# Expected Performance of the ATLAS Experiment

## Detector, Trigger and Physics

**The ATLAS Collaboration**

A detailed study is presented of the expected performance of the ATLAS detector. The reconstruction of tracks, leptons, photons, missing energy and jets is investigated, together with the performance of *b*-tagging and the trigger. The physics potential for a variety of interesting physics processes, within the Standard Model and beyond, is examined. The study comprises a series of notes based on simulations of the detector and physics processes, with particular emphasis given to the data expected from the first years of operation of the LHC at CERN.

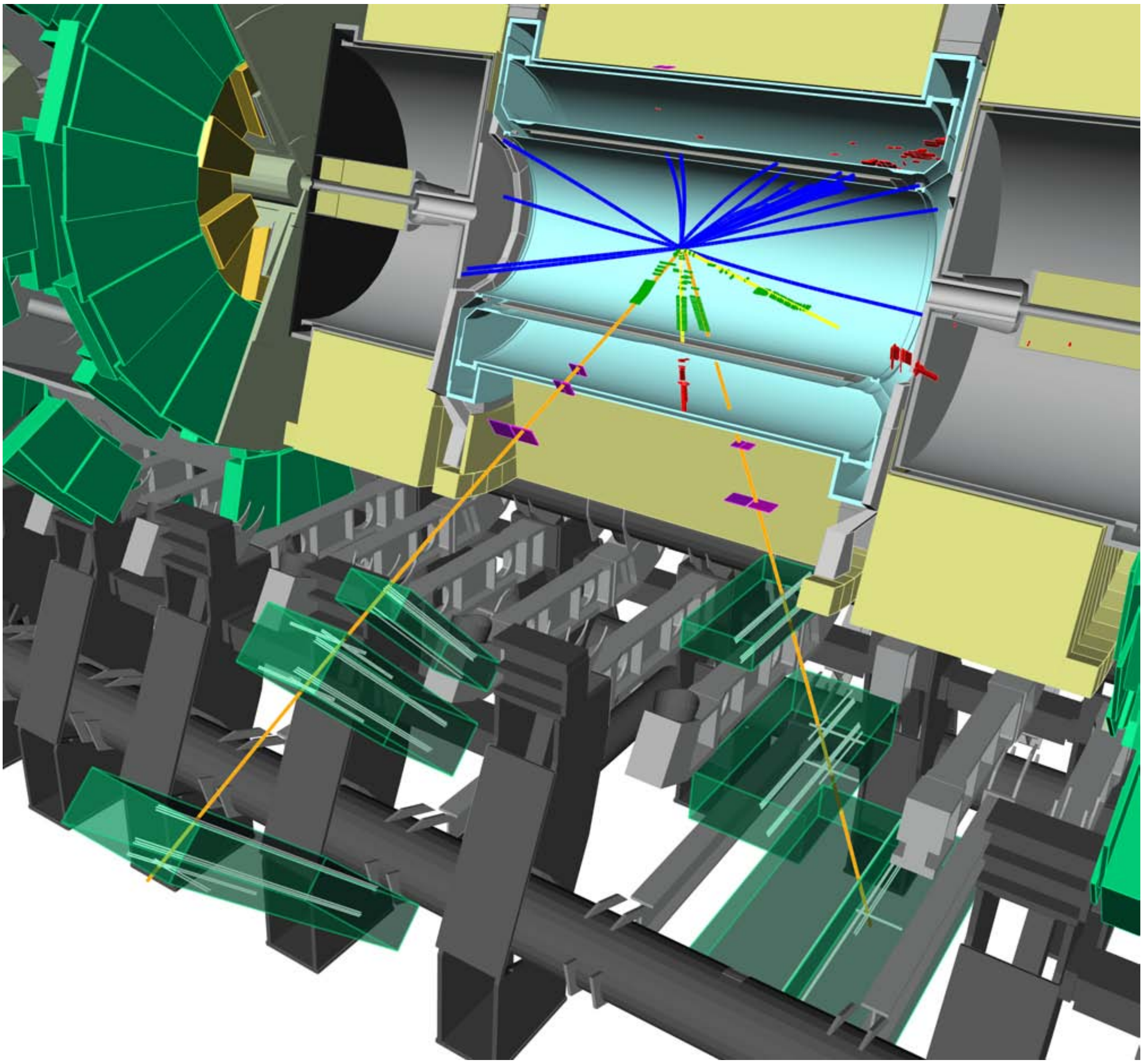

Display of a high-$p_T$ $H \rightarrow ZZ^* \rightarrow ee\mu\mu$ decay ($m_H = 130$ GeV), after full simulation and reconstruction in the ATLAS detector. The four leptons and the recoiling jet with $E_T = 135$ GeV are clearly visible. Hits in the Inner Detector are shown in green for the four reconstructed leptons, both for the precision tracker (pixel and silicon micro-strip detectors) at the inner radii and for the transition radiation tracker at the outer radii. The other tracks reconstructed with $p_T > 0.5$ GeV in the Inner Detector are shown in blue. The two electrons are depicted as reconstructed tracks in yellow and their energy deposits in each layer of the electromagnetic LAr calorimeter are shown in red. The two muons are shown as combined reconstructed tracks in orange, with the hit strips in the resistive-plate chambers and the hit drift tubes in the monitored drift-tube chambers visible as white lines in the barrel muon stations. The energy deposits from the muons in the barrel tile calorimeter can also be seen in purple.

# Contents

# The ATLAS Collaboration

G. Aad[81], E. Abat[18,*], B. Abbott[108], J. Abdallah[11], A.A. Abdelalim[48], A. Abdesselam[115], O. Abdinov[10], B. Abi[109], M. Abolins[86], H. Abramowicz[148], B.S. Acharya[158a,158b], D.L. Adams[24], T.N. Addy[55], C. Adorisio[36a,36b], P. Adragna[73], T. Adye[126], J.A. Aguilar-Saavedra[121a], M. Aharrouche[79], S.P. Ahlen[21], F. Ahles[47], A. Ahmad[144], H. Ahmed[2], G. Aielli[130a,130b], T. Akdogan[18], T.P.A. Åkesson[77], G. Akimoto[150], M.S. Alam[1], M.A. Alam[74], J. Albert[163], S. Albrand[54], M. Aleksa[29], I.N. Aleksandrov[63], F. Alessandria[87a,87b], C. Alexa[25a], G. Alexander[148], G. Alexandre[48], T. Alexopoulos[9], M. Alhroob[20], G. Alimonti[87a], J. Alison [117], M. Aliyev[10], P.P. Allport[71], S.E. Allwood-Spiers[52], A. Aloisio[100a,100b], R. Alon[164], A. Alonso[77], J. Alonso[14], M.G. Alviggi[100a,100b], K. Amako[64], P. Amaral[29], C. Amelung[22], V.V. Ammosov[125], A. Amorim[121b], G. Amorós[161], N. Amram[148], C. Anastopoulos[136], C.F. Anders[57a], K.J. Anderson[30], A. Andreazza[87a,87b], V. Andrei[57a], M-L. Andrieux[54], X.S. Anduaga[68], F. Anghinolfi[29], A. Antonaki[8], M. Antonelli[46], S. Antonelli[19a,19b], B. Antunovic[41], F.A. Anulli[129a], G. Arabidze[8], I. Aracena[140], Y. Arai[64], A.T.H. Arce[14], J.P. Archambault[28], S. Arfaoui[29], J-F. Arguin[14], T. Argyropoulos[9], E. Arik[18,*], M. Arik[18], A.J. Armbruster[85], O. Arnaez[4], C. Arnault[112], A. Artamonov[93], D. Arutinov[20], M. Asai[140], S. Asai[150], S. Ask[80], B. Åsman[142], D. Asner[28], L. Asquith[75], K. Assamagan[24], A. Astbury[163], A. Astvatsatourov[51], T. Atkinson[84], G. Atoian[168], B. Auerbach[168], E. Auge[112], K. Augsten[124], M.A. Aurousseau[4], N. Austin[71], G. Avolio[157], R. Avramidou[9], A. Axen[162], C. Ay[53], G. Azuelos[91,a], Y. Azuma[150], M.A. Baak[29], G. Baccaglioni[87a,87b], C. Bacci[131a,131b], H. Bachacou[133], K. Bachas[149], M. Backes[48], E. Badescu[25a], P. Bagnaia[129a,129b], Y. Bai[32,b], D.C. Bailey [152], J.T. Baines[126], O.K. Baker[168], F. Baltasar Dos Santos Pedrosa[29], E. Banas[38], S. Banerjee[163], D. Banfi[87a,87b], A. Bangert[97], V. Bansal[120], S.P. Baranov[92], S. Baranov[5], A. Barashkou[63], T.B. Barber[27], E.L. Barberio[84], D. Barberis[49a,49b], M.B. Barbero[20], D.Y. Bardin[63], T. Barillari[97], M. Barisonzi[41], T. Barklow[140], N.B. Barlow[27], B.M. Barnett[126], R.M. Barnett[14], S. Baron[29], A. Baroncelli[131a], A.J. Barr[115], F. Barreiro[78], J. Barreiro Guimarães da Costa[56], P. Barrillon[112], R. Bartoldus[140], D. Bartsch[20], J. Bastos[121b], R.L. Bates[52], J.R. Batley[27], A. Battaglia[16], M. Battistin[29], F. Bauer[133], M. Bazalova[122], B. Beare[152], P.H. Beauchemin[115], R.B. Beccherle[49a], N. Becerici[18], P. Bechtle[41], G.A. Beck[73], H.P. Beck[16], M. Beckingham[47], K.H. Becks[167], I. Bedajanek[124], A.J. Beddall[18,c], A. Beddall[18,c], P. Bednár[141], V.A. Bednyakov[63], C. Bee[81], S. Behar Harpaz[147], P.K. Behera[140,d], M. Beimforde[97], C. Belanger- Champagne[159], P.J. Bell[80], W.H. Bell[48], G. Bella[148], L. Bellagamba[19a], F. Bellina[29], M. Bellomo[116a], A. Belloni[56], K. Belotskiy[94], O. Beltramello[29], S. Ben Ami[147], O. Benary[148], D. Benchekroun[132a], M. Bendel[79], B.H. Benedict[157], N. Benekos[160], Y. Benhammou[148], G.P. Benincasa[121b], D.P. Benjamin[44], M. Benoit[112], J.R. Bensinger[22], K. Benslama[127], S. Bentvelsen[103], M. Beretta [46], D. Berge[29], E. Bergeaas Kuutmann[142], N. Berger[4], F. Berghaus[163], E. Berglund[48], J. Beringer[14], K. Bernardet[81], P. Bernat[112], R. Bernhard[47], C. Bernius[75], T. Berry[74], A. Bertin[19a,19b], N. Besson[133], S. Bethke[97], R.M. Bianchi[47], M. Bianco[70a,70b], O. Biebel[96], J. Biesiada[14], M. Biglietti[100a,100b], H. Bilokon[46], S. Binet[14], A. Bingul[18,c], C. Bini[129a,129b], C. Biscarat[173], M. Bischofberger[84], U. Bitenc[47], K.M. Black[56], R.E. Blair[5], G. Blanchot[29], C. Blocker[22], J. Blocki[38], A. Blondel[48], W. Blum[79], U. Blumenschein[53], C. Boaretto[129a,129b], G.J. Bobbink[103], A. Bocci[44], B. Bodine [135], J. Boek[167], N. Boelaert[77], S. Böser[75], J.A. Bogaerts[29], A. Bogouch[88], C. Bohm[142], J. Bohm[122], V. Boisvert[74], T. Bold[157], V. Boldea[25a], V.G. Bondarenko[94], M. Bondioli[157], M. Boonekamp[133], C.N. Booth[136], P.S.L. Booth[71,*], J.R.A. Booth[17], A. Borisov[125], G. Borissov[69], I. Borjanovic[70a], S. Borroni[129a,129b], K. Bos[103], D. Boscherini[19a], M. Bosman[11], M. Bosteels[29], H. Boterenbrood[103], J. Bouchami[91], J. Boudreau[120], E.V. Bouhova-Thacker[69], C. Boulahouache[120], C. Bourdarios[112], J. Boyd[29], I.R. Boyko[63], A. Braem[29], P. Branchini[131a], G.W. Brandenburg[56], A. Brandt[7], O. Brandt[115], U. Bratzler[151], J.E. Brau[111], H.M. Braun[167], B. Brelier[91,e], J. Bremer[29], R. Brenner[159], S. Bressler[147], D. Breton[112], N.D. Brett[115], D. Britton[52], F.M. Brochu[27], I. Brock[20], R. Brock[86], E. Brodet[148], F. Broggi[87a,87b], G. Brooijmans[34], W.K. Brooks[31b], E. Brubaker[30], P.A. Bruckman de Renstrom[38],

D. Bruncko[141], R. Bruneliere[47], S. Brunet[41], A. Bruni[19a], G. Bruni[19a], M. Bruschi[19a], T. Buanes[13], F.B. Bucci[48], P. Buchholz[138], A.G. Buckley[75,f], I.A. Budagov[63], V. Büscher[20], L. Bugge[114], F. Bujor[29], O. Bulekov[94], M. Bunse[42], T. Buran [114], H. Burckhart[29], S. Burdin[71], S. Burke[126], E. Busato[33], C.P. Buszello[159], F. Butin[29], B. Butler[140], J.M. Butler[21], C.M. Buttar[52], J.M. Butterworth[75], T. Byatt[75], S. Cabrera Urbán[161], D. Caforio[19a,19b], O. Cakir[3], P. Calafiura[14], G. Calderini[76], R. Calkins[5], L.P. Caloba[23a], R. Caloi[129a,129b], D. Calvet[33], P. Camarri[130a,130b], M. Cambiaghi[116a,116b], D. Cameron[114], F. Campabadal Segura[161], S. Campana[29], M. Campanelli[75], V. Canale[100a,100b], J. Cantero[78], M.D.M. Capeans Garrido[29], I. Caprini[25a], M. Caprini[25a], M. Capua[36a,36b], R. Caputo[144], C. Caramarcu[25a], R. Cardarelli[130a], T. Carli[29], G. Carlino[100a], L. Carminati[87a,87b], B. Caron[2,g], S. Caron[47], S. Carron Montero[152], A.A. Carter[73], J.R. Carter[27], J. Carvalho[121b], D. Casadei[105], M.P. Casado [11], M. Cascella[119a,119b], C. Caso[49a,49b,*], A.M. Castaneda Hernadez[165], E. Castaneda Miranda[165], V. Castillo Gimenez[161], N.F. Castro[121a], G. Cataldi[70a], A. Catinaccio[29], J.R. Catmore[69], A. Cattai[29], G. Cattani[130a,130b], S. Caughron[34], D. Cauz[158a,158c], P. Cavalleri[76], D. Cavalli[87a], M. Cavalli-Sforza[11], V. Cavasinni[119a,119b], A. Cazzato[70a,70b], F. Ceradini[131a,131b], A.S. Cerqueira [23a], A. Cerri[29], L. Cerrito[73], F. Cerutti[46], S.A. Cetin[18,h], F. Cevenini[100a,100b], A.C. Chafaq[132a], D. Chakraborty[5], J.D. Chapman[27], J.W. Chapman[85], E.C. Chareyre[76], D.G. Charlton[17], S.C. Chatterjii[20], S. Cheatham[69], S. Chekanov[5], S.V. Chekulaev[153a], G.A. Chelkov[63], H. Chen[24], T. Chen[32], X. Chen[165], S. Cheng[32], T.L. Cheng[74], A. Cheplakov[52], V.F. Chepurnov[63], R. Cherkaoui El Moursli[132d], V. Tcherniatine[24], D. Chesneanu[25a], E. Cheu[6], S.L. Cheung[152], L. Chevalier[133], F. Chevallier[133], V. Chiarella[46], G. Chiefari[100a,100b], L. Chikovani[50], J.T. Childers[57a], A. Chilingarov[69], G. Chiodini[70a], S. Chouridou[134], D. Chren[124], I.A. Christidi[149], A. Christov[47], D. Chromek-Burckhart[29], M.L. Chu[146], J. Chudoba[122], G. Ciapetti[129a,129b], A.K. Ciftci[3], R. Ciftci[3], V. Cindro[72], M.D. Ciobotaru[157], C. Ciocca[19a,19b], A. Ciocio[14], M. Cirilli[85], M. Citterio[87a], A. Clark[48], W. Cleland[120], J.C. Clemens[81], B. Clement[54], C. Clément[142], D. Clements[52], Y. Coadou[29], M. Cobal[158a,158c], A. Coccaro[49a,49b], J. Cochran[62], S. Coelli[87a,87b], J. Coggeshall[160], E. Cogneras[16], C.D. Cojocaru[28], J. Colas[4], B. Cole[34], A.P. Colijn[103], C. Collard[112], N.J. Collins[17], C. Collins-Tooth[52], J. Collot[54], G. Colon[82], R. Coluccia[70a,70b], P. Conde Muiño[121b], E. Coniavitis[159], M. Consonni[102], S. Constantinescu[25a], C. Conta [116a,116b], F. Conventi [100a,i], J. Cook[29], M. Cooke[34], B.D. Cooper[73], N.J. Cooper-Smith[74], K. Copic[34], T. Cornelissen[29], M. Corradi[19a], F.C. Corriveau[83,j], A. Corso-Radu[157], A. Cortes-Gonzalez[160], G. Costa[87a], M.J. Costa[161], D. Costanzo[136], T. Costin[30], D. Côté[41], R. Coura Torres[23a], L. Courneyea[163], G. Cowan[74], C.C. Cowden[27], B.E. Cox[80], K. Cranmer[105], J. Cranshaw[5], M. Cristinziani[20], G. Crosetti[36a,36b], R.C. Crupi[70a,70b], S. Crépé-Renaudin[54], C.-M. Cuciuc[25a], C. Cuenca Almenar[157], M. Curatolo[46], C.J. Curtis[17], P. Cwetanski[60], Z. Czyczula[35], S. D'Auria[52], M. D'Onofrio[11], A. D'Orazio[97], A. Da Rocha Gesualdi Mello[23a], P.V.M. Da Silva[23a], C.V. Da Via[80], W. Dabrowski[37], T. Dai[85], C. Dallapiccola[82], S.J. Dallison[126], C.H. Daly[135], M. Dam[35], H.O. Danielsson[29], D. Dannheim[29], V. Dao[48], G. Darbo[49a], W.D. Davey[84], T. Davidek[123], N. Davidson[84], R. Davidson[69], A.R. Davison[75], I. Dawson[136], J.W. Dawson[5], R.K. Daya[39], K. De[7], R. de Asmundis[100a], S. De Castro[19a,19b], P.E. De Castro Faria Salgado[29], S. De Cecco[76], N. De Groot[102], P. de Jong[103], E. De La Cruz-Burelo[85], C. De La Taille[112], L. De Mora[69], M. De Oliveira Branco[29], D. De Pedis[129a], A. De Salvo[129a], U. De Sanctis[87a,87b], A. De Santo[74], J.B. De Vivie De Regie[112], G. De Zorzi[129a,129b], S. Dean[75], G. Dedes[97], D.V. Dedovich[63], P.O. Defay[33], J. Degenhardt [117], M. Dehchar[115], C. Del Papa[158a,158c], J. Del Peso[78], T. Del Prete[119a,119b], A. Dell'Acqua[29], L. Dell'Asta[87a,87b], M. Della Pietra[100a,i], D. della Volpe[100a,100b], M. Delmastro[29], N. Delruelle[29], P.A. Delsart[54], S. Demers[140], M. Demichev[63], B. Demirköz[29], W. Deng[24], S.P. Denisov[125], C. Dennis[115], F. Derue[76], P. Dervan[71], K.K. Desch[20], P.O. Deviveiros[152], A. Dewhurst[69], R. Dhullipudi[24,k], A. Di Ciaccio[130a,130b], L. Di Ciaccio[4], A. Di Domenico[129a,129b], A. Di Girolamo[29], B. Di Girolamo [29], S. Di Luise[131a,131b], A. Di Mattia[86], R. Di Nardo[130a,130b], A. Di Simone[130a,130b], R. Di Sipio[19a,19b], M.A. Diaz[31a], E.B. Diehl[85], J. Dietrich[47],

S. Diglio[131a,131b], K. Dindar Yagci[39], D.J. Dingfelder[47], C. Dionisi[129a,129b], P. Dita [25a], S. Dita [25a], F. Dittus[29], F. Djama[81], R. Djilkibaev[105], T. Djobava[50], M.A.B. do Vale[23a], M. Dobbs[83], R. Dobinson [29,*], D. Dobos[29], E. Dobson[115], M. Dobson[29], O.B. Dogan[18,*], T. Doherty[52], Y. Doi[64], J. Dolejsi[123], I. Dolenc[72], Z. Dolezal[123], B.A. Dolgoshein[94], M. Donega[117], J. Donini[54], T. Donszelmann[136], J. Dopke[167], D.E. Dorfan[134], A. Doria[100a], A. Dos Anjos[165], M. Dosil[11], A. Dotti[119a,119b], M.T. Dova[68], A. Doxiadis[103], A.T. Doyle[52], J.D. Dragic[74], Z. Drasal[123], N. Dressnandt[117], C. Driouichi[35], M. Dris [9], J. Dubbert[97], E. Duchovni[164], G. Duckeck[96], A. Dudarev[29], M. Dührssen [47], I.P. Duerdoth[80], L. Duflot[112], M-A. Dufour[83], M. Dunford[30], A. Duperrin[81], H. Duran Yildiz[3,l], A. Dushkin[22], R. Duxfield[136], M. Dwuznik[37], M. Düren[51], W.L. Ebenstein[44], S. Eckert[47], S. Eckweiler[79], K. Edmonds[20], P. Eerola[77,m], K. Egorov[60], W. Ehrenfeld[41,n], T. Ehrich[97], T. Eifert[48], G. Eigen[13], K. Einsweiler[14], E. Eisenhandler[73], T. Ekelof[159], M. El Kacimi[4], M. Ellert[159], S. Elles[4], K. Ellis[73], N. Ellis[29], J. Elmsheuser [96], M. Elsing[29], R. Ely[14], D. Emeliyanov[126], R. Engelmann[144], A. Engl[96], B. Epp[61], A. Eppig [85], V.S. Epshteyn[93], J. Erdmann[97], A. Ereditato[16], D. Eriksson[142], I. Ermoline[86], J. Ernst[1], E. Ernst[24], J. Ernwein[133], D. Errede[160], S. Errede[160], M. Escalier[112], C. Escobar[161], X. Espinal Curull[11], B. Esposito[46], F. Etienne[81], A.I. Etienvre[133], E. Etzion[148], H. Evans[60], L. Fabbri[19a,19b], C. Fabre[29], P. Faccioli[19a,19b], K. Facius[35], R.M. Fakhrutdinov[125], S. Falciano[129a], A.C. Falou[112], Y. Fang[165], M. Fanti[87a,87b], A. Farbin[7], A. Farilla[131a], J. Farley[144], T. Farooque[152], S.M. Farrington[115], P. Farthouat[29], F. Fassi[161], P. Fassnacht[29], D. Fassouliotis[8], B. Fatholahzadeh[152], L. Fayard[112], F. Fayette[76], R. Febbraro[33], P. Federic[141], O.L. Fedin[118], I. Fedorko[29], L. Feligioni[81], C. Feng[32], E.J. Feng[30], A.B. Fenyuk[125], J. Ferencei[141], J. Ferland[91], W. Fernando[106], S. Ferrag[52], A. Ferrari[159], P. Ferrari[103], R. Ferrari[116a], A. Ferrer[161], M.L. Ferrer[46], D. Ferrere[48], C. Ferretti[85], M. Fiascaris[115], F. Fiedler[79], A. Filipčič[72], A. Filippas[9], F. Filthaut[102], M. Fincke-Keeler[163], L. Fiorini[11], A. Firan[39], G. Fischer[41], M.J. Fisher[106], H.F. Flacher[29], M. Flechl[159], I. Fleck[138], J. Fleckner[79], P. Fleischmann[133], S. Fleischmann[20], C.M. Fleta Corral[161], T. Flick[167], L.R. Flores Castillo[165], M.J. Flowerdew[71], F. Föhlisch[57a], M. Fokitis[9], T. Fonseca Martin[74], D.A. Forbush[135], A. Formica[133], A. Forti[80], J.M. Foster[80], D. Fournier[112], A. Foussat[29], A.J. Fowler[44], K.F. Fowler [134], H. Fox[69], P. Francavilla[119a,119b], S. Franchino[116a,116b], D. Francis[29], S. Franz[29], M. Fraternali[116a,116b], S. Fratina[117], J. Freestone[80], R. Froeschl[29], D. Froidevaux[29], J.A. Frost[27], C. Fukunaga[151], E. Fullana Torregrosa[5], J. Fuster[161], C. Gabaldon[78], O.G. Gabizon[164], T. Gadfort[34], S. Gadomski[48,o], G. Gagliardi[49a,49b], P. Gagnon[60], E.J. Gallas[115], M.V. Gallas[29], B.J. Gallop[126], E. Galyaev[40], K.K. Gan[106], Y.S. Gao[140,p], A. Gaponenko[14], M. Garcia-Sciveres[14], C. García[161], J.E. García Navarro[48], R.W. Gardner[30], N. Garelli[49a,49b], H. Garitaonandia[103], V.G. Garonne[29], C. Gatti[46], G. Gaudio[116a], O. Gaumer[48], P. Gauzzi[129a,129b], I.L. Gavrilenko[92], C. Gay[162], G.G. Gaycken[20], J-C. Gayde[29], E.N. Gazis[9], C.N.P. Gee[126], Ch. Geich-Gimbel[20], K. Gellerstedt[142], C. Gemme[49a], M.H. Genest[96], S. Gentile[129a,129b], F. Georgatos[9], S. George[74], P. Gerlach[167], C. Geweniger[57a], H. Ghazlane[132d], P. Ghez[4], N. Ghodbane[33], B. Giacobbe[19a], S. Giagu[129a,129b], V. Giangiobbe[119a,119b], F. Gianotti[29], B. Gibbard[24], A. Gibson[152], S.M. Gibson[115], L.M. Gilbert[115], M. Gilchriese[14], V. Gilewsky[89], A.R. Gillman[126], D.M. Gingrich[2,g], J. Ginzburg[148], N. Giokaris[8], M.P. Giordani [158a,158c], P. Giovannini[97], P.F. Giraud[29], P. Girtler[61], D. Giugni[87a], P. Giusti[19a], B.K. Gjelsten[114], L.K. Gladilin[95], C. Glasman[78], A. Glazov[41], K.W. Glitza[167], G.L. Glonti[63], K.G. Gnanvo[73], J.G. Godfrey[139], J. Godlewski[29], T. Göpfert[43], C. Gössling[42], T. Göttfert[97], V.G. Goggi[116a,116b], S. Goldfarb[85], D. Goldin[39], T. Golling[14], N.P. Gollub[29], A. Gomes[121b], R. Gonçalo[74], C. Gong[32], S. González de la Hoz[161], M.L. Gonzalez Silva[26], S. González-Sevilla[48], J.J. Goodson[144], L. Goossens[29], P.A. Gorbounov[152], H. Gordon[24], I. Gorelov[101], G. Gorfine[167], B. Gorini[29], E. Gorini[70a,70b], A. Gorišek[72], E. Gornicki[38], S.A. Gorokhov[125], S.V. Goryachev[125], V.N. Goryachev[125], B. Gosdzik[41], M. Gosselink[103], M.I. Gostkin[63], I. Gough Eschrich[157], M. Gouighri[132a], D. Goujdami[132a], M. Goulette[29], A.G. Goussiou[135], S. Gowdy[140], C. Goy[4], I. Grabowska-Bold[157], P. Grafström[29], K-J. Grahn[143],

L. Granado Cardoso[121b], F. Grancagnolo[70a], S. Grancagnolo[70a,70b], V. Gratchev[118], H.M. Gray[34,q], J.A. Gray[144], E. Graziani[131a], B. Green[74], Z.D. Greenwood[24,k], I.M. Gregor[41], E. Griesmayer[45], N. Grigalashvili[63], A.A. Grillo[134], K. Grimm[144], Y.V. Grishkevich[95], L.S. Groer[152], J. Grognuz[29], M. Groh[97], M. Groll[79], E. Gross[164], J. Grosse-Knetter[53], J. Groth-Jensen[77], C. Gruse[25a], K. Grybel[138], V.J. Guarino[5], C. Guicheney[33], A.G. Guida[70a,70b], T. Guillemin[4], J. Gunther[122], B. Guo[152], A. Gupta[30], Y. Gusakov[63], P. Gutierrez[108], N.G. Guttman[148], O. Gutzwiller[29], C. Guyot[133], C. Gwenlan[115], C.B. Gwilliam[71], A. Haas[34], S. Haas[29], C. Haber[14], R. Hackenburg[24], H.K. Hadavand[39], D.R. Hadley[17], R. Härtel[97], Z. Hajduk[38], H. Hakobyan[48], H. Hakobyan[169], R.H. Hakobyan[2], J. Haller[41,n], K. Hamacher[167], A. Hamilton[48], H. Han[32], L. Han [32], K. Hanagaki[113], M. Hance[117], C. Handel[79], P. Hanke[57a], J.R. Hansen[35], J.B. Hansen[35], J.D. Hansen[35], P.H. Hansen[35], T. Hansl-Kozanecka[134], P. Hansson[143], K. Hara[154], G.A. Hare[134], T. Harenberg[167], R.D. Harrington[21], O.B. Harris[75], O.M. Harris[135], J.C. Hart[126], J. Hartert[47], F. Hartjes[103], T. Haruyama[64], A. Harvey[55], S. Hasegawa[99], Y. Hasegawa[137], K. Hashemi[22], S. Hassani[133], M. Hatch[29], F. Haug[29], S. Haug[16], M. Hauschild[29], R. Hauser[86], M. Havranek[122], R.J. Hawkings[29], D. Hawkins[157], T. Hayakawa[65], H.S. Hayward[71], S.J. Haywood[126], M. He[32], S.J. Head[80], V. Hedberg[77], L. Heelan[28], B. Heinemann[14], F.E.W. Heinemann[115], M. Heldmann[47], S. Hellman[142], C. Helsens[133], R.C.W. Henderson[69], M. Henke[57a], A.M. Henriques Correia[29], S. Henrot-Versille[112], T. Henß[167], A.D. Hershenhorn[147], G. Herten[47], R. Hertenberger[96], L. Hervas[29], N.P. Hessey[103], A. Hidvegi[142], E. Higón-Rodriguez[161], D. Hill[5,*], J.C. Hill[27], K.H. Hiller[41], S.J. Hillier[17], I. Hinchliffe[14], C. Hinkelbein[57b], F. Hirsch[42], J. Hobbs[144], N.H. Hod[148], M.C. Hodgkinson[136], P. Hodgson[136], A. Hoecker[29], M.R. Hoeferkamp[101], J. Hoffman[39], D. Hoffmann[81], M.H. Hohlfeld[20], S.O. Holmgren[142], T. Holy[124], Y. Homma[65], P. Homola[124], T. Horazdovsky[124], T. Hori[65], C. Horn[140], S. Horner[47], S. Horvat[97], J-Y. Hostachy[54], S. Hou[146], M.A. Houlden[71], A. Hoummada[132a], J. Hrivnac[112], I. Hruska[122], T. Hryn'ova[4], P.J. Hsu[168], G.S. Huang[108], J. Huang[157], Z. Hubacek[124], F. Hubaut[81], F. Huegging[20], E.W. Hughes[34], G. Hughes[69], R.E. Hughes-Jones[80], P. Hurst[56], M. Hurwitz[30], T. Huse [114], N. Huseynov[10], J. Huston[86], J. Huth[56], G. Iacobucci[100a], M. Ibbotson[80], I. Ibragimov[138], R. Ichimiya[65], L. Iconomidou-Fayard[112], J. Idarraga[91], P. Iengo[29], O. Igonkina[103], Y. Ikegami[64], M. Ikeno[64], Y. Ilchenko[39], D.I. Iliadis[149], Y. Ilyushenka[63], M. Imori[150], T. Ince[163], P. Ioannou [8], M. Iodice[131a], A. Ishikawa[65], M. Ishino[150], Y. Ishizawa[153a], R. Ishmukhametov[39], T. Isobe[150], V. Issakov[168], C. Issever[115], S. Istin[18], A.V. Ivashin[125], W. Iwanski[38], H. Iwasaki[64], J.M. Izen[40], V. Izzo[100a], J.N. Jackson[71], M. Jaekel[29], M. Jahoda[122], V. Jain[60], K. Jakobs[47], J. Jakubek[124], D. Jana[108], E. Jansen[102], A. Jantsch[97], R.C. Jared[165], G. Jarlskog[77], P. Jarron[29], K. Jelen[37], I. Jen-La Plante[30], P. Jenni[29], P. Jez[35], S. Jézéquel[4], W. Ji[77], J. Jia[144], Y. Jiang[32], G. Jin[32], S. Jin[32], O. Jinnouchi[64], D. Joffe[39], L.G. Johansen[13], M. Johansen[142], K.E. Johansson[142], P. Johansson[136], K.A. Johns[6], K. Jon-And[142], A. Jones[160], G. Jones[80], R.W.L. Jones[69], T.W. Jones[75], T.J. Jones[71], O. Jonsson[29], D. Joos[47], C. Joram[29], P.M. Jorge[121b], S. Jorgensen[11], P. Jovanovic[17], V. Juranek[122], P. Jussel[61], V.V. Kabachenko[125], S. Kabana[16], M. Kaci[161], A. Kaczmarska[38], M. Kado[112], H. Kagan[106], M. Kagan[56], S. Kaiser[97], E. Kajomovitz[147], L.V. Kalinovskaya[63], A. Kalinowski[127], S. Kama[41], N. Kanaya[150], M. Kaneda[150], V.A. Kantserov[94], J. Kanzaki[64], B. Kaplan[168], A. Kapliy[30], J. Kaplon[29], M. Karagounis[20], M. Karagoz Unel[115], K. Karr[5], V. Kartvelishvili[69], A.N. Karyukhin[125], L. Kashif[56], A. Kasmi[39], R.D. Kass[106], M. Kataoka[29], Y. Kataoka[150], E. Katsoufis [9], J. Katzy[41], K. Kawagoe[65], T. Kawamoto[150], M.S. Kayl[103], F. Kayumov[92], V.A. Kazanin [104], M.Y. Kazarinov[63], S.I. Kazi[84], J.R. Keates[80], R. Keeler[163], P.T. Keener[117], R. Kehoe[39], M. Keil[48], G.D. Kekelidze[63], M. Kelly[80], J. Kennedy[96], M. Kenyon[52], O. Kepka[133], N. Kerschen[136], B.P. Kerševan[72], S. Kersten[167], M. Khakzad[28], F. Khalilzade[10], H. Khandanyan[160], A. Khanov[109], D. Kharchenko[63], A. Khodinov[144], A.G. Kholodenko[125], A. Khomich[57a], G. Khoriauli[20], N. Khovanskiy[63], V. Khovanskiy[93], E. Khramov[63], J. Khubua[50], G. Kilvington[74], H. Kim[7], M.S. Kim[2], S.H. Kim[154], O. Kind[15], P. Kind[167], B.T. King[71], J. Kirk[126], G.P. Kirsch[115], L.E. Kirsch[22], A.E. Kiryunin[97], D. Kisielewska[37], T. Kittelmann[120], H. Kiyamura[65], E. Kladiva[141], J. Klaiber-Lodewigs[42], M. Klein[71], U. Klein[71],

K. Kleinknecht[79], A. Klier[164], A. Klimentov[24], R. Klingenberg[42], E.B. Klinkby[44], T. Klioutchnikova[29], P.F. Klok[102], S. Klous[103], E.-E. Kluge[57a], T. Kluge[71], P. Kluit[103], M. Klute[53], S. Kluth[97], N.S. Knecht[152], E. Kneringer[61], B.R. Ko[44], T. Kobayashi[150], M. Kobel[43], B. Koblitz[29], A. Kocnar[110], P. Kodys[123], K. Köneke[41], A.C. König[102], S. König[47], L. Köpke[79], F. Koetsveld[102], P. Koevesarki[20], T. Koffas[29], E. Koffeman[103], Z. Kohout [124], T. Kohriki[64], T. Kokott[20], H. Kolanoski[15], V. Kolesnikov[63], I. Koletsou[4], I. Koletsou[112], M. Kollefrath[47], S. Kolos[157,r], S.D. Kolya[80], A.A. Komar[92], J.R. Komaragiri[139], T. Kondo[64], T. Kono[29], A.I. Kononov[47], R. Konoplich[105], S.P. Konovalov[92], N. Konstantinidis[75], A. Kootz[167], S. Koperny[37], K. Korcyl[38], K. Kordas[16], V. Koreshev[125], A. Korn[14], I. Korolkov[11], V.A. Korotkov[125], O. Kortner[97], V.V. Kostyukhin[49a], M.J. Kotamäki[29], S. Kotov[97], V.M. Kotov[63], K.Y. Kotov [104], Z. Koupilova [123], C. Kourkoumelis[8], A. Koutsman[103], S. Kovar[29], R. Kowalewski[163], H. Kowalski[41], T.Z. Kowalski[37], W. Kozanecki[133], A.S. Kozhin[125], V. Kral[124], V.A. Kramarenko[95], G. Kramberger[72], M.W. Krasny[76], A. Krasznahorkay[29], A.K. Kreisel[148], F. Krejci[124], A. Krepouri[149], P. Krieger[152], G. Krobath[96], K. Kroeninger[53], H. Kroha[97], J. Kroll[117], J. Krstic[12a], U. Kruchonak[63], H. Krüger[20], Z.V. Krumshteyn[63], T. Kubota[150], S.K. Kuehn[47], A. Kugel[57b], T. Kuhl[167], D. Kuhn[61], V. Kukhtin[63], Y. Kulchitsky[88], S. Kuleshov[31b], C.K. Kummer[96], M. Kuna[81], A. Kupco[122], H. Kurashige[65], M.K. Kurata[154], L.L. Kurchaninov[153a], Y.A. Kurochkin[88], V. Kus[122], W. Kuykendall[135], E.K. Kuznetsova[129a,129b], O. Kvasnicka[122], R. Kwee[15], M. La Rosa[84], L. La Rotonda[36a,36b], L. Labarga[78], J.A. Labbe[54], C. Lacasta[161], F. Lacava[129a,129b], H. Lacker[15], D. Lacour[76], V.R. Lacuesta[161], E. Ladygin[63], R. Lafaye[4], B. Laforge[76], T. Lagouri[78], S. Lai[47], M. Lamanna[29], M. Lambacher[96], C.L. Lampen[6], W. Lampl[6], E. Lancon[133], U. Landgraf[47], M.P.J. Landon[73], J.L. Lane[80], A.J. Lankford[157], F. Lanni[24], K. Lantzsch[29], A. Lanza[116a], S. Laplace[4], C.L. Lapoire[81], J.F. Laporte[133], T. Lari[87a], A.V. Larionov [125], C. Lasseur[29], M. Lassnig[29], P. Laurelli [46], W. Lavrijsen[14], A.B. Lazarev[63], A-C. Le Bihan[29], O. Le Dortz[76], C. Le Maner[152], M. Le Vine[24], M. Leahu[29], C. Lebel[91], T. LeCompte[5], F. Ledroit-Guillon[54], H. Lee[103], J.S.H. Lee[145], S.C. Lee[146], M. Lefebvre[163], R.P. Lefevre[48], M. Legendre[133], A. Leger[48], B.C. LeGeyt[117], F. Legger[97], C. Leggett[14], M. Lehmacher[20], G. Lehmann Miotto[29], X. Lei[6], R. Leitner[123], D. Lelas[163], D. Lellouch[164], M. Leltchouk[34], V. Lendermann[57a], K.J.C. Leney[71], T. Lenz[167], G. Lenzen[167], B. Lenzi[133], C. Leroy[91], J-R. Lessard[163], C.G. Lester[27], A. Leung Fook Cheong[165], J. Levêque[81], D. Levin[85], L.J. Levinson[164], M.S. Levitski[125], S. Levonian[41], M. Lewandowska[21], M. Leyton[14], J. Li[7], S. Li[41], X. Li[85], Z. Liang[39], Z. Liang[146], B. Liberti[130a], P. Lichard[29], M. Lichtnecker[96], W. Liebig[103], R. Lifshitz[147], D. Liko[29], J.N. Lilley[17], H. Lim[5], M. Limper[103], S.C. Lin[146], S.W. Lindsay[71], V. Linhart[124], A. Liolios[149], L. Lipinsky[122], A. Lipniacka[13], T.M. Liss[160], A. Lissauer[24], A.M. Litke[134], C. Liu[28], D.L. Liu[146], J.L. Liu[85], M. Liu[32,b], S. Liu[2], T. Liu[39], Y. Liu[32], M. Livan[116a,116b], A. Lleres[54], S.L. Lloyd[73], E. Lobodzinska[41], P. Loch[6], W.S. Lockman[134], S. Lockwitz[168], T. Loddenkoetter[20], F.K. Loebinger[80], A. Loginov[168], C.W. Loh[162], T. Lohse[15], K. Lohwasser[115], M. Lokajicek[122], J. Loken [115], D. Lopez Mateos[34,q], M. Losada[156], M.J. Losty[153a], X. Lou[40], K.F. Loureiro[106], L. Lovas[141], J. Love[21], A. Lowe[60], F. Lu[32,b], J. Lu[2], H.J. Lubatti[135], C. Luci[129a,129b], A. Lucotte[54], A. Ludwig[43], I. Ludwig[47], J. Ludwig[47], F. Luehring[60], L. Luisa[158a,158c], D. Lumb[47], L. Luminari[129a], E. Lund[114], B. Lund-Jensen[143], B. Lundberg[77], J. Lundquist[35], A. Lupi[119a,119b], G. Lutz[97], D. Lynn[24], J. Lys[14], E. Lytken[29], H. Ma[24], L.L. Ma[152], M. Maaßen[47], G. Maccarrone [46], A. Macchiolo[97], B. Maček[72], R. Mackeprang[29], R.J. Madaras[14], W.F. Mader[43], R. Maenner[57b], T. Maeno[24], P. Mättig[167], C. Magass[20], C.A. Magrath[102], Y. Mahalalel[148], K. Mahboubi[47], A. Mahmood[1], G. Mahout[17], C. Maidantchik[23a], A. Maio[121b], G.M. Mair[61], S. Majewski[24], Y. Makida[64], N.M. Makovec[112], Pa. Malecki[38], P. Malecki[38], V.P. Maleev[118], F. Malek[54], U. Mallik[140], D. Malon[5], S. Maltezos [9], V. Malychev[104], M. Mambelli[30], R. Mameghani[96], J. Mamuzic[41], A. Manabe[64], L. Mandelli[87a,87b], I. Mandić[72], J. Maneira[121b], P.S. Mangeard[81], I.D. Manjavidze[63], A. Manousakis-Katsikakis[8], B. Mansoulie[133], A. Mapelli[29], L. Mapelli[29], L. March Ruiz[78], J.F. Marchand[4], F.M. Marchese[130a,130b], M. Marcisovsky[122], C.N. Marques[121b], F. Marroquim[23a], R. Marshall[80], Z. Marshall[34,q],

F.K. Martens[152], S. Marti i Garcia[161], A. Martin[73], A.J. Martin[168], B. Martin[29], B. Martin[86], F.F. Martin[117], J.P. Martin[91], M. Martinez Perez[11], V. Martinez Outschoorn[56], A. Martini[46], V. Martynenko[153b], A.C. Martyniuk[80], T. Maruyama[154], F. Marzano[129a], A. Marzin[133], L. Masetti[20], T. Mashimo[150], R. Mashinistov[94], J. Masik[80], A.L. Maslennikov[104], G. Massaro[103], N. Massol[4], A. Mastroberardino[36a,36b], M. Mathes[20], P. Matricon[112], H. Matsumoto[150], H. Matsunaga[150], T. Matsushita[65], J.M. Maugain[29], S.J. Maxfield[71], E.N. May[5], A. Mayne[136], R. Mazini[152], M. Mazzanti[87a,87b], P. Mazzanti[19a], S.P. Mc Kee[85], R.L. McCarthy[144], C. McCormick[157], N.A. McCubbin[126], K.W. McFarlane[55], S. McGarvie[74], H. McGlone[52], R.A. McLaren[29], S.J. McMahon[126], T.R. McMahon[74], R.A. McPherson[163,j], J.M. Mechnich[103], M. Mechtel[167], D. Meder-Marouelli[167], M. Medinnis[41], R. Meera-Lebbai[108], R. Mehdiyev[91], S. Mehlhase[41], A. Mehta[71], K. Meier[57a], B. Meirose [47], A. Melamed-Katz[164], B.R. Mellado Garcia[165], Z.M. Meng[146], S. Menke[97], E. Meoni[36a,36b], D. Merkl[96], P. Mermod[142], L. Merola[100a,100b], C. Meroni[87a], F.S. Merritt[30], A.M. Messina[29], I. Messmer[47], J. Metcalfe[101], A.S. Mete[62], J-P. Meyer[133], J. Meyer[53], T.C. Meyer[29], W.T. Meyer[62], L. Micu[25a], R. Middleton[126], S. Migas[71], L. Mijović[72], G. Mikenberg[164], M. Mikuž[72], D.W. Miller[140], R.J. Miller[86], B.M. Mills[162], C.M. Mills[56], M. Milosavljevic[12a], D.A. Milstead[142], S. Mima[107], A.A. Minaenko[125], M. Miñano[161], I.A. Minashvili[63], A.I. Mincer[105], B. Mindur[37], M. Mineev[63], L.M. Mir[11], G. Mirabelli[129a], S. Misawa[24], S. Miscetti[46], A. Misiejuk[74], J.M. Mitrevski[134], V.A. Mitsou[161], P.S. Miyagawa[80], J.U. Mjörnmark[77], D. Mladenov[22], T. Moa[142], M. Moch[129a,129b], A. Mochizuki[154], P. Mockett[135], P. Modesto[161], S. Moed[56], V. Moeller[27], K. Mönig[41], N. Möser[20], B. Mohn[13], W. Mohr[47], S. Mohrdieck-Möck[97], R. Moles-Valls[161], J. Molina-Perez[29], G. Moloney[84], J. Monk[75], E. Monnier[81], S. Montesano[87a,87b], F. Monticelli[68], R.W. Moore[2], C.M. Mora Herrera[48], A. Moraes[52], A. Morais[121b], J. Morel[4], D. Moreno[156], M. Moreno Llácer[161], P. Morettini [49a], M. Morii[56], J. Morin[73], A.K. Morley[84], G. Mornacchi[29], S.V. Morozov[94], J.D. Morris[73], H.G. Moser[97], M. Mosidze[50], J.M. Moss[106], A. Moszczynski[38], E. Mountricha[9], S.V. Mouraviev[92], E.J.W. Moyse[82], J. Mueller[120], K. Mueller[20], T.A. Müller[96], D.M. Muenstermann[42], A.M. Muir[162], R. Murillo Garcia[157], W.J. Murray[126], E. Musto[100a,100b], A.G. Myagkov[125], M. Myska[122], J. Nadal[11], K. Nagai[24], K. Nagano[64], Y. Nagasaka[59], A.M. Nairz[29], I. Nakano[107], H. Nakatsuka[65], G. Nanava[20], A. Napier[155], M. Nash[75,s], N.R. Nation[21], T. Naumann[41], G. Navarro[156], S.K. Nderitu[20], H.A. Neal[85], E. Nebot[78], P. Nechaeva[92], A. Negri[116a,116b], G. Negri[29], A. Nelson[62], S. Nemecek[122], P. Nemethy[105], A.A. Nepomuceno[23a], M. Nessi[29], S.Y. Nesterov[118], M.S. Neubauer[160], A. Neusiedl[79], R.N. Neves[121b], P. Nevski[24], F.M. Newcomer[117], C. Ng[154], C. Nicholson[52], R.B. Nickerson[115], R. Nicolaidou[133], G. Nicoletti[46], B. Nicquevert[29], J. Nielsen[134], A. Nikiforov[41], N. Nikitin[95], K. Nikolaev[63], I. Nikolic-Audit[76], K. Nikolopoulos[8], H. Nilsen[47], P. Nilsson[7], A. Nisati[129a], R. Nisius[97], L.J. Nodulman[5], M. Nomachi[113], I. Nomidis[149], H. Nomoto[150], M. Nordberg[29], D. Notz[41], J. Novakova[123], M. Nozaki[64], M. Nozicka[41], A.-E. Nuncio-Quiroz[20], G. Nunes Hanninger[20], T. Nunnemann[96], S.W. O'Neale[17,*], D.C. O'Neil[139], V. O'Shea[52], F.G. Oakham[28,a], H. Oberlack[97], A. Ochi[65], S. Odaka[64], G.A. Odino[49a,49b], H. Ogren[60], S.H. Oh[44], T. Ohshima[99], H. Ohshita[137], T. Ohsugi[58], S. Okada[65], H. Okawa[150], Y. Okumura[99], M. Olcese[49a], A.G. Olchevski[63], M. Oliveira[121b], D. Oliveira Damazio[24], J. Oliver[56], E.O. Oliver Garcia[161], D. Olivito [117], A. Olszewski[38], J. Olszowska[38], C. Omachi[65], A. Onea[29], A. Onofre[121b], C.J. Oram[153a], G. Ordonez[102], M.J. Oreglia[30], Y. Oren[148], D. Orestano[131a,131b], I.O. Orlov [104], R.S. Orr[152], E.O. Ortega[127], B. Osculati[49a,49b], C. Osuna[11], R. Otec[124], F. Ould-Saada[114], A. Ouraou[133], Q. Ouyang[32], O.K. Øye[13], V.E. Ozcan[75], K. Ozone[64], N. Ozturk[7], A. Pacheco Pages[11], S. Padhi[165], C. Padilla Aranda[11], E. Paganis[136], F. Paige[24], K. Pajchel[114], A. Pal[7], S. Palestini[29], J. Palla[29], D. Pallin[33], A. Palma[121b], Y.B. Pan[165], E. Panagiotopoulou[9], B. Panes[31a], N. Panikashvili[85], S. Panitkin[24], D. Pantea[25a], M. Panuskova[122], V. Paolone[120], Th.D. Papadopoulou[9], W. Park[24,t], M.A. Parker[27], S. Parker[14], F. Parodi[49a,49b], J.A. Parsons[34], U. Parzefall[47], E. Pasqualucci[129a], G. Passardi[29], A. Passeri[131a], F. Pastore[131a,131b], Fr. Pastore[29], S. Pataraia[97], J.R. Pater[80], S. Patricelli[100a,100b], P. Patwa[24], T. Pauly[29], L.S. Peak[145],

M. Pecsy[141], M.I. Pedraza Morales[165], S.V. Peleganchuk[104], H. Peng[165], R. Pengo[29], J. Penwell[60], M. Perantoni[23a], A. Pereira[121b], K. Perez[34,q], E. Perez Codina[11], V. Perez Reale[34], L. Perini[87a,87b], H. Pernegger[29], R. Perrino[70a], P. Perrodo[4], P. Perus[112], V.D. Peshekhonov[63], B.A. Petersen[29], J. Petersen[29], T.C. Petersen[29], C. Petridou[149], E. Petrolo[129a], F. Petrucci[131a,131b], R. Petti[24,t], R. Pezoa[31b], M. Pezzetti[29], B. Pfeifer[47], A. Phan[84], A.W. Phillips[27], G. Piacquadio[47], M. Piccinini[19a,19b], R. Piegaia[26], S. Pier[157], J.E. Pilcher[30], A.D. Pilkington[80], J. Pina[121b], J.L. Pinfold[2], J. Ping[32], B. Pinto[121b], O. Pirotte[29], C. Pizio[87a,87b], R. Placakyte[41], M. Plamondon[112], W.G. Plano[80], M.-A. Pleier[20], A. Poblaguev[168], F. Podlyski[33], P. Poffenberger[163], L. Poggioli[112], M. Pohl[48], F. Polci[112], G. Polesello[116a], A. Policicchio[135], A. Polini[19a], J.P. Poll[73], V. Polychronakos[24], D.M. Pomarede[133], K. Pommès[29], L. Pontecorvo[129a], B.G. Pope[86], R. Popescu[24], D.S. Popovic[12a], A. Poppleton[29], J. Popule[122], X. Portell Bueso[47], R. Porter[157], G.E. Pospelov[97], P. Pospichal[29], S. Pospisil[124], M. Potekhin[24], I.N. Potrap[97], C.J. Potter[74], C.T. Potter[83], K.P. Potter[80], G. Poulard[29], J. Poveda[165], R. Prabhu[20], P. Pralavorio[81], S. Prasad[56], R. Pravahan[7], T. Preda[25a], K. Pretzl[16], L. Pribyl[29], D. Price[69], L.E. Price[5], M.J. Price[29], P.M. Prichard[71], D. Prieur[126], M. Primavera[70a], K. Prokofiev[29], F. Prokoshin[31b], S. Protopopescu[24], J. Proudfoot[5], H. Przysiezniak [4], C. Puigdengoles[11], J. Purdham[85], M. Purohit[24,t], P. Puzo[112], Y. Pylypchenko[114], M.T. Pérez García-Estañ[161], M. Qi[32], J. Qian[85], W. Qian[126], Z. Qian[81], Z. Qin[41], D. Qing[146], A. Quadt[53], D.R. Quarrie[14], W.B. Quayle[165], F. Quinonez[31a], M. Raas[102], V. Radeka[24], V. Radescu[41], B. Radics[20], T. Rador[18], F. Ragusa[87a,87b], G. Rahal[173], A.M. Rahimi[106], D. Rahm[24], S. Rajagopalan[24], S. Rajek[42], P.N. Ratoff[69], F. Rauscher[96], E. Rauter[97], M. Raymond[29], A.L. Read [114], D.M. Rebuzzi[97], G.R. Redlinger[24], R. Reece [117], K. Reeves[167], E. Reinherz-Aronis[148], I. Reisinger[42], D. Reljic[12a], C. Rembser[29], Z. Ren[146], P. Renkel[39], S. Rescia[24], M. Rescigno[129a], S. Resconi[87a], B. Resende[103], E. Rezaie[139], P. Reznicek[123], A. Richards[75], R.A. Richards[86], R. Richter[97], E. Richter-Was[38,u], M. Ridel[76], S. Rieke[79], M. Rijpstra[103], M. Rijssenbeek[144], A. Rimoldi[116a,116b], R.R. Rios [39], C. Risler[15], I. Riu [11], G. Rivoltella[87a,87b], F. Rizatdinova[109], K. Roberts[160], S.H. Robertson[83,j], A. Robichaud-Veronneau[48], D. Robinson[27], A. Robson[52], J.G. Rocha de Lima[5], C. Roda[119a,119b], D. Rodriguez[156], Y. Rodriguez[156], S. Roe[29], O. Røhne[114], V. Rojo[1], S. Rolli[155], A. Romaniouk[94], V.M. Romanov[63], G. Romeo[26], D. Romero[31a], L. Roos[76], E. Ros[161], S. Rosati[129a,129b], G.A. Rosenbaum[152], E.I. Rosenberg[62], L. Rosselet[48], L.P. Rossi[49a], M. Rotaru[25a], J. Rothberg[135], I. Rottländer[20], D. Rousseau[112], C.R. Royon[133], A. Rozanov[81], Y. Rozen[147], B. Ruckert[96], N. Ruckstuhl[103], V.I. Rud[95], G. Rudolph[61], F. Rühr[57a], F. Ruggieri[131a], A. Ruiz-Martinez[161], V. Rumiantsev[89,*], L. Rumyantsev[63], N.A. Rusakovich[63], D.R. Rust[60], J.P. Rutherfoord[6], C. Ruwiedel[20], P. Ruzicka[122], Y.F. Ryabov[118], V. Ryadovikov[125], P. Ryan[86], A.M. Rybin[125], G. Rybkin[112], S. Rzaeva[10], A.F. Saavedra[145], H.F-W. Sadrozinski[134], R. Sadykov[63], H. Sakamoto[150], G. Salamanna [103], A. Salamon[130a], M. Saleem[108], D. Salihagic[97], A. Salnikov[140], J. Salt[161], B.M. Salvachua Ferrando[5], D. Salvatore[36a,36b], F. Salvatore[74], A. Salzburger[41], D. Sampsonidis[149], B.H. Samset[114], M.A. Sanchis Lozano[161], H. Sandaker [13], H.G. Sander[79], M. Sandhoff[167], S. Sandvoss[167], D.P.C. Sankey[126], B. Sanny[167], A. Sansoni[46], C. Santamarina Rios[83], L. Santi[158a,158c], C. Santoni[33], R. Santonico[130a,130b], D. Santos[121b], J.G. Saraiva[121b], T. Sarangi [165], F. Sarri[119a,119b], O. Sasaki[64], T. Sasaki[64], N. Sasao[66], I. Satsounkevitch[88], G. Sauvage[4], P. Savard[152,a], A.Y. Savine[6], V. Savinov[120], L. Sawyer[24,k], D.H. Saxon[52], L.P. Says[33], C. Sbarra[19a,19b], A. Sbrizzi[19a,19b], D.A. Scannicchio, J. Schaarschmidt[43], P. Schacht [97], U. Schäfer[79], S. Schaetzel[29], A.C. Schaffer[112], D. Schaile[96], R. Schamberger[144], A.G. Schamov [104], V.A. Schegelsky[118], M. Schernau[157], M.I. Scherzer[14], C. Schiavi[49a,49b], J. Schieck[97], M. Schioppa[36a,36b], S. Schlenker[29], J.L. Schlereth[5], P. Schmid[29], M.P. Schmidt[168,*], C. Schmitt[20], M. Schmitz[20], M. Schott[29], D. Schouten[139], J. Schovancova[122], M. Schram[83], A. Schreiner[140,d], M.S. Schroers[167], S. Schuh[29], G. Schuler[29], J. Schultes[167], H-C. Schultz-Coulon[57a], J. Schumacher[43], M. Schumacher[47], B.S. Schumm[134], Ph. Schune[133], C.S. Schwanenberger[80], A. Schwartzman[140], Ph. Schwemling[76], R. Schwienhorst[86], R. Schwierz[43], J. Schwindling[133], W.G. Scott[126], E. Sedykh[118],

E. Segura[11], S.C. Seidel[101], A. Seiden[134], F.S. Seifert[43], J.M. Seixas[23a], G. Sekhniaidze[100a], D.M. Seliverstov[118], B. Selldén[142], M. Seman[141], N. Semprini-Cesari[19a,19b], C. Serfon[96], L. Serin[112], R. Seuster[163], H. Severini[108], M.E. Sevior[84], A. Sfyrla[160], L. Shan[32,b], J.T. Shank[21], M. Shapiro[14], P.B. Shatalov[93], L. Shaver[6], C. Shaw[52], K.S. Shaw[136], D. Sherman[29], P. Sherwood[75], A. Shibata[105], M. Shimojima[98], T. Shin[55], A. Shmeleva[92], M.J. Shochet[30], M.A. Shupe[6], P. Sicho[122], A. Sidoti[15], A. Siebel[167], M. Siebel[29], J. Siegrist[14], D. Sijacki[12a], O. Silbert[164], J. Silva[121b], S.B. Silverstein[142], V. Simak[124], Lj. Simic[12a], S. Simion [112], B. Simmons[75], M. Simonyan[4], P. Sinervo[152], V. Sipica[138], G. Siragusa[79], A.N. Sisakyan[63], S.Yu. Sivoklokov[95], J. Sjölin[142], P. Skubic[108], N. Skvorodnev[22], T. Slavicek[124], K. Sliwa[155], J. Sloper[29], T. Sluka[122], V. Smakhtin[164], S.Yu. Smirnov[94], Y. Smirnov[24], L.N. Smirnova[95], O. Smirnova[77], B.C. Smith[56], K.M. Smith[52], M. Smizanska[69], K. Smolek[124], A.A. Snesarev[92], S.W. Snow[80], J. Snow [108], J. Snuverink[103], S. Snyder[24], M. Soares[78], R. Sobie[163,j], J. Sodomka[124], A. Soffer[148], C.A. Solans[161], M. Solar[124], E. Solfaroli Camillocci[129a,129b], A.A. Solodkov[125], O.V. Solovyanov[125], R. Soluk[2], J. Sondericker[24], V. Sopko[124], B. Sopko [124], M. Sosebee[7], V.V. Sosnovtsev[94], L. Sospedra Suay[161], A. Soukharev[104], S. Spagnolo[70a,70b], F. Spanò[34], P. Speckmayer[29], E. Spencer[134], R. Spighi[19a], G. Spigo[29], F. Spila[129a,129b], R. Spiwoks[29], L. Spogli[131a,131b], M. Spousta[123], T. Spreitzer[139], B. Spurlock[7], R.D. St. Denis[52], T. Stahl[138], R. Stamen[57a], S.N. Stancu[157], E. Stanecka[29], R.W. Stanek[5], C. Stanescu[131a], S. Stapnes[114], E.A. Starchenko[125], J. Stark[54], P. Staroba[122], J. Stastny[122], A. Staude[96], P. Stavina[141], G. Stavropoulos[14], P. Steinbach[43], P. Steinberg[24], I. Stekl[124], H.J. Stelzer[41], H. Stenzel[51], K.S. Stevenson[73], G. Stewart[52], T.D. Stewart[139], M.C. Stockton[17], G. Stoicea[25a], S. Stonjek[97], P. Strachota[123], A. Stradling[7], A. Straessner[43], J. Strandberg[85], S. Strandberg[14], A. Strandlie[114], M. Strauss[108], P. Strizenec[141], R. Ströhmer[96], D.M. Strom[111], J.A. Strong[74,*], R. Stroynowski[39], B. Stugu[13], I. Stumer[24,*], D. Su[140], S. Subramania[60], S.I. Suchkov[94], Y. Sugaya[113], T. Sugimoto[99], C. Suhr[5], M. Suk[123], V.V. Sulin[92], S. Sultansoy[3,v], J.E. Sundermann[47], K. Suruliz[158a,158b], S. Sushkov[11], G. Susinno[36a,36b], M.R. Sutton[75], T. Suzuki[150], Yu.M. Sviridov[125], I. Sykora[141], T. Sykora[123], R.R. Szczygiel[38], T. Szymocha[38], J. Sánchez[161], D. Ta[20], A.T. Taffard[157], R. Tafirout[153a], A. Taga[114], Y. Takahashi[99], H. Takai[24], R. Takashima[67], H. Takeda[65], T. Takeshita[137], M. Talby[81], B. Tali[149], A. Talyshev[104], M.C. Tamsett[74], J. Tanaka[150], R. Tanaka[112], S. Tanaka[128], S. Tanaka[64], G.P. Tappern[29], S. Tapprogge[79], S. Tarem[147], F. Tarrade[24], G.F. Tartarelli[87a], P. Tas[123], M. Tasevsky[122], E.T. Tassi[36a,36b], C. Taylor[75], F.E. Taylor[90], G.N. Taylor[84], R.P. Taylor[163], W. Taylor[153b], F. Tegenfeldt[62], P. Teixeira-Dias[74], H. Ten Kate[29], P.K. Teng[146], S. Terada[64], K. Terashi[150], J. Terron[78], M. Terwort[41,n], R.J. Teuscher[152,j], C.M. Tevlin[80], J. Thadome[167], R. Thananuwong[48], M. Thioye[168], J.P. Thomas[17], T.L. Thomas[101], E.N. Thompson[82], P.D. Thompson[17], R.J. Thompson[80], A.S. Thompson[52], E. Thomson[117], R.P. Thun[85], T. Tic [122], V.O. Tikhomirov[92], Y.A. Tikhonov[104], C.J.W.P. Timmermans[102], P. Tipton[168], F.J. Tique Aires Viegas[29], S. Tisserant[81], J. Tobias[47], B. Toczek[37], T.T. Todorov[4], S. Todorova-Nova[155], J. Tojo[64], S. Tokár[141], K. Tokushuku[64], L. Tomasek[122], M. Tomasek[122], F. Tomasz[141], M. Tomoto[99], D. Tompkins[6], L. Tompkins[14], K. Toms[101], A. Tonazzo[131a,131b], G. Tong[32], A. Tonoyan[13], C. Topfel[16], N.D. Topilin[63], E. Torrence[111], E. Torró Pastor[161], J. Toth[81,w], F. Touchard[81], D.R. Tovey[136], S.N. Tovey[84], T. Trefzger[166], L. Tremblet[29], A. Tricoli[126], I.M. Trigger[153a], S. Trincaz-Duvoid[76], M.F. Tripiana[68], N. Triplett[62], W. Trischuk[152], A. Trivedi[24,t], B. Trocmé[54], C. Troncon[87a], C. Tsarouchas[9], J.C-L. Tseng[115], I. Tsiafis[149], M. Tsiakiris[103], P.V. Tsiareshka[88], G. Tsipolitis[9], E.G. Tskhadadze[50], I.I. Tsukerman[93], V. Tsulaia[120], S. Tsuno[64], M. Turala[38], D. Turecek[124], I. Turk Cakir[3,x], E. Turlay[112], P.M. Tuts[34], M.S. Twomey[135], M. Tyndel[126], D. Typaldos[17], G. Tzanakos[8], I. Ueda[150], M. Uhrmacher[53], F. Ukegawa[154], G. Unal[29], D.G. Underwood[5], A. Undrus[24], G. Unel[157], Y. Unno[64], E. Urkovsky[148], P. Urquijo[48], P. Urrejola[31a], G. Usai [30], L. Vacavant[81], V. Vacek[124], B. Vachon[83], S. Vahsen[14], C. Valderanis[97], J. Valenta[122], P. Valente[129a], S. Valkar[123], J.A. Valls Ferrer[161], H. Van der Bij[29], H. van der Graaf[103], E. van der Kraaij[103], E. van der Poel[103], N. van Eldik[82], P. van Gemmeren[5], Z. van Kesteren[103],

I. van Vulpen[103], R. VanBerg[117], W. Vandelli[29], G. Vandoni[29], A. Vaniachine[5], P. Vankov[71], F. Vannucci[76], F. Varela Rodriguez[29], R. Vari[129a], E.W. Varnes[6], D. Varouchas[112], A. Vartapetian[7], K.E. Varvell[145], V.I. Vassilakopoulos[55], L. Vassilieva[92], E. Vataga[101], F. Vazeille[33], G. Vegni[87a,87b], J.J. Veillet[112], C. Vellidis[8], F. Veloso[121b], R. Veness[29], S. Veneziano[129a], A. Ventura[70a,70b], D. Ventura [135], S. Ventura [46], N. Venturi[16], V. Vercesi[116a], M. Verducci[129a,129b], W. Verkerke[103], J.C. Vermeulen[103], M.C. Vetterli[139,a], I. Vichou[160], T. Vickey[165], G.H.A. Viehhauser[115], M. Villa[19a,19b], E.G. Villani[126], M. Villaplana Perez[161], E. Vilucchi[46], M.G. Vincter[28], V.B. Vinogradov[63], M. Virchaux[133,*], S. Viret[33], J. Virzi[14], A. Vitale [19a,19b], O.V. Vitells[164], I. Vivarelli[119a,119b], R. Vives[161], F. Vives Vaques[11], S. Vlachos[9], M. Vlasak[124], N. Vlasov[20], H. Vogt[41], P. Vokac[124], M. Volpi[11], G. Volpini[87a,87b], H. von der Schmitt[97], J. von Loeben[97], E. von Toerne[20], V. Vorobel[123], A.P. Vorobiev[125], V. Vorwerk[11], M. Vos[161], R. Voss[29], T.T. Voss[167], J.H. Vossebeld[71], N. Vranjes[12a], V. Vrba[122], M. Vreeswijk[103], T. Vu Anh[20], M. Vudragovic[12a], R. Vuillermet[29], I. Vukotic[112], P. Wagner [117], H. Wahlen[167], J. Walbersloh[42], J. Walder[69], R. Walker[153a], W. Walkowiak[138], R. Wall[168], C. Wang[44], J. Wang[32], J.C. Wang[135], S.M.W. Wang[146], C.P. Ward[27], M. Warsinsky[47], P.M. Watkins[17], A.T. Watson[17], G. Watts[135], S.W. Watts[80], A.T. Waugh[145], B.M. Waugh[75], M. Webel[47], J. Weber[42], M. Weber[126], M.S. Weber[16], P. Weber[57a], A.R. Weidberg[115], J. Weingarten[42], C. Weiser[47], H. Wellenstein[22], P.S. Wells[29], M. Wen[46], T. Wenaus[24], S. Wendler[120], T. Wengler[80], S. Wenig[29], N. Wermes[20], M. Werner[47], P. Werner[29], U. Werthenbach[138], M. Wessels[57a], S.J. Wheeler-Ellis[157], S.P. Whitaker[21], A. White[7], M.J. White[27], S. White[24], D. Whiteson[157], D. Whittington[60], F. Wicek[112], D. Wicke[167], F.J. Wickens[126], W. Wiedenmann[165], M. Wielers[126], P. Wienemann[20], C. Wiglesworth[71], A. Wildauer[29], M.A. Wildt[79], I. Wilhelm[123], H.G. Wilkens[29], H.H. Williams[117], W. Willis[34], S. Willocq[82], J.A. Wilson[17], M.G. Wilson[140], A. Wilson [85], I. Wingerter-Seez[4], F.W. Winklmeier[29], L. Winton[84], M. Wittgen[140], M.W. Wolter[38], H. Wolters[121b], B. Wosiek[38], J. Wotschack[29], M.J. Woudstra[82], K. Wraight[52], C. Wright[52], B. Wrona[71], S.L. Wu[165], X. Wu[48], S. Xella[35], S. Xie[47], Y. Xie[32], G. Xu[32], N. Xu[165], A. Yamamoto[64], S. Yamamoto[150], T. Yamamura[150], K. Yamanaka[62], T. Yamazaki[150], Y. Yamazaki[65], Z. Yan[21], H. Yang[85], U.K. Yang[80], Y. Yang[32], Z. Yang[28], W-M. Yao[14], Y. Yao[14], Y. Yasu[64], J. Ye[39], S. Ye[24], M. Yilmaz[3,y], R. Yoosoofmiya[120], K. Yorita[30], R. Yoshida[5], C. Young[140], S.P. Youssef[21], D. Yu[24], J. Yu[7], M. Yu[57b], X. Yu[32], J. Yuan[97], L. Yuan[76], A. Yurkewicz[144], R. Zaidan[81], A.M. Zaitsev[125], Z. Zajacova[29], L. Zanello[129a,129b], P. Zarzhitsky[39], A. Zaytsev[104], M. Zdrazil[14], C. Zeitnitz[167], M. Zeller[168], P.F. Zema[29], C. Zendler[20], A.V. Zenin[125], T. Zenis[141], Z. Zenonos[119a,119b], S. Zenz[14], D. Zerwas[112], Z. Zhan[32], H. Zhang[81,z], J. Zhang[5], Q. Zhang[5], W. Zheng[120], X. Zhang[32], L. Zhao[105], T. Zhao[135], Z. Zhao[85], A. Zhelezko[94], A. Zhemchugov[63], S. Zheng[32], J. Zhong[146], B. Zhou[85], N. Zhou[34], S. Zhou[146], Y. Zhou[146], C.G. Zhu[32,b], H. Zhu[136], Y. Zhu[165], X.A. Zhuang[97], V. Zhuravlov[97], B. Zilka[141], R. Zimmermann[20], S. Zimmermann[47], M. Zinna[116a,116b], M. Ziolkowski[138], R. Zitoun[4], L. Živković[34], V.V. Zmouchko[125,*], G. Zobernig[165], A. Zoccoli[19a,19b], M. zur Nedden[15], V. Zychacek[124].

[1] University at Albany, 1400 Washington Ave, Albany, NY 12222, United States of America
[2] University of Alberta, Department of Physics, Centre for Particle Physics, Edmonton, AB T6G 2G7, Canada
[3] Ankara University, Faculty of Sciences, Department of Physics, TR 061000 Tandogan, Ankara, Turkey
[4] LAPP, Université de Savoie, CNRS/IN2P3, Annecy-le-Vieux, France
[5] Argonne National Laboratory, High Energy Physics Division, 9700 S. Cass Avenue, Argonne IL 60439, United States of America
[6] University of Arizona, Department of Physics, Tucson, AZ 85721, United States of America
[7] The University of Texas at Arlington, Department of Physics, Box 19059, Arlington, TX 76019, United States of America
[8] University of Athens, Nuclear & Particle Physics, Department of Physics, Panepistimiopouli,

Zografou, GR 15771 Athens, Greece
[9] National Technical University of Athens, Physics Department, 9-Iroon Polytechniou, GR 15780 Zografou, Greece
[10] Institute of Physics, Azerbaijan Academy of Sciences, H. Javid Avenue 33, AZ 143 Baku, Azerbaijan
[11] Institut de Física d'Altes Energies, IFAE, Edifici Cn, Universitat Autònoma de Barcelona, ES - 08193 Bellaterra (Barcelona), Spain
[12] [(a)]University of Belgrade, Institute of Physics, P.O. Box 57, 11001 Belgrade; Vinca Institute of Nuclear Sciences[(b)], Mihajla Petrovica Alasa 12-14, 11001 Belgrade, Serbia
[13] University of Bergen, Department for Physics and Technology, Allegaten 55, NO - 5007 Bergen, Norway
[14] Lawrence Berkeley National Laboratory and University of California, Physics Division, MS50B-6227, 1 Cyclotron Road, Berkeley, CA 94720, United States of America
[15] Humboldt University, Institute of Physics, Berlin, Newtonstr. 15, D-12489 Berlin, Germany
[16] University of Bern, Laboratory for High Energy Physics, Sidlerstrasse 5, CH - 3012 Bern, Switzerland
[17] University of Birmingham, School of Physics and Astronomy, Edgbaston, Birmingham B15 2TT, United Kingdom
[18] Bogazici University, Faculty of Sciences, Department of Physics, TR - 80815 Bebek-Istanbul, Turkey
[19] INFN Sezione di Bologna[(a)]; Università di Bologna, Dipartimento di Fisica[(b)], viale C. Berti Pichat, 6/2, IT - 40127 Bologna, Italy
[20] University of Bonn, Physikalisches Institut, Nussallee 12, D - 53115 Bonn, Germany
[21] Boston University, Department of Physics, 590 Commonwealth Avenue, Boston, MA 02215, United States of America
[22] Brandeis University, Department of Physics, MS057, 415 South Street, Waltham, MA 02454, United States of America
[23] Universidade Federal do Rio De Janeiro, Instituto de Fisica[(a)], Caixa Postal 68528, Ilha do Fundao, BR - 21945-970 Rio de Janeiro; [(b)]University of Sao Paolo, address in Sao Paolo, Brazil
[24] Brookhaven National Laboratory, Physics Department, Bldg. 510A, Upton, NY 11973, United States of America
[25] National Institute of Physics and Nuclear Engineering[(a)], Bucharest, P.O. Box MG-6, R-077125; [(b)]West University in Timisoara, Bd. Vasile Parvan 4, Timisoara, Romania
[26] Universidad de Buenos Aires, FCEyN, Dto. Fisica, Pab I - C. Universitaria, 1428 Buenos Aires, Argentina
[27] University of Cambridge, Cavendish Laboratory, J J Thomson Avenue, Cambridge CB3 0HE, United Kingdom
[28] Carleton University, Department of Physics, 1125 Colonel By Drive, Ottawa ON K1S 5B6, Canada
[29] CERN, CH - 1211 Geneva 23, Switzerland
[30] University of Chicago, Enrico Fermi Institute, 5640 S. Ellis Avenue, Chicago, IL 60637, United States of America
[31] Pontificia Universidad Católica de Chile, Facultad de Fisica, Departamento de Fisica[(a)], Avda. Vicuna Mackenna 4860, San Joaquin, Santiago; Universidad Técnica Federico Santa María, Departamento de Física[(b)], Avda. Espãna 1680, Casilla 110-V, Valparaíso, Chile
[32] Institute of HEP, Chinese Academy of Sciences, P.O. Box 918, CN-100049 Beijing; USTC, Department of Modern Physics, Hefei, CN-230026 Anhui; Nanjing University, Department of Physics, CN-210093 Nanjing; Shandong University, HEP Group, CN-250100 Shadong, China
[33] Laboratoire de Physique Corpusculaire, CNRS-IN2P3, Université Blaise Pascal, FR - 63177 Aubiere Cedex, France
[34] Columbia University, Nevis Laboratory, 136 So. Broadway, Irvington, NY 10533, United States of

America
[35] University of Copenhagen, Niels Bohr Institute, Blegdamsvej 17, DK - 2100 Kobenhavn 0, Denmark
[36] INFN Gruppo Collegato di Cosenza[(a)]; Università della Calabria, Dipartimento di Fisica[(b)], IT-87036 Arcavacata di Rende, Italy
[37] Faculty of Physics and Applied Computer Science of the AGH-University of Science and Technology, (FPACS, AGH-UST), al. Mickiewicza 30, PL-30059 Cracow, Poland
[38] The Henryk Niewodniczanski Institute of Nuclear Physics, Polish Academy of Sciences, ul. Radzikowskiego 152, PL - 31342 Krakow, Poland
[39] Southern Methodist University, Physics Department, 106 Fondren Science Building, Dallas, TX 75275-0175, United States of America
[40] University of Texas at Dallas, 800 West Campbell Road, Richardson, TX 75080-3021, United States of America
[41] DESY, Hamburg and Zeuthen, Notkestr. 85, D-22603 Hamburg, Germany
[42] Universitaet Dortmund, Experimentelle Physik IV, DE - 44221 Dortmund, Germany
[43] Technical University Dresden, Institut fuer Kern- und Teilchenphysik, Zellescher Weg 19, D-01069 Dresden, Germany
[44] Duke University, Department of Physics, Durham, NC 27708, United States of America
[45] Fachhochschule Wiener Neustadt; Johannes Gutenbergstrasse 3 AT - 2700 Wiener Neustadt, Austria
[46] INFN Laboratori Nazionali di Frascati, via Enrico Fermi 40, IT-00044 Frascati, Italy
[47] Albert-Ludwigs-Universität, Fakultät für Mathematik und Physik, Hermann-Herder Str. 3, D - 79104 Freiburg i.Br. , Germany
[48] Université de Genève, Section de Physique, 24 rue Ernest Ansermet, CH - 1211 Geneve 4, Switzerland
[49] INFN Sezione di Genova[(a)]; Università di Genova, Dipartimento di Fisica[(b)], via Dodecaneso 33, IT - 16146 Genova, Italy
[50] Institute of Physics of the Georgian Academy of Sciences, 6 Tamarashvili St., GE - 380077 Tbilisi; Tbilisi State University, HEP Institute, University St. 9, GE - 380086 Tbilisi, Georgia
[51] Justus-Liebig-Universitaet Giessen, II Physikalisches Institut, Heinrich-Buff Ring 16, D-35392 Giessen, Germany
[52] University of Glasgow, Department of Physics and Astronomy, Glasgow G12 8QQ, United Kingdom
[53] Georg-August-Universitat, II. Physikalisches Institut, Friedrich-Hund Platz 1, D-37077 Goettingen, Germany
[54] Laboratoire de Physique Subatomique et de Cosmologie, CNRS/IN2P3, Université Joseph Fourier, INPG, 53 avenue des Martyrs, FR - 38026 Grenoble Cedex, France
[55] Hampton University, Department of Physics, Hampton, VA 23668, United States of America
[56] Harvard University, Laboratory for Particle Physics and Cosmology, 18 Hammond Street, Cambridge, MA 02138, United States of America
[57] Ruprecht-Karls-Universitaet Heidelberg, Kirchhoff-Institut fuer Physik[(a)], Im Neuenheimer Feld 227, DE - 69120 Heidelberg; ZITI Ruprecht-Karls-University Heidelberg[(b)], Lehrstuhl fuer Informatik V, B6, 23-29, DE - 68131 Mannheim, Germany
[58] Hiroshima University, Faculty of Science, 1-3-1 Kagamiyama, Higashihiroshima-shi, JP - Hiroshima 739-8526, Japan
[59] Hiroshima Institute of Technology, Faculty of Applied Information Science, 2-1-1 Miyake Saeki-ku, Hiroshima-shi, JP - Hiroshima 731-5193, Japan
[60] Indiana University, Department of Physics, Swain Hall West 117, Bloomington, IN 47405-7105, United States of America
[61] Institut fuer Astro- und Teilchenphysik, Technikerstrasse 25, A - 6020 Innsbruck, Austria
[62] Iowa State University, Department of Physics and Astronomy,Ames High Energy Physics Group,

Ames, IA 50011-3160, United States of America
[63] Joint Institute for Nuclear Research, JINR Dubna, RU - 141 980 Moscow Region, Russia
[64] KEK, High Energy Accelerator Research Organization, 1-1 Oho, Tsukuba-shi, Ibaraki-ken 305-0801, Japan
[65] Kobe University, Graduate School of Science, 1-1 Rokkodai-cho, Nada-ku, JP Kobe 657-8501, Japan
[66] Kyoto University, Faculty of Science, Oiwake-cho, Kitashirakawa, Sakyou-ku, Kyoto-shi, JP - Kyoto 606-8502, Japan
[67] Kyoto University of Education, 1 Fukakusa, Fujimori, fushimi-ku, Kyoto-shi, JP - Kyoto 612-8522, Japan
[68] Universidad Nacional de La Plata, FCE, Departamento de Física, IFLP (CONICET-UNLP), C.C. 67, 1900 La Plata, Argentina
[69] Lancaster University, Physics Department, Lancaster LA1 4YB, United Kingdom
[70] INFN Sezione di Lecce[(a)]; Università del Salento, Dipartimento di Fisica[(b)], Via Arnesano IT - 73100 Lecce, Italy
[71] University of Liverpool, Oliver Lodge Laboratory, P.O. Box 147, Oxford Street, Liverpool L69 3BX, United Kingdom
[72] University of Ljubljana, Jožef Stefan Institute and Department of Physics, SI-1000 Ljubljana, Slovenia
[73] Queen Mary University of London, Department of Physics, Mile End Road, London E1 4NS, United Kingdom
[74] Royal Holloway, University of London, Department of Physics, Egham Hill, Egham, Surrey TW20 0EX, United Kingdom
[75] University College London, Department of Physics and Astronomy, Gower Street, London WC1E 6BT, United Kingdom
[76] Laboratoire de Physique Nucléaire et de Hautes Energies, Université Pierre et Marie Curie (Paris 6), Université Denis Diderot (Paris-7), CNRS/IN2P3, Tour 33, 4 place Jussieu, FR - 75252 Paris Cedex 05, France
[77] Lunds universitet, Naturvetenskapliga fakulteten, Fysiska institutionen, Box 118, SE - 221 00 Lund, Sweden
[78] Universidad Autonoma de Madrid, Facultad de Ciencias, Departamento de Fisica Teorica, ES - 28049 Madrid, Spain
[79] Universitaet Mainz, Institut fuer Physik, Staudinger Weg 7, DE - 55099 Mainz, Germany
[80] University of Manchester, School of Physics and Astronomy, Manchester M13 9PL, United Kingdom
[81] CPPM, Aix-Marseille Université, CNRS/IN2P3, Marseille, France
[82] University of Massachusetts, Department of Physics, 710 North Pleasant Street, Amherst, MA 01003, United States of America
[83] McGill University, High Energy Physics Group, 3600 University Street, Montreal, Quebec H3A 2T8, Canada
[84] University of Melbourne, School of Physics, AU - Parkvill, Victoria 3010, Australia
[85] The University of Michigan, Department of Physics, 2477 Randall Laboratory, 500 East University, Ann Arbor, MI 48109-1120, United States of America
[86] Michigan State University, Department of Physics and Astronomy, High Energy Physics Group, East Lansing, MI 48824-2320, United States of America
[87] INFN Sezione di Milano[(a)]; Università di Milano, Dipartimento di Fisica[(b)], via Celoria 16, IT - 20133 Milano, Italy
[88] B.I. Stepanov Institute of Physics, National Academy of Sciences of Belarus, Independence Avenue 68, Minsk 220072, Republic of Belarus
[89] National Scientific & Educational Centre of Particle & High Energy Physics, NC PHEP BSU, M.

Bogdanovich St. 153, Minsk 220040, Republic of Belarus
[90] Massachusetts Institute of Technology, Department of Physics, Room 24-516, Cambridge, MA 02139, United States of America
[91] University of Montreal, Group of Particle Physics, C.P. 6128, Succursale Centre-Ville, Montreal, Quebec, H3C 3J7 , Canada
[92] P.N. Lebedev Institute of Physics, Academy of Sciences, Leninsky pr. 53, RU - 117 924 Moscow, Russia
[93] Institute for Theoretical and Experimental Physics (ITEP), B. Cheremushkinskaya ul. 25, RU 117 259 Moscow, Russia
[94] Moscow Engineering & Physics Institute (MEPhI), Kashirskoe Shosse 31, RU - 115409 Moscow, Russia
[95] Lomonosov Moscow State University, Skobeltsyn Institute of Nuclear Physics, RU - 119 991 GSP-1 Moscow Lenskie gory 1-2, Russia
[96] Ludwig-Maximilians-Universität München, Fakultät für Physik, Am Coulombwall 1, DE - 85748 Garching, Germany
[97] Max-Planck-Institut für Physik, (Werner-Heisenberg-Institut), Föhringer Ring 6, 80805 München, Germany
[98] Nagasaki Institute of Applied Science, 536 Aba-machi, JP Nagasaki 851-0193, Japan
[99] Nagoya University, Graduate School of Science, Furo-Cho, Chikusa-ku, Nagoya, 464-8602, Japan
[100] INFN Sezione di Napoli[(a)]; Università di Napoli, Dipartimento di Scienze Fisiche[(b)], Complesso Universitario di Monte Sant'Angelo, via Cinthia, IT - 80126 Napoli, Italy
[101] University of New Mexico, Department of Physics and Astronomy,Albuquerque, NM 87131, United States of America
[102] Radboud University Nijmegen/NIKHEF, Department of Experimental High Energy Physics, Toernooiveld 1, NL - 6525 ED Nijmegen , Netherlands
[103] Nikhef National Institute for Subatomic Physics, and University of Amsterdam, Kruislaan 409, P.O. Box 41882, NL - 1009 DB Amsterdam, Netherlands
[104] Budker Institute of Nuclear Physics (BINP), RU - Novosibirsk 630 090, Russia
[105] New York University, Department of Physics, 4 Washington Place, New York NY 10003, USA, United States of America
[106] Ohio State University, 191 West Woodruff Ave, Columbus, OH 43210-1117, United States of America
[107] Okayama University, Faculty of Science, Tsushimanaka 3-1-1, Okayama 700-8530, Japan
[108] University of Oklahoma, Homer L. Dodge Department of Physics and Astronomy, 440 West Brooks, Room 100, Norman, OK 73019-0225, United States of America
[109] Oklahoma State University, Department of Physics, 145 Physical Sciences Building, Stillwater, OK 74078-3072, United States of America
[110] Palacký University in Olomouc, streetname, Czech Republic
[111] 1274 University of Oregon, Eugene, OR 97403-1274, United States of America
[112] LAL, Univ. Paris-Sud, IN2P3/CNRS, Orsay, France
[113] Osaka University, Graduate School of Science, Machikaneyama-machi 1-1, Toyonaka, Osaka 560-0043, Japan
[114] University of Oslo, Department of Physics, P.O. Box 1048, Blindern, NO - 0316 Oslo 3, Norway
[115] Oxford University, Department of Physics, Denys Wilkinson Building, Keble Road, Oxford OX1 3RH, United Kingdom
[116] INFN Sezione di Pavia[(a)]; Università di Pavia, Dipartimento di Fisica Nucleare e Teorica[(b)], Via Bassi 6, IT-27100 Pavia, Italy
[117] University of Pennsylvania, Department of Physics, High Energy Physics Group, 209 S. 33rd Street,

Philadelphia, PA 19104, United States of America
[118] Petersburg Nuclear Physics Institute, RU - 188 300 Gatchina, Russia
[119] INFN Sezione di Pisa[(a)]; Università di Pisa, Dipartimento di Fisica E. Fermi[(b)], Largo B.Pontecorvo 3, IT - 56127 Pisa, Italy
[120] University of Pittsburgh, Department of Physics and Astronomy, 3941 O'Hara Street, Pittsburgh, PA 15260, United States of America
[121] [(a)]Universidad de Granada, Departamento de Fisica Teorica y del Cosmos and CAFPE, E-18071 Granada; Laboratorio de Instrumentacao e Fisica Experimental de Particulas - LIP[(b)], Avenida Elias Garcia 14-1, PT - 1000-149 Lisboa, Portugal
[122] Institute of Physics, Academy of Sciences of the Czech Republic, Na Slovance 2, CZ - 18221 Praha 8, Czech Republic
[123] Charles University in Prague, Faculty of Mathematics and Physics, Institute of Particle and Nuclear Physics, V Holesovickach 2, CZ - 18000 Praha 8, Czech Republic
[124] Czech Technical University in Prague, Zikova 4, CZ - 166 35 Praha 6, Czech Republic
[125] Institute for High Energy Physics (IHEP), Federal Agency of Atom. Energy, Moscow Region, RU - 142 284 Protvino, Russia
[126] Rutherford Appleton Laboratory, Science and Technology Facilities Council, Harwell Science and Innovation Campus, Didcot OX11 0QX, United Kingdom
[127] University of Regina, Physics Department, Canada
[128] Ritsumeikan University, Noji Higashi 1 chome 1-1, JP - Kusatsu, Shiga 525-8577, Japan
[129] INFN Sezione di Roma I[(a)]; Università La Sapienza, Dipartimento di Fisica[(b)], Piazzale A. Moro 2, IT- 00185 Roma, Italy
[130] INFN Sezione di Roma Tor Vergata[(a)]; Università di Roma Tor Vergata, Dipartimento di Fisica[(b)] , via della Ricerca Scientifica, IT-00133 Roma, Italy
[131] INFN Sezione di Roma Tre[(a)]; Università Roma Tre, Dipartimento di Fisica[(b)], via della Vasca Navale 84, IT-00146 Roma, Italy
[132] Université Hassan II, Faculté des Sciences Ain Chock[(a)], B.P. 5366, MA - Casablanca; Centre National de l'Energie des Sciences Techniques Nucleaires (CNESTEN)[(b)], Rabat; Université Mohamed Premier[(c)]LPTPM, Faculté des Sciences, B.P.717. Bd. Mohamed VI, 60000, Oujda ; Université Mohammed V, Faculté des Sciences[(d)], BP 1014, MO - Rabat, Morocco
[133] CEA, DSM/IRFU, Centre d'Etudes de Saclay, FR - 91191 Gif-sur-Yvette, France
[134] University of California Santa Cruz, Santa Cruz Institute for Particle Physics (SCIPP), Santa Cruz, CA 95064, United States of America
[135] University of Washington, Seattle, Department of Physics, Box 351560, Seattle, WA 98195-1560, United States of America
[136] University of Sheffield, Department of Physics & Astronomy, Hounsfield Road, Sheffield S3 7RH, United Kingdom
[137] Shinshu University, Department of Physics, Faculty of Science, 3-1-1 Asahi, Matsumoto-shi, JP - Nagano 390-8621, Japan
[138] Universitaet Siegen, Fachbereich Physik, DE - 57068 Siegen, Germany
[139] Simon Fraser University, Department of Physics, 8888 University Drive, CA - Burnaby, BC V5A 1S6, Canada
[140] SLAC National Accelerator Laboratory, Stanford, California 94309, United States of America
[141] Comenius University, Faculty of Mathematics, Physics & Informatics, Mlynska dolina F2, SK - 84248 Bratislava; Institute of Experimental Physics of the Slovak Academy of Sciences, Dept. of Subnuclear Physics, Watsonova 47, SK - 04353 Kosice, Slovak Republic
[142] Stockholm University, Department of Physics, AlbaNova, SE - 106 91 Stockholm, Sweden
[143] Royal Institute of Technology (KTH), Physics Department, SE - 106 91 Stockholm, Sweden

[144] Stony Brook University, Department of Physics and Astronomy, Nicolls Road, Stony Brook, NY 11794-3800, United States of America
[145] University of Sydney, School of Physics, AU - Sydney NSW 2006, Australia
[146] Insitute of Physics, Academia Sinica, TW - Taipei 11529, Taiwan
[147] Technion, Israel Inst. of Technology, Department of Physics, Technion City, IL - Haifa 32000, Israel
[148] Tel Aviv University, Raymond and Beverly Sackler School of Physics and Astronomy, Ramat Aviv, IL - Tel Aviv 69978, Israel
[149] Aristotle University of Thessaloniki, Faculty of Science, Department of Physics, Division of Nuclear & Particle Physics, University Campus, GR - 54124, Thessaloniki, Greece
[150] The University of Tokyo, International Center for Elementary Particle Physics and Department of Physics, 7-3-1 Hongo, Bunkyo-ku, JP - Tokyo 113-0033, Japan
[151] Tokyo Metropolitan University, Graduate School of Science and Technology, 1-1 Minami-Osawa, Hachioji, Tokyo 192-0397, Japan
[152] University of Toronto, Department of Physics, 60 Saint George Street, Toronto M5S 1A7, Ontario, Canada
[153] TRIUMF[(a)], 4004 Wesbrook Mall, Vancouver, B.C. V6T 2A3; [(b)]York University, Department of Physics and Astronomy, 4700 Keele St., Toronto, Ontario, M3J 1P3, Canada
[154] University of Tsukuba, Institute of Pure and Applied Sciences, 1-1-1 Tennoudai, Tsukuba-shi, JP - Ibaraki 305-8571, Japan
[155] Tufts University, Science & Technology Center, 4 Colby Street, Medford, MA 02155, United States of America
[156] Universidad Antonio Narino, Centro de Investigaciones, Cra 3 Este No.47A-15, Bogota, Colombia
[157] University of California, Irvine, Department of Physics & Astronomy, CA 92697-4575, United States of America
[158] INFN Gruppo Collegato di Udine[(a)]; ICTP[(b)], Strada Costiera 11, IT-34014, Trieste; Università di Udine, Dipartimento di Fisica[(c)], via delle Scienze 208, IT - 33100 Udine, Italy
[159] University of Uppsala , Department of Physics and Astronomy, P.O. Box 516, SE- 75120 Uppsala, Sweden
[160] University of Illinois , Department of Physics, 1110 West Green Street, Urbana, Illinois 61801, United States of America
[161] Instituto de Física Corpuscular (IFIC) Centro Mixto UVEG-CSIC, Apdo. 22085 ES-46071 Valencia, Dept. Física At. Mol. y Nuclear; Univ. of Valencia, and Instituto de Microelectrónica de Barcelona (IMB-CNM-CSIC) 08193 Bellaterra Barcelona, Spain
[162] University of British Columbia, Department of Physics, 6224 Agricultural Road, CA - Vancouver, B.C. V6T 1Z1, Canada
[163] University of Victoria, Department of Physics and Astronomy, P.O. Box 3055, Victoria B.C., V8W 3P6, Canada
[164] The Weizmann Institute of Science, Department of Particle Physics, P.O. Box 26, IL - 76100 Rehovot, Israel
[165] University of Wisconsin, Department of Physics, 1150 University Avenue, WI 53706 Madison, Wisconsin, , United States of America
[166] Julius-Maximilians-University of Würzburg, Physikalisches Institute, Am Hubland, 97074 Wuerzburg , Germany
[167] Bergische Universitaet, Fachbereich C, Physik, Postfach 100127, Gauss-Strasse 20, D- 42097 Wuppertal, Germany
[168] Yale University, Department of Physics, PO Box 208121, New Haven CT, 06520-8121 , United States of America
[169] Yerevan Physics Institute, Alikhanian Brothers Street 2, AM - 375036 Yerevan, Armenia

[170] ATLAS-Canada Tier-1 Data Centre 4004 Wesbrook Mall, Vancouver, BC, V6T 2A3, Canada
[171] GridKA Tier-1 FZK, Forschungszentrum Karlsruhe GmbH, Steinbuch Centre for Computing (SCC), Hermann-von-Helmholtz-Platz 1, 76344 Eggenstein-Leopoldshafen, Germany
[172] Port d'Informaci Cientfica (PIC), Universitat Autnoma de Barcelona (UAB), Edifici D, E-08193 Bellaterra, Spain
[173] Centre de Calcul CNRS/IN2P3, Domaine scientifique de la Doua, 27 bd du 11 Novembre 1918, 69622 Villeurbanne Cedex, France
[174] INFN-CNAF, Viale Berti Pichat 6/2, 40127 Bologna, Italy
[175] Nordic Data Grid Facility, NORDUnet A/S, Kastruplundgade 22, 1, DK-2770 Kastrup, Denmark
[176] SARA Reken- en Netwerkdiensten, Science Park 121, 1098 XG Amsterdam, Netherlands
[177] Academia Sinica Grid Computing, Institute of Physics, Academia Sinica, No.128, Sec. 2, Academia Rd., Nankang, Taipei, Taiwan 11529, Taiwan
[178] UK-T1-RAL Tier-1, Rutherford Appleton Laboratory, Science and Technology Facilities Council, Harwell Science and Innovation Campus, Didcot OX11 0QX, United Kingdom
[179] RHIC and ATLAS Computing Facility, Physics Department, Building 510, Brookhaven National Laboratory, Upton, New York 11973, United States of America
[a] Also at TRIUMF, 4004 Wesbrook Mall, Vancouver, B.C. V6T 2A3, Canada
[b] Also at CPPM, Aix-Marseille Université, CNRS/IN2P3, Marseille, France
[c] Also at Gaziantep University, Turkey
[d] University of Iowa, 203 Van Allen Hall, Iowa City IA 52242-1479, United States of America
[e] Also at Laboratoire de Physique Subatomique et de Cosmologie, CNRS/IN2P3, Université Joseph Fourier, INPG, 53 avenue des Martyrs, FR - 38026 Grenoble Cedex, France
[f] Also at Institute for Particle Phenomenology, Ogden Centre for Fundamental Physics, Department of Physics, University of Durham, Science Laboratories, South Rd, Durham DH1 3LE, United Kingdom
[g] Also at TRIUMF, 4004 Wesbrook Mall, Vancouver, B.C. V6T 2A3, Canada
[h] Currently at Dogus University, Kadik
[i] Also at Università di Napoli Parthenope, via A. Acton 38, IT - 80133 Napoli, Italy
[j] Also at Institute of Particle Physics (IPP), Canada
[k] Louisiana Tech University, 305 Wisteria Street, P.O. Box 3178, Ruston, LA 71272, United States of America
[l] Currently at Dumlupinar University, Kutahya, Turkey
[m] Currently at Department of Physics, University of Helsinki, P.O. Box 64, FI-00014, Finland
[n] Also at Institut für Experimentalphysik, Universität Hamburg, Luruper Chaussee 149, 22761 Hamburg, Germany
[o] Also at H. Niewodniczanski Institute of Nuclear Physics PAN, Cracow, Poland
[p] At Department of Physics, California State University, Fresno, 2345 E. San Ramon Avenue, Fresno, CA 93740-8031, United States of America
[q] Also at California Institute of Technology, Physics Department, Pasadena, CA 91125, United States of America
[r] Also at Petersburg Nuclear Physics Institute, RU - 188 300 Gatchina, Russia
[s] Also at Rutherford Appleton Laboratory, Science and Technology Facilities Council, Harwell Science and Innovation Campus, Didcot OX11 0QX, United Kingdom
[t] University of South Carolina, Dept. of Physics and Astronomy, 700 S. Main St, Columbia, SC 29208, United States of America
[u] Also at Institute of Physics, Jagiellonian University, Cracow, Poland
[v] Currently at TOBB University, Ankara, Turkey
[w] Also at KFKI Research Institute for Particle and Nuclear Physics, Budapest, Hungary
[x] Currently at TAEA, Ankara, Turkey

[y] Currently at Gazi University, Ankara, Turkey
[z] Also at Institute of High Energy Physics, Chinese Academy of Sciences, P.O. Box 918, CN-100049 Beijing, China
[*] Deceased

# Acknowledgements

We are greatly indebted to all CERN's departments and to the LHC project for their immense efforts not only in building the LHC, but also for their direct contributions to the construction and installation of the ATLAS detector and its infrastructure. We acknowledge equally warmly all our technical colleagues in the collaborating Institutions without whom the ATLAS detector could not have been built. Furthermore we are grateful to all the funding agencies which supported generously the construction and the commissioning of the ATLAS detector and also provided the computing infrastructure.

The ATLAS detector design and construction has taken about fifteen years, and our thoughts are with all our colleagues who sadly could not see its final realisation.

We acknowledge the support of ANPCyT, Argentina; Yerevan Physics Institute, Armenia; ARC and DEST, Australia; Bundesministerium für Wissenschaft und Forschung, Austria; National Academy of Sciences of Azerbaijan; State Committee on Science & Technologies of the Republic of Belarus; CNPq and FINEP, Brazil; NSERC, NRC, and CFI, Canada; CERN; NSFC, China; Ministry of Education, Youth and Sports of the Czech Republic, Ministry of Industry and Trade of the Czech Republic, and Committee for Collaboration of the Czech Republic with CERN; Danish Natural Science Research Council; European Commission, through the ARTEMIS Research Training Network; IN2P3-CNRS and Dapnia-CEA, France; Georgian Academy of Sciences; BMBF, DESY, DFG and MPG, Germany; Ministry of Education and Religion, through the EPEAEK program PYTHAGORAS II and GSRT, Greece; ISF, MINERVA, GIF, DIP, and Benoziyo Center, Israel; INFN, Italy; MEXT, Japan; CNRST, Morocco; FOM and NWO, Netherlands; The Research Council of Norway; Ministry of Science and Higher Education, Poland; GRICES and FCT, Portugal; Ministry of Education and Research, Romania; Ministry of Education and Science of the Russian Federation, Russian Federal Agency of Science and Innovations, and Russian Federal Agency of Atomic Energy; JINR; Ministry of Science, Serbia; Department of International Science and Technology Cooperation, Ministry of Education of the Slovak Republic; Slovenian Research Agency, Ministry of Higher Education, Science and Technology, Slovenia; Ministerio de Educación y Ciencia, Spain; The Swedish Research Council, The Knut and Alice Wallenberg Foundation, Sweden; State Secretariat for Education and Science, Swiss National Science Foundation, and Cantons of Bern and Geneva, Switzerland; National Science Council, Taiwan; TAEK, Turkey; The Science and Technology Facilities Council, United Kingdom; DOE and NSF, United States of America.

# Introduction

# Preface

The Large Hadron Collider (LHC) at CERN promises a major step forward in the understanding of the fundamental nature of matter. The ATLAS experiment is a general-purpose detector for the LHC, whose design was guided by the need to accommodate the wide spectrum of possible physics signatures. The major remit of the ATLAS experiment is the exploration of the TeV mass scale where ground-breaking discoveries are expected. In the focus are the investigation of the electroweak symmetry breaking and linked to this the search for the Higgs boson as well as the search for Physics beyond the Standard Model.

In this report a detailed examination of the expected performance of the ATLAS detector is provided, with a major aim being to investigate the experimental sensitivity to a wide range of measurements and potential observations of new physical processes. An earlier summary of the expected capabilities of ATLAS was compiled in 1999 [1]. A survey of physics capabilities of the CMS detector was published in [2].

The design of the ATLAS detector has now been finalised, and its construction and installation have been completed [3]. An extensive test-beam programme was undertaken. Furthermore, the simulation and reconstruction software code and frameworks have been completely rewritten. Revisions incorporated reflect improved detector modelling as well as major technical changes to the software technology. Greatly improved understanding of calibration and alignment techniques, and their practical impact on performance, is now in place.

The studies reported here are based on full simulations of the ATLAS detector response. A variety of event generators were employed. The simulation and reconstruction of these large event samples thus provided an important operational test of the new ATLAS software system. In addition, the processing was distributed world-wide over the ATLAS Grid facilities and hence provided an important test of the ATLAS computing system – this is the origin of the expression "CSC studies" ("computing system commissioning"), which is occasionally referred to in these volumes.

The work reported does generally assume that the detector is fully operational, and in this sense represents an idealised detector: establishing the best performance of the ATLAS detector with LHC proton-proton collisions is a challenging task for the future. The results summarised here therefore represent the best estimate of ATLAS capabilities before real operational experience of the full detector with beam. Unless otherwise stated, simulations also do not include the effect of additional interactions in the same or other bunch-crossings, and the effect of neutron background is neglected. Thus simulations correspond to the low-luminosity performance of the ATLAS detector.

This report is broadly divided into two parts: firstly the performance for identification of physics objects is examined in detail, followed by a detailed assessment of the performance of the trigger system. This part is subdivided into chapters surveying the capabilities for charged particle tracking, each of electron/photon, muon and tau identification, jet and missing transverse energy reconstruction, b-tagging algorithms and performance, and finally the trigger system performance. In each chapter of the report, there is a further subdivision into shorter notes describing different aspects studied. The second major subdivision of the report addresses physics measurement capabilities, and new physics search sensitivities. Individual chapters in this part discuss ATLAS physics capabilities in Standard Model QCD and electroweak processes, in the top quark sector, in b-physics, in searches for Higgs bosons, supersymmetry searches, and finally searches for other new particles predicted in more exotic models.

# Cross-Sections, Monte Carlo Simulations and Systematic Uncertainties

**Abstract**
The studies presented in this volume share several common features, including use of the same event samples for Standard Model processes, and the same detector description and simulation framework for all samples. Common cross-section assumptions were made. These assumptions, and the Monte Carlo generator programs employed, are listed. Information is also given on the different detector configurations and geometries simulated, and on the consistent treatment of systematic uncertainties.

## 1 Introduction

The studies presented in this volume have many shared features starting from a common simulation framework of the ATLAS detector, and the same detector description. They are based on a full simulation of the ATLAS detector using the GEANT4 [1] program and the event samples produced were shared by the various analysis groups. For the simulation of the physics events standard event generators for high energy proton-proton collisions were used and interfaced to the ATLAS simulation framework.

In many searches for new particles at the LHC Standard Model processes represent important backgrounds and the signal significance depends on the precise knowledge of these backgrounds. As discussed in several studies presented in this book, methods were investigated on how to determine these cross-sections from the data themselves. However, this will not be always possible, and reliable theoretical predictions must be used to estimate these backgrounds. In addition, the Standard Model cross-sections are relevant for the estimate of signal rates and consequent measurement precision for Standard Model parameters, or for tests of the Standard Model. All studies presented in this book made common assumptions on the cross-sections for Standard Model processes.

In this introductory note features common to the simulations used in these studies are discussed. After reviewing the cross-section assumptions and models used for different processes, the Monte Carlo generator programs employed are summarised. Information is given, next, on the different detector configurations and geometries simulated. Finally, a common treatment is described of systematic uncertainties which affect many analyses.

## 2 Cross-Sections of Physical Processes

A consistent set of cross-sections for Standard Model processes was used in all studies reported in this volume. Over recent years considerable progress has been made in the calculation of higher-order QCD corrections (often expressed as "K-factors") for many physics processes at the LHC. Wherever these corrections are known for both the signal and the dominant background processes, they were included in the analyses. In case the K-factors are not known for the dominant background processes, the studies have consistently refrained from using K-factors and resorted to Born-level predictions for both signal and backgrounds. In this note we detail the values of cross-sections that were used in the different studies reported in this volume: we do not discuss the uncertainties, such as those from missing higher order corrections.

For the simulation of physics processes both leading order (LO) and next-to-leading order (NLO) Monte Carlo programs were used. For the simulation of several processes tree-level matrix element calculations with parton shower matching were adopted. Unless otherwise stated, all tree-level Monte

Carlo calculations were normalized to the NLO cross-section calculation. In the case of parton shower matching several final state parton multiplicities were simulated for predefined shower matching cuts and the sum of the exclusive cross-sections was normalized to the result of the higher-order calculation. By applying this procedure, it is expected that the shapes of inclusive distributions are reasonably well described. However, large uncertainties are expected in the absolute cross-section predictions in extreme phase-space regions such as, for example, final states with high jet multiplicities.

Table 1: Leading order (LO) and higher order (N)NLO cross-sections for some important Standard Model production processes for pp collisions at a centre-of-mass energy of 14 TeV. In the calculation of all cross-sections the CTEQ6L and CTEQ6M structure function parametrizations have been used. For inclusive $W$ and $Z$ production, the cross-section quoted includes the branching ratio into one lepton generation.

| Process | Comments | Reference | Order in pert. theory | $\sigma$ (nb) |
|---|---|---|---|---|
| Total inelastic pp | | PYTHIA [2] | | $79 \cdot 10^6$ |
| Non Single Diffractive | | PYTHIA [2] | | $65 \cdot 10^6$ |
| Dijet | $p_T^{jet} > 25$ GeV | PYTHIA [2] | LO | $367 \cdot 10^3$ |
| | | NLOJET++ [3, 4] | NLO | $477 \cdot 10^3$ |
| $\gamma$-jet | $p_T^{\gamma} > 25$ GeV | PYTHIA [2] | LO | 180 |
| $b\bar{b} \to \mu + X$ | $p_T^{\mu} > 6$ GeV | PYTHIA [2] | LO | $6.1 \cdot 10^3$ |
| $b\bar{b} \to \mu\mu + X$ | $p_T^{\mu_1/\mu_2} > 6 / 4$ GeV | PYTHIA [2] | LO | 110 |
| $t\bar{t}$ | | | NLO | 0.794 |
| | | Ref. [5] | NLO+NLL | 0.833 |
| Single top production | t-channel | AcerMC [6] | LO | 0.251 |
| | | Ref. [7–9] | NLO | 0.246 |
| | s-channel | AcerMC [6] | LO | 0.007 |
| | | Ref. [7] | NLO | 0.011 |
| | Wt | AcerMC [6] | LO | 0.058 |
| | | Ref. [10–12] | NLO | 0.066 |
| $W \to \ell\nu$ | | FEWZ [13] | LO | 16.8 |
| | | FEWZ [13] | NLO | 20.7 |
| | | FEWZ [13] | NNLO | 20.5 |
| $Z \to \ell\ell$ | $m_{\ell\ell} > 60$ GeV | FEWZ [13] | LO | 1.66 |
| | | FEWZ [13] | NLO | 2.03 |
| | | FEWZ [13] | NNLO | 2.02 |
| WW | $m_{W^{(*)}} > 20$ GeV, $p_T^W > 10$ GeV | MCFM [14] | LO | 0.072 |
| | | MCFM [14] | NLO | 0.112 |
| WZ | $m_{W^{(*)}/Z^{(*)}} > 20$ GeV, $p_T^{W/Z} > 10$ GeV | MCFM [14] | LO | 0.032 |
| | | MCFM [14] | NLO | 0.056 |
| ZZ | $m_{Z^{(*)}} > 12$ GeV | MCFM [14] | LO | 0.0165 |
| | | MCFM [14] | NLO | 0.0221 |
| $\gamma\gamma$ $(qq, qg \to \gamma\gamma)$ | $80 < m_{\gamma\gamma} < 150$ GeV | RESBOS [15] | NLO | 0.0209 |
| $(gg \to \gamma\gamma)$ | $80 < m_{\gamma\gamma} < 150$ GeV | RESBOS [15] | NLO | 0.0080 |

The cross-sections for the most relevant Standard Model production processes are summarised in Table 1. The cross-sections for new physics signal processes are presented in the respective sub-chapters of this book. For the calculation of the leading order cross-sections the CTEQ6L [16, 17] set of structure function parametrizations was used. Processes available at (N)NLO were calculated by using the

CTEQ6M [16, 17] parametrizations. The following comments concern the various cross-sections:

- The total pp cross-section at a centre-of-mass energy of 14 TeV is predicted by PYTHIA [2] to be 102 mb. This is split into elastic (23 mb) and inelastic (79 mb) parts. The total inelastic pp cross-section includes contributions from single and double-diffractive scattering which are estimated to be 14 and 10 mb, respectively. The non-single diffractive cross-section, which is usually also denoted as the *minimum bias* cross-section, is given by $\sigma_{\text{NSD}} = \sigma_{\text{inel.}} - \sigma_{\text{SD}}$ = 65 mb.

- Multijet production via QCD processes is the dominant high-$p_{\text{T}}$ process at the LHC and is an important background in many physics studies. Even if next-to-leading order corrections are partially known, the remaining uncertainties from missing higher-order corrections remain large. We therefore used leading-order estimates in most physics studies and large errors were assigned to cover the uncertainty.

- The pair production of $b$-quarks provides a copious source of leptons at the LHC. The single and dimuon cross-sections from $b\bar{b}$ production were calculated with $p_{\text{T}}$ thresholds as expected at the trigger level. A leading order PYTHIA calculation has been used in the present studies. Even if the higher-order corrections are known [18], large uncertainties remain.

- For the $t\bar{t}$ production cross-section several calculations beyond leading order exist. In the studies presented in this volume the NLO calculation including a next-to-leading log (NLL) resummation [5] was used. The cross-sections for the three relevant sub-processes for single-top production were calculated at NLO.

- The inclusive production cross-sections of $W$ and $Z$ bosons are known at next-to-next-to-leading order (NNLO) and these values were used in the studies. The residual uncertainties from variations of the renormalization and factorization scales are estimated to be at the level of a few percent [13]. In many cases the production of $W$ and $Z$ bosons with jets constitutes an important background to searches. Exclusive $W/Z$ + jet cross sections have in general been calculated with leading order Monte Carlos, such as PYTHIA, or the parton shower matched Monte Carlos ALPGEN or Sherpa. These calculations were normalized to the inclusive NNLO cross-sections. Only in case of the $Wb\bar{b}$ and $Zb\bar{b}$ production were exclusive NLO cross-sections calculated, to which the tree-level Monte Carlo generator results were normalized. The results of these calculations for a few relevant phase space regions are:

| Process | Comments | Reference | Order in pert. theory | $\sigma$ (pb) |
|---|---|---|---|---|
| $Wb\bar{b}$ | $p_{\text{T}}^{b} > 10$ GeV, $\lvert\eta_b\rvert < 2.5$, $\Delta R_{b\bar{b}} >0.7$ | ALPGEN [19] | LO | 68.7 |
| | $m_W^{(*)} > 30$ GeV, $m_{b\bar{b}} > 9.24$ GeV | MCFM [14] | NLO | 176.9 |
| $Zb\bar{b}$ | $p_{\text{T}}^{b} > 10$ GeV, $\lvert\eta_b\rvert < 2.5$, $\Delta R_{b\bar{b}} >0.7$ | AcerMC [6] | LO | 60.7 |
| | $m_Z^{(*)} > 30$ GeV, $m_{b\bar{b}} > 9.24$ GeV | MCFM [14] | NLO | 86.4 |
| $Zb\bar{b}$ | $p_{\text{T}}^{b} > 5$ GeV, $\lvert\eta_b\rvert < 2.5$, $\Delta R_{b\bar{b}} >0.7$ | AcerMC [6] | LO | 27.9 |
| | $m_Z^{(*)} > 60$ GeV, $m_{b\bar{b}} > 9.24$ GeV | MCFM [14] | NLO | 44.8 |

- The cross-sections for diboson production are available at NLO. In addition to the $q\bar{q}$-initiated processes, the $gg$ box-diagram contributions are sizeable, and both have been taken into account in the analyses. For $ZZ$ production the $gg$ box contributions were estimated to be at the level of 30% [20] and the NLO result was scaled accordingly. A re-evaluation of this contribution using the program of Ref. [21] yielded a contribution of 23.8%. For the $\gamma\gamma$ production process the box contribution was calculated using the RESBOS Monte Carlo program [15].

## 3 Monte Carlo Simulation

The samples of fully simulated events were made using a variety of Monte Carlo generators. Interfaces in the ATLAS software framework provided mechanisms to feed the particle-level events generated into the ATLAS simulation software packages. The production of these events was a major effort: a plethora of physics processes were simulated, and over 1300 different data sets were produced. Unless otherwise stated, samples were produced simulating only one proton-proton interaction: the effect of additional interactions was neglected.

The principal general-purpose Monte Carlo generators employed were PYTHIA, HERWIG, Sherpa, AcerMC, ALPGEN, MadGraph/MadEvent and MC@NLO. In addition to these, further generators were used for specific processes: Charybdis, CompHEP, TopReX and WINHAC. The versions of the generators used are summarised in Table 2. Parton-level Monte Carlo generators used either PYTHIA or HERWIG/JIMMYfor hadronisation and underlying event modelling. HERWIG hadronisation was complemented by an underlying event simulation from the JIMMY program [22] (versions 4.2 and 4.31). The underlying event model parameters were tuned, for PYTHIA and HERWIG/JIMMY, to published data from Tevatron and other experiments, as described in Ref. [23] and references therein. For Sherpa, the default parton shower and underlying event modelling was used. Examples of the specific processes generated with each program are given in the Appendix.

Table 2: Monte Carlo event generators used for the production of event samples for the studies reported here. The fourth column shows, for the parton-level event generators, which software was used for the hadronisation and underlying event (UE) simulation.

| Generator | Versions | Reference | Hadronisation+UE |
|---|---|---|---|
| PYTHIA | 6.323-6.411 | [2] | |
| HERWIG | 6.508-6.510 | [24] | JIMMY for UE |
| Sherpa | 1.008-1.011 | [25] | |
| AcerMC | 3.1-3.4 | [26] | PYTHIA,HERWIG |
| ALPGEN | 2.05-2.13 | [19] | HERWIG/JIMMY |
| MadGraph/MadEvent | 3.X-4.15 | [27] | PYTHIA |
| MC@NLO | 3.1-3.3 | [28] | HERWIG/JIMMY |
| Charybdis | 1.001-1.003 | [29] | HERWIG/JIMMY |
| CompHEP | – | [30] | PYTHIA |
| TopReX | 4.11 | [31] | PYTHIA |
| WINHAC | 1.21 | [32] | PYTHIA |

The decay of $\tau$ leptons was normally not treated by the main Monte Carlo generators themselves, but rather via the TAUOLA package [33], version 2.7. The radiation of photons from charged leptons was also treated specially, using the PHOTOS QED radiation package, version 2.15 [34]. These two packages were used for a range of processes and generators: this required implementation of new interfaces for HERWIG and Sherpa. When simulating specific b-hadron decays for B-physics analyses [35], the EvtGen [36] dedicated b-hadron decay package was used in combination with PYTHIA.

The Monte Carlo tools in ATLAS are taken, where available, from the LHC Computing Grid GENSER (generator services) sub-project [37]. These are modified with custom ATLAS software patches when needed. For most Monte Carlo programs more than one version was employed during the long series of simulations: changes in version were motivated by physical model, or technical improvements to the package. Common particle mass definitions were also used where relevant (for example, the top mass was taken to be 175 GeV, unless otherwise stated). The Monte Carlo tools are then either wrapped in-side the ATLAS Athena environment [38], or interfaced via the Les Houches accord event format [39], depending on the implementation simplicity. The latter interfaces were used for the Sherpa, AcerMC, ALPGEN, MadGraph/MadEvent, MC@NLO and CompHEP event generation. These interfaces rely on widespread use of the HepMC C++-based event record format [40]: several improvements were made during the series of event production processings.

LHAPDF, the Les Houches accord PDF interface library [41], was used throughout, and was linked to all Monte Carlo event generators to provide the PDF set values. The PDF sets [16] used were CTEQ6L for leading order (LO) Monte Carlo event generators, and CTEQ6M for the next-to-leading order (NLO) Monte Carlo event generator MC@NLO.

# 4 Detector Description

One important aspect of the Computing Commissioning Challenge was the test of the alignment and calibration procedures with an imperfect, *i.e.* more realistic, description of the ATLAS detector. In particular, misalignments were introduced for the inner detector and additional material was added in the inner detector and in front of the calorimeters. In addition, distorted magnetic field configurations were introduced, where the symmetry axis of the field did not coincide with the beam axis.

The goal was to establish and validate the alignment and calibration procedures and to determine the known distortions. This has a strong physics motivation: for example, a knowledge of the energy scale of the electromagnetic calorimeter with a precision of 0.02%, as required for a precise measurement of the W mass, requires knowledge of the total radiation length of the material in the inner detector with a precision at the level of 1%.

Two different geometries were used in the simulations. In a so-called *as-built geometry* realistic alignment shifts and distortions of the magnetic field were introduced. In the *distorted geometry* additional material was added. The calibration samples were simulated and calibration constants determined with the *as-built geometry*. All physics samples were, however, simulated with the *distorted geometry* and the calibrations constants as determined from the as-built geometry were applied.

**As-built geometry** The *as-built geometry* includes misalignments of the main subdetectors (pixel detector, silicon microstrip tracker (SCT), and transition radiation tracker (TRT)) of the inner detector. The misalignments were introduced as independent translations and rotations at three levels: (i) of the main subdetector parts (pixel detector, SCT barrel, two SCT endcaps, TRT barrel and two TRT endcaps), (ii) of major detector sub-units, like pixel and SCT barrel layers, pixel and SCT endcap disks and TRT barrel modules and (iii) of individual silicon detector modules. The sizes of displacements were chosen to lie within the expected build tolerances. The actual displacements were assigned randomly in most cases.

The shifts described in the following were applied for the levels (i) and (ii) in the global ATLAS coordinate system, defined as a right-handed system with the $x$-axis pointing to the centre of the LHC ring, the $y$-axis in the vertical direction and the $z$-axis along the beam direction. The level (iii) misalignments refer to the local coordinate system of individual detector modules.

At level (i), the whole subdetector parts were displaced in the three spatial coordinates at the level of 1-2 mm followed by rotations around the three axes at the level of 10-50 mrad.

The alignments of the endcap detector sub-units include additional in-plane ($x$-$y$) displacements and rotations around the $z$-axis. They were generated randomly from uniform distributions centred around zero with a width of $\pm 150\ \mu$m and $\pm 1$ mrad, respectively. For the TRT barrel modules the translations are generated randomly from uniform distributions around zero, with widths of $\pm 200\ \mu$m, $\pm 100\ \mu$m and $\pm 300\ \mu$m respectively for modules of the three TRT layers. In addition, a systematic radial shift of +1.0,

-0.5 and +1.5 mm is applied for all modules of the respective layers. No rotations nor displacements were applied to the TRT endcap modules.

For the individual pixel and SCT detector modules individual position displacements were applied randomly from uniform distributions with widths of 30-50 $\mu$m for pixel and 100-150 $\mu$m for SCT modules, followed by rotations around the three axes, also randomly chosen from uniform distributions with widths of $\pm 1$ mrad.

**Distorted geometry** The *distorted geometry* is based on the *as-built geometry* with additional material added in different locations of the inner detector and in front of the electromagnetic calorimeter. Material corresponding to an increase of 1-3% of a radiation length was added just behind the first pixel layer, and just behind the second SCT layer, and in the endcaps adjacent to one of the endcap pixel disks and adjacent to two of the endcap SCT disks. This amount of additional material is considered to be much larger than the uncertainty on the knowledge of the exact amount of the material. Within the active tracking volume the material in regions of service routing was increased by 1-5% of a radiation lenght. For services outside the active tracking volume the material was increased by up to 15% of a radiation length. These increases are also expected to be larger than the uncertainties. It should be noted that for the inner detector the extra material was only added in one half of the azimuthal angle ($0 < \phi < \pi$) to allow for a straightforward study of the difference in calibration and performance with single particles.

Additional material was also added in a $\phi$-asymmetric way in front of the calorimeter. In the region $\eta > 0$ additional material corresponding to 8-11% $X_0$ were added in front of the barrel cryostat, 5% $X_0$ between the barrel presampler and strip layers (in $\pi/2 < \phi < 3\pi/2$), and 7-11% $X_0$ behind the cryostat. In the region $\eta < 0$, additional material corresponding to 5% $X_0$ was added between the barrel presampler and the strip layer in the region $-\pi/2 < \phi < \pi/2$. The density of material in the gap between the barrel and the endcap cryostat was increased by 70%. Again, this is considered to be conservative and larger than the uncertainties on the precise knowledge of the material distribution in this region of the detector.

**Applications in performance of physics studies** Several performance studies were carried out using the as-built and distorted geometries in simulation and the impact is documented elsewhere in this volume. Among the important studies is the impact of the misalignments on the b-tagging performance or on the reconstructed resolution of the Z resonance in muon final states. In addition the impact on the mass resolution and reconstruction efficiencies was studied for $H \to \gamma\gamma$, $H \to ZZ \to 4\ell$ and $Z \to ee$ samples.

## 5 Treatment of systematic uncertainties

The results of the physics and performance studies are affected by systematic uncertainties, some of which are common to many studies. To allow a uniform treatment of these uncertainties across the various analyses, the following effects and prescriptions were applied.

There are detector-related uncertainties, such as those on particle identification efficiencies, on background rejections, and on the precise knowledge of energy scales and resolution functions. These uncertainties can be largely constrained and determined from the data themselves. However, this can only be done with a finite, and integrated luminosity-dependent, accuracy. In the present studies, rough estimates of these uncertainties were used, considering three canonical integrated luminosity values: 0.1, 1 and 10 fb$^{-1}$. Detector-related systematic uncertainties were applied to signal and background samples by varying the energy scale, resolution, or efficiency or rejections.

In addition, uncertainties come from the approximations made in Monte Carlo generators, modelling and from the theoretical calculation of cross-sections. Unless stated otherwise when discussing individual analyses, the following assumptions were applied, for the various systematic uncertainties.

## 5.1 Uncertainties on the detector performance

**Electrons and photons** For electrons and photons, uncertainties on the identification efficiency of 1.0%, 0.5% and 0.2% were assumed for the three values of integrated luminosity, 0.1, 1 and 10 $fb^{-1}$, respectively. These values can be determined from data by applying the so called tag-and-probe methods [42, 43] to known resonance decays, like $Z \to ee$. The uncertainty on the energy scale was assumed to be 1% (0.1%) for integrated luminosities below (above) 1 $fb^{-1}$. The electron and photon resolutions were estimated to be known with precisions of 20%, 10% and 5% at the three values of integrated luminosity. The electron fake rates were assumed to have overall uncertainties of 50%, 20% and 10% at the three integrated luminosity values. All uncertainties were assumed to be independent of $p_{\mathrm{T}}$ and $\eta$.

**Muons** Uncertainties on the identification efficiency of 1%, 0.3% and 0.1% were used for muons with $p_T < 100$ GeV for the three integrated luminosity values. As for electrons, it should be noted that these numbers are expected to be conservative, since the statistical precision that can be obtained from studies of $Z \to \mu\mu$ decays amounts to 0.2% for an integrated luminosity of 0.1 $fb^{-1}$. For higher muon momenta the efficiencies must be estimated using extrapolations based on Monte Carlo and therefore larger values were assumed: for muons with a $p_{\mathrm{T}}$ of 1 TeV, for example, the uncertainties were assumed to be 5%, 3% and 1%, respectively.

The muon energy scale was assumed to be known with precisions of 1%, 0.3% and 0.1% for the three integrated luminosity values. Furthermore, uncertainties of 12%, 4% and 1% were assumed on the muon momentum resolution below 100 GeV, whereas a value of 100% was used for muons with $p_T$ of 1 TeV. All these uncertainties were considered to be independent of $\eta$.

**Jets and Missing $E_T$** Unless otherwise stated, the overall uncertainty on the jet energy scale was assumed to $\pm5\%$ over the pseudorapidity region $|\eta| < 3.2$ and $\pm10\%$ for jets in the forward calorimeters, $3.2 < |\eta| < 4.9$. This scale uncertainty is applied, independently of jet $p_{\mathrm{T}}$, for both light-quark jets and jets from $b$-quarks. In addition, unless stated otherwise, an uncertainty of 10% on the jet energy resolution was considered.

The missing transverse energy, $E_{\mathrm{T}}^{\mathrm{miss}}$, is calculated by summing high-$p_{\mathrm{T}}$ objects like leptons and jets, in addition a component from unclustered energy is added. Part of the uncertainty in $E_{\mathrm{T}}^{\mathrm{miss}}$ is thus correlated with the jet and lepton energy scale uncertainties, but also a wrong calibration of unclustered energy can affect $E_{\mathrm{T}}^{\mathrm{miss}}$.

After identified objects were rescaled or smeared, the $E_{\mathrm{T}}^{\mathrm{miss}}$ was re-calculated with the corrected energies. In most of the studies also the low $p_{\mathrm{T}}$ part of the unclustered energy was modified. In this procedure, the momenta of the leptons and jets with $p_T > 20$ GeV were subtracted first from the $E_{\mathrm{T}}^{\mathrm{miss}}$, a 10% uncertainty on the remainder was applied, and then the effects of the leptons and jets were added back in.

**Heavy-flavour tagging** For the $b$-tagging efficiency a 5% relative uncertainty was assumed, independently of luminosity. This is considered to be a conservative estimate for integrated luminosities of 1 $fb^{-1}$ or higher. For the mistag rate of light and $c$-jets an integrated-luminosity independent uncertainty of 10% was assumed. It is expected that the mistag rates can be measured with this precsion or better from data sets exceeding an integrated luminosity of 0.1 $fb^{-1}$. The efficiency and mistag variations were implemented in analyses by randomly rejecting 5% of the jets tagged as $b$-jets or by randomly changing the tag status of light and $c$-jets.

### 5.2 Uncertainties on cross-sections and Monte Carlo modelling

Several theoretical uncertainties affect the predicted cross-sections. The details and the size of the uncertainty depend on the signal and background processes considered and no general numbers can be quoted. They are therefore usually addressed in the respective studies presented in this volume. The main effects can be classified as follows:

- The theoretical calculations are affected by missing unknown higher-order corrections. These uncertainties are usually estimated by varying the renormalization and factorization scales within factors of two around the nominal scale chosen.
- Despite the normalization of tree-level Monte Carlo programs – with or without parton shower matching – to the (N)NLO cross sections, large uncertainties remain, in particular for exclusive final states in specific phase space regions after the application of cuts. These uncertainties have been estimated either by varying parton-shower matching cuts or by comparisons with different Monte Carlo event generators.
- Uncertainties in the parton distribution functions result in uncertainties on the calculated cross-sections which are typically of the order of 10%. These uncertainties have either been addressed by varying the eigenvalues of the CTEQ parametrization parameters [17] within the suggested values or by comparing the CTEQ and MRST2001 [44] parametrizations.

## Appendix

In the following, additional technical information is given on some of the Monte Carlo event generators employed, together with example processes.

#### PYTHIA

The PYTHIA Monte Carlo event generator [2] was employed for the event simulation of many samples. The new implementation of parton showering, commonly known as $p_{\mathrm{T}}$ -ordered showering, was used, as was the new underlying event model where the phase-space is interleaved/shared between initial-state radiation (ISR) and the underlying event. In addition to the standard processes implemented in PYTHIA, two extensions were implemented containing a chiral lagrangian model [45], and an R-hadron model.

#### HERWIG and JIMMY

HERWIG [24] was used, for example, for simulation of SUSY signal processes [46]. The pre-generated input tables for these processes were provided by ISAJET and ISAWIG [47].

#### Sherpa

The Sherpa Monte Carlo event generator [25], was used for several processes, most notably for the production of electroweak bosons in association with jets: these profited from the implemented CKKW parton-showering and matrix-element matching technique. Some representative processes for which Sherpa was used are: a $W$ or $Z$ produced in association with up to four light jets; Higgs boson production via vector boson fusion; and associated production of $b\bar{b}A$.

### AcerMC

Some processes for which AcerMC [26] was used were: $Zb\bar{b}$ production; $Zt\bar{t}$ production; $t\bar{t}$ production; single top processes; $t\bar{t}b\bar{b}$ production; and $t\bar{t}t\bar{t}$ production. The AcerMC program was used both with PYTHIA and HERWIG hadronisation, to allow tests of systematic uncertainties related to parton shower modelling.

A procedure was developed for combining samples with $t\bar{t}$ production modelled with MC@NLO with samples from the AcerMC $t\bar{t}b\bar{b}$ process. There is an overlap of the two samples since the extra gluon in the NLO $t\bar{t}$ calculation can split into a $b\bar{b}$ pair during parton showering. For studies where this channel was relevant [48], events with additional $b\bar{b}$ pairs in the MC@NLO samples were rejected, since the matrix-element $t\bar{t}b\bar{b}$ generation is expected to describe such events better in the region of the phase space selected by the analysis (especially for relatively large opening angle between the two quarks of the $b\bar{b}$ pair). The corresponding number of events (10% of the total) was also removed from the high jet-multiplicity $t\bar{t}$ sample for normalization purposes.

### ALPGEN

The ALPGEN Monte Carlo event generator [19] was used for several processes, most notably for the production of electroweak bosons in association with jets, in order to profit from the implemented MLM parton-showering and matrix-element matching technique. Some processes for which ALPGEN was used were: $W$ or $Z$ production in association with up to five light jets; $t\bar{t}$ production with up to three additional light jets; $b\bar{b}$ or $c\bar{c}$ production with up to three additional light jets; electroweak boson pair production in association with up to three jets Higgs production via vector boson fusion; and photon pair production in association with up to three jets.

### MadGraph/MadEvent

The MadGraph/MadEvent Monte Carlo event generator [27] was used for a selection of processes, for example for exclusive final states involving multiple electroweak bosons and associated light jets, as well as some Standard Model Higgs boson production channels. Although MadGraph/MadEvent processes in the 4.X versions can be combined with a native version of parton-showering and matrix-element matching technique, this functionality was not used here. Some representative processes for which MadGraph/MadEvent was used are: $W$ or $Z$ production in association with four light partons; $WW$, $WZ$ or $ZZ$ pair production in association with two light partons; electroweak boson production in association with two photons; and photon pair production in association with two additional partons.

### MC@NLO

The MC@NLO event generator [28] is one of the few Monte Carlo tools incorporating full NLO QCD corrections to a selected set of processes in a consistent way. It was used to simulate a number of processes, including: inclusive $W$ or $Z$ production; $t\bar{t}$ production; electroweak boson pair production; and Higgs boson production and decay, for the $W^+W^-$ and $\gamma\gamma$ Higgs boson decay modes.

### Charybdis

The Charybdis Monte Carlo event generator [29] is a special-purpose program implementing production and decay of microscopic black holes in models with TeV-scale gravity.

**CompHEP**

The CompHEP Monte Carlo event generator [30] was used for a small set of processes: excited electron production, $Z' \to e^+e^-\gamma$, and the production of E6 heavy iso-singlet D quarks decaying to $Z$ or $W$ pairs, or to a $ZH$ pair in association with additional quarks.

**TopReX**

The TopReX Monte Carlo event generator [31] was used for top pair or single top production involving flavour-changing neutral current (FCNC) couplings in top quark decays, explicitly: $t\bar{t}$ production where one top quark decays conventionally (to $bW$), and the other to either $q\gamma$ or $qZ$; and single top production and decay to either $q\gamma$ or $qZ$. TopReX was interfaced directly with PYTHIA for parton showering, hadronisation and the underlying event: a point to note is that TopReX single top generation is intimately interfaced with the PYTHIA old (virtuality-ordered) parton showering model and thus cannot be used with the new PYTHIA $p_T$ -ordered showering.

**WINHAC**

WINHAC [32] is a Monte Carlo event generator dedicated to the hadro-production of single $W$ bosons decaying into leptons. Comparisons done within ATLAS have shown that the WINHAC predictions match well the predictions of PHOTOS for radiative corrections to $W$ boson leptonic decays. PHOTOS was used throughout this work.

# Tracking

# The Expected Performance of the Inner Detector

**Abstract**

The ATLAS inner detector will see of the order of 1000 charged particle tracks for every beam crossing at the design luminosity of the CERN Large Hadron Collider (LHC). This paper summarizes the design of the detector and outlines the reconstruction software. The expected performance for reconstructing single particles is presented, along with an indication of the vertexing capabilities. The effect of the detector material on electrons and photons is discussed along with methods for improving their reconstruction. The studies presented focus on the performance expected for the initial running at the start-up of the LHC.

## 1 Introduction

In ATLAS, at the LHC design luminosity of $10^{34}$ cm$^{-2}$s$^{-1}$, approximately 1000 particles will emerge from the collision point every 25 ns within $|\eta| < 2.5$, creating a very large track density in the detector. To achieve the momentum and vertex resolution requirements imposed by the benchmark physics processes, high-precision measurements will be made in the inner detector (ID), shown in Fig. 1. Pixel and silicon microstrip (SCT) trackers, used in conjunction with the straw tubes of the transition radiation tracker (TRT), will make high-granularity measurements. The original performance specifications were set out in 1994 and are detailed in [1] – the focus being on challenging physics channels such as the measurement of leptons from the decays of heavy gauge bosons and the tagging of $b$-quark jets.

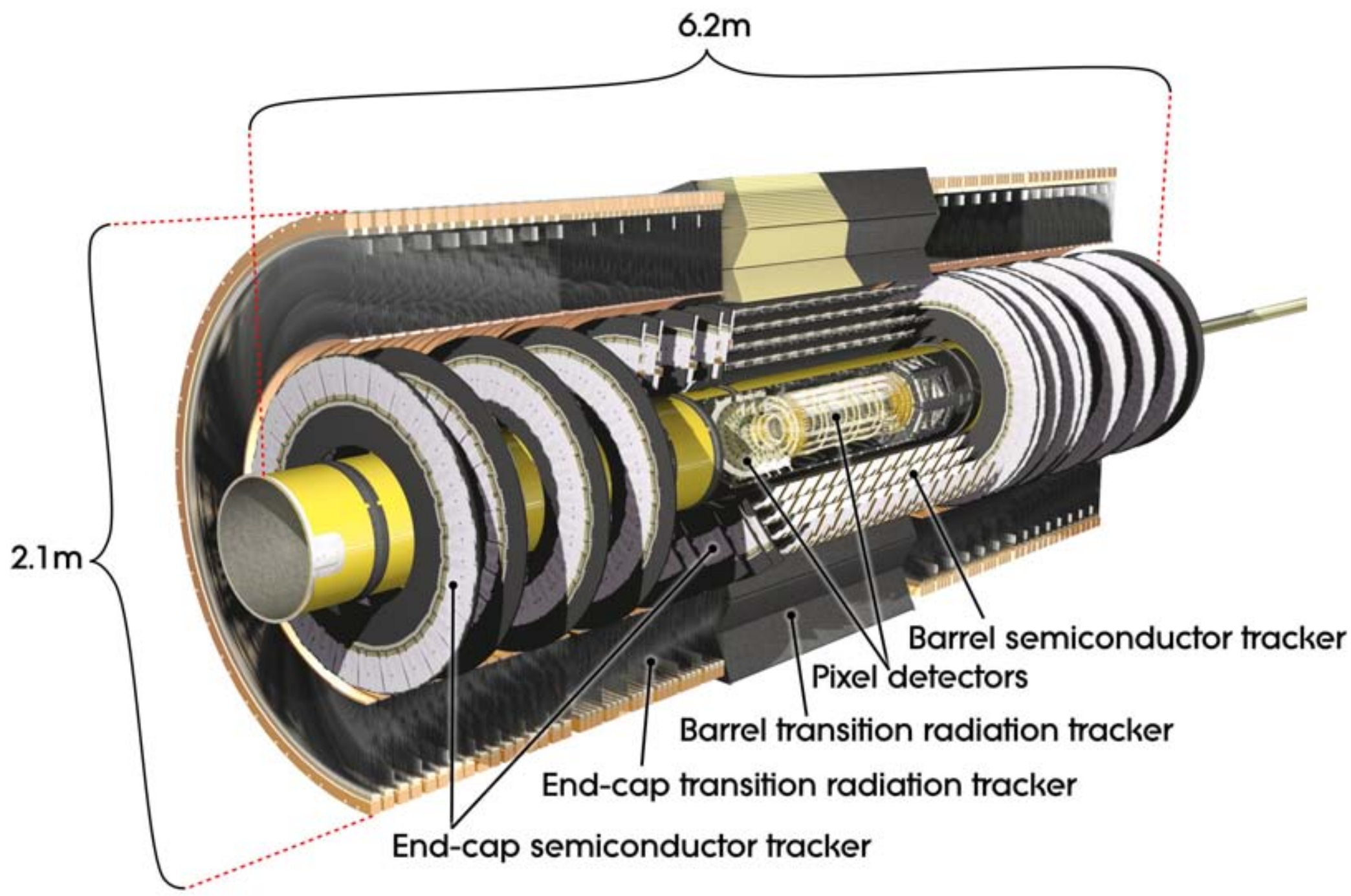

Figure 1: Cut-away view of the ATLAS inner detector.

The ID surrounds the LHC beam-pipe which is inside a radius of 36 mm. The layout of the detector is illustrated in Fig. 2 and detailed in [2]. Its basic parameters are summarised in Table 1. The ID is immersed in a 2 T magnetic field generated by the central solenoid, which extends over a length of 5.3 m with a diameter of 2.5 m.

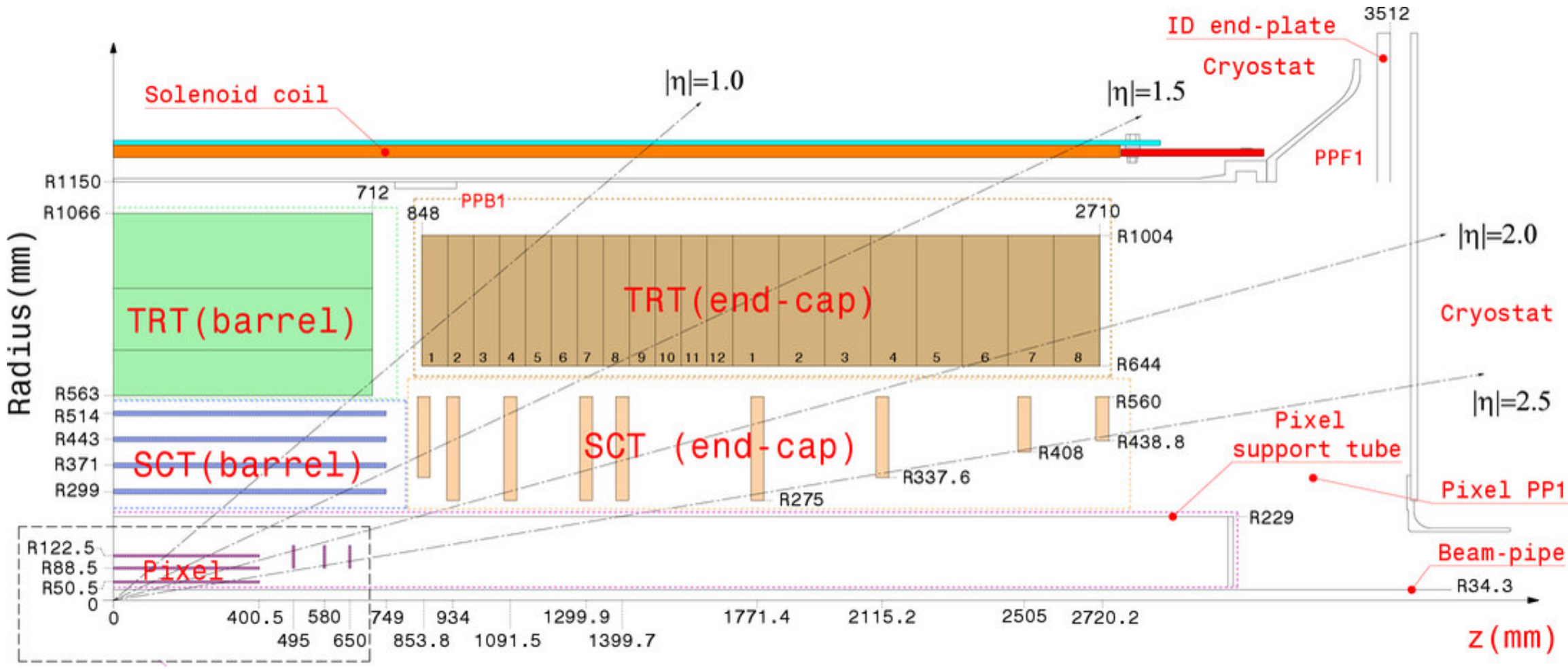

Figure 2: Plan view of a quarter-section of the ATLAS inner detector showing each of the major elements with its active dimensions.

| **Item** | | **Radial extension (mm)** | **Length (mm)** |
|---|---|---|---|
| **Pixel** | Overall envelope | $45.5 < R < 242$ | $0 < \|z\| < 3092$ |
| 3 cylindrical layers | Sensitive barrel | $50.5 < R < 122.5$ | $0 < \|z\| < 400.5$ |
| $2 \times 3$ disks | Sensitive end-cap | $88.8 < R < 149.6$ | $495 < \|z\| < 650$ |
| **SCT** | Overall envelope | $255 < R < 549$ (barrel) | $0 < \|z\| < 805$ |
| | | $251 < R < 610$ (end-cap ) | $810 < \|z\| < 2797$ |
| 4 cylindrical layers | Sensitive barrel | $299 < R < 514$ | $0 < \|z\| < 749$ |
| $2 \times 9$ disks | Sensitive end-cap | $275 < R < 560$ | $839 < \|z\| < 2735$ |
| **TRT** | Overall envelope | $554 < R < 1082$ (barrel) | $0 < \|z\| < 780$ |
| | | $617 < R < 1106$ (end-cap ) | $827 < \|z\| < 2744$ |
| 73 straw planes | Sensitive barrel | $563 < R < 1066$ | $0 < \|z\| < 712$ |
| 160 straw planes | Sensitive end-cap | $644 < R < 1004$ | $848 < \|z\| < 2710$ |

Table 1: Main parameters of the inner detector.

The precision tracking detectors (pixels and SCT) cover the region $|\eta| < 2.5$. In the barrel region, they are arranged on concentric cylinders around the beam axis while in the end-cap regions, they are located on disks perpendicular to the beam axis. The highest granularity is achieved around the vertex region using silicon pixel sensors. All pixel modules are identical and the minimum pixel size on a sensor is $50 \times 400$ $\mu$m$^2$. The pixel layers are segmented in $R-\phi$ and $z$ with typically three pixel layers crossed by each track. The first layer, called the "vertexing layer", is at a radius of 51 mm. The intrinsic accuracies in the barrel are 10 $\mu$m ($R-\phi$) and 115 $\mu$m ($z$) and in the disks are 10 $\mu$m ($R-\phi$) and 115 $\mu$m ($R$). The pixel detector has approximately 80.4 million readout channels.

For the SCT, eight strip layers (four space points) are crossed by each track. In the barrel region, this detector uses small-angle (40 mrad) stereo strips to measure both coordinates, with one set of strips in each layer parallel to the beam direction, measuring $R-\phi$. Each side of a detector module consists of two 6.4 cm long, daisy-chained sensors with a strip pitch of 80 $\mu$m. In the end-cap region, the detectors have a set of strips running radially and a set of stereo strips at an angle of 40 mrad. The mean pitch of the strips is also approximately 80 $\mu$m. The intrinsic accuracies per module in the barrel are 17 $\mu$m ($R-\phi$) and 580 $\mu$m ($z$) and in the disks are 17 $\mu$m ($R-\phi$) and 580 $\mu$m ($R$). The total number of readout channels in the SCT is approximately 6.3 million.

A large number of hits (typically 30 per track, with a maximum of 36, see Fig. 34) is provided by the 4 mm diameter straw tubes of the TRT, which enables track-following up to $|\eta| = 2.0$. The TRT only provides $R-\phi$ information, for which it has an intrinsic accuracy of 130 $\mu$m per straw. In the barrel region, the straws are parallel to the beam axis and are 144 cm long, with their wires divided into two halves, approximately at $\eta = 0$. In the end-cap region, the 37 cm long straws are arranged radially in wheels. The total number of TRT readout channels is approximately 351,000.

| **Item** | **Intrinsic accuracy ($\mu$m)** | **Alignment tolerances ($\mu$m)** | | |
|---|---|---|---|---|
| | | **Radial (R)** | **Axial (z)** | **Azimuth (R-$\phi$)** |
| **Pixel** | | | | |
| Layer-0 | 10 ($R$-$\phi$) 115 ($z$) | 10 | 20 | 7 |
| Layer-1 and -2 | 10 ($R$-$\phi$) 115 ($z$) | 20 | 20 | 7 |
| Disks | 10 ($R$-$\phi$) 115 ($R$) | 20 | 100 | 7 |
| **SCT** | | | | |
| Barrel | 17 ($R$-$\phi$) 580 ($z$)[1] | 100 | 50 | 12 |
| Disks | 17 ($R$-$\phi$) 580 ($R$)[1] | 50 | 200 | 12 |
| **TRT** | 130 | | | 30[2] |

1. Arises from the 40 mrad stereo angle between back-to-back sensors on the SCT modules with axial (barrel) or radial (end-cap) alignment of one side of the structure. The result is pitch-dependent for end-cap SCT modules.
2. The quoted alignment accuracy is related to the TRT drift-time accuracy.

Table 2: Intrinsic measurement accuracies and mechanical alignment tolerances for the inner detector sub-systems, as defined by the performance requirements of the ATLAS experiment. The numbers in the table correspond to the single-module accuracy for the pixels, to the effective single-module accuracy for the SCT and to the drift-time accuracy of a single straw for the TRT.

The combination of precision trackers at small radii with the TRT at a larger radius gives very robust pattern recognition and high precision in both $R-\phi$ and $z$ coordinates. The straw hits at the outer radius contribute significantly to the momentum measurement, since the lower precision per point compared to the silicon is compensated by the large number of measurements and longer measured track length.

The inner detector system provides tracking measurements in a range matched by the precision measurements of the electromagnetic calorimeter [2]. The electron identification capabilities are enhanced by the detection of transition-radiation photons in the xenon-based gas mixture of the straw tubes. The semiconductor trackers also allow impact parameter measurements and vertex reconstruction (“vertexing”) for heavy-flavour and $\tau$-lepton tagging. The secondary vertex measurement performance is enhanced by the innermost layer of pixels, at a radius of about 5 cm.

Charged particle tracks with transverse momentum $p_T > 0.5$ GeV and $|\eta| < 2.5$ are reconstructed and measured in the inner detector and the solenoid field. However, the efficiency at low momentum is limited because of the large material effect in the inner detector (see Fig. 3). The intrinsic measurement performance expected for each of the inner detector sub-systems is summarised in Table 2. This performance has been studied extensively over the years [1], both before and after irradiation of production modules, and also, more recently, during the combined test beam (CTB) runs in 2004 [2,3] and in a series of cosmic-ray tests in 2006 [2,4]. The results have been used to update and validate the modelling of the detector response in the Monte-Carlo simulation. This paper describes the expected performance of the inner detector in terms of tracking, vertexing and particle identification. The alignment of the inner detector is described elsewhere ( [2] and the references therein).

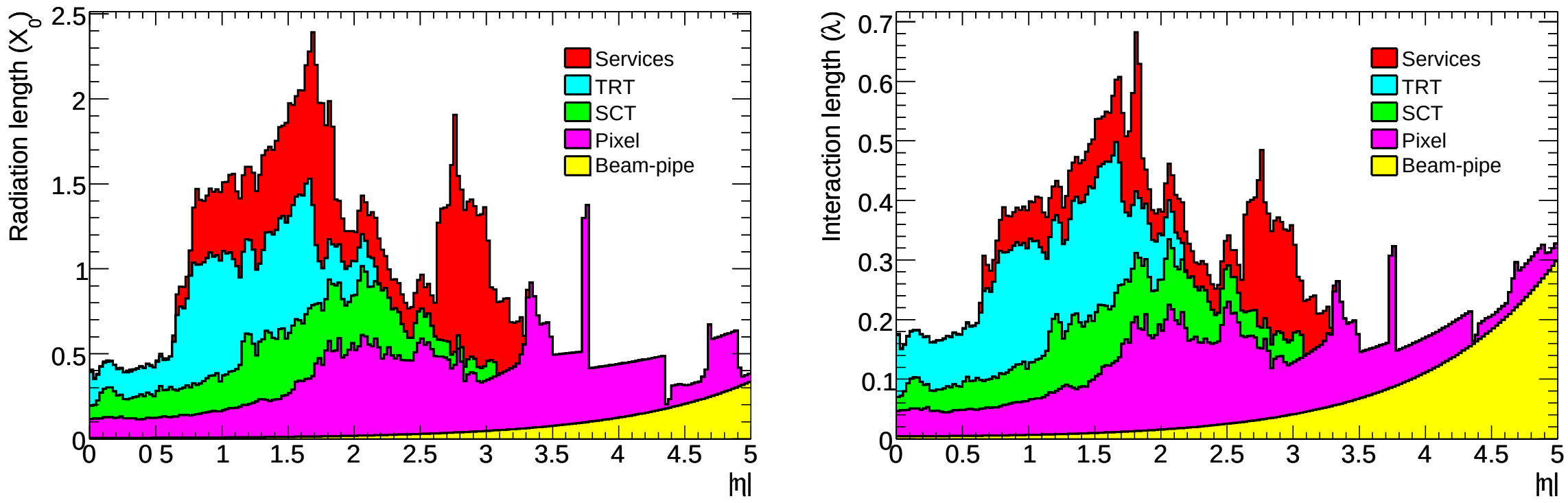

Figure 3: Material distribution ($X_0$, $\lambda$) at the exit of the ID envelope, including the services and thermal enclosures. The distribution is shown as a function of $|\eta|$ and averaged over $\phi$. The breakdown indicates the contributions of external services and of individual sub-detectors, including services in their active volume.

## 2 Track reconstruction

The inner detector track reconstruction software [5] follows a modular and flexible software design, which includes features covering the requirements of both the inner detector and muon spectrometer [2] reconstruction. These features comprise a common event data model [6] and detector description [7], which allow for standardised interfaces to all reconstruction tools, such as track extrapolation, track fitting including material corrections and vertex fitting. The extrapolation package combines propagation tools with an accurate and optimised description of the active and passive material of the full detector [8] to allow for material corrections in the reconstruction process. The suite of track-fitting tools includes global-$\chi^2$ and Kalman-filter techniques, and also more specialised fitters such as dynamic noise adjustment (DNA) [9], Gaussian-sum filters (GSF) [10] and deterministic annealing filters [11]. Optimisation of these tools continues and their performance will need to be evaluated on real data. The tools intended to cope with electron bremsstrahlung (DNA and GSF – see Section 5.1) will be run after the track reconstruction, as part of the electron-photon identification. Other common tracking tools are provided, including those to apply calibration corrections at later stages of the pattern recognition, to correct for module deformations or to resolve hit-association ambiguities.

Track reconstruction in the inner detector is logically sub-divided into three stages:

1. A pre-processing stage, in which the raw data from the pixel and SCT detectors are converted into clusters and the TRT raw timing information is translated into calibrated drift circles. The SCT clusters are transformed into space-points, using a combination of the cluster information from opposite sides of a SCT module.

2. A track-finding stage, in which different tracking strategies [5, 12], optimised to cover different applications, are implemented. (The results of studies of the various algorithms are reported else-where [13].) The default tracking exploits the high granularity of the pixel and SCT detectors to find prompt tracks originating from the vicinity of the interaction region. First, track seeds are formed from a combination of space-points in the three pixel layers and the first SCT layer. These seeds are then extended throughout the SCT to form track candidates. Next, these candidates are fitted, "outlier" clusters are removed, ambiguities in the cluster-to-track association are resolved, and fake tracks are rejected. This is achieved by applying quality cuts. For example, a cut is made on the number of associated clusters, with explicit limits set on the number of clusters shared between several tracks and the number of holes per track (a hole is defined as a silicon sensor crossed by a track without generating any associated cluster). The selected tracks are then extended into the TRT to associate drift-circle information in a road around the extrapolation and to resolve the left-right ambiguities. Finally, the extended tracks are refitted with the full information of all three detectors. The quality of the refitted tracks is compared to the silicon-only track candidates and hits on track extensions resulting in bad fits are labelled as outliers (they are kept as part of the track but are not included in the fit).

A complementary track-finding strategy, called back-tracking, searches for unused track segments in the TRT. Such segments are extended into the SCT and pixel detectors to improve the tracking efficiency for secondary tracks from conversions or decays of long-lived particles.

3. A post-processing stage, in which a dedicated vertex finder is used to reconstruct primary vertices. This is followed by algorithms dedicated to the reconstruction of photon conversions and of secondary vertices.

# 3 Tracking performance

## 3.1 Introduction to performance studies

The expected performance of the tracking system for reconstructing single particles and particles in jets is determined using a precise modelling of the individual detector response (including electronic noise and inefficiencies), geometry and passive material in the simulation. In this paper, a consistent set of selection cuts for reconstructed tracks has been used. Generally, only prompt particles (those originating from the primary vertex) with $p_T > 1$ GeV and $|\eta| < 2.5$ are considered. Standard quality cuts require reconstructed tracks to have at least seven precision hits (pixels and SCT). In addition, the transverse and longitudinal impact parameters at the perigee must fulfil respectively $|d_0| < 2$ mm and $|z_0 - z_v| \times \sin\theta < 10$ mm, where $z_v$ is the position of the primary vertex along the beam and $\theta$ is the polar angle of the track. Stricter selection cuts, called $b$-tagging cuts, are defined by: at least two hits in the pixels, one of which should be in the vertexing layer, as well as $|d_0| < 1$ mm and $|z_0 - z_v| \times \sin\theta < 1.5$ mm. A reconstructed track is matched to a Monte-Carlo particle if at least 80% of its hits were created by that particle. The efficiency is defined as the fraction of particles which are matched to reconstructed tracks passing the quality cuts, and the fake rate is defined as the fraction of reconstructed tracks passing the quality cuts which are not matched to a particle.

## 3.2 Track parameter resolutions

The resolution of a track parameter $X$ can be expressed as a function of $p_T$ as:

$$\sigma_X(p_T) = \sigma_X(\infty)(1 \oplus p_X/p_T) \tag{1}$$

where $\sigma_X(\infty)$ is the asymptotic resolution expected at infinite momentum, $p_X$ is a constant representing the value of $p_T$ for which the intrinsic and multiple-scattering terms in the equation are equal for the parameter $X$ under consideration and $\oplus$ denotes addition in quadrature. This expression is approximate, working well at high $p_T$ (where the resolution is dominated by the intrinsic detector resolution) and at low $p_T$ (where the resolution is dominated by multiple scattering). $\sigma_X(\infty)$ and $p_X$ are implicitly functions of the pseudorapidity. Figures 4, 5 and 6 show the momentum resolution for isolated muons and the transverse and longitudinal impact parameter resolutions for isolated pions[1], all without a beam constraint and assuming the effects of misalignment, miscalibration and pile-up to be negligible. The resolutions are taken as the RMS evaluated over a range which includes 99.7% of the data (corresponding to $\pm 3\sigma$ for a Gaussian distribution). The TRT measurements are included in the track fits for tracks with $|\eta| < 2.0$, beyond which there are no further TRT measurements. Table 3 shows the values of $\sigma_X(\infty)$ and $p_X$ for tracks in two $\eta$-regions, corresponding to the barrel and end-caps. The use of the beam-spot constraint in the track fit improves the momentum resolution for high-momentum tracks by about 5%. The impact parameter resolutions are quoted only for tracks with a hit in the vertexing layer (this requirement has a very high efficiency, as illustrated in Fig. 14 by the small difference between the standard quality and the b-tagging quality tracks). Figure 7 shows the comparison of the impact parameter resolutions for pions and muons. The muon distributions are very close to Gaussian, while those for the pions are slightly broader and have small tails, in addition. The tails are even larger for electrons, and this is discussed in Section 5.

| **Track parameter** | $0.25 < \lvert\eta\rvert < 0.50$ | | $1.50 < \lvert\eta\rvert < 1.75$ | |
|---|---|---|---|---|
| | $\sigma_X(\infty)$ | $p_X$ (GeV) | $\sigma_X(\infty)$ | $p_X$ (GeV) |
| Inverse transverse momentum ($q/p_T$) | 0.34 TeV$^{-1}$ | 44 | 0.41 TeV$^{-1}$ | 80 |
| Azimuthal angle ($\phi$) | 70 $\mu$rad | 39 | 92 $\mu$rad | 49 |
| Polar angle ($\cot\theta$) | $0.7 \times 10^{-3}$ | 5.0 | $1.2 \times 10^{-3}$ | 10 |
| Transverse impact parameter ($d_0$) | 10 $\mu$m | 14 | 12 $\mu$m | 20 |
| Longitudinal impact parameter ($z_0 \times \sin\theta$) | 91 $\mu$m | 2.3 | 71 $\mu$m | 3.7 |

Table 3: Expected track-parameter resolutions (RMS) at infinite transverse momentum, $\sigma_X(\infty)$, and transverse momentum, $p_X$, at which the multiple-scattering contribution equals that from the detector resolution (see Eq. (1)). The momentum and angular resolutions are shown for muons, whereas the impact-parameter resolutions are shown for pions (see text). The values are shown for two $\eta$-regions, one in the barrel inner detector where the amount of material is close to its minimum and one in the end-cap where the amount of material is close to its maximum. Isolated, single particles are used with perfect alignment and calibration in order to indicate the optimal performance.

The consequences of the pseudorapidity variation of the track parameter resolutions can be seen from the reconstructed $J/\psi \to \mu\mu$ masses in the barrel and end-caps. This is shown in Fig. 8 where both muons are either in the barrel or the end-caps.

The determination of the lepton charge at high $p_T$ is particularly important for measuring charge asymmetries arising from the decays of possible heavy gauge bosons ($W'$ and $Z'$). Typically, such measurements require that the charge of the particle be determined to better than $3\sigma$[2]. Whereas the charge of high-energy muons will be measured precisely in the muon system, the charge of high-energy electrons can only be measured by the inner detector. Figure 9 shows the reconstructed values of $q/p_T$ for negatively charged isolated muons and electrons with $p_T = 0.5$ TeV and $p_T = 2$ TeV. The peaks of the distributions are at negative values, reflecting the negative charges of the simulated particles. It can be seen that the shape of the muon distributions is unchanged in going from 0.5 to 2 TeV – at high momen-tum, the resolution of $q/p_T$ is independent of the true momentum of the muon and determined by the intrinsic resolution of the detector.

[1]Muons suffer less from interactions and hence provide the best reference; impact parameter determination is important for vertexing, and this is more commonly required for hadrons, for example when $b$-tagging.

[2]The charge of a particle is considered well measured if it is at least $3\sigma$ from 0 in the variable $q/p$.

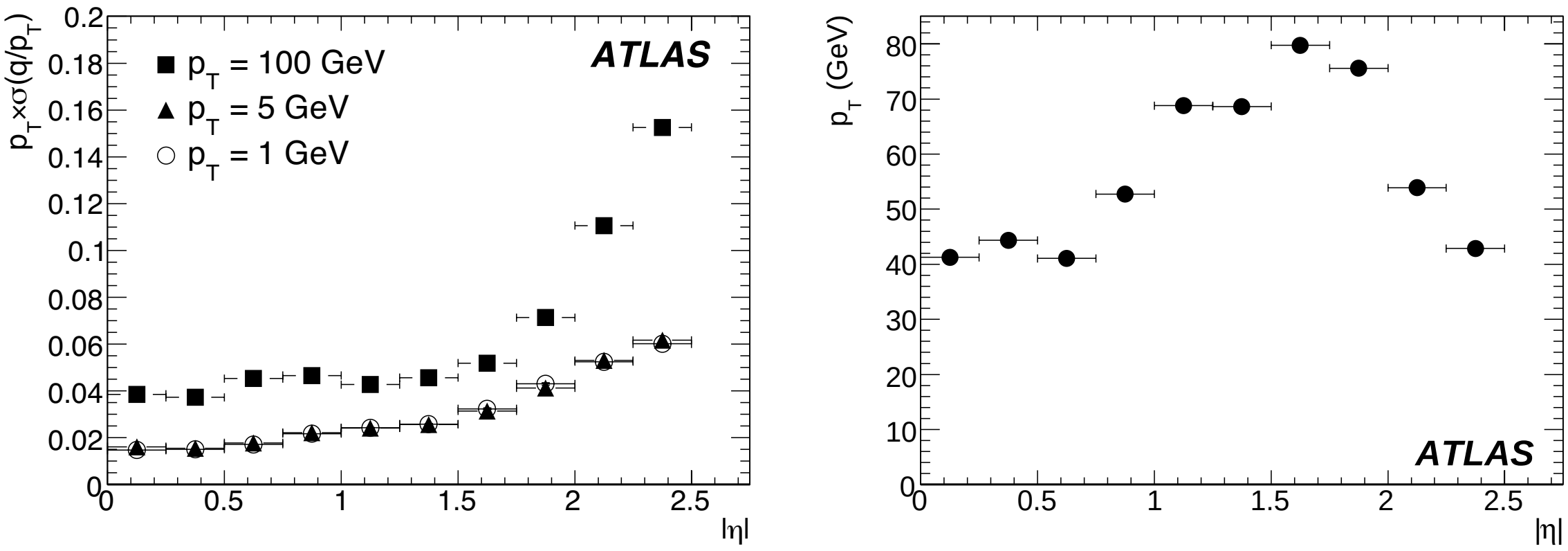

Figure 4: Relative transverse momentum resolution (left) as a function of $|\eta|$ for muons with $p_T$ = 1, 5 and 100 GeV. Transverse momentum, at which the multiple-scattering contribution equals the intrinsic resolution (corresponding to $p_X$ in Eq. (1)), as a function of $|\eta|$ (right).

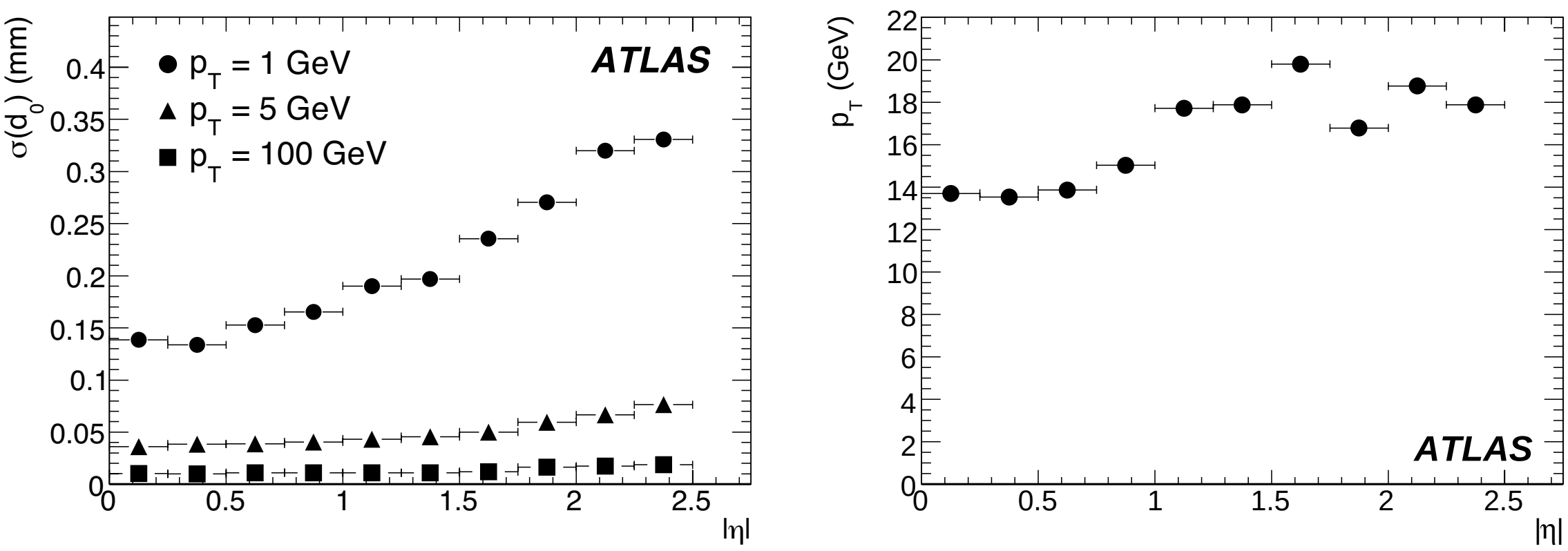

Figure 5: Transverse impact parameter, $d_0$, resolution (left) as a function of $|\eta|$ for pions with $p_T$ = 1, 5 and 100 GeV. Transverse momentum, at which the multiple-scattering contribution equals the intrinsic resolution (corresponding to $p_X$ in Eq. (1)), as a function of $|\eta|$ (right).

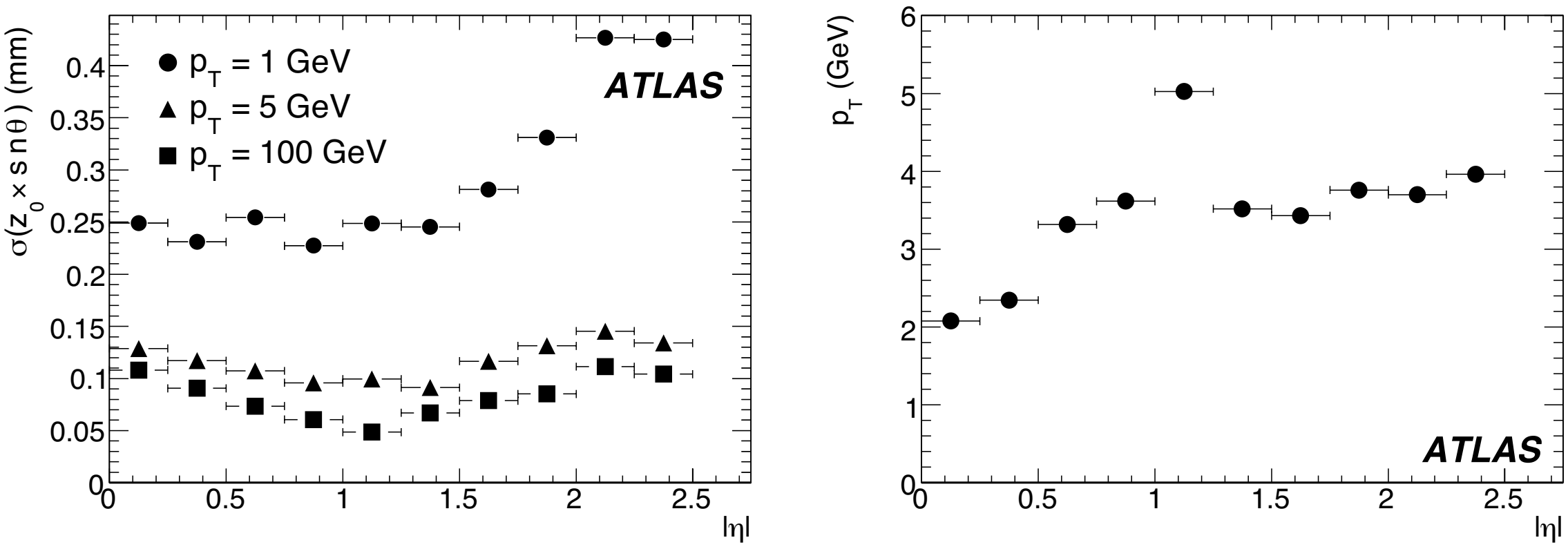

Figure 6: Modified longitudinal impact parameter, $z_0 \times \sin\theta$, resolution (left) as a function of $|\eta|$ for pions with $p_T$ = 1, 5 and 100 GeV. Transverse momentum, at which the multiple-scattering contribution equals the intrinsic resolution (corresponding to $p_X$ in Eq. (1)), as a function of $|\eta|$ (right).

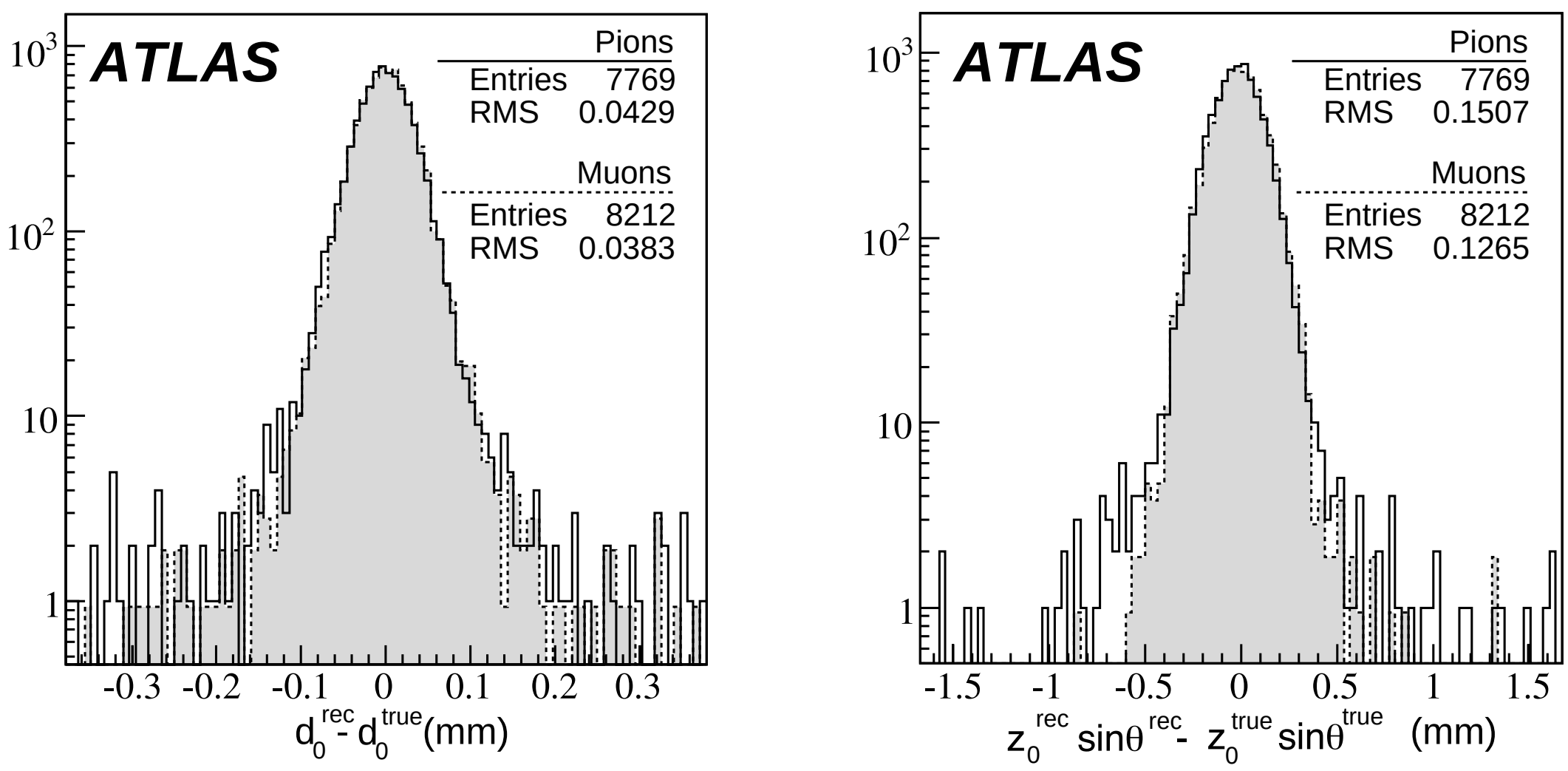

Figure 7: Resolution of the transverse impact parameter, $d_0$ (left) and the modified longitudinal impact parameter, $z_0 \times \sin\theta$ (right) for 5 GeV muons and pions with $|\eta| \leq 0.5$ – corresponding to the first two bins of the previous two figures.

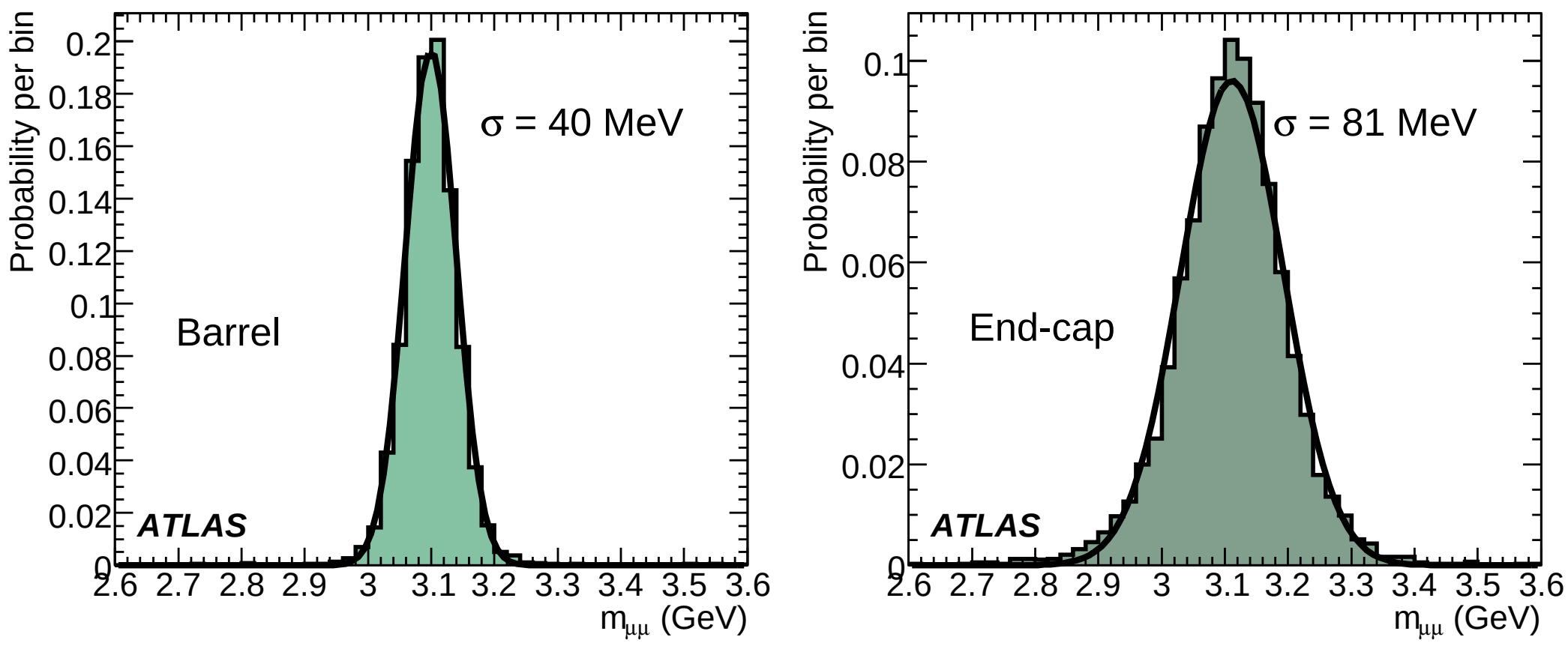

Figure 8: Probability for the reconstructed invariant mass of muon pairs from $J/\psi \to \mu\mu$ decays in events with prompt $J/\psi$ production. Distributions are shown for both muons with $|\eta| < 0.8$ (left) and $|\eta| > 1.5$ (right).

For electrons, things are more complicated. As well as the intrinsic resolution, there are competing effects from bremsstrahlung (which lowers the track momentum and makes the charge easier to measure) and the conversion of bremsstrahlung photons (leading to pattern-recognition problems and degraded charge determination). At 0.5 TeV, the effects of the conversions are significant, causing the electrons to be measured worse than the corresponding muons. However, at 2 TeV, the intrinsic resolution dominates the electron charge misidentification, and this is partially compensated for by the bremsstrahlung. The fractions of muons and electrons for which the sign of the charge is incorrectly determined are shown in Fig. 10. For these plots, perfect alignment has been assumed; any misalignment will degrade the charge sign determination.

### 3.3 Track reconstruction efficiency

Figures 11, 12 and 13 show the efficiencies for reconstructing isolated muons, pions and electrons. In addition to multiple-scattering, pions are affected by hadronic interactions in the inner detector material, while electrons are subject to even larger reconstruction inefficiencies which arise from the effects of bremsstrahlung. As a result, the efficiency curves as a function of $|\eta|$ for pions and electrons reflect the shape of the amount of material in the inner detector (see Fig. 3). As expected, the efficiency becomes larger and more uniform as a function of $|\eta|$ at higher energies.

Previous studies [1] have shown that the reconstruction efficiency is little affected by the "pile-up" of additional minimum bias events at high luminosity ($10^{34}$ cm$^{-2}$s$^{-1}$). A more challenging environment is found in the core of an energetic jet. Figure 14 shows the track reconstruction efficiency for prompt pions (produced before the vertexing layer) and the fake rate for tracks in jets in $t\bar{t}$ events as a function of $|\eta|$. For these events, the mean jet $p_T$ is 55 GeV, and the mean $p_T$ of the accepted tracks which they contain is 4 GeV. The loss of efficiency at $|\eta| = 0$ with the $b$-tagging criteria arises from inefficiencies in the pixel vertexing layer, which are assumed here to be 1%; this improves at higher $|\eta|$, owing to the presence of larger clusters when the track incidence angle decreases. Beyond $|\eta| \sim 1$, the tracking performance deteriorates, mostly because of increased material. As shown in Fig. 15, the fake rate increases near the

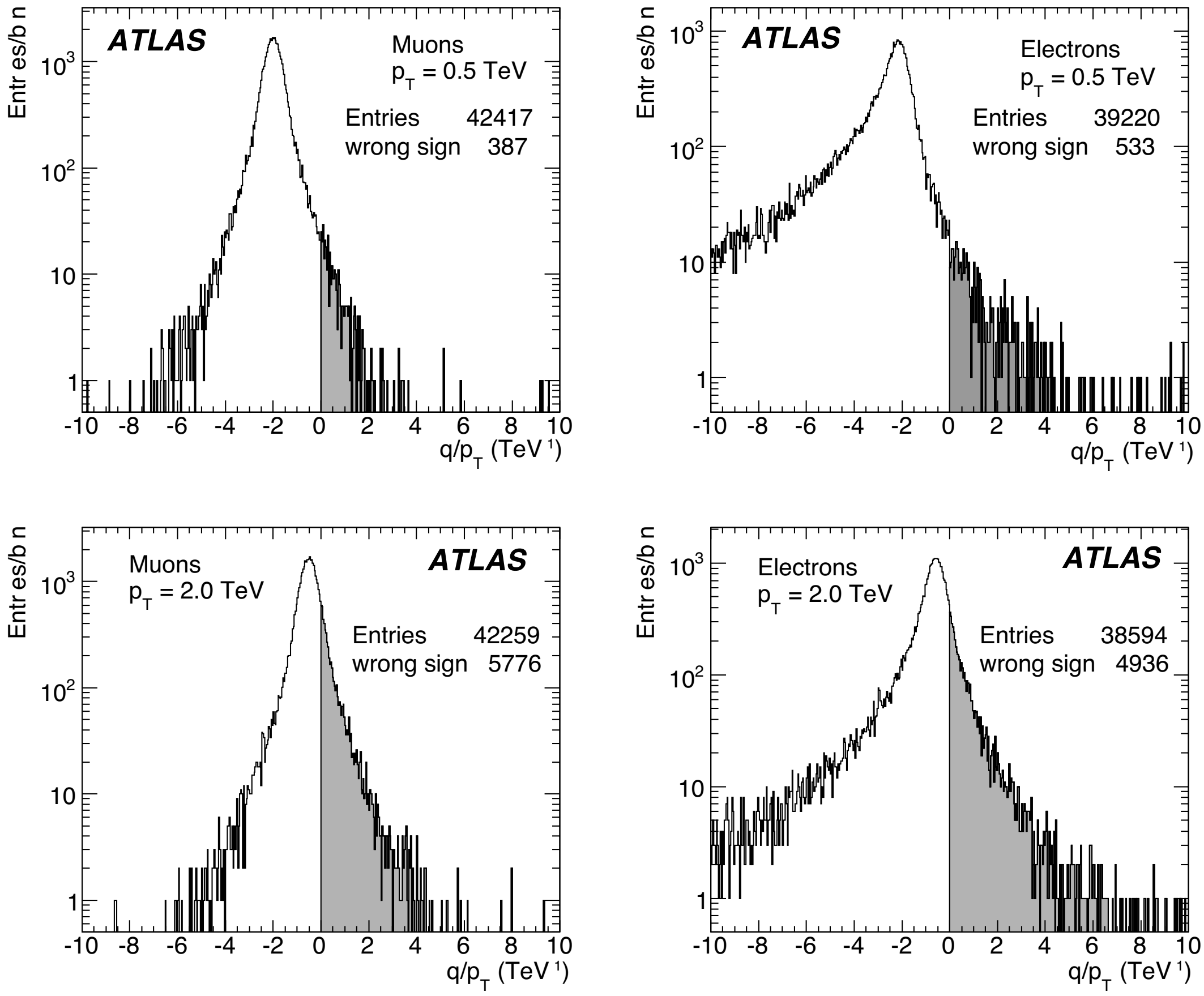

Figure 9: Reconstructed inverse transverse momentum multiplied by the charge for high-energy muons ($\mu^-$) (left) and electrons ($e^-$) (right) for $p_T = 0.5$ TeV (top) and $p_T = 2$ TeV (bottom) and integrated over a flat distribution in $\eta$ with $|\eta| \leq 2.5$. Those tracks which have been incorrectly reconstructed with a positive charge are indicated by the shaded regions. At 2 TeV, the fraction of electrons (muons) whose charge has been misidentified is 12.8% (13.7%).

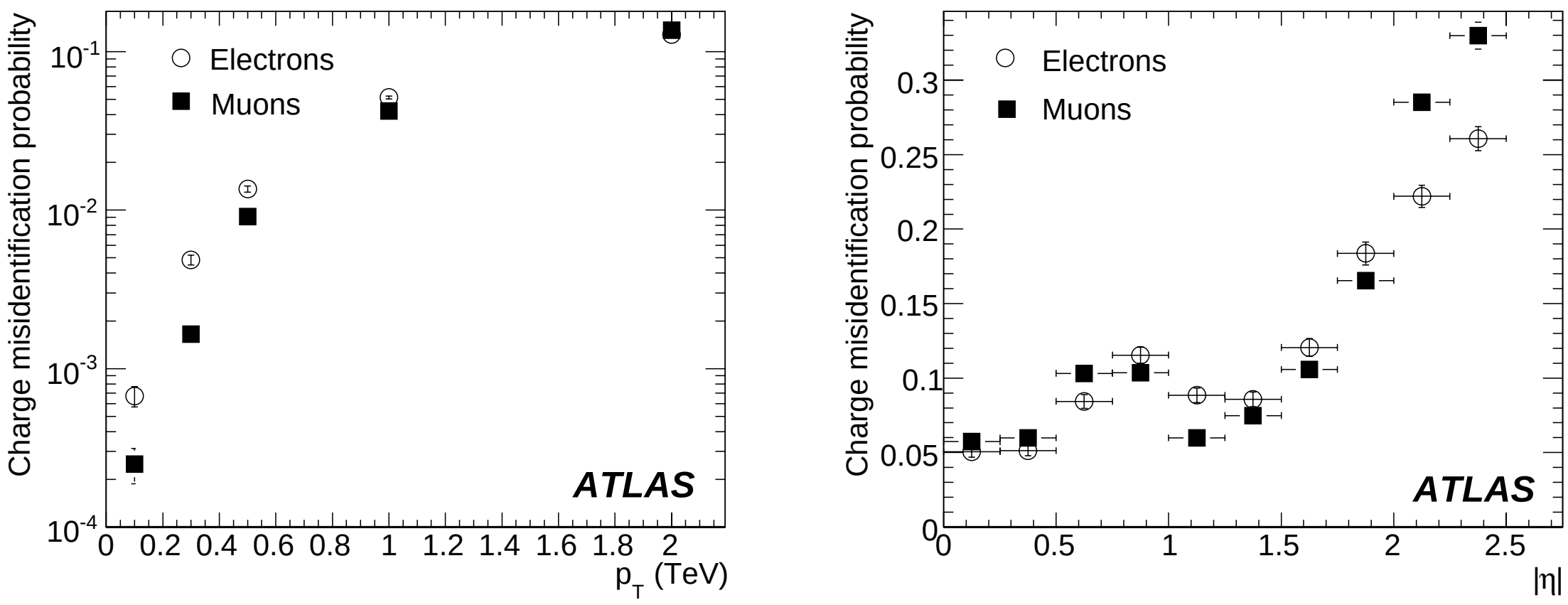

Figure 10: Charge misidentification probability for high-energy muons and electrons as a function of $p_T$ for particles with $|\eta| \leq 2.5$ (left) and as a function of $|\eta|$ for $p_T = 2$ TeV (right).

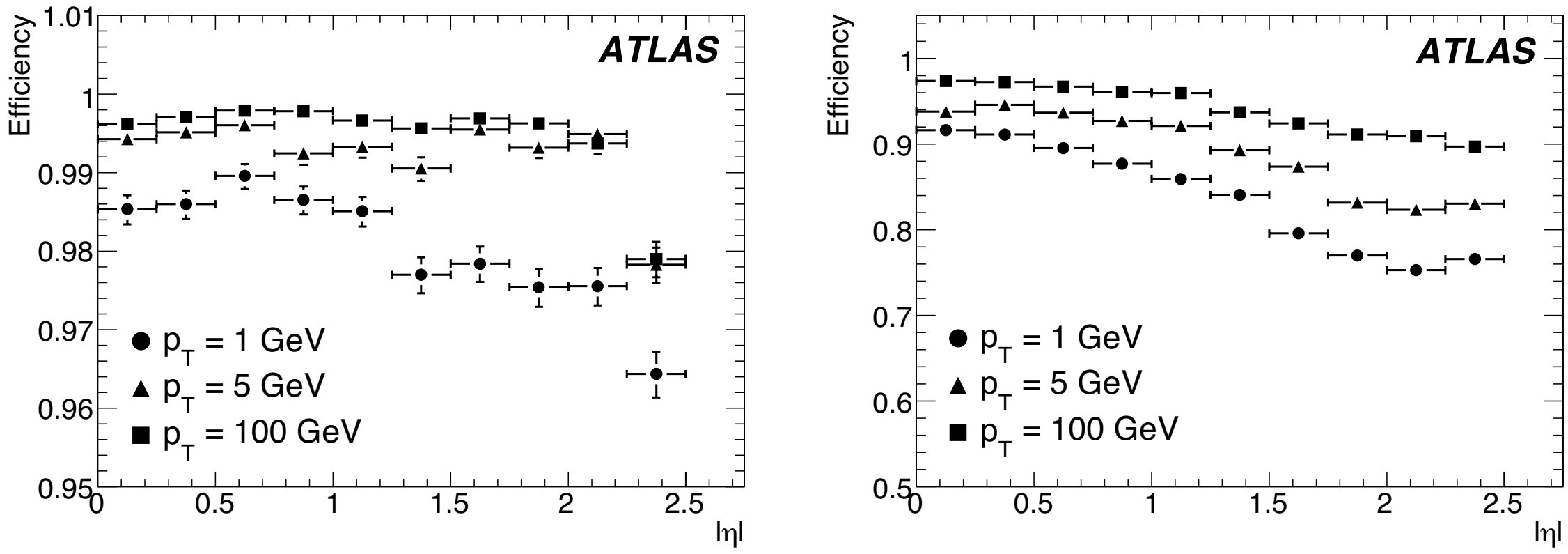

Figure 11: Track reconstruction efficiencies as a function of $|\eta|$ for muons (left) and pions (right) with $p_T$ = 1, 5 and 100 GeV.

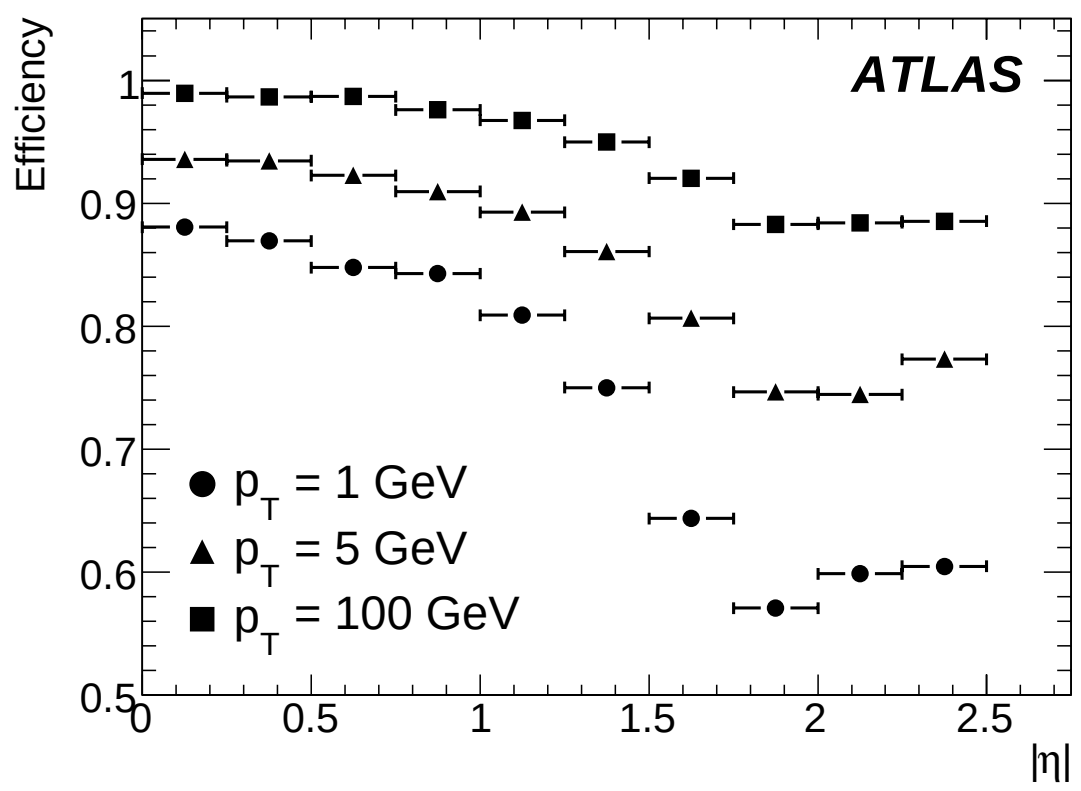

Figure 12: Track reconstruction efficiencies as a function of $|\eta|$ for electrons with $p_T$ = 1, 5 and 100 GeV.

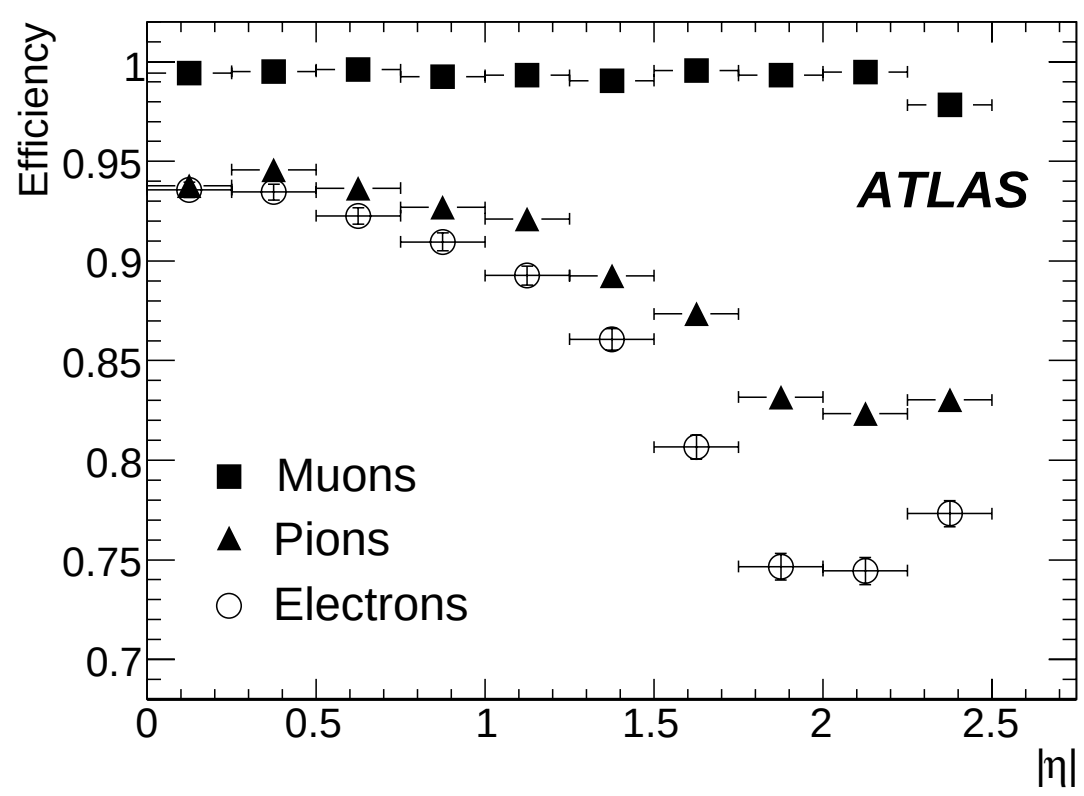

Figure 13: Track reconstruction efficiencies as a function of $|\eta|$ for muons, pions and electrons with $p_T$ = 5 GeV. The inefficiencies for pions and electrons reflect the shape of the amount of material in the inner detector as a function of $|\eta|$.

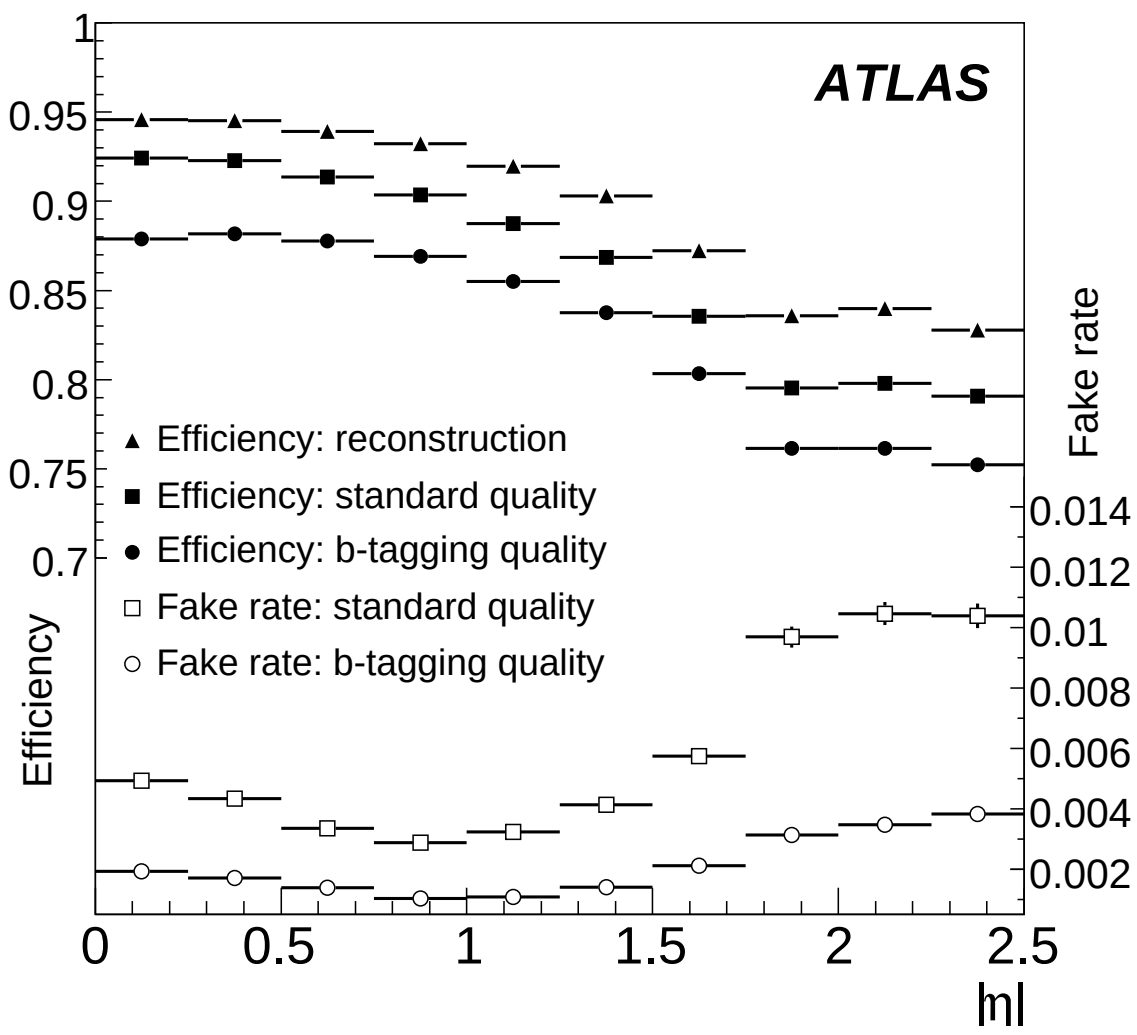

Figure 14: Track reconstruction efficiencies and fake rates as a function of $|\eta|$, for charged pions in jets in $t\bar{t}$ events and for different quality cuts (as described in Section 3.1). "Reconstruction" refers to the basic reconstruction before additional quality cuts.

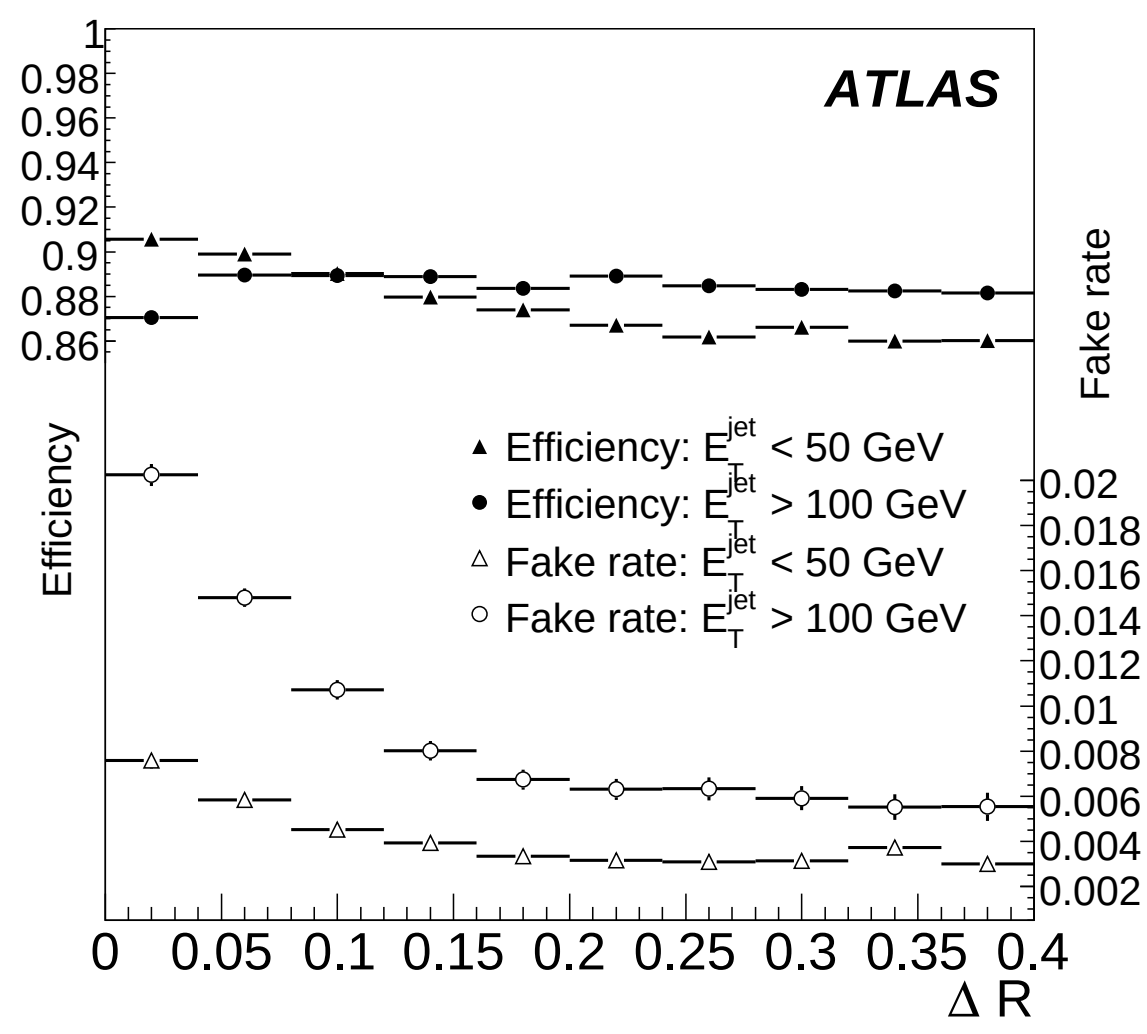

Figure 15: Track reconstruction efficiencies and fake rates as a function of the distance $\Delta R$ (defined as $\Delta R = \sqrt{\Delta\eta^2 + \Delta\phi^2}$) of the track to the jet axis, using the standard quality cuts and integrated over $|\eta| < 2.5$, for charged pions in jets in $t\bar{t}$ events.

core of the jet, where the track density is the highest and induces pattern-recognition problems. This effect increases as the jet $p_T$ increases. Using alternative algorithms, a few percent efficiency can be gained at the cost of doubling the fake rate in the jet core.

The reconstruction described in Section 2 is aimed at tracks with $p_T > 0.5$ GeV. Multiplicity studies in minimum bias events will be among the first analyses undertaken by ATLAS. In these events, the peak of the track $p_T$ spectrum is around 0.3 GeV. The reconstruction of these low-momentum tracks will be difficult because of the high curvature of the tracks, increased multiple scattering, and at very low momentum, reduced numbers of hits, since the tracks may fail to reach the outer layers of the inner detector. To complement the track-finding strategy described in Section 2, an additional strategy is employed in which hitherto unused pixel and SCT hits are used. To further aid the reconstruction, the algorithm for the space-point track seeding is modified to use looser internal cuts and the cut on the number of precision hits is reduced to at least five hits. Tracks are accepted with $p_T > 0.1$ GeV, and in some cases, inefficiencies for $p_T > 0.5$ GeV are recovered. The resulting track reconstruction efficiency is shown in Fig. 16. The distribution of candidate fake tracks is shown in Fig. 17.

# 4 Vertexing performance

## 4.1 Primary vertices

Vertexing tools constitute important components of the higher-level tracking algorithms. The residuals of the primary vertex reconstruction are shown in Fig. 18, as obtained without using any beam constraint, for $t\bar{t}$ events and $H \to \gamma\gamma$ events with $m_H$ = 120 GeV. The results shown here for $H \to \gamma\gamma$ events are based on tracks reconstructed from the underlying event and do not make use of the measurement of the photon direction in the electromagnetic calorimeter. The primary vertex in $t\bar{t}$ events has always a rather large multiplicity and includes a number of high-$p_T$ tracks, resulting in a narrower and more Gaussian

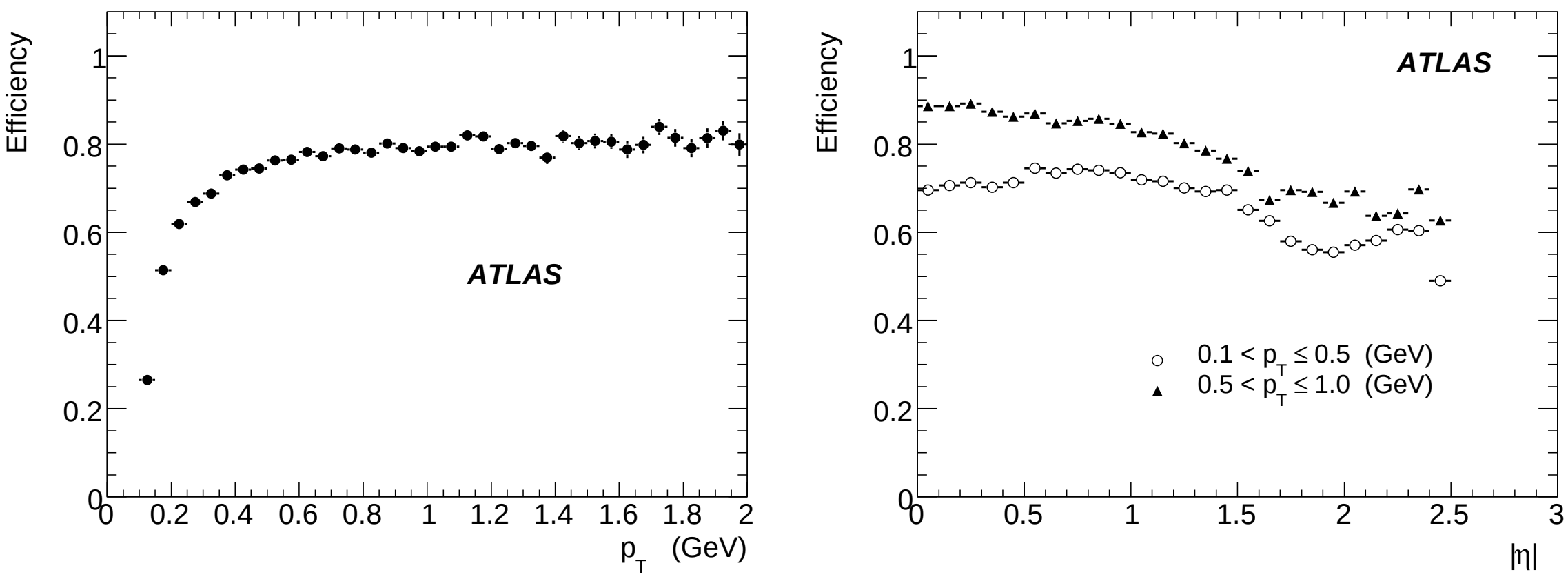

Figure 16: Track reconstruction efficiencies as a function of $p_T$ for $|\eta| < 2.5$ and $p_T > 0.1$ GeV (left) and as a function of $|\eta|$ for two different $p_T$ ranges (right) in minimum bias events (non-diffractive inelastic events).

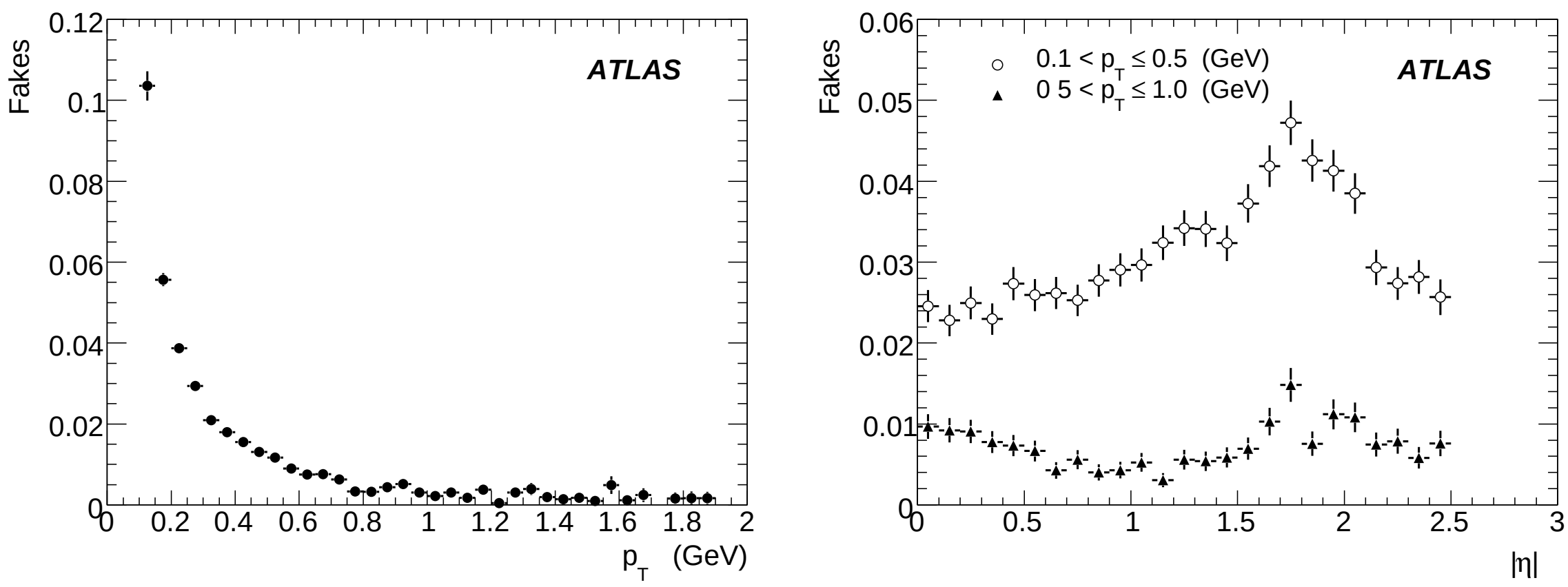

Figure 17: Rate of candidate fake tracks as a function of $p_T$ for $|\eta| < 2.5$ and $p_T > 0.1$ GeV (left) and as a function of $|\eta|$ (right) in minimum bias events (non-diffractive inelastic events). The rate of such tracks is a function of the amount of material, indicating that a large fraction of them are secondaries for which the Monte-Carlo truth information is not kept

distribution than for $H \to \gamma\gamma$ events. Table 4 shows the resolutions of the primary vertex reconstruction in these $t\bar{t}$ and $H \to \gamma\gamma$ events, without and with a beam constraint in the transverse plane, as well as the efficiencies to reconstruct and select correctly (within $\pm 300$ $\mu$m) these primary vertices in the presence of pile-up at a luminosity of $10^{33}$ cm$^{-2}$s$^{-1}$.

| **Event type** | **x-y resolution ($\mu$m)** | **z resolution ($\mu$m)** | **Reconstruction efficiency (%)** | **Selection efficiency (%)** |
|---|---|---|---|---|
| $t\bar{t}$ (no BC) | 18 | 41 | 100 | 99 |
| $t\bar{t}$ (BC) | 11 | 40 | 100 | 99 |
| $H \to \gamma\gamma$ (no BC) | 36 | 72 | 96 | 79 |
| $H \to \gamma\gamma$ (BC) | 14 | 66 | 96 | 79 |

Table 4: Primary vertex resolutions (RMS), without and with a beam constraint (BC) in the transverse plane, for $t\bar{t}$ events and $H \to \gamma\gamma$ events with $m_H$ = 120 GeV in the absence of pile-up. Also shown, in the presence of pile-up at a luminosity of $10^{33}$ cm$^{-2}$s$^{-1}$, are the efficiencies to reconstruct and then select the hard-scattering vertex within $\pm 300$ $\mu$m of the true vertex position in $z$. The hard-scattering vertex is selected as the primary vertex with the largest $\Sigma p_T^2$, summed over all its constituent tracks.

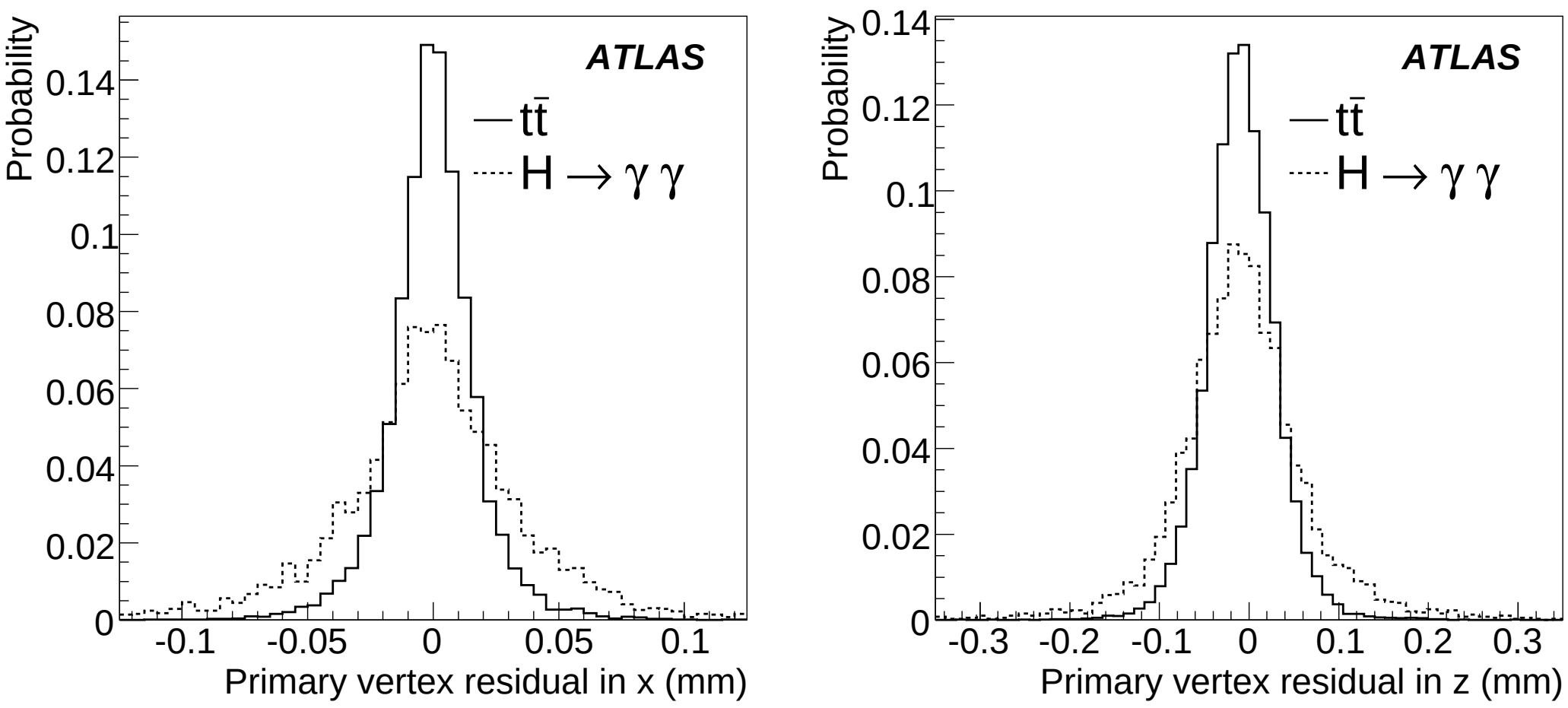

Figure 18: Primary vertex residual along $x$, in the transverse plane (left), and along $z$, parallel to the beam (right), for events containing top-quark pairs and $H \to \gamma\gamma$ decays with $m_H$ = 120 GeV. The results are shown without pile-up and without any beam constraint.

## 4.2 Secondary vertices

The resolution for the reconstruction of the radial position of secondary vertices for $J/\psi \to \mu\mu$ decays in events containing $B$-hadron decays (mean $p_T$ of 15 GeV for the $J/\psi$) is shown in Fig. 19. While there are some tails in the resolution distributions (left-hand plot), these are small. The corresponding distributions for three-prong hadronic $\tau$-decays in $Z \to \tau\tau$ events (mean $p_T$ of 36 GeV for the $\tau$-lepton) are shown in Fig. 20. Because there are three charged tracks in close proximity, the reconstruction of these decays is more challenging: the vertex resolutions are Gaussian in the central region, but have long tails as can be seen from the points showing 95% coverage in right-hand plot.

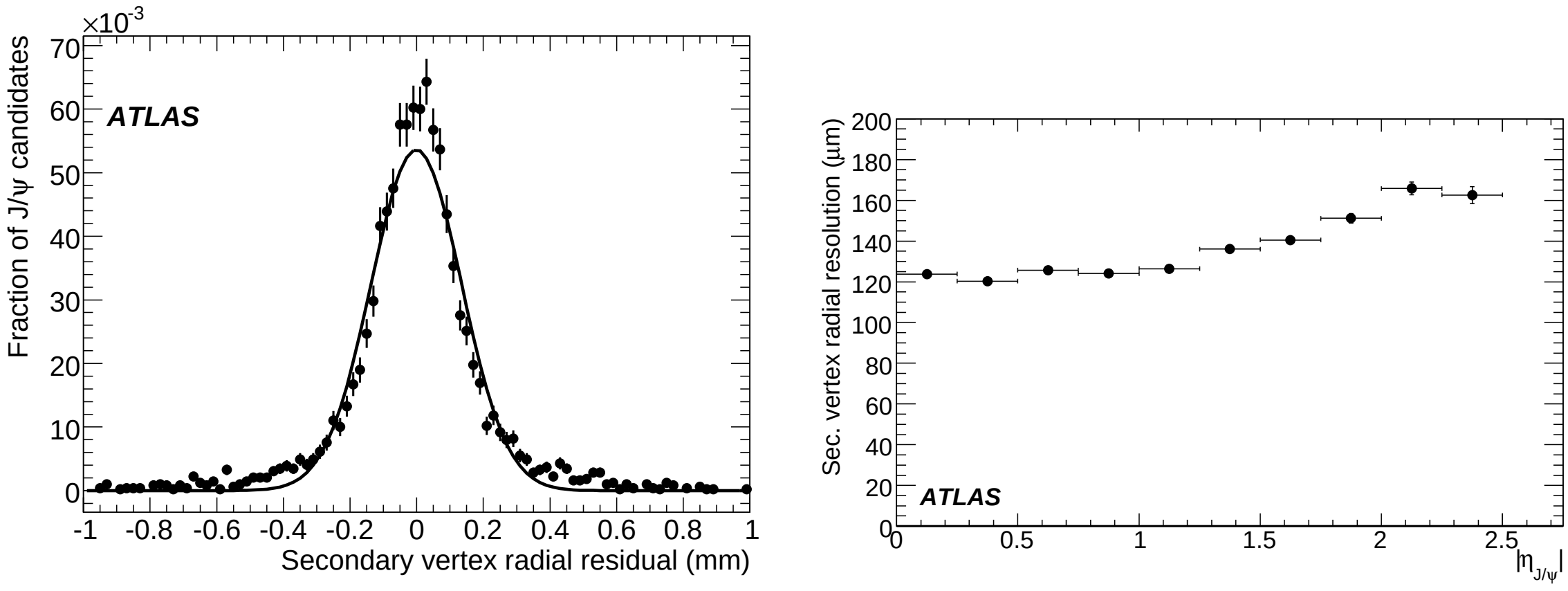

Figure 19: Resolution for the reconstruction of the radial position of the secondary vertex for $J/\psi \rightarrow \mu\mu$ decays in events containing $B$-hadron decays for tracks with $|\eta|$ around 0 (left) and as a function of the pseudorapidity of the $J/\psi$ (right). The $J/\psi$ have an average transverse momentum of 15 GeV.

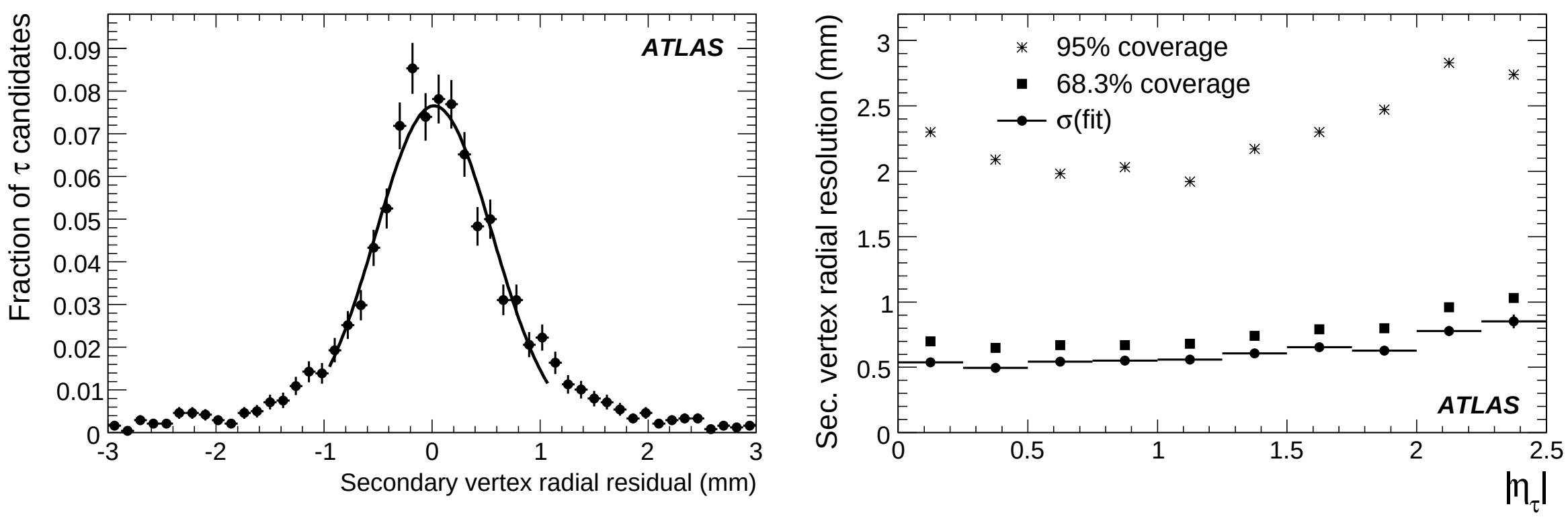

Figure 20: Resolution for the reconstruction of the radial position of the secondary vertex for three-prong hadronic $\tau$-decays in $Z \rightarrow \tau\tau$ events for tracks with $|\eta|$ around 0 (left) and as a function of the pseudorapidity of the $\tau$ (right). In the right-hand plot, the circles with bars correspond to Gaussian fits, as illustrated in the left-hand plot; the points showing 68.3% (95%) coverage show the width of the integrated distribution containing 68.3% (95%) of the measurements (corresponding to $1\sigma$ ($2\sigma$) for a Gaussian distribution). The $\tau$-leptons have an average transverse momentum of 36 GeV.

Finally, Fig. 21 shows the resolution as a function of decay radius for the reconstruction of the radial position of secondary vertices for $K_s^0$ decays (mean $p_T$ of 6 GeV) in events containing $B$-hadron decays. The resolution in each radial slice is determined from a Gaussian fit to the core of the distribution. It can be seen that there are significant tails: just before the barrel layers, the resolution for decays in the barrel region is good, giving rise to the core; while that from the end-caps is variable, depending on the actual position of the decay, giving rise to a broader distribution. The tails can be reduced and the resolutions improved somewhat by tighter cuts on track quality and the reconstructed invariant mass, if desirable. The effect of crossing the three successive pixel layers is clearly visible as well as the degraded resolution for decays beyond the last pixel layer. Figure 22 shows the resolution as a function of decay radius for the reconstruction of the invariant mass of the charged-pion pair for the same $K_s^0 \to \pi^+\pi^-$ decays. Figure 23 shows the efficiency to reconstruct the $K_s^0$ decays. The reconstruction requires 3D information provided by the silicon detectors, and hence the efficiency falls to zero once the decay is beyond the penultimate SCT layers.

# 5 Particle identification, reconstruction of electrons and photon conversions

The reconstruction of electrons and of photon conversions is a particular challenge for the inner detector. The fraction of energy lost by electrons traversing the inner detector is shown in Fig. 24. In the energy range over which the inner detector will measure electrons, the fraction has little dependence on the actual electron energy. Electrons lose on average between 20 to 50% of their energy (depending on $|\eta|$) by the time they have left the SCT, as illustrated in Fig. 25. The probability for photons to convert is fairly independent of their energies for $p_T > 1$ GeV. A histogram of the location of photon conversions in $|\eta| < 0.8$ is shown in Fig. 26 - the radial structure of the detector is clearly visible. Between 10 to 50% of photons have converted into an electron-positron pair before leaving the SCT, as illustrated in Fig. 27.

The TRT plays a central role in electron identification, cross-checking and complementing the electromagnetic calorimeter, especially at energies below 25 GeV [2]. In addition, the TRT contributes to the reconstruction and identification of electron track segments from photon conversions down to 1 GeV and of electrons which have radiated a large fraction of their energy in the silicon layers.

## 5.1 Electron reconstruction

In the absence of bremsstrahlung, the distribution $p_{true}/p_{recon}$ should be Gaussian; but in the presence of bremsstrahlung, this is far from true, as can be seen for the end-cap in Fig. 28 (left-hand plot). By fitting electron tracks in such a way as to allow for bremsstrahlung, it is possible to improve the reconstructed track parameters, as shown in Figs. 28 and 29 for two examples of bremsstrahlung recovery algorithms. These algorithms rely exclusively on the inner detector information and therefore provide significant improvements only for electron energies below $\sim$ 25 GeV. The dynamic noise adjustment (DNA) method extrapolates track segments to the next silicon detector layer. If there is a significant $\chi^2$ contribution, compatible with a hard bremsstrahlung, the energy loss is estimated and an additional noise term is included in the Kalman filter [9]. The Gaussian-sum filter (GSF) is a non-linear generalisation of the Kalman filter, which takes into account non-Gaussian noise by modelling it as a weighted sum of Gaussian components and therefore acts as a weighted sum of Kalman filters operating in parallel [10]. With real data, to improve the fitted track parameters for electrons without deteriorating the fits for non-electrons, it is necessary to assess whether a track is likely to correspond to an electron or not. This can be done to some extent by the algorithms themselves by looking at the fits; additional information can be obtained from the transition radiation in the TRT (see 5.2) and the electromagnetic calorimeter. Ultimately, since information is lost during the bremsstrahlung, there is an unavoidable degradation of

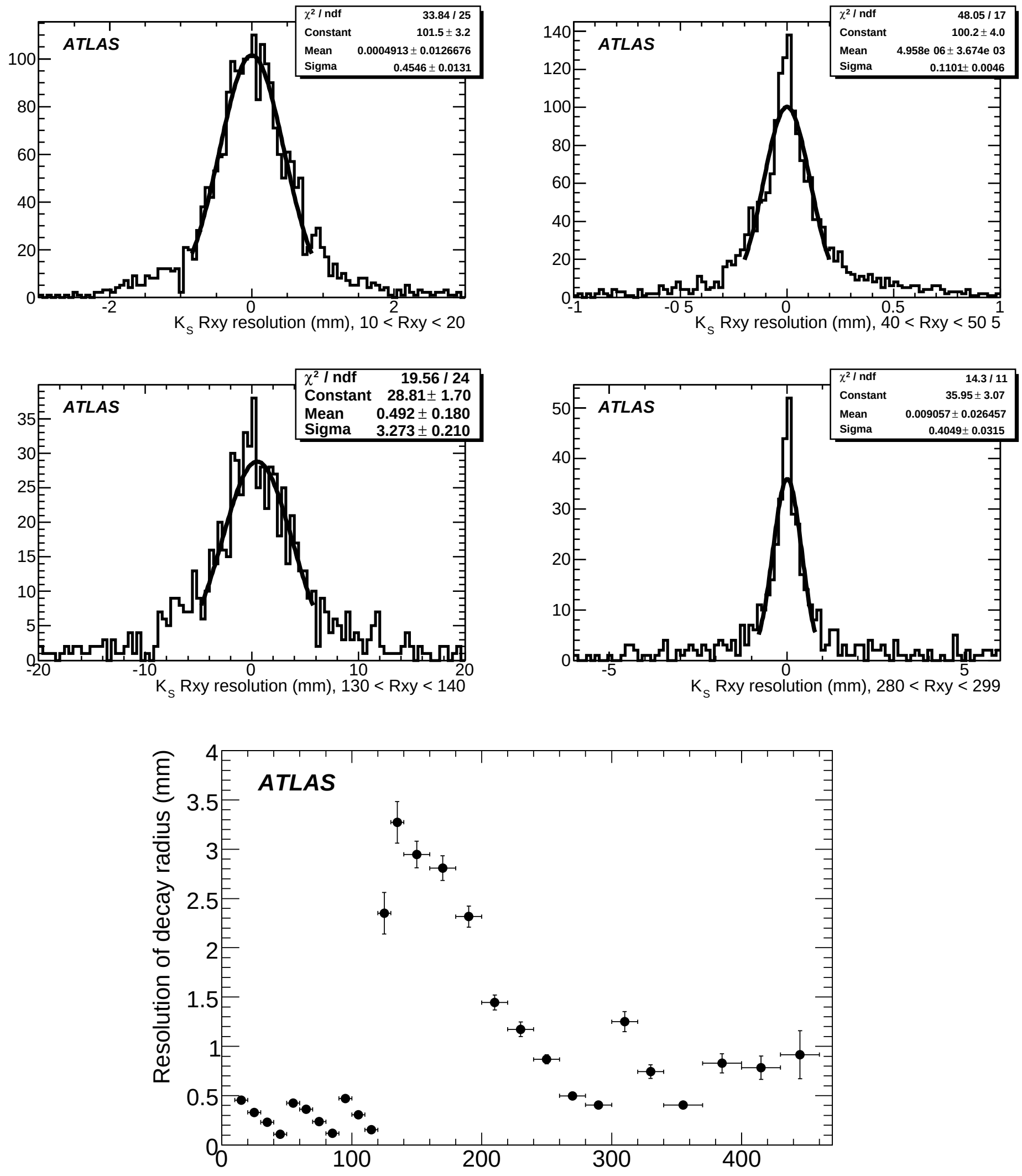

Figure 21: Resolution for the reconstructed radial position of the secondary vertex for $K_s^0 \rightarrow \pi^+\pi^-$ decays in events containing $B$-hadron decays in various radial intervals (upper) and as a function of the $K_s^0$ decay radius (lower). The resolutions are best for decays just in front of the detector layers. The barrel pixel layers are at: 51, 89 and 123 mm; the first two SCT layers are at 299 and 371 mm.

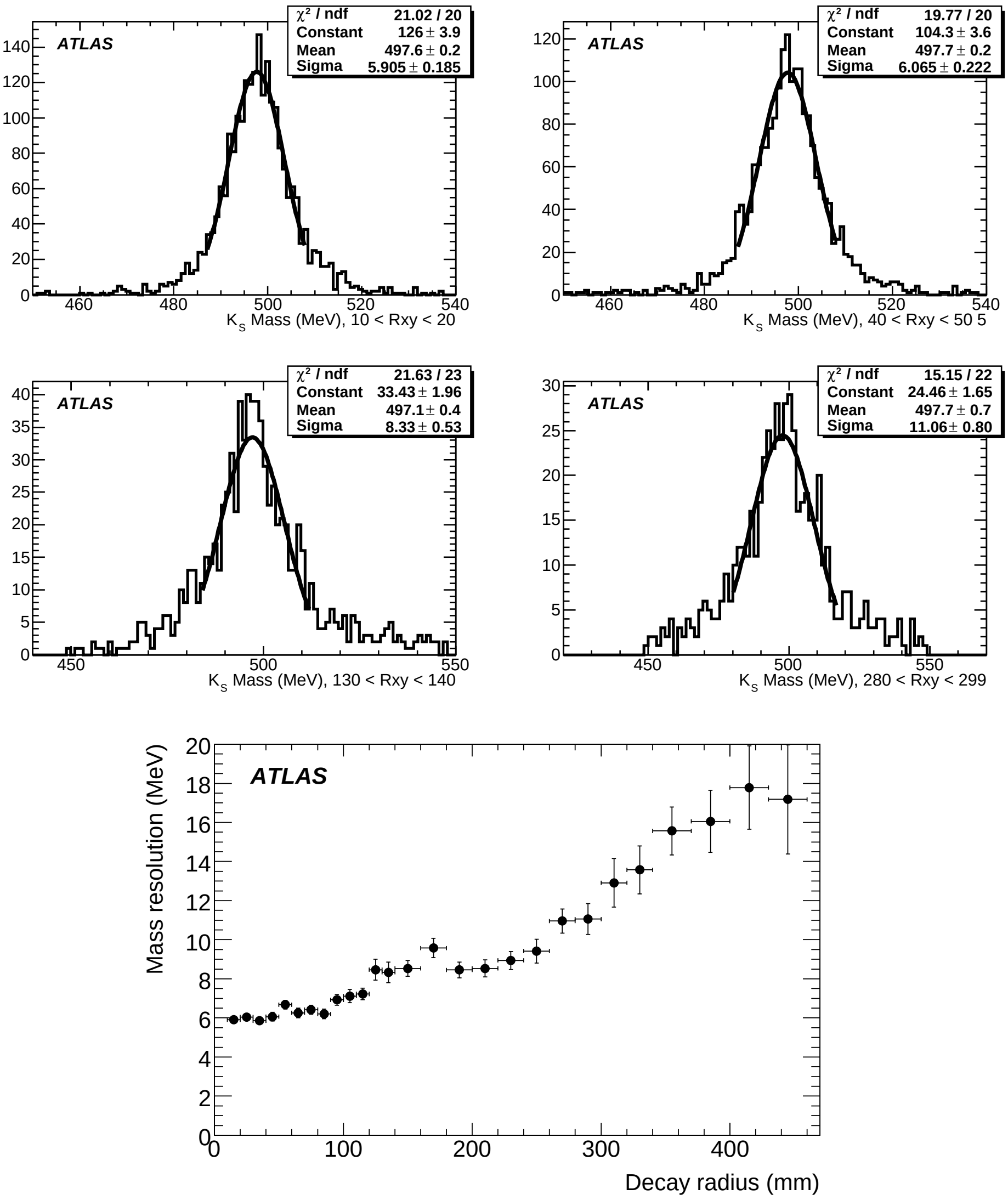

Figure 22: Resolution for the reconstruction of the invariant mass of the charged-pion pair for $K_s^0 \to \pi^+\pi^-$ decays in events containing $B$-hadron decays in various radial intervals (upper) and as a function of the $K_s^0$ decay radius (lower).

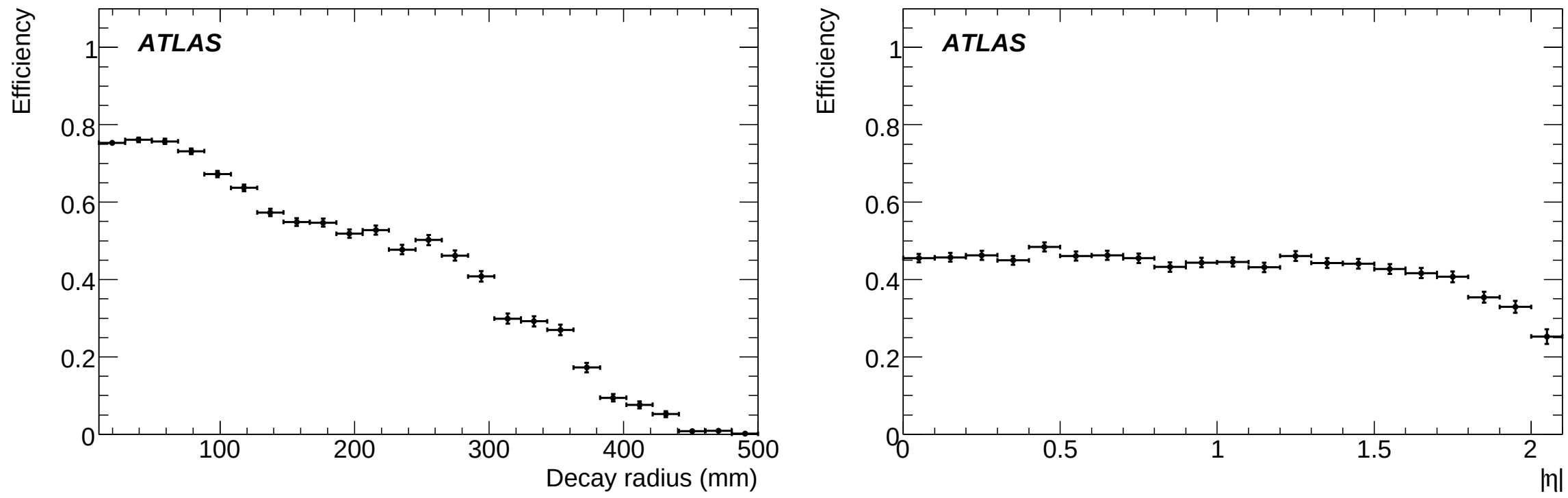

Figure 23: Efficiency to reconstruct charged-pion pairs for $K_s^0 \to \pi^+\pi^-$ decays in events containing $B$-hadron decays as a function of the $K_s^0$ decay radius (left) and as a function of the $|\eta|$ of the $K_s^0$ (right).

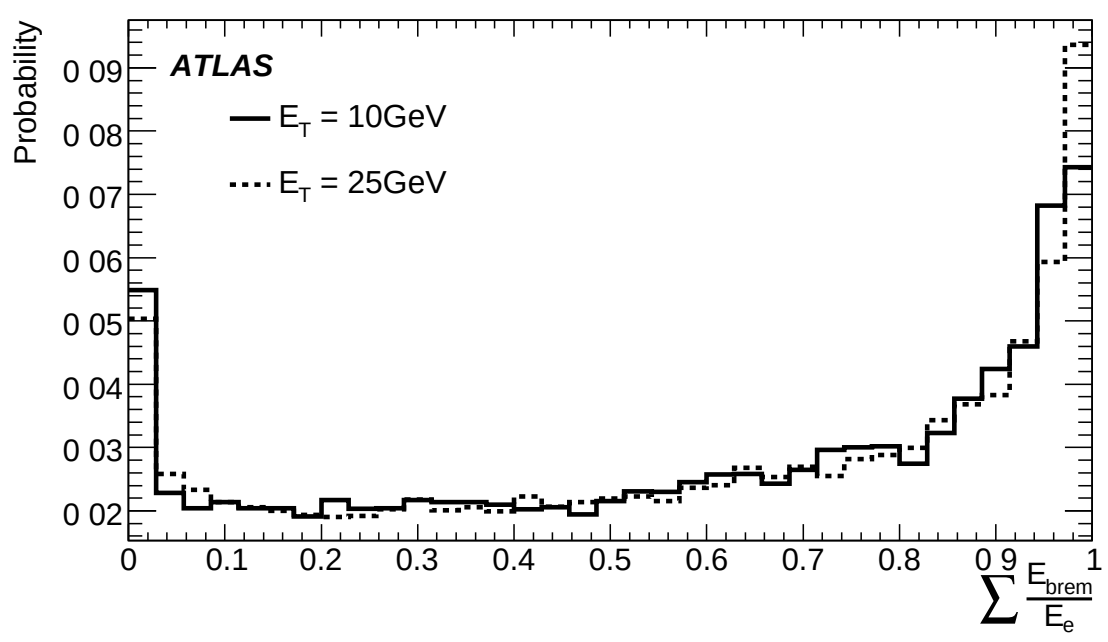

Figure 24: Probability distribution as a function of the fraction of energy lost by electrons with $p_T$ = 10 GeV and 25 GeV (integrated over a flat distribution in $\eta$ with $|\eta| \leq 2.5$) traversing the complete inner detector.

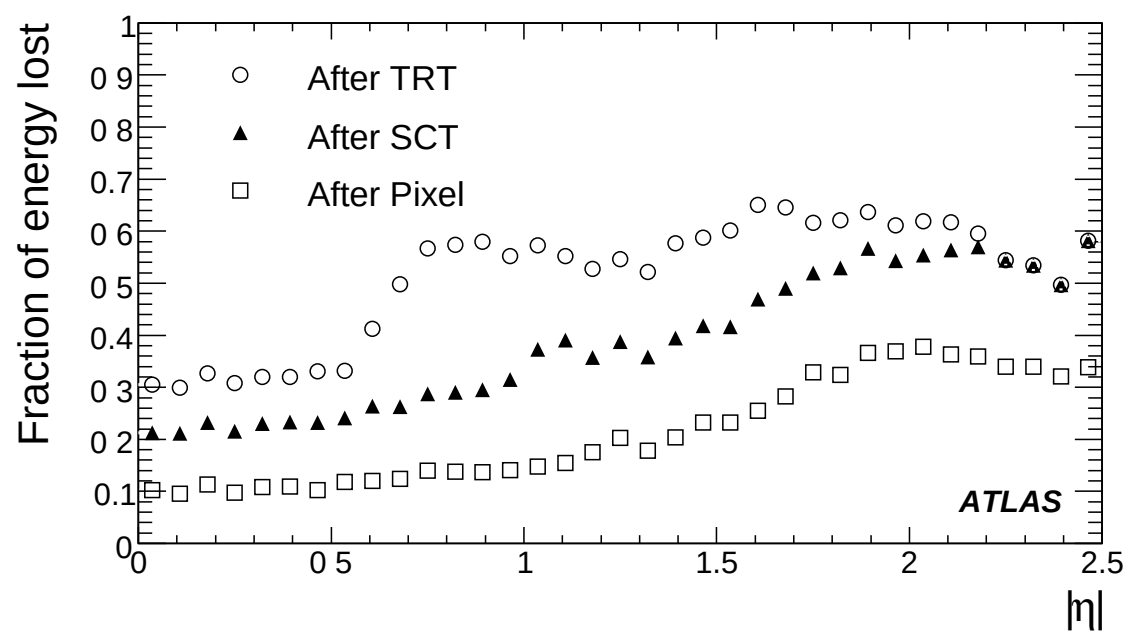

Figure 25: Fraction of energy lost on average by electrons with $p_T$ = 25 GeV as a function of $|\eta|$, when exiting the pixel, the SCT and the inner detector tracking volumes. For $|\eta| > 2.2$, there is no TRT material, hence the SCT and TRT lines merge.

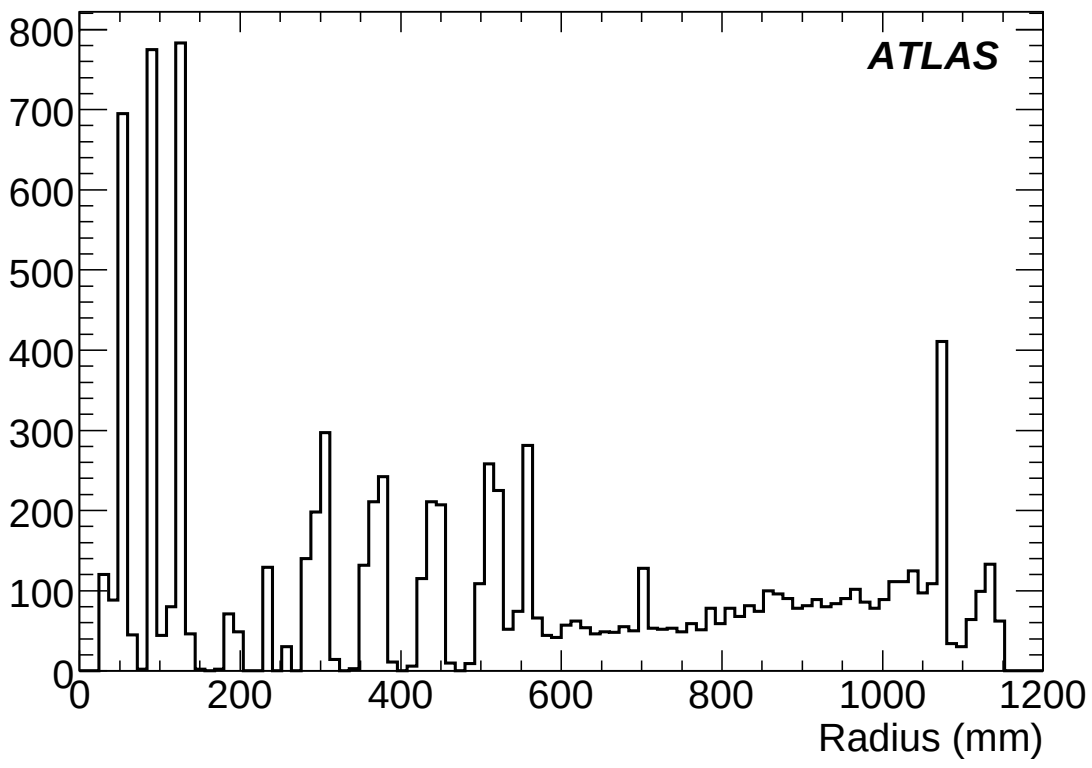

Figure 26: Radial position of photon conversions in the barrel region ($|\eta| < 0.8$) deduced from Monte-Carlo truth information (arbitrary normalisation).

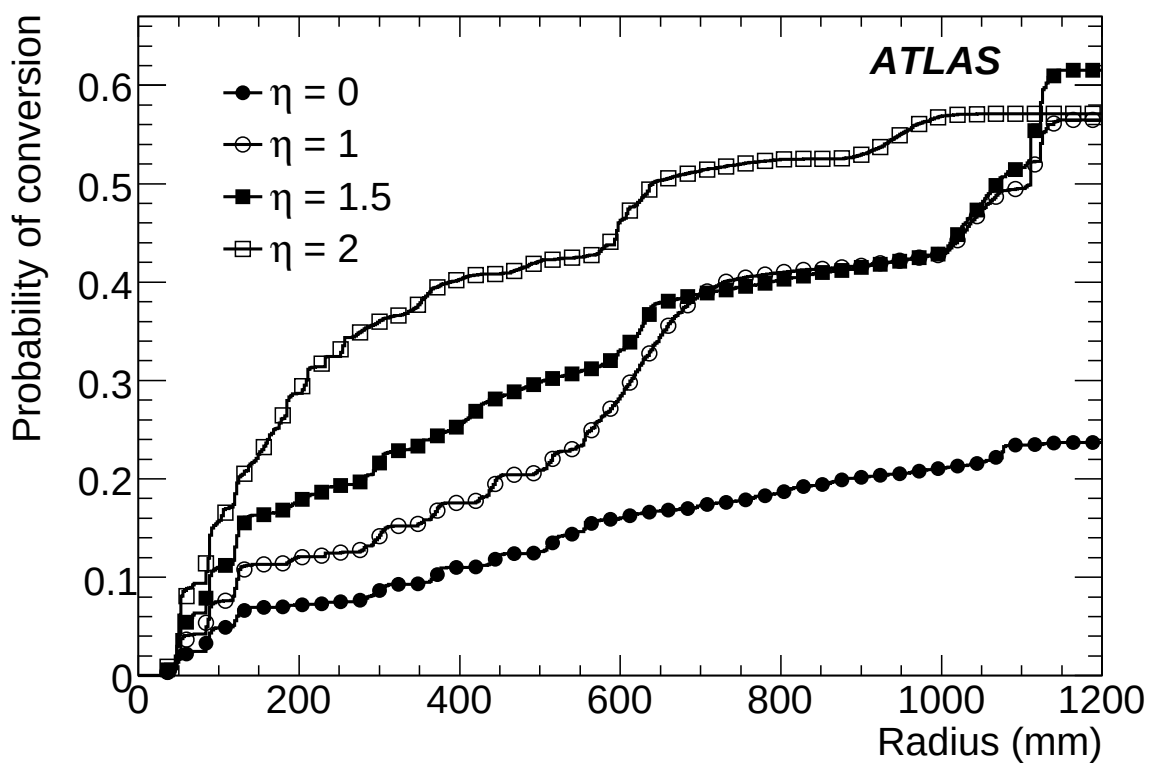

Figure 27: Probability for a photon to have converted as a function of radius for different values of $|\eta|$, shown for photons with $p_T > 1$ GeV in minimum bias events.

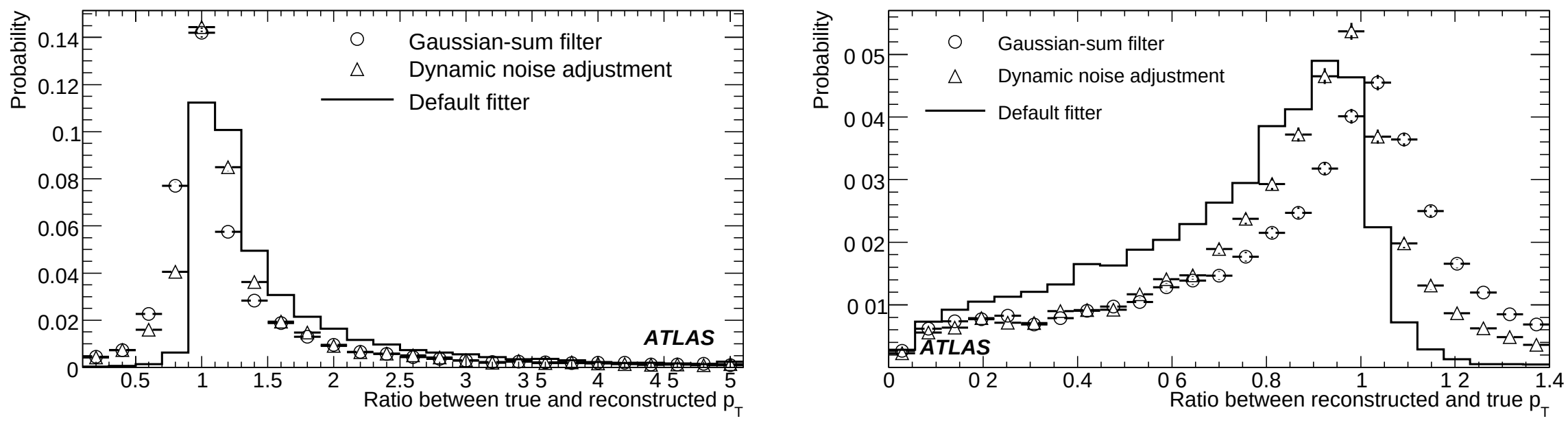

Figure 28: Probability distributions for the ratio of the true to reconstructed momentum (left) and its reciprocal (right) for electrons with $p_T$ = 25 GeV and $|\eta| > 1.5$. The results are shown as probabilities per bin for the default Kalman fitter and for two bremsstrahlung recovery algorithms (see text).

the electron measurement. The algorithms serve to reduce the bias of the track fits caused by the increased track curvature. Only by adding additional information, such at the position of the cluster in the electromagnetic calorimeter [2], is it possible to make a real improvement on the measured momentum.

By allowing for changes in the curvature of the track, the bremsstrahlung recovery algorithms "follow" the tracks better and correctly associate more of the hits, leading to improvements in the reconstruction efficiencies, as can be seen in Fig. 30. GSF has 2-3% greater efficiency than the default reconstruction, since it does not flag hits as outliers, hence a track is less likely to fail the quality cuts on the numbers of hits.

Figure 31 shows the improvements from bremsstrahlung recovery for the reconstructed $J/\psi \rightarrow ee$ mass. Integrating over the complete pseudorapidity acceptance of the ID, and without using any bremsstrahlung recovery, only 42% of events are reconstructed within ±500 MeV of the nominal $J/\psi$ mass, whereas with the use of the bremsstrahlung recovery, this fraction increases to 53% and 56% for DNA and GSF respectively, and the bias of the peak position is reduced. In the inner detector alone, the $J/\psi$ signal in the end-caps is more or less completely lost because of the effects of the increased material compared to that in the barrel. The poor performance in the end-caps arises from the significant fraction of energy lost by electrons (O(30)% by the time they have left the pixels) as well as the change in track direction. These distributions should be contrasted with those for the muonic decays of the $J/\psi$ in Fig. 8.

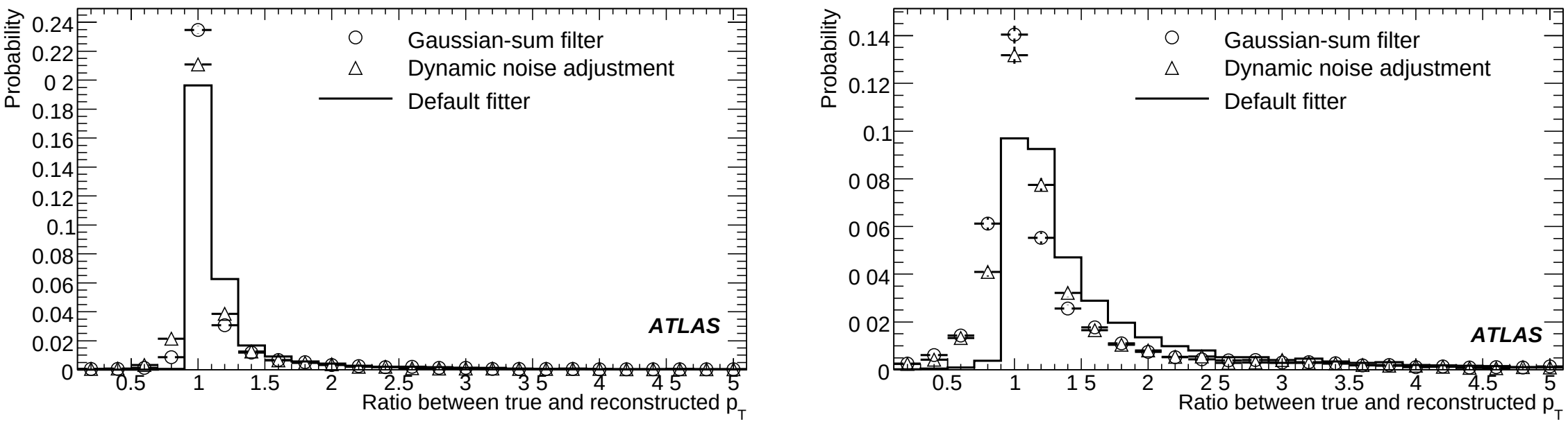

Figure 29: Probability distributions for the ratio of the true to reconstructed momentum for electrons with $p_T$ = 25 GeV and $|\eta| < 0.8$ (left) and $p_T$ = 10 GeV and $|\eta| > 1.5$ (right). The results are shown as probabilities per bin for the default Kalman fitter and for two bremsstrahlung recovery algorithms (see text).

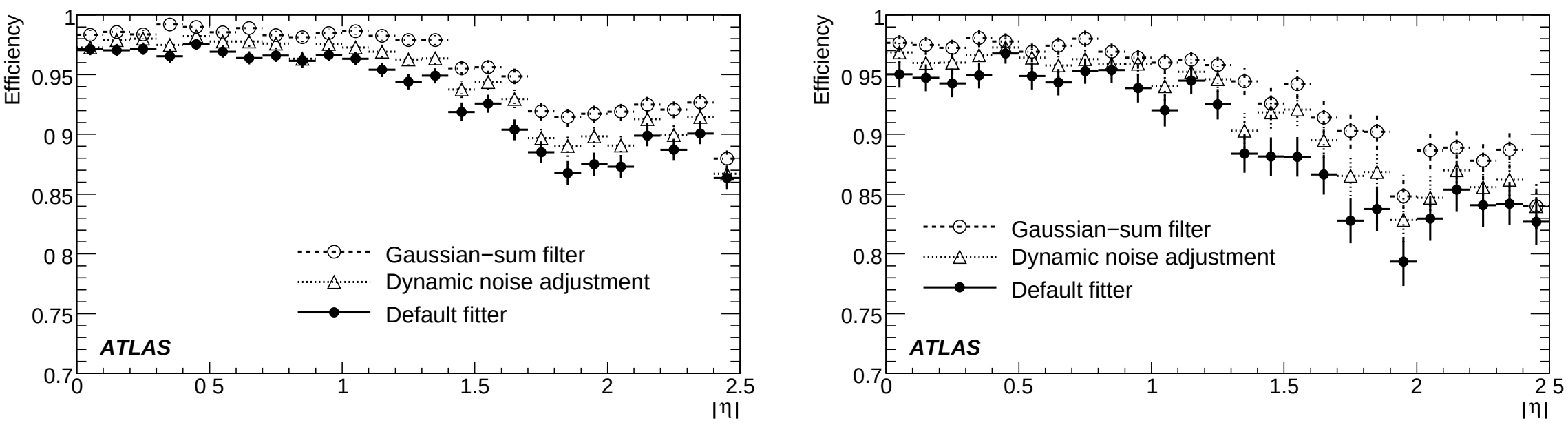

Figure 30: Efficiencies to reconstruct electrons as a function of $|\eta|$ for electrons with $p_T$ = 25 GeV (left) and $p_T$ = 10 GeV (right). The results are shown for the default Kalman fitter and for two bremsstrahlung recovery algorithms (see text).

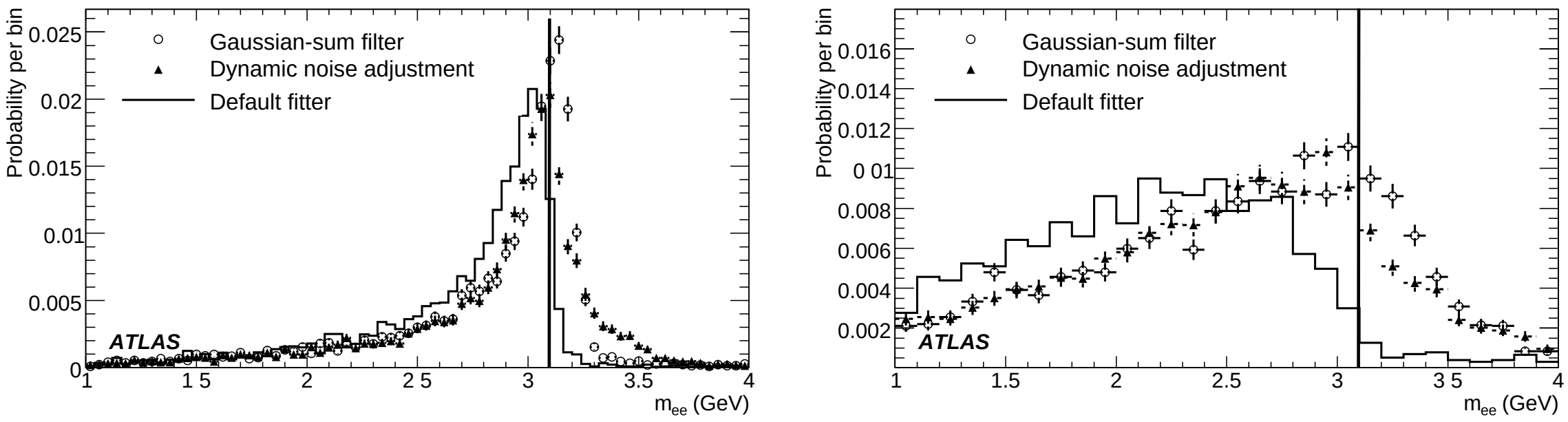

Figure 31: Probability for the reconstructed invariant mass of electron pairs from $J/\psi \to ee$ decays in events with $B_d^0 \to J/\psi(ee)K_s^0$. Distributions are shown for both electrons with $|\eta| < 0.8$ (left) and $|\eta| > 1.5$ (right). The results are shown for the default Kalman fitter and for two bremsstrahlung recovery algorithms (see text). The true $J/\psi$ mass is shown by the vertical line.

To conclude, the material of the inner detector causes a significant amount of bremsstrahlung for electrons, biasing their fitted parameters. This can be partially compensated within the inner detector using the so-called bremsstrahlung recovery procedures, DNA and GSF. These algorithms should be applied to tracks in a way so as to improve electrons and not degrade pions or muons. DNA runs in a time comparable with other simple fitters, while GSF, albeit producing better results, is a factor of twenty slower than DNA. Exactly how these algorithms are used will depend on individual physics analyses.

## 5.2 Electron identification

While the end-cap TRT (discrete radiator foils) is relatively easy to simulate, the barrel TRT (matrix of fibres) is harder and the best indication of the expected performance comes from the test beam (CTB), where a complete barrel TRT module was tested. Using pion, electron and muon samples in the energy range between 2 and 350 GeV, the barrel TRT response has been measured in the CTB in terms of the high-threshold hit probability, as shown in Fig. 32. The measured performance has been used to parametrise the response in the TRT barrel. The transition-radiation X-rays contribute significantly to the high-threshold hits for electron energies above 2 GeV and saturation sets in for electron energies above 10 GeV. Figure 33 shows the resulting pion identification efficiency (probability of pions being misidentified as electrons) for an electron efficiency of 90%, achieved by performing a likelihood evaluation based on the high-threshold probability for electrons and pions for each straw. Figure 33 also shows the effect of including time-over-threshold information, which improves the pion rejection by about a factor of two when combined with the high-threshold hit information. At low energies, the pion rejection (the inverse of the pion efficiency plotted in Fig. 33) improves with energy as the electrons emit more transition radiation. The performance is optimal at energies of $\sim$ 5 GeV, and pion-rejection factors above 50 are achieved in the energy range of 2–20 GeV. At very high energies, the pions become relativistic and therefore produce more $\delta$-rays and eventually emit transition radiation, which explains why the rejection slowly decreases for energies above 10 GeV.

The electron-identification performance expected for the TRT in ATLAS, including the time-over-threshold information, is shown as a function of $|\eta|$ in Fig. 35 in the form of the pion identification efficiency expected for an electron efficiency of 90% or 95%. The shape observed is closely correlated to the number of TRT straws crossed by the track (see Fig. 34), which decreases from approximately 35 to a minimum of 20 in the transition region between the barrel and end-cap TRT, $0.8 < |\eta| < 1.1$, and which also decreases rapidly at the edge of the TRT fiducial acceptance for $|\eta| > 1.8$. Because of its more efficient and regular foil radiator, the performance in the end-cap TRT is better than in the barrel TRT where it consists of radiating fibres [2].

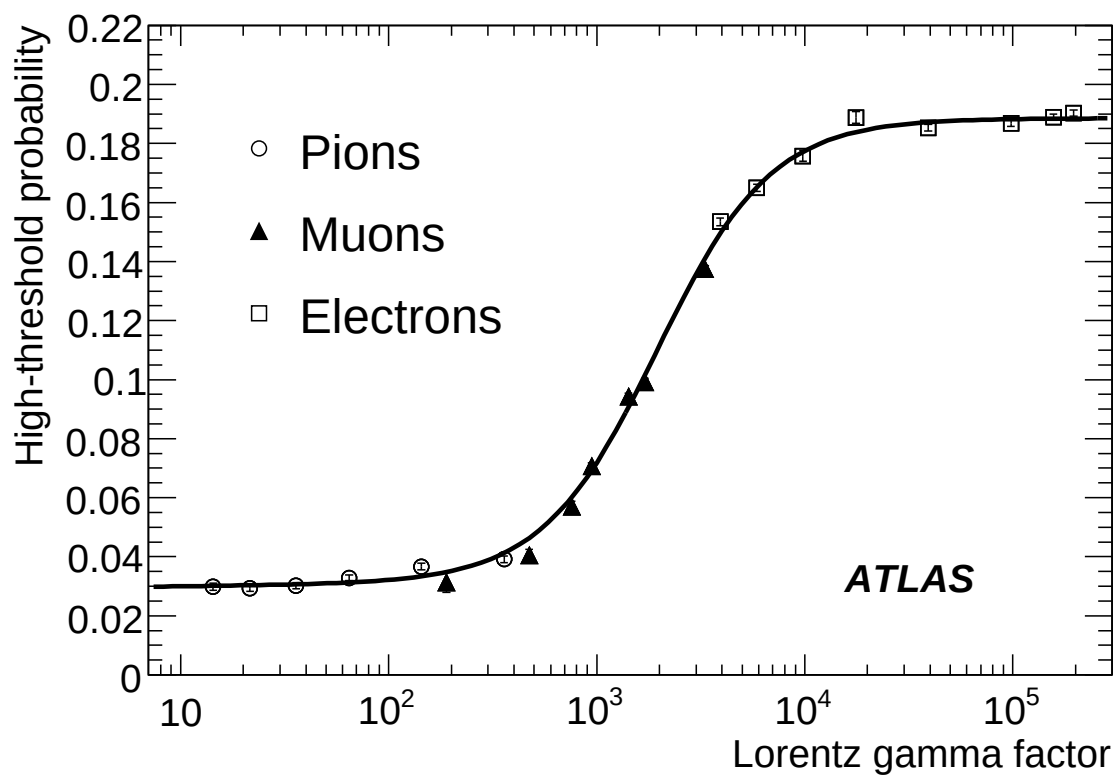

Figure 32: Average probability of a high-threshold hit in the barrel TRT as a function of the Lorentz $\gamma$-factor for electrons (open squares), muons (full triangles) and pions (open circles) in the energy range 2–350 GeV, as measured in the combined test-beam (CTB).

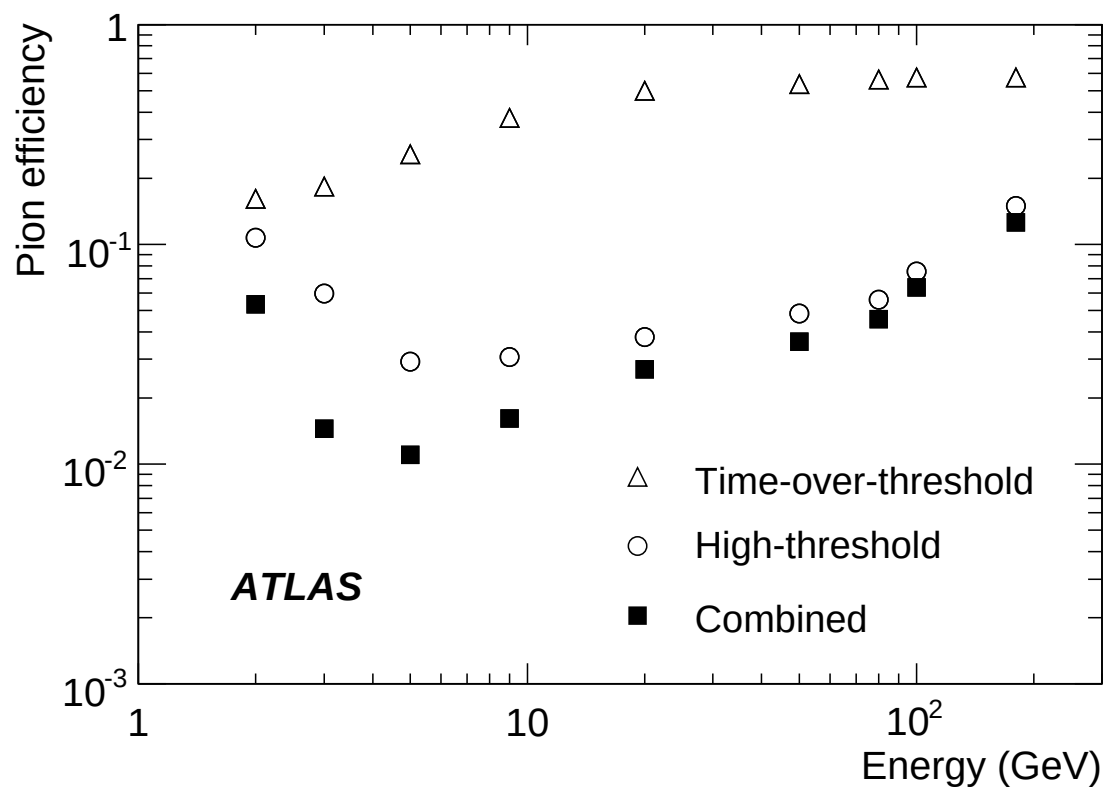

Figure 33: Pion efficiency shown as a function of the pion energy for 90% electron efficiency, using high-threshold hits (open circles), time-over-threshold (open triangles) and their combination (full squares), as measured in the combined test-beam.

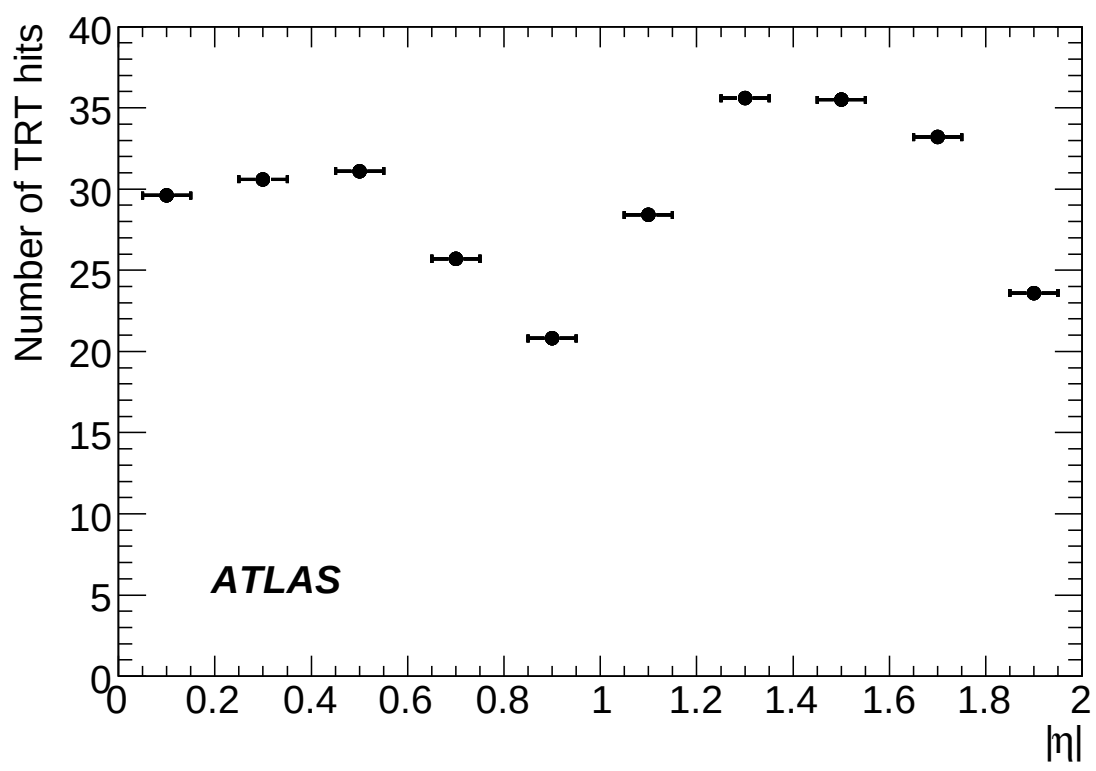

Figure 34: Number of hits on a track as a function of $|\eta|$ for a track crossing the TRT.

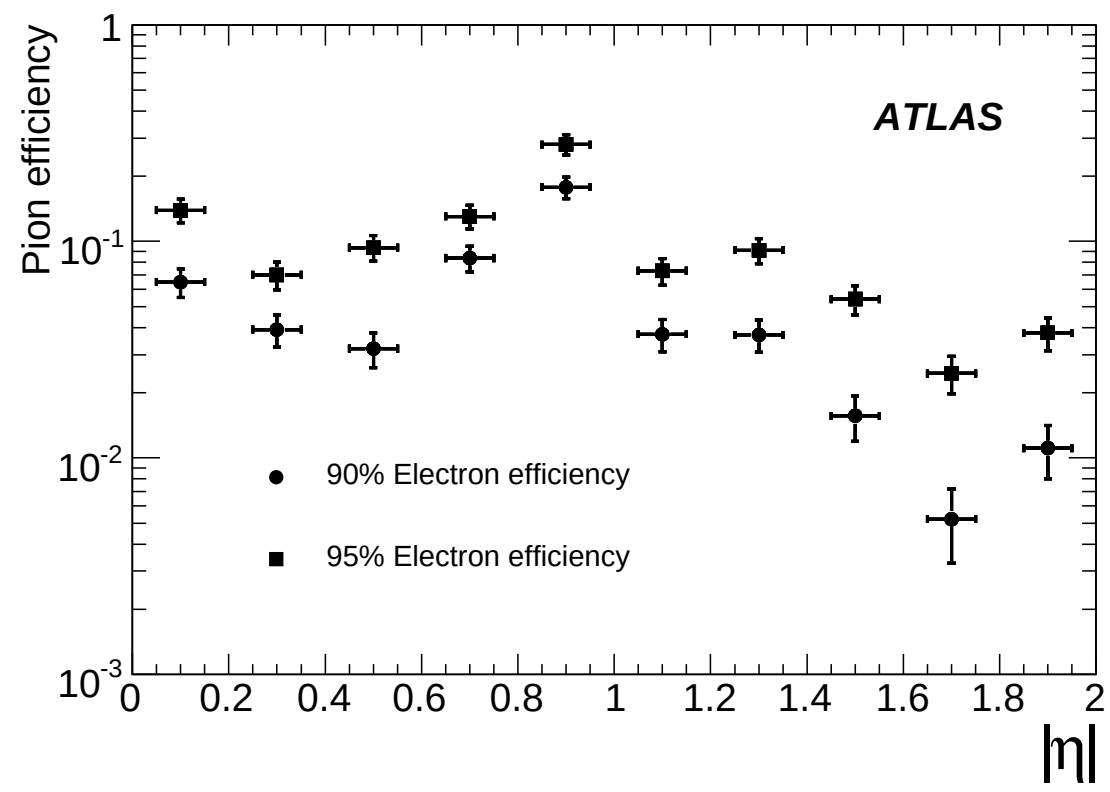

Figure 35: Pion efficiency expected from simulation as a function of $|\eta|$ for an efficiency of 90% or 95% for electrons with $p_T$ = 25 GeV.

## 5.3 Conversion reconstruction

Figure 36 shows the efficiency for reconstructing conversions of photons with $p_T$ = 20 GeV and $|\eta| < 2.1$ as a function of the conversion radius and pseudorapidity, using the standard tracking algorithm combined with the back-tracking algorithm described in Section 2. At radii above 50 cm, the efficiency for reconstructing single tracks drops and that for reconstructing the pair drops even faster because the two tracks are merged. If both tracks from the photon conversion are reconstructed successfully, vertexing tools can be used to reconstruct the photon conversion with high efficiency up to radii of 50 cm. The overall conversion-identification efficiency can be greatly increased at large radii by flagging single tracks as photon conversions under certain conditions. (The identification is distinct from the reconstruction, since with a single electron, the photon conversion cannot be reconstructed.) Only tracks which have no hits in the vertexing layer, which are not associated to any fitted primary or secondary vertex, and which pass a loose electron identification cut requiring more than 9% high-threshold hits on the TRT segment of the track are retained. The resulting overall efficiency for identifying photon conversions is almost uniform over all radii below 80 cm, as shown in Fig. 37.

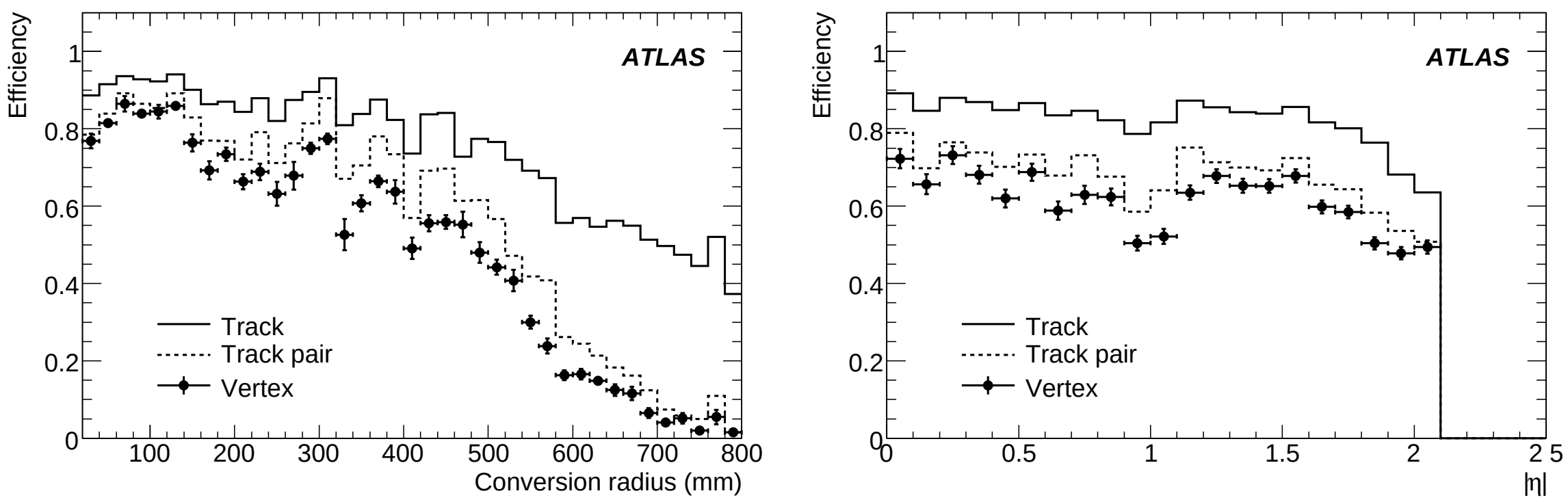

Figure 36: Efficiency to reconstruct conversions of photons with $p_T$ = 20 GeV and $|\eta| < 2.1$, as a function of the conversion radius (left) and pseudorapidity (right). Shown are the efficiencies to reconstruct single tracks from conversions, the pair of tracks from the conversion and the conversion vertex.

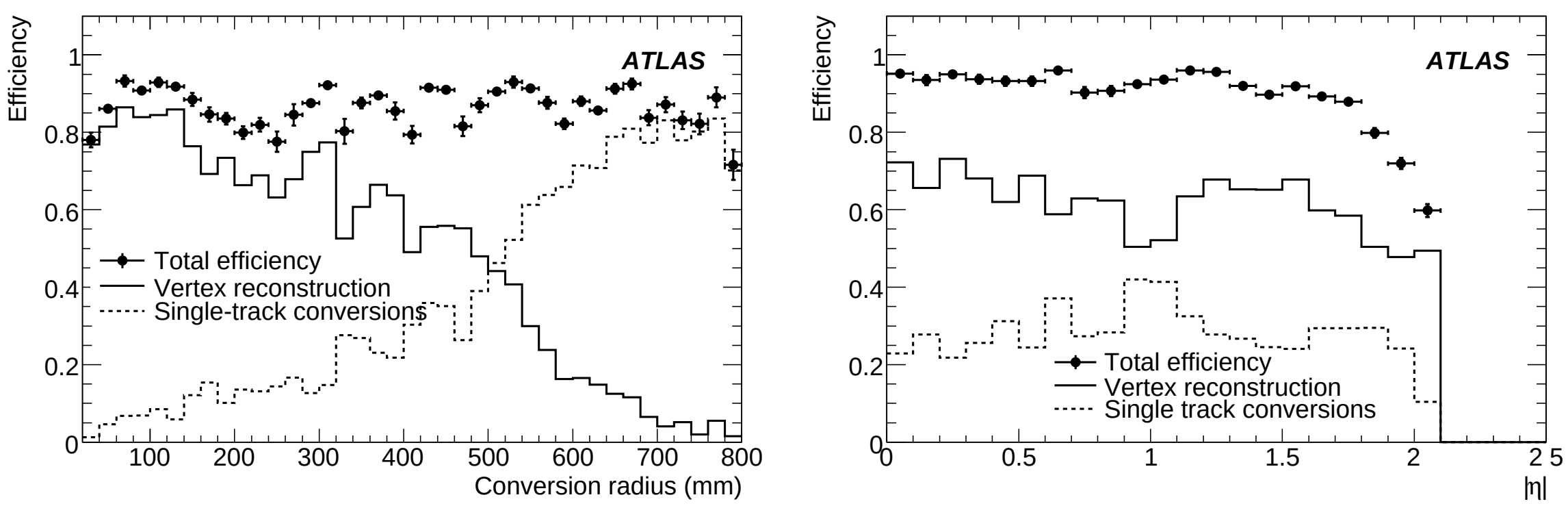

Figure 37: Efficiency to identify conversions of photons with $p_T$ = 20 GeV and $|\eta| < 2.1$, as a function of the conversion radius (left) and pseudorapidity (right). The overall efficiency is a combination of the efficiency to reconstruct the conversion vertex, as shown also in Fig. 36, and of that to identify single-track conversions (see text).

## 6 Conclusions

This paper documents the expected performance for the ATLAS inner detector, focusing on the low-luminosity running at the start-up of the LHC. Most of the performance specifications set out in [1] have been met – it is only at larger values of $|\eta|$, where there are significant amounts of material, that the track-finding efficiencies are less than the targets.

The reconstruction of muons, electrons and pions has been studied in detail as a function of transverse momentum and pseudorapidity. For high-$p_T$ muons in the barrel region, the resolution for $1/p_T$ is expected to be 0.34 TeV$^{-1}$ and the resolution for the transverse impact parameter 10 $\mu$m. The charge of muons and electrons will be measured in the inner detector over the complete acceptance up to 1 TeV with misidentification probabilities on average of no more than a few percent. In the barrel region, muons with $p_T \geq 1$ GeV can be identified with efficiencies in excess of 98%. For high-$p_T$ muons, this rises to $\geq$ 99.5% across the whole acceptance. Electrons and pions suffer from material effects; for tracks around 5 GeV, they are reconstructed with efficiencies between 70 and 95%. The inner detector is able to reconstruct pions down to 0.2 GeV with efficiencies around 50%. Fake rates are low; even in the core of moderate-energy jets (O(50) GeV $E_T$), rates are less than 1%.

Algorithms have been developed to reconstruct primary and secondary vertices, as well as $K_s^0$ (and other V$^0$s) decays and conversions. In the case of $t\bar{t}$ events, primary vertices can be identified with 99% efficiency in the presence of low-luminosity pile-up. $K_s^0$ decays can be reconstructed up to a radius of 400 mm, while conversions can be identified by reconstructing pairs of tracks or tagging single electrons in the TRT with 80% efficiency all the way up to a radius of 800 mm.

Electrons suffer from bremsstrahlung caused by the significant material in the inner detector. Algorithms have been developed to improve the reconstruction of electrons, reducing the bias on the measured momentum. While reasonable electron reconstruction is possible in the inner detector barrel, it is quite difficult in the end-caps because of the increased amount of bremsstrahlung – here, the use of the electromagnetic calorimeter will be essential. Electrons can be identified by their transition radiation in the TRT. For an electron efficiency of 90% at $p_T = 25$ GeV, the pion misidentification probability is of the order of a few percent over most of the acceptance, and the pion rejection will be optimal around 5 GeV.

After many years of preparing the ATLAS inner detector software and having tested it on simulated and test-beam data, we are ready to reconstruct and analyse data from collisions. We now look forward to the first data from the LHC.

# Electrons and Photons

# Calibration and Performance of the Electromagnetic Calorimeter

**Abstract**
This note describes the calibration of electromagnetic clusters, as implemented in current releases of the ATLAS reconstruction program. A series of corrections are applied to calibrate both the energy and position measurements; these corrections are derived from Monte-Carlo simulations and validated using test-beam data. The possibility of obtaining inter-calibration energy corrections from $Z \to ee$ data is also discussed.

## 1 Introduction

In order to realise the full physics potential of the LHC, the ATLAS electromagnetic calorimeter must be able to identify efficiently electrons and photons within a large energy range (5 GeV to 5 TeV), and to measure their energies with a linearity better than 0.5%. The $W$ boson mass measurement, not considered here, will require better precision.

The procedure to measure the energy of an incident electron or photon in the ATLAS electromagnetic (EM) calorimeter has been described in Ref. [1]. Each step of the energy reconstruction has been validated by a series of beam tests over many years, both using only the calorimeter [2, 3] and also combined with representative components from all detector sub-systems. This has allowed considerable refinement of the calorimeter simulation. This simulation is then used to model the behaviour of the full detector.

One of the key ingredients for the description of the detector performance is the amount and position of the upstream material. The understanding of the ATLAS detector geometry has also made progress over the years; an overview of the present knowledge of the detector and its expected performance can be found in [4]. The amount of material in front of the calorimeter for the as-built detector is significantly larger than was initially estimated; this leads to larger energy losses for electrons and to a larger fraction of photons converting (see Figs. 1 and 2).

The standard ATLAS coordinate system is used: the beam direction defines the $z$-axis, and the $x$-$y$ plane is transverse to the beam direction. The azimuthal angle $\phi$ is measured around the beam axis and the polar angle $\theta$ is the angle from the beam axis. The pseudorapidity is defined as $\eta \equiv -\ln(\tan(\theta/2))$.

### 1.1 Electron and photon candidates

The "sliding window" algorithm [5] is used to find and reconstruct electromagnetic clusters. This forms rectangular clusters with a fixed size, positioned so as to maximise the amount of energy within the cluster. An alternate algorithm is available which forms clusters based on connecting neighbouring cells until the cell energy falls below a threshold; this is not used by the default electron and photon reconstruction. The optimal cluster size depends on the particle type being reconstructed and the calorimeter region: electrons need larger clusters than photons due to their larger interaction probability in the upstream material and also due to the fact that they bend in the magnetic field, radiating soft photons along a range in $\phi$. Several collections of clusters are therefore built by the reconstruction software, corresponding to different window sizes. These clusters are the starting point of the calibration and selection of electron and photon candidates.

One of the recent improvements in the calibration procedure is that electron and photon candidates are treated separately. For each of the reconstructed clusters, the reconstruction tries to find a matching track within a $\Delta\eta \times \Delta\phi$ window of $0.05 \times 0.10$ with momentum $p$ compatible with the cluster energy $E$ ($E/p < 10$ [6, 7]). If one is found, the reconstruction checks for presence of an associated conversion. An electron candidate is created if a matched track is found and no conversion is flagged. Otherwise, the candidate is classified as a photon.

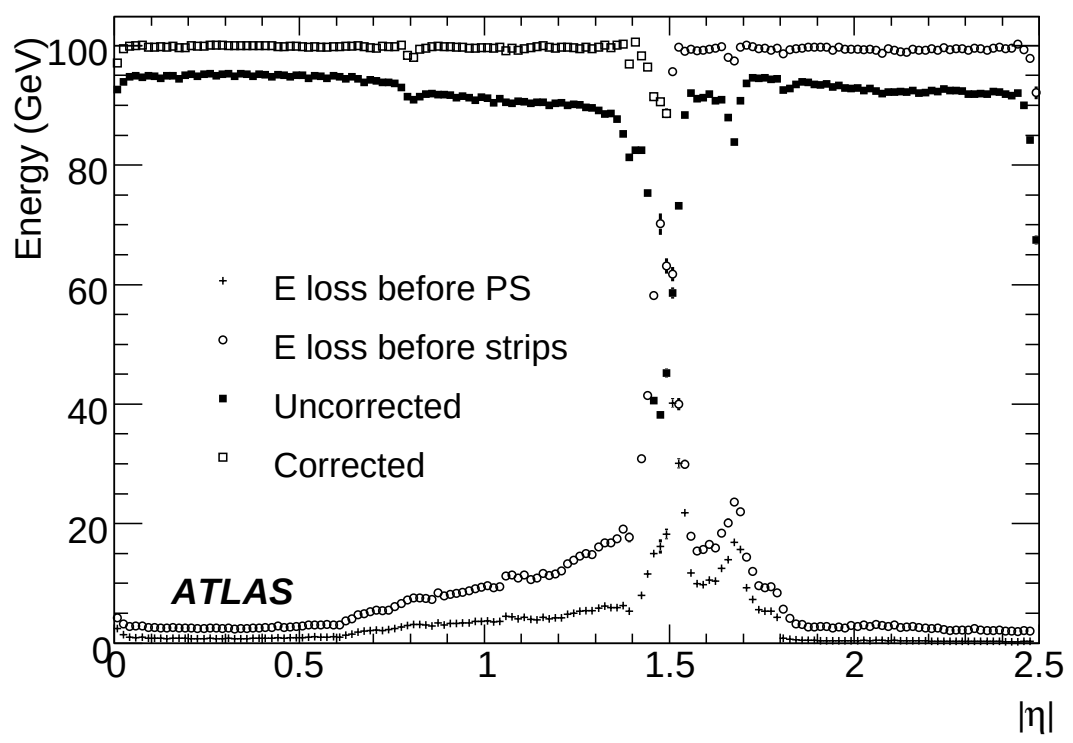

Figure 1: Average energy loss vs. $|\eta|$ for $E = 100$ GeV electrons before the presampler/strips (crosses/open circles), and reconstructed energies before/after (solid/open boxes) corrections.

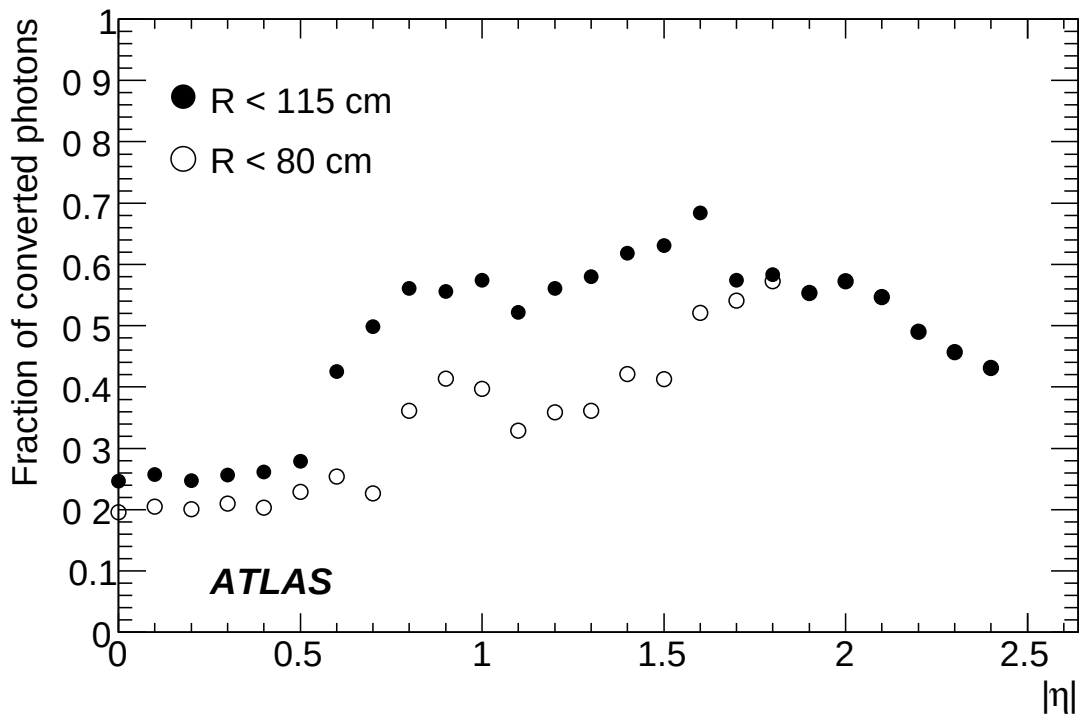

Figure 2: Fraction of photons converting at a radius of below 80 cm (115 cm) in open (full) circles, as a function of $|\eta|$ [4].

This early classification allows applying different corrections to electron and photon candidates. It is the starting point of a more refined identification based largely on shower shapes, described in companion notes [6, 7]. Four levels of electron quality are defined (loose, medium, tight, and tight without isolation). The available photon selection corresponds to the tight electron selection (excluding tracking requirements). The medium and tight selections are used in some parts of the calibration analysis described in this note. But the corrections derived are then applied to all electron and photon candidates.

## 1.2 Calorimeter granularity

The electromagnetic calorimeter (Fig. 3) was designed to be projective in $\eta$, and covers the pseudorapidity range $|\eta| < 3.2$. Precision measurements are however restricted to $|\eta| < 2.5$; regions forward of this are outside of the scope of this note. The calorimeter is installed in three cryostats: one containing the barrel part ($|\eta| < 1.475$), and two which each contain the two parts of the end-cap ($1.375 < |\eta| < 3.2$). Its accordion structure provides complete $\phi$ symmetry without azimuthal cracks. The total thickness of the calorimeter is greater than 22 radiation lengths ($X_0$) in the barrel and greater than $24X_0$ in the end-caps. It is segmented in depth into three longitudinal sections called layers, numbered from 1 to 3 outwards from the beam axis. These layers are often called "front" (or "strips"), "middle," and "back." The $\eta$ granularity of the calorimeter for the front and middle layers is shown in Table 1. The $\phi$ size of cells is 0.025 in layer 2 and 0.1 in layer 1. Layer 3 has a granularity of $\Delta\eta \times \Delta\phi = 0.050 \times 0.025$. For $|\eta| < 1.8$, a presampler detector is used to correct for the energy lost by electrons and photons upstream of the calorimeter. All these regions must be treated separately in deriving the individual corrections.

The effect of the choice of cluster size on electron and photon energy reconstruction has been studied in Refs. [1] and [8]. These results are still the baseline of the present software. For electrons, the energy in the barrel electromagnetic calorimeter is collected over an area corresponding to $3 \times 7$ cells in the middle layer, i.e. $\Delta\eta \times \Delta\phi = 0.075 \times 0.175$. For unconverted photons, the area is limited to $3 \times 5$ cells in the middle layer, whereas converted photons are treated like electrons. The cluster width in $\eta$ increases with increasing $|\eta|$; therefore, an area of $5 \times 5$ cells in the middle layer is used for both electrons and photons in the end-cap calorimeter.

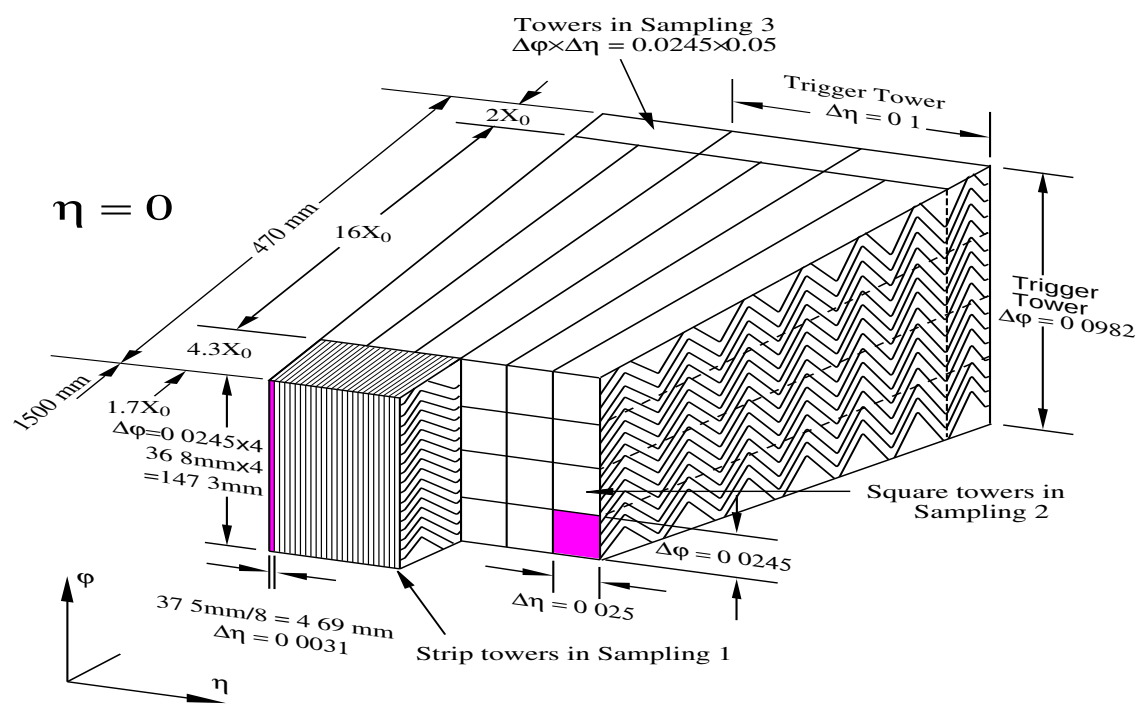

Figure 3: Sketch of the accordion structure of the EM calorimeter [8].

| | $\|\eta\|$ range | Cell $\eta$ size | |
|---|---|---|---|
| | | Layer 1 | Layer 2 |
| Barrel | 0–1.4 | 0.025/8 | 0.025 |
| | 1.4–1.475 | 0.025 | 0.075 |
| end-cap | 1.375–1.425 | 0.05 | 0.05 |
| | 1.425–1.5 | 0.025 | 0.025 |
| | 1.5–1.8 | 0.025/8 | 0.025 |
| | 1.8–2.0 | 0.025/6 | 0.025 |
| | 2.0–2.4 | 0.025/4 | 0.025 |
| | 2.4–2.5 | 0.025 | 0.025 |

Table 1: Calorimeter $\eta$ granularity in layers 1 and 2.

## 1.3 Geometries and data sets

The present knowledge of the detector geometry, resulting from the detector survey, is described in [4] (Sec. 9). But even before the final survey, it was known that the inner detector services located in the crack region would be wider than originally expected, and that the end-cap electromagnetic calorimeter would be shifted by about 4 cm, compared to the nominal (and pointing) geometry described in Ref. [1]. This is taken into account in the simulation, and is treated as a misalignment in the cell calibration procedure described below.

High statistics samples of single electrons and photons, processed with the full detector simulation based on GEANT 4.7 [9], were used to derive and study the corrections. Two detector geometries are available. The first is the “ideal geometry,” which contains the best knowledge of the dead material, but which has no misalignments except for the 4 cm shift of the end-caps. The data sets based on this geometry are used to derive the corrections and for most of the performance studies. The second available geometry is a distorted one, in which extra material is added between the tracking detectors and the calorimeters, and in which misalignments are introduced. For example, the amount of material in the inner detector has increased in some regions by up to 7% of a radiation length for positive $\phi$, and the density of material in the gap between the barrel and end-cap cryostats has increased by a factor of 1.7. The distorted data-sets using this geometry are used to estimate systematic uncertainties and to check the sensitivity of the methods to additional material. In addition to these single-particle data sets, $Z \to ee$ decays are also available.

The standard calorimeter reconstruction for simulated data includes the effects of possible cell-level miscalibrations by smearing the measured energy of each cell (by about 0.7%), therefore increasing the constant term of the energy resolution. (The fractional energy resolution is conventionally parametrised as $\sigma(E)/E = a/E \oplus b/\sqrt{E} \oplus c$, where $a$ is the noise term, $b$ is the sampling term and $c$ is the constant term.) Unless otherwise stated, the results in this note do not include this additional smearing, and therefore correspond to assuming a perfect cell-level calibration.

## 1.4 Energy and position reconstruction

The calibration of the LAr calorimeter is factorised into a channel-by-channel calibration of the electronics readout and an overall energy scale determination.

The first step, often called “electronics calibration”, converts the raw signal extracted from each cell (in ADC counts) into a deposited energy. The method used for this step, which is beyond the scope of this note, was described in Ref. [1]. It was refined and validated when final barrel and end-cap modules were studied in test beams [2, 3]. In the past two years, the experience gained and the algorithm developed were integrated into the standard ATLAS calibration software [10].

The second step deals with clusters. The energies deposited in the cells of each individual layer of a cluster are summed, and an energy-weighted cluster position is calculated for each layer. There are several important effects which must then be understood:

- Due to the accordion geometry, the amount of absorber material crossed by incident particles varies as a function of $\phi$. This produces a $\phi$ modulation of the reconstructed energy.
- The shower is not fully contained in the $\eta$ window chosen for clusters, and the cells have a finite granularity. This introduces a modulation in the energy and a bias in the measured position ("S-shape") which depend on the particle impact point within a cell.
- A perfectly projective particle, coming from the origin of the coordinate system, intersects the calorimeter at the same $\eta$ position in all layers. The luminous region, however, extends significantly in $z$; a particle from a vertex away from the origin intersects the calorimeter at slightly different $\eta$ positions in each layer. Properly combining these $\eta$ measurements requires an accurate parametrisation of the shower depth within each layer.

An early study of these corrections, using both simulation and test beam data, can be found in [11]. The present prediction of these effects and their dependencies on the impact point and energy of the incident particle are described in detail in this note.

The measured energy and position of EM clusters are corrected as described below (see Fig. 4). The required scale of the correction is illustrated by the upper points in Fig. 1, which shows the reconstructed energies of $E = 100$ GeV electrons before and after calibration. It is about 10% over most of the calorimeter, but is larger in the transition region between cryostats.

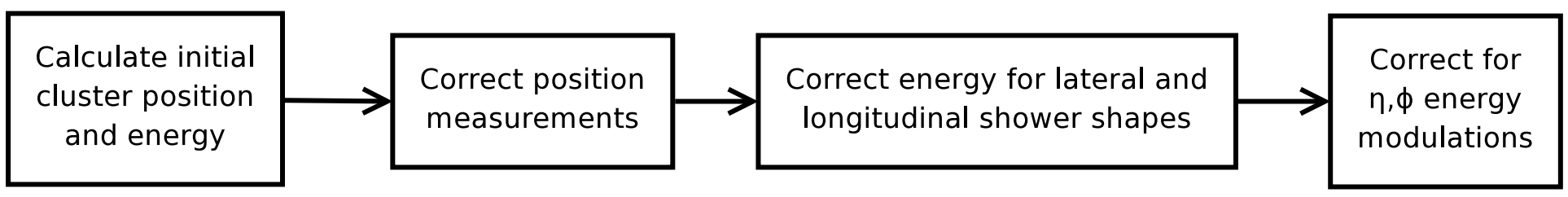

Figure 4: Cluster correction steps.

- To start with, the energies in the cluster cells are summed, and an energy weighted $(\eta,\phi)$ position is calculated for each calorimeter layer. Before applying the cluster corrections, the energy resolution has a constant term of about 0.65% (quoted for photons at $|\eta| = 0.3$).
- As the first step, corrections are applied to the cluster position, measured in each layer. These are described in Sec. 2. The position measurements from the first two layers are then combined to define the shower impact point in the calorimeter, which can then be used for energy reconstruction.
- The next step is to combine the energies deposited in each layer. Two separate procedures have been developed to do this which are described in Secs. 3.1 and 5. In the first one, per-layer energy coefficients, called longitudinal weights, are adjusted to optimise at the same time the energy resolution and the linearity of the response. In the second one, the simulation is used to correct for different types of energy loss one by one, by correlating each of them with measured observables. The corrections are calculated separately for electrons and photons, and determined as a function of $|\eta|$. This reduces the local constant term to about 0.61%.
- The third step, described in Sec. 3.2, uses the shower impact point to correct the total energy for modulations in $\eta$ and $\phi$. This reduces the local constant term to about 0.43%.

In spite of the skill and care put into the detector construction, calibration, and operation, some local or "medium range" inhomogeneities in the calorimeter response have to be expected: localised high-voltage or temperature effects or unexpected additional dead material must be detected and corrected for using data. It is planned to use $Z \to ee$ decays to measure and correct for such effects and to help fix the absolute energy scale. The method developed and the precision expected are described in Sec. 6.

## 2 Cluster position measurement

The position of a cluster is measured in $\eta$ and $\phi$. The positions are first calculated independently for each calorimeter layer as the energy-weighted barycenters of all cluster cells in the layer. (The barrel and end-cap are also treated separately at this stage.) Secondly, the individual layer measurements are corrected for known systematic biases. Finally, the position measurements from layers 1 and 2 are combined to produce the overall cluster position. The position corrections are derived using single-particle electron and photon data samples. Each sample is mono-energetic, and the available samples span the range 5–1000 GeV.

The $\eta$ positions that are calculated at this stage are "detector"-$\eta$, corresponding to the angle that would be made by a particle originating from the origin of the detector coordinate system. In order to properly compare the calculated detector-$\eta$ positions with the $\eta$ of a generated incident particle, which will in general have its production vertex offset in $z$ from the detector origin, one must assume a depth for each calorimeter layer. Here, "depth" refers to the radial distance from the beam axis for the barrel calorimeter, and to the distance from the $x-y$ plane passing through the origin for the end-cap calorimeter. The depths used are those which optimise the $\eta$-position resolution; they are shown in Fig. 5.

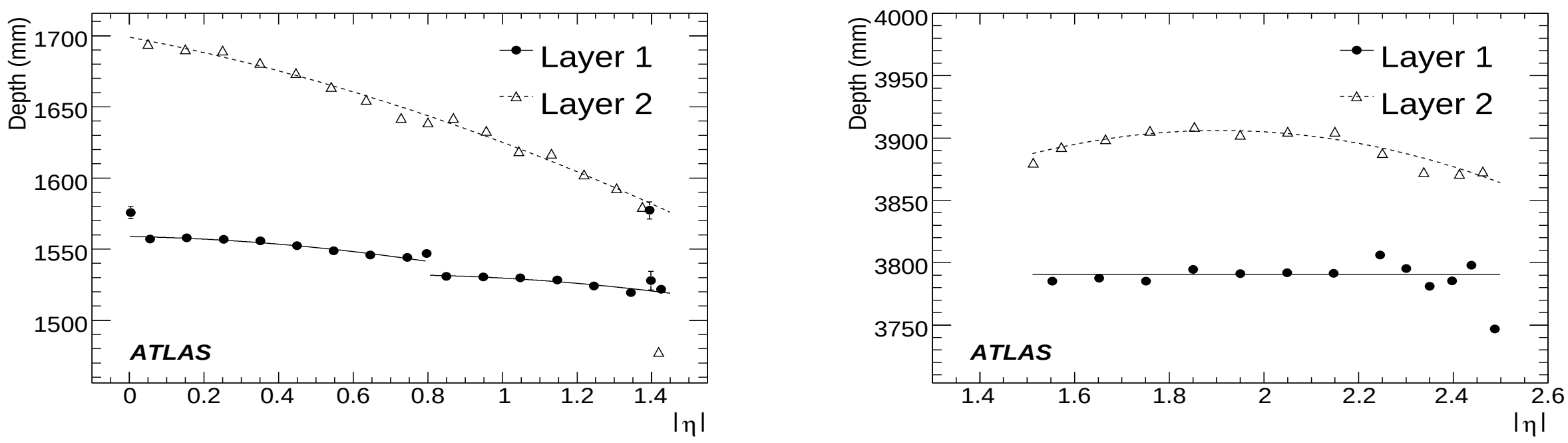

(a) Barrel. (b) End-cap.

Figure 5: Calorimeter depths versus $|\eta|$ for layers 1 and 2 and for 100 GeV photons. The points show the derived optimal depths, and the curves are piecewise polynomial fits to the points. For layer 2 of the barrel, a single curve yielded an adequate fit across $|\eta| = 0.8$; this may be revisited in future versions. From 100 GeV photons.

### 2.1 $\eta$ position correction (S-shape)

The cluster $\eta$ position is first calculated in each layer as the energy-weighted barycenter of the cluster cells in that layer. (In layer 1, only the three strips around the cluster center are used, regardless of the specified cluster size.) Due to the finite granularity of the readout cells, these measurements are biased towards the centers of the cells. For examples, see Fig. 6. This figure plots the difference in $\eta$ between the incident particle and the reconstructed cluster ($\Delta\eta = \eta_{\text{true}} - \eta_{\text{reco}}$) as a function of $v$, the relative $\eta$ offset of the cluster within the cell, which varies from $-1/2 \ldots 1/2$ across the cell. (The sign of $\Delta\eta$ is inverted for negative $\eta$, and in plots it is usually shown as a fraction of the cell $\eta$ width.) The general functional form shown in this figure is often referred to as “S-shape”.

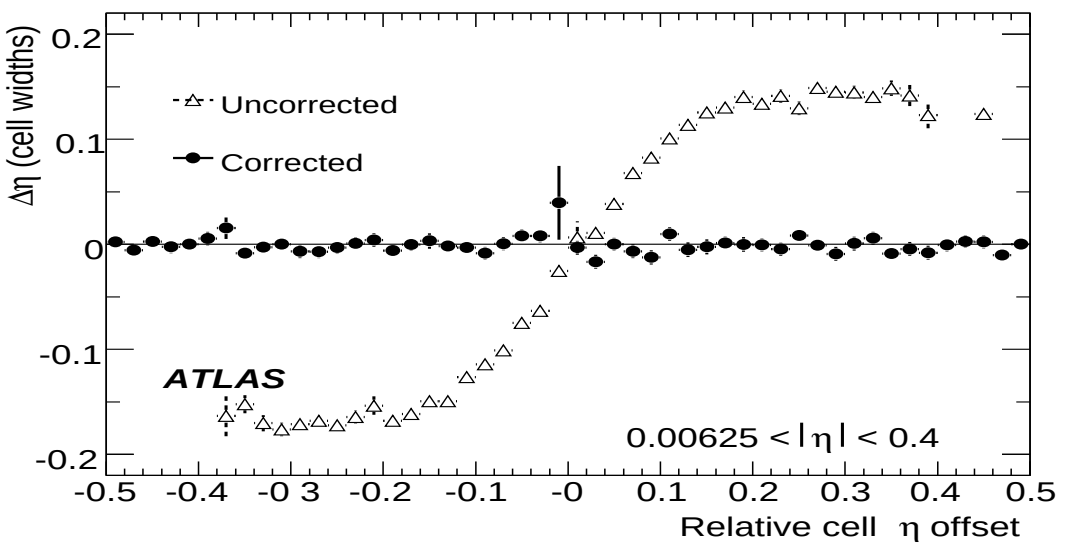

(a) Layer 1, barrel.

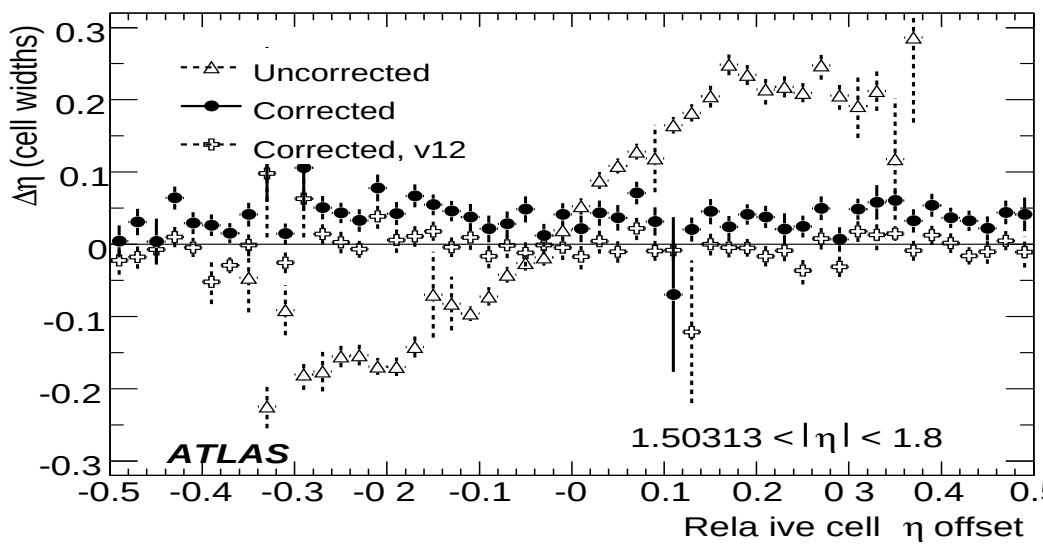

(b) Layer 1, end-cap.

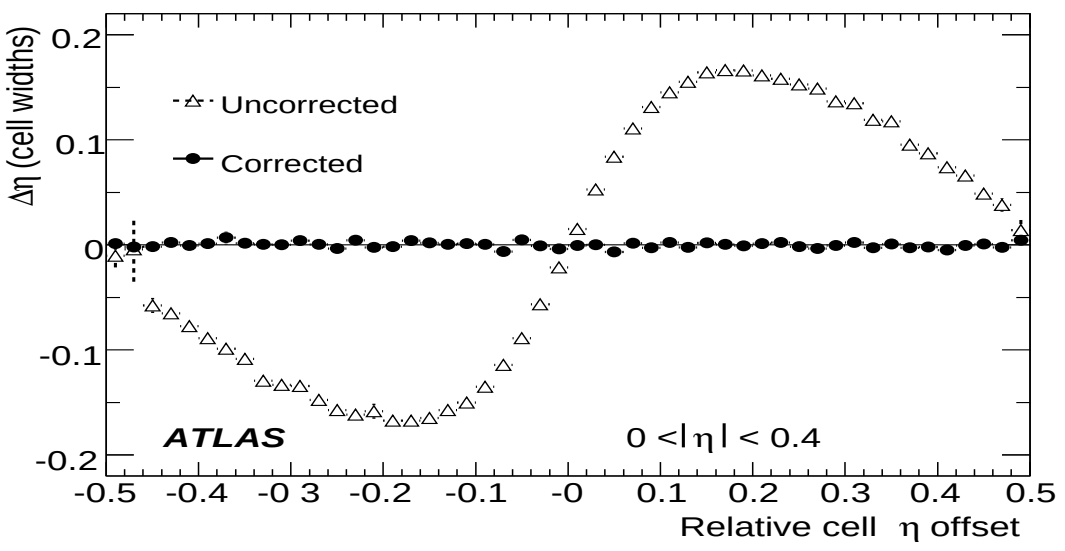

(c) Layer 2, barrel.

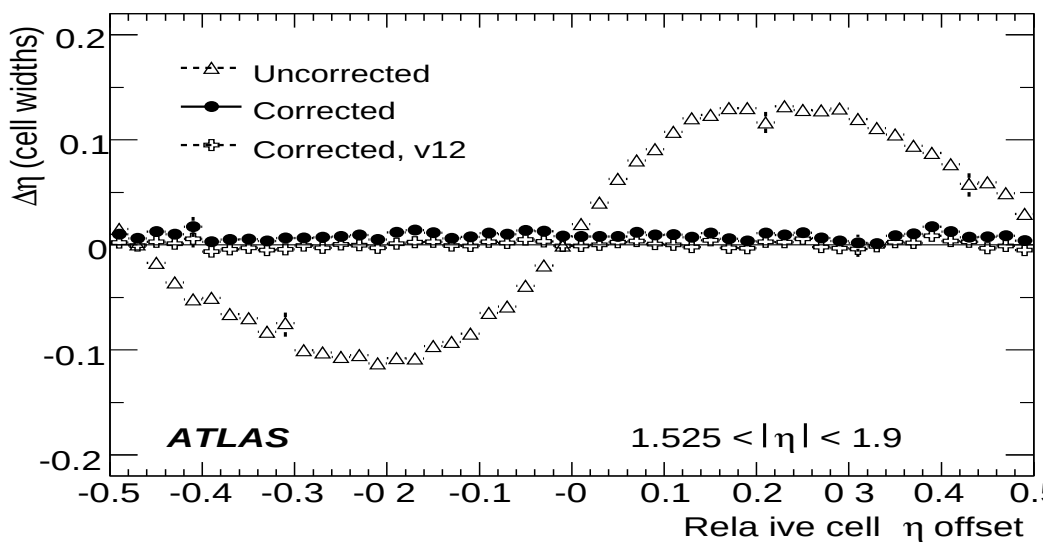

(d) Layer 2, end-cap.

Figure 6: $\Delta\eta$ versus $v$ before and after correction for different regions and for 100 GeV electrons. Note the small systematic offset in the end-cap due to a change in the end-cap geometry since the corrections were derived. For comparison, the “v12” points show results reconstructed using the same geometry as that used to derive the corrections.

Figure 6 shows the correction averaged over an $|\eta|$ range. The actual correction, however, varies continuously over $\eta$, due to changes in the detector geometry (the corrections change to a much greater extent near discontinuities in the calorimeter). For example, the calorimeter cells are not perfectly projective (as the inner and outer cell faces are parallel to the beam-line, rather than being perpendicular to a line from the detector origin); this induces a bias away from the center of the calorimeter. The correction will also depend on the cluster energy, as that affects the average shower depth.

To derive the correction, the calorimeter is divided in $\eta$ into regions based on where the behaviour of the correction changes discontinuously. Within each region, an empirical function is constructed to describe the correction, and an unbinned fit is performed to simulated data for a particular cluster size, type, and energy.

The function used for the empirical fit is of the form

$$f(v) = A\tan^{-1} Bv + Cv + D|v| + E, \tag{1}$$

where $-1/2 \leq v \leq 1/2$ across a cell (for the actual fit, the parameters are redefined to reduce correlations). To turn this into a function of $\eta$, the fit parameters are written as polynomials (usually of second or third degree) in $|\eta|$:

$$A = \sum_i a_i |\eta|^i, \tag{2}$$

and similarly for the other parameters. The fit parameters are then the coefficients $a_i$, $b_i$, etc.

One feature to note about this function is that, in general, $f(-1/2) \neq f(1/2)$, so that it will be discontinuous crossing a cell boundary. For layer 1, this is usually acceptable, since reconstructed cluster positions cluster well away from the cell boundary (Fig. 7(a)). However, in layer 2, the distribution of reconstructed cluster positions remains populated across the cluster boundary (Fig. 7(b)). Therefore, for layer 2, the function is modified so that $f(-1/2) = f(1/2)$.

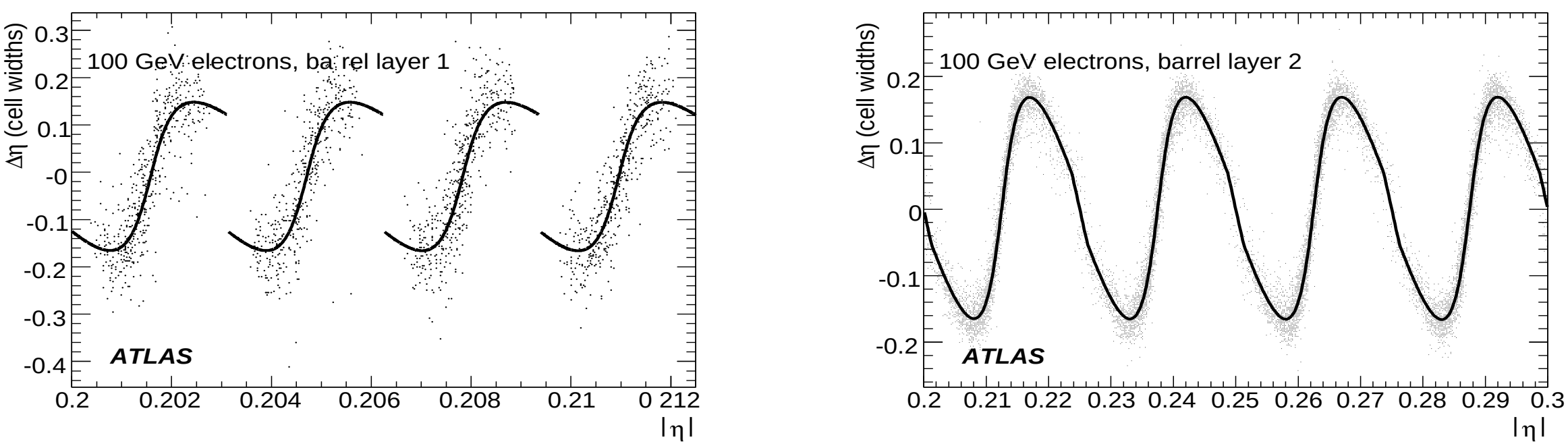

(a) Barrel layer 1

(b) Barrel layer 2

Figure 7: $\Delta\eta$ versus $|\eta|$ in layers 1 and 2 of the barrel, along with the empirical fit function.

In some cases, there is still a significant periodic residual after fitting to this form; in such cases, an additional general trigonometric term is added to the fit:

$$f'(v) = f(v) + \alpha \cos(\beta \pi v + \gamma). \tag{3}$$

Finally, a few regions near the calorimeter edges do not exhibit the S-shape form; a general polynomial is used as the empirical function there.

The correction is evaluated separately for each cluster size and type (electrons, photons). The difference in the correction between electrons and photons is a few percent, and there is about a 10% difference between $5 \times 5$ and $3 \times N$ clusters.

The correction also depends on energy; over the range 25–1000 GeV, the required correction varies by $\sim 20\%$. To apply the correction for a given cluster, the correction is first tabulated for each of the energies for which simulated data samples were available. The final correction is then found by doing a cubic polynomial interpolation within this table. Note a subtlety here: the energies at which the corrections are tabulated are the true cluster energies. However, when the correction is applied, only the reconstructed cluster energy is known. Since the position corrections are done before the energy corrections, the reconstructed cluster energy will be systematically lower than the true energy. If this were used for the interpolation, this would bias the position measurements. So, for the purpose of this interpolation, a crude energy correction is performed by scaling the reconstructed cluster energy by the ratio of the true to reconstructed energy observed in a 100 GeV sample, parametrised as a function of $|\eta|$. This energy correction is used only for the energy interpolation of the position corrections.

Plots of $\Delta\eta$ before and after corrections for several regions are shown in Fig. 6. Note that since the present corrections were derived, the simulated detector geometry was changed slightly in the end-cap, in order to match more closely the as-built detector. This results in a small systematic offset of $O(10^{-4})$ in these regions.

The $\eta$ position resolution for photons versus $|\eta|$ is shown for the two main calorimeter layers (strips and middle) in Fig. 8. The resolution is fairly uniform as function of $|\eta|$ and is $2.5$–$3.5 \times 10^{-4}$ for the strips (which have a size of 0.003 in $\eta$ in the barrel electromagnetic calorimeter) and $5$–$6 \times 10^{-4}$ for the middle-layer cells (which have a size of 0.025 in $\eta$). The regions with worse resolution correspond to the barrel/end-cap transition region and, for the strips, to the region with $|\eta| > 2$, where the strip granularity of the end-cap calorimeter becomes progressively much coarser.

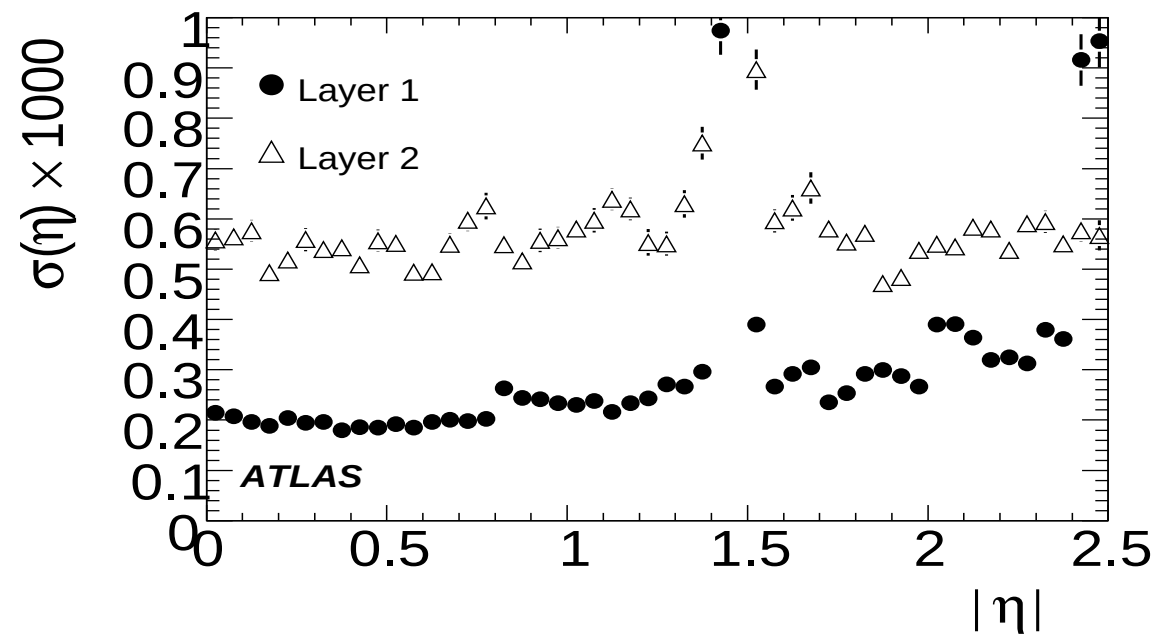

Figure 8: Expected $\eta$ position resolution versus $|\eta|$ for $E = 100$ GeV photons for the two main layers of the barrel and end-cap EM calorimeters [4].

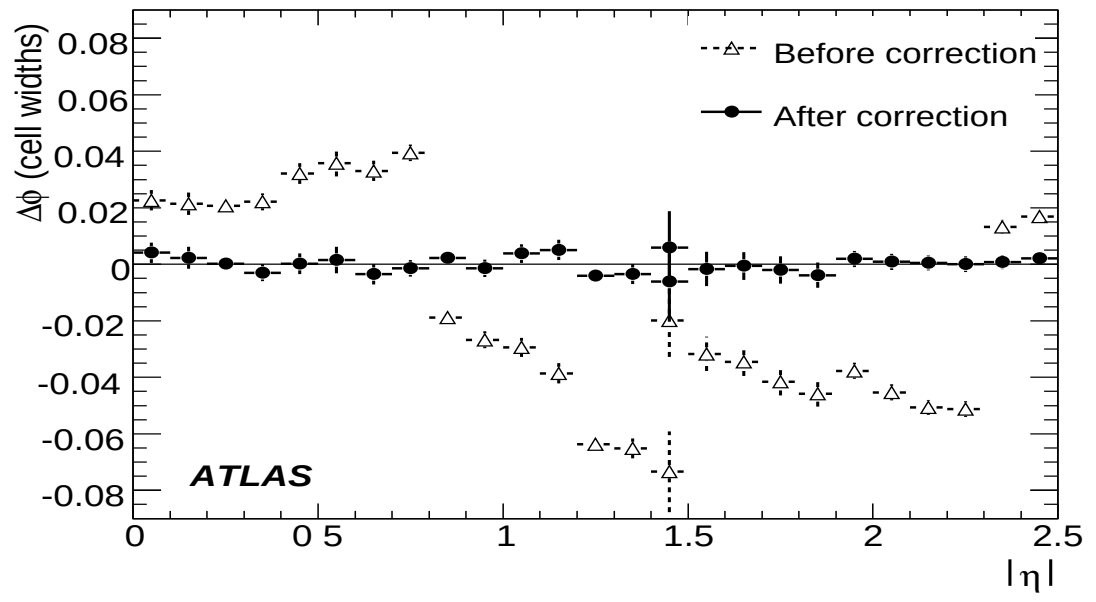

Figure 9: Profile plot of $\Delta\phi$ versus $|\eta|$ before (triangles) and after (circles) correction. For 100 GeV electrons.

## 2.2 $\phi$ position correction

The measurement of the cluster $\phi$ position must also be corrected. These corrections are applied only in calorimeter layer 2 (the $\phi$ granularity is best in this layer). As opposed to the $\eta$ direction, the accordion geometry results in more energy sharing between cells in the $\phi$ direction, which washes out the S-shape in this direction. There is, however, a small bias in the $\phi$ measurement which depends on the average shower depth with respect to the accordion structure (and thus on $|\eta|$). A profile plot of $\Delta\phi = \phi_{\text{true}} - \phi_{\text{reco}}$ before the correction is shown in Fig. 9. (The sign of the offset is flipped for $\eta < 0$, as the two halves of the calorimeter are identical under a rotation.) The discontinuity at $|\eta| = 0.8$, where the absorber thickness and the middle layer depth change, is clearly visible.

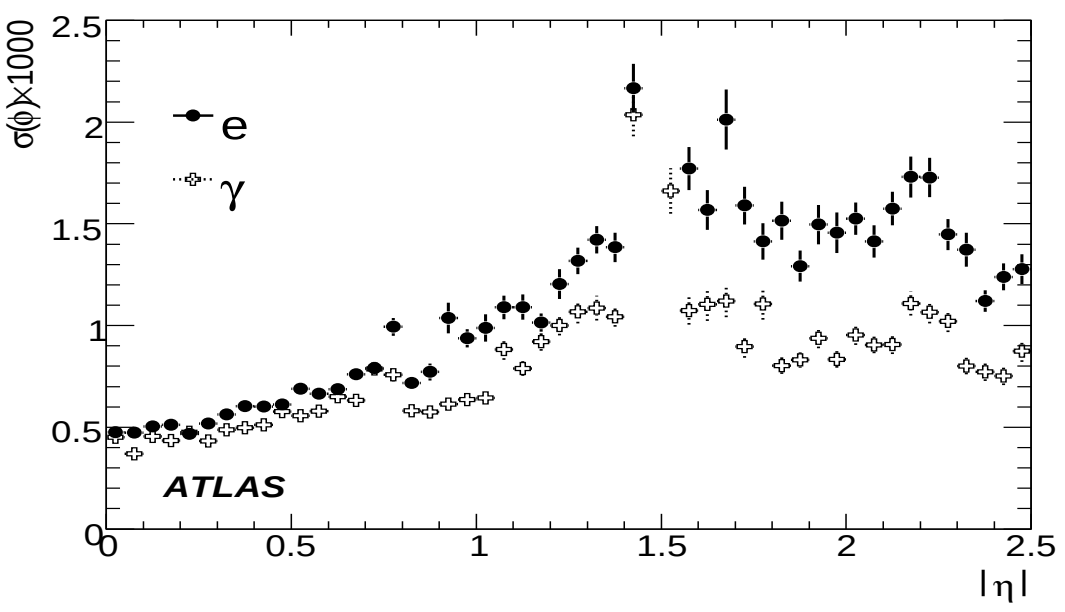

Figure 10: Expected $\phi$ position resolution as a function of $|\eta|$ for electrons and photons with an energy of 100 GeV.

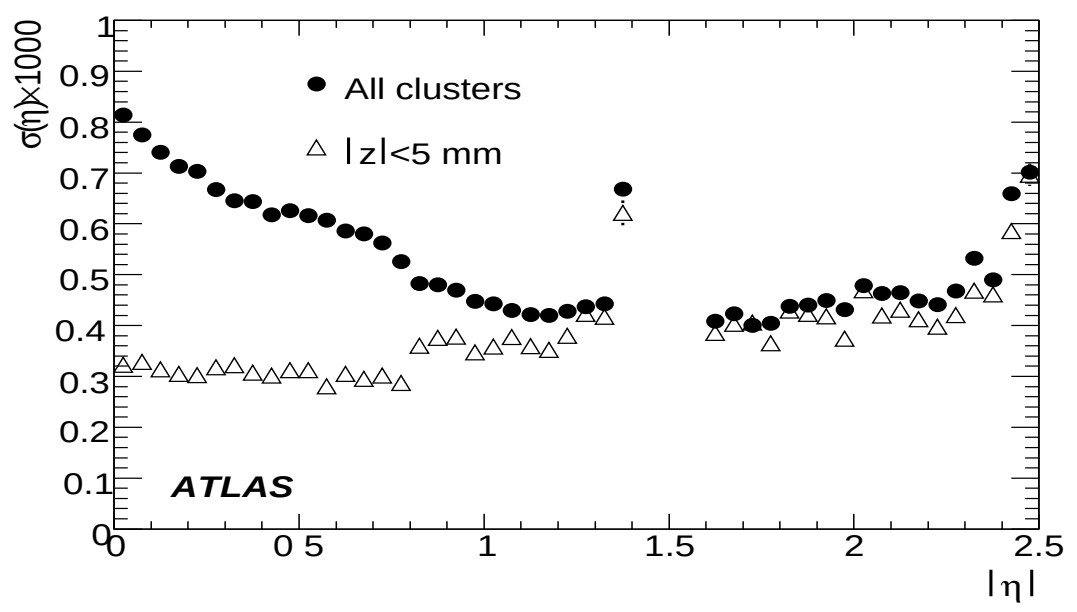

Figure 11: Resolution of $\eta$ position measurement from layers 1 and 2 combined for 100 GeV photons.

The correction derived here is symmetric in $\phi$. In the real detector, the absorbers sag slightly due to gravity, causing a $\phi$-dependent modulation in the $\phi$ offset with a maximum value of about 0.5 mrad [8]. This has not been included in the present simulations, and it is therefore not taken into account in this correction. Studies have shown, however, that the extra smearing of the position measurement from this effect has a negligible contribution to the widths of the invariant mass distributions of $e^+e^-$ pairs. (These studies were performed by generating decays of massive particles using a toy Monte Carlo, smearing the decay products with energy and angular resolutions roughly appropriate to ATLAS, and comparing the widths of the resulting invariant mass distributions before and after shifting the $\phi$ positions by $A\cos\phi$. No significant broadening was observed for $A < 50$ mrad.) The contribution of this effect to the constant term of the energy resolution has not been studied quantitatively, but should also be small.

To produce a correction, the data are binned in $\eta$. The result for one sample is shown in Fig. 9. This function is interpolated in $\eta$; it is then also interpolated in energy as for the $\eta$ position correction.

The $\phi$ position resolution versus $|\eta|$ is shown for calorimeter layer 2 in Fig. 10. Electron clusters, which get smeared in the $\phi$ direction as they radiate while propagating through the magnetic field, have a worse $\phi$ position resolution than do photon clusters. A discontinuous step is seen in the resolution at $|\eta| = 0.8$, where the absorber thickness changes, and the resolution is worst in the transition region between the cryostats.

### 2.3 Position measurement combination

The individual layer $\eta$ and $\phi$ measurements are combined to produce the overall $\eta$ and $\phi$ for a cluster. For $\phi$, only layer 2 is used, so the combination is trivial except in the overlap region, where the energy-weighted average of the barrel and end-cap $\phi$ measurements is used. For $\eta$, both layer 1 and layer 2 are averaged. However, layer 1 is weighted three times as much as layer 2 to roughly take into account the better resolution in layer 1. This prescription, which does not use the actual position resolutions and does not account for correlations, is known to be suboptimal and will be improved in future software versions.

Note that the $\eta$ combination implicitly assumes that the incoming particle is projective. If its production vertex is shifted from the origin, then the combined $\eta$ will be biased. This is illustrated in Fig. 11, which shows the resolution of the combined cluster $\eta$ measurement. Here, the measured cluster $\eta$ is compared to the $\eta$ position of the calorimeter intersected by the true particle track at a depth corresponding to the cluster barycenter. This is shown both for all clusters and for clusters with the $z$ position of the production vertex within 5 mm of the detector center.

### 2.4 Shower direction

At high luminosity, the inner detector cannot accurately determine the interaction vertex due to the large number of additional interactions. This is an issue for the reconstruction of a $H \rightarrow \gamma\gamma$ signal. For this analysis, achieving the best possible resolution on the invariant mass of the photon pair is crucial for separating the signal peak from the continuum background. If the $z$-position of the interaction vertex is unknown, then there will be a large uncertainty in the polar angle of the photons and thus in the pair invariant mass. We can, however, recover information about the incidence angle of the photons by comparing the impact points that are reconstructed in the first and second layers of the EM calorimeter. To do this, we need to know the photon $\eta$ position and the shower barycenter in each of the two layers (Fig. 5). We can then draw a straight line between these two $(\eta, \text{depth})$ points; extending this line to the beam axis gives an estimate of the position of the interaction vertex.

Here, this method is applied to single photons with energies compatible with photons from $H \rightarrow \gamma\gamma$ decays. For $m_H = 120$ GeV, these photons are predominantly in the range $50 - 100$ GeV. Figure 12 shows the resolutions of the photon angle and the interaction vertex measurements as functions of $|\eta|$. Figure 13 shows the same resolution as a function of the photon energy, for $|\eta| < 0.5$.

## 3 Cluster energy measurement

Most of the energy of an electromagnetically interacting particle is deposited in the sensitive volume of the calorimeter, including the lead absorbers and the liquid-argon gaps. A small fraction is deposited in non-instrumented material in the inner detector, the cryostats, the solenoid, and the cables between the presampler and the first EM calorimeter layer. Energy also escapes from the back of the calorimeter.

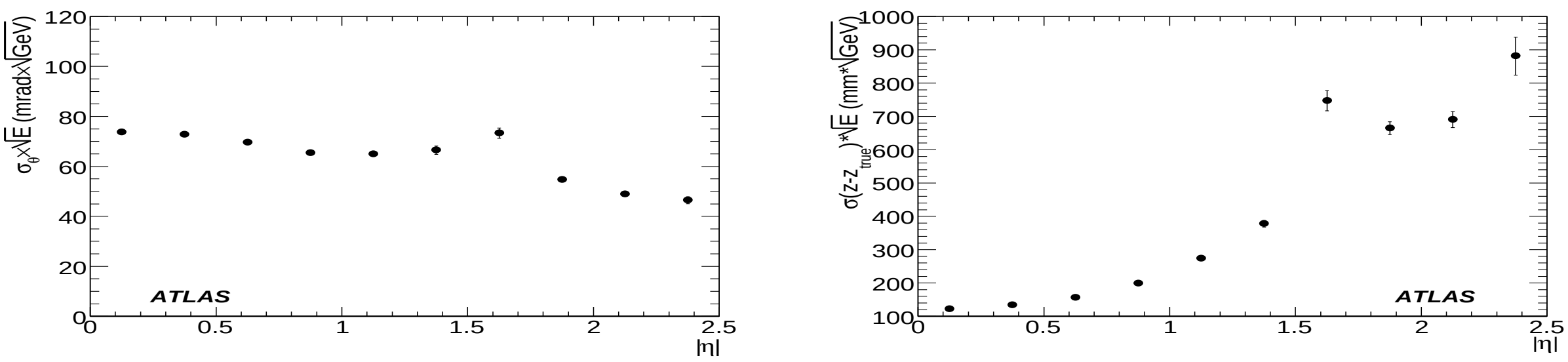

(a) Angular resolution.

(b) Vertex resolution.

Figure 12: Angular and vertex resolution as functions of $|\eta|$ (Gaussian fits), multiplied by $\sqrt{E}$.

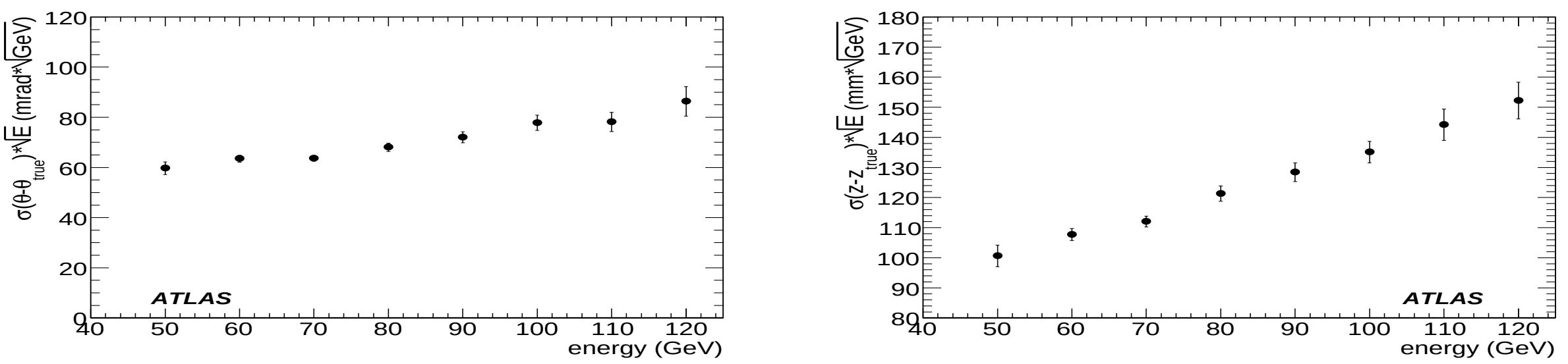

(a) Angular resolution.

(b) Vertex resolution.

Figure 13: Angular and vertex resolution as functions of $E$ (Gaussian fits), for $|\eta| < 0.5$.

The cluster energy is calculated as a linearly weighted sum of the energy in each of the three calorimeter layers plus the presampler. The factors applied to the four energies are called longitudinal weights and their purpose is to correct for the energy losses, providing optimum linearity and resolution.

The ATLAS longitudinal weighting method was first described in Ref. [8]. However, recent ATLAS test beam analyses [2, 3, 12] provided simple extensions of the technique. They also allowed validating this method with real data.

The first section below describes the weighting correction that is performed in current versions of the reconstruction, called the 4-weight method. This is followed by a description of the corrections for $\eta$- and $\phi$-dependent modulations in the energy. A more advanced energy-dependent calibration scheme, called the calibration hit method, is described separately in Section 5.

### 3.1 4-weight method

The weighting method described in this section is is a modification of that described in Ref. [8] and is currently the default in ATLAS reconstruction. The weights used are functions only of $|\eta|$; no energy dependencies are used. The method could be readily extended to include $\phi$- and energy-dependent weights in order to minimise residual non-linearities. The reconstructed energy is given by

$$E_{\text{reco}} = A(B + W_{\text{ps}} E_{\text{ps}} + E_1 + E_2 + W_3 E_3), \tag{4}$$

where $E_{\text{ps}}$ and $E_{1\ldots3}$ are the cluster energies in the presampler and the three layers of the calorimeter (including sampling fractions). The offset term $B$ corrects for upstream energy losses for which the corresponding electron has not reached the presampler (PS). In the limiting case of no energy in the PS, this offset corresponds to the energy an electron loses before it undergoes a hard bremsstrahlung for which the resulting photon passes through the PS without converting (i.e., no energy recorded in the PS). The parameters $A$, $B$, $W_{ps}$, and $W_3$ are calculated by a $\chi^2$ minimisation of $(E_{true} - E_{reco})^2/\sigma(E_{true})^2$ using Monte Carlo single particle samples, where $\sigma(E_{true})$ is a parametrisation of the expected energy resolution. This minimisation is done for separate $|\eta|$ bins, defined by the $\Delta\eta = 0.025$ granularity of the second layer of the calorimeter. Equal-sized samples with energies between 10 and 200 GeV are combined for the fits (the linearity of low energy points could be improved by using more events at those energies.) The fits are done separately for each cluster size and particle type (electron and photon).

A special parametrisation is applied in the gap region between the barrel and end-cap calorimeters ($1.447 < |\eta| < 1.55$), within which the parametrisation of Eq. (4) is not adequate. Moreover, this region is instrumented with scintillator tiles that can be used to recover some of the energy lost in the gap. The parametrisation used in the crack is

$$E_{reco} = A(B + E_b + E_e + W_{scint}E_{scint}), \tag{5}$$

where $E_b$ and $E_e$ are the energies the cluster deposits in the barrel and end-cap calorimeters, respectively. $E_{scint}$ is the scintillator energy, and $W_{scint}$ the weight applied to it. This parametrisation is found to perform significantly better than that used in [1].

The longitudinal weights in Eq. (4) were extracted for electrons and photons and are shown as a function of $|\eta|$ in Fig. 14. In Fig. 14(a) one can see that the overall scale $A$ for electrons (solid) is larger than that for photons. The reason is due to the fact that photons travel on average $9/7X_0$ before they start losing energy. This effect is close to 1% in the middle of the barrel and increases with the increase of upstream material. The offset term $B$ is shown in Fig. 14(b); photons have a very small offset, as expected. (Future versions of the correction will use larger statistics to reduce the scatter observed in the fit results.) The PS weight $W_{ps}$ shown in Fig. 14(c) is the usual factor applied to preshower/presampler energy responses to correct for upstream losses. Finally, in Fig. 14(d), $W_3$ is a weight applied to the last calorimeter layer to correct for energy leakage behind the calorimeter.

Detailed studies have revealed that the physical meaning attributed to these weights is only approximate. For example, the weights compensate for losses after the PS via the minimisation procedure. In addition, the weights have a non-negligible energy dependence. However, this energy dependence does not result in large non-linearities because the weights adjust their values to compensate. These effects are more evident at low energies $E < 15$ GeV, and with large amounts of upstream material. A more rigorous treatment of the longitudinal weighting is presented in Sec. 5.

The performance of this method is shown in Sec. 4

## 3.2 Cluster energy modulation corrections

As the $\phi$ impact position of a particle shifts across the accordion structure of the absorbers, the amount of passive absorber material it encounters and thus the ratio $R \equiv E_{reco}/E_{true}$ varies slightly, with a periodicity equal to that of the absorber spacing. This effect is small, with a maximum value of about a half-percent. Further, at lower energies, the $\phi$ position resolution becomes comparable to the absorber spacing; this contributes to washing out the effect at these energies. The reconstructed energy is corrected for this.

To derive the correction, the calorimeter is binned in $|\eta|$. The binning used is not uniform, but is chosen so as to segregate regions of the calorimeter with non-uniform $R$. Within each $|\eta|$ bin, $R$ is plotted versus the $\phi$ offset of the cluster relative to the absorber. These plots are divided into $\phi$ bins, each bin is fit to a Gaussian, and the means of the fits are plotted. The resulting plot is then normalised to unity and fit to a two-term Fourier series:

$$f(\phi) = 1 + A\left[\alpha\cos(N\phi + C) + (1-\alpha)\cos(2N\phi + D)\right], \tag{6}$$

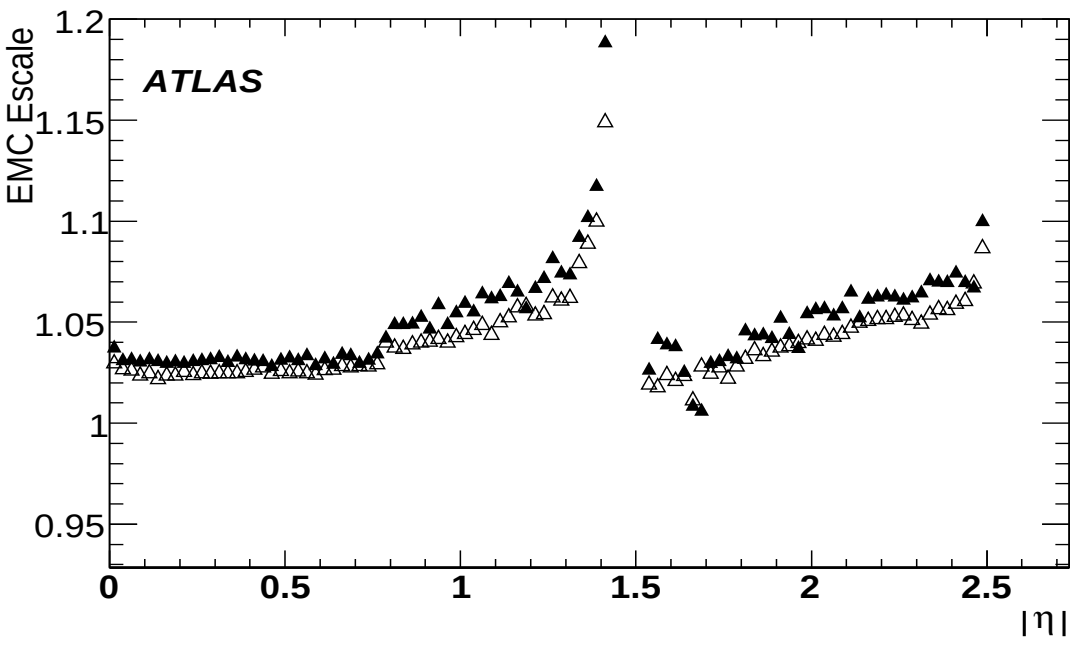

(a) Overall scale $A$.

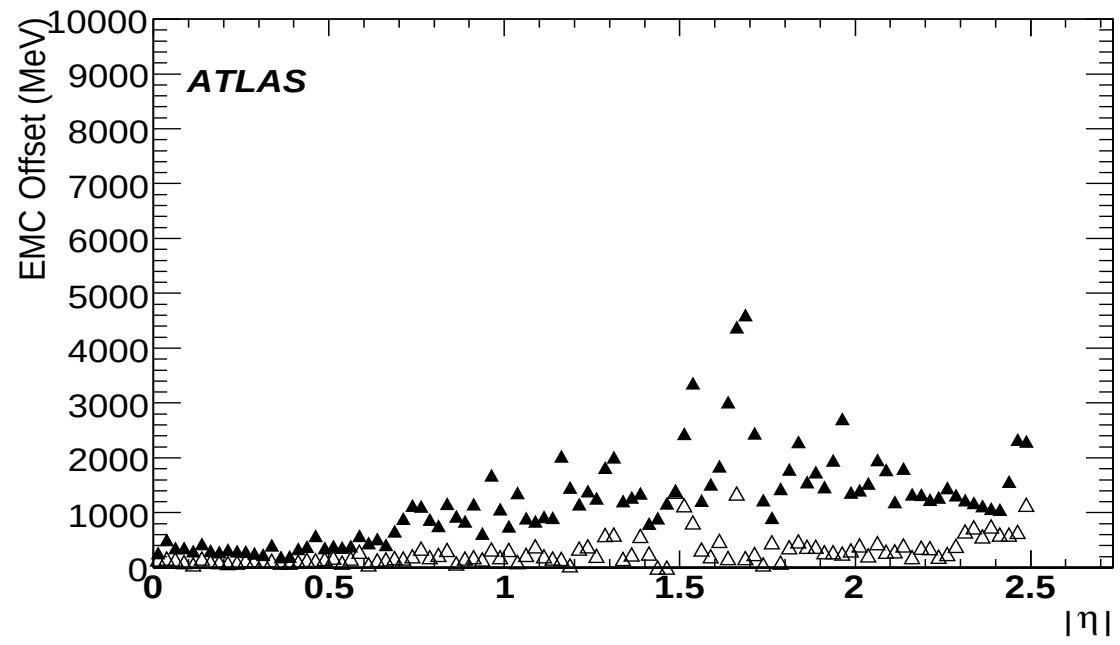

(b) Offset $B$.

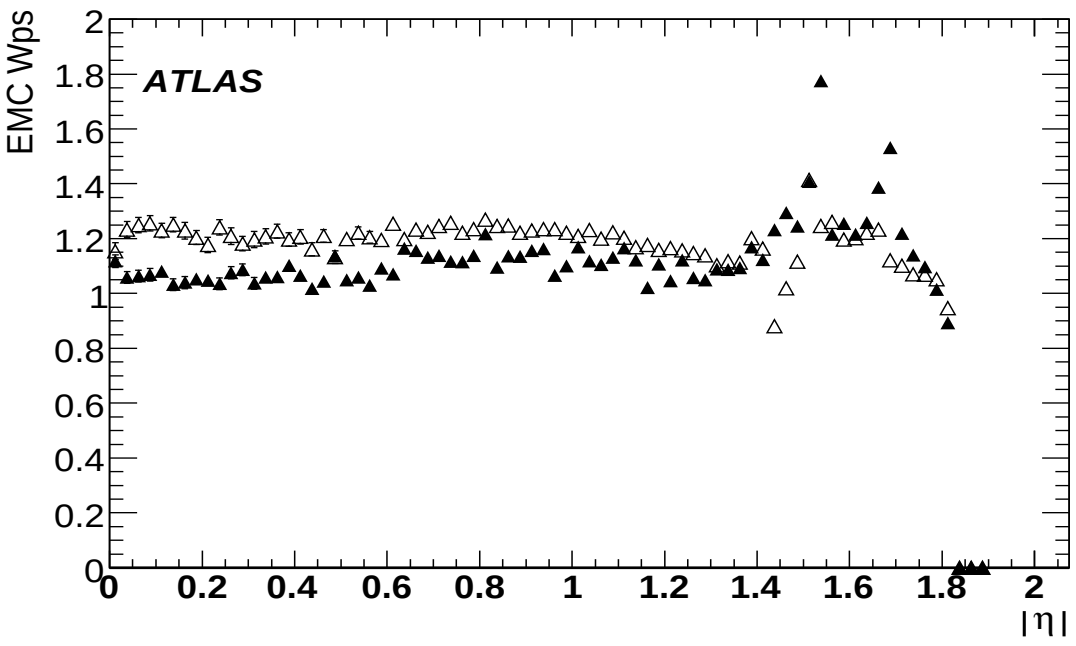

(c) Longitudinal weight $W_{ps}$.

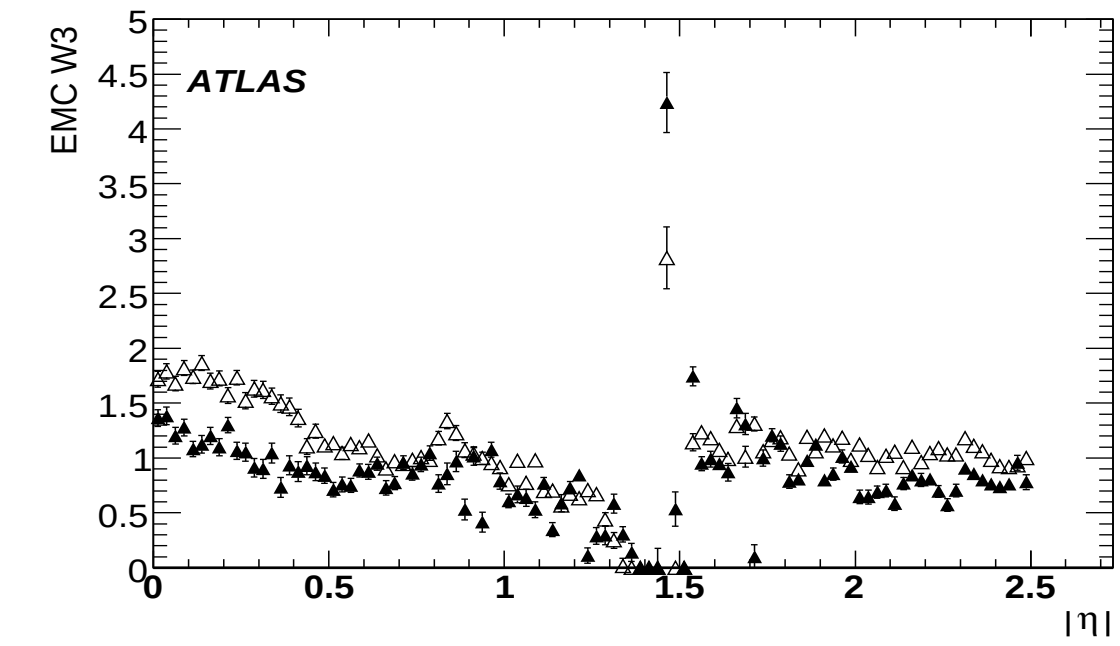

(d) Longitudinal weight $W_3$.

Figure 14: Fitted longitudinal weights for electrons (solid) and photons (open) as functions of $|\eta|$.

for fit parameters $A$, $\alpha$, $C$, and $D$. Parameter $\alpha$ is restricted to the range 0–1. $N$ is the total number of absorbers in $2\pi$ (1024 in the barrel and 768 in the end-capend-cap). An example of such a fit is shown in Fig. 15.

Fits are performed separately for each energy, cluster size, and particle type. To apply the correction, it is calculated for each $\eta$ and energy bin. It is then interpolated both in $\eta$ and in energy. This correction reduces the constant term in the energy resolution (for photons at $|\eta| = 0.3$) from 0.61% to 0.50%.

Energy modulations are also observed along the $\eta$ direction. The energy of a cluster is defined as the energy within a rectangular window of fixed size in $\eta \times \phi$. The window can only shift by an integral number of cells; however, the impact point of a particle may be anywhere within a cell. Thus, on average, a larger fraction of the cluster energy will be contained in the window when the particle hits at the center of a cell than if it hits near an edge. The size of this effect is a few tenths of a percent, and is larger for smaller cluster sizes. The modulation can be fit well with a quadratic; see Fig. 16. Note that this modulation is very small, $< 0.1\%$, in the $\phi$ direction, due to increased energy sharing between the cells; this modulation is not presently corrected. (A larger modulation was seen in the test beam [13], which used $3 \times 3$ clusters.)

The plots to fit are prepared in a similar manner as for the $\phi$ modulations, except that the $x$-axis is taken to be the $\eta$ offset within a cell. The plots from all bins where the detector is mostly uniform are then combined into a single plot; that is, the $|\eta|$ ranges 0.05–0.75, 0.85–1.30, and 1.70–2.50. The resulting plot is then scaled so as to average to unity and fit to a quadratic. The correction is performed separately for each energy, cluster size, and particle type. The final correction is then determined by interpolating in energy. An example fit is shown in Fig. 16. Applying this correction further reduces the constant term to 0.43%. A major contribution to the remaining constant term is from the $\phi$-dependency of the inner detector material distribution. (The present weighting correction is averaged over $\phi$.)

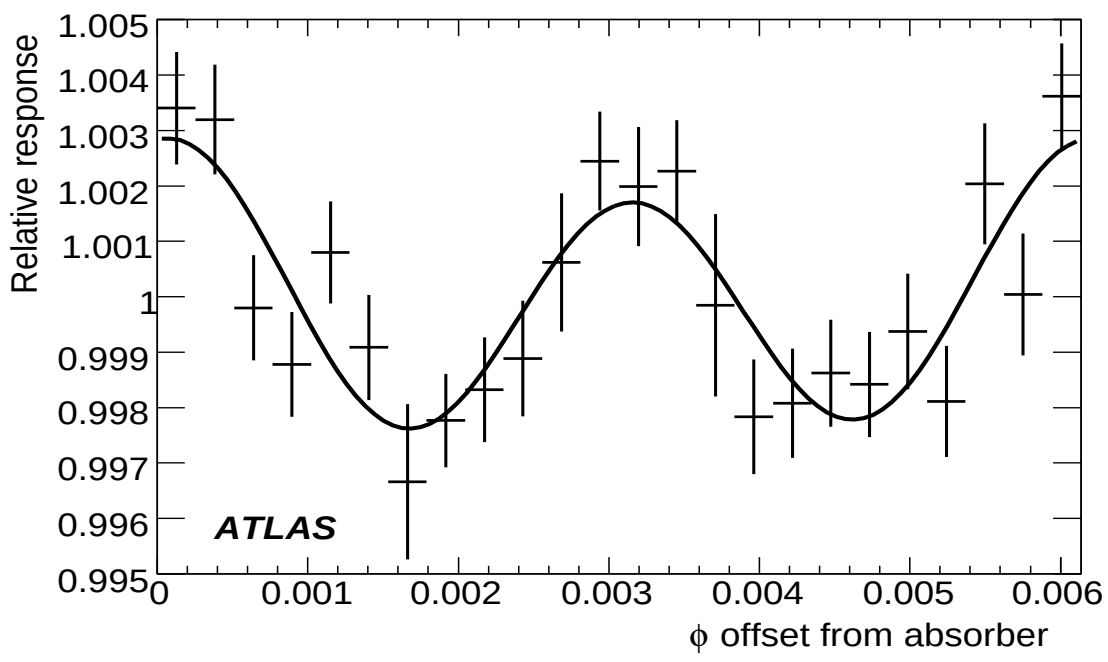

Figure 15: Energy modulation in $\phi$ for 200 GeV $3\times7$ electrons with $0.2 < |\eta| < 0.4$, along with the modulation fit.

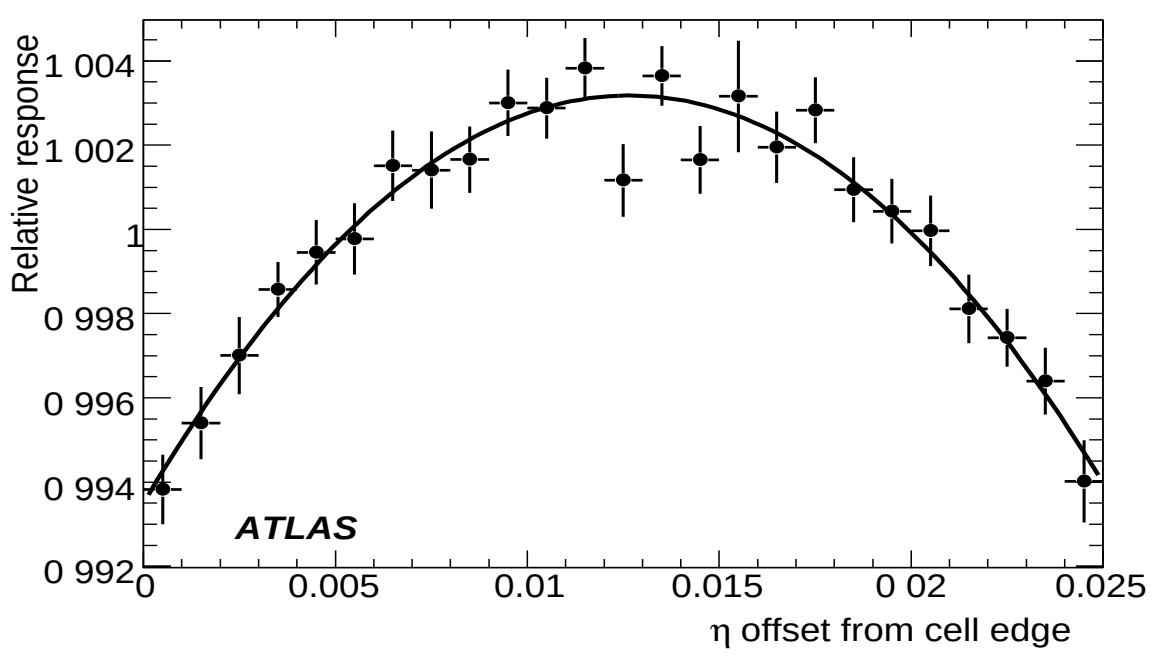

Figure 16: Energy modulation in $\eta$ for 200 GeV $3\times7$ electrons, along with the modulation fit [4].

# 4 Energy calibration performance

This section shows the performance of the calibration chain used in the current version of the ATLAS reconstruction software used for all of the electron and photon reconstruction and identification studies reported here and elsewhere.

## 4.1 Single electrons and photons

In Fig. 17, the energy response, plotted as the difference between measured and true energy divided by the true energy, is shown for electrons with an energy of 100 GeV for two illustrative $\eta$-positions in the barrel electromagnetic calorimeter. The central value of the energy is reconstructed with excellent precision ($\sim 3\times10^{-4}$) if one assumes perfect knowledge of the material in front of the calorimeter. Both the Gaussian core and the non-Gaussian component of the tail of the energy distribution are significantly worse at the point with larger $|\eta|$ due to the larger amount of material in front of the calorimeter. The resolution and non-Gaussian tails are better for photons than for electrons, but are somewhat worse for all photons than for photons that do not convert before leaving the volume of the inner detector.

The linearity (relative difference between the fitted mean energy and the true energy) and resolution are shown in Fig. 18 for electrons and photons. The expected performance is very similar for electrons and photons, with a somewhat larger degradation at larger values of $|\eta|$ in the case of electrons, as expected from the impact of upstream material. For electrons, the linearity is shown for $|\eta| = 0.3$ (barrel) and $|\eta| = 2.0$ (end-cap). The deterioration of the performance seen in the end-cap is attributed to the absence of a presampler ($|\eta| > 1.8$) and the relatively limited statistics of the simulated samples. The resolution shown in Fig. 18(b) is given for three $|\eta|$ points: $|\eta| = 0.3$ (inner barrel), $|\eta| = 1.1$ (outer barrel), and $|\eta| = 2.0$ (end-cap). The resolution drop at larger $|\eta|$ is attributed to the significant increase of upstream material in front of the calorimeter with respect to the small $|\eta|$ region. The extra material causes increased early showering upstream of the calorimeter, which affects the lateral shower shape in the calorimeter. Since Eq. (4) absorbs the corrections for lateral losses into the overall scale constant $A$, an increase in lateral-loss fluctuations will result in a deterioration of the resolution. The fits in Fig. 18(b) give a sampling term of $(10.17\pm0.33)\%$ at small $|\eta|$, and $(14.5\pm1.0)\%$ in the end-cap.

In Fig. 19, the energy resolution for electrons and photons is shown as a function of $|\eta|$. The photon resolution is better than the electron resolution in regions with more material in front of the calorimeter. The extracted constant term of the resolution is shown for photons in Fig. 20 after the weight and modulation corrections. This figure also shows the constant term observed when the standard simulation of cell-level miscalibrations is enabled in the reconstruction program. In Fig. 21, the linearity and resolution

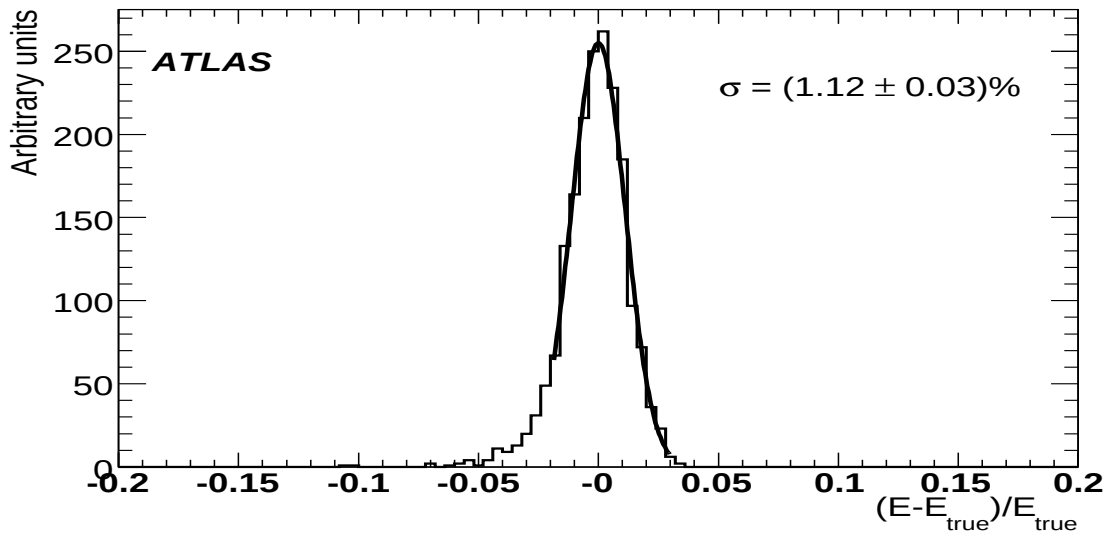

(a) Electrons, $|\eta| = 0.325$.

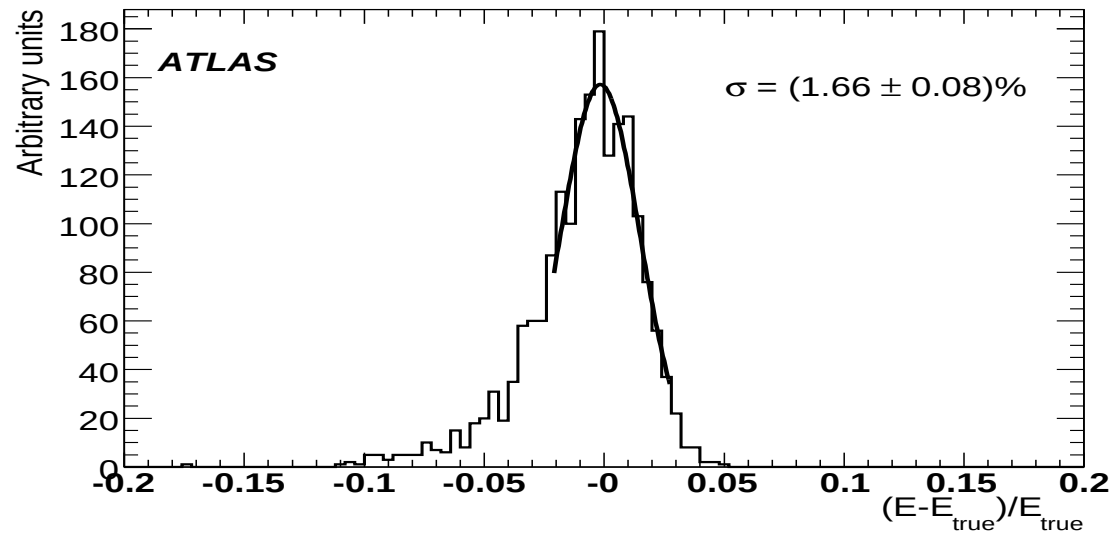

(b) Electrons, $|\eta| = 1.075$.

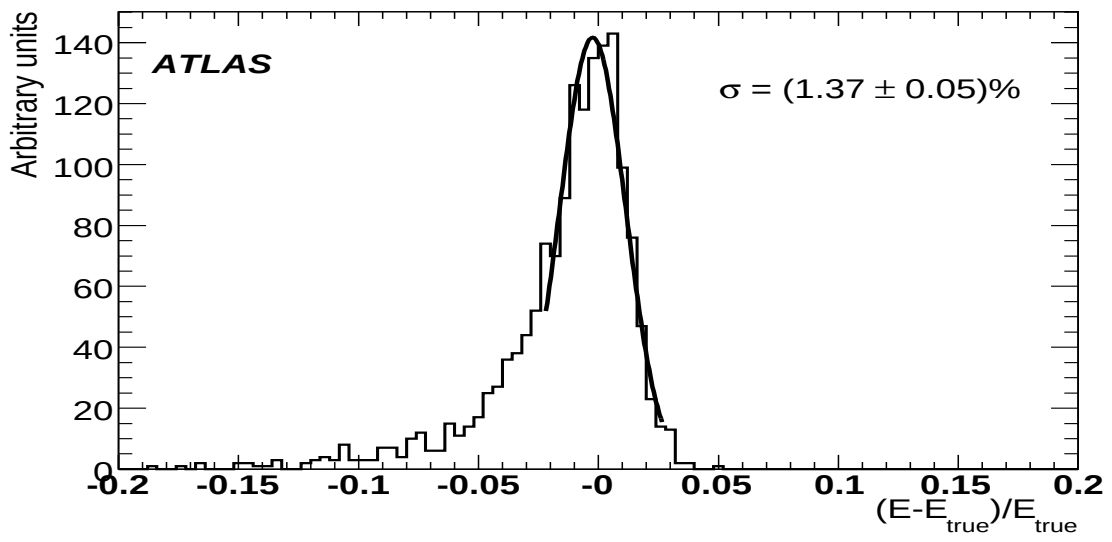

(c) All photons, $|\eta| = 1.075$.

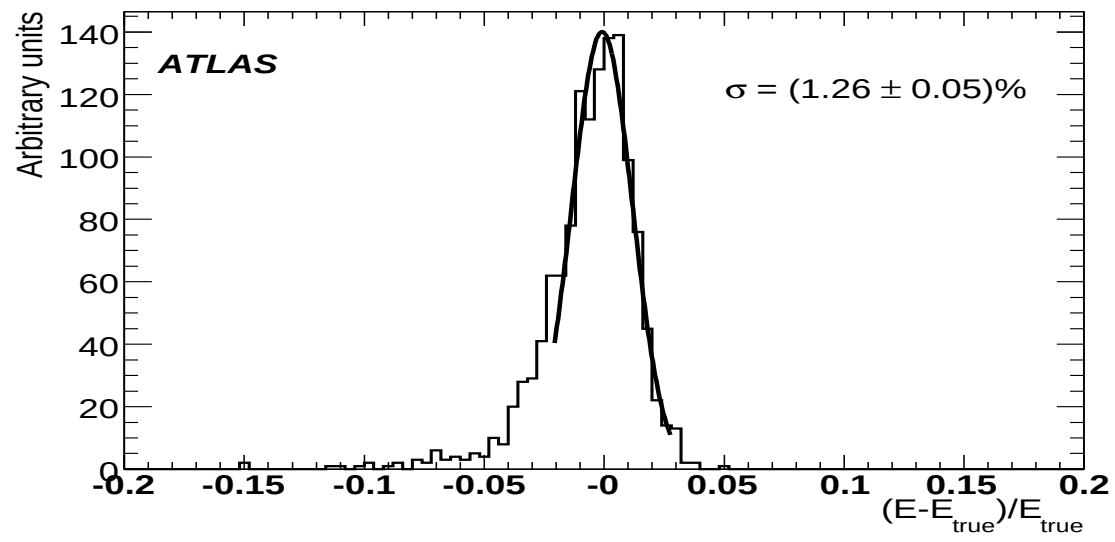

(d) Unconverted photons, $|\eta| = 1.075$.

Figure 17: Difference between measured and true energy normalised to true energy at $E = 100$ GeV.

as a function of $|\eta|$ is shown for a range of energies for single photons.

## 4.2 Mass resolution obtained in $H \to 4e$ and $H \to \gamma\gamma$ final states

Figure 22 shows the reconstructed distribution, after calibration, of the invariant mass of the electrons in $H \to 4e$ decays, with $m_H = 130$ GeV. (Loose electron selection applied, as defined in [6].) A global constant term of 0.7% has been included in the electromagnetic calorimeter resolution for the two plots in this subsection. The central value of the reconstructed invariant mass is correct to $\sim 1$ GeV, corresponding to a precision of 0.7%, and the expected Gaussian resolution is $\sim 1.5\%$. The non-Gaussian tails in the distribution amount to 20% of events lying further than $2\sigma$ away from the peak. They are mostly due to bremsstrahlung, particularly in the innermost layers of the inner detector, but also to radiative decays and to electrons poorly measured in the barrel/end-cap transition region of the electromagnetic calorimeter.

Figure 23 shows the reconstructed photon pair invariant mass for $H \to \gamma\gamma$ decays with $m_H = 120$ GeV (tight photon selection applied and barrel/end-cap transition region excluded). The photon directions are derived from a combination of the direction measurement in the electromagnetic calorimeter described above (see Section 2.4) with the primary vertex information from the inner detector. The central value of the reconstructed invariant mass is correct to $\sim 0.2$ GeV, corresponding to a precision of 0.2%, and the expected resolution is $\sim 1.2\%$. Most of the non-Gaussian tails at low values of the reconstructed photon pair mass are seen to be due to photons which convert in the inner detector. The shift in the means comes from the fact that the corrections to-date do not distinguish between converted and unconverted photons.

## 4.3 Study of systematic effects using $H \to 4e$

The energy linearity for electrons in $H \to 4e$ is shown in Fig. 24(a) for samples based on the ideal (full triangles) and distorted (circles) geometries. The departure from linearity for the distorted geometry is attributed to the presence of extra material in front of the calorimeter. The corresponding resolution is shown in Fig. 24(b) for the distorted geometry.

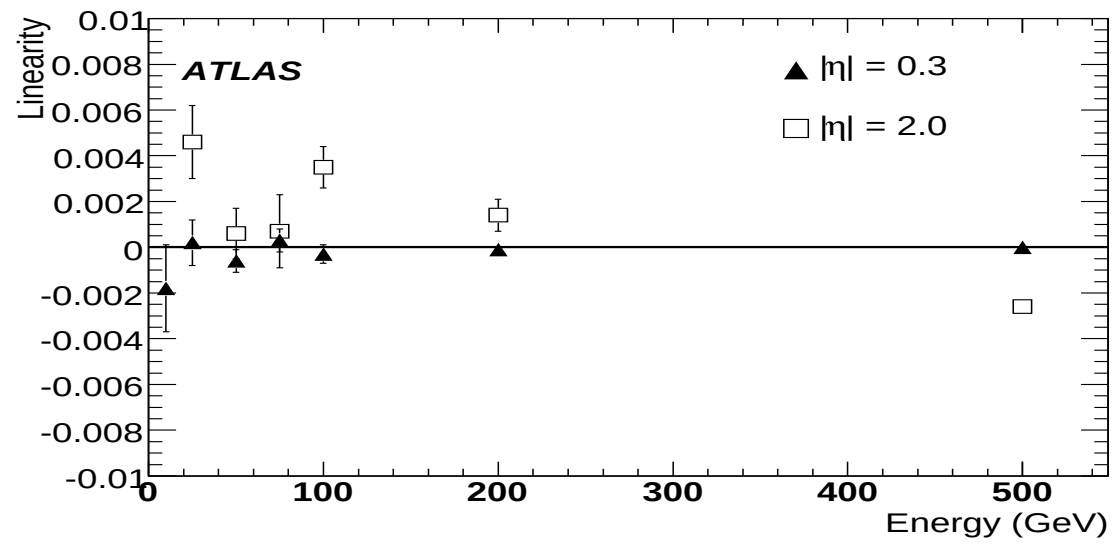

(a) Electron energy linearity.

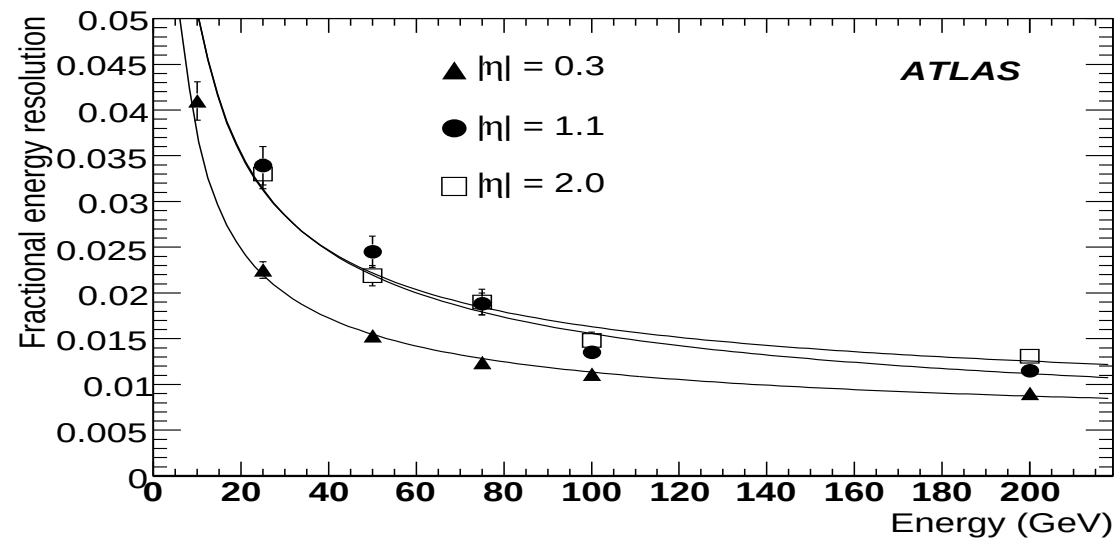

(b) Electron energy resolution.

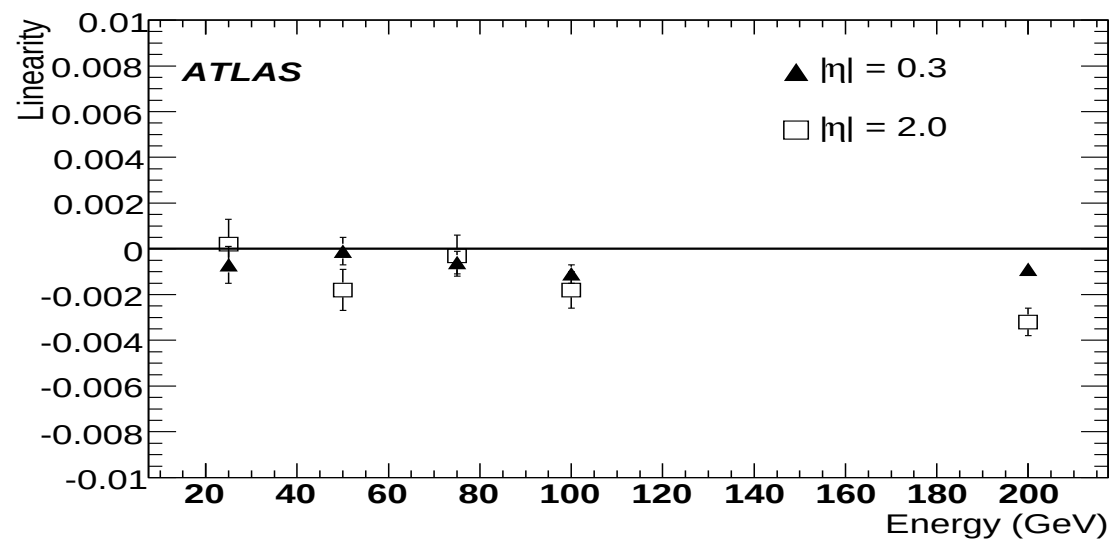

(c) Photon energy linearity.

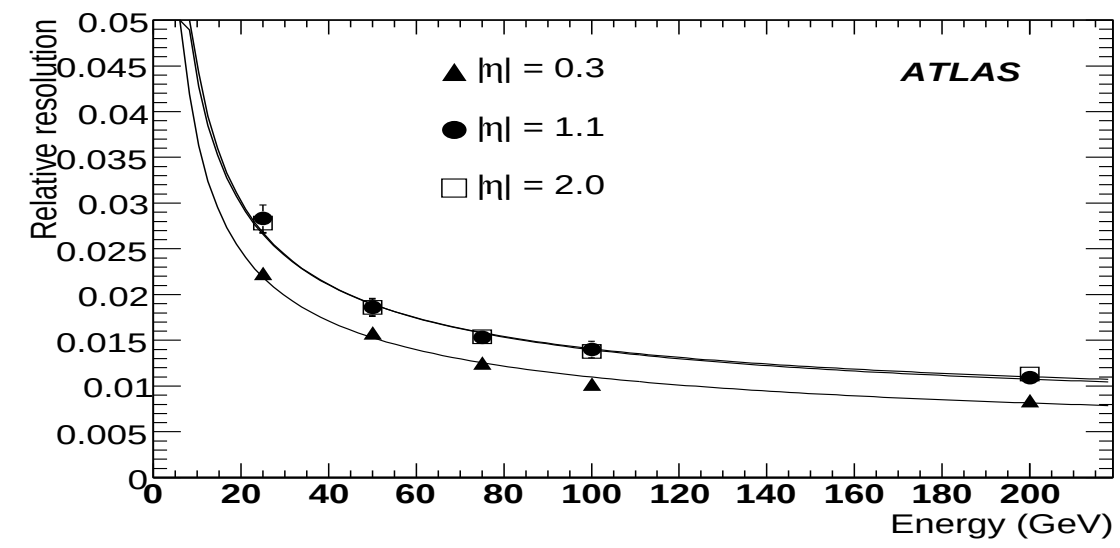

(d) Photon energy resolution.

Figure 18: Energy linearity (left) and resolution (right) for electrons (top) and photons (bottom).

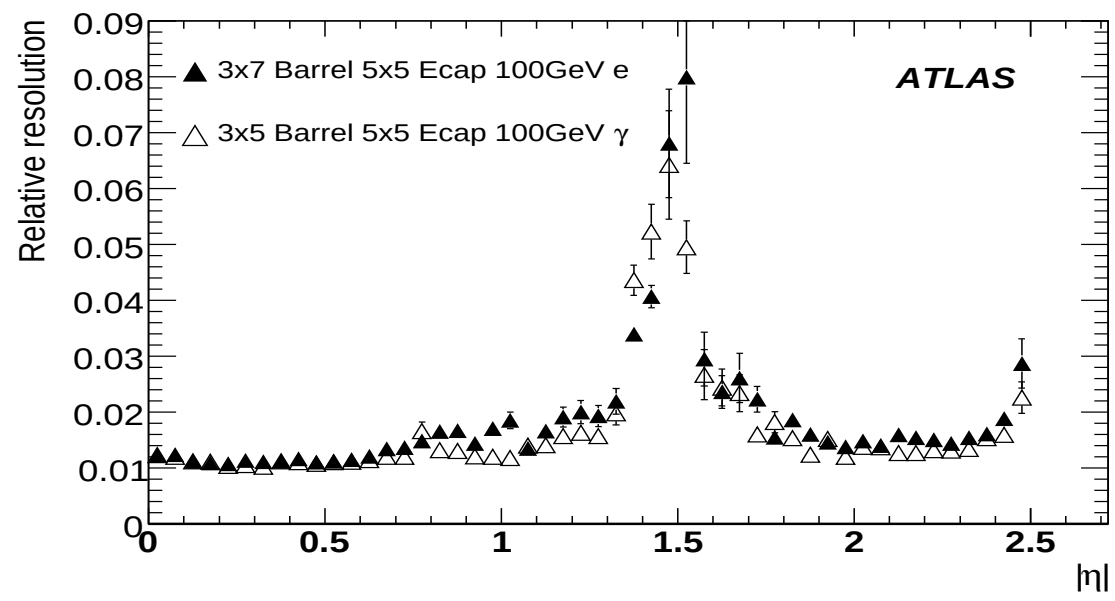

Figure 19: Energy resolution for electrons and photons as a function of $|\eta|$.

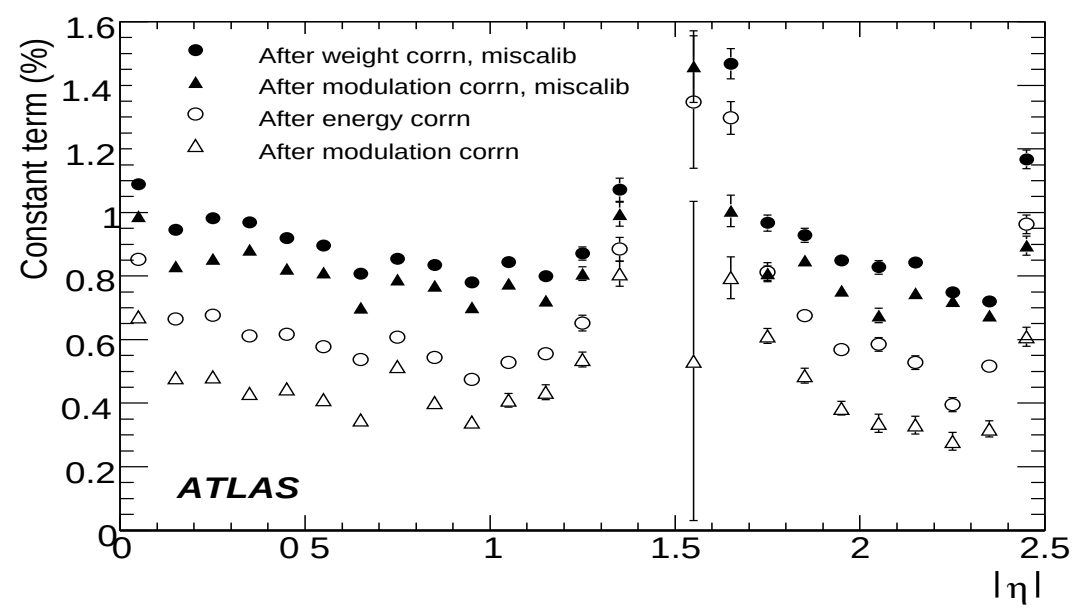

Figure 20: Extracted constant term of the energy resolution for photons, as a function of $|\eta|$, after weight and modulation corrections. Also shown with cell-level miscalibrations enabled.

The uniformity in $\phi$ and $\eta$ observed in this sample is shown in Fig. 25. The non-uniformities seen at higher $|\eta|$ and at positive $\phi$ are due to simulated extra material in these regions. In the $\phi$-uniformity plot (Fig. 25(a)) a residual modulation is observed. This is most likely due to an artefact in the simulation. The longitudinal weights used in the reconstruction depend only on $\eta$, and are averaged over $\phi$. Adding a dependency on $\phi$ as well would make the energy scale along $\phi$ more uniform and also improve the mass resolution of $Z \to ee$.

# 5 Energy correction using calibration hits

This section describes an alternate method for calculating the total energy from the energies in the individual calorimeter layers and the presampler. It is a development of ideas introduced in [14, 15] to analyse test beam data and is described in some detail in [16]. Special simulations are used in which the energy deposited by a particle is recorded in all detector materials, not just the active ones. Through these simulations, the energy depositions in the inactive material can be correlated with the measured quantities. For example, the energy lost in the material in front of the calorimeter (inner detector, cryostat, etc.) can be estimated from the energy deposited in the presampler. The result is a method which provides a modular way to reconstruct the energies of electrons and photons by decoupling all the different corrections. This approach eases comparisons between electrons and photons, and might be particularly useful in the initial stages of the experiment.

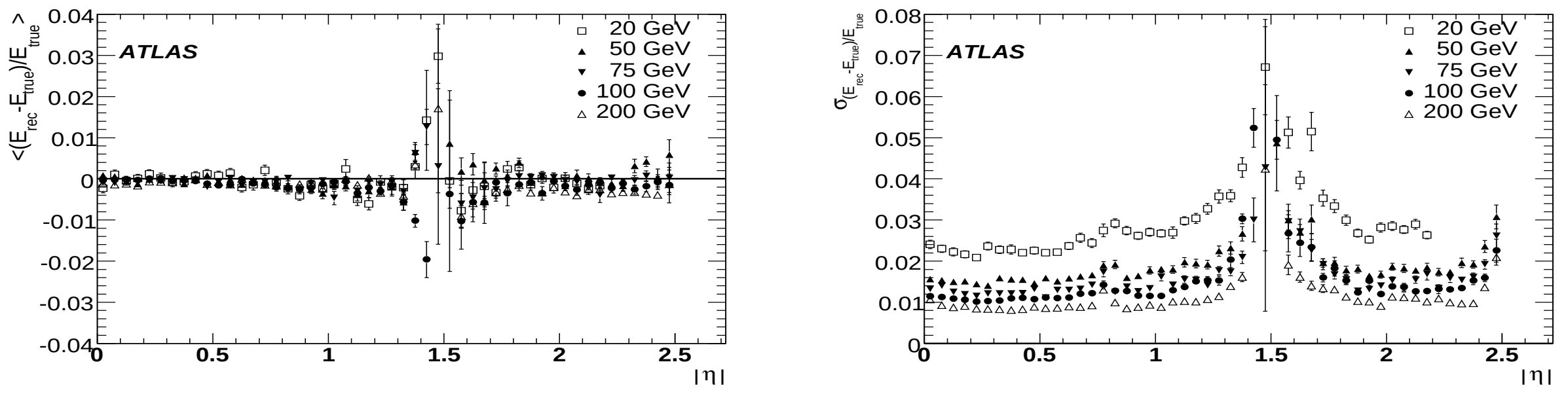

(a) Energy linearity.

(b) Energy resolution.

Figure 21: Energy linearity and resolution for photons ($5 \times 5$ clusters).

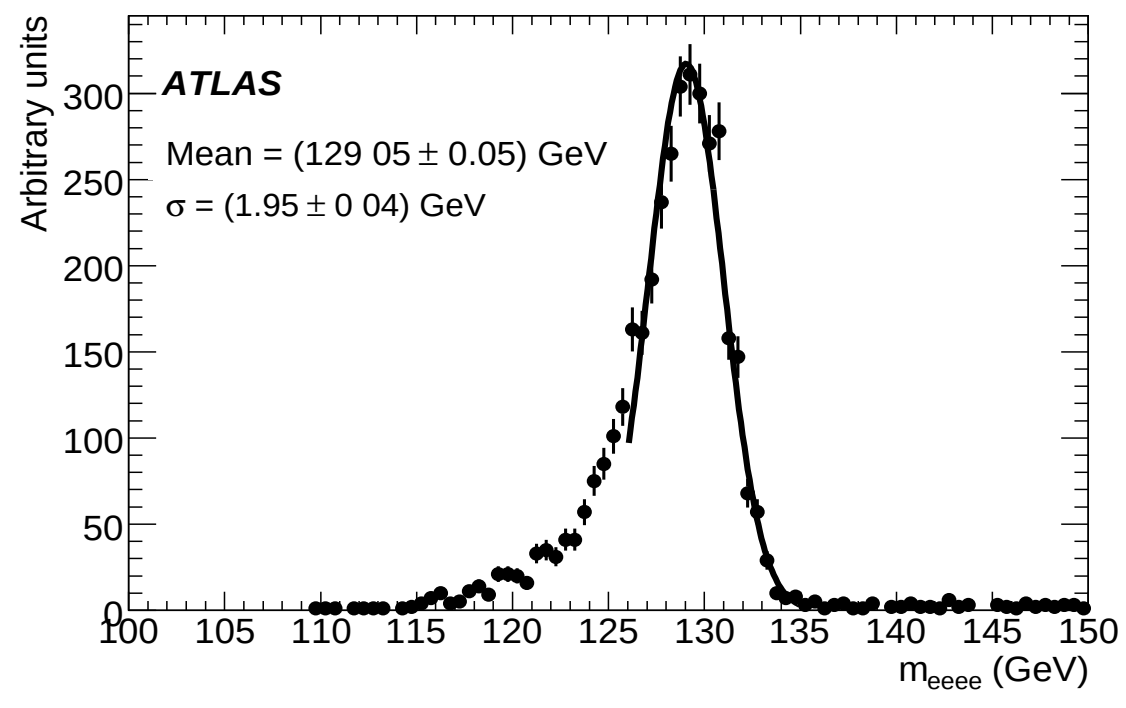

Figure 22: $M(eeee)$ from Higgs boson decays with $m_H = 130$ GeV (energy from calorimeter only, with no $Z$ boson mass constraint).

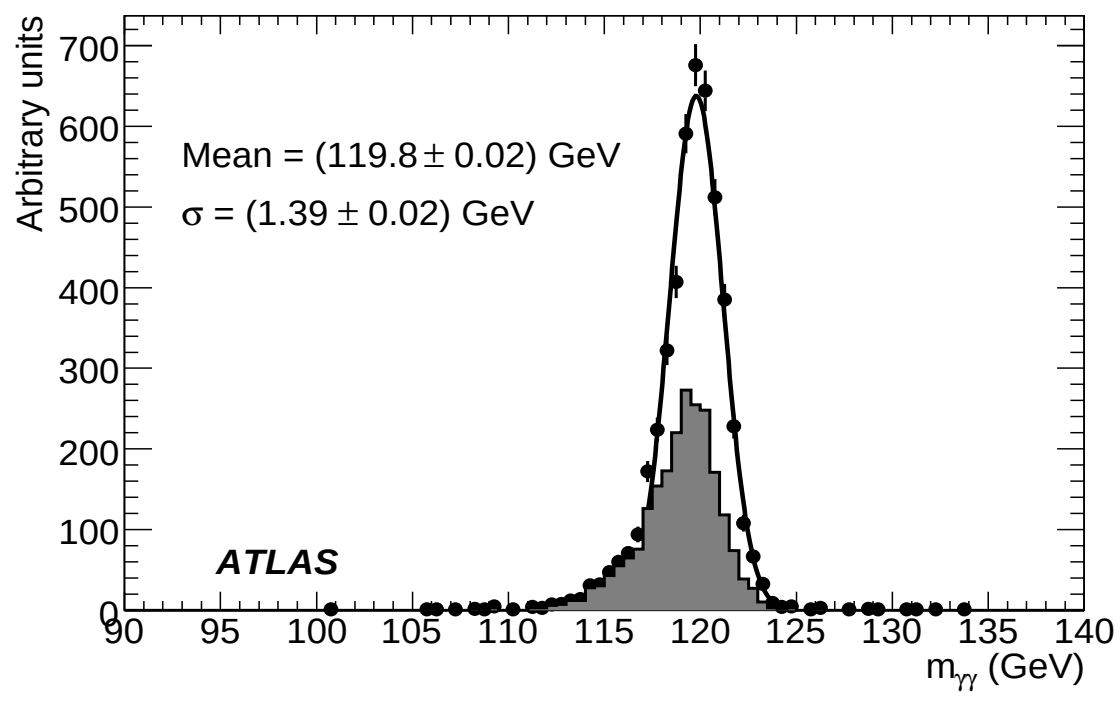

Figure 23: $M(\gamma\gamma)$ from Higgs boson decays with $m_H = 120$ GeV. The shaded plot corresponds to at least one photon converting at $r < 80$ cm.

The cluster energy is decomposed into three pieces, which will be treated separately below:

$$E = E_{\text{cal}} + E_{\text{front}} + E_{\text{back}}, \tag{7}$$

where $E_{\text{cal}}$ is the energy deposited in the electromagnetic calorimeter, $E_{\text{front}}$ is the energy deposited in the presampler and in the inactive material in front of the calorimeter, and $E_{\text{back}}$ is the energy that leaks out the rear of the EM calorimeter.

This analysis uses simulated single-particle, mono-energetic electron and photon samples, with energies ranging from 25 to 500 GeV.

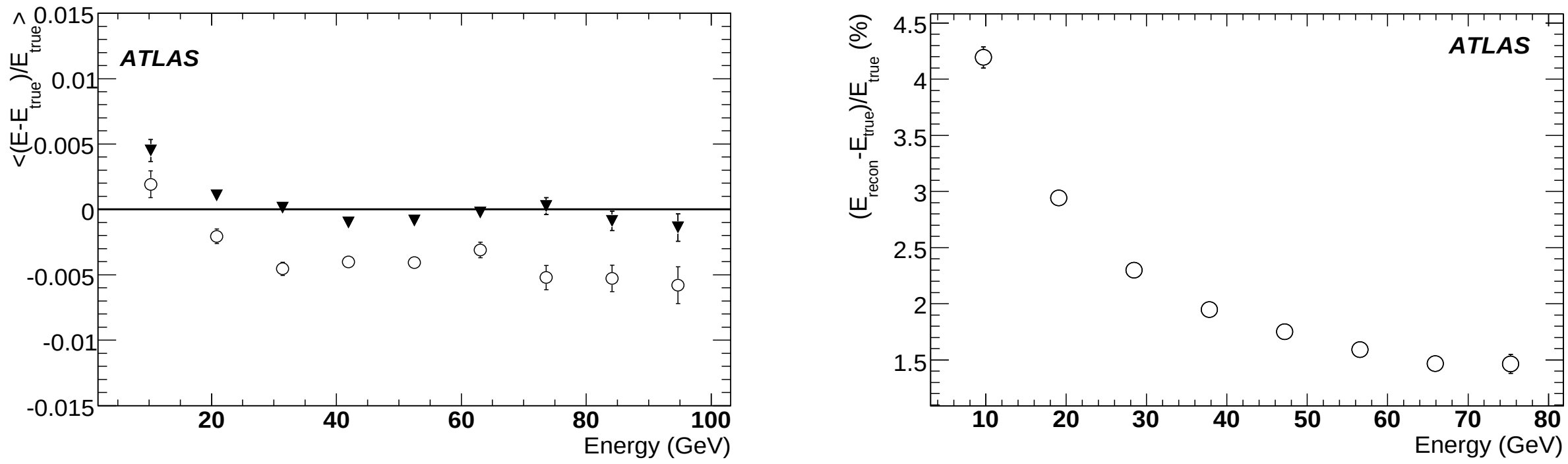

(a) Energy linearity.

(b) Energy resolution.

Figure 24: Electron linearity and resolution in $H \rightarrow 4e$ for the ideal (full triangles) and distorted (circles) geometries.

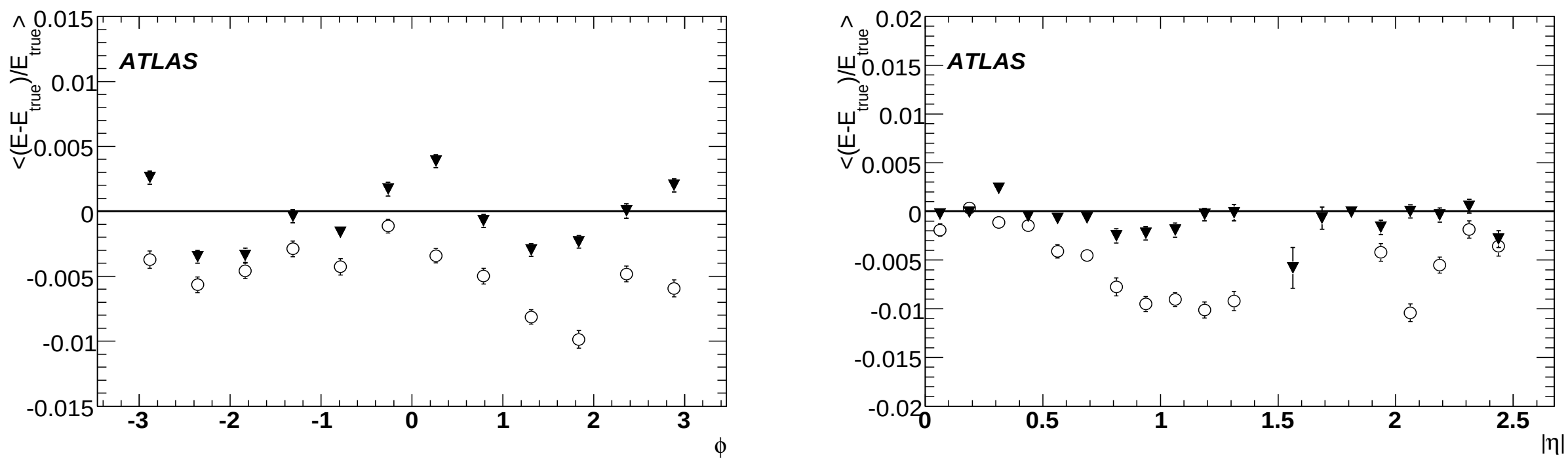

(a) Energy uniformity in $\phi$ integrated over $p_T$ and $\eta$.

(b) Energy uniformity in $\eta$ integrated over $p_T$ and $\phi$.

Figure 25: Electron energy uniformity in $\eta$ and $\phi$, integrated over other kinematic variables, for the ideal (full triangles) and distorted (circles) geometries.

## 5.1 Reconstruction of the energy deposited in the calorimeter

The energy deposited by a particle in the EM calorimeter, $E_{\text{cal}}$, is estimated as

$$E_{\text{cal}} = C_{\text{cal}}(X,\eta)(1+f_{\text{out}}(X,\eta))E_{\text{cl}}, \tag{8}$$

where

- $E_{\text{cl}} = \sum_{i=1}^{3} E_i$, and $E_{1...3}$ are the energies deposited in each of the three calorimeter layers in a given cluster. In the following, $E_{\text{ps}}$ will denote the energy deposited in the presampler. The energies $E_i$ available at this stage of the reconstruction are the energies deposited in the liquid-argon ionisation medium divided by a region-dependent sampling fraction.

- $X$ is the the longitudinal barycentre or shower depth, defined by

$$X = \frac{\sum_{i=1}^{3} E_i X_i + E_{\text{ps}} X_{\text{ps}}}{\sum_{i=1}^{3} E_i + E_{\text{ps}}}, \tag{9}$$

  where $E_i$ is as above and $X_i$ is the longitudinal depth, expressed in radiation lengths, of compartment $i$, computed from the centre of the detector. The $X_i$, which are computed using a geantino[1] scan, are functions of $\eta$.

- $\eta$ is the cluster barycentre, corrected for the S-shape effect (see Sec. 2.1).

- $f_{\text{out}}$ is the fraction of the energy deposited outside the cluster.

- $C_{\text{cal}}(X,\eta)$ is the calibration factor for the energy in the EM calorimeter.

The calibration factor $C_{\text{cal}}$ is defined as the average ratio between the true energy deposited in the EM calorimeter (both absorbers and ionisation medium) and the reconstructed cluster energy $E_{\text{cl}}$. It is within a few percent of unity, and takes into account effects such as the dependence of the sampling fraction on $\eta$ and on the longitudinal profile of the shower. Once the correction factor $C_{\text{cal}}$ is expressed as a function of $X$ it is fairly energy independent. The correction factor averaged over all energies is shown in Fig. 26(a). Its dependence on $X$ is parametrised with a second order polynomial. The fit is performed excluding the bins with less than 0.5% of the total statistics. This criterion is also applied to all the fits performed in the following.

Due to the presence of the magnetic field and bremsstrahlung radiation, the fraction of energy deposited in the calorimeter outside of the cluster is energy dependent. Since only single electrons and photons with no noise or underlying event are simulated, this fraction is easily calculated. The profile of the out-of-cluster energy is asymmetric with the tail on the high side. However the most probable value, obtained with a Gaussian fit around the maximum of the distribution ($-2\sigma$, $+1.5\sigma$), is energy independent when plotted as a function of $X$. The most probable value of the fraction of energy deposited outside the cluster averaged over all energies is shown in Fig. 26(b) for electrons and photons and the two $|\eta|$ values. Electrons and photons behave similarly in the central region but differently in the forward region. This is due to the large difference in the amount of material present in front of the calorimeter ($\sim 2.5X_0$ at $|\eta| = 0.3$ and $\sim 7X_0$ at $|\eta| = 1.65$) combined with the presence of bremsstrahlung and the magnetic field.

[1] A "geantino" is an imaginary non-interacting particle used in the simulation. The properties of the material crossed by the particle are recorded.

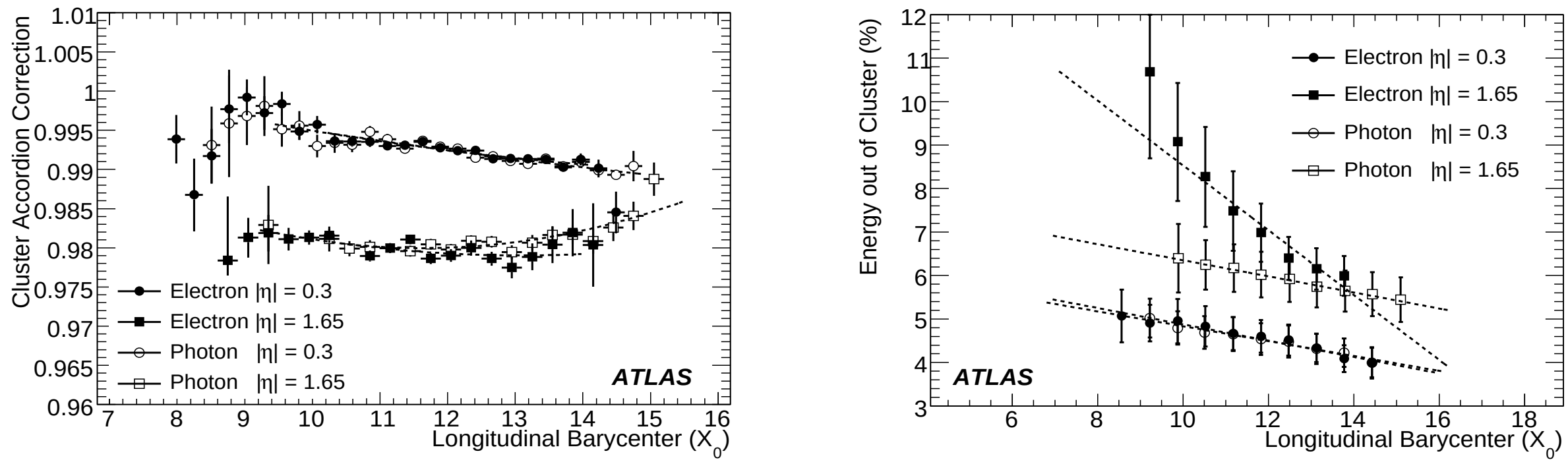

(a) $C_{\text{cal}}$ vs. $X$.

(b) Fraction of out-of-cluster energy.

Figure 26: Correction factor $C_{\text{cal}}$ and fraction of out-of-cluster energy as a function of the shower depth $X$, averaged over all energies, at two representative $|\eta|$ points. The dashed lines show the results of the parametrisation.

## 5.2 Energy deposited in front of the calorimeter

The energy lost in the material in front of the calorimeter (inner detector, cryostat, coil, and material between the presampler and strips) is parametrised as a function of the energy lost in the active material of the presampler ($E_{\text{ps}}$):

$$E_{\text{front}} = a(E_{\text{cal}}, \eta) + b(E_{\text{cal}}, \eta)E_{\text{ps}} + c(E_{\text{cal}}, \eta)E_{\text{ps}}^2. \tag{10}$$

An example of this relation is shown in Fig. 27. All coefficients are parametrised in terms of the energy deposited by a particle in the calorimeter $(E_{\text{cal}})$ and $\eta$. The coefficient $c$ is used only in the end-cap, $1.55 < |\eta| < 1.8$, and is set to zero otherwise. Note explicitly that $E_{\text{front}}$ includes the energy deposited in the presampler and between the presampler and the strips. An alternate form for $E_{\text{front}}$, which depends on the energy in the first calorimeter layer in addition to $E_{\text{ps}}$, was also tried. This did not improve the resolution, so the simpler parametrisation above is retained.

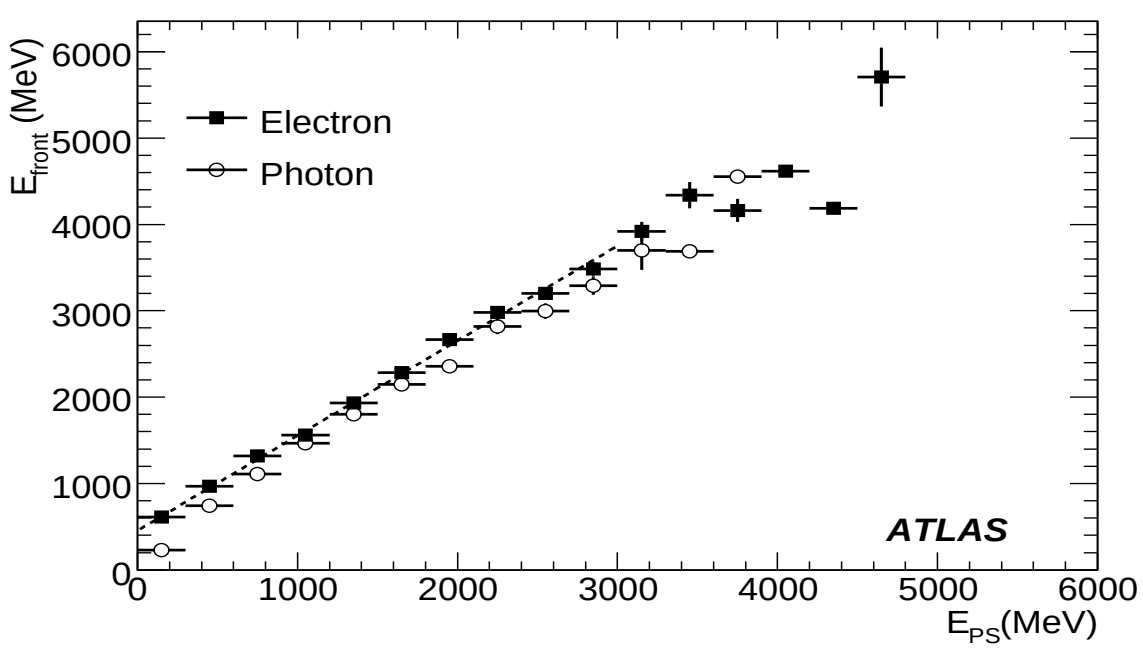

Figure 27: Energy lost in front of the EM calorimeter as a function of the energy measured in the presampler at $|\eta| = 0.3$ for electrons of 100 GeV. The dashed curve shows the parametrisation derived for electrons.

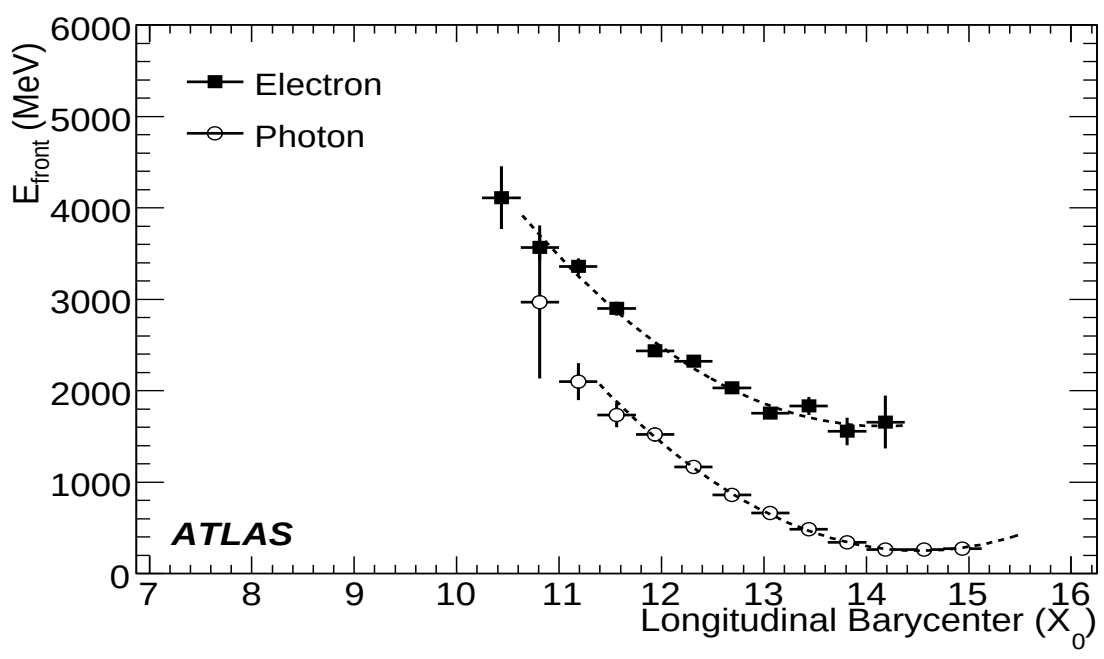

Figure 28: Energy lost in front of the calorimeter as a function of shower depth $X$, for electrons of 100 GeV at $|\eta| = 1.9$, in a region where the calorimeter is not instrumented with the presampler.

In the region $1.8 < |\eta| < 3.2$, not instrumented with the presampler, the energy deposited in front of the calorimeter is parametrised as a function of $X$ with a second degree polynomial. Figure 28 shows this correlation for electrons and photons of 100 GeV at $|\eta| = 1.9$. The coefficients of this polynomial are parametrised in terms of $E_{\text{cal}}$.

### 5.3 Longitudinal leakage correction

The energy deposited by the showers behind the EM calorimeter is computed as a fraction of the energy reconstructed in the calorimeter. This fraction, when parametrised as a function of $X$, is fairly energy independent both for electrons and photons. Averaged over the particle energies, it is parametrised by

$$f_{\text{leak}} \equiv E_{\text{back}}/E_{\text{cal}} = f_0^{\text{leak}}(\eta)X + f_1^{\text{leak}}(\eta)e^X. \tag{11}$$

Figure 29 shows the leakage and the result of the fit for $|\eta| = 0.3$ and 1.65.

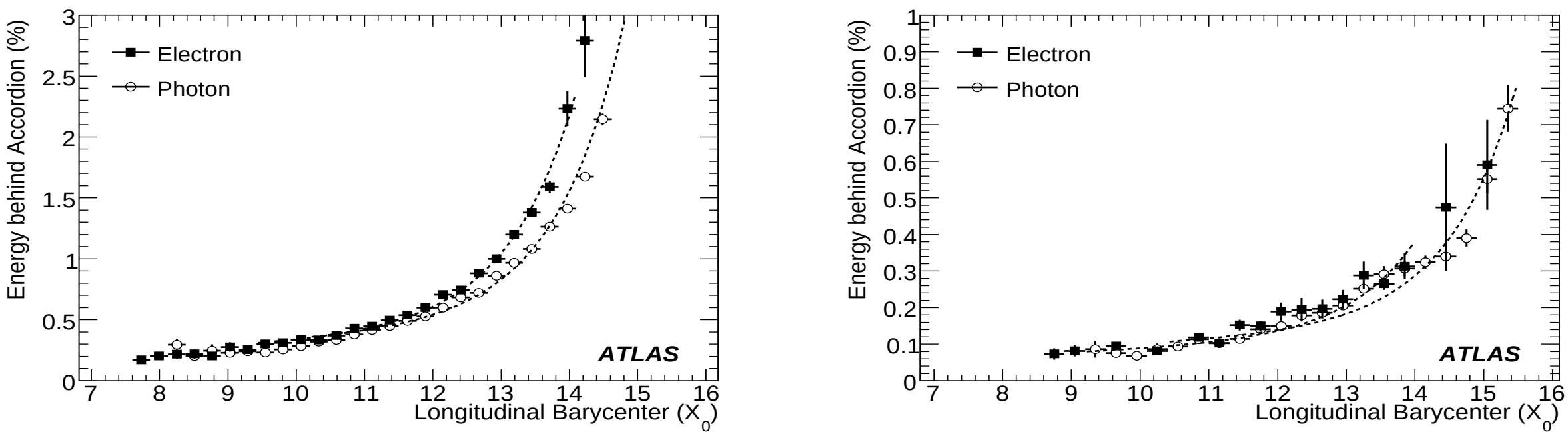

(a) $|\eta| = 0.3$.

(b) $|\eta| = 1.65$.

Figure 29: Fraction of energy deposited behind the calorimeter, averaged over particle energies, as a function of the shower depth $X$. The parametrisation used is superimposed.

### 5.4 Results

The total cluster energy is computed by adding these three contributions. Example distributions of reconstructed energies are shown in Fig. 30. Mean values and standard deviations are found from a fit to a Crystal-Ball function (a Gaussian with a low-side tail of the form $(1-x)^{-n}$).

The resolution is shown in Fig. 31 as a function of the particle energy for electrons and photons at two $|\eta|$ values and in Fig. 32 for various photon energies and all $\eta$ values. The sampling term is shown in Fig. 33 as a function of $|\eta|$ for electrons and photons.

For electrons, the sampling term increases from 8.7% at low $|\eta|$ to 21% at $|\eta| = 1.55$. This worsening of the energy resolution is related to the increase of the material in front of the calorimeter. This effect is much less relevant for photons, which have a maximum sampling term of 12%. The constant term is in general lower than 0.6% and is related to the energy modulation in a cell (see Sec. 3.2), not corrected at this stage. The linearity, the ratio between the fitted mean value and the true particle energy, is shown in Fig. 34. It is better than 0.5% over the full $|\eta|$ range and in the energy interval 25–500 GeV.

The results from the calibration hits correction are comparable in terms of resolution and linearity with the longitudinal weights method. However there are a few differences worth mentioning. The coefficients of the longitudinal weights method are averaged over a range of energies, while the parametrisations of the calibration hits method are energy dependent. This means that it should be easier to extend the calibrated energy range for the calibration hits method without compromising energy linearity. Another important difference is that while the coefficients of the longitudinal weights method have no direct

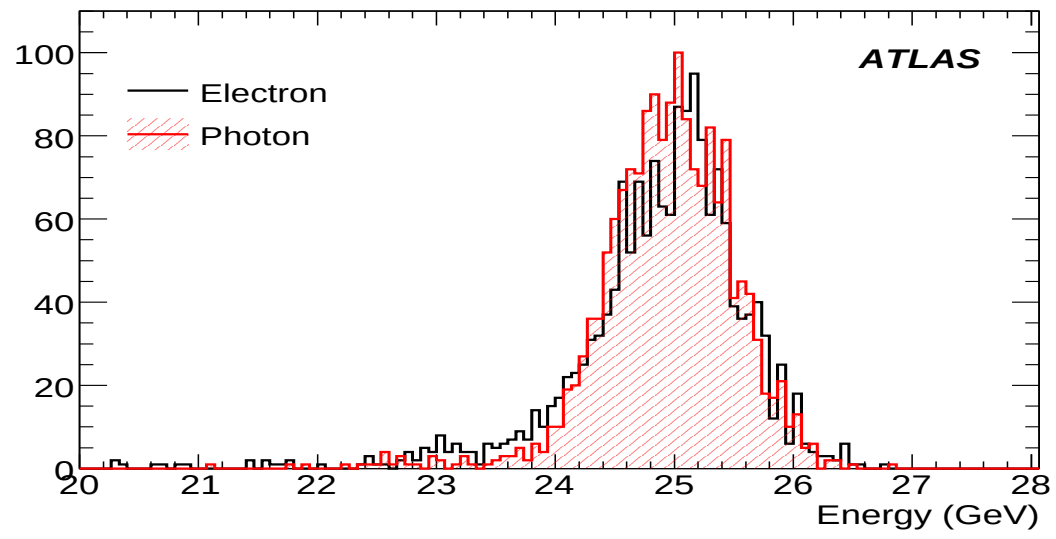

(a) $E = 25$ GeV, $|\eta| = 0.3$.

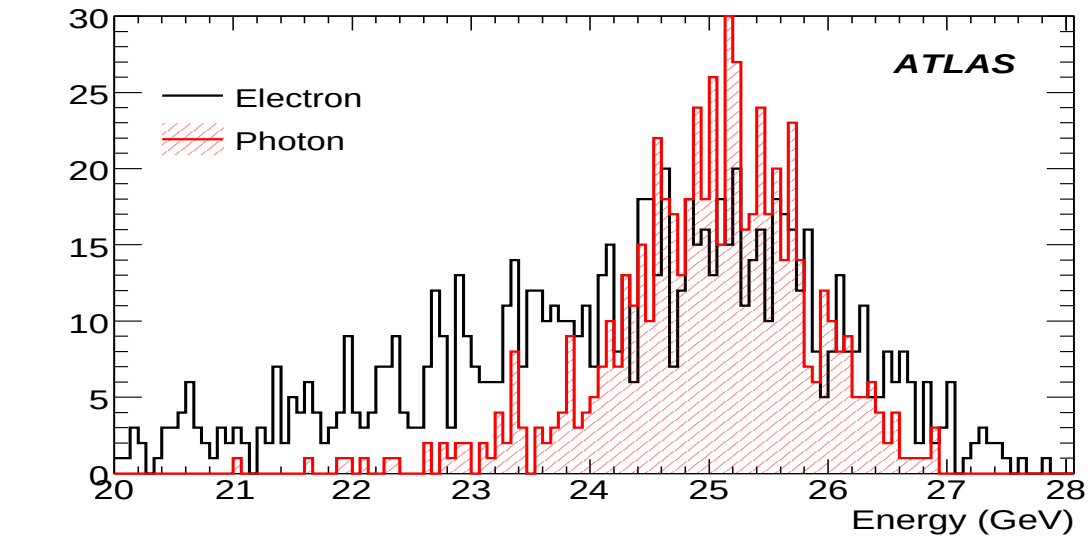

(b) $E = 25$ GeV, $|\eta| = 1.65$.

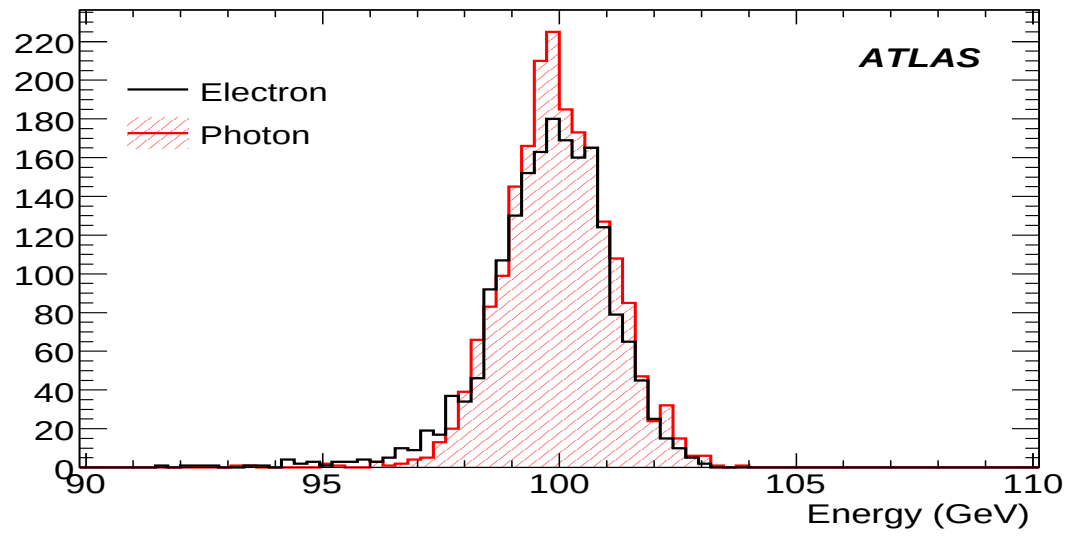

(c) $E = 100$ GeV, $|\eta| = 0.3$.

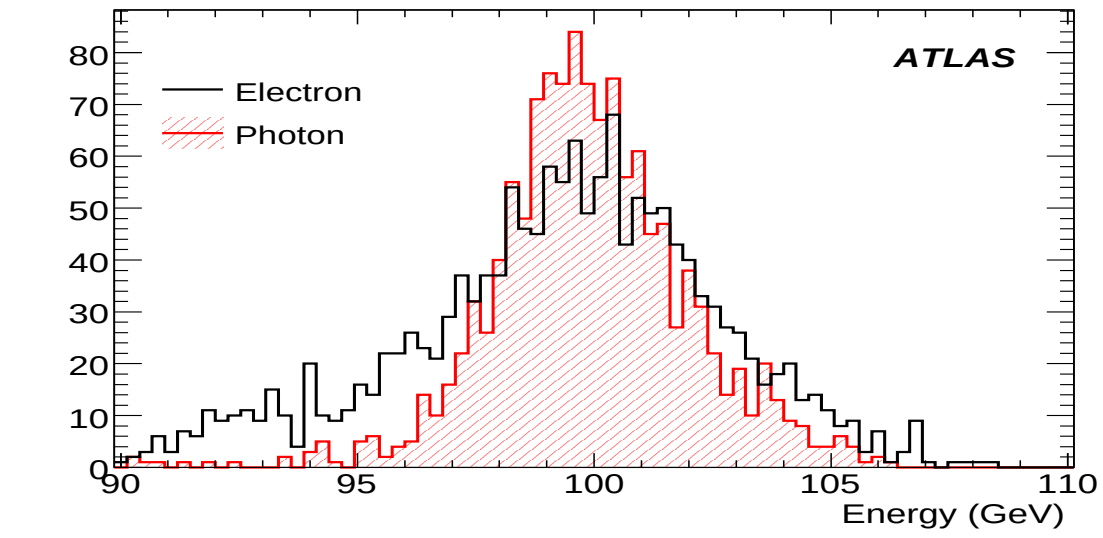

(d) $E = 100$ GeV, $|\eta| = 1.65$.

Figure 30: Total reconstructed energy profiles.

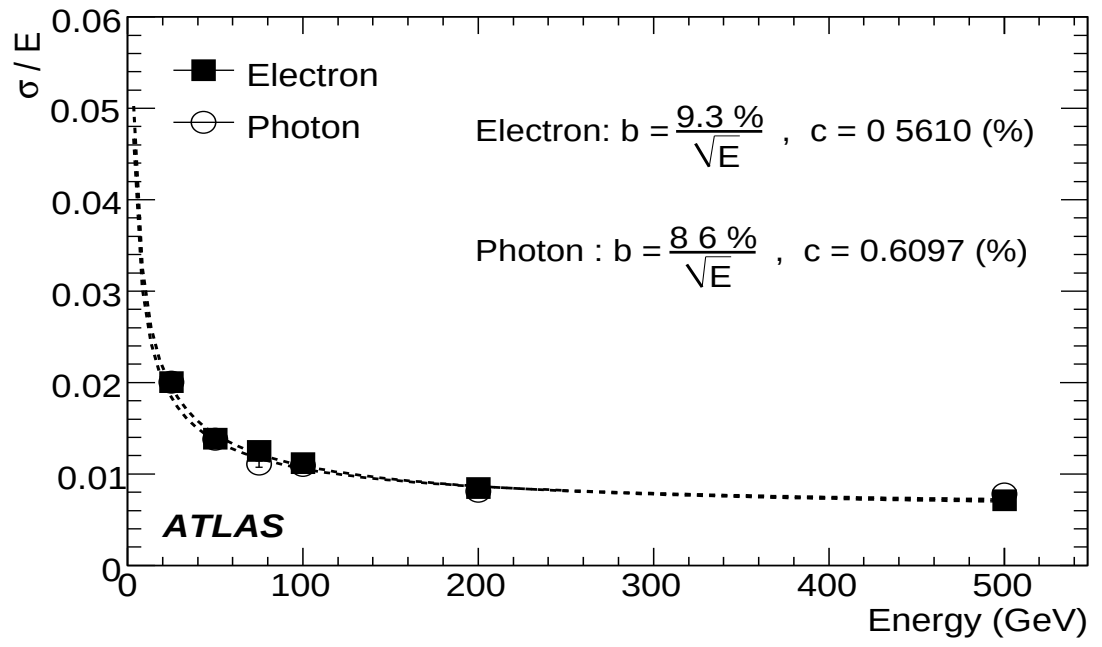

(a) $|\eta| = 0.3$.

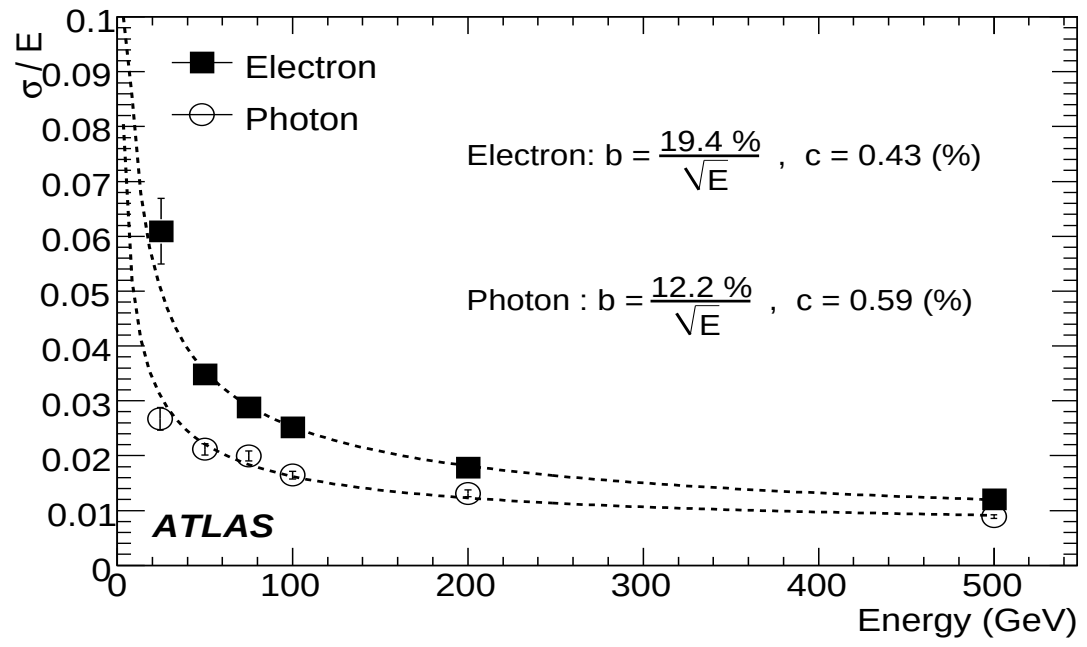

(b) $|\eta| = 1.65$.

Figure 31: Resolution versus particle energy.

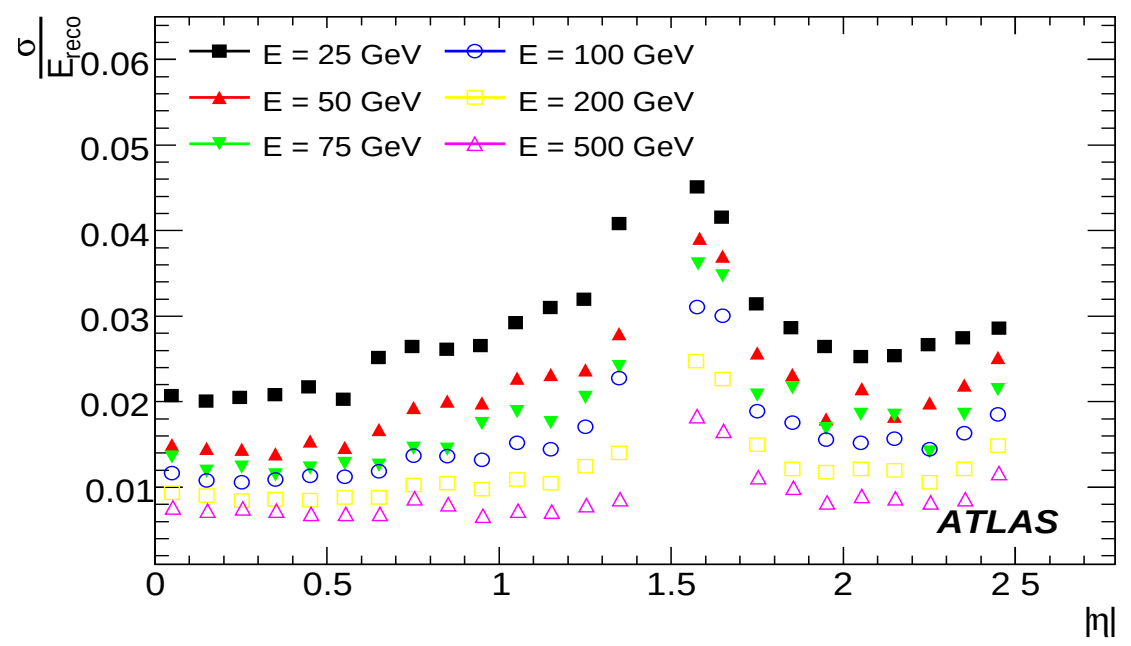

Figure 32: Resolution for various photon energies as a function of $|\eta|$.

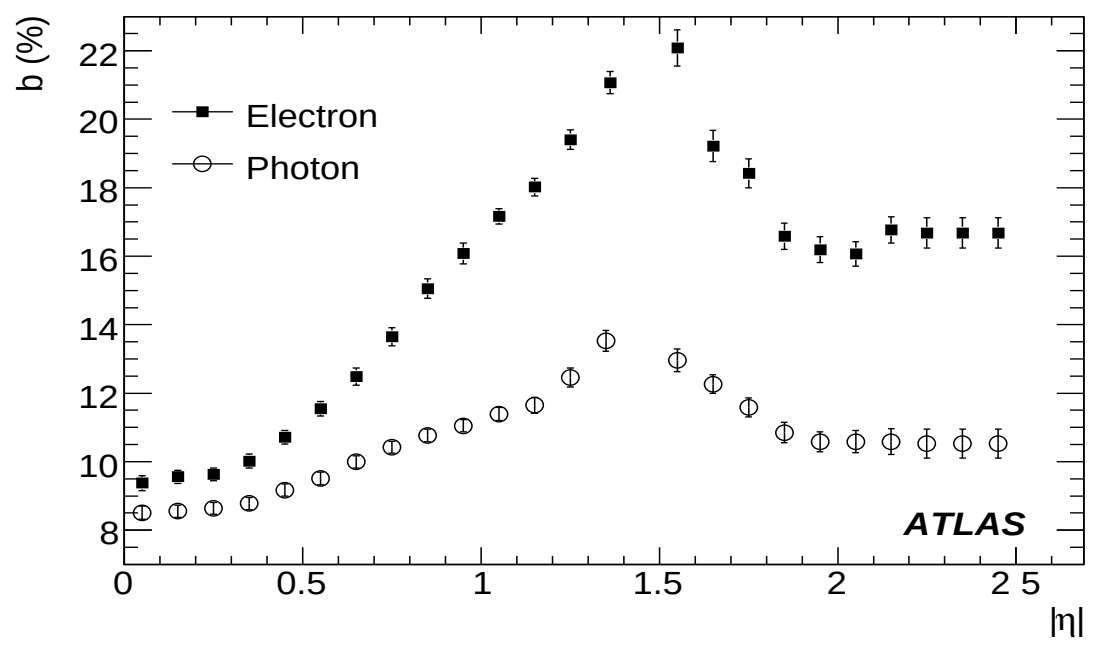

Figure 33: Sampling term as a function of $|\eta|$.

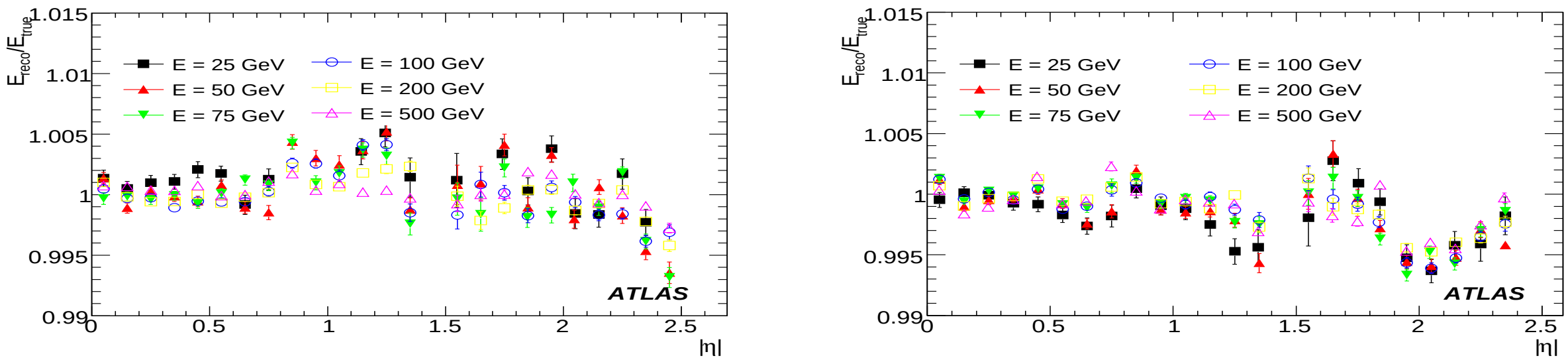

(a) Electrons.

(b) Photons.

Figure 34: Linearity for various particle energies as a function of $|\eta|$.

physical meaning, the parametrisation of the calibration hits method allows isolating the different components of the calibrated cluster energy: that deposited in the calorimeter, inside and outside of the cluster, and in front and behind of it. The knowledge of these separate contributions, which depend on accurate and detailed simulations of the tracker and the calorimeters, could be particularly useful in the early stages of the experiment, for example to disentangle effects such as a miscalibration of the calorimeter or an imperfect knowledge of the inner detector material. It is also worth noting that the estimate of the energy lost in front of the calorimeter is crucial to obtaining a good resolution and linearity; at low energies and large rapidities, a large fraction of the energy of an electron is deposited in front of the calorimeter. The calculation of missing momentum could also benefit from this separation of effects.

# 6 In-situ calibration with $Z \to ee$ events

## 6.1 Motivation

In the EM calorimeter, the construction tolerances and the calibration system ensure that the response is locally uniform, with a constant term $< 0.5\%$ over regions of size $\Delta\eta \times \Delta\phi = 0.2 \times 0.4$. This has been shown with test beam data [13]. Electron pairs from $Z$ boson decays can then be used to intercalibrate the 384 regions of such size within the acceptance of $|\eta| < 2.4$. These regions must be intercalibrated to within 0.5% in order to achieve a desired global constant term of $< 0.7\%$. The basic idea of this calibration method is to constrain the di-electron invariant mass distribution to the well-known $Z$ boson line shape. A second goal of the calibration is to provide the absolute calorimeter electromagnetic energy scale. This must be known to an accuracy of $\sim 0.1\%$ in order to achieve the ATLAS physics goals[2].

## 6.2 Description of the method

Long-range non-uniformities can arise for many reasons, including variations in the liquid argon impurities and temperature, amount of upstream material, mechanical deformations, and high voltage (as localised calorimeter defects may necessitate operating a small number of channels below nominal voltage). For a given region $i$, we parametrise the long-range non-uniformity modifying the measured electron energy as $E_i^{\text{reco}} = E_i^{\text{true}}(1+\alpha_i)$. Neglecting second-order terms and supposing that the angle between the two electrons is perfectly known, the effect on the di-electron invariant mass is:

$$M_{ij}^{\text{reco}} \simeq M_{ij}^{\text{true}}\left(1+\frac{\alpha_i+\alpha_j}{2}\right) = M_{ij}^{\text{true}}\left(1+\frac{\beta_{ij}}{2}\right), \tag{12}$$

where $\beta_{ij} \equiv \alpha_i + \alpha_j$.

[2]Except for the $W$ boson mass measurement, which needs a much better knowledge of the energy scale ($\sim 0.02\%$).

The method to extract the $\alpha$'s is fully described in [17] and is done in two steps. First, the $\beta$'s are determined, then the $\alpha$'s. For a given pair of regions $(i,j)$, the coefficient $\beta_{ij}$ and its associated uncertainty are determined by minimising the following log-likelihood:

$$-\ln L_{\text{tot}} = \sum_{k=1}^{N_{ij}} -\ln L\left(M_k/\left(1+\frac{\beta_{ij}}{2}\right), \sigma_{M,k}\right), \tag{13}$$

where $k$ counts all selected events populating the pair of regions $(i,j)$, $M_k$ is the di-electron invariant mass of event $k$, and $L(M,\sigma_M)$ quantifies the compatibility of an event with the $Z$ boson line shape and is described below. Fits with only one event are removed. Once the $\beta$'s are determined from the minimisation, the $\alpha$'s can be found from the overdetermined linear system given by $\beta_{ij} \equiv \alpha_i + \alpha_j$. This is done using a generalised least squares method, and gives an analytic solution.

The $Z$ boson line shape is modeled with a relativistic Breit-Wigner distribution [18, 19]:

$$\text{BW}(M) \sim \frac{M^2}{(M^2 - M_Z^2)^2 + \Gamma_Z^2 M^4/M_Z^2}, \tag{14}$$

where $M_Z$ and $\Gamma_Z$ are the mass and the width of the $Z$ boson. They were measured precisely at LEP; the values used are, respectively, $91.188 \pm 0.002$ GeV and $2.495 \pm 0.002$ GeV [20]. In proton-proton collisions, the mass spectrum of the $Z$ boson differs from the Breit-Wigner shape of the partonic process cross section. The probability that a quark and antiquark in the interacting $pp$ system produce an object of mass $M$ falls with increasing mass. In order to take this into account, the Breit-Wigner is multiplied by the ad-hoc parametrisation $\mathscr{L}(M) = 1/M^{\beta}$. The parton luminosity parameter $\beta$ is assumed to be a constant and is determined by fitting the $Z$ boson mass distribution obtained with events generated with PYTHIA version 6.403 [21]. Figure 35(a) shows the $Z$ boson mass distribution fitted with a Breit-Wigner with and without the parton luminosity factor. The fitted value of the parameter $\beta$ is $1.59 \pm 0.10$; this will be used in the following. Since the photon propagator and the interference term between the photon and the $Z$ boson were not taken into account in the previous parametrisation, the parton luminosity term also accounts for the effects of these two terms.

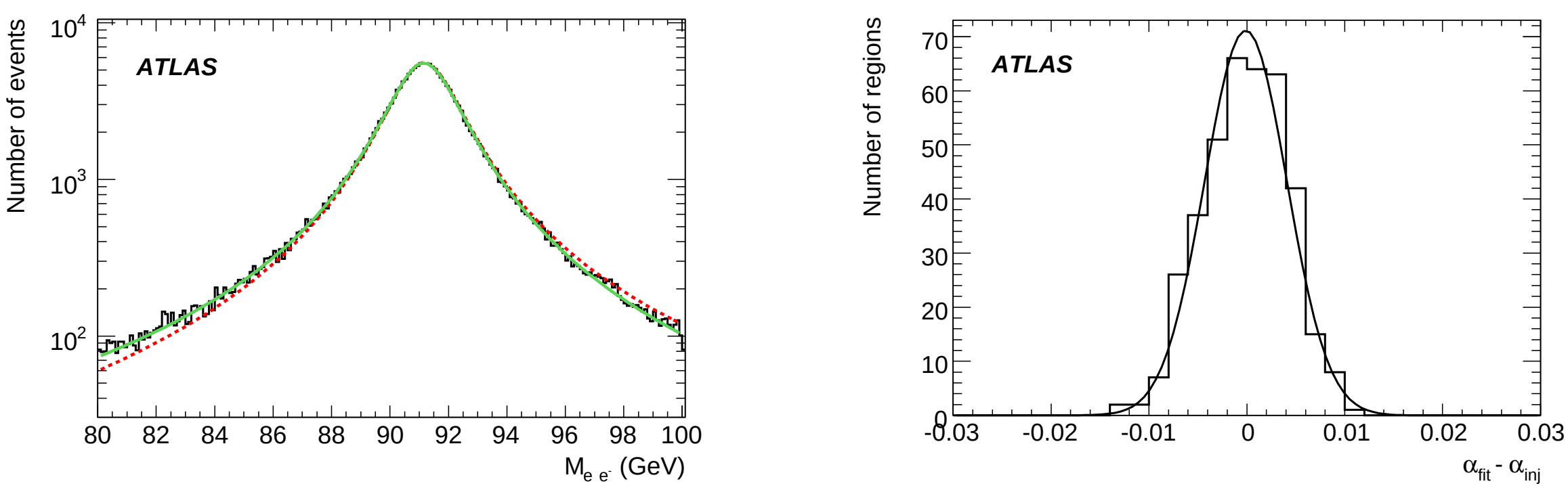

(a) Z line shape

(b) Residual distribution

Figure 35: (a) $Z$ boson mass distribution for PYTHIA events fitted with a Breit-Wigner distribution with (solid line) and without (dashed line) the parton luminosity factor. $\chi^2/N_{\text{DOF}}$ is 1.09 and 3.96, respectively. (b) Residual distribution fitted with a Gaussian.

Finally, in order to take into account the finite resolution of the electromagnetic calorimeter, the Breit-Wigner multiplied by the parton luminosity term is convoluted with a Gaussian:

$$L(M,\sigma_M) = \int_{-\infty}^{+\infty} \text{BW}(M-u)\mathscr{L}(M-u)\,\frac{e^{-u^2/2\sigma_M^2}}{\sqrt{2\pi}\sigma_M}\,du, \tag{15}$$

where $\sigma_M$ is the resolution of the measured mass. It is related to the electron energy resolution via

$$\frac{\sigma_M}{M} = \frac{1}{2}\sqrt{\left(\frac{\sigma_{E_1}}{E_1}\right)^2 + \left(\frac{\sigma_{E_2}}{E_2}\right)^2}. \tag{16}$$

At $|\eta| = 0.3$, the sampling term of the electron energy resolution is equal to 10.0% and increases with increasing $|\eta|$. Technically, the integral is converted to a discrete summation over the convolution parameter $u$ which takes values between $-5\sigma_M$ and $+5\sigma_M$.

### 6.3 Generator-level tests

The method is first tested on generator-level $Z \to ee$ Monte Carlo events. These were generated using PYTHIA 6.403 [21] with $M_Z = 91.19$ GeV and $\Gamma_Z = 2.495$ GeV. Events are required to have at least one electron with $p_T > 10$ GeV and $|\eta| < 2.7$ and a di-electron invariant mass $M_{ee} > 60$ GeV. To simulate the detector resolution, generated electron energies are smeared to obtain $\sigma_E/E = 10\%/\sqrt{E/\text{ GeV}}$.

For each calorimeter region $i$, a bias $\alpha_i$ is generated from a Gaussian distribution with a mean $\mu_{\text{bias}}$ and width $\sigma_{\text{bias}}$. These will be called the "injected" $\alpha$'s, $\alpha_{\text{inj}}$.

For the first tests, $\mu_{\text{bias}}$ is fixed to 0 and $\sigma_{\text{bias}}$ to 2%. The calibration method explained above is applied to 50,000 events after selection. The residual distribution $(\alpha_{\text{fit}} - \alpha_{\text{inj}})$ is shown in Fig. 35(b). The mean value of the residual distribution corresponds to the energy scale, and its width to the energy resolution. Thus it can be seen that the fitting method gives unbiased estimators of the injected $\alpha$'s.

In the case where $\mu_{\text{bias}}$ is different from zero, the mean value of the residual distribution will be different from zero. For example, for $\mu_{\text{bias}} = -3\%$, $\langle\alpha_{\text{fit}} - \alpha_{\text{inj}}\rangle = 0.1\%$. This is a consequence of neglecting the higher-order terms in the Taylor expansion of Eq. (12). Iterating the procedure twice suffices to recover an unbiased estimate of the $\alpha$'s, as shown in Fig. 36(a).

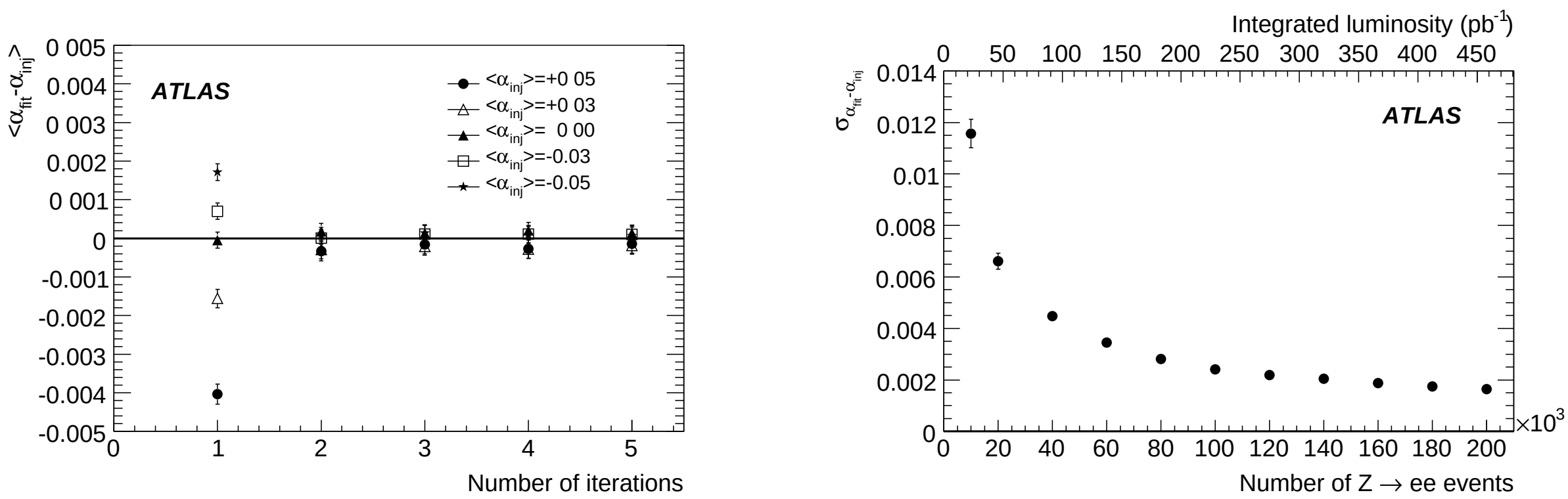

(a) Mean value.

(b) Constant term.

Figure 36: (a) Mean value of the Gaussian fitting the residual distribution as a function of the number of iterations for different mean values of the injected $\alpha$'s; (b) Constant term as a function of the number of events or as a function of the luminosity.

Figure 35(b) also shows the resulting uniformity. After the fit, the RMS of the distribution has been reduced from 2% to 0.4%. The RMS of the residual distribution is a measure of the expected long-range constant term. Figure 36(b) shows the long-range constant term as a function of the number of reconstructed $Z \to ee$ decays or of the integrated luminosity assuming an event selection efficiency of 25%. Therefore, by summing the local constant term of 0.5% with the long-range constant term of 0.4% obtained here, a total constant term of about 0.7% could be achieved with $\sim 100$ pb$^{-1}$. These results assume perfect knowledge of the material in front of the electromagnetic calorimeter.

### 6.4 Results with distorted geometry

The previous section showed results based on generator-level Monte Carlo. The results in this section use PYTHIA events with full detector simulation and reconstruction, using a geometry with additional material in front of the electromagnetic calorimeter.

The number of events available is 349,450 corresponding to an integrated luminosity of $\sim 200$ pb$^{-1}$. Events with at least two reconstructed electrons are kept. The two leading electrons are required to be of at least medium quality [6], to have $p_T > 20$ GeV and $|\eta| < 2.4$, and to be of opposite sign. Finally, the di-electron invariant mass is required to be within $80 < M_{ee} < 100$ GeV. The total selection efficiency is 21.5%; the efficiency for finding two electron candidates within $|\eta| < 2.4$ is 50%.

The calibration method is applied first without injecting any biases ($\alpha_{\text{inj}} = 0$ for all regions). However, the presence of the misalignments and extra material means that there will be some biases intrinsic to the simulation. These "true" biases can be estimated using generator information:

$$\alpha_{\text{true},i} = \frac{1}{N_i} \sum_{k}^{N_i} \frac{p_T^{\text{reco},k} - p_T^{\text{gen},k}}{p_T^{\text{gen},k}}, \tag{17}$$

where $k$ counts over the $N_i$ electrons falling in region $i$, and $p_T^{\text{reco},k}$ and $p_T^{\text{gen},k}$ are the reconstructed and true transverse momenta of electron $k$. The distribution of $\alpha_{\text{true}}$ is shown in Fig. 37(a), as is the results of the fit. The low-end tail corresponds to regions located in the gap between the barrel and end-cap cryostats (Fig. 38(a)), where the density of material has been increased by a factor of 1.7. There is fair agreement between the $\alpha$'s extracted using the data-driven method and those estimated from generator information. Figure 37(b) shows the difference between $\alpha_{\text{fit}}$ and $\alpha_{\text{true}}$; a Gaussian fitted to this distribution has a mean of 0.1% and a width of 0.5%. The distribution of $\alpha_{\text{fit}}$ as a function of $\eta$ and $\phi$ is shown in Fig. 38 for the ideal and distorted geometries. The asymmetry between positive and negative $\phi$ is due to the effect of the extra material in the inner detector at positive $\phi$. The difference between positive and negative $\phi$ values is about 0.6%.

The same exercise is also done by introducing, on top of the non-uniformities due to extra material, a bias $\alpha_{\text{inj}}$ generated from a Gaussian distribution with a mean $\mu_{\text{bias}} = 0$ and width $\sigma_{\text{bias}} = 2\%$. Results are shown in Fig. 39. The Gaussian fitted to this distribution also has a mean of 0.1% and a width of 0.5%.

One can conclude that, using $\sim 87{,}000$ reconstructed $Z \to ee$ events (which corresponds to about 200 pb$^{-1}$), and with an initial spread of 2% from region to region, the long-range constant term should not be greater than 0.5%.[3] This should give an overall constant term $\sim 0.7\%$. The bias on the absolute energy should be small and of the order of 0.2%. If the exercise is repeated with only 100 pb$^{-1}$ of data, the Gaussian fitted to the residual distribution also has a mean of 0.2%, but the width is larger, leading to a long-range constant term of 0.8%.

## 7 Estimation of the systematic uncertainty on the energy scale

The absolute energy scale has been obtained using electrons from $Z \to ee$ decays. It has been determined on events simulated with the misaligned geometry while the longitudinal weights were found with the ideal geometry. On top of the non-uniformities due to extra material, a bias modeling the calorimeter non-uniformities is introduced and is generated from a Gaussian distribution with a mean $\mu_{\text{bias}} = 0$ and width $\sigma_{\text{bias}} = 2\%$. The resulting bias on the energy scale can be assessed by comparing the fitted $\alpha$'s with those from generator information; the bias is equal to 0.2%. This bias is understood and is due to the fact that the model of the $Z$ boson line shape doesn't take into account the effects of bremsstrahlung. Work is ongoing to improve this issue.

[3] Part of the RMS of the residual distribution is also due to uncertainties on the measurement of $\alpha_{\text{true}}$.

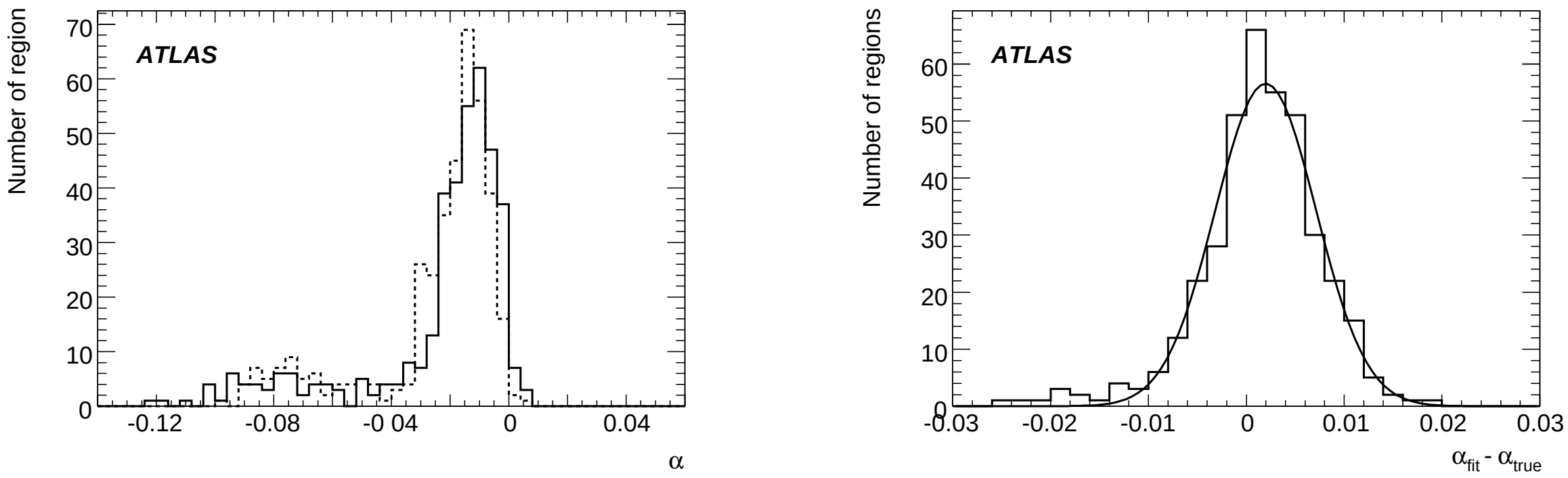

(a) $\alpha_{\text{fit}}$ (solid) and $\alpha_{\text{true}}$ (dashed). (b) Difference between $\alpha_{\text{fit}}$ and $\alpha_{\text{true}}$.

Figure 37: Fit results with distorted geometry and $\alpha_{\text{inj}} = 0$.

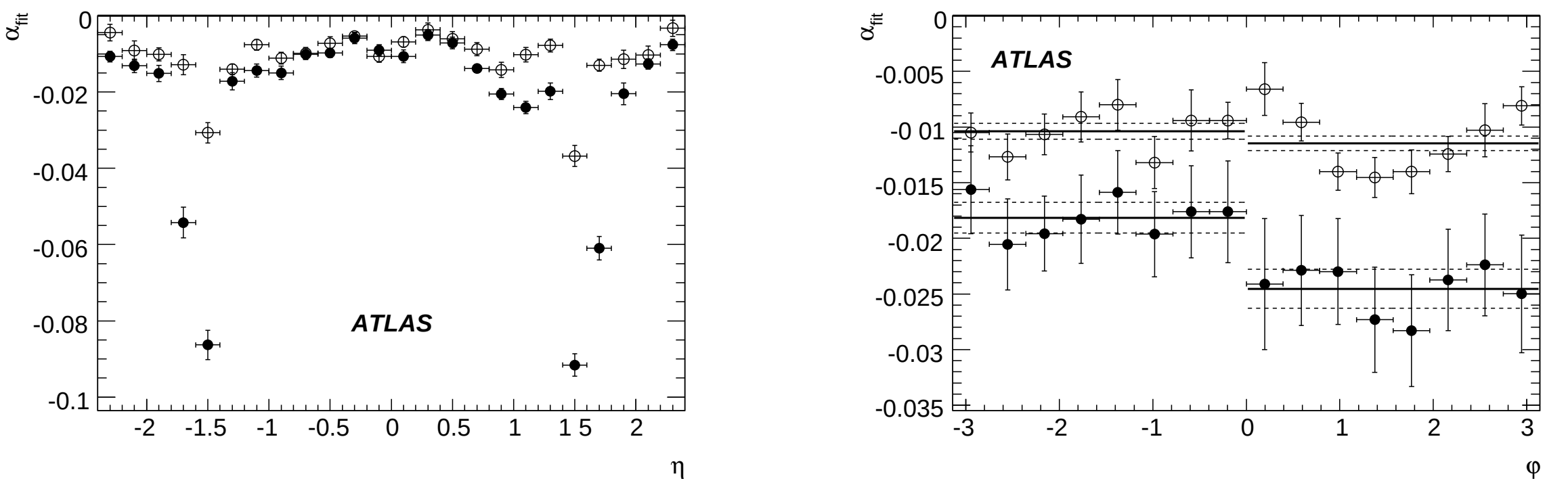

(a) $\alpha_{\text{fit}}$ integrated over $\phi$ as a function of $\eta$. (b) $\alpha_{\text{fit}}$ integrated over $\eta$ as a function of $\phi$, fitted in two separate regions.

Figure 38: $\alpha_{\text{fit}}$ distributions with $\alpha_{\text{inj}} = 0$ and with distorted/ideal (full/open circles) geometry.

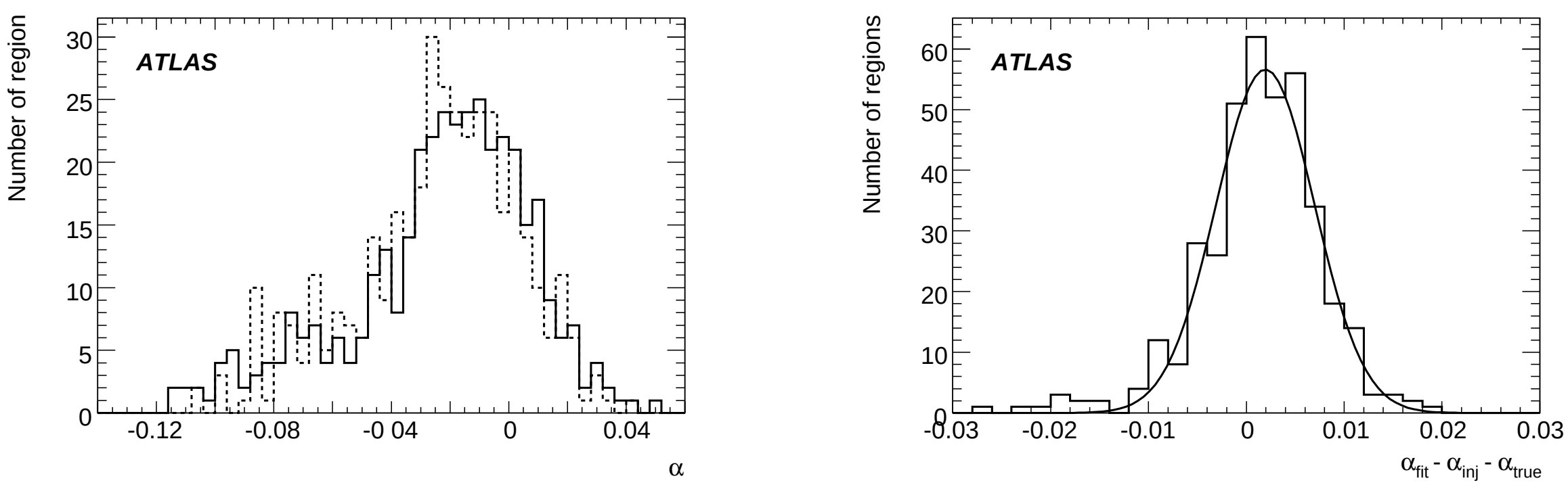

(a) $\alpha_{\text{fit}}$ (solid) and $\alpha_{\text{true}} + \alpha_{\text{inj}}$ (dashed). (b) Difference between $\alpha_{\text{fit}}$ and $\alpha_{\text{true}} + \alpha_{\text{inj}}$.

Figure 39: Fit results with distorted geometry and additional injected biases.

The background has been neglected but it has been checked that the contribution from QCD events where the two jets are misidentified as electrons is small. Thus, it should have a negligible effect on the determination on the energy scale.

Electrons from $Z$ boson decays have a $p_T$ spectrum with a maximum value around 45 GeV. Care will thus have to be taken to extrapolate the calibration obtained from $Z \to ee$ decays to electron energy regions not well populated by these events. Corrections determined with $Z$ boson decays were applied to single electron samples with different generated transverse momenta (20, 40, 120, and 500 GeV) reconstructed with the misaligned geometry. Figure 40 shows $\langle\alpha_{\text{true}}\rangle$ after correction as a function of $p_T$ for four $|\eta|$ bins. In principle, $\langle\alpha_{\text{true}}\rangle$ should be equal to zero. This is true for the 40 GeV electron sample at a level of 0.2% except in the bin ($1.4 < |\eta| < 2.0$) containing the crack region. For central electrons ($|\eta| < 0.6$), the dependence versus $p_T$ is smaller than 0.5%. The effect is worse for non-central electrons. For instance, at $p_T = 120$ GeV, $\alpha_{\text{true}}$ after corrections varies from 1 to 1.6 percent. This non-linearity is due to the presence of extra material in front of the calorimeter.

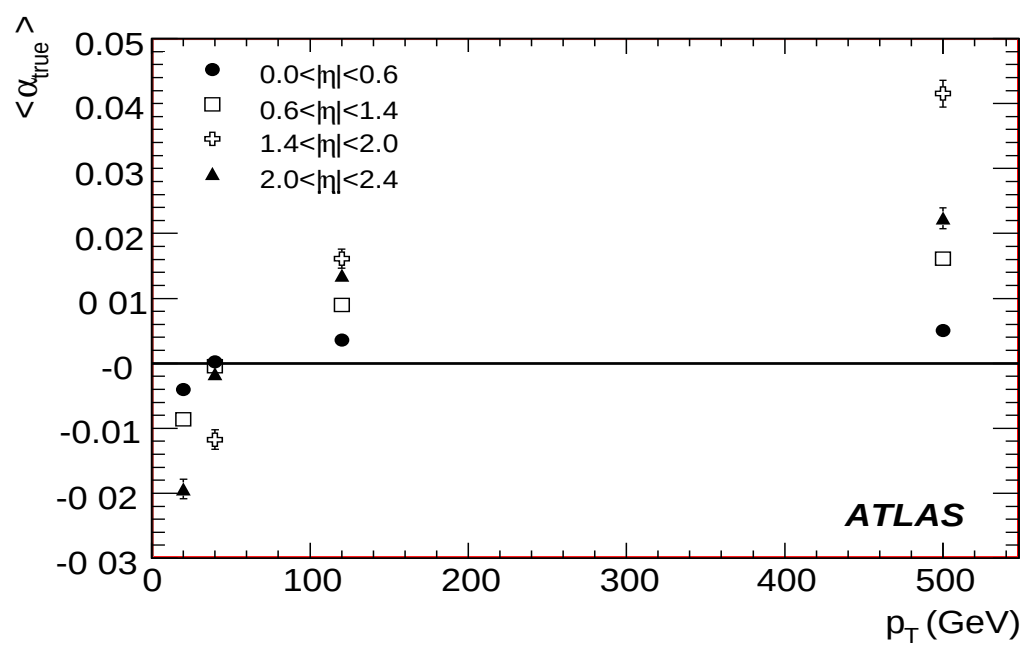

Figure 40: $\langle\alpha_{\text{true}}\rangle$ after correction as a function of $p_T$ for four $\eta$ bins.

To conclude, at the $Z$ boson energy scale, the estimate of the systematic uncertainty is around 0.2%. At other energy scales, the systematic uncertainty is dominated by effects of extra material. For central electrons, corrections can be extrapolated over the full $p_T$ spectrum to a level of 0.5%. The linearity is degraded for non-central electrons at a level of 1 or 2 percent except in the crack region where it is worse. These numbers depend on the amount of extra material added to the misaligned geometry compared to the ideal geometry and will likely be different with real data.

The performance presented here corresponds to our current understanding of the determination of the absolute energy scale. Improvements are expected to achieve systematic uncertainties smaller than 0.5%. For instance, including information from the $E/p$ ratio measured for isolated high-$p_T$ electrons from $W \to e\nu$ decays will compliment the direct calibration of the absolute scale with $Z \to ee$ events. Photon conversions can also help to determine the amount of material in front of the calorimeter.

## Conclusion

The methods and algorithms described in this note were already mentioned in Ref. [1] many years ago. Over the years, they have reached a higher level of stability and maturity, and have been implemented in the ATLAS reconstruction software. It is believed that, given the constraints of the ATLAS detector, in particular the amount of dead material in front of the calorimeter, the performances described here will not evolve much further.

The real challenge at the beginning of data-taking will be the detection and correction for additional inner detector material or calorimeter inhomogeneities which would not have affected the somewhat smaller-scale detectors used in the test beam. Discrepancies between data and simulation will have to be understood prior to the use of the methods described above. The in-situ calibration with $Z \rightarrow ee$ events described in Section 6 will play an important role, and refinements of the method presented here are expected.

# Reconstruction and Identification of Electrons

**Abstract**
This note discusses the overall ATLAS detector performance for the reconstruction and identification of high-$p_T$ electrons over a wide range of transverse energies, spanning from 10 GeV to 1000 GeV.

Electrons are reconstructed using information from both the calorimeter and the inner detector. The reference offline performance in terms of efficiencies for electrons from various sources and of rejections against jets is described. In a second part, this note discusses the requirements and prospects for electrons as probes for physics within and beyond the Standard Model: Higgs-boson, supersymmetry and exotic scenarios. In the last part, this note outlines prospects for electron identification with early data, corresponding to an integrated luminosity of 100 pb$^{-1}$, focusing on the use of the signal from $Z \rightarrow ee$ decays for a data-driven evaluation of the offline performance.

## 1 Introduction

Excellent particle identification capability is required at the LHC for most physics studies. Several channels expected from new physics, for instance some decay modes of the Higgs boson into electrons, have small cross-sections and suffer from large (usually QCD) backgrounds. Therefore powerful and efficient electron identification is needed to observe such signals. Even for standard processes, the signal-to-background ratio is usually less favourable than at past and present hadron colliders. The ratio between the rates of isolated electrons and the rate of QCD jets with $p_T$ in the range 20-50 GeV is expected to be $\sim 10^{-5}$ at the LHC, almost two orders of magnitude smaller than at the Tevatron $p\bar{p}$ collider. Therefore, to achieve comparable performances, the electron identification capability of the LHC detectors must be almost two orders of magnitude better than what has been achieved so far.

Physics channels of prime interest at the LHC are expected to produce electrons with $p_T$ between a few GeV and 5 TeV. Good electron identification is therefore needed over a broad energy range. In the moderate $p_T$ region (20 - 50 GeV), a jet-rejection factor exceeding $10^5$ will be needed to extract a relatively pure inclusive signal from genuine electrons above the residual background from jets faking electrons. The required rejection factor decreases rapidly with increasing $p_T$ to $\sim 10^3$ for jets in the TeV region. For multi-lepton final states, such as possible $H \rightarrow eeee$ in the mass region $130 < m_H < 180$ GeV, a rejection of $\sim 3000$ per jet should be sufficient to reduce the fake-electron backgrounds to a level well below that from real electrons. In this case, however, the electrons have a rather soft $p_T$ spectrum (as low as 5 GeV), resulting in lower reconstruction and identification efficiencies.

Since the publication of the ATLAS physics TDR [1], the ATLAS detector description has been greatly improved, with, in particular, the introduction of a more realistic material description for the inner detector and for the region between the inner detector and the first layer of the electromagnetic calorimeter [2] [3]. This has led to some significant changes in the expected performance. The reconstruction software has also evolved significantly. Each step of the energy reconstruction has been validated by a series of beam tests [4] [5] [6] using prototype modules of the liquid argon electromagnetic calorimeter, and also more recently, combined with prototype modules of the inner detector. At present, two electron reconstruction algorithms have been implemented in the ATLAS offline software, both integrated into one single package and a common event data model.

- The standard one, which is seeded from the electromagnetic (EM) calorimeters, starts from clusters reconstructed in the calorimeters and then builds the identification variables based on information from the inner detector and the EM calorimeters.

- A second algorithm, which is seeded from the inner detector tracks, is optimized for electrons with energies as low as a few GeV, and selects good-quality tracks matching a relatively isolated deposition of energy in the EM calorimeters. The identification variables are then calculated in the same way as for the standard algorithm.

The standard algorithm is the one used to obtain the results presented in this note, while the track-based algorithm is used for low $p_T$ and non-isolated electrons and is the subject of another note [7].

This note is organised as follows. Section 2 discusses the reconstruction and identification of electrons in the fiducial range of the ATLAS detector ($|\eta| < 2.5$), whereas section 3 describes the identification of electrons in the forward region ($2.5 < |\eta| < 4.9$). Section 4 describes some important performance aspects of electron identification in discovery physics processes. Section 5 discusses the strategies for measuring reconstruction and identification efficiencies using a data-driven approach based on $Z \rightarrow ee$ events.

# 2 Calorimeter-seeded reconstruction and identification

In the standard reconstruction of electrons, a seed electromagnetic tower with transverse energy above $\sim$ 3 GeV is taken from the EM calorimeter [3] and a matching track is searched for among all reconstructed tracks which do not belong to a photon-conversion pair reconstructed in the inner detector. The track, after extrapolation to the EM calorimeter, is required to match the cluster within a broad $\Delta\eta \times \Delta\phi$ window of $0.05 \times 0.10$. The ratio, $E/p$, of the energy of the cluster to the momentum of the track is required to be lower than 10. Approximately 93% of true isolated electrons, with $E_T > 20$ GeV and $|\eta| < 2.5$, are selected as electron candidates. The inefficiency is mainly due to the large amount of material in the inner detector and is therefore $\eta$-dependent. As an example, 4% of electron candidates with $p_T = 40$ GeV fail the cut $E/p < 10$ and most of the losses are in the end-cap region. Various identification techniques can be applied to the reconstructed electron candidates, combining calorimeter and track quantities and the TRT information to discriminate jets and background electrons from the signal electrons. A simple cut-based identification procedure is described below together with its expected performance. This is followed by a brief overview of the possibilities offered by more advanced methods, such as a likelihood discriminant.

## 2.1 Electron-jet studies

For the purposes of this note, the electron identification efficiency is defined as

$$\varepsilon = \frac{N_e^{\mathrm{Id}}}{N_e^{\mathrm{truth}}},$$

where $N_e^{\mathrm{Id}}$ is the number of reconstructed and identified candidates and $N_e^{\mathrm{truth}}$ is the number of true electrons selected using the appropriate kinematic cuts at the generator level. A geometrical matching (within a cone of size $\Delta R = 0.2$) between the reconstructed cluster and the true electron is required in the calculation of $N_e^{\mathrm{Id}}$. A classification is applied to define whether a reconstructed electron candidate should be considered as signal or background. This classification is based on the type of the Monte Carlo particle associated to the reconstructed track, as well as that of its non-electron parent particle. As shown in Table 1, candidates are divided into four categories and signal efficiencies are calculated separately for isolated and non-isolated electrons.

For the jet rejection studies, the PYTHIA (version 6.4) [10] event generator has been used to produce the large statistics of jet background samples required to assess both the trigger and offline performance of the electron reconstruction and identification tools described in this note. Two different samples were generated to cover the $E_T$ -range of interest for single electrons (10-40 GeV). The first one, referred to as filtered di-jets, contains all hard-scattering QCD processes with $E_T > 15$ GeV, e.g. $qg \rightarrow qg$, including heavy-flavour production, together with other physics processes of interest, such as prompt-photon production and single $W/Z$ production. The second one, referred to as minimum bias, contains the same processes without any explicit hard-scattering cut-off. A filter was applied at the generator level to simulate the L1 trigger requirements [11], with the goal of increasing in an unbiased way the probability that the selected jets pass the electron identification cuts after GEANT [12] simulation. The summed transverse energy of all stable particles (excluding muons and neutrinos) with $|\eta| < 2.7$ in a region $\Delta\phi \times \Delta\eta = 0.12 \times 0.12$ was required to be greater than a chosen $E_T$ -threshold for an event to be retained. For the filtered di-jet sample, this $E_T$ -threshold is 17 GeV, while for the minimum-bias sample, it is 6 GeV. The filter retains 8.3% of the di-jet events and 5.7% of the minimum-bias events. The total number of events available for analysis after filtering, simulation and reconstruction, amounts to 8.2 million events for the di-jet sample and to 4.1 million events for the minimum-bias sample.

| Category | Type of particle | Type of parent particle |
|---|---|---|
| Isolated | Electron | $Z$, $W$, $t$, $\tau$ or $\mu$ |
| Non-isolated | Electron | $J/\psi$, $b$-hadron or $c$-hadron decays |
| Background electron | Electron | Photon (conversions), $\pi^0/\eta$ Dalitz decays, $u/d/s$-hadron decays |
| Non-electron | Charged hadrons, $\mu$ | |

Table 1: Classification of simulated electron candidates according to their associated parent particle. Muons are included as source because of the potential emission of a Bremsstrahlungs photon.

| $E_T > 17$ GeV | | | $E_T > 8$ GeV | |
|---|---|---|---|---|
| Isolated | Non-isolated | Background | Non-isolated | Background |
| $W$ – 75.0% | $b$-hadrons – 38.7% | $\gamma$-conv. – 97.8% | $b$-hadrons – 39.3% | $\gamma$-conv. – 98.4% |
| $Z$ – 20.9% | $c$-hadrons – 60.6% | Dalitz decays – 1.8% | $c$-hadrons – 59.7% | Dalitz decays – 1.3% |
| t – < 0.1% | $J/\psi$ – 0.7% | $u/d/s$-hadrons – 0.4% | $J/\psi$ – 1.0% | $u/d/s$-hadrons – 0.3% |
| $\tau$ – 4.1% | | | | |

Table 2: Contribution and origin of isolated, non-isolated, and background electron candidates in the two di-jet samples before the identification criteria are applied.

The jet rejections quoted in this note are normalised with respect to the number of particle jets reconstructed using particle four-momenta within a cone size $\Delta R = 0.4$ and derived from a dedicated un-filtered generated sample of di-jets or minimum-bias events. In the di-jet and minimum-bias samples, the average numbers per generated event of such particle jets with $E_T$ above 17 and 8 GeV, respectively, and in the range $|\eta| < 2.47$, are 0.74 and 0.31, respectively.

After reconstruction of electron candidates and before any of the identification cuts are applied, the signal is completely dominated by non-isolated electrons from $b-$ and $c$-hadron decays. The expected signal-to-background ratios for the filtered di-jet ($E_T$ above 17 GeV) and minimum-bias ($E_T$ above 8 GeV) samples are 1:80 and 1:50, respectively. The residual jet background is dominated by charged hadrons. Only a small fraction of the background at this stage consists of electrons from photon conversions or Dalitz decays, namely 6.4% and 9.4%, respectively. Table 2 summarises the relative compositions of the filtered di-jet and minimum-bias samples in terms of the three categories containing electrons described in Table 1.

| Type | Description | Variable name |
|---|---|---|
| Loose cuts | | |
| Acceptance of the detector<br>Hadronic leakage | $\|\eta\| < 2.47$<br>Ratio of $E_T$ in the first sampling of the hadronic calorimeter to $E_T$ of the EM cluster | |
| Second layer of EM calorimeter. | Ratio in $\eta$ of cell energies in $3 \times 7$ versus $7 \times 7$ cells.<br>Ratio in $\phi$ of cell energies in $3 \times 3$ versus $3 \times 7$ cells.<br>Lateral width of the shower. | $R_\eta$<br>$R_\phi$ |
| Medium cuts (includes loose cuts) | | |
| First layer of EM calorimeter. | Difference between energy associated with the second largest energy deposit and energy associated with the minimal value between the first and second maxima.<br>Second largest energy deposit normalised to the cluster energy.<br>Total shower width.<br>Shower width for three strips around maximum strip.<br>Fraction of energy outside core of three central strips but within seven strips. | $\Delta E_s$<br>$R_{\max 2}$<br>$w_{\text{stot}}$<br>$w_{\text{s3}}$<br>$F_{\text{side}}$ |
| Track quality | Number of hits in the pixel detector (at least one).<br>Number of hits in the pixels and SCT (at least nine).<br>Transverse impact parameter (<1 mm). | |
| Tight (isol) (includes medium cuts) | | |
| Isolation | Ratio of transverse energy in a cone $\Delta R < 0.2$ to the total cluster transverse energy. | |
| Vertexing-layer | Number of hits in the vertexing-layer (at least one). | |
| Track matching | $\Delta\eta$ between the cluster and the track ($< 0.005$).<br>$\Delta\phi$ between the cluster and the track ($< 0.02$).<br>Ratio of the cluster energy to the track momentum. | $E/p$ |
| TRT | Total number of hits in the TRT.<br>Ratio of the number of high-threshold hits to the total number of hits in the TRT. | |
| Tight (TRT) (includes tight (isol) except for isolation) | | |
| TRT | Same as TRT cuts above, but with tighter values corresponding to about 90% efficiency for isolated electrons. | |

Table 3: Definition of variables used for loose, medium and tight electron identification cuts. The cut values are given explicitly only when they are independent of $\eta$ and $p_T$ . For a detailed description of the cut variables used for the loose and medium cuts, refer to sections 2.1.1.1 and 2.1.1.2.

### 2.1.1 Cut-based method description

Standard identification of high-$p_T$ electrons is based on many cuts which can all be applied independently. These cuts have been optimised in up to seven bins in $\eta$ and up to six bins in $p_T$ . Three reference sets of cuts have been defined: loose, medium and tight, as summarised in Table 3. This provides flexibility in analysis, for example to improve the signal efficiency for rare processes which are not subject to large backgrounds from fakes.

**2.1.1.1 Loose cuts** This set of cuts performs a simple electron identification based only on limited information from the calorimeters. Cuts are applied on the hadronic leakage and on shower-shape variables, derived from only the middle layer of the EM calorimeter (lateral shower shape and lateral shower width ). This set of cuts provides excellent identification efficiency, but low background rejection.

**2.1.1.2 Medium cuts** This set of cuts improves the quality by adding cuts on the strips in the first layer of the EM calorimeter and on the tracking variables:

- Strip-based cuts are effective in the rejection of $\pi^0 \to \gamma\gamma$ decays. Since the energy-deposit pattern from $\pi^0$'s is often found to have two maxima due to $\pi^0 \to \gamma\gamma$ decay, showers are studied in a window $\Delta\eta \times \Delta\phi = 0.125 \times 0.2$ around the cell with the highest $E_T$ to look for a second maximum. If more than two maxima are found the second highest maximum is considered. The variables used include $\Delta E_s = E_{\mathrm{max2}} - E_{\mathrm{min}}$, the difference between the energy associated with the second maximum $E_{\mathrm{max2}}$ and the energy reconstructed in the strip with the minimal value, found between the first and second maxima, $E_{\mathrm{min}}$. Also included are: $R_{\mathrm{max2}} = E_{\mathrm{max2}}/(1+9\times 10^{-3}E_T)$, where $E_T$ is the transverse energy of the cluster in the EM calorimeter and the constant value 9 is in units of $\mathrm{GeV}^{-1}$; $w_{\mathrm{stot}}$, the shower width over the strips covering 2.5 cells of the second layer (20 strips in the barrel for instance); $w_{\mathrm{s3}}$, the shower width over three strips around the one with the maximal energy deposit; and $F_{\mathrm{side}}$, the fraction of energy deposited outside the shower core of three central strips.
- The tracking variables include the number of hits in the pixels, the number of silicon hits (pixels plus SCT) and the tranverse impact parameter.

The medium cuts increase the jet rejection by a factor of 3-4 with respect to the loose cuts, while reducing the identification efficiency by $\sim 10\%$.

**2.1.1.3 Tight cuts** This set of cuts makes use of all the particle-identification tools currently available for electrons. In addition to the cuts used in the medium set, cuts are applied on the number of vertexing-layer hits (to reject electrons from conversions), on the number of hits in the TRT, on the ratio of high-threshold hits to the number of hits in the TRT (to reject the dominant background from charged hadrons), on the difference between the cluster and the extrapolated track positions in $\eta$ and $\phi$, and on the ratio of cluster energy to track momentum, as shown in Table 3. Two different final selections are available within this tight category: they are named tight (isol) and tight (TRT) and are optimised differently for isolated and non-isolated electrons. In the case of tight (isol) cuts, an additional energy isolation cut is applied to the cluster, using all cell energies within a cone of $\Delta R < 0.2$ around the electron candidate. This set of cuts provides, in general, the highest isolated electron identification and the highest rejection against jets. The tight (TRT) cuts do not include the additional explicit energy isolation cut, but instead apply tighter cuts on the TRT information to further remove the background from charged hadrons.

Figures 1 and 2 compare the distributions expected from $Z \to ee$ decays and from the filtered di-jet sample for a few examples of the basic discriminating variables described above for electron identification.

### 2.1.2 Performance of cut-based electron identification

The performance of the cut-based electron identification is summarised in Tables 4 and 5. Table 4 shows, for each of the background samples, the composition of each of the three categories of electron candidates containing real electrons, as it evolves from reconstruction (no identification cuts) to loose, medium and tight cuts. In the case of non-isolated electrons, there is a strong reduction of the initially dominant component from $c$-hadrons as the identification cuts applied become tighter. In the case of background electrons, there is a significant reduction of the contribution from photon conversions when applying tight cuts, since the vertexing-layer requirement does not much affect electrons from Dalitz decays and $u/d/s$-hadrons. As shown in Table 5, the signal from prompt electrons is dominated by non-isolated electrons from heavy flavours, which are usually close in space to hadrons from the jet fragmentation. The resulting overlap between the electron shower and nearby hadronic showers explains the much lower efficiency observed for these electrons than for isolated electrons from $Z \rightarrow ee$ decays. These non-isolated electrons will nevertheless provide the most abundant initial source of signal electrons and will be used for alignment of the electromagnetic calorimeters and the inner detector, for $E/p$ calibrations, and more generally to improve the understanding of the material of the inner detector as a radiation/conversion source. For tight cuts and an electron $E_T$ of $\sim$ 20 GeV, the isolated electrons from $W$, $Z$ and top-quark decays represent less than 20% of the total prompt electron signal.

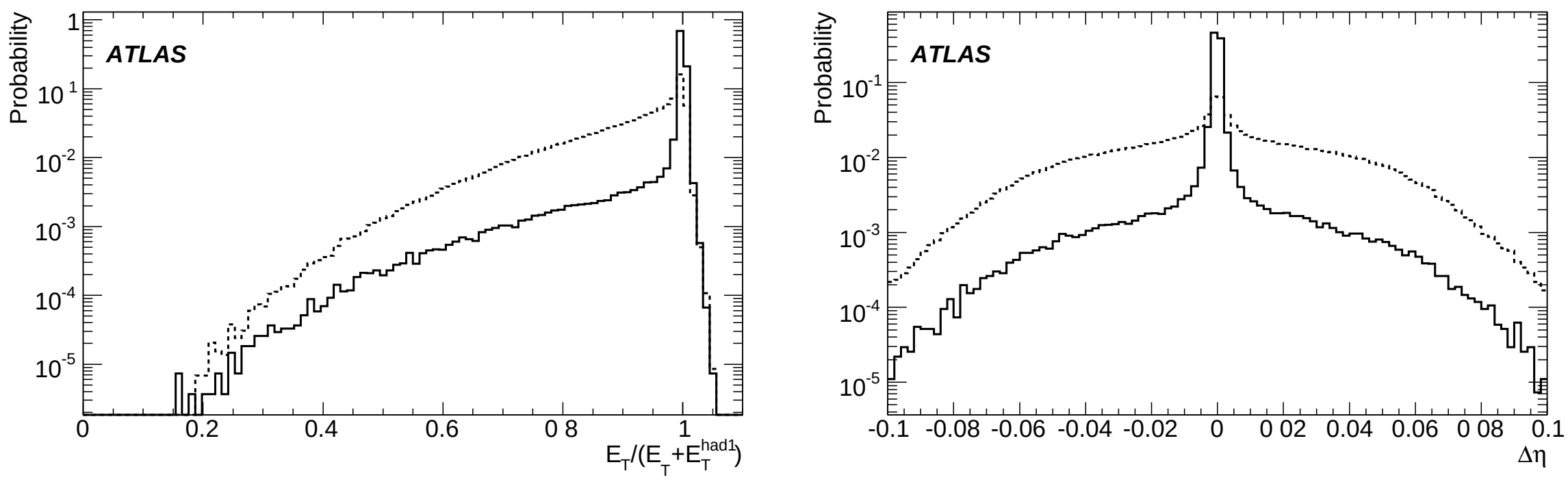

Figure 1: Left: ratio between the transverse energy of the electron candidate and the sum of this transverse energy and that contained in the first layer of the hadronic calorimeter. The distributions are shown for electrons from $Z \rightarrow ee$ decays (solid line) and for filtered di-jets (dotted line). Right: difference in $\eta$ between cluster and extrapolated track positions for electrons from $Z \rightarrow ee$ decays (solid line) and for filtered di-jets (dotted line).

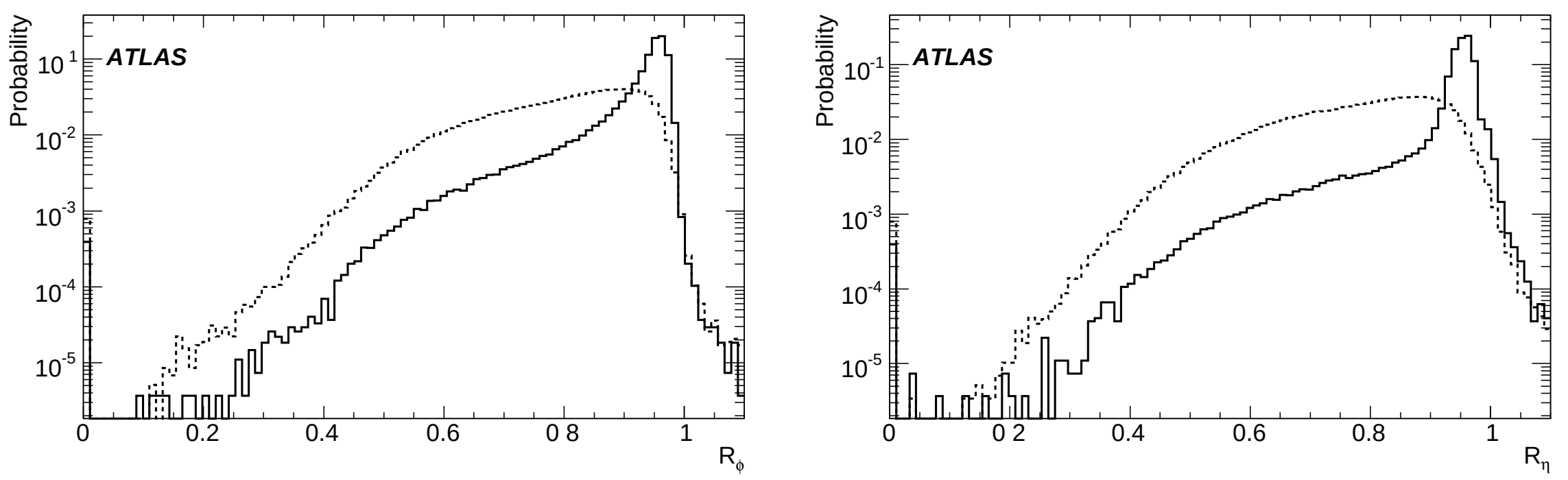

Figure 2: Shower-shape distributions for electrons from $Z \rightarrow ee$ decays (solid lines) compared to those from filtered di-jets (dotted lines). Shown are the energy ratios $R_\phi$ (left) and $R_\eta$ (right) described in Table 3.

For the lower $E_T$ -threshold of 8 GeV, the expected signal from isolated electrons is negligible. Not surprisingly, the tight (TRT) cuts are more efficient to select non-isolated electrons from heavy-flavour decay, while the tight (isol) cuts are more efficient at selecting isolated electrons. After tight cuts, the signal-to-background ratio is close to 3:1, and depends only weakly on the $E_T$ - threshold in the 10-40 GeV $E_T$ -range studied here. The residual background is dominated by charged hadrons, which could be further rejected by stronger cuts (TRT and/or isolation). The initial goal of obtaining a rejection of the order of $10^5$ against jets has been achieved with an overall efficiency of 64% for isolated electrons with

<table>
<tr><td colspan="11">Isolated</td></tr>
<tr><td></td><td colspan="10">$E_T > 17$ GeV</td></tr>
<tr><td></td><td colspan="2">No cut</td><td colspan="2">Loose</td><td colspan="2">Medium</td><td colspan="2">Tight (TRT)</td><td colspan="2">Tight (isol)</td></tr>
<tr><td>$W$</td><td colspan="2">75.0</td><td colspan="2">75.1</td><td colspan="2">74.9</td><td colspan="2">73.9</td><td colspan="2">73.6</td></tr>
<tr><td>$Z$</td><td colspan="2">20.9</td><td colspan="2">20.9</td><td colspan="2">21.1</td><td colspan="2">22.4</td><td colspan="2">22.9</td></tr>
<tr><td>$\tau$</td><td colspan="2">4.1</td><td colspan="2">4.0</td><td colspan="2">4.0</td><td colspan="2">3.7</td><td colspan="2">3.6</td></tr>
<tr><td colspan="11">Non-isolated</td></tr>
<tr><td></td><td colspan="5">$E_T > 17$ GeV</td><td colspan="5">$E_T > 8$ GeV</td></tr>
<tr><td></td><td>No cut</td><td>Loose</td><td>Medium</td><td>Tight (TRT)</td><td>Tight (isol)</td><td>No cut</td><td>Loose</td><td>Medium</td><td>Tight (TRT)</td><td>Tight (isol)</td></tr>
<tr><td>$b$-hadrons</td><td>38.7</td><td>57.6</td><td>71.1</td><td>74.2</td><td>79.1</td><td>39.3</td><td>51.2</td><td>55.2</td><td>57.0</td><td>59.5</td></tr>
<tr><td>$c$-hadrons</td><td>60.6</td><td>41.4</td><td>27.6</td><td>24.4</td><td>19.6</td><td>59.7</td><td>47.6</td><td>43.2</td><td>41.3</td><td>38.6</td></tr>
<tr><td>$J/\psi$</td><td>0.7</td><td>1.0</td><td>1.3</td><td>1.4</td><td>1.3</td><td>1.0</td><td>1.2</td><td>1.6</td><td>1.7</td><td>1.9</td></tr>
<tr><td colspan="11">Background</td></tr>
<tr><td></td><td colspan="5">$E_T > 17$ GeV</td><td colspan="5">$E_T > 8$ GeV</td></tr>
<tr><td></td><td>No cut</td><td>Loose</td><td>Medium</td><td>Tight (TRT)</td><td>Tight (isol)</td><td>No cut</td><td>Loose</td><td>Medium</td><td>Tight (TRT)</td><td>Tight (isol)</td></tr>
<tr><td>$\gamma$-conv.</td><td>97.8</td><td>97.7</td><td>94.9</td><td>88.0</td><td>88.1</td><td>98.4</td><td>98.1</td><td>94.5</td><td>78.5</td><td>83.0</td></tr>
<tr><td>Dalitz decays</td><td>1.8</td><td>1.9</td><td>4.0</td><td>8.5</td><td>8.0</td><td>1.3</td><td>1.4</td><td>3.5</td><td>12.5</td><td>12.4</td></tr>
<tr><td>$u/d/s$-hadrons</td><td>0.4</td><td>0.4</td><td>1.1</td><td>3.5</td><td>3.9</td><td>0.3</td><td>0.5</td><td>2.0</td><td>9.0</td><td>4.6</td></tr>
</table>

Table 4: Percentage contribution and origin of isolated, non-isolated and background electrons in the filtered di-jet and minimum-bias samples. The classification is based on the type of the parent particle of the electron.

$E_T \sim$ 10-40 GeV. The efficiency may be improved with further optimisation of the cuts, as discussed below.

Table 6 shows the efficiencies for prompt electrons and the jet rejections in more detail in the case of medium identification cuts, using a fine binning as a function of $|\eta|$. The efficiency for prompt electrons is significantly worse in the end-cap region ($|\eta| > 1.52$) with a correspondingly higher background rejection. The overlap region region between the barrel and end-cap calorimeters ($1.37 < |\eta| < 1.52$) has both worse efficiency and rejection, as expected because of the large amount of passive material in front of the EM calorimeter. To improve the electron efficiency in the end-cap region, the EM calorimeter cuts in the first layer and the tracking cuts will need to be studied and tuned further.

### 2.1.3 Expected differential rates for inclusive electron signal and background

Figure 3 (left: $E_T > 17$ GeV and right: $E_T > 8$ GeV) show the expected differential cross-sections for electron candidates as a function of $E_T$, for an integrated luminosity of 100 pb$^{-1}$. The different histograms correspond to electron candidates before any identification cuts and after the loose, medium, tight (TRT) and tight (isol) cuts. As illustrated in Table 5, these differential rates are dominated by the jet background except when applying the tight cuts.

The expected differential cross-sections after tight (TRT) cuts are shown in Fig. 4, where they are broken down into their three main components, isolated electrons from $W$, $Z$ and top-quark decays, non-isolated electrons from $b$, $c$ decay, and the residual jet background. The shapes of the spectra for the non-isolated electrons and residual jet background are very similar, whereas the spectrum from isolated electrons exhibits the expected behaviour for a sample dominated by electrons from $W$, $Z$ decay. For an integrated luminosity of 100 pb$^{-1}$, Fig. 4 (right) shows that one may expect approximately ten million reconstructed and identified inclusive electrons from $b$, $c$ decay with $E_T > 10$ GeV, while Fig. 4 (left) shows that for the same integrated luminosity one may expect 500 000 such electrons with $E_T > 20$ GeV, with a dominant contribution from $W$, $Z$ decays for $E_T > 35$ GeV. These large data samples expected for a modest integrated luminosity are an integral part of the trigger menu strategy for early data, as explained in more detail in [11], and will clearly be extremely useful to certify many aspects of the electron identification performance of ATLAS with real data. One example is the understanding of material effects and of inter-calibration between inner detector and EM calorimeter using $E/p$ for a clean subset of the inclusive electrons with $E_T > 10$ GeV. This sample will be complementary to the samples of low-mass electron pairs from $J/\psi$ and $\Upsilon$ decays, discussed in [7]. A second example is the certification of the isolated electron identification using a clean sample of $W \rightarrow e\nu$ decays. Clearly, with more statistics, the large samples of $Z \rightarrow ee$ decays which will be collected will provide the opportunity to refine the understanding of the performance to an extremely high level of accuracy, as discussed in Section 5.

| Cuts | $E_T > 17$ GeV | | | $E_T > 8$ GeV | | |
|---|---|---|---|---|---|---|
| | Efficiency (%) | | Jet rejection | Efficiency (%) | | Jet rejection |
| | $Z \rightarrow ee$ | $b,c \rightarrow e$ | | Single electrons ($E_T$ =10 GeV) | $b,c \rightarrow e$ | |
| Loose | 87.96 ± 0.07 | 50.8 ± 0.5 | 567 ± 1 | 75.8 ± 0.1 | 55.8 ± 0.7 | 513 ± 2 |
| Medium | 77.29 ± 0.06 | 30.7 ± 0.5 | 2184 ± 13 | 64.8 ± 0.1 | 41.9 ± 0.7 | 1288 ± 10 |
| Tight (TRT.) | 61.66 ± 0.07 | 22.5 ± 0.4 | $(8.9 \pm 0.3)10^4$ | 46.2 ± 0.1 | 29.2 ± 0.6 | $(6.5 \pm 0.3)10^4$ |
| Tight (isol.) | 64.22 ± 0.07 | 17.3 ± 0.4 | $(9.8 \pm 0.4)10^4$ | 48.5 ± 0.1 | 28.0 ± 0.6 | $(5.8 \pm 0.3)10^4$ |
| | Fraction of surviving candidates (%) | | | Fraction of surviving candidates (%) | | |
| | Isolated | Non-isolated | Jets | Non-isolated | | Jets |
| Medium | 1.1 | 7.4 | 91.5 (5.5 + 86.0) | 9.0 | | 91.0 (5.0 + 86.0) |
| Tight (TRT) | 10.5 | 63.3 | 26.2 (8.3 + 17.9) | 77.8 | | 22.2 (7.1 + 15.1) |
| Tight (isol) | 13.0 | 58.3 | 28.6 (8.7 + 19.9) | 75.1 | | 24.9 (6.4 + 18.5) |

Table 5: Expected efficiencies for isolated and non-isolated electrons and corresponding jet background rejections for the four standard levels of cuts used for electron identification. The results are shown for the simulated filtered di-jet and minimum-bias samples, corresponding respectively to $E_T$ -thresholds of 17 GeV (left) and 8 GeV (right). The three bottom rows show the fractions of all surviving candidates which fall into the different categories for the medium cuts and the two sets of tight cuts. The isolated electrons are prompt electrons from $W$, $Z$ and top-quark decay and the non-isolated electrons are from $b$, $c$ decay. The residual jet background is split into its two dominant components, electrons from photon conversions and Dalitz decays (first term in brackets) and charged hadrons (second term in brackets). The quoted errors are statistical.

### 2.1.4 Systematic uncertainties on expected performance

To estimate possible systematic uncertainties related to the cut-based electron identification, two shower shape variables have been studied as a function of the amount of material in front of the EM calorimeter. Figure 5 illustrates the impact of additional material, the effect of which has not been included in the EM cluster corrections which are applied as described in [3], for electrons from $H \rightarrow eeee$ decays. The results are shown in two $|\eta|$-ranges for the nominal material and for the case of additional material accounting in total to $\sim 0.1\ X_0$ and $\sim 0.2\ X_0$ (Fig. 5). It is evident that in regions with significant amounts of material the shower is broader (less energy in the core). These differences reduce the electron efficiency; however, the true systematic error on the efficiency due to such effects will depend on how well the inner-detector material can be measured using data.

Figure 6 shows the fraction of energy in the strip layer outside the three core strips and inside the seven-strip window for the same $|\eta|$-ranges. The impact of the additional material is also clearly visible. The estimated change in the electron efficiencies quoted in Table 5 is expected to be less than 2%. It is important to note that the material effects are more pronounced in the strip layer than in the middle layer of the calorimeter. Therefore, one should expect larger uncertainties from this source of systematics for the medium electron cuts than for the loose electron cuts, which rely only on the middle layer of the calorimeter.

| $\|\eta\|$ | $E_T > 17$ GeV | | | $E_T > 8$ GeV | | |
|---|---|---|---|---|---|---|
| | Efficiency (%) | | Jet rejection | Efficiency (%) | | Jet rejection |
| | $Z \rightarrow ee$ | $b,c \rightarrow e$ | | Single electrons ($E_T$ =10 GeV) | $b,c \rightarrow e$ | |
| 0.00 − 0.80 | 88.2 ± 0.1 | 35 ± 1 | 3740 ± 50 | 79.3 ± 0.2 | 51 ± 1 | 1960 ± 30 |
| 0.80 − 1.35 | 83.5 ± 0.1 | 40 ± 1 | 1581 ± 20 | 70.6 ± 0.2 | 52 ± 1 | 914 ± 11 |
| 1.35 − 1.50 | 71.5 ± 0.4 | 41 ± 2 | 444 ± 5 | 49.6 ± 0.5 | 40 ± 3 | 342 ± 5 |
| 1.50 − 1.80 | 63.8 ± 0.2 | 18 ± 1 | 2440 ± 40 | 41.8 ± 0.4 | 24 ± 2 | 890 ± 15 |
| 1.80 − 2.00 | 62.5 ± 0.2 | 12 ± 1 | 9800 ± 450 | 55.1 ± 0.4 | 25 ± 2 | 4660 ± 220 |
| 2.00 − 2.35 | 65.8 ± 0.2 | 16 ± 1 | 8400 ± 300 | 55.0 ± 0.3 | 21 ± 2 | 6000 ± 250 |
| 2.35 − 2.47 | 67.8 ± 0.3 | 14 ± 2 | 4050 ± 170 | 62.5 ± 0.6 | 30 ± 3 | 3980 ± 250 |
| 0.00 − 2.47 | 77.3 ± 0.06 | 31 ± 1 | 2184 ± 13 | 64.8 ± 0.1 | 42 ± 1 | 1288 ± 8 |

Table 6: Expected efficiencies for isolated and non-isolated electrons and corresponding jet background rejections for the medium identification cuts as a function of $|\eta|$. The results are shown for the simulated filtered di-jet and minimum-bias samples, corresponding respectively to $E_T$ -thresholds of 17 GeV (left) and 8 GeV (right). The quoted errors are statistical.

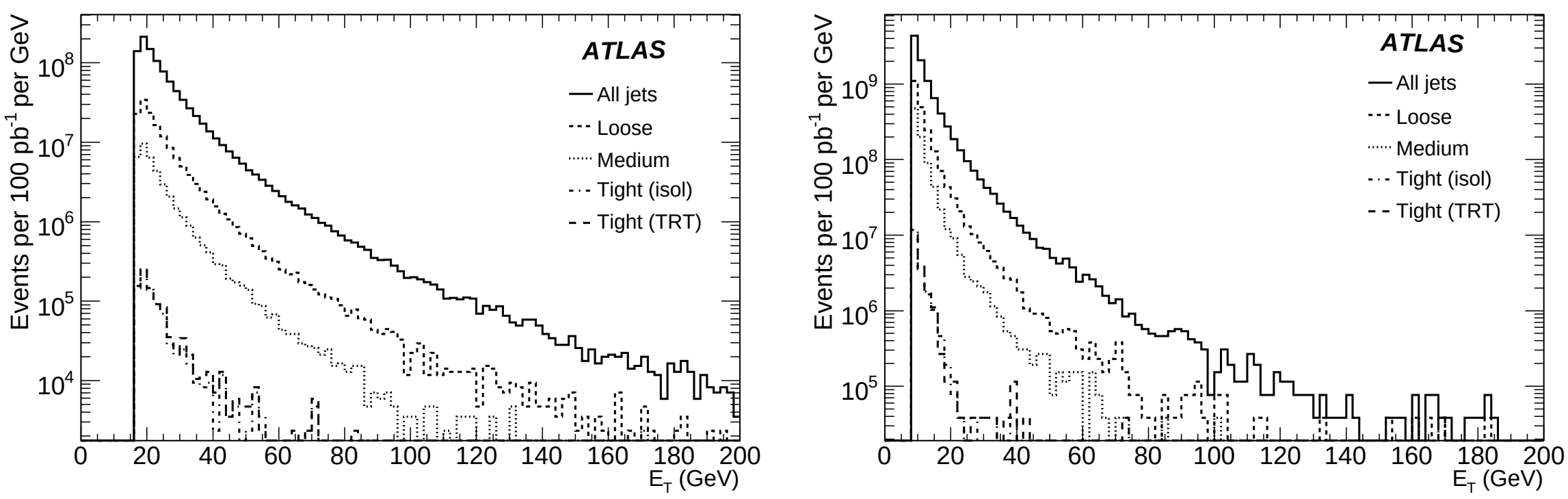

Figure 3: Differential cross-sections as a function of $E_T$ before identification cuts and after loose, medium, tight (TRT) and tight-isol cuts, for an integrated luminosity of 100 pb$^{-1}$ and for the simulated filtered di-jet sample with $E_T$ above 17 GeV (left) and the simulated minimum-bias sample with $E_T$ above 8 GeV (right).

Another important source of systematics affects the jet rejections quoted in Table 5: this arises from the exact $p_T$ -spectrum and mixture of quark and gluon jets, and to a certain extent from heavy flavour jets present in the background under consideration. The numbers quoted in this note are related to the rather low-$p_T$ di-jet background which is relevant for the search for early signals from single electrons. Other background samples relevant to certain physics studies have been shown to display worse rejections, by up to a factor of 3 to 5. This clearly indicates that the fake electron rates will only be better understood with real data.

### 2.1.5 Multivariate techniques

In addition to the standard cut-based electron identification described above, several multivariate techniques have been developed and implemented in the ATLAS software. These include a likelihood discriminant, a discriminant called H-matrix, a boosted decision tree, and a neural network. Table 7 sum-marises the gains in efficiency and rejection which may be expected with respect to the cut-based method by using the likelihood discriminant method. The gains appear to be artificially large in the case of the loose and medium cuts, because these cuts do not make use of all the information available in terms of electron identification, since they were designed for robustness and ease of use with initial data. Nevertheless, they indicate how much the electron efficiency may be improved once all the discriminant variables will be understood in the data.

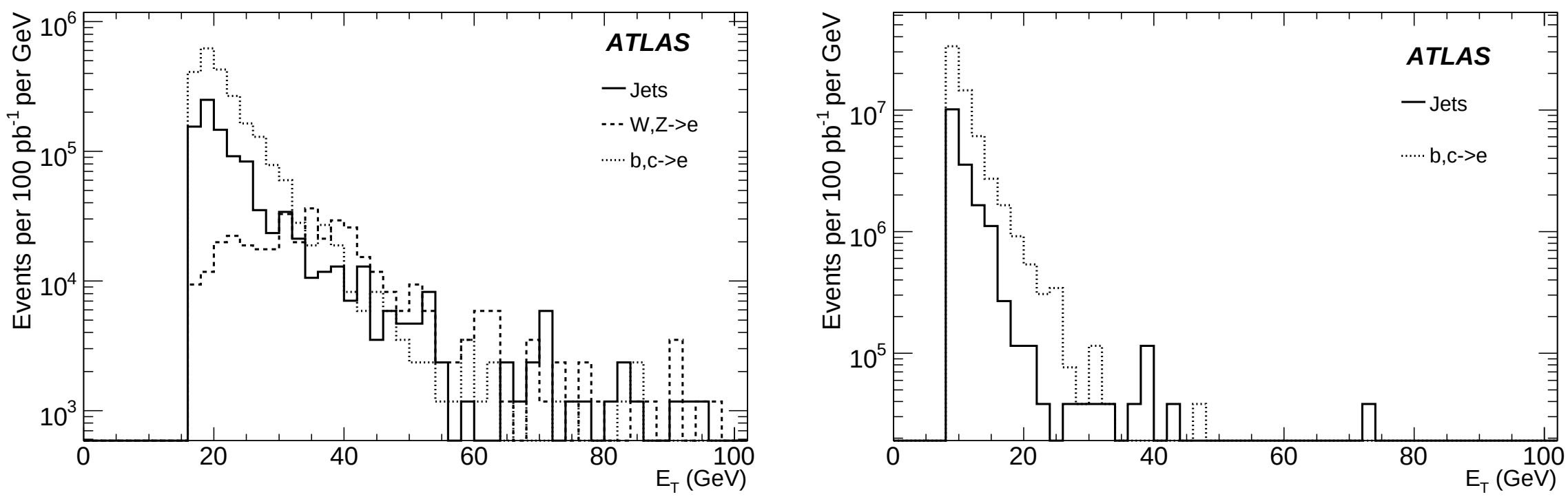

Figure 4: Differential cross-sections as a function of $E_T$ after tight (TRT) cuts, shown separately for the expected components from isolated electrons, non-isolated electrons and residual jet background, for an integrated luminosity of 100 pb$^{-1}$ and for the simulated filtered di-jet sample with $E_T$ above 17 GeV (left) and the simulated minimum-bias sample with $E_T$ above 8 GeV (right).

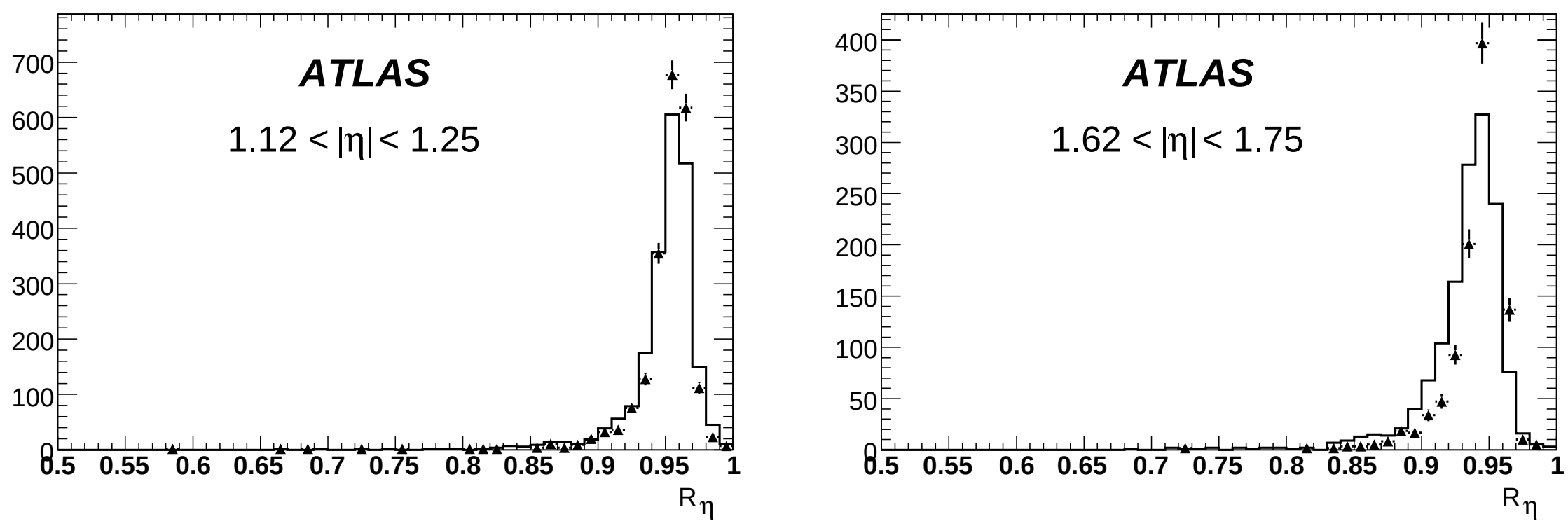

Figure 5: Energy containment, $R_\eta$ (Table 3), for $1.12 < |\eta| < 1.25$ (left) and $1.62 < |\eta| < 1.75$ (right). The symbols correspond to the nominal description and the histogram to the one with additional material.

Figure 7 shows the rejection versus efficiency curve obtained using the likelihood discriminant method, compared to the results obtained for the two sets of tight cuts shown in Table 5. The likelihood discriminant method provides a gain in rejection of about 20-40% with respect to the cut-based method for the same efficiency of 61-64%. Alternatively, it provides a gain in efficiency of 5-10% (tight and medium cuts) for the same rejection. Multivariate methods of this type will of course only be used once the detector performance has been understood using the simpler cut-based electron identification criteria.

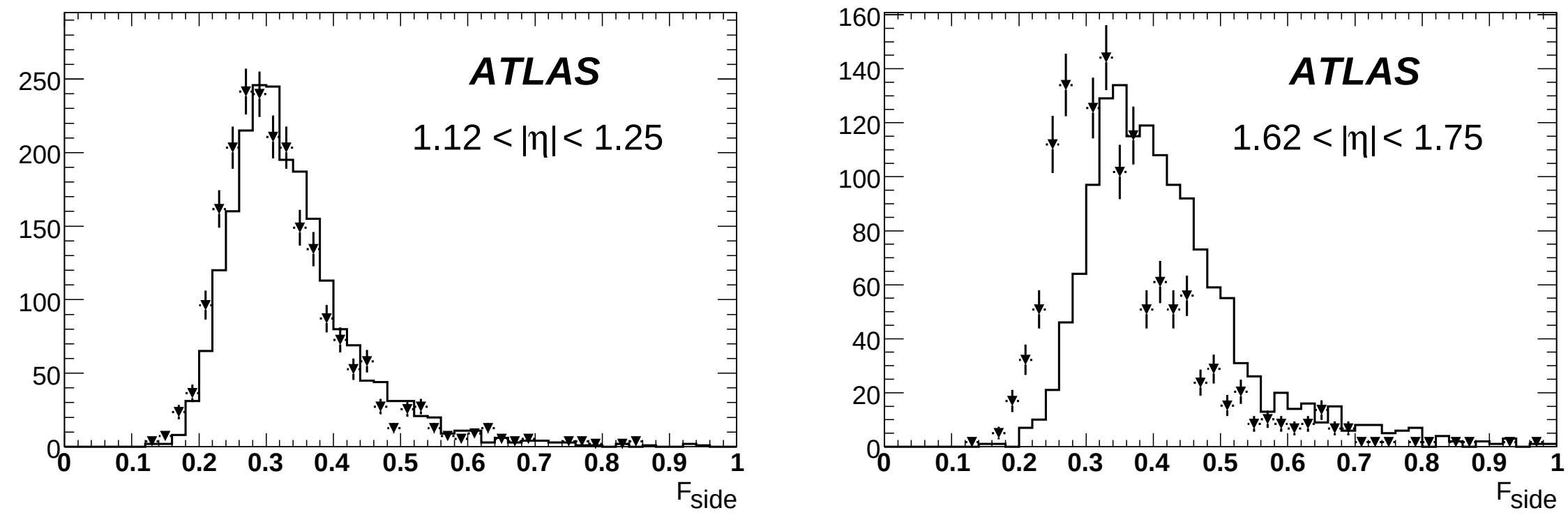

Figure 6: Energy fraction outside a three-strip core, $F_{side}$ (Table 3), for $1.12 < |\eta| < 1.25$ (left) and $1.62 < |\eta| < 1.75$ (right). The symbols correspond to the nominal description and the histogram to the one with additional material.

| **Cuts** | **Cut-based method** | | **Likelihood method** | |
|---|---|---|---|---|
| | Efficiency $\varepsilon_e$ (%) | Rejection $R_j$ | Efficiency (%) at fixed $R_j$ | Rejection at fixed $\varepsilon_e$ |
| Loose | $87.97 \pm 0.05$ | $567 \pm 1$ | $89.11 \pm 0.05$ | $2767 \pm 17$ |
| Medium | $77.29 \pm 0.06$ | $2184 \pm 7$ | $88.26 \pm 0.05$ | $(3.77 \pm 0.08) \times 10^4$ |
| Tight (isol) | $64.22 \pm 0.07$ | $(9.9 \pm 0.2) \times 10^4$ | $67.53 \pm 0.06$ | $(1.26 \pm 0.05) \times 10^5$ |
| Tight (TRT) | $61.66 \pm 0.07$ | $(8.9 \pm 0.2) \times 10^4$ | $68.71 \pm 0.06$ | $(1.46 \pm 0.06) \times 10^5$ |

Table 7: For the loose, medium and tight electron identification cuts, expected electron efficiencies for a fixed jet rejection and jet rejections for a fixed electron efficiency, as obtained from the likelihood discriminant method. The quoted errors are statistical.

## 2.2 Isolation studies

Many physics analyses in ATLAS will be based on final states with isolated leptons from decays of $W$- or $Z$-bosons. These channels usually have the advantage of small background expectation from processes with similar signature, compared to channels with hadronic final states. Nevertheless, they may also suffer from jet background processes, namely if leptons from semi-leptonic heavy-quark decays mimic the isolated leptons of the signal. Therefore, dedicated tools beyond the lepton identification algorithms are needed in order to suppress such sources of background by factors of up to the order of $10^3$. In this section, the performance of a projective likelihood estimator for the separation of isolated electrons from non-isolated electron backgrounds is described. The four variables chosen as input to this isolation likelihood are:

- transverse energy deposited in a small cone of $\Delta R < 0.2$ around the electron cluster;
- transverse energy deposited in a hollow cone of $0.2 < \Delta R < 0.4$ around the electron cluster;
- sum of the squares of the transverse momenta of all additional tracks measured in a cone of $\Delta R < 0.4$ around the electron cluster;
- impact parameter significance of the electron track (with respect to the primary vertex in the transverse plane).

Electrons from $Z \to ee$ decays were used as a clean source of isolated electrons. The reconstructed electrons from this sample were required to be matched to a Monte Carlo electron from $Z$-boson decay and to pass the medium identification cuts in order to be considered as signal electrons. Background electrons were selected from a high-statistics $t\bar{t}$ sample, filtered for a pair of like-sign Monte Carlo electrons, and matched to a Monte Carlo electron from $b/c$-decay.

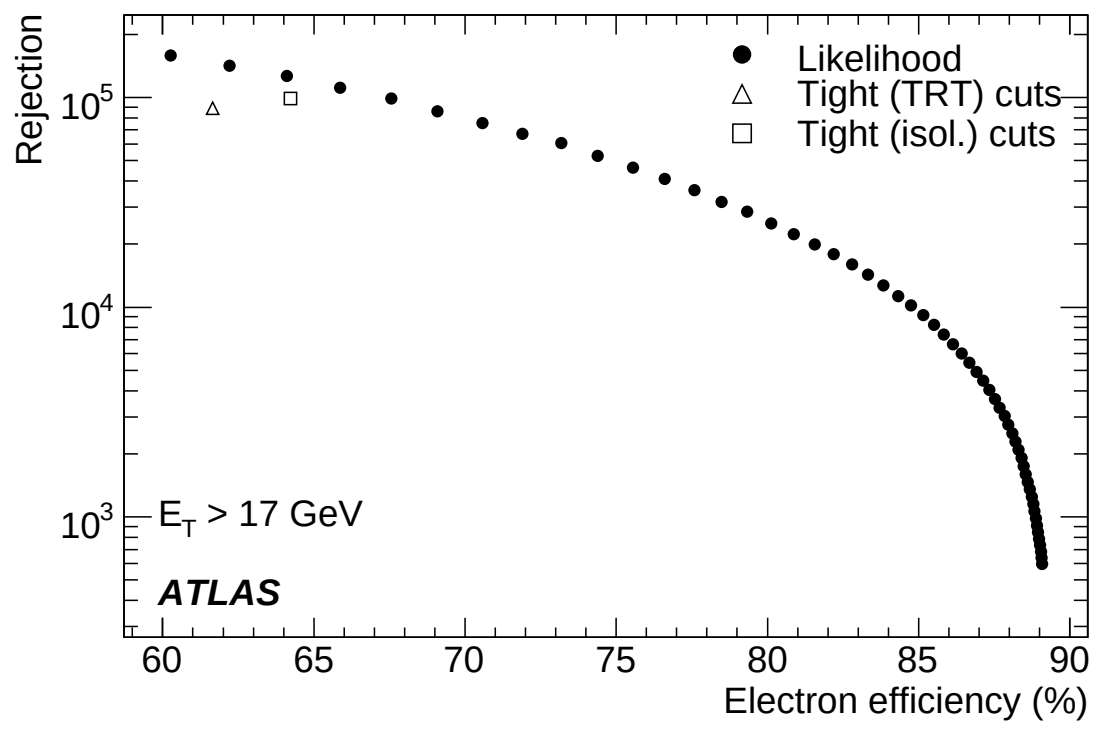

Figure 7: Jet rejection versus isolated electron efficiency obtained with a likelihood method (full circles) compared to the results from the two sets of tight cuts (open triangle and open square).

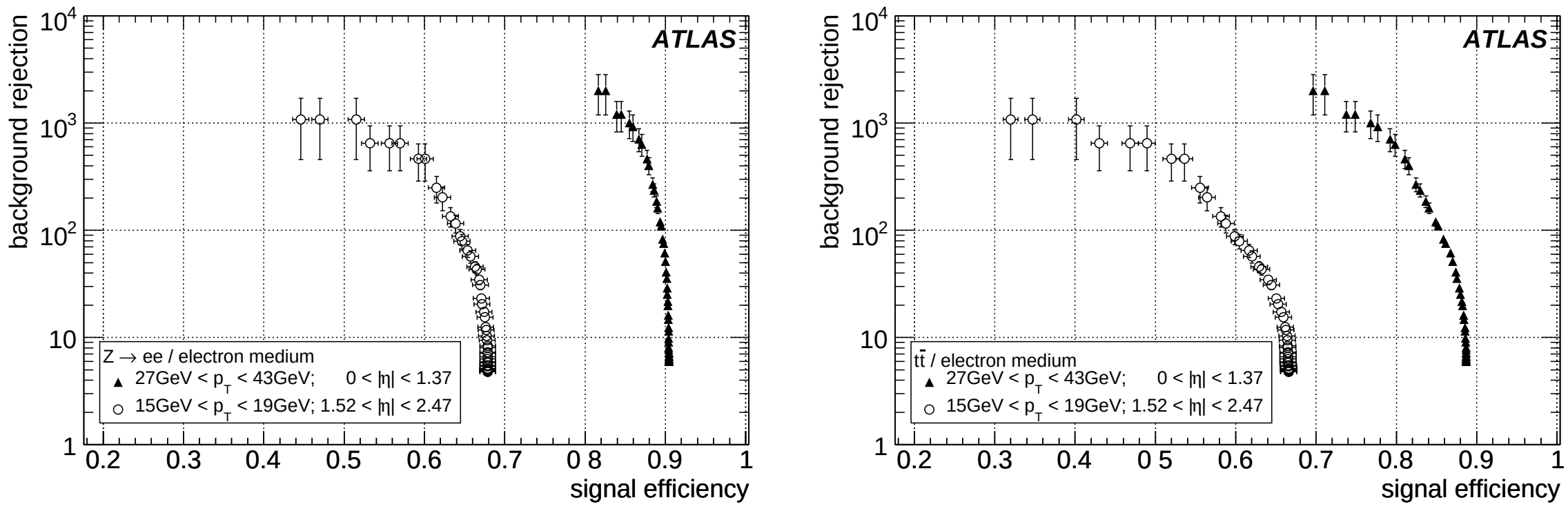

Figure 8: Background electron rejections versus signal efficiencies for electrons in $Z \rightarrow ee$ decays (left) and in $t\bar{t}$ decays (right), for two illustrative bins in $|\eta|$ and $p_T$ .

The results of the performance studies of the isolation likelihood are shown in Fig. 8 for two illustrative bins in $|\eta|$ and $p_T$ . The best results are achieved for high-$p_T$ electrons measured in the barrel region of the EM calorimeter. As can be seen in Fig. 8 left, for electrons with only little hadronic activity in the final state, such as those from $Z \rightarrow ee$ and $H \rightarrow eeee$ decays, the isolation likelihood provides a background rejection of the order of $10^3$, for signal electron efficiencies of 80% (barrel) and 50% (end-caps). The difference observed between barrel and end-caps is mostly due to the $\eta$-dependence of the medium identification cuts shown in Table 6. For comparison, the efficiency for the selection of signal electrons in $t\bar{t}$ events is shown in Fig. 8 right: due to the additional hadronic activity in these final states, the efficiency decreases by 5–10% for the same background rejection, when compared to that quoted for $Z \rightarrow ee$ decays.

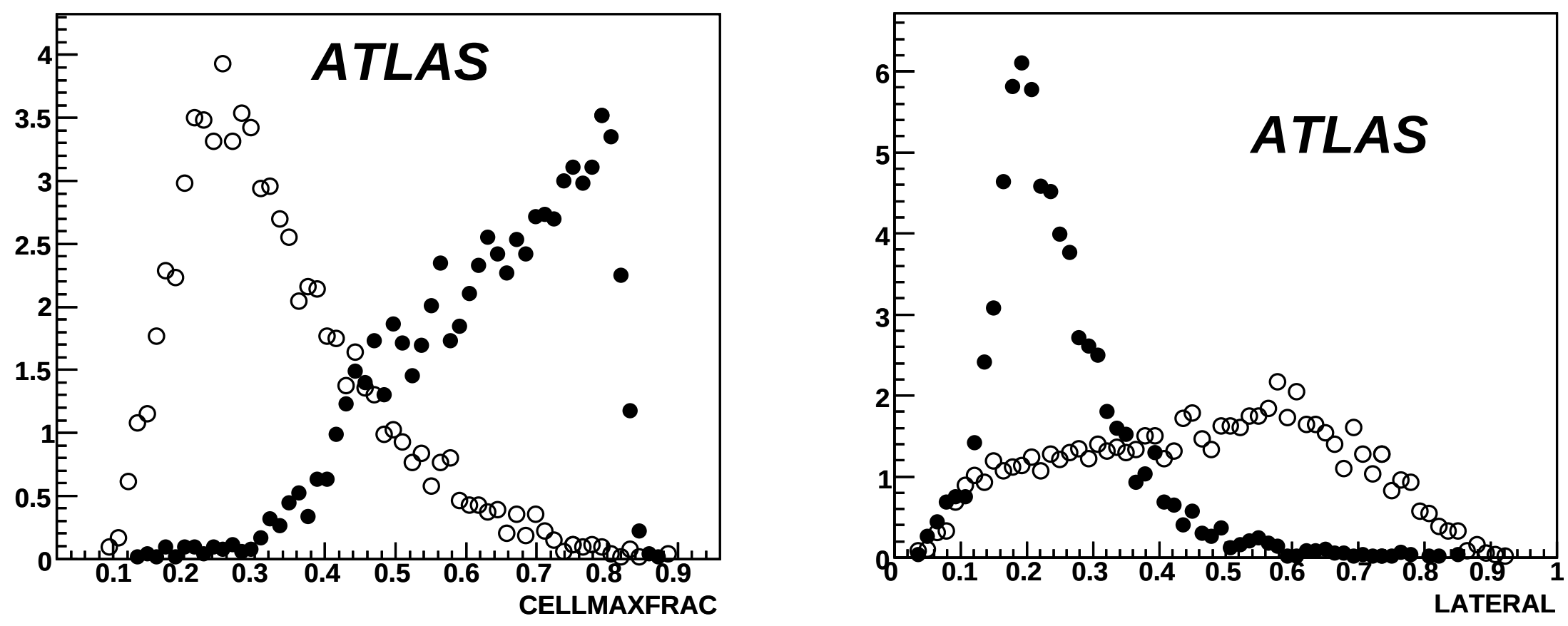

Figure 9: Example of discriminating variables used in the forward region for signal electrons (full circles) and the QCD di-jet background (open circles). Shown in the case of the FCal are the fraction of the total cluster energy deposited in the cell with maximum energy (left) and the relative lateral moment (right).

# 3 Electron identification outside the inner detector acceptance

Electron identification in the forward region ($|\eta| > 2.5$) will be important in many physics analyses, including electroweak measurements and searches for new phenomena. In contrast to the central electrons, forward electron reconstruction can only use information from the calorimeters, since the inner detector covers only $|\eta| < 2.5$. Such electrons can therefore only be identified cleanly above the background in specific topologies, such as $Z \to ee$ or $H \to eeee$ decays.

This section describes the performance of a cut-based method used to identify electrons in the forward region and separate them from the QCD background. The comparison of the performance obtained with a likelihood method is also presented.

Signal electrons are selected from $Z \to ee$ decays and background electrons from a high-statistics sample of QCD di-jet events. Three $|\eta|$-regions are considered: the first one covers the inner wheel of the electromagnetic end-cap, i.e. $2.5 < |\eta| < 3.2$ (the HEC is not used), the second one covers the overlap region between the electromagnetic end-cap and the forward calorimeter (FCal), i.e. $3.2 < |\eta| < 3.4$, and the last region covers the FCal acceptance, i.e. $3.4 < |\eta| < 4.9$. A topological clustering algorithm [13] is used in this analysis and only clusters with $E_T > 20$ GeV are considered. Two examples of the discriminating variables used in these studies are shown in Fig. 9, namely the fraction of the total cluster energy deposited in the cell with maximum energy and the relative lateral moment. The relative lateral moment is defined as $\mathrm{lat}_2/(\mathrm{lat}_2 + \mathrm{lat}_{\mathrm{max}})$, where the lateral moments $\mathrm{lat}_2$ and $\mathrm{lat}_{\mathrm{max}}$ differ in the treatment of the two most energetic cells. Other examples include the first moment of the energy density, the relative longitudinal moment, defined in the same way as the relative lateral moment only with two longitudinal moments, the second moments of the distances of each cell to the shower barycentre and to the shower axis, and the distance of the cluster barycentre from the front face of the calorimeter.

The likelihood discriminant uses the same variables as the cut-based method. Figure 10 shows the performance of the cut-based and likelihood discriminant methods for electrons from $Z \to ee$ decay with $E_T > 20$ GeV. For an electron identification efficiency of 80%, both methods achieve the required goal of $\sim 1\%$ fake rate from the QCD background. This performance is expected to yield, for example, a clean $Z \to ee$ sample with one electron already selected in the central region and one electron in the forward region [14]: the expected background contribution under the $Z$-boson peak is estimated to be below $\sim$ 1%.

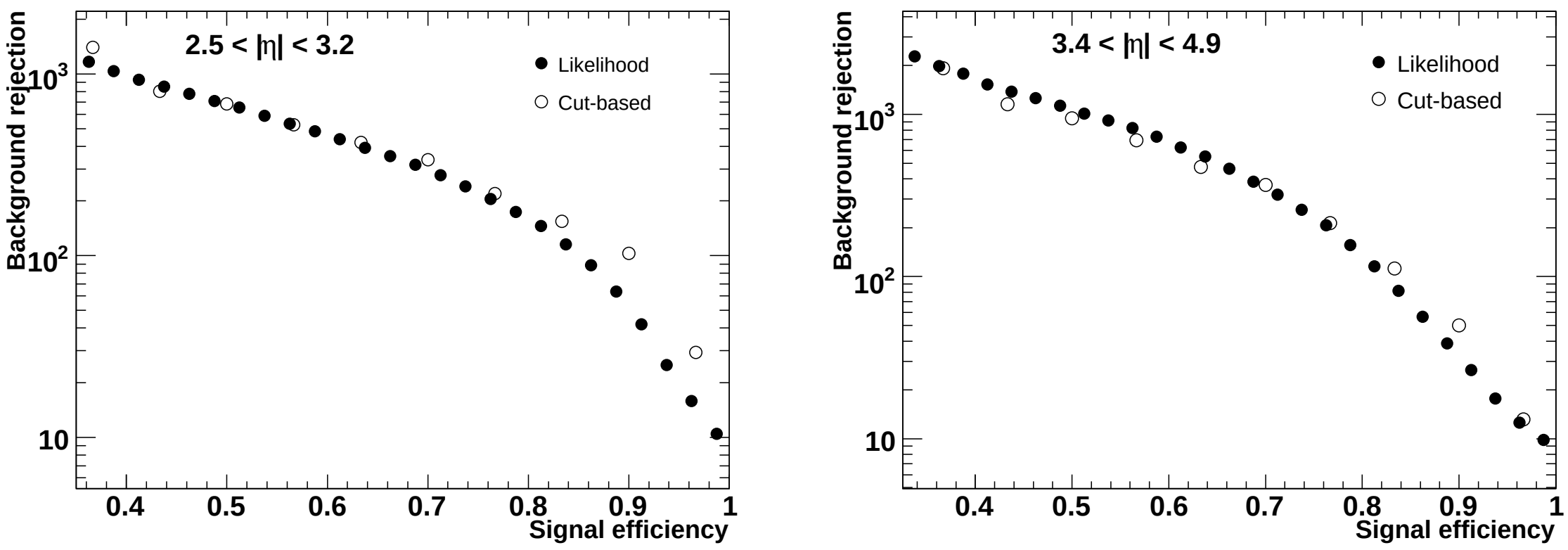

Figure 10: Expected rejection against QCD jets versus efficiency for signal electrons from $Z \rightarrow ee$ decay, for the cut-based and likelihood discriminant methods in the inner wheel of the electromagnetic end-cap (left) and in the FCal (right). The rejection power of the likelihood method is expected to increase when additional variables beyond the minimal set shown here are added.

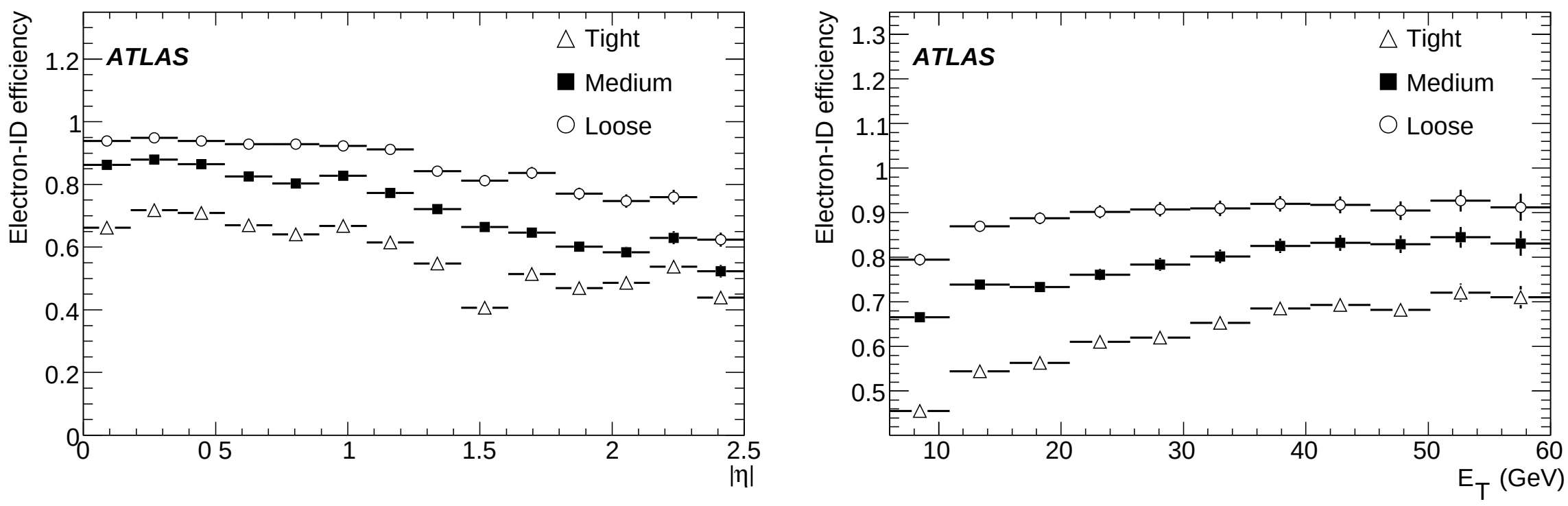

Figure 11: Electron identification efficiency as a function of $\eta$ (left) and $E_T$ (right) for electrons with $E_T > 5$ GeV from $H \rightarrow eeee$ decays.

# 4 Electrons as probes for physics within and beyond the Standard Model

## 4.1 Electrons in Higgs-boson decays

Electrons from the $H \rightarrow eeee$ decay with $m_H < 2m_Z$ are an important benchmark for the evaluation of the performance of the electron reconstruction and identification [15]. Here, only electrons with $|\eta| < 2.5$ and $E_T > 5$ GeV are considered. The electron efficiency as a function of $|\eta|$ and $E_T$ for loose, medium, and tight electron cuts is shown in Fig. 11. The drop in efficiency at low $E_T$ is mainly due to the loss of discrimination power of the shower-shape cuts at lower transverse energies. A loss of efficiency is also visible in the transition region between the barrel and end-cap calorimeters. The results shown here are in quantitative agreement with those obtained for electrons from $Z \rightarrow ee$ decay discussed in Section 2.1.2.

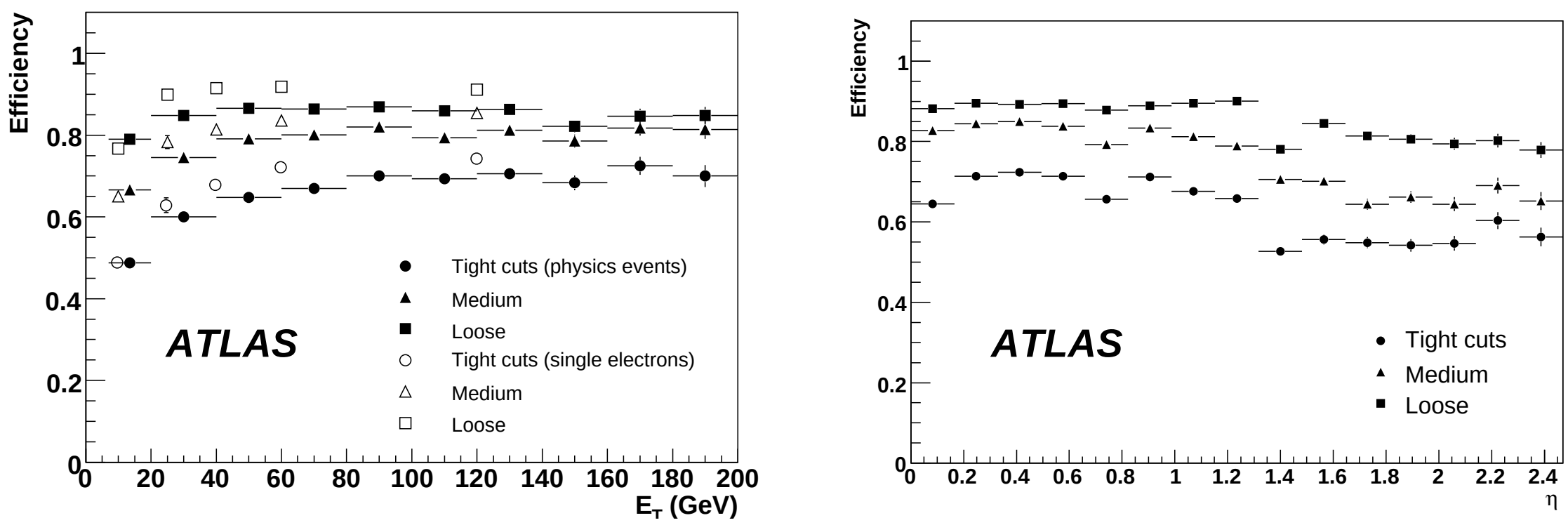

Figure 12: Electron identification efficiency as a function of $E_T$ (left) and $|\eta|$ (right). The full symbols correspond to electrons in SUSY events and the open ones to single electrons of fixed $E_T$. The efficiencies as a function of $|\eta|$ are shown only for electrons with $E_T > 17$ GeV.

## 4.2 Electrons produced in decays of supersymmetric particles

In many supersymmetry (SUSY) scenarios, the most abundantly produced sparticles are squarks (directly or from a gluino decay), which generally decay into a chargino or neutralino and jets. In turn, charginos and neutralinos are very likely to decay into leptons. One interesting mode for SUSY searches is the tri-lepton signal, in which three isolated leptons are expected in the final state. Such SUSY events would feature high-$p_T$ isolated leptons accompanied by a high multiplicity of high-$E_T$ jets. Hence, it is crucial to efficiently identify electrons in such an environment, while preserving the very high jet rejection presented in Section 2. The electron identification efficiency in SUSY events is calculated using the SU3 ATLAS point [16]. In this scenario, a large number of charginos and neutralinos are produced and numerous leptons are expected in the final state.

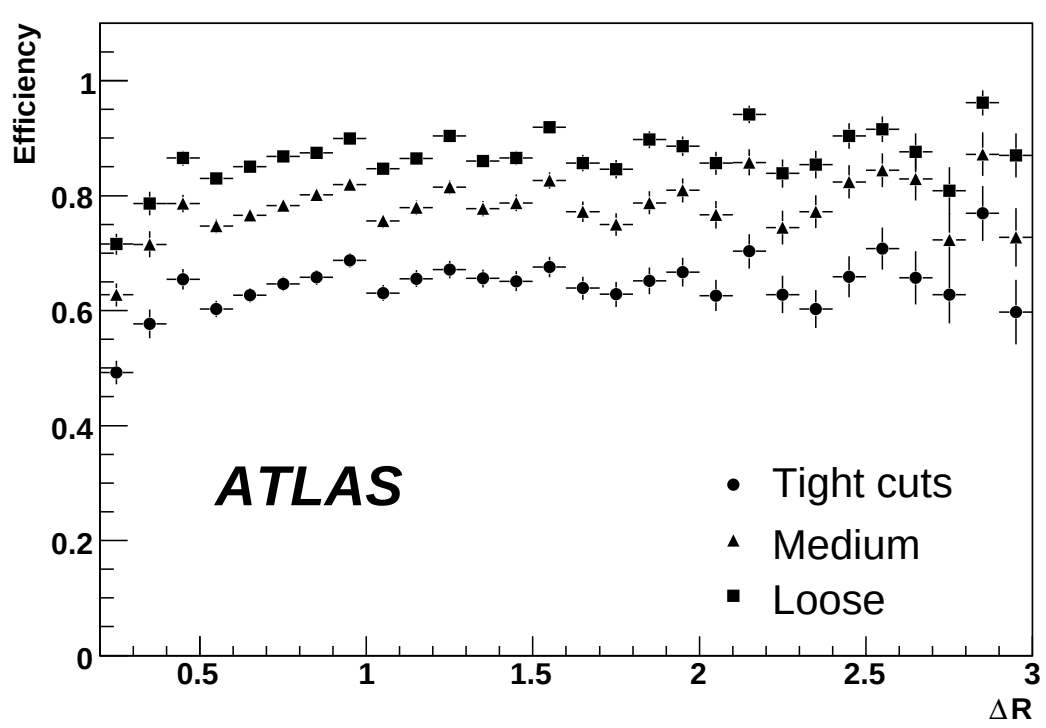

Figure 13: Electron identification efficiency as a function of the distance $\Delta R$ to the closest jet in SUSY events, for electrons with $E_T > 17$ GeV.

Figure 12 shows the identification efficiency of the loose, medium and tight (isol) cuts as a function of $E_T$ and $|\eta|$. The efficiencies shown as a function of $E_T$ are compared with efficiencies for single electrons of $E_T = 10, 25, 40, 60$ and 120 GeV. As expected, single electrons display higher efficiencies than those in SUSY events, because of the large hadronic activity in these events. The efficiencies

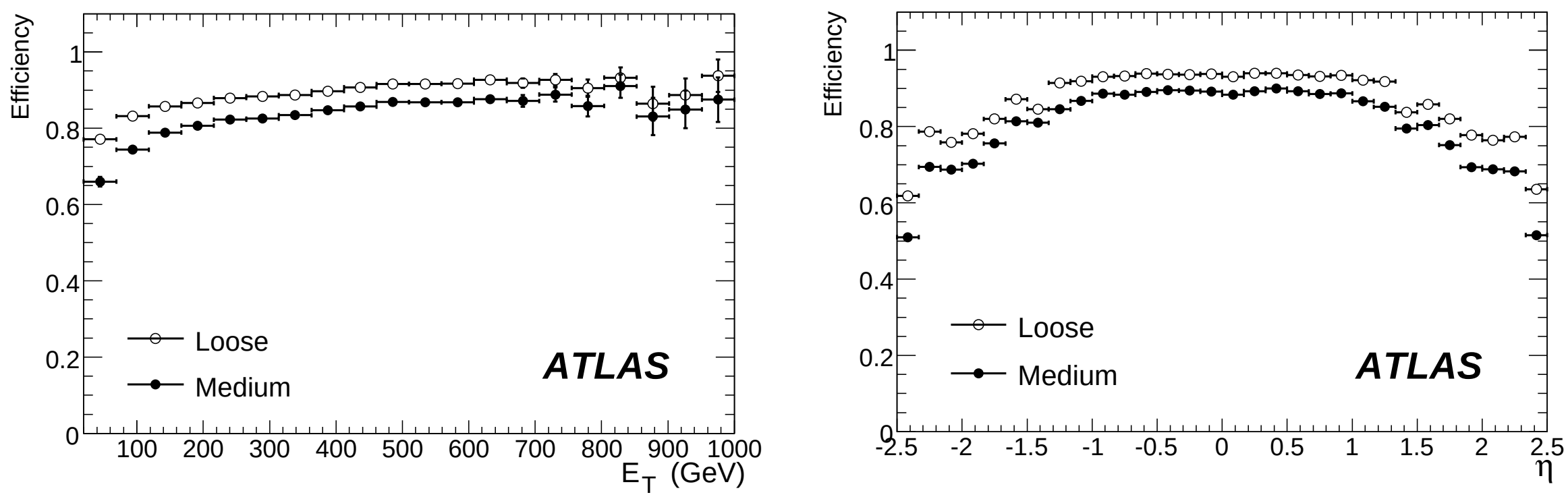

Figure 14: Electron identification efficiency as a function of $E_T$ (left) and $|\eta|$ (right), for electrons from $Z' \to e^+e^-$ decays with $m_{Z'}$ = 1 TeV.

obtained for values of $E_T$ below 20 GeV, are significantly below the plateau values at high $E_T$, for which the cuts were initially optimised.

The efficiencies as a function of $|\eta|$ show the same features as those discussed in Table 6, namely the efficiency in the end-cap region is lower than in the barrel, whereas the jet rejection is significantly higher. Specific drops in efficiency can be seen for $|\eta| \sim 1.35$, which corresponds to the barrel/end-cap transition region, and for $|\eta| \approx 0.8$, which corresponds to the change in the lead thickness between the two types of electrodes in the barrel EM calorimeter.

Figure 13 shows the electron identification efficiency as a function of the distance $\Delta R$ to the closest jet in SUSY events. Jets are reconstructed from topological clusters using a $\Delta R = 0.4$ cone algorithm. For values of $\Delta R > 0.4$, the efficiencies are compatible with those expected for single electrons, whereas for values of $\Delta R < 0.4$, the efficiencies decrease because of the overlap between the hadronic showers from the jet and the electron shower itself.

| Jet $E_T$ -range | 140 – 280 GeV | | 280 – 560 GeV | | 560 – 1120 GeV | |
|---|---|---|---|---|---|---|
| | Efficiency | Rejection | Efficiency | Rejection | Efficiency | Rejection |
| Loose cuts | $86.6 \pm 0.2\%$ | $825 \pm 35$ | $89.6 \pm 0.1\%$ | $620 \pm 25$ | $91.5 \pm 0.4\%$ | $550 \pm 20$ |
| Medium cuts | $80.6 \pm 0.2\%$ | $4000 \pm 370$ | $84.6 \pm 0.1\%$ | $2300 \pm 170$ | $86.7 \pm 0.5\%$ | $1900 \pm 120$ |

Table 8: Electron identification efficiencies and QCD di-jet background rejections obtained for loose and medium identification cuts, including a calorimeter isolation cut (see text), and for three different jet $E_T$ -ranges. The signal electrons are from $Z' \to e^+e^-$ decays with $m_{Z'}$ = 1 TeV and are required to have $E_T$ > 100 GeV.

## 4.3 Electrons in exotic events

High-mass di-electron final states are a promising source of early discovery physics, because of the simplicity and robustness of very high-$p_T$ electron reconstruction, identification and resolution. Very high-$p_T$ electrons refer here to those with transverse momentum ranging from 100 GeV up to several TeV. The backgrounds to very high-$p_T$ electron pairs are expected to be small, and, therefore, only loose or medium identification cuts are considered here. Isolated electrons are required to satisfy the calorimeter isolation cut described in Section 2.

Figure 14 shows efficiencies as a function of $E_T$ and $|\eta|$ for the loose and medium identification cuts, for electrons from $Z' \to e^+e^-$ decays with $m_{Z'}$ = 1 TeV [17]. From these curves, one can note the slow increase in efficiency with $E_T$ before reaching a plateau in the very high-$E_T$ region. Overall efficiencies of $\sim$ 90% and of $\sim$ 85% can be achieved for loose and medium electron cuts, respectively, with a uniform behaviour limited to the barrel region, i.e. $|\eta| < 1.5$.

The QCD background rejection was studied as a function of the jet transverse energy, as shown in Table 8. Using the medium identification cuts, which correspond to an overall efficiency of $\sim$ 85%, a jet rejection factor of several thousand can be achieved for $E_T > 100$ GeV, which should be sufficient to observe the signal in many exotic scenarios.

# 5 Electrons from $Z \to ee$ decays in early data

The experimental uncertainty on the electron identification efficiency is expected to be the source of one of the main systematic errors in many measurements, and in particular in cross-section determinations. In addition, a reliable monitoring of the electron identification efficiency is important in the commissioning phase of the detector and software. The previous sections have shown detailed estimates of the expected electron identification efficiency based on simulated samples. This section focuses on the measurement of electron reconstruction and identification efficiencies using a data-driven approach based on $Z \to ee$ events.

The tag-and-probe method [18] is used in this analysis. It consists of tagging a clean sample of events using one electron, and then measuring the efficiency of interest using the second electron from the $Z$-boson decay. Although more difficult because of trigger-threshold issues and of more severe background conditions, the same approach could be applied to $J/\psi$ and $\Upsilon$ resonances, thus covering the lower end of the $p_T$ spectrum [7].

## 5.1 Tag-and-probe method

The tag condition typically requires an electron identified with tight cuts. Both electrons are also required to be above a $p_T$ threshold consistent with the trigger used. The invariant mass of the lepton pair is then used to identify the number of tagged events, $N_1$ (containing $Z \to ee$ decays), and a sub-sample $N_2$, where the second pre-selected electron further passes a given set of identification cuts. The efficiency for a given signature is given by the ratio between $N_2$ and $N_1$.

To account for background, the lepton-pair invariant mass spectrum is fitted around the $Z$ mass peak using a Gaussian distribution convoluted with a Breit-Wigner plus an exponential function. The dominant background arises from QCD and is estimated using a procedure explained in [18]; its contribution is small in general and its impact on the measurement is therefore very limited.

The probe electron is checked against the selection as an electron candidate (to which only the pre-selection cuts are applied), and as a loose, medium or tight electron. To monitor in detail the efficiency dependence, the results are presented in bins of $\eta$ and $p_T$, at the expense of an increased statistical error in each bin.

A quantitative comparison between the efficiency computed with this tag-and-probe method ($\varepsilon_{TP}$) and the efficiency obtained from the Monte Carlo truth ($\varepsilon_{MC}$) is used to validate the tag-and-probe method.

## 5.2 Electron reconstruction efficiency

The reconstruction and identification of electrons is based on seed-clusters in the electromagnetic calorimeter matched to tracks, as explained in Section 2. The tag electron is a reconstructed electron selected using tight (isol) cuts and also required to pass the trigger EM13i/e15i [11]. The tag electron is also required to be outside the barrel/end-cap transition region ($1.37 < |\eta| < 1.52$). The probe electron is pre-selected by identifying a cluster in the opposite hemisphere, such that the azimuthal difference between tag and probe electrons is $\Delta\phi > 3/4\pi$. Both tag and probe electrons are required to have $E_T$ >15 GeV. The invariant mass of the lepton pair is required to be between 80 and 100 GeV. Figure 15 compares $\varepsilon_{TP}$ and $\varepsilon_{MC}$ as a function of $|\eta|$ and $E_T$ . Table 9 summarises the results obtained for this first step in the reconstruction and identification of the probe electron.

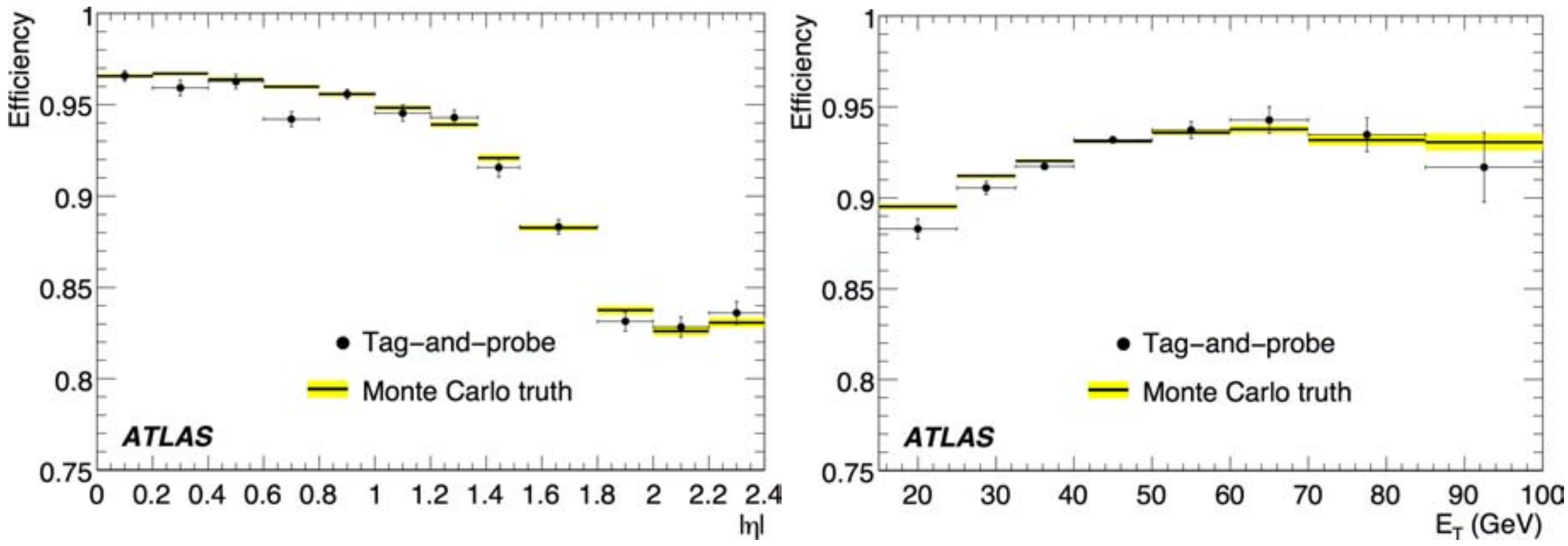

Figure 15: Efficiency of the electron pre-selection as a function of $|\eta|$ (left) and $E_T$ (right) for $Z \rightarrow ee$ decays, using the tag-and-probe method and the Monte Carlo truth information.

| $E_T$−range (GeV) | 15 – 25 | | 25 – 40 | | 40 – 70 | |
|---|---|---|---|---|---|---|
| $\|\eta\|$−range | $\varepsilon_{TP}$ | $\Delta\varepsilon_{TP/MC}$ | $\varepsilon_{TP}$ | $\Delta\varepsilon_{TP/MC}$ | $\varepsilon_{TP}$ | $\Delta\varepsilon_{TP/MC}$ |
| 0 – 0.80 | 96.1±0.4 | 2.0±0.4 | 96.2±0.2 | 0.1±0.2 | 99.0±0.1 | 2.0±0.1 |
| 0.80 – 1.37 | 94.9±0.6 | 1.5±0.6 | 96.0±0.2 | 1.6±0.2 | 95.1±0.2 | -0.5±0.2 |
| 1.52 – 1.80 | 89.0±1.2 | 3.6±1.2 | 88.8±0.6 | 1.3±0.6 | 91.9±0.6 | 1.7±0.6 |
| 1.80 – 2.40 | 83.0±1.0 | 0.6±1.0 | 83.2±0.6 | 0.8±0.6 | 84.9±0.6 | 1.1±0.6 |

Table 9: Efficiency of the electron pre-selection, $\varepsilon_{TP}$, in percent as obtained from the tag-and-probe method, for different ranges of electron $E_T$ and $|\eta|$. The errors quoted for $\varepsilon_{TP}$ are statistical and correspond to an integrated luminosity of 100 pb$^{-1}$ . Also shown is the difference, $\Delta\varepsilon_{TP/MC}$, between this estimate of the pre-selection efficiency and that obtained using the matching to the Monte Carlo electron.

### 5.3 Electron identification efficiency.

In this section, the electron identification efficiency is presented with respect to the reconstructed electrons discussed in Section 5.2. The QCD background was not considered here, since it is less than a few percent below the $Z$-boson mass peak. The reconstructed probe electron was checked against loose, medium and tight selection cuts. Table 10 summarises the results obtained for this second step in the reconstruction and identification of the probe electron. Figure 16 shows as a function of $\eta$ and $p_T$ the comparison between $\varepsilon_{TP}$ and $\varepsilon_{MC}$, for the medium cuts. The losses at high $\eta$ are due to the material in the inner detector, as discussed in Section 2.

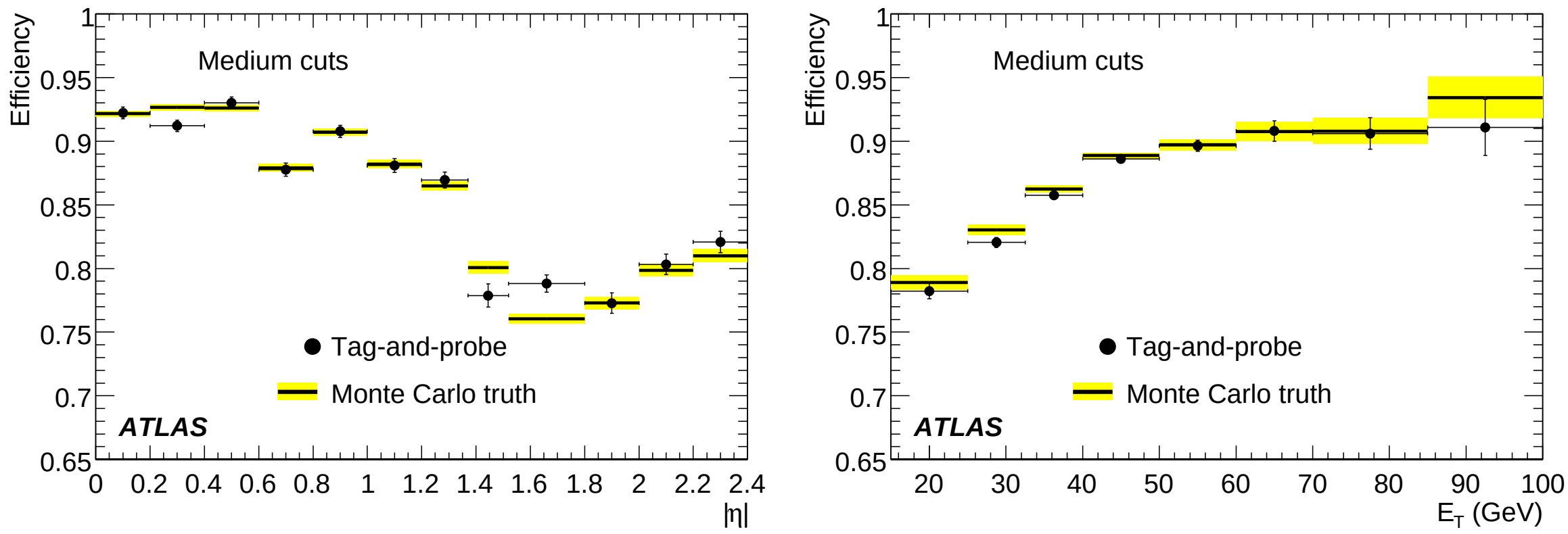

Figure 16: Efficiency of the medium electron identification cuts relative to the pre-selection cuts as a function of $|\eta|$ (left) and $E_T$ (right) for $Z \rightarrow ee$ decays, using the tag-and-probe method and the Monte Carlo truth information.

### 5.4 Statistical and systematic uncertainties

A number of uncertainties may affect these tag-and-probe measurements once the accumulated data will provide high enough statistics to perform similar measurements to those quoted above:

- Differences between $\varepsilon_{TP}$ and $\varepsilon_{MC}$
  The relative difference $\Delta\varepsilon_{TP/MC}$ in regions (in $p_T$ and $|\eta|$), where the efficiency is flat, is less than 0.5%, assuming that the statistical error on $\varepsilon_{MC}$ is negligible. $\Delta\varepsilon_{TP/MC}$ marginally depends on the definition of a true electron and the systematic uncertainty related to this is estimated to be $< 0.1\%$, when varying the cut on the separation in $\eta/\phi$ space ($\Delta R$) between the reconstructed electron candidate and the true electron.

- Statistical uncertainty.
  The size of the available $Z$-boson sample is a source of systematic error. With an integrated luminosity of 100 pb$^{-1}$, the error is expected to be in the range 1-2% for $p_T > 25$ GeV, and $\sim 4\%$ in the low-$p_T$ bin.

- Selection criteria
  Another source of systematic error comes from varying the selection criteria. For instance, uncertainties introduced by varying the cut on the $Z$-boson mass or requiring an isolation criterion for the probe electron were evaluated. The magnitude of the uncertainty introduced is smaller than 0.5% for $p_T > 40$ GeV. At low $p_T$, this uncertainty is estimated to be in the 1-2% range.

- QCD background contribution
  Adding the expected contribution from the QCD background to the signal does not degrade the results, except for $1.52 < |\eta| < 1.8$, a region which is close to the barrel/end-cap transition region and also where the efficiency is not uniform. The contribution from the uncertainties on the residual QCD background is expected to be negligible.

## 6 Conclusion

Excellent electron identification will clearly play an important role at the LHC, since high-$p_T$ leptons will be powerful probes for physics within and beyond the Standard Model. Based on this motivation, various algorithms and tools have been developed to efficiently reconstruct and identify electrons and separate them from the huge backgrounds from hadronic jets.

| *Loose* | 15 – 25 | | 25 – 40 | | 40 – 70 | |
|---|---|---|---|---|---|---|
| $\|\eta\|\backslash p_T$ | $\varepsilon_{TP}$ | $\Delta\varepsilon_{TP/MC}$ | $\varepsilon_{TP}$ | $\Delta\varepsilon_{TP/MC}$ | $\varepsilon_{TP}$ | $\Delta\varepsilon_{TP/MC}$ |
| 0 – 0.8 | $95.2 \pm 2.0$ | $-4.1 \pm 2.0$ | $98.8 \pm 0.3$ | $-0.5 \pm 0.3$ | $99.8 \pm 0.1$ | $0.2 \pm 0.1$ |
| 0.8 – 1.37 | $92.3 \pm 2.1$ | $-6.9 \pm 2.1$ | $98.9 \pm 0.3$ | $-0.7 \pm 0.3$ | $99.6 \pm 0.2$ | $0.0 \pm 0.2$ |
| 1.52 – 1.8 | $100.0 \pm 2.8$ | $1.7 \pm 2.8$ | $99.4 \pm 0.5$ | $0.0 \pm 0.5$ | $99.6 \pm 0.5$ | $0.0 \pm 0.5$ |
| 1.8 – 2.4 | $98.8 \pm 1.6$ | $0.6 \pm 1.7$ | $98.8 \pm 0.5$ | $0.0 \pm 0.5$ | $99.1 \pm 0.4$ | $-0.2 \pm 0.4$ |

| *Medium* | 15 – 25 | | 25 – 40 | | 40 – 70 | |
|---|---|---|---|---|---|---|
| $\|\eta\|\backslash p_T$ | $\varepsilon_{TP}$ | $\Delta\varepsilon_{TP/MC}$ | $\varepsilon_{TP}$ | $\Delta\varepsilon_{TP/MC}$ | $\varepsilon_{TP}$ | $\Delta\varepsilon_{TP/MC}$ |
| 0 – 0.8 | $83.6 \pm 2.3$ | $-4.3 \pm 2.7$ | $89.7 \pm 0.7$ | $-0.8 \pm 0.8$ | $92.6 \pm 0.5$ | $-0.2 \pm 0.6$ |
| 0.8 – 1.37 | $75.6 \pm 2.8$ | $-7.5 \pm 3.4$ | $87.6 \pm 0.9$ | $0.7 \pm 1.0$ | $90.9 \pm 0.8$ | $-0.4 \pm 0.8$ |
| 1.52 – 1.8 | $71.9 \pm 4.4$ | $5.9 \pm 6.5$ | $76.9 \pm 1.9$ | $-2.2 \pm 2.4$ | $83.6 \pm 1.9$ | $0.7 \pm 2.3$ |
| 1.8 – 2.4 | $78.0 \pm 2.7$ | $6.5 \pm 3.7$ | $79.2 \pm 1.4$ | $1.7 \pm 1.8$ | $82.5 \pm 1.4$ | $-1.0 \pm 1.6$ |

| *Tight* | 15 – 25 | | 25 – 40 | | 40 – 70 | |
|---|---|---|---|---|---|---|
| $\|\eta\|\backslash p_T$ | $\varepsilon_{TP}$ | $\Delta\varepsilon_{TP/MC}$ | $\varepsilon_{TP}$ | $\Delta\varepsilon_{TP/MC}$ | $\varepsilon_{TP}$ | $\Delta\varepsilon_{TP/MC}$ |
| 0 – 0.8 | $68.7 \pm 2.6$ | $-5.2 \pm 3.5$ | $73.8 \pm 1.0$ | $-1.2 \pm 1.3$ | $77.0 \pm 0.9$ | $-1.5 \pm 1.1$ |
| 0.8 – 1.4 | $61.8 \pm 3.0$ | $-3.1 \pm 4.7$ | $72.9 \pm 1.2$ | $0.7 \pm 1.7$ | $77.3 \pm 1.1$ | $0.2 \pm 1.5$ |
| 1.5 – 1.8 | $55.7 \pm 4.5$ | $6.8 \pm 8.6$ | $65.9 \pm 2.1$ | $-0.8 \pm 3.1$ | $73.7 \pm 2.2$ | $1.2 \pm 3.1$ |
| 1.8 – 2.4 | $66.2 \pm 3.0$ | $8.5 \pm 4.9$ | $66.0 \pm 1.6$ | $2.6 \pm 2.5$ | $73.4 \pm 1.6$ | $0.7 \pm 2.2$ |

Table 10: Loose, medium and tight electron identification efficiencies relative to the pre-selection efficiencies for different bins in $E_T$ and $|\eta|$. The first error is statistical and corresponds to an integrated luminosity of 100 pb$^{-1}$ . The second error is the difference obtained between $\varepsilon_{TP}$ and $\varepsilon_{MC}$.

Presently, two reconstruction algorithms have been implemented in the ATLAS offline software, both integrated into one single package and a common event model. The first one relies on calorimeter seeds for reconstructing electrons, whereas the second algorithm relies on track-based seeds, is optimised for electrons with lower energies, and relies less on isolation.

The calorimeter based algorithm starts from the reconstructed cluster in the electromagnetic calorimeter, then builds identification variables based on information from the calorimeter and the inner detector. The rejection power with respect to QCD jets comes almost entirely from the identification procedure. Depending on the electron transverse energy and the analysis requirements, rejection factors of 500 to 100 000 can be achieved, for efficiencies of 88% to 64%, using a simple cut-based selection. More refined identification procedures combining calorimeter and track quantities using multivariate techniques provide a gain in rejection of about 20 – 40% with respect to the cut-based method, for the same efficiency of 61 – 64%. Alternatively, they provide a gain of 5 – 10% in efficiency, for the same jet rejection (tight and medium cuts).

Electrons in the forward region can also be identified and separated from the background. A simple cut-based method, exploring the energy depositions in the inner wheel of the electromagnetic end-cap calorimeter and in the forward calorimeter as well as the shower-shape distributions, shows that $\sim$ 99% of the QCD background can be rejected, for an electron identification efficiency of $\sim$ 80%. This performance should be sufficient to select cleanly, for example, $Z \rightarrow ee$ decays with one electron in the forward region.

Studies of the strategies for measuring efficiencies and fake rates in early data show that the tag-and-probe method is a good tool to estimate the electron identification efficiency and to control the reliability of the Monte Carlo simulation. With 100 $pb^{-1}$ , the method is limited by the statistics of the Z sample, whereas its systematic uncertainty is of the order of 1 to 2 %.

The work presented here primarily addresses the description and performance of the offline reconstruction and identification of electrons. However, it also gives an overview of the possible path towards physics discoveries with electrons in Higgs, SUSY, and exotic scenarios.

# Reconstruction and Identification of Photons

**Abstract**
This note presents the description and performance of photon identification methods in ATLAS. The reconstruction of an electromagnetic object begins in the calorimeter, and the inner detector information determines whether the object is a photon - either converted or unconverted - or an electron. Three photon identification methods are presented: a simple cut-based method, a Log-likelihood-ratio-based method and a covariance-matrix-based method. The shower shape variables based on calorimeter information and track information used in all three methods are described. The efficiencies for single photons and for photons from the benchmark $H \to \gamma\gamma$ signal events, as well as the rejection of the background from jet samples, are presented. The performance of the cut-based method on high-$p_T$ photons from a graviton decay process $G \to \gamma\gamma$ is also discussed.

## 1 Introduction

Isolated photons with large transverse momentum, $p_T$, in the final state are distinguishing signatures for many physics analyses envisaged at the LHC. The Higgs particle has been sought over several decades in many high-energy experiments, including those currently running at the Tevatron. It is understood that if the Standard Model Higgs particle exists, and unitarity is not violated, its mass is within the reach of LHC. As described in detail in other parts of this work [1], while the expected cross-section times branching ratio of the Higgs particle decaying into the two photon final state is relatively small, given its distinct signature, isolated high-$p_T$ photons may play a significant role in discovering the Higgs particle in the low mass region. In addition, very high-$p_T$ photons are also signatures of more exotic particles, such as the graviton predicted in Ref. [2], which is expected to have mass larger than 500 GeV. These photons appear as a single, isolated objects with most of their energy deposit in the electromagnetic compartment of the calorimeter. Thus the primary source for background to these photons, namely fake photons, result from jets that fluctuate highly electromagnetic which contain a high fraction of photons from neutral hadron decays, such as $\pi^0 \to \gamma\gamma$.

Since the ATLAS electromagnetic calorimeter [3] is highly segmented with a three-fold granularity in depth and with an $\eta \times \phi$ granularity in the barrel of $0.003 \times 0.1$, $0.025 \times 0.025$, and $0.05 \times 0.025$, respectively, in the front, middle and rear compartments assisted by a pre-sampler in front of the calorimeter, photon identification methods in ATLAS should be much more powerful that those used in past experiments. The experiment also employs elaborate trigger systems that select electrons and photons efficiently, as described in detail in Ref. [4].

This paper presents three ATLAS photon identification methods and their performance for single, isolated photons as well as for photons from physics processes.

## 2 Data samples

The $H \to \gamma\gamma$ ($m_H = 120$ GeV) process is used as the primary signal benchmark sample for medium $p_T$ photons and with the pile-up that corresponds to the instantaneous luminosity $10^{33}$ cm$^{-2}$s$^{-1}$. Rejection studies were conducted using a pre-filtered jet sample (described in details in Ref. [5]), containing all relevant hard-scattering QCD processes with $p_T > 15$ GeV. A filter is applied at the generator level, requiring the summed transverse energy of all stable particles (excluding muons and neutrinos) in a region of $\Delta\phi \times \Delta\eta$ = 0.12×0.12 to be above 17 GeV. A total number of 3 million events were used in rejection studies. Two additional samples with 150 GeV $< p_T <$ 280 GeV (Jet5) and 280 $< p_T <$ 400 GeV (Jet6) were also employed for high-$p_T$ photon rejection studies. Finally, an additional 300,000 event $\gamma$+jet sample has been used for rejection and fake rate studies.

In addition to these signal and background samples, the three identification methods described in this paper were developed using single photon samples - events with no activity except the photon - with full detector simulation in the energy range $10-1000$ GeV with flat pseudorapidity distributions over $|\eta| < 2.5$. For high-$p_T$ photons, graviton samples with masses of 0.5 and 1.0 TeV were employed.

All the samples used in this note were generated using PYTHIA and its fragmentation scheme and were passed through the full detector simulation. Some of the simulations were done with the nominal geometry and material distribution ("ideal") and others with additional material added ("distorted").

In order to maintain the consistency between different studies, the following requirements and definitions are used for efficiencies and rejections.

- Truth match: the reconstructed photons must lie within a cone of radius $\Delta R = \sqrt{\Delta\eta^2 + \Delta\phi^2} < 0.2$ of the true photons in the simulation.
- The reconstructed photons must be within the fiducial volume, pseudorapidity $0 < |\eta| < 1.37$ or $1.52 < |\eta| < 2.47$ to avoid the overlap between the barrel and end-cap calorimeters.

Using the base samples that satisfy the above requirements, the efficiency is defined as follows:

$$\varepsilon = \frac{N_\gamma^{reco}}{N_\gamma^{truth}} \tag{1}$$

where $N_\gamma^{truth}$ is the number of true photons in the simulation that satisfy all the requirements above with the true $E_T$ greater than either 25 GeV or 40 GeV and $N_\gamma^{reco}$ is the number of reconstructed photons that satisfy all the requirements with the true $E_T$ greater than either 25 GeV or 40 GeV and that pass the threshold for one of the three methods.

Similarly, the rejection from the pre-filtered jet sample is computed as follows:

$$R = \frac{N_{jet}}{N_{fake\gamma}} \frac{N_1}{N_2} \frac{1}{\varepsilon_{\gamma-filter}} \tag{2}$$

where $N_{jet}$ is the total number of jets reconstructed in the normalisation sample (same generation as the reconstructed sample but without the filter requirements) using particle four-momenta from the generator hadron level within a cone size $\Delta R$ = 0.4, and $N_2 (= 400,000)$ is the number of events used in this normalisation sample. The values for $N_{jet}/N_2$ in the fiducial volume of $|\eta| < 1.37$ or $1.52 < |\eta| < 2.37$ are 0.226 for jets with $E_T > 25$ GeV and 0.042 for jets with $E_T > 40$ GeV. $N_{fake\gamma}$ is the number of fake photons in the reconstructed (filtered) sample with the candidates that matched to true photons from the hard scatter or from quark bremsstrahlung removed, and $N_1 (= 3,095,900)$ is the number of events analyzed from this sample. Finally, $\varepsilon_{\gamma-filter}$ $(= 0.082)$ is the efficiency of the generator level filter applied to the jet sample.

## 3 Photon identification methods

As discussed in previous sections, three photon identification methods have been developed and are available at present in ATLAS: a simple cut-based identification method, a Log-likelihood-ratio-based identification method (LLR) and the covariance-matrix-based identification method (H-matrix). A partial description of the basic electromagnetic object reconstruction and a detailed presentation of their calibration can be found in Ref. [6].

## 3.1 Characteristic variables and cut-based photon identification

In order to separate real photons from fake photons resulting from jets, several discriminating variables are defined using the information both from the calorimeters and the inner tracking system. Cuts on these variables are developed to maintain high photon efficiency even in the presence of pile-up resulting from the overlapping minimum bias events due to high instantaneous luminosity at the LHC. The discriminating variables used in this study are the same as in previous studies [7–11]. Calorimeter information is used to select events containing a high-$E_T$ electromagnetic shower. The fine-grained first compartment allows to reject showers from photons from $\pi^0$ decays. Track isolation is used to improve the rejection. Only electromagnetic clusters with $E_T > 20$ GeV are used in this study.

### 3.1.1 Variables using calorimeter information

In the electromagnetic calorimeter, photons are narrow objects, well contained in the electromagnetic calorimeter, while fake photons induced from jets tend to have a broader profile and can deposit a substantial fraction of their energy in the hadronic calorimeter. Hence, longitudinal and transverse shower-shape variables can be used to reject jets.

- **Hadronic leakage :** The hadronic leakage is defined as the ratio of the transverse energy in the first layer of the hadronic calorimeter in a window $\Delta\eta \times \Delta\phi = 0.24 \times 0.24$ to the transverse energy of the cluster in order to avoid boundary effects that could result from using readout cells. Real photons are purely an electromagnetic object, therefore they deposit their energy primarily in the electromagnetic compartment of the calorimeter. Fake photons induced from jets contain hadrons that would penetrate deeper into the calorimeter depositing sizable energy beyond the electromagnetic calorimeter.

- **Variables using the second compartment of the ECAL :** Electromagnetic showers deposit most of their energy in the second layer of the electromagnetic calorimeter. For this reason several variables that measure the shape of the shower are available as follows:

  - The real photons deposit most of their energy in a $\Delta\eta \times \Delta\phi = 3 \times 7$ window (in units of middle cells). The lateral shower-shape variables, $R_\eta$ and $R_\phi$, are given by the ratio of the energy reconstructed in $3 \times 7$ middle cells to the energy in $7 \times 7$ cells and the ratio of the energy reconstructed in $3 \times 3$ cells to the energy in $3 \times 7$ cells, respectively. Due to the effect of the magnetic field increasing the width of the converted photon contributions in the $\phi$ direction, $R_\phi$ is less discriminating than $R_\eta$.
  - The lateral width in $\eta$ is calculated in a window of $3 \times 5$ cells using the energy weighted sum over all cells. $w_2 = \sqrt{\frac{\Sigma(E_c \times \eta_c^2)}{\Sigma E_c} - \left[\frac{\Sigma(E_c \times \eta_c)}{\Sigma E_c}\right]^2}$, where $E_c$ is the energy deposit in each cell, and $\eta_c$ is the actual $\eta$ position of the cell represented by the center of the cell in $\eta$ direction. Therefore, $w_2$ is given in units of $\eta$. A correction is applied as a function of the impact point within the cell to reduce the bias from the finite cell size.

- **Variables using the first compartment of the ECAL :** Cuts applied on the variables in the hadronic calorimeter and the second layer of the electromagnetic calorimeter reject jets which contain high-energy hadrons and resulting broad showers. Jets containing single or multiple neutral hadrons such as $\eta$ and $\pi^0$, provide the main contribution which can fake photons. The readout of the first layer of the calorimeter uses strips and provides very fine granularity in pseudorapidity. Thus, the information from this layer can be used to identify substructures in the showers and distinguish isolated photons from the hard scatter and photons from $\pi^0$ decays efficiently. The lateral shower shape in the strips is exploited for $|\eta| < 2.35$ where the strip granularity is sufficiently fine, as long as a 0.5% or larger fraction of the total energy is reconstructed in this layer.

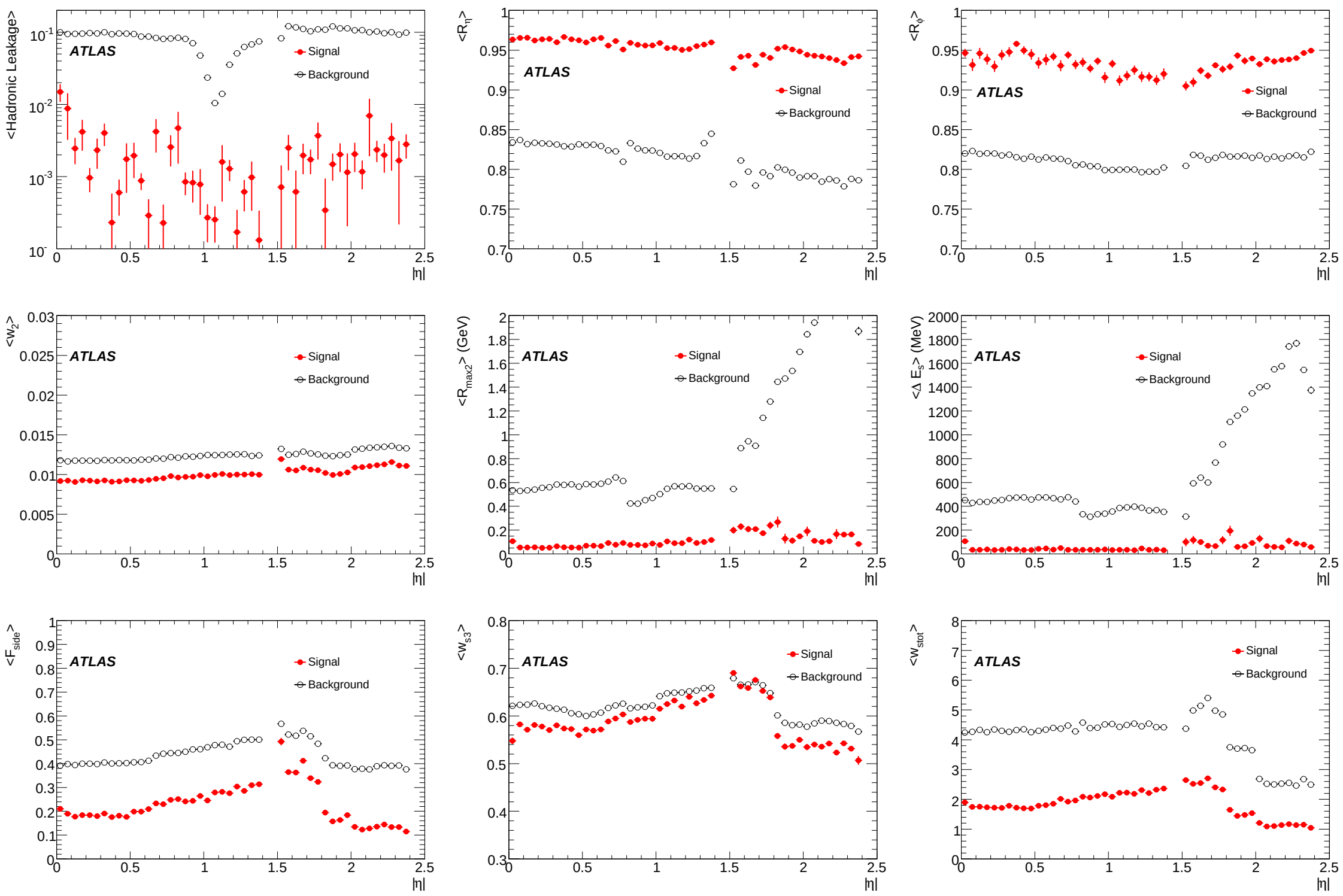

Figure 1: Distributions of the mean of each calorimetric discriminating variable as a function of the pseudorapidity $|\eta|$ for true and fake photons (before cuts) with $20 < E_T < 30$ GeV. The samples have been simulated with the geometry under the realistic alignment scenario and additional material.

- Since the energy-deposit pattern from $\pi^0$'s is often found to have two maxima due to $\pi^0 \to \gamma\gamma$ decay, showers are studied in a window $\Delta\eta \times \Delta\phi = 0.125 \times 0.2$ around the cell with the highest $E_T$ to look for a second maximum. If more than two maxima are found the second highest maximum is considered. The following two variables are constructed using the information from the identified second maximum:
  - $\Delta E_s = E_{\mathrm{max2}} - E_{\mathrm{min}}$, the difference between the energy associated with the second maximum $E_{\mathrm{max2}}$ and the energy reconstructed in the strip with the minimum value, found in between the first and second maxima, $E_{\mathrm{min}}$.
  - $R_{\mathrm{max2}} = E_{\mathrm{max2}}/(1 + 9 \times 10^{-3} E_T/\mathrm{GeV})$, where $E_T$ is the transverse energy of the cluster in the electromagnetic calorimeter. The value of the second maximal energy deposit is corrected as a function of the transverse energy of the cluster to minimise its sensitivity to fluctuations [9, 10].
- $F_{\mathrm{side}} = [E(\pm 3) - E(\pm 1)]/E(\pm 1)$, the fraction of the energy deposited outside the shower core of three central strips. The variable $E(\pm n)$ is the energy deposited in $\pm\, n$ strips around the strip with the highest energy.
- $w_{s3} = \sqrt{\sum E_i \times (i - i_{max})^2 / \sum E_i}$, the shower width over the three strips around the one with the maximal energy deposit. The index $i$ is the strip identification number, $i_{max}$ the identi-fication number of the most energetic strip, and $E_i$ is the energy deposit in strip $i$. $w_{s3}$ is expressed in units of strip cells and corrected for impact point dependence [9].

- $w_{stot}$, the shower width over the strips that cover 2.5 cells of the second layer (20 strips in the barrel for instance). It is expressed in units of strip cells.

Figure 1 shows the average values of the calorimeter-based discriminating variables as a function of the absolute value of pseudorapidity. Features in the plots can be explained by: upstream material thickness which increases with pseudorapidity in the barrel; physical cell-size changes in the end-cap to maintain a constant granularity in $\eta$-$\phi$; and the change in the granularity of the first layer in the end-cap. In particular, the rise of $R_{max2}$ and of $\Delta E_s$ for $|\eta| > 1.5$ stems from a combination of effects from the variation of the quantity of the upstream material and changes in the strip-cell sizes in the end-cap calorimeters. The dip in the hadronic leakage variable near $|\eta| = 1.1$ corresponds to a smaller coverage by the first hadronic layer in this region.

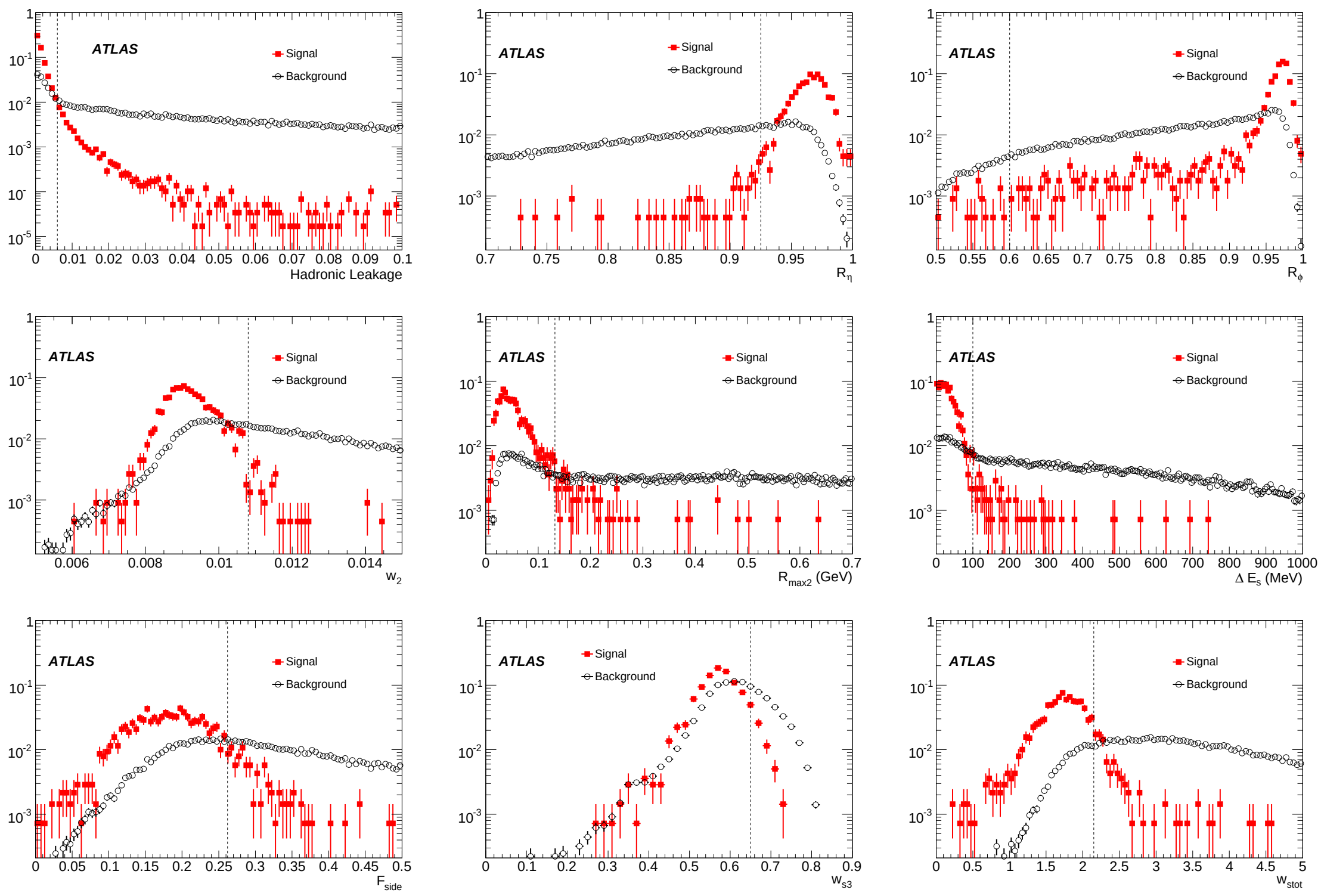

Figure 2: Normalised distributions of the discriminating variable for $|\eta| < 0.7$ for true and fake photons (before cuts) with $20 < E_T < 30$ GeV. The samples have been simulated with the geometry under the realistic alignment scenario.

The cut values are tuned separately in six pseudorapidity intervals in $|\eta| < 2.37$ to reflect the pseudorapidity dependence of these variables. The subdivision is motivated by the varying granularity and material in front of the electromagnetic calorimeter. The quantities calculated using the first compartment can be used only in the regions $|\eta| < 1.37$ and $1.52 < |\eta| < 2.37$ since there are no strips in the crack region or beyond $|\eta| > 2.40$. In addition, up to eight different bins in transverse energy are also used for the cut value adjustment. Figure 2 shows the distributions of the variables in the first $\eta$ bin and in one energy bin. The dashed vertical lines represent the cut values in this bin. The variables are shown for all reconstructed electromagnetic objects before cuts.

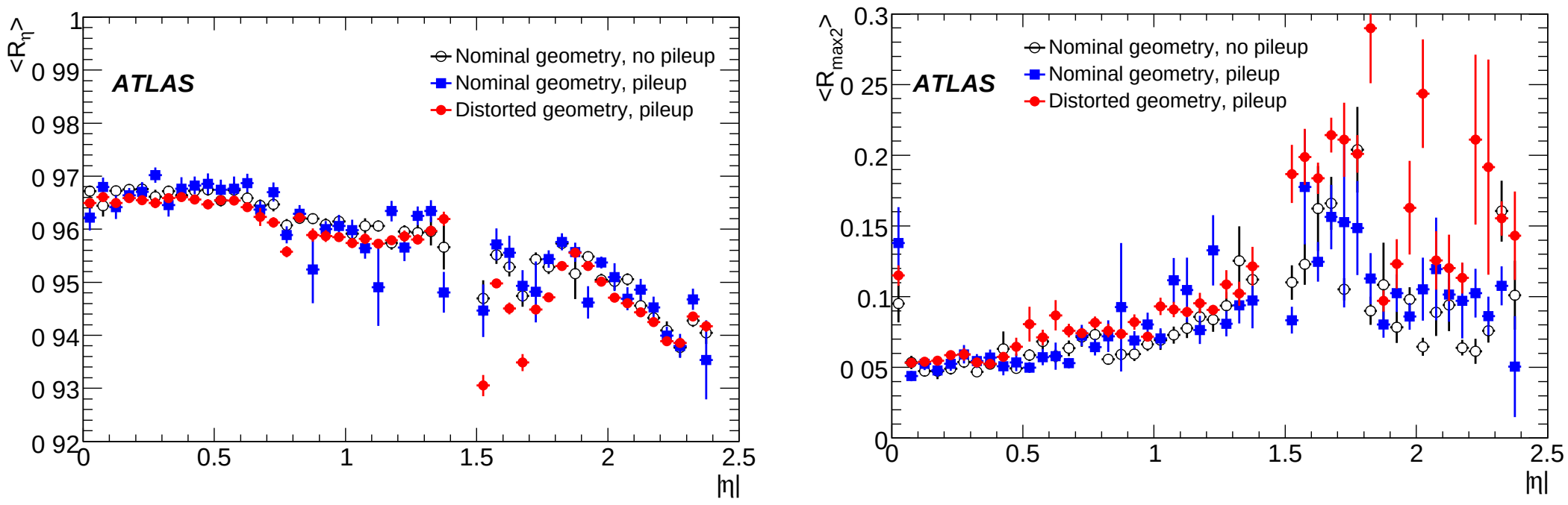

Figure 3: Effect of pile-up and distorted material on mean values of two shower-shape variables for photons from $H \rightarrow \gamma\gamma$ decays: $R_\eta$ (left) and energy of the second maximum in the first layer (right).

Figure 3 shows the impact of pile-up and additional material before the calorimeter on the shower-shapes for photons from Higgs decays. The impact of the large amount of additional material in the transition region, $1.5 < |\eta| < 1.8$, in the realistic alignment geometry can clearly be seen for two shower-shape variables. While pile-up at a luminosity of $10^{33}$ cm$^{-2}$s$^{-1}$ does not change the average shower shape significantly as can be seen for the two variables in Figure 3, it is observed that it does increase RMS of the distributions.

At present, the same cuts are applied for converted and unconverted photons. Studies of the $\gamma$-$\pi^0$ separation, however, have shown that if conversions can be identified efficiently, different cuts can be applied for converted and unconverted photons [12], which could improve rejection by 10-20% while maintaining the same overall photon identification efficiency.

The cuts have been chosen comparing the photons from $H \rightarrow \gamma\gamma$ decays to fake candidates in inclusive jet samples. For this optimisation, samples generated with realistic alignment geometry and pile-up have been used. Some improvement in the performances should be possible at higher $E_T$ for further refinement and optimisation in some of the variables, such as hadronic leakage. The rejection presented in this paper has been estimated on a sample statistically independent from the one used to tune the cuts.

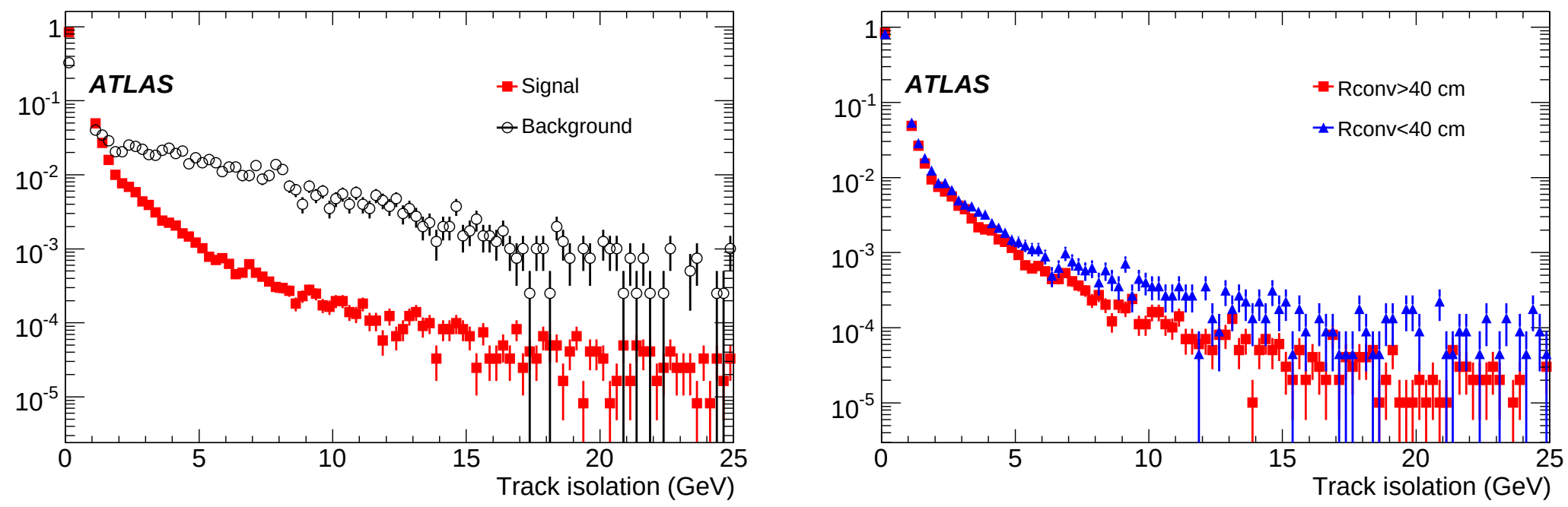

Figure 4: Normalised distribution of the track-isolation variable for events passing the calorimeter selection criteria. Left: comparison of true and fake photons. Right: comparison of early conversions (true conversion radius less than 40 cm) and late conversions (true conversion radius above 40 cm) for photons from $H \rightarrow \gamma\gamma$ decays.

#### 3.1.2 Track isolation

After the calorimeter cuts, the contamination of the inclusive signal from charged hadrons is greatly reduced. The remaining background is dominated by low track multiplicity jets containing high-$p_T$ $\pi^0$ mesons. In order to further remove fake photons from these jets, the track-isolation variable is defined as the sum of the $p_T$ of all tracks with $p_T$ above 1 GeV within $\Delta R < 0.3$, where $\Delta R$ is the $\eta - \phi$ distance between the track position at the vertex and the cluster centroid. Track $p_T > 1$ GeV is imposed to minimise the effect of pile-up and underlying events.

Since the tracks from photon conversions should not be included in computing this variable, some additional selections are applied to tracks within $\Delta R < 0.1$ of the cluster centroid. The impact parameter with respect to the beam line must be less than 0.1 mm. The track $p_T$ must not exceed 15 GeV to remove tracks from very asymmetric conversions, must not be part of a reconstructed conversion vertex and must have a hit in the innermost pixel layer.

The plot on the left in Fig. 4 shows the distribution of the track-isolation variable for true and fake-photon candidates, after the calorimeter shower-shape cuts. An additional rejection of factor 1.5 to 2 is possible for a relatively small efficiency loss. The plot on the right in this figure shows the track-isolation variable for early converted and late converted photons. The difference between the two distributions is rather small, showing that the tracks from conversions have been efficiently removed. At present, a 4 GeV upper cut on the track-isolation variable is applied for this method.

### 3.2 Log-likelihood-ratio-based photon identification

In the Log-likelihood-ratio (LLR)-based method, the distribution of each of the shower-shape variables is normalised to unity to obtain the probability density functions (PDF). The shower-shape variables are pseudorapidity-dependent, so they are separated in four regions of $|\eta|$ and three bins in $p_T$ for this method. The PDF's are obtained using 1.6 million $\gamma$+jet events which provided slightly over 100,000 events in each bin. Since the statistics for the PDF computation is somewhat low in some kinematic phase-space regions, further improvement can be obtained by using tools to smooth the PDF's to compensate for the low statistics [13]. Once the PDF's are established, the Log-likelihood-ratio parameter is defined as:

$$LLR = \sum_{i=1}^{n} \ln\left(L_{si}/L_{bi}\right), \tag{3}$$

where $L_{si}$ and $L_{bi}$ are PDF's of the $i^{th}$ shower-shape variable for the photon and the jet, respectively.

The shower-shape variables used for the LLR method were the same as those used for the cut-based method described previously. Track isolation was also included as a discriminating variable in Equation 3. Figure 5 shows the LLR parameter distribution for photons and for jets. The LLR cut can be tuned over $\eta$ and $p_T$ to obtain an optimal separation between photons and jets.

### 3.3 Covariance-matrix-based photon identification

The shower-shape variables associated with a photon shower in the calorimeter are correlated. The covariance matrix (H-matrix) technique takes advantage of these correlations. The technique was employed successfully in the DØ experiment at the Tevatron and was used to identify electrons [14].

The ten photon shower-shape variables used in the ATLAS H-matrix method are as follows:

- Five longitudinal shower-shape variables: fraction of energy deposited in pre-sampler layer; fractions of energy deposited in sampling layers 1, 2 and 3 separately; and the hadronic leakage, the energy leakage into the first layer of the hadronic calorimeter.

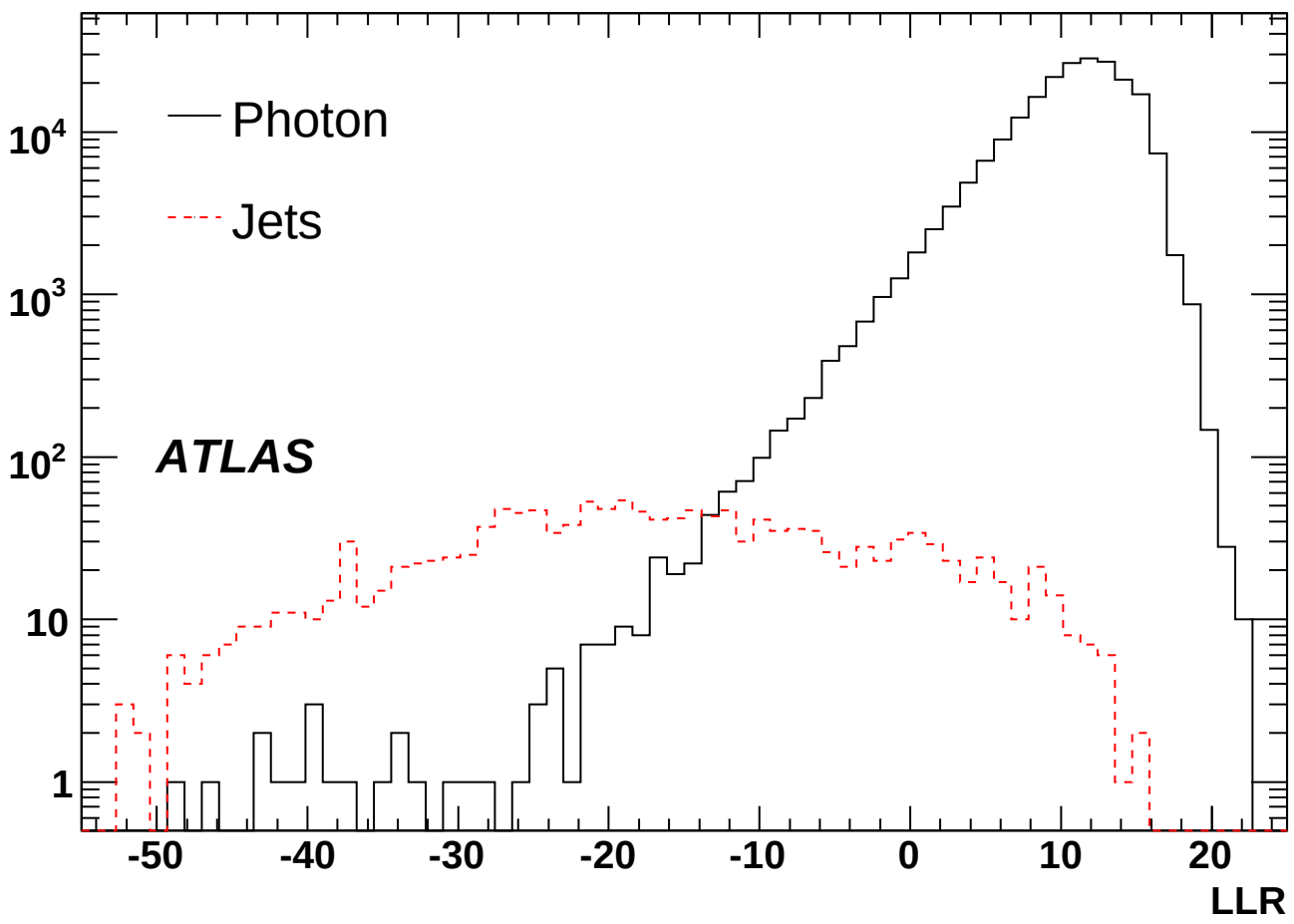

Figure 5: Expected Log-likelihood ratio (LLR) cut-parameter distributions for photons (solid histogram) and for jets (dashed histogram).

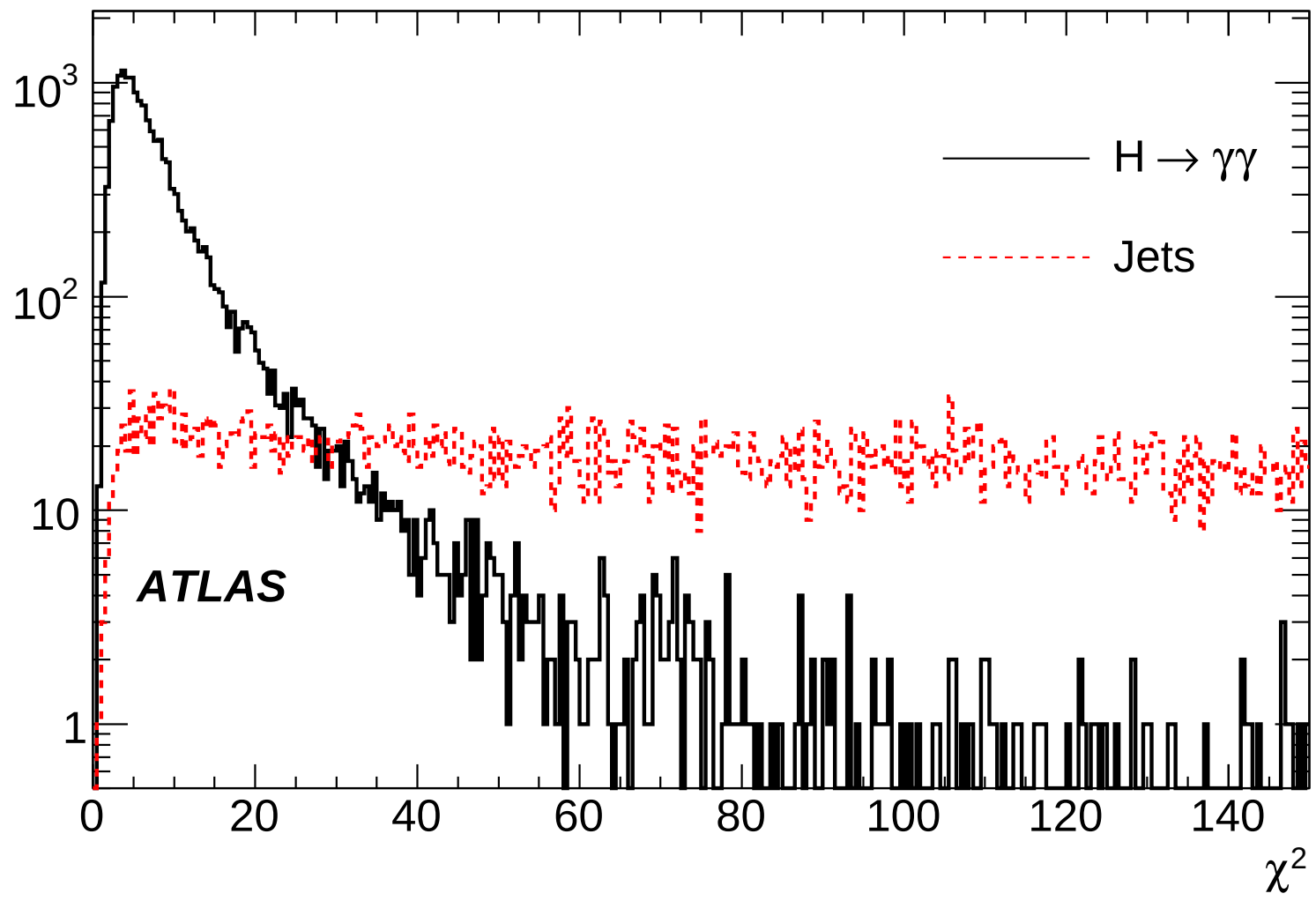

Figure 6: The distributions of H-matrix $\chi^2$ for photons from the $H \to \gamma\gamma$ sample (solid histogram) and for jets from the inclusive jet samples (dashed histogram).

- Five transverse shower-shape variables: the ratio of the energy in $3\times3$ cells to the energy in $7\times7$ in the second sampling layer of the electromagnetic calorimeter; $w_{rms3}$, the corrected width in 3 strips in sampling layer 1; $w_2$, the corrected width in a $3\times5$ window in sampling layer 2; the energy outside of the shower core; $R_\phi$, the ratio of energy in a $3\times3$ to a $3\times7$ window around the cluster centroid.

Using the above variables, a covariance matrix, $M$, is constructed as follows:

$$M_{ij} = \frac{1}{N}\Sigma_{n=1}^{N}(y_i^{(n)} - \bar{y}_i)(y_j^{(n)} - \bar{y}_j), \tag{4}$$

where indices $i$ and $j$ run over the ten variables, $N$ is the total number of photons used in the training sample, $y_j^n$ is the $j^{th}$ variable for the $n^{th}$ photon candidate, and $\bar{y}_j$ is the mean value of $y_j$ variable for the control sample electrons/photons. These matrix elements are constructed for each $\eta$ bin and parametrised for energy dependences. The photon likeness of an object is then measured by the value of the $\chi^2$, defined as follows:

$$\chi^2 = \Sigma_{i,j=1}^{dim}(y_i^{(m)} - \overline{y_i})H_{ij}(y_j^{(m)} - \overline{y_j}) \quad (5)$$

where $H \equiv M^{-1}$, the inverse of the covariance matrix, and the indices $i$ and $j$ run from 1 to the total number of variables (ten) which is the same as the dimension of the matrix, $dim$.

The mean value of the $\chi^2$ is close to the number of dimensions for a photon shower. The shapes of the distributions of the selected shower-shape variables depend on the $\eta$ and the energy of the incident photon. These effects are taken into account in the construction of the H-matrix using single photon samples of energies 10 – 1000 GeV generated flat in $|\eta|$ and parametrising each of the covariance terms in the matrix $M$ of Eq. 4 as a function of the photon energy. The parametrisation as a function of photon energy is obtained in each of the 12 $\eta$ bins. The discrimination power of the H-matrix between real photons and jets is well illustrated in Fig. 6, where the $\chi^2$ distribution of the H-matrix for the jet sample is contrasted to that obtained from photons from $H \rightarrow \gamma\gamma$ decays.

Since the H-matrix implementation at this time does not include the same variables as the other two methods, its performance is currently not directly comparable. Consequently, the performance is not reported here, although the method is decribed for completeness.

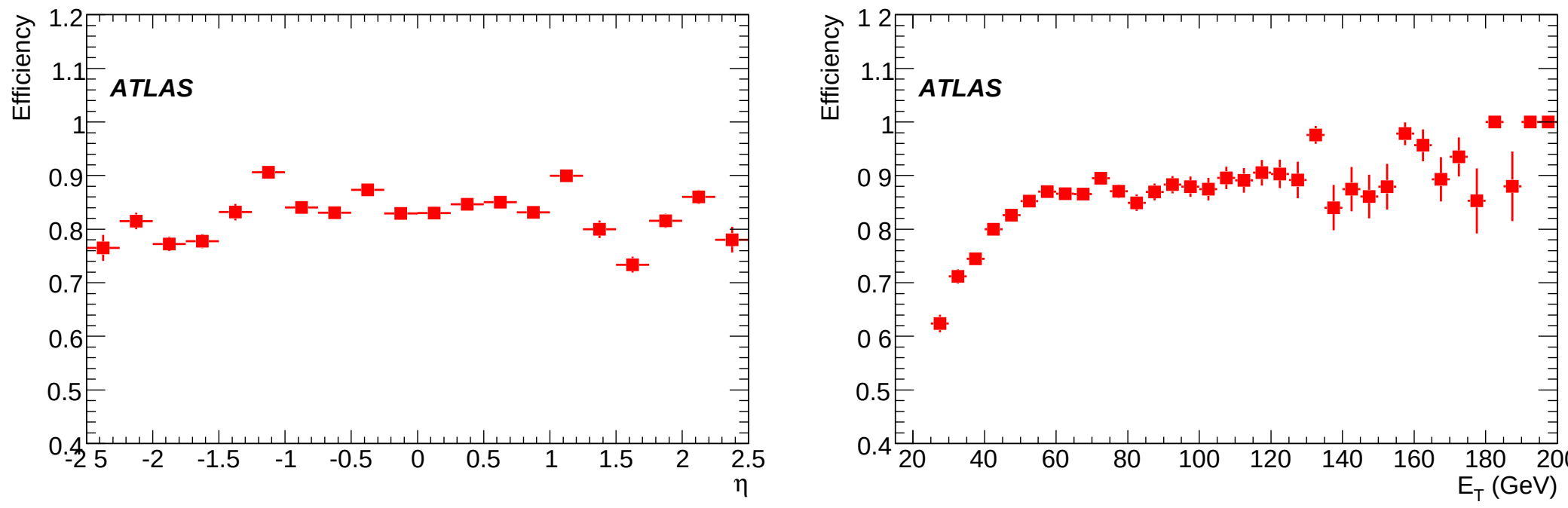

Figure 7: Efficiency of the calorimeter cuts as a function of pseudorapidity (left) and transverse energy (right) of the photons for the distorted geometry.

| Efficiency | $\varepsilon$ (calorimeter cuts) | $\varepsilon$ (track-isolation cut) |
|---|---|---|
| Nominal geometry no pile-up | (87.6±0.2)% | (99.0±0.1)% |
| Nominal geometry with pile-up | (86.6±0.5)% | (98.0±0.2)% |
| Distorted geometry with pile-up | (83.6±0.2)% | (98.1±0.1)% |

Table 1: Overall efficiency for photons from $H \rightarrow \gamma\gamma$ decays for three different simulation choices.

# 4 Photon identification performance for medium-$p_T$ photons

This section describes the performance (efficiencies and rejections) of the cut-based method and the Log-likelihood ratio method on medium-$p_T$ photons, in particular the photons from $H \rightarrow \gamma\gamma$ decays and the jet background samples described in Section 2.

## 4.1 Performance of the cut-based method

In the performance studies presented in this section, all reconstructed electromagnetic objects, including both electron and photon candidates are considered. The efficiency as defined in Section 2 includes both the reconstruction efficiency and the efficiency of the identification cuts.

Figure 7 shows the efficiency of the calorimeter cuts for photons with $E_T > 25$ GeV from $H \to \gamma\gamma$ decay as a function of pseudorapidity (left) and transverse energy (right) for events in the presence of the pile-up expected at a luminosity of $10^{33}$ cm$^{-2}$s$^{-1}$. The optimisation of the cuts for the $H \to \gamma\gamma$ signal has led to an efficiency which is uniform for $E_T > 40$ GeV, but which decreases substantially below 40 GeV because of the much larger fake backgrounds from jets expected at these lower transverse energies. The average efficiencies of the calorimeter and track-isolation cuts are summarised in Table 1.

| | All | Quark jets | Gluon jets |
|---|---|---|---|
| $N$(jet)/$N$(generated events) | 0.23 | 0.056 | 0.177 |
| | Before isolation cut | | |
| $N$(fake)/$N$(filtered events) | $(5.43\pm0.13).10^{-4}$ | $(3.87\pm0.11).10^{-4}$ | $(1.44\pm0.07).10^{-4}$ |
| Rejection | 5070±120 | 1770±50 | 15000±700 |
| | After isolation cut | | |
| $N$(fake)/$N$(filtered events) | $(3.38\pm0.10).10^{-4}$ | $(2.47\pm0.08).10^{-4}$ | $(0.78\pm0.49).10^{-4}$ |
| Rejection | 8160± 250 | 2760±100 | 27500±2000 |

Table 2: Rejection (Equation 2) measured in the inclusive jet sample for $E_T > 25$ GeV

| | All | Quark jets | Gluon jets |
|---|---|---|---|
| $N$(jet)/$N$(generated events) | 0.042 | 0.011 | 0.034 |
| | Before isolation cut | | |
| $N$(fake)/$N$(filtered events) | $(1.16\pm0.06).10^{-4}$ | $(8.3\pm0.5).10^{-5}$ | $(2.8\pm0.3).10^{-5}$ |
| Rejection | 4400±230 | 1610±100 | 15000±1600 |
| | After isolation cut | | |
| $N$(fake)/$N$(filtered events) | $(6.4\pm0.4).10^{-5}$ | $(4.6\pm0.5).10^{-5}$ | $(1.5\pm0.2).10^{-5}$ |
| Rejection | 7800±540 | 2900±240 | 28000±4000 |

Table 3: Rejection (Equation 2) measured in the inclusive jet sample for $E_T > 40$ GeV

The rejection from the pre-filtered jet sample is computed using Equation 2. The rejection is computed separately for all jets, for quark-initiated jets and for gluon-initiated jets. The quark or gluon initiation is defined using the type of the highest $E_T$ parton from the PYTHIA record inside the cone $\Delta R = 0.4$ around the reconstructed jet object. The rejection values are summarised in Table 2 for the three categories of jets. A small fraction ($\approx$ 1-2%) of jet objects are not classified, so the sum of quarks and gluons is slightly smaller than the total. A cut $E_T > 25$ GeV is applied to both reconstructed photons and jets. Table 3 shows the same computation, but for $E_T > 40$ GeV.

Figure 8 shows the fake rate, defined as the inverse of the rejection, as a function of pseudorapidity for all jets with $E_T$ greater than 25 GeV. There is a slight increase of fake rate as a function of pseudorapidity due to the increase in material in front of the calorimeter, which imposes somewhat looser cuts to preserve a constant efficiency. Some additional increase near $|\eta| = 1.1$ is also visible probably coming from the reduced energy in the first layer of the hadronic calorimeter as pointed out previously. This effect, however, gives a less than 10% increase in the overall fake rate.

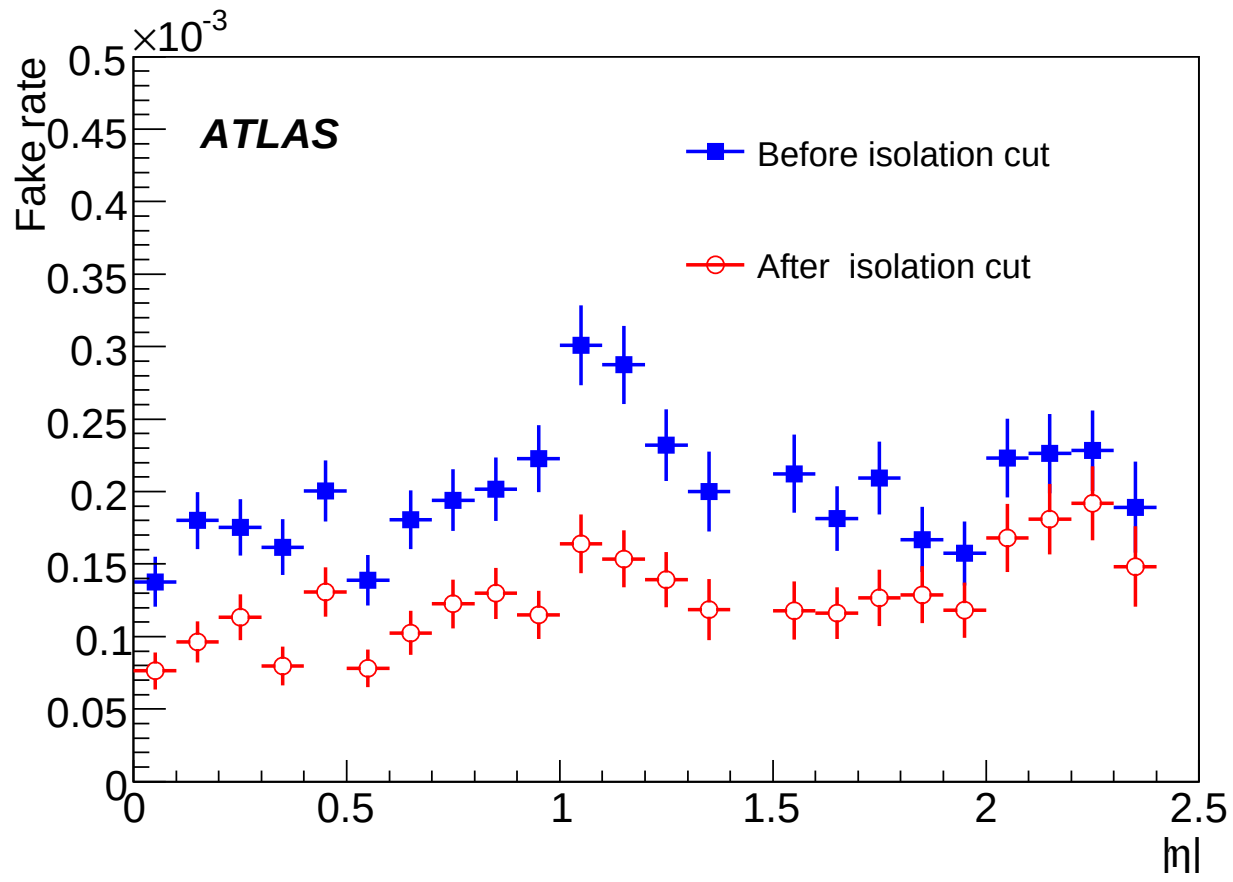

Figure 8: Fake-photon rate as a function of pseudorapidity in the filtered jet sample

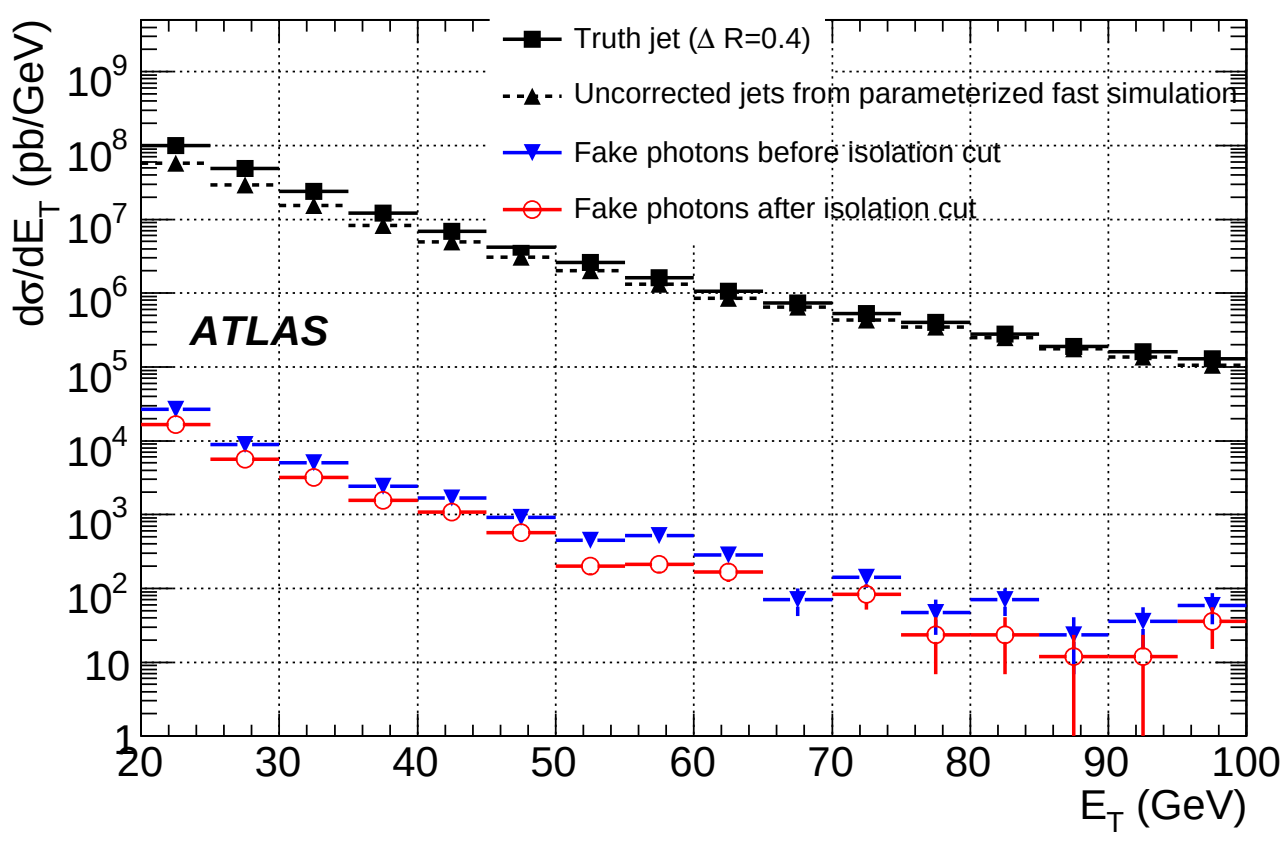

Figure 9: $E_T$ spectra from the inclusive jet sample, for the generated jets (solid squares for full simulation and solid triangles for uncorrected jets from parametrised fast simulation) and the fake-photon candidates before (inverted solid triangles) and after (open circles) the track-isolation cut. The normalisation is that predicted by PYTHIA.

Figure 9 shows the $E_T$ distribution of the jets and of the fake photon candidates before and after the track-isolation cut. This figure also shows that the rejection at 25 GeV is $\approx$ 30% lower if the normalisation is based on the uncorrected parametrised jets from the fast simulation, as was done in Ref. [11].

Figure 10 shows the $\pi^0$ content of the fake-photon candidates at three different cut levels; all reconstructed electromagnetic objects, after the cut on the hadronic leakage and the second layer shower-shape variables (Had+S2) and after all the cuts (Had+S1+S2). A fake photon is defined as coming from a $\pi^0$ if the energy of the leading $\pi^0$ in the cone of 0.2 around the cluster centroid is more than 80% of the reconstructed cluster energy. The figure shows already after the second layer shower-shape cuts, the dominant background contribution comes from $\pi^0$ as expected. After all cuts, the fraction of $\pi^0$ is $\approx$ 70% of the remaining fake-photon candidates.

Figure 11 shows the rejection of the cuts on the first layer variables for candidates from single $\pi^0$'s passing the cuts on the hadronic leakage and the second layer shower-shape variables. As expected, the rejection power against these isolated $\pi^0$'s decreases with energy, as the opening angle between the two photons from $\pi^0$ decays become smaller. The rejection is also better in the central part in the barrel as there is less material than in the higher $\eta$ part of the barrel, and also opening angle is larger than in the end-cap for the same $p_T$ . As a cross-check, Fig. 12 shows the efficiency of the calorimeter cuts for single photons and single $\pi^0$ of $E_T$ = 40 GeV, as a function of pseudorapidity. Again, the rejection is slightly higher than 3 in the central part of the barrel calorimeter and is in reasonable agreement with findings from previous studies [15].

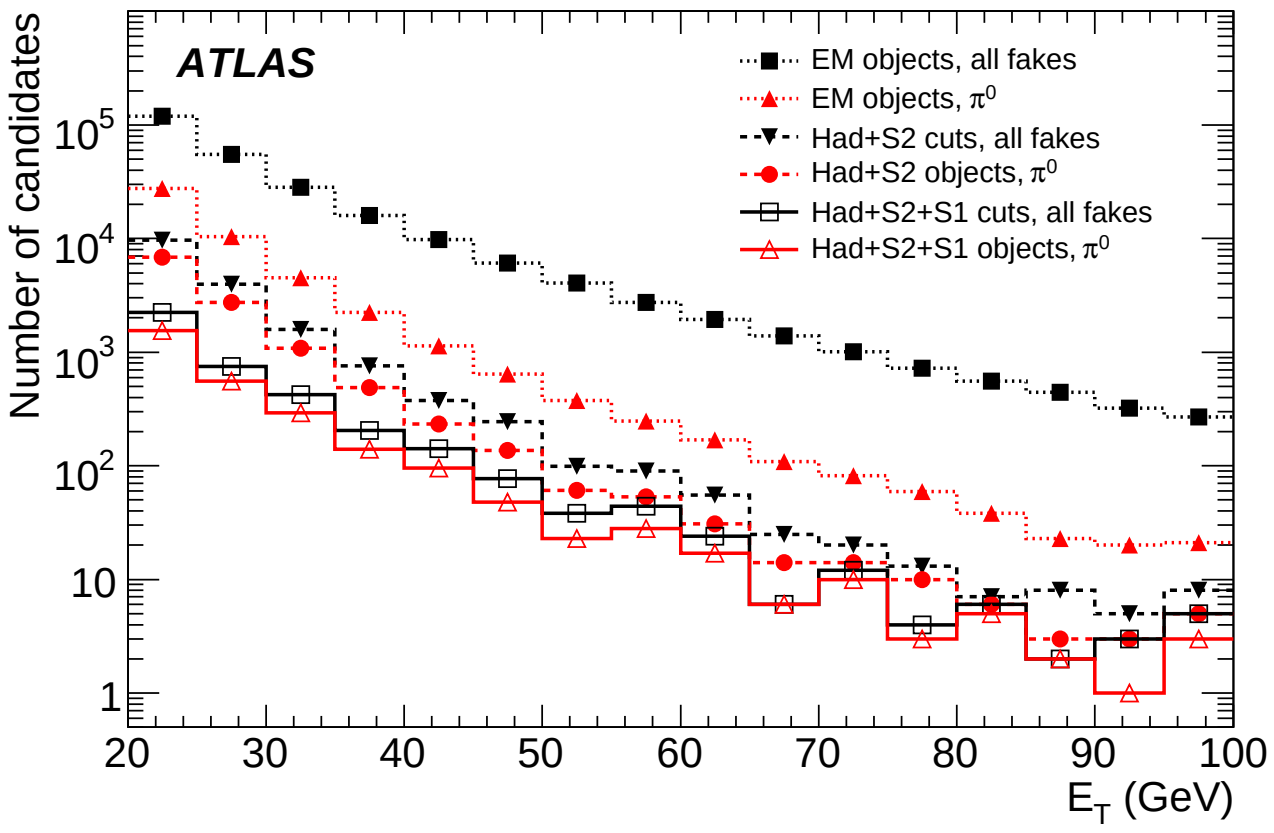

Figure 10: $E_T$ distribution of fake-photon candidates in jets after different level of cuts. The contribution from "single" $\pi^0$ is also shown

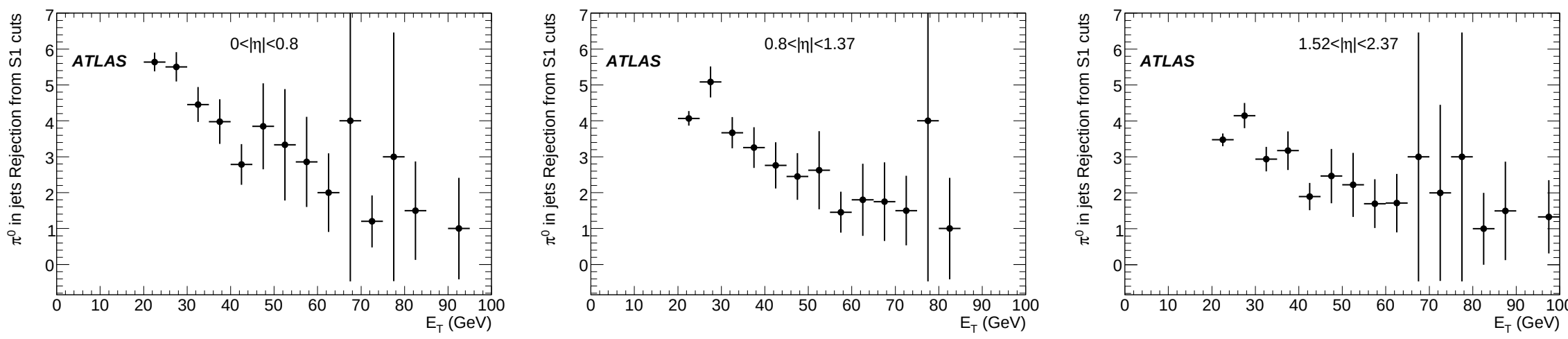

Figure 11: Rejection of the strip-layer cuts against fake photons coming from "single" $\pi^0$ in the jet sample as a function of the transverse energy, for three different pseudorapidity regions.

The rejections measured in these studies have to be taken with care as they rely strongly on the modelling of the fragmentation tail in PYTHIA and the details of the simulation of the detector response. A discussion of the first effect can be found in Ref. [8] from which one would expect an uncertainty of $50-100\%$, and where the uncertainty is larger for gluon initiated jets. In addition, a recent investigation on the differences in fragmentation algorithms in PYTHIA and HERWIG shows appreciable differences in $\pi^0$ production rates. Some differences in rejection are anticipated if the momentum distributions of the $\pi^0$'s from the two fragmentation algorithms differ.

## 4.2 Performance of the Log-likelihood-ratio method

The efficiency for the Log-likelihood-ratio (LLR) method is computed for individual photons from the $H \rightarrow \gamma\gamma$ events generated with the nominal geometry. Figure 13 shows the photon efficiency as a function of $p_T$ (left) and $\eta$ (right) for LLR cut values set at 8, 9 and 10. The overall efficiencies for LLR cuts at 8, 9 and 10 are summarised in Table 4. Jet rejection (left) and photon identification efficiency (right) are shown in Fig. 14 as a function of LLR cut parameter values for three different jet $p_T$ ranges which correspond to the three mean jet $p_T$ values indicated.

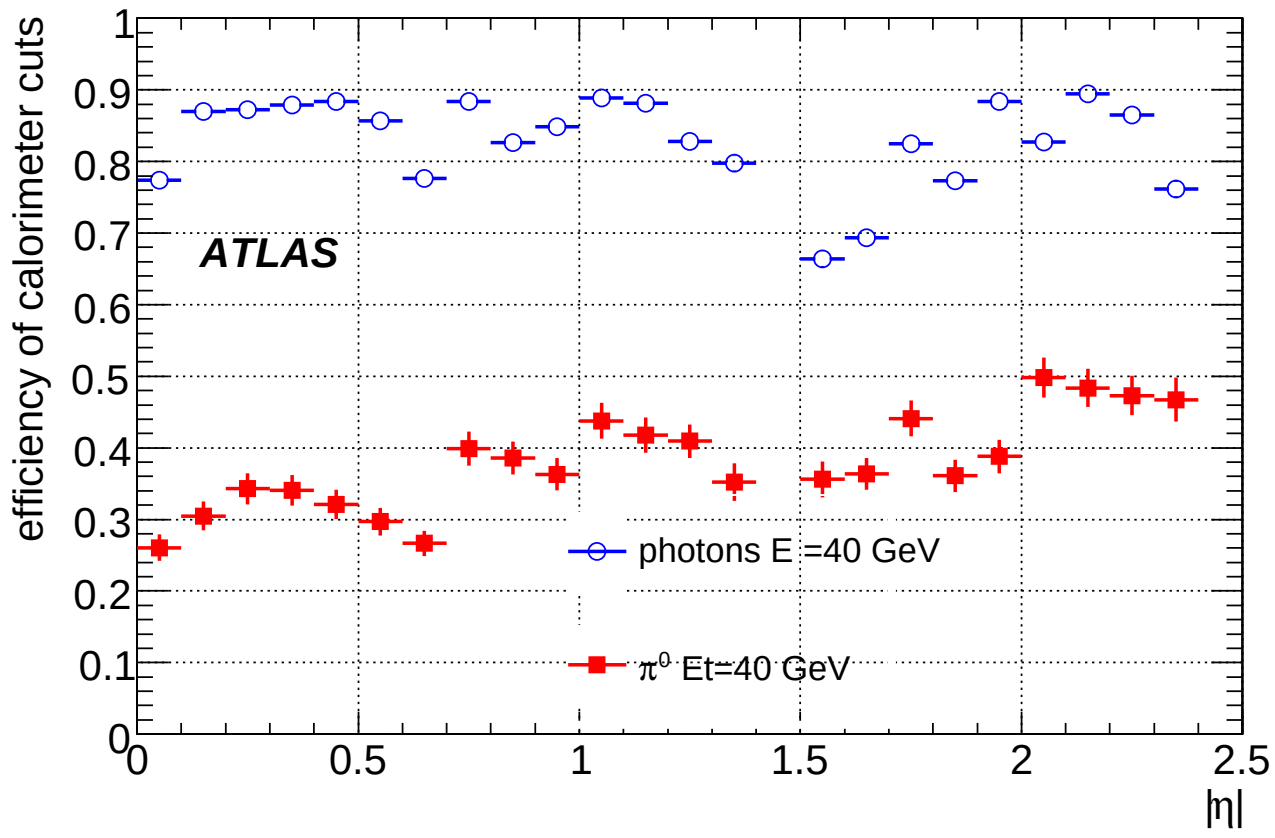

Figure 12: Efficiency of calorimeter cuts versus pseudorapidity for 40 GeV $E_T$ single photons and $\pi^0$ (distorted geometry without pile-up).

| | $E_T > 25$ GeV | | | $E_T > 40$ GeV | | |
|---|---|---|---|---|---|---|
| LLR cut | $LLR > 8$ | $LLR > 9$ | $LLR > 10$ | $LLR > 8$ | $LLR > 9$ | $LLR > 10$ |
| Efficiency(%) | $87.6 \pm 0.3$ | $84.3 \pm 0.2$ | $80.0 \pm 0.2$ | $86.4 \pm 0.3$ | $83.2 \pm 0.2$ | $79.0 \pm 0.2$ |
| Rej.($\gamma$+jet) | $1660 \pm 170$ | $2190 \pm 260$ | $2930 \pm 390$ | $1690 \pm 140$ | $2170 \pm 210$ | $2650 \pm 280$ |
| Rej. (di-jet) | $6820 \pm 440$ | $8930 \pm 650$ | $12430 \pm 1070$ | $6780 \pm 1000$ | $7800 \pm 1230$ | $11550 \pm 2220$ |

Table 4: Overall photon efficiencies and jet rejections for different Log-likelihood ratio (LLR) cut values.

Figure 13 shows the $p_T$-dependence of the photon efficiency. A looser cut on low-$p_T$ photons seems to be beneficial in order to retain a flat photon efficiency as a function of $p_T$. Furthermore, it might also be useful to parametrise the LLR cut values as a function of photon $p_T$ for further optimisation. The jet rejection is also $p_T$-dependent as shown in the plot on the left in Fig. 14. A harder cut on LLR for varying jet $p_T$ can help to keep the rejection constant as a function of $p_T$.

The rejection for jets from $\gamma$+jet and di-jet samples are shown in the fourth and fifth rows in Table 4. The cuts on the photon and jet $p_T$ are 25 GeV and 40 GeV, respectively. The rejection against jets from the di-jet samples is significantly higher than that from the $\gamma$+jet samples. This is largely due to the fact that the jets in $\gamma$+jet events are dominated by quark-initiated jets while those in di-jet events are enriched with gluon-initiated jets.

## 5 Photon identification performance for high-$p_T$ photons

Searches for particles of very high mass decaying to photons, such as the Randall-Sundrum graviton, G, decaying via $G \to \gamma\gamma$ [2], require excellent detector and particle identification performance in a kinematic region very different from the benchmark $H \to \gamma\gamma$ process. The $p_T$-dependent effect caused by differences in kinematics can complicate high-mass graviton searches because they modify the shape of background distributions as a function of the two-photon invariant mass, $m_{\gamma\gamma}$.

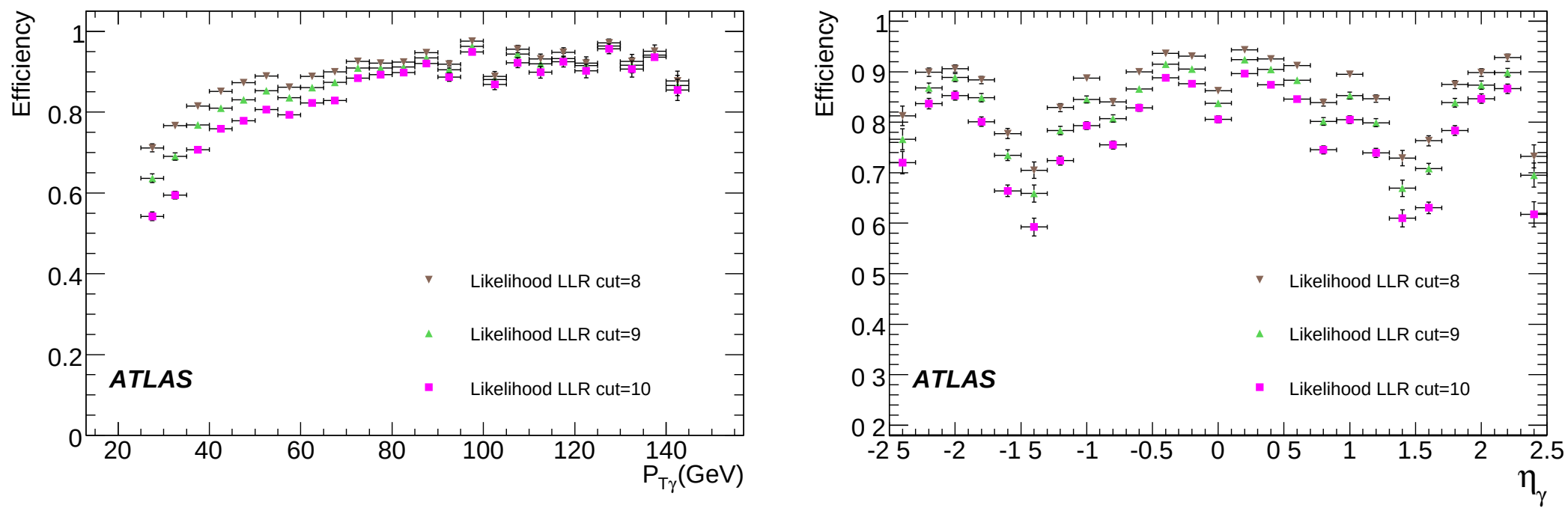

Figure 13: Photon efficiency as a function of $p_T$ and $\eta$ for different Log-likelihood ratio (LLR) cuts. The photons are from $H \rightarrow \gamma\gamma$ decays simulated with the nominal geometry.

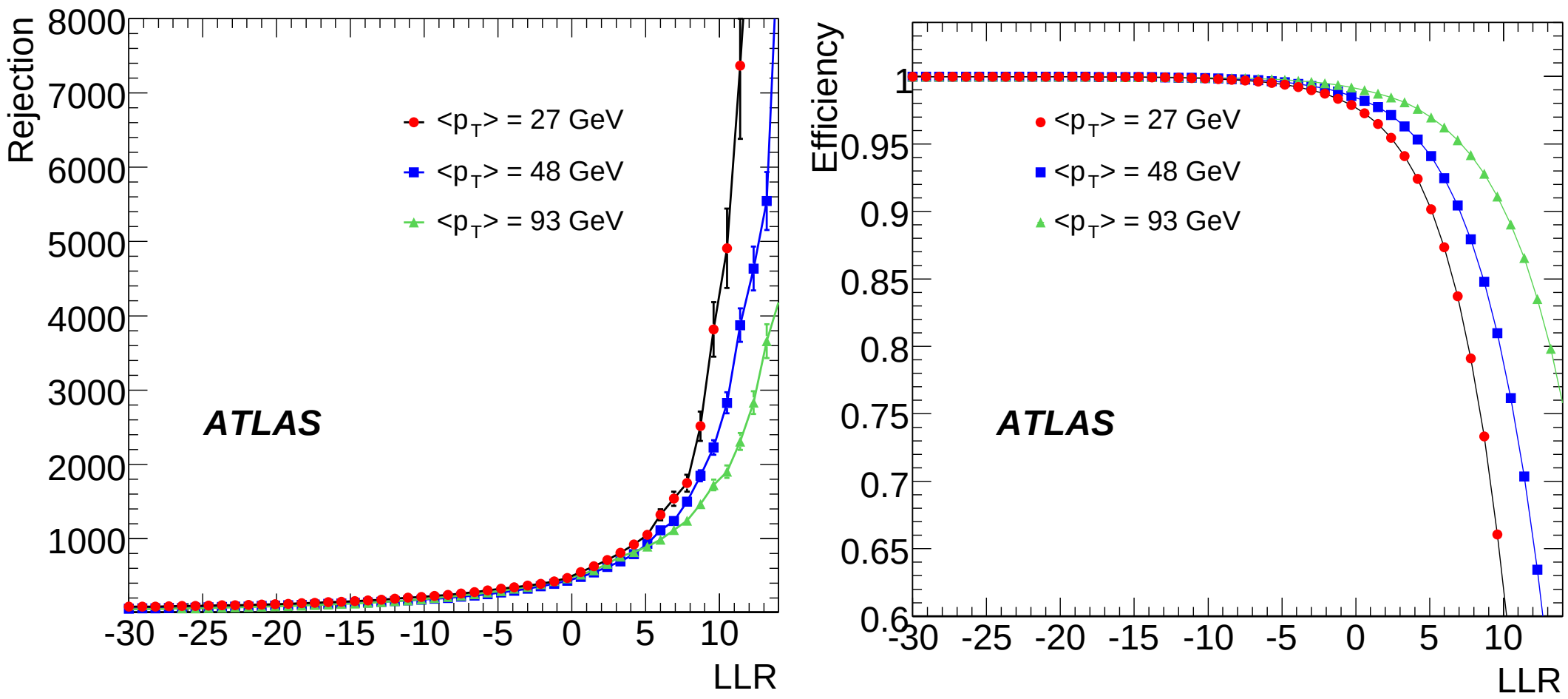

Figure 14: Jet rejection (left) and photon efficiency (right) as a function of Log-likelihood ratio (LLR) cut-parameter values.

The performance of the cut-based identification method for high-$p_T$ photons has been investigated. Studies of the shower characteristics of the photons in the $H \rightarrow \gamma\gamma$ and $G \rightarrow \gamma\gamma$ ($m_G = 500$ GeV) processes found only minor differences in most of the shower-shape variables. In the absence of the track-isolation cut, the photon efficiency as a function of $p_T$ in both the barrel and end-cap calorimeters is approximately constant above $p_T = 50$ GeV. The barrel and end-cap photon electromagnetic reconstruction efficiencies, before applying any identification or isolation cuts, are found to be within 10% of one another for photons from graviton decays. After applying photon identification but no isolation requirements, the efficiencies are $0.829 \pm 0.004$ in the barrel and $0.639 \pm 0.010$ in the end-cap calorimeters for $p_T^{\gamma} > 100$ GeV and $m_G > 500$ GeV.

An isolation variable based on the calorimeter energy in a cone of size $\Delta R = 0.45$ around the cluster centroid was studied. The cut on the calorimeter isolation was observed to produce roughly constant efficiency as a function of $p_T$. A linearly $p_T$-dependent selection cut was determined for barrel and end-cap photons independently. The efficiencies of these $p_T$-dependent cuts for barrel and end-cap calorimeters are shown in Fig. 15 for photons from 500 GeV graviton decays. As can be seen in the figures, these $p_T$-dependent isolation cuts show about a 0.1% reduction in efficiency for photons over the entire $p_T$-range.

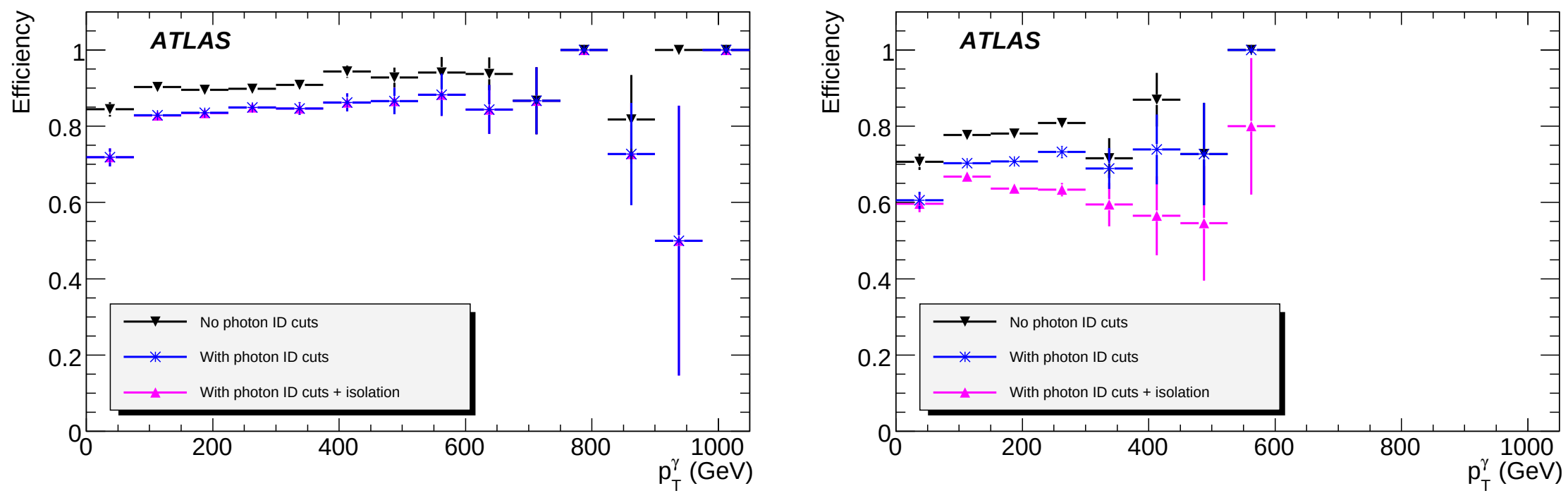

Figure 15: Photon efficiency in the 500 GeV graviton sample as a function of $p_T$ for barrel (left) and end-cap (right) calorimeters.

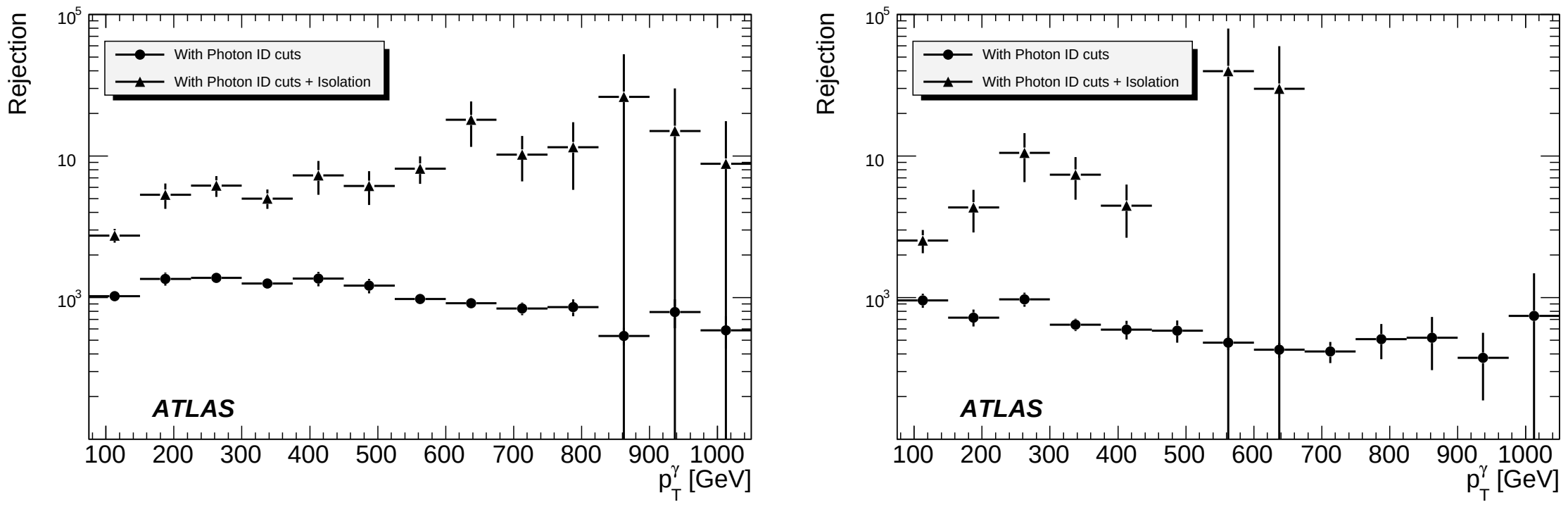

Figure 16: Fake-photon rejection as a function of $p_T$ of the reconstructed photon object for high-$p_T$ binned di-jet samples in the barrel (left) and end-cap (right) calorimeters.

The Jet5 and Jet6 high-$p_T$ jet samples discussed in Section 2 were used for rejection studies. Figure 16 shows the $p_T$ dependence of jet rejection with and without the calorimeter energy isolation cuts. It can be seen that while the efficiency loss is small, employing the isolation cut increases rejection across the full $p_T$ range. In particular, the region below $p_T = 500$ GeV shows a factor $5-10$ increase in rejection. Table 5 provides the measured rejections in the barrel and end-cap calorimeters using these two di-jet samples.

## 6 Comparison of the photon identification methods

Figure 17 shows the rejection and efficiency curves for two of the three currently available photon identification methods - the cut-based method and the Log Likelihood Ratio method - for $\gamma$+jet generated in specific photon momentum bins and the benchmark $H \to \gamma\gamma$ samples. Similarly, Fig. 18 shows the rejection and efficiency curves for these methods for di-jet and $H \to \gamma\gamma$ samples. Tables 6 and 7 provide numerical comparisons of fake-photon rejections for the two methods, for similar photon identification efficiencies and for the $\gamma$+jet and di-jet samples, respectively.

The $\gamma$+jet events, with jets dominated by quark-initiated jets, are the source of the largest background to the $H \to \gamma\gamma$ process. It is apparent from Figs. 17 and 18 that the methods demonstrate significantly reduced rejections for jets from the $\gamma$+jet samples than for those from di-jet samples whose jets are predominantly from gluons. As discussed in previous sections, this difference in rejection can be attributed to the fragmentation differences between the quark and gluon-initiated jets.

Finally, Figs. 17 and 18 also illustrate that, for equal efficiencies, the Log-likelihood ratio method and the cut-based method perform comparably in rejecting jets.

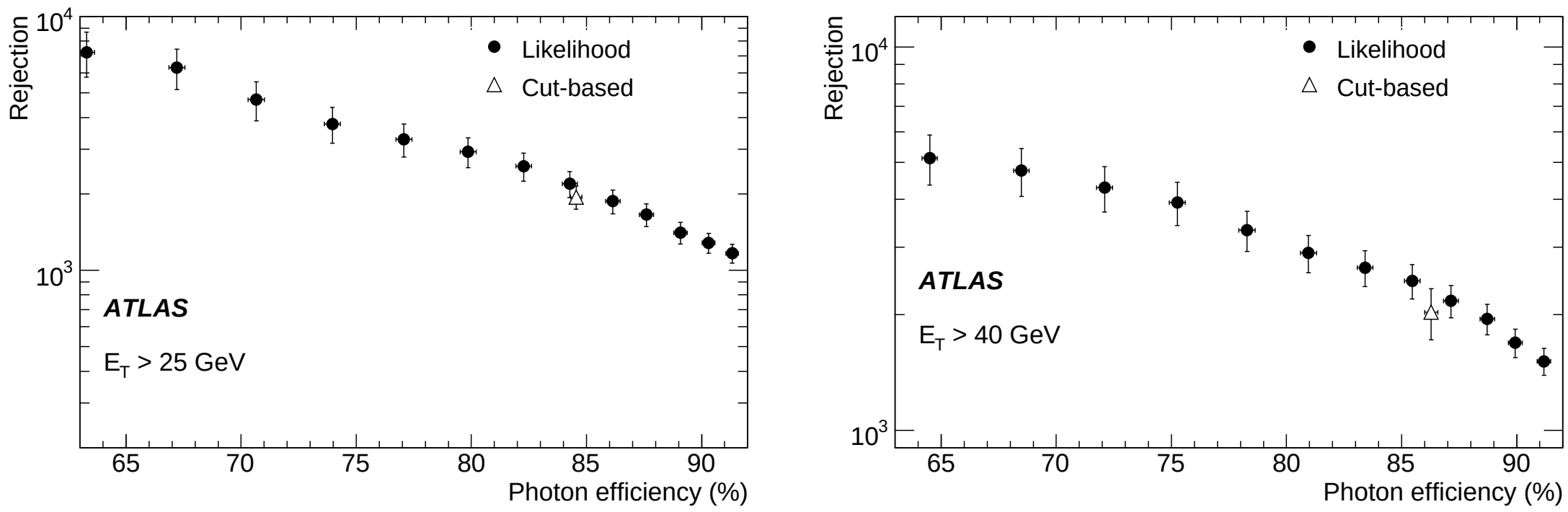

Figure 17: Jet rejection vs photon efficiency for binned $\gamma$+jet and $H \to \gamma\gamma$ benchmark samples for $p_T^{\gamma}, p_T^{jet} > 25$ GeV(left) and $p_T^{\gamma}, p_T^{jet} > 40$ GeV(right).

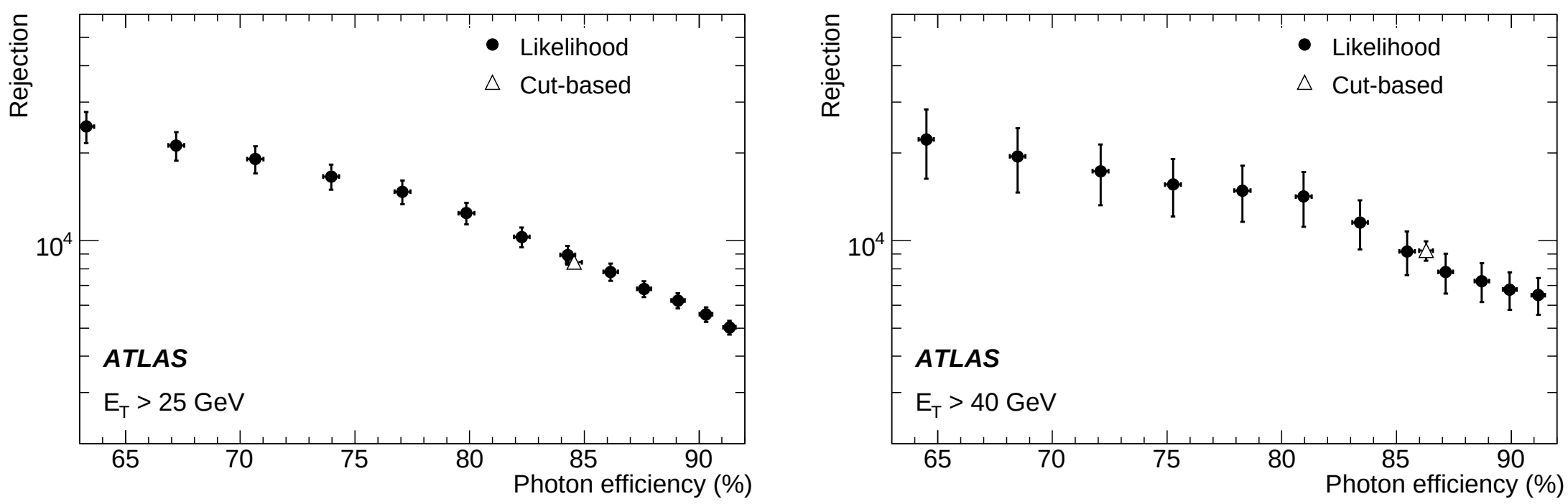

Figure 18: Jet rejection vs photon efficiency of the two methods for filtered di-jet and $H \to \gamma\gamma$ benchmark samples for $p_T^{\gamma},\ p_T^{jet} > 25$ GeV (left) and $p_T^{\gamma},\ p_T^{jet} > 40$ GeV (right).

# 7 Conclusions

This note presents the three photon identification methods developed in ATLAS, the cut-based method, the Log-likelihood ratio (LLR)-based method and the covariance-matrix-based method (H-matrix). The efficiencies and fake-photon rejections of the first two methods have been measured using fully simulated $H \to \gamma\gamma$ ($m_H = 120$ GeV), $\gamma$+jet and filtered electromagnetic di-jet samples. The cut-based and LLR methods show similar rejection factors at equal efficiencies. The strength of the continuous methods such as the LLR and H-matrix is the ability to vary the cuts on LLR or $\chi^2$ values to optimise for specific physics analyses. The performance of the cut-based method for very high-$p_T$ photons from Randall-Sundrum graviton samples has also been studied and, while the cut selection was optimised at low-$p_T$ compared to the signal in the graviton sample, the efficiency remains high. While the currently available photon identification methods perform very well in rejecting background, with high efficiency in retaining photons, it is of critical importance to study the performance of the methods with beam-collision data.

| Region | Rejection($\times 10^3$) | Rejection($\times 10^3$) |
|---|---|---|
| Barrel | $1.55 \pm 0.05$ | $6.59 \pm 0.5$ |
| End-cap | $0.84 \pm 0.04$ | $7.66 \pm 1.1$ |
| Total | $1.32 \pm 0.04$ | $6.79 \pm 0.4$ |

Table 5: Jet rejections obtained using two binned high-$p_T$ di-jet samples, using the cut-based photon identification without (left) and with (right) the track isolation cut.

| | $E_T > 25$ GeV | | $E_T > 40$ GeV | |
|---|---|---|---|---|
| | LLR | Cut-based | LLR | Cut-based |
| Efficiency (%) | $84.3 \pm 0.2$ | $84.5 \pm 0.2$ | $87.1 \pm 0.2$ | $86.3 \pm 0.2$ |
| Rejection | $2190 \pm 250$ | $1940 \pm 230$ | $2170 \pm 210$ | $2030 \pm 190$ |

Table 6: Comparison of jet rejection ($\gamma$+jet sample) versus photon efficiency for the cut-based and LLR methods.

| | $E_T > 25$ GeV | | $E_T > 40$ GeV | |
|---|---|---|---|---|
| | LLR | Cut-based | LLR | Cut-based |
| Efficiency(%) | $84.3 \pm 0.2$ | $84.6 \pm 0.2$ | $85.5 \pm 0.2$ | $86.3 \pm 0.2$ |
| Rejection | $8930 \pm 650$ | $8240 \pm 270$ | $9170 \pm 1570$ | $9240 \pm 710$ |

Table 7: Comparison of jet rejection (di-jet sample) versus photon efficiency with the cut-based method and the Log-likelihood (LLR) method.

# Reconstruction of Photon Conversions

**Abstract**
The reconstruction of photon conversions in the ATLAS detector is important for improving both the efficiency and the accuracy of the detection of particle decays with photon final states, including $H \to \gamma\gamma$. In this note, the performance of the reconstruction of photon conversions for simulated events of different types is described, using both standard inside-out tracking and the more recently implemented outside-in tracking.

## 1 Introduction

Reconstruction of photon conversions in the ATLAS detector is important for a variety of physics measurements involving electromagnetic decay products. In particular, the efficiency of detection of particles with high-mass di-photon final states, such as the Higgs boson or a heavy graviton, is greatly enhanced by efficient conversion reconstruction. Conversion reconstruction will also be used for detector-related studies: mapping the locations of the conversion vertices provides a precise localisation of the material in the ATLAS inner detector.

As photons may convert at any point in the tracker in the presence of material, the ability to reconstruct conversions will depend strongly on the type of tracking algorithm used. Due to the structure of the ATLAS tracker, photons which convert within 300 mm of the beam axis may be reconstructed with a high efficiency with standard (inside-out) Si-seeded tracking, while photons which convert further from the beam pipe may only be reconstructed using (outside-in) tracks, which begin with TRT seeds with or without associated Si hits. Track reconstruction will be discussed in Section 2, while the reconstruction of conversion vertices will be discussed in Section 3, and the overall reconstruction of conversions will be discussed in Section 4. Applications of photon conversion reconstruction in the case of neutral pion decays and low-$p_T$ photons as well as the application of conversion reconstruction to the case of high-$p_T$ physics measurements (such as $H \to \gamma\gamma$), will be found in Section 5. A summary and concluding remarks are found in Section 6.

### 1.1 Theory

The ATLAS detector is designed to measure, among other things, the energies and momenta of photons produced in high-energy proton-proton collisions. The photons which are relevant to physics measurements will have energies in excess of 1 GeV. These photons must pass through the ATLAS tracker before depositing their energy in the Liquid Argon Calorimeter. At photon energies above 1 GeV, the interaction of the photons with the tracker will be completely dominated by $e^+e^-$pair production in the presence of material, otherwise known as photon conversion. All other interactions between the photons and the tracker material, such as Compton or Rayleigh scattering, will have cross-sections which are orders of magnitude below that for the photon conversion, and may thus be safely ignored. The leading-order Feynman diagrams for photon conversions in the presence of material are shown in Figure 1. The presence of the material is required in order for the conversion to satisfy both energy and momentum conservation.

The cross-section for the conversion of photons in the presence of material is both well understood theoretically and thoroughly measured. Work on calculating this cross-section began almost immediately after the discovery of the positron by Anderson in 1932 [1]. Bethe and Heitler first gave a relativistic treatment of photon conversion in 1934 [2] in which the screening of the nuclear Coulomb field was taken into account. A detailed review of the theory regarding photon conversion and the calculation of the conversion cross-section for a variety of materials was given by Tsai in 1974 [3]. A more modern treatment of the topic of conversions, including corrections to the Bethe-Heitler formula for photon energies above 5 TeV was given by Klein in 2006 [4].

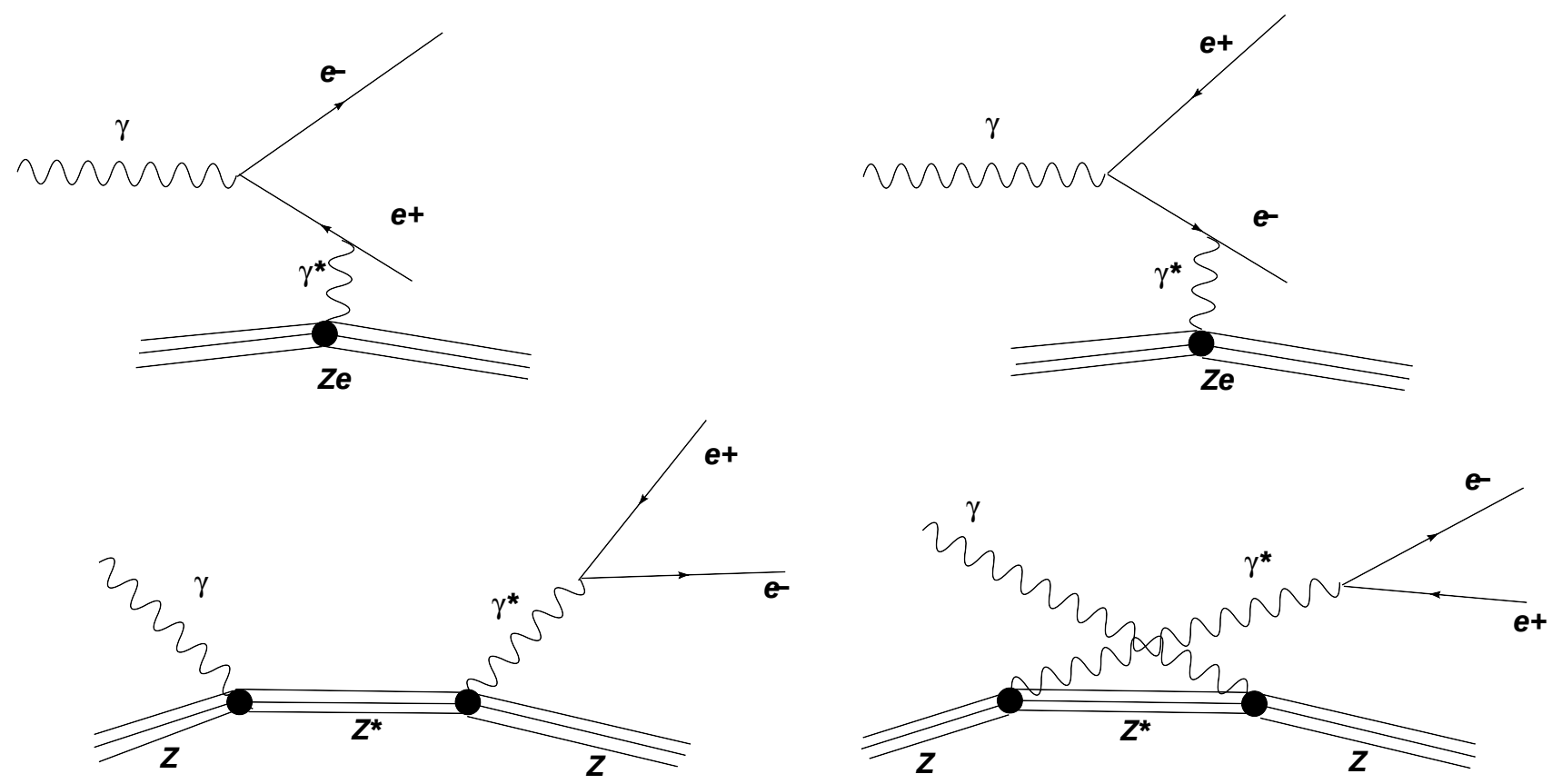

Figure 1: Leading-order Feynman diagrams for photon conversions.

For photon energies used in this study (1 GeV and above) the cross-section for the conversion process is almost completely independent of the energy of the incident photon, and may be given by the following equation [3]:

$$\sigma = \frac{7A}{9X_0N_A}. \tag{1}$$

In this expression $A$ is the atomic mass of the target given in g/mol, and $N_A = 6.022 \times 10^{23}$ is Avogadro's number. $X_0$ is known as the radiation length of the material through which the photon passes, which for elements heavier than helium may be approximated from the atomic mass $A$ and the atomic number $Z$ by the following relation [5]:

$$X_0 = \frac{716.4\,\text{g cm}^{-2} A}{Z(Z+1)\ln(287\sqrt{Z})}. \tag{2}$$

This radiation length is defined such that it is 7/9 of the mean free path for photon conversion. Plots showing the total radiation length traversed by photons in the tracker before reaching the calorimeter may be found in the next section.

The differential cross-section for photon conversions of energies of 1 GeV and above in terms of the quantity $x = (E_{electron}/E_{photon})$ is [4]:

$$\frac{d\sigma}{dx} = \frac{A}{X_0N_A}\left(1 - \frac{4}{3}x(1-x)\right). \tag{3}$$

This cross-section is symmetric in $x$ and $1-x$, the electron and positron energies, and it implies that the momentum of the photon is not simply shared equally between the electron and the positron. Some fraction of the photon conversions will be highly asymmetric, and either the electron or the positron may be produced with a very low energy. If this energy falls below the threshold required to produce a reconstructable track in the ATLAS tracker, then the converted photon will be seen to have only one track, and will be difficult to distinguish from a single electron or positron. This problem is more serious at lower photon energies, as the proportion of conversions which are asymmetric enough to cause the loss of one of the two tracks increases as the photon energy decreases. The difficulties involved in identifying these highly asymmetric single-track conversions will be discussed in a later section.

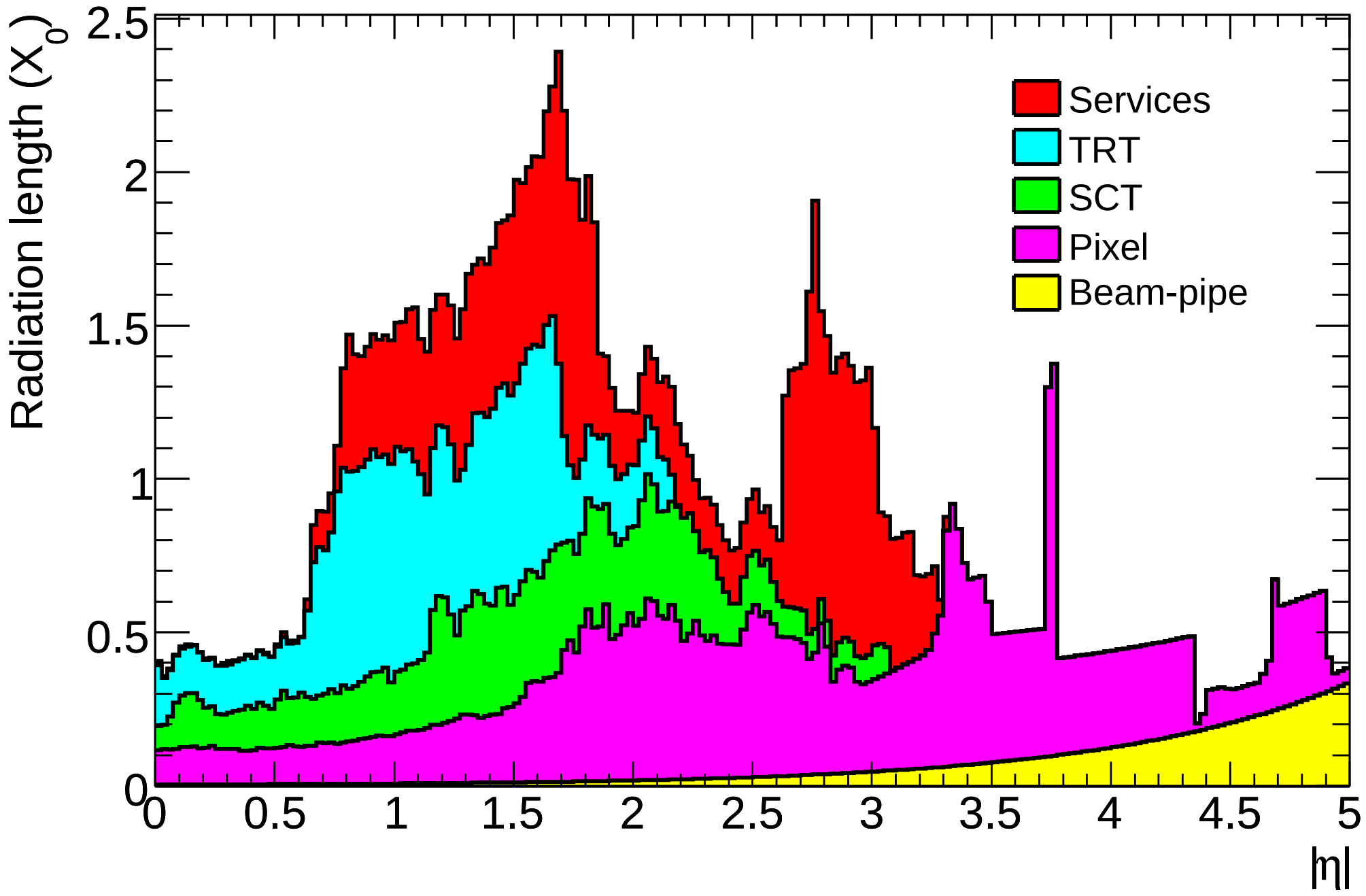

Figure 2: Material in the inner detector as a function of $|\eta|$.

### 1.2 Experimental setup

In this section a very brief description of the ATLAS tracker and electromagnetic calorimeter is included. These are the two sub-systems necessary for the studies relevant to this note. A detailed description of the ATLAS detector can be found in the ATLAS detector paper [6] and references therein.

The ATLAS tracker consists of several co-axial layers immersed in a 2T solenoidal magnetic field. In the so-called barrel region, the innermost of these is a pixel detector consisting of three highly segmented cylindrical layers surrounded by four stereo-pair silicon microstrip (SCT) layers. In addition to the cylindrical layers forming the barrel, both the pixel and the SCT also have end-caps consisting of disk shaped segments used for tracking particles with large pseudorapidities ($|\eta| > 1.5$). The outermost portion of the tracker consists of the Transition Radiation Tracker (TRT), which is comprised of many layers of gaseous straw tube elements interleaved with transition radiation material. The TRT is divided into a barrel detector, covering the small pseudorapidity region $|\eta| < 1$, and two end-cap detectors covering the large pseudorapidity region $1 < |\eta| < 2.1$. The lack of TRT detector elements at higher pseudorapidities is the reason for all the results presented in this note having a cut-off at $|\eta| = 2.1$.

The amount of material in the tracker given in radiation lengths as a function of pseudorapidity can be seen in Fig. 2 [6]. As mentioned earlier, the probability of a photon converting in any given layer is proportional to the amount of material in that layer. Overall, as many as 60 % of the photons will convert into an electron-positron pair before reaching the face of the calorimeter [6]. This number varies greatly with pseudorapidity as can be seen in Fig. 3 [6], for the case of photons with $p_T > 1$ GeV in minimum-bias events. The probability is lowest in the most central region $|\eta| < 0.5$, where the amount of tracker material is at its minimum. A plot showing the true position of photon conversions in the ATLAS tracker, as obtained from a sample of 500,000 simulated minimum-bias events, can be seen in Fig. 4 [6]; the three pixel layers and disks as well as the four barrel SCT layers and their corresponding end-cap layers can be clearly seen.

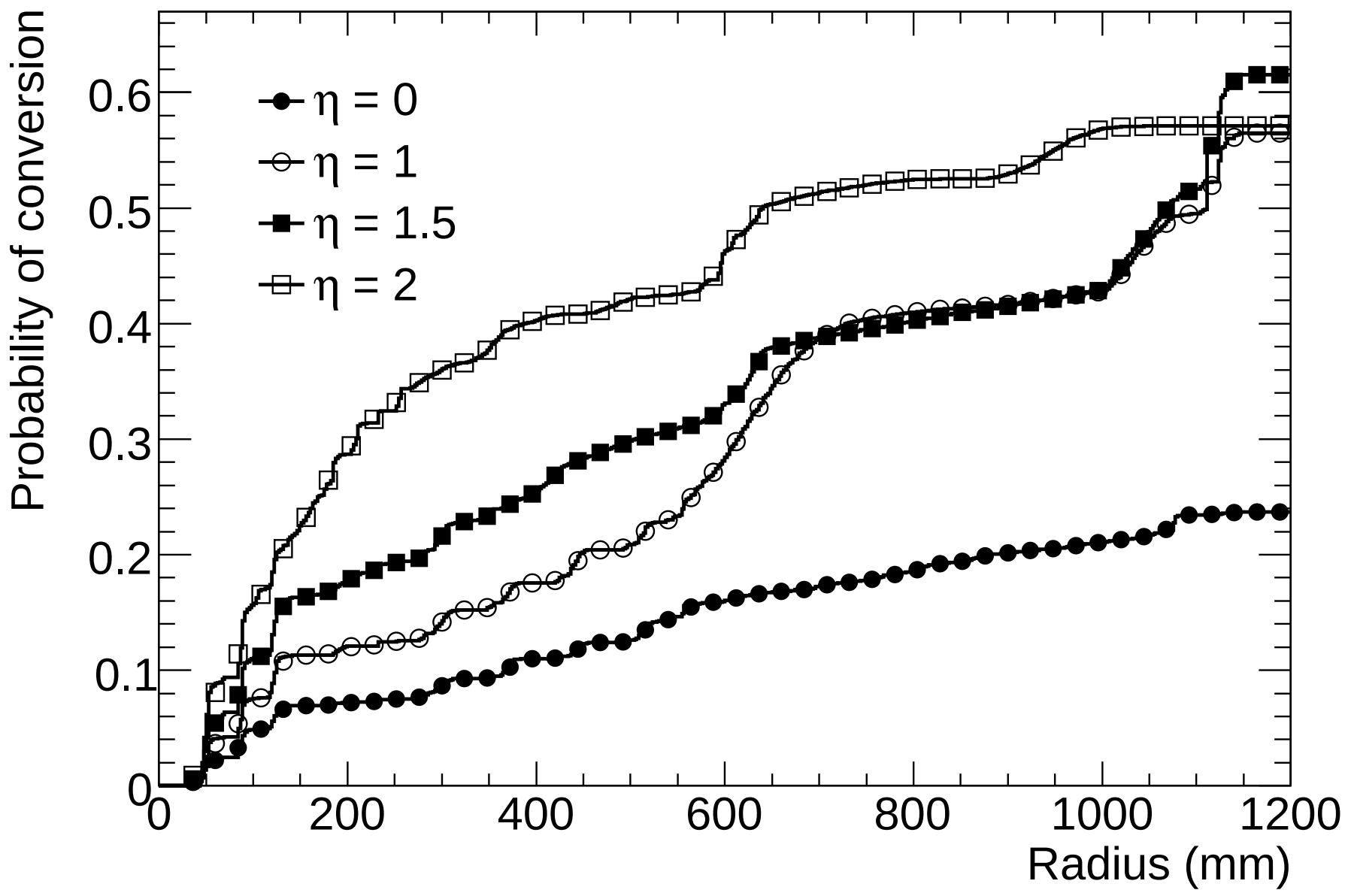

Figure 3: Probability of a photon to have converted as a function of radius for different values of pseudorapidity.

Finally, the energies of the electrons resulting from photon conversions are measured in the electromagnetic calorimeter segments. These are lead-liquid argon detectors with accordion-shaped absorbers and electrodes. Their fine-grained lateral and longitudinal structure, ensures high reconstructed energy resolution for photons with $E_T > 2-3$ GeV, as described in reference [7]. Although the daughter electron tracks and the vertices resulting from the converted photons are reconstructed without any calorimetric information, the latter plays a crucial role later in the reconstruction and particle identification process.

## 2 Track reconstruction

The current track reconstruction process consists of two main sequences, the primary *inside-out* track reconstruction for charged particle tracks originating from the interaction region and a consecutive *outside-in* track reconstruction for tracks originating later inside the tracker. Both methods reconstruct tracks that have both silicon (Si) and transition radiation tracker (TRT) hits and place these tracks in two distinct track collections. A third track category contains those tracks that have only TRT hits and no Si hits; these TRT-only tracks are placed in their own distinct track collection. All three track collections are then examined to remove ambiguities and double counting and are finally merged into a global track collection to be used later during the vertex-reconstruction phase. For a track to be reconstructed by any of these methods, a minimum transverse momentum $p_T > 0.5$ GeV is required throughout. In the following section, brief descriptions of the various tracking algorithms are provided. More detailed descriptions, in particular of the *inside-out* tracking, can be found in reference [8].

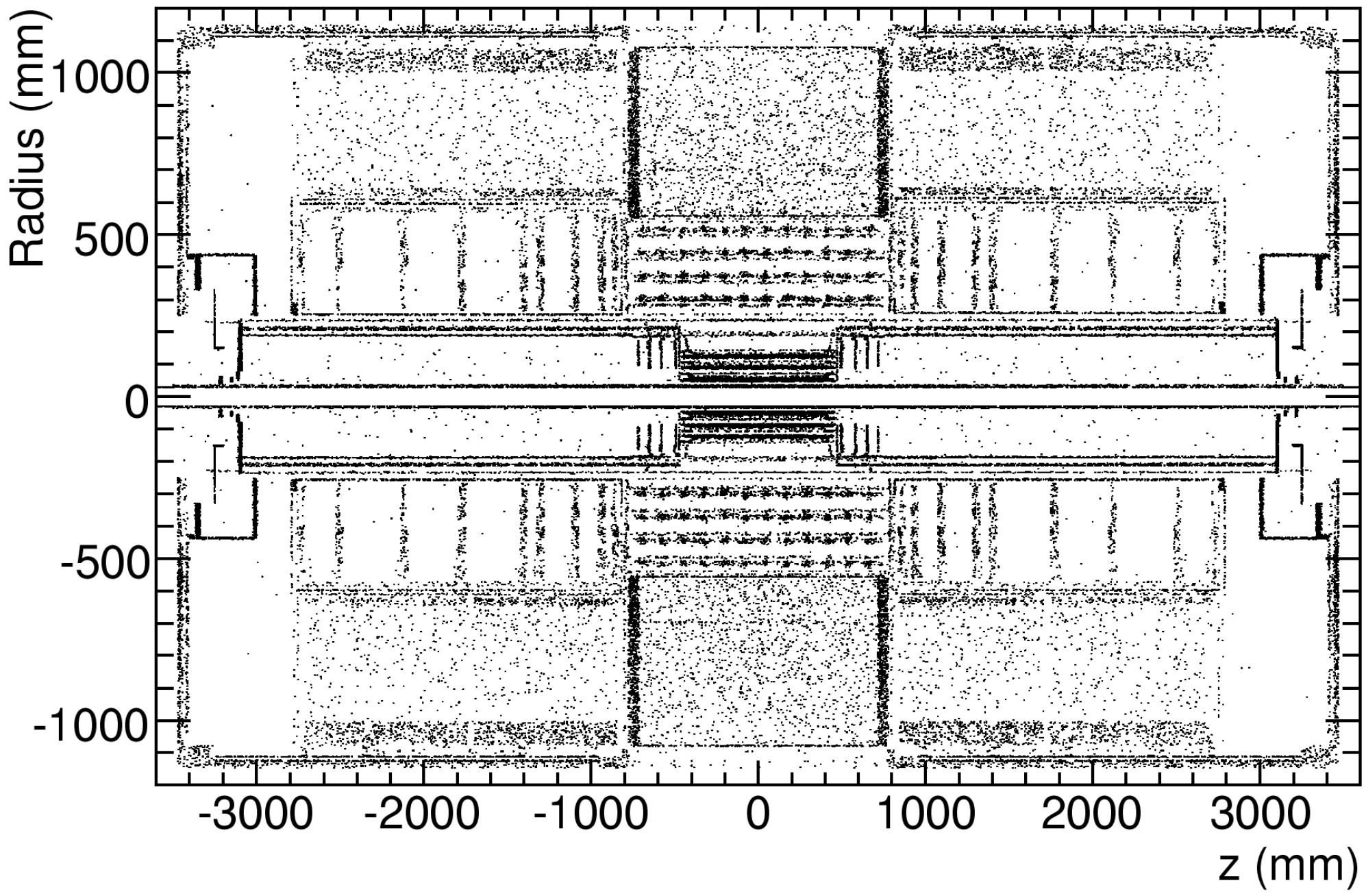

Figure 4: Location of the inner detector material as obtained from the true positions of simulated photon conversions in minimum-bias events.

## 2.1 Inside-out track reconstruction

After the reconstruction of space points inside the pixel and SCT sub-detectors, candidate tracks (seeds) are then formed using three space-point combinations. These seeds are subject to some constraints, such as the curvature, to limit the number of possible combinations. Seeds which pass these constraints then become the starting points for reconstructing tracks. Once a seed has been formed a geometric tool is then invoked in order to provide a list of Si-detector elements that should be searched for additional hits. A combinatorial Kalman-fitter/smoothing formalism is then used to add successive hits to the track. The track information is updated after every step in the search and extraneous outlier hits are efficiently eliminated through their large contribution to the $\chi^2$ of the track fit. Not all space-point seeds coming from Si hits result in a track; the rate at which seeds give rise to a fully reconstructed track is on the order of 10% in a typical $t\bar{t}$ physics event.

A large fraction of the reconstructed track candidates either share hits, are incomplete, or may be fakes resulting from random combinations of hits. It is therefore necessary to evaluate the tracks based on a number of quality criteria and score them accordingly, with the score providing an indication of the likelihood of a specific track to describe a real particle trajectory. Tracks with the highest score are refitted and used as the quality reference for all the remaining tracks. Shared hits are removed in this stage and the remaining part of the track is evaluated again and refitted. Track candidates with too many shared hits are then discarded as well as any other track candidate that fails to comply with any of the quality criteria during evaluation.

At this stage, each one of the resolved track candidates is assigned a TRT extension. First, a geometric extension of the Si track is built inside the TRT and compatible measurements are selected. Possible TRT-track extensions are constructed by combining all such TRT measurements. The full track, including any TRT extension, is then refitted and scored in a way analogous to that during the previous ambiguity-resolving stage. If the new track has a quality score which is higher than that of the original Si track, the TRT extension is kept and added to the Si track, thus creating a "global" Inner Detector track. In other cases only the original Si track is kept, without the TRT extension. The final reconstructed tracks, with or without TRT extensions, are then stored in a dedicated track collection. At this stage they can be classified into three categories:

1. Tracks without TRT extensions (e.g. $|\eta| > 2$);
2. Tracks with extensions which are used in the final fit;
3. Tracks with extensions which are not used in the final fit (outliers).

This last category is characteristic of tracks that have suffered large material interactions as they propagated through the tracker material.

The *inside-out* track reconstruction (as described in the previous section) is a very powerful technique for reconstructing tracks, especially in busy environments where the high granularity of the Si subdetectors (and in particular that of the pixel detector) can provide the necessary resolution for recovering the track-hit pattern. However, it may also lead to fake tracks if not carefully implemented. In order to reduce the number of fake reconstructed tracks, a minimum number of Si hits is required for a track to be reconstructed; in the present implementation of the algorithm this number is seven. This requirement immediately leads to a decreased efficiency in reconstructing tracks that originate late inside the tracker, i.e. in the SCT. Furthermore, tracks which are present only inside the TRT will not be reconstructed at all. These tracks can appear in the cases of secondary decays inside the tracker (e.g. $K_s$ decays) or during photon conversions, the latter being of special interest to this note.

## 2.2 Outside-in track reconstruction

The *outside-in* track reconstruction (also referred to as back-tracking) can offer a remedy to the inefficiency in reconstructing tracks which originate after the pixel detector.

The starting point for this type of track reconstruction is the TRT, where initial track segments are formed using a histogramming technique. The TRT tracker can be divided in two parts, a barrel and an end-cap one, the dividing line being at the $|\eta| = 0.8$ pseudorapidity range. In the $R-\phi$ plane of a TRT barrel sector or the $R-z$ plane of a TRT end-cap sector, tracks which originate roughly at the primary interaction region appear to follow straight lines (this is exactly true in the second case). These straight-line patterns can be characterised by applying the Hough transform [9], which is based on the simple idea that in the $R-\phi(R-z)$ plane, a straight line can be parametrised using two variables: $(\phi_0, c_T)$ or $(\phi_0, c_z)$ respectively, where $c_T$ and $c_z$ are the corresponding azimuthal and longitudinal curvatures and $\phi_0$ is the initial azimuthal angle. As a result, in a two-dimensional histogram formed by these two parameters, TRT straw hits lying on the same straight line will fall within a single cell. Straight lines can therefore be detected by scanning for local maxima in these histograms. To improve the accuracy in the longitudinal direction, the TRT is divided into 13 pseudorapidity slices on either side of the $\eta$=0 plane. The slice size varies, it being smaller around the TRT barrel/end-cap transition region and bigger inside the TRT barrel or end-cap regions. The two-variable approximate track parameters can then be used to define a new set of geometric divisions inside the TRT, within which all straws that could possibly be crossed are included. Using the transformation described in [10], the curved trajectory suggested by the straw hits may be transformed into a straight line in a rotated coordinate system. This is the initial step for a "local" pattern recognition process, in which the best TRT segment may be chosen as the one that crosses the largest number of straws in this straight-line representation. A cut on the minimum number of straw hits necessary to consider the segment as valid is applied during this step. A final Kalman-filter smoother procedure is then applied to determine as accurately as possible the final track parameters of the segment. The above TRT-segment reconstruction procedure has been adopted from the original ATLAS track reconstruction algorithm *xKalman* as described in the references [11].

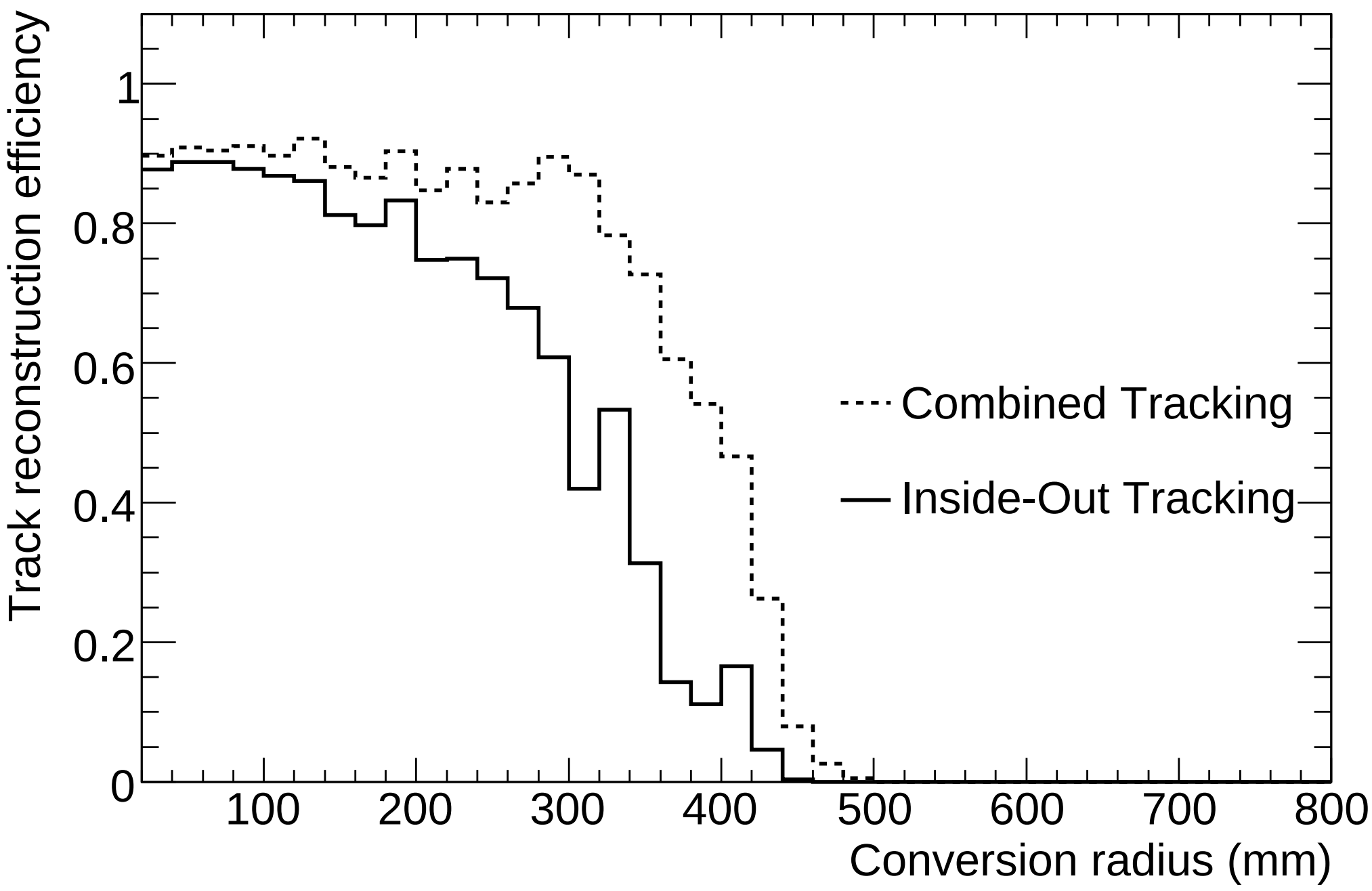

Figure 5: Track reconstruction efficiency for conversions from 20 GeV $p_T$ photons as a function of the conversion radius. The gain in track reconstruction efficiency when tracks reconstructed moving inwards from the TRT are combined with tracks reconstructed by the *inside-out* algorithm, is evident particularly at higher radial distances.

The reconstructed TRT segments are then fed into the second step of the back-tracking algorithm in which extensions are added to them from the Si sub-detectors. Space-point seeds are searched for in narrow $R-\phi$ wedges of the Si tracker, indicated by the transverse TRT-segment track parameters derived in the previous step. A minimum of two space points is required in this case, the search being confined to the last three SCT layers. To reduce the number of space-point combinations cuts on the curvature are then applied, with the third measurement point provided by the first hit in the initial TRT segment. As soon as seeds with pairs of space points are formed, the initial-segment track parameters can then be significantly improved, especially the longitudinal components. A new geometric section through the Si-detector elements is then constructed and a combinatorial Kalman-fitter/smoother technique, as in the case of the *inside-out* tracking, is applied to produce Si-track extension candidates. The Si-track extensions provide a much improved set of track parameters, which can be used to find new TRT extensions to be assigned to every Si-track candidate, thus creating once more a “global” track. Ambiguity resolving and track refitting follow afterwards in the appropriate manner. The final set of resolved tracks from this process is stored in a dedicated track collection. In order to reduce the time required for the reconstruction and minimise double counting, the *outside-in* tracking procedure excludes all the TRT-straw hits and Si-detector space points that have already been assigned to *inside-out* tracks. The enhancement of the track reconstruction efficiency after the *outside-in* reconstructed tracks are included is shown in Fig. 5. Here the track reconstruction efficiency for photon conversions is plotted as a function of the radial distance of the conversion for the case of 20 GeV $p_T$ single photons, before and after the *outside-in* tracking is performed. The bulk of the gain in tracking efficiency is, as expected, at larger radii. The inefficiencies of this method as a function of radius are discussed further in Section 2.3 and again in Sections 4.3 and 4.4. Due to the more limited pseudorapidity coverage of the TRT tracker, the *outside-in* tracking can be used to efficiently reconstruct tracks up to a pseudorapidity value of $|\eta| = 2.1$. All the results presented here have therefore been restricted to within this pseudorapidity range.

### 2.3 Stand-alone TRT tracks and final track collection.

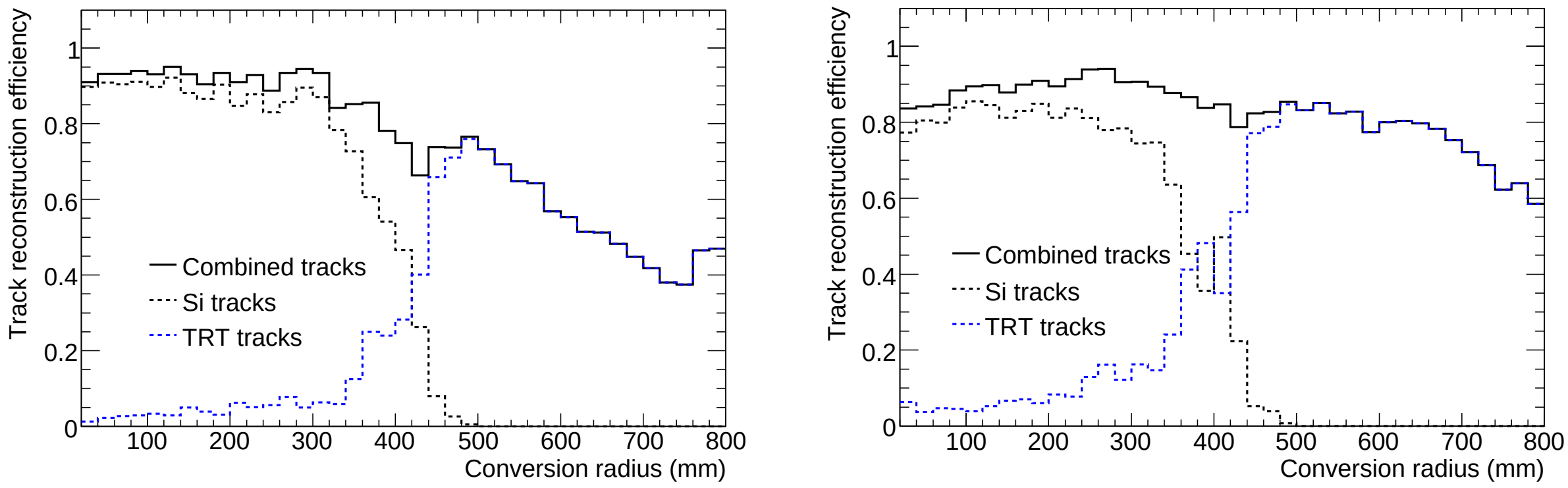

Figure 6: Track reconstruction efficiency for conversions from 20 GeV $p_T$ converted photons (left) and 5 GeV $p_T$ converted photons (right) as a function of conversion radius.

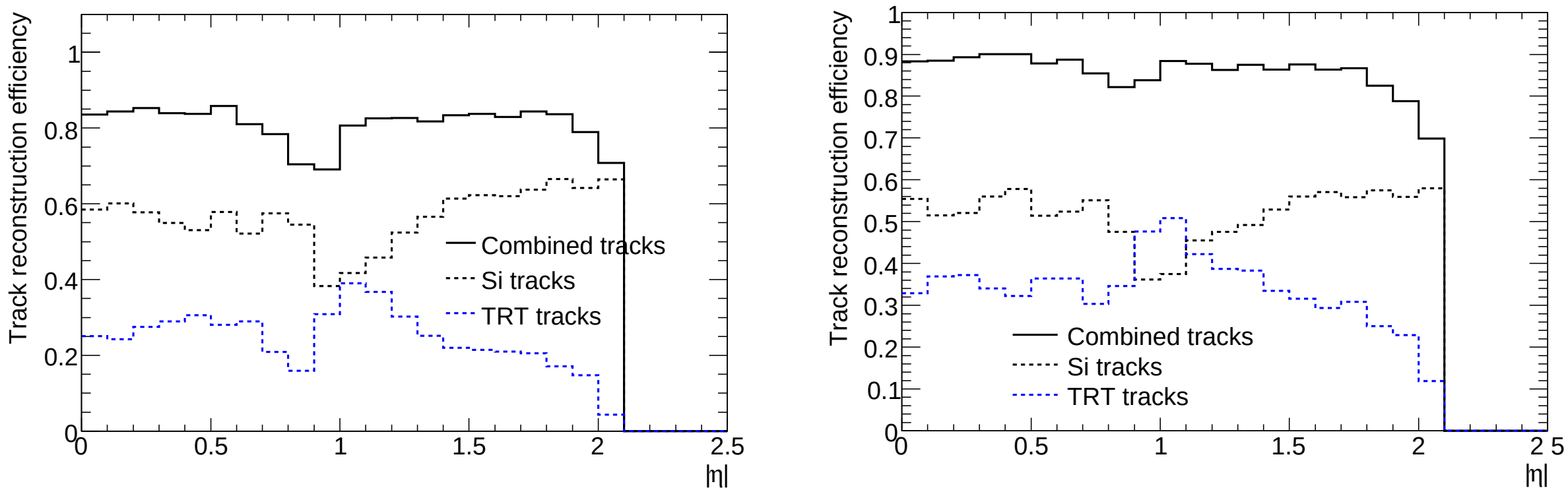

Figure 7: Track reconstruction efficiency for conversions from 20 GeV photons (left) and 5 GeV photons (right) as a function of pseudorapidity.

After the *inside-out* track collection has been formed, all TRT segments that have not been assigned any Si extensions are then used as the basis of one more distinct track collection. These segments are first transformed into tracks, and the segment local parameters are used as the basis for producing the corresponding track parameters assigned to the surface of the first straw hit. Perigee parameters are also computed, but no overall track refitting is performed. These new TRT tracks are then scored and arranged accordingly and a final ambiguity resolving is performed in order to reject any tracks that share too many straw hits. Finally, these stand-alone TRT tracks are then stored in a special track collection.

At the end of the track reconstruction process, and before any primary or secondary vertex fitters are called or other post-processing tasks are executed, the three track collections described above are merged. One last ambiguity resolving is performed in order to select unique tracks from all three collections, although this is mostly for consistency since the straw hits and Si space points associated with the *inside-out* tracks have already been excluded before the *outside-in* track reconstruction. This merged track collection is then used by the photon conversion reconstruction algorithm.

The overall tracking efficiency after all three track collections discussed above are merged, is shown in Fig. 6 for both the case of a 20 GeV $p_T$ single photon sample, and also for a 5 GeV single photon sample, which is more indicative of the case of low track momenta. Two competing effects become apparent as one observes these two plots. The overall track reconstruction efficiency for conversions that happen early inside the tracker, i.e. in $R < 150$ mm, is higher in the case of the 20 GeV $p_T$ photons than that for the 5 GeV $p_T$ ones. This is a clear indication of the larger effect that bremsstrahlung losses have on low $p_T$ tracks, especially on those that originate early inside the tracker. Furthermore it is possible that, depending on the amount of the incurred losses, only part of the track will be reconstructed, i.e. its TRT component, with the pattern recognition failing to recover the corresponding Si clusters. The small fraction of stand-alone TRT tracks that enhance the track reconstruction efficiency from early conversions, is primarily due to this effect. On the other hand the overall track reconstruction efficiency at higher radii is much better for the case of the 5 GeV $p_T$ photons. This is due to the fact that the radius of curvature, being much larger for those tracks, enables them to separate from each other faster as they traverse the tracker under the influence of the applied magnetic field. It is therefore easier in this case to distinguish the two tracks and reconstruct them during the pattern recognition stage. Figure 7 shows the track reconstruction efficiency as a function of pseudorapidity, for both 20 GeV and 5 GeV $p_T$ photons. The overall track reconstruction efficiency is very uniform along the whole pseudorapidity range, starting only to significantly fall off as one approaches the limit of the TRT pseudorapidity extent ($|\eta| = 2.1$). The reduction in efficiency observed around $|\eta| = 1$, is due to the gap at the transition from the barrel to the end cap TRT. The seemingly higher overall tracking efficiency in this plot compared to that in Fig. 6, is due to the fact that the great majority of converted photons originate from the earlier layers of the Si tracker. In this region, as Fig. 6 demonstrates, the converted photon track reconstruction efficiency is very high.

## 3 Vertex fitting

Track finding is only the first step in reconstructing photon conversions; the next step is being able to reconstruct the conversion vertex using the pair of tracks produced by the converted photon. Reconstruction of the conversion vertex is quite different from finding the primary interaction vertex, since for conversions additional constraints can be applied that directly relate to the fact that the converted photon is a massless particle. A specific vertex algorithm, appropriately modified in order to take into account the massless nature of the conversion vertex, has been developed for use by the photon conversion algorithm.

The vertex fit itself is based on the fast-Kalman filtering method; different robust versions of the fitting functional can also be set up in order to reduce the sensitivity to outlying measurements. The vertex fitting procedure uses the full 3D information from the input tracks including the complete error matrices [12].

### 3.1 Algorithm description

The goal of a full 3D vertex fit is to obtain the vertex position and track momenta at the vertex for all tracks participating in the fit as well as the corresponding error matrices. From the input tracks, the helix perigee parameters defining the particle trajectory along with their weight matrix are extracted as described in the references [13, 14]. If one assumes that the particle is created at the vertex $\vec{V}$, then the trajectory parameters $q_i$ may be expressed as a function of the vertex position and the particle momentum at this vertex $q_i = T(\vec{V}, \vec{p}_i)$. A vertex is then obtained by minimising:

$$\chi^2 = \sum_{i=1}^{2} (q_i - T(\vec{V}, \vec{p}_i))^\top w_i (q_i - T(\vec{V}, \vec{p}_i)), \tag{4}$$

where $w_i$ is the $5 \times 5$ weight matrix from the track fit. In order to find the $\vec{V}$ and $\vec{p}_i$ which minimise the above $\chi^2$, equation 4 can be linearised at some convenient point close to the vertex as:

$$\chi^2 = \sum_{i=1}^{2} (\delta q_i - D_i \delta \vec{V} - E_i \delta \vec{p}_i)^\top W_i (\delta q_i - D_i \delta \vec{V} - E_i \delta \vec{p}_i), \tag{5}$$

where $D_i = (\partial T(\vec{V}, \vec{p}_i))/(\partial \vec{V})$ and $E_i = (\partial T(\vec{V}, \vec{p}_i))/(\partial \vec{p}_i)$ are matrices of derivatives. A fast method to find a solution that minimises equation 5 has been proposed in the references [13,14]. It can be shown that this method is completely equivalent to a Kalman-filter based approach [15], where the vertex position is recalculated after every new track addition.

If the initial estimation of the vertex position is far from the fitted vertex, then the track perigee parameters and the error matrix are extrapolated to the fitted point, all derivatives are recalculated and the fitting procedure is repeated. The official tracker extrapolation engine, along with a magnetic field description based on the actual measurement of the ATLAS tracker solenoidal field, is used in this case.

### 3.2 Vertex fit constraints

Constraints are included in the vertex fit algorithm via the Langrange multiplier method. A constraint can be viewed as a function

$$A_j(\vec{V}, \vec{p}_1, \vec{p}_2, ..., \vec{p}_n) = const \tag{6}$$

which is added to the fitting function of equation 4 as

$$\chi^2 = \chi_0^2 + \sum_{j=1}^{N_{const}} \lambda_j \cdot A_j^2 \tag{7}$$

Here $\chi_0^2$ is the function without constraints, $\lambda_j$ is a Lagrange multiplier and $j$ is the constraint number. $A_j^2(...)$ can be linearised around some point $(\vec{V}_0, \vec{p}_{0i})$ to obtain

$$\chi^2 = \chi_0^2 + \sum_{j=1}^{N_{const}} \lambda_j \cdot (A_{j0}^2 + H_j^\top \delta V + \delta V^\top H_j + F_{ij}^\top \delta p_i + \delta p_i^\top F_{ij}) \tag{8}$$

where $H_j = (\partial A_j)/(\partial \vec{V}), F_{ij} = (\partial A_j)/(\partial \vec{p}_i), A_{j0}$ is an *exact* value of $A_j$ at the $(\vec{V}_0, \vec{p}_{0i})$ point, $\delta \vec{V} = \vec{V} - \vec{V}_0$ and $\delta \vec{p}_i = \vec{p}_i - \vec{p}_{0i}$.

The solution of equation 8 then has the form $\vec{V} = \vec{V}_0 + \vec{V}_1$, $\vec{p}_i = \vec{p}_{0i} + \vec{p}_{1i}$, where $\vec{V}_0, \vec{p}_{0i}$ is the solution of the corresponding problem without the constraint $\chi^2 = \chi_0^2$. The second component $\vec{V}_1, \vec{p}_{1i}$ of the above solution is obtained through the normal Lagrange multiplier system of equations. In the case of the conversion vertex, a single angular constraint needs to be implemented. This requires that the two tracks produced at the vertex should have an initial difference of zero in their azimuthal and polar angles $\delta\phi_0, \delta\theta_0 = 0$. This is a direct consequence of having an initial massless particle, but it has the advantage of being much easier to implement.

The right-hand plot in Fig. 8 shows the reconstructed photon inverse transverse momentum after vertex fitting for conversions where neither of the emitted electrons suffered significant bremsstrahlung (less than 20% of the energy of each electron is lost in the inner detector material), while the left-hand plot shows the transverse momentum for the cases where significant bremsstrahlung energy losses occurred. Similarly, the corresponding radial position resolution for conversions with/without significant energy losses due to bremsstrahlung is shown in Fig. 9. Single converted photons with $p_T$ = 20 GeV were used for the plots above, and the emitted electron tracks were required to have at least two silicon space points. The angular constraints $\delta\phi, \delta\theta = 0$, implemented as described earlier, have been used throughout. The overall vertex reconstruction efficiency will be discussed in the following section. It is evident that the presence of bremsstrahlung significantly deteriorates the performance of the vertex fitter.

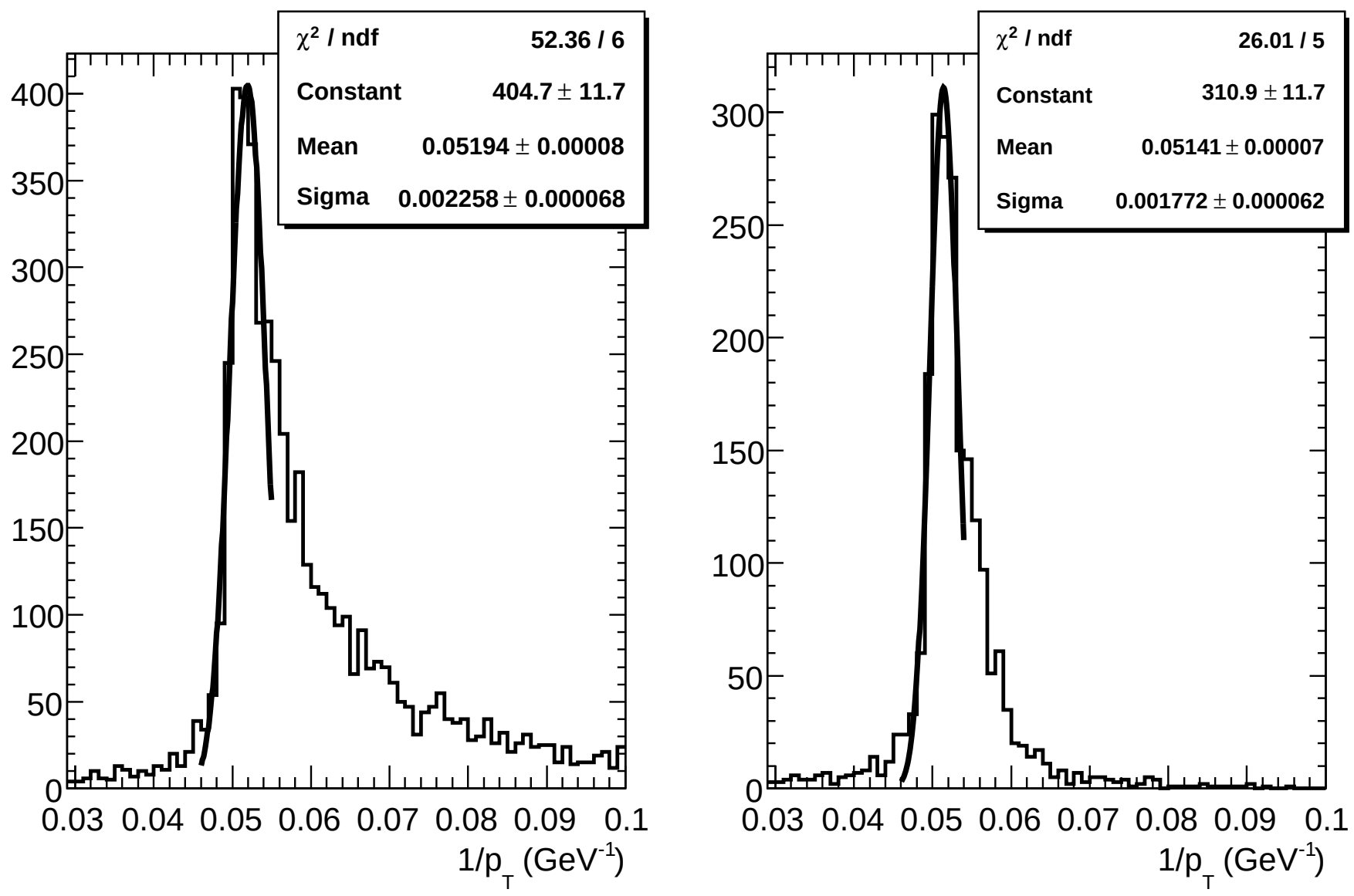

Figure 8: Reconstructed inverse transverse momentum from 20 GeV $p_T$ converted photons with (left) and without (right) significant energy losses due to bremsstrahlung.

As a further check of the performance of the vertex algorithm described in this section, one can apply it to the case of $K_s^0 \to \pi^+\pi^-$ decays. The absence of losses due to bremsstrahlung for the pion tracks, as well as the non-zero opening angle, provide a good test scenario for the constrained vertex fitting. Instead of the angular constraint used in the case of the photon conversions, a straightforward mass constraint is implemented in this case. Figure 10 shows the resolution of both the reconstructed $1/p_T$ and the radial position for 10 GeV $p_T$ $K_s^0$ decays. The absence of a bremsstrahlung-related tail in the left-hand plot compared to those in Fig. 8 is striking.

In a direct comparison to the converted photon case, Fig. 11 shows the relative $1/p_T$ resolution, with and without significant bremsstrahlung losses (20%) respectively, for reconstructed 20 GeV $p_T$ converted photons, together with that for 10 GeV $p_T$ $K_s^0$ decays, as a function of the radial distance from the beam axis. In the case of the $K_s^0$ decays, the reconstructed momentum resolution is better than 2% irrespective of the radial distance from the beam axis, deteriorating only slightly as one moves away from the beam axis. For the case of the photon conversions though, a deterioration in the transverse momentum reconstruction resolution due to the presence of bremsstrahlung losses, is clearly observable when compared to the $K_s^0$ case. Due to the bremsstrahlung losses, the reconstructed $1/p_T$ distribution has a non-gaussian shape, characterised by a tail towards the higher $1/p_T$ ranges, as shown in Fig. 8. As a result, a gaussian fit performed on the core of the $1/p_T$ distribution, will result in a worse overall reconstructed momentum resolution, even in the case of small bremsstrahlung losses, as Fig. 11 demonstrates. The effect is even more significant if one recalls that the reconstructed converted photons have a transverse momentum which is twice that of the $K_s^0$ decays shown in the same figure. The fact that the photon is a massless particle, resulting in an extremely small angular opening of the emitted tracks, makes it also more difficult to reconstruct accurately the position of the conversion vertex, as shown in Fig. 12.

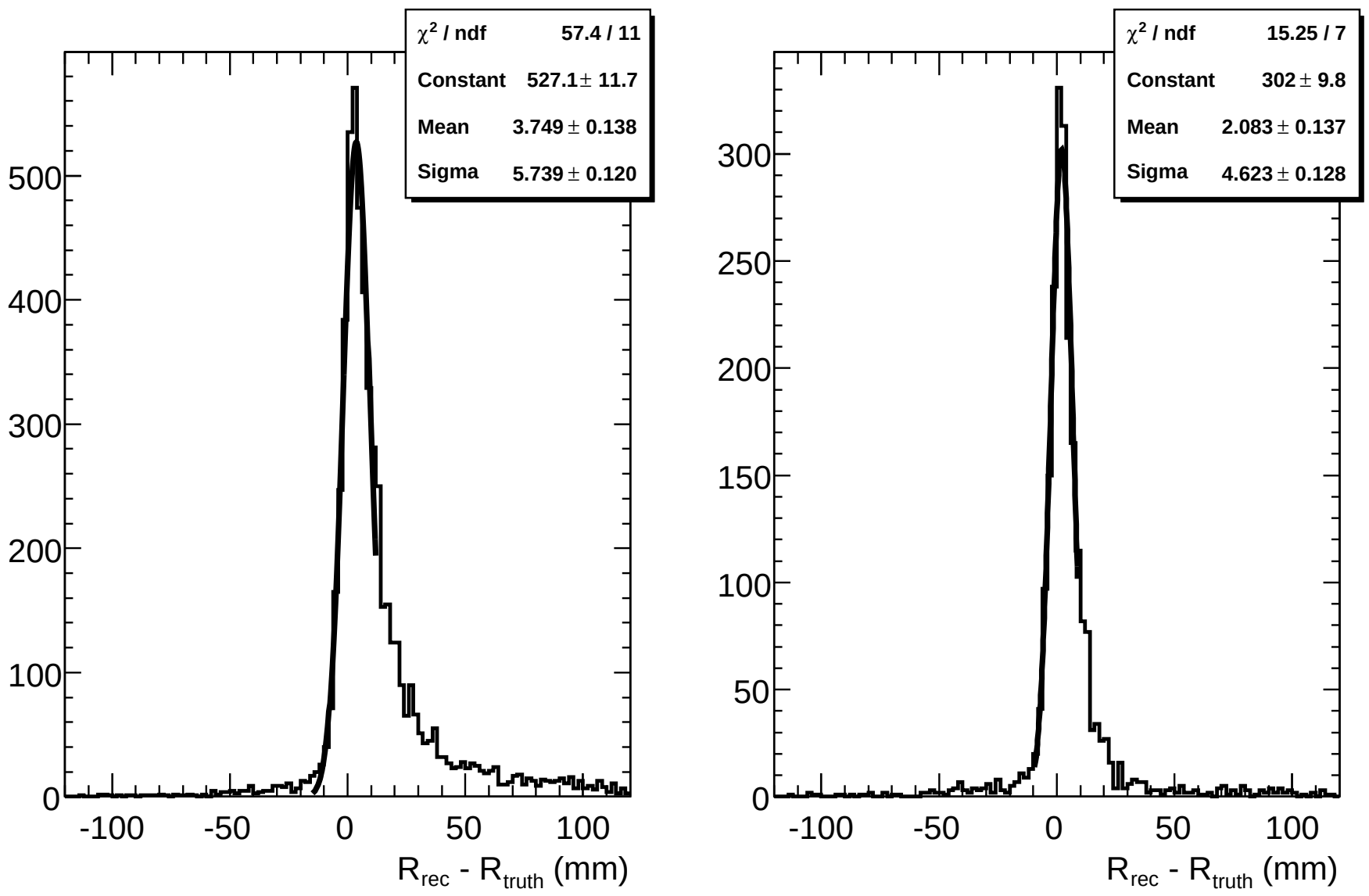

Figure 9: Reconstructed vertex radial positions for 20 GeV $p_T$ converted photons, compared to their true values, with (left) and without (right) significant energy losses due to bremsstrahlung.

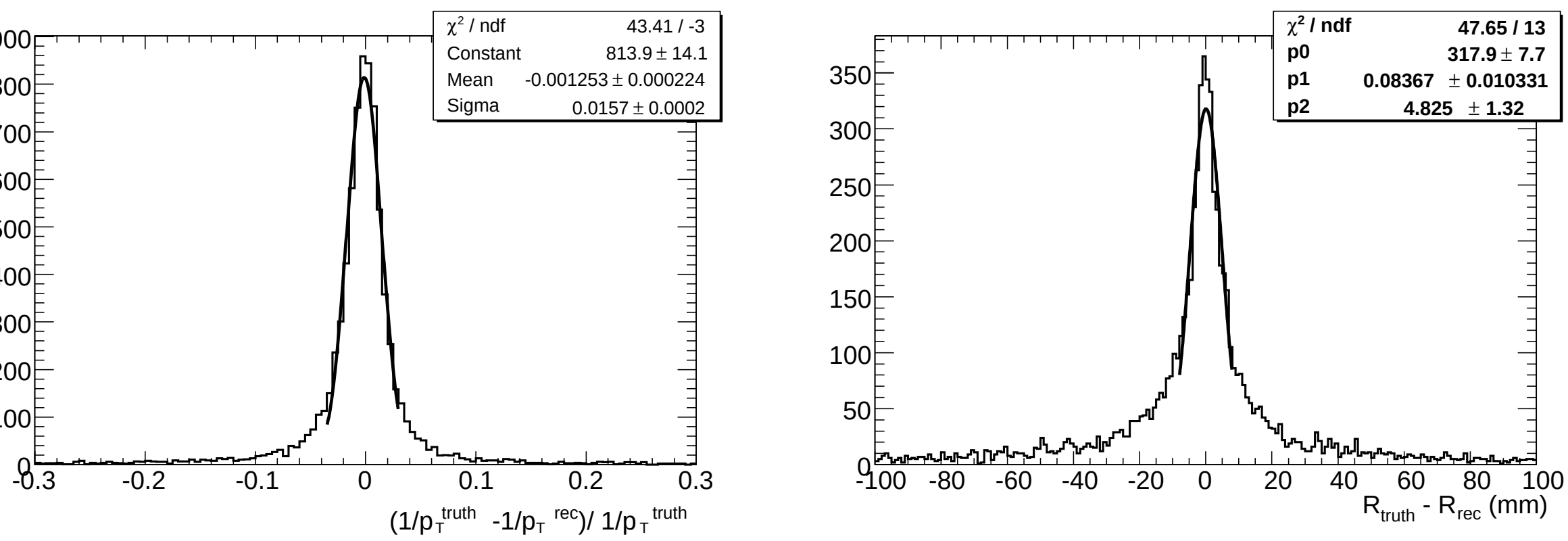

Figure 10: Overall reconstructed relative $1/p_T$ resolution (left) and radial position resolution (right) for $K_s^0$ decays (to charged pions) with $p_T$ = 10 GeV. Only tracks with at least two silicon space points are used.

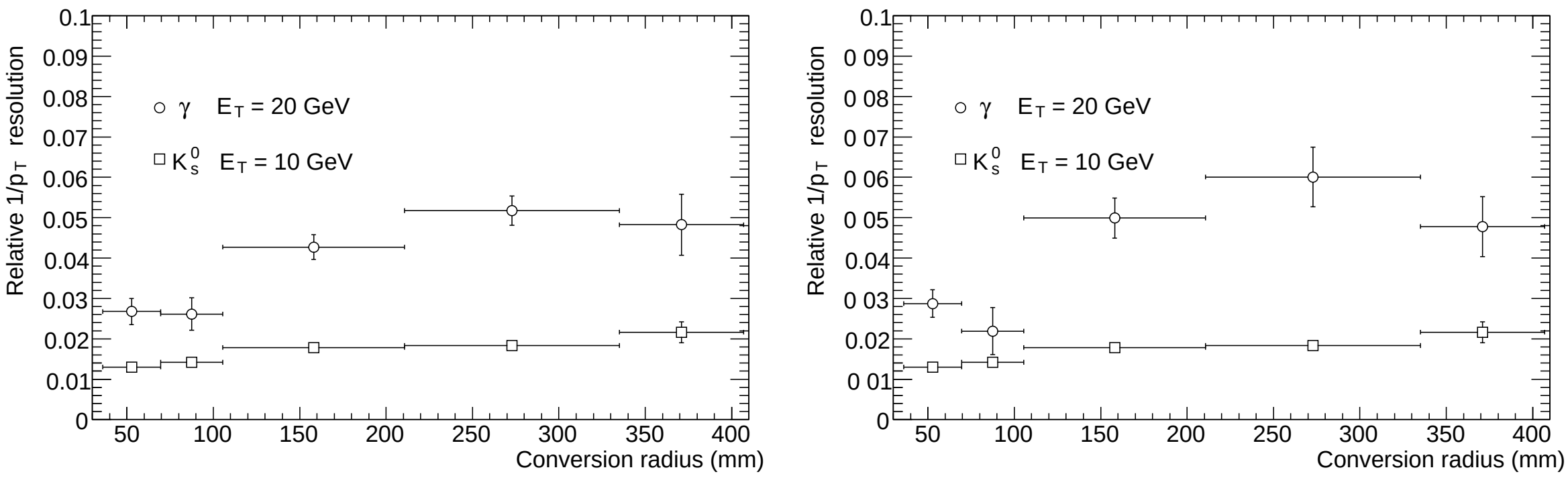

Figure 11: Reconstructed relative $1/p_T$ resolution as a function of radial distance from the beam axis for 20 GeV $p_T$ converted photons and 10 GeV $p_T$ $K_s^0$ decays to charged pions. In the plot on the left, only converted photons, where both of the daughter electrons lost less than 20% of their energy due to bremsstrahlung, are shown. In the plot on the right, all conversions are included.

# 4 Conversion reconstruction

With the three track collections and the vertex fitting algorithm described in the previous two sections, we now have all the necessary tools in place in order to fully reconstruct photons which convert as far as 800 mm away from the primary interaction point. Beyond that radius, the track reconstruction efficiency drops off dramatically due to the lack of a sufficient number of hits in any sub-detector to reliably reconstruct the particle trajectory and accurately predict its track parameters. The conversion reconstruction algorithm is run within the framework of the overall Inner Detector reconstruction software; it is one of the last algorithms run during the post-processing phase. The basic components of the conversion reconstruction are: the track selection and subsequent track classification, the formation of pairs of tracks with opposite charge, the vertex fitting and reconstruction of photon conversion vertex candidates, and finally the reconstruction of single-track conversions. The conversion candidates are then stored in a separate vertex collection, to be retrieved and further classified through matching with electromagnetic clusters during the next level of the event reconstruction. In the results presented in this section, the reconstruction efficiency is estimated for those photon conversions that happen as far as 800 mm away from the primary interaction point, emit daughter electrons with each having at least $p_T$ = 0.5 GeV and are within the $|\eta| = 2.1$ pseudorapidity range. This amounts to $\sim 77\%$ of the total photons converted inside the ATLAS tracker volume in the case of the $H \rightarrow \gamma\gamma$ sample.

## 4.1 Track selection

Only a fraction of the possible track pairs reconstructed by the tracking algorithms and included in the final track collection come from converted photons. Although the wrong-track combinations may be rejected later during the conversion reconstruction process or by physics specific analysis, it is important to remove them as efficiently as possible at an early stage, not least because of the large amount of CPU time involved in processing every possible track pair. Cuts on the perigee impact and longitudinal track parameters, as well as the transverse momentum, are first applied. Tracks that are most probably associated to electrons are then selected by cutting on the probability reconstructed by using the ratio of high-threshold TRT hits over the total number of TRT hits on each track. These cuts have been tuned using $H \to \gamma\gamma$ events, with background present due to the underlying event. All the efficiencies and rejection factors due to track selection cuts which are quoted in this note refer to this physics sample. Table 1 shows the performance of these cuts in accepting tracks produced by converted photons and rejecting non-conversion related tracks. The starting efficiency of $\sim 74\%$ reflects entirely the inefficiency of reconstructing all the conversion related tracks during tracking. After applying these cuts, the surviving tracks are then arranged into two groups with opposite charges.

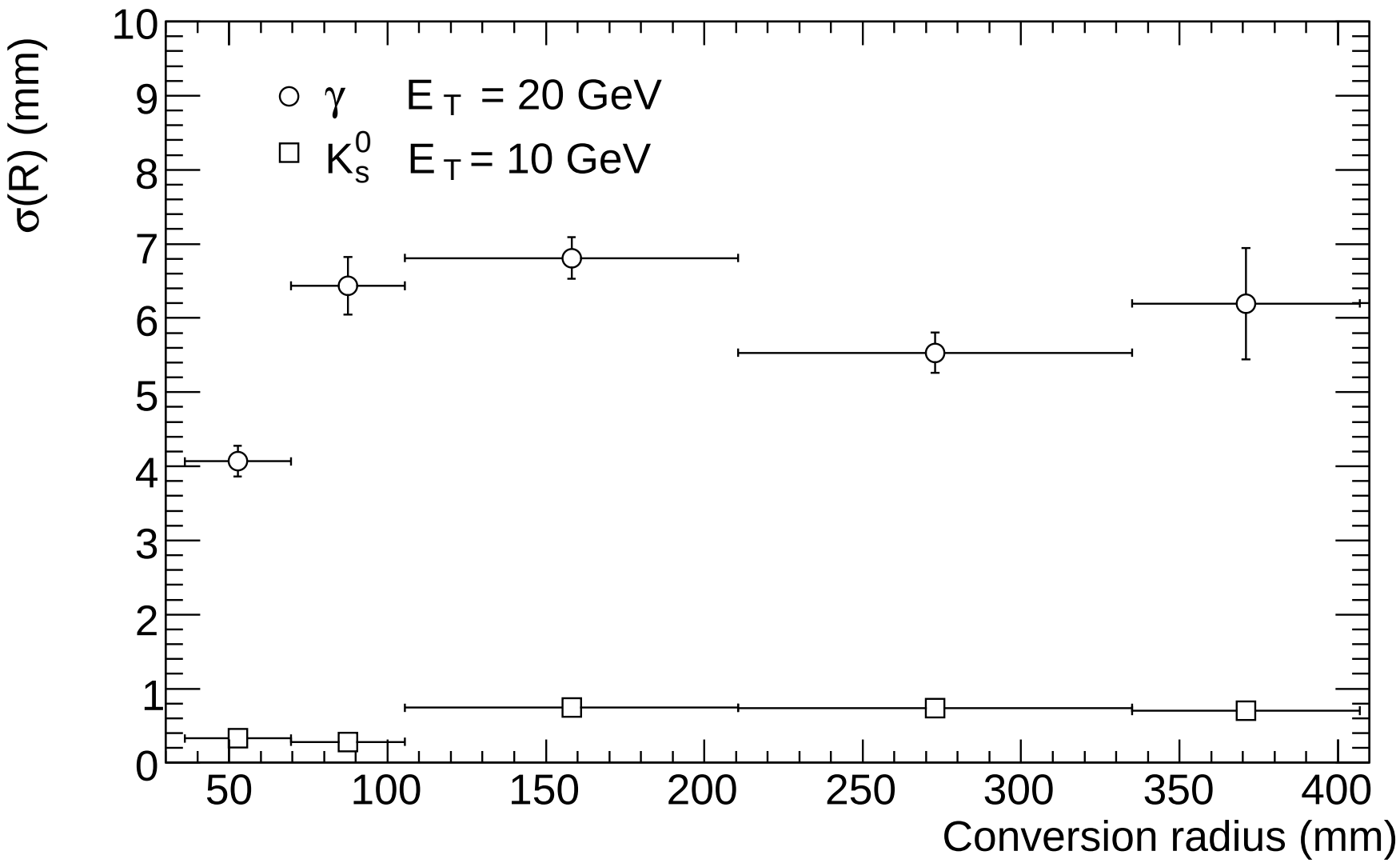

Figure 12: Reconstructed radial resolution as a function of radial distance from the beam axis for 20 GeV $p_T$ converted photons and 10 GeV $p_T$ $K_s^0$ decays to charged pions. In the case of the converted photons, all daughter electrons regardless of bremsstrahlung losses have been included.

| Cut | Efficiency | Rejection |
|---|---|---|
| No Cuts | 0.7378 | 1.00 |
| Impact d0 | 0.7334 | 1.16 |
| Impact z0 | 0.7316 | 1.18 |
| TR ratio | 0.7119 | 2.12 |

Table 1: Track selection cuts: cumulative efficiencies and rejection rates are presented.

## 4.2 Track-pair selection

At this point in the reconstruction process, all possible pairs of tracks with opposite signs are formed and further examined. There are three possible types of track pairs:

1. Pairs in which both tracks have Si hits;
2. Pairs in which one of the two tracks is a stand-alone TRT track;
3. Pairs in which both tracks are stand-alone TRT tracks.

| Cut | Efficiency | Rejection |
|---|---|---|
| Polar angle | 0.7070 | 10.8 |
| Radial distance between first hits | 0.7049 | 12.5 |
| Minimum distance | 0.6970 | 16.5 |
| Vertex radius | 0.6959 | 16.6 |
| Minimum arc length | 0.6935 | 40.3 |
| Maximum arc length | 0.6890 | 111.6 |
| Distance in $z$ | 0.6870 | 111.9 |

Table 2: List of cuts employed during the track-pair selection for the three possible types of track pairs.The cumulative efficiencies and rejection rates are presented (see text for the definition of the cut variables).

In order to reduce the combinatorial background, a series of cuts are applied during the pair formation. These are common to all three track-pair types described above, although their actual values may differ. Table 2 lists those cuts along with the corresponding efficiencies and rejection factors for selecting the correct track pairs and discarding fakes resulting from wrong track combinations. The first criterion for accepting a track pair is that the difference in polar angles between the two daughter tracks in a conversion should be small, based on the fact that the photon is massless. Furthermore, the distance between the first hits of the two tracks in the pair should be reasonably close; this is particularly true in the case where both of them are stand-alone TRT tracks. Finally, the distance of minimum approach between the two tracks in the pair is checked. An iterative method has been implemented that uses the Newton approach to find the set of two points (one on each track) which are closest to each other. The distance of minimum approach between the two tracks is then calculated and a cut is applied to reject those cases where the tracks fail to come within a specified distance from each other.

In order to enhance the performance of the constrained vertex fitter, it is important to begin with a reasonable initial estimate of the vertex position. Using the perigee parameters of the two tracks in the pair, the corresponding radius of curvature and the centre of curvature of the track-helix projection in the $R-\phi$ plane can be derived. As this track-helix projection is circular in the case of a uniform magnetic field such as that of the ATLAS tracker, the estimated vertex position can be identified as either the point of intersection of two circles, or in the case of non-intersecting circles, as the point of minimum approach between two circles. If the two circles do not intersect or approach each other closer than a set minimum distance then the pair is discarded. In principle, two circles may intersect at two points. Since two tracks originating from a conversion vertex (or any vertex for that matter) should also intersect in the $R-z$ plane, the correct intersection point in the $R-\phi$ plane is then chosen to be the one which is closer to the point of minimum approach of the two tracks in the $R-z$ plane. The points of minimum approach both in the $R-\phi$ and the $R-z$ planes should clearly be sufficiently close to each other. If they are separated by more than a set minimum distance, then the track pair is discarded. A cut is also applied on the arc length of the $R-\phi$ plane projection of the two track helices between the line connecting the centres of curvature of the two circles and the actual intersection points. This arc length is required to fall within a specific range which again ideally should tend to be very small. Finally, the distance from the track origin (the candidate conversion vertex location) and the actual points of intersection should also be small. Only track pairs with intersection or minimum-approach points that satisfy the above criteria are further examined. Estimating the initial vertex position allows for a larger number of quality criteria of the track pair to be used in the overall selection process. All the cuts applied during this step have been tested using the 120 GeV $H \to \gamma\gamma$ physics sample; the cuts are tuned so that at least two orders of magnitude of the combinatorial background can be rejected at this point without significant loss in overall conversion reconstruction efficiency. As a consequence, cut values have been intentionally kept fairly loose since even correct track pairs could be characterised by less than optimal selection quantities. This is especially true in cases where at least one of the two tracks involved has only TRT hits resulting in reduced reconstructed track parameter accuracy along the z-axis, or in cases where the tracks have suffered substantial bremsstrahlung losses during their propagation through the ATLAS tracker. In general, the position of the initially estimated vertex falls within a few millimetres of the actual conversion vertex for the correct pair combinations, all deviations being due to the reasons mentioned just before.

| Cut | Efficiency | Rejection |
|---|---|---|
| Fit convergence | 0.6870 | 171.5 |
| Fit $\chi^2$ | 0.6710 | 288.9 |
| Invariant mass | 0.6626 | 353.9 |
| Photon $p_T$ | 0.6625 | 377.1 |

Table 3: Post-vertex fit selection cuts: cumulative efficiencies and rejection rates are presented.

### 4.3 Vertex fitting

The original track perigee assigned during the track reconstruction process is set at the primary interaction point and for the case of photon conversions, especially those that happen far inside the tracker, this is a rather poor assignment. Using the initial estimate for the vertex position described previously, we can redefine the perigee at this point. The new perigee parameters need to be recomputed by carefully extrapolating from the first hit of each track in the pair to this new perigee, taking into account all the material encountered on the way. It is these tracks with their newly computed perigee parameters that are passed to the vertex fitter. This also has the desirable effect of avoiding long extrapolations during the various iterations of the vertex fitting process, which might lead to distortions due to unaccounted-for material effects. At the end of the process the new vertex position along with an error matrix and a $\chi^2$ value for the fit are computed. A vertex candidate is then reconstructed that also contains the track parameters as they are redefined at the fitted conversion vertex. The fit is always successful in the case of the correct track pairs, and it often fails otherwise. After the fit is executed, post-selection cuts on the $\chi^2$ of the fit, on the reconstructed photon invariant mass and on the reconstructed photon $p_T$ can be applied, to reduce even further the wrong pair combinations. These are listed in Table 3.

The track pair selection and the vertex fitting process result in a reduction in the combinatorial background rate by more than two orders of magnitude, with only a rather small loss in overall conversion reconstruction efficiency, amounting to $\sim 8\%$ in the case of $H \rightarrow \gamma\gamma$ decays with $m_H = 120$ GeV. A more quantitative description of the conversion reconstruction efficiency in such decays is presented in Section 5.3. At this stage of the conversion vertex reconstruction, which is still within the tracking software framework, vertices which come from the combinatorial background outnumber the correct conversion vertices by almost a factor of six. The main part of this remaining background consists of reconstructed vertices where at least one of the participating tracks is not an electron at all. This is primarily due to the rather weak particle identification capabilities of the tracker without any access to the electromagnetic calorimeter information. Part of this background can be reduced by some more stringent requirements on the reconstructed conversion vertices after the constrained fit is preformed. But effective improvement is only expected during the subsequent stages of the photon conversion reconstruction, when information from the calorimeter becomes available. Use of the electromagnetic calorimeter should also help to reduce a different type of combinatorial background originating when two electrons from different sources are combined in order to form a track pair. Recent studies indicate significant reduction of both types of the combinatorial background, both by applying tighter vertex selection criteria after the vertex fit is performed and by using the electromagnetic calorimeter information, although they are beyond the scope of this note. The possibility of using the reconstructed photon $p_T$ in order to reduce the number of reconstructed fake vertices, is worth investigating. Figure 13 shows the $p_T$ distribution of the reconstructed conversion vertices along with the distribution for fake vertices resulting from wrong combinations. It is clear that the latter tend to concentrate at the lower $p_T$ region. Nevertheless a final cut on the reconstructed photon $p_T$ will not be as efficient as expected, due to the limited ability at present to correct the reconstructed track momentum for losses due to bremsstrahlung. This is evident in the figure when comparing the reconstructed converted photon $p_T$ distribution with (top row) and without (bottom row) significant bremsstrahlung losses. It becomes even more striking once it is compared to the truth $p_T$ distribution of the converted photon. In the remaining part of this section, the overall performance of the conversion reconstruction software, without utilising the electromagnetic calorimeter information, is examined in the case of single 20 GeV $p_T$ photons, where the combinatorial background is minimal.

Figure 14 shows the track, track-pair, and vertex reconstruction efficiencies for conversions coming from 20 GeV $p_T$ photons as a function of both conversion radius and pseudorapidity. Both the track and track pair efficiencies shown in the figure are measured before any of the selection criteria described above are applied. The large drop in the efficiency at $R > 400$ mm is primarily due to the inefficiency of reconstructing both tracks in the track pair from the photon conversion. It is noteworthy that both the track and the conversion vertex reconstruction efficiency are essentially constant as a function of pseudorapidity. For completeness, Fig. 15 shows the slightly different version of the left-hand plot in Fig. 14 as published in Ref. [6].

Finally, Fig. 16 shows the overall vertex reconstruction efficiency for converted photons with low transverse momenta as a function of conversion radial position. The two competing effects, the bremsstrahlung losses that affect more severely the low $p_T$ tracks, and the higher radii of curvature that result in increased resolving ability of the smaller $p_T$ tracks, that were discussed in Section 2.3, are once more evident here.

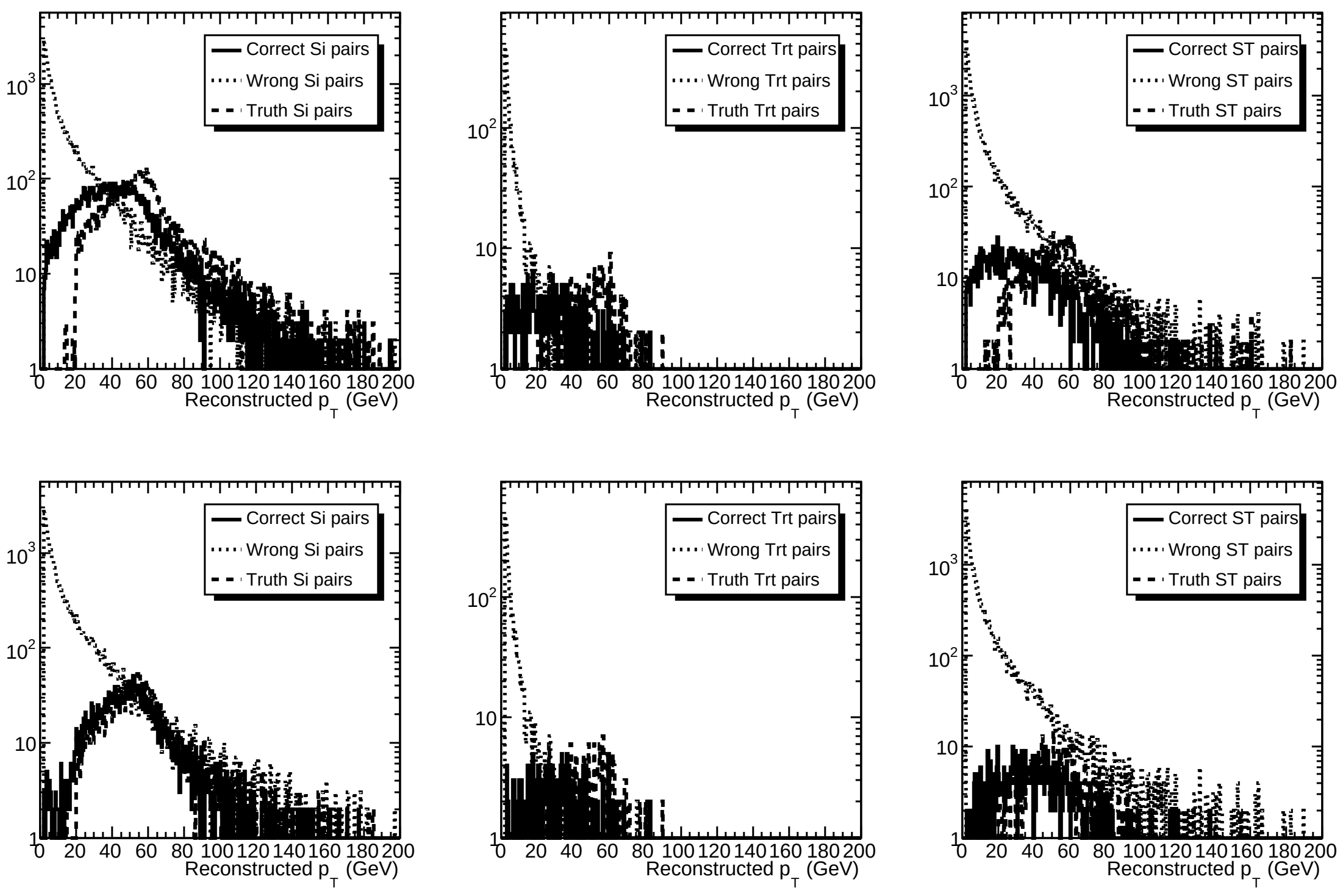

Figure 13: Transverse momentum distribution of reconstructed photon conversions for both correct and wrong track pairs for all three types of pairs: Silicon-Silicon (Si, left column), TRT-TRT (Trt, centre column), and Silicon-TRT (ST, right column). In the top row all electron tracks regardless of bremsstrahlung energy losses are considered for the case of the correct track pairs. In the bottom row only track pairs where both electrons have lost less than 20% of their energy due to bremsstrahlung are shown. For comparison the truth $p_T$ of the converted photon is also shown.

## 4.4 Single-track conversions

Due to conversions which decay asymmetrically (as described in Section 1.1), as well as cases where the conversion happens so late that the two tracks are essentially merged, there are a significant number of conversions where only one of the two tracks from the photon conversion is reconstructed. Depending on the photon momentum scale, these "single-track" conversions become the majority of the cases for conversions that happen late in the tracker and especially inside the TRT. The ability of the TRT to resolve the hits from the two tracks is limited, especially if those tracks do not traverse a long enough distance inside the tracker for them to become fully separated. As a result, only one track is reconstructed, but it will still be highly desirable to recover these photon conversions.

At the end of the vertex fitting process, all of the tracks that have been included in a pair that successfully resulted in a new photon conversion vertex candidate, are marked as "assigned" to a vertex. The remaining tracks are then examined once more on an individual basis in order to determine whether or not they can be considered as products of a photon conversion. For a track to be considered, it should have its first hit beyond the pixel vertexing layer. Furthermore, the track should be electron-like, where again the probability reconstructed by using the ratio of the high threshold TRT hits over the total number of TRT hits (as in the initial track selection described earlier in this section, but requiring a higher value) is used to select likely electron tracks. At the end of this selection tracks wrongly identified as emerging from photon conversions, outnumber the actual photon conversion electron tracks, by almost a factor of two. These are tracks which are not electrons at all, misidentified as such due to the inherent weakness of the particle identification process without the presence of any information from the electromagnetic calorimeter.

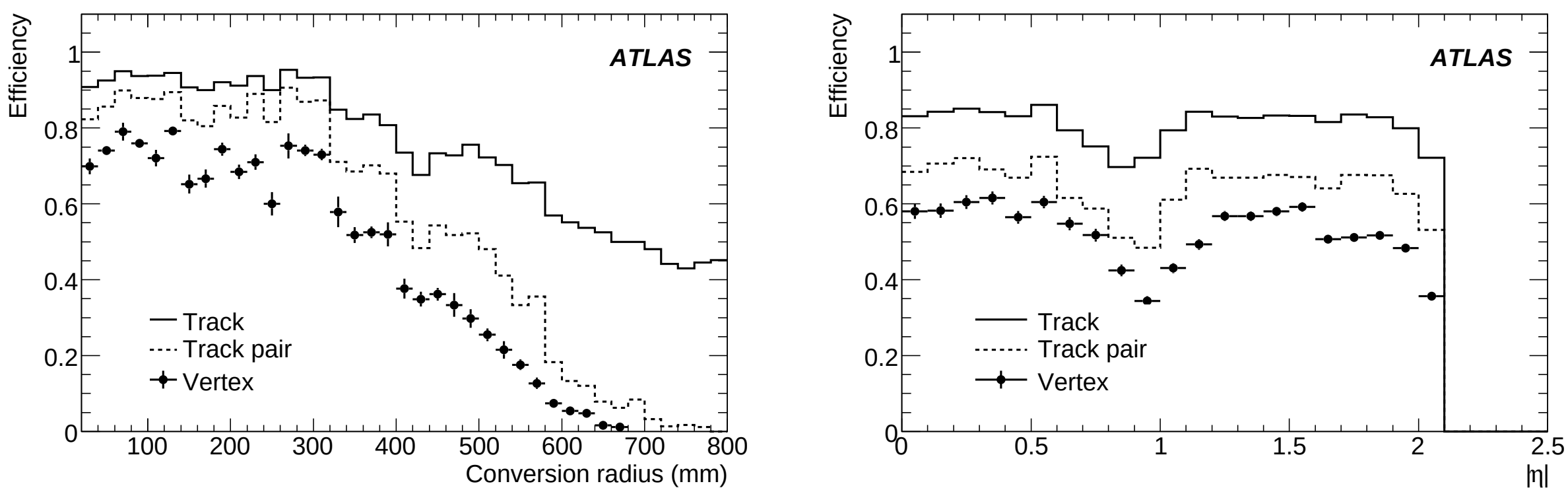

Figure 14: Conversion reconstruction efficiency for conversions from 20 GeV $p_T$ photons as a function of conversion radius (left) and pseudorapidity (right). The solid histograms show the track reconstruction efficiency, the dashed histograms show the track-pair reconstruction efficiency, and the points with error bars show the conversion vertex reconstruction efficiency.

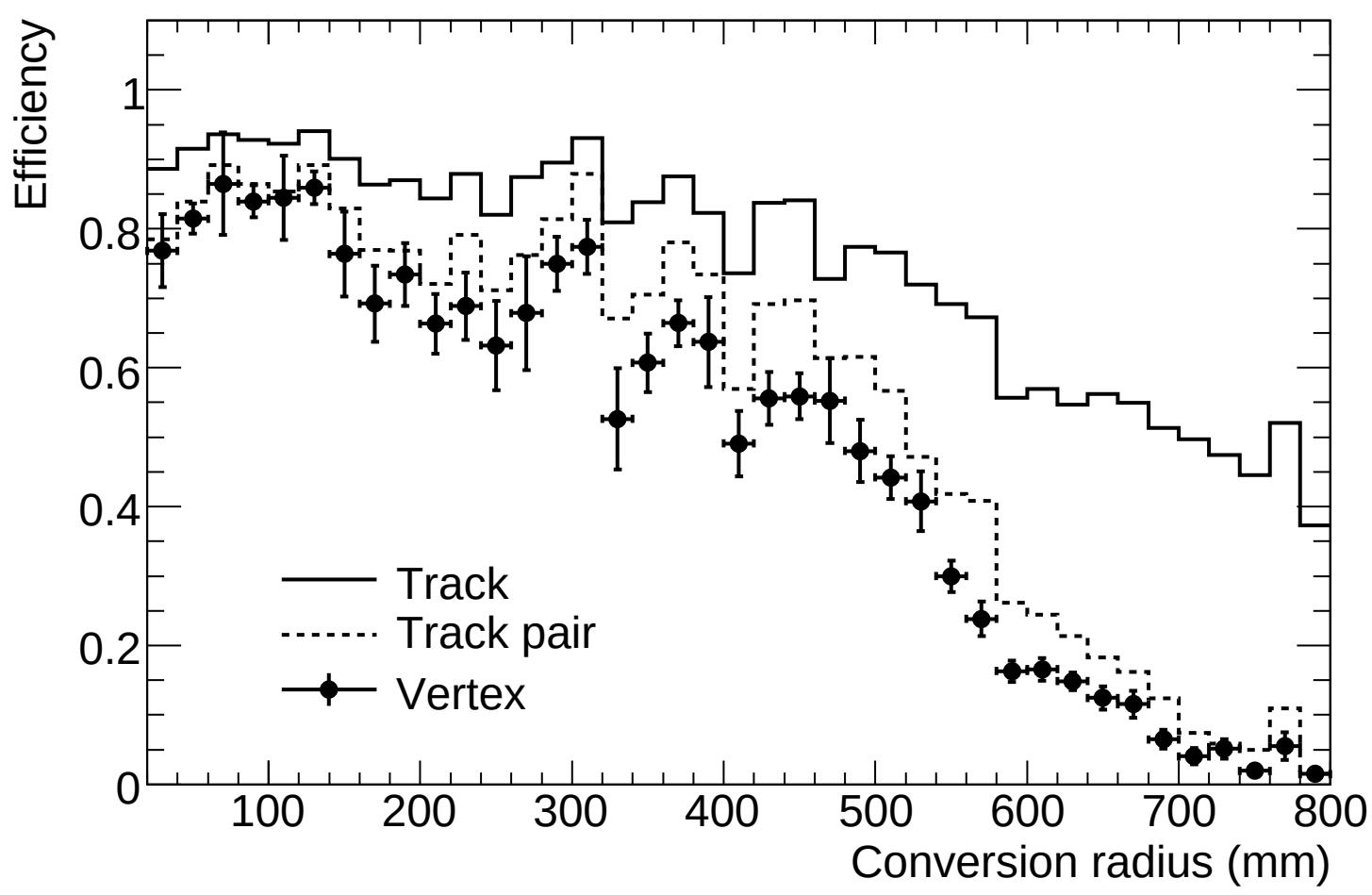

Figure 15: Conversion reconstruction efficiency for conversions coming from 20 GeV $p_T$ photons as a function of conversion radius. The solid histogram shows the track reconstruction efficiency, the dashed histogram shows the track-pair reconstruction efficiency, and the points with error bars show the conversion vertex reconstruction efficiency as published in Ref. [6].

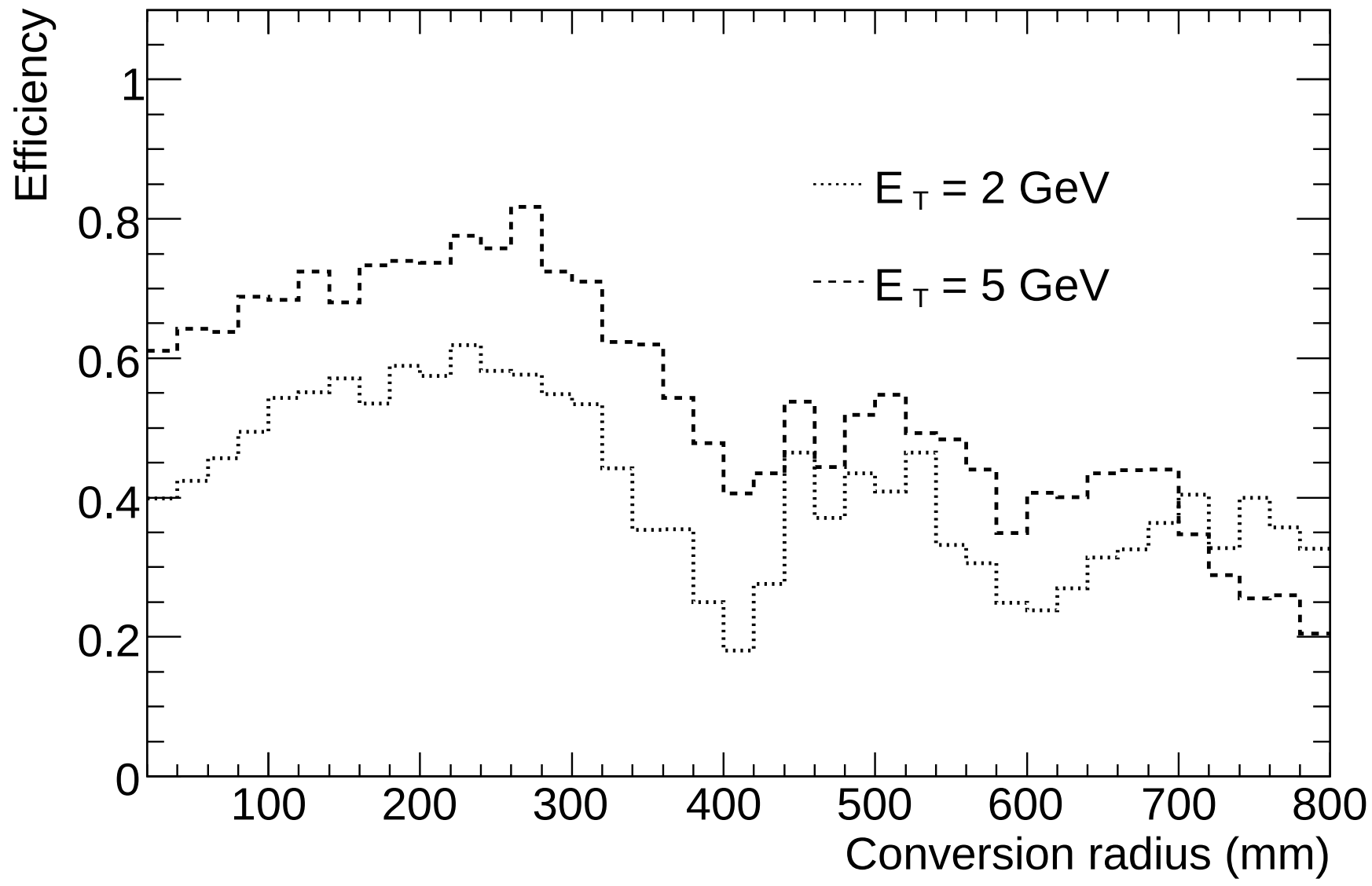

Figure 16: Conversion vertex reconstruction efficiency as a function of conversion radius for photons with transverse energy of 2 and 5 GeV.

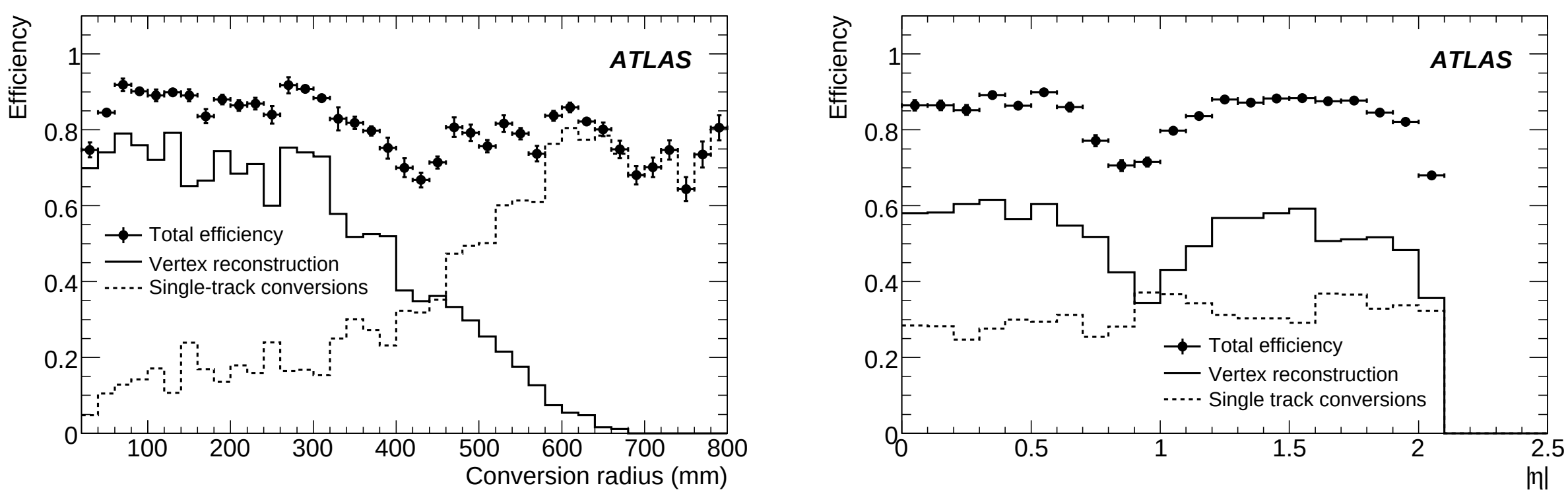

Figure 17: Reconstruction efficiencies for conversions from 20 GeV $p_T$ photons as a function of conversion radius (left) and pseudorapidity (right). The points with error bars show the total reconstruction efficiency, the solid histograms show the conversion vertex reconstruction efficiency, and the dashed histograms show the single-track conversion reconstruction efficiency.

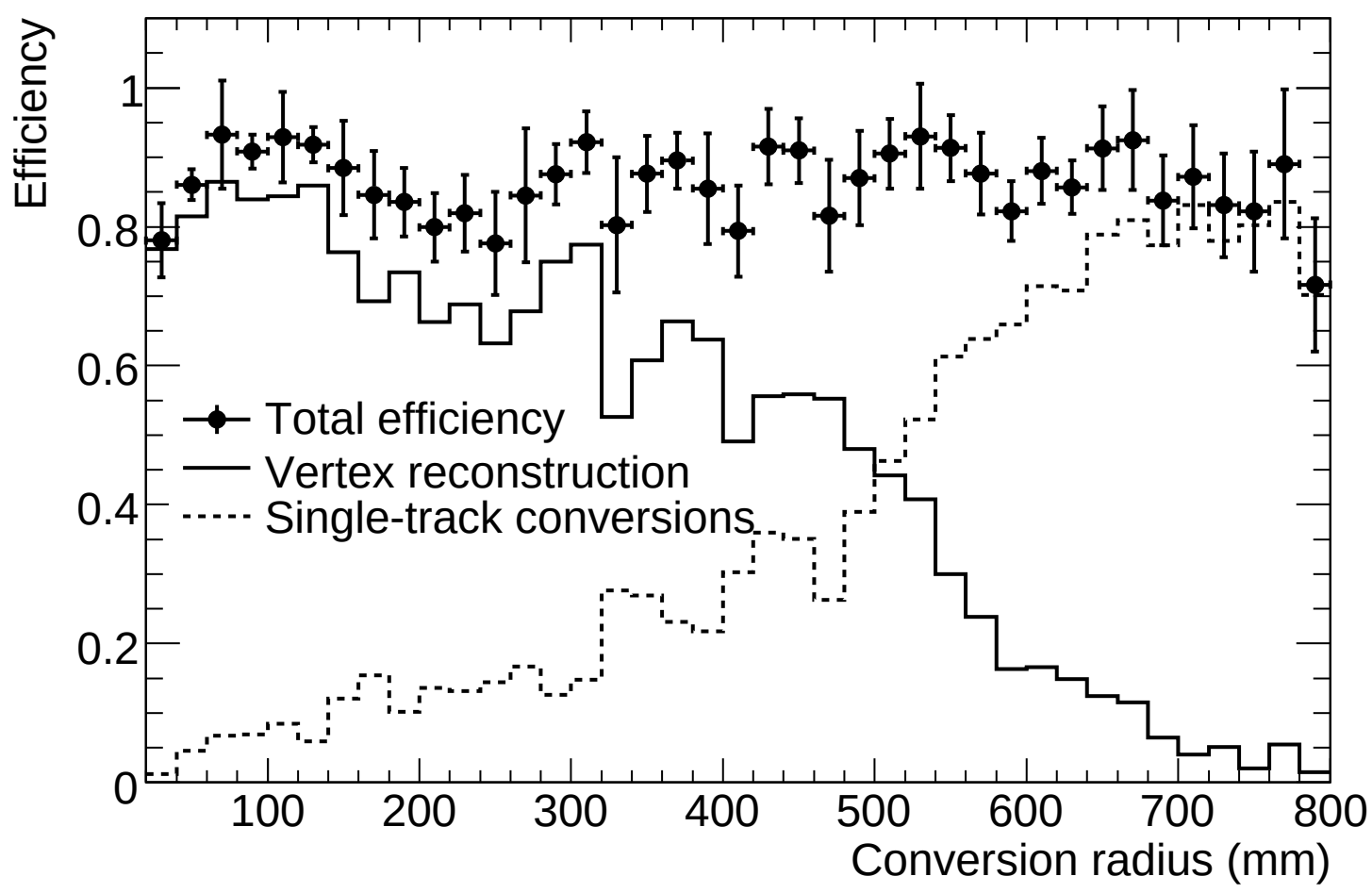

Figure 18: Conversion reconstruction efficiency for conversions coming from 20 GeV $p_T$ photons as a function of conversion radius. The points with error bars show the total reconstruction efficiency, the solid histogram shows the conversion vertex reconstruction efficiency, and the dashed histogram shows the single-track conversion reconstruction efficiency as published in Ref. [6].

A conversion vertex candidate is then reconstructed at the position of the first track hit. It is clear that, especially in the case where the first hit is inside the Si part of the tracker, the position of the conversion vertex reconstructed in this way can be off by as much as a detector layer. This discrepancy is normally much smaller in the case of a vertex inside the TRT due to the higher straw density. On the technical side, this type of reconstruction requires a careful transformation of the local track parameters and error matrix into global ones that are directly assigned to the newly defined vertex. A new vertex candidate is then stored, identical in structure to the one derived from a vertex fit with the important difference that it has only one track assigned to it. The effect of including the single-track conversions into the overall conversion reconstruction efficiency is significant as is shown in Fig. 17. The plot shows the conversion reconstruction efficiency for 20 GeV $p_T$ photons as a function of both radius and pseudorapidity. As expected, the single-track conversions become more and more dominant at higher radial positions, and single-track conversions are fairly uniformly distributed across the full pseudorapidity range. For completeness, Fig. 18 shows the slightly different version of the left-hand plot in Fig. 17 as published in Ref. [6]. While it is not possible to reconstruct the two merged tracks in these single-track conversions, it should be possible to separate such cases from very asymmetric conversions with the lower-energy partner of the pair not reconstructed: the transition radiation information should correspond on average to that expected from two electrons and the drift-time information should be inconsistent with that expected from a single track (resulting in a significant fraction of unused drift circles in the track fit).

# 5 Physics applications: low-$p_T$ conversions, $\gamma/\pi^0$ separation, $H \rightarrow \gamma\gamma$

In this section some interesting applications of the usage of the photon conversions are presented. Only results from photon conversions where both of the daughter electron tracks have been reconstructed are included. It needs to be stressed at this point, that everything that is presented here is meant only as an application example and that nofull-scale analysis has been made.

## 5.1 Low-$p_T$ photon conversions

Of particular interest during initial data taking is the use of the reconstruction of converted photons as a tool to obtain a measurement of the amount of material inside the ATLAS tracker, including passive material. The abundance of low-$p_T$ neutral pions in minimum bias events represents a very rich source of photons and makes this approach particularly promising. The number of photon conversions measured on a detector volume of known $x/X_0$ can be used as a normalisation point to extract the amount of material at any other location inside the detector by counting the relative number of conversions occurring in that portion. To obtain an unbiased map of the tracker material it is necessary to correct the measured number of conversions by the conversion reconstruction efficiency. Several methods are being investigated to measure this efficiency from data, e.g. embedding Monte Carlo photon conversions in data or extracting it from the measure of decays with similar topology like $K_s^0 \rightarrow \pi^+\pi^-$.

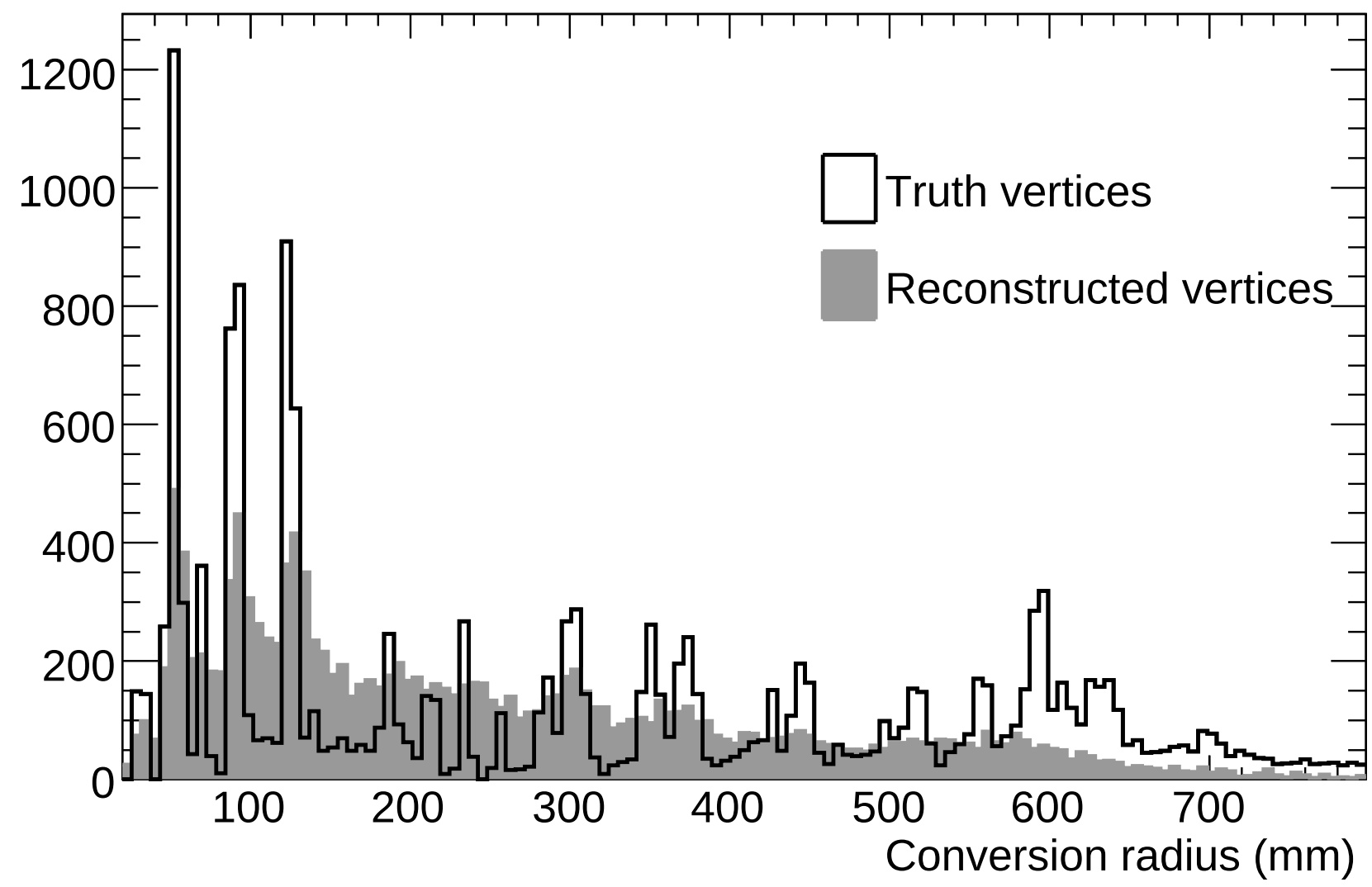

Figure 19: Reconstructed radial positions for conversions of 5 GeV $p_T$ photons. The black histogram shows the truth radial position of the conversion vertices, and the gray histogram shows the radial positions of the reconstructed vertices, regardless of the bremsstrahlung losses of their daughter electrons.

Figure 19 shows the reconstructed radial positions of photon conversions with 5 GeV $p_T$. A few structures may be identified: the initial peak caused by the beampipe, the three layers of the pixel detector and then with lower resolution and significance the SCT layers and the TRT. The observed smearing of the reconstructed position of the conversion vertex is mainly due to bremsstrahlung effects. The position resolutions (in the radial direction) of the reconstructed conversion vertex, for photon conversions produced by the decay of neutral pions with various energies, are shown in Fig. 20 as a function of the distance from the beam axis. All conversions regardless of the amount of energy lost due to bremsstrahlung by the daughter electrons, have been used. In the case of the lower $p_T$ neutral pions, more relevant in the case of minimum bias events, the radial position resolution improves somewhat, as might be expected from the larger angular separation between the produced electrons. On the other hand the use of low $p_T$ tracks can be limited by the lower tracking efficiency caused by multiple scattering and especially bremsstrahlung.

In order to be able to determine the amount of material at a given position, it is necessary to compare the number of reconstructed converted photons at that position with the number of conversions reconstructed at the position of some reference point. This necessitates being able to resolve the position of the reference, which may not be trivial.

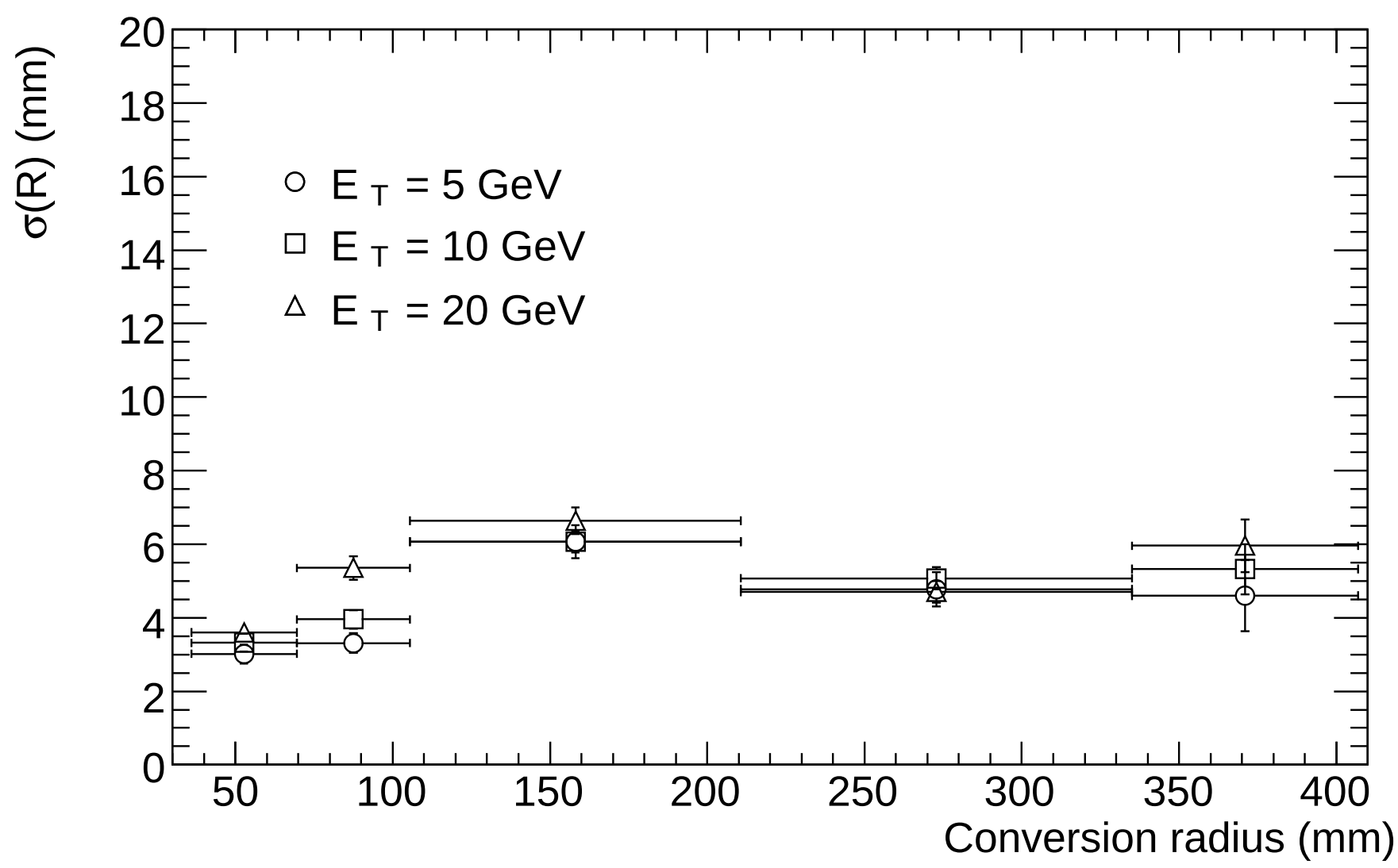

Figure 20: Reconstructed radial position resolution for converted photons produced by the decay of neutral pions with various energies.

## 5.2 $\gamma/\pi^0$ Separation

Another application of conversion reconstruction is the possibility of using the converted photons to identify, and subsequently remove, neutral pions in which at least one of the photons resulting from the decay of the pion has converted. Low multiplicity pions constitute the dominant background to the photon signal after all the calorimeter-specific cuts have been applied during photon identification [16]. In the case of converted photons from $\pi^0$ decays, additional handles could be derived as soon as their reconstructed transverse momentum is made available. About 30% of the neutral pions will have at least one of their daughter photons converted and subsequently reconstructed as such, thus providing an estimate of their $p_T$.

The transverse momentum reconstruction resolution is important when attempting to use conversions to identify low $p_T$ neutral pions. The ratio of the reconstructed $p_T$ of a converted photon inside the ATLAS tracker to the $E_T$ measured by the electromagnetic calorimeter is different for photons from $\pi^0$ decays and for prompt photons. Figure 21 shows such distributions for the case of converted 20 GeV $p_T$ single photons and converted photons from the decay of a 20 GeV $p_T$ neutral pion. The photon $p_T$ shown is that reconstructed by the conversion algorithm, while the $E_T$ shown is taken from the truth value from the simulation. Three regions in pseudorapidity are shown separately, namely those corresponding approximately to the tracker barrel, barrel/end-cap transition and end-cap regions. The top row of plots include all converted photons, irrespective of losses due to bremsstrahlung of their daugter electrons, while the bottom row has only those converted photons where both of their daughter electrons have lost $< 20\%$ of their energy due to bremsstrahlung. Clearly the distinction between conversions from single photons and conversions from photons produced in neutral pion decays is less pronounced in the case of strong bremsstrahlung losses, although an effective bremsstrahlung recovery mechanism should be able to significantly improve the separation between the two distributions. A certain degredation is also evident as we move from the barrel to the end-cap tracker, due to the less accurate reconstruction of the transverse momentum of the daughter electron tracks at higher pseudorapidities. Figure 22 shows the fraction of remaining $\pi^0$ particles as a function of converted photon efficiency, both with and without significant losses due to bremsstrahlung. The overall $\pi^0$ rejection corresponding to a photon acceptance of 90 % for the three different pseudorapidity regions, as described above, is shown in Fig. 23. Again a distinction is made for the cases with and without significant energy losses due to bremsstrahlung. Although reduced, the discriminatory power against $\pi^0$ is significant even when severe losses due to bremsstrahlung are present.

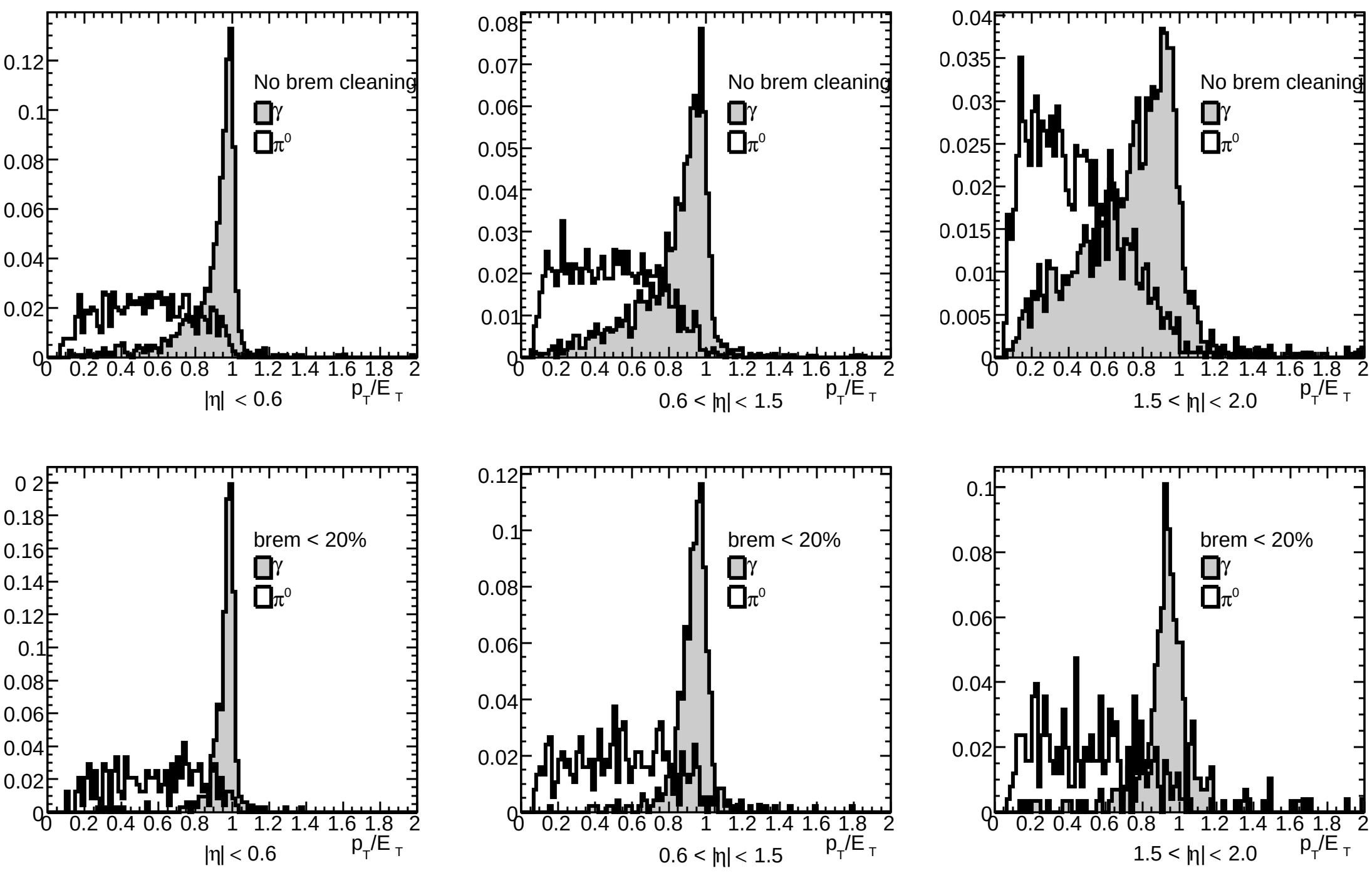

Figure 21: $p_T/E_T$ distribution for 20 GeV $p_T$ converted photons and for photons from a 20 GeV $\pi^0$. The top row shows the distribution for all photons irrespective of the daughter electron energy losses due to bremsstrahlung. The bottom row shows the distribution only for those photon conversions in which the daughter electrons have lost less than 20% of their energy to bremsstrahlung. Three different pseudorapidity ranges are shown, corresponding to the barrel (left), the barrel/end-cap transition (centre) and the end-cap (right) regions.

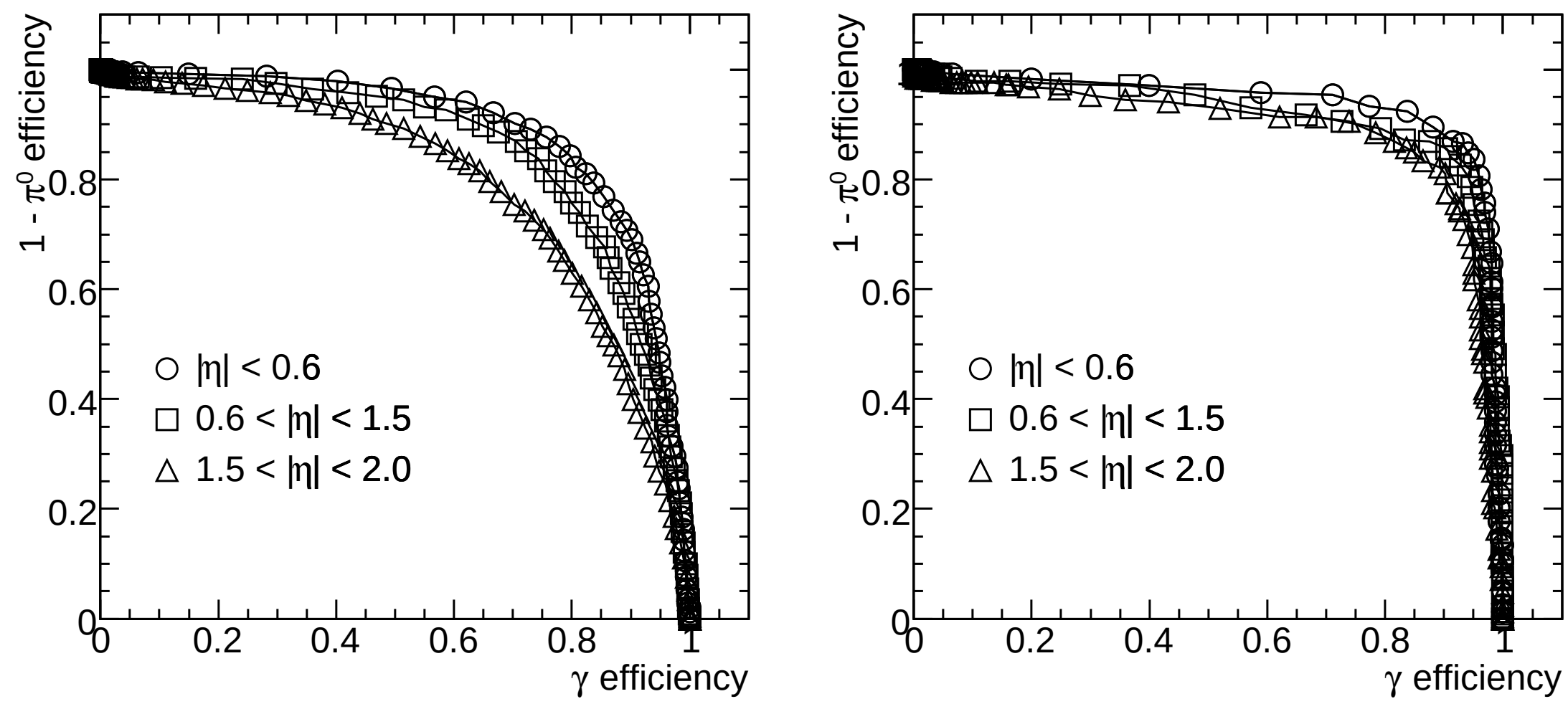

Figure 22: Fraction of remaining $\pi^0$ as a function of converted photon efficiency with (left) and without (right) significant bremsstrahlung losses of the corresponding daughter electrons, for three pseudorapidity regions as described in the text.

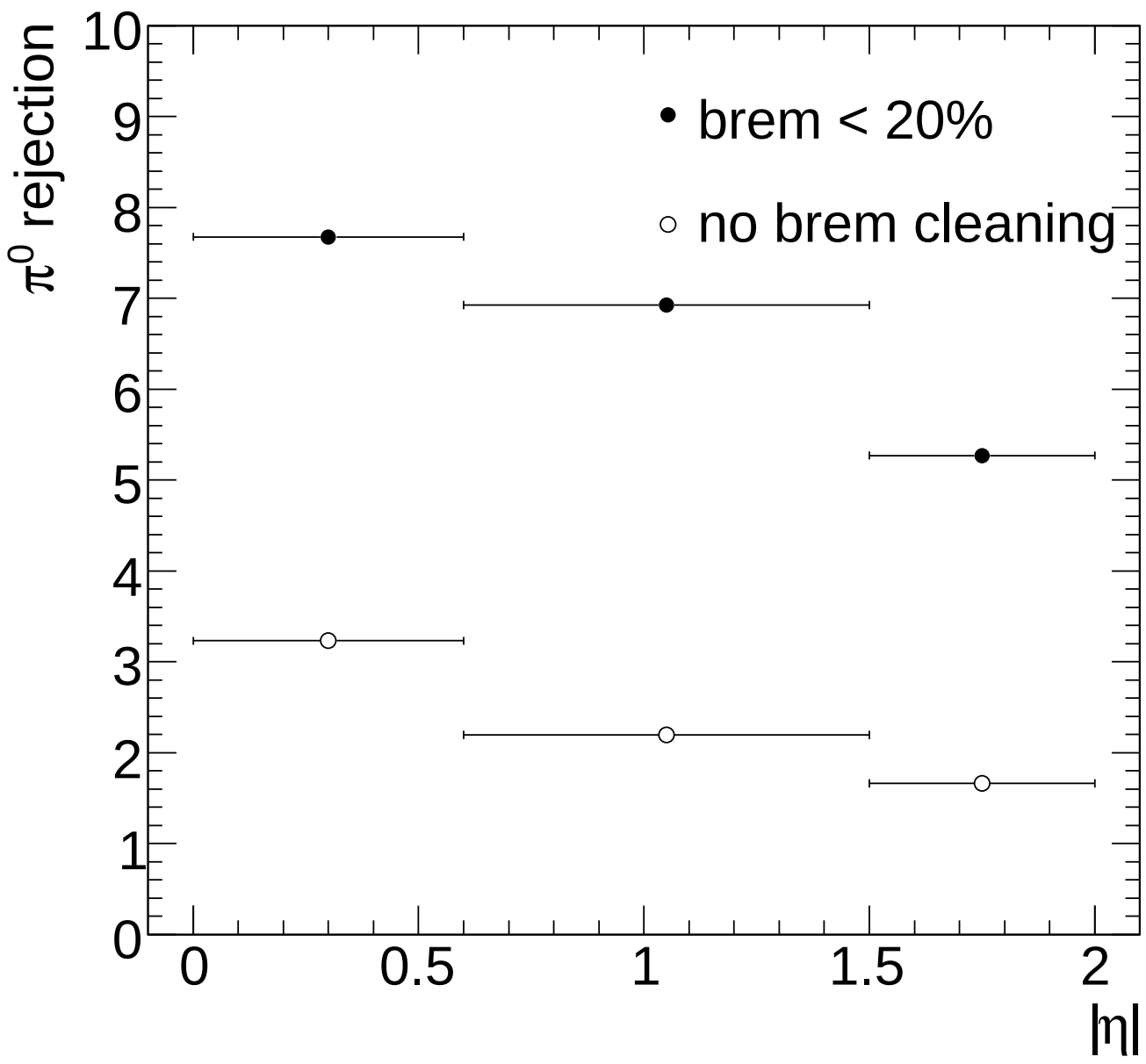

Figure 23: Rejection factors against $\pi^0$ corresponding to photon acceptance efficiencies of 90 %, with and without significant energy losses due to bremsstrahlung for the three pseudorapidity regions described in the text. The results are shown for converted photons and $\pi^0$ with a $p_T$ of 20 GeV.

### 5.3 $H \to \gamma\gamma$ decays

As mentioned in the introduction, the recovery of converted photons is of primary importance in the search for physics processes in which photons are the primary decay product. In particular, accurate reconstruction of the $H \to \gamma\gamma$ process is heavily dependent on the ability to properly reconstruct photon conversions for the following reasons:

1. A significant fraction of photons will convert inside the ATLAS tracker volume. Efficient reconstruction of these photons will enhance the signal statistics for this process.

2. Photon identification, using a combination of inner detector and electromagnetic calorimeter selection criteria, will be improved with effective conversion reconstruction. Even single-track conversions will be useful in this context.

3. The electromagnetic calorimeter calibration will be significantly enhanced when converted photons are identified as such. Again, even single-track conversions will be very useful.

4. The ability to accurately point back to the mother Higgs particle is dramatically enhanced for the case of reconstructed converted photons where both daughter electron tracks are properly recovered.

It is important, therefore, to investigate the performance of the conversion reconstruction strategy in this case, not least because of the higher transverse momenta which characterise the photons produced in $H \to \gamma\gamma$ decays. Decays to photon pairs from a Standard Model Higgs boson with 120 GeV mass have been studied throughout this section.

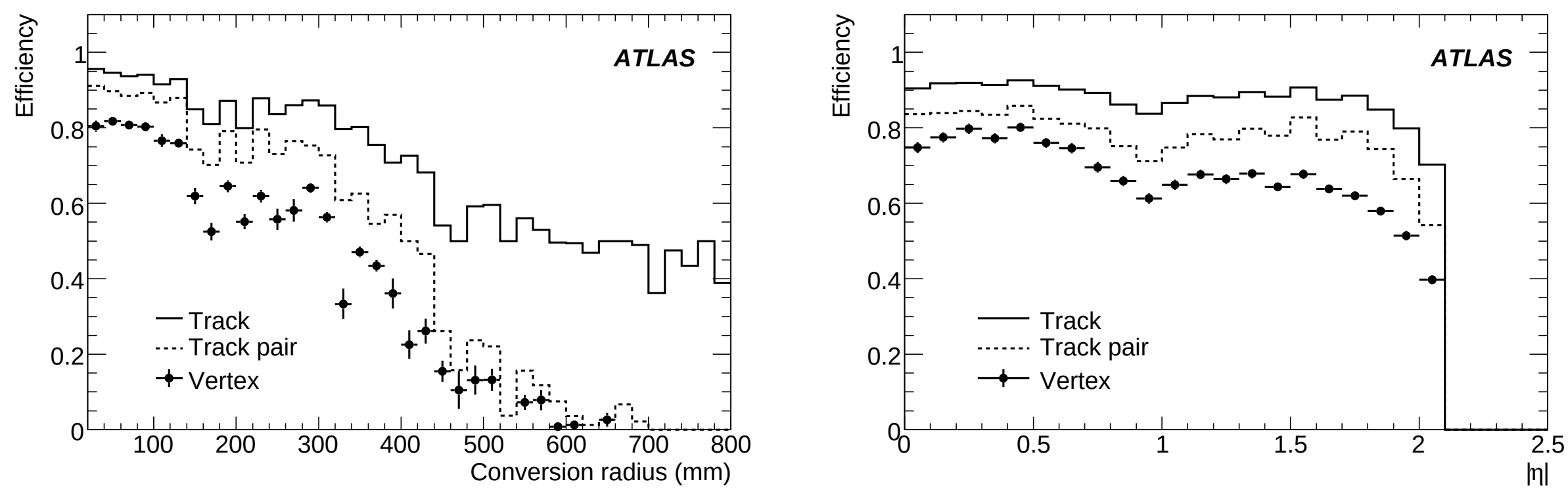

Figure 24: Track, track-pair, and vertex reconstruction efficiencies for converted photons from $H \to \gamma\gamma$ decays with $m_H = 120$ GeV, as a function of radial distance from the beam axis (left) and pseudorapidity (right). The efficiency reduction at $|\eta| \sim 0.8$, is due to the track reconstruction inefficiencies in the gap region between the TRT barrel and end-cap detectors.

Figure 24 shows the converted photon reconstruction efficiency as a function of both radius and pseudorapidity for photons coming from $H \to \gamma\gamma$ decays. The reduced efficiency at higher radii is primarily due to the smaller distance which the produced electron tracks travel inside the magnetic field, reducing the separation between them. The conversion reconstruction efficiency as a function of pseudorapidity is fairly flat, independent of the material distribution inside the ATLAS tracker, as expected. The effect of including the single-track conversions into the overall conversion reconstruction efficiency is also significant for $H \to \gamma\gamma$ decays with large conversion radius and over the full pseudorapidity range, as shown in Fig. 25. The reconstructed conversion vertex radial position resolution is shown in Fig. 26 for reconstructed converted Higgs photon vertices where the participating tracks have lost $> 20\%$ ($< 20\%$) of their energy due to bremsstrahlung, along with all vertices put together. The results are fairly comparable to the ones shown for single photons in Section 4, despite the fact that the resulting photon momenta in this case are on average at least a factor of two bigger and the fact that the presence of the underlying event causes additional complications for the track reconstruction.

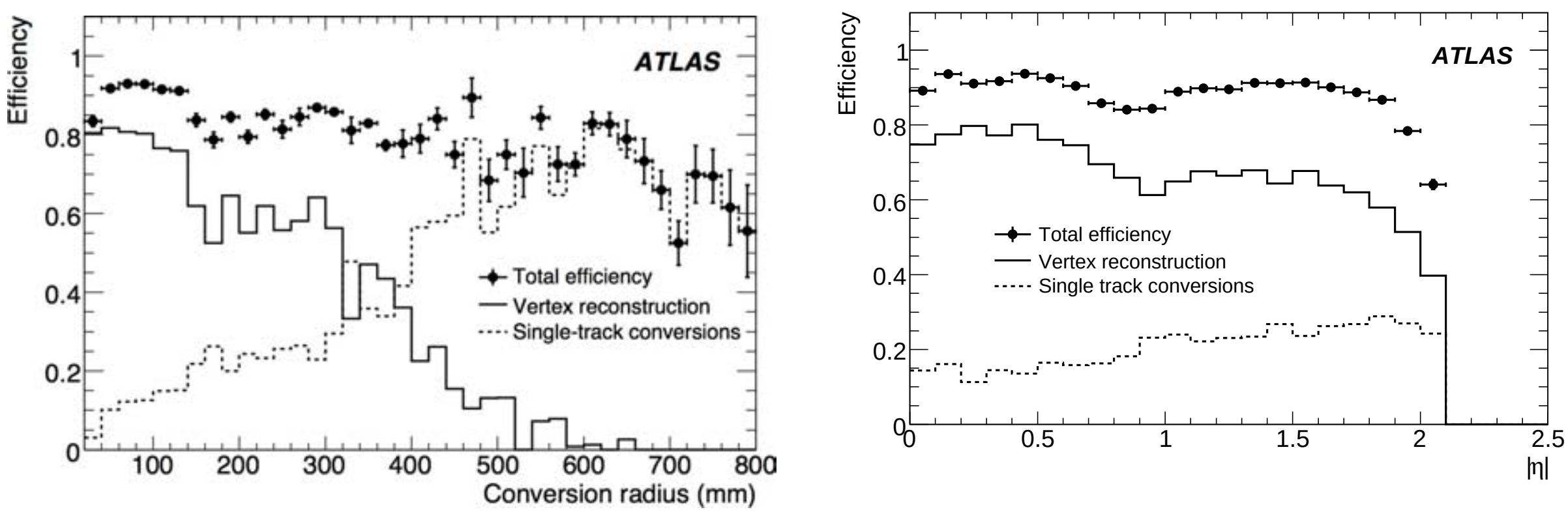

Figure 25: Reconstruction efficiencies for converted photons from $H \to \gamma\gamma$ decays with $m_H = 120$ GeV, as a function of conversion radius (left) and pseudorapidity (right). The points with error bars show the total reconstruction efficiency, the solid histograms show the conversion vertex reconstruction efficiency, and the dashed histograms show the single-track conversion reconstruction efficiency.

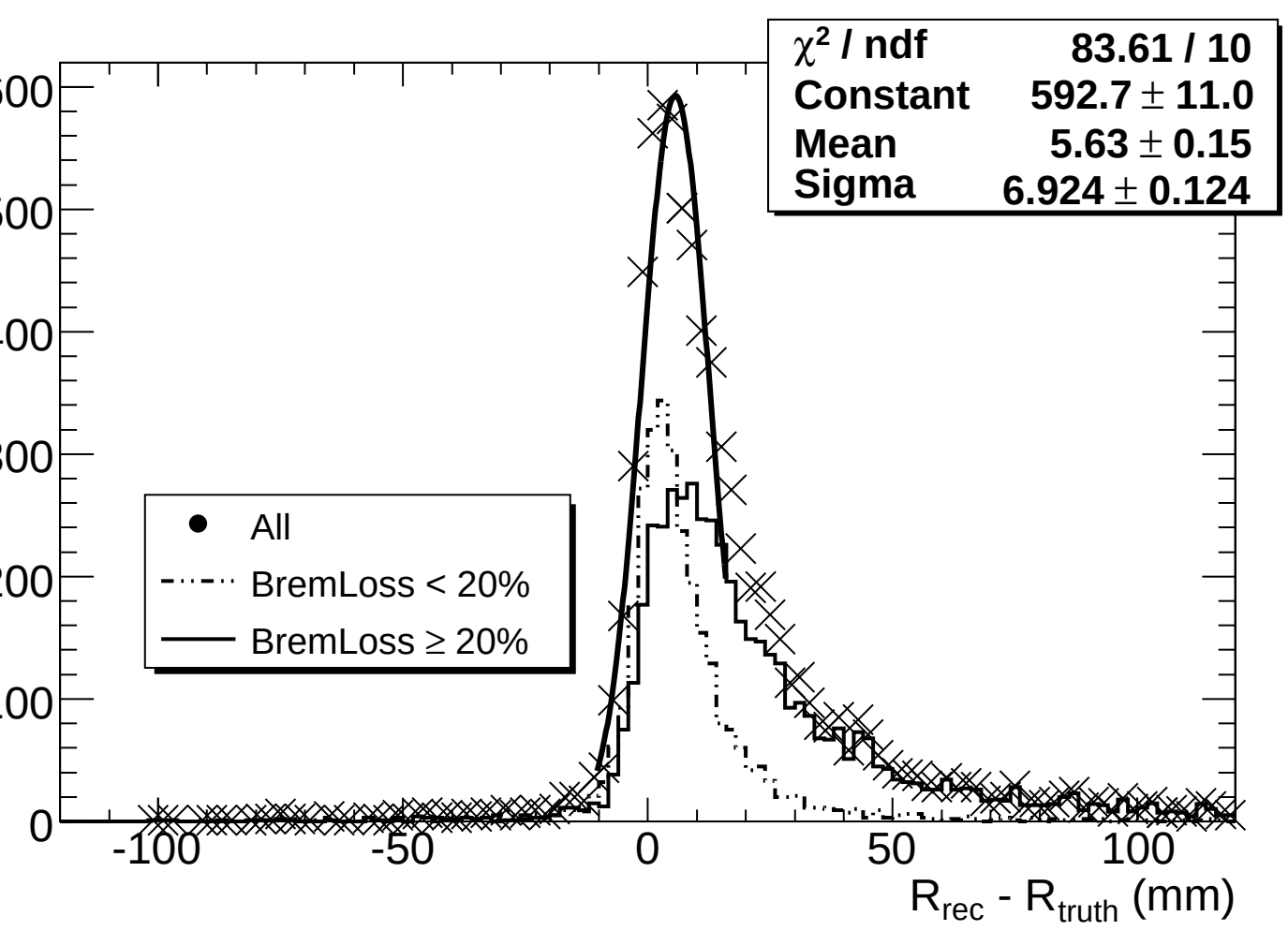

Figure 26: Reconstructed vertex radial position resolution (in mm) for converted photons from $H \to \gamma\gamma$ decays with $m_H = 120$ GeV. For comparison, the two cases where the participating tracks have lost $> 20\%$ ($< 20\%$) of their energy due to bremsstrahlung are also shown separately.

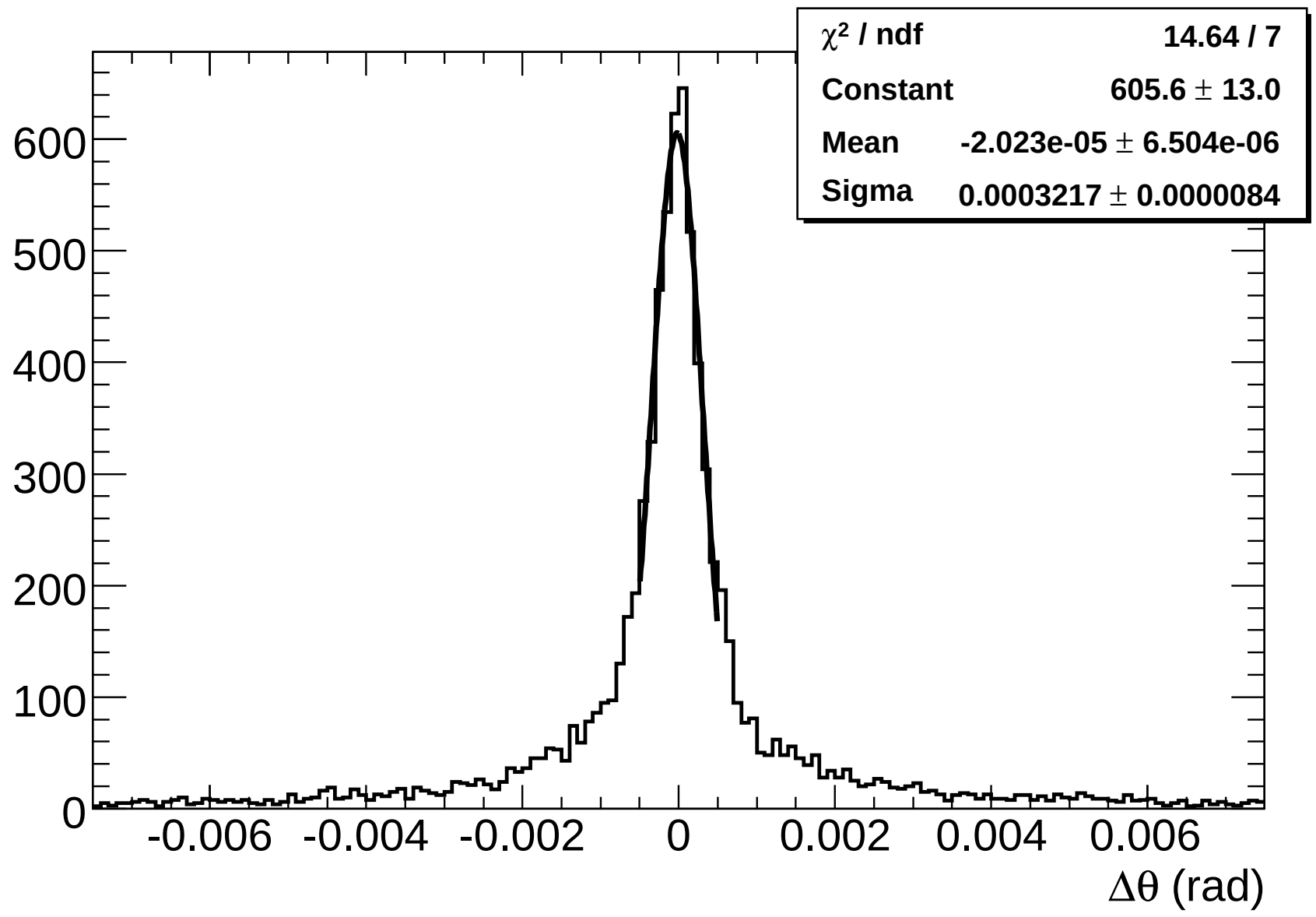

Figure 27: Reconstructed polar angle resolution (in radians) for converted photons from $H \rightarrow \gamma\gamma$ decays with $m_H = 120$ GeV.

Of particular interest for the reconstruction of the Higgs invariant mass is the resolution on the measurement of the polar angle of the reconstructed converted photon. This is shown in Fig. 27 for the case of conversions where both electron tracks have Si hits. These account for $\sim 58\%$ of the reconstructed converted photons inside the ATLAS tracker volume. The resulting resolution is of the order of 0.5 mrad regardless of the transverse momentum of the converted photon. This is an improvement of at least an order of magnitude with respect to the polar angle resolution derived using the electromagnetic calorimeter response [6].

# 6 Summary and conclusions

This note has described and presented a detailed performance evaluation of the conversion reconstruction algorithm which will be used to reconstruct and study early data at the LHC. All three types track collections delivered by the tracking software have been combined and used. A dedicated vertex fit algorithm has been developed for the purpose of reconstructing converted photon vertices. Special care has been given to flagging possible conversions where only one of the produced electron tracks has been reconstructed (or where the two tracks are merged into one in the case of late conversions). Combining all of these tools, a reconstruction efficiency of almost 80% has been achieved for conversions that occur up to a distance of 800 mm from the beam axis. A transverse momentum reconstruction resolution of the order of 5% has been found for converted single photons of various energies. This has also been shown to be valid for the case of photons produced by the decay of a Standard Model Higgs boson with a mass of 120 GeV, as well as for those coming from the decay of low $p_T$ neutral pions. The position resolution is found to be better than 5 mm in the radial direction, making this a promising method for mapping the material inside the ATLAS inner detector. The angular resolution is found to be below 0.6 mrad, giving effective pointing to converted photons from physics processes.

# Reconstruction of Low-Mass Electron Pairs

**Abstract**
This note discusses the reconstruction of $J/\psi$ and $\Upsilon$ decays to electron pairs based on ATLAS Monte Carlo simulated signal and background samples. The possible trigger strategies are described, one geared to select two low-energy electromagnetic objects in direct production, the second one taking advantage of the possible presence of a muon in the final state in $b\bar{b}$ production followed by the decay of one $b$-quark to $J/\psi + X$. The low-energy electrons are reconstructed using a dedicated algorithm seeded by a track reconstructed in the inner detector and identified combining information from the inner detector and the electromagnetic calorimeter. The performance of this algorithm is presented and the potential of using such events for early LHC data studies is investigated.

## 1 Introduction

When switched on, the LHC will produce charm and beauty quarks in abundance which will be collected by the ATLAS experiment [1], even during the low luminosity periods. The number of produced quarkonium states such as J/$\psi$ and $\Upsilon$, important for many physics studies, will be equally numerous. On average, one in every hundred collisions will contain a $b\bar{b}$ pair. The large $b\bar{b}$ cross-section and the high luminosity of the machine give therefore a high rate for $B$-hadrons, making $B$-physics an interesting and competitive subject at the LHC. Low energy resonances, such as J/$\psi$ and $\Upsilon$ will be one of the main sources of isolated electrons in the early data. Both the J/$\psi$ and $\Upsilon$ signal samples are important for understanding the production of prompt quarkonia. But there is another aspect which is the main focus of this note: these samples are ideal to study the performances of trigger and offline reconstruction at low energies as well as being potentially useful for the in-situ calibration of the electromagnetic calorimeter.

This note is organised as follows: Section 2 gives a description of the data-samples used in this note, Section 3 details the trigger selections, and Section 4 describes the offline electron reconstruction and identification procedure. Finally, in Section 5, the physics potential of these channels is explored with initial data, assuming an instantaneous luminosity of $10^{31}$ cm$^{-2}$ s$^{-1}$ and an integrated luminosity of 100 pb$^{-1}$.

## 2 Data samples

The different data samples used in this study are summarised in Table 1. The total cross-sections for charm production at LHC is 7.8 mb and the one for bottom production is 0.5 mb. Quarkonium production was originally described by the colour singlet model which failed to reproduce the direct J/$\psi$ production cross section measured by the CDF experiment [2]. The colour octet model [3] was proposed as a solution to this quarkonium deficit. Direct quarkonia Monte Carlo samples comprise of directly produced J/$\psi$ or $\Upsilon$ in colour singlet and octet states, along with promptly-produced $\chi$'s, which decay into J/$\psi$'s or $\Upsilon$'s [4] [5]. The inclusive production cross sections of J/$\psi$ and $\Upsilon$ are respectively $\sim 90\mu$b and $\sim 0.7\mu$b. A minimum transverse momentum of 3 GeV and a pseudo-rapidity $\eta$ <2.7 are required for the two electrons. The resulting cross sections for the used data samples are respectively $\sim$ 117nb and $\sim$ 47nb. Another sample used in this study is originated from Drell-Yan production. In addition to the electron filter also applied to the J/$\psi$ and $\Upsilon$ samples, the generated di-electron invariant mass $m_{ee}(DY)$ has to be $1 < m_{ee}(DY) < 60$ GeV. Studies also include non-diffractive minimum-bias events with a total assumed cross section of 70 mb.

Table 1: Data samples: process, production cross-section and total number of events available.

| Process | Cross section | Number of events ($\times 10^3$) |
|---|---|---|
| Direct production | | |
| $pp \to J/\psi(e3e3)X$ | 116.3 nb | 160 |
| $pp \to \Upsilon(e3e3)X$ | 47.6 nb | 150 |
| $pp \to$ Drell-Yan$(e3e3)$ | 2.9 nb | 250 |
| minimum-bias | 70 mb | 1,000 |
| $b\bar{b}$ production | | |
| $bB_d \to \mu(6)J/\psi(e2e2)X$ | 0.2 nb | 50 |

For the production via the decay of $b\bar{b}$, only J/$\psi$ events are considered. The signal sample is made of $bB_d \to \mu(6)J/\psi(ee)+X$ events, where the $\mu(6)$ refers to a muon coming from the $b$ quark with a transverse momentum above 6 GeV. A minimum transverse momentum threshold of 2 GeV is applied to the generated electrons.

The simulated data have been produced for the ATLAS Computing System Commissioning [6]. All samples have been generated using the Pythia 6.403 [7] Monte Carlo event generator. More details on the Monte Carlo generators used can be found in [8]. Data have been simulated using GEANT4 [9], with the ATLAS software ATHENA [10], with a realistic geometry including material distortions in front of the electromagnetic calorimeter. Studies presented here correspond to very early data taking, with an initial luminosity of $10^{31}$ cm$^{-2}$ s$^{-1}$. No pile-up has been included. Detailed information about these samples is given in Table 1.

Signal electrons come from $J/\psi$ and $\Upsilon$ decays[1]. The background electrons arise from other direct ($b \to e$, $c \to e$) and cascade ($b \to c \to e$) semileptonic decays of meson with an electron in the final state and $b \to \tau \to e^-$ decays[2]. Other background electrons arise from $\pi^0$ Dalitz decays, $\gamma$-conversions occurring in the inner detector and decays of light hadrons. Distributions of generator level transverse momentum $p_T$ and pseudorapidity $\eta$ for electrons and pions are shown for the $pp \to J/\psi\ (ee)+X$ sample on Fig. 1. The $\eta$ distribution of electrons from conversion reflects the amount of material in front of the electromagnetic calorimeter. Ref. [13] details the reconstruction of such electrons. Table 2 gives the mean $p_T$ for each population in the different samples.

Fig. 2 shows the distance $\Delta R$, at generator-level, between the two signal electrons from $J/\psi$. On average, electrons from direct reconstructed J/$\psi$(e3e3) are separated by $\Delta R = 0.7$, and are restricted from being produced at separations larger than 1.1. Electrons from J/$\psi$ originated from $B$ hadrons on the contrary are on average more collimated, with a mean $\Delta R = 0.6$ and with a larger spread. In comparison, the higher mass of of $\Upsilon$ requires the electrons to have a much larger opening angle, with a broad distribution in $\Delta R$, the two electrons being almost back-to-back.

# 3 Trigger selection

## 3.1 General requirements

ATLAS has a three level trigger system which reduces the 40 MHz bunch crossing rate to about 200 Hz to be recorded. The first level (L1) is a hardware-based trigger which makes a fast decision (in 2.5 $\mu$s) about which events are of interest for further processing, with a rate reduced down to below 40 kHz in its initial implementation. Coarse granularity information from the calorimeter and muon trigger systems are used at this stage of the trigger to identify regions of the detector which contain interesting signals corresponding to, for instance, electrons, muons, taus, and jets. These are called "Regions of Interest" (RoIs) and are used to guide the later stages of the trigger reconstruction. The high level trigger (HLT) is software-based and is split into two levels. At level 2 (L2) the full granularity of the detector is used to confirm the L1 signals and then to combine information from different sub-detectors within the RoIs identified at L1. Fast algorithms are used for the reconstruction at this stage and the rate is reduced to $1-2$ kHz with an average execution time of about 40 ms. Lastly, at the event filter (EF), the whole event is available and "offline-like" algorithms are used along with better alignment and calibration information to form a final decision whether or not an event is accepted. With an execution time of about 4 s, the rate is reduced to 200 Hz.

[1] The corresponding branching ratio [11] is $Br(J/\psi \to ee) = (5.94 \pm 0.06)\%$ and $Br(\Upsilon \to ee) = (2.38 \pm 0.11)\%$.

[2] The corresponding branching ratios [12] are $Br(b \to l^-) = (10.71 \pm 0.22)\%$, $Br(b \to c \to l^+) = (8.01 \pm 0.18)\%$, $Br(b \to \bar{c} \to l^-) = (1.62^{+0.44}_{-0.36})\%$, $Br(b \to \tau \to e^-) = (0.419 \pm 0.055)\%$ and $Br(b \to (J/\psi, \Upsilon) \to ee) = (0.072 \pm 0.006)\%$.

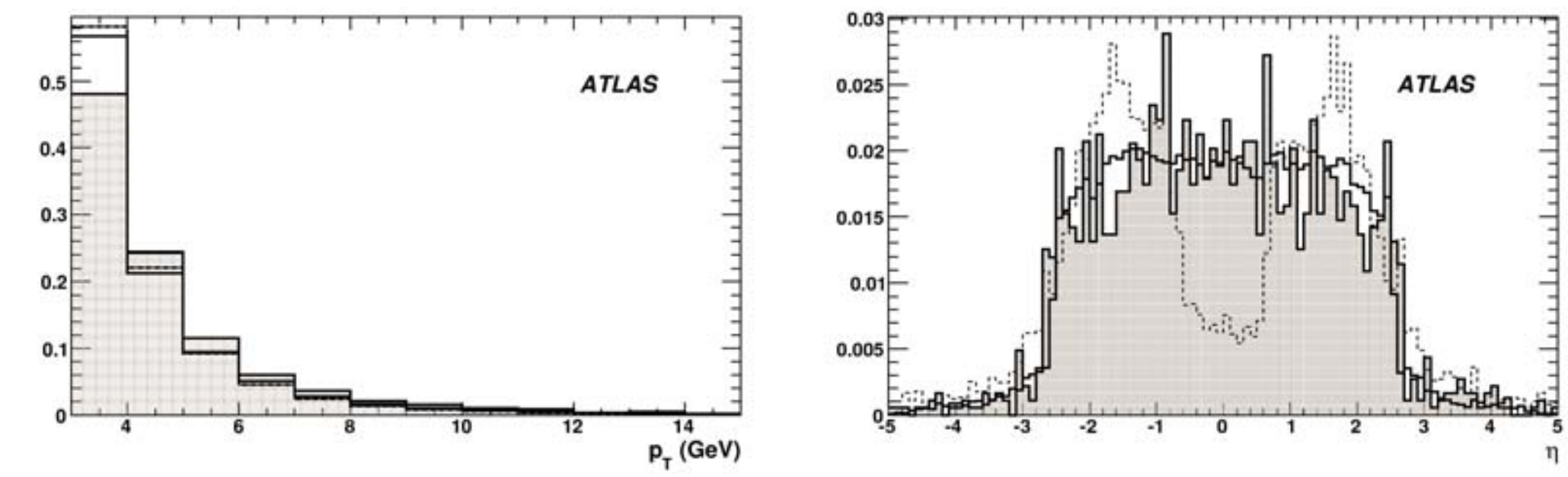

Figure 1: Normalised distributions of generator-level transverse momentum $p_T$ (left) and pseudorapidity $|\eta|$ (right) in the $pp \to J/\psi X$ sample are shown for signal electrons (hatched histograms), electrons from conversions (dotted line histogram), and pions (plain histograms).

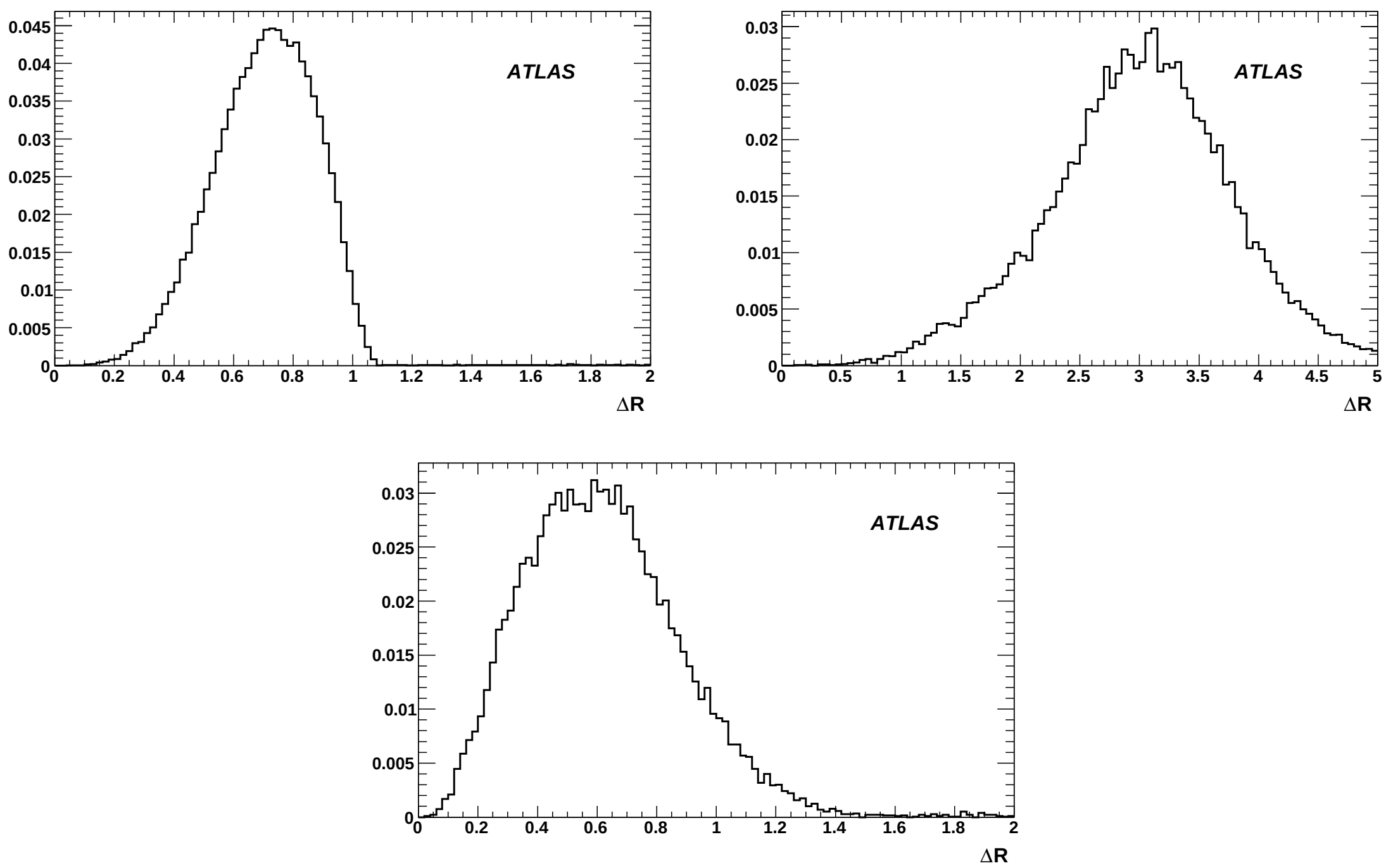

Figure 2: Distance $\Delta R$ at generator-level between the two signal electrons for direct $J/\psi$ events (top left), direct $\Upsilon$ (top right) and $J/\psi$ from b decays (bottom).

Table 2: Mean generator-level $p_T$ (in GeV) for electrons and pions having $p_T > 2$ GeV. Typical RMS on distributions of Fig. 1 is 1.4 GeV.

| sample | electrons | | | pions |
|---|---|---|---|---|
| | signal | $B$ and $D$ hadrons | $\gamma$-conversions and $\pi^0$ Dalitz | |
| $pp \to J/\psi X$ | 4.7 | 4.7 | 4.3 | 4.4 |
| $pp \to \Upsilon X$ | 4.8 | 4.5 | 4.0 | 4.2 |
| $pp \to$ Drell-Yan | - | 5.9 | 4.9 | 4.7 |
| minimum bias | - | 4.2 | 4.2 | 4.2 |
| $bB_d \to \mu(6)J/\psi X$ | 6.3 | 5.5 | 4.9 | 5.1 |

The expected ATLAS trigger performance at an initial luminosity of $10^{31}$ cm$^{-2}$ s$^{-1}$ is studied using the samples described in the previous section. Two trigger menus are considered here: the first is a purely electromagnetic menu which could be used only for early data taking; the second menu relies on the $B$-trigger and could be extended for data taking at low luminosity $10^{33}$ cm$^{-2}$ s$^{-1}$. More details about the overall trigger strategy, in particular for these channels, can be obtained in [14] and [15].

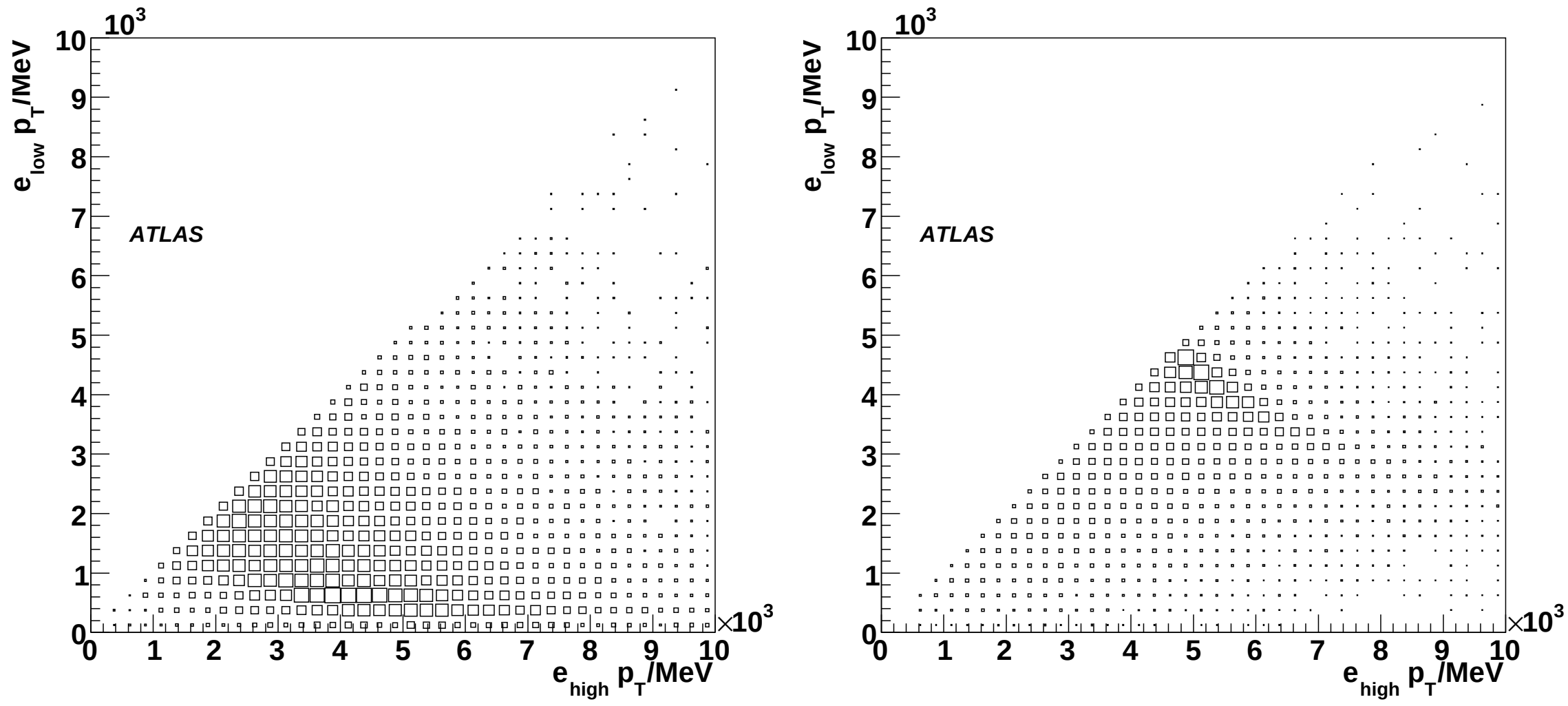

Figure 3: Distribution of the generator-level transverse momentum of the less energetic electron versus the transverse momentum of the most energetic electron in the direct J/$\psi$ (left) and $\Upsilon$ (right) decays.

## 3.2 Electron selection

J/$\psi \to ee$ and $\Upsilon \to ee$ events are very demanding for the trigger system. Due to their relatively low masses, the electrons produced in the J/$\psi$ and $\Upsilon$ decays are very soft. Figure 3 shows the distribution of the transverse momentum of the less energetic electron versus the transverse momentum of the most energetic electron in J/$\psi$ (left) and $\Upsilon$ (right) decays [5]. This poses a huge challenge for the L1 calorimeter trigger. Its performance at the low-energy end is limited by the noise of typically 0.5 GeV per RoI and a 3 GeV threshold is the limit of what is feasible for the L1 trigger. The 6.5 kHz L1 output rate for 2EM3 (corresponding to two L1 electromagnetic clusters greater than 3 GeV ) makes it one of the biggest consumers of the total bandwidth [14]. The strategy to trigger on J/$\psi$ and $\Upsilon \rightarrow ee$ events is based on low $E_T$ L1 electromagnetic RoIs and further electron identification using calorimeter and inner detector information at the HLT. The inner detector tracks are reconstructed in regions of half-size $\Delta\eta \times \Delta\varphi = 0.1 \times 0.1$ around these electromagnetic RoIs.

Figure 4 shows the distribution of the invariant mass of the pairs of electrons for signal and background events after the L1 selection. The J/$\psi$ and $\Upsilon$ samples can be easily recognised by the resonance peaks. Table 3 gives the number of events which are expected to pass the L1 selection. The L1 trigger efficiency is calculated with respect to the number of generated events, which in particular include a requirement on the minimum transverse momentum of 3 GeV on each electron as detailed in Section2. The selected events are in the tail of the J/$\psi$ distribution (see Figure 3). Additionally the requirement of $E_T > 3$ GeV at L1 implies $E_T >= 4$ GeV thus starting to cut into the peak of the $\Upsilon$ distribution. The efficiency of this level is measured to be 27% for J/$\psi$ and $\Upsilon$ events. About 43% of Drell-Yan events pass the L1 in the generated mass range. A total of $4.2 \times 10^6$ J/$\psi \rightarrow$ee, $1.2 \times 10^6$ $\Upsilon \rightarrow$ee and $0.12 \times 10^6$ Drell-Yan events are expected after this level. For the minimum bias sample, the enhancement above

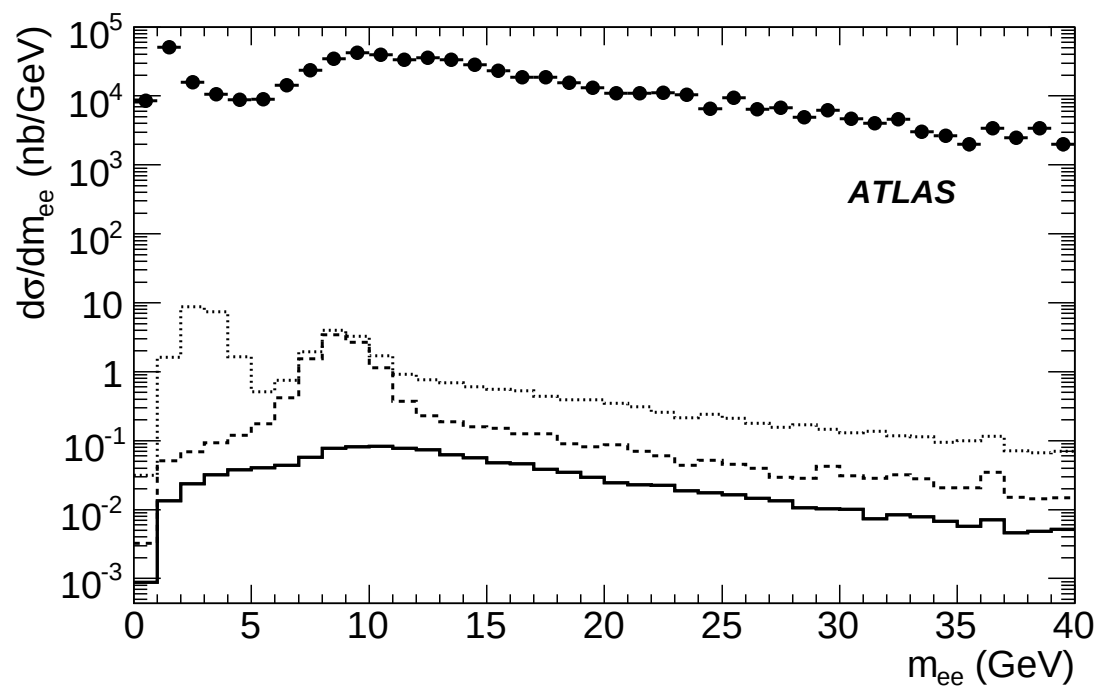

Figure 4: Expected differential cross section for low-mass electron pairs using the 2EM3 trigger menu item after L1 selection for J/$\psi$ decays (dotted histogram), $\Upsilon$ decays (dashed histogram), Drell-Yan production (solid histogram) and expected background (full circles). The invariant mass is reconstructed with calorimeter only information available at L1.

6 GeV arises from the requirement of the presence of two L1 clusters with energy greater than 3 GeV. Studies and implementation of an efficient HLT selection, based on the selection of two electrons with $E_T > 5$ GeV (2e5 menu) is ongoing. Typical rates are expected to be $\sim$ 40 Hz at L2 and 6 Hz at EF.

Table 3: Performance of the 2EM3 trigger at the luminosity of $10^{31}$ cm$^{-2}$ s$^{-1}$ for the direct production of J/$\psi$, $\Upsilon$, Drell-Yan and background events. For signal events the efficiency $\varepsilon$ is given as well as the number of expected events. For background the rate is provided. Quoted errors are statistical only.

| | J/$\psi$ | | $\Upsilon$ | | Drell-Yan | | background |
|---|---|---|---|---|---|---|---|
| | $\varepsilon$ (%) | $10^6$ ev / 100 pb$^{-1}$ | $\varepsilon$ (%) | $10^6$ ev / 100 pb$^{-1}$ | $\varepsilon$ (%) | $10^6$ ev 100 pb$^{-1}$ | Rate (Hz) |
| L1 | 27.4±0.3 | 4.17±0.04 | 27.3±0.3 | 1.22±0.01 | 43.0±0.5 | 0.12±0.001 | 6500±27 |

### 3.3 *B*-physics

The *B*-trigger is expected to account for 5-10% of the total ATLAS trigger resources. The trigger for *B*-physics is initiated by a single- or a di-muon selection at L1. At $10^{31}$ cm$^{-2}$ s$^{-1}$, a threshold $p_T > 4$ GeV will be used, rising to about 6 GeV at $10^{33}$ cm$^{-2}$ s$^{-1}$ to match the rate capabilities of the HLT. For final states such as the $bB_d \rightarrow \mu(6)J/\psi X$ events, inner detector tracks are combined to reconstruct the J/$\psi$ particles. Two different strategies are used for finding the tracks, depending on luminosity [16]. At $10^{31}$ cm$^{-2}$ s$^{-1}$ full reconstruction over the whole inner detector can be performed, since the L1 muon rate is comparatively modest, while at higher luminosities reconstruction will be limited to L1 electromagnetic RoIs with $E_T > 3$ GeV. For the $bB_d \rightarrow \mu(6)J/\psi X$ events the L1 trigger efficiency is $\sim 88\%$. This latter approach has lower efficiency for selecting the signal but requires fewer HLT resources for a fixed L1 rate. If one combines triggers for electromagnetic final states and pre-scaled single muon-triggers needed for trigger efficiency measurements, the overall rate for *B*-physics triggers is approximately 20 Hz at $10^{31}$ cm$^{-2}$ s$^{-1}$.

# 4 Electron reconstruction and identification

The standard electron reconstruction procedure [17], optimised for high energetic electrons, is based on calorimeter clusters to which tracks are associated in a second step. An alternative procedure will be used for the reconstruction of electrons originated from J/$\Psi$ and $\Upsilon$ decays. It takes full advantage of the tracking capabilities of the inner detector as well as the granularity of the electromagnetic calorimeter. The method is seeded by a track which is extrapolated into the electromagnetic calorimeter and allows for efficient reconstruction of electrons in jets for *b*-tagging purpose (cf. [18]) and very low $p_T$ electrons.

### 4.1 Electron reconstruction

The track-based algorithm could handle any charged track particles with a transverse momentum greater than 0.5 GeV. Still, as it will be detailed further, in order to reduce the amount of fake candidates, in particular in jets, only particles with $p_T > 2$ GeV are considered. The inner detector coverage goes to pseudorapidity values up to 2.5, except for the transition radiation tracker (TRT) which extends up to 2. This subdetector being crucial nonetheless in the identification procedure but also to preselect tracks, the electron reconstruction is limited to $|\eta| < 2$. Strict selection criteria, similar to the *b*-tagging ones, are required to have at least nine precision hits (pixel and silicon detectors); at least two hits in the pixel detector and at least one hit in the vertexing layer. The TRT plays a central role in electron identification. Selection criteria are thus required to have at least 20 hits in the TRT and at least one high energy hit (HTR hit) in the TRT detector along the track. After these criteria, only 50% of initial tracks remain. All the tracks that pass these criteria are extrapolated [19] to the second sampling of the electromagnetic calorimeter. Around this position a cluster of size $\Delta\eta \times \Delta\phi = 0.075 \times 0.125$ ($3 \times 5$ in units of cells in the middle sampling) in the barrel and $\Delta\eta \times \Delta\phi = 0.125 \times 0.125$ ($5 \times 5$) in the end-caps is built. The cell with the maximum energy is searched within a small $\eta$ and $\phi$ window, $0.075 \times 0.075$ in the middle layer, around the extrapolation point. Shower shapes are estimated with respect to this position.

Since the algorithm is the same as for the reconstruction of electrons in jets [18], a set of *preselection* criteria are applied to decrease the number of fake candidates:

- the fraction of energy reconstructed in the core of the shower in the first sampling $E_1(\text{core})/E >$0.03;
- the fraction of energy reconstructed in the core of the shower in the third sampling $E_3(\text{core})/E <$0.5;
- the ratio of the energy $E$ reconstructed in the electromagnetic calorimeter over the momentum $p$ reconstructed in the inner detector $E/p >$0.7.

The above selection rejects about 5% of signal electrons but also ensures that the shower shapes in the first sampling are correctly defined. Finally, candidates which are also reconstructed as originating from a conversion [13] are vetoed, corresponding to a loss of 1-3% of signal electrons and pions.

By fitting electron tracks in such a way as to allow for bremsstrahlung, it is possible to improve the reconstructed track parameters, as shown in Fig. 5 on the ratio between the reconstructed and the true momentum for electrons. These algorithms rely exclusively on the inner detector information. The method of dynamic-noise-adjustment extrapolates track segments to the next silicon layer. If it finds a significant $\chi^2$ contribution, compatible with an energy loss by the track due to bremsstrahlung, the fraction of radiated energy is estimated and a corresponding additional noise term is included in the Kalman filter [1] [20].

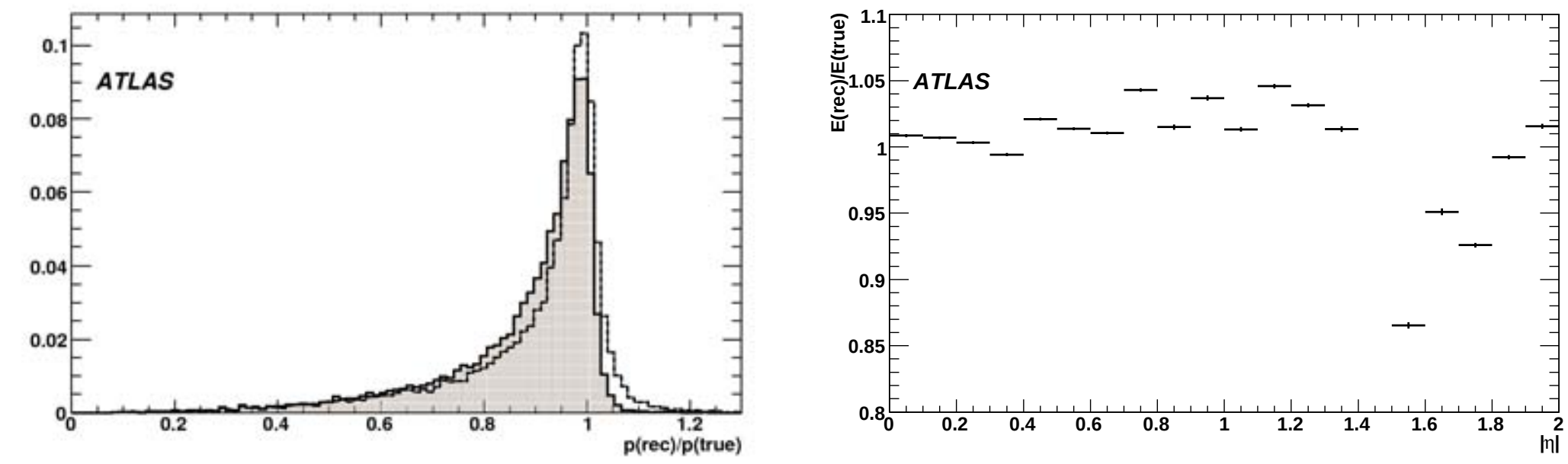

Figure 5: Left : ratio of the reconstructed to true momentum for electrons, for the default Kalman filter (hatched histogram) and for bremsstrahlung recovery algorithm (plain histogram) in the J/$\psi$ samples. Right : ratio of the reconstructed to true energy versus $\eta$ for electrons.

Position and energy corrections are applied in the precise reconstruction of the electromagnetic cluster and are described in [21]. These corrections have been tuned for high energy clusters and are not optimal for low energy electrons. Moreover, they have been determined with electron samples simulated with a detector taken to be perfectly aligned. In Fig. 5 the ratio between the reconstructed and the true energy is shown as a function of $|\eta|$ for signal electrons from $J/\psi$ samples. It can be seen that these corrections over-estimate the electron energy except in the crack region where the effect of extra-material in front of the calorimeter is important. Work is on going to improve the energy reconstruction at low energy.

By default the four-momentum of an electron is defined as the energy reconstructed in the calorimeter, whereas direction is taken from the associated track. As in this note the main physics processes result in electron transverse momenta of less than 15 GeV, the tracker momentum is used instead of the energy unless stated otherwise. Future developments in ATLAS will ensure an optimal combination of calorimeter and tracker measurements in the energy definition.

## 4.2 Electron identification

The most common background processes for producing electron-like showers in the calorimeters were described in section 2. Because the development of showers is different for electrons and hadrons, the electron identification algorithm incorporates variables that describe the shower shapes, quality of the match between the track and its corresponding cluster and the fraction of high threshold hits in the transition radiation tracker.

#### 4.2.1 Identification of isolated electrons

The identification for isolated electrons is based on cuts on the shower shapes, on information from the reconstructed tracks and on the combined reconstruction [17]. To be consistent with the trigger selection, only particles having a transverse energy $E_T > 5$ GeV are considered in the following. Three levels of selection are available:

- "loose", consisting of simple shower-shape cuts (longitudinal leakage, shower shape in the middle layer of the electromagnetic calorimeter) and very loose matching cuts between reconstructed tracks and calorimeter clusters;
- "medium", which adds shower shape cuts making use of the important information contained in the first layer of the electromagnetic calorimeter and track-quality cuts; and
- "tight", with tighter track matching criteria and the cut on the energy-to-momentum ratio. This selection also explicitly requires the presence of a vertexing-layer hit on the track (to further reject photon conversions) and a large ratio of high-threshold to low-threshold hits in the TRT detector (to further reject the background from charged hadrons). Additionally, further isolation of the electron may be required by using calorimeter energy isolation beyond the cluster itself. Two sets of tight selection cuts are used to estimate the overall performance of the electron identification. They are labeled as "tight(TRT)", in the case where a TRT cut with approximately 90% efficiency for electrons is applied, and as "tight(isol)", in the case where a TRT cut with approximately 95% efficiency is applied in combination with a calorimeter isolation cut.

The discriminating variables show a significant dependence on the pseudorapidity and a less pronounced one on the transverse momentum. In $\eta$ the dependence corresponds to varying granularities, lead thickness and material in front of the electromagnetic calorimeter. The separation between the distributions obtained for electrons and pions can vary also with $\eta$. The thresholds applied for cuts have been optimised in five $\eta$ bins, (0,0.8), (0.8,1.37), (1.37,1.52), (1.52,1.8), (1.8,2.0), and for transverse energies below 7.5 GeV, between 7.5 and 15 GeV and above 15 GeV.

The electron identification efficiency is defined as $\varepsilon_e = N_e^t/N_e$, where $N_e$ is the number of signal electron tracks, which pass the track cuts and $N_e^t$ is the number of signal electrons which pass identification cuts. The charged pion rejection is defined as $R_\pi = N_\pi/N_\pi^t$, where $N_\pi$ is the number of *good quality* pion tracks and $N_\pi^t$ is the number of *good quality* pion tracks misidentified as signal electrons. Table 4 shows the electron identification efficiency and pion rejection factor after loose, medium, tight with no isolation requirement and tight selections for the different data samples. For tight selection the efficiency is $\varepsilon_e \sim 65\%$ for direct J/$\psi$ production. Performance is similar for the $\Upsilon$ sample, despite the higher average momentum of the signal electrons, due to the cut on $E_T > 5$ GeV. Figure 6 shows in more detail the overall reconstruction and identification performance: the $p_T$ and $\eta$ dependences of the efficiencies for electrons. There is still an important $\eta$ dependence due to a few discriminating variables and work is ongoing to improve it.

#### 4.2.2 Identification of electrons from $b$ quark

Another identification procedure is optimised for non-isolated electrons and is thus particularly useful for $b\bar{b}$ events. The trigger anticipated for these events is based on a muonic decay mode of either the $b$ or the $\bar{b}$ quark. All good quality tracks are considered above a transverse momentum $p_T >$2 GeV. When possible we use the same variables as for the isolated electron identification but some variables - like the hadronic leakage by the fraction of energy reconstructed in the third sampling - are replaced or only the core of the electromagnetic shower is used. In addition to the traditional cut-based analysis, multivariate techniques have been developed, based on the similar variables, and in particular a likelihood

Table 4: Expected efficiencites $\varepsilon_e$ for electrons from J/$\psi$ and $\Upsilon$ decay for the four standard levels of cuts used for isolated electron identification. Only electrons with $E_T > 5$ GeV, corresponding to the HLT threshold, are considered. The crack region in the electromagnetic calorimeter, between $1.37 < |\eta| < 1.52$ is removed. The quoted errors are statistical only.

| Selection | $pp \to J/\psi X$ | | $pp \to \Upsilon X$ | |
|---|---|---|---|---|
| | $\varepsilon_e$ (%) | $R_\pi$ | $\varepsilon_e$ (%) | $R_\pi$ |
| Loose | $84.3 \pm 0.1$ | $36 \pm 3$ | $83.7 \pm 0.4$ | $32 \pm 7$ |
| Medium | $78.4 \pm 0.1$ | $72 \pm 9$ | $78.4 \pm 0.4$ | $49 \pm 13$ |
| Tight(TRT) | $71.4 \pm 0.1$ | $109 \pm 17$ | $71.3 \pm 0.4$ | $57 \pm 16$ |
| Tight (isol) | $65.5 \pm 0.1$ | $900 \pm 400$ | $66.1 \pm 0.5$ | $740 \pm 300$ |

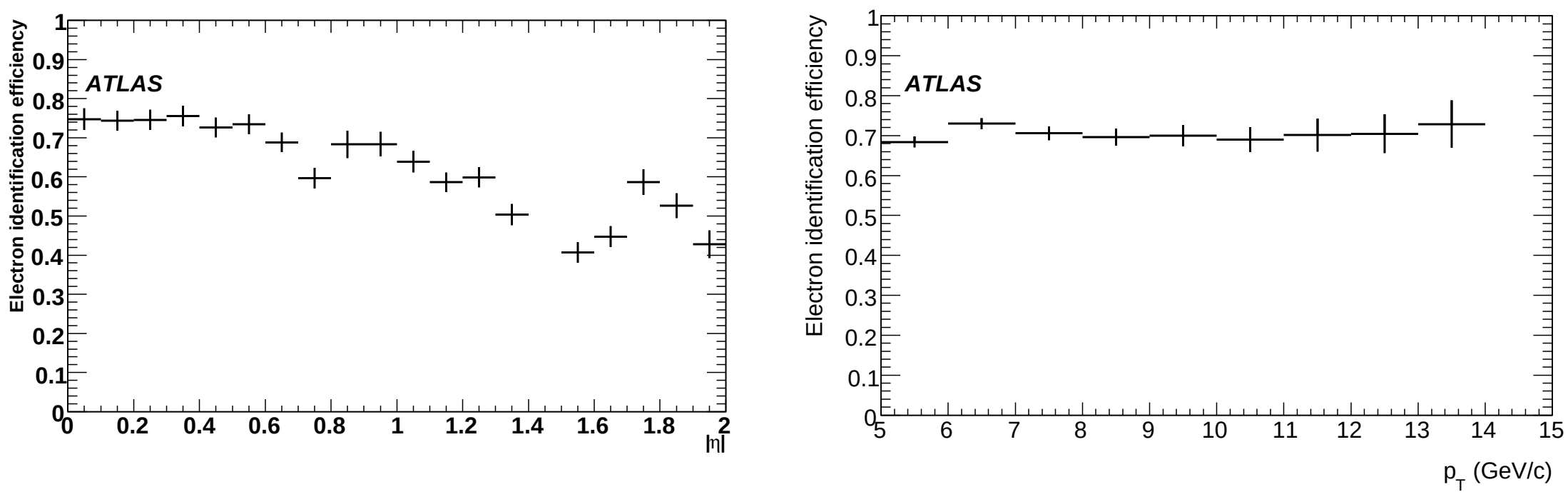

Figure 6: Electron identification efficiency with "Tight(TRT)" cuts level as a function of the pseudorapidity (left) and the transverse momentum (right) in direct J/$\psi$ events.

technique can also be used. Figure 7 shows the obtained pion rejection curve as a function of the electron identification efficiency. In the following the working point is an electron identification efficiency of

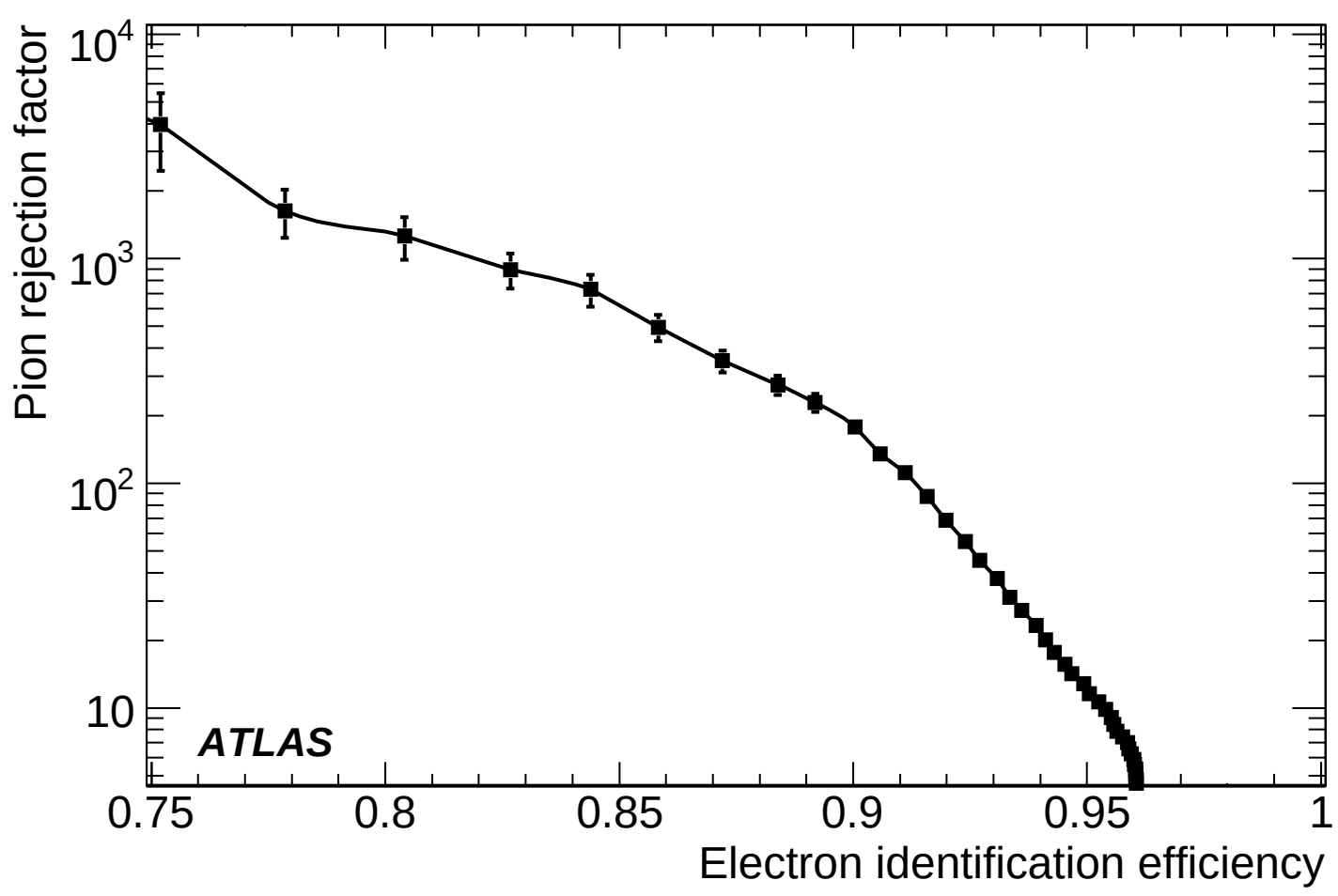

Figure 7: Pion rejection as a function of the electron identification efficiency, in $bB_d \to \mu J/\psi X$ sample.

80%, corresponding to a pion rejection factor of $\sim$ 1300. Figure 8 shows the overall reconstruction and identification performance in more details: the $p_T$ and $\eta$ dependencies of the efficiencies are shown for electrons.

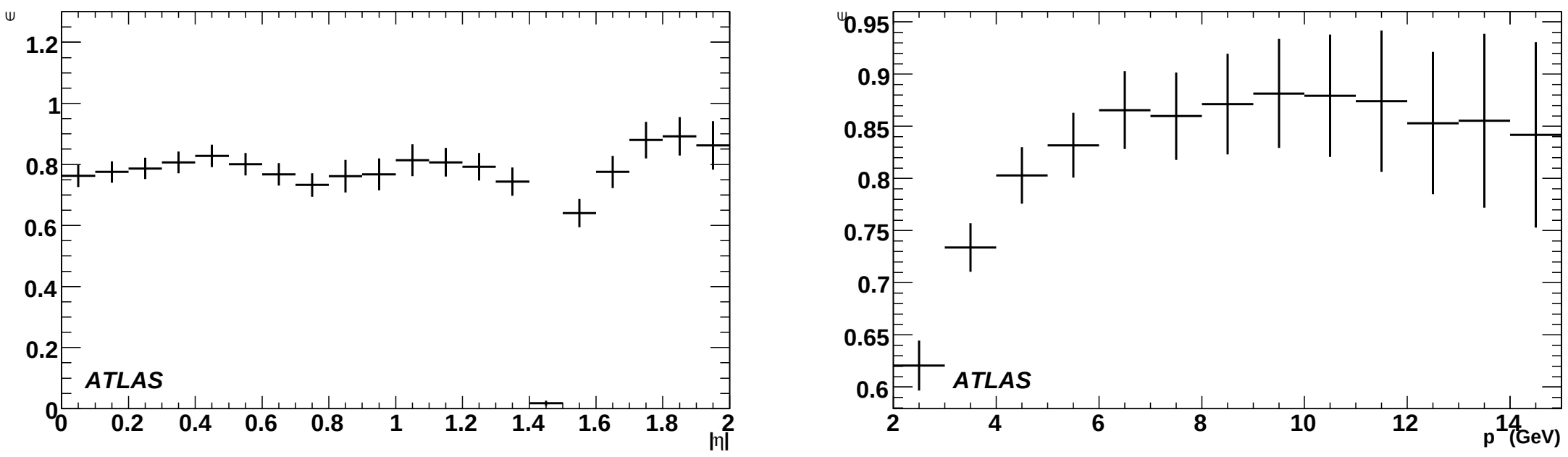

Figure 8: Electron identification efficiency in $bB_d \rightarrow \mu(6)J/\psi X$ sample as a function of the pseudorapidity (left) and the transverse momentum (right). The mean electron identification efficiency is $\varepsilon_e = 80\%$.

# 5 Expected physics studies for early data

## 5.1 Number of expected events

As described in section 3, for an initial luminosity of $10^{31}$ cm$^{-2}$ s$^{-1}$, the trigger seelction of two low energy electrons (2EM3 menu at level 1) should provide good statistics for J/$\psi \rightarrow ee$ and $\Upsilon \rightarrow ee$ decays. Fig. 9 shows the expected differential cross-section for low-mass electron pairs using the 2EM3 trigger menu item and the offline selection in linear (left) and log (right) scale. The invariant mass is reconstructed with direction taken from the inner detector and energy from the electromagnetic calorimeter which allows a better reconstruction of the invariant mass than using calorimeter only information as done at level 1. The signal-to-background ratio obtained is greater than one at the J/$\psi$ and $\Upsilon$ peaks. With an integrated luminosity of 100 pb$^{-1}$ and an efficient identification and reconstruction of these low-mass pairs, approximately two hundred thousand J/$\psi$ decays could be extracted (see table 5).

Table 5: Number of expected events for direct production of J/$\psi$, $\Upsilon$ and Drell-Yan events passing the 2EM3 trigger and offline analysis. Numbers are given for an integrated luminosity of 100 pb$^{-1}$ with early data taking at $10^{31}$ cm$^{-2}$ s$^{-1}$. Quoted errors are statistical only.

| | J/$\psi$ | $\Upsilon$ | Drell-Yan |
|---|---|---|---|
| | $10^3$ ev / 100 pb$^{-1}$ | $10^3$ ev / 100 pb$^{-1}$ | $10^3$ ev / 100 pb$^{-1}$ |
| offline + $E_T > 5$ GeV | 256±9 | 45±5 | 13.9±0.3 |
| offline + $E_T > 5$ GeV + L1 | 230±9 | 43±5 | 13.3±0.3 |

Moreover, the standard $B$-physics trigger, using a single muon above a threshold of $p_T > 4$ GeV, can give access to a sample of $J/\psi$ events originating from the $b\bar{b}$ production, without possible bias on the selection of electromagnetic objects. Due to its lower cross-section, the expected number of events is much less, around $2.3 \times 10^3$ after offline selection and $\sim 1.9 \times 10^3$ after trigger and offline selection, but without any selection on the electrons themselves. A better estimation of this number requires combining a single muon trigger with a trigger for electromagnetic final states as described in section 3.3.

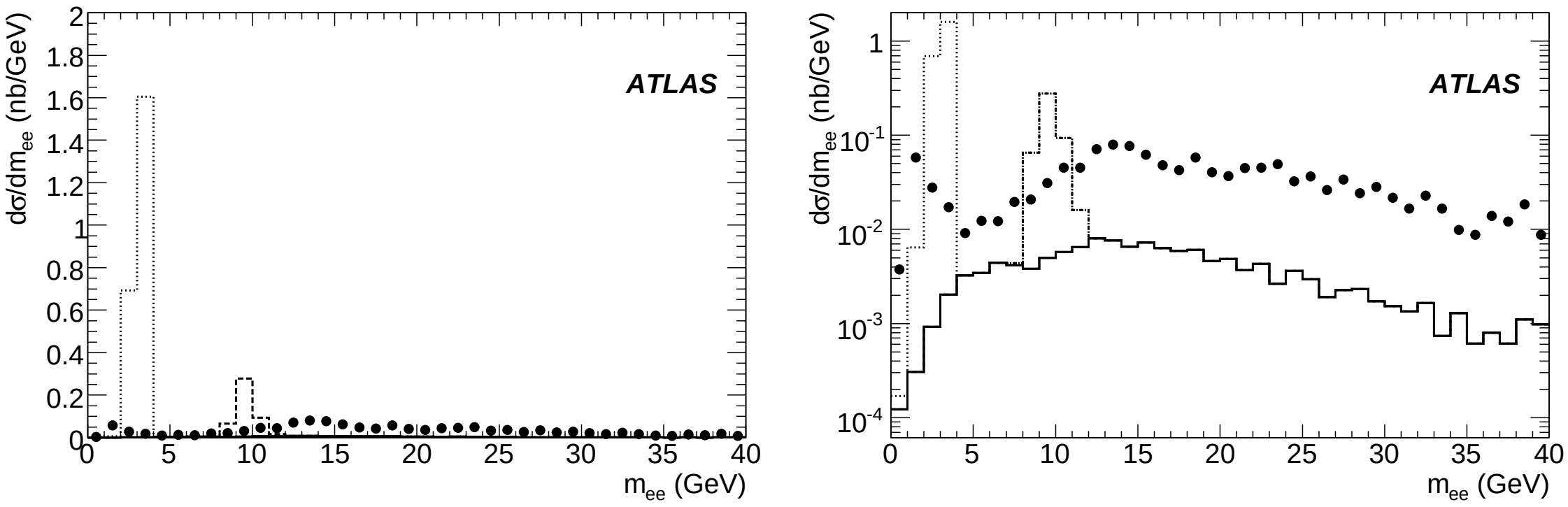

Figure 9: Expected differential cross section for low-mass electron pairs using the 2EM3 trigger menu item and the offline selection in linear (left) and log (right) scale. Shown is the invariant di-electron mass distribution reconstructed using tracks for J/$\psi \rightarrow ee$ decays (dotted histogram), $\Upsilon \rightarrow ee$ decays (dashed histogram) and Drell-Yan production (full histogram). Also shown is the expected background (full circles). The invariant mass is reconstructed with direction taken from the inner detector and energy from the electromagnetic calorimeter.

## 5.2 Quality of the mass reconstruction with initial data

In this section, we study the offline reconstruction of the J/$\psi$ and $\Upsilon$ particles from their decay products. After a short description of the algorithm, we study the performance of the reconstruction for $J/\psi$s originating from the $b\bar{b}$ decays. The invariant mass has been reconstructed with the inner detector only, combining information from the inner detector and the electromagnetic calorimeter, and using only the latter information. For the reconstruction with inner detector information we present results with and without the bremsstrahlung recovery procedure included. For direct production of J/$\psi$ and $\Upsilon$, we only show results using mass reconstruction with the inner detector.

### 5.2.1 Reconstruction of $J/\psi$ and $\Upsilon$ events

The identification of electrons is performed using the electron reconstruction algorithm described above. Electrons are identified with either the "tight" cuts for isolated electrons, or based on the likelihood method tuned to an electron identification efficiency of 80%. Pairs of electrons are thus selected. These pairs define the overall detection efficiency of J/$\psi$ (or $\Upsilon$) events which is the product of the losses due to the removal of clusters located in the crack in the electromagnetic calorimeter, the track quality cuts, and the electron identification efficiency.

Pairs of reconstructed opposite-charge tracks are fitted to a common vertex. Only events with a quality of the fit with $\chi^2$ per degree of freedom $< 6$ are retained Fig. 10 shows the distribution of the reconstructed transverse decay length $L_{xy}$ for direct J/$\psi$ events and events originated from $B$ hadrons decay. It is defined as: $L_{xy} = \frac{\vec{D}\cdot\vec{p}_T(J/\psi)}{||\vec{p}_T(J/\psi)||}$ , where $D$ is the distance between the primary and secondary vertices and $p_T(J/\psi)$ is the J/$\psi$ reconstructed transverse momentum. It is used to distinguish between the prompt J/$\psi$, which have a pseudo-proper time of zero ($L_{xy} < 0.4$ mm) , and $B$-hadron decays into J/$\psi$+X having an exponentially decaying pseudo-proper time distribution due to the non-zero lifetime of the parent $B$-hadrons ($L_{xy} > 0.25$ mm).

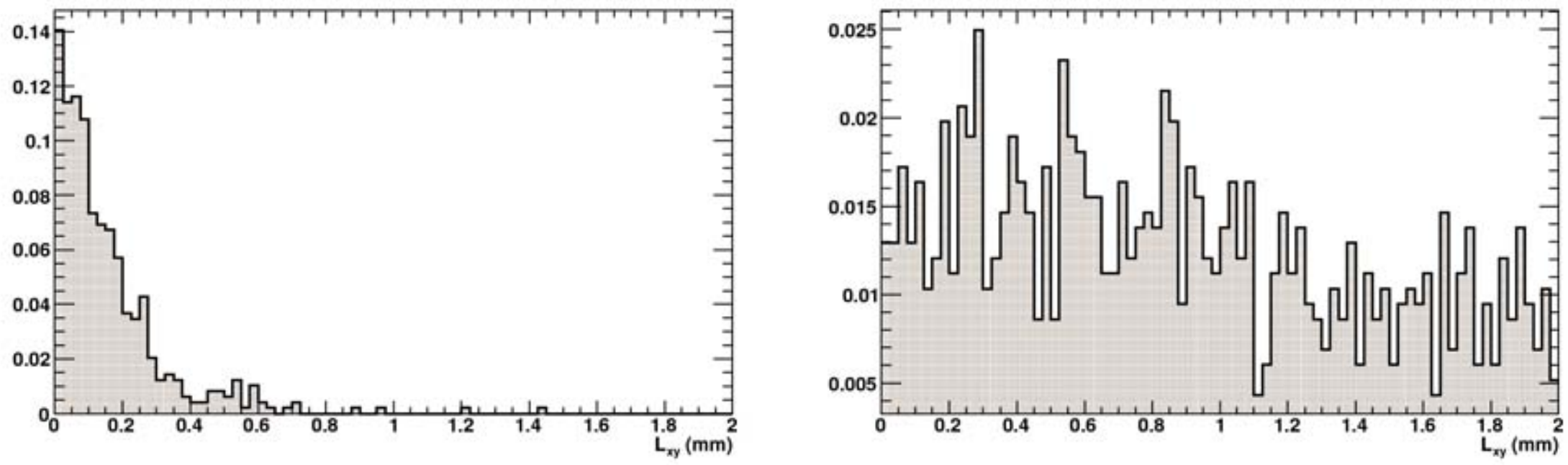

Figure 10: Distributions of the reconstructed transverse decay length direct J/$\psi$ events (left) and J/$\psi$ events originated from $B$ hadrons decay (right).

### 5.2.2 Reconstruction of $J/\psi$ from $b\bar{b}$ decays

After selection, only ~2000 events are reconstructed. This statistics is scaled to $1.9 \times 10^3$ events, corresponding to the expected statistics for an integrated luminosity of 100 $pb^{-1}$.

**Reconstruction in the inner detector:**

Fig. 11 shows the electron pair invariant mass distribution using only the inner detector information for signal events. The fitted function behaves as a Breit-Wigner distribution $\sim \Gamma/(\Delta m_0^2 + (\Gamma/2)^2)$ to the left of the peak $m_0$, and as a Gaussian of width $\sigma_{\rm right}$ to the right, as shown in Fig. 11. The parameter $\sigma_{\rm right}$ characterises the effective resolution in the invariant mass distribution of the pair, while $\Gamma$ is a measure of the intensity of energy loss by the electrons due to the bremsstrahlung. $\Delta m_0 = m_0 - M_{J/\psi}$, where $M_{J/\Psi} = 3096$ MeV is the nominal J/$\psi$ mass. The fitted values of the parameters are shown in Table 6. The J/$\psi$ reconstruction performance is assessed separately for the three cases: TRT barrel, when both electrons have their track pseudorapidity $|\eta| < 0.7$, the TRT end-caps, when at least one electron has $|\eta| > 0.7$, and the full $\eta$ range. In general the quality of the fit is not very high, in particular we see difficulties with correctly reproducing the peak. Table 6 shows the results of the fit of the invariant mass. A shift in the reconstructed mass is measured around 77 MeV, larger in the end-caps than in the barrel. As mentioned in [5], such mass shifts may be due to detector alignment, material effects, magnetic field scale and its stability. The CDF collaboration extensively and successfully used this method but it took many years at the Tevatron to collect sufficient statistics to allow for the disentanglement of various detector effects [22]. The parameter $\Gamma$ is around 550 MeV. The Gaussian width, estimated from the right part of the distribution is around 67 MeV. One can also notice the improvement in the mass reconstruction from bremsstrahlung recovery. Without any bremsstrahlung recovery, only 47% of events are reconstructed within $\pm$ 200 MeV of the nominal $J/\psi$ mass, whereas with the use of the bremsstrahlung recovery, this fraction increases to approximately 55% for the dynamic-noise-adjustment algorithm.

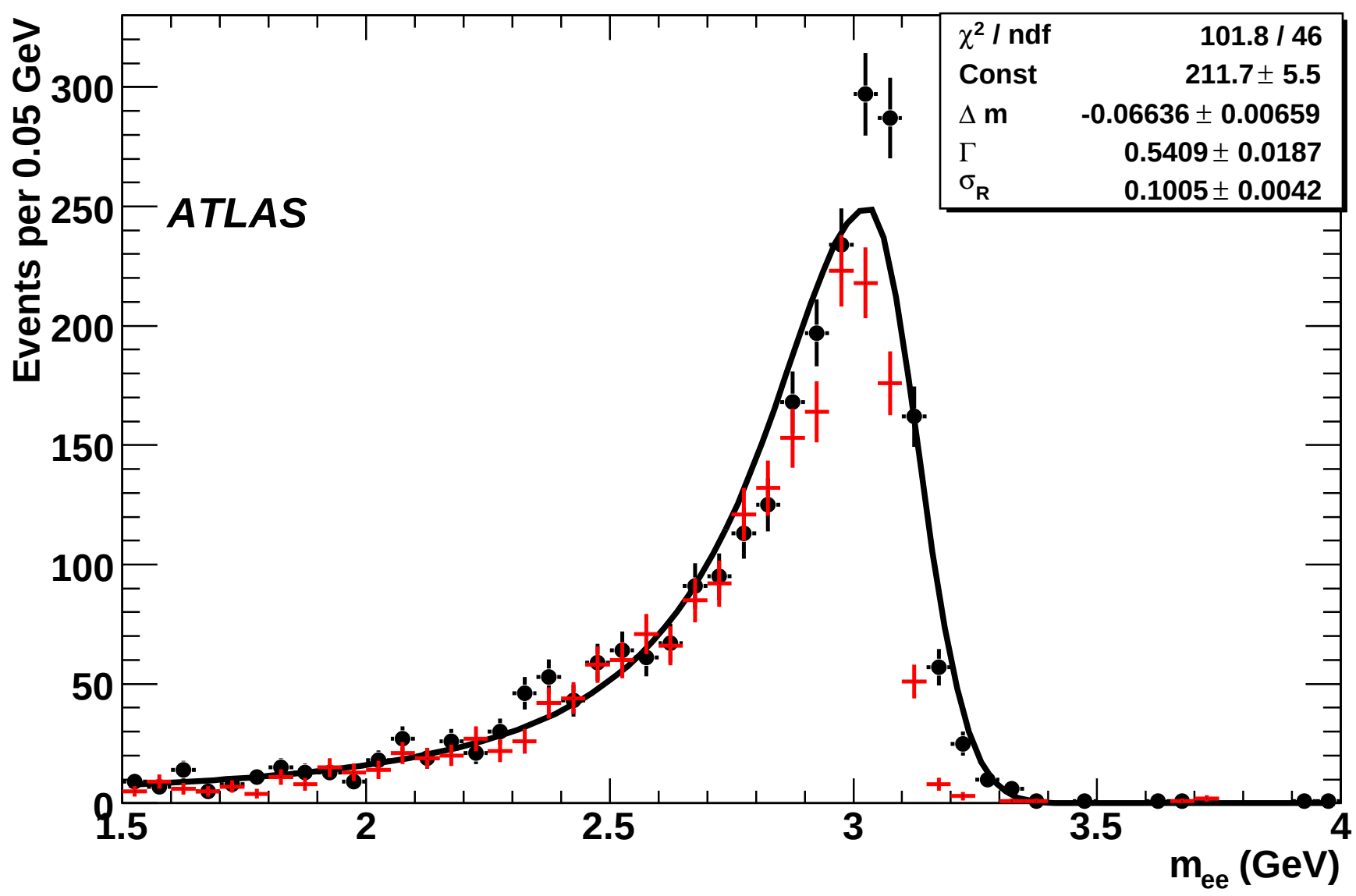

Figure 11: The electron pair invariant mass distribution for $bB_d \to \mu(6)J/\psi X$ events. The energy and direction information are taken from the inner detector. An asymmetric fit is performed with a function which behaves as a Breit-Wigner distribution to the left of the peak and as a Gaussian to the right of the peak. Results are shown without (crosses) and with (bullets) bremsstrahlung recovery included. Selection of events includes L1 trigger and offline and number of events is scaled to 100 $pb^{-1}$.

Table 6: Results of an asymmetric fit to the invariant mass distributions for $bB_d \to \mu(6)J/\psi X$ events, with a function that behaves as a Breit-Wigner distribution to the left of the peak and as a Gaussian to the right of the peak. The direction and energy information are taken from the inner detector.

| brem fit | $\eta$ range | $\Delta m_0$ (MeV) | $\Gamma$ (MeV) | $\sigma_{\text{right}}$ (MeV) |
|---|---|---|---|---|
| No | all | $-77 \pm 7$ | $557 \pm 20$ | $67 \pm 4$ |
| | Barrel | $-71 \pm 7$ | $393 \pm 19$ | $65 \pm 4$ |
| | End-caps | $-178 \pm 17$ | $688 \pm 43$ | $123 \pm 11$ |
| Yes | All | $-66 \pm 6$ | $540 \pm 18$ | $99 \pm 4$ |
| | Barrel | $-45 \pm 7$ | $417 \pm 19$ | $77 \pm 5$ |
| | End-caps | $-128 \pm 12$ | $657 \pm 33$ | $155 \pm 8$ |

**Combined reconstruction:**

The J/$\psi$ mass can be also determined combining information from the inner detector and the electromagnetic calorimeter. The energy is taken from the electromagnetic calorimeter and the direction from the more accurate measurements provided by the inner detector, taking into account the bremsstrahlung recovery procedure. Figure 12 shows the di-electron invariant mass distribution obtained from the signal sample only. An asymmetric gaussian function is fitted, with different width, $\sigma_{\text{left}}$ and $\sigma_{\text{right}}$, either side of the fitted peak mass $m_0$. It is performed in a narrow mass interval, between 2.5 and 3.6 GeV. The parameter $\sigma_{\text{right}}$ characterises the effective resolution in the invariant mass, while $\sigma_{\text{left}}$ is a measure of the deterioration of this resolution due to bremsstrahlung. The fitted values of the parameters are shown in Table 7. Performance is assessed separately for the three cases: TRT barrel, when both electrons have their track pseudorapidity $|\eta| < 0.7$, the TRT end-caps, when at least one electron has $|\eta| > 0.7$, and the full $\eta$ range. The resolution obtained is highly asymmetric, $\sim$ 387 MeV on the left and $\sim$ 189 MeV on the right. It can be also noticed that the quality of the fit is rather poor.

**Reconstruction in the electromagnetic calorimeter:**

Finally it is interesting to investigate the performance if we rely only on the information from the electromagnetic calorimeter. Fig. 13 shows the electron candidates invariant mass distribution obtained from the signal sample only. The same function defined for the combined reconstruction is used to fit the distributions. The fitted values of the parameters are shown in Table 8. Performance is assessed separately in three cases: the barrel region of the electromagnetic calorimeter, when both electrons have their pseudorapidity $|\eta| < 1.4$; the end-cap region, when at least one electron has $|\eta| > 1.4$; and for the full $\eta$ range. The resolution obtained from the width of the Gaussian is $\sim 550$ MeV.

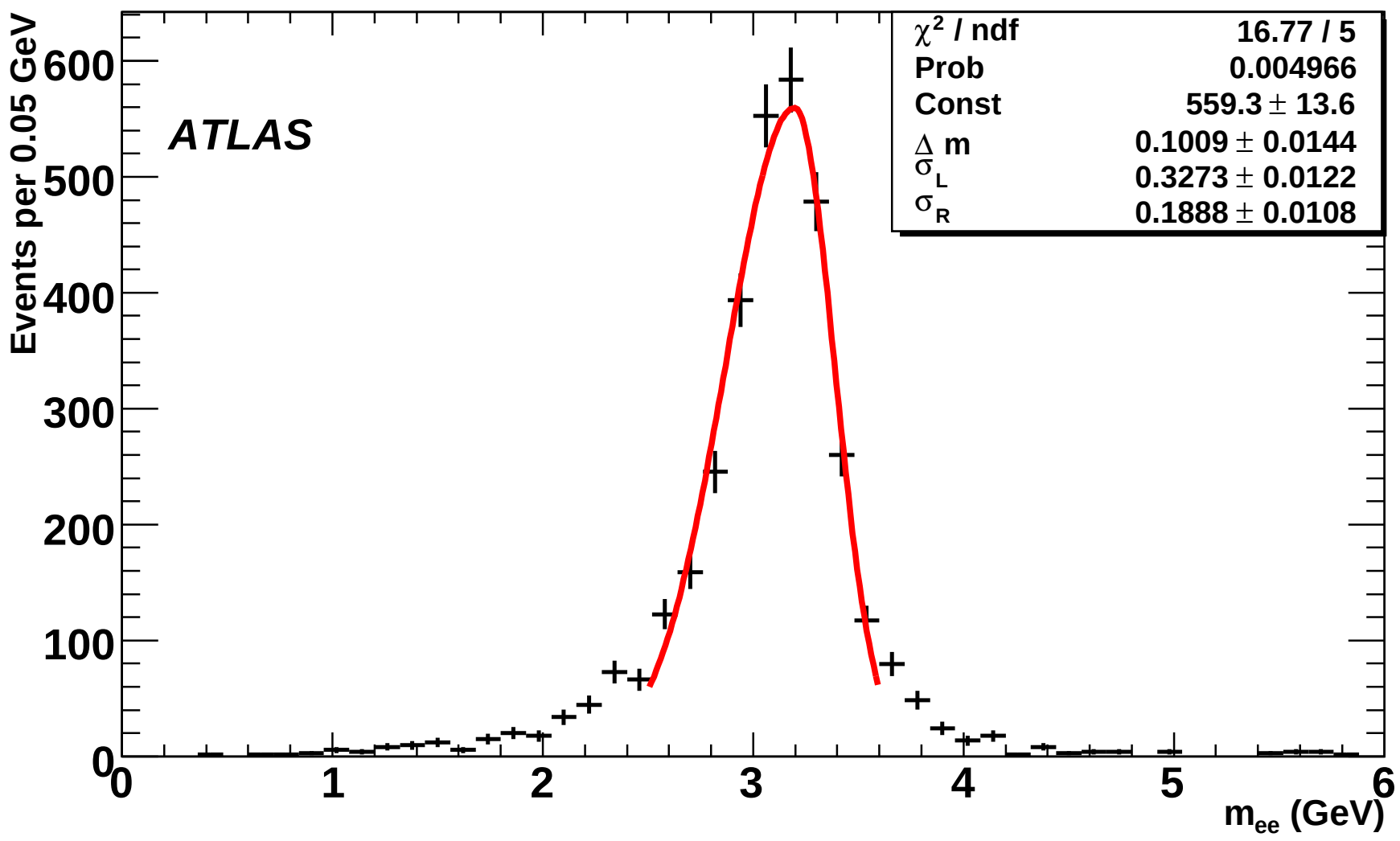

Figure 12: The electron pair invariant mass for $bB_d \to \mu(6)J/\psi X$ events. The energy is taken from the electromagnetic calorimeter and the direction from the inner detector (including bremstrahlung recovery). Selection of events includes L1 trigger and offline and number of events is scaled to 100 $pb^{-1}$.

Table 7: Asymmetric Gaussian fit results for $bB_d \to \mu(6)J/\psi X$ events. The energy is taken from the electromagnetic calorimeter and the direction from the inner detector.

| $\eta$ | $\Delta m_0$ (MeV) | $\sigma_{\rm left}$ (MeV) | $\sigma_{\rm right}$ (MeV) |
|---|---|---|---|
| All | $101 \pm 14$ | $327 \pm 12$ | $189 \pm 11$ |
| Barrel | $94 \pm 16$ | $285 \pm 12$ | $183 \pm 12$ |
| End-caps | $113 \pm 28$ | $385 \pm 27$ | $191 \pm 21$ |

Table 8: Asymmetric Gaussian fit results for $bB_d \to \mu(6)J/\psi X$ events. The energy and direction information is taken from the electromagnetic calorimeter only.

| $\eta$ | $\Delta m_0$ (MeV) | $\sigma_{\rm left}$ (MeV) | $\sigma_{\rm right}$ (MeV) |
|---|---|---|---|
| all | $-17 \pm 54$ | $567 \pm 46$ | $541 \pm 53$ |
| barrel | $-9 \pm 62$ | $558 \pm 50$ | $560 \pm 60$ |
| end-cap | $-68 \pm 102$ | $629 \pm 125$ | $414 \pm 74$ |

### 5.2.3 Reconstruction of direct J/$\psi$ and $\Upsilon$ events

After selection, only ~4000 events are reconstructed. This statistics is scaled to $2.3 \times 10^5$ events, corresponding to the expected statistics for an integrated luminosity of 100 $pb^{-1}$. Figure 14 shows the electron pair invariant mass distribution using only the inner detector information. The same function as defined previously is fitted. The fitted values of the parameters $\Delta m_0$, $\Gamma$ and $\sigma_{\rm right}$ are shown in Table 9. The fitted mass value is shifted by about 100 MeV, the $\Gamma$ factor is $\sim$ 300 MeV and the resolution term is $\sim$ 70 MeV.

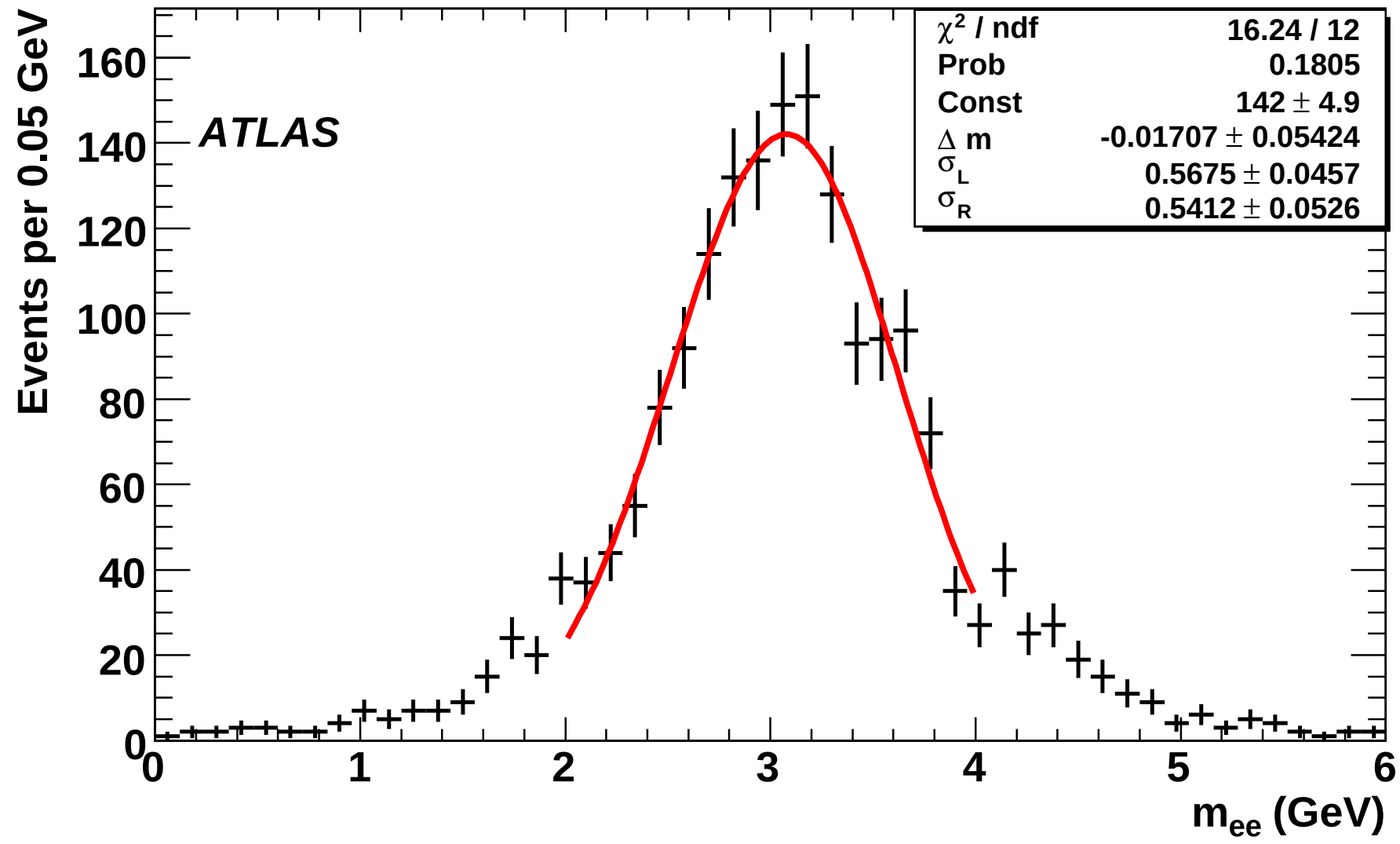

Figure 13: The electron pair invariant mass for $bB_d \to \mu(6)J/\psi X$ events. The energy and direction are taken from the electromagnetic calorimeter. Selection of events includes L1 trigger and offline and number of events is scaled to 100 $pb^{-1}$.

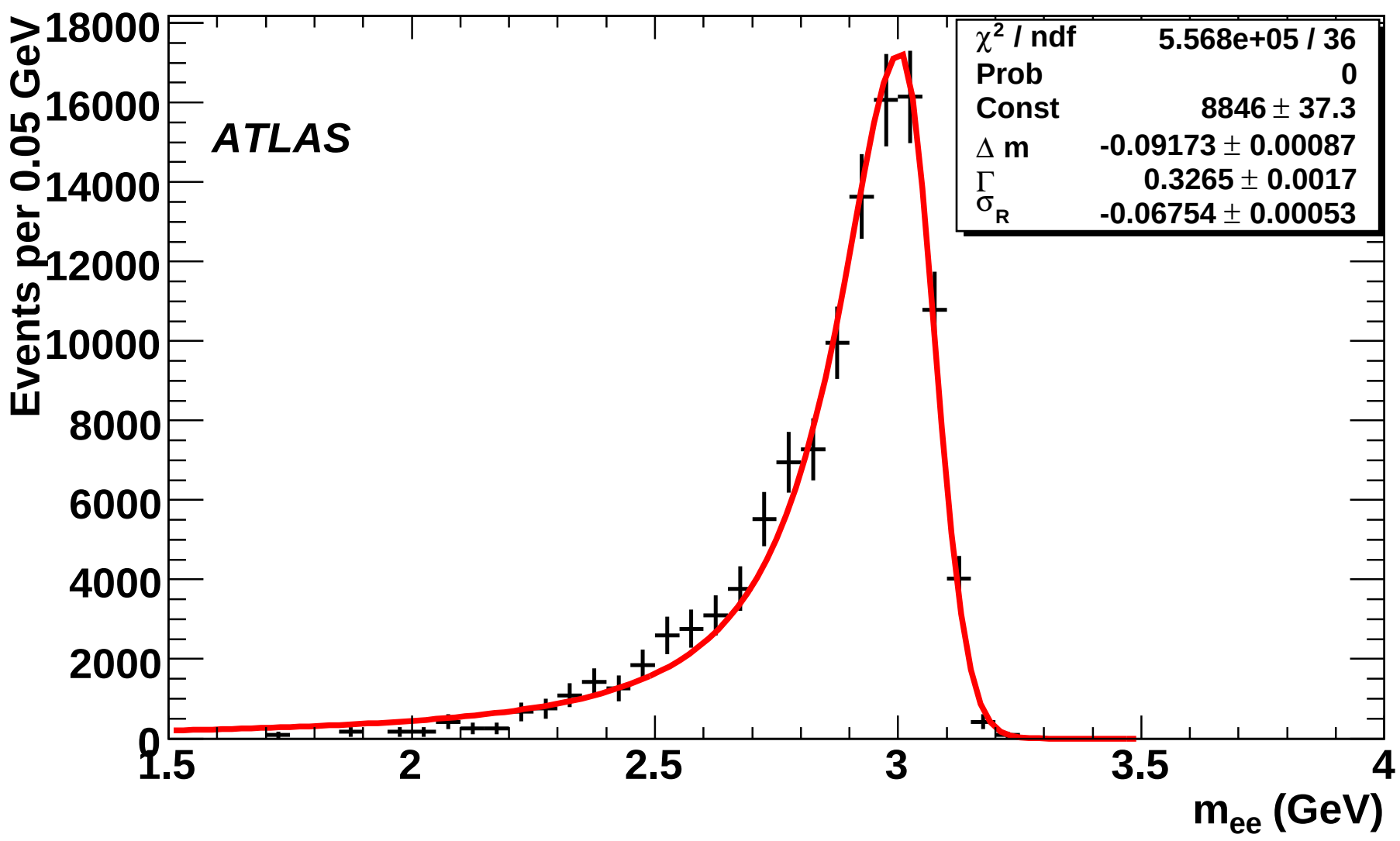

Figure 14: The electron pair invariant mass distribution for $pp \to J/\psi X$ events. The energy and direction information are taken from the inner detector. An asymmetric fit is performed with a function which behaves as a Breit-Wigner distribution to the left of the peak and as a Gaussian to the right of the peak. Selection of events includes L1 trigger, offline and a cut on $E_T > 5$ GeV for each electron to mimic the HLT. The number of events is scaled to 100 pb$^{-1}$.

Table 9: Results of an asymmetric fit to the invariant mass distributions for $pp \to J/\psi X$ events, with a function that behaves as a Breit-Wigner distribution to the left of the peak and as a Gaussian to the right of the peak. The direction and energy information are taken from the inner detector.

| $\eta$ | $\Delta m_0$ (MeV) | $\Gamma$ (MeV) | $\sigma_{\text{right}}$ (MeV) |
|---|---|---|---|
| All | -98 $\pm$ 1 | 298 $\pm$ 2 | 71 $\pm$ 1 |
| Barrel | -77 $\pm$ 1 | 255 $\pm$ 2 | 62 $\pm$ 1 |
| End-caps | -142 $\pm$ 2 | 354 $\pm$ 3 | 87 $\pm$ 2 |

### 5.2.4 $\Upsilon$ reconstruction

After selection, only $\sim$1000 events are reconstructed. This statistics is scaled to $4.3 \times 10^4$ events, corresponding to the expected statistics for an integrated luminosity of 100 pb$^{-1}$. Fig. 15 shows the electron pair invariant mass distribution from the inner detector information. The fitted values of the parameters are shown in Table 10. The fitted mass value is shifted by about 180 MeV, the $\Gamma$ factor is $\sim$ 1 GeV and the resolution term is $\sim$ 140 MeV.

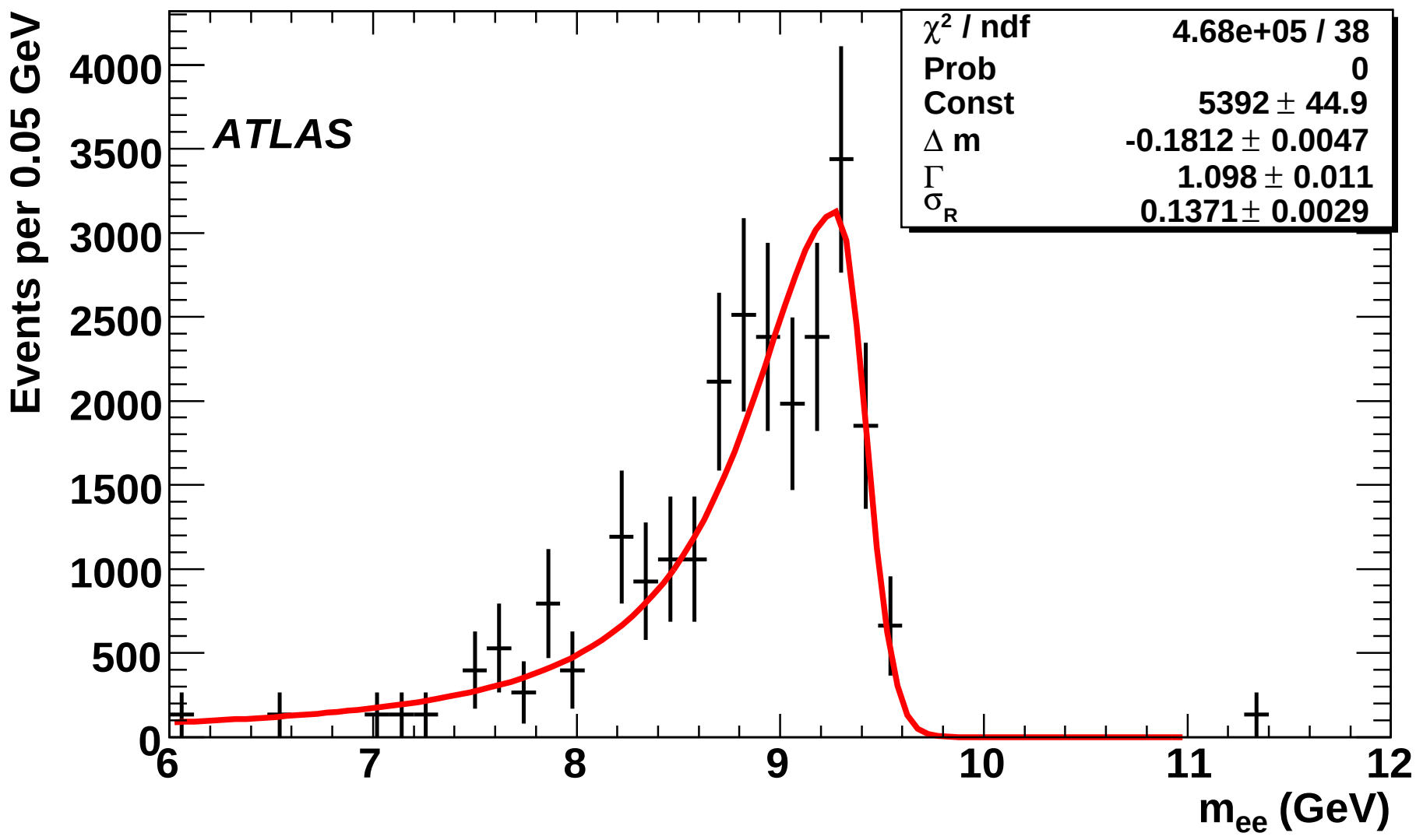

Figure 15: The electron pair invariant mass distribution for $pp \to \Upsilon X$ events. The energy and direction information are taken from the inner detector. An asymmetric fit is performed with a function that behaves as a Breit-Wigner distribution to the left of the peak and as a Gaussian to the right of the peak. Selection of events includes L1 trigger, offline and a cut on $E_T > 5$ GeV for each electron to mimic the HLT. Number of events is scaled to 100 $pb^{-1}$.

## 5.3 Assessment of performance in situ with initial data

Initial studies have been performed for the J/$\psi \to ee$ tag-and-probe method briefly outlined below, using events satisfying a single electron trigger with $E_T > 5$ GeV. Due to too high rate at L1 (40 kHz) it has to be pre-scaled by a factor of 60, which reduces the final statistics. Those events are used to look for an opposite-charge electron pair identified by the offline electron reconstruction with an invariant mass near the $J/\psi$ peak. Using the second electron as the probe which was not required to pass any trigger selection, the efficiency (relative to the offline selection) of a given trigger signature can be measured. We expect to collect of the order of $\approx 20 \times 10^3$ $J/\psi$ signal events after the pre-scale with an integrated luminosity of 100 $pb^{-1}$. Similar studies could be performed to study the offline electron selection.

Table 10: Results of an asymmetric fit to the invariant mass distributions for $pp \rightarrow \Upsilon X$ events, with a function which behaves as a Breit-Wigner distribution to the left of the peak and as a Gaussian to the right of the peak. The direction and energy information are taken from the inner detector.

| $\eta$ range | $\Delta m_0$ (MeV) | $\Gamma$ (MeV) | $\sigma_{\text{right}}$ (MeV) |
|---|---|---|---|
| All | -181 $\pm$ 5 | 1098 $\pm$ 11 | 137 $\pm$ 3 |
| Barrel | -177 $\pm$ 5 | 930 $\pm$ 12 | 137 $\pm$ 3 |
| End-caps | -252 $\pm$ 11 | 1335 $\pm$ 25 | 166 $\pm$ 7 |

One important ingredient in the calibration strategy for the electromagnetic calorimeter is the use of large statistics samples of $Z \rightarrow ee$ decays to perform an accurate inter-calibration of regions with a fixed size of $\Delta\eta \times \Delta\varphi = 0.2 \times 0.4$. To cross-check the calibration obtained from the $Z^0$ decays and also to check the linearity of the calorimeter, it is important to have calibration coefficients for a lower electron energy range, which can be obtained using the J/$\psi \rightarrow ee$ and $\Upsilon \rightarrow ee$ decays as shown in [23]. With the expected statistics, a statistical precision of $\sim 0.6\%$ can be expected on the inter-calibration of the electromagnetic calorimeter based on 100 $pb^{-1}$. Still, more studies are needed in particular to improve the energy reconstruction and to disentangle effects of inter-calibration with the distribution of material in front of the electromagnetic calorimeter.

More generally, these electron samples will allow us to study the performance of both the reconstruction of tracks in the inner detector and clusters in the electromagnetic calorimeter, as well as the alignment between these two detectors. All these studies are crucial for the very first measurements (such as, for example cross-section measurements) to be performed by the ATLAS experiment on the early data.

# 6 Conclusion

In this note, the strategy to reconstruct J/$\psi$ and $\Upsilon$ particles, decaying into electron-positron pairs, has been investigated. The possible trigger strategies have also been described. For initial luminosities of $10^{31}$ cm$^{-2}$ s$^{-1}$, a trigger on low-energy di-electron pairs (2EM3 at L1) should provide good statistics for the direct production of these particles. Moreover the standard $B$-physics trigger, using a single muon above a certain $p_T$ threshold can give access to these events through the $b\bar{b}$ production, without biasing the selection of electromagnetic objects. For these studies, the electron reconstruction seeded by a track in the inner detector has been used. Compared to previous studies, the main improvement comes from the identification procedure, which can either use the standard cut-based analysis, with thresholds tuned at low energy, or a dedicated identification developed for non-isolated electrons. The signal-to-background ratio obtained is larger than one at the J/$\psi$ and $\Upsilon$ peaks, but the extraction of the Drell-Yan signal requires further studies. With an integrated luminosity of 100 $pb^{-1}$ and an efficient identification and reconstruction of these low-mass pairs, approximately two hundred thousand J/$\psi$ decays could be isolated for detailed studies of the electron identification and reconstruction performance, in particular in terms of matching energy and momentum measurements at a scale quite different from that of the more commonly used $Z \rightarrow ee$ decays.

# Muons

# Muon Reconstruction and Identification: Studies with Simulated Monte Carlo Samples

**Abstract**
The strategy and performance for muon identification and reconstruction in ATLAS are described. Performance metrics include efficiency, fake rates and momentum resolution. Results are based on data simulated and reconstructed in 2007.

## 1 Introduction

The ATLAS experiment will detect particles created in 14 TeV proton-proton collisions produced by the CERN LHC (Large Hadron Collider). Only a tiny fraction of these collisions will correspond to interesting standard model processes and an even smaller fraction to new physics. Muons, especially those with high-$p_T$ (transverse momentum) and those that are isolated (from other activity in the detector), will be much more common in these interesting events than in the background, and thus provide important means to identify such events and to determine their properties. The ATLAS detector has been designed to be efficient in the detection of muons and to provide precise measurement of their kinematics up to one TeV.

In parallel with the construction of the detector, software has been developed to reconstruct these muons, i.e., for each recorded event, to identify muons and measure their position, direction and momentum. Here we describe the strategies being pursued for this reconstruction and the current performance characterized in terms of efficiency, fake rate and precision and accuracy of measurement. The results reported here are based on simulation data generated and reconstructed in 2007.

We begin with descriptions of the detector, the reconstruction algorithms and the means by which we measure the performance. These are followed by sections documenting this performance for each of the various reconstruction strategies and finally a section summarizing results and commenting on future developments.

## 2 Detector

The ATLAS detector [1] has been designed to provide clean and efficient muon identification and precise momentum measurement over a wide range of momentum and solid angle. The primary detector system built to achieve this is the muon spectrometer, shown in Figure 1. The spectrometer covers the pseudorapidity range $|\eta| < 2.7$ and allows identification of muons with momenta above 3 GeV/c and precise determination of $p_T$ up to about 1 TeV/c.

The muon spectrometer comprises three subsystems:

- Superconducting coils provide a toroidal magnetic field whose integral varies significantly as a function of both $\eta$ and $\varphi$ (azimuthal angle). The integrated bending strength (Figure 2) is roughly constant as a function of $\eta$ except for a significant drop in the transition between the barrel and endcap toroid coils ($1.4 \lesssim |\eta| \lesssim 1.6$).

- Precision detectors are located in three widely-separated *stations* at increasing distance from the collision region. Each station includes multiple closely-packed layers measuring the $\eta$-coordinate, the direction in which most of the magnetic field deflection occurs. Monitored drift tubes provide these measurements everywhere except in the high-$\eta$ ($|\eta| > 2.0$) region of the innermost station where cathode strip chambers are used. The measurement precision in each layer is typically better than 100 $\mu$m. The cathode strip chambers additionally provide a rough (1 cm) measurement of the $\varphi$-coordinate.

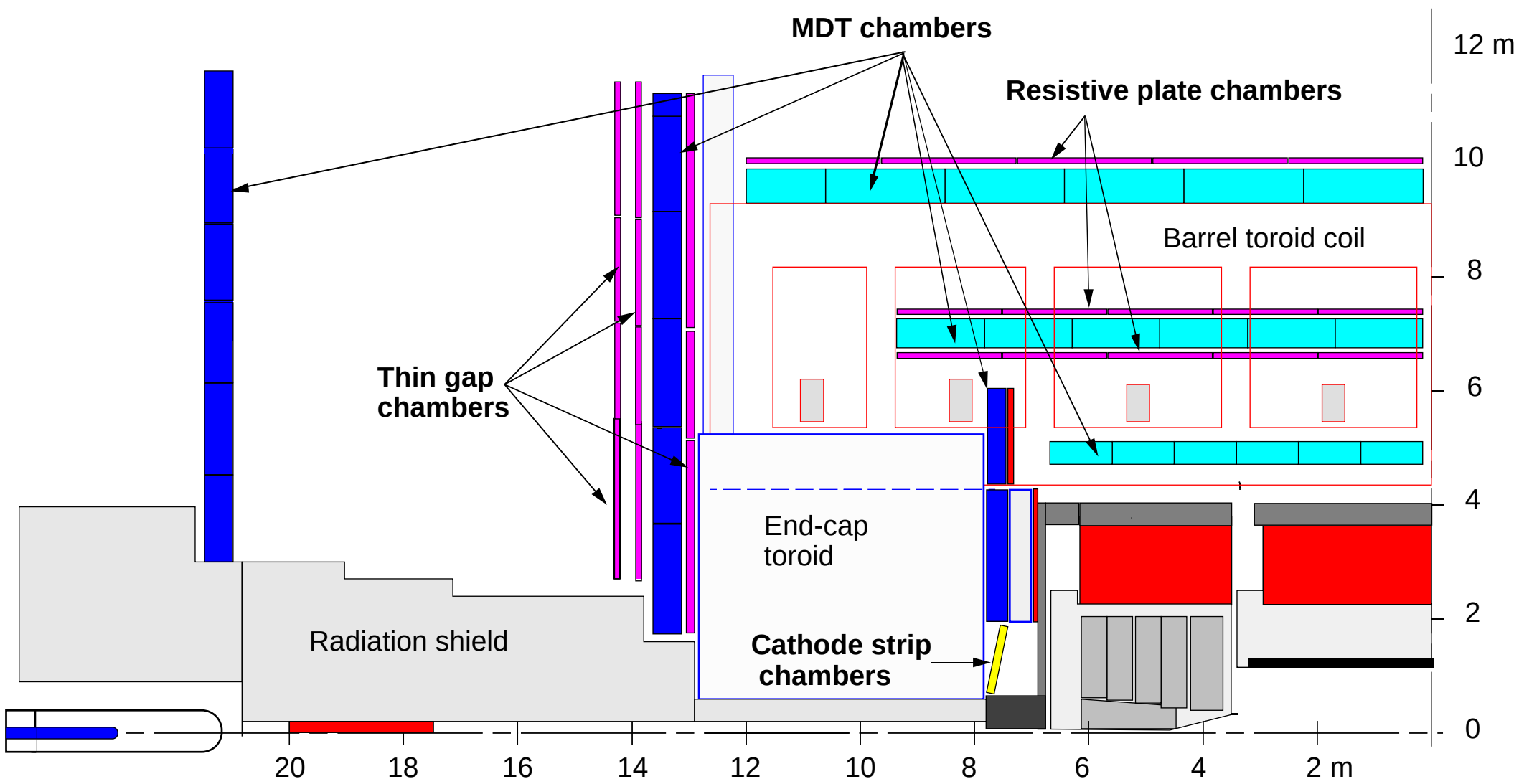

Figure 1: The ATLAS muon spectrometer.

- Resistive plate and thin gap chambers provide similarly rough measurements of both $\eta$ and $\varphi$ near selected stations.

High-$p_T$ muons typically traverse all three stations but there are $\eta$-$\varphi$ regions where one, two or all three stations do not provide a precision measurement, e.g. those regions with support structures or passages for services. There are also regions where overlaps allow two measurements from a single station. Figure 3 shows the number of station measurements as function of $\eta$ and $\varphi$. The resolution and efficiency are degraded where one or more stations do not provide a measurement.

Figure 4 shows how contributions to the muon spectrometer momentum resolution vary as a function of $p_T$. At low momentum, the resolution is dominated by fluctuations in the energy loss of the muons traversing the material in front of the spectrometer. Multiple scattering in the spectrometer plays an important role in the intermediate momentum range. For $p_T > 300$ GeV/c, the single-hit resolution, limited by detector characteristics, alignment and calibration, dominates.

The other ATLAS detector systems also play important roles in achieving the ultimate performance for muon identification and measurement. The calorimeter, with a thickness of more than 10 interaction lengths, provides an effective absorber for hadrons, electrons and photons produced by proton-proton collisions at the center of the ATLAS detector. Energy measurements in the calorimeter can aid in muon identification because of their characteristic minimum ionizing signature and can provide a useful direct measurement of the energy loss [2].

A tracking system inside the calorimeters detects muons and other charged particles with hermetic coverage for $|\eta| < 2.5$, providing important confirmation of muons found by the spectrometer over that $\eta$ range. This *inner detector* has three pixel layers, four stereo silicon microstrip layers, and, for $|\eta| < 2.0$, a straw-tube transition radiation detector that records an average of 36 additional measurements on each track. A 2 Tesla solenoidal magnet enables the inner detector to provide an independent precise momentum measurement for muons (and other charged particles). Over most of the acceptance, for $p_T$ roughly in the range between 30 and 200 GeV/c, the momentum measurements from the inner detector

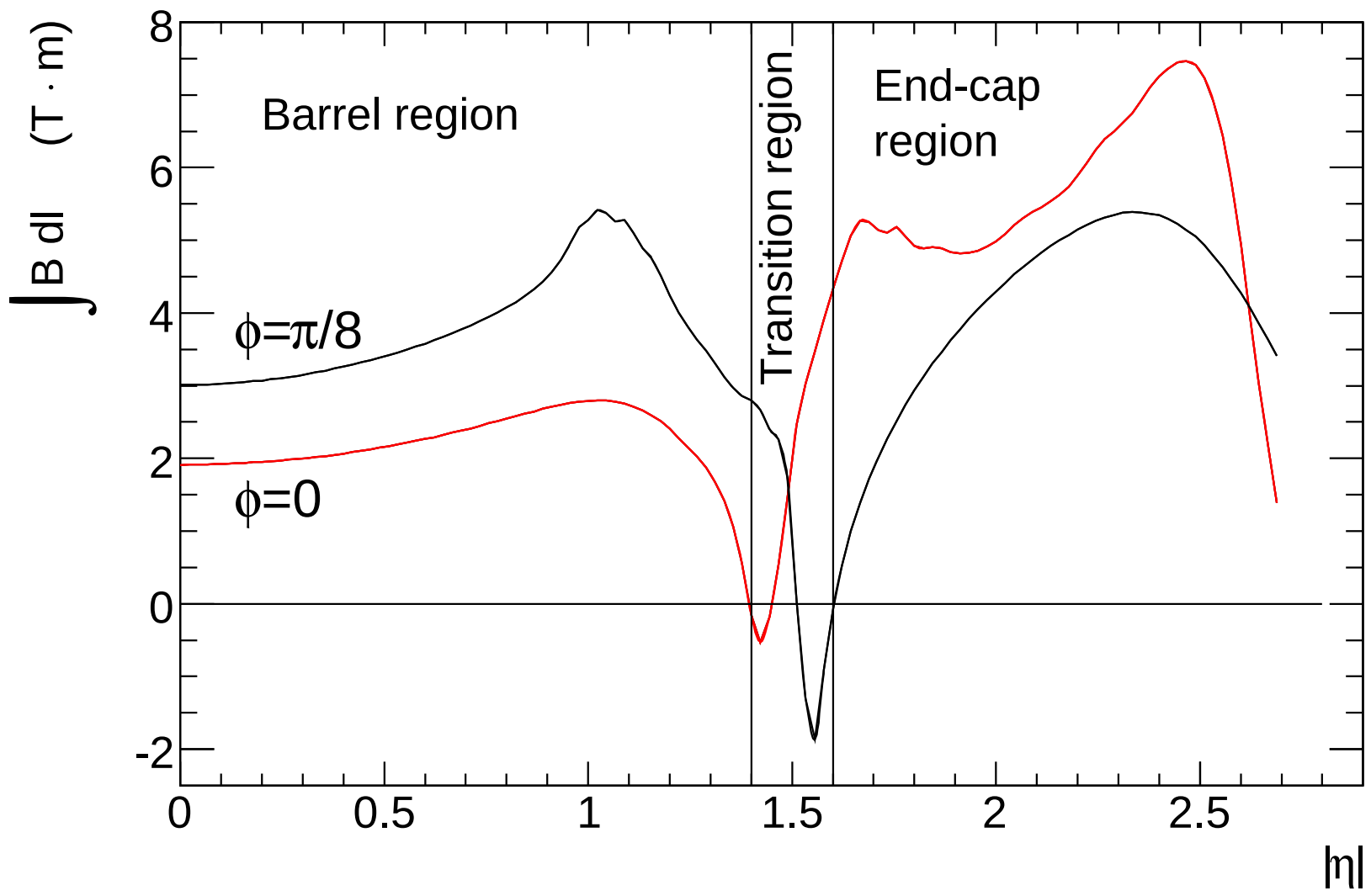

Figure 2: ATLAS muon spectrometer integrated magnetic field strength as a function of $|\eta|$.

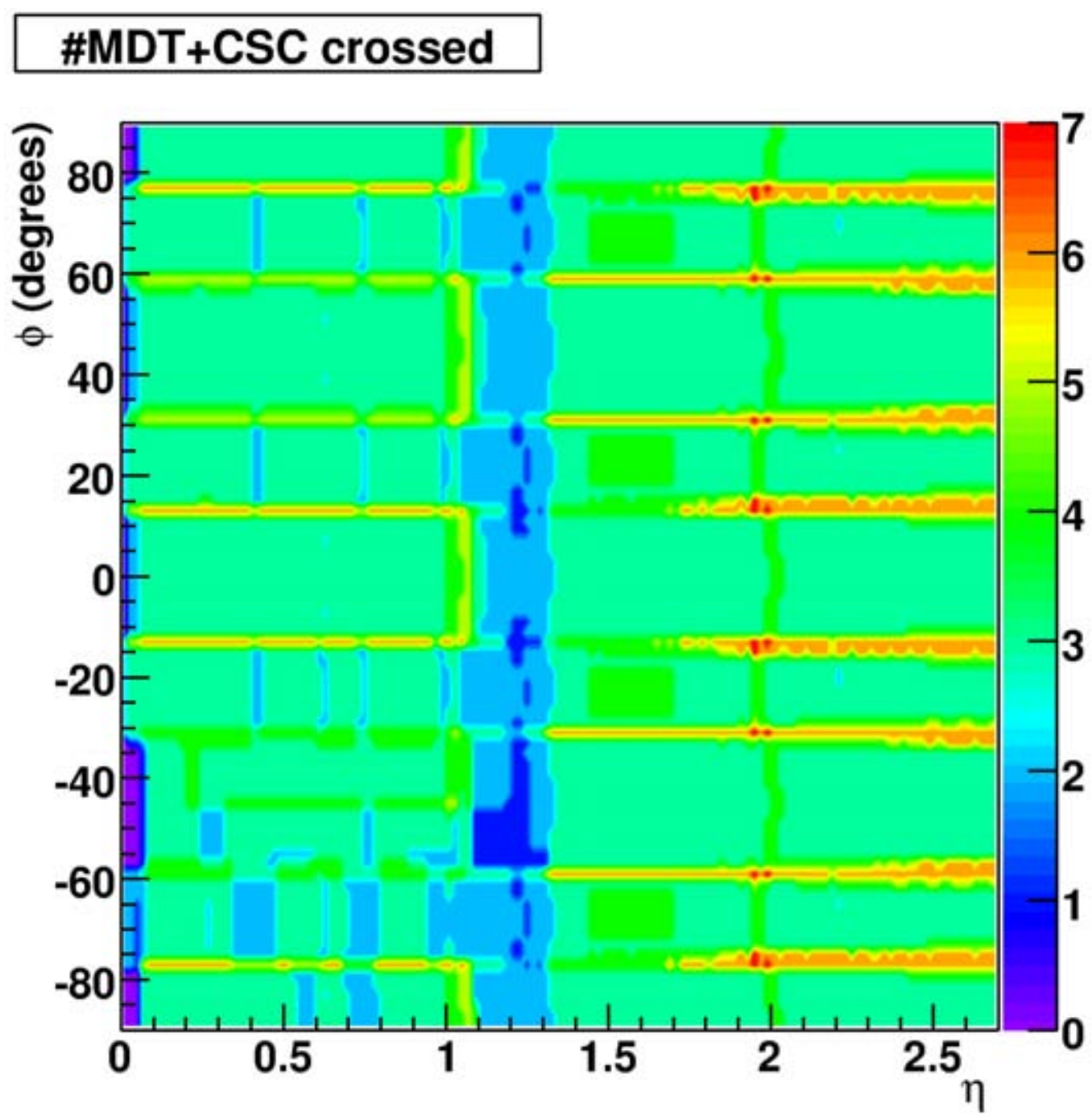

Figure 3: Number of detector stations traversed by muons passing through the muon spectrometer as a function of $|\eta|$ and $\varphi$.

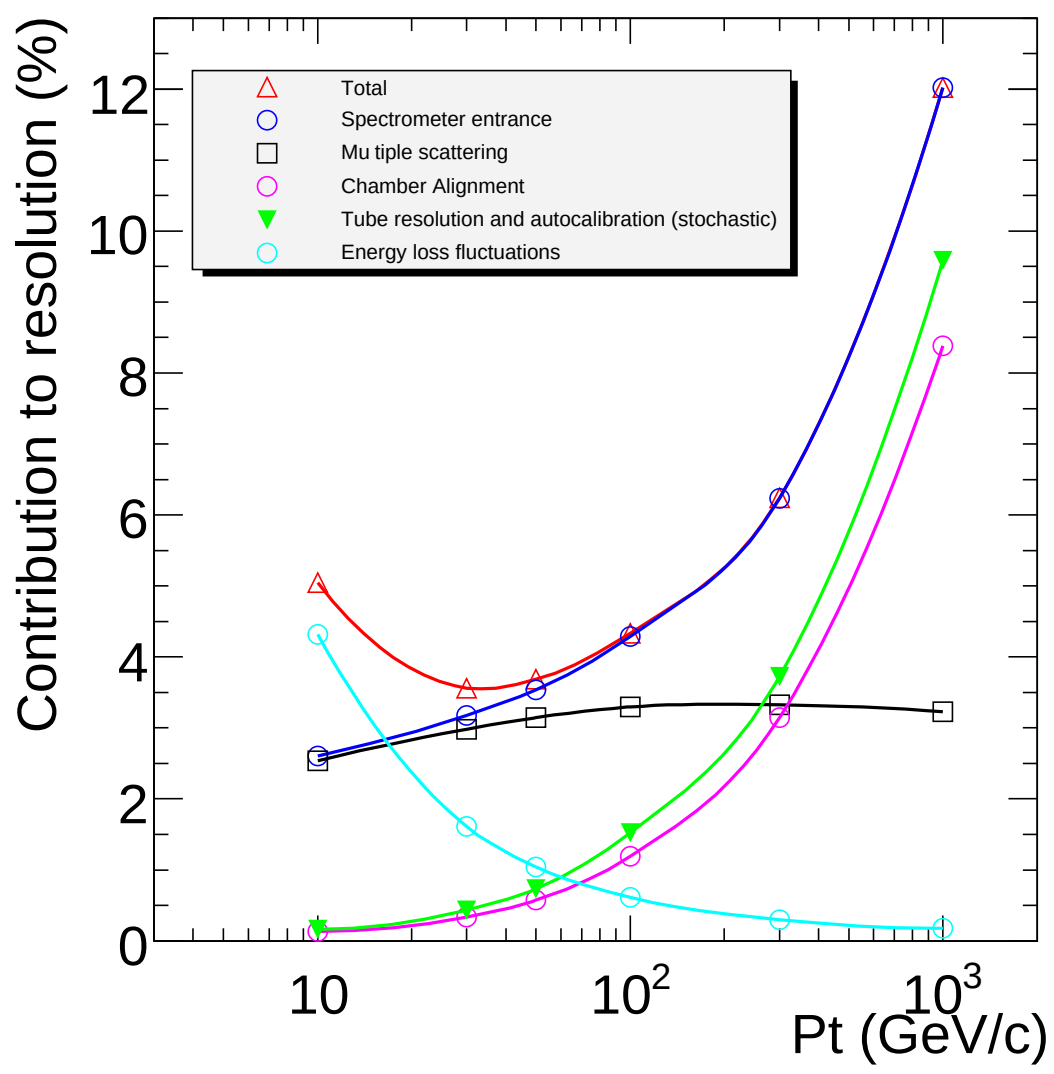

Figure 4: Contributions to the momentum resolution for muons reconstructed in the Muon Spectrometer as a function of transverse momentum for $|\eta| < 1.5$. The alignment curve is for an uncertainty of 30 $\mu$m in the chamber positions.

and muon spectrometer may be combined to give precision better than either alone. The inner detector dominates below this range, and the spectrometer above it.

## 3 Overview of reconstruction and identification algorithms

ATLAS employs a variety of strategies for identifying and reconstructing muons. The direct approach is to reconstruct *standalone* muons by finding tracks in the muon spectrometer and then extrapolating these to the beam line. *Combined* muons are found by matching standalone muons to nearby inner detector tracks and then combining the measurements from the two systems. *Tagged* muons are found by extrapolating inner detector tracks to the spectrometer detectors and searching for nearby hits. Calorimeter tagging algorithms are also being developed to tag inner detector tracks using the presence of a minimum ionizing signal in calorimeter cells. These were not used in the data reconstruction reported here and their performance is documented elsewhere [2].

The current ATLAS baseline reconstruction includes two algorithms for each strategy. Here we briefly describe these algorithms. Later sections describe their performance.

The algorithms are grouped into two families such that each family includes one algorithm for each strategy. The output data intended for use in physics analysis includes two collections of muons—one for each family—in each processed event. We refer to the collections (and families) by the names of the corresponding combined algorithms: Staco [3] and Muid [4]. The Staco collection is the current default for physics analysis.

### 3.1 Standalone muons

The standalone algorithms first build track segments in each of the three muon stations and then link the segments to form tracks. The Staco-family algorithm that finds the spectrometer tracks and extrapolates them to the beam line is called Muonboy [3]. On the Muid side, Moore [5] is used to find the tracks and the first stage of Muid performs the inward extrapolation.

The extrapolation must account for both multiple scattering and energy loss in the calorimeter. Muonboy assigns energy loss based on the material crossed in the calorimeter. Muid additionally makes use of the calorimeter energy measurements if they are significantly larger than the most likely value and the muon appears to be isolated [6].

Standalone algorithms have the advantage of slightly greater $|\eta|$ coverage—out to 2.7 compared to 2.5 for the inner detector—but there are holes in the coverage at $|\eta|$ near 0.0 and 1.2 (see figure 3). Very low momentum muons (around a few GeV/c) may be difficult to reconstruct because they do not penetrate to the outermost stations.

Muons produced in the calorimeter, e.g. from $\pi$ and K decays, are likely to be found in the standalone reconstruction and serve as a background of "fake" muons for most physics analyses. There are a few exotic channels for which charged particles appearing in the calorimeter are a signal of interest.

## 3.2 Inner detector

The primary track reconstruction algorithm for the inner detector is described in Ref. [7]. Space points are identified in the pixel and microstrip detectors, these points are linked to form track seeds in the inner four layers, and tracks are found by extending these seeds to add measurements from the outer layers. This strategy is expected to give very high detection efficiency over the full detector acceptance, $|\eta| < 2.5$.

## 3.3 Combined muons

Both of the muon combination algorithms, Staco and Muid, pair muon-spectrometer tracks with inner-detector tracks to identify combined muons. The match chi-square, defined as the difference between outer and inner track vectors weighted by their combined covariance matrix:

$$\chi^2_{match} = (\mathbf{T_{MS}} - \mathbf{T_{ID}})^{\mathrm{T}} \, (\mathbf{C_{ID}} + \mathbf{C_{MS}})^{-1} \, (\mathbf{T_{MS}} - \mathbf{T_{ID}}) \tag{1}$$

provides an important measure of the quality of this match and is used to decide which pairs are retained. Here **T** denotes a vector of (five) track parameters—expressed at the point of closest approach to the beam line—and **C** is its covariance matrix. The subscript ID refers to the inner detector and MS to the muon spectrometer (after extrapolation accounting for energy loss and multiple scattering in the calorimeter).

Staco does a statistical combination of the inner and outer track vectors to obtain the combined track vector:

$$\mathbf{T} = (\mathbf{C_{ID}^{-1}} + \mathbf{C_{MS}^{-1}})^{-1} \, (\mathbf{C_{ID}^{-1}} \, \mathbf{T_{ID}} + \mathbf{C_{MS}^{-1}} \, \mathbf{T_{MS}}) \tag{2}$$

Muid does a partial refit: it does not directly use the measurements from the inner track, but starts from the inner track vector and covariance matrix and adds the measurements from the outer track. The fit accounts for the material (multiple scattering and energy loss) and magnetic field in the calorimeter and muon spectrometer.

## 3.4 Tagged muons

The spectrometer tagging algorithms, MuTag [3] and MuGirl [8], propagate all inner detector tracks with sufficient momentum out to the first station of the muon spectrometer and search for nearby segments. MuTag defines a tag chi-square using the difference between any nearby segment and its prediction from the extrapolated track. MuGirl uses an artificial neural network to define a discriminant. In either case, if a segment is sufficiently close to the predicted track position, then the inner detector track is tagged as corresponding to a muon.

At present, both algorithms simply use the inner-detector track to evaluate the muon kinematics, i.e. the inner track and spectrometer hits are not combined to form a new track. This is not very important in the low-$p_T$ regime that these algorithms were originally intended to address. Both algorithms are being further developed to allow extrapolation to other and multiple stations and add the capability to include the spectrometer measurements in a track refit.

There is an important difference in the way these algorithms are run in the standard reconstruction chain. MuGirl considers all inner-detector tracks and redoes segment finding in the region around the track. MuTag only makes use of inner-detector tracks and muon-spectrometer segments not used by Staco. Thus MuTag serves only to supplement Staco while MuGirl attempts to find all muons. Obviously, MuTag is part of the Staco family and most sensibly used in that context. MuGirl muons are recorded as part of the Muid family.

### 3.5 Merging muons

The muon finding efficiency (and fake rate) may be increased by including muons found by multiple algorithms but care must be taken to remove overlaps, i.e. cases where the same muon is identified by two or more algorithms. To a large extent, this is done when the collections are created. Standalone muons that are successfully combined are not recorded separately. In those cases where a standalone muon is combined with more than one inner-detector track, exactly one of the muons is flagged as "best match." In the Staco collection, the tagged and combined muons do not overlap by construction. In the Muid collection, overlaps between MuGirl and Muid muons are removed by creating a single muon when both have the same inner detector track.

Analysts wishing to merge standalone and tagged muons or muons from different collections may make use of a muon selection tool to remove overlaps. It requires muons have different inner-detector tracks and merges standalone muons that are too close to one another. Closeness is defined by $\eta$-$\varphi$ separation with a default limit of 0.4.

## 4 Tools for performance evaluation and classification of tracks

Simulation samples were created in the ATLAS framework by running an event generator (PYTHIA [9] or MC@NLO [10, 11]) and using GEANT4 [12] to propagate the final-state particles using ATLAS-specific code to describe the geometry and response of the detector. The data were then reconstructed using the software based on the algorithms described in the previous chapter.

### 4.1 Truth matching and track classification

Muon reconstruction performance is evaluated for each event by comparing selected reconstructed muons with the true muons, i.e. those in the Monte Carlo truth record. The latter include muons created in the initial event generation as well as secondaries produced during propagation through the tracking volume. Muons produced in the calorimeter or muon spectrometer are not included in the truth record. True muons with transverse momentum below 2 GeV/c are also excluded to avoid spurious matches with candidates we do not expect to be able to reconstruct.

For each event, a one-to-one matching is performed between the selected reconstructed muons and the true muons. The matching makes use of two distance metrics: $D_{ref}$ is the *reference* distance measured from true muon to the reconstructed muon:

$$D_{ref} = \sqrt{\left(\frac{\varphi_{reco} - \varphi_{true}}{0.005}\right)^2 + \left(\frac{\eta_{reco} - \eta_{true}}{0.005}\right)^2 + \left(\frac{\Delta p_T/p_T}{0.03}\right)^2} \tag{3}$$

and $D_{eva}$ is the *evaluation* distance measured from the reconstructed muon to the true muon:

$$D_{eva} = \sqrt{(\mathbf{T_{reco}} - \mathbf{T_{true}})\,\mathbf{C_{reco}^{-1}}\,(\mathbf{T_{reco}} - \mathbf{T_{true}})} \tag{4}$$

In the first equation, $\Delta p_T/p_T$ is the fractional momentum resolution:

$$\frac{\Delta p_T}{p_T} = \frac{1/p_{Treco} - 1/p_{Ttrue}}{1/p_{Ttrue}} = \frac{p_{Ttrue} - p_{Treco}}{p_{Treco}} \tag{5}$$

Here $p_T$ is signed (i.e. carries the charge sign), but elsewhere in the text it denotes the magnitude of the transverse momentum. In the second distance equation, **T** again denotes the vector of (five) track parameters (expressed at the distance of closest approach to the beam line) and **C** the associated covariance matrix. Note that $D_{eva}^2$ is a chi-square with five degrees of freedom.

There is a maximum allowed value for each of these distances. For $D_{eva}$ the maximum value is 1000, a very loose cut. The limit for $D_{ref}$ is 100 and we see from equation 3 this implies the matched muons must be within a distance of 0.5 in $\eta$ and $\varphi$ and have the same charge sign with $p_{Treco} > 0.25\,p_{Ttrue}$ or opposite sign with $p_{Treco} > 0.50\,p_{Ttrue}$.

The matching is carried out by first examining each reconstructed muon and assigning it to the nearest true muon using the evaluation distance. The reconstructed muon is left unmatched if no distance is less than the maximum allowed value. The reference distance is evaluated for each match and the match is discarded if it exceeds the threshold for that quantity. If more than one match remains for any true muon, then only the match with the smallest reference distance is retained.

True muons that are matched are said to be *found* and those left unmatched are *lost*. Found muons are classified as *good* if they have $D_{eva} < 4.5$ corresponding to a chi-square probability above 0.0011.

Reconstructed muons are said to be *real* if they are matched and *fake* if unmatched. Note that these fakes may correspond to true muons produced outside the tracking volume (e.g. in the calorimeter) and hence not included in the truth record.

### 4.2 Performance measures

Our performance measures include efficiency, fake rate, resolutions and resolution tails. The efficiency or *finding efficiency* is defined to be the fraction of true muons that are found and is typically evaluated for some kinematic selection (applied after matching). The *good efficiency* is the fraction of true muons that are found and classified as good (as defined in the previous section). The *good fraction* is the fraction of found muons that are classified as good. In the sections that follow, we present the overall efficiency for various physics samples and the efficiency as a function of $\eta$ for the primary benchmark sample.

The fake rate is defined to be the mean number of fake muons per event and it is presented for a variety of $p_T$ thresholds corresponding to the values that might be chosen for different physics analyses.

Five kinematic variables characterize a track, but here we examine only the measurement of the transverse momentum. The precision and accuracy of the direction measurements are typically much better than that required for any physics analysis. The measurement of the initial position of the track (e.g. at the distance of closest approach to the beam line or vertex) is discussed in another note [13]. For the momentum, we use the fractional residual, $\Delta p_T/p_T$, defined in equation 5. This distribution is fitted with a Gaussian and the resolution is defined to be the sigma of this fit. The tails in the distributions are often more important than the core resolution and we characterize these by evaluating the fraction of found muons in five tail categories. The first three are those for which the magnitude of this residual exceeds 5%, 10% or 30%. The last category is the fraction for which the charge sign is incorrectly measured. Finally there is an intermediate category in which either the sign is incorrect or the magnitude of the measured momentum is more than two times larger than the true value.

### 4.3 Monte Carlo samples

Our primary benchmark sample is a collection of $t\bar{t}$ events requiring the presence of at least one lepton (electron, muon or tau). The initial inclusive sample was produced using MC@NLO in conjunction with Herwig [14]. This sample provides a variety of mechanisms for producing muons and we present results for two: *direct* muons which do not have any quarks in their ancestry and *indirect* muons whose ancestry includes a heavy quark (b or c) but not a tau. In this sample, the former are produced directly in the leptonic decay of a W-boson.

Performance metrics are plotted as a function of $\eta$ for $t\bar{t}$ direct muons. In addition, we tabulate efficiencies and fake rates for these muons, for $t\bar{t}$ indirect muons, and for muons in separate low- and high-$p_T$ samples. The low-$p_T$ sample is taken from direct PYTHIA $J/\psi$ production with the $J/\psi$ forced to decay to two muons and a filter selecting only those events where both muons have $|\eta| < 2.5$ and $p_T > 4$ GeV/c. Muons produced by other processes in these events are suppressed by restricting the analyzed sample to muons that have a c-quark in their ancestry. The high-$p_T$ sample consists of direct muons in PYTHIA production of $Z' \to \mu\mu$ with a $Z'$ mass of 2 TeV. The generation also includes $Z/\gamma$ and interference but a dimuon mass cut ($m_{\mu\mu} > 500$ GeV/c) ensures that the average muon $p_T$ is above 500 GeV/c.

At design luminosity, ATLAS will have many interactions in each beam crossing (pileup) and there will be significant background in the muon chambers from low-energy photons and neutrons (cavern background). To get an estimate of the effect this will have on our reconstruction algorithms, we processed a $t\bar{t}$ sample overlaid with the backgrounds expected for a reference luminosity of $10^{33}$ cm$^{-2}$s$^{-1}$. The cavern background was included with a safety factor of 2.0, i.e. at twice the value expected for this luminosity. In the following, this sample is called the high-luminosity $t\bar{t}$ sample. Low luminosity refers to samples without any pileup or cavern background.

There is considerable uncertainty in the estimate of the cavern background and active development is underway to improve reconstruction in this environment, and so the results presented here provide only a rough indication of the performance we expect at high luminosity.

Figure 5 shows the $p_T$, $\eta$ and isolation energy distributions for the true muons in the samples studied in this note. The isolation energy was obtained by summing the calorimeter transverse energy in an $\eta$-$\varphi$ cone of radius 0.2 about the muon. The most probable value for the muon energy loss (as discussed in reference [2]) is subtracted from these values.

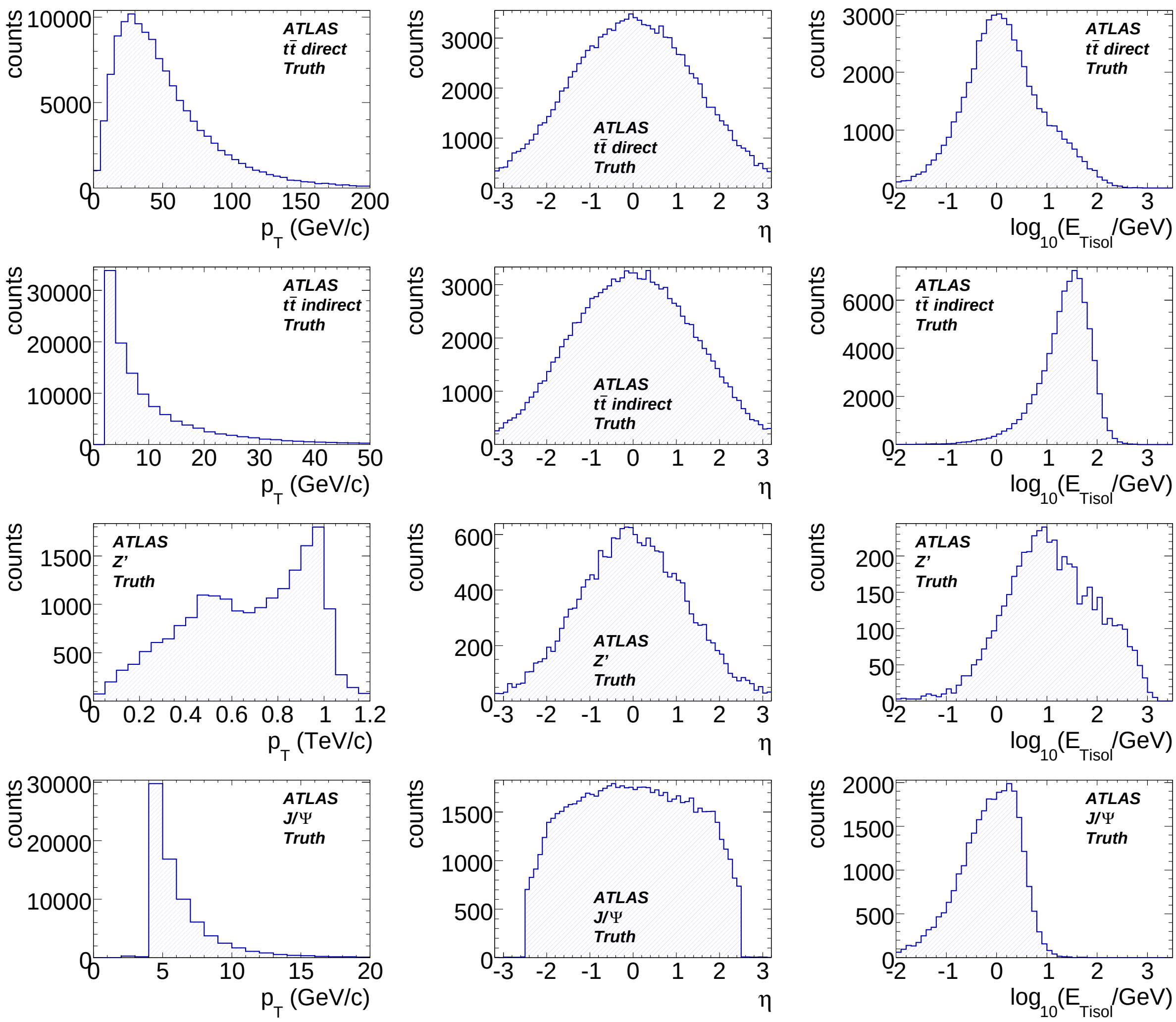

Figure 5: True $p_T$ (left), $\eta$ (center) and isolation (right) distributions for the $t\bar{t}$ direct muons (top), $t\bar{t}$ indirect muons (second from top), $Z'$ (mass 2 TeV) direct muons (third from top) and $J/\psi$ muons (bottom). Note that the $p_T$ range is different in each of the plots of that variable.

# 5 Standalone muon performance

## 5.1 Efficiencies and fake rates

Figure 6 shows the standalone $t\bar{t}$ direct muon efficiencies and fake rates as functions of $\eta$ at low luminosity (i.e. without any pileup or cavern background) and at our reference luminosity ($10^{33}$ /cm$^2$/sec with cavern background safety factor 2.0). Table 1 gives the integrated efficiencies and fake rates for these and other samples.

| | Efficiency | | Fakes/(1000 events) above $p_T$ limit (GeV/c) | | | |
|---|---|---|---|---|---|---|
| Sample | found | good | 3 | 10 | 20 | 50 |
| Muonboy | | | | | | |
| $t\bar{t}$ direct<br>$t\bar{t}$ indirect | 0.951 (1)<br>0.949 (1) | 0.812 (1)<br>0.783 (2) | 24.0 (3) | 4.4 (1) | 1.69 (7) | 0.52 (4) |
| hi-$\mathscr{L}$ $t\bar{t}$ direct | 0.950 (2) | 0.809 (3) | 53 (1) | 8.2 (4) | 3.9 (2) | 1.9 (2) |
| $Z'$ direct | 0.914 (2) | 0.781 (3) | 141 (4) | 79 (3) | 61 (3) | 37 (2) |
| $J/\psi$ | 0.959 (3) | 0.764 (6) | 51 (1) | 5.0 (4) | 1.6 (2) | 0.6 (1) |
| Moore/Muid | | | | | | |
| $t\bar{t}$ direct<br>$t\bar{t}$ indirect | 0.943 (1)<br>0.920 (2) | 0.861 (1)<br>0.838 (2) | 19.8 (3) | 3.9 (1) | 1.44 (6) | 0.47 (4) |
| hi-$\mathscr{L}$ $t\bar{t}$ direct | 0.932 (2) | 0.836 (3) | 984 (4) | 301 (2) | 156 (2) | 61 (1) |
| $Z'$ direct | 0.887 (2) | 0.769 (3) | 168 (4) | 102 (3) | 75 (3) | 43 (2) |
| $J/\psi$ | 0.830 (5) | 0.723 (6) | 6.7 (4) | 1.1 (2) | 0.5 (1) | 0.13 (6) |

Table 1: Muonboy and Moore/Muid efficiencies and fake rates for various samples (section 4.3). Efficiencies are presented both for all found muons and for those with a good truth match ($D_{eva} < 4.5$). Both are calculated for true muons with $|\eta| < 2.5$ and $p_T > 10$ GeV/c. Fake rates are presented for a variety of $p_T$ thresholds.

Comparing with Figure 3, we see most of the efficiency loss occurs in regions where the detector coverage is poor, i.e. for $|\eta|$ around 0.0 and 1.2. Otherwise, the $t\bar{t}$ muon efficiency is close to 100% for Muonboy and around 99% for Moore/Muid. The Muid good fraction is significantly higher than for Muonboy, presumably because of better handling of the material in the calorimeter. The algorithms have similar fake rates at low luminosity. At the higher luminosity, the Staco rate increases significantly (by a factor of 2-4) while the Moore/Muid rate increases dramatically (factor of 100). In the high-$p_T$ $Z'$, the efficiency falls by a few percent for both algorithms. For the low-$p_T$ (and non-isolated) $J/\psi$ muons, the Moore/Muid efficiency degrades significantly while Muonboy remains high.

## 5.2 Resolution

Figure 7 shows the $p_T$ resolutions and tails as functions of $\eta$ and $p_T$. The resolution is degraded at intermediate pseudorapidity ($1.2 < |\eta| < 1.7$) because of the reduced number of measurements (figure 3), the low field integral in the overlap between barrel and endcap toroids (figure 2), and the material in the endcap toroid (figure 1). The average resolution is very similar for the two algorithms. Despite having a lower good fraction, Muonboy has fewer muons for which the charge sign is incorrectly measured. This suggests that, at least in the tails, Moore/Muid provides a better estimate of the momentum error while Muonboy provides a better estimate of its value. The Moore/Muid tails are likely due to the assignment of incorrect hits to spectrometer tracks.

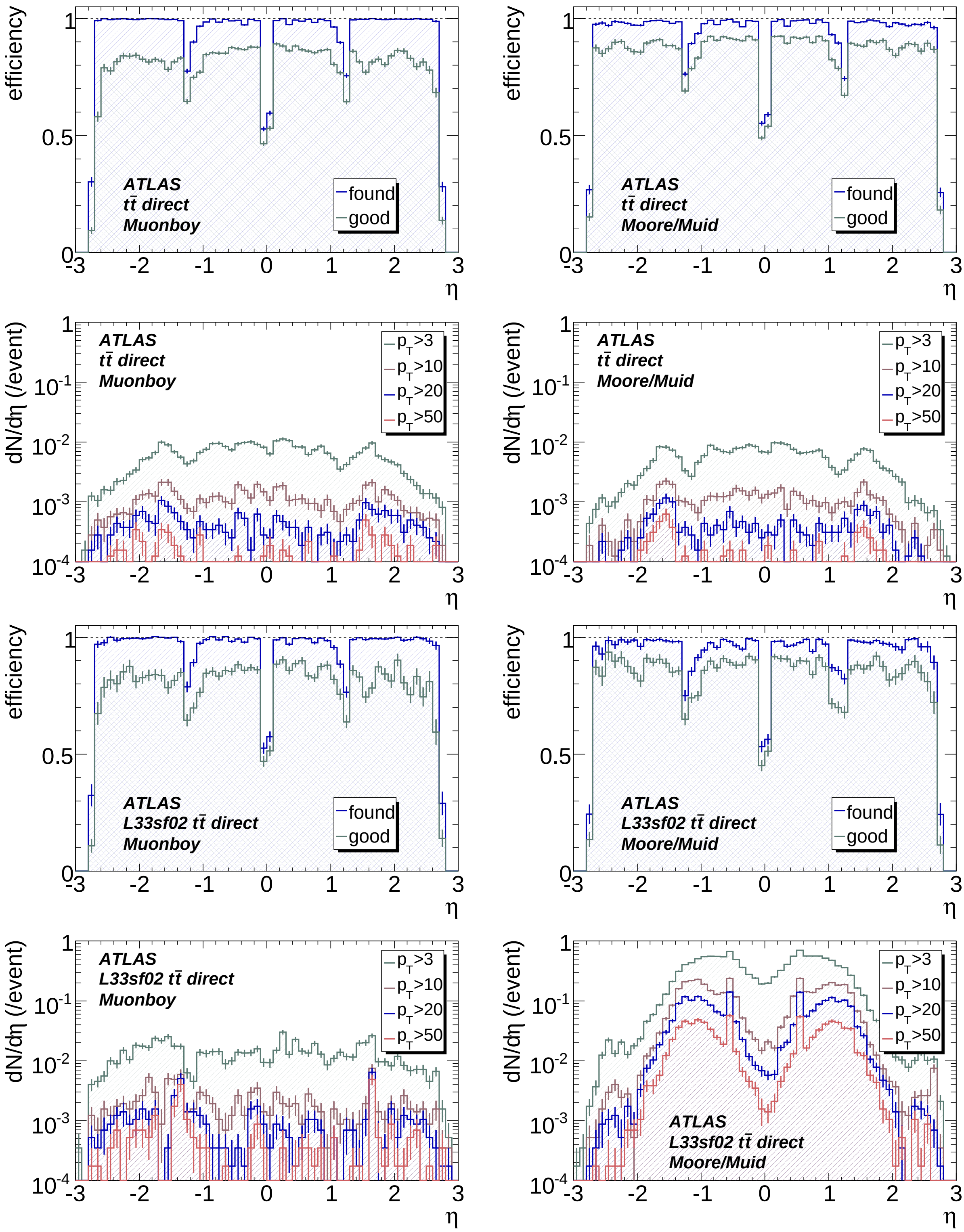

Figure 6: Standalone efficiency and fake rate as functions of true $\eta$ for Muonboy (left) and Moore/Muid (right) for direct muons in $t\bar{t}$ at low (top) and high (bottom) luminosity. In the efficiency plots, the upper curve (blue) is the efficiency to find the muon while the lower curve (green) additionally requires a good match ($D_{eva} < 4.5$) between reconstructed and true track parameters. Fake rates are shown for a variety of $p_T$ thresholds.

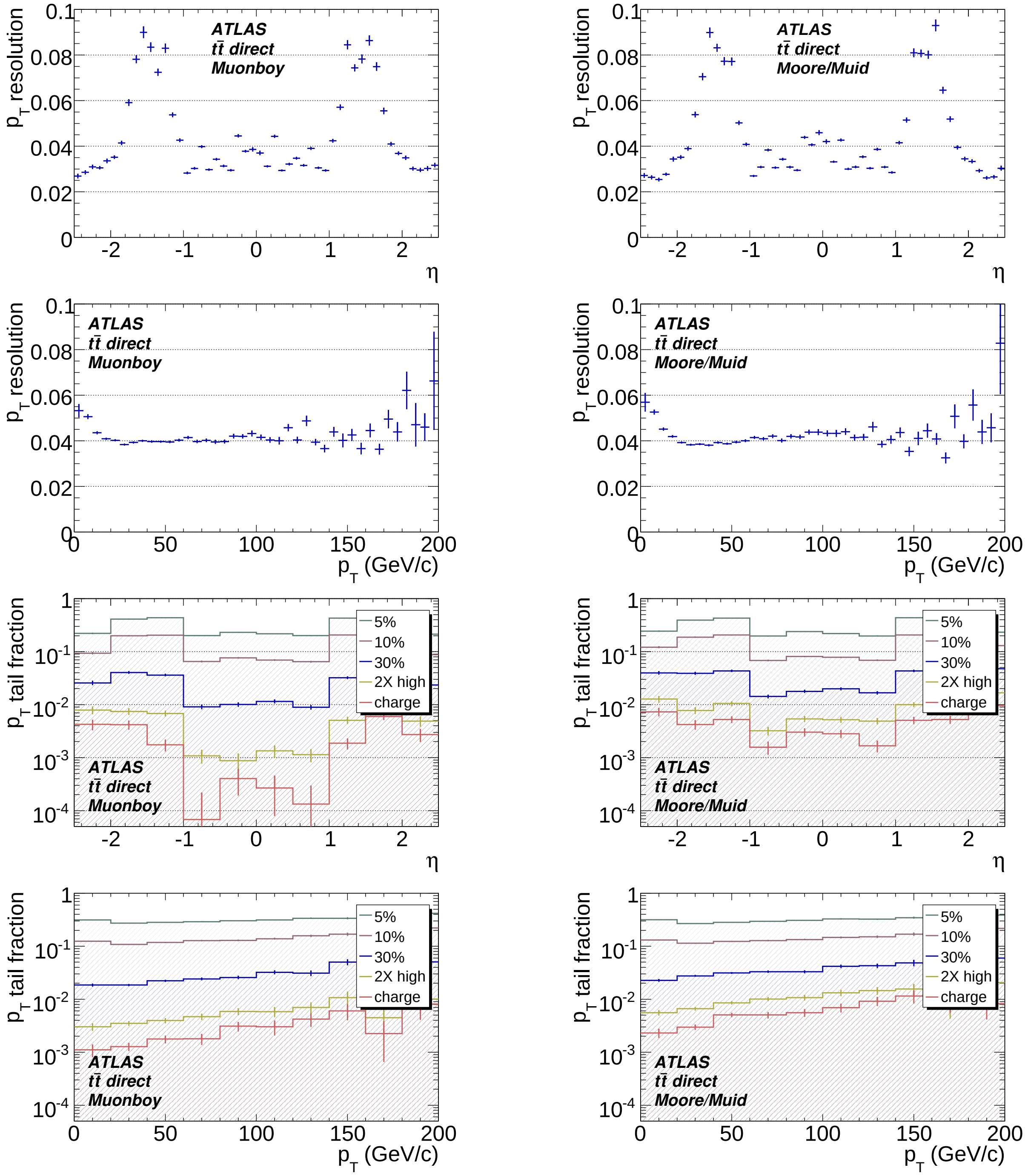

Figure 7: Standalone fractional momentum resolution ($\Delta p_T/p_T$) as function of $\eta$ (top) and $p_T$ (2nd row) and tails in that parameter also as functions of $\eta$ (3rd row) and $p_T$ (bottom). All are for both Muonboy (left) and Moore/Muid (right). The tail is the fraction of reconstructed muons with magnitude of $\Delta p_T/p_T$ outside a range and is shown for a wide range of values. The last tail curve (red, “charge”) includes only muons reconstructed with the wrong charge sign. The 4th tail curve (yellow, “2X high”) includes these and those with momentum magnitude more than two times the true value.

# 6 Inner detector performance

Figure 8 shows the efficiency for $t\bar{t}$ direct muons and Table 2 gives the integrated efficiencies for all of the samples. The efficiency is high for all $\eta$ (within the acceptance) and all samples. There is no evidence of degradation when pileup is added.

The inner detector momentum resolution is the same as that for tagged muons, reported later: see Figure 13 in section 8.

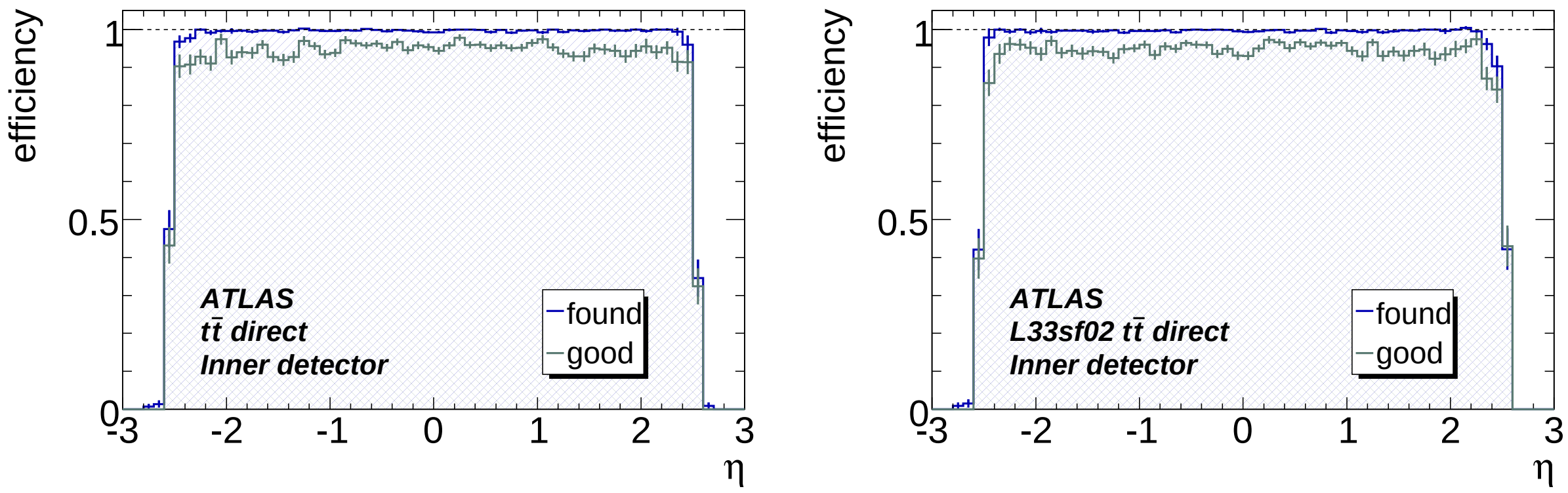

Figure 8: Inner detector $t\bar{t}$ direct muon efficiency as a function of true $\eta$ at low (left) and high (right) luminosity. In each figure, the upper curve (blue) is the efficiency to find the muon while the lower curve (green) additionally requires a good match ($D_{eva} < 4.5$) between reconstructed and true track parameters. The efficiency is for $p_T > 10$ GeV/c.

| | Efficiency | |
|---|---|---|
| Sample | found | good |
| $t\bar{t}$ direct | 0.996 (1) | 0.950 (2) |
| $t\bar{t}$ indirect | 0.997 (1) | 0.833 (5) |
| hi-$\mathscr{L}$ $t\bar{t}$ direct | 0.995 (1) | 0.947 (2) |
| Zprime direct | 0.993 (1) | 0.966 (1) |
| $J/\psi$ | 0.995 (1) | 0.941 (3) |

Table 2: Inner detector efficiencies. The samples and algorithms are described in the text. Efficiencies are presented both for all found muons and for those with a good truth match ($D_{eva} < 4.5$). Efficiencies are calculated for true muons with $|\eta| < 2.5$ and $p_T > 10$ GeV/c.

# 7 Combined muon performance

## 7.1 Efficiencies and fake rates

Figure 9 shows the combined $t\bar{t}$ direct muon efficiency and fake rates for each algorithm as a function of $\eta$ for both low and high luminosity. Compared with the performance for standalone muons (figure 6), Staco shows a small drop in efficiency with little reduction of the fake rate except for the lowest $p_T$ threshold at high luminosity. In fact, the high-$p_T$ fake rates increase at either luminosity because low-$p_T$ standalone muons are matched to high-$p_T$ inner-detector tracks. At low luminosity, Muid $t\bar{t}$ shows a small decrease in both efficiency and fake rate. When background is added, the dramatic increase in fakes for Moore standalone is not observed in Muid combined, i.e. the matching suppresses most of the fakes and the Muid high-$p_T$ fake rates are lower than those of Staco. However, the high-luminosity $t\bar{t}$ Muid efficiency is significantly worse than that of Staco.

When matching inner detector and muon spectrometer tracks, both Staco and Muid calculate $\chi^2_{match}$ (section 3.3) which serves as a discriminant for separating real and fake muons. The fakes include pion or kaon decays in or near the calorimeter. Figure 10 shows the $\chi^2_{match}$ distributions for both direct found muons and fakes. We see that with a cut on this quantity, e.g. $\chi^2_{match} < 100$, many of the Staco high-$p_T$ fakes can be suppressed with only a modest loss in efficiency. The higher Staco fake rates come from looser cuts during reconstruction and, if the $\chi^2_{match}$ cuts are adjusted to give the same efficiencies, the Staco fake rate is lower.

Table 3 shows the integrated Staco and Muid muon efficiencies and fake rates for all samples including an entry showing the effect of the above cut on $\chi^2_{match}$.

| | Efficiency | | Fakes/(1000 events) above $p_T$ limit (GeV/c) | | | |
|---|---|---|---|---|---|---|
| Sample | found | good | 3 | 10 | 20 | 50 |
| Staco | | | | | | |
| $t\bar{t}$ direct<br>$t\bar{t}$ indirect | 0.943 (1)<br>0.933 (1) | 0.875 (1)<br>0.767 (2) | 22.0 (3) | 9.6 (2) | 3.4 (1) | 0.62 (4) |
| $t\bar{t}$ direct cut | 0.924 (1) | 0.865 (1) | 14.8 (2) | 3.1 (1) | 0.39 (3) | 0.01 (1) |
| hi-$\mathscr{L}$ $t\bar{t}$ direct | 0.941 (2) | 0.871 (3) | 25.9 (7) | 11.2 (4) | 4.3 (3) | 0.7 (1) |
| $Z'$ | 0.910 (2) | 0.824 (3) | 14 (1) | 8.4 (9) | 5.2 (7) | 3.4 (6) |
| $J/\psi$ | 0.943 (3) | 0.873 (4) | 0.9 (2) | 0.24 (8) | 0.11 (5) | 0.0 (0) |
| Muid | | | | | | |
| $t\bar{t}$ direct<br>$t\bar{t}$ indirect | 0.926 (1)<br>0.888 (2) | 0.877 (1)<br>0.748 (3) | 15.4 (2) | 2.36 (9) | 0.48 (4) | 0.05 (1) |
| $t\bar{t}$ direct cut | 0.917 (1) | 0.871 (1) | 14.0 (2) | 1.96 (8) | 0.33 (3) | 0.03 (1) |
| hi-$\mathscr{L}$ $t\bar{t}$ direct | 0.904 (2) | 0.854 (3) | 35.5 (8) | 5.0 (3) | 1.1 (1) | 0.24 (6) |
| $Z'$ direct | 0.872 (2) | 0.811 (3) | 11 (1) | 4.5 (7) | 3.1 (6) | 2.7 (5) |
| $J/\psi$ | 0.793 (5) | 0.741 (6) | 0.8 (1) | 0.03 (3) | 0.0 (0) | 0.0 (0) |

Table 3: Staco and Muid efficiencies and fake rates. The samples and algorithms are described in the text. Algorithm names are followed by “cut” to indicate that reconstructed muons are required to have $\chi^2_{match} < 100$ for both efficiency and fake calculations. Efficiencies are presented both for all found muons and for those with a good truth match. Both are calculated for true muons with $|\eta| < 2.5$ and $p_T > 10$ GeV/c. The fake rates are presented for a variety of $p_T$ thresholds.

## 7.2 Resolution

Figure 11 shows the $t\bar{t}$ direct muon $p_T$ resolutions and tails as functions of $\eta$ and $p_T$. Comparing with the same for standalone reconstruction (figure 7), we see, as expected, the combined resolution is significantly better especially in the overlap region ($|\eta|$ around 1.5) and for $p_T$ below 100 GeV/c. There are also significant reductions in the tails of momentum residuals. Misreconstruction and charge misidentification rates are around 0.01% for the combined muons instead of 0.1% for the standalone.

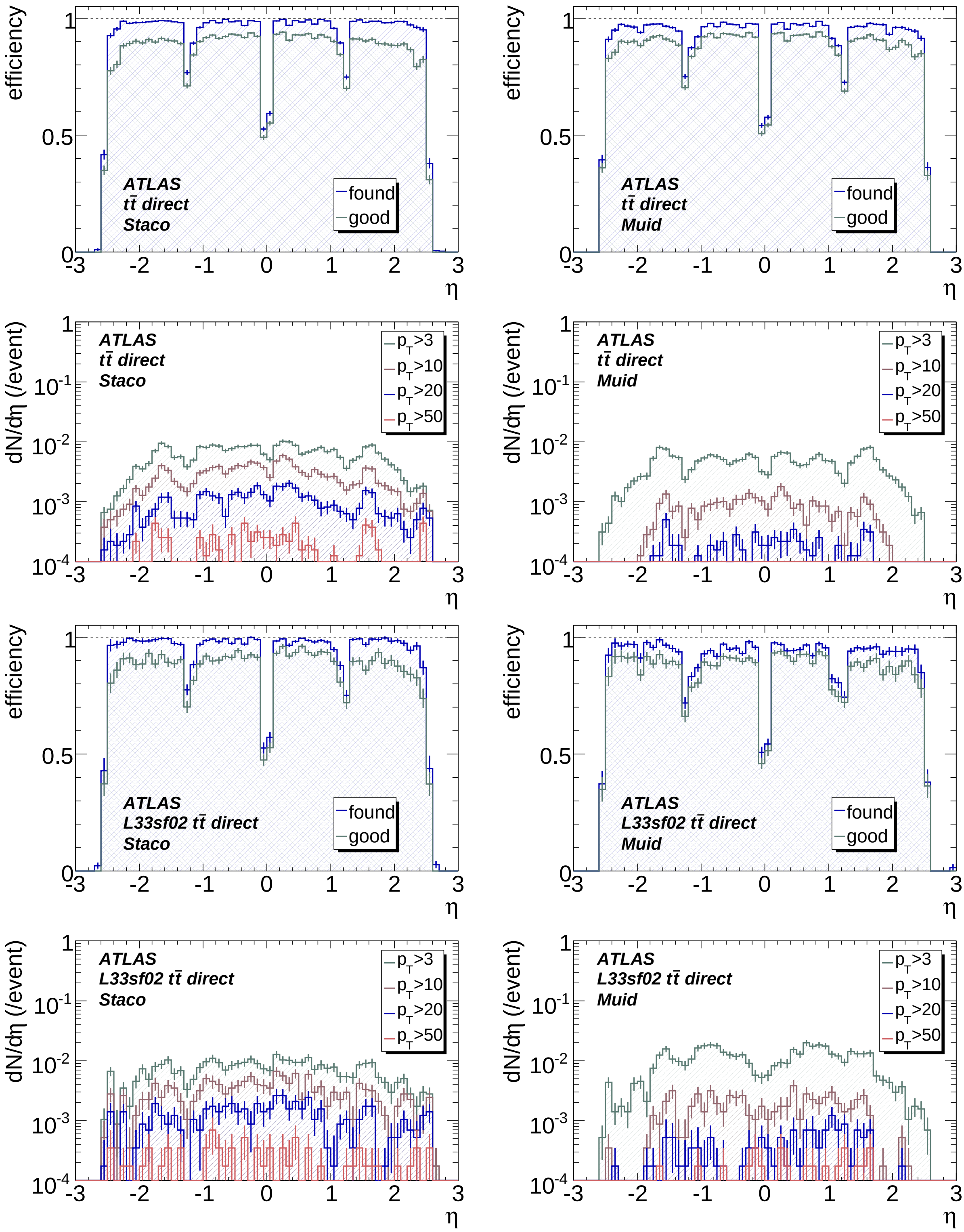

Figure 9: Combined muon efficiency and fake rate for Staco (left) and Muid (right) as functions of true $\eta$ for direct muons in $t\bar{t}$ at low (top) and high (bottom) luminosity. In each efficiency plot, the upper curve (blue) is the efficiency to find the muon while the lower curve (green) additionally requires a good match ($D_{eva} < 4.5$) between reconstructed and true track parameters. The efficiencies are for $p_T > 10$ GeV$/c$. Fake rates are shown for a variety of $p_T$ thresholds.

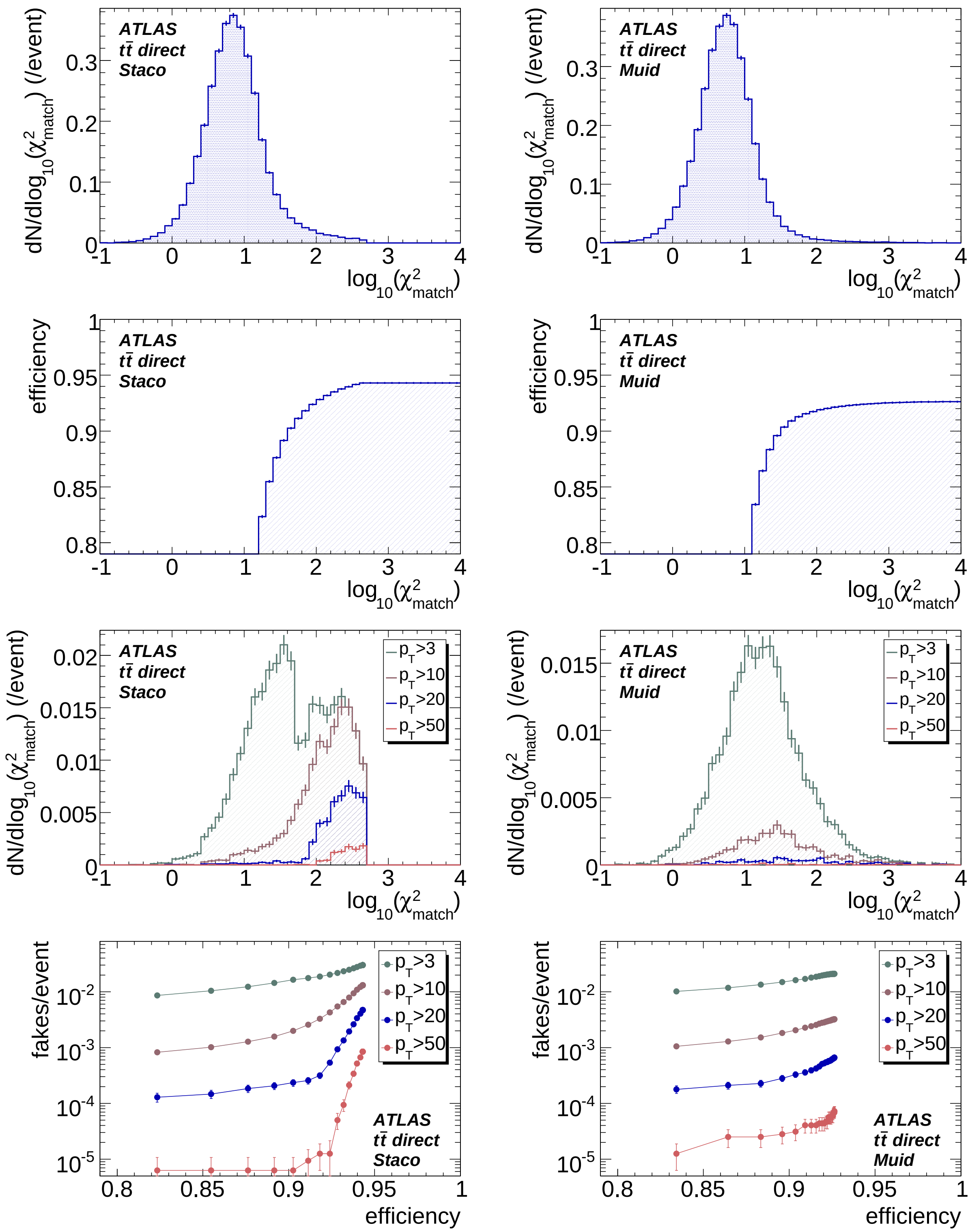

Figure 10: Distributions of $\chi^2_{match}$ for direct muons (top) and fakes (third from top). The fakes are shown for a variety of $p_T$ thresholds. The second row shows the efficiency as function of $\chi^2_{match}$ when muons above that value are rejected. The bottom row shows the fake rates as a function of efficiency as that threshold is varied. All are shown for both Staco (left) and Muid (right). The sharp drops in the Staco $\chi^2_{match}$ distribution come from cuts on that quantity made during reconstruction, i.e. before filling the output muon collection.

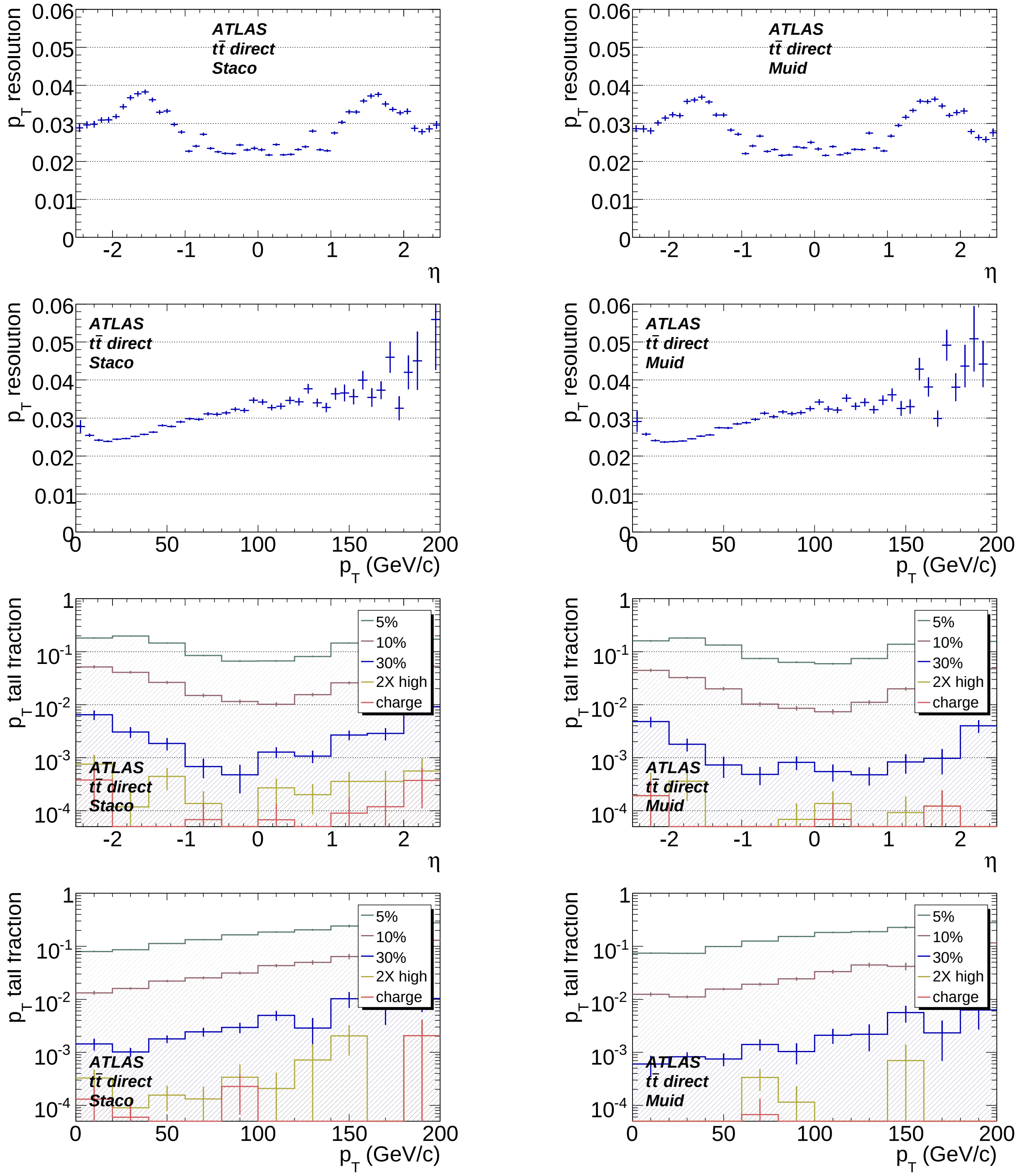

Figure 11: Combined muon fractional momentum resolution ($\Delta p_T/p_T$) as function of $\eta$ (top) and $p_T$ (2nd row) and tails in that parameter also as functions of $\eta$ (3rd row) and $p_T$ (bottom). All are for both Staco (left) and Muid (right). The tail is the fraction of reconstructed muons with magnitude of $\Delta p_T/p_T$ outside a range and is shown for a wide range of values. The last tail curve (red, "charge") includes only muons reconstructed with the wrong charge sign. The 4th tail curve (yellow, "2X high") includes these and those with momentum magnitude more than two times the true value.

# 8 Tagged muon performance

## 8.1 Efficiencies and fake rates

ATLAS runs two tagging algorithms but only MuGirl attempts to find all muons. MuTag is run in a manner to complement Staco and the performance of the combination of these two is reported in the following section.

Figure 12 shows the MuGirl direct muon efficiency and fake rates as a function of $\eta$ in $t\bar{t}$ at low and high luminosity. Table 4 gives the MuGirl integrated efficiencies and fake rates for all our samples.

Comparing with the combined muon results (figure 9 and table 3), we see that MuGirl has lower efficiency and a substantially higher fake rate. We also observe that its performance degrades faster when luminosity background is added. MuGirl has higher efficiency than Muid for reconstructing the low-$p_T$ muons in the $J/\psi$ sample.

## 8.2 Resolution

Figure 13 shows the MuGirl $p_T$ resolution and tail as functions of $\eta$ and $p_T$. MuGirl does not refit the tracks and so this is just the resolution of the inner detector. Comparing with the standalone (figure 7) and combined (figure 11), we see how the standalone and inner measurements complement one another to give high precision over the full $\eta$ and $p_T$ range of the $t\bar{t}$ sample.

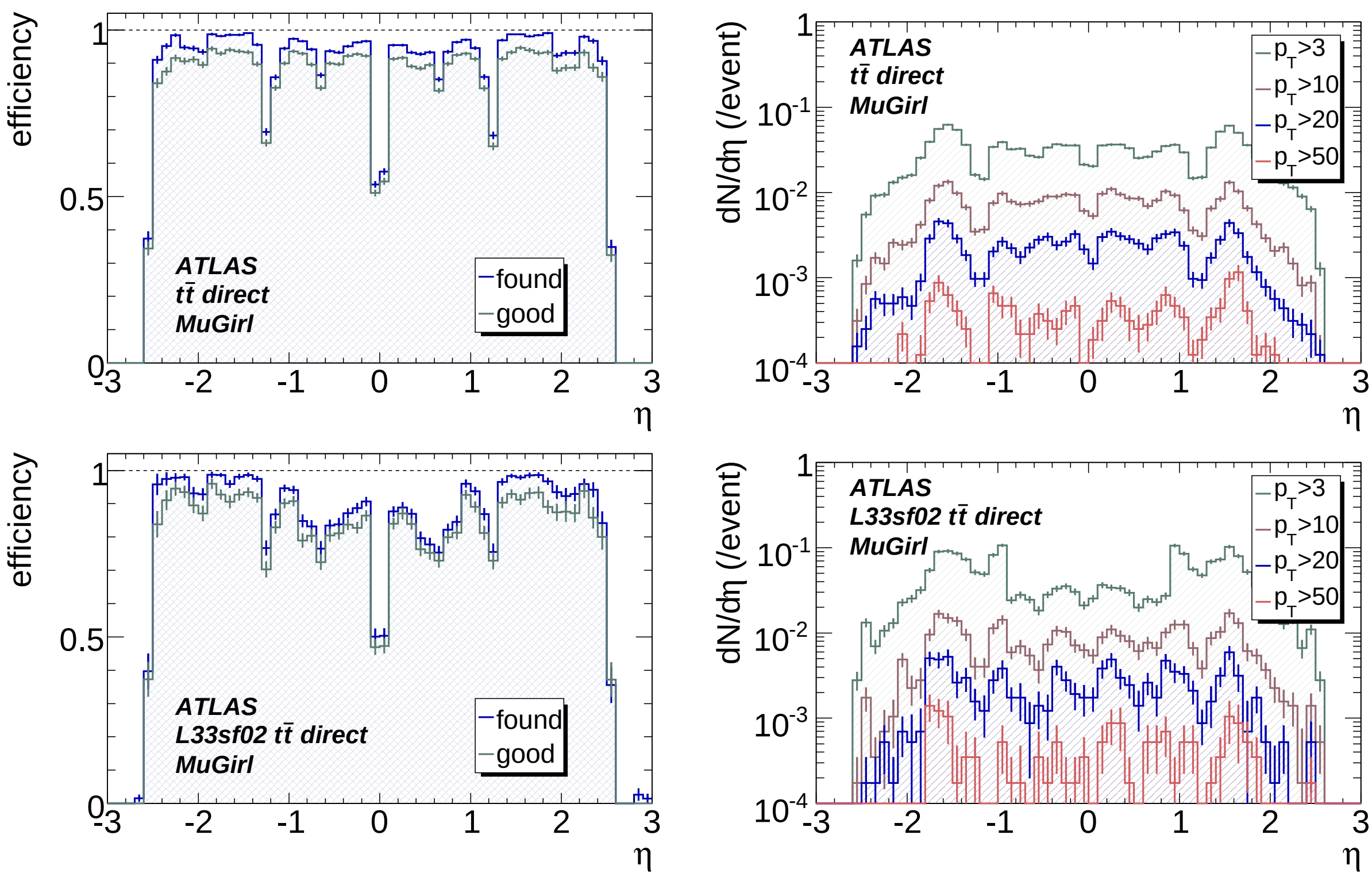

Figure 12: MuGirl efficiency (left) and fake rates (right) as a function of true $\eta$ in $t\bar{t}$ at low (top) and high (bottom) luminosity. In each efficiency plot, the upper curve (blue) is the efficiency to find the muon while the lower curve (green) additionally requires a good match ($D_{eva} < 4.5$) between reconstructed and true track parameters. The efficiency is for muons with true $p_T > 10$ GeV/c. Fake rates are presented for a variety of $p_T$ thresholds.

| Sample | Efficiency found | Efficiency good | Fakes/(1000 events) above $p_T$ limit (GeV/c) 3 | 10 | 20 | 50 |
|---|---|---|---|---|---|---|
| $t\bar{t}$ direct<br>$t\bar{t}$ indirect | 0.911 (1)<br>0.899 (2) | 0.870 (1)<br>0.748 (3) | 105.0 (6) | 23.7 (3) | 7.3 (2) | 1.14 (6) |
| hi-$\mathscr{L}$ $t\bar{t}$ direct | 0.866 (3) | 0.825 (3) | 154 (2) | 26.1 (7) | 7.6 (4) | 1.2 (1) |
| $Z'$ direct | 0.802 (3) | 0.781 (3) | 57 (2) | 26 (2) | 15 (1) | 5.9 (8) |
| $J/\psi$ $c$-quark | 0.888 (4) | 0.839 (5) | 4.4 (3) | 0.11 (5) | 0 (0) | 0 (0) |

Table 4: MuGirl efficiencies and fake rates. The samples and algorithms are described in the text. Efficiencies are presented both for all found muons and for those with a good truth match. Both are calculated for truth muons with $|\eta| < 2.5$ and $p_T > 10$ GeV/c. The fake rates are presented for a variety of $p_T$ thresholds.

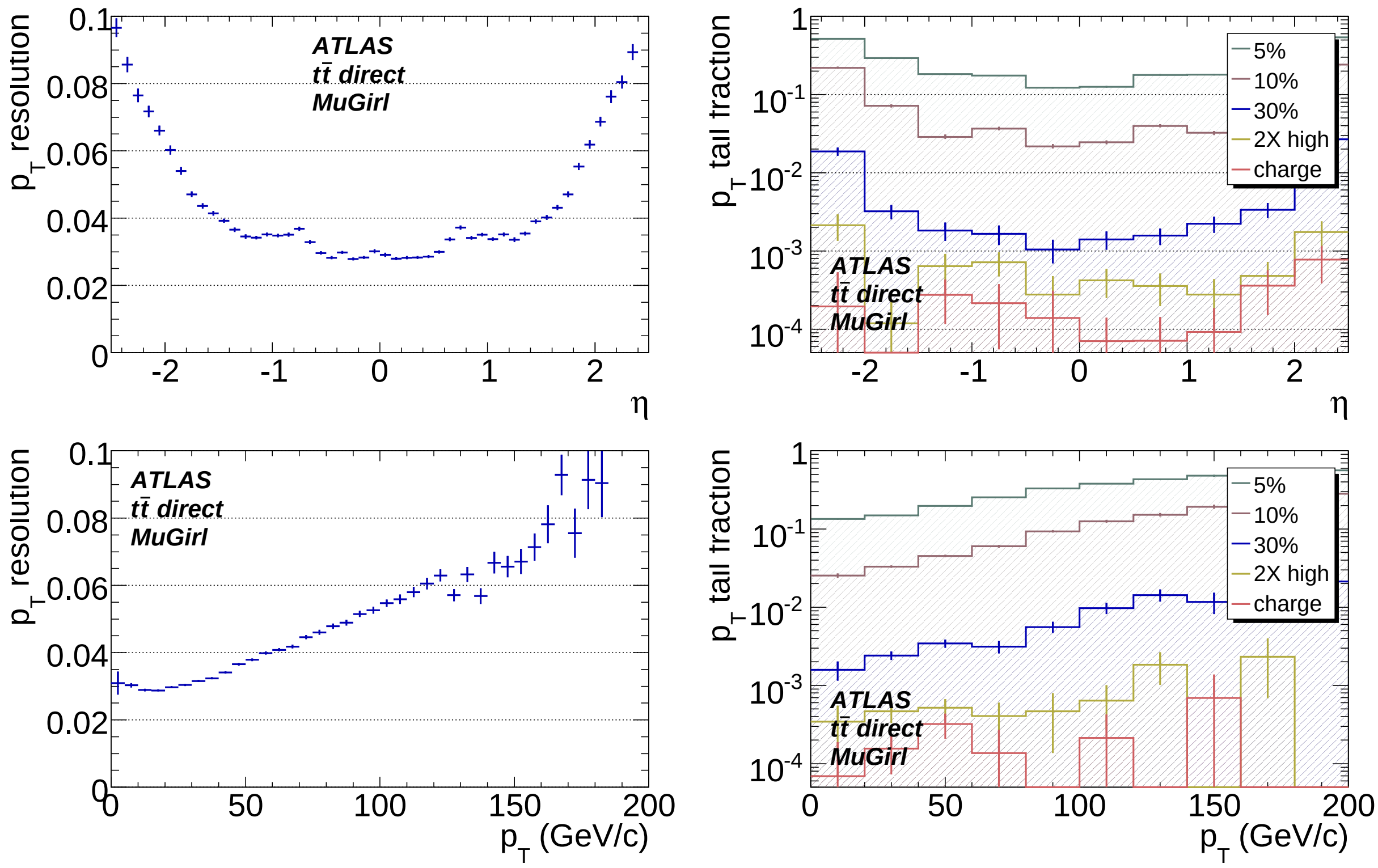

Figure 13: MuGirl fractional momentum resolution ($\Delta p_T / p_T$) as a function of $\eta$ (top) and $p_T$ (bottom). Both the distribution (left) and tails (right) are shown for each. The tail is the fraction of reconstructed muons with magnitude of residual greater than a threshold and results are shown for a variety of thresholds. The last tail curve (red, "charge") includes only muons reconstructed with the wrong charge sign. The 4th tail curve (yellow, "2X high") includes these and those with momentum magnitude more than two times the true value.

# 9 Merged muon performance

Finally we consider merging the muons produced by different algorithms. There are many possible combinations but we restrict ourselves to two simple but important cases: merging the combined and tagged muons separately within each collection (family), i.e. we examine Staco+MuTag and Muid+MuGirl.

Figure 14 shows the corresponding direct muon efficiencies and fake rates in $t\bar{t}$ at low and high luminosity. The integrated efficiencies and fake rates for all samples are summarized in table 5. One of the primary goals of the tagging algorithms is to reconstruct low-$p_T$ muons which the standalone reconstruction misses because the energy loss in the calorimeter leaves these muons with very little momentum in the muon spectrometer. Figure 15 shows the low-$p_T$ efficiency as function of $p_T$ for combined alone and combined supplemented with tagged for each of the collections.

| Sample | Efficiency | | Fakes/(1000 events) above $p_T$ limit (GeV/c) | | | |
|---|---|---|---|---|---|---|
| | found | good | 3 | 10 | 20 | 50 |
| Staco+MuTag | | | | | | |
| $t\bar{t}$ direct | 0.948 (1) | 0.879 (1) | 49.0 (4) | 14.4 (2) | 4.8 (1) | 0.86 (5) |
| $t\bar{t}$ indirect | 0.940 (1) | 0.772 (2) | | | | |
| hi-$\mathscr{L}$ $t\bar{t}$ direct | 0.946 (2) | 0.876 (3) | 58 (1) | 16.6 (5) | 6.1 (3) | 1.1 (1) |
| $Z'$ direct | 0.931 (2) | 0.844 (3) | 32 (2) | 14 (1) | 7.1 (9) | 4.2 (7) |
| $J/\psi$ | 0.954 (3) | 0.883 (4) | 2.5 (3) | 0.3 (1) | 0.11 (5) | 0 (0) |
| Muid+MuGirl | | | | | | |
| $t\bar{t}$ direct | 0.955 (1) | 0.903 (1) | 113.1 (6) | 24.9 (3) | 7.6 (2) | 1.17 (6) |
| $t\bar{t}$ indirect | 0.946 (1) | 0.790 (2) | | | | |
| hi-$\mathscr{L}$ $t\bar{t}$ direct | 0.952 (2) | 0.898 (2) | 181 (2) | 29.8 (7) | 8.4 (4) | 1.2 (2) |
| $Z'$ direct | 0.929 (2) | 0.866 (3) | 61 (3) | 28 (2) | 16 (1) | 7.5 (9) |
| $J/\psi$ | 0.946 (3) | 0.885 (4) | 4.7 (4) | 0.11 (5) | 0 (0) | 0 (0) |

Table 5: Staco+MuTag and Muid+MuGirl efficiencies and fake rates. The samples and algorithms are described in the text. Efficiencies are presented both for all found muons and for those with a good truth match. Both are calculated for truth muons with $|\eta| < 2.5$ and $p_T > 10$ GeV/c. The fake rates are presented for a variety of $p_T$ thresholds.

The merge provides only a small improvement in the Staco efficiencies and a substantial increase in the fake rates (factor of about four). For Muid, the efficiency gains are more substantial: the indirect $t\bar{t}$ efficiency increases by 6% and the $J/\psi$ by 15%. The fake rates are increased by a factor of five, i.e. slightly above the MuGirl rates. Overall, the Muid+MuGirl performance is very similar to that of Staco+MuTag. In both cases, we see the tagging algorithms do provide the significant efficiency improvement for $p_T$ below 10 GeV/c.

# 10 Summary

## 10.1 Present status

The starting point for most ATLAS analyses are the combined muons, i.e. those muons constructed by combining tracks found independently in the inner detector and muon spectrometer. Their momentum resolution and fake rate (with appropriate quality cuts) are both significantly better than muons reconstructed from either the spectrometer alone or muons identified by tagging inner detector tracks. In $t\bar{t}$ events, for muons from $W \to \mu\nu$ with $|\eta| < 2.5$, the Staco combined muon efficiency is 94% with most of the loss coming from regions of the spectrometer where the detector coverage is thin. The efficiency falls by a few percent when the muon transverse momentum reaches the TeV scale where it is much more likely that a muon will radiate a substantial fraction of its energy. The $t\bar{t}$ rate for fakes is a few per thousand events for $p_T > 20$ GeV/c and this can be reduced by an order a of magnitude (with a 2% loss in efficiency) by cutting on the muon quality ($\chi^2_{match}$). The performance of the Muid algorithm is only slightly worse for $t\bar{t}$ but it is significantly less robust, losing additional efficiency at low-$p_T$ and high-$p_T$ and when luminosity background is added.

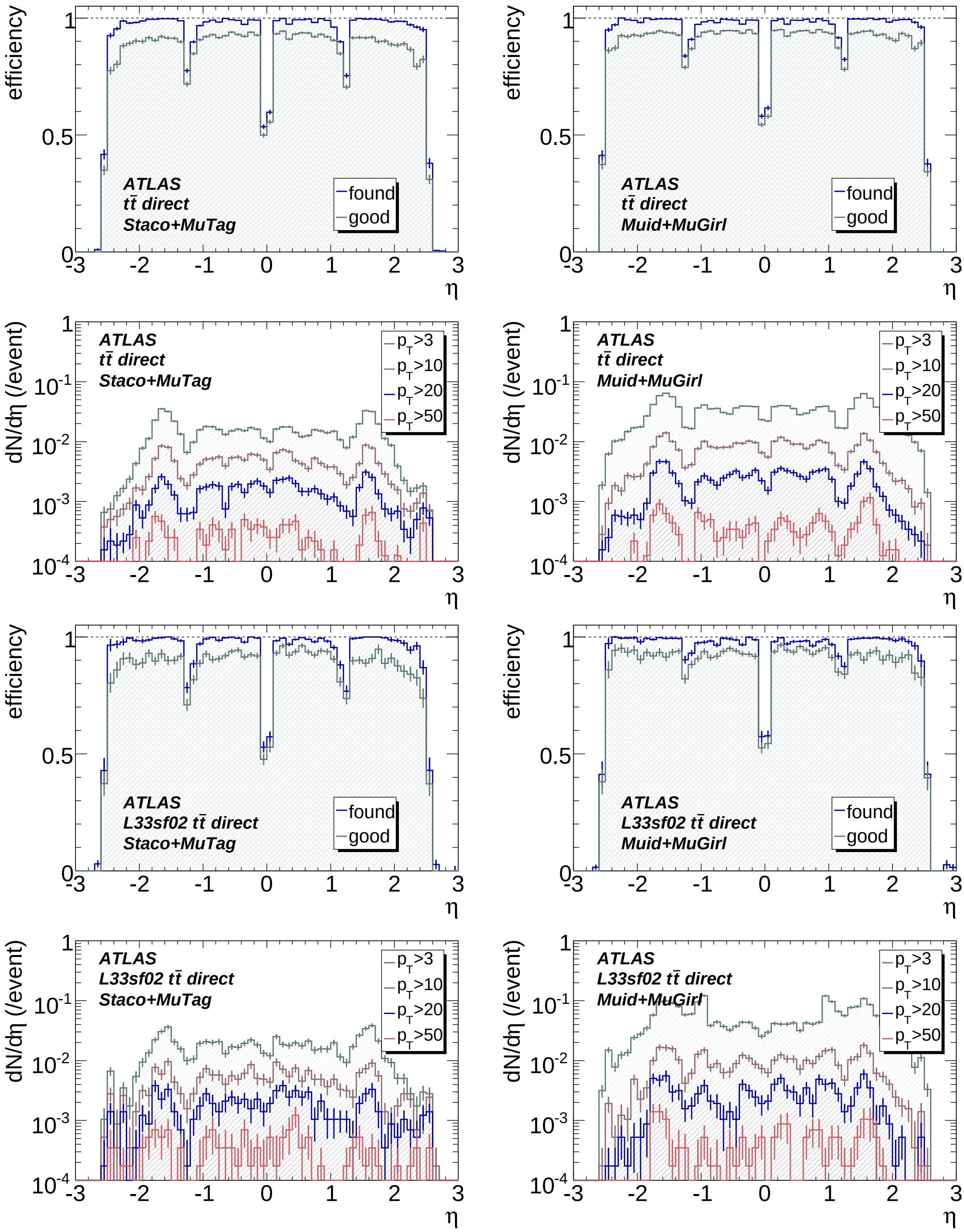

Figure 14: Muon efficiencies and fake rates for Staco+MuTag (left) and Muid+MuGirl (right) as functions of true $\eta$ in $t\bar{t}$ at low (top) and high (bottom) luminosity. In each efficiency plot, the upper curve (blue) is the efficiency to find the muon while the lower curve (green) additionally requires a good match ($D_{eva} < 4.5$) between reconstructed and true track parameters. The muon selection is described in the text. The efficiency is calculated for true $p_T > 10$ GeV/c. The fake rates are presented for a variety of $p_T$ thresholds.

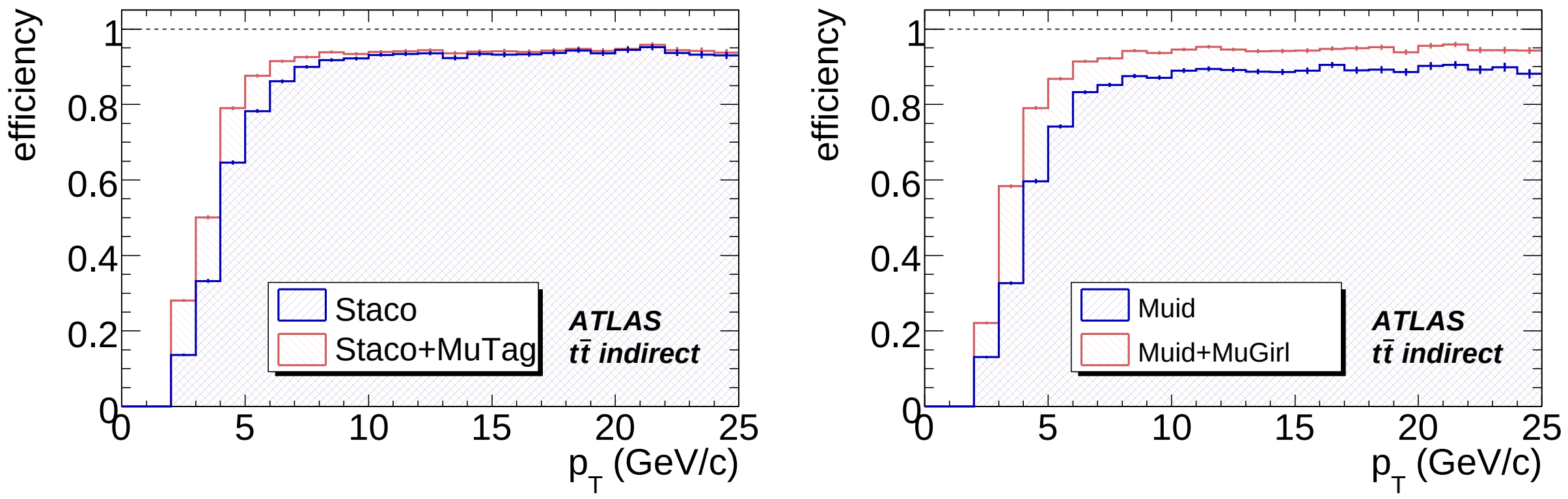

Figure 15: Low-$p_T$ muon finding efficiencies for combined muons alone and combined plus tagged for the Staco (left) and Muid (right) collections. Results are show for the $t\bar{t}$ indirect selection. The other samples show similar behavior but have much poorer statistics at low-$p_T$. The efficiency is calculated for muons with $|\eta| < 2.5$.

The combined muons can be supplemented with the standalone muons to extend the $\eta$ coverage to 2.7 and to recover the percent or so efficiency loss in combination. We do not report on this merge but it is clear from the standalone results that the fake rates will increase significantly especially when luminosity background is present. In the case of Moore, the fake rate is likely intolerable.

We find that merging with MuTag provides only slight improvement to the Staco efficiency with a significant increase in fakes. This may reflect the success of Staco more than deficiencies in MuTag. MuGirl is able to improve the Muid efficiency, so that the merge Muid+MuGirl has performance similar to Staco or Staco+MuTag. By itself, the MuGirl efficiency is somewhat less than that of Staco especially for high-$p_T$ muons, and the fake rates are substantially higher.

## 10.2 Future

The results presented here reflect the status of the ATLAS software used to reconstruct (Monte Carlo) production data in 2007. Work continues both to improve the algorithms described here and to add new ones. The high-luminosity fake rate for Moore is being addressed by introducing timing cuts and investigating alternative approaches to the pattern recognition. The latter also has the goal of reducing the number of false hit assignments. Combined muons with large $\chi^2_{match}$ are being studied to see if a second stage of pattern recognition can reduce the efficiency loss or resolution tails. Efforts are underway to improve or replace the existing spectrometer-tagging algorithms; in particular, code is already in place to extrapolate to additional stations enabling recovery of much of the standalone/combined efficiency loss near $|\eta| = 1.2$. Two calorimeter-tagging algorithms have been developed and offer the possibility of recovering much of the efficiency loss near $\eta = 0$. Improvements in modularity will make it possible to mix components from the different algorithms, (e.g. to use Muid to combine Muonboy muons) and enable algorithms to share common tools such as those being developed to calculate energy loss, refit muon tracks, and repair muons with poor fit quality.

# Muons in the Calorimeters: Energy Loss Corrections and Muon Tagging

**Abstract**
The muon spectrometer is the outermost subdetector of the ATLAS detector, beginning after a muon has traversed 100 radiation lengths of material. Muon momentum measurements must be corrected for energy loss in the calorimeters and the inert material before the muons reach the muon spectrometer. Energy lost in the calorimeters can be estimated from parameterizations or from a measurement of the energy deposited in the calorimeters. In addition, the muon energy loss measurement can be used to tag muons not reconstructed in the muon spectrometer due to inefficiencies, spectrometer acceptance or their low momenta.

In this document we discuss different algorithms developed to perform the energy loss correction in the muon reconstruction. We compare the performance of the muon reconstruction algorithms before and after the energy loss correction is applied. In addition, we describe the muon tagging algorithms, based on measurements obtained in the calorimeters, and contrast their performance in different simulated data samples.

## 1 Introduction

Muons traverse the inner detector and the calorimeters in the ATLAS experiment before reaching the muon spectrometer. The material thickness traversed by the muons before reaching the muon spectrometer is over 100 radiation lengths ($X_0$) (see Figure 1). By passing through this material, muons undergo electromagnetic interactions which result in a partial loss of their energy. As over 80% of this material is in the instrumented areas of the calorimeters, the energy loss can be measured. Understanding how this energy loss happens, its magnitude and how to measure it is essential to obtain the best performance in muon reconstruction and identification.

In this document we discuss the aspects of muon reconstruction and identification that make use of all available energy loss information in the ATLAS software. The Muonboy [1] and Muid [2] algorithms for muon reconstruction take into account internally the calorimeter material effects for tracks already found in the muon spectrometer. Algorithms that calculate the energy loss and transport the track anywhere in the detector are also available. The detailed computation of this correction is the main focus of Sections 2 and 3, while Section 4 is devoted to the use of the energy loss information for muon identification. This note gives an overview of the current algorithms and techniques which will be used for the reconstruction of the first data.

## 2 Algorithmic Treatment of Material Effects

When a muon traverses the detector material, it undergoes successive deflections and a loss of energy. The total angular deflection is an accumulation of many small angle deflections, referred to as multiple (Coulomb) scattering; and it is well approximated by a gaussian distribution that is centered at a zero mean value. The expected root mean square of the projected scattering angle can be described by the formula of Highland [4]:

$$\sigma_{ms}^{proj} = \frac{13.6\ \mathrm{MeV}}{\beta c p}\sqrt{t}[1 + 0.038 \ln t], \qquad (1)$$

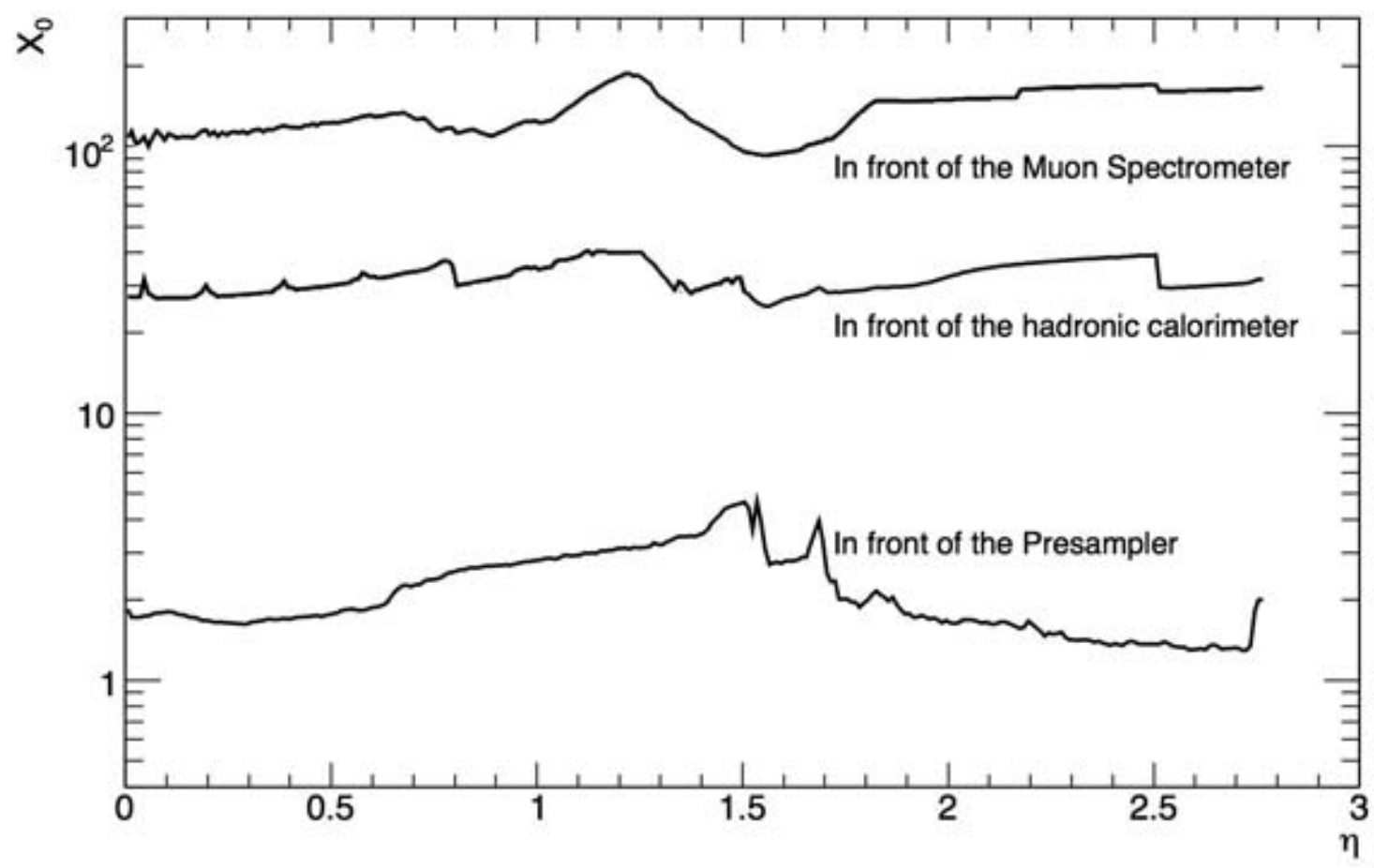

Figure 1: Material distribution before the muon spectrometer in ATLAS as a function of $\eta$ [3]. The material is expressed in radiation lengths ($X_0$).

where $t$ is the thickness of the traversed material in units of the radiation length $X_0$. The energy loss, on the other hand, is non-gaussian. Throughout this document, we will study the energy loss of muons going through the ATLAS detector in detail. The discussion of multiple scattering, however, will be limited to this section, because it is simpler and it will be based on the Highland formula shown above. The thickness in the formula above is calculated from the geometry description for all algorithms. However, there are small differences in how the multiple scattering information is used in the track fitting. These differences are explained below, as the different track fitting strategies are discussed.

In ATLAS track reconstruction applications, two main track fitting strategies are deployed: the classical least squares method and the progressive method that corresponds to the Kalman filter formalism [5].

**The least squares fit:** In the global fitting technique, most material effects are directly integrated into the $\chi^2$ function (the energy loss may or may not be fitted). This is done by introducing the deflection angles and, possibly, energy losses as additional parameters to the fit.

The contribution of the fitted scattering angles to the $\chi^2$ function has to be regulated by the expected range of the scattering process in the traversed material. Scattering effects are applied to the muon on two surfaces along its trajectory, because the scattering effects from material bulk can be accurately described by two scattering centers. The Muonboy algorithm iterates its calculation of the muon trajectory in a complex geometry with many scattering centers. The number of scattering centers is reduced to two after the iteration. The iteration allows for a calculation of the material traversed after the trajectory has been modified to account for the energy loss. On the other hand, the Muid algorithm and the ATLAS tracking global-$\chi^2$ fitter [6] currently use a map of the material from the Monte Carlo on two surfaces. The $\eta$ coordinate of the track on these two surfaces is then used to calculate the amount of material traversed by the muon.

The least squares fit with the calorimeter energy loss as a fitted variable can only be performed in a combined fit including measurements of both the muon spectrometer segments and the inner detector hits. If no inner detector hits exist, the treatment of the energy loss effects is fundamentally equivalent for both a least-squares-inspired algorithm and a Kalman-filter-inspired algorithm.

To minimize the number of degrees of freedom in the least-squares fit, the number of fitted variables must be minimized. In particular, one energy loss variable in a track fit is preferable. This does not mean that the trajectory cannot be affected smoothly by the energy loss, because an extended set of material layers can be calculated using a detailed detector description as in Figure 2 and the fitted energy loss distributed proportionally among these layers. This is done for the purpose of transporting the track through the calorimeters inside the Muonboy algorithm. An alternative approach is currently taken in the Muid and ATLAS tracking global-$\chi^2$ fitter. These algorithms apply the energy loss to the track on one surface inside the calorimeters hence approximate the rate of change of curvature within the calorimeter volume (i.e.: they assume the momentum of the muon changes only at one place along its trajectory).

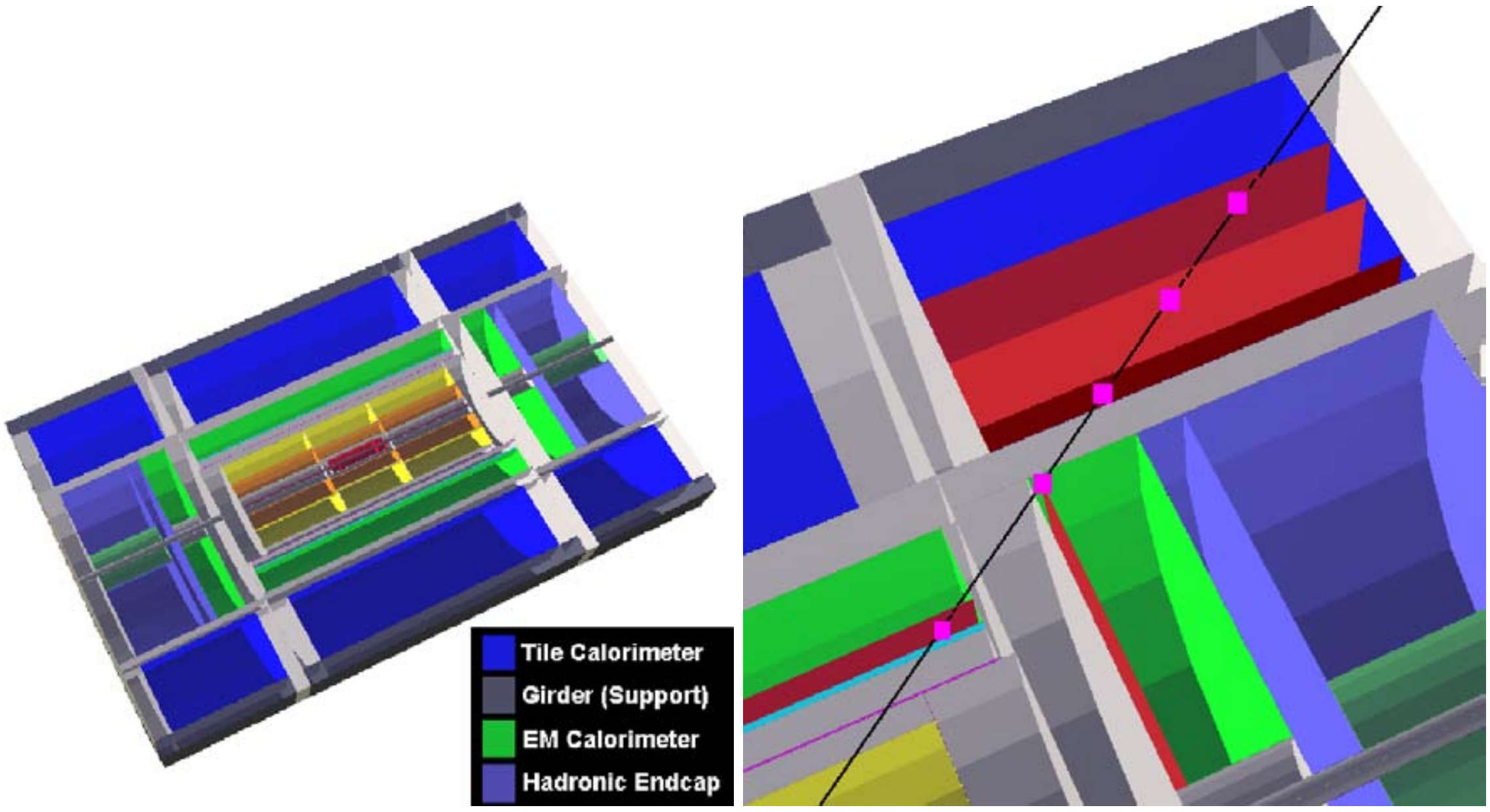

Figure 2: Left: 3-D view of the tracking geometry up to the muon spectrometer. Right: Example set of energy loss update layers (shown as additional surfaces with respect to the figure on the left; update positions shown as squares) created during the extrapolation of a track (black line) through the calorimeter.

The effect of this simplification on the muon combined reconstruction is expected to be small, because the energy loss only affects the trajectory of the track if the track is bending. However, the area where most of the energy loss happens (the calorimeter) has a small magnetic field. A quantitative estimate of the effect of the simplification can be obtained by comparing the multiple scattering effects on the track and the bending that the track undergoes from its entrance in the calorimeters to its exit. The bending is shown in Figure 3. Equation 1 indicates that a 10 GeV (100 GeV) muon going through the calorimeters scatters following a gaussian distribution with RMS $\approx$14-20 (1.4-2) milliradians (with $X_0 = 100$-200 from Figure 1). Figure 3 shows that the deviation of the track due to the magnetic field is comparable to the deviation expected from multiple scattering at least in the $\phi$ direction. Algorithms that use one surface to apply the energy loss correction approximate the mean trajectory inside the calorimeter with a systematic offset that increases with depth to a maximum value at the calorimeter centre. The magnitude of this offset is proportional to the magnetic bending scaled by the fraction of energy loss to muon energy. It thus remains small with respect to the uncertainties caused by Coulomb scattering.

**Progressive fitting techniques:** In progressive fitting techniques, the particle-detector interaction is part of the transport process of the track to the next surface where a hit may exist (measurement surface). The transported track can then be compared (and updated) with the measurement obtained on the next measurement surface. In this transport process magnetic field and material effects (multiple scattering and energy loss) are applied to the parameterization of the track. Multiple scattering is applied by increasing the uncertainties of the angular direction variables, while energy loss effects are taken into account in two ways. A mean energy loss is applied to the track parameterization, and then an uncertainty is added to the corresponding covariance matrix term to account for the stochastic behavior of the energy loss. The resulting increased covariance terms degrade the track prediction for the subsequent measurement surface.

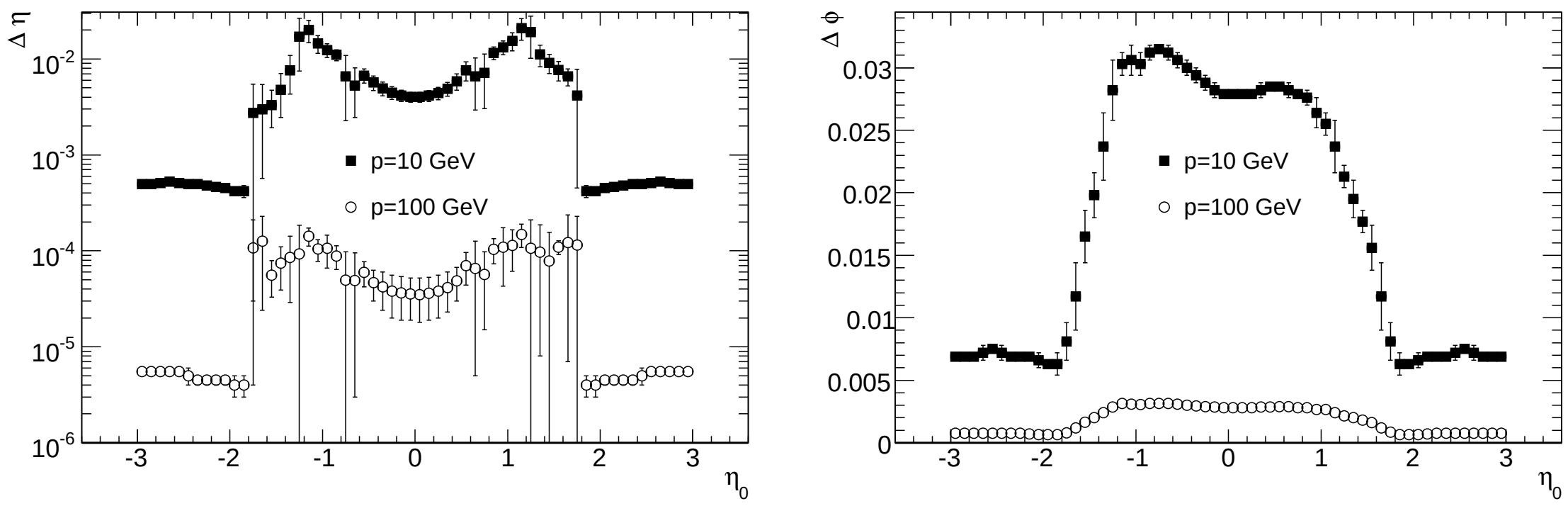

Figure 3: Calculated difference between the calorimeter entrance and exit coordinates ($\Delta\eta$, left, and $\Delta\phi$, right) for 10 GeV (solid squares) and 100 GeV muons as a function of $\eta_0$ of the muon at the interaction point. The lack of mirror symmetry is due to the combined effect of the return flux of the solenoid (unidirectional) and the toroidal magnetic field (symmetric around the $z$ axis).

Progressive fitting tools rely, therefore, on a precise description of the detector material and magnetic field. An example is shown in Figure 2. The illustration on the right of Figure 2 shows material layers that are calculated dynamically during the extrapolation process into the calorimeter active volumes.

In ATLAS, the stand-alone muon reconstruction algorithms (MOORE [7] and Muonboy) use exclusively the least-squares formalism to fit tracks in the muon spectrometer. On the other hand, the inner detector reconstruction uses by default the progressive fitting techniques. In combined muon reconstruction, when the hits in the inner detector are used in combination with the muon spectrometer, the Muonboy-based algorithm (STACO) combines tracks reconstructed in the inner detector and muon spectrometer independently, therefore being a mixture of the tracking fit and the least-squares fit carried out by Muonboy. On the other hand, the MOORE-based algorithm (Muid) performs a least-squares fit in both subsystems.

# 3 Corrections for the Energy Loss from the Beam Pipe to the Muon Spectrometer

In this section we describe how the energy loss is calculated from GEANT4 [8] based parameterizations and/or measurements of the energy loss by the calorimeters. Muon isolation is also discussed in this context. Finally, the energy loss corrections are validated as part of the muon reconstruction algorithms.

## 3.1 Parameterizations of the Energy Loss

Relativistic muons going through matter lose energy mostly through electromagnetic processes: ionization, $e^+e^-$ pair-production, and bremsstrahlung. Ionization energy loss dominates for muons of momenta $\lesssim$100 GeV. Bremsstrahlung and $e^+e^-$ pair-production energy losses are often jointly referred to as radiative energy losses. Higher energy muons lose energy mostly through radiative energy losses.

However, when passing through materials made of high-$Z$ elements the radiative effects can be already significant for muons of energies ≈10 GeV [9].

Ionization energy losses have been studied in detail, and an expression for the mean energy loss per unit length as a function of muon momentum and material type exists in the form of the Bethe-Bloch equation [10]. Other closed-form formulae exist to describe other properties of the ionization energy loss. Bremsstrahlung energy losses can be well parameterized using the Bethe-Heitler equation. However, there is no closed-form formula that accounts for all energy losses. Nevertheless, theoretical calculations for the cross-sections of all these energy loss processes do exist. With these closed-form cross-sections, simulation software such as GEANT4 can be used to calculate the energy loss distribution for muons going through a specific material or set of materials.

The fluctuations of the ionization energy loss of muons in thin layers of material are characterized by a Landau distribution. Here "thin" refers to any amount of material where the muon loses a small percentage of its energy. Once radiative effects become the main contribution to the energy loss, the shape of the distribution changes slowly into a distribution with a larger tail. Fits to a Landau distribution still characterize the distribution fairly well, with a small bias that pushes the most probable value of the fitted distribution to values higher than the most probable energy loss [11]. These features are shown for the energy loss distributions of muons going from the beam-pipe to the exit of the calorimeters in Figure 4.

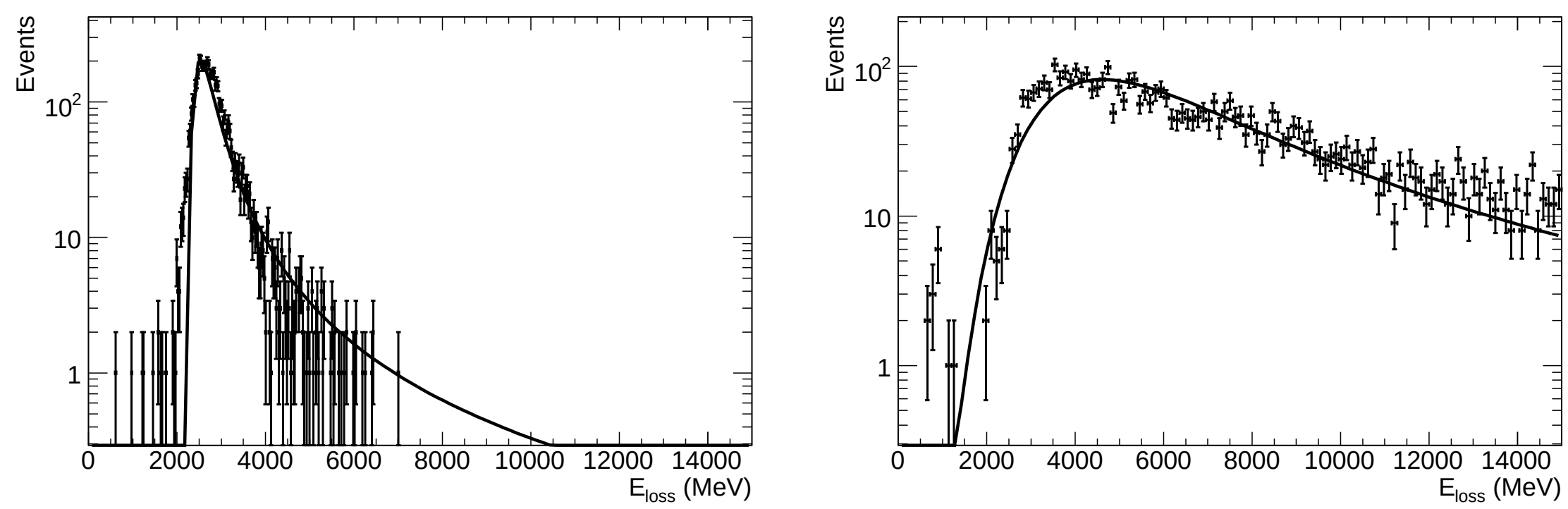

Figure 4: Distribution of the energy loss of muons passing through the calorimeters ($|\eta| < 0.15$) as obtained for 10 GeV muons (left) and 1 TeV muons (right) fitted to Landau distributions (solid line).

As can be seen in Figure 4 the Landau distribution is highly asymmetrical with a long tail towards higher energy loss. For track fitting, where most of the common fitters require gaussian process noise, this has a non-trivial consequence: in general, a gaussian approximation has to be performed for the inclusion of material effects in the track fitting [12].

In order to express muon spectrometer tracks at the perigee, the total energy loss in the path can be parameterized and applied to the track at some specific position inside the calorimeters. As the detector is approximately symmetric in $\phi$, parameterizations need only be done as a function of muon momentum and $\eta$. The $\eta$-dependence is included by performing the momentum parameterizations in different $\eta$ bins of width 0.1 throughout the muon spectrometer acceptance ($|\eta| < 2.7$). The dependence of the most probable value of the energy loss, $E_{\text{loss}}^{\text{mpv}}$, as a function of the muon momentum, $p_\mu$, is well described by

$$E_{\text{loss}}^{\text{mpv}}(p_\mu) = a_0^{\text{mpv}} + a_1^{\text{mpv}} \ln p_\mu + a_2^{\text{mpv}} p_\mu, \tag{2}$$

where $a_0^{\text{mpv}}$ describes the minimum ionizing part, $a_1^{\text{mpv}}$ describes the relativistic rise, and $a_2^{\text{mpv}}$ describes the radiative effects. The width parameter, $\sigma_{\text{loss}}$, of the energy loss distribution is well fitted by a linear function $\sigma_{\text{loss}}(p_\mu) = a_0^\sigma + a_1^\sigma p_\mu$. Some of these fits are illustrated in Figure 5. This parameterization is used as part of the Muid algorithm for combined muon reconstruction [3].

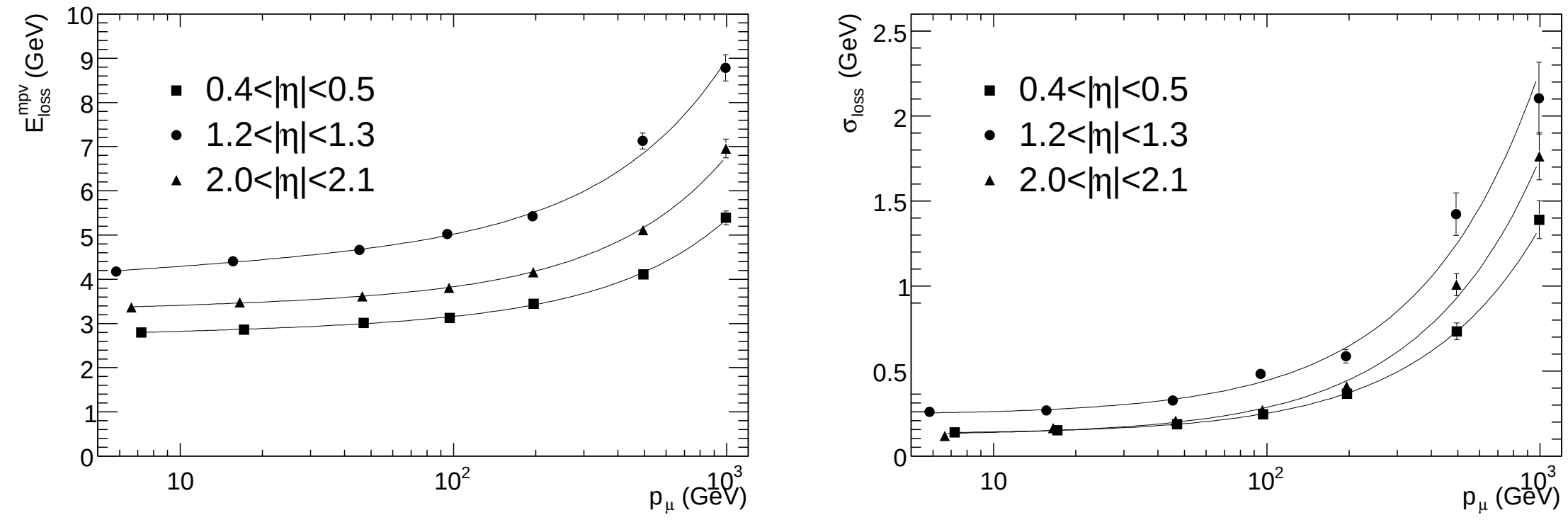

Figure 5: Parameterization of the $E_{\text{loss}}^{\text{mpv}}$ (left) and $\sigma_{\text{loss}}$ (right) of the Landau distribution as a function of muon momentum for different $\eta$ regions. One sees a good agreement between the GEANT4 values and the parameterization.

An alternative approach exists in the ATLAS tracking. In this approach, the energy loss is parameterized in each calorimeter or even calorimeter layer. The parameterization inside the calorimeters is applied to the muon track using the detailed geometry described in Section 2.

The most probable value and width parameter of the Landau distribution are not affected by radiative energy losses in thin materials in the muon energy range of interest ($\sim$5 GeV to a few TeV). This justifies treating energy loss in non-instrumented material, such as support structures, up to the entrance of the muon spectrometer as if it was caused by ionization processes only. The most probable value of the distribution of energy loss by ionization can be calculated if the distribution of material is known [13]. Since material properties are known in each of the volumes in the geometry description used, it is easy to apply this correction to tracks being transported through this geometry.

For the instrumented regions of the calorimeters, a parameterization that accounts for the large radiative energy losses is required. To provide a parameterization that is correct for the full $\eta$ range and for track transport inside the calorimeters, a study of energy loss as a function of the traversed calorimeter thickness, $x$, was performed. Two parameters that characterize fully the pdf of the energy loss for muons were fitted satisfactorily using several fixed momentum samples as

$$E_{\text{loss}}^{\text{mpv},\sigma}(x,p_\mu) = b_0^{\text{mpv},\sigma}(p_\mu)x + b_1^{\text{mpv},\sigma}(p_\mu)x\ln x. \quad (3)$$

The momentum dependence of the $b_i(p_\mu)$ parameters was found to follow the same form as in Equation 2. Fits for some of the absorber materials are shown in Figure 6. These parameterizations have been validated over the $\eta$ range from -3 to 3. A direct comparison of the most probable energy loss in GEANT4 simulation and in the geometry of the ATLAS tracking algorithms is shown in Figure 7 for muons propagating from the beam-pipe to the exit of the electromagnetic calorimeters and to the exit of the hadronic calorimeters.

## 3.2 Measurements of the Energy Deposited in the Calorimeters

In this section the measurement of the muon energy loss in the calorimeters is discussed. Understanding this measurement is important because it allows for an improvement in the energy loss determination. This section provides a basic description of the ATLAS calorimeters and their measurements which is important for understanding the topics discussed in Sections 3.3, 3.4 and 4.

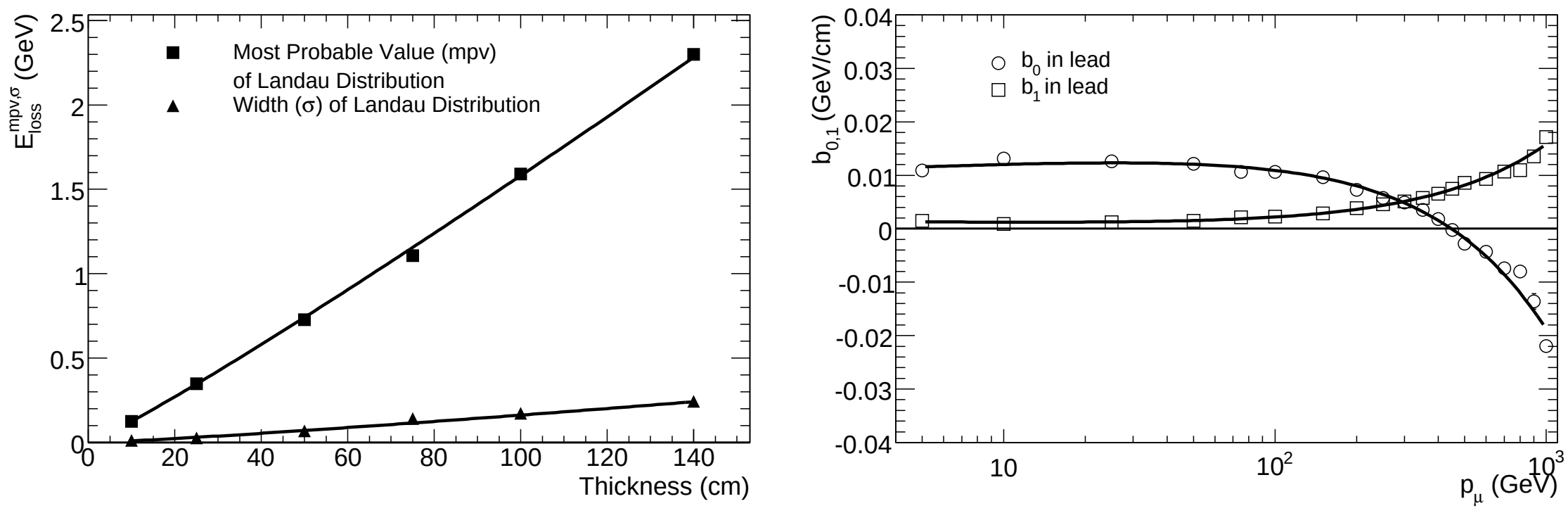

Figure 6: Left: Fit to the most probable value and width of the Landau distribution as a function of thickness of iron for muons of momentum 200 GeV. The fitting function has the form $b_0 x + b_1 x \ln x$. Right: Fit to the parameters $b_0$ and $b_1$ for the most probable value of the energy loss in lead as a function of muon momentum.

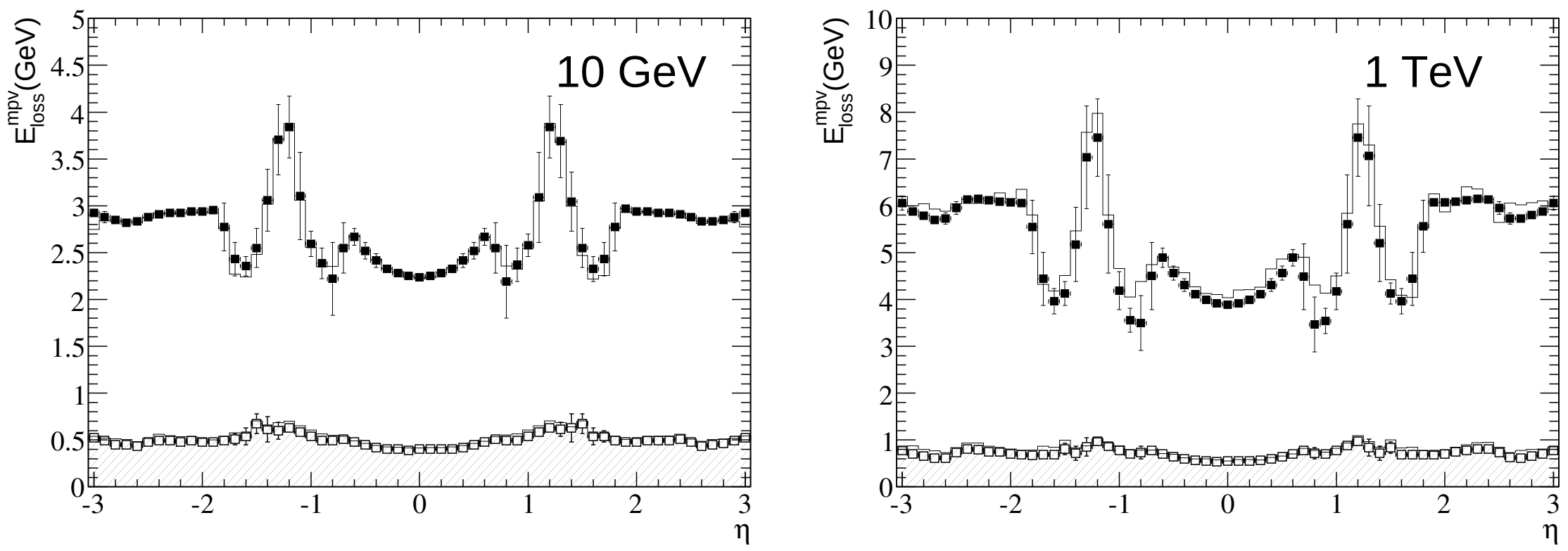

Figure 7: Most probable value of the energy loss as parameterized in the geometry of the ATLAS tracking (points) and in GEANT4 for muons of momentum 10 GeV (left) and 1 TeV (right) as a function of pseudorapidity. The solid line and points correspond to the energy loss of muons propagating from the beam pipe to the exit of the hadronic calorimeters. The filled histogram and hollow points correspond to the energy loss of muons propagating from the beam pipe to the entrance of the hadronic calorimeters.

### 3.2.1 Muons in the Liquid Argon Calorimeters

The electromagnetic calorimeter is a lead-liquid argon sampling calorimeter with accordion shaped absorbers and electrodes, covering the $|\eta|$ range up to 3.2.

The hadronic end-cap calorimeter, also based on liquid argon technology, covers the $|\eta|$ range from 1.5 to 3.2. The absorbers are made of parallel plates of copper. The total thickness of the hadronic end-cap calorimeters is 10 interaction lengths ($\lambda_{\text{int}}$). The measurement of a muon signal in the hadronic end-cap is complicated because the noise levels are high compared to the muon signal itself [14].

The detailed geometrical description of the LAr calorimeters is presented in [15]. Only the aspects relevant for muon studies will be recalled. Both barrel and end-cap calorimeters possess up to three longitudinal samplings (called strip, middle and back). They are completed by a liquid argon presampler detector to estimate the energy lost in upstream material. The signal and noise distributions in two longitudinal calorimeter samples in the barrel of the electromagnetic calorimeter are shown in Figure 8 for 150 GeV muons. Further discussion of how these distributions were calculated from the combined test beam data can be found in Section 4. The signal can be separated from the noise, especially in the middle sampling. In addition, comparisons between GEANT4 simulation and test beam data show that, despite the high noise in the first sampling, the electromagnetic calorimeter can measure reliably the energy lost by muons traversing it.

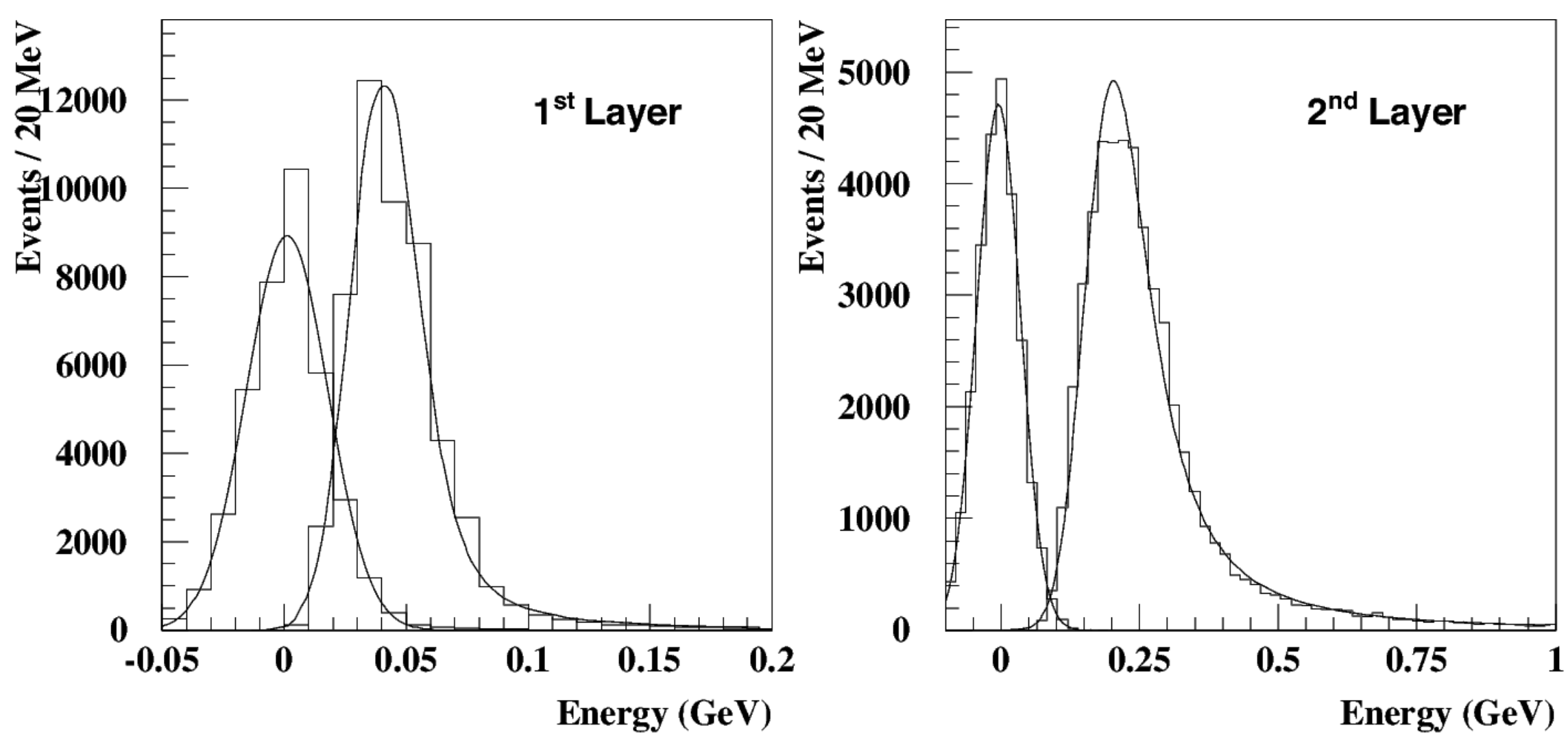

Figure 8: Distribution of the muon energy deposited in one electromagnetic calorimeter cell by 150 GeV muons, fitted to a Landau function convolved with a gaussian [16]. The gaussians on the left of each plot are the distributions of the noise. Left (right): energy deposit in a cell belonging to the first (middle) longitudinal sampling traversed by the muon. The energy is the sum of the energies of the (up to two) cells belonging to the muon cluster (see Section 4). The data were collected in the 2004 Combined Test Beam.

#### 3.2.2 Muons in the Tile Calorimeter

The Tile Calorimeter (TileCal) [17] is a plastic scintillator/steel sampling calorimeter, located in the region $|\eta| < 1.7$; it is divided into three cylindrical sections, referred to as the barrel and extended barrels. It extends from an inner radius of 2.28 m to an outer radius of 4.25 m. Modules are segmented in $\eta$ and in radial depth. In the direction perpendicular to the beam axis, the three radial segments span 1.5, 4.1 and 1.8 $\lambda_{\text{int}}$ in the barrel and 1.5, 2.6, 3.3 $\lambda_{\text{int}}$ in the extended barrels. The resulting typical cell dimensions are $\Delta\eta \times \Delta\phi = 0.1 \times 0.1$ ($0.2 \times 0.1$ in the outermost layer). This segmentation defines a quasi-projective tower structure.

The TileCal response to high-energy muons follows a Landau-type distribution with characteristically long tails at high energies caused by radiative processes and energetic $\delta$–rays. This response has been extensively studied in test beams with 180 GeV muons incident at projective angles. The peak values of the muon signals vary by more than a factor of two in projective geometry. An example of the muon signal, expressed in units of collected charge (pC), is shown in Figure 9, both for the whole tower and the last radial compartment. The signal is well separated from the noise, with a signal-to-noise ratio ($S/N$) of $\sim 44$ and $\sim 18$ respectively. The muon response was shown to be uniform in $\eta$ to within 1.9 % over all modules tested. The energy deposition spectrum observed in the TileCal test beams is within a few percent of the GEANT4 prediction.

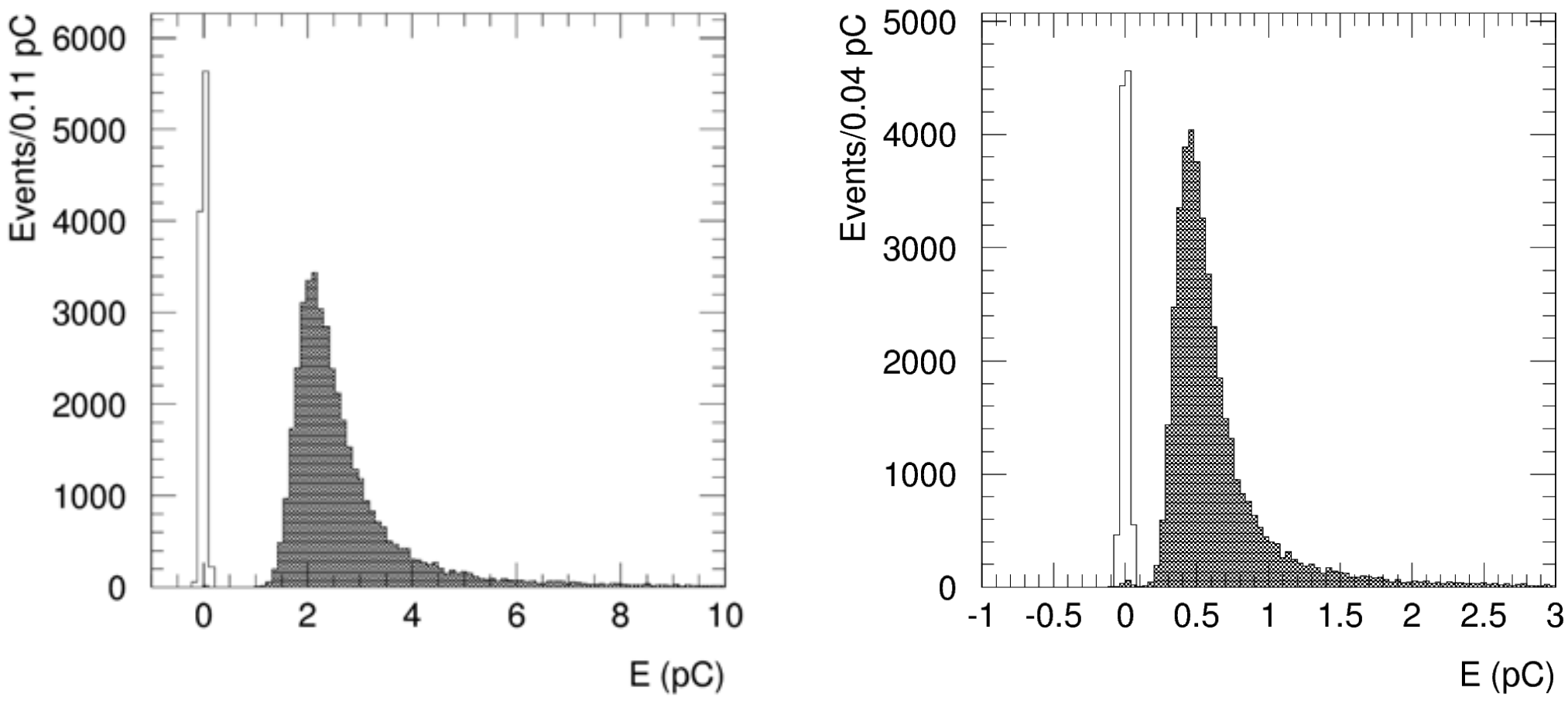

Figure 9: Example of the isolated muon signal as measured at $\eta = 0.35$ in the whole tower (left) and in the last radial compartment (right). The narrow peaks represent the corresponding noise. The energy is measured in units of collected charge. For a muon 1 pC corresponds to roughly 1 GeV, yielding a noise width of roughly 40 MeV for the last radial compartment. The data were collected in test beams in 2002 and 2003.

### 3.2.3 Measurements in Muon Algorithms

The previous sections discussed the reconstruction of energy depositions at the cell level. To provide estimates of muon energy loss and muon isolation, several cells need to be used along the muon trajectory. In addition, muon calibration factors such as the $e/\mu$ ratio for minimum ionizing muons, need to be adjusted in order to find the correct energy deposition.

The classical method for measuring the energy loss of muons in calorimeters is based on the concept of a calorimeter tower, where the muon is assumed to follow a straight trajectory inside the calorimeters. A tower is defined as all calorimeter cells within a cone of fixed radius $\Delta R = \sqrt{\Delta\eta^2 + \Delta\phi^2}$ centered around the muon trajectory. Motivated by this concept, but with a few muon-specific changes, the *Straight Line* method has been developed as part of the Muid algorithm for muon reconstruction. The Straight Line method calculates the coordinates of the relevant track at half the depth of the calorimeter by transporting the track to that position. These coordinates are used to calculate the calorimeter cells included in the measurement cone.

In the *Track Update* method, the muon trajectory through the calorimeters is extrapolated either from inner detector tracks or muon spectrometer tracks. Given this trajectory, the center of the measurement cone is recalculated at each calorimeter layer. Figure 10 shows a qualitative comparison between the Straight Line method and the Track Update method.

Figure 3 illustrated the quantitative differences in the muon trajectories from the two methods. The difference between the Straight Line and Track Update methods can be estimated by comparing the coordinates of the muon at the entrance of the electromagnetic calorimeters and at the exit of the hadronic calorimeters. The difference is clearly negligible for muons of $p_T > 100$ GeV, even though it can be as big as a third of a hadronic cell width for 10 GeV muons.

In Figure 11, a comparison between the measured energy and the true energy loss is shown for the Track Update method. The average measured transverse energy loss in a cone of 0.2 around the muon trajectory for single muons of momentum 10, 100 and 300 GeV is shown as a function of $\eta$. In the same $\eta$ bins the average true (as obtained from the GEANT4 full simulation) transverse energy lost between the interaction point and the entrance of the muon spectrometer is shown. The energy lost by the muons is well estimated by measurements in the calorimeters. The region around $|\eta| = 1$ corresponds to the crack in the TileCal. Consequently, the measurement underestimates the energy loss in that region.

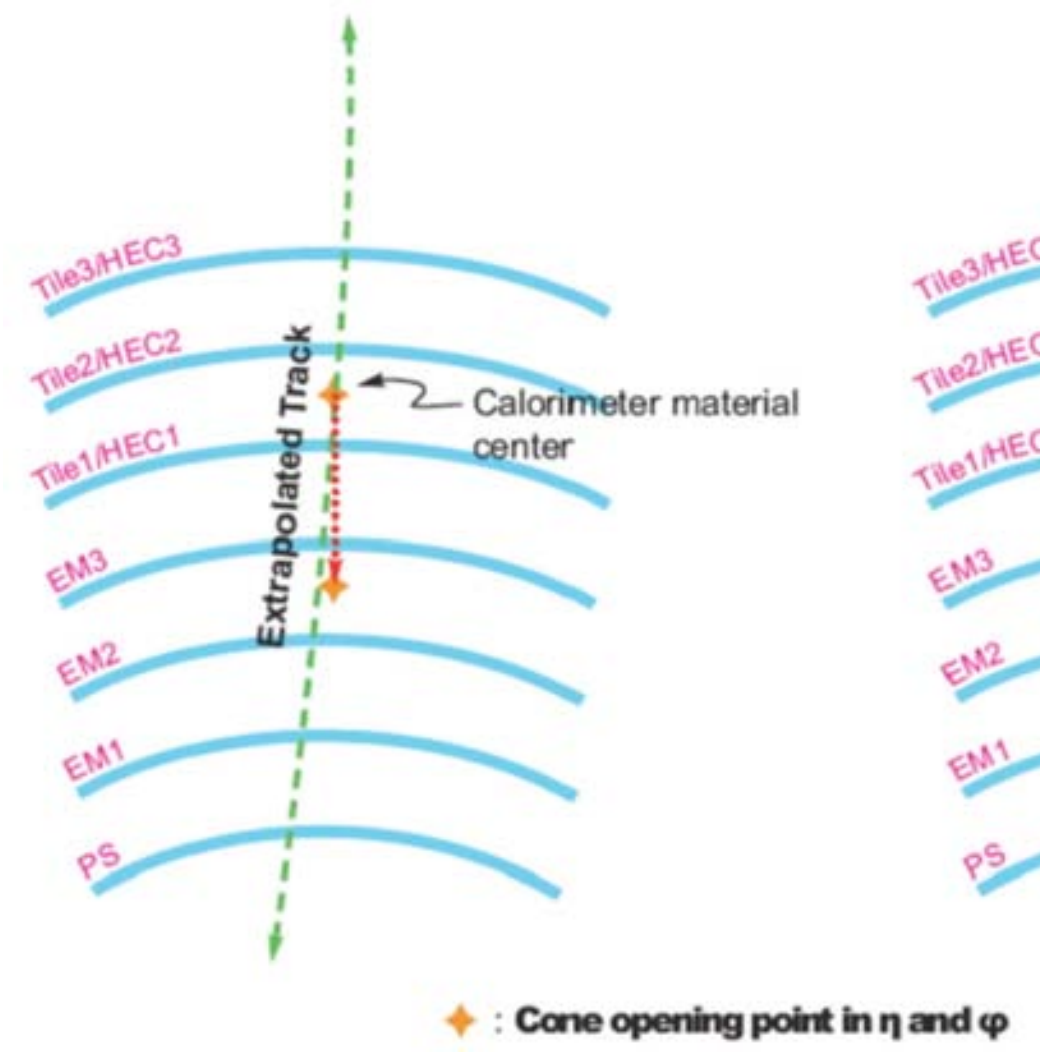

Figure 10: Illustration of the Straight Line (left) and Track Update (right) concepts.

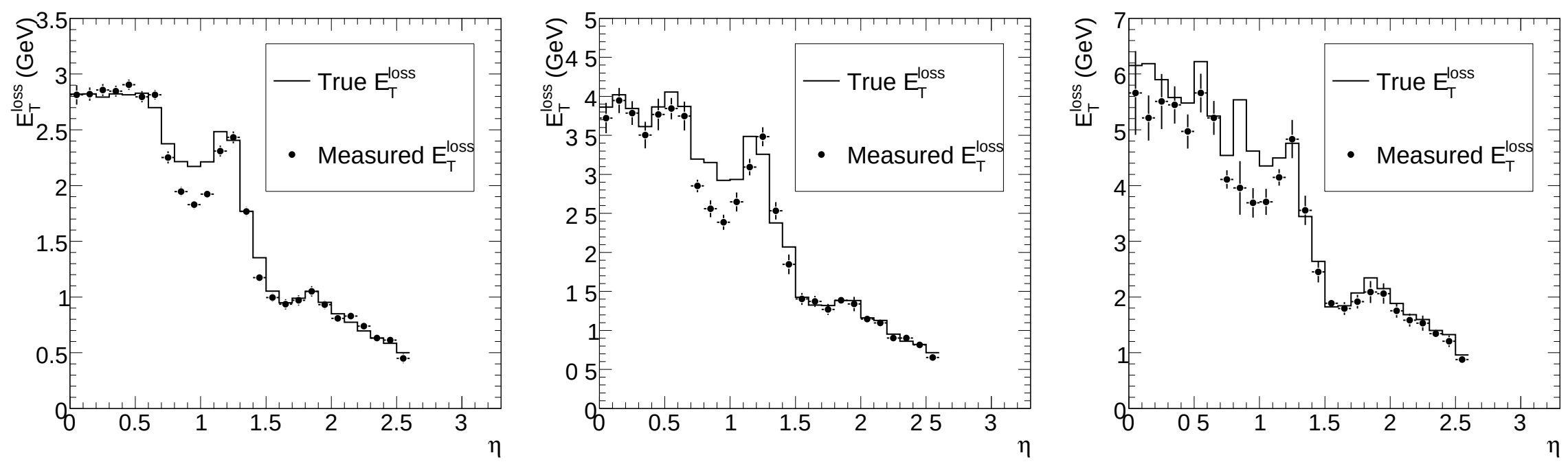

Figure 11: Comparison between the average measured transverse energy deposition (points) and true energy lost between the beam-pipe and the muon spectrometer (line) for muons of momentum 10 GeV (left), 100 GeV (center) and 300 GeV (right). The errors shown are statistical only.

## 3.3 Muon Isolation

The previous studies demonstrate the capabilities of the calorimeters to measure the energy lost by muons. However, these studies were all performed with single muon samples. In real physics samples, muons do not reach the calorimeters alone, but are often accompanied by additional particles that deposit energy in the cells around the muon trajectory and contaminate the muon energy loss measurement. Therefore, in order to determine the energy loss of such muons, isolation criteria must also be defined and optimized for maximum reliability in the energy measurement.

Isolation criteria can be divided into two categories: calorimeter-based and track-based. In muon reconstruction an area is defined around the muon trajectory with a minimum and maximum radius for the purpose of determining calorimeter isolation. This achieves the purpose of excluding the cells where the muon deposits its energy. The size of this inner radius needs to be optimized to collect most of the energy lost by the muon but as little energy as possible from other particles. The energy deposited in the annulus between the inner radius and the outer radius, where the muon deposits little energy, is what the following paragraphs refer to as *isolation energy*. The optimal radii that define this annulus depend on the underlying event and luminosity. However, the muon shower is contained in a small cone of radius $\approx 0.1$ [18]. Therefore, a choice of an inner radius much bigger than 0.1 does not achieve the purpose of collecting the energy deposited by the muon, and it adds noise to the measurement.

A study to determine possible isolation criteria [3] has been performed on a fully simulated $t\bar{t}$ sample, where the $W$ bosons were forced to decay into a muon and a neutrino. Muons produced by the semi-leptonic decays of $b$ quarks tend to be non-isolated, while those from $W$ decays tend to be isolated. In Figure 12 the distribution of the isolation energy for muons originating from quarks and $W$s is shown for the electromagnetic and hadronic calorimeters. The isolation energy inner and outer radii are 0.075 (0.15) and 0.15 (0.30) for the electromagnetic (hadronic) calorimeters, respectively. This reflects their different granularities. The electromagnetic isolation energy is a more powerful discriminant for the annuli radii chosen.

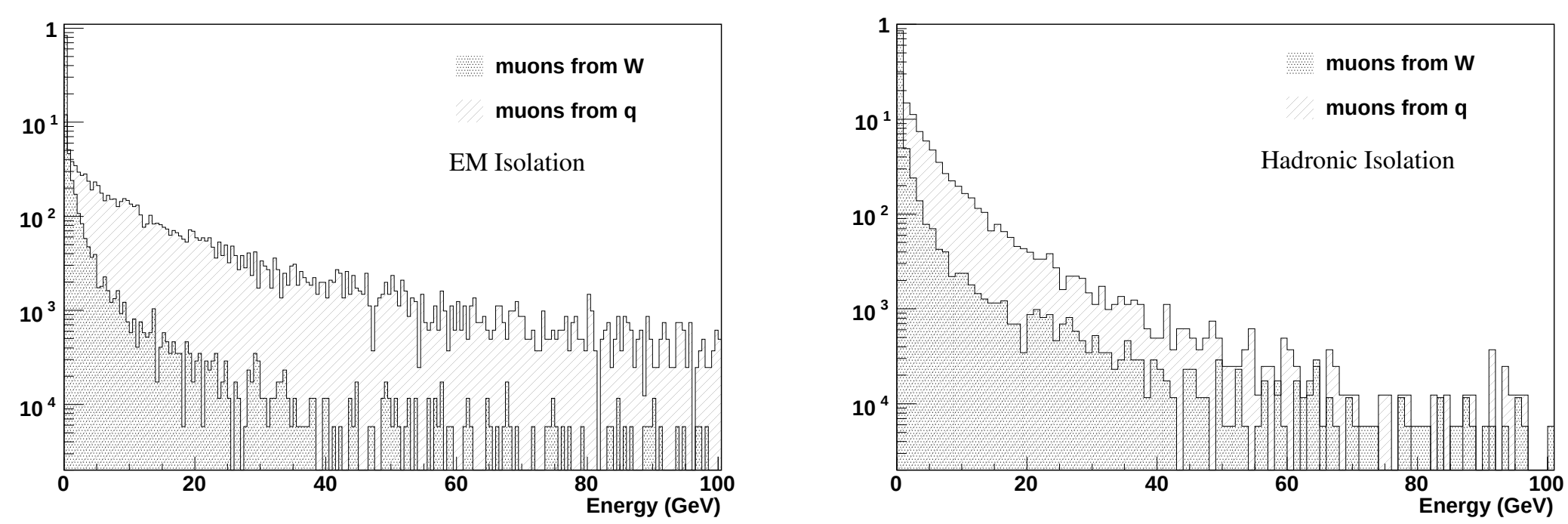

Figure 12: Distribution of the isolation energy in the electromagnetic ($0.075 < \Delta R < 0.15$) (left) and hadronic calorimeters ($0.15 < \Delta R < 0.30$) (right) in muons from a $t\bar{t}$ sample without pile-up.

Based on this figure, for the purpose of the rest of the studies in this section, a cut of 2 GeV on electromagnetic isolation was used to discriminate isolated muons from non-isolated muons. An additional cut of 10 GeV in hadronic isolation was used, even though this cut does not help rejecting non-isolated muons in the vast majority of events. These cuts were relaxed slightly with increasing muon $p_T$ to account for a possible slight increase of the transverse radius of the shower caused by muons in the calorimeters.

Tracking-based criteria can be used to determine isolation cuts independent of calorimeter-based criteria. If used together, they can help eliminate non-isolated muons belonging to highly-collimated jets that escape being identified by calorimeter-based criteria. In Figure 13 the number of inner detector tracks around the muon are plotted for muons from quarks and $W$s in the same samples used for Figure 12. These distributions were obtained for muons that passed the calorimeter-based isolation cuts mentioned above.

The production rate of low-$p_T$ non-isolated muons from $b$-quark decays is expected to be very significant. At the same time, a typical muon from a $W$ or $Z$ decay, will have a $p_T$ of about 40 GeV. Thus, it is unlikely that low momentum muons will be isolated in a $t\bar{t}$ sample. For this reason, all muons with a $p_T$ of less than 15 GeV were automatically tagged as non-isolated and are excluded. In most cases, the muon originating from a $W$ boson is not accompanied by other tracks in the inner detector. In fact, the only track found in the track isolation cone is, essentially, the muon itself as reconstructed in the inner detector. In contrast, several inner detector tracks can be found close to muons originating from quarks. For an isolation cone of $\Delta R$=0.2, the most probable value is three tracks, including the muon itself, but it can be much larger. A cut on tracking isolation has been applied that complements the cut on calorimeter isolation. This cut constrains an isolated muon track to be accompanied by at most one extra track inside the tracking isolation cone. Using the criteria described above (electromagnetic and hadronic calorimeter isolation, track isolation and $p_T^{\mu} > 15$ GeV) approximately 0.2% of the muons originating from $b$ quarks have an energy loss overestimated by more than 6 GeV. On the other hand, 80% of the muons originating from $W$s are tagged as isolated.

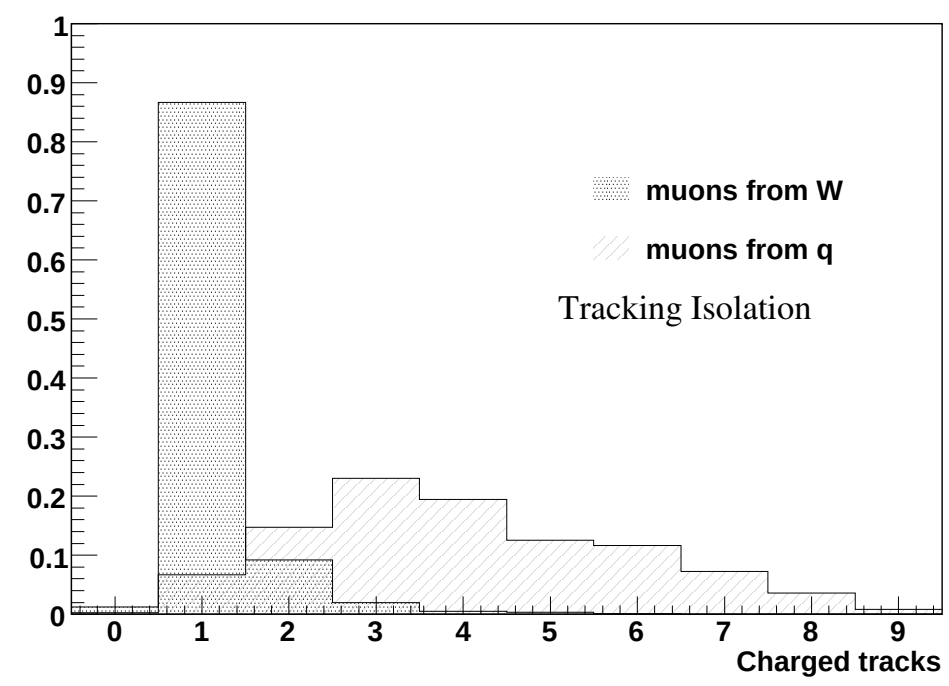

Figure 13: Distribution of the number of inner detector tracks (including the muon track) with $\Delta R < 0.2$ around the muon spectrometer track, after the calorimeter isolation and $p_T$ threshold cuts are applied to muons in a $t\bar{t}$ sample.

All the cuts mentioned above are used by default as isolation criteria in the Muid muon reconstruction algorithm, to establish the contamination of the calorimeter measurement. If the cuts were not chosen tightly enough, non-isolated muons would exhibit an artificial increase in energy loss. If this measurement was then used in muon reconstruction, the reconstructed momentum at the interaction vertex would be artificially increased. This could significantly deteriorate the momentum measurement. A calorimeter measurement of the energy loss will then only make sense if the muon is tagged as isolated. These cuts were considered conservative enough that they could be used by default without inducing biases in the momentum reconstruction [3]. These cuts have not been studied with pile-up or in other samples with an important source of non-isolated muons, like high-$p_T$, $b\bar{b}$ samples. Studies of this type are important in order to set conservative isolation cuts as default for muon reconstruction involving calorimeter measurements.

In addition, it is worth discussing the relationship between muon isolation in reconstruction and muon isolation in physics analyses. While both concepts represent an attempt to determine whether a muon is inside a jet, analysis cuts are also decided on the basis of criteria such as efficiency or fake rate that are not necessarily important for the momentum reconstruction. It is, however, important to emphasize that the optimization of the cuts on reconstruction isolation for specific analyses is possible. It requires, nevertheless, a refitting of the track with analysis-specific cuts; and it should, therefore, only be attempted when the recovery of the Landau energy loss tails is crucial for the analysis and the standard treatment is not adequate.

For muon tagging, however, isolation criteria can overlap with the criteria derived for specific analyses. Isolation studies are necessary to provide a reliable muon tag and do not affect the momentum reconstruction, because the tracking algorithms are independent of the tagger. Default isolation criteria can, for example, be relaxed based on information from the physics sample and the specific analysis. As an example, Figure 14 shows the rejection on the $Zb\bar{b}$ background versus the Standard Model $H(130\text{ GeV}) \rightarrow ZZ^* \rightarrow \mu^+\mu^-\mu^+\mu^-$ efficiency using the calorimeter isolation cuts [19]. At this stage, after a preselection procedure, the four muon candidates have already been selected. Both absolute and normalized (with respect to muon $p_T$) isolation are presented. In these analyses, the isolation energy was defined using a cone of fixed radius. The isolation energy of an event was then defined as the isolation energy of the least isolated muon of the event. The optimum cone radius depends on signal efficiency, with $\Delta R$ of 0.2 being an efficient choice.

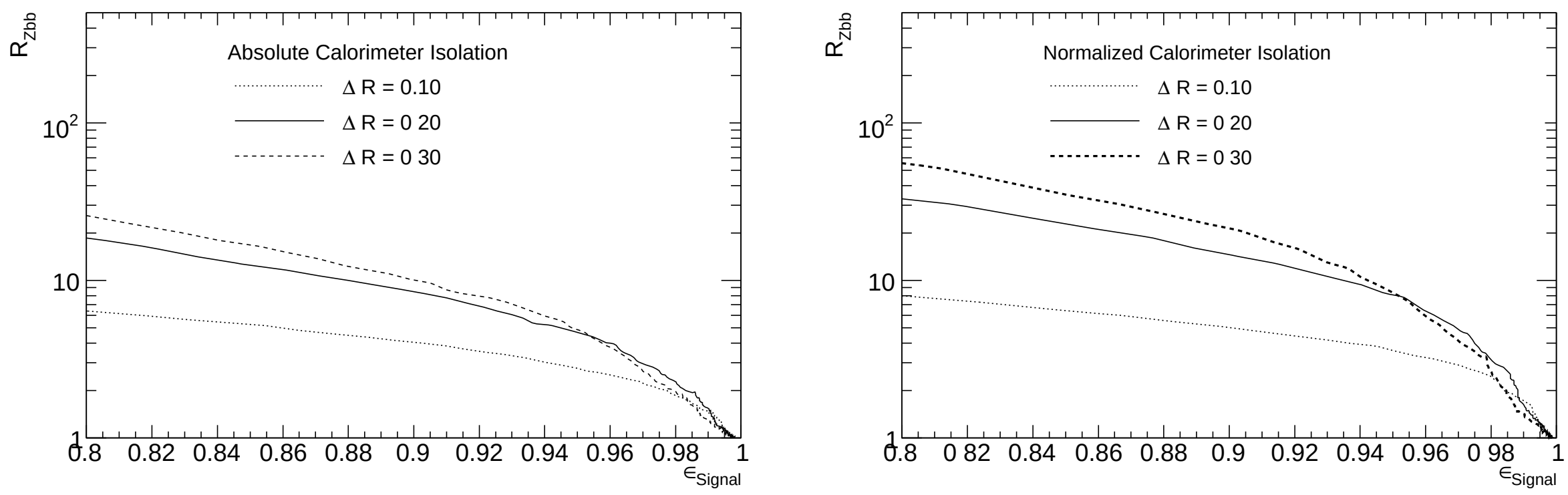

Figure 14: Rejection of the $Zb\bar{b}$ background as a function of the $H(130\text{ GeV}) \rightarrow 4\mu$ signal efficiency. Different radii ($0.1 < \Delta R < 0.3$) are compared for absolute, left, and normalized (with respect to muon $p_T$), right, calorimeter isolation. No pile up events were simulated.

### 3.4 Measurement/Parameterization Combination Methods

To integrate energy loss in tracking algorithms, the energy loss is assumed to be gaussian. If the parameterization is used exclusively to correct for the energy loss, any event in which muons undergo a large energy loss will be incorrectly reconstructed. There is, thus, an advantage in using the parameterizations together with measurements in the calorimeters to optimize the energy loss reconstruction. Here we describe two algorithms developed to use the calorimeter information as well as the energy loss parameterizations for the muon reconstruction algorithms: the *Hybrid Method* [3] and the *Bayesian Method*. The Hybrid Method is used by default after isolation cuts as part of Muid. The Bayesian Method is used if specified by the user as part of Muonboy. By default, Muonboy uses a parameterization of the energy loss only.

The Hybrid Method consists in fully separating the two regions of the Landau distribution: the peak region and the tail region. The calorimetric energy loss measurement is used when the energy deposition is significantly larger than the most probable value (tail region); otherwise the parameterization is used (peak region). The transition point between the two regions is taken as $E_{\text{mpv}} + 2\sigma_{\text{Landau}}$.

The Bayes Method is based on a statistical combination of the parameterization and the measurement in the calorimeters. This combination is performed using Bayes' theorem. This method uses information from the calorimeters, even when the measurement falls in the peak region. When the calorimeter measurement falls in the tail, the results provided by this method are similar to those obtained by the Hybrid Method. If the measurements falls in the peak region, the measurement still constrains the energy loss pdf, improving the energy loss reconstruction resolution. In addition, this method generalizes the selection procedure of the hybrid method, with an event-by-event optimization and for each calorimeter subsystem. This generalization allows also, in principle, an automatic improvement in the energy loss reconstruction as the calorimeter calibration improves.

The full validation of these methods in muon reconstruction with all muon reconstruction effects is shown in the next section. The most significant of these effects are the intrinsic resolutions of the inner detector and muon spectrometer. However, to validate the methods standalone, it is necessary to do it in a model free of these reconstruction effects.

The Hybrid Method has been validated through the energy loss reconstruction distributions in the ATLAS full simulation using the muon kinematics from the simulation. The ratio of the energy loss resolution for the Hybrid Method, $\sigma_{\text{hybrid}}$, with respect to the parameterization alone, $\sigma_{\text{param}}$, is presented in Figure 15. The resolution is defined as the square root of the variance of the energy loss resolution. For low-$p_T$ values the ratio is close to unity, as expected due to the smaller fraction of events in the Landau tail. For increasing $p_T$ values the ratio decreases, approaching 30% at $p_T = 1$ TeV. Thus, using the Hybrid Method results in a significant improvement in the energy loss estimation with respect to the parameterization alone.

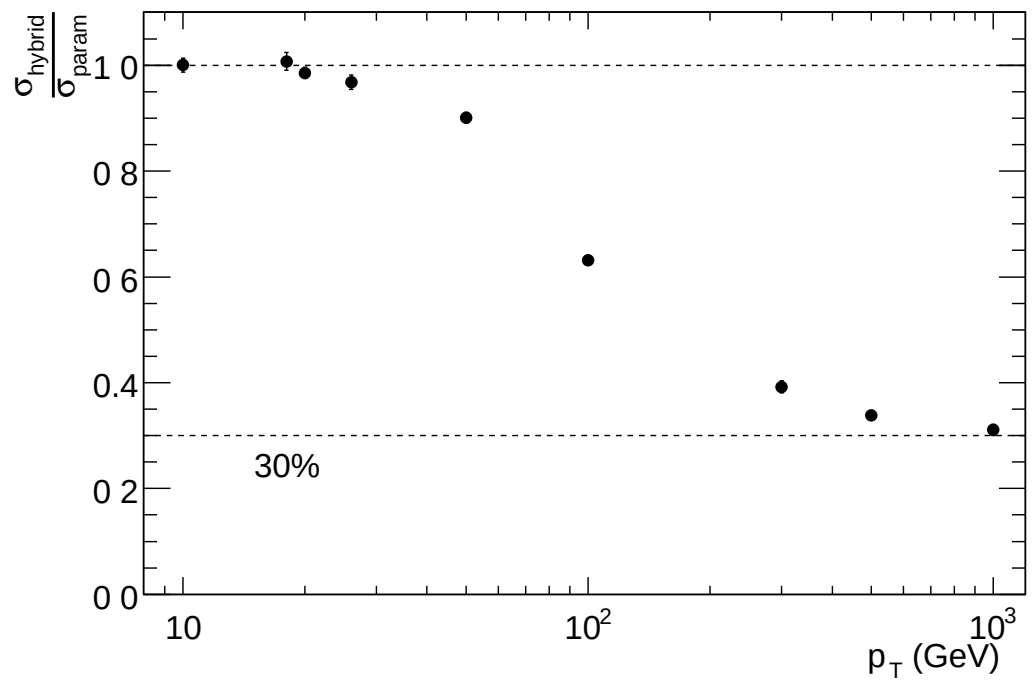

Figure 15: Ratio of the energy loss resolution for the Hybrid Method with respect to the parameterization alone for single muons.

In addition, the performance of the Bayesian Method has been studied in a toy model under the assumption that the calorimeter calibration is understood. Muons of 1 TeV were shot through a block of matter representative of one of the samplings of the hadronic calorimeter. A perfect muon spectrometer was assumed as it is done for the Hybrid Method above. Figure 16 shows some results from this study that demonstrate the potential of the Bayesian Method to reconstruct energy loss. The use of the measurement by itself biases the energy loss reconstruction. The Bayesian Method, on the other hand, shows no biases. Incidentally, there are no biases in the energy loss reconstruction if the muon spectrometer measurement is coupled to the energy loss reconstruction in these studies. In addition, the $E_{\text{loss}}^{\text{true}} - E_{\text{loss}}^{\text{reco}}$ distributions show that the resolution obtained with the Bayesian Method is better than that obtained using the parameterization or the measurement only.

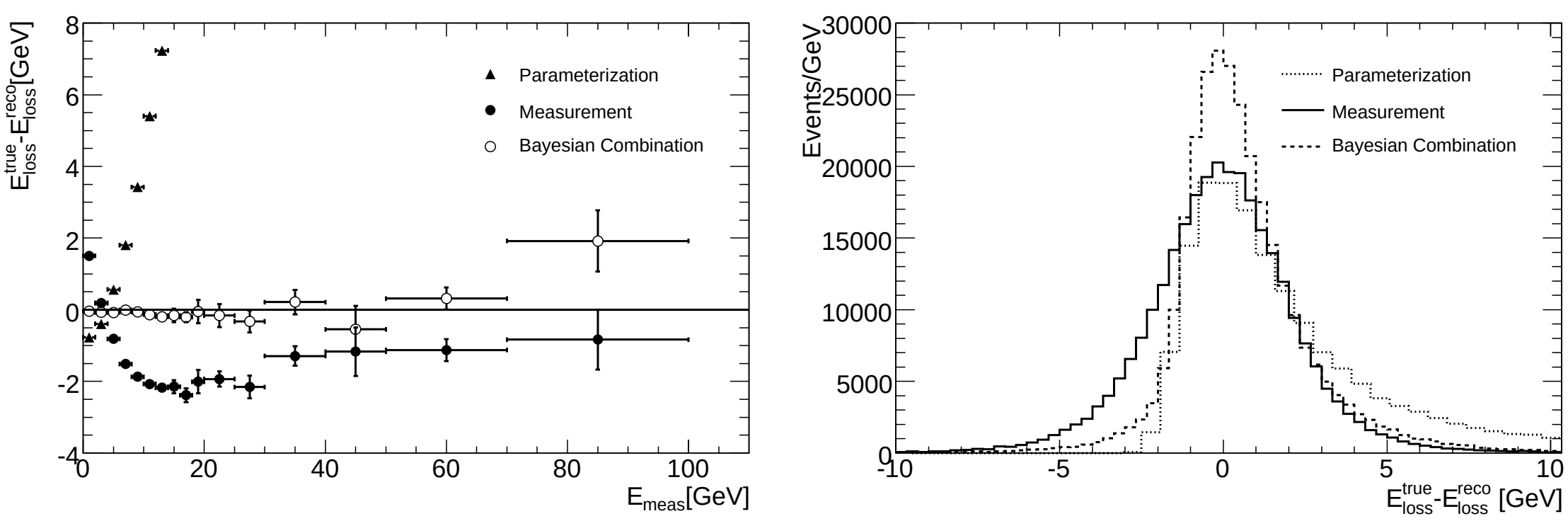

Figure 16: Demonstration of the potential of the statistical method to reconstruct the energy loss. The left plot shows the bias in the energy loss reconstruction, while the right plot shows the $E_{\text{loss}}^{\text{true}} - E_{\text{loss}}^{\text{reco}}$ distribution. Both plots compare the energy loss reconstruction using the parameterization only (triangles, dotted line), the measurement only (filled circles, solid line) and both statistically combined through Bayes' theorem (open circles, dashed line).

These studies prove the potential of these two methods. Now their performance is analyzed when they are used as part of the reconstruction software.

## 3.5 Impact of the Energy Loss Corrections in Reconstruction

Energy loss estimates must be validated as part of the muon reconstruction algorithms. Effects such as the resolution of muon reconstruction, biases intrinsic to the track transport or the effects of gaussian assumptions will be coupled to the energy loss reconstruction. However, the studies shown here are still important for understanding the energy loss correction. They also allow for an investigation of which data samples are most sensitive to an incorrect estimation of the energy loss.

In Figure 17, two parameters that characterize the gaussian distributions $1/p_T^{\rm reco} - 1/p_T^{\rm true}$ are shown. In these plots, the label "MS" corresponds to tracks from fits in the muon spectrometer only. The label

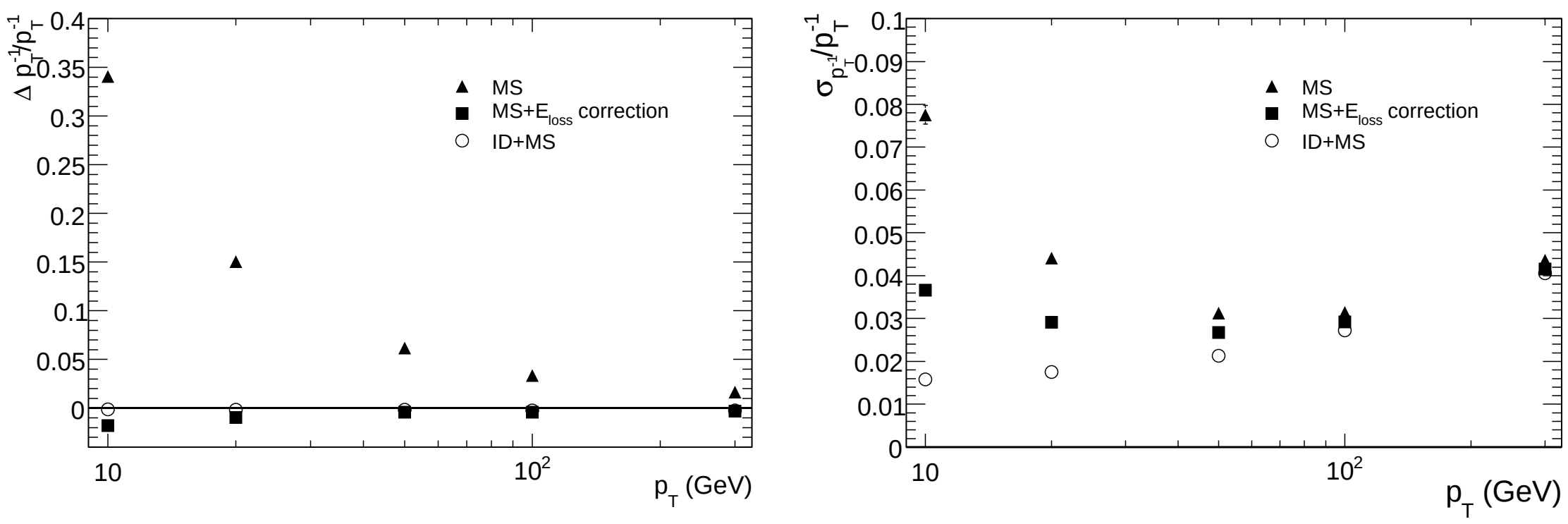

Figure 17: Left: Muon reconstruction bias for different algorithms as a function of muon $p_T$. Right: Muon reconstruction resolution for different algorithms as a function of muon $p_T$. These plots were produced with the Muonboy/STACO algorithms for muon reconstruction [1], but similar performance is obtained with the MOORE/Muid algorithms [2, 7].

"MS+$E_{\rm loss}$ correction" refers to tracks reconstructed at the muon spectrometer and transported to the interaction point (IP), applying an energy loss correction (parameterized or hybrid). For Figure 17 the energy loss correction was calculated using the Muonboy parameterization only. The label "MS+ID" refers to tracks reconstructed with the muon spectrometer and the inner detector. To obtain these combined tracks, the energy loss correction needs to be considered in the fit. The distributions are calculated in $1/p_T$-space because the muon spectrometer reconstruction and inner detector reconstruction have gaussian fluctuations in $1/p_T$. These plots refer to muons reconstructed in the barrel ($|\eta| < 1.0$) of the muon spectrometer. In the left plot, the bias in the reconstruction is shown, defined as the mean of the distribution

$$\frac{\delta p_T^{-1}}{p_T^{-1}} = p_T^{\rm true,IP}\left(\frac{1}{p_T^{\rm reco}} - \frac{1}{p_T^{\rm true,IP}}\right), \quad (4)$$

with $p_T^{\rm true,IP}$ being the true $p_T$ of the muon at the IP. Clearly, in the absence of an energy loss correction, this reconstruction can be highly biased. Such a bias also causes a degradation in the resolution. The application of an energy loss correction reduces this bias and improves the resolution. Further bias reduction and resolution improvement is obtained by using the inner detector together with the muon spectrometer to reconstruct the muon track. However, the combination of inner detector tracks and muon spectrometer tracks is only possible if a proper energy loss correction exists.

Figure 18 shows the invariant mass resolution ($M_{\text{inv}}^{\text{reco}} - M_{\text{inv}}^{\text{true}}$) for $Z \rightarrow \mu\mu$ and $Z'(1000 \ \text{GeV}) \rightarrow \mu\mu$ samples. The events include a generation cut that requires the $p_T$ of both decay muons to be above 7 GeV for $Z$ decays and above 20 GeV for $Z'$ decays. Inner detector and muon spectrometer tracks were required for both reconstructed muons. The effect of the energy loss correction is most significant in the $Z$-mass reconstruction. If the energy loss is not included a shift of about 7 GeV in the mass peak and a significant deterioration in the resolution are visible. These effects are much less pronounced in the reconstruction of the $Z'$ peak. This happens because the more energetic muons from the $Z'$ lose a smaller fraction of their total energy as they pass through the calorimeters.

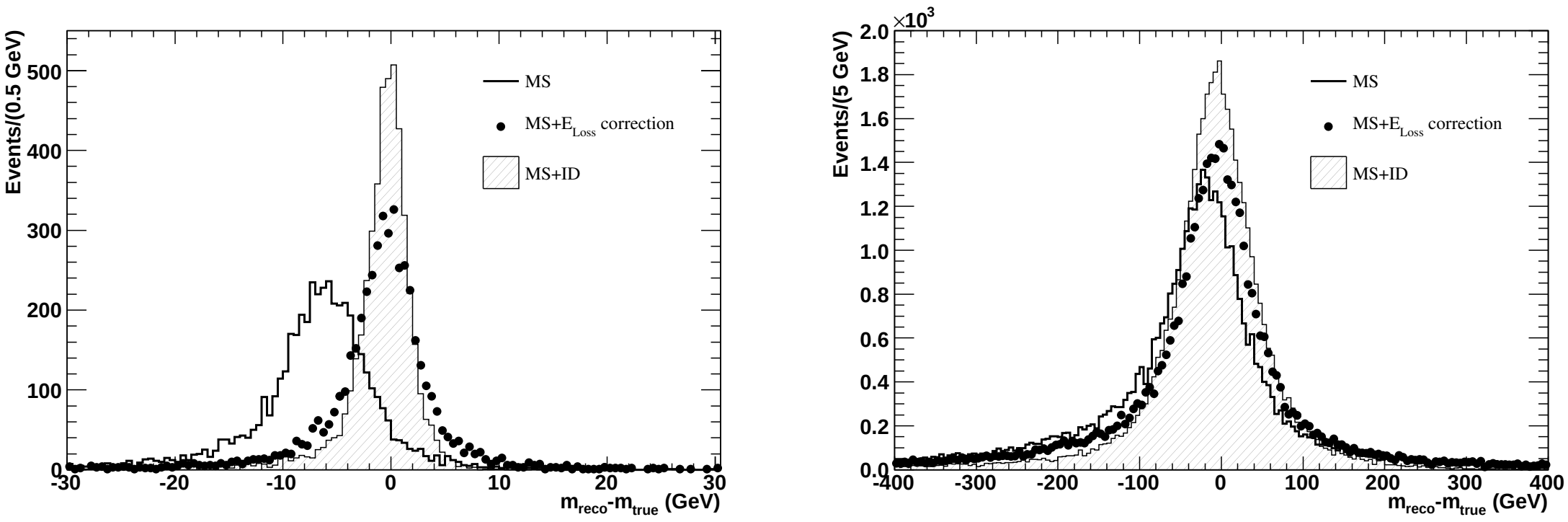

Figure 18: Left: Reconstruction resolution of the $Z$ peak for different algorithms. Right: Reconstruction resolution of the $Z'$ peak for a $Z' \rightarrow \mu\mu$ of mass 1 TeV for different algorithms. These plots were produced with the MOORE/Muid algorithms for muon reconstruction [2, 7], but similar performance is obtained with the Muonboy/STACO algorithms [1].

An additional improvement in the $Z$ and $Z'$ resolution is possible if the energy loss correction uses the calorimeter measurement [20]. This is demonstrated in Figure 19, where a comparison is shown for the muon spectrometer reconstruction with energy loss correction with and without the calorimeter measurement. The same samples as for Figure 18 were used. A few events are recovered from the tails and populate the peak region. This results in $\approx 8\%$ resolution improvement when the inner detector hits are not used. If the inner detector hits are used the resolution improves by $\approx 4\%$, showing that the combined fit is less sensitive to the energy loss correction. These plots were produced with the MOORE/Muid algorithms for muon reconstruction using the Hybrid Method (see Section 3.4). A similar performance is expected using the Bayesian Method, implemented in the ATLAS tracking.

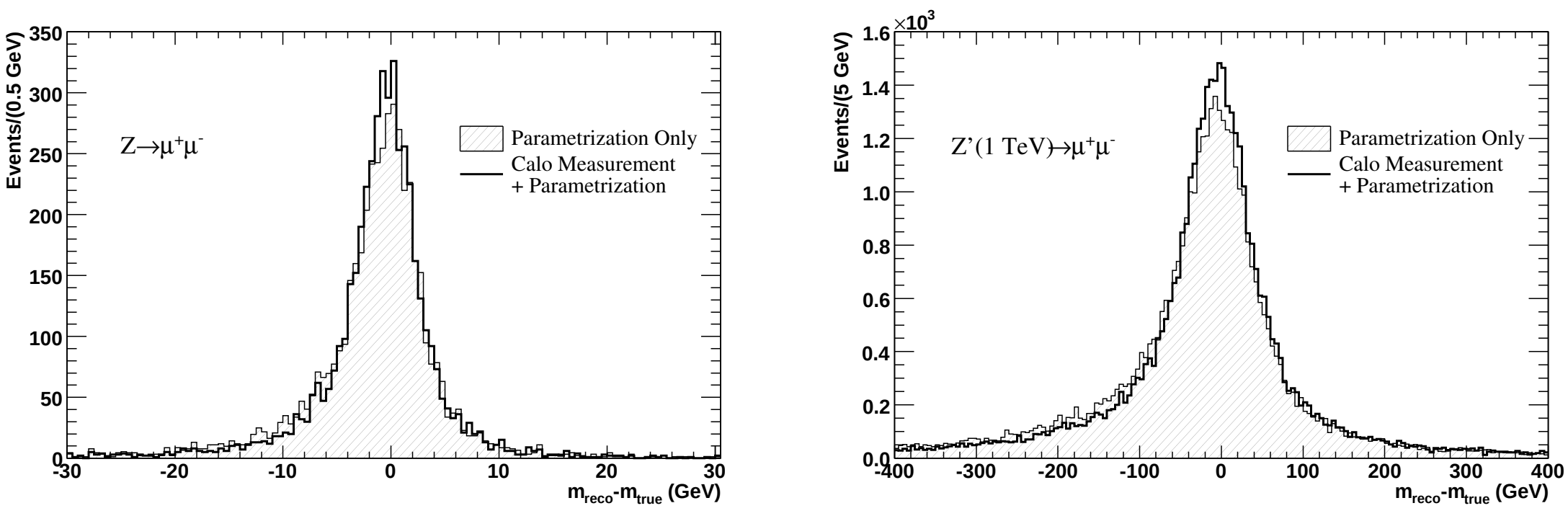

Figure 19: Left: Reconstruction resolution of the $Z$ peak for an algorithm using muon spectrometer standalone tracks and the parameterized energy loss correction (filled histogram) and an algorithm using a combination of a parameterization and the calorimeter measurement for the energy loss correction (empty histogram). Right: Reconstruction resolution of the $Z'$ peak for a $Z' \rightarrow \mu\mu$ of mass 1 TeV for the same algorithms.

# 4 Tagging of Muons in the Calorimeters

In this section the different calorimeter-based muon identification algorithms are described. Section 4.1 provides a detailed description of the algorithm that is currently part of the standard reconstruction; and Section 4.2 illustrates its performance in different physics samples.

These algorithms have been developed with the main goal of complementing the muon spectrometer in two ways: recovering muons with low transverse momentum ($p_T = 2$-5 GeV), and in the regions of limited spectrometer acceptance (especially in the $\eta \sim 0$ region). For completeness, algorithms used for commissioning or triggering are also discussed.

There are two types of calorimeter-based muon tagging algorithms. Their main difference lies in how they initiate the muon search. The calorimeter-seed algorithms search for muons looking at the measured energy. Cells with energy depositions inside some energy range are used as seeds. The lower limit of this range is known as the *initiation* threshold. The cluster of cells used to identify the muon is then built up by adding cells around the seed cell whose energy is above a second lower threshold, so called *continuation* threshold. At the end of the clustering, the $\eta$ and $\phi$ directions of the reconstructed cluster can be used to match a track in the inner detector. There are two algorithms that correspond to this description:

- `LArMuID` is based on a topological clustering algorithm used by ALEPH [21]. A topological clustering algorithm groups neighboring cells whose energy is above a given threshold. Therefore, the resulting clusters have a variable number of cells. This algorithm was used to find muons using the electromagnetic calorimeter data during the test beam and the cosmic commissioning data analysis. It builds the cluster from a seed cell in the middle sampling of the electromagnetic calorimeter. Then, it creates the cluster adding another cell (if any) adjacent in $\phi$ above the continuation threshold. Due to the accordion structure of the electromagnetic calorimeters there are no more than two adjacent cells in $\phi$ that can share the muon signal, thus the clusters consist of at most two cells. The efficiency was measured with a muon beam during the combined test beam as the fraction of events with a reconstructed muon cluster. The efficiency and the probability to generate a fake muon from noise fluctuations were evaluated as a function of both thresholds. The clustering algorithm inherently biases the energy reconstruction, so the lower the thresholds the better the estimation of the reconstructed energy for the muon. The spectrum of energies collected has been compared to and shown agreement with GEANT4.

- `TileMuId` is simple and fast and is used for triggering purposes. Its clustering methods are similar to those of `LArMuID`. This algorithm, however, runs by default as part of the reconstruction software. It starts with a search for a "candidate" muon in the cells belonging to the last TileCal sampling, where the muon gives the clearest signature, due to the screening effect of the two previous samplings. If the measured energy is between a lower and and upper threshold, it uses the $\eta$ and $\phi$ coordinates of the cell to look for another cell with energy within the same thresholds in the central sampling. If both searches are successful it looks for a third cell with energy within the thresholds in the first sampling. When cells with energies within the thresholds are found in all three samplings, the candidate is confirmed to be a muon. Performance studies of `TileMuId` can be found in [22].

There are two track-seed algorithms that extrapolate inner detector tracks through the calorimeter identifying those matching the energy deposition pattern of a muon. The track-seed algorithms use no tracking information from the muon spectrometer. The first track-seed algorithm, `CaloMuonTag`, will be described in detail in the next section.

The second track-seed algorithm, `CaloMuonLikelihoodTool`, builds a likelihood ratio to discriminate muons from pions. The likelihood discriminant is built out of different energy ratios in order to capture the global features of the energy depositions. The discrimination power of these ratios varies both as a function of the momentum of the particles considered, and as a function of $\eta$. Therefore three bins in $\eta$ (barrel, crack, end-cap) and three bins in momentum (0-10 GeV, 10-50 GeV, 50-100 GeV) are used, and a different set of ratios is selected for each of the 9 regions. The likelihood ratio for the muon candidate is then defined as

$$\mathscr{L}(x_1,...,x_N) = \prod_{i=1}^{N} \frac{P_i^{\mu}(x_i)}{P_i^{\mu}(x_i)+P_i^{\pi}(x_i)}. \tag{5}$$

Where $P_i^{\mu}$ and $P_i^{\pi}$, $i = 1,...,N$ are the pdf for the energy ratios. The performance of this algorithm is still being studied, however, the first results show that it is comparable to the performance of `CaloMuonTag` discussed in Section 4.2.

## 4.1 CaloMuonTag

`CaloMuonTag` extrapolates inner detector tracks through the calorimeters, collecting the energy in the cell closest to the extrapolated track for each traversed sampling. The muon can deposit energy in more than one cell in the hadronic calorimeter, but the probability of this happening is rather low, as illustrated in Figure 20. For the purpose of this algorithm, it is enough to assume that all the energy deposited by the muon is localized in the central cell. This also minimizes the electronic noise, which is particularly large in the HEC.

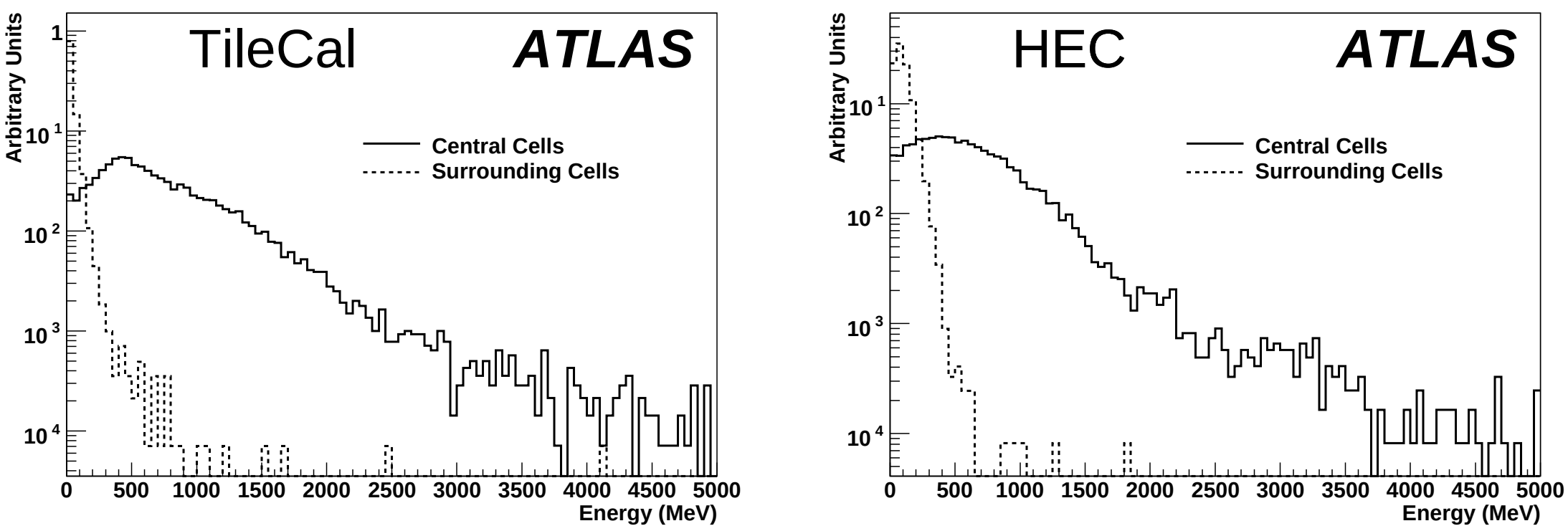

Figure 20: Energy found in the cell traversed by the extrapolated track (solid line) and the surrounding cells (dashed line) in the TileCal (left) and in the HEC (right). Distributions obtained for momentum 100 GeV muons.

A track preselection is made to reduce the number of fakes in the output of the algorithms. In addition, this preselection reduces the time needed by the algorithm to run on events with high track multiplicity. The following cuts in $p_T$, and transverse isolation energy ($E_T^{\text{iso}}$) inside a cone of 0.45 are applied:

- $p_T > 2$ GeV and $E_T^{\text{iso}} < 10$ GeV for tracks pointing to the barrel ($\eta < 1.6$).
- $p_T > 3$ GeV and $E_T^{\text{iso}} < 8$ GeV for tracks pointing to the end-cap.

Tracks are rejected if any of the collected energies are above veto values defined for each sampling. As most fakes are seeded by low-$p_T$ tracks, more stringent cuts can be set for low-$p_T$ ($<$ 10 GeV) track candidates. These cuts can be relaxed for track candidates of higher $p_T$.

Once calorimeter cells along the muon trajectory have been identified, the algorithm determines the lower threshold energy cut that should be used for the tagging as a function of $\eta$:

- $E_{\text{th}} = \frac{E_0^{\text{barrel}}}{\sin^2\theta}$ for $|\eta| < 1.7$,
- $E_{\text{th}} = \frac{E_0^{\text{end-cap}}}{(1-\sin\theta)^2}$ for $|\eta| > 1.7$,

where $\theta$ is the polar angle. The values of $E_{\text{th}}$ from these two equations roughly follow the shape of the measured energy distributions, which increases with the path length of the muon in the cell.

Energy depositions in the last sampling of the calorimeters give the most reliable muon signals. However, due to the gap between the TileCal barrel and extended barrel modules, and the transition region between the TileCal and the end-cap (HEC) calorimeters, it is necessary to look in the two previous samplings to obtain a good efficiency throughout $\eta$. For this reason, if the energy in the last sampling, or one of the two previous samplings depending on the $\eta$ of the track, is above the threshold cut, $E_{\text{th}}$, the track is tagged as a muon. A different tag is given depending on which sampling passes the threshold cut.

## 4.2 Performance

In this section the performance of `CaloMuonTag` is analyzed. The performance of `CaloMuonTag` is studied in some relevant physics samples:

- $pp \to J/\psi \to \mu\mu$. A direct production of a $J/\psi$ decaying to two muons with the following cuts at generation level: one muon with $p_T > 6$ GeV and the other with $p_T > 4$ GeV.
- $H \to ZZ^* \to 4\ell$. A Higgs generated with an invariant mass of 130 GeV is forced to decay into two $Z$'s (one of them offshell) that decay leptonically. Only events with four muons are used for reconstructing the Higgs peak, but all events are used for the efficiency/fake rate calculation.
- $t\bar{t}$ . A sample of pair produced top quarks with all decay channels allowed.
- $Zbb \to 4\ell$. A $Z$ is produced in association with 2 $b$ quarks, and is forced to decay into two charged leptons. The $b$ quarks are also forced to decay into electrons or muons.

All samples were generated with pile-up with a safety factor of 5, i.e. five times nominal value expected at a luminosity of $10^{33} cm^{-2}s^{-1}$. Except for the $pp \to J/\psi \to \mu\mu$ sample, where a safety factor of 2 was used.

Figure 21 shows the performance of the calorimeter muon tagger algorithm on selected samples. The vertical axis on the left shows the efficiency (top distributions). The vertical axis on the right (in red) shows the fake rate (the number of misidentified tracks per event), represented by the shaded distribution at the bottom of the plots. The efficiency (fake rate) is defined as the fraction of muons that are found by the algorithm and (not) matched with a true MC muon. Muons from minimum-bias interactions were not included in the efficiency calculation.

For the $t\bar{t}$ ($Zbb \to 4\ell$) sample, the algorithm performs well in identifying isolated muons from $W$'s ($Z$'s). However, due to the isolation and veto cuts applied to reduce the number of fakes, the efficiency for non-isolated muons from $b$ quarks is very poor, affecting the overall efficiency.

To reduce the fake rate due to the low-$p_T$ tracks in the end-caps, some efficiency in that region needs to be sacrificed. The "peaks" in the fake rate in $\eta$ match the regions where the acceptance of the last calorimeter sampling is limited, and the two previous samplings are used for muon identification. Due to the higher electronic noise, `CaloMuonTag` presents a higher fake rate in the HEC than in the TileCal. The results for these samples, and for dijet samples generated with different $p_T$ transfers at the hard scattering interaction are summarized in Table 1.

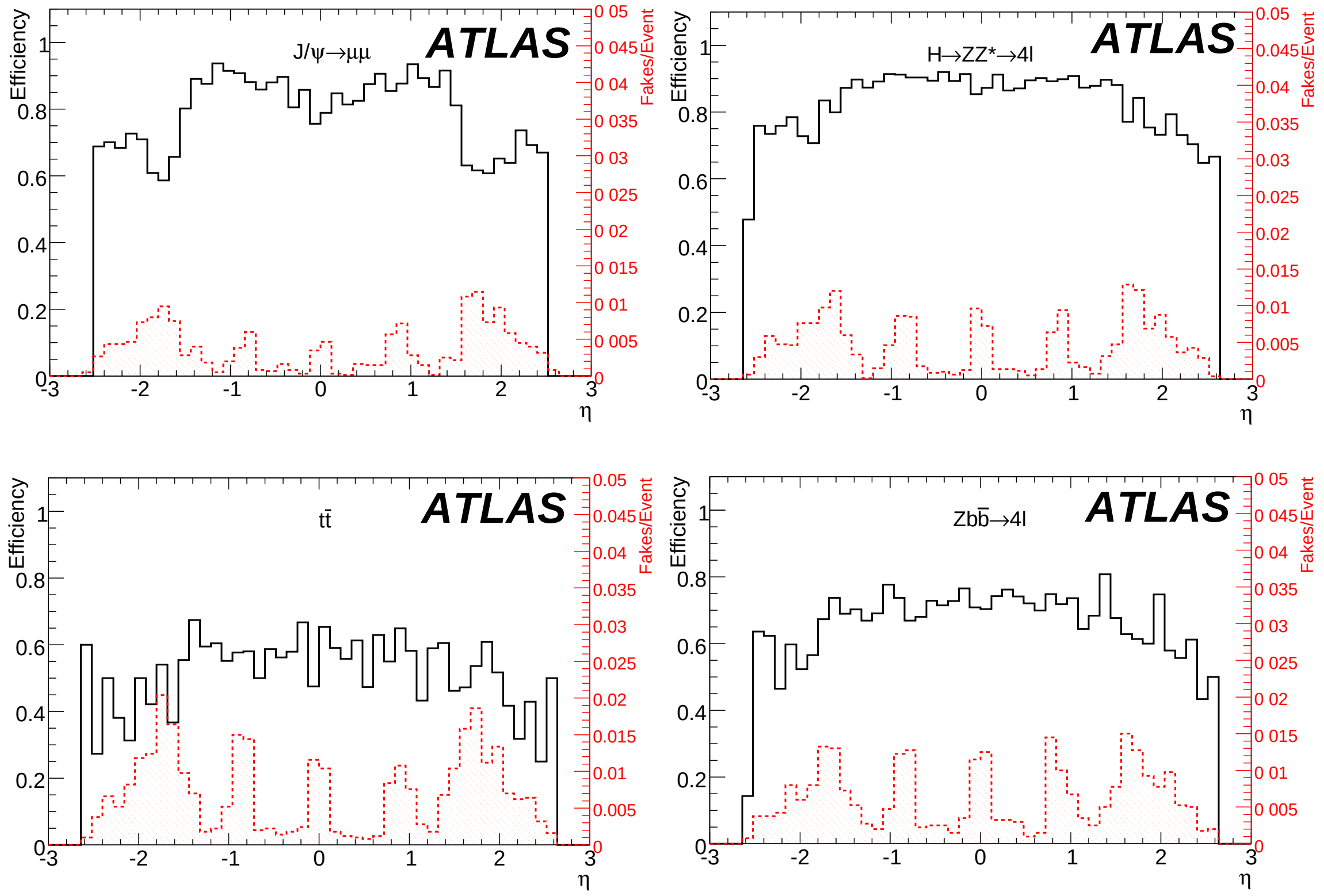

Figure 21: Efficiency (and fakes per event, right axis in red and shaded histograms) vs $\eta$ for different samples. Top left: $pp \to J/\psi \to \mu\mu$. Top right: $H(130) \to ZZ^* \to 4\ell$. Bottom left: $t\bar{t}$. Bottom right: $Zbb \to 4\ell$.

| | $J/\psi \to \mu\mu$ | $H \to 4\ell$ | $t\bar{t}$ | $Zbb \to 4\ell$ | 1120-2240 | 560-1120 | 280-560 | 70-140 | 17-35 |
|---|---|---|---|---|---|---|---|---|---|
| Eff. | 0.80 | 0.86 | 0.54 | 0.69 | - | - | - | - | - |
| f/e | 0.05 | 0.09 | 0.16 | 0.14 | 0.11 | 0.12 | 0.12 | 0.13 | 0.12 |

Table 1: Summary of the efficiencies and fakes per event (f/e) for different physics processes and dijet samples (top numbers show the ranges of $p_T$ transfers at the hard scattering interaction, in GeV).

Finally, Figures 22 and 23 are used to evaluate the performance improvement when using calorimeter muons to reconstruct the Higgs and the $J/\psi$ mass. The plots on the left show the invariant mass reconstructed with muons from a combined muon reconstruction algorithm, which makes use of both inner detector and muon spectrometer tracks. The plots on the right show the invariant mass including muons tagged by `CaloMuonTag` with momenta reconstructed by the inner detector.

To obtain the plots in Figure 22 the same set of cuts were used as in the $H(130) \to ZZ^* \to 4\ell$ stud-ies [19]. The increase on the number of reconstructed events was achieved by adding an extra muon found by the `CaloMuonTag` algorithm during the muon preselection. The extra muon was requested to be found in the last sampling of the TileCal and within the $|\eta| < 0.1$ region. No loss in mass resolution or shift in the mean of the mass peak are observed between the two selected muon samples. The acceptance gap around $|\eta| < 0.1$ represents 4% of the $|\eta| < 2.5$ region covered by the combined muon reconstruction. Since, four muons are recontructed in this analysis, the total efficiency loss due to the gap is 16%. This results shows that calorimeter identification can recover almost all of the lost events (14.9%).

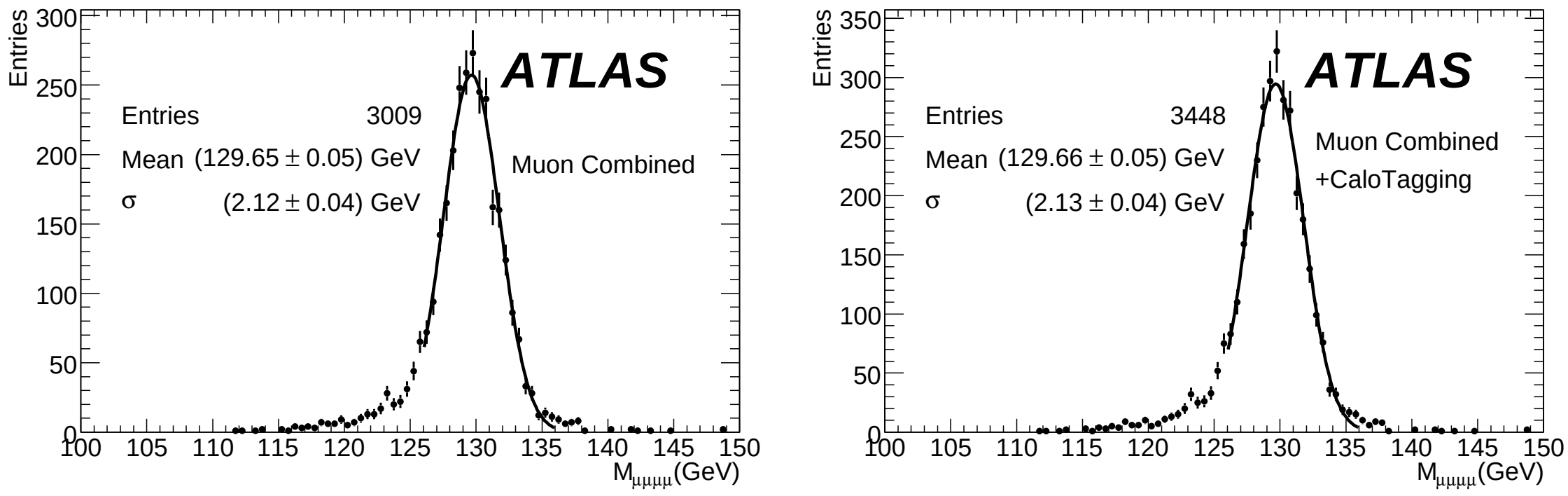

Figure 22: Reconstructed Higgs peak in the $H \to 4\ell$ invariant mass reconstruction for the standard combined muons (left) and for combined muons together with inner detector muons tagged by `CaloMuonTag` in the $\eta$ region $|\eta| < 0.1$ (right).

For the plots shown in Figure 23 the muons used to reconstruct the invariant mass peak were matched to the Monte Carlo truth. In this case, all muons identified by the calorimeter were added to the combined reconstruction muons. Again, no loss in mass resolution or shift in the mean of the mass peak are observed between the two selected muon samples.

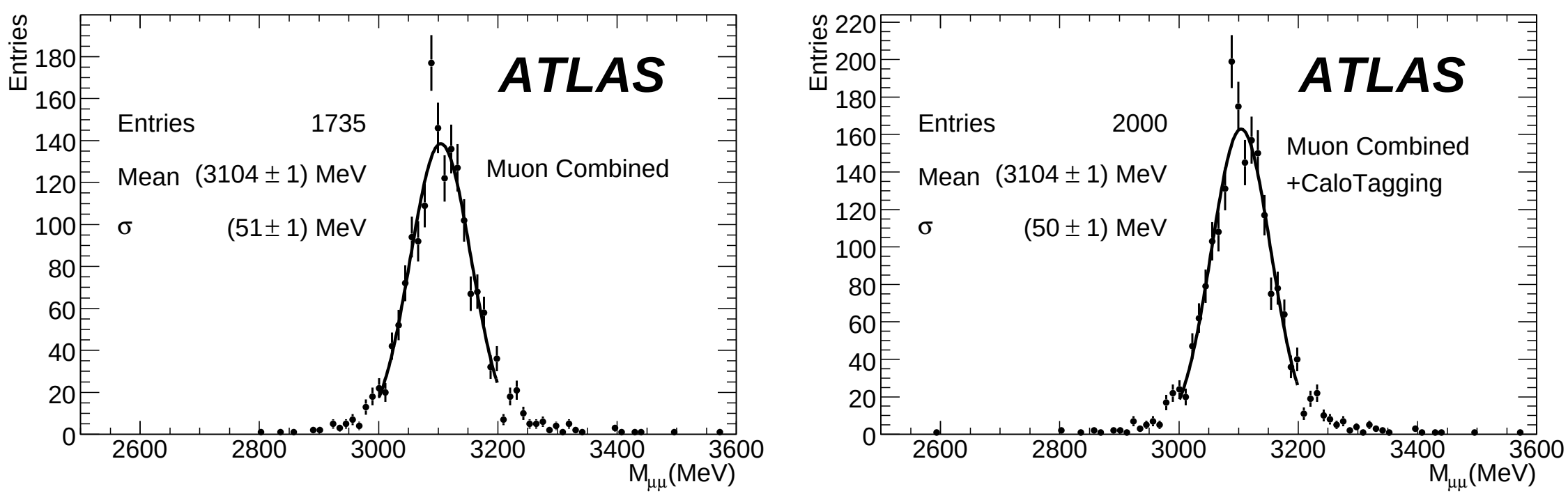

Figure 23: Reconstructed $J/\psi$ peak in the $J/\psi \to \mu\mu$ invariant mass reconstruction for standard combined muons (left) and for combined muons together with inner detector muons tagged by `CaloMuonTag` in the $\eta$ region $|\eta| < 0.1$ (right).

`CaloMuonTag` shows very good efficiency and acceptable fake rate for a high-$p_T$ analysis like $H \to ZZ^* \to 4\mu$. An additional 15% of events were reconstructed when muons identified by `CaloMuonTag` were added to the muons found by the standard reconstruction algorithm. In low-$p_T$ analyses, such as $J/\psi \to 2\mu$, tighter selection cuts need to be applied to keep an acceptable fake rate. This reduces the efficiency of the tagging algorithms. However, an additional 15% of events were reconstructed when muons from CaloMuonTag were also used.

## 5 Conclusion

This document reviews the current status of the understanding of muon energy loss in the ATLAS calorimeters. Although energy losses and their distribution along the muon track have a very small impact on muon reconstruction inside the muon system, they play an important role in the transport of a reconstructed muon track to the beam pipe. During this backtracking, the muon momentum can be corrected using the energy measured in the calorimeter cells traversed. This procedure is justified for muons that have a catastrophic energy loss. In most cases, the muon momentum can be corrected using a parameterization of energy loss, estimated from the reconstructed momentum and the amount and the nature of the material traversed by the reconstructed trajectory. Techniques that attempt to combine the measurement with the parameterization to improve the energy loss estimate have been developed and validated. A performance improvement is achieved in some important analyses through the use of these techniques.

Muon tagging algorithms that use calorimeter measurements and track information have been presented. These algorithms have been developed with the main goal of complementing the muon spectrometer in two ways: recovering muons with very low transverse momentum, and in the regions of limited spectrometer acceptance, especially in the $\eta \sim 0$ region. The performance of the calorimeter muon tagger algorithm in a few relevant data samples has been shown.

# In-Situ Determination of the Performance of the Muon Spectrometer

**Abstract**

The ATLAS muon spectrometer consists of three layers of precision drift-tube chambers in a toroidal magnetic with a field integral between 2.5 and 6 Tm. Muon tracks are reconstructed with 97% efficiency and a momentum resolution between 3% and 4% for 10 GeV< $p_T$ <500 GeV and better than 10% for transverse momenta up to 1 TeV. In this note, the performance of a perfectly calibrated and aligned muon spectrometer will be reviewed and the impact of deteriorations of the magnetic field, the calibration and misalignment of the muon chambers on the performance will be discussed. The main part of the note describes how the performance of the muon spectrometer can be determined using dimuon decays of $Z$ bosons and $J/\psi$ mesons.

## 1 Introduction

Muons with a transverse momentum[1] $p_T$ greater than 3 GeV are detected in the ATLAS muon spectrometer, which is designed to measure muon momenta with a resolution between 3% and 4% for a range of transverse momenta of 10 GeV< $p_T$ <500 GeV and better than 10% for $p_T$'s up to 1 TeV. The muon spectrometer consists of a system of superconducting air-core toroid coils producing a magnetic field with a field integral between 2.5 and 6 Tm [1]. Three layers of chambers are used to precisely measure muon momenta from the deflection of the muon tracks in the magnetic field (see Figure 1). Three layers of trigger resistive-plate chambers (RPC) in the barrel and three layers of fast thin-gap chambers (TGC) in the end caps of the muon spectrometer are used for the muon trigger. The trigger chambers measure the muon tracks in two orthogonal projections with a spatial resolution of about 1 cm. The precision measurement of the muon trajectory is performed by three layers of monitored drift-tube (MDT) chambers in almost the entire muon spectrometer and by cathode-strip chambers (CSC) in the innermost layer of the end caps at $|\eta| > 2.2$. The precision muon chambers provide track points with 35 $\mu$m resolution in the bending plane of the magnetic field. The goal of a momentum resolution better than 10% up to 1 TeV scale requires the knowledge of the chamber positions with an accuracy better than 30 $\mu$m in addition to the high spatial resolution of the chambers. This is achieved by a system of optical alignment monitoring sensors [1].

In the first part of this note, we review the performance of a perfectly calibrated and aligned muon spectrometer and discuss the dependence of the performance on the knowledge of the following quantities: the magnetic field, the calibration of the chambers, the alignment of the chambers, and the accuracy of the determination of the energy loss of the muons in the calorimeters. In the second and main part, we describe how the performance of the muon spectrometer can be determined by means of dimuon decays of $Z$ bosons and $J/\psi$ mesons.

## 2 Performance of a perfect and deteriorated spectrometer

### 2.1 Muon reconstruction

The muon spectrometer measures the momenta of charged particles at the entrance of the muon spectrometer. The energies lost by the muons on the passage through the calorimeters have to be added to the energy measured at the entrance of the muon spectrometer in order to obtain the muon momentum at the primary vertex. This reconstruction strategy is called *stand-alone* muon reconstruction. In order to correct for the energy loss, the expected average energy loss is used as a first estimation; in a second step the energy deposition measured in the calorimeters is used to account for the large energy losses of highly energetic muons due to bremsstrahlung and direct $e^+e^-$ pair production. One speaks of *combined* muon reconstruction when the momentum measurement of the inner detector is combined with the stand-alone reconstruction. In this note, muon momenta will always be given at the $pp$ interaction point.

[1]The transverse momentum is defined as the components of momentum in the transverse plane.

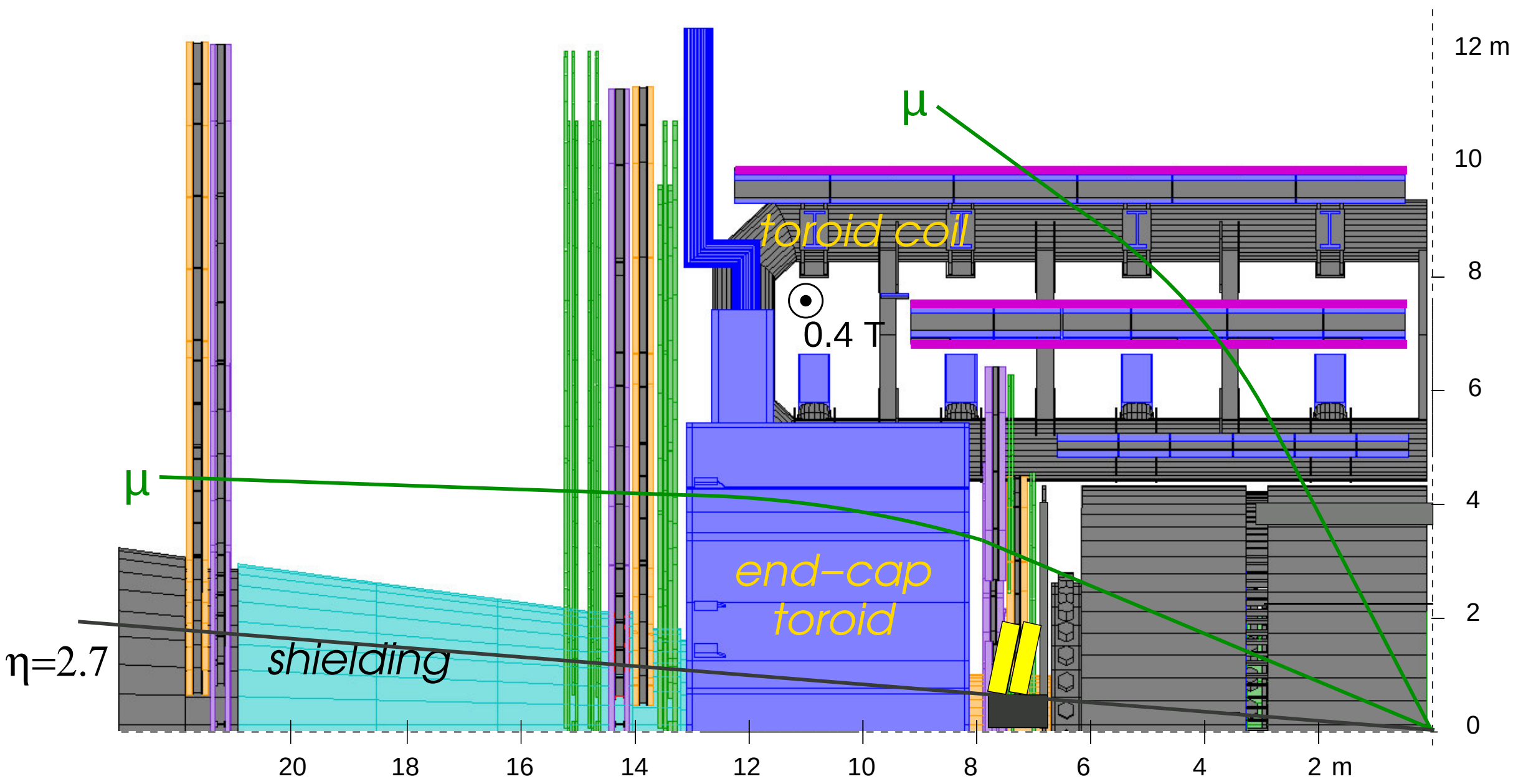

Figure 1: Sketch of a quadrant of the ATLAS muon spectrometer.

The muon reconstruction is described in detail in [2,3]. The focus of this note is the measurement of the performance of the stand-alone reconstruction from real data.

## 2.2 Definitions

The performance of the muon spectrometer is characterized in terms of *efficiency* and *momentum resolution*. In the analysis of simulated data, let $\eta_{rec}$ and $\eta_{truth}$ denote the pseudorapidities and $\phi_{rec}$ and $\phi_{truth}$ denote the azimuthal angles of the reconstructed and generated muons. The distance $\Delta R = \sqrt{(\eta_{rec}-\eta_{truth})^2+(\phi_{rec}-\phi_{truth})^2}$ of a reconstructed and generated muon is shown for a Monte Carlo sample with muons of $p_T = 50$ GeV at the $pp$ interaction point in Figure 2. More than 99.7% of all reconstructed muons have a distance $\Delta R < 0.05$. We therefore define the muon reconstruction *efficiency* as the fraction of generated muons which can be matched to a reconstructed muon within a cone of $\Delta R < 0.05$.

The momentum resolution is measured by comparing the deviation of the reconstructed inverse transverse momentum from the generated inverse transverse momentum:

$$\rho = \frac{\frac{1}{p_{T,truth}} - \frac{1}{p_{T,rec}}}{\frac{1}{p_{T,truth}}} \tag{1}$$

$\rho$ would be normally distributed for a muon spectrometer uniform in $\eta$ and $\phi$. The momentum resolution is not independent of $\eta$ and $\phi$ due to the nonuniformity of the magnetic field and the nonuniformity of the material distribution in $\eta$ and $\phi$. This leads to non-Gaussian tails in the $\rho$ distribution when integrated over $\eta$ and $\phi$, as illustrated in Figure 3. In order to minimize the effect of tails, the momentum resolution is determined in the following way throughout this note: In the first step, a Gaussian $g_0$ is fitted to the distribution. In the next step $i$ a Gaussian $g_i$ is fitted to the data between the $x_{m,i-1} \pm 2\sigma_{i-1}$, where $\sigma_{i-1}$ is the fitted width of $g_{i-1}$ and $x_{m,i-1}$ its fitted mean. The iterative procedure is terminated when the fit relative change of the fit parameters from one to the next iteration is less than 0.1%. The standard deviation of the final fit curve is taken as a measure for the *momentum resolution*. The mean of final fit is referred to as the *momentum scale*, which is a measure for systematic shifts of measured muon momenta with respect to the correct values.

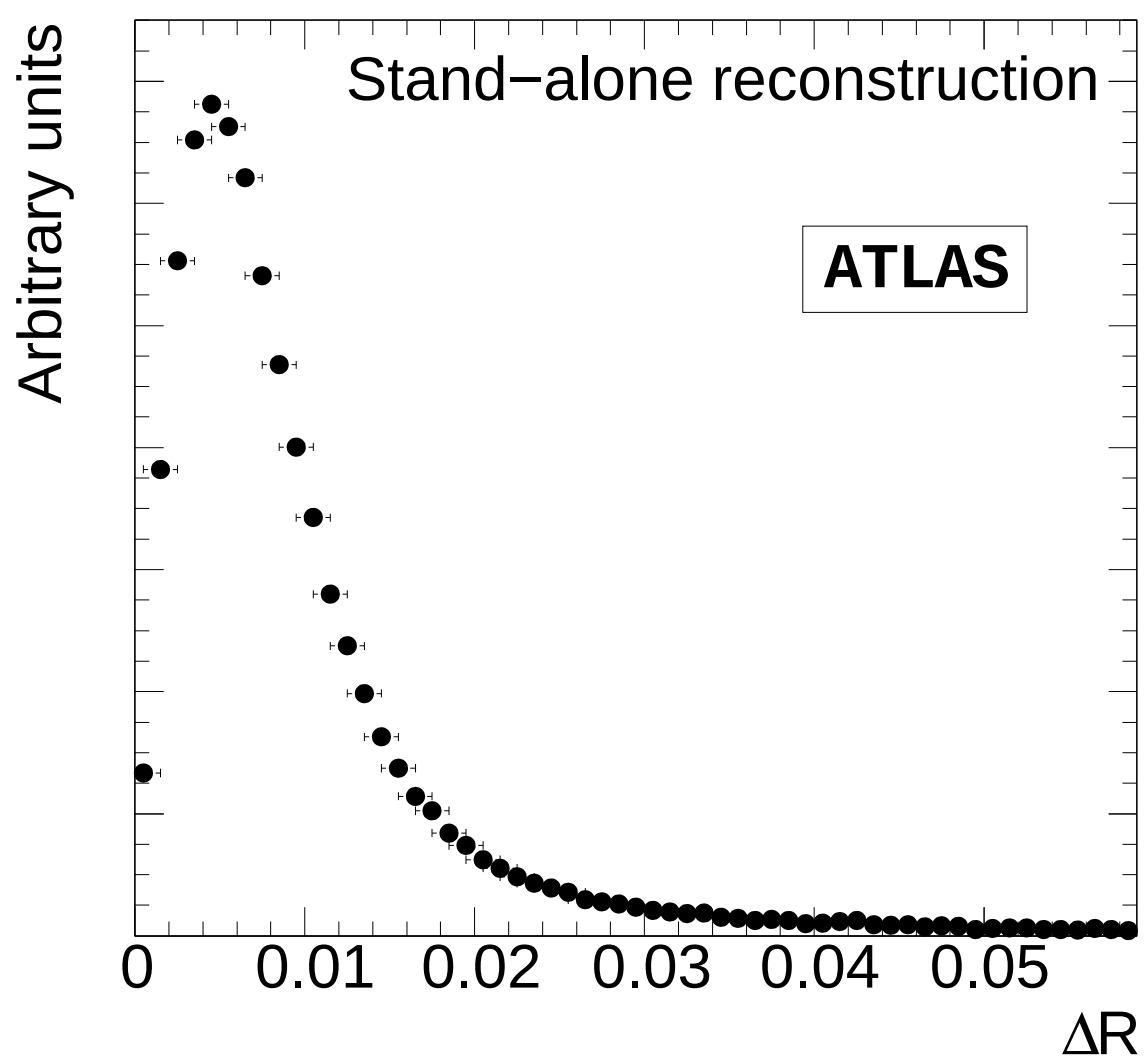

Figure 2: Distribution of distance $\Delta R$ of reconstructed from generated muons in a 50 GeV single muon Monte Carlo sample.

### 2.3 Performance of a perfect muon spectrometer

We briefly review the performance of a perfectly calibrated and aligned muon spectrometer. We refer to [1] and [2] for a more detailed discussion of the performance.

Figure 4(a) shows the reconstruction efficiency for muons with $p_T$=50 GeV as a function of $\eta$ and $\phi$. The efficiency is close to 100% in most of the $\eta$-$\phi$ plane. It drops significantly in the acceptance gaps of the muon spectrometer. The inefficiency near $|\eta| = 0$ is caused by the gap for services of the calorimeters and the inner tracking detector. The inefficiency near $|\eta| = 1.2$ will disappear after the installation of additional muon chambers in the transition region between the barrel and the end caps which will not be present in the initial phase of the LHC operation. The inefficiencies at $\phi \approx 1.2$ and $\phi \approx 2.2$ for $|\eta| < 1.2$ are related to acceptance gaps in the feet region of the muon spectrometer.

The stand-alone reconstruction efficiency is presented as function of $p_T$ in Figure 4(b). It rises from 0 to its plateau value of 95% between $p_T = 3$ GeV and 10 GeV.

The $p_T$-resolution is independent of $\phi$ apart from the feet region where it is degraded due to the material introduced by the support structure of the detector. The resolution also depends on pseudorapidity. It is almost constant in the barrel part of the spectrometer ($|\eta| < 1.05$). It is up to three times worse in the transition region between the barrel and the end caps for $1.05 < |\eta| < 1.7$ mainly due to the

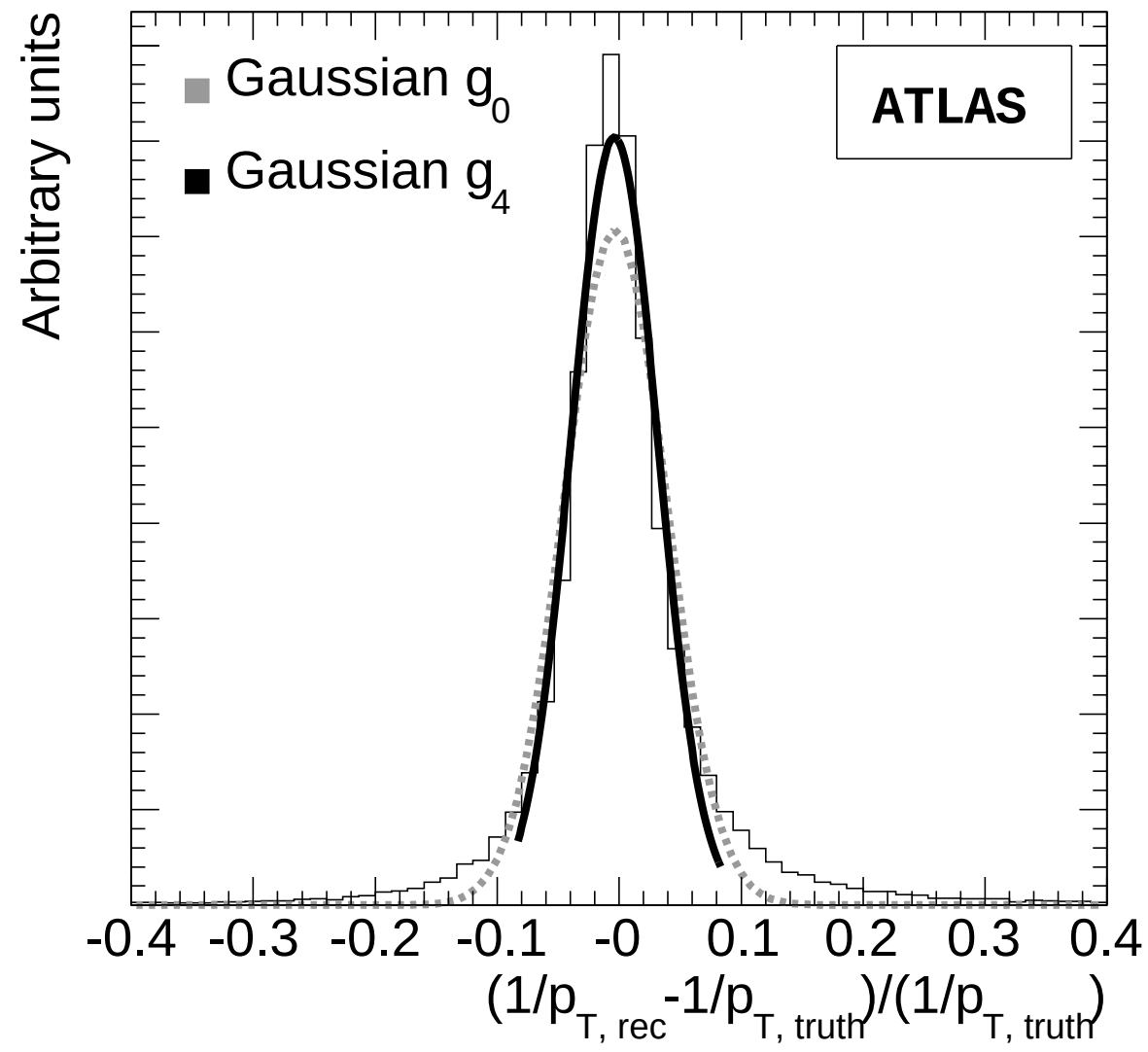

Figure 3: Illustration of the iterative fit of normal distributions to the fractional deviation of the reconstructed inverse momentum from the generated inverse momentum. $g_0$ is the fitted Gaussian of iteration step 0. $g_4$ is the fitted Gaussian of final iteration step 4.

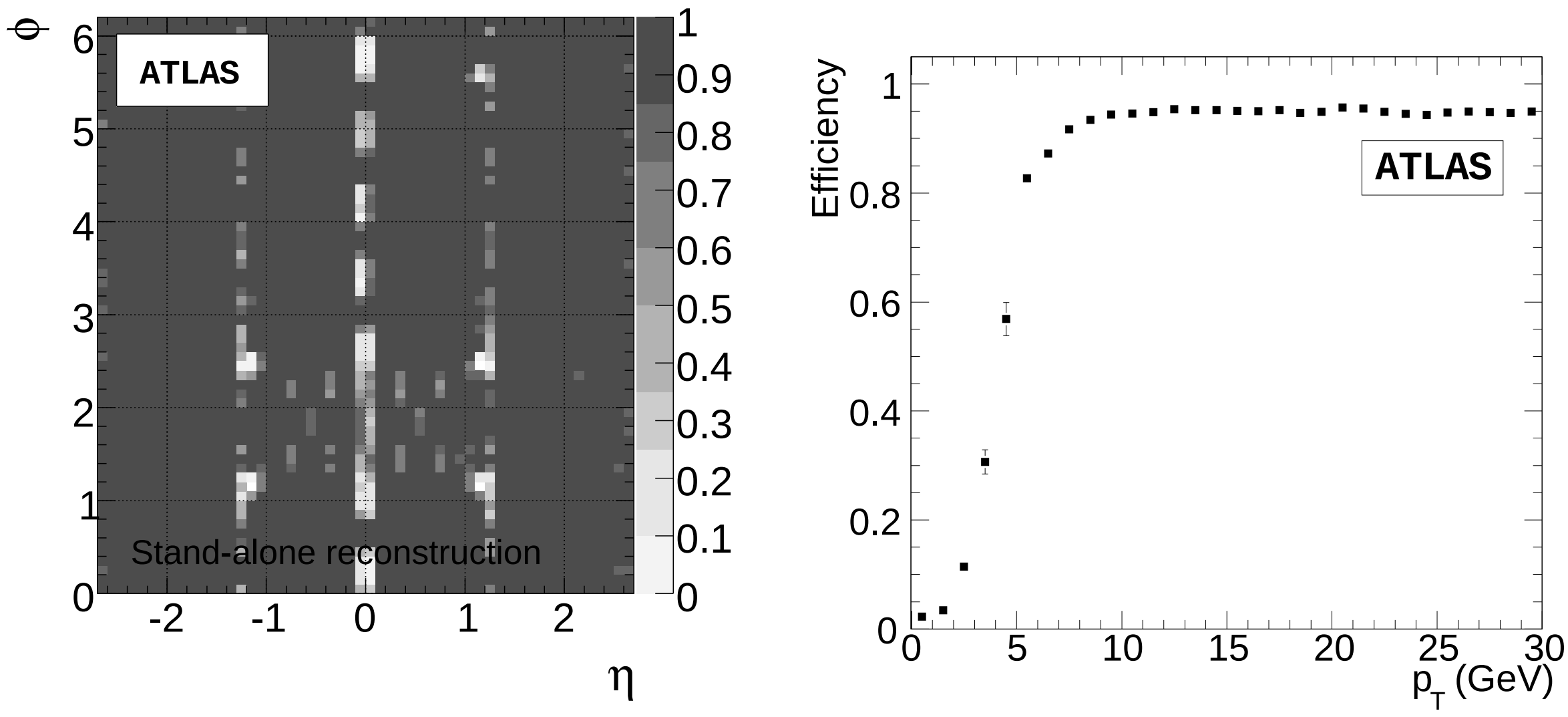

(a) Reconstruction efficiency vs. $\eta$ and $\phi$ for muons of $p_T$=50 GeV.

(b) Reconstruction efficiency vs. $p_T$ integrated over $\eta$ up to $|\eta| < 2.7$ and $\phi$.

Figure 4: Efficiencies of the reconstruction of tracks in the muon spectrometer.

small integral of the magnetic field in this region. The momentum resolution becomes uniform again for $|\eta| > 1.7$.

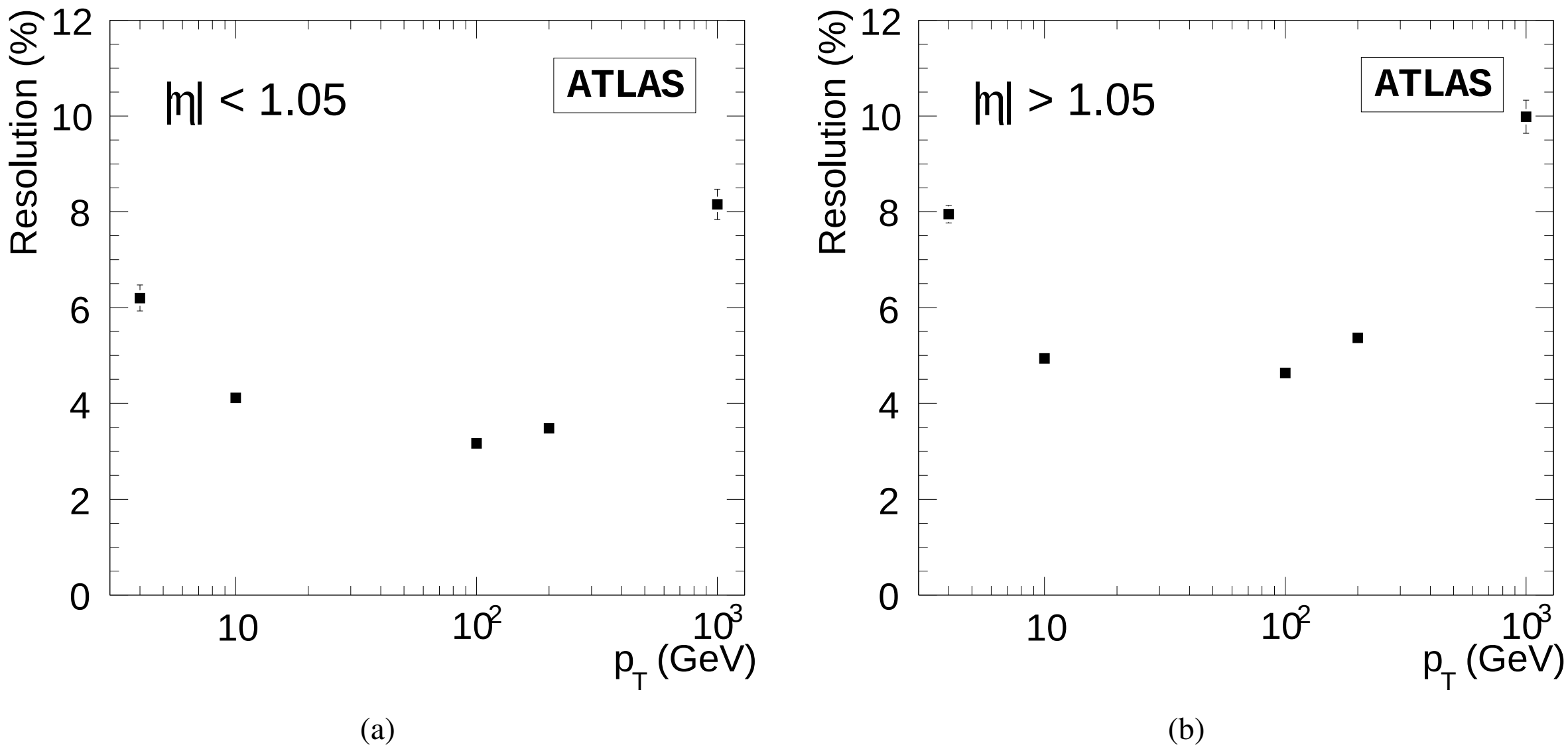

Figure 5: Stand-alone momentum resolution integrated over $\eta$ and $\phi$ as a function of $p_T$ for the barrel (5(a)) and the end-cap region (5(b)).

The stand-alone momentum resolution varies with $p_T$ (see Figure 5). The momentum resolution in the barrel is dominated by fluctuations of the energy loss in the calorimeters for $p_T < 10$ GeV where it is about 5% at $p_T = 6$ GeV. It is best, 2.6% (4%) in the barrel (end cap), for $p_T \approx 50$ GeV where it is dominated by multiple scattering in the muon spectrometer. The momentum resolution at high momenta is limited by the spatial resolution and the alignment of the precision chambers and approaches 10% at $p_T = 1$ TeV.

## 2.4 Deterioration of the performance

The performance of the stand-alone muon reconstruction is affected by the limited knowledge of the magnetic field in the muon spectrometer, the limited knowledge of the material distribution along the muon trajectory required for the calculation of the energy loss, the calibration of the position measurements by the monitored drift-tube chambers, and the alignment of the muon chambers.

The magnetic field will be known with a relative accuracy better than $5 \times 10^{-3}$ based on the measurements of 1840 magnetic field sensors which are mounted on the muon chambers. As a consequence the relative impact on the momentum resolution is less than 3% [1].

Studies for the technical design report of the muon spectrometer [4], which have been confirmed by studies in the context of this note, show that the space-drift-time relationship $r(t)$ of the MDT chambers must be determined with 20 $\mu$m accuracy in order to give a negligible contribution to the momentum resolution up to $p_T = 1$ TeV. A strategy to calibrate $r(t)$ with muon tracks with the required accuracy has been worked out and is described in detail in [5].

The muon chambers are installed with a positioning accuracy of 1 mm in the muon spectrometer with respect to global fiducials in the ATLAS cavern. The studies for the technical design report, however, showed that the muon chambers must be aligned with an accuracy better than 30 $\mu$m in the bending plane. A bias of 30 $\mu$m on the sagitta of a 1 TeV muon corresponds to a systematic shift of the measured momentum of 60 GeV. The alignment of the muon spectrometer is based on a system of optical alignment sensors monitoring relative movements of the chambers on the level of a few micrometers. Muon tracks are used for the absolute calibration of the optical sensor with 30 $\mu$m accuracy. The optical system does not cover the whole muon spectrometer. The positions of the end caps with respect to the barrel must be measured with muon tracks traversing the overlap between the barrel and the end-cap part of the spectrometer. There are also chambers in the transition region between the barrel and end caps whose positions are not monitored by the optical system. These chambers will be aligned with the rest of the muon spectrometer by muon tracks. The alignment of the muon spectrometer is discussed in [6].

The expectation of the muon energy loss in the calorimeters can be checked by comparing the muon momentum as measured by the inner detector and the muon momentum at the entrance of the muon spectrometer, for instance. We shall not discuss the measurement of the muon energy loss in this article and refer the reader to [7].

The initial misalignment will be the dominant source of performance degradation. We shall show in the next section that $Z \to \mu^+\mu^-$ will lead to a clearly visible resonance peak in the dimuon mass distribution even in the case of the initial misalignment. It will therefore be possible to measure the muon performance of a misaligned muon spectrometer with $Z \to \mu^+\mu^-$ events.

**Impact of misalignment on the performance**

In order to study the impact of the initial misalignment of the muon spectrometer on the performance, the simulated data were reconstructed with a different geometry from the one used in the simulation. In the reconstruction geometry, the chambers were randomly shifted from the nominal positions with Gaussian distribution centred at 0 and a standard deviation of 1 mm and rotated randomly with Gaussian distribution centred at 0 and a standard deviation of 1 mrad. Deformations of the chambers which are monitored by an optical system mounted on the chambers were not considered in our studies.

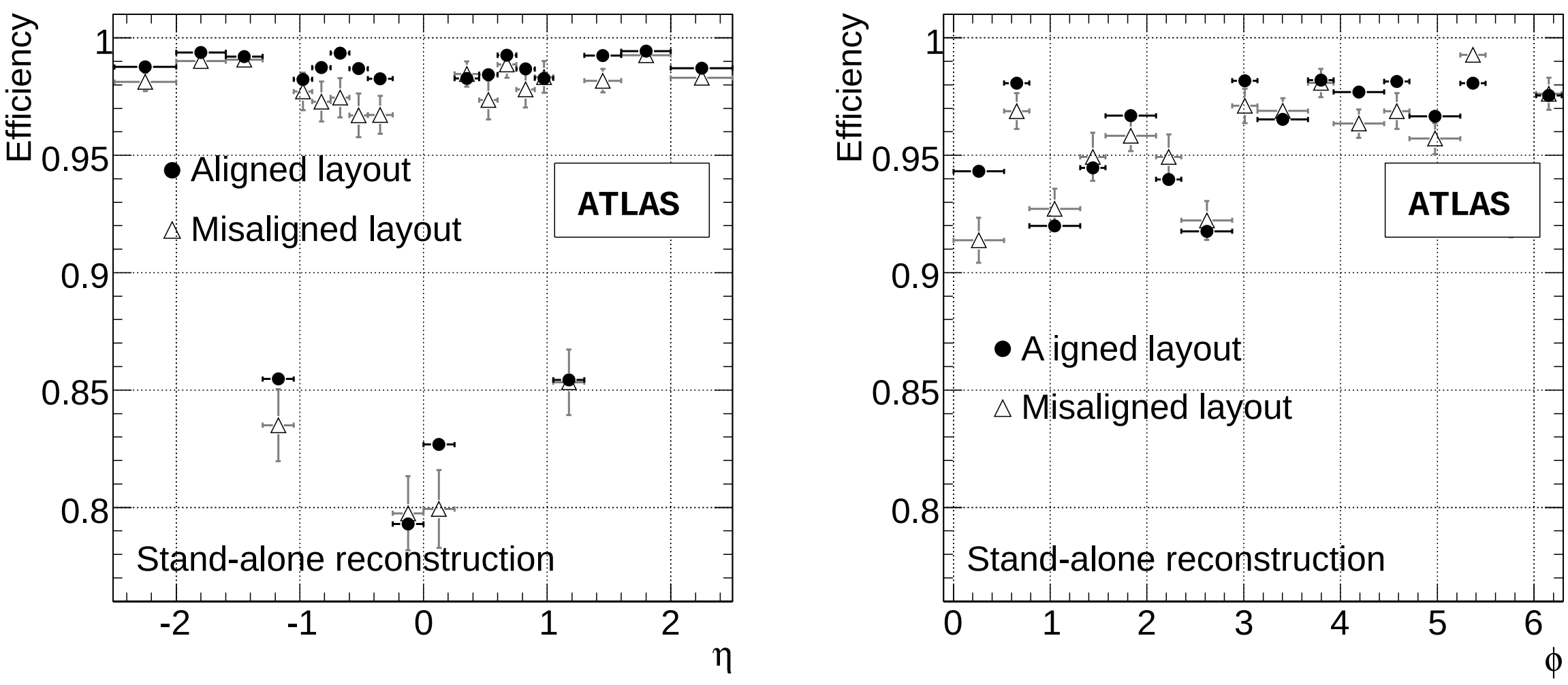

(a) Efficiency vs. $\eta$ integrated over $\phi$ for $p_T$=50 GeV.

(b) Efficiency vs. $\phi$ integrated over $\eta$ for $p_T$=50 GeV.

Figure 6: Comparison of reconstruction efficiency for an aligned muon spectrometer and a misaligned muon spectrometer with a average positioning uncertainty of 1 mm for a simulated single muon sample.

Figure 6 illustrates the comparison of the stand-alone track reconstruction efficiency for 50 GeV muons in the aligned and the misaligned case. Only a small decrease in the reconstruction efficiency can be observed for muons with a momentum of 50 GeV, a momentum typical for muons originating from $W$ or $Z$ bosons. The relatively small decrease in the reconstruction efficiency is mainly due to the fact that the used definition of efficiency is based on a simple $\eta$ and $\phi$ matching and does not take into account the measured transverse momentum of the muons. The reconstruction efficiency could be increased in the misaligned case by applying softer cuts in the pattern recognition stage of the track reconstruction.

Figure 7(a) and 7(b) show the impact of a misaligned muon spectrometer on the fractional transverse momentum resolution; the resolution is highly degraded. The overall observed fractional muon spectrometer resolution $\sigma_{tot}$ can be expressed as the quadratic sum of the intrinsic fractional $p_T$-resolution at the ideal geometry ($\sigma_{ideal}$) and the fractional resolution due to the misaligned geometry ($\sigma_{Alignment}$).

$$\sigma_{tot} = \sqrt{\sigma^2_{Alignment} + \sigma^2_{ideal}}$$

This leads to $\sigma_{Alignment} \approx 0.14$ for muons with $p_T \approx 50$ GeV as expected from the relationship between sagitta and momentum. The effect on the momentum scale is relatively small for the overall muon spectrometer, since random misalignments cancel to a certain extent. In physics signatures, such as the decay of a $Z$ boson into two muons, the impact on the average momentum scale is even less, since a misaligned geometry has the opposite effect for opposite charged muons to first order.

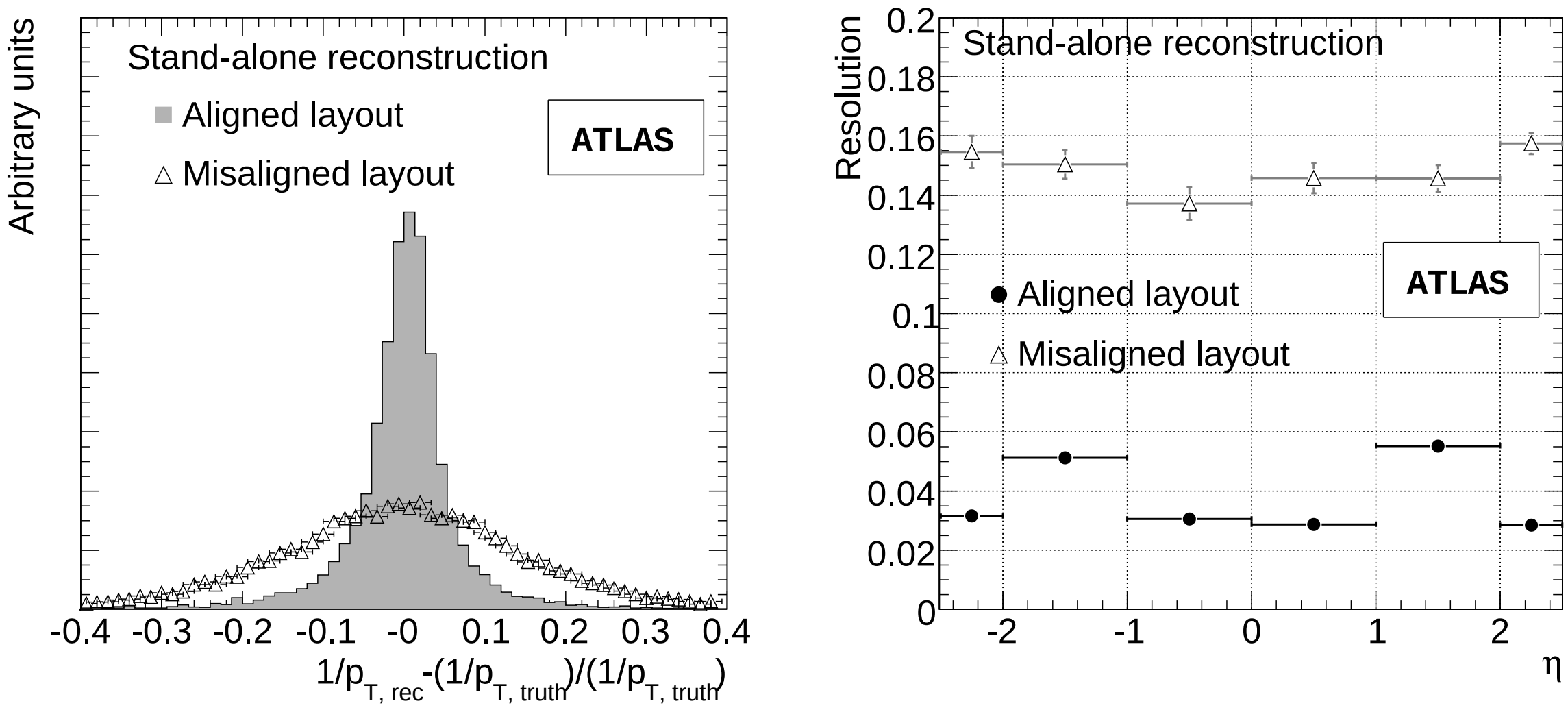

(a) Overall $p_T$-resolution

(b) $p_T$-resolution vs. $\eta$

Figure 7: Comparison of the fractional $p_T$-resolution for an aligned muon spectrometer and a misaligned muon spectrometer.

The impact of initial misalignment of the muon spectrometer on the $Z$ resonance is shown in Figure 8. It is expected that the mean of the invariant mass distribution does not change significantly, since the momentum scale of the reconstructed muon $p_T$ is hardly affected by misalignment. On the other hand a large broadening of the distribution due to the degradation of the $p_T$-resolution of the muons is expected, which is shown in Figure 8. The dependence of the reconstructed width of the $Z$ boson mass distribution on the size of the misalignment is shown in Figure 9. $\sigma_m^{scale}$ is a scaling factor applied to the initial misalignment of 1 mm and 1 mrad. The observed dependence is the basis for the determination of the muon spectrometer resolution with data, which is discussed in section 4. A more detailed discussion of misalignment impacts on the muon spectrometer performance can be found in [8].

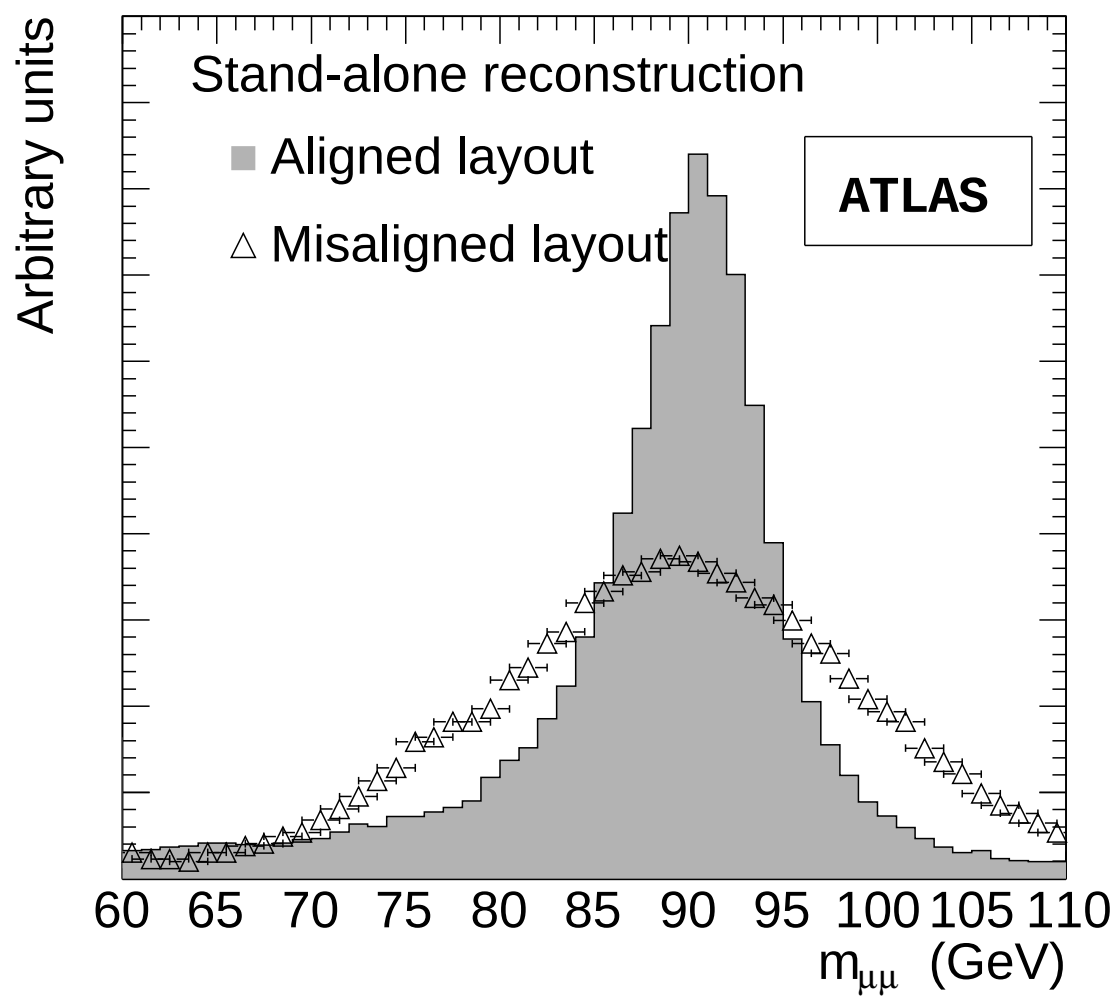

Figure 8: Reconstructed $Z$ boson mass distributions for an aligned and a misaligned ($\sigma_m^{scale} = 1$) muon spectrometer layout.

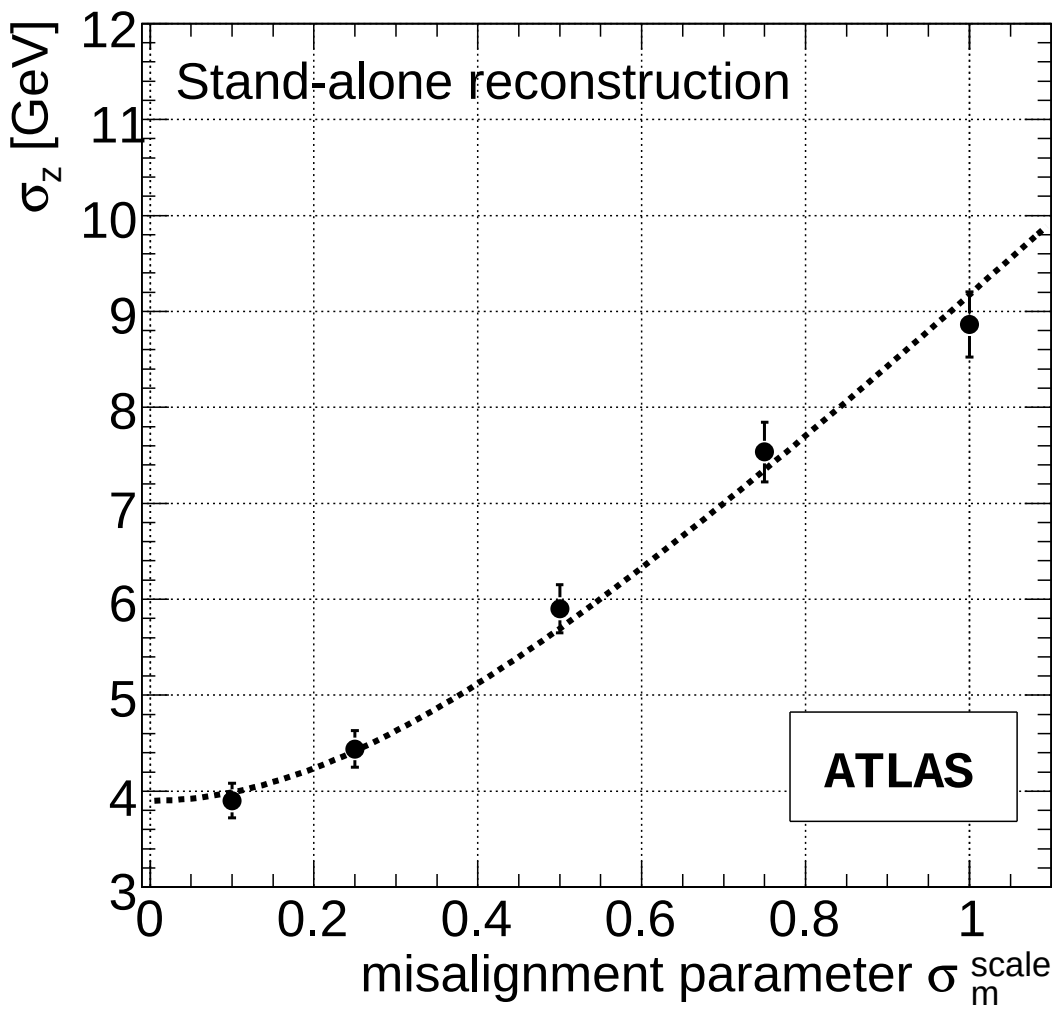

Figure 9: Width of the $Z$ resonance peak including the natural width of the $Z$ vs. misalignment parameter $\sigma_m^{scale}$.

# 3 Measurement of the reconstruction efficiency from $pp$ collision data

The simulation of the ATLAS detector is still under development and is not expected to reproduce the actual performance of the detector in all details at the beginning of the LHC operation. Therefore it is necessary to determine all efficiencies with data in order not to rely on the simulation.

## 3.1 Reconstruction efficiency from dimuon decays of the $Z$ boson

### 3.1.1 Tag-and-probe method

The so-called "tag-and-probe" method can be used to determine the muon spectrometer reconstruction efficiencies from $pp$ collision data. Muons from $Z$ decays will be detected by the inner tracking detector and the muon spectrometer in the common acceptance range of $|\eta| < 2.5$. The measurements of the inner detector and the muon spectrometer are independent, though not necessarily uncorrelated. We require two reconstructed tracks in the inner detector, at least one associated track in the muon spectrometer, and the invariant mass of the two inner-detector tracks to be close to the mass of the $Z$ boson. The last requirement ensures that the reconstructed tracks are the tracks of the decay muons of the $Z$ boson. Moreover, the two inner tracks are required to be isolated to reject possible OCD background. The inner track which could be associated to the track in the muon spectrometer is therefore a muon and is called the *tag muon*. It is also required that the tag muon fired the 20 GeV single-muon trigger in order to ensure that the event is recorded. This selection ensures that a $Z \rightarrow \mu^+\mu^-$ decay has been detected. The second inner track must then be a muon, too, which is called the *probe muon* (see Figure 10). In the analysis of dimuon events from $pp$ collisions, the probe muon plays the role of the generated muon in the determination of the efficiency with simulated data.

The tag-and-probe technique is not restricted to the measurement of the stand-alone reconstruction efficiency. It can, for instance, be used to measure the muon reconstruction efficiency of the inner detector or the trigger efficiency [9].

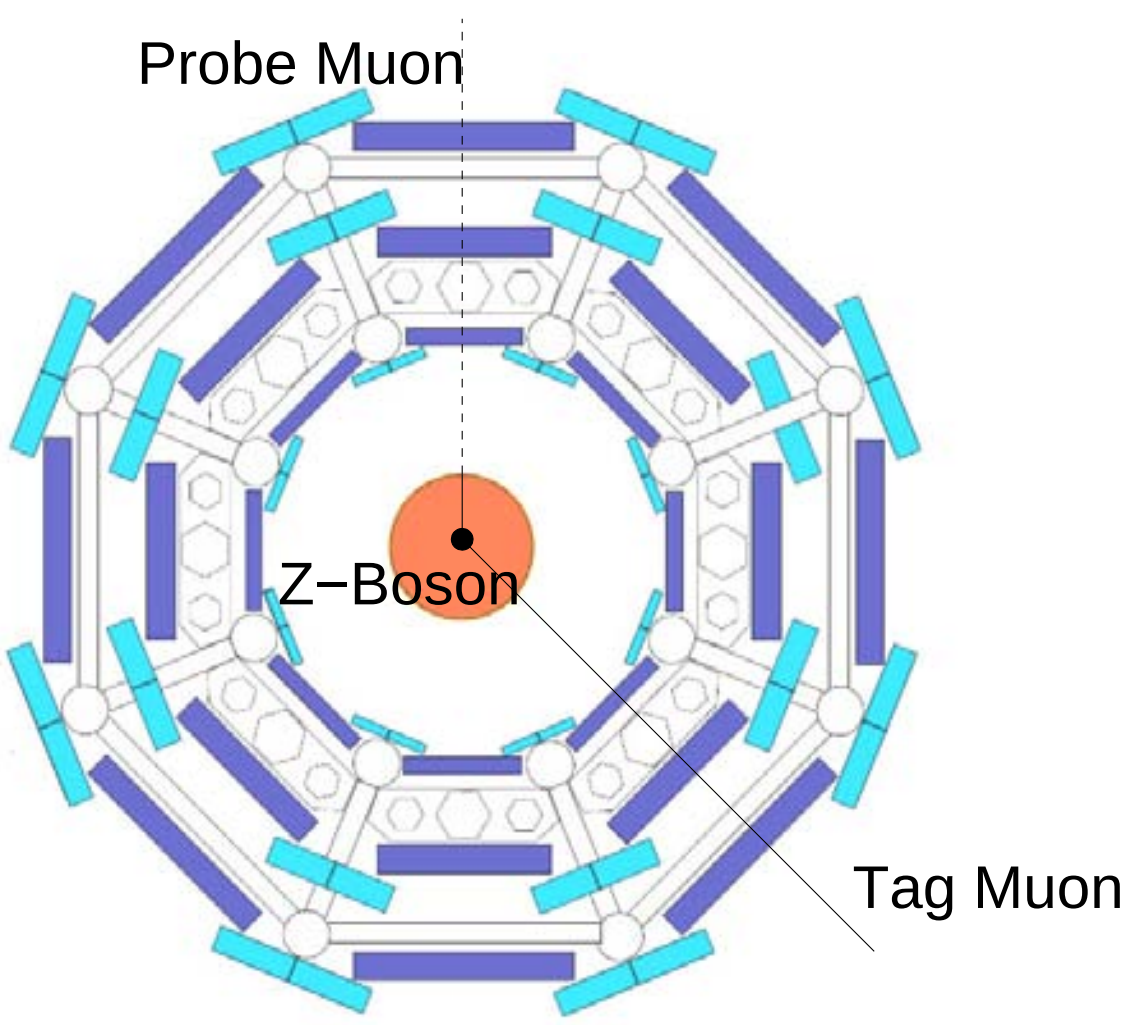

Figure 10: Schematic illustration of the tag and probe method.

Our studies show that the acceptance gaps of the muon trigger which are reflected in uncovered $\eta$-$\phi$ regions of the tag muon do not create uncovered $\eta$-$\phi$ regions of the probe muon. The tag-and-probe method therefore allows us to determine the efficiency over the full $\eta$ and $\phi$ coverage of the inner detector.

Some systematic uncertainties of the tag-and-probe method must be considered. Muons from $Z \rightarrow \mu^+\mu^-$ decays usually fly in opposite directions in the plane transverse to the proton beam axis. Hence inefficiencies which are symmetric in $\Delta\phi \approx \pi$ may not be detected with this method.

The topology of $pp \rightarrow Z/\gamma^* \rightarrow \mu^+\mu^-$ events is characterized by two highly energetic and isolated muons in the final state. A significant QCD-background contribution is expected due to the huge cross section of QCD processes. Moreover, the decay of a $W^\pm$ boson into one highly energetic muon and a neutrino plus an additional muon from a QCD jet and the process $Z \rightarrow \tau^+\tau^- \rightarrow \mu^+\bar{\nu_\tau}\nu_\mu\mu^-\nu_\tau\bar{\nu_\mu}$ were studied as possible background processes in our analysis.

Because of the high collision energy of the LHC, the production of top quark pairs has a cross section of the order of the signal cross section. Top quarks mostly decay into a $W$ boson and bottom quark. The $W$ boson and the bottom quark can decay into muons or electrons, which also might fake the signal process.

The cross section of QCD processes is far too large to be simulated within a full Monte Carlo simulation of the ATLAS detector. Hence it is assumed that the dominant contribution from highly energetic muons is due to the decay of $b$-mesons. A more detailed discussion of the selection of $pp \rightarrow Z \rightarrow \mu^+\mu^-$ events and the background processes which must be considered can be found in [9].

### 3.1.2 Selection of candidate tracks

Figure 11 shows the invariant dimuon mass and the transverse momenta of the selected muon track candidates for the signal and the chosen background processes.

The following cuts have been applied to get a clean track selection. Tracks of opposite charge and a difference in their $\phi$ coordinates greater than 2.0 rad are selected. The rapidity of the tracks is limited to a rapidity coverage of the inner detector of $|\eta| < 2.5$. Each of the selected muon candidate tracks is required to have $p_T > 20$ GeV. The invariant mass $M_{\mu\mu}$ of the two muon candidate tracks must agree with the $Z$ mass within $\pm 10$ GeV, i.e.$|M_{\mu\mu} - 91.2\,GeV| < 10$ GeV.The following isolation cuts are applied to the selected tracks:

- number of reconstructed tracks in the inner detector within a hollow cone around the candidate muon: $N^{\text{ID Tracks}}_{r_1<r<r_2} < 5$
- sum of the $p_T$'s of reconstructed tracks in the inner detector within a hollow cone around the candidate muon: $\sum\limits_{r_1<r<r_2} p_T^{\text{ID Tracks}} < 8$ GeV
- sum of reconstructed energy in the cells of the calorimeter within a hollow cone around the candidate muon: $\sum\limits_{r_1<r<r_2} E_T < 6$ GeV
- energy of a possible reconstructed jet within a hollow cone around the candidate muon: $E^{\text{Jet Energy}}_{r<r_2} <$ 15 GeV

These isolation variables are defined within a hollow cone in the $\eta-$ and $\phi-$plane of the reconstructed muon track,

$$r_1 < \sqrt{\left(\eta_\mu - \eta_{ic}\right)^2 + \left(\phi_\mu - \phi_{ic}\right)^2} < r_2 \tag{2}$$

where $r_1$ and $r_2$ are the inner and the outer radius of the cone. The index $\mu$ stands for the reconstructed muon track while the index $ic$ labels the isolation criteria. The smaller radius is set to $r_1 = 0.05$ and is introduced to exclude the candidate muon track from the calculations of the isolation quantities. The specific value of the outer radius $r_2$ has only a minor effect on the signal and background separation, as long it is large enough to contain a significant amount of data for the definition of isolation variables,

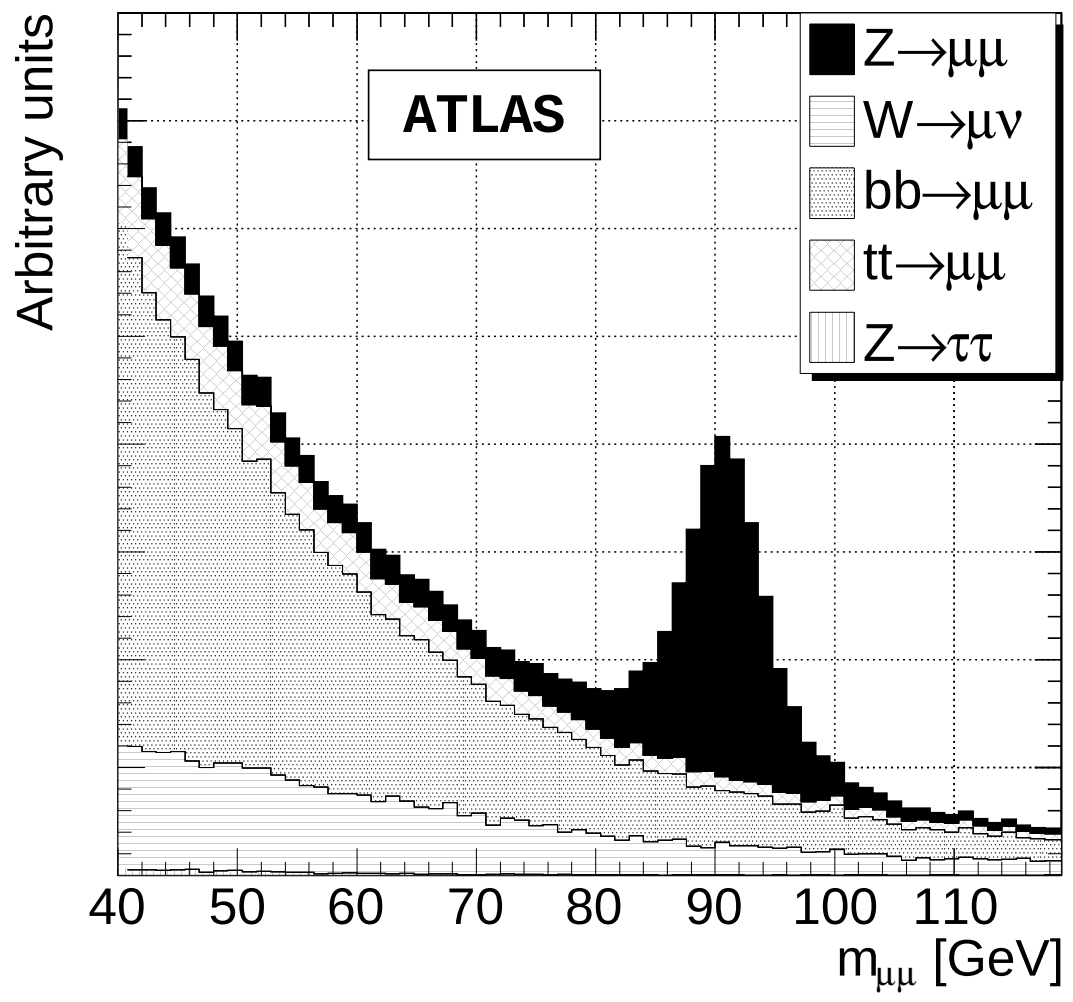

(a) Invariant mass of $Z$ candidates.

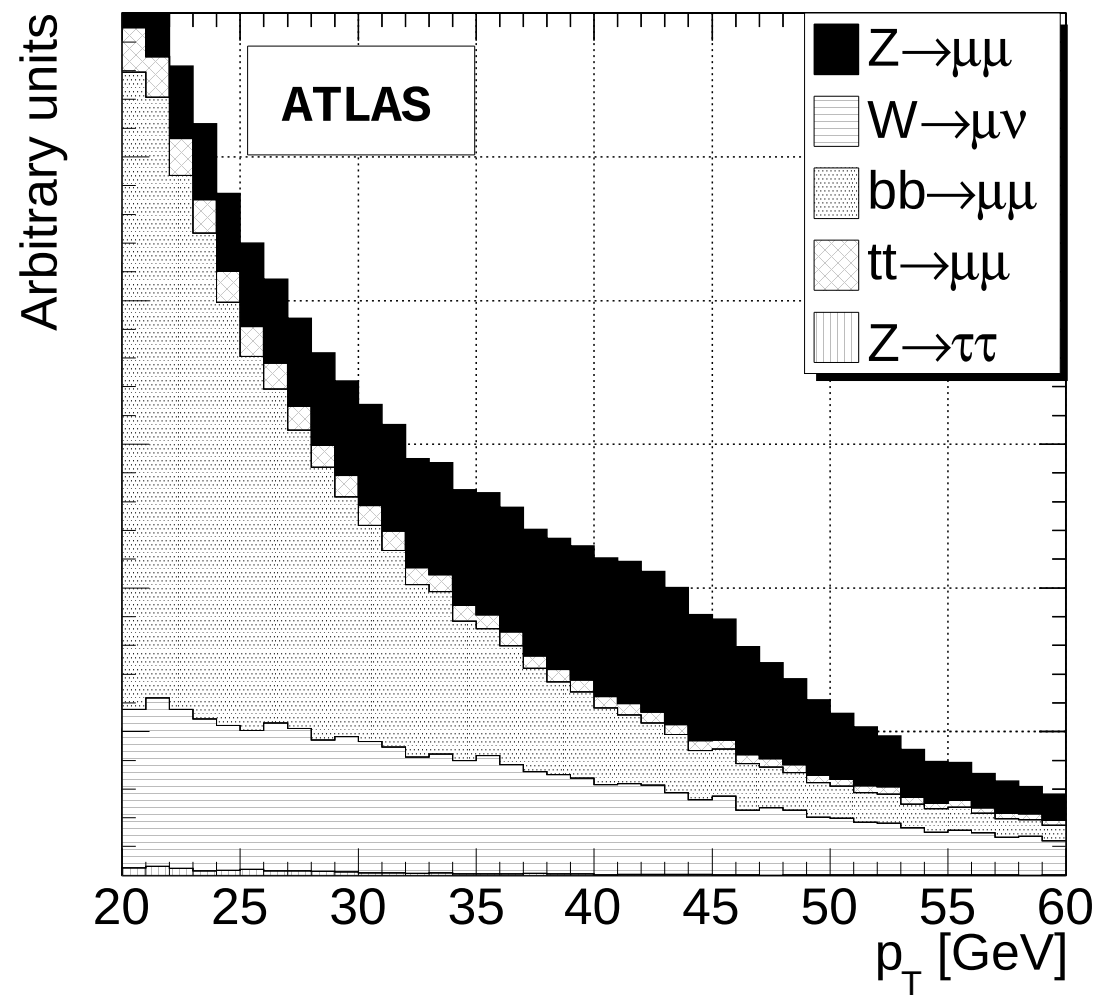

(b) Transverse momentum distribution.

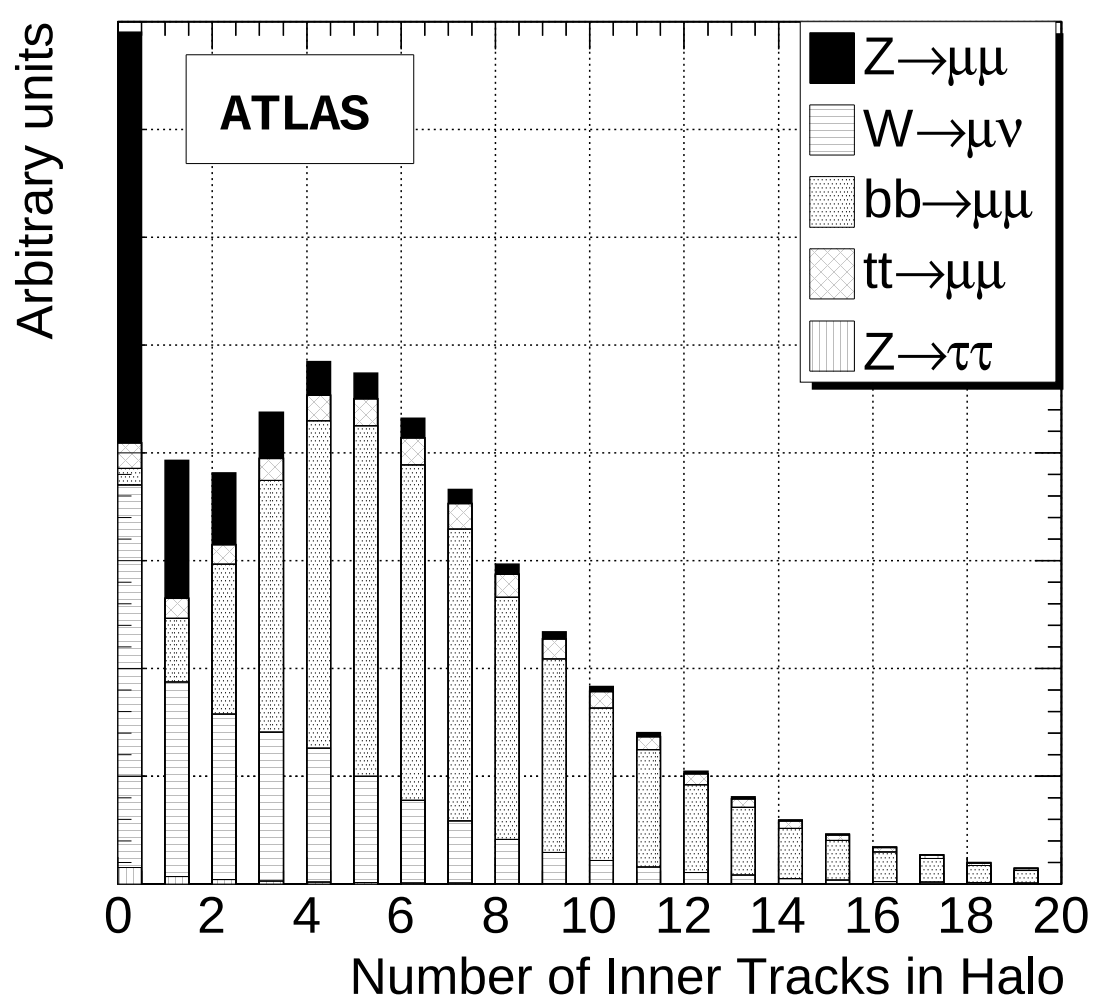

(c) Number of reconstructed tracks within a cone of $\Delta R = 0.5$ around the candidate track.

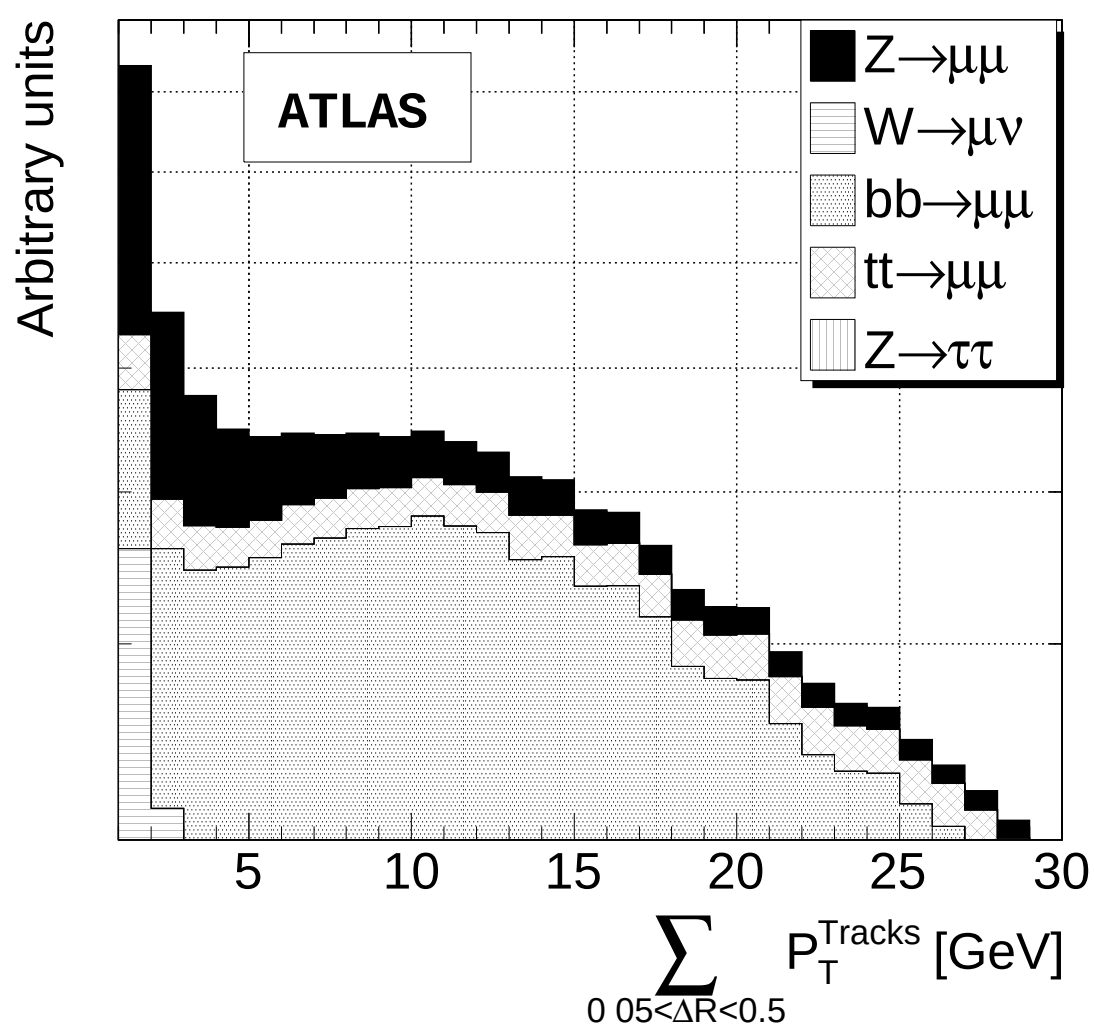

(d) Sum of transverse momenta of all tracks within a cone of $\Delta R = 0.5$ around the candidate track.

Figure 11: Reconstructed quantities for $Z$ candidate events only using inner detector tracks with a transverse momentum above 6 GeV and no further cuts for signal and background processes.

i.e. $r_2 > 0.3$. Our choice of $r_2 = 0.5$ is the same as used in the measurement of the cross section of the process $pp \to Z \to \mu^+\mu^-$ (see [9]). The isolation criteria listed here are optimized for events without pile-up of inelastic $pp$ collisions in a selected event. Pile-up of inelastic $pp$ collisions is expected for the operation of the LHC at a luminosity of $10^{33}$ cm$^{-2}$ s$^{-1}$ and will lead to more energy in a cone around the muons. It was checked that the efficiency of our event selection is reduced by less than 5% in the presence of pile-up and that the purity of our selected samples is not affected by the presence of pile-up.

The distributions of the first two isolation variables for signal and background processes normalized to their cross sections are presented in Figure 11(c) and 11(d) in absence of pile-up. The selection of isolated high-$p_T$ muons allows for a substantial suppression of the background.

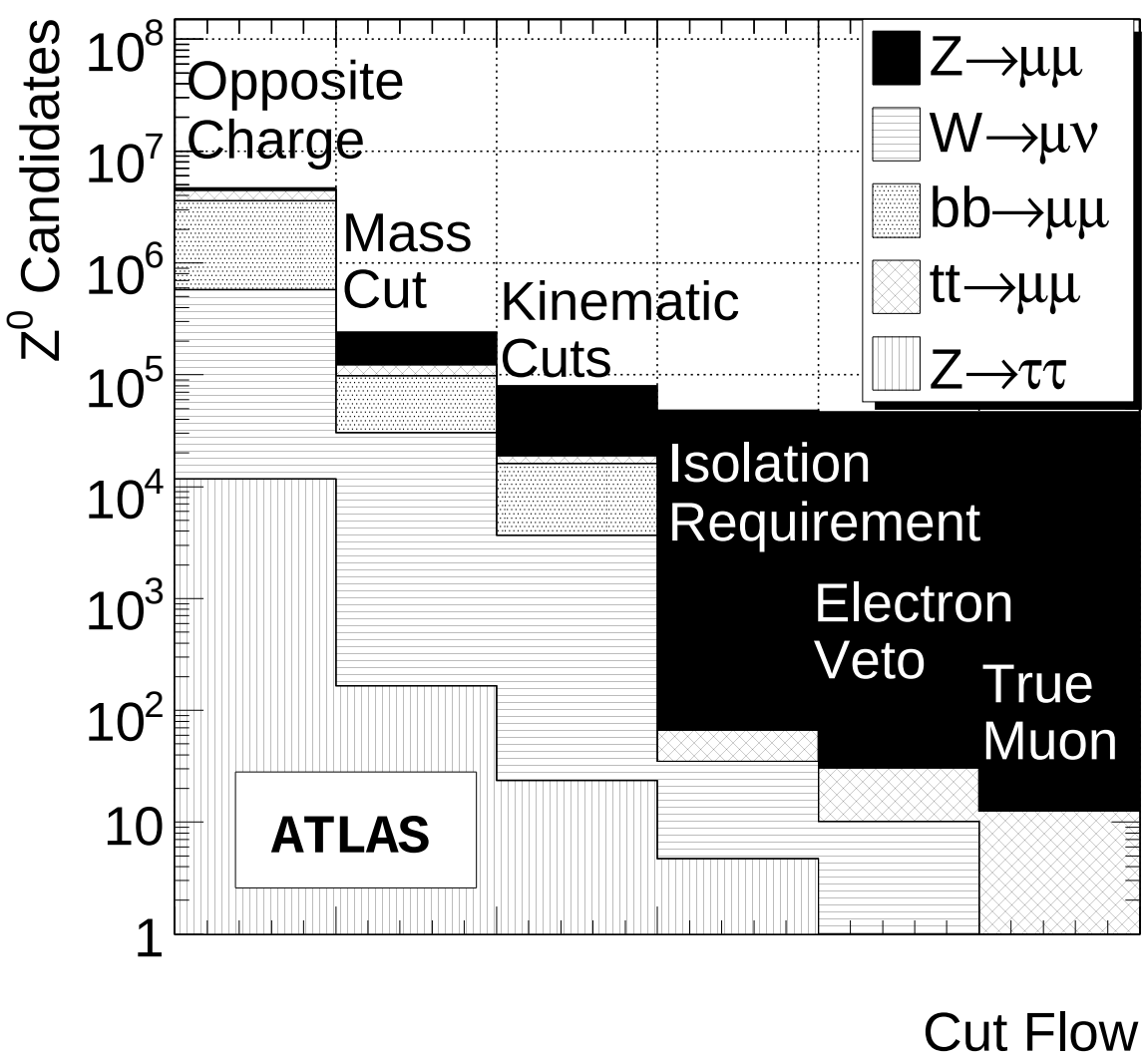

Figure 12: Cut-flow diagram for probe muon tracks: (0) opposite charge requirement, (1) invariant mass requirement, (2) kinematic cuts, (3) isolation requirements, (4) electron veto, (5) found at least one track in the muon spectrometer.

The cut-flow diagram for probe muons is shown in Figure 12. The QCD background can be rejected with isolation cuts. More problematic in this selection is the $W \to \mu\nu$ background, and those $t\bar{t}$-events in which at least one $W$ boson decays into a muon and a neutrino. These processes produce one highly energetic isolated muon track which passes all selection cuts for a tag muon. A further track in the inner detector which passes the other cuts and is not a muon will decrease the measured efficiency. Such a track is most likely caused by an electron, since it is expected that electrons also appear as isolated tracks in the inner detector. Therefore it is required that no reconstructed electromagnetic jet in the electromagnetic calorimeter can be matched to an inner track as an additional selection requirement. This applies especially for probe tracks stemming from a $t\bar{t}$-event. Here, again, one has to distinguish between inner tracks, which result from the decay of the bottom quark or simple QCD-interactions and those, which result from the decay of the $W$ boson. The first case is suppressed by the isolation requirement and can be neglected. The second case can lead to a highly energetic isolated electron, stemming from the decay of the second $W$ boson. These electrons are expected to be vetoed. The cut-flow diagram also shows that the probe muon candidates from the background processes can also be associated to a muon spectrometer track and hence have no negative effect on the efficiency determination.

An overview of the remaining background expected from Monte Carlo is shown in Table 1; there we have assumed at least three events surviving the cuts as a systematic uncertainty in order not to underestimate the background contribution. After all selection cuts, the purity of our sample is high: less than 0.1% of the selected dimuon events are from background processes.

Table 1: Fractional background contribution in % based on Monte Carlo prediction including estimated systematic and statistical uncertainties.

| $b\bar{b} \to \mu\mu$ | $W^{\pm} \to \mu^{\pm}\nu$ | $Z/\gamma^{*} \to \tau\tau$ | $t\bar{t} \to W^{+}bW^{-}b$ | Overall |
|---|---|---|---|---|
| $\approx 0 + 0.03$ | $\approx 0 + 0.06$ | $\approx 0$ | $\approx 0.02 \pm 0.01$ | $0.02 \pm 0.1$ |

Our results are stable against variations of the track matching distance $\Delta R$ from 0.05 to 0.3. The larger track matching cut of $\Delta R$=0.3 takes account for possible misalignment effects in the first phase of LHC. The robustness of our results against the $\Delta R$ matching cut indicates that our selected data sample will allow an efficiency determination which is not significantly affected by background processes even with a possible misalignment of the muon spectrometer.

### 3.1.3 Determination of the stand-alone reconstruction efficiency

The stand-alone reconstruction efficiency depends on $p_T$, $\eta$ and $\phi$ of the muons. Hence, one should determine the efficiency in appropriate bins in these quantities. The lower value of the $p_T$-binning is given by the selection cut of 20 GeV. The highest value is set to 70 GeV and 10 bins are used to ensure high enough statistics within each bin. For larger statistics also values above 100 GeV can be considered.

A natural binning in $\eta$ and $\phi$ is given by the geometry of the muon spectrometer. The muon spectrometer consists of 16 sectors in the $\phi$ plane, small and large MDT chambers sequentially ordered as illustrated in Figure 13(a). Therefore 16 bins in $\phi$ are used. The same geometrical argument applies to the $\eta$-plane of the detector. Three MDT-chambers which are projective to the interaction point define one tower. Twenty towers are defined in $\eta$ which are the basis for the chosen binning (Figure 13(b)). In total 320 regions are defined in the $\eta - \phi$ plane.

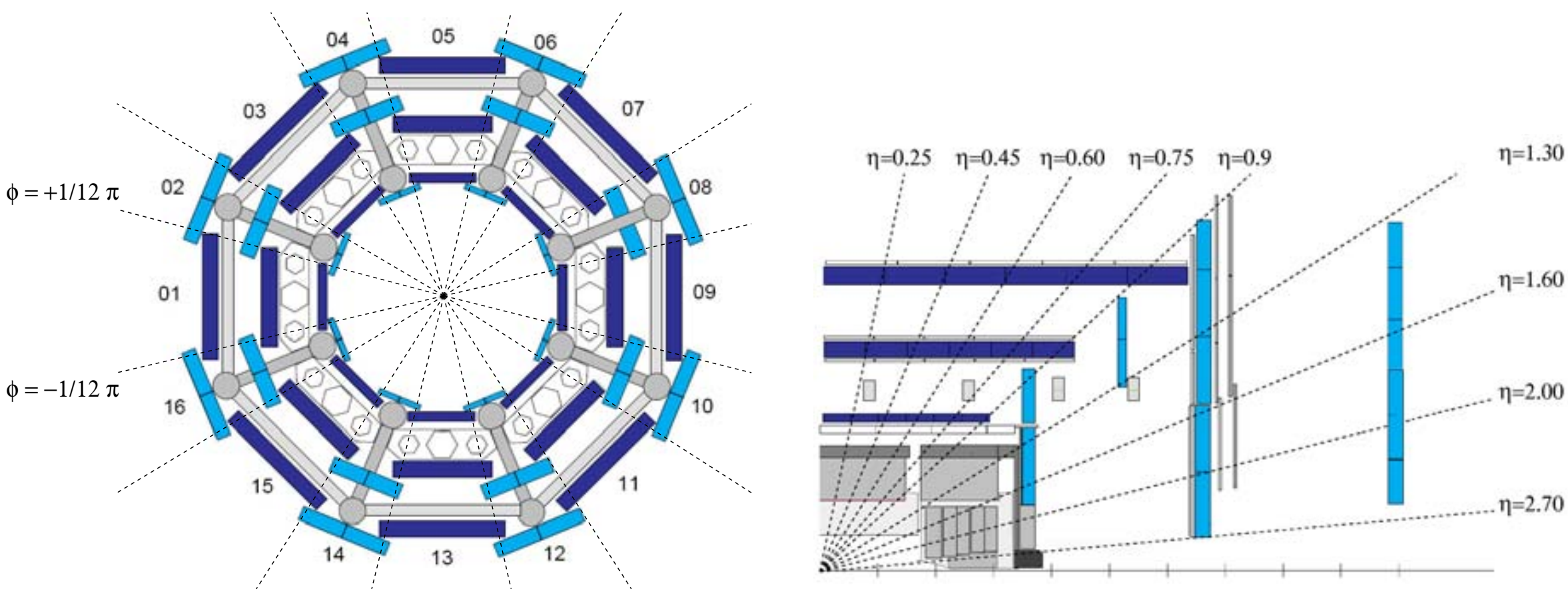

(a) $\phi$-binning (b) $\eta$-binning

Figure 13: Illustration of the choosen $\phi$ and $\eta$-binning of the muon spectrometer

It is important to note that the dominant effect of losing reconstruction efficiency is the acceptance gap due to the absence of MDT chambers. Hence it is a pure geometrical effect mainly in the $\eta$-direction. Therefore different physics samples with different $\eta$- and to a certain extent also different $\phi$- and $p_T$- distributions will lead to different overall reconstruction efficiencies. An overview of the overall reconstruction efficiencies for different physics samples and track matching distances is shown in Table 2. Hence the in-situ determined efficiencies must be applied in an appropriate binning for different physics samples.

Table 2: Overall reconstruction efficiencies for different physics processes. Efficiencies with respect to the Monte Carlo truth information are quoted for the sample of events that pass the single muon trigger.

| Sample | $\Delta R = 0.05$ | $\Delta R = 0.075$ | $\Delta R = 0.15$ |
|---|---|---|---|
| $Z \to \mu^+\mu^-$ | 0.952 | 0.956 | 0.958 |
| $W^\pm \to \mu^\pm \nu$ | 0.953 | 0.958 | 0.960 |
| $t\bar{t} \to W^+W^-b\bar{b}$ | 0.943 | 0.948 | 0.950 |
| $b\bar{b} \to \mu^+\mu^-$ | 0.930 | 0.944 | 0.952 |

The comparison of the efficiencies determined with Monte Carlo truth information and the trag-and-probe method is shown in Figure 14 for $\eta$ and $p_T$, assuming an aligned muon spectrometer. A track matching distance of $\Delta R < 0.075$ was chosen. The efficiencies determined in both ways coincide within their statistical uncertainty for an integrated luminosity of 100 pb$^{-1}$. This proves that possible correlations between tag and probe muons are small and can be neglected to a good extent.

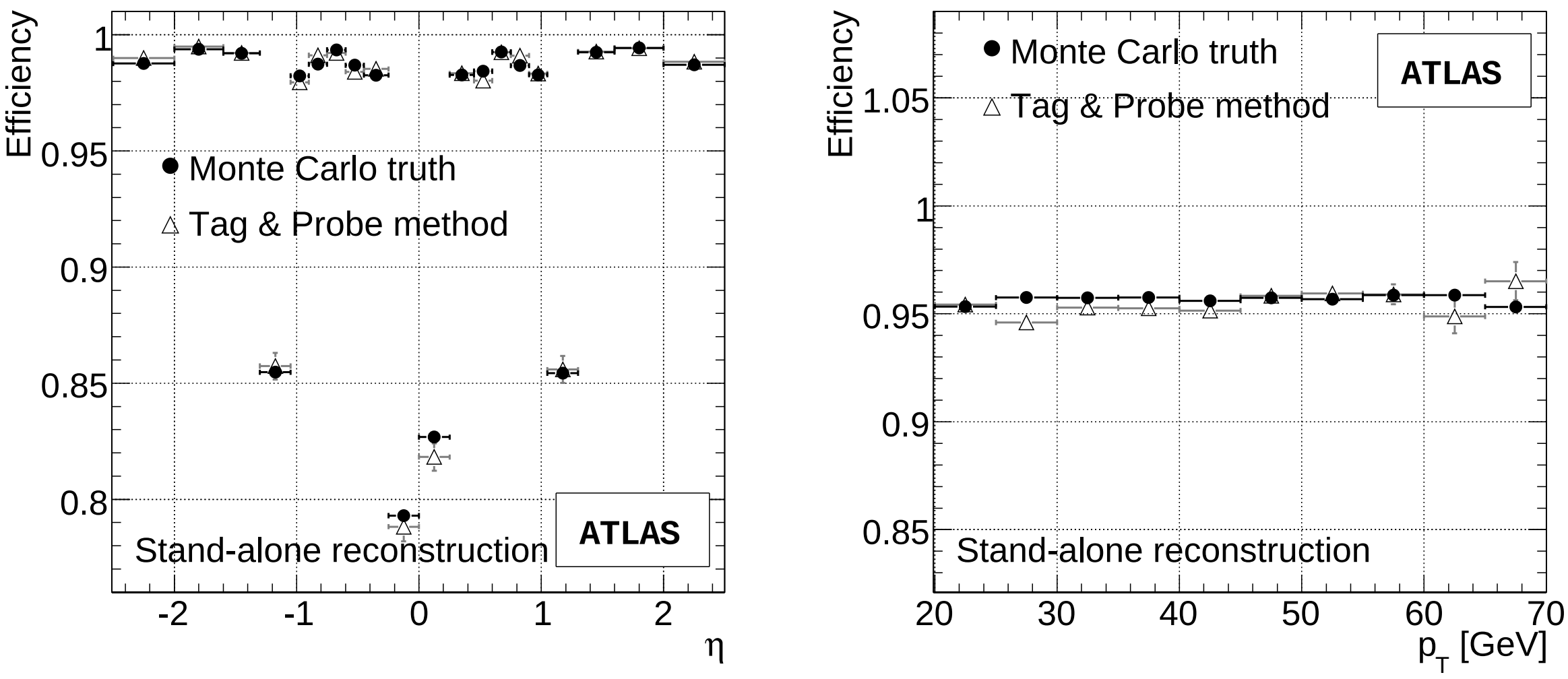

(a) Efficiency integrated over $\phi$ and $p_T$ in $Z \to \mu^+\mu^-$ events vs. $\eta$

(b) Efficiency integrated over $\phi$ and $\eta$ in $Z \to \mu^+\mu^-$ events vs. $p_T$

Figure 14: Comparison of the muon reconstruction efficiency of the muon spectrometer vs. $\eta$ and $p_T$ determined by the tag and probe method and via the Monte Carlo truth information.

The statistical error on the reconstruction efficiency $\varepsilon$ can be calculated (for large $N$) by

$$\Delta\varepsilon = \sqrt{\frac{\varepsilon(1-\varepsilon)}{N}}, \tag{3}$$

where $N$ is the number of tag muons. Note that both muons can, and will, be chosen as tag muons in most cases, as the muon spectrometer is expected to have a reconstruction efficiency of 95% on average. Figure 15 shows the distribution of the in-situ determined efficiencies for all 320 regions. The overall reconstruction efficiency can be determined to a high statistical precision even for relatively low integrated luminosities. A statistical precision of 1% of the overall muon spectrometer reconstruction efficiency can be reached with less than 1 $pb^{-1}$. Figure 16 illustrates the statistical uncertainty averaged over all 320 regions versus the integrated luminosity.

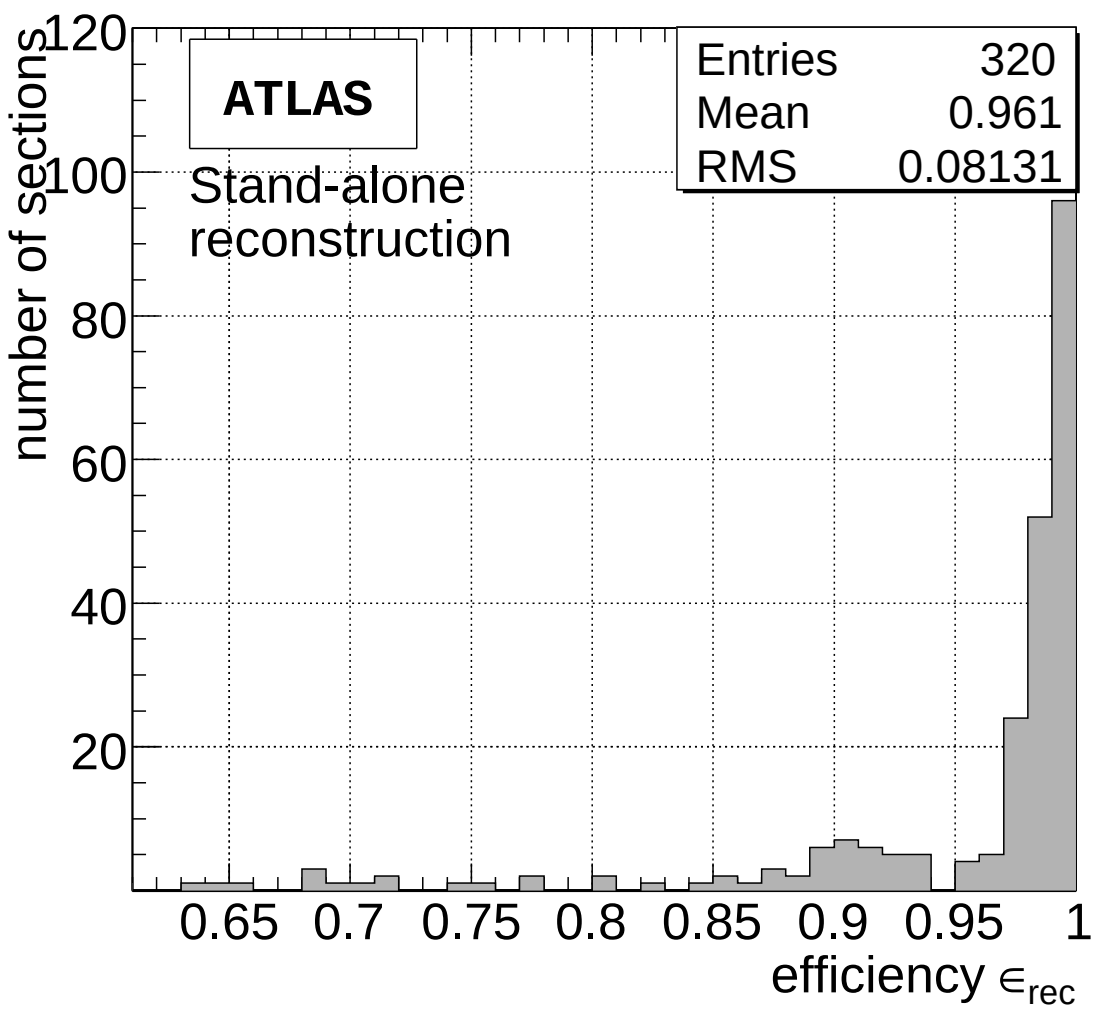

Figure 15: Distribution of muon reconstruction efficiency of the 320 muon spectrometer regions.

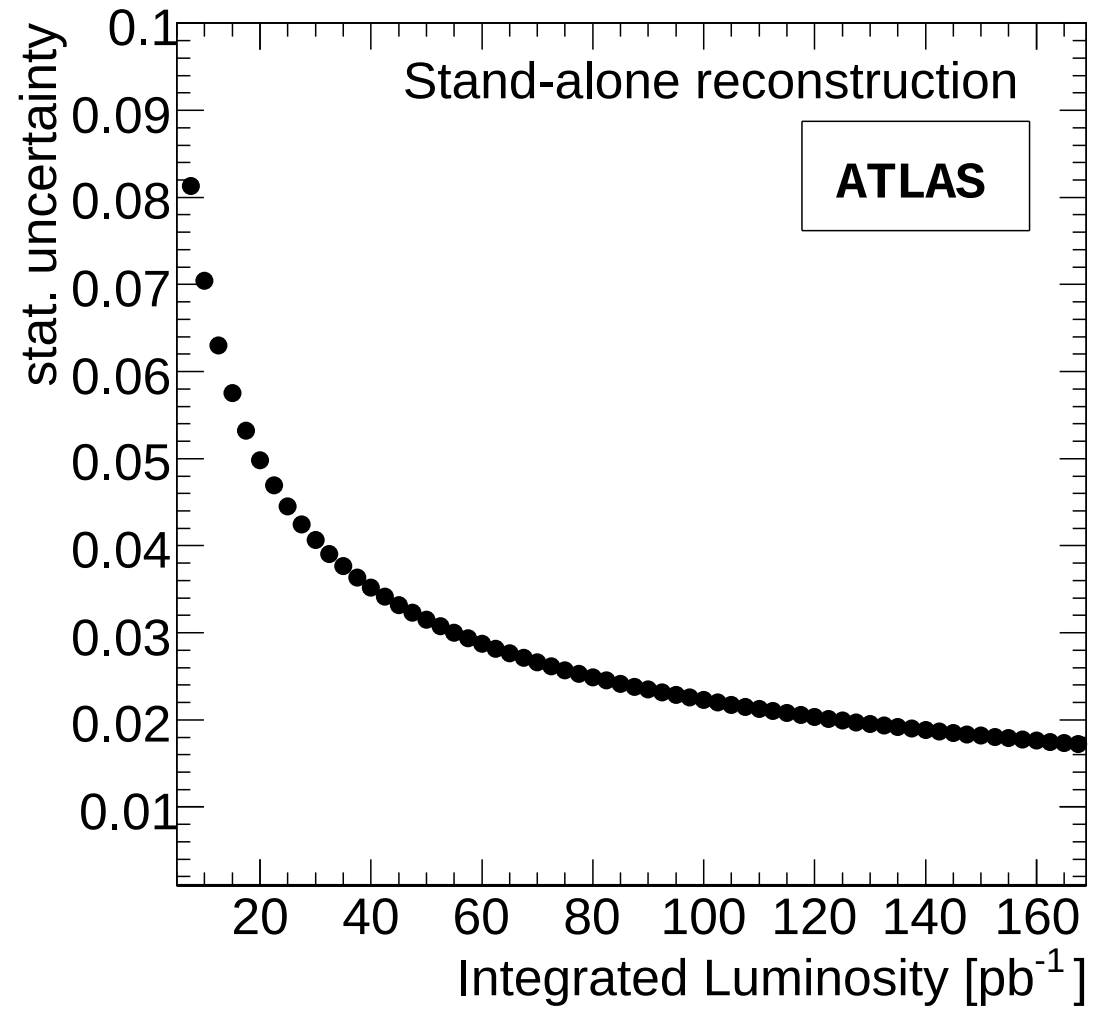

Figure 16: Average statistical error of reconstruction efficiency of the 320 regions vs. integrated luminosity.

A possible correlation between tag and probe muons could be caused by the trigger. The probability of reconstructing a muon is significantly higher if it was triggered, as shown in Figure 17. Hence, it might be suspected that this correlation implies also a correlation in real data, since data events must contain at least one muon which has been triggered. This is not a problem as long as the trigger requirement is only applied on the tag muon.

In Section 3.1.1 it was already mentioned that the tag and probe approach has problems in detecting inefficiencies which have a $\phi \approx \pi$ symmetry. Dividing the data sample in two parts differing in the angle $\Delta\Phi$ could overcome this problem. One part contains reconstructed tag and probe muons with $\Delta\Phi < 2.8$ rad the second sample with $\Delta\Phi > 2.8$ rad. The chosen value of 2.8 rad leads to roughly equally sized samples. Applying the tag and probe method on both sub-samples will lead to different efficiency distributions in case of $\phi$-symmetric inefficiencies. Monte Carlo studies showed that for the presently simulated detector layout we expect only small differences (Fig. 18).

Table 3 summarizes statistical and systematic uncertainties of the in-situ determined stand-alone reconstruction efficiency for two different integrated luminosities. The difference in $|\varepsilon_{in-situ} - \varepsilon_{true}|$ is calculated via

$$|\varepsilon_{in-situ} - \varepsilon_{true}| = \sum_{i=1}^{N} \frac{1}{N} |\varepsilon^i_{in-situ} - \varepsilon^i_{true}| \tag{4}$$

where the index $i$ runs over all bins in $\eta$-direction. This is treated as primary source of systematic uncertainty. One should note that the given systematic error has a strong statistical component from the Monte Carlo statistics which is reflected in the large decrease of the systematic uncertainty in Table 3.

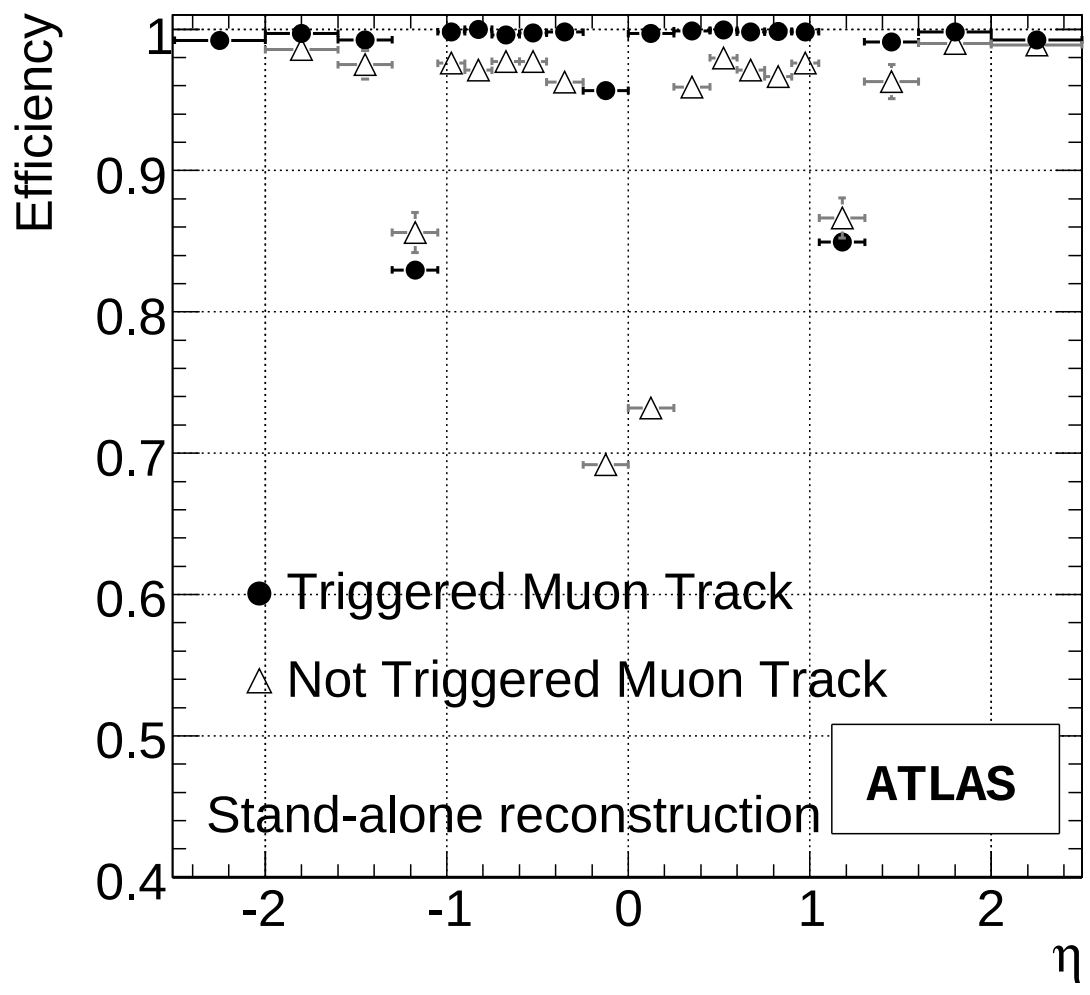

Figure 17: Reconstruction efficiency of the muon spectrometer for muon tracks which have been triggered and muon tracks which have not been triggered.

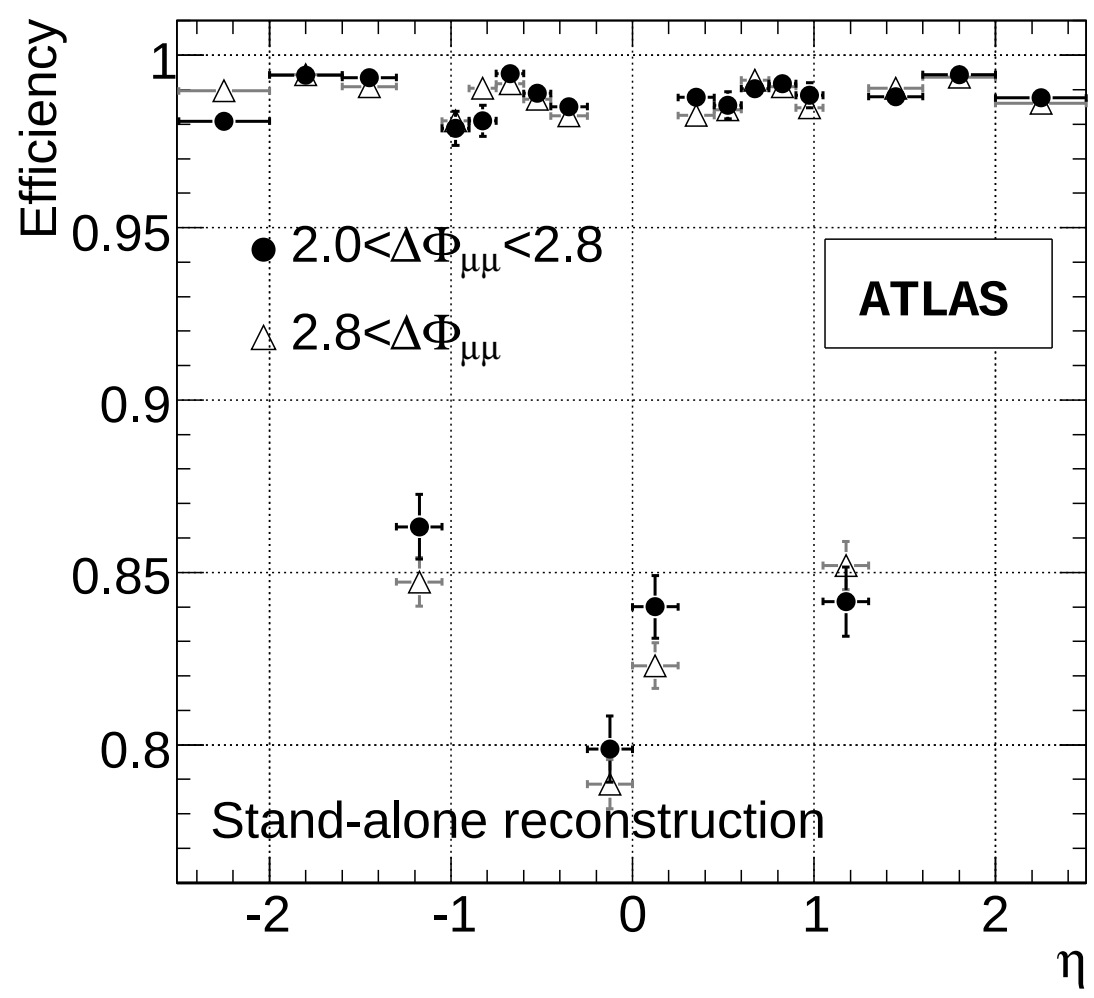

Figure 18: Comparison of muon reconstruction efficiencies determined via tag and probe approach for two sets of muons differing by $\Delta\phi$.

Table 3: Estimated uncertainties of in-situ determined muon spectrometer reconstruction efficiencies for muons in a $p_T$-range between 20 GeV and 70 GeV and within an $\eta$-range smaller than 2.5 from a $Z \to \mu\mu$ decay.

| $\int \mathscr{L}$ | Statistical Uncertainty | $\lvert\varepsilon_{in-situ} - \varepsilon_{true}\rvert$ | Background Contribution | Overall Systematic |
|---|---|---|---|---|
| 100 pb$^{-1}$ | 0.08% | 0.9% | 0.02% | $\approx$ 1% |
| 1 fb$^{-1}$ | 0.03% | 0.1% | 0.02% | $\approx$ 0.1% |

We take the difference between the efficiency obtained for the misaligned layout and the efficiency obtained for the aligned layout as a conservative estimate of the precision which can be achieved with the tag-and-probe method in case of small unresolved misalignments. The difference in both efficiencies for the different $\Delta\phi$-sample is comparable within its statistical uncertainties. The background contribution is only estimated by the Monte Carlo prediction and treated as a systematic uncertainty.

The Gaussian sum of the two systematic uncertainties, namely $|\varepsilon_{in-situ} - \varepsilon_{true}|$ and the background contribution, is defined as the overall systematic uncertainty.

The given uncertainty estimation assumes that nearly all MDT chambers work and $\varepsilon_{true} \approx 96\%$. A lower value of $\varepsilon_{true}$ will lead to an increase of the statistical uncertainty via Equation (3) and also to a higher systematic uncertainty. For real data a conservative estimate of the systematic uncertainty would be the difference of the Monte Carlo prediction for the efficiency and the efficiency determined with collision data. Moreover, it should be noted that the given uncertainties apply for muons in a $p_T$-range between 20 GeV and 60 GeV and within an $\eta$-range smaller than 2.5.

### 3.1.4 Alternative approach

The tag-and-probe analysis presented above uses isolation cuts to reject background events and assumes that a negligable background contribution remains. In this section, we explore the possibility of determining the reconstruction efficiency from collision data without isolation cuts and determine the background contribution directly in data. We apply only cuts on the transverse momenta, e.g. $p_T > 10$ GeV. This leads to a dominant background contribution in the lower invariant dimuon-mass region.

In this approach, a tag muon is defined as a muon spectrometer and inner detector combined muon track, with $p_T > 10$ GeV. A probe muon is defined as any inner detector track, also with $p_T > 10$ GeV. An invariant mass is then calculated from every combinatoric tag and probe pair with opposite charges. The size of this sample is denoted as $N$ in the following (Figure 19(b)). Finally, we select a subsample requiring that the probe muon also be a combined muon track. The size of this sample is denoted as $n$ (Figure 19(a)).

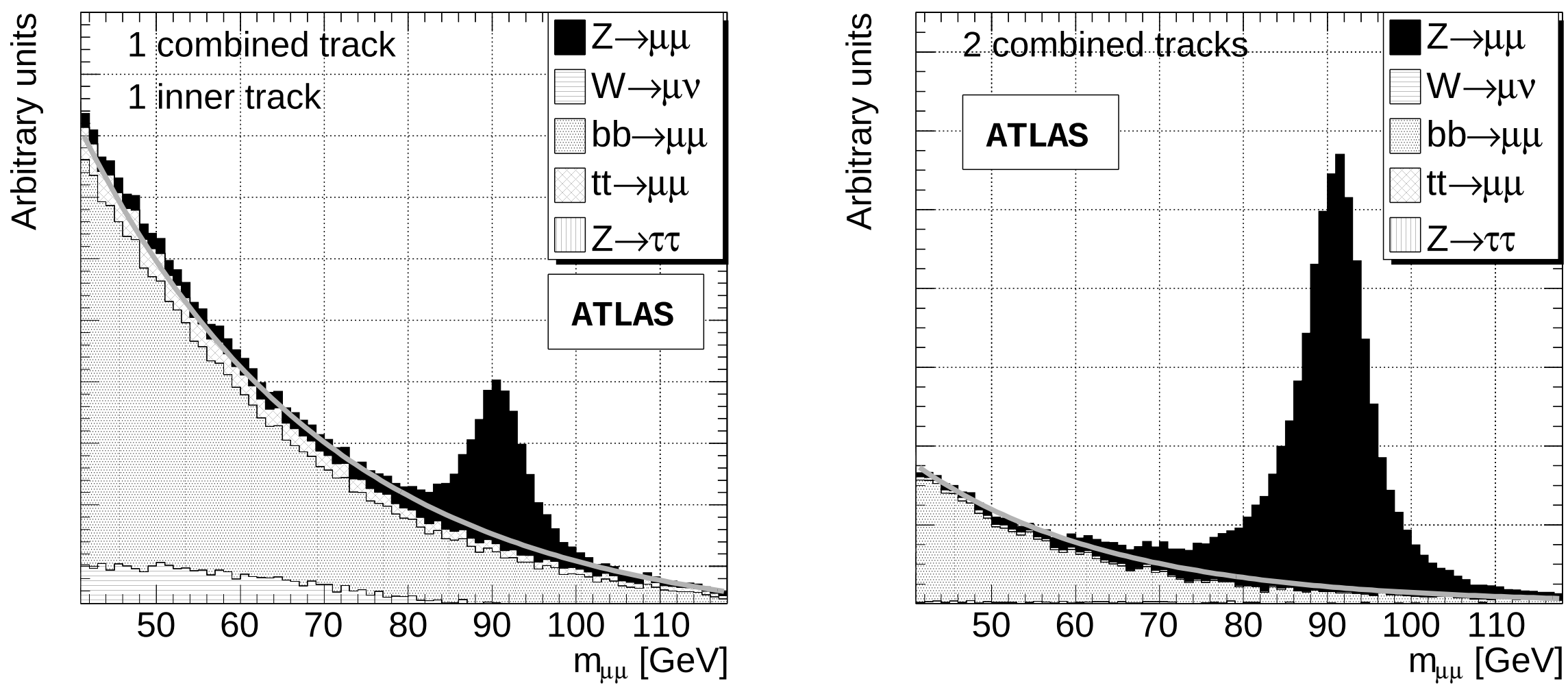

(a) At least one of the muons is matched to a muon spectrometer track.

(b) Both muons are matched to muon spectrometer tracks.

Figure 19: Expected invariant Masses $M_{\mu\mu}$ resulting from two inner tracks where both muons must be matched to a muon spectrometer track (a) or at least one of the muons must be matched to a muon spectrometer tracks (b).

The track-finding effciency of the muon spectrometer, $\varepsilon$, is then defined as $n/N$. Missing tracks in the muon spectrometer will result in $n < N$ and thus effciency loss. The main difference from the approach presented in Section 3.1 is that no isolation cuts are used for the background rejection but instead the background is directly estimated from data via side band subtraction. In this approach an exponential function is fitted to the invariant mass region between $\sim 40$ GeV to $\sim 60$ GeV, where it is assumed that the background contribution is dominating. The exponential function is then extrapolated to the invariant mass region between $\sim 81$ GeV to $\sim 101$ GeV and used for subtraction of the background in this region. The remaining number of events between $\sim 81$ GeV to $\sim 101$ GeV define $n$ and $N$, respectively. In this way, the background contribution is accounted for implicitly in data and no further assumptions on the Monte Carlo predictions are made. The disadvantage of this procedure are the systematic uncertainties of the fitting procedure and the choice of the fitting function. One possible improvement with higher statistics of the background sample would be that the Monte Carlo prediction of the shape of the background distribution could be used to obtain a better fit function than the pure exponential for the side band subtraction.

The systematic uncertainty of this method is again estimated by the residual difference of the in-situ determined efficiency and the true efficiency. For a simulated data sample corresponding to an integrated luminosity of $\int L dt = 100$ pb$^{-1}$ it is expected to determine the efficiency with this approach up to a precision of

$$\Delta\varepsilon = \pm 0.05(sys.) \tag{5}$$

The relative large systematic uncertainty arises mainly from the limited available statistics of background Monte Carlo samples which has a direct impact on goodness of applied fit. Hence, further improvements are likely to be achieved in future studies.

### 3.2 Determination of the reconstruction efficiency with $J/\Psi$ events

The reconstruction efficiency for muons with transverse momenta less than 20 GeV is not determined from $Z \to \mu^+\mu^-$ events due to the cuts on the transverse momenta of the muons. Muons from $J/\psi \to \mu^+\mu^-$ decays populate the momentum range below 20 GeV. We explored the possibility of using the tag-and-probe method on $J/\psi \to \mu^+\mu^-$ events for the measurement of the reconstruction efficiency at low transverse momenta. The method works well on signal events. Yet the huge QCD background contaminates the selected dimuon data sets so much that a reliable efficiency measurement becomes very difficult. Studies using muon isolation techniques have started. The muon reconstruction efficiency of low-$p_T$ muons must therefore be extracted from Monte Carlo simulations and not be determined easily from data.

## 4 Measurement of the momentum resolution and momentum scale

The muon momentum measurement will be affected by the limited knowledge of the magnetic field, the uncertainty in the energy loss of the muons, and the alignment of the muon spectrometer as discussed in Section 2.4.

The analysis of the measurements of the optical alignment sensors and the collision data with the switched-off toroid coils will provide the position of the muon chambers with an accuracy better than 100 $\mu$m at the start-up of the LHC [1]. A systematic error of 100 $\mu$m on the sagitta corresponds to an additional systematic error in the muon momentum of about 0.1 TeV$^{-1} \cdot p^2$ which amounts to 250 MeV for $p$=50 GeV.

Muons with energies below 100 GeV lose on average about 3 GeV of their energy on their passage through the calorimeters almost independently of their energy. The material distribution of the ATLAS detector is modelled in the detector simulation with an accuracy better than a few percent [1]. A 5% uncertainty in the amount of the material traversed by the muons would reflect in a 5% uncertainty of the energy loss, that is an uncertainty of the average energy loss of $\pm$150 MeV.

The uncertainty in the bending power of the toroidal field will lead to a momentum uncertainty which is significantly smaller than the energy loss uncertainty and the impact of the misalignment on the momentum measurement. It can therefore be neglected with respect to energy loss uncertainties and the misalignment of the spectrometer.

A bias in the measured muon momentum translates into a bias in the measurement of the dimuon mass in $Z \to \mu^+\mu^-$ decays. An $\eta$, $\phi$, and momentum dependent bias will also broaden the dimuon mass peak. The shape of the dimuon invariant mass distribution for $Z \to \mu^+\mu^-$ decays can therefore be used to measure the accuracy of the momentum measurement with collision data.

As the momentum bias caused by misalignment is of the same magnitude, but of opposite sign for $\mu^+$ and $\mu^-$ leptons while the energy loss uncertainty has the same sign and magnitude for $\mu^+$ and $\mu^-$ leptons, it is possible to disentangle the effect of misalignment and the effect of energy loss errors on the reconstructed $Z$ mass. The the sensitivities of the dimuon invariant mass spectrum to misalignment and errors in the energy-loss correction were therefore studied separately to get a first insight.

### 4.1 Determination of the energy-loss uncertainty with $Z \to \mu^+\mu^-$ events

We begin with the determination of the energy-loss uncertainty with $Z \to \mu^+\mu^-$ events. We assume that the detector is aligned and that the magnetic field is known with the expected accuracy such that its impact on the momentum scale can be neglected. We allow for an error in the energy-loss and, therefore, correct the reconstructed muon energy in each of the 320 spectrometer towers by a tower-dependent constant $\delta E_{rec,tower}$:

$$E_{rec,tower} \to E_{rec,tower} + \delta E_{rec,tower}. \tag{6}$$

We determine the 400 constants $\delta E_{rec,tower}$ by minimizing

$$\chi^2 = \sum_{dimuon\ pairs\ k} \frac{[(p_{corr,+,k} + p_{corr,-,k})^2 - M_Z^2]^2}{\sigma_k^2} \tag{7}$$

where $p_{corr,\pm,k}$ denotes the corrected measured $\mu^\pm$ momentum and $\sigma_k$ the expected dimuon mass resolution. To estimate the sensitivity to the energy-loss correction, we applied this fit to 40,000 simulated $Z \to \mu^+\mu^-$ events (corresponding to an integrated luminosity of 50 pb$^{-1}$). The fit gives $\delta E_{rec,tower}$ with a bias of 100 MeV and a stastitical error of the same size. Studies to improve the check of the energy-loss correction with collision data are ongoing.

### 4.2 Determination of the momentum scale and resolution for a misaligned spectrometer

In a second step, we assume that the energy-loss correction is right and consider the misalignment of the muon spectrometer as the only source of a deterioration of the momentum measurement.

If the Monte Carlo simulation describes the detector correctly, it also predicts the shape of the reconstructed dimuon mass spectrum for $Z \to \mu^+\mu^-$ events correctly. The misalignment of the muon chambers causes a deviation of the measured from the predicted shape of the invariant dimuon mass spectrum. In order to match the Monte Carlo prediction with the experimental measurement, the reconstructed simulated muon momenta must be smeared and shifted. The following procedure was adopted in our analysis: A random number $\delta p$ normally distributed around 0 with standard deviation $\sigma_{res}$ was added to the reconstructed simulated muon momenta $p_{rec,MC}$ and multiplied by a scale factor $\alpha$:

$$p_{corr} = \alpha(p_{rec,MC} - \delta p). \tag{8}$$

The inclusion of $\delta p$ corrects for an underestimation of the momentum resolution. The scale factor $\alpha$ takes care of systematic shifts between the reconstructed momenta in the experiment and the simulation. $\alpha$ and $\sigma_{res}$ are determined by a fit of the corrected simulated invariant dimuon mass spectrum to the experimentally measured spectrum.

To test this approach, the existing $Z \to \mu^+\mu^-$ Monte Carlo data set was divided into two subsamples of equal size corresponding to an integrated luminosity of 50 pb$^{-1}$. The one sample serves as Monte Carlo reference for an aligned muon spectrometer, the other plays the role of the experimental data set. Two scenarios were investigated:

1. The Monte Carlo reference sample and the experimental sample were simulated and reconstructed with the same (aligned) geometry. $\sigma_{res}$ was fixed to 0 in the analysis of this scenario. Separate scale factors $\alpha_B$ and $\alpha_E$ were applied to muon in the barrel ($|\eta| < 1$) and the end-cap region ($1 \leq |\eta| < 2.7$).

2. The Monte Carlo reference sample was simulated and reconstructed with the same (aligned) geometry. But the experimental sample was reconstructed with a different geometry misaligned as described in Section 2.4. In this scenario, two scale factors $\alpha_B$ and $\alpha_E$ for the barrel and end-cap parts of the muon spectrometer and a global standard deviation $\sigma_{res}$ were used as fit parameters.

Table 4: Fit results for the scale and resolution parameters for an integrated luminosity of 50 pb$^{-1}$.

| Layout | $1-\alpha_B$ | $1-\alpha_E$ | $\sigma_{res}$ |
|---|---|---|---|
| Aligned | $(4\pm14)10^{-4}$ | $(1\pm13)10^{-4}$ | – |
| Misaligned | $(6\pm2)10^{-3}$ | $(5\pm2)10^{-3}$ | $(11.6\pm0.3)$ % |

The results of the tests are summarized in Table 4. In the ideal case in which the reference and the experimental sample are statistically independent, but equivalent otherwise, the fit gives factors $\alpha_B$ and $\alpha_E$ equal to 1 within the statistical errors as expected. In the second scenario, the uncorrected misalignment in the experimental sample leads to a systematic shift of the reconstructed momenta, hence $\alpha_B$ and $\alpha_E$ differ from 1 slightly, but significantly, and a large degradation of the momentum resolution from 3.5% to 12% is observed which is consistent with the degradation presented in Section 6. A large $Z \to \mu^+\mu^-$ sample would clearly allow for a finer segmentation than the division in barrel and end-cap parts for the scale factors and lead to smaller values of $\sigma_{res}$.

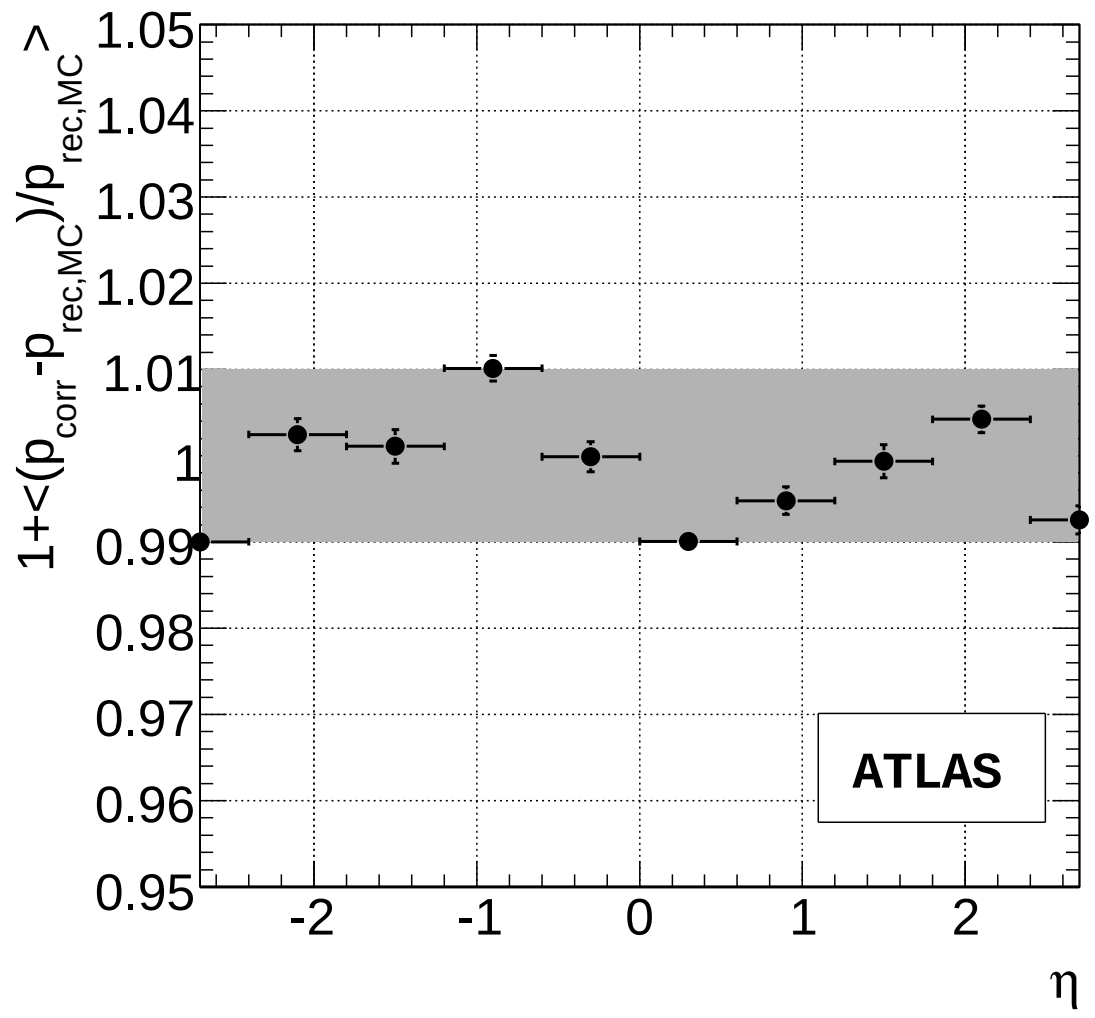

Figure 20: Dependence of $< p_{corr} - p_{rec,MC} > /p_{rec,MC}$ on $\eta$ integrated over $p_T$ and $\phi$ for $Z \to \mu^+\mu^-$ events in the second scenario of a misaligned detector.

The mean value of $1+\frac{p_{corr}-p_{rec,MC}}{p_{rec,MC}}$ is presented in Figure 20 as a function of $\eta$ for the second scenario. The mean values are spread around 1 with a standard deviation of 0.3%. The maximum deviation from 1 is less than 1%. This results indicates that the $Z$-mass distribution permits the detection of imperfections in the momentum reconstruction. Studies which use a more refined parametrization of the momentum correction and take into account energy-loss and alignment corrections at the same time are in progess.

We conclude from the studies in this section that it should be possible to control the muon momentum and energy scale on the level of 0.5 GeV for 50 GeV muons with 40,000 $Z \to \mu^+\mu^-$ events corresponding to an integrated luminosity of 50 $pb^{-1}$.

# 5 Conclusions

The performance of the ATLAS muon spectrometer can be predicted by Monte Carlo simulations. The performance of the spectrometer will, however, differ from the prediction due to the initial misalignment of the muon chambers and imperfections in the corrections of the muon energy-loss. It is therefore important to measure the performance with collision data.

We showed in the present article that it is possible to measure the muon reconstruction efficiency with $Z \to \mu^+\mu^-$ events with an accuracy better than 1% with an integrated luminosity of 100 $pb^{-1}$. Selection cuts and the $p_T$ spectrum of the $Z$ decay muons limit the momentum measurement to range of 20 GeV$< p_T <$70 GeV. The efficiency measurement can be extended to higher momenta with increased luminosity when the tails of the $p_T$ spectrum get populated.

We explored the possibility of measuring the efficiency at low transverse momenta with $J/\psi \to \mu^+\mu^-$ events. Our studies show that a reliable efficiency measurement will be difficult due to large irreducible QCD background.

We finally addressed the question of how the momentum and energy scale can be measured with $Z \to \mu^+\mu^-$. According to our feasibility study it will be possible to control the energy-loss correction on the level of 100 MeV and the momentum scale on the level of 1% for an integrated luminosity of about 100 $pb^{-1}$. More detailed studies are needed to obtain a better estimate of the achievable precision.

# Tau Leptons

# Reconstruction and Identification of Hadronic $\tau$ Decays

**Abstract**
In this note the overall performance of the ATLAS detector is discussed for the identification of and measurements with hadronic decays of $\tau$ leptons in a wide dynamic range of transverse energies, spanning from 10-15 GeV up to at least 500 GeV. In general, hadronically decaying $\tau$ leptons are reconstructed by matching narrow calorimetric clusters with a small number of tracks. Two complementary approaches, a calorimeter-seeded and a track-seeded algorithm, have been developed to efficiently reconstruct these decays while providing the necessary large rejection against jets from QCD processes. The performance of these algorithms in terms of efficiency and rejection against jets is discussed. In addition, the prospects for the determination of fake $\tau$ rates as well as the extraction of $\tau$ lepton signals from $W$ and $Z$ boson decays and from $t\bar{t}$ events in early ATLAS data corresponding to an integrated luminosity of 100 pb$^{-1}$ are discussed.

## 1 Introduction

Tau leptons, and particularly their hadronic decays, will play an important role at the LHC. They will provide an excellent probe in searches for new phenomena: the Standard Model Higgs boson at low masses, the MSSM Higgs boson or Supersymmetry (SUSY). Therefore, understanding their selection efficiencies and the cross-sections at which they will be produced is essential for discovering new physics.

Tau leptons are massive particles with a measurable lifetime undergoing electroweak interactions only. The production and the decay of $\tau$ leptons are well separated in time and space ($\Gamma_\tau/m_\tau \sim 10^{-11}$), providing potential for unbiased measurements of the polarisation, spin correlations, and the parity of the resonances decaying into $\tau$ leptons. The excellent knowledge of $\tau$ decay modes from low energy experiments indeed makes this an ideal signature for the observations of new physics.

The interesting transverse momentum range of $\tau$ leptons spans from below 10 GeV up to at least 500 GeV. Experiments at the LHC will thus have to identify them in a wide momentum range. The low energy range should be optimized for analyses related to W and Z boson observability with $\tau$ decays and also to Higgs boson searches and SUSY cascade decays. The higher energy range is mostly of interest in searches for heavy Higgs bosons in MSSM models and for extra heavy W and Z gauge bosons. For illustration, Fig. 1 shows the transverse energy spectrum of the visible decay products of $\tau$ leptons from different processes of interest normalized to the predicted cross-section with which they will be produced at the LHC and to an integrated luminosity of 10 fb$^{-1}$.

The reconstruction of $\tau$ leptons is usually understood as a reconstruction of the hadronic decay modes, since it would be difficult to distinguish leptonic modes from primary electrons and muons. Despite a strong physics motivation for exploring data with $\tau$ leptons in the final state, their reconstruction at hadron colliders remains a very difficult task in terms of distinguishing interesting events from background processes dominated by QCD multi-jet production. Another related challenge is providing efficient triggering for these events while keeping trigger rates at manageable levels.

The availability of various decay modes makes $\tau$ leptons a rich but not totally unique signature. Hadronically decaying $\tau$ leptons[1] are distinguished from QCD jets on the basis of low track multiplicities contained in a narrow cone, characteristics of the track system and the shapes of the calorimetric showers. Isolation from the rest of the event is required both in the inner detector and the calorimeter. From this information, a set of identification variables is built, to which either a traditional cut-based selection or multi-variate discrimination techniques are applied.

[1]We will often use notation $\tau_{had}$ in this note when discussing objects reconstructed from the visible part of the hadronic decay products of a $\tau$ lepton.

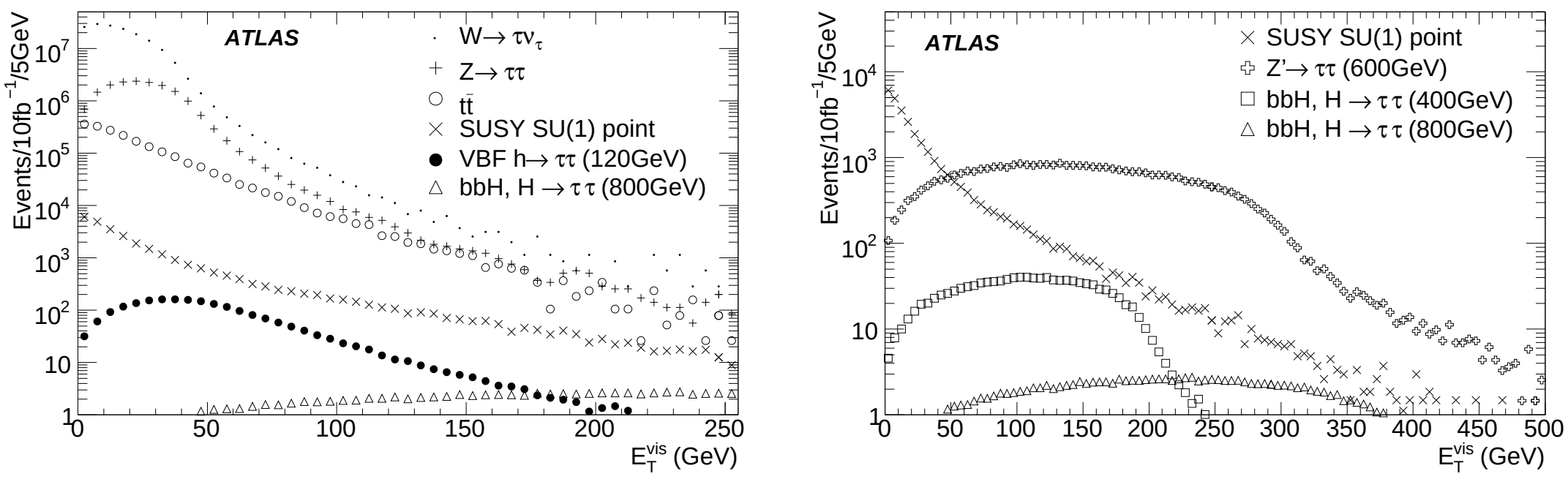

Figure 1: The visible transverse energy of $\tau$ leptons from different physics processes: top quark decays, W/Z production, Standard Model vector boson fusion Higgs boson production for $m_H = 120$ GeV with $H \to \tau\tau$, for $\tau$ leptons from low energy Supersymmetry with a light stau (SU1 sample), heavy $Z'$ bosons, and heavy Higgs bosons from $bbH$ production in the MSSM with $\tan\beta = 20(45)$ for masses of 400 GeV (800 GeV).

The inner detector provides information on the charged hadronic track or the collimated multi-track system reconstructed in isolation from the rest of the event. These tracks should neither match track segments in the muon spectrometer nor reveal features characteristic of an electron track (e.g. high threshold hits in the Transition Radiation Tracker). In the case of a multi-track system, they should be well collimated in $(\eta, \phi)$ space and the invariant mass of the system should be below the $\tau$ lepton mass. The charge of the decaying $\tau$ lepton can be directly determined from the charge(s) of its decay product(s).

Calorimetry provides information on the energy deposit from the visible decay products (i.e. all decay products excluding neutrinos). Hadronically decaying $\tau$ leptons are well collimated (with an opening angle limited by the ratio $m_\tau/E_\tau$ ) leading to a relatively narrow shower in the electromagnetic (EM) calorimeter with, for single-prong decays with one or few $\pi^0$'s, a significant pure electromagnetic component. On average in this case about 55% of the energy is carried by $\pi^0$s present among the decay products.

The calorimeter and tracking information should match, with narrow calorimeter cluster being found close to the track(s) impact point in the calorimeter. Furthermore, the invariant mass of the cluster should be small and the cluster should be isolated from the rest of the event.

The algorithms for the reconstruction of hadronically decaying $\tau$ leptons are considered higher level reconstruction as they use components provided by algorithms specific to different subdetectors like track reconstruction in the inner detector or topological clustering of the energy deposits in the calorimeter. At present, two complementary algorithms have been implemented into the ATLAS offline reconstruction software.

- The calorimetry-based algorithm starts from clusters reconstructed in the hadronic and electromagnetic calorimeters and builds the identification variables based on information from the tracker and the calorimeter.

- The track-based algorithm starts from seeds built from few (low multiplicity) high quality tracks collimated around the leading one. The energy is calculated with an energy-flow algorithm based only on tracks and the energy in the electromagnetic calorimeter. All identification variables are built using information from the tracker and the calorimeter.

A short overview of the features of $\tau$ lepton decays is included in Section 2. In Section 3 selected topics on the performance of the detector directly relevant to the reconstruction and identification of the hadronically decaying $\tau$ leptons are discussed. Offline reconstruction algorithms and performance results are described in Section 4. In the remaining part of the note strategies for analyses using $\tau$ leptons with the first 100 pb$^{-1}$ of data are presented.

## 2 Topology of $\tau$ leptons in LHC collisions

The transverse momentum range of interest spans from below 10 GeV up to 500 GeV. $\tau$ leptons decay hadronically in 64.8% of all cases, while in $\sim$ 17.8% (17.4%) of the cases they decay to an electron (muon) [1]. From the detection point of view, hadronic modes are divided by the number of charged $\pi$s among the decay products into single-prong (one charged $\pi$) and three-prong (three charged $\pi$s) decays. The small fraction (0.1%) of five-prong decays is usually too hard to detect in a jet environment. The $\tau \to \pi^{\pm}\nu$ mode contributes 22.4% to single-prong hadronic decays and the $\tau \to n\pi^0\pi^{\pm}\nu$ modes 73.5%. For three-prong decays, the $\tau \to 3\pi^{\pm}\nu$ decay contributes 61.6%, and the $\tau \to n\pi^0 3\pi^{\pm}\nu$ mode only 33.7%. In general, one- and three-prong modes are dominated by final states consisting of $\pi^{\pm}$ and $\pi^0$. There is a small percentage of decays containing $K^{\pm}$ which nevertheless can be identified using the same technique as for states with $\pi^{\pm}$ from the ATLAS detector point of view. A small percentage of states with $K^0_S$ cannot be easily classified as belonging to either the single-prong or three-prongs categories as the number of registered prongs depends on the actual $K^0_S$ interaction within the detector. Unless specific studies are done, other multi-prong hadronic modes can be safely neglected.

The lifetime of the $\tau$ lepton ($c\tau = 87.11\mu m$) in principle allows for the reconstruction of its decay vertex in the case of three-prong decays. The flight path in the detector increases with the Lorentz boost of the $\tau$ lepton, but at the same time the angular separation of the decay products decreases. A resulting transverse impact parameter of the $\tau$ decay products can be used to distinguish them from objects originating from the production vertex.

The incorporation of spin effects in $\tau$ lepton decays is often of importance. This was done within the framework of the ATLAS Monte Carlo simulation and events were generated using `PYTHIA` [2] interfaced with `TAUOLA` [3]. The generation process has correctly included full spin correlations in production and decays of the $\tau$ leptons. Tau leptons from the decay of gauge bosons, Higgs bosons or in SUSY cascade decays will carry information on the polarisation of the decaying resonance and in the case of pair production also some information on the spin correlations. Tau leptons from $W \to \tau\nu$ and $H^{\pm} \to \tau\nu$ will be 100% longitudinally polarised, with $P_{\tau}$ = +1.0 and $P_{\tau}$ = -1.0 respectively, resulting in different distributions of the charged to total visible energy for single-prong decays in the center-of-mass system of the decaying resonance. At the LHC this effect can be used to suppress $W \to \tau\nu$ background and to increase the $H^{\pm} \to \tau\nu$ observability [4]. The $\tau$ polarisation could also be used as a tool to discriminate between MSSM versus Extra Dimension scenarios [5]. The longitudinal polarisation of $\tau$ leptons from neutral Higgs boson decays will be democratic with 50% probability, thus $\tau$ leptons from Higgs boson decays are effectively not polarized. The polarisation of $\tau$ leptons from Z boson decays will be a more complicated function of the center-of-mass energy of the system and the angle of the decay products [6]. In the cleaner environment of the ILC and also perhaps at the sLHC, building variables sensitive to the longitudinal and transverse spin correlations may lead to a CP measurement of the Higgs boson [7, 8].

## 3 Performance of the ATLAS detector for $\tau$ identification

Hadronic $\tau$ decays can be efficiently reconstructed and identified using information from the inner detector and from the calorimeter. Reconstruction is done only for the visible part of the decay products, however, for specific analyses like $H \to \tau\tau$ the complete invariant mass of the $\tau\tau$ system may be reconstructed using the collinear approximation [9] (neutrino momenta parallel to that of the visible decay products). A few selected topics related to the performance of the detector are discussed below before turning to the reconstruction algorithms.

## 3.1 Tracking and vertexing

The reconstruction of tracks from charged pion decays is an important ingredient of the $\tau_{had}$ reconstruction algorithms. The track-based algorithm is seeded by one or more good quality tracks which allow for the calculation of the $\tau_{had}$ energy with the so called energy-flow scheme. Both the calo-based and the track-based algorithm determine the charge of the $\tau_{had}$ candidate by summing up the charge(s) of the tracks reconstructed in the $\tau_{had}$ core region[2]. The tracking information is further used to identify hadronically decaying $\tau$ leptons and to discriminate them against the background from hadronic jets by considering the track multiplicity, the impact parameter and the transverse flight path in the case of multi-track candidates. The track selection should therefore ensure high efficiency and quality of the reconstructed tracks over a broad momentum range from 1 GeV to a few hundred GeV.

### 3.1.1 Reconstruction efficiency and track quality

The efficiency for track reconstruction in $\tau$ decays is defined as the probability for a given charged $\pi$ from a $\tau$ decay to be reconstructed as a track. With respect to the reference tracking performance of the detector established for single muons in the low $p_T$ range a degradation due to hadronic interactions (a charged $\pi$ interacting with the material of the inner detector) is expected. In the higher $p_T$ range a degradation is caused by the strong collimation of the multi-track system for three-prong decays.

Good quality tracks reconstructed with $p_T$ as low as 1 GeV are required by the track-based algorithm, while the calorimeter-based algorithm accepts any track with $p_T > 2$ GeV. A standard quality selection has been defined in Ref. [10]. However, for the reconstruction of $\tau$ leptons a somewhat stricter selection has been applied. Good quality tracks are required to satisfy $\chi^2/\text{n.d.f} < 1.7$, to have a number of pixel and SCT hits $\geq 8$ and transverse impact parameters $d_0 < 1$mm. For the leading track in addition the number of low threshold TRT hits has to be larger than 10 in a pseudorapidity $\eta$ range up to 1.9, while for the second or third track the presence of a B-Layer hit and ratio of the of high-to-low threshold hits of smaller than 0.2 are required. Both requirements were added to minimize the number of accepted tracks from conversions. A dedicated veto against electron tracks being used as leading tracks is not applied at the reconstruction level. This will be taken care of separately as part of the identification procedure.

Figure 2 shows the reconstruction efficiency for $p_T = 1 - 50$ GeV using the standard quality selection as defined in Ref. [10]. Adding the additional quality criteria as described above, the overall efficiency for reconstructing good quality tracks from $\tau$ lepton hadronic decays is reduced to $82 - 83\%$. The reconstruction efficiency is slightly higher for tracks from single prong decays compared to three-prong decays, where tracks could be very collimated particularly for boosted $\tau$ leptons.

### 3.1.2 Charge misidentification

The charge of the $\tau$ lepton is calculated as the sum of the charges of the reconstructed tracks. For the leading track, which is required (e.g. by the track-based algorithm) to have a transverse momentum [3] larger than 9 GeV, charge mis-identification is limited to $\sim 0.2\%$ using the quality cuts described above.

[2]The core region for the track-based (calo-based) algorithm is understood here as $\Delta R < 0.2(0.3)$ cone in $(\eta, \phi)$ around the reconstructed direction of the visible decay products.

[3]The threshold $p_T > 9$ GeV on the leading track was used for results presented here, while it was lowered to 6 GeV in the more recent software releases.

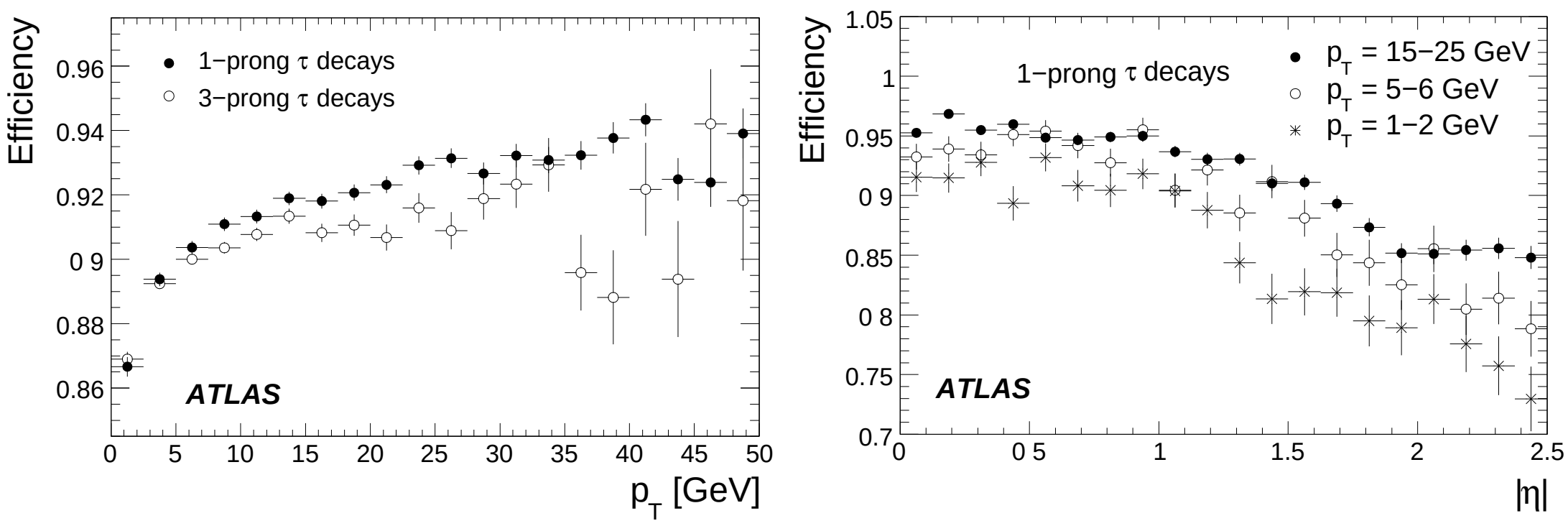

Figure 2: Reconstruction efficiency for tracks from charged $\pi$s for one- and three-prong hadronic $\tau$ decays from $W \to \tau\nu$ and $Z \to \tau\tau$ signal samples as a function of the transverse momentum of the track (left) and of the pseudorapidity for three different ranges of track $p_T$ (right).

The overall charge mis-identification probability for the hadronically decaying $\tau$ lepton is however dominated by combinatorial effects: single-prong decays may migrate to the three-prong category due to photon conversions or the presence of additional tracks from the underlying event. A three-prong decay might be reconstructed as a single-prong decay due to inefficiencies of the track reconstruction and selection. This overall charge mis-identification is estimated to be below $\sim 3.6\%$ without requiring additional quality cuts. In fact, the $\tau$ charge misidentification is dominated by a combination of effects, but the contributions from the charge misidentification of the individual tracks should not be neglected.

Table 1 shows the percentage of contamination for one- and three-prong candidates using the aforementioned quality criteria for tracks in the core region. For the roughly 3.9% contamination of the single-track candidates from three-prong decays, about 85% are due to hadronic interactions. A 3.8% contamination of three-track candidates from one-prong decays is observed with 70% of them being due to conversions. The percentage of the overall charge misidentification is also shown. Requiring at least one B-Layer hit reduces the charge misidentification both in the case of electron tracks from conversions and in the case of hadronic interactions at low radii. However, this happens at the expense of an additional loss in efficiency in particular for three-prong decays.

### 3.1.3 Tracks from conversions

Photons from $\pi^0$ decays might convert in the material of the inner detector and then contribute additional tracks to the core or isolation region of the $\tau_{had}$ candidate. This could result in one-prongs being reconstructed as three-prong candidates, in an inefficiency of the reconstruction and identification criteria and in a degradation of the energy resolution as calculated from the energy-flow algorithm.

A large fraction of reconstructed $\tau_{had}$ candidates are accompanied by conversions. In 1.5% of the cases a conversion electron is reconstructed as the leading track of the one-prong candidate, while 5.7% of the three-prong candidates contain one reconstructed track coming from a conversion electron. In Table 1 the effects of charge misidentification and contamination from photon conversions are quantified.

### 3.1.4 Impact parameter

The mean proper lifetime of the $\tau$ lepton is about 0.29 ps. Although the lifetime of the $\tau$ lepton is about five times shorter than that of the $b$-quark, the transverse impact parameters of its decay products are still useful for $\tau$ identification. The impact parameters, $d_0$ and $z_0 \sin(\theta)$, have been studied for one-prong candidates reconstructed by the track-based algorithm. The transverse impact parameter $d_0$ is defined as the smallest distance in the transverse plane between the track and the reconstructed primary vertex. The impact parameter $z_0$ is given by the distance in $z$-direction between the reconstructed primary vertex and the point of closest approach in the transverse plane of the track multiplied by $\sin(\theta)$ to obtain the component transverse to the track direction. Tracks assigned to the $\tau_{had}$ candidate are not used in the primary vertex fit.

Table 1: Percentage of one- and three prong $\tau$ lepton hadronic decays within reconstructed one-, two- and three-prong $\tau_{had}$ candidates by the track-based algorithm, matched to true $\tau$ decays. Tracks in a cone of $\Delta R = 0.2$ around the leading good quality track are considered. A transverse momentum of $p_T > 9$ GeV is required for the leading track. An estimate for electron contamination and charge misidentification is given in addition. Separately specified are results for a subsample where no hadronic secondary interaction of primary charged $\pi$ was recorded inside the inner detector volume. Events from $Z \to \tau\tau$ and $W \to \tau\nu$ samples were used.

| Seeds for track-based $\tau_{had}$-candidates | Reconstructed as single-prong | Reconstructed as three-prong | Reconstructed as two-prong |
|---|---|---|---|
| Electron contamination (from conversion) | 1.5% | 5.7% | 2.9% |
| $\tau \to \pi^{\pm} n\pi^0 \nu$ | 96.1% | 3.8% | 23.8% |
| $\tau \to 3\pi^{\pm} n\pi^0 \nu$ | 3.9 % | 96.2% | 76.2% |
| Charge misid. | 1.7% | 3.6% | |
| (no had. interact.) | 0.4% | 2.1% | |

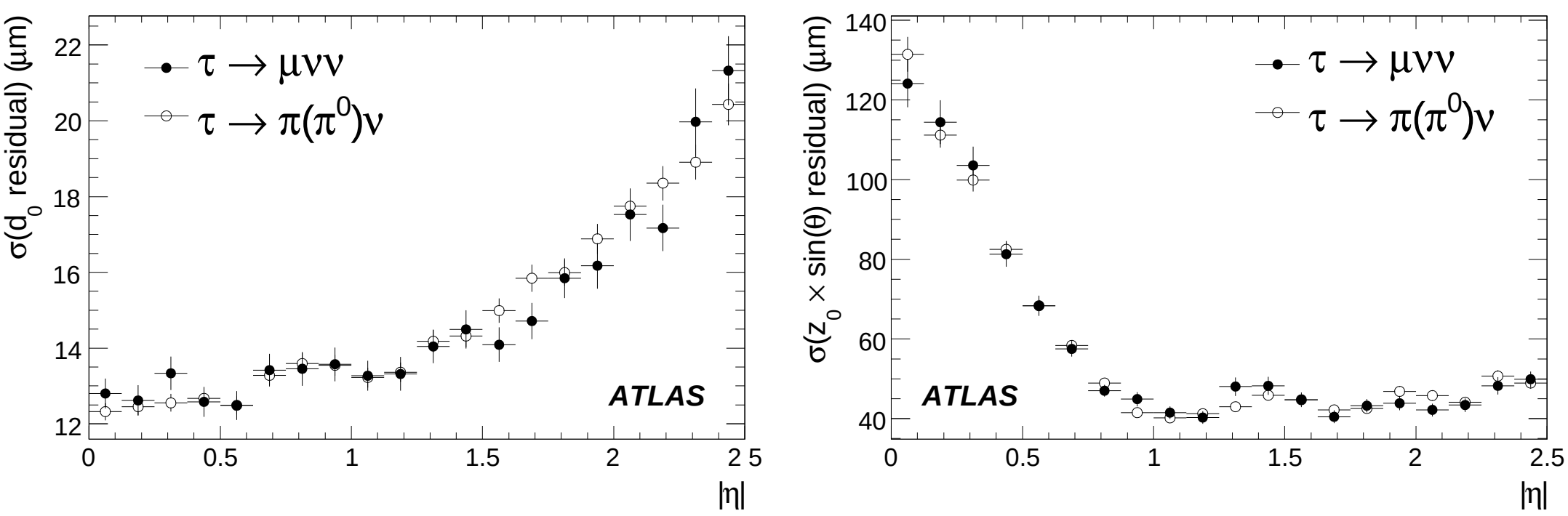

Figure 3: Transverse (left) and longitudinal (right) impact parameter resolution as a function of $|\eta|$ from a one-prong $Z \to \tau\tau$ sample. The open (full) circles are from $\tau \to \pi(\pi^0)\nu$ ($\tau \to \mu\nu\bar{\nu}$) events.

In Fig. 3 the resolution of the transverse (left) and longitudinal (right) impact parameters are shown as a function of $|\eta|$. The resolution for final state muons and pions from $\tau \to \mu\nu\nu$ and $\tau \to \pi(\pi^0)\nu$ decays are similar: about 13 $\mu$m for $|\eta| < 1.0$ and about 50 $\mu$m for $|\eta| > 1.0$ for the transverse and longitudinal impact parameters, respectively. No degradation related to hadronic interactions for $\tau \to \pi\nu$ decays is observed. This has been verified by studying events with elastic interaction in the Inner Detector (hadronic interaction), defined as events where the outgoing $\pi^{\pm}$ carries more than 90 % of the transverse energy of the incoming $\pi^{\pm}$. This observation is consistent with that presented in Ref. [11].

In Fig. 4 distributions of the significances of the impact parameters are presented. The significance is defined as the impact parameters divided by its estimated error. The distribution shows a moderate discrimination power between one-prong candidates reconstructed from hadronic $\tau$ decays and fake one-prong candidates. Due to the limited resolution of the longitudinal impact parameter, the separation

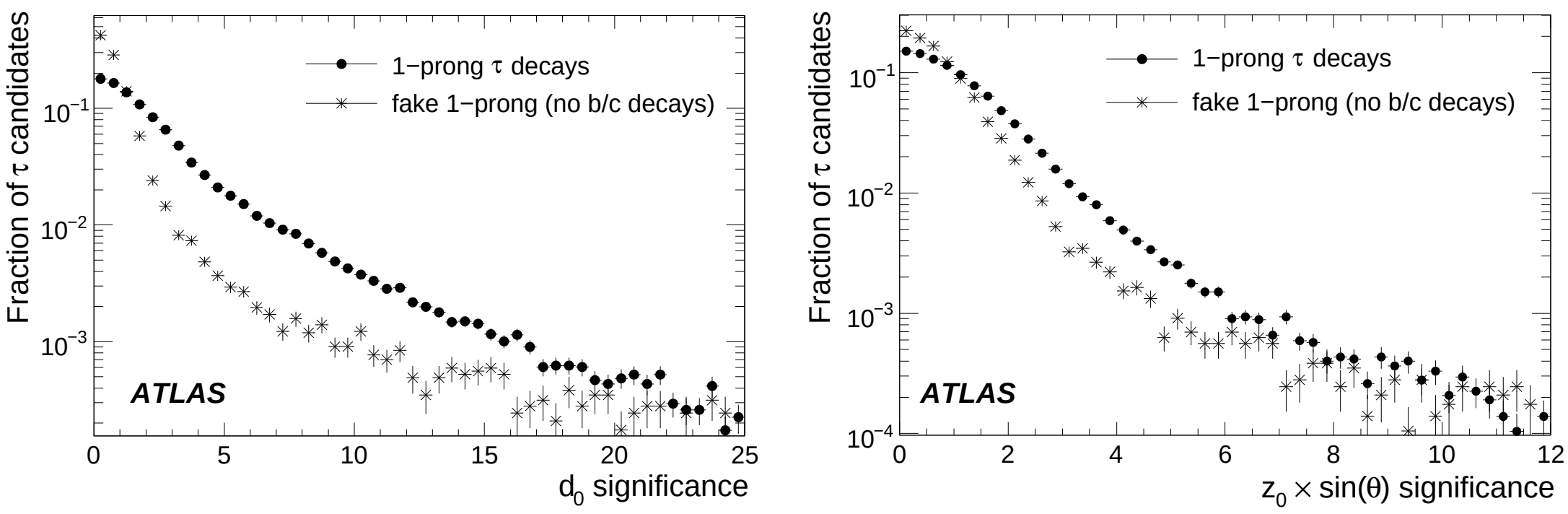

Figure 4: Significances of the impact parameters $d_0$ (left) and $z_0 \sin(\theta)$ (right) for 1-prong $\tau_{had}$ candidates reconstructed by the track-based algorithm. Distributions are shown for $\tau_{had}$ candidates reconstructed from $\tau$ decays and for fake candidates which do not originate from the decays of *b*- or *c*-hadrons.

between the two classes is less significant in that case.

### 3.1.5 Secondary vertex reconstruction and transverse flight path

The significant lifetime of the $\tau$ lepton ($c\tau = 87.11\mu$m) allows the reconstruction of its decay vertex for three-prong decays. Currently, five vertex fitting algorithms [12–14] are implemented in the ATLAS reconstruction framework. Among these the adaptive vertex fitter [13], an iterative re-weighted fit which down-weights tracks according to their weighted distance to the vertex, was found to give the optimal performance.

To estimate its performance, secondary vertex fits were performed using tracks associated with $\tau_{had}$ candidates from $Z \to \tau\tau$ and $W \to \tau\nu$ events. The quality criteria applied in the track-based reconstruction were required to be met by the tracks. The $\tau_{had}$ candidates associated with a true hadronic $\tau$ decay were divided into two classes. Candidates with three tracks successfully matched to true particles coming from the same true three-prong hadronic $\tau$ decays were used as a reference. These candidates are denoted hereafter as fully-matched. The second class is composed of the remaining candidates with at least two tracks of which at least one is matched to a true particle coming from a hadronic $\tau$ decay. These candidates are denoted hereafter as partially matched.

The resolution of the secondary vertex position varies strongly if measured in the perpendicular or parallel direction with respect the momentum of $\tau_{had}$ candidate. The resolution on the position of the secondary vertex calculated in the plane perpendicular to the momentum of the $\tau_{had}$ candidate is expected to be better than in the parallel direction due to the collimation of tracks. To estimate the resolution in the transverse plane, the residuals of the vertex position in the direction perpendicular to both momentum of a $\tau_{had}$ candidate and the beam axis were calculated. The distributions were approximated by a double Gaussian fit. For fully matched three-prong $\tau_{had}$ candidates there is no significant difference between the distributions obtained with different fitters. The distributions of residuals of the secondary vertex position obtained with the adaptive fitter, parallel and perpendicular to the direction of flight of the $\tau_{had}$ candidate are presented in Fig. 5. Shown in Table 2 are the resolution[4], the mean values of the fit and 68.3 % and 95.0 % coverages[5] for the fully matched, partially matched and combined samples. As expected, the transverse resolution ($\sigma \sim 10\ \mu m$) is far more accurate than the parallel one ($\sigma \sim 600\ \mu m$). The non-Gaussian tails are significant in both cases, but far more important in the case of the parallel component. A precise reconstruction of the transverse flight path is therefore possible, which is important for further rejection of the QCD background. It may also be possible to obtain a competitive measurement of the $\tau$ lifetime, which requires a measurement of the flight path and momentum of the $\tau$ lepton.

[4]In the case of a double Gaussian fit, the width of the central Gaussian will be quoted as the resolution hereafter.
[5]The coverage is the half-width of a symmetric interval covering a given percentage of the distribution.

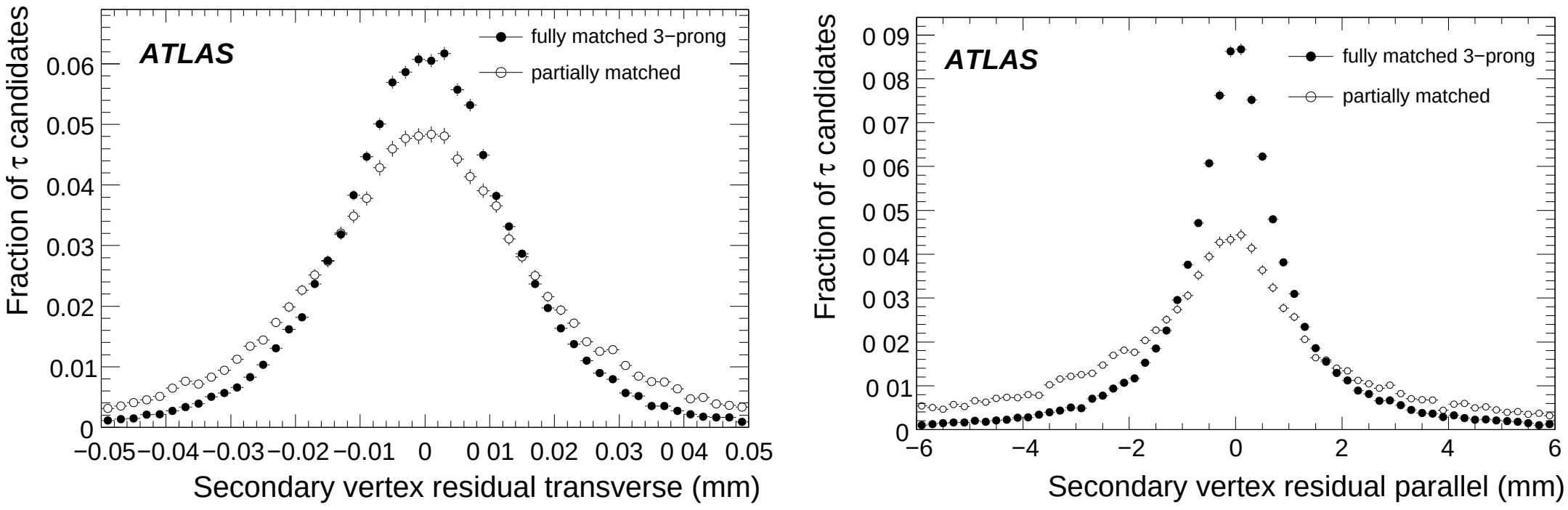

Figure 5: Residuals of the secondary vertex position parallel and perpendicular to the direction of flight of the $\tau_{had}$ candidate using the adaptive vertex fitter. Fully (solid) and partially (open) matched three-prong $\tau_{had}$ candidates reconstructed with the track-based algorithm from $Z \to \tau\tau$ and $W \to \tau\nu$ processes are used.

Table 2: Resolution and mean of the distribution of residuals of the secondary vertex position in the directions parallel and transverse to that of the reconstructed momentum vector of the $\tau_{had}$ candidate as obtained from the adaptive vertex fitter. Candidates with up to three associated tracks reconstructed by the track-based algorithm were used. The resolution quoted is the $\sigma$ of the core Gaussian of a double Gaussian fit in the range $[-4\,\text{mm}, 4\,\text{mm}]$ in the parallel direction and $[-50\,\mu\text{m}, 50\,\mu\text{m}]$ in the transverse direction. The 68.3% and 95% coverages are also quoted.

| | Resolution | Mean | 68.3% | 95% |
|---|---|---|---|---|
| | Parallel | | | |
| Fully matched 3-prong | $0.593 \pm 0.008$ mm | $0.006 \pm 0.006$ mm | 1.27 mm | 5.33 mm |
| Partially matched | $0.703 \pm 0.030$ mm | $-0.035 \pm 0.020$ mm | 3.83 mm | $> 15$ mm |
| Combined | $0.613 \pm 0.008$ mm | $0.004 \pm 0.006$ mm | 1.89 mm | 11.37 mm |
| | Transverse | | | |
| Fully matched 3-prong | $10.1 \pm 0.2\,\mu$m | $0.2 \pm 0.1\,\mu$m | $14.4\,\mu$m | $36.9\,\mu$m |
| Partially matched | $11.3 \pm 0.5\,\mu$m | $-0.1 \pm 0.2\,\mu$m | $20.9\,\mu$m | $72.2\,\mu$m |
| Combined | $10.5 \pm 0.2\,\mu$m | $0.1 \pm 0.1\,\mu$m | $16.4\,\mu$m | $48.1\,\mu$m |

Shown in Fig. 6 is the resolution on the transverse flight path as a function of the transverse momentum and the pseudorapidity of the $\tau_{had}$ candidates. The resolution was obtained from a Gaussian fit to a central interval covering 80% of distributions of residuals of a transverse flight path for the adaptive vertex fitter for fully matched three-prong candidates. In addition the 68.3% and 95% coverages of distributions of residuals are presented.

Distributions of the significance of the transverse flight path, for different classes of three-prong candidates are shown in Fig. 7. This distribution might be used to discriminate between true $\tau_{had}$ candidates and fake candidates from light jets. The discrimination in the case of *b*- and *c*- jets seems however to be difficult.

An efficient rejection of tracks coming from photon conversions, decays of long-lived particles and hadronic interactions in the material reduce the number of one-prong candidates wrongly reconstructed as two- or three-prong candidates and improve the separation between correctly reconstructed three-prong candidates and candidates from light jets. It is a subject of further studies currently in progress.

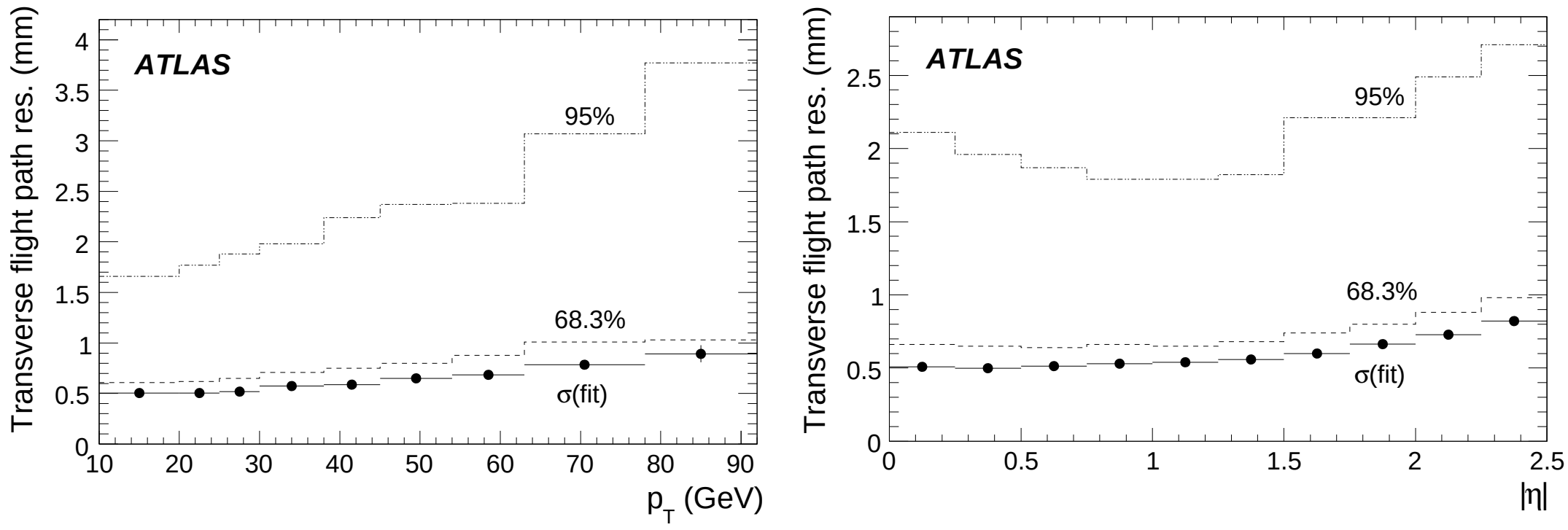

Figure 6: Resolution on the transverse flight path reconstructed with the adaptive vertex fitter for fully matched three-prong $\tau_{had}$ candidates as a function of the transverse momentum (left) and the pseudorapidity (right). Standard deviations of Gaussians fitted to central intervals covering 80% of the residual distributions are shown (black points). In addition the 68.3 % and 95 % coverages of the distributions of residuals of the secondary vertex position are shown (dashed and dot-dashed lines).

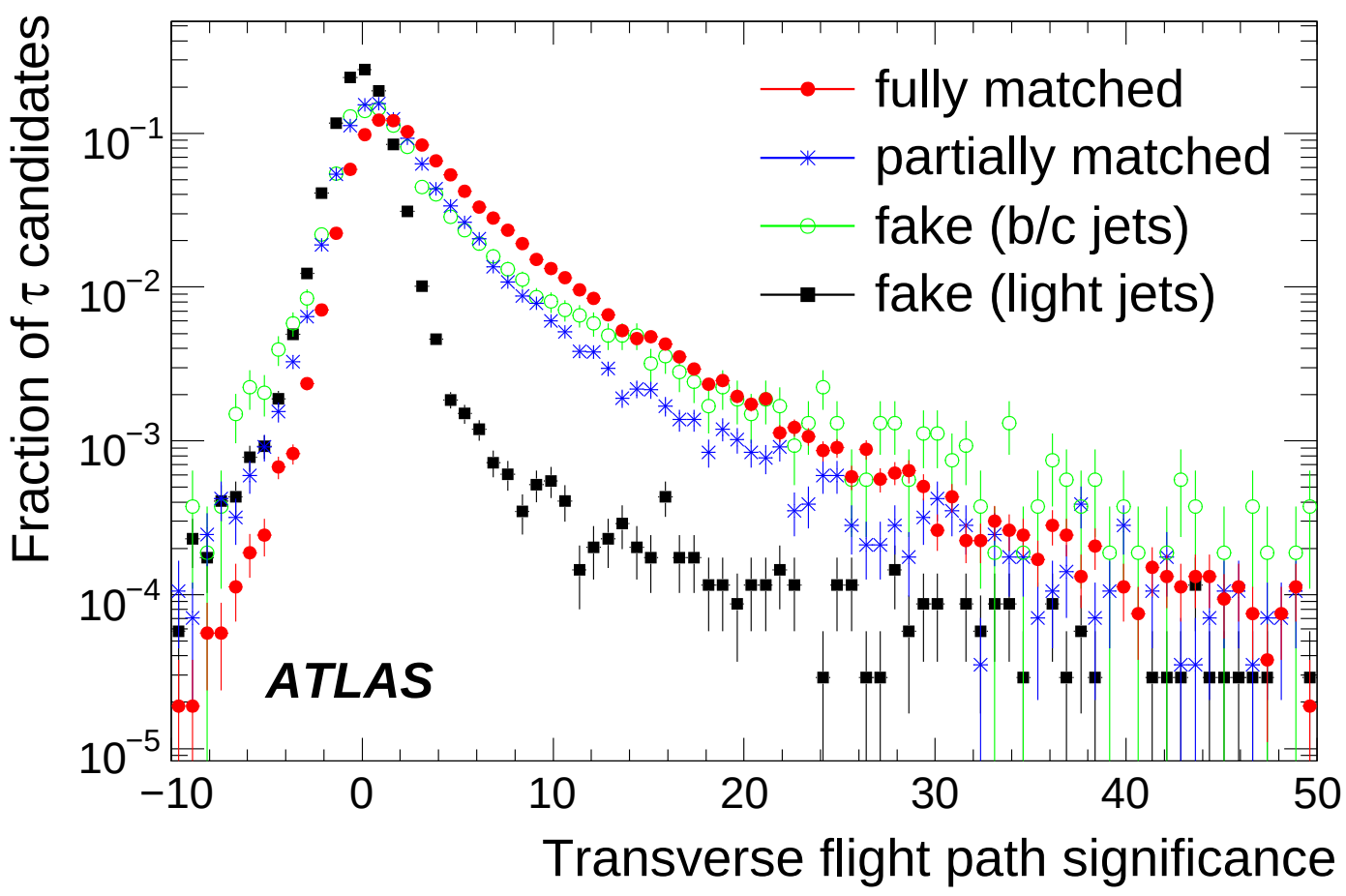

Figure 7: Significance of the transverse flight path for fully matched and partially matched three-prong and for fake candidates with and without hadrons containing b or c quarks (the contribution from semileptonic decays of b/c jets into $\tau$ leptons was not subtracted).

Table 3: Single prong candidates: fractions with zero, one and two or more reconstructed $\pi^0$ subclusters.

| decay mode | no $\pi^0$ subclusters | 1 $\pi^0$ subcluster | $\geq 2$ $\pi^0$ subclusters |
|---|---|---|---|
| all $\tau \to$ had$\nu$ | 32% | 35% | 33% |
| $\tau \to \pi\nu$ | 65% | 20% | 15% |
| $\tau \to \rho\nu$ | 15% | 50% | 35% |
| $\tau \to a_1(\to 2\pi^0\pi)\nu$ | 9% | 34% | 57% |

### 3.2 Reconstruction of $\pi^0$ subclusters

The high granularity of the electromagnetic calorimeter in ATLAS allows for the identification of isolated subclusters from $\pi^0$s inside the core region of the reconstructed $\tau$ lepton hadronic decays.

Studies have been performed based on the topological clustering algorithm [15] with only the middle layer of the calorimeter used for finding primary maxima, and the strip layer used for finding the secondary maxima. The clustering was performed based on cells in a region $\Delta R < 0.4$ around the direction of the leading track satisfying $p_T > 9$ GeV and only subclusters with center within $\Delta R < 0.2$ were taken. A subtraction procedure was applied first to reduce the impact from energy deposits of nearby $\pi^{\pm}$'s and of energy double-counting when adding the latter (track + $\pi^0$ clusters) to reconstruct the visible $\tau_{had}$ energy. Namely, before clustering procedure, cells being closest to the impact point of the track ($\Delta R < 0.0375$) were removed. The subtraction was stopped when the subtracted energy exceeded 70% of the track momentum. In the case of coincidence of large energy deposits in the hadronic calorimeter (above 40% of the track momentum) and in the presampler+strip layer close to the track, indicating superposition of $\pi^0$ and $\pi^{\pm}$ showers, cells were subtracted only from the middle layer up to the point where the transverse energy of the remaining cells exceeded $2.5 \cdot E_T$ collected in the presampler+strip layer (always counted in $\Delta R < 0.0375$ from the track impact point).

Reconstructed subclusters were required to have $E_T > 1$ GeV and be separated by $\Delta R > 0.0375$ from the impact point of the track in the middle layer. In addition, subclusters were accepted if their reconstructed energy in the strip+presampler layers exceeded 10% of their total energy. These requirements efficiently removed about 50% of satellite clusters from charged $\pi$s in the case of $\tau \to \pi\nu$ decays. Table 3 summarizes the results in terms of the fraction of one-prong candidates reconstructed with a given multiplicity of $\pi^0$ subclusters.

Reconstructing the track and $\pi^0$ subclusters for single-prong decays allows for the definition of the energy and visible mass of the hadronic $\tau$ decays from the vector sum of both components. The procedure was evaluated for one-prong decays from $W \to \tau\nu$ events, i.e. for $\tau$ leptons with visible transverse momenta below 50 GeV. Figure 8 shows the response and resolution obtained by this algorithm for reconstructing the visible energy in decays of type $\tau \to \rho\nu$ from $W \to \tau\nu$ events in case at least one $\pi^0$ subcluster is reconstructed. A Gaussian fit to the core region of the distribution yields a resolution of 4.6% with an effective shift of $-2.4\%$, dominated by the calibration of the electromagnetic calorimeter not being optimal for the $\pi^0$ subcluster reconstruction.

As a final benchmark for the quality of the $\pi^0$ cluster reconstruction discussed above, the invariant mass of $\tau \to \rho\nu \to \pi^0\pi\nu$ decays is reconstructed from the track + $\pi^0$ subcluster system which is more difficult than the reconstruction of the transverse energy only since the resolution is dominated by the precision of the reconstruction of the angle between the charged and the neutral pion. Figure 8 (right) shows the reconstructed visible mass for $\tau \to \rho\nu$, $\tau \to a_1(\to 2\pi^0\pi)\nu$, and $\tau \to \pi\nu$ decays. The relative contributions are proportional to the branching fractions convoluted with the experimental efficiencies of the algorithm applied to inclusive hadronic decays of the $\tau$ lepton. If more than one $\pi^0$ subcluster is reconstructed the energy weighted barycenter of the cluster system is taken.

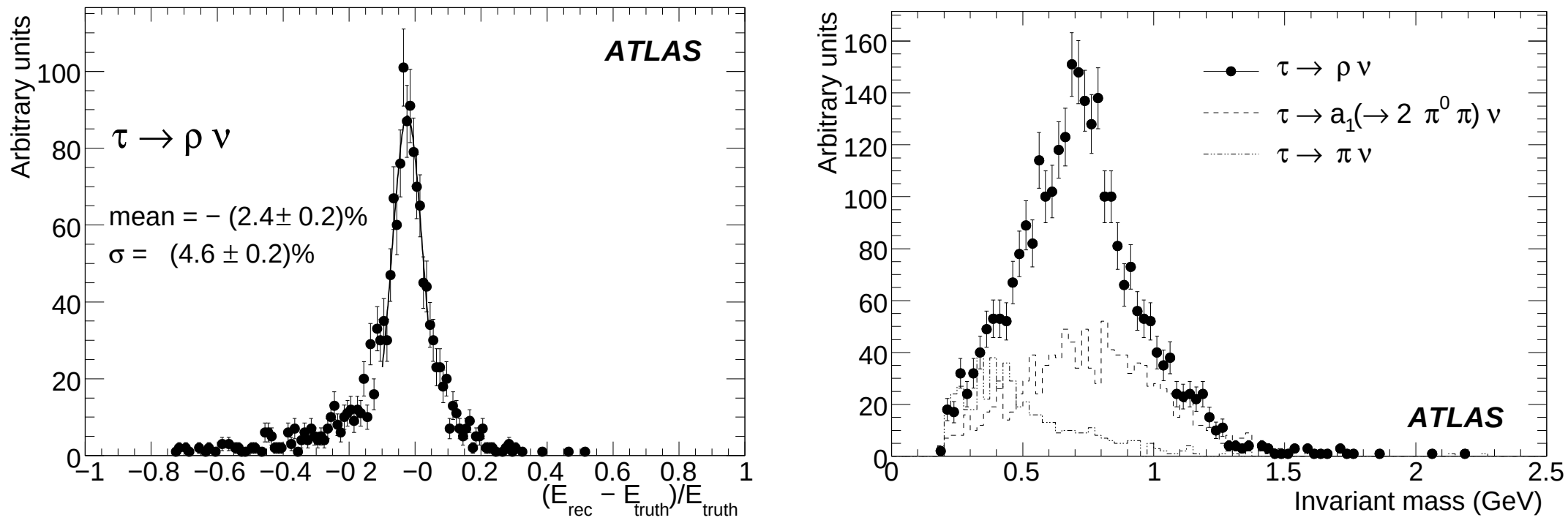

Figure 8: The energy response obtained for the visible energy from $\tau \to \rho\nu$ events using candidates with one $\pi^0$ subcluster (left). The invariant mass of the visible decay products for hadronic single-prong $\tau \to \rho\nu$, $\tau \to a_1(\to 2\pi^0\pi)\nu$, and $\tau \to \pi\nu$ decays using candidates from $W \to \tau\nu$ events with at least one $\pi^0$ subcluster reconstructed (right).

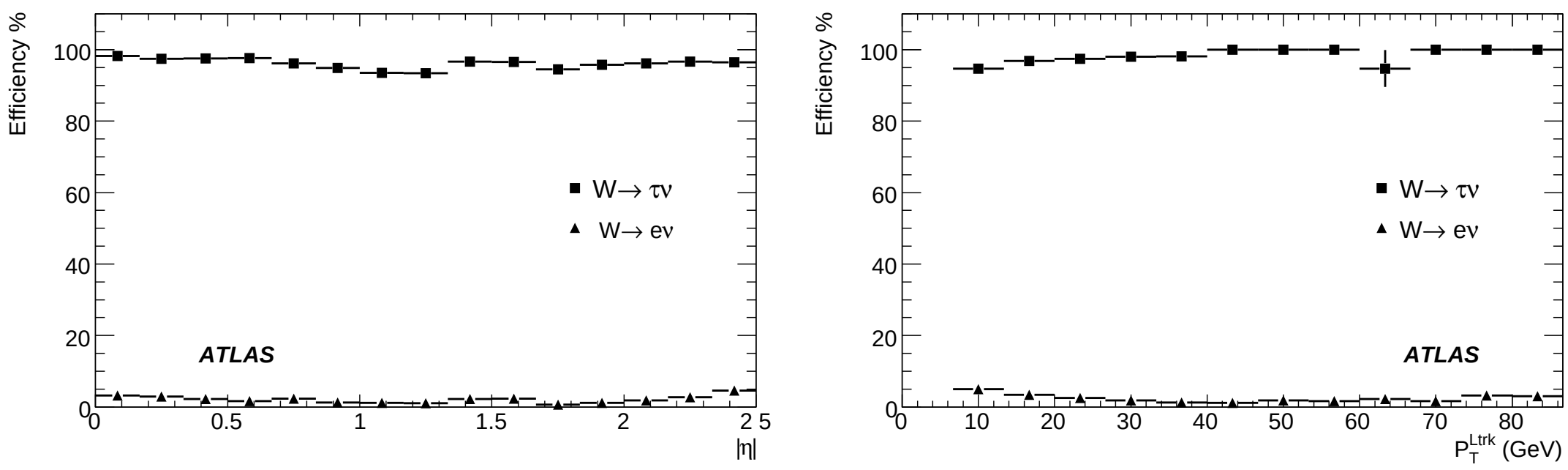

Figure 9: The efficiency of the electron veto algorithm for $W \to \tau\nu$ (rectangles) and $W \to e\nu$ (triangles) events as a function of $|\eta|$ and $p_T$ of the leading track.

### 3.2.1 Combined veto on electron tracks

An efficient rejection of tracks originating from isolated electrons is important for rejecting backgrounds for example from $W \to e\nu$ and $Z \to ee$ events. One possibility would be to reject tracks that have been identified as good electron candidates by the standard electron reconstruction algorithm. With the so called tight selection this algorithm is found to reject $\sim 85\%$ of all electrons from $W \to e\nu$ events with a loss of efficiency for true hadronic $\tau$ decays with $p_T^{\text{track}} > 9$ GeV of less than 1%.

To achieve a more stringent selection, a dedicated algorithm to veto electrons has been developed aiming at a higher rejection rate while, at the same time, retaining a high fraction of hadronic $\tau$ decays. It is based on the following variables:

- The energy deposited in the hadronic part of the calorimeter ($E^{\text{HCAL}}$).
- The energy not associated with a charged track in the strip compartment of the electromagnetic

Table 4: Efficiency for hadronically decaying $\tau$ leptons and true electrons from $W \to \tau\nu$ for passing the electron veto algorithm. The numbers given are normalized to true electrons with $p_T > 9$ GeV and $|\eta| < 2.5$ (vs. true e) and to reconstructed one-prong or three-prong candidates with the leading track being matched to a $\pi$ from $W \to \tau\nu$ events (vs. reconstructed $\tau_{had}$). The probability that an electron from $W \to e\nu$ events with $p_T > 9$ GeV and $|\eta| < 2.5$ is reconstructed as one-prong (three-prong) candidate is $\sim 70\%$ ($\sim 0.7\%$). In addition the performance of the standard algorithm for electron reconstruction [16] is shown. The statistical uncertainty on the numbers presented here is at the level of $0.1 - 0.5\%$.

| Candidates | Reconstructed as single-prong | Reconstructed as three-prong | Overall |
|---|---|---|---|
| Electron-veto algorithm | | | |
| $\tau$ from $W \to \tau\nu$ (vs reconstructed $\tau_{had}$) | 94.1% | 96.2% | 94.9% |
| Electron from $W \to e\nu$ (vs true e) | 1.5% | < 0.1% | 1.6% |
| Standard algorithm (tight selection) | | | |
| $\tau$ from $W \to \tau\nu$ (vs reconstructed $\tau_{had}$) | 99.9% | 99.9% | 99.9% |
| Electron from $W \to e\nu$ (vs true e) | 15.6% | 0.4% | 16.4% |
| Standard algorithm (medium selection) | | | |
| $\tau$ from $W \to \tau\nu$ (vs reconstructed $\tau_{had}$) | 90.6% | 95.1% | 92.1% |
| Electron from $W \to e\nu$ (vs true e) | 4.2% | 0.2% | 4.6% |

part of the calorimeter ($E_{\max}^{\text{strip}}$).

- The ratio in the transverse plane of the associated energy in the electromagnetic calorimeter and the track momentum ($E_T/p_T$).
- The ratio of the number of high threshold to low threshold hits (including outliers) in the TRT ($N_{HT}/N_{LT}$).

The first two variables are used to divide tracks into categories in which discrimination is provided by fixed cuts on the remaining two variables.

The algorithm yields a rejection factor of 60 against electrons from $W \to e\nu$ events at the expense of losing 5% of the signal from $W \to \tau\nu$ events. The efficiency of the algorithm [6] as a function of $|\eta|$ and $p_T$ is shown in Fig. 9 and its performance is summarized in Table 4. For completeness results from the standard electron reconstruction algorithm are also shown. The dedicated electron-veto described here gives much better efficiency for rejecting isolated electrons from W decay than the standard electron reconstruction algorithm for comparable loss in accepting true hadronic $\tau$ decays.

## 4 Offline algorithms for $\tau$ reconstruction

Two complementary algorithms for the reconstruction of hadronic $\tau$ decays have been implemented in the ATLAS offline reconstruction software. Each algorithm is discussed separately below and their performance is compared.

[6] For hadronic decays of $\tau$ leptons the efficiency is defined w.r.t. the reconstructed $\tau_{had}$ candidates and for electrons w.r.t. to all electrons inside $|\eta| \leq 2.5$ with $p_T > 9$ GeV.

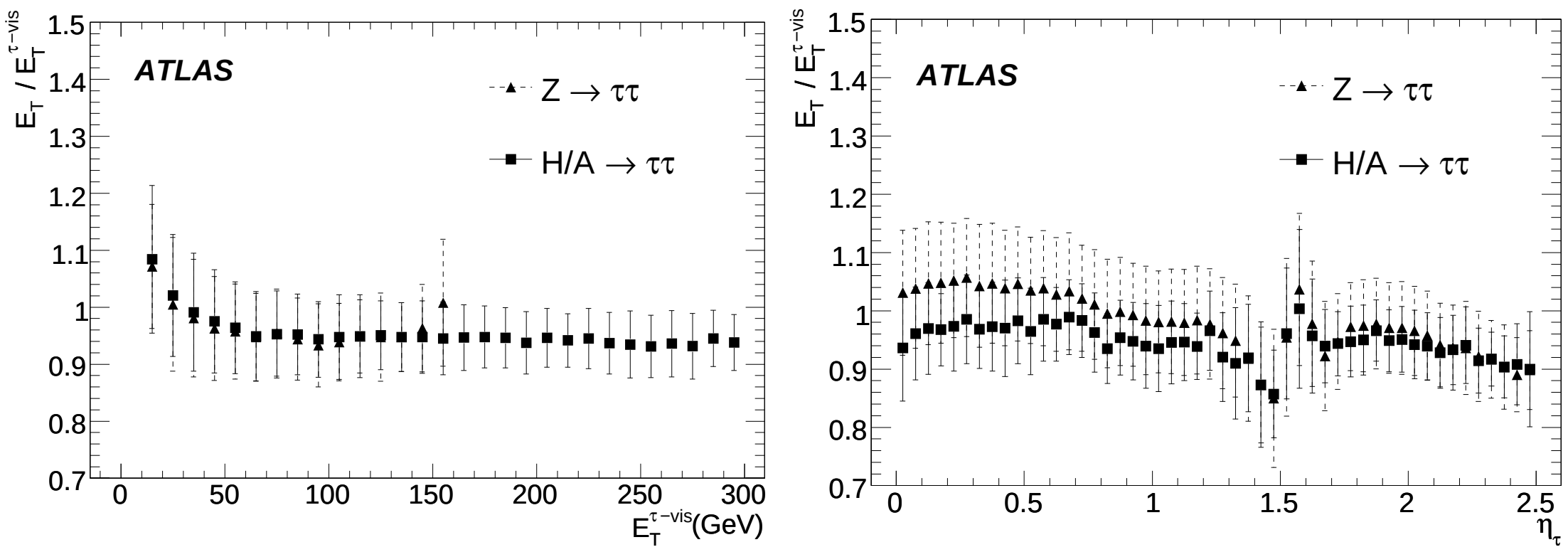

Figure 10: The ratio of the reconstructed ($E_T$) and the true ($E_T^{\tau-vis}$) transverse energy of the hadronic $\tau$ decay products is shown as a function of the visible true transverse energy $E_T^{\tau,\mathrm{vis}}$ (left), calculated in $|\eta| < 2.5$ and $|\eta|$ (right) for taus from $Z \to \tau\tau$ (triangles) and $A \to \tau\tau$ with $m_A = 800$ GeV (squares) decays. The ordinate value is the mean and the error bars correspond to the sigma of the Gaussian fit performed in the range $0.8 < E_T/E_T^{\tau,\mathrm{vis}} < 1.2$. The results are obtained after applying the loose likelihood selection, see below.

## 4.1 The calorimeter-based algorithm

In this approach [17], hadronically decaying $\tau$ candidates are reconstructed using calorimeter clusters as seeds. They are obtained from a sliding window clustering algorithm applied to so called calorimeter towers which are formed from cells of all calorimeter layers on a grid of size $\Delta\eta \times \Delta\phi = 0.1 \times 2\pi/64$. The energy and position are calculated from the clusters, while all cells with the full granularity of the corresponding calorimeters are used to calculate the quantities involved in $\tau$ identification as described in the following. Only clusters with a transverse energy $E_T > 15$ GeV are used. The probability for a true $\tau$ to be reconstructed as a cluster increases from 20% to 68% over the visible $\tau$ transverse energy range from 15 to 20 GeV and saturates at 98% for $E_T > 30$ GeV.

All cells within $\Delta R < 0.4$ around the barycenter of the cluster are then calibrated with an H1-style calibration [18]. The cell weights are a function of the cell energy density, $\eta$ and the calorimeter region. These weights have been optimized for jets [18] and only approximately for hadronic $\tau$ decays. The mean and sigma of a Gaussian fit to the ratio of the reconstructed and the generated energy of the visible $\tau$ decay products, $E_T^{\tau-vis}$, in the range from 0.8 to 1.2 is shown in Fig. 10 as a function of $E_T^{\tau-\mathrm{vis}}$ and $\eta$. The resolution is of the order of 10% and an offset in the range from +5 to -7% is observed in the $\tau$ energy range from 20 to 50 GeV, while at larger energies the offset is of the order of -3 to -5%.

Several quantities that exploit the $\tau$ lepton properties have been combined in a likelihood function to discriminate hadronic $\tau$ decays from fake candidates originating from QCD jets. These quantities are described in the following:

- **The electromagnetic radius $R_{em}$:**

  To exploit the smaller transverse shower profile in $\tau$ decays, the electromagnetic radius $R_{em}$ is used, defined as

$$R_{\mathrm{em}} = \frac{\sum_{i=1}^{n} E_{T,i} \sqrt{(\eta_i - \eta_{\mathrm{cluster}})^2 + (\phi_i - \phi_{\mathrm{cluster}})^2}}{\sum_{i=1}^{n} E_{T,i}}, \qquad (1)$$

  where $i$ runs over all cell in the electromagnetic calorimeter in a cluster with $\Delta R < 0.4$. The quantities $\eta_i$, $\phi_i$, and $E_{T,i}$ denote their position and transverse energy in cell $i$. Cells may have different sizes depending on the layer and their $\eta$ value. The size varies from $\Delta\eta \times \Delta\phi = 0.003 \times$ 0.1 in the $\eta$-strip region of the barrel to $0.025 \times 0.025$ for the second calorimeter layer. This leads to a dependence of the performance on $\eta$. This variable shows good discrimination power at low $E_T$ but becomes less effective at higher $E_T$.

- **Isolation in the calorimeter**:

  Clusters built from hadronic $\tau$ decays are well collimated and therefore rather tight isolation criteria can be used. Here a ring of $0.1 < \Delta R < 0.2$ was chosen as the isolation region and the quantity

$$\Delta E_{\mathrm{T}}^{12} = \frac{\sum_i E_{T,i}}{\sum_j E_{T,j}}, \tag{2}$$

  is calculated, where the indices $i$ and $j$ run over all electromagnetic calorimeter cells in a cone around the cluster axis with $0.1 < \Delta R < 0.2$ and $\Delta R < 0.4$, respectively, and $E_{T,i}$ and $E_{T,j}$ denote the transverse cell energies.

  Like $R_{\mathrm{em}}$, the $\Delta E_T^{12}$ distribution shows an $E_T$ dependence and becomes narrower with increasing $E_T$. This variable also depends on the event type and is expected to be less effective for events with higher hadronic activity, like e.g. $t\bar{t}$ events.

- **Charge of the $\tau$ candidate**:

  The charge of a $\tau$ candidate is defined as the sum over the charge(s) of the associated track(s). The misidentification of the charge on the level of a few percent shows almost no $E_T$ dependence.

- **Number of associated tracks**:

  The number of tracks, $N_{tr}$, associated with a given cluster within $\Delta R < 0.3$. The tracks are required to have $p_T > 2$ GeV and no specific requirements on the quality of the track reconstruction is made. A significant fraction of events with zero, two, and even four tracks is observed for true hadronic $\tau$ decays.

- **Number of hits in the $\eta$ strip layer**:

  The number of hits in $\eta$ direction in the finely segmented strip detector, $N_{strip}$, in the first layer of the electromagnetic barrel calorimeter is also used in the likelihood discrimination. Cells in the $\eta$ strip layer within $\Delta R < 0.4$ around the cluster axis are counted as hits if the energy deposited exceeds 200 MeV. In contrast to jets, a significant fraction of $\tau$ leptons deposit nearly no energy in the $\eta$ strip layer ($\tau \to \pi\nu$ decays) and the number of corresponding hits is small.

- **Transverse energy width in the $\eta$ strip layer**

  The transverse energy width $\Delta\eta$ is defined as

$$\Delta\eta = \sqrt{\frac{\sum_{i=1}^{n} E_{Ti}^{\mathrm{strip}} (\eta_i - \eta_{\mathrm{cluster}})\, 2}{\sum_{i=1}^{n} E_{Ti}^{\mathrm{strip}}}}. \tag{3}$$

  where the sum runs over all strip cells in a cone with $\Delta R < 0.4$ around the cluster axis and $E_{Ti}^{\mathrm{strip}}$ is the corresponding strip transverse energy. Like $R_{\mathrm{em}}$ it is a powerful discriminator at low $E_T$ but loses discrimination power with increasing $E_T$ for higher collimated high $E_T$ jets.

- **Lifetime signed pseudo impact parameter significance**:

  At present only a 2-dimensional impact parameter, also called the pseudo impact parameter, is used. It is defined as the distance from the beam axis to the point of closest approach of the track in the plane perpendicular to the beam axis. From this information and from the jet axis, a quantity denoted as lifetime signed pseudo impact parameter significance, defined as $\mathrm{sig}_{d_0} = d_0/\sigma^2_{d_0}$ where $\sigma$ is the impact parameter resolution, is calculated.

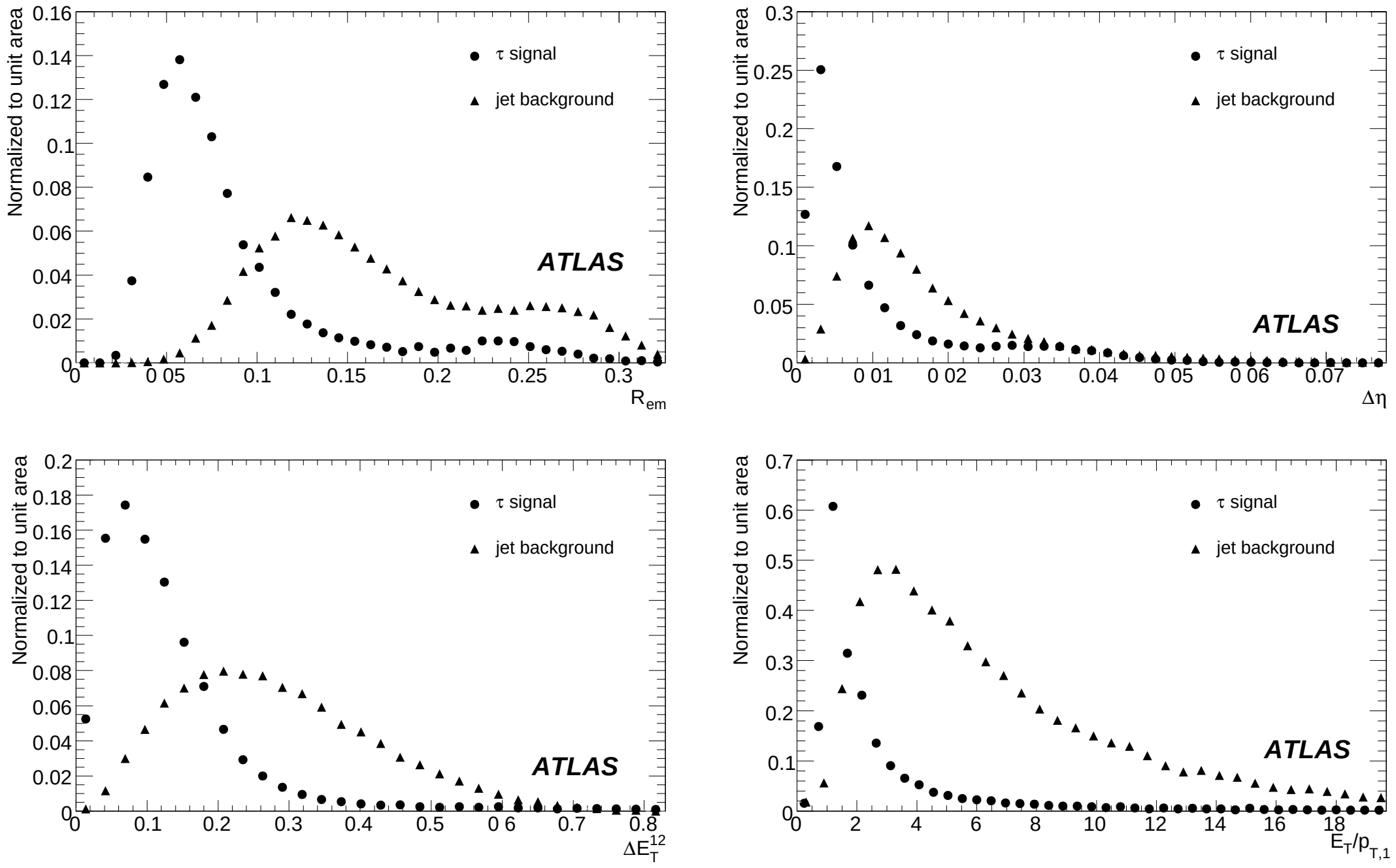

Figure 11: The distributions of a few discriminating variables (electromagnetic radius, energy isolation, transverse energy width in the $\eta$ strip layer and $E_T$ over $p_{T1}$ of the leading track) used in the calorimeter-based tau identification for true tau decays and jets with visible transverse cluster energies $E_T$ in the range from 40 to 60 GeV and track multiplicities between 1 and 3.

- **$E_T$ over $p_T$ of the leading track:** $E_T/p_{T1}$ :

  For $\tau$ decays a large fraction of the energy is expected to be carried by the leading track and the ratio of the cluster energy $E_T$ to the momentum of the leading track $p_{T1}$ is expected to be large, close to 1. This provides another discrimination against QCD jets, which are expected to have a more uniform distribution of $p_T$ among the tracks. They are also expected to have more additional neutral particles. Values above one are also expected from $\tau$ decay modes involving additional $\pi^0$s and for three-prong decays. The $E_T$ dependence is rather modest for $\tau$ decays but more pronounced for QCD jets, which tend to become more signal like with higher $E_T$.

In Fig. 11 the distributions of a few discriminating variables are shown for signal and backgrounds for transverse cluster energies $E_T$ in the range between 40 and 60 GeV and for candidates with 1 or 3 tracks.

For the calorimeter-based algorithm the $\tau$ identification is based on a one-dimensional likelihood ratio constructed from three discrete variables ($N_{\mathrm{tr}}$, $N_{\mathrm{strip}}$ and the charge of the $\tau$ lepton) and five continuous variables ($R_{\mathrm{em}}$, $\Delta E_T^{12}$, $\Delta\eta$, $\mathrm{sig}_{d_0}$, and $E_T/p_{T,1}$). For the discrete variables the ratios are directly taken from the reference histograms. For the continuous variables fits of appropriate functions to each variable for all $E_T$ bins have been performed. The distribution of the likelihood for taus and jets are shown in Fig. 12. Despite any limitation from using only one-dimensional distributions it shows a good separation power.

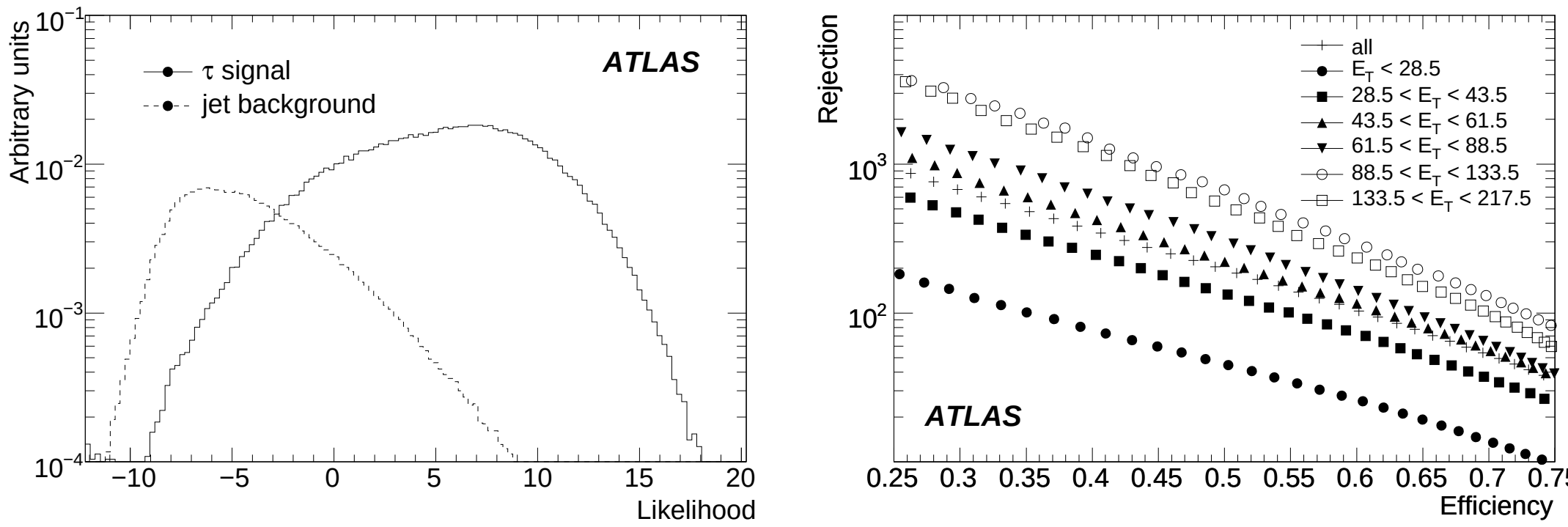

Figure 12: Left: The log likelihood (LLH) distribution for $\tau$ leptons (solid) and jets from QCD production (dashed). The likelihood is applied after a preselection on the number of associated tracks, i.e. requiring $1 \leq N_{tr} \leq 3$. (Candidates with LLH $< -10$ had variables outside the boundaries of histograms used when obtaining the PDFs for the likelihood calculation). Right: Efficiency for $\tau$ leptons and rejection against jets for different $E_T$ ranges, achieved with the likelihood selection.

It should be noted that the $\tau$ identification efficiency chosen to keep enough signal events and to achieve the necessary rejection against background depends on the physics channel. Despite the use of $E_T$ bins the likelihood discrimination shows a residual $E_T$ dependence. Therefore, a fixed cut on the likelihood value neither will result in a generally flat efficiency, nor will it be optimal.

## 4.2 The track-based algorithm

In this approach [19], the visible part of the hadronically decaying $\tau$ lepton is seen as a very well collimated object consisting of charged and neutral pions, with the charged component being the leading one, *i.e.* reproducing well the direction of the visible decay products and having significant transverse momentum. This assumption is followed by the requirement of a low multiplicity of tracks reconstructed in the region considered the core of the $\tau_{had}$ candidate, and the requirement of only minimal energy deposit in the isolation region around the core. The energy-scale of the object and the calorimetric variables used in the identification are built following this picture. For most of the analyses only one-track and three-track candidates should be used. Including candidates with two tracks helps to recover a large fraction of lost three-prong candidates, however it also significantly increases the background from QCD events, in particular for $\tau$ leptons with visible transverse momentum below 30 GeV. Candidates with track multiplicities larger than three should be used for monitoring the level of fake candidates only.

The reconstruction step consists of identifying and qualifying a leading hadronic track[7] which becomes a seed for building the $\tau$ candidate. Then up to six additional tracks are allowed in the core region. The $(\eta, \phi)$ position of the candidate is taken from the direction of the track at the vertex or the track-$p_T$ weighted bary-center in the case of multi-track candidates and the energy of the candidate is calculated from the energy flow method. In addition charge $\pm 1$ or 0 is required in the case of multi-prong candidates. The identification step consists of calculating calorimetric and tracking quantities and then providing a decision either based on selection with cuts or a discriminating variable based on multi-variate techniques.

[7]The threshold $p_T > 9$ GeV on the leading track was used for results presented here, while it was lowered to 6 GeV in the more recent software releases.

### 4.2.1 The energy - flow approach

The energy scale of the hadronic $\tau_{had}$ candidate is defined using an energy flow algorithm. The energy deposit in cells is divided into categories.

- The pure electromagnetic energy, $E_T^{\text{emcl}}$:

  The energy is seeded by an isolated electromagnetic cluster which is isolated from the good quality tracks and which has no substantial hadronic leakage. The energy is collected in a narrow window around the seed. Only presampler, strip and middle layers are used.

- The charged electromagnetic energy, $E_T^{\text{chrgEM}}$, $E_T^{\text{chrgHAD}}$:

  The energy is seeded by the impact point of the track(s) in each layer and the energy is collected in a narrow window around it.

- The neutral electromagnetic energy, $E_T^{\text{neuEM}}$:

  The energy is seeded by the $(\eta, \phi)$ of the track at the vertex and in each layer the closest cell is searched for. The energy is collected from not yet used cells in a cone of $\Delta R = 0.2$ with respect to the cell closest to the impact point. Only presampler, strip and middle layers are used.

In the energy-flow approach the charged energy deposits $E_T^{\text{chrgEM}} + E_T^{\text{chrgHAD}}$ are replaced by the track(s) momenta (no hadronic neutrals) in order to define the energy scale of the $\tau_{had}$. The contribution from $\pi^0$'s is included in $E_T^{\text{emcl}}$ and $E_T^{\text{neuEM}}$; the effects of $\pi^0$ and $\pi^\pm$ depositing energy in the same calorimeter cells or charged energy leakage outside a narrow cone around the track are corrected by adding two terms: $\sum \text{res}E_T^{\text{chrgEM}}$ and $\text{res}E_T^{\text{neuEM}}$. The complete definition for the energy scale $E_T^{\text{eflow}}$ reads as follows:

$$E_T^{\text{eflow}} = E_T^{\text{emcl}} + E_T^{\text{neuEM}} + \sum p_T^{\text{track}} + \sum \text{res}E_T^{\text{chrgEMtrk}} + \text{res}E_T^{\text{neuEM}}. \tag{4}$$

The fractional energy response, calculated as $(E_{rec} - E_{truth})/E_{truth}$, for one and three prong candidates is shown in Fig. 13.

The evident advantage from using the above approach for defining the energy scale comes from the fact that while performing well for true hadronic decays of $\tau$ leptons, it significantly underestimates the nominal energy of fake $\tau_{had}$s from jets. This effect is rather obvious since a cone of $\Delta R = 0.2$ is too narrow to efficiently collect the energy of a QCD jet (particularly with low transverse momentum) and also since a large fraction of the neutral hadronic component is largely omitted in the definition itself (as the energy deposit in the hadronic calorimeter does not contribute to the energy calculations). This leads to a faster falling background spectrum as a function of $E_T$ compared to that using calibrated calorimetric clusters as implemented in the calorimeter-based algorithm (Section 4.1). This method leads however to more non-Gaussian tails in the fractional energy response than the more conventional energy estimates from calorimetry only.

### 4.2.2 Identification with calorimetric and tracking variables

Several calorimetric and tracking variables are used to discriminate a narrow, low track multiplicity $\tau_{had}$ cluster from a hadronic cluster originating from quarks or gluons. If not stated otherwise, the calorimetric and tracking identification quantities are calculated from cells/tracks within a core cone of $\Delta R = 0.2$ around the seed. The isolation criteria used here are checked in an isolation cone $\Delta R = 0.2 - 0.4$. Please note, that although some definitions are very similar for the calorimeter-based and track-based algorithms, in case of the latter the narrower core cone is often used for the calculation of calorimetric quantities and a more explicit distinction between core and isolation cone is made.

Not all discriminating quantities discussed in Section 3 have been already implemented in the identification procedure. In particular transverse impact parameter, transverse flight path and categorizing single-prong candidates using $\pi^0$ subclusters have been added only in the current releases of the reconstruction software. Therefore they are not used for the results presented below.

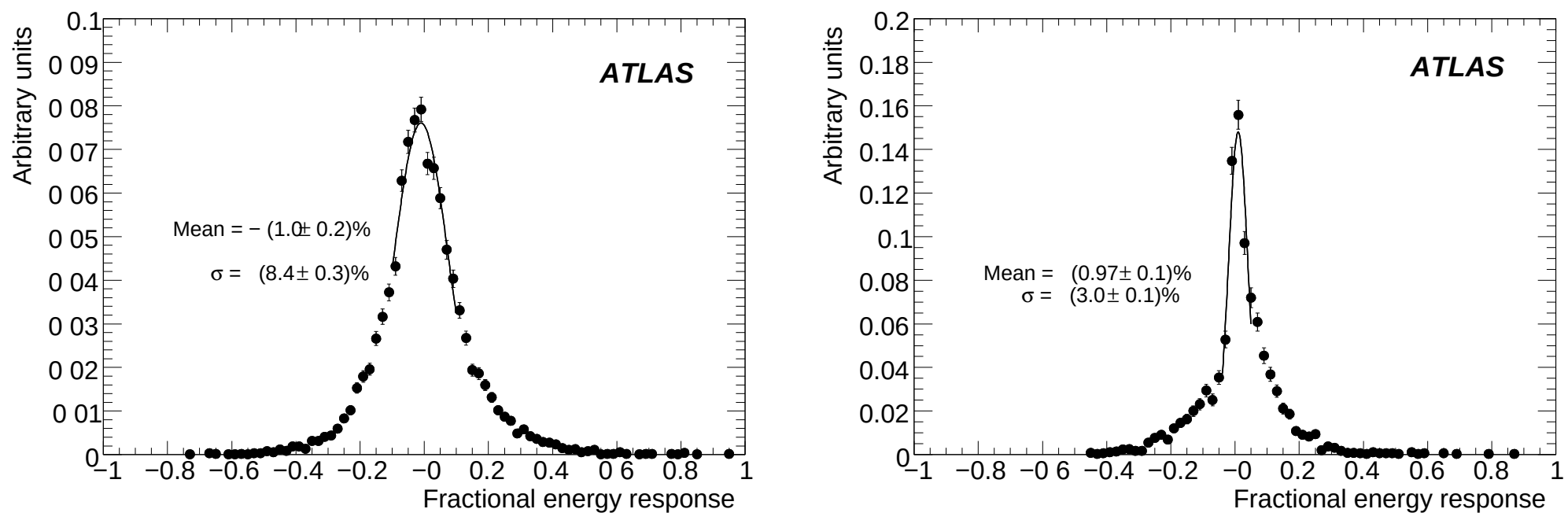

Figure 13: The fractional energy response for single-prong (left) and three-prong (right) true $\tau_{had}$ candidates reconstructed with the track-based algorithm. Events from a $W \to \tau\nu$ sample are shown.

- **Tracking quantities**
  - The variance $W^{\tau}_{\text{tracks}}$ (for multi-prong candidates only), defined as

$$W^{\tau}_{\text{tracks}} = \frac{\sum(\Delta\eta^{\tau,\text{track}})^2 \cdot {p_T}^{\text{track}}}{\sum {p_T}^{\text{track}}} - \frac{(\sum \Delta\eta^{\tau,\text{track}} \cdot {p_T}^{\text{track}})^2}{(\sum {p_T}^{\text{track}})^2}. \tag{5}$$

  - The invariant mass of the tracks system (for multi-track candidates), $m_{trk3p}$,
  - The number of tracks in the isolation cone.

- **Calorimetric quantities**
  - The electromagnetic radius of the $\tau_{had}$ candidate, $R^{\tau}_{\text{em}}$, as defined in Eq. (1) for the calorimeter-based algorithm, but calculated from cells around the seed belonging to the first three samplings of the electromagnetic calorimeter only (presampler, strips and middle layer).
  - The number of $\eta$ strips, $N^{\tau}_{\text{strips}}$, with energy deposits above a certain threshold.
  - The width of the energy deposit in the strips, as defined in Eq. (3) for the calorimeter based algorithm but calculated in the core cone only.
  - The fraction of the transverse energy, $\text{frac}{E_T}^{R12}$, deposited in a cone of radius $0.1 < \Delta R < 0.2$ with respect to the total energy in a cone of $\Delta R = 0.2$. Cells belonging to all layers of the calorimeter are used:

$$\text{frac}E_T^{R12} = \frac{\sum E_T^{\text{cell}}(R^{\tau,\text{cell}} < 0.2) - \sum E_T^{\text{cell}}(R^{\tau,\text{cell}} < 0.1)}{\sum E_T^{\text{cell}}(R^{\tau,\text{cell}} < 0.2)}. \tag{6}$$

  - The transverse energy, $E_T^{\text{core}}$, at the EM scale deposited inside the core cone.
  - The transverse energy, $E_T^{isol}$ and $E_T^{\text{isolHAD}}$, at the EM scale, deposited inside the isolation cone.

- **Tracking and calorimetric quantities**
  - The ratio of transverse energy deposited in the hadronic calorimeter in the core region (at the EM scale), $E_T^{\text{chrgHAD}}$, with respect to the sum of the transverse momenta of the tracks.

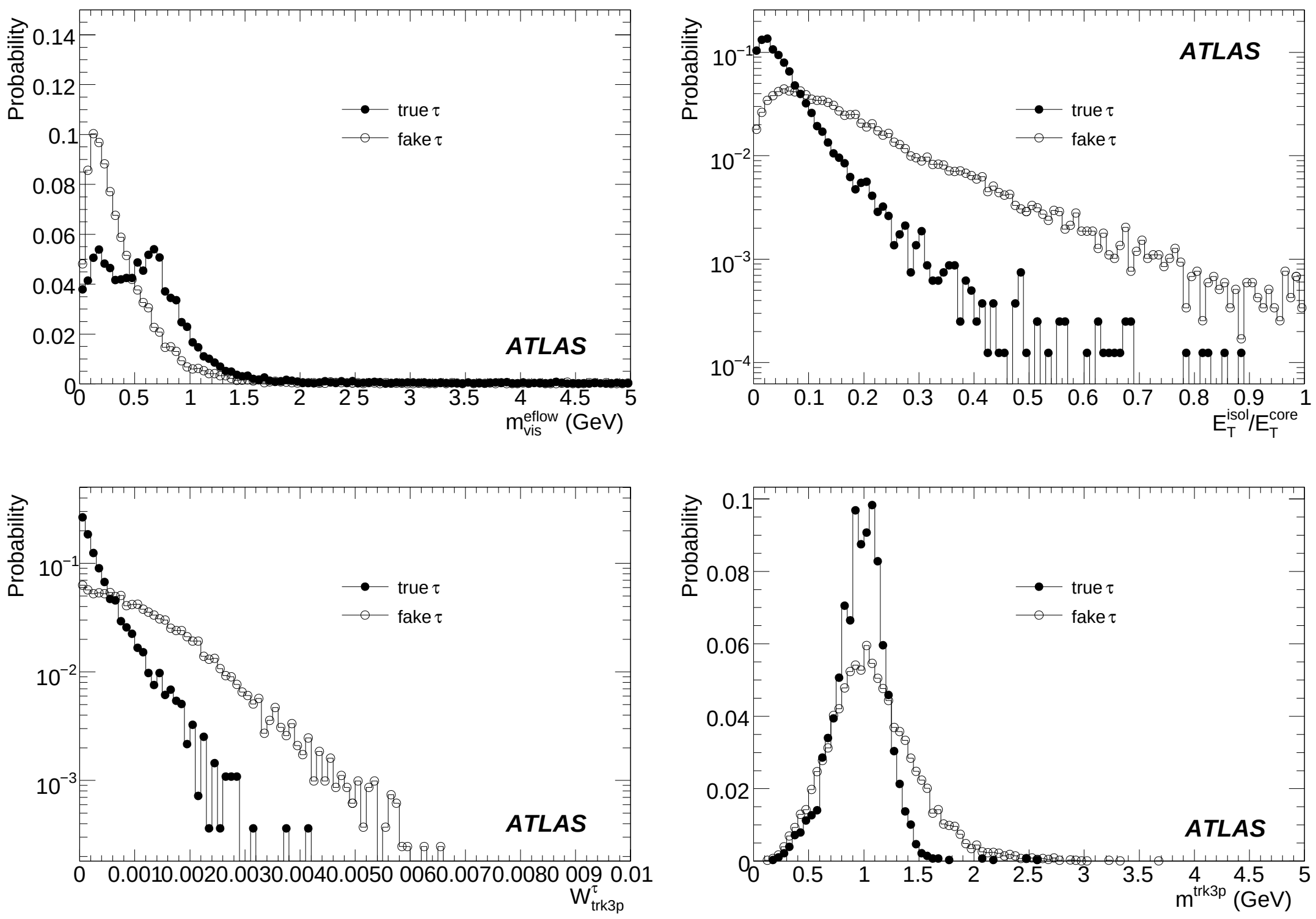

Figure 14: The distributions for signal and backgrounds for the visible mass $m_{vis}^{eflow}$ and ratio of the transverse energy in the isolation and core region $E_T^{isol}/E_T^{core}$ for single-prong candidates, and variance $W_{\text{tracks}}^{\tau}$ and invariant mass of the track system $m^{trk3p}$ for three-prong candidates. Distributions are shown for the candidates in the transverse energy range $E_T = 20-40$ GeV.

- The visible mass $m_{vis}^{eflow}$ calculated from cells used for the energy-flow calculation and tracks. In case of multi-prong candidates, where this mass is smaller than that calculated from the four-momenta of the tracks, the invariant mass of the track system is taken instead.

In Fig. 14 as an example the distributions for signal and backgrounds for the $m_{vis}^{eflow}$, the ratio $E_T^{isol}/E_T^{core}$ for single-prong candidates, the variance $W_{\text{tracks}}^{\tau}$ and the invariant mass $m_{trk3p}$ for three-prong candidates are shown. Note that the $m_{vis}^{eflow}$ distribution shows a double peak structure coming from $\tau \to \pi^{\pm}\nu$ and $\tau \to \rho(a1)\nu$ decays. The separation for candidates with and without $\pi^0$ clusters was not done for the distribution shown.

### 4.2.3 Overall efficiency and rejection

The identification step is done by calculating discriminants using basic cut methods, cut methods optimized by the TMVA package [20], multi-variate analyses based on neural network technique, and from PDRS discrimination [21].

The rejection power expected from the identification step only is quite modest, given that a quite good rejection is already achieved in the reconstruction step. The overall performance is summarized in Table 5. For an efficiency of about 30% with respect to all hadronic decays in the energy range $10-30$ GeV, rejection rates of 200/360 for one-prong/three-prong hadronic $\tau$ decays can be achieved with the cut based selection and of 500/700 with multi-variate selection techniques.

Table 5: Efficiencies and rejection rates for different discrimination techniques for the track-based algorithm for fixed efficiencies. The efficiencies are normalized to all hadronic $\tau$ decays. The rejection rates are calculated with respect to jets reconstructed from true particles in the Monte Carlo. Events from $Z \to \tau\tau$ signal samples and QCD dijets were used. The errors given are statistical only.

| Selection | Efficiency | Rejection cuts | Rejection TMVA cuts | Rejection NN | Rejection PDRS |
|---|---|---|---|---|---|
| | $E_T$ = 10-30 GeV | | | | |
| one-prong | 0.33 | 225 ± 10 | 435 ± 30 | 510 ± 40 | 460 ± 40 |
| three-prong | 0.28 | 360 ± 25 | 470 ± 40 | 740 ± 70 | 670 ± 60 |
| | $E_T$ = 30-60 GeV | | | | |
| one-prong | 0.42 | 140 ± 10 | 170 ± 10 | 440 ± 40 | 320 ± 30 |
| three-prong | 0.45 | 60 ± 2 | 9 0 ± 10 | 160 ± 10 | 130 ± 10 |

Table 6: Rejection against jets from Monte Carlo true particles for a 30% efficiency and separately for the one-prong (1p) and three-prong (3p) candidates. The efficiencies are normalized to true hadronic $\tau$ decays. For the signal $Z \to \tau\tau$ events and events from $bbH$, $H \to \tau\tau$ with $m_H = 800$ GeV were used; for the background QCD dijet-samples were used. The errors given are statistical only.

| Algorithm | $E_T$ = 10-30 GeV | $E_T$ = 30-60 GeV | $E_T$ = 60-100 GeV | $E_T > 100$ GeV |
|---|---|---|---|---|
| Track-based (neural network) | 1p: 740 ± 70<br>3p: 590 ± 50 | 1p: 1030 ± 160<br>3p: 590 ± 70 | | |
| Calo-based (likelihood) | | 1p: 1130 ± 50<br>3p: 187 ± 3 | 1p: 2240 ± 140<br>3p: 310 ± 7 | 1p: 4370 ± 280<br>3p: 423 ± 8 |

### 4.3 Comparison of the two algorithms

Figure 15 shows the expected performance of the two algorithms, illustrated as curves describing the jet rejection versus the efficiency, separately for one and three-prong hadronic $\tau$-decays and for different ranges of the visible transverse energy. The jet rejections are computed with respect to jets reconstructed from true particles in the Monte Carlo. The rejections obtained are between a factor of two and ten higher for one-prong decays than for three-prong decays, depending on the algorithm and on the transverse energy range considered. For an efficiency of 30% for one-prong decays, the rejection against jets is typically between 500 and 5000, as illustrated more quantitatively and as a function of the visible transverse energy in Table 6.

Figure 16 shows the normalized track-multiplicity spectra for hadronic $\tau$ candidates with visible transverse energies above 20 GeV, from $Z \to \tau\tau$ decays and from jets, as reconstructed by the track-based algorithm. The distributions are shown after the reconstruction step, after a cut-based identification algorithm and finally after applying a neural network discrimination. The track multiplicity in the jet sample is quite different from that in the signal sample, independently from the cuts applied.

At the same time, Figure 16 indicates that the purity of one-prong and three-prong $\tau_{had}$ candidates improves in the signal sample since the expected fractions of single versus three prongs are reproduced. For one-prong (three-prong) candidates, the purity improves from 87% (74%) after reconstruction to 91% (86%) after cut-based identification and to 92% (93%) after applying the neural-network discrimination technique.

Figure 16 also shows that using candidates with track multiplicities above three to normalize the QCD background will allow a reasonably precise calibration of the performance using real data, provided the rejection against QCD jets is proven to be sufficient to extract a clean signal in the one-prong and three-prong categories. The sensitivity of such a method can be enhanced by also studying the track multiplicity outside the narrow cone used for $\tau$-identification and combining this information with that presented in Fig. 16.

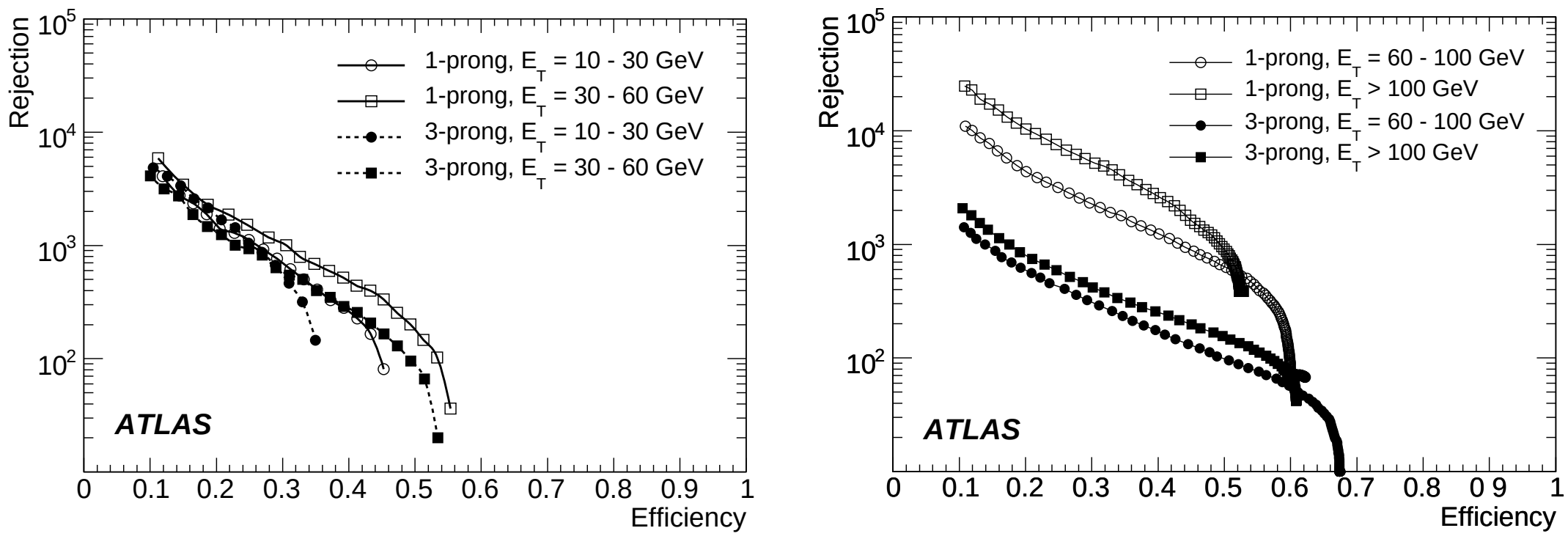

Figure 15: Expected performance for the track-based algorithm with a neural-network selection (left) and the calorimeter-based algorithm with the likelihood selection (right). The rejection rates against jets from Monte-Calo particles as a function of the efficiency for hadronic $\tau$ decays for various ranges of the visible transverse energy are shown. For signal events $Z \to \tau\tau$ and $bbH, H \to \tau\tau$ with $m_H = 800$ GeV were used, for the background QCD dijet samples were used.

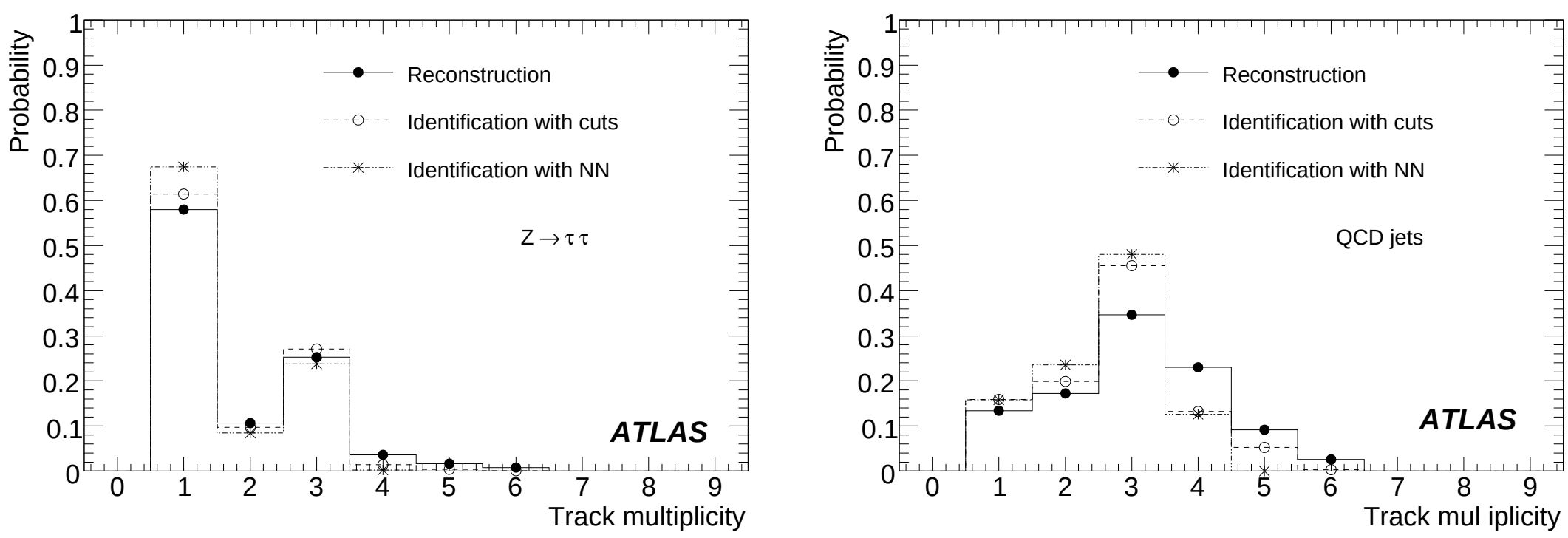

Figure 16: Track multiplicity distributions obtained for hadronic $\tau$ decays with a visible transverse energy above 20 GeV and below 60 GeV using the track-based $\tau$ identification algorithm. The distributions are shown after reconstruction, after cut-based identification and finally after applying the neural network (NN) discrimination technique for an efficiency of 30% for the signal (left) and the background (right).

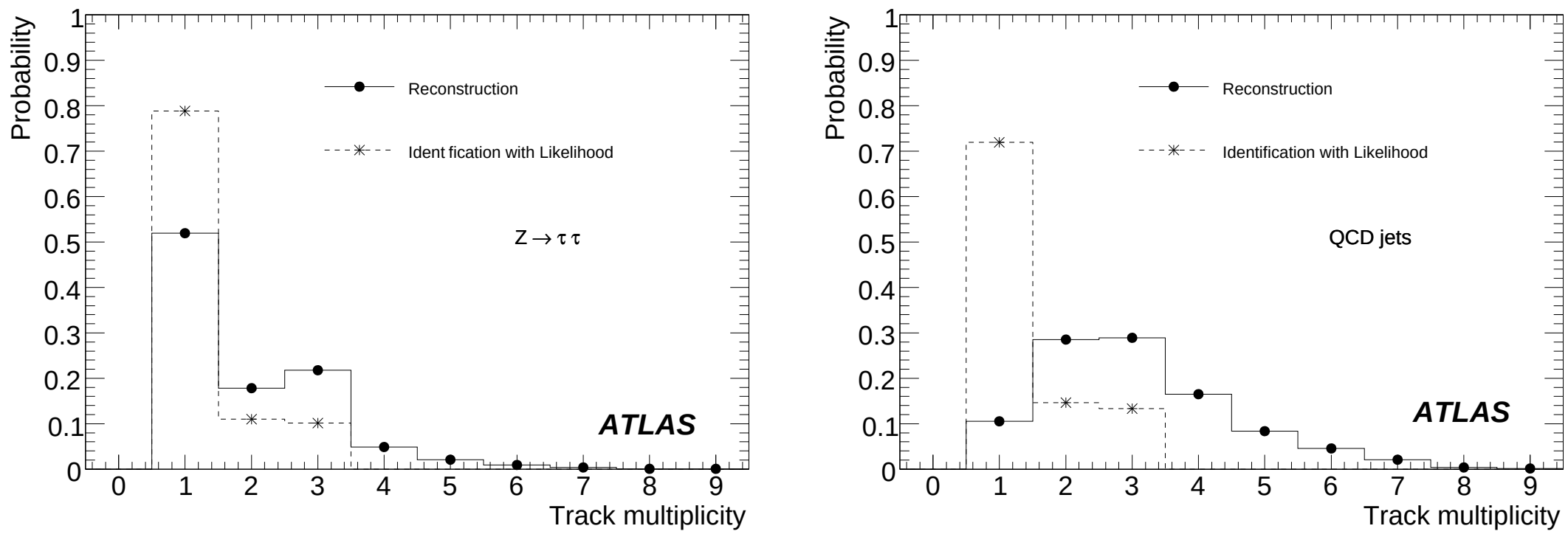

Figure 17: Track multiplicity distributions obtained for hadronic $\tau_{had}$-decays with visible transverse energy above 20 GeV and below 60 GeV using the calorimeter-based $\tau$ identification. The distributions are shown after reconstruction and after applying the likelihood discrimination technique (medium selection) for the signal (left) and the background (right).

The corresponding spectra for the calorimeter-based algorithm are shown in Figure 17 after reconstruction and after applying the likelihood discriminant. The optimization of the likelihood results in a lower efficiency for accepting three-prong decays which biases the track multiplicity spectra. One should note that candidates with 1-3 tracks are accepted as good $\tau_{had}$ candidates.

The application of these $\tau$ identification algorithms to extract $\tau$ signatures from physics processes and to reject the large backgrounds from QCD processes is discussed in Section 6.

# 5 Fake-rates from QCD di-jet samples

This study demonstrates a simple and generic method to determine the $\tau_{had}$ fake rate from jets in early data. Since fake rates are expected to be in the range of $10^{-3}$ to few $10^{-2}$ for low $p_{\mathrm{T}}$ $\tau$ leptons, and since the expected rate of QCD jets far exceeds the rate of hadronically decaying $\tau$ leptons, a precise measurement of fake rates is expected to be crucial for many analyses, and also for a further optimization of the $\tau_{had}$ identification algorithms. In the following, the method and the results are described and statistical and systematic uncertainties are discussed.

The method proposed here uses a very clean sample of QCD jets, with no significant contamination from true $\tau$ leptons. This is achieved by selecting dijet events with two jets having similar $p_{\mathrm{T}}$ and being back-to-back in $\phi$ (see Fig. 18). One of the two jets is randomly chosen as the so-called 'tag jet', for which a cut on the number of tracks ($n_{\mathrm{Trk}} \geq 4$ for $p_{\mathrm{T}} \leq 50$ GeV + 1 track for each additional 50 GeV interval in $p_{\mathrm{T}}$) ensures that it is not a true hadronic $\tau_{had}$ decay. If this cut is fulfilled, the other jet, called 'probe jet', can be used to measure the fake rate from QCD jets [8]. This is performed both for the calorimeter-based and the track-based $\tau_{had}$ reconstruction algorithm, with identification according to the medium likelihood selection and cut discriminant, respectively. Note also that, in order to avoid a direct dependence on the trigger, the probe jet should be required to not have caused the event to be triggered.

[8] The fake rate is determined as the number of probe jets identified as $\tau_{had}$ divided by the number of probe jets.

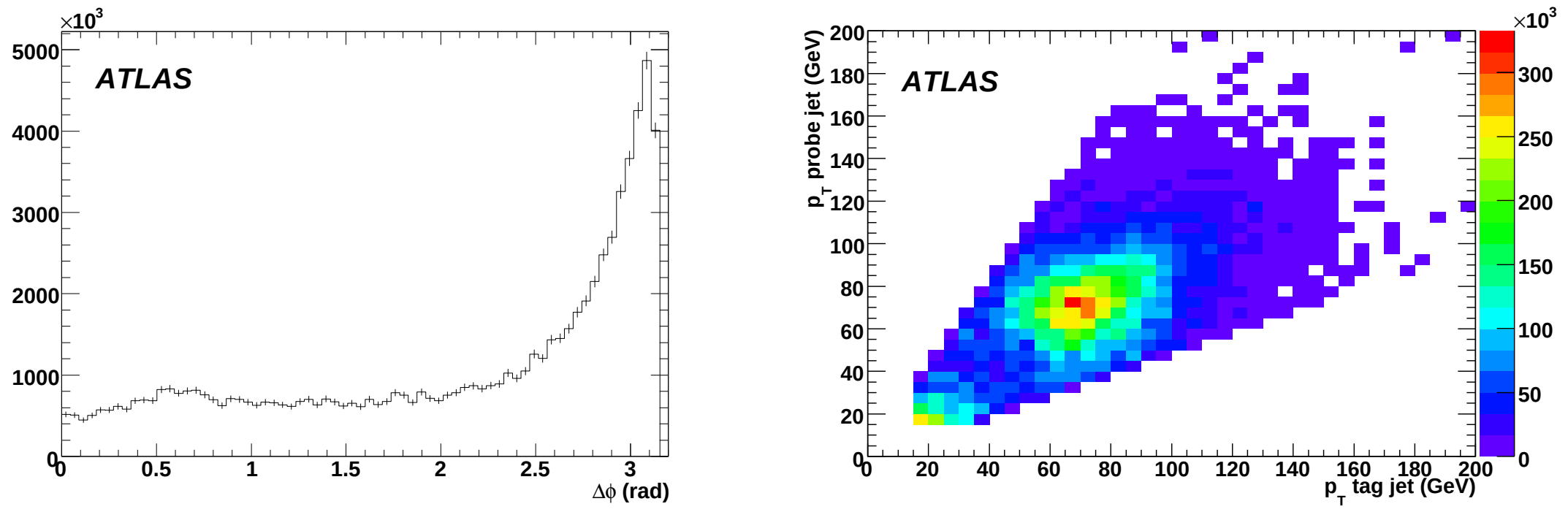

Figure 18: Example of selections on a MC dijet sample, generated with $70 \le p_{\mathrm{T}} \le 140$ GeV. The two jets have to fulfill $\Delta\phi \ge (\pi - 0.3)$ in order to be back to back in $\phi$ (left) and have similar $p_{\mathrm{T}}$ values (right).

Table 7: The $\tau_{had}$ fake rate from QCD jets and its statistical uncertainty for the available Monte Carlo statistics and for expected $100\,\mathrm{pb}^{-1}$ of data in bins of $p_T$ for both $\tau_{had}$ reconstruction algorithms.

| | Calorimeter-based algorithm | | Track-based algorithm | |
|---|---|---|---|---|
| $p_T$ range (GeV) | MC stat. (%) | Expected stat. error for 100 $\mathrm{pb}^{-1}$ (%) | MC stat. (%) | Expected stat. error for 100 $\mathrm{pb}^{-1}$ (%) |
| 15-40 | $2.3 \pm 0.3$ | $\pm\, 0.02$ | $2.5\pm0.5$ | $\pm\, 0.02$ |
| 40-80 | $5.2 \pm 2.2$ | $\pm\, 0.01$ | $6.7\pm2.2$ | $\pm\, 0.01$ |
| 80-120 | $0.5 \pm 0.2$ | $\pm\, 0.001$ | $1.8\pm0.6$ | $\pm\, 0.002$ |
| 120-160 | $0.2 \pm 0.2$ | $\pm\, 0.002$ | $1.4\pm0.6$ | $\pm\, 0.004$ |

This method, which achieves low statistical uncertainties even for small datasets, only relies on the dijet and tag jet selection to acquire a clean QCD jet sample and it does not depend, to first order, on the number of true $\tau$ leptons present in the sample nor on the $\tau_{had}$ efficiencies. Also, the selection can be easily adapted to select probe jets in an environment similar to the one of any given physics analysis using $\tau_{had}$ candidates.

To perform these studies, Monte Carlo dijet samples (generated in various $p_{\mathrm{T}}$ ranges) and samples containing true $\tau$ leptons ($Z \to \tau\tau$, $W \to \tau\nu$) were used. Proper weighting of the samples, including trigger prescales for running at $\mathscr{L} = 10^{31}\,\mathrm{cm}^{-2}\mathrm{s}^{-1}$, has been applied. To cross-check the method, it has been verified that there is very good agreement between the values obtained with all selected jets and those coming from jets matched to Monte Carlo particles jets only, which indicates a very high jet purity.

The numerical results of the fake rate determination using the available Monte Carlo statistics for the two algorithms can be found in Table 7. Systematic uncertainties (discussed below) are not included. Note that the uncertainty in the range $40 < p_{\mathrm{T}} < 80$ GeV, for both algorithms using available Monte Carlo statistics, is dominated by one dijet event (with large weight) in one of the Monte Carlo samples. More interesting are the expected statistical uncertainties in data, which are at the percent or sub-percent level for 100 $\mathrm{pb}^{-1}$, and even for 10 $\mathrm{pb}^{-1}$ of integrated luminosity. This demonstrates the relevance of this method for very early data.

Given the statistical precision of the fake rate results, this measurement of the systematics of other measurements is going to be limited by its own systematic uncertainties. In the following, an outline of the necessary systematic studies is given, including results where possible.

The presence of true hadronically decaying $\tau$ leptons in the selected Monte Carlo sample is not statistically significant enough to alter the results within their uncertainties. When more statistics are available, tighter cuts on the tag side can be applied in order to remove more efficiently these types of events.

A slight tendency was observed for the jet samples with lower hard scattering transverse momenta, to show higher fake rates. This can be explained by the fact that the jet characteristics depend on the degree of parton showering. Also, jets from gluons are wider and on average have higher tracks multiplicity than jets from quarks. Then, the fake rate of gluon-jets should be smaller than that of quark-jets. Data triggered with the jet-triggers, that has more gluon-jets than quark-jets, and with single photon triggers, where the quark-jets very likely will be dominant, can be used to understand this effect. However, this type of systematic uncertainty will be small as long as the distributions of observable quantities (such as the number of tracks and the jet isolation) of the probe jets studied for the fake rate is comparable with the properties of the jets faking hadronically decaying $\tau$ leptons in the physics analysis.

In case there is a physical correlation between the properties of the tag and the probe jet, the selection of the tag jet would directly influence the properties of the probe jet, and hence distort the results. As an example of a possible observable correlation, the number of tracks per jet was studied and it has been found that the correlations were in the sub-percent range. Hence it is concluded that given the current knowledge from Monte Carlo, this source of uncertainty can be neglected. For data, this can be revisited.

# 6 Tau leptons in Standard Model processes

The goals for early $\tau$ physics in ATLAS include acquiring a sample of $\tau$ leptons from data with a purity as high possible so that the $\tau_{had}$ identification efficiency can be measured and the simulation can be tuned. Collecting data with an integrated luminosity of 100 pb$^{-1}$ at an instantaneous luminosity of $10^{31}\text{cm}^{-2}\text{s}^{-1}$ will provide a unique opportunity to access and understand statistically significant $\tau$ samples from Standard Model processes at relatively low transverse momenta. Processes, like the production of W and Z bosons and top quark pairs with their huge cross section will lead to samples of a few hundred to a few thousand identified hadronic $\tau$ decays. Hadronically decaying $\tau$ leptons will then become a well understood probe for discovery physics like searches for Higgs bosons, SUSY, or unexpected phenomena. Below we present feasibility studies for analyses which can be envisaged with an integrated luminosity of 100 pb$^{-1}$.

## 6.1 The $W \to \tau\nu$ inclusive production

The $W \to \tau\nu$ signal will be produced with $\sigma \times BR = 1.7 \cdot 10^4$ pb, and will be dominated by events with low $p_T$ of the W-boson resulting in soft $\tau$ leptons with low missing transverse energy. The expected cross-section of the dominant background from hadronic jets is $\sim 10^{10}$ pb (calculated for hard-scattering $p_T^{hard} > 8$ GeV), about 6 orders of magnitude larger than the signal production.

The analysis is very sensitive to the performance of the hadronic $\tau$-trigger [22]. These events will have to be triggered with a $\tau$+$E_T^{miss}$ trigger, with a configuration adequate to fit into the allowed budget for trigger rates. Given the fact that the physics motivation is to get signal events with $\tau_{had}$ candidates at low transverse momenta, the present base-line configuration for this analysis is to use $\tau$20i+EFxE30[9], with the $E_T^{miss}$ trigger applied only at the Event Filter level. The present trigger optimization gives an overall $\sim 70\%$ trigger efficiency with respect to off-line analysis.

The signal will be extracted requiring one identified hadronic $\tau$ decay with transverse energy $E_T = 20-60$ GeV and observing the characteristic track multiplicity spectrum of identified hadronic $\tau$ decays. In the present study the track-based algorithm was used with the medium identification, corresponding to rejection of 700-1000 for the single- and 600 for the three-prong $\tau$ selection against jets and 30% efficiency for true hadronic $\tau_{had}$ decays (see Table 6). The overwhelming background from QCD events will be further suppressed by vetoing events with an additional isolated electrons or muons and by increasing the threshold on $E_T^{miss}$. An additional handle is to select only events with the topological configuration of the $\tau_{had}$, $E_T^{miss}$ or additional jets optimal for suppressing events with large fake $E_T^{miss}$, *i.e.* excluding those where the $E_T^{miss}$ direction is close to the direction of the identified $\tau_{had}$ or an additional jets (see Ref. [23]).

[9] In this notation, $\tau$20i+EFxE30 denotes a trigger which requires at least one $\tau_{had}$ candidate with a transverse energy above 20 GeV and "EFxE30" is short for $E_T^{miss} > 30$ GeV at the Event Filter level.

The QCD background has been estimated with a mixture of full and fast simulation, taking into account the necessary corrections for the different slopes in the $E_T^{miss}$ distribution in the full and fast simulations. It should be noted also that predictions for this overwhelming background are subject to large uncertainties.

An important background is also expected from $W \to e\nu$ events, where an electron passes the hadronic $\tau$-trigger selection criteria. This channel, with an initial production cross-section of the same order as the signal, but contributing almost exclusively to the single-prong mode, will exceed the signal rates by some factor, before a dedicated electron veto is applied to the single-prong candidates. With the expected performance of such an algorithm, as discussed in Section 3.2.1, this background can be efficiently suppressed, also providing a control channel for the topology of the hadronic part of the $W \to \tau\nu$ events passing the trigger and offline selections. Backgrounds from $Z \to \tau\tau$, $t\bar{t}$, $Z \to ee$, $W \to \mu\nu$ events were also considered and it was estimated that they will be suppressed with the offline selection below a few percent of the signal.

Table 8 summarises the expected event yield at the various stages of the selection. With an $E_T^{miss}$ threshold of 50 GeV, the expected signal-to-background ratio is 1:1 and about 3240 signal events would be observed. Increasing the threshold to 60 GeV would reduce the number of accepted signal events to 1550 but increase the signal-to-background ratio to 3:1. Figure 19 shows the expected track multiplicity spectrum after the final selection. The optimisation of this selection can be performed using control samples, i.e. $W \to e\nu$ events extracted from the same filter stream, hence modeling of the W-recoil part of the event will be possible directly from the data. The final optimisation of the chosen efficiency/rejection and offline threshold on $E_T^{miss}$ will have to be tuned with data, given the large uncertainties on Monte Carlo predictions for the background from hadronic jets. The control on the QCD background normalization will be possible with fake $\tau_{had}$ with track multiplicity above 3, i.e in the signal-free region.

Table 8: Expected number of events in 100 pb$^{-1}$ of data for signal and background after subsequent steps of the selection. The track-based algorithm has been used for $\tau_{had}$ reconstruction. The QCD background has been estimated combining fast and full simulation. Given are the expected number of events of track multiplicity one to three, i.e. contributing to signal region only.

| Selection | $W \to \tau\nu$ | $W \to e\nu$ | $W \to \mu\nu$ | QCD dijet | $t\bar{t}, Z \to ee, Z \to \tau\tau$ |
|---|---|---|---|---|---|
| Trigger $\tau 20i$+EFxE30 | $8.8 \cdot 10^4$ | $6.1 \cdot 10^5$ | $3.2 \cdot 10^4$ | $4.8 \cdot 10^8$ | $3.0 \cdot 10^5$ |
| Identified $\tau + E_T^{miss} > 30$ GeV | $2.0 \cdot 10^4$ | 2600 | 200 | $3.0 \cdot 10^6$ | 1600 |
| $E_T^{miss} > 50$ GeV | 4200 | 530 | 90 | $5.0 \cdot 10^4$ | 550 |
| Veto fake $E_T^{miss}$ topology | 3600 | 500 | 80 | $1.8 \cdot 10^4$ | 150 |
| Require jet $p_T > 15$ GeV | 3240 | 450 | 60 | 3200 | 80 |
| Increase to $E_T^{miss} > 60$ GeV | 1550 | 150 | 25 | 500 | 30 |

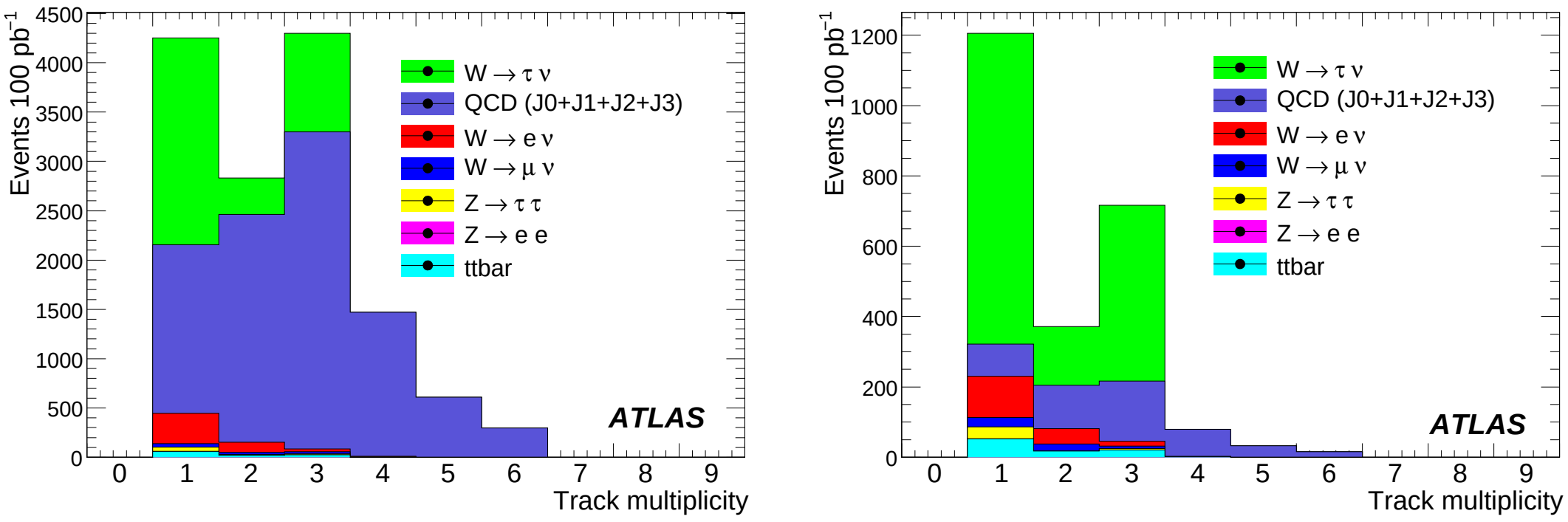

Figure 19: The track multiplicity spectrum of accepted $\tau_{had}$ candidates after selection as described in the text with thresholds respectively $E_T^{miss} > 50$ GeV (left) and $E_T^{miss} > 60$ GeV (right). The expected event numbers are given for an integrated luminosity of 100 $pb^{-1}$.

## 6.2 The $Z \to \tau\tau$ inclusive production

The inclusive $Z \to \tau\tau$ process will provide a ten times lower rate compared to $W \to \tau\nu$, but will have more robust prospects for analysis. It will be possible to cross-check channels with $e\ \tau_{had}$ and $\mu\ \tau_{had}$ final states to control the background, comparing the number of events observed in the same-sign and opposite-sign samples. Moreover, events will be primarily triggered with lepton triggers providing an unbiased sample of hadronic $\tau$ decays which could also serve to understand efficiencies of the hadronic $\tau$-trigger. The measured cross-section for the $Z \to \tau\tau$ process will be an excellent check on the $\tau$ identification efficiencies, while the lepton identification and trigger efficiencies will be measured first from $Z \to \ell\ell$ channels. A measurement of the visible mass of the $\ell\ \tau_{had}$ system at low background levels will have sensitivity to the energy scale of the reconstructed $\tau_{had}$s.

The analysis presented here is designed to select in the first 100 $pb^{-1}$ of data a sufficient number of $Z \to \tau\tau \to \ell\nu\nu\ \tau_{had}\nu$ events with very low background, which then can be used to determine the $\tau_{had}$ energy scale from the reconstructed $\ell\tau_{had}$ visible mass and to determine the $E_T^{miss}$ scale [23] from the reconstructed complete invariant mass of the $\tau\tau$ pair (including neutrinos). Events, which have been selected by the single electron or single muon trigger stream are analysed and as a first step an isolated lepton (electron or muon) with $p_T^{\ell} > 15$ GeV is required. Then, the set of basic selection cuts is applied. It requires a missing transverse energy $E_T^{miss} > 20$ GeV (to suppress $Z \to \ell\ell$ and QCD backgrounds), transverse mass of the lepton and $E_T^{miss}$ system $m_T^{\ell,\ E_T^{miss}} < 30$ GeV (against $W \to \ell\nu$ background), total transverse energy deposited in the calorimeter $\Sigma E_T^{calo} < 400$ GeV (against $t\bar{t}$ and QCD backgrounds) and finally no identified b-jet (against $t\bar{t}$ and QCD backgrounds). In the next step, events with an identified $\tau_{had}$ with $p_T > 15$ GeV are selected. The track multiplicity of identified $\tau_{had}$ is required to be one or three. In addition the angular separation between the isolated lepton and $\tau$ is imposed, requiring $\Delta\phi(\ell, \tau_{had})$ to be in the ranges between 1.0 - 3.1 or 3.2-5.3 .

The analysis was performed using $\tau_{had}$ reconstructed with the calorimeter-based algorithm and the identification with the likelihood discriminant. The thresholds on the likelihood discriminant were optimized for this analysis and correspond to an overall efficiency of $\sim 35\%$ with respect to all hadronic $\tau$ decays. The QCD background has been estimated with a mixture of full and fast simulation which assumes uncorrelated probabilities for a jet to produce an isolated lepton candidate (predominantly leptons from heavy flavor decays with a small contribution from fakes in the electron case) and for the second jet to produce a fake $\tau_{had}$ candidate.

Table 9 gives the expected number of signal and background events passing the selection criteria for a data sample of 100 pb$^{-1}$. About 520 signal events are expected in the visible mass $m_{\ell\tau_{had}}$ window between $37-75$ GeV. The expected background levels are 10% from $W\to\ell\nu$ events and about 5% background from QCD events. The reconstructed visible mass of the $\ell\tau_{had}$ pair is shown in Figure 20 for opposite-sign events.

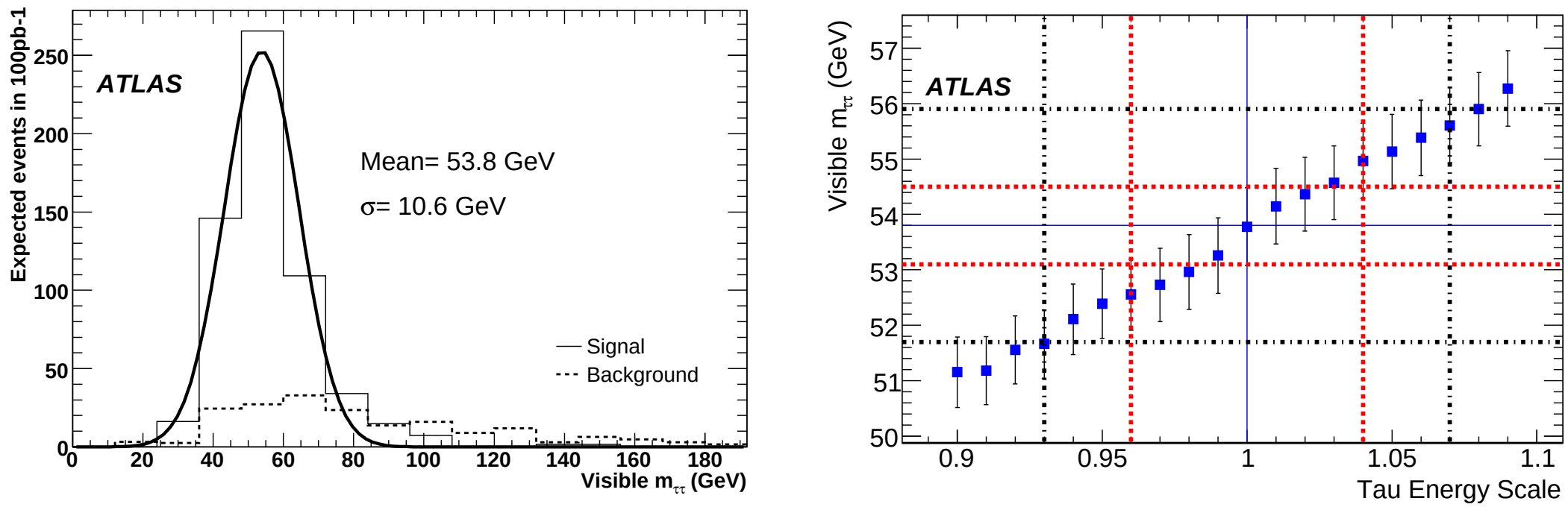

Figure 20: Left: The reconstructed visible mass of the $(\ell\tau_{had})$ pair for $Z\to\tau\tau$ decays (solid line) and QCD, $W\to\ell\nu$, $Z\to\ell\ell$ backgrounds (dashed line). Right: The reconstructed visible mass of the $(\ell\tau_{had})$ pair from $Z\to\tau\tau$ decays as a function of the $\tau_{had}$ energy scale (right). The dashed lines correspond to $\pm1\sigma$ and $\pm3\sigma$ with respect to the reconstructed peak position. The results were obtained with the calorimeter-based algorithm.

Table 9: Expected number of events in 100 pb$^{-1}$ of data for signal and background after reconstruction of the $\tau$ candidate with the calorimeter-based algorithm and after application of the selection cuts for the $Z\to\tau\tau$ channel. The QCD background has been estimated combining fast and full simulation.

| Selection | $Z\to\tau\tau$ | $W\to\ell\nu$ | QCD dijet | $t\bar{t}$ | $Z\to\ell\ell$ |
|---|---|---|---|---|---|
| Isolated lepton | $1.5\cdot10^4$ | $16.7\cdot10^5$ | $1.1\cdot10^7$ | $2.6\cdot10^4$ | $2.2\cdot10^5$ |
| $E_T^{miss} > 20$ GeV | 4750 | $14.3\cdot10^5$ | $3.2\cdot10^5$ | $2.4\cdot10^4$ | $1.0\cdot10^4$ |
| $m_T^{\ell,E_T^{miss}} < 30 GeV$ | 3200 | $2.6\cdot10^4$ | $1.8\cdot10^5$ | 3650 | 3200 |
| $\Sigma E_T < 400$ GeV | 3000 | $2.4\cdot10^4$ | $1.7\cdot10^5$ | 1280 | 2800 |
| b-jet veto | 2780 | $2.4\cdot10^4$ | $2.7\cdot10^4$ | 135 | 2600 |
| $\tau_{had}$-id + $\Delta\phi(\ell\tau_{had})$ cuts | 630$\pm$30 | 210 $\pm$10 | 74$\pm$11 | 10$\pm$2 | 30$\pm$5 |
| OS events, $m^{\ell,\tau_{had}}$ = 37-75 GeV | 520$\pm$30 | 45 $\pm$5 | 29$\pm$5 | $<5$ | 10 $\pm$5 |

This selection for $Z\to\tau\tau$ events will provide access to signal-suppressed and signal-enriched samples. The same-sign events (e $\tau_{had}$ and $\mu$ $\tau_{had}$) will be essentially signal-free. This will allow a study of $\tau_{had}$ identification and mistagging efficiencies. The mistagging efficiency will be estimated from the ratio of accepted to all candidates in different categories (one-prong with and without $\pi^0$ subclusters, multi-prong). Then, this estimate can be used to predict the background component in opposite-sign events and to tune the Monte Carlo predictions for the identification variables. It will allow a confirmation of the overall consistency and an estimate of the relative error on the background predictions in the signal enriched sample. Finally, a measurement of the cross-section for $Z\to\tau\tau$ events relative to the $Z\to ee,\mu\mu$ channels will provide a cross-checks on the associated efficiencies.

Once the lepton energy scale is determined with the very first data, the selected $Z\to\tau\tau$ events can be used to determine the $\tau_{had}$ energy scale in-situ. Subtracting the estimated background from opposite-sign events, as measured with the same-sign events, will allow for better estimates on the energy scale from the shape of the distribution of the visible mass. In Fig. 20 the sensitivity of the measured visible $Z$ boson mass, as obtained from the reconstructed $\tau$ pairs, on the absolute $\tau$ energy scale is shown, assuming only signal events.The statistics correspond to 100 pb$^{-1}$ of data. Taking into account only the statistical uncertainties, the $\tau_{had}$ energy scale could be determined with a precision of $\sim 3\%$.

Table 10: Expected number of events in 100 pb$^{-1}$ of data for $t\bar{t} \to W(\ell\nu)W(\tau_{had},\nu_\tau)b\bar{b}$ signal and background after subsequent steps in the selection. The track-based algorithm has been used for $\tau_{had}$ reconstruction.

| Selection | $t\bar{t}(\ell,\tau_{had})$ | $W \to \ell\nu + 3\,jets$ | single t | Z$\to \ell\ell$ + 2 jets |
|---|---|---|---|---|
| Isolated lepton $p_T > 20$ GeV | 1300 | $3.9 \cdot 10^5$ | 4300 | 630 |
| Identified $\tau_{had}$ $p_T > 15$ GeV | 190 | 22000 | 210 | 120 |
| 1st jet $E_T > 50$ GeV, 2nd jet $E_T > 30$ GeV | 170 | 4000 | 170 | 35 |
| $E_T^{miss} > 25$ GeV | 150 | 3400 | 150 | 15 |
| $\Sigma E_T > 250$ GeV | 150 | 1750 | 130 | 10 |
| Opposite-sign events | 130 | 850 | 54 | $< 10$ |
| 1 b-jet tag | 67 | 28 | 20 | |

### 6.3 The $\tau$ leptons from $t\bar{t}$ production

With a cross-section of 833 pb [24], about 16500 events are expected in 100 pb$^{-1}$ with a W boson decaying into a $\tau$ lepton. Due to the increased center of mass energy available at the LHC the cross section for $t\bar{t}$ production increases by nearly two orders of magnitude over what is available at the Tevatron. The $t\bar{t}$ channel is discussed here as an additional source for $\tau$ leptons from the SM processes, supplementing samples expected from $W \to \tau\nu$ and $Z \to \tau\tau$ process.

The decay chain $t\bar{t} \to W(qq')W(\tau_{had}\nu)b\bar{b}$ requires events triggered using $\tau + E_T^{miss}$ triggers, $\tau$ triggers and multi–jets triggers. In the latter case this will lead to an unbiased sample of $\tau_{had}$ candidates. The event is required to have at least two light quark jets, two $b$-tagged jets, and an identified hadronic $\tau$ decay. If the event has more than two light quark jets, the pair with the invariant mass closest to the nominal mass of the $W$ bosons is chosen which is then combined with the closest b jet to constitute the hadronically decaying top quark. With 100 pb$^{-1}$ of data about 300 $t\bar{t} \to W(qq')W(\tau_{had}\nu)b\bar{b}$ signal events with S:B of 20:1 are expected. These events can be used to study the $\tau_{had}$ reconstruction and identification performance and to commission the $\tau$ trigger. The $p_T$ range of identified $\tau$ leptons will be complementary to that available from the inclusive W and Z boson production. A more detailed description of the analysis is included in Ref. [25].

The decay chain $t\bar{t} \to W(e\nu_e,\mu\nu_\mu)W(\tau_{had},\nu_\tau)b\bar{b}$ is also interesting for both its physics potential and the possibility of using this channel to understand $\tau_{had}$ identification. These events will be triggered with single lepton triggers and the main background will come primarily from $W(\to \ell\nu)$+jets, single top production and from $Z(\to \tau\tau)$+jets production. The analysis requires an isolated lepton and identified $\tau_{had}$. To suppress backgrounds from W+jets and Z+jets events it requires two additional high $E_T$ jets, significant energy deposition in the calorimeter $\Sigma E_T > 250$ GeV and $E_T^{miss} > 25$ GeV. Additional background suppression can be achieved by requiring that one or two jets are b-tagged. Table 10 summarizes cut flow of the analysis.

In the first 100 pb$^{-1}$ of data, $54 \pm 4$ signal events in the $e\tau_{had}$ channel are expected, with a signal-to-background ratio (S:B) 1:10. The use of b-tagging rejects considerably the dominant $W + 3$ jets background (see Fig. 21). If at least one tight b-tag jet is required the expected number of signal events decreases to $28 \pm 3$ with S:B improving to 1:1. When requiring a jet in the event that passes the tight b tagging criteria, the dominant source of background is still $W(\to \ell\nu)$+jets, however the composition of the background changes and single top production starts contributing significantly, with quark or gluon jets faking $\tau_{had}$ and true b-jets from b-quark fragmentation.

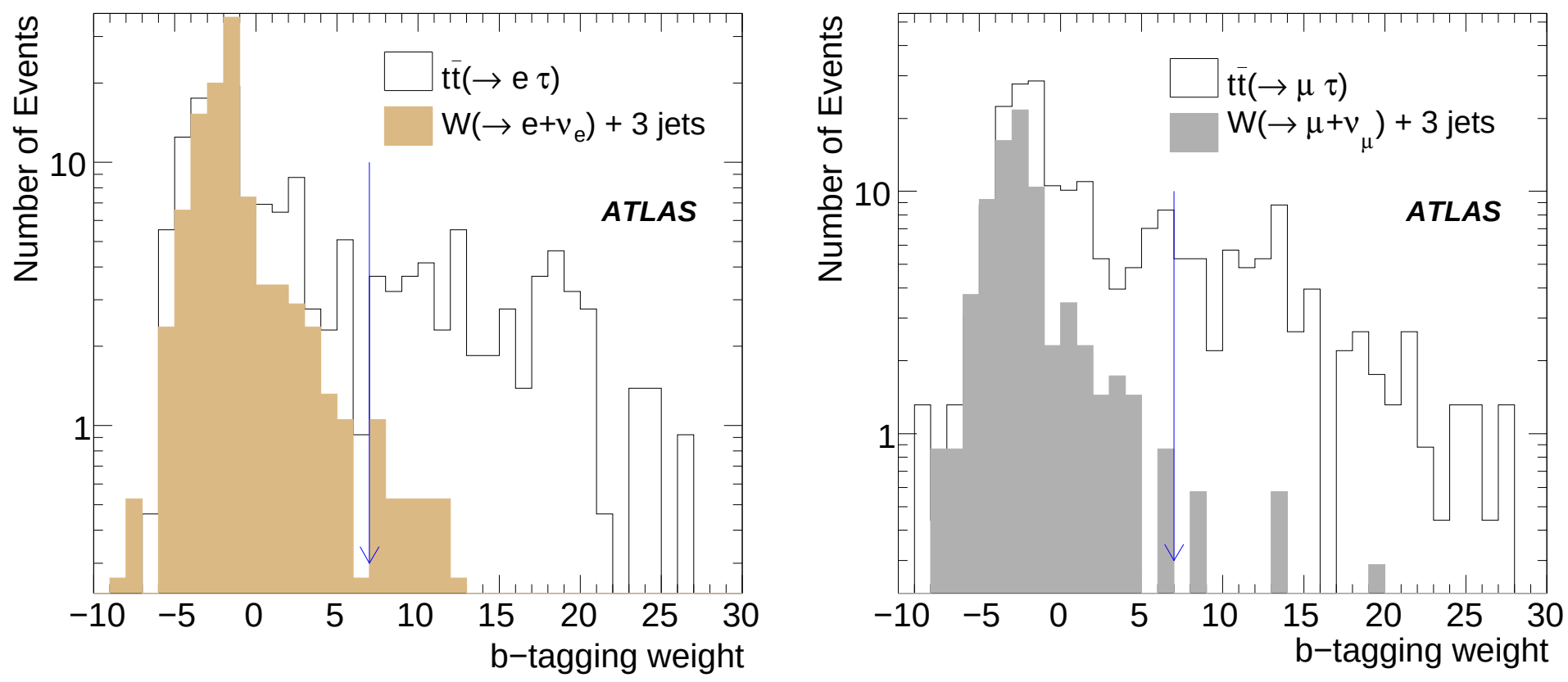

Figure 21: Combined $b$-tagging weights using impact parameter and secondary vertex information for the first two leading $E_T$ jets, both in $t\bar{t} \to W(e\nu_e, \mu\nu_\mu)W(\tau_{had}\nu_\tau)b\bar{b}$ and $W + 3$ *jets* background. The $e\ \tau$ ($\mu\ \tau$) channel is shown on the left (right). The cut value of 7 on the b-tagging weight is indicated with the arrows. An integrated luminosity of 100 pb$^{-1}$ of data is assumed.

## 7 Summary

Two complementary algorithms for the identification of hadronic $\tau$ decays in the ATLAS experiment have been developed. The first one (calorimeter based) is seeded from a reconstructed cluster in the calorimeter, the second one (track based) relies on seeds built from reconstructed tracks in the inner detector. Several discrimination methods have been established, including a simple cut-based selection as well as multivariate selections based on likelihood, neutral network, and probability range search techniques. Rejection factors against jets from QCD processes of a few hundred up to a few thousand can be achieved for a $\tau$ efficiency of 30% in the $p_T$ range between 10 to 60 GeV. In addition, a dedicated algorithm has been developed to reject electrons that pass the $\tau$ identification criteria. In the low energy range, rejection factors of the order of 50 and higher against electrons from $W$ and $Z$ bosons decays are achieved at the expense of a 5% efficiency loss for hadronic $\tau$ decays.

It has also been estimated that the expected performance in the ATLAS experiment will be adequate to extract $\tau$ signals in early LHC data from $W \to \tau\nu$ and $Z \to \tau\tau$ decays. These signals are important to establish and calibrate the $\tau$ identification performance with early data. The study of dijet events from QCD processes will allow a determination of $\tau$ fake rates. It is expected that such rates can be measured with a statistical precision at the percent level or better already with data corresponding to an integrated luminosity of 100 pb$^{-1}$.

# Jets and Missing Transverse Energy

# Jet Reconstruction Performance

**Abstract**
This section summarizes the general aspects of jet reconstruction with the ATLAS detector. General but brief descriptions of the available jet algorithms are provided, together with a discussion of the performance expectations for the various algorithm configurations in different detector regions and for different physics environments. The emphasis is on realistic estimates for the initial jet reconstruction performance, determined in the absence of experimental data. The corresponding expectations for important jet reconstruction parameters like signal linearity and uniformity, the relative energy resolution, and the jet reconstruction efficiency and purity, are presented.

## 1 Introduction

High quality and highly efficient jet reconstruction is an important tool for almost all physics analyses to be performed with the ATLAS experiment at the Large Hadron Collider (LHC) at CERN. The requirements especially for the absolute precision on the jet energy scale often exceed the corresponding achieved performance in previous experiments. Typically, an absolute systematic uncertainty of better than 1% is desirable for precision physics like the measurement of the top quark mass, and the reconstruction of some SUSY final states.

The principal detector for jet reconstruction is the ATLAS calorimeter system, with its basic components depicted in Fig. 1. It provides near hermetic coverage in a pseudorapidity range $-4.9 < \eta < 4.9$. The technology choices are well suited for high quality jet reconstruction in the challenging environment of the proton-proton ($pp$) collisions at $\sqrt{s} = 14$ TeV at the LHC. The electromagnetic liquid ar-gon/lead calorimeters feature an accordion geometry homogeneous in azimuthal coverage for $|\eta| < 3.2$. The hadronic calorimeters surrounding them are iron with scintillating tile readout in the central region ($|\eta| \lesssim 1.7$) and parallel plate liquid argon/copper in the endcap region ($1.7 \lesssim |\eta| \lesssim 3.2$). The forward region is covered by liquid argon/copper and liquid argon/tungsten calorimetry with a tubular electrode readout accommodating the high ionization rates expected at LHC at design luminosity. The readout of the calorimeters is highly granular for the electromagnetic devices, with typically three longitudinal segments with varying lateral cell sizes, e.g. $\Delta\eta \times \Delta\phi = 0.025 \times 0.025$ in the second segment containing the electromagnetic shower maximum. The hadronic calorimeters are coarser, with typically $\Delta\eta \times \Delta\phi = 0.1 \times 0.1$, but have also at least three longitudinal segments. The total number of readout cells in the ATLAS calorimeter system is close to $200,000$. The total thickness of the ATLAS calorimeter system for hadrons is at least 10 absorption lengths over the whole acceptance region. More details on the calorimeters, and any other detectors in ATLAS, can be found in Ref. [1].

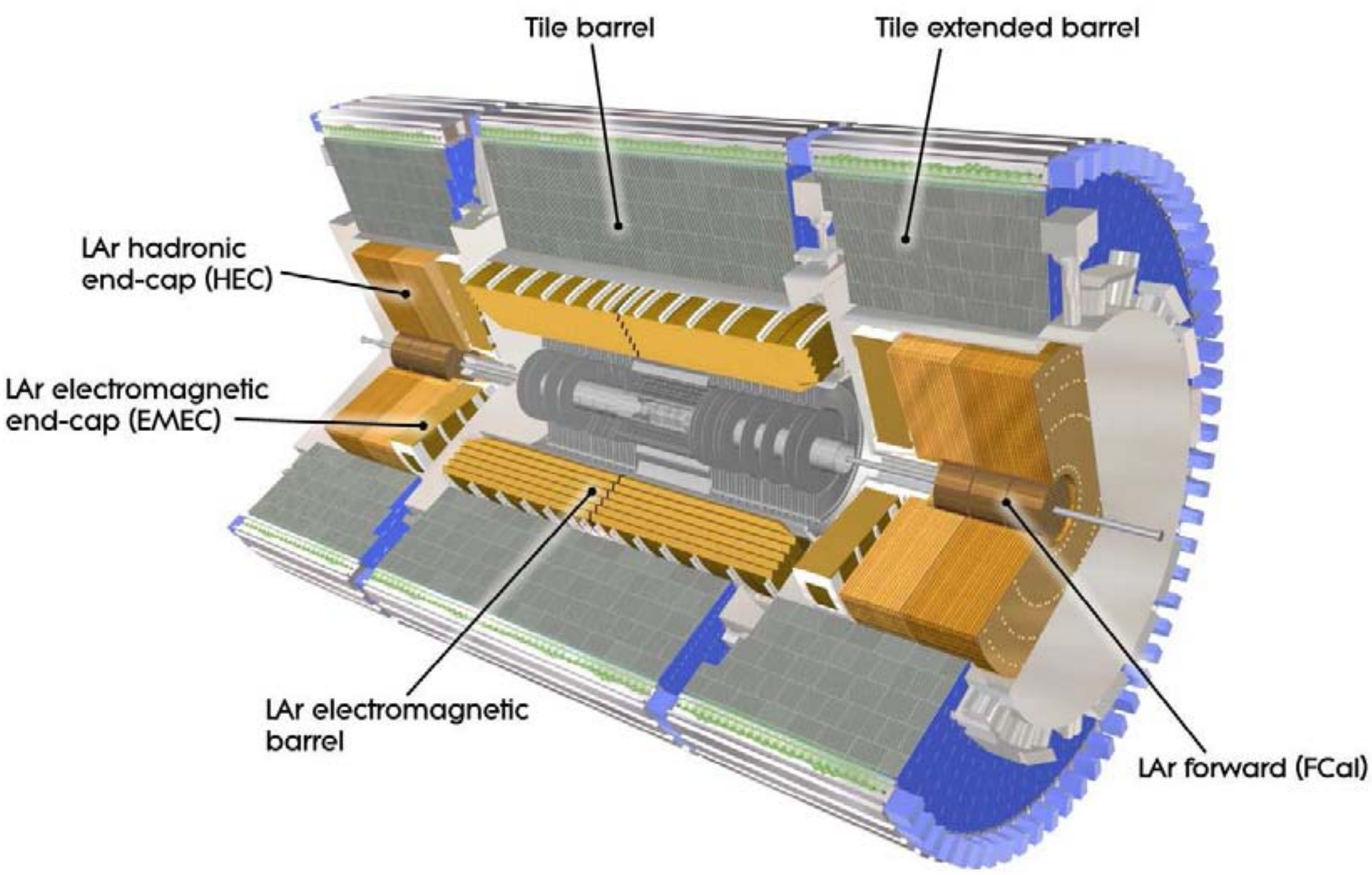

Figure 1: The ATLAS calorimeter system.

In this note the approaches used by the ATLAS collaboration to achieve the challenging performance goals are discussed. First, the most commonly used jet finders are briefly introduced in Section 2, together with the theoretical and experimental guidelines for the ATLAS implementations. Then, the expectations for the performance of these algorithms using different detector signals are shown in Section 3. As the focus is on the upcoming initial data taking period of ATLAS, some distortions in the detector alignment and material distributions have been included but not corrected in the results presented here. Finally, special challenges to jet reconstruction like forward going jets and jets in minimum bias events are discussed in Section 4, followed by conclusions in Section 5.

# 2 Jet algorithms in ATLAS

In general an attempt is made to provide implementations of all relevant jet finding algorithms in ATLAS. These include fixed sized cone algorithms as well as sequential recombination algorithms and an algorithm based on event shape analysis. This approach is a response to the fact that there is no universal jet finder for the hadronic final state in all topologies of interest. For example, for the measurement of the inclusive QCD jet cross-sections wider jets are typically preferred to capture the hard scattered parton kinematics, including possible small angle gluon radiation, completely. On the other hand, to reconstruct a $W$ boson decaying into two jets or to find jets in very busy final states like $t\bar{t}$ production or possible SUSY signatures, narrow jets are preferred.

The common feature of all jet finder implementations in ATLAS is full four-momentum recombination whenever the constituents of a jet change, either through adding a new constituent, or by removing one, or by changing the kinematic contribution of a given constituent to the jet. Also, in the ATLAS reconstruction software framework ATHENA, the same jet finder code can be run on objects like calorimeter signal towers, topological cell clusters in the calorimeters, reconstructed tracks, and generated particles and partons.

In this section the basics of the present default jet algorithms used in ATLAS are discussed after a brief summary of the theoretical and experimental guidelines for jet algorithm implementation. In addition, some features of jet finders not included in the more comprehensive presentation of jet reconstruction performance in Section 3 are shown.

## 2.1 Guidelines for jet reconstruction

The basic guidelines for jet reconstruction in ATLAS have been extracted from Ref. [2]. They also reflect the concept of *jet definition* discussed in Ref. [3], which is an attempt to provide a common understanding between experiments and theory on how a given jet finding strategy should be specified to assure the highest level of comparability between the results from various sources.

### 2.1.1 Theoretical guidelines

The major theoretical guidelines for jet reconstruction are:

**Infrared safety:** The presence of additional soft particles between two particles belonging to the same jet should not affect the recombination of these two particles into a jet. In the same sense, the absence of additional particles between these two should not disturb the correct reconstruction of the jet. Generally, any soft particles not coming from the fragmentation of a hard scattered parton should not effect the number of jets produced.

**Collinear safety:** A jet should be reconstructed independent of the fact that a certain amount of transverse momentum is carried by one particle, or if a particle is split into two collinear particles.

**Order independence:** The same hard scattering should be reconstructed independently at parton-, particle- or detector level.

Note that from the perspective of experimental data the particles mentioned in these guidelines can, to a point, be replaced by four-momentum type objects reconstructed from detector signals, see e.g. Section 2.5 and Section 2.6 below.

### 2.1.2 Experimental guidelines

Additional aspects of jet reconstruction in ATLAS include features also reflected in the design of the detector. They can be divided into three classes.

**Detector technology independence:** The reconstructed jet and its kinematic variables should not dependent on the signal source, i.e. all detector specific signal characteristics and inefficiencies must be calibrated out or corrected as much as possible.

- **Detector resolution:** contributions from the finite spatial and energy resolution must be at a minimum;
- **Detector environment:** effects from the detector environment like electronics noise, signal losses in un-instrumented (inactive) materials and cracks between detectors must be at a minimum;
- **Stable signals:** the detector signal reconstruction and calibration must provide a stable input signal to jet reconstruction.

**Environment independence:** The jet reconstruction environment is characterized by the additional activity in the collision event due to multiple interactions and pile-up, the source of the jet, the underlying event activity, and other features of the *pp* collisions at LHC.

- **Stability:** a jet should be found and reconstructed safely even in the case of changing underlying event activity and changing instantaneous luminosity, thus changing number of multiple interactions;
- **Efficiency:** all physically interesting jets from energetic partons must be identified with high efficiency.

**Implementation:** The jet algorithm implementation must be fully specified in that the jet definition, which consists of the jet finder and its configuration together with the choice of kinematic recombination given in Ref. [3], must be complete. Also included must be all selections and, if important for the measured jet, the signal choices. In addition, the implementation of the jet reconstruction must make efficient use of computing resources, i.e. it must be sufficiently fast and avoid excessive memory consumption.

The most commonly used jet finder implementations in ATLAS are a seeded fixed cone finder with split and merge (see below), and a $k_T$ algorithm [4, 5] implementation, with an initial implementation as described in Ref. [6], but later replaced by a faster implementation similar to the one in the FASTJET package [7]. It is anticipated that for the first experimental collision data all implementations of the FASTJET library ($k_T$, anti-$k_T$, Cambridge flavor $k_T$ [8]) will be available, as well as the seedless infrared-safe cone algorithm SISCONE [9]. As there are no complete systematic evaluations of these algorithms available for this note, they are excluded from further discussions here.

## 2.2 Fixed cone jet finder in ATLAS

The ATLAS implementation of the iterative seeded fixed-cone jet finder follows the algorithm description of Ref. [2]. First, all input is ordered in decreasing order in transverse momentum $p_T$. If the object with the highest $p_T$ is above the seed threshold, all objects within a cone in pseudorapidity $\eta$ and azimuth $\phi$ with $\Delta R = \sqrt{\Delta\eta^2 + \Delta\phi^2} < R_{\mathrm{cone}}$, where $R_{\mathrm{cone}}$ is the fixed cone radius, are combined with the seed. A new direction is calculated from the four-momenta inside the initial cone and a new cone is centered around it. Objects are then (re-)collected in this new cone, and again the direction is updated. This process continues until the direction of the cone does not change anymore after recombination, at which point the cone is considered stable and is called a jet. At this point the next seed is taken from the input list and a new cone jet is formed with the same iterative procedure. This continues until no more seeds are available. The jets found this way can share constituents, and signal objects contributing to the cone at some iteration maybe lost again due to the recalculation of the direction at a later iteration.

This algorithm is not infrared safe, which can be (at least) partly recovered by introducing a split and merge step after the jet formation is done. Jets which share constituents with more than a certain fraction $f_{\mathrm{sm}}$ of the $p_T$ of the less energetic jet are merged, while they are split if the amount of shared $p_T$ is below $f_{\mathrm{sm}}$, with $f_{\mathrm{sm}} = 0.5$ in ATLAS. Other important parameters of the ATLAS cone jet finder are a seed threshold of $p_T > 1$ GeV, and a narrow ($R_{\mathrm{cone}} = 0.4$) and a wide cone jet ($R_{\mathrm{cone}} = 0.7$) option.

From a theoretical standpoint this particular cone jet finder is by design only meaningful to leading order for inclusive jet cross-section measurements and final states like $W/Z+1$ jet, but is not meaningful at any order for 3-jet final states, $W/Z+2$ jets, and for the measurement of the dijet invariant mass in 2 jets $+X$ final states [10].

## 2.3 Sequential recombination algorithms

The default implementation of a sequential recombination jet finder in ATLAS is the $k_T$ algorithm. Here all pairs ij of input objects (partons, particles, reconstructed detector objects with four-momentum representation) are analyzed with respect to their relative transverse momentum squared, defined by

$$d_{ij} = \min(p_{T,i}^2, p_{T,j}^2)\frac{\Delta R_{ij}^2}{R^2} = \min(p_{T,i}^2, p_{T,j}^2)\frac{\Delta\eta_{ij}^2 + \Delta\phi_{ij}^2}{R^2},$$

and the squared $p_T$ of object $i$ relative to the beam $d_i = p_{T,i}^2$. The minimum $d_{\mathrm{min}}$ of all $d_{ij}$ and $d_i$ is found. If $d_{\mathrm{min}}$ is a $d_{ij}$, the corresponding objects $i$ and $j$ are combined into a new object $k$ using four-momentum recombination. Both objects $i$ and $j$ are removed from the list, and the new object $k$ is added to it.

If $d_{\mathrm{min}}$ is a $d_i$, the object $i$ is considered to be a jet by itself and removed from the list. This procedure is repeated for the resulting new sets of $d_{ij}$ and $d_i$ until all objects are removed from the list. This means that all original input objects end up to be either part of a jet or to be jets by themselves. Contrary to the cone algorithm described earlier, no objects are shared between jets. The procedure is infrared safe. As it does not use seeds, it is also collinear safe. The distance parameter $R$, which is the only free parameter besides the choice of recombination scheme in this inclusive implementation of the $k_T$ algorithm, allows some control on the size of the jets. Default configurations in ATLAS are $R = 0.4$ for narrow and $R = 0.6$ for wide jets.

Table 1: Default jet finder configurations used in ATLAS.

| Algorithm | Main parameter | Clients |
|---|---|---|
| Seeded fixed cone | $R_{cone} = 0.4$ | $W \to jj$ in $t\bar{t}$, SUSY |
| (seed $p_T > 1$ GeV) | $R_{cone} = 0.7$ | inclusive jet cross-section, $Z' \to jj$ |
| $k_T$ | $R = 0.4$ | $W \to jj$ in $t\bar{t}$, SUSY |
| | $R = 0.6$ | inclusive jet cross-section, $Z' \to jj$ |

Table 1 summarizes the algorithms and configurations which have been used by ATLAS for basically all pre-collision physics studies. Thus, these are the base for most predictions related to the performance of the hadronic final state reconstruction for all studied physics channels available to date.

## 2.4 Alternative jet finders

As jet finders others than the "default" algorithms and configurations discussed above can be more appropriate for the precision analysis of specific final states, additional jet algorithms are available in ATLAS for application at the analysis stage. Those are the mid-point variant of the fixed cone algorithm originally introduced by CDF in Ref. [2, 11], and the "optimal jet finder" discussed in Ref. [12]. Both algorithms have been studied for ATLAS and in general provide a very similar performance when compared to the default seeded cone and $k_T$ implementations. This can be seen in Fig. 2 for the mid-point algorithm, which, as already said, is a flavour of the fixed cone algorithm with the modification that the seeds are placed between two particles with significant $p_T$, rather than just using an individual particle $p_T$ as seed directly. Besides a slightly smaller efficiency in the central region, there are no significant differences between the studied jet finders observed in this non-comprehensive investigation with simulated $t\bar{t}b\bar{b}$ events.

The optimal jet finder is a departure from the traditional approach of reconstructing each jet rather independent from previously reconstructed jets in the same event. Here the basic scheme is to calculate a weight for each particle reflecting the contribution to any jet by minimizing a test function event by event. This function actually includes weights for contributions to transverse momentum not clustered into any jet at all, thus using the overall event shape when reconstructing the jets. Parameters of the algorithm are a jet cone size and a threshold for the test function. A more exclusive mode is available where the number of jets to be reconstructed can be fixed beforehand. In general this jet finder works well in busy final states like full hadronic top-quark decays in $t\bar{t}$ production. Predictions for the relative top mass resolution, for example, are basically identical for the default $k_T$ implementation with $R = 0.6$ and the optimal jet finder for the same jet size.

## 2.5 Calorimeter jets

The most important detectors for jet reconstruction are the ATLAS calorimeters. In this section the input signals and default calibration schemes for calorimeter jets are briefly described.

The ATLAS calorimeter system [13] has about $200,000$ individual cells of various sizes and with different readout technologies and electrode geometries. For jet finding it is necessary to first combine these cell signals into larger signal objects with physically meaningful four-momenta. The two concepts available are calorimeter *signal towers* and *topological cell clusters*.

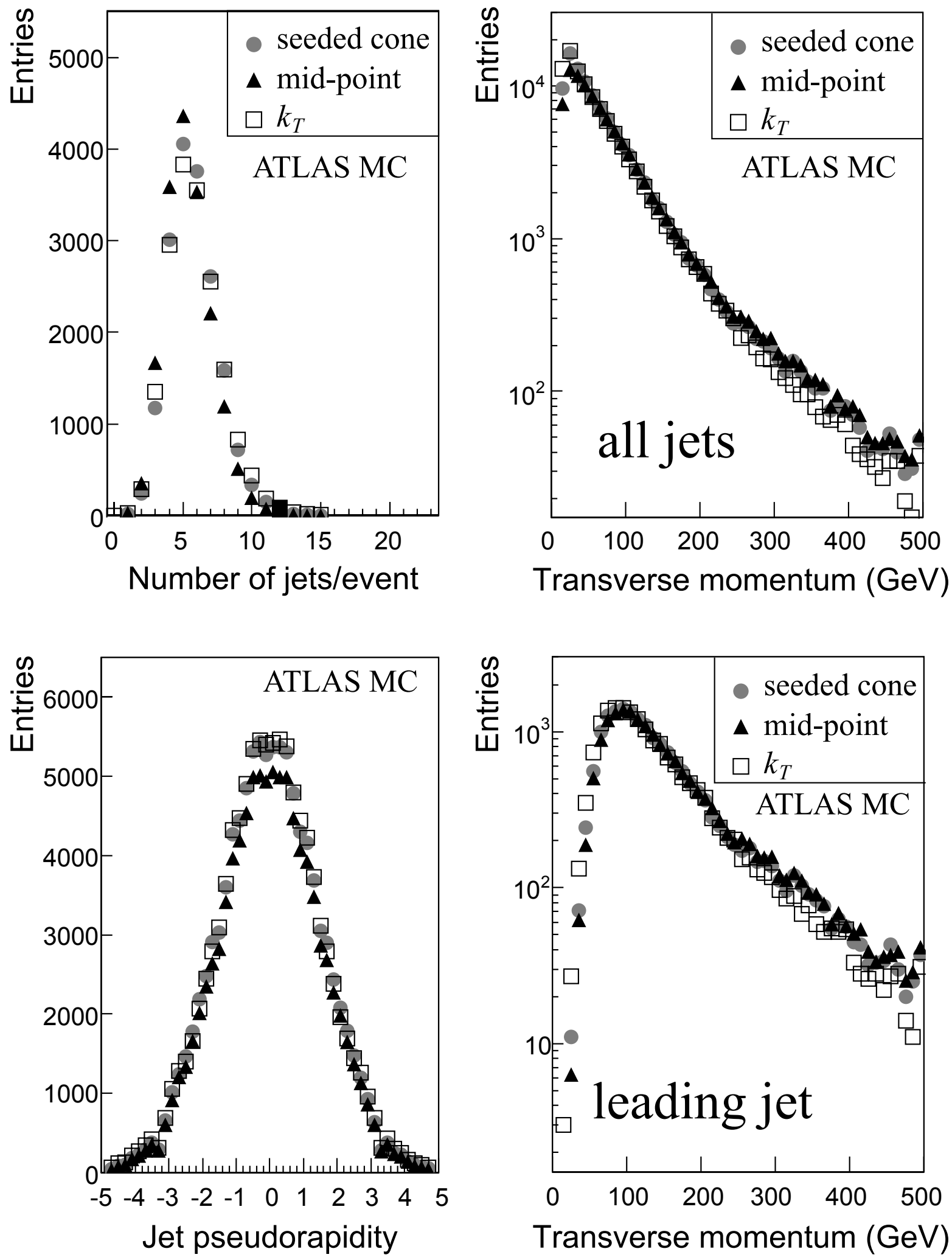

Figure 2: Performance comparison of the ATLAS seeded cone, $k_T$, and the mid-point seeded cone for $t\bar{t}b\bar{b}$ events, as calculated from simulations. Shown are the distributions for jet multiplicity (top left), the $p_T$ spectrum for all jets (top right), the rapidity distribution (bottom left) and the $p_T$ spectrum for the leading jets (bottom right).

### 2.5.1 Calorimeter tower signals

In case of the towers, the cells are projected onto a fixed grid in pseudorapidity ($\eta$) and azimuth ($\phi$). The tower bin size is $\Delta\eta \times \Delta\phi = 0.1 \times 0.1$ in the whole acceptance region of the calorimeters, i.e. in $|\eta| < 5$ and $-\pi < \phi < \pi$ with $100 \times 64 = 6,400$ towers in total. Projective calorimeter cells which completely fit inside a tower contribute their total signal, as reconstructed on a basic electromagnetic energy scale[1], to the tower signal. Non-projective cells and projective cells larger than the tower bin size contribute a fraction of their signal to several towers, depending on the overlap fraction of the cell area with the towers, see Fig. 3 for illustration.

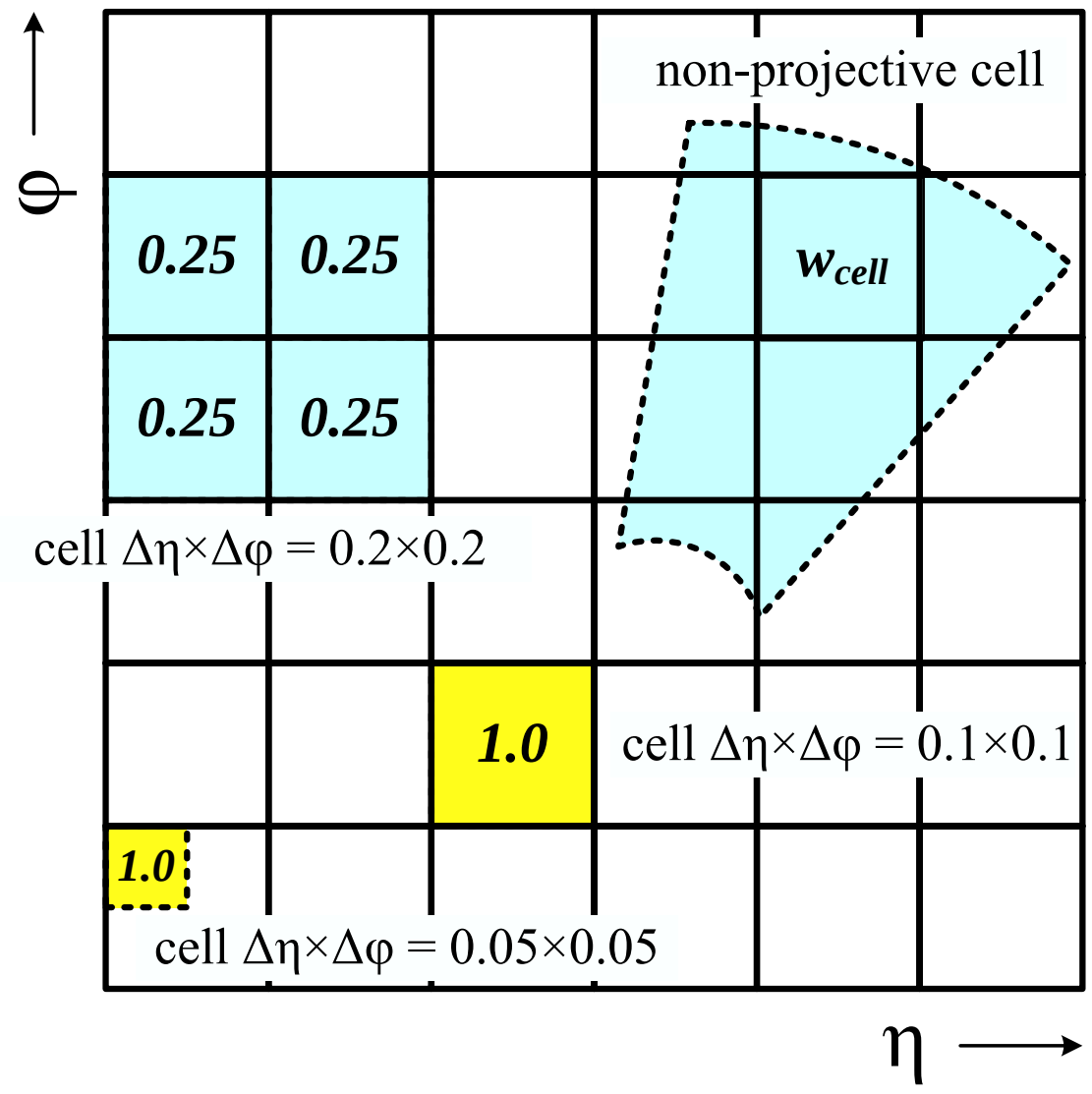

Figure 3: Calorimeter cell signal contributions to towers on a regular $\Delta\eta \times \Delta\phi = 0.1 \times 0.1$ grid, for projective and non-projective cells. The signal contribution is expressed as a geometrical weight and is calculated as the ratio of the tower bin area over the projective cell area in $\eta$ and $\phi$.

Thus, the tower signal is the nondiscriminatory sum of possibly weighted cell signals (all cells are included). As the cell signals are on the basic electromagnetic energy scale, the resulting tower signal is on the same scale. No further corrections or calibrations are applied at this stage.

### 2.5.2 Topological cell clusters

The alternative representation of the calorimeter signals for jet reconstruction are topological cell clusters, which are basically an attempt to reconstruct three-dimensional “energy blobs” representing the showers developing for each particle entering the calorimeter. The clustering starts with seed cells with a signal-to-noise ratio, or signal significance $\Gamma = E_{\text{cell}} / \sigma_{\text{noise,cell}}$, above a certain threshold $S$, i.e. $|\Gamma| > S = 4$. All directly neighbouring cells of these seed cells, in all three dimensions, are collected into the cluster. Neighbours of neighbours are considered for those added cells which have $\Gamma$ above a certain secondary threshold $N$ ($|\Gamma| > N = 2$). Finally, a ring of guard cells with signal significances above a basic threshold $|\Gamma| > P = 0$ is added to the cluster. After the initial clusters are formed, they are analyzed for local signal maximums by a splitting algorithm, and split between those maximums if any are found [15].

[1] This is the raw signal from the ATLAS calorimeters. The nomenclature indicates that this scale has been derived from electron signals, but it lacks all corrections applied in high precision electron or photon reconstruction as described in Ref. [14]. It typically includes all electronic corrections and the geometrically motivated corrections for high voltage problems, like inactive electrode sub-gaps and similar.

Figure 4 shows the average number of particles in Monte Carlo generated jets from QCD dijet production together with the number of topological cell clusters in the corresponding simulated calorimeter jets. The figure indicates that the clustering algorithm resolves the particle content of the jet in the pseudorapidity range $1.5 \lesssim |\eta| \lesssim 2.5$. The shower overlap between the particles in the jet cannot be resolved as well in the central region $|\eta| \lesssim 1.5$, where the calorimeter cell sizes are a bit larger on the scale of the hadronic shower. Here the ratio of number of particles per cluster is approximately 1.6. Similarly, in the forward region $|\eta| \gtrsim 2.5$ both the increase in shower overlap, due to the decreasing linear distance between jet particles, and the increase of the cell sizes reduce the resolution power of the clustering algorithm for individual particle showers.

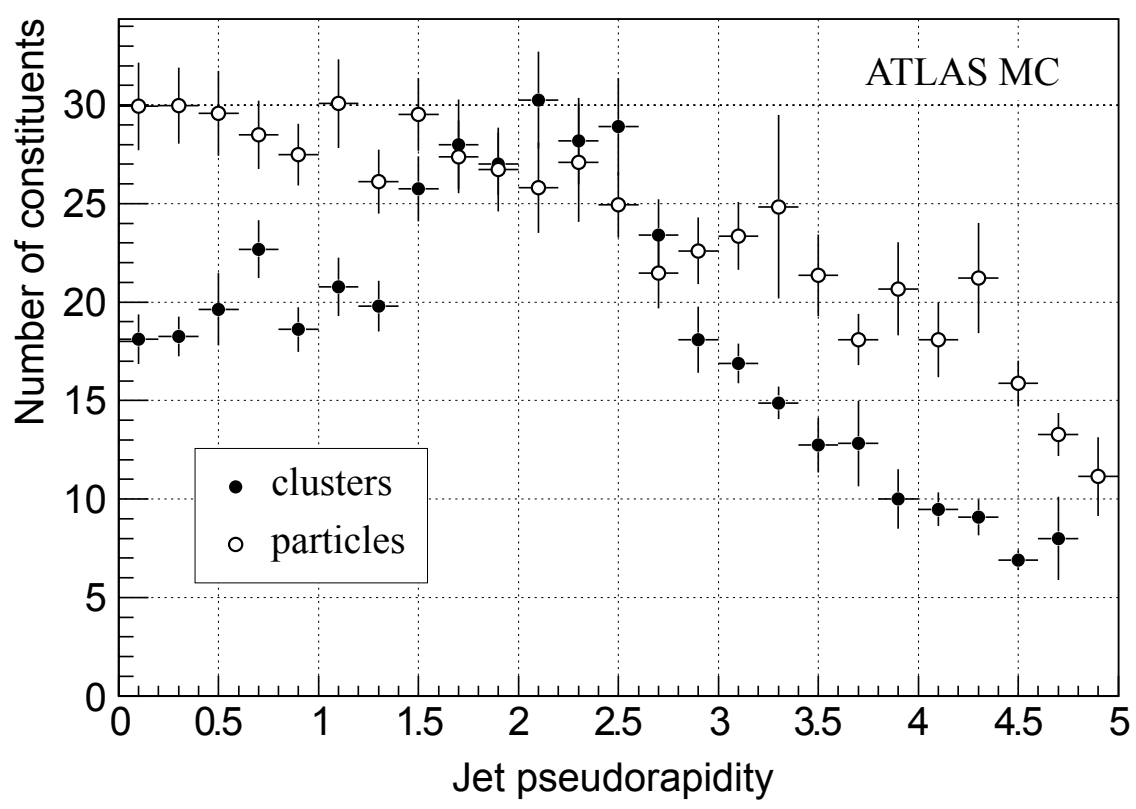

Figure 4: Estimates for the average number of particles in seeded fixed size cone jets with $R_{\rm cone} = 0.7$ in fully simulated QCD dijet production, shown as function of the pseudorapidity $\eta$ of the jet. Also shown is the corresponding average number of topological clusters in matching jets found with the same algorithm in the ATLAS calorimeters.

Like towers, clusters are initially formed using the basic electromagnetic energy scale cell signals. These clusters can already be used for jet reconstruction. In addition, clusters can be calibrated to a local hadronic energy scale. This calibration starts with a classification step characterizing clusters as electromagnetic, hadronic, or noise, based on their location and shape. After that, cell signals inside hadronic clusters are weighted with functions depending on cluster location, energy, and the cell signal density. Then, a correction for energy losses in inactive materials close to or inside the cluster is applied. Finally, a correction for signal losses due to the clustering itself (out-of-cluster correction) is applied. Note that all calibrations and corrections for topological clusters are derived from single particle simulations and do not use the jet context.

### 2.5.3 Characteristics of calorimeter input to jet finding

There are attempts in ATLAS to go beyond the high quality reconstruction of the total jet signal with the best possible resolution. Especially the reconstruction of jet structure, including the lateral and longitudinal signal distributions, can be useful to apply jet energy scale corrections jet by jet, or reconstruct the origin of a given jet. The ability to reconstruct this structure depends on the choice of the calorimeter signal definition used in jet finding, as is qualitatively indicated in the simulated high $p_T$ QCD event in Fig. 5. A given calorimeter signal definition like clusters may reproduce the jet shape at particle level better in certain regions of the calorimeters than in others. For example, from the depicted event in this figure the cluster signals represent the transverse energy flow of particles inside a jet better than the tower jets in the central and endcap regions, while in the forward region the clusters cannot resolve individual showers anymore and thus cannot reproduce the jet shape very well. Here the towers still reflect some spatial structure of the incoming particles.

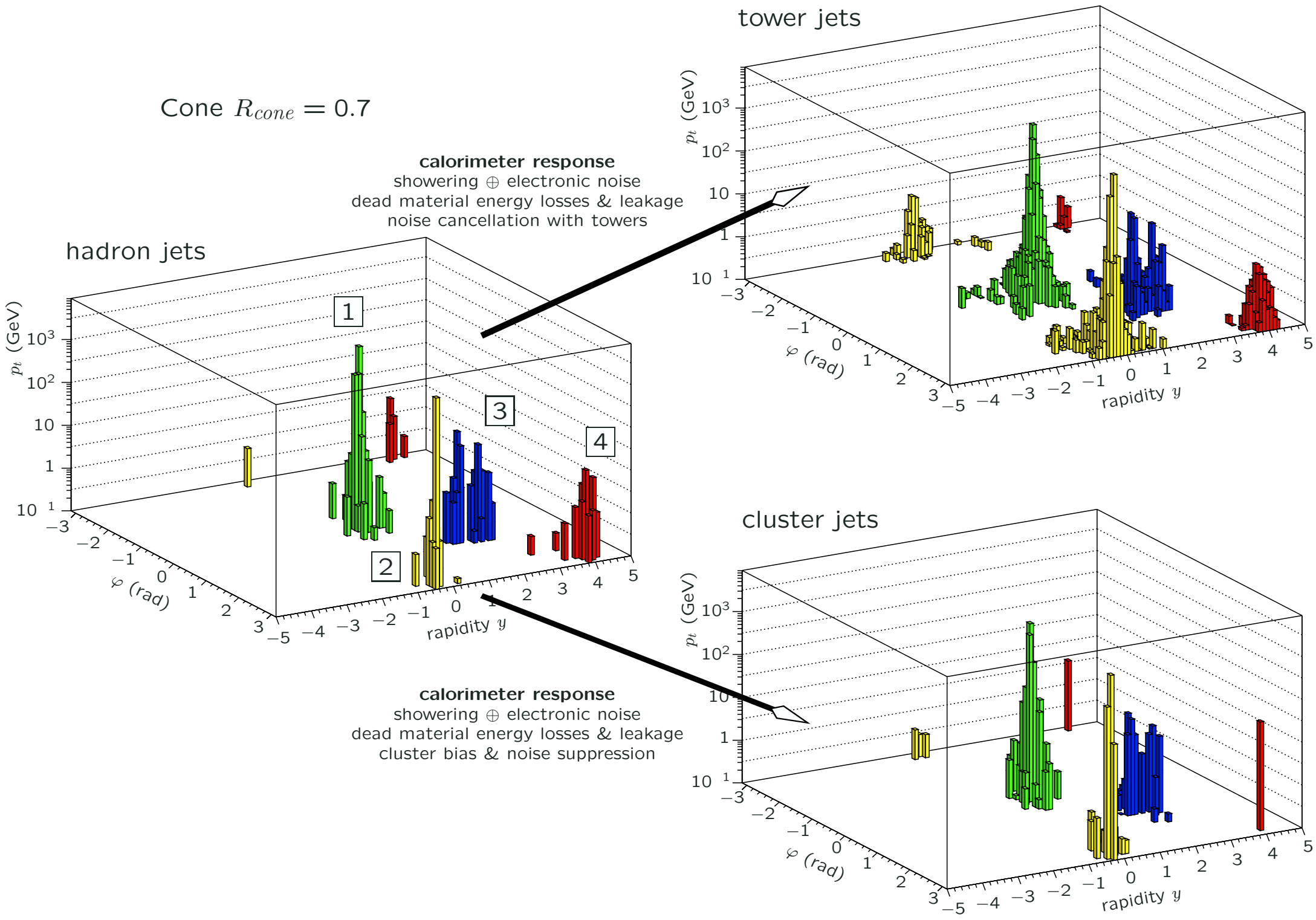

Figure 5: A simulated QCD dijet event with four jets in the final state, as seen at particle level and in the ATLAS calorimeters when using towers or clusters (extracted from Ref. [16]).

Another important difference between tower and cluster jets is the number of calorimeter cells used in the jet. While towers include all cells of the ATLAS calorimeters, topological clustering actually applies noise suppression due to the cell signal significance cuts used. This means that many fewer cells contribute to jets in case of clusters, and that the noise contribution per jet is also smaller for cluster jets than for tower jets, see Section 3.3.1 for further discussion.

Jet finding needs physical four-momenta on input. Thus, both towers and clusters are defined as massless pseudo-particles with a four-momentum $(E, \vec{p})$, reconstructed from the reconstructed energy $E$ (either electromagnetic or hadronic scale, see above), and the directions $\eta$ and $\phi$:

$$
\begin{aligned}
E &= |\vec{p}| = \sqrt{p_x^2 + p_y^2 + p_z^2} \\
p_x &= p \cdot \frac{\cos\phi}{\cosh\eta} \\
p_y &= p \cdot \frac{\sin\phi}{\cosh\eta} \\
p_z &= p \cdot \tanh\eta \,.
\end{aligned}
$$

The directions are fixed to the bin center in the $(\eta, \phi)$ grid for each tower, while they are reconstructed from the energy-weighted barycenter for topological clusters.

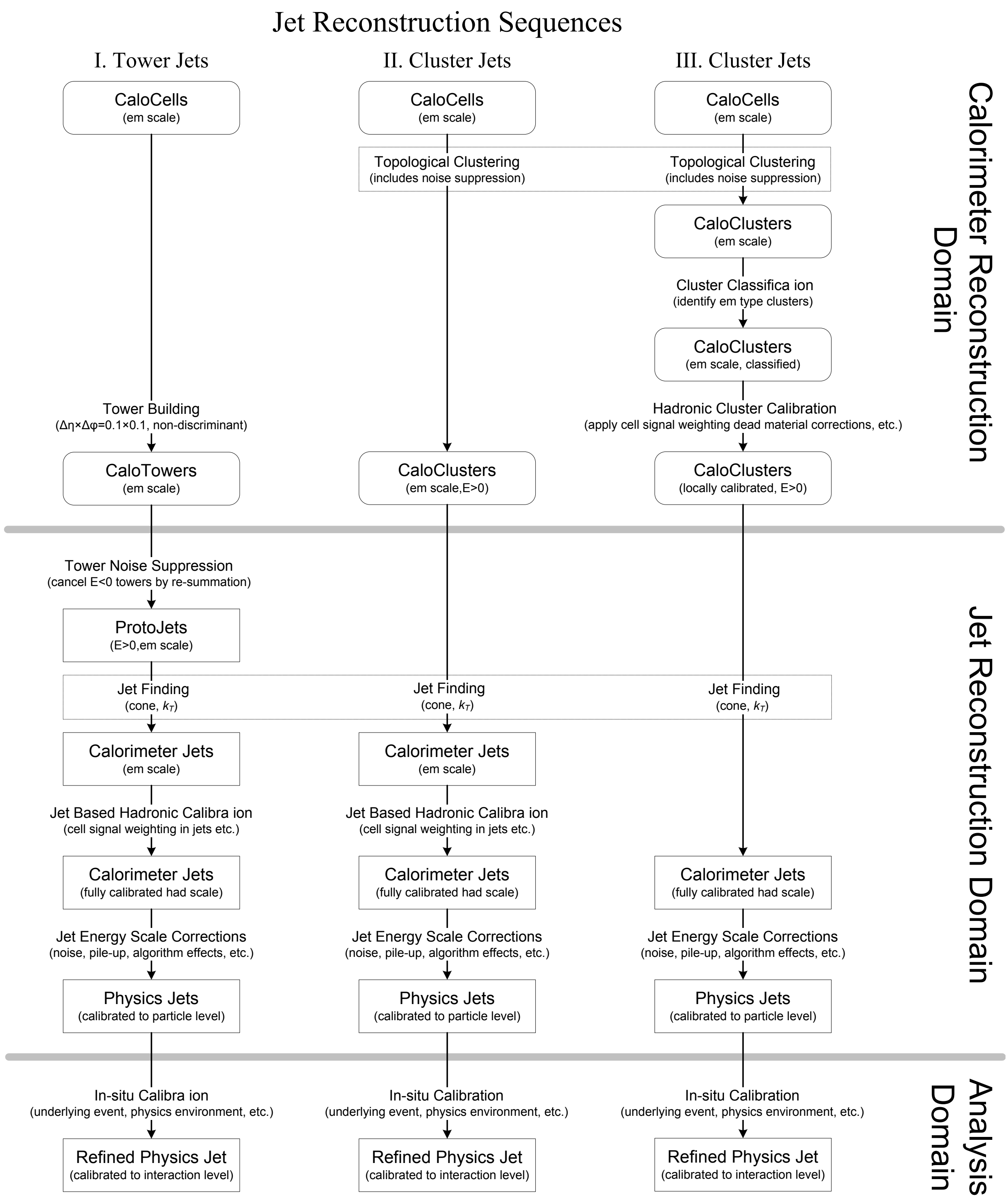

Figure 6: Schematic view on the reconstruction sequences for jets from calorimeter towers (left), uncalibrated (center) and calibrated (right) topological calorimeter cell clusters in ATLAS. The reconstruction (software) domains are also indicated.

#### 2.5.4 Reconstruction flow for calorimeter jets

The general reconstruction algorithm flows for calorimeter jets are summarized in Fig. 6. They reflect the differences between tower and cluster signals. As already discussed, tower signals are on the electromagnetic energy scale while topological clusters are either on this scale or are calibrated on a local hadronic energy scale. Also, tower signals do not include noise suppression, while topological clustering has noise suppression built in.

**Reconstruction sequence I: tower jets** Jet reconstruction from calorimeter towers (left diagram in Fig. 6) starts with a re-summation step, which addresses a possible unphysical four-momentum due to a negative (net) tower signal $E_{\text{tower}} < 0$. This can be generated by signal fluctuations from noise (electronics and physics from pile-up) in the cells entering into the corresponding towers. Simply ignoring the negative signal towers enhances the contribution of positive noise fluctuations, but combining negative signal towers with nearby positive signals such that the combined four-momentum is physical with $E_{\text{tower}} > 0$, leads to cancellations of some of the noise fluctuations and avoids signal biases. Only negative signal towers without nearby positive signals are completely dropped.

The resulting "protojets" represent either one or a few towers, and have all physically valid four-momenta. They are the input to the actual jet finding algorithm like seeded fixed cone or $k_T$. The outputs of the jet finder are then jets with energies on the electromagnetic energy scale. Their constituents are the original calorimeter towers. They are subjected to a cell signal based calibration discussed below in Section 2.5.5. After calibration, jets with $p_T < 7$ GeV are discarded.

More refined corrections are needed to calibrate the tower jets to the particle level. Those include corrections for residual non-linearities in the jet response due to algorithm effects, like missing energy from the jet, or adding energy not belonging to the jet, in the jet clustering procedure. Other corrections include suppression of signal contributions from the underlying event and/or pile-up. Most of these can only be addressed in the context of a specific physics analysis.

**Reconstruction sequence II: cluster jets** When topological clusters on electromagnetic energy scale are used for jet reconstruction, the reconstruction flow is rather similar to the tower jet reconstruction, see center diagram in Fig. 6. The main difference is the treatment of negative signals. Due to the symmetric noise cut applied in the cell selection in the clustering step, some clusters may have net negative signal as well. These can be ignored for jet reconstruction without significantly biasing the jet signal by positive noise contributions, because the noise suppression applied by the cell clustering already severely reduces any noise contribution. Some additional average cancellation is achieved by the symmetric noise cut, which allows inclusion of some negative cell signals even into positive (physical) clusters.

The cluster jets are initially on the electromagnetic energy scale as well. The same cell signal weighting functions used for tower jets are applied to initially calibrate these jets, with some additional corrections for the fact that these calibration functions have not been optimized for the cluster signals, see Section 2.5.5 for more details. Like in reconstruction sequence I, all jets with $p_T < 7$ GeV after calibration are discarded.

**Reconstruction sequence III: locally calibrated cluster jets** In this sequence the input objects to jet finding are already calibrated to the local hadronic energy scale [17]. This means that after the jets are formed, they are also calibrated on this scale. Additional corrections needed are related to the fact that all calibrations and corrections for this particular scale have been derived from single pion response only. Additional jet energy losses due to loss of particles in the magnetic field in ATLAS, or in inactive material without leaving any signal above clustering threshold in the calorimeters, have to be corrected in the jet context itself, in addition to the corrections for the physics environment contributions already discussed for the tower jets. See the right hand side diagram in Fig. 6 for a schematic overview. Fully calibrated jets from this sequence with $p_T < 7$ GeV are again discarded.

#### 2.5.5 Calorimeter jet calibration

The long standing calibration scheme for calorimeter jets in ATLAS is based on cell signal weighting. It can be applied to both tower and cluster jets from the reconstruction sequences I and II, respectively. The basic idea behind this approach, which was originally developed for the CDHS experiment [18] and further refined for the H1 experiment Ref. [19], is that low signal densities in calorimeter cells indicate a hadronic signal in a non-compensating calorimeter and thus need a signal weight for compensation of the order of the electron/pion signal ratio $e/\pi$, while high signal densities are more likely generated by electromagnetic showers and therefore do not need additional signal weighting.

To apply the cell signal weighting, first all calorimeter cell signals contributing to a jet are retrieved. This is possible even if the jets have not directly been reconstructed from these cells. The signal in each cell $i$ in the jet is weighted by a function depending on the cell location $\vec{X}_i$ and the cell signal density $\rho_i = E_i/V_i$, with $E_i$ being the electromagnetic energy signal of the cell, and $V_i$ being its volume. The weighting factor is $\approx 1$ for high density signals and rising up to 1.5, the typical $e/\pi$ for the ATLAS calorimeters, with decreasing cell signal densities. The weighting functions are universal in that they do not depend on any jet feature or variable. The calibrated jet four-momentum $(E_{\text{jet,calo}}, \vec{p}_{\text{reco}})$ is then recalculated from the weighted cell signals, which are treated as massless four-momenta $(E_i, \vec{p}_i)$ with fixed directions:

$$\left(E_{\text{jet,calo}}, \vec{p}_{\text{calo}}\right) = \left( \sum_i^{N_{\text{cells}}} w(\rho_i, \vec{X}_i) E_i, \sum_i^{N_{\text{cells}}} w(\rho_i, \vec{X}_i) \vec{p}_i \right) . \quad (1)$$

The signal weighting functions have been determined using seeded fixed size cone jets ($R_{\text{cone}} = 0.7$) in fully simulated QCD dijet events by fits of reconstructed calorimeter tower jet energies to matching Monte Carlo truth particle jet energies. Residual non-linearities (as function of $p_T$) and non-uniformities (as function of $\eta$) are corrected by an additional calibration function parametrized in both variables. These corrections have also been calculated for other standard jet finding configurations and calorimeter signals. The calibration has been determined using the ideal, non-distorted detector geometries. See Ref. [17] for more details.

### 2.6 Track jets

Finding jets in reconstructed inner detector tracks is useful to recover possible inefficiencies of the calorimeter signals, especially for jets pointing to transition or crack regions. Even though the scheme of seeding calorimeter jets with track jets in these regions has not been completely evaluated, some recovery may be possible. At least tagging of suspicious event topologies using track jets without matching calorimeter jet, or matching a poorly reconstructed calorimeter jet, allows a suppression of events with significant fake missing transverse momentum in the hadronic final state reconstruction, thus improving the quality of any given sample for physics analysis.

Matching the track jets with calorimeter jets is another promising approach to refine the jet energy measurement jet by jet, and the hadronic final state reconstruction in general. First, the $p_T$ fraction carried by tracks, defined as

$$f_{\text{trk}} = \frac{p_{\text{T,track}}}{p_{\text{T,calo}}} , \quad (2)$$

with $p_{\text{T,track}}$ being the transverse momentum from tracks and $p_{\text{T,calo}}$ being the one reconstructed by the calorimeter, can be measured for each jet within the inner detector acceptance $|\eta| < 2.5$. It can then be used to (relatively) improve the jet energy measurement, as indicated in Fig. 7. The figure shows results obtained from simulations of QCD dijet processes of jets within $|\eta| < 0.7$ and $40 < p_T < 60$ GeV. Even though the calorimeter jet calibration performs well on average, there are residual dependencies of the individual jet signal reconstruction quality on the jet fragmentation, i.e. the particle composition of the jet, for which $f_{trk}$ provides a handle for a relative correction which can be applied jet by jet. This fragmentation dependency can be understood in that jets with large $f_{trk}$ have a larger amount of their energy carried by charged hadrons, which tend to generate a smaller signal in the non-compensating ATLAS calorimeters. The standard jet calibration based on the calorimeter cell signals alone cannot completely recover the corresponding signal loss. This is certainly one of the promising techniques for a relative jet energy scale correction in the context of a refined jet calibration, with the immediate goal of improving the relative jet energy resolution.

Jet finders usually cluster four-momentum in two dimensions, like azimuth $\phi$ and pseudorapidity $\eta$. Inner detector tracks in ATLAS have a reconstructed vertex associated, thus the $z_{vtx}$ coordinate can be added as a third dimension to jet finding from these tracks, see Fig. 8. Jet clustering in $\phi$, $\eta$, and $z_{vtx}$ then allows the assignment of a vertex to a matching calorimeter jet, which is especially of interest in events with multiple interactions from pile-up. In this case jets not associated with the primary vertex can be tagged as such, and removed from the final state in jet counting experiments, e.g. a $W + n$ jets analysis.

### 2.7 Energy flow jets

In Section 2.6 some aspects of the use of jets from charged tracks reconstructed with the inner detector are discussed. Another approach to improve the jet energy measurement is to match the calorimeter response in towers or clusters with a charged track pointing to it, and use the track kinematics if the track momentum resolution is better than the calorimeter energy resolution for the matched cluster, which for hadrons in ATLAS is the case up to $p_T \approx 140$ GeV at $\eta = 0$. The principal method is referred to as *energy flow* reconstruction, and was pioneered at LEP [20] and is in use in hadron colliders; for example, the CDF application is described in Ref. [21].

The most important feature of energy flow reconstruction is the removal of the calorimeter signal generated by an accepted track. In ATLAS this has been studied using a track-cluster match approach. Figure 9 shows the relative variation of the $p_T$ resolution for cone jets from energy flow objects and topological clusters in QCD dijet production, both without final energy scale corrections. From this there are indications that for jets with $p_T \lesssim 80$ GeV in this environment, which is characterized by typically low event activity in general, the jet energy resolution can be improved by the energy flow technique. The expected gain is most significant at lower jet $p_T$, e.g. about 15% (relative) at $p_T = 40$ GeV.

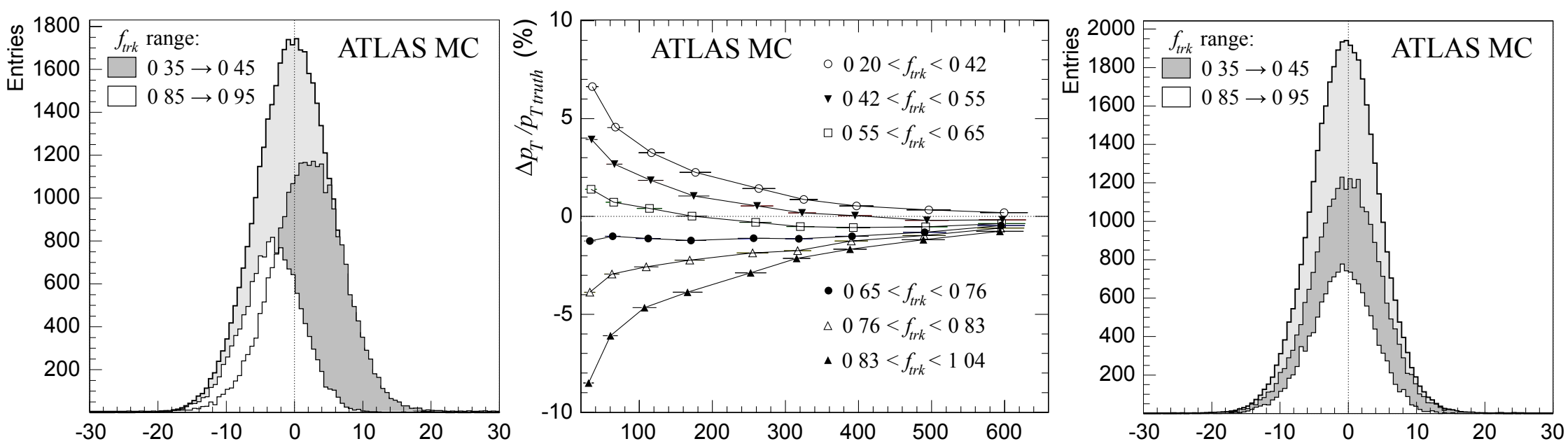

Figure 7: The difference between jet $p_T$ reconstructed from the calorimeter and the matching truth particle jet $p_T$, for all jets with $|\eta| < 0.7$ and $40 < p_T < 60$ GeV, and jets from this sample in two regions of the track $p_T$ fraction $f_{trk}$ defined in Eq.(2) (left). The plot in the center shows the relative difference $\Delta p_T/p_{T,truth}$ for various bins of $f_{trk}$ and as function of the calorimeter jet $p_T$. The distributions of $\Delta p_T$ after applying these corrections are shown in the right figure. The lightly shaded area indicates the distribution for all jets, without any selection based on $f_{trk}$.

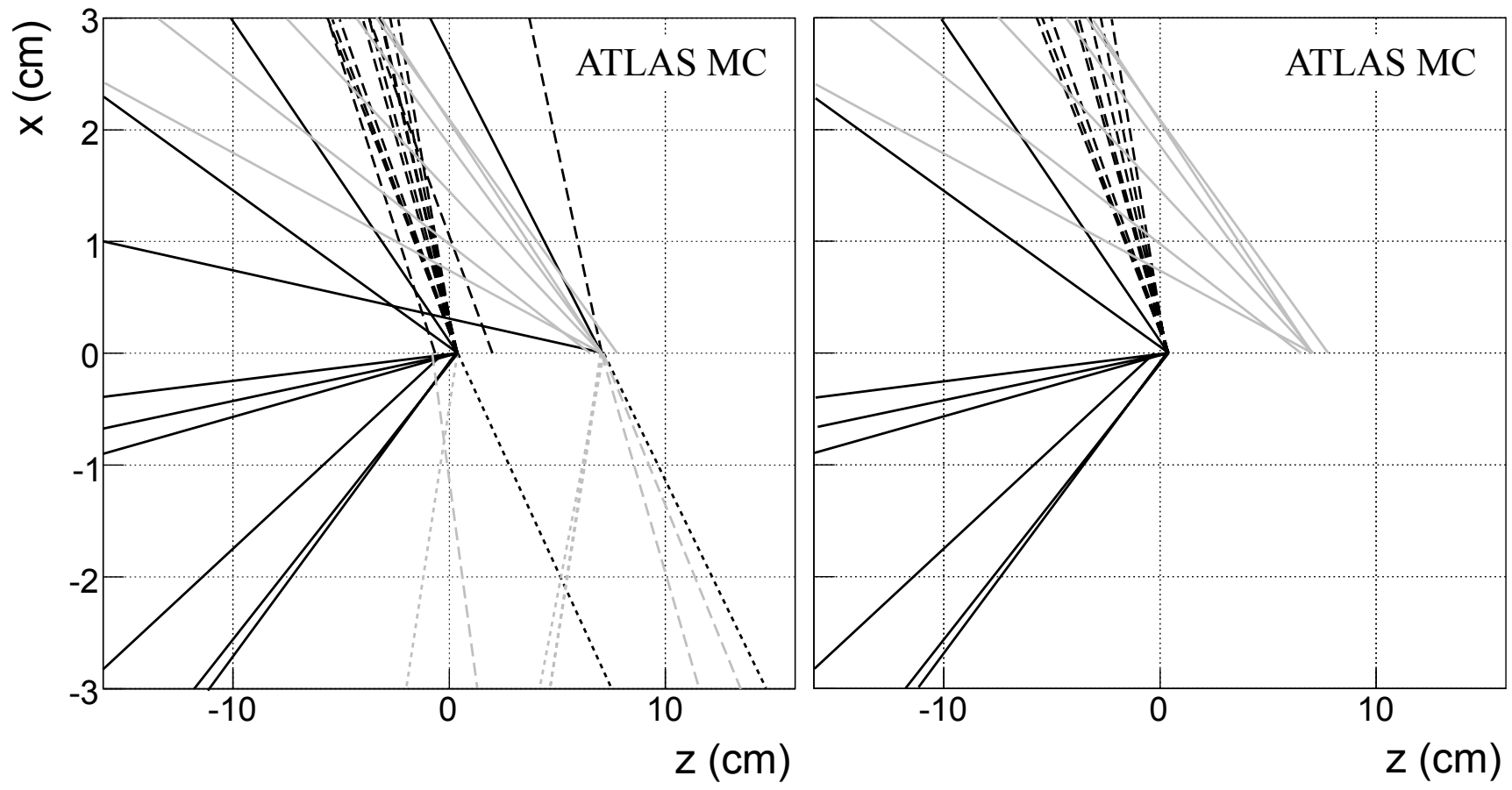

Figure 8: Track jets in the horizontal $(x,z)$ plane of a simulated event with multiple vertices. The left figure shows tracks assigned to cone jets with $p_T > 5$ GeV using the usual two-dimensional clustering algorithm in $\eta$ and $\phi$, with $R_{cone} = 0.4$ (same line patterns and shades indicate same jet). Clearly tracks from different vertices can end up in the same jet. The right figure shows cone jets from three-dimensional clustering for the same event, explicitly including the track vertices ($R_{cone} = 0.4$, $\Delta z = 10$mm).

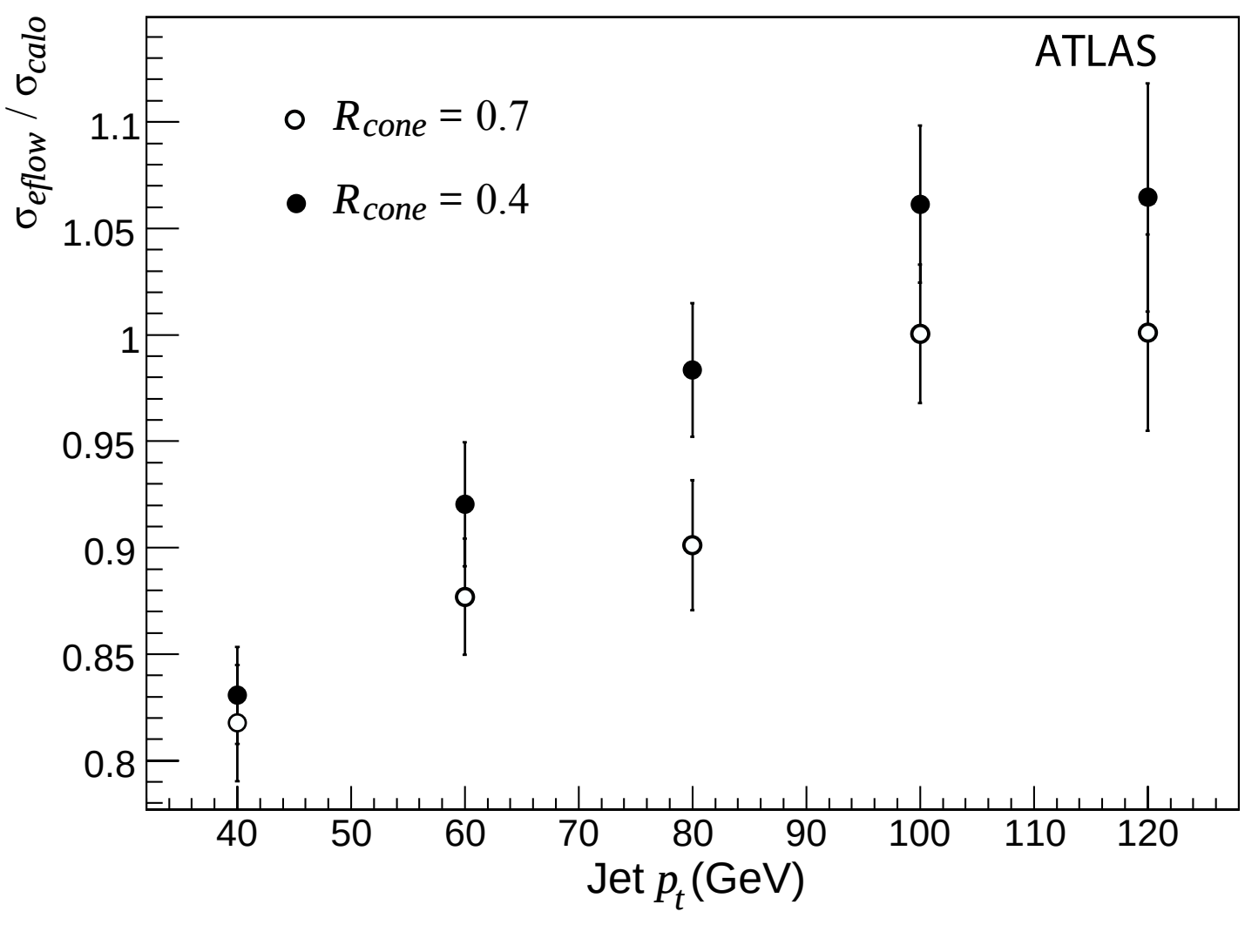

Figure 9: The ratio of the relative $p_T$ resolutions for energy flow and cluster cone jets $\sigma_{\text{eflow}}/\sigma_{\text{calo}}$, as determined with fully simulated dijet events, as function of the jet $p_T$. Results for both narrow ($R_{cone} = 0.4$) and wide ($R_{cone} = 0.7$) jets within $|\eta| < 1.8$ are shown.

### 2.8 Particle jets

Particle jets are only available in simulated events. They are built from stable particles produced by the fragmentation model in the physics generator. Stable particles in ATLAS are those with a laboratory frame lifetime of about 10 picoseconds or more, thus typically including electrons, muons, photons, charged pions, kaons, protons, neutrons, neutrinos, and their corresponding antiparticles. These particles represent the "truth" reference of a hard scattering process for performance studies and simulation based calibration approaches. The particle jets are therefore referred to as *truth particle jets*, or *truth jets*, in the following sections of this note.

Neutrinos and muons generated in the collision are excluded from these truth jets, as they have their own observables, i.e. missing transverse momentum for neutrinos and explicitly reconstructed tracks for muons. Jet finding with generated particles uses the same code as calorimeter reconstruction, obviously excluding the signal preparation for towers in reconstruction sequence I and all calibration steps.

## 3 Jet reconstruction performance in ATLAS

The performance of the ATLAS detector for jet reconstruction has been evaluated within the present day limitations of physics generators, mostly PYTHIA [22], and detector response simulations, all performed with GEANT4 [23, 24]. All results presented here should therefore be considered to be expectations and of preliminary nature.

### 3.1 Preliminaries

The comparisons of calibrated jet features discussed in this section uses calorimeter jets to which the cell signal based calibration has been applied, in addition to the overall scale correction for different jet finder configurations and calorimeter signal choices, i.e. jets from reconstruction sequence I and II introduced in Section 2.5.4, with calibrations applied as described in Section 2.5.5. As already discussed in that section, the parameters of the corresponding calibration functions have been determined using simulations of dijets from QCD processes with the ideal detector geometry, meaning no misalignment between detectors or detector elements, no detector shape distortions, and assuming a perfect knowledge of the material distribution from supports, services, cryostats, etc., in the complex ATLAS geometry.

The simulations performed for the evaluation of the jet reconstruction performance use the same generated physics, meaning the same events at particle level, but include small shifts in relative detector positioning and small changes to the amount (increased) and location (more realistic asymmetric distribution in pseudorapidity and azimuth) of dead material in the detector description. As a consequence, the calorimeters do not respond optimally for this evaluation with respect to jet signals, but the estimated performance is likely closer to the initial one in ATLAS, with some a priori unknown distortions, misalignments, and other imperfections.

Final corrections can be derived once experimental data from the detector become available with sufficient statistics by e.g. following the strategies lined out in Ref. [25]. For the performance expectations presented here these corrections, which could have been derived rather straight forwardly by using the misaligned detector geometry in the simulations for calibration as well, are intentionally not applied to probe some of the initial systematic uncertainties to be expected for the initial running of ATLAS. Note that none of the results presented here include pile-up.

### 3.2 Comparing jet reconstruction algorithm performances

Jet reconstruction performance is typically expressed in terms of expected or measured signal linearity, i.e. flatness of the detector response to particle jets over the whole kinematic range of interest at the LHC (from $p_{\mathrm{T}} \approx 10$ GeV to a few TeV), signal uniformity in pseudorapidity $\eta$ and azimuth $\phi$ over the whole detector system coverage, and the achievable energy resolution. Additional features of jet reconstruction performance are the efficiency to find jets, and the purity of the found jet sample, which is of course related to the number of fake jets reconstructed.

Other jet features studied with the ATLAS detector are the possibility to reconstruct the original source of a given jet. This is particularly interesting for heavily boosted heavy particle decays like for top quarks, where the final state will likely be reconstructed as just one (narrow) jet. The measurement of the jet mass and a substructure analysis are experimental tools which could address this question.

The quality of the reconstructed variables depends on the choice of the calorimeter signal (towers or clusters), the choice of the jet finder and its configuration (wide/narrow jets), and the ability to unfold as much as possible the physics environment, like reflected in underlying event and pile-up contributions. A high precision analysis of a given event topology may require different configurations than offered in default jet reconstruction to optimize the signal. In this section the expected effect of the choices discussed above on the performance variables is shown for selected configurations.

In most cases the performance is evaluated using a truth reference provided in the simulation by jets reconstructed at particle level, see Section 2.8. The detector and truth jets are associated through directional matching. The following list of variables is used:

**Jet directions** are given by pseudorapidity $\eta$ and azimuth $\phi$. As $\eta$ is directly related to the polar angle $\theta$, and thus useful (even for massive jets) to understand the variations of the detector response, the rapidity $y$ can be used for physics analysis motivated selections. The variables are defined as:

$$\eta = -\frac{1}{2}\ln\left(\frac{p+p_z}{p-p_z}\right) = -\ln\left[\tan\left(\frac{\theta}{2}\right)\right] \tag{3}$$

$$y = -\frac{1}{2}\ln\left(\frac{E+p_z}{E-p_z}\right) \tag{4}$$

$$\phi = \arctan\left(\frac{p_y}{p_x}\right). \tag{5}$$

**Matching radius** $R_m$ is defined by the directional distance between truth and detector jets, i.e.

$$R_m = \sqrt{\Delta\eta^2 + \Delta\phi^2}. \tag{6}$$

Two jets are matched if the radial distance between them is $R_m \leq 0.2$, if not stated otherwise. Only one match is allowed for each reference (truth) jet. In case of two or more nearby jets, the one closest to the reference is taken for efficiency and purity studies. For signal linearity and uniformity studies the reference and the calorimeter jets are omitted in this case.

**Signal linearity** can be determined by the ratio $\lambda$ of the energy reconstructed for the calorimeter jet $E_{\mathrm{jet,calo}}$ and the matched truth jet energy $E_{\mathrm{jet,truth}}$,

$$\lambda = \frac{E_{\mathrm{jet,calo}}}{E_{\mathrm{jet,truth}}}, \tag{7}$$

when $\lambda$ can be calculated as function of $E_{\mathrm{jet,truth}}$ or $E_{\mathrm{jet,calo}}$.

**Signal uniformity** is measured by the variation of the signal as function of the jet direction in the detector frame, as given by $\eta$ in Eq.(3). The variable $\lambda$ defined in Eq.(7) can be used to estimate the uniformity from simulations, if calculated as function of $\eta$.

**Relative energy resolution** is given by the width of the distribution of the relative difference between $E_{\mathrm{jet,calo}}$ and $E_{\mathrm{jet,truth}}$:

$$\frac{\sigma}{E} = \sqrt{\left\langle \left( \frac{E_{\mathrm{jet,calo}} - E_{\mathrm{jet,truth}}}{E_{\mathrm{jet,truth}}} \right)^2 \right\rangle - \left\langle \frac{E_{\mathrm{jet,calo}} - E_{\mathrm{jet,truth}}}{E_{\mathrm{jet,truth}}} \right\rangle^2} . \tag{8}$$

**Jet reconstruction efficiency** $\varepsilon$ is defined by the following ratio:

$$\varepsilon(R_m) = \frac{\text{\# matches of truth particle jets with reconstructed jets}}{\text{\# truth particle jets}} = \frac{N_m^{\mathrm{jets}}(R_m)}{N_{\mathrm{truth}}^{\mathrm{jets}}} . \tag{9}$$

It depends on the matching radius $R_m$ and is determined as function of the true jet energy, $p_\mathrm{T}$, or $\eta$.

**Jet reconstruction purity** $\pi$ is given by

$$\pi(R_m) = \frac{\text{\# matches of truth particle jets with reconstructed jets}}{\text{\# reconstructed jets}} = \frac{N_m^{\mathrm{jets}}(R_m)}{N_{reco}^{\mathrm{jets}}} \tag{10}$$

and also depends on the choice for $R_m$. The fake rate $f(R_m)$ is then given by $f(R_m) = 1 - \pi(R_m)$. Purity is calculated as function of the reconstructed jet energy, $p_\mathrm{T}$, and/or direction.

Additional variables reconstructed from jets, like substructure and shape measures, are discussed in Section 3.6.

## 3.3 Comparisons of basic jet signal features

The main data source for the evaluation of the performance of the ATLAS calorimeters for jet reconstruction are fully simulated QCD dijet events with at least one of the hard scattered partons having a $p_\mathrm{T}$ within a certain bin. Eight bins are defined, with incrementing delimiters approximately following a $2^n$ power law: $17 \rightarrow 35$ GeV, $35 \rightarrow 70$ GeV, etc., up to the final bin $p_\mathrm{T} > 2240$ GeV with the upper limit set by the kinematic limit introduced by the parton direction and the 14 TeV center-of-mass energy in the $pp$ collisions at LHC.

### 3.3.1 Signal linearity and resolution

Signal linearity has been studied in detail for the evaluation of the various calorimeter jet calibration schemes under discussion in ATLAS [17]. The general expectation from these studies with the distorted detector is a response flat within $\pm 1\%$ for jets with $p_\mathrm{T} \gtrsim 50$ GeV, and a slightly larger deviation from linearity for jets with lower transverse momentum down to $\sim 10$ GeV [17]. As the calibration functions are determined using one jet finder configuration and one simulated calibration sample in a perfect detector model (see discussion in Section 2.5.5), one can estimate the shift from a response flat within the margins introduced by the distorted detector when other configurations and/or a different calorimeter signal basis are considered. This shift can be expressed as the ratio of $\lambda_{alt}$, as given in Eq.(7), from a given alternative jet reconstruction to $\lambda_{ref}$, the reference from the same jet reconstruction configuration used to derive calibration functions in the ideal detector:

$$\xi = \frac{\lambda_{\mathrm{alt}}}{\lambda_{\mathrm{ref}}} = \frac{E_{\mathrm{jet,calo}}^{\mathrm{alt}} \Big/ E_{\mathrm{jet,truth}}^{\mathrm{alt}}}{E_{\mathrm{jet,calo}}^{\mathrm{ref}} \Big/ E_{\mathrm{jet,truth}}^{\mathrm{ref}}} . \tag{11}$$

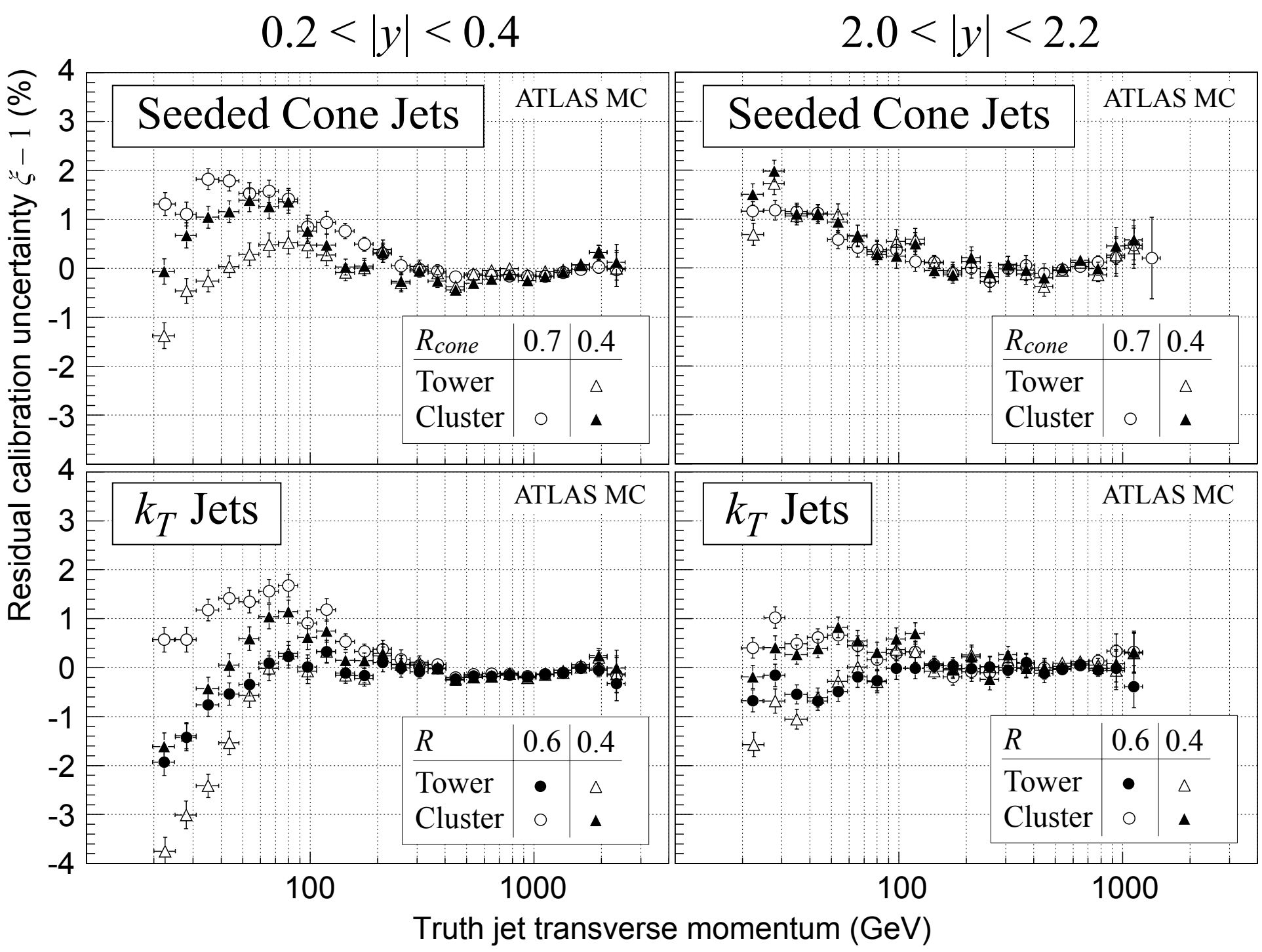

Figure 10: Residual calibration uncertainties for reconstructed jets from various jet finder configurations and the two calorimeter signals, as function of the truth jet $p_T$ and in two bins of jet rapidity $y$. The reference configuration is seeded fixed cone tower jets with $R_{\text{cone}} = 0.7$. The residual calibration uncertainty is given by $\xi$, as defined in Eq.(11).

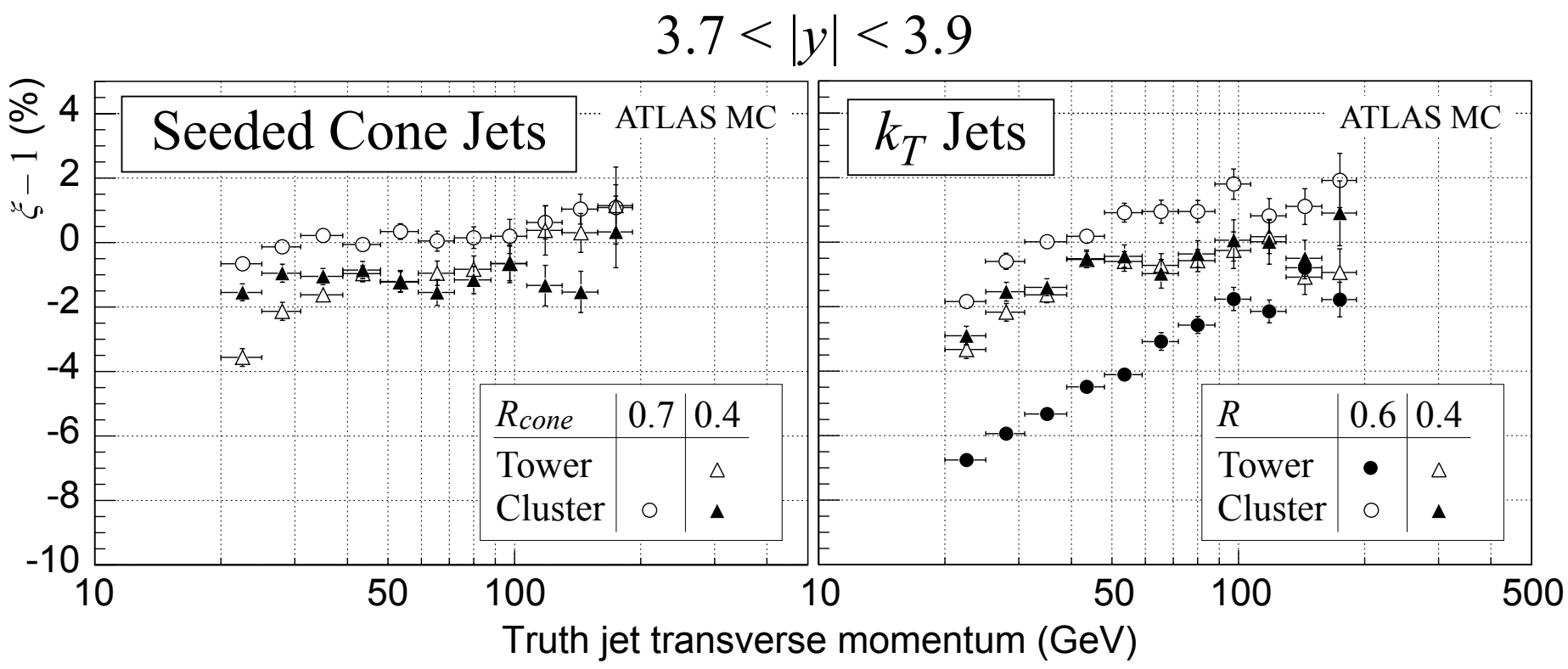

Figure 11: Residual calibration uncertainty $\xi$, calculated relative to the fixed seeded cone ($R_{\text{cone}} = 0.7$) tower jet calibration as described by Eq.(11), as function of the truth jet $p_T$ in the forward direction of ATLAS.

The quantity $\xi$ can be viewed as a measure of the residual calibration uncertainty with respect to the best calibrated jet reconstruction configuration, and is thus an estimate of one of the systematic uncertainty contributions in the general jet reconstruction. Note that

$$\xi \approx \frac{E^{\rm alt}_{\rm jet,calo}}{E^{\rm ref}_{\rm jet,calo}}$$

when the alternative reconstruction uses the same jet finder with the same parameters as the reference, because $E^{\rm alt}_{\rm jet,truth} \approx E^{\rm ref}_{\rm jet,truth}$ in this case. The prime example here is the comparison of seeded fixed cone tower jets with $R_{\rm cone} = 0.7$ (the reference) with cluster jets found with the same configuration.

Figure 10 shows expectations for $\xi$ as function of the jet $p_T$ in two different regions of jet rapidity. From this simulation based study it can be concluded that the cell signal based jet calibration, with the additional overall scale corrections discussed in Section 2.5.5 applied, is universal for the distorted detector at a level of about 2% for the studied QCD jets with $p_T > 20$ GeV and the particular choice of distortions implemented in the detector description of the simulation program. As these distortions include additional inactive material between the electromagnetic liquid argon and the hadronic tile calorimeter (about 10% increase in nuclear absorption length), the effect is in particular emphasized for low $p_T$ jets in the central region, which mostly occupy the $0.4 < |y| < 0.4$ bin[2]. Here the sensitivity to the jet algorithm (seeded cone or $k_T$), its configuration (narrow or wide jets), and the choice of calorimeter signals (clusters or tower) is also largest, see Fig. 10.

Sensitivities to signal and jet finder choices are stronger for jets within $3.7 < |y| < 3.9$. The calorimeter jet shape in the corresponding forward region is dominated by (lateral) hadronic shower extension rather than the particle flow in a cone in $(\eta, \phi)$, meaning that a considerable part of the calorimeter signal can be outside of a chosen jet cone and therefore be lost for the total jet signal. Jets from each of the jet finder configurations and calorimeter signal choices have been individually corrected for these signal losses using simulations with the ideal detector geometry. The effect of the distorted detector, which includes a change of the relative $z$ position of the endcap and forward calorimeters in ATLAS and thus a change of the aspect ratio of the particle level jet shape to the calorimeter jet shape and consequently different energy losses, can be significant especially for the $k_T$ jets, as indicated in Fig. 11. Again using the seeded fixed-size cone tower jets with $R_{\rm cone} = 0.7$ as a reference, the cluster jets reconstructed with the same algorithm show a similar response, indicated by $|\xi - 1| < 2\%$ in the whole kinematic range studied here. The narrow cone cluster jets lose about 1% of their energy, rather independent of their $p_T$, which can be expected due to the lateral hadronic shower size varying only slightly with energy. The larger effects of the distorted detector on the $k_T$ jets, like the prediction of a nearly linear rise of $\xi$ with $\log p_T$ for $p_T \lesssim 100$ GeV for all considered $k_T$ jets, but most pronounced for wide $k_T$ tower jets, require more investigation of the distortion effects and their reflection in towers and clusters on the more complex dynamics driving the $k_T$ algorithm.

The jet energy resolution is the other important contribution to precision jet reconstruction. It has been evaluated in the distorted detector geometry with the same QCD sample. The results are again discussed in more detail in Ref. [17]. A typical relative energy resolution, calculated as described by Eq.(8), and achieved without particular corrections for the distorted detector, has a stochastic term of about $60\%/\sqrt{E(\text{ GeV})}$ and a high energy limit of about 3% in the central region of ATLAS, all determined with these simulated events.

It is expected that the relative energy resolution depends on the calorimeter signal choice, the jet algorithm, the underlying event and pile-up activity, and the general particle density and flow of the physics environment. Using QCD dijet simulations, with electronics noise included in the detector simulation but without any pile-up activity, the difference in resolution between tower and cluster jets can be estimated with the test variable $\psi_\sigma$, which uses the fractional difference $\Delta_\sigma$ in the energy resolution

$$\Delta_\sigma = \left(\frac{\sigma}{E}\right)^2_{\text{cluster}} - \left(\frac{\sigma}{E}\right)^2_{\text{tower}},$$

with the relative resolutions $\sigma/E$ as given in Eq.(8). $\psi_\sigma$ can then be defined as

$$\psi_\sigma = \begin{cases} \sqrt{\Delta_\sigma} & \text{for} \quad \Delta_\sigma > 0 \\ -\sqrt{-\Delta_\sigma} & \text{for} \quad \Delta_\sigma < 0 \end{cases} . \tag{12}$$

[2] Note that for the QCD jets under consideration here the jet mass $m$ is generally small, in particular $m \ll p$, i.e. $y \approx \eta$.

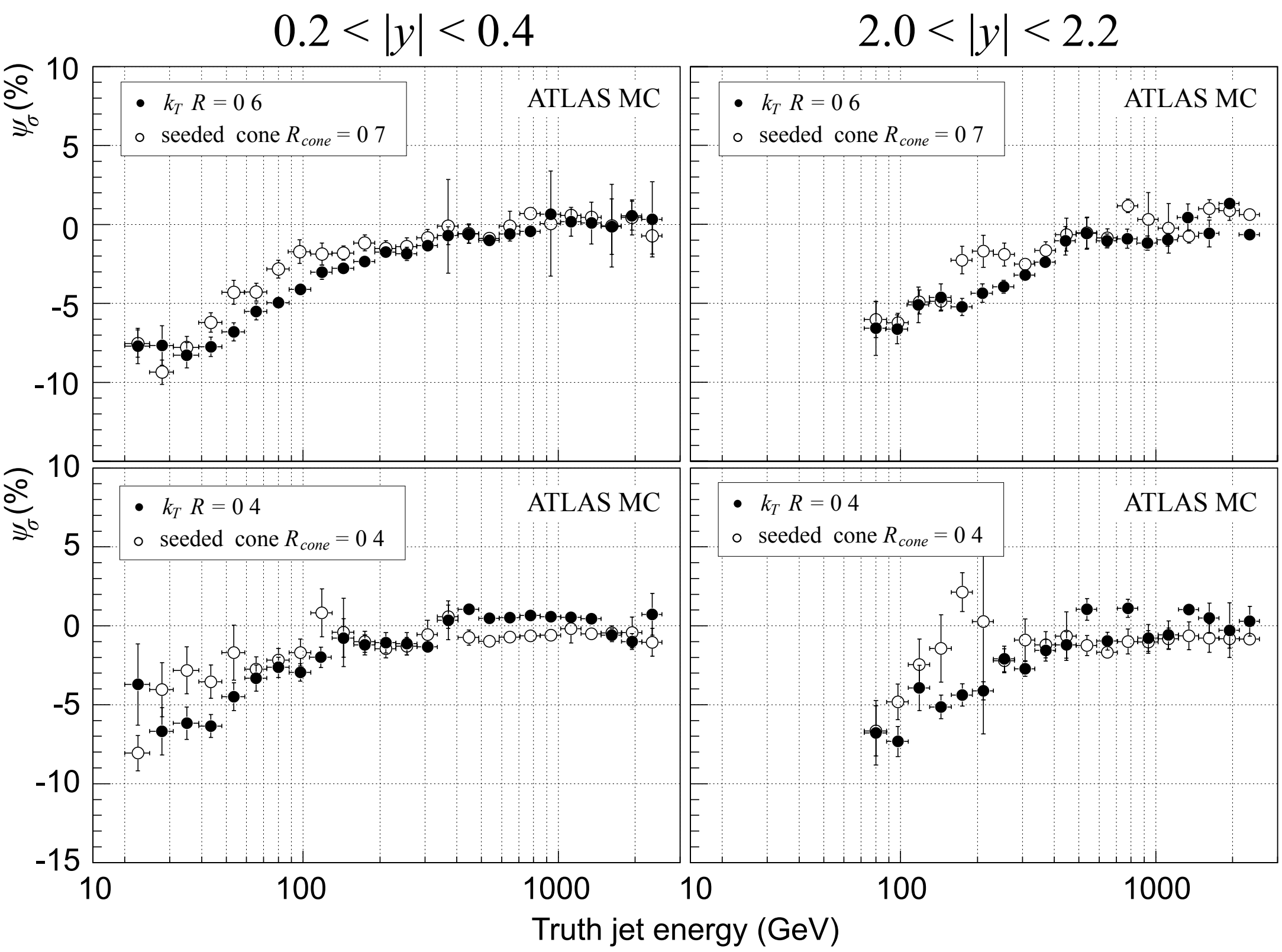

Figure 12: The difference in relative energy resolution $\psi_\sigma$ (see Eq.(12)) between seeded cone cluster and tower jets, and $k_T$ cluster and tower jets, respectively, as function of the matching particle jet energy, and in two different regions of jet rapidity $y$. A negative value for $\psi_\sigma$ indicates a better resolution for cluster jets.

Figure 12 shows the prediction for $\psi_\sigma$ for various jet configurations in two different kinematic regimes defined by the jet rapidity. As this variable mainly folds differences in the stochastic and noise contribution to the jet energy resolution, the effect of noise suppression implicit for cluster jets is particularly visible at low energies and for wider jets, where more calorimeter cells with noise contribute to the tower jets than in narrow jets. At higher jet energies, the energy resolution contribution introduced by signal fluctuations from electronics noise is significantly reduced and $\psi_\sigma$ is comparable with zero in both kinematic regimes.

The relative difference $\psi_\sigma$ can be transformed into an energy-equivalent difference in a straight forward manner:

$$\Delta\sigma_{\text{abs}}(E) = \psi_\sigma(E) \cdot E . \tag{13}$$

Figure 13 shows predictions for this difference for the QCD jets discussed here. The general flat (energy independent) behaviour indicates that the differences between cluster and tower jets indeed are mostly related to electronics noise. The average noise contribution $\langle \Delta\sigma_{\rm abs} \rangle$ has been determined by straight line fits to the flat regions. Seeded cone tower jets with $R_{\rm cone} = 0.7$ in the lower rapidity regime $0.2 < |y| < 0.4$ yield $|\langle \Delta\sigma_{\rm abs} \rangle| \approx 2.5$ GeV more noise than the corresponding cluster jets. This contribution is estimated to be higher for $k_T$ jets with $R = 0.6$, at approximately 4 GeV. $|\langle \Delta\sigma_{\rm abs} \rangle|$ increases in the higher rapidity region $2.0 < |y| < 2.2$ to about 6 GeV for cone and 8 GeV for $k_T$ jets, respectively. This observation already reflects the larger electronics noise in the endcap calorimeters, where most of the jets in this rapidity range go. The evaluation of $\psi_\sigma$ as function of the jet direction, as shown in Fig. 15 in Section 3.4, explores this observation further.

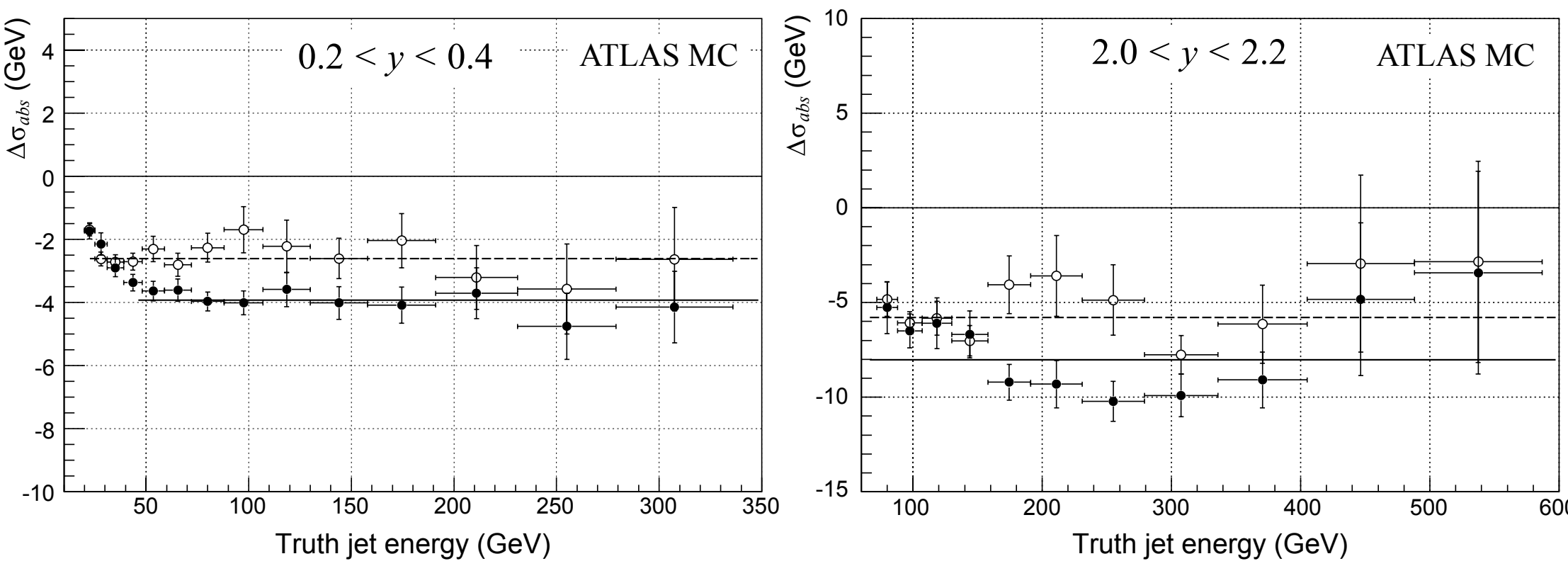

Figure 13: Estimates for the absolute difference in resolution, as defined in Eq.(13) between seeded cone cluster and tower jets ($R_{\rm cone} = 0.7$, open circles ∘). The filled circles (•) show the same difference for $k_T$ cluster and tower jets, with $R = 0.6$. The solid and dashed lines show results and the range of fits to the data points. Note the linear energy scale and the limited energy range, compared to Fig. 12.

Figure 13 also suggests that for $E < 50$ GeV, $k_T$ jets made from clusters in the lower rapidity regime are subjected to fluctuations from other sources than electronics noise, which partly cancel the noise reduction as $|\Delta\sigma_{\rm abs}|$ gets smaller with decreasing energy. These fluctuations are not present in $k_T$ jets made from towers and in seeded cone jets reconstructed from towers or clusters in the same kinematic region, and need further exploration.

### 3.4 Jet signal uniformity

The variation of the jet response as function of the jet direction is a measure of the uniformity of the jet signals and more powerful to diagnose possible detector inefficiencies than the linearity studies discussed in the previous section. Figure 14 shows the variation of the response to wide and narrow seeded cone and $k_T$ tower jets as function of the jet pseudorapidity[3] for simulated QCD dijet events. The additional inactive material introduced between the electromagnetic and hadronic calorimeters in the central region, which is part of the distorted detector geometry model, especially affects the response to the lower energy jets. The drop in the jet response is also more enhanced for lower energetic jets in the crack region around $|\eta| \approx 1.5$, with up to 10% of signal loss, again due to more inactive material in this region. The response to the higher energetic jets is more uniform, as expected, with a signal drop in the cracks around 4%, which is also observed for the crack at $|\eta| \approx 3.2$.

[3]Contrary to the preceding discussion of the signal linearity and energy resolution, where the rapidity $y$ is used to categorize different kinematic regimes for jets, in the discussion of the signal uniformity the pseudorapidity $\eta$ is preferred, as it directly reflects a direction in ATLAS (see Eq.(3) in Section 3.2) and thus is more appropriate to characterize the response in a certain calorimeter region.

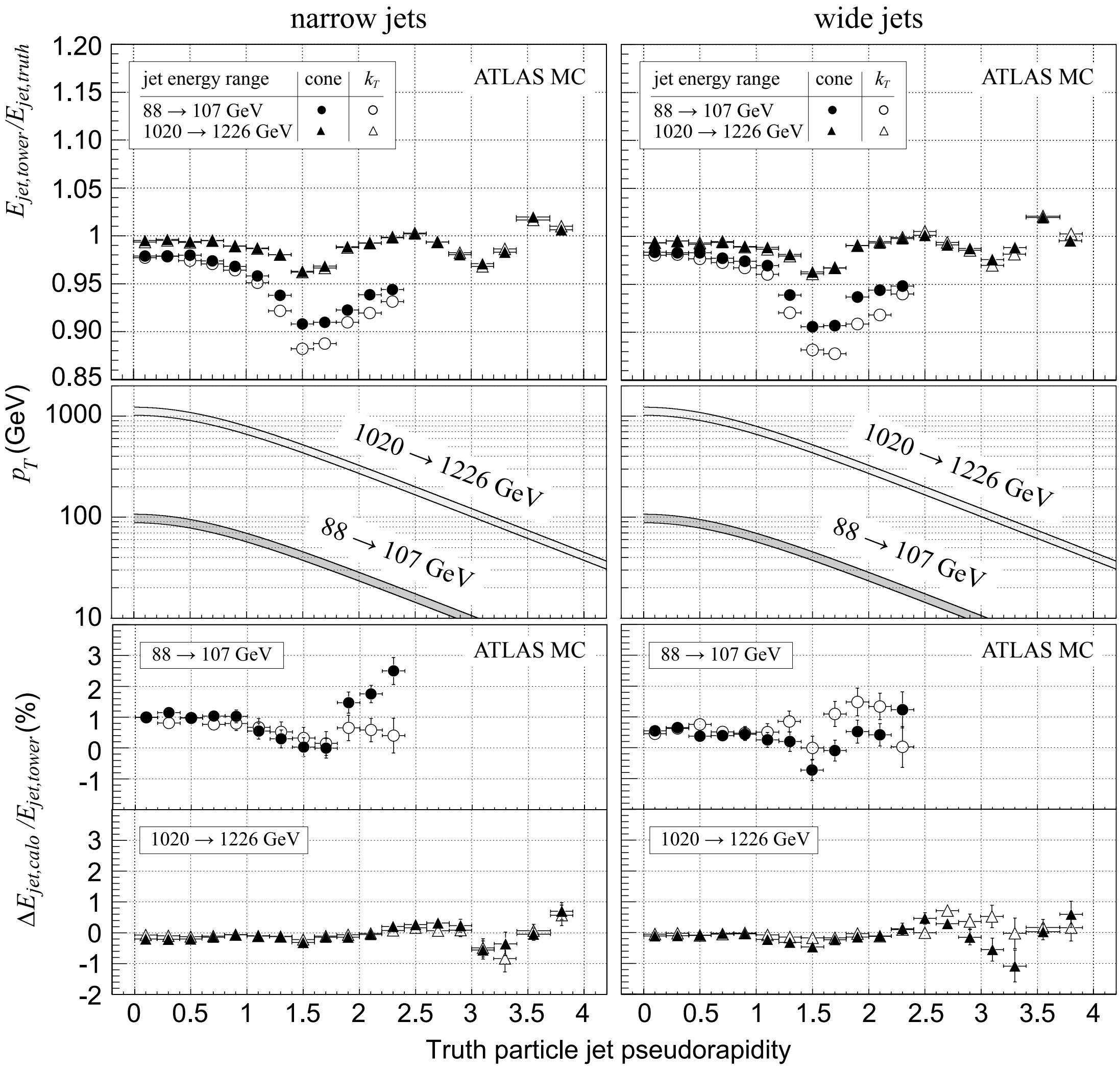

Figure 14: The uniformity of the ATLAS calorimeter response for wide ($R_{\mathrm{cone}} = 0.7$) and narrow ($R_{\mathrm{cone}} = 0.4$) seeded fixed sized cone jets, and for wide ($R = 0.6$) and narrow ($R = 0.4$) $k_{\mathrm{T}}$ jets. The upper two plots show the response variation as function of the jet pseudorapidity $\eta$ for tower jets in two different jet energy ranges, while the lower plots show the signal difference between cluster and tower jets $\Delta E_{\mathrm{jet,calo}} = E_{\mathrm{jet,cluster}} - E_{\mathrm{jet,tower}}$ as function of $\eta$, again in two jet energy bins. The $p_{\mathrm{T}}$ variation for the two energy bins is shown in the middle section, for comparisons.

There are only relatively small difference between the seeded cone and $k_T$ jet signal uniformity, which are more pronounced at lower jet energies. Using clusters instead of towers as the calorimeter signal basis does not affect the signal uniformity for the higher energetic jets, but is expected to recover some signal losses, of the order of 1%, for the lower energetic jets.

Predictions for the dependency of the relative energy resolution on the jet direction is shown in Fig. 15, for the same QCD jet sample. From this study there are significant differences expected between the seeded cone and the $k_T$ algorithm for lower jet energies. Both for wide and narrow jets the cone algorithm seems to perform better, especially in the endcap calorimeter region. For higher energetic jets these differences become insignificant.

Testing the difference between cluster and tower jet energy resolution has been done using the relative resolution difference $\psi_\sigma$, as defined in Eq.(12). For the lower energetic jet sample the noise-like contribution to $\psi_\sigma$ already observed in Fig. 12 is confirmed, with the additional observation that in the endcap calorimeters clusters seem to reduce these fluctuations even more than in the central region. High energetic cluster and tower jets perform very similarly. From the absolute differences between the resolutions shown in Fig. 13, which is about $2.5 - 8$ GeV depending on the kinematic regime and the chosen jet finder and its configuration, the expected value for $\psi_\sigma$ for higher jet energies is about $0.2 - 0.7\%$, which is in agreement with the predictions for the high energy jet sample shown in Fig. 15. The larger and positive $\psi_\sigma$ at higher pseudorapidities observed especially for narrow $k_T$ jets indicates signal fluctu-ations introduced by clustering in the forward region. A more detailed investigation of the $k_T$ algorithm performance when configured for narrow jets in this region is under way.

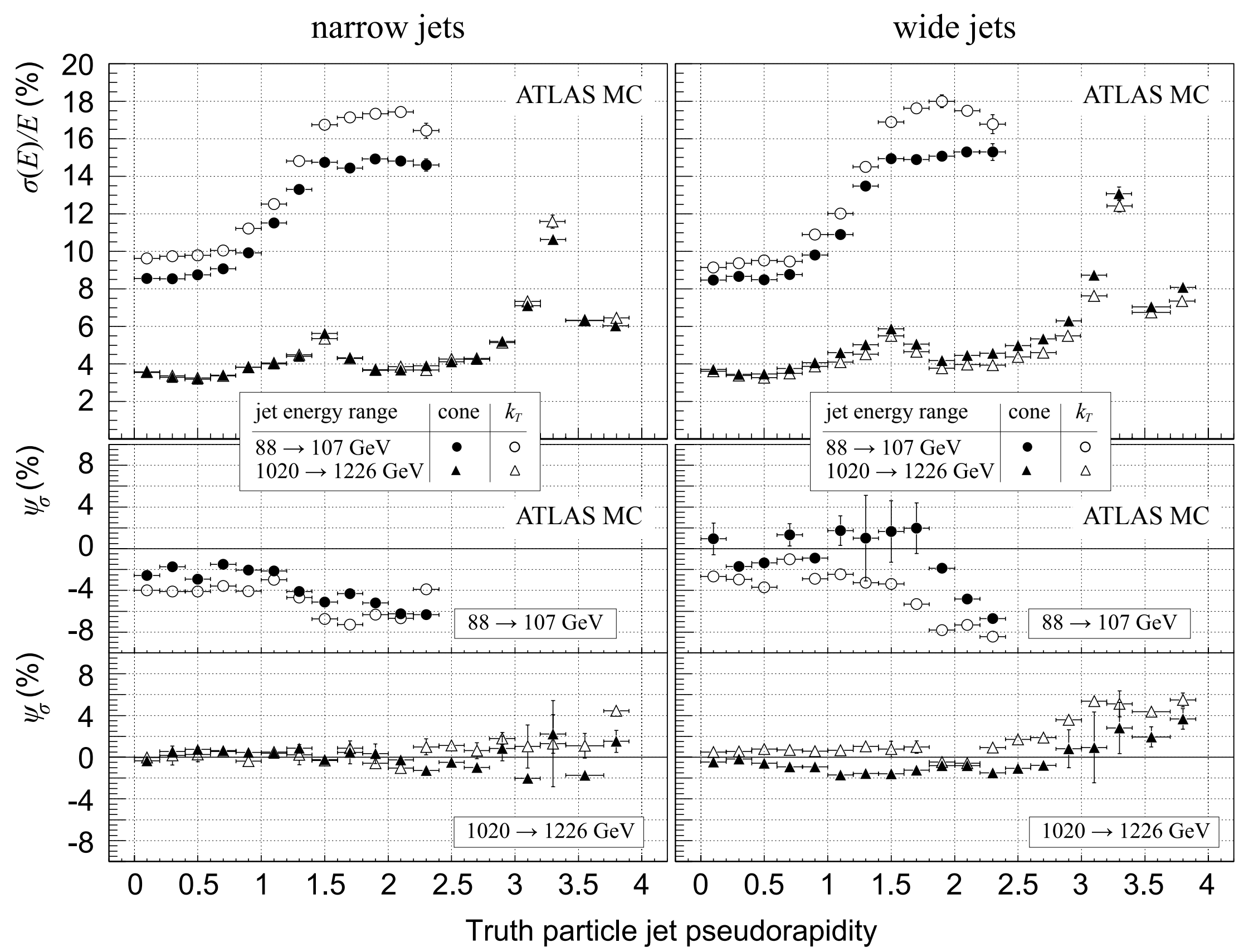

Figure 15: Predictions for the direction dependence of the relative jet energy resolution for seeded cone (wide, $R_{\mathrm{cone}} = 0.7$, left, and narrow, $R_{\mathrm{cone}} = 0.4$, right) and wide ($R = 0.6$) and narrow ($R = 0.4$) $k_T$ tower jets in two different jet energy bins (upper plots). The relative resolution difference to cluster jets $\psi_\sigma$, as defined in Eq.(12), is shown for the same jet samples in the bottom plots.

### 3.5 Efficiency and purity

The efficiency of the jet reconstruction in ATLAS depends not only on the jet $p_T$ but also on the jet direction and the physics environment. For basic performance evaluation it has been studied using QCD dijet production. Figures 16, 17, and 18 show the expectations for seeded cone jets with $R_{cone} = 0.7$ and 0.4, and $k_T$ jets with $R = 0.4$ and 0.6, respectively, in various kinematic regimes defined by the jet rapidity $y$. It is noticeable that the efficiency at lower $p_T$ is rather low, with the wide $k_T$ jets performing better than the corresponding wide cone jets. Both jet algorithms have similar efficiencies for narrow jets for $0.2 < |y| < 0.4$, with increasingly better efficiencies predicted for $k_T$ jets at higher rapidities.

Figures 16 through 18 also show the estimated differences in the jet reconstruction efficiency between tower and cluster jets, defined as $\Delta\varepsilon = \varepsilon_{cluster} - \varepsilon_{tower}$, with $\varepsilon_{cluster}$ and $\varepsilon_{tower}$ defined in Eq.(9). In general cluster jets are expected to be reconstructed with higher efficiency in the low $p_T$ regime, especially in case of wide jets and in the forward direction. Narrow seeded cone jets are reconstructed with basically the same efficiency for both calorimeter signal choices, except at $p_T \lesssim 50$ GeV in the forward region, where cluster jets are again more efficient.

The predictions for $\Delta\varepsilon$ for wide $k_T$ indicate similar behaviour as the seeded cone jets in the central calorimeter region, but a much less pronounced difference between the tower and cluster jets in the forward region. There are also indications the narrow $k_T$ jet reconstruction efficiency is slightly improved when using clusters, see again Fig. 16 through Fig. 18. No significant variation of the jet reconstruction efficiency as function of the jet direction is expected, especially not for higher $p_T$ jets.

The fake rate $f = 1 - \pi$ (see also Eq.(10)) for jet reconstruction in QCD dijet events has also been studied. The rate drops quickly from more than 50% at jet $p_T$ of 10 GeV to around 1% for jets with $p_T > 50$ GeV. There are indications that wide jets reconstructed from topological clusters have a lower number of fakes at low $p_T$, especially in the central region. Also, the $k_T$ algorithm seems to generate fewer wide jets than the cone algorithm in this region, independent of the calorimeter signal choice. These differences are expected to be much smaller for narrow jets.

### 3.6 Jet composition and mass

The reconstruction of jet masses and substructure has gained interest at LHC, especially because of the expected production of heavy particles with masses of $\mathcal{O}(100)$ GeV and transverse momenta of several hundred GeV. These may decay hadronically into very collimated final states. An important standard model example is the $W$ boson [26]. Other examples include SUSY [27], exited heavy quarks [28], and in exotic final states involving extra dimensions [29].

In any case, the complete final state of a heavy and boosted particle may be reconstructed into one jet, which in the example of a fully hadronically decaying top quark actually contains three highly collimated jets. The mass of the reconstructed jet is then one of the indicators of its origin, in addition to a possible substructure reconstruction reflecting the three “internal” jets.

The reconstruction of the jet mass is inherently difficult from calorimeter signals, as the shower development washes out the directions and energies of individual particles in the jet. In addition, the true jet mass is of course best reconstructed if all of its original particles can be measured at high precision. The mass reconstruction is thus disturbed by undetected particles, like the ones curling in the solenoidal magnetic field in the inner detector cavity of ATLAS, and the ones losing too much energy in upstream inactive material to generate a signal above threshold in the calorimeters.

Figure 19 shows the expected composition of $k_T$ jets with $R = 0.6$ built from particles, towers, and clusters in three different rapidity regions. The number of constituents shown in this figure is of course depending on the calorimeter signal choice, i.e. towers or clusters. The variation in the number of clusters between the rapidity regions reflects the changing spatial resolution power of the calorimeter with respect to resolving individual showers inside a jet. The observation that the number of particles inside jets seems to drop at highest $p_T$ indicates a change in the origin of the jets. Most jets at lower $p_T$ in the studied QCD sample are gluon jets, while for higher $p_T$ a significant fraction of jets is produced by quarks. As gluons have a larger probability to radiate off other gluons than quarks have, one can expect more particles inside gluon jets, and even more jets in the final state in case of gluons.

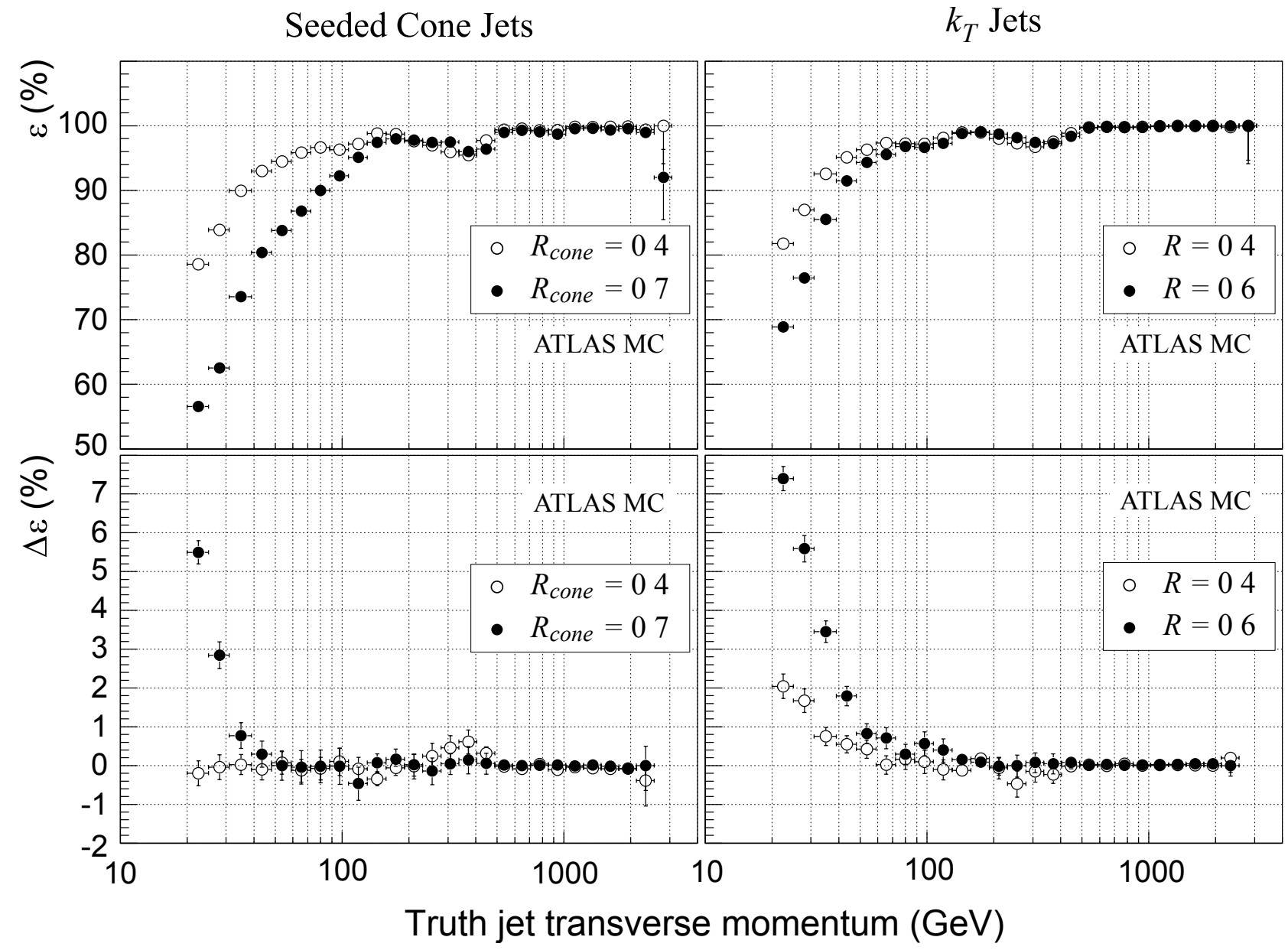

Figure 16: The calorimeter jet reconstruction efficiency $\varepsilon$ in ATLAS, calculated from simulations using Eq.(9), for seeded cone tower jets with $R_{\text{cone}} = 0.7$ and $R_{\text{cone}} = 0.4$, as function of the truth particle jet $p_T$, in $0.2 < |y| < 0.4$ (top left). The plot on the top right shows predictions for wide ($R = 0.6$) and narrow ($R = 0.4$) $k_T$ tower jet reconstruction efficiencies in the same kinematic regime. The difference in efficiency between cluster and tower jets, defined as $\Delta\varepsilon = \varepsilon_{\text{cluster}} - \varepsilon_{\text{tower}}$, is shown in the lower left plot for cone and in the lower right plot for $k_T$ jets.

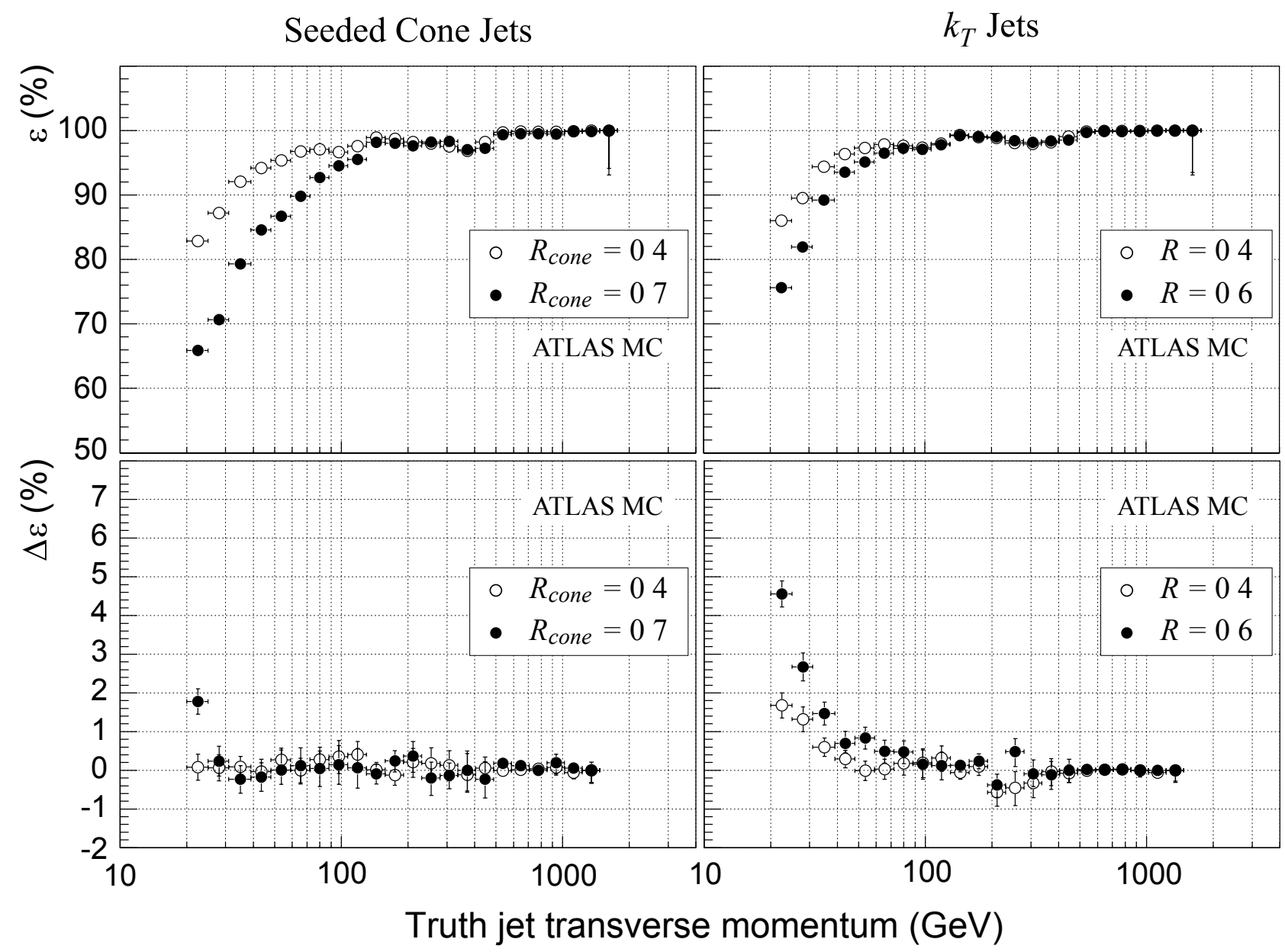

Figure 17: Predictions for the calorimeter jet reconstruction efficiency in ATLAS for wide and narrow seeded cone (top left) and $k_T$ (top right) tower jets with $2.0 < |y| < 2.2$, shown together with corresponding differences in efficiency between cluster and tower jets (bottom plots), again as function of the truth particle jet $p_T$ (see caption of Fig. 16 for more details).

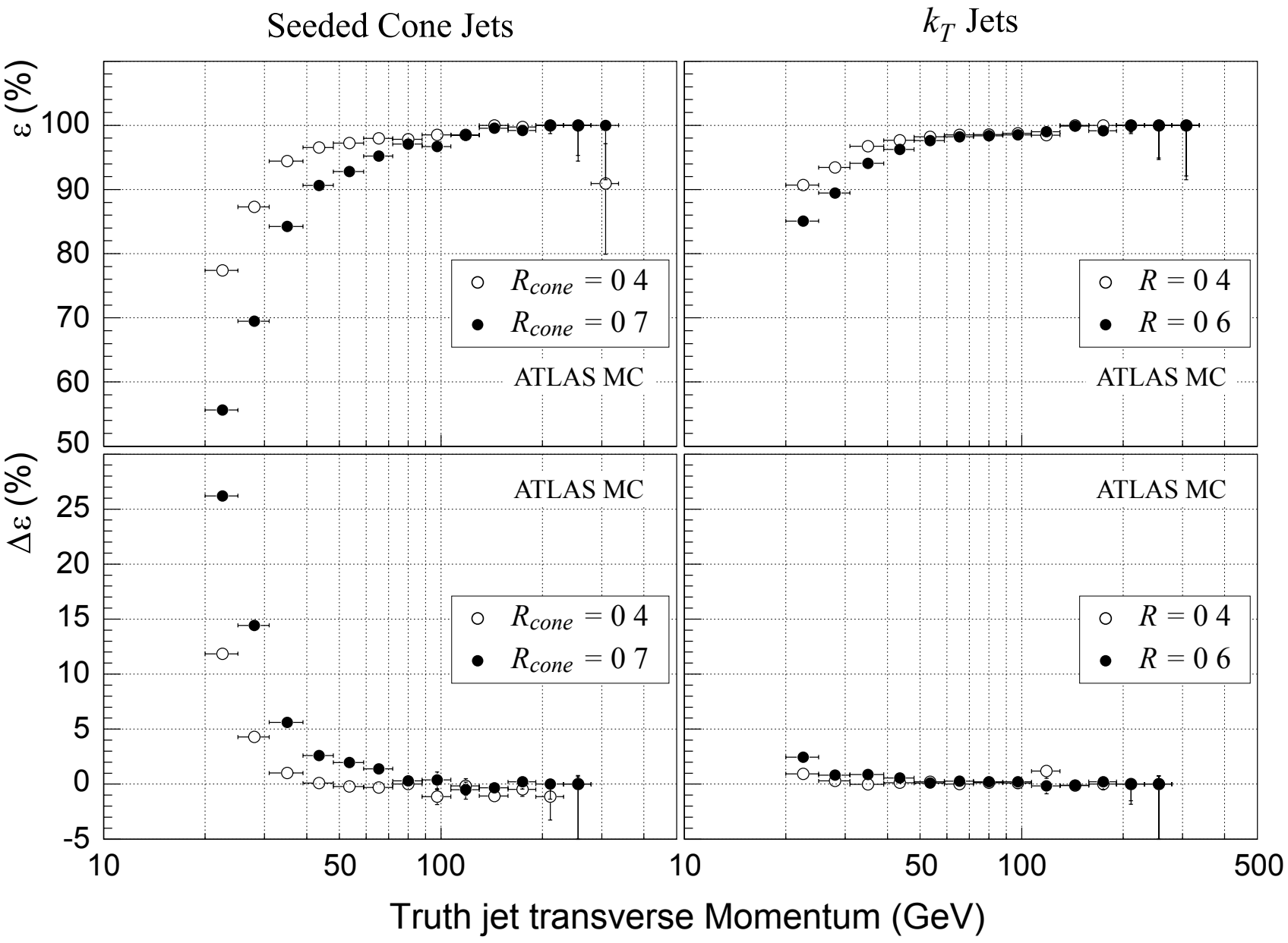

Figure 18: Jet reconstruction efficiencies $\varepsilon$ as function of the truth particle jet $p_T$, for wide and narrow seeded cone and $k_T$ tower jets in ATLAS (top), estimated with simulations of QCD dijet processes, in the jet rapidity range $3.7 < |y| < 3.9$. The corresponding difference in efficiency between cluster and tower jets is shown in the bottom plots. See caption of Fig. 16 for more details.

The relative sensitivity of the jet mass reconstruction to low signal contributions has been studied for cluster, tower and particle jets with QCD dijet simulations. The relative change in the jet mass if $p_T$ thresholds are applied to the constituents, is defined as

$$\Delta m_{\rm rel}(p_T^{\rm min}) = \frac{m(p_T > p_T^{\rm min})}{m(p_T > 0)} . \tag{14}$$

Here $m(p_T > p_T^{\rm min})$ means the jet mass recalculated using only jet constituents with a transverse momentum above $p_T^{\rm min}$, while $m(p_T > 0)$ is the jet mass using all constituents. Figure 20 shows the expectations for the variation of $\Delta m_{\rm rel}$ as a function of the reconstructed $m(p_T > 0)$, for different $p_T^{\rm min}$. Especially for high mass jets the cluster jets show a similar behaviour as the particle jets, indicating that clusters are probably better reflecting the particle composition in these jets than the tower jets. Note that $p_T > 400$ MeV is around the threshold for charged particles to reach the calorimeter in the magnetic field of the inner detector, and is also close to the general signal threshold for the calorimeters in the central region of ATLAS.

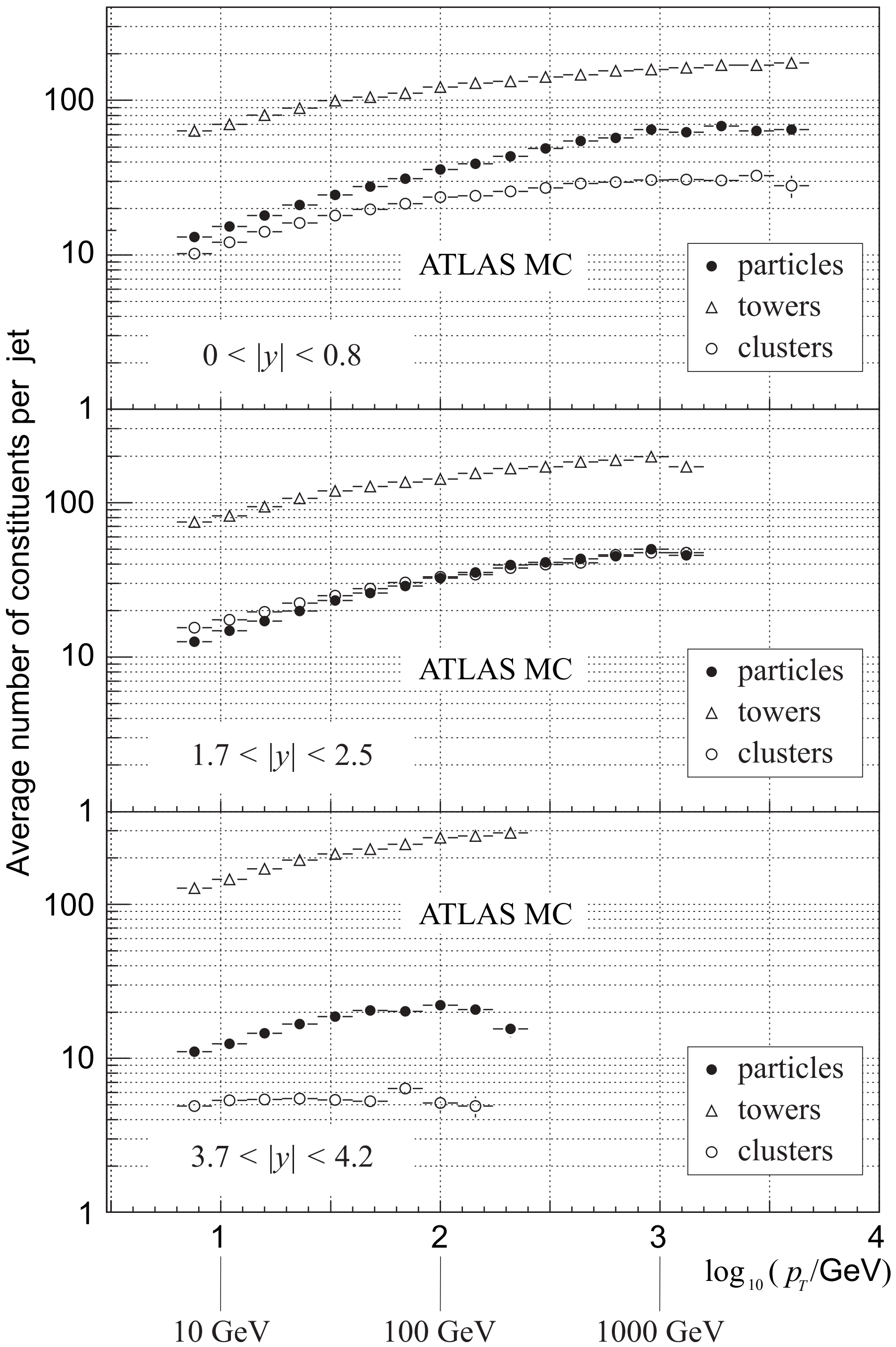

Figure 19: The average number of constituents of $k_T$ jets with $R = 0.6$ as function of the jet $p_T$, in three different regions of ATLAS, as predicted by simulations of QCD dijet events (figure adapted from Ref. [16]).

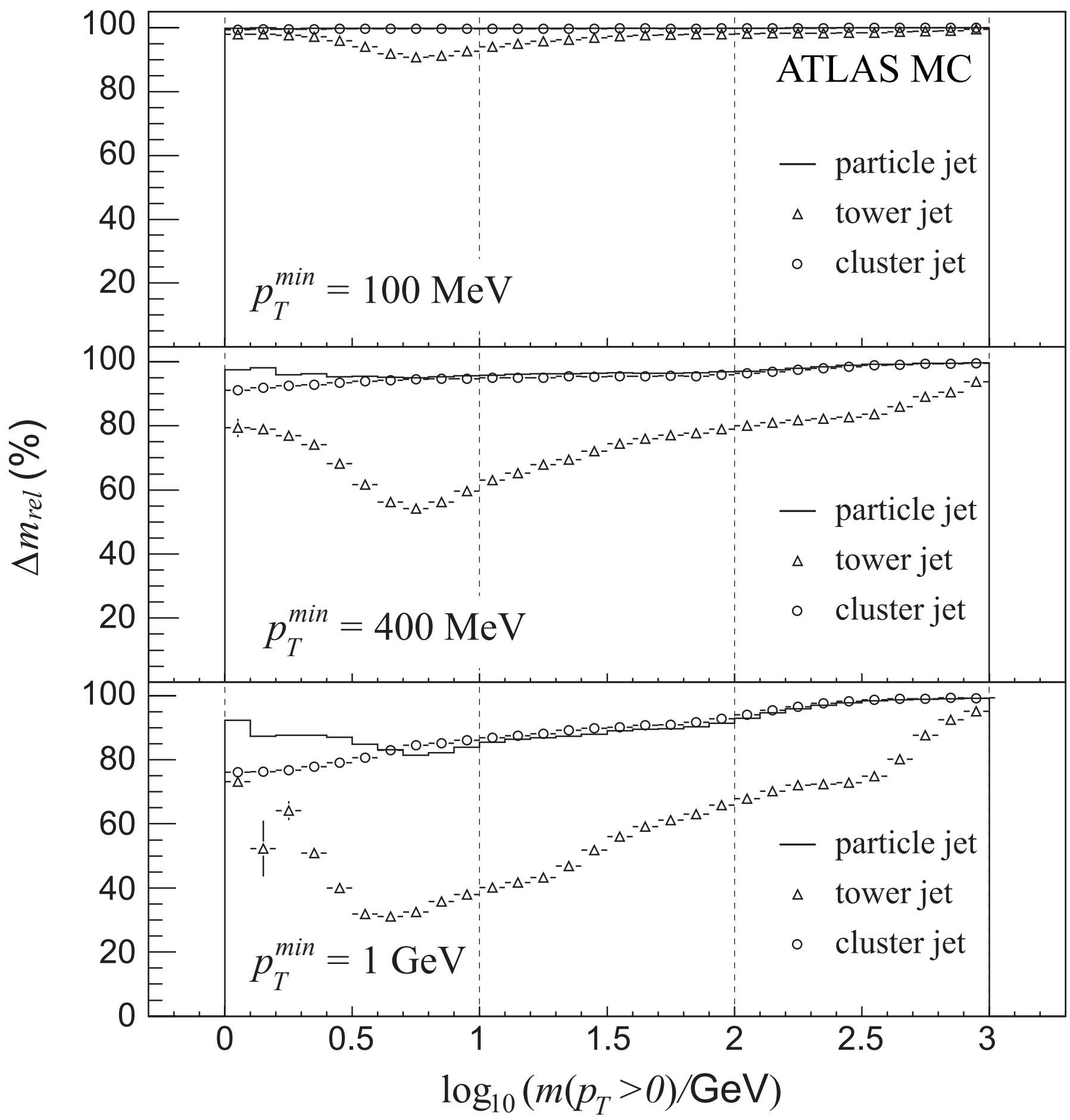

Figure 20: The variation of the jet mass re-reconstructed from its constituents with different levels of biases in $p_{\mathrm{T}}$, as function of the least-biased mass using all constituents, for simulated tower, cluster, and particle level jets in QCD dijet events in ATLAS (figure adapted from Ref. [16]).

### 3.7 Jet substructure

The sensitivity in the mass reconstruction observed in Fig. 20 suggests the need for jet variables less sensitive to the soft particle contribution and response, yet providing sensitivity to the possible origin of the jet. One of the interesting variables is the characteristic scale $y_{\text{scale}}$ for sub-jet splitting in $k_T$ jets, i.e. the $p_T$ scale at which the $n$ last recombinations of the $k_T$ algorithm are undone. It can be defined using $y_n$ such that

$$y_{\text{scale}}^2 = p_T^2 \times y_n \,.$$

For example, $n = 2$ means splitting into two sub-jets, identical to undoing the last recombination in the $k_T$ algorithm. $y_{\text{scale}}$ is logarithmically below the jet $p_T$ in gluon and quark jets, due to the strongly ordered (in $k_T$) QCD evolution, see e.g. Ref. [30]. In case of a strongly boosted $W$ boson decaying into quark and anti-quark, on the other hand, $y_{\text{scale}}$ is closer to the $W$ mass $m_W$. The left plot in Fig. 21 shows a qualitative comparison of the $y_{\text{scale}}$ spectrum for jets with masses $m_{\text{jet}} > 40$ GeV in simulated QCD dijet processes with the one for jets from simulated boosted $W$ boson decays with the same mass cut applied.

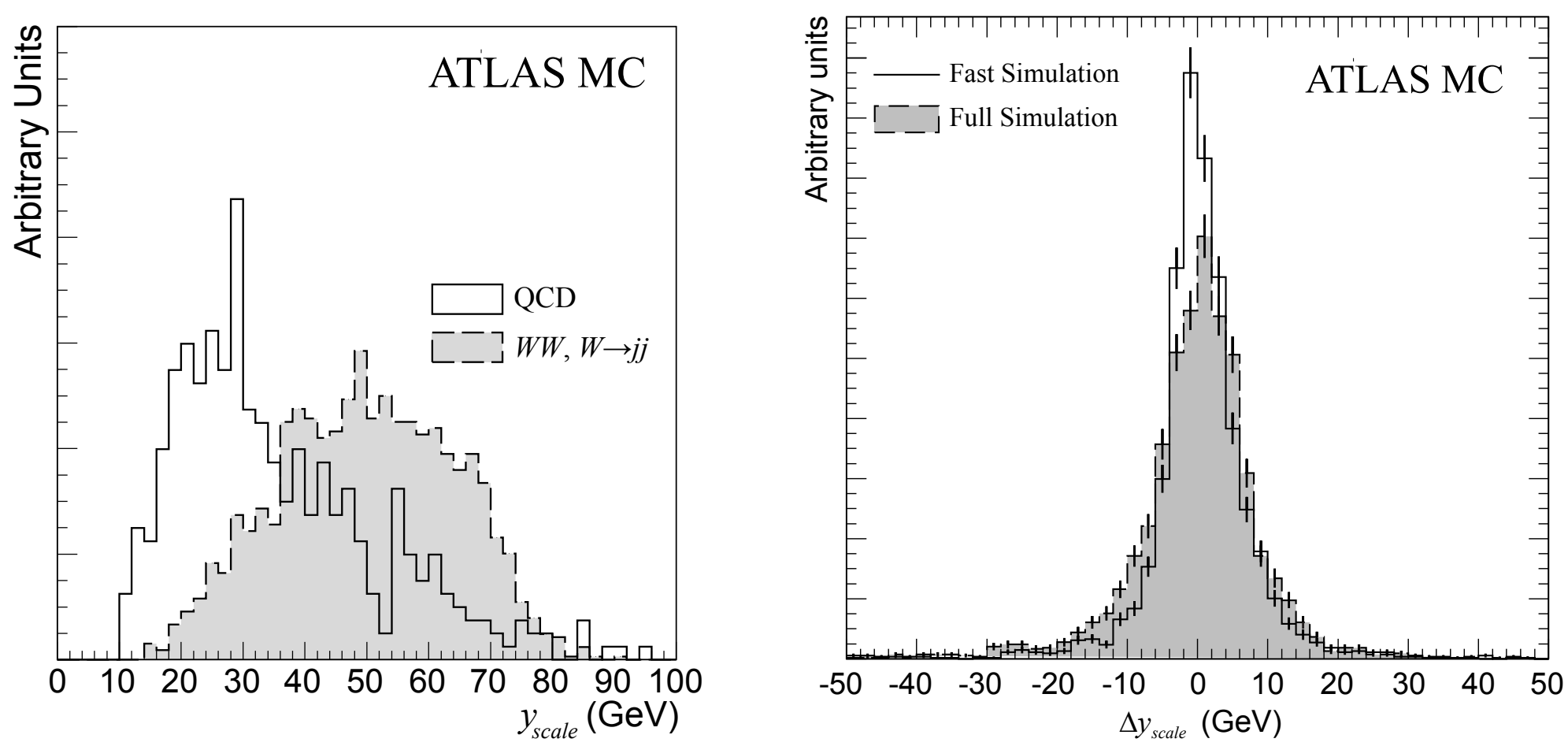

Figure 21: Distributions of the scale variable $y_{\text{scale}}$ indicating the threshold for splitting a given jet with mass $m_{\text{jet}} > 40$ GeV into two sub-jets, for simulated QCD dijet processes and boosted $W$ bosons decaying hadronically (left figure). The right figure shows predictions of the resolution power for $y_{\text{scale}}$ for the boosted $W$ bosons. The filled area distribution shows the spectrum for $\Delta y_{\text{scale}} = y_{\text{scale,reco}} - y_{\text{scale,truth}}$ obtained with fully detailed simulations, while the solid distribution shows the optimistic estimate from a smeared particle-level calculation. Both distributions are normalized to unity.

The resolution power of ATLAS for a measurement of $y_{\text{scale}}$ has been studied for splitting into two jets, i.e.:

$$y_{\text{scale}} = \sqrt{y_2} \times p_T \,,$$

with a sample of simulated highly boosted $W$ decaying hadronically. For these the corresponding $y_{\text{scale}}$ distribution approximately peaks at $m_W/2$, as expected, see Fig. 21. The resolution power for $y_{\text{scale}}$ can be evaluated by comparing this variable reconstructed from the calorimeter jet with the one from the matching particle level jet. The right plot in Fig. 21 shows the resolution for full simulation and for a smeared particle-level fast simulation. Pending a full scale evaluation in the context of a physics analysis, the present observation based on the the similarity of both distributions is that the reconstruction of $y_{\text{scale}}$ does not seem to be too sensitive to details affecting the calorimeter signal, like showering, limited acceptance for low energetic particles, and similar. The prediction for the experimental $y_{\text{scale}}$ resolution from this figure is about 12%, while the optimistic simulation based on four-momentum smearing predicts about 9%.

# 4 Forward jet reconstruction and jets in minimum bias events

The ATLAS detector provides near hermetic coverage within pseudorapidities of approximately $-4.9 < \eta < 4.9$. The forward region is particularly challenging for jet reconstruction, as the jets of interest often have rather low transverse momentum (at high energy) and are thus closer to the fluctuations introduced by pile-up at design luminosity at LHC. For example, in a jet cone with $R_{\mathrm{cone}} = 0.7$ one expects fluctuations in transverse momentum of the order of 12 GeV [31], meaning that the minimum $p_T$ which can safely be reconstructed is around 40 GeV. The estimates for jet performance in the forward region presented here have been calculated without pile-up, i.e. only electronics noise is folded into the reconstructed signals.

Events with depleted hadronic activity in the central region of the $pp$ collisions are important signatures for discoveries, including but not limited to leptonically decaying Higgs bosons produced in vector boson fusion (VBF) events, like $WW$ scattering. To reconstruct this signal with significant efficiency a central jet veto can be applied. The effectiveness of this veto can be understood from the efficiency to reconstruct jets in minimum bias events without hard scattering, and the rate for fake jet reconstruction. This has been studied with simulations for ATLAS and the results are presented in Section 4.2.

## 4.1 Forward jets

Recent studies indicate that the detection of forward going (light quark) jets with pseudorapidities $2.7 \leq |\eta| \leq 4.9$ helps to significantly increase the discovery potential not only for heavy Higgs bosons [13], but also for intermediate mass Higgs bosons produced in VBF [32]. This region in the ATLAS detector features a complex calorimeter geometry in the transition region from the end-cap to the forward calorimeters. This leads to a loss of precision from the changing readout geometries. Some aspects of the performance of the ATLAS detector for these forward jets are discussed in this section.

The VBF events have a specific topology in that on average two hard *tag jets* are produced by the two quarks radiated off the vector boson. Figure 22 shows predictions for the jet multiplicity distributions in these events, both over the whole detector acceptance ($|\eta| < 4.9$), where most often the two tag jets are found, and in the forward direction ($2.7 < |\eta| < 4.9$) only. From this the expectation is in most events only one of the jets is going into the forward direction, at least for the particular Higgs boson mass considered here ($m_H = 120$ GeV). This observation is independent of the jet size for seeded cone and $k_T$ jets.

The relative transverse momentum resolution for both forward seeded fixed cone and $k_T$ jets in VBF produced $H \to \tau\tau \to \mu\mu$ events is estimated at about 9% for $20 < p_T < 120$ GeV. Signal linearity is expected to be within $\pm 2\%$ for these jets, at least for $p_T \gtrsim 30 - 40$ GeV, see Fig. 23. The precision achieved in the reconstruction of the jet kinematics for even lower $p_T$, an attempt only realistic in low luminosity running at LHC, seems to be limited from this study to $\approx 5\%$.

Predictions for the efficiency and purity of forward jets reconstructed with the seeded cone ($R_{\mathrm{cone}} = 0.4$) finder are shown in Fig. 24. Here the indication is that using topological cell clusters as calorimeter signals for jet reconstruction makes jet reconstruction more efficient at the likely less relevant lower end of the jet $p_T$ spectrum ($p_T \lesssim 30 - 40$ GeV), but generates more fake jets in the same $p_T$ range, i.e. has a lower purity in this region for this event sample. As both calorimeter signals reconstruct $k_T$ jets with basically the same efficiency and purity in the whole kinematic range of interest for these VBF events, there is an indication that the splitting of cell signals to fill towers, which is predominant in the forward region, generates less seeds than the more integrating cell clustering, where cell signals are summed up rather than split and thus create more likely signal objects above the seed threshold. See also text and Fig. 3 in Section 2.5.1 for tower formation, and Section 2.5.2 for clustering.

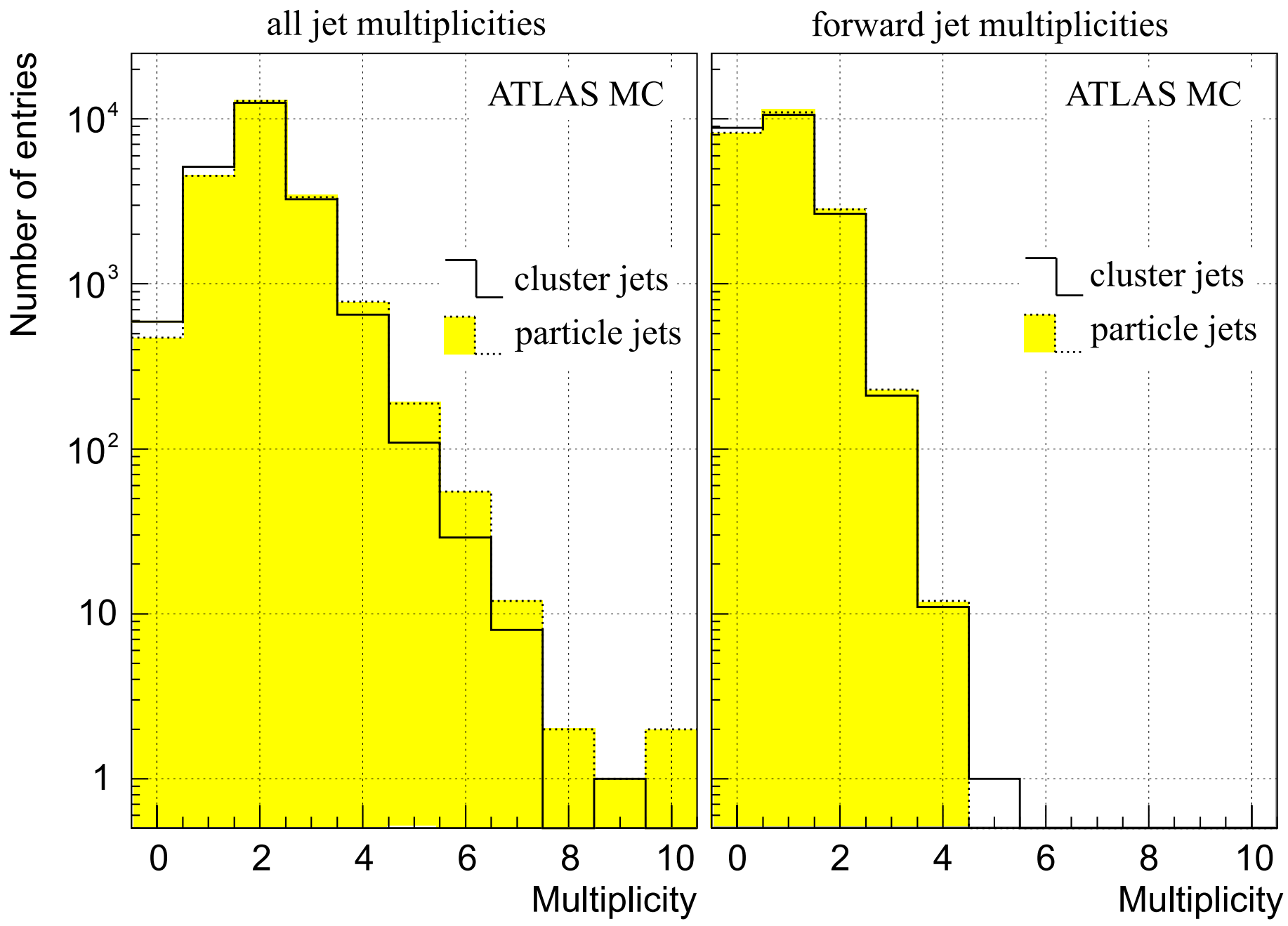

Figure 22: Simulated jet multiplicities for seeded cone jets ($R_{\text{cone}} = 0.4$) with $p_T > 20$ GeV, built from topological cell clusters and generated particles, respectively, in VBF produced Higgs boson events (left). The Higgs boson decays as $H \to \tau\tau \to \mu\mu$. The right plot shows the multiplicities of forward going jets with rapidities $2.7 \leq |y| < 4.9$.

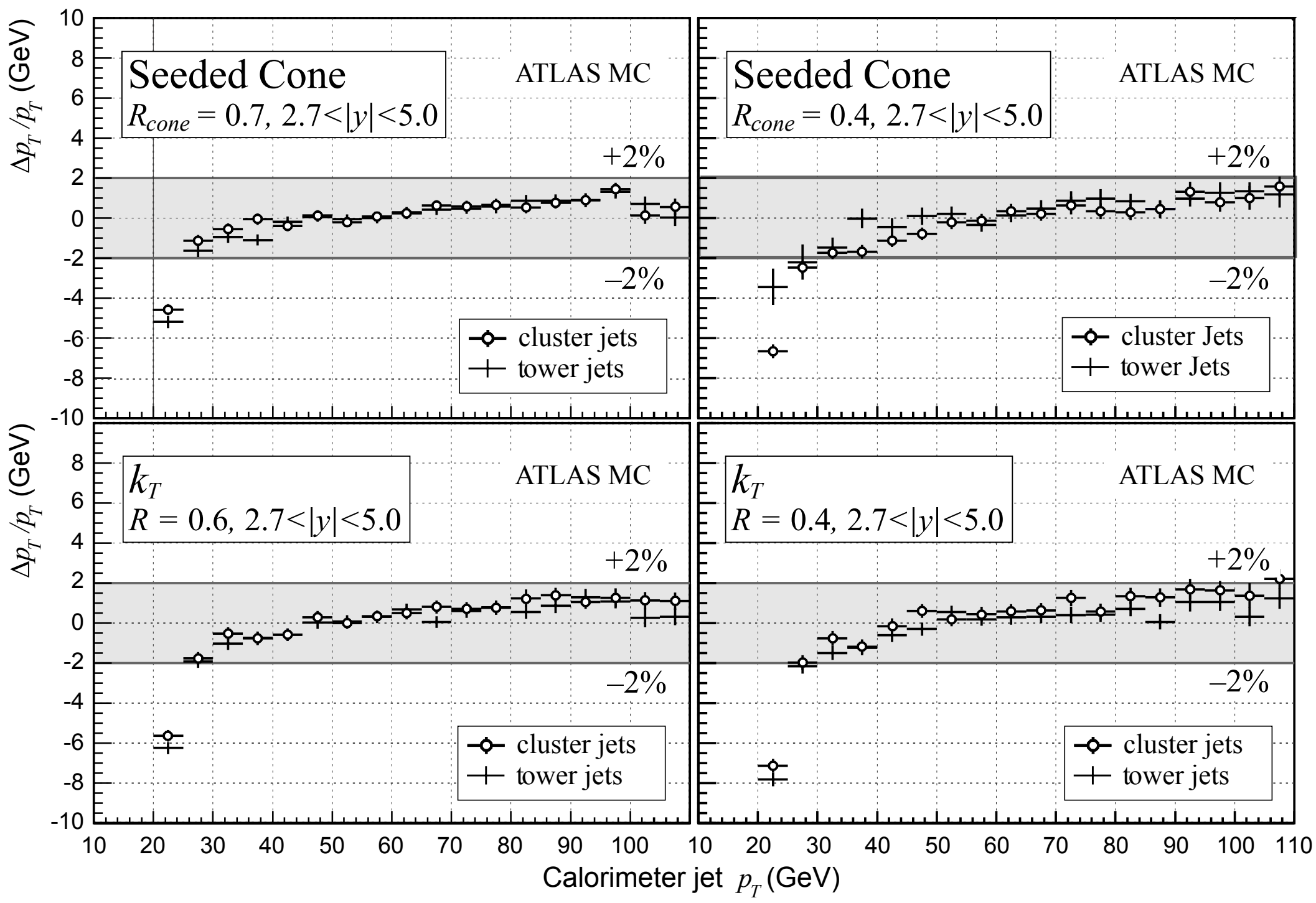

Figure 23: Relative deviation of reconstructed transverse momentum $p_T$ from calorimeter jets and the $p_T$ from the matched particle jet in VBF Higgs boson production, for various jet finder configurations and the two calorimeter signal definitions (towers and clusters).

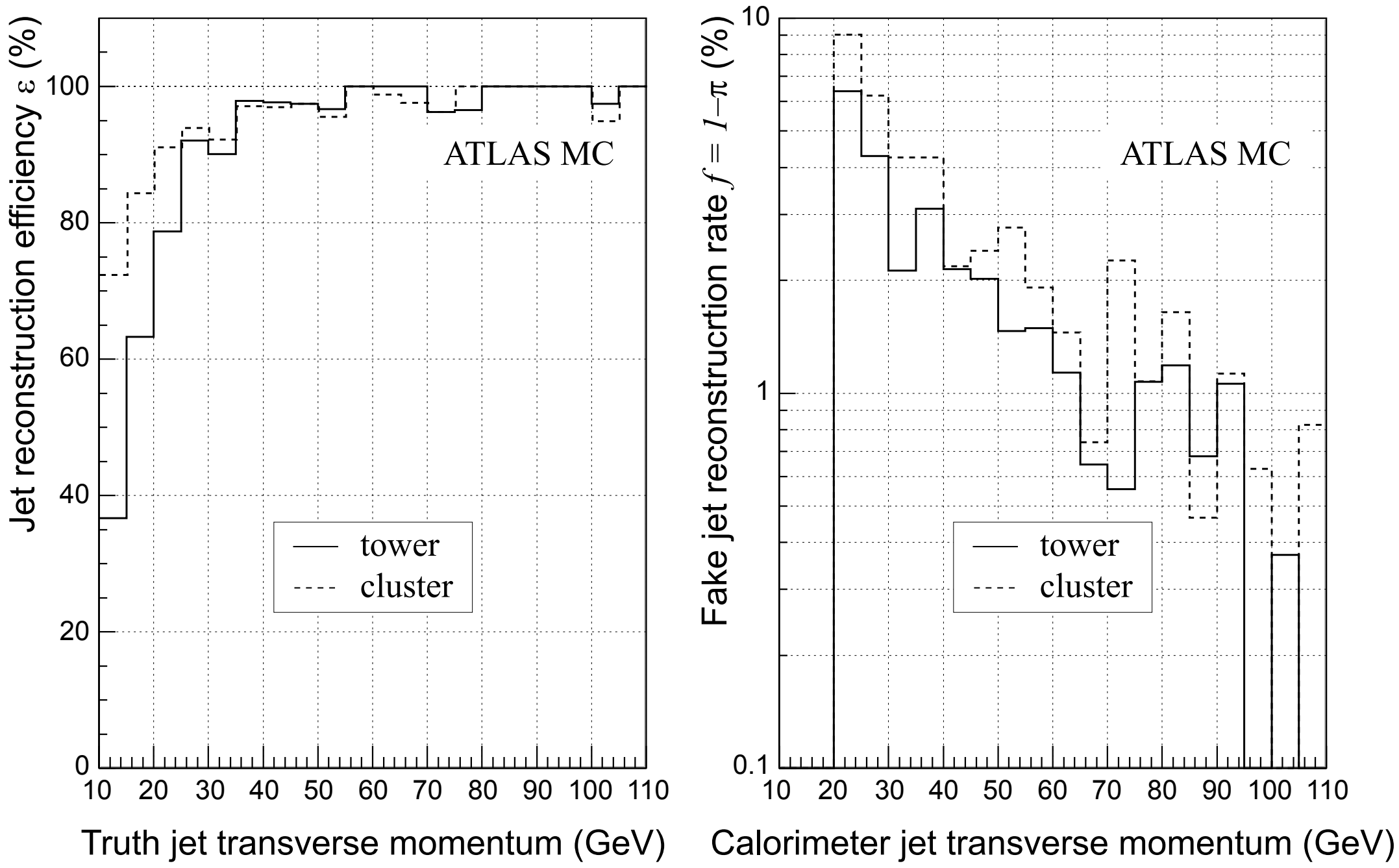

Figure 24: Estimated efficiency to reconstruct narrow seeded cone tower and cluster jets ($R_{\mathrm{cone}} = 0.4$) generated in VBF Higgs boson production in the forward direction ($2.7 < |\eta| < 4.9$), as function of the truth jet transverse momentum (left). The plot on the right shows the corresponding fake reconstruction rate as function of the transverse momentum of the calorimeter jet.

## 4.2 Jets in minimum bias events

Soft underlying physics, as generated by the underlying event in hadron colliders and the multiple soft interactions at high luminosity is an important source of jet production not directly related to the triggered hard scattering process of interest. For efficient application of a central jet veto, which is an important tool in background suppression in VBF produced Higgs boson events, it is crucial to understand the jet rate from soft interactions in this region, and the particular characteristics of these jets. The latter point is subject to ongoing studies, but first estimates on the jet multiplicity and rate from simulated single minimum bias events are available.

Figure 25 shows the expected average number of jets in these events as function of the $p_T$ threshold applied in the final jet selection, for various calorimeter signal definitions and the most commonly used jet finder configurations. The $k_T$ jet multiplicity for narrow ($R = 0.4$) jets is rather independent of the calorimeter signal choice, while for wider jets ($R = 0.6$) the cluster jets have a lower average multiplicity, i.e. are less problematic for a central jet veto. For seeded cone jets, the wider ($R_{\mathrm{cone}} = 0.7$) tower and cluster jets have very similar multiplicities, while here the narrow tower jets ($R_{\mathrm{cone}} = 0.4$) have a lower multiplicity than the cluster jets.

Predictions for the probability $P\left(N_{\mathrm{jet}} \geq 1, p_T > 20\ \mathrm{GeV}, \eta_{\mathrm{range}}\right)$ of reconstructing at least one jet per single minimum bias event with $p_T > 20$ GeV within a pseudorapidity range of $|\eta| < \eta_{\mathrm{range}}$ is shown in Fig. 26. Narrow jets from towers and clusters behave rather similarly for both the seeded cone and the $k_T$ algorithm. Wider jets are more often found in calorimeter tower jets than in topological cluster jets. In general the $k_T$ algorithm is less likely to reconstruct wider jets with $R = 0.6$ than the seeded cone is with $R_{\mathrm{cone}} = 0.7$. Here $P\left(N_{\mathrm{jet}} \geq 1, p_T > 20\ \mathrm{GeV}, \eta_{\mathrm{range}}\right)$ has been calculated including the occasional jet from the minimum bias (soft or semi-hard) interaction as well as “true” fake jets from calorimeter signal fluctuations due to noise. Both lead to efficiency losses in an analysis selecting final states with no hadronic activity in a given region $|\eta| < \eta_{range}$. Note that the expectations for the jet veto efficiency discussed here are estimated by $1 - P\left(N_{jet} \geq 1, p_T > 20\text{ GeV}, \eta_{range}\right)$ under the assumption that the minimum bias events generate a similar underlying activity as can be expected in signal events. An additional loss of signal events is associated with the multiple soft interactions from pile-up. One expects $\sim 25$ minimum bias interactions per bunch crossing at the design luminosity of $\mathscr{L} = 10^{34}\text{cm}^{-1}\text{s}^{-1}$. Experimental effects introduced by the calorimeter readout can increase this number by another factor of 2, giving jet veto efficiencies in the order of $95(50)\%$ for narrow jets with $p_T > 20$ GeV at $\mathscr{L} = 10^{33}\text{cm}^{-2}\text{s}^{-1}(10^{34}\text{cm}^{-2}\text{s}^{-1})$.

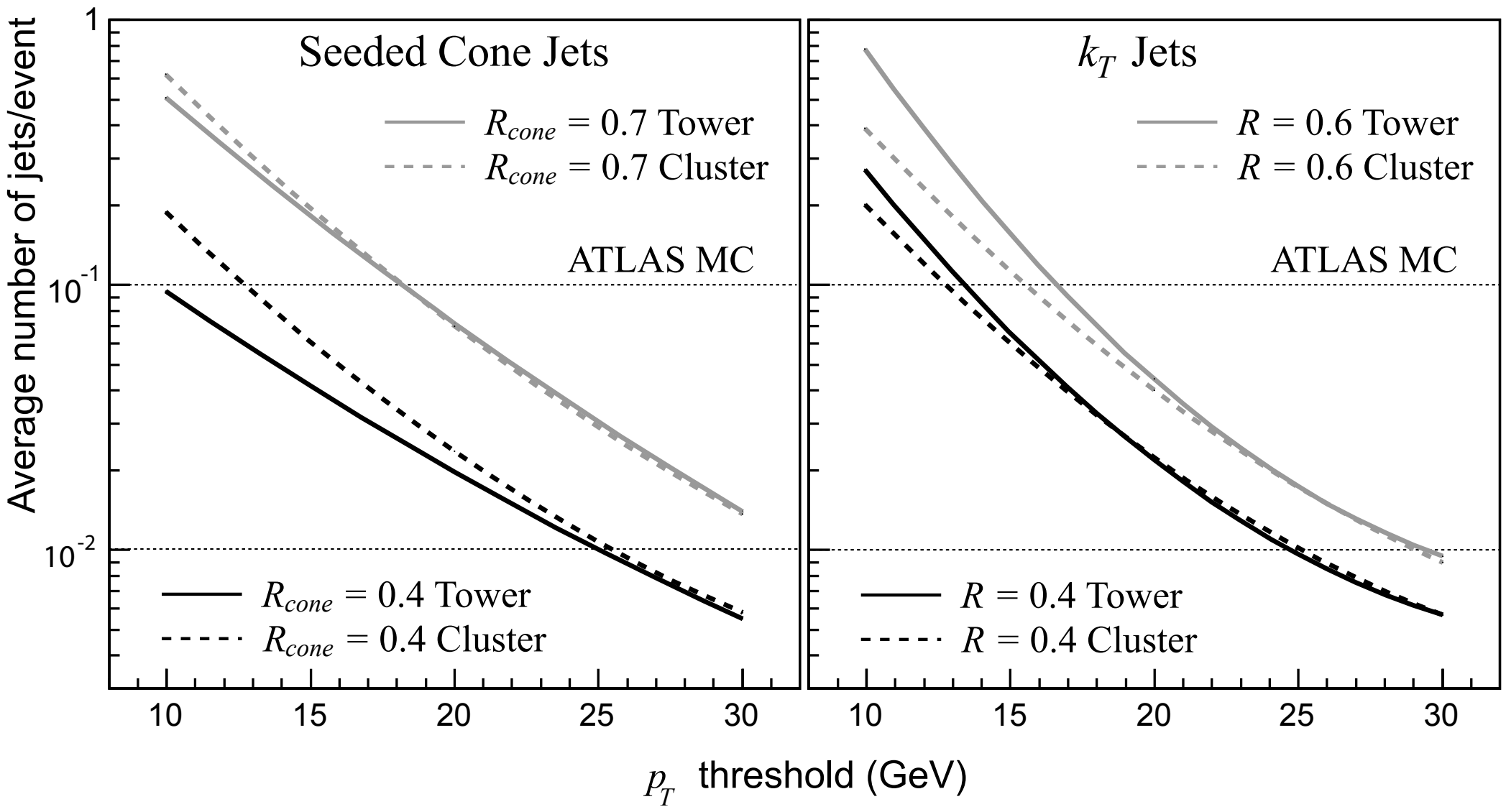

Figure 25: Estimates for the average number of jets per minimum bias event, as function of the $p_T$ threshold applied to the final jet. Shown are results for wide and narrow cluster and tower jets in minimum bias simulations.

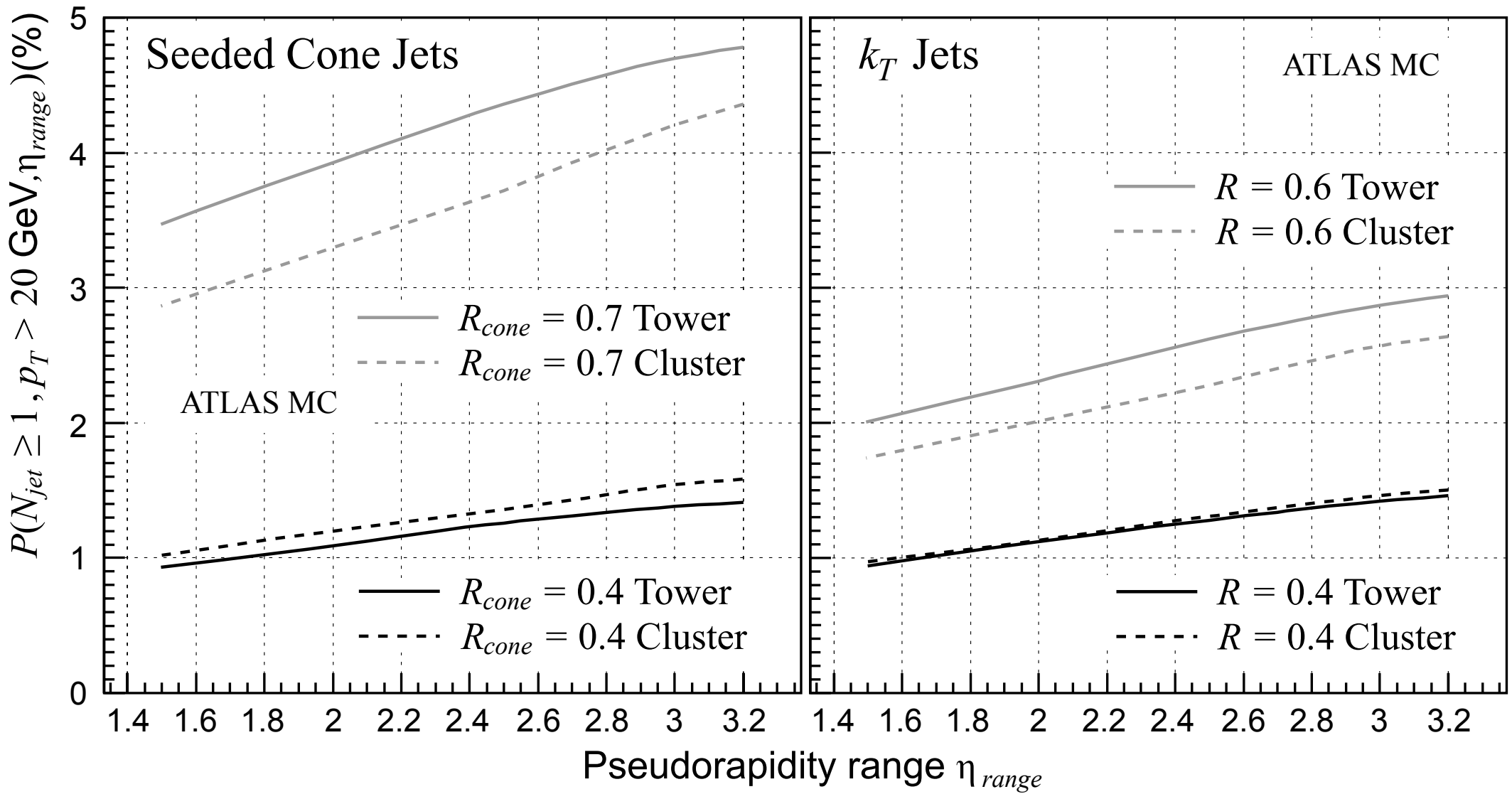

Figure 26: Prediction from full simulations for the probability $P\left(N_{\text{jet}} \geq 1, p_{\text{T}} > 20 \text{ GeV}, \eta_{\text{range}}\right)$ to reconstruct at least one jet with $p_{\text{T}} > 20$ GeV within a pseudorapidity range $|\eta| < \eta_{\text{range}}$ considered for jet reconstruction in single minimum bias events, as function of $\eta_{range}$. The jet veto efficiency in the region $|\eta| < \eta_{\text{range}}$ can be estimated from these curves as $1 - P\left(N_{\text{jet}} \geq 1, p_{\text{T}} > 20 \text{ GeV}, \eta_{\text{range}}\right)$.

Reconstructing jets in minimum bias events can provide important information for the modeling of soft physics and on the underlying event for hard scattering processes if correlation effects are neglected. For example, the pseudorapidity distributions of $k_T$ jets with $p_T > 10$ GeV are shown in Fig. 27. Most jets above threshold are centrally produced with $|\eta| < 1$ and reconstructed with some limited efficiency, both for towers and clusters. The fake rate for jet reconstruction at this $p_T$ threshold can be considerable ($\mathscr{O}(30\%)$ in the same central region), but decreases quickly with increasing jet $p_T$, thus likely restricting the accessible experimental phase space to test jet production in soft or semi-hard proton collisions with precision on production rates or cross section.

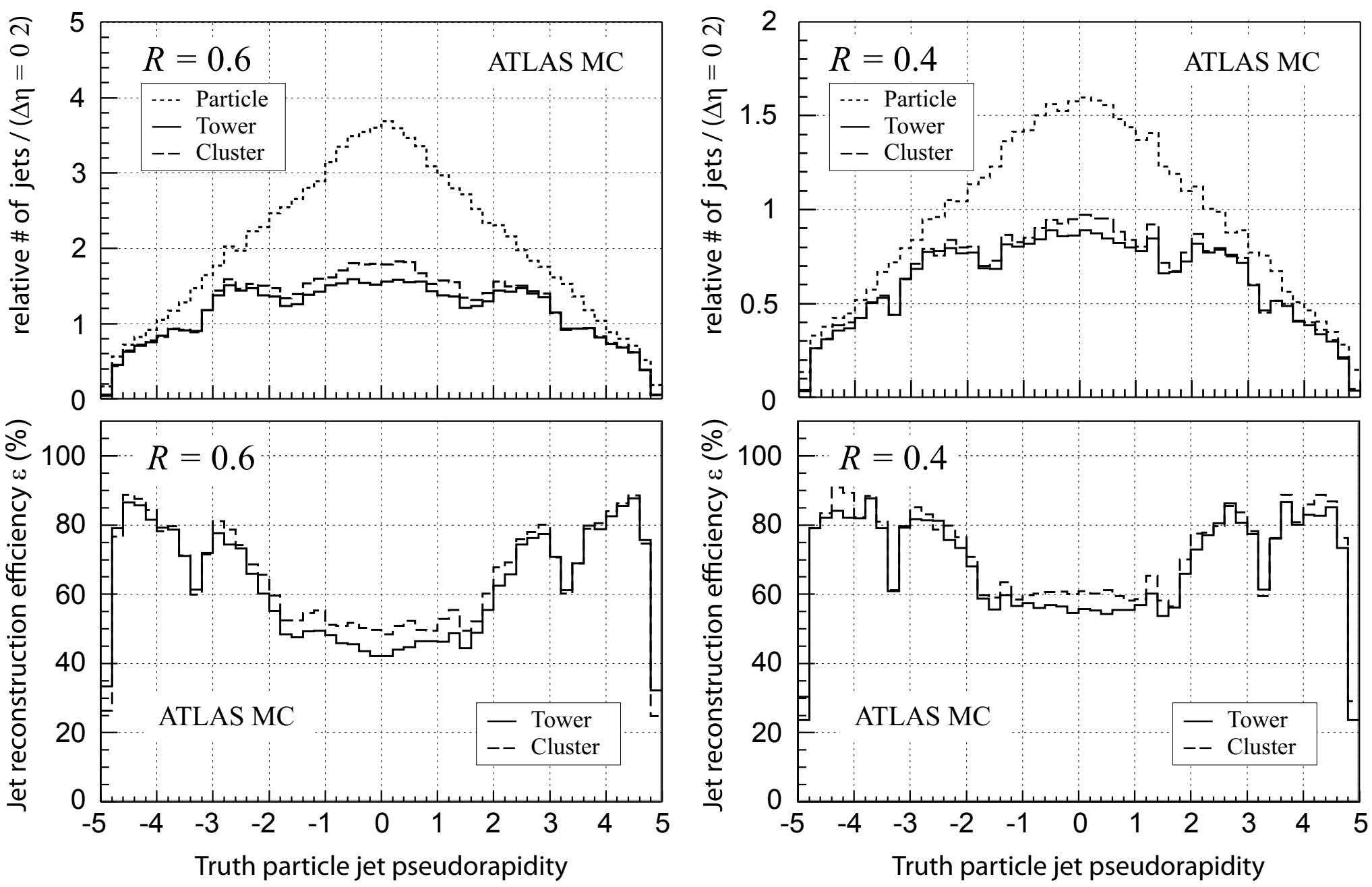

Figure 27: The pseudorapidity distribution of jets with $p_T > 10$ GeV in single minimum bias events in ATLAS (top plots for $k_T$ with $R = 0.6$ and $R = 0.4$, respectively). The distributions for the calorimeter jets exclude fake jets, i.e. each reconstructed calorimeter jet is matched with a truth particle jet. The efficiency for jet reconstruction in these events is shown in the bottom plots.

## 5 Conclusions and outlook

ATLAS supports a highly configurable and flexible jet reconstruction framework, which can easily be adapted to accommodate new jet algorithms or signal definitions from the detectors. At the same time, a default set of configurations for jet finding strategies is provided: a seeded fixed cone algorithm with split and merge, and the $k_T$ algorithm, both with two different parameters controlling the size of the reconstructed jet. For the seeded cone, cone radii of $R_{cone} = 0.7$ and $R_{cone} = 0.4$ are available. Similarly, the default choices for narrow and wide $k_T$ jets are distance parameters $R = 0.4$ and $R = 0.6$, respectively. For the detector jets, which are reconstructed from projective calorimeter towers or topological cell clusters, a calorimeter cell signal weighting based calibration is applied, with calibration weights derived from simulations with an ideal ATLAS detector geometry model, together with an additional jet energy scale correction parametrized in pseudorapidity and transverse momentum.

Performance predictions for a slightly misaligned and distorted ATLAS detector system have been extracted mostly from PYTHIA generated QCD dijet processes simulated through the detector with GEANT4. The emphasis in these studies was to estimate possible deviations from signal linearity and uniformity, the deterioration of the relative jet energy resolution, the effect on jet reconstruction efficiencies and the fake jet reconstruction rate for the first collisions, when the inactive material distributions and the alignment of detector components may not be well known. The corresponding pre-collision physics data estimates have been derived from Monte Carlo for the default jet configurations for the two considered calorimeter signal definitions (tower and cluster). A preliminary conclusion from these studies is that the signal linearity can likely be controlled at the level of $2-3\%$, depending on the detector region. The effect of the combination of a particular calorimeter signal choice with a given jet reconstruction configuration has been found to be consistent with expectations, i.e. the noise suppression intrinsic to clusters can be observed. In general there are strong indications that the effect of uncertainties in the knowledge of the detector geometry only leads to a modest degradation of the jet reconstruction performance.

Special challenges to jet reconstruction in the forward direction or jet vetoes in the central region, have been evaluated and found to be of acceptable performance for physics analyses like vector-boson fusion produced Higgs boson events. Pile-up at high LHC luminosities leads to considerable degradation especially of the jet veto efficiency. Due the large uncertainties in the modeling of soft physics underlying and overlapping the *pp* collisions, a precise quantitative evaluation of this degradation needs to be postponed until experimental minimum bias data and other event topologies become available with sufficient data quality and statistics.

In preparation for the experimental data ATLAS is now focusing on "data only" calibration approaches using tools like $p_T$ balance in prompt photon production and in QCD dijets. Additional efforts are concentrating on the subtraction of the pile-up contribution to jets, which can be estimated measuring the (transverse) energy scattered into a given area in pseudorapidity and azimuth in minimum bias events.

# Detector Level Jet Corrections

**Abstract**
The jet energy scale is proven to be an important issue for many different physics analyses. It is the largest systematic uncertainty for the top mass measurement at Tevatron, it is one of the largest uncertainties in the inclusive jet cross section measurement, whose understanding is the first step towards new physics searches. Finally, it is an important ingredient of many standard model analyses. This note discusses different strategies to correct the jet energy for detector level effects.

## 1 Introduction

The jet calibration process can be seen as a two-step procedure. In the first step, the jet reconstructed from the calorimeters is corrected to remove all the effects due to the detector itself (nonlinearities due to the non-compensating ATLAS calorimeters, the presence of dead material, cracks in the calorimeters and tracks bending in/out the jet cone due to the solenoidal magnetic field). This calibrates the jet to the particle level, i.e. to the corresponding jet obtained running the same reconstruction algorithm directly on the final state Monte Carlo particles. The second step, is the correction of the jet energy back to the parton level, which will not be discussed in this section.

There are currently several calibration approaches studied in the ATLAS collaboration based on the calorimeter response on the cell level or layer level and either in the context of jets or of clusters.

The first part of the section describes a possible approach for the calibration to the particle jet. The energy of the jet is corrected using cell weights. The weights are computed by minimizing the resolution of the energy measurement with respect to the particle jet. The performance of the calibration in terms of jet linearity and resolution is assessed in a variety of events (QCD dijets, top-pairs and SUSY events). The different structure of these events (different color structure, different underlying event) will manifest itself as a variation in the quality of the calibration. This method has been the most widely used so far in the ATLAS collaboration.

Other methods have also been studied. Here we discuss one alternative global calibration approach, which makes use of the longitudinal development of the shower to correct for calorimeter non-compensation. The jet energy is corrected weighting its energy deposits in the longitudinal calorimeter samples. Although the resolution improvement is smaller with respect to other methods, this method is simple and less demanding in terms of agreement between the detector simulation predictions and real data.

The second part of this section describes the concept of local hadronic calibration. First clusters are reconstructed in the calorimeters with an algorithm to optimize noise suppression and particle separation. Shower shapes and other cluster characteristics are then used to classify the clusters as hadronic or electromagnetic in nature. The hadronic clusters are subject to a cell weighting procedure to compensate for the different response to hadrons compared to electrons and for energy deposits outside the calorimeter. In contrast to the cell weights mentioned above no minimization is performed and the actual visible and invisible energy deposits in active and inactive calorimeter material as predicted by Monte Carlo simulations are used to derive the weights. One of the advantages of this method is that the jet reconstruction runs over objects which have the proper scale (in contrast to the global approach, where the scale corrections are applied after the jet is reconstructed from uncorrected objects).

The third part of the note describes refinements of the jet calibration that can be done using the tracker information: the residual dependence of the jet scale on the jet charged fraction can be accounted for improving the jet resolution. An algorithm to correct the $b$-jet scale in case of semileptonic decays will also be discussed.

## 2 The calibration to the truth jet

According to the perturbative QCD, jets are the manifestations of scattered partons (quarks and gluons). After undergoing fragmentation, a collimated collection of hadrons emerges and its energy is measured in the calorimeter system. In addition to this hard scattering, the final state also contains energy coming from multiple proton–proton (pile–up) interactions and the underlying event.

The typical output of an event generator will provide theoretical predictions about the particle content and spectra at this stage, the so called particle level. Jets resulting from the application of a jet reconstruction algorithm at the particle level are thus relevant as “truth”, since they represent the final state jets that ideally must be reconstructed starting from the detector level. In the following we refer to them using the expression “truth jets”.

Since jet fragmentation functions are independent of jet energy, the fraction of the total jet energy carried by the different particle types in a jet is basically independent of energy. Figure 1 shows the relative contribution of the different particle types to the jet energy as a function of the jet $E_T$. About 40% of the total energy is carried by charged pions, 25% is carried by photons (mainly coming from the $\pi^0$ decay), another 20% is accounted for by kaons, nearly 10% by protons and neutrons. Therefore, 25% of the energy deposits in the calorimeters come directly from pure electromagnetic showers. The right plot of Fig. 1 shows the average fractional energy deposit in the different calorimeter samplings with respect to the true jet energy in the central calorimeter regions ($|\eta| < 0.7$). Most of the energy (about 2/3 of the reconstructed energy) is measured by the electromagnetic calorimeter. The total reconstructed energy differs significantly from the true jet energy. This is because of a number of detector effects:

- if the calorimeters are non-compensating (as in ATLAS), their response to hadrons is lower than that to electrons and photons, and is non-linear with the hadron energy.
- part of the energy is lost because of dead material, cracks and gaps in the calorimeters, and is also non-linear with hadron energy.
- The solenoidal magnetic field will bend low energy charged particles outside the jet cone.

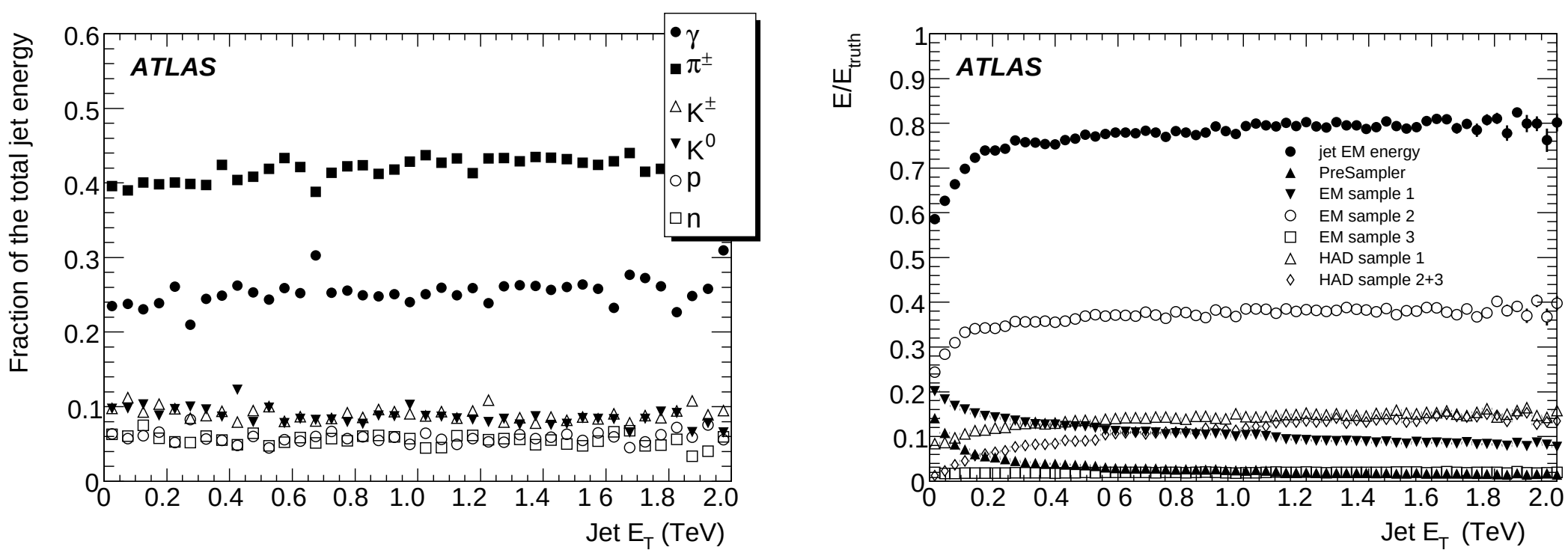

Figure 1: Left: fractional energy carried by different particle types as a function of the jet energy. Right: fraction of true energy deposited in the different calorimeter samplings for a jet in the central ($|\eta| < 0.7$) calorimeter region as a function of its true energy.

The reconstructed jet energy must be corrected for these effects to obtain the best estimator for true jet energy.

In the following we will discuss two possible strategies. The first one (referred to as global calibration) aims to provide calibration coefficients at jet level; the second one, the local calibration, provides calibration constants at the jet constituent level. The performance of both the approaches will be discussed in detail. Both methods use simulated events to obtain the calibration coefficients.

# 3 An energy density based cell calibration

The shower produced by a jet impinging on the calorimeters is composed of an electromagnetic and a hadronic component. The electromagnetic component is characterized by a compact, highly dense, energy deposit, while the hadronic one is broader and less dense. This fact can be used to correct the energy measurement to recover for the non-linear calorimeter response to the hadrons.

After jet reconstruction using calorimeter cells calibrated at the electromagnetic scale, the total energy of a jet is reconstructed by summing the energies of its constituent cells multiplied by a weight which depends on the energy density of the cell itself. We thus define the EM scale jet energy as:

$$E_{\mathrm{em}} = \sum_{i=\mathrm{cells}} E_i \tag{1}$$

where $E_i$ is the energy in the cell $i$ for the considered jet. We then define a jet weighted 4-vector

$$E = \sum_{i=\mathrm{cells}} w_i E_i \qquad \vec{P} = \sum_{i=\mathrm{cells}} w_i \vec{P}_i \tag{2}$$

where $E_i$, $\vec{P}_i$ are the $i$-th cell energy and momentum (whose direction is defined by the position in the calorimeter and whose magnitude is equal to $E_i$), and $w_i$ are correction factors that need to be determined. They depend on the cell energy density $E_i/V_i$, where $V_i$ is the volume of the $i$-th cell.
In order to reduce the number of weights to be computed, the following steps are done:

- The energy density distributions of the cells are divided into different bins with width increasing logarithmically with the cell energy density.
- The calorimeters are subdivided into several regions $k$. The longitudinal segmentation is partially exploited. Broad pseudorapidity bins are also defined. Table 1 shows the defined regions.
- The weight in the $k$-th calorimeter region, in the $j$-th energy density bin is defined to be:

$$w_i^{(k,j)} = \sum_{m=0}^{N_p-1} a_m^{(k)} \log^m (E/V)_j \tag{3}$$

where $N_p$ (the number of parameters used in the fit) is a number which depends on the region $k$ considered. The value of $\log(E/V)_j$ is defined at the lower edge of the $j$-th bin.

With this procedure, the number of independent parameters to be determined is significantly reduced (see Table 1). The pre-sampler and the strip-layer of the EM calorimeter have a single weight, constant with respect to the density of the energy deposits. The last three rows of the table refer to three energy terms which are also corrected with a single multiplicative factor: they are the cryostat term, the scintillator term and the gap term. The cryostat term is computed as the geometrical average of the energy deposited in the back of the electromagnetic barrel and the first layer of the TileCal barrel. It was in fact found in the past [1] that this gives a good estimate of the energy loss in the cryostat. The scintillator and gap terms correspond to the energy deposited in the scintillation counters in the region between the Tile barrel and extended barrel [2]. The weights applied to these terms are meant to recover for the presence of a large amount of dead material.

The parameters have been determined considering QCD dijet events, simulated with PHYTHIA6.4 [3] with the ATLAS settings [4], and the detector simulated using GEANT4. The events have been generated in bins of the partonic $p_T$, as illustrated in Table 2. Approximately 10k events have been used for each bin. The jets have been reconstructed using calorimeter towers as input. The jet reconstruction algorithm used is a seeded cone algorithm with a seed threshold of $E_T = 1$ GeV, and a cone size $R_{\text{cone}} = 0.7$. Jets with a reconstructed axis lying close to the gap region ($1.3 < |\eta| < 1.5$), or to the crack region ($3.0 < |\eta| < 3.5$), or in the very forward region ($|\eta| > 4.4$), are excluded from the minimization.

Table 1: Definition of regions defined for the minimization that determines the cell weights. The third column shows the number of parameters used in the minimization.

| Region name | Longitudinal Sample | Number of parameters $N_P$ |
|---|---|---|
| wemb0 | Barrel pre-sampler | 1 |
| wemb1 | Barrel EM strips | 1 |
| weme0 | End-Cap pre-sampler | 1 |
| weme1 | End-Cap EM strips | 1 |
| emb0 | Barrel middle and back sample, $\|\eta\| < 0.8$ | 4 |
| emb1 | Barrel middle and back sample, $\|\eta\| > 0.8$ | 4 |
| eme0 | Endcap middle and back sample, $\|\eta\| < 2.5$ | 4 |
| eme1 | Endcap middle and back sample, $\|\eta\| > 2.5$ | 4 |
| til0 | Barrel | 4 |
| til1 | Extended Barrel | 4 |
| hec0 | Hadronic End-Cap, $\|\eta\| < 2.5$ | 4 |
| hec1 | Hadronic End-Cap, $\|\eta\| > 2.5$ | 4 |
| fem | FCal first layer | 3 |
| fhad | FCal second and third layer | 3 |
| cryo | Cryostat term | 1 |
| scint | Scintillator term | 1 |
| gap | Gap term | 1 |
| Total | | 45 |

Table 2: List of the QCD dijet events used to compute the calibration constants listed in the text.

| Sample Tag | $p_T$ cut |
|---|---|
| J1 | 17 GeV $< p_T <$ 35 GeV |
| J2 | 35 GeV $< p_T <$ 70 GeV |
| J3 | 70 GeV $< p_T <$ 140 GeV |
| J4 | 140 GeV $< p_T <$ 280 GeV |
| J5 | 280 GeV $< p_T <$ 560 GeV |
| J6 | 560 GeV $< p_T <$ 1120 GeV |
| J7 | 1120 GeV $< p_T <$ 2240 GeV |
| J8 | $p_T >$ 2240 GeV |

Reconstructed jets are associated to the nearest truth jet (in $\phi - \eta$ space), obtained, as discussed in Section 2, running the same reconstruction algorithm on the final state particles from the event generator.

The following quantity is minimized using MINUIT:

$$\chi^2 = \sum_e \left( \frac{E^{(e)} - E^{(e)}_{\text{truth}}}{E^{(e)}_{\text{truth}}} \right)^2, \tag{4}$$

where the sum runs over the considered events $e$, $E^{(e)}$ is defined in equation (2) and $E_{\text{truth}}^{(e)}$ is the energy of the matched truth jet.

It should be noted that this approach partially absorbs effects that are not purely calorimetric into the weights. In particular, the energy smearing introduced by the central solenoidal magnetic field, which bends low $p_T$ particles in and out of the jet cone, is not treated separately, but the effect is proven to be small for the cone $R_{\text{cone}} = 0.7$ jets used for the minimization.

In order to correct for residual non–linearities in the jet response, a further, reconstruction algorithm dependent, correction function is introduced. The final 4-vector of a jet is thus defined as

$$E_\delta = \rho_\delta(E_T, \eta)E \qquad \vec{p}_\delta = \rho_\delta(E_T, \eta)\vec{P} \tag{5}$$

where $\delta$ indicates the dependence on the jet reconstruction algorithm and $E_T$ and $\eta$ are the transverse energy and the pseudorapidity of the 4-vector $p^\mu$. The *scale factor* $\rho_\delta(E_T, \eta)$ is obtained fitting the ratio $E_T/E_{T,\text{truth}}$ in 44 bins of $\eta$ as a function of $E_T$ with the following function:

$$f(E_T) = \sum_{i=0}^{3} c_i \log^i E_T \tag{6}$$

and, for a given $\delta$ and $\eta$ bin,

$$\rho = 1/f. \tag{7}$$

The scale factor corrects for the residual non-linearity introduced by the cracks and gaps in the calorimeter and for differences introduced by the use of different reconstruction algorithms, finally recovering the truth jet scale. The size of this final correction is at the level of few percent (up to 5%) in the crack and gap calorimeter region, while it is of the order of 1-2% (depending on the jet algorithm) in the rest of the $p_T$–$\eta$ phase space.

Therefore, the complete set of calibration parameters for a given reconstruction algorithm includes the cell energy density dependent weights obtained with cone $R_{\text{cone}} = 0.7$ jets, plus specific scale factors.

## 3.1 Results on dijet events

All the jet corrections computed as described in the previous section (scale factor included) have been applied to dijet events. The parameters that are considered in order to assess the quality of the calibration are the jet linearity, defined as the ratio between the reconstructed jet energy and the corresponding truth jet energy (as defined in Section 2) and the energy resolution.

The matching between the reconstructed jets and the truth jets is done considering their separation in a $\eta - \phi$ plane, defined as

$$R_{\text{cone}} = \sqrt{\Delta\phi^2 + \Delta\eta^2} \tag{8}$$

A truth jet is matched with a reconstructed jet if $R_{\text{cone}} < 0.2$.

Once the matching is done, the $E - \eta$ phase space of the truth jets is subdivided into bins. For each bin in energy and pseudorapidity, a histogram is filled with the ratio between the reconstructed energy and the truth energy $E_{\text{rec}}/E_{\text{truth}}$. The resulting histogram is first fitted with a gaussian in the whole histogram range. This provides two estimates ($\mu_{\text{raw}}$, $\sigma_{\text{raw}}$) for the mean value and the width of the distribution. The fit is then repeated in the range $\mu_{\text{raw}} \pm 2\sigma_{\text{raw}}$. This provides the final values ($\mu$, $\sigma$) that are used in the summary plots. Two examples of such histograms are shown in Fig. 2 for two different pseudorapidity and energy bins.

Figure 3 shows the dependence of $\langle E_{\text{rec}}/E_{\text{truth}} \rangle$ (linearity, in the following) on the truth jet energy $E_{\text{truth}}$ for jets reconstructed from calorimeter towers. The plot on the left refers to jets reconstructed with a cone algorithm with radius of 0.7 while the one on the right is for $k_T$ algorithm with $R = 0.6$ [5].

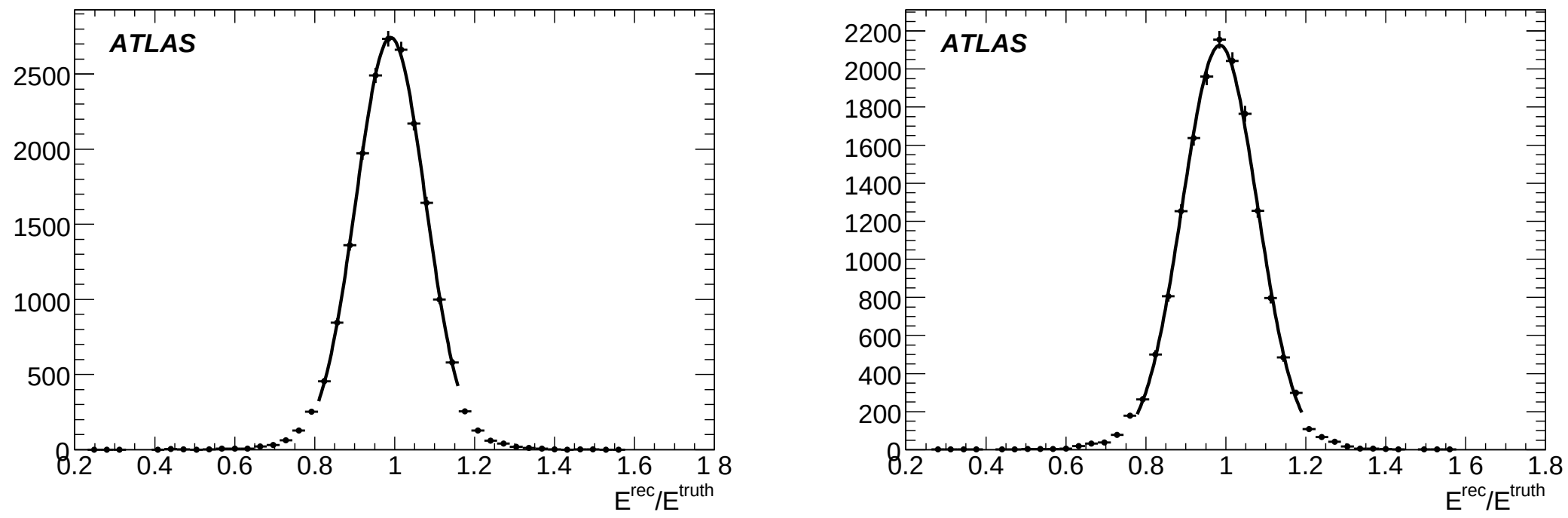

Figure 2: Two example histograms of $E_{rec}/E_{truth}$. On the left, the histogram is done for $88\text{ GeV} < E_{truth} < 107\text{ GeV}$ and $|\eta| < 0.5$, on the right for $158\text{ GeV} < E_{truth} < 191\text{ GeV}$ and $1.0 < |\eta| < 1.5$.

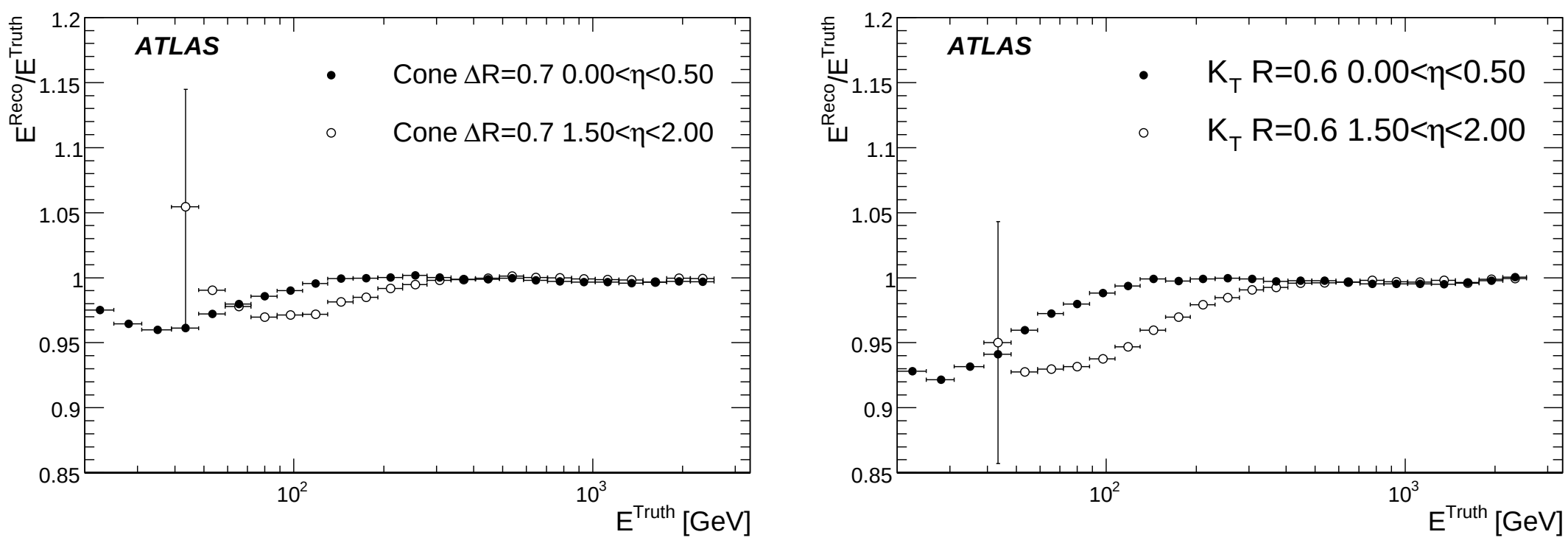

Figure 3: Dependence of the ratio $E_{rec}/E_{truth}$ on $E_{truth}$ for jets reconstructed with a cone algorithm with $R_{cone} = 0.7$ and with a $k_T$ algorithm with $R = 0.6$. The black (white) dots refer to jet with $|\eta| < 0.5$ $(1.5 < |\eta| < 2)$. An ideal detector geometry has been used to simulate the events.

The results show that the linearity is recovered over a wide energy range, both in the central ($|\eta| < 0.5$) and in the intermediate ($1.5 < |\eta| < 2$) regions. For the cone algorithm, at low energy ($E = 20-30\,\mathrm{GeV}$), the linearity differs by up to 5% from 1 in the central region. At low energy, there is a 5% residual non linearity, not fully recovered by the parametrization chosen for the scale factor.

Concerning the intermediate pseudorapidity region, we can see a similar behavior around 100 GeV (note that in this region $E \sim 100\,\mathrm{GeV}$ corresponds to $E_T = E/\cosh\eta \sim 35\,\mathrm{GeV}$).

The linearity plot for the $k_T$ algorithm shows a more pronounced deviation from 1 at low energy ($\langle E_{rec}/E_{truth}\rangle = 5\%$ at 50 GeV, 8% at 30 GeV). The linearity is fully recovered above $\sim 100\,\mathrm{GeV}$ in the central region, $\sim 300\,\mathrm{GeV}$ in the intermediate region.

The uniformity of the response over pseudorapidity is also satisfactory. Figure 4 shows the dependence of the ratio $E_T^{rec}/E_T^{truth}$ on the pseudorapidity of the matched truth jet for three different transverse energy bins. Again, the left plot refers to cone 0.7 jets, while the right one refers to $k_T$ jets with $R = 0.6$. We can observe that for the lowest considered transverse energy bin, the ratio increases with the pseudorapidity. This is a consequence of the fact that energy increases with $\eta$ at fixed $E_T$ and that the linearity improves with increasing energy.

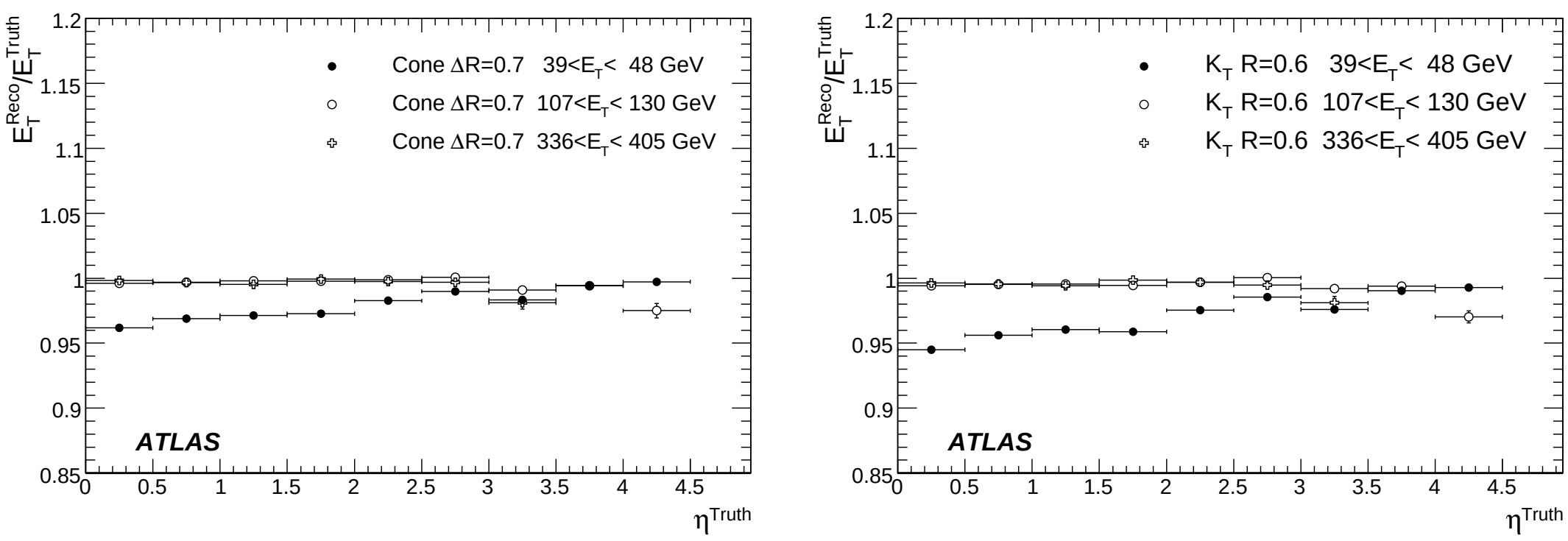

Figure 4: Dependence of the ratio $E_T^{rec}/E_T^{truth}$ on the pseudorapidity for the cone algorithm with $R_{cone} = 0.7$ (on the left) and for the $k_T$ algorithm with $R = 0.6$ (on the right). An ideal detector geometry has been used to simulate the events.

We can also observe (in particular for the $k_T$ algorithm) the remnants of the calorimeter structure, which is not completely corrected by the procedure. There is a first, small dip at $|\eta| \sim 1.5$, in correspondence with the gap between the TileCal barrel and extended barrel [2]. A second dip is observed in correspondence with the calorimeter crack between the End-Cap and the forward calorimeters.

Even if the effect is smaller when higher $E_T$ bins are considered, it is still present in the crack region. Jets with $E_T \sim 400\,\mathrm{GeV}$ still show a slight $\eta$ dependence in their response. As a last indicator of the quality of the correction factors, we consider the energy resolution $\sigma(E_{rec})/E_{rec}$. The dependence of the energy resolution on the jet energy is shown in Fig. 5 for the cone (left) and $k_T$ (right) algorithms in two pseudorapidity bins. The fit to the data is done considering three terms contributing independently to the resolution:

$$\frac{\sigma}{E} = \frac{a}{\sqrt{E\ (\mathrm{GeV})}} \oplus b \oplus \frac{c}{E}. \tag{9}$$

The sampling term ($a$) is due to the statistical, poissonian, fluctuations in the energy deposits in the calorimeters. The constant term ($b$) reflects the effect of the calorimeter non-compensation and all the detector non-uniformities involved in the energy measurement. The noise term ($c$) is introduced to describe the noise contribution to the energy measurement. Although the physics origin of the different terms is quite clear, it should be kept in mind that many different ways of combining them have been used in literature. In particular, the three parameters are correlated, and their values depend on the particular functional form used to parameterize the resolution.

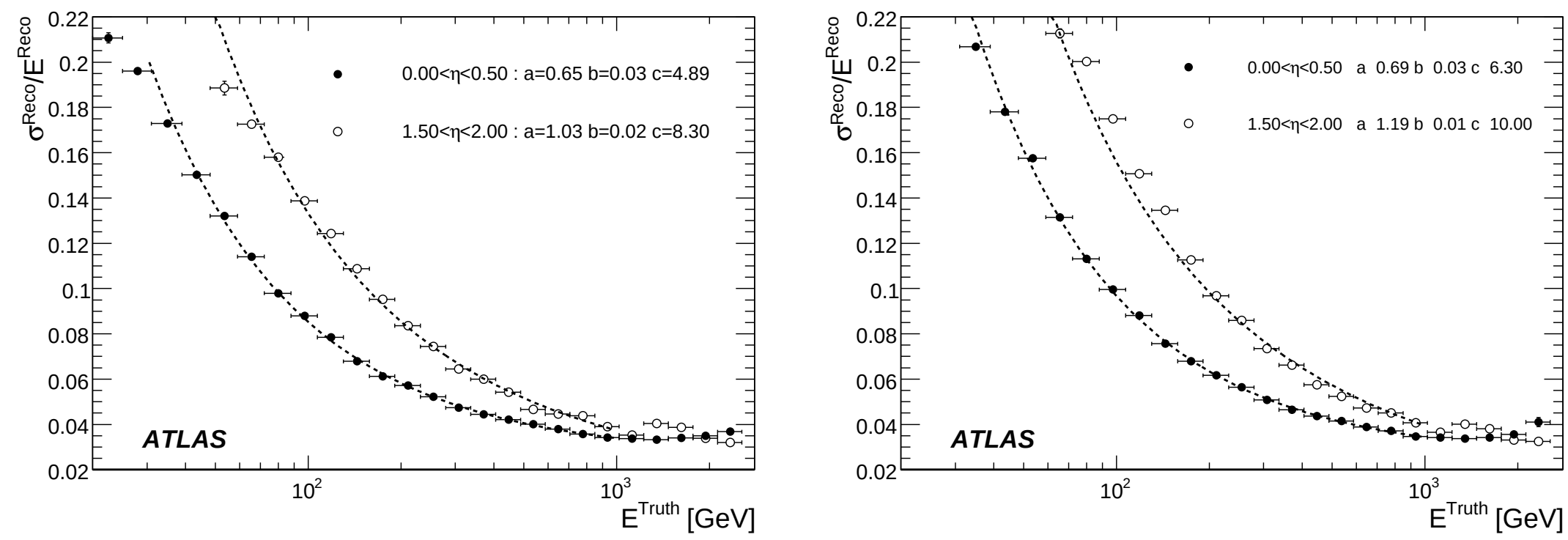

Figure 5: Energy resolution as a function of the jet energy for the cone algorithm with $R_{\text{cone}} = 0.7$ (on the left) and for the $k_T$ algorithm with $R = 0.6$ (on the right). The black (white) dots refer to jets with $|\eta| < 0.5$ ($1.5 < |\eta| < 2$). The smooth curves correspond to a fit done using the parametrization of Eq. 9. An ideal detector geometry has been used to simulate the events.

The fit is performed between 30 GeV and 1 TeV. In the central region, the resolution of the $k_T$ algorithm is significantly worse than that of the cone algorithm at low energy, while it is similar in the high energy region. The dependence of the resolution on the pseudorapidity is shown in Fig. 6. Again, the effect of the gap and crack regions are visible in particular for the lowest $E_T$ bin, where a clear worsening of the resolution is present in particular around $\eta = 3.2$.

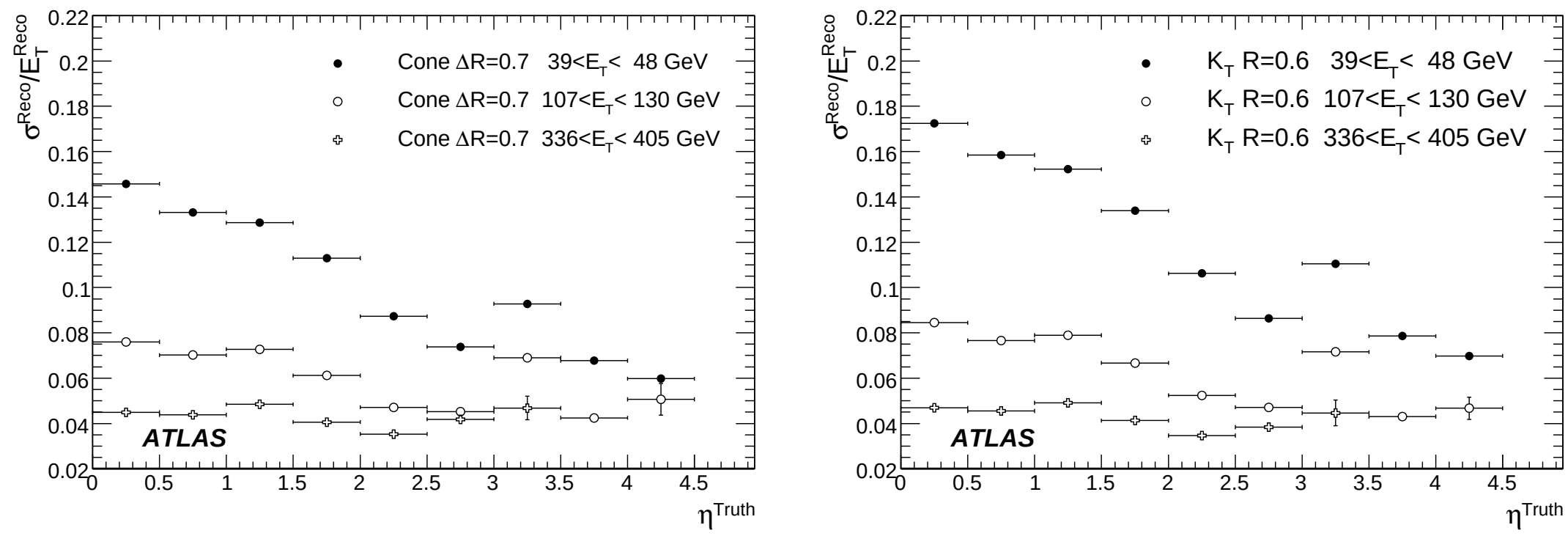

Figure 6: Energy resolution as a function of the pseudorapidity for the cone algorithm with $R_{\text{cone}} = 0.7$ (on the left) and for the $k_T$ algorithm with $R = 0.6$. An ideal detector geometry has been used to simulate the events.

The analysis has been repeated also for jets obtained using the uncalibrated topological clusters (see Section 5) as input for the jet reconstruction algorithms, obtaining very similar results for the linearity and the uniformity. The values of the parameters obtained with the fit of the function expressed in equation (9) to the simulation results are given in Table 3.

First, at fixed pseudorapidity, the sampling term is almost constant for the different reconstruction algorithms and for the different inputs. The sampling term in the central calorimeter region is about 65%, in rough agreement with what has been found previously with similar calibration approaches [6]. The constant term is also independent of the reconstruction algorithm and input type. The noise term on the other hand shows a slight variation between jets reconstructed from towers or from topological clusters, and is a significant contribution to the jet energy resolution for energies below 100 GeV. In particular, the noise term is lower if topoclusters are used. A significant difference is also observed between the cone and the $k_T$ algorithm, the latter showing a larger noise value.

Table 3: Parameters of the resolution parametrization described in the text as obtained with the fit of the resolution curves of Fig.9.

| Reconstruction Algorithm | $0 < \eta < 0.5$ | | | $1.5 < \eta < 2.5$ | | |
|---|---|---|---|---|---|---|
| | a (%) | b (%) | c (GeV) | a (%) | b (%) | c (GeV) |
| Cone $R_{cone} = 0.7$ Tower | $64 \pm 4$ | $2.6 \pm 0.1$ | $4.9 \pm 0.5$ | $103 \pm 10$ | $2.6 \pm 0.8$ | $8 \pm 1$ |
| $k_T R = 0.6$ Tower | $68 \pm 5$ | $2.5 \pm 0.2$ | $6.3 \pm 0.5$ | $110 \pm 1$ | $1 \pm 1$ | $12.2 \pm 2.5$ |
| Cone $R_{cone} = 0.7$ Topo | $63 \pm 4$ | $2.7 \pm 0.1$ | $4.2 \pm 0.5$ | $107 \pm 8$ | $1 \pm 1$ | $6.5 \pm 1.5$ |
| $k_T R = 0.6$ Topo | $64 \pm 5$ | $2.7 \pm 0.2$ | $5.4 \pm 0.5$ | $112 \pm 4$ | $1 \pm 1$ | $10.0 \pm 1.5$ |

## 3.2 Results on $t\bar{t}$ and SUSY events

We now verify the quality of the jet calibration on two radically different physics samples. We consider first a $t\bar{t}$ sample, generated with MC@NLO [7]. The event generator used to produce the final state particles is HERWIG interfaced with JIMMY [8], whose fragmentation model is different from the PHYTHIA one considered so far. Moreover, most of the jets in $t\bar{t}$ events are produced by the fragmentation of quarks, while the QCD dijet events contain mostly gluon jets for moderate $E_T$. The simulation layout used for the $t\bar{t}$ events is the same distorted layout discussed in [9], i.e. a geometry with increased dead material in particular in the gap region.

We consider a cone algorithm with $R_{cone} = 0.4$ since it is the one most widely used for analyses concerning the top quark in ATLAS. The performance on QCD dijet events has been checked and the results are very similar to those discussed in Section 3.1 and in [9] for the cone with $R_{cone} = 0.7$

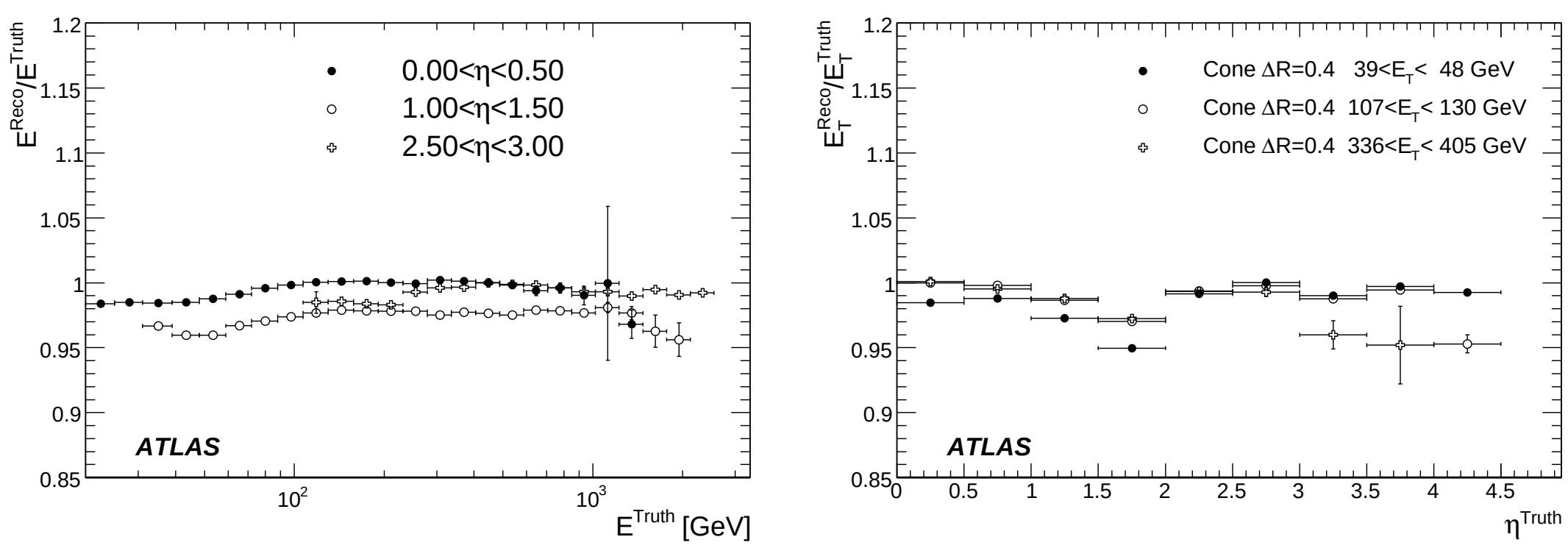

Figure 7: Linearity as a function of energy for three pseudorapidity regions (on the left) and of the pseudorapidity for three transverse energy bins (on the right) for cone 0.4 tower jets in $t\bar{t}$ events.

The linearity as a function of the energy and the pseudorapidity is shown in Fig. 7. Despite the fundamentally different event structure, an overall acceptable linearity is found also in top events.

We finally consider SUSY events. The sample has been generated with HERWIG/JIMMY and simulated with a distorted geometry. Such events are characterized by a high multiplicity of quark jets (the SUSY point chosen is discussed in [10]). They are thus useful to asses the performance of the jet calibration in busy events.

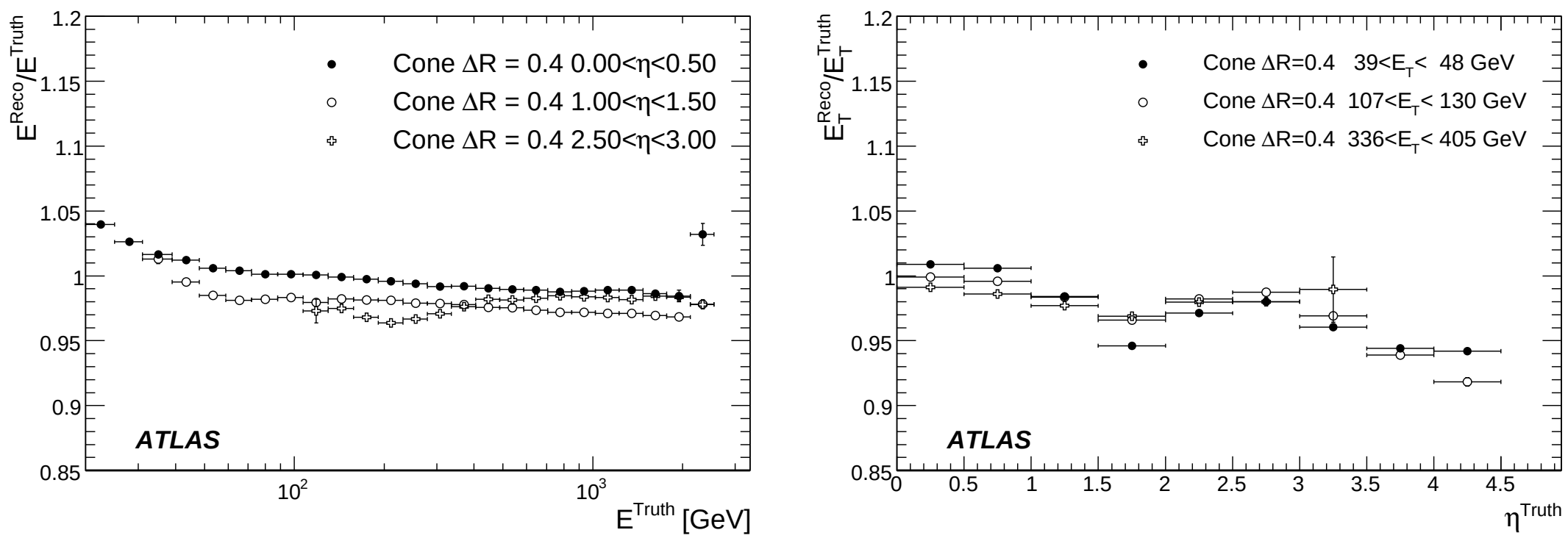

Figure 8: Linearity as a function of energy for three pseudorapidity regions (on the left) and of the pseudorapidity for three transverse energy bins (on the right) for cone 0.4 tower jets in SUSY events.

The linearity (Fig. 8) is overall good. In the central region, a deviation from 1 of maximum 4% is observed at low jet energy. Apart from the expected dip at $\eta = 1.5$, we observe, also, a good uniformity of the linearity as a function of the pseudorapidity. At large pseudorapidity ($|\eta| > 3.5$) , the linearity is off by 5-6%.

### 3.3 A check of the systematics with real data

We applied the method discussed so far to single pions from the ATLAS combined test beam of the year 2004 [11]. The weights have been computed on events fully simulated with GEANT4 using the test beam geometry.

Positively charged beams of different energy impinging with an incident angle of 20 degrees on the calorimeters surface have been considered. The beams are composed of pions, protons, positrons and muons. Signals from scintillators present upstream and downstream of the calorimeters are used as vetoes to reject early showering particles and muons, respectively. To reject the electrons we required an energy deposit in the first two layers of the electromagnetic calorimeter of less than a certain threshold (75% - 90% of the beam energy). A fraction of protons equal to that expected for the chosen beam line has been added to the simulated sample.

The energy distributions obtained for each energy point at the electromagnetic scale and after the calibration are fitted with a Gaussian function. The Gaussian mean values ($\langle E \rangle$) are used to evaluate the calibration procedure.

In Fig. 9 (left) the ratios $\langle E \rangle / E_{\text{beam}}$ are shown as a function of the beam energy. The black dots refer to the Monte Carlo at the electromagnetic scale. The black squares refer to the Monte Carlo after the weighting. The procedure restores the linearity at the 2% level for the simulation.

In the same figure the results on the real data are also shown (gray markers). In this case the linearity is also restored to within a few percent.

To evaluate these differences, we define the ratio $R = \langle E_{\text{data}} / E_{\text{MC}} \rangle$. Fig. 9 (right) shows the double ratio $R_{\text{HAD}} / R_{\text{EM}}$ of the points of Fig. 9 (left). This plot is showing the effect of the calibration procedure on the agreement between data and simulation. The agreement is worse after the application of the calibration procedure by maximum 4%.

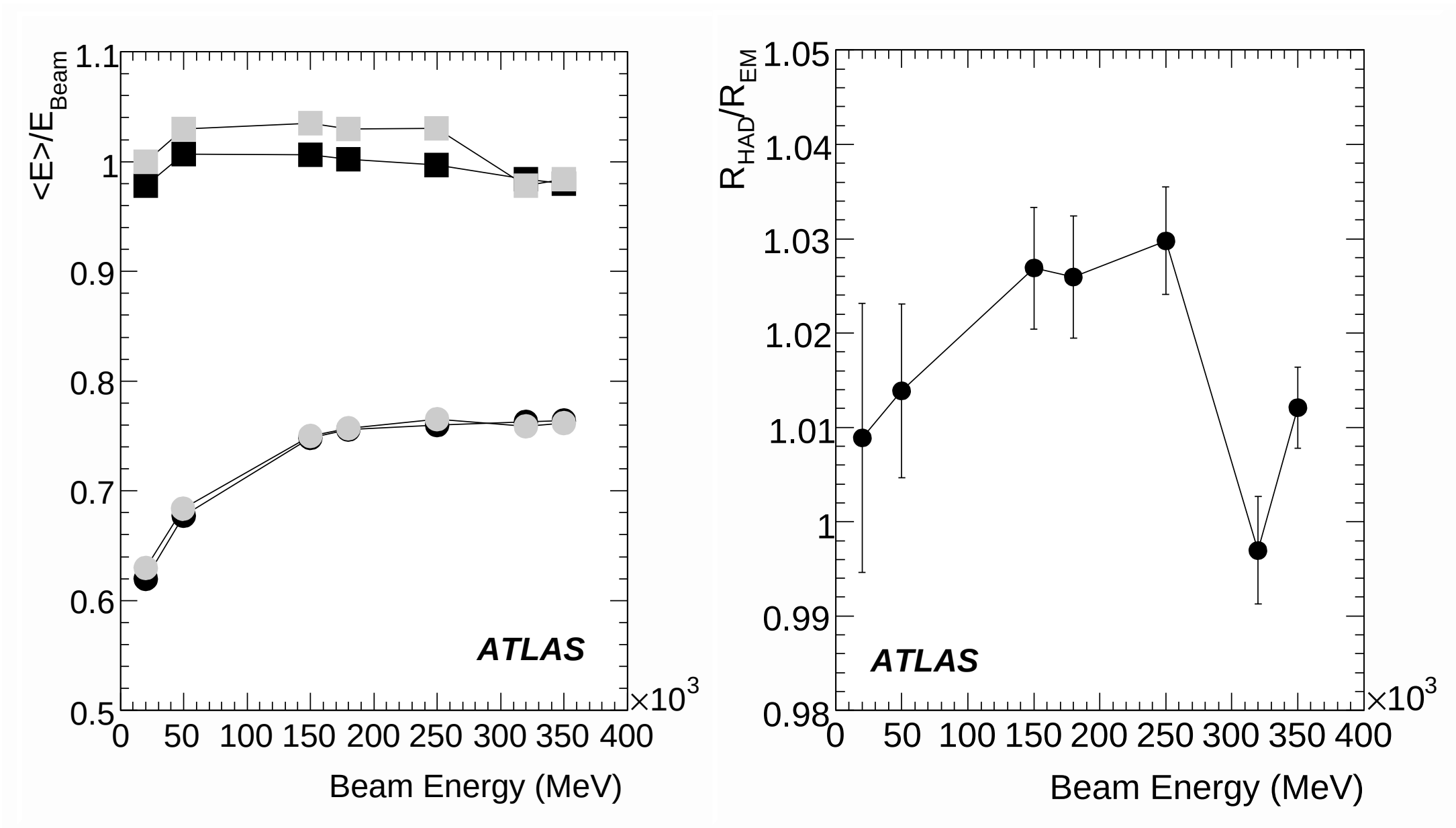

Figure 9: On the left: $\langle E\rangle/E_{\text{beam}}$ for simulated (black points) and real (gray points) data at the EM (dots) and calibrated (squares) scales. On the right: Double ratio $R_{\text{HAD}}/R_{\text{EM}}$.

### 3.4 Summary

The tests discussed in this Section are meant to demonstrate the robustness of the jet corrections computed as discussed at the beginning of the present Section. Summarizing, we can say that the discussed strategy is able to recover the linearity of the jet energy measurement over a wide energy range, invoking a relatively low number of parameters (of the order of 50) constrained by a fit on QCD dijet events. The fact that the jet corrections can be applied with success to events generated with different showering models, with a very different quark-gluon jet ratio and different topologies, gives confidence in the correction strategy as a method to remove the detector effects.

# 4 Alternative global calibration methods

Although the discussed calibration scheme is the most widely used at present in the ATLAS collaboration, it is not the only one that has been investigated.

## 4.1 Longitudinal shower development

On average, the early part of a hadron shower is dominated by electromagnetic energy deposited by neutral pions and the ratio of visible to invisible energy is large. In the deeper part of the shower this ratio becomes smaller and more of the hadron shower goes undetected. This can be seen in the a quantitative study that was carried out by the ATLAS TileCal collaboration in the 1996 test-beam [12]. It shows that in the first interaction length of the calorimeter approximately 70% of the energy of the hadron shower is deposited as visible electromagnetic energy. The fraction falls off with depth in the calorimeter and at $6\lambda$ only 25% the energy of the hadron shower is deposited as electromagnetic energy. Therefore a longitudinal weighting of energy deposition as a function of depth can provide improved resolution and linearity [13]. Figure 1 (right) shows the fraction of energy deposited by a hadronic jet at different depths in the calorimeter. The layers used in this weighting scheme are defined below based on the properties and geometry of different calorimeters.

The above motivation for longitudinal weighting is based on the average shower behavior. Hadron showers fluctuate event-by-event and in a jet, the incoming particle type and energy also varies depending on how the jet fragmentation proceeds. Figure 1 (left) shows the average energy carried by different particle types in a jet. To better account for these differences in shower fluctuation and electromagnetic content of a hadronic jet, the longitudinal weighting is performed in bins of the fraction of energy deposited in the LAr calorimeter. Furthermore, the Atlas calorimeter has a significant variation in geometry as a function of pseudo-rapidity and we therefore fit the parameters in independent bins of jet $\eta$.

As shown below, a longitudinal weighting based on the above properties of hadron shower development and jets shows a significant improvement in jet energy resolution and linearity with respect to uncorrected jet energy.

### Longitudinal weighting method

In general the choice of energy layers to be weighted are motivated from the following. In the barrel LAr calorimeter, the first three depths (presampler, EMB1, EMB2) provide a total of $24X_0$. This gives 99% containment for photons with energies up to 140 GeV [14]. For hadronic jets, the energy in these layers is expected to be predominately from neutral pions. The presampler and EMB1 are defined as a single layer for longitudinal weighting purposes. The weight for this layer is expected to be sensitive to energy losses in the inner detector. EMB2, which has $18X_0$ is weighted alone. The EMB3 is thin and has small energy deposit. Therefore the energy in this layer is added to the energy in the first layer of the Tile calorimeter. This allows for simulations to provide an average correction for energy loss in the cryostat. Depending on the jet pseudorapidity, a different number of calorimeter layers are used in the fitting. Up to a pseudorapidity of 1.5 the jet energy is fitted in four layers in the calorimeter defined as follows:

$$\begin{aligned} E_0 &= E_{\mathrm{presampler}} + E_{\mathrm{EMB1}} \\ E_1 &= E_{\mathrm{EMB2}} \\ E_2 &= E_{\mathrm{EMB3}} + E_{\mathrm{Tile1}} \\ E_3 &= E_{\mathrm{Tile2}} + E_{\mathrm{Tile3}} + E_{\mathrm{HCAL}}. \end{aligned}$$

For a jet $\eta$ between 1.5 and 3.2 the jet is fitted for two layers in the calorimeter defined as follows.

$$E_0 = E_{\mathrm{presampler}} + E_{\mathrm{LArEM1}} + E_{\mathrm{LArEM2}}$$

$$E_1 = E_{\text{LArEM3}} + E_{\text{Tile1}} + E_{\text{Tile2}} + E_{\text{Tile3}} + E_{\text{HCAL}} + E_{\text{FCAL}}$$

Beyond $\eta$ of 3.2 up to 4.4 the jet is not divided into calorimeter layer segments and the full jet energy is fitted.

The general strategy of deriving weights for different layers is to minimize the function:

$$S = \sum_n \left[ \left( E_n^{\text{Ref}} - E_n^{\text{rec}} \right)^2 + \lambda \left( E_n^{\text{Ref}} - E_n^{\text{rec}} \right) \right] \tag{10}$$

with

$$E_n^{\text{rec}} = \sum_i w_i E_i \tag{11}$$

where the $w_i$ are weights assigned to the elements $E_i$ of a calorimeter layer in a jet. $E_n^{\text{Ref}}$, the true energy to which we want to calibrate, is defined as the energy of all the MC generated particles contained in the cone of the reconstructed jet. The Lagrange multiplier $\lambda$ constrains the minimization such that $\langle E^{\text{Ref}} - E^{\text{rec}} \rangle = 0$. The minimization is performed separately for jets classified in bins of eta (44 eta bins of size 0.1), three fractional energies ($f_{\text{em}}$) deposited in the EM calorimeter and two energy bins. The fractional energy $f_{\text{em}}$ is defined as

$$f_{\text{em}} = (E_{\text{Presampler}} + E_{\text{LArEM1}} + E_{\text{LArEM2}})/E^{\text{rec}}. \tag{12}$$

Three bins in $f_{\text{em}}$ are chosen such that each bin has roughly the same statistics. At high energies the three bins are (small: 0.0-0.65), (mid: 0.65-0.75) and (large: 0.75-1.0). At low energies they are (0.0-0.75), (0.75-0.85) and (0.85-1.0). The bin size varies with energy since more energetic jets deposit more energy in the deeper part of the calorimeter. Bins in $f_{em}$ are used only for jets with $\eta < 3.2$.

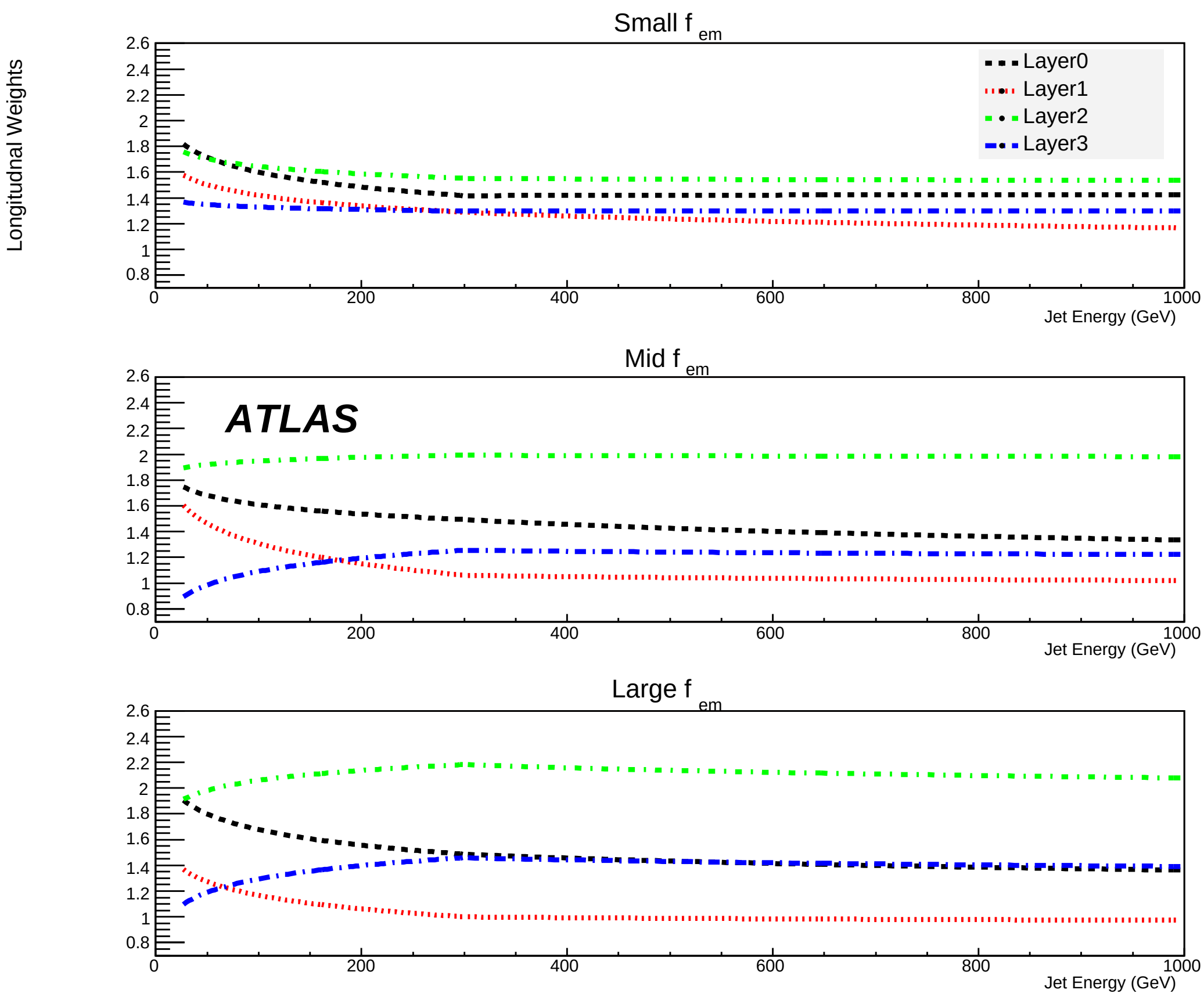

Figure 10: The longitudinal weights as a function of jet energy for four layers in three $f_{\text{em}}$ bins and for central jet $\eta$.

For each layer, weights are chosen to have the following dependence on the true jet energy:

$$w = a + b\log(E/E_{Cut}), \tag{13}$$

where $a$ and $b$ are the parameters to be determined by minimization. When applying these weights to jets, the uncorrected jet energy is used instead of the true jet energy. The result is iterated until a stable value of the corrected jet energy is obtained.

In the above equation, $E_{Cut}$ is an arbitrary energy chosen according to the following criteria. Since the energy range covered is quite large (25 GeV - 2 TeV) the fit is performed in two independent energy bins. For $|\eta| < 1.2$, $E_{Cut} = 300$ GeV, for $1.2 < |\eta| < 3.2$, $E_{Cut} = 450$ and for $|\eta| > 3.2$ it is set to $35 \cdot \cosh(\eta)$ GeV. By choosing $E_{Cut}$ to be the bin boundary one forces the weight to be equal to the value of $a$ at the boundary. The energy range below $E_{Cut}$ is fitted with a fixed value of $a$. This ensures a smooth behavior of weights across $E_{Cut}$ and reduces the lower energy fit from a two parameter to a one parameter fit. The same smoothness in the behavior of weights across $\eta$ and $f_{em}$ is not strictly imposed in the present fit, although the weights do not have strong variation across the bin boundaries.

The fitting procedure was applied to a fully simulated and reconstructed QCD dijet sample. To suppress noise, topological clusters were used. The jet algorithm (cone $R_{cone} = 0.7$) is run on calorimeter towers which contain only cells which are included in the reconstructed topological clusters. Half of the events in the sample were used to determine the layer weights. These weights were then applied to the other half of the events to determine the effect of the weights on jet energy linearity and resolution.

Figure 10 shows the behavior of the weights in bins of $f_{em}$. A common feature in all the weight distributions is a small variation with respect to the jet energy, especially for high energies. This ensures insensitivity to the use of the true jet energy in equation 13. Layer 2, which measures the bulk of the jet energy has a weight close to 1 when $f_{em}$ is large i.e when the jet is predominantly electromagnetic in nature. When $f_{em}$ is large i.e when the jet is predominantly hadronic, the layer 2 weights are around 1.4. Layer 1 acquires a generally higher weight due to losses in the inner detector, even though it is within the early part of the jets. Layer 3 and 4 get weights larger than 1 corresponding to jets being predominantly hadronic in these layers. Layer 3 gets larger weights than layer 2 since it also corrects for energy lost in the cryostat.

Figure 11 shows the jet energy scale linearity as a function of jet energy (left). The corrected jet energy scale is linear to about 2% with the largest non-linearity coming from low energies where the uncorrected non-linearity is approximately 30%. Figure 11 (right) shows the corresponding linearity as a function of detector pseudo-rapidity for jets of 1000 GeV in energy. The typical non-uniformity is about 1%, increasing to about 2% in the region of $\eta \sim 3.0$. Jet energy resolutions as a function of the jet energy scale is shown in Fig. 12 for two different jet eta regions. At high energy the jet resolution approaches about 4%.

In terms of jet energy linearity, the longitudinal weighting and H1-style weighting scheme (Section 3) have comparable performance, although the H1-style weighting scheme shows a slightly better resolution. This is expected since H1-style weighting uses local cell energy density to discriminate between EM and non-EM like energy deposits. In contrast, longitudinal weighting is less sensitive to local energy fluctuations, which may be an advantage in the early data taking period, when the simulation of energy deposition in the calorimeter may not accurately reproduce the data.

## 5 Local hadron calibration

In contrast to the global calibration method just described, where first jets are made from towers or clusters on the electromagnetic scale and the calibration is applied after jet-making on cell or sampling level, the local hadron calibrated jets are made from clusters which are already calibrated to the hadronic scale.

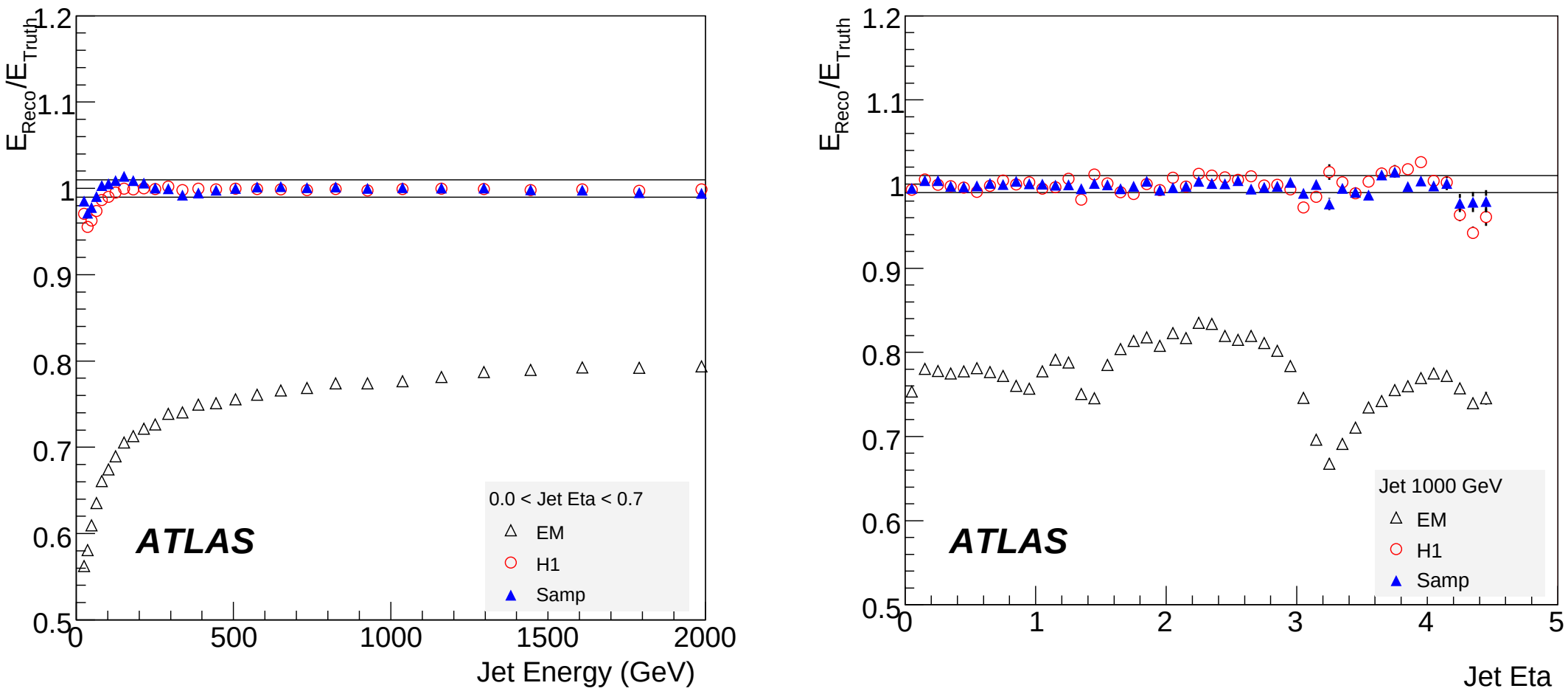

Figure 11: Jet energy linearity as a function of jet energy (left), and as a function of jet pseudorapidity (right). The points are for jets reconstructed at the electromagnetic scale (EM), for the global weighting scheme described here (Samp) and for the H1-style calibration described in the previous Section. The jets have a cone radius of $R_{\text{cone}} = 0.7$.

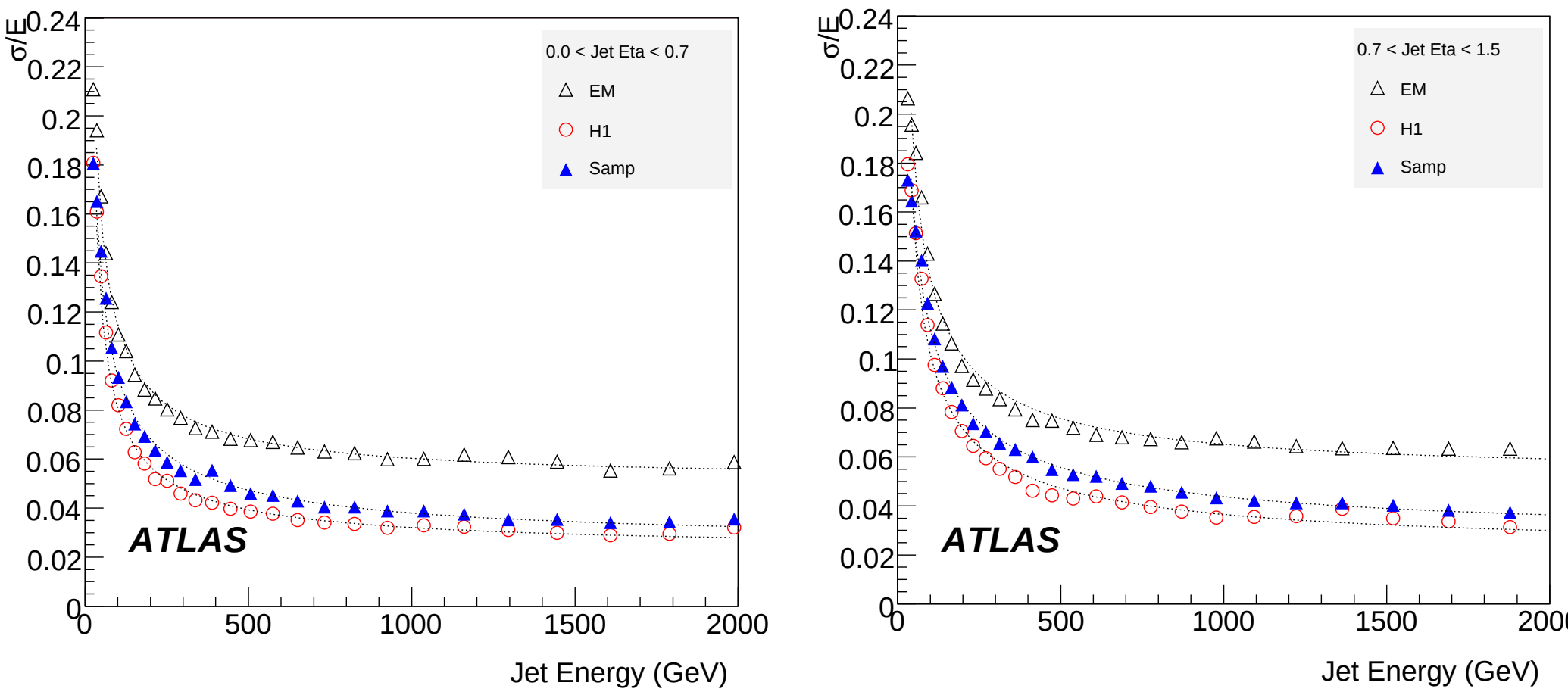

Figure 12: Jet energy resolution for jets with a cone radius of 0.7 for two regions in pseudorapidity. The three sets of point show the resolution at the detector (EM) scale, after H1-style and longitudinal weighting.

### 5.1 Topological clusters

The cluster algorithm used is described in detail in Ref. [15]. Clusters grow dynamically around seed cells based on noise thresholds and are re-grouped in a second splitter step around local maxima.

The aim of the clustering step before the actual jet making is two-fold:

1. To suppress noise from electronics and pile-up by reducing the number of cells included in the jets via noise-driven clustering thresholds.

2. To improve the correspondence between clusters and particles. Due to the dynamic nature of the cluster growing, individual clusters correspond better to stable particles than towers or cells and the jet constituents can serve to further study the substructure of jets.

To illustrate the effect of noise reduction by using topological clusters as input to jets the amount of noise at the electromagnetic scale and the number of cells per jet for cone jets with $R_{\mathrm{cone}} = 0.7$ is compared in Fig. 13 for jets from dijet simulations with towers as input and with topological clusters as input. The noise reduction is a direct consequence of selecting fewer cells with topological clusters. The effect is largest for low energetic jets since the size and number of signal clusters becomes small. The number of cells per jet for tower jets does not depend on the energy since no threshold for the towers is applied. Subsequently the noise changes only if a cell included in a tower jet switches to a lower gain. For the displayed energies this effect is visible in the forward region only. For jets from topological clusters the noise increases with energy and at transverse energies of 150 GeV it is typically a factor of 2 lower than for a corresponding tower jet except for the very forward region where the signals are so dense that the topological clusters again include almost all cells and the noise level reaches that of the tower jets.

Figure 14 shows the correspondence of clusters and stable truth particles in a dijet simulation. The sample shown is a PYTHIA QCD dijet sample with the transverse energy of the leading jet between 140 GeV and 280 GeV. Roughly 1.6 truth particles correspond to each of the $65 \pm 30$ clusters for a cut at 1 GeV in transverse energy, but the ratio does not depend on the cut. It is close to the expected ratio of $\sim 4/3$ since about $1/2$ of the stable particles in jets are photons from $\pi^0$-decays, which usually merge to one cluster.

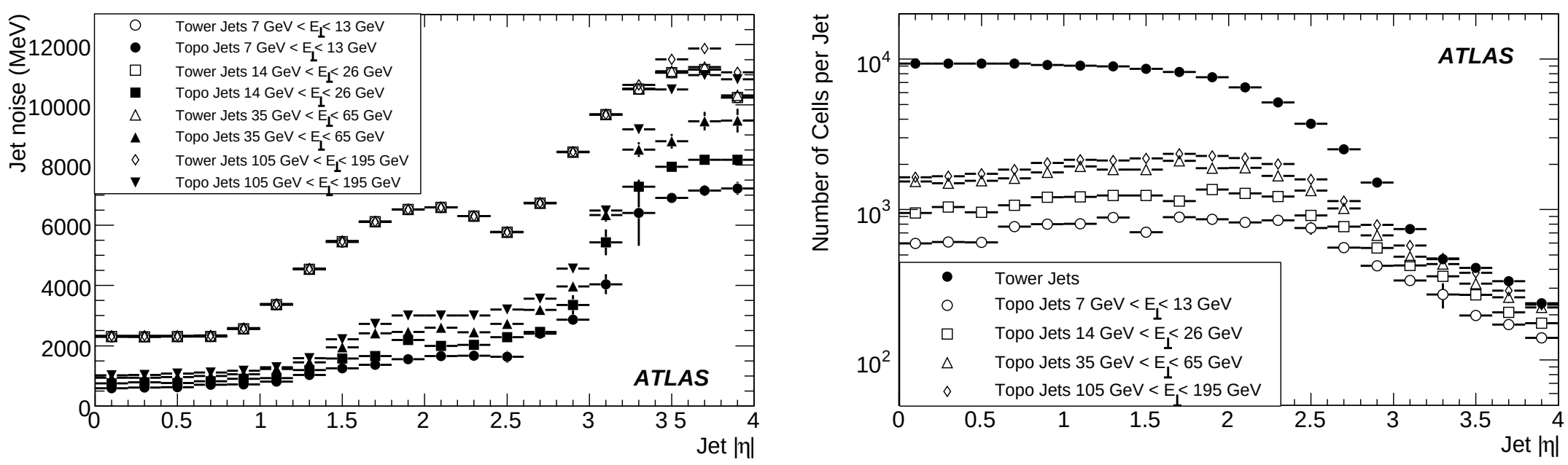

Figure 13: Noise contents (left) and number of cells per jet (right) of Cone jets with $R_{\mathrm{cone}} = 0.7$ for different energies for towers as input (open symbols) and topological clusters as input (filled symbols).

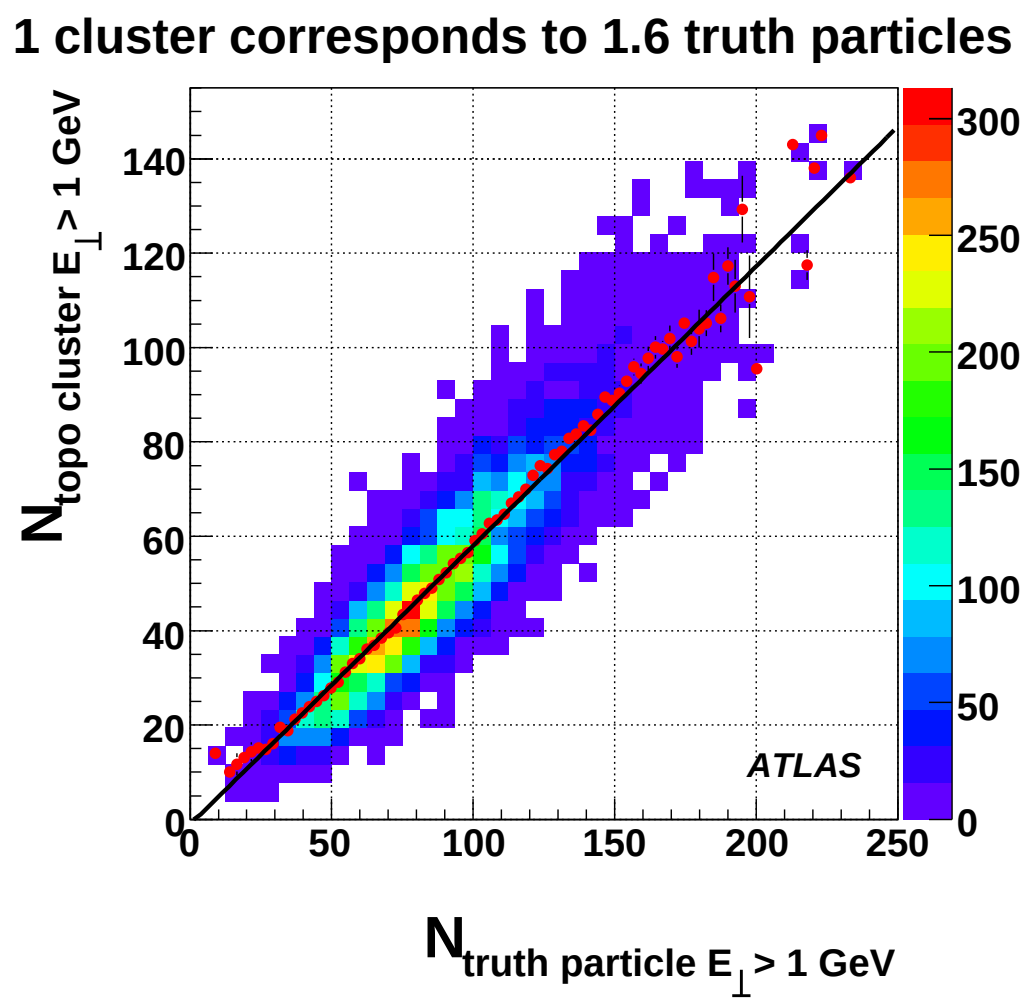

Figure 14: Number of topo clusters with $E_{\perp} > 1\,\mathrm{GeV}$ vs. number of stable truth particles with $E_{\perp} > 1\,\mathrm{GeV}$ from a QCD dijet simulation.

## 5.2 Cluster Calibration

The local hadron calibration of topological clusters is described in detail in Ref. [16]. The calibration starts by classifying clusters as mainly electromagnetic, hadronic, or unknown depending on cluster shape variables, moments derived from the positive cell contents of the cluster and the cluster energy. The classification is based on predictions from GEANT4 [17, 18] simulations for charged and neutral pions. The expected phase space population in logarithmic bins of the cluster energy, cluster depth in the calorimeter, and average cell energy density and linear bins in $|\eta|$ from neutral and charged pions with a ratio of $1:2$ is converted to a classification weight, reflecting the a-priori assumption that $2/3$ of the pions should be charged.

Roughly 90% of the energy of charged pions is classified as hadronic by this procedure for all energies, while for neutral pions 90% of their energy is classified as electromagnetic on average beyond 100 GeV and the performance drops with the logarithm of the pion energy to about 50% at 10 GeV. The ideal fraction of 100% is not reached for the charged pions as sometimes the shower is split into more than one cluster with one of them being predominantly electromagnetic in nature. At low energies neutral pion clusters occupy the same phase space as charged pion clusters and the a-priori precedence for charged pions makes the classification as electromagnetic less likely. This leads to the high fraction of neutral pion energy classified as hadronic at low energies which is still acceptable, since the weights applied here are close to 1. Clusters classified as hadronic receive cell weights derived from detailed GEANT4 simulations of charged pions with so-called calibration hits in active and inactive calorimeter materials, which contain the energy from ionization losses and also from invisible processes, such as nuclear excitation, and from escaping particles, such as neutrinos. Cells in individual calorimeter samplings are treated in 0.2-wide $|\eta|$-bins. The weights are binned logarithmically in cluster energy and cell energy density. A flat distribution in the logarithm of the particle energy was used to generate the single pion events.

Out-of-cluster (OOC) corrections are applied to correct for energy deposits inside the calorimeter but outside calorimeter clusters due to the noise thresholds applied during cluster making. These corrections depend on $|\eta|$, cluster energy and the cluster depth in the calorimeter.

Dead material (DM) corrections are applied to compensate for energy deposits in materials outside of the calorimeters. For deposits in upstream material like the inner wall of the cryostat the presampler signals are found to be highly correlated with the lost energy and the corrections are derived from the sum of calibration hit energies in the upstream regions and the presampler signal.

The correction for energy deposited in the outer cryostat wall between the electromagnetic and hadronic barrel calorimeters is based on the geometrical mean of the energies in the samplings just before and just beyond the cryostat wall. Corrections for other energy deposits without clear correlations to cluster observables are obtained from lookup tables binned in cluster energy, $|\eta|$, and shower depth.

### 5.3 Performance for jets

The aim in this section is to evaluate the degree of completeness of the local hadron calibration when applied to jets. The performance of the local hadronic calibration scheme was evaluated using the dijet samples listed in Table 2 and two methods: by comparison to particle jets as described above in section 3.1, and by comparison to the calibration hits in the GEANT4 record. Since no truth matching occurs in the derivation of the calibration constants genuine jet-level effects are expected to be visible once the reconstructed jet is compared to the matching jet made of stable truth particles. The main sources of remaining energy corrections are:

**Misclassification** Hadronic energy deposits which are treated as electromagnetic lead to a lower energy response, while electromagnetic energy deposits wrongly treated as hadronic lead to a higher energy response. The effect of energy underestimation dominates and is roughly 3% for $p_\perp \simeq 150\,\text{GeV}$.

**Lost Particles** Low energetic particles might be bent outside the acceptance cone of the reconstructed jet or reach the calorimeter inside the acceptance cone but leaving a signal below threshold for the clustering. Both effects are estimated to add up to 5% for $p_\perp \simeq 150\,\text{GeV}$, with 3% stemming from low energy deposits not included in the clusters and 2% from particles bent outside the acceptance cone.

Jets formed from topological clusters, calibrated using the local hadronic calibration scheme, were compared to the truth jets as done with the previous calibration methods. Figure 15 shows the linearity for Cone $R_{\text{cone}} = 0.7$ and $k_T$ $R = 0.6$ dijets for 3 different $|\eta|$ regions as a function of the jet energy. The performance in the forward region is especially low because of a scale error of 10% introduced in the simulations[1]. The forward scale error highlights another strength of the calibration hits – they can in fact reveal that there is a problem in the predicted reconstructed energy. All calibration methods discussed in this note would yield an overestimation of the jet energy in the forward region in real data if this simulation problem would not be fixed in the samples needed to derive the calibration constants.

In the other pseudo-rapidity regions the linearity is rising from 80% at 30 GeV to over 95% at 1 TeV. Figure 16 shows the linearity as a function of the true jet $|\eta|$ for 4 different jet energies. Dips in the linearity are clearly visible for the transition regions between the calorimeter systems at the gap region ($1.3 < |\eta| < 1.5$) and the crack region ($3.0 < |\eta| < 3.5$). For the forward region, the mentioned scale error is very clearly visible: the linearity cannot be recovered and the scale is off by 10%. The dependency of the linearity on the jet energies can also be observed in these plots: while the linearity for jets with about 100 GeV can be recovered up to about 90% (85% in the gap region), high energy jets (of about 1 TeV) show a linearity of about $95-97\%$ which is compatible with the 3% loss due to misclassification as discussed above. Although a simple scale function, such as given in Eq. (5) on the jet level, would restore the linearity and give a comparable performance as the global calibration, our goal is to understand and correct for these effects in order to recover the linearity instead by correction functions based on the jet constituents.

[1] The assumed sampling fraction did not correspond to the actual sampling fraction in the FCal region and thus the assumption that electromagnetic showers can remain un-scaled leads to an underestimation of the energy.

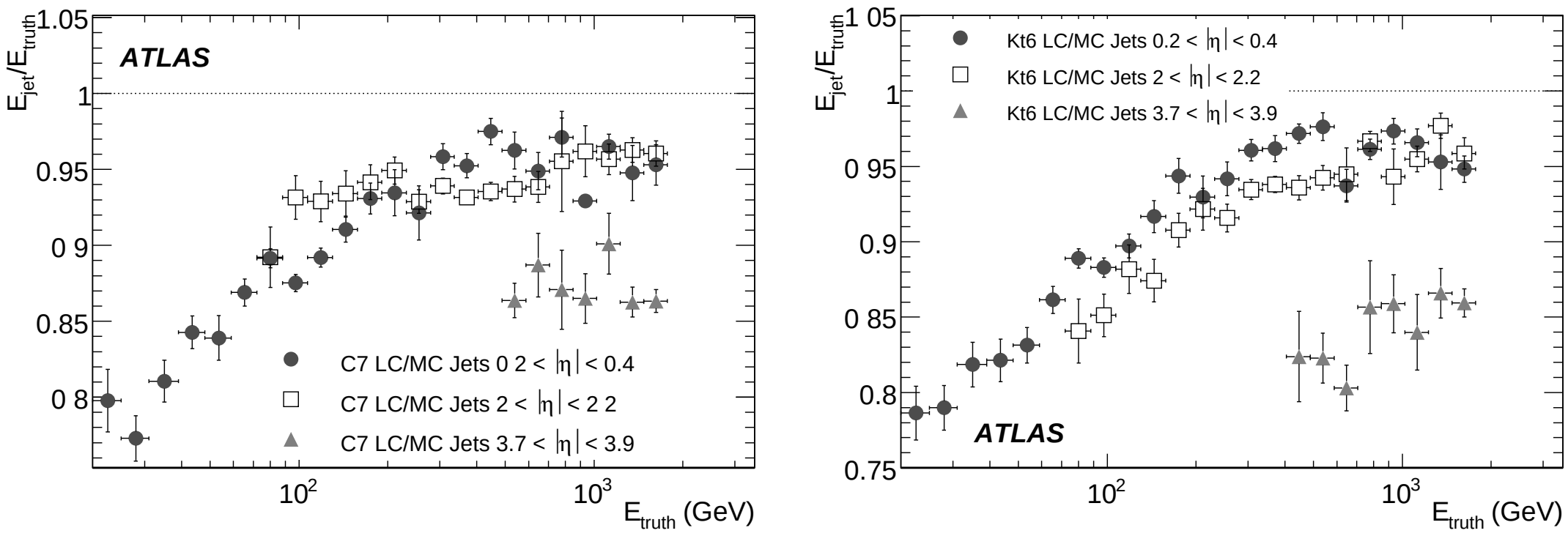

Figure 15: Linearity for Cone jets with $R_{\mathrm{cone}} = 0.7$ (left) and Kt jets with $R = 0.6$ (right), both calibrated with the local hadron calibration method (LC), using truth particle jets (MC) as reference. The linearity is shown as a function of the matched truth jet energy.

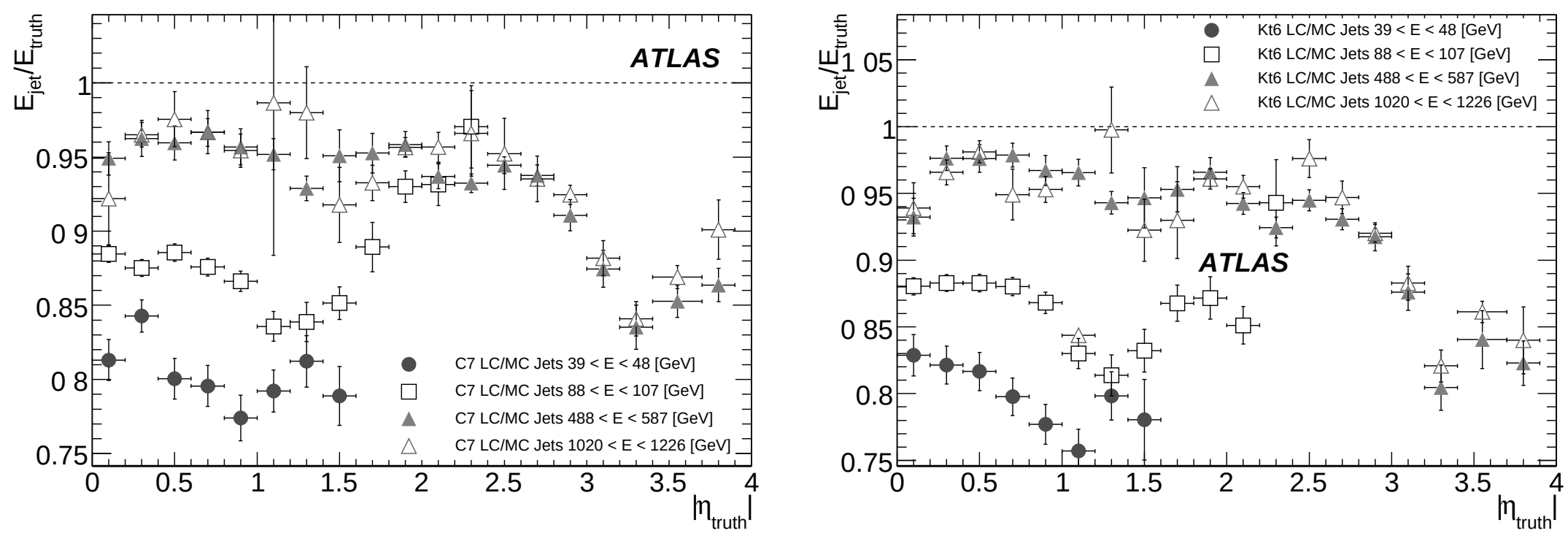

Figure 16: Linearity for Cone jets with $R_{\mathrm{cone}} = 0.7$ (left) and Kt jets with $R = 0.6$ (right), both calibrated with the local hadron calibration method, using truth particle jets as reference. The linearity is shown as a function of the matched truth jet $|\eta|$.

The jet energy resolution is shown in Fig. 17 as a function of the true jet energy. Table 4 shows the parameterised resolution obtained using this method as a function of energy and rapidity. It's performance is typically 20% or more above that obtained using the global calibration method. We discuss some possible improvements to the local hadronic calibration method below.

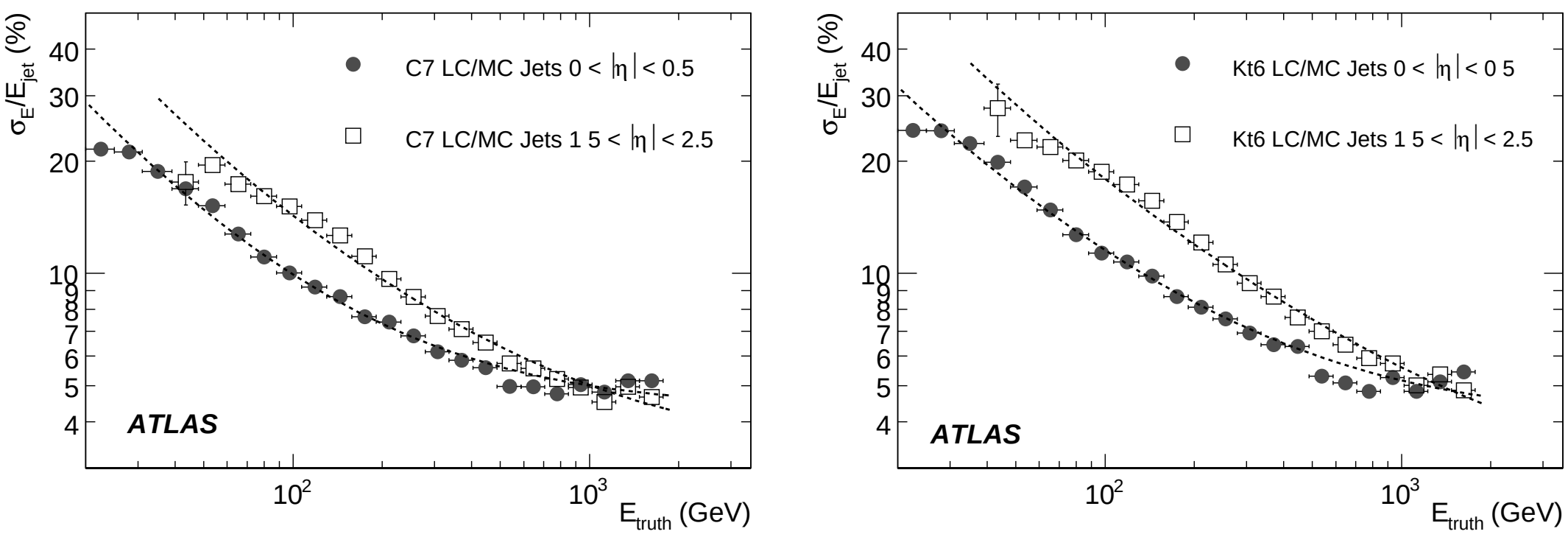

Figure 17: Resolution for Cone jets normalized to the reconstructed jet energy with $R_{\mathrm{cone}} = 0.7$ (left) and Kt jets with $R = 0.6$ (right), both calibrated with the local hadron calibration method, using truth particle jets as reference. The resolution is shown as a function of the matched truth jet energy.

Table 4: Resolution as function of $E_{\mathrm{true}}$ for jet with the local hadron calibration applied.

| Reconstruction Algorithm | $0 < \lvert\eta\rvert < 0.5$ | | | $1.5 < \lvert\eta\rvert < 2.0$ | | |
|---|---|---|---|---|---|---|
| | a (%) | b (%) | c (GeV) | a (%) | b (%) | c (GeV) |
| Cone $R_{\mathrm{cone}} = 0.7$ LC | $78 \pm 8$ | $3.5 \pm 0.8$ | $2.3 \pm 0.9$ | $98 \pm 14$ | $7.7 \pm 1.7$ | $3.3 \pm 0.7$ |
| $k_T$ $R = 0.6$ LC | $79 \pm 8$ | $4.7 \pm 0.7$ | $2.4 \pm 0.6$ | $117 \pm 15$ | $9.7 \pm 1.9$ | $1.2 \pm 2.3$ |

A detailed analysis of the performance of the local calibration when applied to jets is also presented in Ref. [19], where different local calibration approaches are compared to the performance of the H1 global calibration.

## 5.4 Further Improvement

As seen in the previous section several jet-level corrections need to be applied in order to bring jets made of local hadron calibrated topological clusters to the truth particle scale. However, the global method is seen to exhibit somewhat superior performance indicating that further improvements should be possible since both methods use shower development as their fundamental basis.

Figure 18 (left) shows the ratio of the corrected energy as obtained from the reconstructed calorimeter cells to the energy obtained from the GEANT4 calibration hits in the clusters and in dead material. A significant deficit is seen at low energy. Figure 18 (right) shows the ratio of the the energy obtained from the GEANT4 calibration hits in the clusters and in dead material to the energy of the nearest truth particle jet. In this case we see roughly unity at high energy and only a 10% deficit at low energy. Since this comparison is to the truth particle jet, we attribute this deficit to out of cone energy. We therefore conclude that the dominant effects on the non-linearity at low energies seen in Fig. 18 stem from particles lost in the dead material upstream of the calorimeters. These low energy pions deposit most of their energy in upstream materials and often do not leave a sufficiently large signal in the calorimeters to cause a cluster to be formed. At present, no good observable on the local cluster level aids in recovering this lost energy and the local calibration method can not account for it. Corrections for these effects are currently being studied. A scaling function like Eq. (5) which is used in the global method would help to restore the linearity in Fig. 15 but would not improve the resolution. The generalization of cluster shape variables to the jet level (number of low energetic constituent clusters, energy distribution of the constituent clusters, etc.) might help in order to obtain correction procedures that depend only indirectly on the used jet algorithm, restore the linearity and improve the resolution. The missing energy content can for example be estimated by extrapolating the actual distribution of constituent cluster energies to zero GeV to recover the lost contributions from low energetic particles. The in-situ methods as discussed in Ref. [20] can be used to validate the corrections obtained and to possibly compensate residual non-linearities.

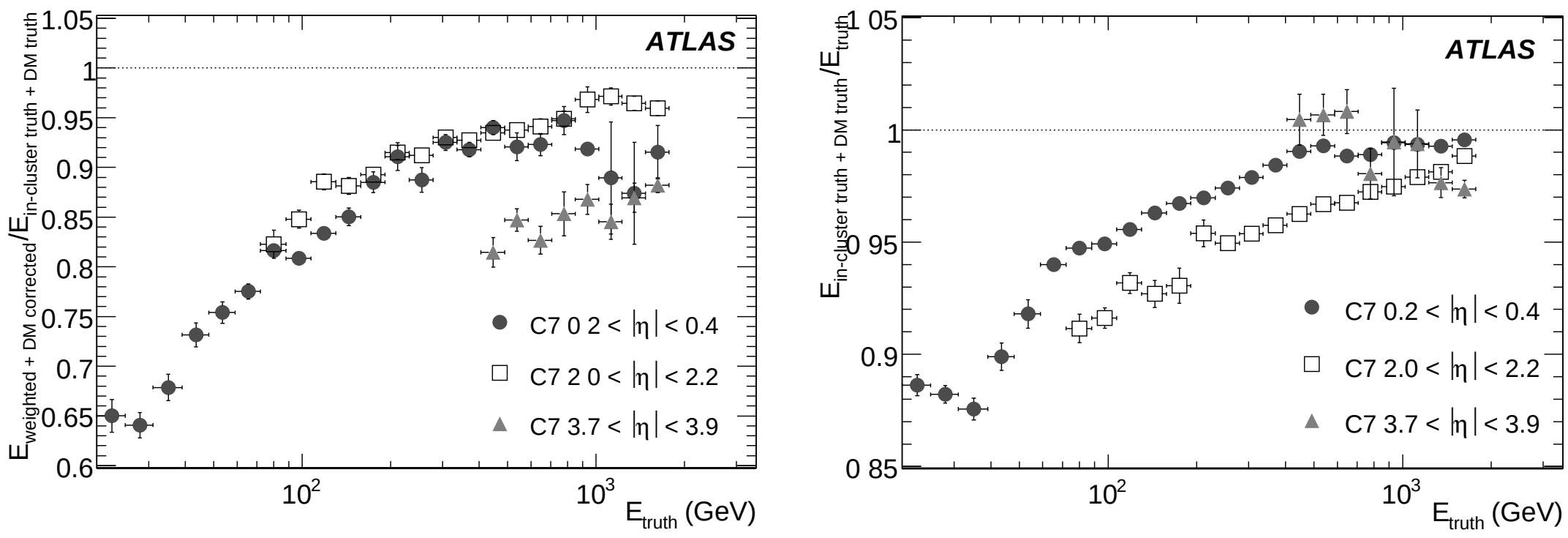

Figure 18: $E_{\text{weighted + DM corrected}}/E_{\text{in cluster truth + DM truth}}$, the reconstructed weighted and dead-material corrected energy over the the predicted true energy inside clusters and associated dead material regions (left) and $E_{\text{in cluster truth + DM truth}}/E_{\text{truth}}$, the ratio of the predicted true energy inside clusters and associated dead material regions (the denominator in the left plot) over the energy of the matched truth particle jet (right) as function of the matched truth jet energy for cone jets with $R_{\text{cone}} = 0.7$.

# 6 Track-based improvement in the jet energy resolution

We present a track-based method for improving the jet energy resolution in ATLAS. Unlike energy-flow techniques reference, information is added to the reconstructed jet, after the global jet energy scale corrections have been implemented, and the track-based correction is applied based on the fraction of jet momentum carried by charged tracks associated with the jet. Using this correction, a $\sim 20\%$ improvement in jet energy resolution at low energy is achieved.

In this chapter we describe a technique that uses tracks in jets to extract information from the jet topology and fragmentation in order to improve the jet energy resolution. The approach is conceptually different from more traditional energy flow methods, where precise track momentum measurements replace calorimeter clusters. In the proposed technique, tracks are used to correct the response of jets as a function of the jet particle composition, specifically using the ratio of track to calorimeter transverse momentum ($f_{\text{trk}} = \frac{p_T^{\text{tracks}}}{p_T^{\text{calorimeter}}}$). Using $f_{\text{trk}}$ provides an improvement in jet energy resolution without changing the jet energy scale applied during reconstruction.

In general, jets are composed primarily of neutral and charged pions. Charged particles leave tracks in the detectors, and so one might naively expect that approximately two-thirds of the jet energy will be carried by tracks associated with that jet. Monte Carlo QCD dijet samples show Gaussian $f_{\text{trk}}$ distributions centered around 0.66, with small tails extending above 1. The tails are more prominent at low energies and include, for example, jets with a true $f_{\rm trk}$ near one and one or more tracks with incorrectly measured momenta.

The fractional jet energy resolution, $\frac{\sigma\left(p_T^{\rm reco}-p_T^{\rm true}\right)}{p_T^{\rm true}}$, is proportional to the width of the jet energy response in bins of transverse energy, normalized to the average jet energy in a bin. If the response of these jets varies significantly with $f_{\rm trk}$, the transverse jet energy resolution will be artificially broadened, as shown in Fig. 19. One sees that the total measured transverse jet energy resolution is considerably wider than either of the constituents corresponding to jets with different charged particle fractions. By correcting the jet response as a function of jet $p_T$ and $f_{\rm trk}$ we reduce the overall broadening of the energy distribution and, hence, improve the jet energy resolution.

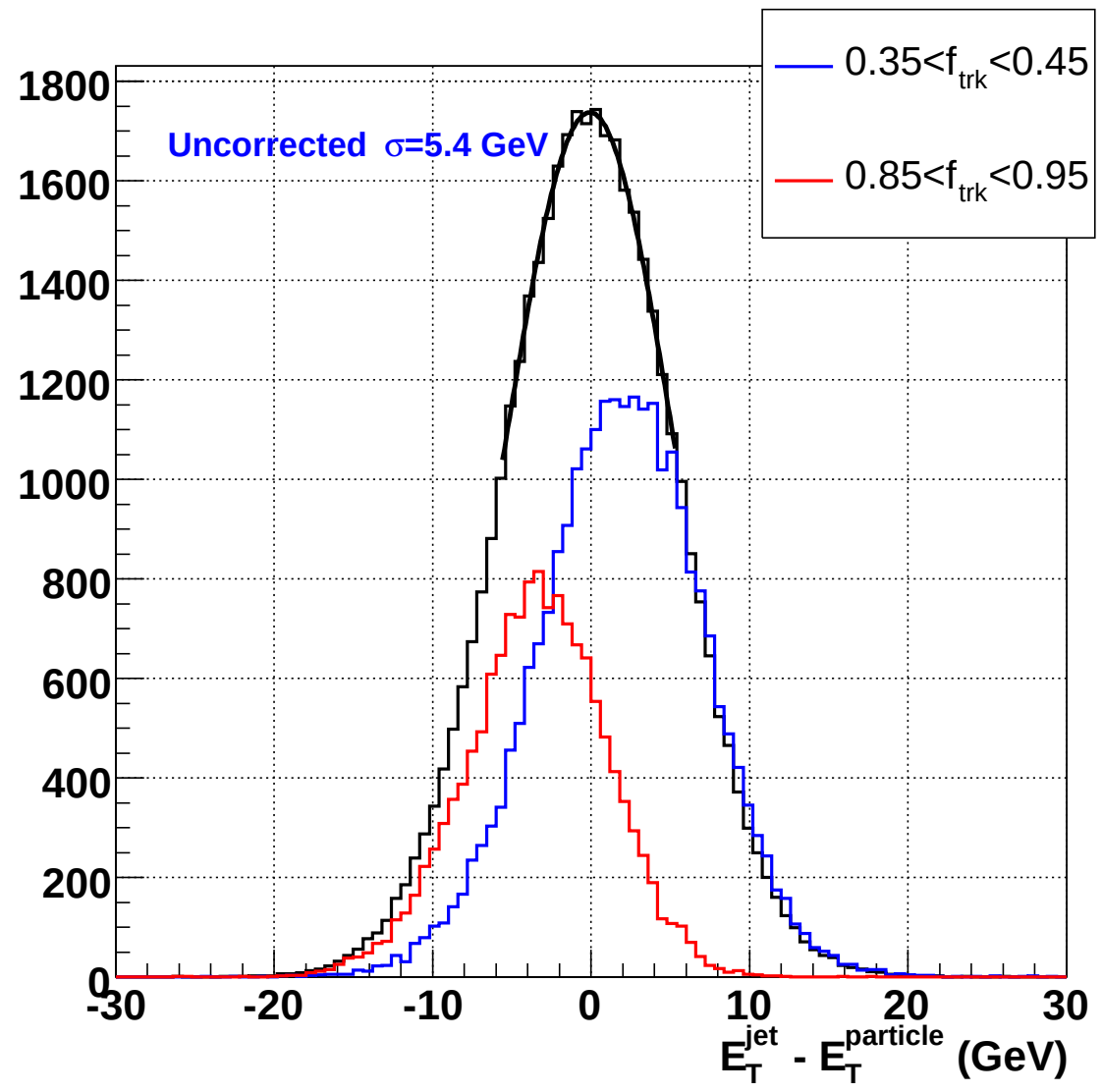

Figure 19: Black: difference between reconstructed and truth jet transverse energy for jets with $0 < |\eta| < 0.7$ and 40 GeV $< p_T < 200$ GeV. The mean (width) of this distribution is proportional to the jet energy response (resolution). Since jets with different $f_{\rm trk}$ have different responses, the transverse energy resolution is artificially broadened because of the offset of the distributions for each $f_{\rm trk}$ bin. The normalization is arbitrary.

### 6.1 Monte Carlo samples and event selection

Track based jet corrections were determined using QCD dijet events. The MC events used are the same J1-J4 samples described in Table 2. The reconstructed jets in the samples ranged in $p_T$ from 7 GeV to 280 GeV. Only jets with energies above 45 GeV were used in the fits.

Reconstructed 0.4 cone, tower-seeded jets were selected from the event and separated into bins of pseudorapidity. Fits were formed for the central ($|\eta| < 1.2$), transition ($1.5< |\eta| <1.8$) and end cap ($1.9< |\eta| <2.2$) regions of the calorimeters. Because the ATLAS tracker acceptance ends at $|\eta|$=2.5, jets beyond $|\eta| = 2.0$ were not considered for fits.

Tracks within a cone of radius 0.4 in $\eta - \phi$ around a jet axis were included in the calculation of $f_{\rm trk}$. In order to remove jets with a single poorly measured track, jets were required to be associated with at least two tracks, and each track was required to have $\chi^2/DoF < 3.0$. Less than 1% of the tracks and jets were rejected by these cuts. The requirement of two tracks was only applied to derive the correction and not to evaluate the performance. In order to be as inclusive as possible, all calorimeter jets were considered. To find the track-based energy corrections, jets were also required to be isolated in a cone of 0.8 in $\eta - \phi$ space, to avoid jet-jet contamination which skew $f_{\text{trk}}$ considerably for low energy jets. After cuts, each jet contained typically 5 to 8 tracks.

Jets were also required to have a truth jet matched within 0.1 in $\eta - \phi$ space, and were required not to be matched to a truth $b$-quark. Jets with $b$-quarks could have different energy responses compared to light quark jets and were excluded from the fits. Incorporating a track-based $b$-jet energy correction is a topic for future studies.

### 6.2 Track-based jet energy response parameterization

Fits to the jet energy response are made as functions of $f_{\text{trk}}$ and $p_{\text{T}}$. The fits were also binned in regions with flat response as a function of pseudorapidity. Three such regions were identified: one in the barrel calorimeter, one in the endcap calorimeter, and one in the transition region between the two. The fits are then extended to nearby regions so that all jets are eventually corrected. Since the transitions between the calorimeters are discontinuous, the transitions between the corrections were discontinuous for this first version of track corrections.

One dimensional fits of the response as a function of jet $p_{\text{T}}$ were performed in bins of $f_{\text{trk}}$ with the function $R^{f_{\text{trk}}}(p_{\text{T}}) = E_T^{\text{jet}}/E_T^{\text{true}} = a\left(1 - e^{b-c*p_{\text{T}}}\right)$. Sample fits are shown in Fig. 20 (left). The bins in $f_{\text{trk}}$ were adjusted so that each bin contained approximately the same number of jets, and no single bin spanned a large range of $f_{\text{trk}}$. The $p_{\text{T}}$ used for each point of the fit was taken to be the average $p_{\text{T}}$ of jets in that bin, and the $f_{\text{trk}}$ recorded for the bin was the average $f_{\text{trk}}$ of all jets in that bin.

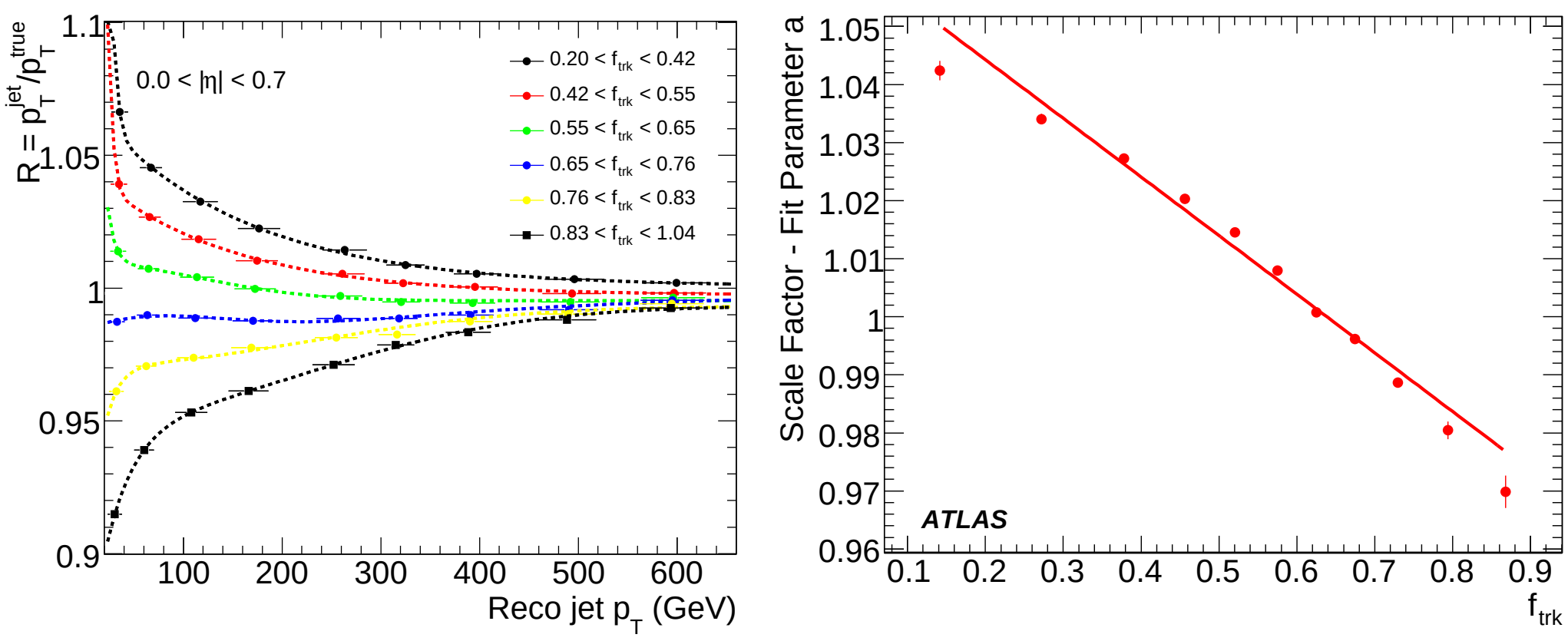

Figure 20: On the left, fits as a function of reconstructed jet $p_{\text{T}}$ for central ($|\eta| < 0.7$) jets in bins of $f_{\text{trk}}$. On the right, straight line fit of one of the parameters determined by $p_{\text{T}}$ fits, as a function of $f_{\text{trk}}$.

The jet $p_{\text{T}}$ fit parameters were then extracted and fitted with a straight line function of $f_{\text{trk}}$, as shown in Fig. 20 (right) for parameter $a$. The 2-dimensional track-jet energy corrections, $R(p_{\text{T}}, f_{\text{trk}})$ can be applied to reconstructed jets to improve their average response.

### 6.3 Algorithm performance

The track-jet response correction $R(p_{\text{T}}, f_{\text{trk}})$ described previously was first applied to jets selected for determining the parameterization. The jet energy response was then checked as a function of $f_{\text{trk}}$. Figure 21 (left) shows the dependence of the response on $f_{\text{trk}}$ before and after applying the track-jet correction. Track-jet corrections flatten the response, $R^{f_{\text{trk}}}(p_{\text{T}}) = E_T^{\text{jet}}/E_{\text{T}}^{\text{true}}$, to a mean value of 1 for all $f_{\text{trk}}$. An overcorrection at high $f_{\text{trk}}$ appears. As the overlaid distribution of $f_{\text{trk}}$ demonstrates, very few jets lie in this region and so the total energy resolution is not altered.

As described earlier and shown in Fig. 19, the jet resolution before the $R(p_T, f_{trk})$ correction can be thought of as a convolution of distributions consisting of the sum of several offset Gaussians with different $f_{trk}$. By re-centering the underlying distributions we improve the jet transverse energy resolution. Fig. 21 (right) shows the overlapping distributions after re-centering.

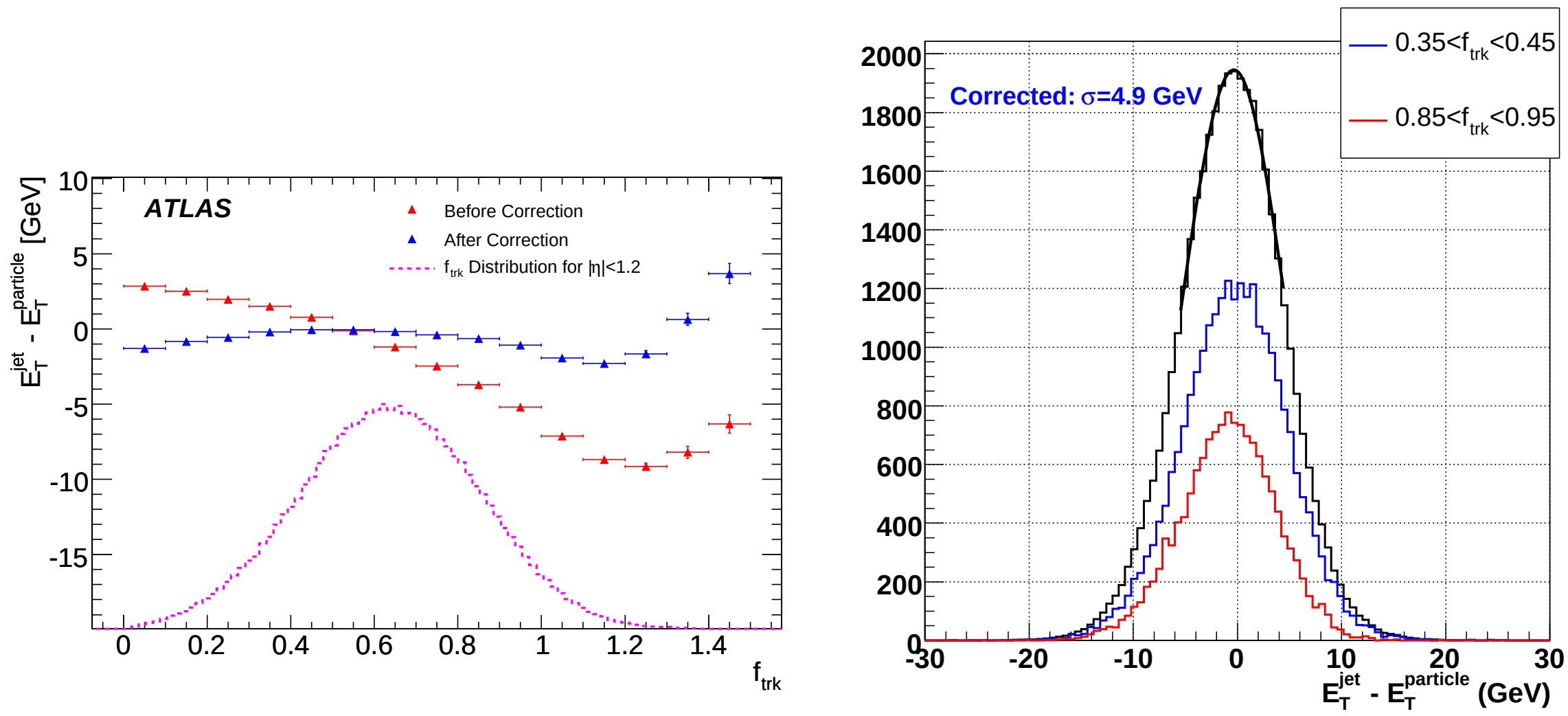

Figure 21: On the left: absolute jet energy response as a function of $f_{trk}$, before and after applying the track-jet response correction. The distribution of $f_{trk}$ has been overlaid to show the jet distribution. All jets used in the fits are included. On the right: jet transverse energy response after the track-jet response correction for jets with $0 < |\eta| < 0.7$ and $40$ GeV $< p_T < 200$ GeV. The underlying Gaussian distributions are now overlapping, and the measured jet transverse energy resolution has been reduced.

The fitted corrections were applied to all jets above 40 GeV in the dijet samples. The following studies show the improvement of jet transverse energy resolution and $E_T^{miss}$ distributions after the corrections are applied.

### Jet transverse energy resolution

The transverse jet energy resolution is considerably improved at low jet $p_T$, as demonstrated in Fig. 22. Fits are shown both before and after the corrections are applied. A $\sim 15\%$ improvement in energy resolution is achieved at 60 GeV.

### $E_T^{miss}$ resolution

Track-based jet corrections can also be used to improve the scale of the missing transverse energy. The track-corrected $E_T^{miss}$ is computed as

$$\vec{E_T}^{corr} = \vec{E_T} - \sum(\vec{p_T}^{trk,corr} - \vec{p_T}) \tag{14}$$

where the sum refers to the two leading jets, and $\vec{p_T}^{trk,corr} = p_T/R(p_T, f_{trk})$ is the jet transverse momentum corrected as described in Section 6.2.

Figure 23 (left) shows the mean value of the $E_T^{miss}$ as a function of the $f_{trk}$ difference between the two leading jets in a dijet sample. Both leading jets were required to have $|\eta| < 1.2$ for this study. Figure 23 (left) shows a large imbalance of energy when the two jets have large differences in $f_{frk}$. When $\Delta f_{trk} < 0$, $f^2_{trk} > f^1_{trk}$ and $p^2_T$ is underestimated resulting in a positive bias on $E_T^{miss}$. Similar argument explains a negative $E_T^{miss}$ bias for $\Delta f_{trk} > 0$.

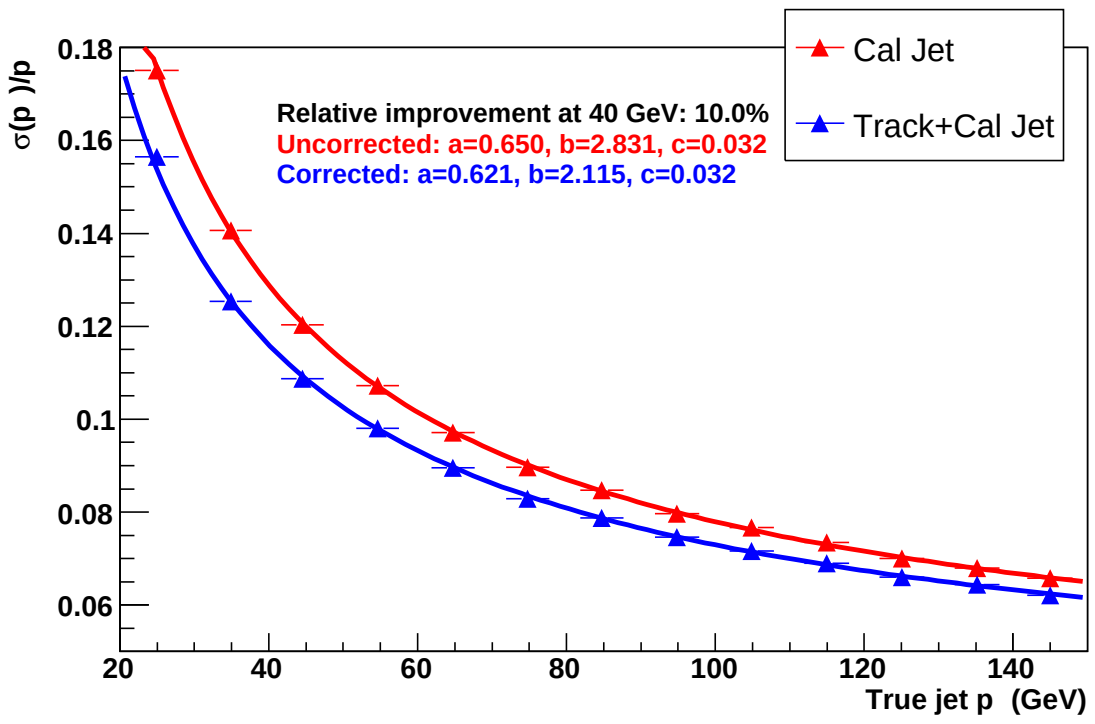

Figure 22: Jet transverse momentum resolution as a function of jet $p_T$ before and after correcting for $f_{trk}$.

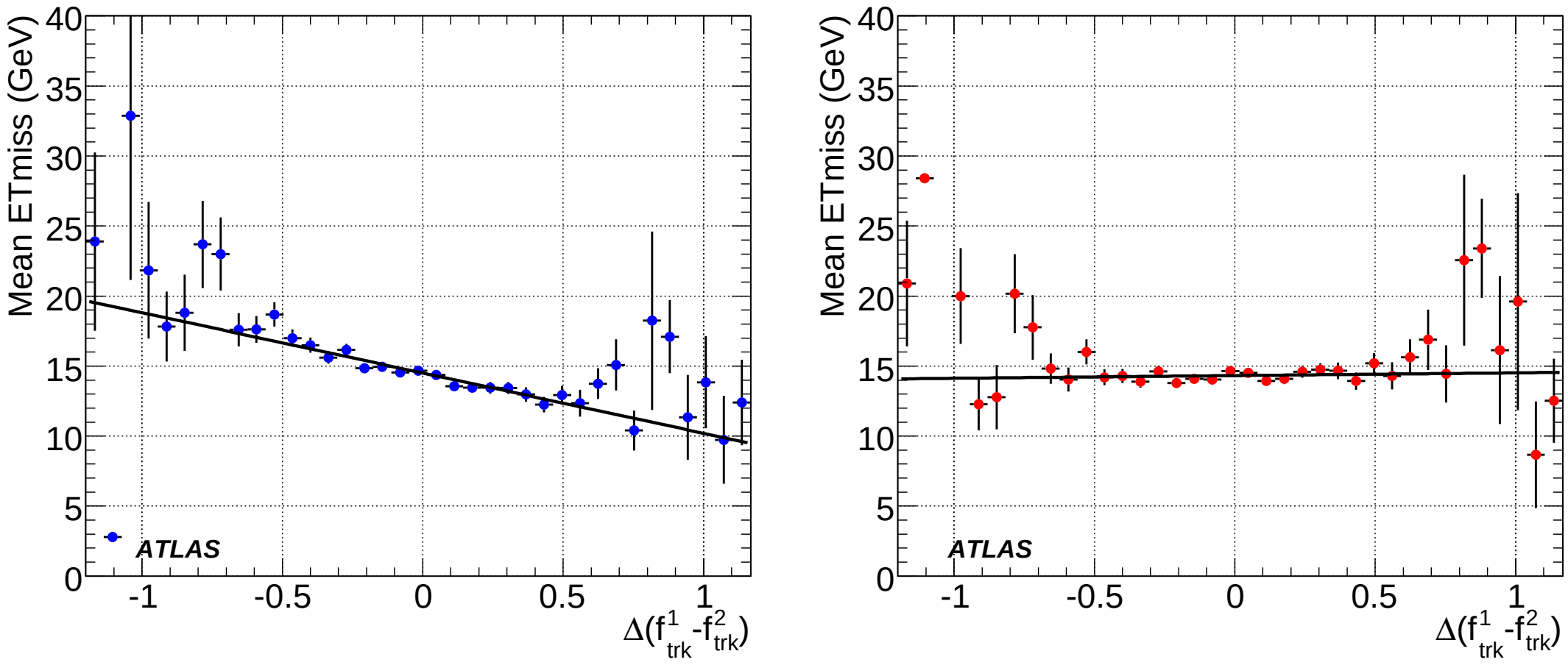

Figure 23: Average missing transverse energy as a function of the $f_{trk}$ difference between the two leading jets in a dijet sample before (left) and after (right) track-based jet energy corrections. The width and tails are improved.

Figure 23 (right) shows that the $E_T^{miss}$ scale is properly corrected after applying the track-based response correction to the leading two jets, and the $E_T^{miss}$ bias has been removed.

## 6.4 Conclusions and future studies

Although the corrections in this section were calculated only for cone jets with $\Delta R = 0.4$, they can be trivially extended to any other jet collection, including $k_T$ jets.

We introduced a track-based method for correcting the response of jets in ATLAS that provides a $\sim 20\%$ improvement in jet energy resolution at 50 GeV. The corrections also improve missing energy distributions. These corrections do not require new jet energy scale corrections and can be applied after the standard reconstruction. By systematically adding information from the tracker to jets already reconstructed based on calorimeter information, considerable improvements can be made.

There are several additions being explored to further improve the jet energy resolution using this technique and other similar track-based variable methods. This technique will be expanded to correct $b$-jets in the same way that light quark jets have already been corrected. The radius used for track-jet association may be adjusted to improve the performance. Track-based response corrections will be expanded to include additional variables such as the fraction of transverse momentum carried by the leading track ($f^1_{trk}$) and track multiplicity ($n_{trk}$).

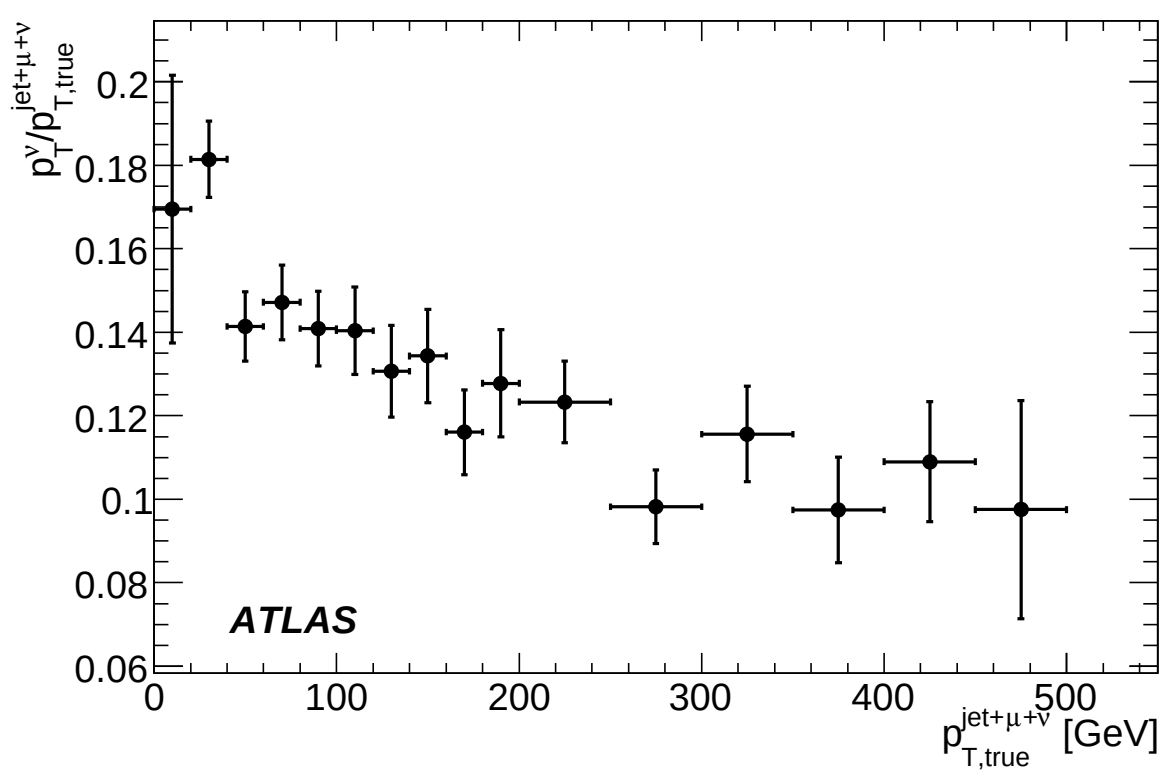

Figure 24: Fraction of $p_T$ carried by the neutrino in $b$ jets decaying semileptonically ($b \to \mu X$ or $b \to c \to \mu X$). The abscissa corresponds to the total transverse component for the jet (all interacting particles except muons), muon and neutrino momenta. We use $b$ jets from QCD dijet samples as described in Section 7.1.

# 7 Jet energy scale corrections to semileptonic $b$ jets

In this section, we discuss a possible strategy to correct the b-jet energy in case of semileptonic decays of the $b$ quark.

The decay of $b$ quarks usually produces a $c$ quark, which subsequently decays to a $d$ quark. The $b$ quark decays into a muon and a neutrino $\approx 10\%$ of the time. As a result, a $b$ jet is accompanied by a neutrino and a muon $\approx 19\%$ of the time and by two neutrinos and two muons $\approx 1\%$ of the time. These neutrinos carry away a fraction of the jet energy, introducing a systematic underestimation of the energy of such jets. In this document, we concentrate on $b$ jets that contain only one neutrino inside. The neutrino from the semileptonic $b$ jet decay carries over 10% of the total jet $p_T$.

However, these jets can be tagged by the presence of a muon, if the muon is energetic enough to reach the Muon Spectrometer. Upon a successful tag, the jet energy scale can be corrected through a parameterization of the energy carried by the neutrino. In the following sections, a correction of the jet energy scale as a function of jet and muon $p_T$ for semileptonic $b$ jets is presented and validated.

## 7.1 Monte Carlo event selection

For the studies in this document, two data samples were used (250 k $t\bar{t}$ and the dijet samples described in Table 2)

In addition, for the present studies, semileptonic $b$ jets were required to be tagged by the soft-$b$ tagger [21] and be contained within $|\eta| < 1.2$. The $\eta$ cut is required because the jet response changes for higher $\eta$. Studies in larger $\eta$ regions were not possible due to a lack of statistics. For the dijet sample, only events that had two $b$ jets with $\Delta\phi > 1.0$ were used. This provides a sample composed mostly of $b\bar{b}$ events as well as a few $gg$ events where one of the gluons decays to $b\bar{b}$.

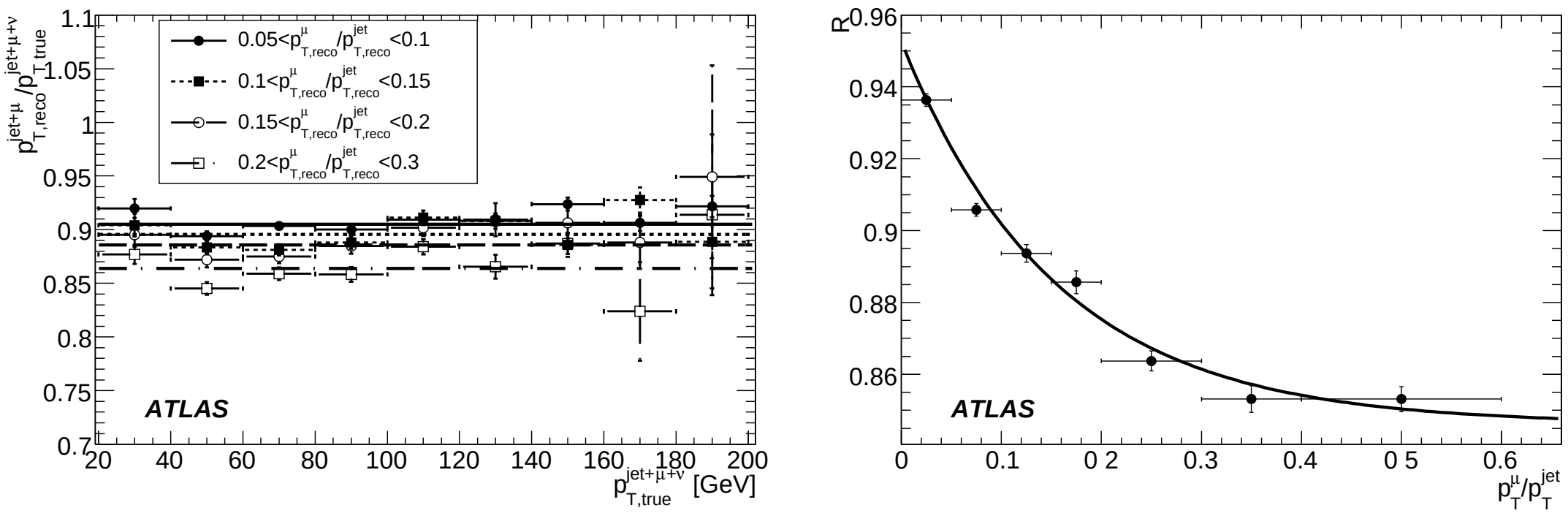

Figure 25: Left: Jet response as a function of jet $p_T$ for semileptonic $b$ jets containing one muon for different values of $x$ (equation 15). Here $p_{T,\text{true}}^{\text{jet}+\mu+\nu}$ refers to the transverse component of the vector addition of the true jet momentum (all interacting particles except muons included), the true muon momentum and the neutrino momentum. Right: Value of the results from straight line fits to the points in the left plot as a function of $x$. The fit function $C(x) = a + be^{-cx}$ is also shown.

## 7.2 Derivation of the jet energy scale correction

The jet energy scale correction was derived using semileptonic $b$ jets from the $t\bar{t}$ sample. The jet response was studied as a function of jet and muon $p_T$. In particular, it was found that a correlation exists between the jet response and the quantity

$$x = p_{T,\text{reco}}^{\mu} / p_{T,\text{reco}}^{\text{jet}}, \tag{15}$$

where $p_{T,\text{reco}}^{\mu}$ is the reconstructed muon $p_T$, and $p_{T,\text{reco}}^{\text{jet}}$ is the reconstructed jet $p_T$, which does not include the muon contribution. The jet responses for samples with different values of $x$ are shown in the left plot of Figure 25. The responses are shown together with the constant fits used to determine the correlation between the jet response and the quantity $x$. The fits do well in correcting for the missing neutrino energy, and, therefore, no further dependence on jet $p_T$ has been considered so far. A better modeling of the dependence of the response as a function of jet $p_T$ in the different samples could help improve the correction. The values of the constant fits ($R$, in the figure above) were then used to parameterize the jet response as a function of $x$. The values of $R$ for samples with different values of $x$ are shown on the right plot of Figure 25. These values were fit to the function $C(x) = a + be^{-cx}$. This fit is shown on the same plot and resulted in the following values for the parameters: $a = 0.846$, $b = 0.11$ and $c = 6$.

This function was used to correct the jet energy scale. In particular, the corrected $p_T$ of the jet was calculated as:

$$p_{T,\text{corr}}^{\text{jet}+\mu} = [C(x)]^{-1} p_{T,\text{reco}}^{\text{jet}+\mu}. \tag{16}$$

The momentum of the jet was increased accordingly along the direction of the vector addition of the muon and jet momenta.

## 7.3 Validation

The jet response before and after applying the correction for the $t\bar{t}$ sample is shown in the left plot of Figure 26. The jet response is corrected appropriately with an uncertainty of $\approx 2\%$ on the correction. The right plot shows the jet response in the dijet samples. The performance of the correction on this sample is comparable to that shown in the $t\bar{t}$ sample. The relative jet $p_T$ resolution, $\sigma_r/r$ where $r = p_{T,\text{reco}}^{\text{jet}+\mu} / p_{T,\text{reco}}^{\text{jet}+\mu+\nu}$, was also studied for these two samples before and after the correction, but it is not shown. The effect of the correction on this quantity is not noticeable within the statistical uncertainties of these studies.

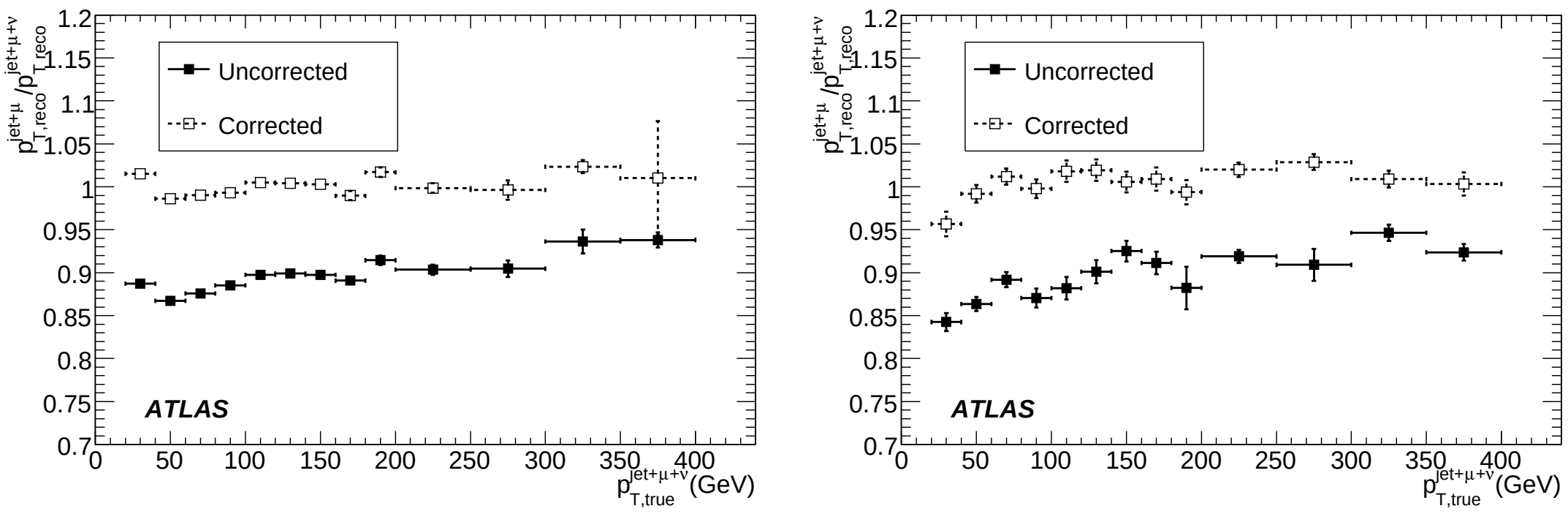

Figure 26: Response of semileptonic $b$ jets before and after applying the neutrino correction to jets from a $t\bar{t}$ sample (left) and a $b\bar{b}$ sample (right).

### 7.4 Conclusion on the corrections to semileptonic $b$ jets

In this section, a procedure for correcting the jet energy scale of semileptonic $b$ jets decaying to a muon has been presented. The procedure has been validated on semileptonic $b$ jets from two data samples showing a noticeable improvement in the jet response, while the relative $p_{\mathrm{T}}$ resolution remains unchanged. To conclude, it should be emphasized that the correction of semileptonic $b$ jets is strongly coupled to our ability to tag them. For this reason, studies in specific physics analyses need to be done to set the operating point of the soft-b tagger. For example, in a $b\bar{b}$ sample, where there are no light jets, we can benefit from this correction the most, while in a $t\bar{t}$ sample the operating point of the soft $b$-tagger might need to be adjusted and could have too low an efficiency for the correction to be noticeable. Further studies on this subject are required in order to determine when the correction is desirable.

## Conclusion

In the first part of the article, we discussed two different ways of using the full ATLAS simulation to calibrate jets. The global method is proven to recover the linearity of the energy measurement while improving the resolution in a wide energy range on Monte Carlo samples. We proved its robustness over different quark content, shower model, and event complexity. The local hadron calibration has been shown to almost fully recover the linearity with respect to the jet calorimeter energy deposits. Even though a correction step to go back to the truth jet scale is missing, the method is promising. In the last part of the article, we discussed the possible use of the tracker information of the jet to improve the jet resolution. After the application of the global calibration, the method is able to further improve the jet energy resolution especially at low $E_T$. Finally we discussed a possible way of recovering the neutrino energy in semileptonic $b$-quark decays.

The calibration methods discussed in this note will be used to provide jet corrections for the ATLAS detector. We stress that the validation of the corrections heavily relies on in–situ measurements as discussed in [20].

# $E/p$ Performance for Charged Hadrons

**Abstract**
The precise measurement of the momentum of hadrons in the tracking detectors, compared to the energy $E$ measured in the calorimeters can provide a validation of the hadronic calibration of the response of the calorimeter to pions at the percent level. The $E/p$ ratio allows the validation of the absolute energy scale calibration over the pseudorapidity range covered by the tracking detectors. Since the selected hadrons are mainly pions, the results obtained for pion response from test beam data are almost directly comparable. The single hadron $E/p$ performances are studied using minimum bias events.

## 1 Introduction

The tracking system can be used to verify the hadronic calorimeter calibration in the energy range between 400 MeV and 200 GeV. The inner tracking will be able to reach a 1% precision on the transverse momentum, $p_T$ , with a few months of data taking at low luminosity. The energy over momentum, $E/p$, ratio allows the validation of the absolute energy scale calibration over the pseudorapidity range of the tracking detectors which covers the Barrel Tile Calorimeter ($|\eta|<1$), the Extended Barrel Tile Calorimeter ($1<|\eta|<1.7$) and the Liquid Argon Hadronic Endcap Calorimeter up to $|\eta|$=2.5. Furthermore, since the selected hadrons are almost exclusively pions, the results obtained for pion response from test beam data are almost directly comparable.

The energy range for which the hadronic calibration must be checked can be determined by jet fragmentation studies at the generator level. It is instructive to investigate the lowest energy needed to obtain a given fraction of the jet energy, denoted $x$. For this reason, it is useful to define the lowest energy jet particles (LEJPs). Using the highest $p_T$ jet in QCD dijet events, the lowest energy jet particles (LEJPs) are determined by ordering in energy the stable particles ($c\tau > 10$ mm) in the jet. Starting with the highest energy, the particle energies are summed up until a fraction $x$ of the jet energy is reached. The particle that pushes this sum over the threshold is the LEJP. Figure 1 shows the mean LEJP energy for a value of $x = 0.95$ as a function of the jet energy for central ($|\eta| < 1.5$) and endcap ($1.5 < |\eta| < 3.2$) jets. To reconstruct 95% of the jet energy in the central region we have to measure particle energies down to 800 MeV. To reconstruct 99% of the jet energy we need to reach energies down to 200 MeV. The energy of the highest energy particle in the jet ranges from 10 GeV to 550 GeV.

Any hadronic calibration scheme is required to be robust over the energy range 200 MeV to 550 GeV to ensure the correct jet energies scale. Furthermore a large fraction of particles in jets have energy below 10 GeV and a large proportion of them are charged pions. Hence it is important to check the hadronic calibration at low particle energies using the $E/p$ method. The robustness of the hadronic calibration, the validity of the hadronic shower model in Monte Carlo, and the single charged hadron $E/p$ performance can be studied in several data samples which cover different energy ranges. The minimum bias sample can reach down to a momentum of 400 MeV and the remainder of this note deals with studies of $E/p$ in the minimum bias sample.

## 2 Study of $E/p$ using Minimum Bias events

As discussed above, low energy hadrons carry a large portion of the energy in a jet. It is therefore important to study the $E/p$ performance at low energies to obtain an ultimate calibration of the jet energy scale to 1% precision. In early data, when a precise hadronic calibration will not be available, this study will be performed using the electromagnetic scale. Subsequently, each step of the local hadronic calibration can be cross-checked and improved. These steps include removal of noise via the use of topological clusters, cluster classification as electromagnetic or hadronic, and calibration of hadronic clusters to the local hadronic energy scale.

Minimum bias samples were used to determine the feasibility of using charged hadrons in the $p_T$ range between 1 to 10 GeV to cross-check the single hadron energy scale for topological clusters calibrated to the local hadronic energy scale [1]. The simulation used for this study replicates data collected over only a few days at a trigger bandwidth of 10 Hz. A year of data-taking at low luminosity will reduce the statistical uncertainty to less than 1%. The data collected in one year should allow the local hadronic calibration to be checked as a function of both $\eta$ and $\phi$.

The calibration of topological clusters calibrated to the local hadronic energy includes corrections for invisible and escaped energy in the hadronic shower as well as energy lost in dead material, such as the material of the inner detector. It does not include corrections for energy lost in cells outside the cluster (out-of-cluster). In the very low energy regime considered by this study, there are significant energy losses in dead material and from out-of-cluster effects. Even with cluster calibration, it is not possible to completely recover these losses, since many low energy particles do not leave sufficient energy in the calorimeter to meet the cluster reconstruction thresholds. As a consequence, we expect the calibrated value of $\langle E/p \rangle$ to be below 1. However, as $p_T$ increases, these energy losses diminish and therefore $\langle E/p \rangle$ approaches 1. Measuring the $E/p$ distributions for single hadrons can provide a way of determining the size of these energy losses in-situ. It can also be directly compared to the single pion Monte Carlo used to derive the weights and cluster classification used as part of the local hadronic calibration.

## 2.1 Isolated pions in Minimum Bias events

To eliminate fake tracks and ensure an accurate momentum measurement, only tracks satisfying the following criteria were considered in this study:

- At least one hit in the B Layer of the pixel detector.
- No more than one missing hit in the other pixel and SCT layers.
- Good quality track fit satisfying $\chi^2/\text{ndof} < 1.5$.

Approximately 76% of all reconstructed tracks pass these criteria. The $p_T$ range for pions surviving the track quality constraints is given by the dashed curve in Fig. 2. This shows that minimum bias events can provide a source of pions to check the hadronic energy scale from 500 MeV, the nominal lower energy required for a reconstructed track, up to energies of the order 10 GeV.

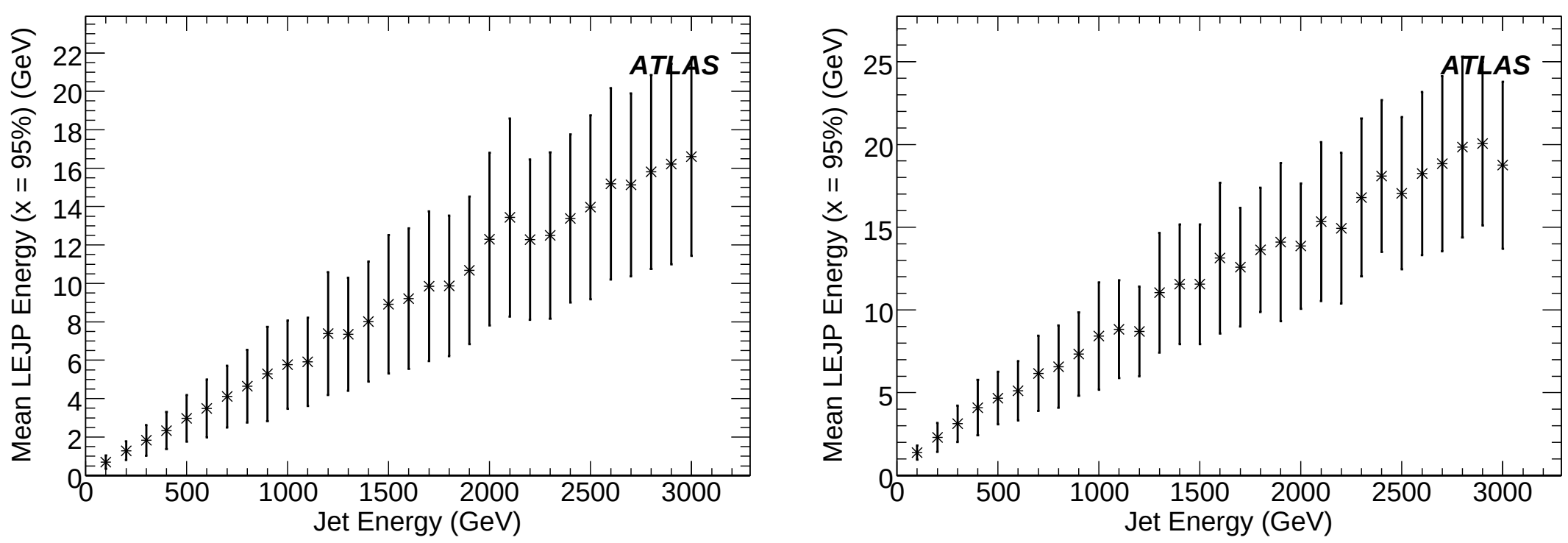

Figure 1: Mean LEJP energy ($x = 0.95$) as a function of the jet energy for jets in central (left) and endcap (right) regions. The error bars represent the Full-Width-Half-Maximum of the LEJP energy spectrum.

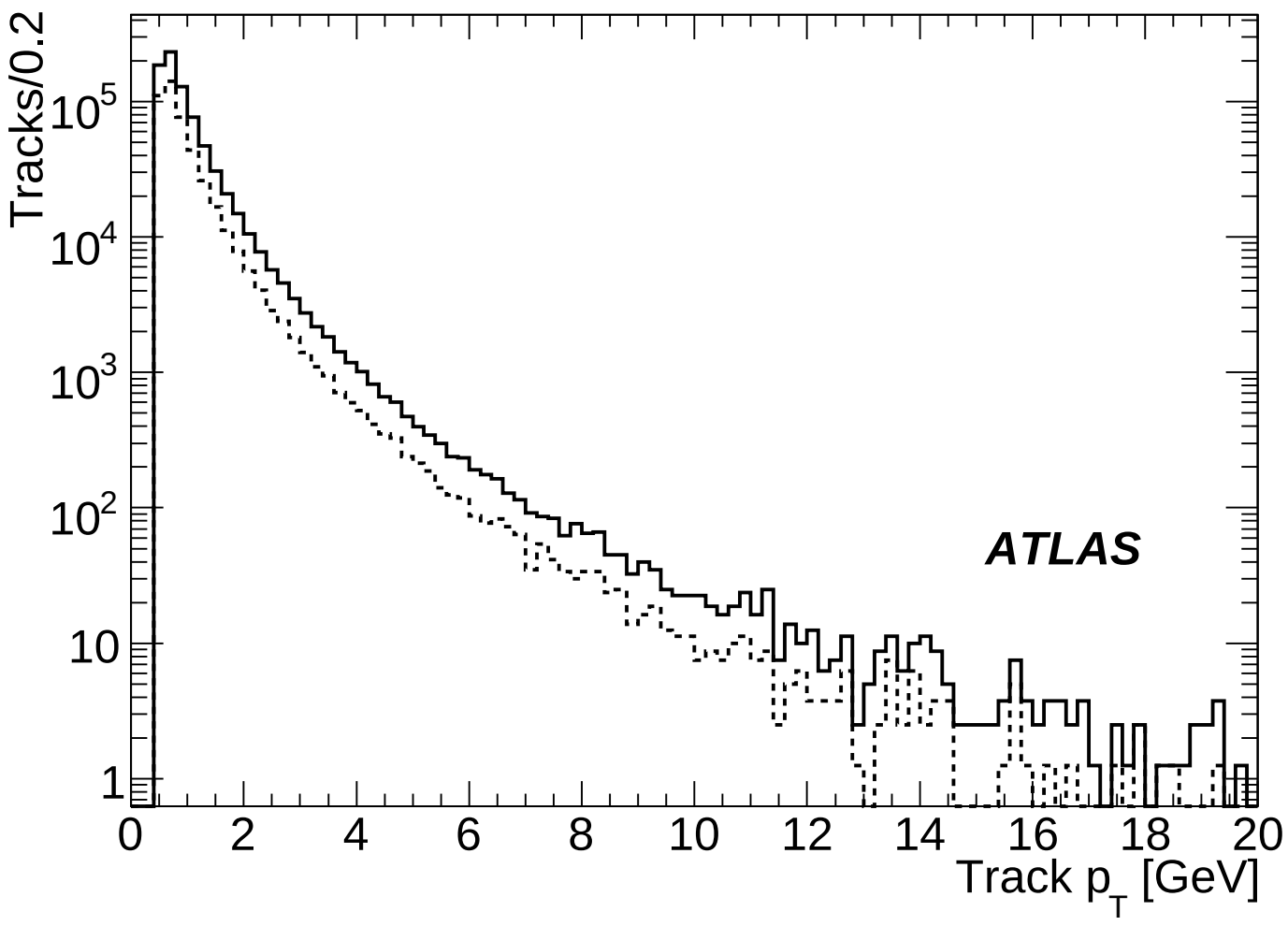

Figure 2: Distribution of $p_T$ of all tracks (solid line) and selected tracks (dashed line) in minimum bias events, scaled to $1\,\mu b^{-1}$.

The pion energy deposited in the calorimeter is found by extrapolating the track direction to the second sampling layer of the EM calorimeter. From this position a cone is defined, $\Delta R_{\text{cone}} = 1.0$, and the energy of each charged hadron is calculated as the sum of all topological clusters within this cone. A topological cluster is considered to be inside the cone if its barycenter is within this $\Delta R_{\text{cone}} = 1.0$ around the extrapolated track position.

In minimum bias events, energy from the underlying event inside the region defined by $\Delta R_{\text{cone}}$ considerably biases the measured pion energies. Photons originating from the decay of neutral pions are the predominant source of background. Charged particles also contribute to the background as the reconstruction efficiency of tracks below 5 GeV can be lower than 80% [2]. Selection criteria were used to reduce the contamination from the underlying event. These criteria differ slightly for charged hadrons with $p_T \leq 3$ GeV and $p_T > 3$ GeV due to the low multiplicity of higher $p_T$ tracks in the event and the different hadronic shower shapes in the calorimeter. The selection criteria used to minimise the amount of energy from the underlying event are:

- Isolated from other tracks by at least $\Delta R = 0.4$. The track positions were taken at the second layer of the EM calorimeter.
- The charged hadron must be one of the harder particles in the event.
  - $p_T / \sum_{\text{all tracks}} p_T > 0.1$ for $p_T \leq 3$ GeV.
  - $p_T / \sum_{\text{all tracks}} p_T > 0.3$ for $p_T > 3$ GeV.
- The energy in the hadronic calorimeter must be isolated by requiring the energy in the outer region $0.4 < \Delta R_{\text{cone}} < 1.0$ of the cone to be small.
  - $E_{\text{HAD}_{1.0-0.4}} < 0.01 \times p_{\text{track}}$ for $p_T \leq 3$ GeV.
  - $E_{\text{HAD}_{1.0-0.4}} < 0.05 \times p_{\text{track}}$ for $p_T > 3$ GeV.

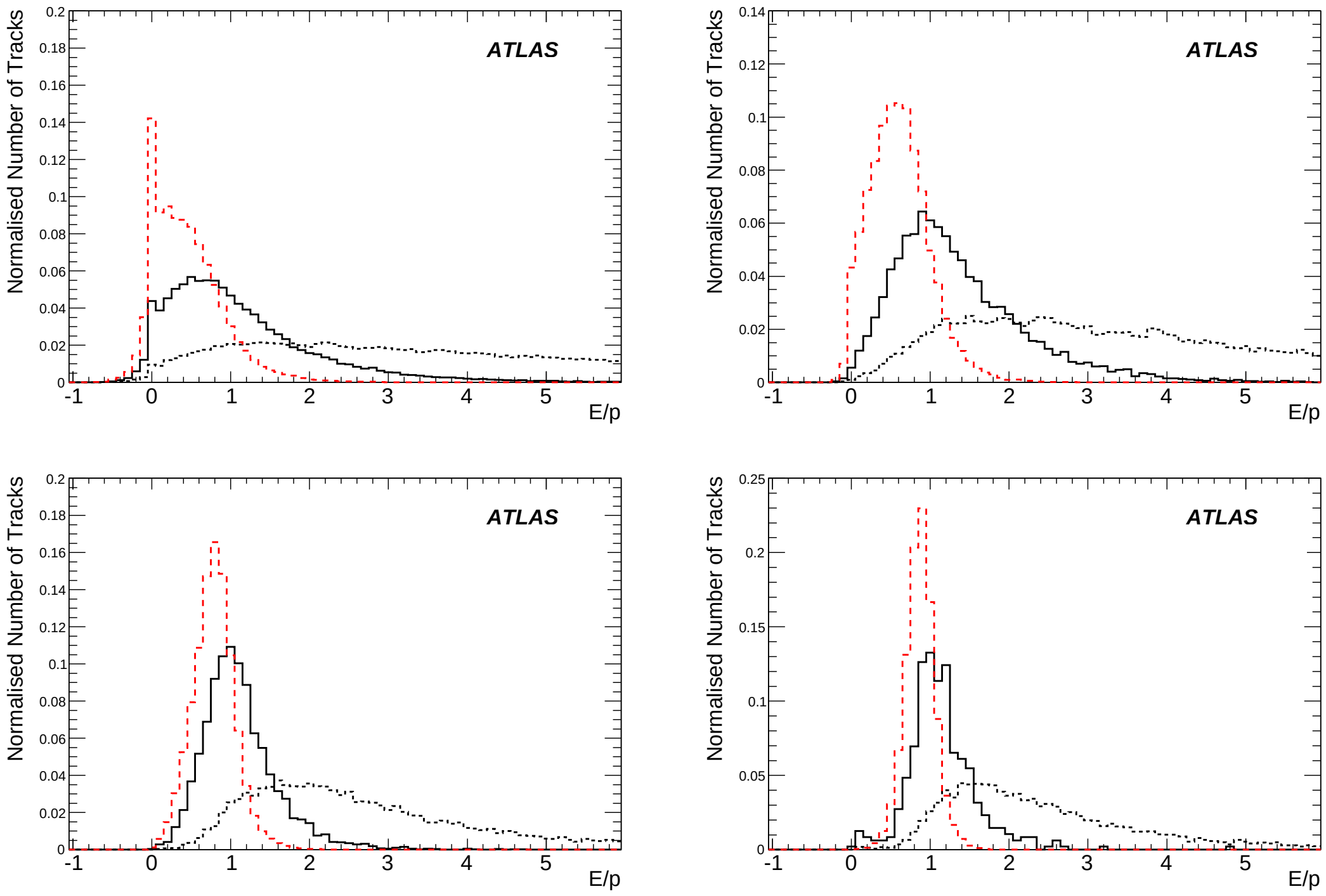

Figure 3: The $E/p$ distributions for pions from minimum bias events before cuts (dotted line), after cuts (dark solid line) and for a reference sample of single particles (dashed line) for tracks in the $p_T$ range $0.8-1.2$ GeV (top left), $1.6-2.4$ GeV (top right), $4-6$ GeV (bottom left) and $8-12$ GeV (bottom right). All distributions are normalised to unity. The negative $E/p$ ratios are a consequence of clusters with negative energy resulting from noise.

Figure 3 shows the $E/p$ distribution in minimum bias events before and after cuts, and compares it to single pions. All results shown here and in Section 2.2 are for tracks associated with real pions (studies of the impact of non-pion tracks are in progress). These selection criteria approximately halve the shift in $\langle E/p \rangle$ caused by other particles in the cone.

The shift of $\langle E/p \rangle$ caused by the selection is primarily due to the isolation cut in the hadronic calorimeter. The shift was estimated using the pion from single particle Monte Carlo and was found to change $\langle E/p \rangle$ by 0.006 for 1 GeV pions, 0.002 for 2 GeV pions, negligible for 5 GeV pions and 0.001 for 10 GeV pions. The remaining background, due to energy from the underlying event, cannot be removed with cuts. This is justified by the absence of isolated charged hadrons in minimum bias events.

## 2.2 Data-driven unfolding procedure

The remaining background, present inside the region defined by $\Delta R_{\text{cone}}$, was estimated using a data-driven method described in Ref. [3]. This was then unfolded from the pion $E/p$ distribution. Most of the energy coming from the underlying event is deposited in the electromagnetic calorimeter. Charged pions which deposit most of their energy in the hadronic calorimeter were therefore selected in order to estimate the energy contamination in the electromagnetic calorimeter. These are labelled as late showering pions.

The energy from the underlying event was studied by dividing $\Delta R_{\text{cone}}$ into the following two regions:

- A region where very little background energy is deposited, consisting of:
  - The hadronic calorimeter and

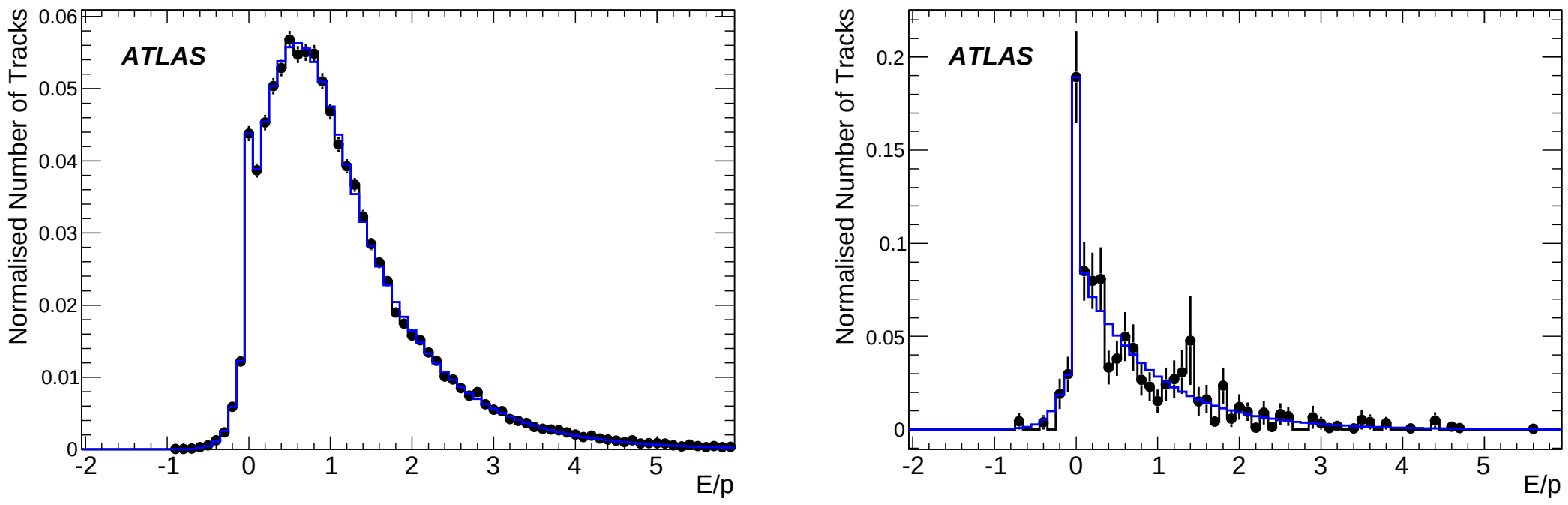

Figure 4: The total, $(E/p)^{\text{meas}}$, (left) and underlying event, $(E/p)^{\text{contam}}$, (right) distributions used to extract the $E/p$ of isolated charged hadrons in the range 0.8 - 1.2 GeV. Error bars show the statistical error on the distribution before fitting (solid line).

- A core region close to the track:
  - $\Delta R_{\text{core}} = 0.1$ for $p_T \leq 3$ GeV.
  - $\Delta R_{\text{core}} = 0.05$ for $p_T > 3$ GeV.

- A ring surrounding the assumed trajectory of the pion through the electromagnetic calorimeter contains energy of the underlying event, $E_{\text{outer-cone}}$, defined to be $\Delta R_{\text{core}} < \Delta R < 1$ in the electromagnetic calorimeter.

We use the energy in the $E_{\text{outer-cone}}$ of late-showering pions to estimate the underlying event energy. For this study we assume that the energy in $E_{\text{outer-cone}}$ for late-showering charged particles is zero as they act as minimum ionising particles (mips) in the electromagnetic calorimeter.[1]

Late-showering charged pions were identified by two criteria based on the energy in a narrow cone of $\Delta R < 0.05$ around the track:

- Energy in the electromagnetic calorimeter $< 0.5 \times p_{\text{track}}$.
- Energy in the hadronic calorimeter $> 0.5 \times p_{\text{track}}$ .

The background estimated using late-showering pions is applied to all pions (late and early showering ones). Figure 4 shows the underlying event energy obtained by this method for tracks in the $p_T$ range $0.8 - 1.2$ GeV. The underlying event energy divided by the track momentum is defined as $(E/p)^{\text{contam}}$.

The $(E/p)^{\text{meas}}$ distributions shown by the solid curves in Fig. 3 are a convolution of the $E/p$ distributions for isolated pions, $(E/p)^{\text{iso}}$, and the underlying event energy. By deconvoluting the background from the measured $E/p$, we recovered $(E/p)^{\text{iso}}$.

The convolution can be written in terms of the number of entries in each bin $i$ in the measured $E/p$ histogram:

$$(E/p)_i^{\text{meas}} = \sum_j P_{ij} \times (E/p)_j^{\text{iso}}, \tag{1}$$

where the elements of the matrix $P_{ij}$ represent the probability of background contamination shifting the energy from bin $j$ to bin $i$.

Each element of $P_{ij}$ is taken from the energy deposited in $E_{\text{outer-cone}}$ (shown in Fig. 4 for $p_T = 1$ GeV pions). The matrix elements are defined as $P_{ij} = (E/p)^{\text{contam}}_{(i-j)}$, i.e. $P_{ij}$ is given by the $i^{\text{th}} - j^{\text{th}}$ bin content of the normalised histogram in the region $E_{\text{outer-cone}}$. The probability that each bin contributes to the $(E/p)^{\text{meas}}$ value is given by $P_{ij}/\det(P_{ij})$. The diagonal elements of the normalised matrix give the probability that a measurement is free from contaminating energy.

[1]We assumed that the mips penetrating depth was uncorrelated with the background.

The distribution $(E/p)^{\text{iso}}$ is derived by solving the set of linear equations described above via matrix inversion. The unfolding method is described in Ref. [4] and Ref. [5]. The background is fitted with two exponential functions: one above zero and one below. The measured $E/p$ is fitted with an exponential function below zero, a 7th order polynomial function above zero up to half the maximum height, and an exponential function for the high $E/p$ tail. The result of these fits for 1 GeV pions is shown in Fig. 4. The final distributions obtained for $(E/p)^{\text{iso}}$ are shown in Fig. 5. The subtraction:

$$\langle E/p\rangle^{\text{iso}} = \langle E/p\rangle^{\text{meas}} - \langle E/p\rangle^{\text{contam}} \tag{2}$$

is used to examine the shift caused by any remaining contamination or by the method itself. Results are shown in Table 1. The $E/p$ values are consistent with the single particle mean, showing that this procedure is promising. However, larger statistics are required to properly assess the ultimate precision of this method.

We compared charged hadrons in minimum bias Monte Carlo events of $p_T = 0.8 - 1.2$ GeV, $1.6 - 2.4$ GeV, $4 - 6$ GeV and $8 - 12$ GeV with single pion samples of 50000 events generated with $p_T = 1$ GeV, 2 GeV, 5 GeV and 10 GeV. As both minimum bias and single particle simulations are done with the same detector geometry and showering model, the $E/p$ distributions are identical, providing a way to test how well we can retrieve the single pion $E/p$ distribution in minimum bias events.

Both positive and negative pions were used, and the results were combined to improve the statistical precision. Due to limited Monte Carlo statistics, here we bin only in $p_T$ . Ultimately, the $E/p$ response will be studied in bins of both energy and pseudorapidity, because the dead material distribution varies dramatically across $|\eta|$. In minimum bias events, the mean charged particle multiplicity is relatively flat in $\eta$ for a given $p_T$ . Single particle samples are weighted to have the same $|\eta|$ distribution as minimum bias tracks.

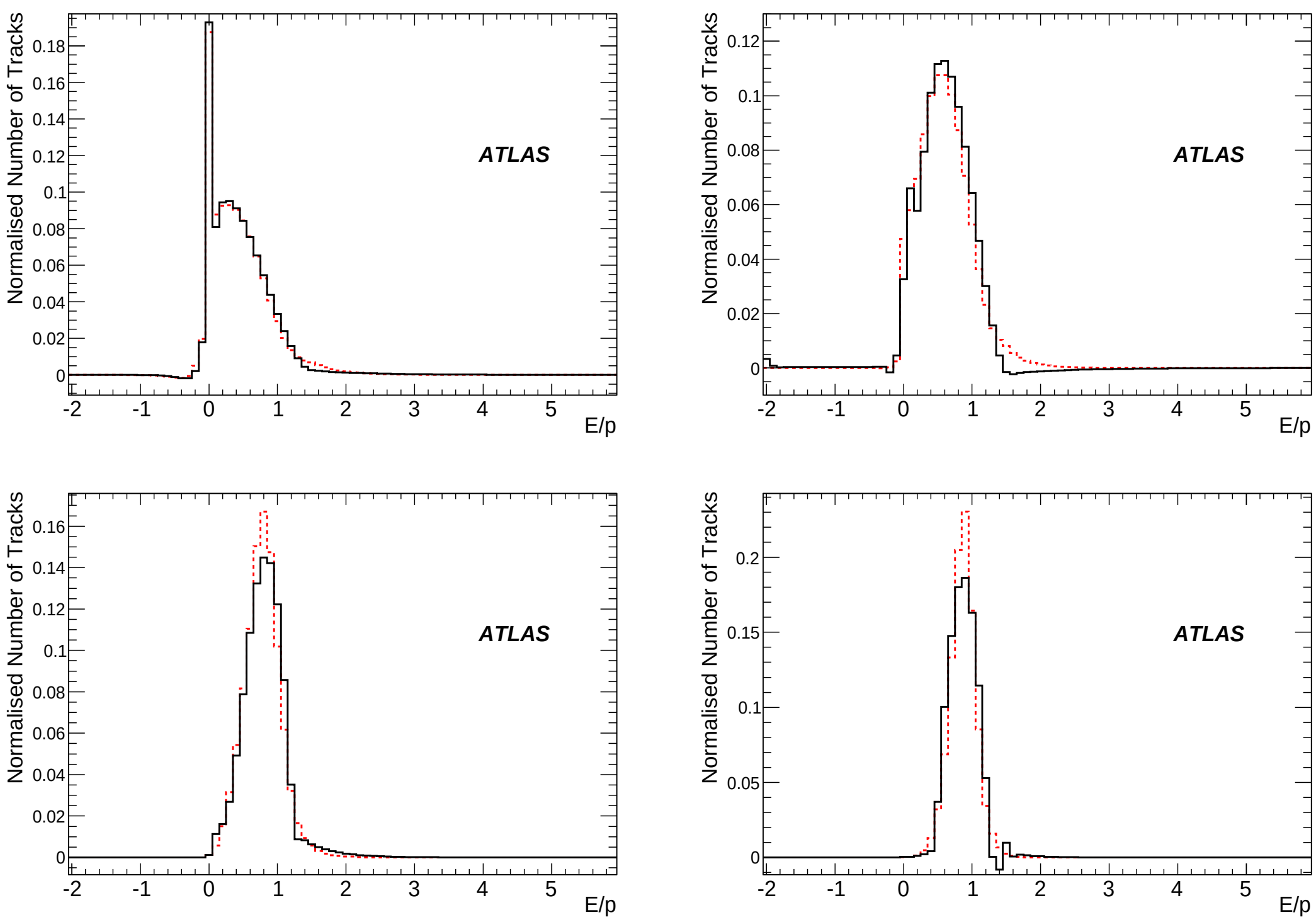

Figure 5: $E/p$ distributions obtained from deconvolution for pions in minimum bias events (solid line) and single pions (dashed line) for the track momentum ranges 0.8 GeV $< p_T <$ 1.2 GeV (top left), 1.6 GeV $< p_T <$ 2.4 GeV (top right), 4 GeV $< p_T <$ 6 GeV (bottom left) and 8 GeV $< p_T <$ 12 GeV (bottom right). The deconvolution also removes the effect of 'fake' clusters in $E_{\text{outer–cone}}$, reconstructed from electronics noise.

Table 1: $\langle E/p \rangle$ for pions at $p_T = 1, 2, 5$ and 10 GeV compared to pions in single particle events.

| Pion $p_T$ Range | Minimum Bias | Single Particles |
|---|---|---|
| $0.8 - 1.2$ GeV | $0.429 \pm 0.052$ | $0.449 \pm 0.003$ |
| $1.6 - 2.4$ GeV | $0.622 \pm 0.043$ | $0.609 \pm 0.003$ |
| $4 - 6$ GeV | $0.818 \pm 0.015$ | $0.786 \pm 0.002$ |
| $8 - 12$ GeV | $0.870 \pm 0.036$ | $0.869 \pm 0.001$ |

### 2.3 Sources of systematic uncertainties

We studied a number of effects which biased the $E/p$ distribution. In our study we assumed that all charged hadrons were pions. In minimum bias events, about 75% of all charged particles are pions, with a small number of kaons and protons. Less than 2% of the tracks are due to heavier hadrons, electrons, muons or are fake. However, the mix of particle types is different for late-showering hadrons. Biases arising from this different particle mixture were studied by comparing the $\langle E/p \rangle$ calculated using all track types with the $\langle E/p \rangle$ derived frompion tracks alone. Within the available statistical precision, we can put an upper limit of 10% on the shift. We use the energy in the $E_{\text{outer–cone}}$ of late-showering pions to estimate the contaminating energy. For this study we assume that the energy in $E_{\text{outer–cone}}$ for late-showering charged particles is zero as they act as mips through the electromagnetic calorimeter. This assumption translates to a shift on $\langle E/p \rangle$ of less than 4%. The energy from the underlying event in the hadron calorimeter and in the $\Delta R_{\text{core}}$ region cannot be measured by this in-situ method. The shift of the $\langle E/p \rangle$ due to this background is estimated using single pions and was found to be less than 4%. The uncertainty in the energy scale is dominated by the unmeasured contaminating energy and the effect of non-pion tracks. All the above systematic uncertainties are statistically dominated.

## 3 Conclusions

Jet fragmentation studies show that low energy hadrons carry a large portion of the energy in a jet. This study shows that isolated charged hadrons produced in minimum bias events can be used to study the $E/p$ performance in the $p_T$ range from 1 to 10 GeV to obtain an ultimate calibration of the jet energy scale. This sample will provide an in-situ test of the cluster level hadronic calibration down to 1 GeV. Within a year at low luminosity, it will be possible to reduce the statistical uncertainty to less than 1%.

With the available Monte Carlo statistics we obtained a statistical uncertainty of 10% at 2 GeV which goes down to 3% at 10 GeV, mostly dominated by the statistics of the late showering pion control sample. Systematic biases from several sources have been studied. Effects, such as unpredicted detector inhomogeneities, will be studied once data are available.

# Jet Energy Scale: In-situ Calibration Strategies

**Abstract**
This note outlines procedures for the in-situ determination of the jet energy scale and resolution of the ATLAS calorimeters. The jet energy scale is evaluated using energy-balanced processes: $\gamma/Z+$jet, dijet, and multijet events. The jet energy resolution is obtained from dijet events. Generators using leading-order multiparton matrix element calculations merged with parton showers along with leading-logarithmic Monte Carlo models are compared. Effects on the energy-balance due to hadronization effects are also studied.

## 1 Introduction

Precise reconstruction of the jet energy is a fundamental ingredient of many physics analyses at the LHC, such as the determination of the top quark mass, the reconstruction of dijet resonances, and the measurement of the inclusive jet cross-section. Moreover the performance obtained on jet reconstruction has a direct impact on the quality of the measurement of the missing transverse energy which will play a decisive role in many searches for new physics at the LHC.

The ultimate goal of the jet energy measurement is, in most cases, the reconstruction of the initial parton momentum. On the other hand, the measurement in the ATLAS detector starts from signals recorded in the calorimeter cells which have been calibrated at the *electromagnetic* (EM) scale. This scale is set in test beams and is defined to reproduce correctly the electron energy in the beams.

A subsequent software jet calibration procedure is performed in two major steps. First, corrections are made for detector effects, in particular calorimeter non-compensation, noise, losses in dead material and cracks, longitudinal leakage and particle deflection in the magnetic field. The procedure is described in Ref. [1]. After this step the *hadronic scale* which provides the jet energy at the *particle level* is obtained, i.e. it should correspond to the jet energy obtained after running the same jet algorithm over all true momenta of the final state particles in the event. Throughout this note, we will refer to these jets as "particle-level jets" or "truth jets". At the second step physics effects, such as clustering, fragmentation, initial and final state radiation (ISR and FSR), underlying event (UE) and pile-up are considered. After that the *final scale* is reached which corresponds to the energy at the *parton level*. These effects can depend on the type of interaction, e.g. they can differ for quark and gluon jets, can depend on the parton momentum scales and multiparton interactions, so that an individual study may be necessary for each data analysis involving jets.

The validation of the whole jet calibration has to be performed *in-situ* using suitable physics processes. In the course of in-situ validation (also called in-situ calibration) the systematic uncertainty of the hadronic energy scale is determined, and the final tuning of this scale is possibly performed. Furthermore, the level of physics effects affecting the final scale is estimated. In addition, the jet energy resolution can be determined in-situ.

The validation procedure will start with QCD dijet events, which allow us to check the uniformity of the calibration as a function of azimuth $\phi$ and of pseudo-rapidity $\eta$. The uniformity in $\phi$ can be checked by studying jet rates; the uniformity in $\eta$ can be validated using $p_T$ balance between the jets. The dijet events also open two ways of determining the jet energy resolution.

After a uniform detector response is obtained, the absolute hadronic energy scale will be studied using $\gamma$ or $Z+$jet events, in which the $Z$ boson is reconstructed via the $Z \to e^+e^-$ or $\mu^+\mu^-$ decay. The $p_T$ balance between the jet and the boson in such events will be used to relate the hadronic scale of the jets to the well understood energy of electromagnetic objects.

At very high jet energies, the statistics of $\gamma/Z+$jet events vanishes. After the validation of the absolute jet energy scale (JES) with these events in a limited $p_T$ range, the scale at higher $p_T$ will be validated using QCD multijet events by balancing the momentum of the leading jet to the momentum sum of the other jets. Alternatively, it can be inferred from the opening angle between the two leading charged particle tracks associated with the jet.

These methods are discussed in the present note. Expectations of several Monte Carlo programs are compared for various jet algorithms. The statistical and systematic uncertainties of the methods are assessed with an emphasis on the first $100\,\mathrm{pb}^{-1}$ and $1\,\mathrm{fb}^{-1}$ of ATLAS data. It is important to note that one missing contribution is the precision with which the Monte Carlo is tuned to and reproduces the data. This, naturally, can only be determined once data are available.

Several related studies are covered in other notes. The determination of JES from the invariant mass of the $W$ boson in $W \to qq$ decays is covered in Ref. [2]. The study of jet fragmentation and of the underlying event using tracks associated with the jets and in the complete event is discussed in Ref. [3]. The measurement of $E/p$ for isolated charged particle tracks in order to study the response of the calorimeter is considered in [4].

This note is organized as follows. First, the Monte Carlo programs used for the present studies are briefly described. Then, the standard ATLAS jet reconstruction and calibration procedure is outlined. Afterward, we discuss the basic features of $\gamma$ and $Z+$jet events and of the $p_T$ balance calibration technique, the physics effects influencing the calibration procedure, the limitations of the underlying models and the expected systematic and statistical uncertainties of the jet energy scale. Subsequently, we consider the calibration techniques in QCD dijet and multijet events, followed by the methods of determination of the jet energy resolution. A summary is given at the end of the note.

# 2 Monte Carlo event generators

The general-purpose Monte Carlo programs HERWIG, PYTHIA and ALPGEN are used here to model the final states at the LHC energies. The event generation relies on phenomenological approaches to describe the processes which occur at all levels apart from calculation of the matrix elements. This is done by factorizing the event generation into several stages:

- hard subprocess at a fixed order of perturbation theory,
- initial and final state QCD radiation using parton shower models,
- multiple parton interactions (MI) contributing to the underlying event (UE),
- hadronization (fragmentation).

PYTHIA and HERWIG include leading order ($2 \to 2$) matrix elements for generating $\gamma/Z+$jet and QCD dijet events. Higher order QCD effects are modeled using parton showers, which can be insufficient to describe hard QCD radiation. These effects may significantly affect the event kinematics and thus spoil the determination of the jet energy scale and resolution. They are checked using ALPGEN which includes leading order matrix elements for generating multiparton final states. ALPGEN simulations are also used in addition to PYTHIA to study multijet events at very high transverse momenta, where such events provide the main means to determine the jet energy scale.

## 2.1 PYTHIA

PYTHIA 6.4 [5] is used with the on-shell leading order (LO) matrix element to model the final states. Higher-order QCD effects are simulated in the leading-logarithmic approximation with initial- and final-state radiation following the DGLAP evolution [6]. Coherence effects from soft-gluon interference are included. The underlying event has been tuned to reproduce the CDF data. The parton density functions (PDF) CTEQ6ll [7] are used for the proton.

The initial-state shower is based on backward evolution to the shower initiators. Initial and final state showers are matched to each other by maximum emission cones. The Lund string fragmentation model is used to produce the final state hadrons. The string model is based on linear confinement, where (anti-)quarks or other color (anti-)triplets are located at the ends of the string, and gluons are energy momentum carrying kinks on the string. When the invariant mass of a string piece gets small enough, it is identified as a hadron, thus the whole system eventually evolves into hadrons.

### 2.2 HERWIG

HERWIG 6.510 [8] is used in conjunction with the underlying event generator JIMMY 4.31 [9]. As with PYTHIA, the CTEQ6ll set is used for the proton PDFs. HERWIG uses the parton shower approach for initial and final state QCD radiation, including color coherence effects and azimuthal correlations both within and between the jets. It includes the angular ordered parton shower algorithm which re-sums both soft and collinear singularities. HERWIG uses the cluster fragmentation model [10, 11], where all the outgoing gluons are first split into quark/anti-quark or diquark/anti-diquark pairs. Then, quarks are combined with their nearest neighbor (in the color field) anti-quark or diquark to form color singlet clusters. These clusters have mass and spatial distributions peaked at relatively low values. For large cluster masses, the $q-$distributions fall rapidly and are asymptotically independent of the hard subprocess scale. If a cluster is too light to decay into two hadrons, it is allowed to become the lightest hadron of the relevant flavor. A similar cluster model is also used to model soft and underlying hadronic events.

### 2.3 ALPGEN

ALPGEN 2.06 [12] is used with the HERWIG parton shower and the JIMMY underlying event model and with the subsequent HERWIG cluster fragmentation. Similarly to PYTHIA and HERWIG simulations, the CTEQ6ll set is used for the proton PDFs. Each final state parton multiplicity is generated individually by ALPGEN. A matching using a given scheme between a generated parton with the parton shower is performed in order to avoid double counting of the jet multiplicity. We use the MLM matching scheme.

## 3 Jet reconstruction in ATLAS

Several jet collections are built during event reconstruction in ATLAS, varying the input to the jet finder and the jet finding algorithm. The inputs considered for the jet finder are “calorimeter towers” (calotowers) and “topological clusters” (topoclusters). In addition, “truth particles” are used for simulated events. Calorimeter towers are built from all calorimeter cells contained in a region of $(d\eta \times d\phi)$ of size $(0.1 \times 0.1)$. The initial cell energy is calibrated at the EM scale. Topological clusters are built according to criteria that identify significant energy deposits in topologically connected cells. Three different levels of signal significance are applied to the seed, the neighboring and peripheral cells. Currently, the settings of $(4,2,0)$ in units of sigma of noise are used [1].

The jet finder methods used are cone and $k_T$ algorithms. The seeded cone algorithm is used with radii of $R = 0.4$ or 0.7, $p_T$ of seeds $= 1$ GeV, and a split/merge fraction of 0.5. The inclusive $k_T$ algorithm (fast version) is run with a $D$ parameter of 0.4, 0.6 or 1.0 [3].

Two calibration approaches are developed in ATLAS, the “global” and the “local” schemes [1]. The global scheme uses H1-style weights [1] to correct calorimeter cells after jet finding. The weights are based on the energy density in a cell. This calibration is specific to each type of jet finder. In the local scheme, jets are built from pre-calibrated topoclusters, which already include hadronic calibration, as well as dead material and out-of-cluster corrections. In this note, we report on the "global" scheme[1].

For most of the studies, a minimum transverse momentum $p_T$ of at least 10 GeV is required for a jet to be considered in the data analysis.

# 4 In-situ studies using $\gamma/Z$ + jet events

## 4.1 Electromagnetic final state in $\gamma/Z$ + jet events

The $Z$ boson is observed via its $Z \to ee$ or $Z \to \mu\mu$ decay products. The good lepton identification capability of the ATLAS detector allows reconstruction of $Z$ decays with very low background [13].

Before comparing the jet energy scales in data and in simulation using $\gamma/Z$ + jet events, basic $p_T$ and $\eta$ spectra of the vector bosons have to be checked. These spectra are affected mainly by the efficiency of photon or lepton identification, and by the trigger efficiency. In the case of $\gamma$ production, significant background due to misidentified jets in QCD dijet events remains despite the rejection power of the detector and of the reconstruction algorithms. The theoretical systematic uncertainties associated with the vector bosons are expected to be small at all levels. The effects on transverse momentum due to additional photon radiation at the hadron level are predicted to be negligible.

## 4.2 Use of $p_T$ balance

In leading order of perturbation theory the final state of $\gamma/Z$ + jet events can be considered as a two-body system in which the transverse momentum of the jet $p_{T,\text{jet}}$ is exactly balanced by the transverse momentum of the vector boson $p_{T,\gamma}$ or $p_{T,Z}$. The $p_T$ balance can thus be defined as

$$B_1 = \frac{p_{T,\text{jet}}}{p_{T,\gamma/Z}} - 1 \,. \quad (1)$$

[1]The local scheme approach was not yet fully developed when studies reported here were done; this approach will be studied and compared to the global scheme in the future.

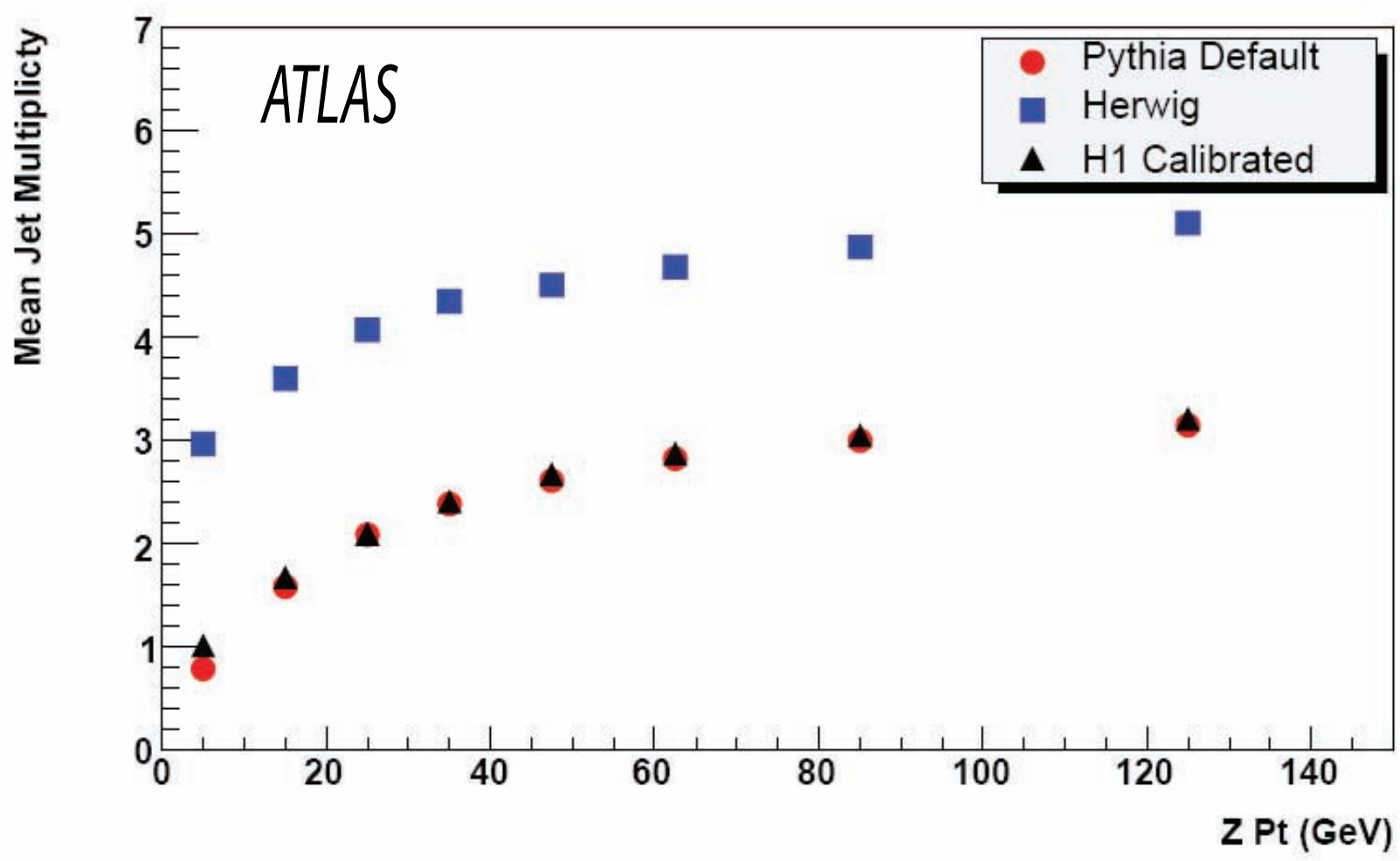

Figure 1: Mean jet multiplicity for jets with $p_T > 10$ GeV as a function of $p_T$ of the $Z$ boson in $Z$ + jet events.

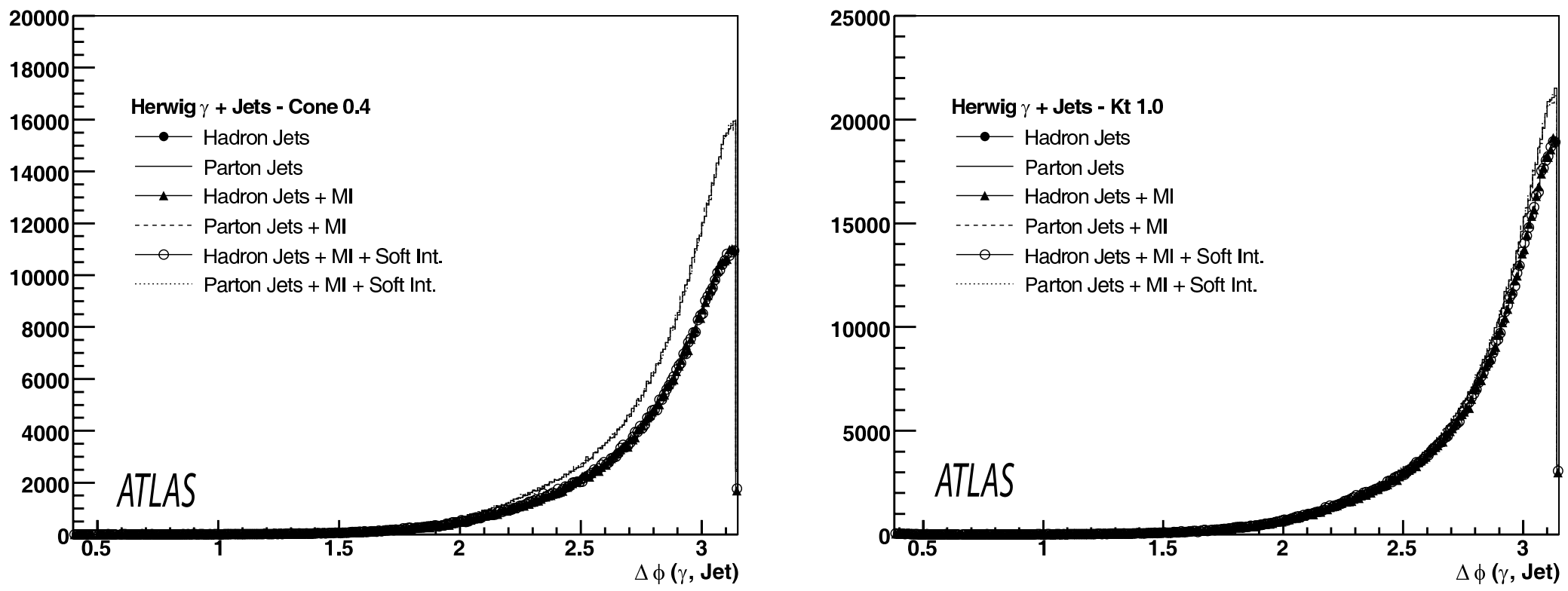

Figure 2: $\Delta\phi$ between the photon and the jet for a) cone algorithm with $R = 0.4$, and b) $k_T$ algorithm with $D = 1$.

However in reality, these events contain more than one jet in the final state at parton, particle and detector level, so that considering only the leading (highest $p_T$) jet is in general too crude an approximation of the event kinematics. This is one of the major issues in this study and requires a detailed understanding of jet multiplicities, as well as their $p_T$ and angular spectra. The Monte Carlo models will have to be tuned to reproduce these jet properties and their energy dependence as observed with the data. A striking example of the current uncertainty in modeling these quantities is shown in Fig. 1, in which the average multiplicity of jets in $Z+$jet events, generated using PYTHIA and HERWIG, is depicted as a function of the transverse momentum of the $Z$ boson. Cone jets reconstructed with $R = 0.7$ at the particle level with $p_T > 10$ GeV in the pseudorapidity range $|\eta| < 5$ are used. HERWIG predicts a larger number of jets than PYTHIA mainly due to its mechanism of cluster formation and the assumptions used to model soft and underlying hadronic events. Further extensive studies have been performed in order to understand the differences in the basic jet distributions at the parton and the particle level for various jet algorithms, and to quantify the influence of initial and final state radiation and of the underlying event model.

A very useful quantity to consider is the azimuthal angular difference $\Delta\phi$ between the boson and the leading jet. It provides information about the event topology and is sensitive to additional physics effects, like initial state radiation and final state radiation. Figure 2 shows the $\Delta\phi$ distribution at parton and at hadron level for three jet algorithms in $\gamma+$jet events with multiple interactions switched on and off, respectively. Although the hadronization effects are apparent for cone jets with $R = 0.4$, the resulting leading jet in most cases is in a back-to-back topology with the photon. In order to reduce the topological bias due to additional radiation, the boson and jets in both $\gamma$ and $Z+$jet events are typically required in our studies to be back-to-back within $\Delta\phi$ of $\pm 0.2$.

In principle, one can consider two extreme ways of taking the physics effects into account:

- Select only those events where the $Z$ or $\gamma$ is back-to-back to only one jet in the event. Require any other jet to have a small transverse momentum. The jet energy correction factors can then be determined from the requirement $B_1 = 0$ as a function of energy and pseudorapidity. The strong back-to-back cut and the requirement against other jets in the event will severely cut the statistics in such an analysis.

- The other extreme is to take all jets in the event that pass loose selection cuts and balance the sum of their momenta to the $Z$ or $\gamma$ momentum. The $p_T$ balance can then be defined as

$$B_\Sigma = \frac{|\sum_{\text{jets}} \vec{p}_T|}{p_{T,\gamma/Z}} - 1 \ , \qquad (2)$$

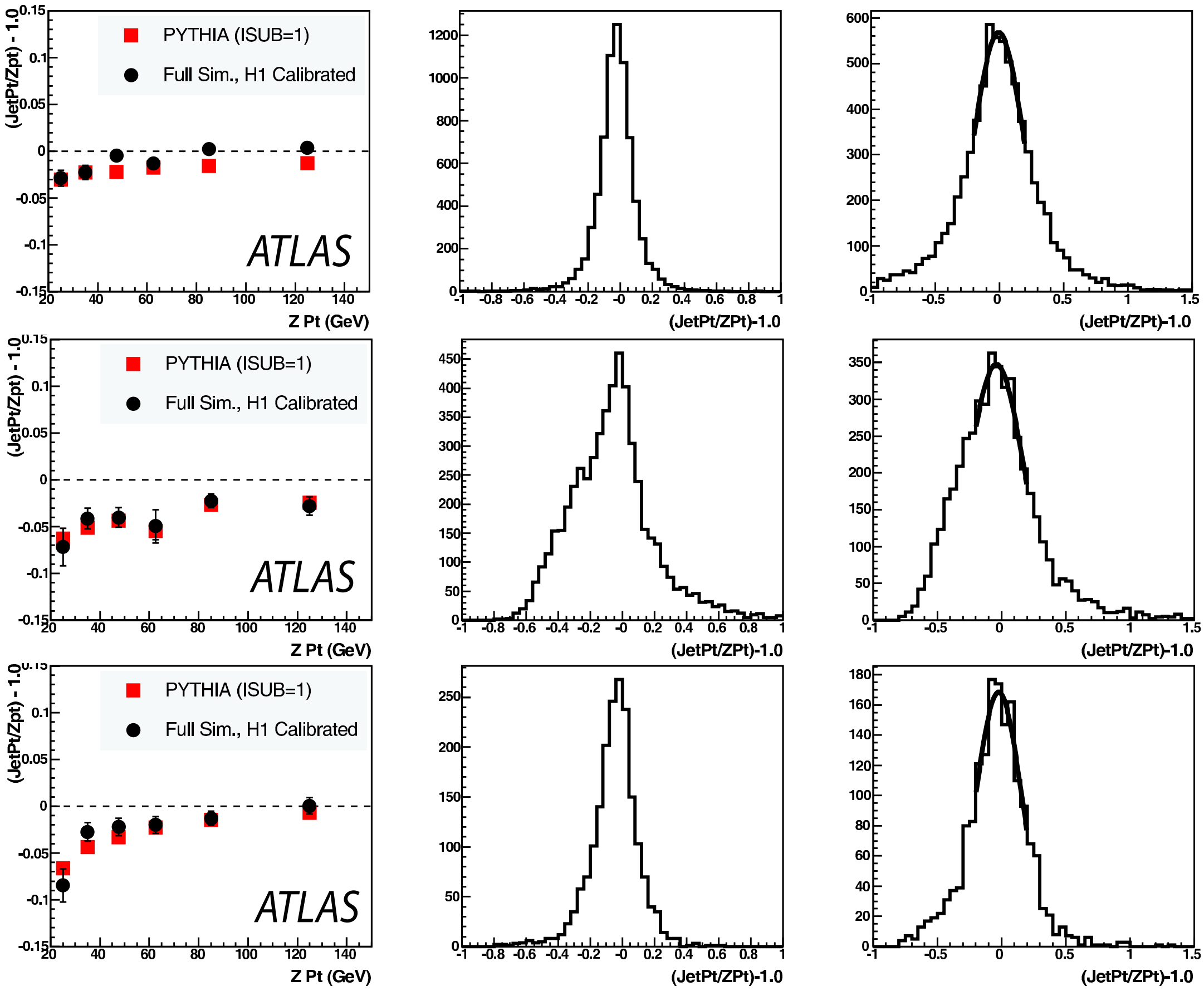

Figure 3: Left in all rows: the mean value of the fitted $p_T$ balance $B_\Sigma$ as a function of $p_{T,Z}$ in $Z$+jet events. Particle level jets (squares) and jets reconstructed from detector signals (circles) are shown. Middle in all rows: $B_\Sigma$ distribution for $p_{T,Z} \sim 50$ GeV for truth jets. Right in all rows: $B_\Sigma$ distribution for $p_{T,Z} \sim 50$ GeV for reconstructed jets. Upper row: all jets with $p_T > 1$ GeV are taken into account. Middle row: only jets with $p_T > 10$ GeV are used and the requirement $|\pi - \Delta\phi| < 0.2$ is imposed. Lower row: in addition, no further jet with $p_T > 10$ GeV is allowed.

This approach should be less sensitive to issues related to physics modeling. However, it is more difficult in this case to relate directly the measured balance to the energy scales of specific jets.

Any kinematic selection cut will affect the global $p_T$ balance between the boson and the hadronic system. This is illustrated in Fig. 3. The central figure in the upper row shows the momentum balance $B_\Sigma$ obtained by summing all particle level jets with $p_T > 1$ GeV for events with $p_{T,Z} \sim 50$ GeV. The actual balance fluctuates around a mean value close to zero. The right plot shows the corresponding distribution for reconstructed (H1) jets which is broader due to the detector resolution effects.

The left plot shows the $p_T$ balance obtained by fitting a Gaussian to the $p_T$ balance distribution in the various $p_{T,Z}$ bins[2]. The fitted mean shows a small negative bias (of the order of 1-3%) which decreases with increasing $p_{T,Z}$. Adding the $p_T$ of neutrinos and muons to that of the jets reduces the bias by about 1%, so this cannot be the origin of this systematic effect, which is further discussed later on.

The effect of selection cuts is studied further, as shown in the middle row of Fig. 3, which presents the

[2] Throughout this note we use the mean value of a Gaussian distribution fitted within a limited sigma range (typically $\pm 1\sigma$) to characterize the $p_T$ balance. We refer to it also as the most probable value.

$p_T$ balance for particle jets with $p_T > 10$ GeV. In addition, the leading jet is required to be back-to-back with the $Z$ within $|\Delta\phi| = (\pi \pm 0.2)$. The bias increases and the spread deteriorates after increasing the $p_T$ threshold for jets. The effects are still strong even though the $\Delta\phi$ cut is imposed. They become less significant as $p_{T,Z}$ increases. The main effect is due to fluctuations of the fragmentation combined with out-of-cone losses and with jet splitting. In addition, residual ISR/FSR contributes significantly to the spread. The underlying event may also contribute particles to the clustered jet. Studies performed with PYTHIA $\gamma$+jet events [14, 15] indicate that the underlying event contributes $\sim 20$ MeV on average to the transverse energy $E_T$ at truth particle level in each $(\Delta\eta \times \Delta\phi)$ region of $(0.1 \times 0.1)$ which corresponds to one calorimeter tower. This results in $\sim 1$ GeV contribution per cone jet with $R = 0.4$ and $\sim 3$ GeV per cone jet with $R = 0.7$. The lower row of Fig. 3 shows the same distributions after an additional requirement to have no further jet with $p_T > 10$ GeV in the event. The spread of the $p_T$ balance distributions is reduced. The negative bias is also reduced but remains significant, especially at low $p_{T,Z}$ values.

Furthermore, these effects strongly depend on the chosen jet algorithm. In Fig. 4 the $p_T$ balance for the leading jet relative to the photon is depicted as a function of photon $p_T$ for cone jets with $R = 0.4$, $R = 0.7$ and for $k_T$ jets with $D = 1$. The jets are reconstructed at the truth particle level and at the parton level after parton showering in HERWIG $\gamma$+jet events and after applying the $\Delta\phi$ cut. The $k_T$ jets on parton level reveal an essentially perfect balance, as expected from theory. Cone jets are subject to out-of-cone losses due to parton showering. The relative losses increase with decreasing $p_T$, and are bigger for cone jets with smaller radii. They are much larger for particle level cone jets due to the lateral spread of the fragmentation. On the contrary, the $k_T$ algorithm tends to include particles originating from other partons, which leads to a significant positive bias at low $p_T$.

A further important issue in the use of the $p_T$ balance is the quantity used as a reference to define the $p_T$ ranges. This is illustrated in Fig. 5 in which PYTHIA $\gamma$+jet events are used which were generated with a minimum $p_T$ of 30 GeV for the hard scattering process. The scatter plot shows the photon $p_T$ versus the $p_T$ of the parton as produced in the hard interaction. The events are distributed around the diagonal as a result of ISR. The cut at the generator stage is visible. If we choose $p_{T,\gamma}$ bins to study $p_T$ balance distributions (e.g. the set of events with 30 GeV $< p_{T,\gamma} <$ 40 GeV contained between the two horizontal dashed lines), we will observe a bias towards lower $p_{\mathrm{T}}$ balance values, since in this case we populate the distribution with significantly more events from the left side of the diagonal than from the right side. The events on the left side are generated on average in interactions with smaller center-of-mass energy, hence with higher cross-section. The cross-section decreases quickly with energy and the negative bias of the $p_{\mathrm{T}}$ balance is significant[3]. One should note also that the bias is proportional to the spread around the diagonal. Hence it decreases with $p_{\mathrm{T}}$. It also decreases if one requires the parton and the $\gamma$ to be back-to-back in azimuth within some $\Delta\phi$ limit, since the ISR spread is then reduced [14]. If we were able to select events according to $(p_{\mathrm{T},\gamma} + p_{\mathrm{T,parton}})/2$ (with lines perpendicular to the diagonal), then the symmetry would be restored.

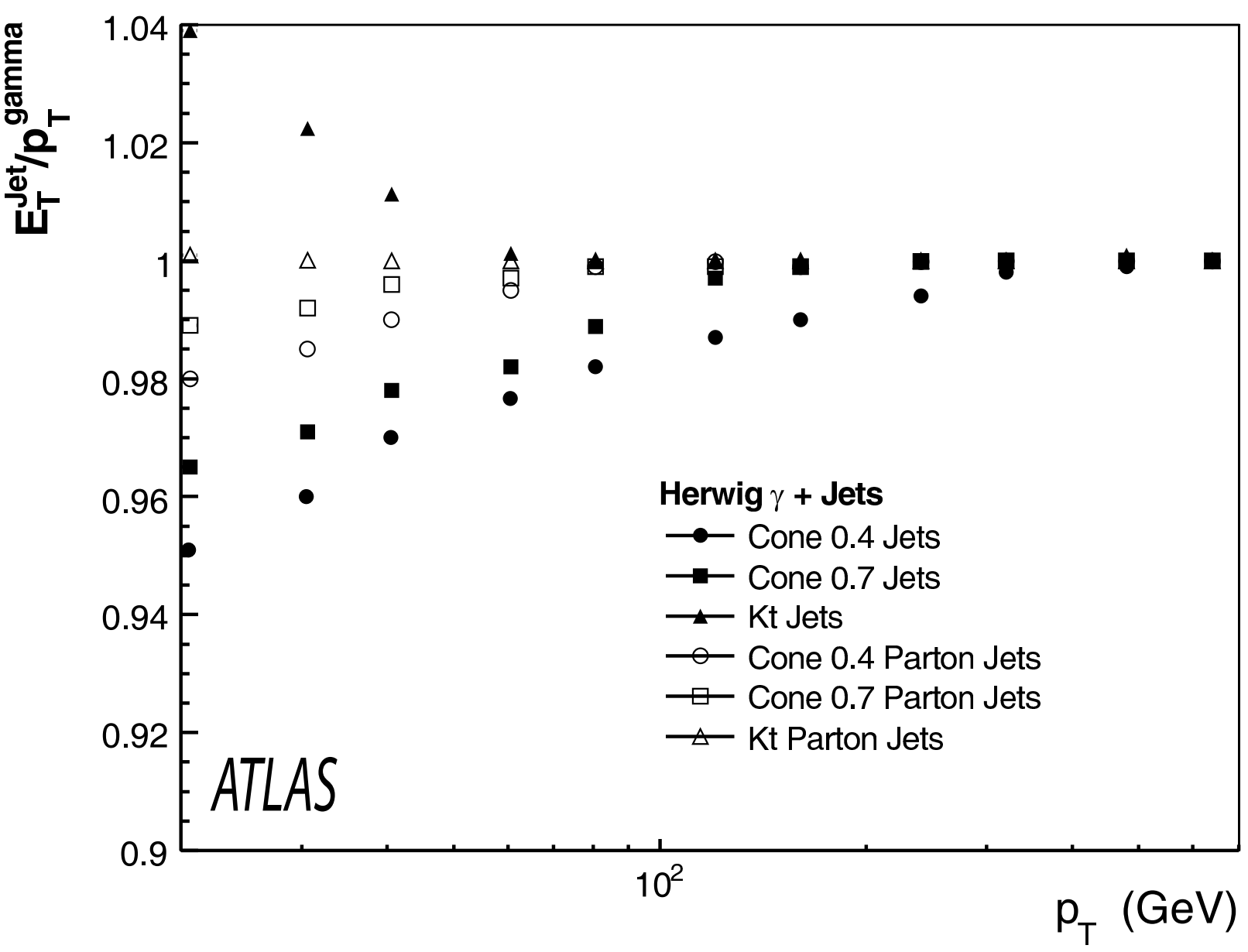

Figure 4: Mean value of the fitted $p_T$ balance ($B_1$ + 1) as a function of $p_{T,\gamma}$ in $\gamma$+jet HERWIG events for various jet algorithms. The points correspond to particle level and parton level jets.

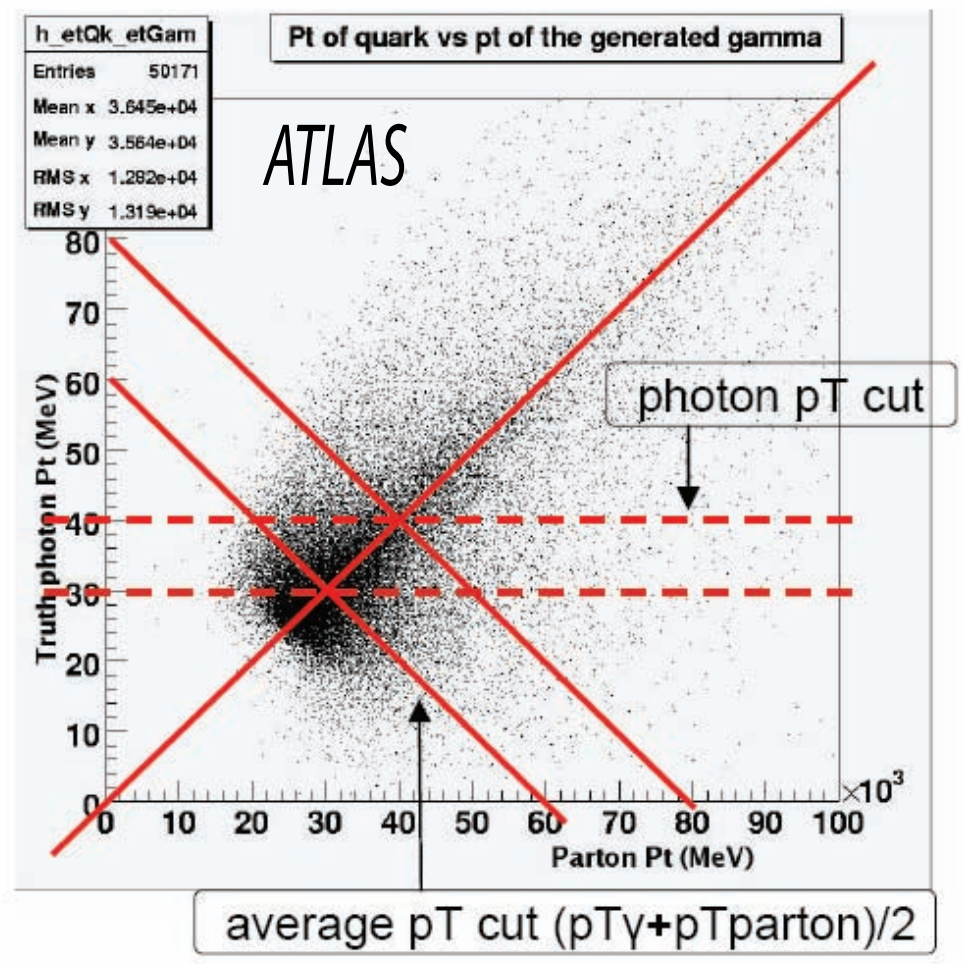

Figure 5: The $p_{\mathrm{T}}$ of the parton versus the $p_{\mathrm{T}}$ of the photon as produced in the hard interaction in $\gamma$+jet events.

This can be seen in Fig. 6. The $p_{\mathrm{T}}$ balance $B_1$ for events with $101 < p_{\mathrm{T}} < 152$ GeV is shown in the left panel of the figure. The solid line shows the balance when the $p_T$ is taken as the average $p_T$ of the photon and the jet[4] and the triangles when it is taken as the photon $p_T$. The difference in the shape is clearly visible with a more pronounced low tail for the $p_{T,\gamma}$ reference, because the cross-section favors the case when ISR has reduced the leading jet $p_T$, as discussed above. The circles show the balance with the photon $p_T$ reference when the photon and the jet are required to be back-to-back within 0.2 to reduce ISR. The resulting distribution is indeed more symmetric.

[3]Throughout this note, we will generically refer to such bias as “bin migration effect”.

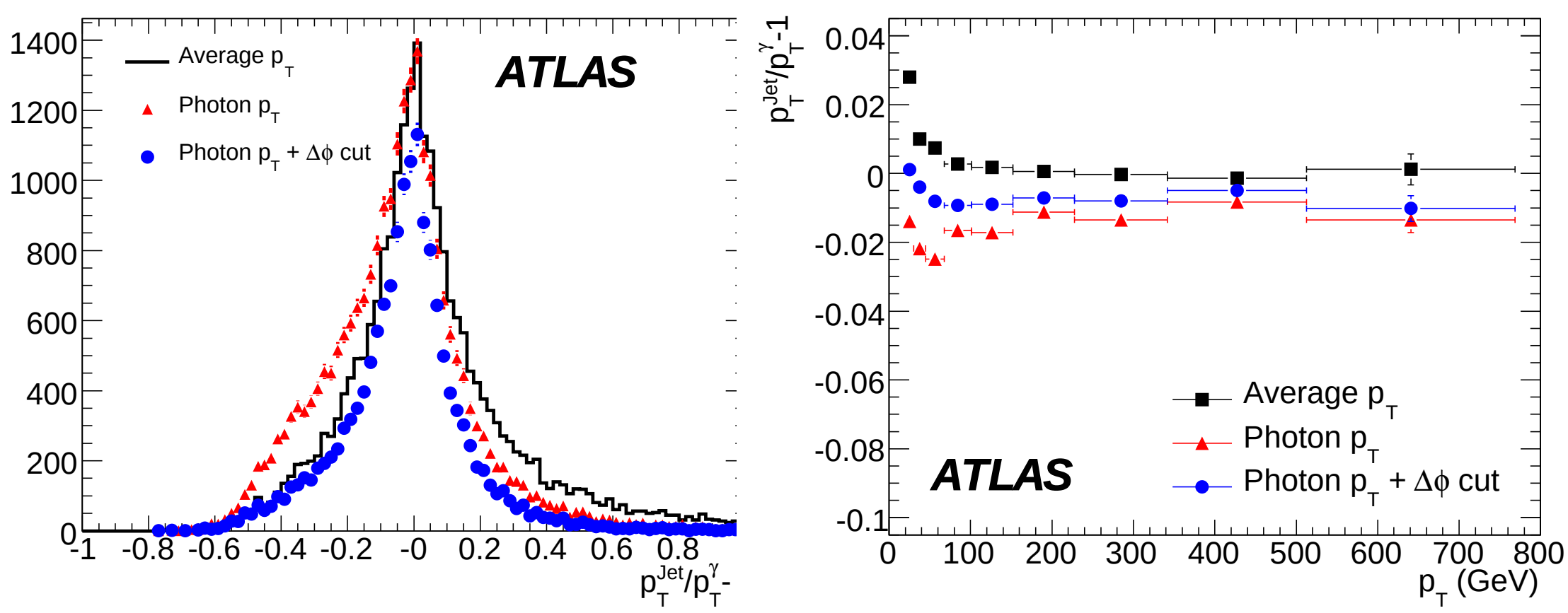

Figure 6: Left: $p_{\mathrm{T}}$ balance at particle level for events with $101 < p_{\mathrm{T}} < 152$ GeV. The solid line shows the balance when the $p_{\mathrm{T}}$ reference for binning is taken as the average $p_{\mathrm{T}}$ of the photon and the jet; the triangles when it is taken as the photon $p_{\mathrm{T}}$. The circles show the balance when the photon $p_{\mathrm{T}}$ is used and the photon and the jet are required to be back-to-back within 0.2. Right: the $p_{\mathrm{T}}$ dependence of the most probable value of the particle level jet balance for these three cases.

The right panel of the figure shows the most probable value of the particle level jet balance for these three cases. The effect of varying the quantity used to define the $p_T$ bins can be as large as 3% for low $p_T$. The $\Delta\phi$ cut helps reducing the effect to the percent level. More details on this issue can be found in Ref. [14]. A disadvantage of using the average $p_T$ to define the energy bins is to introduce a coupling with the jet energy scale that one is trying to measure. Hence, we decided to use $p_{T,\gamma}$ and $p_{T,Z}$ in the baseline study. The average $p_T$ will be used as a cross-check.

The above studies show that the level of imbalance is determined in first order by "out-of-cone" losses related to the jet algorithm under consideration. In addition ISR/FSR and UE influence the measured balance. ISR/FSR effects can be reduced by suitable cuts like a "back-to-back" requirement. UE can be estimated separately by looking at the energy outside jets and can be subtracted, if necessary. The imbalance, depending on the algorithm and on the choice of the reference scale, is up to $5-10\%$ at 20 GeV and becomes smaller than 1% around $100-200$ GeV. A careful tuning of Monte Carlo simulations will be necessary to reproduce and disentangle the effects that cause the residual imbalance.

In the following, the specific issues of $\gamma$+jet and $Z$+jet data analyses are discussed. In particular, backgrounds, event rates and the kinematic reach for each channel are considered.

### 4.3 Analysis of $\gamma$+jet events

The data used in this section are $\gamma$+jet events generated using PYTHIA. They are generated in intervals of $p_T$ of the hard scattered partons to provide adequate statistical coverage over a wide range of $p_T$. The signal has been generated via the annihilation process $q\bar{q} \to g\gamma$ (ISUB = 15) [5], and the QCD Compton process $qg \to q\gamma$ (ISUB = 30). The Compton process dominates over the whole $p_T$ range.

The main background comes from QCD dijet events, where one of the jets is misidentified as a photon in the calorimeter. For background estimation, QCD jets are simulated with the same version of PYTHIA and similar intervals of $p_T$. Table 1 gives the cross-sections for the various $p_T$ intervals for $\gamma$+jet and QCD dijet events.

In this study, the leading jet $p_T$ balance method is used, which consists of first selecting the leading photon in the event and then selecting the leading jet in the opposite hemisphere. The balance is investigated at the reconstruction level using full detector simulation. The cone jet algorithm with $R = 0.7$ is used. Only jets with $|\eta| < 2.5$ are considered.

[4] The jet $p_T$ is used as an approximation to the parton $p_T$.

Table 1: $\gamma$+jet and QCD dijet cross-sections for the various $p_T$ intervals. The total inelastic cross-section is $8 \times 10^{11}$ pb.

| $p_T$ interval | $\gamma$+jet $\sigma$ (pb) | Dijet $\sigma$ (pb) | $\sigma$ ratio (Dijet / $\gamma$+jet) |
|---|---|---|---|
| $17 < p_T < 35$ GeV | $2.61 \times 10^5$ | $1.38 \times 10^9$ | $5.29 \times 10^3$ |
| $35 < p_T < 70$ GeV | $2.76 \times 10^4$ | $9.33 \times 10^7$ | $3.38 \times 10^3$ |
| $70 < p_T < 140$ GeV | $2.59 \times 10^3$ | $5.88 \times 10^6$ | $2.27 \times 10^3$ |
| $140 < p_T < 280$ GeV | $1.99 \times 10^2$ | $3.08 \times 10^5$ | $1.55 \times 10^3$ |
| $280 < p_T < 560$ GeV | $1.17 \times 10^1$ | $1.25 \times 10^4$ | $1.07 \times 10^3$ |
| $560 < p_T < 1120$ GeV | $4.90 \times 10^{-1}$ | $3.60 \times 10^2$ | $7.35 \times 10^2$ |

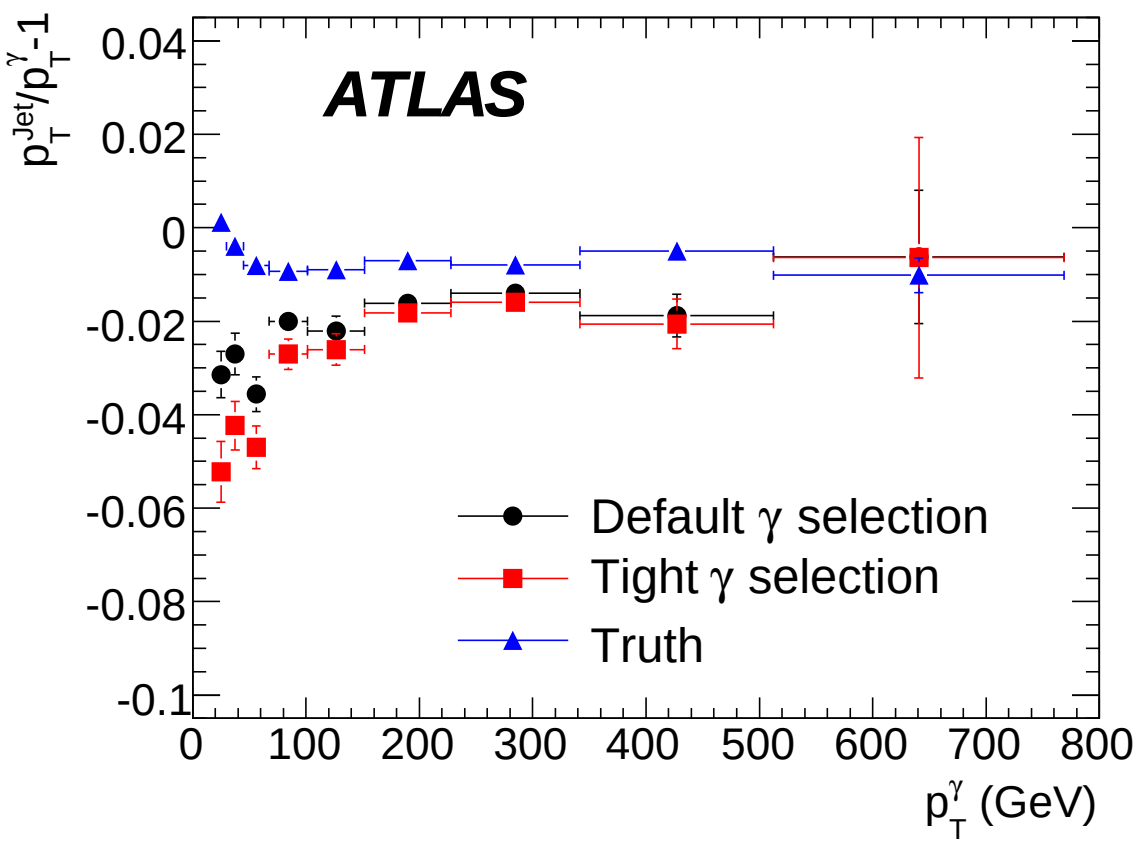

Figure 7: The most probable value of the balance at reconstruction level for cone jets with $R = 0.7$. Black and dots are for default and tight selection, respectively, and the points show the truth level balance. The back-to-back $\Delta\phi$ cut is applied.

Two different photon selections have been considered. The tight photon selection adds isolation criteria in the calorimeter and tracker to provide further jet rejection, as discussed later in this section. A calorimeter isolation criterion is applied on the relative transverse isolation energy in a cone with half-opening angle 0.2, requiring that $E_T(\text{cone})/p_{T,\gamma} < 0.05$. The track isolation criterion requires that the number of tracks pointing to any jet around the photon is less than three. Here, track parameters are measured at the origin. The tight selection maintains a good efficiency of $> 60\%$ above 100 GeV while it is significantly reduced at low $p_T$. These tight selection cuts have not been optimized. Their purpose in this note is to show that additional purity can be obtained, without too much loss of efficiency, at least for the higher $p_T$ range. The photon energy calibration, $p_T(\text{reconstructed})/p_T(\text{truth})$, has been checked and is between 0.995 and 1 over most of the $p_T$ range. No photons impinging in the calorimeter crack regions with $1.37 < |\eta| < 1.52$ are used.

The average response of jets over the various calorimeter regions was checked and a residual miscalibration of about 1% was found for all $p_T$ values. Jets close to the crack regions are not well calibrated, therefore jets with $1.3 < |\eta| < 1.8$ are excluded from further analysis. Eventually we will rely on dijet balance to correct for such $\eta$-dependent effects in the calibration (see Section 5).

Figure 7 shows the measured $p_T$ balance. The photon $p_T$ is used as the reference and the standard back-to-back $\Delta\phi$ cut is applied. Table 2 shows the fitted mean, the statistical uncertainty and the integrated luminosity for each $p_T$ interval for the tight photon selection. The $p_T$ balance above 80 GeV is flattening at the level of -0.02. At low $p_T$ the balance with the tight photon selection is a few percent below the balance measured with the default cuts. The photon reconstruction efficiency is low, and the stringent isolation cuts likely bias the sample composition rejecting events with strong ISR or underlying event. The match between the leading reconstructed jet and the leading truth jet is better than 98.5% for jets above 70 GeV; it degrades to 85% for jets above 17 GeV. The reconstruction level $p_T$ balance is one percent below the truth level $p_T$ balance. This is the result of the residual miscalibration of the jets in the particular release of the ATLAS event reconstruction software, as checked by comparing the reconstructed jet $p_T$ with the truth particle level jet $p_T$.

The PYTHIA QCD jet samples described in Table 1 were used to estimate the influence of the dijet background. The jet rejection factors for the standard photon selection and the tight photon selection were studied as a function of jet $p_T$. With the default photon selection criteria, the probability of identifying a jet as the leading photon is of the order of $10^{-3}$ and has some $p_T$ dependence. The resulting signal to background ratio is estimated to be of the order of 1 above 100 GeV and degrades to about 0.1 for lower $p_T$. The tight photon selection improves the signal to background reduced by a factor of three or more.

Table 2: Fitted mean $p_T$ balance, error and integrated luminosity of the analysed samples for the various $p_T$ intervals for the tight photon selection. The last column shows the precision expected for 10 pb$^{-1}$ obtained by scaling the error according to recorded integrated luminosity.

| $p_T$low edge | Bin width | Fitted balance | Integrated luminosity, pb$^{-1}$ | Error for 10 pb$^{-1}$ |
|---|---|---|---|---|
| 20 GeV | 10 GeV | $-0.052 \pm 0.007$ | 0.67 | 0.2% |
| 30 GeV | 15 GeV | $-0.042 \pm 0.005$ | 0.67 | 0.2% |
| 45 GeV | 22.5 GeV | $-0.047 \pm 0.005$ | 9.1 | 0.4% |
| 67.5 GeV | 33.5 GeV | $-0.027 \pm 0.003$ | 9.1 | 0.4% |
| 101 GeV | 51 GeV | $-0.026 \pm 0.003$ | 47 | 0.7% |
| 152 GeV | 76 GeV | $-0.018 \pm 0.002$ | 47 | 0.4% |
| 228 GeV | 114 GeV | $-0.016 \pm 0.002$ | 535 | 1.7% |
| 342 GeV | 171 GeV | $-0.021 \pm 0.005$ | 535 | 4% |
| 513 GeV | 256 GeV | $-0.006 \pm 0.026$ | 535 | 19% |

Figure 8 left shows the $p_T$ balance distribution for QCD background events passing the default photon selection in the interval $140 < p_{T,\gamma} < 280$ GeV. The back-to-back $\Delta\phi$ cut has been applied. The fitted most probable value of the distribution with the default photon selection and the $\Delta\phi$ cut is 0.13. Applying the tight selection reduces the number of background events, but a few events still remain in the tail. It will thus be desirable to maintain a good signal to background ratio to avoid a bias that would be difficult to control. For the tight photon selection, the mean of the Gaussian fit to the signal shown in Fig. 8 (right) for the interval $140 < p_{T,\gamma} < 280$ GeV changes by less than 1% when the background is added in.

Thus, the background studies have shown that above $\sim 80$ GeV, the signal to background ratio is good enough to avoid bias. A more detailed and higher statistics study of the background rejection at lower $p_T$ should be carried out, to understand if the range of high precision could be extended to lower $p_T$.

Table 2 shows that a good statistical precision is obtained for a relatively small integrated luminosity. For example, 100 pb$^{-1}$ is sufficient to reach a precision of $1-2\%$ for jets in the range $300 < E_T < 500$ GeV. We expect that the threshold of the un-prescaled photon trigger will be 60 GeV for early running. Hence above that value the full statistics should be recorded while prescale factors will have to be applied for lower $p_T$.

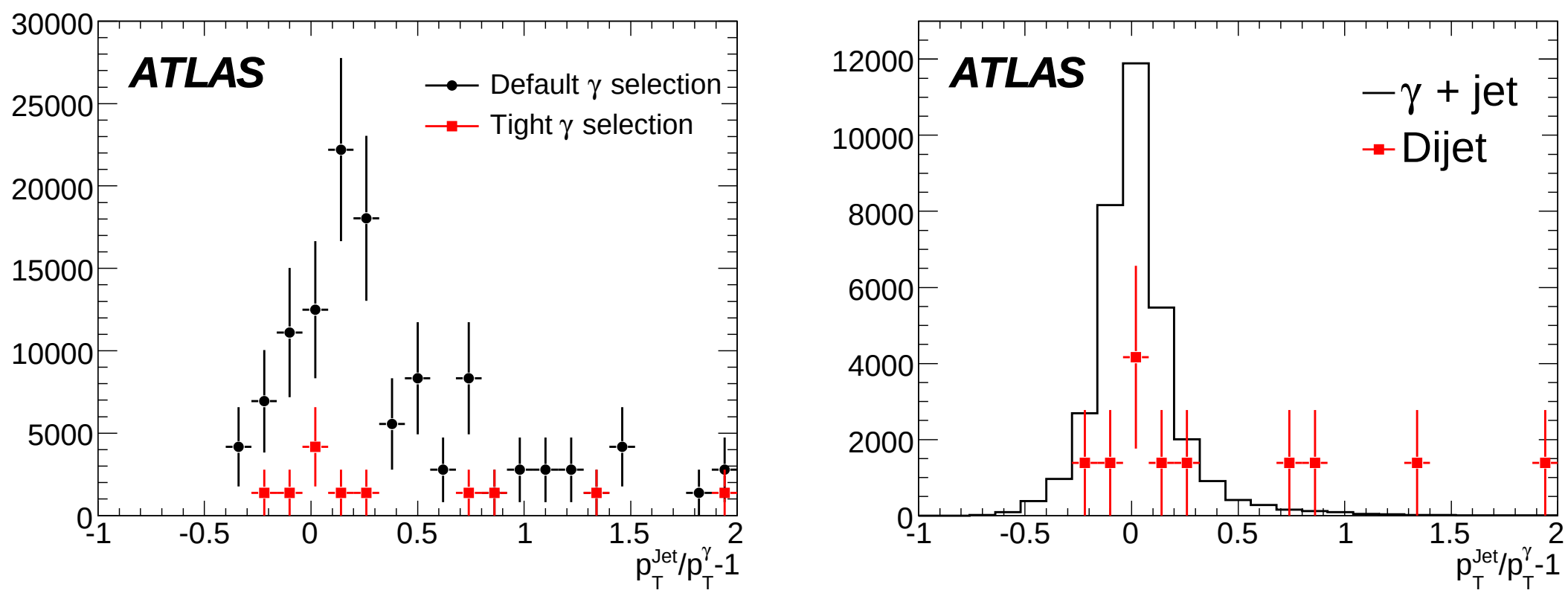

Figure 8: Left: $p_T$ balance for the background sample of $140 < p_T < 280$ GeV for the default and tight photon selection. Right: $p_T$ balance for the signal and background sample in the interval $96 < p_{T,\gamma} < 224$ GeV for tight photon selection.

### 4.4 Analysis of $Z+$jet events

A sample of inclusive $Z$ boson events corresponding to an integrated luminosity of $500\,\text{pb}^{-1}$ has been produced using ALPGEN. The MLM matching cut is imposed at $p_T > 20$ GeV and $|\eta| < 6$, and the factorization scale is set at $m_Z^2 + p_{T,Z}^2$ where $m_Z$ is the $Z$ boson mass. The invariant mass is required to lie between 40 and 200 GeV. A generator-level filter requires one truth cone jet with $R = 0.4$, with $p_T > 20$ GeV and $|\eta| < 5.0$ and two electrons/muons with $p_T > 10$ GeV and $|\eta| < 2.7$ in the event.

Another sample of inclusive $Z$ events corresponding to $120\,\text{pb}^{-1}$ , as well as the relevant background samples have been generated using PYTHIA. The signal has been generated via the Drell-Yan process (ISUB = 1) [5]. As for $\gamma+$jet events, the $qg$ Compton process dominates over the whole $p_T$ range. The Drell-Yan process contributes due to higher order QCD radiation which may reproduce the leading jet.

The $Z \to ee$ and $W \to e\nu$ events are produced with a generator-level filter requiring one truth electron with $p_T > 10$ GeV and $|\eta| < 2.7$. The $Z \to \tau\tau$ events are generated with a filter requiring two electrons/muons with $p_T > 5$ GeV and $|\eta| < 2.8$. For $Z \to \tau\tau$ and $Z \to ee$ the dilepton mass is required to be larger than 60 GeV. For more details on the data samples see Ref. [16].

The fully simulated events are required to pass the isolated single-electron trigger with a $p_T$ threshold of 25 GeV or the isolated di-electron trigger with the $p_T$ threshold of 15 GeV at trigger level 1 and 2. Electrons are required to have $p_T > 25$ GeV, $|\eta| < 2.5$ and to be outside the calorimeter cracks defined as $1.37 < |\eta| < 1.52$. Jets are reconstructed with the cone tower algorithm with $R = 0.7$. They are required to have a distance of $\Delta R > 0.4$ from a reconstructed electron and $p_T > 30$ GeV.

The $Z$ bosons are selected by requiring two reconstructed electrons with invariant mass in the range $m_Z \pm 20$ GeV. If more than two electrons are reconstructed, the pair with the invariant mass closest to $m_Z$ is chosen. The combined distribution of the invariant mass for the signal and background events is shown in the left panel of Fig. 9 for ALPGEN $Z+$jet events where jets with $p_T > 40$ GeV are reconstructed using the cone tower algorithm with $R = 0.7$. The main backgrounds are from QCD multijets and from top production, while $Z \to \tau\tau$ and $W \to e\nu$ events are an order of magnitude smaller. The background level is reasonably low. Requiring the jet and the $Z$ to be back-to-back in azimuth within 0.2 reduces the background to a negligible level.

The right panel of Fig. 9 shows the measured balance for cone jets with $R = 0.7$ as predicted by ALPGEN for $500\,\text{pb}^{-1}$ . The precision with which the $p_T$ balance can be measured experimentally depends on the efficiency for reconstructing the $Z$ in its electron or muon decay channels, on the trigger efficiency and on the experimental width of the $p_T$ balance distribution. Table 3 shows the fitted most probable value of the $p_T$ balance and the statistical precision for an integrated luminosity of $500\,\text{pb}^{-1}$ obtained with the $Z \to ee$ sample. Below 200 GeV, the precision is of the order of 0.8%, while above 200 GeV, it gets worse. The $Z \to \mu\mu$ decay mode can also be used. A similar efficiency and background level can be achieved as shown in [16] doubling the available statistics. A statistical precision of 1% can be achieved below 200 GeV with an integrated luminosity of $100\,\text{pb}^{-1}$ , while above the precision is at the level of 2%. This assumes also that the complete $\eta$ range can be used for the measurement, relying on techniques like dijet balancing to ensure uniformity across $\eta$ to the desired precision.

The measurement is also affected by systematic uncertainties. The balance is sensitive to the correct modeling of higher order radiation. Comparing the predictions from different Monte Carlo simulation programs gives an indication of the size of such effects. This is illustrated in Fig. 10, in which ALPGEN and PYTHIA samples of $Z+$jet events corresponding to $120\,\text{pb}^{-1}$ and $500\,\text{pb}^{-1}$ (ALPGEN only) are compared after applying the $\Delta\phi$ cut, for cone jets with $R = 0.7$. PYTHIA predicts a more negative average $p_T$ balance at low $p_T$ than ALPGEN. This effect is related to broader distributions and larger tails in the balance distributions resulting from low $p_T$ activity in PYTHIA induced by higher order radiation. The differences between the two generators can be tested with $\sim 100\,\text{pb}^{-1}$ of data for $p_T < 100$ GeV.

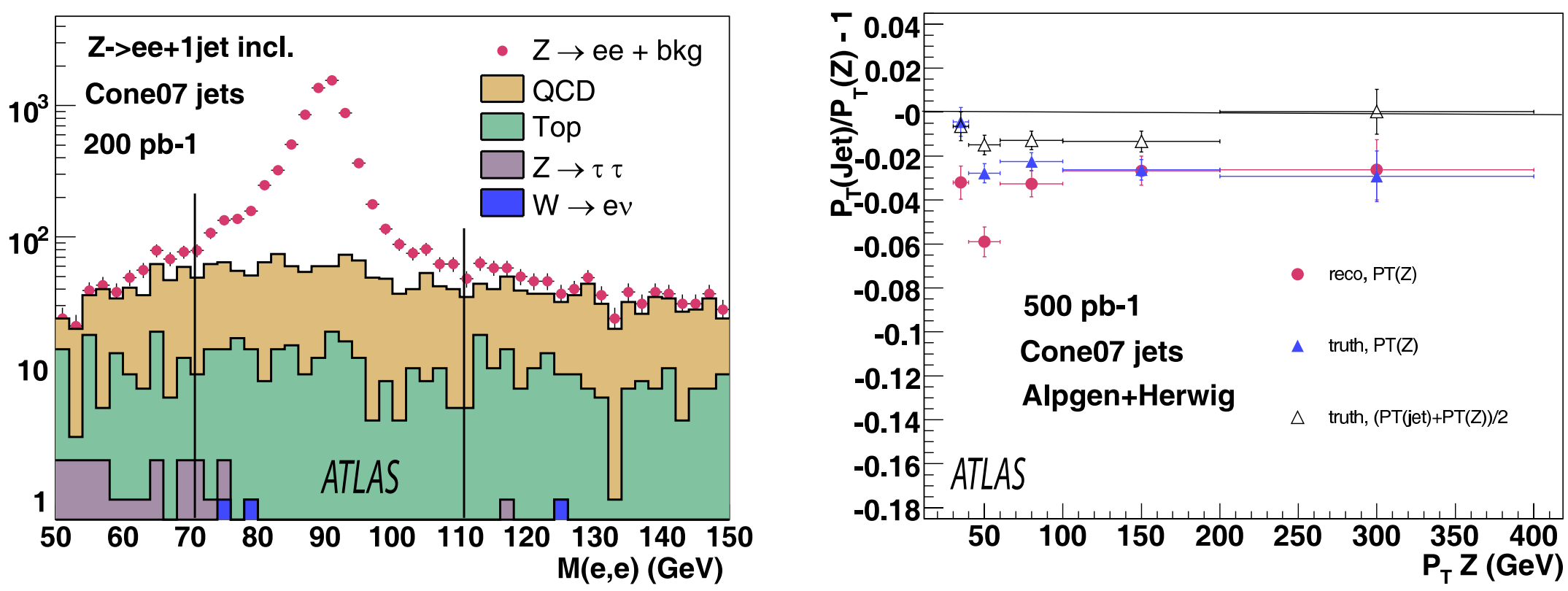

Figure 9: Left: Distribution of the dielectron mass for $Z \to ee+ \geq 1$jet events and the relevant background in a simulated event sample corresponding to an integrated luminosity of $200\,\text{pb}^{-1}$ with cone jets with $R = 0.7$. Right: The $p_T$ balance for an integrated luminosity of $500\,\text{pb}^{-1}$ of cone jets with $R = 0.7$ in events generated with ALPGEN in 5 bins of $p_{T,Z}$. The red dots are for reconstructed jets, solid triangles for truth jets and open triangles for truth in bins of average $p_{T,Z}$ and jet $p_T$.

Table 3: The fitted most probable value of the $p_T$ balance and the statistical precision for an integrated luminosity of $500\,\text{pb}^{-1}$ obtained with the $Z \to ee$ + jets sample for reconstructed cone jets with $R = 0.7$ and $R = 0.4$.

| $p_T$ interval | cone 0.7 jet balance | error | cone 0.4 jet balance | error |
|---|---|---|---|---|
| $30 < p_T < 40$ GeV | −0.032 | 0.008 | −0.150 | 0.007 |
| $40 < p_T < 60$ GeV | −0.059 | 0.007 | −0.162 | 0.008 |
| $60 < p_T < 100$ GeV | −0.032 | 0.006 | −0.113 | 0.007 |
| $100 < p_T < 200$ GeV | −0.027 | 0.007 | −0.081 | 0.008 |
| $200 < p_T < 400$ GeV | −0.026 | 0.014 | −0.060 | 0.016 |

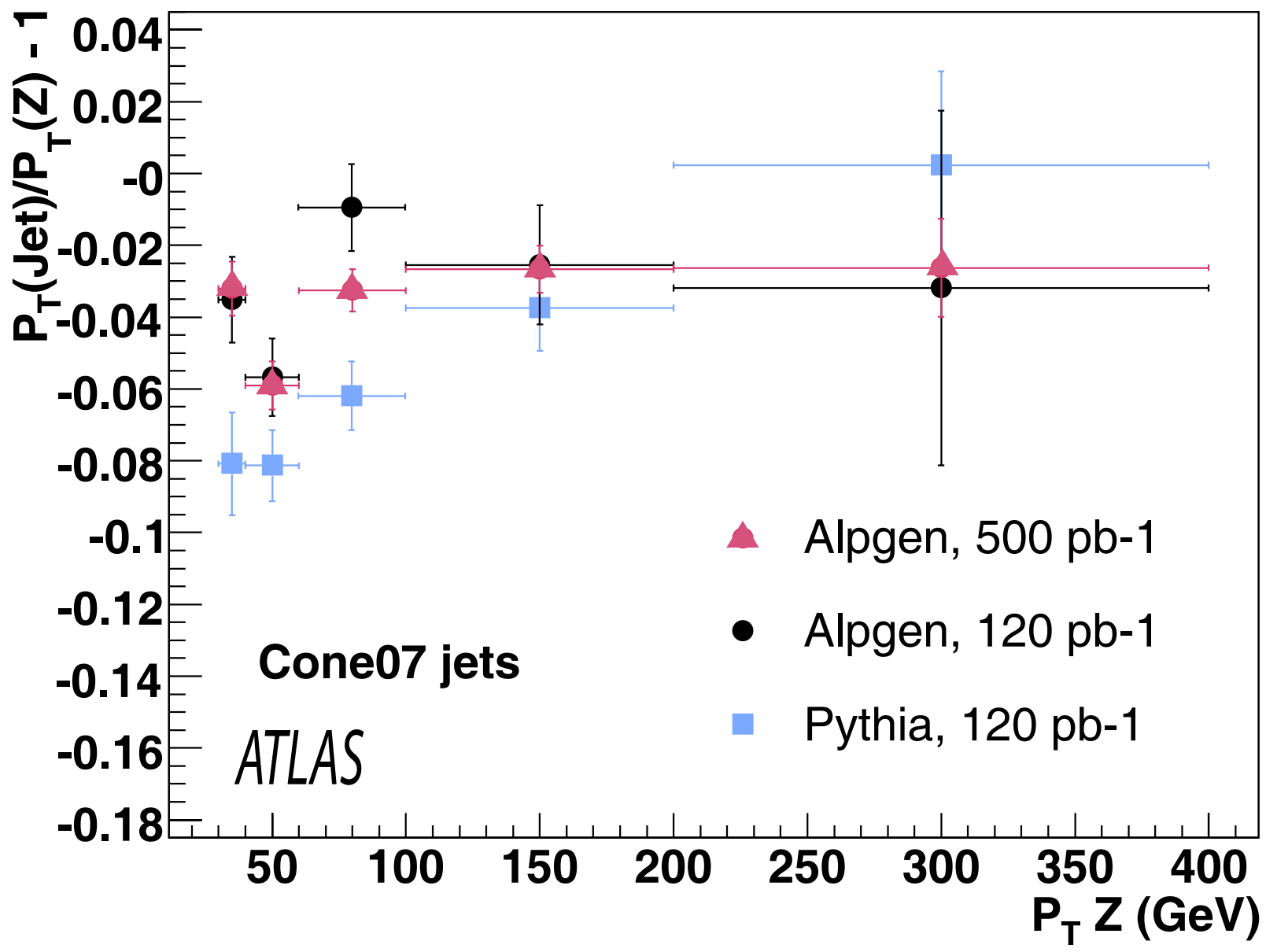

Figure 10: The $p_T$ balance for an integrated luminosity of $120\text{pb}^{-1}$ and $500\text{pb}^{-1}$ in events generated with ALPGEN (dots and triangles, respectively) and for $120\text{pb}^{-1}$ in events generated with PYTHIA (squares) in bins of $p_{T,Z}$ for cone jets with $R = 0.7$.

### 4.5 The missing $E_T$ projection method

The missing $\vec{E}_T$ projection method is an alternative approach to test the jet energy scale and has been used successfully by the D0 experiment to determine the response of their calorimeter to hadronic jets [17]. It is explained here for the example of $\gamma$+jet events, but it can also be used for $Z$+jet events. Details about the study presented in this section can be found in Ref. [15].

At leading order, momentum conservation between the photon and the jet gives

$$\vec{E}_{T,\gamma} + \vec{p}_{T,\text{parton}} = 0\ . \tag{3}$$

Neglecting parton showering and hadronization effects, this can be rewritten at particle level as

$$\vec{E}_{T,\gamma} + \vec{E}_{T,\text{jet}} \approx 0\ , \tag{4}$$

where momentum has been replaced by energy, which is a good approximation for light quark jets.

The systematic effects of parton showering and hadronization, including initial state radiation (ISR) and final state radiation (FSR), can be reduced by requiring that the leading jet and the photon are back-to-back within 0.2 in azimuth in a similar way to what was shown in Fig. 2.

At the calorimeter level, the balance equation is modified to

$$e\vec{E}_{T,\gamma} + j(E_\text{jet})\vec{E}_{T,\text{jet}} = -\vec{E}_T^\text{miss}\ . \tag{5}$$

where $e$, $j$ are the electromagnetic and hadronic responses, $E_\text{calo}/E_\text{particle}$, of the calorimeters, respectively. It is implicit in Eq. 5 that energy deposited in the calorimeters by the underlying event and pileup is symmetric in $\phi$. These assumptions have been tested and their effect is small. Anticipating that the electromagnetic scale will be well calibrated, for example by using the electrons from the decay of a Z boson, we can assume that $e \approx 1$. The quantities can then be projected in the direction of the photon, yielding

$$E_{T,\gamma} + j(E_{jet})\vec{E}_{T,jet} \cdot \hat{n}_\gamma = -\hat{n}_\gamma \cdot \vec{E}_T^{miss} , \tag{6}$$

where $\hat{n}_\gamma$ is the unit vector in the direction of the photon. Using the relation

$$\vec{E}_T^{miss} = -\vec{E}_{T,\gamma} - \sum{}' \vec{E}_T , \tag{7}$$

where $\sum'$ indicates a sum over all activity in the calorimeter other than the $\gamma$, the response can be further simplified as

$$j = \frac{\sum' \vec{E}_T \cdot \hat{n}_\gamma}{\vec{E}_{T,jet} \cdot \hat{n}_\gamma} = \frac{\sum' \vec{E}_T \cdot \hat{n}_\gamma}{E_{T,\gamma}}, \tag{8}$$

where Eq. 4 was used in the last step. In this form it is clear that this method of measuring the response is independent of the underlying event, since the hadronic activity outside the $\gamma$+jet system is approximately $\phi$-symmetric and its contribution to the sum cancels out. It is also (mostly) independent of the jet algorithm.

However, the jet response depends on the jet energy because the relative fraction of EM energy in the calorimeter level jet increases with increasing energy, hence a correction for non-compensation must be applied. Thus, the response has to be measured as a function of the jet energy. If we bin the response as a function of the measured jet energy ($E_{jet}$), it would be biased towards lower $E_{T,\gamma}$ due to the bin migration described in Section 4.2. Instead, the jet energy is taken as $E' = E_{T,\gamma} \cdot \cosh(\eta_{jet})$, which uses the much better measured energy of the photon and projects it in the jet direction; this procedure is based on the momentum balance of Eq. 4. The EM scale jet energy distribution is fitted with a Gaussian in each bin of reference energy $E'$, and the jet energy scale is then calculated. The correspondence between $E'$ and the measured jet energy is also determined. The resulting jet energy scale can then be plotted as a function of $E_{jet}$.

The response function is then plotted for each algorithm as a function of the corresponding mean jet energy and parameterized by

$$j(E) = b_0 + b_1 \ln \frac{E}{E_{scale}} + b_2 \ln^2 \frac{E}{E_{scale}} \tag{9}$$

with $E_{scale}$ set to 200 GeV and the $b$ parameters fitted, as shown in Fig. 11.

The fit function is used to correct the jet energy at the EM scale $E_T^{meas}$. The corrected energy $E_T^{calib} = E_T^{meas}/j(E)$ is then compared to the truth information $E_T^{MC}$ in the Monte Carlo simulation. Figure 12 is a plot of the ratio $E_T^{MC}/E_T^{calib}$ as a function of jet energy, which shows a linear response within $\approx 2\%$ over the range $50-900$ GeV.

This procedure accounts for the fact that the ATLAS calorimeters are non-compensating and that the electromagnetic content of a hadronic shower is energy dependent. Further adjustments, such as out-of-cone corrections, must be made to obtain the final jet energy scale. It is only at this point that jet algorithm dependent corrections are made and is therefore a useful method as a cross-check of our understanding of the energy scale with different sensitivity to systematic effects.

## 4.6 Conclusions of the studies of $\gamma/Z$ + jet events

The use of the in-situ $\gamma/Z$+jet processes allows us to propagate the knowledge of the electromagnetic scale characteristic of the $\gamma$ or the $Z \to \ell\ell$ decay to the hadronic recoil system. Typically the leading jet is required to be back-to-back in azimuth with the $\gamma$ or the $Z$ boson, and the $p_T$ balance between them is used to connect the two scales.

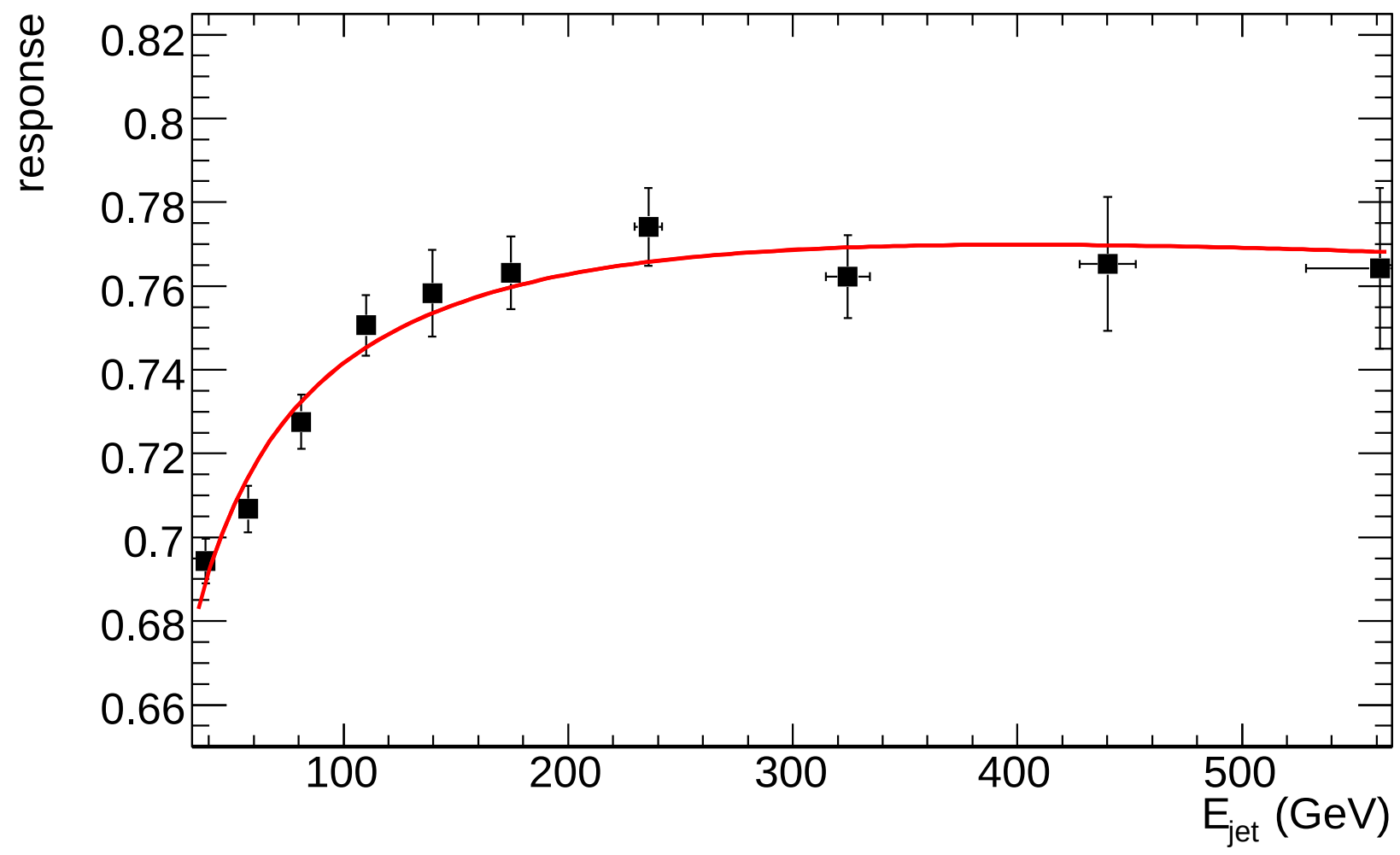

Figure 11: The energy dependence of the jet response for cone jets with $R = 0.4$. The solid line corresponds to the fit using Eq. 9.

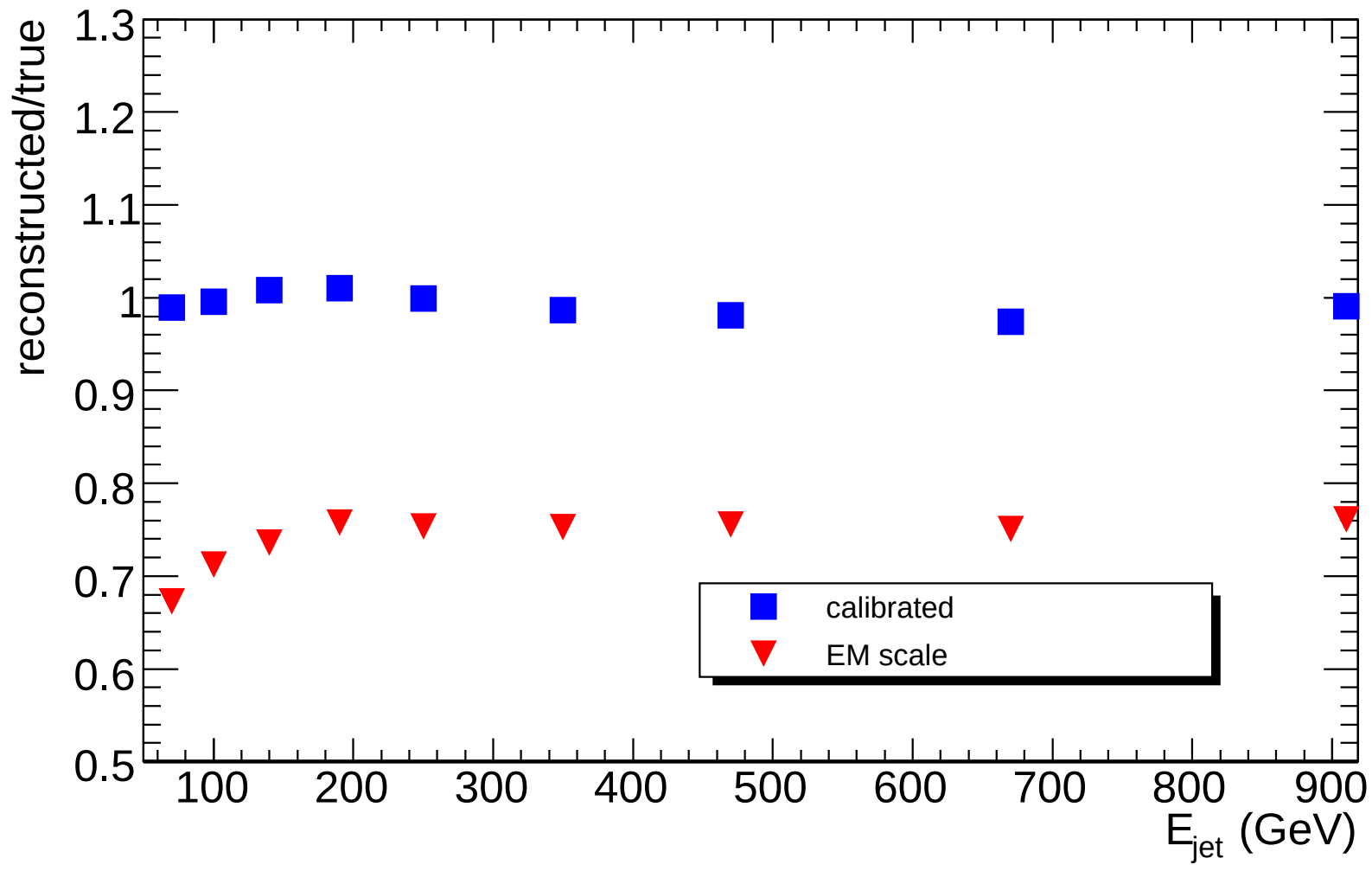

Figure 12: The ratios $E_T^{MC}/E_T^{calib}$ (triangles) and $E_T^{MC}/E_T^{meas}$ (squares) for jets reconstructed using the cone algorithm with $R = 0.4$. See the text for an explanation of the symbols.

The balance is affected by various physics effects which systematically limit the precision of the in-situ validation procedure. These effects can be as large as $5-10\%$ at 20 GeV and tend to decrease to the percent level at about 100 GeV.

The $\gamma$+jet process has the advantage of higher statistics compared to $Z$+jet. However, at low $p_T$ it may be seriously affected by background from QCD jets. Hence it may turn out to be most useful above $\sim 50-100$ GeV where the signal to background ratio is more favorable and where an inclusive single photon trigger is available. A statistical precision better than a percent is achieved already with 10pb$^{-1}$ at 100 GeV, and in the range $300-500$ GeV the percent level is reached with about 100pb$^{-1}$. The $Z$+jet process allows us to measure the absolute energy scale in the low $p_T$ range with a statistical precision of 1% with $\sim 300\text{pb}^{-1}$ of data. It provides also the highest $p_T$ reach with percent level precision in the range $300-500$ GeV with $\sim 200\text{pb}^{-1}$. Above $\sim$ 500 GeV, other methods will be used as discussed further on.

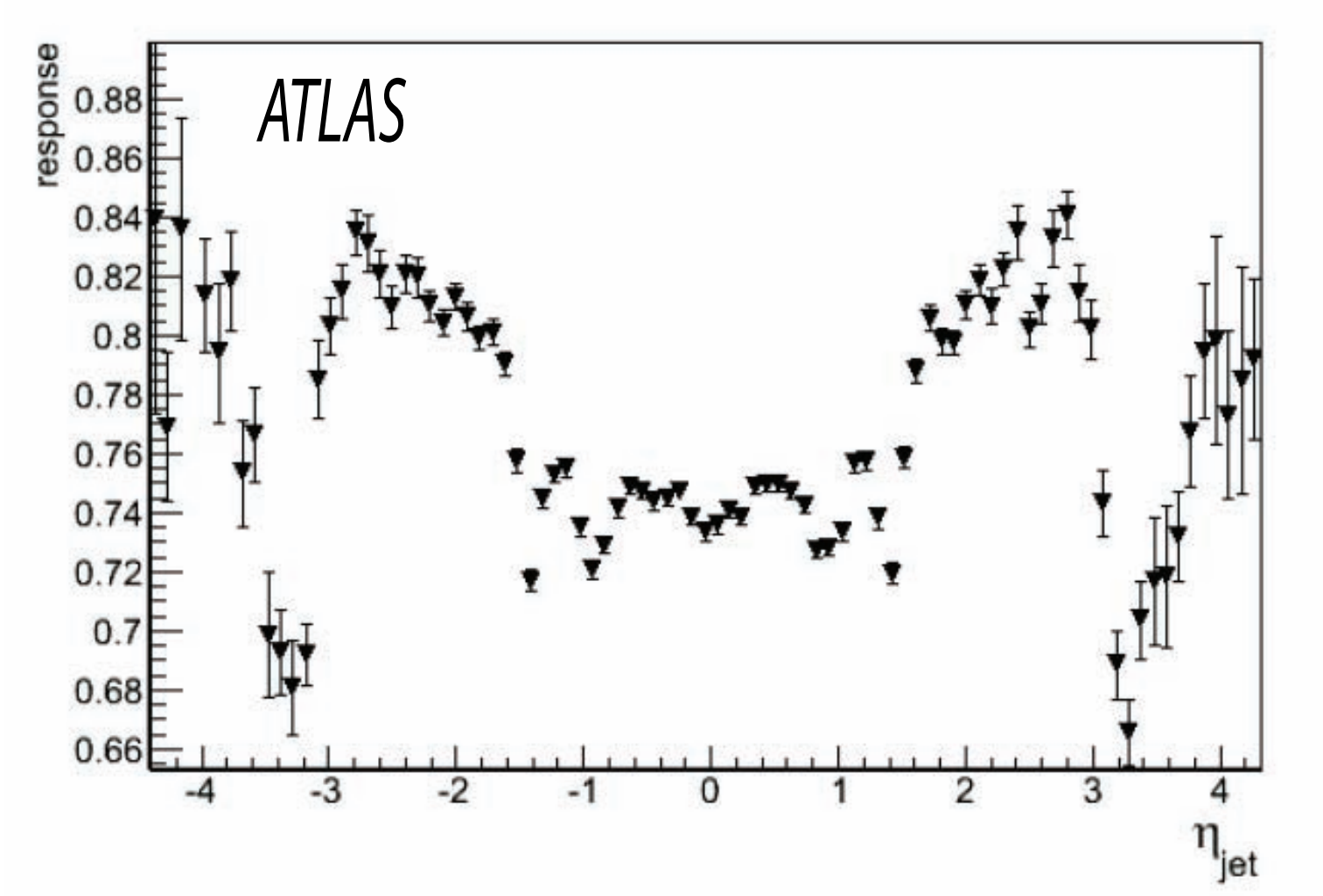

Figure 13: The jet response $p_T$(reconstructed)$/p_T$(truth) at the EM scale versus the jet pseudorapidity $\eta$ .

# 5 In-situ studies using QCD jet events

## 5.1 Calorimeter intercalibration in QCD dijet samples

The calorimeter response $p_T$(reconstructed)$/p_T$(truth) for jets at the EM scale reveals significant variations with pseudorapidity $\eta$ , as shown in Fig. 13. The dips correspond to known cracks and dead material regions. The response may also be non-uniform in azimuth $\phi$ for the same $\eta$ due to the variations in the basic calorimeter response and due to the non-uniform distribution of the dead material. The response in $\phi$ and $\eta$ should become flat at the hadronic scale. This has to be checked in-situ with high precision, and if necessary, an intercalibration of different regions of the calorimeter has to be performed. For this study we use PYTHIA QCD jet events, generated in intervals of $p_T$ of the hard scattered partons, as described in Section 4.3.

### 5.1.1 Intercalibration in azimuth

The intercalibration in $\phi$ can be checked using the high rates of QCD jet events. As the scattering cross section is constant in $\phi$, the jet rates above a fixed $p_T$ threshold have to be nearly equal in different $\phi$ sectors, as shown in Fig. 14 (left) for 64 $\phi$ sectors in the $\eta$ range $|\eta| < 0.1$ which corresponds to the segmentation of the ATLAS hadronic calorimeter in one calorimeter wheel. Strongly deviating rates in particular sectors would point to dead calorimeter cells, losses in dead material, or noise.

For a perfectly calibrated calorimeter the spread of rates around the mean rate value $N$ should be $\sqrt{N}$ following the Poissonian distribution. Thus, collecting on average e.g. $\sim$1000 events per sector, one would expect the spread of $\sim$30 events, i.e. 3% of the average rate. A significantly larger spread would point to a relative miscalibration of different sectors. The spread can be obtained by filling a histogram with the rates and fitting it with a gaussian. Due to the strongly falling jet $p_T$ distribution the rates are extremely sensitive to the jet energy scale: a $1.5-2\%$ shift in the energy scale would result in $\sim 6-9\%$ change of the rate, which is a $2-3\sigma$ effect for the above example with 1000 events per sector. Thus, the statistics of 1000 events per sector should allow us to control the jet energy scale to within $1.5-2\%$. The luminosity necessary to collect such statistics in 64 $\phi$ sectors within the $\eta$ interval $|\eta| < 0.1$ is shown for different $p_T$ thresholds in Fig. 14 (right). A study of the systematics for this method is on the way.

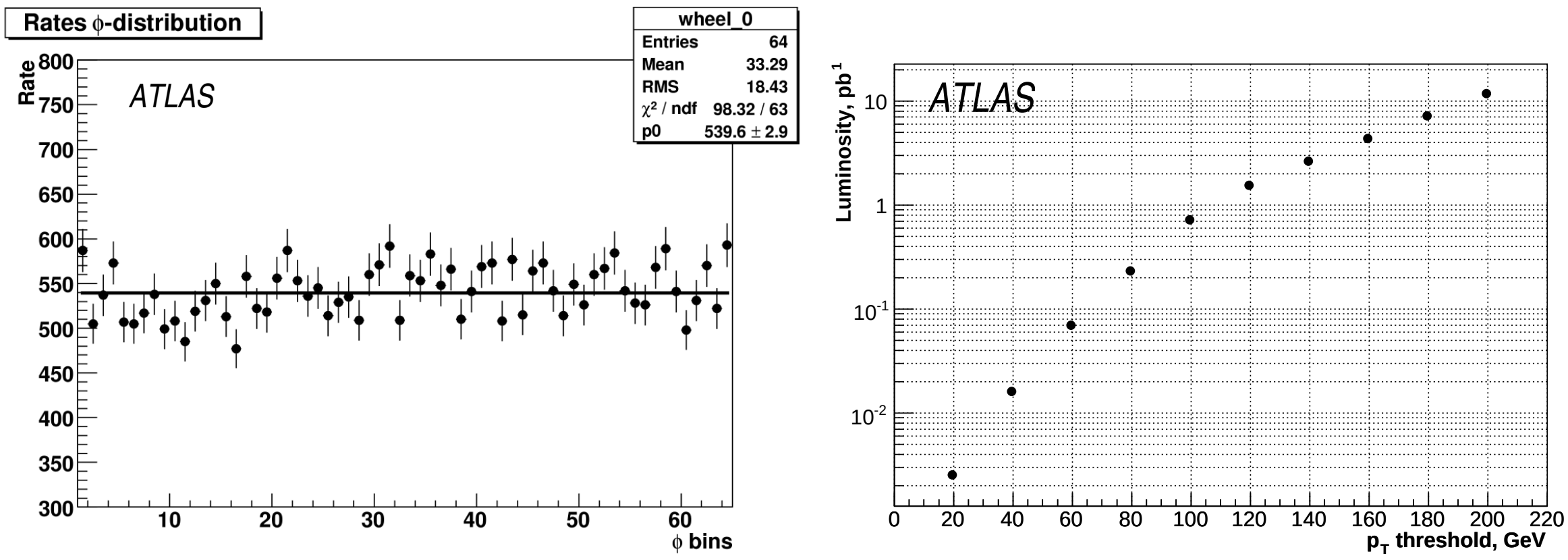

Figure 14: Left: The jet rate as a function of $\phi$ for jets with the transverse momentum above a certain threshold. Right: Integrated luminosity required to collect 1000 events with jets above the given $p_T$ thresholds in each of the 64 $\phi$ sectors in the region $|\eta| < 0.1$.

The trigger menu for the initial running period at the luminosity of $10^{31}$ cm$^{-2}$s$^{-1}$ is currently under development, and the effect of trigger prescaling has not yet been taken into account in this plot. The highest $p_T$ threshold for the inclusive jet trigger is expected to be at $100-150$ GeV, above which no prescaling will be applied. The prescale factors will steeply rise for lower thresholds, such that the recorded event rate should be roughly flat in $p_T$. Scaling the shown luminosity values for the lowest two $p_T$ bins down with the currently foreseen prescale factors shows that the rate of recorded events will be still sufficient to reach the statistical precision of $1.5-2\%$ with $\lesssim 10\,\text{pb}^{-1}$ of data.

However, the rates may be affected by the calibration of the level 1 calorimeter trigger. To avoid a bias, the $p_T$ thresholds chosen for this analysis should be significantly higher than the respective trigger thresholds. This will lead to a loss of statistics, and thus a higher integrated luminosity will be necessary for the same statistical precision.

### 5.1.2 Intercalibration in pseudorapidity

The intercalibration in $\eta$ can be done based on the $p_T$ balance in QCD dijet events. The roughly three orders of magnitude higher cross section of dijet events, as compared to $Z/\gamma+$jet events, allows an intercalibration of the calorimeter with higher granularity, higher precision and for higher energy ranges than can be achieved in $Z/\gamma+$jet events for the same integrated luminosity. Applying the calibration based on the $p_T$ balance, one corrects for the relative difference in calorimeter response for jets, the noise and pile-up in different regions of pseudorapidity $\eta$ . The goal is to achieve a flat response in $\eta$ and to extend the previous calibrations for $|\eta| < 2.5$ to higher pseudorapidities.

The method is straightforward: one calorimeter region in pseudorapidity $\eta$ is chosen as the reference region and jets in other $\eta$ regions are calibrated relative to jets in this region. Normally, one calorimeter $\eta$ wheel in the central range, far from the crack regions, will be chosen as the reference region. Other considerations are the existence of dead or noisy cells in a particular part of the calorimeter, the uniformity of the response in $\phi$ within one wheel etc. The width of the region will also depend on the available statistics.

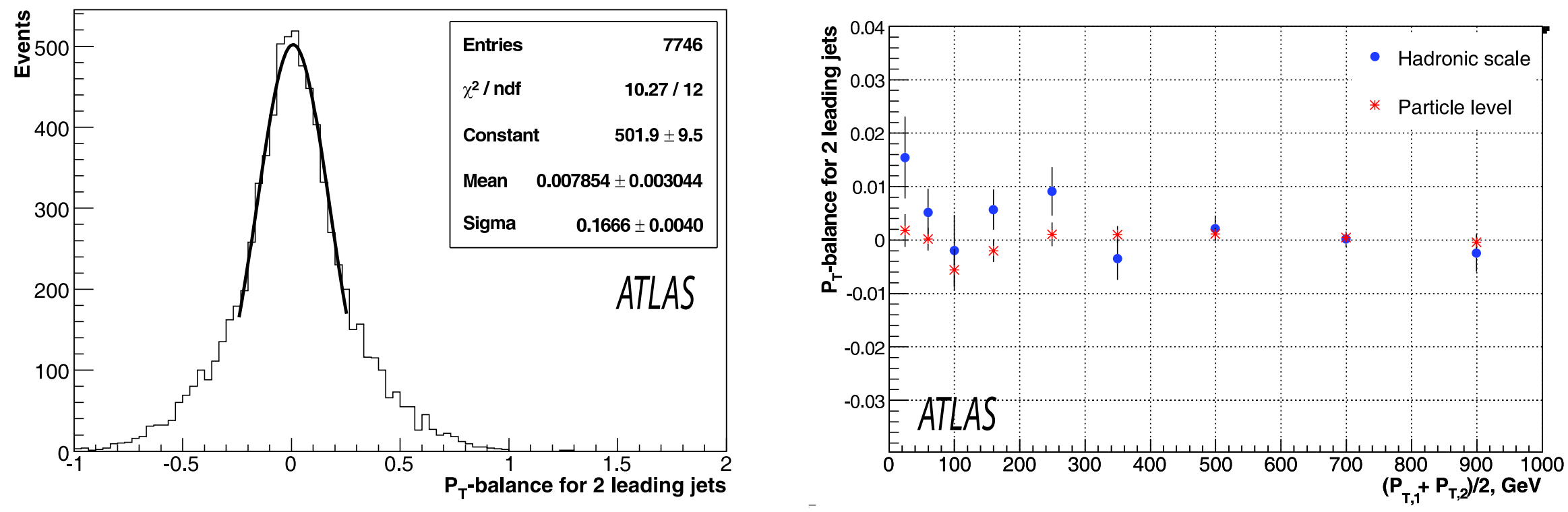

Figure 15: Left: The asymmetry $A$ as measured with both jets in the central region $|\eta| < 0.7$ as defined in Eq. 10. Right: The mean asymmetry obtained from gaussian fits, plotted as a function of the half scalar sum of $p_T$ of both jets at the reconstruction level (closed circles) and at the truth particle level (stars).

We choose to characterize the $p_T$-balance in dijet events by the asymmetry $A$ given by

$$A = \frac{p_T^{\text{probe}} - p_T^{\text{ref}}}{(p_T^{\text{probe}} + p_T^{\text{ref}})/2}\,, \tag{10}$$

where $p_T^{\text{ref}}$ is the transverse momentum of the jet in the reference region, and $p_T^{\text{probe}}$ is that of the jet in the region to be calibrated. Using this form, the $p_T$ balance is symmetric, as shown in Fig. 15 (left) for reconstructed cone jets with $R = 0.7$. It thus provides better properties than the simple definition $p_T^{\text{probe}}/p_T^{\text{ref}}$ which is intrinsically asymmetric.

To limit ISR/FSR and thereby reduce the width of the asymmetry distribution, a cut on the azimuthal angle between the two jets, $\Delta\phi > 3$, is applied. The fitted mean value of the asymmetry as a function of $p_T$ is displayed in the right panel of Fig. 15 and shows that for this definition there is very little bias both for truth particle jets and for reconstructed jets at the hadronic scale.

Further cuts can be applied to obtain a cleaner dijet sample and thus reduce further the width of the $p_T$ balance distributions. For example, the total number of reconstructed jets with $p_T > 10$ GeV can be limited to $N_{\text{jet}} < 4$ or even more strongly to $N_{\text{jet}} = 2$. However, these additional cuts reduce the statistics of the sample, as demonstrated in Fig. 16, which shows the integrated luminosity required to reach a statistical precision of 0.5% of the $p_T$ balance fit mean value in the probe range $0.7 < \eta < 0.8$, where $0 < |\eta| < 0.7$ is taken as the reference region. The luminosity values are shown for different sets of cuts. For the case of applying only the back-to-back cut on the angle between the two leading jets, 0.5% precision can be reached in this $\eta$ region with 10 pb$^{-1}$ of data for $p_T \lesssim 300$ GeV, and with 100 pb$^{-1}$ for $p_T \lesssim 500$ GeV.

Similar plots were produced for the other $\eta$ regions, which show that an order of magnitude more luminosity is necessary for $|\eta| > 2.5$, compared to the central region. On the other hand, the size of the reference region can be increased sequentially, i.e. as soon as one $\eta$ has been checked or recalibrated, it can be added to the reference region to study the next $\eta$ range.

As in the luminosity estimation for the $\phi$ intercalibration in Section 5.1.1, the trigger prescaling has not yet been taken into account in Fig. 16. As in Sect. 5.1.1, scaling the shown luminosity values for the lowest two $p_T$ bins down with the currently foreseen prescale factors shows that the rate of recorded events will be sufficient to reach 0.5% precision with $\lesssim 10$ pb$^{-1}$ of data.

The calibration method is further tested in the full $\eta$ region for different $p_T$ ranges. For each $\eta$ region $i$ in each $p_T$ range $j$, the asymmetry $A$ defined in Eq. 10 is fitted and the mean value $A_{ij}$ is used to calculate a correction factor $c_{ij}$ via

$$c_{ij} = \frac{2-A_{ij}}{2+A_{ij}} \; . \tag{11}$$

In the case of an imperfect original calibration, factors deviating from 1 are obtained. Each jet $p_T$ can then be multiplied by the appropriate factor to restore the balance of fully simulated jets. More details of this study can be found in Ref. [18].

## 5.2 High $p_T$ jet calibration

The calibration of the jet energy scale for high $p_T$ jets (with $p_T \gtrsim 500$ GeV) is a challenge at hadron collider experiments. In this energy range, methods developed for lower $p_T$ jets, normally using well defined objects as reference, start to fail. This difficulty will be further increased at the early stage of data taking, where both our understanding of the detectors and the statistics of such high $p_T$ jets are usually limited. Our goal is to develop a calibration technique for jets with $p_T > 500$ GeV based on data collected with an integrated luminosity of 0.1–1 fb$^{-1}$. The basic idea is to determine the energy scale of high $p_T$ jets from lower $p_T$ jets, for which the energy scale will be obtained by the techniques described earlier in this note. Two calibration techniques are discussed in the following sections: 1) the correlation between different $p_T$ jets in momentum balance in the transverse plane, and 2) the angle between particles in the jets.

### 5.2.1 Multijet $p_T$ balance method

This method is based on selecting events with multiple ($>2$) jets and calibrating the energy scale of the highest $p_T$ jet by requiring a $p_T$ balance between the leading jet and the system of remaining lower $p_T$ jets. The absolute jet energy scale of the non-leading jets, which is expected to be obtained from photon+jets, $Z$+jets and/or $W \rightarrow$ jet jet studies, is thus propagated to the JES of the higher $p_T$ jets using $p_T$ balance methods. The large QCD jet cross-section allows for a very high reach in $p_T$. This method has been studied with samples generated with both ALPGEN and PYTHIA using the fast and full detector simulation. Two different analysis techniques were applied as described in the following.

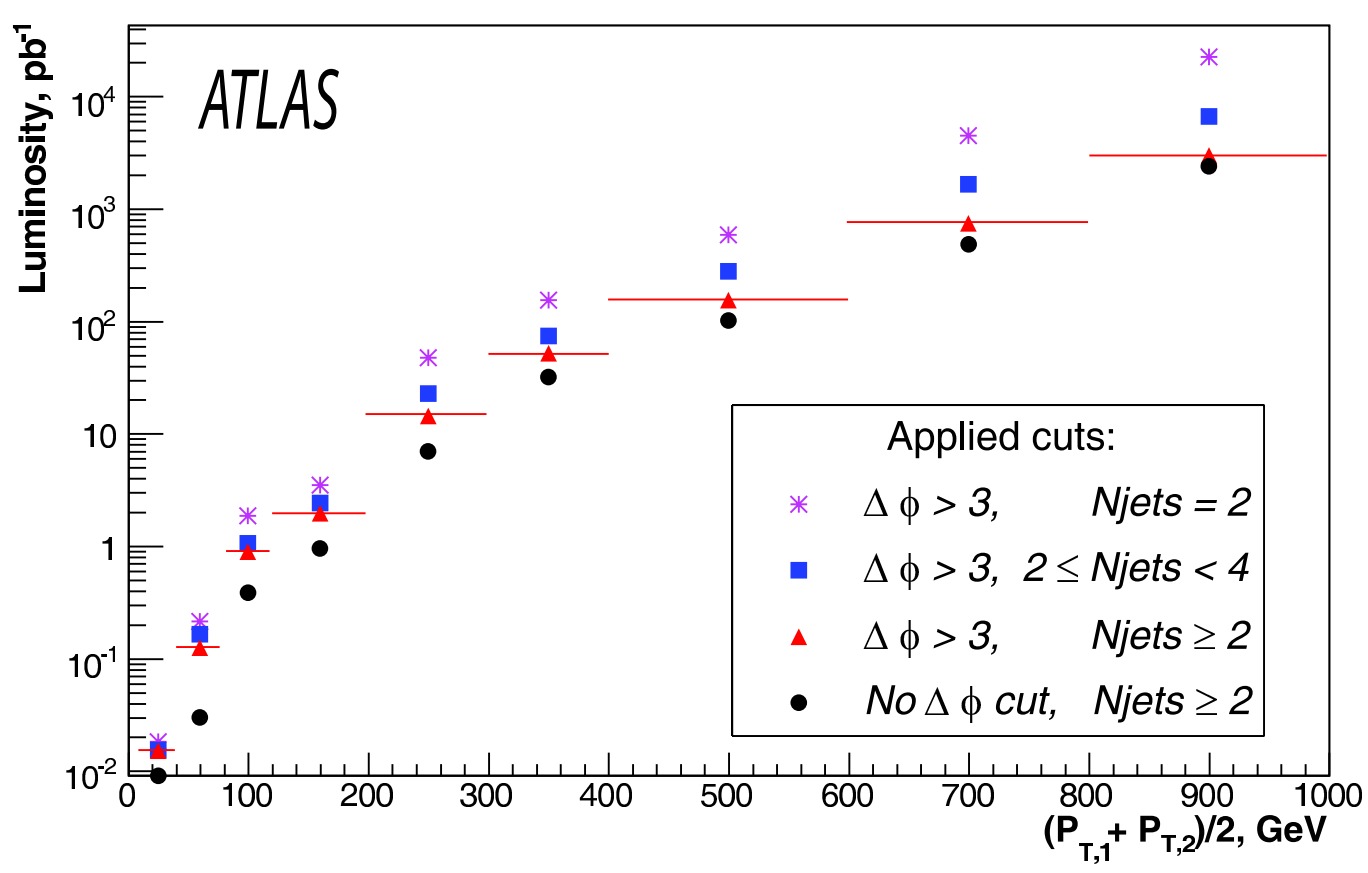

Figure 16: Integrated luminosity required to reach 0.5% precision for various $p_T$ ranges in the region $0.7 < \eta < 0.8$ with different sets of selection cuts: all PYTHIA dijet events (circles), requiring $\Delta\phi > 3$ between the two leading jets (triangles), requiring in addition less than 4 reconstructed jets in an event (squares), requiring exactly two reconstructed jets (stars). The reference region is $0 < |\eta| < 0.7$.

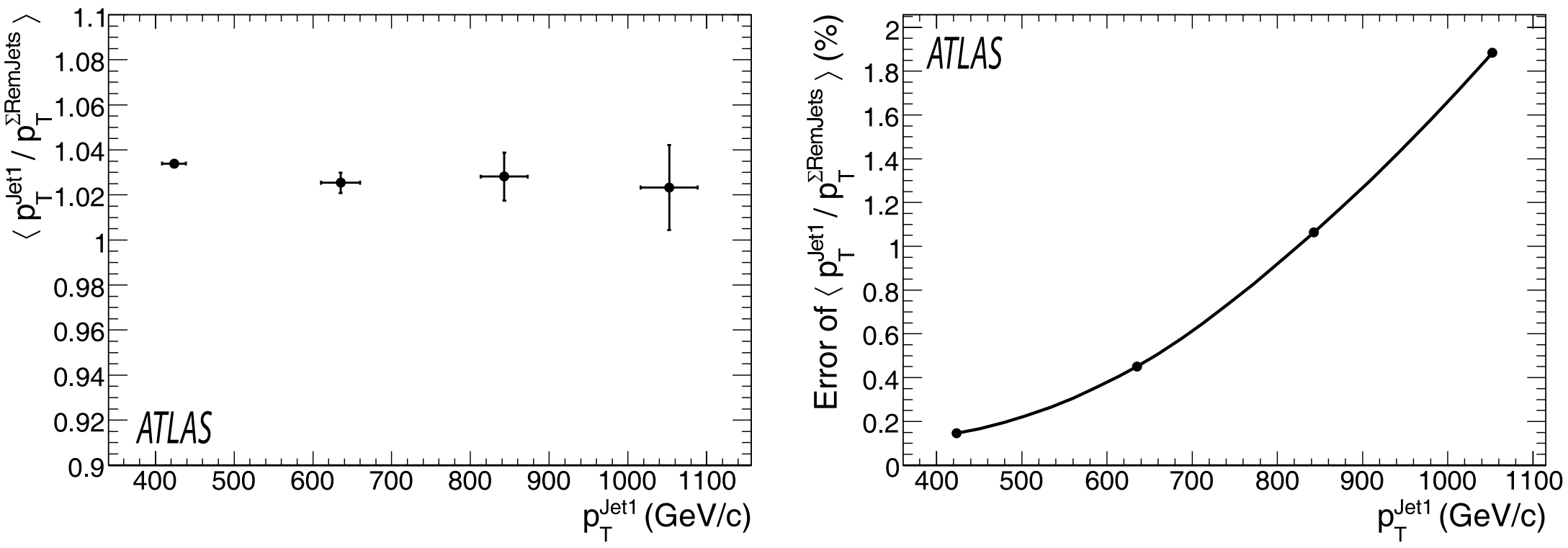

Figure 17: Energy scale (left) and energy scale uncertainty (right) of high $p_T$ jets relative to lower $p_T$ remnant jets as a function of jet $p_T$, obtained by multijet $p_T$ balance method at an integrated luminosity of 1 fb$^{-1}$. The error bars shown are statistical only.

### ALPGEN study

For this study QCD multijet samples with 4, 5 and 6 partons were generated using ALPGEN interfaced with JIMMY. Four data samples were formed by requiring the leading parton $p_T$ to be larger than 400, 600, 800, or 1000 GeV, while the $p_T$ of the next-to-leading parton was required to be smaller than 200, 300, 400, or 500 GeV, respectively. The total number of partons ($p_T > 40$ GeV) can be 4, 5 and 6 in each of the above samples. All the partons are required to be in the range $|\eta| < 3$. These samples, referred to as PT400, PT600, PT800, and PT1000 hereafter, were processed through the full detector simulation.

At least four cone jets with $R = 0.7$ with $p_T > 40$ GeV must be reconstructed in each event. The leading jet $p_T$ in the samples PT400, 600, 800, and 1000 must be in the interval $400 < p_T^{jet1} < 460$ GeV, $600 < p_T^{jet1} < 700$ GeV, $800 < p_T^{jet1} < 920$ GeV, and $1000 < p_T^{jet1} < 1140$ GeV, respectively, while the $p_T$ of the next-to-leading jet must correspondingly respect the bound $p_T^{jet2} < 190$ GeV, $p_T^{jet2} < 280$ GeV, $p_T^{jet2} < 370$ GeV, and $p_T^{jet2} < 470$ GeV. In addition the leading jet and the vector sum of the remaining lower $p_T$ jets with $p_T > 40$ GeV (called remnant jets) are required to be back-to-back in azimuth within $\pm 20°$. After all the cuts, about 3200, 310, 40, and 11 events remain in the PT400, 600, 800, and 1000 samples, respectively, for an integrated luminosity of 1 fb$^{-1}$.

The distribution of the $p_T$ balance, defined as

$$B'_{\Sigma} = \frac{p_T^{jet1}}{\text{non}-\text{leading jets} \ |\sum p_T|} \ , \tag{12}$$

is plotted for each sample and fitted by a Gaussian. The mean values of the Gaussian fits with their uncertainties for 1 fb$^{-1}$ are shown in Fig. 17 as a function of $p_T^{jet1}$. The figure indicates that $B'_{\Sigma}$ could be determined with a statistical accuracy of 2% or better for high $p_T$ jets with $400 < p_T < 1100$ GeV and that it shows a constant offset of 2-3% over the $p_T^{jet1}$ range, as also observed in the PYTHIA study.

Various potential sources of systematic effect have been studied. Low $p_T$ jets will not contribute to remnant jet system because of the 40 GeV $p_T$ threshold applied to the jets. If the threshold is varied by $\pm 20$ GeV, $B'_{\Sigma}$ varies by $\pm 1\%$ at most. The effect is smaller, for the same range of thresholds, than in the PYTHIA analysis because of the back-to-back requirement imposed on the system. A sizable fraction of the soft radiation is still missed is the event. This is the main cause of the positive $B'_{\Sigma}$ offset. It could be reduced by requiring the overall event $p_T$ to be balanced, e.g. by requiring that $E_T^{miss} < 40$ GeV. In that case, $B'_{\Sigma}$ decreases by 1–2% and becomes close to 1. However, an $E_T^{miss}$ cut should be used with caution as fake $E_T^{miss}$ could cause biases. The effect of the underlying event on $B'_{\Sigma}$ has also been evaluated by subtracting the $E_T$ measured in a cone ($R = 0.7$) placed randomly in a region of azimuth within $60^\circ < \Delta\phi < 120^\circ$ from the leading jet and not overlapping with any of the remnant jets. After subtraction, $B'_\Sigma$ increases by 1–2%.

The precision on the JES for high $p_T$ jets is affected by the JES uncertainty of the remnant jets. If we assume the uncertainty to be $\pm7\%$ for low $p_T$ jets with $|\eta| < 3.2$, which is the current linearity performance of the H1 weight calibration method used in this analysis [1], vary the scale of the remnant jets by $\pm7\%$, and apply again all the event selection, $B'_\Sigma$ changes by $\pm7\%$ independently of $p_T^{\rm jet1}$. This provides an estimate of the JES uncertainty for high $p_T$ jets. The uncertainty on $B'_\Sigma$ is increased since more events fail the next-to-leading jet $p_T$ cut that is included in the overall JES uncertainty.

Adding the statistical and above systematic uncertainties in quadrature, the JES uncertainty obtained from multijet balance method is estimated to be about 8% in the $p_T$ range of $400 < p_T < 1100$ GeV and is totally dominated by the absolute JES uncertainty for lower $p_T$ remnant jets. If the absolute JES uncertainty is not included in the systematics, the relative JES uncertainty is 3%.

### PYTHIA study

In this study, PYTHIA samples with full and fast detector simulation were used. The fully simulated sample was generated with kinematic cuts of $280 < p_T < 560$ GeV for the hard scattering process. It corresponds to an integrated luminosity of about $25\,\mathrm{pb}^{-1}$. Cone jets with $R = 0.7$ were reconstructed based on calorimeter towers, topoclusters and on Monte Carlo truth particles.

The fast simulation package ATLFAST was used for a high statistics study. A cut on the hard scattered parton at $p_T > 280$ GeV was applied. The sample represents an integrated luminosity of about $0.8\,\mathrm{fb}^{-1}$. Cone jets with $R = 0.4$ were used for this study.

Events are selected with a minimum of three reconstructed jets, the leading jet being above a fixed $p_T$ threshold, up to which the JES is assumed to be calibrated. All non-leading jets are required to be below that value. A further event selection cut is $|\eta| < 2$ for the leading jet. Generally non-leading jets have been reconstructed up to $|\eta| < 5$ and down to $p_T > 10$ GeV. For specific studies of their systematic influences all these cuts have been varied.

Figure 18 presents results of the method for the sample using full detector simulation. The event selection cut on non-leading jets is $p_T < 300$ GeV. On the left, the $p_T$ balance $B'_\Sigma$ is shown for leading jets with $370 < p_T < 380$ GeV, together with a Gaussian fit. While non-Gaussian tails are visible, which could be partly reduced by a stricter event selection, they do not influence the fit and it was thus decided to use the maximum statistics.

To reduce the bias of the $p_T$ balance due to migration effects, the division of the $p_T$ region into bins is done based on the average $p_T$ of the sum of the non-leading jets and the leading jet of each event. This reduces the bias to about 1% at truth level. This is similar to what is studied in the $Z/\gamma$+jet analyses (Section 4.2), where the average $(p_{T,\gamma/Z} + p_{T,\rm jet})/2$ is an option for the binning.

On the right of Fig. 18 the JES was studied from 350 GeV to 550 GeV. The result is reasonably flat but shows a systematic offset of about 2%, as also observed in the ALPGEN study. There are various effects contributing to this imbalance, mostly originating from soft radiation affecting the low $p_T$ jets in the hemisphere opposite the leading jet. At truth jet level, changing the minimum jet $p_T$ from 5 to 10/20/40 GeV introduces a negative bias of 0.2/1/3%, respectively. The $p_T$ threshold value for real data should be chosen taking the noise, pile-up conditions and jet finding efficiency at low $p_T$ into account. Out of cone losses also play a role: changing the cone size from 0.7 to 0.4 at truth level, affects the measured balance by $\sim 1\%$. Furthermore, imperfect calibration of low $p_T$ jets at the level of few percent [1] may be a cause of bias.

In a high statistics study, based on ATLFAST, the non-leading jets were restricted to a $p_T$ value below 350 GeV during event selection. Figure 19 shows the result on the left. The $p_T$ balance distribution is lower than for cone jets with $R = 0.7$, as discussed above. Truth and ATLFAST jets give closer results in this case, since the latter are truth jets smeared for resolution effects.

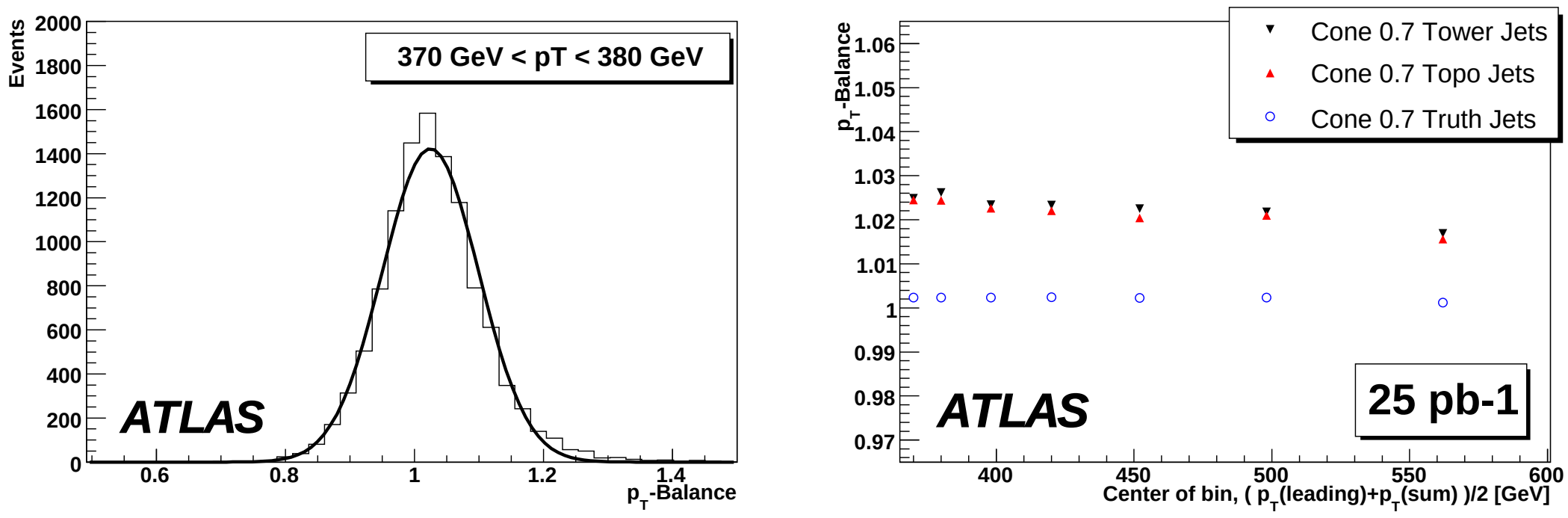

Figure 18: Left: the ratio of the absolute value of the vector sum of the non-leading jet $p_T$ to the leading jet $p_T$ for the $p_T$ bin 370–380 GeV fitted by a Gaussian. Right: this ratio as a function of jet $p_T$. The mean and the error of the mean of the Gaussian fits are shown. The average of the leading jet $p_T$ and of the total $p_T$ of the non-leading jets is used for the binning.

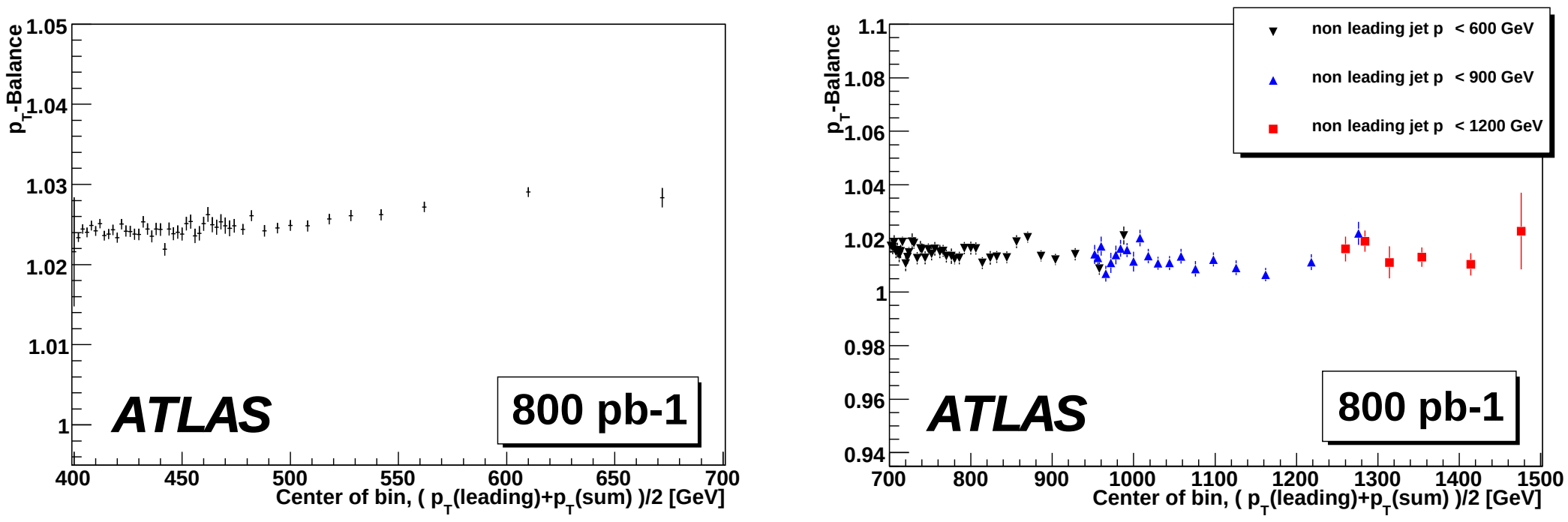

Figure 19: Results using ATLFAST with cone jets with $R = 0.4$. Left: The fitted balance as a function of the average leading and non-leading jets $p_T$. Right: Iterations of the method using the $p_T$ range checked by one iteration as the reference region for the next.

The reach in $p_T$ of the method can be further increased by iterating the procedure. Apart from the QCD jet cross-section, the statistics are mainly limited by the cut on the $p_T$ of the non-leading jets during event selection, essentially requiring that one parton in the hard scattering lost a significant part of its energy due to gluon radiation. For higher $p_T$ ranges this cut should be moved to higher values to obtain sufficient statistics. The procedure is thus, after applying the method once, up to a certain $p_T$, to use the range up to this $p_T$ as reference region for a next iteration. Results using this approach based on the ATLFAST sample are shown in Fig. 19 on the right hand side. For the studied high $p_T$ region, the multijet statistics predicted by PYTHIA may be an underestimation, as the PYTHIA parton shower model is not perfectly tuned for this $p_T$ range.

A crucial feature for this method, as observed in PYTHIA simulations, is that the imbalance value remains essentially constant for one iteration if the tested JES is correct. The balance can differ between the iterations. In general, it is slightly closer to 1 for higher $p_T$ ranges, as the system of non-leading jets becomes harder and thus less sensitive to low energy effects. To ensure that the JES is correct, we require that the beginning of the test region for a given iteration overlaps with the end of the region verified in the previous iteration, or at the very beginning verified by another method.

A complete understanding of the feature that the balance should remain constant for all $p_T$ within one iteration, is necessary in order to rely on it in data. In this case, in addition to the absolute JES uncertainty of the non-leading jets, the systematic uncertainty of this method is given by deviations of the balance from a constant value as a function of $p_T$.

To determine this additional systematic uncertainty, the deviations are split into a global slope of a linear fit and into local deviations around this fit. As possible sources of systematic effects, the absolute JES of the remnant system, the choice of the $p_T$ binning, the lower and upper $p_T$ thresholds for the non-leading jets, the effect of the underlying event and of the noise, and the effect of the varying jet multiplicity in the remnant system are studied. To estimate the deviations quantitatively, the absolute JES is varied by 10%; the lower jet $p_T$ threshold is varied between 10 and 50 GeV; the upper threshold for each iteration is varied by 100 GeV; the effect of the underlying event and of the noise is estimated by adding $\pm 5$ GeV to each jet in the event; and the analysis is performed separately for events with exactly 4, 5 and 6 jets.

The slope of the fit is compatible with zero within statistical uncertainties for all variations of the sources. The corresponding systematic uncertainties for each source is estimated to be $\lesssim 0.1\%$. The estimation is limited by the available statistics. All local deviations from a linear fit are of the order of the statistical uncertainties. The width of their distribution is found to be less than 0.2% above the statistical expectation for each of the above influences. Adding all uncertainties quadratically results in a very small total systematic uncertainty of 0.5%. Thus, the main contribution to the JES uncertainty is expected to be the accuracy of the calibration in the reference region. The JES can then be calibrated up to $p_T \gtrsim 1.5$ TeV, with the specific uncertainties introduced by this method being below 1% and requiring less than $1\text{fb}^{-1}$ of integrated luminosity.

### 5.2.2 Track angle method

The idea behind this method is that the invariant mass of two particles in a jet is approximately constant (given by the scale of $\Lambda_{QCD}$), which leads to a $p_T^{-1}$ behavior of the $\eta$-$\phi$ distance between two particles in a jet $\Delta R$, where $p_T$ is the transverse momentum of the jet. The dependence of $\Delta R$ on $p_T^{-1}$ of the jet could provide a means of determining the energy scale of high $p_T$ jets once the JES of lower $p_T$ jets is calibrated with some accuracy using other in-situ techniques. The procedure to determine the JES of high $p_T$ jets is as follows: 1) measure $\Delta R$ for low $p_T$ jets with the calibrated JES in both data and Monte Carlo samples, 2) calibrate the simulation so that $\Delta R$ in the simulation matches $\Delta R$ in data in the low $p_T$ range, then 3) measure the $\Delta R$ for a sample of given high $p_T$ jets in data, and look for the Monte Carlo jet $p_T$ scale corresponding to the measured $\Delta R$ value on the curve of $\Delta R(p_T^{-1})$.

The studies were performed on PYTHIA dijet samples generated in intervals of $p_T$. The jets selected have to have $p_T > 20$ GeV and should not overlap with other physics objects within $R < 0.4$. Afterward all tracks are selected which fall in a cone of $R = 0.4$ around reconstructed jet axis of the jet with the highest $p_T$. If fewer than two tracks are found for the jet, the event is rejected. The $\Delta R$ of two tracks is then calculated for all combinations from two up to the five highest $p_T$ tracks in the jet. Figure 20 shows the mean of the $\Delta R$ values (one entry per event), normalized to an integrated luminosity of $1\text{fb}^{-1}$, for leading jets with $140 < p_T^{\text{truth}} < 160$ GeV (top-left) and $1120 < p_T^{\text{truth}} < 1280$ GeV (top-right). The two histograms correspond to the $\Delta R$ obtained from leading two and five tracks. When more low $p_T$ tracks are involved, the $\Delta R$ distribution broadens as a result of jet fragmentation.

The bottom plots in Fig. 20 show the mean (left) and most probable value (MPV) obtained from a Landau fit to the peak (right) of the $\Delta R$ distributions as a function of the leading jet truth $p_T$. Both mean and MPV are well fit by a function of the form $p_0/x + p_1$ as shown by the curves. Figure 21 (top-left) shows the uncertainty of the jet $p_T$ scale obtained by evaluating the jet $p_T$ range on the curves covered by the statistical uncertainties of the data points in the bottom of Fig. 20 and dividing the $p_T$ range by the jet truth $p_T$. The $p_T$ scale uncertainty varies significantly at high $p_T$ with different choices of $\Delta R$ values. In general, the MPV is more stable than the mean value as it suffers less from fluctuations in jet hadronization. The MPV-based scale uncertainty seems less dependent on the leading track multiplicity in PYTHIA-only comparisons, but the situation changes when PYTHIA is compared with HERWIG as in Fig. 21 (top-right). Apparently the choice of the leading two tracks is more robust against different jet fragmentation models, and therefore the $\Delta R$ obtained from the MPV of leading two tracks is used hereafter as a central value.

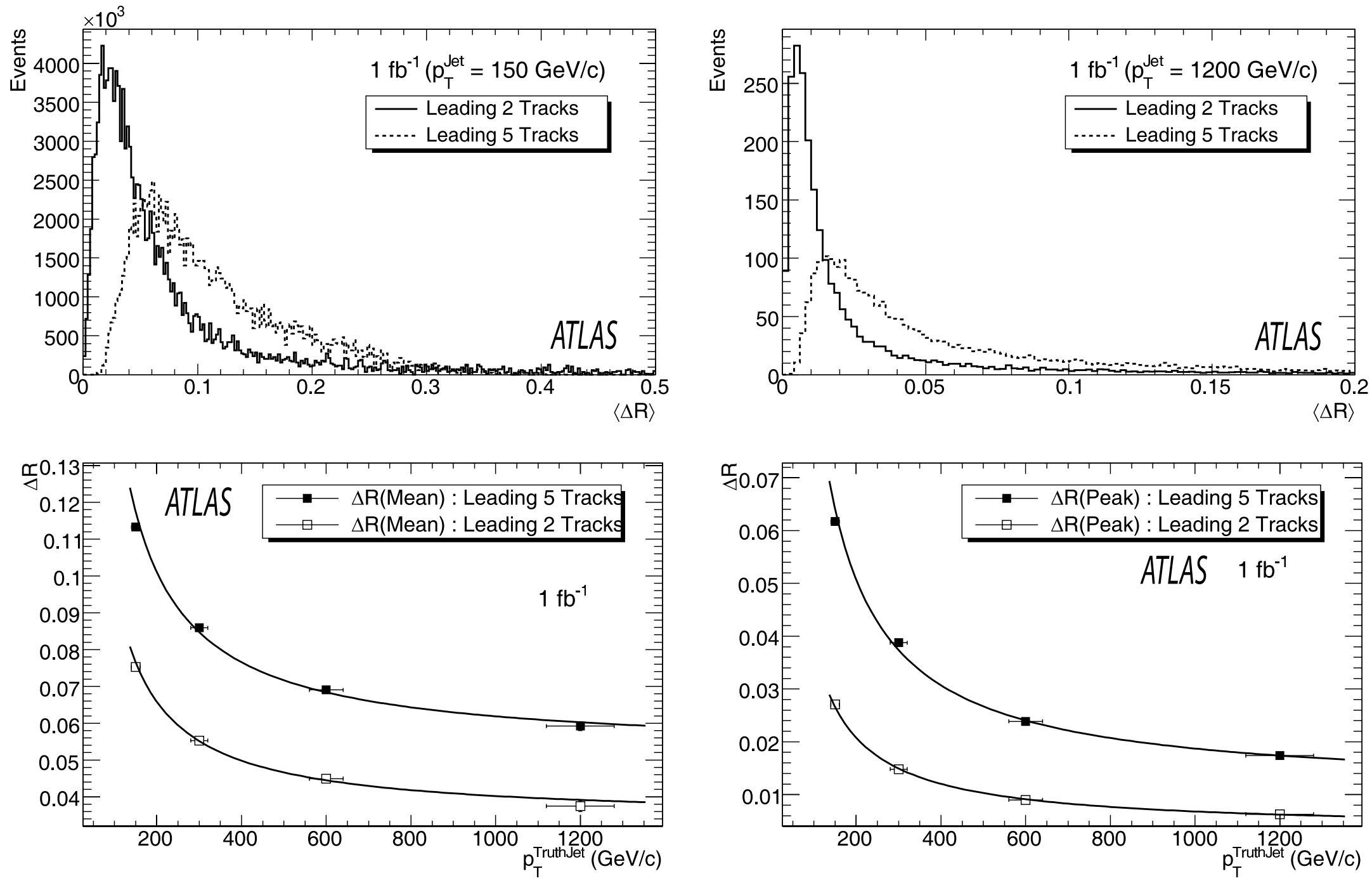

Figure 20: Top: Distributions of the mean of the $\Delta R$ values for the leading two (solid histogram) and five (dashed histogram) tracks in jets with $140 < p_T^{\rm truth} < 160$ GeV (left) and $1120 < p_T^{\rm truth} < 1280$ GeV (right) for an integrated luminosity of $1 \rm fb^{-1}$. Bottom: Mean (left) and most probable value obtained from a Landau fit to the peak (right) of the $\Delta R$ distributions as a function of the leading jet truth $p_T$ for the leading two (solid points) and five (open points) tracks. The curves represent fits with a function of the form $p_0/x + p_1$.

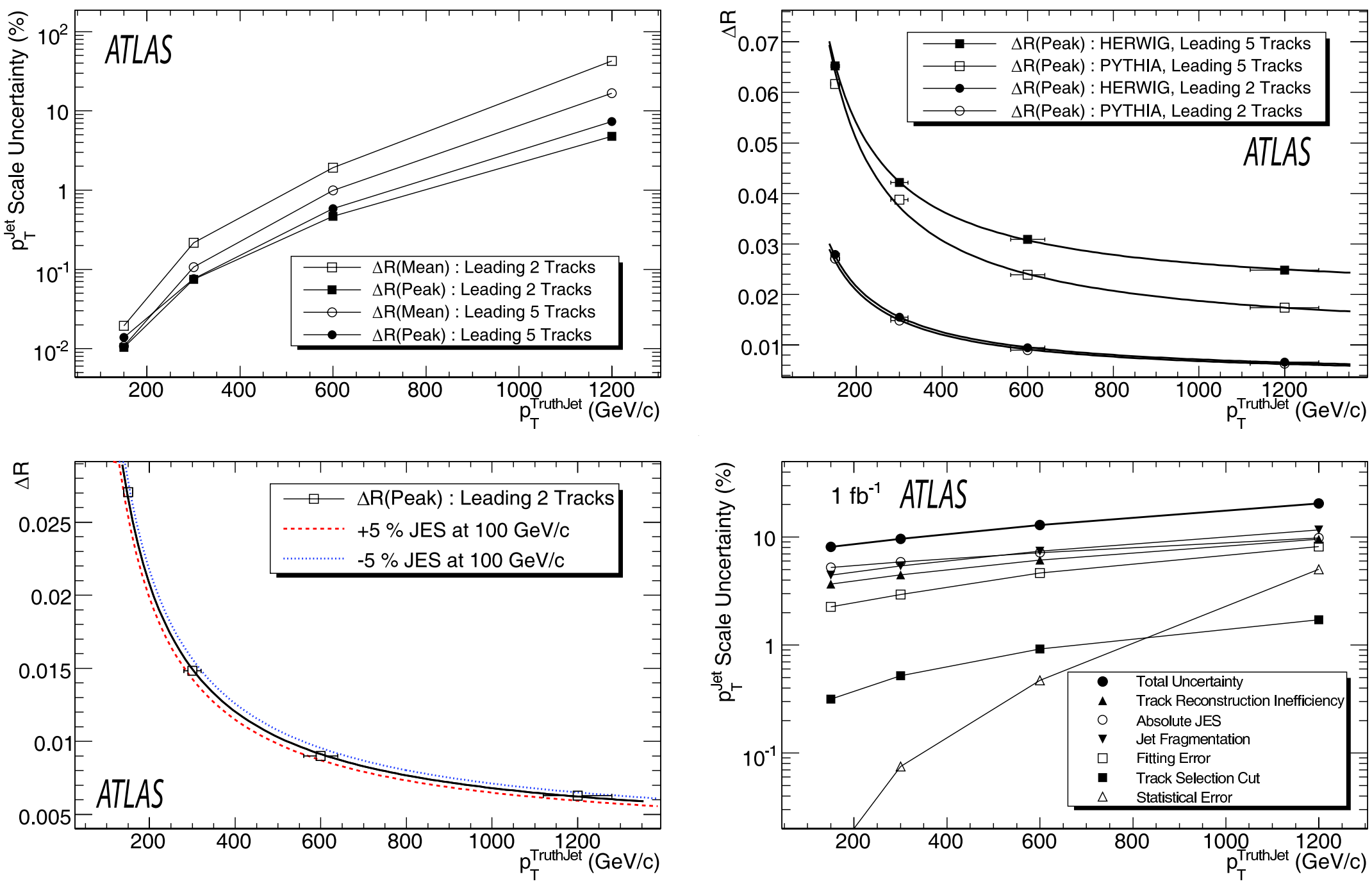

Figure 21: Top-Left: Jet $p_T$ scale uncertainty (statistical uncertainty only) as a function of jet truth $p_T$ obtained for different choices of $\Delta R$ values and track multiplicities. Top-Right: The most probable $\Delta R$ of the leading two and five tracks as a function of the jet truth $p_T$ in PYTHIA (open points) and HERWIG (solid points). Bottom-Left: Default fit (solid curve) to the most probable $\Delta R$ of the leading two tracks as a function of the truth jet $p_T$ and the curves corresponding to $\pm 5\%$ JES variations at $p_T^{\text{jet}} = 5$ GeV (dashed and dotted). Bottom-Right: Total and individual systematic and statistical uncertainties as a function of the truth jet $p_T$ expected to be obtained from the track angle method for an integrated luminosity of $1\text{fb}^{-1}$.

The following sources of systematic uncertainties are considered: 1) the jet fragmentation model as discussed above, 2) the track selection cuts and reconstruction inefficiencies, 3) absolute JES at low energies, and 4) the uncertainty associated with the fit. The systematic uncertainty on the jet $p_T$ scale is evaluated by obtaining the $\Delta R$ versus jet $p_T$ distributions shifted by $\pm 1\sigma$ (or the equivalent amount specified) of the source, and looking at the variation in the jet $p_T$ at the measured $\Delta R$ values. For 1), the difference between PYTHIA and HERWIG is assigned. For 2), the deviated distributions are obtained by applying different selection cuts on tracks and assuming an extra 5% reconstruction inefficiency for tracks in jets. The JES uncertainty for low $p_T$ jets is assumed to be $\pm 5\%$ at $p_T^{\text{jet}} = 100$ GeV, and the deviated distributions are obtained by estimating the corresponding variations in $\Delta R$ at $p_T^{\text{jet}} = 100$ GeV using the default fit function and transferring the variations to uncertainties on the parameters of the function. The JES deviated and default distributions are shown in Fig. 21 bottom-left. The total uncertainty obtained by adding all the systematic uncertainties and the statistical uncertainty in quadrature, and individual systematic uncertainties as a function of the jet truth $p_T$ are shown in the bottom-right of the figure. With the assumption of $\pm 5\%$ JES uncertainty at $p_T^{jet} = 100$ GeV, the JES of 150, 300, 600, and 1200 GeV $p_T$ jets can be potentially measured with an accuracy of 8, 10, 13, and 21%, respectively, with an integrated luminosity of $1\text{fb}^{-1}$.

## 5.3 Obtaining jet energy resolution in QCD dijet sample

A precise measurement of the jet energy resolution is necessary to model and control the systematic effects associated with it on jet observables and the quantities used in physics measurements. A key issue in estimating the jet energy and momentum resolution is the need to disentangle detector effects from those associated with physics such as the underlying event, and initial and final state radiation.

Two data-based techniques are presented which allow the jet energy resolution to be estimated, and they were compared where appropriate to the resolution obtained using particle jets: the dijet balance method, used by the DØ collaboration, and the $k_T$ balance technique, developed by UA2 and used by CDF.

The QCD dijet events for this analysis were generated using PYTHIA in the $p_T$ range from 17 GeV up to 1120 GeV. Jets were reconstructed using the cone algorithm with $R = 0.7$.

To select dijet events the following basic cuts are applied:

- One primary vertex;
- Exactly two back-to-back leading jets with $|\Delta\phi| < 0.3$ and $p_T > 10$ GeV;
- Both jets in the same $\eta$ region $|\eta| < 1.2$.

The $\Delta\phi$ and minimum $p_T$ cuts are varied for some systematic studies.

### 5.3.1 Dijet balance method

The determination of the jet $p_T$ resolution in the dijet balance technique is based on energy conservation in the transverse plane. It assumes that there are only two jets in the event which have similar $p_T$. If one defines the asymmetry distribution $A$ for the two jets as:

$$A \equiv \frac{p_{T,1} - p_{T,2}}{p_{T,1} + p_{T,2}} , \tag{13}$$

then for jets in the same rapidity region, which therefore have the same resolution, the fractional transverse jet energy resolution can be expressed as a function of $\sigma_A$:

$$\frac{\sigma_{p_T}}{p_T} = \sqrt{2}\sigma_A , \tag{14}$$

where $p_T = (p_{T,1} + p_{T,2})/2$.

The asymmetry distribution $A$ for two representative $p_T$ bins obtained by using cone jets with $R = 0.7$ are shown in Fig. 22. The distributions have been symmetrized by computing either $p_{T,1} - p_{T,2}$ or $p_{T,2} - p_{T,1}$ randomly for each event. The resolution function is well described by a Gaussian fit, although it is worth mentioning that previous studies done by the DØ Collaboration suggest using a double Gaussian fit for large $p_T$ bins, to absorb effects of non-Gaussian tails.

The simple dijet balance method does not return the true fractional transverse energy resolution associated purely with detector effects, since the two leading jets of an event could be imbalanced due to undetected jets in the event with $p_T < 10$ GeV such as from soft radiation. This radiation will broaden the resolution function and a correction may be derived by studying the effect of the cut on the $p_T$ of any third jet in the event as described in the next section.

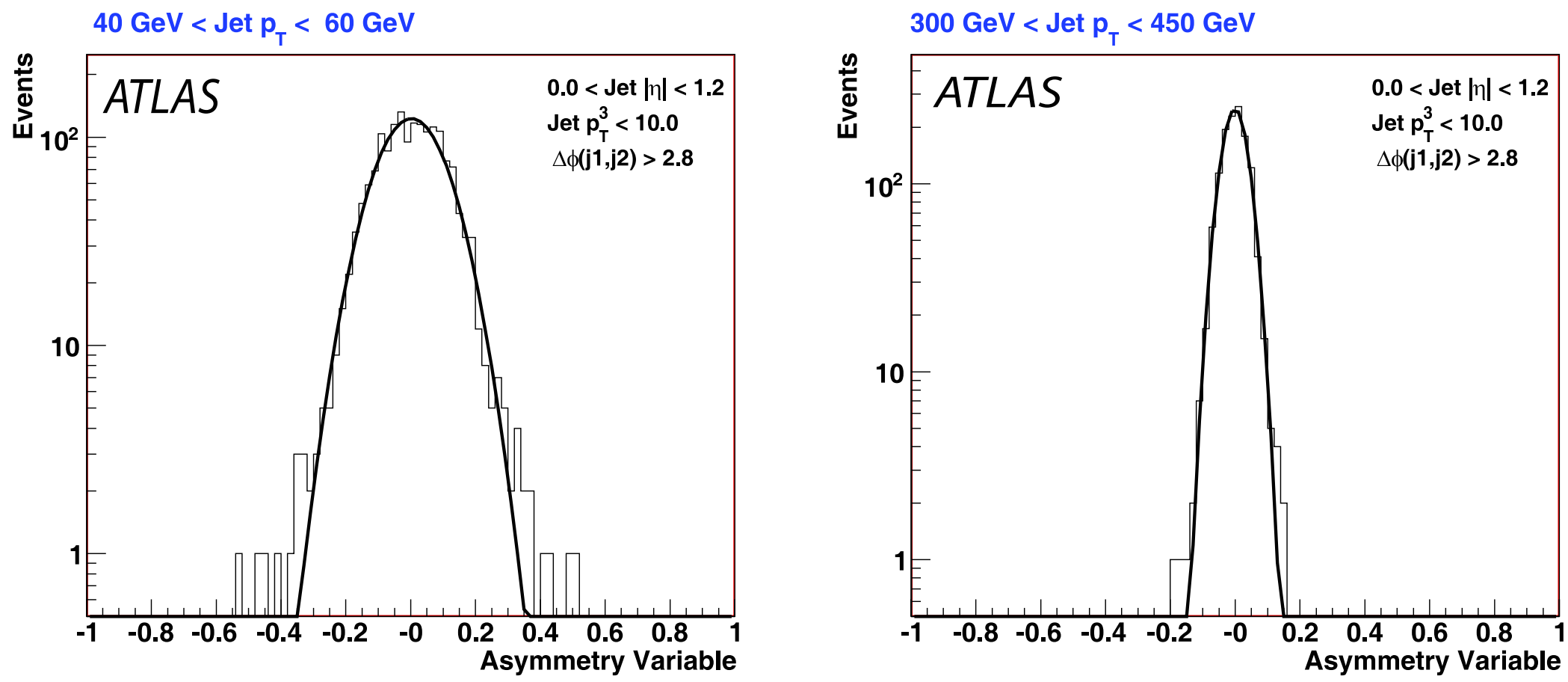

Figure 22: Asymmetry distributions of two jets for two representative $p_T$ bins. Cone jets with $R = 0.7$ in the pseudorapidity region $|\eta| < 1.2$ are used. The distributions were fitted with a single Gaussian function.

### 5.3.2 Soft radiation correction

In our basic event selection, both the $\Delta\phi$ cut and the cut requiring exactly two jets with $p_T > 10$ GeV are implemented in order to reduce the broadening of the $p_T$ balance distribution due to soft radiation effects. However, the effects are not fully suppressed, as further jets with $p_T < 10$ GeV may remain in the selected events. To account for this effect, the jet resolution is computed by allowing a third jet with transverse momentum $p_{T,3}$ up to some threshold $\varepsilon$, and varying the threshold values between $\varepsilon = 12.5$ and 27.5 GeV. As before, the resolution distributions in the jet $p_T$ bins are fitted using single Gaussian functions. For each jet $p_T$ bin, the set of resolutions obtained from the different $p_{T,3}$ thresholds are fitted with a straight line and extrapolated to $\varepsilon = 0$. We define this value as:

$$\left(\frac{\sigma_{p_T}}{p_T}\right)^{p_{T,3}\to 0}, \tag{15}$$

which would be the resolution that we would have measured from an ideal dijet sample with $\varepsilon = 0$. Examples of the linear fits and the extrapolations for two $p_T$ bins are presented in Fig. 23.

After fitting the resolution for the different third jet thresholds, we calculate a correction factor, $K(p_T)$, as follows:

$$K(p_T) = \left(\frac{\sigma_{p_T}}{p_T}\right)^{p_{T,3}\to 0} \Big/ \left(\frac{\sigma_{p_T}}{p_T}\right)^{\varepsilon=10\text{ GeV}}. \tag{16}$$

Finally, correcting by this factor, the unbiased fractional transverse energy resolution is obtained via

$$\left(\frac{\sigma_{p_T}}{p_T}\right)_{\text{corrected}} = K(p_T) \times \left(\frac{\sigma_{p_T}}{p_T}\right)^{\varepsilon=10\text{ GeV}}_{\text{uncorrected}}. \tag{17}$$

Since the soft radiation bias should be larger at small transverse energies, and negligible at high $p_T$ we parametrize $K(p_T)$, with the function:

$$K(p_T) = 1 - \exp(a - b p_T). \tag{18}$$

This parametrization can be used to correct the resolution for each $p_T$ bin and determine the unbiased resolution associated purely with detector effects. The jet energy resolutions with and without the soft radiation correction are shown in Fig. 24. As we will see below, by comparing this result to that obtained by other approaches we can establish an estimate of the systematic uncertainty of the jet energy resolution.

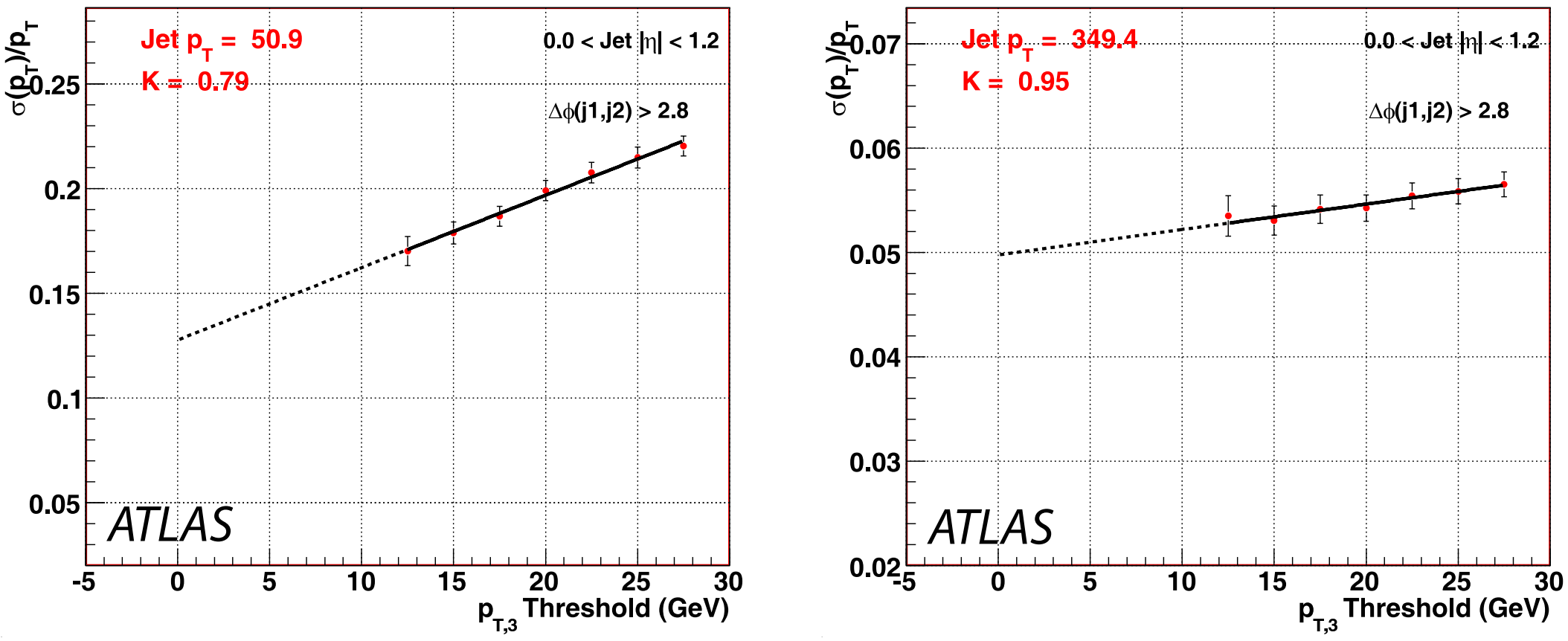

Figure 23: Resolution versus the $p_{\mathrm{T},3}$ threshold cut for different $p_{\mathrm{T}}$ bins. The line corresponds to the linear fit applied while the dashed-line shows the extrapolation to $p_{\mathrm{T},3} = 0$, which corresponds to an ideal dijet sample ($\varepsilon = 0$).

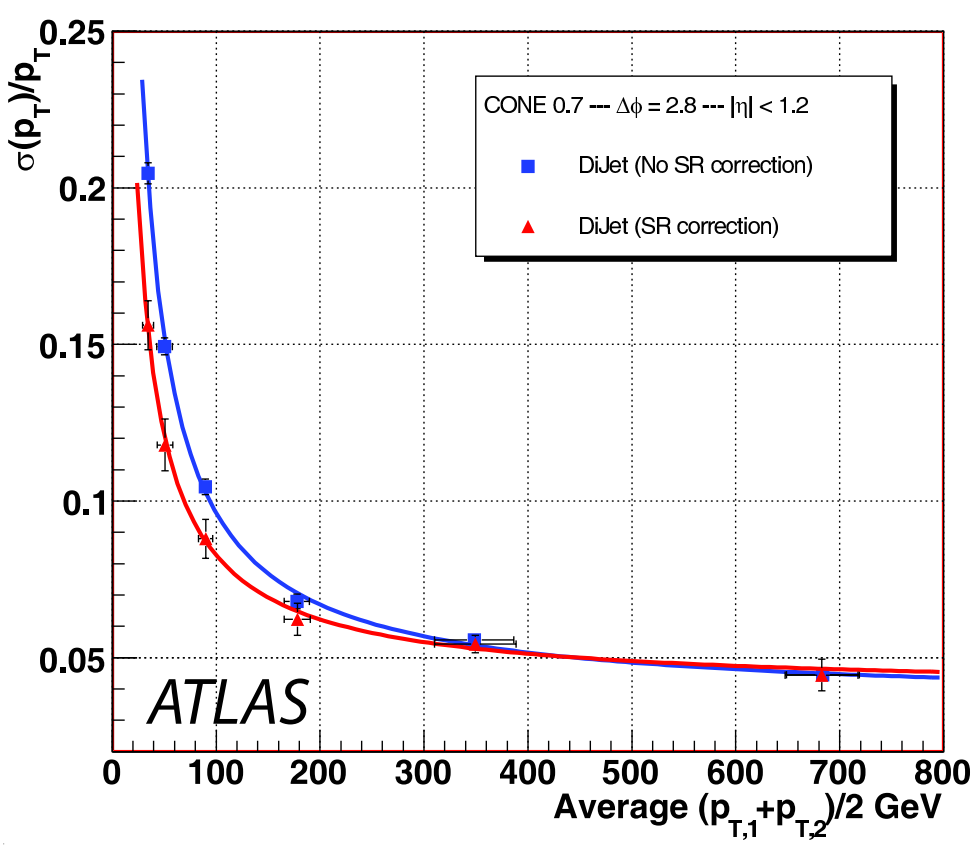

Figure 24: Jet energy resolution for cone jets with $R = 0.7$ in the pseudorapidity range $|\eta| < 1.2$. The results are obtained by using dijet balance techniques with and without applying the soft radiation correction.

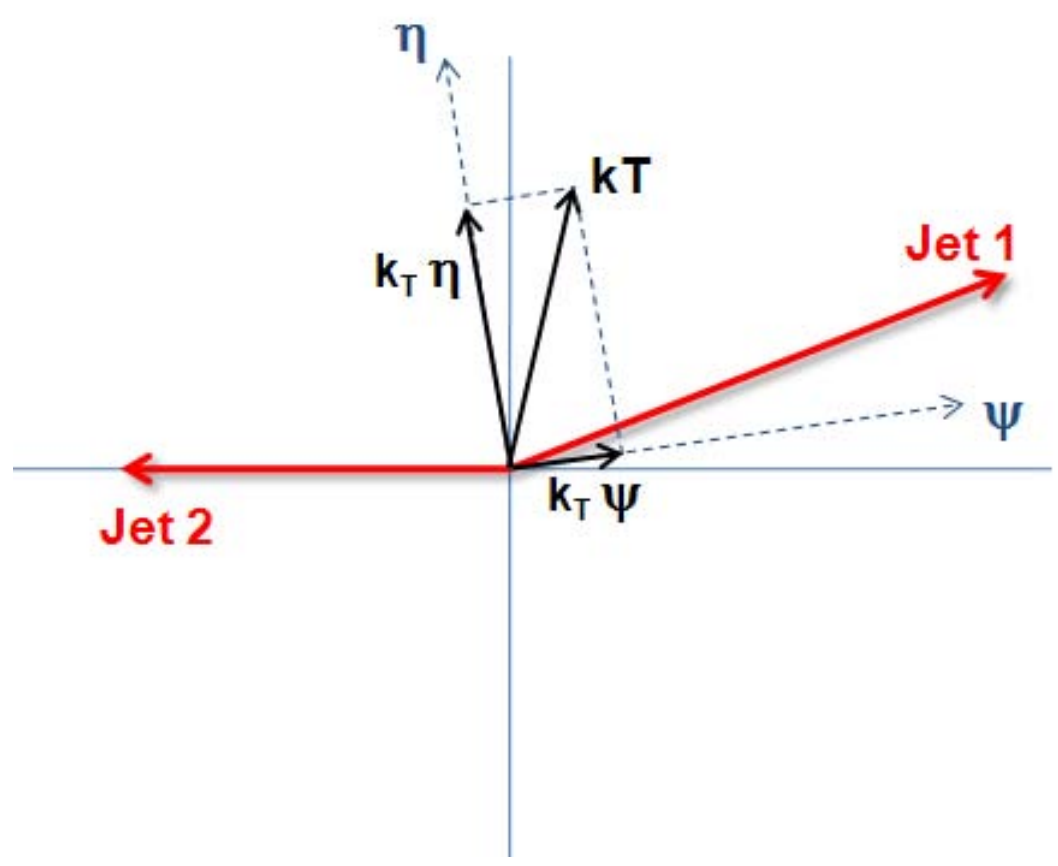

Figure 25: Sketch of the $k_{\mathrm{T}}$ balance technique. The $\eta$ axis corresponds to the azimuthal angular bisector of the dijet system while the $\psi$ axis is defined as being orthogonal to the $\eta$ axis.

### 5.3.3 The $k_{\mathrm{T}}$ balance technique

The $k_{\mathrm{T}}$ balance technique was developed by UA2 and studied at CDF. To extract the contributions of detector effects on to the jet energy resolution, the imbalance vector $\vec{K}_T = \vec{p}_{T,1} + \vec{p}_{T,2}$ is projected onto two components ($\psi$,$\eta$) as shown in Fig. 25.

The $\eta$ axis corresponds to the azimuthal angular bisector of the dijet system while the $\psi$ axis is defined as being orthogonal to the $\eta$ axis. The two components, $K_{T,\psi}$ and $K_{T,\eta}$, are sensitive to different effects. The calorimeter energy resolution represents the main source for $\sigma_\psi$, the spread of $K_{T,\psi}$. Gluon radiation effects are smaller but affect this component as well. On the other hand, $\sigma_\eta$, the spread of $K_{T,\eta}$, is significantly affected by gluon radiation. In addition, there are other smaller effects such as jet angular resolution, underlying event and out-of-cone fluctuations. In order to reduce the hard gluon radiation effects, events with $p_{\mathrm{T},3} > 11$ GeV are rejected. If we define the contributions from the calorimeter resolution, $\sigma_{\mathrm{res}}$, and from soft radiation, $\sigma_{\mathrm{SR}_\parallel}$ and $\sigma_{\mathrm{SR}_\perp}$, then we can write

$$\begin{aligned} \sigma_\psi^2 &= \sigma_{\mathrm{res}}^2 + \sigma_{\mathrm{SR}_\parallel}^2 \\ \sigma_\eta^2 &= \sigma_{\mathrm{SR}_\perp}^2 . \end{aligned}$$

Assuming

$$\sigma_{\mathrm{SR}_\parallel}^2 = \sigma_{\mathrm{SR}_\perp}^2 , \tag{19}$$

the soft radiation effects are removed by subtracting in quadrature $\sigma_\eta$ from $\sigma_\psi$:

$$\sigma_{\mathrm{res}} = \sqrt{\sigma_\psi^2 - \sigma_\eta^2} . \tag{20}$$

The data sample was divided into six $p_{\mathrm{T}}$ regions, and the distributions $K_{T,\eta}$ and $K_{T\psi}$ were fitted with single Gaussians. Since in this method the measured resolution comes from the convolution of the single jet resolution, to obtain the single jet energy resolution, $\sigma_{\mathrm{res}}$ must be scaled by a factor $\frac{1}{\sqrt{2}}$.

Figure 26 shows the resulting $\frac{1}{\sqrt{2}}\sigma_\psi$ and $\frac{1}{\sqrt{2}}\sigma_\eta$ as a function of the square root of the average $p_{\mathrm{T}}$ of both jets. $\sigma_\psi$ has an approximately linear dependence with $\sqrt{\langle p_{\mathrm{T}_{1,2}} \rangle}$, while $\sigma_\eta$ has a flat dependence, especially at high $p_{\mathrm{T}}$, as expected. The effective jet energy resolution after removing soft radiation effects by subtracting in quadrature $\sigma_\eta$ from $\sigma_\psi$ is shown in Fig. 27.

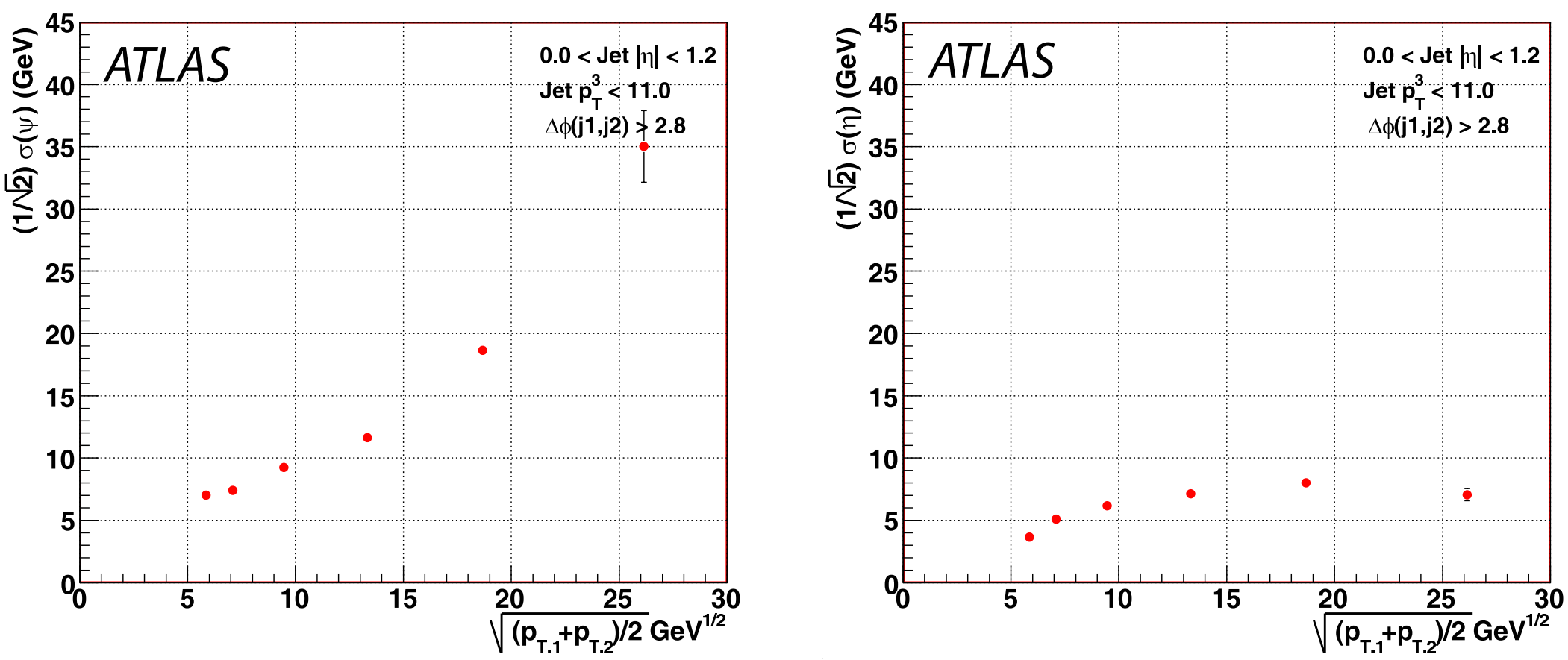

Figure 26: $\frac{1}{\sqrt{2}}\sigma_\psi$ and $\frac{1}{\sqrt{2}}\sigma_\eta$ in dijet events as a function of the square root of the average $p_{\mathrm{T}}$ of both jets.

### 5.3.4 Comparison of dijet and $k_{\mathrm{T}}$ balance techniques

A comparison of the two methods, the dijet balance and the $k_{\mathrm{T}}$ balance technique, allows an estimate of the systematic uncertainty associated with the determination of the resolution associated with detector response. By comparing the reconstructed jets with the jets from Monte Carlo truth we obtain information on its dependence on the jet algorithm and the jet matching. Each of the first two leading calorimeter jets is matched to particle jets, which lie within $\Delta R < 0.1$ w.r.t. the reconstructed jet. After matching, the response $p_T^{calo}/p_T^{truth}$ is determined, and the true resolution is obtained by fitting the response with single Gaussians. This is shown in Fig. 28 together with the results from the two data-driven estimates.

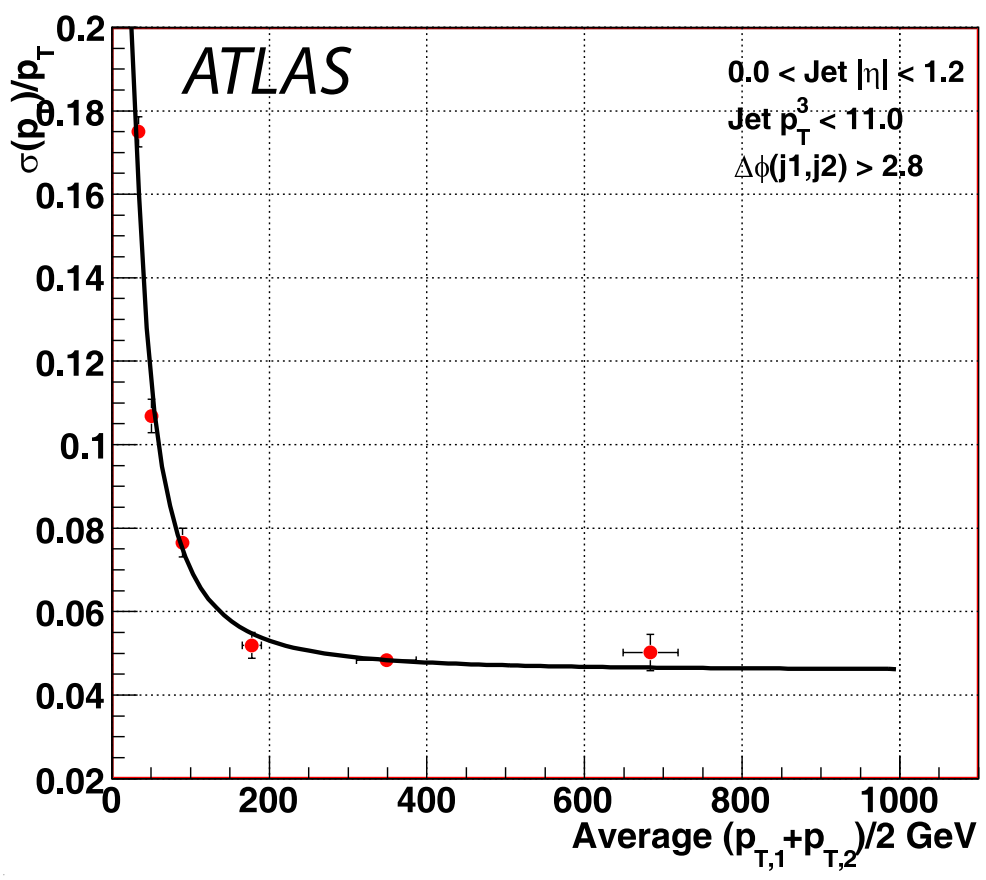

Figure 27: The effective jet energy resolution for cone 0.7 jets after removing soft radiation effects by subtracting in quadrature $\sigma_\eta$ from $\sigma_\psi$.

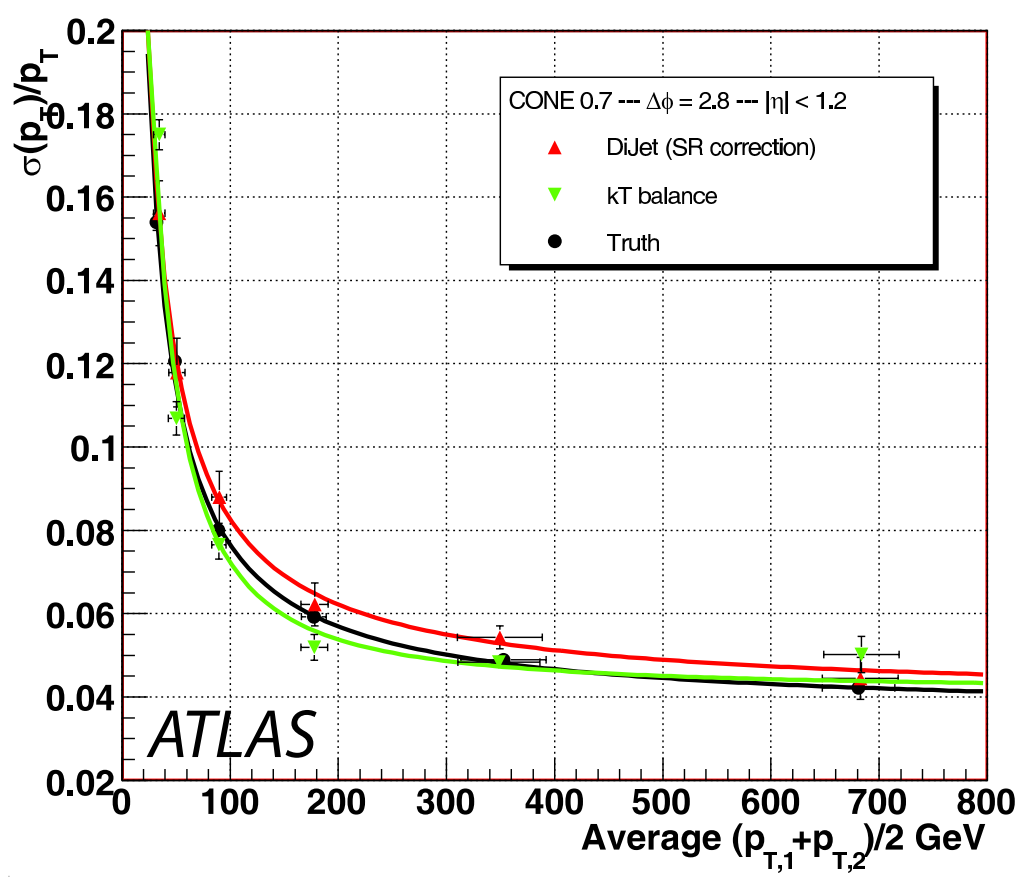

Figure 28: True jet energy resolution for cone 0.7 compared with those obtained using dijet and $k_T$ balance techniques.

The two data-driven approaches agree within uncertainties, though a hint that the $k_T$ method is underestimating the true detector resolution might be visible. The determination of the resolution using reconstructed jets matched to particle jets gives results in between the two data-driven approaches and together with them provides a reasonable basis from which to estimate the systematic uncertainties in the determination of the noise, stochastic and constant terms in the jet energy resolution.

## 6 Summary and conclusions

Precise reconstruction of the jet energy is required by many physics analyses. In-situ processes provide the platform against which our understanding, based on theory and detector simulation, must be tested in order to achieve maximum precision. From the analyses presented in this paper, we conclude that one can characterize three ranges of jet $p_T$, in which the different features of detector response and underlying physics control the precision which can be achieved.

From the reconstruction threshold of 10 GeV up to $\sim 100-200$ GeV, the combination of effects from jet clustering and physics effects like ISR/FSR and underlying event can be as large as $5-10\%$ at low $p_T$, slowly decreasing to the percent level around $100-200$ GeV. With QCD dijet events, we will be able to control the uniformity of the response on the percent level with high granularity in $\phi$ and $\eta$ already with $\sim 10\text{pb}^{-1}$. In turn, $Z$+jet balance is an adequate process to measure the absolute energy scale with a statistical precision of 1% with approximately $300\text{pb}^{-1}$ of data. Hence the precision will rather quickly be dominated by the systematic uncertainty. Validation of the Monte Carlo simulation to ensure that both physics and detector effects are well modeled will be particularly important to control systematic uncertainties. Differences between currently available Monte Carlo generators are on the level of $5-10\%$ in some of the observables, which sets the scale for the challenge in this region of jet $p_T$.

The medium $p_T$ region extends from $100-200$ GeV to about 500 GeV. Physics effects are at the level of $1-2\%$ in this region. The momenta of the particles resulting from jet fragmentation are within the range of the test beam measurement. It will be an important range in which to validate the Monte

Carlo description of the detector response. In addition the dijet balance is still statistically powerful, and 1% precision for an $\eta$ bin size of 0.1 can be achieved with $\sim 100\text{pb}^{-1}$. In $\gamma$+jet events the $p_T$ balance can be measured with negligible bias from QCD background. The statistical precision is better than one percent for $10\text{pb}^{-1}$ at 100 GeV, and in the range $300-500$ GeV the percent level is reached with $\sim 100\text{pb}^{-1}$.

Above 500 GeV, one enters a regime which is beyond any existing jet measurement and extrapolation of the Monte Carlo modeling and comparison to in-situ data is crucial. A much larger integrated luminosity is required for in-situ measurements of similar precision. In addition, for the highest momentum particles which may be produced in these jets, our simulation of the detector no longer has test beam data available as a basis. Techniques such as multijet balancing or extrapolation based on jet particle properties can be used and should allow us to control the scale at the level of a few percent with $1\text{fb}^{-1}$ for jets up to 1 TeV.

# Measurement of Missing Tranverse Energy

**Abstract**
This note discusses the overall ATLAS detector performance for the reconstruction of the missing transverse energy, $\not\!\!E_T$. Two reconstruction algorithms are discussed and their performance is evaluated for a variety of simulated physics processes which probe different topologies and different total transverse energy regimes. In addition, effects of fake $\not\!\!E_T$ resulting from instrumental effects and from false reconstructions are investigated. Finally, studies with first data, corresponding to an integrated luminosity of $100\text{pb}^{-1}$, are suggested which can be used to assess and calibrate the $\not\!\!E_T$ performance at the startup of data taking.

## 1 Introduction

A very good measurement of the missing transverse energy, $\not\!\!E_T$, is essential for many physics studies in ATLAS. Events with large $\not\!\!E_T$ are expected to be the key signature for new physics such as supersymmetry and extra dimensions. A good $\not\!\!E_T$ measurement in terms of linearity and resolution is also important for the reconstruction of the top-quark mass from $t\bar{t}$ events with one top quark decaying semi-leptonically. Furthermore, it is crucial for the efficient and accurate reconstruction of the Higgs boson mass when the Higgs boson decays to a pair of $\tau$-leptons.

This Note describes the overall performance of the $\not\!\!E_T$ measurement in ATLAS. The performance is checked using fully simulated Monte Carlo (MC) samples in different physics channels with differences in event topology and kinematics range. Events with no true $\not\!\!E_T$ as well as events with a true $\not\!\!E_T$, $\not\!\!E_T^{\text{True}}$, due to particles unseen in the detector as neutrinos or lightest supersymmetric particles, ranging from $\sim 20$ to $\sim 500$ GeV are used.

An important requirement on the measurement of $\not\!\!E_T$ is to minimize the impact of limited detector coverage, finite detector resolution, presence of dead regions and different sources of noise that produce fake $\not\!\!E_T$, $\not\!\!E_T^{\text{Fake}}$. The ATLAS calorimeter coverage extends to large pseudorapidity angles to minimize the impact of high energy particles escaping in the very forward direction. Even so, there are inactive transition regions between different calorimeters that produce $\not\!\!E_T^{\text{Fake}}$. Dead and noisy readout channels in the running detector, if present, will also produce $\not\!\!E_T^{\text{Fake}}$. Such $\not\!\!E_T^{\text{Fake}}$ sources can significantly enhance the background from QCD multi-jet events in supersymmetry searches or the background from $Z \to \ell\ell$ events accompanied by high-$p_T$ jets in Higgs boson searches when the Higgs boson decays into two leptons and neutrinos.

The calorimeter plays a crucial role in the $\not\!\!E_T$ measurement and an important first step of the $\not\!\!E_T$ measurement is the suppression of noise in the calorimeter. Section 2 describes the techniques used for noise suppression in calorimeters. The two $\not\!\!E_T$ reconstruction algorithms used in ATLAS, Cell-based and Object-based, are described in detail in Sections 2.2 and 2.3. The overall performance of the $\not\!\!E_T$ measurement in ATLAS is reported in Section 3. Any mis-measurement in the event, due to finite resolution or acceptance of the detector or due to instrumental effects related to dead or noisy channels, will degrade the $\not\!\!E_T$ measurement. Such sources of $\not\!\!E_T$, that can lead to large values of fake $\not\!\!E_T$, are studied in Section 4. Section 5 introduces the $\not\!\!E_T$ algorithm for the ATLAS triggers and describes its performance. Finally, Section 6 describes techniques in different physics channels that can be used with the very first ATLAS data to validate the $\not\!\!E_T$ measurement and determine the $\not\!\!E_T$ scale in-situ.

# 2 The algorithms for $\not{E}_T$ reconstruction in ATLAS

The transverse missing energy in ATLAS is primarily reconstructed from energy deposits in the calorimeter and reconstructed muon tracks. Apart from the hard scattering process of interest, many other sources, such as the underlying event, multiple interactions, pile-up and coherent electronics noise, lead to energy deposits and/or muon tracks. Classifying the energy deposits into various types (e.g. electrons or jets) and calibrating them accordingly is the essential key for an optimal $\not{E}_T$measurement. In addition, the loss of energy in dead regions and readout channels make the $\not{E}_T$ measurement a real challenge.

There are two algorithms for $\not{E}_T$ reconstruction in ATLAS that emphasize different aspects of energy classification and calibration.

The Cell-based algorithm starts from the energy deposits in calorimeter cells that survive a noise suppression procedure. The cells can be calibrated using global calibration weights depending on their energy density. This procedure will be robust already at initial data taking because it does not rely on other reconstructed objects. In a subsequent step, the cells can be calibrated according to the reconstructed object they are assigned to. Corrections are applied for the muon energy and for the energy lost in the cryostats.

The Object-based algorithm starts from the reconstructed, calibrated and classified objects in the event. The energy outside these objects is further classified as low $p_T$ deposit from charged and neutral pions and calibrated accordingly.

The noise suppression in the calorimeter is common for the Cell- and Object-based algorithm and is described below, followed by a detailed description of the algorithms.

## 2.1 Calorimeter noise suppression

The electronics noise alone in the $\approx$ 200k readout channels of the ATLAS calorimeter contributes about 13 GeV to the width of the $\not{E}_T$ distribution. Especially in events that do not have large $\not{E}_T$, such as in $Z \to \tau\tau$ used for an in-situ determination of the $\not{E}_T$ scale (Section 6.2), the noise suppression is of crucial importance.

For the $\not{E}_T$ measurement, two noise suppression methods have been studied so far. Both require knowledge of the width of the noise distribution, $\sigma_{\text{noise}}$, which can be either purely electronics noise or a combination of electronics and pile-up noise.

**Standard Noise Suppression Method.** The first method is based on only using calorimeter cells with energies larger than a threshold, generally corresponding to a certain number of $\sigma_{\text{noise}}$. The threshold is optimized for $\not{E}_T$ resolution, the scale of $\not{E}_T$, the total transverse energy in the calorimeters, $\Sigma\not{E}_T$, and for the highest $p_T$ jet to be close to the case without noise simulation. Two cases are studied: a symmetric threshold ($|E_{\text{cell}}| > n \times \sigma_{\text{noise}}$) and an asymmetric one ($E_{\text{cell}} > n \times \sigma_{\text{noise}}$). A symmetric threshold with $n = 2$ for all calorimeters is generally used.

**Noise Suppression using TopoClusters.** The second method only uses cells in 3-dimensional topological calorimeter clusters [1, 2], hereafter called TopoClusters. A TopoCluster is reconstructed starting from a seed cell with an absolute energy value $|E_{\text{cell}}| > 4\sigma_{\text{noise}}$ to which neighbors with $|E_{\text{cell}}| > 2\sigma_{\text{noise}}$ are added. Finally the cells at the boundary are required to have $|E_{\text{cell}}| > 0\sigma_{\text{noise}}$. The cells that constitute the TopoCluster are hereafter called TopoCells. This set of thresholds, referred to as $4/2/0$, is optimized to suppress electronics noise as well as pile-up from minimum bias events, while keeping the single pion efficiency as high as possible.

As a result of the large energy density of electromagnetic showers, the $\pi^0$ reconstruction efficiency is high (close to 100% for energies $>$ 4 GeV) for the $4/2/0$ configuration. On the other hand, the reconstruction efficiency for charged pions is very sensitive to the parameters of the TopoClusters. For example, changing the cuts on neighbors from $4/2$ to $6/3$, the $\pi^\pm$ efficiency significantly decreases for TopoClusters with $E < 4$ GeV. This sensitivity highlights the importance of a good modeling of the noise level from first data.

In $Z \rightarrow \nu\bar{\nu}$ events simulated with electronics noise, the $\not{E}_{\mathrm{T}}$ resolution degrades by only 3 % for the 4/2/0 configuration as compared to the same events without noise added. Also, the TopoCluster algorithm performs better in terms of linearity and resolution of the $\not{E}_{\mathrm{T}}$ measurement, compared to the standard noise suppression method. Therefore Cell- and Object-based $\not{E}_{\mathrm{T}}$ algorithms apply the noise suppression method based on TopoClusters with configuration $4/2/0$.

## 2.2 Cell-based $\not{E}_{\mathrm{T}}$ reconstruction

The Cell-based $\not{E}_{\mathrm{T}}$ reconstruction includes contributions from transverse energy deposits in the calorimeters, corrections for energy loss in the cryostat and measured muons:

$$\not{E}_{x,y}^{\mathrm{Final}} = \not{E}_{x,y}^{\mathrm{Calo}} + \not{E}_{x,y}^{\mathrm{Cryo}} + \not{E}_{x,y}^{\mathrm{Muon}}. \tag{1}$$

In the following, the three terms in the above equation, referred to as calorimeter, cryostat and muon terms, are described in some detail.

### 2.2.1 The $\not{E}_{\mathrm{T}}$ calorimeter term

As described in the previous section, the first step is to select calorimeter cells that belong to reconstructed TopoClusters to minimize the impact of noise.

The $x$ and $y$ components of the calorimeter $\not{E}_{\mathrm{T}}$ term are calculated from the transverse energies measured in TopoCells:

$$\not{E}_{x,y}^{\mathrm{Calo}} = - \sum_{\mathrm{TopoCells}} E_{x,y}. \tag{2}$$

The total transverse energy in the calorimeters, $\Sigma\not{E}_{\mathrm{T}}$, is calculated from the scalar sum of $E_{\mathrm{T}}$ of all TopoCells:

$$\Sigma\not{E}_{\mathrm{T}}{}^{Calo} = \sum_{\mathrm{TopoCells}} E_{\mathrm{T}}. \tag{3}$$

The straightforward result, obtained by using the electromagnetic calibration for all cells, gives a large shift in the $\not{E}_{\mathrm{T}}$ scale of about 30% with respect to $\not{E}_{\mathrm{T}}^{\mathrm{True}}$ (see Section 3).

This result illustrates the necessity of developing a dedicated calibration scheme to reduce the systematic shift of the $\not{E}_{\mathrm{T}}$ scale and optimize its resolution. This goal is achieved in several steps according to the cell classification. The classification depends on whether the energy deposits in the calorimeter are electromagnetic or hadronic in nature and whether they are associated with high $p_{\mathrm{T}}$ particles.

To classify energy deposits, schemes to calibrate hadronic showers such as 'H1-like' calibration or 'Local-Hadronic' calibration [3] utilize the energy density in a cell. Electromagnetic showers tend to have higher energy densities as compared to hadronic showers. The 'Local-Hadronic' calibration scheme uses further information related to shape and depth of the calorimetric shower to classify a TopoCluster. The next step in the cell-based $\not{E}_{\mathrm{T}}$ reconstruction is to globally calibrate all calorimeter cells using the 'H1-like' or 'Local-Hadronic' calibration schemes. As can be seen in the performance section, this already gives a very good $\not{E}_{\mathrm{T}}$ performance. The final refinement step of the calibration using the association of cells with reconstructed objects is described in Section 2.2.4. It improves the linearity and the resolution particularly for events containing electrons (Section 3).

### 2.2.2 The $\not{E}_T$ muon term

The $\not{E}_T$ muon term is calculated from the momenta of muons measured in a large range of pseudorapidity, defined by $|\eta| < 2.7$:

$$\not{E}_{x,y}^{\text{Muon}} = - \sum_{\text{RecMuons}} E_{x,y}. \quad (4)$$

In the region $|\eta| < 2.5$ only good-quality muons in the muon spectrometer with a matched track in the inner detector are considered. The matching requirement reduces considerably contributions from fake muons, sometimes created from high hit multiplicities in the muon spectrometer in events with very energetic jets. For higher values of the pseudorapidity ($2.5 < |\eta| < 2.7$), outside the fiducial volume of the inner detector, there is no matched track required and the muon spectrometer is used alone.

The muon momentum measured by the muon spectrometer is taken in the two cases. Energy lost in the calorimeter is already included in the calorimeter term. No $p_T$ threshold cut is applied to reconstructed muons. Apart from the loss of muons outside the acceptance of the muon spectrometer ($|\eta| > 2.7$), there is a loss of muons in other regions (see Section 4.1) due to limited coverage of the muon spectrometer. The muons reconstructed from the inner detector and calorimeter energy deposits could be used to recover these events, but they are not yet used here.

As can be seen in the performance section, the $\not{E}_T$ resolution is only marginally affected by the muon term, due to the good identification efficiency and resolution of the ATLAS muon system. However, unmeasured, badly measured or fake muons can be a source of large fake $\not{E}_T$ (see Section 4).

### 2.2.3 $\not{E}_T$ cryostat term

The thickness of the cryostat between the LAr barrel electromagnetic calorimeter and the tile barrel hadronic calorimeter is about half an interaction length where hadronic showers can lose energy. The $\not{E}_T$ reconstruction recovers this loss of energy in the cryostat using the correlation of energies between the last layer of the LAr calorimeter and the first layer of the hadronic calorimeter. A similar correction for the end-cap cryostats is applied. This correction is called the cryostat term when used for jet energy correction [3]. It is defined as follows:

$$\not{E}_{x,y}^{\text{Cryo}} = - \sum_{\text{recJets}} Ejet_{x,y}^{\text{Cryo}}, \quad (5)$$

where all reconstructed jets are summed in the event, and

$$Ejet^{\text{Cryo}} = w^{\text{Cryo}} \sqrt{E_{\text{EM3}} \times E_{\text{HAD}}}, \quad (6)$$

where $w^{\text{Cryo}}$ is a calibration weight (determined together with the cell calibration weights in the H1-like calibration) and $E_{\text{EM3}}$ and $E_{\text{HAD}}$ are the jet energies in the third layer of the electromagnetic calorimeter and in the first layer of the hadronic calorimeter, respectively. The cryostat correction turns out to be non-negligible for high-$p_T$ jets. It contributes at the level of $\sim 5\%$ per jet with $p_T$ above 500 GeV.

### 2.2.4 Refined calibration of the $\not{E}_T$ calorimeter term

The final step is the refinement of the calibration of cells associated with each high-$p_T$ object. Calorimeter cells are associated with a parent reconstructed and identified high-$p_T$ object, in a chosen order: electrons, photons, muons, hadronically decaying $\tau$-leptons, $b$-jets and light jets. Refined calibration of the object is then used in $\not{E}_T$ to replace the initial global calibration cells. The calibration of these objects is known to higher accuracy than the global calibration, enabling to improve the $\not{E}_T$ reconstruction.

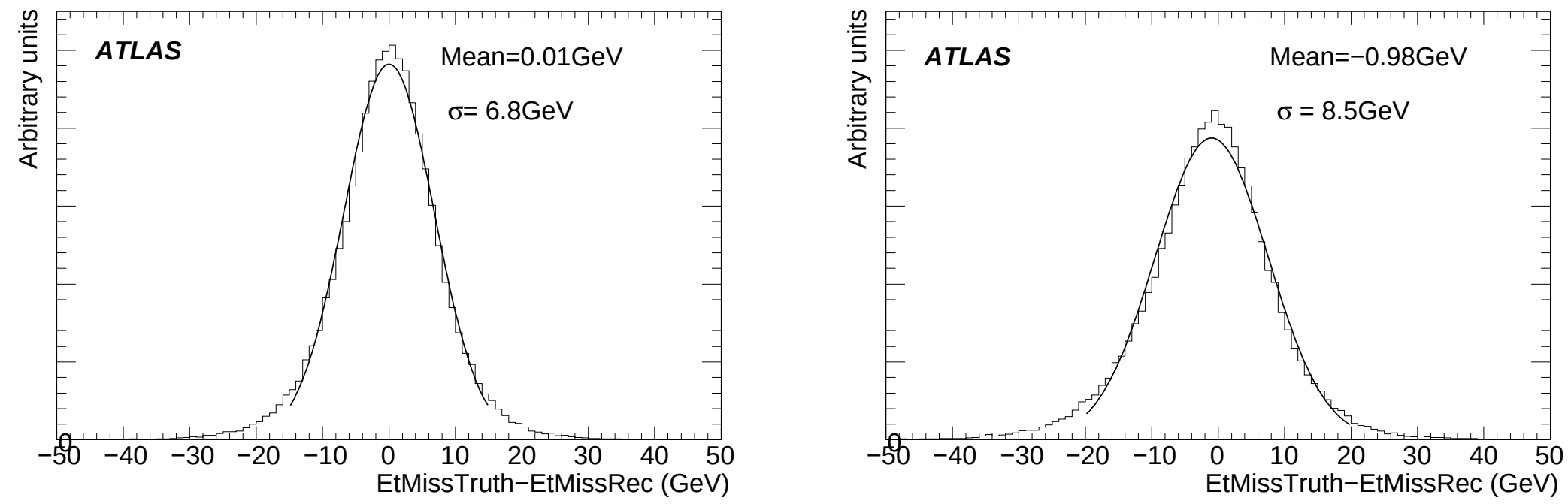

Figure 1: Distribution of the difference between true and reconstructed $\not\!\!E_{\mathrm{T}}$ for $Z \rightarrow \tau\tau$ events (left) including and (right) excluding cells in Topoclusters not associated with reconstructed high-$p_{\mathrm{T}}$ objects.

The calorimeter cells are associated with the reconstructed objects through the use of an association map. This map is filled starting from the reconstructed/identified objects in the chosen order, navigating back to their component clusters and back again to their cells. If a cell belongs to several kinds of reconstructed objects, only the first association is included in the map, i.e. the overlap removal is done at cell level. This avoids double counting of cells in the $\not\!\!E_{\mathrm{T}}$ calculation. If a cell belongs to more than one object of the same kind, all associations are included in the map and the geometrical weight of the cells, accounting for the sharing of energy of cells owned by two different TopoClusters, is also included to avoid double counting.

Attention has to be paid to the calibration of cells inside different objects. For example, for electrons/photons, the final cluster-level calibration (which can be propagated back to the cell-level) corrects for upstream material, longitudinal leakage and out-of-cone energy. The last correction should not be applied in the $\not\!\!E_{\mathrm{T}}$ calculation because the contribution of cells outside objects already accounts for it. In a similar way, for $\tau$ lepton decays and for jets, the overall scale factors which correct the energy for physics effects like final state radiation, fragmentation or the underlying event as well as for the effects due to the clustering algorithm are not applied in the calculation of $\not\!\!E_{\mathrm{T}}$, because they also contain the out-of-cluster correction.

All TopoCells, even if not associated with any high-$p_{\mathrm{T}}$ reconstructed object, are used in the $\not\!\!E_{\mathrm{T}}$ calculation. They are calibrated using the global calibration scheme. The importance of the energy deposits of these low energy particles for the $\not\!\!E_{\mathrm{T}}$ calculation is shown in Fig. 1. The shift in the absolute value of the reconstructed $\not\!\!E_{\mathrm{T}}$ increases by about 1 GeV while the resolution is degraded by a factor $\sim 1.25$.

Once the cells are associated with categories of objects as described above, the contribution to $\not\!\!E_{\mathrm{T}}$ is calculated as follows:

$$\not\!\!E_{x,y}^{\mathrm{Calo}} = \not\!\!E_{x,y}^{\mathrm{RefCalib}} = -(\not\!\!E_{x,y}^{\mathrm{RefEle}} + \not\!\!E_{x,y}^{\mathrm{RefTau}} + \not\!\!E_{x,y}^{\mathrm{Refbjets}} + \not\!\!E_{x,y}^{\mathrm{RefJets}} + \not\!\!E_{x,y}^{\mathrm{RefMuo}} + \not\!\!E_{x,y}^{\mathrm{RefOut}}), \tag{7}$$

where each term is calculated from the negative of the sum of calibrated cells inside a specific object and $\not\!\!E_{x,y}^{\mathrm{RefOut}}$ is calculated from the cells in TopoClusters which are not included in the reconstructed objects. In the following the final $\not\!\!E_{\mathrm{T}}$ calculation obtained from Equation (1) with $\not\!\!E_{x,y}^{\mathrm{Calo}} = \not\!\!E_{x,y}^{\mathrm{RefCalib}}$ will be referred to as $\not\!\!E_{\mathrm{T}}^{\mathrm{RefFinal}}$.

### 2.3 Object-based $\not\!\!E_{\mathrm{T}}$ reconstruction

The motivation of the method is to reliably reconstruct $\not\!\!E_{\mathrm{T}}$ for analyses that are sensitive to low $p_{\mathrm{T}}$ deposits coming mostly from neutral and charged pions, from soft jets, from the underlying event and from pile-up. This is important, for example, in reconstructing the invariant mass of the Standard Model Higgs boson, $m_H$, in the $H \rightarrow \tau^+\tau^-$ final state for masses in the range $115 < m_H < 140$ GeV [4].

The object-based method comprises two main steps:

- Establish a classification between two main types of objects: high $p_T$ (e/$\gamma, \mu, \tau$, jets) and low $p_T$ objects ($\pi^0, \pi^\pm$, unclustered deposits) coming from underlying event, pile-up of multiple $pp$ collisions, initial and final state radiation, and other soft QCD processes.
- Apply the object-based calibration optimized for $\not{E}_T$ calculation.

The $x$ and $y$ components of $\not{E}_T$ and $\sum E_T$ are calculated by adding the contributions from each type of components:

$$\not{E}_{x,y} = -E_{x,y}^{\text{High}} - E_{x,y}^{\text{Low}} \tag{8}$$

$$\sum E_T = \sum E_T^{\text{High}} + \sum E_T^{\text{Low}}, \tag{9}$$

where the indices 'High' and 'Low' correspond to the $\not{E}_T$ and $\sum E_T$ calculated from high $p_T$ and low $p_T$ objects, defined below.

The object-based algorithm uses mostly the calorimeter to reconstruct $\not{E}_T$. Some objects such as electrons and taus also use the inner detector tracking, while the muons use both, inner detector and muon spectrometer information. Tracking is also used for the low $p_T$ deposits of soft objects.

The object-based method uses the TopoClusters $4/2/0$ (Section 2.1) and first calculates all contributions of high $p_T$ objects. Each TopoCluster is allowed to be included only once by the first object that is associated with it. The classification starts with the identification of electrons, photons, muons and $\tau$'s. Once the clusters belonging to these objects are removed from the event record, hadronic jets above a certain threshold are identified. TopoCells not part of any of the above high $p_T$ objects are classified as low $p_T$ deposit. The next subsections discuss the calibration of these objects.

### 2.3.1 Calorimeter objects: electrons, hadronic $\tau$-jets, jets

**Electrons:** The electron objects are taken from the standard electron reconstruction and a matched track is always required. The default calibration of an electron is based on the 'sliding window' EM clusters [5]. To be consistent with the combined $4/2/0$ TopoClustering used for the object-based reconstruction, electrons are reconstructed from TopoClusters and they are calibrated by weighting the energy deposits in the longitudinal calorimeter layers. The electron $p_T$ is required to be at least 8 GeV.

**Tau Leptons:** The object-based method uses the calo-based reconstructed $\tau$-jets [6] and it calibrates them as jets. A cut of 20 GeV on the $\tau$-jet $p_T$ is applied and a $\tau$ likelihood $> 4$ is required.

**Jets:** A cone ($\Delta R = 0.7$) jet algorithm running on the TopoClusters is used for calculating the $\not{E}_T$ contribution from jets. Jets not overlapping with the electron and tau jets are chosen. The jet energy calibration is based on the 'H1-like' hadronic calibration [3], which corrects for the difference in response for e/$\gamma$ and hadrons, followed by a scale correction due to the non-uniformity of the reconstruction in $\eta$. It is based on the $\not{E}_T$ projection method (used by CDF and D0)[7] using a dijet sample where forward jets are corrected with respect to well measured central jets ($|\eta| < 0.8$). The jet $p_T$ is required to be at least 20 GeV .

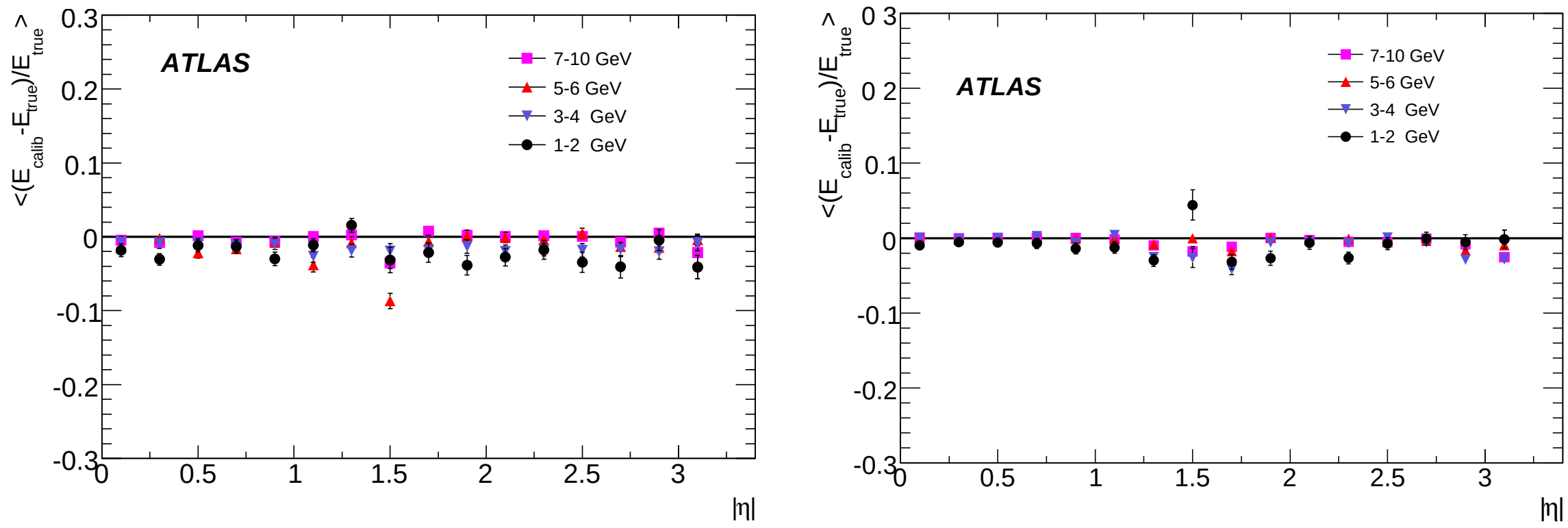

Figure 2: Linearity for single charged $\pi$ (left) and single $\pi^0$ (right) as a function of $\eta$ for the energy bin $1 < E < 10$ GeV).

### 2.3.2 Muons

In general, the combined muons, reconstructed from the inner detector and the muon spectrometer, are used. In the region where the inner detector has no coverage ($2.5 < |\eta| < 2.7$) muons reconstructed from the spectrometer only are used. To reduce the number of fake muons originating from punch through of high $p_T$ jets, strong quality requirements are imposed on the combined track from the inner detector and the muon spectrometer. Additionally, if a muon passes inside a jet and the ratio of measured momenta in the inner detector and in the muon spectrometer is below 0.2, the muon is rejected. If the muon is found inside a jet or the total $p_T$ at the EM scale of the TopoClusters within a cone of 0.2 around the muon is $> 10$ GeV, either the measured muon energy loss in the calorimeter is subtracted from the combined $p_T$ or the muon spectrometer $p_T$ is used. The combined muon $p_T$ is required to be at least 6 GeV.

Inner detector tracks may be used to improve the $\not{E}_T$ reconstruction. They contain the muons not covered by the combination of inner detector and muon spectrometer, especially in the crack regions of the muon system. Only those tracks are retained which are isolated from others by a cone of size 0.3. Those tracks overlapping with electrons, muons or jets that are already used for $\not{E}_T$ are discarded. The tracks should have $p_T > 6$ GeV and should satisfy a muon likelihood criterion based on $E/p$ and the energy in the calorimeter sampling layers. In addition, isolated tracks (mostly pions) with little energy deposits in the calorimeter can be used to improve $\not{E}_T$. These tracks are used only if $p_T$ less than 10 GeV, to avoid tracks with spurious high $p_T$,which deteriorates the $\not{E}_T$ performance.

### 2.3.3 Low $p_T$ depositions: classification and calibration

Only those TopoClusters that have not been assigned to high $p_T$ objects are considered for classification as low $p_T$ objects of either electromagnetic or hadronic nature. For this purpose, clusters are further subdivided into so-called mini-jets. Mini-jets are reconstructed from TopoClusters using a cone algorithm of a relatively small radius of $\Delta R = 0.2$ and a seed cluster of $p_T > 0.5$ GeV. A mini-jet is required to have $p_T > 0.5$ GeV. The mini-jets are next classified as charged and neutral pions.

The separation between charged and neutral pions is done only in $|\eta| < 3.2$. A mini-jet is defined to be a $\pi^0$ if the fractional energy in the hadronic compartments is less than 2% and $p_T > 10$ GeV. The low $p_T$ deposits are calibrated under single charged and neutral pion hypotheses in the full $\eta$ range. A sampling method similar to that used to extract longitudinal weights as implemented for the electron and photon calibration is used. The calorimeter sampling weights are determined separately for neutral and charged pions in different energy and $\eta$ regions. A good linearity is achieved for both neutral and charged pion hypotheses. Figure 2 shows the linearity for charged and neutral pions in the central region, which roughly fluctuates within 5%.

Mini-jets will not saturate all the low $p_{\mathrm{T}}$ calorimeter deposits. There will remain a non-trivial amount of energy left in the calorimeter and clustered in TopoClusters from very low momentum pions, which will not form a mini-jet. These deposits are referred to as unassociated deposits. The energy calibration of these unassociated deposits has been estimated in three $\eta$ regions, depending on the calorimeter region (barrel, end-cap, forward), and has been added to the $\not{E}_{\mathrm{T}}$ calculation.

# 3 Performance of the reconstructed $\not{E}_{\mathrm{T}}$

In this section the performance of the $\not{E}_{\mathrm{T}}$ reconstruction is discussed, focusing on the linearity and resolution of the reconstructed $\not{E}_{\mathrm{T}}$ as a function of the true missing transverse energy, $\not{E}_{\mathrm{T}}^{\mathrm{True}}$. The measurement of the $\not{E}_{\mathrm{T}}$ direction and the dependence on topology are also discussed. Note that the performance of the two $\not{E}_{\mathrm{T}}$ reconstruction methods is very similar, so it is not specified which method has been used to produce each performance plot.

$\not{E}_{\mathrm{T}}^{\mathrm{True}}$ is defined from the sum of all stable and non-interacting particles in the final state (neutrinos and the lightest supersymmetric particles). Comparisons between $\not{E}_{\mathrm{T}}$ and $\not{E}_{\mathrm{T}}^{\mathrm{True}}$ are made for a number of physics processes with different topologies and final states.

The $\not{E}_{\mathrm{T}}$ performance in case of events with large $\not{E}_{\mathrm{T}}^{\mathrm{Fake}}$, which contribute predominantly to the tails of the $\not{E}_{\mathrm{T}}$ distribution, is described in the next section.

## 3.1 Linearity and resolution

The $\not{E}_{\mathrm{T}}$ linearity is defined by the following expression:

$$Linearity = (\not{E}_{\mathrm{T}}^{\mathrm{True}} - \not{E}_{\mathrm{T}})/\not{E}_{\mathrm{T}}^{\mathrm{True}}, \tag{10}$$

where $\not{E}_{\mathrm{T}}$ and $\not{E}_{\mathrm{T}}^{\mathrm{True}}$ are reconstructed and true $\not{E}_{\mathrm{T}}$, respectively. This definition of linearity assumes an $\not{E}_{\mathrm{T}}^{\mathrm{True}}$ value above a threshold and an $\not{E}_{\mathrm{T}}^{\mathrm{Fake}}$ value that is small such that the $\not{E}_{\mathrm{T}}$ angle is well measured.

Figure 3 shows the reconstructed linearity as a function of $\not{E}_{\mathrm{T}}^{\mathrm{True}}$ for a number of physics processes. The following statements summarize the behavior of the linearity distributions:

- The uncalibrated $\not{E}_{\mathrm{T}}$ corresponds to the use of cell energies at the electromagnetic scale and shows a large systematic bias of 30%. In $W \to e\nu$ and $W \to \mu\nu$ decays, the bias is smaller since the hadronic activity on average is smaller.
- The reconstructed $\not{E}_{\mathrm{T}}$ based on globally calibrated cell energies and reconstructed muons gives a linearity to within 5%.
- The reconstructed $\not{E}_{\mathrm{T}}$ including the cryostat correction shows a linearity to within 1% for all processes except for $W \to e\nu$ .
- The refined $\not{E}_{\mathrm{T}}$ calibration, which optimizes the calibration with reconstructed object identity, recovers the linearity for $W \to e\nu$ events to within 1%. The refined calibration also gives the best resolution when compared with the above steps of calibration (see also Section 6).

The linearity for $A \to \tau\tau$ with $m_A = 800$ GeV is shown in Fig. 3 (right) as a function of $\not{E}_{\mathrm{T}}^{\mathrm{True}}$. The bias of linearity at low $\not{E}_{\mathrm{T}}^{\mathrm{True}}$ is due to the finite resolution of the $\not{E}_{\mathrm{T}}$ measurement. The reconstructed $\not{E}_{\mathrm{T}}$ is positive by definition, so the linearity is negative when the true $\not{E}_{\mathrm{T}}$ is near to zero. Excluding the events with $\not{E}_{\mathrm{T}}^{\mathrm{True}} < 40$ GeV, which have a small statistics the observed linearity is found to be within 2%.

The resolution is estimated from the width of the $\not\!\!E_{x,y} - \not\!\!E_{x,y}^{\mathrm{True}}$ distribution in bins of the total transverse energy deposited in the calorimeters ($\Sigma\not\!\!E_{\mathrm{T}}$). The core of each distribution is fitted with a Gaussian shape to estimate the width. Figure 4 shows the $\sigma$ of the fit plotted as a function of $\Sigma\not\!\!E_{\mathrm{T}}$ when refined calibration is applied. The $\not\!\!E_{\mathrm{T}}$ resolution approximately follows a stochastic behaviour as a function of $\Sigma\not\!\!E_{\mathrm{T}}$. Deviations from this simple behaviour are expected, and observed for low values of $\Sigma\not\!\!E_{\mathrm{T}}$ where the contribution of noise is important and for very high values of $\Sigma\not\!\!E_{\mathrm{T}}$ where the constant term in the resolution of the calorimetric energy measurement dominates.

The $\not\!\!E_{\mathrm{T}}$ resolution is fitted with a function $\sigma = a \cdot \sqrt{\Sigma\not\!\!E_{\mathrm{T}}}$ for values of $\Sigma\not\!\!E_{\mathrm{T}}$ between 20 and 2000 GeV. The parameter $a$, which quantifies the $\not\!\!E_{\mathrm{T}}$ resolution, varies between 0.53 and 0.57 (see Fig. 4 left and right, respectively). Refined $\not\!\!E_{\mathrm{T}}$ calibration yields the best results when compared to earlier stages of the calibration as described above. For $W \to e\nu$ decays the $a$ parameter is reduced by 88% and for $Z \to ee$ events it is reduced by 78% with respect to the global calibration with the cryostat correction applied. Figure 5 shows the resolution in the high $\Sigma\not\!\!E_{\mathrm{T}}$ region for QCD jet samples. The jet samples used are generated in parton $p_{\mathrm{T}}$ bins: J1 corresponds to $17 < p_{\mathrm{T}} < 35$ GeV, J2 to $35 < p_{\mathrm{T}} < 70$ GeV, J3 to $70 < p_{\mathrm{T}} < 140$ GeV, J4 to $140 < p_{\mathrm{T}} < 280$ GeV, J5 to$280 < p_{\mathrm{T}} < 560$ GeV, J6 to $560 < p_{\mathrm{T}} < 1120$ GeV, J7 to $1120 < p_{\mathrm{T}} < 2240$ GeV. There is a clear degradation in the performance for the high $p_{\mathrm{T}}$ jet samples (J6 and J7), where the linear term dominates.

The effect of angular calorimeter coverage on the $\not\!\!E_{\mathrm{T}}$ measurement is evaluated by comparing the resolution with and without including the forward calorimeters (FCAL). In $Z \to \tau\tau$ events the $\not\!\!E_{\mathrm{T}}$ resolution is 7.8 and 10.1 GeV, respectively, showing that the ATLAS coverage minimises by design the effect of particles escaping at very large $\eta$ and that the forward calorimeter is very important to guarantee that.

Projections of $\not\!\!E_{\mathrm{T}}$ along suitable axes can be used to check calibration problems and understand topology dependences. The quantity $\not\!\!E_{\mathrm{L}}$ (longitudinal $\not\!\!E_{\mathrm{T}}$ projection) is the $\not\!\!E_{\mathrm{T}}$ projection onto the axis pointing in the direction of the genuine $\not\!\!E_{\mathrm{T}}$ of the event. In events without genuine $\not\!\!E_{\mathrm{T}}$, such as $Z \to \ell\ell$ events this axis is reconstructed from the direction of flight of the $Z$ and in dijet events the projection is done along the dijet thrust axis. The quantity $\not\!\!E_{\mathrm{P}}$ (perpendicular $\not\!\!E_{\mathrm{T}}$ projection) is defined as a projection orthogonal to the $\not\!\!E_{\mathrm{L}}$ direction.

A bias of $\not\!\!E_{\mathrm{L}}$ usually indicates mis-calibration of the object's energy scale (in most cases the hadronic scale). For events with significant values of $\not{E}_{\mathrm{T}}^{\mathrm{True}}$, $\not{E}_{\mathrm{P}}$ is a measure of the $\not{E}_{\mathrm{T}}$ angular resolution. For dijet events and other back-to-back topologies with similarly defined $\not{E}_{\mathrm{P}}$ usually there is no bias in $\not{E}_{\mathrm{P}}$, but the resolution of $\not{E}_{\mathrm{P}}$ is sensitive to soft radiation in the event. One possible source of bias in $\not{E}_{\mathrm{P}}$ comes from a bias in the angular measurement of the objects in the event, which will cause a shift in the same direction on both sides of $\not{E}_{\mathrm{L}}$.

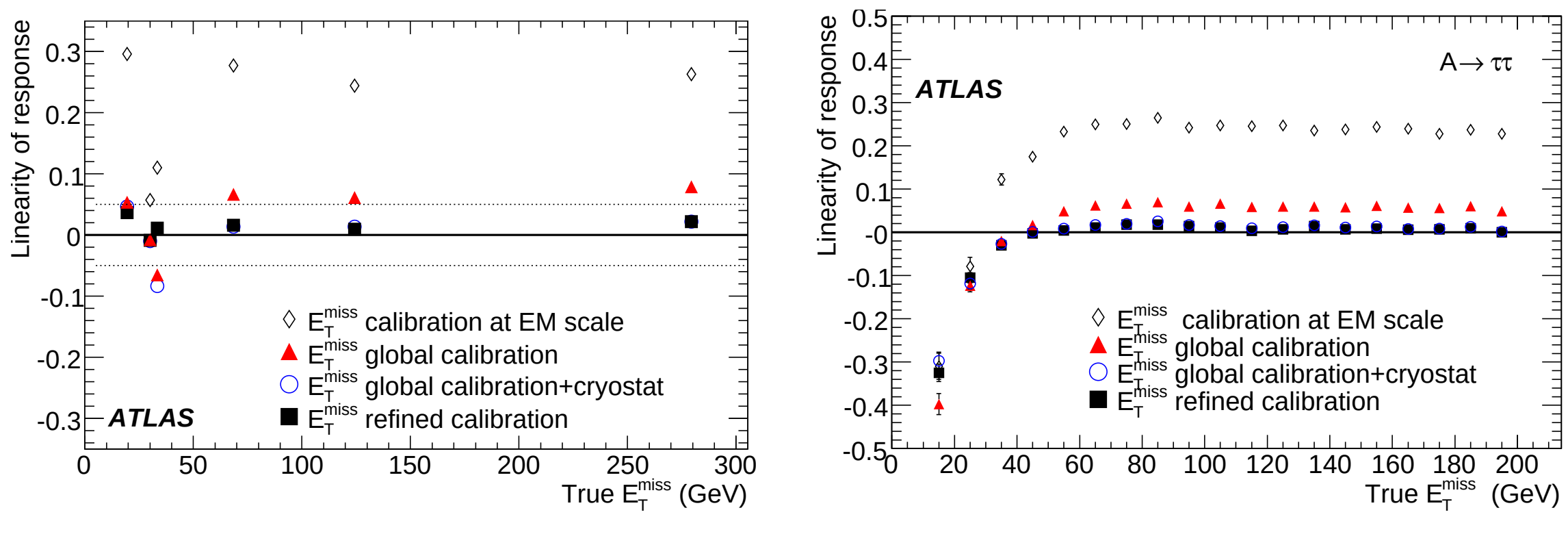

Figure 3: (left) Linearity of response for reconstructed $\not\!\!E_{\mathrm{T}}$ as a function of the average true $\not\!\!E_{\mathrm{T}}$ for different physics processes covering a wide range of true $\not\!\!E_{\mathrm{T}}$ and for the different steps of $\not\!\!E_{\mathrm{T}}$ reconstruction (see text). The points at average true $\not\!\!E_{\mathrm{T}}$ of 20 GeV are from $Z \to \tau\tau$ events, those at 35 GeV are from $W \to e\nu$ and $W \to \mu\nu$ events, those at 68 GeV are from semi-leptonic $t\bar{t}$ events, those at 124 GeV are from $A \to \tau\tau$ events with $m_A = 800$ GeV, and those at 280 GeV are from events containing supersymmetric particles at a mass scale of 1 TeV. (right) Linearity of response for reconstructed $\not\!\!E_{\mathrm{T}}$ as a function of the true $\not\!\!E_{\mathrm{T}}$ for $A \to \tau\tau$ events with $m_A = 800$ GeV.

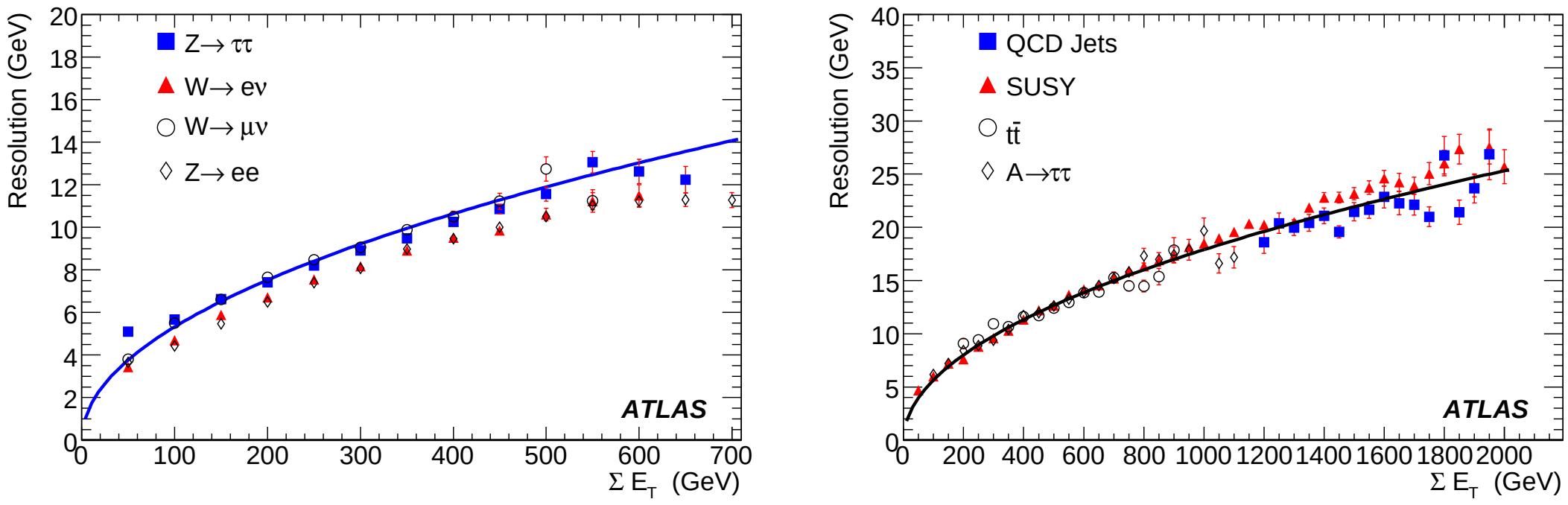

Figure 4: Resolution of the two $\not{E}_T$ components with refined calibration as a function of the total transverse energy, $\Sigma E_T$ for low to medium values (left) and for higher values (right). The curves correspond to the best fits of $\sigma = 0.53\sqrt{\Sigma E_T}$ through the points from $Z \to \tau\tau$ events (left) and $\sigma = 0.57\sqrt{\Sigma E_T}$ through the points from $A \to \tau\tau$ events (right). The points from $A \to \tau\tau$ events are for masses $m_A$ ranging from 150 to 800 GeV and the points from QCD jets correspond to dijet events with $560 < p_T < 1120$ GeV.

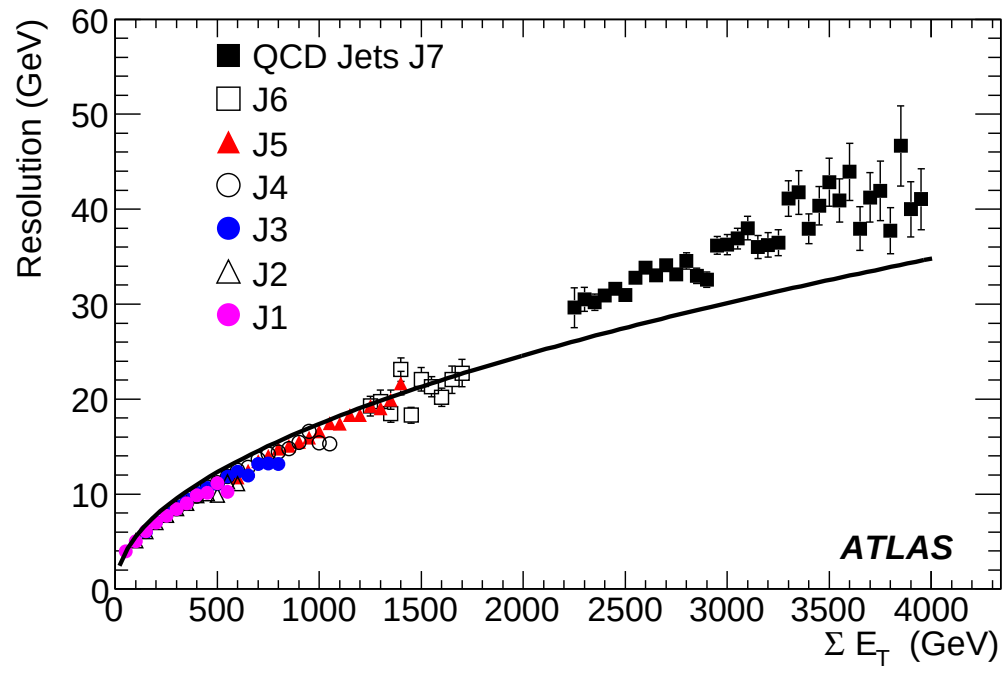

Figure 5: Resolution of the two $\not{E}_T$ components with refined calibration as a function of $\Sigma\not{E}_T$ for QCD dijet samples ($17 < p_T < 2240$ GeV). See text for the definition of samples J1-J7. The curve corresponds to $\sigma = 0.55\sqrt{\Sigma\not{E}_T}$ (combined fit in the low and medium $\Sigma\not{E}_T$ regions).

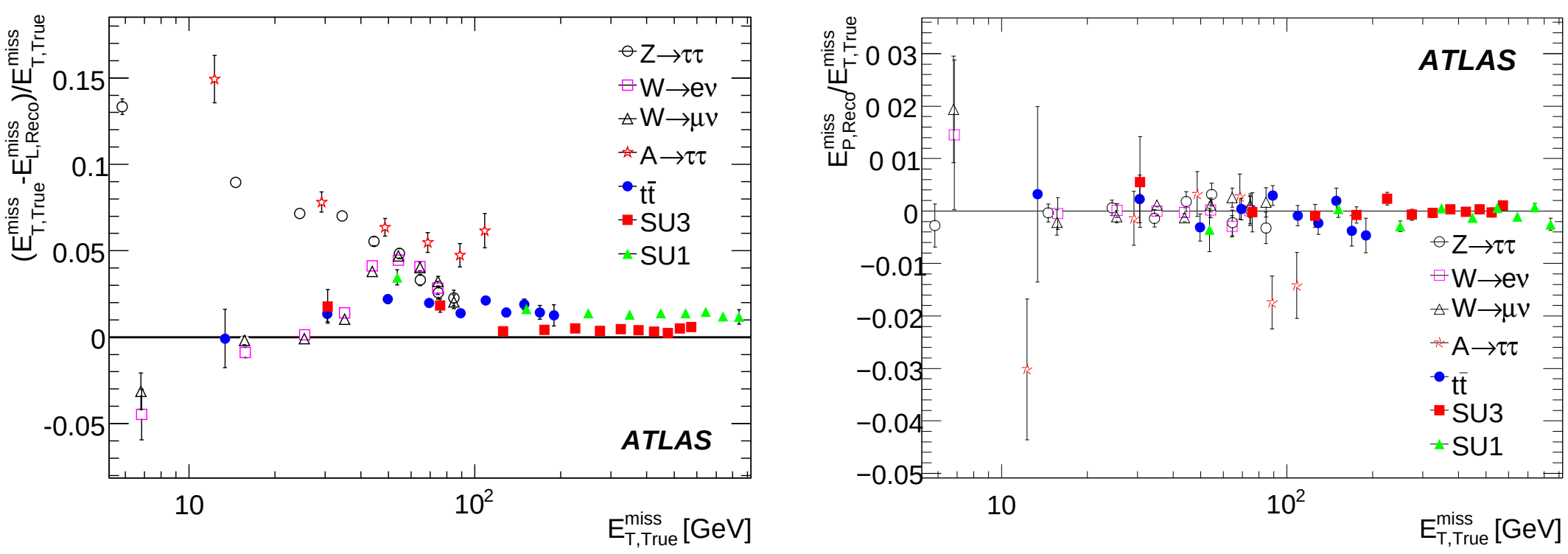

Figure 6: Linearity of response for $\not{E}_L$ (left) and $\not{E}_P$ (right) as a function of the average true $\not{E}_T$ for different physics processes. $\not{E}_L$ and $\not{E}_P$ are defined in the text.

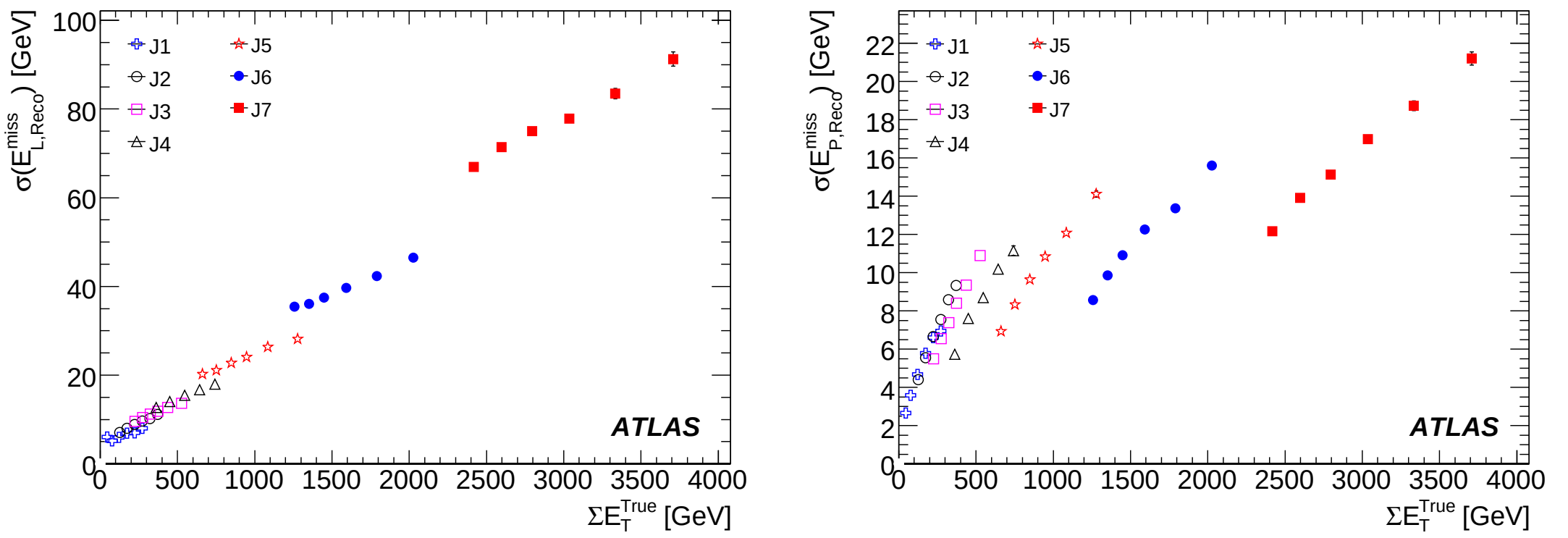

Figure 7: Resolution of $\not{E}_{\mathrm{L}}$(left) and $\not{E}_{\mathrm{P}}$(right) as a function of $\Sigma\not{E}_{\mathrm{T}}$ for QCD jet samples. See text for the definition of samples J1-J7.

Figure 6 shows the $\not{E}_{\mathrm{L}}$ and $\not{E}_{\mathrm{P}}$ linearity as a function of the true $\not{E}_{\mathrm{T}}$ for different physics channels. The $\not{E}_{\mathrm{L}}$ linearity is within 2% above 100 GeV and deviates by 5-10% at lower $\not{E}_{\mathrm{T}}^{\mathrm{True}}$. The $\not{E}_{\mathrm{P}}$ bias is consistent with zero throughout. Figure 7 shows the $\not{E}_{\mathrm{L}}$ and $\not{E}_{\mathrm{P}}$ resolution as a function of the true $\Sigma\not{E}_{\mathrm{T}}$ for the QCD jet samples. As the $\not{E}_{\mathrm{L}}$ for jet events is defined as the $\not{E}_{\mathrm{T}}$ projection onto the thrust axis of the two leading jets which are back-to-back, a much larger resolution in absolute terms is expected with respect to $\not{E}_{\mathrm{P}}$. The discontinuity in J4 ($140 < p_{\mathrm{T}} < 280$ GeV) and J5 ($280 < p_{\mathrm{T}} < 560$ GeV) $\not{E}_{\mathrm{P}}$ resolution is a result of dividing the dijet events into samples based on the parton $p_{\mathrm{T}}$. Events with the same $\Sigma\not{E}_{\mathrm{T}}$ in J4 and J5 can have different energies in the perpendicular direction due to differences in underlying event, initial or final state radiation. This changes the $\not{E}_{\mathrm{P}}$ resolution in the two samples.

A good performance in terms of linearity and resolution may enhance the ability to reconstruct the mass of final states which involve neutrinos. Despite the presence of several neutrinos in the final state, the invariant mass of the $\tau$ pair can also be reconstructed in $Z \rightarrow \tau\tau$ and supersymmetric Higgs boson decays like $A \rightarrow \tau\tau$ under simplifying assumptions [8, 9, 10]. Figure 8 shows reconstructed mass peaks of $Z \rightarrow \tau\tau$ and supersymmetric Higgs boson decays $A \rightarrow \tau\tau$ with $m_A = 450$ GeV. The reconstructed masses are correct to $\sim 2\%$ and the mass resolution is approximately 11%. Nevertheless, significant tails remain in the distributions because of the highly non-Gaussian effects induced by mis-measurements of $\not{E}_{\mathrm{T}}$ and by the approximations used.

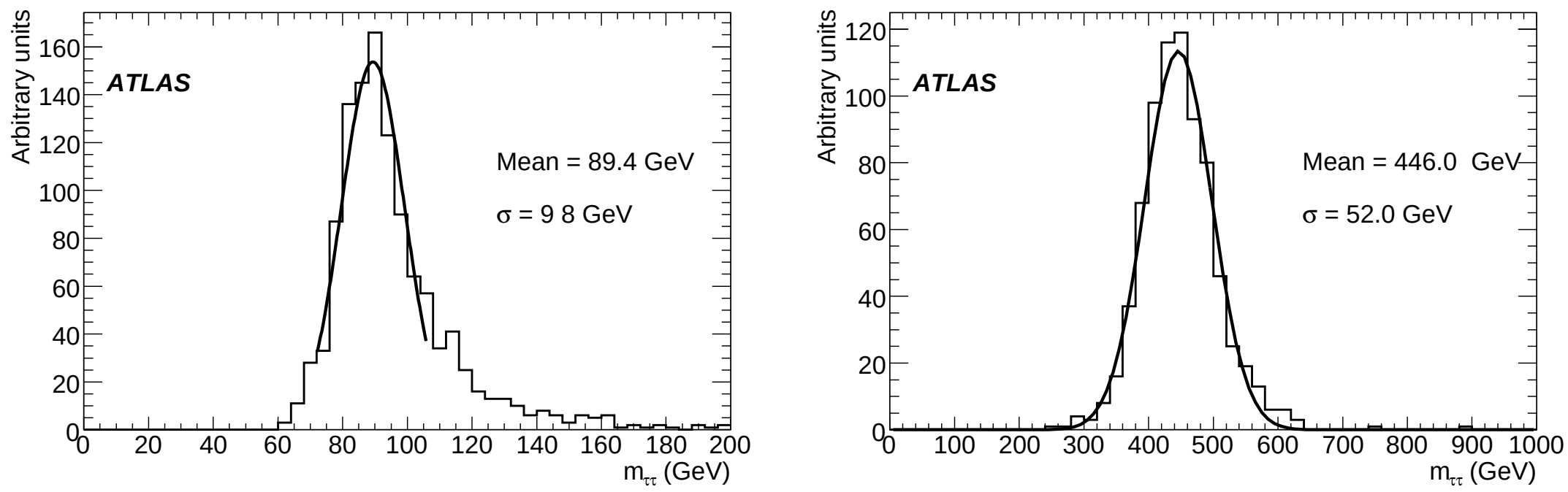

Figure 8: Distributions of the reconstructed invariant mass of $\tau$-lepton pairs with one $\tau$-lepton decaying to a lepton and the other one decaying to hadrons. The results are shown for $Z \rightarrow \tau\tau$ decays (left) and for $A \rightarrow \tau\tau$ decays with $m_A = 450$ GeV (right).

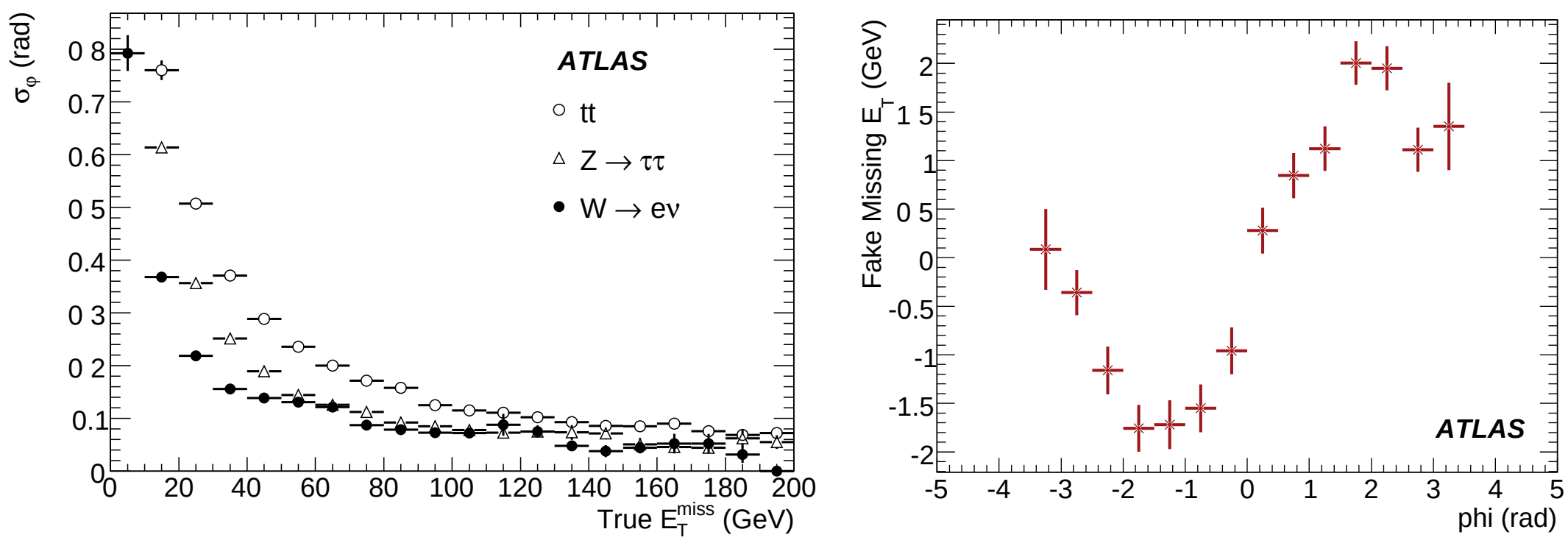

Figure 9: (left) Accuracy of the measurement of the azimuth of the $\not{E}_{\mathrm{T}}$ vector as a function of the true $\not{E}_{\mathrm{T}}$ for three different physics processes: semi-leptonic $t\bar{t}$ events, $Z \rightarrow \tau\tau$ and $W \rightarrow e\nu$ events. (right) $\not{E}_{\mathrm{T}}^{\mathrm{Fake}}$as a function of the reconstructed $\phi(\not{E}_{\mathrm{T}})$ in $t\bar{t}$ events, simulated with extra material in $\phi$.

### 3.2 Measurement of the $\not{E}_{\mathrm{T}}$ direction

Large energy fluctuations in the calorimeter or muon mis-measurements can produce large $\not{E}_{\mathrm{T}}^{\mathrm{Fake}}$. In general, for events with genuine missing transverse energy, the $\not{E}_{\mathrm{T}}$ angular resolution will depend on the relative fraction of $\not{E}_{\mathrm{T}}^{\mathrm{Fake}}$ and on the event topology. Figure 9 shows the $\not{E}_{\mathrm{T}}$ azimuthal angular resolution as a function of the $\not{E}_{\mathrm{T}}^{\mathrm{True}}$ for three different physics processes. The measurement of the $\not{E}_{\mathrm{T}}$ azimuth is clearly more accurate for $W \rightarrow e\nu$ events, which in general contain one high-$p_{\mathrm{T}}$ electron and moderate hadronic activity in addition, compared to $t\bar{t}$ events. For values of $\not{E}_{\mathrm{T}}^{\mathrm{True}}$ below 40 GeV, the accuracy of the measurement of the direction degrades rapidly. In contrast, for high values of $\not{E}_{\mathrm{T}}^{\mathrm{True}}$, azimuthal accuracies below 100 mrad are achieved.

Detector inefficiencies may perturb the radial symmetry of the physics events. Thus, observations of $\phi$ asymmetries in reconstructed variables may be a hint of instrumental problems. Due to the increased material (of $\sim 5\%$ to 10%) in the upper half of the detector that was added artificially into the simulation, a $\phi$ asymmetry is observed in $\not{E}_{\mathrm{T}}$ as seen in Fig. 9 (right). This $\phi$ asymmetry can also be observed in the $\not{E}_{\mathrm{T}}$ computed at the event filter trigger level.

Similarly, problematic $\eta$ regions may be spotted by looking at $\not{E}_{\mathrm{T}}$ correlations with jet pseudorapidity, in particular in QCD events. In fact, in this kind of events, $\not{E}_{\mathrm{T}}$ is mainly due to jet mis-measurements which will affect more the $\not{E}_{\mathrm{T}}$ component parallel to the dijet axis than that perpendicular to it. Detector failures or incorrect calibration sets may be revealed as unexpected peaks in such plots.

## 4 Fake $\not{E}_{\mathrm{T}}$

The reconstructed $\not{E}_{\mathrm{T}}$ has two constituents - one that is produced by particles that interact weakly with the detector ($\not{E}_{\mathrm{T}}^{\mathrm{True}}$) and the other one due to detector inefficiencies and resolution ($\not{E}_{\mathrm{T}}^{\mathrm{Fake}}$). Figure 10 shows the rate of $\not{E}_{\mathrm{T}}^{\mathrm{Fake}}$ and $\not{E}_{\mathrm{T}}^{\mathrm{True}}$ for the QCD sample generated with $560 < p_{\mathrm{T}} < 1120$ GeV, where $\not{E}_{\mathrm{T}}^{\mathrm{Fake}}$ dominates at lower values and also has a larger tail. The same figure also shows these distributions after excluding events with high $p_{\mathrm{T}}$ jets within $17^o$ of the reconstructed $\not\!\!E_{\mathrm{T}}$ in the transverse plane. This considerably lowers the $\not\!\!E_{\mathrm{T}}^{\mathrm{Fake}}$ rate compared to $\not\!\!E_{\mathrm{T}}^{\mathrm{True}}$. For an accurate measurement of $\not\!\!E_{\mathrm{T}}$ it is important to have a good understanding of the sources of $\not\!\!E_{\mathrm{T}}^{\mathrm{Fake}}$ in data. Since the goal here is to study the performance and not the relative contributions of signal and background, comparisons between different physics samples are made for a common number of events rather than for a common luminosity.

The first two subsections discuss the $\not\!\!E_{\mathrm{T}}^{\mathrm{Fake}}$ from muons and from the calorimeter under the assumption that all detector readout channels are functional. The impact of dead-regions in the detector is examined in the subsequent subsections.

## 4.1 Fake $\not\!\!E_{\mathrm{T}}$ from muons

$\not\!\!E_{\mathrm{T}}^{\mathrm{Fake}}$ from muons can be caused either by inefficiencies in reconstructing a high $p_{\mathrm{T}}$ muon or by reconstructing a fake high $p_{\mathrm{T}}$ muon. The latter could be present due to a combination of a lower $p_{\mathrm{T}}$ muon and/or random hits from high $p_{\mathrm{T}}$ jet punch-throughs from the calorimeter to the muon chambers. It can be argued that for reasonable muon identification efficiencies, $\not\!\!E_{\mathrm{T}}^{\mathrm{Fake}}$ from missed muons will only be a small fraction of the $\not\!\!E_{\mathrm{T}}^{\mathrm{True}}$ from neutrinos. For example in QCD samples the neutrino to muon ratio is roughly two. It gets higher in other physics samples depending on the fraction of $\tau$ candidates in the event. On the other hand fake muons that are reconstructed from random hits in the muon chambers can be arbitrarily hard and strongly contribute to $\not\!\!E_{\mathrm{T}}^{\mathrm{Fake}}$. However, the study here shows that $\not\!\!E_{\mathrm{T}}^{\mathrm{Fake}}$ from muons is dominated by missed muons rather than fake muons.

The total $\not\!\!E_{\mathrm{T}}^{\mathrm{Fake}}$ in the MC sample can be defined as the vectorial difference of reconstructed $\not\!\!E_{\mathrm{T}}$ and $\not\!\!E_{\mathrm{T}}^{\mathrm{True}}$, as follows:

$$\not\!\!E_{\mathrm{T}}^{\mathrm{Fake}} = \sqrt{{\not\!\!E_{x}^{\mathrm{Fake}}}^2 + {\not\!\!E_{y}^{\mathrm{Fake}}}^2} \qquad \text{where} \qquad \not\!\!E_{x,y}^{\mathrm{Fake}} = \not\!\!E_{x,y} - \not\!\!E_{x,y}^{\mathrm{True}}. \tag{11}$$

$\not\!\!E_{x,y}$ and $\not\!\!E_{x,y}^{\mathrm{True}}$ are the $x$ and $y$ components of the final reconstructed $\not\!\!E_{\mathrm{T}}$ and of the $\not\!\!E_{\mathrm{T}}^{\mathrm{True}}$ defined in Section 3.

The contribution to the total $\not\!\!E_{\mathrm{T}}^{\mathrm{Fake}}$ from muon mis-measurements can be defined by

$$\not\!\!E_{x,y}^{\mathrm{FakeMuon}} = \not\!\!E_{x,y}^{\mathrm{Muon}} - \not\!\!E_{x,y}^{\mathrm{TrueMuon}}, \tag{12}$$

where $\not\!\!E_{x,y}^{\mathrm{Muon}}$ and $\not\!\!E_{x,y}^{\mathrm{TrueMuon}}$ are calculated by summing the reconstructed and true $x$ and $y$ components from muons in the event.

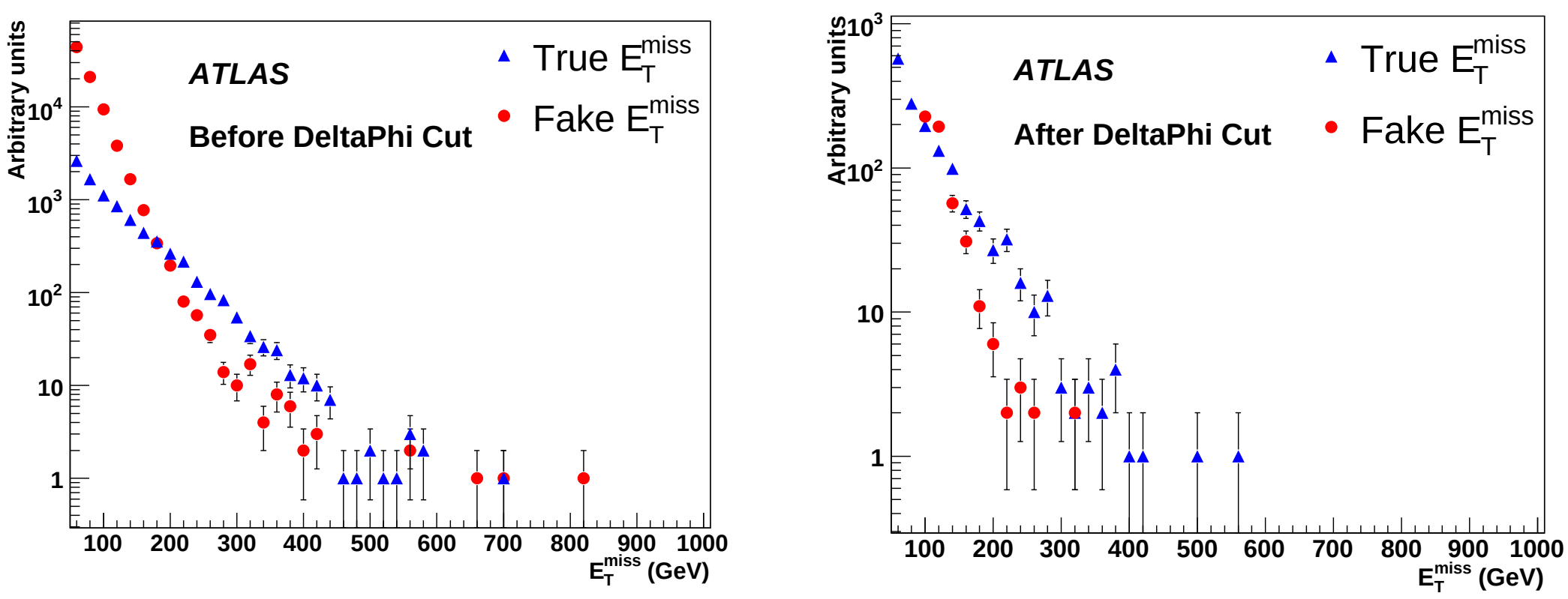

Figure 10: The rates of $\not\!\!E_{\mathrm{T}}^{\mathrm{Fake}}$ and $\not\!\!E_{\mathrm{T}}^{\mathrm{True}}$ in the QCD sample with $560 < p_{\mathrm{T}} < 1120$ GeV: (left) overall rates, (right) requiring a $\Delta\phi$ separation between $\not\!\!E_{\mathrm{T}}$ and the leading high-$p_{\mathrm{T}}$ jet in the event. The $\not\!\!E_{\mathrm{T}}^{\mathrm{Fake}}$ rates can be strongly reduced. It should be noted though that such cuts are very analysis dependent.

Table 1: Number of events with $\not\!\!E_{\mathrm{T}}^{\mathrm{Fake}}$ above various thresholds from muon (top) and calorimeter (bottom) mis-measurements. The J6 (QCD jets in $560 < p_{\mathrm{T}} < 1120$ GeV range), SU3, $t\bar{t}$ , and $Z \to \mu\mu$ samples are normalized to the same number of events (25k).

| | $\not\!\!E_{\mathrm{T}}^{\mathrm{Fake}} > 60$ GeV | $> 90$ GeV | $> 120$ GeV | $> 150$ GeV |
|---|---|---|---|---|
| $\not\!\!E_{\mathrm{T}}^{\mathrm{Fake}}$ from Muon | | | | |
| J6 | 61 | 33 | 20 | 13 |
| SU3 | 195 | 109 | 57 | 42 |
| $t\bar{t}$ | 147 | 64 | 33 | 20 |
| $Z \to \mu\mu$ | 436 | 94 | 37 | 20 |
| $\not\!\!E_{\mathrm{T}}^{\mathrm{Fake}}$ from Calorimeter | | | | |
| J6 | 4273 | 1249 | 351 | 110 |
| SU3 | 1005 | 176 | 56 | 53 |
| $t\bar{t}$ | 104 | 15 | 4 | 2 |

Figure 11 (left) shows the scatter plot of the two quantities defined in Equations (11) and (12) for the QCD sample with $560 < p_{\mathrm{T}} < 1120$ GeV. In order to separate the $\not\!\!E_{\mathrm{T}}^{\mathrm{Fake}}$ from muons with respect to the calorimeter related $\not\!\!E_{\mathrm{T}}^{\mathrm{Fake}}$ the following cuts can be used: (a) $\not\!\!E_{\mathrm{T}}^{\mathrm{FakeMuon}} > \not\!\!E_{\mathrm{T}}^{\mathrm{Fake}}/2$ which selects events with fake $\not\!\!E_{\mathrm{T}}$ coming predominantly from muon mis-measurements, and (b) $\not\!\!E_{\mathrm{T}}^{\mathrm{FakeMuon}} < \not\!\!E_{\mathrm{T}}^{\mathrm{Fake}}/2$ which selects events with fake $\not\!\!E_{\mathrm{T}}$ coming predominantly from other sources like jet mis-measurements in the calorimeter. Figure 11 also shows how these cuts separate events with $\not\!\!E_{\mathrm{T}}^{\mathrm{Fake}}$ from muons and from calorimeter induced effects.

Table 1 shows the number of events above various $\not\!\!E_{\mathrm{T}}^{\mathrm{Fake}}$ thresholds for physics samples (QCD jets in the range $560 < p_{\mathrm{T}} < 1120$ GeV, SU3, $t\bar{t}$ and $Z \to \mu\mu$ ) that are predominantly from muons (top rows) or from the calorimeter (bottom rows). It can be seen that the relative rate of $\not\!\!E_{\mathrm{T}}^{\mathrm{Fake}}$ across samples depends on the muon and calorimeter activities.

In Table 2 different categories of muon mis-measurements and their relative contribution to $\not\!\!E_{\mathrm{T}}^{\mathrm{Fake}}$ are shown[1]. The first two rows are from missed muons and the last two rows are from fake muons reconstructed from random hits in the muon chambers and with a possible match to soft muon tracks in the inner detector. Table 2 shows a smaller contribution to $\not\!\!E_{\mathrm{T}}^{\mathrm{Fake}}$ from fake muons compared to missed muons. Therefore the dominant contribution to $\not\!\!E_{\mathrm{T}}^{\mathrm{Fake}}$ from muons is due to inefficiencies in the muon

[1] Due to technical reasons a small fraction of muon events were not classified in any category.

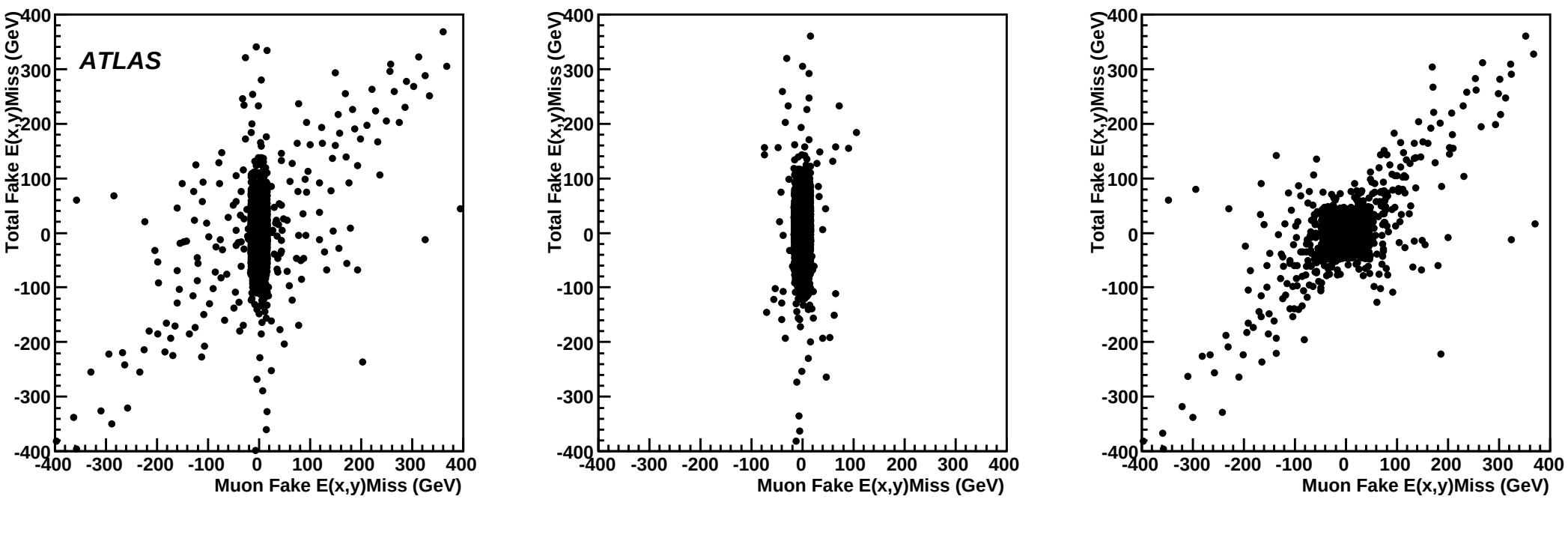

Figure 11: (left) Total $\not\!\!E_{\mathrm{T}}^{\mathrm{Fake}}$ as a function of $\not\!\!E_{\mathrm{T}}^{\mathrm{Fake}}$ from muon mis-measurements. (middle and right) Cuts defined in (b) and (a) (see text) are used to separate calorimeter/jet and muon $\not\!\!E_{\mathrm{T}}^{\mathrm{Fake}}$ components.

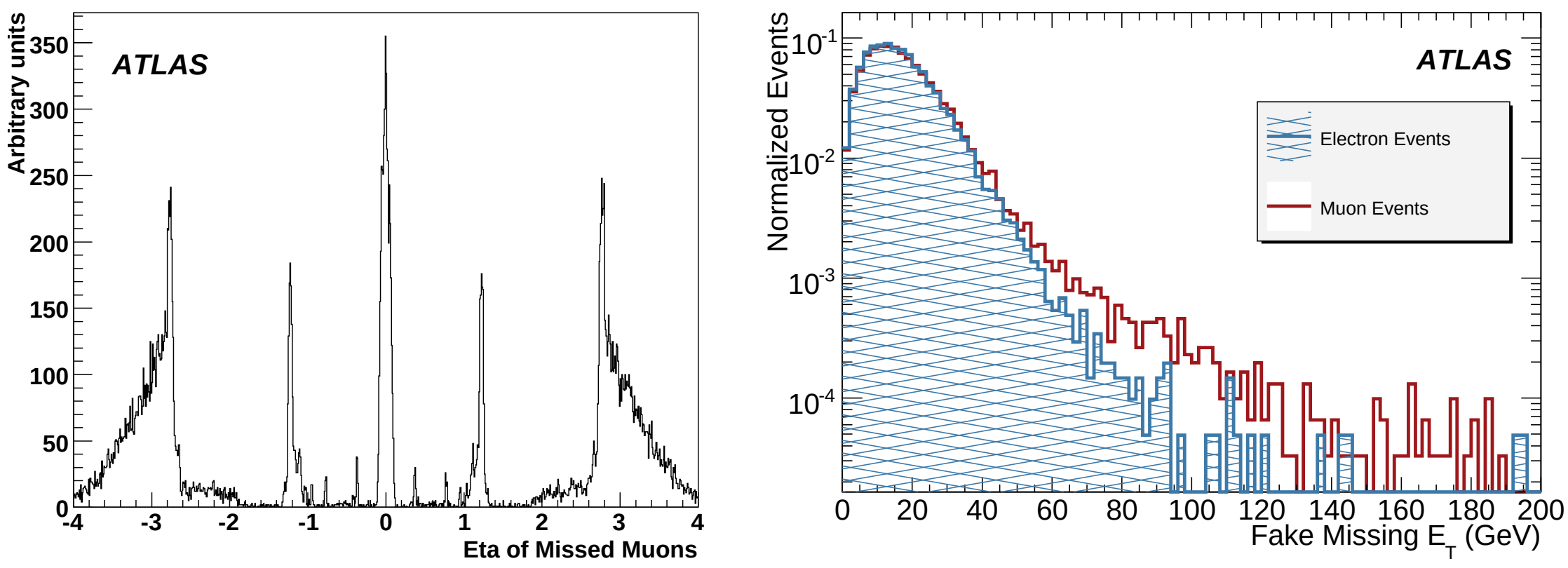

Figure 12: (left) The $\eta$ distribution of true muons that were missed during reconstruction in a $Z \rightarrow \mu\mu$ high $p_T > 100$ GeV sample. (right) $\not{E}_T^{Fake}$ in $t\bar{t}$ events in the electron (hatched) and muon channel.

identification.

Figure 12 (left) shows the $\eta$ distribution of the true muons that were missed at the reconstruction level in the $Z \rightarrow \mu\mu$ sample with $p_T > 100$ GeV. There are missed muons around $\eta = 0$, $|\eta|$=1.2 and at high $\eta$ ($|\eta|$>2.7) where there is no muon coverage. Muon tracks cannot be reconstructed by the muon system around $\eta = 0$ ($-0.05 < \eta < 0.05$), because of service holes required for cables and cryogenics passage to the inner detectors and calorimeters. In the region around $|\eta|$=1 there is a loss of efficiency in muon reconstruction, due to the middle muon station missing for initial data taking. In these studies muons missed due to limitations of muon detector coverage or poor muon reconstruction have not been recovered. In the next software releases algorithms to recover some of these missed muons using energy deposits in the calorimeter and tracks in the inner detector will be used.

Figure 12 (right) shows the $\not{E}_T^{Fake}$ distributions for events in which the leptonically decaying $W$ results in an electron and those resulting in a muon, respectively. The latter distribution clearly contains larger non-Gaussian tails. As discussed above, the sources of these large tails are either missed or fake muons.

## 4.2 Fake $\not{E}_T$ from the calorimeter

In this section it is assumed that all calorimeter readout channels are functional. $\not{E}_T^{Fake}$ in the calorimeter is then produced by mis-measurements of hadronic jets, taus, electrons or photons.

The calorimeter has cracks and gaps in the transition regions, which are also used for service outlets. These regions have poorer resolution and are expected to have larger contributions to $\not{E}_T^{Fake}$ compared to the rest of the calorimeter. There are two gap regions defined in the following $\eta$ ranges: ($1.3 < |\eta| < 1.6$) and ($3.1 < |\eta| < 3.3$). Figure 13 shows the $\eta$ distribution of the worst and the second worst measured jet (defined w.r.t. the closest true jet and their energy difference) in the calorimeter for the QCD sample generated with $560 < p_{\mathrm{T}} < 1120$ GeV. It shows that a large number of the worst measured jets have $\eta$ pointing to $|\eta|$ in 1.3-1.6. The $\eta$ distribution of the second worst measured jet is more flat and peaks around $|\eta|$ in 0.6-0.9, the transition region of the barrel tile calorimeter to the extended barrel tile calorimeter.

Table 2: The sources of mis-measured muons that contribute to $\not{E}_T^{Fake} > 60$ GeV. The columns are normalized to the same number of events (25k).

| Sources | SU3 | J6 | $t\bar{t}$ | $Z \rightarrow \mu\mu$ |
|---|---|---|---|---|
| MC muon not reconstructed | 121 | 17 | 42 | 332 |
| Reconstructed muon failed quality cut | 30 | 18 | 40 | 18 |
| Muon reconstructed with badly measured $p_T$ | 29 | 3 | 28 | 28 |
| Reco muon not close to MC muon (“fake muon”) | 11 | 23 | 0 | 25 |

The above correlation of worst measured jets and their $\eta$ suggests a large correlation between the jet $\eta$ and $\not\!\!E_{\mathrm{T}}^{\mathrm{Fake}}$. However, the $\not\!\!E_{\mathrm{T}}^{\mathrm{Fake}}$ distribution from full simulation samples suggests otherwise. Figure 14 shows the $\not\!\!E_{\mathrm{T}}^{\mathrm{Fake}}$ distribution in the QCD samples generated with $560 < p_{\mathrm{T}} < 1120$ GeV and $140 < p_{\mathrm{T}} < 280$ GeV, when a jet points to the crack/gap region or not. The slope of the distributions suggests no significant correlation between jets pointing to cracks and $\not\!\!E_{\mathrm{T}}^{\mathrm{Fake}}$.

This apparent contradiction between Fig. 13 and Fig. 14 can be understood as follows: even though the worst measured jet contributes strongly to the $\not\!\!E_{\mathrm{T}}^{\mathrm{Fake}}$, it is not the only source of $\not\!\!E_{\mathrm{T}}^{\mathrm{Fake}}$ in the event. The worst measured jet contributes on average about 60% and the second worst measured jet contributes about 20% to $\not\!\!E_{\mathrm{T}}^{\mathrm{Fake}}$. But not all worst measured jets are along the crack region and each event has many jets. In lower $\not\!\!E_{\mathrm{T}}^{\mathrm{Fake}}$ regions there is a stronger correlation between $\not\!\!E_{\mathrm{T}}^{\mathrm{Fake}}$ and jets pointing to cracks. For higher $\not\!\!E_{\mathrm{T}}^{\mathrm{Fake}}$ ( $> 50$ GeV considered here ) there is more than one source contributing to $\not\!\!E_{\mathrm{T}}^{\mathrm{Fake}}$ and the correlation of jets pointing to cracks is smeared out as can be seen in Fig. 14.

### 4.3 Fake $\not\!\!E_{\mathrm{T}}$ from calorimeter leakage

Jet leakage from the calorimeters or fluctuations in large jet energy deposits in non-instrumented regions such as the cryostat between the liquid argon and tile calorimeters can also be a source of $\not\!\!E_{\mathrm{T}}^{\mathrm{Fake}}$. The method used to detect events with potential jet leakage is to look for large energy deposits in the following regions: the outermost layers of the TileCal and the HEC, the outermost LAr barrel layer and the innermost TileCal barrel layer, and in the TileCal gap and crack scintillators.

The following shows an example of selection cuts applied on different variables of the three leading $p_{\mathrm{T}}$ jets with $p_{\mathrm{T}} > 100$ GeV: $\mathrm{E_{Tile2}/E_{Total}} > 0.05$, $\mathrm{E_{Tile10}/E_{Total}} > 0.7$, $\mathrm{E_{Cryo}/E_{Total}} > 0.2$, $\mathrm{E_{Gap}/E_{Total}} > 0.2$ and $\mathrm{E_{HEC3}/E_{Total}} > 0.5$, where $\mathrm{E_{Total}}$ is the total jet energy, $\mathrm{E_{Tile2}}$ is the jet energy in the outermost tile layer, $\mathrm{E_{Tile10}}$ is the jet energy in the first two innermost tile layer, $\mathrm{E_{Cryo}}$ is the energy lost by the jet in the cryostat, $\mathrm{E_{Gap}}$ is the jet energy in the gap scintillators and $\mathrm{E_{HEC3}}$ is the jet energy in the outermost layer of the HEC calorimeters. If any of these cuts is satisfied the event is rejected.

Furthermore, as the tracks found in the inner detector are not affected by the reconstruction, com-plementary information on events with fake $\not\!\!E_{\mathrm{T}}$ can be obtained using $\not\!\!E_{\mathrm{T}_{\mathrm{trk}}}$ from tracks which is the $\not\!\!E_{\mathrm{T}}$ computed only from tracks. A cut on $(\not\!\!E_{\mathrm{T}_{trk}} - \not\!\!E_{\mathrm{T}} > 50\ \mathrm{GeV})$ was chosen for the optimization of the signal significance.

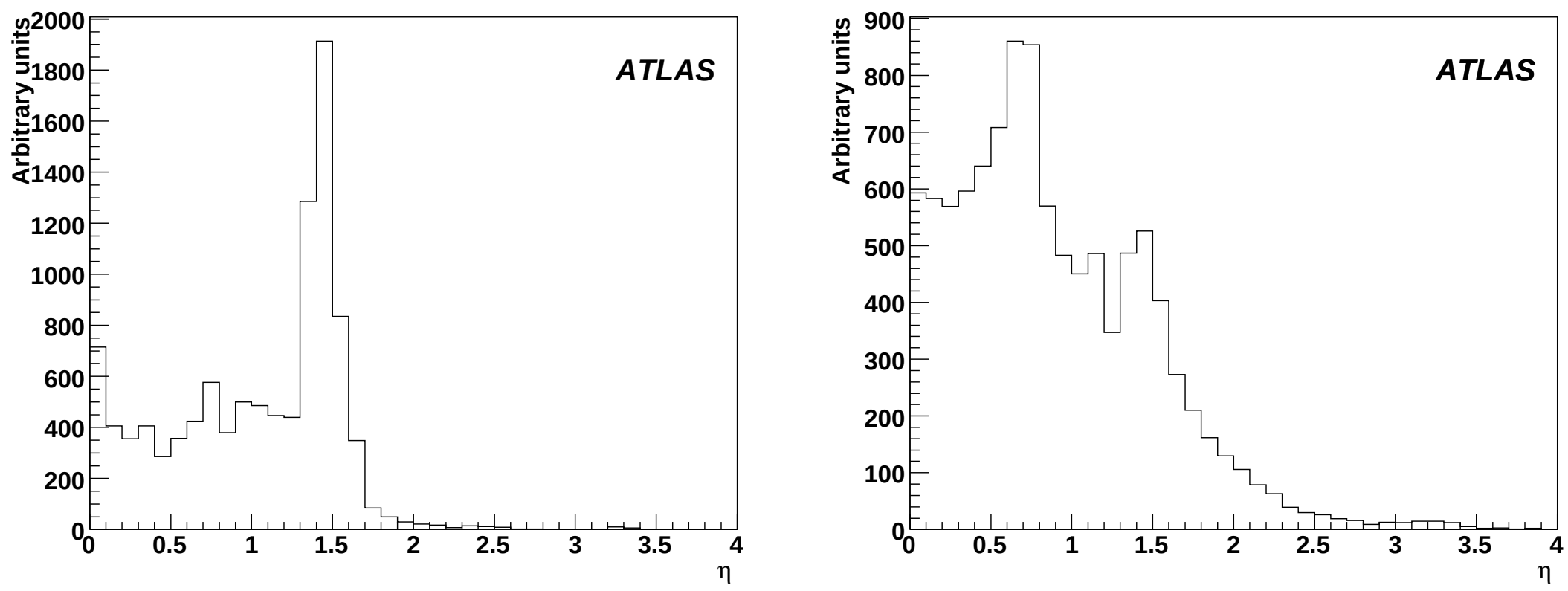

Figure 13: The $\eta$ distribution of the worst (left) and second-worst measured jet (right) in the calorimeter in QCD events generated with $560 < p_{\mathrm{T}} < 1120$ GeV.

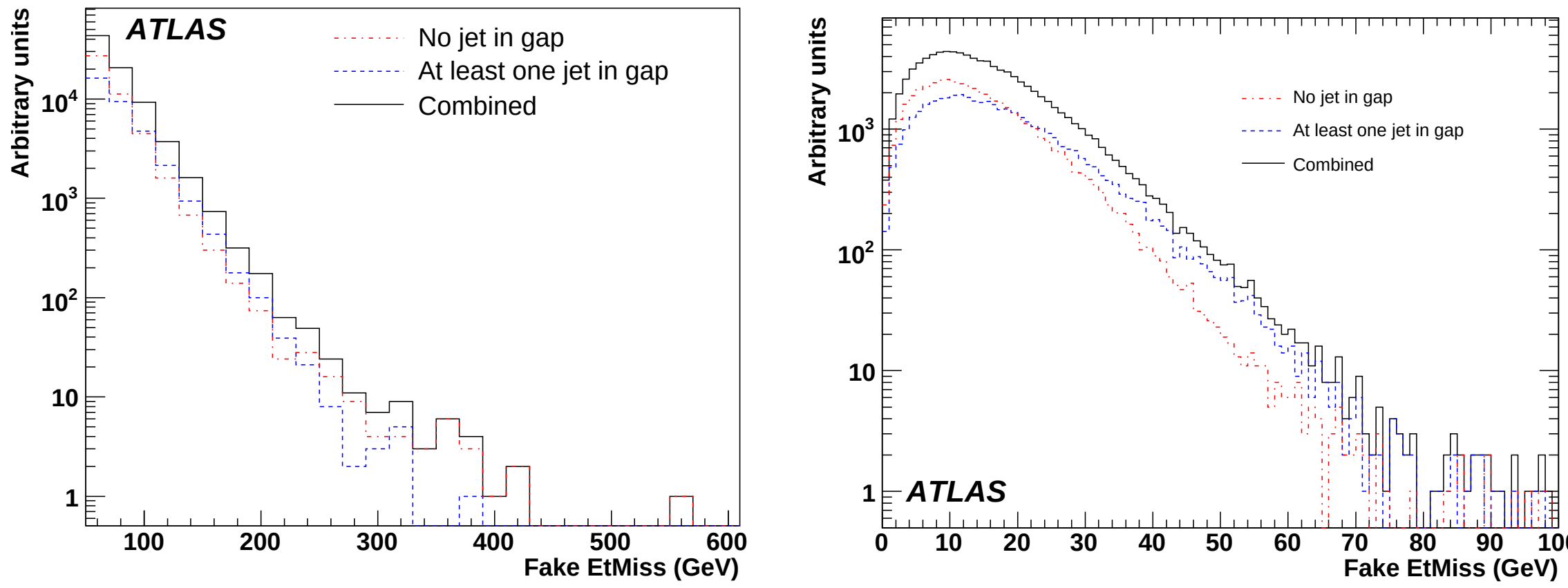

Figure 14: The $\not\!\!E_{\mathrm{T}}^{\mathrm{Fake}}$ rate for QCD sample in $560 < p_{\mathrm{T}} < 1120$ GeV range (left) and QCD sample in $140 < p_{\mathrm{T}} < 280$ GeV (right) due to calorimeter mis-measurements.

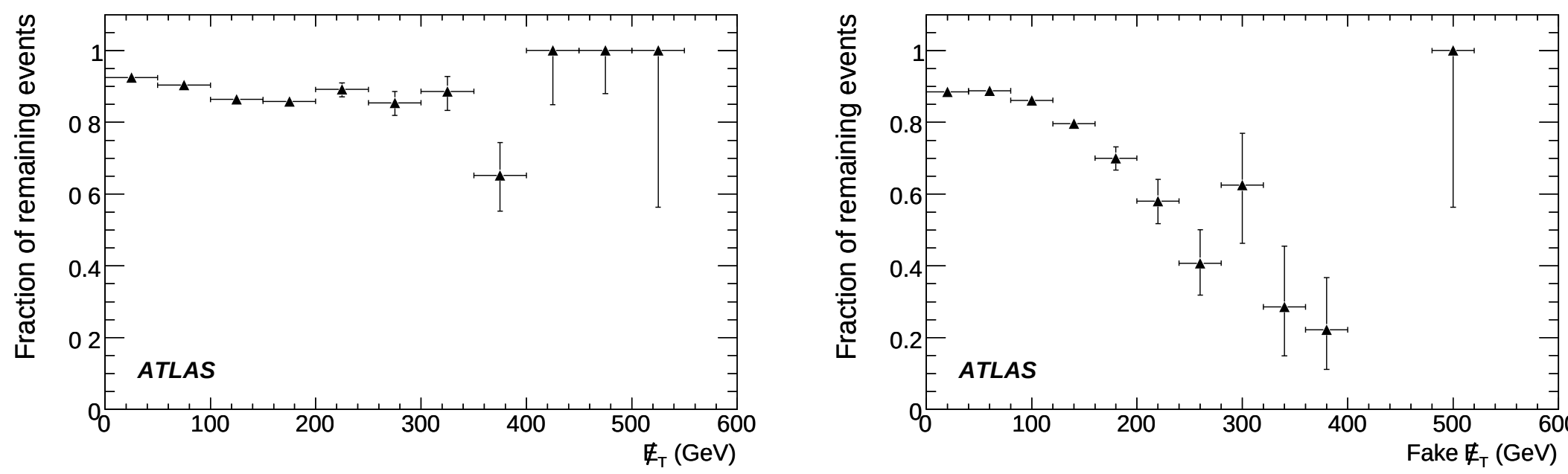

Figure 15: Fraction of events remaining after the cuts discussed in the text as a function of $\not\!\!E_{\mathrm{T}}$(left) and $\not\!\!E_{\mathrm{T}}^{\mathrm{Fake}}$(right) for the QCD sample generated with $560 < p_{\mathrm{T}} < 1120$ GeV.

Figure 15 shows the percentage of events remaining after the cuts described above are applied on QCD sample generated with $560 < p_{\mathrm{T}} < 1120$ GeV. The right plot shows suppression of large fake $\not\!\!E_{\mathrm{T}}$ generated due to high $p_{\mathrm{T}}$ jet leakage. Larger fake $\not\!\!E_{\mathrm{T}}$ are suppressed more strongly. It can also be noticed from the left plot that these cuts, although removing a large fraction of the events dominated by fake $\not\!\!E_{\mathrm{T}}$, are not sensitive to the overall $\not\!\!E_{\mathrm{T}}$ in the event and the fraction of remaining events is fairly constant over $\not\!\!E_{\mathrm{T}}$. The method is of course analysis dependent and the values of the cuts should be chosen taking into account the signal efficiency.

## 4.4 Fake $\not\!\!E_{\mathrm{T}}$ from instrumental effects

In real data there will be sources of $\not\!\!E_{\mathrm{T}}^{\mathrm{Fake}}$ which are not fully modeled in Monte Carlo simulations including, for example, mis-modeling of material distributions and instrumental failures. As the details of these $\not\!\!E_{\mathrm{T}}^{\mathrm{Fake}}$ sources will be understood with time, increasingly refined analyses will be developed to minimize their associated backgrounds while maintaining high selection efficiencies for signals with genuine missing energy. While it is difficult to predict in advance the exact sources of mis-modeled $\not\!\!E_{\mathrm{T}}^{\mathrm{Fake}}$, it is nevertheless possible to insert problems into the Monte Carlo that match potential hardware failures. These include trips in high-voltage channels or readout power supplies of the calorimeter or noise in calorimeter channels or regions. The initial studies presented here are based on samples with simulated dead regions of the calorimeter, so called 'cell-killed' samples: one dead front-end readout crate in the LAr electromagnetic barrel calorimeter and one dead front-end readout crate affecting LAr electromagnetic and endcap calorimeters. Based on the location of these hardware failures the calorimeter is divided into three regions of $\phi(\not\!\!E_{\mathrm{T}})$: 'Region 1' with EM endcap and hadronic endcap problems, 'Region 2' with EM barrel problems, and 'Region 3' with no problems.

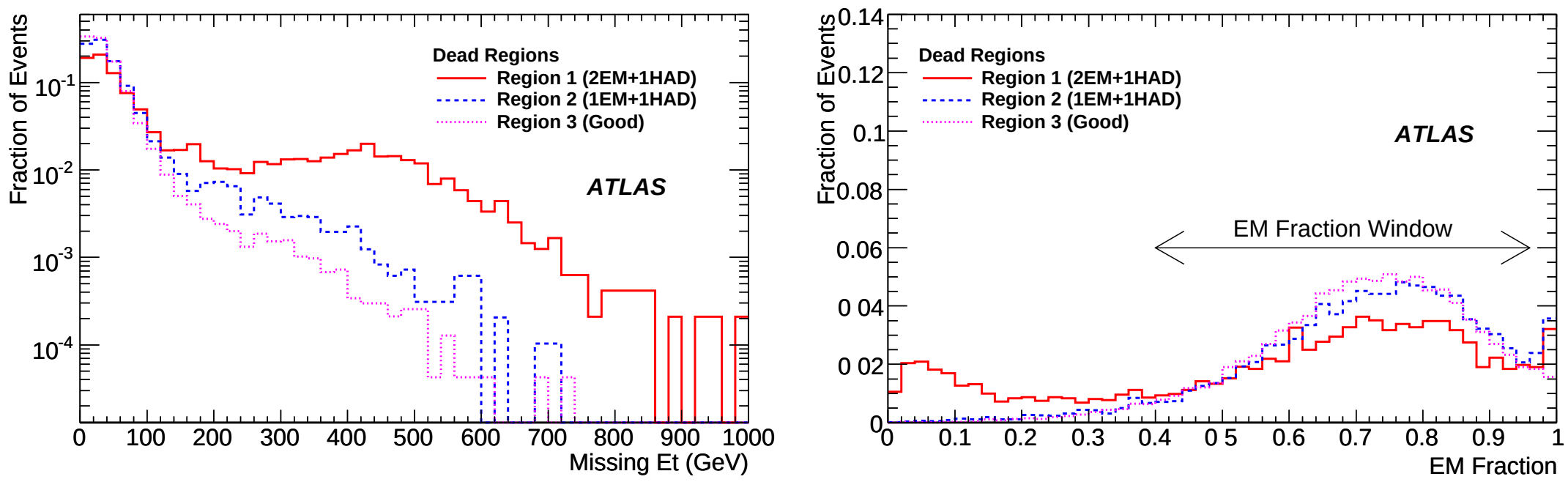

Figure 16: Cell-killed QCD sample in $560 < p_{\mathrm{T}} < 1120$ GeV range: the $\not\!\!E_{\mathrm{T}}$ distribution (left) and the EM fraction (right). The histograms are normalized by area.

These samples were used as references for the development of cuts that reduce the $\not\!\!E_{\mathrm{T}}^{\mathrm{Fake}}$ background while maintaining high efficiencies for potential signal events. Since we will not know the precise location and nature of hardware problems in advance, the cuts are not tuned assuming that knowledge. In real data it will be possible to significantly improve the analysis performance by cutting harder when energy deposits are expected near regions with detector hardware problems, but that is not exploited in the studies so far. Future work will also include using these samples to develop data-driven techniques for predicting the $\not\!\!E_{\mathrm{T}}^{\mathrm{Fake}}$ tails.

The high-$p_{\mathrm{T}}$ ($560 < p_{\mathrm{T}} < 1120$ GeV) $\gamma$+jet MC sample was processed with the cell-killed configuration described above. Events with at least one back-to-back photon-jet pair were selected. Figure 16 shows the $\not\!\!E_{\mathrm{T}}^{\mathrm{Fake}}$ generated in the three regions. Large $\not\!\!E_{\mathrm{T}}$ tails are seen in Regions 1 and 2 where the holes in calorimeter coverage have been introduced. The effect of dead regions can also be seen in Table 3. The columns show the number of events above various $\not\!\!E_{\mathrm{T}}^{\mathrm{Fake}}$ thresholds. The first two rows show the number of events with no dead-regions and with the above dead-regions simulated. A very large increase in high $\not\!\!E_{\mathrm{T}}^{\mathrm{Fake}}$ is seen.

Several methods have been developed to suppress events with large fake $\not\!\!E_{\mathrm{T}}$:

- **EM Fraction Method:** The EM fraction method starts by finding the closest calorimeter jet to the $\not\!\!E_{\mathrm{T}}$ direction vector using $\Delta\phi$ between the calorimeter jet and $\not\!\!E_{\mathrm{T}}$. Figure 16 (right) shows the EM fraction distributions. Small EM fractions are due to a dead LAr EM calorimeter crates, whereas large EM fractions are due to a dead hadron calorimeter crates. The fake $\not\!\!E_{\mathrm{T}}$ generated this way can be suppressed by requiring the EM fraction to be in a window from 0.40 to 0.96. The effect of this cut can be seen in the third row of Table 3 when compared to the increase in $\not\!\!E_{\mathrm{T}}^{\mathrm{Fake}}$ due to dead-regions in row 2. A rejection of 1.5 to 4 is seen from low to high $\not\!\!E_{\mathrm{T}}^{\mathrm{Fake}}$ bins. The selection efficiency can be defined as the ratio of the number of events after and before the cuts when events fall in Region 3 (Region 3 has small $\not\!\!E_{\mathrm{T}}^{\mathrm{Fake}}$). For the EM fraction method the selection efficiency is $\sim 90\%$.

- **Track-Jet Methods:** jets were reconstructed from inner detector tracks using the cone algorithm

Table 3: The number of events from $\gamma$+jet samples above various $\not{E}_T^{\text{Fake}}$ thresholds with no dead-regions, with simulated dead-regions and after applying various suppression techniques (see text). All numbers are normalized to 25k events sample size.

| $\not{E}_T^{\text{Fake}}$ (GeV) | $> 100$ | $> 200$ | $> 300$ | $> 400$ | $> 500$ |
|---|---|---|---|---|---|
| No dead regions | 555 | 18 | 3 | 1 | 0 |
| w/ dead regions | 2482 | 1308 | 864 | 501 | 199 |
| w/ EM fraction method | 1651 | 572 | 287 | 122 | 50 |
| w/ track-jet cluster method | 1786 | 664 | 313 | 129 | 49 |
| w/ comb. track-jet cluster method | 1402 | 392 | 150 | 46 | 14 |

with $\Delta R = 0.4$. A fiducial volume cut of $|\eta| < 2.5$ is applied due to the tracking coverage. Since track-jets use the inner tracking detectors, they provide a complementary identification of events with fake $\not{E}_T$. A number of different methods of exploiting the track-jets were studied. An effective method which only uses information present in the Analysis Object Data (AOD) is to sum the $E_T$ of calorimeter topological clusters within the track-jet $\eta$-$\phi$ cone. The distributions of the $E_T$ ratio of the track-jet to the clusters has tails due to the dead calorimeter regions. A cut is applied requiring $E_T$ ratios larger than 1.0; the value was chosen to maintain a significant efficiency for signal samples, and could be substantially tightened for different physics analyses. The fourth row of Table 3 shows the effectiveness of this cut. Rejections from $\eta = 1.4$ to $\eta = 4$ are achieved from the low to high $\not{E}_T^{\text{Fake}}$ regions, with efficiencies of $\sim 93\%$.

Since the EM fraction method and track-jet methods are largely uncorrelated, they can be combined for better suppression of large $\not{E}_T^{\text{Fake}}$. The last row in Table 3 shows the performance of the combined track-jet method. The selection efficiency for this combined method is $\sim 84\%$. It must be emphasized that the results presented in this table represent the worst case since the cuts do not use the detailed information about detector problems that will be known from the detector control and data quality systems. In a final analysis, much harder cuts can be applied in regions with known detector problems. Of course, this will reduce the acceptance of the detector.

# 5 The $\not{E}_T$ Trigger algorithm performance

This section briefly describes the $\not{E}_T$ algorithms applied at the first level trigger (L1) and the higher-level trigger (HLT). The HLT is a combination of second level trigger (L2), and a third level trigger (or event filter, EF). The L1 algorithm is based on hardware, while the HLT algorithms are software based.

## 5.1 The $\not{E}_T$ at L1

The $\not{E}_T$ L1 calorimeter triggers cover the region $|\eta| < 4.9$, which is the limit of the forward calorimeters [11]. The basic units of L1 $\not{E}_T$ and $\Sigma\not{E}_T$ trigger algorithms are 'jet elements', formed by summing over trigger towers within windows of $0.2 \times 0.2$ in the $(\eta, \phi)$ plane.[2] They are processed by the Jet/Energy modules, which compute $E_x$, $E_y$ and $\Sigma\not{E}_T$ of each jet element. Four thresholds are available on $\Sigma\not{E}_T$ with values up to 2044 counts in steps of 4 (usually, 1 count = 1 GeV). Eight $\not{E}_T$ thresholds are available, up to a maximum threshold value of 504 counts. If an overflow occurs at any point in the $\not{E}_T$ or $\Sigma\not{E}_T$ algorithm, all of the corresponding thresholds are set as passed.

[2] For $2.4 < |\eta| < 3.2$ the granularity is either $\Delta\eta = 0.2$ or $\Delta\eta = 0.3$, while FCAL jet elements extend from $|\eta| = 3.2$ to $|\eta| = 4.9$. The $\phi$ granularity of the FCAL jet elements is $\Delta\phi = 0.4$.

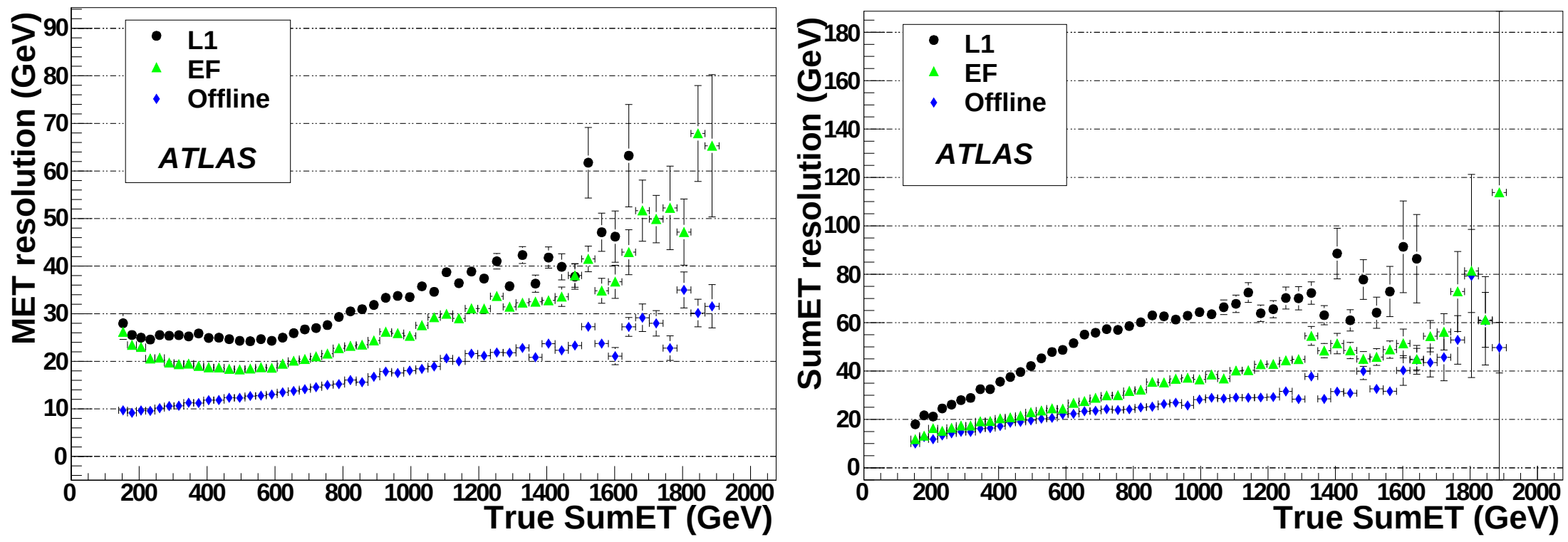

Figure 17: L1, EF and offline results for the $\not\!\!E_{\mathrm{T}}$ (left) and the $\Sigma\not\!\!E_{\mathrm{T}}$ (right) resolution as a function of the true $\Sigma\not\!\!E_{\mathrm{T}}$ in $t\bar{t}$ events.

For each event, the L1 $E_x$, $E_y$ and $\Sigma\not\!\!E_{\mathrm{T}}$ are saved into one object of the RecEnergyRoI class, including overflow flags. If any threshold is passed, the JetEnergy RoI is also produced containing the bit pattern of the thresholds passed.

## 5.2 $\not\!\!E_{\mathrm{T}}$ at HLT

The offline algorithm described in Section 2.2 is too resource intensive (both in terms of memory access and calculations to be done) to be applied in the HLT. The following algorithms are ready for the first data taking[3]:

- The L2 algorithm uses the calorimeter information in RecEnergyRoI provided by the L1 trigger and applies a correction for L2 muon objects.
- The default EF algorithm sums all calorimeter cells and applies a 0-th order hadronic calibration by multiplying $\not\!\!E_{\mathrm{T}}$ and $\Sigma\not\!\!E_{\mathrm{T}}$ by a constant related to the hadronic/electromagnetic calorimeter energy fraction in a jet. Finally, the EF takes the muon contribution into account.

Both at L2 and EF, the muon correction may be switched off independently for each trigger chain.

The decision taken both at L2 and EF is carried out by the same software package. This hypothesis testing code can be configured to accept events based on $\not\!\!E_{\mathrm{T}}$, on $\Sigma\not\!\!E_{\mathrm{T}}$ or on both.

## 5.3 The $\not\!\!E_{\mathrm{T}}$ and $\Sigma\not\!\!E_{\mathrm{T}}$ resolution at trigger level

Figure 17 shows the $\not\!\!E_{\mathrm{T}}$ and $\Sigma\not\!\!E_{\mathrm{T}}$ resolution for the L1, EF and offline algorithms as a function of the true $\Sigma\not\!\!E_{\mathrm{T}}$ for $t\bar{t}$ events. Since the L2 algorithm takes the L1 result and applies relatively small corrections, the present L2 resolution is practically the same as for L1 and therefore not shown. At true $\Sigma\not\!\!E_{\mathrm{T}}$ values of 500 GeV, the L1 resolution on the $\not\!\!E_{\mathrm{T}}$ measurement is about 25 GeV which is a factor of two larger than what is achieved offline. The EF resolution, for both $\not\!\!E_{\mathrm{T}}$ and $\Sigma\not\!\!E_{\mathrm{T}}$ lies in between the values for L1 and offline resolution.

[3] At present, work is in progress to detect fake sources due to detector effects like uninstrumented regions or hardware failure, or physics environment, e.g. beam-halo or cosmic ray tracks.

# 6 $\not{E}_T$ in early data

Validation of the $\not{E}_T$ reconstruction described in the previous sections will be performed with the first LHC data accumulated by ATLAS. For the very first data, the two main issues are controlling instrumental failures and calibration of energy deposits in the calorimeter. At this stage, the overwhelming number of minimum bias events will be used to monitor and diagnose $\not{E}_T$ reconstruction problems.

The development of algorithms for data quality checks is being actively pursued to minimize the impact of such failures and will be optimized when the first data is collected. The 'standard' $\not{E}_T$ calculation, using the calorimeter cells at the electromagnetic scale above a threshold will be always provided as a reference. The $\not{E}_T$ calibration strategy follows the steps described in Section 3. First, a simple global calibration for all the TopoCells will be used. As shown in Section 3 this already gives a good linearity behavior. As the event reconstruction becomes more robust, the event objects (e.g. electrons, jets, taus) will be used to obtain the best $\not{E}_T$ resolution.

Figure 3 already shows how the $\not{E}_T$ linearity improves from a simple to a more refined $\not{E}_T$ calculation. Figure 18 shows how the $\not{E}_T$ resolution improves arriving in steps to the final refined calibration.

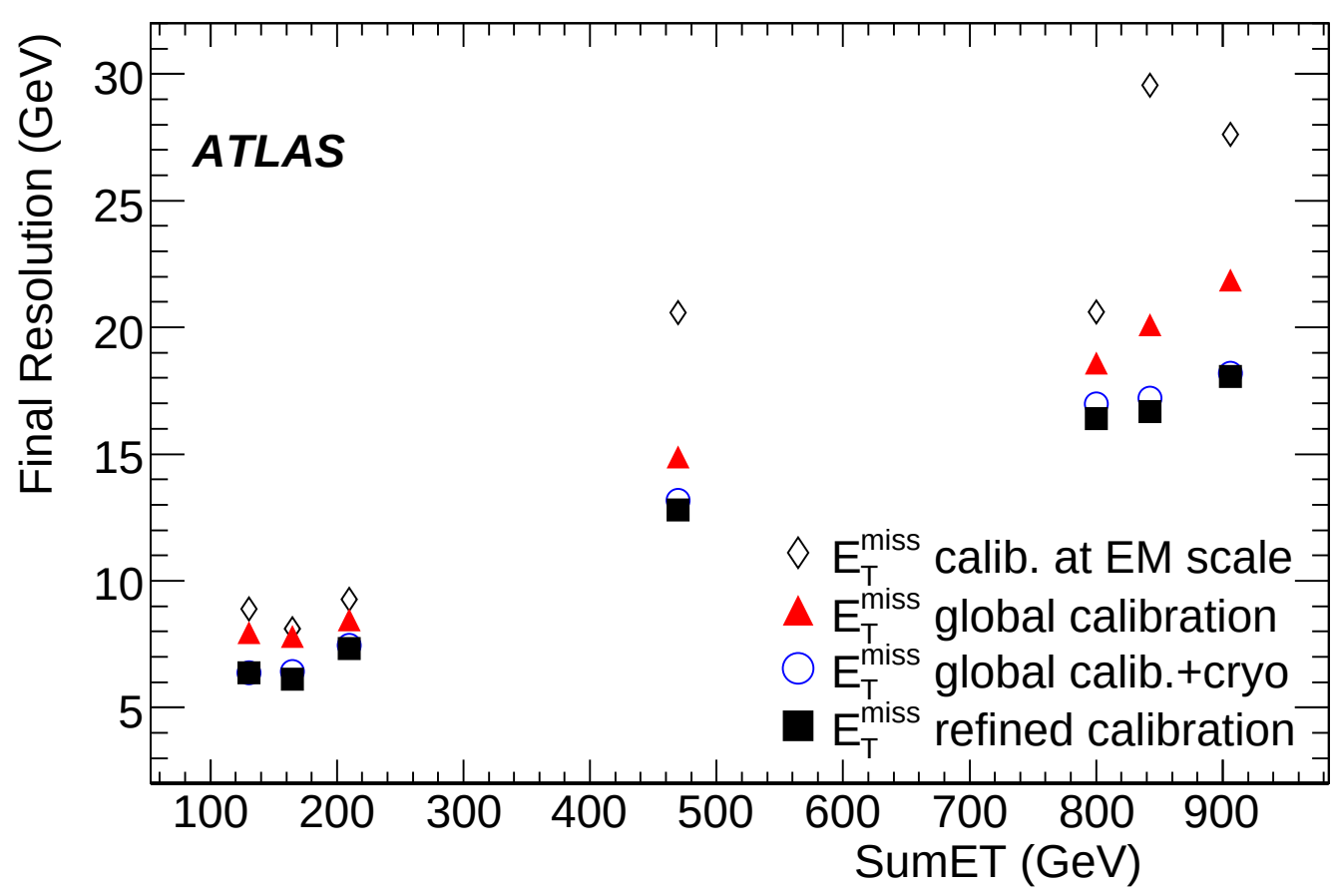

Figure 18: The $\not{E}_T$ resolution as a function of true $\Sigma\not{E}_T$ from the uncalibrated to the refined calibration as indicated for the following channels: $W \to \mu\nu$ corresponding to true $\Sigma\not{E}_T = 130$ GeV, $W \to e\nu$ to 165 GeV, $Z \to \tau\tau$ to 210 GeV, $t\bar{t}$ to 470 GeV, $A/H \to \tau\tau$ to 843 GeV, J5 to 800 GeV and SUSY to 906 GeV.

Once data of the order of 100 pb$^{-1}$ are collected, the $\not{E}_T$ validation focuses on physics channels with relatively large $\not{E}_T$ and/or $\Sigma\not{E}_T$. This section briefly describes a few studies that will be performed with the first data.

The $Z \to \tau\tau$ process, using the $Z$ mass constraint, can be used to determine the $\not{E}_T$ scale in-situ to about 8% accuracy. The $Z \to \ell\ell$ process with decays to electrons and muons does not have any significant $\not{E}_T^{True}$ from neutrinos. The small background to these events will help to test possible $\not{E}_T$ biases, expected to be zero, and the resolution in a straightforward manner. The copiously produced $W \to e\nu$ and $W \to \mu\nu$ events can be used to test the reconstructed $\not{E}_T$ in the $(20-150)$ GeV range. Two methods to use these events are discussed. Semileptonic $t\bar{t}$ events also have genuine $\not{E}_T$ and allow a test of the $\not{E}_T$ reconstruction in an environment relevant for many physics analyses and searches, notably SUSY.

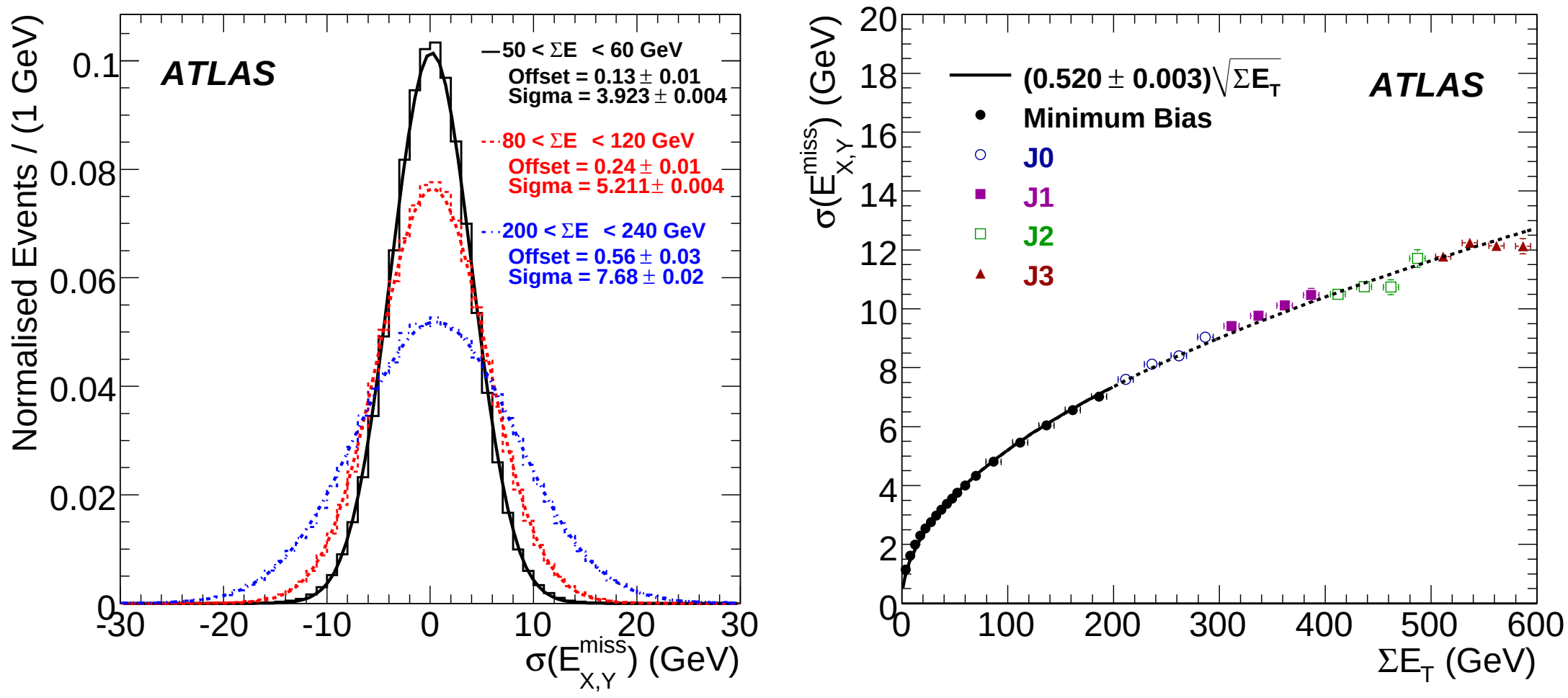

Figure 19: (left) The $\not{E}_{x,y}$ resolution for different $\Sigma\not{E}_{\mathrm{T}}$ regions in minimum bias events. (right) The $\not{E}_{\mathrm{T}}$ resolution in QCD dijet events (J0-J3: see Section 3.1 for definition) is shown together with the $\not{E}_{\mathrm{T}}$ resolution from minimum bias events (black filled circles) as a function of $\Sigma\not{E}_{\mathrm{T}}$. An integrated luminosity of the order of $10^{-5}$ pb$^{-1}$ is used.

## 6.1 Minimum bias events

Minimum bias interactions at the LHC are dominated by soft collisions of the two interacting protons. These events are useful for $\not{E}_{\mathrm{T}}$ commissioning, especially in the early stages of the experiment, due to their large statistics and their comparatively simple event selection. The main background in minimum bias events will originate from empty, beam gas and beam halo events, especially at the beginning of the experiment. Minimum bias events will be used to verify the $\not{E}_{\mathrm{T}}$ reconstruction procedure and estimate the $\not{E}_{\mathrm{T}}$ resolution for low $\Sigma\not{E}_{\mathrm{T}}$ events.

In the early stages of the experiment, minimum bias events will be selected by three types of triggers: randomly selected bunch crossings (MB1), randomly selected bunch crossings together with a SemiConductor Tracker space point trigger (MB2), and minimum bias trigger scintillator (MBTS2). The details of the triggers are described in Ref. [12].

For the study of $\not{E}_{\mathrm{T}}$ in minimum bias events, high signal efficiency and background rejection are required. The relative fraction of non-diffractive, single diffractive and double diffractive events in the sample is not a concern. The selection criteria require at least 20 semiconductor tracker space points to reject empty events and at least one good reconstructed track to reject beam gas and halo events.

A Monte Carlo study predicts an overall trigger efficiency of 96.8%. The offline track selection efficiency (with respect to events passing the trigger) is 80.6% for a total selection efficiency of 78.0%.

The $\not{E}_{\mathrm{T}}$ in minimum bias events is fairly low with a mean of 4.3 GeV. Fake $\not{E}_{\mathrm{T}}$ is caused mainly by calorimeter energy resolution (82%) and acceptance (18%). The true $\not{E}_{\mathrm{T}}$ is 0.06 GeV on average, originating from $K/\pi$ decays-in-flight and from the decay of charm and bottom particles. The true $\Sigma\not{E}_{\mathrm{T}}$ in non-diffractive minimum bias events is typically 64 GeV, while the reconstructed $\Sigma\not{E}_{\mathrm{T}}$ is on average 49 GeV due to the loss of low energy particles which do not reach the calorimeters. Since minimum bias events are dominated by soft (low-$p_{\mathrm{T}}$) interactions, jets are reconstructed with a rate depending on $\Sigma\not{E}_{\mathrm{T}}$. When minimum bias events with $\Sigma\not{E}_{\mathrm{T}}$ of 50 GeV are compared to events with $\Sigma\not{E}_{\mathrm{T}}$ of 250 GeV, the average number of reconstructed jets with $p_{\mathrm{T}} > 7$ GeV increases from below one to about nine with an average jet energy of 10 and 13 GeV, respectively.

The $\not{E}_{\mathrm{T}}$ resolution in minimum bias events is expected to scale as $\sqrt{\Sigma E_T}$ because the stochastic term of the calorimeter resolution is dominant in $\Sigma \not{E}_T$ regions as shown in the left plot of Fig. 19. These distributions are well fitted by Gaussian functions with offsets of zero (in the case of no $\phi$ asymmetry) and resolutions which scale with $\Sigma \not{E}_T$.

The right plot of Fig. 19 shows the comparison of the $\not{E}_T$ resolution evaluated in this study with the higher $\Sigma \not{E}_T$ region ($\Sigma \not{E}_T > 300$ GeV). The $\not{E}_T$ resolution in QCD dijet events matches well the resolution obtained from minimum bias events.

### 6.2 Determining the $\not{E}_T$ scale using $Z \to \tau\tau$ events

At the beginning of ATLAS operation, about 70k events of type $Z \to \tau\tau$ with one leptonic and one hadronic $\tau$-decay will be produced in 100 pb$^{-1}$ of data. Such events can be selected with a lepton trigger. The $Z \to \tau\tau \to \ell h$ are produced with genuine $\not{E}_T$ of typically 20 GeV and $\Sigma \not{E}_T$ of the order of 200 GeV. In these events the peak position of the $\tau\tau$ invariant mass distribution is sensitive to $\not{E}_T$ and can be very useful in determining the $\not{E}_T$ scale [10].

The main backgrounds come from $W \to \ell\nu$+jets events, where one jet fakes a $\tau$ decay, and from QCD events (mainly $b\bar{b}$). The $t\bar{t}$ background has a much lower cross-section; the $Z \to ee$ and $Z \to \mu\mu$ backgrounds are also small and the $WW$ background is negligible. Events with the final state lepton and $\tau$-jet of the same-sign are not expected to come from $Z \to \tau\tau$ events which have opposite-sign. The backgrounds (apart from the $t\bar{t}$ background, which is anyway low) contribute in the same way to the opposite-sign and the same-sign samples. Hence, the effect of backgrounds will be minimized using same-sign events, subtracted from the opposite-sign events.

For each reconstructed event, the leading and isolated lepton (electron or muon) with $p_T^{\ell} > 15$ GeV and $|\eta^{\ell}| < 2.5$ is chosen and a set of basic cuts is applied: $\not{E}_T > 20$ GeV (rejects QCD events), the transverse mass calculated from $\not{E}_T$ and the lepton $< 50$ GeV (suppresses events from semileptonic $W$ decays), and $\Sigma \not{E}_T < 400$ GeV (suppresses QCD). In addition it is required to have no tagged $b$ jets (suppresses $t\bar{t}$ and $b\bar{b}$ events). Then at least one identified $\tau$-jet with $p_T^{\tau-\mathrm{jet}} > 15$ GeV, $|\eta^{\tau-\mathrm{jet}}| < 2.5$, and a track multiplicity of one or three is required. The $\Delta\phi$ between the isolated lepton and the $\tau$-jet is required to be in the range between $1-2.8$, which reduces badly reconstructed events and further rejects backgrounds.

With 100 pb$^{-1}$ of data, 210 signal events (opposite-sign) are expected in the invariant mass range 66 GeV $< m_{\tau\tau} <$ 116 GeV. A total background of 16 events is expected. Figure 20 (left) shows the reconstructed mass peak for $Z \to \tau\tau$ events as well as the small total backgrounds after analysis cuts for opposite-sign and same-sign events.

Figure 20 (right) shows a very good sensitivity of the measured $Z$ mass reconstructed from $\tau$-pairs to the absolute $\not{E}_T$ scale. With an integrated luminosity of 100 pb$^{-1}$, the $Z$ mass can be reconstructed with an uncertainty of $\pm 0.8$ GeV. Taking into account the statistical uncertainty only, the $\not{E}_T$ scale could be determined with a precision of $\sim 3\%$. But systematic effects, such as the subtraction of same-sign events and the stability of the fit will affect the measurement of the reconstructed mass peak. Therefore, assuming a resolution of $\pm 3\sigma$ on the reconstructed $Z$ mass, the $\not{E}_T$ scale can be determined to about $\pm 8\%$.

### 6.3 $Z \to \ell\ell$ events

This analysis uses inclusive $Z \to ee$ and $Z \to \mu\mu$ samples to investigate the scale and resolution of the $\not{E}_T$ reconstruction in the first data. In these samples the transverse momentum of the two leptons from the $Z$ boson decay are balanced by the hadronic recoil and $\Sigma \not{E}_T$ reaches values up to a few hundred GeV.

Events are selected by requiring two well reconstructed, identified and isolated leptons with $p_T > 25$ GeV. They have to have equal or opposite charge and a reconstructed mass, $m_{\ell\ell}$, in the interval $70-100$ GeV. In a sample of 250 pb$^{-1}$ of data, about 400k events are expected.

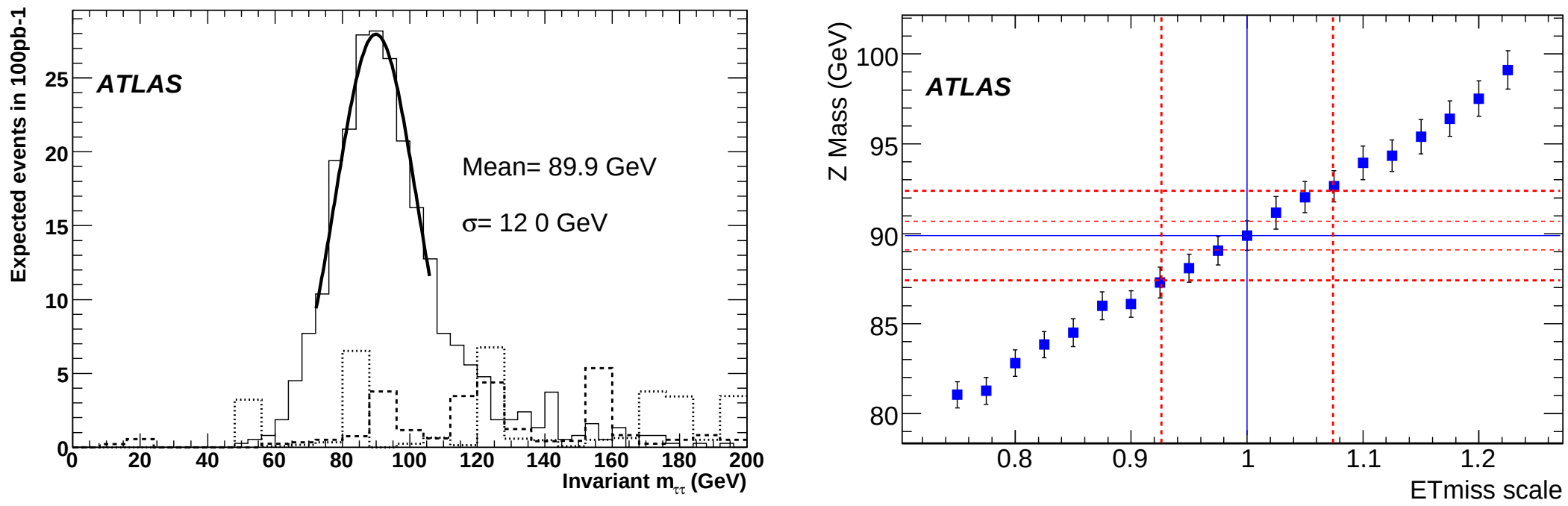

Figure 20: (left) Reconstructed invariant mass of the pair of $\tau$ leptons for $Z \to \tau\tau$ decays and all backgrounds: opposite-sign background (dashed) and same-sign background (dotted). (right) Reconstructed invariant mass of the pair of $\tau$ leptons for $Z \to \tau\tau$ decays as a function of the $\not{E}_T$ scale. The horizontal lines correspond to $\pm 1\sigma$ and to $\pm 3\sigma$ w.r.t. the $Z$ peak position. The analysis is based on an integrated luminosity of 100 pb$^{-1}$ of data.

Backgrounds from $Z \to \tau\tau$ and $W \to \ell\nu$ events are negligible. The background from QCD events in which two leptons are falsely identified is expected to be small but has to be carefully evaluated when data are available. In the present study, these backgrounds are not considered as they are expected to have negligible impact.

In Section 3, projections of $\not{E}_T$, called $\not{E}_L$ and $\not{E}_P$, were introduced. This analysis aims at optimizing the principle of using projections by resolving the missing transverse momentum along the so called 'longitudinal axis' which is defined by the combined direction of flight of the two leptons. The perpendicular axis is also defined in the transverse plane which is orthogonal to the longitudinal axis. The axes as reconstructed from the measured angles of the leptons and their measured energies are thus not used at this point, which would fully exploit the good angular resolution of the ATLAS detector. In general, the longitudinal axis points in the direction of flight of the $Z$ boson and away from the hadronic recoil.

Figure 21 (left) shows, for $Z \to ee$ events, the average $\not{E}_T$ resolved along both axes as a function of the transverse momentum of the lepton system resolved along the longitudinal axis.

The results for the longitudinal axis exhibit a negative offset of up to $\sim 4$ GeV at high values of $p_T$ of the lepton system, while the results for the perpendicular axis are consistent with zero. For $Z \to \mu\mu$ events similar results are obtained. It has been verified that this offset is not caused by real neutrinos in the event. If the event topology is considered, it is clear that this is suggesting that the magnitude of the hadronic recoil $p_T$ is underestimated. The resolution of the $\not{E}_T$ projection on both axes as a function of the total scalar sum of the activity in the hadronic calorimeter, $\sum E_{T,\text{cluster}}$, is shown in Figure 21(right) for $Z \to ee$ events. The fitted curve is of the form $\sigma(\not{E}_T) = P_0\sqrt{\sum E_{T,\text{cluster}}} + P_1$ and illustrates the stochastic behavior of the calorimetric energy measurement.

These results on $Z \to ee$ and $Z \to \mu\mu$ events demonstrate that potential problems of the $\not{E}_T$ reconstruction can be located with high accuracy using the first data.

### 6.4 $W \to \ell\nu$ events

Events with $W \to e\nu$ and $W \to \mu\nu$ will be copiously produced at the LHC. Tens of thousands of events with an excellent signal-to-background ratio can be collected per pb$^{-1}$ . With these events, a good understanding of the $\not{E}_T$ reconstruction can be achieved up to $\not{E}_T$ values of few hundred GeV. The average $\Sigma\not{E}_T$ is of the order of 150 GeV.

In order to isolate $W \to \ell\nu$ events, two basic selection cuts are required: existence of one high $p_T$

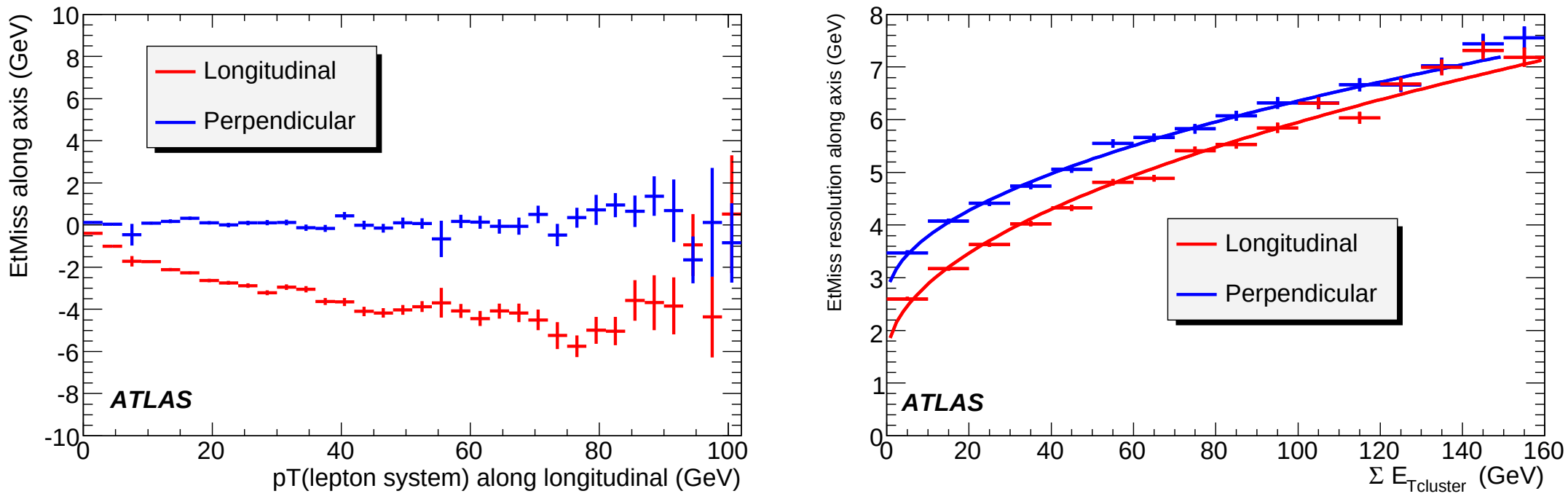

Figure 21: For $Z \rightarrow ee$ events: (left) $\not\!\!E_{\mathrm{T}}$ projected onto the longitudinal and perpendicular axes as explained in the text as function of the $p_{\mathrm{T}}$ of the lepton system resolved along the longitudinal axis. (right) The width $\sigma(\not\!\!E_{\mathrm{T}})$ as a function of $\sum E_{\mathrm{T,cluster}}$. Both plots use a sample corresponding to 250 pb$^{-1}$ of data.

charged lepton with $|\eta| < 2.5$ and $\not\!\!E_{\mathrm{T}} > 20$ GeV. Possible backgrounds are from $t\bar{t}$ production, $W \rightarrow \tau\nu$, $Z \rightarrow \ell\ell$, and QCD events.

Two methods have been investigated to check the $\not\!\!E_{\mathrm{T}}$ reconstruction using these events. The first is based on the fact that the average $p_{\mathrm{T}}$ of the charged lepton and the neutrino are the same, so the ratio $R = p_{\mathrm{T},\nu}/p_{\mathrm{T},\ell}$ has been studied. This variable should be $\sim 1$, but its distribution is distorted by the kinematic and acceptance requirements on the charged leptons. The method and related systematics have been checked with the fast ATLAS simulation. It is expected to be sensitive to values of $\not\!\!E_{\mathrm{T}}$ up to 60 GeV even with 1pb$^{-1}$ of data. Full simulation studies are in progress.

The second method, based on the shape of the reconstructed transverse mass of the $W$ boson, is sensitive to both the $\not\!\!E_{\mathrm{T}}$ resolution and the scale. The transverse mass, $m_{\mathrm{T}}^{W}$, is reconstructed under the hypothesis that $\not\!\!E_{\mathrm{T}}$ is completely due to $p_{\mathrm{T}}^{\nu}$. An example of an $m_{\mathrm{T}}^{W}$ distribution is shown in the next section for $t\bar{t}$ events, which have high values of $\Sigma\not\!\!E_{\mathrm{T}}$ of typically 500 GeV. The focus in this section however is on a dedicated analysis of the corresponding distribution for Drell-Yan events at lower $\Sigma\not\!\!E_{\mathrm{T}}$ as statistically required.

The $m_{\mathrm{T}}^{W}$ distribution (for Drell-Yan events) is fitted in a binned log-likelihood fit that uses template histograms. To minimize the dependence on the kinematics of the $W$ boson, e.g. on its transverse momentum, the fit is restricted to values of $m_{\mathrm{T}}^{W}$ in the range of 65 to 90 GeV. The template histograms of the $m_{\mathrm{T}}^{W}$ distributions are generated by convolving the true transverse mass distribution with the $\not\!\!E_{\mathrm{T}}$ response: $\not\!\!E_{x,y} = \alpha p_{\mathrm{T}}^{\nu}(x,y) \oplus \mathrm{Gauss}(0,\sigma)$ where parameters $\alpha$ and $\sigma$ are the $\not\!\!E_{\mathrm{T}}$ scale and resolution (in GeV), respectively. Since the $\not\!\!E_{\mathrm{T}}$ resolution strongly depends on the activity in the calorimeter, the analysis is performed in several $\Sigma\not\!\!E_{\mathrm{T}}$ intervals.

Figure 22 (left) shows the resolution for $W \rightarrow \mu\nu$ events. The results of the fit agree well with the expectations using truth information labeled as 'pseudo-data'. In Fig. 22 (right) the result for the $\not\!\!E_{\mathrm{T}}$ scale is shown. The scale is measured at the 1% level over a large range of $\Sigma\not\!\!E_{\mathrm{T}}$, confirming the excellent performance of this technique. The template fitting performs well in the low $\Sigma\not\!\!E_{\mathrm{T}}$ region, while a small discrepancy is observed in the high $\Sigma\not\!\!E_{\mathrm{T}}$ region.

This method described so far is applicable to W$\rightarrow \mu\nu$ events, whereas in $W \rightarrow e\nu$ events it has to be modified because the electron is included in the $\Sigma\not\!\!E_{\mathrm{T}}$ calculation, while the muon is not. This leads to a significant correlation between $\Sigma\not\!\!E_{\mathrm{T}}$ and the shape of the transverse $W$ mass distribution when the template is made including the $\Sigma\not\!\!E_{\mathrm{T}}$ dependence. Similar results for the $\not\!\!E_{\mathrm{T}}$ scale and resolution are obtained, but systematic uncertainties have not yet been estimated.

Also if the backgrounds, including the QCD background, cannot be efficiently suppressed by the selection cuts, their influence could be non-negligible. Work is in progress on that.

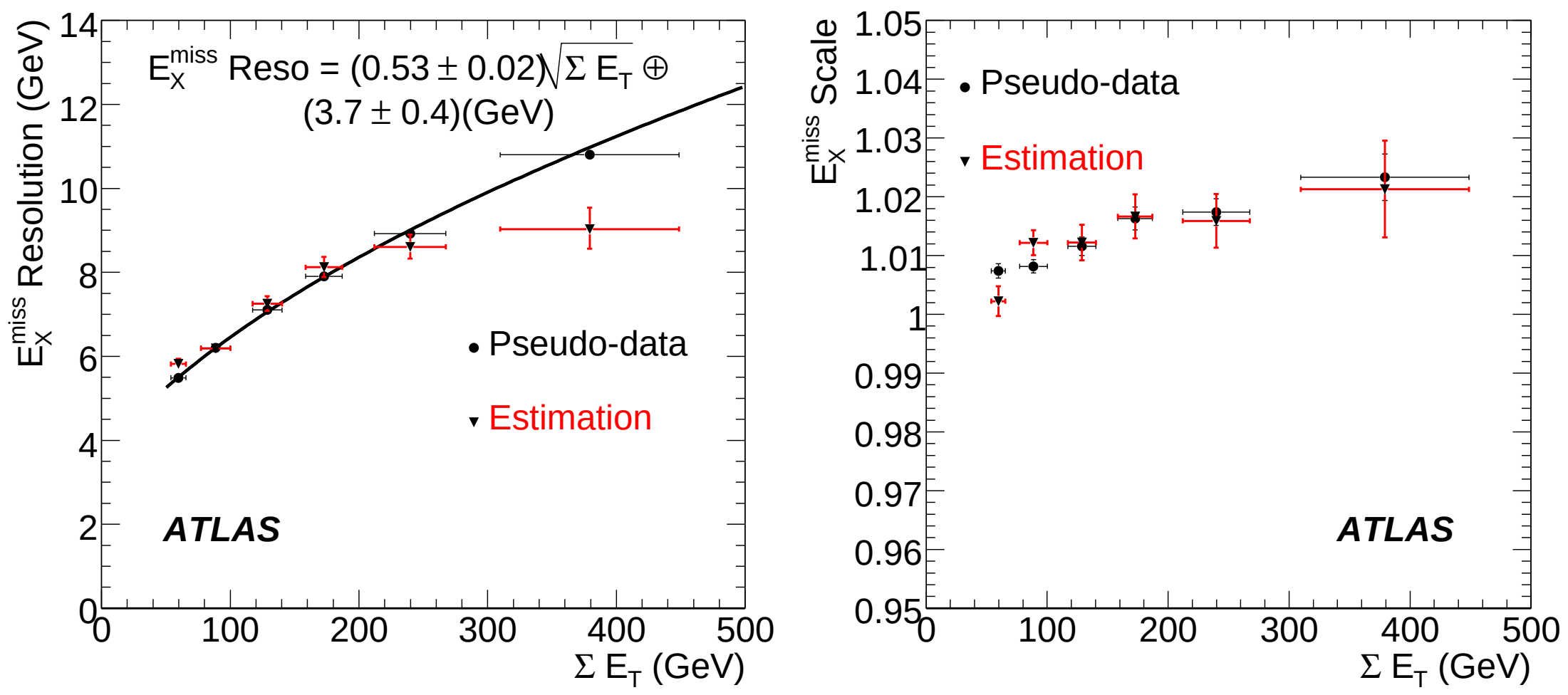

Figure 22: For W→ $\mu\nu$ events: $\not\!\!E_T$ resolution (left) and scale (right) as a function of $\Sigma\not\!\!E_T$, using the second method described in the text. The circular dots represent the value calculated from pseudo-data and the triangular markers represent the estimation.

### 6.5 Semileptonic $t\bar{t}$ events

Semileptonic $t\bar{t}$ events have an interesting multi-jet topology. With genuine $\not\!\!E_T$ in the range from 20 GeV to 100 GeV and a total transverse energy of typically 500 GeV, they are representative of other physics channels such as SUSY. This section shows that the reconstructed transverse $W$ mass as well as a kinematic fit that exploits all mass constraints in $t\bar{t}$ events without requiring $b$ jet tagging, will be useful to investigate possible problems of the $\not\!\!E_T$ measurement in early data. Both methods are sensitive to the scale of $\not\!\!E_T$ to the level of a few percent (statistically) when a sample with an integrated luminosity of 200 pb$^{-1}$ is used. These methods are affected differently by jet energy scales and background and can provide complementary information.

About 7k events survive the selection requirements, which are: at least 3 jets with $p_T \geq 40$ GeV, at least one more jet with $p_T \geq 20$ GeV, $\not\!\!E_T \geq 20$ GeV, and one isolated lepton ($e$ or $\mu$), with $p_T \geq 20$ GeV. The requirements strongly suppress the background from QCD events, which is expected to have no effect and is ignored. The QCD background is expected to be $< 10\%$ in $t\bar{t}$ events, so, after the requirements for the kinematic fit, this assumption should be safe. The background from $W$+jets events is at the level of 20% and is included in this study.

In $t\bar{t}$ analyses the usual assumption is that the $\not\!\!E_T$ in an event can be assigned to the neutrino from the leptonically decaying $W$. With this assumption, the transverse mass $m_T^W$ can be reconstructed from the $\not\!\!E_T$ vector and the transverse momentum of the charged lepton. Figure 23 (left) shows that the shape of the $m_T^W$ distribution is distinctly different for various ranges of fake $\not\!\!E_T$[4], illustrating the power of these events to locate problems. To demonstrate that the transverse $W$ mass distribution can be used to check the $\not\!\!E_T$ scale in early data, two additional event samples with the true $\not\!\!E_T$ scaled by 0.8 and 1.2, respectively have been produced and reconstructed. The samples are analysed by fitting a Gaussian shape to the core of the peak in the transverse mass distributions. The peak position shifts by -7 and +7 GeV for the sample with scale of 0.8 and 1.2 respectively, both with an statistical uncertainty of 0.5 GeV, indicating a sensitivity

[4]Here fake $\not\!\!E_T$ is defined as the scalar difference of reconstructed $\not\!\!E_T$ and true $\not\!\!E_T$.

to the $\not\!\!E_T$ scale at the level of 2%.

Note that backgrounds other than $W$+jets events are not considered in this study. Background from SUSY events can have a severe impact on the distribution by shifting the peak and thus mimicking a $\not\!\!E_T$ scale calibration offset. The existing knowledge of $t\bar{t}$ events can be combined in a kinematic fit to improve the measured quantities and to investigate the scale of the $\not\!\!E_T$ measurement. The following mass constraints are available: $m_W^{had} = m_W^{lep} = 80.4$ GeV, where $m_W^{had}$ is the reconstructed mass of two light jets of the hadronically decaying $W$ and $m_W^{lep}$ is the reconstructed mass of the lepton and the neutrino of the leptonically decaying $W$. The reconstructed mass of the leptonically decaying top quark and that of the hadronically decaying top quark are assumed to be $m_{top}^{had} = m_{top}^{lep} = 175$ GeV. The neutrino's transverse momentum is set equal to $\not\!\!E_T$ and its longitudinal momentum is analytically calculated from the $m_W^{lep}$ constraint.

The $\chi^2$ function of the fit is built using the energy of the four leading jets, with fit parameters to scale the corrected jet energies, a constraint on the product of the fit parameters to the a-priori known or assumed overall jet energy scale and with the implementation of the four mass constraints. All twelve possible permutations of assigning jets to the two top quarks and $W$ bosons respectively are considered. Finally, only the permutation with the lowest $\chi^2$ is selected in each event.

It is found that the $\not\!\!E_T$, re-calculated after the fit, is not significantly improved. However, using a cut on $\chi^2$ improves the resolution on $\not\!\!E_T$, so it is possible to use the $\chi^2$ of a kinematic fit to classify $t\bar{t}$ events with a relatively good $\not\!\!E_T$ measurement without using $b$ tagging. In the first data this classification helps to locate possible detector problems.

The kinematic fitting procedure can be utilized to check the $\not\!\!E_T$ scale in early data. Background events are expected to be incompatible with the constraints used in the fit and thus be reduced by a cut on $\chi^2$. Therefore, in contrast to the study using the transverse $W$ mass as described above, this method suffers significantly less from backgrounds.

After applying the fit to the events and requiring $\chi^2 < 10$ and in additional $\not\!\!E_T > 40$ the background from $W$+jets event is reduced to the 1% level. A robust estimator of the $\not\!\!E_T$ scale is the measured transverse momentum difference of the (anti-) top quark mother of the leptonically and hadronically decaying $W$ respectively: $\Delta p_T = p_T^{lep} - p_T^{had}$ where $p_T^{lep}$ is the combined transverse momentum of the measured charged lepton, $\not\!\!E_T$ and one $b$ jet, while $p_T^{had}$ is the momentum of the jets of the hadronically decaying $W$ and the other $b$ jet. Of course, the assignment of the jets to the correct top quark is not guaranteed. Nevertheless, this quantity is remarkably sensitive to the $\not\!\!E_T$ scale, as can be seen in Fig.

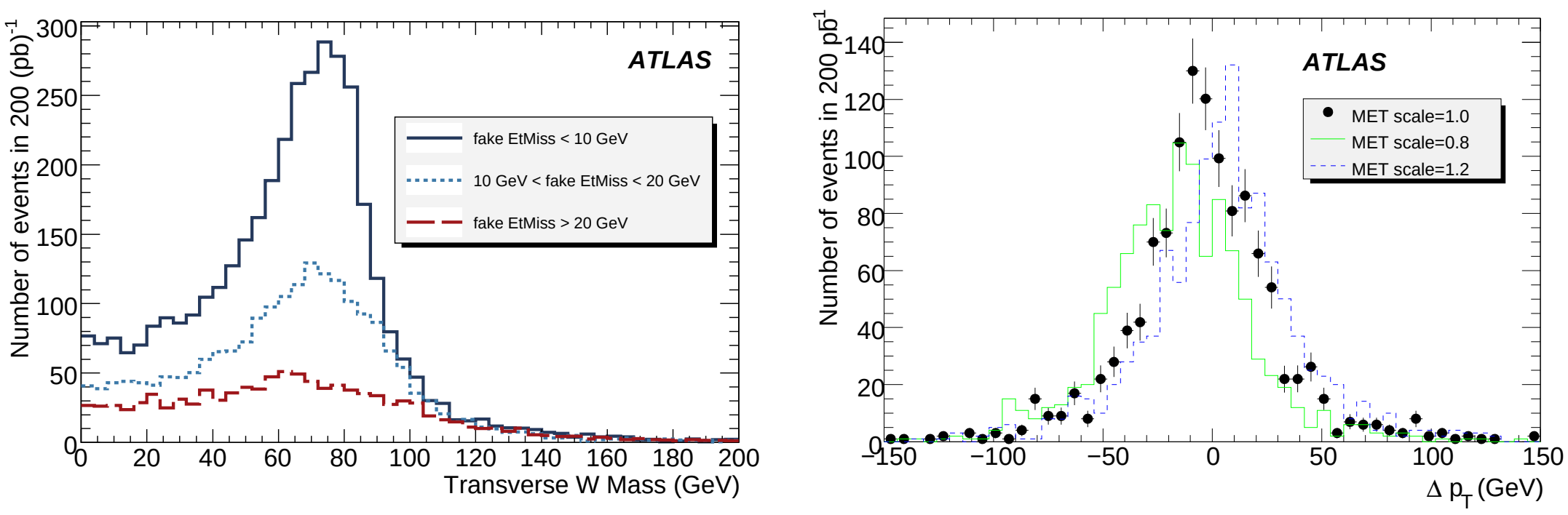

Figure 23: In semileptonic $t\bar{t}$ events, (left) the reconstructed transverse $W$ mass in various ranges of fake $\not\!\!E_T$, (right) the distribution of the measured momentum difference of the two top quarks, $\Delta p_T$ for (true) $\not\!\!E_T$ scales of 0.8, 1.0 and 1.2 as indicated. The analysis is based on an integrated luminosity of 200 pb$^{-1}$ of data.

23(right). The mean values of the distributions are $-16.4\pm0.9$ GeV , $-4.5\pm0.8$ GeV, and $4.5\pm0.9$ GeV for scales of 0.8, 1.0, and 1.2, respectively. This implies a sensitivity on the $\not{E}_T$ scale at the level of 2%. The systematic variation due to a shift of the top quark mass of 2.5 GeV is about 2%.

## 7 Summary

The $\not{E}_T$ in ATLAS is calculated from the energy in the calorimeter and from the reconstructed muons. The energy in the calorimeter is classified and calibrated according to the reconstructed objects to which it belongs. Two algorithms for reconstruction and calibration are presently implemented in the ATLAS software, one Cell-based, where the $\not{E}_T$ reconstruction and calibration is done starting from the energy deposited in calorimeter cells, and the other one Object-based, where the $\not{E}_T$ reconstruction is done from the reconstructed, classified and calibrated objects and from the energy outside of them. The performance of the two is similar.

The $\not{E}_T$ performance has been checked on a large variety of events with physical $\not{E}_T$ such as $Z\to\tau\tau$ , $A/H\to\tau\tau$, SUSY, and $t\bar{t}$, as well as events with no physical $\not{E}_T$ like minimum bias events, events with QCD jets, and $Z$+jets processes. The resulting linearity of the response is within 5%, even for low true $\not{E}_T$ values of the order of 40 GeV. The $\not{E}_T$ resolution, $\sigma$, follows an approximate stochastic behaviour over a wide range of values of the total transverse energy deposited in the calorimeters. A simple fit to a function $\sigma = a\cdot\sqrt{\Sigma E_T}$ yields values between 0.53 and 0.57 for the parameter $a$, for $\Sigma E_T$ values between 20 and 2000 GeV. Deviations from this simple behaviour are expected and observed for low values of $\Sigma E_T$ where noise is an important contribution, and for very high values of $\Sigma E_T$ where the constant term in the jet energy resolution dominates. For values of the true $\not{E}_T$ below 40 GeV, the accuracy of the measurement of the direction of the $\not{E}_T$ vector for small values of $\not{E}_T$ degrades rapidly. In contrast, for high values of the true $\not{E}_T$, azimuthal accuracies better than 100 mrad can be achieved. This accuracy of the measurement of the $\not{E}_T$ direction allows an isolation cut on $\not{E}_T$, which can efficiently suppress events with a badly measured jet and the resulting $\not{E}_T$ pointing in the jet direction.

A dedicated study of fake $\not{E}_T$ shows that instrumental effects like hot/dead/noisy cells (regions) in calorimeters, as well as beam-gas scattering or other machine backgrounds, or displaced vertices, are very important, and that their understanding will be crucial in the first days of data taking. Different methods can be used to clean events and to correct/recover the $\not{E}_T$ measurement. Mis-measurements in the detector itself, due to high-$p_T$ muons escaping from the fiducial acceptance or from large losses of deposited energy in cracks or inactive materials, might also effectively limit the performance of the $\not{E}_T$ reconstruction and have therefore been studied in detail.

With the first 100pb$^{-1}$ the algorithms for $\not{E}_T$ reconstruction and calibration can be checked studying the $\not{E}_T$ linearity and resolution in minimum bias events and in Standard Model processes like $Z$ and $W$ decays and in $t\bar{t}$ processes. Complementary methods for the determination of the $\not{E}_T$ scale in-situ have been studied and it has been shown that it will be possible to determine the $\not{E}_T$ scale with a precision of at least 8%.

# *b*-Tagging

# $b$-Tagging Performance

**Abstract**
The ability to identify jets stemming from the fragmentation and hadronization of $b$ quarks is important for the high-$p_T$ physics program of ATLAS: top physics, Higgs boson searches and studies, new phenomena. After an overview of the reconstruction of the key ingredients for $b$-tagging, the tagging techniques are described. The performance of $b$-tagging algorithms is then detailed, as well as the impact on performance of several factors and new promising directions. Finally, expected performance in the first data and the anticipated uncertainty with which it can be measured are briefly discussed.

## 1 Introduction

This note discusses the identification of jets stemming from the hadronization of $b$ quarks, or $b$-tagging. The ability to identify jets containing $b$-hadrons is important for the high-$p_T$ physics program of a general-purpose experiment at the LHC such as ATLAS. This is in particular useful to select very pure top samples, to search and/or study Standard Model or supersymmetric (SUSY) Higgs bosons which couple preferably to heavy objects or are produced in association with heavy quarks, to veto the large dominant $t\bar{t}$ background for several physics channels and finally to search for new physics: SUSY decay chains, heavy gauge bosons, etc.

The large majority of these studies requires good $b$-tagging performance for jets with a transverse momentum ranging from 20 to 150 GeV. However, for super-symmetric processes, jets of $p_T$ as high as 500 GeV may have to be tagged [1], and for exotic phenomena $b$-jets of up to a few TeV can be produced. For top studies, the signal rates are very high at the LHC and therefore a moderate $b$-tagging efficiency ($> 50\%$) is acceptable, while a fraction of light jets mis-identified as $b$-jets below a few per mille suppresses most of the $W$+jets background (see for instance Ref. [2]). One of the most demanding channels for $b$-tagging is the production of a light Standard Model Higgs boson in association with a top-antitop pair [3]: $t\bar{t}H(H \to b\bar{b})$. Four $b$-jets have to be tagged with very high efficiency ($\varepsilon_b \approx 70\%$) since the signal cross-section is low, and the mis-tagging rate must be kept below 1% to fight the large $t\bar{t}$+jets background.

The identification of $b$-jets takes advantage of several of their properties which allow us to distinguish them from jets which contain only lighter quarks. First the fragmentation is hard and the $b$-hadron retains about 70% of the original $b$ quark momentum. In addition, the mass of $b$-hadrons is relatively high ($> 5$ GeV). Thus, their decay products may have a large transverse momentum with respect to the jet axis and the opening angle of the decay products is large enough to allow separation. The third and most important property is the relatively long lifetime of hadrons containing a $b$ quark, of the order of 1.5 ps ($c\tau \approx 450\mu$m). A $b$-hadron in a jet with $p_T = 50$ GeV will therefore have a significant flight path length $\langle l \rangle = \beta\gamma c\tau$, traveling on average about 3 mm in the transverse plane before decaying. Such displaced vertices can first be identified inclusively by measuring the impact parameters of the tracks from the $b$-hadron decay products. The transverse impact parameter, $d_0$, is the distance of closest approach of the track to the primary vertex point, in the $r-\varphi$ projection. The longitudinal impact parameter, $z_0$, is the $z$ coordinate of the track at the point of closest approach in $r-\varphi$. The tracks from $b$-hadron decay products tend to have rather large impact parameters which can be distinguished from tracks stemming from the primary vertex. The other more demanding option is to reconstruct explicitly the displaced vertices. These two approaches of using the impact parameters of tracks or reconstructing the secondary vertex will be referred to later on as spatial $b$-tagging. Finally, the semi-leptonic decays of $b$-hadrons can be used by tagging the lepton in the jet. In addition, thanks to the hard fragmentation and high mass of $b$-hadrons, the lepton will have a relatively large transverse momentum and also a large momentum relative to the jet axis. This is the so-called soft lepton tagging (the lepton being soft compared to high-$p_T$ leptons from $W$ or $Z$ decays).

The tagging methods relying on the impact parameter of tracks are detailed in this note. Only a summary and the main results of the other methods are given. The techniques employed to reconstruct either a single inclusive vertex or to attempt to resolve the complex topologies with a secondary $b$-hadron vertex and a tertiary $c$-hadron vertex are discussed in Ref. [4], as well as the reconstruction of the primary vertex. The tagging with soft muons or electrons from $b$-hadron decays is detailed respectively in Ref. [5] and Ref. [6]. The expected performance of the $b$-tagging algorithms in ATLAS, and the impact of several factors, are explained in detail in this note. However, the assessment of the impact of residual misalignments on the performance is just starting and first results are available in Ref. [7]. While a large effort is put into having a very accurate Monte Carlo simulation, the $b$-tagging performance must be measured in data. Several studies aiming at measuring the $b$-tagging efficiency in dijet events (Ref. [8]) or in $t\bar{t}$ events (Ref. [9]) have been performed. The studies to measure the mis-tagging rates are just starting and are not discussed. Finally, the high-level trigger of ATLAS has the capability to select $b$-jets. This is particularly interesting for channels with several $b$-jets where jet thresholds can be lowered at the first level thanks to the $b$-tagging applied at the second and event-filter levels. The high-level trigger $b$-tagging performance and strategies are discussed in Ref. [10].

The layout of this note is as follows: in Section 2, the reconstruction of the key objects for $b$-tagging is briefly explained and the performance summarized. Since the definition of the flavour of a jet is not unambiguous in Monte Carlo, the estimators used to assess the performance are defined in Section 3. Section 4 is intended to be a pedagogical approach to the various tagging algorithms available and to the likelihood ratio formalism used by ATLAS. The $b$-tagging performance for various physics processes is described in Section 5, relying on the current state-of-the-art $b$-tagging production software. In Section 6, a few additional studies aiming at better understanding some critical aspects of the $b$-tagging are detailed, while in Section 7 three studies showing new directions to improve the $b$-tagging performance are presented. In both cases the studies are described in a separate section because either they required specific datasets or they relied on software and/or cuts/optimizations which were different from the ones currently in use in the ATLAS software or they even required new software developments. In addition, the anticipated uncertainty with which the $b$-tagging performance may be measured in data is discussed in Section 8. Finally in Section 9, the main findings and the expected performance in the first data are summarized.

# 2 Reconstruction of the key objects

The reconstruction of the various objects needed for $b$-tagging and its performance are summarized in this section.

## 2.1 Charged tracks

The tracks reconstructed in the ATLAS Inner Detector [11] are the main ingredient for $b$-tagging. On average a track consists of 3 pixel hits, 4 space-points in the silicon micro-strip detector and about 36 hits in the Transition Radiation Tracker (TRT). The innermost pixel layer (the so-called $b$-layer) is located at a radius of 5 cm, while the TRT extends up to a radius of 1 m. The tracker is immersed in a 2 T magnetic field generated by the central solenoid. The intrinsic measurement accuracy of the pixels is around 10 $\mu$m in $r\phi$ and 115 $\mu$m in $z$. All these allow the tracker to measure efficiently and with good accuracy the tracks within $|\eta| < 2.5$ and down to $p_T \sim 500$ MeV. For a central track with $p_T = 5$ GeV, which is typical for $b$-tagging, the relative transverse momentum resolution is around 1.5% and the transverse impact parameter resolution is about 35 $\mu$m. Further details can be found in Refs. [11, 12].

Most of the results in this note are based on the default pattern-recognition and fitting algorithm, NewTracking. Its performance is described in Ref. [12]. When relevant, some comparisons are made with an alternate algorithm, iPatRec.

### 2.1.1 Baseline track selection

The track selection for *b*-tagging is designed to select well-measured tracks and reject fake tracks and tracks from long-lived particles ($K_s, \Lambda$ or other hyperon decays) and material interactions (photon conversions or hadronic interactions).

Two different quality levels are used. For the standard quality level, at least seven precision hits (pixel or micro-strip hits) are required. The transverse and longitudinal impact parameters at the perigee must fulfil $|d_0| < 2$ mm and $|z_0 - z_{pv}| \sin\theta < 10$ mm respectively, where $z_{pv}$ is the longitudinal location of the primary vertex. Only tracks with $p_T > 1$ GeV are considered. For the *b*-tagging quality, the extra requirements are: at least two hits in the pixel detector of which one must be in the *b*-layer, as well as $|d_0| < 1$ mm and $|z_0 - z_{pv}| \sin\theta < 1.5$ mm. This selection is used by all the tagging algorithms relying on the impact parameters of tracks, while slightly different selections are used by the secondary vertex algorithms as discussed in Ref. [4].

### 2.1.2 Tracking efficiency

The *b*-tagging performance strongly depends upon the tracking efficiency. The tracking performance inside jets, where the track density may be high, is discussed in the following. The tracking performance for single tracks is discussed in Ref. [12].

Figure 1 shows the tracking efficiency and fake rate for tracks in $t\bar{t}$ events as a function of the track pseudo-rapidity. For the efficiency denominator, only charged primary pions[1] produced well before the *b*-layer ($|x - x_{pv}| < 10$ mm, $|y - y_{pv}| < 10$ mm) and with $p_T > 1$ GeV and $|\eta| < 2.5$ are considered. The first level of the efficiency corresponds to the basic reconstruction efficiency, where a track matched to a Monte Carlo particle is found. The fake rate is defined as the fraction of reconstructed tracks which do not pass the matching criteria used for the efficiency, i.e. less than 80% of their hits are coming from the same Monte Carlo particle. At high pseudo-rapidities, the tracking performance deteriorates mostly because of increased material and more ambiguous measurements.

Figure 2 shows the tracking efficiency and fake rate for tracks in $t\bar{t}$ events as a function of their distance $\Delta R = \sqrt{\Delta\phi^2 + \Delta\eta^2}$ to the axis of the closest jet, for tracks fulfilling the *b*-tagging quality cuts. The tracking performance degrades near the core of the jet where the track density is the highest and induces pattern-recognition problems. This is especially visible for high-$p_T$ ($> 100$ GeV) jets.

Finally, in Figure 3 the tracking efficiency and fake rates obtained with the default algorithm and with iPatRec are compared. The first plot shows the comparison for several bins in the track $p_T$ for all jets, while the second plot is as a function of the distance to the jet axis for jets with $E_T > 100$ GeV. It is interesting to note that the two algorithms have a different working point: the default algorithm maintains a low level of fakes at the price of losing in efficiency, while the complementary choice was taken for iPatRec. This difference in treatment will lead to different *b*-tagging performance for jets with high momentum, as discussed in particular in Section 5.6. The features seen in these plots are specific to the pattern-recognition inside jets: for instance the decrease of the NewTracking efficiency at high track $p_T$ is not visible for isolated tracks; it is here correlated with the local track density since high-$p_T$ tracks are more likely to originate from denser high-$p_T$ jets.

[1] For the tracking studies only pions were considered, but similar results are expected for charged kaons, protons, etc. which are all used for *b*-tagging.

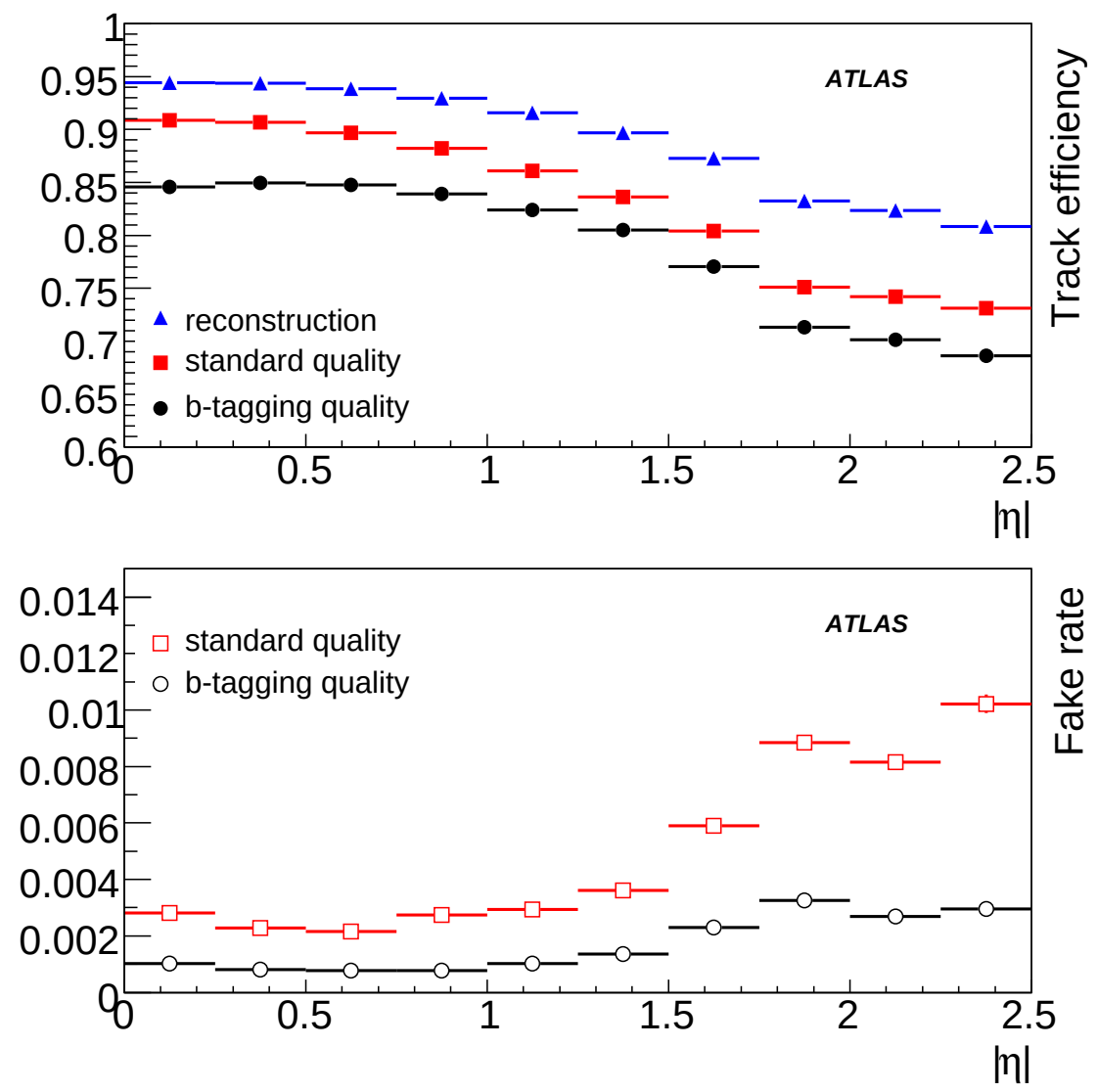

Figure 1: Tracking efficiency (top plot) and fake rate (bottom plot) versus track pseudo-rapidity, for three levels of track selection: matching (blue triangles), standard quality cuts (red squares) and *b*-tagging quality cuts (black circles), in $t\bar{t}$ events.

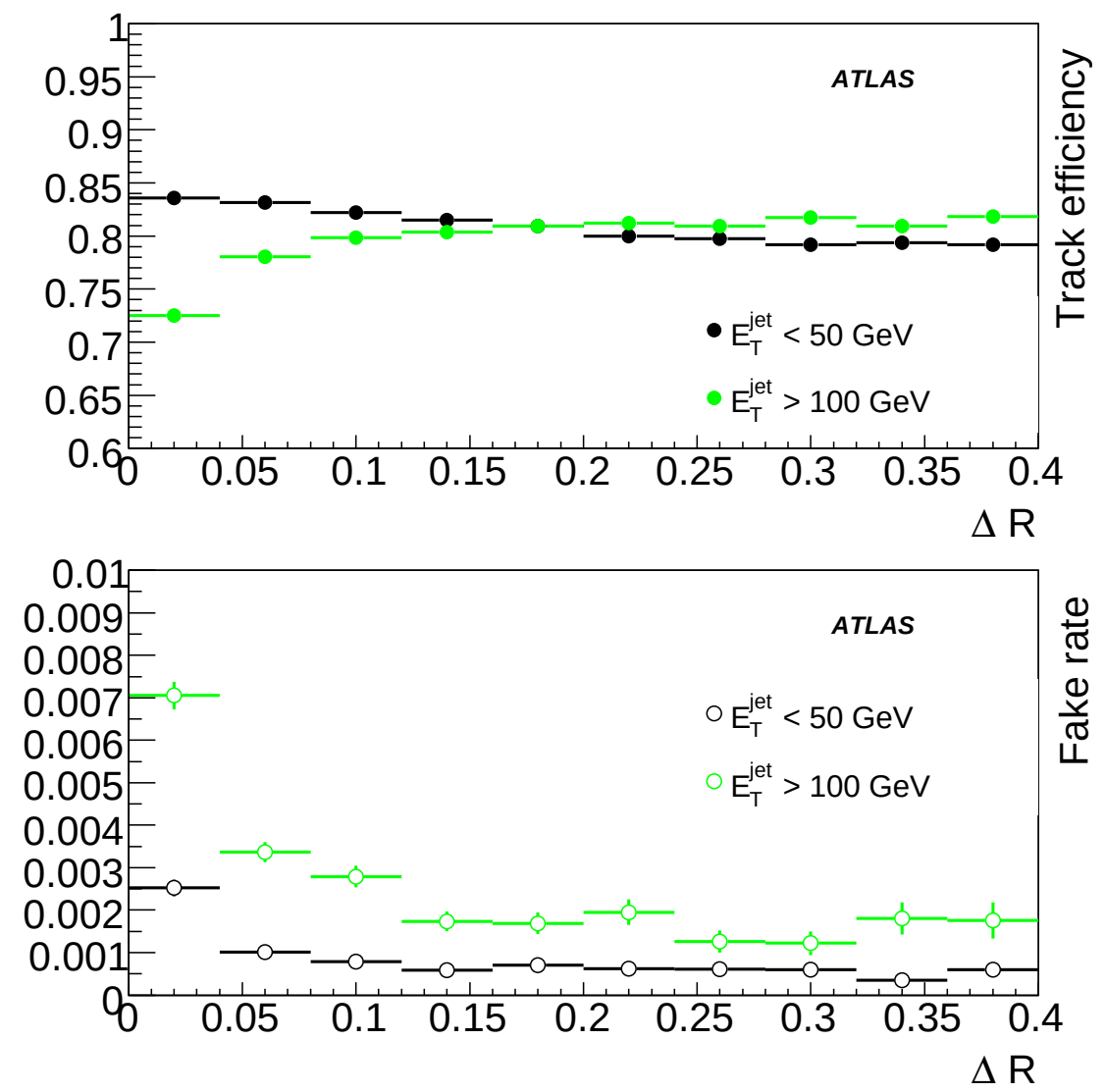

Figure 2: Tracking efficiency (top plot) and fake rate (bottom) versus distance to jet axis, for tracks fulfilling the *b*-tagging quality cuts and associated to low-$p_T$ jets (black symbols) or high-$p_T$ jets (green symbols), in $t\bar{t}$ events.

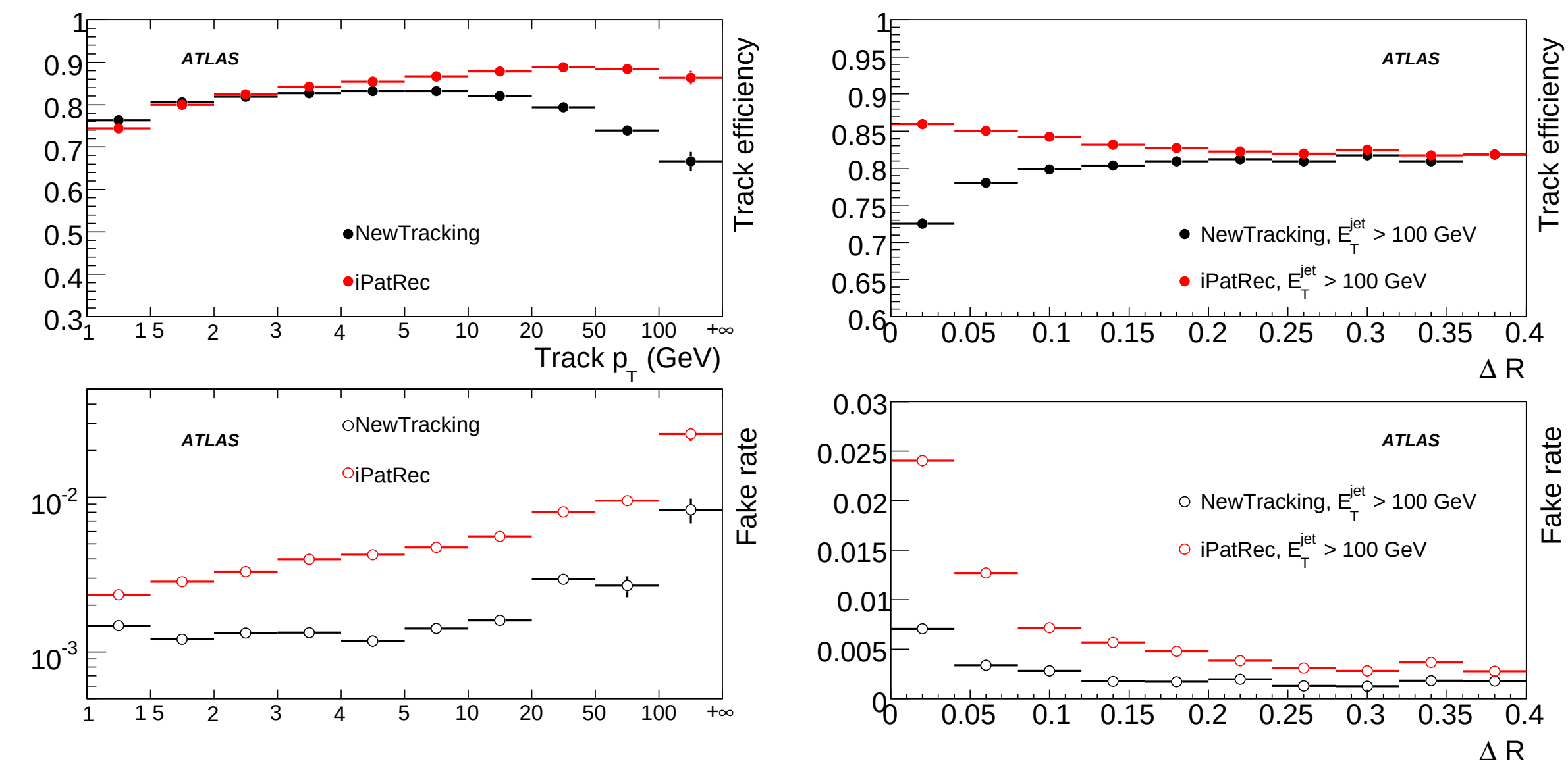

(a) for several bins of track $p_T$ (b) versus the distance to the jet axis

Figure 3: Tracking efficiency (top plots) and fake rate (bottom plots) in $t\bar{t}$ events after the *b*-tagging quality cuts, for two tracking algorithms: default NewTracking (black symbols) and iPatRec (red symbols).

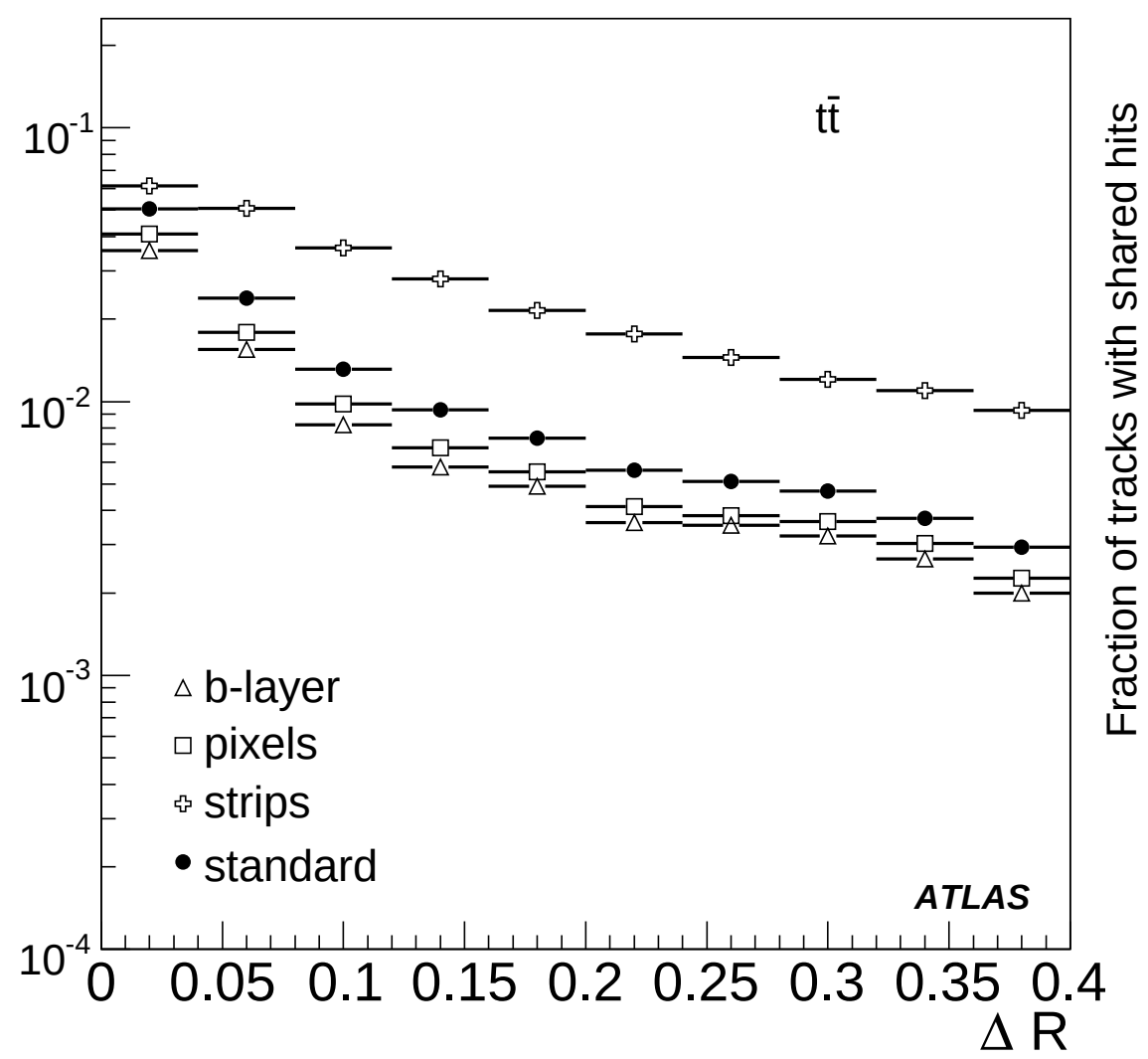

Figure 4: Fraction of tracks with shared hits versus distance to the jet axis. Tracks fulfilling the $b$-tagging quality cuts, and with at least one shared hit in the silicon systems are shown. The standard definition of shared hits (see text) is shown as well.

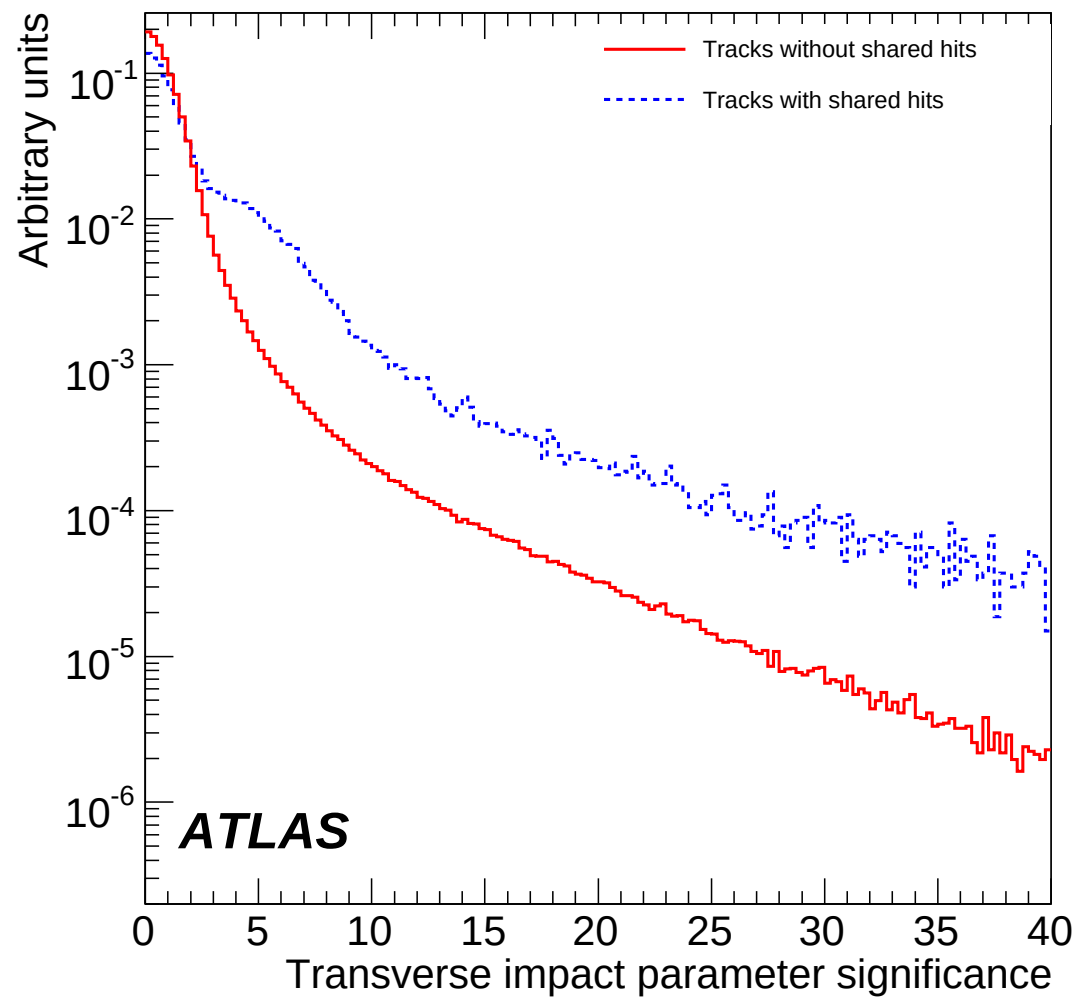

Figure 5: Transverse impact parameter significance $d_0/\sigma_{d_0}$ for tracks in light jets. Two categories of tracks are used: regular ones (red plain curve) and tracks with shared hits (blue dashed). Both distributions are normalized to unity.

### 2.1.3 Tracks with shared hits

Tracks originating from the same point and passing the track selection cuts will not necessarily have the same impact parameter distributions. First of all, even using the track parameters normalized to their error will not compensate for all resolution effects, such as non-Gaussian tails. In addition, the pattern-recognition process itself can produce tracks of variable quality depending on their hit contents. Those tracks require a special treatment to be flagged appropriately. The most significant subset of such tracks is formed by the tracks which are sharing some of their hits with other tracks.

Figure 4 shows the fraction of tracks which are sharing at least one hit with another reconstructed track versus the distance of the track to the jet axis, for jets originating from $t\bar{t}$ events. Currently for $b$-tagging purposes, a track is defined as a track with shared hits if it has at least one shared hit in the pixels or two shared hits in the strips. As expected, the fraction of tracks with shared hits increases with the local track density, and is therefore higher for high-$p_T$ jets and in the core of the jets. In $t\bar{t}$ events the average $p_T$ for taggable (*i.e.* $p_T > 15$ GeV and $|\eta| < 2.5$) $b$-jets and light jets are respectively 74 and 55 GeV. The fraction of tracks with shared hits is about 2%. For jets with a transverse momentum of about 140 GeV ($WH$ events with $m_H$=400 GeV, see below), this fraction is twice as high. In both cases the fraction is roughly similar for NewTracking and iPatRec. In an extreme case, for $Z' \to b\bar{b}$ events with $m_{Z'} = 2$ TeV, a majority of tracks have shared hits and the fraction depends significantly on the reconstruction algorithm (cf. Section 5.6). Even when the overall level of shared hits is relatively low, it has been demonstrated that those tracks should be treated appropriately since their impact on the $b$-tagging performance is significant. Indeed, the impact parameter significances, defined as the ratios $d_0/\sigma_{d_0}$ and $z_0/\sigma_{z_0}$ of the impact parameters to their measured error, for tracks in light jets exhibit a very different behavior depending on whether the track is a regular one or a track with shared hits, as shown in Figure 5. It is clear that tracks with shared hits can mimic lifetime more easily.

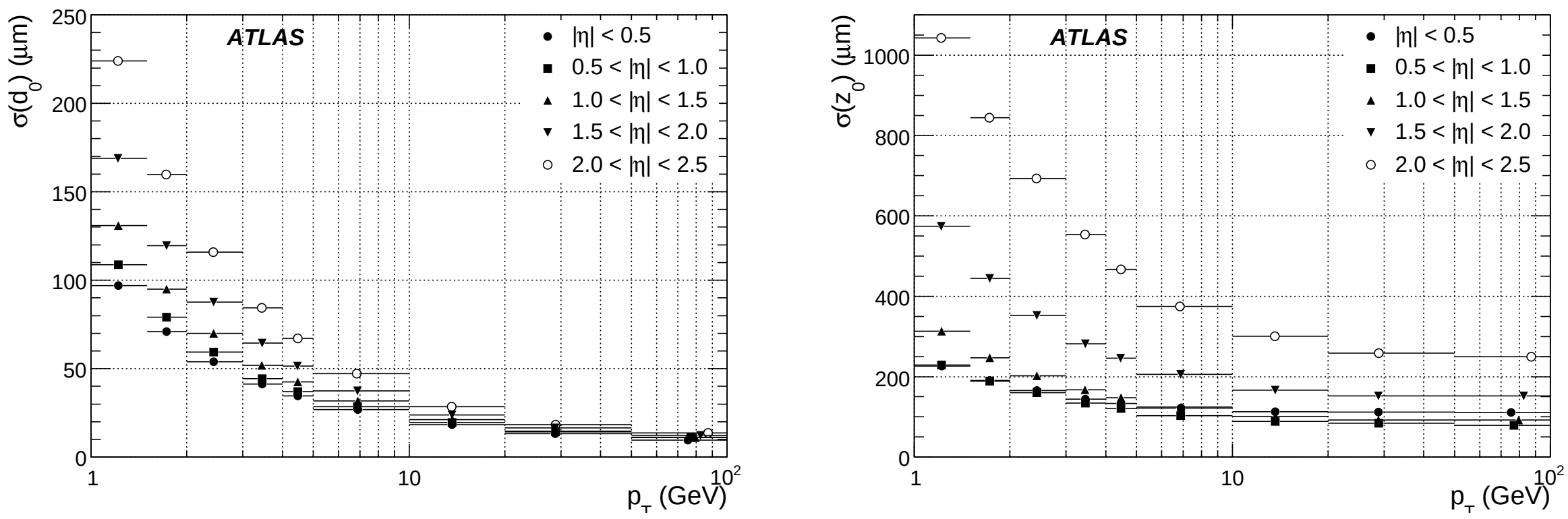

(a) Transverse impact parameter

(b) Longitudinal impact parameter

Figure 6: Track impact parameter resolution versus track $p_T$, for several bins in the track pseudo-rapidity.

### 2.1.4 Impact parameter resolution

The resolution of the track impact parameter is a crucial ingredient to be able to discriminate tracks coming from long-lived hadrons and prompt tracks. To estimate it, all the reconstructed tracks in $t\bar{t}$ events fulfilling the $b$-tagging quality cuts and matched to a good Monte-Carlo track as defined in Section 2.1.2 were used. The difference between the reconstructed and the true impact parameter within a bin was fitted with a single gaussian, whose $\sigma$ is reported on Figures 6(a) and 6(b), for respectively the transverse and longitudinal impact parameters. For a central track with $p_T = 5$ GeV, which is typical for $b$-tagging, the transverse impact parameter resolution is about 35 $\mu$m.

## 2.2 Primary vertex finding

Another key ingredient for $b$-tagging is the primary vertex of the event. The impact parameters of tracks are recomputed with respect to its position and tracks compatible with the primary vertex are excluded from the secondary vertex searches. At LHC the beam-spot size will be $\sigma_{xy} = 15$ $\mu$m and $\sigma_z = 5.6$ cm: therefore the primary vertex is especially important for the $z$ direction, while in the transverse plane only the beam-line could be used. The strategies to find the primary vertex and their performance are explained in Ref. [4]. The efficiency to find the primary vertex is very high in the high-$p_T$ events of interest, and the resolution on its position is around 12 $\mu$m in each transverse direction and 50 $\mu$m along $z$. With pile-up, the presence of additional minimum bias vertices makes the choice of the primary vertex less trivial: at a luminosity of $2 \times 10^{33}$ cm$^{-2}$s$^{-1}$ (on average 4.6 minimum bias events per bunch-crossing) a wrong vertex can be picked up as the primary vertex in about 10% of the cases [4], thus causing a deterioration in the $b$-jet tagging efficiency.

## 2.3 Jet algorithms

The baseline jet algorithm for the studies in this note is a seeded cone algorithm using the calorimeter towers with a cone size of $\Delta R = 0.4$, and where the cells were calibrated using the H1 method (see Ref. [13] for details). The impact on $b$-tagging performance of using other jet algorithms is discussed in Section 6.4.

For $b$-tagging purposes, only the jet direction is relevant. In the first place, this direction is used to define which tracks should be associated with the jets. The actual tagging is done on this subset of tracks. Currently tracks within a distance $\Delta R < 0.4$ of the jet axis are associated to the jet. A given track is associated to only one jet (the closest in $\Delta R$). This is the case for actually any jet collections, regardless of the cone size of the jet. The jet direction is also used to sign the impact parameters of the tracks in the jet as explained in Section 4.1.2.

Except when stated otherwise, there was no attempt to remove from the reconstructed jet collection the jets which are composed of only electrons. In the $t\bar{t}$ sample (semi-leptonic and di-leptonic channels), about 5% of the taggable reconstructed jets are electrons. There is no dedicated treatment for muons. Isolated muons are very unlikely to fake jets, at least for the common processes under consideration in this note where $p_T(\mu) < 100$ GeV. Muons in jets, stemming from *b*/*c*-hadron semi-leptonic decays and measured in the muon spectrometer, deposit on average about 3 GeV in the calorimeter but their momentum as measured in the inner detector and the muon system is not used to refine the kinematics of the jet, which remain purely calorimeter-based.

Only jets fulfilling $p_T > 15$ GeV and $|\eta| < 2.5$ are deemed taggable and considered in the performance studies.

### 2.4 Soft lepton reconstruction

Leptons arising from semi-leptonic decays of *b*-hadrons or subsequent *c*-hadrons can be used to tag *b*-jets.

Soft muons are reconstructed [5] using two complementary reconstruction algorithms. A *combined muon* corresponds to a track fully reconstructed in the muon spectrometer that matches a track in the inner detector. Low-momentum muons (below $p \sim 5$ GeV) which cannot reach the muon middle and outer stations are identified by matching an inner detector track with a segment in the muon spectrometer inner stations. Muons satisfying some basic requirements ($p_T > 3$ GeV, $|d_0| < 4$ mm) are associated to the closest jet provided that $\Delta R < 0.5$. Finally, the kinematic properties of the jet-muon system are used in order to reject the background caused by punch-throughs and decays-in-flight in light jets.

Reconstructing soft electrons [6] in the calorimeter inside a jet is more difficult. This is achieved by matching an inner detector track to an electromagnetic cluster. For a given track, only the energy contained in a small window around the track extrapolation is used. The contribution of neighbouring hadronic showers is therefore reduced. The identification procedure takes full advantage of the tracking capabilities of the inner detector as well as of the granularity of the electromagnetic calorimeter: a likelihood ratio combines inner detector information such as transition radiation hits with shower shape variables from the calorimeter. The performance is, however, highly dependent on the track density in jets as well as the quantity of matter in front of the electromagnetic calorimeter.

## 3 Performance estimators

### 3.1 Labelling

To define *b*-tagging performance, the Monte Carlo event history is used to know the type of parton from which a jet originates. This *labelling* procedure is not unambiguous and is not strictly identical for different Monte Carlo generators. For the results presented here, a quark labelling has been used: a jet is labelled as a *b*-jet if a *b* quark with $p_T > 5$ GeV is found in a cone of size $\Delta R = 0.3$ around the jet direction. The various labelling hypotheses are tried in this order: *b* quark, *c* quark and $\tau$ lepton. When no heavy flavour quark nor $\tau$ lepton satisfies these requirements, the jet is labelled as a light-jet. No attempt is made to distinguish between *u*, *d*, *s* quarks and gluon since such a label is even more ambiguous.

### 3.2 Efficiency and rejection

For performance studies, only jets fulfilling $p_T > 15$ GeV and $|\eta| < 2.5$ are considered and refered to as taggable jets. In the following, jets for which no track passed the $b$-tagging quality cuts are still counted in the performance estimators. However, events where the primary vertex could not be reconstructed are ignored. In addition, $b$-jets were not categorized according to the nature of the $b$-hadron decay: $b$-jets with semi-leptonic decays behave quite differently from jets with hadronic decays, even when using purely spatial methods, but in the following no distinction was made.

The tagging efficiency is naturally defined as the fraction of taggable jets labelled as $b$-jets (see previous section) which are actually tagged as $b$-jets by the tagging algorithm under study. The mis-tagging rate is the fraction of taggable jets not labelled as $b$ which are actually tagged as $b$-jets. For historical reasons the jet rejection is used instead: this is simply the inverse of the mis-tagging rate.

### 3.3 Purification

A difficulty arises as soon as the jet multiplicity is high and various jet flavours are present in a single event: a jet with $\Delta R(\text{jet} - b) = 0.31$ is labelled as a light jet, although tracks from $b$-hadron decay with high lifetime content are likely to be associated to it.

This leads to a decrease of the estimated performance, not related to the $b$-tagging algorithm itself but to the labelling procedure which strongly depends on the activity of the event. In order to obtain a more reliable estimation of $b$-tagging performance, a purification procedure has been devised: light jets for which a $b$ quark, a $c$ quark or a $\tau$ lepton is found within a cone of size $\Delta R = 0.8$ around the jet direction are not used to compute the rejection.

The performance estimated after purification represents the intrinsic power of the $b$-tagging algorithms and should be similar for different kinds of hard event, whereas results obtained for the complete light jet sample are more dependent on the event type. On the other hand, the latter is more representative of the actual $b$-tagging power for a given physics analysis. This is illustrated in Figure 7: the light jet rejection in simple $WH$ events is similar without or with purification (left plot), while for busier $t\bar{t}$ events (right plot) the two curves differ in the region where lifetime content as opposed to resolution effects dominates (*i.e.* for $\varepsilon_b < 80\%$). In the following, jets fulfilling the purification procedure will be referred to as purified or pure jets, the ones failing this procedure will be called non-pure jets, while all the jets will be called raw jets.

## 4 b-tagging algorithms

In this section the various algorithms used in ATLAS to tag $b$-jets are explained. The spatial algorithms, built on tracks and subsequently vertices, are the most powerful ones. Most of them are based on a likelihood ratio approach, but simpler and more robust tagging algorithms are also available. Soft lepton tagging algorithms are also very important, in particular since the correlation with the previous ones is minimal. Their performance is summarized in section 4.3.

### 4.1 Spatial algorithms based on likelihood ratio

All tracks in the jet fulfilling the $b$-tagging quality cuts described in 2.1.1 are considered for the spatial $b$-tagging algorithms. In typical $t\bar{t}$ events, the average number of those tracks per light ($b$-) jet is 3.7 (5.5) and their average $p_T$ is 6.6 (6.3) GeV, respectively.

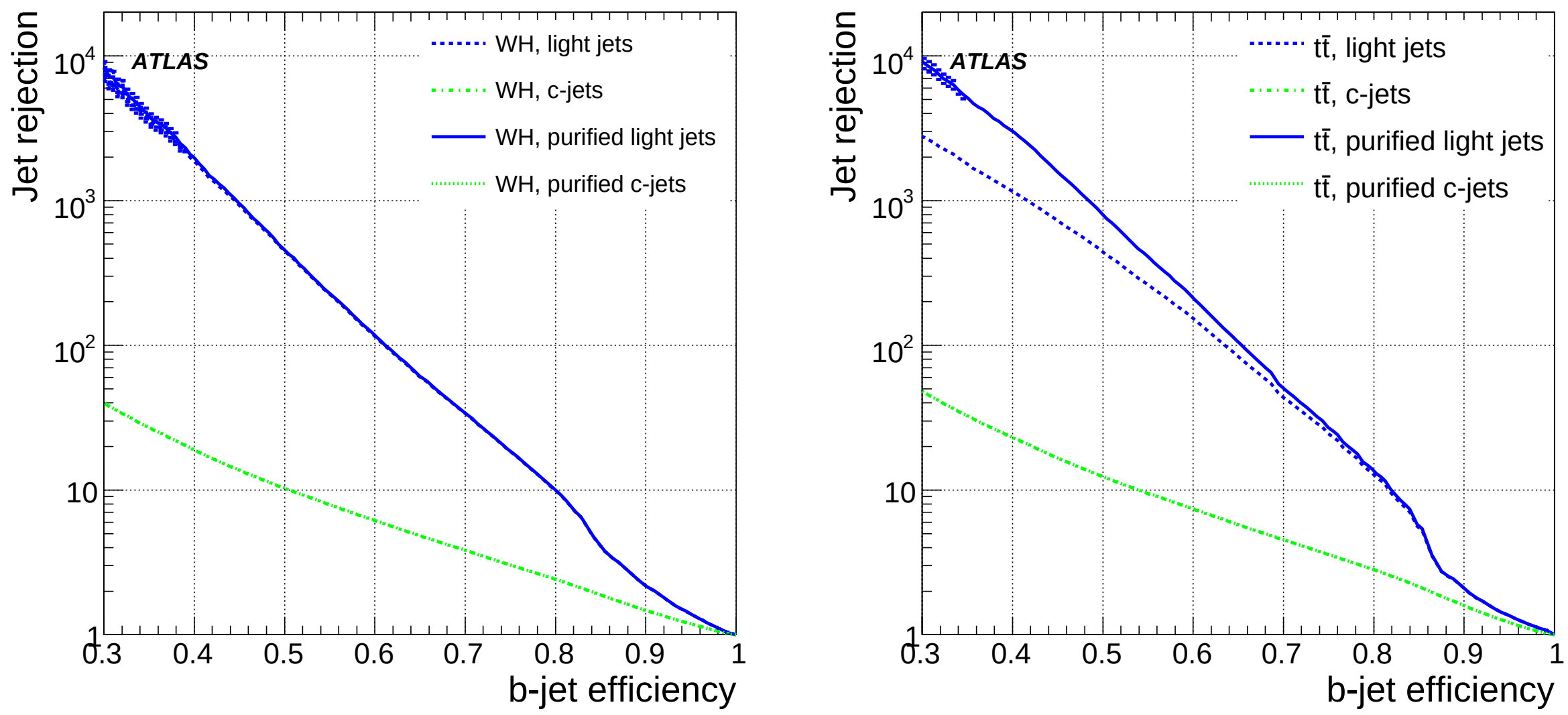

Figure 7: Rejection of light jets and *c*-jets with and without purification versus *b*-jet efficiency for $WH$ ($m_H$ =120 GeV) and $t\bar{t}$ events, using the tagging algorithm based on 3D impact parameter and secondary vertex.

#### 4.1.1 $V^0$ and secondary interactions rejection

The preselection cuts on impact parameters reject a large fraction of long-lived particles and secondary interactions. Among the remaining tracks, the ones identified by the secondary vertex search (section 4.1.3) as likely to come from $V^0$ decays are rejected (they amount to between 1% and 3% of the tracks in light and *b*-jets respectively). To do so, the search starts by building all two-track pairs that form a good vertex. The mass of the vertex is used to reject the tracks which are likely to come from $K_s, \Lambda$ decays and photon conversions. The radius of the vertex is compared to a crude description of the innermost pixel layers to reject secondary interactions in material. The cuts and performance of this selection are described in Ref. [4].

#### 4.1.2 Impact parameter tagging algorithms

For the tagging itself, the impact parameters of tracks are computed with respect to the primary vertex (cf. section 2.2). On the basis that the decay point of the *b*-hadron must lie along its flight path, the impact parameter is signed to further discriminate the tracks from *b*-hadron decay from tracks originating from the primary vertex. The sign is defined using the jet direction $\vec{P}_j$ as measured by the calorimeters (cf. section 2.3), the direction $\vec{P}_t$ and the position $\vec{X}_t$ of the track at the point of closest approach to the primary vertex and the position $\vec{X}_{pv}$ of the primary vertex:

$$\text{sign}(d_0) = (\vec{P}_j \times \vec{P}_t) \cdot \left(\vec{P}_t \times (\vec{X}_{pv} - \vec{X}_t)\right)$$

The experimental resolution generates a random sign for the tracks originating from the primary vertex, while tracks from the *b*/*c* hadron decay tend to have a positive sign. The sign of the longitudinal impact parameter $z_0$ is given by the sign of $(\eta_j - \eta_t) \times z_{0t}$ where again the $t$ subscript refers to quantities defined at the point of closest approach to the primary vertex.

The distribution of the signed transverse impact parameter $d_0$ is shown on Figure 8, left plot, for tracks coming from *b*-jets, *c*-jets and light jets. The right plot shows the significance distribution $d_0/\sigma_{d_0}$ which gives more weight to precisely measured tracks. Combining the impact parameter significances of all the tracks in the jet is the basis of the first method to tag $b$-jets. Three tagging algorithms are defined in this way: IP1D relies on the longitudinal impact parameter, IP2D on the transverse impact parameter and finally IP3D which uses two-dimensional histograms of the longitudinal versus transverse impact parameters, taking advantage of their correlations.

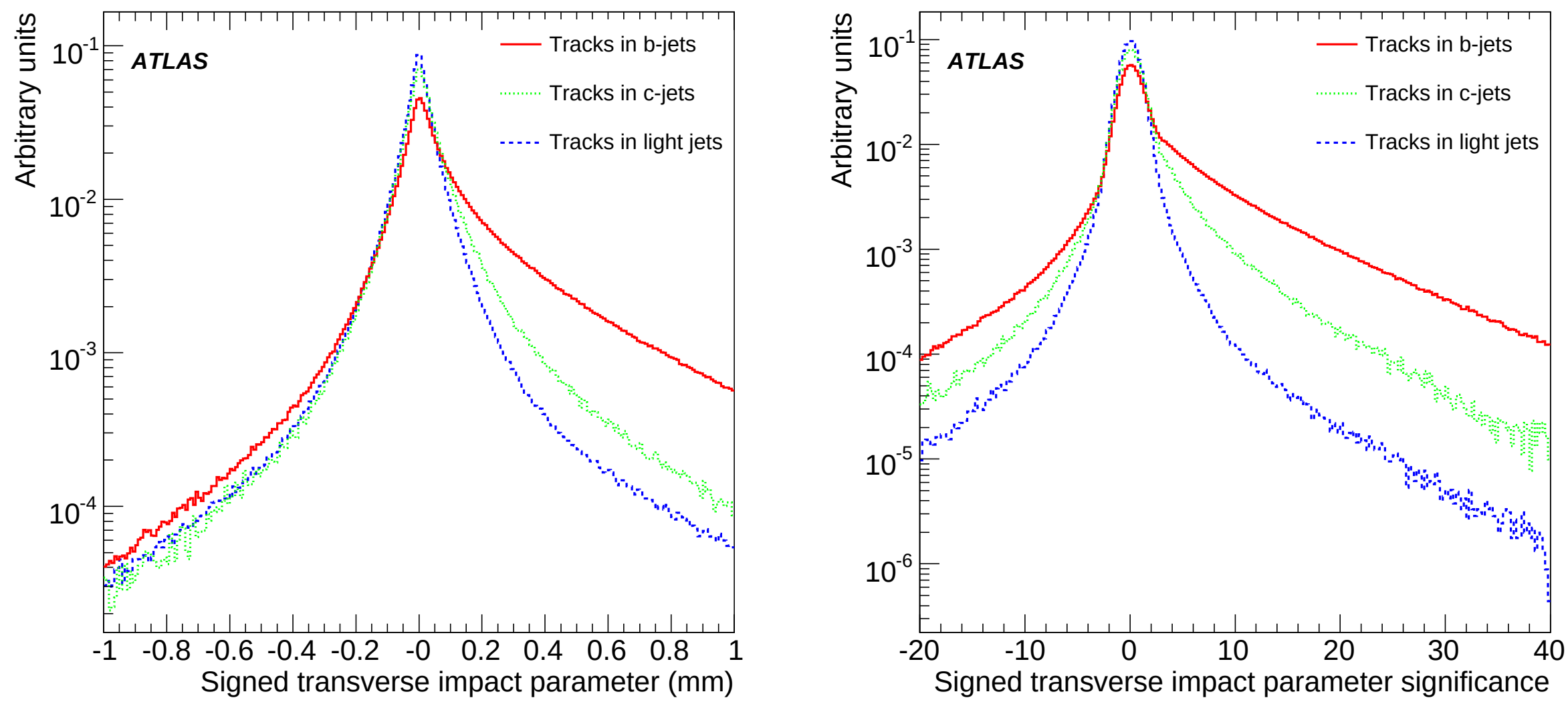

Figure 8: Signed transverse impact parameter $d_0$ distribution (left) and signed transverse impact parameter significance $d_0/\sigma_{d_0}$ distribution (right) for $b$-jets, $c$-jets and light jets.

### 4.1.3 Secondary vertex tagging algorithms

To further increase the discrimination between $b$-jets and light jets, the inclusive vertex formed by the decay products of the bottom hadron, including the products of the eventual subsequent charm hadron decay, can be sought. The reader is referred to Ref. [4] for all details. The search starts by building all two-track pairs that form a good vertex, using only tracks far enough from the primary vertex ($L_{3D}/\sigma_{L_{3D}} > 2$ where $L_{3D} \equiv \|\vec{X}_{pv} - \vec{X}_t\|$ is the three dimensional distance between the primary vertex and the point of closest approach of the track to this vertex). Vertices compatible with a $V^0$ or material interaction are rejected. All tracks from the remaining two-track vertices are combined into a single inclusive vertex, using an iterative procedure to remove the worst track until the $\chi^2$ of the vertex fit is good. Three of the vertex properties are exploited: the invariant mass of all tracks associated to the vertex, the ratio of the sum of the energies of the tracks participating to the vertex to the sum of the energies of all tracks in the jet and the number of two-track vertices. These properties are illustrated in Figure 9 for $b$-jets and light jets. The so-called SV tagging algorithms make different use of these properties: SV1 relies on a 2D-distribution of the two first variables and a 1D-distribution of the number of two-track vertices, while SV2 is based on a 3D-histogram of the three properties which requires quite some statistics. The secondary vertex finding efficiency depends in particular on the event topology, but the typical efficiency $\varepsilon_b^{SV}$ is higher than 60% in $b$-jets. The SV taggers require an a priori knowledge of $\varepsilon_b^{SV}$ and $\varepsilon_u^{SV}$.

A completely new algorithm, JetFitter, is also available, which exploits the topological structure of weak $b$- and $c$-hadron decays inside the jet. A Kalman filter is used to find a common line on which the primary vertex and the beauty and charm vertices lie, as well as their position on this line approximating the $b$-hadron flight path. With this approach, the $b$- and $c$-hadron vertices are not merged, even when only a single track is attached to each of them. The discrimination between $b$-, $c$- and light jets is based on a likelihood using similar variables to the SV tagging algorithm above, and additional variables such as the flight length significances of the vertices. This algorithm and its performance are also described in detail in Ref. [4].

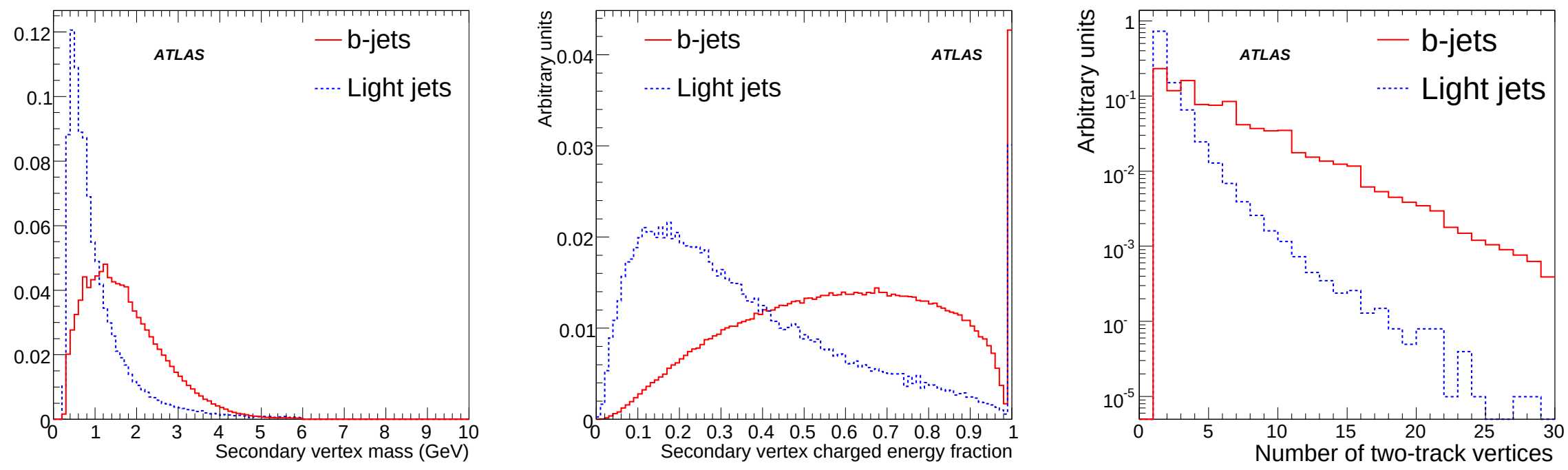

Figure 9: Secondary vertex variables: invariant mass of all tracks in vertex (left), energy fraction vertex/jet (center) and number of two-track vertices (right) for $b$-jets and light jets.

### 4.1.4 Formalism of likelihood ratio

For both the impact parameter tagging and the secondary vertex tagging, a likelihood ratio method is used: the measured value $S_i$ of a discriminating variable is compared to pre-defined smoothed and normalized distributions for both the $b$- and light jet hypotheses, $b(S_i)$ and $u(S_i)$. Two- and three-dimensional probability density functions are used as well for some tagging algorithms. The ratio of the probabilities $b(S_i)/u(S_i)$ defines the track or vertex weight, which can be combined into a jet weight $W_{Jet}$ as the sum of the logarithms of the $N_T$ individual track weights $W_i$:

$$W_{Jet} = \sum_{i=1}^{N_T} \ln W_i = \sum_{i=1}^{N_T} \ln \frac{b(S_i)}{u(S_i)} \tag{1}$$

The distribution of such a weight is shown in Figure 10 for $b$-, $c$- and light jets for two different tagging algorithms: IP2D and the sum of the weights from IP3D and SV1. When no vertex is found, the SV taggers return a weight of $\ln \frac{1-\varepsilon_b^{SV}}{1-\varepsilon_u^{SV}}$. To select $b$-jets, a cut value on $W_{Jet}$ must be chosen, corresponding to a given efficiency. The relation between the cut value and the efficiency depends on the jet transverse momentum and rapidity, and therefore is different for different samples.

### 4.1.5 Likelihood ratio and track categories

As seen already, tracks may exhibit different behavior even after the track selection, such as the tracks with shared hits (Figure 4). One idea to take advantage of the different properties of tracks is to arrange all tracks into various categories and use dedicated probability density functions for each category. The likelihood ratio formalism permits to incorporate such categories in a straightforward way. After the division of the tracks into disjoint categories $j$, where every category has its own set of reference histograms $b_j$ and $u_j$, the jet weight can simply be written as the sum over all tracks in each category $N_T^j$ and all categories $N_C$:

$$W_{Jet} = \sum_{j=1}^{N_C} \left( \sum_{i=1}^{N_T^j} \ln \frac{b_j(S_i)}{u_j(S_i)} \right) \tag{2}$$

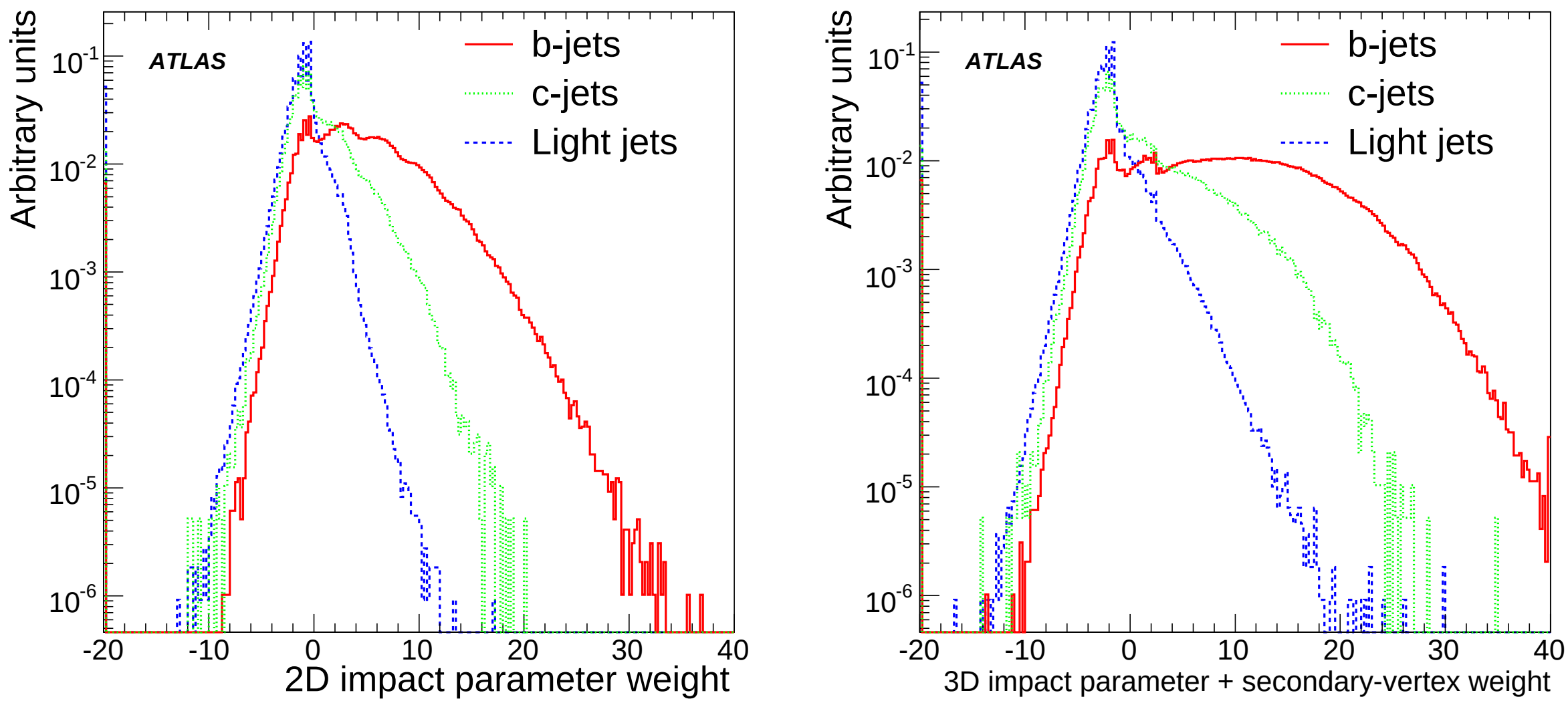

Figure 10: Jet $b$-tagging weight distribution for $b$-jets, $c$-jets and purified light jets. The left plot is for the IP2D tagging algorithm. The right plot corresponds to the IP3D+SV1 tagging algorithm.

Currently in the $b$-tagging software, two track categories are used: the *Shared* tracks (tracks with shared hits), and the complementary subset of tracks called *Good* tracks. These track categories are only used for the time being for the IP1D, IP2D and IP3D tagging algorithms.

## 4.2 Other spatial algorithms

The spatial algorithms based on likelihood ratios require an a-priori knowledge of the properties of both $b$-jets and light jets. Methods to measure them in data are being devised for the $b$-jets [8, 9] but will require at least about 100 pb$^{-1}$. In addition, there is no clear way to extract a pure enough sample of light jets, and Monte Carlo simulation will probably have to be used once a thorough validation against data has been performed. A few other spatial tagging algorithms, less powerful, are therefore developed, which have less reliance on Monte Carlo and are expected to be easier and faster to commission with the first real data.

The simplest approach that could be used, at least at the beginning, is the counting of tracks with large impact parameter or large impact parameter significance. Requiring a few of these tracks provides a sample enriched in $b$-jets. The performance of such a tagging algorithm is not discussed in this note because it is not yet fully implemented in ATLAS. Such a simple tagger may also be very useful at the trigger level.

Another approach is to combine the impact parameter of all the tracks in the jet. JetProb is an implementation of the ALEPH tagging algorithm [14], used extensively at LEP and later at the Tevatron. The signed impact parameter significance $d_0/\sigma_{d_0}$ of each selected track in the jet is compared to a resolution function $\mathscr{R}$ for prompt tracks, in order to measure the probability that the track $i$ originates from the primary vertex (Figure 11(a)):

$$\mathscr{P}_i = \int_{-\infty}^{-|d_0^i/\sigma_{d_0}^i|} \mathscr{R}(x)dx \tag{3}$$

The resolution function can be measured in data using the negative side of the signed impact parameter distribution (cf. section 6.5.1), assuming there is no contribution from heavy-flavour particles which is not strictly true.

The individual probability of each of the $N$ tracks associated to the jet are then combined to obtain a jet probability $\mathscr{P}_{jet}$ which discriminates $b$-jets against light jets (Figure 11(b)):

$$\mathscr{P}_{jet} = \mathscr{P}_0 \sum_{j=0}^{N-1} \frac{(-ln\mathscr{P}_0)^j}{j!} \tag{4}$$

where

$$\mathscr{P}_0 = \prod_{i=1}^{N} \mathscr{P}'_i \quad \text{and} \begin{cases} \mathscr{P}'_i = \frac{\mathscr{P}_i}{2} & \text{if } d_0^i > 0 \\ \mathscr{P}'_i = \left(1 - \frac{\mathscr{P}_i}{2}\right) & \text{if } d_0^i < 0 \end{cases} \tag{5}$$

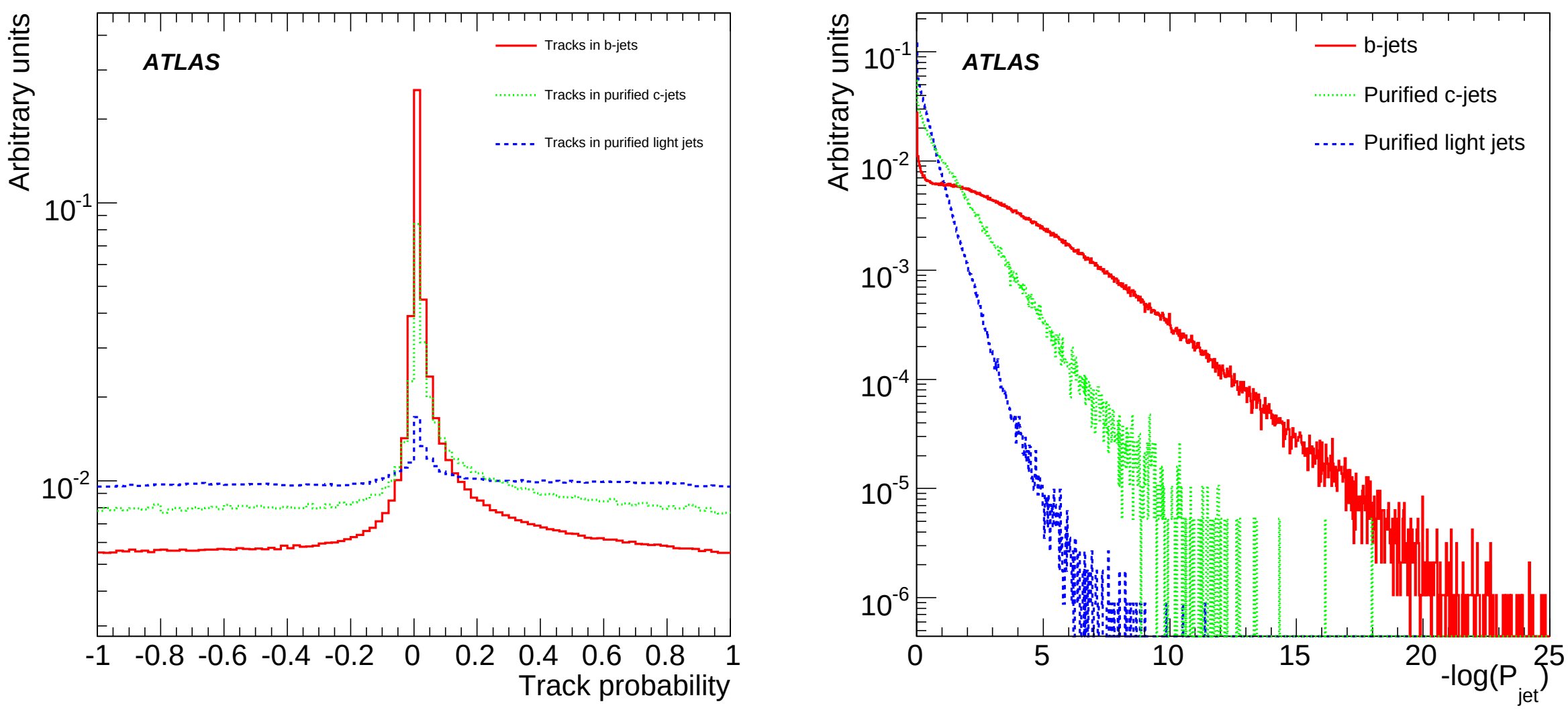

(a) Individual track probability $\mathscr{P}_i \times \mathrm{Sign}(d_0^i)$ (b) Jet probability $\mathscr{P}_{jet}$

Figure 11: Distributions of the probability of compatibility with the primary vertex for individual tracks (left plot) and for all tracks in the jet (right plot) as defined for JetProb. The cases of $b$-jets (red plain), $c$-jets (green dashed) and light jets (blue dotted line) are shown.

### 4.3 Soft lepton algorithms

Soft lepton tagging relies on the semi-leptonic decays of bottom and charm hadrons. Therefore it is intrinsically limited by the branching ratios to leptons: at most 21% [15] of $b$-jets will contain a soft lepton of a given flavour, including cascade decays of bottom to charm hadrons. However, tagging algorithms based on soft leptons exhibit very high purity and low correlations with the track-based tagging algorithms, which is very important for checking and cross-calibrating performance in data (see for instance Ref. [8]).

#### 4.3.1 Soft muons

Once a reconstructed muon is associated to a jet as explained briefly in Section 2.4, a likelihood permits to discriminate light jets from $b$-jets. The algorithm and its performance are detailed in Ref. [5] and will not be discussed further in this note. To summarize, a light jet rejection of about 300 can be achieved for a $b$-tagging efficiency of 10%. Those numbers include the semi-leptonic branching ratio, the detector acceptance, the reconstruction efficiency as well as the jet-muon association efficiency. This was estimated in $t\bar{t}$ events, including a simulation of the cavern background (low-energy neutrons and photons stemming from the interaction of forward particles with the shielding) which reduces the rejection level by about 30%. The performance is relatively steady in the jet $p_T$ range 15-100 GeV and in pseudo-rapidity.

#### 4.3.2 Soft electrons

A likelihood ratio is also used for soft electrons. The algorithm and its performance are detailed in Ref. [6] and will not be discussed further in this note. A light jet rejection of about 100 can be achieved for a $b$-tagging efficiency of 7%. The efficiency of the soft electron identification is high, since two-thirds of the true $b$-jets containing a real soft electron are tagged by the soft electron algorithm. However, about 25% of light jets are mis-tagged by real electrons from photon conversions and Dalitz decays. This was estimated in $WH$ ($m_H = 120$ GeV) events without pile-up. Based on previous study [16], a further degradation by 10% (30%) is expected when on average 4.6 (23) minimum-bias events are added. While the performance is constant in jet $p_T$ in the range 15-100 GeV, it degrades quickly with the jet pseudo-rapidity because of the higher amount of dead material, the poor performance in the transition region between the barrel and end-cap cryostats of the electromagnetic calorimeter ($1.37 < |\eta| < 1.52$) and the absence of the TRT beyond $|\eta| > 2$.

### 4.4 Combining tagging algorithms

Currently only the likelihood-based tagging algorithms have been combined, since the formalism is easy in this case: the weights of the individual tagging algorithms are simply summed up. The most commonly used tagging algorithm, IP3D+SV1, is actually such a combination. It should be noted that the SV tagging algorithms have been optimized to work in conjunction with the IP ones. Another one combines IP3D and JetFitter. Multivariate approaches to combine all tagging algorithms, including the soft lepton ones, have not received much attention so far. There are, however, some new studies and the use for instance of boosted decision trees is discussed in Section 7.3.

### 4.5 Calibration of tagging algorithms

The likelihood-based tagging algorithms require knowledge of the probability density functions of the discriminating variables for both the $b$- and light jet hypotheses: this is called the calibration of the tagging algorithms, or their reference histograms. In the following, those functions have been derived from a large sample of jets coming from $t\bar{t}$ and $t\bar{t}jj$ events. Several issues about the calibration and its impact on $b$-tagging performance are discussed in Section 6.5.

## 5 Performance for various physics processes

In this section, the $b$-tagging performance is reviewed for several physics channels of interest. Several spatial tagging algorithms are considered: JetProb and IP2D which are best suited for the initial period, IP3D and then IP3D+SV1 for regular operations once the secondary vertexing is understood and finally IP3D+JetFitter for the ultimate performance.

### 5.1 Dependence on jet transverse momentum and pseudo-rapidity

The spatial $b$-tagging performance depends strongly on the jet momentum and rapidity: the $p_T$ and $\eta$ dependencies of the $b$-tagging efficiency and light jet rejection for a given cut on the $b$-tagging weight are shown in Figure 12. At high $p_T$ or at high $|\eta|$, the $b$-jet tagging performance is poor, regardless of which tagging algorithm is used. At low $p_T$, maintaining a reasonable $b$-jet efficiency is possible only by loosening the cut on the weight, at the price of a very low rejection of light jets. The strong dependence, especially in $p_T$, makes the extraction of the $b$-jet efficiency from data complicated and means that more integrated luminosity will be required, since several bins are needed.

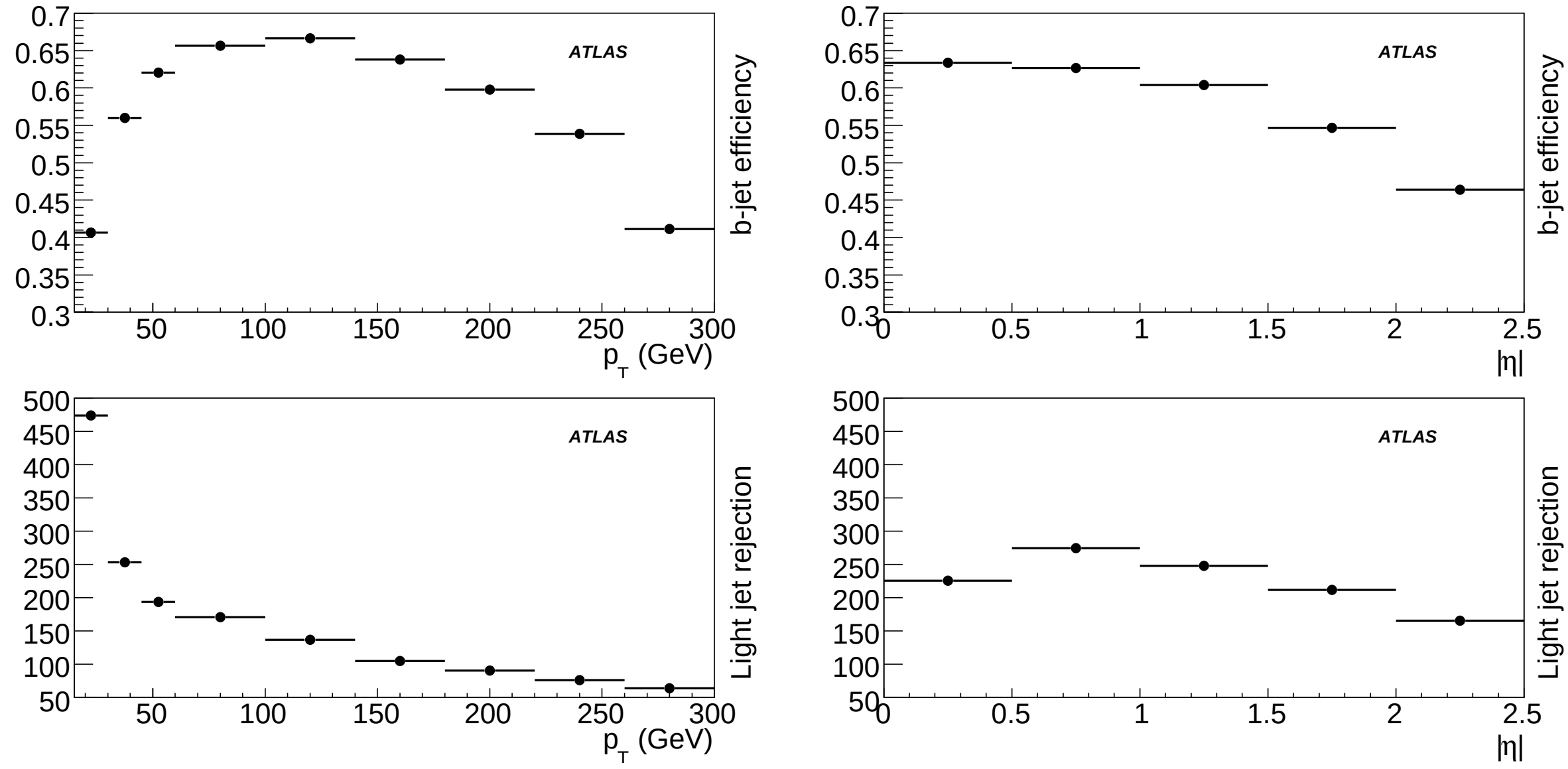

(a) versus jet $p_T$ (b) versus jet $|\eta|$

Figure 12: $b$-tagging efficiency and purified light jet rejection obtained with the IP3D+SV1 tagging algorithm operating at a fixed cut of 4 on the $b$-tagging weight, for $t\bar{t}$ events.

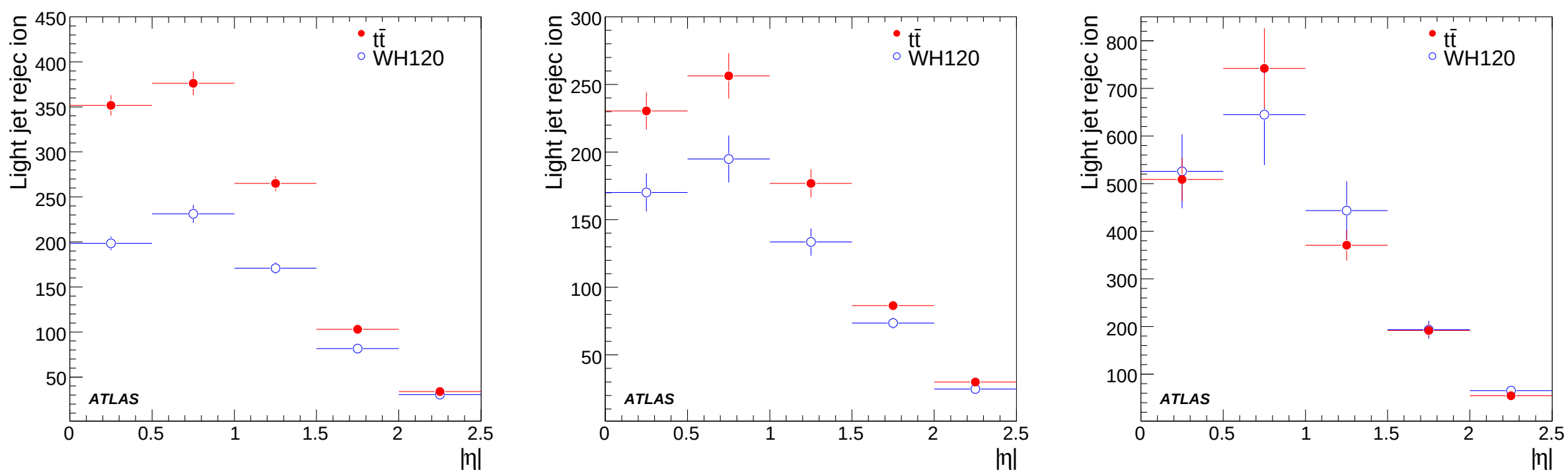

(a) All jet $p_T$ (b) Jets with $30 < p_T < 45$ GeV (c) Jets with $60 < p_T < 100$ GeV

Figure 13: Rejection of light jets with purification versus jet $\eta$ for the IP3D+SV1 tagging algorithm and for two different physics channels: jets from $t\bar{t}$ events and from $WH$ ($m_H = 120$ GeV) events, for a fixed 60% tagging efficiency in each bin.

Because of these strong $p_T$ and $\eta$ dependencies, and since various samples have very different spectra, it is not straightforward to compare between channels the integrated rejection numbers shown in the following. It is worth noting that this dependence is really a two-dimensional one, thus Figure 12 is useful for illustrative purposes but the $(p_T, |\eta|)$ spectrum of the jets in the sample considered is not properly factorized out. This is important for instance to parametrize the $b$-tagging performance. This is further illustrated in Figure 13: the rejections achieved in two different samples become more similar as a function of $\eta$ when looking in bins of $p_T$. The remaining differences are mostly because the binning in $p_T$ is still too large, but also because of other minor differences between the samples: for example they have been generated with different Monte Carlo generators (cf. section 6.6).

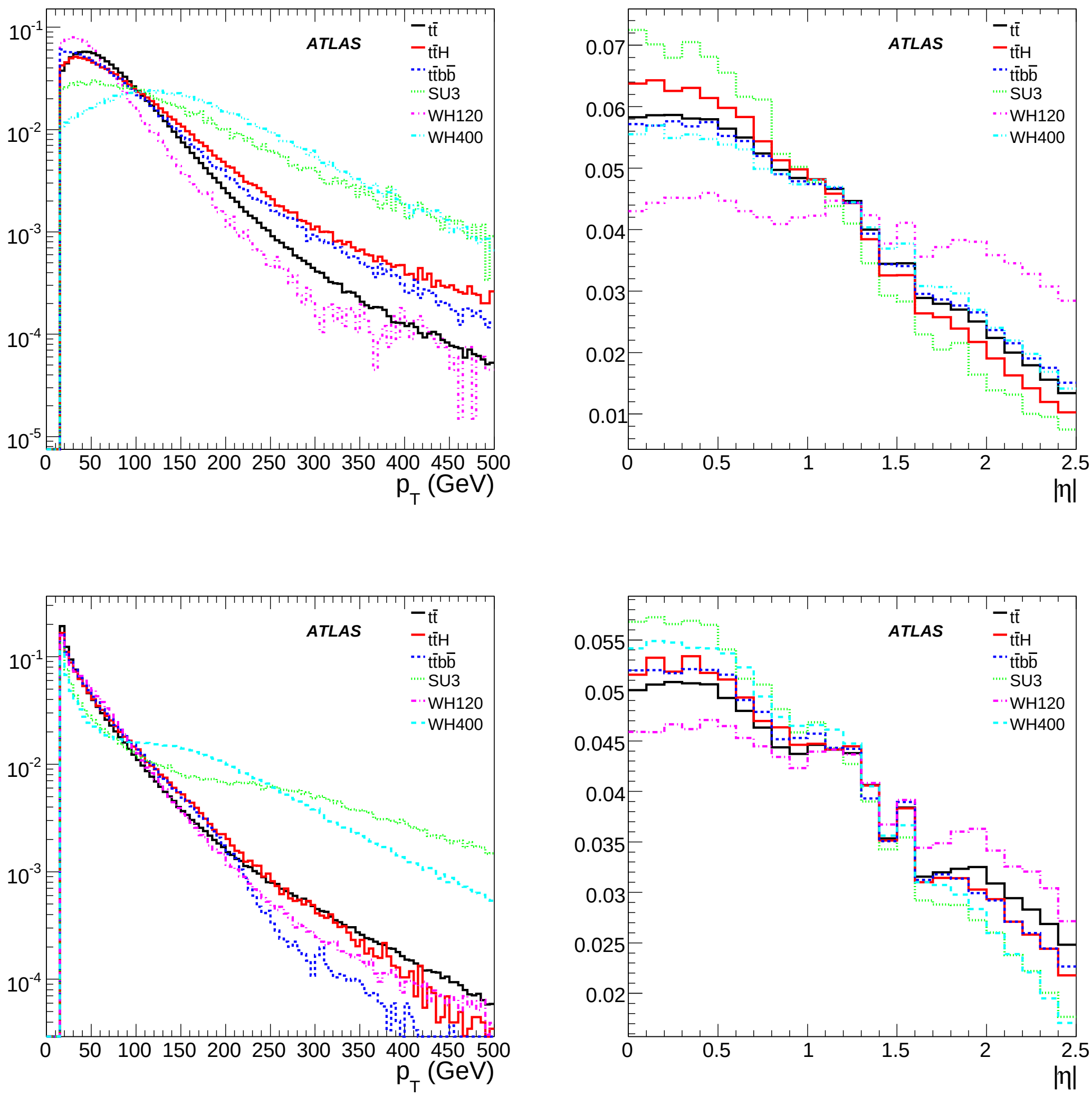

Figure 14: $p_T$ and $|\eta|$ spectra of b (upper plots) and light (lower plots) jets for the various channels considered in this section.

For reference, the $p_T$ and $|\eta|$ spectra of $b$ and light jets in the various samples used in the following are shown on Figure 14. They affect the integrated rejections for the various channels.

## 5.2 Simple topologies: WH channels

This first class of events illustrates the performance obtained on simple event topologies where the jet multiplicity is very low. As discussed in Section 3.3, purification is not an issue in this case.

Events from Higgs boson production in association with a $W$ boson are interesting in this respect and are a benchmark for $b$-tagging performance, even though the channel itself is no longer thought to be very promising at the LHC (see however Ref. [17] for a recent re-investigation). The $W$ decays leptonically, and there are only two jets coming from the hard process, originating from the $H$ decay. To study the $b$-tagging efficiency the decay $H \to b\bar{b}$ is simulated while here for the rejection of charmed and light jets the Higgs boson is forced to decay to $c\bar{c}$ or to the unlikely $u\bar{u}$ channel respectively.

The $b$-tagging performance obtained on this kind of events and for $m_H = 120$ GeV is shown in Table 1 for the tagging algorithms considered. Two typical $b$-tagging efficiencies were considered: 50% and 60%. For each tagging algorithm, the cuts on the weight required to achieve these efficiencies were determined over the whole sample, and then applied to estimate the rejections.

To study more energetic jets, similar physics processes have been considered for a different Higgs boson mass, $m_H$=400 GeV (again such a choice is unphysical since a 400 GeV Higgs boson would not decay to $b\bar{b}$ but is useful for these studies). The results are shown in Table 1 for the light jet rejection and in Table 2 for the $c$-jet rejection. The differences in performance between the two mass cases are the result of the different $p_T$ and $\eta$ spectra: the jets for $m_H = 400$ GeV are more energetic and explain most of the discrepancy, this effect being only slightly balanced by the fact that jets for $m_H = 120$ GeV are more forward. For this channel, the gain obtained with JetFitter is more visible.

## 5.3 Multi-jets channels: the top case

The jet multiplicity in pair-produced top quark events is much higher. In the following, the channels where at least one of the $W$ bosons decays to leptons are considered. For the dominant lepton+jets channel, there are usually at least four jets from the hard process and extra jets from radiation. Several flavours of jets are present at the same time in the event: two $b$-jets from the top quarks, light jet(s) and often a $c$-jet from the $W$ decaying hadronically. This increases the likelihood of having light jets contaminated with heavy flavour and also makes the labelling of jets even more ambiguous as discussed previously. The benchmark curves of jet rejection versus $b$-tagging efficiency are shown on Figures 15(a) and 15(b), for light jets and for several tagging algorithms. The jets of the various flavours were taken from the same sample in this case, unlike for events of the $WH$ channels. Table 1 shows the light jet rejection achieved in $t\bar{t}$ events and in $t\bar{t}jj$ events, both samples being generated with MC@NLO+HERWIG. The latter events are $t\bar{t}$ events which were filtered in order to have at least six jets, of which four are taggable. Since the performance in those two samples is similar, they have been merged. For light jets, both the raw (without purification) and purified rejections are shown. For $c$-jets and $\tau$-jets the purification does not make any significant difference. The rejection power of $c$-jets (Table 2 and Figure 15(c)) is naturally very limited because of the lifetime of $c$-hadrons and is almost independent of the physics process. Without any optimization, the $b$-tagging algorithms also prove to be useful for the identification of $\tau$-jets, as shown on Figure 15(d). The small discontinuities in the curves on Figure 15, visible notably for the IP3D tagger, are due to the conjunction of a coarser binning of the underlying probability density functions for this tagger and the presence of single-track jets (notably electrons faking jets). This effect is more pronounced for the $\tau$-jets (Figure 15(d)) where single-prong decays are abundant.

The impact on the light jet rejection of electrons faking jets can be seen in the fourth block of Table 1. Electron jets are seldom mis-tagged as $b$-jets, since they have usually a single prompt high-$p_T$ track which is well-measured. In this sample, the high-$p_T$ electrons are coming from $W \to e\nu$ or indirectly from $W \to \tau\nu$. In the fourth part of Table 1, a jet $j$ is considered as an electron faking a jet, and therefore discarded, if it matches with a reconstructed electron candidate $e$: $\Delta R(e, j) < 0.1$ and $E_T(e)/E_T(j) > 0.75$.

It is interesting to notice that, despite the more complex topology of these $t\bar{t}$ events, the integrated light-jet rejection achieved is higher than for the $WH$ ($m_H$=120 GeV) case. This is mostly because jets in $t\bar{t}$ events are more central than the ones in $WH$ ($m_H$=120 GeV) production (cf. Figure 14), and the $b$-tagging performance degrades quickly at large pseudo-rapidities as seen already.

Table 1 shows the light jet rejection achieved in even more complex topologies with at least six jets. Those channels are relevant for the Higgs discovery channel $t\bar{t}H(b\bar{b})$ which requires a high $b$-tagging efficiency since four jets are $b$-tagged and the cross section is low: therefore the more typical working points of $\varepsilon_b$ around 60-70% are shown. As shown in Figure 14, the $p_T$ spectrum for the $b$-jets in these samples is harder than for $b$-jets in $t\bar{t}$ events, explaining partly the differences in performance. In addition the high $b$-jet multiplicity in $t\bar{t}H$ events leads to some lifetime contamination in the few light jets available in this sample. All these samples are based on PYTHIA Monte Carlo, unlike the $t\bar{t}jj$ sample which is a background for this channel as well but is based on MC@NLO+HERWIG Monte Carlo and was kept separate for this reason.

Table 1: Integrated rejection of light jets (with and without purification when it applies), for various event types and for several tagging algorithms. For each case, the cut on the $b$-tagging weight is chosen to lead to the quoted average $b$-tagging efficiency $\varepsilon_b$ over the sample considered. The quoted errors are statistical only.

| | JetProb | IP2D | IP3D | IP3D+SV1 | IP3D+JetFitter |
|---|---|---|---|---|---|
| $WH$ ($m_H = 120$ GeV) events | | | | | |
| $\varepsilon_b = 50\%$ | 83±1 | 116±2 | 190±3 | 458±13 | 555±17 |
| $\varepsilon_b = 60\%$ | 30±0 | 42±0 | 59±1 | 117±2 | 134±2 |
| $WH$ ($m_H = 400$ GeV) events | | | | | |
| $\varepsilon_b = 50\%$ | 73±1 | 163±3 | 179±3 | 298±7 | 396±11 |
| $\varepsilon_b = 60\%$ | 27±0 | 56±1 | 58±1 | 96±1 | 123±2 |
| $t\bar{t}$ and $t\bar{t}jj$ events | | | | | |
| Raw, $\varepsilon_b = 50\%$ | 91±0 | 146±1 | 232±2 | 456±4 | 635±7 |
| Purified, $\varepsilon_b = 50\%$ | 97±0 | 186±1 | 310±3 | 789±10 | 924±13 |
| Raw, $\varepsilon_b = 60\%$ | 28±0 | 46±0 | 67±0 | 154±1 | 189±1 |
| Purified, $\varepsilon_b = 60\%$ | 28±0 | 51±0 | 76±0 | 206±1 | 224±2 |
| $t\bar{t}$ and $t\bar{t}jj$ events, once electrons faking jets are removed | | | | | |
| Raw, $\varepsilon_b = 50\%$ | 92±0 | 142±1 | 219±1 | 423±4 | 593±6 |
| Purified, $\varepsilon_b = 50\%$ | 99±0 | 181±1 | 293±2 | 732±10 | 863±12 |
| Raw, $\varepsilon_b = 60\%$ | 31±0 | 49±0 | 67±0 | 144±1 | 180±1 |
| Purified, $\varepsilon_b = 60\%$ | 33±0 | 56±0 | 76±0 | 194±1 | 213±2 |
| $t\bar{t}H$ events | | | | | |
| Raw, $\varepsilon_b = 60\%$ | 23±0 | 35±0 | 49±1 | 90±2 | 113±2 |
| Purified, $\varepsilon_b = 60\%$ | 25±0 | 48±1 | 72±1 | 188±5 | 188±5 |
| Raw,$\varepsilon_b = 70\%$ | 10±0 | 14±0 | 18±0 | 32±0 | 31±0 |
| Purified, $\varepsilon_b = 70\%$ | 11±0 | 17±0 | 22±0 | 46±1 | 37±1 |
| $t\bar{t}bb$ events | | | | | |
| Raw, $\varepsilon_b = 60\%$ | 23±0 | 34±0 | 50±1 | 100±2 | 123±2 |
| Purified, $\varepsilon_b = 60\%$ | 24±0 | 41±0 | 64±1 | 156±4 | 166±4 |
| Raw, $\varepsilon_b = 70\%$ | 10±0 | 13±0 | 18±0 | 32±0 | 28±0 |
| Purified, $\varepsilon_b = 70\%$ | 10±0 | 15±0 | 20±0 | 40±0 | 31±0 |
| SUSY SU3 events | | | | | |
| Raw, $\varepsilon_b = 50\%$ | 66±1 | 140±4 | 162±5 | 246±9 | 328±14 |
| Purified, $\varepsilon_b = 50\%$ | 68±1 | 161±5 | 183±6 | 290±13 | 375±19 |
| Raw, $\varepsilon_b = 60\%$ | 24±0 | 50±1 | 55±1 | 89±2 | 110±3 |
| Purified, $\varepsilon_b = 60\%$ | 25±0 | 53±1 | 58±1 | 99±3 | 117±3 |

Table 2: Integrated rejection of $c$- and $\tau$-jets, for various event types and for several tagging algorithms. For each case, the cut on the $b$-tagging weight is chosen to lead to the quoted average $b$-tagging efficiency $\varepsilon_b$ over the sample considered.

| | JetProb | IP2D | IP3D | IP3D+SV1 | IP3D+JetFitter |
|---|---|---|---|---|---|
| $c$-jet rejection for $WH$ ($m_H = 400$ GeV) events | | | | | |
| $\varepsilon_b = 50\%$ | 7.9±0.1 | 9.7±0.1 | 10.7±0.2 | 12.4±0.2 | 12.7±0.2 |
| $\varepsilon_b = 60\%$ | 4.7±0.0 | 5.7±0.1 | 6.1±0.1 | 6.8±0.1 | 7.3±0.1 |
| $c$-jet rejection for $t\bar{t}$ and $t\bar{t}jj$ events | | | | | |
| $\varepsilon_b = 50\%$ | 8.4±0.0 | 9.5±0.0 | 10.6±0.0 | 12.4±0.1 | 12.3±0.1 |
| $\varepsilon_b = 60\%$ | 5.1±0.0 | 5.8±0.0 | 6.5±0.0 | 7.4±0.0 | 7.4±0.0 |
| $\tau$-jet rejection for $t\bar{t}$ and $t\bar{t}jj$ events | | | | | |
| $\varepsilon_b = 50\%$ | 10.2±0.1 | 13.9±0.1 | 20.3±0.2 | 45.2±0.8 | 36.9±0.6 |
| $\varepsilon_b = 60\%$ | 5.1±0.0 | 6.4±0.0 | 8.0±0.1 | 24.6±0.3 | 19.3±0.2 |

## 5.4 High-$p_T$ jets: SUSY

Events from the SUSY bulk region (SU3 point, see Ref. [1]) were considered. In these events, the average taggable jet multiplicity is about 5.3 and a large number of $\tau$-leptons are produced in the decay chain of charginos and neutralinos. There are on average 0.6 $b$-jets per event, with a relatively hard $p_T$ spectrum as shown on Fig. 14: the average $p_T$ is 144 GeV. On average about 0.6 taggable jets per event are labelled as $\tau$, compared to 0.2 in the semi-leptonic $t\bar{t}$ channel. They are not considered as light jets. The results are shown in Tables 1 and 2: because the $p_T$ of the jets is quite high, the light jet rejection is similar to that achieved for $WH$ ($m_H = 400$ GeV) events but significantly worse than for the other channels.

## 5.5 Degradation of performance at low and high $p_T$

At low $p_T$, performance is degraded mostly because multiple scattering is increased. This also holds for the high $|\eta|$ region, where the amount of material in the tracking region increases very significantly, inducing more secondary interactions. There is currently no rejection of secondary interactions found in the pixel disks, unlike in the barrel (cf. Section 4.1.3). More importantly, the increase of the extrapolation distance from the $b$-layer to the primary vertex at large pseudo-rapidities significantly degrades the $z_0$ resolution as seen in Figure 6(b).

Several effects conspire to reduce the $b$-tagging performance as the jet $p_T$ increases above 120 GeV. First of all, the fraction of fragmentation tracks increases with the parton transverse momentum, as shown in Figure 16, while the jet is collimated into a narrower cone: since a fixed-size cone is currently used to associate tracks to the jet this leads to a dilution of the discriminating power in $b$-jets. The density of tracks in the core of energetic jets challenges the pattern-recognition ability of the software and of the inner detector itself, leading to either a reduced tracking efficiency or a high level of fakes as shown in Figure 3 for jets with $E_T > 100$ GeV. Finally, for very energetic $b$-hadrons the Lorentz boost leads to a much enhanced decay length. The typical $c\tau$ ($\sim 450\mu$m) of $b$-hadrons is thus scaled by a factor $\gamma \sim |p_B|/m_B$ which can be large. For high $p_T$ jets, the $b$-hadron can decay at a rather large radius $R_B$, as illustrated in Table 3: close to the inner radius of the pixel detector, leading to more tracking ambiguities in the first detection layers, or even after the first pixel layer. In the latter case, the current requirement (for the IPnD and JetProb tagging algorithms) of a hit on the $b$-layer actually kills the signal. At very high $p_T$ (above 500 GeV), these effects become so critical that a dedicated strategy has to be devised, as discussed in the next section. In the current simulation, the $b$-hadrons decaying at a radius larger than the beam-pipe or the $b$-layer radii do not interact with these objects, while in real events this will even more reduce the performance.

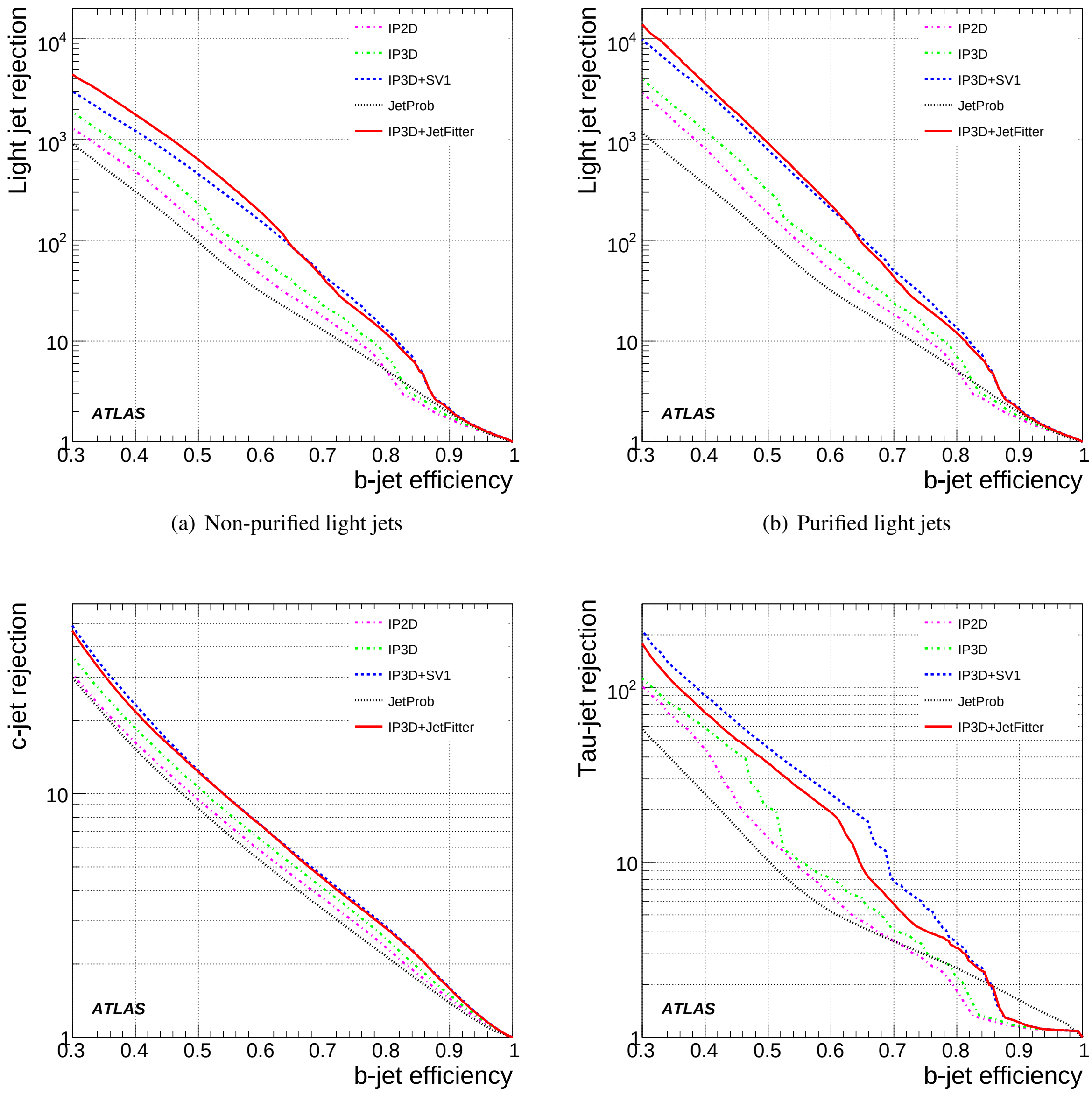

(a) Non-purified light jets

(b) Purified light jets

(c) $c$-jets

(d) $\tau$-jets

Figure 15: Rejection of light jets, $c$- and $\tau$-jets versus $b$-jet efficiency for $t\bar{t}$ and $t\bar{t}jj$ events and for all tagging algorithms: JetProb, IP2D, IP3D, IP3D+SV1, IP3D+JetFitter.

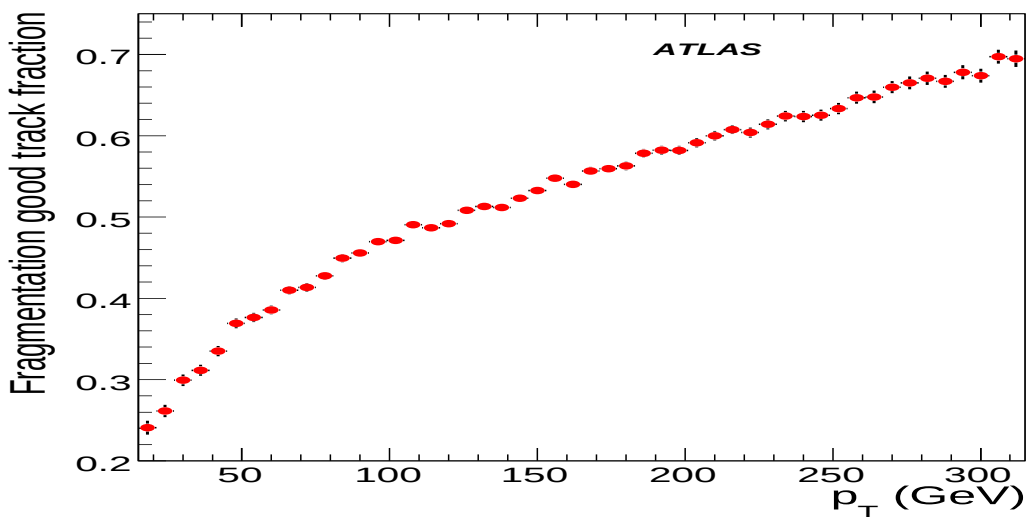

Figure 16: Fraction of selected tracks which are not from B/D decays versus jet $p_T$, in $b$-jets from $WH$ ($m_H = 400$ GeV) events.

Table 3: Fraction of $b$-jets in $WH$ ($m_H = 400$ GeV) events for which the $b$-hadron decays beyond the beam-pipe vacuum (first column) or beyond the $b$-layer (second column).

| | $R_B > 2.9$ cm | $R_B > 5.1$ cm |
|---|---|---|
| all $E_T$ | 9.0% | 2.8% |
| $E_T > 100$ GeV | 12.2% | 3.9% |
| $E_T > 200$ GeV | 21.1% | 7.9% |

### 5.6 The case of very high-$p_T$ jets: exotic physics

The very high $p_T$ range is defined as jets exceeding a transverse energy of 500 GeV. Identification of such very high $p_T$ $b$-jets is required for the search for heavy resonances with (predominantly) hadronic decays. A large number of exotic physics models presents signatures with very high $p_T$ $b$-jets, up to a few TeV. An example is the decay $Z_H \to Zh$ in the little Higgs model [18], where $Z \to e^+e^-$ and $h \to b\bar{b}$.

In the following, three samples corresponding to the process $Z' \to q\bar{q}$, where $q$ denotes $u$, $b$ and $c$ quarks respectively were used. The $Z'$ mass is chosen to be 2 TeV, so that the primary partons have transverse momenta in the range from 300 GeV to 1 TeV.

The $b$-tagging algorithms rely particularly on the determination of the jet axis as an approximation of the $b$- and $c$-hadron flight direction. The difference between the jet pseudo-rapidity and the true $b$-hadron direction exhibits a narrow Gaussian core. The widths of the core range are $\sigma_\eta \approx 0.025$ and $\sigma_\phi \approx 0.010$ mrad, with a moderate dependence on jet $E_T$ and pseudo-rapidity. Non-gaussian tails give rise to large RMS: RMS $_{\eta,\phi} \approx 0.050$. The particles from the decay of the highly boosted $b$-hadron are emitted at very small angles. Up to half of the tracks from the $b$-hadron decay lie within the azimuthal angle between the $b$-hadron and the jet. In these cases the sign of the impact parameter cannot be determined accurately.

The reconstruction of tracks in high $p_T$ jets presents a series of specific challenges, as already explained in Section 5.5. Particles are associated to reconstructed jets using a $\Delta R < 0.4$ criterion. To highlight reconstruction effects only true pions reaching the outer radius of the tracker are taken into account. Particles are moreover required to originate from a well-defined vertex: either the primary vertex of the event or the $b$/$c$-hadron decay vertex. Only the first level of the efficiency, i.e. the matching to hits (Section 2.1.2), is considered. With this definition the efficiency on a reference sample of low $p_T$ jets is close to 100% and essentially independent of the jet energy and of the distance to the jet axis. In Figure 17 the track efficiency in the $Z'$ sample is plotted versus the jet transverse energy, for the default track reconstruction - NewTracking - and for iPatRec.

The algorithmic reconstruction efficiency for prompt tracks is only slightly degraded. Even in the very harsh environment of a 1 TeV jet, the efficiency is approximately 90%. The degradation is most pronounced in the core of the jet ( $\Delta R < 0.1$ ). For pions originating in $b/c$-hadron decays a much more significant degradation of the efficiency towards high jet $E_T$ is observed. For 1 TeV jets the efficiency is approximately 50%. The efficiency shows a strong dependence on the decay vertex radius. It is worth mentioning that NewTracking and iPatRec assign very different errors to the positions defined by large pixel or SCT clusters arising in the inner layers with such dense jets: the former assigns a very small error assuming only one particle was involved in the cluster, while the latter is assuming the opposite and assigns the maximal error (cluster width/$\sqrt{12}$).

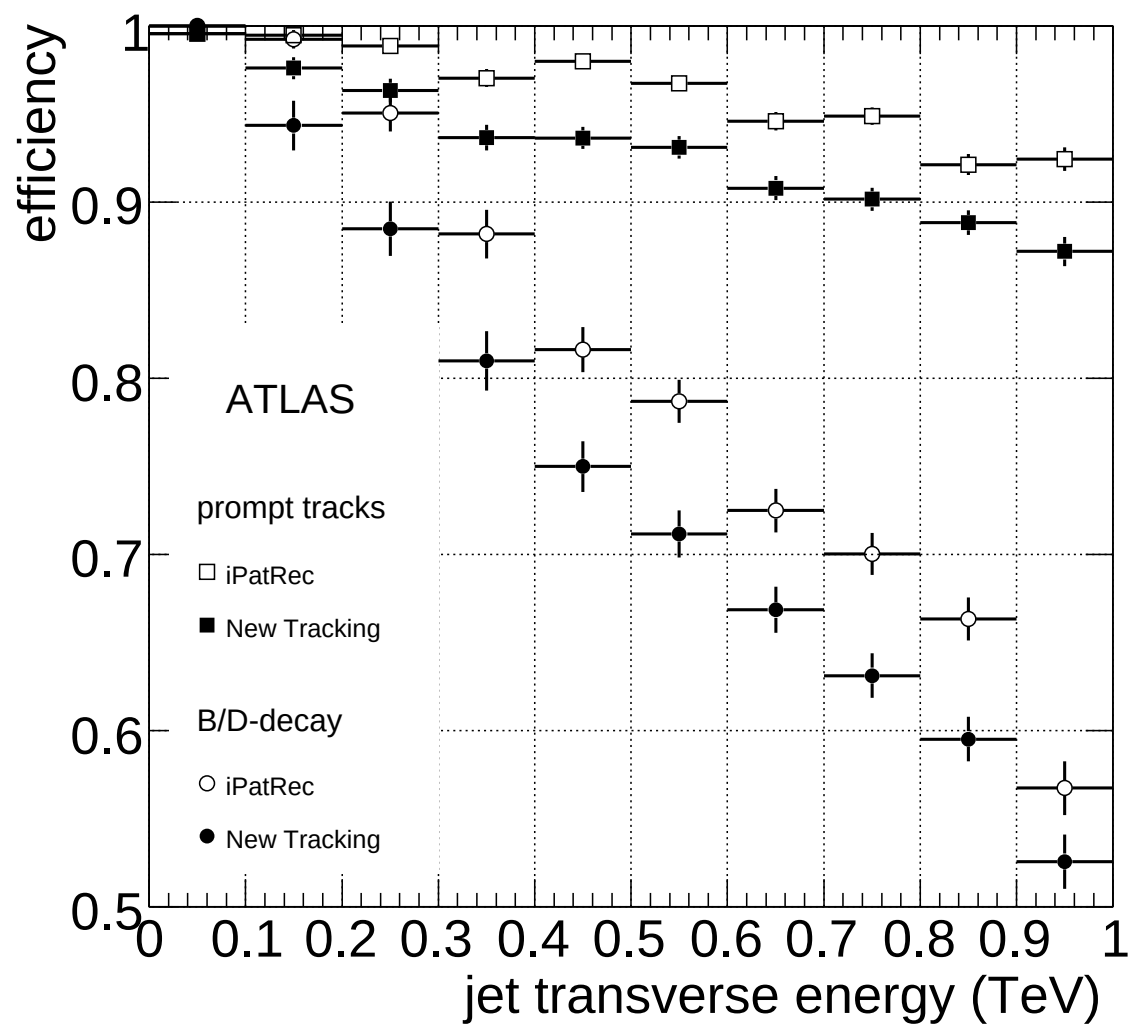

Figure 17: Algorithmic tracking efficiency for long-lived pions as a function of the jet transverse energy, for prompt tracks or tracks from $b/c$-hadron decays, and for two reconstruction algorithms: NewTracking and iPatRec.

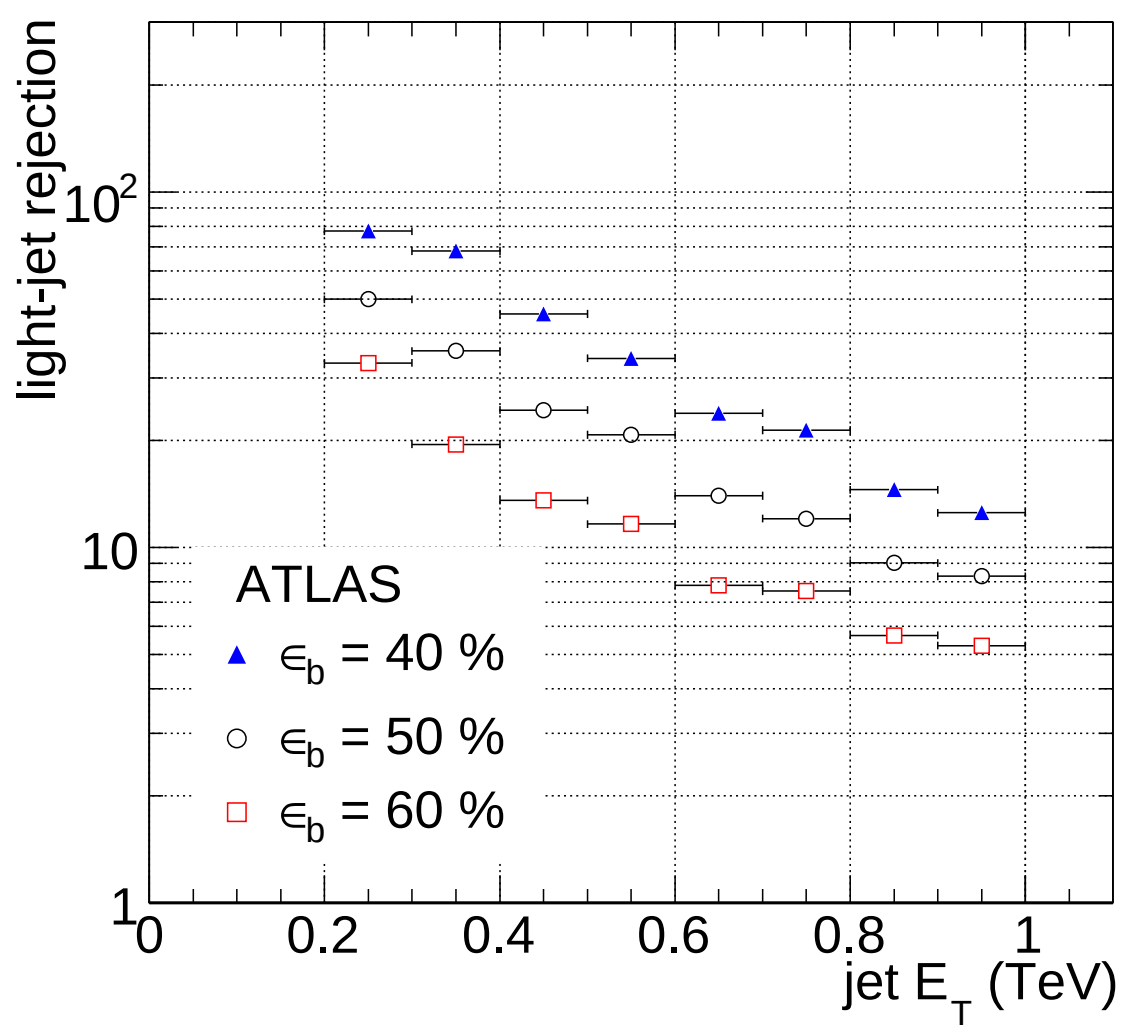

Figure 18: Raw light jet rejection for jets from $Z' \to q\bar{q}$ with $m_{Z'} = 2$ TeV, versus jet transverse energy, for the IP3D+SV1 tagging algorithm (with tracks from iPatRec). and for three $b$-tagging efficiencies.

In very high $p_T$ jets the probability that two or more tracks share a hit is very high ($> 10\%$), unlike what was seen for low and moderate $p_T$ jets (cf. Figure 4). The number of shared hits per track - evaluated for individual sub-detectors, or for the complete silicon tracker - is a factor 2-3 larger in iPatRec. For particles from very displaced $b/c$-hadron decay vertices reconstructed with iPatRec, tracks with a shared hit in the $b$-layer actually outnumber the tracks with an unambiguous assignment. Again the two tracking algorithms have made opposite choices for their working point: NewTracking considers shared hits are stemming from pattern-recognition errors and tries to assign them to the best track, which is not necessarily meaningful when the cluster is really originating from several near-by particles, while iPatRec does not try to resolve the ambiguity.

The high multiplicity of fragmentation tracks, the degraded tracking efficiency, the ambiguities in hit assignment particularly in the innermost layer and the uncertainty in the impact parameter sign, all render high $p_T$ jets a harsh environment. This may be improved though by the use of dedicated reconstruction algorithms. A priori, the current simulation should also be updated to transport properly the $b$-hadrons through the material. While studies have started, they are beyond the scope of this note. For the time being, a re-optimization of the default tagging algorithm parameters has been performed with the aim of improving the $b$-tagging performance over a large jet $E_T$ range (from 200 GeV to 1 TeV). The cone size for the jet-to-track association was reduced to 0.2 (see Section 7.2 for a possible improvement of the current treatment), the $p_T$ cut on tracks raised to 5 GeV and the use of tracks without a hit in the $b$-layer was allowed (see also Section 7.1).

The tagging performance on iPatRec tracks is found to be on average 60% better than for the default algorithm. The choice of iPatRec to maintain high tracking efficiency inside dense jets (cf. Figure 3), even at a price of higher fake rates, seems to be instrumental in achieving better performance here. The resulting $b$-tagging performance is presented in Figure 18. Given the modest level of rejection achieved, a tagging efficiency of 40% is considered here. Without any tuning, the rejection level of the IP3D+SV1 tagging algorithm would be about three times worse for the same $b$-tagging efficiency.

In this study, the standard (low $p_T$) reference histograms were used for the tagging algorithms. So far, none of the methods developed to extract the calibration histograms from data has been shown to work for these very high $p_T$ jets. Reference histograms for very high $p_T$ jets may be extracted from Monte Carlo simulation, provided it reliably describes the data. Doing so, a modest improvement of the performance (up to 50% higher light jet rejection for the same b-tagging efficiency compared to the results shown here) can be achieved.

To conclude, $b$-tagging for very high $p_T$ jets faces a series of specific difficulties. This study demonstrated that a rejection between 10 and 70 for jets with $p_T > 500$ GeV can be achieved by tuning the current algorithms. Further improvements require dedicated treatments at the clustering level (with probably a second-pass approach to break down large clusters coming from near-by particles) and at the pattern-recognition stage of the track reconstruction. On small preselected datasets, retracking with a specially optimized pattern-recognition algorithm should be possible.

# 6 Specific studies to characterize $b$-tagging performance

In this section, a few additional studies aimed at better understanding some critical aspects of the $b$-tagging performance are detailed. Those studies are described in a separate section because either they required specific datasets or they rely on software and/or cuts/optimizations which are different from the ones currently in use in the ATLAS software.

## 6.1 Impact of residual misalignments

All the studies in this note do not take into account the effect of residual misalignments. While all the samples studied were simulated with misalignments, they were reconstructed assuming a perfect knowledge of those misalignments. However, detailed studies, discussed in Ref. [7], are in progress on this subject. Two different approaches were used: residual misalignment sets and actual realignment. In the former, the events simulated with misalignments are reconstructed using the knowledge of the misalignments, but the true detector elements positions are shifted and/or rotated slightly from their actual position to mimic residual misalignments. The individual pixel modules were shifted by about 10 $\mu$m in $x$ and 30 $\mu$m in $y$ and $z$, and rotated by about 0.3 mrad. The pixel layers, disks and the whole detector were displaced by slightly smaller amounts. In this case, the light jet rejection drops by a factor 2 for the same $b$-tagging efficiency. For residual misalignments about half as big, the drop in light jet rejection is degraded by 40% compared to the ideal case. The reduction of residual misalignments relies on the actual alignment procedure, which is performed to obtain the new positions of the detector elements. This is the most realistic case considered so far, and includes many (but not all) systematic deformations caused by the alignment procedure itself. In this case, the light jet rejection is at most 25% lower for the same $b$-tagging efficiency.

### 6.2 Impact of the tracker material on performance

A major effort has been invested in describing accurately the material in the tracking volume of ATLAS. However, some underestimation is possible. To assess the impact of extra material on the *b*-tagging performance, results with different geometries were compared. Extra material was added, mostly beyond the *b*-layer, increasing the thickness in radiation lengths by about 8% (15%) at $|\eta| \approx 0(1)$. The first noticeable effect is the degradation of the impact parameter resolution. The other effect is an increased fraction of particles undergoing interactions in the matter of the detector and producing secondary particles which can, directly or through pattern-recognition problems, fake non-prompt tracks. At a 60% *b*-tagging efficiency, the extra material decreases by 10% the light jet rejection power. About 60% of the loss of rejection is explained by the worsening of the impact parameter resolution, and about 40% by extra secondaries.

### 6.3 Impact of the pixel detector conditions

The pixel detector and notably the innermost *b*-layer are critical for achieving good *b*-tagging performance. The detector efficiency clearly affects the tracking performance but is also explicitly a key ingredient for *b*-tagging since the *b*-tagging quality cuts require that each track have at least two pixel hits of which one is in the *b*-layer. These pixel hit requirements are made in order to maintain the highest resolution on the impact parameter of tracks.

A single pixel inefficiency of 5% has been used to simulate the events. Measurements on the pixel staves before the detector integration gave a single pixel inefficiency below $\sim 0.3\%$ (and below $\sim 0.1\%$ for the *b*-layer for which the highest quality components were used). The effect of this inefficiency is especially relevant at small $|\eta|$ where half of the pixel clusters contain only one pixel. The impact of varying the fraction of randomly distributed dead pixels was studied for three tagging algorithms on a large statistics (600k events) $t\bar{t}$ sample and the results are shown in Table 4. When decreasing the fraction of dead pixels from 5% to 1%, the tracking efficiency for tracks fulfilling the *b*-tagging quality cuts improves by up to 2.5% absolute (around $\eta \sim 0$), leading to a relative gain in rejection of about 10%.

Table 4: Reference light jet rejection for several tagging algorithms and the relative change with various configurations of the pixel system (see text) for a 60% *b*-tagging efficiency in $t\bar{t}$ events. The used reference histograms were produced with the respective samples.

| | IP2D | IP3D | IP3D+SV1 |
|---|---|---|---|
| Reference rejection (5% of dead pixels) | $54 \pm 1$ | $77 \pm 2$ | $229 \pm 10$ |
| Relative change with 1% of dead pixels | +9% | +10% | +17% |
| Relative change with a dead half-stave on *b*-layer | -8% | -8% | -10% |
| Relative change with a dead bi-stave on *b*-layer | -34% | -34% | -28% |
| Relative change with a dead half-stave on external pixel layer | $< 1\%$ | $< 1\%$ | -1% |
| Relative change with a dead bi-stave on external pixel layer | -1% | -1% | -4% |

More global problems, such as chip and module inefficiencies were not considered in the studies. Their effect has been studied in detail in Ref. [19] and can be very important. However, the latest measurements made right before the detector integration indicate that only one module (in the middle layer) out of 1744 is dead and that fewer than 5 chips are dead out of 27904. Besides single module failures, more dramatic failures might happen. During the pixel operation, it is thought that the two most likely sources of potential failures could be an opto-board failure, leading to half a stave (at most 7 modules) not functioning and a cooling problem implying that a whole bi-stave (26 modules) could not be used. To study those scenarios, the pixel digitization was modified to disable the corresponding modules, in either the $b$-layer or the external pixel layer. The impact on $b$-tagging performance of these two scenarios is shown in Table 4. In the case of well-identified module failures, it is clear that some recovery strategies can be used, either directly in the tracking code or at least in the $b$-tagging algorithm, for instance by not requiring a hit on a dead module.

## 6.4 Jet algorithms

For $b$-tagging purposes, an accurate knowledge of the jet direction is relevant. In the first place, this direction is used to define which tracks should be associated to the jets. Then it is used to sign the impact parameter of tracks.

As mentioned earlier, all $b$-tagging results are given for jets reconstructed with a cone algorithm of size $\Delta R = 0.4$. Since a given physics analysis may opt for a different jet algorithm, the impact of this choice on the $b$-tagging performance is tested in this section. In all cases, as it is the default in the $b$-tagging software, only the tracks within a distance $\Delta R < 0.4$ of the jet axis were used for the tagging, even for jet algorithms which could benefit from a less geometric track-jet association such as the $k_T$ algorithm.

Several cone sizes $\Delta R$ and size parameters $R$ were studied for the cone algorithm and the $k_T$ algorithm: from 0.2 to 0.8 in steps of 0.1. Finally, the $b$-tagging performance of the mid-point algorithm [20], an alternate jet algorithm addressing the infrared sensitivity of cone algorithms, was also checked, for two different cone size of $\Delta R = 0.4$ and 0.7. In this study, electrons faking jets were removed.

Figure 19 shows the rejection of light jets versus the $b$-tagging efficiency obtained with the IP3D+SV1 tagging algorithm, for several jet algorithms run on $t\bar{t}$ events. Results with and without purification are very different. With purification (Figure 19(b)) there is no significant difference in performance in the relevant range of $b$-tagging efficiency ($40\% < \varepsilon_b < 75\%$).

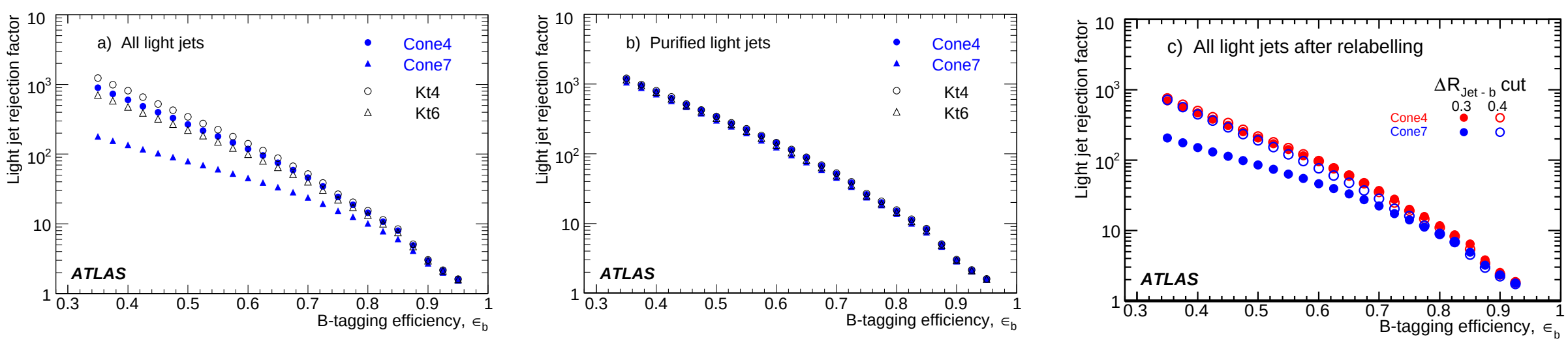

Figure 19: Rejection of light jets versus $b$-tagging efficiency for the IP3D+SV1 tagging algorithm applied on jets reconstructed with different algorithms: cone algorithm with size $\Delta R = 0.4, 0.7$ or $k_T$ algorithm with parameter $R = 0.4, 0.6$. See text for the last plot.

This stability is the anticipated behavior, since only the jet direction is meaningful for $b$-tagging purposes and it does vary with the jet algorithm but not drastically for moderate jet $p_T$: the mean of the distance $\Delta R(b, jet)$ for a 50 GeV jet is 0.081 for a jet of size $\Delta R = 0.4$ and 0.095 for a jet of size $\Delta R = 0.7$ (see Ref. [9] for more details). For completeness, it should be mentioned that no differences were found between tower-based and topological cluster-based jets.

Without the purification procedure (Figure 19(a)), the results are different for different jet definitions. However, the interpretation is not straightforward. In principle broader jets could be more easily contaminated by neighbouring tracks originating from distinct partons whose showers could not be resolved: light jets for instance could be contaminated by heavy-flavour decay products. However, this effect should be marginal since the maximum track-jet distance for association is kept to $\Delta R = 0.4$ in all cases.

A better explanation is linked to the ambiguity and arbitrariness of the labelling procedure: with fewer, broader jets the assignment of partons to the jets is more ambiguous and more likely to have $\Delta R(b, jet) > 0.3$, and thus more jets wrongly labelled as light jets. Therefore, for example, the rejection at $\varepsilon_b = 50\%$ with a cone radius $\Delta R = 0.7$ appears to be three times less than with a cone of radius 0.4, by relabelling the jets with a $\Delta R$ cut of 0.4 instead of 0.3 this difference can be virtually eliminated as seen in Figure 19(c). Thus although it appears to be more difficult to define the true flavour of a broader jet this does not necessarily exclude the choice of a cone size of 0.7 for a given analysis.

## 6.5 Sensitivity to the calibration of tagging algorithms

Most of the ATLAS tagging algorithms make use of an a priori knowledge to discriminate *b*-jets from light jets, which comes in various forms. The simplest of these ingredients is the transverse impact parameter resolution function used by the JetProb tagging algorithm to measure the compatibility of tracks with the primary vertex. The likelihood ratio tagging algorithms rely on several such distributions, with the further complication that they must be known for both the light and the *b*- hypothesis. In this section, the way this knowledge may affect the performance is studied. It is not currently possible to know if these settings are a good representation, both in nature and amplitude, of the differences data/Monte Carlo that will be observed, but at least they give information about the robustness of the tagging algorithms. All the results are given for the $t\bar{t}$ sample.

### 6.5.1 JetProb tagging algorithm

JetProb is expected to be one of the first tagging algorithms to be commissioned in ATLAS. To perform well, the resolution function (cf. Section 4.2) must be measured in data to avoid possible short-comings of the Monte Carlo simulation (resolutions and non-gaussian tails mostly). One of the major advantages of this tagging algorithm is that a priori any track from any physics process, e.g. the tracks from the first minimum-bias events, could be used to calibrate the resolution function, provided the contamination of non-prompt tracks can be kept at a very low level.

The sensitivity to this last point was studied by checking two different scenarios to select tracks in order to build the resolution functions. In all cases, only reconstructed tracks with negative impact parameter significance and fulfilling the *b*-tag quality cuts are used. This was done on $t\bar{t}$ events but similar or better (because of less heavy flavour contamination) results are expected for minimum-bias events. In the ideal case, the tracks are required to match to a true particle whose true origin is at the primary vertex of the event. In the realistic case, this requirement was not enforced. The distributions of the negative impact parameter significance $d_0/\sigma_{d_0}$ of tracks obtained in the two cases exhibit significant differences in the tails: in the ideal case, 0.6% of the tracks have $|d_0| > 5\sigma_{d_0}$ (the RMS of the distribution is 1.3) while this fraction is 3.2% for the realistic case (RMS is 2.1). These distributions are then used as resolution functions to measure the *b*-tagging performance on the same events: at a 50% *b*-tagging efficiency, the light jet rejection with the realistic scenario is 15% lower than for the ideal case.

### 6.5.2 Likelihood-based tagging algorithms

For the likelihood-based tagging algorithms, the probability density functions for all the variables (cf. Section 4.1) are built for the *b*-jet and light jet hypotheses using Monte Carlo. The $t\bar{t}$ channel can be used in data to isolate a sample of pure *b*-jets from which the various distributions can be derived, using the methods described in [9]. However, more than a few hundreds of $pb^{-1}$ of data are needed. For the light jets, it seems very difficult to isolate a pure enough sample in data. In this section, the sensitivity to the calibration is estimated using reference histograms obtained from Monte Carlo with very different settings.

First the impact of using a different tracking algorithm for the calibration (iPatRec) and for the performance measurement (NewTracking) was assessed and led to a very small variation of the rejection power, below 10%. Another study consisted of using different detector descriptions for calibrating and testing: the two geometries compared were relatively similar, with a relative difference in the amount of material in the tracking volume of 8% (15%) at $\eta \sim 0$ (1). At most a 5% change in rejection power is seen in this case. Another issue is the sample composition of the reference histograms: using only $t\bar{t}$ events or a mix of $t\bar{t}$, $WH$ ($m_H = 120, 400$ GeV) and SUSY events does not change significantly ($< 5\%$ relative change on rejection power) the *b*-tagging performance on a $t\bar{t}$ sample. However, larger effects are expected when *b*-tagging is run on a very different sample from the one used for the calibration.

A possible bias when building the calibration distributions and looking at performance on the same event sample was studied by dividing the sample into two. The bias on the resulting rejection factors was found to be usually negligible, and in all cases below 10%.

Finally, it was checked what statistics are needed to define the underlying histograms for the various tagging algorithms in order for results to be stable. This was checked on semi-leptonic $t\bar{t}$ events by halving a 600k event sample, building calibrations on 10k, 50k, 100k and 300k events from the first half-sample and checking the performance on the other half. To obtain a rejection level stable within 3%, 50k events are needed for all the IP and SV tagging algorithms when used in a regime where $\varepsilon_b \geq 50\%$.

## 6.6 Sensitivity to the Monte Carlo modelling

In the Monte Carlo modelling, several parameters can affect the ability to tag *b*-jets. Any effect that can change the lifetimes of the produced particles, the multiplicity of the charged tracks or the momenta of these tracks can potentially change the tagging efficiency. This modelling is not necessarily a good description of data, and in addition it is also performed differently across generators.

### 6.6.1 Fragmentation

First of all, various fragmentation models, describing the non-perturbative process in which quarks hadronize into colorless hadronic states, are implemented in the Monte Carlo generators. For heavy-flavour quarks, two options are available in the PYTHIA generator: the Lund-Bowler model and the Peterson fragmentation model. While the former has been found to give reasonable agreement with experimental data at LEP, SLC and HERA, the latter is currently the default in ATLAS PYTHIA productions. The impact of these various fragmentation models was investigated with six different $t\bar{t}$ samples. For the Peterson fragmentation, three samples were produced with different values for the $\varepsilon_b$ parameter: 0.003, 0.006 (default), 0.012. For the Lund-Bowler model, the $r_Q$ parameter was varied: 0.50, 0.75, 1.0 (default). The maximum relative discrepancy in the *b*-tagging efficiency was found to be around 6%, comparing the Peterson model with $\varepsilon_b = 0.012$ to the Lund-Bowler model for $r_Q = 0.5$. However, this choice of parameters is a bit extreme: the difference between the default Peterson model and the Lund-Bowler model with $r_Q = 0.75$ (which was found to fit best the OPAL and SLD data) leads to an uncertainty on the *b*-tagging efficiency of 1.1% for a fixed *b*-tagging cut leading to a *b*-tagging efficiency of 72%.

### 6.6.2 Heavy flavour production

The production fraction of various *b*/*c*-hadron species can also lead to different *b*-tagging efficiencies since they have different lifetimes and decay modes. The measured *b*-hadron fractions [15] and their values in the generators are shown in Table 5. For HERWIG, the defaults have been changed following the CDF tuning [21] by setting the CLPOW parameter to 1.2, in order to obtain a $b$-baryon fraction in agreement with the PDG and with PYTHIA. Another ingredient is the production of excited bottom and charm states which can give rise to soft charged pions or kaons, affecting the topology of the events: this has not been studied yet.

Table 5: Fraction (in %) of $b$-hadron species from the PDG (assuming $f(B_d) = f(B^{\pm})$) and in PYTHIA, the default HERWIG and HERWIG tuned for ATLAS (CLPOW=1.2).

| | $B_d$ | $B^{\pm}$ | $B_s$ | Baryons |
|---|---|---|---|---|
| PDG | $39.8 \pm 1.0$ | $39.8 \pm 1.0$ | $10.4 \pm 1.4$ | $9.9 \pm 1.7$ |
| PYTHIA (ATLAS) | 39.7 | 39.2 | 12.1 | 9.1 |
| HERWIG | 44.3 | 44.8 | 10.8 | 0.0 |
| HERWIG (tuned) | 39.4 | 39.9 | 10.4 | 10.3 |

The various production fractions of each type of the $b$-mesons were varied according to the measured errors from PDG 2006 [15], and the impact on a simulated PYTHIA $t\bar{t}$ sample was studied by a re-weighting technique. This source of systematics can be safely neglected since the net effect is an uncertainty below the per mil level on the tagging efficiency.

### 6.6.3 $b$-hadron lifetimes and decays

The uncertainty on the lifetime of the various b hadrons was also studied and found to give rise to an uncertainty of 0.3% for a $b$-tagging efficiency of 72%.

The uncertainty in the charged track multiplicity of b hadron decays was estimated by comparing PYTHIA with measurements from LEP [22]. The resulting uncertainty on the $b$-tagging efficiency was found to be 0.9%.

### 6.6.4 Heavy flavour decays with EvtGen

The two event generators used in ATLAS to fragment and decay particles, PYTHIA and HERWIG, implement different algorithms to simulate the decays of generated particles, using their own decay tables to specify decay modes and branching fractions. The sophistication of the decay simulation and the scope of decay tables vary considerably between generators. For B meson decays, arguably the most detailed simulation is currently provided by EvtGen [23].

Since the $b$-tagging performance on Monte-Carlo samples may depend on details of simulated particle decays, such as the charged particle multiplicity or the spatial distribution of secondary decay vertices in B decays, the impact of using EvtGen as a decayer instead of PYTHIA was studied on $t\bar{t}$ events. For this study, a decay file for inclusive decays was assembled based on the latest (as of summer 2005) version of the decay files used by the BaBar and CDF experiments. For decay channels where experimental data is available, branching fractions were taken from PDG [15], while the remaining decays are simulated generically with JETSET.

For this study, two specific samples were generated: the first $t\bar{t}$ sample was generated by PYTHIA and the decays were handled by PYTHIA. The second one was generated in the same way except that particle decays were simulated by EvtGen. As expected, changing the particle decay simulation leads to small differences in some distributions of generator–level quantities. For example, the mean multiplicity of charged pions in decays of $B^{\pm}$ and $B^0$ mesons, not including decays of long-lived weakly decaying strange particles such as $K_s^0$ and $\Lambda$, is 3.89 (RMS 2.19) with PYTHIA and 3.62 (RMS 2.13) with EvtGen. The average multiplicity obtained by EvtGen agrees well with the experimental value of $3.58 \pm 0.07$ [15].

The $b$-tagging weight distributions obtained with the two generators are thus slightly different. For a fixed $b$-tagging weight cut, the $b$-jet efficiency varies by about 1%. Tuning the cut to keep the same $b$-tagging efficiency in both samples leads to non-negligible changes in the light jet rejection: it decreases by about 5% to 15% when EvtGen is used, depending on the tagging algorithm and chosen $b$-jet efficiency.

# 7 Specific studies for improving $b$-tagging performance

In this section, three studies aiming at improving the $b$-tagging performance are presented. Most of them rely on specific software developments.

In the first part, new track categories are defined to make a better use of the slight differences in e.g. impact parameter resolution that such tracks may exhibit. A second study evaluates the potential gain by varying the way tracks are associated to jets. The last study shows the improvement obtained when combining several tagging algorithms in a multivariate approach. Those studies were done independently and no attempt was made yet to combine them. It is worth noting that some approaches advocated in the first two studies are expected to be highly correlated.

## 7.1 Improving performance with track categories

Grouping the tracks used for $b$-tagging in several categories has been discussed in Section 4.1.5. Dedicated treatment for *Shared* tracks (cf. 2.1.3) is already implemented and used in the current $b$-tagging software. Using dedicated probability density functions for the *Shared* tracks improves the light jet rejection by 23% (7%) for a $b$-jet tagging efficiency of 50% (60% respectively) in $t\bar{t}$ events. This is the default treatment in the software. This effect is sample-dependent and is more important for samples with high jet multiplicities and energetic jets, which tend to be more collimated.

Using additional track categories to improve the $b$-tagging performance is being further investigated. A possible interest of the track categories is to try to loosen the track quality cuts and therefore gain in efficiency without diluting the discrimination power of the good tracks. As discussed in Section 2.1.2, requiring each track to have a hit on the $b$-layer leads to an absolute loss in efficiency of about 2.5%, which is not negligible for $b$-tagging purposes where few tracks are available. Thus one attempt consisted of trying to keep tracks with no $b$-layer hit in a special category. About 4% of the tracks in jets from $t\bar{t}$ events which pass the rest of the $b$-tagging selection fall in this category.

The categories could also be used to deal with the non-Gaussian resolution tails and the imperfect treatment of the matter in the tracking error estimation process. A priori some fraction of these effects would be better treated from first principles directly in the tracking, but the experience with previous experiments shows that this is difficult in practice and therefore ad-hoc treatments may be justified. The natural variables to partition tracks are $p_T$ (actually $p$ for multiple-scattering) and pseudo-rapidity (since the material is very non-uniform in $\eta$).

Finally, another potential use of track categories was studied: in $b$-jets, the fragmentation tracks accompanying the $b$-hadron decay products are prompt and should therefore have different distributions of the discriminating variables. Tracks with $p_T(\text{track})/p_T(\text{jet}) < 0.04$ were defined as fragmentation tracks since those are in principle softer and were put in a special category. In typical $b$-jets from the $t\bar{t}$ sample, this cut selects 13% of the tracks. Some correlation with the treatment in $p_T$ bins (previous case above) is expected.

The use of these categories brings some improvement to the light jet rejection, as high as 60% for the binning in track $p_T$ which is the most powerful way of partioning tracks. The improvement for some categories depends on the sample: the dedicated category for fragmentation tracks for example is more helpful for the $WH$ ($m_H$ =400 GeV) sample where the actual decay products of the $b$-hadron are very collimated and the fixed-size cone for associating tracks to the jets brings in a larger fraction of prompt tracks.

Various ways to combine all the new categories were investigated. The gain in rejection brought by the best combinations, using 11 partitions formed with the aforementioned categories, ranges from 20% to 70% depending on the sample and the tagging algorithm, for a 60% $b$-tagging efficiency.

### 7.2 Optimizing the track-to-jet association

As the jet transverse momentum increases, its particles are collimated into a narrower cone. But currently all tracks within $\Delta R < 0.4$ of the jet axis are associated with the jet, regardless of its momentum. For a 300 GeV $b$-jet, only 30% of the tracks associated to the jet comes from the $b$-hadron decay products, as shown in Figure 16. Therefore at high $p_T$ the $b$-tagging discriminating power is diluted since a larger fraction of the tracks in the jet may be picked up from environmental contamination: underlying event, pile-up or neighbouring jets in busy events. An alternative track-to-jet association has been studied, with a $\Delta R$ cut varying with the jet $p_T$: $\Delta R < f(p_T)$. Based on the distribution of $\Delta R(\mathrm{jet},\mathrm{track})$ for tracks originating from $b$-hadron decays in $b$-jets, a functional form $f(p_T)$ has been chosen which ensures that 95% of the tracks from $b$-hadron decays in these events are associated to the jet for any jet $p_T$ in the range $[15,500]$ GeV.

The impact on $b$-tagging performance of using this association instead of the standard one is checked on $t\bar{t}$ events. The relative improvement on the overall raw light jet rejection is 46% (7%) for respectively a $b$-tagging efficiency of 50% (60%). This improved treatment actually affects only the non-pure jets (22% of the light jets in this sample), for which the rejection triples for $\varepsilon_b = 50\%$ and doubles for $\varepsilon_b = 60\%$. Obviously the fraction of non-isolated jets and therefore the possible gain with this method are sample-dependent.

### 7.3 Combining tagging algorithms with boosted decision trees

Several multi-variant techniques exist that can combine different $b$-tagging algorithms into a single classifier for discriminating $b$-jets from light jets. We investigated boosted decision trees (BDT).

BDT can be applied to any classification problem and their use for combining several $b$-tagging algorithms into a single classifier for discriminating $b$-jets from light jets was investigated [24]. In this study, a BDT classifier was optimized on a training sample containing $b$-jet and light jet patterns extracted from $WH$ ($m_H = 120$ GeV, $H \to b\bar{b}$ or $H \to u\bar{u}$) and $t\bar{t}$ samples. The following input variables were used: the weight from the IP3D tagging algorithm, the three variables on which the SV tagging algorithms are based (cf. Figure 9), the number of tracks associated with the secondary vertex, the weights of the soft muon and soft electron tagging algorithms, the largest transverse and longitudinal impact parameter significances and transverse momentum of the tracks in the jet, the jet transverse momentum and the number of tracks in the jet. For a fair comparison with IP3D+SV1, a BDT classifier with only the first four variables was also studied. The predictive power of the classifier was estimated using a distinct sample of $b$-jet and $u$-jet patterns (test sample).

Table 6: Rejection of light jets given by IP3D+SV1 and the boosted decision tree for $WH$ (with $m_H$=120 GeV) and $t\bar{t}$ events, for fixed $b$-tagging efficiencies of 50% and 60% in each sample.

| | IP3D+SV1 | | BDT 4 variables | | BDT 12 variables | |
|---|---|---|---|---|---|---|
| | $\varepsilon_b = 50\%$ | $\varepsilon_b = 60\%$ | $\varepsilon_b = 50\%$ | $\varepsilon_b = 60\%$ | $\varepsilon_b = 50\%$ | $\varepsilon_b = 60\%$ |
| $WH$ ($m_H$=120 GeV) | 529±40 | 155±6 | 682±79 | 189±12 | 762±93 | 201±13 |
| $t\bar{t}$ | 393±16 | 143±3 | 484±34 | 161±6 | 563±42 | 187±8 |
| $t\bar{t}$ after purification | 720±42 | 205±6 | 808±73 | 226±11 | 1021±103 | 278±15 |

Table 6 compares the rejection of light jets given by the two BDT configurations and the likelihood ratio weight IP3D+SV1 for $WH$ events and $t\bar{t}$ events. It shows that with the same variables the BDT outperforms IP3D+SV1 by 10 to 30%. When using additional information, including the soft lepton tagging, the light jet rejection on both event topologies increases by about 50% with respect to IP3D+SV1. For these results, the training and test samples were based on similar events ($WH$ or $t\bar{t}$). When training the BDT on $WH$ events and using $t\bar{t}$ events for the test sample, the gain in rejection compared to IP3D+SV1 is lower but still interesting ($> 20\%$).

# 8 Measuring $b$-tagging performance in data

For analyses using $b$-tagging, the estimation of the backgrounds from well-known Standard Model processes requires knowledge of the tagging and mis-tagging efficiency for the various flavours of jets with high accuracy. The quality of Monte Carlo simulation of these properties is unknown and therefore strategies must be developed to measure the tagging and mis-tagging efficiencies directly in data.

## 8.1 $b$-tagging efficiency

Several strategies for measuring the $b$-tagging efficiency directly in data are investigated in detail. The relative precision they permit has been estimated for a typical $b$-tagging efficiency of 60%.

The first approach, described in Ref. [8], relies on a sample of the abundantly produced dijet events, in which one of the jets contains a muon. A muon+jet trigger has been conceived and proposed for this purpose. Two methods, also employed at the Tevatron, are used to estimate the $b$-content of the dijet sample. The $p_{T_{rel}}$ method is based on templates of the muon $p_T$ relative to the jet axis. The templates are derived from Monte Carlo events for the three types of jets: bottom, charm and light (the latter will eventually be derived from data). The so-called System 8 method uses two samples of differing bottom quark content and two uncorrelated tagging algorithms, typically the soft muon one and the tagging algorithm to be calibrated, to form a system of equations from which the $b$-tagging efficiency can be extracted. Using 50 pb$^{-1}$ of data, a detailed $p_T$- or $\eta$-dependent calibration curve could be derived with the $p_{T_{rel}}$ method and with System 8. Since it is expected that the systematic uncertainties will dominate rapidly the total error for these methods, a careful study of systematics errors has to be done which was not fully completed for this note: the systematic errors studied so far indicate that it should be possible to control the absolute error on $\varepsilon_b$ to 6%. Currently the two methods are proven to work well for jets below a $p_T$ of 80 GeV.

The second approach, discussed in Ref. [9], makes use of $t\bar{t}$ events and is complementary: a little more data is needed but the tagging efficiency of higher $p_T$ jets can be measured. Two distinct ways are described: by counting the number of selected $t\bar{t}$ events with one, two or more $b$-tagged jets, or by studying directly the output distributions of $b$-tagging algorithms on samples of $b$-jets pre-selected by several methods (topological, kinematic or likelihood selection). The counting method allows measurement of the integrated $b$-tagging efficiency with a relative precision of $\pm 2.2$(stat.)$\pm 3.5$(syst.)% in the lepton+jets channel and $\pm 3.7$(stat.)$\pm 2.7$(syst.)% in the di-lepton channel for 100 pb$^{-1}$ of data. For the topological selection, 200 pb$^{-1}$ of data are needed and allow a relative precision of $\pm 7.7$(stat.)$\pm 3.2$(syst.)%.

## 8.2 Mis-tagging rates

Measuring the mis-tagging rate in data is more difficult and under study. The main approach is to use the negative tags (either in impact parameter or decay length) which describe the effects of a limited resolution on a priori prompt tracks and then to correct for long-lived particles ($K_s$, $\Lambda$, etc), material interactions and heavy flavour jets which are negatively tagged. So far no study in ATLAS has estimated the accuracy with which fake rates can be measured. However, based on the Tevatron experience, it seems that a 10% relative error could be achievable with 100 $pb^{-1}$.

### 8.3 Extracting reference distributions from data

For the likelihood tagging algorithms, reference distributions for light and $b$-jets are needed. In the case of $b$-jets, they could in principle [25] be measured in data from a pure sample of $b$-jets using the techniques developed for the $b$-tagging efficiency estimation in $t\bar{t}$ events. As shown in Ref. [9], the various distributions can be checked in data with a few hundred $pb^{-1}$. However, much more data is needed to extract multi-dimensional likelihoods. For light jets, it seems difficult to extract from data a sample with sufficient purity. In any case, a Monte Carlo accurately describing the data is also needed to extrapolate those reference distributions to ranges where they certainly can not be measured, the very high-$p_T$ regime for instance.

## 9 Conclusions and outlook: realistic performance in first data

The first tagging algorithm to be commissioned with real data is expected to be (besides the track counting method) JetProb, using the tracks from any kind of events to define its resolution function. For a $b$-tagging efficiency of 60%, a light jet rejection of around 30 could be achieved with this tagging algorithm but further improvements are expected. The soft lepton tagging algorithms will be commissioned at the same time, leading to higher rejection levels when considering semi-leptonic $b$-jets. Once the quality of the Monte Carlo simulation is checked and better understood with data, a tagging algorithm relying on Monte Carlo templates for $b$ and light jet hypotheses such as IP3D can be used, perhaps doubling the rejection power. The commissioning of the tagging algorithms relying on secondary vertexing may take more time, but tagging algorithms like SV1 should quadruple the initial rejection level, bringing it above 100. Finally the combination of the ultimate JetFitter tagging algorithm and the various improvements described in this note should permit a rejection of 200, or more interestingly to maintain a rejection of 100 at a higher $b$-tagging efficiency, around 70%.

Those estimates do not take into account the impact of residual misalignments in the tracker. However, a first study has been performed in which the actual alignment procedure was run. This is the most realistic study so far, and includes many systematic deformations caused by the alignment procedure itself. It concludes that the mis-tagging rate is at most 30% lower for the same $b$-tagging efficiency with a realistic early detector alignment.

The impact of several other effects on the $b$-tagging performance has been studied in this note. In the simulation used, the fraction of dead pixels is overestimated. Using 1% instead of 5% dead pixels improves the light jet rejection by about 10%. Relative improvement in the light jet rejection, from 10% in typical $t\bar{t}$ events to 50% for high-$p_T$ samples, could be achieved with a tuning of the tracking in jets. The sensitivity to the accuracy of the passive material description in the simulation could be quite dramatic. However, a large effort has been made in describing accurately the material in the tracking volume. If an 8% to 15% discrepancy between Monte Carlo and data would remain, the impact on the $b$-tagging performance is a 10% relative change in light jet rejection power.

New ideas to improve the performance have been studied: the generalization of the track categories could bring a 10% to 60% improvement, optimizing the track-to-jet association could help significantly in busy events and finally the use of multivariate techniques such as BDT could lead to gains in the range 10-50%. However, all these potential gains only apply to some tagging algorithms, some $p_T$ region, etc. Furthermore some correlations are expected among them. Therefore it will be interesting to assess the net impact of these improvements when they are all available, used simultaneously and optimized.

Finally, detailed studies have shown that the $b$-tagging efficiency can be measured directly in data using dijet or $t\bar{t}$ events. With 100 pb$^{-1}$, a relative precision of about 5% can be achieved for $b$-jet efficiency. The accuracy with which the mis-tagging rates can be measured deserves more study, however, a 10% precision seems feasible based on the Tevatron experience.

# Vertex Reconstruction for *b*-Tagging

**Abstract**

Tagging of *b*-quark jets, "*b*-tagging", is an important ingredient for Standard Model analyses as well as for searches for new physics. A property of *b*-quark jets exploited by *b*-tagging algorithms is the presence of secondary *b*- and *c*-hadron decay vertices. In this note, methods for the explicit reconstruction of secondary and/or tertiary decay vertices of *b*- and *c*-hadron decays are presented. The performance of the secondary vertex based *b*-tagging algorithms and the dependences on the event topology and jet kinematics are studied. The efficient reconstruction of the primary interaction vertex is also crucial for *b*-tagging, especially in the presence of pile-up interaction vertices at LHC luminosity. The ATLAS primary vertex reconstruction strategies and performances are presented in this note as well.

## 1 Introduction

Identification of *b*-quark jets relies on the properties of the production and weak decay of *b*-hadrons. The most important one is their relatively large lifetime of about 1.5 ps ($c\tau \approx 450\mu$m). The resulting *b*-hadron flight path $< l >= \beta\gamma c\tau$ leads to a signature of one or more displaced secondary vertices (e.g. inside a jet originating from a *b*-quark with transverse momentum of 50 GeV *b*-hadrons travel on average about 3 mm in the transverse plane before their decay).

Numerous methods can be used for the identification of *b*-quark jets (usually called *b-jet tagging* or just *b-tagging*), based e.g. on the presence of leptons inside jets due to semileptonic *b*-hadron decays, kinematical properties of jets or explicit *b*- or *c*-hadron reconstruction. The most efficient tagging methods are based on the presence of displaced vertices inside jets. Such vertices can be detected either by the presence of tracks incompatible with the primary event vertex or by explicit reconstruction of those secondary vertices. Many track based and vertex based *b*-tagging methods are known together with combined methods using both track impact parameters and vertex reconstruction simultaneously.

The reconstruction of secondary *b*-hadron decay vertices inside jets is challenging for several reasons. The multiplicity of charged particle tracks belonging to the vertex is, contrary to the reconstruction of exclusive decay modes, a-priori not known. In cases of less than two reconstructed charged particles, a well defined secondary vertex cannot be reconstructed. This can happen either because of the charged particle multiplicity produced in the decay or limitations in the track reconstruction efficiency, as imposed by the geometrical acceptance of the inner tracking detectors or interactions in the detector material. Secondary vertex reconstruction efficiencies thus show an upper limit. In addition, weak *b*-hadron decays almost always lead to one or more charm hadrons which subsequently decay through weak interaction. Since *c*-hadrons also have significant lifetimes, the resulting topology is a set of charged particle tracks either stemming from the primary event vertex, the secondary *b*-hadron or tertiary *c*-hadron decay vertices. The resolution of the ATLAS tracking system does not resolve this decay topology in all cases. Vertex reconstruction inside jets for *b*-jet tagging thus has to be done in an inclusive way. It should be targeted towards highest efficiency to detect secondary vertices and identify its topology as well as possible. Kinematical features of reconstructed vertices like e.g. the invariant mass may be used in combination with spatial features (track-vertex and vertex-vertex distances, etc.) to increase the *b*-jet identification power of the algorithms.

Another important ingredient for tagging *b*-quark jets is the reconstruction of the primary event vertex. The size of the beamspot in the transverse plane (about 15 $\mu$m) is sufficiently small to allow the application of *b*-tagging algorithms if only information as defined in the transverse plane is used, once the position of the beam is known with high precision. Explicit reconstruction on event by event basis does not improve significantly the resolution of the primary event vertex in the transverse plane. In the longitudinal direction along the beam, however, the a-priori knowledge of the interaction point is only poor (several cm). Explicit reconstruction is thus mandatory if *b*-tagging algorithms are not only based on track and vertex information in the transverse plane. Furthermore, at nominal running of the LHC, additional minimum bias events produce additional so-called pile-up vertices which have to be distinguished from the hard primary interaction. The main task of the primary vertex reconstruction for *b*-tagging is thus to reconstruct and identify the signal event vertex out of all interaction vertices along the beam.

At the LHC, *b*-tagging has to be applied to jets covering a wide kinematic range both in transverse momentum and pseudorapidity. *b*-tagging has to be applied over the full acceptance of the tracking detectors of about $|\eta| < 2.5$. The track reconstruction performance, and thus also the vertex reconstruction and *b*-tagging performance, varies strongly with transverse momenta and pseudorapidities of charged particle tracks.

This note focuses on aspects of primary and secondary vertex reconstruction relevant for *b*-tagging. Other aspects, also more technical ones, and performance issues are addressed in notes dedicated to primary [1] and secondary vertex reconstruction [2, 3]. The note is structured as follows. In Section 2 some information on definitions, physics input objects and data sets used in this note is given. Section 3 describes the reconstruction of the primary event vertex, its performance and impact on *b*-tagging. Algorithms designed for the inclusive reconstruction of secondary decay vertices inside jets based on two different methods are described in Section 4. The performance of these algorithms in terms of secondary vertex related quantities is discussed in Section 5. Their combination with the pure track impact parameter based *b*-tagging algorithms is described in Section 6. The *b*-tagging performance of the algorithms is presented in Section 7. Section 8 gives a summary of the note together with an outlook towards possible future improvements.

## 2 Definitions, Reconstruction Details and Datasets

This section gives definitions and technical details of data reconstruction and analysis procedures used in this note. More detailed information can be found in [4].

The efficiency to tag a jet of flavour $q$ as $b$-jet, $\varepsilon_q$, is defined as:

$$\varepsilon_q = \frac{\text{Number of jets of real flavour q tagged as b}}{\text{Number of jets of real flavour q}}. \tag{1}$$

Usually $\varepsilon_b$ is called *tagging efficiency* and $\varepsilon_{udsc}$ *mistagging rate*. The inverse of the mistagging rate $r_{udsc} = 1/\varepsilon_{udsc}$ is called *b-tagging rejection power* or simply *rejection*. The assignment of a certain flavour to a jet in the Monte Carlo simulation is not unambiguously defined. The following definition has been introduced: the jet flavour is the flavour of the heaviest quark after gluon radiation and splitting within a cone of some size around the jet direction. The default cone size used in ATLAS and in the current note is $\Delta R < 0.3$. To test the pure algorithmic performance of the algorithms, a procedure called "purification" may be applied. Here, a light quark jet is only considered if there is no heavy parton (*b*- or *c*-quark) or $\tau$ lepton within a cone of 0.8 around the jet direction.

Several jet reconstruction algorithms are available in ATLAS [5]. The *b*-tagging performance may depend on the jet algorithm, but the studies of this dependence are outside the scope of this note (see [4] for such studies). For the studies in this note, an iterative cone algorithm with a cone size of $\Delta R = 0.4$ using combined calorimeter towers as input, has been used. Charged particle tracks have been reconstructed with a Kalman filter based algorithm [6].

Offline ATLAS jet reconstruction and track reconstruction are independent. An assignment of tracks to jets is thus necessary. The current ATLAS method used also in this note is also based on a geometrical cone around the jet direction. All charged tracks within a certain cone (the default value is $\Delta R < 0.4$) around the jet axis are assigned to a jet.

The following fully simulated Monte Carlo samples have been used for the performance studies described in this note:

- Higgs boson production in association with a $W$ boson. The $W$ boson was forced to decay into a muon and its anti-neutrino, $W \rightarrow \mu \overline{\nu}_\mu$, the Higgs boson was forced to decay into pairs of $b$-, $c$- or $u$-quarks: $\mathrm{H} \rightarrow \mathrm{b\overline{b}}$, $\mathrm{H} \rightarrow \mathrm{c\overline{c}}$, $\mathrm{H} \rightarrow \mathrm{u\overline{u}}$. This ensures that the jets of different quark flavour have very similar kinematics. To cover a wide range of jet transverse momenta, samples with Higgs boson masses of $m_H = 120$ GeV and $m_H = 400$ GeV have been produced.
- Events with pairs of top quarks, $t\bar{t}$, where one or both $W$ bosons decayed leptonically.
- Events with pairs of top quarks and at least two additional jets: $t\bar{t}jj$. These events show a large hadronic activity with a high probability of overlapping jets.

During the first few years the LHC will operate at a luminosity of $2 \cdot 10^{33}$ cm$^{-2}$s$^{-1}$ (low luminosity mode), reaching $10^{34}$ cm$^{-2}$s$^{-1}$ (high luminosity mode) at later stages. Each signal event triggered and reconstructed in the ATLAS detector will be overlayed by several low-$p_T$ proton-proton interactions, commonly denoted as minimum bias interactions. The average number of minimum bias interactions per bunch crossing is 4.6 and 23 for the low and high luminosity modes, respectively. To study the influence of these additional so-called *Pile-Up* interactions, fully simulated Monte-Carlo samples for the low luminosity mode have also been generated for the WH ($m_H = 120$ GeV) data set.

To parametrize the trajectory of a charged particle in a magnetic field, so-called "perigee" parameters are used [6]. The ATLAS track based $b$-tagging algorithms use $d_0$ and $z_0$, the transverse and longitudinal impact parameters at the point of closest approach of the trajectory to the primary vertex in the transverse plane. To reduce the dependence on the track parameter resolutions, the corresponding variables are divided by their errors. Each impact parameter *significance* also gets a sign based on the track and jet directions. The sign is positive if the track crosses the jet axis in front of the primary vertex and negative if behind. Most of the tracks produced in $b$-hadron decays have a positive sign whereas tracks originating from the primary vertex have both signs with equal probabilities [4]. The distance between the primary and secondary vertices may also have a sign based on the jet direction.

# 3 Primary Vertex Reconstruction

In this section different strategies for the reconstruction of primary vertices currently implemented in the ATLAS reconstruction software are described. The evaluation of the algorithms is performed in the *single collision* and low luminosity pile-up modes.

The reconstruction of primary vertices can generally be subdivided into two problems: vertex finding and vertex fitting. The primary vertex finding deals with the association of reconstructed tracks to a particular vertex candidate. The task of the vertex fitting is to reconstruct the position of the primary vertex and its covariance matrix, recalculate the parameters of the incident tracks using the vertex constraint and provide a measure of goodness of the fit, such as for example the $\chi^2$ of the fit [7]. In addition, the $\chi^2$ of the refit of track parameters with the knowledge of the reconstructed vertex is a criterion of track to vertex compatibility.

A brief description of the mathematical properties of the algorithms for primary vertex reconstruction and evaluation of their performance with respect to $b$-tagging are presented below. For a more detailed description of the algorithms the reader is referred to [1].

## 3.1 Algorithms for Vertex Fitting

Several strategies for vertex fitting are currently implemented in the ATLAS reconstruction software. All implemented vertex fitters are based on the minimization of a $\chi^2$ function with respect to a position of the vertex and parameters of incident tracks at this position. In addition, all fitters are currently using the exact analytical solution for the dependence of the track parameters on the vertex position and track parameters at the vertex. Indeed, the full solution with respect to the ATLAS version of perigee parametrization has a simple form, which is also computationally economical [1]. The approach to the minimization of the $\chi^2$, treatment of components of track momenta in the vicinity of the vertex candidate and reweighting of tracks during the iterative process, however, differs from algorithm to algorithm.

### 3.1.1 The Billoir Full and Fast Vertex Fitting Algorithms

Two algorithms, the *Billoir Full* and the *Billoir Fast Vertex Fitter* were implemented, following the schema presented in [8]. Here, the inversion of the full $(3n+3)\times(3n+3)$ matrix[1] of parameters of the fit, where $n$ is the number of tracks, is replaced with $n$ inversions of $(3\times 3)$ matrices. Compared to the global $\chi^2$ minimization, this approach leads to a significant gain in CPU time due to the inversion of smaller matrices. In addition to the reconstruction of the vertex position and its covariance matrix, the *Billoir Full Vertex Fitter* also refits the parameters of the incident tracks with the knowledge of the vertex.

The *Billoir Fast Vertex Fitter* is a simplified version of the vertex fit, where the trajectories of charged particles are approximated with straight lines in the vicinity of the vertex and their momentum components $p = (\theta, \phi_v, q/p)$ are considered to be constant. While the precision of this approach is only a little smaller than the one of the *Full Billoir Vertex Fitter*, the fit itself is significantly faster due to the reduction of the size of covariance matrices to be inverted. It should also be noted, that the *Fast Vertex Fit* does not refit the momenta of the incident tracks. Due to its reasonable resolution and high CPU performance, this is the default vertex fitter to be used with *InDetPriVxFinder*, as explained in Section 3.2.1.

### 3.1.2 The Sequential and Adaptive Vertex Fitting Algorithms

The *Sequential Vertex Fitter* implements the conventional Kalman Filter for vertex fitting as described in [7]. The vertex estimate is updated iteratively using the information from a single track at a time. The *SequentialVertexSmoother* allows the refit of parameters of incident trajectories with the knowledge of the reconstructed vertex position. In addition, the smoother allows for a calculation of the smoothed $\chi^2$ of a track, which is a good criterion of compatibility of a track to a vertex.

A robust version of the above algorithm is the *AdaptiveVertexFitter*. It is an iterative reweighted Kalman Vertex Fitter, where each track is down-weighted according to its compatibility to the actual vertex position [9]. The dependence of the weighting factor on the iteration number of the fit is determined by a thermodynamic annealing procedure. The assignment of tracks to a vertex candidate thus becomes stronger with iterations and the outliers are efficiently discarded.

### 3.1.3 The TrkVKalVrtFitter Algorithm

The *TrkVKalVrtFitter* is a universal vertex fitter with constraints. It uses the Billoir method [8] to estimate the local vertex position and calculate refit track parameters with the requirement of passing through the vertex position. The possibility of applying a beam spot constraint when reconstructing the primary vertex is also provided. Neutral and charged particles can be used simultaneously in the fit.

[1] In the vicinity of the vertex, the polar angle $\theta$ and the momentum $p$ of the trajectories are considered to be constant. The number of parameters of the fit therefore reduces to $3n+3$: 3 remaining parameters for each of $n$ tracks and 3 vertex coordinates.

In the presence of badly measured tracks the precision and stability of the fit may be improved by the optional use of robust functionals. In each iteration step the error matrix for each track is recalculated based on an eigenvector decomposition and downweighting of badly measured directions in the track parameter space. Several functionals for downweighting known as *M-type estimators* are implemented [2].

### 3.2 Vertex Finding Strategies

Three different strategies for primary vertex finding are currently implemented. These are the *InDetPriVxFinder*, the *InDetAdaptiveMultiPriVxFinder* and the *VKalVrtPrim*. In the *InDetPriVxFinder* the process of primary vertex finding is decoupled from the vertex fitting and the maximal number of reconstructed vertices is therefore defined at the seeding step. The latter two algorithms exhibit a *finding through fitting* approach, where the number of vertex candidates changes according to the results of the previous iteration of the fit.

#### 3.2.1 The InDetPriVxFinder Algorithm

The reconstruction strategy of the Inner Detector Primary Vertex Finder consists of three steps. It starts with the selection of good quality tracks originating from the beam crossing area (the detailed selection cuts can be found in [1]).

After this initial track preselection, clusters in $z$ are searched for in the resulting set of tracks. A simple sliding window algorithm is used for this purpose. First, the tracks are ordered in ascending order of their $z_0$ impact parameter. The full range of $z_0$ impact parameters is then scanned and clusters of fixed length are formed. In the low luminosity scenario, the default maximal cluster length was chosen to be 3 mm. The resulting clusters of tracks are considered as independent primary vertex candidates and corresponding vertices are reconstructed with the provided vertex fitting algorithm.

After the reconstruction of the initial vertex candidates, the tracks least compatible with a given vertex candidate are rejected and the vertex candidate is refitted (the *Billoir Fast Vertex Fitter* described in 3.1.1 is used by default) using one of the following procedures:

- All tracks, with a $\chi^2$ contribution greater than a predefined value (by default, the threshold of $\chi^2 = 5$ per track is chosen) are discarded and the vertex candidate is reconstructed again using remaining tracks.
- The least compatible tracks are removed from the vertex candidate iteratively one by one until no track with a $\chi^2$ contribution greater than a predefined value is left or the number of remaining tracks is too little to continue. The vertex candidate is refitted at each iteration.

At present, the first algorithm is used as the default one. At the last stage of the primary vertex reconstruction, the obtained vertex candidates are sorted in descending order according to the sum of transverse momenta of their tracks. The vertex with the highest $\sum p_T$ of the tracks is tagged as the signal one. It should be noted that the maximum number of reconstructed primary vertices is fully defined by the output of the cluster finding in the beginning of the algorithm. The iterative refit of the vertex candidates followed by the rejection of incompatible tracks serves for the refinement of the knowledge of the vertex positions only. The minimum number of tracks to form a vertex candidate is one in the case when the beam position is used as constraint, and two otherwise.

#### 3.2.2 The InDetAdaptiveMultiPriVxFinder Algorithm

The *InDetAdaptiveMultiPriVxFinder* is an example of a *finding-through-fitting* approach to the primary vertex reconstruction. The reconstruction starts with the preselection of tracks originating from the beam crossing region (the detailed selection cuts can be found in [1]). A single vertex seed is created out of the preselected set of tracks. A vertex candidate is then reconstructed using the Adaptive Multi Vertex Fitter [10], as described in 3.1.2. The tracks which were considered to be outliers during the first fit are used to create a new vertex seed. At the next iteration a simultaneous fit of two vertices is performed. Each track is down-weighed with respect to the two vertices. The number of vertex candidates is growing with iterations and the vertices are competing with each other in order to attain more tracks. An annealing procedure is applied to this process: the assignment of tracks to vertices is becoming harder with iterations. The result of the fit is a set of reconstructed vertices with an almost solid track assignment. It has to be noted that for the adaptive fitters the starting point of the fit is of great importance.

#### 3.2.3 The VKalVrtPrim Algorithm

In this algorithm, primary vertex reconstruction starts with the selection of a subset of tracks close to the beamspot position. A sliding window type algorithm (with window size $\approx 4.5$ mm) selects a position along the beam which maximizes the vertex candidate quality value. This position becomes a seed for the vertex fitter. It accepts all tracks close to the found vertex candidate position and tries to fit a single vertex. If the quality of the vertex is not satisfactory, one or several tracks with the biggest $\chi^2$ contribution are rejected and the fit is repeated. The algorithm works until a vertex with good quality is obtained. Best results may be achieved with the combination of a robust fitting functional and the rejection of tracks during the fit. After a successful fit, the participating tracks are removed from the set of selected tracks and the sliding window algorithm starts a search of the next vertex candidate for the remaining tracks. *VKalVrtPrim* runs until no more vertex candidates can be obtained in the remaining track sample.

The vertex candidate quality value used in the search is a sum of the track transverse momenta, the transverse mass and $p_T$ nonuniformity in the transverse plane. The vertex fitter uses the beam spot position as constraint in the fit.

### 3.3 Performance of Primary Vertex Reconstruction

In this section, the performance of different primary vertex reconstruction algorithms is evaluated. The studies are performed in single collision and low luminosity pile-up modes.

As mentioned above, the main tasks of primary vertex finding algorithms are:

- To find all primary vertices in a bunch crossing and reconstruct their positions with highest possible accuracy.
- To identify the hard scatter signal primary vertex among the set of reconstructed vertices with highest possible efficiency.

These two aspects will be investigated in the following. In this section the performance of five strategies for primary vertex reconstruction are compared: *VKalVrtPrim*, *InDetAdaptiveMultiPriVxFinder* and *InDetPriVxFinder* used with *SequentialVertexFitter*, *Billoir Fast* and *Billoir Full Vertex Fitters*. Monte Carlo samples of $WH(120) \to \mu\nu b\bar{b}$ and $WH(120) \to \mu\nu u\bar{u}$ events with and without low luminosity pile-up are used for this study.

#### 3.3.1 Efficiency of Hard Scatter Signal Primary Vertex Identification

As mentioned above, an average of 4.6 minimum bias events are expected to overlay the signal event during low luminosity running of the LHC. The number of simulated primary vertices per bunch crossing, corresponding to the low luminosity mode of the LHC is shown in Figure 1 (left), together with the number of primary vertices reconstructed with different algorithms. All primary vertex finders return containers of reconstructed vertices ordered according to the probability of a given vertex to be produced in a signal collision. The identification of a signal vertex (first in the list) is performed in different ways depending on the algorithm. The *InDetPriVxFinder* orders the vertices according to the sum of transverse momenta of incident tracks, while the output of the *InDetAdaptiveMultiPriVxFinder* is sorted according to the sum of squares of transverse momenta. In all cases, the first vertex in the list is considered to be the primary signal vertex and is used by the $b$-tagging algorithms. In order to estimate the efficiency to find the correct signal vertex from the list of primary vertices two methods are used. In the first method the signal vertex is considered as reconstructed correctly if it is closer than 500 $\mu$m in the $z$ direction from the truth: $|z_{rec} - z_{truth}| < 500\,\mu$m. The second method is based on the purity of reconstructed vertices. The definition of the primary vertex purity is based on the association of tracks fitted to a primary vertex to truth particles:

$$P = \frac{\sum w_{signal}}{\sum w_{signal} + \sum w_{pileup} + \sum w_{noCorres}} \tag{2}$$

where $\sum w_{signal}$ is the sum of weights of all tracks associated with the signal primary vertex, $\sum w_{pile-up}$ is the sum of weights of all tracks associated with pile-up events and $\sum w_{noCorres}$ is the sum of weights of all tracks which were not matched to truth particles. It can be noted that for vertex algorithms which do not assign lower weights to tracks with respect to the vertex candidate during the fit, the sum of the weights of all tracks fitted to the vertex is equal to the track multiplicity. If the value of the purity for a reconstructed primary vertex candidate is above 0.5, it is assumed that the main contribution to the reconstructed vertex stems from signal tracks and therefore the signal primary vertex is correctly identified.

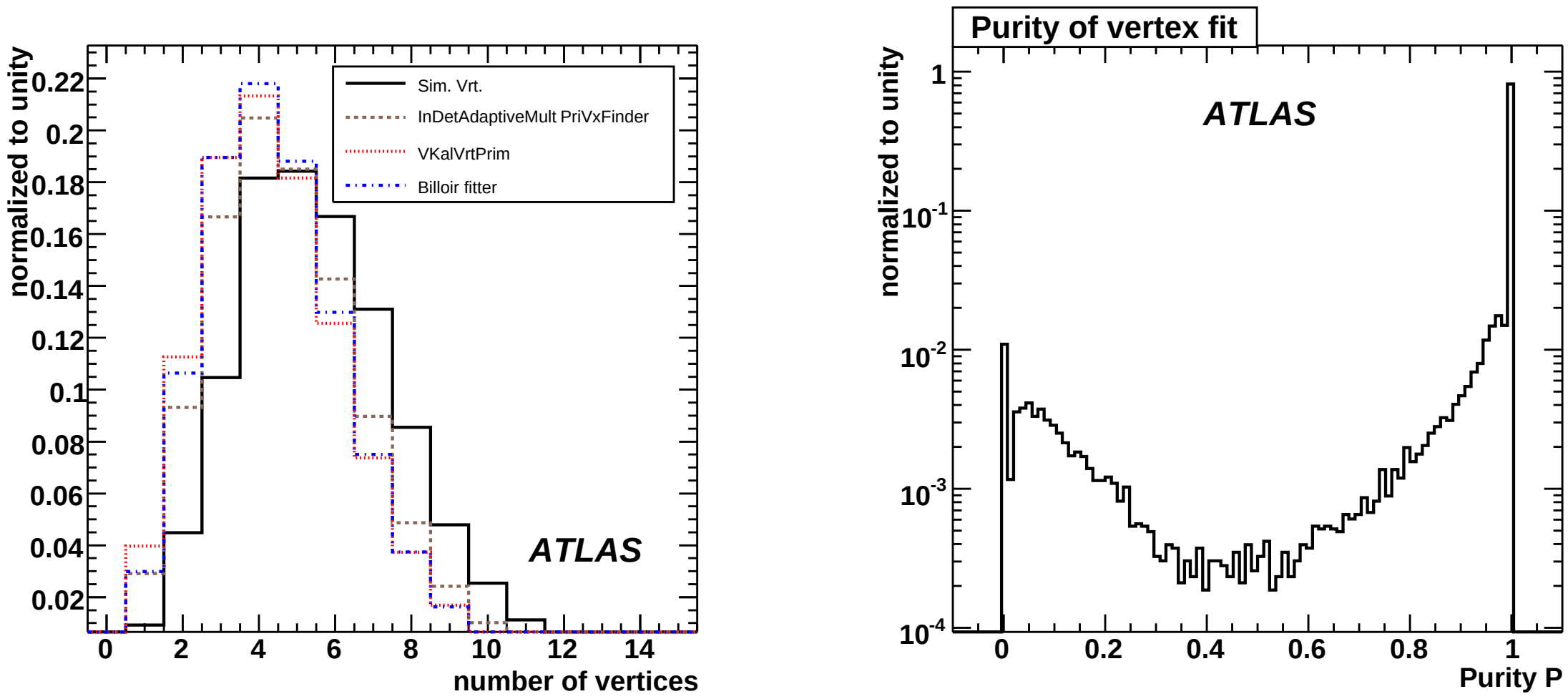

Figure 1: Left: Number of simulated (black line) and reconstructed vertices per event for different reconstruction algorithms (dotted lines); right: distribution for the *InDetAdaptiveMultiPriVxFinder* of the primary vertex purity as defined in Eq. (2). The $WH(m_H = 120\text{GeV}) \rightarrow \mu\nu b\bar{b}$ event sample has been used for these studies.

In Figure 1 (right) the distribution of the purity of the first vertex in the output list of the *InDetAdaptiveMultiPriVxFinder* for $WH(120) \rightarrow \mu\nu b\bar{b}$ events is shown. It can be noted that in the majority of cases the purity of a tagged signal primary vertex is close to unity.

In Table 1 the fraction of events with correctly identified vertices for both distance based and purity based definitions are shown for different reconstruction algorithms. The $WH(120) \rightarrow \mu\nu b\bar{b}$ and $WH(120) \rightarrow \mu\nu u\bar{u}$ Monte Carlo samples with low luminosity pile-up were used for this study. It can be noted that the best performance is shown by the *InDetAdaptiveMultiPriVxFinder* and by *VKalVrtPrim*.

Table 1: Fractions of correctly identified primary vertices for different primary vertex reconstruction algorithms. The fractions are calculated according to the purity based (1) and distance based (2) definitions of the efficiency. The fraction of events where the signal vertex is in the list of reconstructed vertices (column denoted *in list*) is calculated according to the distance based definition.

| | $WH \to \mu\nu b\bar{b}$ | | | $WH \to \mu\nu u\bar{u}$ | | |
|---|---|---|---|---|---|---|
| Algorithm | (1) [%] | (2) [%] | in list [%] | (1) [%] | (2) [%] | in list [%] |
| *InDetAdaptiveMultiPriVxFinder* | $93.7 \pm 0.1$ | $93.8 \pm 0.1$ | $99.1 \pm 0.1$ | $96.1 \pm 0.1$ | $96.2 \pm 0.1$ | $99.3 \pm 0.1$ |
| *VKalVrtPrim* | $94.4 \pm 0.1$ | $94.5 \pm 0.1$ | $99.1 \pm 0.1$ | $95.1 \pm 0.1$ | $95.2 \pm 0.1$ | $99.4 \pm 0.1$ |
| *InDetPriVxFinder (Billoir Full)* | $89.8 \pm 0.2$ | $89.3 \pm 0.2$ | $97.6 \pm 0.1$ | $94.0 \pm 0.2$ | $93.8 \pm 0.2$ | $98.2 \pm 0.1$ |
| *InDetPriVxFinder (Billoir Fast)* | $89.8 \pm 0.2$ | $89.3 \pm 0.2$ | $97.6 \pm 0.1$ | $94.0 \pm 0.2$ | $93.8 \pm 0.1$ | $98.2 \pm 0.1$ |

Also no significant difference is observed between purity based and distance based definitions of the efficiency. In addition, the observation is made that in the single-collision mode, the rate of wrongly identified vertices is essentially zero for all algorithms.

The misidentification of the signal vertex leads to a wrong estimate of the longitudinal impact parameters of tracks. This wrong estimate may lead to an incorrect identification of *b*- and light quark jets and a decrease of the *b*-tagging performance. A dependence of the *b*-tagging performance on the type of the primary vertex reconstruction algorithm is therefore expected.

### 3.3.2 Pointing Lepton

In the events containing isolated leptons (electrons or muons), or even a lepton inside a jet coming possibly from semileptonic decays of heavy hadrons, the properties of these leptons can be used to further increase the efficiency to identify correctly the primary vertex. The trajectory of a reconstructed lepton can be extrapolated to the interaction region. Each reconstructed vertex, for which the lepton has a longitudinal impact parameter $z_0 < 3\,\text{mm}$, gets an additional weight and can then be tagged as the signal vertex after reordering the list of reconstructed vertices. In Table 2 the fractions of events with correctly identified primary vertices after reordering according to the pointing lepton information are shown for different primary vertex finding algorithms. To estimate these fractions, the distance based definition of efficiency is used. Comparing to Table 1, it can be noted that the use of information from a reconstructed lepton increases the efficiency of correctly identifying the signal primary vertex in case of the *Billoir fitters*, which use the sum of squares of transverse momenta of tracks as sorting criteria. In most cases of misidentified vertices, the direction of the isolated lepton from the $W$ boson decay is not in the kinematical acceptance of the inner detector, e.g. for the *InDetAdaptiveMultiPriVxFinder* $(73 \pm 1)\%$ of the outliers have the isolated muon with $|\eta_\mu| > 2.5$. Events with a well reconstructed isolated lepton, as it is the case in many analyses, thus show a much lower fraction of misidentified primary vertices.

Table 2: Fraction of correctly identified primary vertices for different algorithms using the distance based definition with reordering due to a lepton matched to the primary vertex.

| | $WH \to \mu\nu b\bar{b}$ | $WH \to \mu\nu u\bar{u}$ |
|---|---|---|
| *InDetAdaptiveMultiPriVxFinder* | $94.7 \pm 0.1$ | $96.0 \pm 0.1$ |
| *VKalVrtPrim* | $94.6 \pm 0.1$ | $95.2 \pm 0.1$ |
| *InDetPriVxFinder (Billoir Full)* | $93.7 \pm 0.1$ | $95.0 \pm 0.1$ |
| *InDetPriVxFinder (Billoir Fast)* | $93.7 \pm 0.1$ | $95.0 \pm 0.1$ |

### 3.4 Primary Vertex Resolution

In presence of the pile-up vertices spatial resolutions of the reconstructed primary vertex may degrade due to the partial acceptance of tracks from the superposed events. Tables 3 and 4 show the resolutions for the $x$ and $z$ coordinates of reconstructed primary vertices for different reconstruction algorithms.

Table 3: Resolutions on the $x$ coordinate of primary vertices reconstructed with different strategies in $WH \to \mu\nu b\bar{b}$ and $WH \to \mu\nu u\bar{u}$ signal only and pile-up scenarios. The numbers shown are the widths of a fit with a single Gaussian to the corresponding distributions of residuals.

| | $\sigma_x$ [$\mu$m] in $WH \to \mu\nu b\bar{b}$ | | $\sigma_x$ [$\mu$m] in $WH \to \mu\nu u\bar{u}$ | |
|---|---|---|---|---|
| Algorithm | signal only | pile-up | signal only | pile-up |
| *InDetAdaptiveMultiPriVxFinder* | $11.46 \pm 0.05$ | $11.66 \pm 0.05$ | $10.13 \pm 0.05$ | $10.34 \pm 0.05$ |
| *VKalVrtPrim* | $11.44 \pm 0.05$ | $11.59 \pm 0.05$ | $10.04 \pm 0.05$ | $10.25 \pm 0.06$ |
| *InDetPriVxFinder (Billoir Full)* | $12.22 \pm 0.05$ | $12.51 \pm 0.05$ | $11.01 \pm 0.06$ | $11.17 \pm 0.05$ |
| *InDetPriVxFinder (Billoir Fast)* | $12.23 \pm 0.05$ | $12.50 \pm 0.05$ | $11.01 \pm 0.06$ | $11.17 \pm 0.06$ |

Table 4: Resolutions on the $z$ coordinate of primary vertices reconstructed with different strategies in $WH \to \mu\nu b\bar{b}$ and $WH \to \mu\nu u\bar{u}$ signal only and pile-up scenarios. The numbers shown are the widths of a fit with two Gaussians to the corresponding distribution of residuals ($\sigma_{z,N}$ for the narrow and $\sigma_{z,T}$ for the broad component, respectively). *Core Fraction* is the fraction of events contained in the narrow Gaussian.

| | | $WH \to \mu\nu b\bar{b}$ | | | $WH \to \mu\nu u\bar{u}$ | | |
|---|---|---|---|---|---|---|---|
| Algorithm | | $\sigma_{z,N}$, [$\mu$m] | $\sigma_{z,T}$, [$\mu$m] | Core Fraction [%] | $\sigma_{z,N}$, [$\mu$m] | $\sigma_{z,T}$, [$\mu$m] | Core Fraction [%] |
| *InDetAdaptiveMulti-PriVxFinder* | signal only | $41.3 \pm 0.5$ | $98 \pm 2$ | 76.3 | $35.3 \pm 0.5$ | $79 \pm 2$ | 75.8 |
| | pile-up | $40.9 \pm 0.5$ | $95 \pm 2$ | 75.1 | $34.9 \pm 0.5$ | $77 \pm 2$ | 75.5 |
| *VKalVrtPrim* | signal only | $40.2 \pm 0.5$ | $92 \pm 2$ | 71.1 | $36.0 \pm 0.5$ | $82 \pm 2$ | 78.1 |
| | pile-up | $41.2 \pm 0.5$ | $98 \pm 2$ | 74.1 | $34.3 \pm 0.5$ | $76 \pm 2$ | 73.2 |
| *InDetPriVxFinder (Billoir Full)* | signal only | $49.3 \pm 0.5$ | $132 \pm 3$ | 76.3 | $40.4 \pm 0.6$ | $99 \pm 3$ | 76.4 |
| | pile-up | $48.6 \pm 0.5$ | $127 \pm 3$ | 76.7 | $41.0 \pm 0.5$ | $103 \pm 3$ | 79.1 |
| *InDetPriVxFinder (Billoir Fast)* | signal only | $49.3 \pm 0.5$ | $132 \pm 3$ | 76.2 | $40.3 \pm 0.6$ | $98 \pm 3$ | 76.0 |
| | pile-up | $48.6 \pm 0.5$ | $127 \pm 3$ | 76.5 | $40.9 \pm 0.5$ | $103 \pm 3$ | 79.0 |

The numbers shown are the widths of a fit with a single Gaussian for the $x$ coordinate and two Gaussians for the $z$ coordinate, respectively, to the corresponding distribution of the residuals. In addition the ratio of the narrow Gaussian and the sum of the two Gaussians is given for the $z$ coordinate. It can be noted that the best results are obtained with the *VKalVrtPrim* and *InDetAdaptiveMultiPriVxFinder* and that the values of coordinate resolutions for all algorithms are very close to those obtained for the non-pile-up case. It can therefore be concluded, that in the low-luminosity conditions the degradation of spatial resolution appears to be very small.

### 3.5 Impact on $b$-Tagging Performance

The beam interaction region in ATLAS has an approximately gaussian transverse profile with $\sigma = 15\mu m$. This size is smaller than the typical transverse impact parameter ($d_0$) [6] resolution of charged particle tracks. $b$-tagging algorithms based on transverse impact parameter significances only (IP2D method [4]) are thus only slightly sensitive to the details of the primary vertex reconstruction and the presence of pile-up vertices. This is demonstrated in the first column of Table 5. Since the vertex reconstruction accuracy in the longitudinal direction is also better than the typical longitudinal impact parameter resolution ($z_0$ [6]), the $b$-tagging performance is not influenced significantly by the choice of the primary vertex algorithm in the case where only the signal vertex is present. This can be seen by comparing the rows labelled *signal only* in Table 5 for the different primary vertex reconstruction algorithms.

Table 5: Light jet rejection for WH($m_H = 120$ GeV) events (signal only and pile-up scenarios) using different primary vertex reconstruction algorithms. Only jets with at least one track associated are considered in these performance studies. $IP2D, IP3D$ and $IP3D + SV1$ $b$-tagging methods are described in details in [4]

| algorithm | | IP2D 50% | IP2D 60% | IP3D 50% | IP3D 60% | IP3D + SV1 50% | IP3D + SV1 60% |
|---|---|---|---|---|---|---|---|
| *InDetAdaptiveMulti-PriVxFinder* | signal only | $133\pm6$ | $46\pm1$ | $226\pm14$ | $67\pm2$ | $438\pm36$ | $119\pm5$ |
| | pile-up | $135\pm6$ | $45\pm1$ | $160\pm8$ | $52\pm1$ | $230\pm14$ | $82\pm3$ |
| | ratio [%] | $102\pm6$ | $98\pm3$ | $71\pm8$ | $78\pm4$ | $53\pm10$ | $69\pm6$ |
| *VKalVrtPrim* | signal only | $131\pm6$ | $46\pm1$ | $234\pm14$ | $69\pm2$ | $486\pm42$ | $129\pm6$ |
| | pile-up | $141\pm6$ | $47\pm1$ | $136\pm6$ | $50\pm1$ | $220\pm12$ | $80\pm3$ |
| | ratio [%] | $108\pm6$ | $102\pm2$ | $58\pm7$ | $72\pm4$ | $45\pm10$ | $68\pm6$ |
| *InDetPriVxFinder (Billoir Fast)* | signal only | $123\pm5$ | $46\pm1$ | $225\pm13$ | $68\pm2$ | $449\pm38$ | $113\pm5$ |
| | pile-up | $124\pm5$ | $41\pm1$ | $140\pm6$ | $48\pm1$ | $220\pm12$ | $71\pm2$ |
| | ratio [%] | $101\pm6$ | $89\pm3$ | $62\pm7$ | $71\pm3$ | $49\pm10$ | $62\pm3$ |
| *InDetPriVxFinder (Billoir Full)* | signal only | $123\pm5$ | $46\pm1$ | $225\pm13$ | $68\pm2$ | $450\pm38$ | $114\pm5$ |
| | pile-up | $124\pm5$ | $41\pm1$ | $140\pm6$ | $48\pm1$ | $221\pm12$ | $71\pm2$ |
| | ratio [%] | $101\pm6$ | $89\pm3$ | $62\pm7$ | $71\pm3$ | $49\pm10$ | $62\pm3$ |

The presence of pile-up vertices, however, changes the situation for the longitudinal coordinate considerably. The $z$ position resolution is not affected significantly by the presence of additional vertices as is shown in Table 4, but the longitudinal size of the beam interaction region is much bigger ($\sigma_z \approx 5.6$ cm) than the transverse one. Any misidentification of the primary interaction vertex thus produces artificially large $z_0$ track impact parameters. Consequently, most of the tracks are rejected by the $b$-tagging track selection procedure in these cases and the jets to be tagged do not have any tracks associated. Such jets can not be tagged as $b$-quark jets anymore. Since the vertex misidentification probability is not negligible in some event topologies, as can be seen from Table 1, the performance of $b$-tagging algorithms using information from the longitudinal coordinate is appreciably affected. The degradation of the $b$-tagging performance in the presence of pile-up vertices for these algorithms is shown in the second and third columns of Table 5 for the *IP3D* and *IP3D+SV1* algorithms which use this information.

An additional complication for $b$-tagging with pile-up originates from the rather large amount of soft jets produced in pile-up interactions. These jets do not have reconstructed tracks inside at all in many cases and are thus considered as light quark jets by the $b$-tagging algorithms. This, however, causes an unphysical change of the $b$-tagging performance not related to the performance of the algorithms themselves, because a different set of jets is used to estimate the performance. To minimize this effect, the results in Table 5 are obtained for jets with at least one associated track. This approach is different from the one normally used for $b$-tagging performance studies. The numbers in Table 5 thus have to be taken with care when comparing with other $b$-tagging results. The observed degradation of the $b$-tagging performance confirms the results obtained in a previous study [11], where a more dramatic decrease of the $b$-tagging performance was demonstrated for the case of the nominal LHC luminosity ($10^{34}$cm$^{-2}$s$^{-1}$) when each beam crossing produces 24 pile-up events on average.

There are several possibilities to improve the $b$-tagging performance in pile-up scenarios. The pres-ence of jets without any tracks associated to them (especially if the requirement of a small longitudinal impact parameter is not fulfilled for most of the tracks) is an indication that the correct primary vertex has not been reconstructed in this particular event. In such cases, one could either try to use the second vertex from the list of primary vertex candidates (according to the ordering of vertices as applied by the primary vertex reconstruction algorithm) or drop the information in the longitudinal direction, thus falling back to a $b$-tagging algorithm working in the transverse plane only. The reconstruction of the primary event vertex, as currently implemented, is a universal procedure not depending on the topology of the event. Adding information specific to a particular analysis could help in some cases to increase the fraction of correctly identified primary vertices, e.g. reconstructing the primary vertex only from prompt isolated high $p_T$ leptons for channels where these are present. These options have to be studied in the future.

## 4 Inclusive Secondary Vertex Reconstruction in Jets

A $b$-jet originates from a $b$-quark, which produces a $b$-hadron in the fragmentation. The $b$-hadron then decays due to electroweak interactions, which cause the transition of the $b$-quark preferably into a $c$-quark ($|V_{cb}|^2 \gg |V_{ub}|^2$), which then subsequently also undergoes a weak decay. As a result, the typical topology of the particles in a $b$-jet seen in the detector is a decay chain with two vertices, one stemming from the $b$-hadron decay and at least one from $c$-hadron decays. The eventual intermediate presence of excited $b$- or $c$-hadron states does not change this picture because their strong or electromagnetic decays do not cause measurable lifetimes. A unique feature of jets originating from the fragmentation of $b$-quarks is thus the presence of $b$- and $c$-hadron decay vertices.

An exclusive reconstruction of these decays cannot be done with high efficiency. The large $b$-hadron masses lead to a huge number of possible decay modes with very small branching ratios, many of them involving neutral particles (which cannot be used for vertex reconstruction). The reconstruction of secondary $b$- and $c$-hadron decay vertices in jets thus has to be done in an inclusive way, where the number of charged particle tracks originating from $b$- and $c$-hadron decays is not known a-priori.

Trying to resolve the $b$- and $c$-hadron vertices of the decay cascade separately is very difficult for the following reasons:

- The probability to have at least two reconstructed charged particle tracks both from the $b$- and $c$-hadron decays is much less than 100%. This is both because of the charged particle multiplicities involved in these decays as well as the limited track reconstruction efficiency, mainly because of material interactions in the detector (as discussed in Section 5).
- The resolutions of the relevant track parameters, especially at low transverse momenta, are not sufficient to separate the two vertices efficiently.

Two different approaches for the inclusive reconstruction of secondary vertices are presented in Sections 4.1 and 4.2. The first one is based on a classical approach of fitting a single geometrical vertex. As explained, this is strictly speaking not the correct hypothesis, however, for the reasons given above, this is an approximation that works well for a large fraction of cases. The second algorithm is based on a kinematical approach, assuming that the primary event vertex and the $b$- and $c$-hadron decay vertices lie approximately on the same line, the flight path of the $b$-hadron. These underlying assumptions are discussed in more detail in the corresponding sections.

The most important issues in inclusive vertex reconstruction are to reconstruct the decay vertices with high efficiency and to associate efficiently to the vertices the charged particle tracks coming from the corresponding decays of $b$- and $c$-hadrons.

Another important issue is the detection and removal of sources of tracks with large impact parameters which do not have any relation to the $b$-quark content of a jet. $K^0$ and $\Lambda^0$ decays, $\gamma$-conversions and hadronic interactions in the detector material must be removed as efficiently as possible before applying *b*-tagging.

Combining the information from secondary vertices and track impact parameters allows the optimization of the *b*-tagging performance and is decribed in Section 6.

## 4.1 The BTagVrtSec Algorithm

The main purpose of the *BTagVrtSec* algorithm is efficient *b*-jet tagging based on the detection of a secondary vertex inside a jet. It reconstructs secondary vertices due to *b*- and/or *c*-hadron decays inside a jet with high efficiency and calculates a jet weight, a discriminating variable which may be combined with similar variables from other tagging algorithms (see Section 6). This algorithm is based on the VKalVrt [2] vertex reconstruction package.

### 4.1.1 BTagVrtSec Vertex Reconstruction

As stated above, to separate *b*-quark jets from jets produced by light quarks and gluons, it is sufficient to detect unambiguously *b*- and *c*-hadron decay vertices inside the jet. The idea of the *BTagVrtSec* algorithm is to maximize the *b*/*c*-hadron vertex detection efficiency, keeping at the same time the probability to find a fake vertex inside light jets low. The default version of the algorithm constructs a single secondary vertex from the *b*- and *c*-hadron decay products. As justification of such an approach it should be noted, that the ability to reconstruct *b*- and *c*-hadron decay vertices separately is quite limited as already explained previously. Moreover, the precise reconstruction of the decay chain is less important for *b*-tagging than the detection of the secondary decay itself. The single vertex approximation allows a high secondary vertex detection efficiency, necessary for powerful *b*-tagging at the price of an imprecise kinematical reconstruction in cases when the distance between the *b*-and *c*-hadron vertices is big. Another advantage of the single vertex approximation is a more simple algorithm that might be easier to tune and calibrate on data.

The *BTagVrtSec* algorithm starts with a selection of tracks inside a jet. Tracks are selected with the same quality cuts as for the primary vertex reconstruction algorithm *VKalVrtPrim*, except for a relaxed cut on the transverse track impact parameter ($d_0 \leq 3.5$ mm) and no requirement of the presence of a hit in the first layer (b-layer) of the pixel detector. This is done in order to maximize the efficiency to reconstruct $V^0$ decays and material interactions with subsequent removal of the corresponding tracks from the *b*-tagging procedure. Tracks are selected in a cone around the jet axis. The size of the cone is a tunable parameter (the default value is $\Delta R = 0.4$) which currently does not depend on the jet reconstruction algorithm.

The vertex search itself starts with a determination of all track pairs which form good ($\chi^2 < 4.5$) two-track vertices inside the jet. In addition, each track of the pair must have a three dimensional distance from the primary vertex[2] divided by its error higher than 2.0 and the sum of these two significances must be higher than 6.0. In order to decrease the fake rate an additional requirement $(\vec{V}_{2tr} - \vec{V}_{primary}, \vec{P}_{jet}) > 0$ is used.

Some of the reconstructed two-track vertices stem from $K_s^0$ and $\Lambda^0$ decays, $\gamma \to e^+e^-$ conversions and hadronic interactions in the detector material. The corresponding distributions are shown in Figure 2. Figures 2 a) and b) show the invariant $\pi^+\pi^-$ and $p\pi^-$ mass spectra for accepted two particle vertices with peaks due to $K_s^0$ and $\Lambda^0$ decays. Figure 2 c) shows the distance between the primary and secondary vertices in the transverse plane with peaks due to interactions in the material of the beam pipe and pixel detector layers. Charged particle tracks coming from such vertices are marked as *bad* and do not participate further in the following *b*-tagging procedure for the given jet.

[2] The distance between the primary vertex and the point of closest approach of the track to this vertex.

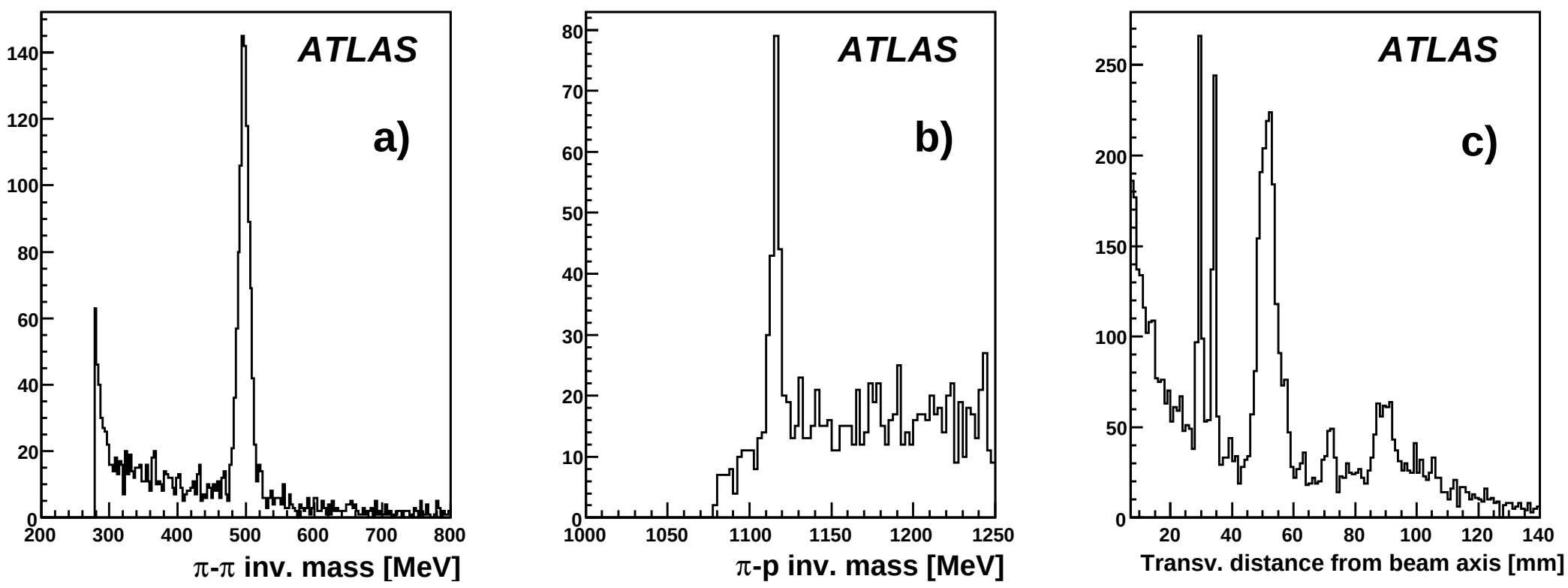

Figure 2: Some distributions for reconstructed two track vertices:
a) the $\pi^+\pi^-$ invariant mass spectrum with a peak of $K^0$ decays; b) the $p\pi$ invariant mass spectrum with a peak of $\Lambda^0$ decays; c) the distance in the transverse plane between the primary and secondary vertices with the peaks due to interactions in the beam pipe (two walls at R≈30 mm) walls and pixel layers (R=50.5 mm and R=88.5 mm).

In the next step of the algorithm all tracks inside the jet from accepted two-track vertices except for marked $V^0$ decays and material interactions are combined into one *secondary track* list and the vertex fitting procedure from the *VKalVrt* package tries to fit a single secondary vertex out of all these tracks. If the resulting vertex has an unacceptable $\chi^2$, the track with the highest contribution to the vertex $\chi^2$ is deleted from the *secondary track* list and the vertex fit is redone. This procedure iterates until a good $\chi^2$ of the vertex fit is obtained or all tracks from the *secondary track* list have been removed.

Secondary vertices in light quark jets mainly stem from tracks coming from the primary vertex, typically with a bad measurement of their track parameters and errors.

Some *b*-tagging efficiency and rejection calibration algorithms require *negative tail* vertices. For the use in those algorithms the requirement for two-track vertices $(\vec{V}_{2tr} - \vec{V}_{prim}, \vec{P}_{jet}) > 0$ may be dropped. In this condition in some cases *BTagVrtSec* reconstructs a final single secondary vertex *behind* the primary vertex: $(\vec{V}_{sec} - \vec{V}_{prim}, \vec{P}_{jet}) < 0$. Such secondary vertices are mostly fake vertices but give the necessary reference for calibration.

### 4.1.2 *b*-Tagging with BTagVrtSec

Each *b*-tagging algorithm is based on variables which show significantly different behaviour for *b*-jets and light jets. For secondary vertex based algorithms the first variable is the presence of a reconstructed vertex in the jet itself, whose probability is big for *b*-quark jets and small for light quark jets. The vertex reconstruction procedure can, however, provide much more information which also may be used in a *b*-tagging algorithm to increase the efficiency. In order not to have a too complex procedure which is still easy to calibrate and to control, only the most sensitive variables should be taken into account.

Although *BTagVrtSec* can be used as a standalone *b*-tagging algorithm, it was developed primarily for working in combination with track impact parameter based algorithms [4]. The distance between primary and secondary vertices is thus not used because this kind of lifetime information is already contained in the track impact parameters. Three additional variables have been chosen for the *BTagVrtSec* tagging algorithm:

1. The mass of the reconstructed secondary vertex: $M$.

2. The ratio of the energy of charged particle tracks included in the secondary vertex and the total energy of all charged particle tracks in the jet: $R$;
3. The number of good (excluding identified $V^0$ decays or material interactions) two-track vertices in the jet: $N$.

The distributions of the first two variables are shown in Figure 9 in Section 5 for light, charm and $b$-jets. The distributions of the number of two-track vertices can be found in [4]. In standalone mode (without combination with track impact parameter based tagging algorithms) the three dimensional distance between the primary and secondary vertices may also be used.

To be reliable any $b$-tagging algorithm must be calibrated on data. To facilitate the calibration and to reduce the necessary amount of data the chosen variables have been transformed (for details see [12]):

- Invariant mass: $M' = \dfrac{M}{M+1}$;
- Energy ratio: $R' = R^{0.7}$;
- Number of good two-track secondary vertices: $N' = \log N$.

Due to the efficiency to reconstruct a secondary vertex inside a jet not reaching 100%, the probability density functions (PDF) of the vertex based variable have to contain a $\delta$–function [12].

$$PDF = (1-\varepsilon)\cdot\delta(M',F',N') + \varepsilon\cdot ASH(M',F',N')$$

with $\varepsilon$ being the efficiency to reconstruct a secondary vertex inside a jet. The continuous probability density function of the vertex variables is constructed from multidimensional calibration histograms using the ASH smoothing method [13].

Two slightly different taggers based on the *BTagVrtSec* algorithms are available in ATLAS, denoted $SV1$ and $SV2$. They use exactly the same variables but handle them in a different way. $SV1$ treats $M'$ and $R'$ jointly and adds $N'$ as independent variable (2+1 decomposition), whereas $SV2$ uses joint three-dimensional probability density functions.

The only criterion for the selection of variables and tuning of the *BTagVrtSec* algorithm was $b$-tagging performance. Other quantities like the purity of the reconstructed secondary vertices were not considered. Although the achieved quality of the $b$-hadron decay vertex is quite good (see Section 5), for applications other than $b$-tagging *BTagVrtSec* may require a different tuning.

## 4.2 The JetFitter Algorithm

A new inclusive secondary vertexing algorithm which exploits the topological structure of weak $b$- and $c$-hadron decays inside a jet was recently developed.

### 4.2.1 Reconstruction of the Decay Chain Topology

As already stated, the fragmentation of a $b$-quark results in a decay chain with two vertices, one stemming from the $b$-hadron decay and at least one from $c$-hadron decays.

As described in Section 4.1, the *BTagVrtSec* algorithm relies on a vertexing algorithm in which displaced tracks are selected and an inclusive single vertex is obtained using a Kalman based $\chi^2$ fit (as sketched in Figure 3).

The algorithm described here, called *JetFitter*, is based on a different hypothesis. It assumes that the $b$- and $c$-hadron decay vertices lie on the same line defined through the $b$-hadron flight path. All charged particle tracks stemming from either the $b$- or $c$-hadron decay thus intersect this $b$-hadron flight axis. There are several advantages to this method:

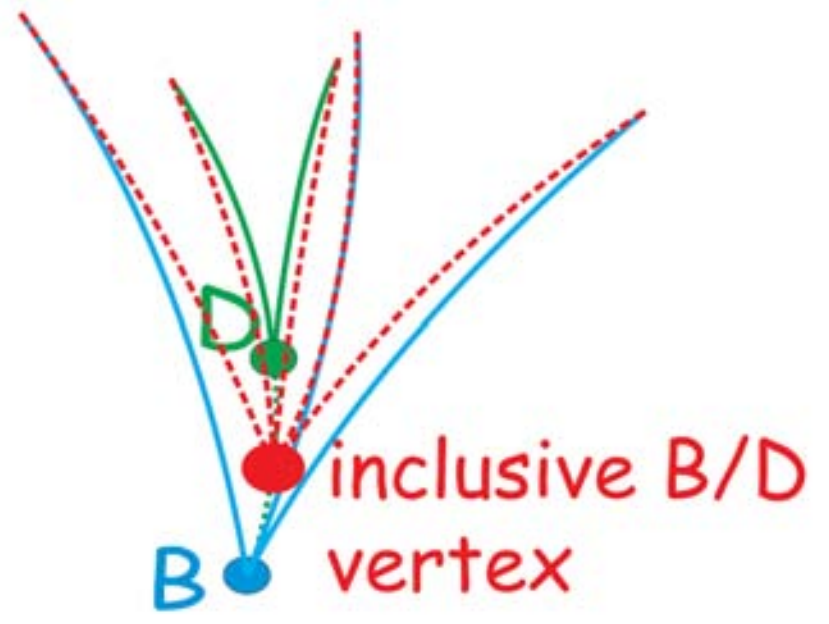

Figure 3: *BTagVrtSec* fits all displaced tracks to an inclusive vertex.

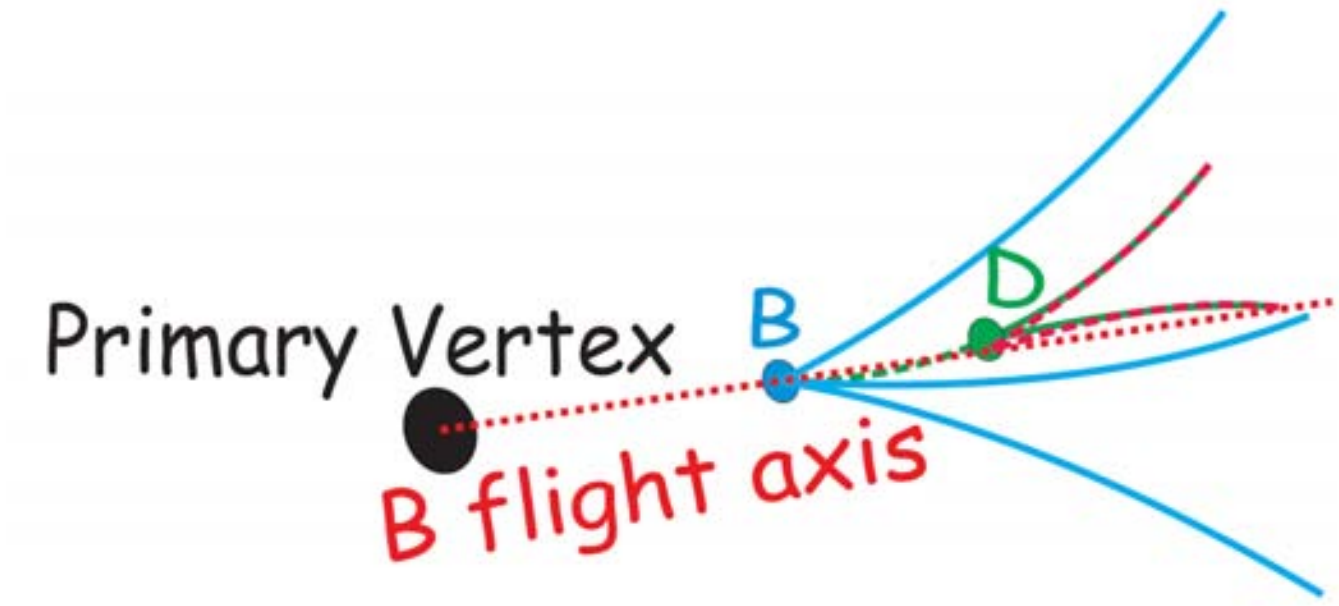

Figure 4: JetFitter performs a multi-vertex fit using the *b*-hadron flight direction as constraint.

- Incomplete topologies can also be reconstructed (in principle even the topology with a single track from the *b*-hadron decay and a single track from the *c*-hadron decay is accessible).
- The fit evaluates the compatibility of the given set of tracks with a *b*-*c*-hadron like cascade topology, increasing the discrimination power against light quark jets.
- Constraining the tracks to lie on the *b*-hadron flight axis reduces the degrees of freedom of the fit, increasing the chance to separate the $b/c$-hadron vertices.

From the physics point of view this hypothesis is justified through the kinematics of the particles involved as defined through the hard *b*-quark fragmentation function and the masses of *b*- and *c*-hadrons. The lateral displacement of the *c*-hadron decay vertex with respect to the *b*-hadron flight path is small enough not to violate significantly the basic assumption within the typical resolutions of the tracking detector (see Figure 4).

This hypothesis, extensively used in the *JetFitter* algorithm, was explored for the first time in the *ghost track algorithm* developed by the SLD Collaboration [14], where the already defined *b*-hadron flight axis is substituted by a *ghost track* and where a numerical global $\chi^2$ minimisation procedure was used to perform the multi-vertex fit.

### 4.2.2 The JetFitter Vertex Reconstruction Algorithm

In JetFitter the vertexing task is mathematically implemented as an extension of the Kalman Filter formalism for vertex reconstruction [7] and the decay chain is described through the determination of the following variables:

$$\vec{d} = (x_{PV}, y_{PV}, z_{PV}, \phi, \theta, d_1, d_2, ..., d_N), \tag{3}$$

with:

- $(x_{PV}, y_{PV}, z_{PV})$: the primary vertex position.
- $(\phi, \theta)$: the azimuthal and polar directions of the *b*-hadron flight axis.
- $(d_1, d_2, ..., d_N)$: the distances of the fitted vertices, defined as the intersections of one or more tracks and the *b*-hadron flight axis, to the primary vertex position along the flight axis (N representing the number of vertices).

Before starting the fit, the variables are initialised with their prior knowledge:

- The primary vertex position (with covariance matrix), as provided by the primary vertex finding algorithm.
- The $b$-hadron flight direction, approximated by the direction of the jet axis, the error being provided by the convolution of the jet direction resolution with the average displacement of the jet axis relative to the $b$-hadron flight axis, as determined from Monte Carlo simulations.

The fit is then performed, resulting in the minimization of the $\chi^2$ containing the weighted residuals of all tracks with respect to their vertices on the $b$-hadron flight axis. The charged particle tracks to be used in the determination of the decay chain are selected according to the $\Delta R$ matching criterion explained in Section 2 and a further track selection is applied, in order to reduce the amount of fake tracks.

After the primary vertex and the $b$-hadron flight axis have been initialised, a first fit is performed under the hypothesis that each track represents a single vertex along the $b$-hadron flight axis, until $\chi^2$ convergence is reached, obtaining a first set of fitted variables $(\phi, \theta, d_1, d_2, ..., d_N)$.

A clustering procedure is then performed, where all combinations of two vertices (picked up among the vertices lying on the $b$-hadron flight axis plus the primary vertex) are taken into consideration, filling a *table of probabilities*. After the table of probabilities is filled, the vertices with the highest compatibility are merged, a new complete fit is performed and a new table of probabilities is filled. This procedure is iterated until no pairs of vertices with a probability above a certain threshold remain.

The result of this clustering procedure is a decay topology with a well defined association of tracks to vertices along the $b$-hadron flight axis, with at least one track for each vertex.

More details about the JetFitter vertex reconstruction algorithm can be found in [3].

### 4.2.3 The JetFitter based $b$-Tagging Algorithm

The $b$-tagging algorithm implemented as a first application of *JetFitter* is based on separating $b$-jets from $c$- and light-quark ($u$,$d$,$s$) jets by means of the definition of a likelihood function.

The decay topology is described by the following discrete variables:

1. Number of vertices with at least two tracks.
2. Total number of tracks at these vertices.
3. Number of additional single track vertices on the $b$-hadron flight axis.

while the vertex information is condensed in the following variables:

1. Mass: the invariant mass of all charged particle tracks attached to the decay chain.
2. Energy Fraction: the energy of these charged particles divided by the sum of the energies of all charged particles associated to the jet.
3. Flight length significance $\frac{d}{\sigma(d)}$: the weighted average vertex position divided by its error.

The use of these variables allows the definition of a likelihood function of the form:

$$L^{b,l,c}(x) = \sum_{cat} coeff(cat) \cdot PDF_{cat}(mass) \cdot PDF_{cat}(energyFraction) \cdot PDF_{cat}\left(\frac{d}{\sigma(d)}\right), \quad (4)$$

which has to be parametrised separately for each of the three different flavours.

The information about the decay topology of the jet as reconstructed by *JetFitter* is represented by the category (the coefficient $coeff(cat)$ representing how probable it is to find a certain topology for a given flavour), while the vertex information is contained in the probability distribution functions (*PDFs*).

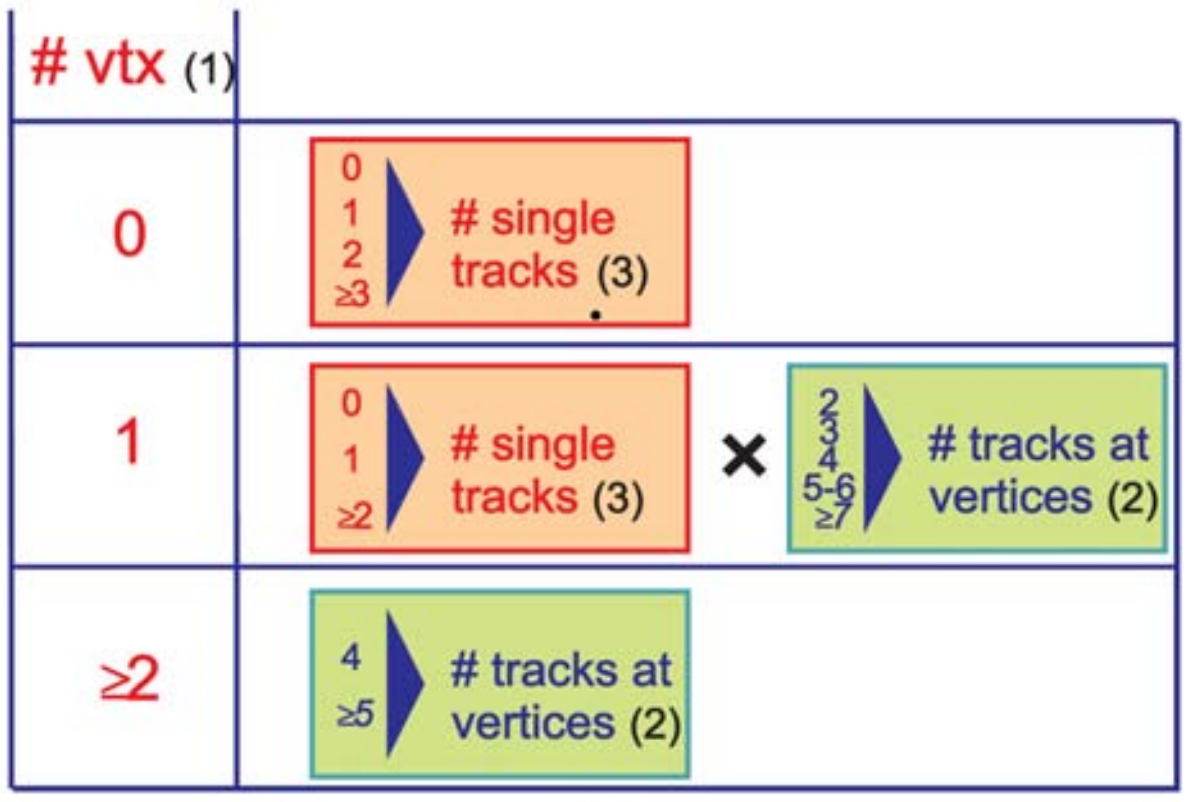

Figure 5: 13 different topologies are defined, combining the three discrete variables in such way as to reduce their correlations. For the case of one single vertex with at least two tracks (1), the discrete *PDFs* for both the other two variables, total tracks at vertices (2) and single additional tracks (3), are used, but they are then considered as uncorrelated, so that their corresponding coefficients are just multiplied.

The discrete variables describing the decay topology are combined according to the scheme of Figure 5 into 13 category coefficients.

In order to reduce the correlations between the decay topology and vertex related variables and to increase the discrimination power, the *PDFs* are made category dependent. This splitting of *PDFs* is done only when strictly needed.

The templates for the vertex variables are shown in Table 6 for all three flavours. Each *PDF* was split independently into the categories it was found to be most correlated with, in order to maintain the number of split *PDFs* to be determined on Monte Carlo simulated events as low as possible. They were determined on the WH($m_H = 120$ GeV) Monte Carlo events with $H \to b\bar{b}$, $H \to c\bar{c}$ and $H \to u\bar{u}$ and on $t\bar{t}$ Monte Carlo events.

The *JetFitter* based $b$-tagging algorithm can be either used as a stand-alone algorithm or in combination with pure impact parameter based algorithms (see Section 6).

## 5 Secondary Vertex Reconstruction Performance

In this section, the performance of the inclusive secondary vertex reconstruction algorithms as described in Sections 4.1 and 4.2 is discussed.

Table 6 shows the rate of reconstructed secondary vertices in $b$-quark and light quark jets in bins of jet transverse momentum and pseudorapidity for both secondary vertex reconstruction algorithms, as obtained on the $t\bar{t}$ and $t\bar{t}jj$ samples.

Above a certain transverse momentum, the efficiency to identify the secondary vertex in a $b$-jet for both algorithms approaches approximately a constant value around $75-80\%$, while the number of vertices reconstructed in light-jets consistently increases with higher jet transverse momentum.

Both effects have to do mainly with the ability to separate real or fake secondary tracks from the primary vertex: tracks from secondaries are selected according to the impact parameter significance of a track with respect to the primary vertex or to the three dimensional flight length significance of the secondary vertex. Therefore, these quantities are strictly related and, above a certain threshold, they are nearly invariant regardless of the boost and thus the flight length of the displaced $b$- or $c$-hadron produced. For low transverse momenta, the charged particle tracks from $b$-hadron decays suffer from larger multiple scattering, thus showing less significant displacements from the primary event vertex and reducing the secondary vertex reconstruction efficiency.

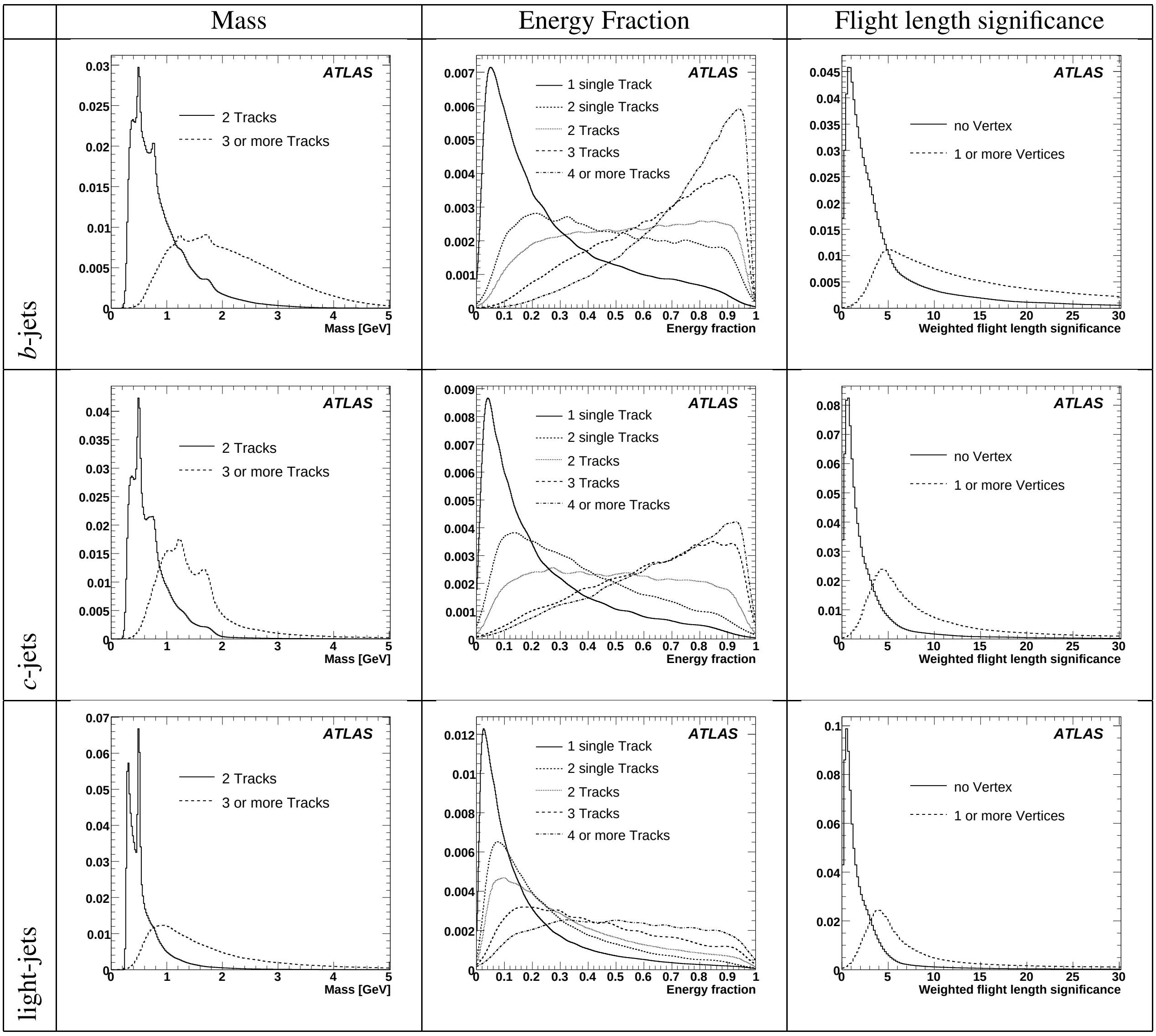

Figure 6: The *PDFs* for the mass, the energy fraction and the flight length significance are shown, separately for the three different jet-flavours and split according to the decay chain topology found by *JetFitter*.

The rising amount of fake vertices in light quark jets with increasing jet transverse momenta is partly due to the nature of the fragmentation process in light quark jets, which produces a larger number of tracks coming from the primary interaction vertex the larger the transverse jet momentum is. Assuming the probability for a reconstructed track to represent an outlying measurement is approximately constant to first approximation, the probability of having tracks which fake secondaries increases roughly linearly with the number of primary tracks. The real secondary vertices produced in light quark jets, like photon conversions and $V^0$ decays represent an additional source of vertices not related to heavy hadron decays. Furthermore, the pattern recognition during the track reconstruction phase is more difficult in dense jets as it is the case for higher jet transverse momenta (see [4]).

Table 6: The fraction of jets with at least one reconstructed secondary vertex passing the selection criteria to be used by the *b*-tagging algorithms in *b*-, *c*- and light quark jets for *BTagVrtSec* (top) and *JetFitter* (bottom). These numbers, given in percent, have been obtained on the $t\bar{t}$ and $t\bar{t}jj$ samples, applying the purification procedure.

| | $0<\|\eta\|<0.5$ | | | $0.5<\|\eta\|<1.0$ | | | $1.0<\|\eta\|<1.5$ | | | $1.5<\|\eta\|<2.0$ | | | $2.0<\|\eta\|<2.5$ | | |
|---|---|---|---|---|---|---|---|---|---|---|---|---|---|---|---|
| $p_T$[GeV] | b | c | l | b | c | l | b | c | l | b | c | l | b | c | l |
| 15–30 | 56.9 | 21.2 | 3.0 | 56.3 | 20.5 | 2.6 | 55.3 | 18.5 | 2.5 | 51.7 | 16.7 | 2.9 | 41.7 | 14.1 | 3.2 |
| | 54.4 | 19.6 | 2.3 | 53.4 | 18.8 | 2.1 | 52.2 | 17.6 | 2.0 | 49.5 | 15.3 | 2.4 | 41.4 | 13.8 | 3.0 |
| 30–50 | 72.4 | 29.6 | 5.3 | 72.1 | 28.9 | 4.5 | 70.7 | 28.1 | 4.6 | 66.5 | 24.0 | 5.1 | 58.5 | 21.0 | 5.6 |
| | 69.6 | 27.1 | 4.1 | 68.5 | 25.7 | 3.6 | 66.8 | 24.5 | 3.7 | 62.9 | 21.9 | 4.4 | 56.8 | 21.1 | 5.6 |
| 50–80 | 78.2 | 35.1 | 7.5 | 78.3 | 33.9 | 6.4 | 76.5 | 32.0 | 6.7 | 72.7 | 29.5 | 7.3 | 65.8 | 27.5 | 7.9 |
| | 74.7 | 31.1 | 5.4 | 73.9 | 29.4 | 4.9 | 71.9 | 28.2 | 5.2 | 67.7 | 25.7 | 6.2 | 63.1 | 25.2 | 7.9 |
| 80–120 | 80.3 | 39.3 | 10.4 | 80.2 | 38.5 | 9.2 | 78.7 | 36.8 | 9.2 | 74.6 | 34.3 | 9.6 | 67.4 | 31.7 | 10.9 |
| | 76.3 | 33.9 | 6.9 | 75.4 | 32.6 | 6.2 | 73.4 | 30.9 | 6.3 | 69.1 | 28.5 | 7.9 | 64.2 | 28.6 | 10.1 |
| 120–200 | 78.4 | 42.6 | 14.7 | 78.0 | 39.8 | 13.3 | 76.9 | 39.9 | 13.5 | 71.8 | 36.5 | 14.1 | 64.2 | 32.1 | 14.8 |
| | 76.3 | 36.8 | 8.9 | 74.5 | 32.2 | 8.1 | 72.4 | 32.5 | 8.6 | 67.0 | 30.6 | 10.0 | 59.9 | 27.3 | 13.5 |

The dependence of the vertex reconstruction efficiency on the jet pseudorapidity is mainly a consequence of the different resolutions which can be achieved in different regions of the Inner Detector: for increasing rapidities, the track resolutions close to the Interaction Point get worse, the track reconstruction efficiency starts to decrease and the number of fake tracks rises (a detailed quantitative explanation of these effects can be found in [6] and [4]).

It can also be concluded that the *BTagVrtSec* algorithm is slightly more efficient in reconstructing secondary vertices inside b quark jets, at the cost of a slightly higher rate of fake vertices produced in light quark jets compared to the *JetFitter* algorithm, at least in the barrel region of the Inner Detector.

Table 7 shows the population of some reconstructed topologies for *b*-quark, *c*-quark and light quark jets for *JetFitter*, as a function of the jet transverse momentum and pseudorapidity. As stated in Sec-tion 4.2, the key feature of *JetFitter* is the ability to distinguish several decay chain topologies, in addition to recognizing the presence of a single inclusive decay vertex. The efficiencies and mistagging rates given above for the different categories show that most of the achievable gain is due to the 1 $Vertex + 1$ $Track$ and 2 $Vertices$ categories, where a gain in rejection against light quark jets of a factor $\sim 4$ and $\sim 16$ can be obtained on approximately $\sim 10\%$ and $\sim 5\%$ of the reconstructed $b$-jets, respectively. The dependence on $p_T$ and $\eta$ follows the same pattern as already described for the inclusive vertex reconstruction.

Table 7: Population of the different topologies of the vertices reconstructed by *JetFitter* in *b*-,*c*- and light quark jets , shown as a function of the jet transverse momentum $p_T$ (top) and the jet pseudorapidity $\eta$. These numbers, given in percent, have been obtained on the $t\bar{t}$ sample, applying the purification procedure.

| | $15<p_T<30$ | | | $30<p_T<50$ | | | $50<p_T<80$ | | | $80<p_T<120$ | | | $120<p_T<200$ | | |
|---|---|---|---|---|---|---|---|---|---|---|---|---|---|---|---|
| | b | c | l | b | c | l | b | c | l | b | c | l | b | c | l |
| Nothing | 30.7 | 66.0 | 87.9 | 18.5 | 56.7 | 83.0 | 13.7 | 51.3 | 79.7 | 11.7 | 46.8 | 76.8 | 10.8 | 43.5 | 71.9 |
| 1 Single Track | 15.3 | 15.2 | 9.4 | 11.4 | 16.5 | 12.3 | 9.9 | 17.4 | 13.6 | 9.3 | 18.0 | 14.7 | 9.7 | 19.1 | 16.5 |
| 2 Single Tracks | 2.8 | 1.3 | 0.4 | 3.7 | 2.1 | 0.7 | 4.5 | 2.6 | 1.0 | 5.5 | 3.5 | 1.4 | 6.8 | 4.3 | 2.2 |
| 1 Single Vertex | 42.8 | 16.2 | 2.2 | 50.3 | 22.3 | 3.9 | 49.6 | 25.1 | 5.2 | 46.3 | 26.8 | 6.4 | 42.2 | 26.9 | 8.1 |
| 1 Vertex + 1 Track | 6.7 | 1.2 | 0.09 | 11.9 | 2.2 | 0.2 | 15.9 | 3.1 | 0.4 | 19.2 | 4.2 | 0.7 | 21.7 | 5.3 | 1.1 |
| 2 Vertices | 1.7 | 0.1 | 0.01 | 4.2 | 0.3 | 0.02 | 6.3 | 0.5 | 0.04 | 8.0 | 0.6 | 0.08 | 8.8 | 0.9 | 0.2 |

| | $0.0<\|\eta\|<0.5$ | | | $0.5<\|\eta\|<1.0$ | | | $1.0<\|\eta\|<1.5$ | | | $1.5<\|\eta\|<2.0$ | | | $2.0<\|\eta\|<2.5$ | | |
|---|---|---|---|---|---|---|---|---|---|---|---|---|---|---|---|
| | b | c | l | b | c | l | b | c | l | b | c | l | b | c | l |
| Nothing | 15.2 | 54.8 | 88.8 | 16.2 | 56.4 | 88.5 | 17.6 | 57.6 | 87.5 | 20.5 | 59.5 | 85.7 | 25.5 | 62.5 | 84.6 |
| 1 Single Track | 10.1 | 16.1 | 8.0 | 10.2 | 16.1 | 8.4 | 11.1 | 16.4 | 9.2 | 12.9 | 17.2 | 10.4 | 14.8 | 16.7 | 10.9 |
| 2 Single Tracks | 4.2 | 2.2 | 0.5 | 4.4 | 2.2 | 0.5 | 4.7 | 2.4 | 0.5 | 5.1 | 2.5 | 0.7 | 5.0 | 2.3 | 0.7 |
| 1 Single Vertex | 46.8 | 23.3 | 2.5 | 47.3 | 22.1 | 2.4 | 46.9 | 21.1 | 2.5 | 44.7 | 18.3 | 3.0 | 41.1 | 16.2 | 3.5 |
| 1 Vertex + 1 Track | 16.8 | 3.1 | 0.2 | 15.6 | 2.7 | 0.2 | 14.3 | 2.3 | 0.2 | 12.8 | 2.2 | 0.2 | 10.5 | 2.0 | 0.3 |
| 2 Vertices | 6.9 | 0.5 | 0.03 | 6.3 | 0.4 | 0.02 | 5.5 | 0.3 | 0.03 | 4.1 | 0.3 | 0.02 | 3.1 | 0.3 | 0.02 |

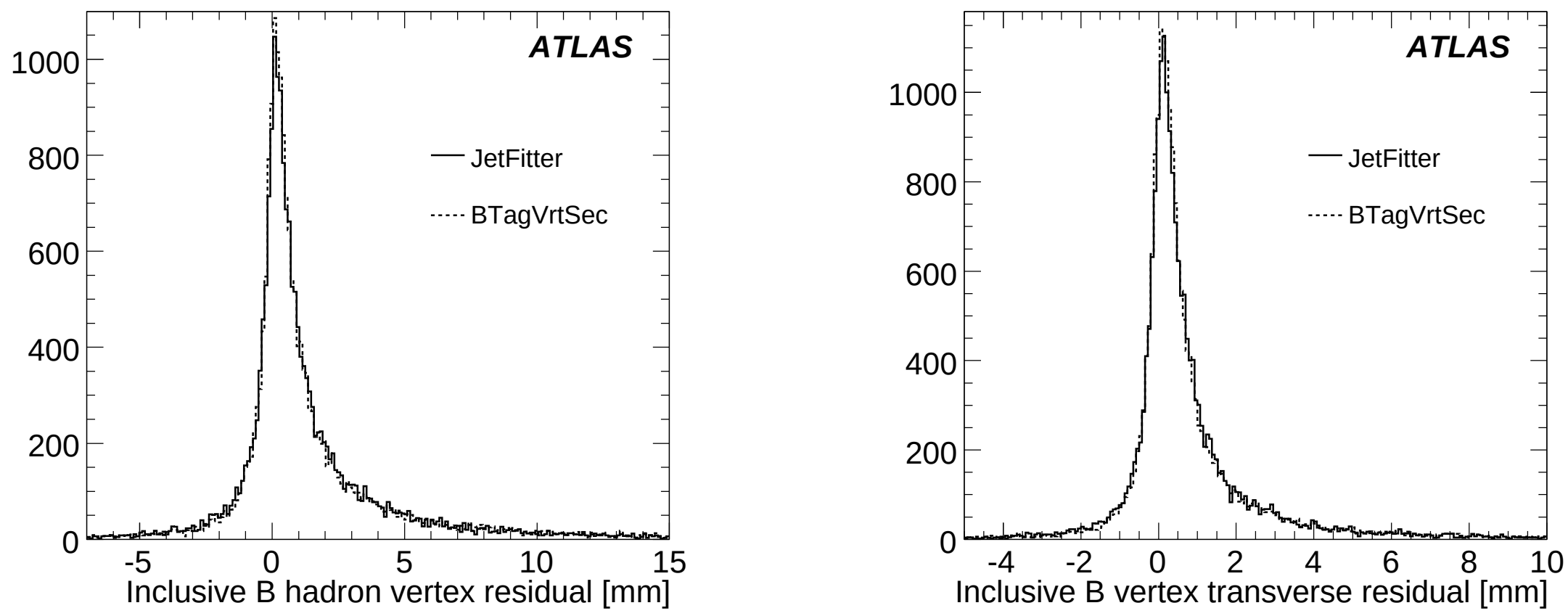

Figure 7: Residuals of the reconstructed three dimensional (left) and transverse (right) flight length of the inclusive secondary vertex with respect to the true $b-$hadron position for both vertexing algorithms.

Figure 7 shows the resolution achieved on the inclusively reconstructed $b$-hadron decay vertex with respect to the true $b$-hadron position. A core can be seen, corresponding to the cases where most of the reconstructed tracks really stem from the $b$-hadron vertex, approaching the intrinsic resolution of the vertex reconstruction, while a very large tail to higher flight lengths can be observed, due to tracks coming from the decay of the charmed hadron of the $b \to c$-hadron cascade.

Another criterion to estimate the algorithmic performance of the secondary vertex finders is the fraction of charged particles arising from the decays of $b$- or $c$-hadrons and reconstructed as tracks in the Inner Detector that are correctly associated to a displaced vertex (*efficiency*) and –at the same time– which fraction of the tracks assigned by the secondary vertex finders to displaced vertices really stem from real $b$- or $c$-hadron decays (*purity*).

The average charged particle multiplicities at the secondary and tertiary decay vertices as obtained from the Monte Carlo generator PYTHIA was analyzed, where a minimum transverse momentum of 500 MeV was required for the charged particles. The decay products of a strong or electromagnetic $b$-hadron decay (e.g. pions from $B^{**}$ decays) are not counted as coming from the secondary vertex, because of the lifetime of these states the decay particles emerge from the primary vertex. The decay products of a strongly or electromagnetically decaying $c$-hadron (e.g. $D^*$ or $D^{**}$) are considered as stemming from the $b$-hadron vertex, while only the decay products of weakly decaying $c$-hadrons are considered as originating from the $c$-hadron vertex.

The average charged particle multiplicity of the inclusive $b/c$-hadron vertex is $\sim 4.7$, as shown in Figure 8, distributed in $\sim 2.3$ charged particles stemming from the $b$-hadron vertex and $\sim 2.4$ charged particles originating from the $c$-hadron vertex. In the same figure the decay multiplicity at generator level (left) is compared with the number of tracks coming from the $b$- and $c$-hadron vertices which have been reconstructed as tracks in the Inner Detector (right). Around 80% of their charged decay products are correctly reconstructed as tracks and pass the standard track quality selection cuts. In order to strongly reduce the contamination of reconstructed secondary vertices by other charged particles originating from the primary interaction point, tracks originating from the $b$- or $c$-hadron vertices, but not distinguishable from primary tracks, are also suppressed to some extent by the track selection cuts. This and other selection criteria further reduce the $b$- and $c$-hadron decay products reconstruction efficiency, resulting in a compromise between the highest possible *efficiency* and a reasonable *purity*. In Table 8 the track association efficiencies and purities of the displaced tracks stemming from the $b$- or $c$- hadrons for the *BTagVrtSec* and *JetFitter* algorithms are shown, the efficiencies being normalized to the number of tracks from $b/c$-hadron decays which are reconstructed by the tracking detector. The fit of an inclusive $b/c$-hadron decay vertex, as performed by *BTagVrtSec*, allows to obtain a very high purity, but at the cost of starting to loose some tracks when the distance between the $b$- and $c$-hadron vertices starts to be relevant. *JetFitter* is able to recover a good part of this inefficiency, thanks to the ability of reconstructing more complex decay chain topologies, at the cost of a slightly lower purity.

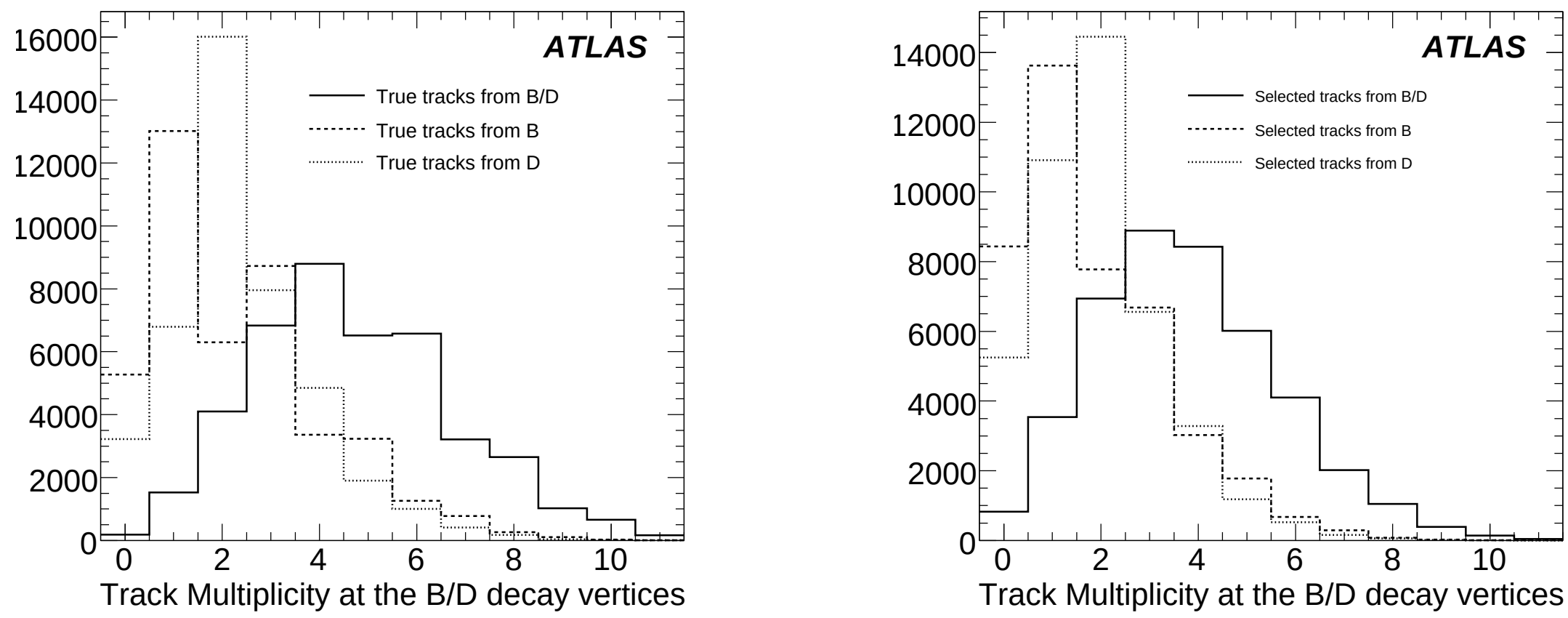

Figure 8: Average charged particle decay multiplicity of the $b$- and/or $c$-hadrons at generator level (left), compared with the number of charged particles coming from the $b$- and/or $c$-hadron vertices reconstructed as tracks in the Inner Detector and passing some standard quality criteria (right). The WH($H \to b\bar{b}$) sample has been used.

Table 8: Efficiency (fraction of reconstructed tracks from the decays of $b$- or $c$-hadrons that are associated to the secondary vertices) and purity (fraction of tracks fitted to the secondary vertices that stem from real $b$- or $c$-hadron decays) for the two secondary vertex finders, separately for different topologies in the case of *JetFitter*. The WH($H \to b\bar{b}$) sample has been used.

| Algorithm | Topology | Track efficiency | Track purity |
|---|---|---|---|
| BTagVrtSec | 1 inclusive $B/D$ vertex | 69 % | 92 % |
| JetFitter | 1 vertex | 74 % | 91 % |
| | 1 vertex + 1 track | 80 % | 85 % |
| | 2 vertices | 85 % | 89 % |

Several topological and kinematical variables related to the reconstructed secondary vertices are used by the $b$-tagging algorithms, as described in sections 4.1 and 4.2. The invariant mass of charged particles associated to the vertex and the fraction of charged energy at the vertex relative to that of the jet, are used by both algorithms. The distributions of these variables are shown in Figure 9 for different regions of jet transverse momenta and pseudorapidities for the *BTagVertSec* algorithm. The flight distance significance, defined as the distance between the primary and secondary vertices divided by its error, is also shown there. The observed dependence of the secondary vertex reconstruction performance on the jet kinematics will directly impact the performance of the *b*-tagging algorithms. This will be discussed in Section 7.

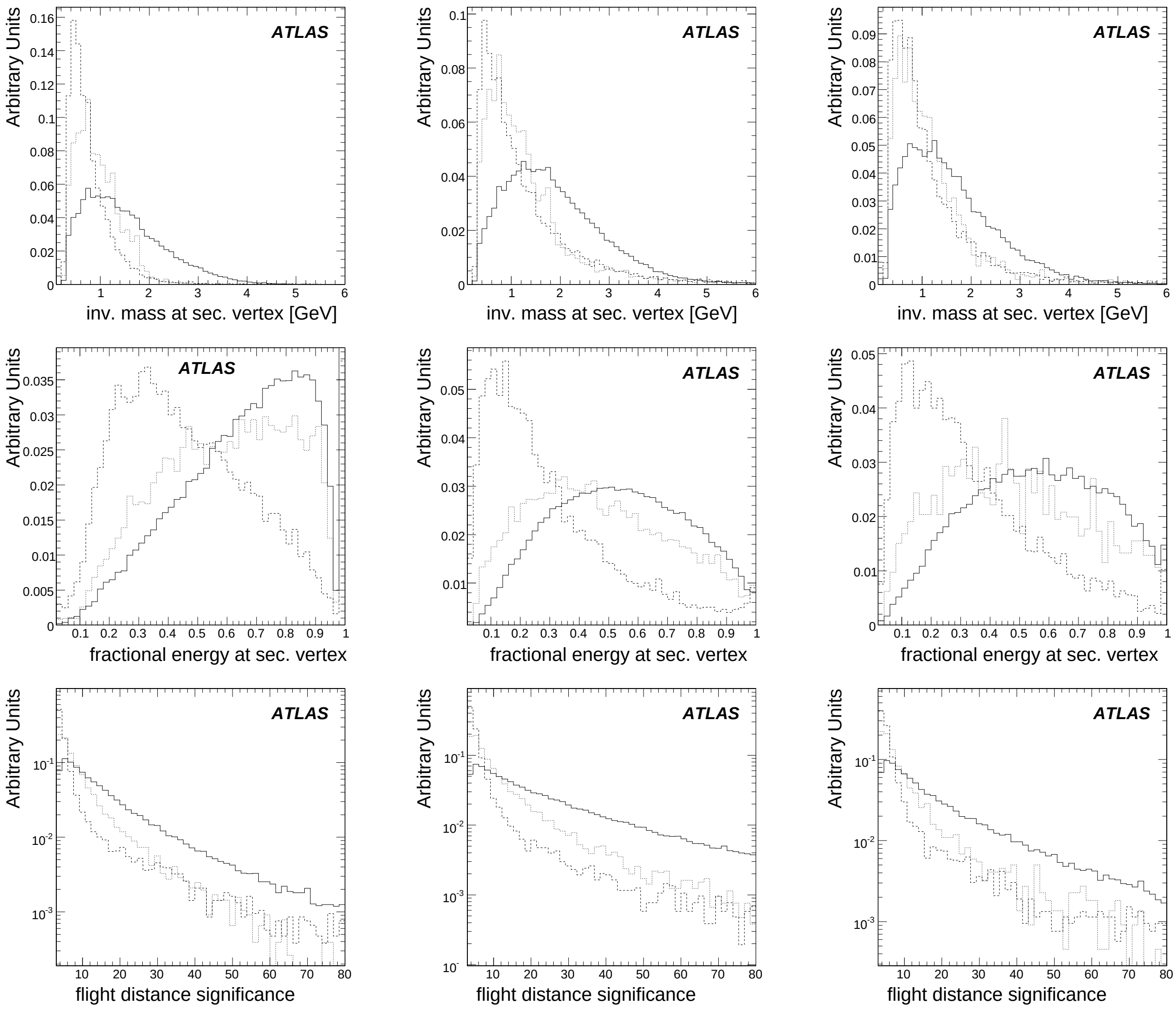

Figure 9: Variables related to the properties of reconstructed secondary vertices as computed by the *BTagVrtSec* algorithm for different jet transverse momenta and pseudorapidities for *b*-quark jets (solid), *c*-quark jets (dotted) and light quark jets (dashed). Left: $0 < |\eta| < 0.5$, $15 < p_T < 30$ GeV; middle: $0 < |\eta| < 0.5$, $80 < p_T < 120$ GeV; right: $2 < |\eta| < 2.5$, $80 < p_T < 120$ GeV. The top row shows the invariant mass of charged particle tracks associated to the reconstructed secondary vertices, the middle row the energy of charged particle tracks associated to the reconstructed secondary vertices divided by the energy of all charged particles in the jet, and the bottom row the flight distance significance as defined in the text.

Essentially all *b*-tagging algorithms use the direction of the jet, as reconstructed from calorimeter towers, to assign e.g. the lifetime sign to the impact parameters of charged particle tracks. The jet direction is a reasonably good approximation of the *b*-quark direction, however, for *b*-tagging it is mainly the direction of the $b$-hadron that matters. The direction of the $b$-hadron flight path can be estimated using information from reconstructed primary and secondary vertices, e.g. for *BTagVrtSec* the line joining the primary and secondary vertices can be used as the $b$-hadron direction if a secondary vertex is present in the jet. The more sophisticated *JetFitter* algorithm delivers by construction an estimate for the $b$-hadron flight direction using information from both the calorimeters and vertices of charged particle tracks. Figure 10 shows a comparison of the resolutions achieved for the azimuth angle $\phi$ for the different approaches. Using an improved $b$-hadron direction may decrease the contribution of charged particles from the decays of $b$-hadrons to the negative tail of the impact parameter distribution. This contribution mainly comes from a wrong assignment of the lifetime sign if the $b$-hadron direction has not been reconstructed sufficiently precisely. This will facilitate the calibration and improve the $b$-tagging performance.

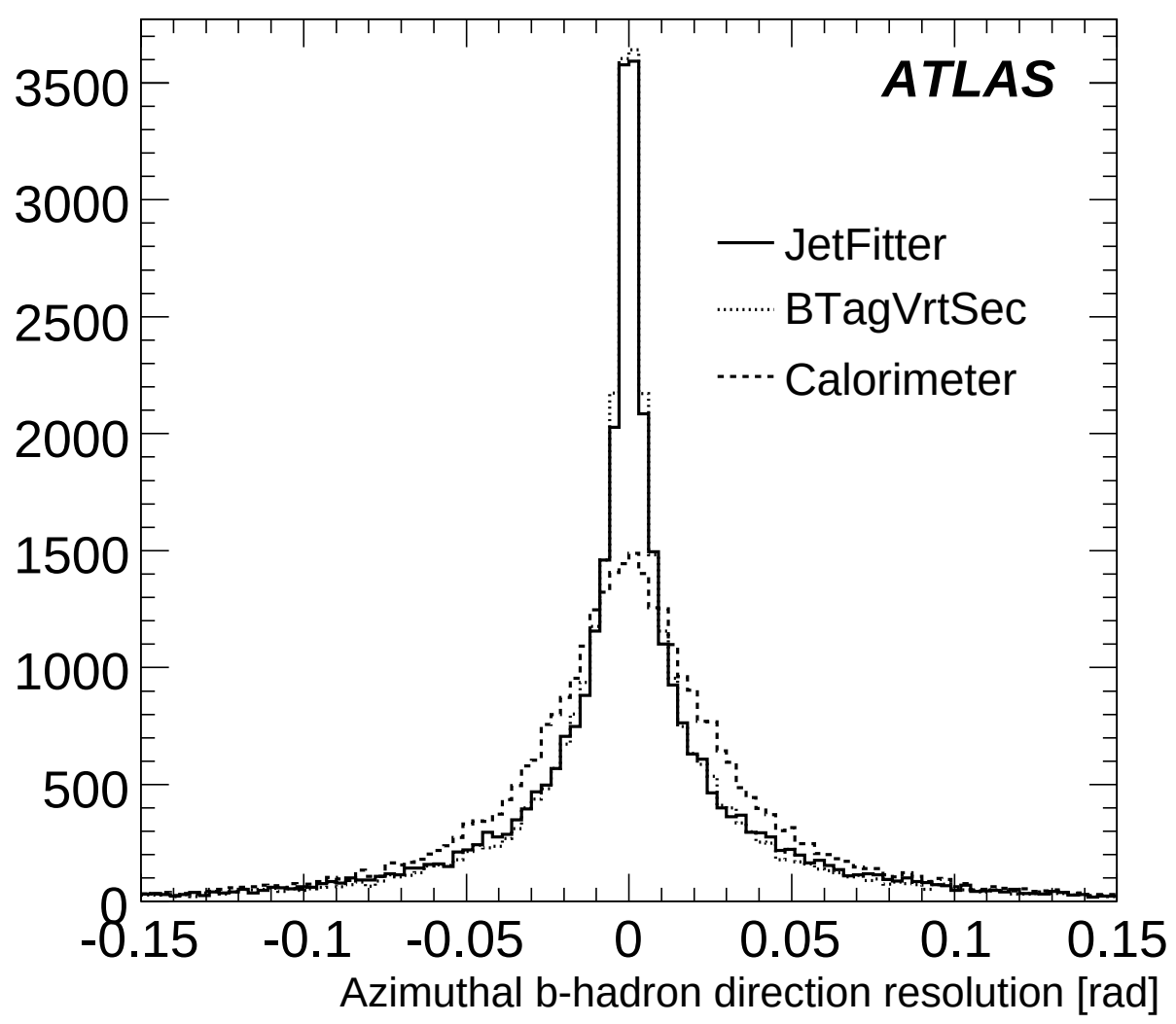

Figure 10: Angular resolution of the $b$-hadron flight direction in the azimuth angle $\phi$ as reconstructed with the *JetFitter* and *BTagVrtSec* algorithms (for the latter, the line joining the primary and secondary vertices has been used in the case where a secondary vertex is present), compared with the corresponding resolution as obtained from calorimeter jets. The $WH$ ($H \to b\bar{b}$; $m_H$=120 GeV) sample was used for this study.

## 6 Combination with Impact Parameter based $b$-tagging Algorithms

The $b$-tagging algorithms presented in the previous sections provide all information needed to be used as stand-alone algorithms. Secondary vertex based $b$-tagging algorithms are, however, limited by the efficiency to reconstruct a secondary vertex inside a jet. To obtain maximum performance, the secondary vertex based algorithms can be combined with other algorithms. $b$-tagging algorithms purely based on the impact parameter significances of charged particle tracks do not have this limitation. They do not offer, however, the amount of topological and kinematical information as the secondary vertex based algorithms.

In ATLAS, the impact parameter information can be used either in two ($r\phi$ plane only) or three dimensions. The corresponding algorithms and their performance are described in detail in [4]. They calculate a likelihood ratio of the probability density functions for observed track impact parameter significances $S$ for tracks coming from $b$-quark jets and light quark jets: $w_{track} = b(S)/u(S)$. The jet likelihood ratio (called *jet weight*) is the sum of track ratios: $w_{jet}^{IP} = \sum_{i=1}^{N_{track}} \ln w_{track,i}$. Assuming that the tracks are independent, the jet likelihood ratio is an optimal variable for making a decision about the jet origin.

Secondary vertex algorithms were designed in a way facilitating combination with track only based algorithms. They provide likelihood ratios of various parameters for $b$-jets and light jets, which may be simply summed with track likelihood ratios. To preserve an optimal performance, secondary vertex variables have been chosen to be maximally independent on track impact parameters (i.e. the distance between primary and secondary vertex is strongly correlated with track impact parameters and then should be used with care). The combined *jet weight* is then computed as sum of the likelihood ratios of the different algorithms: $W_{jet}^{combined} = W_{jet}^{SV} + W_{jet}^{IP}$.

The likelihood ratio approach used in ATLAS makes $b$-tagging an easily scalable procedure. Any new algorithm or new variable can be added in the same way to already existing tagging information. Choice of independent variables (or at least weakly correlated ones) guarantees an optimality of combined procedure.

# 7 $b$-Tagging Performance

In this section, the performance of the algorithms described in this note is discussed. The data samples described in Section 2 have been used for these studies. Jets had to be within the acceptance of the tracking detectors ($|\eta| < 2.5$) and their reconstructed transverse momenta had to exceed 15 GeV: $p_T^{jet} >$ 15 GeV. The performance is shown in terms of $b$-tagging efficiency, $\varepsilon_b$, and light quark rejection, $r_{u,c}$ as defined in Section 2. A certain working point is chosen by placing a cut on the weight as computed by the algorithm.

The performance was studied both using only the information as delivered by the secondary vertex based algorithms (denoted *BTagVrtSec* and *JetFitter* as in the previous sections) and after combination with a tagging algorithm based on a combination of transverse and longitudinal impact parameter significances as described in Section 6. The $b$-tagging algorithm purely based on the information from impact parameter significances of charged particle tracks showing the best performance is called *IP3D* and is used in the following studies. Details about this algorithm can be found in [4]. The results from the combined algorithms are denoted *BTagVrtSec+IP3D* and *JetFitter+IP3D*, respectively. Tables 9 and 10 show the rejection rates against light quark and charm quark jets for different tagging algorithms and data samples.

Table 9: The rejection against light quark jets for fixed $b$-tagging efficiencies of 50% and 60% for different data samples without and with applying the purification procedure, denoted as *raw* and *purified*, respectively.

| Sample | $\varepsilon_b$ | BTagVrtSec | JetFitter | BTagVrtSec+IP3D | JetFitter+IP3D |
|---|---|---|---|---|---|
| WH(120) | 50% raw | $97 \pm 1$ | $156 \pm 3$ | $454 \pm 13$ | $545 \pm 17$ |
| | 60% raw | $37 \pm 0$ | $52 \pm 0$ | $116 \pm 2$ | $133 \pm 2$ |
| WH(120) | 50% purified | $98 \pm 1$ | $156 \pm 3$ | $462 \pm 13$ | $554 \pm 17$ |
| | 60% purified | $38 \pm 0$ | $52 \pm 1$ | $118 \pm 2$ | $134 \pm 2$ |
| WH(400) | 50% raw | $55 \pm 1$ | $101 \pm 1$ | $285 \pm 6$ | $379 \pm 10$ |
| | 60% raw | $25 \pm 0$ | $45 \pm 0$ | $93 \pm 1$ | $121 \pm 2$ |
| WH(400) | 50% purified | $55 \pm 1$ | $101 \pm 1$ | $296 \pm 7$ | $390 \pm 11$ |
| | 60% purified | $25 \pm 0$ | $46 \pm 0$ | $95 \pm 1$ | $123 \pm 2$ |
| $t\bar{t}$ +$t\bar{t}jj$ | 50% raw | $110 \pm 1$ | $176 \pm 1$ | $456 \pm 4$ | $633 \pm 7$ |
| | 60% raw | $48 \pm 0$ | $66 \pm 0$ | $154 \pm 1$ | $190 \pm 1$ |
| $t\bar{t}$ +$t\bar{t}jj$ | 50% purified | $130 \pm 1$ | $189 \pm 1$ | $791 \pm 11$ | $924 \pm 14$ |
| | 60% purified | $53 \pm 0$ | $68 \pm 0$ | $206 \pm 1$ | $226 \pm 2$ |

Table 10: The rejection against charm quark jets for fixed $b$-tagging efficiencies of 50% and 60% for different data samples. The purification procedure has not been applied here.

| Sample | $\varepsilon_b$ | BTagVrtSec | JetFitter | BTagVrtSec+IP3D | JetFitter+IP3D |
|---|---|---|---|---|---|
| WH(400) | 50% raw | 8.1 ± 0.1 | 9.8 ± 0.2 | 12.4 ± 0.2 | 12.6 ± 0.2 |
| | 60% raw | 4.8 ± 0.0 | 6.1 ± 0.1 | 6.8 ± 0.1 | 7.3 ± 0.1 |
| $t\bar{t}$ +$t\bar{t}jj$ | 50% raw | 9.9 ± 0.0 | 10.3 ± 0.0 | 12.4 ± 0.1 | 12.3 ± 0.1 |
| | 60% raw | 5.9 ± 0.0 | 6.2 ± 0.0 | 7.4 ± 0.0 | 7.4 ± 0.0 |

Figure 11 shows the rejection of light quark (u,d,s) and gluon jets, $r_u$, versus the $b$-tagging efficiency $\varepsilon_b$ for the pure secondary vertex based algorithms and a comparison with the most performant tagging algorithm purely based on three dimensional impact parameter information, *IP3D*. It can be seen that the maximum achievable $b$-tagging efficiency of the vertex based algorithms is limited by the efficiency to reconstruct a displaced vertex or $b$-hadron decay topology. It is higher in the case of *JetFitter* which imposes only very soft requirements on the topology. The performance of the impact parameter based tagging algorithm *IP3D* is better over almost the full range of $b$-tagging efficiencies. There is also a significant difference between *BTagVrtSec* and *JetFitter*, the latter one showing better performance. It has to be noted, however, that *JetFitter* explicitely uses lifetime information (the flight distance significance), which is not used by *BTagVrtSec* to decorrelate the algorithm better from *IP3D*. The right plot of Figure 11 exhibits some steep structures at $b$-tagging efficiencies of about 50% and 70%. These can be related to sudden drops in the rejection rate of the *IP3D* algorithm for these $b$-tagging efficiencies, as can be seen in the left part of Figure 11. This behaviour can be explained mainly by the presence of jets with only a single track associated to them and the finite binning of the reference distributions for the probability density functions of the impact parameter significances as used by the *IP3D* $b$-tagging algorithm (see [4] for details).

Figure 12 shows the rejection against charm quark jets. The rejection against charm quark jets is significantly lower compared to light quark jets because of the real lifetime of charm hadrons and thus similar topology. Here, the vertex based algorithms show a performance that is much closer to the impact parameter based algorithm.

Figures 13 and 14 show the rejections against light quark jets and charm quark jets for the secondary vertex based tagging algorithms after combination with the *IP3D* algorithm, again compared with *IP3D*. This combination results in a significant increase in rejection power, especially for light quark jets.

The dependence of the performance on the jet kinematics is of particular importance. Figures 15 and 16 show the rejection against light quark jets for a fixed $b$-tagging efficiency of 50% versus the jet transverse momentum and jet pseudo rapidity, respectively. Table 11 shows the rejection against light quark jets for a fixed $b$-tagging efficiency of 50% for several regions of jet transverse momenta and pseudorapidities.

The reasons for the observed behaviour are manifold. For larger pseudorapidities, the particles pass through more material and thus the detector resolution is worse. Another reason for degradation of the longitudinal impact parameter at high pseudorapidity is the dramatic increase of the extrapolation distance from b-layer hit to the vertex. The degradation for high jet transverse momenta is caused by an increased fragmentation multiplicity, resulting in an increased combinatorics in the secondary vertex finding stage when trying to find the tracks stemming from the $b$-hadron decay. Furthermore, the pattern recognition in the track reconstruction becomes more difficult in the very dense environment of jets with very large transverse momenta. The steep fall for low jet transverse momenta is mainly due to the strongly enhanced multiple scattering of low momentum charged particle tracks, leading to significantly degraded impact parameter resolutions. The observed dependence of the $b$-tagging performance on the jet kinematics can be related to the peformance of the secondary vertex reconstruction as discussed in Section 5. It can be seen there, that the secondary vertex reconstruction efficiency in $b$-quark jets drops in the same kinematical regions as the resulting $b$-tagging performance, with an increased rate of (fake) vertices in light quark jets.

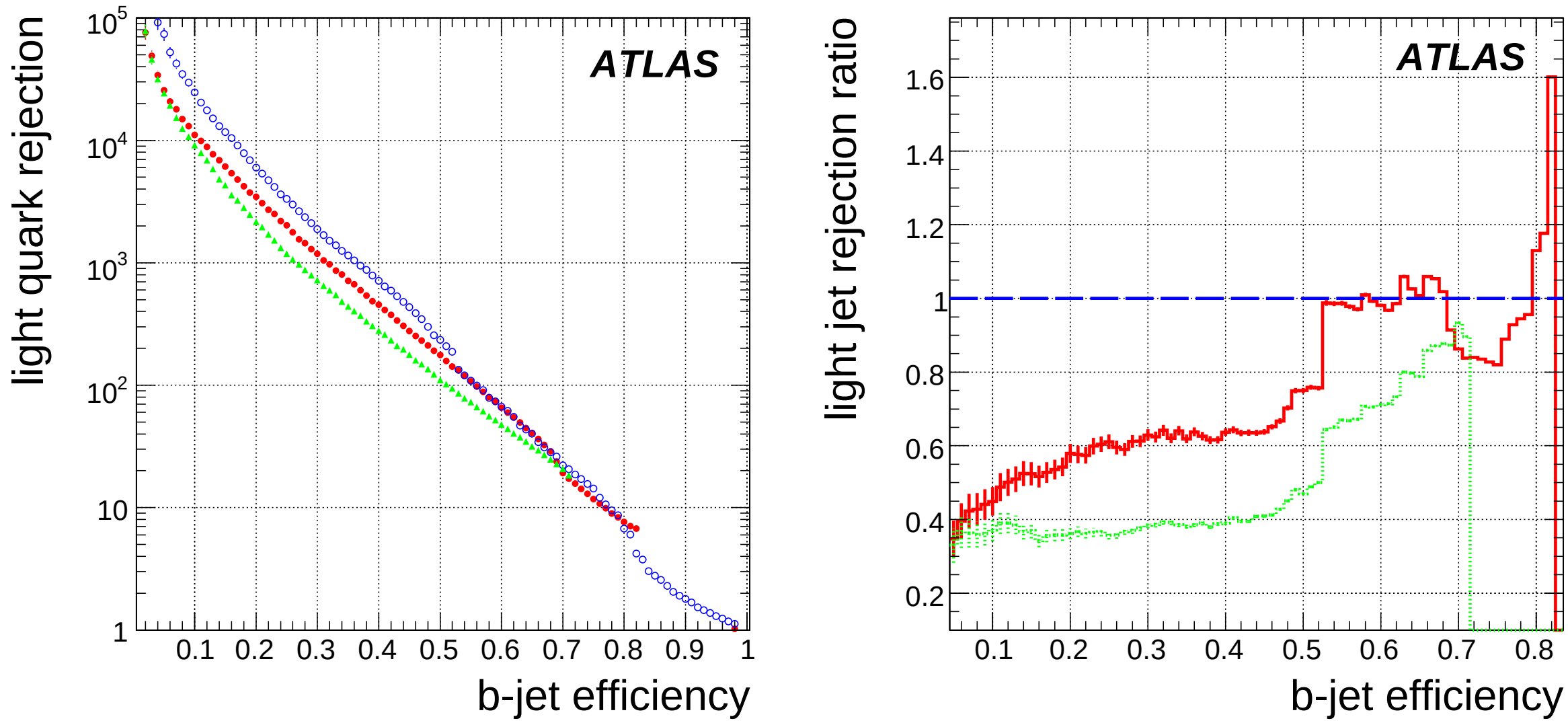

Figure 11: Left: Light jet rejection versus $b$-tagging efficiency for *BTagVrtSec* (triangles, green) and *JetFitter* (full circles, red). The pure impact parameter based algorithm *IP3D* is also shown for comparison (open circles, blue); right: The ratio with respect to *IP3D* for *BTagVrtSec* (dashed line, green) and *JetFitter* (full line, red). These results have been obtained on the $t\bar{t}$ and $t\bar{t}jj$ samples. No purification of light quark jets (see Section 2) has been applied.

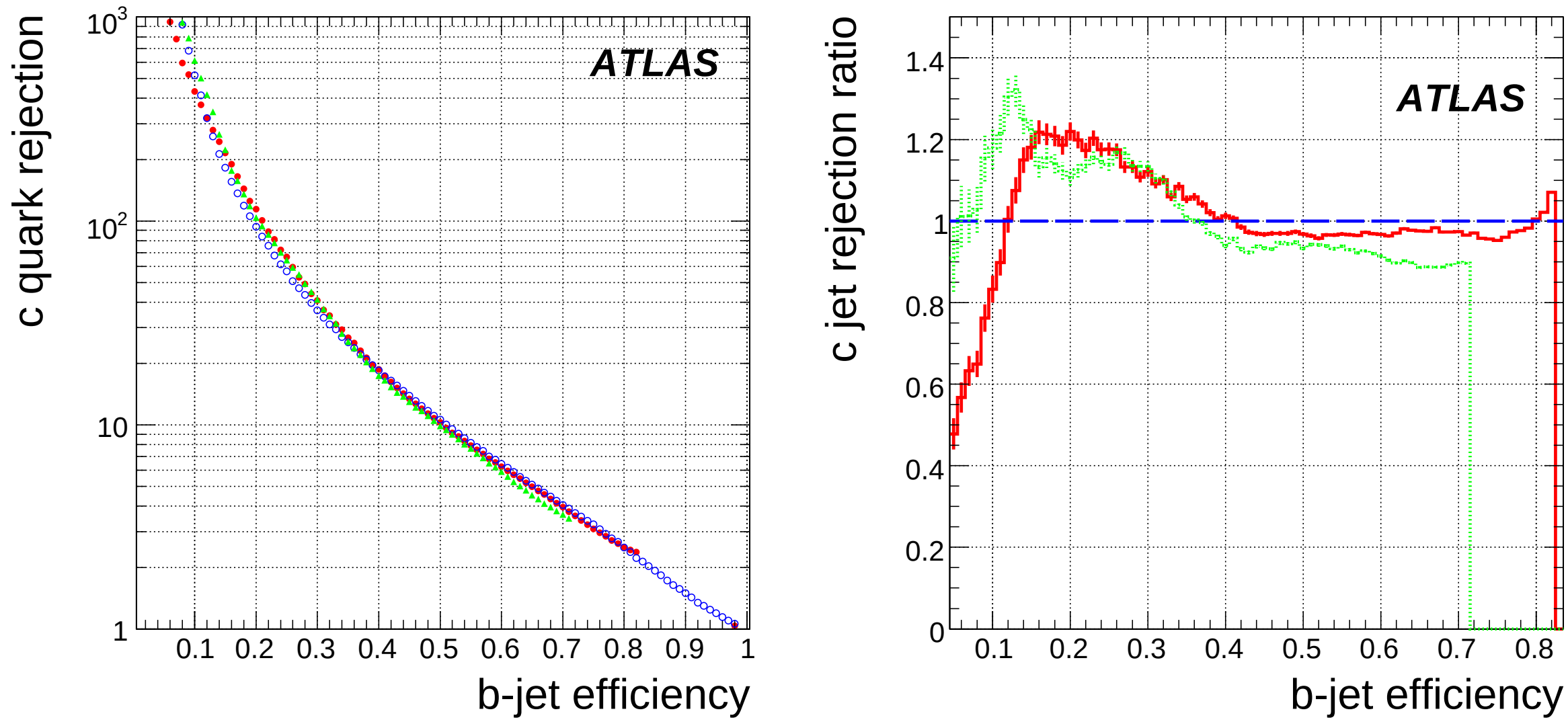

Figure 12: Left: Charm jet rejection versus $b$-tagging efficiency for *BTagVrtSec* (triangles, green) and *JetFitter* (full circles, red). The pure impact parameter based algorithm *IP3D* is also shown for comparison (open circles, blue); right: The ratio with respect to *IP3D* for *BTagVrtSec* (dashed line, green) and *JetFitter* (full line, red). These results have been obtained on the $t\bar{t}$ and $t\bar{t}jj$ samples.

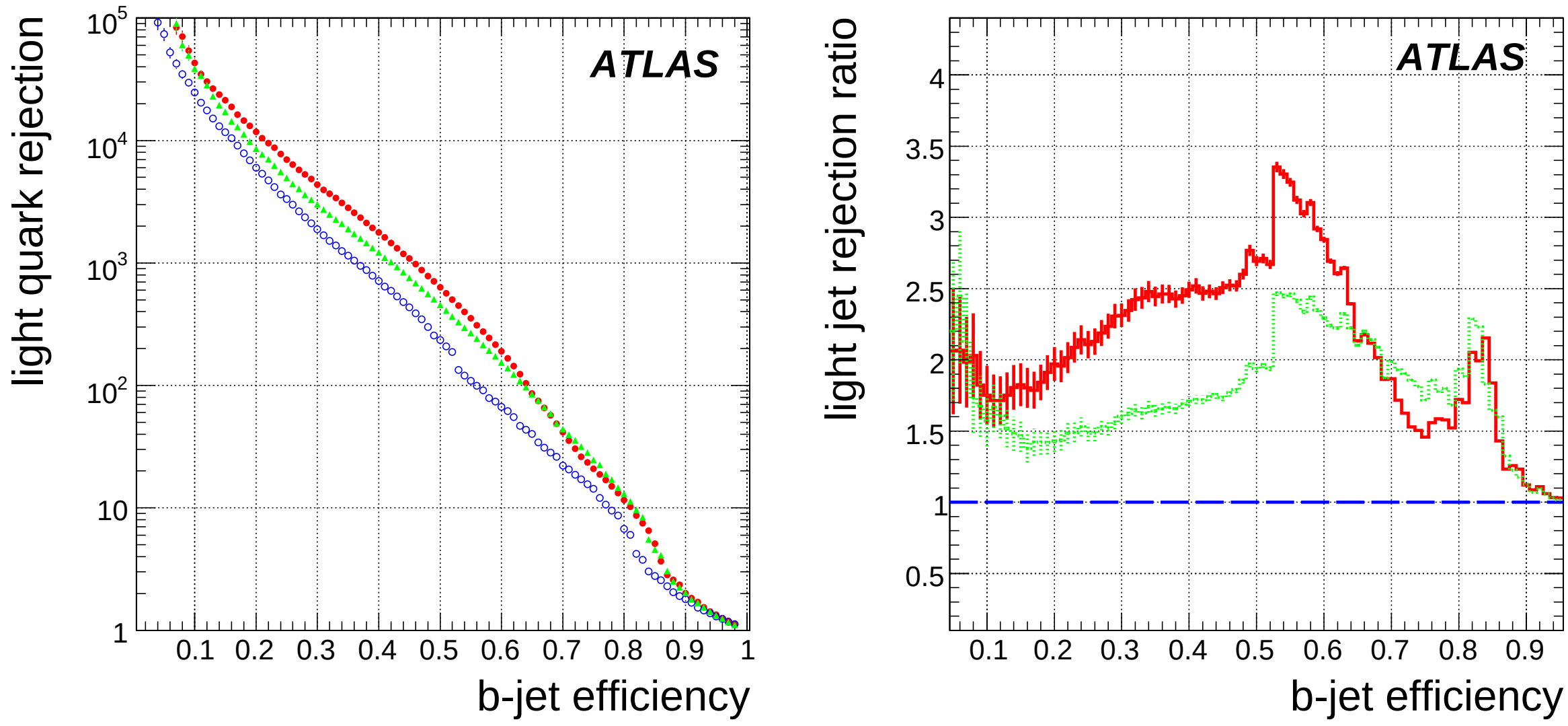

Figure 13: Left: Light jet rejection versus $b$-tagging efficiency for *BTagVrtSec* (triangles, green) and *JetFitter* (full circles, red) after combination with *IP3D*. The pure impact parameter based algorithm *IP3D* is also shown for comparison (open circles, blue); right: The ratio with respect to *IP3D* for *BTagVrtSec* (dashed line, green) and *JetFitter* (full line, red) combined. These results have been obtained on the $t\bar{t}$ and $t\bar{t}jj$ samples. No purification has been applied.

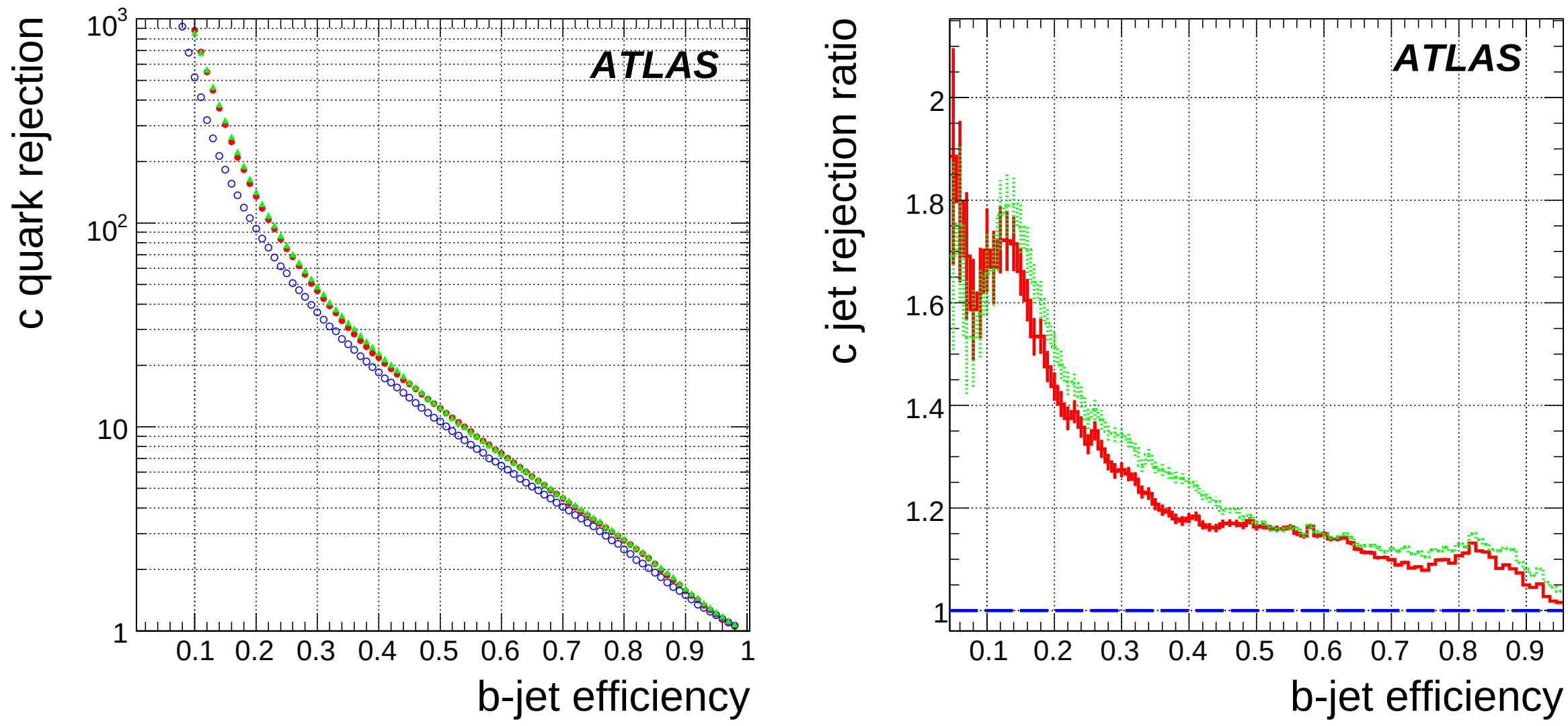

Figure 14: Left: Charm jet rejection versus $b$-tagging efficiency for *BTagVrtSec* (triangles, green) and *JetFitter* (full circles, red) after combination with *IP3D*. The pure impact parameter based algorithm *IP3D* is also shown for comparison (open circles, blue); right: The ratio with respect to *IP3D* for *BTagVrtSec* (dashed line, green) and *JetFitter* (full line, red) combined. These results have been obtained on the $t\bar{t}$ and $t\bar{t}jj$ samples.

Table 11: The rejection against light quark jets (with purification) for a fixed $b$-tagging efficiency of 50% for the $t\bar{t}$ and $t\bar{t}jj$ samples in regions of jet transverse momenta and pseudorapidities for the secondary vertex based $b$-tagging algorithms (top: *BTagVrtSec*; bottom: *JetFitter* ) after combination with the pure impact parameter based tagging algorithm, *IP3D*.

| | $0<\|\eta\|<0.5$ | $0.5<\|\eta\|<1.0$ | $1.0<\|\eta\|<1.5$ | $1.5<\|\eta\|<2.0$ | $2.0<\|\eta\|<2.5$ |
|---|---|---|---|---|---|
| 15 GeV$< p_T <$30GeV | $177 \pm 4$ | $180 \pm 4$ | $158 \pm 4$ | $74 \pm 1$ | $28 \pm 0$ |
| | $200 \pm 5$ | $183 \pm 4$ | $149 \pm 3$ | $74 \pm 1$ | $25 \pm 0$ |
| 30 GeV$< p_T <$50GeV | $1170 \pm 79$ | $1140 \pm 79$ | $782 \pm 48$ | $375 \pm 18$ | $92 \pm 2$ |
| | $1269 \pm 90$ | $1286 \pm 93$ | $957 \pm 65$ | $409 \pm 21$ | $111 \pm 3$ |
| 50 GeV$< p_T <$80GeV | $1534 \pm 132$ | $2613 \pm 306$ | $1380 \pm 127$ | $536 \pm 36$ | $149 \pm 6$ |
| | $2354 \pm 251$ | $2415 \pm 272$ | $1678 \pm 170$ | $677 \pm 51$ | $195 \pm 9$ |
| 80 GeV$< p_T <$120GeV | $1698 \pm 203$ | $2050 \pm 281$ | $1293 \pm 152$ | $559 \pm 50$ | $143 \pm 8$ |
| | $2286 \pm 317$ | $3293 \pm 573$ | $1311 \pm 156$ | $715 \pm 73$ | $196 \pm 12$ |
| 120GeV$< p_T <$200GeV | $1016 \pm 128$ | $1116 \pm 152$ | $736 \pm 86$ | $235 \pm 17$ | $63 \pm 3$ |
| | $1231 \pm 171$ | $1370 \pm 206$ | $1194 \pm 178$ | $299 \pm 25$ | $67 \pm 3$ |

Two of the most discriminating variables, for both algorithms, are the invariant mass of charged particle tracks associated to the secondary vertex and the fraction of the charged energy at the secondary vertex divided by the total charged energy in the jet. Figure 9 in Section 5 shows these variables for $b$-quark, $c$-quark and light quark jets for various jet transverse momenta and pseusorapidities. The loss of discrimination power in the regions where the degradation of the $b$-tagging performance is observed, is clearly visible. Apart from the kinematic dependence, the $b$-tagging performance also depends critically on the contamination of the light quark jets by displaced tracks stemming from nearby $b$- or $c$-quark jets, as can already be concluded comparing the $b$-tagging performance before and after application of the purification procedure. It is however worth looking at this dependence in more detail, analyzing the $b$-tagging performance as a function of the distance $\Delta R$ of the light quark jets to the nearest $b$- or $c$-quark or $\tau$ lepton.

Table 12 shows, both for the pure secondary vertex based algorithms and after combination with the *IP3D* algorithm, that the more the light quark jets are contaminated by $b$- or $c$-hadron decay products (smaller values of $\Delta R$), the stronger the degradation of the $b$-tagging performance is. At very small angles of $\Delta R$ the light quark-jets can be barely distinguished from the nearby heavy flavour jet. A small kinematical bias (slightly different $p_T$ and $\eta$ distribution of the jets for different $\Delta R$ intervals) should also be taken into account when interpreting these results.

There is a noticeable difference in the behaviour of *JetFitter* compared to *BTagVrtSec*, the second being less robust against building up vertices which catch up contributions from nearby heavy flavour jets. As stated in Section 4.2.2, *JetFitter* uses the jet direction as reconstructed by the calorimeter as a constraint in the fit of the hypothetical $b$-hadron flight axis, thus being more efficient in discarding tracks coming from nearby jets.

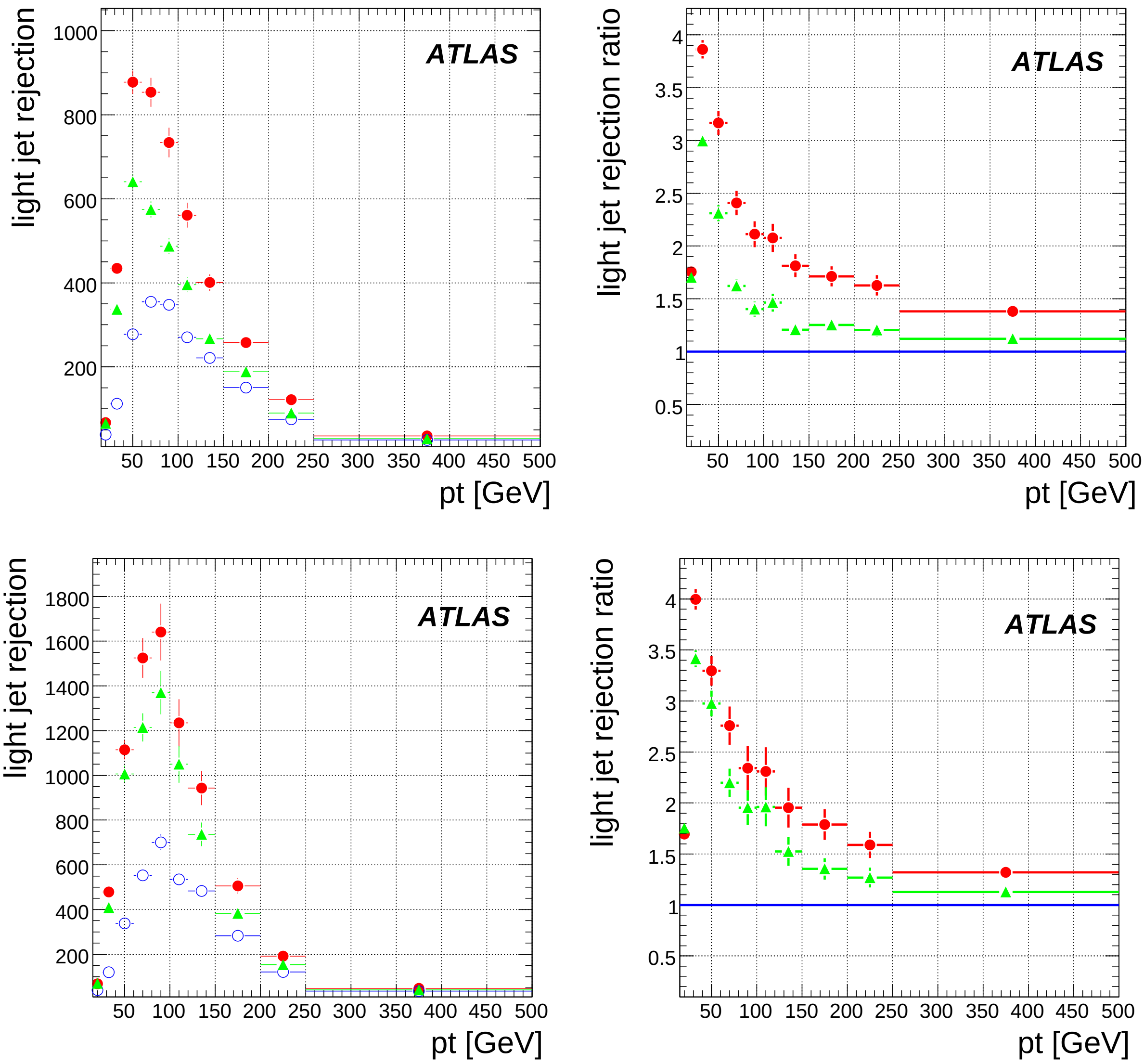

Figure 15: Left: Light jet rejection for a fixed $b$-tagging efficiency of 50% versus the jet transverse momentum for *BTagVrtSec* (triangles, green) and *JetFitter* (full circles, red) after combination with *IP3D*. The pure impact parameter based algorithm *IP3D* is also shown for comparison (open circles, blue); right: The ratio with respect to *IP3D* for *BTagVrtSec* (triangles, green) and *JetFitter* (full circles, red) combined. The plots in the top (bottom) row show the performance without (with) applying the purification procedure. These results have been obtained on the $t\bar{t}$ and $t\bar{t}jj$ sample.

## 8 Summary and Outlook

In this note, an overview of vertex reconstruction algorithms used in ATLAS, both for the reconstruction of the primary event vertex and the inclusive reconstruction of secondary decay vertices inside jets, has been given. The focus has been put on applications to the tagging of $b$-quark jets.

Several primary vertex reconstruction algorithms are available in ATLAS. The performance of these algorithms has been studied both for the case where only the hard signal interaction is present and for a luminosity of $2 \times 10^{33}\mathrm{cm}^{-2}\mathrm{s}^{-1}$, when 4.6 additional pile-up vertices are present on average. It was demonstrated that the presence of additional vertices does not degrade the primary vertex reconstruction precision significantly. A more severe problem is the misidentification of a pile-up vertex as the signal vertex. For the luminosity of $2 \times 10^{33}\text{cm}^{-2}\text{s}^{-1}$ and the event topologies studied in this note, this misidentification rate can be as high as 5%. This misidentification leads to a significant degradation of the performance of the $b$-tagging algorithms that use information of charged particle tracks in the longitudinal direction. Possible procedures to recover a large part of this performance loss have been discussed and will be studied an implemented in the future.

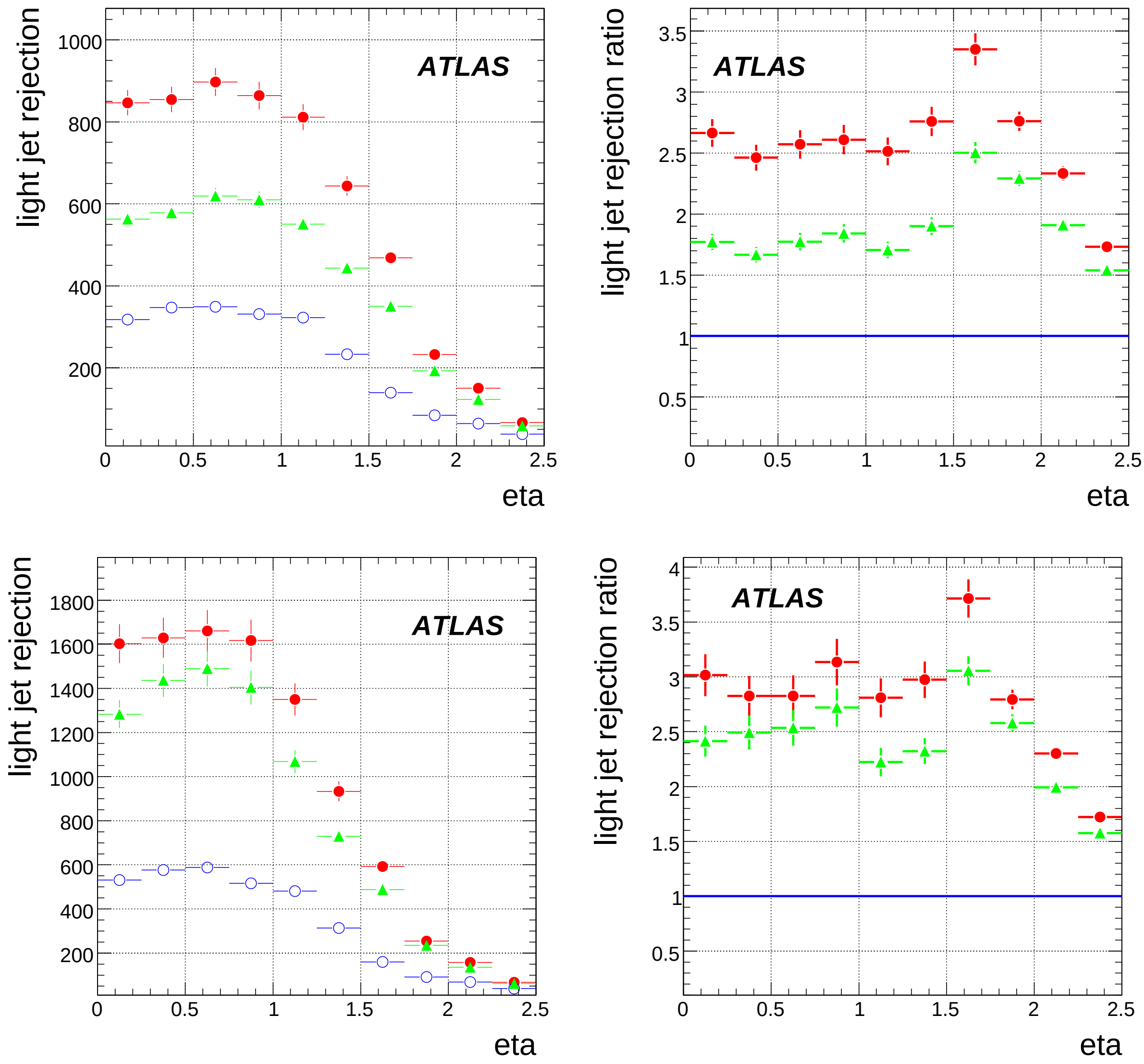

Figure 16: Left: Light jet rejection for a fixed $b$-tagging efficiency of 50% versus the jet pseudorapidity for *BTagVrtSec* (triangles, green) and *JetFitter* (full circles, red) after combination with *IP3D*. The pure impact parameter based algorithm *IP3D* is also shown for comparison (open circles, blue); right: The ratio with respect to *IP3D* for *BTagVrtSec* (triangles, green) and *JetFitter* (full circles, red) combined. The plots in the top (bottom) row show the performance without (with) applying the purification procedure. These results have been obtained on the $t\bar{t}$ and $t\bar{t}jj$ sample.

Table 12: The rejection against light quark jets for different $b$-tagging efficiencies ($\varepsilon_b$) as function of the difference in $\Delta R$ of the light quark jets to the nearest $b$-, $c$- or $\tau$-jet (top row: *BTagVrtSec*; bottom row: *JetFitter*), for the secondary vertex based $b$-tagging algorithms used stand-alone (upper table) and after combination with the impact parameter based tagging algorithm *IP3D* (lower table).

| $\varepsilon_b$ | $0.3 < \Delta R < 0.35$ | $0.35 < \Delta R < 0.4$ | $0.4 < \Delta R < 0.45$ | $0.45 < \Delta R < 0.5$ | $0.5 < \Delta R < 0.6$ | $0.6 < \Delta R < 0.7$ |
|---|---|---|---|---|---|---|
| 50% | $12.4 \pm 0.3$ | $28 \pm 1$ | $80 \pm 5$ | $177 \pm 14$ | $231 \pm 11$ | $244 \pm 10$ |
| | $29 \pm 1$ | $89 \pm 6$ | $244 \pm 27$ | $363 \pm 43$ | $345 \pm 20$ | $327 \pm 16$ |
| 60% | $7.3 \pm 0.1$ | $14 \pm 0.3$ | $8.7 \pm 1$ | $68 \pm 3$ | $91 \pm 3$ | $91 \pm 3$ |
| | $17.3 \pm 0.4$ | $43 \pm 2$ | $30 \pm 6$ | $113 \pm 7$ | $128 \pm 5$ | $119 \pm 4$ |

| $\varepsilon_b$ | $0.3 < \Delta R < 0.35$ | $0.35 < \Delta R < 0.4$ | $0.4 < \Delta R < 0.45$ | $0.45 < \Delta R < 0.5$ | $0.5 < \Delta R < 0.6$ | $0.6 < \Delta R < 0.7$ |
|---|---|---|---|---|---|---|
| 50% | $18.4 \pm 0.5$ | $51 \pm 2$ | $179 \pm 17$ | $467 \pm 62$ | $939 \pm 92$ | $951 \pm 79$ |
| | $31 \pm 1$ | $141 \pm 12$ | $501 \pm 80$ | $793 \pm 138$ | $1133 \pm 121$ | $1077 \pm 95$ |
| 60% | $10.1 \pm 0.2$ | $23 \pm 1$ | $66 \pm 4$ | $143 \pm 11$ | $265 \pm 14$ | $266 \pm 12$ |
| | $17.0 \pm 0.4$ | $50 \pm 2$ | $158 \pm 14$ | $222 \pm 20$ | $309 \pm 17$ | $302 \pm 14$ |

Many interesting physics events at the LHC have a lepton with large transverse momentum in the final state, so one may try to decrease the misidentification rate by selecting the reconstructed interaction vertex closest to this lepton. This strategy, however, does not provide a significant improvement in comparison with the standard primary vertex selection algorithm since the presence of a well reconstructed high $p_T$ lepton increases the weight of the corresponding vertex and then is taken into account automatically.

Two algorithms are available in ATLAS for the inclusive reconstruction of secondary decay vertices in jets, following different approaches. The first algorithm, *BTagVrtSec*, reconstructs explicitly geometrical secondary vertices inside the jet. The second algorithm, *JetFitter*, is based on the specific kinematics of the $b$- and $c$-hadrons decay chain. The performance of the algorithms has been investigated thoroughly. The $b$-tagging performance of both algorithms has been studied using only the information from the secondary vertex algorithms as well as after combination with $b$-tagging algorithms based on charged track impact parameters. A strong dependence of the vertex reconstruction and thus $b$-tagging performance on the jet pseudorapidity and transverse momentum was demonstrated. As explained in the text, this is caused by physics or instrumental effects.

Several new developments and further improvements are planned for the future. A dedicated reconstruction of $V^0$ decays and material interactions will also be available in *JetFitter* soon. Together with other improvements, like a modified seeding procedure and the application of multivariate techniques, a significant improvement can be expected for the *JetFitter* based $b$-tagging algorithm in the near future. An extension of the *BTagVrtSec* algorithm is able to reconstruct several vertices in a jet. The definition of variables related to the additional information to be used for $b$-tagging will be done in the near future and increase the $b$-tagging performance. The implementation of another inclusive vertex reconstruction algorithm, the so-called *Topological Vertex Finder* is in progress. It is based on an algorithm originally developed by the SLD collaboration [15].

In the inclusive secondary vertex reconstruction algorithms presented in this note, different cuts are applied at different stages during vertex finding and fitting. Currently, these cuts do not depend on the jet parameters. Due to the strong dependence of the $b$-tagging performance on them, however, it seems more appropriate to optimize these parameters depending on the kinematics of the jet under consideration. One quantity worth investigating is the size of the cone around the jet axis within which charged particle tracks are associated to the jet and thus used by the vertex reconstruction and $b$-tagging algorithms. For large jet transverse momenta, the cone that contains most of the particles from the decays of heavy $b$- and $c$-hadrons as well as from fragmentation, will be significantly smaller than for lower transverse momenta. It is thus desirable to choose the size of this track association cone depending on the transverse momentum of the jet. Since this is not restricted to secondary vertex based $b$-tagging algorithms, this point is addressed in detail in [4]. To give an idea of what can be expected from such a parameter tuning, one parameter of the *BTagVrtSec* algorithm has been optimized in bins of the jet transverse momentum. This parameter is a cut on the significance of the displacement of vertex candidates built from pairs of tracks from the primary event vertex. Table 13 shows the values chosen after the optimization procedure.

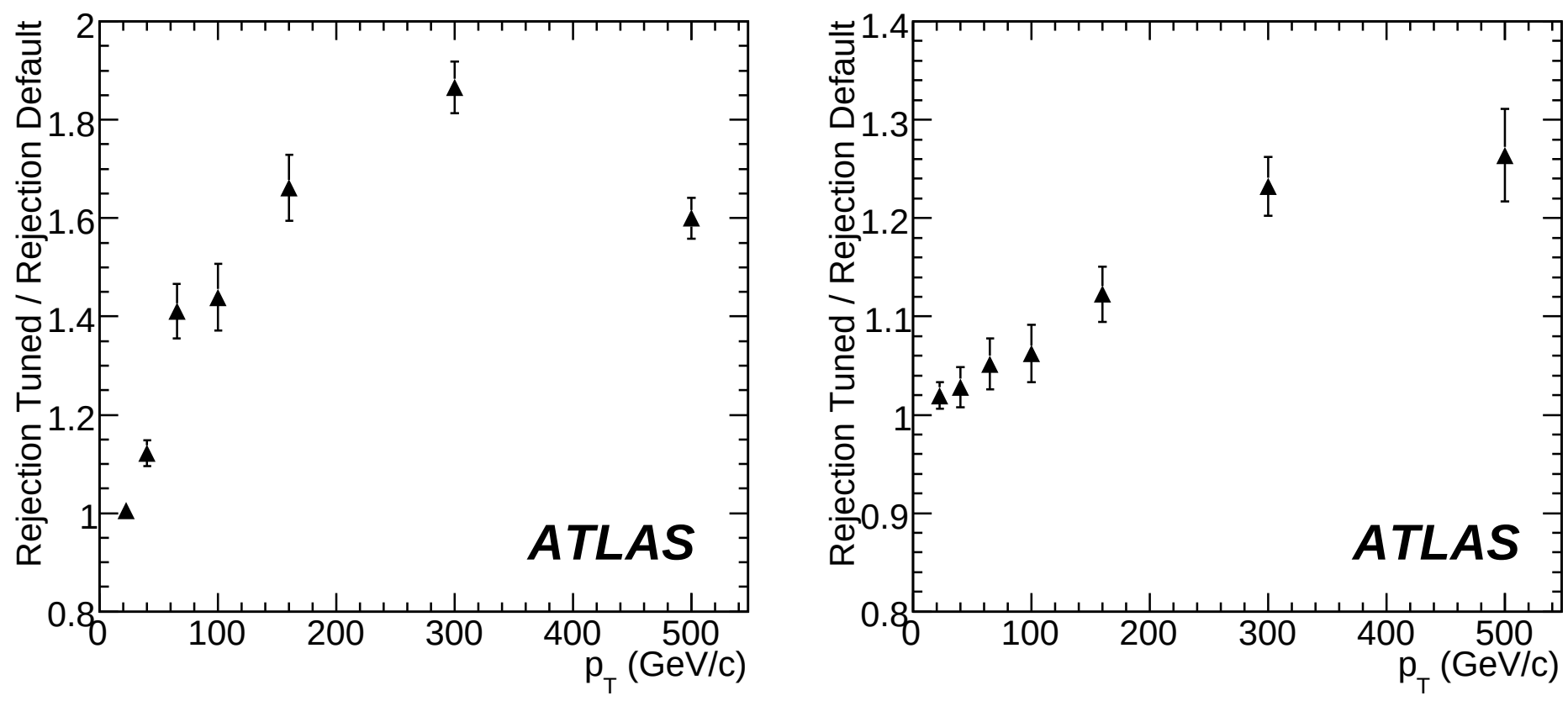

Figure 17: Ratios of the tuned rejections to the default rejection as a function of the jet transverse momentum for a $b$-tagging efficiency of 60%. The left hand side plot shows this ratio for light quark jet rejection, while the right hand side plot shows the ratio for charm jet rejection.

It can be seen, that the optimal value varies strongly with the jet transverse momentum. Figure 17

Table 13: Optimal values for the significance cut on the displacement of two track vertex candidates from the primary event vertex for the different jet $p_T$ bins.

| Jet $p_T$ [GeV] | 15-30 | 30-50 | 50-80 | 80-120 | 120-200 | 200-400 | 400-1000 |
|---|---|---|---|---|---|---|---|
| Optimal Cut | 3.5 | 4.5 | 5.0 | 6.0 | 6.5 | 7.0 | 7.0 |

shows the gain in $b$-tagging performance that is achieved by the parameter tuning. It can be seen that the rejection against light quark jets improves by a factor of about two in the region of large jet transverse momenta, whereas the gain in charm jet rejection is more moderate. This parameter tuning will be continued in the future and will be included in the $b$-tagging algorithms.

# Effects of Misalignment on $b$-Tagging

**Abstract**
This note investigates the effects of misalignment on $b$-tagging performance using Monte Carlo simulations. Four different alignment sets were considered, two with known random misalignments, one produced with the ATLAS alignment procedures and one with a perfectly aligned detector. Error tuning was investigated to compensate for the larger effective hit errors caused by misalignment. In addition the effects of misalignment on the tracking and vertexing performance were evaluated.

## 1 Introduction

The ATLAS detector has been built to provide high precision tracking and vertexing which are essential for good $b$-tagging performance. Misalignment of the detector will degrade the tracking resolution and consequently the performance of the $b$-tagging is expected to be sensitive to the alignment of the detector. As well as random misalignments of modules which give an effective smearing of each hit, systematic distortions introduced by the structure of the detector and by the alignment algorithms can have unexpected consequences.

The effects of misalignment on $b$-tagging have been studied with Monte Carlo simulations using a number of different alignment sets. These include a set that perfectly aligns the detector, two hand-made sets with known levels of random misalignment and an alignment set produced with the actual alignment algorithms to be used to align the ATLAS detector.

The assignment of correct errors for the hits is important for proper track and vertex reconstruction. Because module misalignments add to the intrinsic error of the module, the errors assigned to the hits need to be adjusted depending on the level of misalignment. Samples were investigated with and without this hit error adjustment.

The note is organized as follows: Section 2 describes the four alignment sets. The error tuning procedure and resulting scale factors used to adjust the hit errors are presented in Section 3. Studies were made with both $t\bar{t}$ and $WH(m_H = 120\ \text{GeV})$ samples and a brief description of these samples is given in Section 4. Section 5 describes the effects of misalignment on the tracking and vertexing performance. The effects of misalignment on the $b$-tagging performance, as measured by comparing the $b$-jet efficiency versus light jet rejection for the different alignment sets, are presented and discussed in Section 6.

## 2 Residual misalignment sets

In order to study the effects of misalignment, a number of different alignment scenarios were considered. The Monte Carlo simulation used in this investigation includes misalignments introduced at the simulation stage. The level of misplacement is representative of the amount of misalignment expected before any attempt to align the detector. The misalignments are of the order of 10–100 $\mu$m at the level of individual modules and assembly structures such as layers and disks and misalignments of the order of a few mm at the whole subsystem level. The level of these misalignment was based on known fabrication precisions and survey measurements. [1]. The misalignments introduced are too large to allow for reasonable reconstruction. What is desired is to reconstruct the resulting data sets with alignment corrections that are typical of what is expected in the real detector, after which only small misalignments should remain. Four alignment sets were used in this study:

- **Perfect:** This is the ideal case where the same set of alignments used in the simulation are used in the reconstruction and so one does not see any misalignment.

- **Aligned:** This uses an alignment set produced using the actual track based alignment algorithms developed for the ATLAS detector. It is expected to include any systematic deformations that the alignment procedure itself causes. While some systematic effects were included in the misalignments introduced in the simulation, such as clocking effects where each subsequent layer was rotated by increasing amounts, it does not contain all the systematic deformations which are expected. In particular large scale structures such as layers and discs were treated as rigid objects without any internal deformations such as a twist. Also pixel stave bows which are known to occur were not introduced. So it is possible that this set is still optimistic. This set is a first attempt at the full scale alignment of the inner detector and so should not be considered the final word on what will be seen in the real detector. However, it is considered to be the most realistic case studied here.

- **Random10:** This is a hand-made alignment set that takes the misalignment set used in simulation and randomly shifts the module positions by small amounts. These residual misalignments were introduced at different levels in the hierarchy. Random shifts and rotations were made to individual modules, and whole layers and disks. A small shift and rotation was also made to the whole pixel structure. Since the degradation of the $b$-tagging performance is expected to be dominated by the alignment of the pixel system, only pixel residual misalignments were introduced, The SCT and TRT were corrected perfectly as in the perfect alignment case. Due to movements of higher level structures in this set, some systematic effects may exist. The levels of misalignment are given in Table 1. The axis definitions for the module level uses a local frame where $x$ and $y$ are the $r\phi$ and $\eta$ measurement directions respectively and $z$ is out of the plane. For higher levels they correspond to the global frame with $z$-axis along the beam direction. $RotX$, $RotY$, $RotZ$ are rotations around the corresponding axes. The module level shifts in the $r\phi$ measurement direction are around 10 $\mu$m. The set attempts to emulate the level of misalignments expected during the early running period. It is not well known what levels of misalignments are expected after certain running periods so this is just an indication rather than being a firm prediction of what is expected at start up. Comparison with the real alignments (“Aligned” set) shows this to be a rather pessimistic scenario.

- **Random5:** As with “Random10”, but with levels of misalignment better by about a factor of 1.5 to 2. This is an estimate of what might be expected after several years of running. Like “Random10”, this set introduces misalignments at the three levels of hierarchy with levels of misalignment given in Table 2.

Table 1: Residual misalignment for “Random10”. Random misalignments were generated with a Gaussian distribution with $\sigma$ as tabulated. Shifts are in $\mu$m and rotations are in mrad.

| Level | $x$ | $y$ | $z$ | $RotX$ | $RotY$ | $RotZ$ |
|---|---|---|---|---|---|---|
| Module | 10 | 30 | 30 | 0.3 | 0.5 | 0.2 |
| Layer | 10 | 10 | 15 | 0.05 | 0.05 | 0.1 |
| Disk | 10 | 10 | 30 | 0.2 | 0.2 | 0.1 |
| Whole pixel | 10 | 10 | 15 | 0.1 | 0.1 | 0.1 |

Table 2: Residual misalignment for "Random5". Random misalignments were generated with a Gaussian distribution with $\sigma$ as tabulated. Shifts are in $\mu$m and rotations are in mrad.

| Level | $x$ | $y$ | $z$ | $RotX$ | $RotY$ | $RotZ$ |
|---|---|---|---|---|---|---|
| Module | 5 | 15 | 15 | 0.15 | 0.3 | 0.1 |
| Layer | 7 | 7 | 10 | 0.02 | 0.02 | 0.05 |
| Disk | 7 | 7 | 20 | 0.1 | 0.1 | 0.05 |
| Whole pixel | 7 | 7 | 10 | 0.05 | 0.05 | 0.05 |

# 3 Error scaling

## 3.1 Error scaling procedure

The intrinsic error of a hit will depend on a number of factors such as the cluster width and track direction. These factors are taken into account when calculating the intrinsic error of the hit. In the case of a perfectly aligned detector, if these intrinsic errors are properly determined one expects the pull distribution (the distribution of the hit residuals divided by the calculated intrinsic error) to have a width close to one.

The differences between the real positions of individual hits and those recorded by a misaligned detector lead to an additional error term that must be added in quadrature to the intrinsic error of the hits.

The errors on the hits directly affect whether a hit is associated to a track, the track propagation and track parameter errors and the objects that use tracks as input, such as vertices. Of particular importance to $b$-tagging is the precision of the impact parameter and the vertexing performance. It is therefore necessary to have accurately assigned hit errors.

In this section hits will refer to clusters in the silicon detectors (pixel and SCT) and drift circles in the TRT. To correct the hit errors the diagonal elements of the error matrix are modified using two parameters $a$ and $c$:

$$\sigma'^2 = a^2 \cdot \sigma^2 + c^2 \tag{1}$$

where:

- $\sigma$ is the original error assigned to the hit which is a function of the cluster size and track angle. This should normally be close to the intrinsic resolution if properly determined,
- $a$ is a multiplicative factor on the error, which is meant to compensate for inaccuracies in the intrinsic error determination,
- $c$ is a constant added in quadrature to the error. This is meant to correct effects attributed purely to residual misalignments.

Since each detector component can have significantly different behaviour, the granularity of each detector component has to be taken into account, and therefore different sets of ($a$, $c$) have to be computed separately for the barrel and endcap regions for each detector technology, as well as for the different $r\phi$ and $\eta$ measurement directions in the case of the pixel detector.

For the derivation of the ($a$, $c$) pairs, the distributions of hit residuals and their pull distributions are analyzed, and in particular the deviations of the pull widths from the ideal value of 1 are investigated.

Since the scale factor $a$ is intended to correct the intrinsic resolutions, this is most easily obtained with a perfectly aligned geometry. Naturally, this is not possible with real data, where more in depth studies will be needed to determine if the assigned intrinsic errors are appropriate. Currently the factor $a$ is needed as the errors used in the reconstruction do not match those observed in the simulation. It is assumed, however, that the best knowledge from test-beam and simulation will be put into the determination of the intrinsic error such that $a$ will be close to 1 and any remaining differences would be absorbed into the parameter $c$.

The widths of the resulting pull distributions can be used directly as the scaling factors $a$. This is iterated a few times, applying the correction, rerunning reconstruction and then determining new values of $a$. The iterations are necessary due to correlations between detector components. The factor $c$ is set to zero when determining the $a$ factor.

The resulting factors $a$ are then kept constant when used for the misaligned detector. Several iterations (apply ($a$, $c$) factors, reconstruct sample, analyze pulls) are performed using a misaligned detector, in order to determine the $c$ factor. It is computed using the formula:

$$c_i^2 = (p_{obs}^2 - 1)a^2\sigma_0^2 + p_{obs}^2 c_{i-1}^2 \quad (2)$$

where $c_i$ and $c_{i-1}$ are the values of the $c$ factor obtained in the iteration $i$ and $i-1$, respectively, $p_{obs}$ is the hit residual pull width observed at step $i$, and $\sigma_0$ is the average intrinsic detector resolution.

The determination of $c$ does not rely on any information about the actual detector positions and the procedure can be applied to real data. For this study a sample of high energy single muons was used, while in practice one would need to study the feasibility of extracting the error tuning with a more realistic event sample and track selection.

## 3.2 Error scaling parameters

The resulting parameters after the tuning are shown in Table 3. The values of $a$ are seen to be well below one for the SCT and TRT. This is due to an overestimate of the intrinsic errors. This is being improved. The value for $c$ gives some indication of the level of residual misalignment. For the “Random5” and “Random10” sets, the values of $c$ are higher than what was input for the module shifts. This is possibly due to a larger error being needed to compensate for the effects of the layer and disc movements. It can be seen that the real alignment results in small values of $c$ compared to the hand-made sets. In the pixel $r\phi$ measurement direction one gets 3 $\mu$m and in the $\eta$ measurement direction one gets around 15 $\mu$m for the “Aligned” set.

Table 3: Error scaling parameters for the different alignment scenarios. The parameter $a$ was tuned using the “Perfect” case and used for all alignment scenarios.

| | All $a$ | Perfect $c(\mu m)$ | Random10 $c(\mu m)$ | Random5 $c(\mu m)$ | Aligned $c(\mu m)$ |
|---|---|---|---|---|---|
| Pixel barrel $r\phi$ | 1.03 | 0 | 31 | 13 | 3 |
| Pixel barrel $\eta$ | 0.97 | 0 | 71 | 34 | 13 |
| Pixel endcap $r\phi$ | 1.05 | 0 | 30 | 14 | 3 |
| Pixel endcap $\eta$ | 1.08 | 0 | 43 | 11 | 15 |
| SCT barrel | 0.78 | 0 | 0 | 2 | 7 |
| SCT endcap | 0.86 | 0 | 6 | 5 | 8 |
| TRT barrel | 0.82 | 0 | 11 | 3 | 37 |
| TRT endcap | 0.77 | 0 | 11 | 10 | 19 |

Table 4: Efficiency of track reconstruction.

| Setup | efficiency (%) |
|---|---|
| Perfect | $97.09 \pm 0.02$ |
| Random10 | $95.50 \pm 0.03$ |
| Random10 + error scaling | $97.22 \pm 0.02$ |
| Aligned | $97.07 \pm 0.02$ |
| Aligned + error scaling | $97.04 \pm 0.02$ |

# 4 Monte Carlo Samples

The performance of $b$-tagging was investigated with $t\bar{t}$ and $WH(m_H = 120\ \text{GeV})$ samples which are standard samples used in $b$-tagging performance studies in ATLAS [2]. The main difference between these two samples is that the $WH$ events have lower jet multiplicities.

The $t\bar{t}$ sample includes semi-leptonic and di-lepton channels and this sample is used for measuring both $b$-jet and light jet efficiencies. The $WH(120)$ sample contains two sub samples. $WH(120) \rightarrow \mu\nu b\bar{b}$ is used to measure $b$-jet efficiencies and $WH(120) \rightarrow \mu\nu u\bar{u}$ for light jet efficiencies. The samples contained no pile-up.

# 5 Effects of misalignment on tracking and vertexing

The tracking and vertexing performance was studied with the $WH(120) \rightarrow \mu\nu b\bar{b}$ sample, although the other samples could equally have been chosen. A sample size of 27,000 events was used. This sample was reconstructed using three of the alignment sets described in Section 2: “Perfect”, “Random10” and “Aligned”. All other settings were kept the same. For the “Random10” and “Aligned” sets, samples were investigated with and without error scaling.

## 5.1 Tracking performance

The track reconstruction efficiency was computed for each scenario by comparing true tracks to corresponding reconstructed tracks. Tracks with $p_T > 1\ \text{GeV}$ and $|\eta| < 2.5$ were selected. A true track was considered to match a reconstructed track if the true track was the source of at least 50% of the hits associated to the reconstructed track. Next, the efficiency was computed as the ratio of the number of matched tracks to the number of all true tracks. The results for the efficiency for each of the three scenarios are shown in Table 4. The presence of residual misalignment in the “Random10” set causes a loss of about 2% in the efficiency, while the introduction of error scaling completely recovers the loss of performance. The “Aligned” set shows no significant change with respect to the “Perfect” case, with or without error scaling.

The number of fake tracks was also investigated in a similar manner to the track reconstruction efficiency calculation. A track was labeled as “fake” if it had fewer than 50% of its hits from a single true track. The percentage of fake tracks from the total accepted tracks is shown in Table 5 for the different alignment scenarios. The misalignments result in more fakes, and as with the efficiency, this is recovered with the introduction of error scaling.

## 5.2 Vertexing performance

The performance of the primary vertex finding algorithm was also investigated for the same five scenarios of alignment and error scaling. The efficiency for the primary vertex finding was computed as the ratio between the total number of reconstructed vertices to the total number of true vertices. It was found that this remains constant, at a value of $99.68 \pm 0.04\%$, irrespective of the misalignment scenario considered.

Table 5: Ratio of fake tracks to the total number of accepted tracks.

| Setup | fake tracks (%) |
|---|---|
| Perfect | $2.33 \pm 0.02$ |
| Random10 | $2.46 \pm 0.02$ |
| Random10 + error scaling | $2.29 \pm 0.02$ |
| Aligned | $2.34 \pm 0.02$ |
| Aligned + error scaling | $2.27 \pm 0.02$ |

The primary vertex resolution was evaluated by looking at the difference between the reconstructed and the true vertex position. The resulting distributions for $x$ and $z$ directions are displayed in Fig. 1. The results for the $y$ direction were similar to that in the $x$ direction. The introduction of residual misalignment causes the distributions to become wider as would be expected with a degradation of the hit resolutions. The shift for the hand-made sets is consistent with the shift of the entire pixel detector that was introduced. For the "Aligned" set a shift in $z$ of about 90 $\mu$m is apparent. The alignment procedures do not fully constrain the six degrees of freedom of the whole detector and no attempt was made to correct to the average primary vertex position in the $z$ direction. Because of this, the alignment procedure can easily result in such a shift when comparing with truth information.

The values for the resolution are computed as the width of a Gaussian fit to the distributions in Fig. 1 and are shown in Table 6. The resolution is degraded by residual misalignment, for both $x$ and $z$ directions. Error scaling helps to partially recover the loss of performance for the "Random10" scenario.

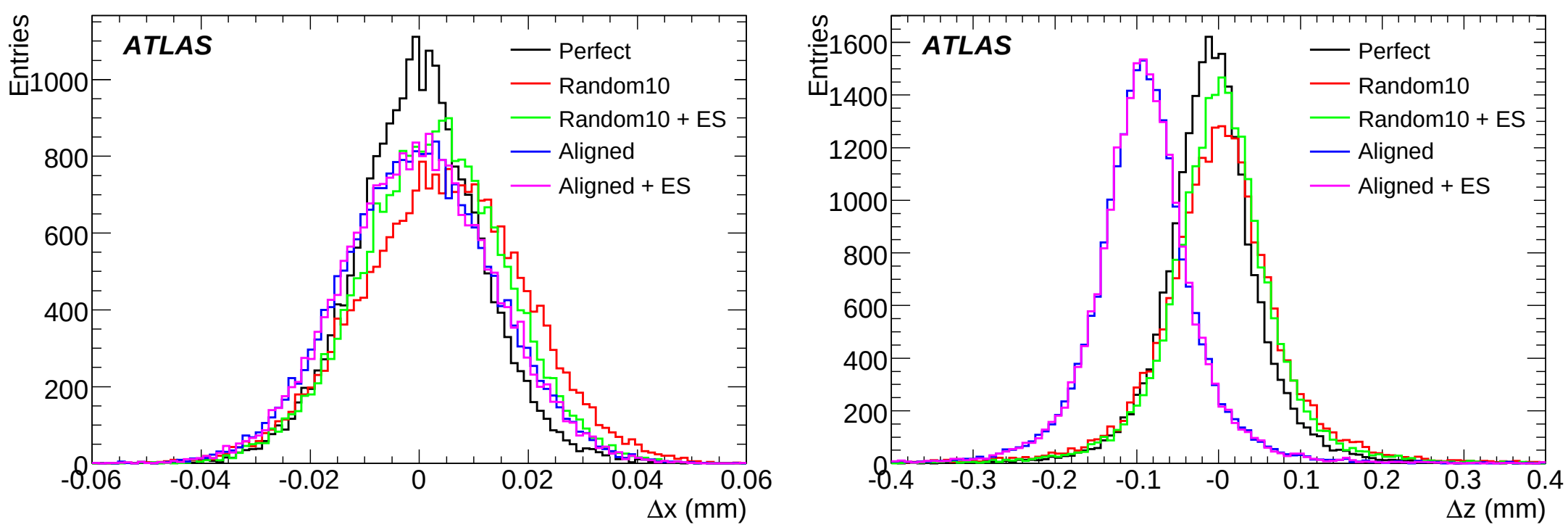

(a) x-direction

(b) z-direction

Figure 1: Primary vertex resolution, shown for the $x$ direction (a) and $z$ direction (b) for the various misalignment scenarios. ES denotes error scaling.

The number of primary vertex outliers was also investigated. Since the shift observed when looking at the resolution should not affect reconstruction, the true vertex position must be corrected by this shift. In the following, a primary vertex was flagged as "outlier" if the distance between the reconstructed vertex and the corrected true vertex position was greater than three sigma (30 $\mu$m in the $x$ direction and 150 $\mu$m in the $z$ direction).

The percentage of outliers is shown in Table 7, which shows that residual misalignment introduces additional outliers, and therefore indicates a degradation in the primary vertex finding. The number of outliers is however partially diminished by the application of error scaling for "Random10". For the "Aligned" scenario the corresponding scaling factors are much smaller than for "Random10" and the effect of error scaling on the primary vertex performance is negligible.

Table 6: Primary vertex resolution.

| Setup | res. in $x$ ($\mu$m) | shift in $x$ ($\mu$m) | res. in $z$ ($\mu$m) | shift in $z$ ($\mu$m) |
|---|---|---|---|---|
| Perfect | $11.4 \pm 0.1$ | $-0.13 \pm 0.07$ | $51.1 \pm 0.4$ | $-8.2 \pm 0.3$ |
| Random10 | $15.1 \pm 0.1$ | $4.2 \pm 0.1$ | $63.0 \pm 0.4$ | $1.4 \pm 0.4$ |
| Random10 + error scaling | $13.2 \pm 0.1$ | $2.6 \pm 0.1$ | $56.6 \pm 0.4$ | $2.3 \pm 0.4$ |
| Aligned | $13.9 \pm 0.1$ | $-0.18 \pm 0.09$ | $53.7 \pm 0.4$ | $-91.5 \pm 0.4$ |
| Aligned + error scaling | $13.8 \pm 0.1$ | $-0.15 \pm 0.09$ | $55.4 \pm 0.4$ | $-91.6 \pm 0.4$ |

Table 7: Fraction of primary vertex outliers.

| Setup | outliers in $x$ (%) | outliers in $z$ (%) |
|---|---|---|
| Perfect | $1.7 \pm 0.1$ | $4.1 \pm 0.1$ |
| Random10 | $5.5 \pm 0.2$ | $8.3 \pm 0.2$ |
| Random10 + error scaling | $2.8 \pm 0.1$ | $6.1 \pm 0.2$ |
| Aligned | $3.2 \pm 0.1$ | $8.2 \pm 0.2$ |
| Aligned + error scaling | $3.2 \pm 0.1$ | $8.0 \pm 0.2$ |

# 6 Effects of misalignment on $b$-tagging performance

For the study of the impact of the residual misalignment on the $b$-tagging performance, several data sets were produced with $WH(m_H = 120\ \text{GeV})$ and $t\bar{t}$ events. Eight cases were considered corresponding to each specific scenario of residual misalignment ("Perfect", "Aligned", "Random10" and "Random5") with and without error scaling as described in Sections 2 and 3. The performance of the $b$-tagging has been assessed by looking at the rejection rate of light quarks at $b$-jet efficiencies of 50% and 60% using various tagging algorithms: IP2D, IP3D, SV1 and the combined tagger IP3D+SV1. A description of the different taggers can be found in Ref. [2].

For each of the scenarios using $WH(120)$ samples, 45,000 $WH(120) \to \mu\nu b\bar{b}$ events and 175,000 $WH(120) \to \mu\nu u\bar{u}$ events were used. The $t\bar{t}$ samples contained 50,000 events each with the exception of the "Perfect" scenario without error scaling which had 570,000 events and the "Aligned" scenario with error scaling which had 500,000 events.

## 6.1 Results

Figure 2 shows the $b$-jet efficiency versus light jet rejection for $t\bar{t}$ and $WH(120)$ samples for the four misalignment sets with and without error scaling. Rejections for the IP3D and IP3D+SV1 tagger at $b$-tag efficiencies of 50% and 60% are tabulated in Table 8 for $WH(120)$ and in Table 9 for $t\bar{t}$.

The results for the IP3D+SV1 are also summarized in Fig. 3. As expected, the larger the misalignment, the greater the degradation of the $b$-tagging performance. In the case of "Random10", which represents the highest amount of misalignment (shifts of 10 $\mu$m in the pixel $r\phi$ measurement direction), there is almost a factor 2 drop in performance. For "Random5", where the level of misalignment is lower (shifts of the order of 5 $\mu$m), the decrease of the light jet rejection rates is lower than in the previous case at around 30% degradation. For the "Aligned" set the loss in performance is around 10 to 20% and lies somewhere between the “Perfect” alignment and the “Random5” set. This is consistent with the level of misalignment suggested by the parameter $c$ in the error tuning which is 3 $\mu$m in the pixel $r\phi$ measurement direction.

Table 8: Light jet rejection rates computed for $b$-jet efficiencies of 50% and 60% for the IP3D and IP3D+SV1 taggers for the various misalignment scenarios with and without error scaling (ES) for $WH(120)$ events.

| | Rejection rate | | | |
|---|---|---|---|---|
| Setup | IP3D (50%) | IP3D (60%) | IP3D+SV1 (50%) | IP3D+SV1 (60%) |
| Perfect | $211 \pm 4$ | $67 \pm 1$ | $399 \pm 11$ | $104 \pm 2$ |
| Perfect + ES | $215 \pm 5$ | $67 \pm 1$ | $372 \pm 11$ | $98 \pm 2$ |
| Random10 | $51 \pm 1$ | $23 \pm 1$ | $49 \pm 1$ | $21 \pm 1$ |
| Random10 + ES | $80 \pm 1$ | $29 \pm 1$ | $166 \pm 3$ | $49 \pm 1$ |
| Random5 | $144 \pm 3$ | $49 \pm 1$ | $165 \pm 3$ | $53 \pm 1$ |
| Random5 + ES | $182 \pm 7$ | $53 \pm 1$ | $311 \pm 16$ | $80 \pm 2$ |
| Aligned | $193 \pm 4$ | $62 \pm 1$ | $300 \pm 8$ | $84 \pm 1$ |
| Aligned + ES | $190 \pm 4$ | $62 \pm 1$ | $306 \pm 8$ | $87 \pm 1$ |

Table 9: Standard light jet rejection rates computed for $b$-jet efficiencies of 50% and 60% for the IP3D and IP3D+SV1 taggers for the various misalignment scenarios with and without error scaling (ES) for $t\bar{t}$ events.

| | Rejection rate | | | |
|---|---|---|---|---|
| Setup | IP3D (50%) | IP3D (60%) | IP3D+SV1 (50%) | IP3D+SV1 (60%) |
| Perfect | $238 \pm 11$ | $68 \pm 2$ | $480 \pm 30$ | $166 \pm 6$ |
| Perfect + ES | $244 \pm 11$ | $70 \pm 2$ | $474 \pm 30$ | $161 \pm 6$ |
| Random10 | $86 \pm 2$ | $32 \pm 1$ | $95 \pm 3$ | $38 \pm 1$ |
| Random10 + ES | $71 \pm 2$ | $25 \pm 0$ | $242 \pm 11$ | $77 \pm 2$ |
| Random5 | $192 \pm 7$ | $56 \pm 1$ | $290 \pm 14$ | $95 \pm 3$ |
| Random5 + ES | $133 \pm 4$ | $46 \pm 1$ | $360 \pm 20$ | $116 \pm 4$ |
| Aligned | $234 \pm 10$ | $67 \pm 2$ | $442 \pm 27$ | $143 \pm 5$ |
| Aligned + ES | $206 \pm 8$ | $62 \pm 1$ | $428 \pm 24$ | $138 \pm 5$ |

Table 10: Purified light jet rejection rates computed for $b$-jet efficiencies of 50% and 60% for the IP3D and IP3D+SV1 taggers for the various misalignment scenarios with and without error scaling (ES) for $t\bar{t}$ events.

| | Rejection rate | | | |
|---|---|---|---|---|
| Setup | IP3D (50%) | IP3D (60%) | IP3D+SV1 (50%) | IP3D+SV1 (60%) |
| Perfect | $331 \pm 19$ | $80 \pm 2$ | $914 \pm 86$ | $243 \pm 12$ |
| Perfect + ES | $332 \pm 19$ | $80 \pm 2$ | $872 \pm 79$ | $234 \pm 11$ |
| Random10 | $97 \pm 3$ | $34 \pm 0$ | $106 \pm 3$ | $41 \pm 1$ |
| Random10 + ES | $76 \pm 2$ | $26 \pm 0$ | $316 \pm 17$ | $89 \pm 3$ |
| Random5 | $250 \pm 12$ | $62 \pm 2$ | $387 \pm 23$ | $113 \pm 4$ |
| Random5 + ES | $154 \pm 6$ | $50 \pm 1$ | $558 \pm 41$ | $148 \pm 6$ |
| Aligned | $321 \pm 18$ | $77 \pm 2$ | $714 \pm 59$ | $190 \pm 8$ |
| Aligned + ES | $273 \pm 14$ | $70 \pm 2$ | $706 \pm 56$ | $180 \pm 7$ |

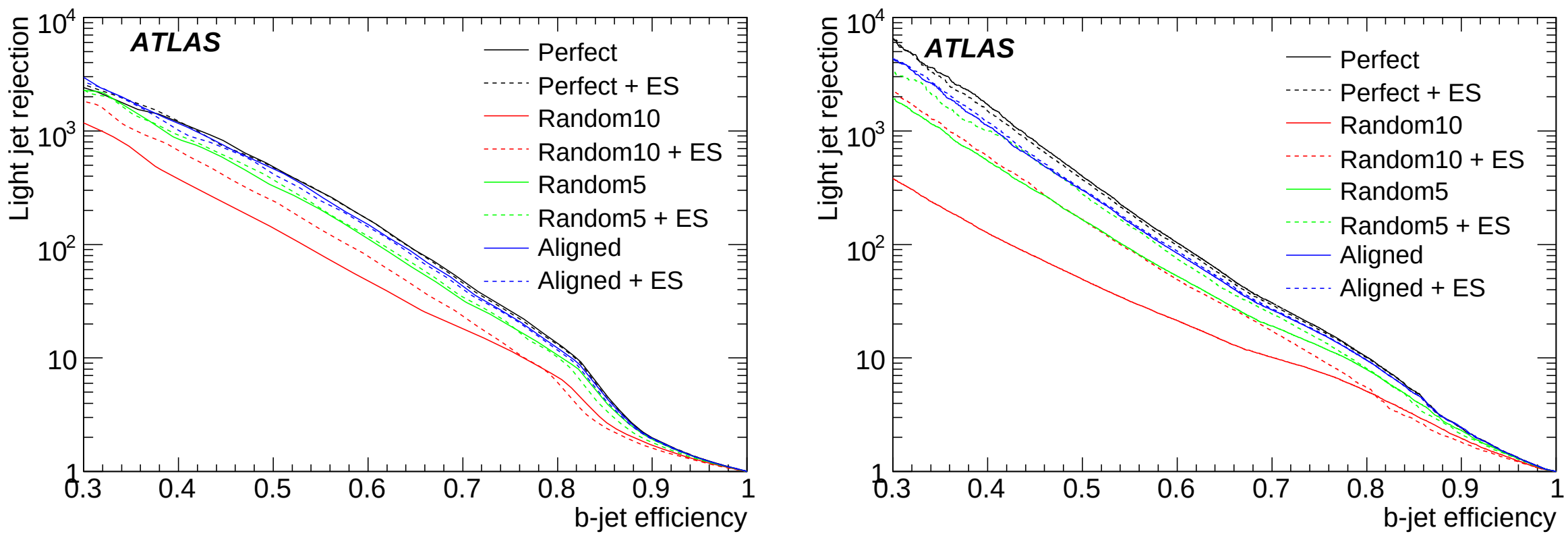

Figure 2: Light jet rejection versus $b$-tagging efficiency for the four different alignment sets for IP3D+SV1 for $t\bar{t}$ (left) and $WH(120)$ (right). ES denotes error scaling.

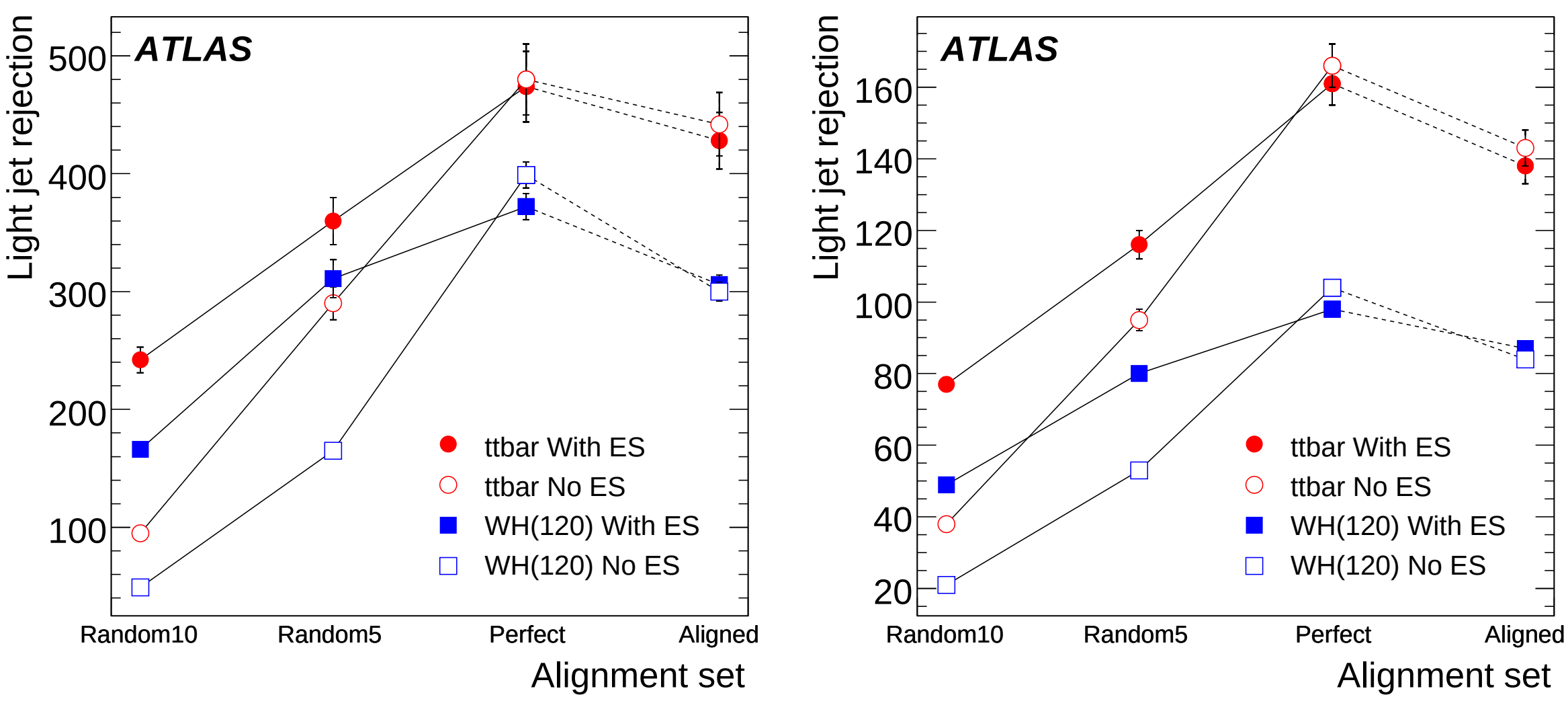

Figure 3: Light jet rejections using IP3D+SV1 tagger for the four misalignment scenarios at $b$-tagging efficiency working points of 50% (left) and 60% (right). Results are shown before and after error scaling (ES).

The rejections for $WH(120)$ are systematically lower than that for $t\bar{t}$ as observed in other studies [2], however, in general they both show similar trends with misalignment. Some differences are observed with the effects of error scaling which are discussed below. Results are shown mainly for $t\bar{t}$, although similar conclusions are reached for both samples.

Figure 4 shows the $b$-tag weight for IP3D+SV1 tagger for the different alignment scenarios. The differences between the “Aligned” and “Perfect” sets are difficult to see in such plots but for the larger misalignments (“Random10” and “Random5”) it is seen that the light jets have slightly larger weights while the $b$-jets have lower weights resulting in the loss of discrimination.

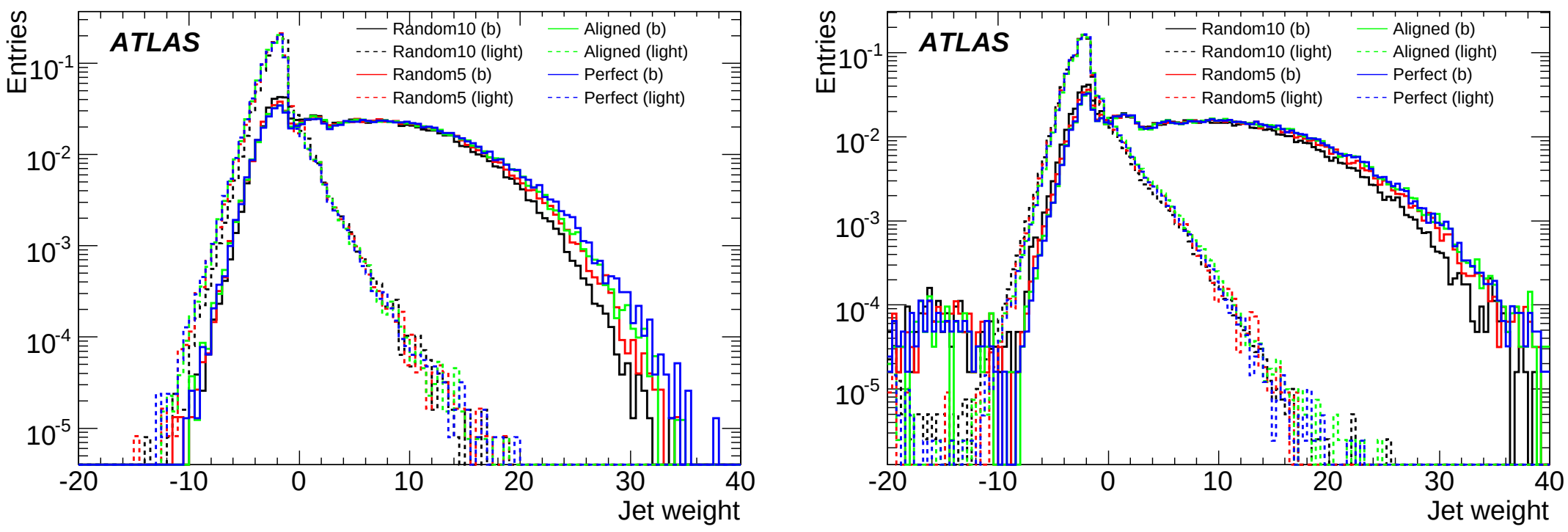

(a) $t\bar{t}$

(b) $WH(120)$

Figure 4: Jet weight distributions for the IP3D+SV1 tagger for the different alignment scenarios with error scaling for $t\bar{t}$ (a) and $WH(120)$ (b).

## 6.2 Effects of error scaling

It is observed in Fig. 3 that for the larger misalignment scenarios (“Random10” and “Random5”) the error scaling gives a significant improvement, while for the “Aligned” and “Perfect” case the impact of error scaling is small.

Figure 5 compares the $b$-tag weights with and without error scaling. Only the “Random10” results are shown. For the other scenarios the differences were less pronounced. The $t\bar{t}$ events show some differences in behaviour for the error scaling as compared with the $WH$ events. For $t\bar{t}$, the error scaling results in only a small difference for the light jets while for the $b$-jets the differences are more evident. This is in contrast with the $WH$ events where the light jets show more differences and the $b$-jets are less affected.

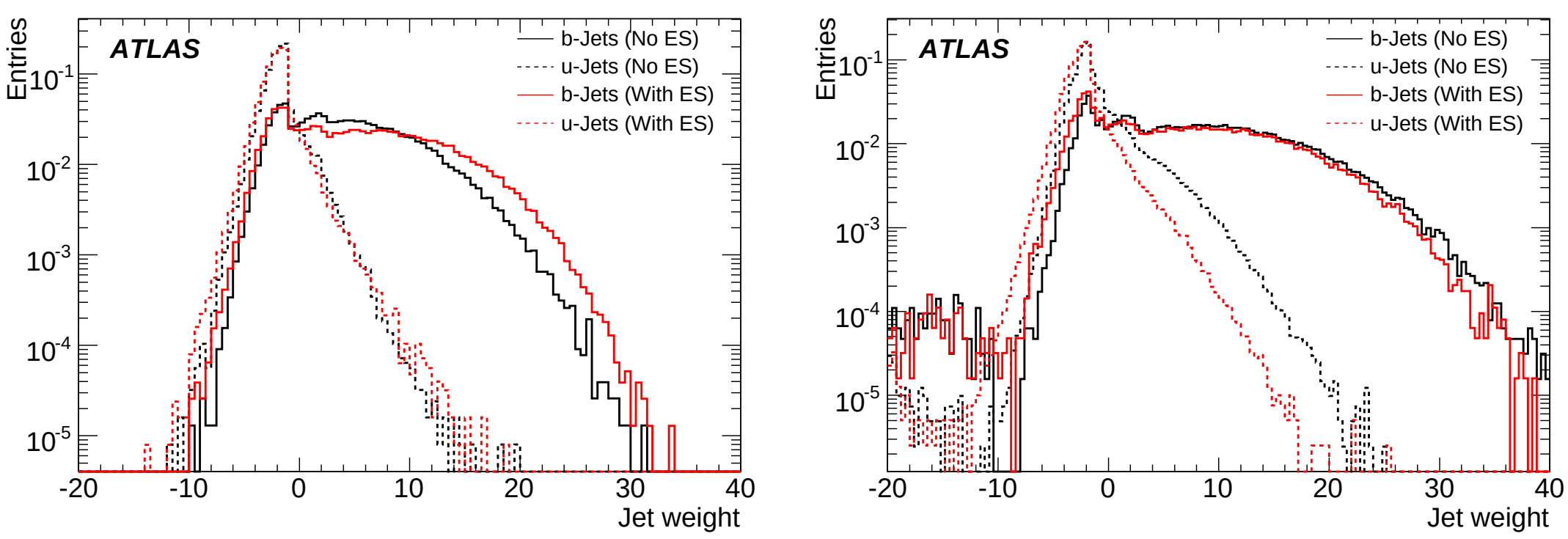

(a) “Random10”, $t\bar{t}$

(b) “Random10”, $WH(120)$

Figure 5: Jet weight distributions for the IP3D+SV1 tagger comparing with and without error scaling (ES) for “Random10” scenario for $t\bar{t}$ (a) and $WH(120)$ (b).

The differences are thought to be associated with the $p_T$ spectra of light jets and $b$-jets which are different from each other and different for the two event types. The effectiveness of the error scaling is expected to have some $p_T$ dependence since for lower $p_T$, the multiple scattering will dominate and differences in hit errors will be less important. There was insufficient statistics in the samples however to verify if this was indeed the case.

In some cases the use of error scaling results in worse performance. This is discussed further in Section 6.4.

## 6.3 Purified jets

For comparison with other studies [2] purified jets were also studied. Purification excludes labelling jets as light jets when there is a $b$-quark within a cone of 0.8. This gives a more physics independent measure of the performance (although differences will still be seen between samples because of the different $p_T$ and $\eta$ distributions of the jets contained in the samples). Table 10 shows the results for purified jets for $t\bar{t}$. For $WH$ events only standard jets were considered as other studies [2] show similar results with and without purification. Figure 6 compares standard and purified jets for $t\bar{t}$ events and it can be seen that the rejections are higher for purified jets but the trends are similar for the different alignment scenarios. The degradation of the "Aligned" case with respect to the "Perfect" alignment is more pronounced after purification (19 – 23% degradation) than for the standard jets (10 – 14% degradation). The effects of error scaling showed similar behaviour for both standard and purified jets.

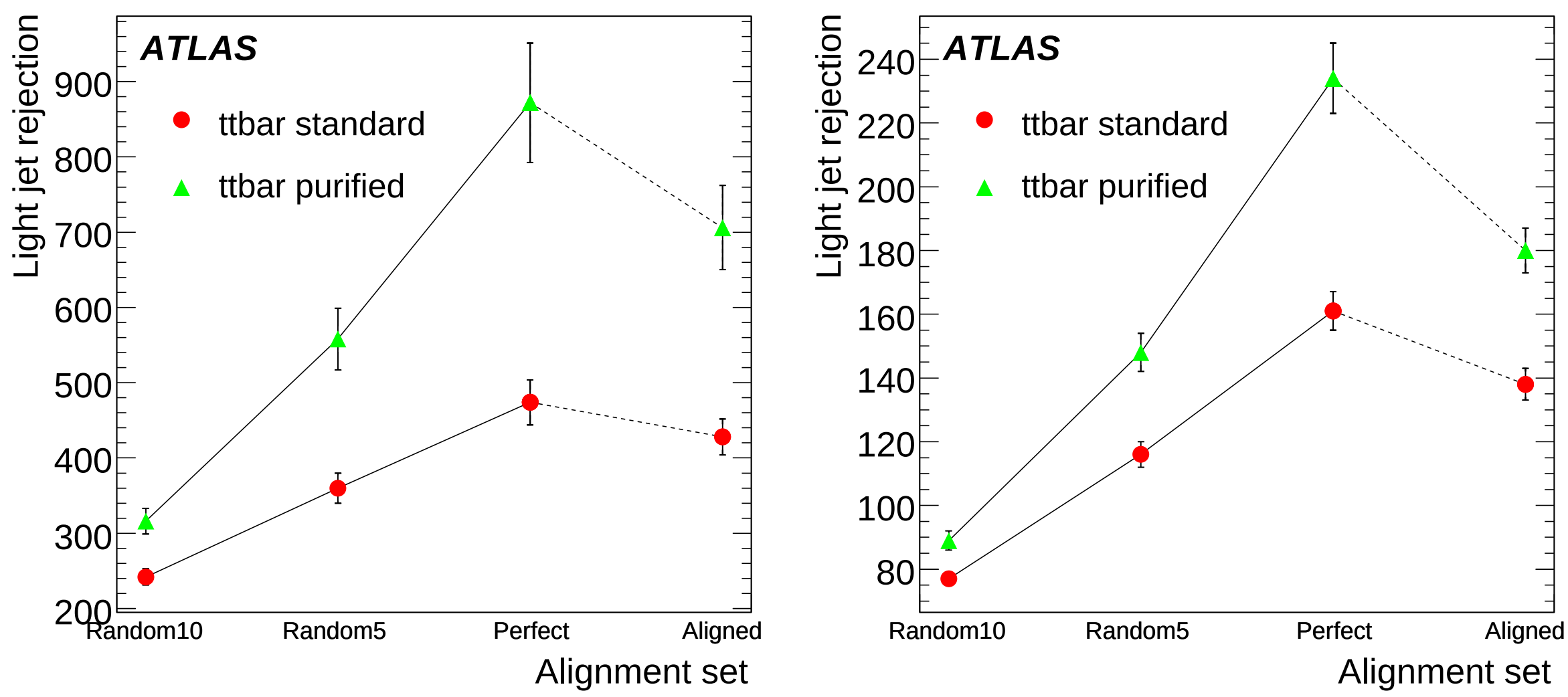

Figure 6: Comparison of light jet rejections using IP3D+SV1 tagger for standard and purified jets in $t\bar{t}$ events. Left plot: 50% $b$-tag efficiency. Right plot: 60% $b$-tag efficiency.

## 6.4 Comparison of the different taggers

Figure 7 shows the rejections for different taggers for standard jets. The impact parameter based taggers are the most affected by misalignment with the "Aligned" set showing 10 – 20% lower rejections than the "Perfect" case and up to a factor 3 degradation for the largest misalignment. After error scaling, the SV1 tagger shows rather uniform performance for all scenarios considered, with the "Aligned" set giving 10% degraded performance with respect to the "Perfect" case.

Without error scaling (see Fig. 8) the SV1 tagger shows significant differences for the different alignment scenarios. The ratio between rejections with error scaling to those without error scaling is shown in Fig. 9. It can be seen that the error scaling has the most beneficial effect with the larger misalignments ("Random10" and "Random5") for the SV1 performance. For the "Aligned" scenario the error scaling has little effect while for the "Perfect" case it actually degrades the performance slightly. For the impact parameter based taggers the error scaling has a smaller effect on the $b$-tagging performance and even degrades the performance in $t\bar{t}$ events.

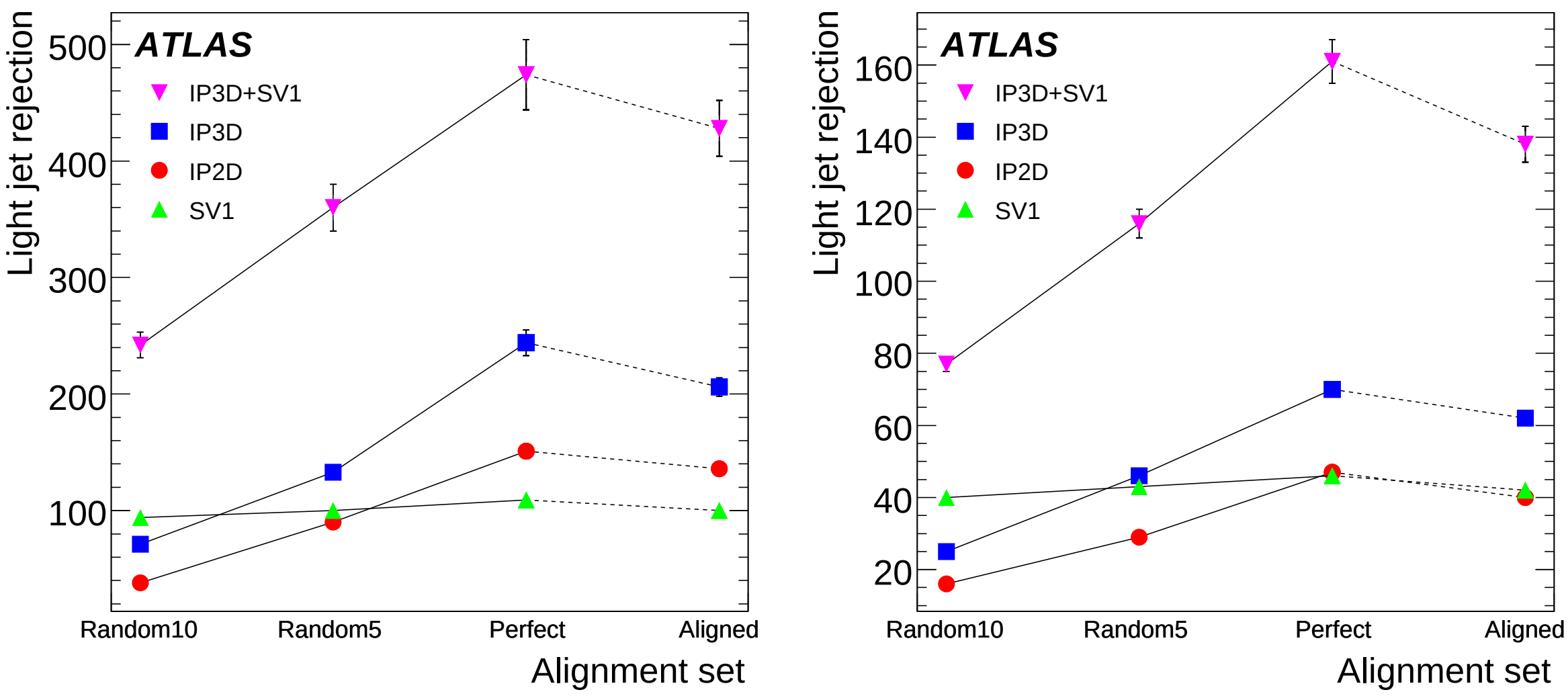

Figure 7: Comparison of the light jet rejections for the different taggers, IP2D, IP3D, SV1 and the combined tagger IP3D+SV1. Left plot: 50% $b$-tag efficiency. Right plot: 60% $b$-tag efficiency. Results are with error scaling using $t\bar{t}$ events.

Due to the beneficial effect of the error scaling on the secondary vertexing, the overall performance of the combined tagger (IP3D+SV1) is also improved with error scaling in the case of the "Random10" and "Random5" sets. The error scaling parameter $c$ is zero for the "Perfect" alignment and small for the "Aligned" set and consequently for both cases the effect of error scaling on the performance of $b$-tagging for the combined tagger is also small. For the "Aligned" case, while the relative difference is smaller than the hand-made sets, the error scaling degrades the performance slightly.

The reason for loss of performance with error scaling for some cases can be explained by the following: since the error scaling will generally increase the errors, it will reduce the impact parameter significance. This is desirable for light jets as it will make them more compatible with zero impact parameter. For $b$-jets, however, it also reduces the significance and so will reduce the $b$-tagging efficiency for a given weight cut, or in other words, one needs a lower weight cut to obtain the same efficiency and hence results in lower rejection for light jets for a given $b$-jet efficiency. The overall effect depends on these two competing effects and so can potentially lead to a loss in performance. While a decrease in performance with error scaling was observed in $t\bar{t}$ events for the IP2D and IP3D taggers, the opposite was observed for $WH$ events. As was already discussed above for the $b$-tag weight in Fig. 5, the error scaling affects the light jets and $b$-jets in the two physics samples differently and this is thought to lead to the differences in the error scaling behaviour seen here.

The effect of error scaling on the secondary vertex tagger will be to recover some secondary vertices which would have otherwise failed quality cuts due to an underestimated error. One will lose some true secondary vertices that are close to the primary vertex as the larger error will make them compatible with the primary vertex, however for similar reasons it will result in fewer fake secondary vertices close to the primary vertex.

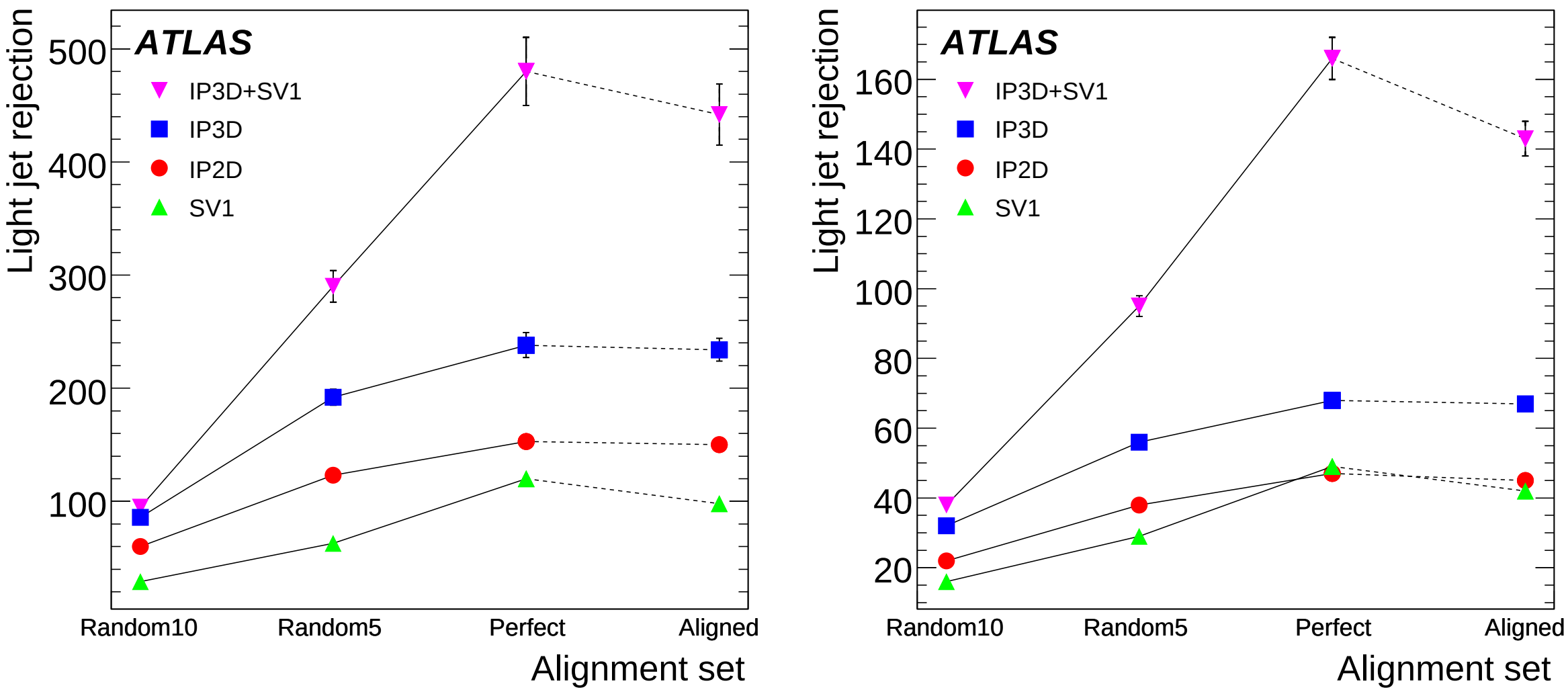

Figure 8: As in Fig. 7 but without error scaling.

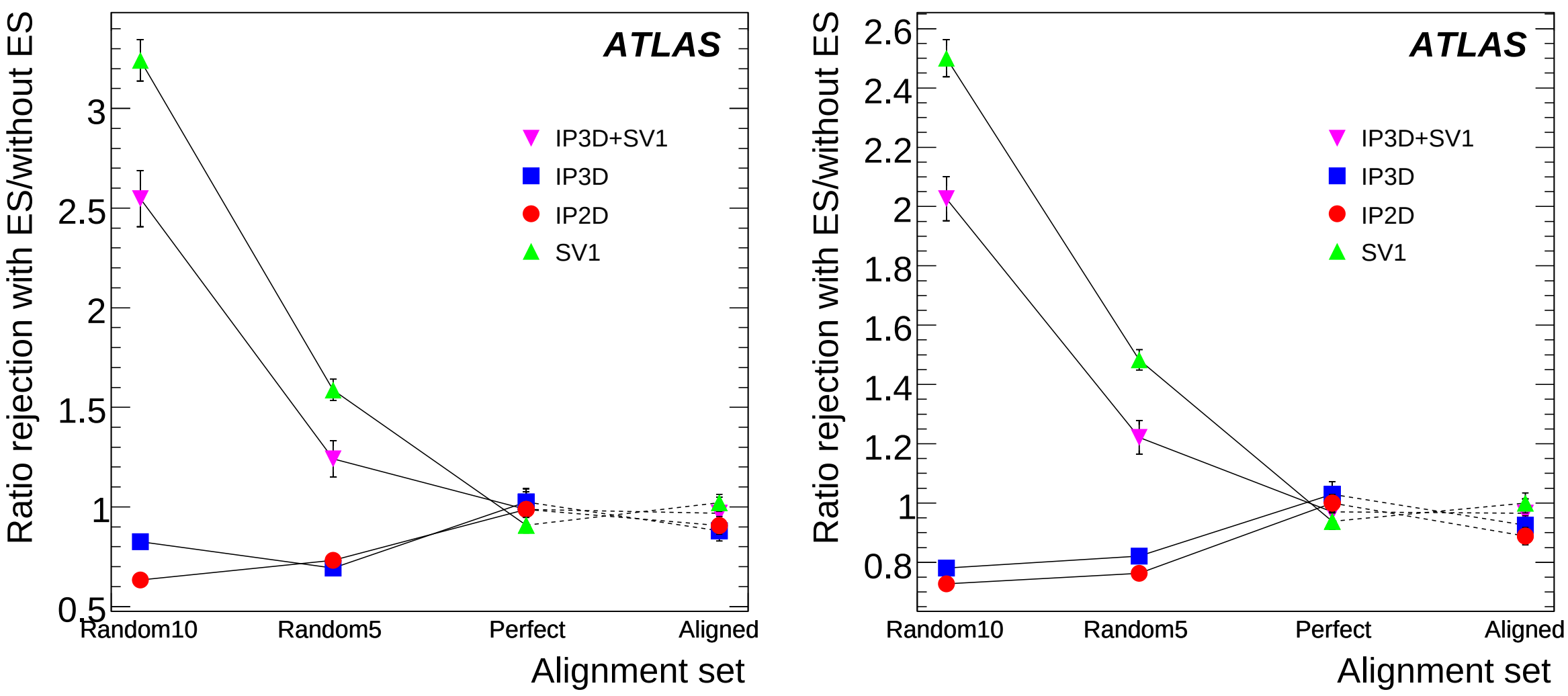

Figure 9: Ratio of rejections with error scaling to rejections without error scaling. Left plot: 50% $b$-tag efficiency. Right plot: 60% $b$-tag efficiency.

### 6.5 Effects of recalibration

The taggers require probability distribution functions for light jets and $b$-jets as described in Ref. [2] and the process of obtaining the set of these reference distributions is known as calibration. The results presented here use the same set of calibrations as used in Ref. [2]. Since misalignments will alter these distributions, it is possible that one can obtain some more discriminating power by recalibrating using the misaligned sample. Methods for obtaining these calibrations from real data are explored in Ref. [3].

To investigate whether recalibrating results in any gain in performance, a new set of reference distributions was obtained for each sample with and without error scaling. In practice one should use independent samples to calibrate and to test the performance, however, here the reference distributions were obtained with the same or a subset of the sample used to measure the performance.

As seen in Fig. 10 recalibration gives better performance, although after error scaling the difference between the fixed calibration and recalibration is only marginal.

One might expect that recalibration may compensate any miscalculation of the errors in a similar way to the error scaling. This is only partially the case as can be seen in Fig. 11, where the relative improvement with error scaling is reduced when recalibrating.

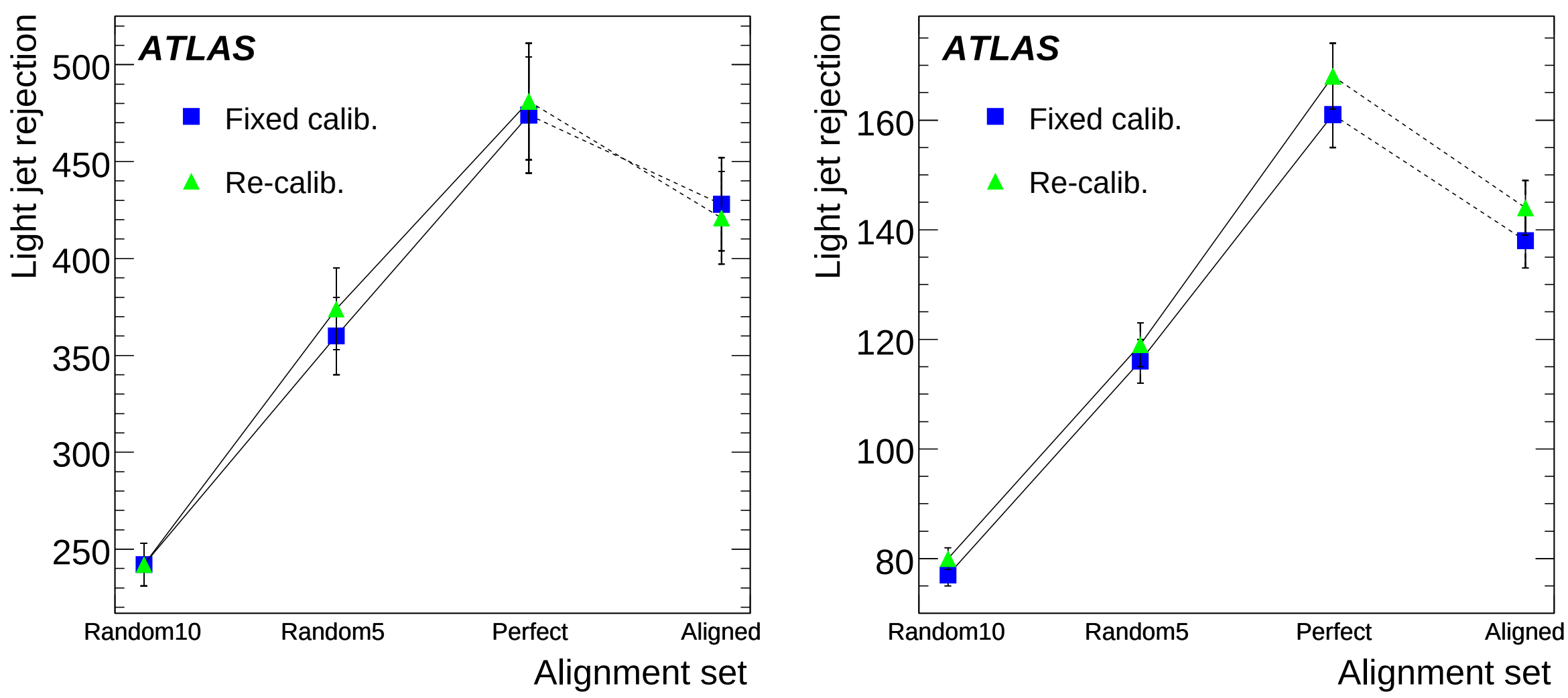

Figure 10: Comparison of the light jet rejection obtained for the IP3D+SV1 tagger using a fixed calibration or recalibrating for each separate sample. Left plot: 50% $b$-tag efficiency. Right plot: 60% $b$-tag efficiency.

## 7 Conclusion

The effect of misalignment on the tracking performance, as measured by tracking efficiency and fake rates, was small, and error scaling recovered the performance almost to the level that was seen with the perfect alignment. The degradation of the primary vertex was more significant, with resolutions about 2.5 $\mu$m degraded and an increase in the number of outliers. The "Aligned" set showed similar performance to the "Random10" despite better $b$-tagging performance.

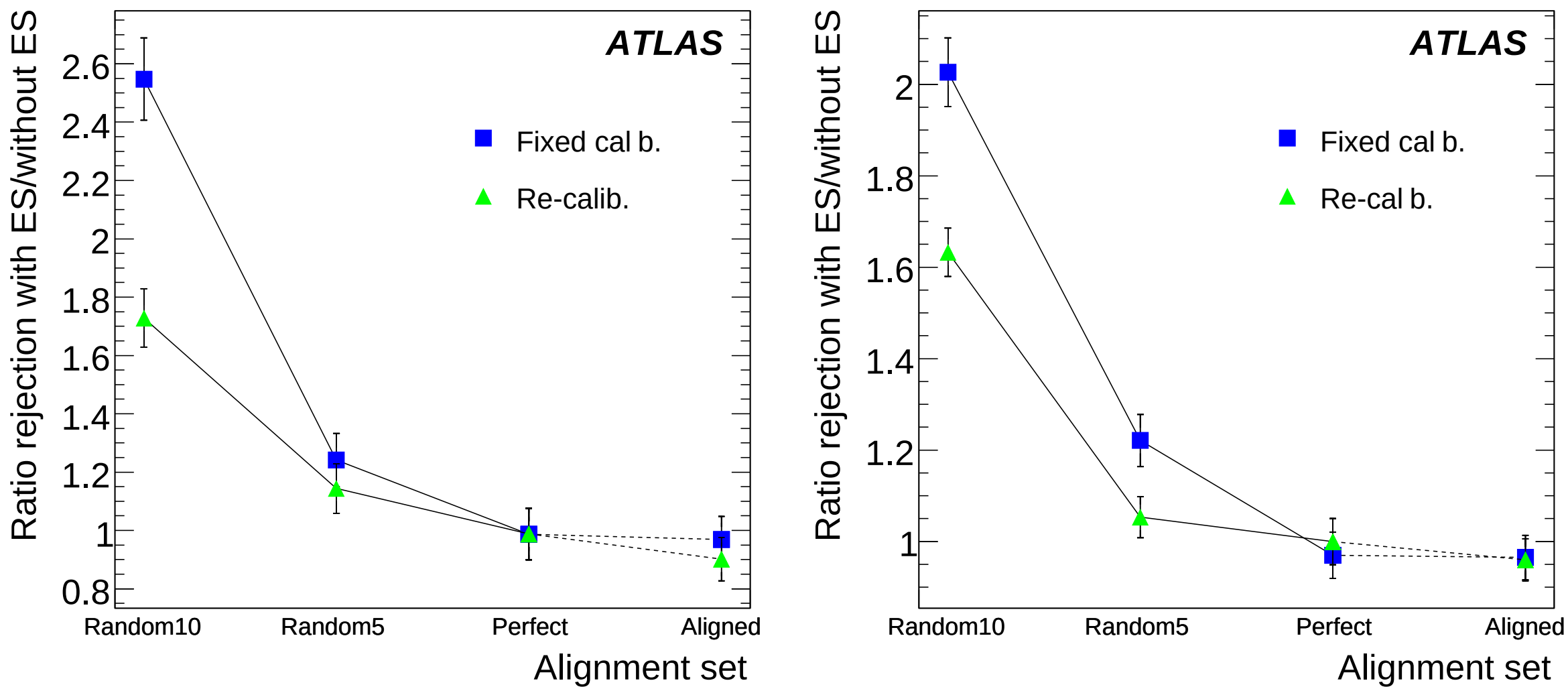

Figure 11: Ratio of rejections with error scaling to rejections without error scaling. Results are shown for the IP3D+SV1 tagger. Left plot: 50% $b$-tag efficiency. Right plot: 60% $b$-tag efficiency. Compares using a fixed calibrations and recalibrating for each separate sample.

The performance of $b$-tagging was clearly degraded with misalignment and the amount of degradation was found to be roughly proportional to the amount of random displacement of modules. Systematic effects are expected to also play an important role, however, the random displacement was the only aspect that was quantified in this study by the parameter $c$ obtained in the error scaling procedure. In order to disentangle the contributions from random and systematic effects it would be necessary to create dedicated residual misalignment sets with known systematic distortions.

The impact parameter based taggers were observed to be the most affected by misalignment and the introduction of error scaling brought little or even negative benefit to the $b$-tagging performance in the case of $t\bar{t}$ events. Error scaling was important for the performance of the secondary vertex finding, and without it, for larger misalignments the degradation for the secondary vertex based tagger was significant. With error scaling most of the degradation was recovered and the secondary vertex tagger showed uniform performance for all alignment scenarios considered. The behaviour of the combined tagger, IP3D+SV1, follows what one might conclude from the behaviour of the separate taggers, that is, it benefits from error scaling but shows a degradation with misalignment even after error scaling.

The "Aligned" set was the most realistic misalignment scenario studied here and the results were encouraging with rather moderate degradation in the $b$-tagging performance. In purified jets from $t\bar{t}$, the degradation was more evident with 19% loss of rejection at 50% $b$-tagging efficiency. However, in a more realistic environment, as seen by looking at standard jets, the amount of degradation was only 10%. At 60% $b$-tagging efficiency, the loss of rejection was slightly larger with 23% degradation for purified jets and 14% degradation for standard jets. For WH events the loss of rejection was similar with around 18% degradation at 50% $b$-tagging efficiency and 11% degradation at 60% $b$-tagging efficiency.

The amount of residual misalignment remaining after applying the actual alignment procedures resulted in only a small loss of performance and so misalignments are not expected to cause a major problem for doing $b$-tagging in ATLAS.

# Soft Muon $b$-Tagging

**Abstract**

$b$-jets can be identified by taking advantage of the presence of a muon coming from the semi-leptonic decay of the $b$ hadrons. This note describes a soft muon tagging algorithm and its performance when applied to different Monte Carlo physics samples in ATLAS. A $b$-jet tagging efficiency of about 10% (including the inclusive $b \rightarrow \mu\nu X$ branching ratio of $\approx 20\%$) is achieved for a light jet rejection of better than 300. The effect of pile-up events and cavern background on the performance is also considered and is found to be significant but manageable.

## 1 Introduction

Soft-lepton tagging relies on the semi-leptonic decays of $b$ and $c$ hadrons. Indeed the presence of a muon is enhanced in $b$-jets thanks to the significant semi-leptonic decay branching ratio of $b$ hadrons ($BR(b \rightarrow \mu\nu X) \approx 11\%$), and of $c$ hadrons produced by the $b$ hadron decay (sequential semi-leptonic decay, $BR(b \rightarrow c \rightarrow \mu\nu X) \approx 10\%$). A soft muon tagger, while intrinsically limited by the small semi-leptonic branching ratio, offers a good alternative or complement to the more performant lifetime taggers. Thanks to its good purity and low correlation with lifetime taggers, it can be used to do cross-calibrations of the two types of tagger [1], or to enhance the lifetime tagger performance by combining the two. Finally, it permits derivation of specific jet energy corrections [2].

The properties of the soft muons and of the various backgrounds are shown in Section 2. The algorithm is then described in detail in Section 3 while its performance applied to different Monte Carlo samples is discussed in Section 4.

## 2 Properties of muons from semi-leptonic decays and their backgrounds

There are three sources of background from particles within light jets: muons coming from the decay of light hadrons (mostly pions), hadrons managing to go through the calorimeter and reaching the muon spectrometer ("punch-through"), and hits caused by the neutron gas that will be surrounding the detector during data taking, producing fake tracks in the muon spectrometer ("cavern background"). $c$-jets are also a source of background as far as $b$-tagging is concerned. Figures 1, 2, 3, and 4 show, respectively, the distance to the closest jet $\Delta R = \sqrt{\eta^2 + \phi^2}$, the muon transverse momentum, the impact parameter, and the transverse momentum relative to the closest jet axis $p_T^{rel}$, for muons from (direct) $b$ decays, sequential $b \rightarrow c \rightarrow \mu\nu X$ decays, $c$ decay, light hadron decays, and fake muons, in jets of $E_T > 15$GeV and $|\eta| < 2.5$ in $t\bar{t}$ events. $p_T^{rel}$ is defined as the muon momentum in the plane orthogonal to the jet axis, where the jet axis is corrected for the presence of a muon by adding the muon momentum to the jet momentum. Contributions from true muons are estimated at the generator level while fake muons are estimated with the full simulation and defined as a reconstructed muon associated with a jet while no muon of momentum larger than 2 GeV was found at the generator level within $\Delta R = 0.6$ of that same jet. Note that the distinction between fake muons and muons from light hadron decays is somewhat arbitrary since the latter can also be produced in the calorimeter as part of the hadronic shower; such muons are present in the full simulation but are not distinguished from fake muons because long-lived particles and showers are taken care of by GEANT and the full GEANT information was not available in the present study. Muons produced by direct $b$ hadron decays have a significantly larger transverse momentum than the various backgrounds, especially muons from light hadron decays; additionally, since light hadrons such as charged pions and charged kaons have a long lifetime, those muons tend to have a large impact parameter. Those two properties will be used to reject this important background efficiently. Thanks to the $b$ hadrons high mass, muons from direct $b$ decays tend to be more boosted in a plane transverse to the jet axis, *i.e.* have a larger $p_T^{rel}$ than the backgrounds. This property is used in the tagger to further separate $b$-jets from light jets through a likelihood technique. It is important to note that muons from sequential $b$ decay are much more difficult to separate from background since they have a much softer spectrum both in $p_T$ and $p_T^{rel}$ than those from direct $b$ decay. Finally, Fig. 5 shows that the $p_T^{rel}$ distribution in $t\bar{t}$ and $WH$ events is very similar. Indeed the $p_T^{rel}$ variable is not very correlated with the jet $E_T$ or pseudo-rapidity. Thus it is not very dependent on the process producing the $b$-jets.

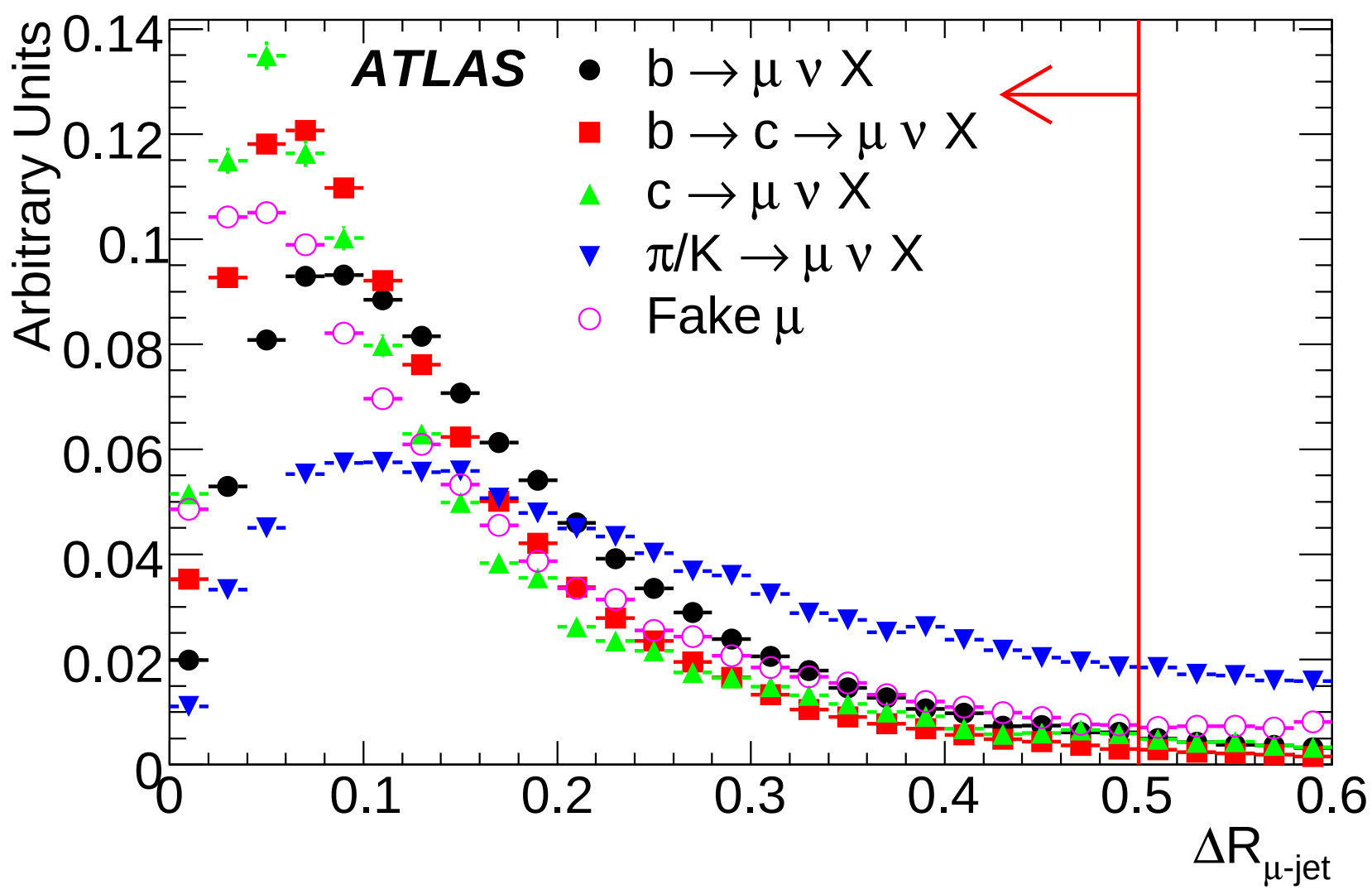

Figure 1: Distribution of $\Delta R = \sqrt{\Delta\eta^2 + \Delta\phi^2}$ between closest jet and muons from $b$ hadron decays, $c$ hadron decays, light hadron decays, and fakes, in jets of $E_T > 15$GeV and $|\eta| < 2.5$ in $t\bar{t}$ events. The red line shows the cut applied for the basic selection described in Section 3. All histograms in Fig. 1 to 5 are normalized to unity.

## 3 Description of the algorithm

The algorithm works in three steps, described in the following sections:

- The standard muon reconstruction algorithms are used to identify muons.
- Muons satisfying some basic requirements are then associated to jets.
- Finally, a 1-dimensional likelihood ratio using the $p_T^{rel}$ variable discriminates further signal from background.

### 3.1 Muon reconstruction

Two complementary muon reconstruction algorithms are used by the soft-muon tagger: so-called "combined" muons, which correspond to a track fully reconstructed in the muon spectrometer and matched with a track in the inner detector; and "tagged" muons, which cannot reach the muon middle and outer stations and are identified by matching an inner detector track with a segment in the muon inner stations only. The ATLAS muon spectrometer is able to fully reconstruct muons of momentum larger than about 5 GeV, providing a precise measurement of its momentum, allowing – in combination with the inner detector track – a very good rejection of the combinatoric background in the dense environment of jets. Muon spectrometer tracks that fail to be combined with an inner detector track are not used by the soft-muon tagger because of the contamination from fakes and muons from light hadron decays. Tagged muons are less pure than combined muons but allow one to recover some efficiency in the crucial low momentum range. Thus both combined muons and tagged muons are used by the soft-muon tagger. For more details of muon reconstruction and identification, see Ref. [3].

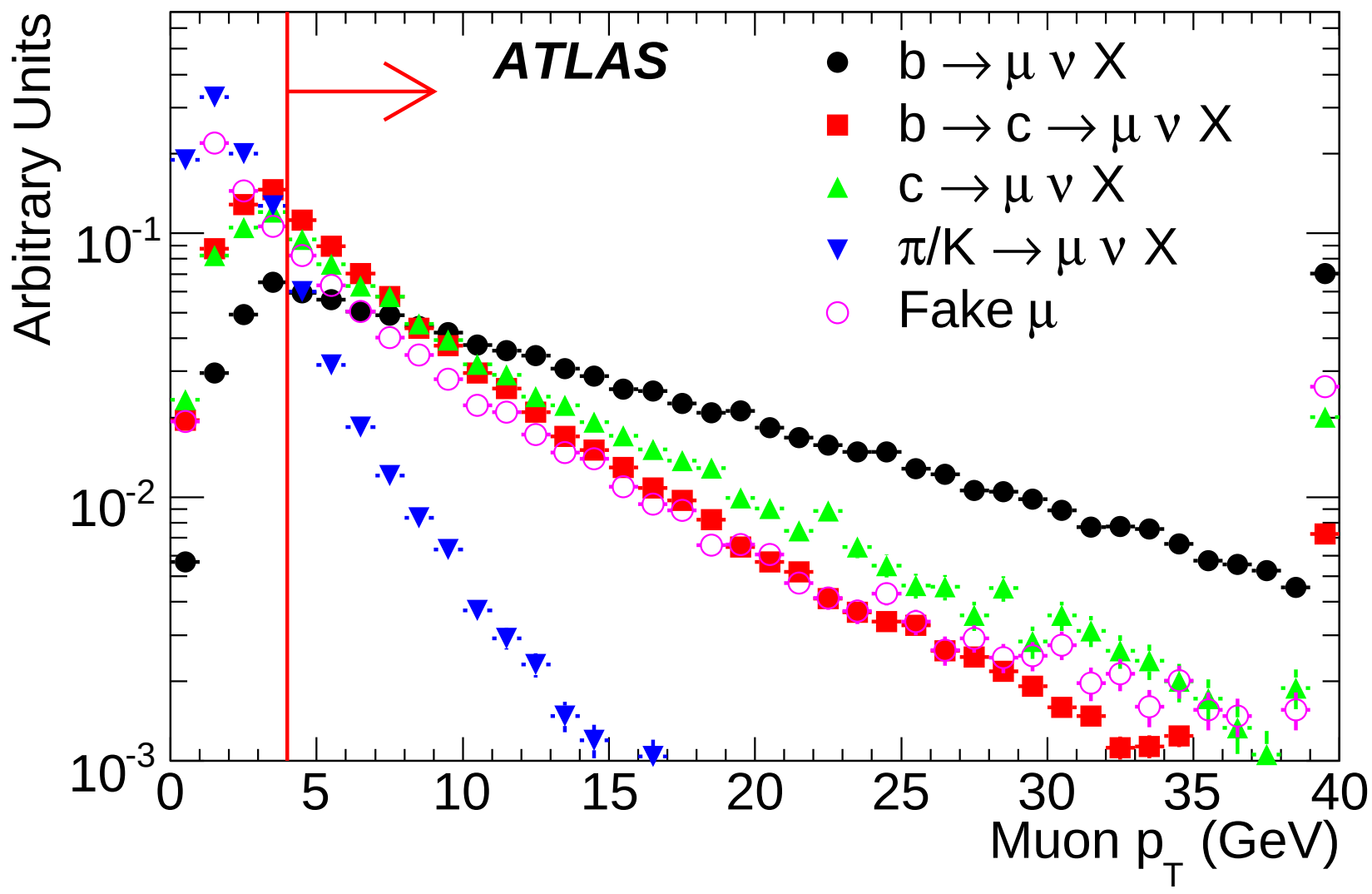

Figure 2: Transverse momentum distribution of muons from $b$ hadron decays, $c$ hadron decays, light hadron decays, and fakes, in jets of $E_T > 15$GeV and $|\eta| < 2.5$ in $t\bar{t}$ events. The last bin includes overflows. The red line shows the cut applied for the basic selection described in Section 3.

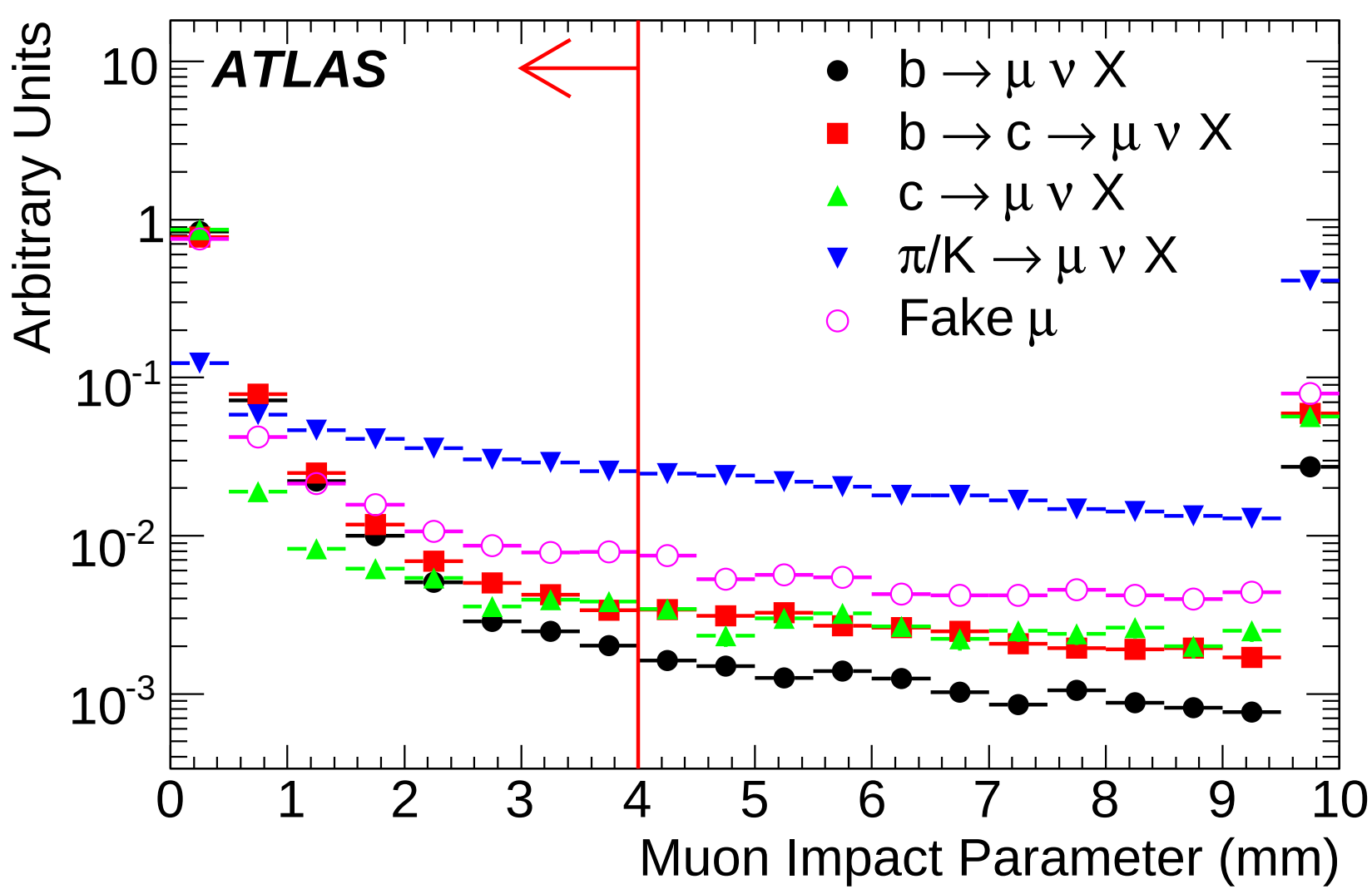

Figure 3: Impact parameter distribution of muons from $b$ hadron decays, $c$ hadron decays, light hadron decays, and fakes, in jets of $E_T > 15$GeV and $|\eta| < 2.5$ in $t\bar{t}$ events. The last bin includes overflows. The red line shows the cut applied for the basic selection described in Section 3.

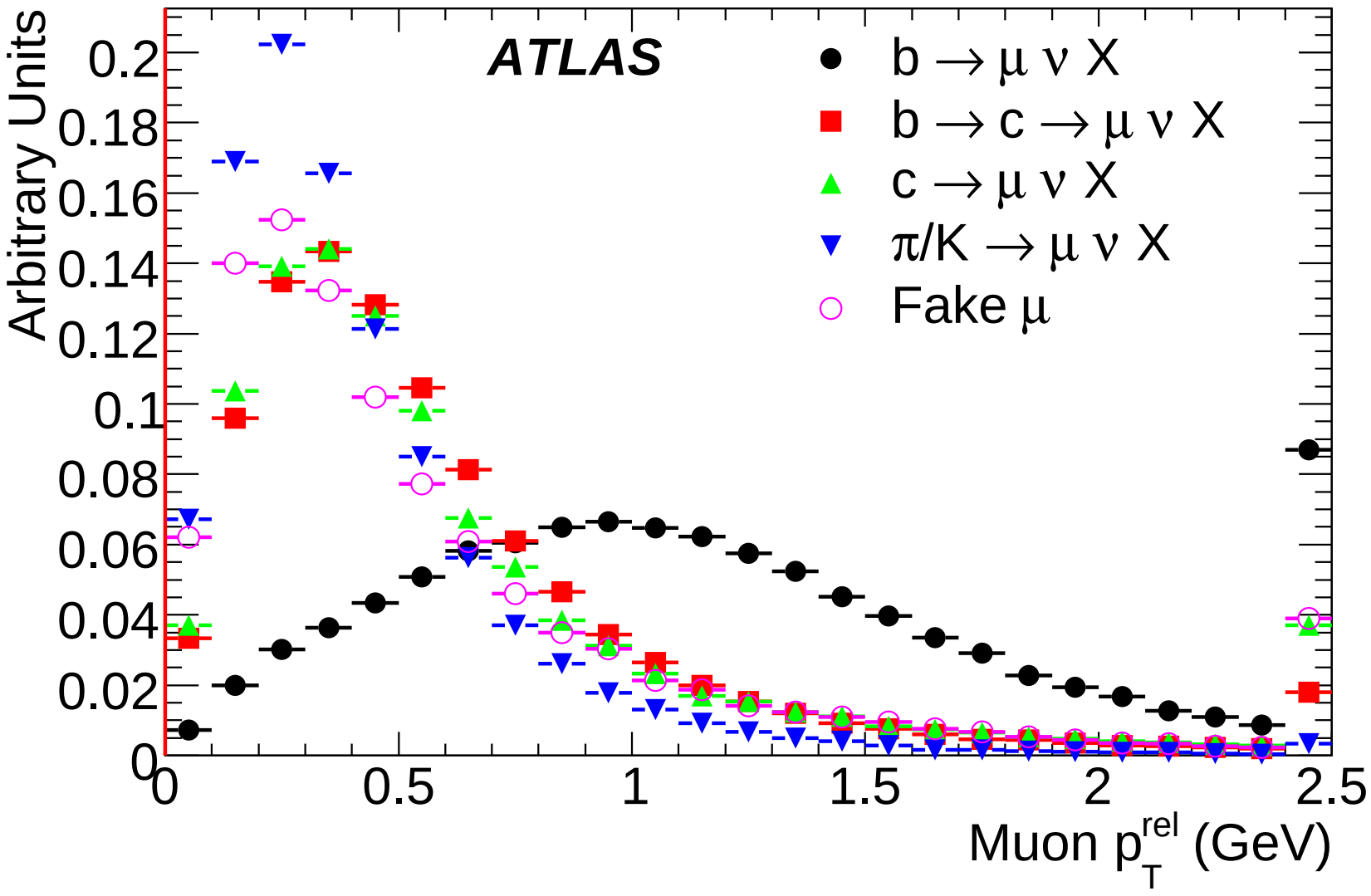

Figure 4: Distribution of the transverse momentum relative to the jet axis of muons from $b$ hadron decays, $c$ hadron decays, light hadron decays, and fakes, in jets of $E_T > 15$GeV and $|\eta| < 2.5$ in $t\bar{t}$ events. The last bin includes overflows.

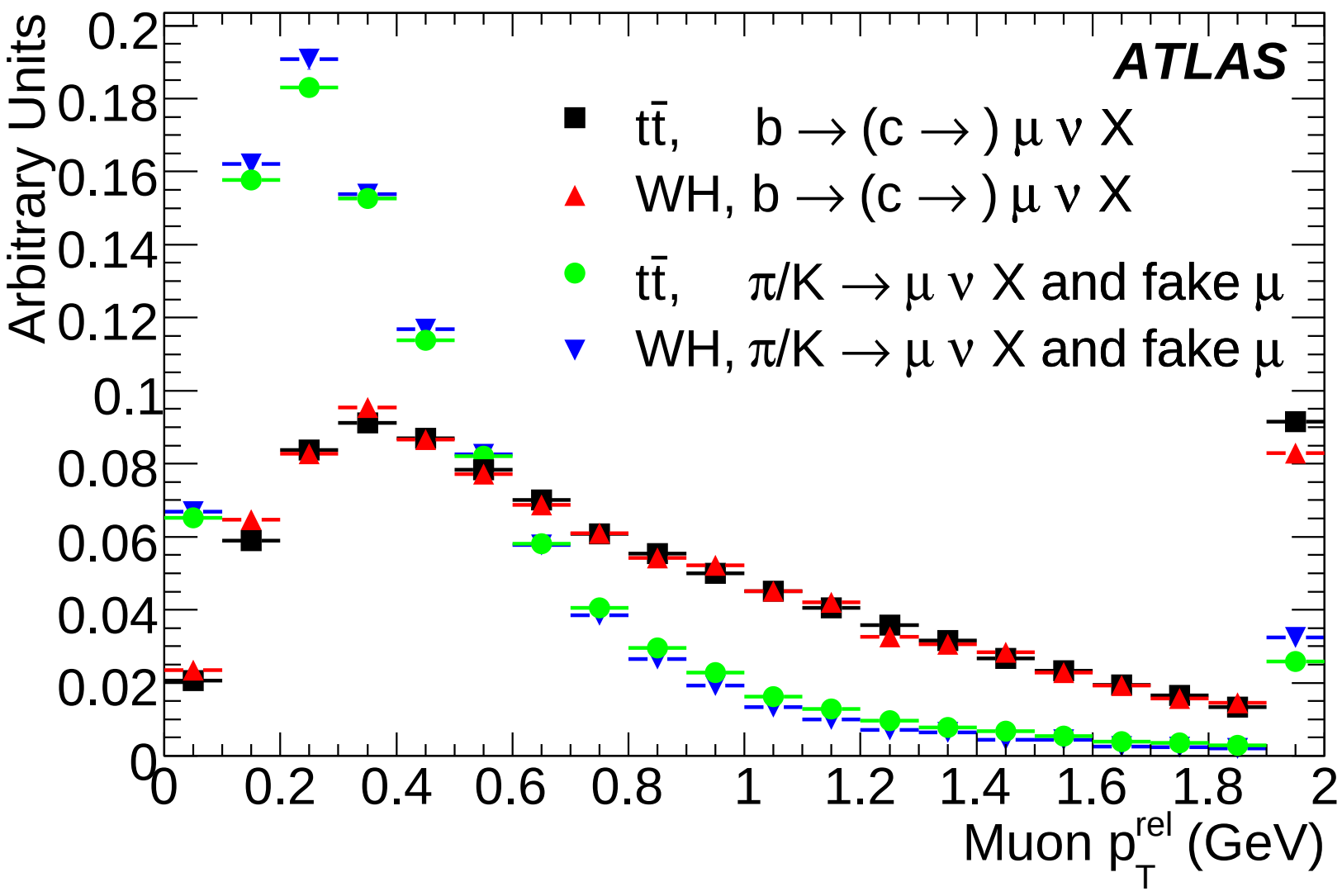

Figure 5: Distribution of the transverse momentum relative to the jet axis of muons from $b$ hadron decays and from background sources, in jets of $E_T > 15$GeV and $|\eta| < 2.5$ for $t\bar{t}$ events and $WH$ events. The last bin includes overflows.

### 3.2 Jet-muon association and basic selection

Muons are associated with the closest jet (in $\eta \times \phi$) in the event (only jets with $E_T > 15$ GeV and $|\eta| < 2.5$ are considered) and the jet-muon pair is required to satisfy $\Delta R$(jet-muon)$< 0.5$. Additionally, muons are required to satisfy the following requirements:

- Impact parameter with regard to the primary vertex $|d_0| < 4$ mm
- $p_T > 4$ GeV
- Matching between the muon spectrometer and inner detector tracks of $\chi^2/dof < 10$

### 3.3 Likelihood ratio

Muons from semi-leptonic $b$-decays are further separable because of their particular kinematics. As shown in Section 1, the muon transverse momentum relative to the jet ($p_T^{rel}$) is a good discriminant variable, and it is used in a likelihood ratio to discriminate between the $b$ and light jet hypotheses. No attempt was made to specifically discriminate between $b$- and $c$-jets. The $p_T^{rel}$ probability density functions were estimated on fully simulated $t\bar{t}$ and $WH \to \mu\nu b\bar{b}/WH \to \mu\nu u\bar{u}$ samples of several hundred thousand events and are shown in Fig. 6 and 7 for tagged and for combined muons separately. Here pile-up and cavern background are omitted. The normalized variable $p_T^{rel}/(p_T^{rel} + 0.5\,GeV)$ is used instead of $p_T^{rel}$ in order to facilitate the smoothing of the probability density functions. The likelihood is found to be less discriminating for tagged muons than for combined muons because muons at low momentum tend to be produced by $b \to c \to \mu X$ sequential decays, which also tend to yield a smaller $p_T^{rel}$ as shown in the previous section. This is visible in the probability density function as a second peak at a value of $\approx 0.5$ (compared to $\approx 0.7$ for direct $b \to \mu X$ decays). Each likelihood is normalized with the probability (estimated on the same fully simulated samples) that a reconstructed muon be associated with a jet of a given flavor. For each type of reconstructed muon (tagged or combined), the likelihood ratio $Q$ can be written as follows:

$$Q = \frac{\varepsilon_b^0 \times L(p_T^{rel}|b)}{\varepsilon_l^0 \times L(p_T^{rel}|l)}$$

where $\varepsilon_b^0$ ($\varepsilon_l^0$) is the fraction of $b$-jets (light jets) that contain a muon satisfying the basic selection and $L(p_T^{rel}|b)$ ($L(p_T^{rel}|l)$) is the probability density function for a $b$-jet (light jet). It was found that $\varepsilon_b^0/\varepsilon_l^0 \approx 6.6$ for tagged muons and $\varepsilon_b^0/\varepsilon_l^0 \approx 44.4$ for combined muons, reflecting the fact that combined muons are purer and contribute to a large fraction of the efficiency.

Several muons can be associated with a jet. In such a case, the muon with highest transverse momentum is considered. At this point, no attempt has been made to use the information given by the presence of an additional muon in the jet.

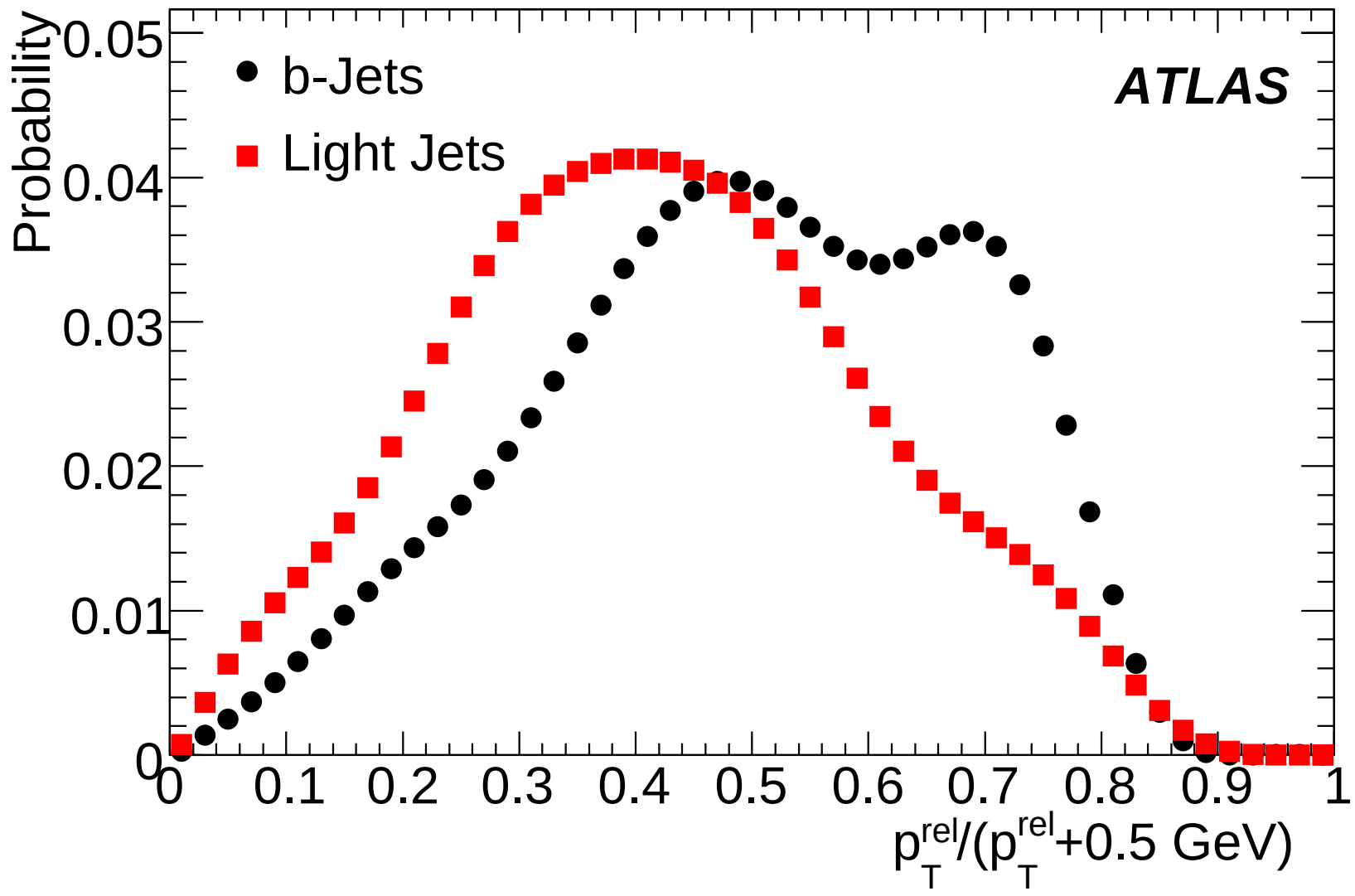

Figure 6: Probability density function used in the algorithm likelihood for the $p_T^{rel}$ variable, for the tagged muons.

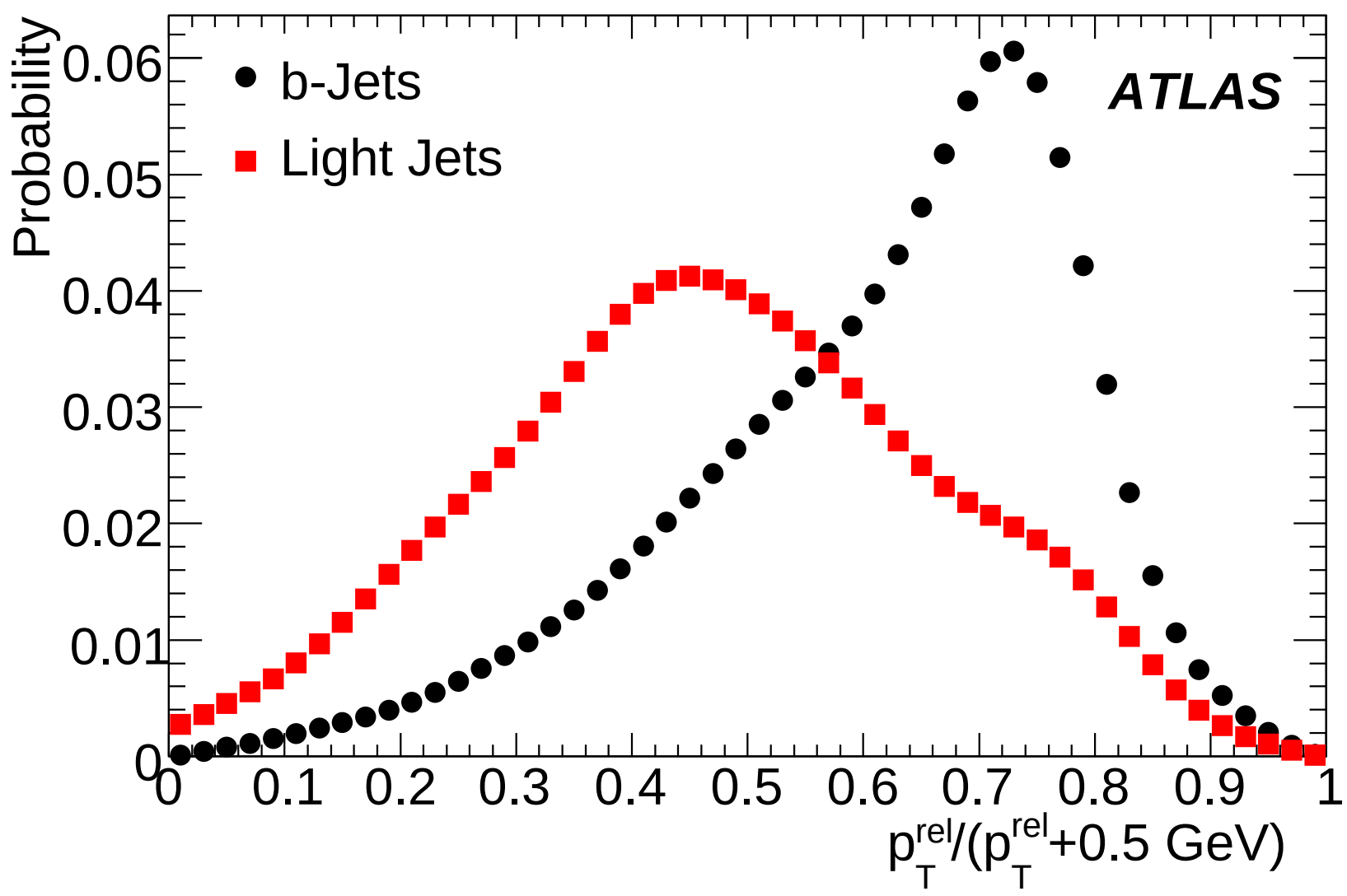

Figure 7: Probability density function used in the algorithm likelihood for the $p_T^{rel}$ variable, for the combined muons.

## 4 Performance

The algorithm performance was estimated on $t\bar{t}$ and $WH \rightarrow \mu\nu b\bar{b}/WH \rightarrow \mu\nu u\bar{u}$ Monte Carlo samples of several hundred thousand events. The jet flavor is determined at the generator level by matching reconstructed jets with quarks considered after final state radiation (for details, see Ref. [4]). Figures 8 and 9 show the $b$-tagging efficiency and the light jet tagging rate, respectively, with and without a cut on the likelihood ratio, as a function of jet $E_T$ and pseudorapidity in $t\bar{t}$ events. An average $b$-tagging efficiency of 10% is reached for a cut on the likelihood ratio of $lnQ > 3.05$, which corresponds to rejecting all tagged muons and selecting combined muons with $p_T^{rel} > 360$ MeV. No requirement on the $b$ decay is made, so that the $b$-tagging efficiency includes the semi-leptonic branching ratios as well as the jet-muon association efficiency, the detector acceptance, and the muon reconstruction efficiency and selection. Likewise, the light jet tagging rate includes both light hadron decay muons and fake muons. The $b$-tagging efficiency tends to decrease as the jet $E_T$ increases, mostly because of a drop in the tracking efficiency of the inner detector in highly collimated jets. Not surprisingly, the light jet tagging rate increases significantly with the jet $E_T$. After the basic selection stage, light hadron decays account for 8% (1.6%) of combined muons (tagged muons) present in tagged light jets in $t\bar{t}$ events.

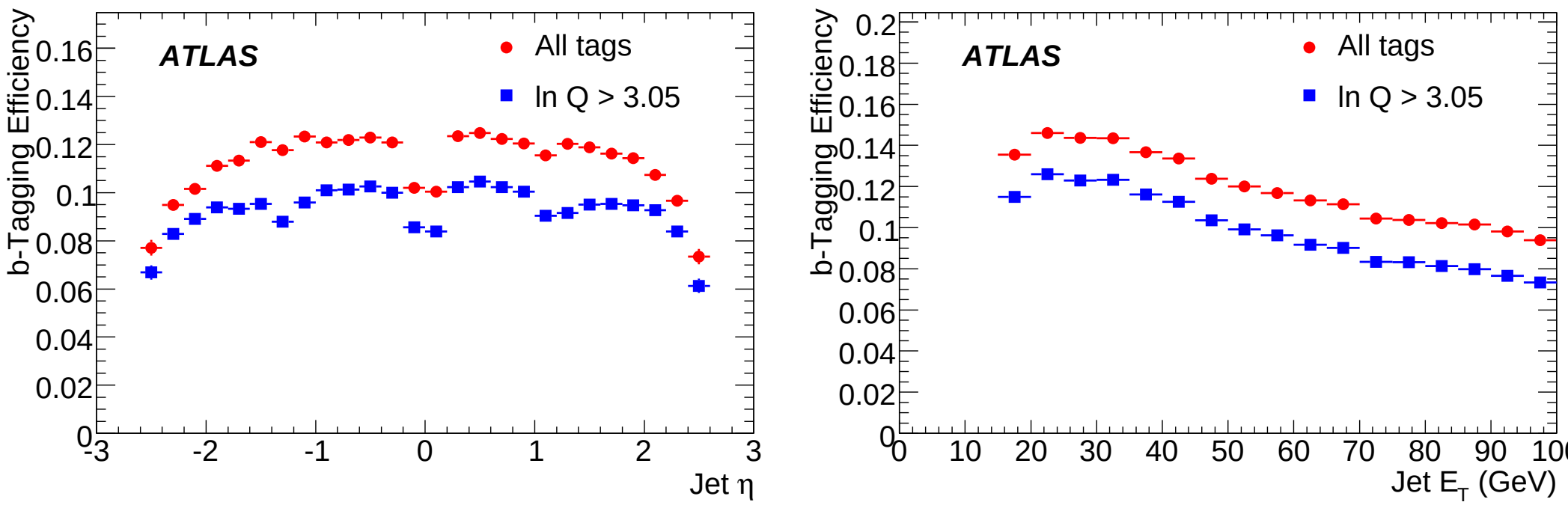

Figure 8: Probability for a $b$-jet to be tagged by the soft-muon tagger as a function of jet pseudorapidity (left) and transverse energy (right: the last bin includes overflows) without and with a requirement on the likelihood ratio corresponding to an average $b$-tagging efficiency of 10%.

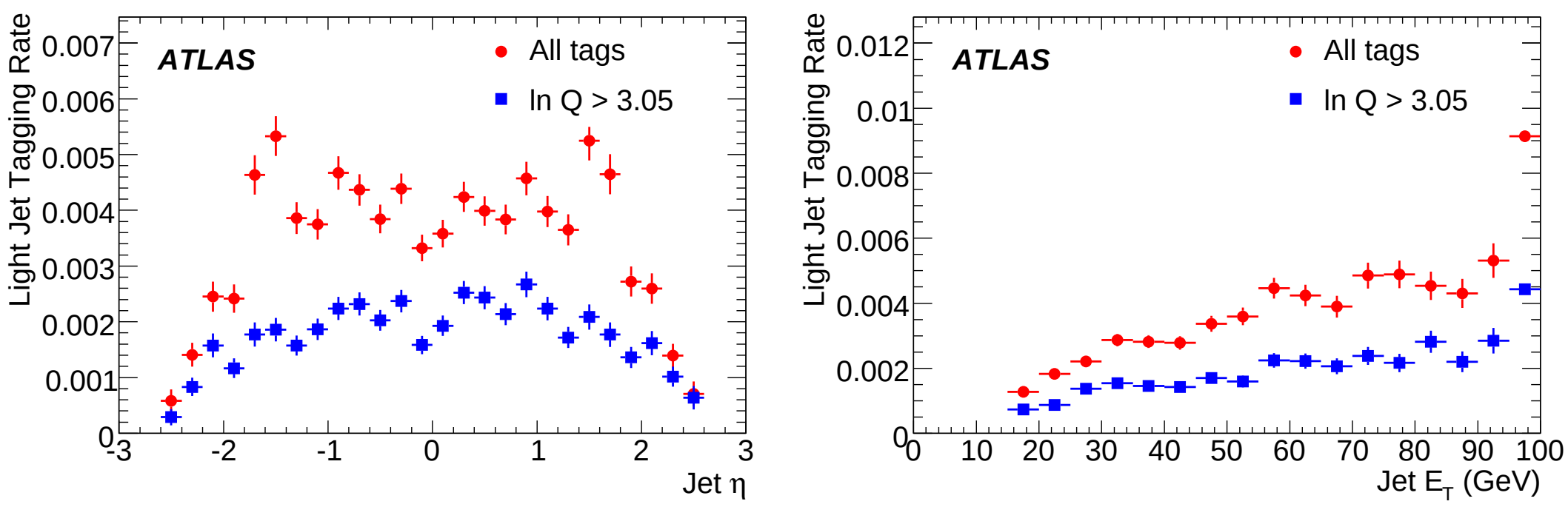

Figure 9: Probability for a light jet to be tagged by the soft-muon tagger as a function of jet pseudorapidity (left) and transverse energy (right, the last bin includes overflows) without and with a requirement on the likelihood ratio corresponding to an average $b$-tagging efficiency of 10%.

Figures 10 and 11 show the $b$-tagging efficiency vs light jet rejection (the inverse of the light jet tagging rate) for different cuts on the likelihood ratio. Figure 10 compares the performance in $t\bar{t}$ and $WH$ events. The rejection for a given efficiency is slightly better in $t\bar{t}$ events than in $WH$ events, due to the fact that jets in $t\bar{t}$ events are more central, while the tagger performance is degraded in the forward region of the detector. The soft-muon tagger reaches a rejection of about 300 in $WH$ events and 380 in $t\bar{t}$ events for an efficiency of 10%.

Figure 11 shows the effect of pile-up (the superposition of several events occuring within the same bunch crossing) and cavern background on the tagger performance. Here, only jets matched to the hard-scatter process quarks are considered in order to make a fair comparison of the algorithm performance with and without cavern background and pile-up. Indeed, soft jets from pile-up events tend to have a lower light jet tagging rate and would bias the comparison. For a luminosity of $2 \times 10^{33}\text{cm}^{-2}\cdot\text{s}^{-1}$, pile-up and cavern background decrease the rejection by about 15% (for a given efficiency of 10%), which is significant but not dramatic.

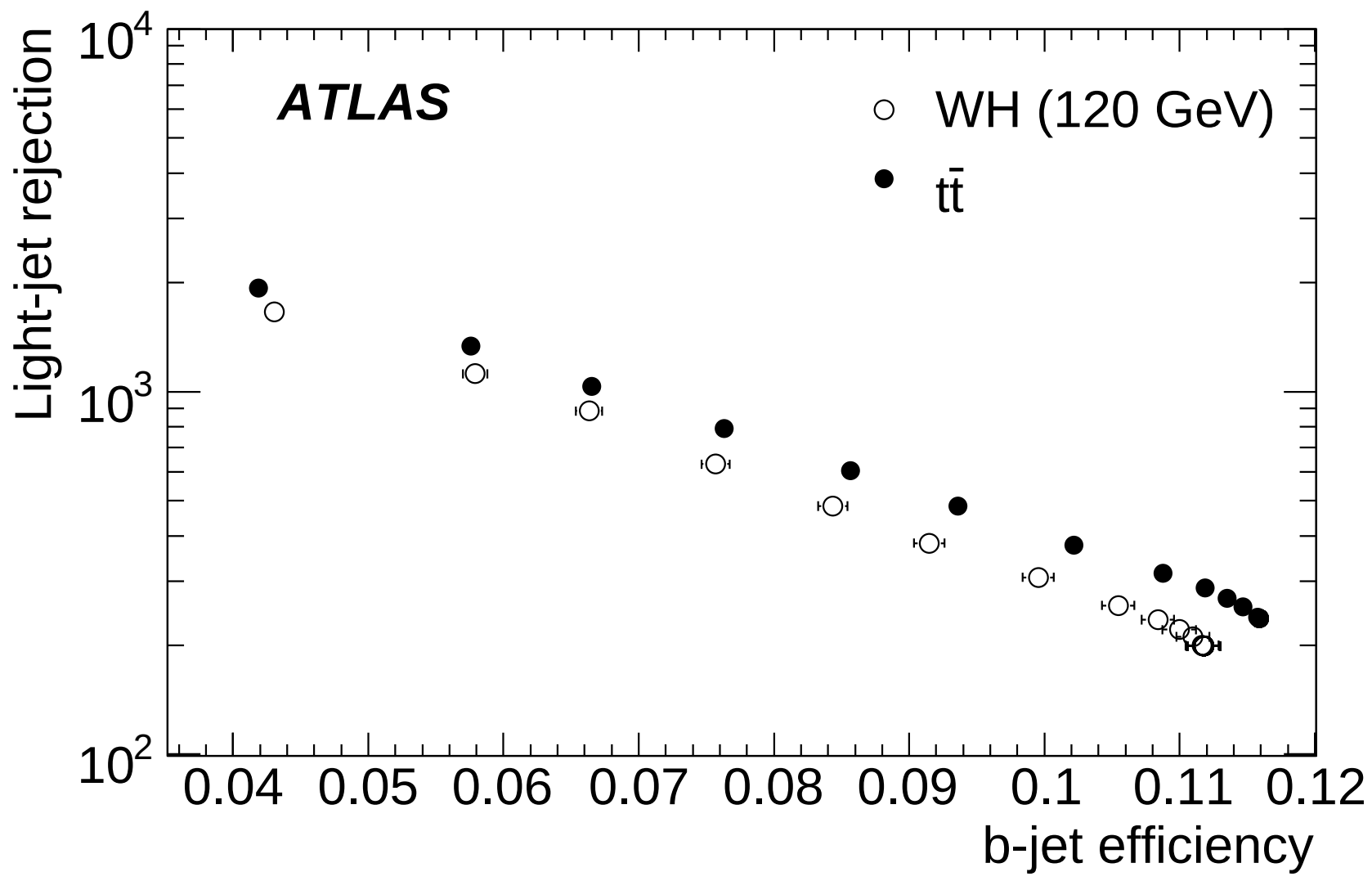

Figure 10: $b$-tagging efficiency vs light jet rejection estimated in $t\bar{t}$ and $WH$ (without pile-up/cavern background).

Performance will have to be evaluated in data, using techniques similar to those developed at the Tevatron [5]. The muon reconstruction efficiency can be measured using a tag-and-probe method applied to $J/\psi$ and $Z$ samples. The light jet tagging rate may be measured using jet events, although disentangling background from heavy-flavor muons is a complex issue. Jet samples may also be used to test the likelihood probability distribution functions against data. Those issues have not been studied in the context of soft muon tagging in ATLAS yet.

# 5 Conclusion

The soft-muon tagger developped for the ATLAS detector shows excellent performance across a large jet $E_T$ spectrum, with a light jet rejection of more than 300 for a $b$-tagging efficiency of 10%. The event pile-up and the cavern background appear to have a significant but not dramatic effect on the performance, with a decrease of about 15% on the light jet rejection at a luminosity of $2 \times 10^{33}\text{cm}^{-2}\cdot\text{s}^{-1}$.

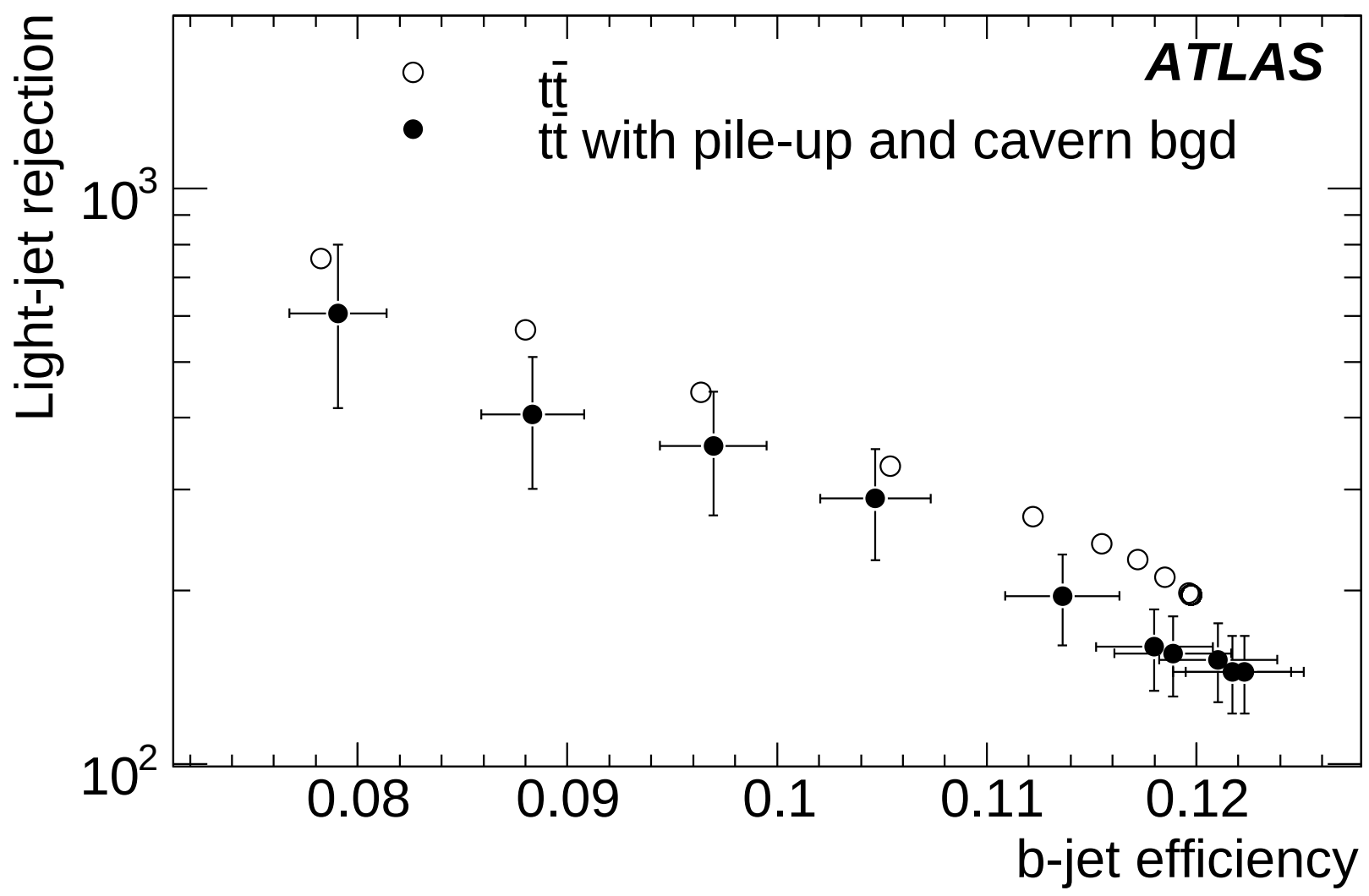

Figure 11: $b$-tagging efficiency vs light jet rejection estimated on a $t\bar{t}$ sample with and without pile-up/cavern background.

# Soft Electron $b$-Tagging

**Abstract**

The presence of an electron coming from the semi-leptonic decay of a $B$ hadron can be used to tag $b$-jets. This note describes a soft electron tagging algorithm and its performance when applied to ATLAS Monte Carlo simulation data. Jets are built from the energy reconstructed in the electromagnetic and hadronic calorimeters with a cone algorithm. Electrons, reconstructed by a track-seeded algorithm, are sought among charged tracks associated to the jets. The tagging of jets originating from $b$ quarks is based on kinematics and on the identification capabilities of the inner tracker and the electromagnetic calorimeter. The performance of the electron identification and $b$-tagging procedure are presented. A $b$-jet tagging efficiency of about 7% (including the inclusive $b \to e\nu X$ branching ratio of $\sim$ 20%) is achieved for a light jet rejection of better than 100.

## 1 Introduction

A variety of interesting physics processes at LHC, such as the $H \to b\bar{b}$ decay for an intermediate Higgs boson mass range, top physics or searches for new physics require efficient identification of $b$-quarks. The performance of $b$-tagging algorithms in ATLAS are studied in Ref. [1]. The semi-leptonic decays of heavy quarks provide a clean signature used to identify the flavour composition of jets. The semi-leptonic decay modes are $b \to \ell$, $c \to \ell$ and the cascade decay $b \to c \to \ell$. Electrons produced in $b$ decays (through direct and cascade decays) can be detected using the electromagnetic calorimeter and the inner detector. Since these electrons are non-isolated and with low transverse momentum, excellent electron/hadron separation capability is required. As a $b$-flavour tag, the lepton tag is not competitive with the lifetime tag because the branching ratio is low for such decays (about 20% for $B$-meson decays to leptons, including cascade decays, per lepton family) and is limited by the efficiency to reconstruct and identify electrons within jets. Still this method can be used with the vertex algorithms to provide a complement to the overall $b$-tagging performance and for cross-checks and calibration.

The note is organised as follows. Monte Carlo data samples used in this analysis are described in Section 2. The jet and track selections are described in Section 3. The electron identification algorithm is described in Section 4. The $b$-tagging algorithm and its performance are described in Section 5.

## 2 Monte Carlo samples

The samples used in this study contain events with electrons in jets from Higgs boson associated production $WH$, with $W \to \mu\nu$. We take the mass of the Higgs boson to be $m_H = 120$ GeV. The signal sample consists of $H \to b\bar{b}$ events and the background sample of $H \to u\bar{u}$ events. Another sample of $H \to b\bar{b}$ events is enriched in true electrons with a filter at the generation level. This filter is defined as follows: both $b$ quarks (before final state radiation, FSR) are required to have a transverse momentum $p_T^b > 15$ GeV and a pseudorapidity $|\eta_b| < 2.5$; at least one true electron with $p_T^e > 1$ GeV is required to be found in a cone $\Delta R < 0.4$ around each $b$-quark direction (before FSR). This sample will be used only in Section 4 to model signal electrons with enough statistics.

All samples have been generated using the PYTHIA 6.403 [2] Monte Carlo event generator. More details on the Monte Carlo generators used can be found in [3]. Data have been passed through a full detector simulation based on GEANT4. No pile-up has been included but its effect on performance, based on previous studies, is mentioned below where relevant.

The signal electrons come from direct ($b \to e$) and cascade ($b \to c \to e$) semi-leptonic decays of $B$-hadrons. Apart from the previous sources, there are also some other sources[1], mainly $b \to \tau \to e$ and $b \to (J/\psi, \psi') \to e^+e^-$. The background electrons arise from $\pi^0$ Dalitz decays, $\gamma$-conversions occurring in the inner detector and decays of light hadrons. The fraction of electrons from photon conversions becomes substantial at large pseudo-rapidities and large transverse momenta.

Distributions of true transverse momentum, $p_T$, and pseudo-rapidity, $\eta$, for electrons and pions are shown in Fig. 1 for the samples considered. Table 1 shows the mean values of the true transverse momentum distributions for signal and background electrons and pions in the two samples.

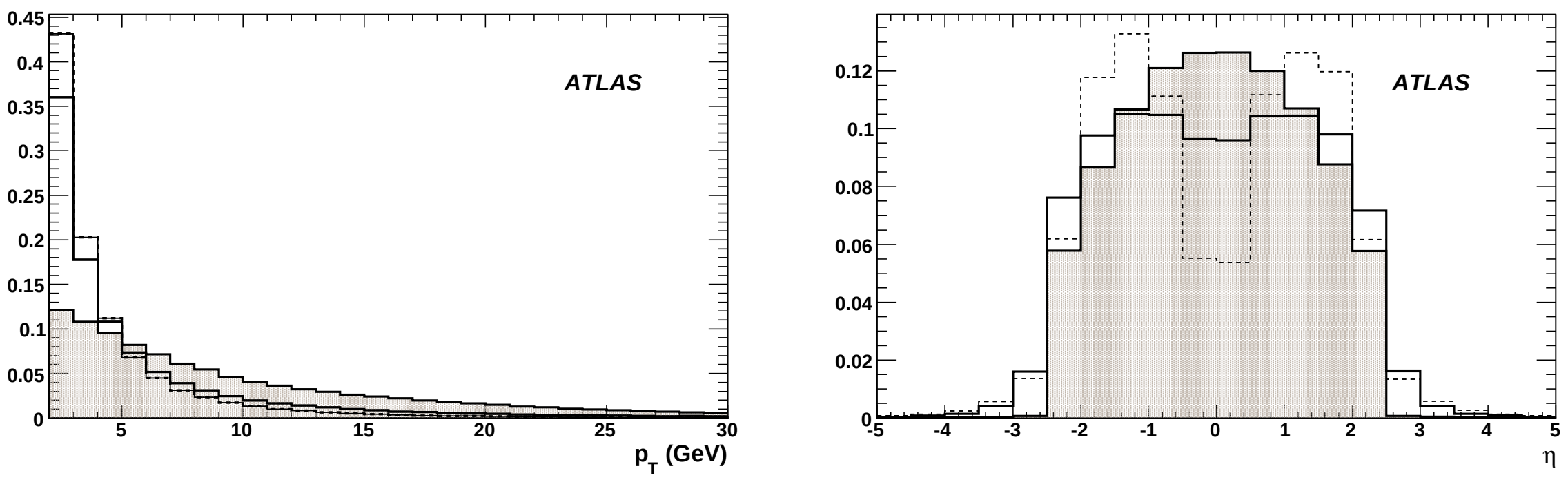

Figure 1: Normalized distributions of true transverse momentum $p_T$ (left) and pseudo-rapidity $\eta$ (right) are shown for signal electrons (hatched histograms), electrons from conversions (dotted line histograms) and pions (plain histograms).

Table 1: The mean true $p_T$ (in GeV) for electrons and pions in various data samples.

| sample | electrons | | | | pions |
|---|---|---|---|---|---|
| | $B$ hadrons | $D$ hadrons | $\gamma$-conversions and $\pi^0$ Dalitz | other sources | |
| $H \to bb$ | 10.8 | 11.9 | 4.9 | 7.3 | 5.9 |
| $H \to u\bar{u}$ | - | - | 5.2 | 4.2 | 7.1 |

# 3 Jet and track selection

The reconstruction of the various objects needed for $b$-tagging and the estimation of its performance are summarized in this section. The reconstruction of soft electrons will be detailed in the next section.

We follow the default jet reconstruction and labelling procedure described in Ref. [1]. Jets are reconstructed in the calorimeters using a standard cone algorithm with a size of $\Delta R = 0.4$. Jets with $p_T > 15$ GeV and $|\eta| < 2.5$ are considered for $b$-tagging. To assess quantitatively the $b$-tagging performance, Monte-Carlo information is used to determine the type of parton (a quark-based labelling) from which a jet originates. Jets are labelled as $b$-jets if a $b$ quark with $p_T > 5$ GeV (after FSR) is found in a cone $\Delta R = 0.3$ around the jet direction. The labelling for $c$-jets (and $\tau$-jets) is done in the same way. By light jets we denote all other reconstructed jets.

[1] The corresponding branching ratios [4] are $Br(b \to \ell^-) = (10.71 \pm 0.22)\%$, $Br(b \to c \to \ell^+) = (8.01 \pm 0.18)\%$, $Br(b \to \bar{c} \to \ell^-) = (1.62^{+0.44}_{-0.36})\%$, $Br(b \to \tau \to e) = (0.419 \pm 0.055)\%$ and $Br(b \to (J/\psi, \psi') \to e^+e^-) = (0.072 \pm 0.006)\%$.

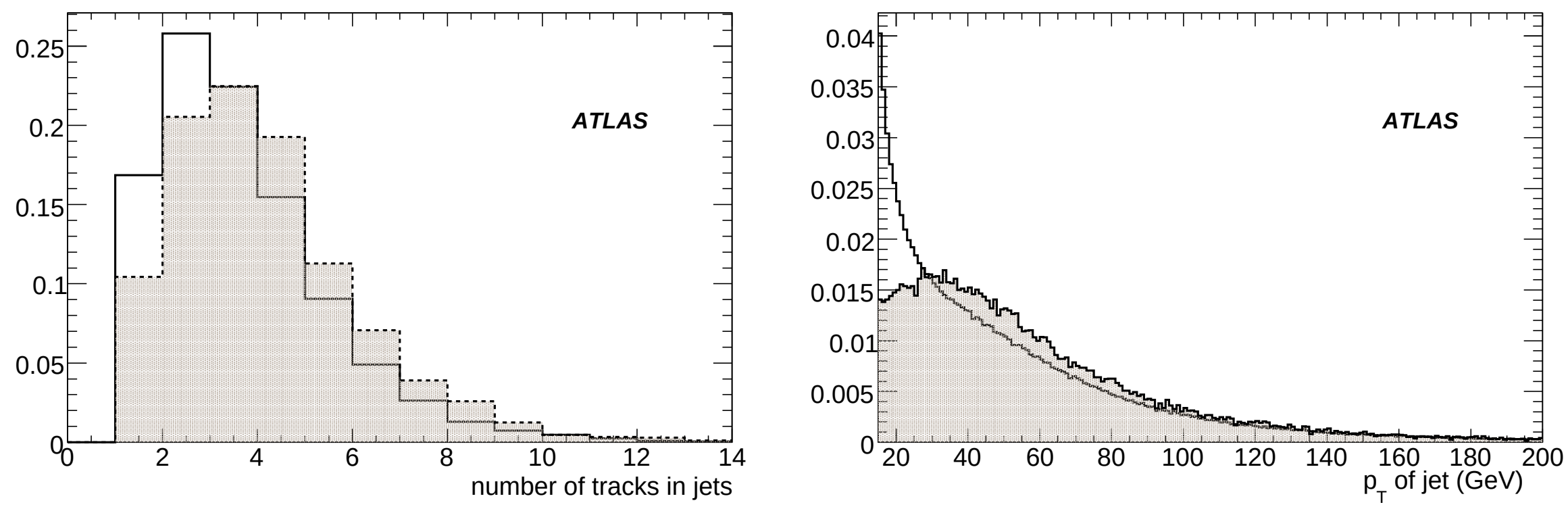

Figure 2: Track multiplicity in jets (left) and jet transverse momentum (right) for $b$ jets (hatched histograms) and light jets (solid line). Only jets having at least one *good quality track* with $p_T > 2$ GeV are considered.

For the soft electron $b$-tagging algorithm, a track-based electron reconstruction algorithm is used and will be described in the next section. In order to be able to identify track candidates as electrons, strict track selection criteria are applied and called hereafter *good track quality cuts*. To reduce the number of fake candidates in jets, only tracks with $p_T > 2$ GeV are considered. The inner detector coverage extends to pseudo-rapidity values of $\pm 2.5$, except for the transition radiation tracker (TRT) which extends up to $\pm 2$. This sub-detector is crucial in the identification procedure, so the track selection requires $|\eta| < 2$. Tracks are required to have at least nine precision hits (pixels and SCT), at least two hits in the pixel detector, at least one hit in the vertexing layer (the so-called $b$-layer) and an unsigned transverse impact parameter at the perigee smaller than 1 mm. The TRT may record either of two types of hits, 'low-energy' hits which are used for track reconstruction and 'high-energy' hits which are used for electron identification. Selection criteria thus require at least 20 low-energy hits and at least one high-energy hit in the TRT detector along the track. After this selection, only about 50% of initial tracks remain. About 50% of actual $b$-jets and 60% of light jets, in the WH sample, have no *good quality track* associated. Figure 2 shows the multiplicity of *good quality track* inside jets and the jet transverse momenta. For jets having at least one *good quality track* the mean multiplicity is 3.8 (3.3) for $b$-jets (light jets). Light jets, which include ISR, FSR and underlying events, have a softer transverse momentum than $b$ jets, which are originated from the $H$ decay, with a mean $p_T$ of 51 GeV against 57 GeV, but with a large spread, an RMS of $\sim$35 GeV.

The fraction of jets containing electrons with a *good quality track* is given in Table 2 for signal and background samples.

Table 2: Fraction of jets (in %) with an electron at the generator level that also have a *good quality track*, for the signal and background samples.

| Sample | $\gamma$-conversions and $\pi^0$ Dalitz | $B$ hadrons | $D$ hadrons | Other sources |
|---|---|---|---|---|
| $H \to b\bar{b}$ | 1.6 | 3.4 | 2.2 | 0.3 |
| $H \to u\bar{u}$ | 1.3 | $< 10^{-4}$ | $< 10^{-4}$ | 0.1 |

# 4 Electron reconstruction and identification

The standard electron reconstruction procedure [5] is based on calorimeter clusters, with a subsequent association to tracks. While this method is efficient for high-energy isolated electrons, such as those arising from $W$ or $Z$ decays, it is not effective for electrons inside hadronic jets, such as those from semi-leptonic decays. Indeed hadron and electron showers tend to overlap in collimated jets so that electron cluster characteristics are obscured. An alternative procedure takes full advantage of the tracking capabilities of the inner detector as well as the granularity of the electromagnetic calorimeter. The method relies on the extrapolation of reconstructed charged particle trajectories into the electromagnetic calorimeter. The most common background processes for producing electron-like showers in the calorimeters were described in Section 2. Because the development of showers is different for electrons and hadrons, the electron identification algorithm incorporates variables that describe the shower shapes, quality of the match between the track and its corresponding cluster, and the fraction of high-energy hits in the transition radiation tracker. This algorithm is used also for the reconstruction of low $p_T$ electrons in J/$\psi$ events [6].

## 4.1 Electron reconstruction

All the tracks that pass the *good track quality cuts*, described in the previous section, are extrapolated to the second (also known as middle) sampling layer of the electromagnetic calorimeter. Around this position a cluster of size $\Delta\eta \times \Delta\phi = 0.125 \times 0.125$ ($5 \times 5$ cells in this sampling layer) is built. The cell with the maximum energy is sought within a small $\eta$ and $\phi$ window ($0.075 \times 0.075$) around the extrapolation point. Shower shapes are estimated with respect to this cell. The contribution of neighbouring hadronic showers is therefore reduced.

A set of *preselection* criteria are applied to decrease the number of fake candidates per jet:

- The ratio of energy reconstructed in the core of the shower in the first sampling layer to the total shower energy reconstructed in the core of the cluster, must fulfil $E_1(\text{core})/E(\text{core}) > 0.03$. $E_1(\text{core})$ is the energy reconstructed in a window of size $\Delta\eta \times \Delta\phi = 0.009375 \times 0.1$ in the first sampling layer ($3 \times 1$ cells). $E(\text{core})$, the core energy in the cluster, is computed in the following windows: $0.075 \times 0.3$ ($3 \times 3$ cells) in the presampler, $0.046875 \times 0.2$ ($15 \times 2$ cells) in the strips, $0.125 \times 0.125$ ($5 \times 5$ cells) in the middle, $0.15 \times 0.125$ ($3 \times 5$ cells) in the back. This ratio tends to be larger for electrons than for hadrons due to the different development of hadronic and electromagnetic showers. This quantity is discussed in detail in the following section.

- Similarly, the fraction of energy reconstructed in the core of the shower in the third sampling layer to the energy reconstructed in the core of the cluster, $E_3(\text{core})/E(\text{core})$, must be smaller than 0.5. $E_3(\text{core})$ is the energy reconstructed in a window of size $\Delta\eta \times \Delta\phi = 0.15 \times 0.075$ in the third sampling layer ($3 \times 3$ cells). This ratio is larger for hadrons than it is for low-energy electrons which almost never reach this layer and is discussed in the following section.

- The ratio of the energy $E$ reconstructed in the electromagnetic calorimeter over the momentum $p$ of the track reconstructed in the inner detector is required to fulfil $E/p > 0.7$.

These *preselection* criteria cut out less than 5% of signal electrons. Finally, candidates which are also reconstructed as originating from a conversion are vetoed, corresponding to a loss of 9% of all signal and background tracks.

Position and energy corrections are applied in the precise reconstruction of the electromagnetic cluster and are described in [7]. These corrections have been tuned for high-energy clusters and are not optimal for low-energy electrons. In Fig. 3 the ratio between the reconstructed and the true electron energies is shown as a function of $|\eta|$ for signal electrons from the $H \to b\bar{b}$ sample. It can be seen that the corrections over-estimate the electron energy. Moreover, as electrons are embedded in jets, there is a contamination from other hadrons in the energy determination which is not visible with the isolated electrons from a J/$\psi$ sample [6]. Work is on going to improve the energy reconstruction at low energy.

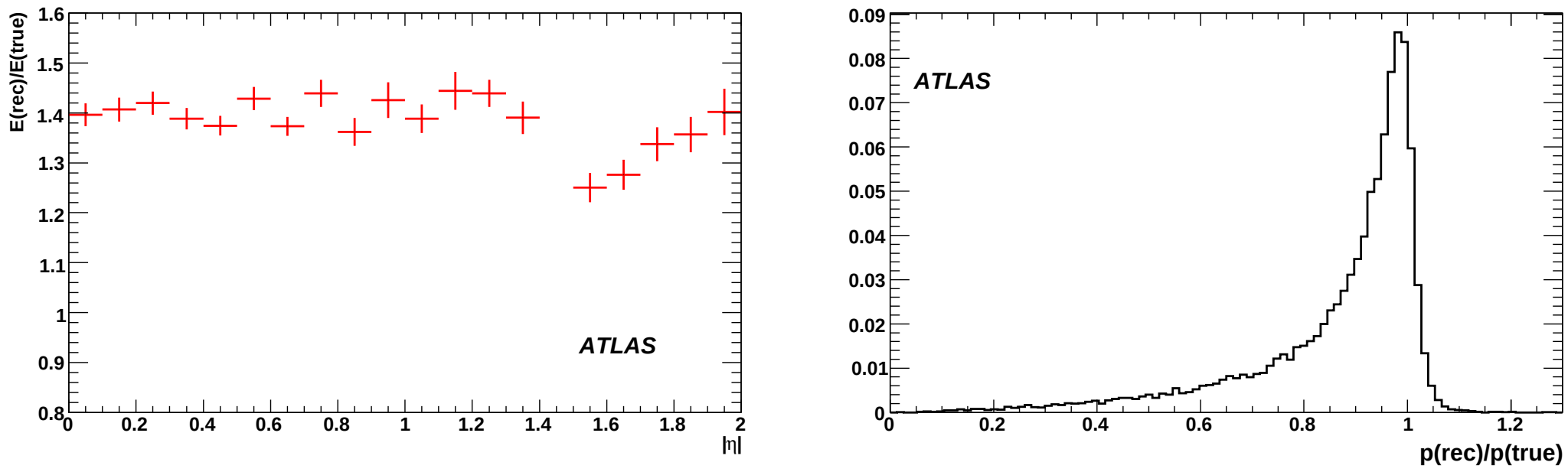

Figure 3: Ratio between reconstructed and true energy as a function of the electron pseudo-rapidity $|\eta|$ (left) and ratio of the reconstructed to true momentum for electrons (right).

By default the four-momentum of an electron is defined as the energy reconstructed in the calorimeter, whereas the direction is taken from the associated track. Since in this note the main physics processes lead to electron transverse momenta of less than 15 GeV, the momentum measured in the tracker is used instead of the energy unless stated otherwise. The reconstructed momentum of the electron is shown in Fig. 3. The track reconstructs the momentum of the electron close to its true value for most candidates. The small downward shift of the peak from unity and the tail towards lower values are due to photon bremsstrahlung. Future developments in ATLAS will ensure an optimal combination of calorimeter and tracker measurements in the energy definition.

## 4.2 Electron identification variables

The hadronic calorimeter has a granularity of $\Delta\eta \times \Delta\phi = 0.1 \times 0.1$ which is too large to disentangle the energy deposit of an electron inside a jet. In the electromagnetic calorimeter, electrons are narrow objects while jets tend to have a broader profile, allowing a discrimination. For this purpose, we use the finer granularity ($\Delta\eta \times \Delta\phi = 0.05 \times 0.025$) of the third sampling layer of the electromagnetic calorimeter. The ratio $E_3(\mathrm{core})/E(\mathrm{core})$, defined above, is larger for pions than for electrons and is clearly discriminating as can be seen in Fig. 4.

Electromagnetic showers deposit most of their energy in the second sampling layer of the electromagnetic calorimeter. The following variables are used:

- The lateral shower shape $R_\eta$ (cf. Fig. 5 on left) is given by the ratio of the energy reconstructed in a window of size $0.075 \times 0.175$ ($3 \times 7$ cells of the middle sampling layer) to the energy reconstructed in $0.175 \times 0.175$ ($7 \times 7$ cells).
- The lateral width $\omega_{\eta 2} = \sqrt{\frac{\sum E_c \times \eta^2}{\sum E_c} - \left(\frac{\sum E_c \times \eta}{\sum E_c}\right)^2}$ (cf. Fig. 5 on right) is calculated in a window of size $0.075 \times 0.125$ ($3 \times 5$ cells of the middle sampling layer) using the energy weighted sum over all cells, which depends on the particle impact point inside the cell.

The first layer, with its very fine granularity in pseudo-rapidity, can be used to detect sub-structures within a shower and thus isolated $\pi^0$'s and $\gamma$'s can be discriminated against efficiently. For all variables computed in the first sampling layer, two cells in $\phi$ are summed.

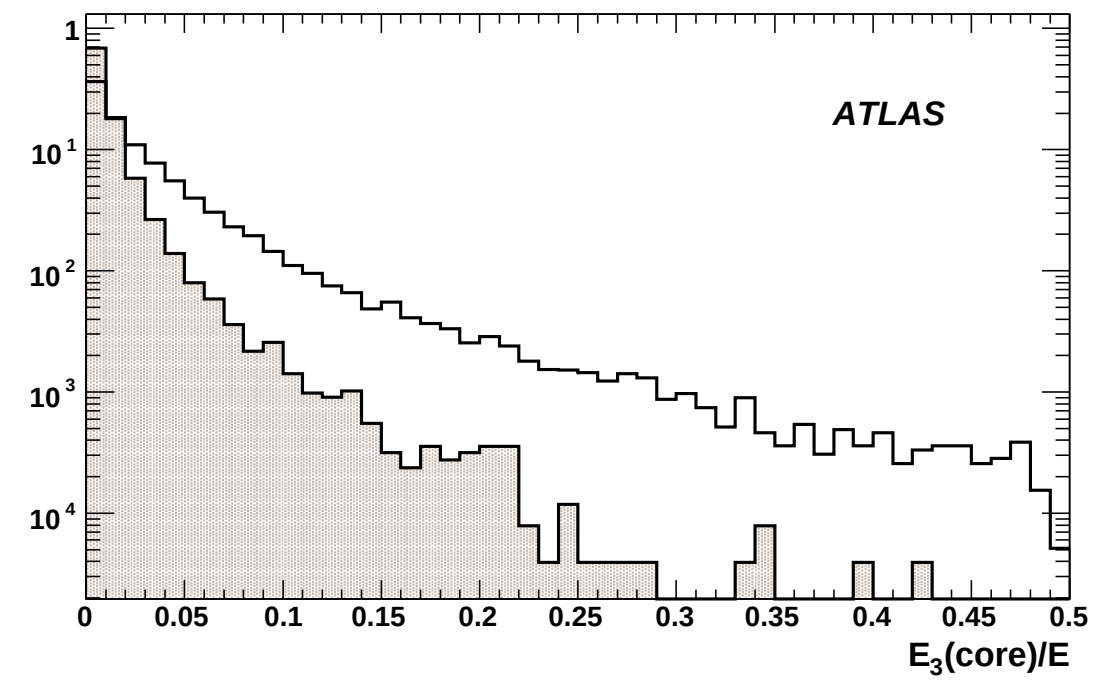

Figure 4: Ratio $E_3(\text{core})/E(\text{core})$ (see text) for electrons in the $H \to b\bar{b}$ sample (hatched histogram) and for charged pions in the $H \to u\bar{u}$ sample (solid line). The distributions are normalized to unit area.

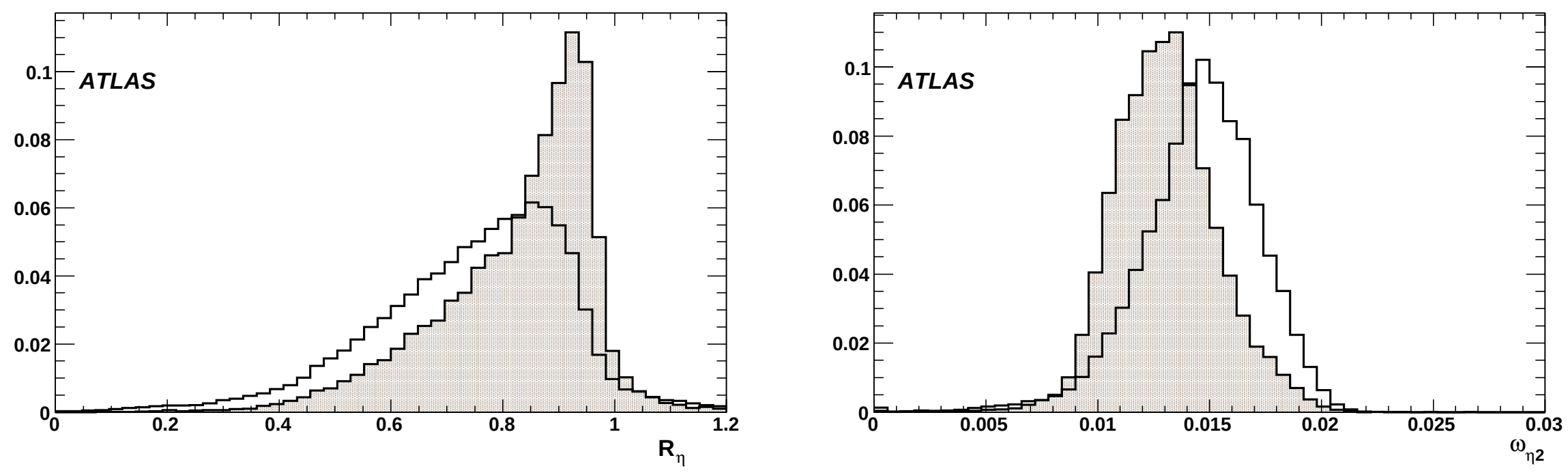

Figure 5: Lateral shower shape $R_\eta$ (left) and lateral width $\omega_{\eta 2}$ (right) in the second layer of the electromagnetic calorimeter (see text for details). The distributions are shown for electrons in the $H \to b\bar{b}$ sample (hatched histograms) and for charged pions in the $H \to u\bar{u}$ sample (solid lines). The distributions are normalized to unit area.

The ratio $E_1(\text{core})/E(\text{core})$, defined above, is larger for electrons than for hadrons as can be seen in Fig 6.

The lateral shower shape in the strips is now exploited.

- The total shower width in strips $\omega_{stot}$ is determined in a window $\Delta\eta = 0.0625$ (corresponding to 20 strips in the barrel for instance). It is calculated from: $\omega_{stot} = \sqrt{\sum E_i \times (i - i_{max})^2 / \sum E_i}$, where $i$ is the strip number and $i_{max}$ the strip number of the first local maximum. This width is shown for electrons and pions in Fig. 7.

- The shower width using three strips around the one with the maximal energy deposit is shown in Fig. 7 on the right. It is given by the following formula: $\omega_{s3} = \sqrt{\sum E_i \times (i - i_{max})^2 / \sum E_i}$, where $i$ is the number of the strip and $i_{max}$ the strip number of the most energetic one.

The pion and jet rejection can be significantly improved by ensuring consistency between the electromagnetic calorimeter and the inner detector information.

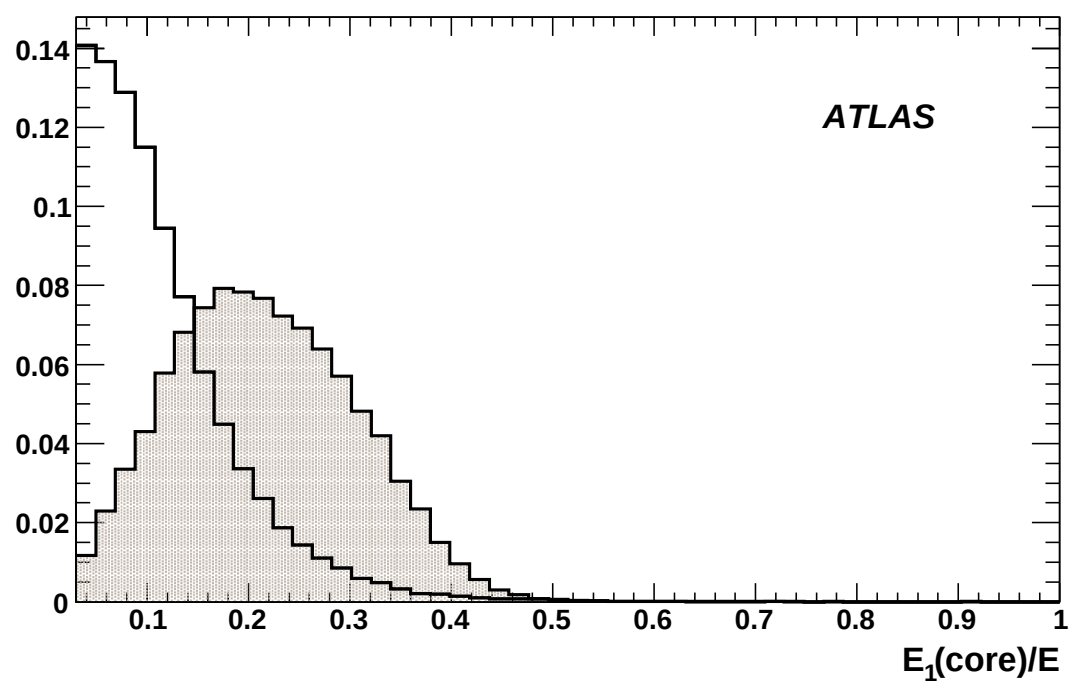

Figure 6: Ratio $E_1(\text{core})/E(\text{core})$ (see text) for electrons in the $H \to b\bar{b}$ sample (hatched histogram) and for pions in the $H \to u\bar{u}$ sample (solid line). The distributions are normalized to unit area.

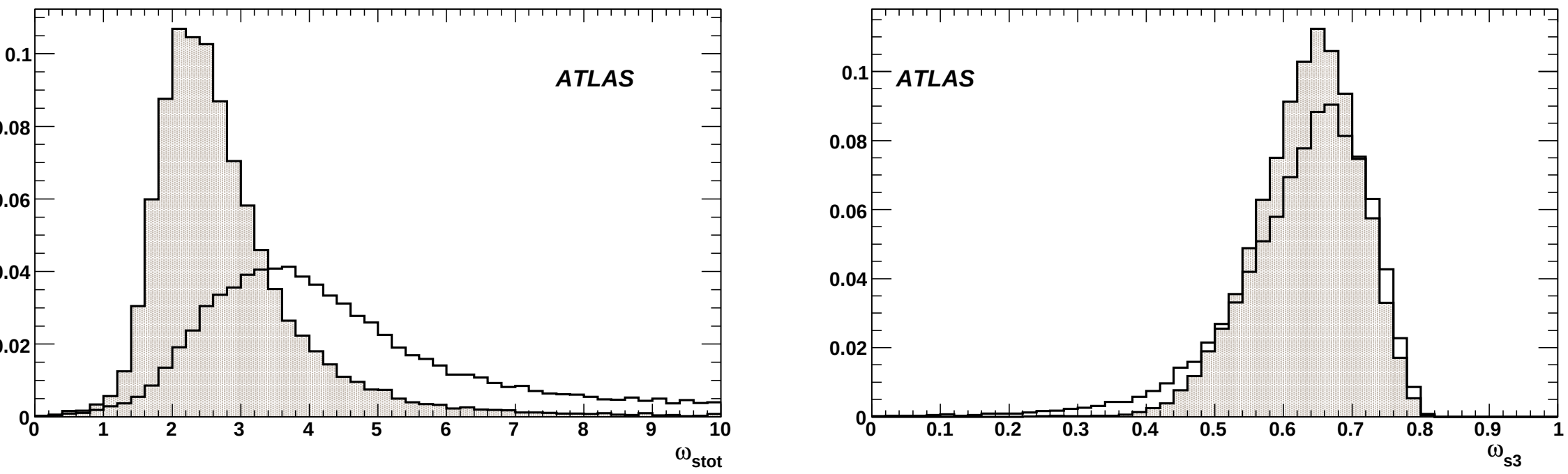

Figure 7: Total shower width $\omega_{stot}$ (left) and shower width in three strips $\omega_{s3}$ (right) in the first layer of the electromagnetic calorimeter. The distributions are shown for electrons in the $H \to b\bar{b}$ sample (hatched histograms) and pions in the $H \to u\bar{u}$ sample (solid lines). The distributions are normalized to unit area.

First, the angular matching between the track and the electromagnetic cluster is checked (cf. Fig. 8): $|\Delta\eta| = \sum_{i=i_m-7}^{i=i_m+7} E_i \times (i - i_m) / \sum_{i=i_m-7}^{i=i_m+7} E_i$, which gives the difference between the track and the shower positions measured in units of distance between the strips, where $i_m$ is the impact cell for the track reconstructed in the inner detector and $E_i$ is the energy reconstructed in the $i$-th cell in the $\eta$ direction, at constant $\phi$ given by the track parameters.

Subsequently, the energy $E$ measured in the electromagnetic calorimeter is compared to the momentum $p$ measured in the inner detector (cf. Fig. 9 on the left). In the case of an electron, the momentum should match the energy. The large tails at high values of the ratio in the signal distribution are due to soft bremsstrahlung.

A further reduction of the charged hadron contamination is obtained by rejecting tracks having a low fraction of high-energy hits in the TRT. Figure 9 (right) shows the ratio $N_{\text{HTR}}/N_{\text{straw}}$ between the number of high threshold hits $N_{\text{HTR}}$ and the total number of TRT hits $N_{\text{straw}}$.

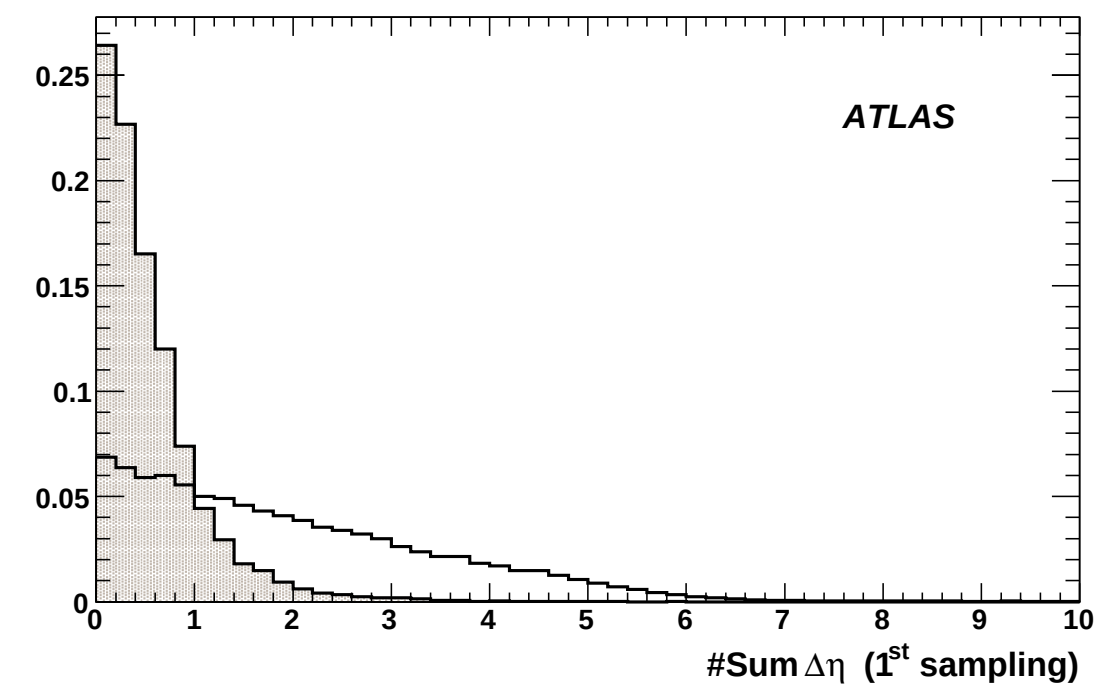

Figure 8: Angular matching between charged tracks extrapolated to the electromagnetic calorimeter and electromagnetic clusters in pseudo-rapidity ($|\Delta\eta|$). The distributions are shown for electrons in the $H \to b\bar{b}$ sample (hatched histogram) and for pions in the $H \to u\bar{u}$ sample (solid line). The distributions are normalized to unit area.

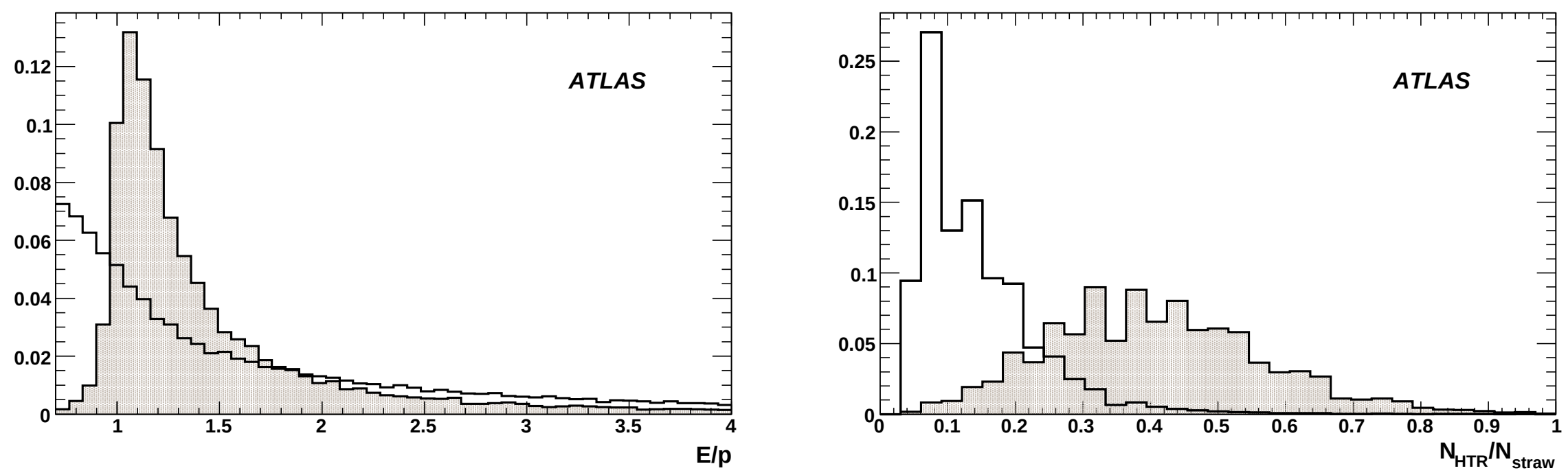

Figure 9: Ratio $E/p$ between the energy of the electromagnetic clusters and the momentum of reconstructed charged tracks (left) and fraction $N_{\mathrm{HTR}}/N_{\mathrm{straw}}$ of high-energy hits in the TRT (right). The distributions are shown for electrons in the $H \to b\bar{b}$ sample (hatched histograms) and for pions in the $H \to u\bar{u}$ sample (solid lines). The distributions are normalized to unit area.

### 4.3 Rejection of electrons from $\gamma$-conversions and Dalitz decays

A significant source of low-$p_T$ electron tracks in jets are photon conversions and Dalitz decays. Such tracks might be identified by the algorithm as signal electron tracks. The *good quality track* cuts help in part to suppress that type of background. In case of $\gamma$-conversions and Dalitz decays, $e^+e^-$ pairs with small invariant mass are observed in the detector. To find them the conversion finding algorithms can be used. Ref. [8] details the last developments on the reconstruction of such electrons. Unfortunately, these were not available for the electron reconstruction algorithms at the time of this study. As detailed in Section 4.1, about 9% of signal electrons are mis-identified as conversions. Based on previous studies [9], for an electron efficiency $\varepsilon_e = 80\%$, the rejection factor of electrons from conversions and Dalitz decays is $\sim 3$ with the conversion package and $\sim 2$ without.

# 5 $b$-tagging algorithm

The variables described in the previous section have been shown to be efficient in distinguishing electron tracks from non-electron tracks or electron tracks from $\gamma$-conversions and Dalitz decays.

In order to construct a $b$-tagging algorithm we combine these variables with additional variables to exploit the specific features of $b$-jets. First we take advantage of the fact that the electron is coming from a $b$ quark and thus can have a significant transverse impact parameter $d_0$. Figure 10 shows the corresponding distributions. In addition, because of the $B$ hadron's high mass, electrons from direct bottom decays tend to be more boosted in a plane transverse to the jet axis, *i.e.* have a larger $p_T^{rel}$ than the backgrounds, as shown in Fig. 11. $p_T^{rel}$ is defined as the electron momentum in the plane orthogonal to the jet axis.

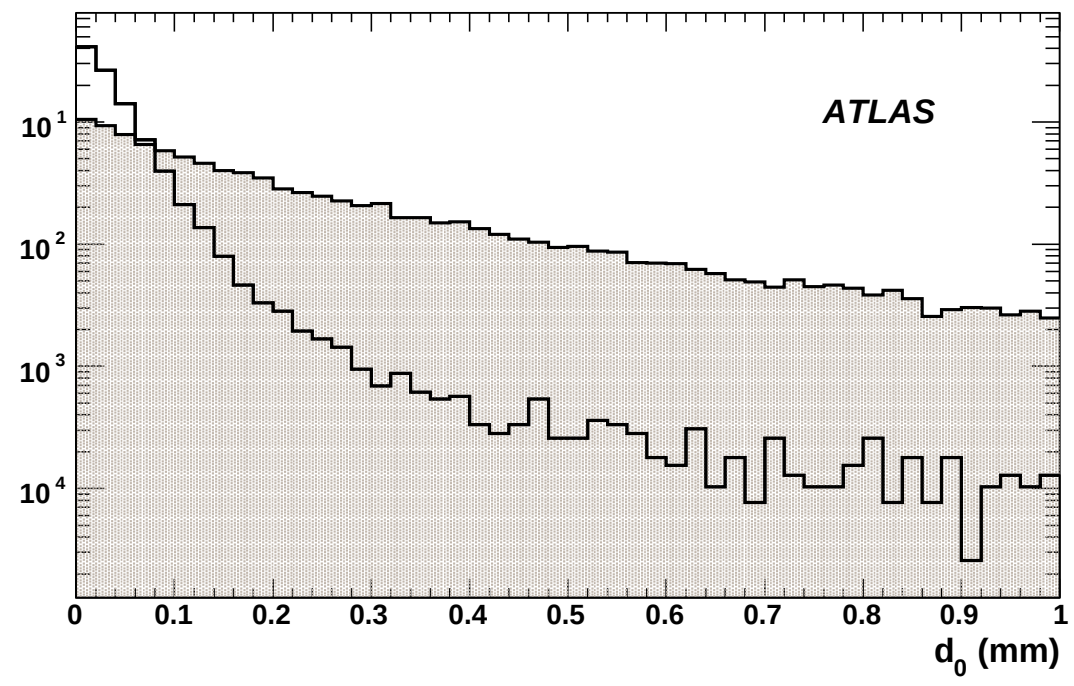

Figure 10: Transverse impact parameter for electrons in the $H \to b\bar{b}$ sample (hatched histogram) and for pions in the $H \to u\bar{u}$ sample (solid line). The distributions are normalized to unit area.

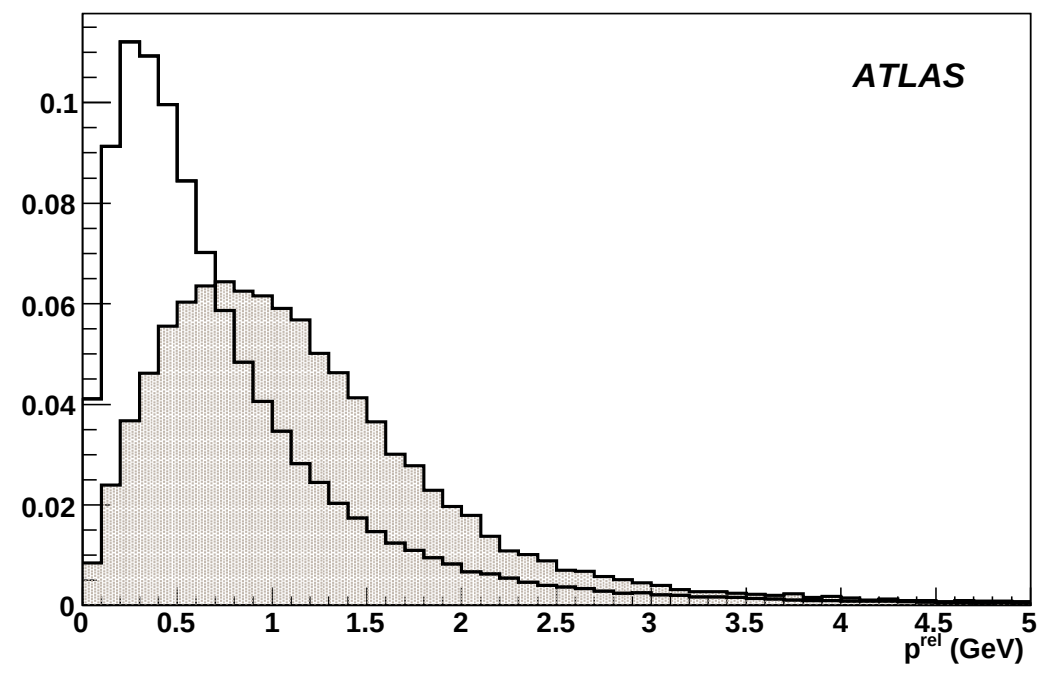

Figure 11: Distribution of the track transverse momentum $p_T^{rel}$ relative to the jet axis for signal electron tracks in $b$ jets (hatched histogram) and for pion tracks in light jets (solid line).

The combination of all variables is performed in two steps; first for variables independent of the jet and second including a jet dependent variable as explained in Section 5.1. Finally the performance of the algorithm will be detailed in Section 5.2.

### 5.1 b-tagging procedure

A likelihood-ratio method is used to combine the variables described in the previous section along with the electron track impact parameter $d_0$. This likelihood ratio uses information only on the characteristics of the electron and is independent of the parameters of the jet. The discriminating variables are compared to pre-defined normalized distributions (probability density functions) for both the electron and the pion hypotheses. We use the signal sample of $H \rightarrow b\bar{b}$ filtered for electrons and part of the $H \rightarrow u\bar{u}$ background samples. The variables show a significant dependence on pseudo-rapidity and a less pronounced one on transverse momentum. In $\eta$ the changes correspond to varying granularities, lead thickness and material in front of the electromagnetic calorimeter. The separation between the distributions obtained for electrons and pions can vary also with $\eta$. Therefore probability density functions are defined in five $|\eta|$ bins: (0-0.8), (0.8-1.37), (1.37-1.52), (1.52-1.8), (1.8-2).

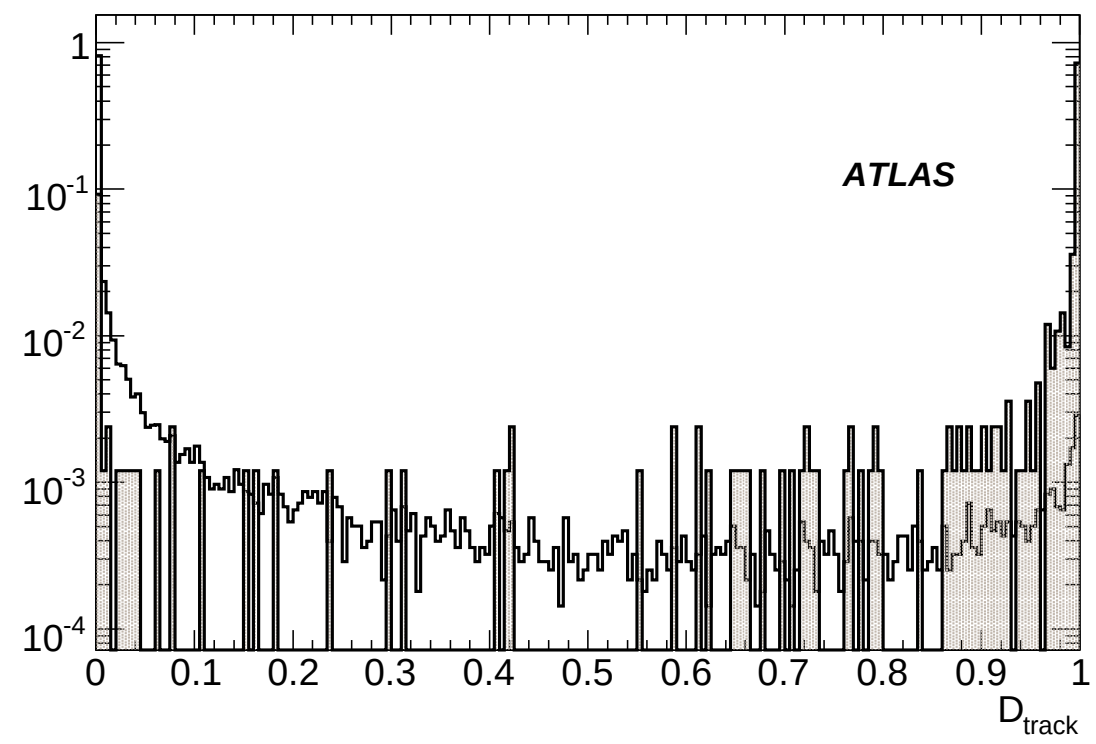

Figure 12: Discriminating function $D_{\text{track}}$ for electrons in the $H \rightarrow b\bar{b}$ sample (hatched histogram) and for pions in the $H \rightarrow u\bar{u}$ sample (solid line). The distributions are normalized to unit area.

For each *good quality* track the discriminating function $D_{\text{track}}$ is calculated as:

$$D_{\text{track}} = \frac{\prod_i P_e(x_i)}{\prod_i P_e(x_i) + \prod_i P_h(x_i)}, \tag{1}$$

where $x_i$ denotes the value of the $i$-th variable for a given track, $P_e(x_i)$ is the probability obtained from the single variable $x_i$ that the track originates from a signal electron, $P_h(x_i)$ is the probability that the track originates from a hadron, and $i$ runs through all the variables used by the algorithm. Distributions of $D_{\text{track}}$ obtained for signal electrons and background pions are shown in Fig. 12. $D_{\text{track}}$ tends to 1 for electrons and tends to 0 for pions. The identification of a candidate track as originating from a signal electron track is based on the value of $D_{\text{track}}$. Those tracks for which $D_{\text{track}}$ is below a given threshold are rejected. Typically, for an electron identification efficiency of 80%, a pion rejection factor of about 200 can be achieved in $H \rightarrow u\bar{u}$ events.

For each *good quality* track in the jet the value of the discriminating function for electron identification, $D_{\text{track}}$, defined previously in Eq. 1, has been calculated. For each jet, the track with the highest value $\max(D_{\text{track}})$ is chosen and this single track will now be used to estimate the discriminating function $D_{\text{jet}}$ for the jet. In addition to the variables used for electron identification based on the inner detector and electromagnetic calorimeter information, the track transverse momentum $p_T^{rel}$ relative to the jet axis, shown in Fig. 11 is included. For each jet the discriminating function $D_{\text{jet}}$ is calculated using the selected track as:

$$D_{\text{jet}} = \max(D_{\text{track}}) \times \frac{P_e(p_T^{rel})}{P_e(p_T^{rel}) + P_h(p_T^{rel})}, \tag{2}$$

where $P_e(p_T^{rel})$ is the probability obtained from the single variable $p_T^{rel}$ that the track originates from a $b$-jet and $P_h(p_T^{rel})$ is the probability that the track originates from a light jet. Distributions of $D_{\text{jet}}$ obtained for $b-$ and light jets are shown in Fig. 13. For a given threshold $D_{\text{jet}}^{\text{thr}}$, a jet with $D_{\text{jet}} > D_{\text{jet}}^{\text{thr}}$ is tagged as a $b$-jet.

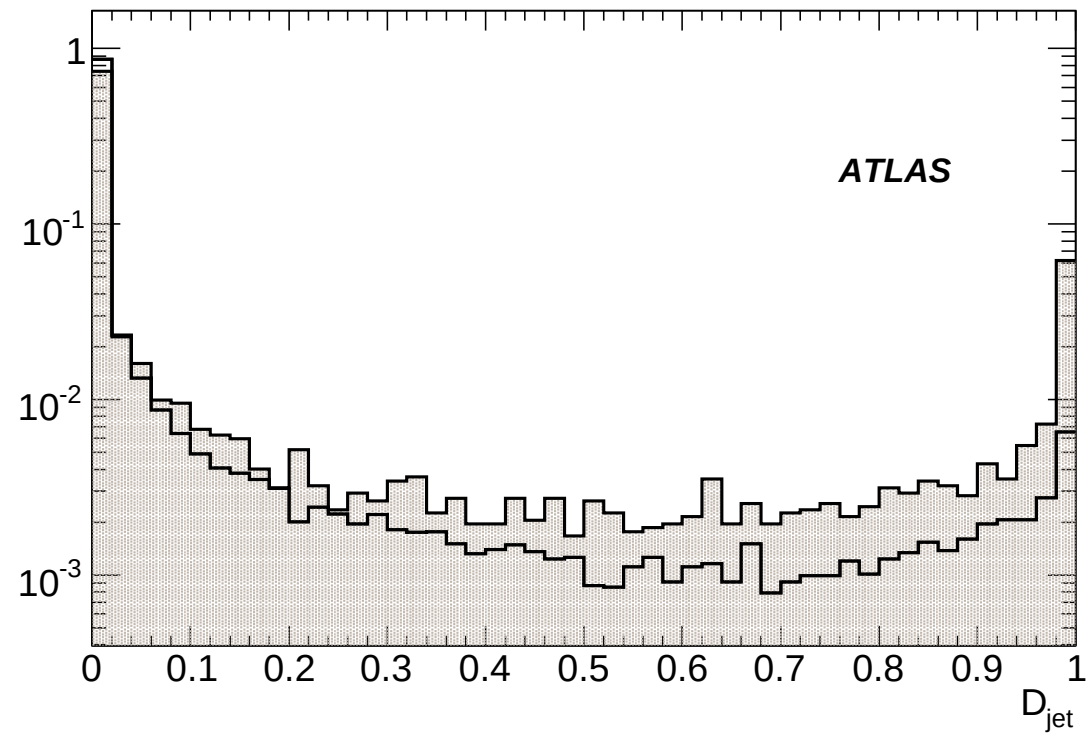

Figure 13: Discriminating function $D_{\text{jet}}$ for $b$-jets in the $H \rightarrow b\bar{b}$ sample (hatched histogram) and light jets in the $H \rightarrow u\bar{u}$ sample (solid line). The distributions are normalized to unit area.

## 5.2 Performance of the $b$-tagging algorithm

The $b$-tagging efficiency is defined as $\varepsilon_b = N_b^t/N_b$ where $N_b^t$ is number of the tagged $b$-labelled jets and $N_b$ is the total number of jets labelled as $b$-jets. The definition includes the semi-leptonic branching ratios as well as the detector acceptance and the electron reconstruction and identification efficiencies. The jet rejection factor is calculated as $R_{\text{light jet}} = N_j/N_j^t$, where $N_j$ is the number of light jets and $N_j^t$ is the number of light jets tagged by mistake as $b$-jets.

Figure 14 shows the rejection of light jets as a function of the $b$-tagging efficiency $\varepsilon_b$. The rejection factors achieved against light jets are presented in Table 3. For a $b$-tagging efficiency $\varepsilon_b = 7\%$ on inclusive $b$-jets, the light jet rejection factor is about 110. This operating point corresponds to a 61% $b$-tagging efficiency on semi-electronic $b$-jets.

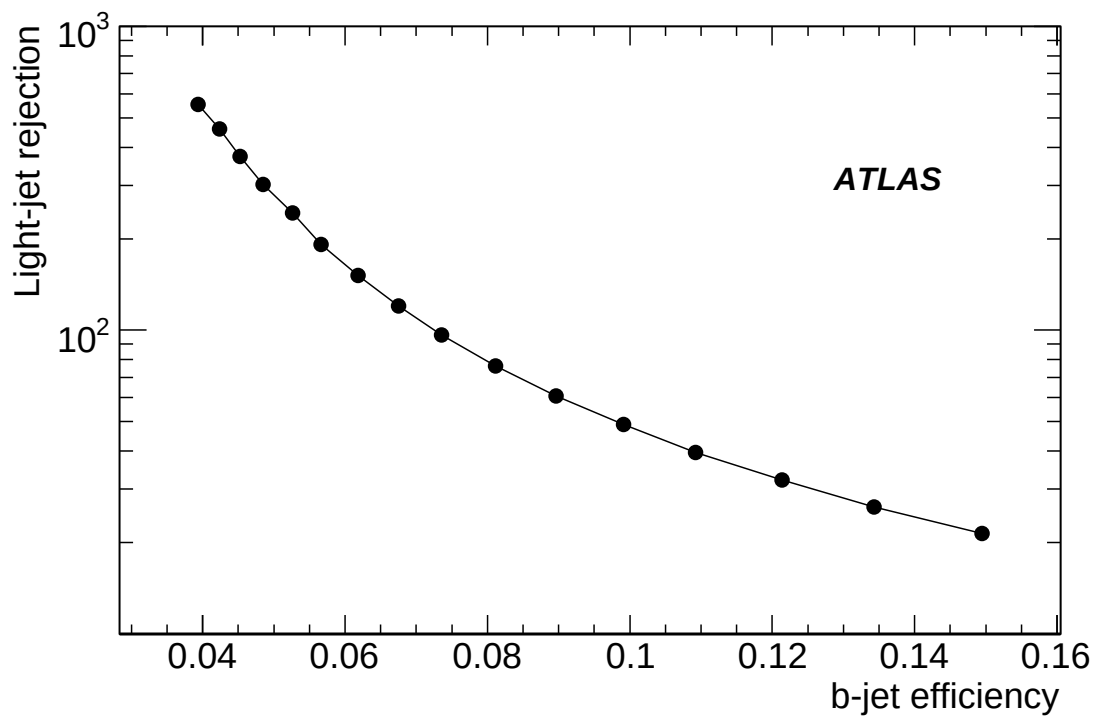

Figure 14: Rejection factor of light jets $R_{\text{light jet}}$ versus $b$-tagging efficiency $\varepsilon_b$.

Table 3: Light jet rejection factors $R_{\text{light jet}}$ for various $b$-tagging efficiencies $\varepsilon_b$ on inclusive jets. Errors are statistical only.

| $\varepsilon_b$ (%) | $R_{\text{light jet}}$ |
|---|---|
| 8 | $80 \pm 1$ |
| 7 | $110 \pm 2$ |
| 6 | $160 \pm 3$ |

This study has been performed on Monte Carlo simulated data which do not include pile-up effects. Based on previous study [10], a further degradation of the light jets rejection factor by 10% (30%) is expected when on average 4.6 (23) minimum-bias events per beam-crossing are added.

Figure 15 shows the light jet rejection as a function of the jet $p_T$ and $|\eta|$ for a total $b$-tagging efficiency $\varepsilon_b = 7\%$. Performance are estimated in five bins in $|\eta|$ and in seven bins in $p_T$. The rejection of light jets does not depend significantly on jet $p_T$ over the range 30-80 GeV. For lower $p_T$, jets are wider and an electron can escape outside the cone of $\Delta R = 0.4$ and this leads to lower rejections for the same $b$-tagging efficiency. For higher jet energies, the jet is narrower and showers coming from the particles in the jet overlap more and thus it is more difficult to separate the electrons, leading to a drop in rejection. A loss in the jet rejection factor can also be observed in the region $1.37 < |\eta| < 1.52$ corresponding to the crack between the barrel and the end-caps of the electromagnetic calorimeter. Even though soft electrons are reconstructed only within $|\eta| < 2$, some sensitivity is retained for jet axes just beyond the cut-off as shown in Fig.15.

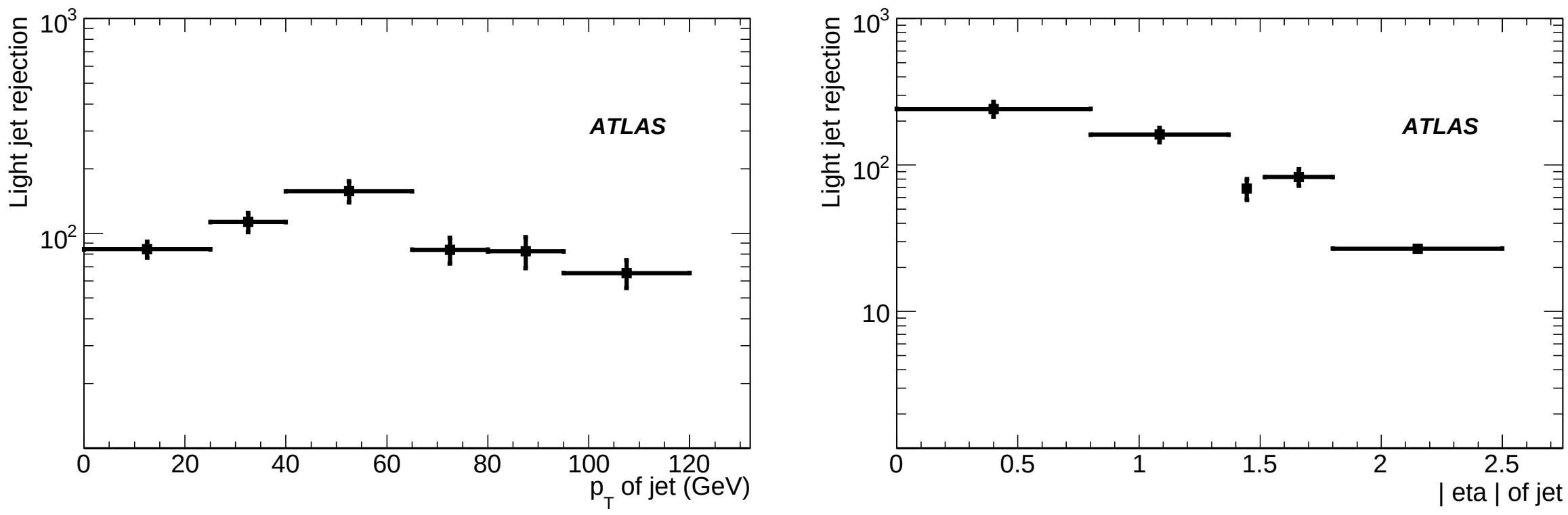

Figure 15: Light jet rejection factor as a function of jet $p_T$ (left) and jet $|\eta|$ (right) for a $b$-tagging efficiency $\varepsilon_b = 7\%$.

Table 4 shows the fraction of jets tagged of a given type of track for $\varepsilon_b = 7\%$. Most $b$-jets are tagged by true electron tracks; in particular by signal electrons in ~65% of cases. Light jets are tagged in ~25% of cases by electrons from $\gamma$-conversions and Dalitz decays. Light jets are also tagged in ~60% of cases by pions.

# 6 Conclusion

Soft electron tagging relies on the semi-leptonic decays of bottom and charm hadrons. It is therefore intrinsically limited by the branching ratios to electrons: about 20% of $b$-jets will contain a soft electron, including cascade decays of bottom to charm hadrons. However, when a signal electron is present the algorithm is quite efficient. In addition, tagging algorithms based on soft leptons have low correlations with the track-based $b$-tagging algorithms, which is very important for checking and cross-calibrating performance with data.

Table 4: Fraction of jets (%) in the $WH$ sample tagged by a specified type of track for $\varepsilon_b = 7\%$. The "*e from **" column corresponds to electrons from conversions and Dalitz decays. Statistical errors on these numbers are negligible.

| Jet type | Fraction of jets tagged by a specified type of track [%] | | | | | | |
|---|---|---|---|---|---|---|---|
| | all $e$ | $e$ from $b$ | $e$ from $c$ | $e$ from * | other $e$ | $\pi$ | others |
| $b$ | 67.6 | 42.0 | 19.4 | 3.6 | 2.6 | 23.9 | 8.5 |
| light | 27.2 | 0.2 | 0.4 | 25.5 | 1.1 | 58.7 | 14.1 |

The study presented here was based on a sample of *WH* events which have two high $p_T$ and well separated jets. A $b$-jet tagging efficiency of about 7% (including the inclusive $b \to e\nu X$ branching ratio of $\sim$ 20%) is achieved for a light jet rejection better than 100. The efficiency of the soft electron identification is high, since two-thirds of the $b$-jets are tagged by the true electron. Work is ongoing to prepare similar analyses with other Monte Carlo simulated samples with different topologies as well as to prepare the measurement with real ATLAS data.

# $b$-Tagging Calibration with $t\bar{t}$ Events

**Abstract**

This note describes various studies of $b$-jet reconstruction and $b$-tagging in simulated ATLAS $t\bar{t}$ events. The performance of several jet algorithms on $t\bar{t}$ event $b$-jets is studied. Techniques for measuring the $b$-tagging efficiency directly from data $t\bar{t}$ events, using both tag counting and explicit $b$-jet selection, are described and compared. Finally, the optimisation of $b$-tagging for $t\bar{t}$ events, using both Monte Carlo and data samples, is discussed. Both semileptonic and dilepton $t\bar{t}$ decay channels are considered, for an initial integrated luminosity of $100\,\text{pb}^{-1}$.

## 1 Introduction

The identification of $b$-jets is of crucial importance to many physics analyses at the LHC, including top physics and the search for new particles including Higgs bosons. The performance of $b$-tagging algorithms will have to be understood using real data, as Monte Carlo will only give an approximate description of the tracking and vertexing performance of the detector, at least initially. Light jet rejection factors can be measured in inclusive jet samples (where the heavy flavour content is small), but pure samples of $b$-jets are required to measure the $b$-jet tagging efficiency. One solution, already widely used at the Tevatron, exploits dijet events in which one of the jets is $b$-tagged using a soft lepton tag [1].

At the LHC, the large $t\bar{t}$ production cross-section offers an alternative source of $b$-jets, in a distinctive topology which is relatively easy to trigger on and isolate, providing at least one of the $W$-bosons from the top decays leptonically. Providing the top quark is assumed to have Standard Model properties (in particular that $\text{Br}(t \to Wb) = 1$), each $t\bar{t}$ event has two $b$-jets, together with at least one high-$E_T$ lepton, missing energy and additional high-$E_T$ jets from hadronically decaying $W$-bosons. The environment (high jet multiplicity, high-$E_T$ b-jets) is also much closer to that which $b$-tagging efficiency measurements are typically needed than in dijet events.

The $b$-tagging efficiency can be extracted from $t\bar{t}$ events in two ways—either by counting events with different numbers of tagged jets, or by reconstructing the $t\bar{t}$ decay topology in order to identify a pure sample of $b$-jets. Both techniques are explored in this note, which is organised as follows. Section 2 briefly describes the Monte Carlo simulation samples and common event selection which is used throughout, and Section 3 discusses reconstruction of $b$-jets in $t\bar{t}$ events, as an essential preliminary to studies of $b$-jet tagging. The counting and $b$-jet selection methods are discussed in Sections 4 and 5. Finally, Section 6 discusses the possibilities for optimising $b$-tagging performance specifically for $t\bar{t}$ events, using both Monte Carlo and data.

## 2 Datasets and event selection

All studies were performed using ATLAS Monte Carlo production samples as discussed in more detail in [2] and [3]. Semileptonic and dileptonic $t\bar{t}$ decays were generated with MC@NLO with Herwig hadronisation. AcerMC with Pythia hadronisation was used as an alternative. The background in the semileptonic channel is dominated by $W$+multijet production, which was simulated with ALPGEN (including $Wb\bar{b}$ and $Wc\bar{c}$), and single top production, simulated with AcerMC. In the dilepton channel (used only for the counting method $b$-tagging efficiency determination), additional backgrounds from $Z$+jets (simulated with ALPGEN) and diboson production (simulated with Herwig) are also important. Background coming from QCD multijet events has not been studied, as this background is difficult to simulate and samples were not available. The normalisation of such background (if significant) will eventually have to be extracted from data.

| | Electron channel | Muon channel | | Dilepton channel |
|---|---|---|---|---|
| $t\bar{t}$ (except all hadronic) | 45232 | 45232 | | 45232 |
| Trigger | 10430 | 12116 | | 21417 |
| Lepton ID | 8350 (80.1 %) | 11268(93.0 %) | | |
| Isolation | 8063 (96.6 %) | 9909 (87.9 %) | Lepton pair selection | 1549 (7.23 %) |
| Missing $E_T > 20$ GeV | 6362 (78.9 %) | 7834 (79.1 %) | Missing $E_T$ cut | 1355 (87.5 %) |
| $\geq 4$ jets $E_T > 20$ GeV | 3651 (57.4 %) | 4555 (58.1 %) | $\geq 2$ jets $E_T > 20$ GeV | 1160 (85.6 %) |
| $\geq 4$ jets $E_T > 30$ GeV | 2329 (63.8 %) | 2927 (64.3 %) | $\geq 2$ jets $E_T > 30$ GeV | 1000 (86.2 %) |
| W mass window cut | 1958 (84.1 %) | 2487 (85.0 %) | - | - |
| top mass window cut | 1378 (70.4 %) | 1773 (71.3 %) | - | - |

Table 1: Numbers of events from $t\bar{t}$ production expected with 100 pb$^{-1}$ in the lepton+jets and dilepton channels. The cut efficiency w.r.t. the previous line is shown in parenthesis. The fully hadronic decay $t\bar{t}$ contribution is not included but is expected to be small. Contributions from non-$t\bar{t}$ background are not included.

Systematic errors have been evaluated following the prescriptions discussed in detail in [2] and [4] wherever possible. The sensitivity to incorrect $b$-tagging of non-$b$ jets was assessed by doubling and setting to zero the corresponding jet tagging efficiencies, separately for light quark and charm jets. The jet energy scale was varied by $\pm 5$ %. The Monte Carlo event generation sensitivity was assessed by using AcerMC instead of MC@NLO, including samples with modified ISR, FSR and parton shower cutoff $Q^2$. The backgrounds from $W$+multijet events and single top production were independently doubled and set to zero. In all cases, the effect of these variations was computed without adjusting parameters of the analysis (*e.g.* acceptance factors and background estimates) to compensate for changes made to the input Monte Carlo samples.

The analyses make use of isolated high $E_T$ electrons and muons (with $E_T > 20$ GeV) and jets (with $E_T > 20$ GeV or higher depending on the particular analysis), selected according to standard ATLAS object definitions and fiducial cuts [2]. In the semileptonic channel, events are required to have exactly one high $E_T$ lepton, at least four jets and missing transverse energy of at least 20 GeV. In the dilepton channel, two leptons of opposite charge and a missing transverse energy of at least 20 GeV are required; in the $e^+e^-$ and $\mu^+\mu^-$ channels, the $Z$+jets background was further reduced by vetoing events with dilepton mass between 81 and 101 GeV and by requiring a missing transverse energy of at least 35 GeV. In both cases, at least one lepton in the event was required to pass the ATLAS level-1, level-2 and event filter triggers, using trigger signatures appropriate for the 20 GeV offline lepton $E_T$ cuts. For electrons, the signatures e25i and e60 were used (the e60 trigger improves the efficiency at high lepton $E_T$), and for muons, the signature mu20i was used. Further studies of the ATLAS trigger performance on $t\bar{t}$ events can be found in [5].

The resulting estimated event yields at various stages of the event selection are shown in Table 1 for 100 pb$^{-1}$. Non-$t\bar{t}$ backgrounds are not included. The individual analyses apply further specific selection cuts as discussed below. The effects of the tighter selections used only by the tag counting analysis described in Section 4 are also shown in Table 1. The hadronic $W$ mass window cut was applied by selecting the pair of jets with invariant mass closest to the $W$ mass and requiring this to be within $\pm 20$ GeV of the nominal value; a top mass cut was applied similarly by adding the jet giving a reconstructed top mass closest to the nominal, and requiring it to be within $\pm 30$ GeV of the top mass.

All studies were performed with the default ATLAS $b$-tagging algorithm (IP3D+SV1), which uses a likelihood weight $w$ constructed from the results of the IP3D impact parameter and SV1 secondary vertex-based taggers [6]. This likelihood weight is large for $b$-jets and small for light and $c$-quark jets. For jets from top decay with $E_T > 20\,\text{GeV}$, a cut $w > 4$ gives a $b$-jet tagging efficiency of around 70 % per jet and a light jet rejection of 50 while a $w > 7$ cut gives an efficiency of 60 % and a rejection of 200, though these values are somewhat dependent on jet $E_T$ and $\eta$. Detailed studies of the expected $b$-tagging performance can be found in [6] (*e.g.* in Table 1 therein). For the purposes of comparison, $b$-tagging efficiency measurements have been done at reference efficiencies of 50, 60 and 70 %. Various other cut values are used internally by the selections in the individual analysis methods. When considering Monte Carlo truth information, jets have been labelled $b$, $c$ and light using the standard ATLAS jet matching and labelling procedures [6].

## 3 $b$ jet reconstruction in $t\bar{t}$ events

The choice of jet reconstruction algorithm is an important issue for top physics analysis in ATLAS, where good resolution for the reconstruction of the original quark energy and direction is mandatory. These questions are explored below, using various jet algorithms and parameter choices. The analysis uses a sample of 250k semileptonic $t\bar{t}$ events generated with MC@NLO and passed through the standard top selection as discussed in Section 2. The choice of jet algorithm also affects the $b$-tagging performance, as studied in detail elsewhere [6].

The standard reconstruction outputs jet collections made with the Cone (with $\Delta R = 0.4$ and $\Delta R = 0.7$) and $k_T$ (with $D = 0.4$ and $D = 0.6$) jet algorithms, each based on either Precluster Projective Towers ('Tower') or Topological Cell Clusters ('Topo') (see [7] for more information on jet reconstruction in ATLAS). For this study, these standard jet reconstruction algorithms and parameter choices have been complemented by additional Cone and $k_T$ jets with a larger variety of $\Delta R$ and $D$ parameters. To compare all these jet algorithm and parameter choices in a controlled environment, re-reconstruction of jets was performed at the AOD level, using 'Topological Cell Clusters' calibrated by 'Local Hadron Calibration' (see [8] and [9] for details about the different strategies used to exploit and calibrate calorimeter information). All parameters except for the $D$ parameter ($k_T$) or $\Delta R$ (Cone) were left unchanged with respect to the standard jet algorithms.

It should be noted that these re-reconstructed jets do not contain the calorimeter energy corrections applied in the standard calibration procedure described in [7] and [8]. Therefore a comparison of the overall performance between the different jet algorithms with different parameters is valid, but no direct comparison between corrected and non-corrected jets should be made.

### 3.1 Jet energy and angle resolution

Reconstructed jets are matched to quarks from the $t\bar{t}$ decay following the standard procedure [6]. The results are shown in Fig. 1.

The top plots of Fig. 1 show the width of the distributions $(E_{quark} - E_{jet})/E_{quark}$ (*i.e.* the energy resolution) as a function of $E_{quark}$, for both $k_T$ and Cone. To calculate the width, a Gaussian is fitted to the individual distributions. The distributions show the expected behaviour. The smallest width is obtained for the jet sizes 0.4, 0.5 and 0.6, for both Cone and $k_T$. However, the differences between the different jet algorithms are small, so no strong conclusions can be drawn from them.

The two lower plots of Fig. 1 show an investigation of the angular resolution. The mean distance between the initial quark and the resulting jet in $\Delta R$ is plotted as function of $E_{quark}$ for Cone and $k_T$. The expected behaviour is that for bigger jet sizes the deviations between the jet axis and the flight direction of the initial quark should increase; this behaviour is observed in Fig. 1. It can also be seen that for larger jets and higher quark energies, the $k_T$ jet algorithm performs better than Cone.

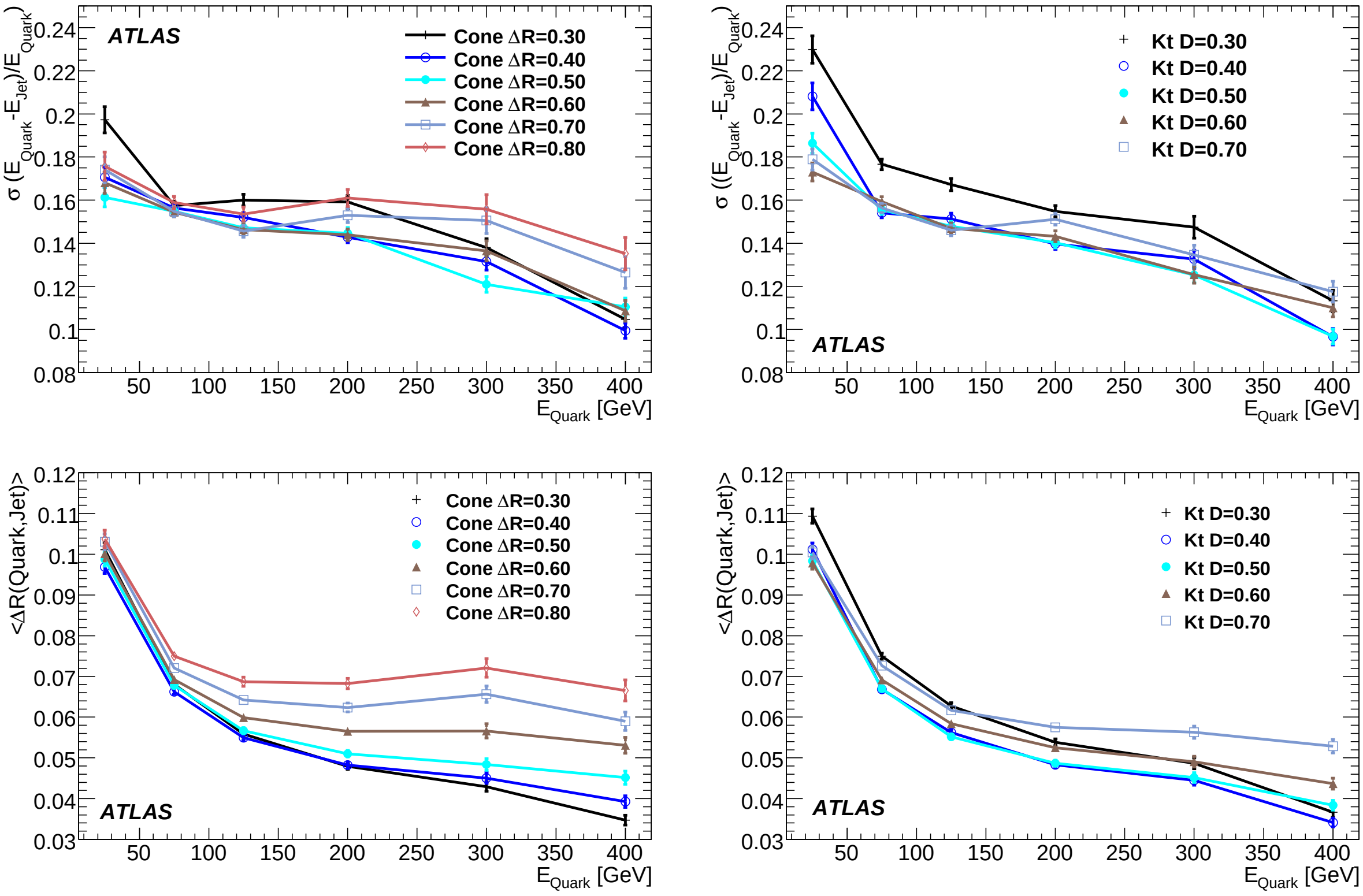

Figure 1: Studies of $b$-jet resolution: width of $(E_{quark} - E_{jet})/E_{quark}$ and average of ΔR(Quark,Jet) for Cone and $k_T$ with different jet sizes.

The choice of the Cone algorithm with $\Delta R = 0.4$ that was driven by the need to reconstruct properly W hadronic decays [10] has been found to be an appropriate choice for the $t\bar{t}$ signal. Background studies that will be performed in the near future will complete the picture.

# 4 $b$-tagging efficiency measurement via tag counting

Conceptually, in the absence of background, the simplest way to determine the $b$-tagging efficiency in $t\bar{t}$ events is to count the number of events with different numbers of $b$-tagged jets; this allows both the $b$-tagging efficiency and the $t\bar{t}$ production cross-section to be measured simultaneously. This method is discussed in detail below, with an emphasis on the $b$- (and $c$-) tagging efficiency measurements. More information on $t\bar{t}$ cross-section measurement can be found elsewhere [4].

In the following analysis, the event selection has been slightly tightened from that used elsewhere, in order to reduce the background. Jets are required to have $E_T > 30$ GeV. In the lepton+jets channel, the single top background is significantly reduced by applying a cut on the $W$ and top reconstructed masses, as described in Section 2.

## 4.1 Method

Top pair-production events give rise to two $b$-jets in the final state. Assuming for pedagogical purposes that every selected event contains two $b$-jets in the detector acceptance and that only $b$-jets can be tagged, then it is clear that the expected number of events with one $b$-tagged jet is proportional to $N2\varepsilon_b(1-\varepsilon_b)$ while the expected number of events with two $b$-tagged jets is proportional to $N\varepsilon_b^2$, where:

- $\varepsilon_b$ is the $b$-tagging efficiency, i.e. the probability to tag a $b$-jet
- $N$ is the number of selected $t\bar{t}$ events prior to any $b$-tagging requirement, which is proportional to the $t\bar{t}$ production cross-section.

It can be seen that it is possible to estimate both the $b$-tagging efficiency and the $t\bar{t}$ production cross-section from the observed number of events with one and two $b$-tagged jets.

In reality, $c$-jets and light jets (either from the hadronic $W$ decay or from ISR/FSR) are present and contribute to the number of tagged jets in the event. Moreover not all $b$-jets coming from the top decays end up being selected, whilst a small number of $b$-jets are produced through gluon radiation. To take these effects into account, the event flavour content is estimated from Monte Carlo with a large simulation $t\bar{t}$ sample. The factors $F_{ijk}$ are defined as the fractions of selected events (prior to any $b$-tagging requirement) with $i$ $b$-jets, $j$ $c$-jets, and $k$ light jets ($i,j,k = 0,1,2,3,...$). The *expected* number of events with $n$ tagged jets $< N_n >$ can be written as the sum over all possible combinations of $i$ $b$-jets, $j$ $c$-jets, and $k$ light jets, as a function of $b$-, $c$-, and light jet tagging efficiency:

$$< N_n >= (L\cdot\sigma_{t\bar{t}}\cdot A_{pre-tag})\cdot\sum_{i,j,k}F_{ijk}\sum_{i'+j'+k'=n}A_i^{i'}\cdot\varepsilon_b^{i'}\cdot(1-\varepsilon_b)^{i-i'}\cdot A_j^{j'}\cdot\varepsilon_c^{j'}\cdot(1-\varepsilon_c)^{j-j'}\cdot A_k^{k'}\cdot\varepsilon_l^{k'}\cdot(1-\varepsilon_l)^{k-k'} \quad (1)$$

where $A_i^{i'}$ is the number of arrangements $i!/(i'!\cdot(i-i')!)$, the prime subscript corresponding to the number of *tagged* jets of a given flavour; $\sigma_{t\bar{t}}$ is the total production cross-section; $A_{pre-tag}$ is the acceptance for $t\bar{t}$ events prior to any $b$-tagging requirement (including trigger efficiency, lepton reconstruction and ID efficiency, *etc.*) and $L$ is the integrated luminosity. The assumption that there is no correlation between tags in a given event is discussed later and is treated as a systematic uncertainty.

Finally, the following likelihood can be written:

$$L = \Pi(Poisson(N_n, < N_n >))) \quad (2)$$

where $N_n$ is the *observed* number of events with $n$ tags. In practice only events with one, two, or three tags in the lepton+jets channel and one or two tags in the dilepton channel are taken into account in the likelihood. Events with no tag suffer from significant background, whilst there are few events with more than three (two in the dilepton channel) tags. In the lepton+jets channel, both $b$- and $c$-tagging efficiencies are allowed to fluctuate in the fit together with the $t\bar{t}$ cross-section (hence three variables and three constraints); the light jet tagging efficiency is fixed in the fit and must be measured elsewhere. In the dilepton channel, the $c$-jet tagging efficiency is also fixed (hence two variables and two constraints).

Tables 2 and 3 show the $F_{ijk}$ values for the lepton+jets and dilepton channels, estimated from a sample of 265k $t\bar{t}$ events. It is interesting to note that the 'nominal' combination (two $b$-jets and two light jets for the lepton+jets channel; two $b$-jets for the dilepton channel) represents only about one third of the events, due to the presence of ISR/FSR light jets and $c$-jets from $W$ decays. The significant presence of $c$-jets in the lepton+jets channels allows the $c$-jet tagging efficiency to also be measured.

Figures 2 and 3 show the expected yield of events in the lepton+jets and dilepton channels for an integrated luminosity of $100\,\mathrm{pb}^{-1}$, as a function of the number of tagged jets. A $b$-jet is considered tagged if its $b$-tagging weight $w$ is greater than 7.

The method was applied to a sample of $t\bar{t}$ events corresponding to about 400 $\mathrm{pb}^{-1}$ of data and independent of those used to estimate the acceptance factors. Table 4 compares the measured efficiencies and cross-section to their true value for three different $b$-tagging purity levels. No bias is visible within the available statistics. The small deviation for the cross-section measurement in the dilepton channel is explained by the lack of statistics in the sample used to estimate the $F_{ijk}$ factors; this issue is discussed later together with other systematic uncertainties. With 100 pb$^{-1}$, a statistical precision of 2.7% (4.2%) can be achieved on the $b$-tagging efficiency, and 2.4% (4.8%) on the $t\bar{t}$ cross-section, respectively in the lepton+jets channel and dilepton channel. In the lepton+jets channel, the statistical uncertainty on the $c$-tagging efficiency is rather large, because this measurement is mostly determined by the number of events with three tagged jets.

| Number of light jets | 1 | 2 | 3 | 4 | any |
|---|---|---|---|---|---|
| 2 $b$-jets, 1 $c$-jet | 15.9 % | 10.6 % | 3.00 % | 0.85 % | 30.5 % |
| 2 $b$-jets, 0 $c$-jet | - | 24.4 % | 13.8 % | 3.74 % | 42.8 % |
| 1 $b$-jet, 1 $c$-jet | - | 6.46 % | 2.38 % | 0.74 % | 9.75 % |
| 1 $b$-jet, 0 $c$-jet | - | - | 7.60 % | 3.21 % | 11.6 % |
| 0 $b$-jet, 1 $c$-jet | - | - | 0.38 % | 0.11 % | 0.49 % |
| 0 $b$-jet, 0 $c$-jet | - | - | - | 0.46 % | 0.65 % |

Table 2: Fractions of selected events with a certain number of $b$, $c$ and light jets in the lepton+jets channel for the counting method. Only the dominant contributions are shown.

| Number of light jets | 0 | 1 | 2 | 3 | any |
|---|---|---|---|---|---|
| 2 $b$-jets, 1 $c$-jet | 1.02 % | 0.73 % | 0.27 % | <0.1 % | 2.08 % |
| 2 $b$-jets, 0 $c$-jet | 35.2 % | 20.9 % | 6.64 % | 1.39 % | 64.3 % |
| 1 $b$-jet, 1 $c$-jet | 1.68 % | 1.08 % | 0.46 % | 0.22 % | 3.54 % |
| 1 $b$-jet, 0 $c$-jet | - | 18.9 % | 6.26 % | 1.66 % | 27.2 % |
| 0 $b$-jet, 1 $c$-jet | - | 0.29 % | 0.11 % | <0.1 % | 0.40 % |
| 0 $b$-jet, 0 $c$-jet | - | - | 1.11 % | 0.31 % | 1.50 % |

Table 3: Fractions of selected events with a certain number of $b$, $c$ and light jets in the dilepton channel for the counting method. Only the dominant contributions are shown.

## 4.2 Backgrounds

In the lepton+jets channel, the main backgrounds are $W$+jets and single top. Other sources of background are $Z$+jets (where one lepton fails to be identified), $WW/WZ/ZZ$+jets, and QCD processes (which has not been studied, as discussed in Section 2). Figure 2 shows the total background due to $W/Z$+jets, single top, and diboson+jets production and Fig. 4 (left) shows the expected signal over background ratio as a function of the number of tagged jets. Events with one, or more than one, tagged jets have good purity, with signal over background ratios of 14.4 and 26.8 respectively, but the background does need to be taken into account. The estimated background is subtracted from each sub-sample.

| | | $\varepsilon_b$ (%) | | $\varepsilon_c$ (%) | | $\sigma_{t\bar{t}}$ (pb) |
|---|---|---|---|---|---|---|
| | | true | meas. | true | meas. | |
| Lepton+jets | $w > 4$ | 72.1 | 71.7±0.7 | 22.3 | 21.9±1.5 | 841±9 |
| | $w > 7$ | 60.4 | 59.8±0.8 | 12.8 | 13.8±1.3 | 844±10 |
| | $w > 10$ | 48.1 | 47.4±0.9 | 6.7 | 8.2±1.4 | 832±13 |
| Dilepton | $w > 4$ | 72.9 | 72.9±1.0 | - | - | 882±17 |
| | $w > 7$ | 61.1 | 60.5±1.2 | - | - | 883±19 |
| | $w > 10$ | 48.4 | 47.9±1.3 | - | - | 883±25 |

Table 4: Counting method: tagging effciencies and cross-section measured on a control sample and compared to their true value for three levels of $b$-tagging purity, for the lepton+jets and dilepton channels. The $t\bar{t}$ production cross-section was assumed to be 833 pb. The uncertainties are statistical only and correspond to about 400 pb$^{-1}$ of data.

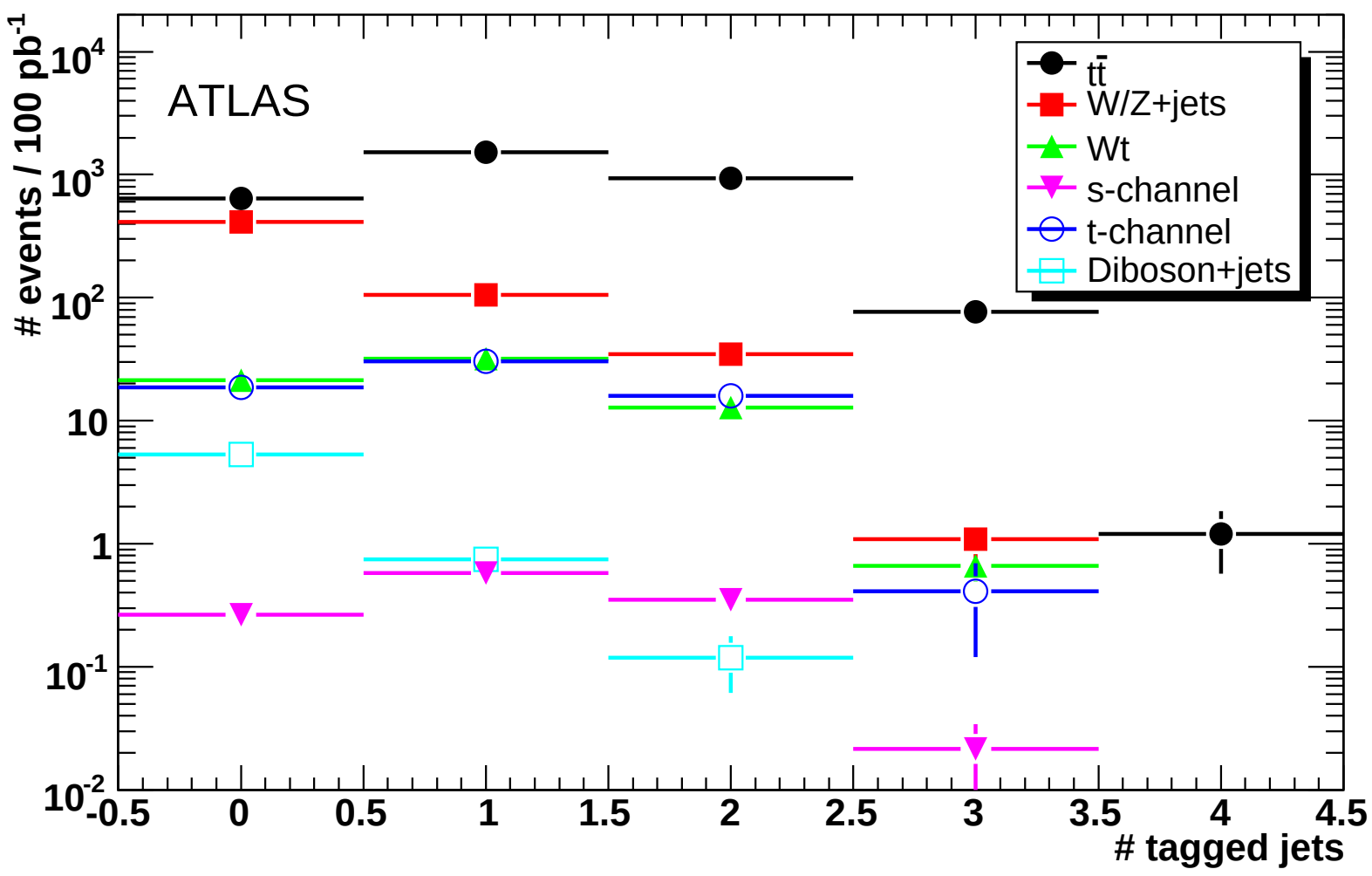

Figure 2: Yield expected for an integrated luminosity of $100\,\mathrm{pb}^{-1}$ in the lepton+jets channel as a function of the number of tagged jets. The expected background from $W/Z$+jets, single top, and diboson production is also shown.

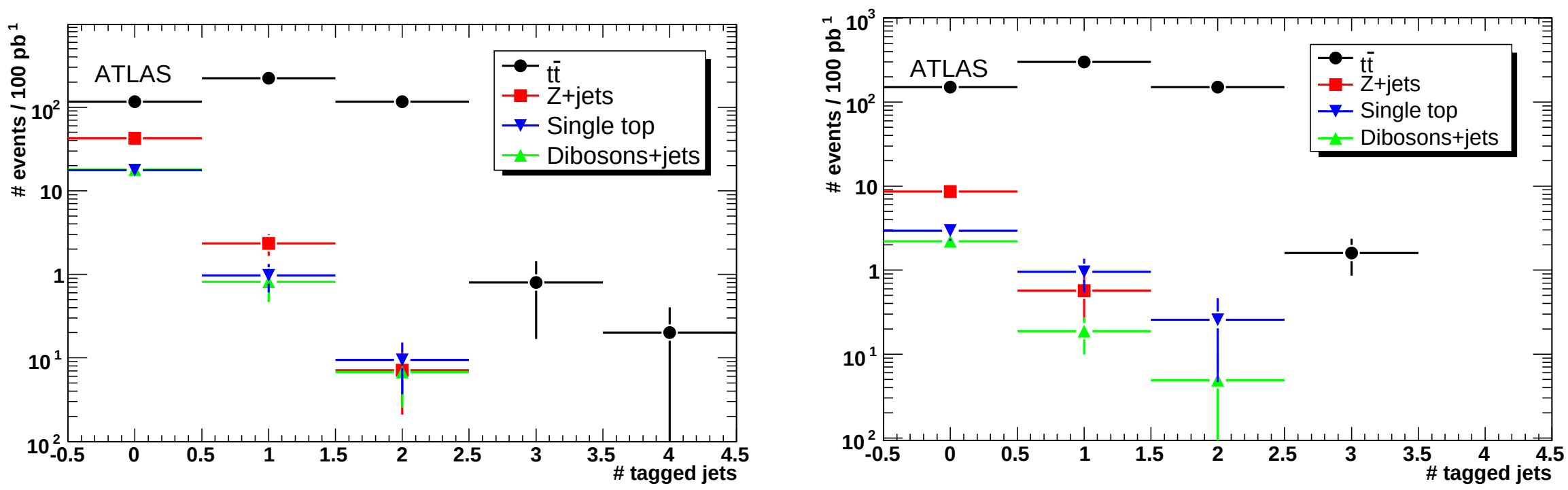

Figure 3: Yield expected for an integrated luminosity of $100\,\mathrm{pb}^{-1}$ in the dilepton+jets channels (left: $ee/\mu\mu$; right: $e\mu$) as a function of the number of tagged jets. The expected background from $Z$+jets and single top is also shown.

In the dilepton+jets channel, the main source of background is $Z$+jets production (but in the $e\mu$ channel, only $Z \to \tau\tau \to e\mu\nu\nu$ is significant). $WW/WZ/ZZ$+jets and single top also contribute. Figure 3 shows the total background due to $Z$+jets and diboson+jets production. The purity of the dilepton+jets sample (especially of the $e\mu$ channel) is good enough for the background estimate not to be an issue. Figure 4 shows that the expected signal over background ratio for events with exactly one tagged jet is 80 in the $ee$ and $\mu\mu$ channels and 175 in the $e\mu$ channel; the background is completely negligible for events with two or more tagged jets. It is remarkable that the $e\mu$ 0-tagged jet sub-sample is also quite pure and could be used in the fit to improve its statistical power, provided a reliable background estimate is available.

| Channel | Lepton+Jets | | | Dilepton | |
|---|---|---|---|---|---|
| Source | $\varepsilon_b$ | $\varepsilon_c$ | $\sigma_{t\bar{t}}$ | $\varepsilon_b$ | $\sigma_{t\bar{t}}$ |
| Light & $\tau$ jets | $<0.1$ | 18 | $<0.1$ | 0.7 | 0.3 |
| $c$-jets | - | - | - | 0.8 | 0.8 |
| $b$-jet labelling | 1.4 | 12 | 0.1 | 1.4 | 0.1 |
| tag correlation | $<0.2$ | $<0.2$ | $<0.2$ | $<0.2$ | $<0.2$ |
| Jet energy scale | 0.9 | 2.9 | $^{+6.8}_{-9.9}$ | 0.5 | $^{+1.3}_{-3}$ |
| $b$-jet energy scale | $<0.1$ | 1.5 | 0.8 | 0.2 | $<0.1$ |
| (MC statistics | 0.5 | 7 | 0.5 | 3 | 3) |
| Background | 1.2 | 3.5 | 4.8 | 0.3 | 0.4 |
| AcerMC vs MC@NLO | $<0.1$ | 9 | (4.9) | 2 | 6 |
| ISR/FSR | 2.7 | 12.5 | 8.9 | 2 | (4) |
| top quark mass | 0.3 | - | 2.2 | 0.5 | 2 |
| Luminosity | - | - | 5 | - | 5 |
| Lepton ID/trigger/pdf's | - | - | 2.8 | - | 2.8 |
| Total | 3.4 | 27 | $^{+12.4}_{-14.4}\pm 2.8\pm 5$ | 3.5 | $^{+6.6}_{-7.2}\pm 2.8\pm 5$ |
| Statistical (100 pb$^{-1}$) | 2.7 | 18 | 2.4 | 4.2 | 4.8 |

Table 5: Relative systematic and statistical uncertainties (in percent) for the lepton+jets channel for a $b$-tagging efficiency of $\varepsilon_{\text{true}} = 0.6$ and jet $E_T > 30$ GeV.

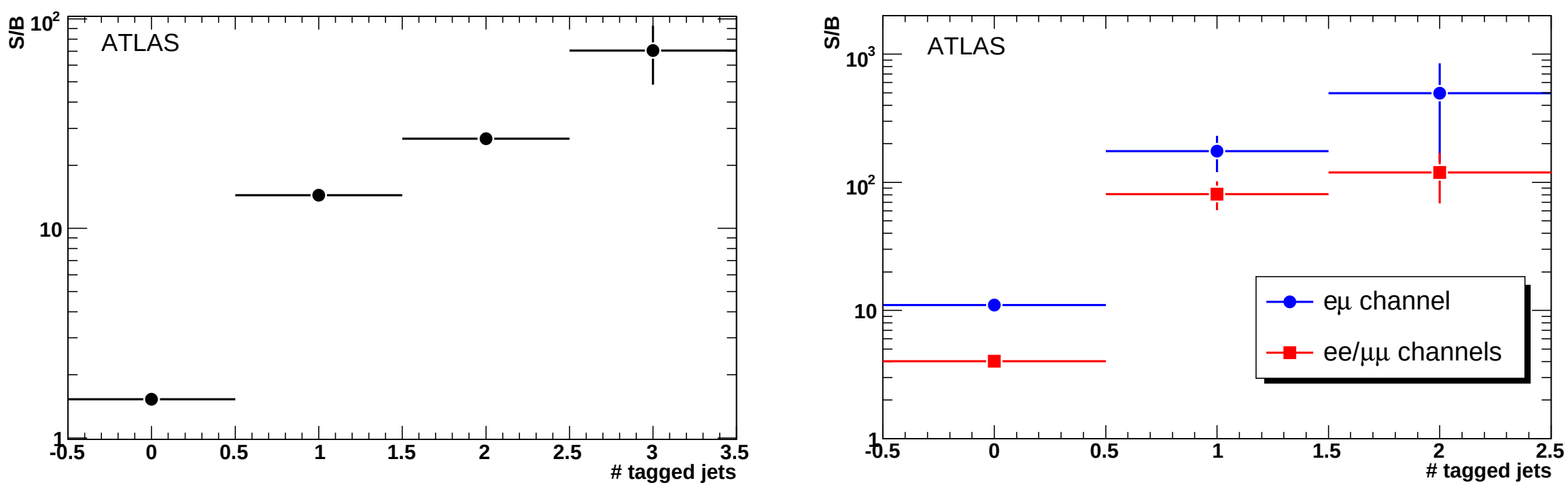

Figure 4: Signal over background ratio vs the number of tagged jets in the lepton+jets channel (left) and in the dilepton $ee/\mu\mu$ and $e\mu$ dilepton+jets channels (right).

## 4.3 Systematic uncertainties

Table 5 summarizes the resulting systematic uncertainties for the counting method in the lepton+jets and the dilepton channels. Some effects specific to this method are discussed in detail below.

The acceptance factors $F_{ijk}$ depend upon the definition of jet flavour, which is arbitrary to some ex-tent. By default, reconstructed jets are matched to the closest heavy quark (after FSR) in $\Delta R = \sqrt{\eta^2 + \phi^2}$; the jet is attributed the quark flavour if $\Delta R < 0.3$ (in case of ambiguity, $b$ quarks have priority over $c$ quarks). Jets that are not associated to any $b$ or $c$ quark are considered light jets. Another possibility is to match jets with hadrons, following the same procedure. It was checked that the choice of matching hadrons or quarks has a negligible effect on the definition of jet flavour. To assess the systematic uncertainty, the cut was shifted from 0.2 to 0.5 (nominal cut: 0.3) and the $F_{ijk}$ factors were re-estimated for each value. The shift observed on the same pseudo-experiments was taken as systematic effect.

The counting method assumes that there is no correlation between $b$-tags within an event, while some detector effects or the reconstructed primary vertex might induce such a correlation. To check this assumption, events with two reconstructed $b$-jets in the detector acceptance were selected. The covariance between the tagging of the two $b$-jets is:

$$cov(tag_1, tag_2) = \frac{N^{2b-tags}}{N} - \varepsilon_b^2$$

(where $N^{2b-tags}$ is the number of events with two tagged $b$-jets, $N$ is the total number of events, and $\varepsilon_b$ is the true $b$-tagging efficiency). It was found that $cov(tag_1, tag_2) = (8.8 \pm 27) \cdot 10^{-4}$, which is consistent with 0; the statistical uncertainty corresponds to a bias on the measured efficiency of 0.2 %, which is conservatively taken as systematic uncertainty.

Backgrounds are subtracted in each sub-sample and a 100% uncertainty is assumed. In the lepton+jets sample, for 100 pb$^{-1}$, 106 events from $W/Z$+jets are expected to populate the lepton+jets 1-tag sub-sample and 34.6 the 2-tag sub-sample. In the dilepton channel, 4.4 events with one tagged jet are expected for all three channels (mostly in the $ee$ and $\mu\mu$ channels), resulting in a small uncertainty.

A 5% uncertainty on the jet energy scale, a 1% additional uncertainty on the $b$-jet energy scale, and a 2 GeV uncertainty on the top mass, are also taken into account.

The statistical uncertainty on the $F_{ijk}$ factors was estimated by comparing the results given by two different training samples. It is not negligible in the dilepton+jets channel (3 %). However it could easily be reduced by employing a larger $t\bar{t}$ sample (1M events would suffice); thus it was not included in the total uncertainty.

The cross-section measurement is very sensitive to the jet multiplicity, which is driven mostly by ISR and FSR. This uncertainty appears also in the comparison of the two MC generators since the hadronisation models (Herwig in one case, Pythia in the other) need to be tuned in data and display important differences. In order to avoid to count twice the uncertainty on jet multiplicity, only the largest of the two estimates (ISR/FSR and MC generators) was considered in the total systematic uncertainty.

Systematic uncertainties on $\sigma_{t\bar{t}}$ that are not specific to this analysis are discussed elsewhere [4]. Uncertainties relative to integrated luminosity and event selection prior to $b$-tagging (namely, lepton identification efficiency, trigger efficiency and pdf's) amount to 5 % and 2.8 %, respectively.

## 4.4 Tag counting results

With 100 pb$^{-1}$ of data, the counting method allows the $b$-tagging efficiency at a working point of $\varepsilon_{\text{true}} = 0.6$ to be measured with a relative precision of $\pm 2.7$(stat.)$\pm 3.4$(syst.) % in the lepton+jets channel, and $\pm 4.2$(stat.)$\pm 3.5$(syst.) % in the dilepton channel. A better understanding of ISR/FSR could significantly reduce the systematic uncertainty.

The uncertainty on the $t\bar{t}$ production cross-section is somewhat larger in both channels, with $\pm 2.4$(stat.)$^{+12.7}_{-14.7}$(syst.)$\pm 5$(lum.) % in the lepton+jets channel, and $\pm 4.8$(stat.)$^{+7.2}_{-7.7}$(syst.)$\pm 5$(lum.) % in the dilepton channel, mostly because of the uncertainties on the jet multiplicity (ISR/FSR), jet energy scale, and background estimate, uncertainties that can be improved in the long term and have been assessed conservatively here.

The lepton+jets channel also allows the $c$-tagging efficiency to be measured. Unfortunately it is very sensitive to light jet contamination, and the uncertainty on light jet rejection and jet multiplicity makes any precision measurement impossible. Isolating a purer $c$-jet sample through some kinematic selection might help but has not been attempted here.

# 5 Selecting $b$ jet samples in data

Going beyond the $b$-tagging efficiency measurement from the counting method discussed above, the kinematics and topology of semileptonic $t\bar{t}$ events also allow pure samples of $b$-jets to be identified directly. Providing this selection is done without biasing any properties of the selected $b$-jets, they can then be used to measure the $b$-tagging efficiency, simply by studying the distribution of the tagging discriminant variable(s) in the selected jets. With enough statistics, this technique should also allow the measurement of the tagging efficiency as a function of other variables (e.g. jet $E_T$ and $\eta$).

Obtaining a pure sample of identified $b$-jets requires full reconstruction of the $t\bar{t}$ decay chain, assigning four jets in the event to the two $b$-jets coming directly from the $t \to Wb$ decays and the jets from hadronic $W \to qq$ decay. For any given event, many different assignments are possible, especially when spectator jets (not from the $t\bar{t}$ decay) are also present. The correct combination can be determined by making use of kinematic information, including the reconstruction of the hadronic and leptonic top masses for each considered combination, and the classification can be further improved by requiring a $b$-tag for *one* of the jets assigned to the $t \to Wb$ decay, providing the other presumed $b$-jet is left unbiased.

Three similar techniques have been explored for making this jet identification, all of which consider each combinatorial jet assignment and try to determine which one is correct:

- A 'topological' selection based on the reconstructed masses for each combination.
- A 'likelihood' selection, exploiting reconstructed top mass, jet momentum and angular information.
- A 'kinematic' selection based on applying a kinematic fit to each combination and choosing the one with the best $\chi^2$ value.

All of these methods start from the selection of semileptonic $t\bar{t}$ events discussed in Section 2, *i.e.* events with a high-$E_T$ lepton, significant missing energy and at least four jets. None of the methods result in a sample which is 100 % pure in $b$-flavoured jets, so background subtraction techniques are needed to derive pure samples on a statistical basis. The three methods, their associated background subtraction techniques, and their application to the measurement of the $b$-tagging efficiency for the standard IP3D+SV1 tagger, are discussed in Sections 5.1, 5.2 and 5.3 below. Comparisons of their performance in terms of $b$-jet sample selection and $b$-tagging efficiency measurement are then given in Sections 5.4 and 5.5.

## 5.1 Topological selection

The topological selection of $b$-jets from the basic sample has two stages: reconstruction of $t\bar{t}$ pairs, followed by fitting the top mass distributions to extract the $b$-jet sample and estimate the remaining background. To improve the purity, an explicit $b$-tag is required on the jet from the hadronic top decay, whilst the $b$-jet from the leptonic top is left unbiased, and is used to form the sample of $b$-jets for measuring the tagging efficiency.

An attempt is made to reconstruct both hadronic and leptonic top decays in each event. First, the invariant mass of every pair of jets is calculated to look for combinations consistent with coming from the decay $W \to jj$. Combinations satisfying $60 < m_{jj} < 100$ GeV are retained, and combined with a third jet to form a candidate $t \to bW \to bjj$ decay. One jet from the W decay and the candidate $b$-jet must have jet $E_T > 40$ GeV, and the second $W$ decay jet must have $E_T > 20$ GeV. To reduce background, this candidate $b$-jet is required to have a $b$-tagging weight of $w > 3$, corresponding to an efficiency of about 74 % for genuine $b$-jets from top decay. Additionally, the two $W$ jets are both required to have $b$-tagging weights of $w < 3$, to reduce the probability that the $b$-jet from the top decay is included in the $W$. No mass requirements are placed on the hadronic top candidate at this stage.

A leptonic top is then reconstructed from one of the other jets and a reconstructed leptonic $W$ decay. The leptonic $W$ is formed from the identified lepton and a neutrino, whose transverse momentum ($p_x$, $p_y$) is taken to be equal to the missing transverse momentum vector of the event. The longitudinal momentum of the neutrino is inferred using the constraint of the known $W$ mass, which leads to a quadratic equation for $p_z$ which has either two or no solutions. In the case of two solutions, the one with the smaller $p_z$ is chosen. If there is no solution, the measured missing $E_T$ is scaled down, preserving its direction, until a solution is possible. The reconstructed $W$ is then combined with another jet which was not used in making the hadronic $W$, which is assumed to be the $b$-jet from the leptonic top decay. This jet is required to have $E_T > 20$ GeV. No requirements are placed on the $b$-tagging weight for this jet, which is instead enhanced in $b$-flavour due to the topological reconstruction of the $t\bar{t}$ event.

This procedure is performed for all possible combinations of jets in each event. In an event with four jets, several assignments of jets to the hadronic top $b$-jet, leptonic top $b$-jet and W decay jets are possible, constrained by the requirements made on the $b$-tagging weights for the jets assigned to the hadronic top decay. In events with more than four jets, some jets will be left unassigned, and are assumed to be spectator jets from initial or final state radiation or the underlying event. As the number of jets in the event increases, the number of combinations satisfying all the requirements increases very rapidly, from typically one or two combinations in events with four jets to an average of nine combinations for events with six or more jets. The combination with the largest scalar sum of the $p_T$ of the two tops is retained for further analysis. In 100 pb$^{-1}$ of fully-simulated $t\bar{t}$ events plus $W$+jets background, 3028 events are selected, of which 80 % are semileptonic $t\bar{t}$ events with $t \to be\nu$ or $b\mu\nu$, with the background being dominated by $t\bar{t}$ with $b \to b\tau\nu$ (11 %) and $W$+jets (6 %), together with smaller contributions from dilepton and single top events.

### 5.1.1 Mass fitting

After all these selections, only one assignment of jets and leptons to the hadronic and leptonic top decays remains. The resulting distributions of reconstructed hadronic and leptonic top masses are shown in Fig. 5. Clear top mass peaks are seen in both distributions, though with significant combinatorial background and a small contribution from $W$+jet and single top background events. Contributions from non-$t\bar{t}$ events (denoted class 1), events where both tops are incorrectly reconstructed (class 2), events where one top (hadronic or leptonic) is correctly reconstructed and the other is not (classes 3 and 4), and events where both tops are correctly reconstructed (class 5) are shown separately. For these purposes, a top decay is considered correctly reconstructed if the associated $b$-jet is actually labelled as a $b$-jet by the Monte Carlo truth analysis, and the direction of the reconstructed top decay is within $\Delta R$ of 0.5 (2.0) for the hadronically (leptonically) decaying top. The hadronic top mass distribution in Fig. 5 shows peaks for classes 3 and 5, and the leptonic top mass shows peaks for classes 4 and 5, though with a considerably worse mass resolution. Both distributions also show some 'peaking' structure in the combinatorial backgrounds.

The analysis aims to extract a pure sample of $b$-jets from the reconstructed leptonic top decay. The leptonic top is chosen since it does not suffer from combinatorial background within the top decay itself—by contrast, in the hadronic top decay with three jets combined to give the top mass, a mis-assignment of jets to $W$ decay and $b$ will give the same top mass, with only the reconstructed hadronic $W$ mass changing. Since the resolution on the latter quantity is relatively poor, many reconstructed $bj$ combinations are compatible with the $W$ mass, and the $W$ mass requirement gives little discrimination. In the leptonic top decay, the $W$ has no decay jets, and this problem does not arise.

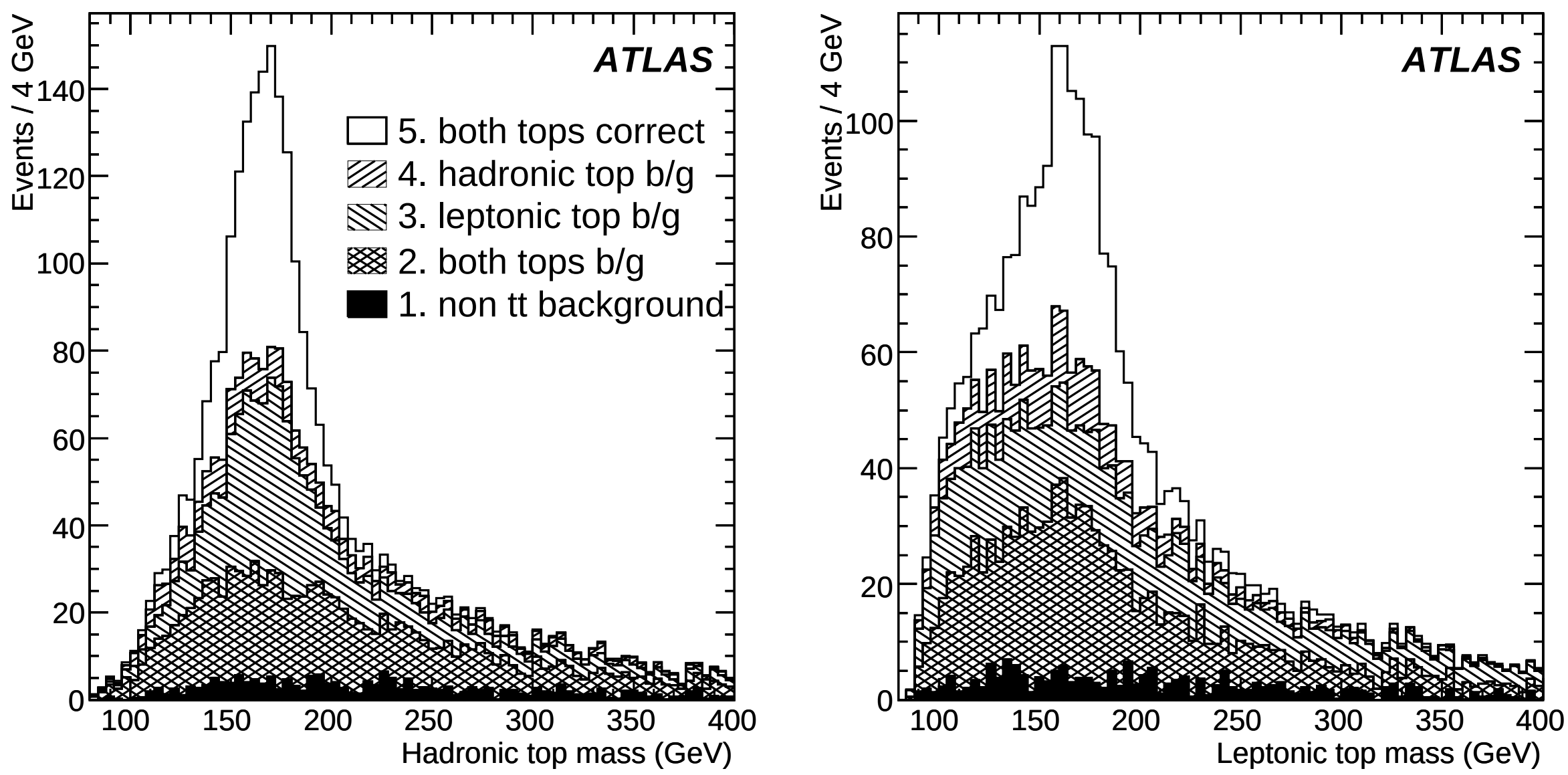

Figure 5: Reconstructed hadronic (left) and leptonic (right) top masses for the selected jet combination, showing the contributions from correctly reconstructed $t\bar{t}$ events, combinatorial and non-$t\bar{t}$ background, normalised to $100\,\mathrm{pb}^{-1}$. The numbers refer to the classes discussed in the text.

The resolution on the reconstructed leptonic top mass is poor, and there is a large combinatorial background. This is reduced by first requiring that the hadronic top mass is in the signal region ($140 < m_{\mathrm{jjj}} < 190\,\mathrm{GeV}$), and then dividing the sample into five sub-samples according to the reconstructed $E_T$ of the leptonic top $b$-jet, as the level of background depends strongly on this jet energy. The resulting leptonic top mass distributions can be seen in Fig. 6, where it can be seen that the top signal is significantly enhanced, especially for high leptonic top jet $E_T$. Events from classes 4 and 5 have correctly reconstructed leptonic tops, where the leptonic top jet is pure $b$ flavour, whilst the other events are combinatorial background, with a mixture of jet flavours (they can include $b$-jets from the hadronic top, and jets which are not unambiguously from a single parton, containing some particles from the decay of $B$ hadrons).

In order to correct for this background of event classes 1–3 under the leptonic top mass peak, both its size and flavour composition must be determined. This is done on a statistical basis using the sideband region with high $m_{b\ell\nu}$ to normalise the background contribution, and a control sample to determine the shape of its mass distribution. The control sample is generated from events where the reconstructed hadronic top mass satisfies $200 < m_{\mathrm{jjj}} < 400\,\mathrm{GeV}$, and where the leptonic top $b$-jet satisfies a cut on the $b$-tagging weight of $w < 4$. These samples are dominated by events in classes 2 and 3, and provide a reasonable model of the background shapes in the signal sample.

The amount of background under the signal is then extracted using a simultaneous fit to both signal and control sample leptonic top mass distributions. The control sample is described using a fit function $F_b(m_{b\ell\nu})$ given by:

$$F_b(m_{b\ell\nu}) = E((m_{b\ell\nu} - c_0)/400)\,, \tag{3}$$

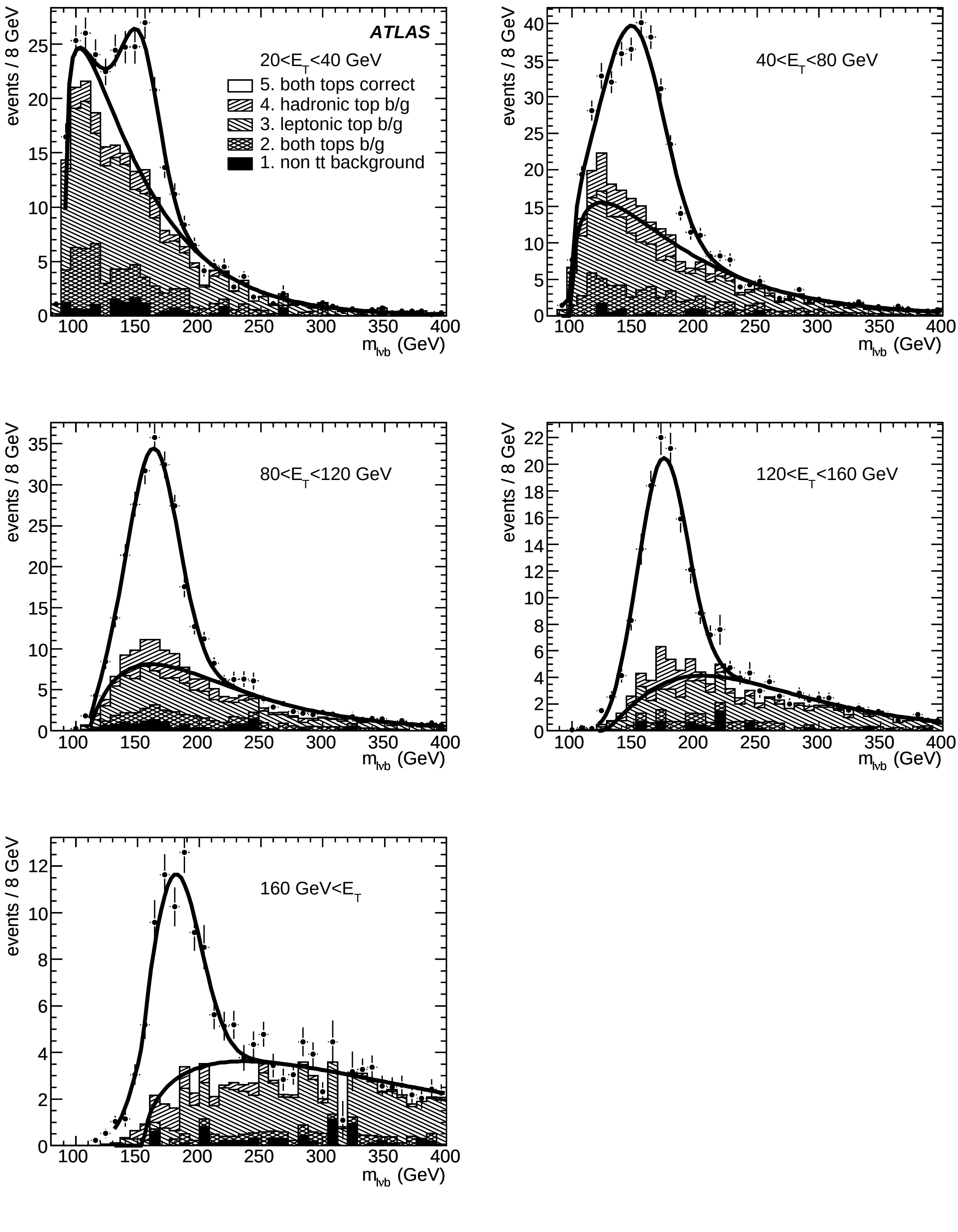

Figure 6: Reconstructed leptonic top mass distributions for different regions of the leptonic top $b$-jet $E_T$, with hadronic top mass in the range $140 < m_{\mathrm{jjj}} < 190$ GeV. The points with error bars show all events, with background contributions indicated by the hatched histograms. The error bars show the full Monte Carlo statistics ($948\,\mathrm{pb}^{-1}$) whilst the event counts indicate the number of events expected for $100\,\mathrm{pb}^{-1}$. The lines show fits to the signal and estimated background, and are discussed further in the text.

where $E(x)$ is given by:

$$E(x) = \begin{cases} c_1 x^{c_2} \exp(-c_3 x) & \text{if } x > 0 \\ 0 & \text{if } x < 0 \end{cases} . \tag{4}$$

The signal is described by a fit function $F_s(m_{b\ell\nu})$ consisting of a scaled background function plus a Gaussian representing the signal, given by:

$$F_s(m_{b\ell\nu}) = c_4 F_b(m_{b\ell\nu}) + c_5 G((m_{b\ell\nu} - c_6)/c_7) . \tag{5}$$

The parameter $c_4$ represents the ratio of class 1–3 background contributions in signal to control samples, and $c_5$ the amount of top signal with fitted mean $c_6$ and RMS width $c_7$. All eight parameters are extracted in a simultaneous $\chi^2$ fit to the signal and control sample mass distributions, thus determining the shape of the background under the signal peak from the control sample, and the normalisation of the background from the sideband of the signal distribution at high $m_{b\ell\nu}$. Separate fits are performed in each leptonic top jet $E_T$ bin, and the results are shown in Fig. 6. The fits are performed over the mass range $b_1$ and $b_2$, where $b_1$ is set to 90, 90, 110, 120, and 130 GeV for the 20–40, 40–80, 80–120, 120–160 and 160-200 GeV jet $E_T$ bins, reflecting the differing kinematic cut-off in each jet mass bin.

### 5.1.2 Background subtraction

The sample of leptonic top $b$-jets is divided using the reconstructed leptonic top mass into a signal region defined by $s_1 < m_{b\ell\nu} < s_2$ where $s_1 = c_6 - 2c_7$ and $s_2 = c_6 + 2c_7$ (*i.e.* $m_{b\ell\nu}$ within $\pm 2\sigma$ of the fitted top mass peak position) and a sideband region with $m_{b\ell\nu}$ outside this range. The distribution of the $b$-tagging weight in the signal region contains contributions both from pure $b$-jets (events of classes 4 and 5), and from the mixture of flavours forming the combinatorial background under the signal peak. This latter mixture is well described by the jets in the sideband region, so its effect is corrected for by subtracting the $b$-tagging weight distribution in the sideband region, scaled according to the expected normalisation derived from the mass fit discussed in Section 5.1.1. The scale factor $S$ is given by:

$$S = \frac{\int_{s_1}^{s_2} F_b(m_{b\ell\nu})\, dm_{b\ell\nu}}{\int_{b_1}^{s_1} F_b(m_{b\ell\nu})\, dm_{b\ell\nu} + \int_{s_2}^{b_2} F_b(m_{b\ell\nu})\, dm_{b\ell\nu}} \tag{6}$$

and varies between around 4 for the 40-80 GeV bin to 0.3 for the 160-200 GeV bin. Once this scaled background has been subtracted, the resulting $b$-tagging weight distribution in the signal region is statistically compatible with that expected from pure $b$-jets. The uncertainty on the amount of background to subtract in each bin of the $b$-tagging weight distribution has two components: the uncertainty on the number of background events in each bin in the sideband distribution, scaled by $S$, and the uncertainty on $S$ itself, which is correlated across all bins. The latter is calculated using equation 6 together with the full correlation matrix for the parameters $c_0$ to $c_6$ which are used in the background function $F_b(m_{b\ell\nu})$.

In order to check that this selection does not introduce a bias, the $b$-tagging weight distribution of true $b$-jets in the selected signal region was checked, and found to be compatible with that of all $b$-jets within the $\eta$ and $E_T$ acceptance. Hence this selected sample of $b$-jets can be safely used to measure the $b$-tagging efficiency.

### 5.1.3 Topological selection results

The results of applying this analysis procedure to full simulation events (including background) are shown as the ‘Topological’ entries in Table 6 in Section 5.4. For each leptonic top jet $E_T$ bin, the number of jets selected in the window $s_1 < m_{b\ell\nu} < s_2$, the number remaining after subtracting the scaled sideband background, and the purity of the selected sample are given, together with the background-subtracted sample $b$-flavour purity, *i.e.* the fraction of the background-subtracted sample which is made up of $b$-flavour jets according to the Monte Carlo truth information. It can be seen that in all cases the flavour purity is compatible with unity within errors, showing that the background subtraction procedure is working well and producing a statistically pure sample of $b$-jets. Table 6 also shows the results integrated from 40-200 GeV. The sample size and purity in the 20–40 GeV jet $E_T$ bin is not sufficient to make a sensible measurement of the $b$-tagging efficiency, and this bin is not considered in what follows.

Using this $b$-jet selection, the $b$-tagging efficiency corresponding to any cut on the $b$-tagging weight can be obtained by integration of the $b$-tagging weight distribution, *i.e.* calculating the fraction of the background-subtracted weight distribution above the cut value. This efficiency is shown as a function of $b$-tagging weight cut for each $b$-jet $E_T$ bin in Fig. 7 as the points with error bars. The solid histogram shows the same distribution for an unbiased sample of pure $b$-jets in $t\bar{t}$ events without any event selection, derived using Monte Carlo truth information. If the method is correct, the two distributions should agree within errors. The statistical errors on the measured $b$-tagging efficiency (which are highly correlated between bins) have two components: the binomial errors due to the fluctuations in numbers of events passing the cut for signal and background, and the uncertainty due to the background subtraction scale factor $S$. Figure 7 also shows the difference $\delta = \varepsilon_{\text{meas}} - \varepsilon_{\text{true}}$ between the estimated and true efficiency, which should be compatible with zero if the method is working well. This is seen to be the case except for very high $b$-tagging efficiencies in the two highest jet $E_T$ bins, where the statistics are limited. The combined efficiency curve for $b$-jet $E_T > 40$ GeV is shown in Fig. 8(a). These results are derived by weighting the individual $b$-tagging efficiencies measured in each energy bin by the $b$-jets $E_T$ distribution in $t\bar{t}$ events.

The method has been validated using ensemble tests based on sets of Monte Carlo subsamples of various integrated luminosities, derived from a sample of 1793 pb$^{-1}$. Studies of the distributions of fit results, estimated uncertainties and pulls show that the uncertainty on the $b$-tagging efficiency returned from the fit underestimates the true uncertainty by about 20 %. They also show that the fit shows significant biases and does not always converge correctly for samples of 100 pb$^{-1}$. With 200 pb$^{-1}$ of data, the fit results are unbiased and the fit converges in 98 % of the test samples; hence this is considered the minimum amount of data for which this technique can be used.

Figure 8(b) shows the absolute uncertainty returned by the tagging measurement as a function of the efficiency itself, scaled to an integrated luminosity of 200 pb$^{-1}$. At $\varepsilon_{\text{true}} = 0.6$, the statistical error is about 0.038, corresponding to a relative error of $\sigma\varepsilon_{\text{true}}/\varepsilon_{\text{true}}$ =6.4 %.

Systematic uncertainties have been evaluated as discussed in Section 4.3, with the exception that no uncertainties were evaluated for $b$-jet labelling, Monte Carlo statistics and the top quark mass, which are not relevant for this analysis. The systematic errors are summarised in Table 7.

## 5.2 Likelihood-based selection

The likelihood selection uses templates based on event and jet kinematic variables to determine the best assignment of jets to $t\bar{t}$ decay products within an event. These same likelihoods are also used to select events for which the jet assignments are most likely to be correct.

Templates are constructed using several kinematic variables to provide discrimination between 'correct' and 'incorrect' permutations. To maximize discrimination, templates are constructed separately for each jet multiplicity. All possible permutations of assignments of jets to final-state quarks are considered, and, for each permutation within each event, the leptonic and hadronic top quark and hadronic $W$ boson are reconstructed, using the reconstructed momenta. The permutation is labelled 'correct' if the jets used to reconstruct the particle ($t$, $W$, $b$-quark, or light quark from $W$ decay) are correctly matched to the quarks arising from the hard-scatter vertex. To limit the number of jet-parton assignment combinations, events with more than six jets are rejected.

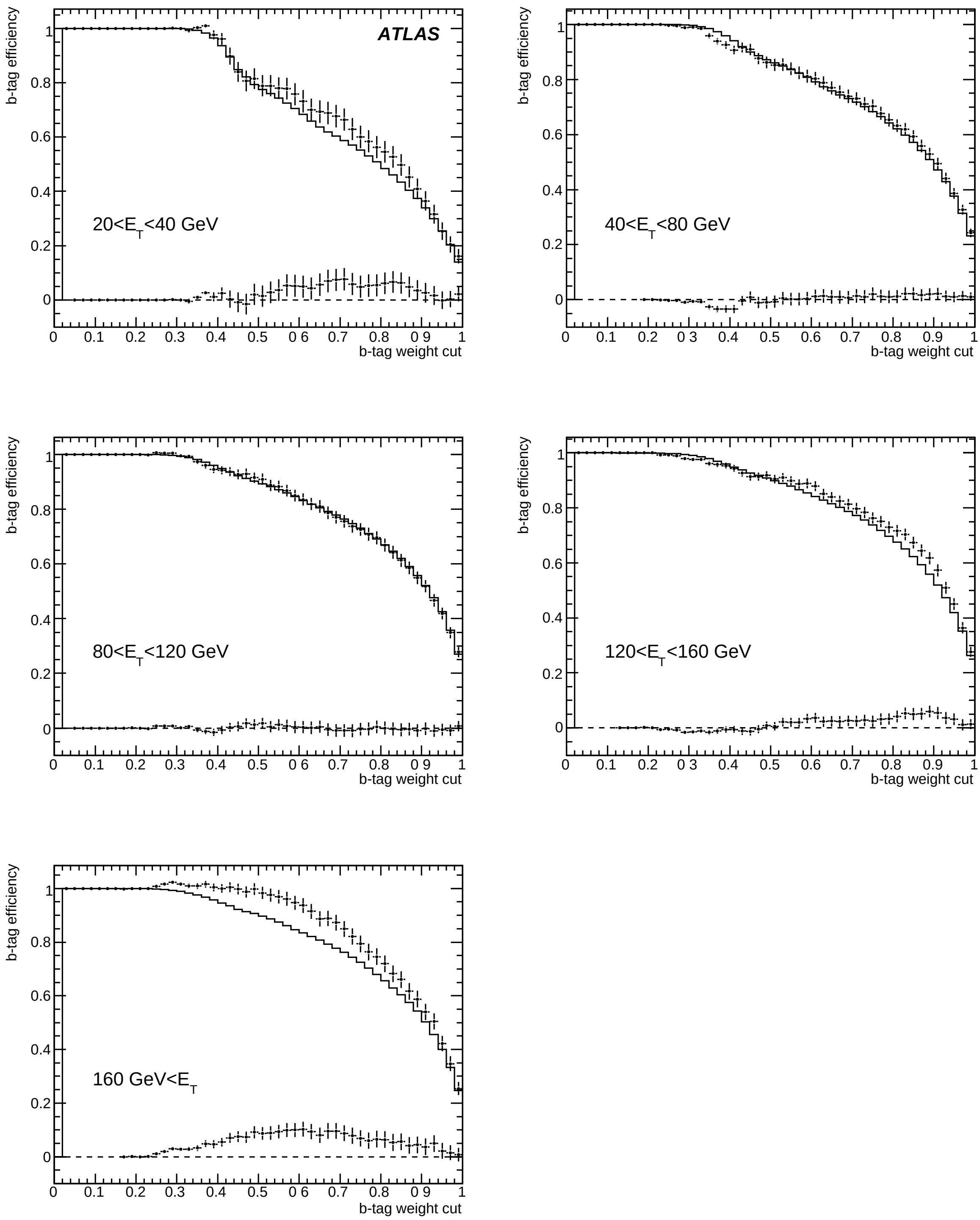

Figure 7: $b$-tagging efficiency *vs.* weight cut for different $b$-jet $E_T$ ranges, as measured by the topological $b$-jet analysis (upper points with error bars, statistics of full simulation sample), and compared to the true efficiency from unbiased $b$-jets (histogram). The difference between the two is shown as the lower points with error bars.

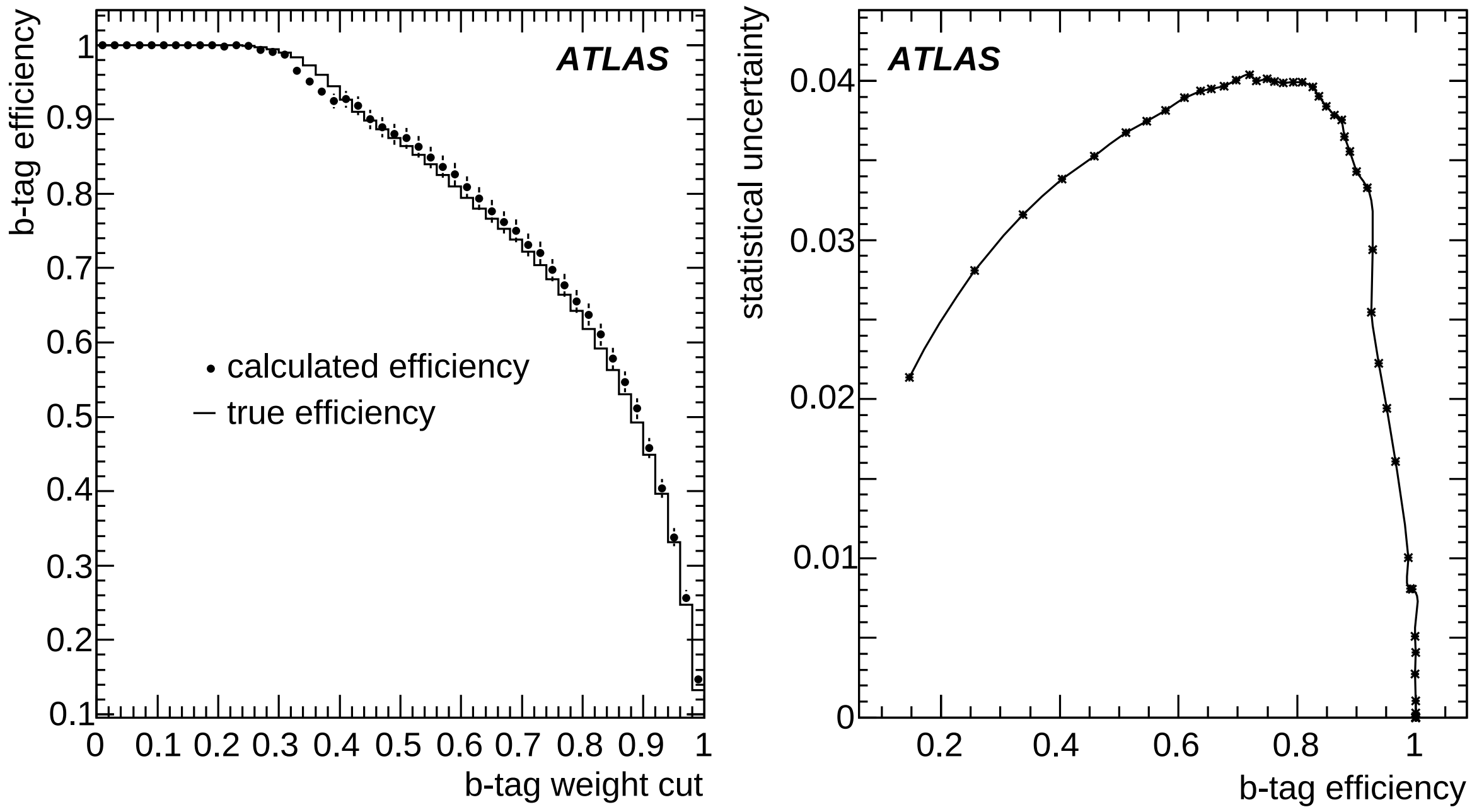

Figure 8: (a) $b$-tagging efficiency *vs.* cut on $b$-tagging weight $w$, as measured from the topological $b$-jet selection (points with error bars), and derived from Monte Carlo truth information in all $b$-jets in $t\bar{t}$ events (histogram), for $948\,\mathrm{pb}^{-1}$ of simulated $t\bar{t}$ plus background events; (b) Estimated statistical uncertainty on the measured $b$-tagging efficiency as a function of tagging efficiency, for $200\,\mathrm{pb}^{-1}$.

The 18 variables used in the likelihood are described below. The two templates based on the hadronic t-quark are shown in Fig. 9.

**1-4. hadronic $W$ boson and top quark $p_T$ and mass.** The two light quark four-momenta are added together to obtain the hadronic $W$ boson four-momentum. The $b$-quark four-momentum is then added to obtain the hadronic top quark momentum.

**5-6. leptonic top quark $p_T$ and mass.** The transverse momentum of the neutrino is taken to be the negative of the total transverse momentum of the event. The $z$-component of the neutrino momentum is unknown, so is determined by assuming the mass of the $W$ boson (80.403 GeV) and solving the resulting quadratic equation. Since there are two solutions, the smaller solution is chosen. If the root is imaginary, the real part is used. The energy of the neutrino is taken from the neutrino momentum assuming the neutrino is massless. The leptonic top quark four-momentum is then determined from the lepton, neutrino and associated $b$-quark jet.

**7-18. hadronic and leptonic $b$-jets and 2 jets from $W$-decay: rank, $E_T$, and $\Delta\phi(b_{had},$lep$)$.** For each of the four jets assigned to the hadronic $W$ quarks and the two $b$-quark, kinematic variables are used to construct additional templates. The jets are ranked by $E_T$, with rank one denoting the jet with the highest $E_T$. This jet rank, along with jet $E_T$ and the separation in $\phi$ between the jet and the lepton ($\Delta\phi$(jet,lep)) are used as additional templates.

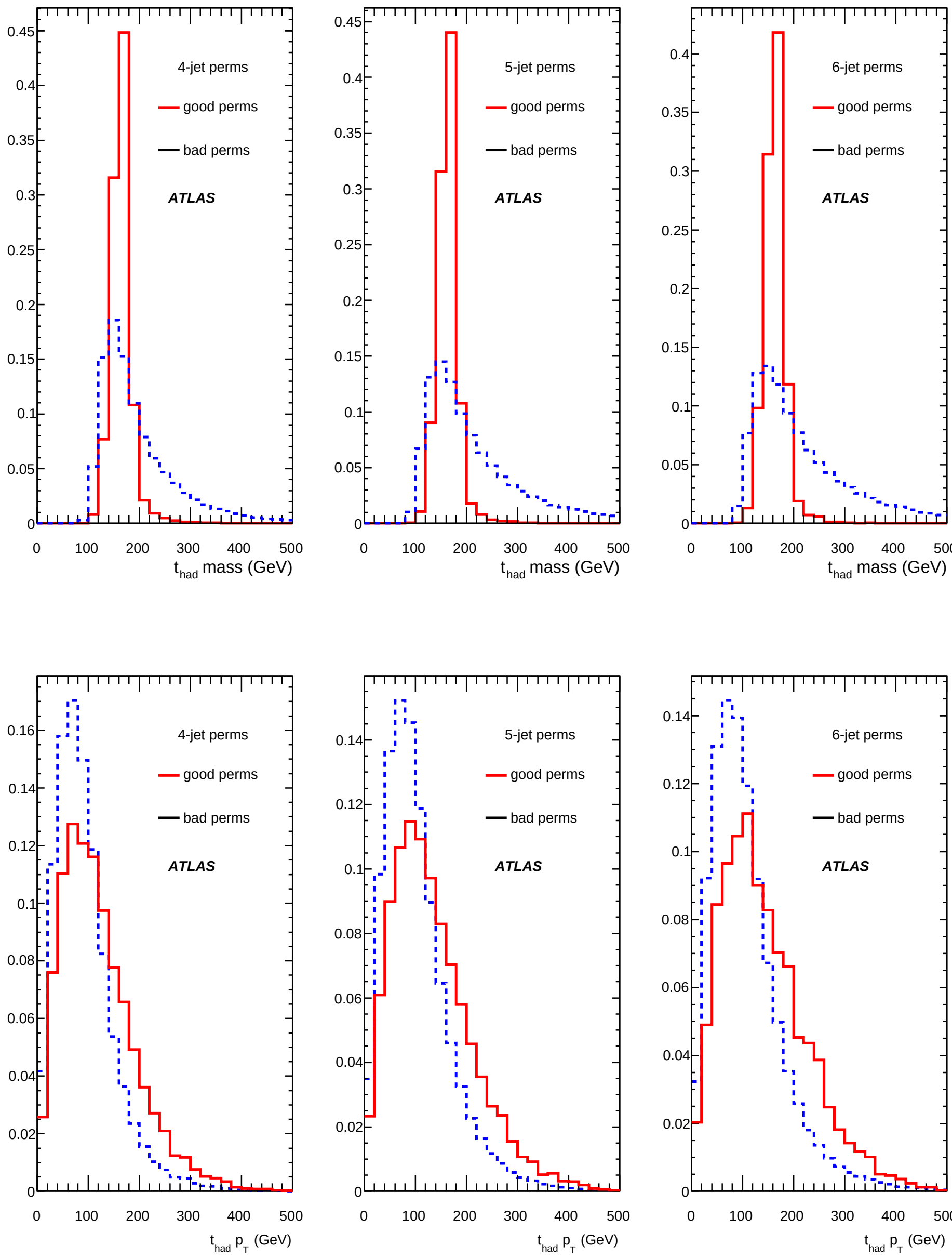

Figure 9: Templates for the likelihood-based selection, using (a) hadronic $t$-quark mass and (b) hadronic $t$-quark $p_T$.

### 5.2.1 Event selection

The $b$-purity of the basic sample selected as discussed in Section 2 is enhanced by requiring one jet to be $b$-tagged, with an IP3D+SV1 weight $w > 6$. Then a discriminant value $\mathscr{D}$ is calculated for each jet combination in the event using the templates, which are taken to be likelihood distribution functions for each variable. The 'correct combination' templates are used to get the '$s$' values for the $i^{th}$ variable, and

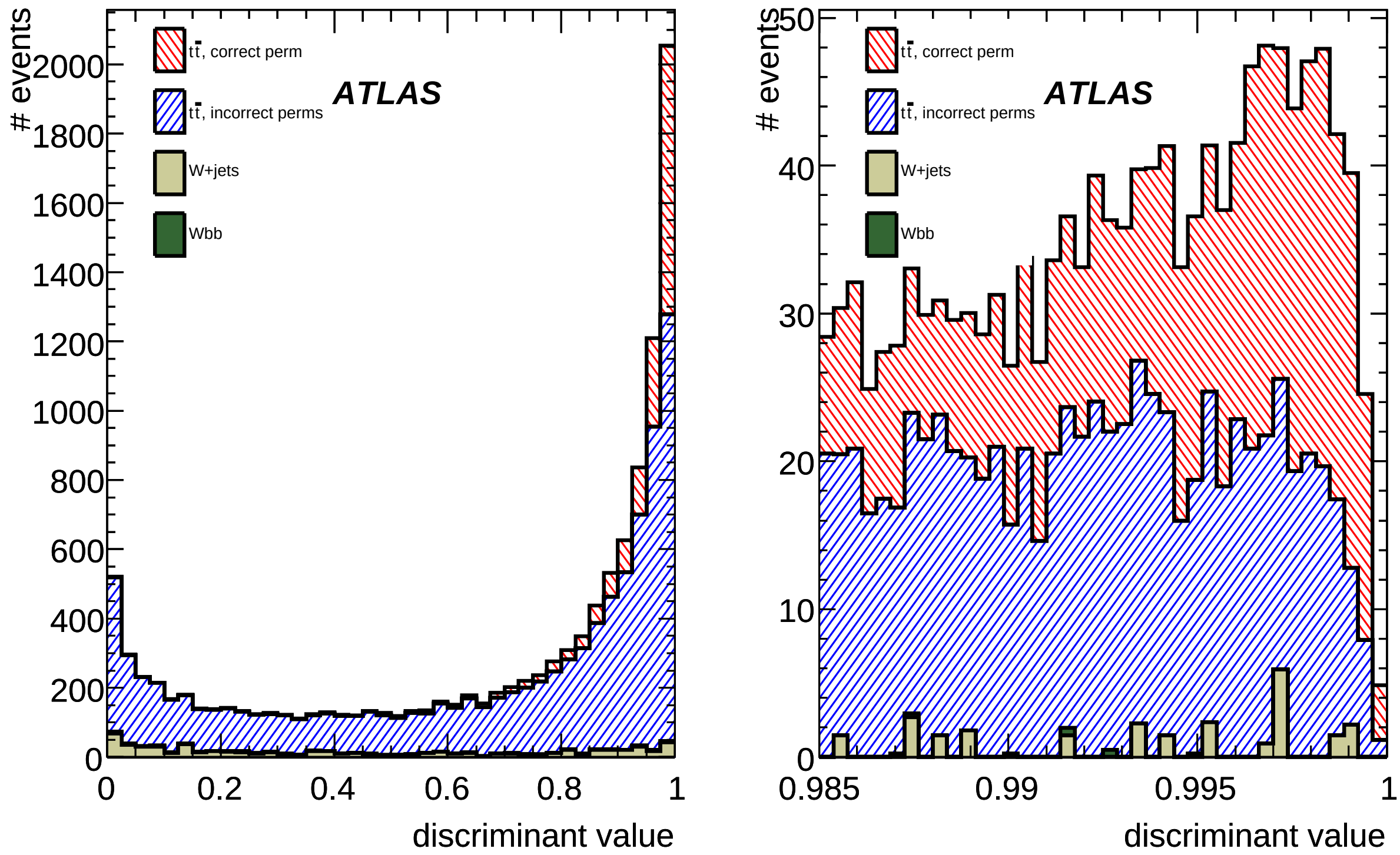

Figure 10: Discrimination between correct and incorrect permutations, best permutation chosen for each event, for entire range of $\mathscr{D}$, and for $\mathscr{D} \geq 0.985$. Events are chosen which have at least 1 $b$-tagged jet.

the 'incorrect combination' templates are used to get the '$b$' values. They are combined according to:

$$\mathscr{D} = \frac{\prod_i (s_i/b_i)}{1 + \prod(s_i/b_i)}. \tag{7}$$

In each event, the combination with the largest value of $\mathscr{D}$ is retained. The resulting distributions are shown in Fig. 10, normalised to 100 pb$^{-1}$. The contributions from the correct combination, incorrect combinations and non-$t\bar{t}$ events are shown separately. As can be seen from the right-hand plot, very high discriminant values are required to select a sample with high purity.

The $b$-jet purity is shown *vs.* the number of selected events in Fig. 11, for several ranges of jet $E_T$. Higher purity samples can be obtained by tightening the cut on $\mathscr{D}$, but at the expense of selection efficiency. A cut of $\mathscr{D} > 0.985$ was chosen for these studies assuming 100 pb$^{-1}$ samples, which could be raised to improve the purity once the data sample size allows.

### 5.2.2 Background subtraction

Assuming the cuts used to discriminate between correct and incorrect jet-parton assignments reduce the background from non-$t\bar{t}$ events to a negligible level, the primary background to be considered arises from jets incorrectly assigned to $b$-quarks. These jets can be from light, $c$, or $s$-quarks from $W$ decay, and from ISR/FSR jets. To estimate the effect of the background on the $b$-tagging efficiencies, it is necessary to determine the tagging rate of jets in background events, as well as the fraction of signal (and hence background) events in the data sample. Using Monte Carlo truth information to determine the true flavour of the jet assigned to the leptonic top decay (and assumed to be a $b$-jet), signal ($S$) and background ($B$) templates are formed, binned by discriminant value. Using these templates, a likelihood is constructed to find the fraction $f_{sgn}$ of signal events. The likelihood formed from data ($D$) and templates is:

$$\ln \mathscr{L}(f_{sgn}) = \sum_{i}^{\#bins} -D_i \, \ln(f_{sgn}S_i + (1 - f_{sgn})B_i)\,, \tag{8}$$

where $D_i$ is the number of data events in the $i^{th}$ bin, and $S_i$ and $B_i$ are the values of the $i^{th}$ bins of signal and background templates. The associated background $b$-tagging efficiencies are determined using events passing a looser discriminant cut ($\mathscr{D} > 0.9$), since the tag rate for jets in the background do not depend strongly on the value of this cut. These are shown for several ranges of jet $E_T$ in Fig. 12. The average of the $b$-tagging efficiencies of non-$b$-jets over all jet $E_T$ bins is used, giving $\varepsilon_{btag}^{\text{bkgd}} = 0.021 \pm 0.004$. Once the $b$-tagging efficiencies for background events and $f_{sgn}$ are obtained, the following equation is used to solve for $\varepsilon_{\text{btag}}^{b\text{-jets}}$:

$$\varepsilon_{\text{btag}}^{b\text{-jets}} = \frac{1}{f_{sgn}} \left( \varepsilon_{\text{btag}}^{\text{all jets}} - (1 - f_{sgn})\varepsilon_{\text{btag}}^{\text{bkgd}} \right) \tag{9}$$

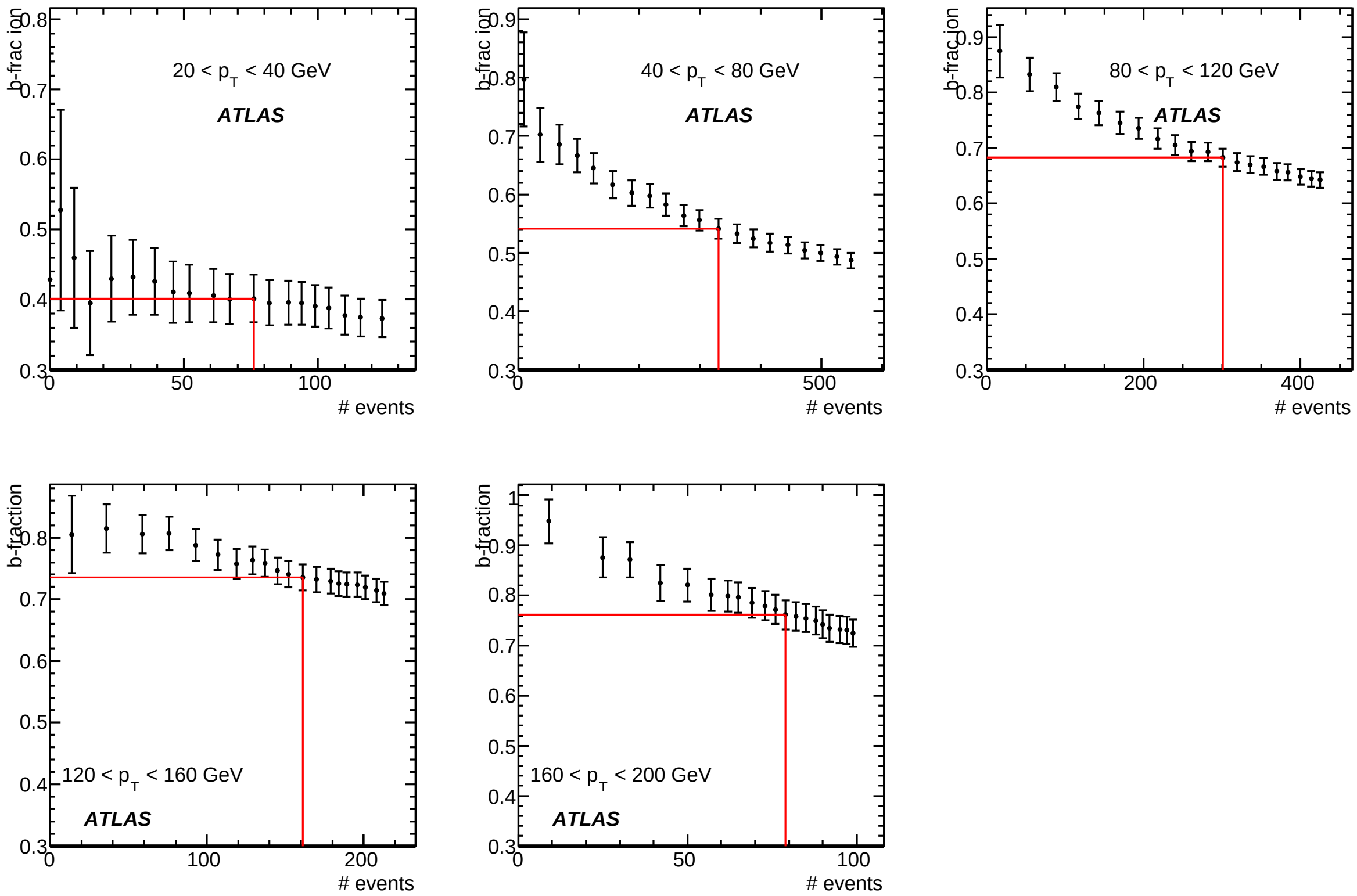

Figure 11: Likelihood selection: $b$-jet purity vs. number of selected events (discriminant cut value ranging from 0.975 to 1) for each $b$-jet $E_T$ bin.

### 5.2.3 Results for $b$-tagging efficiency measurement

The $b$-tagging efficiencies are determined for events binned in $b$-jet $E_T$, with bin sizes chosen to match the kinematic and topological methods. To simulate the $100\,\text{pb}^{-1}$ sample, histograms have been scaled to the number of events expected in the sample. The sample purities are shown for the $E_T$-binned sample in Table 6 for comparison with the other methods, and the $b$-tagging efficiencies are shown in Fig. 13. The average of $b$-tagging efficiencies for jet $E_T > 40\,\text{GeV}$ is $0.674 \pm 0.086$, consistent with the actual value of 0.658.

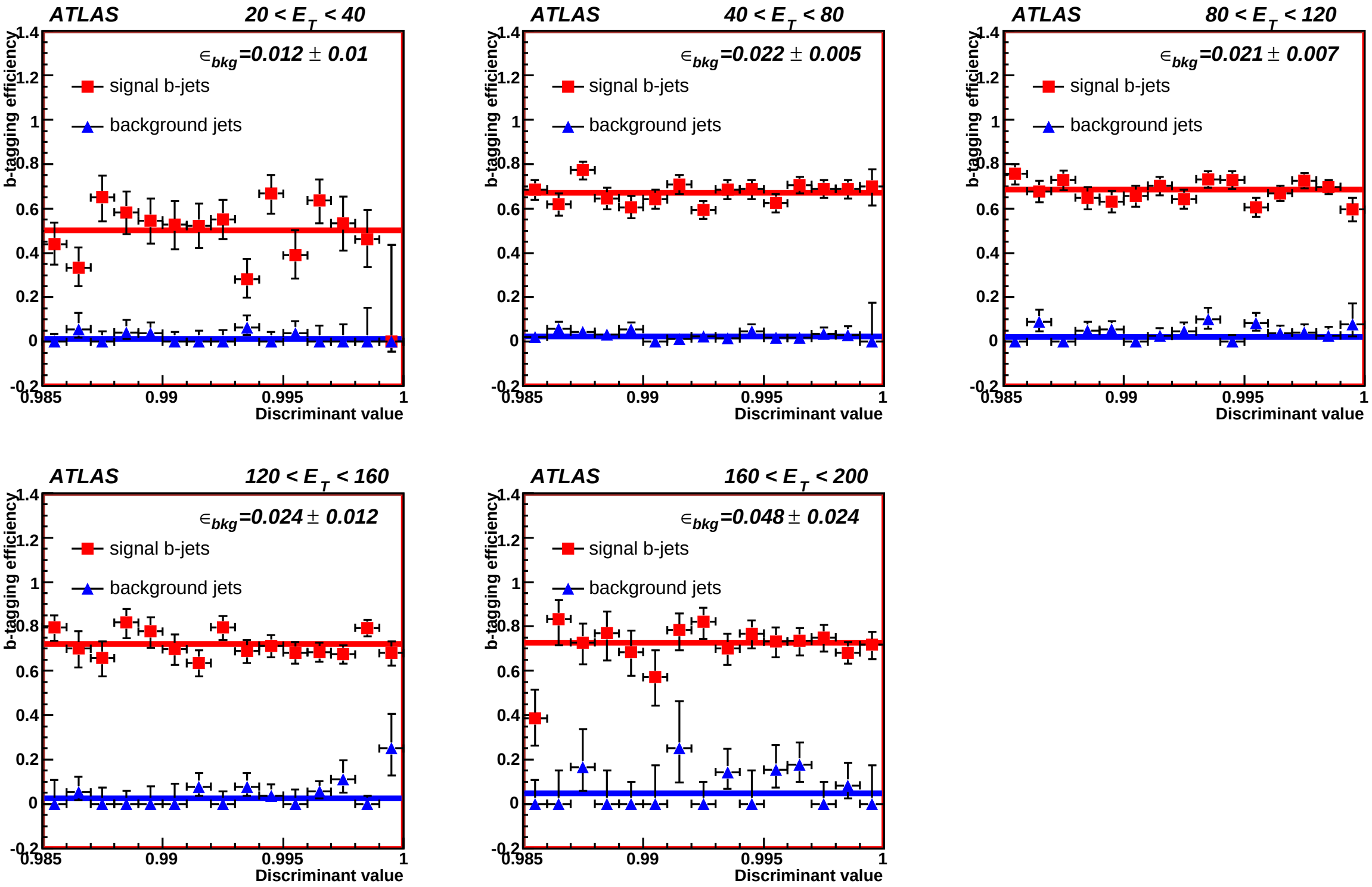

Figure 12: $b$-tagging efficiencies for signal ($b$-jets), and tag rate for light and $c$-jets (background) for events passing a discriminant cut of 0.9. Each plot was done for a different range of jet $E_T$.

Ensemble testing was done using 100 pseudo-experiments, verifying the same central value and error. The pull distribution of the 100 pseudo-experiments has a width of 0.60 standard deviations, so the error (0.086) is scaled by 0.60 to get an expected statistical error 0.05 for the $100\,\text{pb}^{-1}$ sample. The low pull width arises because, for some pseudo-experiments in the ensemble, likelihood fits of signal fractions are close to 0 or 1. Failing to determine asymmetric errors properly for these pseudo-experiments results in an overestimation of the error.

Systematic uncertainties are determined for the effects discussed in Section 4.3, using the full available MC statistics for all except the effect of background events in the sample. Ensemble testing was used to evaluate this error, using double the expected number of events and no background events. In general, the systematic errors are large for the cuts chosen for the $100\,\text{pb}^{-1}$ sample. These are expected to improve once data samples are available and the discriminants are optimised, especially once the sample size allows tighter jet $E_T$ and discriminant cuts to be used in order to reduce the background from non-$t\bar{t}$ and wrong combinations.

## 5.3 Kinematic selection

This section describes the selection of a high purity $b$-jet sample through the use of a kinematic fit and cuts based on the resulting fit $\chi^2$. Other selection criteria, including the use of a $b$-tag on the hadronic top

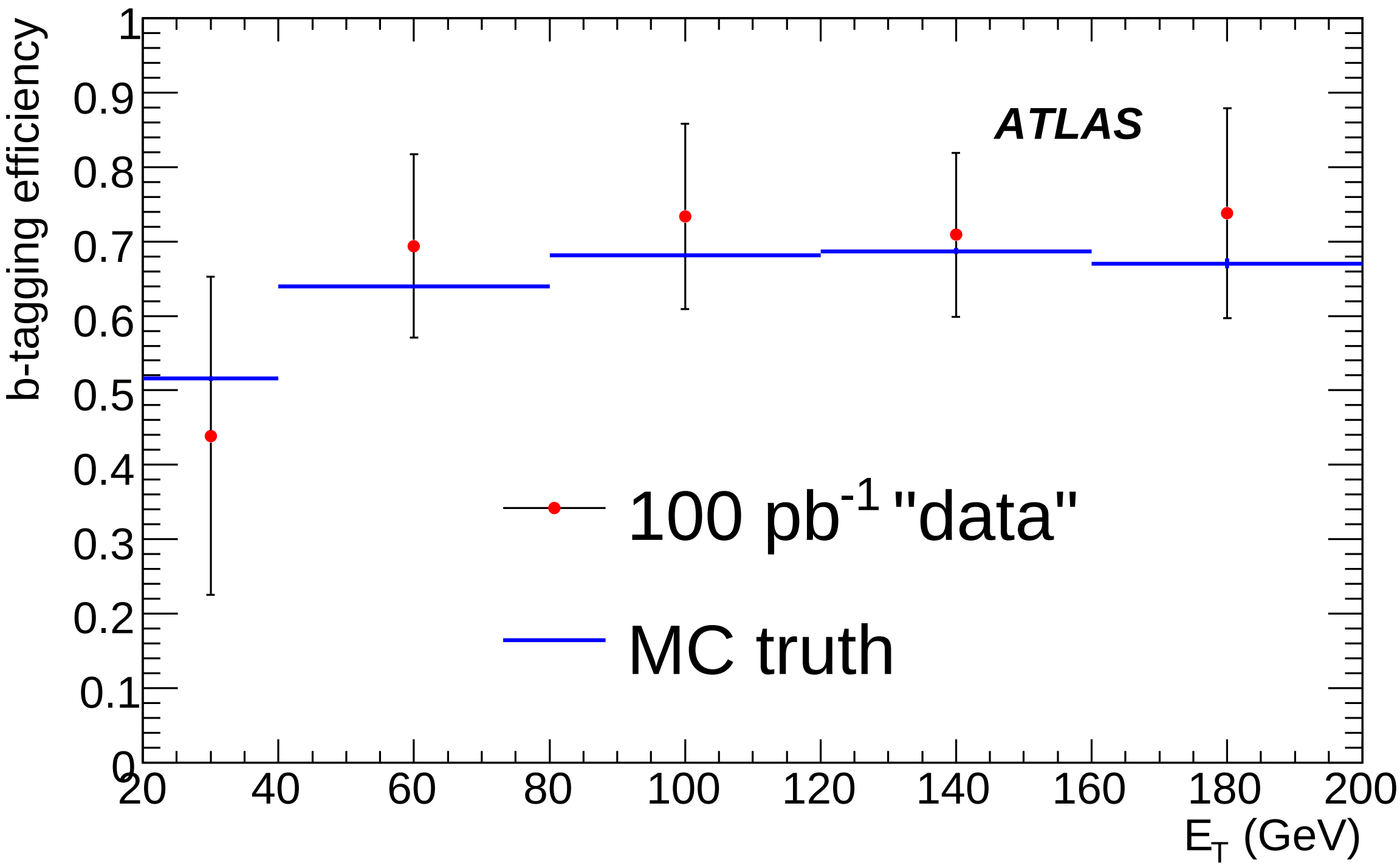

Figure 13: Likelihood method: simulation of $b$-tagging efficiencies vs. $E_T$ for a 100 $\text{pb}^{-1}$ sample using the best permutation for events selected with a discriminant cut. The solid line shows the true $b$-tagging efficiencies from the full MC sample using truth information. ($b$-tag weight=6.0, discriminant cut 0.985)

$b$-jet are also used to enhance the purity. As with the other methods described above, it is difficult to get high purity whilst maintaining sufficient statistics, and a data-based background subtraction procedure is employed to derive an effective pure $b$-jet sample.

#### 5.3.1 Kinematic fit and performance

The analysis uses the kinematic fit HITFIT [11], originally developed at DØ. A $W$ mass constraint is included and the two reconstructed tops are constrained to have the same mass. The top mass can also be constrained. For this study the mean top mass obtained with the unconstrained fit (172 GeV) was used as a constraint in subsequent fits. The mean top mass obtained from the fit was 172 GeV. Both neutrino solutions are considered and the one with the lower $\chi^2$ is used. The fit requires calibrations of the jet energy to the parton energy and resolutions of the jets and leptons, which were determined by comparing reconstructed quantities with the Monte Carlo truth information.

Only the four largest $E_T$ jets in the event were used in the kinematic fit, giving rise to 12 permutations which were each fitted. The correct combination is not always in the four leading jets. In 49 % of the events passing the basic semileptonic event selection discussed in Section 2, the four jets from the $t\bar{t}$ system are reconstructed and pass the jet selection cuts. Out of the events where all four jets are reconstructed 44 % contain the ttbar jets in the four leading jets, rising to 92 % of events if the first six jets are used. However, this leads to a large increase in combinatorial background and was not considered further.

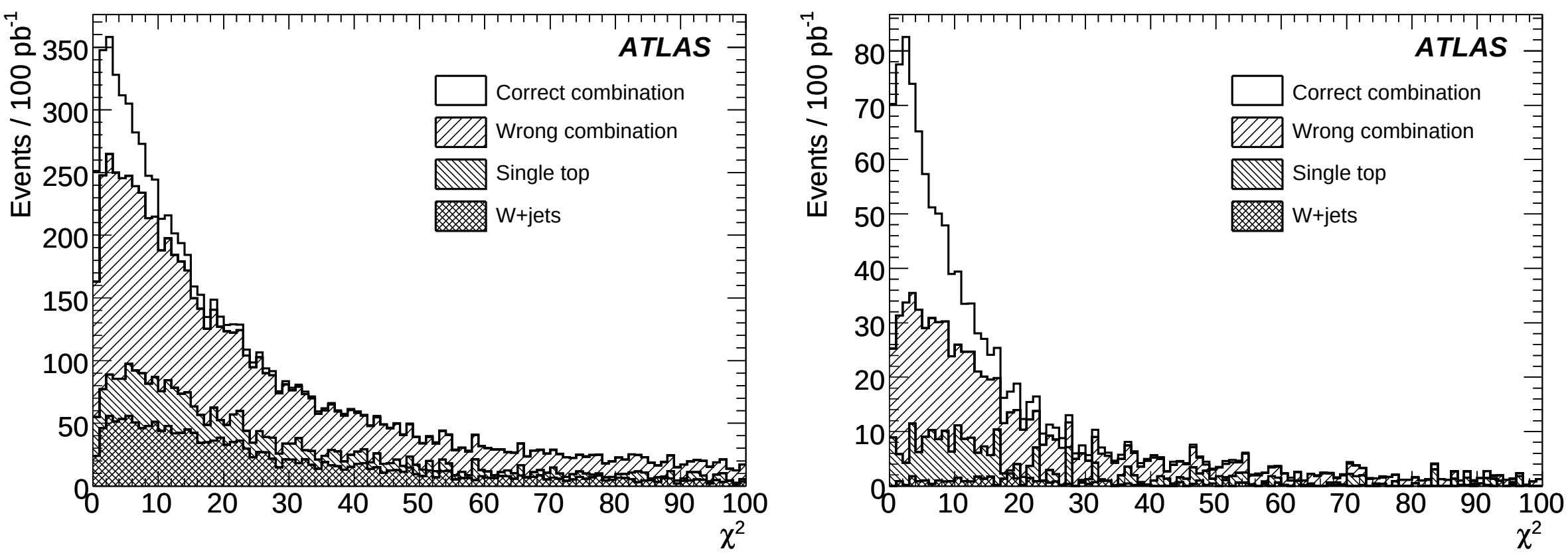

Figure 14: Kinematic selection: Distribution of the $\chi^2$ for the permutation with the minimum $\chi^2$ showing contributions of the signal and backgrounds with standard selection cuts (left plot), and with additional requirements according to selection $S_3$ described in the text (right plot).

From the four leading jets, the permutation with the lowest $\chi^2$ was chosen, and events where this $\chi^2$ was below a particular cut value were retained. If the four leading jets are four jets from the $t\bar{t}$ system, then the correct permutation is the one with the lowest $\chi^2$ in 58 % of the cases. Figure 14 (left) shows the $\chi^2$ distribution for the selected combination, including the contributions from correct and wrong combinations in $t\bar{t}$ events, and from $W$+jets and single top backgrounds. It can be seen that the fraction of correct combinations is not high, even for low $\chi^2$ values. Several additional selections beyond the 'standard cuts' $S_0$ defined above were tried in order to enhance the purity as follows. Selection $S_1$ requires a $b$-tag with weight $w > 5$ on the jet assigned as the hadronic $b$-jet; the effect of this is mainly to reduce the $W$+multijet background. Selection $S_2$ adds a $b$-tag veto ($w < 5$) on both jets assigned to the hadronically decaying $W$. Selection $S_3$ also requires there are six or less jets with $E_T > 20$ GeV, and selection $S_4$ requires that the hadronic top $p_T$ satisfies $p_T > 150$ GeV. The relative reduction of wrong combinations and background can be seen in the $\chi^2$ distribution after selection $S_3$ as shown in Fig. 14 (right).

Figure 15 shows the results of these additional selections, in terms of the fraction of jets assigned as the leptonic top $b$-jet which are really $b$-jets, and the number of selected events for 100 pb$^{-1}$. Purities of up to 90 % are reachable, but at the cost of low selection efficiency. To retain a reasonable number of events, a signal sample is defined using selection $S_3$ with a cut $\chi^2 < 10$, which leaves around 600 events for 100 pb$^{-1}$.

### 5.3.2 Background subtraction

As for the topological method, a background control sample is used to estimate the composition of the remaining background in the sample of selected jets from data. The use of the kinematic fit excludes most events with masses outside the top mass window, so it is not possible to use top mass sidebands to estimate the background composition. However, events with high $\chi^2$ are predominantly wrong combinations (see Fig. 14). The wrong-combination fraction can be enhanced by selecting events that pass the standard selection $S_0$ and requiring that one of the jets assigned to the $W$ is tagged as a $b$-jet ($w > 5$). The $\chi^2$ distribution of this control sample follows closely that of the wrong combination background in the selected jet sample. The shape of the $\chi^2$ distribution is used to predict the amount of combinatorial background in the signal sample, normalising the two samples in the region $30 < \chi^2 < 100$. Any test distribution (*e.g.* the $b$-tagging weight) extracted from the signal sample can then be corrected for this contribution of background events, taking the shape of the test distribution in background from the region of the control sample with $\chi^2 > 30$. This background subtraction method is very similar to that discussed in Section 5.1, with the high $\chi^2$ region playing the role of the top mass sideband, and the control sample with a $b$-tagged jet from the $W$ decay corresponding to the sample with high hadronic top mass in the topological selection.

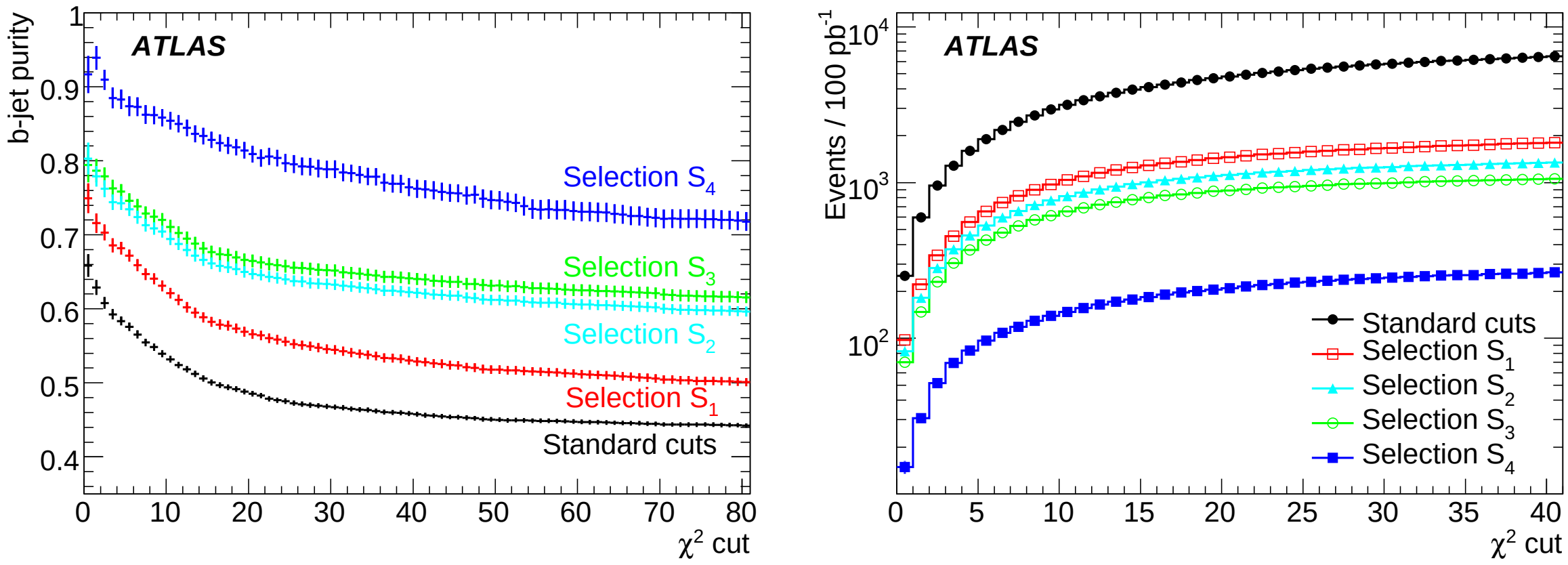

Figure 15: Left: The purity (fraction of true $b$ jets) of the jet assigned as the $b$-jet on the leptonic side as a function of the $\chi^2$ cut. Right: The corresponding number of events. The different selection criteria are described in the text.

#### 5.3.3 Results for $b$-tagging efficiency measurement

The results of applying this selection to simulated data including background and scaling the event count to $100\,\text{pb}^{-1}$ are shown in Table 6. The purity in the 20-40 GeV jet $E_T$ bin is rather low, and the background from the significant contribution of single top is not well-estimated. As for the topological analysis, this bin is not used for calculating the $b$-tagging efficiency. The combined $b$-tagging weight distribution for the other bins ($E_T > 40\,\text{GeV}$) is shown in Fig. 16. The $b$-tagging efficiency corresponding to any given cut can be calculated by integration, and the corresponding statistical errors for various tag working points can be seen in Table 7.

### 5.4 Comparison of selection methods

The performance of the different $b$-jet selection methods are compared in Table 6, which shows the number of jets selected as a function of jet $E_T$, the effective number of jets after background subtraction procedures have been performed, the corresponding purity of the original sample, and the $b$-jet fraction of the final background-subtracted sample, which should be compatible with one.

The different selections have roughly similar overall performance, with the topological selection giving relatively more jets at high $E_T$, and the kinematic selection more at low $E_T$. All selections allow a sample of several hundred $b$-jets to be selected in $100\,\text{pb}^{-1}$ of data, although the selections of low $E_T$ $b$-jets (20–40 GeV) suffer from large backgrounds.

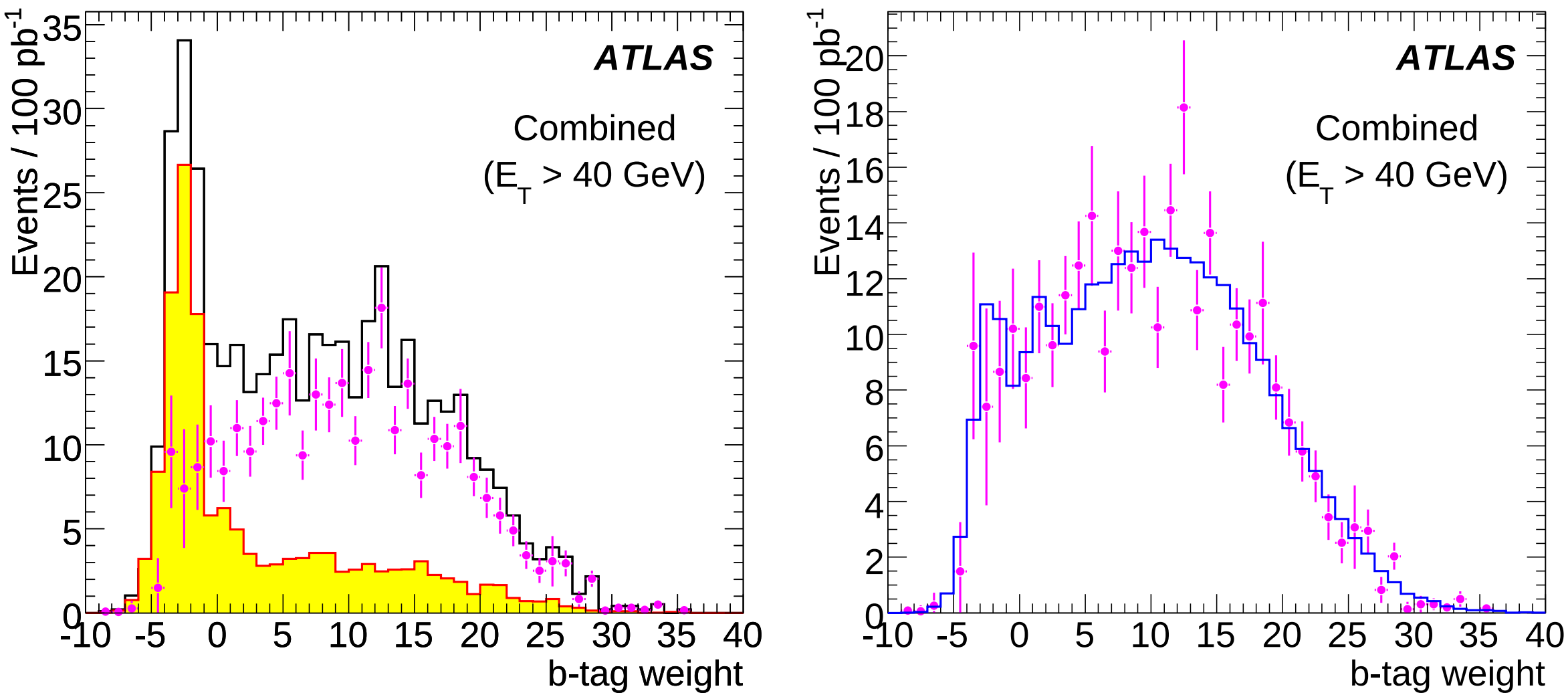

Figure 16: Kinematic selection: (Left): The $b$-tag weight distribution for the uncorrected sample (unfilled histogram), for the estimated background sample (filled histogram) and the corrected distribution calculated from the difference (data points). Right: The $b$-tag weight distribution for the corrected sample (data points) compared with the distribution for true $b$-jets (histogram). Both plots are normalised to $100\,\mathrm{pb}^{-1}$, but use $967\,\mathrm{pb}^{-1}$ of simulated data.

### 5.5 Comparison of efficiency measurements

The performance of the different $b$-tagging efficiency measurements is compared in Table 7, which shows the statistical and significant systematic errors for a true tagging efficiency of 0.6 for all methods. The systematic errors have been evaluated as discussed in [2] and Section 2. The differences in systematic uncertainties reflect the different uses made of Monte Carlo information by the various methods. The expected statistical errors for other tagging efficiencies, and the simpler IP2D tagging algorithm, are shown (where available) in Table 8.

With $100\,\mathrm{pb}^{-1}$ of data, the counting method can measure the overall $b$-tagging efficiency to a precision of better than 5 %, and, particularly as the luminosity increases through $200\,\mathrm{pb}^{-1}$ several alternative approaches are possible that also enable the efficiency dependence on other variables (*e.g.* jet $E_T$) to be studied.

## 6 Optimisation of $b$-tagging for $t\bar{t}$ events

The ATLAS multivariate $b$-tagging algorithms rely heavily on likelihood reference distributions for light- and $b$-quark jets derived from Monte Carlo simulation. The default reference distributions are produced using a mixture of physics processes. In principle, using reference histograms from Monte Carlo $t\bar{t}$ events alone could lead to an improvement in tagging performance on $t\bar{t}$ events. The selection of a pure sample of $b$-jets from $t\bar{t}$ data offers an opportunity to tune the $b$-tagging algorithms directly on data, reducing the dependence on Monte Carlo simulation. Both these topics are discussed below.

| Jet $E_T$ range (GeV) | Selected jets | Effective jets | Purity (%) | Estimated $b$ flavour purity (%) |
|---|---|---|---|---|
| Topological | | | | |
| 20–40 | 144 | 51 | 54 | $99 \pm 8$ |
| 40–80 | 340 | 189 | 67 | $92 \pm 3$ |
| 80–120 | 223 | 138 | 78 | $98 \pm 3$ |
| 120–160 | 122 | 92 | 84 | $98 \pm 3$ |
| 160–200 | 75 | 54 | 86 | $108 \pm 4$ |
| 40–200 | 819 | 515 | 60 | - |
| Likelihood | | | | |
| 20–40 | 52 | 24 | 47 | $86 \pm 6$ |
| 40–80 | 225 | 130 | 58 | $94 \pm 3$ |
| 80–120 | 204 | 143 | 70 | $97 \pm 3$ |
| 120–160 | 110 | 91 | 83 | $89 \pm 3$ |
| 160–200 | 54 | 40 | 73 | $104 \pm 5$ |
| 40–200 | 593 | 403 | 68 | - |
| Kinematic | | | | |
| 20–40 | 164 | 102 | 53 | $74 \pm 6$ |
| 40–80 | 284 | 182 | 73 | $93 \pm 4$ |
| 80–120 | 113 | 78 | 88 | $107 \pm 5$ |
| 120–160 | 38 | 30 | 90 | $103 \pm 5$ |
| 160–200 | 17 | 13 | 94 | $107 \pm 6$ |
| 40–200 | 451 | 302 | 79 | $98 \pm 3$ |

Table 6: Summary of $b$-jet selection method performance, showing the number of selected $b$-jets, the number of jets after background subtraction, the selected sample purity, and estimated $b$-jet purity of the final sample after background subtraction, for each selection method. The results correspond to $100\,\mathrm{pb}^{-1}$ of MC@NLO $t\bar{t}$ Monte Carlo plus backgrounds from $W$+jets, $Wb\bar{b}$, $Wc\bar{c}$ and single top production.

| Systematic | Counting | | Topological | Likelihood | Kinematic |
|---|---|---|---|---|---|
| | lepton+jet | dilepton | | | |
| Light jets and $\tau$ | 0.1 | 0.7 | 0.5 | 5.2 | 0.6 |
| Charm jets | 0.0 | 0.8 | 0.7 | 4.6 | 2.2 |
| Jet energy scale | 0.9 | 0.5 | 0.5 | 2.5 | 1.1 |
| $b$-jet labelling | 1.4 | 1.4 | - | - | - |
| MC generators | 0.1 | 2 | 0.2 | 5.9 | 5.5 |
| ISR/FSR | 2.7 | 2 | 1 | 2.2 | 0.5 |
| $W$+jet background | 1.2 | 0.3 | 2.8 | 9.6 | 0.3 |
| Single top background | 0.1 | 0.1 | 1.2 | - | 1.2 |
| Top quark mass | 0.3 | 0.5 | - | 4.1 | - |
| Total systematic | 3.4 | 3.5 | 3.4 | 14.2 | 6.2 |
| Statistical ($100\,\mathrm{pb}^{-1}$) | 2.7 | 4.2 | - | 5.0 | 7.7 |
| Statistical ($200\,\mathrm{pb}^{-1}$) | 1.9 | 3.0 | 6.4 | 4.4 | 5.5 |

Table 7: Summary of systematic and statistical uncertainties on the measurement of the $b$-tagging efficiency at a true efficiency of $\varepsilon_{\mathrm{true}} = 0.6$, for the counting method in lepton+jets and dilepton channels, and the topological, likelihood and kinematic jet selection methods. The uncertainties are expressed as relative errors (in %).

| Tag | $\varepsilon_{true}$ | Counting lepton+jet | Counting dilepton | Topological | Likelihood | Kinematic |
|---|---|---|---|---|---|---|
| IP3D+SV1 | 0.5 | 2.8 | 4.0 | 6.2 | 8.8 | 6.1 |
| IP3D+SV1 | 0.6 | 1.9 | 3.0 | 6.4 | 5.2 | 5.5 |
| IP3D+SV1 | 0.7 | 1.4 | 2.0 | 6.7 | 5.1 | 4.9 |
| IP2D | 0.5 | | | 7.4 | 4.7 | 6.2 |
| IP2D | 0.6 | | | 6.5 | 5.1 | 5.5 |
| IP2D | 0.7 | | | 5.6 | 5.1 | 5.0 |

Table 8: Expected statistical errors (relative errors in %) on the $b$-tagging efficiency for 200 pb$^{-1}$ measured using each technique, for IP3D+SV1 and IP2D taggers at various working points.

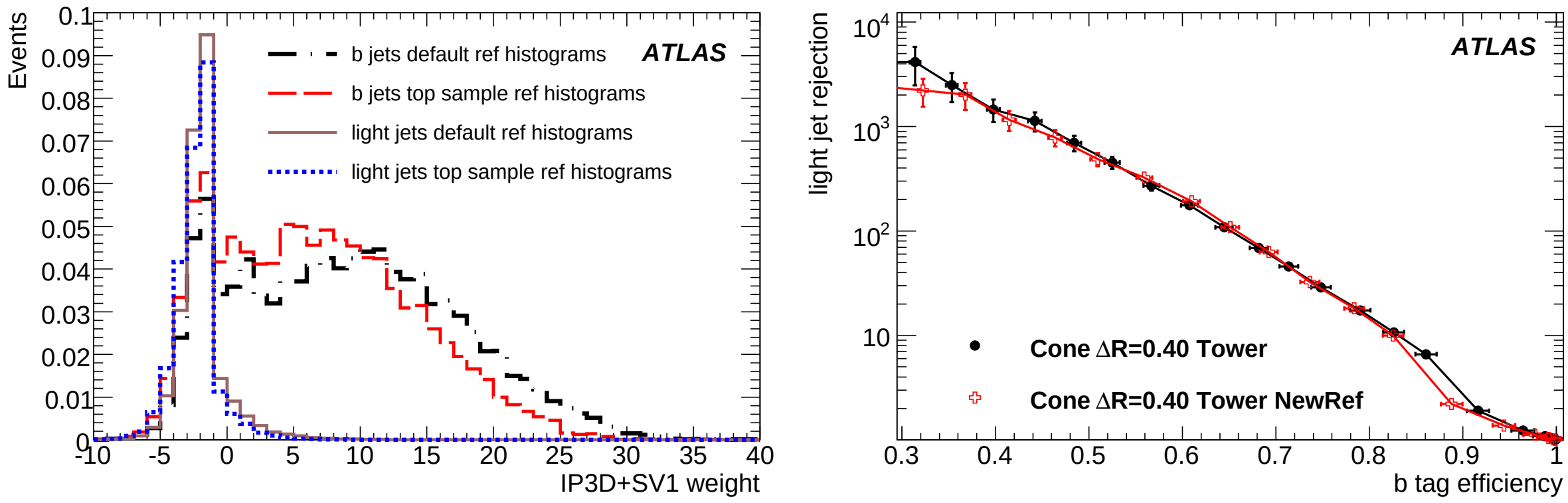

Figure 17: Comparison of b tagging weights and b tag performance using default and special top sample reference histograms.

## 6.1 Using $t\bar{t}$ specific reference histograms

For this study, reference histograms based only on $t\bar{t}$ events (without the light jet purification procedure which discards light jets biased by nearby $b$-jets [6]) were created using a total of 250k events. By contrast, the default reference histograms were produced using a variety of physics processes, light jet purification, as well as different versions of the detector geometry, simulation, and reconstruction. The following comparison does not attempt to disentangle the various effects of those differences; for more details, see [6]. The resulting $t\bar{t}$ event $b$-tagging weights are compared to those calculated using the default reference histograms, for the IP3D+SV1 $b$-tagging algorithm with the default tower-based Cone jet algorithm with $\Delta R = 0.4$. The results are shown in Fig. 17. The left plots shows the $b$-tagging weights for $b$-jets and light jets calculated with both sets of reference histograms.

The weights for both $b$-jets and light quark jets calculated with the top sample reference histograms are on average smaller than those from the default reference histograms. However, because the $b$-tagging weights are smaller for both $b$-jets and light jets, the performance should not change significantly. This is confirmed by the right plot in Fig. 17, showing the $b$-tag efficiency *vs* the light jet rejection. Within the statistical uncertainty there is no difference in the performance. So there appears to be no significant advantage in using special dedicated $t\bar{t}$ reference histograms when applying $b$-tagging to $t\bar{t}$ events.

### 6.2 Calibrating $b$-tagging using data

In principle, the $b$-jet selection methods discussed in Section 5 should allow the reference distributions for $b$-jets in $t\bar{t}$ events to be determined directly from data, removing the need for Monte Carlo $b$-jets completely. However, none of the methods produce a pure $b$-jet sample without the need for background subtraction, and hence do not identify a pure sample of $b$-jets from which multiple variables (and their correlations) can be extracted simultaneously on a jet-by-jet basis. Instead, background-subtracted distributions are produced. These can be used directly as reference distributions only for one-dimensional likelihoods with a single input variable. For $n$-dimensional likelihoods (for example the standard tagging weight which is a two-dimensional combination of IP3D and SV1 taggers), an $n$-dimensional background-subtracted distribution would be needed in order to properly model both the two input distributions and their correlation. For higher-dimensional distributions, the required data statistics would quickly become prohibitive.

At least for initial data, a more feasible approach would be to use the background-subtracted $b$-jet samples to check each of the likelihood input variables individually, to determine how well the Monte Carlo simulation models the data. Monte Carlo would then be used to form the likelihood reference distributions as before (including correlations between input variables), with the comparison with data being used to assess systematic uncertainties, *e.g.* by reweighting Monte Carlo events to more precisely follow the data distributions if discrepancies are seen.

This technique is illustrated in Fig. 18, which shows background-subtracted distributions of five variables related to the $b$-tagging weight computation, derived from the $b$-jet sample selected using the topological technique of Section 5.1, compared to the true distributions from Monte Carlo. The integrated luminosity is 948 $\mathrm{pb}^{-1}$. The first two variables are IP3D, the weight derived from the 3D-impact parameters of all tracks in the jet; and SV1, the weight derived from the SV1 secondary vertex reconstruction algorithm. The others are all input variables used to form the SV1 secondary vertex weight: N2Track, the number of two-track vertices found; SVMass, the mass of the reconstructed secondary vertex, and SVEFrac, the energy fraction of the jets associated to the vertex. The spikes around zero for these last three variables correspond to jets where no secondary vertex was found. The variables have been transformed linearly onto the range [0,1] to simplify the subtraction procedure.

It can be seen that the background-subtracted distributions are consistent with the Monte-Carlo truth, demonstrating the validity of the method. It would be straightforward to construct reweighting factors from these distributions, to explore possible systematic variations. However, it should also be noted that the techniques discussed in Section 5.5 already allow the $b$-tagging efficiency to be measured with minimal dependency on the Monte Carlo—the main use of this technique would then be to understand how to improve the Monte Carlo simulation of $b$-jets, which could help to increase the $b$-tagging performance in $t\bar{t}$ events by giving each variable its optimal weight in the likelihood.

## 7 Conclusions

The studies of jet reconstruction presented above show that the $b$-jet energy resolution in $t\bar{t}$ events is not very sensitive to the jet algorithm and parameters chosen. The jet angular resolution is more sensitive, degrading in particular with large jet cone sizes.

With 100 $\mathrm{pb}^{-1}$ of data, several methods are available to make useful measurements of the $b$-tagging efficiency in $t\bar{t}$ events. The tag counting method can reach an overall relative precision of around 5 %, in both lepton+jets and dilepton channels. However, this method only gives an 'integrated' measurement, valid for the $E_T$ and $\eta$ distribution of $b$-jets selected by the $t\bar{t}$ event selection. Monte Carlo techniques would have to be used to extrapolate this measurement for other jet $E_T$ and $\eta$ distributions, perhaps also incorporating information derived from studies of dijet samples [1].

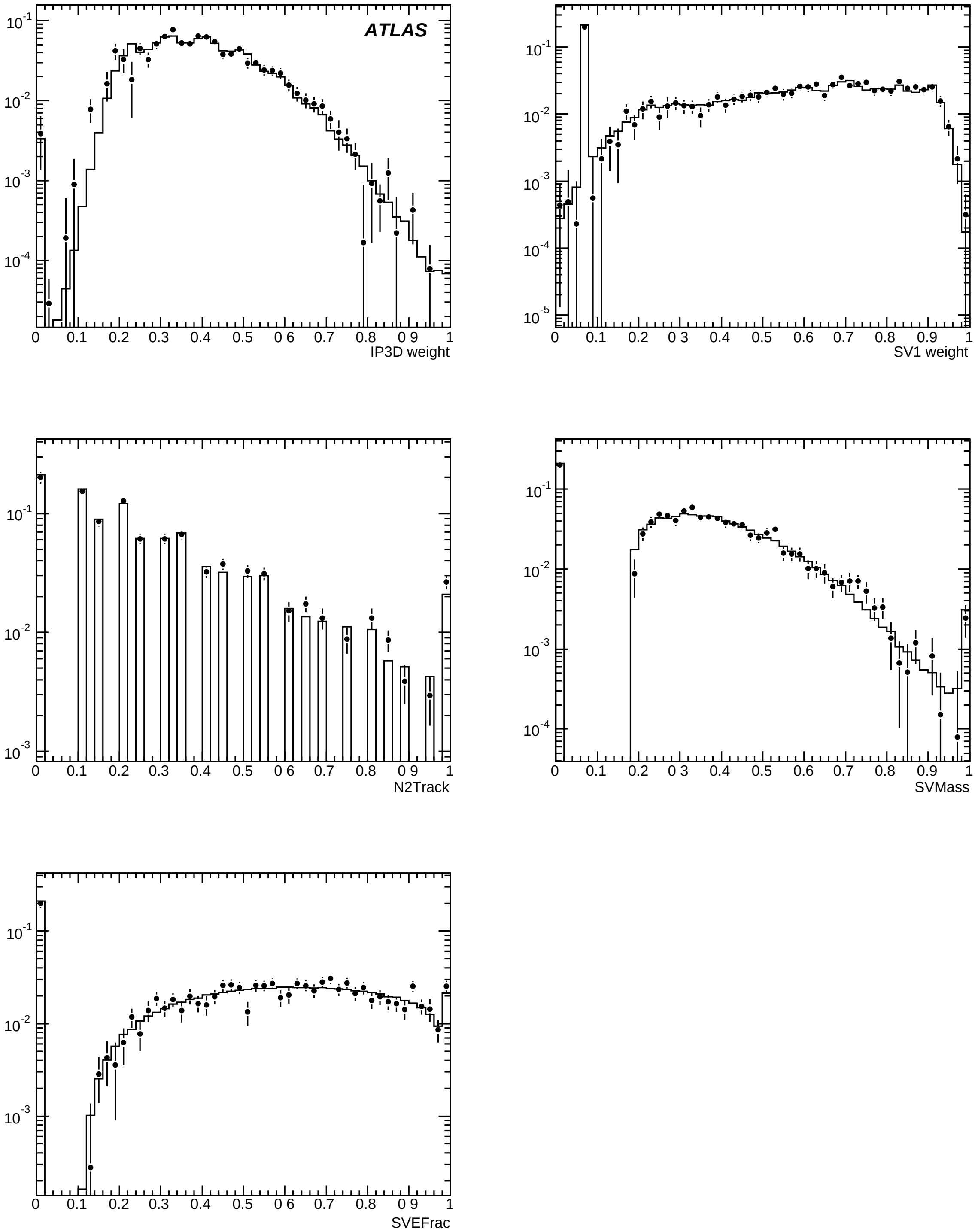

Figure 18: Background-subtracted $b$-tagging variable distributions derived from the $b$-jet sample selected by the topological method, with $948\,\mathrm{pb}^{-1}$ of simulated $t\bar{t}$ plus background data. The derived distributions are shown by the points with error bars, and the Monte Carlo truth for an unbiased sample of $b$-jets is shown by the solid histograms.

Other techniques, based on the topology and kinematics of $t\bar{t}$ events, can be used to select a sample of jets enriched in $b$-flavour, which can then be used to measure the $b$-tagging efficiency as a function of other variables. This requires that the remaining non-$b$ background in the selected jet samples be estimated and removed, and a variety of techniques are available for doing that, based either on data or Monte Carlo. These techniques are less statistically powerful than the tag counting method, achieving overall $b$-tagging uncertainties of around 10 % with $100\,\mathrm{pb}^{-1}$, but will become increasingly powerful as more data becomes available.

Calibrating the $b$-tagging likelihood reference distributions from Monte Carlo $t\bar{t}$ events rather than the generic event mixture used by default does not significantly affect the performance of the standard algorithms, which demonstrates the robustness of the $b$-tagger. On the other hand, the $b$-jet selection methods discussed above also allow reference distributions from $t\bar{t}$ data to be obtained; with moderate integrated luminosity these are likely to be more useful to cross-check the Monte Carlo description of likelihood input variables rather than as a direct input to $b$-tag calibration.

# $b$-Tagging Calibration with Jet Events

**Abstract**

This note describes two strategies for data-based measurement of the semi-leptonic $b$-tagging efficiency of ATLAS lifetime $b$-tagging algorithms using jet data. The $p_{T,rel}$ method uses templates of the muon $p_T$ relative to the jet+muon axis for bottom, charm, and light jets to estimate the $b$-quark content of a jet sample. The $b$-tagging efficiency is obtained by measuring the $b$-quark content before and after tagging a sample. The second method, System 8, uses two samples of jets of differing $b$-quark content and two uncorrelated tagging algorithms to form a system of 8 equations and 8 unknowns, one of which is the $b$-tagging efficiency. Both methods use a Monte Carlo scale factor to convert the measured semi-leptonic efficiency to an inclusive efficiency. Both methods give results binned in $p_T$ and $\eta$. Good agreement was found between the $b$-tagging efficiencies determined by both techniques and true Monte Carlo for $15 < p_T < 80$ GeV, however above 80 GeV both methods have difficulty. Initial systematics studies have been performed which indicate that it should be possible to control the absolute error on $b$-tagging efficiency to 6% for both methods. This error does not include the error associated with converting the semi-leptonic efficiency to the inclusive efficiency.

## 1 Introduction

Many analyses at the LHC will rely on the presence of $b$-quarks. Top production, Standard Model Higgs boson searches, and searches for physics beyond the Standard Model, for example, all have final states involving $b$-quarks. To estimate the backgrounds from well known Standard Model processes after applying $b$-tagging algorithms it is necessary to know the tagging efficiency for $b$-jets, for light-jets (fake rate or mistagging rate), and for $c$-jets (charm-tagging efficiency). All three efficiencies should be measured as accurately as possible: systematic errors in the $b$-tagging calibration can translate to large errors in an analysis. Although the fake rate and $c$-tagging rate are important, only the $b$-tagging efficiency is considered in this note.

In order to measure the $b$-tagging efficiency we must know the bottom quark content ('$b$-content') of a calibration sample well. Unfortunately, it is difficult to select a pure $b$-jet sample in the data. Two independent methods to calibrate the $b$-tagging algorithms are currently under study in ATLAS. The first method uses a high purity sample of top quark events. Full kinematic reconstruction enables the proper identification of the jet resulting from the fragmentation of the $b$-quark from the top quark decay. A tagging algorithm can be run on that sample of jets and the efficiency measured [1]. Similarly, one can count tags in a high purity sample of top events and estimate both the tag rate and the cross section [1].

This note describes the determination of the $b$-tagging efficiency in QCD jet data. Unfiltered QCD jet data have a very small fraction of $b$-jets at small jet energy and small effective $\sqrt{s}$, however. In order to increase the fraction of $b$-jets we require that jets in the sample contain a muon. Though muons come from other sources, a major source is the semi-leptonic decay of $b$-quarks or $c$-quarks resulting from the $b$-quark decays. A consequence is that only the lifetime tagging efficiency in semi-leptonic decays of $b$-jets is measured by the techniques described here. The semi-leptonic efficiency must be corrected to obtain an inclusive $b$-tagging efficiency. The scaling is determined from Monte Carlo; this note only briefly touches on the determination of this scale factor.

Two standard lifetime-based $b$-tagging algorithms, IP2D and IP3D+SV1 [2], are used to demonstrate the calibration methods and estimate systematic errors. Briefly, IP2D uses reconstructed tracks and their transverse impact parameters to look for jets with displaced tracks inconsistent with light-jets. IP3D+SV1 is a combined tagger: IP3D is similar to IP2D, but also takes the longitudinal impact parameter into account; SV1 reconstructs a secondary vertex from tracks near a jet. SV1 uses a likelihood made up of the invariant mass of the tracks associated with the secondary vertex, the ratio of the energy of tracks in the secondary vertex to the energy of tracks associated to the jet, and the total number of two-track vertices reconstructed in a jet. Methods that measure the *b*-tagging efficiency that work for these particular taggers are expected to work for other lifetime taggers as well.

Two separate calibration techniques are described in this note:

- The $p_{T,rel}$ method, described in Section 3, uses Monte Carlo-derived templates, for *b*-, *c*-, and light-jets, of the relative $p_T$ of a muon with respect to the jet+muon axis. The *b*-content of a jet data sample can then be determined by fitting the $p_{T,rel}$ distribution of the data with these templates before and after the lifetime tagging algorithms are applied. The *b*-tagging efficiency is derived from the changing *b*-fraction.

- The System 8 method, described in Section 4, employs two samples with different *b*-content and two uncorrelated tagging algorithms to construct a system of 8 nonlinear equations and 8 unknowns. One of the unknowns is the *b*-lifetime tagging efficiency.

Both of these techniques require a large sample of jets with muons. Section 2 describes a dedicated trigger that will be used to collect this sample along with the Monte Carlo samples and selection cuts used in this study.

The performance of the tagging algorithms varies with both jet $p_T$ and $\eta$ [2]. This is caused by both geometrical acceptance effects and variations in the track reconstruction efficiency. Both methods described in this note measure the efficiency as a function of $p_T$ and $\eta$. The dependence on other variables could also be measured if required.

The efficiency measured from jet data is largely uncorrelated with that measured from the $t\bar{t}$ data, since the data sets and the techniques are uncorrelated. Thus the results obtained with the two methods can be combined to further improve the understanding of *b*-tagging and reduce associated systematic errors. At low jet energies the $t\bar{t}$ method is rather sensitive to background contamination, a region where the jet method will perform best. At higher jet energies the $t\bar{t}$ method will have a better accuracy.

## 2 Samples and selection

Several sets of Monte Carlo QCD jet events were generated specially for this study. The jet samples were generated with the standard dijet process using the PYTHIA Monte Carlo generator and full ATLAS detector simulation and detector reconstruction software. ATLAS splits generation of its QCD samples by requiring the hard scatter parton $p_T$ to be within a certain range: between 17 and 35 GeV, between 35 and 70 GeV, between 70 and 140 GeV, and between 140 and 280 GeV. Approximately 100,000 events were generated in each range. After generation the samples are combined for the analysis.

The *b*-jet statistics in the QCD sample are not sufficient to test the methods discussed here. To increase statistics we also generated *muon+jet samples*, which further required in each event a muon from any source with a true $p_T > 3$ GeV and $|\eta| < 2$. The $p_T$ cut was chosen as a representative lower limit for muons that can accurately be reconstructed and identified (we make a reconstructed muon $p_T > 4$ GeV cut). The requirement of a muon has a dramatic effect on the flavor composition of the sample, increasing the *b*-jet content by about a factor of 10, and increasing the *c*-jet content by about a factor of 5.

Identification of electrons in jets is more challenging than that of muons, and so in the studies reported here we restrict ourselves to the muon channel.

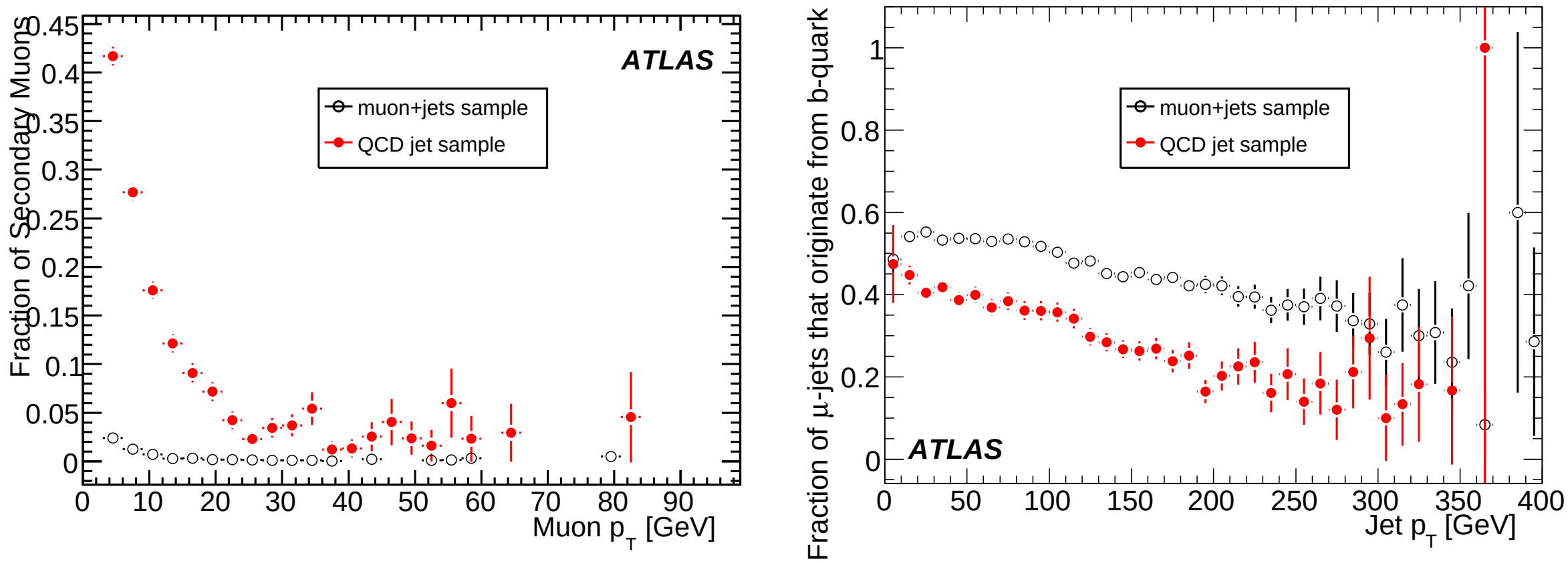

Figure 1: The left plot shows the fraction of all muons with reconstructed $p_T > 4$ GeV that are secondary muons vs muon $p_T$ in the QCD samples (filled circles) and in the muon+jet samples (open circles). The difference is caused by the generator-level filtering of the $\mu$ samples before ATLAS simulation has a chance to create the secondary muons. The right plot shows the fraction of muon-tagged jets that are due to a $b$-quark as identified by the default Monte Carlo labeling algorithm. In both plots, the black open circles represent muon-tagged jets found in the muon+jet sample and the red filled circles represent muon-tagged jets found in the QCD jet sample.

## 2.1 Selection cuts

This analysis depends, primarily, on two reconstructed objects: jets and muons. Jets are found as $\Delta R < 0.4$ cone jets with a $p_T > 15$ GeV [3]. Jets are calibrated with a standard jet energy scale calibration (and the jet $p_T$ includes the muon $p_T$) [4]. Muons are reconstructed from tracks in the outer muon detectors and must match an inner detector track with a fit $\chi^2 < 10$. The muons must have a reconstructed $p_T > 4$ GeV [5]. A *muon-tagged jet* is a jet with a muon contained within a cone of $\Delta R < 0.4$ of the jet axis.

The *flavor* of each jet must be determined ($b$-, $c$-, or light-jet) in Monte Carlo. The ATLAS flavor labeling algorithm [2] is used. A jet is labeled as a $b$-jet if a $b$-quark with $p_T > 5$ GeV is found in a cone of size $\Delta R = 0.3$ around the jet direction. If no $b$-quark is present, but instead a $c$-quark is found then the jet is labeled as a charm jet. Remaining jets are labeled as light-jets. Light-jets include jets that originate from a gluon as well as a $u$-, $d$-, or $s$-quark. A $b$- or $c$-quark that originates from gluon splitting will be classified as a $b$- or $c$-jet as long as the $b$- or $c$-quark is within $\Delta R < 0.3$ of the jet axis.

## 2.2 Monte Carlo biases

Requiring that a true muon from $b$ or $c$ decay is present as imposed on the muon+jet Monte Carlo sample means that almost all muons from decays of $\pi$/K, or from material interactions, are missing. These latter sources are here denoted *secondary muons.* They are present in the inclusive jet sample, of course.

The QCD jet sample predicts the secondary muon rate shown by the filled circles in Fig. 1 (left). The muon+jet samples have a very different secondary muon average fraction, especially at low muon $p_T$, as seen by the open circles in Fig. 1 (left). Fig. 1 (right) shows the fraction of jets with selected muons which are $b$-jets in the two samples.

The adequacy of the modeling of the secondary muons was checked, firstly by checking that the secondary muon rates were consistent separately for $b$-jets, $c$-jets, and light-jets between the inclusive jet and muon+jet samples. Secondly, we compared the lifetime tagging rates of $b$-jets and light-jets in the two samples as a function of $p_T$ after requiring a reconstructed muon. The results are shown in Fig. 2. The $b$-jet tag rates (a tag is defined as a jet having an IP3D+SV1 weight greater than 4) in the QCD jet and muon+jet samples are consistent within statistics. The light-jet lifetime-tag rate is a factor of about three higher in the muon+jet sample than in the muon-tagged inclusive jet sample. This arises from semi-leptonic heavy flavor decays in nearby jets which contaminate the light-jets with a muon and a displaced vertex. Muon-tagged jets in the QCD jet sample are less likely to suffer from this contamination because the muon is more likely to be a secondary, and thus not be associated with an event containing a $b$-quark. Real data is expected to look more like the QCD jet sample with the lower lifetime-tagger tag rate. We have checked that neither tagging calibration technique is affected by changes in the relative mix of light-jet muon sources as far as current Monte Carlo statistics allow.

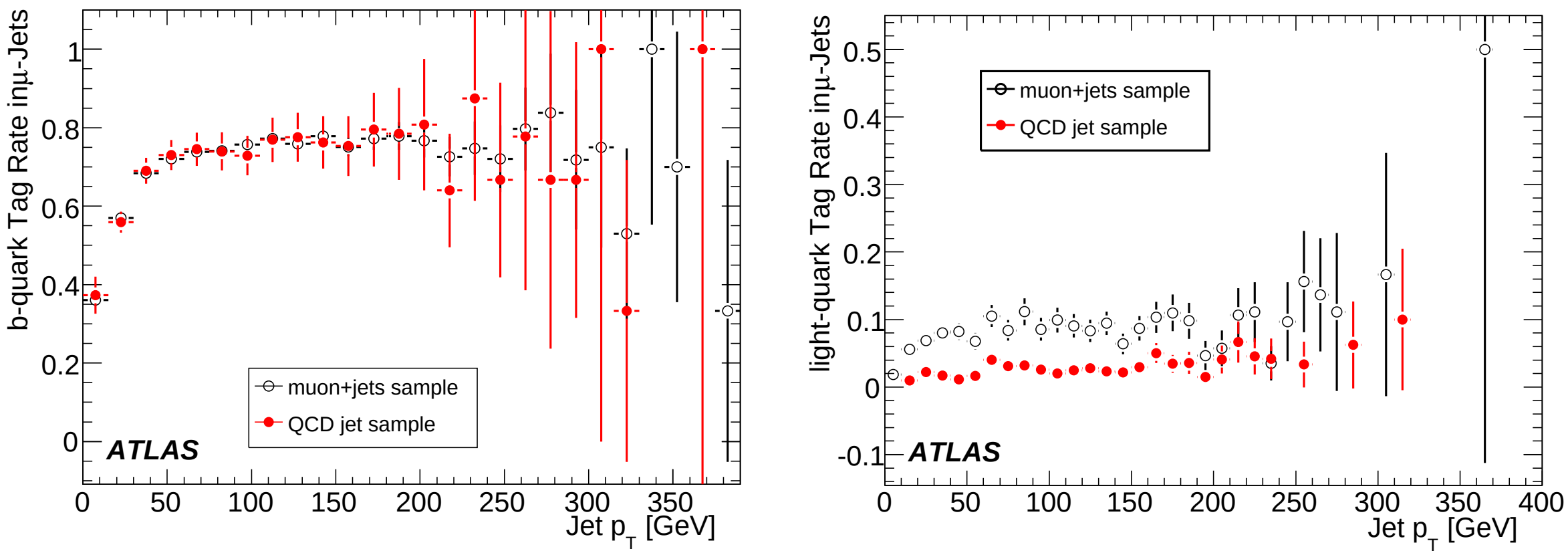

Figure 2: Tag rates (IP3D+SV1 weight > 4) for jets containing a reconstructed muon. The left plot are jets labeled as $b$-jets and the right plot are jets labeled as light-jets. In both plots, the open circles represent muon-tagged jets found in the muon+jet sample and the filled circles represent muon-tagged jets found in the QCD jet sample.

## 2.3 Trigger

Dedicated trigger signatures are necessary to obtain the required number of muon+jet events. We consider the statistics obtainable as a function of jet $p_T$. We also discuss a more sophisticated trigger to increase statistics at high jet $p_T$ compared to a single jet trigger. A full discussion of the operation of the ATLAS jet and muon trigger systems can be found elsewhere [6, 7].

The most obvious choice for a muon+jet trigger is a simple coincidence of lepton and jet triggers requiring no geometrical correlation. The rate for such a trigger is high, so that it must be prescaled, especially at low $p_T$. The flexible multi-level trigger system of ATLAS allows geometrical correlations to be used, explicitly requiring that the muon is close to the triggering jet direction.

In the current trigger menu foreseen for running at luminosities around $10^{31}$ cm$^{-2}$s$^{-1}$ [8], various muon and jet trigger thresholds are available at Level-1 (L1). Example combinations of these thresholds which are promising for the studies reported here are the two triggers L1_MU4_J10 and L1_MU6_J10, which are L1 signatures with a muon with $p_T > 4$ or 6 GeV and a jet with $p_T > 10$ GeV.

At Level-2 (L2) the muon selection is refined [7]: the selection sequence is started by the *muFast* algorithm which confirms L1 muon candidates and makes a more precise muon momentum measurement using muon spectrometer hit information. A fast track combination algorithm, *muComb*, matches tracks found by muFast with tracks found in the inner detector. For the studies reported here, no further selection is made at the third trigger level, the event filter (EF).

Since the trigger rate is likely to be limited by bandwidth considerations at low $p_T$ rather than the physics rate, the key parameter in considering the trigger performance is the purity of the sample in terms of the final selected events which are desired: the offline muon-tagged jets. We define the purity of the triggered sample as the fraction of events passing the trigger that contain such an offline muon-tagged jet. For the purpose of this study, the offline muon-tagged jet candidate is as defined in Section 2 ($p_T > 4$ GeV matching a reconstructed jet with $p_T > 15$ GeV). The purities of jet samples selected by the different trigger requirements discussed above are shown in Table 1.

The second and subsequent rows of Table 1 show that a further refinement is possible by asking for angular matching between the trigger muon and jet directions (taken here to be within $\Delta R = \sqrt{\Delta\phi^2 + \Delta\eta^2} < 0.4$). The choice between L1_MU4_J10 and L1_MU6_J10 will be driven essentially by the required muon $p_T$ acceptance. Overall, purities relative to the offline selected sample of around 80% are attainable. This refined algorithm will be implemented for data-taking.

Table 1: Purities of muon-tagged jet events in the triggered sample for different trigger selections. The purity is shown for four triggers with two L1 muon trigger thresholds ($X$). Errors are purely Monte Carlo statistics.

| Signature | $X = 4$ GeV | $X = 6$ GeV |
|---|---|---|
| L1_MU$X$_J10 | $20 \pm 1\%$ | $40 \pm 3\%$ |
| L1_MU$X$_J10 ($\mu$-jet matching) | $51 \pm 3\%$ | $79 \pm 7\%$ |
| L2_mu$X$(muFast)_J10 ($\mu$-jet matching) | $70 \pm 5\%$ | $82 \pm 8\%$ |
| L2_mu$X$(muComb: $\mu$+ID)_J10 ($\mu$-jet matching) | $78 \pm 6\%$ | $84 \pm 8\%$ |

A further concern is to provide a good coverage in jet $p_T$ extending from low $p_T$ as discussed above to much higher jet $p_T$, where the highest jet threshold will run unprescaled. The strategy adopted, similar to the one used for the inclusive jet trigger, consists of building up a set of muon-jet triggers with different jet trigger thresholds with prescale factors that diminish as $p_T$ rises. Figure 3 (left) shows the $p_T$ distribution of the jet belonging to the muon-jet candidate selected using the signature L2_mu4_J10. Using the set of signatures L2_mu4_J10, L2_mu4_J18, L2_mu4_J23, L2_mu4_J35, L2_mu4_J42 with prescale factors 50/15/12/12/1 a more uniform jet $p_T$ distribution is obtained as shown in Fig. 3 (right).

Under the assumption that the rate budget for the muon+jet trigger is 1 Hz, 100 000 muon+jet events are expected for around 30 hours of running time, corresponding to 1 pb$^{-1}$ of data at a luminosity of $10^{31}$ cm$^{-2}$s$^{-1}$. The combined muon-jet sample used in this study is therefore equivalent to roughly 5 pb$^{-1}$ of data, assuming that the $p_T$ acceptance of the trigger is comparable to that generated in the Monte Carlo samples.

# 3 The $p_{T,rel}$ method

The $p_{T,rel}$ method is based on the different relative transverse momentum distributions of muons in $b$-jets, $c$-jets, and light-jets. This arises because the muon typically originates from the semi-leptonic decay of a heavy hadron in heavy-flavor jets. The variable $p_{T,rel}$ is defined as the $p_T$ of the muon with respect to the jet+muon axis. As can be seen Fig. 4, $p_{T,rel}$ has good discrimination between $b$-jets and $c$-jets and light-jets.

The fraction of $b$-jets in the muon-tagged jet sample is estimated by performing a fit to the $p_{T,rel}$ distribution of muons using templates describing the $p_{T,rel}$ shape from $b$-jets, $c$-jets and light-jets. The first two templates are determined from Monte Carlo using muons originating either from $b$- or $c$-hadron decays. The template for light-jets is obtained by picking random tracks from jets with no heavy flavor in close proximity, assuming a uniform probability that such tracks fake a muon. We plan on using the same procedure to determine the light-jet templates from data once data is available, although we will still use Monte Carlo for the $b$- and $c$-templates. A systematic uncertainty will arise from the presence of non $b$- and $c$-jets in the real data sample, as discussed in Section 3.3.

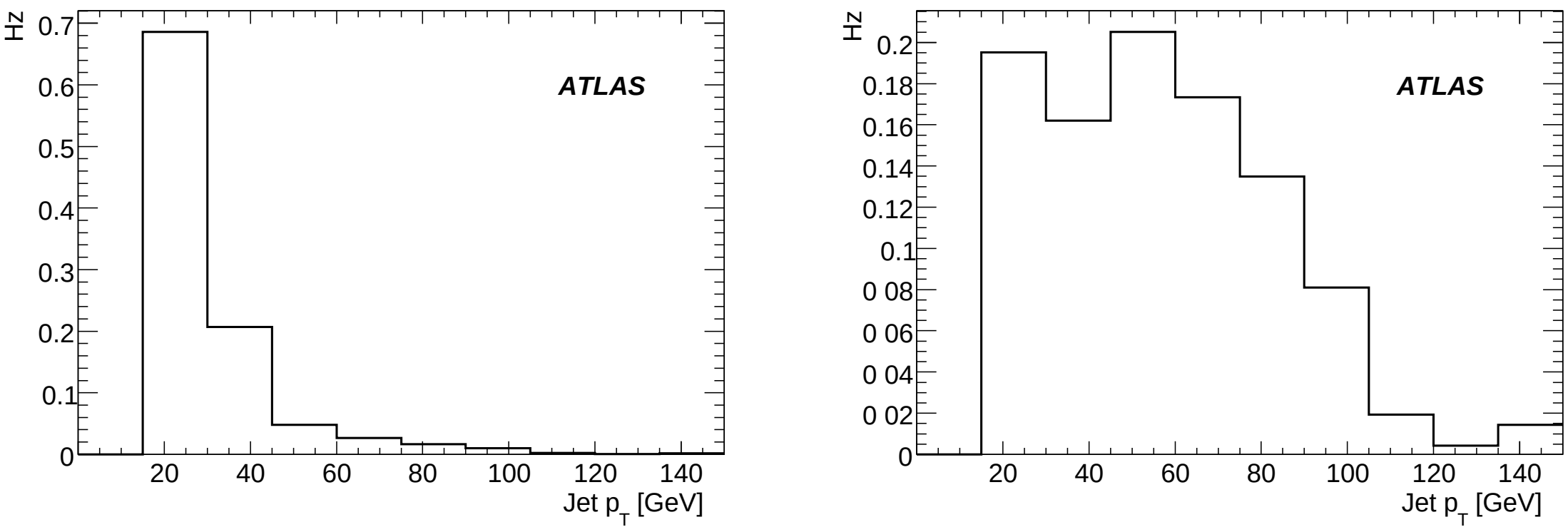

Figure 3: Jet $p_T$ distribution for trigger L2_mu4_J10 (left plot) and the jet $p_T$ distribution for the sum of triggers L2_mu4_J10, L2_mu4_J18, L2_mu4_J23, L2_mu4_J35, L2_mu4_J42 (right plot). In the case of the sum of triggers, the triggers were relatively prescaled by factors 50/15/12/12/1. In both plots the muon is confirmed at L2 by the muComb algorithm.

The $p_{T,rel}$ templates are determined in bins of jet $p_T$ and $\eta$ to allow the $b$-tagging efficiency to be measured as a function of these variables. The $p_{T,rel}$ distribution is fitted by allowing the normalization of the three templates to vary and minimizing a likelihood that also accounts for the template statistics [9]. The fit results are expressed in terms of fit fractions of the $b$-, $c$-, and light-jet $p_{T,rel}$ templates, $F_b$, $F_c$ and $F_{\text{light}}$. The $b$-tagging efficiency is obtained from the fit results by:

$$\varepsilon_b^{data,i} = \frac{N_{\mu-jet}^{tag,i} F_b^{tag,i}}{N_{\mu-jet}^{i} F_b^{i}} \tag{1}$$

where $N_{\mu-jet}^{i}$ and $N_{\mu-jet}^{tag,i}$ are, respectively, the number of $\mu$-jets in the $i$'th jet $p_T$ and $\eta$ bin before and after tagging.

As described in Section 5, a Monte Carlo-based correction is needed to correct the efficiency for $b$-jets with a muon to inclusive $b$-jets. In this section we discuss only the lifetime-tagging efficiency measurement for $b$-jets with muons.

### 3.1 Measuring the $b$-tagging efficiency using $p_{T,rel}$

The full muon+jet sample was split into two equal parts. The first half was used to obtain the $p_{T,rel}$ templates for light-, $c$- and $b$-jets, while the second half was used to measure the tagging performance.

As mentioned above, the light-jet $p_{T,rel}$ templates were built from the $p_{T,rel}$ of all tracks within a cone of $\Delta R < 0.4$ and $p_T > 4$ GeV of any reconstructed jet labeled as 'light'. The candidate jet was further required to be at least $\Delta R$ away from any other non-light-labeled jet.

Representative templates are shown in Fig. 4. The $p_{T,rel}$ distributions depend on jet $p_T$, especially for $b$-jets, and the templates are therefore derived in several bins of jet $p_T$. For high-$p_T$ jets, the separation power of the $p_{T,rel}$ variable significantly decreases: both the shape and peak value of the $p_{T,rel}$ distribution for $b$-jets approaches that of $c$- and light-jets. This can be seen in the right plot of Fig. 4.

The ROOT TFractionFitter [9] algorithm is used to fit the templates to the data (the second half of the Monte Carlo sample for this study). TFractionFitter uses a standard likelihood fit that takes into account both the template and data statistics. It includes the constraint that the templates must sum to the data distribution. The histograms in Fig. 5 show the results of the fit applied to muon-tagged jet samples before applying the IP3D+SV1 tagger (left) and after (right). Since the shape of the templates for $c$- and light-jets do not differ greatly, the relative rate of $c$- and light-jets is not so well determined. The shape of the template for $b$-jets does differ significantly, and so the fraction of $b$-jets can be reliably determined. The $b$-tagging efficiency is then determined directly using Equation (1).

As a first step, an inclusive $b$-tagging efficiency was measured, averaged over the $p_T$ and $\eta$ spectra of the jets. The results for the inclusive $b$-tagging efficiency, for various cuts on the tagger weight $w$, are presented in Table 2. Good agreement is observed between the true Monte Carlo $b$-tagging efficiency and that measured with the $p_{T,rel}$ method, within 1-2% statistical precision.

Table 2: True Monte Carlo efficiencies and efficiencies obtained with the $p_{T,rel}$ method for 2 taggers using the combined muon+jet samples. The errors are statistical only.

| Tagger | Weight cut | $\varepsilon_{true}$ | $\varepsilon_{meas}$ |
|---|---|---|---|
| IP3D+SV1 | $w > 4$ | 0.748 | $0.758 \pm 0.018$ |
| | $w > 7$ | 0.627 | $0.630 \pm 0.013$ |
| | $w > 10$ | 0.489 | $0.475 \pm 0.010$ |
| IP2D | $w > 2$ | 0.731 | $0.715 \pm 0.017$ |
| | $w > 3$ | 0.640 | $0.631 \pm 0.013$ |
| | $w > 4$ | 0.550 | $0.541 \pm 0.009$ |

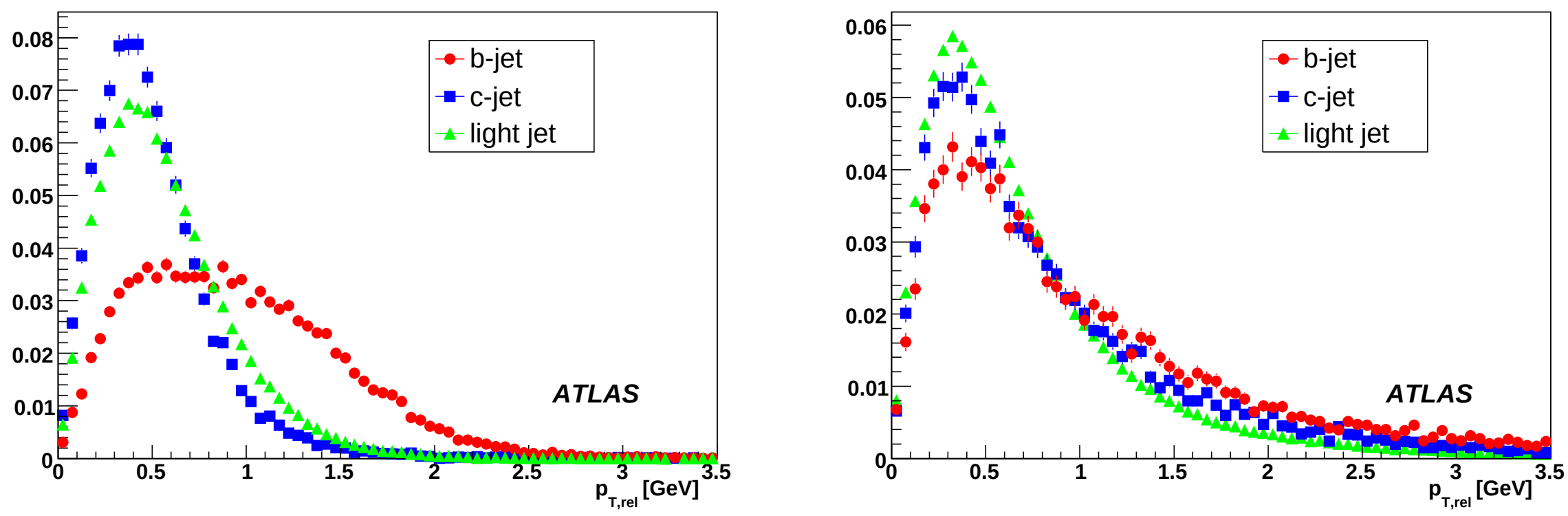

Figure 4: $p_{T,rel}$ templates obtained at low $p_T^{jet}$ ($15 < p_T^{jet} < 28$ GeV) (left) and high $p_T^{jet}$ ($163 < p_T^{jet} < 300$ GeV) region. Intermediate $p_T^{jet}$ ranges have distributions lying between these extremes.

The stability and sensitivity of the algorithm was tested by varying the fraction of $b$-jets in the Monte Carlo data-like test sample. This can be done in two ways: by decreasing the number of $b$-jets or by adding more light-jets to the initial sample. We have chosen the second approach.

We varied the input fraction of light-jets in the Monte Carlo test sample (and thus the fraction of $b$-jets) and re-measured the efficiency. The true input fractions and those obtained from fits to templates are shown in Table 3. We find agreement within statistical errors.

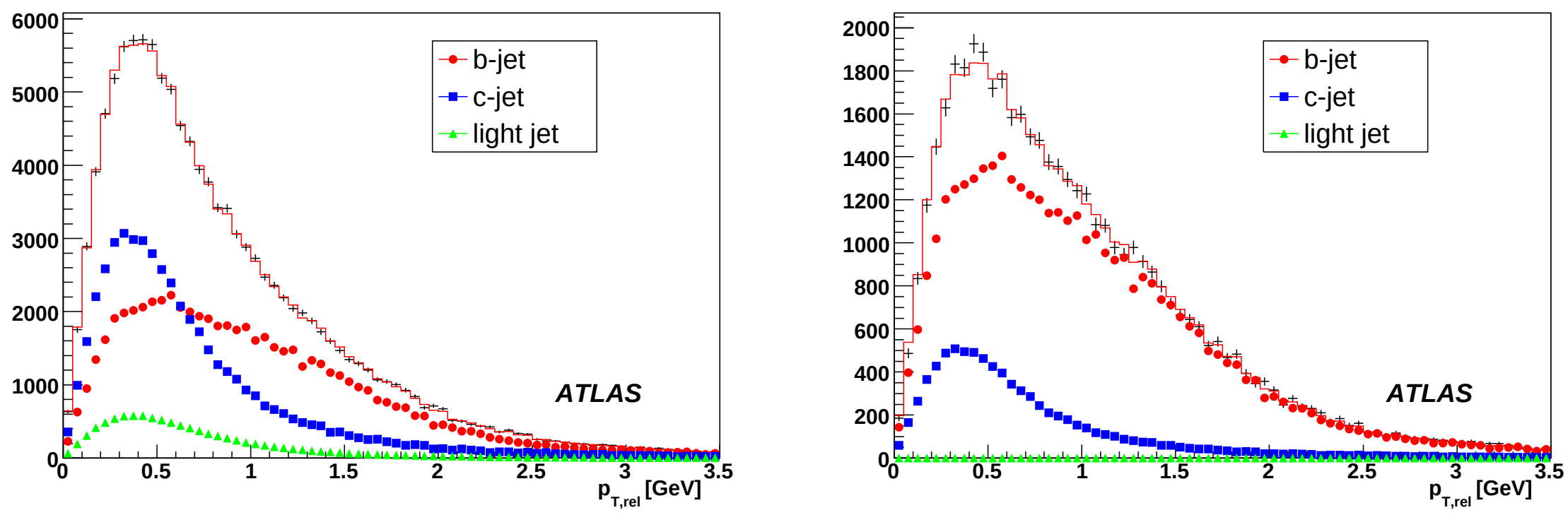

Figure 5: A fit of the $p_{T,rel}$ templates to the test sample. The test sample (black error bars) was fit with the $p_{T,rel}$ templates obtained from QCD jet Monte Carlo samples (green triangle: light-jet, blue square: $c$-jet, and red dot: $b$-jet). The red histogram is the result of the fit. The left plot shows the fit results for all muon-tagged jets, and the right shows that obtained after tagging.

Table 3: The stability of the $p_{T,rel}$ method as a function of changing $b$-, $c$-, and light-jet fractions. The fraction of $b$-jets was varied by altering the number of light-jets as described in the text. The first two columns show the true and fitted fractions of $b$-jets in the muon-tagged jet sample ($f_B$ - true and $f_B$ - result of fit). The remaining columns show the efficiency for two taggers (IP3D+SV1 with $w > 7$ and IP2D with $w > 3$) as measured directly in Monte Carlo ($\varepsilon_{true}$, and as determined by the $p_{t,rel}$ method ($\varepsilon_{meas}$).

| $f_B$ - true | $f_B$ - result of fit | $\varepsilon_{true}^{wIP3D+SV1>7}$ | $\varepsilon_{meas}^{wIP3D+SV1>7}$ | $\varepsilon_{true}^{wIP2D>3}$ | $\varepsilon_{meas}^{wIP2D>3}$ |
|---|---|---|---|---|---|
| 0.259 | $0.260 \pm 0.005$ | 0.627 | $0.630 \pm 0.024$ | 0.640 | $0.632 \pm 0.014$ |
| 0.298 | $0.298 \pm 0.005$ | 0.627 | $0.632 \pm 0.013$ | 0.640 | $0.633 \pm 0.013$ |
| 0.347 | $0.349 \pm 0.006$ | 0.627 | $0.630 \pm 0.015$ | 0.640 | $0.631 \pm 0.014$ |
| 0.398 | $0.397 \pm 0.006$ | 0.627 | $0.634 \pm 0.015$ | 0.640 | $0.635 \pm 0.012$ |
| 0.457 | $0.463 \pm 0.010$ | 0.627 | $0.625 \pm 0.016$ | 0.640 | $0.625 \pm 0.016$ |

## 3.2 $b$-tagging efficiency as a function of jet $p_T$ and $\eta$

The $b$-tagging efficiency is strongly dependent on jet $p_T$ and $\eta$ [2]. The above Monte Carlo test is an average and is only applicable for physics analysis that have the same kinematic properties as the muon+jet sample. In order to properly account for such effects the $b$-tagging efficiency should be parameterized and measured as a function of at least these two variables. Since the Monte Carlo statistics are limited, the method is tested by binning separately in $p_T$ and $\eta$. With more statistics a two-dimensional $(p_T, \eta)$-binning will be used. For the first case, the $p_{T,rel}$ templates are derived in several jet $p_T$ bins. Event samples in each $p_T$ bin are split in half as before: the first half was used to determine the $b$, $c$, and light jet $p_{T,rel}$ templates, while the second half was used as a data-like test sample to measure the $b$-tagging efficiency. The efficiency obtained in this way is integrated over the whole jet $\eta$ spectrum. The procedure was repeated to measure the efficiency dependence on jet $|\eta|$. In this case, the efficiency was integrated over the $p_T$ spectrum of jets.

The measured $p_T$ dependence of the $b$-tagging efficiency is shown in Fig. 6 (left) for the tagger IP3D+SV1. We observe a good agreement between measured and true efficiencies in this low jet $p_T$ region. For jets with $p_T > 80$ GeV, not shown, the statistical error from the fit becomes large and the results become unreliable. At present, therefore, we conclude that the efficiency can be reliably determined only in the low jet $p_T$ region ($p_T < 80$ GeV). This is related to the poor separation of $b$-jet templates from $c$- and light-jet templates in the high jet $p_T$ region (compare the plots in Fig. 4 for an example). In view of this, the $|\eta|$-dependence of the $b$-tagging efficiency was studied only for jets with $p_T < 80$ GeV (Fig. 6 right). Good agreement over the whole $|\eta|$ range is observed.

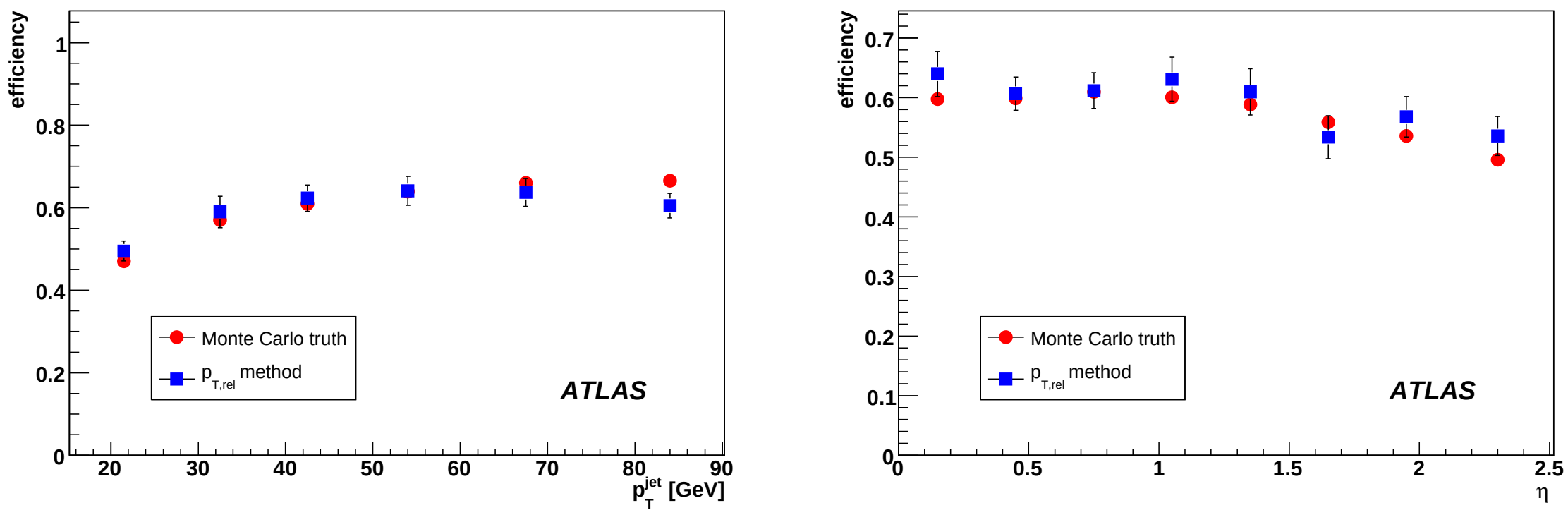

Figure 6: Efficiency as a function of jet $p_T$ (left) and jet $|\eta|$ (right) for the tagger IP3D+SV1 as measured using the $p_{T,rel}$ method. The dots are the true value as measured in the Monte Carlo, and the squares (with error bars) are determined from the $p_{T,rel}$ method. The lines are parameterizations to the measured $p_{T,rel}$ points. At high jet $p_T$ the $p_{T,rel}$ measurement technique fails and so we do not attempt to measure $b$-tagging performance above 80 GeV. The $|\eta|$ plot (right) includes only jets with $p_T < 80$ GeV.

### 3.3 Expected Errors

The high yield of muon-tagged jets from the muon-jet trigger means that the statistical error for determining efficiencies with the $p_{T,rel}$ method should rapidly become small: for 100 pb$^{-1}$ of data, for example, the overall statistical error would already be well below 1%.

The study of the systematic errors in the $p_{T,rel}$ method is in progress, and will require real data for detailed evaluation. Sources of systematic error being considered include:

- The major uncertainty of this method is the use of Monte Carlo to model the $p_{T,rel}$ templates for $b$- and $c$-jets. A change in the fragmentation function, for example, will likely change the muon $p_{T,rel}$ spectrum. We estimated the contribution of these effects by scaling the $p_{T,rel}$ shape of the $b$ and $c$ templates, resulting in an change in $\varepsilon_{\text{meas}}$ of 5%.

- The light-jet templates will be drawn from data and thus will have some heavy flavor contamination. QCD Monte Carlo predicts a jet sample will contain 2.5% $b$-jets and 5% $c$-jets. We expect little error to be introduced by $c$-jets as their $p_{T,rel}$ distribution is so similar to light-jets. We tested the effect of 2.8% $b$-jets in the pool of jets that was used to determine the light-jet template and re-ran the $p_{T,rel}$ fit. We observed a systematic shift to smaller $b$-jet efficiencies by 3%. We add $\pm 3\%$ as a systematic error due to heavy flavor contamination of the light jet templates. There is evidence we can mitigate the size of this error by rejecting jets from the light-jet template that satisfy a loose lifetime tagging requirement.

- Detector modeling has also been examined as a source of further systematic error. Jet $p_T$ resolution, the muon $p_T$ resolution, and the jet and muon direction resolution are all potential contributors. Studies indicate the error due to this are of order a few percent. Detector modeling of muon systematics have not been studied as it is assumed that the Monte Carlo modeling errors will be larger.

- We have an implicit assumption that all tracks in jets correctly model the $p_{T,rel}$ distribution for light jets. A further systematic will exist because $p_{T,rel}$ distribution for fake muons in light jets is not the same as the tracks associated with light jets.

These preliminary studies to date indicate that the systematic error sources should be controllable at the level of 6% or better on the lifetime tagging efficiency.

# 4 The System 8 method

The System 8 technique is designed to measure the $b$-tagging efficiency with reduced dependence on Monte Carlo [10], [11]. This method uses two data samples with different $b$-fractions and two uncorrelated tagging algorithms. A system of 8 nonlinear equations involving known quantities - like the total number of jets tagged in each sample - and 8 unknown quantities - like the $b$-tagging efficiency - can be written down and solved.

Before writing down the system of 8 equations we define the taggers and two samples more fully. The two taggers used must be as uncorrelated as possible. For this study we use the muon-in-jet tagger algorithm (SMT) [5] as one tagger. The SMT algorithm forms a 1-dimensional likelihood using the $p_{T,rel}$ of the muon. In the present analysis an event is said to be tagged by the SMT tagger if the likelihood value is greater than 1.4. The second tagger, the one whose efficiency we want to measure is either the IP3D+SV1 or the IP2D algorithm. We abbreviate this second algorithm as “LT”, short for “lifetime tagger”.

The two samples are denoted the “n-sample” and the “p-sample.” The n-sample is the full sample of jets containing a muon described in Section 2. The p-sample is a subset of the n-sample selected to have an enhanced $b$-content. A muon-tagged jet in the p-sample is required to have at least one back-to-back ($\Delta\phi > 2.5$) lifetime-tagged (IP3D+SV1 weight $> 3$) jet. This selection criterion increases the $b$-fraction of the p-sample relative to the n-sample by about a factor of two. The cuts for the p-sample were chosen to keep the statistics of the p-sample as large as possible and also to ensure the stability of the method.

## 4.1 The 8 equations

Four numbers are measured in each sample: the number of jets before tagging ($n$ in the n-sample, $p$ in the p-sample), the number of jets tagged by the LT algorithm ($n^{LT}$, $p^{LT}$), the number of jets tagged by the SMT algorithm ($n^{SMT}$, $p^{SMT}$), and the number of jets tagged by both algorithms ($n^{both}$, $p^{both}$).

While the total number of muon-tagged jets in each sample is an experimental observable, their flavor composition is not. We denote the number of $b$-jets in each sample as ($n_b$, $p_b$) and $c$- and light-jets as ($n_{cl}$, $p_{cl}$). The tagging efficiencies of the algorithms on the selected $b$-jets are ($\varepsilon_b^{LT}$, $\varepsilon_b^{SMT}$) and non-$b$-jets tagging efficiencies are ($\varepsilon_{cl}^{LT}$, $\varepsilon_{cl}^{SMT}$) — which are also unknown. Unless otherwise stated, each efficiency is that which would apply to tagging of the n-sample.

These 8 quantities can be related by 8 equations. These equations also contain parameters encoding the extent to which the following assumptions are valid: that the efficiency of each tagger is the same on the n- and p-sample, and that the two tagging algorithms are uncorrelated. The parameters introduced are $\alpha_i$ as follows:

$$\varepsilon_b^{both} = \alpha_1\, \varepsilon_b^{LT}\, \varepsilon_b^{SMT} \quad (2)$$

$$\varepsilon_{cl}^{both} = \alpha_2\, \varepsilon_{cl}^{LT}\, \varepsilon_{cl}^{SMT} \quad (3)$$

$$\varepsilon_{cl}^{SMT} \text{ (on p-sample)} = \alpha_3\, \varepsilon_{cl}^{SMT} \text{ (on n-sample)} \quad (4)$$

$$\begin{aligned}
\varepsilon_{cl}^{LT} \text{ (on p-sample)} &= \alpha_4\, \varepsilon_{cl}^{LT} \text{ (on n-sample)} && (5)\\
\varepsilon_{b}^{SMT} \text{ (on p-sample)} &= \alpha_5\, \varepsilon_{b}^{SMT} \text{ (on n-sample)} && (6)\\
\varepsilon_{b}^{LT} \text{ (on p-sample)} &= \alpha_6\, \varepsilon_{b}^{LT} \text{ (on n-sample)} && (7)
\end{aligned}$$

The $\alpha_1$ and $\alpha_2$ coefficients measure how correlated the two taggers are on $b$-jets and non-$b$-jets. The $\alpha_3$, $\alpha_4$, $\alpha_5$, and $\alpha_6$ are sensitive to any tag rate differences caused by the selection of the p-sample. The method is constructed in such a way that the $\alpha_i$ should each be approximately unity. The values of the $\alpha_i$ must be determined from Monte Carlo and possible differences of each $\alpha_i$ from that in data must be included in the systematic error. The current method used to evaluate this systematic is described in Section 4.3.

The eight jet counts enumerated above can be related by the following eight equations, including the $\alpha_i$:

$$\begin{aligned}
n &= n_b + n_{cl} && (8)\\
p &= p_b + p_{cl} && (9)\\
n^{LT} &= \varepsilon_b^{LT}\, n_b + \varepsilon_{cl}^{LT}\, n_{cl} && (10)\\
p^{LT} &= \alpha_6\, \varepsilon_b^{LT}\, p_b + \alpha_4\, \varepsilon_{cl}^{LT}\, p_{cl} && (11)\\
n^{SMT} &= \varepsilon_b^{SMT}\, n_b + \varepsilon_{cl}^{SMT}\, n_{cl} && (12)\\
p^{SMT} &= \alpha_5\, \varepsilon_b^{SMT}\, p_b + \alpha_3\, \varepsilon_{cl}^{SMT}\, p_{cl} && (13)\\
n^{both} &= \alpha_1\, \varepsilon_b^{LT}\, \varepsilon_b^{SMT}\, n_b + \alpha_2\, \varepsilon_{cl}^{LT}\, \varepsilon_{cl}^{SMT}\, n_{cl} && (14)\\
p^{both} &= \alpha_1\, \alpha_5\, \alpha_6\, \varepsilon_b^{LT}\, \varepsilon_b^{SMT}\, p_b + \alpha_2\, \alpha_3\, \alpha_4\, \varepsilon_{cl}^{LT}\, \varepsilon_{cl}^{SMT}\, p_{cl} && (15)
\end{aligned}$$

This system of eight equations (giving the method its name) is well specified if the value of each $\alpha_i$ is known. The $\varepsilon_b^{LT}$ is the unknown that is of most interest — most of the others are not directly usable for other analysis. The System 8 method is based on a technique used by the D0 experiment at the Tevatron [10]. The CDF experiment has also used jet events to calibrate their lifetime $b$-tagging efficiency [12] using a different technique.

## 4.2 Solving System 8, statistical error, and stability

Solving the system of 8 equations is in general straightforward with tools like Mathematica (analytical solution) or MINUIT (numerical solution). However, the nonlinearity of the 8 equations makes evaluating the statistical errors nontrivial. As a result Monte Carlo methods are employed. The System 8 observables, the jet counts $\{n, p, n^{LT}, p^{LT}, n^{SMT}, p^{SMT}, n^{both}, p^{both}\}$ are correlated; from them 8 uncorrelated jet counts $\{x_1, .., x_8\}$ are defined by dividing the n-sample into non-overlapping subsets. These parameters are varied according to Gaussian distributions with standard deviation $\sqrt{x_i}$. The variation of all $x_i$ was performed simultaneously assuming no correlations between them. The counts were varied in this way 100,000 times and the 8 equations were solved each time. The resulting distribution of $\varepsilon_b^{LT}$ was used to determine the one-sigma statistical error. Statistical errors are shown for current Monte Carlo statistics in Table 5.

The System 8 solution is not well constrained when the equations are close to being linearly-dependent. As the equations near linear dependence, the System 8 solution becomes increasingly sensitive to statistical fluctuations, resulting in unreliable values of $\varepsilon_b^{LT}$. The term ‘stability’ will be used to describe the sensitivity of System 8 to statistical variations in the data.

The 8 equations are linearly independent as long as the $b$-content of the n- and p-samples differ, and as long as $\varepsilon_b \neq \varepsilon_{cl}$ for both the SMT and LT tagging algorithms. In practice the first condition is easily met by construction of the p-sample (which requires a back-to-back $b$-tagged jet), but in the case of the SMT the second stability requirement ($\varepsilon_b^{SMT} \neq \varepsilon_{cl}^{SMT}$) becomes increasingly difficult to satisfy at high jet $p_T$.

The SMT algorithm is highly correlated with the $p_{T,rel}$ distribution of muons to distinguish between $b$-, $c$- and light-jets. These distributions become very similar at high jet $p_T$ (see Fig. 4), leading to the poor ability of the SMT to separate $b$ and non-$b$-jets at high momentum. It is for this reason we restrict the analysis to $p_T < 80$ GeV.

### 4.3 Correlation systematic error

It is important to determine how well the Monte Carlo describes the $\alpha_i$ in the data. Possible deviations should be reflected in the systematic error. We evaluate this contribution as the change in $\varepsilon_b^{LT}$ obtained by shifting each $\alpha_i$ from its value in Monte Carlo to unity. The total contribution to the systematic error, formed by adding the individual shifts in quadrature, is abbreviated as the correlation error.

Each $\alpha_i$ is measured from Monte Carlo and the results are shown for the IP3D+SV1 tagger ($w > 4$) in Table 4. The errors on each $\alpha_i$ come only from Monte Carlo statistics. We have checked other tagging algorithms and weight cuts and the results are similar. It can be seen that only $\alpha_3$ differs by much more than one sigma from unity, and the others lie within 3% of unity.

Table 4: The $\alpha$ coefficients measured on Monte Carlo for jet $15 < p_T < 80$ GeV, for the IP3D+SV1 tagger weight $w > 4$ and SMT cut of 1.4 ($\alpha_i^{meas}$). The last two columns show how the $b$-tagging efficiency ($\varepsilon_b$) is affected by different variations in $\alpha_i$: a 1% variation ($\alpha_i = 1 \pm 0.01$) and the measured offset of each $\alpha_i$ from unity ($\alpha_i = 1 \pm (\alpha_i^{meas} - 1)$). The change in $\varepsilon_b$ is labeled $\Delta_{\varepsilon_b}$. The latter column is added in quadrature to determine the total systematic error for the correlation error (see text).

| Assumption | Value ($\alpha_i^{meas}$) | $\Delta_{\varepsilon_b}$ for $\alpha_i = 1 \pm 0.01$ | $\Delta_{\varepsilon_b}$ for $\alpha_i = 1 \pm (\alpha_i^{meas} - 1)$ |
|---|---|---|---|
| $\alpha_1$ | $1.016 \pm 0.012$ | 0.027 | 0.043 |
| $\alpha_2$ | $1.018 \pm 0.024$ | 0.002 | 0.004 |
| $\alpha_3$ | $1.052 \pm 0.019$ | 0.006 | 0.031 |
| $\alpha_4$ | $1.028 \pm 0.020$ | $< 0.001$ | $< 0.001$ |
| $\alpha_5$ | $1.011 \pm 0.011$ | $< 0.001$ | $< 0.001$ |
| $\alpha_6$ | $1.005 \pm 0.010$ | 0.016 | 0.008 |

We vary each $\alpha_i$ by $\pm 0.01$ to give an idea of how each $\alpha_i$ contributes to the error in $\varepsilon_b^{LT}$. Table 4 shows that variations in $\alpha_2$, $\alpha_4$ and $\alpha_5$ should have rather less impact on the $b$-tagging efficiency measurement than $\alpha_1$, $\alpha_3$ and $\alpha_6$. The measured values of the coefficients $\alpha_1$ and $\alpha_6$ are statistically consistent with unity and show little statistically-significant dependence on jet $p_T$ or SMT cut. The coefficient $\alpha_3$, on the other hand, is not consistent with unity and shows a $p_T$ dependence, and has a pronounced effect on the measured efficiency at high jet $p_T$. The non-unity $\alpha_3$ is understood to be primarily due to the different relative fractions of charm and light in the n- and p-samples.

In order to work out the correlation error we evaluate the shift in $\Delta_{\varepsilon_b^{LT}}$ for a variation in $\alpha_i$ of $\pm(\alpha_i - 1)$, as shown in Table 4. Adding the errors in quadrature gives an estimate of the correlation error of 6%. Larger Monte Carlo samples will help improve our understanding of this error and help better understand which $\alpha_i$ really aren't unity.

### 4.4 Tests on Monte Carlo

We evaluated the System 8 method on the muon+jet Monte Carlo sample, for the following standard ATLAS taggers: IP2D, and the default combination of IP3D+SV1 for jets with $p_T < 80$ GeV.

To derive $p_T$ and $\eta$ dependent efficiency curves, the System 8 method was applied in bins of $p_T$ and $\eta$. Due to lack of Monte Carlo statistics it is not currently possible to bin in both $p_T$ and $\eta$ simultaneously; instead, System 8 was tested in $p_T$ bins for $|\eta| < 2.5$, and separately in $\eta$ bins for $15 < p_T < 80$ GeV. With larger Monte Carlo statistics and much larger data statistics (such as with 50 pb$^{-1}$) we should have enough for four bins in $p_T$ and four in $|\eta|$ simultaneously for jet $p_T < 80$ GeV. The exact binning configuration will be optimized to give the best possible resolution at low jet $p_T$ where the $b$-tagging efficiency is changing most rapidly. For this study, three bins in $p_T$ (15-30 GeV, 30-50 GeV, 50-80 GeV), and four bins in $|\eta|$ (0.0-0.5, 0.5-1.0, 1.0-1.5, 1.5-2.5) were considered.

Table 5 shows the results of calibrating the IP3D+SV1 tagger ($w$ >4) and IP2D tagger ($w$ >3) in bins of $p_T$ and $\eta$. The statistical error is shown. Recall the correlation error is an additional 6% which should be added in quadrature.

Table 5: Measured efficiencies, the statistical error, and their deviation from the true efficiencies in bins of $p_T$ and $\eta$, as described in the text. The statistical error is for the muon+jet Monte Carlo statistics. The correlation error, estimated to be $\pm 0.06$, must be added as well.

| Tagger | Bin | Measured $\varepsilon_b$ | $\|\Delta_{meas.,true}\|$ |
|---|---|---|---|
| IP3D+SV1 | $p_T$: 15-30 | $0.672 \pm 0.036$ | 0.040 |
| | $p_T$: 30-50 | $0.723 \pm 0.026$ | 0.004 |
| | $p_T$: 50-80 | $0.755 \pm 0.029$ | 0.017 |
| | $\|\eta\|$: 0.0-0.5 | $0.748 \pm 0.041$ | 0.008 |
| | $\|\eta\|$: 0.5-1.0 | $0.734 \pm 0.045$ | 0.029 |
| | $\|\eta\|$: 1.0-1.5 | $0.736 \pm 0.038$ | 0.017 |
| | $\|\eta\|$: 1.5-2.5 | $0.707 \pm 0.043$ | 0.001 |
| IP2D | $p_T$: 15-30 | $0.520 \pm 0.030$ | 0.002 |
| | $p_T$: 30-50 | $0.619 \pm 0.027$ | 0.003 |
| | $p_T$: 50-80 | $0.671 \pm 0.033$ | 0.011 |
| | $\|\eta\|$: 0.0-0.5 | $0.663 \pm 0.043$ | 0.003 |
| | $\|\eta\|$: 0.5-1.0 | $0.656 \pm 0.052$ | 0.016 |
| | $\|\eta\|$: 1.0-1.5 | $0.606 \pm 0.036$ | 0.034 |
| | $\|\eta\|$: 1.5-2.5 | $0.587 \pm 0.046$ | 0.018 |

To conclude, System 8 works well for different types of lifetime tagging algorithms for jet $p_T < 80$ GeV. It demonstrates stable results for different cuts on $b$-tagging weights.

## 5 The inclusive $b$-tagging efficiency

This note describes two methods which measure the lifetime-tagging efficiency for $b$-jets containing muons. In this section we briefly discuss how to convert from the measured tag efficiency to a lifetime tagging efficiency relevant to inclusive jet samples.

Fig. 7 shows the ratio of the muon-jet lifetime tag rate to the inclusive-jet tag rate for the default IP3D+SV1 tagger ($w > 4$) as a function of $p_T$. The difference between the tag rates in the two samples is primarily due to about a 50% difference in the $p_T$ of the highest $p_T$ track and in the average number of tracks per jet.

We define a scale factor, $S_b$, to describe the difference in tag rates. The scale factor is defined in

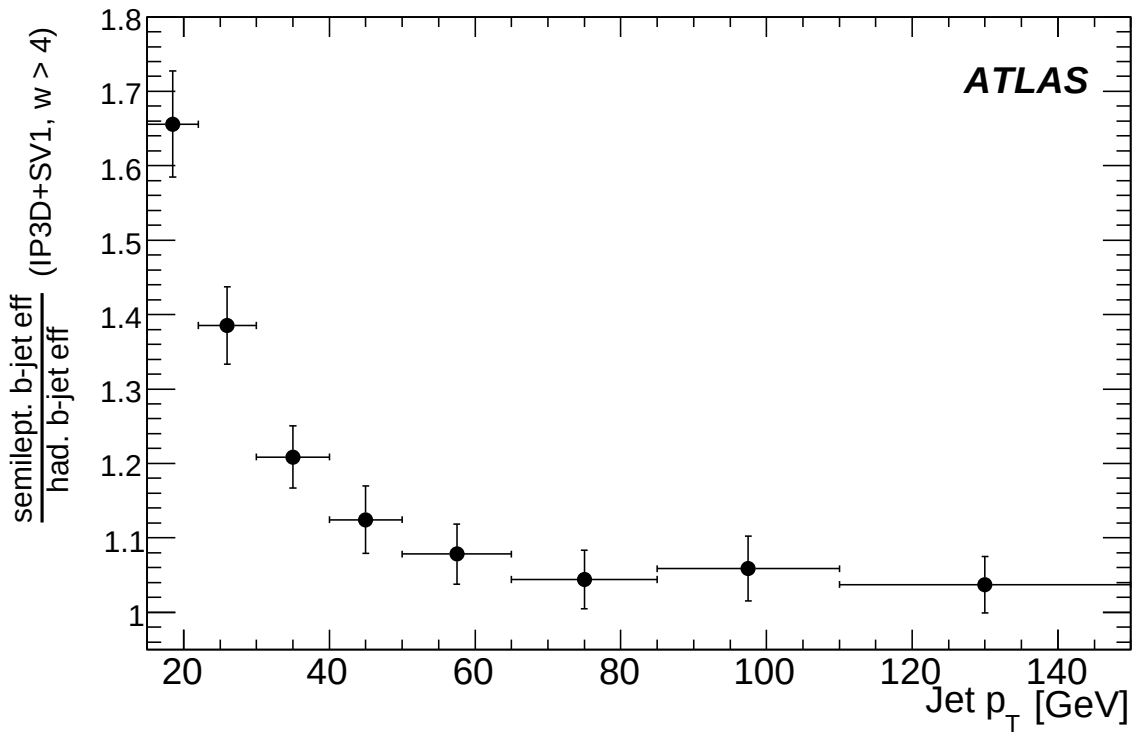

Figure 7: The right hand plot shows the ratio of efficiencies for semi-leptonic jet tagging and hadronic jet tagging in the QCD jet sample.

Monte Carlo:

$$\varepsilon^{MC}_{b\to\mathrm{had}}(p_T,\eta) = S_b(p_T,\eta)\varepsilon^{MC}_{b\to\ell\nu X}(p_T,\eta) \tag{16}$$

and thus, given a hadronic Monte Carlo $b$-jet, one calculates its calibrated lifetime tagging efficiency as follows:

$$\varepsilon_{b\to\mathrm{had}}(p_T,\eta) = \varepsilon_b(p_T,\eta)S_{b\to\mathrm{had}}(p_T,\eta) \tag{17}$$

where $\varepsilon_b$ is the tagging efficiency measured with System 8 or $p_{T,rel}$ in muon-tagged jets.

To understand the systematic error on the scale factor we need a better understanding of the causes of the differences in tag rate. This work is in progress.

## 6 Conclusions

We have presented two methods to calibrate lifetime $b$-tagging algorithms using dijet data. Trigger studies were performed in order to assure that there will be enough data for the $b$-tagging calibration. Detailed $b$-tagging performance can be obtained with data corresponding to 50 pb$^{-1}$ of integrated luminosity.

The suggested methods, $p_{T,rel}$ and System 8, were studied using Monte Carlo jet events, at least one of which has an associated muon. Good agreement is observed between true semi-leptonic $b$-tagging efficiency and that measured by $p_{T,rel}$ and System 8. Both methods were verified in different jet $\eta$ and $p_T$ regions. We found that the current version of the $p_{T,rel}$ method can be used for jets below 80 GeV in entire $\eta$ region. System 8 was also proved to work well up to 80 GeV in the entire $\eta$ region.

The total error of both methods is expected to be rapidly dominated by systematic uncertainties. The systematic errors studied so far indicate that both methods should be able to control the absolute error on $\varepsilon_b$ to 6%. The additional systematic uncertainties associated with converting the semi-leptonic efficiency to the inclusive efficiency have not yet been determined The two methods are complementary to those discussed in [1], in that they are most useful in measuring the turn-on of the $b$-tagging efficiency at low jet $p_T$.

# Trigger

# The Trigger for Early Running

**Abstract**
The ATLAS trigger and data acquisition system is based on three levels of event selection designed to capture the physics of interest with high efficiency from an initial bunch crossing rate of 40 MHz. The selections in the three trigger levels must provide sufficient rejection to reduce the rate to 200 Hz, compatible with offline computing power and storage capacity. The LHC is expected to begin its operation with a peak luminosity of $10^{31}$ cm$^{-2}$ s$^{-1}$ with a relatively small number of bunches, but quickly ramp up to higher luminosities by increasing the number of bunches, and thus the overall interaction rate. Decisions must be taken every 25 ns during normal LHC operations at the design luminosity of $10^{34}$ cm$^{-2}$ s$^{-1}$, where the average bunch crossing will contain more than 20 interactions. Hence, trigger selections must be deployed that can adapt to the changing beam conditions while preserving the interesting physics and satisfying varying detector requirements. In this paper, we provide a menu of trigger selections that can be deployed during the startup phase at each trigger level and show its evolution to higher luminosities. The studies presented in this paper are based on simulated data.

## 1 Introduction

The ATLAS trigger [1, 2] is composed of three levels of event selection: Level 1 (L1) [3] which is hardware-based using ASICs and FPGAs, the Level 2 (L2) and Event Filter (EF) (collectively referred to as the High Level Trigger or HLT [4]) based on software algorithms analyzing the data on large computing farms. The three levels of the ATLAS trigger system must reduce the output event storage rate to $\sim$200 Hz (about 300 MB/s) from an initial LHC bunch crossing rate of 40 MHz. It is evident from Fig. 1 that large rejection against QCD processes is needed while maintaining high efficiency for low cross section physics processes that include searches for new physics.

During the ATLAS startup phase, where low luminosity conditions ($10^{31}$ cm$^{-2}$ s$^{-1}$) are expected to prevail, the focus of the trigger selection strategy will be to commission the trigger and the detector and to ensure that established Standard Model processes are observed. It is therefore important to deploy loose selection criteria at each stage. In early operations many triggers will operate in pass-through mode, which entails executing the trigger algorithms but accepting the event independent of the algorithmic decision. This allows the trigger selections and algorithms to be validated to ensure that they are robust against the varying beam and detector conditions that are hard to predict before data-taking. As the luminosity increases, the use of higher thresholds, isolation criteria and tighter selections at HLT become necessary to reduce the background rates while achieving selection of interesting physics with high efficiency.

This note describes the possible triggers that can be deployed during the initial low luminosity running and discusses the strategy for triggering as the LHC ramps up to its design luminosity of $10^{34}$ cm$^{-2}$ s$^{-1}$. The performance of the various trigger algorithms at each of the three trigger levels is described in a set of additional accompanying notes. It should be emphasized that the rate estimates discussed in this note are based on simulations and are subject to several sources of uncertainty which include lack of knowledge of the exact cross-sections, detector performance, and beam related background conditions. Observations with early data will allow validation of these estimated rates and extrapolations to higher luminosities.

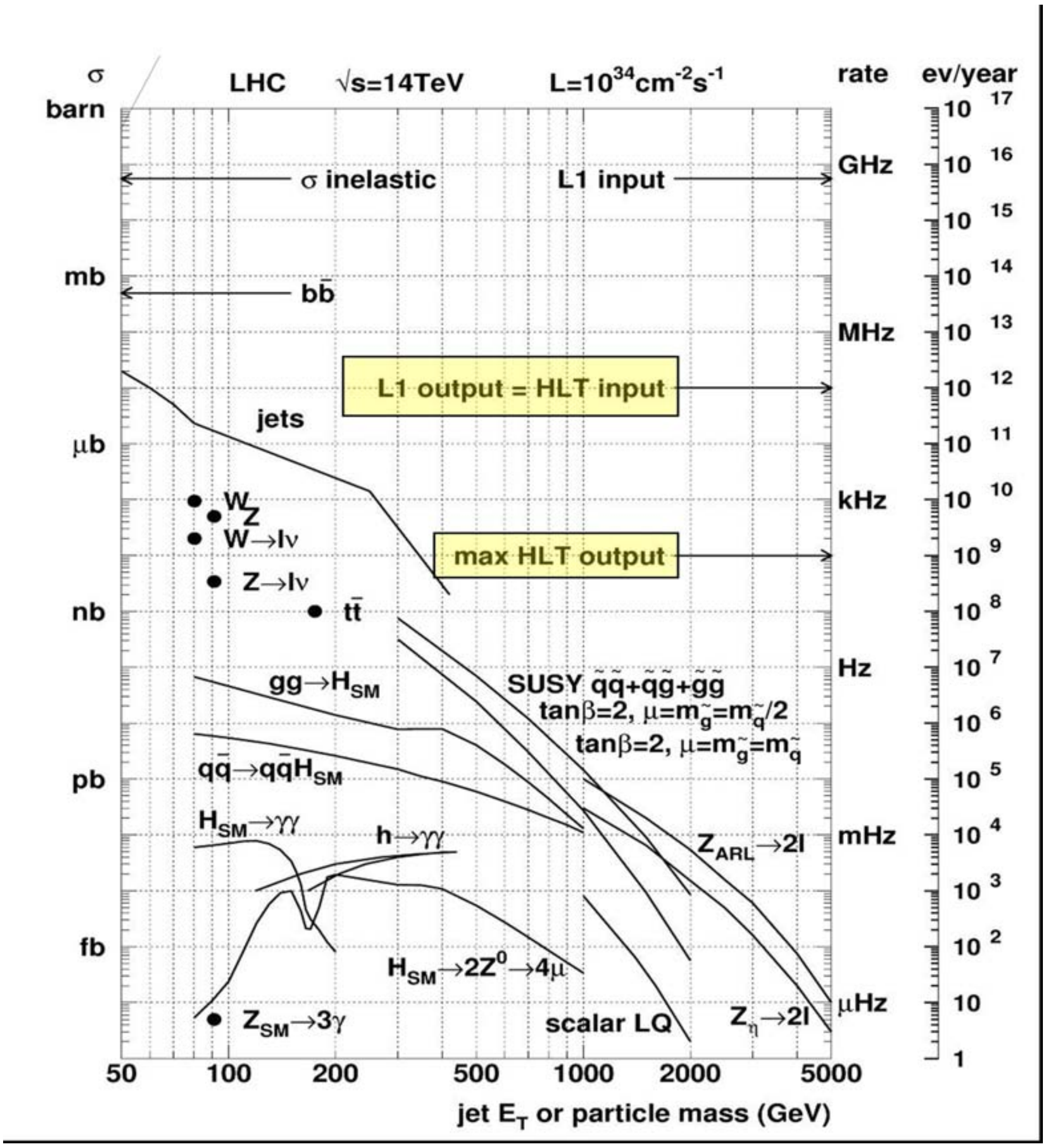

Figure 1: Expected event rates for several physics processes at the LHC design luminosity.

## 2 Level 1 trigger

The Level 1 trigger system receives data at the full LHC bunch crossing rate of 40 MHz and must make its decision within 2.5 $\mu$s to reduce the output rate to 75 kHz (∼40 kHz during ATLAS start-up). The L1 trigger has dedicated access to data from the calorimeter and muon detectors. The L1 calorimeter trigger [5] decision is based on the multiplicities and energy thresholds of the following objects observed in the ATLAS Liquid Argon [6] and Tile [7] calorimeter sub-system: Electromagnetic (EM) clusters, taus, jets, missing transverse energy ($\not{E}_T$), scalar sum $E_T$ ($\sum E_T$) in calorimeter, and total transverse energy of observed L1 jets ($\sum E_T$(jets)). These objects are computed by the L1 algorithms using the measured $E_T$ values in trigger towers of $0.1 \times 0.1$ granularity in $\Delta\eta \times \Delta\phi$. The L1 muon trigger [8] uses measurement of trajectories in the different stations of the muon trigger detectors: the Resistive Plate Chambers [9] (RPC) in the barrel region and the Thin Gap Chambers [8] (TGC) in the endcap region. The input to the trigger decision is the multiplicity for various muon $p_T$ thresholds.

There are a limited number of configuration choices that are available at L1. The most common difference between configuration choices is the amount of transverse energy or momentum required, so we refer to these configurations as "thresholds," but note that in addition to the $E_T$ threshold condition, three different isolation criteria can be applied for L1 EM and tau objects, and three different window sizes can be specified for L1 jet objects. Table 1 gives the number of these so-called thresholds that can be set for each object type. The total number of thresholds allowed for EM and tau objects is 16, where 8 are dedicated to be EM objects and 8 can be configured to be either EM or tau objects. The forward jets have four thresholds that can be set independently in each of the detector arms.

| Object | EM | Taus | Jets | For. Jets | $\not{E}_T$ | $\sum E_T$ | $\sum E_T(jets)$ | $\mu_{\leq 10\ \text{GeV}}$ | $\mu_{>10\ \text{GeV}}$ |
|---|---|---|---|---|---|---|---|---|---|
| # of thresholds | 8 - 16 | 0 - 8 | 8 | 4+4 | 8 | 4 | 4 | 3 | 3 |

Table 1: Number of L1 thresholds that can be set for each L1 object type at any given time (see text for details).

The total number of allowed L1 configurations (also called L1 items) that can be deployed at any time is 256. Each of these L1 items, programmed in the Central Trigger Processor (CTP) [10], is a logical combination of the specified multiplicities of one or more of the configured L1 thresholds. As an example L1_EM25i and L1_EM25 (A single L1 EM object with $E_T > 25$ GeV with and without isolation respectively) uses two L1 EM thresholds while L1_2EM25i (Two L1 isolated EM object with $E_T > 25$ GeV) uses the same L1 threshold as the L1_EM25i item. Furthermore, for each of the 256 L1 items, a prescale factor N can be specified (where only 1 in N events is selected and passed to the HLT for further consideration). As the peak luminosity drops during a fill, the L1 prescale value can be adjusted to keep the output bandwidth saturated without stopping and restarting a data-taking run, if desired. A given data-taking run is sub-divided into time intervals of the order of one minute. These sub-divisions, called luminosity blocks [11], provide the smallest granularity at which various data will be monitored and available for physics analysis. The trigger configuration, including the L1 prescale settings, remains unchanged within this luminosity block and adjustments to L1 prescale factors will be made on luminosity block boundaries.

## 3 Level 2

The L2 trigger is software-based, with the selection algorithms running on a farm of commodity PCs. The selection is largely based on regions-of-interest (RoI) identified at L1 and uses fine-grained data from the detector for a local analysis of the L1 candidate. A seed is constructed for each trigger accepted by L1 that consists of a $p_T$ threshold and an $\eta$-$\phi$ position. The L2 algorithms use this seed to construct an RoI window around the seed position. The size of the RoI window is determined by the L2 algorithms depending on the type of triggered object (for example, a smaller RoI is used for electron triggers than for jet triggers). The L2 algorithms then use the RoI to selectively access, unpack and analyze the associated detector data for that $\eta$-$\phi$ position. The ability to move, unpack, and analyse the local data only around the seed position greatly reduces both the processing times and the required data bandwidth.

The L2 algorithms provide a refined analysis of the L1 features based on fine-grained detector data and more optimal calibrations to provide results with improved resolution. They provide the ability to use detector information that is not available at L1, most notably reconstructed tracks from the Inner Detector. The information from individual sub-systems can then be matched to provide additional rejection and higher purity at L2. For each L1 RoI, a sequence of L2 algorithms is executed which compute event feature quantities associated with the RoI. Subsequently, a coherent set of selection criteria is applied on the derived features to determine if the candidate object should be retained.

The L2 farm will consist of around 500 quad-core CPUs. On average, the L2 can initiate the processing of a new event every 10 $\mu$s. The average processing time available for L2 algorithms is 40 ms, which includes the time for data transfers. The L2 system must provide an additional rejection compared to L1 of about 40 to reduce the output rate down from $\sim$75 (40) kHz to $\sim$2 (1) kHz during nominal (startup) operations.

## 4 Event Filter

The final online selection is performed by software algorithms running on the Event Filter (EF), a farm of processors that will consist of 1800 dual quad-core CPUs. The EF receives events accepted by L2 at a rate of 2 kHz (1 kHz) during nominal (startup) operations and must provide the additional rejection to reduce the output rate to $\sim$200 Hz, corresponding to about 300 MB/s. An average processing time of 4 s per event is available to achieve this rejection. The output rate from the Event Filter is limited by the offline computing budget and storage capacity.

As in L2, the EF works in a seeded mode, although it has direct access to the complete data for a given event as the EF selection is performed after the event building step. Each L2 trigger that has been accepted can be used to seed a sequence of EF algorithms that provide a more refined and complete analysis. Unlike L2, which uses specialized algorithms optimized for timing performance, the EF typically uses the same algorithms as the offline reconstruction. The use of the more complex pattern recognition algorithms and calibration developed for offline helps in providing the additional rejection needed at the EF.

## 5 Trigger rate estimation

Trigger rates have been estimated using a sample of simulated events. The design of a specific trigger menu often requires several iterations of selection optimization to ensure that the output rate is within allowed bandwidths and that interesting physics is triggered with high efficiency.

The first step in approximating trigger rates is to choose an appropriate input simulation sample. Most trigger selections are dominated by backgrounds from common processes, so samples with large physics cross-sections are generally used. However, these typically contain very few events that satisfy the trigger criteria and hence a very large number of events are required to obtain adequate statistical uncertainties on the estimated rates. In order to design the trigger menu for a luminosity of $10^{31}$ cm$^{-2}$ s$^{-1}$, a minimum-bias dataset containing seven million non-diffractive events with a cross-section of approximately 70 mb was used. To estimate the trigger rates with comparable statistical uncertainties for higher luminosities would require prohibitively large generated samples, hence other approaches are being pursued. These include using a combination of QCD and minimum bias event samples or alternatively using the so-called enhanced bias sample. The enhanced bias sample is a loosely filtered minimum bias sample requiring the lowest L1 $p_T$ thresholds for muon, EM, or jet to have been fulfilled. Only events that pass the filtering process are reconstructed, resulting in a much more effective use of the computing resources. In addition to QCD processes, other high cross-section physics processes, such as $W$ and $Z$ boson production, need to be considered for estimating trigger rates at very high luminosities. Although such simulated samples provide a reasonable starting point to establish a data taking menu, these trigger menus will evolve as our understanding of the detector and trigger evolve, and as the physics requirements mature. Once data taking operations begin, dedicated data samples for further menu optimization and rate estimations will be collected.

In order to compute the initial trigger rates, the full trigger simulation (L1, L2, and EF) is executed on a minimum bias data sample generated with PYTHIA [12] and simulated with realistic detector effects in Geant4 [13]. The rate estimates presented in this paper are based on an analysis of seven million minimum-bias events. At a luminosity of $10^{31}$ cm$^{-2}$ s$^{-1}$ each unprescaled event corresponds to a rate of about 0.1 Hz. These raw rates are subsequently weighted for any applied prescale factors, which allows more accurate rate measurements as it makes use of all the events in the data sample. For each level, the individual trigger rates are computed using the equation

$$\mathrm{R} = \mathscr{L} \times \frac{\mathrm{n_{accepted}}}{\mathrm{n_{total}}} \times \frac{1}{\prod_{\mathrm{l=lowest\ level}}^{\mathrm{current\ level}} \mathrm{P_l}} \times \sigma \tag{1}$$

where $R$ is the rate at the current level, $\mathscr{L}$ is the instantaneous luminosity, $P_l$ is the prescale factor applied at a specific level, $n_{accepted}$ is the number of events accepted after the specific trigger level studied (and hence all the previous lower levels as well), and $n_{total}$ is the total number of events in the dataset and $\sigma$ is the cross section.

Trigger rates are estimated by assigning a probability for an event to have been accepted. The computation of rates based on probabilities makes maximal use of the available set of simulated events from a statistical point of view. The probability $Pr_i$ of an event being accepted by a trigger item $i$ is given by:

$$\mathrm{Pr_i(event)} = \mathrm{D_i(event)/P_i}, \tag{2}$$

where $D_i$ is the decision probability accepting an event when no prescale factor is applied, and $P_i$ is the overall prescale associated with the trigger item.

In order to compute the overall acceptance rate of a specific menu, the overlap in acceptance from different trigger items needs to be correctly taken into account. The probability that two triggers accept an event simultaneously is then given by:

$$\mathrm{Pr_{12}(evt)} = \mathrm{Pr_1(evt)} \times \mathrm{Pr_2(evt)}. \tag{3}$$

The overall probability of accepting an event in a menu of two triggers, including their correlations, is thus given by

$$\mathrm{Pr_{menu}(evt)} = \mathrm{Pr_1(evt)} + \mathrm{Pr_2(evt)} - \mathrm{Pr_1(evt)} \times \mathrm{Pr_2(evt)} \tag{4}$$

This computation, although simple in the case of two triggers, becomes increasingly complex as the number of items in the menu increases. Fortunately, this problem can be solved recursively. Dedicated tools employing these methods have been developed to compute trigger acceptance rates and overlaps for various trigger menus.

# 6 Trigger menu for a luminosity of $10^{31}$ cm$^{-2}$ s$^{-1}$

The initial LHC startup luminosity is expected to be approximately $10^{31}$ cm$^{-2}$ s$^{-1}$ with a low number of bunches in the machine. These conditions will be ideal for commissioning the trigger and detector systems, as well as for the initial data taking, which will be dedicated to high cross section Standard Model signatures. Hence, the trigger selections deployed during this early running phase will primarily be a combination of low $p_T$ thresholds and loose selection criteria. Triggers at higher selection stages will be operated in pass-through mode wherever possible.

Trigger menus are tables of triggers that incorporate the signatures for various physics objects at each of the three trigger levels. These signatures fully specify the thresholds and selection criteria at each level, providing a recipe for triggering on various physics processes. Each signature is considered carefully addressing its physics goals, the efficiency and background rejection it provides for meeting these goals and the trigger bandwidth it consumes. Trigger menus also must include additional triggers for trigger validation, monitoring, calibration and measuring the performance of the physics triggers. With these considerations, a trigger menu for the startup phase has been developed. The rates have been estimated using the previously described simulated minimum-bias events.

The following notation is used to label the different trigger items: e (electron), g (photon), EM (electromagnetic), J (jets), FJ (forward jets), XE (Missing $E_T$), TE (Total scalar sum $E_T$), JE (Scalar sum of jet $E_T$), MU (muons), and tau (tau leptons). A typical example of a trigger item is 2e15i (two isolated electrons with a $p_T$ greater than 15 GeV) or tau20i_XE30 (an isolated tau decaying hadronically with visible $p_T$ above 20 GeV and $\not{E}_T$ above 30 GeV). A prefix to the item name is used to specify the trigger level at which the item is deployed. If the presence of several trigger object types are required, the AND of these multiple object types is shown using an "_", for example, tau25i_XE30 requires an isolated tau lepton with a $p_T$ above 25 GeV AND a Missing ET above 30 GeV (when objects are combined with no qualifier a simpler notation can be used, for example, $e+\mu$ for a trigger with an electron and a muon).

## 6.1 L1 items foreseen for a luminosity of $10^{31}$ cm$^{-2}$ s$^{-1}$

All rates given in this section are measured using simulated events with a luminosity of $10^{31}$ cm$^{-2}$ s$^{-1}$. Table 2 shows a potential set of EM L1 signatures, their prescale factors and estimated rates that could be deployed during start-up for an assumed luminosity of $10^{31}$ cm$^{-2}$ s$^{-1}$. It shows the eight L1 EM thresholds and the multi-EM object thresholds that are direct combinations of the single EM thresholds. At low luminosities, a single non-isolated trigger of 7 GeV can be used without the application of prescale factors at the first trigger level with a rate of about 5 kHz. All multi-object EM triggers at L1 have sufficiently low trigger rates and can be deployed without the application of any prescale factors. The total L1 output rate out for the EM objects defined in Table 2 is about 10 kHz after correctly accounting for events that have passed multiple L1 EM object signatures. The L1 EM object triggers are then used to seed both electron and photon signatures at the HLT.

| Trigger Item | EM3 | EM7 | EM13 | EM13I | EM18 | EM18I | EM23I | EM100 |
|---|---|---|---|---|---|---|---|---|
| Prescale | 60 | 1 | 1 | 1 | 1 | 1 | 1 | 1 |
| Rate (Hz) | 674 | 4900 | 950 | 480 | 369 | 143 | 53 | 1.5 |
| Trigger Item | 2EM3 | 2EM7 | 2EM13 | 2EM13I | 2EM18 | 2EM18I | 2EM23I | 3EM7 |
| Prescale | 1 | 1 | 1 | 1 | 1 | 1 | 1 | 1 |
| Rate (Hz) | 6500 | 534 | 108 | 8 | 47 | 2 | 0.6 | 53 |

Table 2: L1 trigger items and estimated rates at $10^{31}$ cm$^{-2}$ s$^{-1}$ for electromagnetic objects.

Several higher threshold signatures are deployed with and without isolation, even though lower thresholds can be deployed without the application of prescale factors at low luminosities. This allows the validation of the higher thresholds and trigger items that will be needed at high luminosities when the lower threshold triggers can only be deployed with prescale factors due to rate considerations.

The inclusive L1 jet trigger items and corresponding prescale factors, as shown in Table 3, are chosen to given an approximately flat trigger rate across the steeply falling jet $E_T$ spectrum. At a luminosity of $10^{31}$ cm$^{-2}$ s$^{-1}$, a single jet trigger with a threshold $E_T$ of 120 GeV can be deployed without prescale factors and has a L1 output rate of about 8 Hz. Figure 2 shows that triggered jet $E_T$ spectrum is fairly flat up to the threshold value of 120 GeV and then falls with the jet cross section. This strategy provides sufficient statistics across the $E_T$ spectrum for the measurement of the differential cross sections and for measuring the efficiencies and performance of different algorithms.

At a luminosity of $10^{31}$ cm$^{-2}$ s$^{-1}$, the HLT algorithms for the inclusive jet triggers are run in pass-through mode with the rate controlled only by using L1 prescale factors. This allows validation of the HLT algorithms for jet triggers that can then be deployed at higher luminosities.

| Trigger Item | J10 | J18 | J23 | J35 | J42 | J70 | J120 | 3J10 | 3J18 | 4J10 | 4J18 | 4J23 |
|---|---|---|---|---|---|---|---|---|---|---|---|---|
| Prescale | 42000 | 6000 | 2000 | 500 | 100 | 15 | 1 | 150 | 1 | 30 | 1 | 1 |
| Rate (Hz) | 4 | 1 | 1 | 1 | 4 | 4 | 9 | 40 | 140 | 40 | 20 | 7 |

Table 3: L1 trigger items and estimated rates at $10^{31}$ cm$^{-2}$ s$^{-1}$ for jet objects.

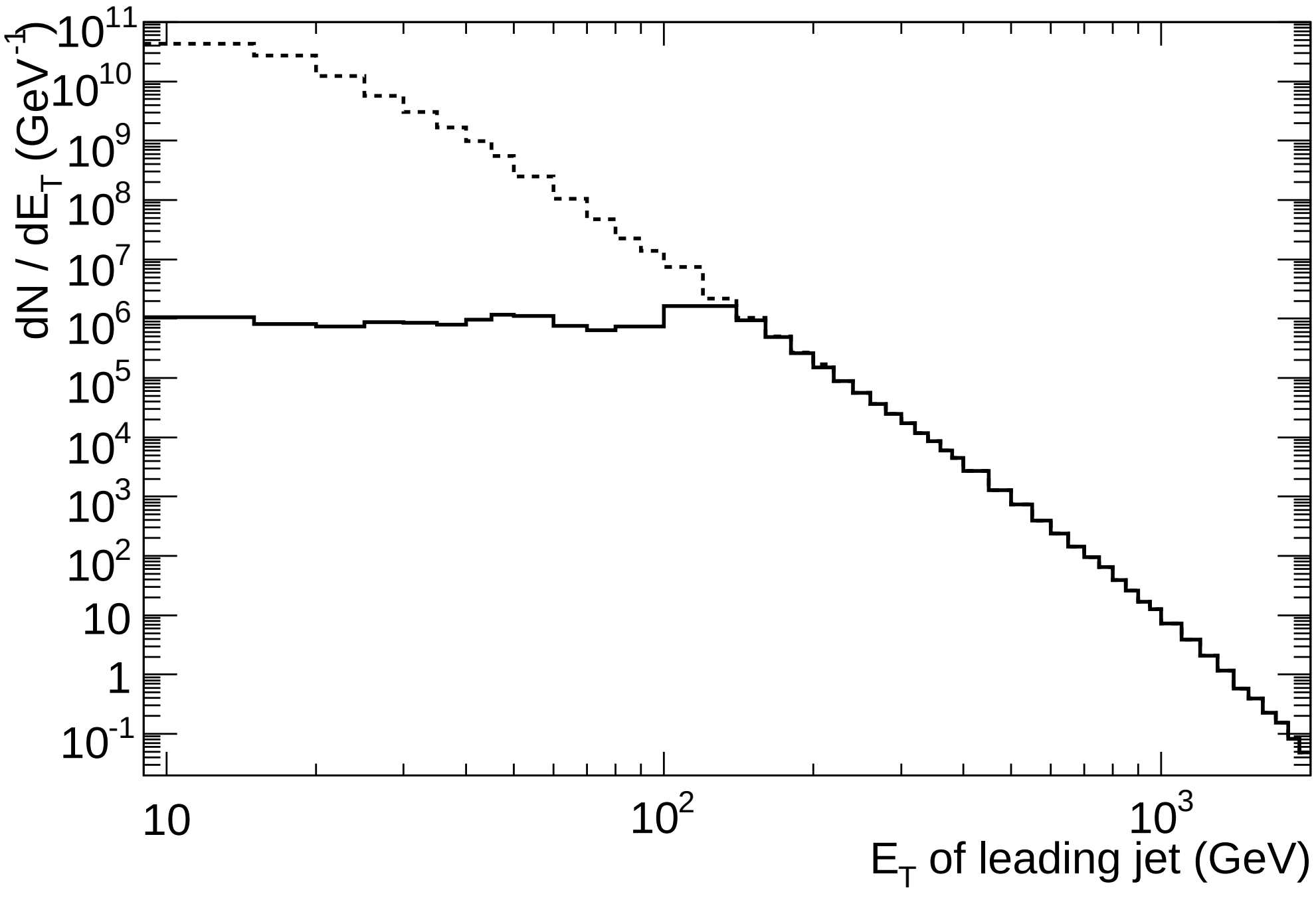

Figure 2: Jet $E_T$ spectrum at $10^{31}$ cm$^{-2}$ s$^{-1}$ before (dashed) and after (solid) pre-scaling at L1.

Unlike single jet triggers, where the output rates are controlled by applying prescale factors at L1, jet triggers with higher multiplicities (multi-jets) have a larger allocated L1 output bandwidth to allow the use of additional selection algorithms only applicable at the HLT (e.g. b-tagging). Their total output rate to disk storage is subsequently controlled at the HLT with either a combination of jet algorithms and additional prescale factors (leading to normal jet signatures) or additional highly rejective selections, (e.g. b-tagging algorithms leading to b-jet signatures).

The prescale factors and rates for forward jet triggers are shown in Table 4. The HLT algorithms at low luminosity are executed in pass-through mode with rates controlled using L1 thresholds and prescale factors.

| Trigger Item | FJ18 | FJ35 | FJ70 | FJ120 | 2FJ18 | 2FJ35 |
|---|---|---|---|---|---|---|
| Prescale | 7000 | 700 | 20 | 1 | 100 | 1 |
| Rate (Hz) | 1 | 1 | 1 | 1 | 1 | 2 |

Table 4: L1 trigger items and estimated rates at $10^{31}$ cm$^{-2}$ s$^{-1}$ for forward jet objects.

Table 5 shows the rates for the suite of single muon and dimuon signatures; these triggers can all be deployed at $10^{31}$ cm$^{-2}$ s$^{-1}$ without L1 prescale factors with further selection at HLT to control the output rates. Three of the six available muon thresholds must be established below $p_T$ of 10 GeV and are based on a coincidence of only two of the inner three stations. The lowest possible muon threshold of 4 GeV is set by opening the coincidence window in the two stations to the maximum allowed size. The remaining three high $p_T$ threshold L1 muon triggers ($p_T > 10$ GeV) require a coincidence in all three muon trigger stations. The largest contribution to the rates shown in Figure 3 come from b, c quark decays and in-flight decays of pions and kaons. The L1 muon trigger is highly efficient (99%) for $p_T$ above the threshold values within the fiducial acceptance of the detector.

| Trigger Item | MU4 | MU6 | MU10 | MU15 | MU20 | MU40 |
|---|---|---|---|---|---|---|
| Rate (Hz) | 1730 | 640 | 360 | 30 | 20 | 10 |
| Trigger Item | 2MU4 | MU4_MU6 | 2MU6 | 2MU10 | 2MU20 | 3MU6 |
| Rate (Hz) | 70 | 45 | 14 | 7 | 0.2 | 0.7 |

Table 5: L1 trigger items and estimated rates at $10^{31}$ cm$^{-2}$ s$^{-1}$ for muon objects.

Some of the signatures and estimated rates for single and double tau triggers at $10^{31}$ cm$^{-2}$ s$^{-1}$ are shown in Table 6. The triggers at low luminosity are chosen to collect large statistics of $W$ and $Z$ boson decays to tau leptons. For $W$ boson decays, additional cuts on $\not{E}_T$ help reduce the rates. However, reliance on $\not{E}_T$ is limited during startup as it is very sensitive to several detector and acceptance effects, and will take time to validate. Alternative approaches that use $\not{E}_T$ only in the Event Filter seeded with a single tau signature at L1 are being studied. In addition to the $2\tau$ triggers at L1, $\tau + e$ and $\tau + \mu$ triggers have been implemented at L1 to trigger on $Z \to \tau\tau$ decays with one of the taus decaying leptonically.

The eight $\not{E}_T$ thresholds likely to be deployed at the L1 stage are shown in Table 7. The strategy here is similar in nature to that of the jet triggers, with L1 prescale factors tuned to provide a flat rate across the $\not{E}_T$ spectrum. The reliance on inclusive $\not{E}_T$ trigger will be small especially during the early running period as it is sensitive to various detector effects that will require time to understand. Most of the $\not{E}_T$ thresholds are expected to be used in combination with other signatures.

In addition to the $\not{E}_T$ trigger thresholds, there are four thresholds available for the scalar sum $E_T$ (TE) and the scalar sum $E_T$ of observed L1 jets (JE). As for the $\not{E}_T$ triggers, reliance on such triggers will be limited in the early running period, but could prove to be valuable to ensure that very high $E_T$ events are recorded. Table 8 shows the selections, the pre-scale factors necessary to achieve the desired rate reduction, and estimated L1 output rates for the TE and JE triggers that could be deployed at startup.

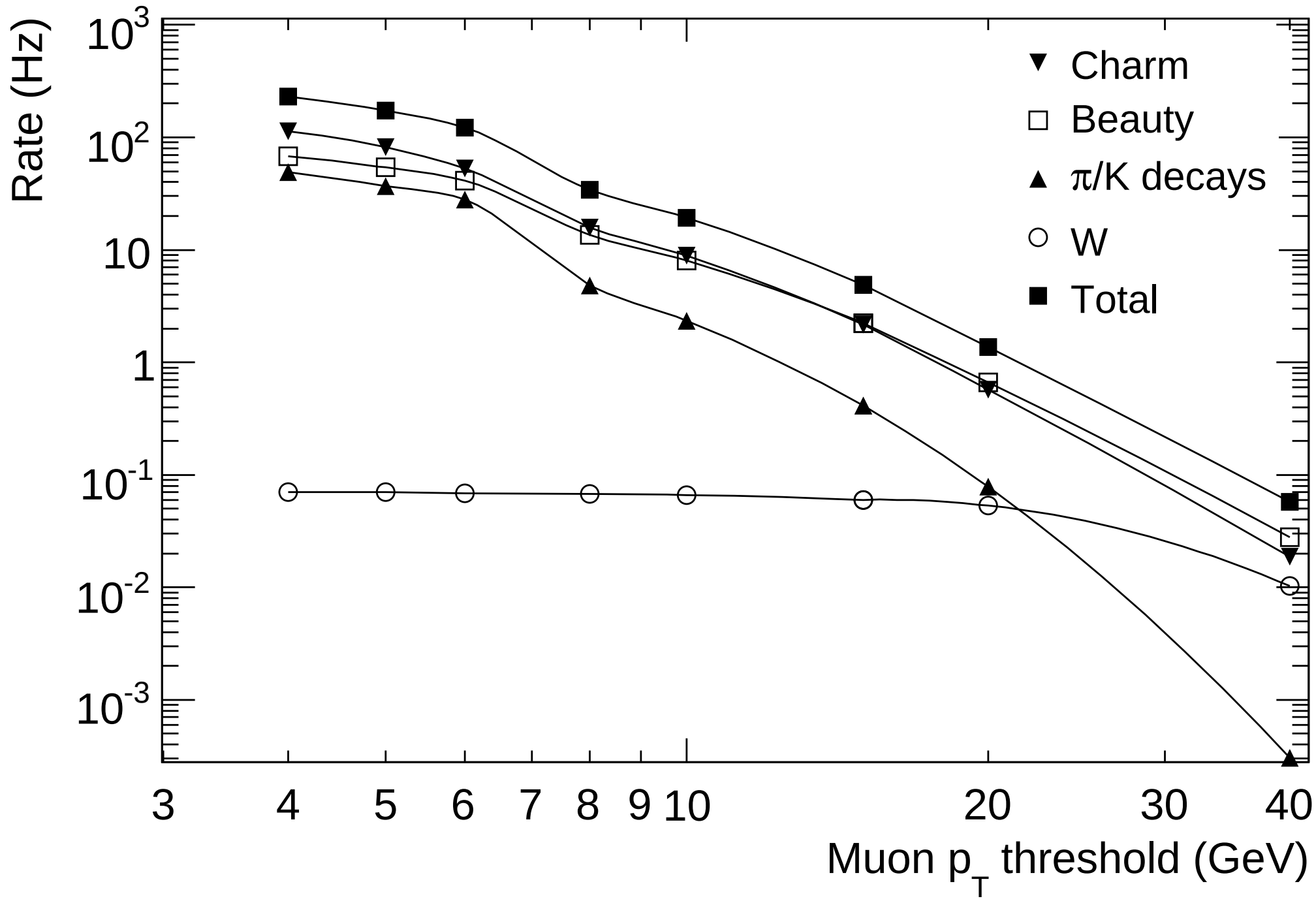

Figure 3: Estimated muon trigger rate for a luminosity of $10^{31}$ cm$^{-2}$ s$^{-1}$. Shown are the total rates and various contributions.

| Signature | tau6 | tau9I | tau11I | tau16I | tau25 | tau25I | tau40 |
|---|---|---|---|---|---|---|---|
| Prescale | 750 | 300 | 1500 | 10000 | 20 | 10 | 1 |
| Rate (Hz) | 19 | 16 | 2 | $< 0.1$ | 16.1 | 25 | 83 |
| Signature | 2tau6 | 2tau9I | 2tau16I | tau6_tau16I | tau9I_EM13I | tau9I_MU6 | tau9I_XE30 |
| Prescale | 100 | 1 | 1 | 10 | 1 | 1 | 1 |
| Rate (Hz) | 19 | 413 | 65 | 46 | 100 | 25 | 160 |

Table 6: L1 trigger items and estimated rates at $10^{31}$ cm$^{-2}$ s$^{-1}$ for tau objects.

| Trigger Item | XE15 | XE20 | XE25 | XE30 | XE40 | XE50 | XE70 | XE80 |
|---|---|---|---|---|---|---|---|---|
| Prescale | 30000 | 7000 | 1500 | 200 | 20 | 2 | 1 | 1 |
| Rate (Hz) | 2.5 | 3 | 4 | 7.5 | 7.5 | 14 | 2 | 1 |

Table 7: L1 trigger items and estimated rates at $10^{31}$ cm$^{-2}$ s$^{-1}$ for Missing $E_T$ objects.

| Trigger Item | TE150 | TE250 | TE360 | TE650 | JE120 | JE220 | JE280 | JE340 |
|---|---|---|---|---|---|---|---|---|
| Prescale | 100k | 1100 | 40 | 1 | 150 | 10 | 2 | 1 |
| Rate (Hz) | 2 | 3 | 1 | 0.5 | 0.5 | 0.5 | 0.5 | 0.1 |

Table 8: L1 trigger items and estimated rates at $10^{31}$ cm$^{-2}$ s$^{-1}$ for the scalar $\sum E_T$ (TE) and $\sum E_T^{jet}$ (JE) triggers.

In addition to L1 triggers that rely on thresholds and multiplicities of a single object type (EM, Jet, muon etc.), several triggers are formed by combining multiple L1 object types (e.g: EM + muon, Jet + missing $E_T$ etc.). Table 9 shows some of the typical combined signatures and their estimated rates. The rates for these triggers are generally small because they combine objects of different types at L1 and can thus be executed without prescale factors. Combining object types at L1 provides a mechanism to control the rates while maintaining low enough thresholds to meet the physics goals of the trigger. Even though single object triggers may suffice for low luminosity running, it is necessary to deploy the multi-object triggers at low luminosity to validate them and ensure their reliability as the LHC moves to high luminosity operations.

| Trigger Item | EM13_XE20 | EM7_MU6 | MU11_XE15 | MU10_J18 |
|---|---|---|---|---|
| Rate (Hz) | 225 | 10 | 13 | 33 |
| Trigger Item | 2J42_XE30 | 4J23_EM13I | 4J23_MU11 | EM13I_J42_XE30 |
| Rate (Hz) | 13 | 6.5 | 1 | 6.5 |

Table 9: A representative list of L1 trigger items and estimated rates at $10^{31}$ cm$^{-2}$ s$^{-1}$ for triggers combining several object types. The "_" notation is used to show the AND between two object types.

## 6.2 HLT signatures foreseen for a luminosity of $10^{31}$ cm$^{-2}$ s$^{-1}$

During low luminosity running, most of the triggers at Level 2 and Event Filter are either executed in pass-through mode or with loose selections. Table 10 shows some of the lowest threshold trigger items that can be executed without applying prescale factors and their estimated rates for $10^{31}$ cm$^{-2}$ s$^{-1}$. Higher threshold triggers, which will become important during high luminosity running are also deployed so that they can be validated with early data.

| Trigger Item | e12 | 2e5 | g20 | tau60 | tau25i_XE30 | MU10 | 2MU4 | e10_MU6 | J120 | 4J23 | 2b23 |
|---|---|---|---|---|---|---|---|---|---|---|---|
| Rate (Hz) | 19 | 7 | 7 | 10 | 3.5 | 18 | 2.3 | 0.5 | 9 | 7 | 3 |

Table 10: Examples of low threshold trigger terms executed without prescale factors and estimated rates that can be deployed at $10^{31}$ cm$^{-2}$ s$^{-1}$.

Figure 4 shows a graphical summary of the EF output rates for each trigger group and the cumulative rates, which provide a running total of the rates. The sum of the rates for all the trigger groups is more than the cumulative rates due to overlaps between the groups. The grouping is done as follows: single and multi-object triggers of the same object type are grouped together, hence "Electrons" refers to the total rates estimated for all single and multi-electron triggers including triggers executed with prescale factors and in pass-through mode. "*B*-Physics and Topological" refers primarily to *B*-physics triggers and other triggers where invariant mass cuts have been applied during the selection process, such as in selection of $J/\psi$ , $\Upsilon$ , and $Z$ decays. The "Other Topological" triggers require two or more object types, such as $e + jets$, $\tau + \not\!\!E_T$ etc.

The trigger grouping and associated rates are shown in finer detail in Table 11 for each of the trigger levels. The rates for each trigger grouping accounts for overlaps between signatures in that group, but not across groups. The "Total" row gives the cumulative rates for this trigger menu, accounting for overlaps between the trigger groups as well. The total output rates for each trigger level, for the proposed trigger menu at a luminosity of $10^{31}$ cm$^{-2}$ s$^{-1}$, is estimated to be within the available bandwidth, although there are large uncertainties inherent in the simulation. The estimated rate out of L1/L2 is 12 kHz/620 Hz well below their respective targets of 40 kHz and 1 kHz available during the LHC startup phase. The selections have been tuned to yield the targeted EF output rate of 200 Hz, but it is evident that this preliminary trigger list will need to be optimized based on early experience with real data.

## 7 Data streams

ATLAS has adopted an inclusive streaming model whereby raw data events can be streamed to one or more files based on the trigger decision. A proposed initial streaming configuration consists of four raw data streams called egamma, jetTauEtmiss, muons, and minbias. Each stream consists of events that pass one or more trigger signatures. The stream names indicate the type of trigger signatures they will contain, for example, events passing electron or photon triggers will be written to the egamma stream. Events passing certain topological triggers could be written to more than one stream. Two examples, $e + \not\!\!E_T$ and $e + \mu$, can be used to demonstrate the two possible modes for streaming events that pass a topological trigger. For the $e + \not\!\!E_T$ signature, events are only written to the egamma stream unless they also pass an inclusive $\not\!\!E_T$ signature after pre-scaling, in which case they are also written to the jetTauEtmiss stream. For the $e + \mu$ signature, events are written to both egamma and muon stream regardless of whether they pass the inclusive single electron or single muon trigger.

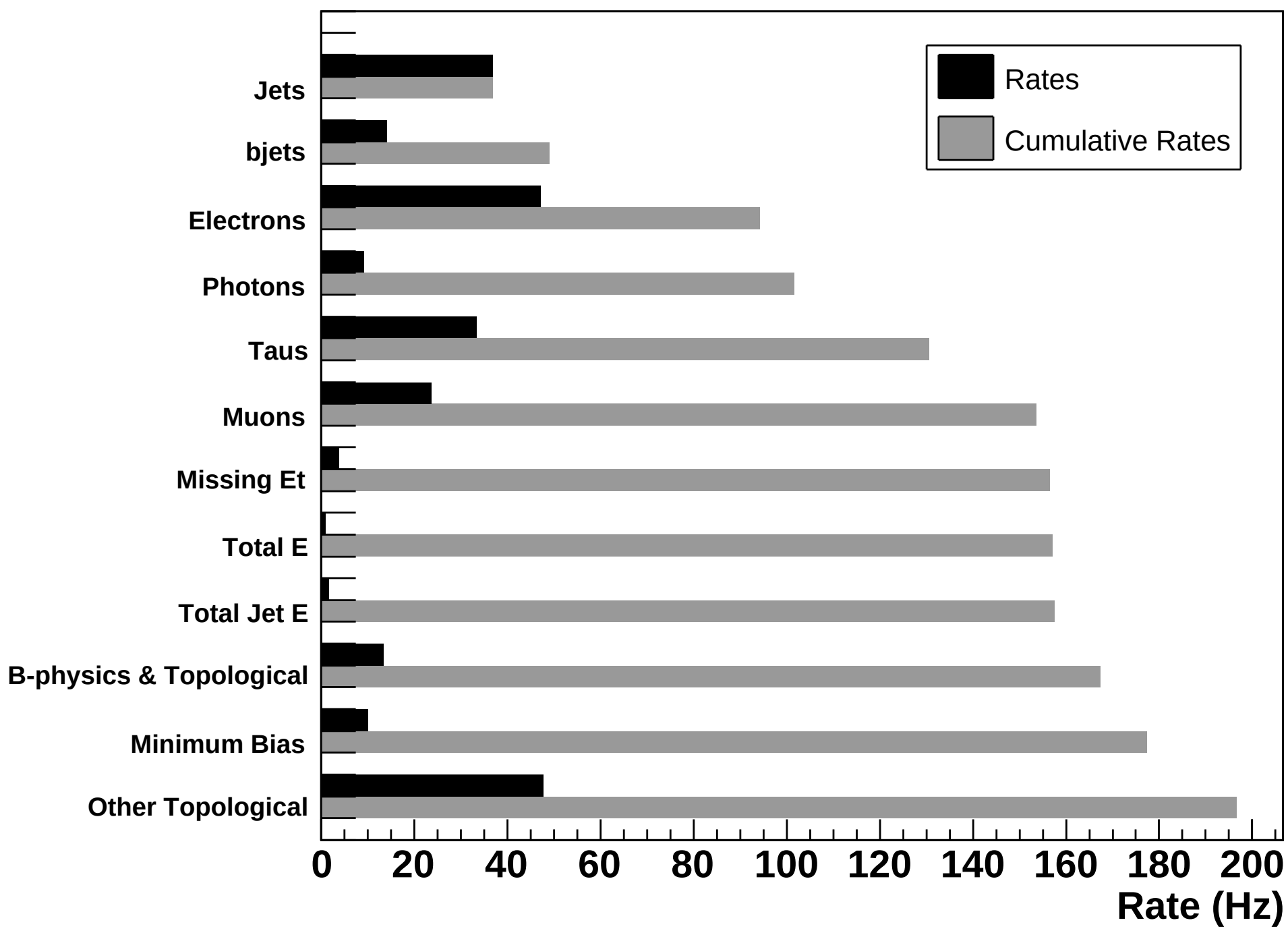

Figure 4: HLT unique (black) and cumulative (gray) estimated rates at $10^{31}$ cm$^{-2}$ s$^{-1}$ for different trigger groups as described in the text.

| Object | L1 (Hz) | L2 (Hz) | EF (Hz) |
|---|---|---|---|
| Single-electrons | 5580 | 176 | 27.3 |
| Multi-electrons | 6490 | 41.1 | 6.9 |
| Multi-photons | common | 2.9 | $< 0.1$ |
| Single-photons | common | 33.4 | 9.1 |
| Multi-Jets | 221 | 7.9 | 7.9 |
| Single-Jets | 24.4 | 24.4 | 24.4 |
| Multi-Fjets | 2.7 | 2.7 | 2.7 |
| Single-Fjets | 3.7 | 3.7 | 3.7 |
| Multi-bjets | common | 12.9 | 2.6 |
| Single-bjets | common | 11.6 | 11.6 |
| Multi-taus | 465 | 14.5 | 12.4 |
| Single-taus | 148 | 32.9 | 22.3 |
| Multi-muons | 68.6 | 5.8 | 2.3 |
| Single-muons | 1730 | 204 | 21.8 |
| Missing $E_T$ | 37.9 | 31. | 3.8 |
| Total $E_T$ | 6.3 | 6.3 | 1 |
| Total Jet $E_T$ | 1.6 | 1.6 | 1.6 |
| BPhysics | common | 25 | 13 |
| Muti-Object | 5890 | 134 | 48 |
| Minimum Bias | 1000 | 10 | 10 |
| **Total** | **12000** | **620** | **197** |

Table 11: L1, L2, and EF estimated rates for several groups of trigger items at a luminosity of $10^{31}$ cm$^{-2}$ s$^{-1}$. The total rate accounts for overlaps between the groups. The L1 objects labeled "common" have the same L1 triggers as other object types and hence do not require any additional L1 bandwidth.

Streams are chosen to have approximately the same proportion of events and to keep the total overlap (event duplication across streams) to less than 10%. The final optimization of these streams can only be achieved with an understanding of the overlaps observed with real data. Furthermore, the number and type of raw data streams may be optimized for use at different luminosity settings.

Table 12 shows the total and the unique rates of the proposed early stream configuration. The unique rate reflects the number of events written solely to the specified stream, hence the difference between the total rate and the unique rate is the rate of replicated events in each stream.

As shown in the Table, in addition to the raw data streams, an express stream and a calibration stream have also been defined. An express stream containing a subset of triggers can be used to provide rapid feedback on the quality of the data. It is therefore reconstructed first and any relevant knowledge of the quality of data is incorporated into the reconstruction of the remaining streams. Events in express stream are primarily intended for monitoring and not for physics analysis. Hence, by definition, events appearing in express stream would also appear in one of the primary raw data streams.

Additional triggers required for detector calibration can be run in a parasitic mode during data taking operations. These include triggers that provide data needed for detector alignment and energy scale determination. Such data can be written to their own raw data stream with the advantage of being processed early and the extracted calibration constants used as part of the bulk reconstruction of the primary raw data streams.

## 8 Evolution to higher luminosities

Experience with early running will allow further optimization of the trigger algorithms and menus and improve the ability to estimate rates in the high luminosity regime. As the LHC ramps up to its design luminosity, complex trigger signatures with multiple observables, higher $p_T$ thresholds and tighter selections will be deployed to maintain the output Event Filter rates at about 200 Hz.

As noted in Section 2, there are a limited number of L1 thresholds available for each object type. The intent is to keep the L1 thresholds as stable as possible. With increasing luminosity, higher L1 thresholds need to be introduced at the expense of some of the lower thresholds. However, many of the thresholds will be retained providing common points of comparison across luminosity regimes. At high luminosity, the luxury of running in pass-through mode or with loose selections will not be possible, and tighter HLT selections will be implemented to achieve the required rejection. For example, jet triggers that used only the L1 prescale to control the rates at low luminosity, will enable the HLT at high luminosity to obtain the additional rejection needed to control the EF output rates. In addition, topological triggers that include jets at lower thresholds than the inclusive jet triggers but in combination with other objects or requirements to achieve reduction in output rate, are deployed to increase the physics acceptance.

At high luminosities, the trigger software and selection must be robust against high detector occupancies, pile-up effects and cavern backgrounds. Pile-up effects becomes significant with increasing luminosities with an average of more than 20 interactions per crossing expected at a luminosity of $10^{34}$ cm$^{-2}$ s$^{-1}$. The trigger should ensure coverage of the full physics programme, including searches for new physics and precision measurements of Standard Model parameters. The signatures include lepton, photon, and jet triggers, but with higher thresholds and tighter selection criteria employed to control the rates. Additional requirements that operate in pass-through mode at low luminosities, such as isolation, large $\not{E}_T$ and other complex criteria such as flavour tagging, must also be deployed to achieve the necessary rate reduction.

Table 13 shows a representative sample of L1 and HLT trigger items that can be expected to be deployed without prescale factors at a luminosity of $2\times10^{33}$ cm$^{-2}$ s$^{-1}$. A comparison of this menu to the one at a luminosity of $10^{31}$ cm$^{-2}$ s$^{-1}$ illustrates the evolution of the rates and thresholds as a function of luminosity. The evolution is not linear as some of the triggers also have employed tighter selection conditions at higher luminosity. While Table 13 gives a flavour of some of the primary triggers and expected rates, the full physics trigger menu will consist of many additional signatures for precision measurements and the discovery program, as well as triggers required for calibration and background studies.

| Stream | Total Rate (Hz) | Unique Rate (Hz) |
|---|---|---|
| egamma | 55 | 48 |
| muon | 35 | 29 |
| jetTauEtmiss | 104 | 89 |
| minBias | 10 | 10 |
| express | 18 | 0 |
| calibration | 15 | 13 |

Table 12: Total and unique rates at $10^{31}$ cm$^{-2}$ s$^{-1}$ for a selected raw data stream configuration.

| L1 item | Rate (kHz) |
|---|---|
| EM18I | 12.0 |
| 2EM11I | 4.0 |
| MU20 | 0.8 |
| 2MU6 | 0.2 |
| J140 | 0.2 |
| 3J60 | 0.2 |
| 4J40 | 0.2 |
| J36_XE60 | 0.4 |
| tau16I_XE30 | 2.0 |
| MU10_EM11I | 0.1 |
| Others | 5.0 |

| HLT item | Rate (Hz) |
|---|---|
| e22i | 40 |
| 2e12i | $< 1$ |
| g55i | 25 |
| 2g17i | 2 |
| MU20i | 40 |
| 2MU10 | 10 |
| J370 | 10 |
| 4J90 | 10 |
| J65_XE70 | 20 |
| tau35i_XE45 | 5 |
| 2MU6 for *B*-physics | 10 |

Table 13: Subset of trigger items from two illustrative trigger menus at L1 (left) and at the HLT (right) for a luminosity of $2\times10^{33}$ cm$^{-2}$ s$^{-1}$.

Preliminary rate studies suggest that about 30% of the 200 Hz bandwidth will be available for electron and photon triggers, 25% for muon triggers, 15% for jet triggers, and 15% for triggers involving taus and $\not{E}_T$. About 5% of the bandwidth is allocated to *B*-physics related triggers which will involve low $p_T$ di-muon signatures with additional mass cuts to select $J/\psi$ and other rate *B*-meson decays, with the balance of the bandwidth used for calibration and background triggers. This proposed distribution, will of course evolve with experience, and will be tuned to ensure full coverage for discovery and precision physics at all luminosities.

# 9 Summary

The LHC is expected to begin its operation at a low luminosity of about $10^{31}$ cm$^{-2}$ s$^{-1}$ and ramp up to the design luminosity of $10^{34}$ cm$^{-2}$ s$^{-1}$ over the first few years of operation. The three levels of the ATLAS trigger have been designed to handle the high rates and occupancies at high luminosity. The trigger items and their performance have been studied in detail in both the low and high luminosity regimes and a comprehensive trigger menu has been developed for the LHC startup phase. Details of the triggers comprising this menu have been discussed in this note and primarily consist of low $p_T$ thresholds and loose selections that would allow for rapid commissioning and preparation for the high luminosity regime. The rates estimated for a trigger menu that will likely be deployed during the initial phase of the LHC run are within the limits of the TDAQ bandwidths, but are subject to fairly large uncertainties due to the use of simulations that extrapolate from the 2 TeV center-of-mass energy of the Tevatron to the 14 TeV expected for the LHC. The trigger menu in the high luminosity regime will use high $p_T$ thresholds and complex triggers involving multiple objects to provide efficient selection of physics processes and high background rejection. A flavour of such signatures has been discussed in this note. The initial running will allow further optimization of the trigger menus, which will evolve as the luminosity increases over three orders of magnitude to the design luminosity.

# HLT Track Reconstruction Performance

**Abstract**
This note reviews the tracking algorithms used at the L2 and Event Filter stages of the High Level Trigger of ATLAS. The tracking performance (efficiency, resolution) is studied for different topologies (single tracks, high and low $p_T$ jets) using simulated data. Detailed information on the execution time of the algorithms is also given.

## 1 Introduction

The aim of this note is to describe the tracking algorithms used at the L2 and Event Filter stages of the High Level Trigger of ATLAS and to study their performance.

The definitions of the relevant quantities (efficiency, fake rate, and resolution) given in this note can differ with the ones used in the different selection algorithms of specific trigger objects (e/$\gamma$, muon, taus, $b$-jet and $B$-physics): here the purpose is to define a common language to study and compare the different tracking algorithms.

A detailed description of the complete ATLAS detector and its performance can be found in [1]. In addition to a description of the trigger system, it also contains relevant information on the ATLAS tracking detectors, the Inner Detector (ID) and its subsystems (Pixel, SCT and TRT).

## 2 Track reconstruction at L2

### 2.1 Data preparation

Detector data should be converted before it can be used by tracking algorithms. The conversion process includes the bytestream decoding, the cluster formation in the Pixel and SCT detectors [1], and their conversion in spatial coordinates (space points).

### 2.2 IDScan

IDScan is a set of algorithms for fast pattern recognition and track reconstruction at the second level trigger, using space points provided by the tools described in Section 2.1. These algorithms first determine the $z$-position of the interaction point along the beam axis and then perform combinatorial tracking only inside groups of space points that point back to that determined position.

The first algorithm, aiming to determine the $z$-position of the primary vertex, divides the region-of-interest (RoI) into many equally-sized $\phi$ slices, whose width is tuned according to the individual RoI type, based on the lowest track momenta desired and the level of background hits in the detector. While tracks of high momenta produce most of their hits in a small number of neighbouring slices, hits from lower-momentum, curved tracks populate several different slices. Every space point is paired (or optionally every space point from the innermost three silicon layers) in each slice to the other space points in that slice and in a few neighboring slices, and each pair is used to calculate a $z$-position by linear extrapolation to the beam line. (This exploits the fact that helical trajectories of charged particles in a solenoidal magnetic field are straight lines in the $\rho$-$z$ projection.) A one-dimensional histogram accumulates all the calculated $z$ values, and the peak(s) in this histogram provide the rest of the IDScan algorithms with the $z$-coordinate of the $pp$ interaction point. The correct position is identified in more than 98% of the RoIs, with a resolution between 150 and 200 $\mu$m (depending on the type of RoI) for the central RoIs.

Using the $z$-position previously reconstructed, the second algorithm computes the pseudorapidity for all the space points in the RoI and fills a two-dimensional histogram in $(\eta, \phi)$. Since all hits from a given (sufficiently) high-$P_T$ track tend to be contained in a small solid angle (with its apex at the origin for the track), the space points from each track that originates from the computed $z$-position on the beam axis form a cluster of neighbouring bins in this histogram. When the bin size is small enough, the occupancy for each bin is low and each cluster, called a group, often contains the space points of a single track. Fake candidates are reduced by keeping track of which detector layers contribute space points to each bin and requiring that at least four out of an expected seven layers to have contributed to a given bin or its immediate neighbours, before that bin is included in a group.

After the groups have been identified random space points and/or space points from multiple tracks in each group are separated. This is achieved by considering all possible triplets of space points within a group and making use of the fact that any three hits from a track can be used to extract the same track parameters $\phi_0$ and $1/P_T$ in the transverse plane. The algorithm fills a two-dimensional histogram with extracted $(\phi_0, 1/P_T)$ values and considering combinations containing space points from at least four different silicon layers.

Finally the cleaned groups are subjected to the clone removal algorithm, which identifies groups sharing at least a certain number of space points (currently 2 or 3 depending on the RoI type) and removes all but the one with the highest number of space points. Furthermore, a group is removed if it shares more than 45% of its space points with others. This step significantly reduces the number of fake groups that contain a few random space points in addition to a small number of space points from an actual track.

After all these steps, the remaining groups are passed on to a track fitter. The default fitter used by IDScan is described in Section 2.5.

## 2.3 SiTrack

The SiTrack L2 algorithm adopts a combinatorial pattern recognition approach to reconstruct tracks starting from space points formed in the ID silicon detectors.

In order to perform space point combinations, these are first of all grouped into sets from which the entries of each combination will then be extracted; the grouping is implemented in SiTrack, using the idea of "logical layers". These correspond to a list of physical detector layers, i.e. barrel layers and end-cap disks, and are labeled with increasing numbers moving away from the beam line. The same physical layer can be included in more logical layers, to increase the robustness of the track finding process. To provide a tangible example, the first logical layer adopted for the reconstruction of high-$p_T$ isolated leptons includes the innermost two pixel barrel layers and the innermost pixel end-cap disk.

Once the space points have been associated to the logical layers they belong to, the track reconstruction algorithm proceeds through the following five steps:

- formation of track seeds;
- optional primary vertex reconstruction along the beam line;
- extension of track seeds;
- merging of extended seeds;
- clone removal.

The formation of track seeds corresponds to a combinatorial pairing of space points coming from the innermost two logical layers. For each seed, the extrapolation to the beam line is evaluated, using a straight line approximation; this process is depicted in Fig. 1. At this point a cut on the transverse impact parameter is applied. This cut, meant to reduce the number of seeds to be further processed, is particularly important, as it fixes the lowest reconstructible track $p_T$ value.

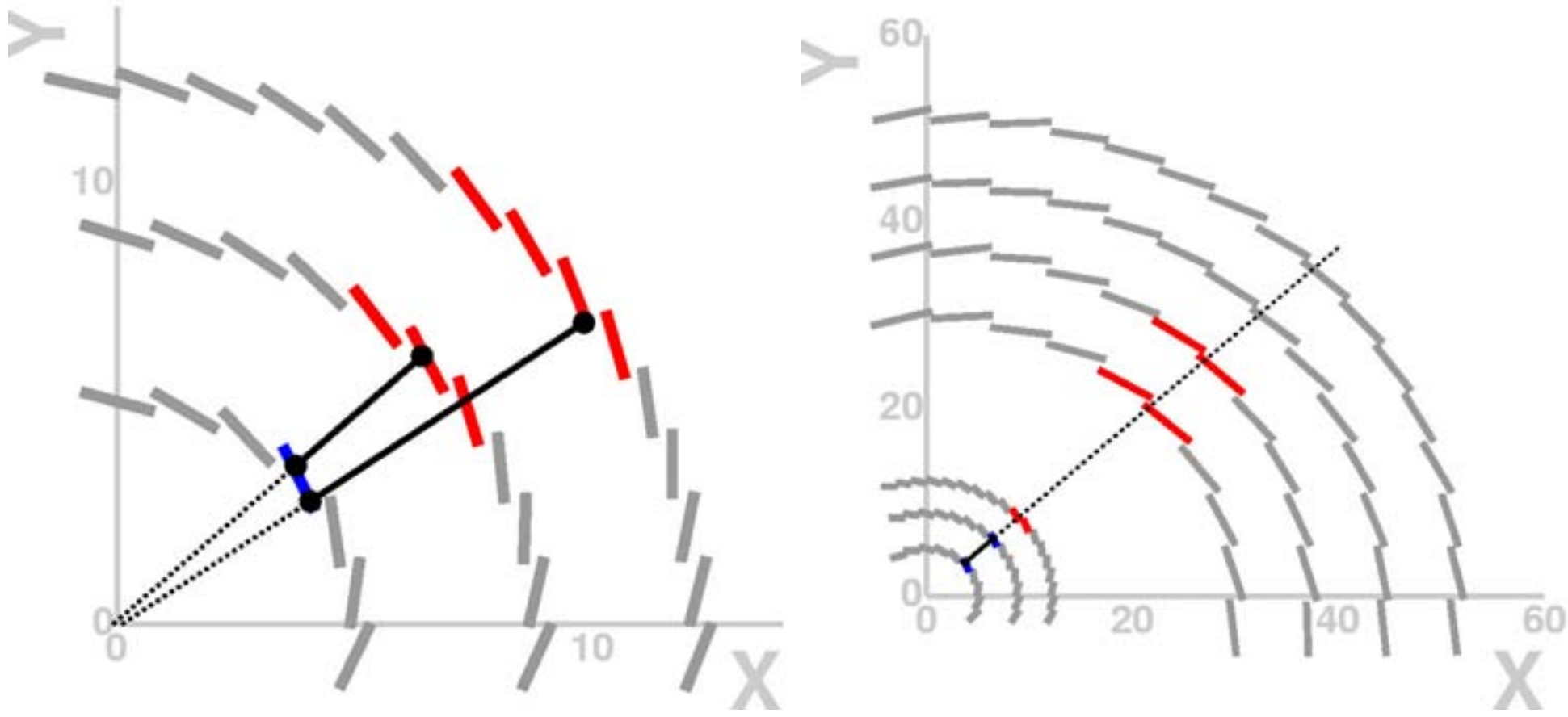

Figure 1: Pictorial scheme of the SiTrack combinatorial strategy for track seeds formation (left) and track seeds extension (right).

The subsequent step is the reconstruction of the position of the primary interaction vertex along the beam line, used to reject tracks not coming from the primary interaction. The vertex reconstruction is performed filling a histogram with the longitudinal impact parameter of the seeds and searching for histogram maxima; more vertex candidates can be retained and seeds not pointing to any of the reconstructed vertexes are discarded. This optional step is useful for high track multiplicity topologies like jets, but is typically skipped in the case of low multiplicity event topologies, e.g. for the reconstruction of single isolated leptons. Each retained seed is extended, as depicted in Fig. 1, extrapolating it to the outer logical layers and forming one or more space point triplets for each seed; extensions are selected applying a cut on the distance between the outer space point and the extrapolated seed. Each extended seed is then fitted by a straight line in the longitudinal plane and parametrized as a circle in the transverse plane.

All the extensions found for each seed must then be merged into a single full track, grouping the triplets having similar track parameters after the fit. The full track is thus formed by the union of the space points from all the merged extensions. All the triplets not involved in the merging process are discarded, while track parameters are re-evaluated for the full track, fitting it with a straight line in the longitudinal plane and a circle in the transverse one.

Two full tracks obtained from different track seeds may still share most of their space point; these tracks are defined as clones. To eliminate these ambiguous cases, only the clone track containing the largest number of space points is retained; in case more clone tracks contain the same number of space points, the one with the lowest $\chi^2$ value prevails. The retained full tracks are finally refit using one of the available common fit tools.

### 2.4 TRT tracking

Information from the TRT part of Inner Detector can be used as the basis of a L2 tracking algorithm. The core of the algorithm is a set of utilities from the offline reconstruction package xKalman [2] for the reconstruction of tracks in the TRT detector. It is based on the Hough-transform (histogramming) method. At the initialization step of the algorithm, a set of trajectories in the $\phi - R(Z)$ space is calculated for the barrel and endcap parts of the TRT. The value of the local magnetic field is taken into account at each straw position and coordinate along it when calculating the trajectories. After initialization, a histogram (with a size of 500 bins in $\phi$ and 70 bins in curvature) is filled for each event with the TRT hit positions. The track candidates can be identified from peaks in the histogram. Bins with at least eight hits are considered as track candidates. These track candidates should satisfy some quality parameters like the number of unique hits and the ratio of hits to number of straws crossed by the trajectory. For each track candidate, the parameters are tuned so that the track lies on the maximum number of drift circle positions. It is at this stage that drift information is taken into account to further improve the resolution of the track parameters.

## 2.5 Track fitting tools

The track fitting procedure used by the L2 ID algorithms in Pixel and SCT detectors is based on a Kalman filtering technique.

At first, space points are dissolved into clusters and a filtering node is created for each cluster. The filtering nodes encapsulate implementations of the Kalman filter algorithm for various measurement models. The fitter object uses the filtering nodes to update a track state described as a 5-dim vector of track parameters (local $x$, local $y$, angles $\phi$ and $\theta$ given in the global coordinate system, and track inverse momentum $Q$) and corresponding covariance matrix.

The track state update consists of three steps. First, the track state is extrapolated using a simple parabolic approximation of a trajectory in uniform magnetic field. The material-related corrections (multiple scattering, energy losses) are added to the covariance matrix during this step. The extrapolated track state is used to *validate* the next hit: if the $\chi^2$ distance between the hit and state is less then a predefined cut for this filtering node the track state is updated. These “extrapolate-validate-update” steps are repeated for every node. After that, a standard backward smoother is applied.

There exists another tool which performs track fit and simultaneous pattern recognition in the TRT. The implementation of this tool is based on a distributed approach. More details on the L2 TRT track extension tool can be found in Ref. [3].

## 2.6 Vertex Fitting tools

An essential part of the L2 event selection (e.g. $B$-physics event selection) is vertex finding and fitting using tracks reconstructed by the L2 tracking algorithms as input. Due to the L2 timing constraints a vertex fitting algorithm for the L2 application has to be fast. An additional requirement stems from the L2 track reconstruction which provides input track parameter errors in form of a covariance matrix. In contrast, vertex fitting algorithms proposed in literature assume uncertainties of the input track parameters to be described by weight (inverse covariance) matrices. However, if only track covariance matrices are available, these algorithms requires them to be inverted beforehand thus resulting in substantial computing time overhead.

To alleviate this drawback a fast vertex fitting algorithm capable of using track covariance matrices directly (i.e. without time-consuming inversion) has been developed. The specific feature of the algorithm is that track momenta at perigee points rather than “at-vertex” momenta are selected as the fit parameters. Such a choice of fit parameters makes it possible to apply a decorrelating measurement transformation so that the transformed measurement can be partitioned into two uncorrelated vectors – measured momenta and its linear combination with measured track coordinates at the perigee. This linear combination comprises a new 2-dim measurement model while the measured momenta and the corresponding blocks of the input track covariance matrices are used to initialise a vertex fit parameter vector and covariance matrix of the Kalman filter. This approach provides a mathematically correct and numerically stable initialisation of the vertex fit. A reduced size (2-dim instead of the usual 5-dim) of the measurement model makes the proposed Kalman filter very fast and therefore suitable for an online application in the ATLAS Level 2 Trigger. A detailed description of the L2 vertex fitting algorithm can be found in Ref. [4].

## 3 Track reconstruction at EF

ID reconstruction at the EF is performed using ATLAS "New Tracking" software [5]. A common approach between offline and online is possible thanks to the modular New Tracking design which allows the replacement of time-critical components and full-featured offline modules by trigger-specific implementations. New Tracking currently covers two sequences, the main inside-out track reconstruction (track finding starts from the Silicon and then is extended to the TRT) and outside-in tracking (from the TRT to the Silicon). The primary pattern search concepts for both sequences have been to a large extent adopted from the already existing ATLAS ID reconstruction program xKalman [2], but integrated and incorporating additional components in the common New Tracking approach. In the following note, only the inside-out reconstruction of tracks is described since is it the only one used in the EF ID online reconstruction. In the future, an outside-in approach is intended to be used in the trigger in cases where photon conversions are present.

The EF ID reconstruction runs for many different signatures, such as electrons, muons, taus, and *b*-jets. Each of these triggers contains a very similar algorithm sequence as the ID inside-out tracking and is followed, depending on a given object, by dedicated event reconstruction algorithms including vertex finding, *b*-tagging or electron processing. In the EF realization of New Tracking dedicated algorithms steer the underlying tools with RoI-seeded input collections [1]. The tools used are directly taken from the offline reconstruction chain but operated in a RoI-seeded mode, where the trigger signature defines a width of the RoI.

The EF ID algorithm sequence is divided into three stages defined as pre-processing, inside-out track finding and post-processing. The pre-processing stage is responsible for building clusters and drift circles in the Silicon and TRT detectors, respectively, and the creation of space points as three-dimensional representations of the Silicon detector measurements.

The inside-out track finding starts from the Pixel and SCT to find track seeds and creates track candidates based on the seeds primarily found. The seeded track finding results in a very high number of track candidates, that have to be resolved before an extension into the TRT detector can be done. Many of these tracks share hits, are incomplete, or describe fake tracks, hence ambiguity resolution is necessary. The track extension from the Silicon to the TRT is divided into two modules. First, tracks found in the Silicon detector are used as an input to find a compatible set of TRT measurements. Then each extended track is evaluated with respect to the original Silicon track. A track scoring mechanism is then used to compare the original track with the one after refitting, and the best track is chosen.

The last stage, post-processing, starts from a primary-vertex search, which is based on the Billoir fitting method [6]. A track object is created which is a representation of the track reconstruction results aimed for analysis applications.

Currently several different fitting techniques are implemented in New Tracking and can be chosen at a configuration level:

- Kalman Fitter as a straightforward implementation of the Kalman filter technique [7] that has been adopted for the track fitting in high energy physics experiments. For the ATLAS Silicon detector, the Kalman Fitter has a dedicated extension for fitting of tracks from electrons, that lose stochastically a significant part of their energy due to bremsstrahlung effects. In that case, an assumption about purely Gaussian noise is far from being optimal. A special Dynamic Noise Adjustment tech-nique has been developed [8]. It still uses a Gaussian error assumption but modifies the applied variance based on the amount of traversed material.

[1] The FullScan operation is an exception which assumes track reconstruction in the entire detector.

- Deterministic Annealing Filter is a deterministic annealing technique [9] which combines the standard Kalman filter formalism with a probabilistic description of the measurement assignment to a track.
- Gaussian Sum Filter is a special multi-Gaussian extension of the standard Kalman fitter [10], dedicated to reconstruction of electron tracks. In the GSF approach, the highly non-gaussian probability density function of electron energy loss is modeled by a mixture of several Gaussians.
- Alignment Kalman Filter is an extended version of the Kalman filter [11] that integrates an update of the detector surface orientation and position into an intrinsic measurement update of a Kalman Fitter step.
- Global Chi2 Fitter [12] is a track fit through a minimization of $\chi^2$ value. Given purely Gaussian process noise, the minimization of the $\chi^2$ value that is built from hit residuals at every measurement surface gives the best set of estimators of the track trajectory. Material effects enter the $\chi^2$ function as additional fitting parameters.

The Kalman Fitter approach without the Deterministic Annealing Filter extension is a default fitter at the EF ID. Other fitters described above can be chosen during the configuration step.

# 4 Performance

## 4.1 Timing measurement

This section presents a summary of CPU timing measurements for various steps of data preparation and track reconstruction in L2 ID and EF ID algorithms.

### 4.1.1 L2 ID

The L2 ID timing has been measured on a quad-core 3GHz Woodcrest CPU machine. Timers provided by the standard trigger monitoring framework have been used for these measurements.

The data preparation timing measurements obtained on $t\bar{t}$ data for the Pixel and SCT are shown in Table 1 for e/$\gamma$, muon, and tau triggers. The average cluster collection and space point multiplicities per RoI are presented in Table 2.

The track reconstruction timing measurements are summarized in Table 3 (note that SiTrack fit timing is included in the pattern recognition time). The average track multiplicity is presented in Table 4.

Tables 5,6, 7, and 8 present a comparison between RoI-based and FullScan running for $B-$physics triggers on $b\bar{b} \to \mu X$ data. These show, respectively, the time of data formation, average space point multiplicity, time of L2 ID track reconstruction, and average track multiplicity for the two cases.

### 4.1.2 EF ID

The EF ID timing measurements are shown in Table 9. They were done on a 3 GHz Intel Xeon 5160 CPU with 8GB of memory. The EF ID software was run in the emulator of the event filter processing and execution times for the algorithms were obtained as mean values from the timing histograms provided by the HLT framework. The results are representative in terms of algorithm execution times but do not account for data collection times, which are platform dependent. Two event samples were used in the tests, one corresponding to a single electrons with 100GeV, the other was a simulation of $t\bar{t}$ production.

| Data preparation step | Mean time [ms] | | |
|---|---|---|---|
| | $\mu$ | $e/\gamma$ | $\tau$ |
| RegionSelector | 0.23 | 0.26 | 0.32 |
| RobDataProvSvc | 0.02 | 0.02 | 0.02 |
| Cluster IDCs retrieval | 0.03 | 0.03 | 0.02 |
| BS-to-clusters, Pixel | 0.63 | 0.73 | 1.02 |
| BS-to-clusters, SCT | 0.60 | 0.69 | 0.95 |
| Pixel space point formation | 0.24 | 0.27 | 0.35 |
| SCT space point formation | 0.31 | 0.37 | 0.40 |
| Total time | 2.11 | 2.45 | 3.23 |

Table 1: The timing measurements of the data formation per RoI.

| Data preparation step | Multiplicity | | |
|---|---|---|---|
| | $\mu$ | $e/\gamma$ | $\tau$ |
| Pixel cluster coll. | 46.2 | 45.6 | 104.4 |
| SCT cluster coll. | 82.8 | 83.5 | 245.8 |
| Pixel space points | 66.4 | 65.9 | 131.6 |
| SCT space points | 36.5 | 38.8 | 87.1 |

Table 2: Average cluster/space point multiplicities per RoI.

The most CPU demanding operation is the track finding in the silicon detectors with typical execution times of 30 ms per RoI ( 300 ms in FullScan mode). Next is the processing of TRT extensions and resolution of ambiguities each with a CPU cost of about 20 ms per RoI (approximately 170 ms in FullScan). Another important contribution to the processing time is the data preparation step which takes about 20 ms per RoI for all detectors (and about 300 ms in FullScan mode). The total time to process $t\bar{t}$ events in full-scan mode is about 1s.

## 4.2 Efficiency and resolution definition

In order to evaluate the performance of any tracking algorithm, three kinds of information have to be provided: track reconstruction efficiency, the percentage of fake reconstructed track candidates, and the resolution of the track parameters.

| Track reconstruction step | Mean time [ms] | | |
|---|---|---|---|
| | $\mu$ | $e/\gamma$ | $\tau$ |
| IDScan : Pattern recognition | 0.60 | 0.70 | 1.45 |
| IDScan : Track Fit | 0.28 | 0.29 | 0.49 |
| IDScan : TRT data preparation | 1.14 | 1.32 | – |
| IDScan : TRT tracking | 2.96 | 2.87 | – |
| SiTrack : Pattern recognition | 0.62 | 0.66 | – |
| SiTrack : TRT data preparation | 1.15 | 1.39 | – |
| SiTrack : TRT tracking | 2.85 | 3.66 | – |

Table 3: The timing measurements of the L2 ID track reconstruction with IDScan and SiTrack.

| L2 ID algorithm | Tracks/RoI | | |
|---|---|---|---|
| | $\mu$ | $e/\gamma$ | $\tau$ |
| IDScan | 1.92 | 1.63 | 4.50 |
| SiTrack | 2.05 | 2.66 | – |

Table 4: The average track multiplicity per RoI for L2 ID track reconstruction with IDScan and SiTrack.

| Data preparation step | Mean time [ms] | |
|---|---|---|
| | RoI $0.75 \times 0.75$ | FullScan |
| RegionSelector | 0.67 | 4.76 |
| RobDataProvSvc | 0.02 | 0.41 |
| Cluster IDCs retrieval | 0.03 | 0.06 |
| BS-to-clusters, Pixel | 2.13 | 11.7 |
| BS-to-clusters, SCT | 1.89 | 13.1 |
| Pixel space point formation | 0.78 | 5.27 |
| SCT space point formation | 0.68 | 5.65 |
| Total time | 6.67 | 44.5 |

Table 5: The timing measurements of the data formation for RoI-based and FullScan running of $B-$physics triggers.

All the results shown in this section refer to reconstructed Monte Carlo (MC) simulated data samples, where the track parameters for both the reconstructed and the simulated tracks are available. In addition, each space point used to build a given track can be traced back to the MC particle that generated the corresponding charge deposit. This information is used by the ID trigger analysis packages to evaluate two additional quantities for each reconstructed track, which prove fundamental for the efficiency definition:

- the number of space points which trace back to the same MC track;
- the link between a reconstructed track and the MC track that generated most of its space points.

In this context we define:

- reconstructible MC particle: a particle passing a set of geometrical selection cuts (being contained in one of the processed RoIs, pointing to the primary vertex) and a $p_T$ cut;

| Data preparation step | Multiplicity | |
|---|---|---|
| | RoI $0.75 \times 0.75$ | FullScan |
| Pixel cluster coll. | 266.2 | 1744 |
| SCT cluster coll. | 824.7 | 8176 |
| Pixel space points | 246.2 | 1591.5 |
| SCT space points | 119.6 | 910.3 |

Table 6: The average cluster/space point multiplicities for RoI-based and FullScan running of $B-$physics triggers.

| Track reconstruction step | Mean time [ms] | |
|---|---|---|
| | RoI $0.75 \times 0.75$ | FullScan |
| IDScan : Pattern recognition | 3.00 | 34.6 |
| IDScan : Track Fit | 0.55 | 2.19 |
| IDScan : TRT data preparation | 2.05 | 9.50 |
| IDScan : TRT tracking | 5.43 | 19.5 |

Table 7: The timing measurements of the L2 ID track reconstruction for RoI-based and FullScan running of $B-$physics triggers.

| L2 ID algorithm | Tracks/RoI | |
|---|---|---|
| | RoI $0.75 \times 0.75$ | FullScan |
| IDScan | 4.26 | 16.9 |

Table 8: The average track multiplicity for RoI-based and FullScan running of the $B-$physics triggers.

- good track: a reconstructed track linked to a reconstructible particle by the majority of its space points; the geometrical and $p_T$ cuts used for the MC particles are applied to the reconstructed tracks too;
- best track: for each reconstructible particle, more than one good track can be available; the best track is defined as the one sharing the largest percentage of space points with the linked particle;
- fake track: a reconstructed track (passing geometrical and $p_T$ cuts) which is not a good track;

The most natural choice for the definitions of efficiency, fake fraction and resolution is then:

- efficiency: ratio between the best tracks and reconstructible particles;
- fake fraction: ratio between fake tracks and all the reconstructed tracks passing the applied cuts;
- resolution: difference between the track parameter of a good track and that of the linked reconstructible MC particle;

These definitions are used to produce the set of plots used in the following subsections to summarize the performance of a given tracking algorithm.

The track reconstruction efficiency is shown as a function of the absolute value of $\eta$ and $p_T$ of the reconstructible MC particle while the fake fraction is shown as a function of the absolute value of $\eta$ and $p_T$ of the reconstructed track. The track parameter resolutions are shown as a function of the absolute value of $\eta$ and $p_T$ of the reconstructible MC particle: the resolutions on $\phi$, $1/p_T$, transverse impact parameter ($d_0$) and longitudinal impact parameter ($z_0$) are shown as a function of the absolute value of $\eta$ while resolutions on $d_0$ and $1/p_T$ are shown as a function of $p_T$. The track parameter resolutions are evaluated with a gaussian fit to the resolution distribution for each parameter.

### 4.3 Results with isolated electrons

The tracking performance for electrons was evaluated with a data sample of single electrons uniformly distributed over a transverse momentum range of 7 to 80 GeV. Figure 2 shows the reconstruction efficiency and the fake fraction as a function of $\eta$ and $p_T$ for SiTrack, IDScan and EF tracking algorithms. Figure 3 summarizes, for the same algorithms, the track parameters resolutions. As shown in the figures, the efficiency is typically 95% or greater, except for low $p_T$ or high $\eta$ where it drops to about 90%. The fake fractions tend to be below 2% except at low $p_T$ or high $\eta$ where they can exceed 6%. Resolutions are observed to be quite good and are also somewhat degraded at low $p_T$ or high $\eta$.

| Algorithm | | Mean time [ms] | | | | |
|---|---|---|---|---|---|---|
| | | single e 100GeV | $t\bar{t}$ | | | |
| | | Electron | Electron | Muon | Tau | FullScan |
| Data preparation | Pix | 2 | 2 | 2 | 3 | 61 |
| | SCT | 8 | 8 | 10 | 10 | 85 |
| | TRT | 8 | 6 | 8 | 9 | 140 |
| Space point finder | | 1 | 1 | 1 | 2 | 30 |
| Track finding in Si | | 6 | 29 | 6 | 31 | 310 |
| Ambiguity solving | | 4 | 15 | 5 | 11 | 135 |
| TRT track extensions | | 1 | 3 | 1 | 1 | 31 |
| TRT extension processing | | 5 | 19 | 8 | 13 | 170 |
| Vertex finding | | 0.1 | 1 | 1 | - | 23 |
| Particle creation | | 0.4 | 2 | 1 | 1 | 24 |
| Total | | 35 | 86 | 43 | 81 | 1009 |

Table 9: Timing of EF reconstruction steps per RoI in two samples of events, single $e^-$ of 100 GeV and $t\bar{t}$ events. Electron, muon, and tau times were measured in the corresponding triggers, FullScan mode comes from the execution of B-physics triggers. The RoI sizes were $0.1 \times 0.1$ ($\Delta\phi \times \Delta\eta$) for the electron and muon triggers, $0.2 \times 0.2$ for tau triggers.

### 4.4 Results with isolated muons

The tracking performance for muons was evaluated with a data sample of single muons with transverse momenta of 6, 9, 21 and 30 GeV. Figure 4 shows the reconstruction efficiency and the fake fraction as a of function $\eta$ and $p_T$ for SiTrack, IDScan and EF tracking algorithms. Figure 5 summarizes, for the same algorithms, the track parameter resolutions. As shown in the figures, the muon performance is even better than for electrons, with close to 100% efficiency throughout the kinematic range and an extremely low fake rate. Resolutions are slightly better than for electrons, and show similar $p_T$ and angular dependence.

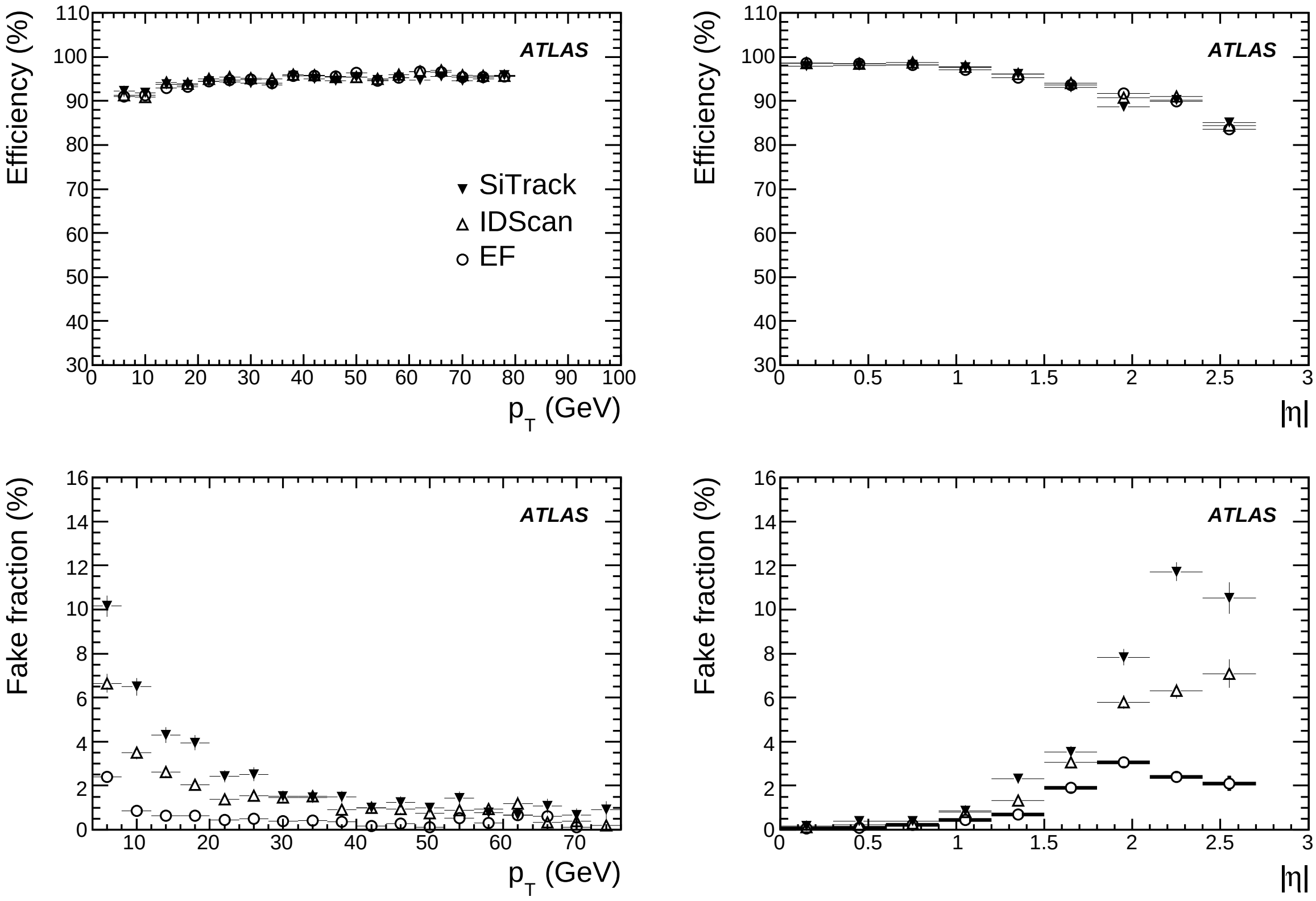

Figure 2: Electron track reconstruction efficiency (top) for single electrons as functions of $p_T$ and $\eta$ for SiTrack (full triangles), IDScan (empty triangles) and EF tracking (empty circles). Bottom plots show fake fraction as a function of the reconstructed $p_T$ and $\eta$.

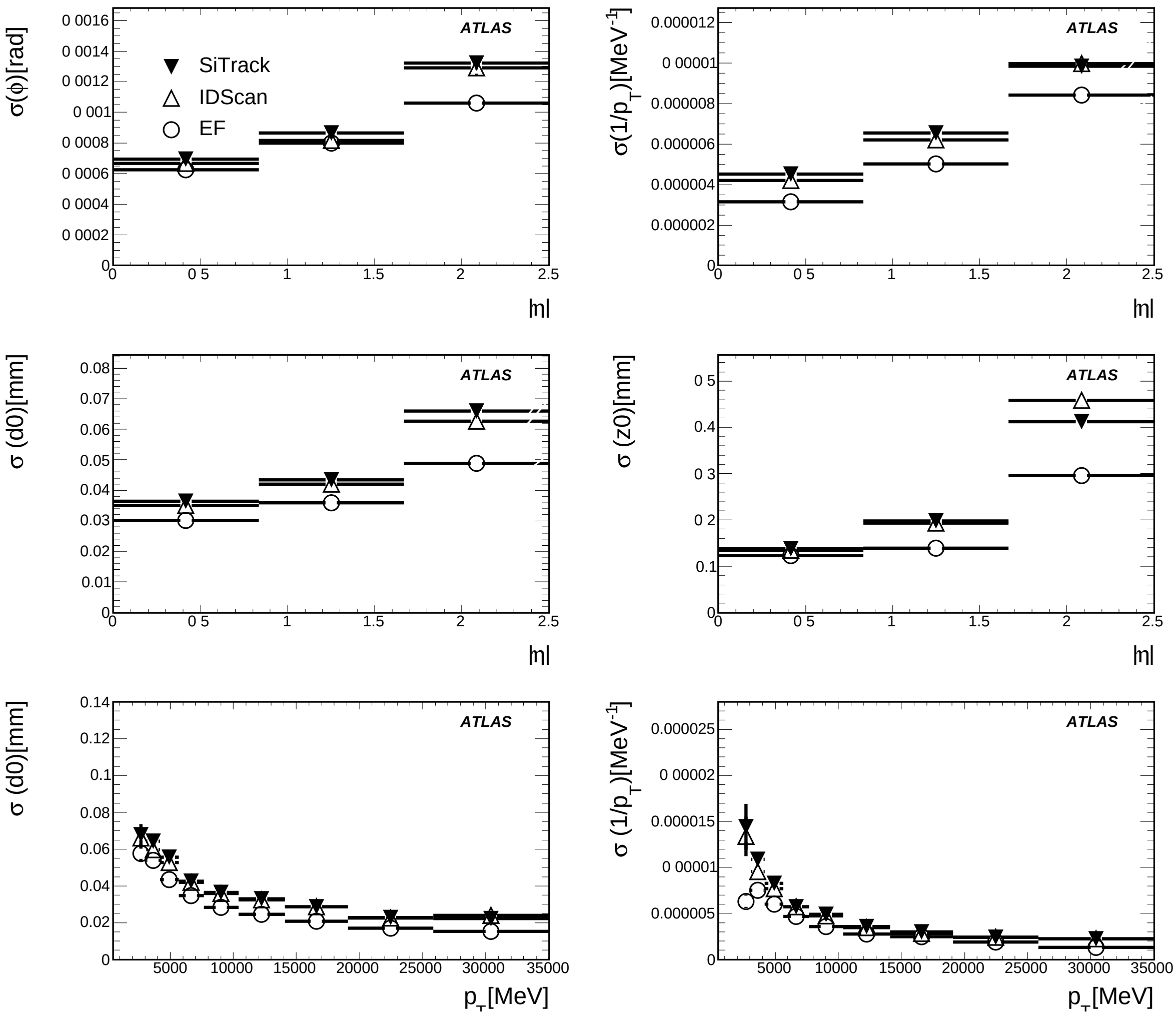

Figure 3: Track parameter resolutions for single electrons as a function of $\eta$ and $p_T$ for SiTrack (full triangles), IDScan (empty triangles) and EF tracking (empty circles).

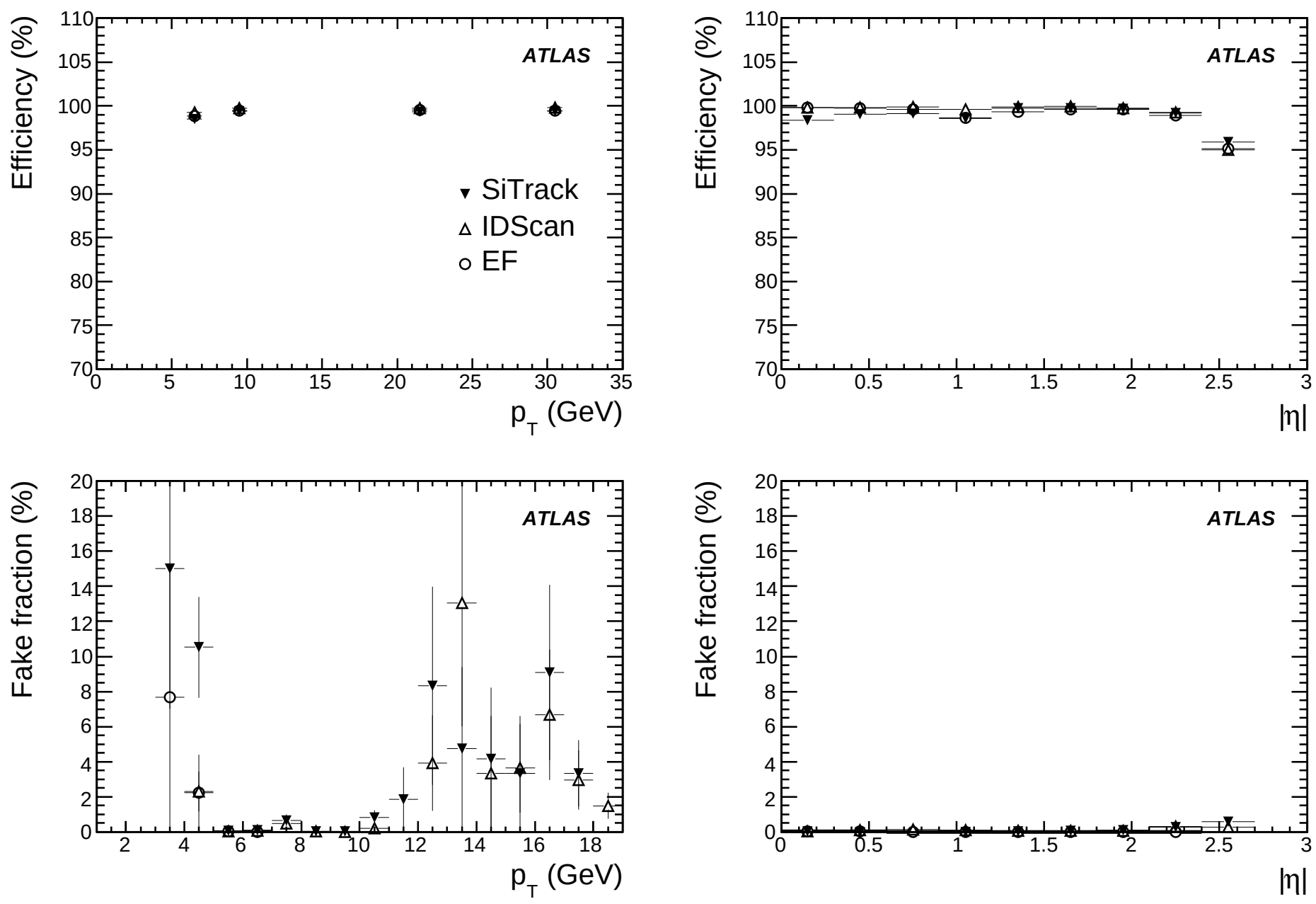

Figure 4: Muon track reconstruction efficiency (top) for single muons as functions of $p_T$ and $\eta$ for SiTrack (full triangles), IDScan (empty triangles) and EF tracking (empty circles). Bottom plots show fake fraction as a function of the reconstructed $p_T$ and $\eta$.

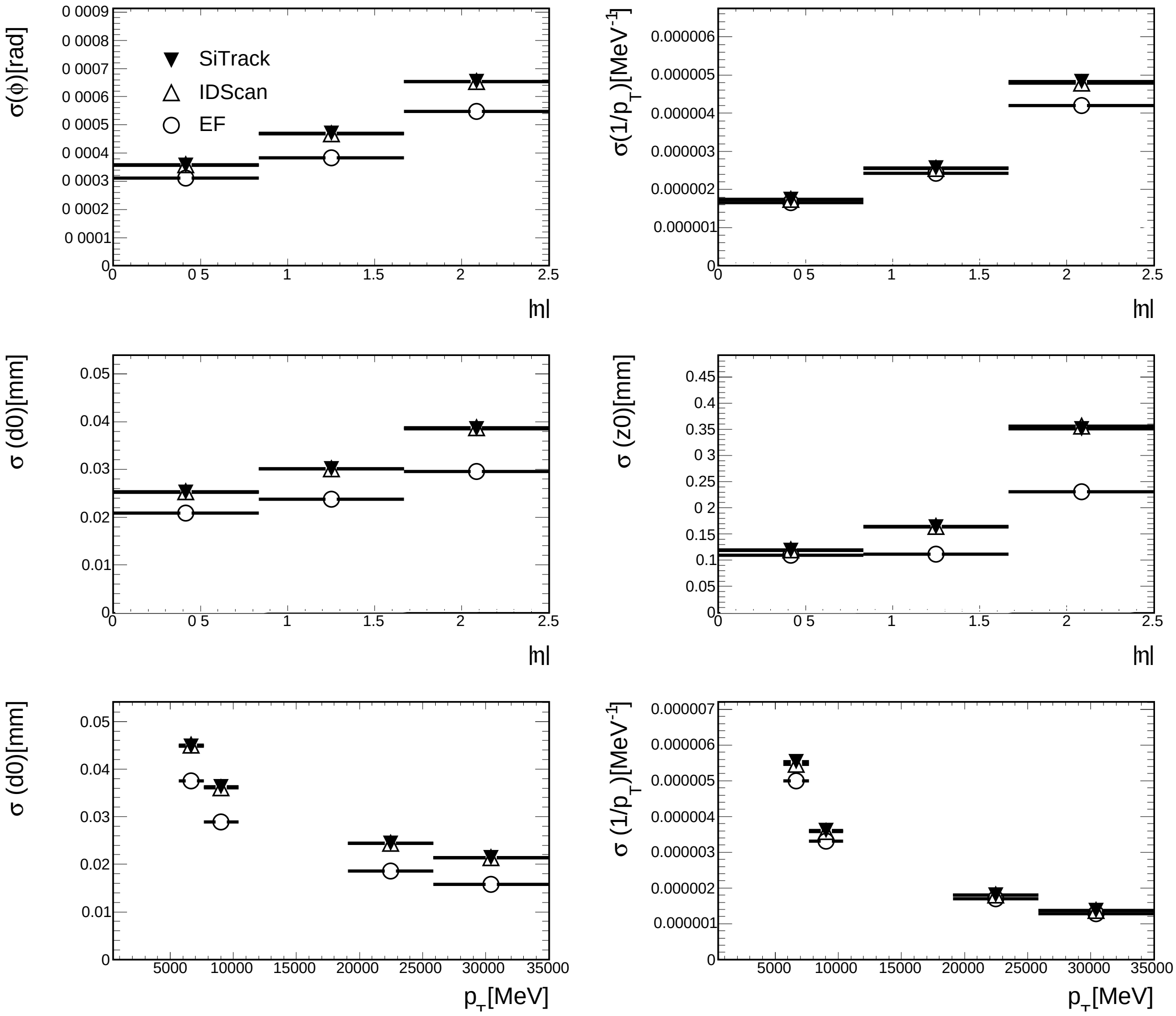

Figure 5: Track parameter resolutions for single muons as a function of $\eta$ and $p_T$ for SiTrack (full triangles), IDScan (empty triangles) and EF tracking (empty circles).

### 4.5 Results with jets

The tracking performance has also been evaluated using a sample of $b$-jets produced in the decay of a Higgs boson (mass 120 GeV) produced in association with a leptonically-decaying $W$ boson. This is the benchmark sample for $b$-tagging selection. Only tracks formed with at least four space points are considered.

Figure 6 shows the reconstruction efficiency and the fake fraction as a function of $\eta$ and $p_T$ for SiTrack, IDScan and EF tracking algorithms. Figure 7 summarizes, for the same algorithms, the track parameter resolutions. Not surprisingly, the efficiency for this sample is a bit lower (80 to 90 %) than the single electron and single muon samples, and has a bit higher fake rates. Resolutions are comparable or slightly worse.

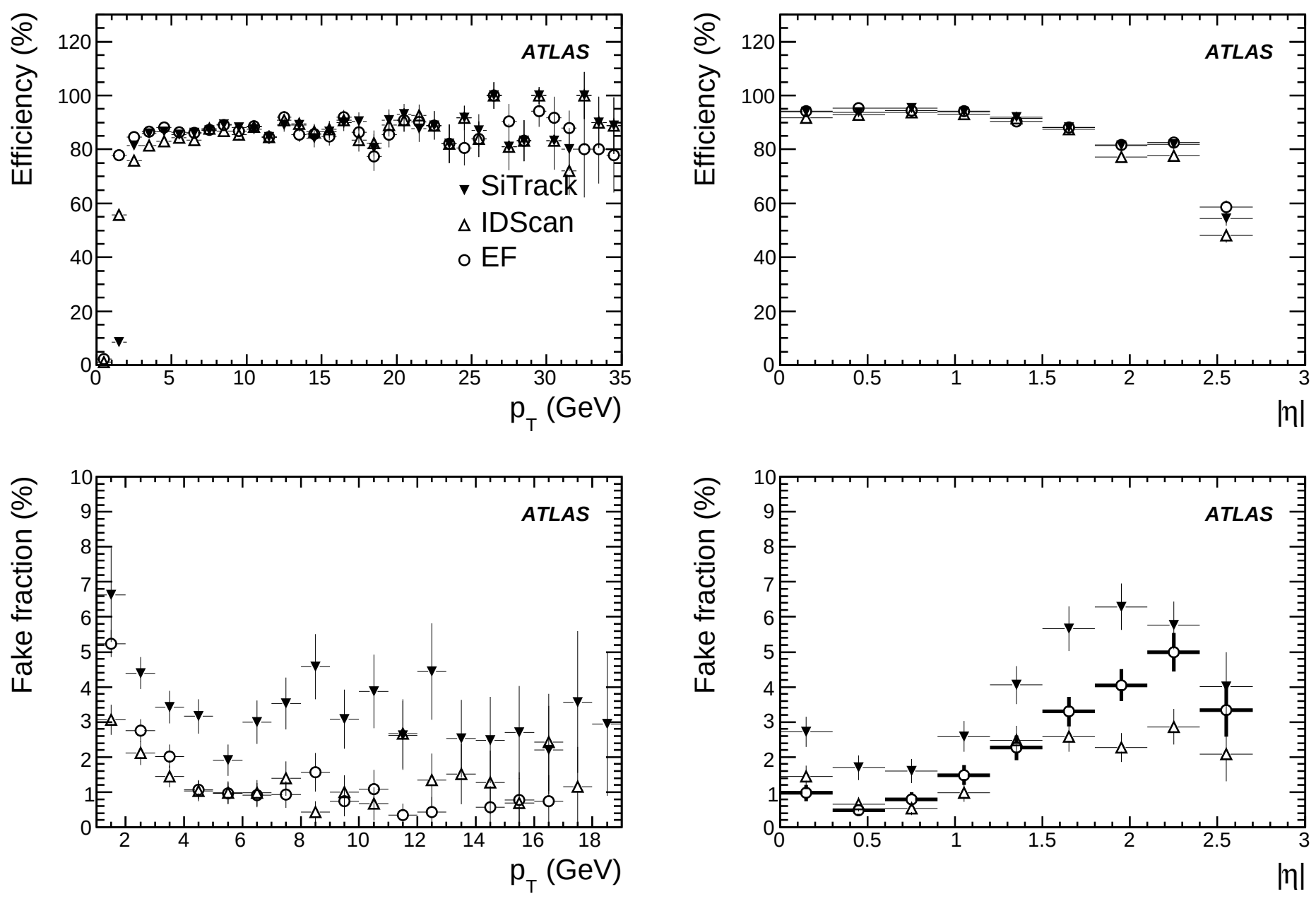

Figure 6: Reconstruction efficiency (top) for tracks in $b$-jets yielded by Higgs decay as functions of $p_T$ and $\eta$ for SiTrack (full triangles), IDScan (empty triangles) and EF tracking (empty circles). The efficiency as a function of $\eta$ is computed for tracks with $p_T > 3\ GeV$. Bottom plots show fake fraction as a function of the reconstructed $p_T$ and $\eta$.

### 4.6 Results with $\pi$ and $K$ in B-physics events

Using $B_s$ decay to $\phi\pi \rightarrow KK\pi$ the reconstruction efficiency for both kaons and pions have been evaluated.

Figure 8 shows the reconstruction efficiency as a function of $\eta$ and $p_T$ for SiTrack, IDScan and EF tracking algorithms. The efficiency is typically close to 95% dropping somewhat at low $p_T$ or high $\eta$.

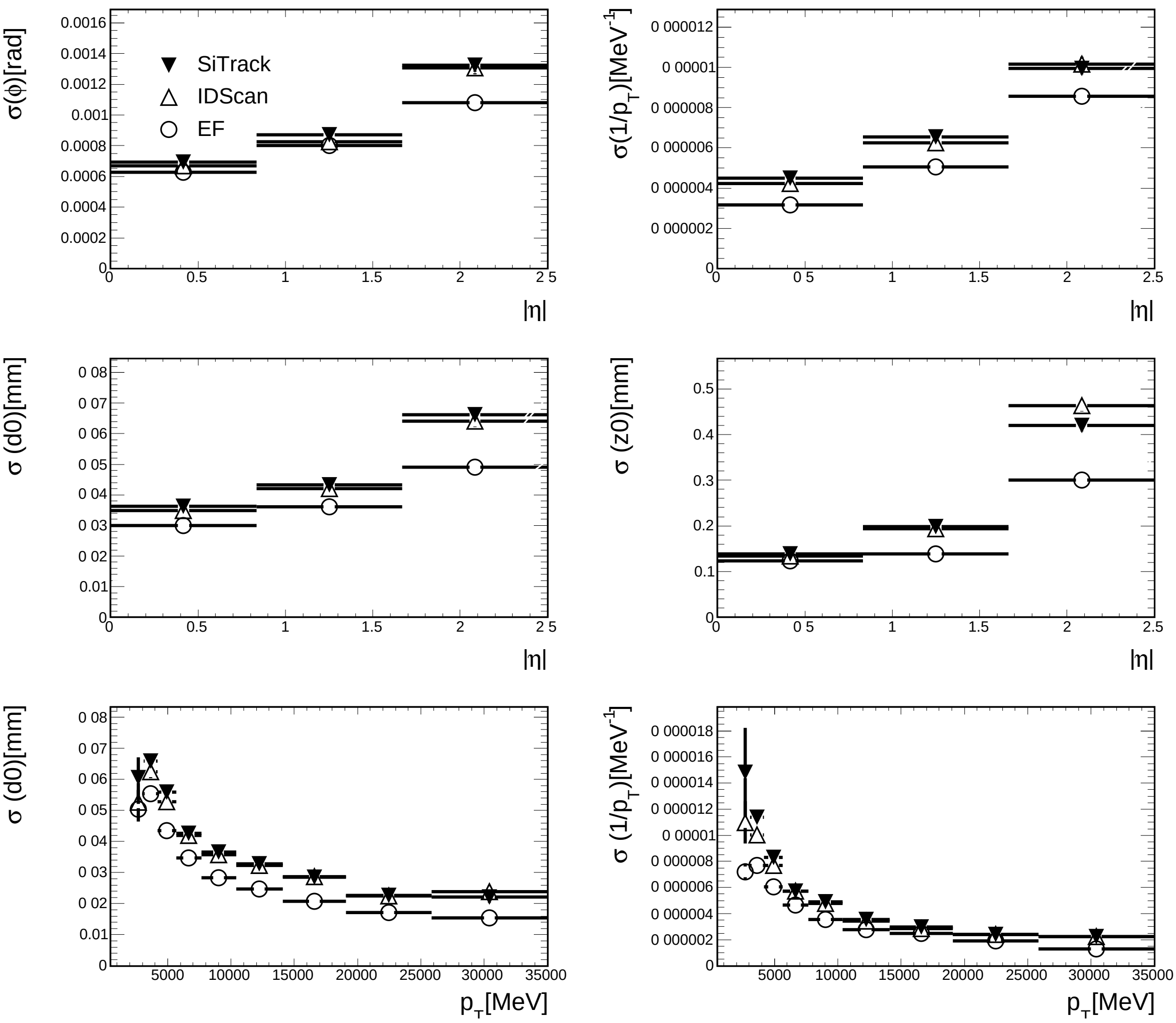

Figure 7: Track parameter resolutions for tracks in $b$-jets yielded by Higgs decay as a function of $\eta$ and $p_T$ for SiTrack (full triangles), IDScan (empty triangles) and EF tracking (empty circles).

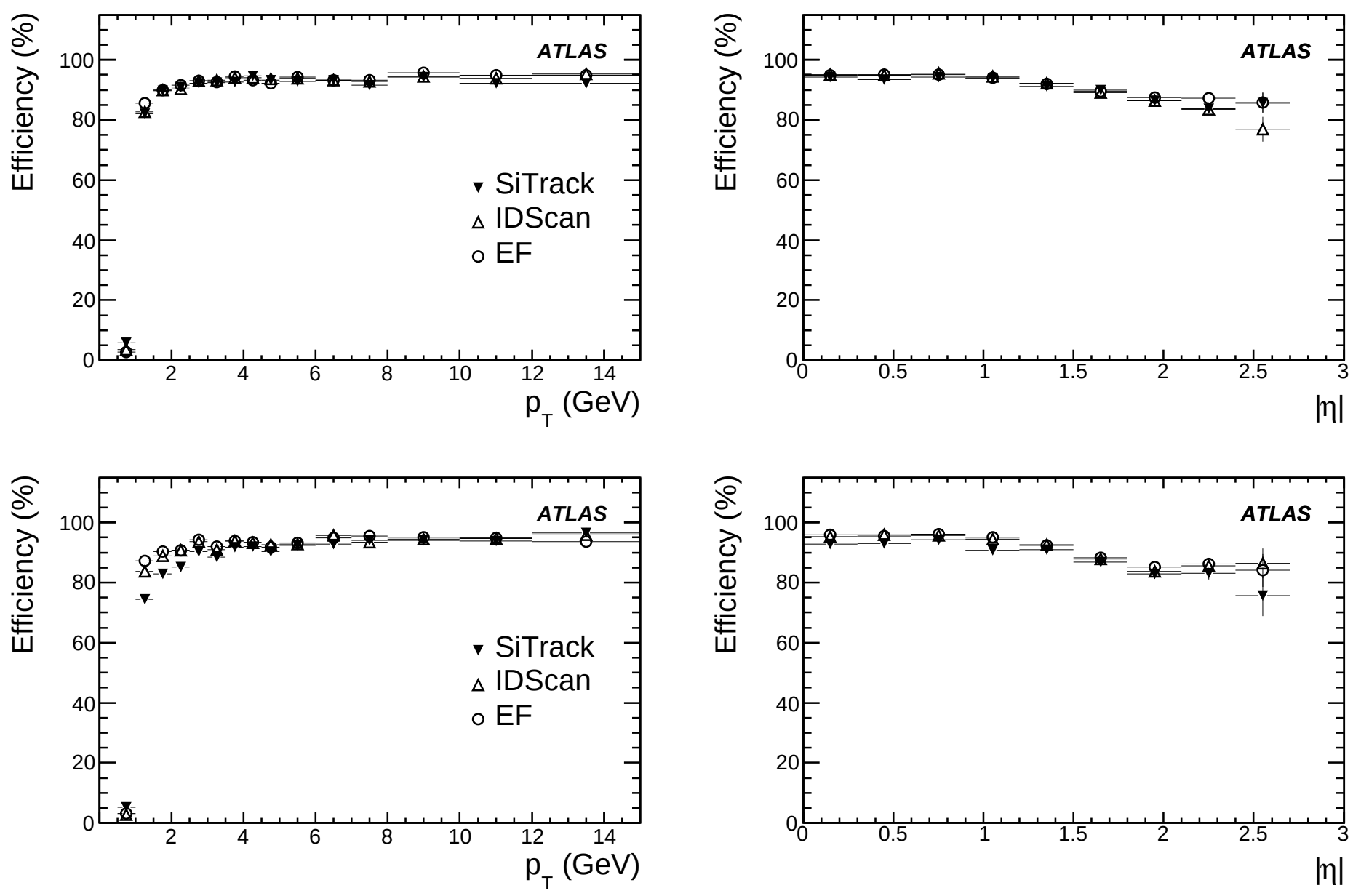

Figure 8: Reconstruction efficiency for kaons (top plot) and pions (bottom plot) yielded from a decay of the $B_s$ meson to $\phi\pi \rightarrow KK\pi$ (sample 16701 with misaligned geometry), as functions of $p_T$ and $\eta$ for SiTrack (full triangles), IDScan (empty triangles) and EF tracking (empty circles).

### 4.7 Comparison between the EF and the offline tracking performance

Since the track reconstruction in the EF is based on the same software as used for offline reconstruction, the EF tracking performance was studied under conditions equivalent to those used offline. This study was done using release 13.0.30.4 with perfectly aligned detector geometry.

To factorize out the bare EF performance, fake L1 RoIs were produced and passed through the L2 to seed the EF tracking, based on the MC truth information from particles with a $p_T > 1$ GeV and a $|\eta| < 3$.

The same selection as normally used for offline studies was applied, where only particles within $|\eta| < 2.5$, $|d0| < 2$ mm and $|z_0 - z_v| \times \sin\theta < 10$ mm were taken into account. The accepted tracks had to pass the requirement that at least 80% of the their hits were caused by the matched MC particle as well as the quality requirement of having at least 7 hits in the Pixel or SCT detectors together.

Figure 9 (left) shows the reconstruction efficiency as a function of $\eta$, for single electrons and muons with a $p_T = 5$ or 100 GeV. The efficiency is defined as the fraction of particles that produce an accepted track and the average efficiencies based on all electrons (muons) were found to be $83.9 \pm 0.2\%$ ($99.50 \pm 0.03$) for a $p_T = 5$ GeV and $92.9 \pm 0.1\%$ ($99.52 \pm 0.03$) for a $p_T = 100$ GeV. Figure 9 (right) together with Fig. 10 shows the muon resolution of $1/p_T$, $\phi$ and $d_0$ for accepted tracks as functions of $\eta$, where the inverse $p_T$ resolution is scaled by the $p_T$ value for easier comparison. In accordance with the offline studies, the resolution is determined from the root-mean-squared of the tracks within a region containing 99.7% of the distributions, i.e. within 3 standard deviations from the mean of a Gaussian distribution. All results are shown together with the offline results presented in [1], which is represented in the plots by the superimposed lines. The results agrees well, however, small deviations are seen due to slightly different software setup with respect to Ref. [1].

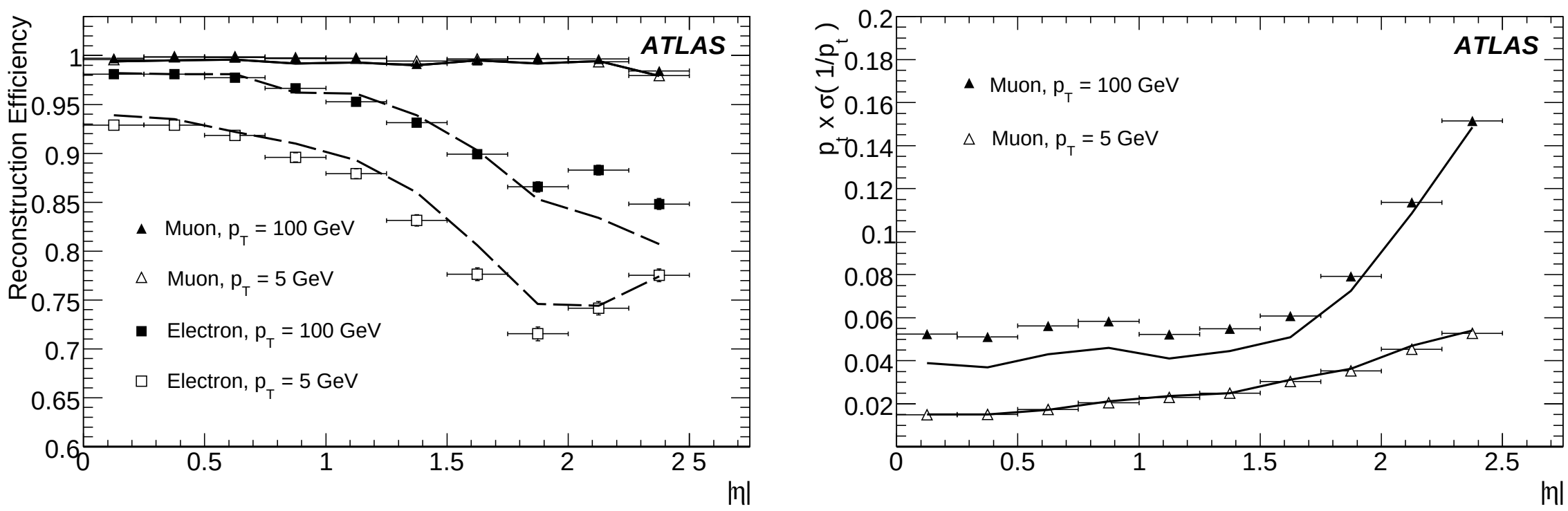

Figure 9: Reconstruction efficiency (left) and scaled $1/p_T$ resolution (right) as functions of $|\eta|$. The superimposed lines represent results from the offline track reconstruction.

## 5 Summary and conclusions

We have reviewed the track reconstruction algorithms used at the L2 and Event Filter stages of the High Level Trigger of ATLAS.

The algorithms performance and timing have been studied running on simulated data different chains of trigger selection.

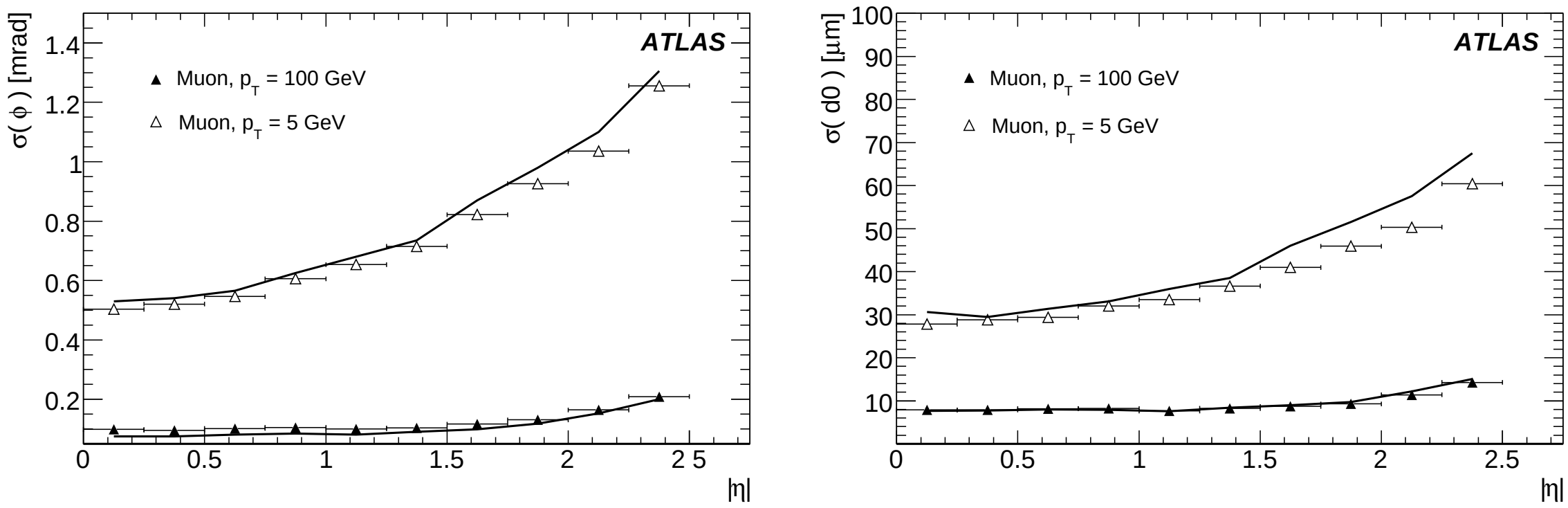

Figure 10: $\phi$ (left) and $d_0$ (right) resolution as functions of $|\eta|$. The superimposed lines represent results from the offline track reconstruction.

# Data Preparation for the High-Level Trigger Calorimeter Algorithms

**Abstract**
This note describes the data preparation necessary to enable the ATLAS High Level Trigger Calorimeter Algorithms. An overview of the infrastructure, which provides the transition from the calorimeter electronics to data reconstruction and trigger algorithm implementation, is given. This infrastructure is detailed as a separate note since it is relevant to all trigger algorithms requiring calorimeter information (electrons, photons, taus, jets, missing $E_T$ and muons).

## 1 Introduction

The calorimeters and part of the muon system were designed to participate in the ATLAS first level hardware-based trigger (L1) [1–4], while all sub-detectors participate in the software-based high level trigger (HLT) [5], comprised of Level 2 (L2) and the Event Filter (EF). One important phase for any trigger software algorithm is the data preparation step which provides the conversion of the bytes of data produced by the detector electronics into a convenient form for the trigger algorithms. In the case of the calorimeters, the digital information provided by the detector must be converted into calorimeter cells as input to the reconstruction algorithms. A good data preparation step will provide the input to the trigger software in an organized manner, so that access to the prepared data is optimized. This note describes this step for the calorimeter trigger software in the HLT. The same software data preparation layer is used in algorithms that are used to identify electrons, photons, taus, jets and muons [6].

### 1.1 Calorimeter readout

The fundamental LAr calorimeter readout unit is the calorimeter cell. The cell electrodes receive the current due to the drift electrons in the liquid argon and form a triangular shaped signal [1]. The shaping and readout of this signal is performed by the Front-End Electronics. To preserve the dynamic range and the energy resolution, the signal is shaped with three possible gains. The Front-End Boards (FEBs) save analog samples of the signals coming from the detector at the bunch crossing rate (25 ns). Each FEB can process up to 128 LAr calorimeter cells.

The signals are converted by the FEBs to a digital format if the event is accepted by the L1 Trigger. The digital information is sent to the ReadOut-Drivers (RODs). These are Digital Signal Processor (DSP) based machines, fast enough to deal with a number of input channels (2 FEBs feed one ROD DSP). From the pulse shape digitized at the FEB, the energy deposited in any cell can be calculated.

Data from each ROD are sent to a ReadOut Buffer (ROB). The ROBs keep this data fragment until it is requested by L2 or the Event Builder (EB). While L2 only requests a limited amount of data fragments, the EB will request fragments from the whole detector for events approved by L2 for subsequent processing in the EF computer nodes.

In the Tile Calorimeter [2], the photons produced in the scintillators are measured by photomultipliers, which produce a negative shaped pulse. The digitized electronic signal is saved into an on-detector memory waiting for the accept signal from the L1 trigger. For each of the 256 Tile Calorimeter modules, a so-called drawer (inserted in the back of the calorimeter structure) contains up to 48 photomultipliers and all the readout electronics.

The analog signals from the detector cells are also summed up by dedicated hardware by detector regions in depth. Trigger Towers (TT) are coarse granularity combinations of the detector cells and can be provided in analog mode to the hardware L1 processing. Except for the very forward regions, the TT size is $0.1 \times 0.1$ in $\eta \times \phi$. The L1 hardware algorithm uses some minimal TT energy and isolation quantities to define a possible L1 calorimeter candidate. A pointing to the found candidate $\eta \times \phi$ position is sent as a seed for software trigger processing. This seed is used to open a region (usually defined in terms of TT coordinates) called Region of Interest (RoI), where the full detector granularity ca be accessed by the reconstruction algorithms.

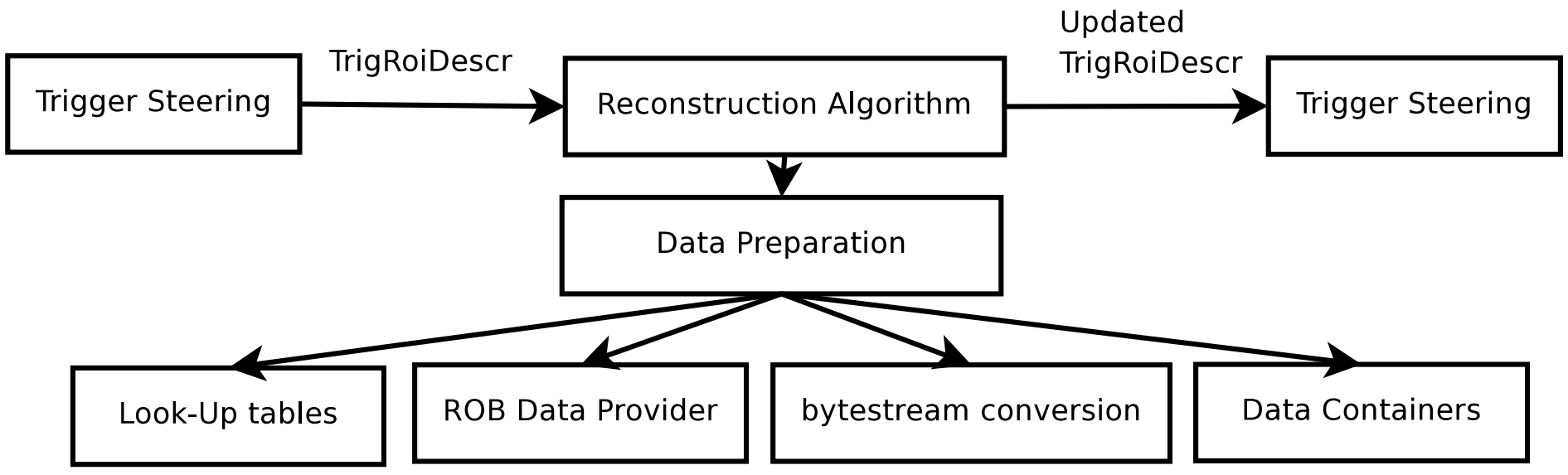

Figure 1: Different parts of the data preparation processing and their relation to the calorimeter algorithm at the L2. For details, see text.

## 2 Data preparation

From a general point of view the preparation of the LAr and Tile Calorimeters data is similar. Figure 1 depicts the global scope of the data preparation for the L2 calorimeter algorithm as described below. The EF structure is similar and is discussed briefly later in the note (see Ref. [7] for additional details of the full HLT data preparation).

The L2 software component called Steering receives the L1 information on the acceptance of an event along with the $\eta$ and $\phi$ coordinates corresponding to the L1 triggered object. The reconstruction algorithm gathers a list of ROB identifiers which contain data for a given RoI. Each ROB may partially contain data from TTs not pertaining to the RoI (ROB data access is not usually defined by the RoI, but rather by the hardware cabling). An optimal way to map cells to the towers and to the addresses of the ROB must be provided.

The mapping of the ROBs and TT are part of the geometry description and are also used in the offline reconstruction software framework. The access to the offline detector description databases which maps any physical position into a set of identifiers is typically very slow, as the full detector description is comprised of a great amount of data. In order to provide faster access, compatible with the L2 speed requirements, a look-up table called the Region Selector is prepared in the initialization phase of the trigger software.

The ROB identifiers are translated to network addresses of the ROB machines and the data are subsequently requested. The processing of the trigger selection algorithms is inhibited while the network acquires the data. Operating in a multiprocessing environment, as forseen for the ATLAS trigger, reduces this dead time [5, 8]. The ROB data provider receives the list of ROB identifiers and returns the detector data to the algorithms.

When data are received, pointers to the beginning of the different fragments are passed to the detector-specific bytestream conversion code. Bytestream conversion is the decoding of the data format produced by the detector RODs, and packaging of the data into an accessible format for the algorithms, in this case calorimeter cells. The last part of the data preparation is to provide the cells in a manner organized for the reconstruction algorithms. For instance, cells are provided by detector layer.

### 2.1 Data processing in the Read-Out Drivers

The LAr Digital Signal Processors (DSPs) are able to prepare data in different formats, the most important one is termed "physics mode." In this mode the DSPs process the nominal 5 samples per cell provided by the front-end electronics. These samples are used to compute the energy deposited in the cell by the particles using an optimal filtering (OF) [9]. This processing is a simple weighted sum of the samples. The weights include the noise autocorrelation, electronics calibration constants, and a normalization factor that converts ADC counts to MeV. For cells with energy above a programmable threshold the timing of the signal and the quality of the pulse shape compared to the expectation are calculated. Finally, for each cell, the choice of electronic gain applied to the analog signal in the FEB is recorded.

Beyond the cell-based data, the DSP can also extract global information at the FEB or TT level, which can be used to improve the L2 and EF processing speed. The DSP can sum up the energy in a given region in space providing $E_x$, $E_y$, and $E_z$ sums for these regions. The cell energies are added within these regions using cell-position-based projection coefficients loaded in the DSP from a database. Zero suppression is applied for cells below a given threshold. FEB's or TT's can be used to reconstruct jets or missing $E_T$ at L2 and EF if unpacking the full detector is too time consuming. Currently FEB's are being used to provide the energy sums, however, using the TT information instead is under evaluation to improve jet and missing $E_T$ resolutions.

The pulses from the Tile Calorimeter photomultipliers are also sampled and digitized by 10-bit ADCs. During a physics data taking, 7 samples (175ns) of the signal pulses are acquired and transmitted to the RODs. The information is processed using DSPs which also apply optimal filtering for cell energy reconstruction [10].

### 2.2 Region selector

As mentioned earlier, part of the information is stored in lookup tables for fast access to the detector description. In the LAr calorimeter case, the information unit to be correlated to the L1 position is the Trigger Tower. The $\eta \times \phi$ minimum and maximum and the ROD identifier for each TT is arranged in a large matrix. Multiple tables corresponding to the different calorimeter layers are available. For the transition region from barrel to endcap calorimeter the data of the fiducial volume covered by a TT may be provided by more than one ROD. In the Tile Calorimeter case, the geometry information is associated with the calorimeter module identifier and, again, the ROB identifiers. The Look-Up tables with geometry information for LAr and Tile are prepared by accessing the relevant conditions database.

### 2.3 Data containers

The data structure of a calorimeter cell includes a part common to LAr and Tile, and parts specific to both subdetectors. In the software these cells are organized in vectors, called collections. For the Liquid Argon Calorimeter each cell collection holds data for a LAr ROD, corresponding to two FEBs or, at most 256 cells. In the case of the Tile cell collection, there are either 23 cells (in the Barrel) or 13 cells (in the Extended Barrel) per collection. Data for four cell collections are associated with a single ROD. A Tile Calorimeter ROD has data for at most 92 Tile cells. Finally, the collections are organized in a vector which is called a container.

The containers for LAr and Tile are stored permanently in memory and the cells and collections are never deleted. This way, on-the-fly memory allocation, which is typically a slow operation in a computing system, is avoided. One problem with reusing collections is that the container must keep track of which collections have already been decoded in a given event. This information is provided by the tools that access the container. If requested subsequently in the same event, the collection will not be decoded again.

### 2.4 Bytestream conversion

The ROD fragments, containing the energy encoded information are provided to the appropriate HLT bytestream conversion code. Based on the ROD fragment identifier, the corresponding cell collection is located by the proper container (LAr or Tile containers). Subsequently, subdetector specific code is used to perform the data unpacking.

The LAr bytestream conversion code automatically identifies the fragment type using the ROD version encoded in the bytestream itself. Depending on the detected format, the corresponding internal infrastructure is selected.

The bytestream conversion software unpacks the energy information using the DSP physics output format as described in Section 2.1. The conversion provides the cell energy, hardware gain, pulse peak time, and pulse-fit quality information (if available) for each of the ROD fragment channels. The channel number is used as an index to the cell position in the cell collection, so that each LAr channel is associated to a single predefined cell object in the collection. Each cell is updated with the current values of energy, time, quality and hardware gain. In the unpacking step, typically more cells are requested than those contained in the RoI as data from one FEB may extend over several TTs. Furthermore, the trigger reconstruction algorithms require data access on a layer-by-layer basis. This results in a very complex operation with many checks of cell layer and position. Maps between TT identifiers and the associated groups of cells are prepared prior to algorithm execution, to speed up the process. Using the TT identifier list obtained from the Region Selector, a chain of cells for those TTs can be obtained, simplifying the algorithm code.

The bytestream conversion software for the Tile calorimeter data also checks the ROD format, ensuring that the correct method of unpacking the data is chosen. The data is decoded and the energy values are stored in a pre-allocated raw data structure. This is again used to avoid online memory allocation. The energy, time, and quality are stored together with the ADC identifier for each cell in a Tile Calorimeter drawer. This raw data is copied into the cell structure. The mapping of raw data to cells is the same for every drawer in a given calorimeter sector. To speed up the processing a mapping is built to the indices of the cells that correspond to each raw data.

Each Tile drawer is unpacked into a cell collection. The data providing in this case is much simpler than in the LAr case. Algorithms are able to iterate through the whole collection after the unpacking is done.

### 2.5 Data preparation in the EF

The Data Preparation tools in the EF make use of the same data unpacking approach that is used by L2, dumping this information into an offline cell container. As the EF has a larger time budget, however, more sophisticated algorithms and tools developed for offline reconstruction are used to process the cells stored in this container.

For each of the subdetectors (EM, HEC, FCal and Tile), the EF cell container is filled. This provides the possibility of unpacking only selected calorimeter sections, as needed. Once the container has been filled with the corresponding calorimeter cells, a set of software tools are executed to organize and check the container, and to perform cell-based calibrations.

## 3 Algorithm performance

In this section, the performance of representative HLT algorithms (e/$\gamma$ for L2 and missing $E_T$ for EF) and data preparation is studied. The results are based on bytestream files prepared with a format similar to the ATLAS raw output data.

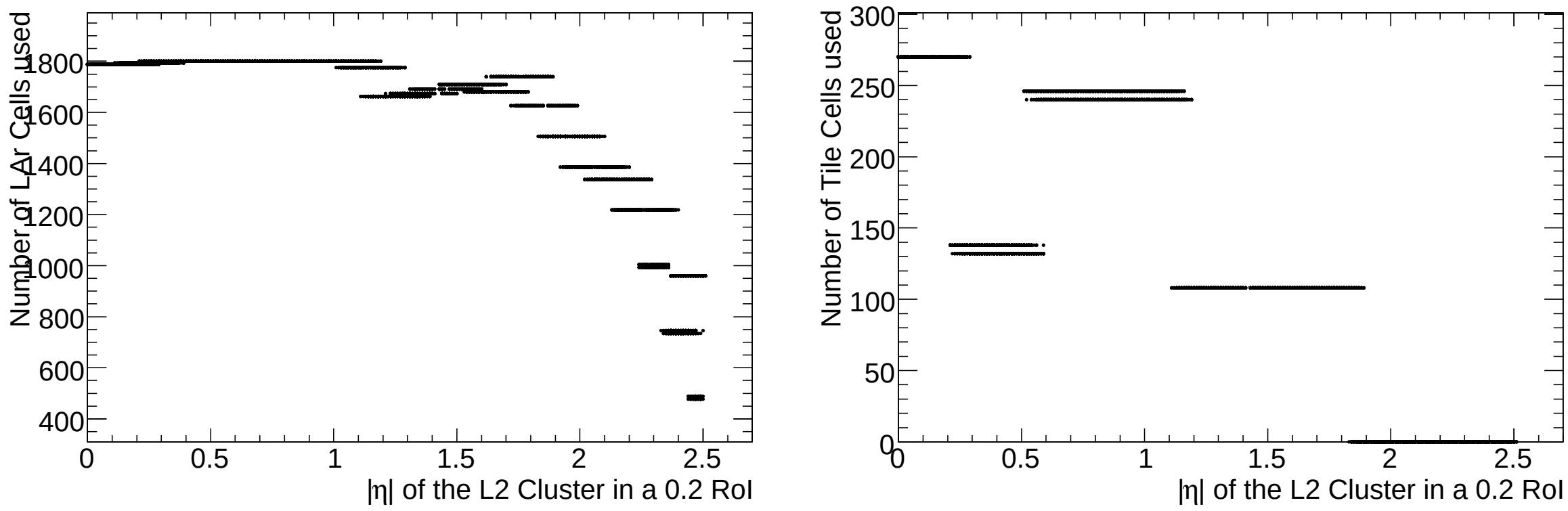

Figure 2: Number of cells used in the L2 e/$\gamma$ selection algorithm as a function of $\eta$ for the LAr (EM, EM endcaps and HEC) Calorimeters (left) and for the Tile Calorimeter (right). The RoI size was $0.4 \times 0.4$ in $\eta \times \phi$.

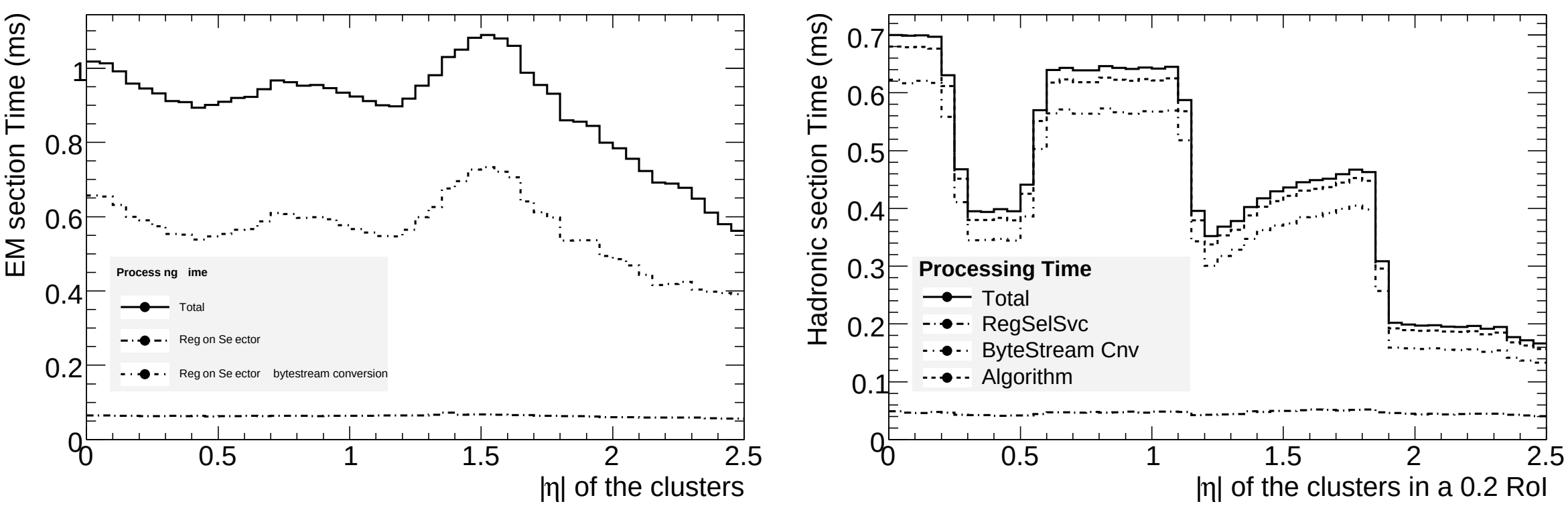

Figure 3: Cumulative time spent in the different phases (Region Selector, bytestream conversion, and algorithm) of the L2 e/$\gamma$ selection as a function of $\eta$ for the electromagnetic part (left) and for the hadronic part (right) for an RoI size of $0.4 \times 0.4$ in $\eta \times \phi$ (a 2.3 GHz machine was used to perform these measurements).

The primary performance issue is processing time. The processing time depends on the number of cells required, which is a function of $\eta$. Figure 2 shows the number of cells separately for LAr (left) and Tile (right) calorimeters; the overlapping bins in the figure are due to variable $\phi$ segmentation in transition regions between different calorimeter modules. The distribution on the left shows that the number of active cells in the barrel is quite uniform; since the endcap granularity is smaller, the number of unpacked cells decreases with increasing $\eta$. The distribution of the number of cells unpacked for the Tile Calorimeter depends on the number of drawers to be unpacked. In the very central region ($|\eta| < 0.4$), data from negative and positive rapidities must be accessed to complete the RoI, doubling the amount of data to be unpacked. A similar effect is observed in the region between the TileCal Barrel and Extended Barrel.

For the standard L2 e/$\gamma$ selection based on a RoI size of $0.4 \times 0.4$ in $\eta \times \phi$, the execution time for each of the processing steps was measured. The results, based on a sample of about 15,000 single electron events are shown in Fig. 3 separately for the EM (left: LAr) and for the hadronic (right: Tile and HEC) sections. It is important to stress that the processing time per RoI does not depend on event type, since the cluster sizes are constant for a given RoI. It can be seen that the Region Selector comprises a small portion of the total time, and that the bytestream conversion is the dominant source of time, with the algorithm itself taking only about 35% (10%) of the total time for the EM (Hadronic) calorimeters. Even though fewer cells are used in the EM calorimeter crack region (around $\eta = 1.5$ as shown in Fig. 2), these cells are distributed in two ROBs (one from the Barrel and another from the EM endcap), resulting in an overall increase in the processing time. Finally, we note that the conversion times are especially large in the regions covered by the Tile calorimeter and in proportion to the number of Tile calorimeter modules accessed. Work is ongoing to reduce these large processing times. The timing results averaged over $\eta$ are also summarized in Table 1. ROB data retrieval times are not included, since they can only be evaluated during real data taking.

| Reco. step | Region Selector | bytestream conversion | Algorithm | Total |
|---|---|---|---|---|
| EM $2^{nd}$ layer | $29\mu s$ | $169\mu s$ | $146\mu s$ | $347\mu s$ |
| EM $1^{st}$ layer | $13\mu s$ | $171\mu s$ | $113\mu s$ | $301\mu s$ |
| EM other layers | $21\mu s$ | $158\mu s$ | $56\mu s$ | $243\mu s$ |
| Hadronic | $46\mu s$ | $334\mu s$ | $43\mu s$ | $438\mu s$ |
| Total | $109\mu s$ (8%) | $833\mu s$ (63%) | $358\mu s$ (27%) | 1.33 ms |

Table 1: Processing time for different algorithm steps and for different actions. Improvements for the Tile calorimeter data preparation are envisaged. Time measurement excludes ROB data retrieval time (a 2.3 GHz machine was used).

These time measurements indicate that the preparation of the Tile calorimeter data needs to be improved. Even though about six times fewer cells are accessed for the barrel region, the data preparation time to run the hadronic part is comparable to the EM part.

Other trigger algorithms such as tau or jet identification need larger RoI sizes and consequently require more time. As an example, a jet algorithm which uses a $1.0 \times 1.0$ RoI size takes 10 to 12 ms.

### 3.1 EF missing $E_{\mathrm{T}}$ performance

The EF missing $E_T$ reconstruction algorithm accesses data from all the calorimeters and computes the missing $E_T$ with its $E_x$, $E_y$ components as well as the total scalar energy sum. In addition, corrections due to energy deposits from muons can be taken into account by including the results from the EF muon reconstruction.

To access the calorimeter data the algorithm uses the same data preparation layer used by L2. Since the ATLAS calorimeters contains about 200,000 cells, the access to every single cell can become too time consuming at the trigger level. A faster option is to use energy sums at the FEB level (discussed earlier in Section 2.1). FEB unpacking has only been implemented for LAr data, where the impact on the unpacking time is most significant.

In Fig. 4 the processing time of the EF missing $E_T$ algorithm is shown for full unpacking and FEB-based unpacking. On average, the time to process the whole calorimeter is dramatically reduced from 57 ms for the cell method to 2.4 ms for the FEB method.

The total scalar sum and the Missing $E_T$ calculations were performed using the two unpacking methods. The missing $E_T$ calculation does not depend on the method, while the scalar sum is systematically reduced in the FEB calculation due to the effect of the zero suppression. However, due to the drastic improvement in speed, the FEB algorithm is a valid option for the missing $E_T$ reconstruction.

In addition to the timing studies detailed here, a thorough study of the memory usage and initial-ization time was performed. A substantial fraction of the initialization time is taken by the detector geometry preparation, including the filling of the cell coordinates and the Region Selector tables. This initialization step requires access to databases containing information on detector conditions, and possibly files with complementary information. It was determined that the initialization time is acceptable and does not inhibit the running of any of the desired algorithms. Furthermore, the memory usage of the algorithms was measured to be stable and within acceptable operating limits.

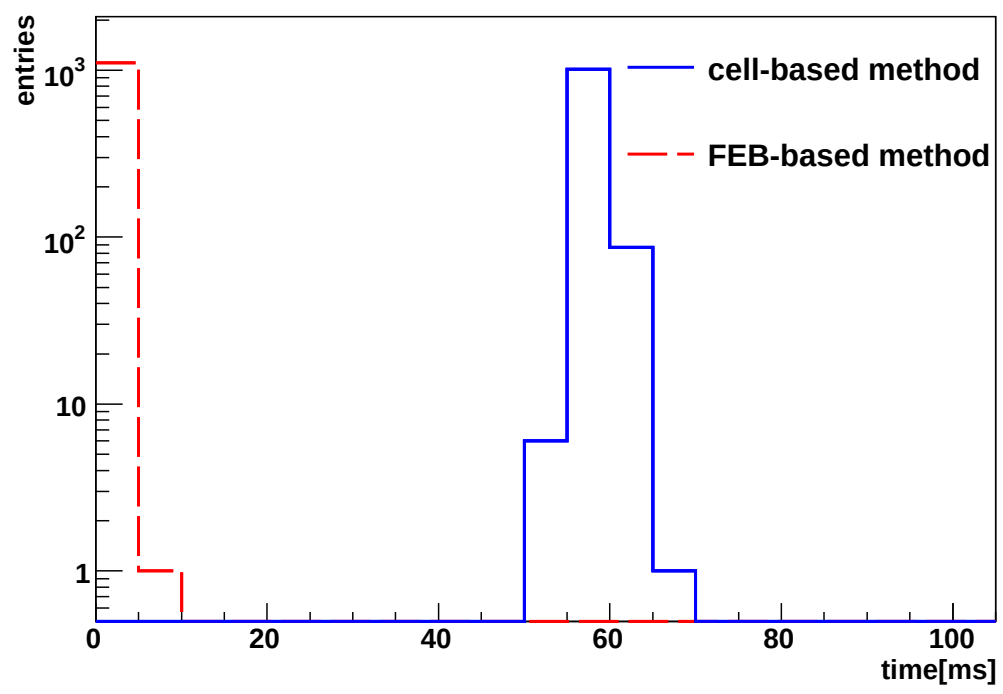

Figure 4: Processing time of the EF missing $E_T$ algorithm. The timing distributions for the cell and FEB method are shown. A 2GHz machine was used for this test.

## 4 Data preparation summary

This note describes the implementation of the whole data preparation step for the HLT calorimeter trigger from the detector electronics up to the reconstruction level. It is fundamental that a data preparation layer is efficient and fast, leaving time for the real physics algorithms. The High-Level Trigger Calorimeter tools described here have been used extensively with simulated data to commission the ATLAS trigger. A unique interface provides access to detector physics quantities (calorimeter cells) obtained with complex computations from the readout data. Knowledge of the detector details is, of course, a fundamental input into optimizing the strategy to be followed in this unpacking procedure. The critical performance issue for calorimeter data preparation is that it be accomplished within the online time budget, and this goal has been achieved with the current system. Even for special algorithms, like the missing $E_T$ which process cells from the whole detector, the data preparation performance is still within the required processing interval restrictions. Whenever FEB summary information can be used, significant timing reductions can be achieved. Further optimization studies are still in progress.

In addition to studies with simulated data, the tools and algorithms discussed here have been applied to commissioning runs of the ATLAS detector using cosmic rays. Until the LHC begins taking data, this is the only exercise that can approximate the real trigger usage in LHC conditions. Many trigger objects, such as taus, jets, and missing $E_T$ are being successfully debugged in this manner, providing important feedback to the algorithm developers.

# Tau Trigger: Performance and Menus for Early Running

**Abstract**
The selection of events with handronically decaying tau leptons is challenging due to high background rates at the LHC. On the other hand, efficient selection of events with tau leptons increases the discovery potential of ATLAS in many physics channels, notably Standard Model or Supersymmetric (SUSY) Higgs boson production. In this note we describe the ATLAS tau trigger system, focusing on the early data taking period, and present results from studies based on simulated events, including trigger rates and the acceptance of tau leptons from $W$ and $Z$ boson decays, Higgs Boson decays, and SUSY processes. In order to cope with the rate and optimize the efficiency of important physics channels, the results of the current simulation studies indicate that ATLAS tau triggers should include either relatively high transverse momentum single tau signatures, or low transverse momentum tau signatures in combination with other signatures, such as missing transverse energy, leptons, or jets.

## 1 Introduction

Tau triggers are designed to select hadronic decays of tau leptons, which mainly consist of one or three charged pions accompanied with a neutrino and possibly neutral pions. Leptonic tau decays are typically selected by electron [1] or muon [2] triggers. Tau triggers are an important part of the ATLAS trigger system, a fundamental component of the ATLAS detector [3].

ATLAS will collect data at different luminosities, starting from $10^{31}\,\mathrm{cm}^{-2}\ \mathrm{s}^{-1}$. At the lowest luminosity the focus of tau triggers is to collect samples that are useful for understanding the detector and the tau reconstruction software. Tau signatures combined with missing transverse energy signatures are essential to provide data samples enriched in W$\rightarrow \tau\nu$ events, which provide an important sample of real taus needed to refine tau identification algorithms. Additionally, single tau triggers with large prescale factors will provide samples for tau fake rates studies.

At higher luminosities, tau triggers will cope with the event rate increase by using higher $E_T$ threshold requirements, more restrictive identification requirements, or demanding a combination of different signatures, such as missing transverse energy, jets, or leptons. At high luminosity, tau triggers will be essential to enable the collection of data samples for searches based on single tau lepton final states, like Minimal Supersymmetric Standard Model (MSSM) $H^{\pm} \rightarrow \tau\nu$ [4] decays. They will also be used for final states with more than one tau lepton, like SM Higgs boson [5], MSSM neutral Higgs boson [6], or $Z'$ boson [7] decays.

The results presented in this paper are based on events fully simulated with GEANT4. The text is organized as follows. An overview of the three-level trigger selection for single tau triggers is presented in Section 2, while the performance in terms of resolution, rates, and efficiencies is described in Section 3. Timing studies of tau trigger algorithms are presented in Section 4. The current tau trigger menu, which includes various single and combined tau triggers, is described in Section 5, with a focus on the early data taking period corresponding to a luminosity of $10^{31}\,\mathrm{cm}^{-2}\ \mathrm{s}^{-1}$. Finally, a summary is presented in Section 6.

# 2 Overview of single tau trigger selection

## 2.1 Level 1 selection

The L1 tau trigger selection is closely related to the L1 electron/photon trigger ($e/\gamma$), and is fully documented in [8]. It is a hardware trigger based on electromagnetic (e.m.) and hadronic calorimeter information, and uses trigger towers of approximate size $\Delta\eta \times \Delta\phi = 0.1 \times 0.1$, with a coverage up to $|\eta|$ $< 2.5$ (given by the inner-detector coverage and the high-granularity e.m. calorimetry).

The algorithm considers a rectangular Region of Interest (RoI) of $4 \times 4$ towers ($0.4 \times 0.4$ in $\eta \times \phi$) in both the e.m. and hadronic calorimeters, and makes use of different elements, each formed by summing $E_T$ over a group of towers. The algorithm uses the following quantities:

- the central $2 \times 2$ *core* cluster is the energy measured in the central 2×2 e.m. and hadronic towers
- the *TauCluster* is the energy defined by the two most energetic neighboring central towers in the e.m. calorimeter plus the central $2 \times 2$ towers of the hadronic calorimeter
- *EmIsol* is the energy in the e.m. isolation ring (the region between 2×2 and 4×4 towers in the e.m. calorimeter).
- *HadIsol* is the energy in the hadronic isolation ring (region between 2×2 and 4×4 towers in the hadronic calorimeter).

The L1 tau trigger candidate is accepted if the core cluster is a local $E_T$ maximum and also satisfies additional conditions on *TauCluster*, *EmIsol* and *HadIsol* [8]. Its position is taken as the center of the 4×4 tower RoI, and its energy is calibrated using a procedure derived for jets (see Section 2.1.1). A maximum of eight trigger thresholds are available at L1 for taus. Each threshold is a combination of requirements of $E_T$ thresholds for *TauCluster*, *EmIsol* and *HadIsol*. A L1 tau trigger candidate passing the requirements is then passed to L2 for further examination.

### 2.1.1 Level 1 Calorimeter Calibration

Trigger towers are formed by the analogue summation of calorimeter cells. Calibration of the trigger consists of adjusting the overall gains of the towers. The e.m. towers are calibrated to optimize the e.m. trigger response. The gains of the hadronic towers are adjusted to provide a uniform jet response in $\eta$ using a jet sample with an $E_T$ range 50-100 GeV and making $\eta$-dependent adjustments to the trigger thresholds.

## 2.2 Level 2 selection

The L2 tau trigger selection uses the full calorimeter granularity and the inclusion of tracking information from the Inner Detector to refine the L1 selection. The selection is designed to further reject QCD jet backgrounds by exploiting more of the characteristics of a hadronic tau decay, such as collimation and low track multiplicity.

### 2.2.1 Level 2 calorimeter selection

At this stage, the selection of the single tau triggers is based on the calorimeter shower shape variables and the energy of a reconstructed cluster within the RoI to enrich the sample of tau candidates. Shape variables are calculated using only the second e.m. sampling layer (out of four). The second sampling is where most of the e.m. energy is deposited, therefore it provides the most information about the

e.m. shower shape. The cluster energy is calculated with all available e.m. and hadronic layers. Three different rectangular windows are defined, centered on a *seed* cluster, with areas of $\eta \times \phi = 0.1 \times 0.1$ (narrow, *Nar*), $0.2 \times 0.2$ (wide, *Wid*), and $0.3 \times 0.3$ (normal, *Nor*) [1]. The *Nor* window is equivalent to the RoI used at L2.

The algorithm consists of several steps. First, a *seed* cluster is found, using the second e.m. sampling. The algorithm unpacks cells in the *Nor* window centered around the L1 RoI position, and finds the cell with the highest energy deposition. In a *Wid* window around the most energetic cell, the cluster position is defined as the energy weighted mean position of the cells. Then, shape variables are reconstructed using the different windows around the *seed*, as described in the following. At the same time, the total energy in all calorimeter samplings is computed. Finally, the total energy is corrected with a simple sampling calibration.

The variables used by the L2 Calorimeter selection are the following:

- *EMRadius* is the energy weighted squared radius of the *seed*, which is obtained from the sum of the individual energy weighted squared cell distances from the *seed*. It is calculated in a *Nor* window around the *seed*, in the second sampling of the e.m. calorimeter, i.e.

$$EMRadius = \frac{\sum_{Nor} E_{cell} \cdot R_{cell}^2}{\sum_{Nor} E_{cell}}. \tag{1}$$

- *IsoFrac* is the difference in energy between the *Nar* and *Wid* window, normalized to the *Wid* window. It is calculated in the second sampling of the e.m. calorimeter. The definition is

$$isoFrac = \frac{\sum_{Wid} E_{cell} - \sum_{Nar} E_{cell}}{\sum_{Wid} E_{cell}}. \tag{2}$$

- *StripWidth* is the width of the energy deposition, defined as the energy weighted standard deviation in $\eta$. It is calculated in a *Nor* window around the *seed*, in the second sampling of the e.m. calorimeter. The formula is

$$stripWidth = \sqrt{\frac{\sum_{Nor} \eta_{cell}^2 \cdot E_{cell}}{\sum_{Nor} E_{cell}} - \left[\frac{\sum_{Nor} \eta_{cell} \cdot E_{cell}}{\sum_{Nor} E_{cell}}\right]^2}. \tag{3}$$

- *EtCalib* is the calibrated total transverse energy, calculated in the e.m. and hadronic calorimeters, in a *Nor* region around the seed.

The distributions of *EMRadius*, *IsoFrac* and *stripWidth* are shown in Fig. 1 for tau trigger candidates from generated tau leptons decaying hadronically and for QCD jets for two different $E_T$ regions. In the top row distributions for low $E_T$ tau leptons from $W \rightarrow \tau\nu$ decays are compared to background QCD jets distributions, while in the bottom row high $E_T$ tau leptons from Supersymmetric neutral Higgs decays (with a mass of 800 GeV) are compared to QCD background. The top row shows the difficulty in triggering on low $E_T$ taus as the separation between signal and background is not so dramatic; the separation is much better for high $E_T$ taus.

[1] The window sizes are currently subject of optimization studies.

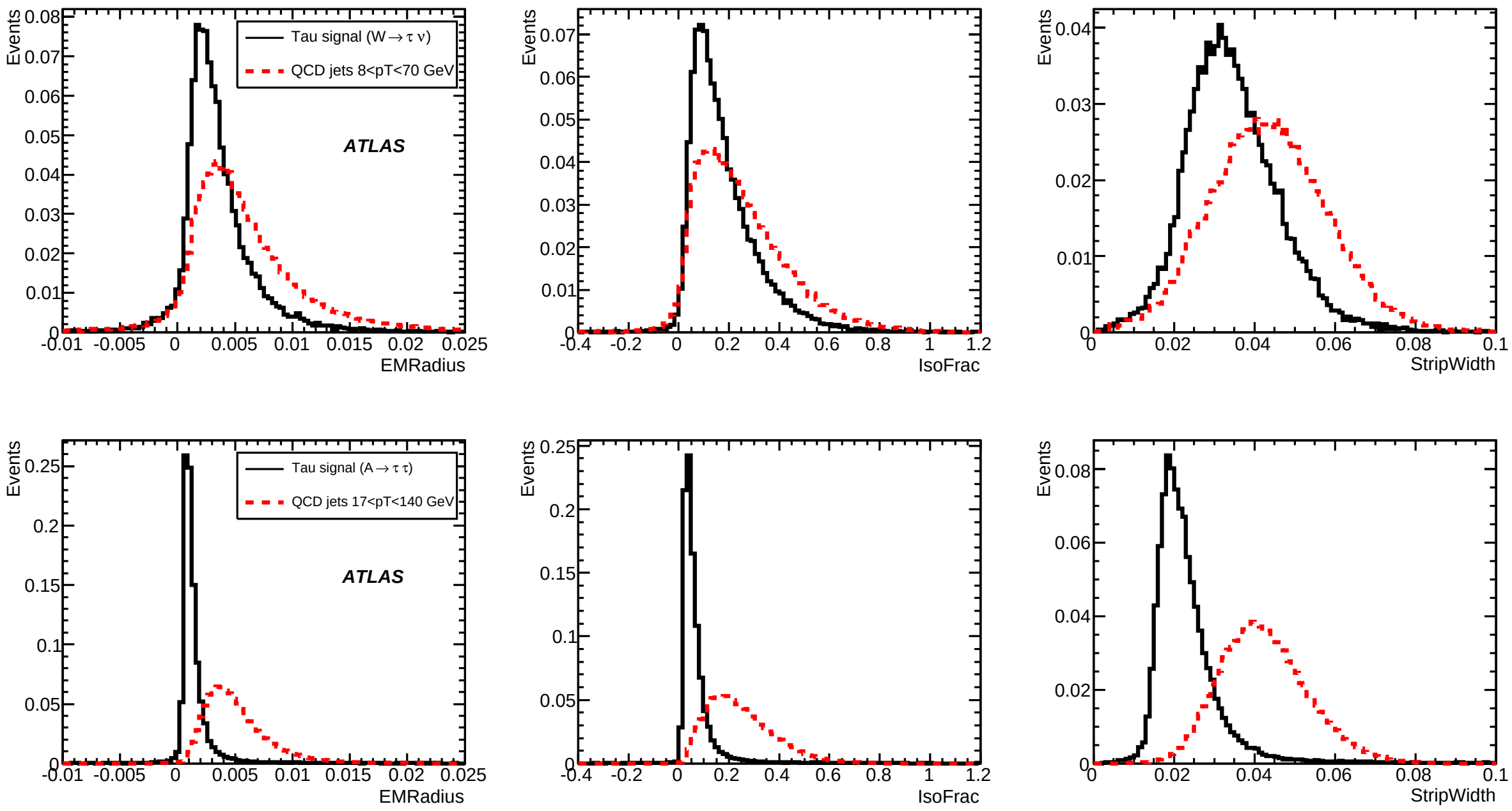

Figure 1: L2 Calorimeter variables to distinguish QCD jets from low $E_T$ taus from $W \to \tau\nu$ decays (top) and from high $E_T$ taus from Supersymmetric Higgs $A \to \tau\tau$ decays (bottom).

### 2.2.2 Level 2 calorimeter calibration

A simple *Sampling* calibration is used in the L2 calorimeter algorithm for reconstruction of the tau energy. This method is fast ($\approx 7\,\mu$s) because it applies a global weight to the energy of each sampling to compute a total calibrated energy, as opposed to other slower algorithms, such as the offline calibration [9], which use a cell-by-cell correction. In the current implementation only two layers are considered (e.m. and hadronic compartments) and the weights are energy and $\eta$ dependent.

### 2.2.3 Level 2 tracking selection

The standard L2 track reconstruction [10] is used in the L2 tau triggers.To keep the L2 execution time within budget, data from the Transition Radiation Tracker is not used.

Tracks with $p_T > 1.5\,$GeV are reconstructed in a rectangular RoI of size $\eta \times \phi = 0.6 \times 0.6$ centered on the L2 Calorimeter *seed*. The output of the L2 Tracking algorithm is a list of tracks found in the RoI. Within this region, two selection cones are defined, corresponding to a distance $\Delta R = \sqrt{(\Delta\eta)^2 + (\Delta\phi)^2}$ of 0.15 (*Core*) and 0.3 (*Nor*) with respect to the direction of the highest $p_T$ track found in the RoI. An isolation ring (*Iso*) with $\Delta R$ between 0.15 and 0.3 is also defined.

The following selection variables are then calculated from the track list:

- *Pt leading* is the $p_T$ of the track with the highest $p_T$. By requiring a minimum $p_T$ this criterion also effectively requires that at least one track is found in the RoI.
- *Pt Iso/Core* is the ratio of the scalar sum of $p_T$s of all tracks in the *Core* and *Iso* region $\sum p_T^{\rm iso} / \sum p_T^{\rm core}$.
- *N Slow tracks* is the number of slow tracks found in the *Core* region. A slow track is defined as a track with $p_T$ below a certain threshold, typically 7.5 GeV/c. The rejection power of this variable might depend on the pile-up conditions.

- *Charge* is defined as the absolute value of the sum of charges of all tracks found in the *Nor* region.
- *N Tracks* is the total number of tracks found in the *Nor* region.

The L2 tracking selection places requirements on these five variables. The set of requirements can be different for different single tau signatures. The three most important criteria are requiring that at least one track be found with a minimum *Pt leading*, an isolation cut on the maximum amount of energy deposited in the *Iso* region, and a cut on the maximum number of slow tracks. The requirements on charge and total tracks are very loose, due to the higher number of fake tracks found by the L2 tracking algorithm compared to the more sophisticated offline reconstruction. In addition, the offline reconstruction of single tau lepton final states (e.g. from $W \to \tau\nu$) relies on an unbiased track distribution to estimate backgrounds and extract the number of signal events. The distributions of *Pt leading*, *N Slow tracks* and *Pt Iso/Core* are shown in Fig. 2 for tau trigger candidates from generated tau leptons decaying hadronically and for QCD jets for two different $E_T$ regions. QCD jet background is compared to low $E_T$ tau leptons from $W \to \tau\nu$ decays (top row), and high $E_T$ tau leptons from Supersymmetric neutral Higgs decays (bottom row). Tau trigger candidates are required to pass different L1 and L2 Calorimeter selection criteria based on their $E_T$. A L2 tau trigger candidate passing the calorimeter and tracking requirements is passed to the EF for further consideration.

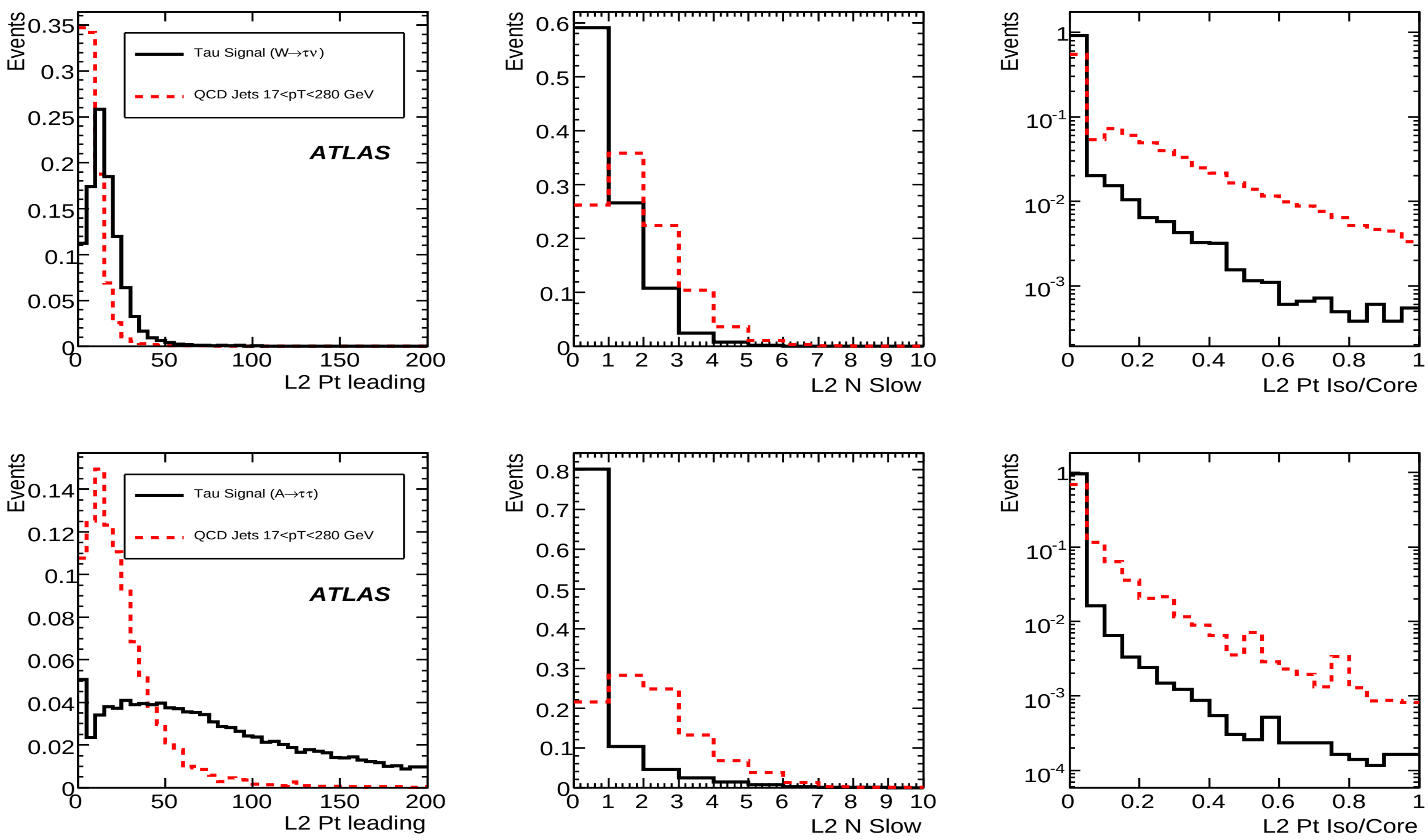

Figure 2: L2 Tracking variables to distinguish QCD background from low $E_T$ taus from $W \to \tau\nu$ decays (top) and from high $E_T$ taus from Supersymmetric Higgs $A \to \tau\tau$ decays (bottom).

## 2.3 Event Filter selection

At the EF level, the selection follows the offline reconstruction procedure as closely as possible. The EF calorimeter algorithm collects cells in a rectangular RoI of size $\eta \times \phi = 0.6 \times 0.6$ centered around the L2 tau trigger candidate. The EF tracking reconstruction is described in [10]. Tracking is performed in a rectangular RoI of size $\eta \times \phi = 0.4 \times 0.4$ centered around the L2 tau trigger candidate.

To create an EF tau trigger candidate, the EF single tau triggers execute the calorimeter based identification algorithm of the offline reconstruction [11] in the following way. The cells collected in the RoI are used to reconstruct the direction of the EF tau trigger candidate. Additionally, some very loose criteria are applied to tracks reconstructed in the RoI, and if more than one track is found a secondary vertex reconstruction is attempted. The transverse energy and the calorimeter shower shape variables are built from the cells collected in the RoI using the newly reconstructed direction. Unlike the offline reconstruction algorithm, no noise suppression is applied at the EF. The addition of electronic noise in the calorimeter cells gives a generally small change in the calorimeter shower shape variables. Finally, an overall hadronic calibration [9] is applied to all cells, and a tau specific jet calibration is applied to the tau trigger candidate. This final calibration is derived from simulated samples of high and low $p_T$ tau lepton sources. A procedure using taus from $Z$ bosons decays and exploiting the correlation between the tau absolute energy scale and the visible reconstructed $Z$ boson mass will be applied to determine calibration constants from real data [11].

The following variables are used for the EF selection:

- *EMRadius* is an energy weighted radius, which exploits the small transverse shower profile of tau leptons. The mathematical expression used is

$$EMRadius = \frac{\sum_i E_{T,cell} \cdot \Delta R_{cell}}{\sum_{Nor} E_{T,cell}}. \tag{4}$$

  where the sum includes not only the second sampling layer of the e.m. calorimeter (as in L2) but also all other samplings except the last one.

- *IsoFrac* is the energy deposited in an annular region between $0.1 < \Delta R < 0.2$ divided by the energy deposited in $\Delta R < 0.3$. A similar expression is given in Eq. 2, where the square regions should be replaced by the annuli.

- *N Tracks* is the number of associated tracks with $p_T > 2\,\mathrm{GeV}$ found in a $\Delta R < 0.2$ region around the trigger candidate from L2.

- *Pt leading track* is the highest $p_T$ track among the tracks found in a $\Delta R < 0.2$ region around the trigger candidate from L2.

- *EtCalib* is the energy calculated in all e.m. and hadronic cells found in the $\Delta R < 0.3$ region around the trigger candidate from L2, and calibrated with the procedure described in [9] and an additional tau specific jet calibration.

Some of the variables used in the EF selection, namely *EtCalib*, *IsoFrac* and *EMRadius*, are shown in Fig. 3 for tau trigger candidates from generated tau leptons decaying hadronically and for QCD jets for two different $E_T$ regions. As for previous figures, QCD background is compared to standard tau signal samples. Events with a tau candidate that pass the EF thus pass the trigger and are recorded for offline analysis.

# 3 Performance of single tau triggers

## 3.1 Performance overview

This section describes in detail the expected trigger efficiencies and background event rates obtained from the various single tau triggers available in the ATLAS trigger menu, while combined triggers including other types of signatures is given in Section 5. In addition, the optimization procedure applied to the selection at each level is introduced. The goal is to obtain at least 90% efficiency at each trigger level with sufficient rejection of background.

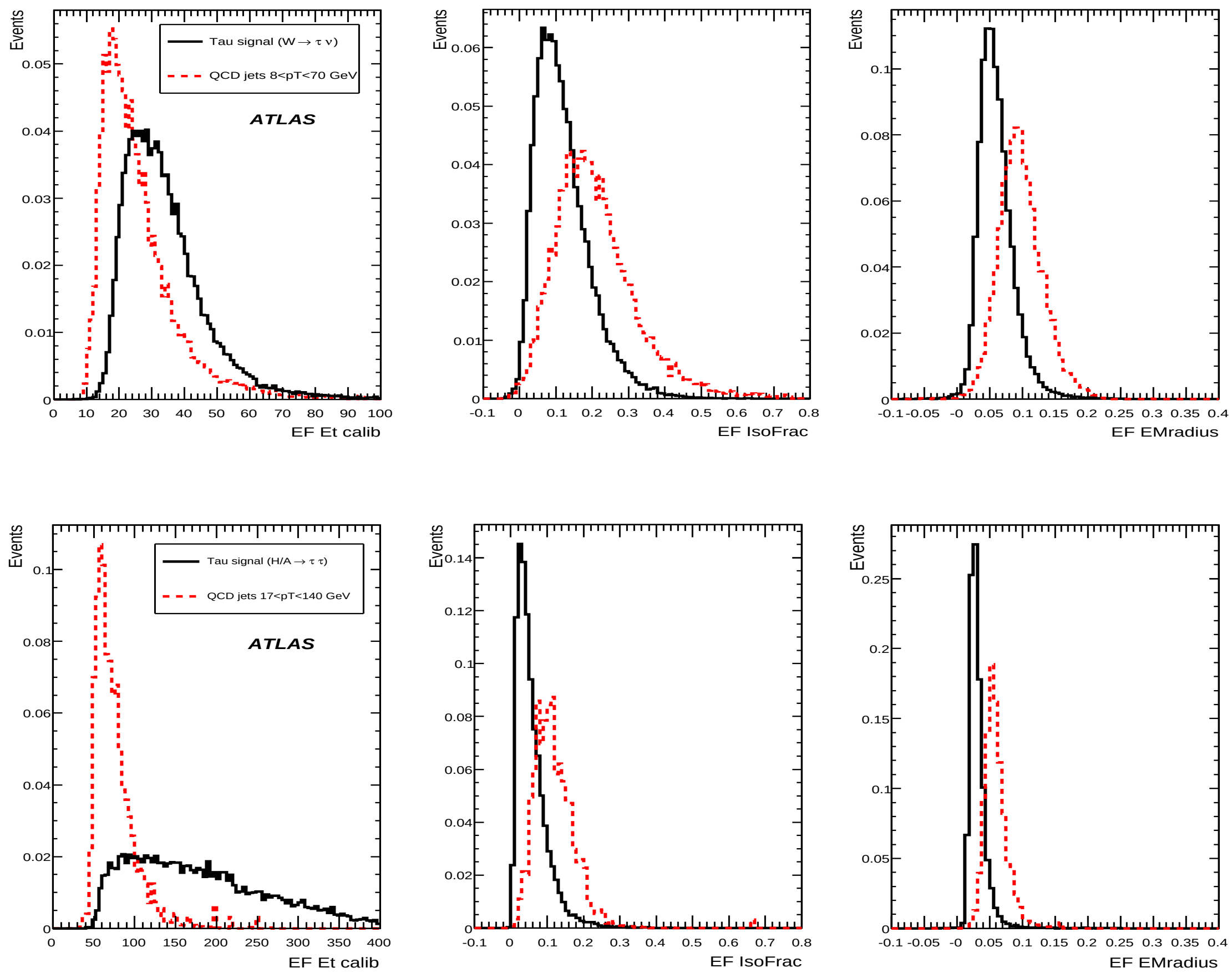

Figure 3: EF variables to distinguish QCD background from low $E_T$ taus from $W \to \tau\nu$ decays (top) and from high $E_T$ taus from Supersymmetric Higgs $A \to \tau\tau$ decays (bottom).

The various types of samples of GEANT4 simulated events used in this note together with the total number of events and the cross section are summarized in Table 1. It is appropriate to note at this point that the rates reported in this note are clearly affected by large uncertainties, because of the limited statistics available and because the simulation has yet to be validated with real data.

| Sample | Events | $\sigma$ (pb) |
|---|---|---|
| Min.Bias | 2900 k | 7.0E+10 |
| QCD dijet 8$< p_T <$ 17 GeV | 742 k | 1.74E+10 |
| QCD dijet 17 $< p_T <$ 35 GeV | 395 k | 1.38E+9 |
| QCD dijet 35 $< p_T <$ 70 GeV | 824 k | 9.33E+7 |
| QCD dijet 70 $< p_T <$ 140 GeV | 272 k | 5.88E+6 |
| $W_{\tau\rightarrow hX}$ (filtered $p_T >$ 12 GeV) | 24000 | 5.54 nb |
| $Z_{\tau\tau}$ | 23000 | 0.246 nb |
| $H_{\tau\tau\rightarrow \ell hX}(120)$ | 22924 | 0.145 |
| $H_{\tau\tau\rightarrow hhX}(120)$ | 40950 | 0.073 |
| $A_{\tau\tau}(800)$ | 26250 | 10 |
| SU3 | 22250 | 18.59 |
| $t\bar{t}$ | 41700 | 800 |

Table 1: GEANT4 simulated data samples used in the note.

### 3.1.1 Naming convention

In the ATLAS trigger menu, different single tau triggers are implemented, corresponding to different $E_T$ threshold requirements [2]: `tau10i`, `tau15i`, `tau20i`, `tau25i`, `tau35i`, `tau45i`, `tau60`. For `tau10i` and `tau15i` the isolation criteria are only applied at the L2 and EF level. For some $E_T$ thresholds, additional triggers are defined using looser isolation criteria (e.g. `tau10`). Hadronic isolation at L1 is not applied in any selection. Due to the $E_T$ resolution at the three trigger levels, the actual cut applied on $E_T$ at each stage is generally lower than the nominal $E_T$ to maintain high efficiency.

### 3.1.2 Trigger efficiency definition

The tau trigger efficiency is optimized with respect to generated tau leptons decaying hadronically, where the visible tau momentum ($p^{\alpha}_{\text{vis}} = p^{\alpha}_{\tau} - p^{\alpha}_{\nu}$) is in the sensitive region of the detector ($|\eta| \leq 2.5$) and is greater than the nominal $E_T$ threshold requirement for a given signature. Furthermore, the efficiency is optimized to select those tau leptons which are likely to be selected by the tau identification algorithms of the offline reconstruction software. This constitutes what we subsequently call the tau trigger reference.

More specifically, a geometrical match of a trigger candidate to a generated tau lepton within a cone region $\Delta R < 0.2$ is requested. Generated tau leptons are considered for calculating signal efficiencies only if the tau lepton is also selected by either of the two tau offline reconstruction algorithms [11]: the calorimeter based or the track based algorithm. It should be noted here, that detailed optimization with respect to 1-prong (1 charged pion) and 3-prong (3 charged pions) tau lepton decays has not been performed for this study. This results in different efficiencies for these classes of events, as pointed out in the following sections. Total trigger efficiencies shown in this section are estimated on various physics processes which cover different kinematic ranges. The samples used are therefore specified in the text.

[2]The first symbol of the signature represents the particle type, the following number is the $E_T$ threshold and the "i" indicates that an isolation requirement is applied.

### 3.2 Energy and angular resolution

The relative resolutions (in %) on $E_T$, $\eta$, and $\phi$ as a function of visible $E_T$ and $\eta$ for tau trigger candidates at the different trigger levels are shown in Fig. 4 for `tau20i`. The visible four-momentum of the tau lepton is reconstructed from all decay products except neutrinos. The resolutions are rather flat as a function of the generated visible $E_T$, however, some dependence on the distributions as a function of the generated visible $\eta$ is observed. In particular a degraded resolution in $E_T$ and $\eta$ in the transition region between the barrel and endcap calorimeters is clearly visible. Due to the inclusion of track information, the angular resolution at L2 is better than that of L1. The $\eta$ and $\phi$ of the tau trigger candidate at the end of the EF algorithm execution is set according to the calorimeter based tau reconstruction algorithm, hence the EF angular resolution is slightly worse than at L2 where tracking information is used.

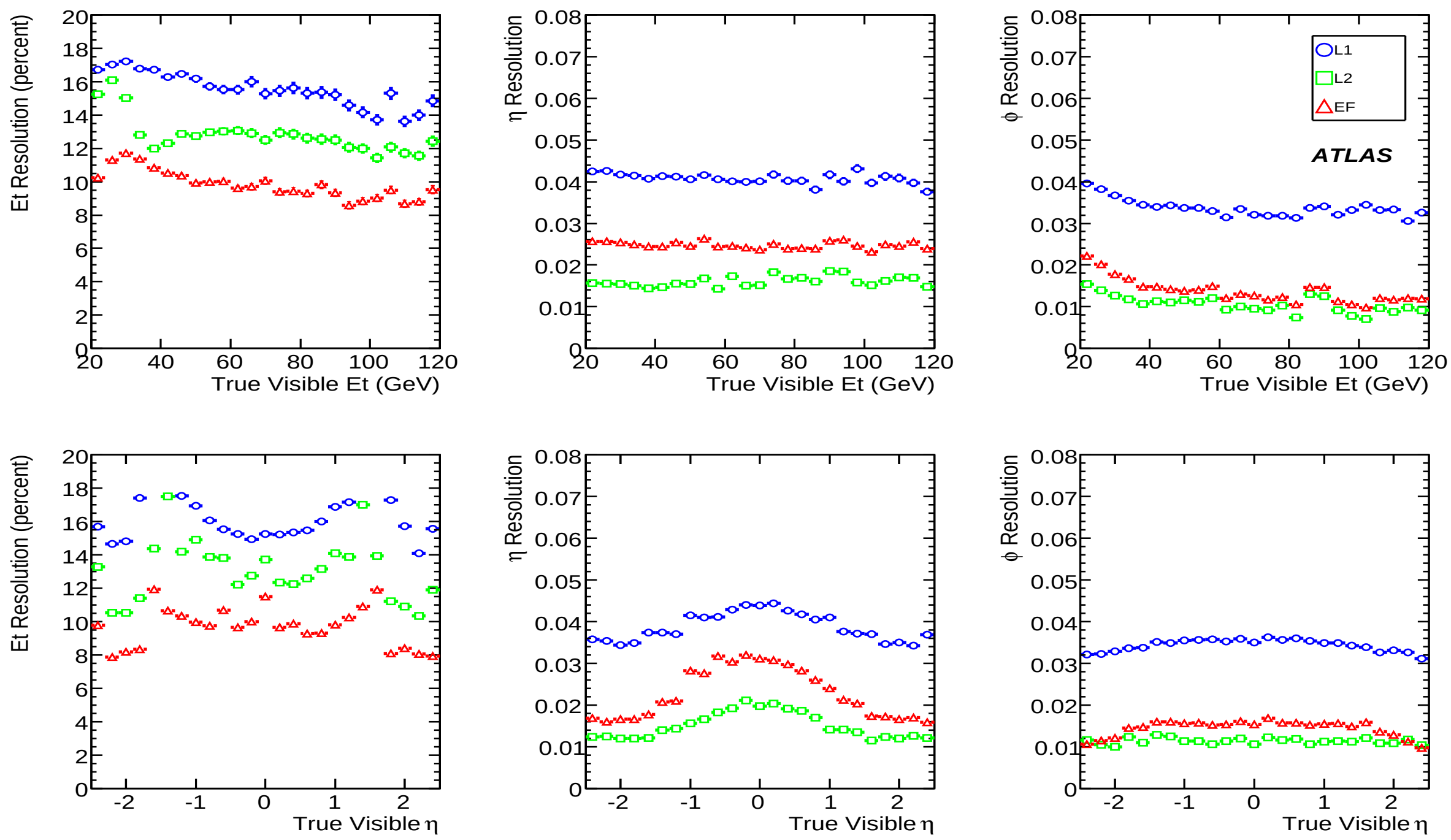

Figure 4: Relative angular and energy resolution for $\tau$ candidates as a function of generated visible $E_T$ and $\eta$. Resolution (in %) is calculated using a variety of simulated tau lepton samples passing the `tau20i` trigger.

### 3.3 L1 performance studies

The standard reconstruction of the visible energy of L1 tau candidates (*TauCluster*) reveals a serious limitation when trying to obtain a high efficiency above a given $E_T$ threshold at L1. A systematic shift of about 30% and a relative resolution of 33% are observed for the reconstructed energy with respect to the generated values. Therefore other possible reconstruction methods have been explored. The energy reconstructed with the standard algorithm (*TauCluster*) has been compared with other reconstruction methods: (1) E4×4, defined as the sum over all towers (e.m. and hadronic part) in the $4 \times 4$ RoI region,

(2) Jet4 $\times$ 4, obtained as the total energy in the $4 \times 4$ RoI region calculated with the L1 Jet algorithm (which is described in Ref. [8]), and (3) Jet6$\times$6. An improvement of 10% in resolution can be achieved with E4$\times$4 and Jet4$\times$4 algorithms with respect to *TauCluster*. However, a comparison of the efficiencies of the different methods, shown in Fig. 5, indicates that the *TauCluster* algorithm efficiency is systematically the highest. While these other algorithms that more fully contain the $E_T$ of the tau lepton give better resolution than *TauCluster*, this gain in resolution is accompanied by a loss in efficiency and discrimination, motivating the continued use of *TauCluster* as the default tau algorithm.

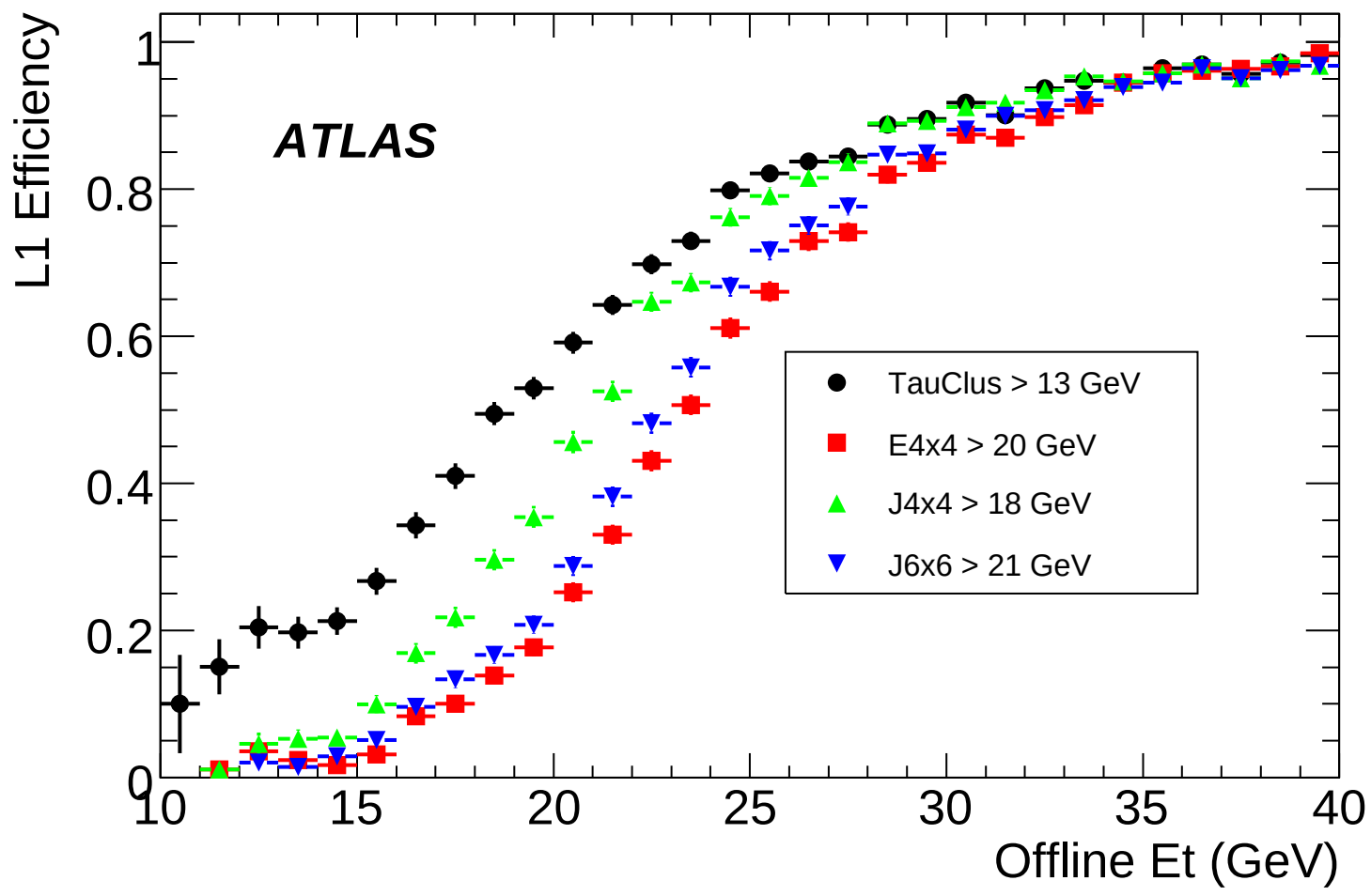

Figure 5: Efficiency curves as a function of the offline $E_T$ for different L1 reconstruction methods; threshold requirements are chosen to give the same rate among the different cases.

Using *TauCluster*, the rates for a luminosity of $10^{31}\,\mathrm{cm}^{-2}\mathrm{s}^{-1}$ are evaluated by using simulations of the relevant QCD backgrounds (see Table 2). The expected rate including all physical processes (minimum bias rate) is shown as a function of the cut value applied to *TauCluster* in Fig. 6. The effect of the isolation cut can also be seen in Fig. 6.

The L1 selection used in tau trigger implementation is presented in Table 2; the cuts have been tuned to maintain a high efficiency, and the signatures are all observed to have an efficiency $> 95\%$ [3]. The efficiency has been obtained using simulated samples of $W \to \tau\nu$ and 800 GeV Supersymmetric Higgs $A \to \tau\tau$ events, while background is evaluated using QCD events with two jets with hard parton $8 < p_T < 140$ GeV. In Fig. 7, the efficiency is plotted as a function of the generated visible $E_T$ for different tau signatures. The rather slow increase of the efficiency curves limits the overall performance of tau triggers.

### 3.4 L2 performance studies of the calorimeter selection

At L2, the full granularity calorimeter and tracking information is available, allowing a more sophisticated selection of tau leptons. As shown in Section 3.2, the angular and energy resolutions are improved with respect to L1.

The L2 calorimeter based selection of tau trigger candidates is optimized with respect to generated tau leptons that pass the L1 selection and are successfully reconstructed by the offline reconstruction software (offline details are given in Section 3.1.2). The optimization is obtained using simulated samples of $W \to \tau\nu$ and 800 GeV Supersymmetric Higgs $A \to \tau\tau$ events, and simulated background samples of QCD events with two jets with hard parton $8 < p_T < 140$ GeV.

[3] The efficiency definition is described in Section 3.1.2

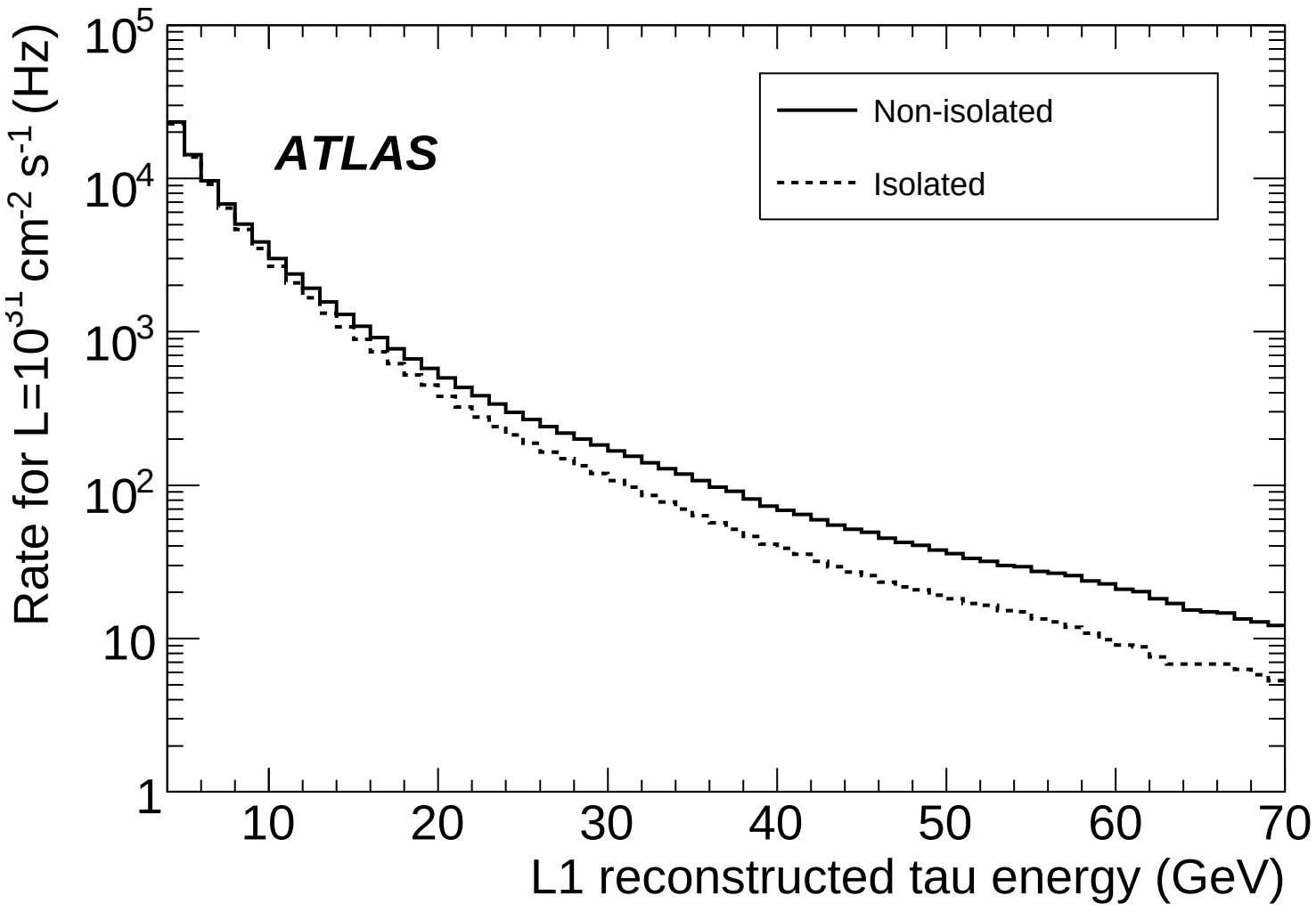

Figure 6: Expected rate including all physical processes at a luminosity of $\mathscr{L} = 10^{31}\,\mathrm{cm}^{-2}\mathrm{s}^{-1}$, as a function of the cut applied to *TauCluster* and with and without the isolation cut *EmIsol*.

| Signature | $E_T$ threshold (GeV) | EmIsol (GeV) | Eff.(%) | Rate (Hz) |
|---|---|---|---|---|
| tau10i | 6 | - | 97 | 13945 |
| tau15i | 6 | - | 98 | 13945 |
| tau20i | 9 | 6 | 96 | 4823 |
| tau25i | 11 | 6 | 96 | 2822 |
| tau35i | 16 | 6 | 96 | 990 |
| tau45i | 25 | 6 | 96 | 263 |
| tau60 | 40 | - | 97 | 97 |

Table 2: Single tau signatures, associated L1 thresholds, and efficiency and rates determined from signal and background samples as described in the text ($\mathscr{L}=10^{31}\,\mathrm{cm}^{-2}\,\mathrm{s}^{-1}$).

The L2 variables based on calorimetry introduced in Section 2.2 are optimized in the following way:

- Since some correlation among the calorimeter shower shape variables (*EMRadius, isoFrac* and *stripWidth*) is expected, the three variables are simultaneously optimized. The thresholds are determined by scanning the possible cut values for each variable and choosing the set of cuts giving the lowest background rate for a requested minimum signal efficiency.

- The threshold for the variable *EtCalib* is determined by requiring that the efficiency becomes flat starting at the nominal $E_T$ of the trigger (see Fig. 8).

The optimized L2 calorimeter requirements for tau signatures and the corresponding efficiencies and rates are presented in Table 3. The cut value on *EtCalib* increases for higher $E_T$ signatures, as we aim to select higher $E_T$ taus. The shower shape variables, a measure of the narrowness of the shower, are tighter for higher $E_T$ signatures, since higher $E_T$ taus are more boosted and their decay products are more collimated. The three shower shape variables requirements have been optimized to give an efficiency of $\approx 95\%$, although the observed efficiency values are a little lower due to the additional *EtCalib* requirement.

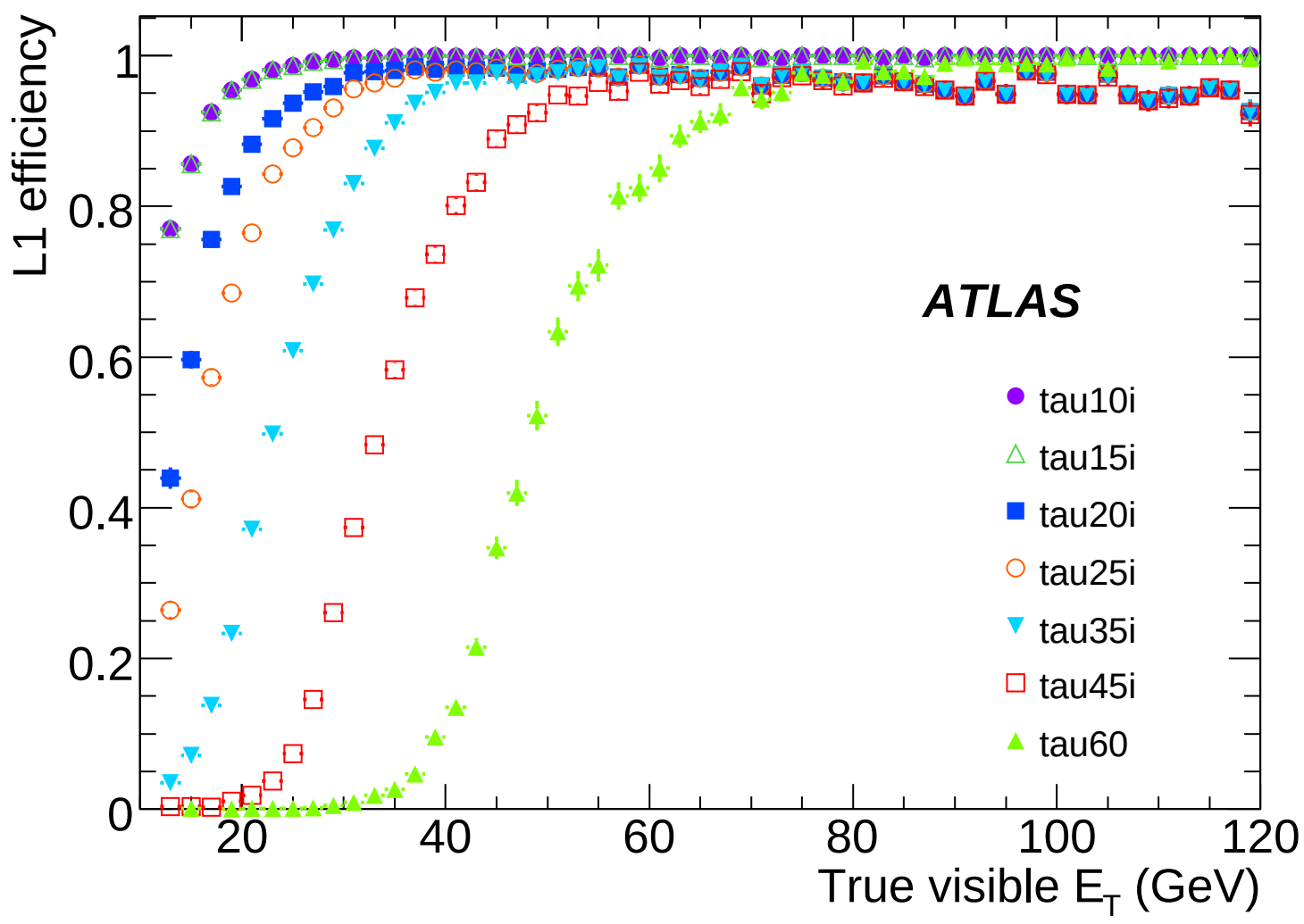

Figure 7: L1 efficiency curves for different tau signatures.

The efficiency is detailed further in Table 4 for two typical tau signatures, one with low and one with high $E_T$. For each signature, efficiency is recalculated using as a reference different tau lepton samples reconstructed offline. Either the sample from each tau algorithm is considered separately, or the sample reconstructed by either of them ("OR"), or by both simultaneously ("AND"), is considered. The errors quoted are statistical only. In order to achieve the needed rejection of QCD backgrounds, the calorimeter-based algorithm of the offline reconstruction biases the distribution of decays towards 1-prong decays using calorimeter information. A similar rejection and efficiency is achieved in a different manner using the track-based algorithm, which requires a large transverse momentum for the leading track in the tau lepton decay. Therefore, the L2 calorimeter selection has a high and uniform efficiency for the sample reconstructed offline by the calorimeter-based algorithm, while it is lower on the sample reconstructed using the track-based algorithm in the 3-prong category.

| Signature | *EtCalib* (GeV) | *EMRadius* | *IsoFrac* | *StripWidth* | Eff.(%) | Rate(Hz) | L1/L2 rate |
|---|---|---|---|---|---|---|---|
| tau10i | 8.0 | 0.023 | 0.74 | 0.058 | 95 | 9251 | 1.51 |
| tau15i | 9.7 | 0.022 | 0.71 | 0.057 | 94 | 5933 | 2.35 |
| tau20i | 12.2 | 0.019 | 0.66 | 0.055 | 94 | 3070 | 1.57 |
| tau25i | 17.0 | 0.016 | 0.65 | 0.051 | 92 | 1224 | 2.31 |
| tau35i | 26.5 | 0.015 | 0.6 | 0.048 | 91 | 351 | 2.82 |
| tau45i | 33.5 | 0.00807 | 0.43 | 0.0465 | 93 | 119 | 2.21 |
| tau60 | 44.9 | 0.00345 | 0.35 | 0.04 | 93 | 47 | 2.06 |

Table 3: Single tau signatures, associated L2 calorimeter thresholds and efficiency on $W \to \tau\nu$ sample and rates on QCD background samples for $\mathscr{L}$=$10^{31}$ cm$^{-2}$ s$^{-1}$.

| | tau20i | | | tau60 | | |
|---|---|---|---|---|---|---|
| Offline Identification method | all taus | 1-prong | 3-prong | all taus | 1-prong | 3-prong |
| Calo-based | 97.6±0.1 | 97.5±0.1 | 98.3±0.4 | 97.8±0.1 | 97.8±0.1 | 96.8±0.5 |
| Track-based | 92.5±0.2 | 95.2±0.2 | 88.4±0.4 | 94.9±0.2 | 95.7±0.2 | 91.7±0.5 |
| Calo- OR Track-based | 94.2±0.2 | 96.3±0.2 | 88.9±0.4 | 96.6±0.1 | 97.3±0.1 | 92.8±0.4 |
| Calo- AND Track-based | 97.2±0.2 | 97.0±0.2 | 98.2±0.4 | 96.6±0.2 | 96.6±0.2 | 96.1±0.9 |

Table 4: Tau signatures and corresponding efficiencies (in %) for different reference samples of tau leptons identified by the offline reconstruction. Efficiencies have been estimated on low and high $E_T$ generated samples of tau leptons. Errors are statistical only. Efficiencies are for L2 calorimeter selection only.

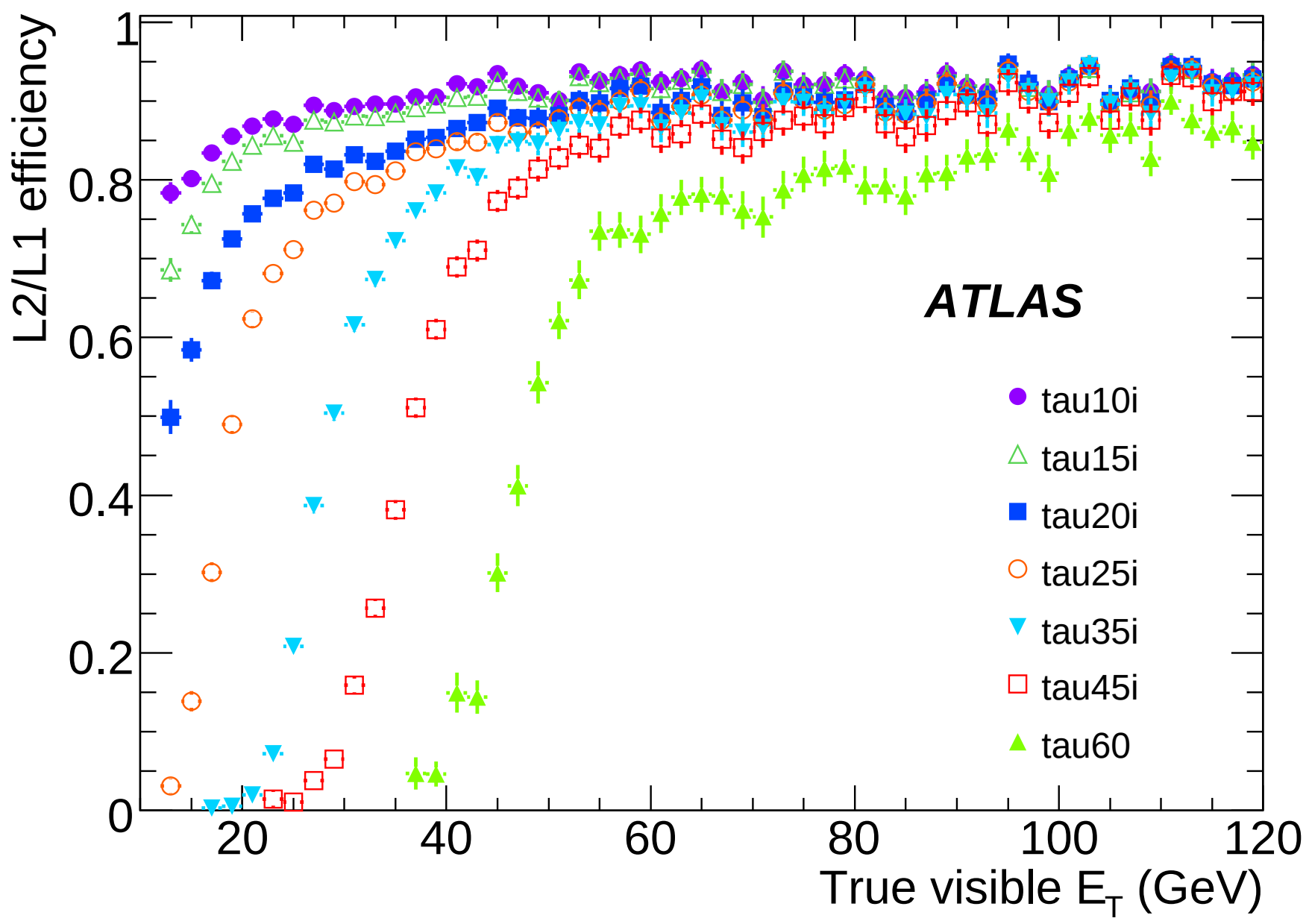

Figure 8: Efficiency curves of L2 selection relative to L1 for different tau signatures.

### 3.5 L2 performance studies of the tracking selection

The addition of tracking information at L2 allows an improvement in the discrimination between hadronic decays of tau leptons and QCD jet backgrounds, although at the cost of a non-negligible loss of trigger reconstruction efficiency. This loss is most noticeable for low $E_T$ tau signatures, mainly due to the requirement of one leading track above a fixed $p_T$ threshold. Overall, the L2 tracking algorithm is about 95% efficient for isolated tracks with $E_T$ above 5 GeV (for example from $W \to \tau\nu$). In $W \to \tau\nu$ events, however, only 90% of generated tau leptons that are identified by the offline reconstruction and pass the L2 calorimeter selection have at least one track with $p_T > 5$ GeV reconstructed at L2.

The L2 tracking selection of tau trigger candidates is optimized with respect to generated tau leptons that pass the L1 and L2 calorimeter selection and are found by the offline reconstruction software. Details of the offline reference and efficiency definition are described in Section 3.1.2. The optimization is obtained with the standard samples described previously.

The efficiency and rejection power of some of the L2 tracking variables can be seen in Fig. 9, where the integrated efficiency for tau samples and background are shown as a function of the applied cut value.

For low $E_T$ signatures, the minimum requirement on *Pt leading* along with the upper limit on *Pt Iso/Core* are the two most important criteria. For higher $E_T$ signatures, the upper limit on *N Slow tracks* becomes more important, as background QCD jets tend to have a higher multiplicity of soft tracks compared to multi-track hadronic decays of tau leptons. The multiplicity requirement, *N Track*, is potentially a useful discriminant for higher $E_T$ signatures, although a tight cut tends to bias the multiplicity distribution of the tau leptons found by offline reconstruction. To avoid such bias a very soft cut is applied.

The optimized L2 tracking requirements for tau signatures and the corresponding efficiencies and rates are presented in Table 5. Aside from the *Pt leading* requirement, which is tightened with increasing $E_T$ threshold, the requirements on *Pt Iso/Core*$<0.1$, *N Slow tracks*$\leq 2$, *Charge*$\leq 2$ and $1\leq$*N Track*$\leq 7$ are kept constant at values which work reasonably well for all signatures, and therefore are not shown in the Table.

The efficiency is detailed in Table 6 for two typical tau signatures, one with low and one with high $E_T$. For each signature, efficiency is recalculated using different reference samples based on offline reconstruction as in the previous section. Again, the errors quoted are statistical only. For the high $E_T$ signatures, there is not a strong efficiency dependence on the particular reference sample chosen. For the low $E_T$ signatures, however, the different selections used in the offline algorithms, particularly in selecting 1-prong tau lepton decays, are apparent. Since the track-based algorithm also requires a track with some minimum $p_T$ when applied to these events the L2 tracking selection is quite efficient. The L2 tracking selection is considerably less efficient when applied to events selected with the calorimeter based algorithm, where no such requirement on the tracks is made.

The overall L2 trigger efficiency for various single tau signatures, using the optimized selection described in Sect. 3.4- 3.5, is shown in Fig. 8.

| Signature | Pt lead (GeV) | Eff. (%) | Rate (Hz) | L2Calo/L2Track |
|---|---|---|---|---|
| `tau10i` | 1.5 | 86 | 3980 | 2.32 |
| `tau15i` | 2.5 | 85 | 2900 | 2.04 |
| `tau20i` | 5.0 | 81 | 947 | 3.24 |
| `tau25i` | 5.0 | 74 | 351 | 3.48 |
| `tau35i` | 5.0 | 75 | 107 | 3.28 |
| `tau45i` | 5.0 | 80 | 39 | 2.01 |
| `tau60` | 5.0 | 79 | 15 | 3.13 |

Table 5: Single tau signatures, associated L2 tracking thresholds and efficiencies for $W \rightarrow \tau\nu$ sample and rates for QCD background samples for $\mathscr{L}=10^{31}\,\mathrm{cm}^{-2}\,\mathrm{s}^{-1}$.

| | `tau20i` | | | `tau60` | | |
|---|---|---|---|---|---|---|
| Offline Identification method | all taus | 1-prong | 3-prong | all taus | 1-prong | 3-prong |
| Calo-based | 87.1±0.3 | 86.6±0.3 | 91.8±0.8 | 97.5±0.1 | 97.9±0.1 | 90.3±0.9 |
| Track-based | 95.5±0.2 | 96.5±0.2 | 93.9±0.3 | 95.8±0.2 | 97.3±0.2 | 89.3±0.6 |
| Calo- OR Track-based | 88.9±0.2 | 87.5±0.3 | 92.8±0.4 | 96.5±0.1 | 97.5±0.1 | 89.3±0.5 |
| Calo- AND Track-based | 97.3±0.2 | 97.4±0.2 | 96.6±0.6 | 97.9±0.1 | 98.2±0.1 | 92.4±1.3 |

Table 6: Tau signature efficiencies (in %) for different reference samples of tau leptons identified by the offline reconstruction. Efficiencies have been estimated on low and high $E_T$ generated samples of tau leptons. Errors are statistical only. Efficiencies are for L2 tracking only.

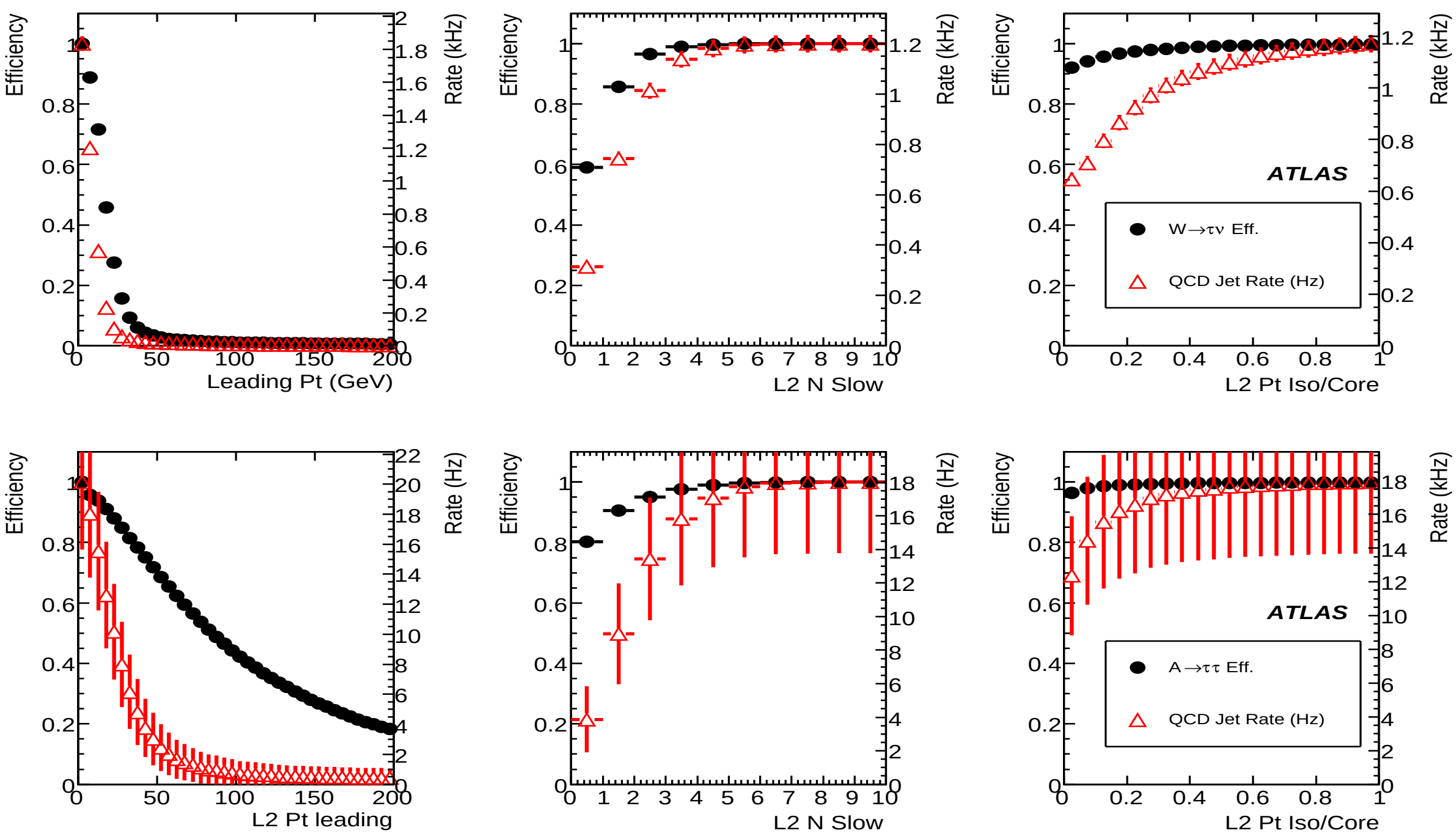

Figure 9: Performance of L2 tracking variables to distinguish QCD background jets from low $E_T$ (top) and high $E_T$ (bottom) tau leptons. For each plot, the closed circle data points show the cumulative efficiency (left-hand scale) for tau leptons as a function of cut value, while the open triangle data points show the QCD jet rates (right-hand scale) for a luminosity of $\mathscr{L} = 10^{31}\,\mathrm{cm}^{-2}\mathrm{s}^{-1}$.

### 3.6 EF performance studies

As described in Section 2.3, the tau identification algorithm developed for offline reconstruction is adapted for EF use. The resolution in $E_T$ and direction achieved at the EF are shown in Section 3.2, and provide a clear improvement in $E_T$ determination with respect to L2.

The EF selection of tau trigger candidates is optimized with respect to generated tau leptons that pass the L1 and L2 selection and are found by the offline reconstruction software. Details of the offline reference and efficiency definition are described in Section 3.1.2. The optimization is performed using the standard samples described previously.

The cuts on calorimeter shape variables and $E_T$ are optimized separately, in analogy to the L2 procedure previously described. First, the correlation between the $E_T$ of the tau lepton and the $p_T$ of the leading track associated with the tau lepton is studied in simulated samples, and then a combined requirement on these two variables is applied to each tau signature. Finally, the calorimeter shower shape variables are studied, and a further requirement is applied on these variables.

In Table 7 the optimized threshold requirements and corresponding efficiencies and background rates are presented. As for L2, the shower shape variables are tightened and the cut value on *EtCalib* and $p_T$ *lead* are raised as a function of the $E_T$ of the tau signature. The two shower shape variables requirements have been optimized to give an overall efficiency of $\approx 90\%$. The requirement on *nTracks* is not stringent, so as not to bias the distribution for this variable, which is a key offline variable for the evaluation of tau lepton purity. Fig. 10 shows the efficiency for various single tau signatures, using the optimized selection summarized in Table 7.

The efficiency is detailed in Table 8 as in previous sections. At EF clearly the efficiency is highest for the tau leptons identified by the calorimeter based algorithm of the offline reconstruction.

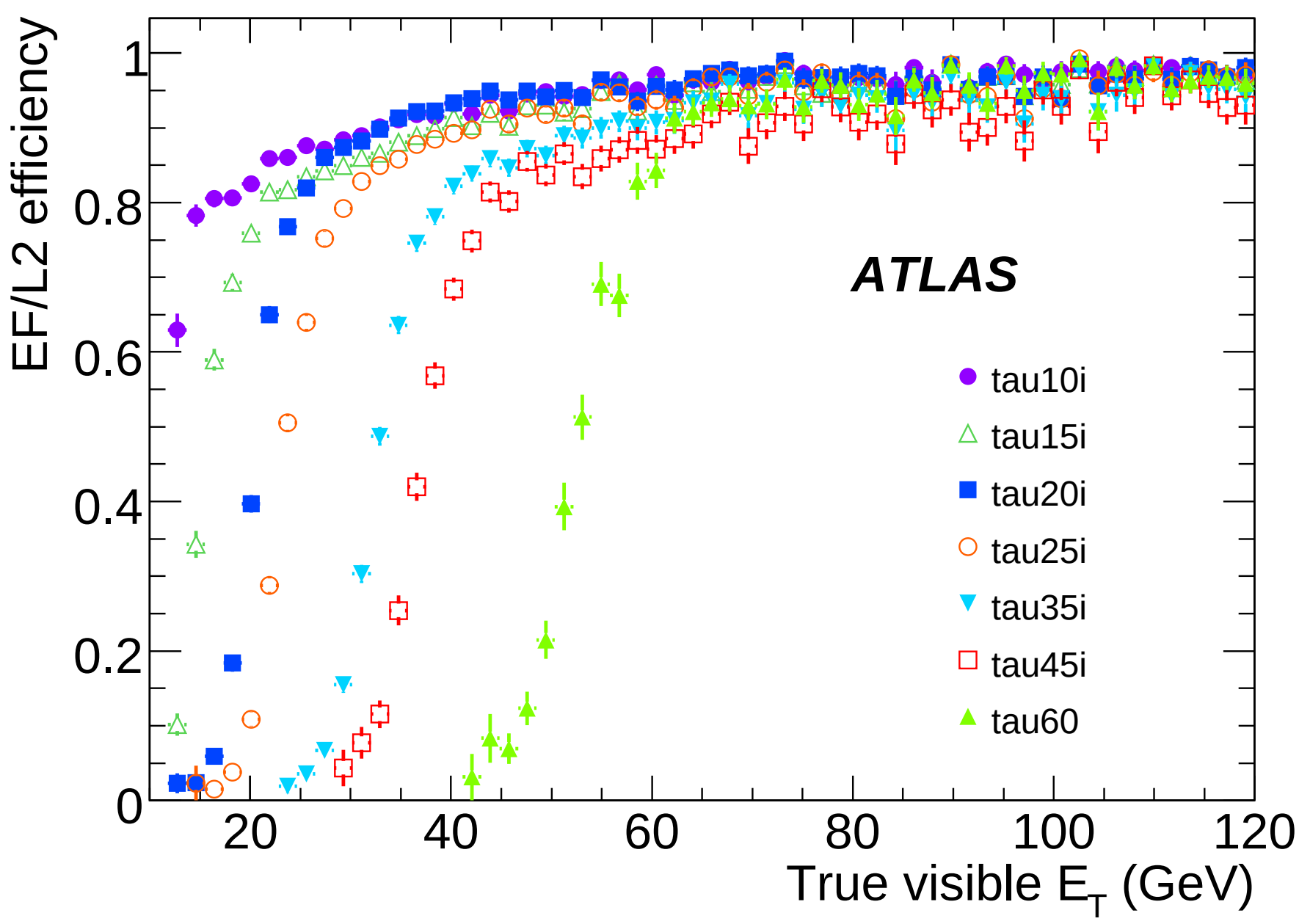

Figure 10: Efficiency curves of EF selection relative to L2 for different tau signatures.

| Signature | *EtCalib* (GeV) | $p_T$ lead. (GeV) | *EMRadius* | *IsoFrac* | *nTracks* | Eff.(%) | Rate (Hz) |
|---|---|---|---|---|---|---|---|
| `tau10i` | 10. | 5. | 0.13 | 0.38 | 1-8 | 90 | 1611 |
| `tau15i` | 14. | 6. | 0.13 | 0.32 | 1-8 | 85 | 897 |
| `tau20i` | 19. | 6. | 0.11 | 0.33 | 1-8 | 86 | 349 |
| `tau25i` | 22. | 7. | 0.1 | 0.3 | 1-8 | 86 | 175 |
| `tau35i` | 31. | 8. | 0.09 | 0.24 | 1-8 | 87 | 64 |
| `tau45i` | 36. | 8. | 0.08 | 0.19 | 1-8 | 92 | 24 |
| `tau60` | 51. | 8. | 0.09 | 0.24 | 1-8 | 93 | 8 |

Table 7: Single tau signatures, associated EF thresholds and efficiencies for the $W \to \tau\nu$ sample and rates for the QCD background sample for $\mathscr{L}=10^{31}\,\mathrm{cm}^{-2}\,\mathrm{s}^{-1}$.

| | `tau20i` | | | `tau60` | | |
|---|---|---|---|---|---|---|
| Offline Identification method | all taus | 1-prong | 3-prong | all taus | 1-prong | 3-prong |
| Calo-based | 90.9±0.3 | 90.3±0.3 | 95.5±0.6 | 98.5±0.1 | 98.5±0.1 | 98.7±0.4 |
| Track-based | 84.7±0.3 | 88.9±0.4 | 77.4±0.6 | 97.3±0.1 | 97.7±0.1 | 95.8±0.4 |
| Calo- OR Track-based | 85.0±0.3 | 87.9±0.3 | 77.9±0.6 | 98.0±0.1 | 98.2±0.1 | 96.4±0.3 |
| Calo- AND Track-based | 93.7±0.3 | 93.2±0.3 | 97.1±0.5 | 98.2±0.1 | 98.2±0.1 | 98.6±0.6 |

Table 8: Tau signature efficiencies (in %) for different reference samples of tau leptons identified by the offline reconstruction. Efficiencies have been estimated on low and high $E_T$ generated samples of tau leptons. Errors are statistical only. Efficiencies are for EF only.

### 3.7 Combined performance of single tau triggers

After evaluating each trigger level separately, we now present the overall combined performance (L1 + L2 + EF) of the single tau triggers. Figures 11 and 12 show the dependence of the efficiency on $E_T$. The shape of the efficiency curves is determined by the $E_T$ resolution, which is different for the different trigger levels. The $E_T$ threshold requirement of the signature is selected such that the efficiency curve reaches a plateau at the visible $E_T$ of the generated tau lepton greater than this threshold. Since the efficiency turn-on is slowest at L1, the cut on $E_T$ at L1 is set significantly lower than for L2 and EF.

The efficiency is detailed in Table 9 for two typical tau triggers as in the previous sections. In addition to the observations of the individual trigger level sections, one can see that the overall performance for the low $E_T$ tau trigger is about 70% with respect to the reference, while at high $E_T$ it is about 90%. This is due to relatively tighter cuts needed to control the large backgrounds present for the low $E_T$ signature. One can also see that the efficiency at low $E_T$ is low in particular for 3-prong decays when the reference is the sample of tau leptons identified by the track-based algorithm. This reflects the previous observation of Section 3.4, and points to the need for the execution of the track-based algorithm at the EF as well as at L2.

## 4 Trigger timing studies

### 4.1 Online setup

Prior to first LHC collisions, it is desirable to test the trigger software and data acquisition system in conditions resembling real data taking. There are two type of tests which can be performed: *cosmic runs*, where the detector signals left by cosmic rays activate the L1 trigger, and *technical runs* where simulated L1 trigger signals are used. These runs allow the monitoring system to be assessed, the algorithm timing to be measured (although the composition of the input events is not representative of the expected L1 trigger output when real data are taken) and the interplay between triggers to be studied. All studies are performed on CPUs dedicated to timing studies in order to minimize the possible influence of other users on timing results. Each of these machines is a dual-core Intel(R) XEON(TM) CPU 2.20 GHz machine.

| | tau20i | | | tau60 | | |
|---|---|---|---|---|---|---|
| Offline Identification package | all taus | 1-prong | 3-prong | all taus | 1-prong | 3-prong |
| Calo-based | 76.5±0.4 | 75.5±0.4 | 85.2±1.0 | 93.8±0.2 | 94.3±0.2 | 86.1±1.0 |
| Track-based | 70.9±0.4 | 79.5±0.4 | 58.3±0.7 | 88.4±0.3 | 90.9±0.3 | 78.1±0.8 |
| Calo- OR Track-based | 68.3±0.3 | 72.5±0.4 | 58.6±0.6 | 91.3±0.2 | 93.1±0.2 | 79.6±0.7 |
| Calo- AND Track-based | 87.9±0.4 | 87.4±0.4 | 91.1±0.9 | 92.8±0.3 | 93.1±0.3 | 87.1±1.6 |

Table 9: Tau trigger efficiencies (in %) for different reference samples of tau leptons identified by the offline reconstruction. Efficiencies have been estimated on low and high $E_T$ generated samples of tau leptons. Errors are statistical only. Efficiencies is for L1 + L2 + EF selection.

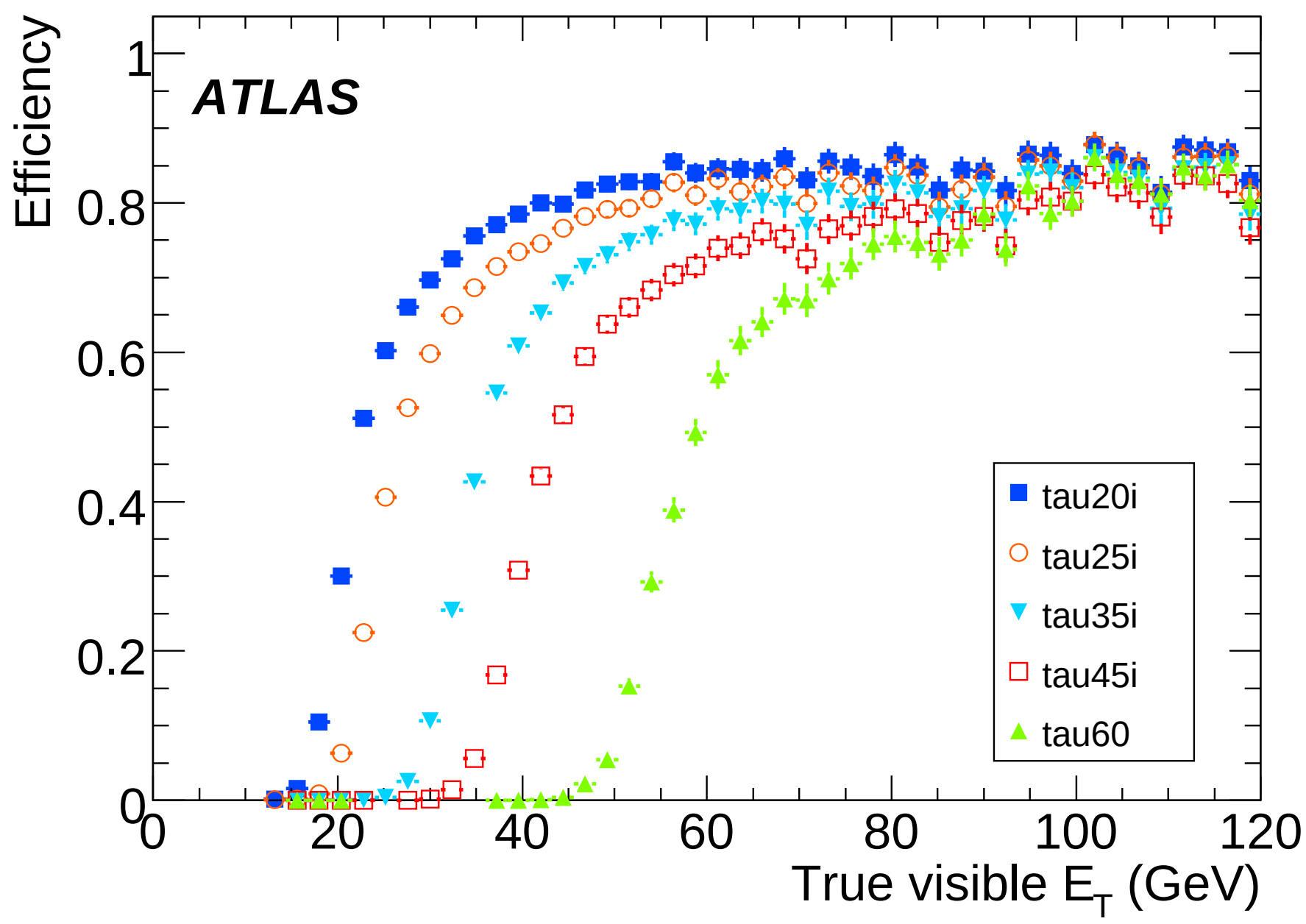

Figure 11: Overall trigger efficiency (L1 + L2 + EF) for different tau triggers.

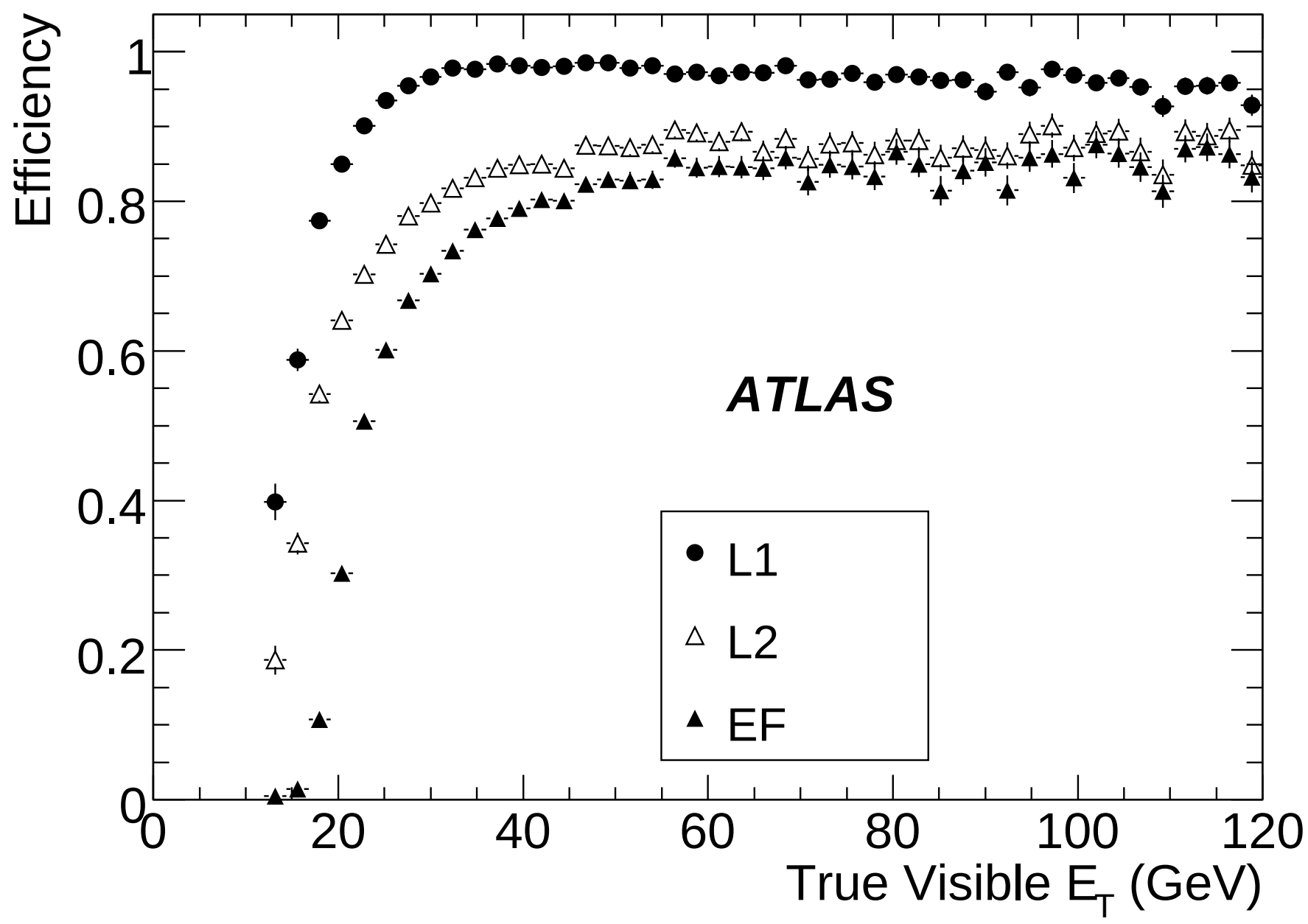

Figure 12: Efficiency for `tau20i` trigger.

### 4.2 Timing results

The CPU time performance of the trigger is a crucial ingredient in the optimization of the trigger. Current design constrains the total execution time per event to 40 ms at L2 and 1 s at EF for the whole trigger. Recent timing studies performed on tau triggers include:

- tests of individual triggers to gauge the impact of increasing threshold levels on the total execution time of the trigger.

  During this test, each of the tau triggers is run individually (and no other triggers are run at the same time). Although the average execution time of each algorithm remains roughly equal between triggers (see Table 10), the average total time per event decreases as a result of the lower number of RoIs per event for high energy triggers. The results shown in Table 10 indicate that the time performance of the tau trigger should be well within the constraint for total execution time at L2 and EF.

- a verification of caching of data from the liquid argon calorimeter in L2. This test requires running the tau algorithms and other calorimeter-based algorithms (electron algorithm, for example) for the same event. In case data in a given RoI is needed by more than one signature, then the caching procedure ensures that data are requested only once. Recent tests verify that caching is correctly implemented and provides roughly a factor of four time savings in the L2 calorimeter part.

| | Threshold Signature | | | | |
|---|---|---|---|---|---|
| Algorithm | tau10i | tau15i | tau20i | tau25i | tau35i |
| L2 Calo | 8.1 | 8.0 | 8.0 | 8.1 | 8.1 |
| L2 Tracking | 15.4 | 15.5 | 15.0 | 14.9 | 14.7 |
| L2 Combined | 1.9 | 1.9 | 2.0 | 2.0 | 2.2 |
| L2 TotalTime | 41.6 | 35.9 | 19.7 | 14.1 | 7.9 |
| EF Calo | 12.3 | 12.5 | 13.4 | 13.0 | 14.0 |
| EF Tracking | 289.7 | 297.7 | 269.5 | 268.4 | 247.8 |
| EF Combined | 77.0 | 76.9 | 80.7 | 77.8 | 78.9 |
| EF TotalTime | 149.1 | 133.6 | 67.5 | 51.2 | 24.6 |

Table 10: Mean algorithm execution time for each of the tau triggers. All times are given in ms and per RoI, except the L2 and EF total times are given per event, in ms. The measurements are performed on a simulated sample of limited statistics, 950 QCD background events with hard parton $35 < p_T < 70$ GeV.

- a timing comparison between the calorimeter-approach and tracking-approach in L2. Currently, the default order of L2 algorithms is that calorimeter data are analysed first and tracking data afterward. The alternative approach should also be evaluated. For the *tau15* signature alone, using a simulated sample of limited statistics (950 QCD background events with hard parton $35 < p_T < 70$ GeV), the calorimeter and tracking approaches have been evaluated and show roughly equal total execution times, 34.1 and 34.0 ms respectively. On a simulated sample of 400 QCD background events with hard parton $p_T > 2$ TeV, for the same tau trigger, the tracking approach is slightly faster than the calorimeter approach, 299 versus 330 ms respectively. The longer execution time on high energy jet samples with respect to low energy ones is mainly due to the longer execution time of the L2 tracking algorithm. This result shows that the L2 tracking selection provides greater background rejection power than the L2 calorimeter one against high energy jets, and therefore suggests that the tracking-approach is advisable for high $E_T$ tau triggers.

# 5 Tau trigger menu

The single tau signatures are the basic element in the ATLAS trigger menu for collecting hadronic decays of tau leptons. Therefore, Section 2 and 3 have been devoted to the selection and performance of the single tau signatures. In this section we describe the full tau trigger menu, with an emphasis on triggers that combine tau signatures with other signatures and the physics goals they address. Such combined triggers are very important for collecting samples of tau leptons of moderate $E_T$ from several SM or even beyond the SM processes. As shown in Sect. 3, the minimal requirement on $E_T$ for a single tau trigger must be high, even at low luminosity, to be allowed to run without prescale factors. A combined requirement of a moderate $E_T$ tau signature with other signatures provides a way to achieve the necessary rejection against backgrounds and avoid prescale factors.

## 5.1 Combined tau triggers

The following triggers aiming at collecting hadronic decays of tau leptons are currently implemented in the ATLAS trigger menu, in addition to the single tau triggers already introduced:

- **tau+missing $E_T$ (tau+$E_T^{miss}$)**. This type of trigger covers a wide spectrum of physics channels. At low luminosity, when the trigger rejection can be relaxed, the selection of events with $W \rightarrow \tau\nu$ is the priority. $t\bar{t}$ events with tau leptons in the final state are also selected by this trigger. Such events are characterized by relatively soft $E_T$ range of tau leptons as well as low $E_T^{miss}$. The tau+$E_T^{miss}$ triggers at design luminosity are intended for SM or SUSY Higgs (neutral or charged) searches as well as for searches of new exotic particles like $Z'$. The $E_T^{miss}$ trigger [14] uses the same threshold requirement at all trigger levels. The turn-on of the $E_T^{miss}$ efficiency curve is mainly limited by the L1 missing $E_T$ resolution. Since the average $E_T^{miss}$ in $W \to \tau\nu$ events is low, and the EF $E_T^{miss}$ resolution is better, additional triggers where the $E_T^{miss}$ requirement is applied only at EF are under study.

- **tau+$\ell$(+jets)**, $\ell = e, mu$. This type of trigger aims at selecting events with two relatively soft tau leptons in the final state. Two tau leptons are found in events with $Z$ boson, neutral SM or SUSY Higgs. In addition, the tau+$\ell$ combination selects events with multiple leptons like $t\bar{t}$ or lepton flavor violating processes. The combination of two trigger signatures allows the use of lower threshold requirements than for the case of the single tau trigger. In case of excessive rates for this type of trigger at design luminosity, the additional requirement of jets or $E_T^{miss}$ can be introduced.

- **tau+tau(+jets)**. This type of trigger records events where both tau leptons decay hadronically. While the rejection rate is less favorable than the tau+$\ell$ case, the sample collected is complementary to the above and both increases statistics and allows the reduction of systematics uncertainties due to lepton identification. This trigger is highly relevant for searches of Higgs boson or new exotic particles like $Z'$ and will also be beneficial for SUSY double tau end point analyzes.

- **tau+jets, tau+*b*-jets**. This type of trigger is an interesting alternative for $t\bar{t}$ studies. At low luminosity, it allows the study of events with low jet $E_T$ thresholds, while at high luminosities it is necessary to reduce QCD and multiple interaction background events.

## 5.2 Commissioning triggers

In addition to single tau triggers there are single *track* triggers, highly optimized to select RoIs with one track only. The main purpose of this type of trigger is alignment of the tracker or hadronic calibration. These triggers work in *parasitic* mode, taking all L1 RoIs passing 6 GeV or 9 GeV with isolation requirements in a given event. In this manner, a sufficient amount of tracks for commissioning can be obtained.

Furthermore, additional triggers like `tau15i_PT` and `2tau25i_PT` are included in the ATLAS trigger menu (PT stand for pass-through). For pass-through triggers, all events that are accepted by the L1 algorithm are recorded for offline study of the HLT algorithms. At L2 and EF the algorithms are executed and their result is stored in the event record, however, their decision is not considered in the global trigger decision. As the output rate of these triggers is equal to the L1 rate, only a small fraction can be recorded, typically $\sim 0.1$ Hz after prescale factors.

## 5.3 Tau trigger menu performance

In this section we summarize tau trigger efficiencies for various physics signals, see Table 11, and the corresponding rates estimated on minimum bias samples for $100\,pb^{-1}$, see Table 12. For the results shown, the physics signal efficiencies are calculated per event, using the following references:

- trigger with one (two) tau leptons: at least one (two) tau leptons at generator level with visible $E_T$ greater than the tau trigger threshold(s) are required. Furthermore, the generated tau leptons are identified by the offline reconstruction algorithms.

- trigger with tau+*e* and tau+*mu*: besides the tau lepton requirement (see item above), the additional lepton is required at generator level to have an $E_T$ greater than the chosen lepton trigger threshold. The generated lepton additionally is identified by the "loose" offline reconstruction algorithm [1], [2].
- trigger tau+$E_T^{miss}$: besides the tau lepton requirement (see item above), the missing $E_T$ at generator level as well as the missing $E_T$ reconstructed by the offline reconstruction algorithms [14] are required to be above the chosen trigger threshold.

Introducing the triggers described above is important for most physics signals with tau leptons in the final state because:

- allows increased statistics
- it is a robust approach against failure or inefficiency of a particular trigger (e.g. due to detector problems)
- allows reduction of systematic uncertainty by comparing results of the same analysis repeated on samples selected with different triggers.

In Table 11 and Table 12 the symbol `tau+xe` stands for the tau+$E_T^{miss}$ triggers, and `j` stands for the jet triggers.

## 5.4 Tau trigger menus for different luminosity periods

According to the foreseen LHC start-up plans, ATLAS will commence data taking at a low luminosity of about $10^{31}\,\text{cm}^{-2}\,\text{s}^{-1}$. Over the course of the first year or so, it is expected that the luminosity will increase to a rather high luminosity of about $10^{33}\,\text{cm}^{-2}\,\text{s}^{-1}$. While the input rate rapidly increases during this period, the maximal output rate of the complete trigger is expected to be constant at about 200 Hz (due to limitations on storage capacities and recording speed), putting a significant burden on the trigger system.

As mentioned in Section 1, the low luminosity period will be used to commission the detector and trigger system. During this period, the focus of the trigger selection is Standard Model physics and events necessary for calibration and efficiency studies. The total sample collected is expected to be of the order of $100\,\text{pb}^{-1}$. Concerning hadronic decays of tau leptons, the absolute scale of tau leptons and the characteristics and rate of QCD jets misidentified as tau leptons are estimated from simulated data studies only, and need to be verified and understood with collider data. To address this, several triggers are proposed, and outlined in the next section. For each trigger, a method of measuring the efficiency on real data should also be proposed.

The transition to higher luminosity is foreseen to be done smoothly, dropping triggers of the low luminosity period menu which were used for commissioning or with unacceptably large rate, and keeping or introducing triggers without any prescale factors which focus on discovery physics.

The high luminosity menu will be obtained either by tightening the threshold requirements and cuts of triggers present in the low luminosity menu, or adding new triggers, especially combinations of several signatures. Only some ideas for a tau trigger menu for a luminosity of $\mathscr{L}=10^{33}\,\text{cm}^{-2}\,\text{s}^{-1}$ currently exist.

Given the current background rates, shown in Section 3, the high luminosity menu will include single high $E_T$ signatures with threshold requirements higher than what shown in Section 3 (most probably `tau100`). Furthermore, the high luminosity menu will be composed mostly of combined triggers like `tau35i_xe45`, `tau25i_e25i`, `tau25i_mu20`, `2tau35i`, `tau20i_j150` and `tau20i_4j50`.

| Trigger Item | $W_{\tau\to hX}$ | $Z_{\tau\tau}$ | $t\bar{t}$ | $A_{\tau\tau}(800)$ | SU3 | $H_{\tau\tau\to\ell hX}$ | $H_{\tau\tau\to hhX}$ |
|---|---|---|---|---|---|---|---|
| tau10 | 82.6±0.3 | 91.9±0.2 | 93.6±0.4 | 97.1±0.1 | 94.3±0.3 | 93.8±0.2 | 96.9±0.1 |
| tau10i | 78.7±0.4 | 89.8±0.2 | 91.1±0.4 | 96.4±0.2 | 92.0±0.4 | 92.1±0.3 | 95.8±0.1 |
| tau15 | 78.2±0.4 | 88.7±0.3 | 91.9±0.4 | 96.5±0.1 | 92.4±0.4 | 91.6±0.3 | 95.4±0.1 |
| tau15i | 74.1±0.4 | 86.0±0.3 | 89.1±0.5 | 96.1±0.2 | 90.4±0.4 | 90.0±0.3 | 94.1±0.1 |
| tau20i | 68.5±0.5 | 79.9±0.3 | 83.2±0.6 | 89.8±0.2 | 82.5±0.5 | 85.0±0.4 | 89.8±0.2 |
| tau25i | 66.5±0.6 | 76.0±0.4 | 80.1±0.7 | 89.0±0.3 | 79.7±0.6 | 82.0±0.4 | 87.1±0.2 |
| tau35i | 65.8±0.9 | 70.0±0.6 | 77.4±0.9 | 87.5±0.3 | 76.9±0.7 | 78.2±0.5 | 82.0±0.2 |
| tau45 | 82.7±1.3 | 78.7±0.8 | 88.0±0.9 | 94.9±0.2 | 89.6±0.6 | 86.2±0.5 | 88.5±0.2 |
| tau45i | 72.1±1.5 | 68.5±0.9 | 76.0±1.2 | 86.1±0.3 | 75.1±0.9 | 75.8±0.7 | 78.5±0.3 |
| tau60 | 77.5±2.6 | 74.4±1.5 | 74.7±1.7 | 91.4±0.2 | 78.2±1.1 | 76.1±0.9 | 77.5±0.4 |
| tau100 | 83.9±6.6 | 78.2±4.1 | 80.2±3.5 | 90.0±0.3 | 81.7±1.9 | 79.1±1.6 | 80.7±0.7 |
| 2tau25i | 0.0±0.0 | 47.2±1.5 | 60.0±11.0 | 62.6±1.2 | 62.6±2.7 | 61.5±6.7 | 59.3±0.6 |
| 2tau35i | 0.0±0.0 | 43.1±3.1 | 57.1±18.7 | 60.6±1.3 | 62.0±3.8 | 50.0±9.1 | 55.6±0.9 |
| tau15i_xe20 | 56.3±0.6 | 48.4±0.8 | 80.1±0.7 | 92.7±0.2 | 89.5±0.4 | 80.8±0.5 | 80.2±0.3 |
| tau20i_xe30 | 45.4±0.9 | 38.6±1.2 | 70.1±0.9 | 84.5±0.3 | 81.4±0.6 | 73.9±0.6 | 73.2±0.4 |
| tau25i_xe30 | 44.2±1.0 | 38.0±1.3 | 67.5±1.0 | 83.8±0.3 | 78.7±0.6 | 71.4±0.7 | 71.1±0.4 |
| tau35i_xe20 | 55.1±1.2 | 42.0±1.2 | 68.7±1.1 | 84.5±0.3 | 76.3±0.7 | 70.8±0.7 | 69.5±0.4 |
| tau35i_xe30 | 47.7±1.5 | 38.5±1.7 | 63.3±1.2 | 82.3±0.3 | 76.2±0.7 | 69.2±0.8 | 66.9±0.4 |
| tau35i_xe40 | 42.1±2.5 | 39.2±2.4 | 58.0±1.4 | 80.8±0.4 | 75.6±0.8 | 68.3±1.0 | 65.3±0.5 |
| tau45_xe40 | 54.7±3.6 | 48.3±3.0 | 67.4±1.6 | 87.6±0.3 | 88.2±0.7 | 76.5±1.0 | 71.0±0.6 |
| tau45i_xe20 | 60.1±2.1 | 43.8±1.7 | 67.3±1.4 | 83.2±0.3 | 74.7±0.9 | 69.0±0.9 | 66.7±0.4 |
| tau20i_e10 | 0.0±0.0 | 68.1±1.2 | 73.7±2.9 | 79.6±0.8 | 77.1±1.7 | 73.2±0.8 | 0.0±0.0 |
| tau20i_mu6 | 0.0±0.0 | 72.0±0.9 | 81.4±2.4 | 80.3±0.7 | 83.3±1.3 | 79.4±0.7 | 0.0±0.0 |
| tau20i_j70 | 70.8±1.4 | 78.8±0.7 | 81.6±0.7 | 89.4±0.2 | 82.7±0.6 | 65.6±0.7 | 72.6±0.3 |
| tau20i_3j23 | 61.1±2.1 | 80.2±0.8 | 82.0±0.7 | 89.9±0.2 | 83.2±0.6 | 83.7±0.7 | 80.7±0.3 |
| tau20i_4j23 | 58.2±4.1 | 82.7±1.6 | 80.5±0.9 | 90.3±0.3 | 84.5±0.6 | 86.9±1.7 | 83.9±0.7 |

Table 11: Physics signal efficiency without prescale factors. The requirements at generator level are summarized in Section 5.3. The Higgs boson mass for the last two columns is 120 GeV. Errors are statistical only.

| Trigger Selection | Level 1 Events | Level 1 Rate (Hz) | Level 2 Events | Level 2 Rate (Hz) | Event Filter Events | Event Filter Rate (Hz) |
|---|---|---|---|---|---|---|
| tauNoCut | 79802 | 23342 | 79252 | 23181 | 77231 | $22590 \pm 81$ |
| tau10 | 48854 | 14290 | 16235 | 4749 | 7748 | $2266 \pm 26$ |
| tau10i | 48854 | 14290 | 14066 | 4114 | 5413 | $1583 \pm 21$ |
| tau15 | 48854 | 14290 | 10472 | 3063 | 4054 | $1186 \pm 19$ |
| tau15i | 48854 | 14290 | 9764 | 2856 | 2953 | $864 \pm 16$ |
| tau15i_PT | 48854 | 14290 | 10237 | 2994 | 10237 | $2994 \pm 30$ |
| tau20i | 16298 | 4767 | 3322 | 972 | 1229 | $359 \pm 10$ |
| tau25i | 9404 | 2751 | 1565 | 458 | 631 | $185 \pm 7$ |
| tau35i | 3149 | 921 | 473 | 138 | 215 | $63 \pm 4$ |
| tau45 | 1062 | 311 | 321 | 94 | 255 | $75 \pm 5$ |
| tau45i | 750 | 219 | 154 | 45 | 79 | $23 \pm 3$ |
| tau60 | 262 | 77 | 35 | 10.2 | 25 | $7.3 \pm 1.5$ |
| tau100 | 262 | 77 | 7 | 2.0 | 4 | $1.2 \pm 0.6$ |
| tau15i_xe20 | 5625 | 1645 | 1329 | 389 | 198 | $58 \pm 4$ |
| tau20i_xe30 | 525 | 154 | 139 | 40.7 | 23 | $6.7 \pm 1.4$ |
| tau20i_xe30_PT | 525 | 154 | 142 | 41 | 142 | $41 \pm 4$ |
| tau25i_xe30 | 385 | 113 | 83 | 24.3 | 17 | $5.0 \pm 1.2$ |
| tau35i_xe20 | 784 | 229 | 129 | 37.7 | 42 | $12.3 \pm 1.9$ |
| tau35i_xe30 | 203 | 59 | 41 | 12.0 | 10 | $2.9 \pm 0.9$ |
| tau35i_xe40 | 58 | 17 | 15 | 4.4 | 3 | $0.9 \pm 0.5$ |
| tau45_xe40 | 47 | 14 | 18 | 5.3 | 6 | $1.7 \pm 0.7$ |
| tau45i_xe20 | 278 | 81 | 54 | 15.8 | 19 | $5.6 \pm 1.3$ |

Table 12: Minimum bias event rates without prescale factors. The cross section value used is $\sigma = 70\,\text{mb}$ and the peak luminosity used is $\mathscr{L} = 10^{31}\,\text{cm}^{-2}\text{s}^{-1}$. Errors are statistical only.

### 5.4.1 Running at $\mathscr{L}$=$10^{31}$ cm$^{-2}$ s$^{-1}$

The trigger menu for a luminosity of $\mathscr{L}$=$10^{31}$ cm$^{-2}$ s$^{-1}$ covers largely SM physics channels ($W \to \tau\nu$, $Z \to \tau\tau$, $t\bar{t}$). A fair amount of the bandwidth is given to commissioning triggers ($<$ 1 Hz each), which are needed to monitor trigger rates and variables used for tau lepton identification in different $E_T$ ranges.

Table 13 gives the trigger menu proposed for initial data taking, including expected prescale factors and corresponding trigger rates, with overlap between triggers taken into account. The prescale factors are necessary to restrict the total rate from low $E_T$ tau triggers, since there are many physics topics that need to share limited bandwidth. The initial prescale values have been determined from simulation, and will be optimised based on experience with early data. Event yields for several tau-based physics topics are shown in Table 14. In this table the results of only one typical trigger are reported for a set of triggers, corresponding to the one currently considered as the most suitable for the physics signals targeted at low luminosity. However, as physics simulation studies are progressing and as background rates are verified on data, it is likely that the trigger menu will be modified.

## 6 Summary

This note describes in detail the ATLAS tau trigger system, which is dedicated to the selection of hadronic decays of tau leptons. These results are obtained from simulations prior to LHC operations. In the selection at L1, energy deposition in the e.m. and hadronic calorimeters are used, while at L2 and EF calorimeter shower shape and energy information is combined with tracking information from the Inner Detector. At design luminosity, tau triggers are envisaged to be used without prescale factors to select single tau trigger candidates with $E_T$ above 100 GeV. The events with tau trigger candidates with lower $E_T$ will be selected using combined triggers which employ tau signatures in conjunction with missing $E_T$ electron, muon, jet or another tau signatures. The largest contribution to the rate will be misidentified low $E_T$ QCD jets, that are narrow and contain few particles mimicking hadronic decays of tau leptons.

In the first months of LHC operation a low luminosity of $10^{31}$ cm$^{-2}$ s$^{-1}$ is anticipated, and the data will primarily be used for commissioning the detector and trigger system. At that time the main focus of tau triggers will be on Standard Model physics such as $W \to \tau\nu$, $Z \to \tau\tau$ and $t\bar{t}$ events with hadronic decays of tau leptons in the final state. The typical tau trigger efficiency at low luminosity for generated tau leptons reconstructed by the algorithms of the offline reconstruction is expected to be around 80% for a wide range of physics channels, resulting in a total tau trigger rate of 28 Hz.

| Trigger | Prescale | Rate (Hz) | Cumulative Rate (Hz) |
|---|---|---|---|
| `tau100` | 1 | 2.4±0.5 | 2.4±0.5 |
| `tau60` | 1 | 10.7±1.0 | 10.7±1.0 |
| `tau45i` | 10 | 2.5±0.2 | 12.6±1.1 |
| `tau45` | 20 | 4.1±0.1 | 16.0±1.4 |
| `tau20i_xe30` | 1 | 4.9±0.7 | 20.1±1.4 |
| `tau20i_e10` | 1 | 1.2±0.4 | 21.2±1.5 |
| `tau20i_mu6` | 1 | 2.8±0.5 | 23.7±1.6 |
| `2tau25i` | 1 | 2.6±0.5 | 25.3±1.6 |
| `tau20i_4j23` | 1 | 0.1±0.1 | 25.4±1.6 |
| `tau20i_3j23` | 1 | 0.9±0.3 | 25.8±1.6 |
| `tau20i_j70` | 1 | 7.1±0.9 | 28.6±1.7 |

Table 13: Tau trigger menu for $\mathscr{L}$=$10^{31}$ cm$^{-2}$ s$^{-1}$.

| Trigger Item | $W_{\tau\to hX}$ | $Z_{\tau\tau}$ | $t\bar{t}$ | $A_{\tau\tau}(800)$ | SU3 | $H_{\tau\tau\to\ell hX}(120)$ | $H_{\tau\tau\to hhX}(120)$ |
|---|---|---|---|---|---|---|---|
| tau15i | 128.4 | 9.53 | 2.56 | 0.37 | 0.002 | 0.003 | 0.004 |
| tau45 | 816 | 121.4 | 63.9 | 26.2 | 0.17 | 0.11 | 0.17 |
| tau45i | 1406 | 211.4 | 110.3 | 47.5 | 0.30 | 0.20 | 0.31 |
| tau60 | 4399 | 670 | 568 | 484 | 3.07 | 1.18 | 1.87 |
| tau100 | 699 | 84 | 116 | 406 | 2.58 | 0.31 | 0.48 |
| 2tau25i | 0 | 544 | 13 | 37 | 0.24 | 0.02 | 0.68 |
| 2tau35i | 0 | 114 | 4 | 35 | 0.22 | 0.01 | 0.33 |
| tau20i_xe30 | 29005 | 668 | 2097 | 429 | 2.72 | 2.30 | 2.01 |
| tau35i_xe40 | 3379 | 178 | 831 | 374 | 2.37 | 1.03 | 1.00 |
| tau20i_e10 | 0 | 1191 | 182 | 83 | 0.53 | 1.36 | 0.00 |
| tau20i_mu6 | 0 | 2003 | 227 | 96 | 0.61 | 1.94 | 0.00 |
| tau20i_j70 | 8672 | 1278 | 2457 | 2.96 | 320 | 2.04 | 0.00 |
| tau20i_3j23 | 3852 | 1019 | 2697 | 2.31 | 288 | 1.30 | 0.00 |
| tau20i_4j23 | 957 | 232 | 1803 | 1.31 | 226 | 0.23 | 0.00 |

Table 14: Number of events expected in 100 pb$^{-1}$at $10^{31}$ cm$^{-2}$ s$^{-1}$ for different physics signals. Cross-sections are given in Table 1. The requirements imposed at generator level are summarized in Section 5.3.

# Physics Performance Studies and Strategy of the Electron and Photon Trigger Selection

**Abstract**
This note gives an overview of the implementation and performance of the electron and photon selection by the ATLAS trigger system. Trigger menus for commissioning as well as for the first physics run are presented together with the strategy for the early data taking phase. The physics performance in terms of selection efficiency and background rejection has been estimated using Monte Carlo simulations for various luminosity scenarios. An example of a method to determine trigger efficiency from real data using $Z \rightarrow ee$ events is discussed.

## 1 Introduction

This note gives an overview of the implementation and performance of the electron and photon selection by the ATLAS trigger system. Electron and photon trigger baseline signatures and menus for LHC commissioning and first physics run are presented. The principles that have determined their design are:

- Coverage of the physics needed for commissioning the trigger and detector systems as well as the offline reconstruction.
- Coverage of the physics channels that allow standard model studies and searches for new physics.
- Keeping the rates within the allowed bandwidth.

Events with electrons and photons in the final state are important signatures for many physics analyses envisaged at the LHC. A good selection by the electron/photon (e/$\gamma$) triggers will be important for many analyses from searches for new physics, such as the Higgs boson, SUSY, $Z'$ boson to standard model (SM) precision physics such as top quark and $W$ boson mass measurement, rare $B$ decays, etc. In the early running processes such as $Z \rightarrow ee$, $J/\psi \rightarrow ee$, $W \rightarrow e\nu$ and $\gamma$-jet events will be crucial for the understanding of the detector. These decays are important benchmark channels for the calibration, alignment and monitoring of the detector performance. The e/$\gamma$ trigger needs to cover the transverse energy range between a few GeV and several TeV. An overview of some relevant physics channels with electrons and photons in the final state, classified according to the corresponding transverse energy thresholds is given in Table 1.

To achieve this physics reach the trigger algorithms have to be optimized in terms of physics performance (signal efficiency, background rejection) and system performance (execution time, data requirements, etc). The trigger menu has to ensure a good selection of the above physics channels within the allocated rate for the various luminosities during LHC running.

In the following sections the implementation, performance and selection strategy are described in detail. Section 2 explains how e/$\gamma$ candidates are reconstructed and selected at the different trigger levels. Section 3 presents the selection strategy for LHC start-up, describing the schema currently envisaged for: commissioning, first menu for an initial luminosity of L$\sim 10^{31}$ cm$^{-2}$ s$^{-1}$ and the subsequent development towards higher luminosity scenarios. In Section 4 the performance of the various electron and photon triggers for a startup luminosity of L$\sim 10^{31}$ cm$^{-2}$ s$^{-1}$ and a luminosity scenario of L$\sim 10^{33}$ cm$^{-2}$ s$^{-1}$ is discussed. Examples of trigger efficiencies for physics processes with electrons and photons in the final state over the transverse energy ($E_T$) spectrum from a few GeV up to several TeV are included. A discussion of the robustness of the trigger selection follows in Section 5. Section 6 explains how the trigger efficiency will be determined from real data using $Z \rightarrow ee$ events.

| Momentum range | Examples of some important processes |
|---|---|
| low $p_T \sim$ 5-15 GeV | $B_d \to J/\psi K_s^0 \to ee\pi\pi$<br>$B_s \to K^*\gamma$<br>$J/\psi \to ee$, Drell-Yan |
| high $p_T \sim 20-100$ GeV | $H \to \gamma\gamma$ (for m(H)<130GeV)<br>$H \to ZZ^{(*)} \to eeee, ee\mu\mu$ (for $130 <$ m(H) $< 700$ GeV)<br>top physics, $Z \to ee, W \to e\nu$,<br>direct photon production |
| very high $p_T \sim 100-1000$ GeV | $Z' \to ee$, $W' \to e\nu$<br>$G \to \gamma\gamma, G \to ee$,<br>$pp \to ee^* \to ee\gamma$ |

Table 1: Examples of some important processes requiring a good electron and photon trigger selection

## 2 Electron and photon trigger selection

In this section the electron and photon trigger reconstruction and selection are described. The reconstruction and selection variables for each of the trigger levels are summarized in separate subsections.

### 2.1 L1 selection

At L1 trigger information from the electromagnetic (EM) and hadronic calorimeter system in the form of so-called trigger towers is used. A trigger tower has a dimension $\Delta\eta \times \Delta\phi \sim 0.1 \times 0.1$. In this region all the cells are summed over the full depth of either the electromagnetic or hadronic calorimeter. The L1 selection algorithm for electromagnetic clusters is based on a sliding $4 \times 4$ window of trigger towers which looks for local maxima [1].

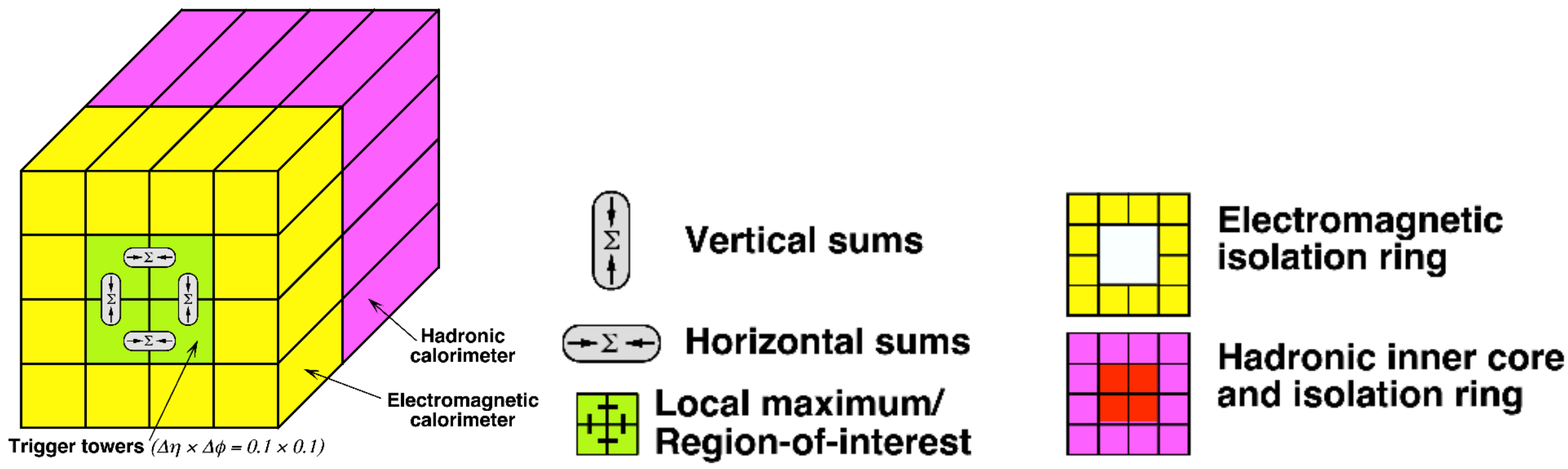

Figure 1: L1 calorimeter trigger schema, showing how trigger towers (each spanning a $0.1 \times 0.1$ $\eta \times \phi$ region) are used to determine the energy for the electromagnetic cluster as well as for the electromagnetic isolation, hadronic core and hadronic isolation.

The trigger object is considered to contain an electron or photon candidate if the following requirements are satisfied:

- The central $2 \times 2$ 'core' cluster consisting of both EM and hadronic towers is a local $E_T$ maximum This requirement prevents double counting of clusters by overlapping windows.

- The most energetic of the four combinations of two neighbouring EM towers passes the electromagnetic cluster threshold.

Figure 1 shows the L1 trigger tower schema used to determine the L1 selection variables. Isolation requirements can be imposed if required to control the rate:

- $E_{isol}^{EM}$: The total $E_T$ in the 12 EM towers surrounding the $2 \times 2$ core cluster is less than the electromagnetic isolation threshold.
- $E_{core}^{HAD}$: The total $E_T$ in the 4 towers of the hadronic calorimeter behind the $2 \times 2$ core cluster of the electromagnetic calorimeter is less than the hadronic core threshold.
- $E_{isol}^{HAD}$: The total $E_T$ in the 12 towers surrounding the $2 \times 2$ core cluster in the hadronic calorimeter is less than the hadronic isolation threshold.

The distributions of these isolation variables for signal and background are shown in Fig. 2 from Monte Carlo simulations. For signal, single electrons with an $E_T$ between 7 and 80 GeV with a flat distribution are used (solid line, hatched histogram). In comparison, background candidates from a simulated sample of QCD background (referred to as dijets) with $E_T > 17$ GeV are shown (dashed line, hollow histogram). Most of the samples discussed in this note, unless otherwise specified, use the Pythia 6.403 [2] Monte Carlo event generator. The ATLAS detector Monte Carlo simulation is based on GEANT4 [3]. The Monte Carlo simulations store details of the generated particles including the type and kinematic variables, this so-called *MC-Truth* information can then be used as a control in subsequent analyses of the simulated data. In this case the truth information has been used to guarantee that no real electrons are included in the background distributions in Fig. 2. The distributions shown have been normalized to unit area. It can be seen that isolation, in particular L1 EM isolation, provides a good handle to reduce jet background rate, typically needed at low transverse energy thresholds. As isolation depends on the topology and energy distribution of the event, as well as on other parameters such as luminosity and beam conditions, isolation cuts need to be well understood and applied carefully during data taking.

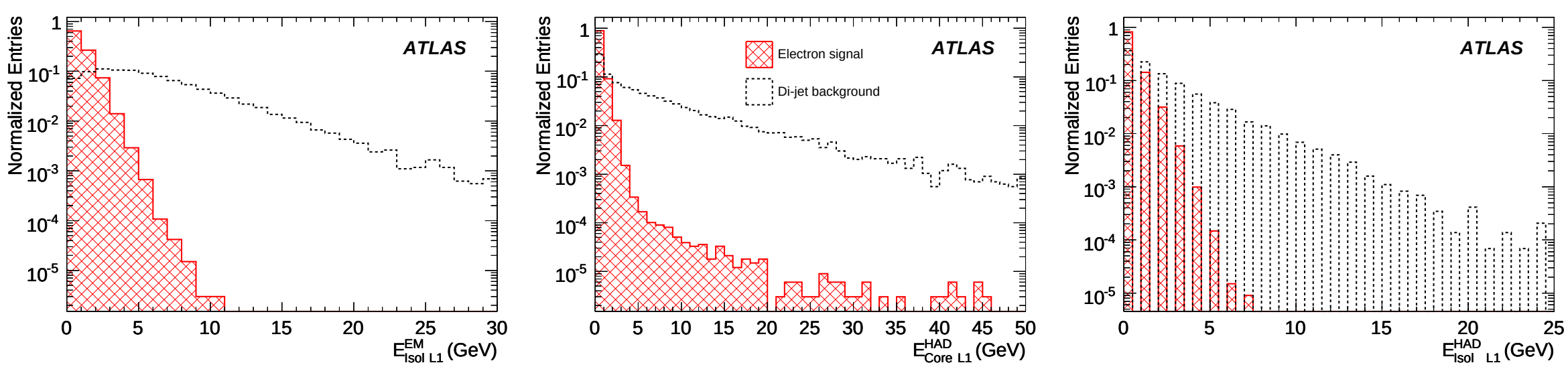

Figure 2: L1 isolation variables for single electrons with an $E_T$ between 7 and 80 GeV with a flat distribution (solid line, hatched histogram). In comparison, background candidates from the $E_T > 17$ GeV dijet sample are shown (dashed line, hollow histogram). For the background, only clusters that do not match to a true electron within a $\Delta R$ cone of 0.1 are considered. The distributions for electromagnetic isolation (left), hadronic core energy (middle), and hadronic isolation (right) are shown.

## 2.2 L2 selection

L2 is seeded by the L1 EM Region of Interest (RoI). Thus L2 receives the reconstructed L1 object with the $\eta$ and $\phi$ positions and the transverse energy thresholds passed. L2 accesses a subsample of the detector data around the given $\eta$ and $\phi$ position and applies trigger specific reconstruction algorithms characterized for their speed and robustness. Both photon and electron selection use the full granularity, full precision calorimeter information now available in the first selection step. The transverse cluster energy and various shower shape variables calculated in the different layers of the EM calorimeter are used to identify e/$\gamma$ candidates. The electron selection uses in addition inner detector information. Tracks are reconstructed in the inner detector and matched to the calorimeter energy clusters. Thus track finding and track-cluster matching variables can be used to select electrons.

### 2.2.1 Calorimeter based electron and photon selection

L2 calorimeter reconstruction is seeded by the $\eta$ and $\phi$ positions provided by L1. Calorimeter cells in a window of size $\Delta\eta \times \Delta\phi = 0.4 \times 0.4$ are retrieved (for more details on the data preparation see [4]). At the L2 trigger the cluster building algorithm scans the cells in the second layer of the EM calorimeter and searches for the cell with highest $E_T$. Subsequently, a cluster of $0.075 \times 0.175$ in $\eta \times \phi$ is built around this seed cell. The larger cluster size in $\phi$ reduces the low-energy tails due to photon conversion and electron bremsstrahlung. Electrons and photons deposit nearly all of their energy in the EM calorimeter and deposit typically less than 1% of their energy into the hadronic calorimeter. In addition, showers from electrons and photons are typically smaller in the plane transverse to its direction than showers from jets. These quantities are used to select a low-background sample of electrons and photons.

In detail, the L2 electron and photon calorimeter algorithms select events base on the following quantities:

- Transverse energy of the EM cluster ($E_T^{EM}$): Due to the energy dependence of the jet cross-section, a cut on $E_T^{EM}$ provides the best rejection against jet background for a given high $p_\mathrm{T}$ signal process.
- Transverse energy in the first layer of the hadronic calorimeter ($E_T^{Had}$): This is required to be below a given threshold. This cut is relaxed for high $E_T$ triggers (90 GeV and above) as the leakage into the hadronic calorimeter increases with energy.
- Shower shape in $\eta$ direction in the second EM sampling: The ratio of the energy deposit in $3 \times 7$ cells (corresponding to $0.075 \times 0.175$ in $\Delta\eta \times \Delta\phi$) over that in $7 \times 7$ cells is calculated: $R_{core} = E_{3x7}/E_{7x7}$. Photons and electrons deposit most of their energy in $3 \times 7$ cells and thus the corresponding ratio is typically larger than 80 %.
- Search for a second maximum in the first EM sampling: After applying the cuts in the hadronic calorimeter and the second sampling of the EM calorimeter, only jets with very little hadronic activity and narrow showers in the calorimeter remain. The fine granularity in rapidity in the first sampling of the EM calorimeter allows checks to be made for substructures within a shower for a further rejection of background such as single or multiple $\pi^0$s or $\eta$s decaying to photons. The energy deposit in a window $\Delta\eta \times \Delta\phi = 0.125 \times 0.2$ is examined. The shower is scanned for local maxima in the $\eta$-direction. The ratio of the difference between the energy deposited in the bin with highest energy $E_{1st}$ and the energy deposited in the bin with second highest energy $E_{2nd}$ divided by the sum of these two energies is calculated: $R_{strips} = (E_{1st} - E_{2nd})/(E_{1st} + E_{2nd})$. This ratio tends to one for isolated electrons and photons, and tends to zero for photons coming for example from $\pi^0$ decay.

Figure 3 shows typical distributions of the above shower shape variables for a signal sample of Higgs decaying into $\gamma\gamma$ and a the dijet background sample. A clear rejection power against background of the $E_T$ is seen for a cut above 20 GeV. The shower shape variables in the first and second EM sampling of the calorimeter provide a good discrimination power above 0.8. These shower shape variables will have a narrower distribution for photons compared to electrons. The high granularity of the first EM sampling of the ATLAS detector permits efficient photon identification using only calorimeter information, tracking information is not used at all in the selection.

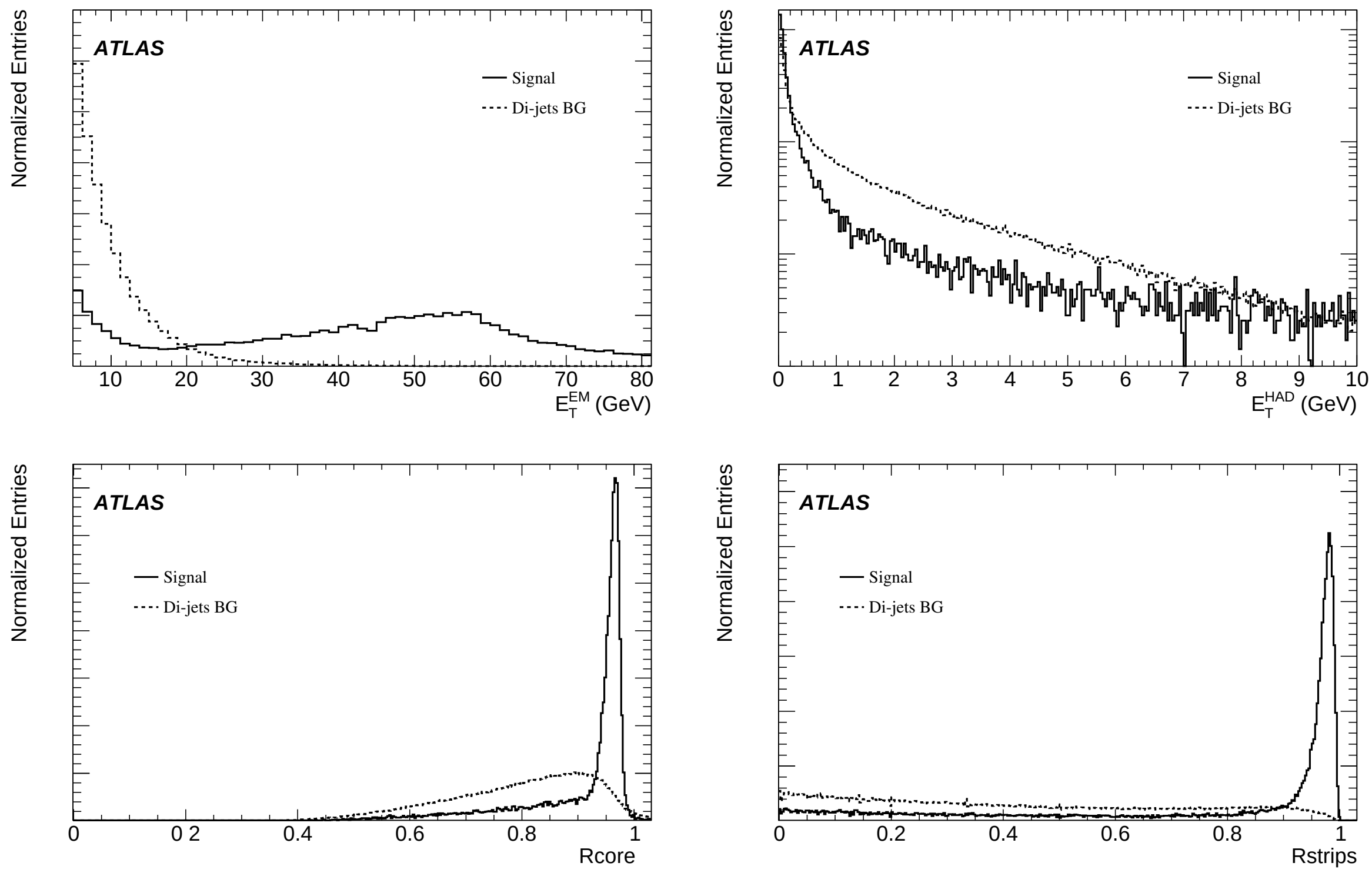

Figure 3: Selection variables for a L2 calorimeter energy cluster. The distributions are shown for signal candidates from a simulated $H \rightarrow \gamma\gamma$ sample (dashed line) and for dijet background candidates that do not have a photon or electron matched within a $\Delta R$ cone of 0.1 and that have at least 1 jet with $E_T > 17$ GeV (black solid line). Both distributions have been normalized to unity. The plots show the transverse energy of the EM cluster (top left), transverse energy deposited in the first layer of the hadronic calorimeter (top right), shower shape in the $\eta$ direction in the second EM sampling ($R_{core}$) (bottom left), and the search for a second maximum in the first electromagnetic sampling ($R_{strips}$) (bottom right).

### 2.2.2 Inner detector electron selection

If all the criteria of the calorimeter based electron selection are fulfilled, a search for tracks is performed in front of the cluster, electron trigger candidates are identified by the presence of a matching reconstructed tracks [5].

### 2.2.3 Combined calorimeter and inner detector based electron selection

A further rate reduction, while maintaining a high electron efficiency, can be achieved by combining the calorimeter and inner detector information. The background rate can be reduced by cutting on $\eta$ and $\phi$ between the EM cluster and the extrapolated track into the calorimeter. As shown in Fig. 4 these distributions are narrower for electrons than for jets. Another quantity useful for electron selection is the ratio of the cluster $E_T$ and the track $p_{\mathrm{T}}$. This quantity is affected by bremsstrahlung effects which cause a tail in the $E_T/p_T$ distribution towards high values (see Fig. 4), so it is not intended to use an upper cut on $E_T/p_T$ in the early data-taking.

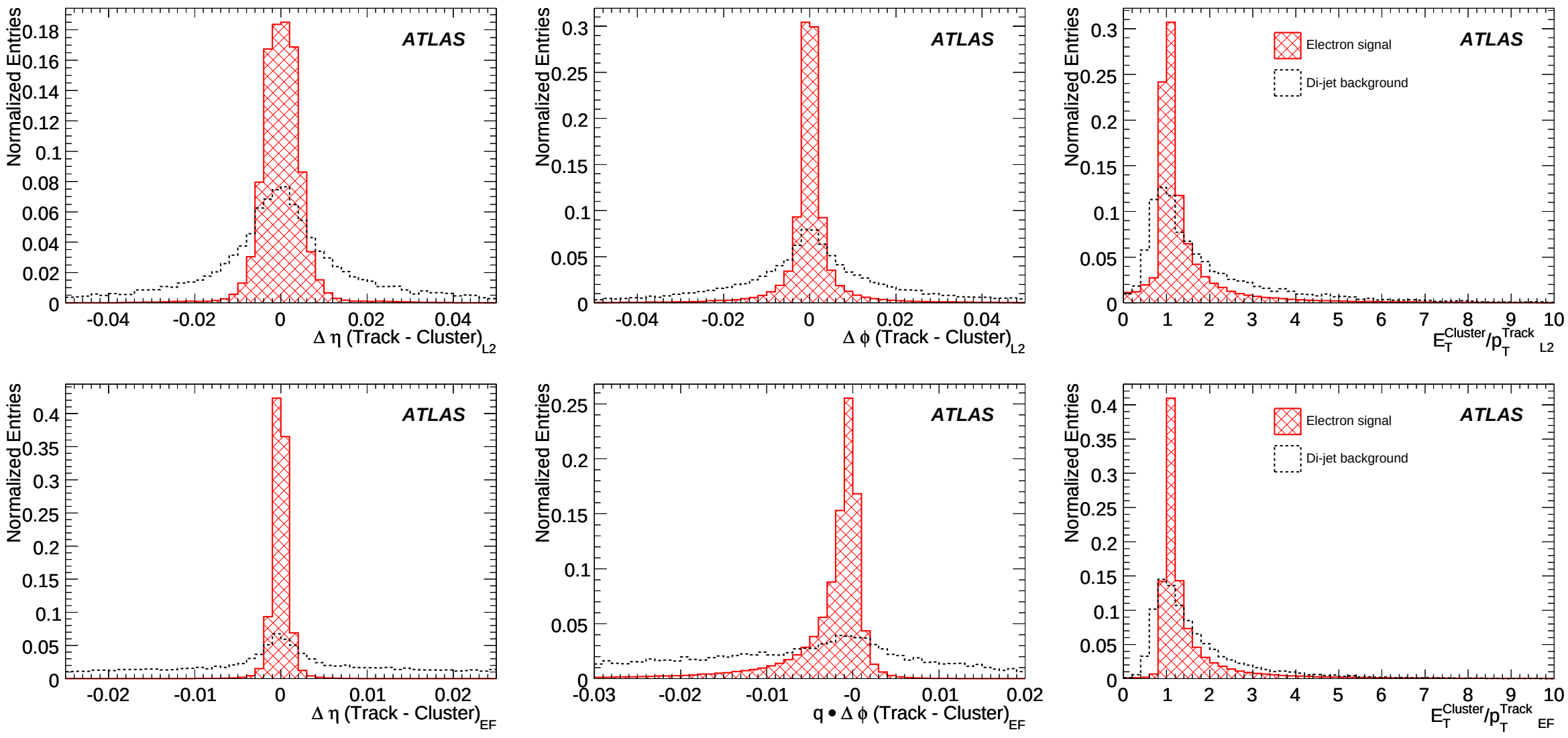

Figure 4: L2 (top) and EF (bottom) electron selection variables based on the combined calorimeter and inner detector information. From left to right the following distributions are shown: the difference in $\eta$ (left) and $\phi$ (middle) between the cluster and track (extrapolated to calorimeter) position. and are shown: ratio of the $E_T$ of the EM cluster and the $p_\mathrm{T}$ of the reconstructed tracks (right). Distributions are shown for signal (solid line, hatched histogram) and background (dashed line, hollow histogram). The reconstructed electrons come from a $7 < E_T < 80$ GeV sample. The background candidates come from a filtered dijet simulated sample, only candidates with no match to a truth electron within a $\Delta R$ cone of 0.1 rad are selected.

## 2.3 EF selection

At the EF trigger level offline reconstruction algorithms and tools are used as much as possible. An important difference, however, between the offline and the EF reconstruction is that the offline reconstruction is run once per event accessing the whole detector, while the EF uses a seeded approach; runs several times per event, once for each RoI given by L2, accessing only the corresponding subsample of the detector.

Currently, in the photon trigger selection only calorimeter information is used. EM clusters are searched for and reconstructed in RoIs of size $\Delta\eta \times \Delta\phi = 0.4 \times 0.4$. The EF calorimeter clustering algorithm searches for a local energy maximum with calorimeter trigger tower granularity. For electron and photon reconstruction only the data from the EM calorimeter is used, in contrast with L2 where the hadronic energy is also computed. The clusters should have an $E_T$ above a given threshold. The default cluster size used is $0.125 \times 0.125$ in $\eta \times \phi$ (whilst the offline reconstruction algorithms perform clustering with different window sizes, a single clustering option is used in the EF ). Once found by the clustering algorithm the cluster parameters (position, energy, etc.) are computed and further refined by a set of cluster correction (position and energy calibration) tools [5]. Also, corrections for the transition region between barrel and end-cap calorimeters are possible using the information from a set of scintillators.

For electron triggers, tracks are subsequently reconstructed in the inner detector. The EF tracking currently implements an inside-out track reconstruction (track finding starts from the inner silicon de-tectors and then is extended to the transition radiation tracker). In the future, an outside-in approach is intended to be used in the trigger in cases where photon conversions are present. There is the possibility to attempt bremsstrahlung recovery using offline tools. This option is not foreseen to run in the electron triggers for start-up but might be applied when running at higher luminosities with tighter selections. At $L\sim10^{31}$ cm$^{-2}$ s$^{-1}$ bremsstrahlung effects do not affect the trigger efficiency since the selection cuts are sufficiently loose to be insensitive to the performance improvements.

Electron and photon identification in the EF is very similar to the offline [6]. Calorimeter shower shapes, leakage into the hadronic calorimeter and the $E_T$ of the EM cluster are used for the calorimeter based selection for electrons and photons. Compared to L2 more shower shape variables are used. Together with improved calibrations this results in a further rate reduction. For electrons track-cluster matching variables, track quality cuts, transverse impact parameter and for high luminosity running potentially transition radiation information could be used to further reduce the rate.

As an example a loose electron EF selection will use the following selections: longitudinal leakage, shower shapes in the middle layer of the EM calorimeter, and very loose track-cluster matching cuts. Tighter selections might also use the shower shapes in the first EM calorimeter layer, information on the transverse impact parameter and on the track quality (number of hits in the pixels and strip silicon detectors and number of hits in the first pixel layer). Distributions for the track-cluster matching variables at EF are shown in Fig. 4. The distributions are very similar to those of the L2 track-cluster matching, but more refined algorithms are available at this stage, with more up-to-date calibration and alignment information.

## 3 Electron and photon trigger selection strategy

In this section the foreseen trigger selection strategy from the start-up phase up to physics running at nominal low luminosity is discussed. First the various commissioning steps at start-up are explained. This is followed by a discussion of a possible electron and photon trigger menu for the first physics run assuming a luminosity of $L\sim10^{31}$ cm$^{-2}$ s$^{-1}$.

When LHC turns on the trigger menus will need to provide data for commissioning and for analysis. In these paragraphs analysis is used to mean standard model measurements and searches for new physics. Typically these two objectives require very different trigger selection criteria. Commissioning usually demands loose selections to allow for more basic understanding of the detector and trigger, but loose selections accept more background events. Analysis is favoured with tight trigger selections that maximize the signal to background ratio in the allocated bandwidth. At start-up commissioning tasks would be prioritized, though signatures that provide data for a physics analysis will be included in the menu whenever possible. Following the progress on commissioning, loose trigger selections will be substituted by tighter ones that increase the analysis capabilities. The main challenge for the ATLAS trigger is the big difference (approximately six orders of magnitude) between the collision rate and the rate at which data can be stored. This imposes tight constraints on the trigger menus.

These principles used to determine the trigger menus, are developed in more detail in the following subsections, which also include early trigger menu tables.

The naming conventions for the various e/$\gamma$ triggers used in the following sections are as follows. The name for L1 electromagnetic candidates is EM. This is followed by the $E_T$ threshold applied for this triggers and if isolation criteria are applied an “i” is added at the end of the name. The number preceding EM indicates the object multiplicity. For example 2EM13I requires that at least two isolated electron or photon candidates are identified at L1 passing a $E_T = 13$ GeV threshold. The naming conventions for L2 and EF are very similar. For example 2$\gamma$17i denotes a trigger which selects events in which two isolated photons are found passing a $E_T$ threshold cut of 17 GeV at EF level. Similarly a possible e20_xe15 trigger selects events in which at least one electron candidate with $E_T\ (EF) > 20$ GeV is identified, in addition, the missing transverse energy at EF exceeds 15 GeV.

## 3.1 Commissioning

At LHC start-up an initial luminosity of L$\sim10^{31}$ cm$^{-2}$ s$^{-1}$ is expected. This will allow lower $E_T$ thresholds compared to those at the nominal LHC luminosity. The start-up menu has to provide the data samples needed to commission the trigger and detectors and at the same time provide useful data to be used for physics analysis. Therefore, the trigger has to guarantee the selection of the following standard model channels: $W \to e\nu$, $J/\psi \to ee, \Upsilon \to ee$, Drell-Yan, and direct photon production. For example: $Z \to ee$, $J/\psi \to ee$ and $\Upsilon \to ee$ will provide input for the electromagnetic calibration, alignment, efficiency measurements, etc. Electrons from bottom and charm quark decays will be useful for studies of $E/p$. Direct photon production will provide input for the jet calibrations using $\gamma$-jet events where the photon and jet are back-to-back. For comparison, in the first 100 pb$^{-1}$ of data we expect 235k of $J/\psi \to ee$, 40k of $\Upsilon \to ee$, 10k of Drell-Yan events, 10M $b,c \to e$, 100k direct photons and 250k of $W \to e\nu$.
The strategy is to apply L1 selections and run the High-Level Trigger (HLT) in pass-through mode (the selection criteria are tested and the trigger decision is recorded but no event is rejected). Table 2 shows the e$\gamma$ triggers foreseen. This will provide the first data samples for the low and high-$p_T$ spectrum and the rates for the various e/$\gamma$ triggers. Fig. 5 shows the expected L1 e/$\gamma$ rate for a luminosity of L$\sim10^{31}$ cm$^{-2}$ s$^{-1}$ with and without isolation criteria applied as a function of the transverse energy threshold.

| **Trigger** | **L1 item** | **Rate w/o prescale [Hz]** | **Pre-scale** | **HLT Rate [Hz]** |
|---|---|---|---|---|
| em5_passHLT | EM3 | 40000 | 20000 | 2 |
| em10_passHLT | EM7 | 5000 | 1300 | 4 |
| em15_passHLT | EM13 | 800 | 200 | 4 |
| em15i_passHLT | EM13I | 390 | 100 | 4 |
| em20_passHLT | EM18 | 280 | 70 | 4 |
| em20i_passHLT | EM18I | 100 | 25 | 4 |
| em25i_passHLT | EM23I | 41 | 10 | 4 |
| em105_passHLT | EM100 | 1 | 1 | 1 |
| 2em5_passHLT | 2EM3 | 6500 | 1600 | 4 |
| 2em15_passHLT | 2EM13 | 80 | 20 | 4 |
| 2em20_passHLT | 2EM18 | 35 | 10 | 4 |

Table 2: 'L1-only selection' trigger menu items for the LHC start-up including prescale factors. Depending on the rate, prescale factors might be readjusted. The L1 selection is applied and HLT selection is run in pass-through mode.

This will provide the input, based on real data, to devise a trigger menu best suited for the first commissioning and physics run with HLT selection enabled. Note that current rate estimates are affected by an uncertainty factor of two or three coming from the theoretical uncertainties in the jet cross-section. Though the L1 pass-through triggers will mainly select background events they are useful samples and can be used to look for clusters and tracks in the whole $\eta - \phi$ space and check that known dead and noisy channels and/or disconnected regions are correctly flagged. In addition the distributions of the e/$\gamma$ selection variables from data can be compared to those from Monte Carlo simulations. For example, signal over background distributions will be studied for variables with good discrimination power (e.g. Rcore).

L2 and EF performance can be checked and studies undertaken to evaluate the tracking performance of the different L2 tracking algorithms.

In the next phase of the commissioning the HLT selection will be progressively enabled. Several monitoring triggers will be kept, though with increased prescale factors applied. For example, L2 and EF in pass-through mode or loose $\gamma$ triggers (to monitor tighter calorimeter cuts as well as tracking efficiency for the electron triggers).

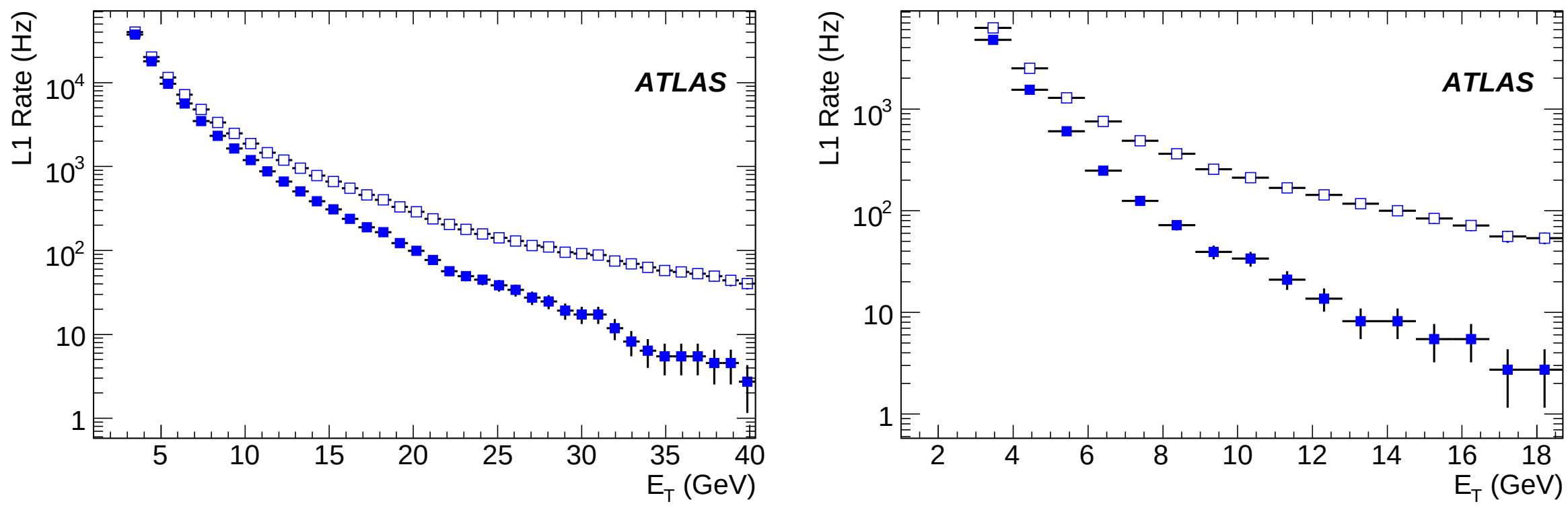

Figure 5: L1 rate for single (left) and double (right) e/$\gamma$ triggers for a luminosity of L$\sim 10^{31}$ cm$^{-2}$ s$^{-1}$. The open correspond to non-isolated triggers. Errors are statistical only.

## 3.2 First physics run

After the commissioning phase of the detector the first physics run is foreseen. At the physics run, each of the trigger menu signatures including the HLT will be enabled as soon as the understanding of the detector and trigger allow. Table 3 gives an overview of the main electron and photon physics triggers. The aim is to select events with at least one electron above $\sim$ 10 GeV or one photon above $\sim$ 20 GeV, in addition to the relevant double object triggers, e.g. for selecting $J/\psi$, $\Upsilon$, and Z events. $J/\psi \to ee$ and $\Upsilon \to ee$ events are particularly demanding events for the trigger system. The trigger rates errors for each trigger menu are statistical. Due to their relatively low masses, the electrons produced in the $J/\psi$ and $\Upsilon$ decays are very soft (with an average transverse momenta of less that 5 GeV). This poses a huge challenge to the L1 calorimeter trigger. Its performance at the low-energy end is limited by the noise of typically 0.5 GeV per RoI. A 3 GeV threshold is the limit for the L1 trigger. The 6.5 kHz L1 output rate for 2e5 makes it one of the biggest consumers of the total bandwidth.

To keep the rate at an acceptable level selections for the electron triggers at low-$p_T$ (e.g. 2e5) have to apply tighter HLT selections compared to the higher-$p_T$ triggers. A prescaled e5 trigger will allow measurement of the efficiencies and optimization of the selection cuts. A range of signatures is foreseen in the trigger menu to adapt to running conditions. If the trigger rate should prove to be too high back-up items with higher thresholds are included and/or prescale factors can be adjusted. Note, the uncertainties related to the detector response and the theoretical uncertainties on the jet cross-section. To ensure the selection of important physics channels, redundant triggers are present. As an example Table 4 shows the various triggers which will be useful for triggering on $W \to e\nu$, these include triggers which combine electrons and missing transverse energy.

| **Signature** | **L1 item** | **EF selection** | **L1 Rate** | **Pre-scale** | **HLT Rate** | **Motivation** |
|---|---|---|---|---|---|---|
| e10 | EM7 | medium | 5.0 kHz | 1 | 21 Hz | $e^{\pm}$ from b,c decays, E/p studies |
| $\gamma$20 | EM18 | loose | 0.3 kHz | 1 | $5.4 \pm 0.2$ Hz | direct photon production, jet calibration using $\gamma$-jet events, high-$p_{\mathrm{T}}$ physics |
| e20 | EM18 | loose | 0.3 kHz | 1 | $4.3 \pm 0.2$ Hz | high-$p_{\mathrm{T}}$ physics, $Z \rightarrow ee, W \rightarrow e\nu$ |
| em105_passHLT | EM100 | | 1 Hz | 1 | $1.0 \pm 0.1$ Hz | New physics, check for possible problems |
| 2e5 | 2EM3 | medium | 6.5 kHz | 1 | 6 Hz | $J/\psi \rightarrow ee$, $Y \rightarrow ee$, Drell-Yan production |
| 2$\gamma$10 | 2EM7 | loose | 0.5 kHz | 1 | $< 0.1$ Hz | di-photon cross-section |
| 2e10 | 2EM7 | loose | 0.5 kHz | 1 | $0.4 \pm 0.2$ Hz | $Z \rightarrow ee$ |

Table 3: Summary of the main electron and photon triggers envisaged for the first physics run at L$\sim 10^{31}$ cm$^{-2}$ s$^{-1}$. In this table the main physics triggers are listed including their expected rates and their physics motivation.

| **Signature** | **L1 item** | **EF selection** | **Pre-scale** | **Rate [Hz]** | **Motivation** |
|---|---|---|---|---|---|
| e20 | EM18 | loose | 1 | $4.3 \pm 0.2$ | main physics trigger |
| $\gamma$20 | EM18 | loose | 1 | $5.4 \pm 0.2$ | redundancy, check of tracking eff. and performance |
| e15_xe20 | EM13_XE20 | loose | 1 | $1.0 \pm 0.4$ | access to lower $p_{\mathrm{T}}$ -range |
| e10_xe30 | EM7_XE30 | medium | 1 | $0.3 \pm 0.3$ | access to lower $p_{\mathrm{T}}$ -range |
| e20i | EM18I | loose | 1 | $2.8 \pm 0.1$ | backup if rate too high |
| e25i | EM23I | loose | 1 | $1.4 \pm 0.1$ | backup if rate too high |
| e20_xe15 | EM18_XE15 | loose | 1 | $1.6 \pm 0.1$ | backup if rate is too high |
| $\gamma$20_xe15 | EM18_XE15 | loose | 1 | $1.9 \pm 0.2$ | backup if rate is too high, check tracking eff and performance |

Table 4: Summary of the main and redundant triggers for selecting $W \rightarrow e\nu$ events foreseen for the L$\sim 10^{31}$ cm$^{-2}$ s$^{-1}$ trigger menu.

### 3.3 Electron and photon trigger menus

This section collects detailed examples of the foreseen trigger menus. Table 5 shows a summary of the trigger menu for the first physics run assuming a luminosity L$\sim 10^{31}$ cm$^{-2}$ s$^{-1}$. Each row corresponds to a different trigger signature. The convention to interpret the name of a trigger signature is explained in Section 3. For each signature the table has three blocks of information. The first block of three columns gives L1 information:

- Name of the L1 trigger item. Which includes the transverse energy threshold required and an "I" if isolation cuts (as described in Section 2.1) are applied.
- Prescale factor applied after L1 selection.
- Corresponding rate after prescale factors applied.

The second block of three columns summarizes the EF:

- Tightness of the EF selection cuts.
- Prescale factor.
- Corresponding HLT rate.

The last column illustrates some relevant channels that the corresponding trigger would collect.

Tables 6 and 7 have the same structure described above. Table 6 gives examples of trigger signatures with tighter selection cuts, to be used if the rates given by the ones presented in Table 5 are too high.

A summary of the main trigger signatures foreseen for a higher luminosity scenario of L$\sim 10^{33}$ cm$^{-2}$ s$^{-1}$, is shown in Table 7.

## 4 Electron and photon trigger physics performance

In this section the performance of the different signatures is presented with trigger efficiency plots as a function of transverse energy $E_T$ and pseudo-rapidity $\eta$.

The trigger efficiency optimization of a given electron/photon trigger menu is a compromise between several factors: trigger efficiency for signal, QCD background rate which depends on the luminosity (constrained by allowed HLT bandwidth) and constrains of the average execution time at each trigger level. The performance of the main electron and photons triggers has been evaluated for a start-up luminosity of L$\sim 10^{31}$ cm$^{-2}$ s$^{-1}$and for a higher luminosity of L$\sim 10^{33}$ cm$^{-2}$ s$^{-1}$. In sections 4.2 and 4.3 the performance of these triggers is discussed for electrons and photons respectively. For these studies two types of events have been used, the so called ideal and the misaligned detector geometry. The misaligned detector geometry simulation contains expected distortions of the detector and in addition contains extra material coming from a more accurate description of cooling, powering and cabling services. This increases significantly the amount of material in the region $1.4 < |\eta| < 1.8$ compared to the ideal geometry, thus affecting the calorimeter energy calibration which was extracted using the ideal geometry. In section 4.2 a comparison is given for the performance based on these two layouts for one of the electron triggers.

### 4.1 Data Samples

The signal samples used for electron and photon trigger performance studies in this note are summarized in Table 8. For the electron and photon trigger optimization and turn-on curves samples with an energy range of $7 < E_T < 80$ GeV (flat energy spectrum) are used.

| Signature | Level-1 Item | Level-1 Pre-scale | Level-1 Rate [kHz] | HLT Sel-ection | HLT Pre-scale | HLT Rate [Hz] | Motivation |
|---|---|---|---|---|---|---|---|
| e5 | EM3 | 60 | 0.7 | medium | 1 | $4.8 \pm 0.2$ | $J/\Psi \to ee$, $Y \to ee$, Drell-Yan |
| 2e5 | 2EM3 | 1 | 6.5 | medium | 1 | 6 | $J/\psi \to ee$, $Y \to ee$, Drell-Yan |
| Jpsiee | 2EM3 | 1 | 6.5 | medium | 1 | 1 | $J/\psi \to ee$, $Y \to ee$ |
| e10 | EM7 | 1 | 5.0 | medium | 1 | 21 | $e^{\pm}$ from b,c decays, E/p studies |
| $\gamma$10 | EM7 | 1 | 5.0 | medium | 100 | $0.6 \pm 0.1$ | $e^{\pm}$ direct photon cross-section, e-no-track trigger |
| e10_xe30 | EM7_ XE30 | 1 | 0.2 | medium | 1 | $0.3 \pm 0.3$ | access low $p_T$-range for $W \to e\nu$ |
| 2$\gamma$10 | 2EM7 | 1 | 0.5 | loose | 1 | $< 0.1$ | di-photon cross-section |
| 2e10 | 2EM7 | 1 | 0.5 | loose | 1 | $0.4 \pm 0.2$ | $Z \to e^+e^-$ |
| Zee | 2EM7 | 1 | 0.5 | loose | 1 | $< 0.1$ | $Z \to e^+e^-$ |
| 2e12i_L33 | 2EM7 | 1 | 0.5 | tight | 1 | $< 0.1$ | trigger for L$\sim 10^{33}$ cm$^{-2}$ s$^{-1}$ |
| $\gamma$15 | EM13 | 1 | 0.7 | medium | 10 | $1.3 \pm 0.1$ | $e^{\pm}$ direct photon cross-section |
| e15_xe20 | EM13_ XE20 | 1 | 0.2 | loose | 1 | $1.0 \pm 0.4$ | access low $p_T$-range for $W \to e\nu$ |
| 2g17i_L33 | 2EM13I | 1 | 0.1 | tight | 1 | $< 0.1$ | trigger for L$\sim 10^{33}$ cm$^{-2}$ s$^{-1}$ |
| $\gamma$20 | EM18 | 1 | 0.3 | loose | 1 | $5.4 \pm 0.2$ | direct photons, jet calibration using $\gamma$-jet events, high-$p_T$ physics,check tracking eff. |
| e20_ passL2 | EM18 | 1 | 0.3 | loose | 200 | $< 0.1$ | check L2EF performance |
| e20 passEF | EM18 | 1 | 0.3 | | 125 | 0.1 | check L2EF performance |
| em20_ passEF | EM18 | 1 | 0.3 | | 750 | $0.5 \pm 0.1$ | check HLT performance |
| em20i_ passEF | EM18I | 1 | 0.1 | | 300 | $0.5 \pm 0.1$ | check L1 isolation |
| e22i_L33 | EM18I | 1 | 0.1 | tight | 1 | $1.2 \pm 0.1$ | trigger for L$\sim 10^{33}$ cm$^{-2}$ s$^{-1}$ |
| $\gamma$55_L33 | EM18 | 1 | 0.3 | tight | 1 | $1.2 \pm 0.1$ | trigger for L$\sim 10^{33}$ cm$^{-2}$ s$^{-1}$ |
| em105_ passHLT | EM100 | 1 | 1 | | 1 | $1.0 \pm 0.1$ | New physics, check for possible problems |
| $\gamma$150_ passHLT | EM100 | 1 | 1 | | 1 | $< 0.1$ | check for possible problems in express stream |

Table 5: Summary of triggers for the first physics run assuming a luminosity of L$\sim 10^{31}$ cm$^{-2}$ s$^{-1}$. For each signature rates and the motivation for this trigger are given.

| | Level-1 | | | Event Filter | | | |
|---|---|---|---|---|---|---|---|
| **Signature** | **Item** | **Pre-scale** | **Rate [kHz]** | **Sel-ection** | **Pre-scale** | **Rate [Hz]** | **back-up for trigger** |
| e5_e7 | 2EM3 | 1 | 6.5 | medium | 1 | $4.3 \pm 0.2$ | 2e5 |
| e5_e10 | EM3_EM7 | 1 | 6.5 | medium | 1 | $4.3 \pm 0.2$ | 2e5 |
| 3g10 | 3EM7 | 1 | $< 0.1$ | tight | 1 | $< 0.1$ Hz | 2g10, 2e10 |
| e20_xe15 | EM18_XE15 | 1 | 0.1 | loose | 1 | $1.6 \pm 0.1$ | e20 for $W \rightarrow e\nu$ selection |
| $\gamma$20_xe15 | EM18_XE15 | 1 | 0.1 | loose | 1 | $1.9 \pm 0.2$ | g20 for $W \rightarrow e\nu$ selection |
| e20i | EM18I | 1 | 0.1 | loose | 1 | $2.8 \pm 0.1$ | e20 |
| e25 | EM18I | 1 | 0.1 | loose | 1 | $2.4 \pm 0.1$ | e20 |
| e25 | EM18I | 1 | 0.1 | loose | 1 | $2.4 \pm 0.1$ | e20 |
| e25i | EM23I | 1 | $< 1$ | loose | 1 | $1.4 \pm 0.1$ | e20 |
| $\gamma$105_ | EM100 | 1 | $<< 1$ | | 1 | $< 0.1$ | em105_passHLT |
| e105_ | EM100 | 1 | 1 | | 1 | $< 0.1$ | em105_passHLT |

Table 6: Summary of backup triggers defined in case the rate is too high for a start-up luminosity of L$\sim 10^{31}$ cm$^{-2}$ s$^{-1}$. For each signature rates and the motivation for this trigger are given.

| | Level-1 | | | HLT | | | |
|---|---|---|---|---|---|---|---|
| **Signature** | **Item** | **Pre-scale** | **Rate [kHz]** | **Sel-ection** | **Pre-scale** | **Rate [Hz]** | **Motivation** |
| 2e12i | 2EM7 | 1 | 0.5 | tight | 1 | 1 | $Z \rightarrow ee$ |
| 2$\gamma$17i | 2EM13I | 1 | 0.1 | tight | 1 | $\sim 1$ | new physics e.g. $H \rightarrow \gamma\gamma$ |
| e22i | EM18I | 1 | 10 | tight | 1 | 120 | high-$p_T$ electron physics |
| $\gamma$55 | EM18 | 1 | 30 | tight | 1 | $20 \pm 3$ | high-$p_T$ photon physics |
| $\gamma$105 | EM100 | 1 | 0.1 | loose | 1 | 1 | new very high-$p_T$ physics |
| e105 | EM100 | 1 | 0.1 | loose | 1 | $< 1$ | new very high-$p_T$ physics |

Table 7: Summary of triggers assuming a luminosity of L$\sim 10^{33}$ cm$^{-2}$ s$^{-1}$. For each signature rates and the motivation for this trigger are given.

| Physics | $E_T$(GeV) | Geometry |
|---|---|---|
| single photons | 60 | misaligned |
| single photon scan | 7-80 | misaligned–ideal |
| single photons | 20 | misaligned |
| single electrons | 25 | misaligned |
| single electron scan | 7-80 | misaligned–ideal |
| Z$\rightarrow ee$ | – | misaligned–ideal |
| direct $J/Psi$ | – | misaligned |
| H$\rightarrow \gamma\gamma$ | 120 | misaligned |
| W$\rightarrow e\nu$ | | misaligned |
| G$\rightarrow ee$ | 500 | misaligned |
| Z$^{'} \rightarrow ee$ | 1000 | misaligned |
| G$\rightarrow \gamma\gamma$ | 500 | misaligned |
| Photon+Jet1 | 17-35 | misaligned |
| Photon+Jet2 | 35-70 | misaligned |
| Photon+Jet3 | 70-140 | misaligned |
| Photon+Jet4 | 140-280 | misaligned |
| Photon+Jet5 | 280-560 | misaligned |
| Photon+Jet6 | 560-1120 | misaligned |

Table 8: Main electron/photon signal and physics samples.

| $p_T$(hard) | jet filter | $\sigma$ after filtering |
|---|---|---|
| 6GeV | default | 4.214 mb |
| 7GeV | loose | 6.531 mb |
| 7GeV | default | 3.788 mb |
| 17GeV | loose | 0.373 mb |
| 17GeV | default | 0.191 mb |
| 35GeV | loose | 0.039 mb |
| 35GeV | default | 0.021 mb |

Table 9: Main background samples and their cross sections.

Background samples and their expected cross-sections are summarized in Table 9. The minimum bias Monte Carlo sample corresponds to our best knowledge of the inclusive expected backgrounds (including both hard and soft processes). Unfortunately most of the events in it are at low energies $< 7$ GeV and it is not very practical for high-statistics studies above this energy. To increase our efficiency for the events with electrons and photons we use a filter which is based on the combination of the EM cluster $E_T$ cut and the area of the EM cluster. Two types of the filter, "default" and "loose" are used to check for the potential biases in the L1 trigger rate which could occur if the area required by the filter is too small or the $E_T$ cut too high for the trigger threshold studied.

For the trigger rate studies in the intermediate energy range $6-17$ GeV we use minimum bias events with a transverse energy above 6 GeV and default filter (a rejection factor of 16.61). Above 17 GeV the dijet processes become the most prominent background for electrons and photons. Dijet samples with a threshold cut of 15 GeV and different level of filters (default and loose respectively) are used to provide background estimates in that area (the physics processes such as W, Z, direct photons have been added). For the very high energy studies dijet samples with a threshold cut of 35 GeV are used.

For this note two detector geometries are used:

- "ideal" which uses the detector geometry to our best knowledge.
- "misaligned" where the detector has misalignments and material distortions.

The misaligned detector geometry has been generated to test the robustness of our reconstruction and trigger algorithms with respect to incorrect alignment and calibration of the detector due to unexpected excess of material. Extra material was added in the misaligned detector geometry with respect to the ideal:

- For the inner detector, extra thin layers of material were added in the azimuthal angle range of $0 < \phi < \pi$ only. The amount of material added varies in the $z$ direction, from a few percent of $X_0$ on the active detector elements up to 1 $X_0$ in the areas occupied by services.
- For the electromagnetic calorimeter, more material was added in the barrel cryostat ($\sim 8-11\%$ of $X_0$), between the barrel presampler and first calorimeter sampling ($\sim 5\%$ $X_0$, always for positive $\phi$), and in the gap between the barrel and endcap cryostats (factor 1.7 increase of material density).

## 4.2 Electron trigger performance

The performance of the electron trigger has been evaluated for the trigger menus foreseen for the luminosity expected at early running L$\sim 10^{31}$ cm$^{-2}$ s$^{-1}$ and for a higher luminosity of L$\sim 10^{33}$ cm$^{-2}$ s$^{-1}$. The trigger efficiencies are quoted with respect to electrons identified with offline particle identification cuts [6]. The trigger efficiencies have been evaluated using simulated single electron samples with a transverse energy spectrum of $7 < E_T < 80$ GeV and other physics samples such as $J/\psi \to ee$, $Z \to ee$, etc. Rates have been estimated using simulated dijet events or minimum bias background depending on the trigger thresholds.

In the following the performance for the start-up trigger items e5, e10, e20, e105 and for a possible e22i trigger for higher luminosity running are discussed.

Figure 6 shows the efficiencies as a function of $E_T$ and $|\eta|$ of the low-$p_\mathrm{T}$ e5 trigger for the three trigger levels L1, L2 and EF. The e5 trigger menu is a trigger with prescale factor at L$\sim 10^{31}$ cm$^{-2}$ s$^{-1}$and the main trigger is the 2e5 trigger, which applies the same selection cuts. The efficiencies are obtained for electrons from a misaligned detector geometry $J/\psi \to ee$ sample. In the offline analysis these electrons typically will be identified via the "medium set" of offline "soft-electron" selection cuts (see [7] for more details). Therefore, the efficiencies shown in Fig. 6 are normalized with respect to such an offline selection. As seen in the right figure, the efficiencies drop significantly in the transition region between the barrel and end-cap calorimeter. A better optimization of the trigger selection cuts in this region might help to recover part of the inefficiencies. Figure 6 shows L1, L2 and EF efficiencies as a function of $E_T$ and of $|\eta|$ of the e10 trigger, which is the lowest $p_{\mathrm{T}}$ unprescaled single electron trigger foreseen for running at L$\sim10^{31}$ cm$^{-2}$ s$^{-1}$. The e10 trigger selects events with at least one good electron candidate with $E_T > 10$ GeV. The efficiencies are obtained for single electrons using ideal detector geometry and are normalized with respect to the medium set of offline electron cuts as discussed in Section 2.3. The efficiency reaches a plateau value for $E_T$ approximately above 15 GeV and is quite uniform as a function of $|\eta|$, except for a $10-20\%$ dip in the transition region between the barrel and end-cap calorimeters. In Fig. 7 e10 trigger efficiencies are compared for samples that use ideal and misaligned detector simulation. The biggest effects are observed at lower $E_T$ for the turn-on curve (left) and in the region $1.4 < |\eta| < 1.8$ (right) where a significant increase of material in the so-called misaligned geometry with respect to the ideal one was introduced. Trigger efficiency plots as a function of $E_T$ and $|\eta|$ for the electron triggers e20 and e105 are shown in Fig. 6. Compared to the e10 trigger signature, the e20 trigger applies looser electron identification cuts. The e105 trigger is aimed at selecting very high $p_{\mathrm{T}}$ electrons with very loose selection cuts for L$\sim10^{32}$ cm$^{-2}$ s$^{-1}$. At L$\sim10^{31}$ cm$^{-2}$ s$^{-1}$this trigger will only apply the L1 selection.

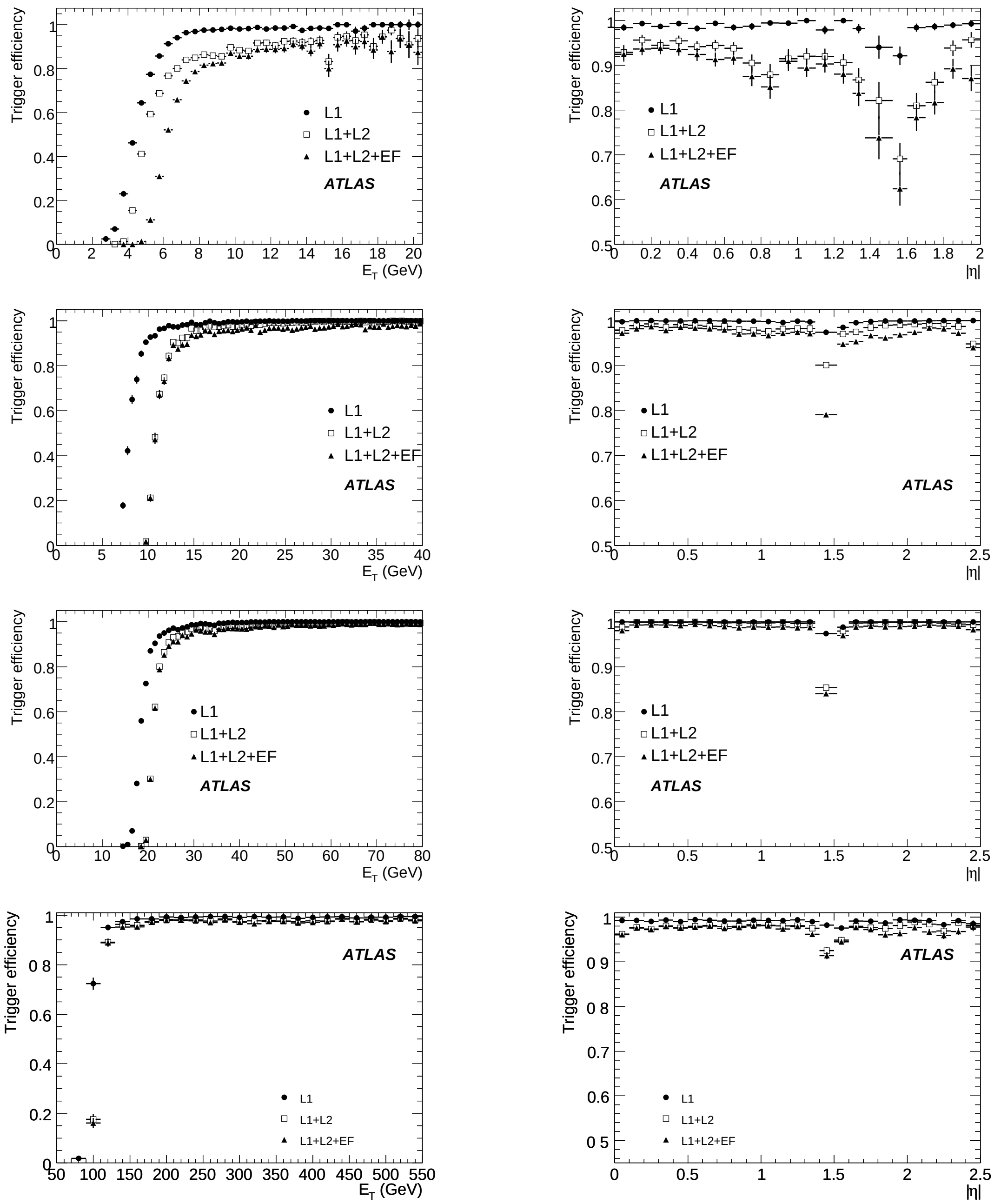

Figure 6: Trigger efficiencies at L1 (solid circles), L2 (open squares) and EF (solid triangles) as a function of true electron $E_T$ (left) and $|\eta|$ (right) for the e5 (top), e10 (second from top), e20 (third from top) e105 (bottom) menu items. The efficiencies are obtained from the following Monte Carlo simulated samples: $J/\psi \rightarrow ee$ decays simulated with misaligned detector geometry for e5 trigger item, $Z' \rightarrow ee$ (1TeV) for e105 and single electrons simulated with ideal detector geometry for e10 and e20. Trigger efficiencies are normalized with respect to the medium set of offline soft-electron cuts for e5, with respect to the medium set of offline electron cuts for e10 and with respect to the loose set of offline electron cuts for e20 and e105. For trigger efficiency versus $|\eta|$ plots, an $E_T$ cut according to the corresponding menu item has been applied: for e5 $E_T > 10$ GeV, for e10 $E_T > 15$ GeV, for e20 $E_T > 30$ GeV and for e105 $E_T > 130$ GeV. For e5 trigger item no data is shown for electrons for $|\eta| > 2$ as this is beyond the coverage of the transition radiation tracker whose information is used for the offline electron selection. Errors are statistical only.

For several physics channels a global trigger efficiency with respect to the offline loose selection is shown in Table 10 for each trigger level. An average trigger efficiency of $\sim 98\%$ is obtained after the EF level trigger. An efficiency close to 100% is expected for very high $p_{\mathrm{T}}$ electrons, therefore the selection has been optimized towards this goal.

Figure 8 (left) shows L1, L2 and EF efficiencies as a function of $E_T$ of the signature e22i, the menu item selecting an electron with $E_T > 22$ GeV. The efficiencies are obtained for single electrons using ideal detector geometry are normalized with respect to a loose set of offline electron cuts as discussed in [6]. The efficiency reaches a plateau value for $E_T$ approximately above 25 GeV.

Figure 8 (right) shows an example of trigger efficiency dependency on the offline electron identification cuts. Trigger efficiency has been determined with respect to loose, medium and tight offline electron identification selections as described in [6] for e15i. This menu item applies medium electron identification cuts at EF, therefore its efficiency with respect to loose offline reconstructed electrons is significantly lower than with respect to medium or tight offline electrons.

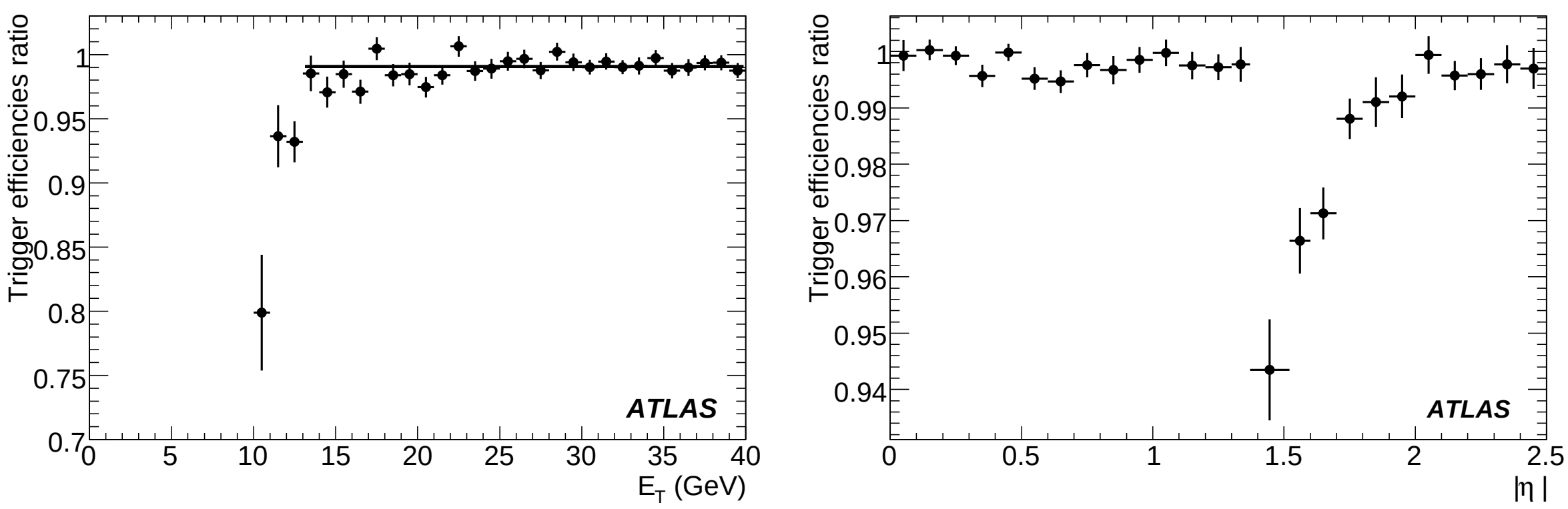

Figure 7: Ratio of trigger efficiencies for single electrons reconstruction in misaligned and ideal detector geometry as a function of true electron $E_T$ (left) and $|\eta|$ (right) for the e10 menu item. Events in the $|\eta|$ plot are required to verify $E_T > 15$ GeV. Errors are statistical only.

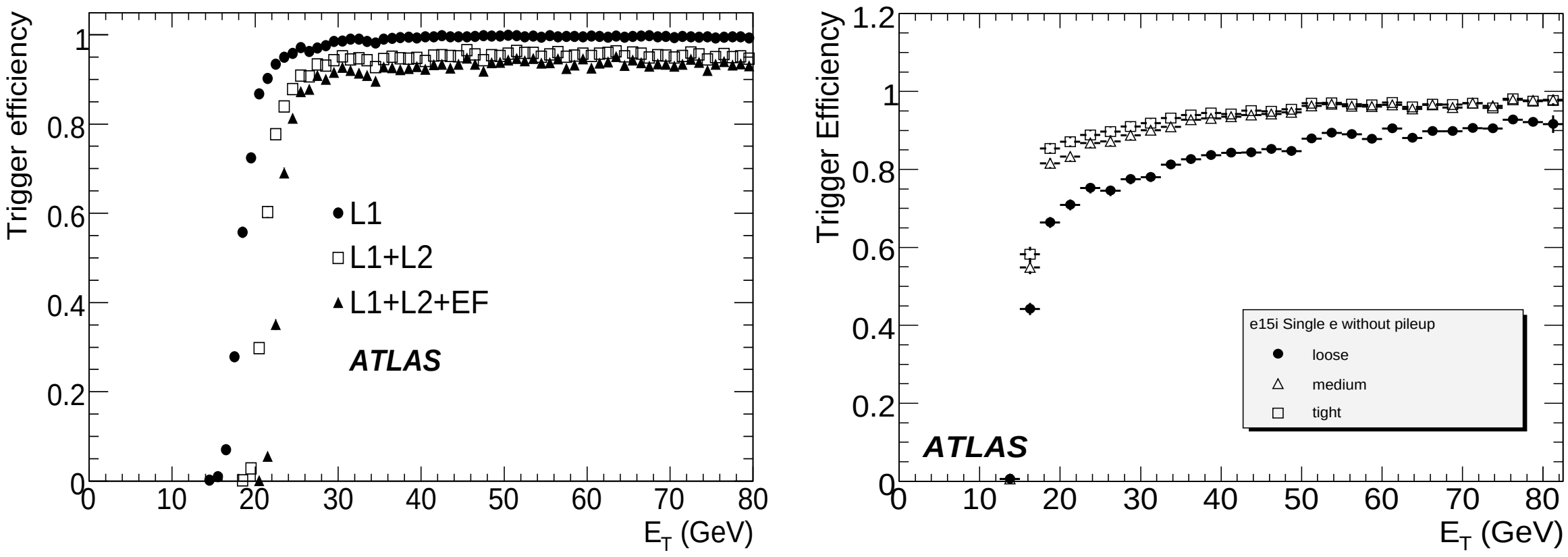

Figure 8: Trigger efficiencies at L1, L2 and EF as a function of true electron $E_T$ for the e22i menu item (left). The efficiencies are obtained for single electrons using ideal detector geometry and are normalized with respect to loose set of offline electron cuts. Trigger efficiency dependency on the offline electron identification (right). Trigger efficiency for the e15i signature is determined with respect to loose, medium and tight offline electron identification selection (described in [6]). Errors are statistical only.

### 4.3 Photon trigger performance

The physics performance of the photon trigger menus presented in Section 3 has been evaluated for early running at L$\sim 10^{31}$ cm$^{-2}$ s$^{-1}$and for a luminosity of L$\sim 10^{33}$ cm$^{-2}$ s$^{-1}$. The performance has been evaluated in terms of trigger efficiency after each trigger level using simulated single photons with transverse energies between 7 and 80 GeV. Rate estimates were calculated using dijet background simulations.
A $\gamma 20$ trigger with loose selection has been defined for the first physics run at L$\sim 10^{31}$ cm$^{-2}$ s$^{-1}$. At this stage the selections still need to be understood and the photon trigger is also used to check the tracking part of the electron selection. Therefore, the same calorimeter selection cuts are used as for electrons.

Figure 9 (top right) shows the relative rate for each trigger level as a function of $|\eta|$. Photons in the transition region between barrel and end-cap calorimeters ($1.37 < |\eta| < 1.52$) are not well measured and are excluded in the physics analysis.
The single photon trigger at L$\sim 10^{33}$ cm$^{-2}$ s$^{-1}$ selects photons above $E_T$ =55 GeV. To keep the rate under control, this trigger has to apply very tight L2 and EF calorimeter selections.

Figure 9 shows the trigger efficiencies as a function of $E_T$ and $|\eta|$ for the $\gamma 55$ trigger using single photons. Only photons with $|\eta| < 1.37$ or $1.52 < |\eta| < 2.45$ are considered. With respect to loosely selected offline photons the $\gamma 55$ trigger is 95.5% efficient.

The double object trigger $2\gamma 17i$ is another of the main photon triggers for running at L$\sim 10^{33}$ cm$^{-2}$ s$^{-1}$. Since the single photon dataset only contains one photon per event, the $2\gamma 17$i trigger efficiency was estimated as the square of the single photon trigger efficiency (Eff$_{2\gamma 17i}$= Eff$^2_{\gamma 17i}$). The overall trigger efficiency estimated from this sample is $93.9 \pm 0.2\%$.

Figure 9 shows the $\gamma$ 17i turn-on curve and efficiency as a function of $|\eta|$ using single photon events. The trigger efficiencies shown for the turn-on are normalized with respect to tight offline photon selection criteria. In addition, the photons are required to be within the region $|\eta| < 1.37$ or $1.52 < |\eta| < 2.45$. This plot shows that the $\gamma 17$i trigger is well set up and is fully efficient at 25 GeV.

The photon trigger performance has been studied not only for single photon samples but also for several physics channels. Table 11 shows a summary of photon trigger efficiencies for simulated samples of standard model direct photon decays, Higgs decaying into two photons (low mass range 120 GeV)

| Dataset (geometry) | Trigger Level | Trigger efficiency (%) |
|---|---|---|
| $W \to e\nu$ (misaligned) | L1 | $98.2 \pm 0.1$ |
| (e20) | L1 + L2 | $96.0 \pm 0.2$ |
| | L1 + L2 + EF | $94.3 \pm 0.2$ |
| $Z' \to ee$ (misaligned) | L1 | $99.7^{+0.3}_{-0.7}$ |
| (e105) | L1 + L2 | $98.9 \pm 0.7$ |
| | L1 + L2 + EF | $98.8 \pm 0.7$ |
| electrons 500 GeV (misaligned) | L1 | $99.1 \pm 0.6$ |
| (e105) | L1 + L2 | $97.6 \pm 0.5$ |
| | L1 + L2 + EF | $97.5 \pm 0.5$ |

Table 10: Global trigger efficiencies for the trigger items e20 and e105. The signal efficiencies are determined from several different signal samples: $W \to e\nu$ $Z' \to ee$ and a sample of single electrons of fixed transverse energy of 500 GeV. The simulation used misaligned detector geometry. Trigger efficiencies are determined with respect to loose offline electron selection. In the $W \to e\nu$ case the offline reconstructed electron is required to be within the $|\eta| < 2.5$ region and to have a transverse energy above 25 GeV. Efficiencies for the e105 item are determined in the kinematic region $|\eta| < 2.5$, the transition region between barrel and end-cap calorimeters removed ($1.37 < |\eta| < 1.52$), and with a minimum true transverse energy of 130 GeV.

| Dataset (luminosity) | Trigger Level | Trigger efficiency (%) |
|---|---|---|
| $\gamma$+Jet (L$\sim 10^{31}$ cm$^{-2}$ s$^{-1}$) | L1 | $100.0 \pm 0.0$ |
| ($\gamma$20) | L1 + L2 | $99.8 \pm 0.1$ |
| | L1 + L2 + EF | $95.0 \pm 0.3$ |
| G $\to \gamma\gamma$ (L$\sim 10^{32}$ cm$^{-2}$ s$^{-1}$) | L1 | $99.5 \pm 0.1$ |
| ($\gamma$105) | L1 + L2 | $98.8 \pm 0.1$ |
| | L1 + L2 + EF | $98.7 \pm 0.2$ |
| H $\to \gamma\gamma$ (L$\sim 10^{33}$ cm$^{-2}$ s$^{-1}$) | L1 | $95.6 \pm 0.2$ |
| (2$\gamma$17i) | L1 + L2 | $93.6 \pm 0.2$ |
| | L1 + L2 + EF | $91.2 \pm 0.2$ |

Table 11: Trigger efficiencies after each trigger level, normalized with respect to the loose $\gamma$ selection, for different physics channels covering a wide range in $E_T$. In the case of $H \to \gamma\gamma$ standard kinematical cuts are applied in addition. For each physics process the efficiencies for the main physics trigger to select these events for a given luminosity scenario are shown.

and an exotic signature: graviton decaying into a pair of photons G$\to \gamma\gamma$ (500 GeV).

Trigger efficiencies after each level are given. Efficiencies are determined with respect to a loose offline selection using the main physics trigger for a given luminosity to select these events. In the case of $H \to \gamma\gamma$ standard kinematical cuts are also applied. A more detailed study of trigger efficiencies for different physics channels can be found in [8] and [9].

# 5 Trigger robustness studies

The electron/photon HLT selection must be robust against detector effects such as mis-calibration, misalignment, dead and noisy read-out cells or sectors, luminosity and beam conditions changes etc. This

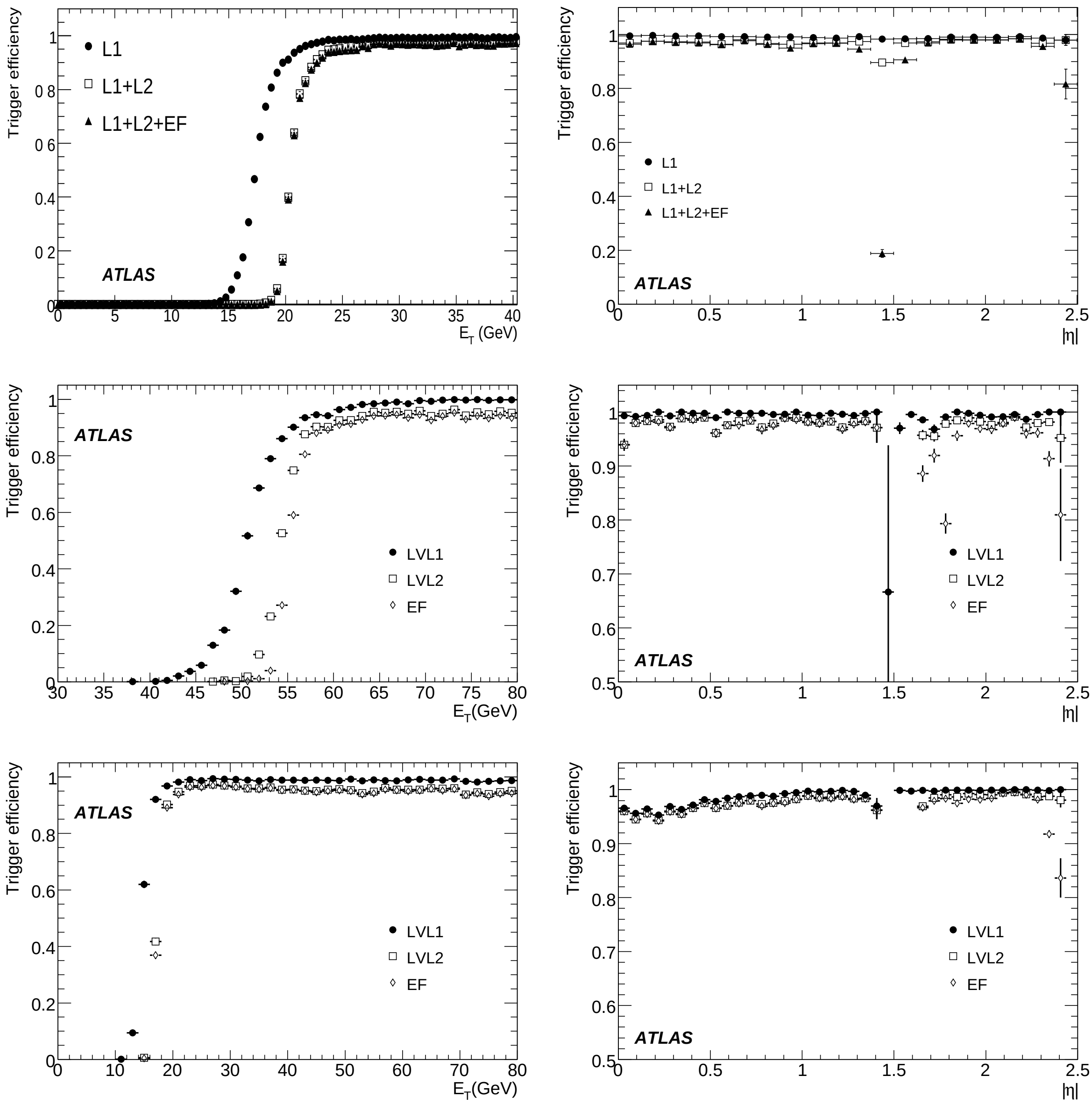

Figure 9: Trigger efficiencies at L1, L2 and EF as a function of the generated photon $E_T$ (top left)and $|\eta|$ (top right) for the $\gamma$20 trigger. The efficiencies are obtained for single photons simulated with ideal detector geometry and are normalized with respect to the loose set of offline photon cuts. Note, the $|\eta|$ plot includes an additional cut of $E_T > 23$ GeV. Trigger efficiencies at L1, L2 and EF as a function of the generated photon $E_T$ (middle left) and $|\eta|$ (middle right) for the $\gamma$55 trigger. The efficiencies are normalized with respect to photons with $E_T > 55$ GeV passing the loose set of offline photon cuts. Trigger efficiencies at L1, L2 and EF as a function of the generated photon $E_T$ (bottom left) and $|\eta|$ (bottom right) for the $\gamma$17i trigger. The efficiencies are normalized with respect to the tight set of offline photon cuts. Errors are statistical only.

will be especially important during early running. Depending on the bunch structure of the LHC, the effect of pile–up might already be important even at a luminosity of L$\sim 10^{32}$ cm$^{-2}$ s$^{-1}$. More than

| Trigger Level | e12i no pile-up | e12i with pile-up | e22i no pile-up | e22i with pile-up |
|---|---|---|---|---|
| L1 | $94.8 \pm 0.1\%$ | $94.8 \pm 0.3\%$ | $96.0 \pm 0.1\%$ | $95.5 \pm 0.3\%$ |
| L1 + L2 | $86.8 \pm 0.2\%$ | $86.7 \pm 0.4\%$ | $90.2 \pm 0.2\%$ | $89.4 \pm 0.4\%$ |
| L1 + L2 + EF | $81.5 \pm 0.2\%$ | $81.3 \pm 0.5\%$ | $88.7 \pm 0.2\%$ | $88.0 \pm 0.4\%$ |

Table 12: The e12i and e22i trigger efficiency from a sample of electrons of $7 < E_T < 80$ GeV, with and without pile-up at a luminosity of L$\sim$2$\times$ $10^{33}$ cm$^{-2}$ s$^{-1}$

one proton-proton interaction might occur per bunch crossing, the proton-proton interactions that are not interesting for analysis but happen in the same bunch crossing that the interesting one are typically denoted as 'pile-up'. At L$\sim$2$\times$$10^{33}$ cm$^{-2}$ s$^{-1}$ around 4.6 minimum bias events are expected per bunch crossing. In this section only the effects of pile-up and mis-calibrations due to additional detector material in front of the calorimeter are discussed.

## 5.1 Effect of pile-up

The robustness of several electron triggers was studied with simulated data which included overlapping events ("pile-up") for a luminosity L$\sim$2$\times$ $10^{33}$ cm$^{-2}$ s$^{-1}$. The presence of pile-up might have effects in reconstruction efficiency, (mostly for tracks but also for calorimeter energy clusters), as well as in the identification of isolated electrons and photons, both at the trigger and offline reconstruction level.

Table 12 shows the effect of pile-up on the e12i and e22i trigger for single electron events. As can be seen the effect of pile-up on the trigger efficiency at L$\sim$2$\times$ $10^{33}$ cm$^{-2}$ s$^{-1}$ is $\sim$ 3%. There is a $\sim$ 2% effect due to the isolation criteria at L1 when tight isolation cuts are applied (e12i) and a $\sim$ 1% loss at the HLT level. This results show that the trigger efficiency is only slightly more affected by pile-up effects than offline electron reconstruction. The effect of pile-up on trigger performance determined with respect to *MC-Truth* electrons as well as on the background rate is being studied.

## 5.2 Effect of additional detector material

A simulated data sample consisting of single electrons as described above was used to study the effect of additional inactive material in the detector. Despite the efforts to accurately describe the detector in Monte Carlo simulation it is not expected to be perfect. The biggest and more problematic differences between simulation and reality are expected to come from the description of inactive material (for example pipes for cooling, power and signal cables, etc). Trigger robustness against those possible distortions has to be tested. Methods to identify and correct those effects have to be developed. The detector simulation used to produce this data sample included distorted material distributions in both the inner detector volume and the electromagnetic calorimeter.

A detailed description of the material added in the Monte Carlo sample is given in Section 4.1 The amount of extra material with respect to the non-distorted simulation, grows from a few percent of a radiation length at $\eta = 0$ up to $\sim 1X_0$ at $1.5 < |\eta| < 1.8$, and then decreases towards higher values of $|\eta|$. The amount of material added in front of the active elements was larger than the uncertainty on the material distribution.

The effect of the extra inactive material on the electron trigger was studied by comparing the trigger efficiency for $\phi > 0$, where extra material was added in the detector simulation, and for $\phi < 0$, where no extra material was added. The resulting efficiencies are shown in Fig. 10. The efficiency is plotted as a function of the kinematic variables of the electron candidate reconstructed offline. A loose offline electron selection was used for normalization.

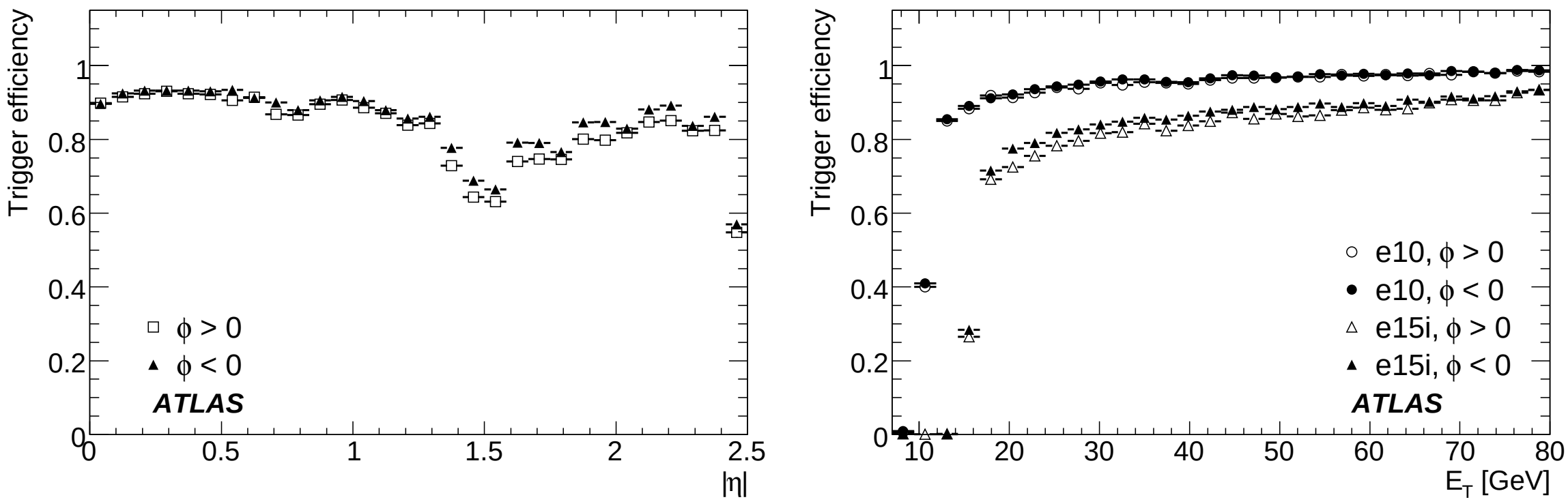


Figure 10: Effect of additional inactive material in the detector on the electron trigger efficiency. The trigger efficiency is compared for the nominal material distribution (at $\phi < 0$) and for increased inactive material (at $\phi > 0$) for the electron triggers e10 and e15i. The efficiency is plotted as a function of $|\eta|$ (left) and $E_T$ (right) of the electron candidate reconstructed offline. The left histograms correspond to the e15i trigger only. Errors are statistical only.

It can be seen that the effect of the added material is more pronounced for the e15i signature than for e10. The e15i signature applies tighter selection cuts than the e10. In particular e15i requires the so-called medium electron identification cuts in EF, this explains the lower trigger efficiency for e15i. If trigger efficiency was computed with respect to medium or tight offline electron identification selection, the trigger efficiency would be higher, as shown in Fig. 8.

# 6 Trigger efficiency determination from real data

The trigger efficiency will need to be determined from data, reducing the dependence on Monte Carlo simulation as much as possible. Before data-taking starts some of the methods to study the trigger performance without relying on the *MC-Truth* information are being developed, an example is given in this section.

The so-called "tag and probe" method has been studied using a Monte Carlo simulation sample of $Z \to ee$. In the first subsection the basis of this method is explained, followed by the detailed explanation of the selection criteria chosen for this study, the trigger efficiency computation and its comparison with *MC-Truth* information. In the following subsection the performance of the method is presented for some loose trigger selections that will be used in early running.

## 6.1 Efficiency extraction method

The so-called "tag and probe" method uses offline identification of $Z \to ee$ decays to select a clean sample of electrons, which are then used to determine the electron trigger efficiency. During data-taking a data sample to perform an analysis will be characterized by a given trigger signature being satisfied. In the study summarized in this section the data sample is defined by a given single electron trigger signature. The electron candidate that has satisfied the trigger is reconstructed and identified offline, it is the so-called "tag" electron. $Z \to ee$ decays are selected requiring a second electron to be identified offline together with some identification conditions on the $Z$ particle. This second electron is the so-called "probe" and it is used to study the trigger performance, since it is know to be a "good electron" it can be used to verify if electron trigger selection cuts are efficient to identify electrons.

The discrimination criteria applied to identify offline the tag and probe electrons and the $Z$ can vary. In the following paragraphs we specify the identification criteria chosen to perform the study presented in this section.

Selection of the tag electron involves tight electron identification cuts [6] in the offline reconstruction to reduce background, and this electron must also trigger the event on its own, through a single electron trigger chain. As trigger efficiencies are measured with respect to one of the established offline electron definitions (the so-called "loose", "medium" or "tight" electron identification set of cuts, described in [6]), this selection is the one initially applied to the probe electron.

Additionally, both offline reconstructed electrons must pass kinematic cuts, *i.e.* $E_T > E_T^{cut}$ and $|\eta| < 2.4$, and not lie in the region between the barrel and endcap calorimeters ($1.37 < |\eta| < 1.52$ region is excluded). The value of $E_T^{cut}$ is chosen to be where the efficiency reaches its high-energy plateau in curves that show trigger efficiency versus transverse energy (such as the one shown in Fig. 11 ). When a trigger efficiency is plotted as a function of $E_T$ or $\eta$, the relevant kinematic selection is relaxed. Finally, there are topological constraints, namely that the two electrons have opposite charge, and that their combined invariant mass lies in the range $70 < m_{\mathrm{ee}} < 110$ GeV.

At this point, the pair of electrons (tag and probe) may be considered for analysis. The trigger efficiency is defined by the frequency with which the probe electron in this sample passes the relevant trigger selection.

It is perfectly possible for the probe electron to satisfy the tag selection as well. In this case, the roles of tag and probe may be swapped, the tag becoming the probe and vice versa. Because the tag selection is at least as tight as the probe selection at every stage, the new probe passes the trigger selection automatically. Thus, tag and probe events fall into one of three categories:

- $N_{1f}$ events where only one electron passes the tag cuts and the probe fails the trigger selection
- $N_{1p}$ events where only one electron passes the tag cut and the probe passes the trigger selection
- $N_{2p}$ events where both electrons pass the tag selection

Counting these events, the measured efficiency may be expressed as

$$\varepsilon_{\text{“tag and probe''}} = \frac{N_{1p} + 2N_{2p}}{N_{1f} + N_{1p} + 2N_{2p}} = \frac{N_{\mathrm{p}}}{N_{\mathrm{T}}} \tag{1}$$

where $N_{\mathrm{p}}$ and $N_{\mathrm{T}}$ are defined by this equation. Assuming that the uncertainty in each variable is simply $\sqrt{N}$, the statistical error on $\varepsilon$ is

$$\sigma_{\varepsilon} = \frac{1}{N_{\mathrm{T}}}\sqrt{\left[(1-2\varepsilon)N_{\mathrm{p}} + \varepsilon^2 N_{\mathrm{T}} + (1-\varepsilon)^2 \cdot 2N_{2p}\right]}. \tag{2}$$

Equation 2 may be generalized to allow for a more comprehensive uncertainty in $N_{\mathrm{p}}$ and $N_{\mathrm{T}}$. Equations 1 and 2 can be extended to be used for differential trigger efficiency measurements.

When using Monte Carlo simulated events, the truth record (*MC-Truth*) can be used to provide another estimate of the trigger efficiency. To understand the systematic uncertainty of the tag and probe method, the fractional efficiency difference between the trigger efficiency determined using tag and probe method ($\varepsilon_{\text{“}tag\ and\ probe''}$) and the trigger efficiency determined with respect to the *MC-Truth* information ($\varepsilon_{MC-Truth}$) can be used:

$$\text{Fractional Efficiency Difference} = \frac{\varepsilon_{\text{“tag and probe''}} - \varepsilon_{MC-Truth}}{\varepsilon_{MC-Truth}}. \tag{3}$$

To reduce kinematic bias, the *MC-Truth* events have been required to satisfy the same kinematic cuts as the tag and probe events, although some bias can remain from the detector resolution. To reduce this, and

estimate just the bias from the tag and probe method itself, reconstructed quantities ($E_T$ and $\eta$) are used to determine the acceptance. Non-reconstruction of a true electron is not a problem here, as the trigger efficiency is always measured relative to some level of offline reconstruction.

| Trigger Level | *w.r.t.* loose (%) (truth) | *w.r.t.* medium (%) (truth) | *w.r.t.* tight(%) (truth) |
|---|---|---|---|
| L1 | 99.995 ± 0.005 (99.994 ± 0.002) | 99.995 ± 0.005 (99.995 ± 0.002) | 99.997 ± 0.005 (99.998 ± 0.001) |
| L2 | 98.74 ± 0.07 (98.67 ± 0.03) | 99.59 ± 0.04 (99.54 ± 0.02) | 99.68 ± 0.04 (99.62 ± 0.02) |
| EF | 98.66 ± 0.07 (98.59 ± 0.03) | 99.15 ± 0.06 (99.12 ± 0.02) | 99.96 ± 0.01 (99.96 ± 0.06) |
| L1 + L2 + EF | 97.41 ± 0.09 (97.28 ± 0.04) | 98.74 ± 0.07 (98.65 ± 0.03) | 99.63 ± 0.04 (99.57 ± 0.02) |

Table 13: Single object tag and probe efficiencies for the e10 selection used in the 2e10 signature, with comparison to efficiencies derived using truth information from simulation. Efficiencies are given with respect to the previous trigger level(s), as well as for the whole trigger (rows) and the given offline electron identification selection (columns). The errors given are statistical only. Errors on tag and probe values are scaled up, to correspond to 50 $pb^{-1}$of data. For this table, the invariant mass cut is $70 < m_{ee} < 100$ GeV. A signal sample of $Z^0 \to ee$ Monte Carlo simulation without background was used to obtain these results.

## 6.2 Efficiency extraction for early running

Early running at ATLAS will be vital for understanding the performance of the detector as well as the trigger and the offline reconstruction. There will be several single object triggers with low thresholds and loose selections that will be impossible to use later on in the experiment. Table 13 shows example results for the e10 trigger selection used to build the 2e10 signature. The single object efficiency is shown, as the tag electron has been required to pass the tighter single electron trigger e10. Both trigger signatures are listed in Table 3. Table 13 displays the relative trigger efficiency for each trigger level as well as the overall one, comparing results obtained from tag and probe method and from simulation truth information, that show a good agreement within the statistical error. Figure 11 shows these results differentially, along with the fractional efficiency difference defined in Equation 3. This shows good agreement between the two methods, and that the loose e10 trigger has high efficiency for $Z^0 \to ee$ electrons. The results presented in this Section have been obtained from a sample of $Z^0 \to ee$ Monte Carlo simulation without adding possible background. Studies performed adding Monte Carlo simulated background have shown that the presence of background could degrade the resolution of the method to a few percent level. The use of a low $E_T$ threshold and the loose electron identification cuts on the probe offline selection means that backgrounds will be significantly higher than in standard data-taking at higher luminosities, in which tighter selection cuts would be applied. The $E_T$ low threshold will, however, mean that background will be present in the invariant mass distribution of the two electrons for values smaller than the $Z$ mass, allowing both sidebands to be used for background subtraction below the $Z$ peak. Even with higher $E_T$ thresholds, where only one sideband can be used, it is possible to extract the correct signal with a small additional systematic error [10].

The e20 signature foreseen for early running phase has looser electron identification cuts than the other single electron signature present in this menu, e10. Even though no isolation to other electromagnetic or hadronic activity in the calorimeters is required, the threshold of the transverse energy is

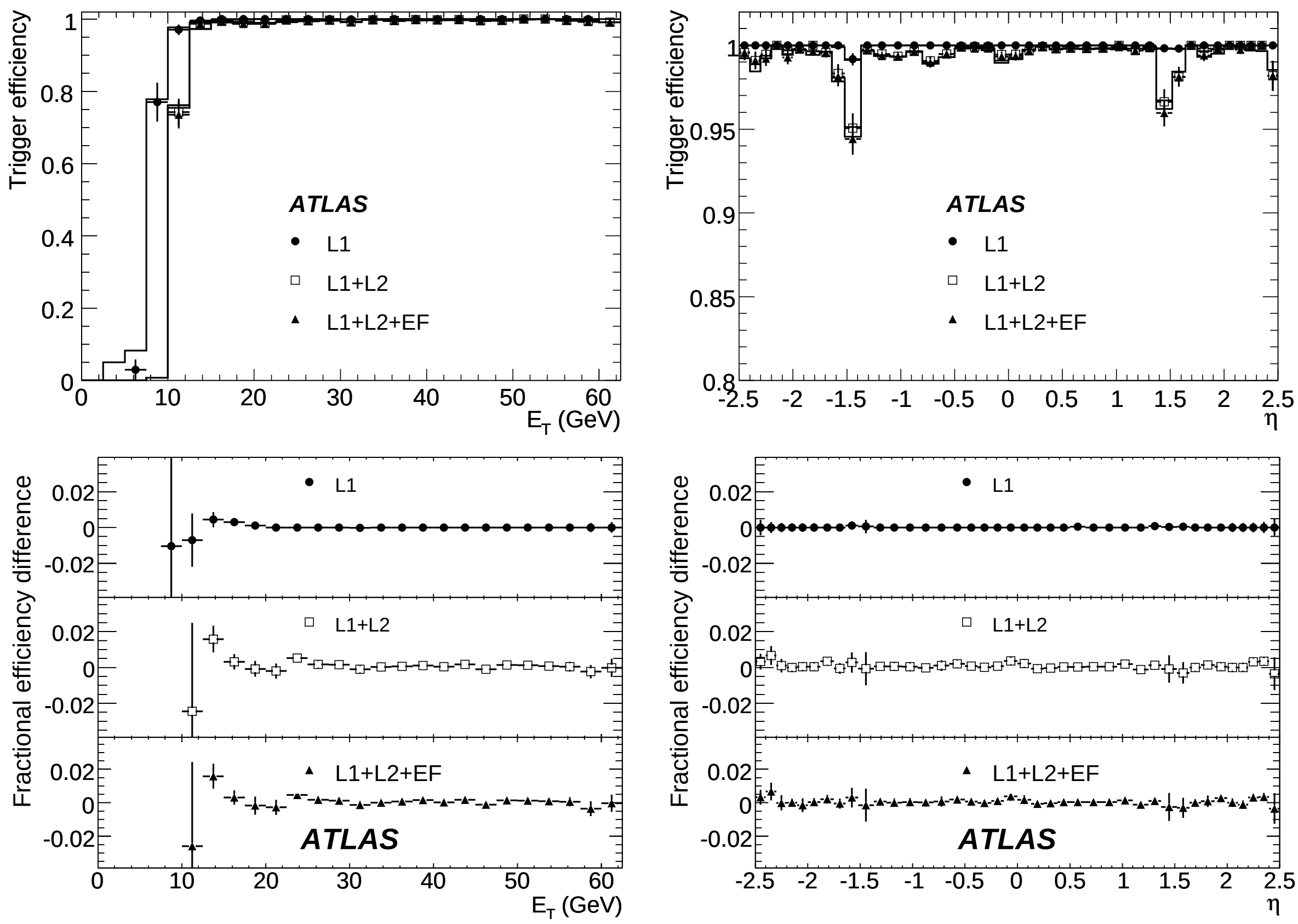


Figure 11: Single object tag and probe efficiencies for the e10 selection of the 2e10 trigger signature. The efficiencies shown are relative to a tight offline electron identification selection as described in [6], as a function of the reconstructed $E_T$(left) and $\eta$ (right). The tag and probe method (points) is compared to MC truth (lines). The lower two plots show the fractional efficiency difference (see Equation 3) between the two. The number of events used corresponds to 50 pb$^{-1}$. For this figure, the invariant mass cut is $70 < m_{ee} < 100$ GeV. Errors are statistical only. A signal sample of $Z^0 \rightarrow ee$ Monte Carlo simulation without background was used to obtain these results.

sufficient to sustain a manageable rate. The trigger efficiencies for the different trigger levels with respect to the offline electron identification selections for the probe electron are shown in Table 14, both for the "tag and probe" method and with respect to *MC-Truth*, showing good agreement within the statistical error, and high trigger efficiencies. In Fig. 12 the trigger efficiency for e20 signature as a function of $E_T$ and $\eta$ is presented. It can be seen that the L2 efficiency is close to 100% with respect to the L1 efficiency already at very low transverse energy, the EF is almost 100% efficient with respect to L2. The optimization for this trigger item made to minimize the loss of efficiency in these areas, especially not losing efficiency in the HLT with respect to L1 at large transverse energy. In the bottom part of the same figure the fractional efficiency difference between the "tag and probe" and *MC-Truth* method as defined in Eq. 3 is shown, good agreement between both methods is observed.

| Trigger Level | *w.r.t.* loose (%) (truth) | *w.r.t.* medium (%) (truth) | *w.r.t.* tight(%) (truth) |
|---|---|---|---|
| L1 | 99.74 ± 0.03 (99.84 ± 0.01) | 99.83 ± 0.03 (99.93 ± 0.01) | 99.88 ± 0.02 (99.95 ± 0.01) |
| L2 | 98.55 ± 0.07 (98.48 ± 0.03) | 99.48 ± 0.04 (99.44 ± 0.02) | 99.58 ± 0.04 (99.53 ± 0.02) |
| EF | 98.67 ± 0.07 (98.60 ± 0.03) | 99.16 ± 0.06 (99.13 ± 0.02) | 99.96 ± 0.01 (99.959 ± 0.06) |
| L1 + L2 + EF | 97.0 ± 0.01 (96.95 ± 0.04) | 98.48 ± 0.08 (98.51 ± 0.03) | 99.41 ± 0.05 (99.44 ± 0.02) |

Table 14: Tag and probe trigger efficiencies for the e20 signature, with comparison to efficiencies derived using truth information from simulation. Efficiencies are given with respect to the previous trigger level(s), as well as for the whole trigger (rows) and the given offline electron identification selection (columns). The errors given are statistical only. Errors on tag and probe values are scaled up, to correspond to 50 $pb^{-1}$of data. For this table, the invariant mass cut is $70 < m_{ee} < 100$ GeV. A signal sample of $Z^0 \rightarrow ee$ Monte Carlo simulation without background was used to obtain these results.

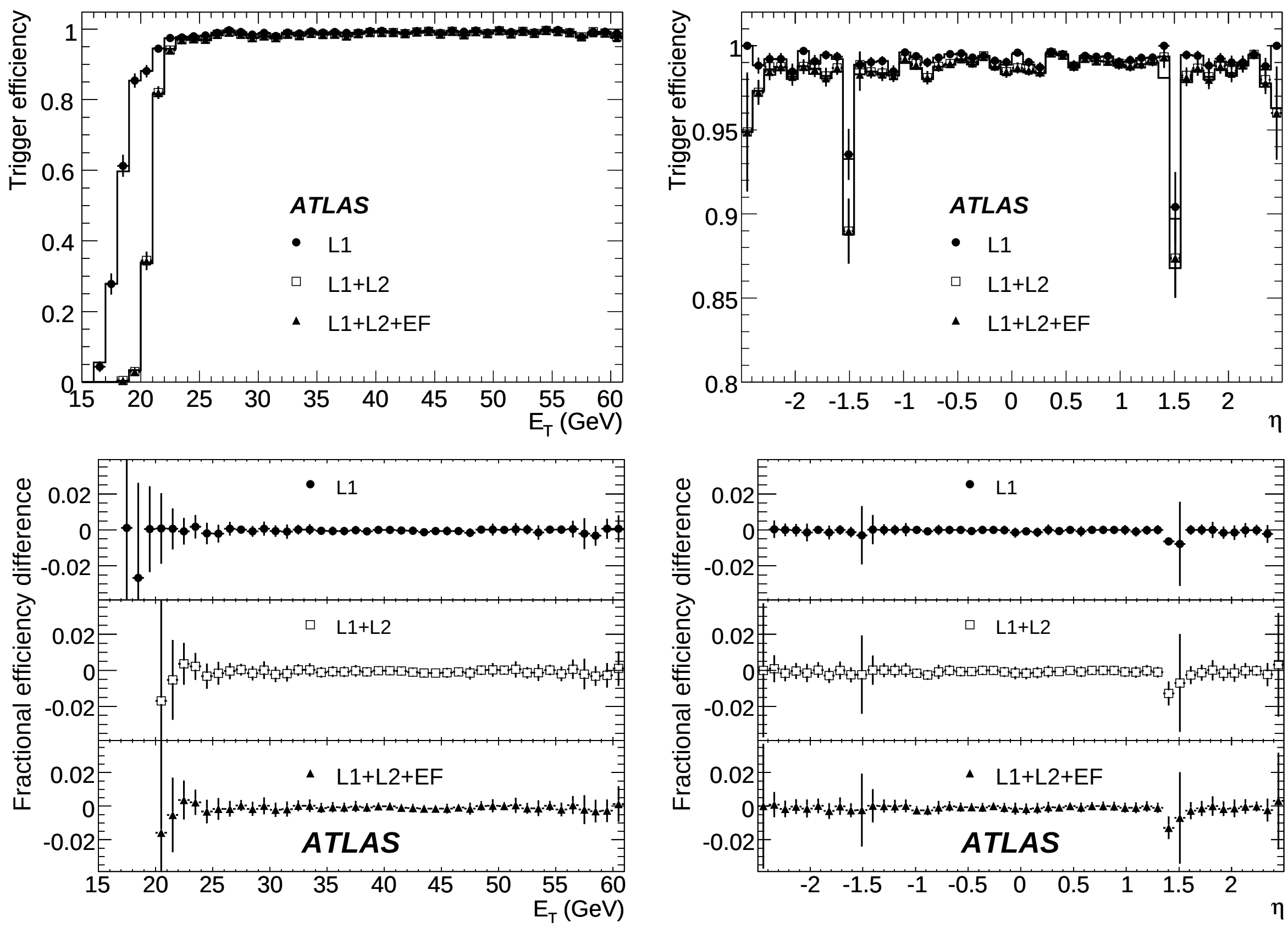


Figure 12: Trigger efficiency from the "tag and probe" method with $Z \rightarrow ee$ for the e20 trigger signature. The efficiencies are shown *w.r.t.* a tight offline electron selection as described in [6], as a function of the reconstructed $E_T$ (left) and the reconstructed $\eta$ (right). The "tag and probe" method (points) is compared with the *MC-Truth* method (solid line) for all three trigger levels, L1 (solid circles), L2 (open triangles), and the EF (solid squares), The fractional efficiency difference (see Eq. 3) between the "tag and probe" and *MC-Truth* methods is shown in the bottom two plots. The number of $Z \rightarrow ee$ events used corresponds to $100\,\text{pb}^{-1}$. Errors are statistical only. A signal sample of $Z^0 \rightarrow ee$ Monte Carlo simulation without background was used to obtain these results.

## 7 Summary

An electron and photon trigger baseline for LHC commissioning and the first physics run have been presented. Many studies and tests have been summarized, without finding any problems that could compromise the successful start-up of ATLAS data-taking. Electron and photon trigger performance have been studied in detail in a wide energy range using single electron and single photon simulated samples. Trigger efficiency dependencies on transverse energy ($E_T$) and pseudo-rapidity ($\eta$) have been observed and studies are ongoing to explore the possibility to minimize them. Trigger efficiencies for the e/$\gamma$ selection have been evaluated for simulations of different physics samples such as $H \to \gamma\gamma$, $Z \to ee$, $W \to e\nu$ and some exotic channels. The electron and photon trigger menus for these selections have proven to be efficient above the threshold of the corresponding transverse energy cut with respect to offline reconstruction. The corresponding background rates have been estimated to be within the allocated bandwidth. It should be noted that due to current uncertainty in the cross-sections used for background estimations, rates could vary significantly. This could require the use of tighter selections with corresponding loss of signal efficiency. Some examples of backup signatures for such situations have been shown. Trigger robustness studies have been started, and are currently being extended. Trigger reconstruction and selection efficiencies do not show significant drops with respect to offline reconstructed electrons and photons. A method of trigger efficiency determination from data using $Z \to ee$ decays has been studied in depth. This results compare well to trigger efficiency computed with respect to Monte Carlo truth. The extension of this method to other channels and samples are being developed.

# Performance of the Muon Trigger Slice with Simulated Data

**Abstract**
The overall functionality and performance of the muon trigger system with respect to data produced as part of the ATLAS Computing System Commissioning effort is described. The physics performance in terms of trigger efficiency and accepted rates is studied for the muon inclusive signatures for different luminosity scenarios. Dedicated studies on physics samples with single and double muon final states are also performed in order to evaluate the trigger efficiencies on realistic data and background rejection capabilities. Methods to evaluate muon trigger efficiencies from real data are discussed. Furthermore, strategies to use the ATLAS calorimeters to tag and select isolated muons are presented.

## 1 Introduction

Triggering and identifying muons will be crucial for many LHC physics analyses. In accordance with the ATLAS general trigger scheme, the muon trigger system has three distinct levels: L1, L2, and the Event Filter (EF). The paper discusses the software tools used for muon trigger reconstruction and the algorithm selection strategy and trigger configuration. Next, the resolution and selection efficiencies of the various muon triggers are presented, followed by a discussion of the trigger rates for various luminosities. Subsequently, the rejection of background from in-flight meson decays and selection of isolated muons using calorimeter information is discussed. Finally, the trigger performance on the di−muon final states $Z \rightarrow \mu\mu$and $Z' \rightarrow \mu\mu$ is presented along with a description of plans to determine the trigger efficiency from collider data.

## 2 Detector simulation and data samples

The samples used in this paper were produced using a full GEANT4 based simulation of the ATLAS detector. The trigger simulation options included both standard and $B$-physics trigger simulation configurations, which correspond to the standard and the low trigger thresholds (see Section 3). The deterioration of efficiency due to the geometrical acceptance and the limited size of coincidence window are taken into account in the L1 simulation and trigger logic emulator.

Large samples of single prompt muons, simulated uniformly in $\eta - \phi$, with fixed $p_T$ ranging from 2 GeV to 1 TeV, have been used to study the muon trigger performance. One of the main backgrounds for the muon trigger selection comes from in-flight decays of charged kaons and pions. This has been evaluated using samples of minimum bias events and single pions, where the mesons are forced to decay inside the Inner Detector cavity in order to facilitate the production of a sizable sample of in-flight $\pi/K$ decays. Muon trigger rates were determined using both single muons and minimum bias events. The selection of muons using the Tile Calorimeter has been studied using low $p_T$ (4 and 6 GeV) single muons and semi-inclusive $b$ quark decays, $b\bar{b} \rightarrow \mu X$. Muon trigger studies on high-$p_T$ dimuon final states and the determination of trigger efficiency from data have been performed using $Z \rightarrow \mu\mu$and $Z' \rightarrow \mu\mu$ as signal processes and $B \rightarrow \mu\mu$, W boson decays, $Z \rightarrow \tau\tau$ and top-pair events as background processes.

## 3 Muon trigger algorithms and configuration

The L1 muon trigger selects active RoIs, in the event using Resistive Plate Chambers (RPC) [1] in the barrel ($|\eta| < 1.05$) and Thin Gap Chambers (TGC) [1] in the endcaps ($1.05 < |\eta| < 2.4$). The

trigger algorithms look for hit coincidences within different RPC or TGC detector layers inside the programmed geometrical windows which define the transverse momentum region. A coincidence is required in both $\eta$ and $\phi$ projections. The information about muon candidates in both the barrel and the end-cap is transmitted to the Muon to Central Trigger Processor Interface (MuCTPI) [1], which calculates the number of L1 muon candidates for six different $p_T$ thresholds and takes overlaps between the trigger sectors into account by using look-up-tables (LUT). There are several L1 signatures each corresponding to a different $p_T$ threshold:

- mu0, mu5, mu6, mu8, mu10 for the low $p_T$ selection;
- mu11, mu20, mu40 for the high $p_T$ selection.

The integer numbers after the “mu” symbolize the required $p_T$ threshold. L1 also provides the coordinates in $\eta$ and $\phi$ of the selected RoIs. The mu0 threshold represents a L1 configuration with completely open coincidence windows; it is also called the “Cosmic” threshold as it can be used to trigger on cosmic rays during the detector commissioning phase and between the LHC fills. Similar thresholds, labeled with “muXX”, have been defined for L2 and EF.

The muon HLT runs L2 and EF algorithms. It starts from the RoI delivered by the L1 trigger and applies trigger decisions in a series of steps, each refining the existing measurement by acquiring additional information from the ATLAS detectors. A list of physics signatures, implemented in the event reconstruction and selection algorithms, are used to build signature and sequence tables for all HLT steps. This stepwise and seeded processing of events is controlled by the trigger steering. The reconstruction progresses by calling feature extraction algorithms. These typically request detector data from within the RoI and attempt to identify muon features. Subsequently, a hypothesis algorithm determines whether the identified feature meets the criteria necessary to continue. The decision to reject the event or continue is based on the validity of signatures, taking into account prescale and pass-through factors. Thus, events can be rejected after an intermediate step if no signatures remain viable.

The main algorithm of the muon L2 system, muFast, runs on full granularity data within the RoI defined by L1. An optimized strategy is used to avoid heavy calculations and access to external services to reduce the execution time of the algorithm. After pattern recognition driven by the trigger hits which selects Monitored Drift Tubes (MDT) regions crossed by the muon track, a track fit is performed using MDT drift time precision measurements. The $p_T$evaluation is performed using LUT. Reconstructed tracks in the Inner Detector can be combined with the tracks found by muFast by a fast track combination algorithm called muComb.

The L2 algorithm (muIso) is used to discriminate between isolated and non-isolated muon candidates by examining energy depositions in the electromagnetic and hadronic calorimeters. The algorithm is seeded by muons selected by muFast or muComb and decodes LAr and Tile Calorimeter quantities in cones centered around the muon direction. For the muon selection two different concentric cones are defined: an internal cone chosen to contain the energy deposit deposited by the muon itself, and an external cone, containing energy only from detector noise, pile-up and jet particles.

A strategy for tagging muons at L2 in the TileCal is implemented in the TileMuId algorithm. It can provide additional redundancy and robustness to the muon trigger, as well as enhance the efficiency in the low $p_{\mathrm{T}}$ region. The search starts from the outermost calorimeter layer, which contains the cleanest signals, and once a deposited energy is compatible with a muon, the algorithm checks the energy deposition in the neighboring cells for the internal layers. Candidates are considered tagged muons when muon compatible cells are found following a $\eta$-projective pattern in all the three TileCal layers. There are two different variants of this algorithm : one (TrigTileLookForMuAlg) is fully executed on the L2 Processing Unit (L2PU) while the other (TrigTileRODMuAlg) has a core part executed on the Readout Driver (ROD).

The EF accesses the full event with its full granularity. Given the larger admissible latency, it is possible to adapt algorithms developed for the off-line reconstruction to the on-line framework, an approach that minimises algorithm development. The EF processing starts by reconstructing tracks in the Muon Spectrometer around the muons found by L2 and is done by three instances of the EF algorithm; the first instance reconstructs tracks inside the Muon Spectrometer, starting with a search for regions of activity within the detector, and subsequently performing pattern recognition and full track fitting. The second step extrapolates muon tracks to their production point. Finally the information from the first two steps is combined with the reconstructed tracks from the Inner Detector.

The hypothesis algorithms define a set of HLT trigger thresholds by applying cuts on the $p_T$ of the muon candidate. The muon trigger efficiency is defined as

$$\frac{\text{The number of events with a triggered muon}}{\text{The number of events with a muon}} \tag{1}$$

The effective trigger thresholds are obtained in such a way that at the nominal threshold value the efficiency is 90% of the corresponding efficiency without cuts. For this reason effective thresholds are slightly lower than nominal thresholds.

# 4 L1 performance

## 4.1 Barrel muon trigger performance

As mentioned earlier, studies of muon trigger performance were conducted using simulated samples of single muon events generated over a large $p_T$ and angular range. L1 selection algorithms show an efficiency greater than 99% for muons with $p_T$ above threshold. The overall acceptance (82% low-$p_T$, 78% high-$p_T$) is due exclusively to geometrical regions of the Muon Spectrometer not covered by the RPC. Figure 1 shows the inefficient regions corresponding to the magnet support structures ($-2.3 \leq \phi \leq -1.7$ and $-1.4 \leq \phi \leq 0.9$) and the spectrometer central crack at $\eta \sim 0$, not covered by RPCs. The overall loss in geometrical acceptance due to the $\eta = 0$ crack is approximately 7% while the loss due to the support structure is about 5%. Moreover smaller inefficiency patterns are clearly visible which are due to magnetic ribs in small trigger sectors. The geometrical acceptance effects are visible also in Fig. 2 where the L1 efficiency above threshold is shown with respect to $\eta$ and $\phi$.

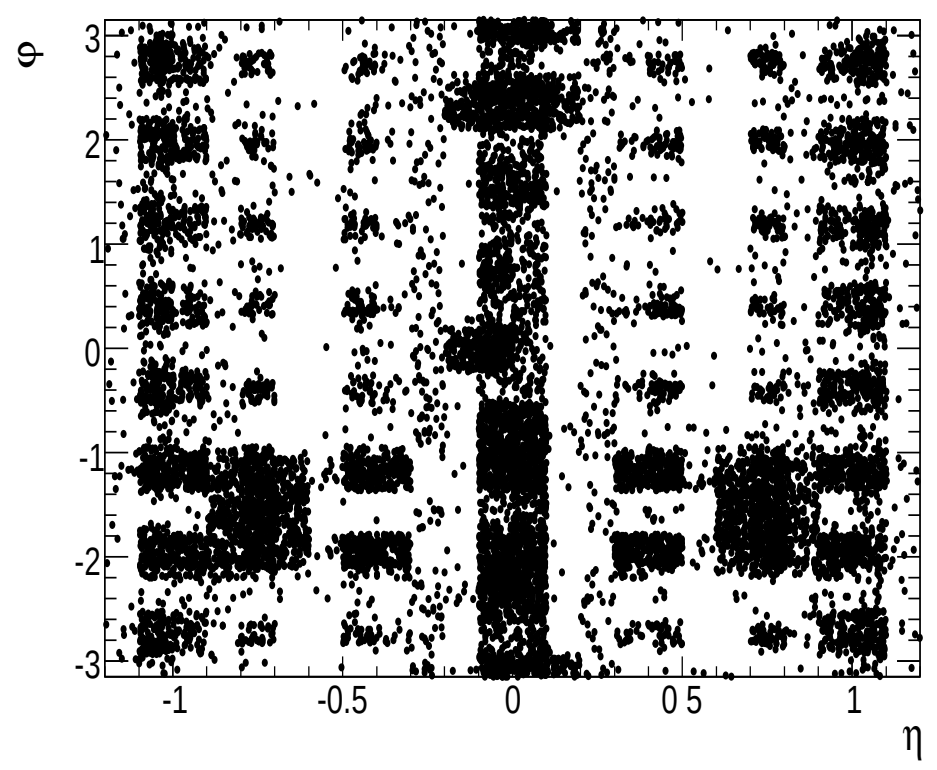


Figure 1: L1 geometrical acceptance in the $\eta$-$\phi$ plane.

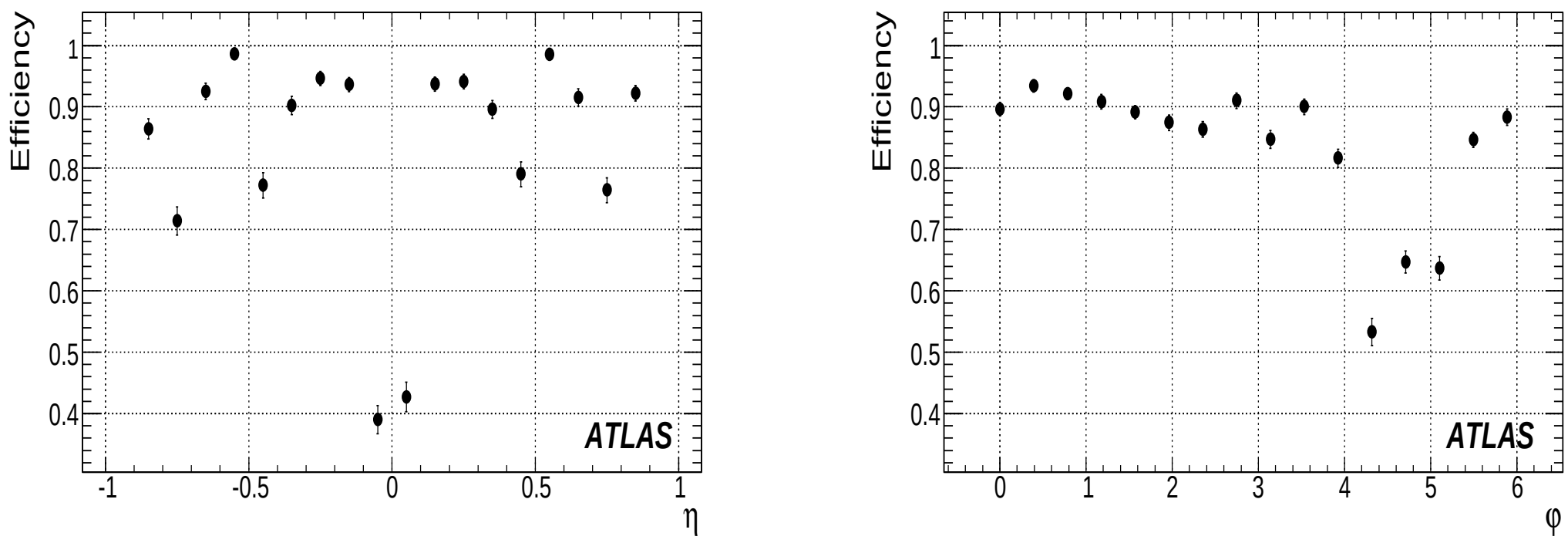


Figure 2: $\eta$ and $\phi$ dependence of barrel trigger efficiency for single muons with a $p_T$=75 GeV.

| Threshold | Plateau Efficiency | Effective Threshold (GeV) | Sharpness(GeV) |
|---|---|---|---|
| mu6 | 0.82 | 5.3 | 2.2 |
| mu8 | 0.82 | 6.1 | 1.9 |
| mu10 | 0.82 | 6.7 | 2.2 |
| mu11 | 0.78 | 10.9 | 3.7 |
| mu20 | 0.78 | 15.3 | 7.1 |
| mu40 | 0.78 | 27.8 | 19.7 |

Table 1: Plateau efficiencies, effective thresholds, and sharpness for L1 signatures. Sharpness is defined as the difference of $p_T$ corresponding to 90% and 10% of the plateau efficiency.

Figure 3 shows turn-on curves for low-$p_T$ and high-$p_T$ thresholds; the efficiencies at plateau and effective thresholds are summarized in Table 1.

Muon tracks are deflected in the $r-\eta$ plane under the action of the toroidal magnetic field. Their trajectories are symmetrical under charge exchange and reflection with respect to the plane $z=0$, but the layout of the Muon Spectrometer is not. This asymmetry, could, in principle, produce a bias in the trigger efficiency calculation. From the single muon data sample, it was found that for muons with $p_T$ greater than the L1 threshold the asymmetry in the efficiency is quite small ($< 1\%$).

Particular attention was devoted to the study of muons with $2 < p_T < 3.5$ GeV in the barrel region ($|\eta| <1.05$). Given the large inclusive cross-section with muons in the final state, this very low-$p_T$ region represents the most significant contribution to the total expected muon rate. Table 2 shows the fraction of these events that produce hits in the RPCs and the L1 barrel efficiency for passing the mu6 trigger. Most of the low $p_T$muons that pass L1 have $|\eta| \simeq 1$.

| Muon $p_T$ (GeV) | Percentage of events with hits in RPC | L1 Efficiency |
|---|---|---|
| 2 | 0.15% | $(1.4 \pm 0.1) \cdot 10^{-3}$ |
| 2.5 | 0.35% | $(3.4 \pm 0.1) \cdot 10^{-3}$ |
| 3 | 0.48% | $(4.8 \pm 0.1) \cdot 10^{-3}$ |
| 3.5 | 3.36% | $(33.4 \pm 0.3) \cdot 10^{-3}$ |

Table 2: L1 RPC efficiency for very low-$p_T$ single muon events with $|\eta| <1.05$.

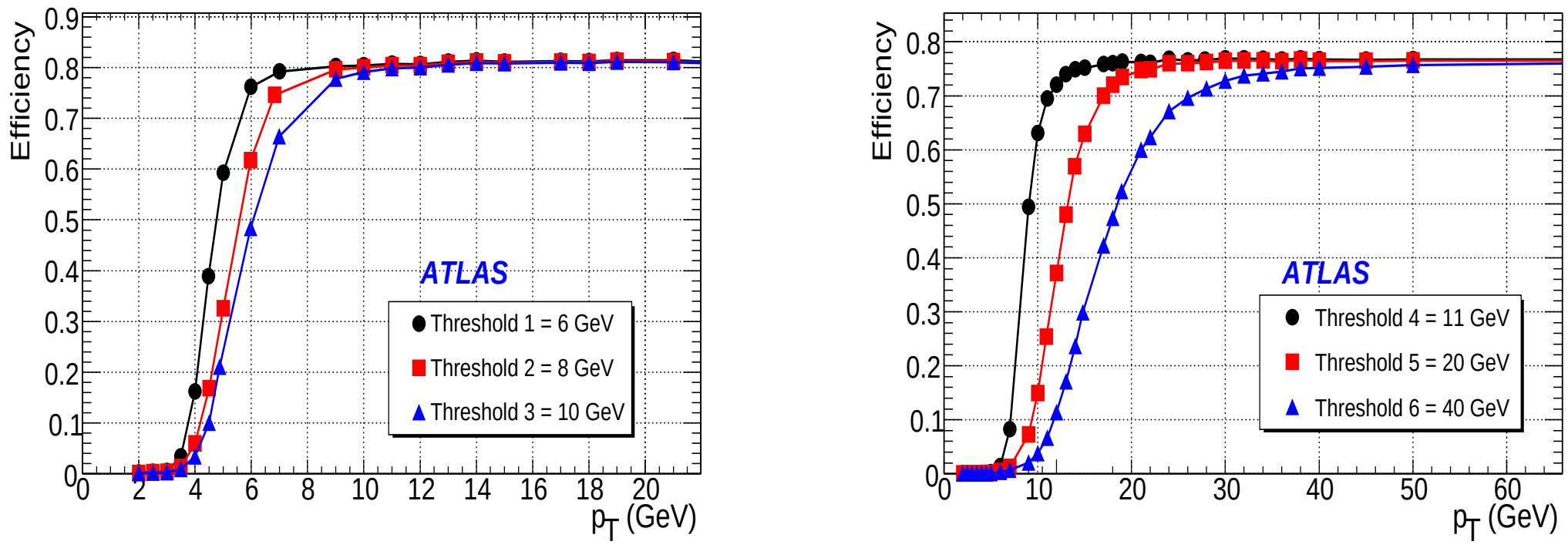


Figure 3: L1 barrel efficiency as a function of $p_T$ for low-pt (left) and high-pt (right) thresholds.

Very low-$p_T$ muons produced in the acceptance of the TGC ($|\eta| > 1.05$) sometimes give hits in the RPC ($|\eta| < 1.05$) because they are strongly deflected by the magnetic field. The fraction of muons with $|\eta| > 1.05$ that give RPC hits ranges from 46% at $p_T = 2$ GeV to about 10% at 3.5 GeV, and is negligible above 4 GeV. The overall efficiency for such muons to pass the mu6 trigger is about $10^{-3}$.

Figure 4 shows L1 end-cap efficiency curves for low $p_T$ (left) and high $p_T$ thresholds (right). Efficiencies at the threshold and plateau are summarized in Table 3. The efficiency of mu6 at threshold is 77%, relatively lower than other cases. This is due to the limited window-size of the three-station coincidence for muons having $p_T$ of 6 GeV. The $\eta$ dependence of the mu6 and mu20 efficiency are

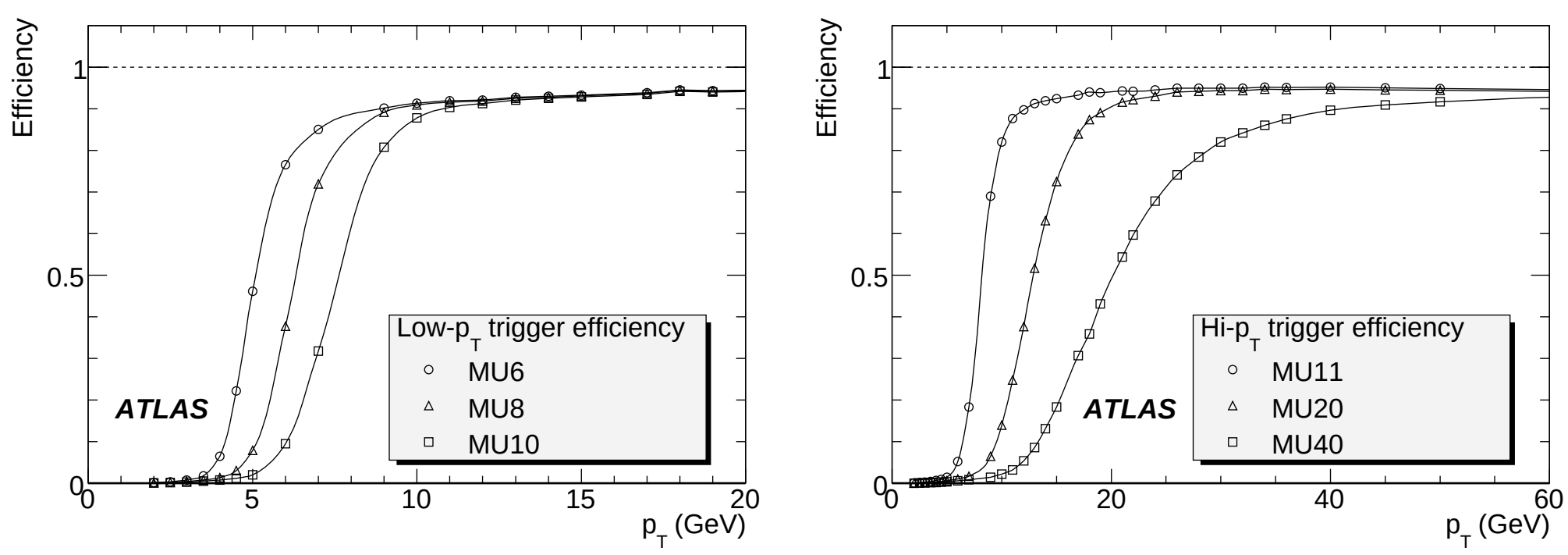


Figure 4: The end-cap trigger efficiency curves for each $p_T$ threshold. The left plot shows the low-$p_T$ thresholds of 6, 8, and 10 GeV and the right plot shows the high-$p_T$ thresholds of 11, 20 and 40 GeV.

| $p_T$ threshold (GeV) | 6 | 8 | 10 | 11 | 20 | 40 |
|---|---|---|---|---|---|---|
| Threshold | 77% | 84% | 88% | 88% | 92% | 90% |
| Plateau | 95% | 95% | 95% | 95% | 94% | 93% |

Table 3: Trigger efficiencies at threshold and plateau for various muon $p_T$ thresholds.

shown in Fig. 5 (at threshold) and Fig. 6 (at plateau) with respect to the sign of charge of muon $q \times \eta$. Because the two muon end-cap stations are made as mirror images, $\mu^{-(+)}$ with $\eta > (<)\ 0$ behaves the same as $\mu^{+(-)}$ with $\eta < (>)\ 0$. The difference of efficiency between the two signs of $q \times \eta$ is large at the geometrical boundary for muons near the $p_T = 6$ threshold as shown in Fig. 5. One more point

worth noting is the dip at $\eta$=2 for the mu6 signature. Muons in the dip region pass through chambers which belong to different trigger sectors, consequently the requirement of a three-station coincidence is not satisfied and trigger efficiency is reduced. Figure 7 shows the $\phi$ dependence of mu6 (left) and mu20 (right) trigger efficiencies at the plateau and at threshold. The effect of octant symmetry of the magnetic field is seen in the plot of the mu6 efficiency for threshold muons. For the mu6 signature at plateau and for the mu20 signature, this effect is not observed, resulting in an approximately uniform efficiency.

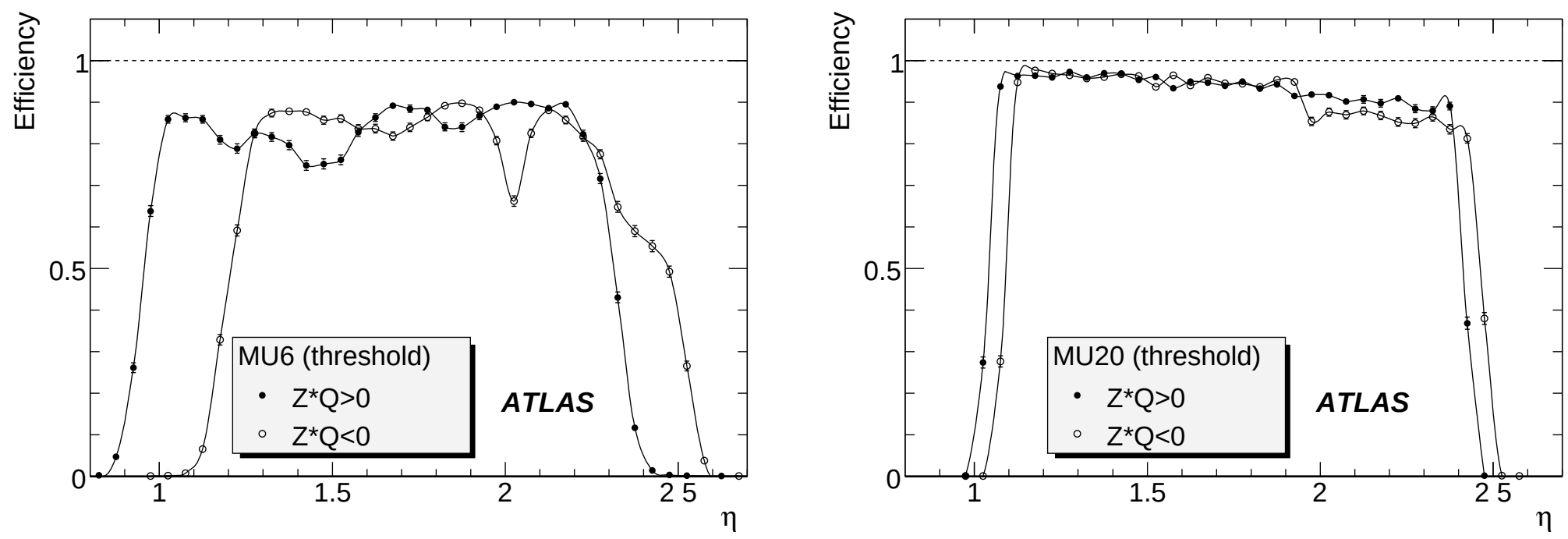


Figure 5: $\eta$ dependence of end-cap trigger efficiency for mu6 (left) and mu20 (right). The solid circles represent $q \times \eta > 0$, the open circles represent $q \times \eta < 0$.

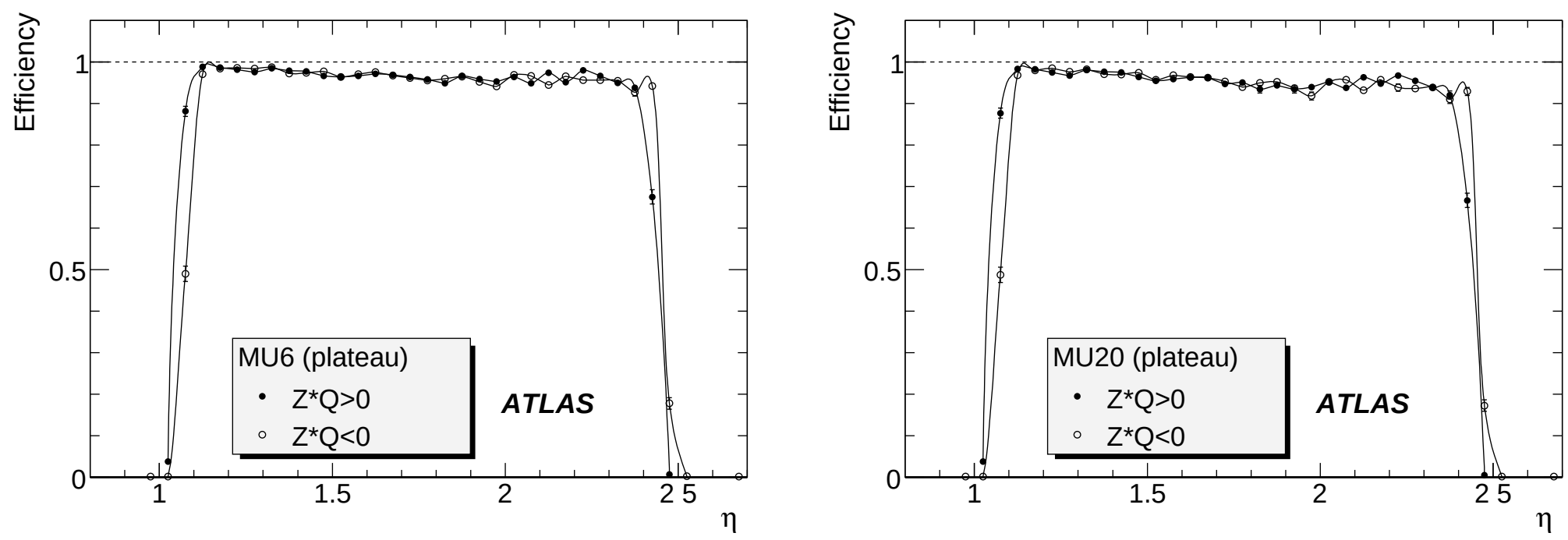


Figure 6: $\eta$ dependency of end-cap trigger efficiency at plateau for mu6 (left) and mu20 (right). The solid circles represent $q \times \eta > 0$, the open circles represent $q \times \eta < 0$.

## 5 Performance of L2 muon algorithms

As described in Section 2, algorithm performance is evaluated on samples of single muons generated with different transverse momenta. The resolution of the inverse of the measured momentum with respect to the generated transverse momentum is studied. Due to the non-uniform magnetic field in the Muon Spectrometer it is divided into four regions according to the pseudorapidity of the muon candidate: the Barrel region with $|\eta| < 1.05$, and three end-cap regions with $1.05 < |\eta| < 1.5$, $1.5 < |\eta| < 2.0$, and $2.0 < |\eta| < 2.4$.

For the Muon Spectrometer standalone reconstruction (muFast), the resolution of inverse $p_T$ as a function of the muon transverse momentum is shown in Fig. 8(left). The degradation in the resolution

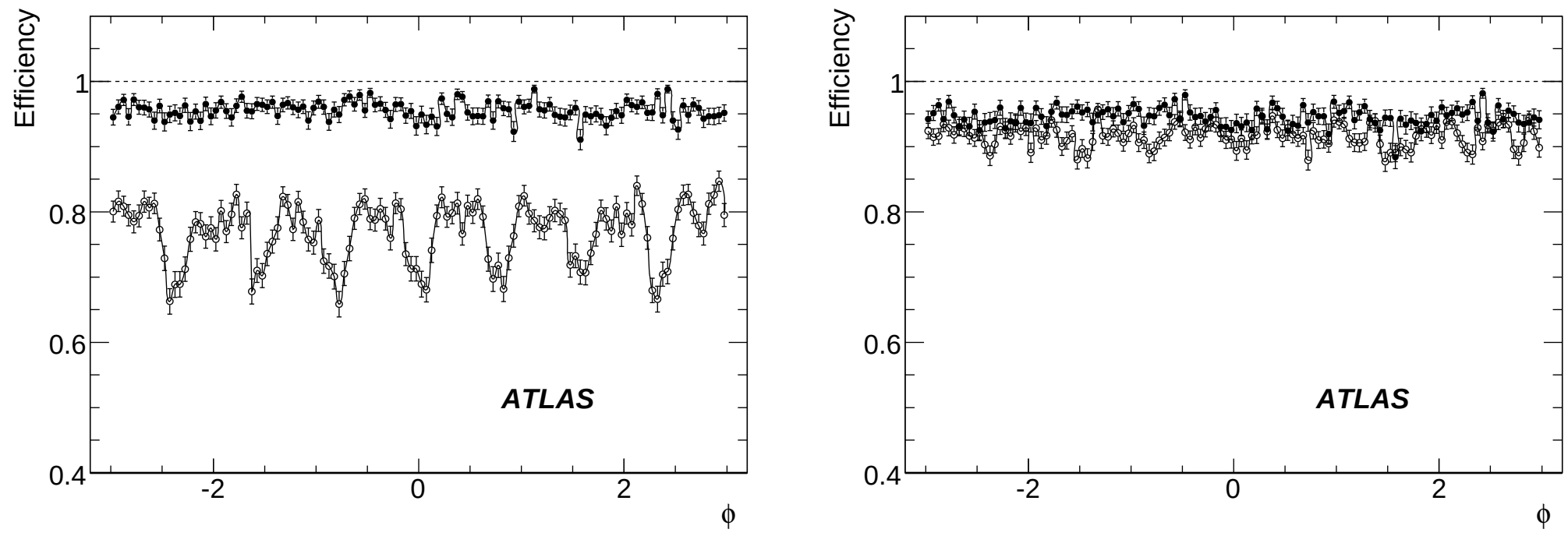


Figure 7: $\phi$ dependence of end-cap trigger efficiency for mu6 (left) and mu20 (right) for muons with $p_T = 45$ GeV. The open circles show the efficiency at threshold and the solid circles show the efficiency at plateau.

with respect to previous results [2] is caused by the realistic geometry misalignment introduced in the muon simulation. Resolution as function of $\eta$ and $\phi_{Loc}$[1] is shown in Fig. 8(right). The degradation of the resolution in the endcap regions is evident. The muFast efficiency with respect to L1 selection as a

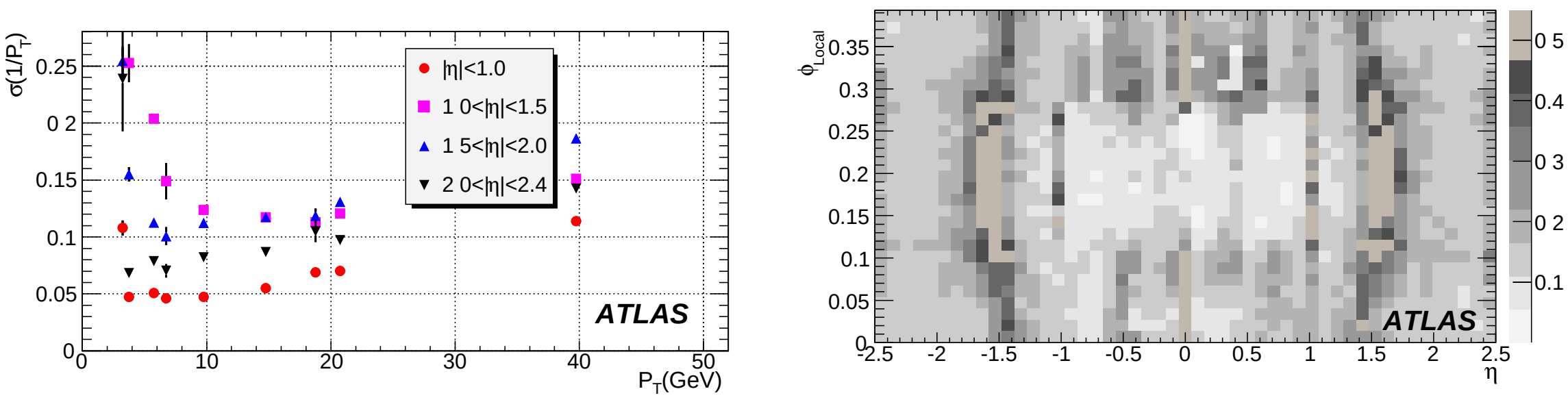


Figure 8: $1/p_T$ resolution (Muon Spectrometer StandAlone) as a function of $p_T$ (left) and $\eta - \phi_{Loc}$ (right).

function of muon momentum are shown for mu4, mu6, and mu20 selections in Fig. 9. The low rejection at small momentum ($2.5 < p_T < 4.5$ GeV), in particular in the barrel region, is caused by candidate tracks not pointing to the nominal interaction vertex due to large scattering angles.

Efficiencies of the MS standalone reconstruction (muFast) (mu6 trigger selection) with respect to L1 selection as a function of $\eta$ and $\phi_{Loc}$ for muons with $P_T = 6$ GeV are shown in Fig. 10.

The combination of a Muon Spectrometer standalone muon candidate with an Inner Detector track found by the L2 tracking algorithms is performed by muComb. For muons with $p_T < 50$ GeV the Inner Detector measurement has a better resolution than the Muon Spectrometer standalone measurement. Therefore, the combination of the two measurements gives better resolutions in the low-$p_T$ range. Figure 11 shows the $1/p_T$ resolution as a function of $p_T$ and also the resolution as function of $\eta$ and $\phi_{Loc}$. The muComb efficiency with respect to muFast as a function of $p_T$ for mu4, mu6, and mu20 is shown in Fig. 12. The problem of low rejection for low-$p_T$ muons is partially solved when the Muon Spectrometer candidates are combined with Inner Detector tracks.

[1] $\phi_{Loc}$ is the azimuthal angle folded up in $[0, \pi/16]$ such to cover half of an odd MS sector ($[0, \pi/32]$) and half of an even MS sector $[\pi/32, \pi/16]$.

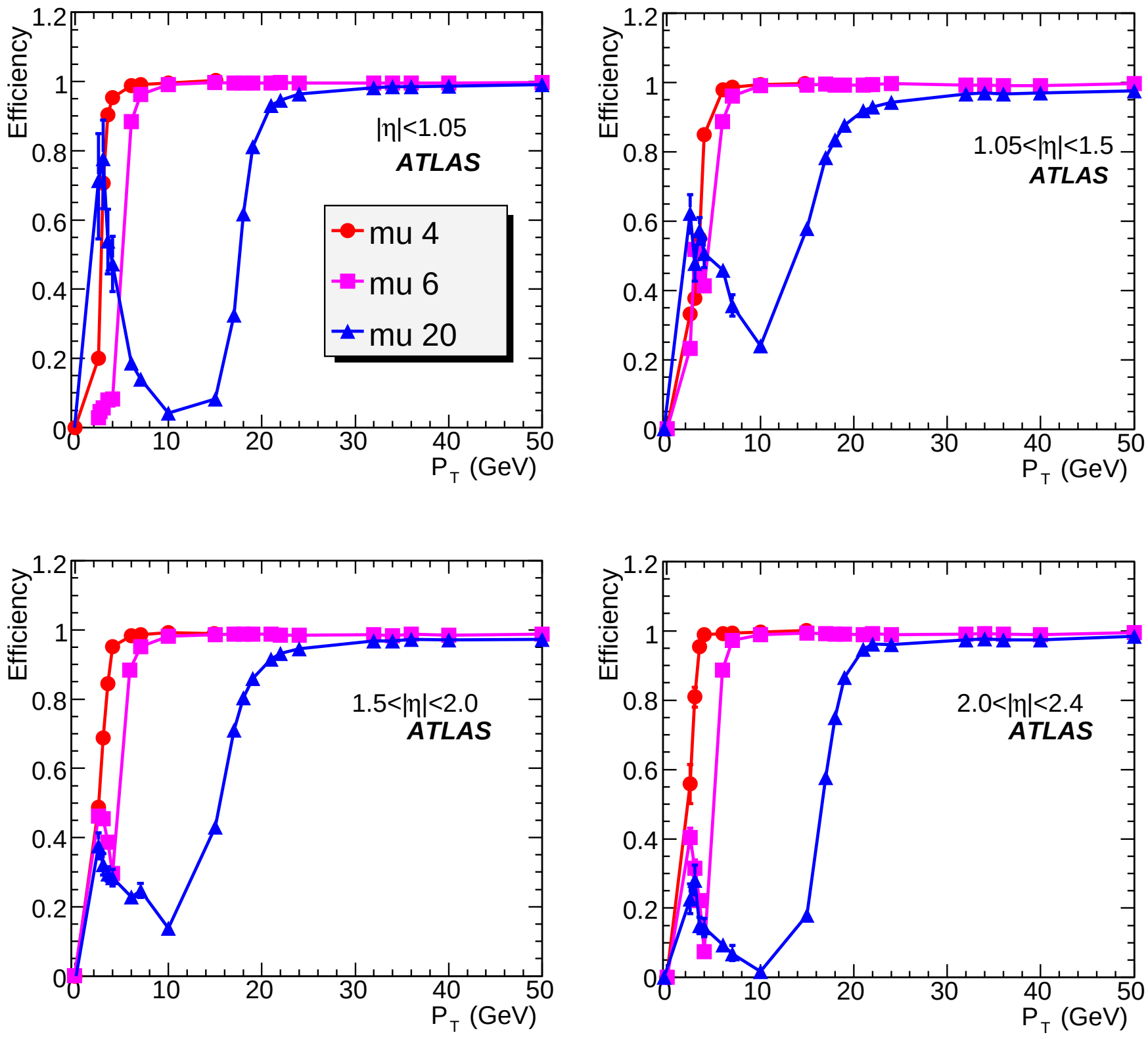


Figure 9: muFast efficiency with respect to L1 selection for the mu4, mu6, and mu20 triggers for different $\eta$ regions.

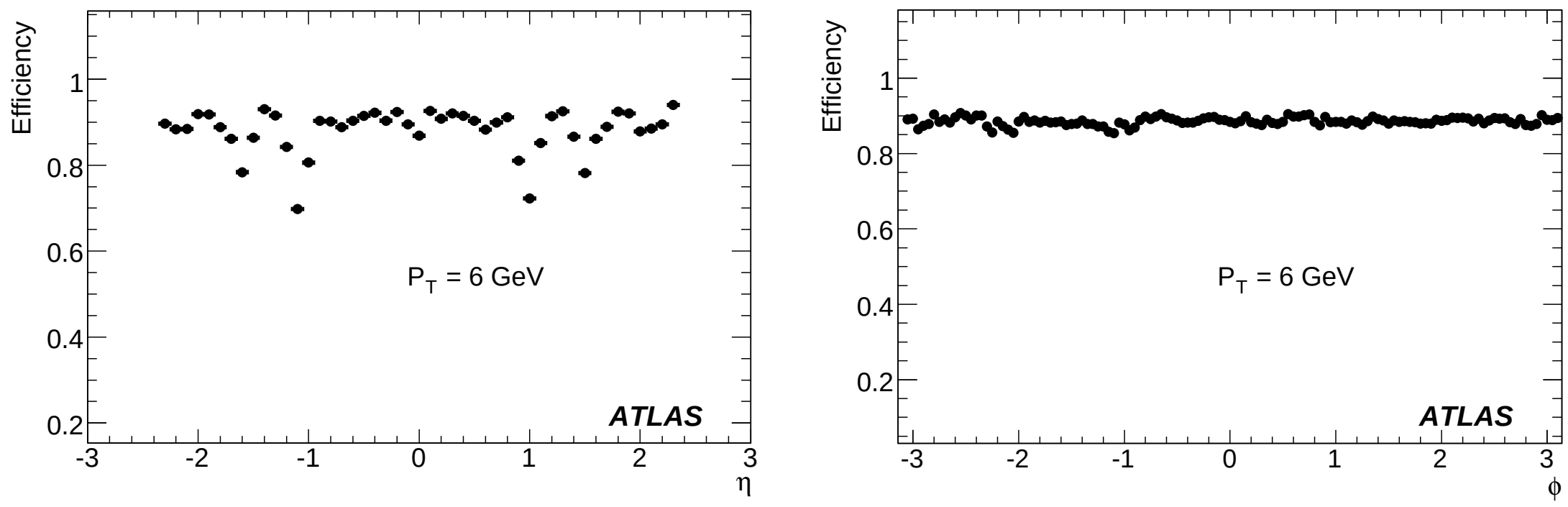


Figure 10: Efficiency of muFast as a function of $\eta$ (right) and $\phi_{Loc}$ (left) for muons with $p_T = 6$ GeV.

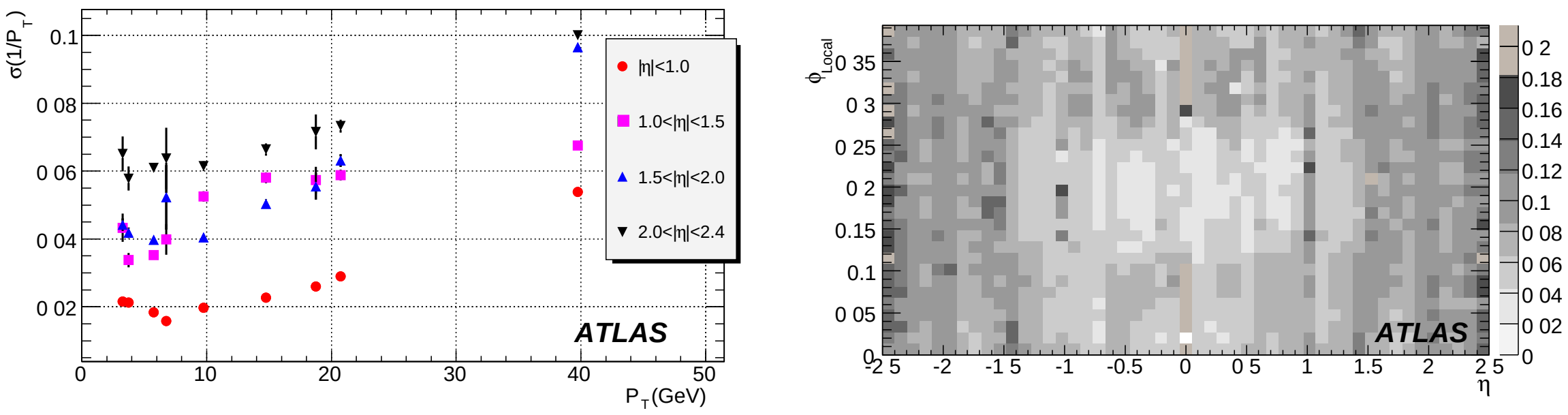


Figure 11: The muon combined $1/p_T$ resolution as a function of $p_T$ (left) and as a function of $\eta - \phi_{Loc}$ (right).

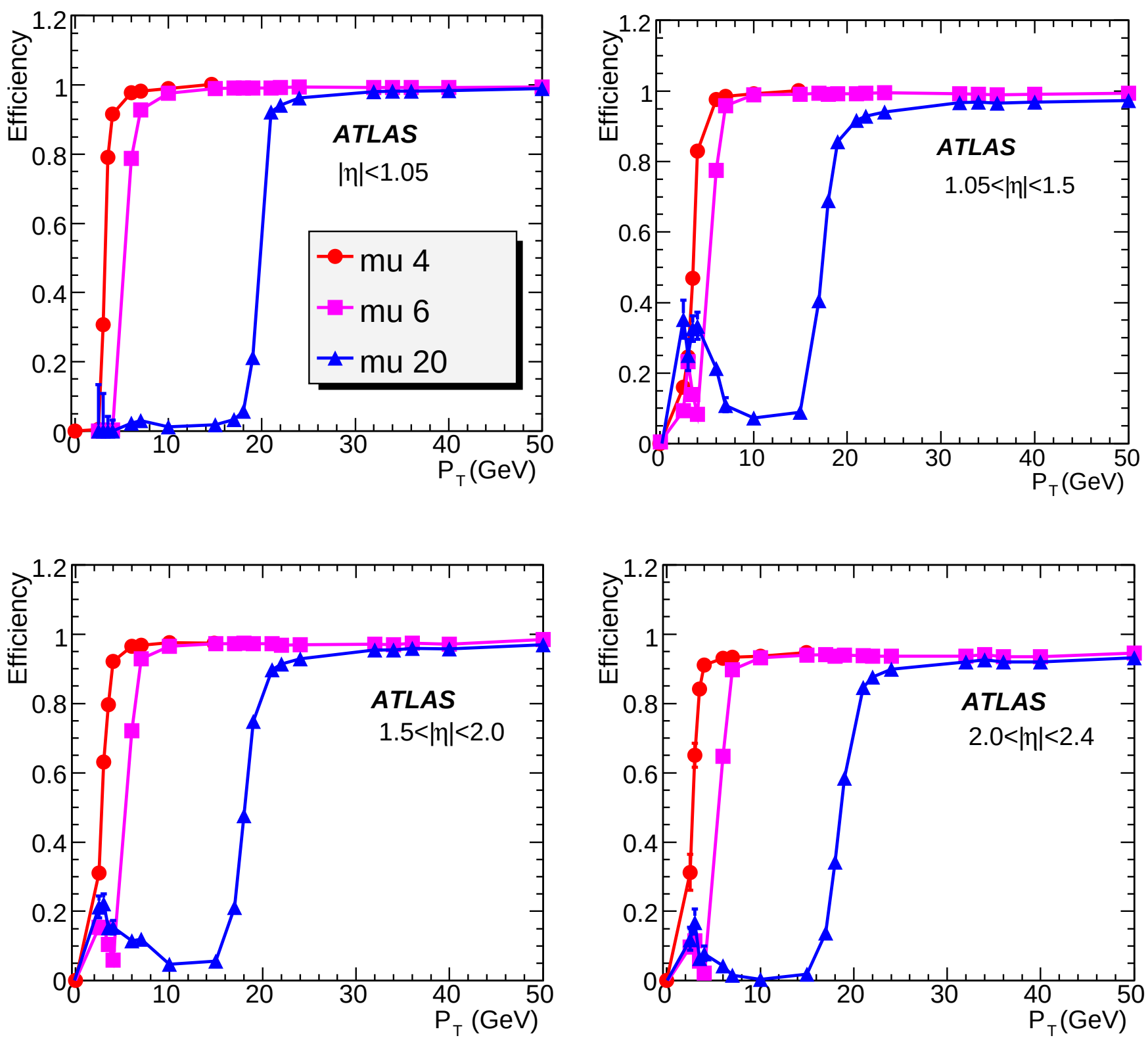


Figure 12: The muComb algorithm efficiency with respect to mu4, mu6, and mu20 triggers for the different $\eta$ regions.

Efficiencies of muComb (mu6 trigger selection) with respect to muFast selection as a function of $\eta$ and $\phi_{Loc}$ for muons with $p_T = 6$ GeV are shown in Fig. 13.

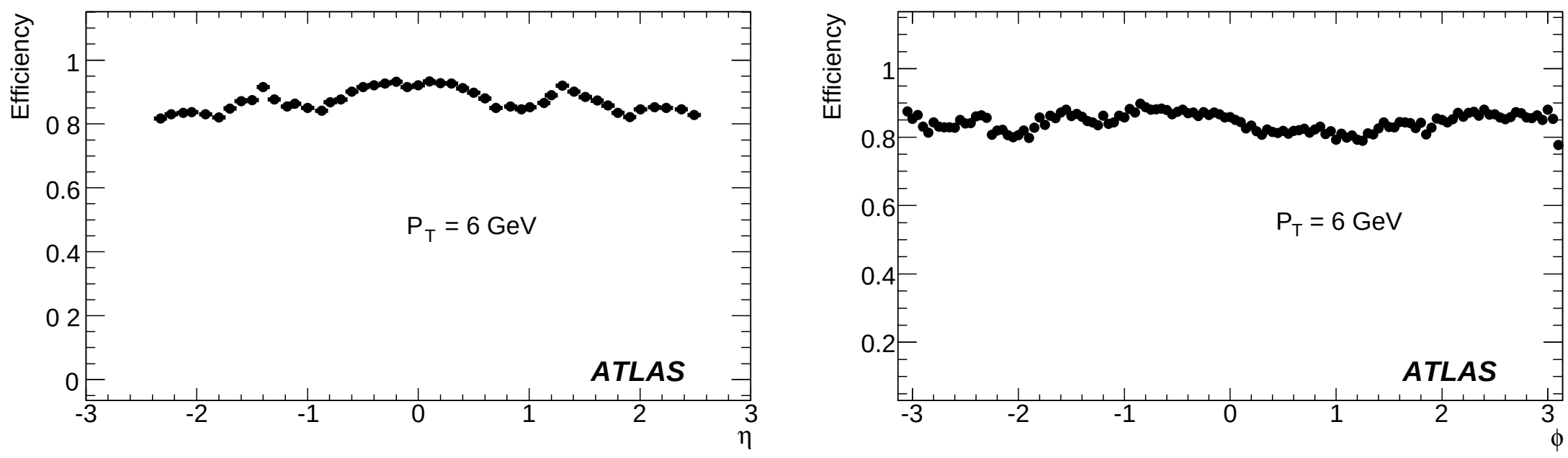


Figure 13: Efficiency of muComb as a function of $\eta$ (right) and $\phi_{Loc}$ (left) for single muons with $p_T = 6$ GeV.

# 6 Event filter performance

The full reconstruction in the Muon EF has been executed on the simulated samples described in Section 2. The reconstruction in the Muon Spectrometer is carried out by the MOORE algorithm, the extrapolation to the vertex of the muon track found in the Muon Spectrometer is performed by the MuId standalone algorithm and the combination of the tracks found in the Muon Spectrometer and in the Inner Detector by the MuId Combined algorithm.

Efficiency for single muon events is defined as the ratio of events with a reconstructed track at the EF after the execution of each reconstruction step to all events which have passed L1 and L2. The efficiency with respect to L2 as a function of muon $p_T$ is shown in Fig. 14 for all three EF algorithms. The efficiency is defined on an event basis, and counts only once events having L2 muon-feature or EF

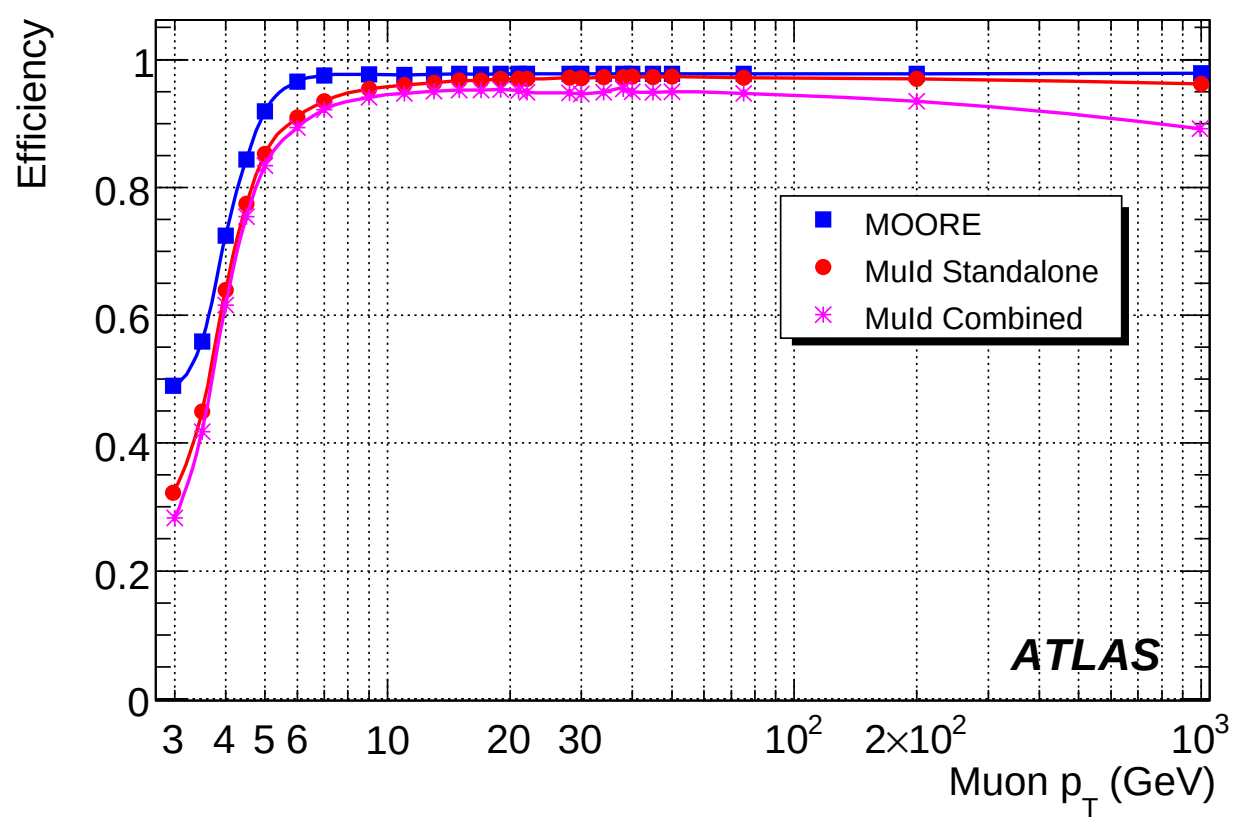


Figure 14: Efficiency as a function of muon $p_T$ for MOORE, MuId Standalone and MuId Combined.

track multiplicity greater than 1. According to this definition, the efficiency to trigger an event with more than one muon is expected to be higher with respect to what is estimated here.

The efficiencies are lower for $3 < p_T < 6$ due to multiple scattering and energy loss fluctuation effects. Moreover, in the case of MuId Combined, at very high $p_T$ the increasing probability of muon showering is responsible for a small loss in efficiency. The efficiencies as a function of $\eta$ and $\phi$ show a structure, especially at low momentum, explainable with some residual dependence in $\eta$ and $\phi$ on the Muon Spectrometer geometrical acceptance and on the magnetic field inhomogeneities which affect less previous levels. It can be seen in Fig. 15 where the efficiency is shown for 6 GeV muons. In Fig. 16

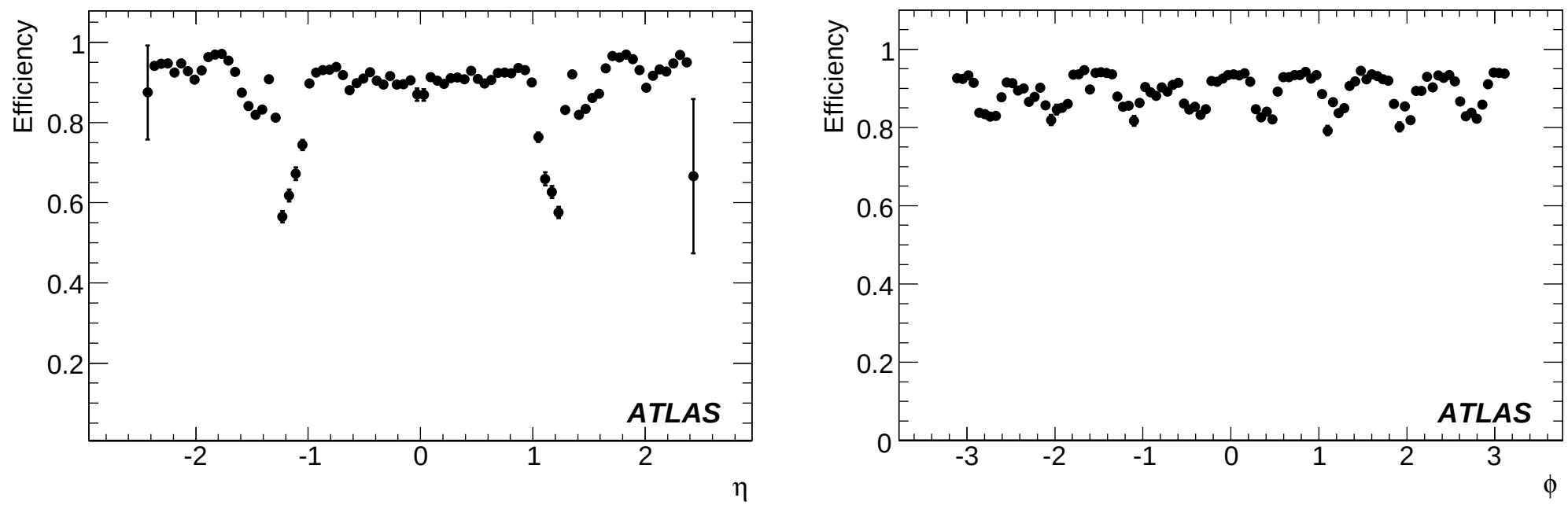


Figure 15: Efficiency of MuId Combined as a function of $\eta$ (left) and of $\phi$ (right) for 6 GeV muons.

the MuId combined efficiency with respect to L2 for different thresholds is shown on the left, while the overall trigger efficiency (L1 + L2 + EF) with respect to the generated muons is shown on the right. All efficiency values are averaged over the whole $|\eta| < 2.4$ range.

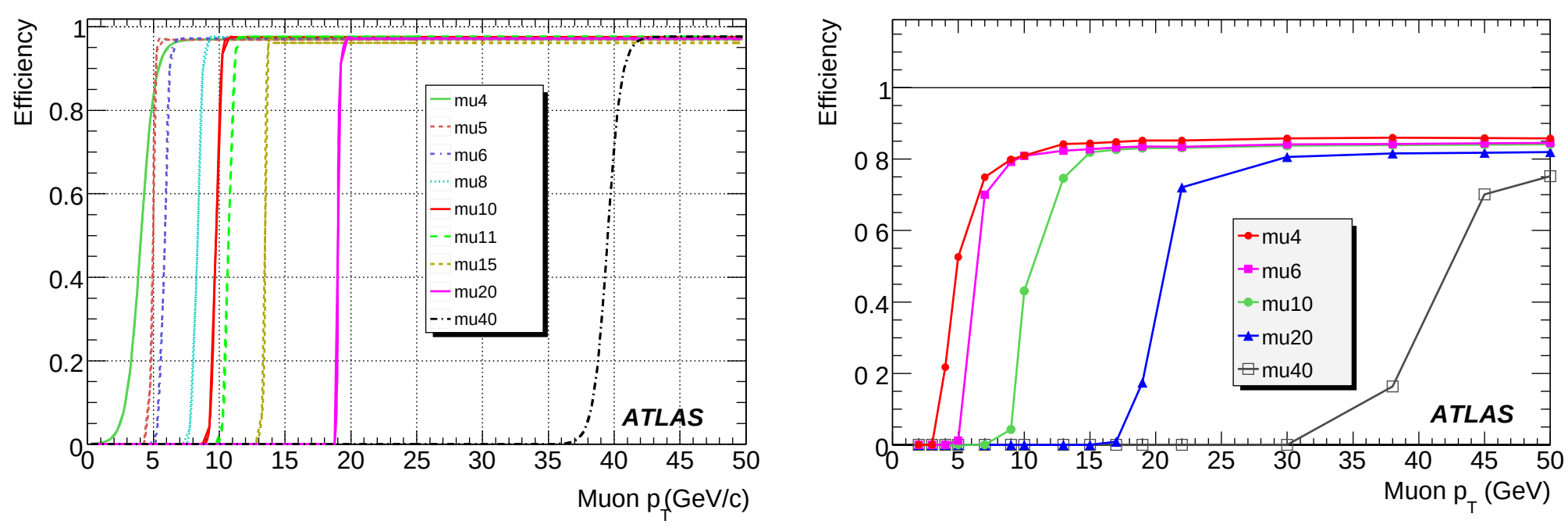


Figure 16: MuId combined efficiencies for various $p_T$ thresholds with respect to L2 (Left) and with respect to generated truth muons (Right).

In Fig. 17, the $1/p_T$ resolution is shown as a function of muon $p_T$ for all EF algorithms. For a muon with $p_T$ below 50 GeV the Inner Detector dominates the reconstruction precision so the combination of measurements greatly improves the resolutions. For a muon with $p_T$ above 100 GeV, the Muon System dominates the measurement of the muon combined transverse momentum. $1/pt$ resolution as a function of $\eta$ is shown in Fig. 18 for 20 GeV muons. The worsening of the resolution in the region $1.0 < |\eta| < 1.5$ can be attributed to the highly inhomogeneous magnetic field in the transition regions of the Muon Spectrometer. This effect is recovered by means of the combined reconstruction which exploits the Inner Detector performance.

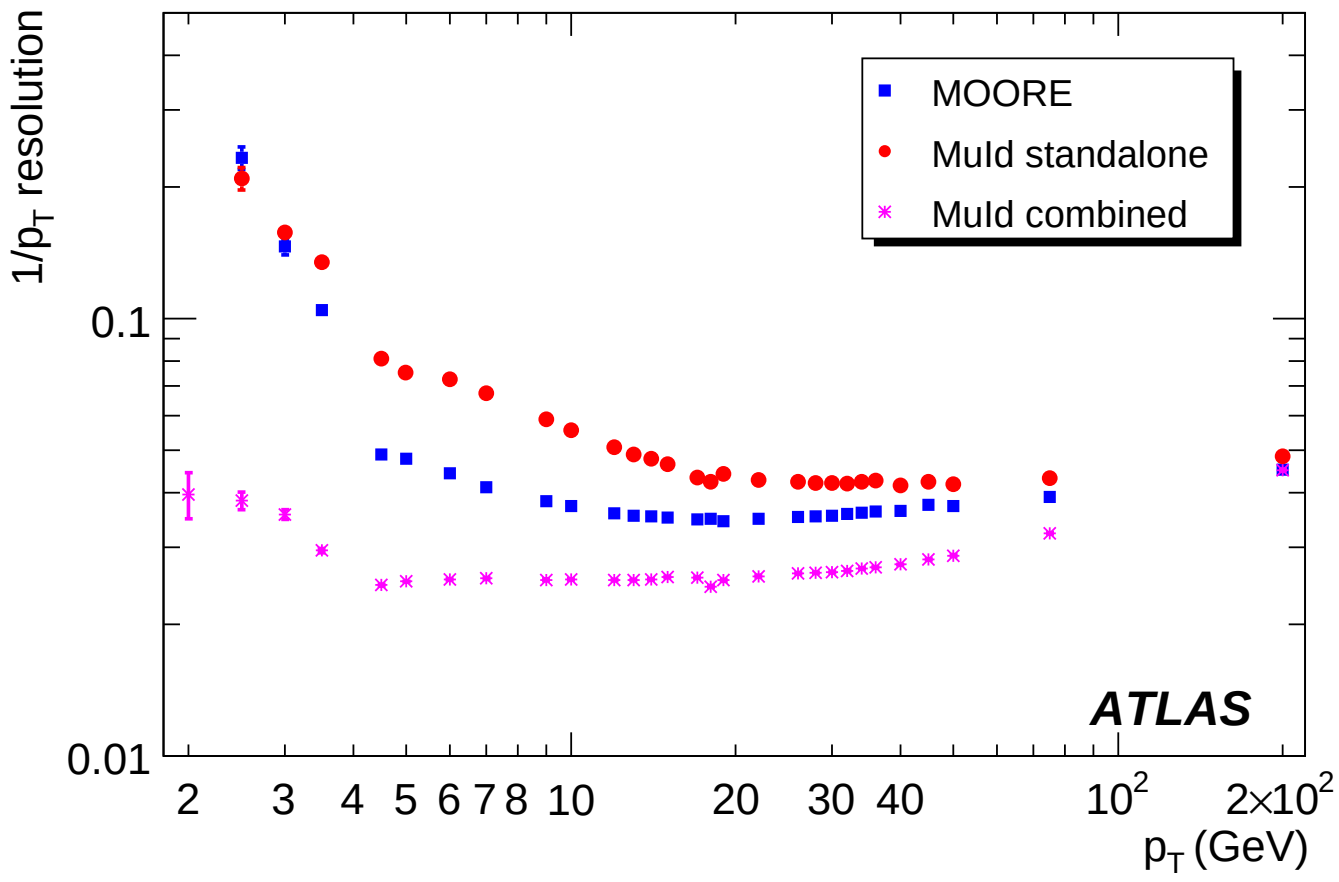


Figure 17: $1/p_T$ resolution as a function of $p_T$ for MOORE, MuId Standalone and MuId Combined.

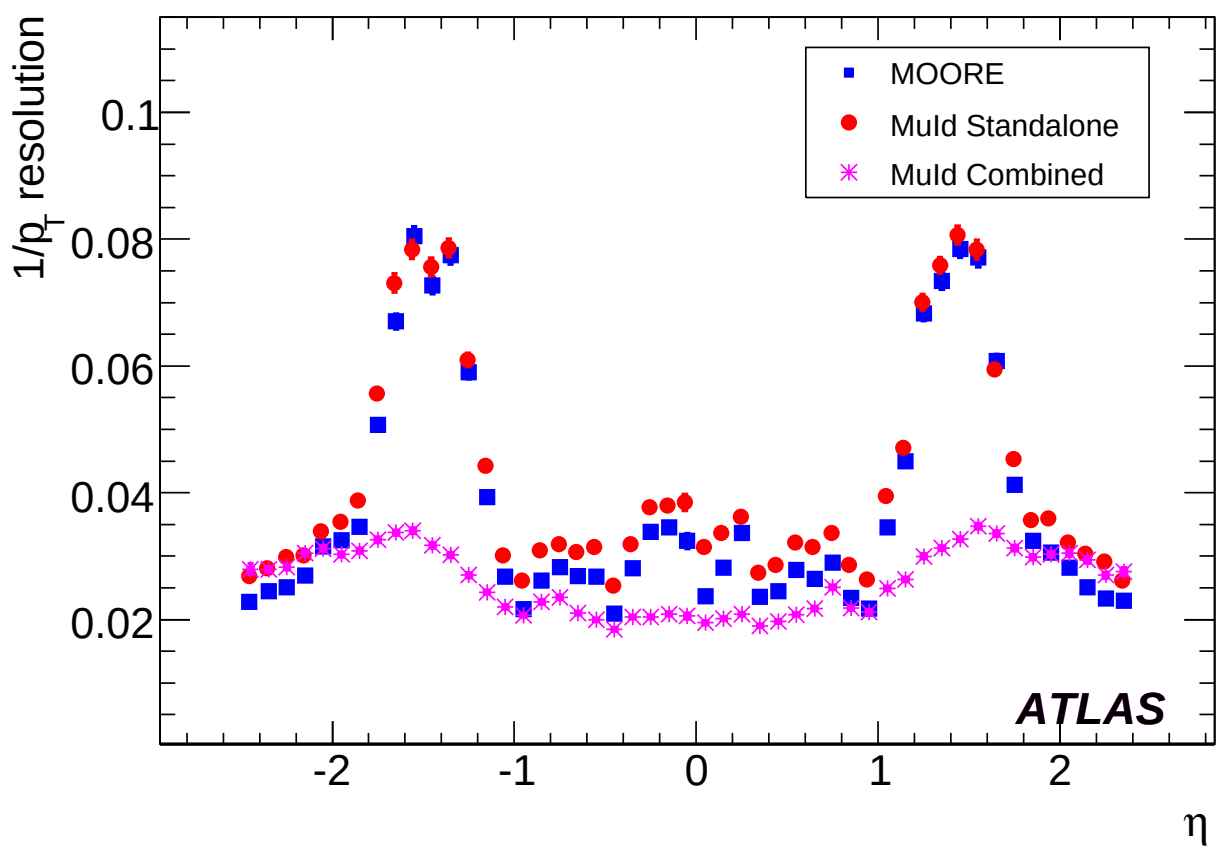


Figure 18: $1/p_T$ resolution for MuId Combined as a function of $\eta$ for muons with $p_T = 20$ GeV.

## 7 Muon trigger rates

The trigger rates for single muon event originating from all the physical processes expected in ATLAS were obtained using the EF MuiD combined algorithm (the events must of course also pass the L1 and L2 muon algorithms). Various luminosity scenarios expected during LHC operation (from $\mathscr{L} = 10^{31}$ cm$^{-2}$ s$^{-1}$ to $10^{34}$ cm$^{-2}$ s$^{-1}$) were considered. Trigger rates were typically computed by convolving, over a given $p_T$ range, the estimated efficiencies with the cross-sections of processes representing the main muon sources at LHC. For the i-th process with cross-section $\sigma_i$, the rate is

$$R_i = \mathscr{L} \int \frac{d\sigma_i}{dp_T} \varepsilon(p_T) dp_T \tag{2}$$

where $\mathscr{L}$ is the instantaneous luminosity and $\varepsilon(p_T)$ is the muon trigger efficiency for a given $p_T$ value. In order to take into account the $\eta$ dependence, separate estimates for different $\eta$ regions have been considered. An 0.5 GeV step has been applied in the numerical integration. The inclusive muon cross-sections at the LHC for $b \to \mu$ and $c \to \mu$ decays have been parameterized by using PYTHIA 6.403 [3]

| L1 muon trigger rates | | | | | | |
|---|---|---|---|---|---|---|
| | $\mathscr{L} = 10^{31}$ cm$^{-2}$ s$^{-1}$ | | $\mathscr{L} = 10^{33}$ cm$^{-2}$ s$^{-1}$ | | $\mathscr{L} = 10^{34}$ cm$^{-2}$ s$^{-1}$ | |
| | Barrel (Hz) | Endcaps (Hz) | Barrel (Hz) | Endcaps (Hz) | Barrel (Hz) | Endcaps (Hz) |
| | "Cosmic" | | 6 GeV | | 20 GeV | |
| $\pi/K$ | 454 | 199 | 8600 | 5300 | 1100 | 5200 |
| beauty | 85 | 74 | 4400 | 5100 | 2500 | 3300 |
| charm | 124 | 104 | 6100 | 6900 | 2800 | 4400 |
| top | <0.1 | <0.1 | <0.1 | <0.1 | 0.3 | 0.5 |
| $W$ | <0.1 | <0.1 | 3.0 | 4.4 | 26 | 41 |
| TOTAL | 663 | 377 | 19100 | 17300 | 6400 | 12900 |
| | 5 GeV | | 8 GeV | | 40 GeV | |
| $\pi/K$ | 162 | 81 | 2200 | 3800 | 470 | 1900 |
| beauty | 54 | 53 | 2900 | 4000 | 1100 | 1300 |
| charm | 76 | 73 | 3800 | 4700 | 1200 | 1400 |
| top | <0.1 | <0.1 | <0.1 | <0.1 | 0.3 | 0.3 |
| $W$ | <0.1 | <0.1 | 4 | 4.5 | 23 | 33 |
| TOTAL | 292 | 207 | 8900 | 12500 | 2800 | 4600 |

Table 4: Single muon trigger rates at L1, for various low and high $p_T$ thresholds, at $\mathscr{L} = 10^{31}$ cm$^{-2}$ s$^{-1}$, $\mathscr{L} = 10^{33}$ cm$^{-2}$ s$^{-1}$ and $\mathscr{L} = 10^{34}$ cm$^{-2}$ s$^{-1}$.

| L2 muon standalone trigger rates | | | | | | |
|---|---|---|---|---|---|---|
| | $\mathscr{L} = 10^{31}$ cm$^{-2}$ s$^{-1}$ | | $\mathscr{L} = 10^{33}$ cm$^{-2}$ s$^{-1}$ | | $\mathscr{L} = 10^{34}$ cm$^{-2}$ s$^{-1}$ | |
| | Barrel (Hz) | Endcaps (Hz) | Barrel (Hz) | Endcaps (Hz) | Barrel (Hz) | Endcaps (Hz) |
| | 4 GeV | | 6 GeV | | 20 GeV | |
| $\pi/K$ | 190 | 140 | 4300 | 3700 | 410 | 1800 |
| beauty | 50 | 67 | 3000 | 3900 | 540 | 1500 |
| charm | 70 | 94 | 4000 | 5200 | 520 | 1700 |
| top | <0.1 | <0.1 | <0.1 | <0.1 | 0.2 | 0.4 |
| $W$ | <0.1 | <0.1 | 3 | 4 | 24 | 38 |
| TOTAL | 310 | 301 | 11300 | 12800 | 1494 | 5038 |
| | 5 GeV | | 8 GeV | | 40 GeV | |
| $\pi/K$ | 82 | 120 | 840 | 1500 | 200 | 690 |
| beauty | 37 | 59 | 1000 | 2200 | 87 | 280 |
| charm | 49 | 81 | 1300 | 2900 | 83 | 290 |
| top | <0.1 | <0.1 | <0.1 | <0.1 | 0.1 | 0.2 |
| $W$ | <0.1 | <0.1 | 3 | 4 | 17 | 23 |
| TOTAL | 168 | 260 | 3143 | 6604 | 387 | 1283 |

Table 5: Single muon trigger rates at L2 muon standalone, for various low and high $p_T$ thresholds, at $\mathscr{L} = 10^{31}$ cm$^{-2}$ s$^{-1}$, $10^{33}$ cm$^{-2}$ s$^{-1}$ and $10^{34}$ cm$^{-2}$ s$^{-1}$. The large expected rate in particularly in the endcap, caused by the relatively low rejection of low $p_T$ muons, can be reduced by improving the selection algorithm.

| L2 muon combined trigger rates | | | | | | |
|---|---|---|---|---|---|---|
| | $\mathscr{L} = 10^{31}$ cm$^{-2}$ s$^{-1}$ | | $\mathscr{L} = 10^{33}$ cm$^{-2}$ s$^{-1}$ | | $\mathscr{L} = 10^{34}$ cm$^{-2}$ s$^{-1}$ | |
| | Barrel (Hz) | Endcaps (Hz) | Barrel (Hz) | Endcaps (Hz) | Barrel (Hz) | Endcaps (Hz) |
| | 4 GeV | | 6 GeV | | 20 GeV | |
| $\pi/K$ | 130 | 124 | 3500 | 2600 | 68 | 890 |
| beauty | 48 | 66 | 2700 | 3400 | 320 | 830 |
| charm | 66 | 91 | 3800 | 4400 | 280 | 840 |
| top | <0.1 | <0.1 | <0.1 | <0.1 | 0.2 | 0.4 |
| $W$ | <0.1 | <0.1 | 3 | 4 | 22 | 35 |
| TOTAL | 244 | 281 | 10000 | 11000 | 690 | 2590 |
| | 5 GeV | | 8 GeV | | 40 GeV | |
| $\pi/K$ | 44 | 55 | 400 | 530 | 6 | 310 |
| beauty | 31 | 45 | 660 | 1100 | 31 | 92 |
| charm | 41 | 61 | 780 | 1300 | 26 | 99 |
| top | <0.1 | <0.1 | <0.1 | <0.1 | 0.1 | 0.1 |
| $W$ | <0.1 | <0.1 | 3 | 4 | 7 | 12 |
| TOTAL | 116 | 161 | 1840 | 2900 | 70 | 513 |

Table 6: Single muon trigger rates at L2 muon combined, for various low and high $p_T$ thresholds, at $\mathscr{L} = 10^{31}$ cm$^{-2}$ s$^{-1}$, $10^{33}$ cm$^{-2}$ s$^{-1}$, and $10^{34}$ cm$^{-2}$ s$^{-1}$.

| Event Filter muon trigger rates | | | | | | |
|---|---|---|---|---|---|---|
| | $\mathscr{L} = 10^{31}$ cm$^{-2}$ s$^{-1}$ | | $\mathscr{L} = 10^{33}$ cm$^{-2}$ s$^{-1}$ | | $\mathscr{L} = 10^{34}$ cm$^{-2}$ s$^{-1}$ | |
| | Barrel (Hz) | Endcaps (Hz) | Barrel (Hz) | Endcaps (Hz) | Barrel (Hz) | Endcaps (Hz) |
| | 4 GeV | | 6 GeV | | 20 GeV | |
| $\pi/K$ | 125 | 119 | 1890 | 1230 | 46 | 40 |
| beauty | 44 | 56 | 1870 | 2190 | 260 | 380 |
| charm | 60 | 76 | 2390 | 2780 | 220 | 330 |
| top | < 0.1 | < 0.1 | < 0.1 | < 0.1 | 0.2 | 0.3 |
| $W$ | < 0.1 | < 0.1 | 2.9 | 3.9 | 21 | 31 |
| TOTAL | 229 | 251 | 6150 | 6200 | 550 | 780 |
| | 5 GeV | | 8 GeV | | 40 GeV | |
| $\pi/K$ | 36 | 25 | 290 | 260 | 0.14 | 0.2 |
| beauty | 27 | 33 | 550 | 800 | 10.5 | 16.3 |
| charm | 36 | 43 | 640 | 930 | 7.1 | 11.1 |
| top | < 0.1 | < 0.1 | < 0.1 | < 0.1 | < 0.1 | < 0.1 |
| $W$ | < 0.1 | < 0.1 | 2.8 | 3.8 | 3.9 | 6.1 |
| TOTAL | 99 | 101 | 1480 | 1990 | 21.7 | 33.7 |

Table 7: Single muon trigger rates at EF muon combined, for various low and high $p_T$ thresholds, at $\mathscr{L} = 10^{31}$ cm$^{-2}$ s$^{-1}$, $10^{33}$ cm$^{-2}$ s$^{-1}$, and $10^{34}$ cm$^{-2}$ s$^{-1}$.

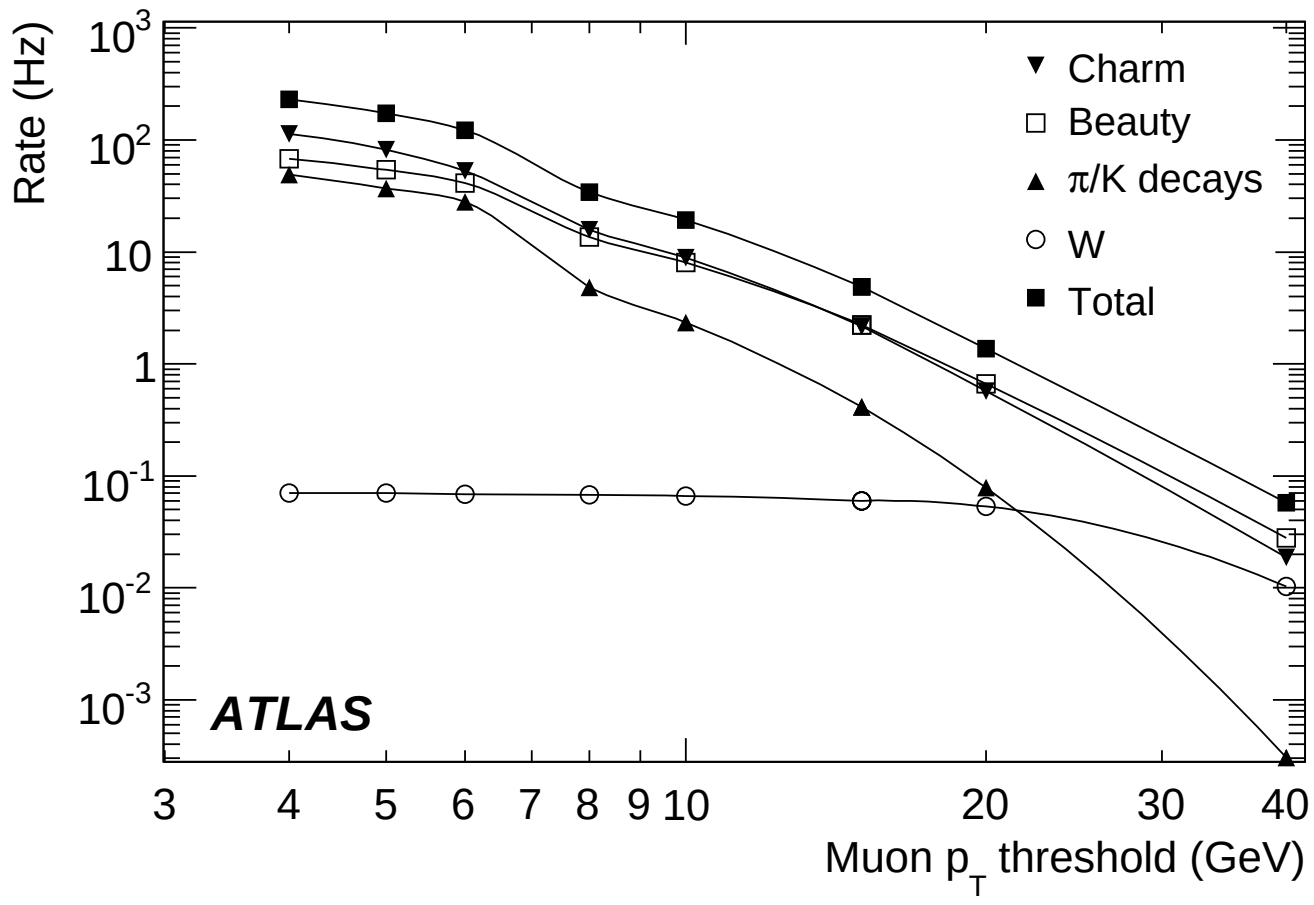


Figure 19: Expected EF rates at $\mathscr{L} = 10^{31}$ cm$^{-2}$ s$^{-1}$ for single muon processes as a function of muon $p_T$ threshold integrated over $|\eta| < 2.4$.

which produces conservative estimates since it predicts cross-sections about 2 to 3 times higher than previous descriptions [4]. Top quark and W/Z decays were simulated using PYTHIA 5.7 [5]. Rates of muon in-flight decays from $\pi/K$ mesons have been computed using the DPMJET Monte Carlo program [6].

To verify the results obtained with this method and to understand the systematics, an alternative approach, relying on event counting, has been applied to the minimum bias events (*counting method*). The convolution and counting methods give EF final rates which are in good agreement, within statistical errors due to the limited size of the minimum bias sample, starting from $p_T$ threshold of 6 GeV. The values obtained with the counting method for lower $p_T$ thresholds (4 and 5 GeV) are a factor of two to four less for muons from $\pi/K$ decays with respect to the convolution (provided by DPMJET) of Eq. 2.

The rates obtained for some low and high $p_T$ thresholds in the barrel and in the endcaps after L1, L2 muFast, L2 muComb and EF selection are shown in Tables 4, 5, 6 and 7.

In Fig. 19 the total (barrel+endcaps) EF rates at L = $10^{31}$ cm$^{-2}$ s$^{-1}$ are shown as a function of the $p_T$ threshold. In this figure, to keep uniformity among the rate results, mostly provided by PYTHIA 6.403, it has been chosen to report for the 4 and 5 GeV thresholds the EF rates obtained with the counting procedure.

## 7.1 Fake dimuon trigger rate

A single muon can be detected in multiple muon trigger sectors, causing such events to erroneously satisfy dimuon triggers. Such fake triggers can be suppressed by the overlap handling implemented in the MuCTPI . For the single muon samples used in the analysis, the fake dimuon trigger probability is defined as

$$P_{\text{fake}} = \frac{\text{Number of events with more than one muon triggered}}{\text{Number of events with a triggered muon}} \tag{3}$$

Four sources of fake double-counts have been considered :

- Barrel-Barrel double counts (BB): When a single muon is detected by two overlapping RPC sectors.

- Barrel-Endcap double counts (BE): When a single muon is detected by an overlapping RPC-TGC sector pair.
- Endcap-Endcap double counts (EE): When a single muon is detected by two overlapping "Endcap" TGC sectors.
- Forward-Forward double counts (FF): When a single muon is detected by two overlapping "Forward" TGC sectors.

The probabilities that a single muon would cause any of these fake double counts have been calculated separately. The effect of the MuCTPI overlap handling can be seen in Fig. 20, where the left plot shows the BE fake dimuon trigger probabilities for the 6 $p_T$ thresholds in the 2 to 50 GeV range without using the overlap handling, while the right plot shows the probabilities after applying the MuCTPI overlap handling.

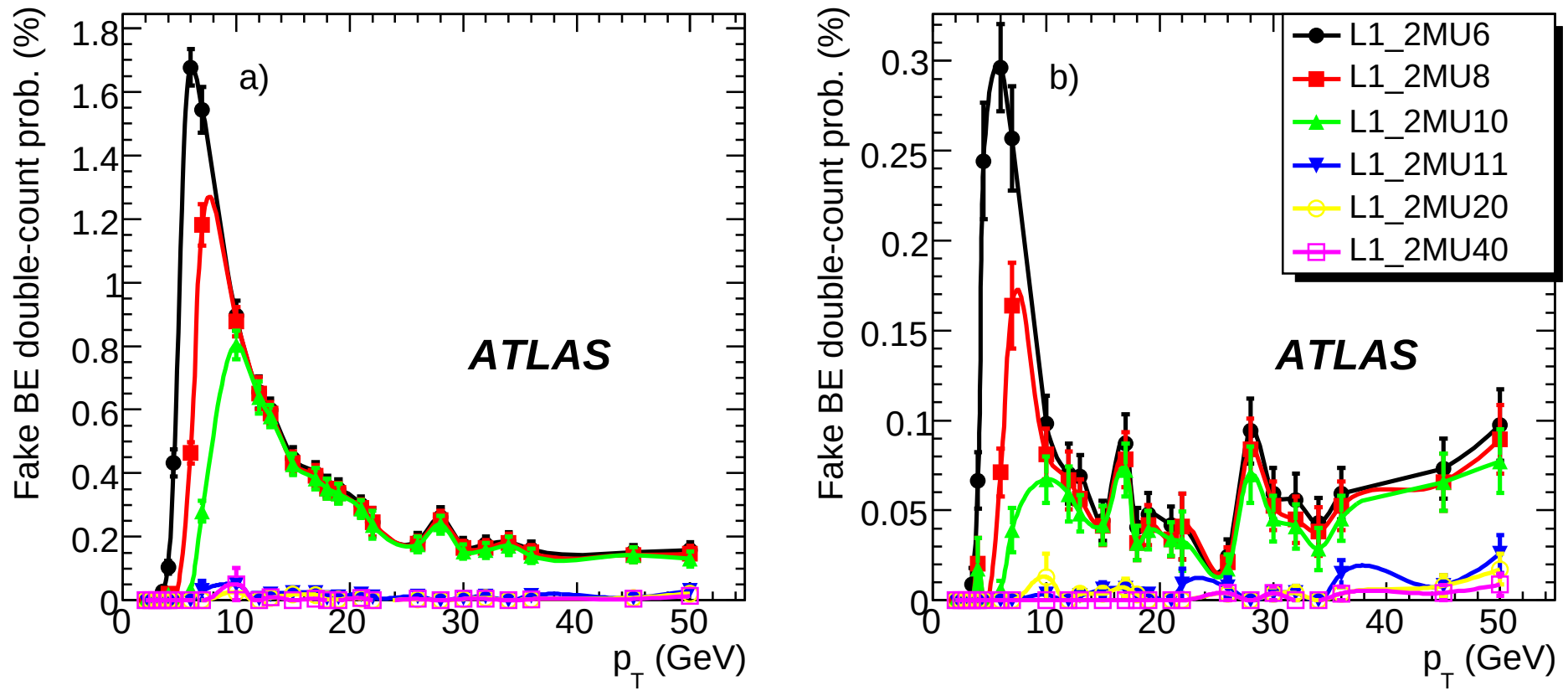


Figure 20: Barrel-Endcap fake dimuon trigger probabilities without (a) and with (b) using the overlap handling of the MuCTPI.

The probabilities for 6 and 20 GeV single muons to produce a fake dimuon trigger if they caused a single-muon trigger, for all available L1 muon thresholds, can be seen in Table 8.

The fake probabilities can be used to calculate the single-muon trigger rates according to

$$R_{\text{fake}} = \mathscr{L} \int_{p_T^{cutoff}}^{p_T^{inf}} \sigma_p(p_T)\varepsilon(p_T)P_{\text{fake}}(p_T)dp_T \tag{4}$$

where $\mathscr{L}$ is the instantaneous luminosity of the accelerator, $\sigma_p$ is the inclusive muon production cross-section at LHC and $\varepsilon$ is the L1 trigger efficiency. The fake dimuon trigger rates are presented in Table 9.

## 8 Rejection of muons from $\pi/K$ decays

Despite the large theoretical uncertainties on the rates of the muon production processes at low transverse momentum, it is clear that in flight decays of pions and kaons are a significant source of single muons and, therefore, a strategy must be developed to reject these events in the trigger. Rejection of muons from $\pi$ and $K$ decays at the EF is described below. A study describing rejection at L2 can be found in [7].

| Trigger item | $p_T$[GeV] | BB prob. [%] | BE prob. [%] | EE prob. [%] | FF prob. [%] |
|---|---|---|---|---|---|
| 2mu4 | 6.0 | 1.56 ± 0.07 | 1.39 ± 0.08 | 1.00 ± 0.07 | 0.81 ± 0.06 |
| | 20.0 | 1.43 ± 0.06 | 0.13 ± 0.02 | 0.49 ± 0.05 | 0.55 ± 0.05 |
| 2mu5 | 6.0 | 1.14 ± 0.06 | 1.17 ± 0.07 | 0.40 ± 0.05 | 0.56 ± 0.06 |
| | 20.0 | 1.43 ± 0.06 | 0.13 ± 0.02 | 0.49 ± 0.05 | 0.55 ± 0.05 |
| 2mu6 | 6.0 | 1.11 ± 0.05 | 0.97 ± 0.06 | 0.39 ± 0.04 | 0.55 ± 0.05 |
| | 20.0 | 1.43 ± 0.06 | 0.13 ± 0.02 | 0.49 ± 0.05 | 0.55 ± 0.05 |
| 2mu8 | 6.0 | 0.87 ± 0.05 | 0.38 ± 0.06 | 0.31 ± 0.05 | 0.58 ± 0.07 |
| | 20.0 | 1.33 ± 0.06 | 0.10 ± 0.02 | 0.42 ± 0.05 | 0.45 ± 0.05 |
| 2mu10 | 6.0 | 0.68 ± 0.05 | 0.12 ± 0.08 | 0.21 ± 0.09 | 0.54 ± 0.13 |
| | 20.0 | 1.26 ± 0.06 | 0.10 ± 0.02 | 0.36 ± 0.04 | 0.36 ± 0.04 |
| 2mu11 | 6.0 | 0.43 ± 0.21 | 0.00 ± 0.00 | 0.32 ± 0.15 | 0.42 ± 0.16 |
| | 20.0 | 0.86 ± 0.05 | 0.00 ± 0.00 | 0.33 ± 0.04 | 0.32 ± 0.04 |
| 2mu20 | 6.0 | 0.28 ± 0.28 | 0.00 ± 0.00 | 0.48 ± 0.34 | 0.00 ± 0.00 |
| | 20.0 | 0.75 ± 0.05 | 0.00 ± 0.00 | 0.24 ± 0.03 | 0.18 ± 0.03 |
| 2mu40 | 6.0 | 0.42 ± 0.42 | 0.00 ± 0.00 | 0.00 ± 0.00 | 0.00 ± 0.00 |
| | 20.0 | 0.49 ± 0.04 | 0.00 ± 0.00 | 0.08 ± 0.03 | 0.08 ± 0.03 |

Table 8: Probabilities that single muons with transverse momenta 6 and 20 GeV which caused a single muon trigger, to also passes a fake dimuon signature at the same threshold.

| Trigger item | BB rate [Hz] | BE rate [Hz] | EE rate [Hz] | FF rate [Hz] | Total fake rate [Hz] |
|---|---|---|---|---|---|
| 2mu4 | 1846.6 ± 119.2 | 271.6 ± 14.1 | 136.2 ± 24.5 | 69.2 ± 12.3 | 2323.7 ± 123.1 |
| 2mu5 | 243.9 ± 13.0 | 203.1 ± 10.6 | 35.3 ± 10.5 | 33.3 ± 6.5 | 515.5 ± 20.8 |
| 2mu6 | 193.9 ± 12.4 | 82.6 ± 7.1 | 24.6 ± 6.0 | 24.7 ± 4.7 | 325.7 ± 16.2 |
| 2mu8 | 114.1 ± 9.8 | 16.1 ± 2.1 | 9.7 ± 3.0 | 12.4 ± 3.3 | 152.3 ± 11.0 |
| 2mu10 | 79.2 ± 8.0 | 4.9 ± 1.2 | 4.8 ± 2.2 | 5.5 ± 2.0 | 94.4 ± 8.6 |
| 2mu11 | 11.7 ± 1.8 | 0.1 ± 0.1 | 3.9 ± 2.0 | 4.5 ± 1.8 | 20.1 ± 3.2 |
| 2mu20 | 2.4 ± 0.4 | 0.1 ± 0.0 | 2.4 ± 1.9 | 0.7 ± 0.5 | 5.5 ± 2.0 |
| 2mu40 | 0.8 ± 0.1 | 0.0 ± 0.0 | 1.7 ± 1.7 | 0.1 ± 0.1 | 2.6 ± 1.7 |

Table 9: Rates of various fake dimuon triggers at $\mathscr{L} = 10^{33}$ cm$^{-2}$ s$^{-1}$.

## 8.1 Data samples and their validation

Minimum bias samples would be the most suitable for studies involving pion and kaon decays. However, the probability that pions or kaons produced in low or moderate $p_T$ QCD scattering would decay before interacting hadronically in the calorimeters is low, between 0.1% and 1% depending on the meson $p_T$. In order to enhance the number of charged pion and kaon decays in the sample, the simulation of events without any charged meson with $p_T$ above a given threshold is aborted and one $\pi^{\pm}$ or $K^{\pm}$ is forced to decay in the Inner Detector cavity.

The samples produced are:

- $10^6$ single pions with $p_T > 2.5$ *GeV* and kinematics ($p_T \times \eta$) generated according to a double differential cross-section of primary pions in minimum bias events
- $10^5$ minimum bias events, where one charged $\pi$ or K with $p_T > 2$ *GeV* per event is forced to decay;

In order to estimate cross-sections or trigger rates, the abundance of forced decays must be re-weighted on an event by event basis according to meson decay probability. In addition to the above samples, standard minimum bias events have been used as a reference to cross check the results obtained from these dedicated productions.

The muon $p_T$ spectra observed in minimum bias events and single pions, forced to decay, were found to be consistent, after appropriate re-weighting, with each other and in agreement with previous predictions and unforced minimum bias events.

## 8.2 Rejection strategy at the event filter

The fraction of in flight decay muons retained at the EF, normalised to the L2 efficiency, for the mu6 trigger item, has been measured as a function of the muon $p_T$. There is a very poor rejection capability ($\lesssim 90\%$) for muons coming from pion decays, which demonstrates that the standard muon identification procedures are not very sensitive, as expected, to the small kink between the pion and muon tracks. The kinematics of charged kaon two-body decays, which are the dominating kaon contribution to the muon rate, is much more favorable toward rejection due to the larger average value of the angle between the kaon and the muon tracks. In order to improve the rejection capability, additional measured parameters providing some discriminating power between background and primary muons have been identified:

- the impact parameter, $d_0$, of the track reconstructed in the inner tracker;
- the number of hits associated to the Inner Detector track in the Pixel Detector ($N_{hits}$(Pixel)), in the pixel B-layer ($N_{hits}(B_{layer})$) and in the Silicon Tracker ($N_{hits}$(SCT));
- the ratio $p_{T_{ID}}/p_{T_{MS}}$ between the transverse $p_T$ in the Inner Detector and in the Muon Spectrometer, after back-extrapolation to the interaction point and correction for the measured energy loss in the calorimeters;
- the $\chi^2_{\text{matching}}$ of the matching between the track parameters as reconstructed in the Muon Spectrometer and in the Inner Detector.

The discrimination power of each variable has been studied by measuring the fraction of accepted events as a function of the cut applied for both isolated muons and fake muons above a given $p_T$ threshold. The results, shown in Fig. 21, are based on the simulations of single muons and single pions with forced decays. For each variable, the fraction of events retained after the cut is normalised to the number of events passing the EF reconstruction before the application of any hypothesis algorithms. From the analysis of the exclusive rejection power of the individual variables, the set of cuts listed below have been defined. These cuts try to minimize the efficiency loss for prompt muons while reducing the background:

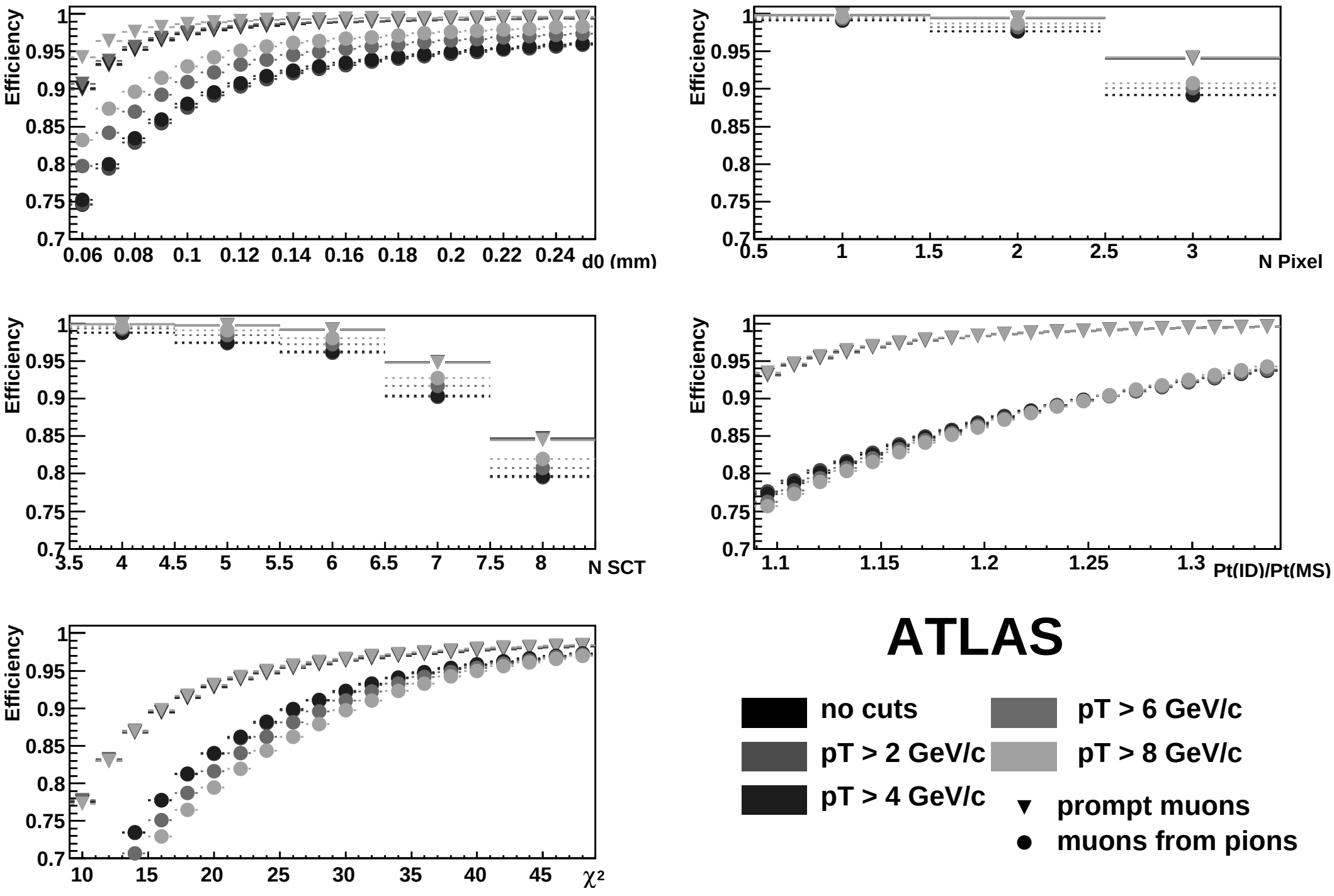


Figure 21: Efficiency for prompt muons and muons from pion decays as a function of the cut on some discriminating variables.

- $|d_0| < 0.15$ mm, $N_{hits}(B_{layer}) \geq 1$, $N_{hits}(Pixel) \geq 3$, $N_{hits}(SCT) \geq 6$,
- $p_{T_{ID}}/p_{T_{MS}} < 1.25$, $\chi^2_{matching} \leq 26$.

In particular, these values have been chosen by considering efficiency and background rejection at $p_T = 4\ GeV$. It is assumed that cuts will be optimized for each muon item in the trigger menu.

From the application of these cuts on the reference sample of events accepted at the EF, the efficiency for prompt muons and in flight decay muons shown in Fig. 22 have been obtained. A loss of efficiency between 25% at the 4 GeV threshold and 10% at 20 GeV correspond to a reduction in background of 65% and 75%, respectively. The rejection achieved for kaon decays is slightly better than that achieved for $\pi$ decays, as expected from the different decay kinematics. These results are derived from nominal detector performance and algorithm resolutions. However, they demonstrate that cuts can be adjusted to obtain reasonable trigger rate at the very low $p_T$ threshold of 4 GeV which is reached mostly by reducing uninteresting events at the cost of some efficiency loss for prompt muons. An optimization of the cuts, with specific tuning for each trigger element, will eventually further improve the signal to background ratio.

# 9 Muon isolation

## 9.1 Optimization procedure

The L2 isolation algorithm is seeded by either muFast or muComb. The algorithm decodes LAr and Tile Calorimeter quantities (i.e. transverse energy deposit or sums of calorimetric cells above a predefined energy threshold) in cones centered around the muon direction. The geometrical definition of these

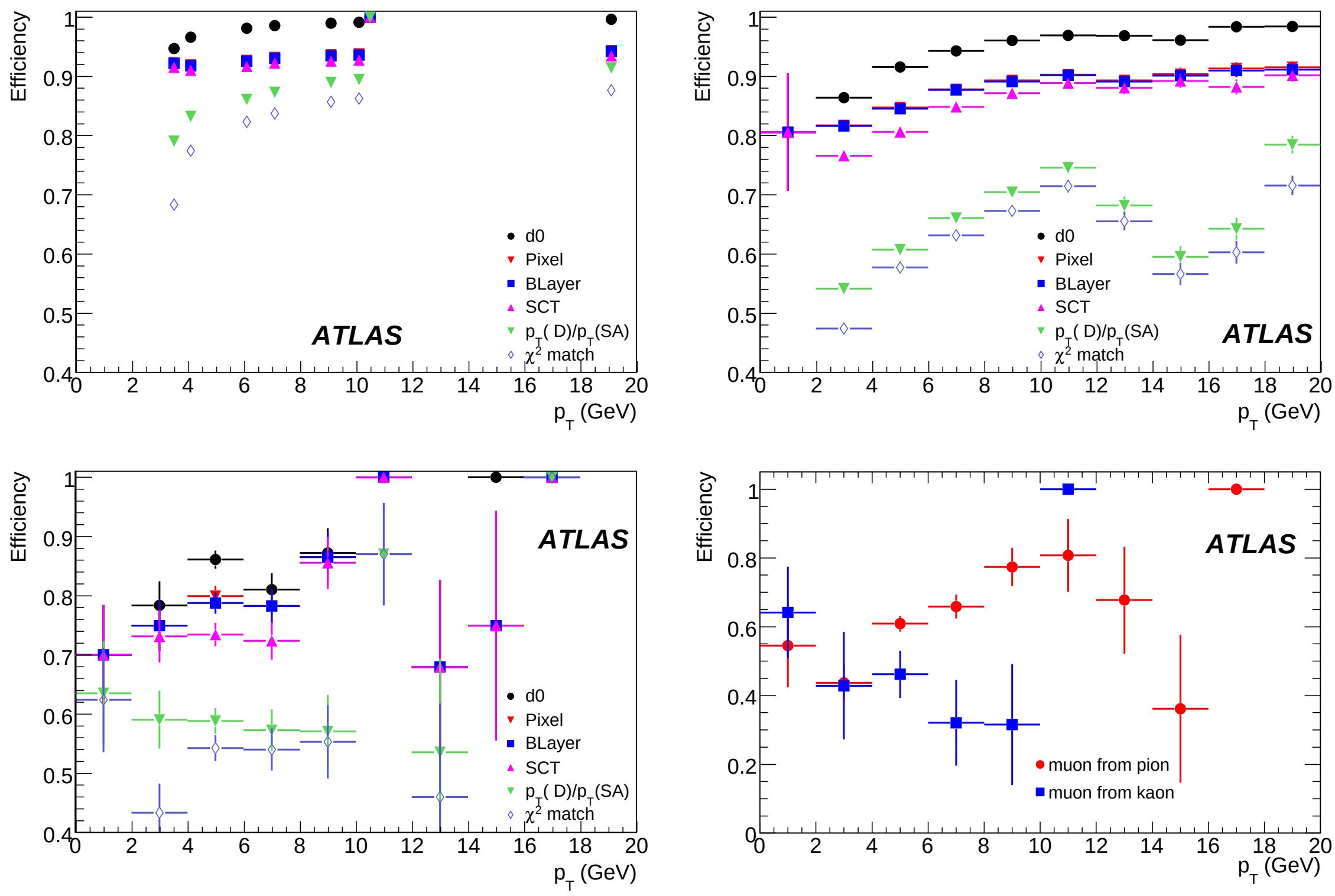


Figure 22: EF efficiency as a function of $p_T$ for different rejection cuts for prompt muons (a), muons from single pion decays (b), minimum bias events (c and d). In (c) $\pi/K$ contribution has been separated. Each efficiency curve shows the data reduction obtained by the addition of the corresponding cut to the overall selection procedure. The specific values of the cuts are discussed in the text.

cones is given by the condition $\Delta R < \Delta R_{MAX}$, where $\Delta R = \sqrt{(\Delta\eta^2 + \Delta\phi^2)}$, and $\Delta\eta$, $\Delta\phi$ are the distances in pseudorapidity and azimuthal angle between the calorimetric cell and the cone axis. Because the muon itself contributes to the energy deposit inside the cone, to improve the discriminating power of the isolation algorithm, two different concentric cones are defined: an internal cone chosen to contain the energy deposit released by the muon itself, and an external one, supposed to include contributions only from detector noise, pile-up and jet particles if present. The optimization of the muon isolation algorithm consists of determining the optimal size of the inner and outer cone radius, the values of the cell energy thresholds, used to compute the transverse energy and number of cells sums, and the isolation requirements.

Table 10 summarizes the samples that have been used to optimize the algorithms and to measure their performance. Half of the events in the samples number 1 and 2 in the Table have been used as signal and background, respectively, in the optimization of the algorithm parameters, the remaining events for sample 1 and 2 and the other samples listed in the Table have instead been used to estimate the algorithm performances.

Only the parameters relative at the muon trigger in the barrel region ($|\eta| < 1.05$) have been studied. Simulation of the electronic readout noise for both LAr and Tile Calorimeters has been also included.

A muon track passing through the calorimetry will deposit energy in the cells which immediately surround it. The deposited energy can be contained within some cone of radius $R_{Inner}$, where $R = \sqrt{\Delta\eta^2 + \Delta\phi^2}$. If the muon is isolated, there will be little energy deposited in cells which lie in an outer

| | Process | Generator | Number of events |
|---|---|---|---|
| 1 | $Z \to \mu^+\mu^-$ | `Pythia` | 1 $10^4$ |
| 2 | $bb \to \mu(15)X$ | `Pythia` | 1.5 $10^4$ |
| 3 | $bb \to \mu(6)X$ | `Pythia` | 1 $10^4$ |
| 4 | $q\bar{q} \to \mu X$ | `Pythia` | 2.1 $10^4$ |
| 5 | Single-$\mu$($p_T$=100 GeV) | `Single-Mu gun` | 2 $10^5$ |
| 6 | Single-$\mu$($p_T$=38 GeV) | `Single-Mu gun` | 2 $10^5$ |
| 7 | Single-$\mu$($p_T$=19 GeV) | `Single-Mu gun` | 2 $10^5$ |

Table 10: Data samples used in the muon isolation algorithm optimization.

annulus around this ($R \in [R_{Inner}, R_{Outer}]$). The radius of the inner cone (i.e. the cone fully containing the muon) has been determined from the distribution of the summed transverse energy contained within a cone of increasing radius around the muon direction from $Z \to \mu\mu$, as shown in Figure 23. The value of $R$ corresponding to the inner cone radius is visible as a change in the slope of the curve. Once the radius for which all the muon energy is contained in the cone is reached, for each further increase of the cone radius only noise will be summed, resulting in a reduction of the slope of the energy sum curve. The reduction in the slope depends on the level of electronic readout noise per cell, as shown in Fig. 23, where curves for several values of the threshold cut on the calorimetric cell energy is shown, ranging from 40 to 90 MeV. The effect of the electronic noise is only relevant for the LAr calorimeter. From the two figures it can be seen that a cone of radius 0.1 (one readout cell), is sufficient to contain the muon energy deposition in the hadronic calorimeter, while a radius of about 0.07 (one to three readout cells, depending on position), is sufficient for the electromagnetic calorimeter, due to the finer readout granularity. The value of the outer cone radius is instead constrained by timing requirements. Increasing the outer cone radius requires a larger fraction of the calorimeter to be read out and decoded. Because the readout step of the algorithm dominates the execution time ($>90\%$ of the overall algorithm time) the requirement to keep the overall timing below $O(10)$ $ms$ constrains the maximum outer cone radius to be below about 0.4. We have verified that optimal background rejection is obtained by keeping the outer cone radius at is maximum value.

An analysis has been performed over all the quantities used in the isolation hypothesis testing, with a goal of minimizing the number of variable used in the optimization step. Each variable used in the optimization is listed in Table 11, together with the respective separation power expressed in term of minimum variance bound [8]. The optimal value of the cell energy cut thresholds, used to compute the transverse energy and number of cell sums, has been obtained by maximizing the background rejection after applying a fixed cut on the isolation variables ($var1$ and $var2$ in Table 11), giving a 95% efficiency for the $Z \to \mu\mu$ signal. A common threshold value of 60 MeV has been obtained with this procedure for both the LAr and Tile calorimeter. Algorithm performances are stable for threshold variations of $\pm 10$ MeV around the optimal values. The distributions of some of the most powerful variables for signal selection and background rejection are shown in Fig. 24.

Optimal cut values for the isolation variables described above have been obtained in a multivariate optimization procedure by simultaneously varying all the cuts in sensible ranges and by minimizing the $b\bar{b} \to \mu X$ background efficiency at fixed $Z \to \mu\mu$ signal efficiency. In the optimization procedure, both background and signal efficiencies are calculated with respect to muons satisfying the L2 muFast mu20 requirement. In Fig. 25 the background rejection,defined as $1/\varepsilon_{BG}$ where $\varepsilon_{BG}$ is the efficiency for the $b\bar{b}$ background sample, and $(1-\varepsilon_{BG})$ as a function of the $Z \to \mu\mu$ signal efficiency, obtained after the optimization procedure, are shown. The chosen working point for the isolation algorithm yields a factor of 10 reduction for the $b\bar{b}$ background at a 95% signal efficiency for muons with $p_T$ >20 GeV.

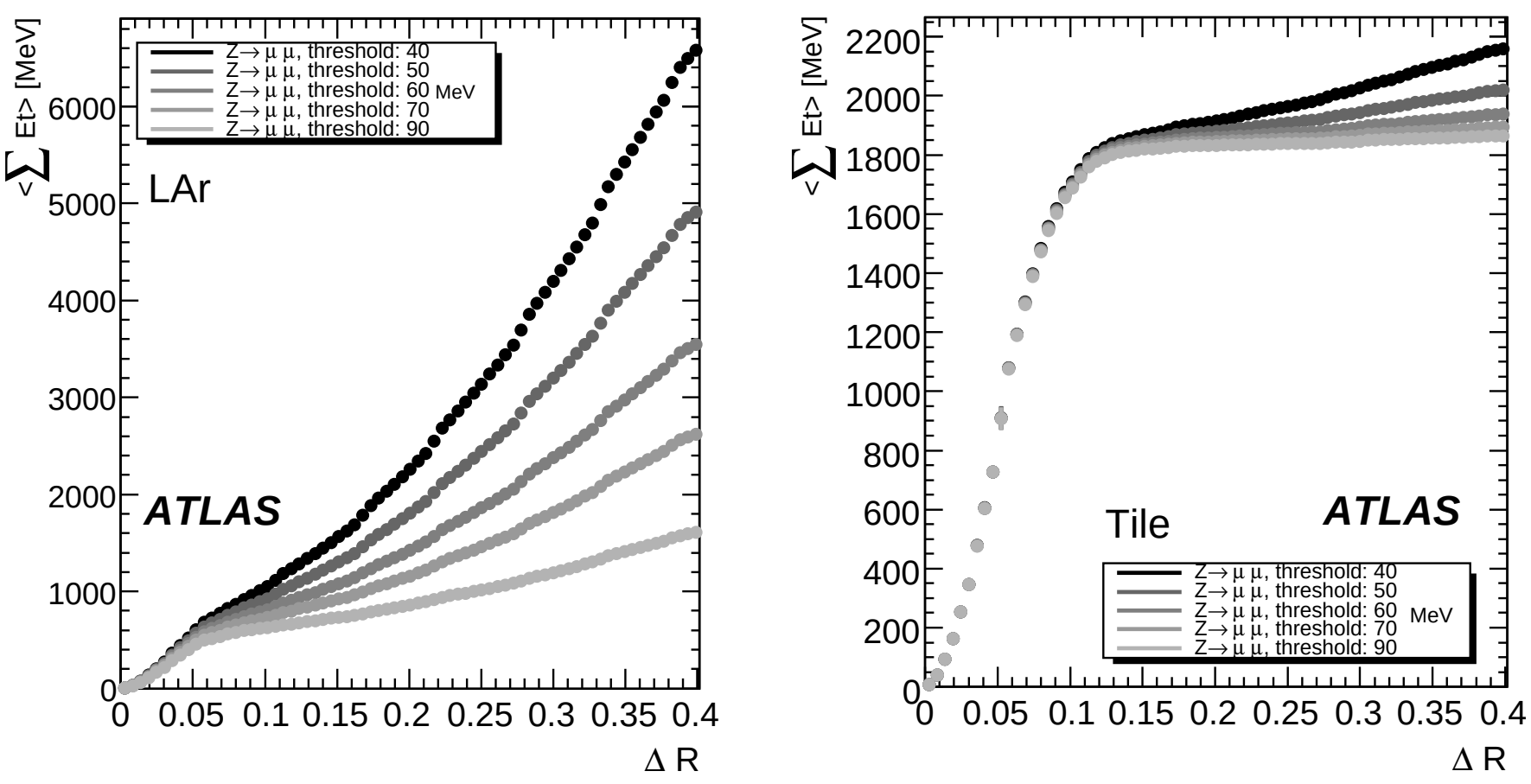


Figure 23: The total transverse energy contained within a cone of increasing radius around the muon candidate track from $Z \rightarrow \mu\mu$ signal events in the LAr calorimeter (left) and Tile calorimeter (right). The different curves on each figure correspond to different thresholds applied on the cell energy.

| Label | variable | Separation |
|---|---|---|
| $var1$ | $Iso_{LAr} = \sum E_T^{\Delta R<0.07} / \sum E_T^{\Delta R<0.4}$ | 0.21 |
| $var2$ | $Iso_{Tile} = \sum E_T^{\Delta R<0.1} / \sum E_T^{\Delta R<0.4}$ | 0.29 |
| $var3$ | $E_{LAr}^{O} = \sum E_T^{\Delta R \in [0.07,0.4]}$ | 0.75 |
| $var4$ | $E_{Tile}^{O} = \sum E_T^{\Delta R \in [0.1,0.4]}$ | 0.40 |
| $var5$ | $E_{LAr}^{I} = \sum E_T^{\Delta R<0.07}$ | 0.23 |
| $var6$ | $E_{Tile}^{I} = \sum E_T^{\Delta R<0.1}$ | 0.06 |
| $var7$ | Number of LAr cells above threshold with $\Delta R \in [0.07, 0.4]$ | 0.72 |
| $var8$ | Number of Tile cells above threshold with $\Delta R \in [0.1, 0.4]$ | 0.34 |
| $var9$ | Number of LAr cells above threshold with $\Delta R < 0.07$ | 0.31 |
| $var10$ | Number of Tile cells above threshold with $\Delta R < 0.1$ | 0.07 |

Table 11: Variable used in the muon isolation optimization procedure. The separation is zero for identical signal and background shapes, and it is one for shapes with no overlap.

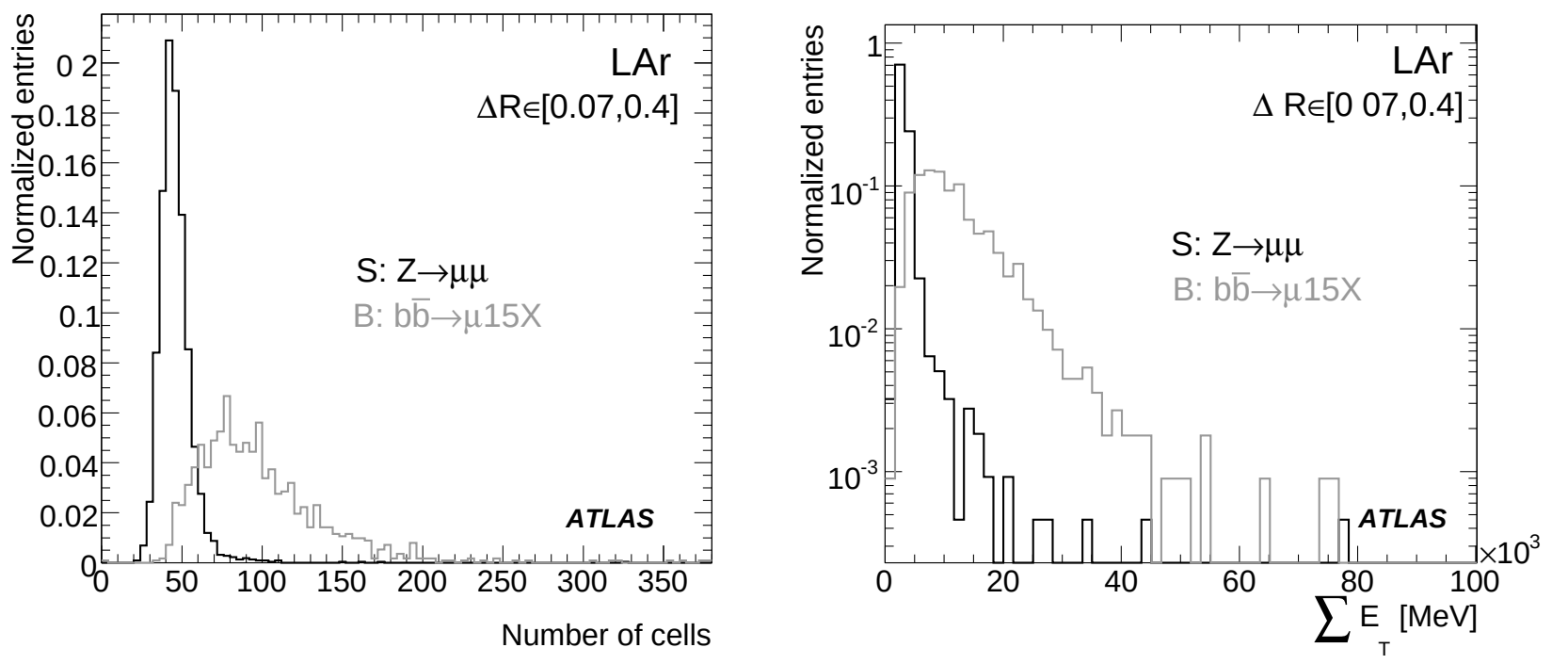


(a) Number of LAr cells above threshold in the outer ring

(b) Transverse energy sum in the outer ring of LAr

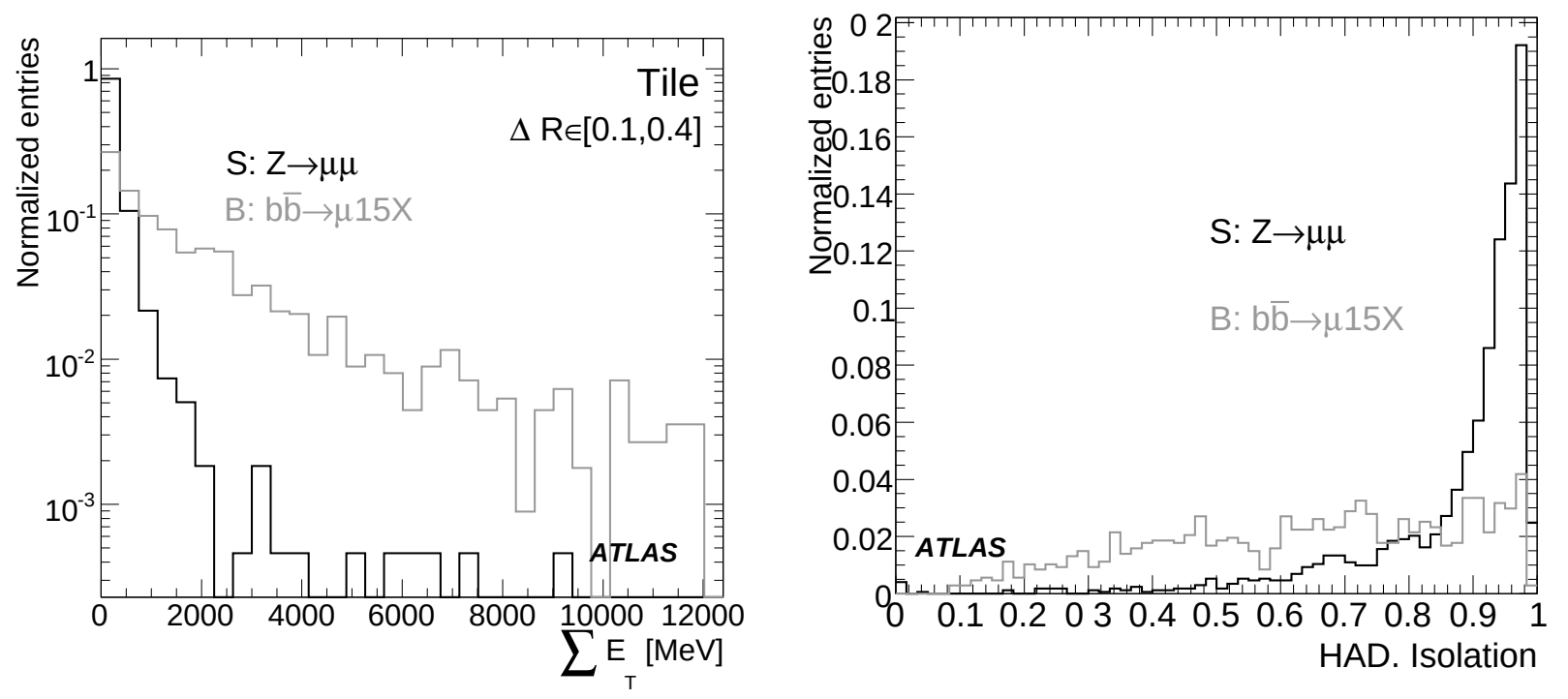


(c) Transverse energy sum in the Tile in the outer ring

(d) Isolation variable in Tile calorimeter

Figure 24: Most powerful variables for calorimetry-based muon isolation.

## 9.2 Performance

The performance of the isolation algorithms in terms of $b\bar{b}$ and dijet background reduction, efficiency of benchmark signal channels and timing is presented below. The performance of isolation algorithms can be affected by the instantaneous luminosity since the pile-up requires higher thresholds for the same nominal efficiency. This is particularly true for calorimetry-based isolation, while for track-based isolation the effect can be reduced by requiring that the contributing tracks come from the same primary vertex as the muon. For this reason, the results of this study should be taken as preliminary and valid only in the framework of the approximations used in the simulated events production for these studies. Possible changes and further development may occur as soon as real data is available.

The effect of isolation algorithms on various sources of non-isolated muons at L2 is shown in Table 12. The quantity $1-\varepsilon_{BG}$ is shown for the isolation requirements corresponding to a working point for the isolation algorithm with a nominal $Z \rightarrow \mu\mu$ signal efficiency of 95%. Results from dijet decays give an estimate of the rejection power of the isolation algorithm for high-$p_T$ muons from $K$ and $\pi$ in flight decays. The rejection power for low-$p_T$ muons from $b\bar{b}$ decays selected by the level 2 mu6 requirement has also been estimated. The reduction in rejection power, from a factor 10 to a factor of about 2, at the low-$p_T$ limit is expected, given the low energy associated with the jets. As already mentioned, calorimetry based isolation algorithms are not effective against these kind of muons, and the track-based isolation is expected to be much more powerful in reducing this kind of background.

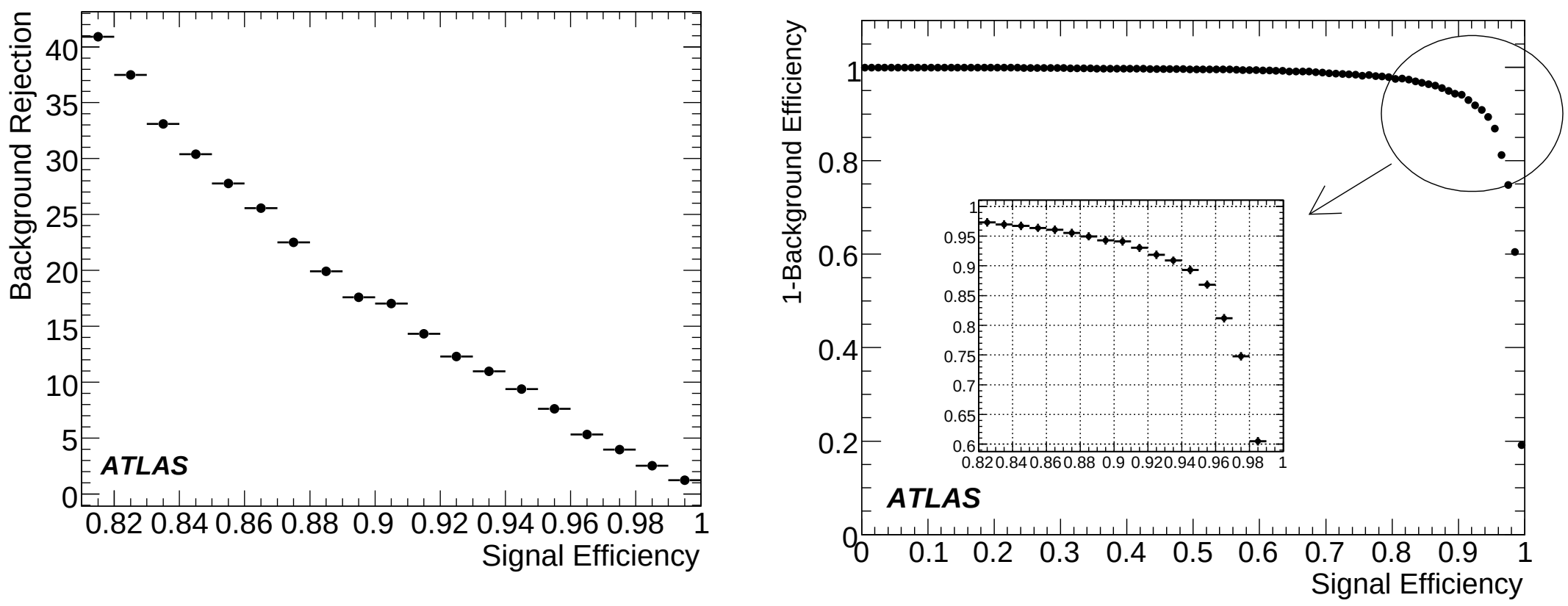


Figure 25: Background rejection ($1/\varepsilon_{BG}$) (left), and $1-\varepsilon_{BG}$ (right), as a function of the signal efficiency as obtained in the muon isolation algorithm optimization procedure.

| Process | Trigger item | Average muon $p_T$ (GeV) | $1-\varepsilon_{BG}$ (%) |
|---|---|---|---|
| $bb \to \mu(15)X$ | mu20 | 25.0 | $89.4 \pm 0.7$ |
| $bb \to \mu(6)X$ | mu6 | 9.0 | $54.6 \pm 0.9$ |
| $q\bar{q} \to \mu X$ | mu20 | 40.0 | $99.6 \pm 0.1$ |
| $q\bar{q} \to \mu X$ | mu6 | 20.0 | $97.3 \pm 0.2$ |

Table 12: Muon isolation algorithm $1-\varepsilon_{BG}$ for muons from several background samples. Efficiencies are calculated with respect to muons passing the level 2 mu20 or mu6 requirements, as specified.

The isolation algorithm has been tuned such that the efficiency is 95% at the chosen working point. The efficiency of the isolation requirement has been studied as a function of $p_T$ using samples of single muons with a $p_T$ of 11, 39, and 100 GeV. No sizable change in the muon efficiency are visible, indicating that radiation effects are small for $p_T$ in this range. Evaluation of the effect of the muon radiation for very high $p_T$ muons (500 to 1000 GeV) is ongoing. The efficiencies for muons from $Z \to \mu\mu$ decays and for single muons are reported in Table 13.

| Process | Trigger path | $\varepsilon$ (%) |
|---|---|---|
| $Z \to \mu^+\mu^-$ | mu20 | $95.5 \pm 0.4$ |
| Single $\mu$ $p_T$=100 GeV | mu20 | $98.68 \pm 0.07$ |
| Single $\mu$ $p_T$=39 GeV | mu20 | $98.97 \pm 0.07$ |
| Single $\mu$ $p_T$=6 GeV | mu6 | $98.54 \pm 0.09$ |

Table 13: Muon isolation algorithm efficiencies for muons from several processes and thresholds.

The time available for running L2 algorithms in the on-line trigger is limited to approximately 20 ms. The CPU processing time is therefore a relevant parameter for the feasibility of algorithms to be included in the trigger chain. The results obtained indicate a typical overall time of less than 10 ms. Further and more detailed timing studies performed on the actual L2 processors are ongoing.

## 10 Muon identification using the tile calorimeter

The muon signatures in the three radial layers of the Tile Calorimeter are well measured quantities with a typical pattern that can be used to identify the muons efficiently down to very low $p_{\mathrm{T}}$. This information can be used to confirm the Muon Spectrometer Triggers (i.e. provide redundancy in noisy/dead regions) or to enhance the selection efficiency for very soft muons typically out of reach for the spectrometer.

The algorithm exploits the radial and transverse calorimeter segmentation. The search starts from the outermost layer, which is the one with the cleanest signal, and once a cell is found with energy compatible with a muon, the algorithm checks the energy deposition in the neighbor cells for the most internal layers. These "candidate patterns" are considered as muons when cells compatible with the typical muon energy deposition are found following a $\eta$-projective pattern in all the three TileCal layers. More details can be found in Ref. [9].

### 10.1 Performance

The performance of the TileMuId algorithms has been studied with MonteCarlo single muons and semi-inclusive muon production ($b\bar{b} \to \mu(4)X$) samples. The effect of minimum-bias pileup at low luminosity ($\mathscr{L} = 10^{33}\mathrm{cm}^{-2}\mathrm{s}^{-1}$) has been investigated as well.

Two algorithms, implementing complementary strategies are described, one (TrigTileLookForMuAlg) is fully executed on the LVL2 Processing Unit (PU) the second (TrigTileRODMuAlg) has a core part executed on the Read Out Driver Digital Signal Processor (ROD-DSP) in order to save time. This allows a very fast processing of the entire detector (full scan) as opposed to the RoI based processing typical of the trigger algorithm running on the LVL2 PUs. Since each ROD-DSP processes a small part of the detector readout the TrigTileRODMuAlg acceptance is lower compared with that of TrigTileLookForMuAlg.

### 10.1.1 Spatial resolution

The spatial resolution of the algorithms can be studied using the distributions of the residuals $\Delta\eta = \eta(\mu_{\text{Tile}}) - \eta(\mu_{\text{Truth}})$ and $\Delta\phi = \phi(\mu_{\text{Tile}}) - \phi(\mu_{\text{Truth}})$ in single muon events with $2 \le p_{\text{T}} \le 15$ GeV. The distributions are well described by a Gaussian and resolution can be defined as $\sigma_\eta$ = 0.05 for TrigTileRODMuAlg ( $\sigma_\eta$ = 0.04 for TrigTileLookForMu) and $\sigma_\phi$ = 0.03 rad. To characterize the performance of the algorithms with MC physics events a matching region with the MC truth will be used. For this analysis a matching region of $\Delta\eta \times \Delta\phi = 0.2 \times 0.12$ is used.

### 10.1.2 Efficiency

The muon-tagging efficiency is defined as the ratio of the number of tagged muons which match a truth muon ($N_{\text{tag}}$) to the number of generated truth muons ($N_{\text{gen}}$). Figure 26 shows the efficiency as a function

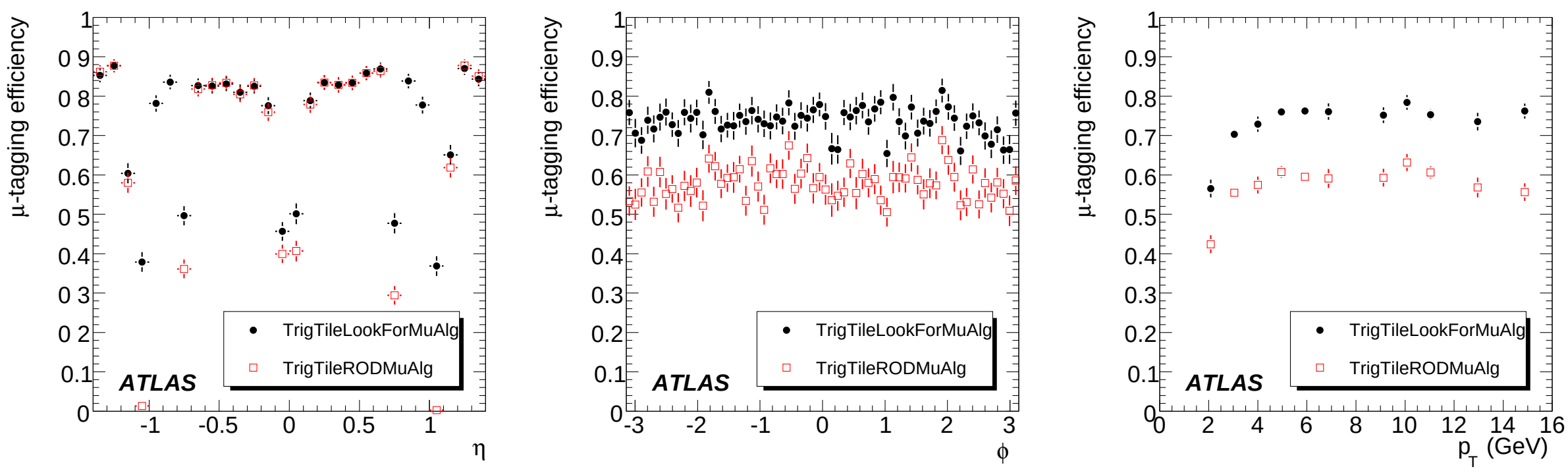


Figure 26: Efficiency as a function of $\eta$ (left), $\phi$ (center) and $p_{\text{T}}$ (right)for TrigTileLookForMu (filled circles) and TrigTileRODMu (open squares) using single muon events.

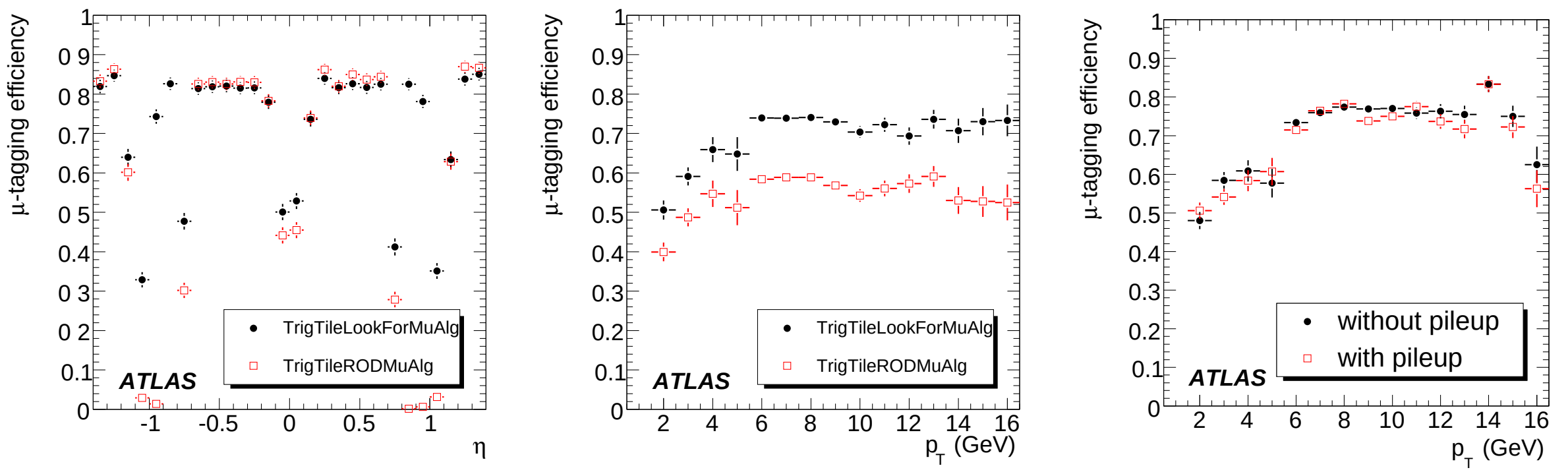


Figure 27: Efficiency as a function of $\eta$ (left) and $p_{\text{T}}$ (center) for TrigTileLookForMu (filled circles) and TrigTileRODMu (open squares) in $b\bar{b} \to \mu(6)X$ events. Right plot show for TrigTileLookforMu the effect of pileup of Minimum Bias events at low luminosity (filled circles) compared with the case without pileup (open squares).

of $\eta$, $\phi$, and $p_{\text{T}}$ of the muon for the two algorithms as obtained using the single muon sample. The efficiency of TileRODMu is lower than that of TileLookForMu. In the region $0.8 \le |\eta| \le 1.1$ the towers

are split between the barrel and the extended barrels, and the cells belonging to different partitions are processed by different ROD DSPs. Similar effects are observed for the boundary at $\eta \sim 0$. Except for these two regions of low geometrical acceptance both algorithms show efficiency $\sim 85\%$ with good agreement. Since TileCal is homogeneous in $\phi$, the efficiency is uniform as a function of $\phi$, see Fig. 26 (center). The efficiency decreases with the muon $p_{\mathrm{T}}$ for $p_{\mathrm{T}} < 3$ GeV and is about 42% at $p_{\mathrm{T}} = 2$ GeV. Most of the muons with $p_T < 2$ GeV stop in the Tile calorimeter. For $p_T > 4$ GeV the efficiency is flat at about 60%. Figure 27 (left) and (center) show the efficiency curves for both algorithms as obtained in $b\bar{b} \to \mu(6)X$ events. These results are in good agreement with the performance obtained using single muons, indicating that the algorithms are not too sensitive to the additional hadronic activity in $b\bar{b}$ events.

To evaluate the performance in a realistic LHC operation scenario a sample of $b\bar{b} \to \mu(6)X$ events simulated with pileup of minimum-bias events at a luminosity $\mathscr{L} = 10^{33}$ cm$^{-2}$s$^{-1}$ was used. As shown in Fig. 27 (right), the efficiencies as a function of $p_{\mathrm{T}}$ for two cases are similar for $p_{\mathrm{T}} > 5$ GeV. The additional muons from minimum-bias events make the efficiency worse in the low $p_{\mathrm{T}}$ region. The average efficiency in the sample with pileup (67.97±0.81)% is slightly lower than the one obtained without pileup (74.25±0.79)%. It can be concluded that the efficiency is not substantially affected by minimum-bias pileup.

### 10.1.3 Fraction of fakes

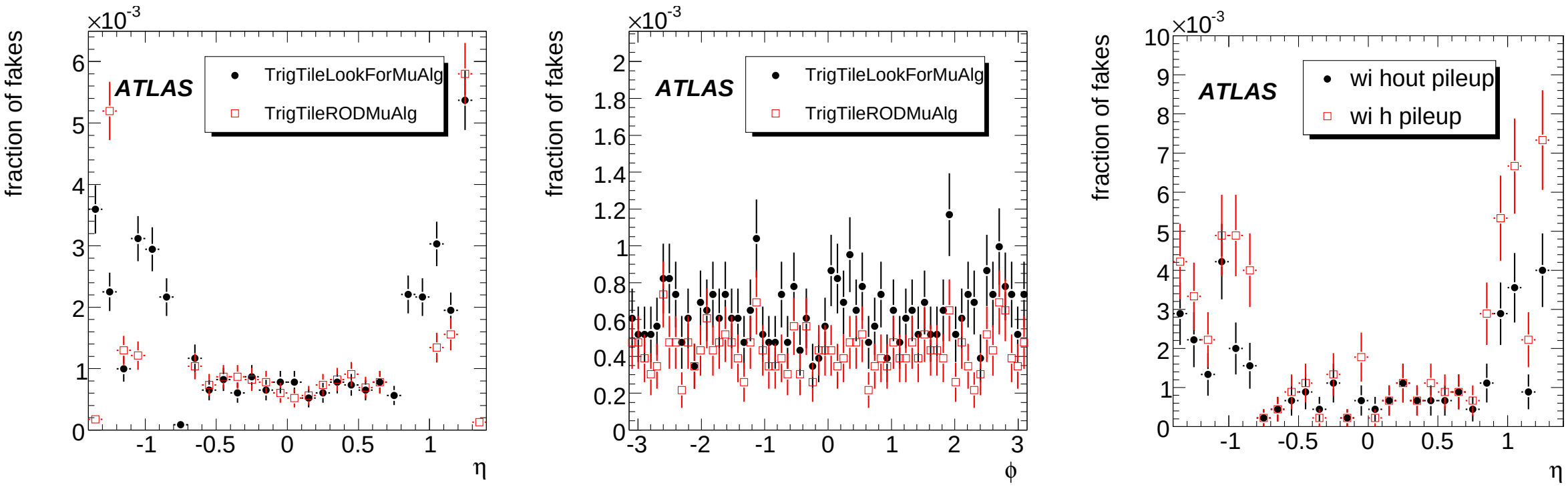


Figure 28: Fraction of fakes as a function of $\eta$ (left) and $\phi$ (center) for TrigTileLookForMu (filled circles) and for TrigTileRODMu (open squares) in $b\bar{b} \to \mu(6)X$ events. The right plot compare performance of TrigTileLookForMu in samples with (filled circles) and without pileup (open squares).

The muon tags which are not matched with truth muons are considered fake. The same $\Delta\eta \times \Delta\phi$ matching cuts are used for the efficiency and fake computation. The fraction of fakes in a given data sample is defined as the ratio of the number of misidentified muons to the total number of events.

The left and center plots of Fig. 28 show the fraction of fakes as a function of $\eta$ and $\phi$ obtained by the two algorithms. Both algorithms show a very small fake rate in the central region (0.12% for $|\eta| < 0.7$). The main contribution of fakes comes from the extended barrel and gap regions, where the cell segmentation is coarse and the projectivity is the worst. The fraction of fakes in the whole range $|\eta| < 1.4$ is $2.7 \pm 0.1\%$ for TrigTileRODMu and $4.1 \pm 0.1\%$ for TrigTileLookForMu. The fraction of misidentified muons as a function of $\phi$ is flat as expected.

Figure 28 (right) shows the performance of TrigTileLookForMu in $b\bar{b} \to \mu(6)X$ events with and without the pileup of minimum-bias events at low luminosity. The fraction of fakes increase from $3.7 \pm 0.1\%$ to $6.0 \pm 0.1\%$ when the minimum bias pileup at $\mathscr{L} = 10^{33}$ cm$^{-2}$s$^{-1}$ is taken into account. The

fake rate increases at larger values of $\eta$ (gap and extended barrel), where the cell granularity is worse and more minimum bias event are expected, compared to the central $\eta$ region.

## 10.2 Combined performance with the inner detector

In order to measure the $p_T$ of the identified muon, the secondary RoI produced by the TileMuId algorithm is used to seed the Inner Detector (ID) track reconstruction algorithm. The size of the ID RoI that requires processing is defined by the Tile algorithm resolution and by the bending in the central solenoid. For $p_{\mathrm{T}}(\mu_{\mathrm{Truth}}) > 2$ GeV, $\Delta\phi = \phi(\mu_{\mathrm{Tile}}) - \phi(\mu_{\mathrm{Track}}) \approx 0.2$ is required. If at least one track is found within the region $\Delta\eta \times \Delta\phi = 0.1 \times 0.2$ and with $p_{\mathrm{T}} > 2$ GeV, the calorimetric tag is confirmed to be a muon and the trigger sequence is successful.

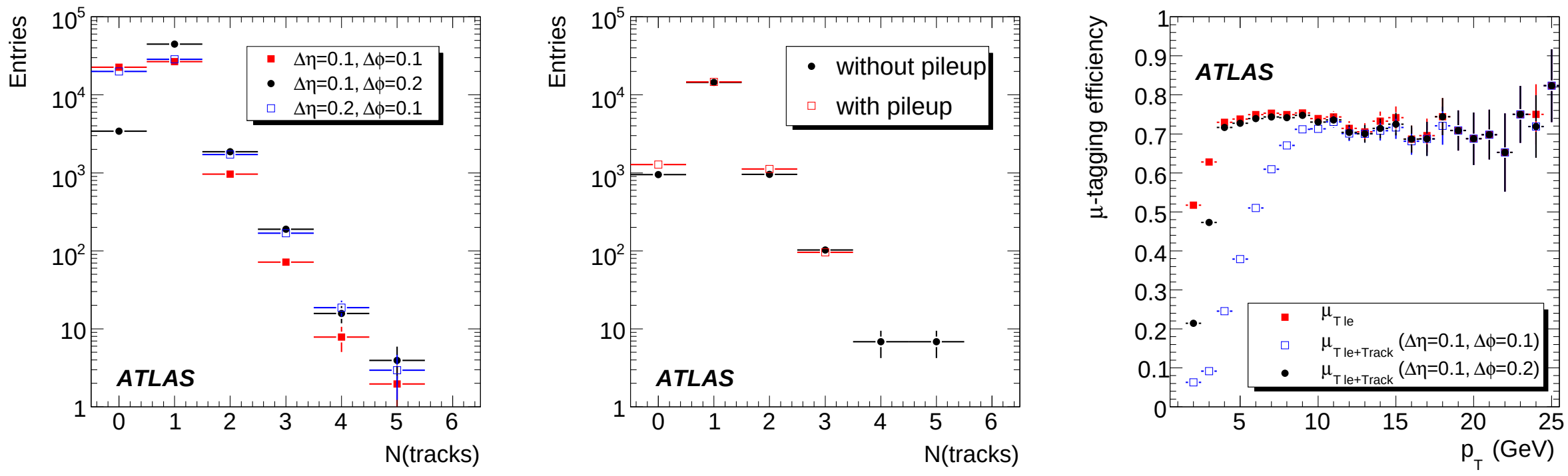


Figure 29: The number of tracks within the given RoI for $b\bar{b} \to \mu(4)X$ (left) and with the fixed RoI size of $\Delta\eta = 0.1$ and $\Delta\phi = 0.2$ for $b\bar{b} \to \mu(6)X$ with/without pileup (center). The right plot shows the efficiency as a function of $p_{\mathrm{T}}$ for the muons tagged by only TileCal (TileLookForMu) and the muons combined with the associated track.

Figure 29 shows the multiplicity of track in the ID RoI; left plot shows that a region with $\Delta\phi = 0.1$ misses the low $p_{\mathrm{T}}$ tracks and results in more events with zero track within the RoI. The RoI with $\Delta\eta = 0.2$ does not give any advantage. The RoI with a size $\Delta\eta = 0.1$ and $\Delta\phi = 0.2$ is a good compromise; the efficiency to reconstruct the muon track is good and the multiplicity of tracks (ambiguity) is acceptable. In the case of reconstruction of multiple tracks, the closest is chosen as the best-matched for the $\mu$ tagged by TileCal, and all tracks are saved since the ambiguity cannot be further resolved at this level. As shown in Figure 29 (center), the multiplicity of tracks within the RoI is not significantly affected by the pileup.

Figure 29 (right) shows the overall combined (TileCal+ID) efficiency for $\Delta\phi = 0.1$ and $\Delta\phi = 0.2$ as a function of muon $p_T$. The combined efficiency obtained with $\Delta\phi = 0.2$ is approximately equal to that of the TileCal stand-alone except for $p_{\mathrm{T}} < 3.5$ GeV. The efficiency from the matched track shows no dependence on $\eta$ or $\phi$. The efficiency, purity and acceptance using the different sizes of RoI are summarized in Table 14. The efficiency and acceptance are significantly improved from $\Delta\phi = 0.1$ to $\Delta\phi = 0.2$. For $\Delta\phi = 0.2$, 97% of tagged muons by TileCal match the associated track. The purity and acceptance of $\Delta\phi = 0.3$ are similar to those of $\Delta\phi = 0.2$. However, the size of $\Delta\eta$ does not affect the efficiency of the matched track with $\mu$ significantly. The differences due to the minimum-bias pileup are observed to be about 2 to 3% due to the small number of events from pileup samples.

| | **TileLookForMu** | **RoI size with $(\Delta\eta, \Delta\phi)$ for matching tracks** | | |
|---|---|---|---|---|
| | | (0.1, 0.1) | (0.1, 0.2) | (0.1, 0.3) |
| Efficiency (%) | $73.08 \pm 0.17$ | $42.02 \pm 0.19$ | $70.91 \pm 0.18$ | $72.06 \pm 0.18$ |
| Unmatched $\mu_{\text{Tile}}$ (%) | | $42.50 \pm 0.36$ | $2.98 \pm 0.08$ | $1.40 \pm 0.05$ |
| Efficiency ($p_{\text{T}} > 4$ GeV) | | $44.09 \pm 0.20$ | $72.94 \pm 0.18$ | $73.06 \pm 0.18$ |
| Purity ($p_{\text{T}} > 4$ GeV ) | | $98.51 \pm 0.88$ | $98.79 \pm 0.67$ | $98.69 \pm 0.67$ |
| Acceptance ($p_{\text{T}} > 4$ GeV) | | $40.78 \pm 0.31$ | $71.93 \pm 0.45$ | $72.01 \pm 0.45$ |

Table 14: Performance with the matched track for $b\bar{b} \to \mu(4)X$.

# 11 Muon trigger performance for $Z \to \mu^+\mu^-$

## 11.1 The "tag and probe" method

The trigger efficiency is a fundamental parameter in physics analyses and therefore it is important to have several independent methods for estimating it. The "Tag and Probe" method is a concrete application of a data-driven technique for performance analysis. This method is based on the definition of a "probe-like" object, used to make the performance measurement, within a properly "tagged" sample of events. Physics processes suitable for this method are generally those characterized by a double-object final state signature. The decay of the Z provides two high-$p_T$ muons that can lead to two trigger tracks in the Inner Detector and Muon Spectrometer and to a combined object. These two measurements are in principle independent, thought not necessarily uncorrelated.

"Tagged" events require one triggered track with $p_T > 20$ GeV and "Probe" objects can be defined as *Inner Detector offline reconstructed tracks (ID-Probe)*, where measurements are referred to the offline Inner Detector reconstruction efficiency (~100%), or as *Muon Spectrometer offline reconstructed tracks (MS-Probe)*, where values are normalized to the offline Muon Spectrometer reconstruction efficiency (standalone or combined with the Inner Detector).

The trigger performance is measured by checking for L1, L2, and EF trigger tracks associated with each probe object. A schematic illustration of the method is shown in Fig. 30. It must be verified that selected tracks come from a Z decay. A background process with two isolated tracks in the Inner Detector, of which only one is a real muon, would introduce a systematic error in the efficiency evaluation. For this reason, cuts have to be applied in order to select a clean signal sample.
A significant background contribution is expected from QCD processes, which have large cross-sections.

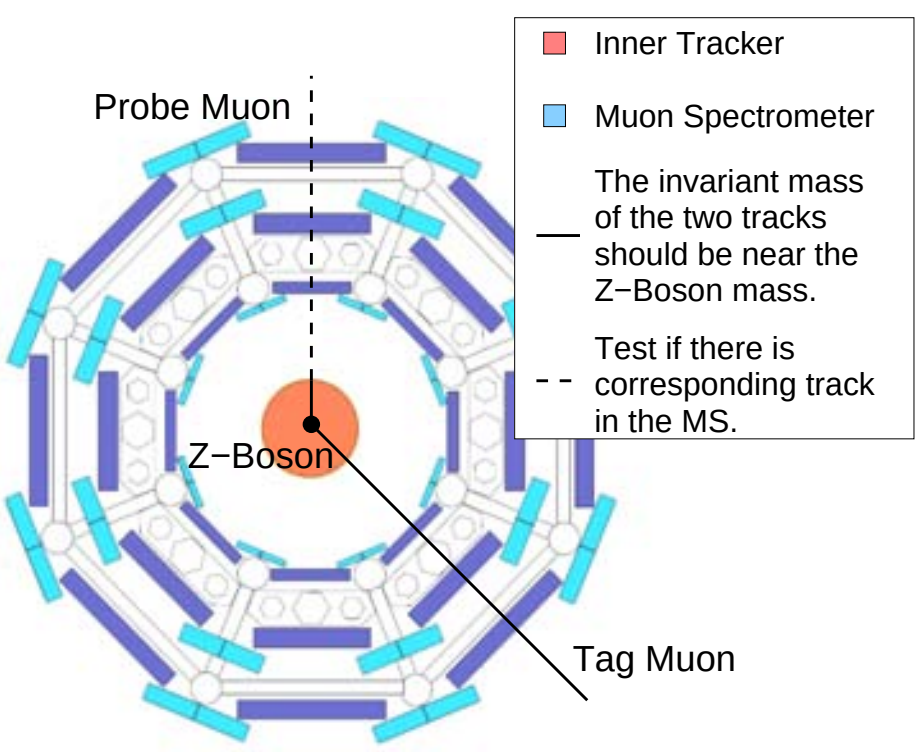


| **Process** | **Generation cuts** | **$\sigma$ [pb]** |
|---|---|---|
| $Z \to \mu^+\mu^-$ | $M_{\mu\mu} > 60$ GeV/$c^2$<br>$1\mu$: $\lvert\eta\rvert < 2.8, p_T > 5$ GeV | 1497 |
| $W \to \mu\nu$ | $1\mu$: $\lvert\eta\rvert < 2.8, p_T > 5$ GeV | 11946 |
| $BB \to \mu\mu X$ | $1\mu$: $\lvert\eta\rvert < 2.5, p_T > 15$ GeV<br>$1\mu$: $\lvert\eta\rvert < 2.5, p_T > 5$ GeV | 4000 |
| $t\bar{t} \to W^+bW^-b$ | only leptonic decay | 461 |
| $Z \to \tau^+\tau^-$ | $\tau\tau \to ll, M_{\mu\mu} > 60$ GeV/$c^2$<br>$1\mu$: $\lvert\eta\rvert < 2.8, p_T > 5$ GeV | 77 |

Figure 30: Illustration of the Tag and Probe method (left) and cross-sections with generation cuts for signal and background processes (right).

This background has been studied by considering the dominant contribution of muons from decays of *B*-meson pairs. Also the muonic W boson decay, which can give a higher energetic muon plus an additional muon from a QCD jet and the $Z \rightarrow \tau^+\tau^- \rightarrow \mu^+\nu_\mu\bar{\nu_\tau}\ \mu^-\bar{\nu_\mu}\nu_\tau$ process have been considered.

Moreover the top-pair production cross-section at LHC is of the same order of magnitude as $Z$ boson cross-section. Top quarks decay with a 99.9% probability into a $W$boson and a $b$ quark. Therefore muons originating from $W$ boson and $b$-quark decays can also give a signal-like signature. Cross-sections and generation cuts of the processes considered are reported in Fig. 30. PYTHIA [5] is used to generate the processes.

Another possible source of background is muons from cosmic-rays. An estimation of cosmic rates in the trigger system has been done in Ref. [10] and shows a negligible effect on trigger performance.

The isolation variables chosen for this analysis are the number of reconstructed tracks in the Inner Detector ($N^{ID}_{cone}$), the sum of $p_T$ of the Inner Detector tracks ($\sum p^{ID}_{T,cone}$), the energy of a jet candidate ($E^{jet}_{cone}$) and the sum of reconstructed energy in the cells of the Calorimeter ($\sum E^{EM}_{cone}$). Muons from QCD processes tend to be produced within a large cascade of other particles and therefore should not appear isolated in the detector. In the case of the decay of top pairs, one highly energetic and isolated muon can come from one $W$ boson decay while the second $W$ boson can decay leptonically into a high-$p_T$ electron which appears as an isolated track in the Inner Detector. In order to not count this as a false probe, electrons are vetoed. The values of the selection cuts applied in this analysis have been defined in [11]. The isolation cuts allow for background rejection of approximately 99% while retaining a signal efficiency of about 76%. After applying the probe selection cuts the signal to background ratio is more than $10^3$. In addition, probe muons selected from background processes can be associated to trigger tracks and hence have no negative impact on trigger efficiency measurements.

## 11.2 Determination of trigger efficiencies

Two measurement scenarios have been studied:

- *Low luminosity* ($\int \mathscr{L} dt \simeq 50\ \text{pb}^{-1}$): in order to not rely on the combined reconstruction based on Inner Detector and Muon Spectrometer matching, only the tracks from the Muon Spectrometer are used. The isolation cuts are also based only on Inner Detector quantities;
- *High luminosity* ($\int \mathscr{L} dt \simeq 1000\ pb^{-1}$): full combined information from Inner Detector and Muon Spectrometer is used and also Calorimeter based cuts are applied to select isolated tracks.

In each scenario both the ID- and MS-Probe methods have been studied. The efficiency dependence on $\phi$ and $\eta$ is determined by the Muon Spectrometer layout. A $p_T$ cut of 20 GeV has been applied on the probe tracks to test the system in its plateau region. The efficiency as a function of $\text{p}_T$ has been also estimated from data in the *high luminosity* scenario.

### 11.2.1 Low luminosity measurements

The relative efficiency as a function of $\eta$, measured at each trigger level, is shown in Fig. 31 using the ID-probe. L1 acceptance losses are related to an incomplete coverage of the trigger detectors due to the presence of support and access structures. The L2 efficiency, with respect to the L1 selection, is about 96% in the barrel region with a small decrease in the endcap, an improvement is expected due to optimization of the TGC cabling in next software releases. The EF shows an $\eta$ efficiency distribution, with respect to L2, close to 99% in the barrel region and a very small decrease from $|\eta| > 2.0$. In the region $1.05 < |\eta| < 1.3$ the absence of the some MDT chambers in the ATLAS initial layout resulted in an efficiency loss of about 10%.[2]

[2] The missing chambers are scheduled to be installed by the end of 2009.

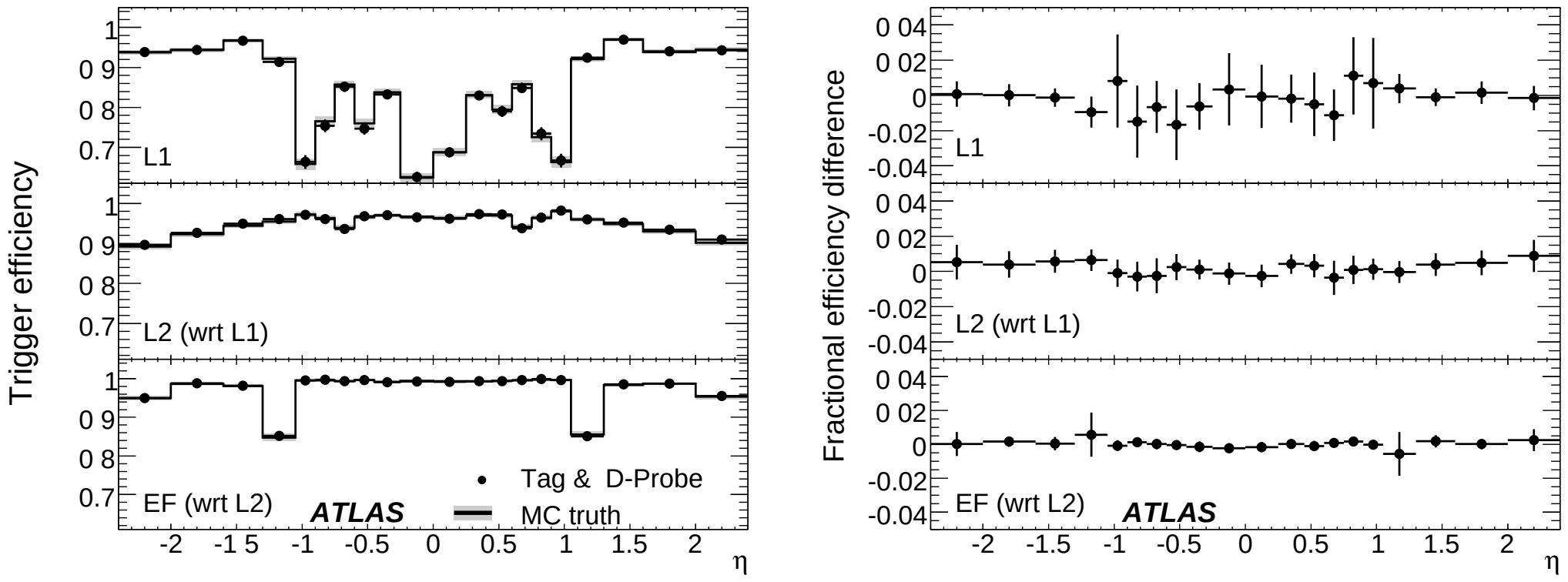


Figure 31: The muon trigger efficiency for each trigger level (left) and fractional efficiency difference (right) as a function of $\eta$ in the *low luminosity* scenario using the ID-Probe. The efficiencies determined with the Tag and Probe method are compared to those calculated in a Monte Carlo truth-based analysis.

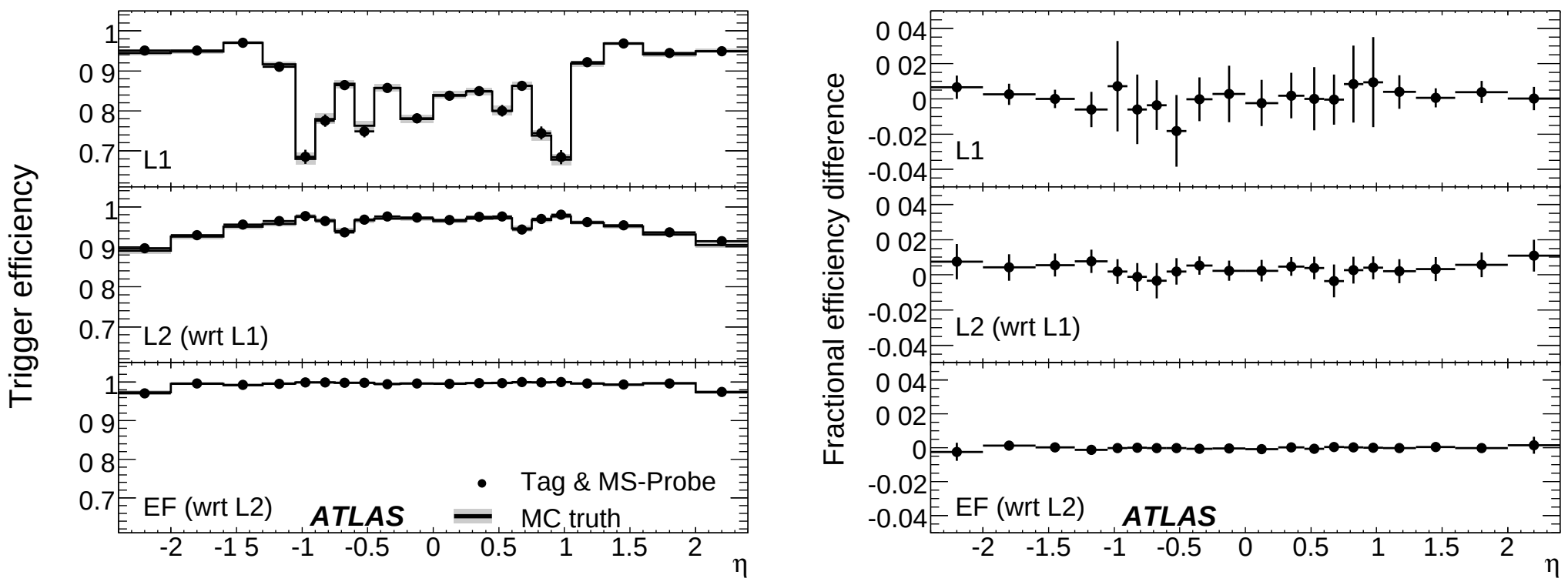


Figure 32: The muon trigger efficiency for each trigger level (left) and fractional efficiency difference (right) as a function of $\eta$ in the *low luminosity* scenario using the MS-Probe. The efficiencies determined with the Tag and Probe method are compared to those calculated in a Monte Carlo truth-based analysis.

The observed agreement with the Monte Carlo truth-based analysis is very good. In order to quantitatively estimate the bin-by-bin differences the "fractional efficiency difference"

$$\frac{\varepsilon_{Tag\&Probe} - \varepsilon_{MC}}{\varepsilon_{MC}} \tag{5}$$

has been computed. This quantity is shown for each trigger level in Fig. 31 as a function of $\eta$.
The agreement between Tag and Probe method and Monte Carlo analysis is very high, more than 99% over all the trigger coverage. The only observed deviations, at the level of 2%, are found in the central crack at $\eta = 0$ and in the transitions from barrel to endcap at $|\eta| = 1.05$. The results obtained by the application of the Tag and Probe method using the standalone MS-Probe are reported in Fig. 32 as a functions of $\eta$. With respect to the values shown in Fig. 31 inefficiencies due to L1 acceptance cracks are partially factorized in the offline muon reconstruction efficiency of the MS-Probe (e.g. the $\eta = 0$ region.). The same effect is clearly evident at the EF level for the efficiency loss at $1.05 < |\eta| < 1.3$ visible in Fig. 31.

Table 15 shows the uncertainties on the overall trigger efficiency in each case, calculated also only in barrel and endcap regions. The statistical uncertainty is reported together with expected systematic

| Detector region | Barrel ($\|\eta < 1.05\|$) | Endcap ($1.05 < \|\eta\| < 2.4$) | Overall ($0 < \|\eta\| < 2.4$) |
|---|---|---|---|
| **Low luminosity - ID probe ($\int \mathscr{L} dt = 50\ pb^{-1}$)** | | | |
| Trigger Efficiency | 71.65 | 83.59 | 77.38 |
| Statistical Uncertainty | 0.42 | 0.36 | 0.28 |
| $\|\varepsilon_{TRUTH} - \varepsilon_{TP}\|$ | 0.23 | 0.40 | 0.10 |
| Expected Background Contribution | 0.57 | 0.17 | 0.40 |
| Overall Systematic Uncertainty | 0.61 | 0.43 | 0.41 |
| **Low luminosity - MS probe ($\int \mathscr{L} dt = 50\ pb^{-1}$)** | | | |
| Trigger Efficiency | 76.94 | 87.83 | 82.13 |
| Statistical Uncertainty | 0.41 | 0.34 | 0.27 |
| $\|\varepsilon_{TRUTH} - \varepsilon_{TP}\|$ | 0.17 | 0.64 | 0.33 |
| Expected Background Contribution | 0.01 | 0.00 | 0.01 |
| Overall Systematic Uncertainty | 0.17 | 0.64 | 0.33 |

Table 15: Estimated uncertainties of in-situ determined muon overall trigger efficiency for the *low luminosity* scenario, using an ID- and an MS-Probe track. Systematic uncertainties are reported for background contribution and absolute difference with Monte Carlo truth-based analysis.

| Detector region | Barrel ($\|\eta < 1.05\|$) | Endcap ($1.05 < \|\eta\| < 2.4$) | Overall ($0 < \|\eta\| < 2.4$) |
|---|---|---|---|
| **High luminosity - ID probe ($\int \mathscr{L} dt = 1000\ pb^{-1}$)** | | | |
| Trigger Efficiency | 73.24 | 86.31 | 79.73 |
| Statistical Uncertainty | 0.10 | 0.08 | 0.06 |
| $\|\varepsilon_{TRUTH} - \varepsilon_{TP}\|$ | 0.02 | 0.72 | 0.58 |
| Expected Background Contribution | 0.05 | 0.01 | 0.03 |
| Overall Systematic Uncertainty | 0.05 | 0.72 | 0.58 |

Table 16: Estimated uncertainties of in-situ determined muon overall trigger efficiency for the *high luminosity* scenario using the ID-Probe. Systematic uncertainties are reported for background contribution and absolute difference with Monte Carlo truth-based analysis.

errors. Two sources of systematic uncertainties are considered: the absolute difference with respect to the value measured in a Monte Carlo truth-based analysis and the background contribution, evaluated by comparing the efficiency calculated with Tag and Probe method using only the signal sample and using a cross-section weighted sum of all processes. Both systematics are less than 0.5%. A greater background contribution is observed when using ID-Probe, since the isolation is based only on Inner Detector quantities.

### 11.2.2 High luminosity measurements

After early data is collected and analyzed, a better understanding of the detector, in terms of calibration and alignment, will allow to use all the available information such as Calorimeter quantities for track isolation and combination of Inner Detector and Muon Spectrometer tracks. The trigger efficiency measurements from data in this scenario are reported using the *high luminosity* dataset of $\int \mathscr{L} dt = 1000\ pb^{-1}$.

As in the *low luminosity* case the measured differences between the Tag and Probe and Monte Carlo analysis are compatible with zero. Small deviations at the level of 1 to 2% are observed in the endcap for the L2 trigger efficiency in the $|\eta| > 1.5$ region. These effects are expected to be reduced by the L2 algorithm optimization. Results are shown in Table 16. With the addition of the calorimeter-based isolation, the background systematic contribution is reduced by a factor of 10 with respect to the low luminosity scenario. The dependence of the trigger efficiency on the $p_T$ shows the typical shape of a turn-on curve. The sharpness of the curve is related to the finite $p_T$ resolution, $\sim$ 30% at L1, $\sim$ 5% at

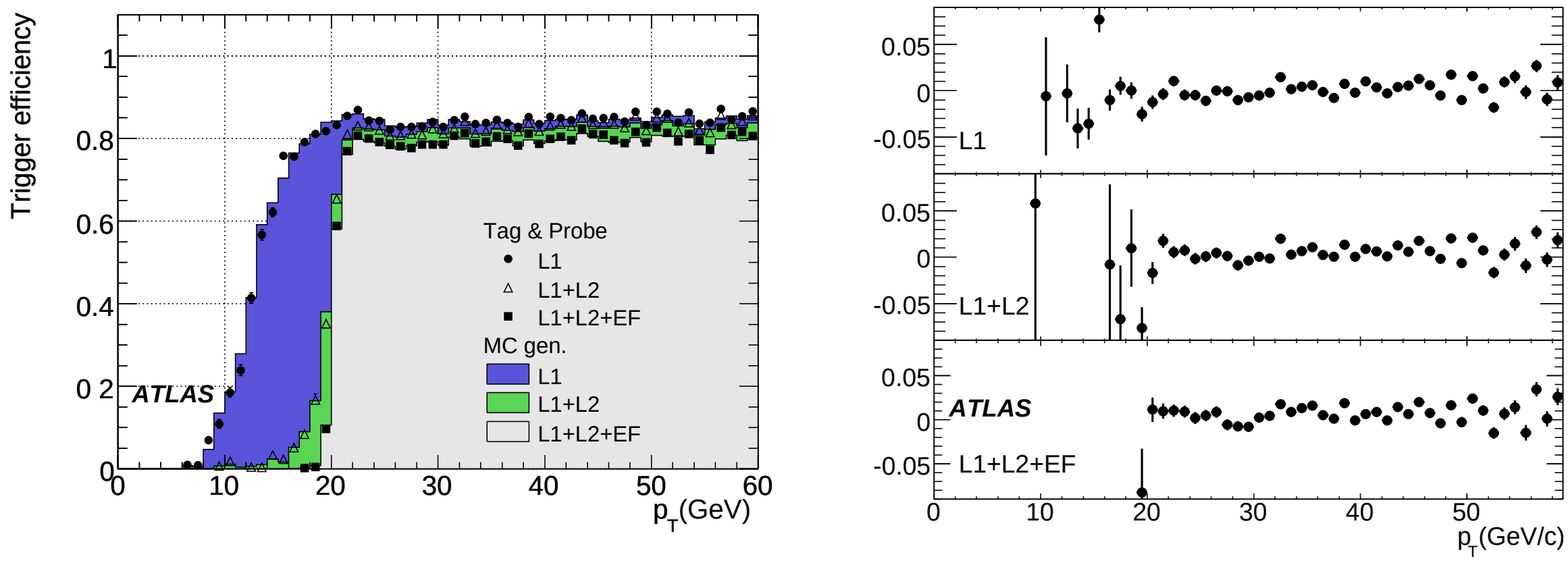


Figure 33: Muon trigger efficiency turn-on curve after each trigger level determined by the Tag and Probe method and by the Monte Carlo truth-based analysis in the *high luminosity* scenario using the ID-Probe (left). In the right plot the fractional efficiency difference is shown.

L2, and $\sim 3\%$ at the EF. Turn-on curves are shown in Fig. 33 using the ID-probe and similar results are obtained with the MS-probe. The turn-on point and the plateau values are correctly reproduced from data. The fractional efficiency difference is shown for each trigger level. The disagreement near the threshold is within 5%, due mainly to resolution effects, while in the plateau region the observed difference is less than $1\%$.

# 12 High p$_T$ Dimuon final states

In principle, the high mass dilepton/diphoton resonance search should have a fairly straightforward trigger strategy as there are very high energy leptons in the event. However, there are several questions that remain: what trigger requirements are optimal for the analysis? What $p_T$ thresholds and object quality selection should be applied? How can one estimate the trigger efficiency from data for such rare (or non-existent) events? Are the same object quality requirements that are appropriate for lower $p_T$ objects appropriate for very high energy objects?

This Section addresses these questions, evaluates the trigger efficiency for the signal samples of interest, and discusses the trigger strategy for the earliest data taking periods. It is expected that during both low and high luminosity periods there will be an unprescaled single muon trigger without an isolation requirement. The 20 or 40 GeV threshold are expected to be highly efficient for a high mass resonance decaying into two muons.

## 12.1 Efficiency estimate

The muon trigger efficiency is estimated using several methods. The first method is to rely on simulation; while this is the simplest and most direct method it is believed to be somewhat more optimistic (better resolution, higher efficiency). Therefore, the trigger efficiency is also estimated using methods which can be applied to real data.

The trigger efficiencies are calculated with respect to the offline event selection. Two combined muons are required to satisfy the cuts: $|\eta| < 2.7$, $p_T > 30$ GeV, track fit $\frac{\chi^2}{\text{D.O.F}} < 10$ and Inner Detector and Muon Spectrometer track match $\frac{\chi^2}{\text{D.O.F}} < 10$. The trigger efficiencies for the dimuon heavy resonance Monte Carlo samples are shown in Table 17. The efficiency as a function of $p_T$ has been fit to

| Sample | L1 % | L2 % | EF % | Total Trigger Efficiency % |
|---|---|---|---|---|
| 400 GeV $\rho_T/\omega_T$ | $97.6 \pm 0.10$ | $98.8 \pm 0.07$ | $99.5 \pm 0.05$ | $96.0 \pm 0.13$ |
| 600 GeV $\rho_T/\omega_T$ | $98.1 \pm 0.08$ | $98.5 \pm 0.08$ | $99.2 \pm 0.06$ | $95.9 \pm 0.13$ |
| 800 GeV $\rho_T/\omega_T$ | $97.6 \pm 0.10$ | $98.7 \pm 0.07$ | $99.2 \pm 0.05$ | $95.6 \pm 0.13$ |
| 1 TeV $\rho_T/\omega_T$ | $97.6 \pm 0.09$ | $98.7 \pm 0.07$ | $99.2 \pm 0.05$ | $95.6 \pm 0.12$ |
| 1 TeV Z' (E6) | $97.8 \pm 0.09$ | $98.9 \pm 0.06$ | $99.5 \pm 0.04$ | $96.3 \pm 0.1$ |
| 2 TeV Z' (SSM) | $97.6 \pm 0.14$ | $98.7 \pm 0.11$ | $98.9 \pm 0.10$ | $95.3 \pm 0.2$ |

Table 17: Trigger efficiencies of dimuon resonance samples. For the meaning of E6 and SSM see [12].

| Trigger Level | $A_0$ | $A_1$ | $A_2$ |
|---|---|---|---|
| L1 | $12.5 \pm 0.3$ | $3.7 \pm 0.4$ | $0.845 \pm 0.02$ |
| L2 | $19.6 \pm 0.2$ | $1.59 \pm 0.19$ | $0.976 \pm 0.02$ |
| EF | $19.5 \pm 0.4$ | $1.56 \pm 0.3$ | $0.931 \pm 0.01$ |

Table 18: Fitted parameter for the L1, L2, and EF of the trigger $p_T$ turn on curves

.

$$f(p_T) = 0.5 \cdot A_2 \cdot (1.0 + erf(\frac{p_T - A_0}{\sqrt{2} \cdot A_1})) \tag{6}$$

where $erf$ is the error function, $A_0$, $A_1$, and $A_2$ are the fit parameters which represent the $p_T$ value at which the efficiency reaches half its maximum value, the slope of the turn-on curve, and the maximum efficiency in the plateau region, respectively.

There are several methods to evaluate the trigger efficiency from the data itself. A possible one is to look at the trigger efficiency for a known experimentally clean signature that is similar to the final state of interest; $Z \rightarrow \mu^+\mu^-$ is one of such signatures. Since the Z is light compared to the total center of mass energy, it can be produced with a significant $p_T$ distribution. The trigger efficiency on the Z can be measured and extrapolated to high $p_T$. The advantage of this method is that it uses data to measure the trigger efficiency which is the most accurate method of measuring the Z trigger efficiency. A disadvantage is that the muon trigger efficiency is being extrapolated to a $p_T$ by a factor of 10 higher than the mean $p_T$ of the muons from the Z decay.

The strategy of evaluating the trigger efficiency from data is as follows. It is first necessary to use one of several methods to estimate the muon trigger efficiency as a function of the muon $p_T$ and its uncertainty. The single object trigger efficiency allow the construction of the probability for an event with N objects to pass the trigger. This probability can be written as:

$$P = 1 - \prod_{i=1}^{N}(1 - P_i) \tag{7}$$

where $P_i$ is the probability for the $i$-th object to pass the trigger.

Two common methods that have been used extensively at the Tevatron are the selection by orthogonal triggers and the 'Tag and Probe' method using $Z \rightarrow \mu\mu$ decay. The 'Tag and Probe' requires two offline muons to have an invariant mass within 12 GeV$^2$ of 91.1 GeV$^2$. The turn on curves as a function of the offline muon $p_T$, obtained using this method, is fit to equation 6. This procedure was repeated for all three trigger levels and the results are summarized in Table 18.

A second possible method of evaluating the trigger efficiency with data is by the method of orthogonal triggers. To obtain a sample of unbiased events we select events that pass one of the calorimeter based triggers, the single 20 GeV jet trigger. We then perform the offline analysis and require that we have a dimuon pair using identical event selection to the 'Tag and Probe' analysis. From this sample we

| *Sample* | L1Mu20 Efficiency % | L2Mu20 Efficiency % | EFMu20 Efficiency | Total Efficiency |
|---|---|---|---|---|
| Z' 1 TeV (SSM) | 97.7 $\pm$ 0.11 | 99.0 $\pm$ 0.07 | 99.6 $\pm$ 0.04 | 96.3 $\pm$ 0.01 |
| $Z \to \mu^+\mu^-$ | 97.83 $\pm$ 0.04 | 98.86 $\pm$ 0.03 | 99.52 $\pm$ 0.02 | 96.26 $\pm$ 0.05 |

Table 19: L1Mu20 trigger efficiencies at L1, L2, and Event Filter w.r.t offline reconstruction using orthogonal trigger selection to record events

.

simply check the fraction of events that pass the L1, L2, and EF trigger conditions for the 20 GeV muon trigger. The results are shown in Table 19 and are in good agreement with the 'Tag and Probe' method and direct emulation of the trigger on the Monte Carlo sample. Unfortunately, in the real experiment a single jet trigger with a threshold of 20 GeV would be very highly prescaled and hence will suffer from poor statistics. Events that passed any calorimeter trigger could be used for this study if biases in the event topology were taken into account, however, such a study is beyond the scope of this note.

We have developed two methods that could be used to evaluate the trigger efficiency from data. Extraction of the muon trigger efficiency as a function of the reconstructed muon kinematics via a tag and probe method and an orthogonal trigger method agree well with the simulated trigger efficiency. These methods will allow us to more accurately estimate the trigger efficiency for LHC data.

## 13 Summary

In this paper Muon trigger baseline performance and rates for initial and standard LHC operation have been presented. Trigger efficiency has been studied in detail in a wide energy range using single muon simulated samples. From efficiencies the muon rates have been evaluated. It should be noted that due to the uncertainties of the inclusive muons cross-sections, rates could vary significantly and different threshold cuts could be adopted. A further rate reduction should come from dedicated strategies to reject muon from in-flight decays of $K$ and $\pi$; in this paper a preliminary analysis is presented at Event Filter. It is demonstrated that a good rejection can be achieved with contained losses of prompt muons.

The possibility to select at the ATLAS second level trigger with high efficiency isolated muons from $W$ and $Z$ decays reducing the ones from heavy quark decays has been studied in depth. Although electronic readout and pileup noise have been simulated, no cavern background has been yet included. A factor of ten reduction on high-$p_T$ muons from heavy-quark decays has been obtained while maintaining a 95% efficiency on $Z \to \mu^+\mu^-$ final state. Next step will be to investigate how much the use of the longitudinal granularity of the calorimeters and inner tracker detector will increase the muon isolation rejection power.

The overall performance of the TileCal muon tagging algorithm has been presented, using MC samples of single muons and inclusive B-Physics processes, including minimum-bias pileup at low luminosity.

We finally addressed the question of how the muon trigger efficiency can be measured with $Z \to \mu^+\mu^-$ and $Z' \to \mu\mu$ using the tag and probe method. This technique shows a very good agreement with results based on Monte Carlo studies.

# HLT $b$-Tagging Performance and Strategies

**Abstract**
The ability to trigger on $b$-jets improves the flexibility and physics performance of the High Level Trigger (HLT), especially for topologies containing more than one $b$-jet. It will be shown that the acceptance for $b$-jets can be increased and background reduced by lowering jet transverse energy thresholds and applying $b$-tagging selections based on the impact parameter of tracks in jets. This note reviews the $b$-jet selection in the HLT and discusses its integration into the ATLAS trigger menu.

## 1 Introduction

Final states containing $b$-jets have been proposed as signatures with substantial discovery potential in a variety of physics channels. The ability to separate $b$-jets from light-quark and gluon jets is thus an important ingredient of the online selection strategy in ATLAS.
One of the most interesting physics cases addressed by such a $b$-jet trigger selection involves events with final states containing four $b$-jets. This event class is relevant for Higgs bosons search in the low mass range, $m_H < 130$ GeV. The most promising channels are the $H \to b\bar{b}$ decay, where the Standard Model Higgs boson is produced by way of the associated production channel $t\bar{t}H$ and, in supersymmetric theories, the channels $b\bar{b}H$, $b\bar{b}A$ with $H/A \to b\bar{b}$ or $H \to hh \to b\bar{b}b\bar{b}$.

The selection of $b$-jets at the trigger level is mainly meant to improve the flexibility of the HLT, extending its physics performance for the topologies described above. This is achieved by increasing the acceptance for signal events, while concurrently reducing the background.

The $b$-jet selection relies on tracking information which is only available starting with the Second Level Trigger (L2). Therefore, the acceptance for signal can only be increased by simultaneously lowering L1 jet thresholds and applying a more discriminating $b$-jet selection in the High Level Trigger (L2 and EF). High rejection power from the $b$-jet trigger is required to compensate for less rejection due to lower L1 thresholds and thereby to comply with L2 and EF output rate limitations.

## 2 Monte Carlo samples

The $b$-tagging performance on single jets, presented in this note, is evaluated on $b$-jets from $H \to b\bar{b}$ decays, where the Higgs boson has a mass of 120 GeV and is produced in association with a $W$ decaying leptonically. The standard background for single-jet studies are the corresponding $u$-jets, obtained by artificially replacing the $b$-quarks from the Higgs decay with $u$-quarks. While these events imprecisely model the real background from light-flavour jets they can be seen as a worst case scenario since the kinematical properties of signal and background are very similar.

Even in this very simple situation, the association between Regions of Interest (RoI), identified by the L1 trigger, and jets is not uniquely defined: a generic $x$-quark in the final state of an interaction or a decay can radiate gluons and, therefore, change its direction. An RoI from $H \to b\bar{b}$ or $H \to u\bar{u}$ is labeled as $x$-jet ($x = b, u$) if an $x$-quark from the original hard process points, after final state radiation, along the RoI direction within an angular distance of $\Delta R = \sqrt{\Delta\eta^2 + \Delta\phi^2} < 0.1$.

In order to evaluate the rate of the $b$-jet trigger menu, the rejection power must be evaluated on a more representative background sample. As for many of the trigger studies in ATLAS, dijet samples are chosen for this purpose since they correctly include all contributions to the $b$-tagging background, including $c$-quarks and taus.

All data samples studied in this note have been generated without pile-up, leaving the influence of pile-up for further studies. The activity due to underlying event is taken into account since it is automatically included in the PYTHIA [1] event generation.

# 3 HLT $b$-jet selection

## 3.1 L1 configuration

The HLT reconstruction starts from the RoIs selected by the L1 trigger [2]. In particular, the $b$-jet trigger starts from a L1 jet-RoI $\Delta\eta \times \Delta\phi = 0.8 \times 0.8$ and performs track and vertex reconstruction in a smaller RoI $\Delta\eta \times \Delta\phi = 0.4 \times 0.4$ in order to reduce data access and consequently processing time.

## 3.2 $b$-jet trigger feature extraction algorithms

The first step in the $b$-jet trigger chain, both at L2 and EF, is the reconstruction of the relevant quantities needed to perform the selection. The $b$-jet RoIs can be separated from light jet RoIs using the impact parameters of the charged tracks, the properties of reconstructed secondary vertices, or soft leptons; all these quantities are related to the $b$-quark lifetime and to its decay properties.

The present $b$-jet trigger implementation relies only on the impact parameters of charged tracks. Primary vertex reconstruction is performed only in the $z$ direction while its coordinates in the transverse plane are assumed to be compatible with the origin.

Track reconstruction algorithms are described, together with their performance, in [3]. The two Inner Detector tracking algorithms available at L2 show equivalent performance when operating on jet samples [3]. Thus to avoid unnecessary comparisons, the results obtained with the SiTrack algorithm are presented. For EF track reconstruction, the algorithm corresponding to that used for offline reconstruction has been adopted (NewTracking).

### 3.2.1 Primary vertex reconstruction

Along the $z$ direction no *a priori* knowledge of primary vertex $z_{vtx}$ is available; consequently, this has to be reconstructed, starting from the tracks available in the RoI. This information is needed for the correct evaluation of the longitudinal parameter of each track with respect to the primary interaction position.

The adopted algorithm, a simple histogramming method based on a sliding window, yields an efficiency of 98(99)% and a resolution on $z_{vtx}$ of about $120(100)\mu$m at L2(EF) as illustrated in Fig. 1.

## 3.3 Tagging variables

The HLT $b$-jet tagging methods are based on the transverse and longitudinal impact parameters of the reconstructed tracks. Since the methods are the same for L2 and EF they will be described using L2 variables only.

### 3.3.1 Transverse impact parameter

The most natural choice is to build the $b$-tagging discriminant variable from the transverse impact parameter $d_0$ of the reconstructed tracks. Since the hadrons containing $b$-quarks have a finite lifetime ($\tau \sim$ 1.6 ps), tracks from their decays are characterized by large $d_0$ values, while tracks from $u$-jets come dominantly from the primary vertex ($d_{vtx} = 0$).

In particular, the significance of the transverse impact parameter $S = d_0/\sigma(d_0)$ is used, where $\sigma(d_0)$ is the error on the impact parameter. The error on the transverse impact parameter at L2 is parametrized

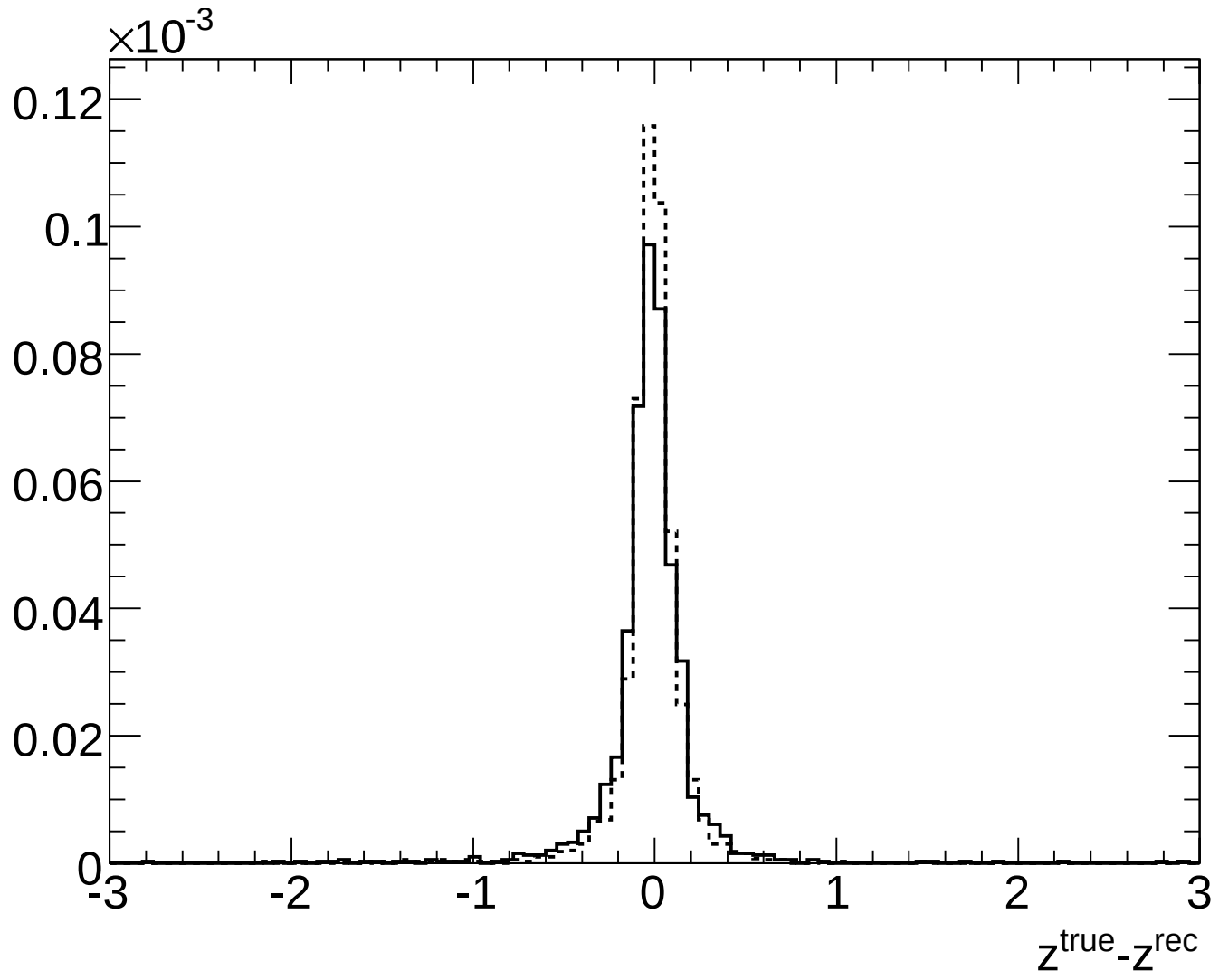


Figure 1: The distribution of the difference between the true and the reconstructed $z$ primary vertex coordinates at L2 (full line) and EF (dashed line). The widths as determined by a fit to the distributions are 120 $\mu$m and 100 $\mu$m, respectively.

as a function of reconstructed $p_T$ as:

$$\sigma(d_0) = \sqrt{p_0^2 + \left(\frac{p_1}{p_T}\right)^{p_2}}$$

where $p_0$ is the asymptotic term, $p_1$ is the term due to multiple scattering, and $p_2$ is the exponent of the multiple scattering contribution (close to two). Although L2 tracking algorithms have recently reached a good level of precision, the above error parametrization is still applicable at L2 during early data, while for the EF the reconstructed error is used.

Figure 2 shows the distributions of the impact parameter significance $d_0/\sigma(d_0)$ for $b$-jets and light jets at L2. The significance has been rescaled according to the function $f(x) = log(1+|x|)$ in order to have a reasonably uniform bin population along the $x$ axis. From these plots it can be guessed that the impact parameter significance is a promising choice for the discriminant variable, since the two distribution are very well separated.

### 3.3.2 Longitudinal impact parameter

The longitudinal impact parameter ($z_0$), i.e. the track's z-intercept, can be adopted, as well as the transverse impact parameter, to discriminate between $b$-jets and light jets. After the primary vertex position has been reconstructed, the $\delta z_0 = z_0 - z_{vtx}$ variable can be used to form a discriminant which can then be used for $b$-jet selection. Figure 3 shows the distributions of the longitudinal impact parameter significance ($\delta z_0/\sigma(z_0)$) of $b$-jets and light jets at L2. The significance has been rescaled as described above for the transverse impact parameter.

As for Fig. 2, the signal and background distributions are different although much less so than for the transverse impact parameter significance. From this comparison, it is clear that most of the discrimination will be provided by the measured transverse impact parameter significance. The resolution of the longitudinal impact parameter significance is not as good due to the coarser resolution of the silicon

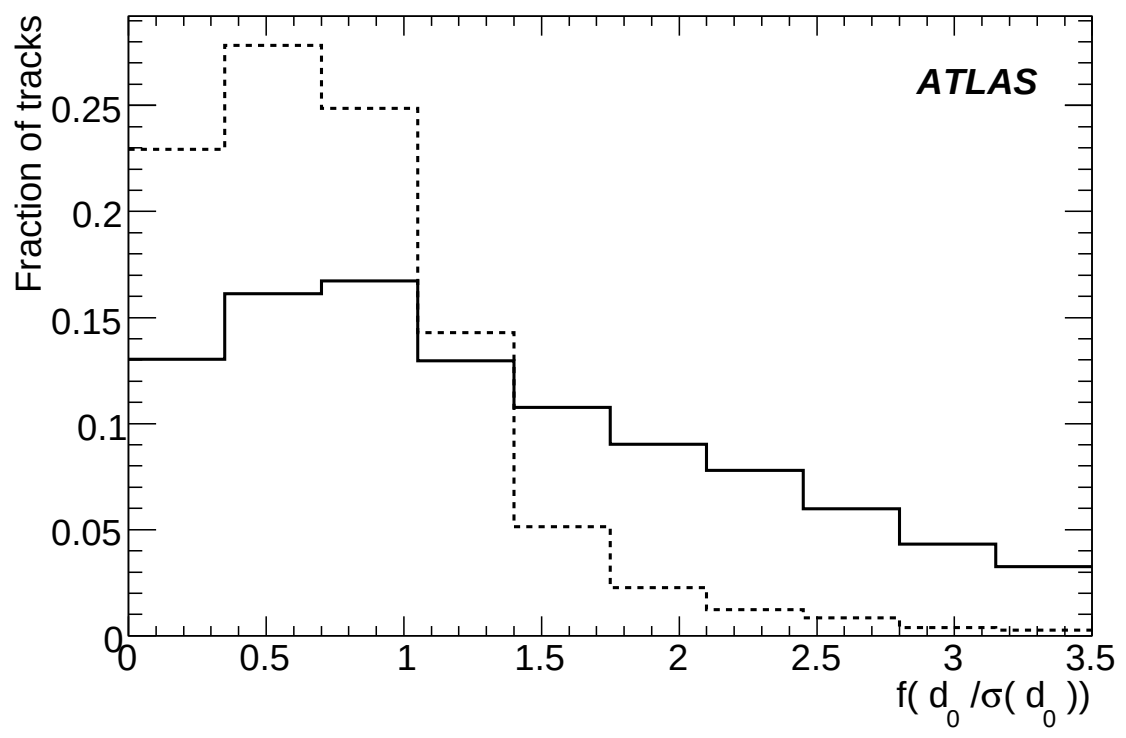


Figure 2: Distribution of the rescaled function (described in the text) of the transverse impact parameter significance for tracks coming from $b$-jets (solid line) and light jets (dashed line) at L2.

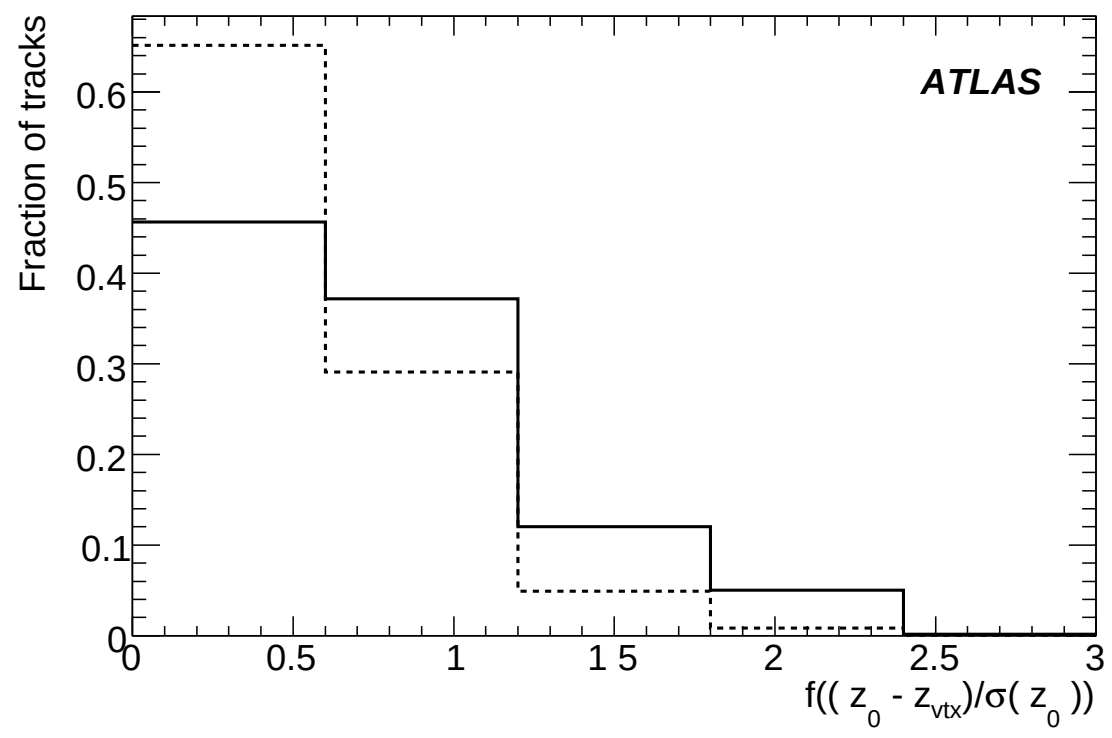


Figure 3: Distribution of the rescaled function (described in the text) of the longitudinal impact parameter significance for tracks coming from $b$-jets (solid line) and light jets (dashed line) at L2.

tracking detectors along the $z$-direction, bigger extrapolation distance from innermost silicon layer hit to primary vertex at high $\eta$, and to the resolution of the reconstructed primary vertex.

### 3.4 HLT $b$-jet tagging methods

In this Section, HLT $b$-tagging methods are described. The likelihood ratio method is quite general and can be applied to different variables while the $\chi^2$ method is essentially designed to test the compatibility of the tracks with respect to the primary vertex using the transverse impact parameter.

The likelihood ratio, using information on the signal and background shape that have to be estimated on real data, is both more powerful and more difficult to tune than the $\chi^2$ method.

#### 3.4.1 The likelihood-ratio method

The likelihood-ratio method is a statistical tool used to separate two or more event classes, and is based on a set of characteristic variables. The likelihood-ratio variable $W$ is evaluated, for a given event, as the ratio between the probability distributions for two alternative hypotheses. In its application to $b$-jet selection, the likelihood-ratio variable is defined as

$$W = S(s)/S(b),$$

where $S(s)$ and $S(b)$ are the probability densities for the signal, the $b$-jets, and the background, represented in this case by the $u$-jets. This variable is widely used to obtain the best possible separation between signal and background, in terms of a single variable, in fits aimed at extracting the fraction of signal events in a given sample. The same variable can be also directly used, as in the $b$-jet selection case, to select signal events, for example by applying a cut on the likelihood-ratio variable itself.

The probability density distributions used in the $b$-tagging application can be functions of some parameter of each track (e.g. the transverse impact parameter $d_0$) or of some collective property of the jet (e.g. its track multiplicity). In the first case, these distributions take the form

$$s(par_1, par_2, par_3, \ldots, par_n),$$
$$b(par_1, par_2, par_3, \ldots, par_n),$$

where the $1,\ldots,n$ indices identify each track belonging to the jet. The corresponding likelihood-ratio variable is thus defined as

$$W = \frac{s(par_1, par_2, par_3, \ldots, par_n)}{b(par_1, par_2, par_3, \ldots, par_n)}$$

Exact evaluation of the $s$ and $b$ functions is very difficult, since it would require an almost infinite amount of simulated data; for example, in order to reasonably populate an $n$-dimensional cube, about 100 entries are needed for each dimension, corresponding to $n^{100}$ tracks; even worse, the number of tracks in a jet is not fixed. However, if we assume that the variables corresponding to different tracks are independent, the ratio between the overall probability densities reduces to the product of the ratios of the single probability densities:

$$W = \prod_{i=1}^{n} \frac{s(par_i)}{b(par_i)},$$

which is much easier to evaluate. In the *b*-tagging case, track parameters have complex correlations which depend on the proper time for the $B$ hadron and on its decay kinematics. Nevertheless, it can be proven that, neglecting these correlations does not invalidate this variable, but results only in a slight reduction of its discriminating power.

The $W$ variable, can take any value between 0 (for the background) and $+\infty$ (for the signal). For practical reasons, it is useful to handle a variable defined on a finite interval; to achieve this, $W$ is usually replaced by another variable

$$X = \frac{W}{1+W},$$

which can only range between 0 and 1.

As an illustration of the method, Fig. 4 and Fig. 5 show the distributions of the discriminant variable $X$ which is based on the combination of the transverse and longitudinal impact parameter for *b*-jets and light jets respectively at L2 and EF. It can be seen that signal events (*b*-jets) accumulate near $X = 1$, while the background (light jets) tends to have $X$ close to 0.

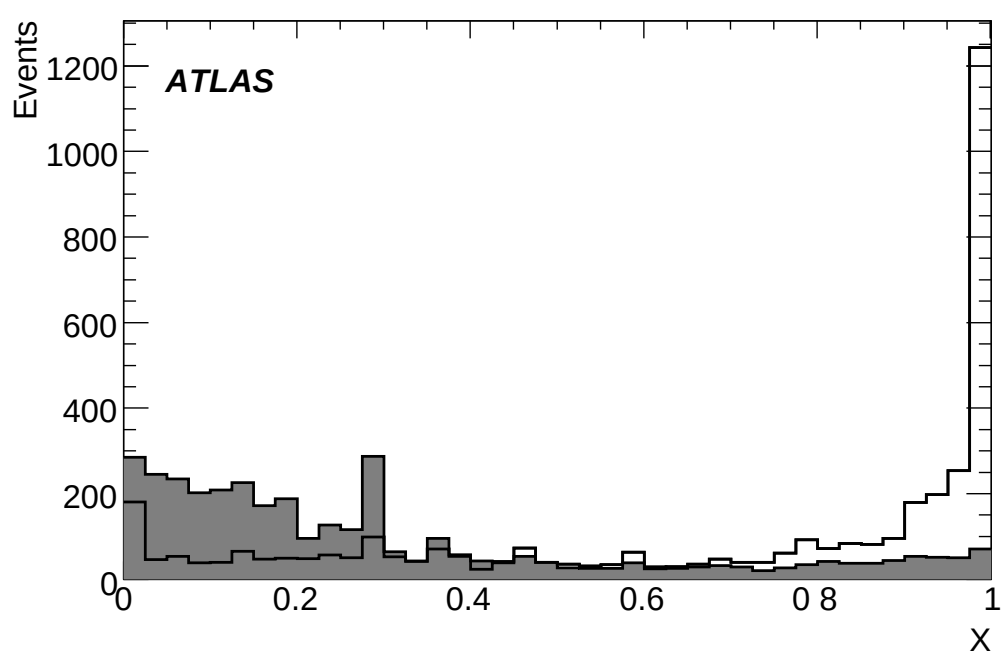


Figure 4: Distribution of the discriminant variable $X$ based on the combination of the transverse and longitudinal impact parameter significances for *b*-jets and *u*-jets (dark area) at L2.

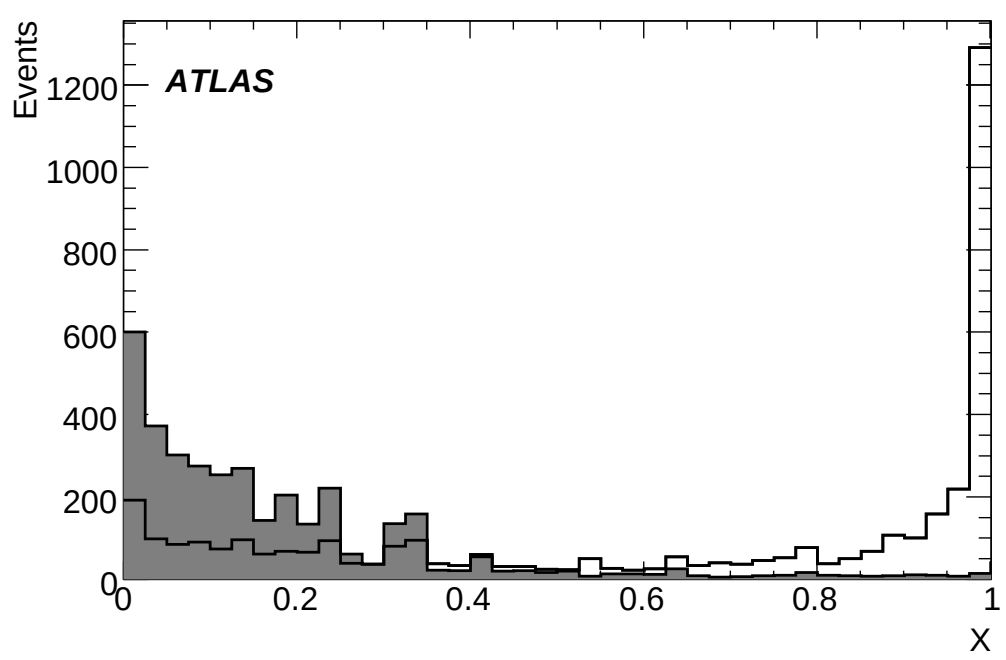


Figure 5: Distribution of the discriminant variable $X$ based on the combination of the transverse and longitudinal impact parameter significances for *b*-jets and *u*-jets (dark area) at EF.

Contrary to the offline *b*-tagging methods based on likelihood ratio, the sign of the impact parameters is currently not used at HLT since the RoI direction does not give a precise estimation of the *b*-jet direction. Future studies will use the impact parameter sign determination respect to a track based cone jet described in the next Section.

### 3.4.2 $\chi^2$ method

The $\chi^2$ method offers an alternative tagging approach that is more robust since it is less dependent on the details of the impact parameter distributions. This method has been studied and characterized only at L2.

The $\chi^2$ method computes the probability for a jet to originate from the primary vertex, based on the *signed* transverse impact parameter significance of tracks pointing to the jet. This technique was originally developed by the ALEPH collaboration and extensively used at LEP and Tevatron experiments [6] [7] [8]. One of the main advantages of this method is that it only relies on the transverse impact parameter significance distribution of prompt tracks in multi-jet events, which can be easily derived completely from real data. On the other hand the performance of this method is limited due to the fact that tracks from beauty and charm particles produce significant tails with negative impact parameters. These negative tails originate from the differences of the direction of the estimated jet and $B$-hadrons and also in the differences of the direction of $B$-hadrons and charmed hadrons in the cascade decays.

The definition of the sign of the transverse impact parameter is based on the angle between the jet axis and the line between the primary vertex and the point of closest approach of the track, such that it is negative when the track appears to originate behind the primary vertex (i.e. when the angle is greater than $\pi/2$) as illustrated in Fig. 6.

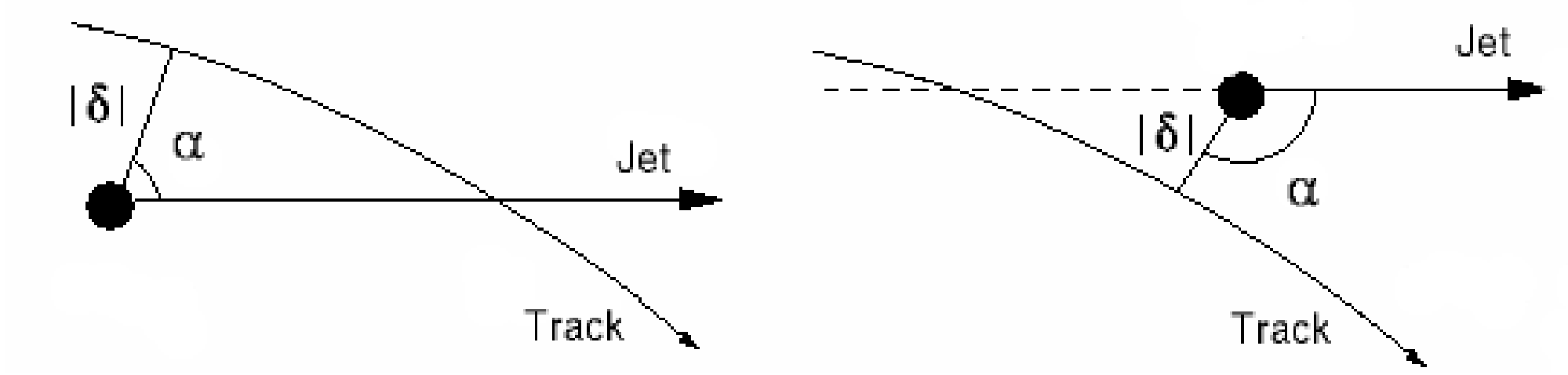


$d_0 = |\delta|$ $d_0 = -|\delta|$

Figure 6: Definition of the sign of the transverse impact parameter. When the angle between the jet axis and the line between the primary vertex and the point of closest approach of the track is lower (greater) than $\pi/2$ the sign is positive (negative).

Tracks from light-quark jets have equal probability to have positive or negative transverse impact parameters, and the width of the signed transverse impact parameter distribution depends on the tracking detector resolution and multiple-scattering effects. The signed transverse impact parameter distribution of tracks from displaced $b$-jets, on the other hand, has a large positive asymmetry due to the fact all that most of the long-lived particles from $b$-hadron decays are produced with positive impact parameters.

Good jet angular resolution is a key to achieving a good $b$/light-quark jet discrimination since the direction of the jet axis enters into the calculation of the sign of the impact parameter. Poor jet angular resolution results in frequent mis-assignment, particularly for tracks with small angle with respect to the jet direction.

In order to improve the resolution of the azimuthal angle ($\phi$) of the jet, a track-based simple cone jet reconstruction algorithm is used instead of the jet-RoI $\phi$ direction. Figure 7 shows that the $\phi$ resolution improves by more than a factor of two when tracks are used to compute the jet direction. The effect of jet $\phi$ resolution on $b$/light quark jet discrimination can be seen in Fig. 8, which shows the distribution of signed transverse impact parameter significance for $b$-jet tracks when $\phi$ is computed using truth, RoI, and track-jet $\phi$ directions.

The negative transverse impact parameter significance is computed using a parameterization for the

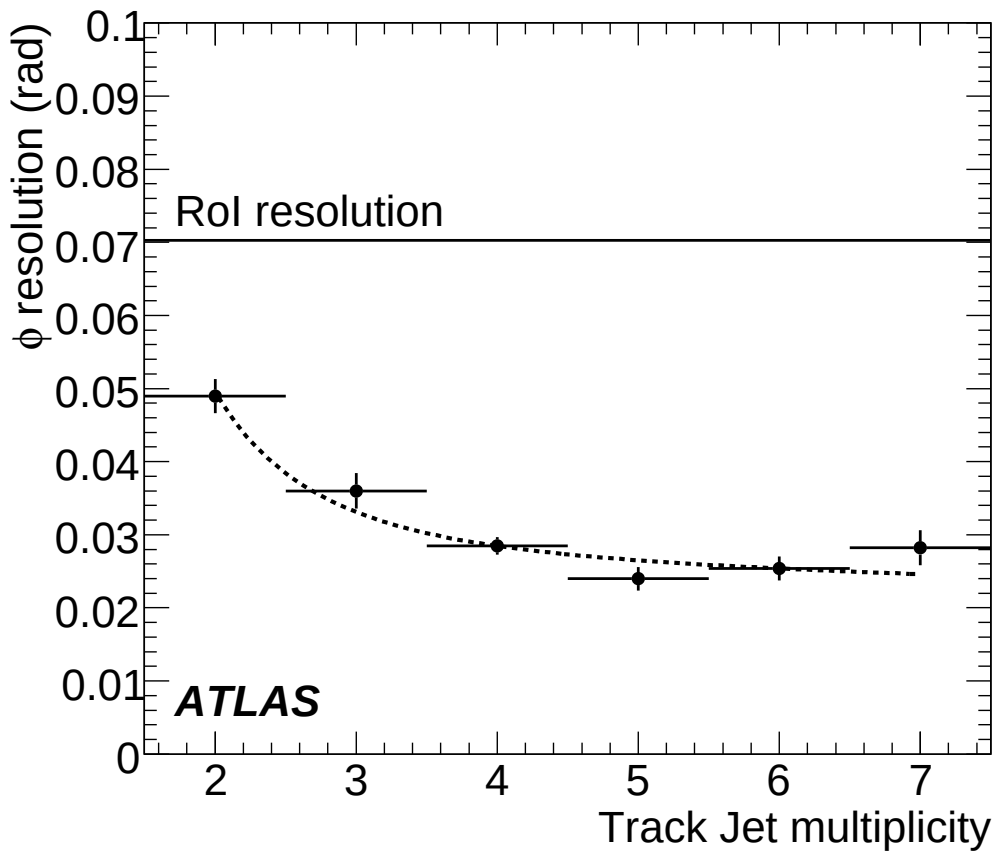


Figure 7: $\Delta\phi$ between the true $b$-quark direction and its estimate using RoIs (solid line) and track-based jets (dashed line), as a function of the track-jet multiplicity.

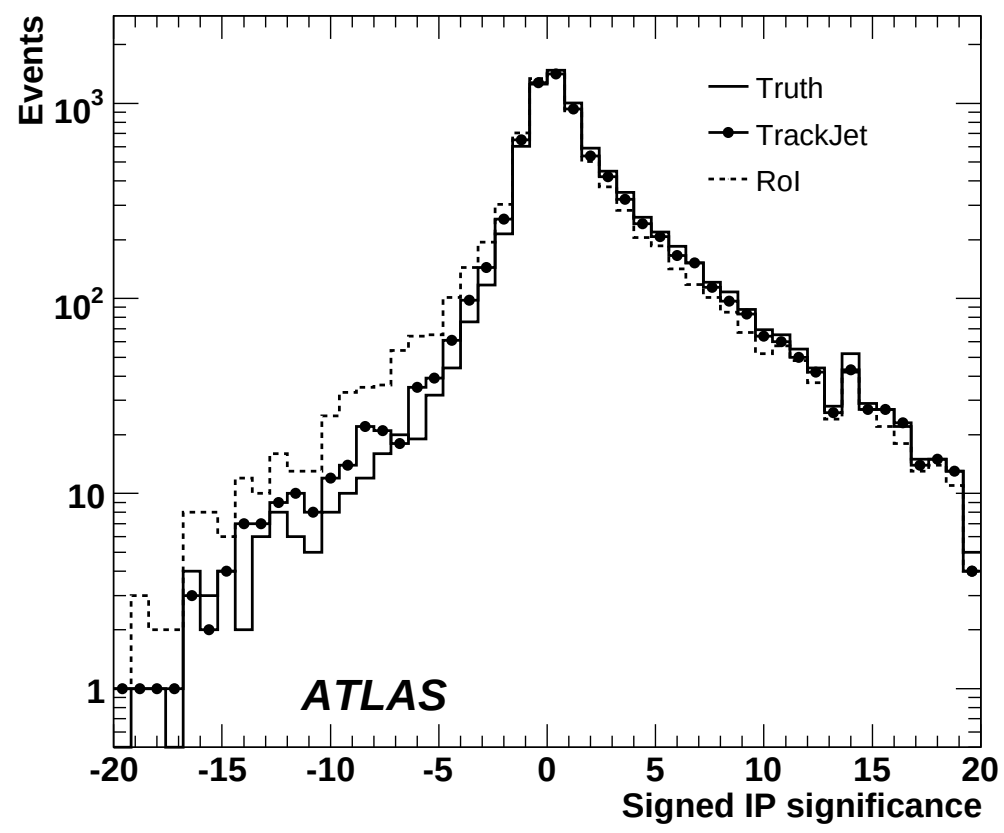


Figure 8: Distribution of the signed impact parameter significance for $b$-jets when the jet $\phi$ direction is computed using truth, track-jet, and RoI information.

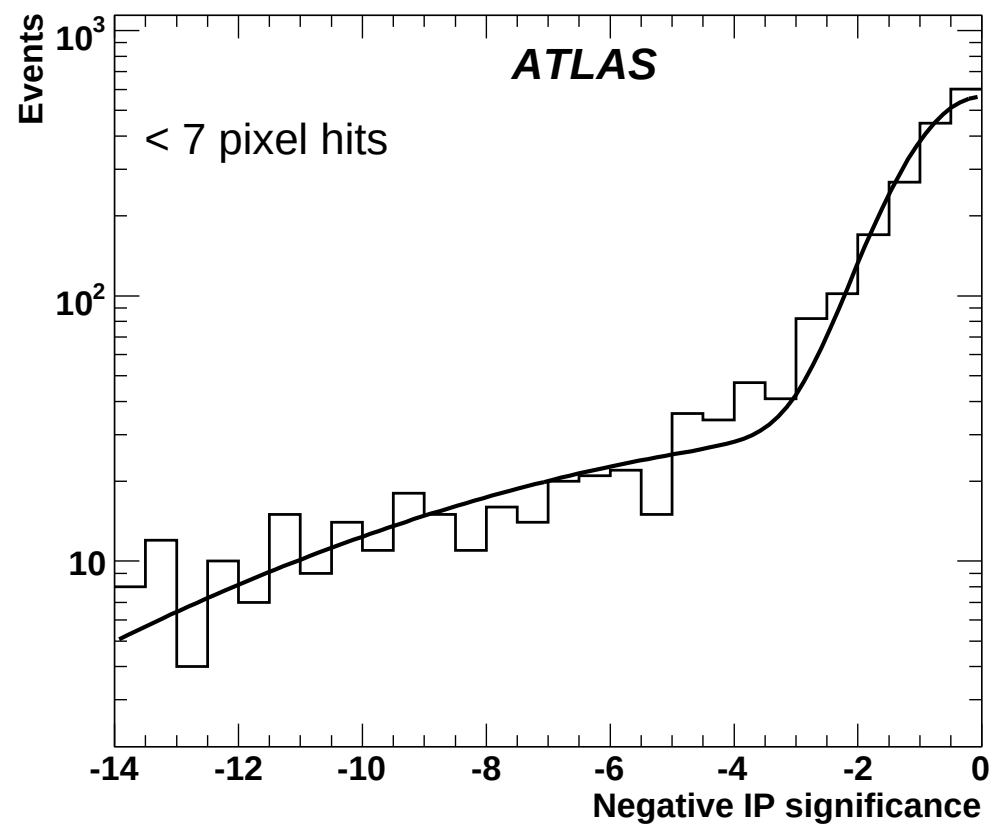


Figure 9: Negative transverse impact parameter significance and the resolution function $R(S)$ for tracks with less than 7 hits in the Silicon detectors.

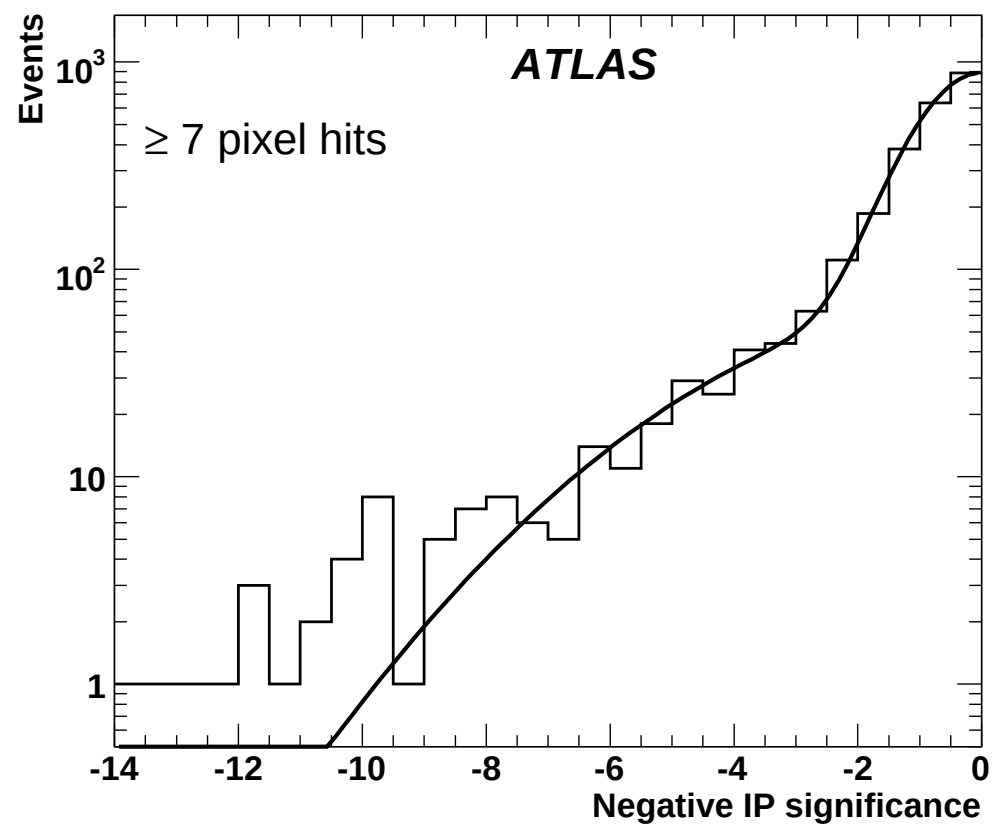


Figure 10: Negative transverse impact parameter significance and the resolution function $R(S)$ for tracks with at least 7 hits in the Silicon detectors.

transverse impact parameter error as a function of $p_{scat} = p \sin\theta^{3/2}$ and the number of hits in the pixel detector. A double-Gaussian fit to the distribution of negative transverse impact parameter significance ($R(S)$) is used to define $p_{trk}(S)$, the probability for a track to originate from a primary vertex:

$$p_{trk}(S) = \frac{\int_{-15}^{-|S|} R(S)dS}{\int_{-15}^{0} R(S)dS} \tag{1}$$

where only tracks with positive impact parameter are used in the calculation. The distribution of negative transverse impact parameter significance and the resolution function $R(S)$ is shown in Fig. 9 and Fig. 10 for tracks with less and more than 7 Silicon detectors hits.

The definition of $p_{trk}$ ensures a uniform distribution between 0 and 1 for tracks originating from the primary vertex. Tracks from displaced $B$ decays result in $p_{trk} \sim 0$.

A $\chi^2$ jet probability is defined by considering the probabilities of all tracks with positive transverse

impact parameter in a jet [6]:

$$p_{jet} = \Pi \sum_{j=0}^{N-1} \frac{(-\log \Pi)^j}{j!} \tag{2}$$

where $\Pi = \prod_1^N p_{trk}(S)$.

Figure 11 shows the $p_{trk}$ distribution for light and *b*-quark jets. The spike at small probability for light quarks tracks is due to tracks from $V^0$ decays, which have positive transverse impact parameter.

Figure 12 shows the $p_{jet}$ distribution for light and *b*-quark jets. Jets are tagged as *b*, if the $p_{jet}$ is below some value, typically between 0.5% and 5%.

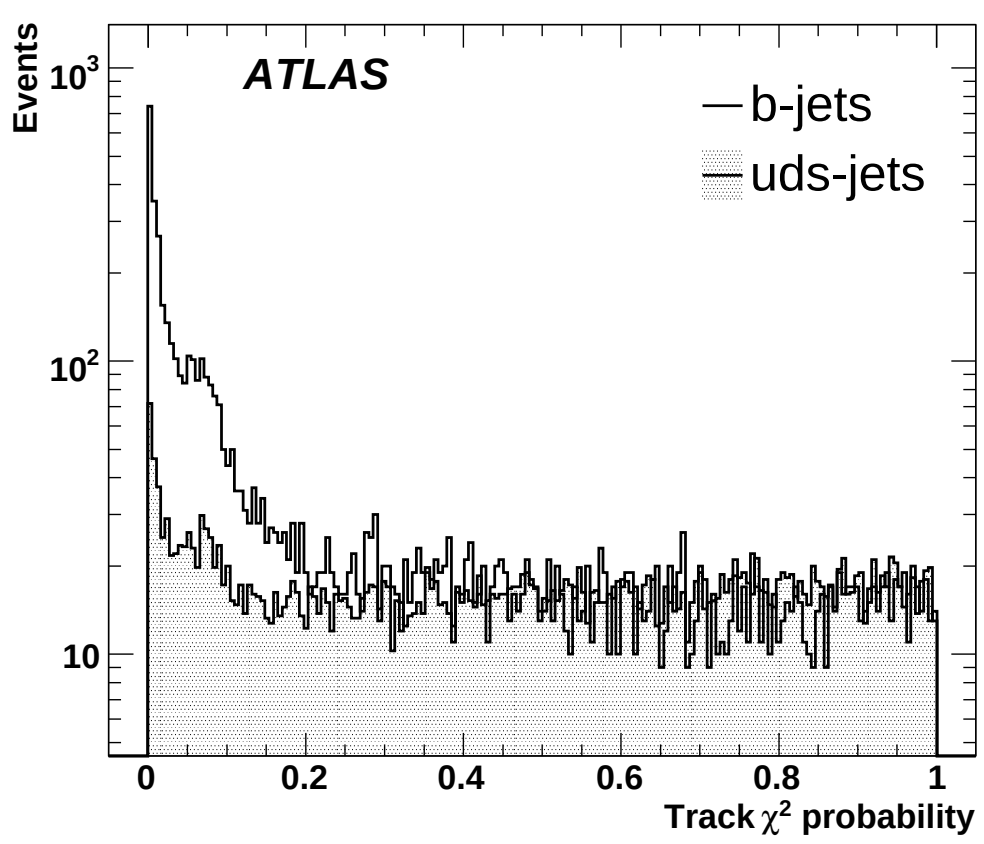


Figure 11: Track $\chi^2$ probability ($p_{trk}$) for *b*-jets (full histogram) and light jets (shaded histogram).

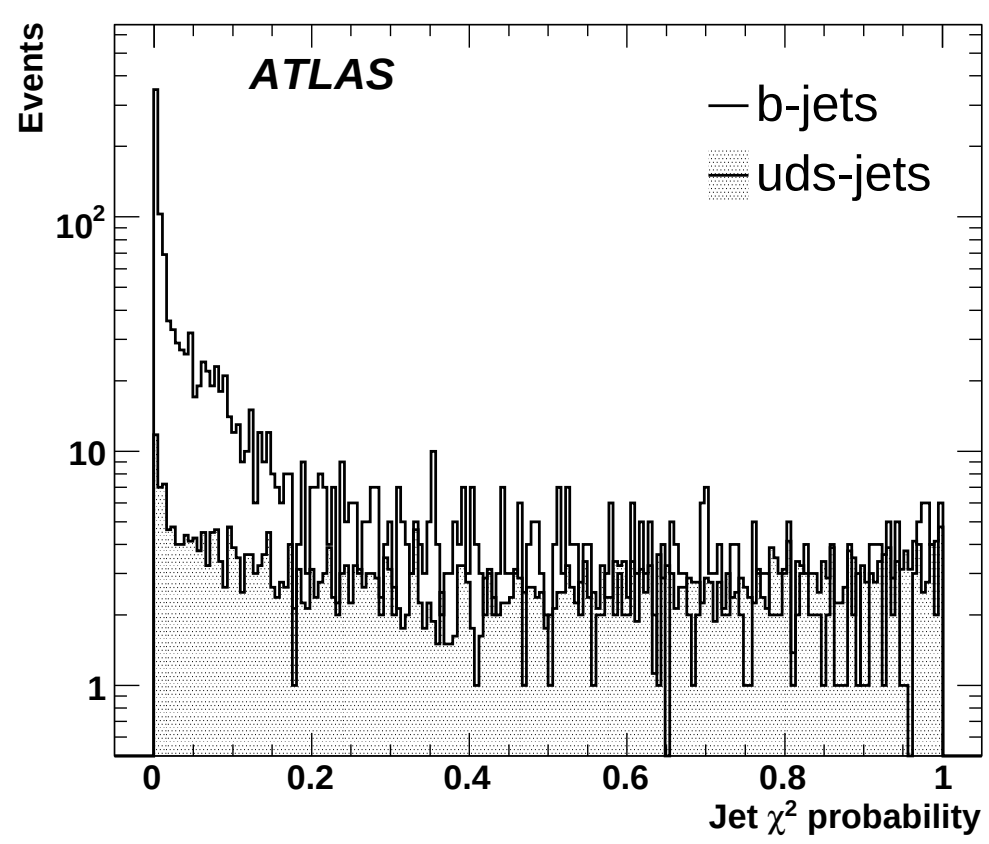


Figure 12: Jet $\chi^2$ probability ($p_{jet}$) for *b*-jets (full histogram) and light jets (shaded histogram).

# 4 HLT *b*-jet selection performance on single jet-RoIs

Every tagging method will be characterized by the curve showing the light-jet rejection versus the efficiency to select *b*-jets ($\varepsilon_b$). The light-jet rejection is defined as the inverse of the efficiency of selecting *u*-jets ($R_u = 1/\varepsilon_u$) where we have assumed that u-jets are representative of light jets in general.

## 4.1 Likelihood ratio method using impact parameters

Figures 13 and 14 show, respectively, the *b*-tagging performance for L2 and EF when the transverse impact parameter significance is used in defining the discriminant variable $X$, while figures 15 and 16 show the *b*-tagging performance curves for L2 and EF, when the significance of the longitudinal impact parameter with respect to the primary vertex is used instead.

Figures 17 and 18 show the *b*-tagging performance curves for L2 and EF when the likelihood ratio method is built on the combination of the transverse and longitudinal impact parameter significances.

## 4.2 $\chi^2$ method

The performance of the $\chi^2$ *b*-tagging algorithm, evaluated as a function of the $\chi^2$ cut is shown in Fig. 19. The limited efficiency of the method is due to the request of at least two reconstructed tracks to define the track-jet. Clearly, an effort should be made to include RoIs having only a single track. Nonetheless,

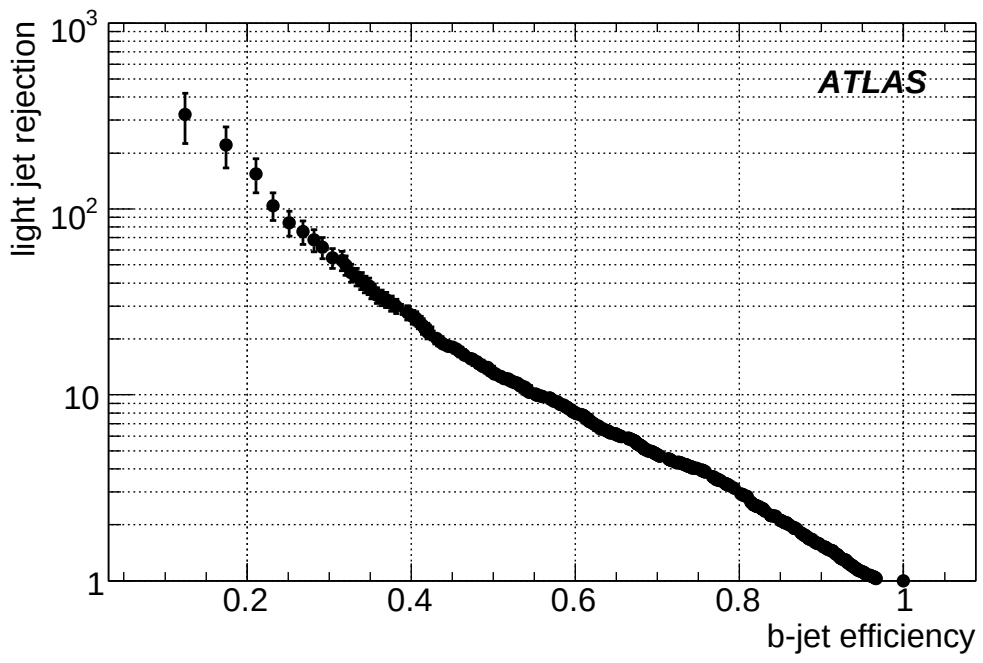


Figure 13: Performance of the *b*-jet selection based on the $d_0$ significance discriminant variable at L2.

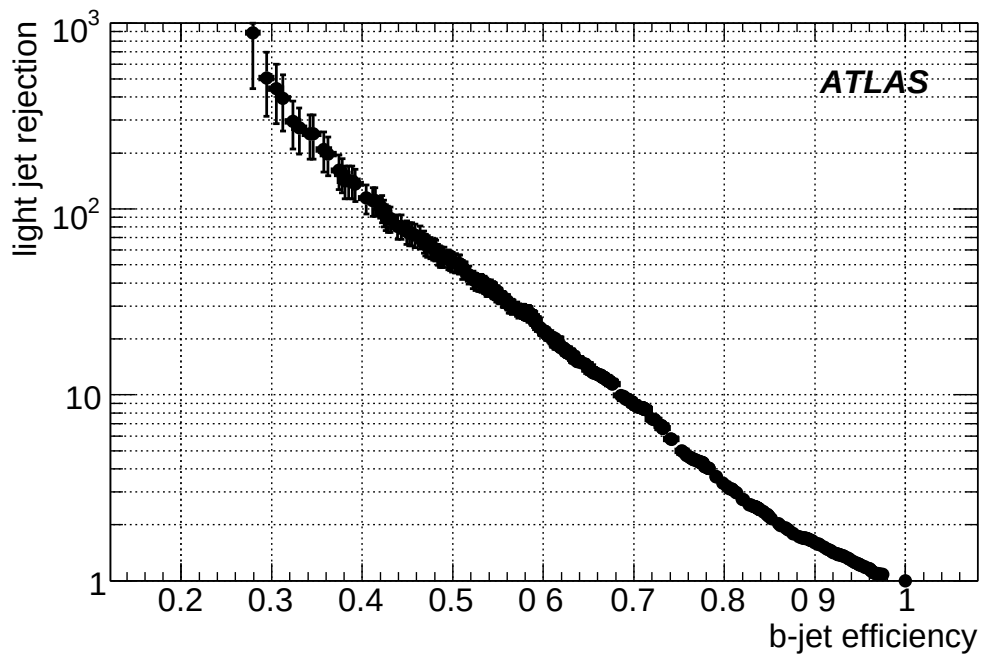


Figure 14: Performance of the *b*-jet selection based on the $d_0$ significance discriminant variable at EF.

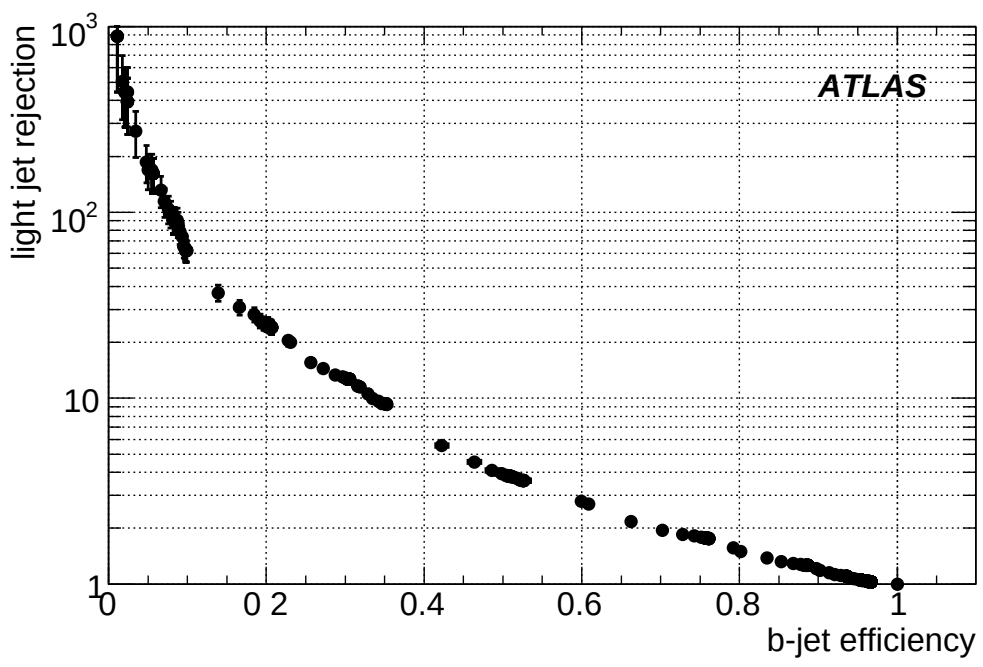


Figure 15: Performance of the *b*-jet selection based on the $\delta z_0$ significance discriminant variable at L2.

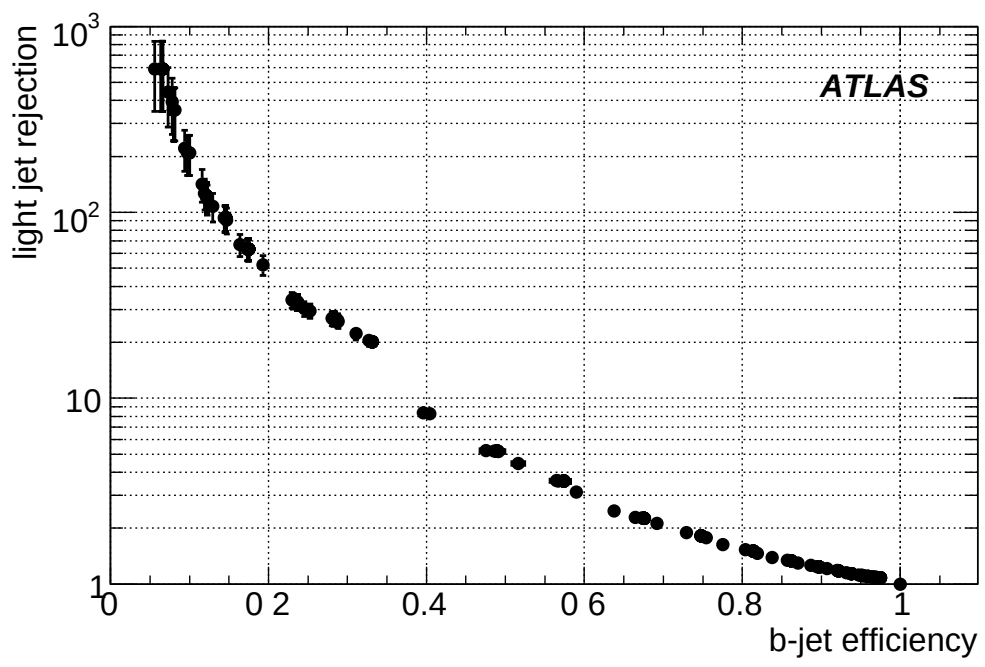


Figure 16: Performance of the *b*-jet selection based on the $\delta z_0$ significance discriminant variable at EF.

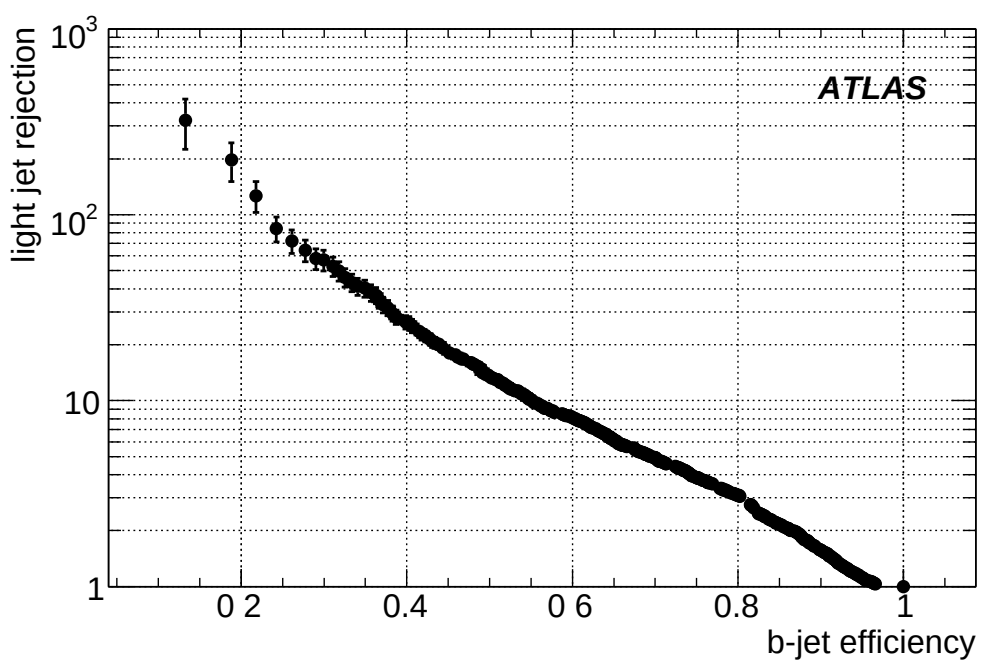


Figure 17: Performance of the *b*-jet selection based on the combination of the transverse and longitudinal impact parameter significances at L2.

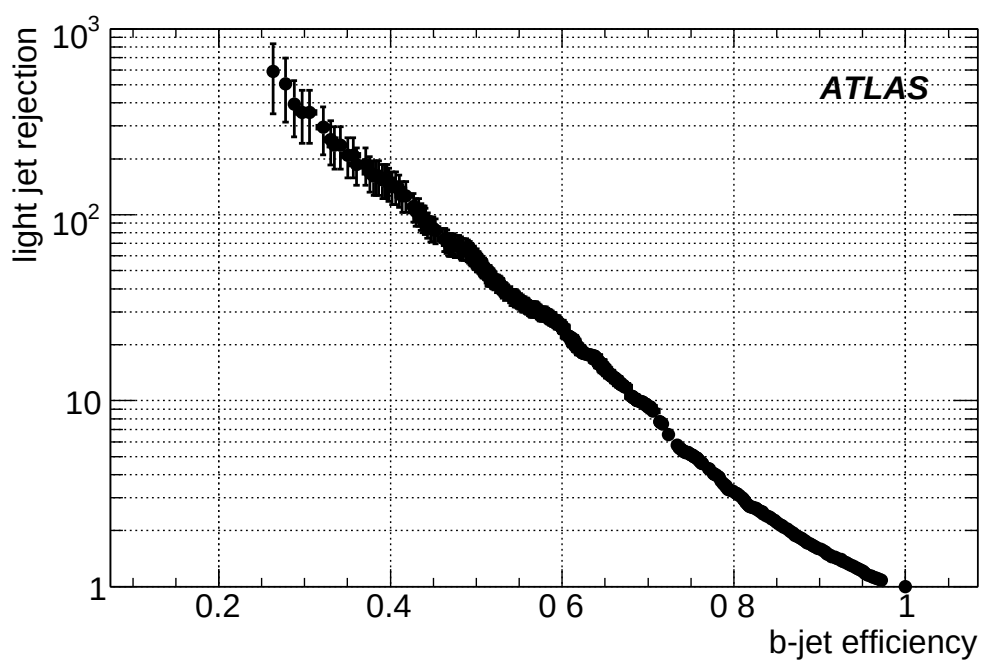


Figure 18: Performance of the *b*-jet selection based on the combination of the transverse and longitudinal impact parameter significances.

we note that the strength of the method lies in its impact intrinsic robustness and this advantage must also be considered when comparing its performance with that of the likelihood method.

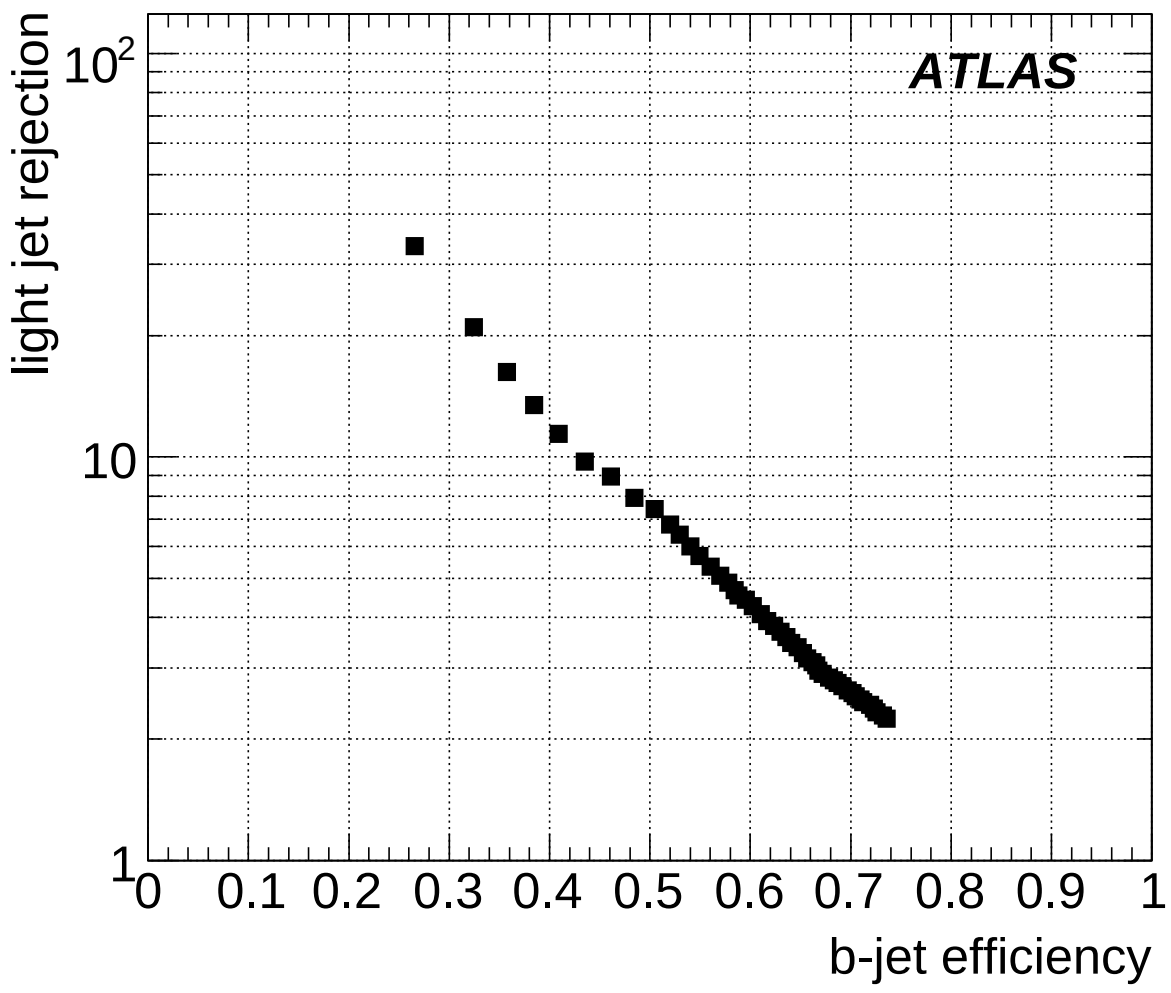


Figure 19: Performance of the $b$-tagging selection based on the jet $\chi^2$ probability variable.

### 4.3 Comparison with the offline selection

To tune the online working points so as to ensure the attainment of the overall (i.e. including offline cuts) efficiency goal of 60% for $b$-jet tagging and avoid biases, it is crucial to evaluate the correlation between the online and offline algorithms.

The performance of the L2 and EF trigger algorithms based on impact parameters in the transverse plane has been compared to that obtained with the corresponding offline algorithm. This choice is motivated by the wish to perform a coherent comparison; more exhaustive comparison studies will be performed on specific physics selections.

Figure 20 demonstrates that the L2, EF and Offline selections are well correlated. In particular it is always possible to recover the full offline performance at a given $b$-jet efficiency if the L2 and EF working points are set at an appropriate higher efficiency. In particular for the trigger menu studies shown in the following, a working point of about 80% efficiency at L2 and about 70% at EF have been chosen in order to ensure full acceptance for the standard offline working point (60%).

Future studies will address remaining differences between EF and offline algorithms.

### 4.4 Execution time at L2 and EF

The execution time needed to reconstruct relevant quantities described in this note and to perform $b$-jet selection was evaluated both at L2 and EF. Results highlight that the timing performance fits design requirements and that the overall time spent is dominated by data preparation and track reconstruction algorithms. Further details are given in [3].

## 5 $b$-tagging trigger strategy

After having defined and characterized the $b$-jet selection algorithm on single $b$-jet RoIs the $b$-jet trigger menu has to be built. Figure 21 illustrates the online $b$-jet selection algorithm's performance as evaluated using high statistics samples. The performance of the L2 algorithm is indicated along with the performance of the EF algorithm on events which are selected by L2 (at the nominal working point of 80% efficiency). Since the $b$-tagging cut is fixed, the $b$-tagging efficiencies vary with $E_T$threshold. Typical

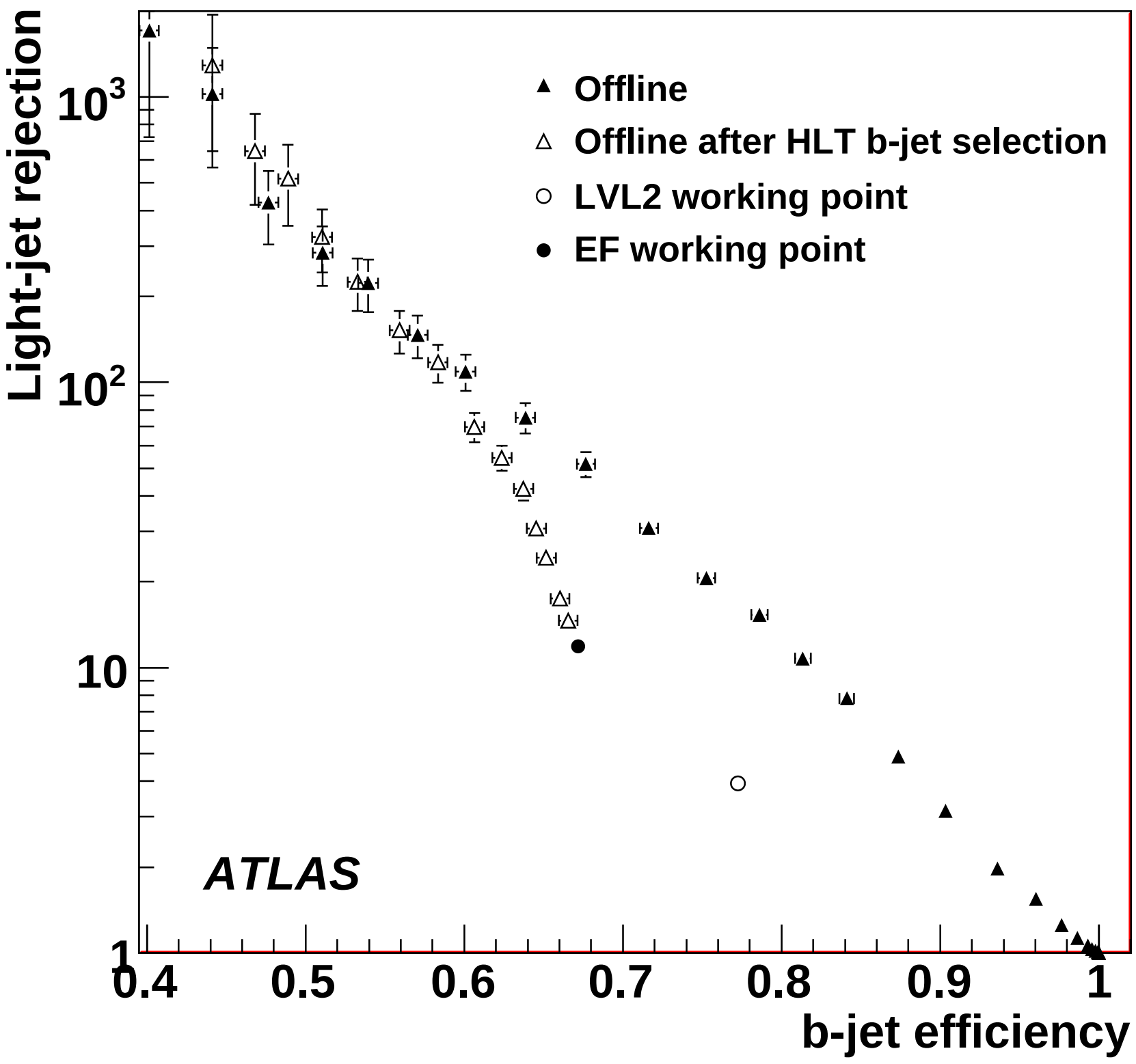


Figure 20: The correlation between L2, EF and offline taggers

L2(EF) *b*-tagging efficiencies are $\varepsilon_b = 76(67)\%$ at $E_T = 18$ $GeV$ and $\varepsilon_b = 80(73)\%$ at $E_T = 70$ $GeV$. This variation of the working point as a function of $E_T$ compensates for the effect of the worsening of the *b*-tagging performance at low $E_T$, so the rate reduction does not change significantly with $E_T$.

It is clear that *b*-jet selection can play an important role especially for events with multiple *b*-jets because the selective filtering of *b*-jets can produce very high rejection and thereby allow a significant decrease of the L1 thresholds while keeping the jet-RoI output rate of L2 and EF almost constant.

## 5.1 *b*-jet trigger menu

The possible *b*-jet signatures initiated by multi-jet L1 signatures with given $E_T$ thresholds can be represented in general as **nb**$E_T$**_mL1J**$E_T$, where n indicates the number of *b*-tagged jets required out of m L1 jets with transverse energy greater than $E_T$. The HLT *b*-tagging working point is the one describe in section 4.3. The rate reduction as a function of the available L1 thresholds is shown in Fig. 22. The EF output rates of different multi *b*-jet signatures at the luminosity of $10^{31}$ cm$^{-2}$s$^{-1}$ are given in Table 1. The rates and uncertainties of these rates have been computed for dijet samples using the relations

$$\begin{aligned} p_i &= N^i_{EF}/N^i_{Total} \\ R &= \mathscr{L}\sum p_i\sigma_i \\ \sigma(R) &= \mathscr{L}\sqrt{\sum \frac{p_i(1-p_i)}{N^i_{Total}}\sigma_i^2} \end{aligned} \quad (3)$$

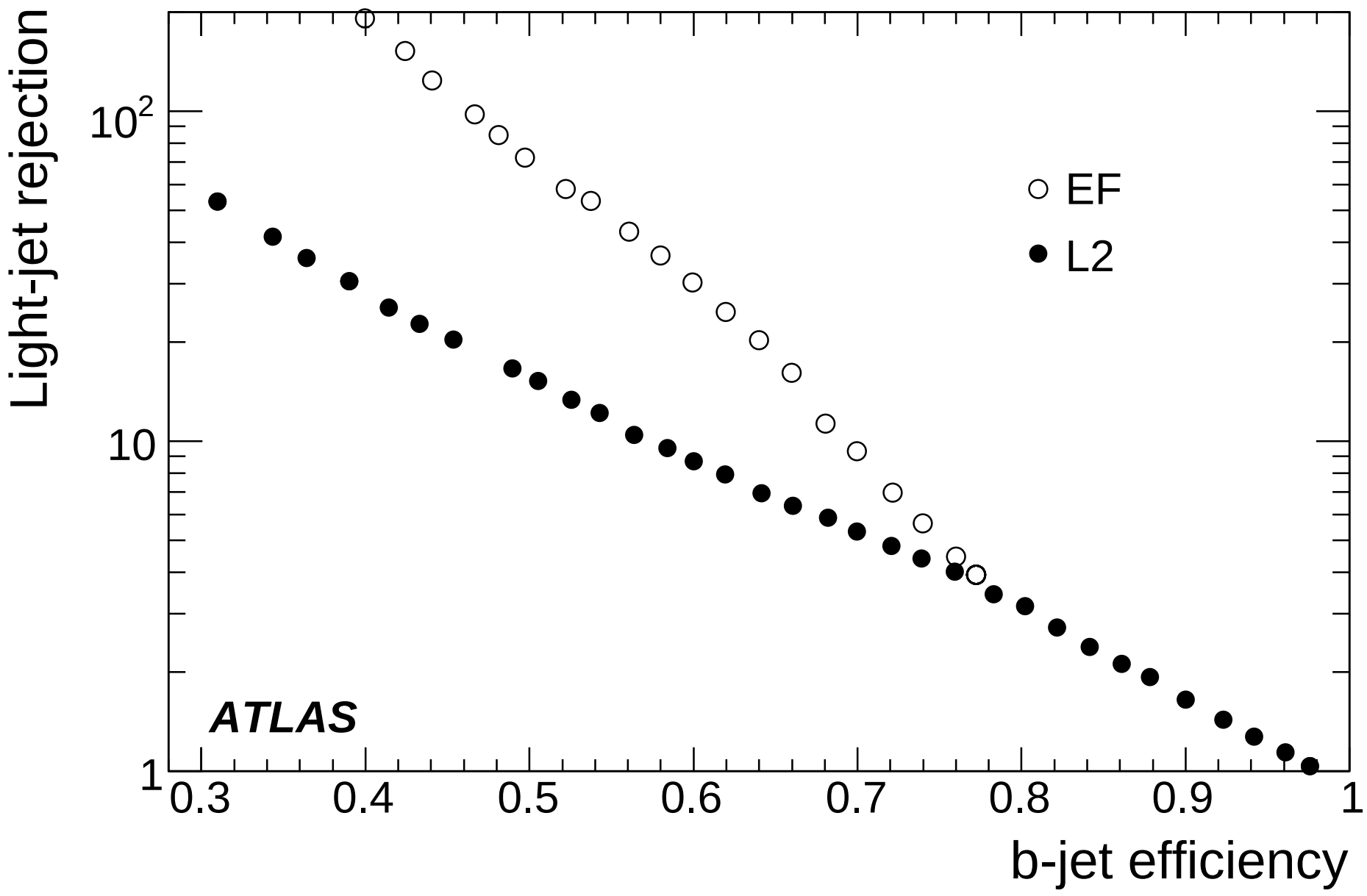


Figure 21: *b*-jet performance based on the combination of the transverse and longitudinal impact parameter (EF selection starts from the chosen L2 working point).

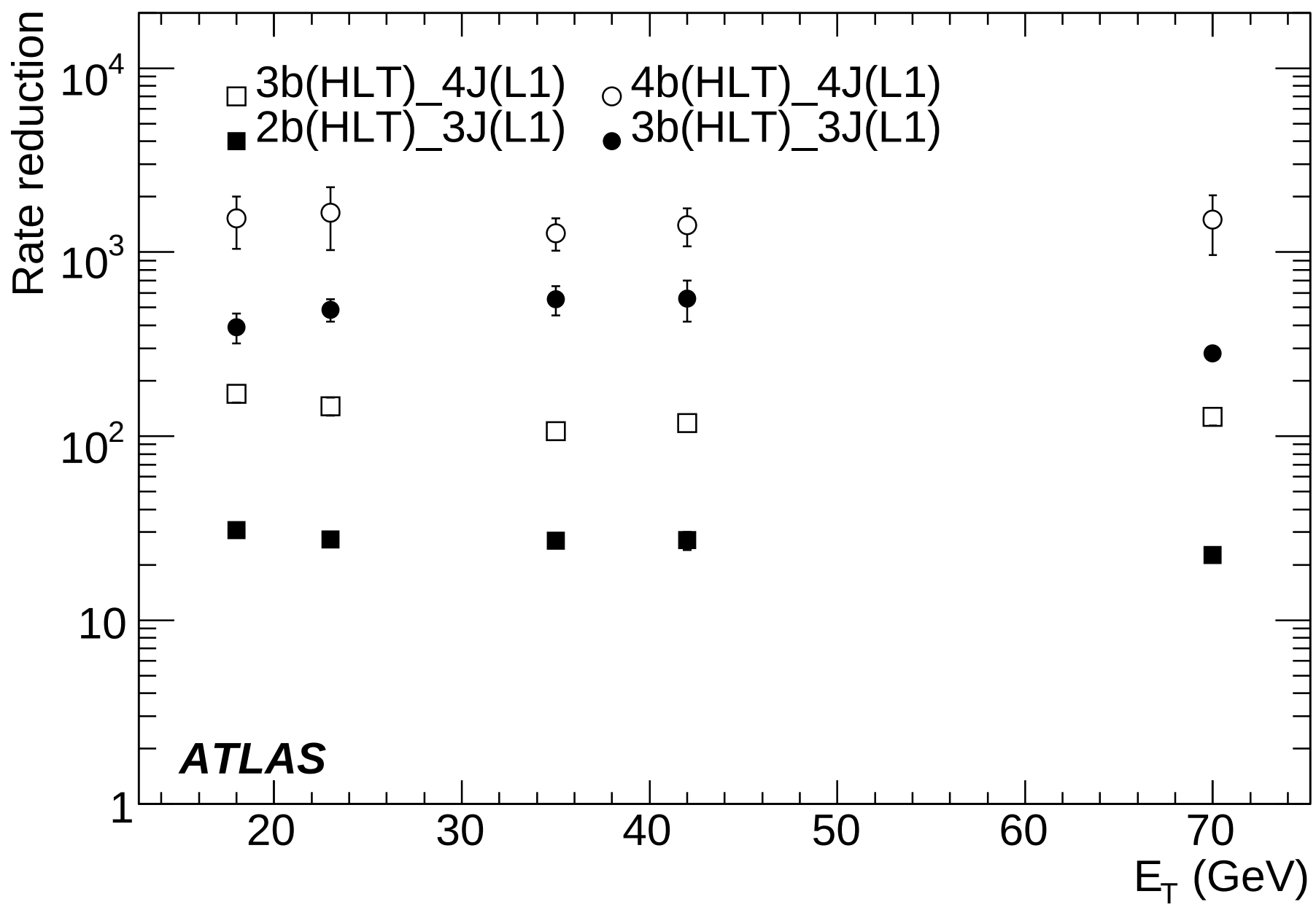


Figure 22: Rate reduction achieved with HLT *b*-jet as a function of the L1 $E_T$ threshold.

where $N^i_{EF}$ and $N^i_{Total}$ are respectively the number of events selected at the end of the trigger chain and the total number of events in the sample $J_i$, $\sigma_i$ is the cross-section of the sample $J_i$ and $\mathscr{L}$ is the luminosity.

The uncertainties in the tables indicate that at high transverse energy, the rate computation is not very

precise. Nevertheless, with the requirement of keeping the EF output rate at a few Hz for each multi *b*-jet signature, trigger menus for different luminosities can be chosen as:

- luminosity $10^{31}$ cm$^{-2}$s$^{-1}$: **3b23_3L1J23**, **3b18_4L1J18**
- luminosity $10^{32}$ cm$^{-2}$s$^{-1}$: **2b42_3L1J42**, **3b35_3L1J35**, **3b23_4L1J23**, **4b18_4L1J18**
- luminosity $10^{33}$ cm$^{-2}$s$^{-1}$: **2b70_3L1J70**, **3b42_3L1J42**, **3b35_4L1J35**, **4b23_4L1J23**

It can be noticed that as the luminosity increases, requiring more *b*-tagged jets is a viable alternative to increasing $E_T$ thresholds.

| Transverse energy | Signature rate [Hz] | | | |
|---|---|---|---|---|
| $E_T$ [GeV] | $2bE_T_3L1JE_T$ | $3bE_T_3L1JE_T$ | $3bE_T_4L1JE_T$ | $4bE_T_4L1JE_T$ |
| 18 | $47 \pm 11$ | $1.5 \pm 0.4$ | $1.0 \pm 0.3$ | $0.2 \pm 0.1$ |
| 23 | $18 \pm 7$ | $0.5 \pm 0.2$ | $0.4 \pm 0.2$ | $0.004 \pm 0.002$ |
| 35 | $1.0 \pm 0.2$ | $0.04 \pm 0.01$ | $0.02 \pm 0.01$ | $0.0007 \pm 0.00006$ |
| 42 | $0.4 \pm 0.1$ | $0.02 \pm 0.01$ | $0.01 \pm 0.01$ | $0.0007 \pm 0.00006$ |
| 70 | $0.01 \pm 0.02$ | $0.0008 \pm 0.0006$ | $0.0007 \pm 0.0006$ | $0.0007 \pm 0.00006$ |

Table 1: EF output rates for the different multiple *b*-jet signatures at $10^{31}$ cm$^{-2}$s$^{-1}$.

The strategy behind the evolution of the *b*-jet signatures is to select more aggressively as luminosity increases and HLT tracking becomes better understood. Before the *b*-jet trigger achieves full performance, a good online resolution of track impact parameters must be achieved. In turn, this requires adequate knowledge of the inner detector alignment and sufficient understanding of the overall detector performance.

### 5.2 Prospects for measuring efficiency and correlation with offline on real data

The HLT *b*-tagging group is working closely with the offline *b*-tagging group to develop a method to measure the *b*-jet efficiency with real data. For an explanation of the method and a discussion of its performance we refer to the *b*-tagging note on dijets [5].

In addition to the "physics" triggers listed in the previous Section , the *b*-jet group has introduced several "technical" triggers in order to study rate and correlation of the online and offline algorithms:

- single *b*-jet signatures: *b*18, *b*23, *b*35, *b*42, *b*70: prescaled to limit their contribution to the EF output to few Hz;
- each multi jet item is duplicated with an identical signature which selects, independently of the HLT *b*-jet result, one over *n* events (where *n* is presently set at 1000 but will be tuned according to the rate allocated to *b*-jet triggers).

## 6 Summary and conclusions

The *b*-jet selection at the L2 and EF stages of the ATLAS High Level Trigger has been described and characterized. An HLT *b*-tagging trigger menu has been implemented which demonstrates the feasibility of increasing the acceptance of events with more than one *b*-jet by decreasing L1 jet $E_T$ thresholds while controlling the output rate by introducing a *b*-jet selection at the HLT.

# Overview and Performance Studies of Jet Identification in the Trigger System

**Abstract**
This note describes in detail the algorithms used to identify jets in the ATLAS trigger system. Results from performance studies of these jet algorithms are presented. An initial trigger menu using jets and proposed strategy to adapt to increases in luminosity are also discussed.

## 1 Introduction

A critical component of the ATLAS trigger system is the ability to efficiently identify hadronic jets in an event. The performance of the jet reconstruction depends on the trigger jet energy resolution and scale. In this note we discuss the algorithms at the different trigger levels and evaluate their performance.

## 2 L1 jet trigger algorithm

A detailed description of the L1 jet trigger algorithm can be found in [1]; however, for completeness, a summary of the relevant features of this algorithm is presented below.

The ATLAS electromagnetic and hadronic calorimeters are segmented into approximately 7200 trigger towers, with granularity of approximately $0.1 \times 0.1$ in $\eta \times \phi$ space. The granularity varies slightly in different sub-detector systems, for further details see [1]. Analog signals from these trigger towers are transmitted directly to the L1 system. The L1 hardware digitises the trigger tower signals, associates them with a bunch crossing and performs pedestal subtraction. The L1 system also applies a noise suppression threshold and transverse energy calibration. The electromagnetic tower $E_T$ response is calibrated at the EM scale and the hadronic tower $E_T$ response is calibrated for jets.

The L1 trigger constructs "jet elements" made of the sum of $2 \times 2$ trigger towers in the electromagnetic (EM) calorimeter added to $2 \times 2$ trigger towers in the hadronic calorimeter which gives a granularity of $0.2 \times 0.2$ in $\eta \times \phi$ space. The jet reconstruction algorithm consists of a sliding window of programmable size that could be either $2 \times 2$, $3 \times 3$ or $4 \times 4$ jet elements. A jet is reconstructed if the total transverse (EM+Hadronic) energy within the window is above a given threshold. The step size for the sliding window is 0.2 in both $\eta$ and $\phi$ which implies significant overlap of the window in neighbouring positions. To prevent the L1 algorithm from identifying overlapping jets, the transverse energy of a cluster, defined as a region spanned by $2 \times 2$ jet elements, is required to be a local maximum within $\pm 0.4$ units in $\eta$ and in $\phi$. The L1 jet algorithm identifies jets within the region of $|\eta| < 3.2$. Figure 1 shows a schematic diagram of the jet reconstruction algorithm at L1.

In contrast to the other calorimeters, the L1 forward calorimeter (FCAL) trigger towers have a granularity of approximately $0.4 \times 0.4$ in $\eta$ and $\phi$. The forward jet trigger electronics was originally designed to only be used for the calculation of missing $E_T$ at L1, and not to identify in addition jets in the forward regions of the detector. As a consequence, limited granularity of the FCAL data is available at L1. A jet element in the FCAL is formed by summing calorimeter towers in $\eta$. Therefore, the FCAL jet elements have a $\phi$ granularity of 0.4 with only a single $\eta$ bin at each end. This has an impact on how the HLT forward jet reconstruction algorithm is implemented, as discussed in Section 5.

In the tentative menu proposed for early data taking, a sliding window size of $4 \times 4$ jet elements has been chosen for almost all thresholds; the exception is the algorithm used to reconstruct b-jets which has a proposed transverse energy threshold of 5 GeV and uses a window size of $2 \times 2$ to minimise the identification of fake jets associated with possible calorimeter noise.

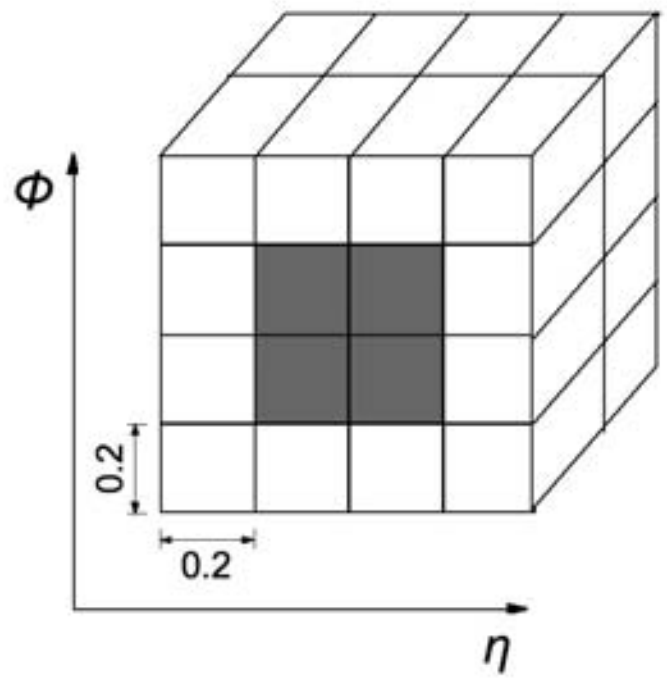


Figure 1: Schematic diagram of the L1 jet algorithm showing a window of $4\times4$ jet elements spanning the electromagnetic and hadronic calorimeter in depth, and a local maximum transverse energy cluster of $2\times2$ jet elements.

Table 1: Simulated dijet event samples used to study the performance of the jet trigger algorithms. The last row gives a summary of the cuts applied on the hard scatter parton in the event.

| Event Sample | J0 | J1 | J2 | J3 | J4 | J5 | J6 | J7 | J8 |
|---|---|---|---|---|---|---|---|---|---|
| Cross-section (mb) | 17.6 | 1.4 | 9.3E-2 | 5.9E-3 | 3.14E-4 | 1.3E-5 | 3.6E-7 | 5.3E-9 | 2.22E-11 |
| $E_T$ Range (GeV) | 8-17 | 17-35 | 35-70 | 70-140 | 140-280 | 280-560 | 560-1120 | 1120-2240 | $>2240$ |

# 3 L1 performance

Dijet events were used to study the performance of the L1 jet trigger algorithm. Table 1 provides a summary of the simulated data samples used along with their respective cuts on the hard scatter parton and cross-section. These simulated dijet event samples, together, span the whole $E_T$ jet spectrum relevant for jet identification in the trigger.

The transverse energy scale of L1 jets is defined as the ratio between the transverse energy measured in L1 divided by the truth jet $E_T$. Each jet identified by the L1 trigger was matched to a truth jet found using the cone algorithm with $R = 0.4$. The matching criterion consists in searching for the closest truth jet in the $\eta-\phi$ plane, where the distance is defined as $\Delta R = \sqrt{(\eta_{L1}-\eta_{Reco})^2+(\phi_{L1}-\phi_{Reco})^2}$. The L1 jet transverse energy scale as function of truth jet $E_T$ is shown in Fig. 2(a). The transverse energy scale at L1 varies from about 70% to 90%, increasing with the transverse energy of the jet. There are several effects that contribute to this behaviour. The first one is the noise suppression mentioned in Section 2. The lower the transverse energy of the jet, the larger is the number of trigger towers that do not satisfy this noise suppression threshold leading to an underestimation of the jet transverse energy. In addition, each L1 trigger tower signal must be associated with a particular Bunch Crossing Identification number. The efficiency of this requirement is approximately 50% for 1 GeV trigger tower signals, reaching nearly 100% for 3 GeV trigger tower signals. Again, this inefficiency particularly affects lower transverse energy jets that will tend to contain a larger fraction of trigger towers with low energy signals. The most important reason for the lower jet transverse energy scale with respect to truth jets comes from the different $e/\pi$ response of EM and hadronic towers. For presentation purposes, a common conversion factor is used to translate the number of counts measured in an EM or hadronic tower into units of transverse energy. This results in an apparent lower jet transverse energy scale compared to truth jets. It is important to note that thresholds in the L1 are applied in terms of the internal L1 quantity "counts" and not in units of transverse energy. In Fig. 2(b), the L1 jet transverse energy scale is also shown as

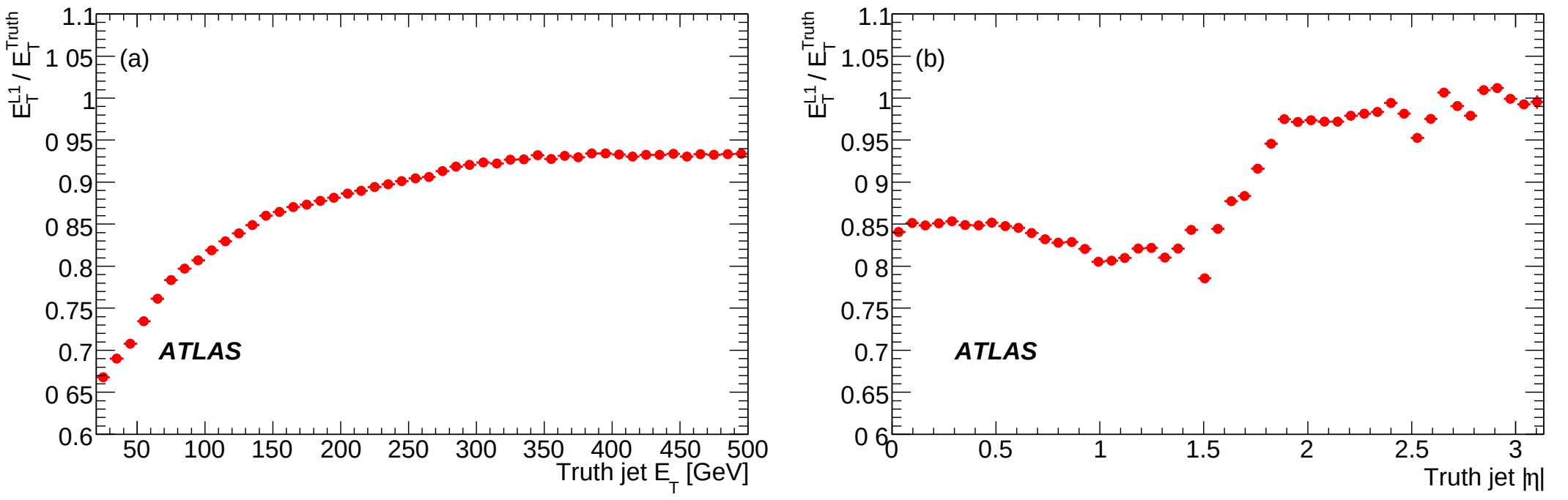


Figure 2: The L1 jet transverse energy scale as function of truth jet transverse energy (a) and pseudo-rapidity (b).

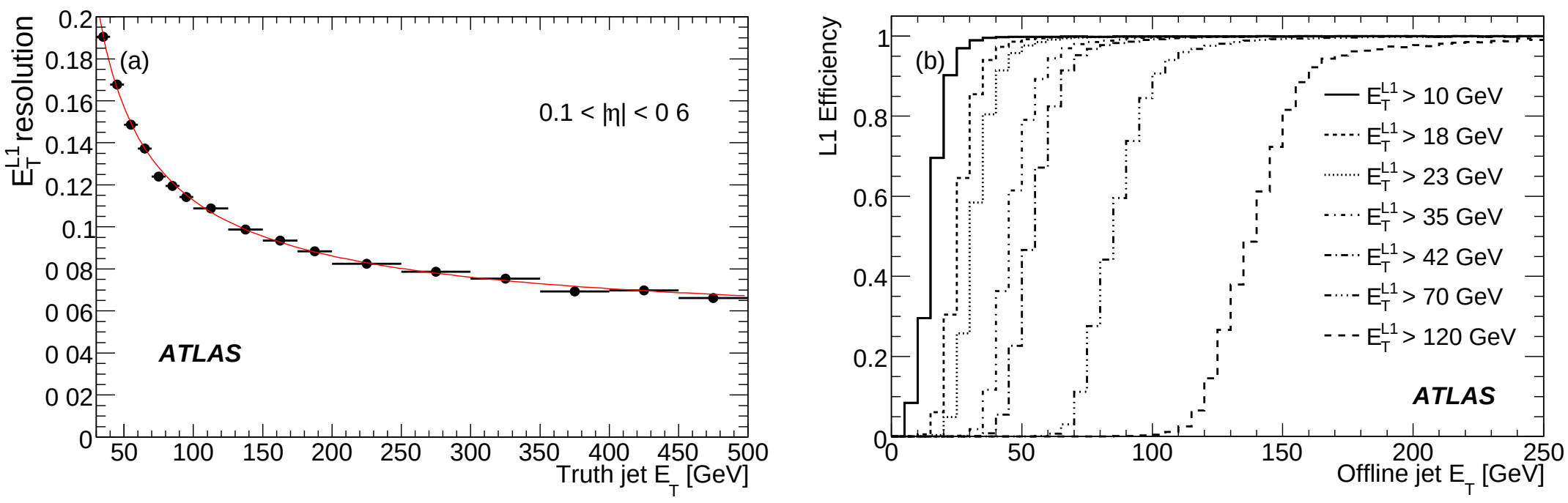


Figure 3: L1 jet transverse energy resolution as a function of truth jet transverse energy (a) and the L1 jet trigger efficiency as function of the offline reconstructed jet $E_T$ for different L1 energy thresholds (b).

a function of the pseudo-rapidity ($\eta$) of the truth jet. The response of the different calorimeter sub-detectors can be identified. Figure 3(a) shows the L1 transverse energy resolution as function of truth jet $E_T$. The transverse energy resolution is defined as the width of the $E_T^{L1} - E_T^{truth}$ distribution in each $E_T^{truth}$ bin, divided by the $E_T^{truth}$ of that bin.

Figure 3(b) shows the L1 jet trigger efficiency as function of offline reconstructed jet $E_T$ for different L1 jet trigger thresholds. The limited jet transverse energy resolution of the L1 system particularly affects the higher L1 energy thresholds.

The effect of pile-up on the performance of the L1 jet trigger reconstruction was also studied. At a luminosity of $10^{33}\mathrm{s}^{-1}\mathrm{cm}^{-2}$, an average of approximately 2 inelastic collisions are expected per bunch crossing. Simulated dijet events including this level of pile-up were generated. The effect of increased occupancy in the calorimeters will have some impact on the reconstructed transverse energy of the jet. Figure 4 shows the impact of this pile-up on the L1 jet transverse energy scale and jet trigger efficiency. The transverse energy scale increases due to the additional energy deposited in the calorimeter from pile-up events. The L1 transverse energy resolution also worsens as seen on Fig. 4(b). The effect is clearly more important for low energy jets where the contribution of pile-up energy can be of the same order of magnitude as the jet energy itself.

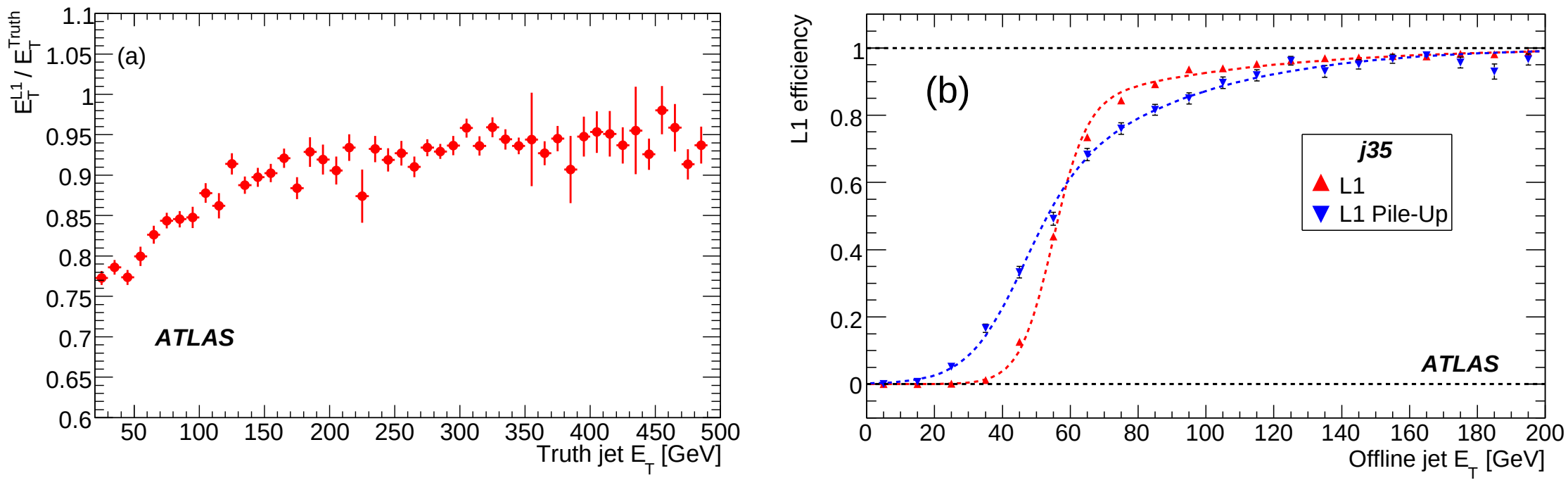


Figure 4: The L1 jet transverse energy scale as function of truth jet transverse energy (a) for simulated dijet events with pile-up. The L1 jet trigger efficiency as function of offline reconstructed jets (b) for a 35 GeV L1 trigger threshold, with and without taking into account pile-up.

## 4 HLT jet algorithms

The HLT object reconstruction is guided by the result of the L1 system. HLT algorithms typically only access data from a limited region of the detector in the vicinity of an RoI provided by L1. The position of this RoI is successively updated and refined by the HLT algorithms.

HLT algorithms are classified in two types:

- "Feature Extraction algorithms": Algorithms that retrieve and unpack detector data and create simple classes composed of useful physics observables. These algorithms consume most of the available time.
- "Hypothesis algorithms": Algorithms that retrieve the physics information produced in the preceding Feature Extraction algorithms, and validate a specific hypothesis (e.g. $E_T$ threshold). These algorithms have a very fast execution time.

This separation between Feature Extraction and Hypothesis algorithms optimises the overall execution time since the data retrieved by a single Feature Extraction algorithm can be used to provide input to several fast Hypothesis algorithms to test various physics signatures, and hence avoids multiple data access and unpacking.

Figure 5 shows a schematic diagram of the sequence of algorithms used to reconstruct jets in the HLT. The Hypothesis algorithms for jets reconstructed at L2 and EF compare the energy of the jet candidates to some predefined $E_T$ thresholds. The next few sections describe in detail each Feature Extraction algorithm appearing in Fig. 5.

## 5 L2 jet algorithm

Standard L2 jets are defined within the $|\eta| < 3.2$ region and are reconstructed using the electromagnetic and hadronic calorimeter data. Forward jets are defined within the range $3.2 < |\eta| < 5$ and are reconstructed using data from the forward calorimeters. As described below, the forward jet reconstruction is different than the standard jet algorithm due to the limited $\eta$ position resolution of L1 jets.

The output of the L2 Feature Extraction algorithm is a reconstructed jet with a given energy and position in $\eta$ and $\phi$. The algorithm contains three distinct parts described below: data preparation, jet finding and calibration.

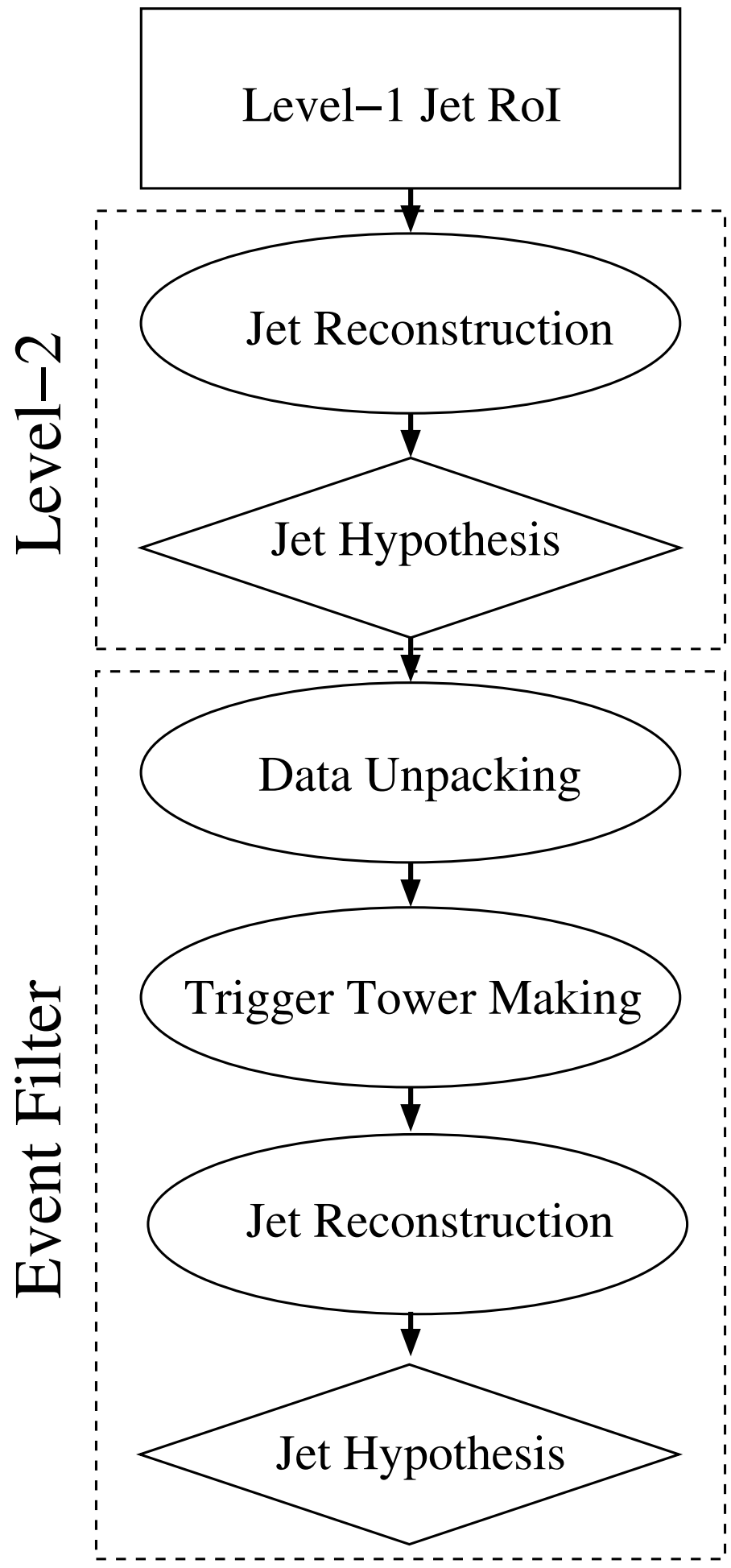


Figure 5: Schematic diagram of the sequence of algorithms used to reconstruct jets in the HLT. The ovals and diamonds represent Feature Extraction algorithms and Hypothesis algorithms, respectively.

### 5.1 L2 jet data preparation

The data preparation for the L2 jet trigger is a critical part of the algorithm chain. It provides the collection of data from the detector readout drivers (ROD) to the L2 processing units and the conversion from the raw data into bytestream files readable by the HLT algorithms. The RODs receive data from the calorimeters front-end boards (FEB) via optical fibres. The FEBs are installed on the detector and contain the electronics for amplifying, shaping, sampling, pipelining, and digitising the signals [2, 3].

The ATLAS calorimeters consist of more than $10^5$ individual readout channels; therefore, in order to meet the L2 timing performance goals of 40 ms total processing time per event, the amount of data unpacked must be kept to a minimum while simultaneously maximising the physics performance of the algorithm.

The L2 jet trigger algorithm accesses calorimeter data that lies in a rectangular region centred around the L1 jet RoI position with a width $\Delta\eta$ and $\Delta\phi$ that can be defined to have any size. The widths $\Delta\eta$ and $\Delta\phi$ are parameters that are specified at trigger configuration time. Figure 6 shows a schematic diagram of the L2 jet reconstruction algorithm.

The position and transverse energy of each detector element that falls into the chosen ($\Delta\eta$, $\Delta\phi$) window is read out by the algorithm. As a result the calorimeter read out region can be regarded as

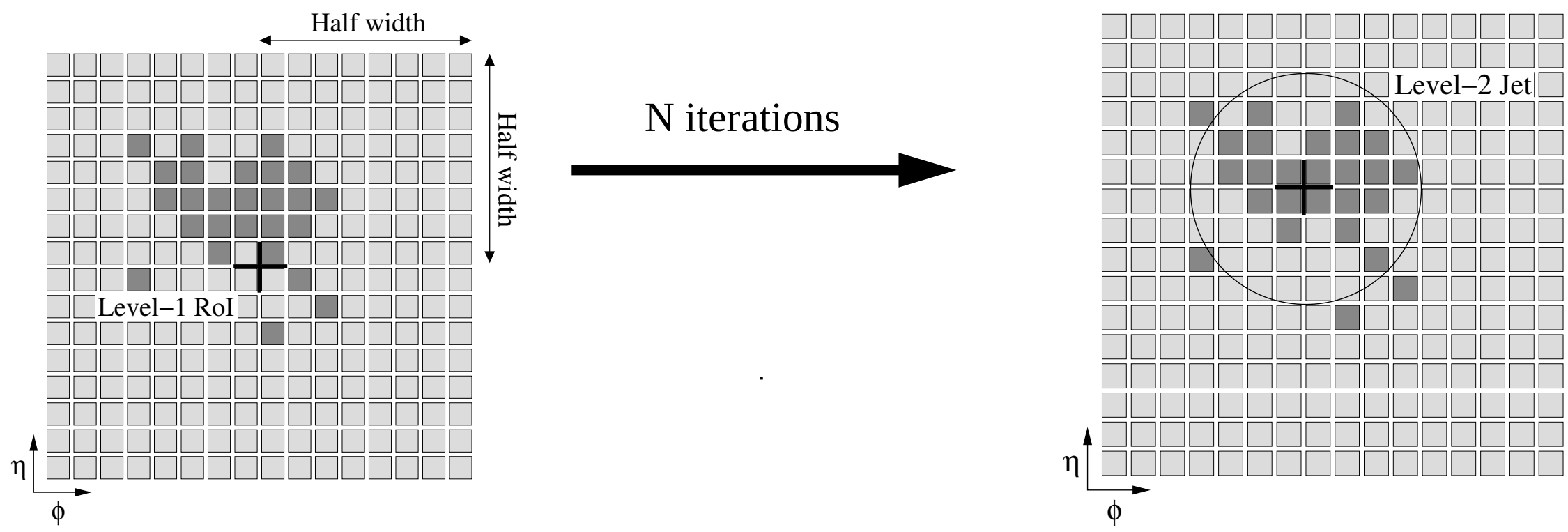


Figure 6: Schematic diagram summarising the L2 jet algorithm. The data unpacking step reads in the necessary calorimeter data within a predefined window size and defines a grid of calorimeter elements each with an associated energy and position. The size of grid elements depends on the calorimeter data unpacking method used. The dark (red) boxes in the diagram represent grid elements with substantial energy deposition. The algorithm is seeded by the L1 position as shown on the left. The final jet is found after a given number of iterations. The position of the jet is calculated as the energy weighted average of the grid elements position within a given cone size. The energy of the jet is calculated as the sum of the energy of each grid elements falling within the given cone size.

partitioned into a grid of elements with associated energies and $(\eta,\phi)$-coordinates as shown in Fig. 6. Therefore, the amount of data accessed is equivalent to the number of grid elements.

Two different data unpacking approaches are implemented and described below. One of the methods, the cell-based approach, has finer granularity and hence produces more accurate energy and transverse energy reconstruction, but is more time consuming. The other method, the front-end board approach, is faster but the coarser granularity produces a less precise reconstruction. Both methods are, as we will see in section 6, reasonably within the L2 time budget limits. The final decision of what approach should be used will be made depending on the final High Level Trigger setup.

### 5.1.1 The cell-based approach

This method uses the full granularity of the calorimeters [1]. Each grid element in Fig. 6 corresponds to a calorimeter cell with a given transverse energy and $(\eta,\phi)$-coordinate provided by the ROD. In the following discussions, jets reconstructed using this data unpacking approach are referred to as "cell-based jets".

### 5.1.2 The front-end board approach

This method uses a coarser granularity than the cell-based data unpacking. Instead of reading out every cell over a specified region of the liquid argon (LAr) calorimeters, only information from the LAr calorimeters front-end boards is used. There are 128 readout channels per FEB and two FEBs are connected to one ROD. For the tile calorimeter, the full cell granularity is still used.

In the RODs, the sums of the Cartesian components of the cell energies are calculated for each FEB:

$$E_{\mathrm{x}} = \sum_{\mathrm{i}=1}^{128} E_{\mathrm{i}} \frac{\cos\phi_{\mathrm{i}}}{\cosh\eta_{\mathrm{i}}} \tag{1}$$

$$E_y = \sum_{i=1}^{128} E_i \frac{\sin\phi_i}{\cosh\eta_i} \quad (2)$$

$$E_z = \sum_{i=1}^{128} E_i \tanh\eta_i \quad (3)$$

where $E_i$ is the energy and $\phi_i$, $\eta_i$ the position of each cell. This sum is computed per FEB and runs over all channels with an energy $E_i > 2 \cdot \sigma_{noise}$. The noise cut value, which is a configuration parameter, has been determined from performance studies with QCD dijet events, such as to give the similar jet energy scale as obtained with the cell-based approach, which will be presented in the next section. The total energy of the cells connected to a FEB is then obtained from the quadratic sum of the three components, $E_{tot} = \sqrt{E_x^2 + E_y^2 + E_z^2}$. The corresponding $\eta$ and $\phi$ coordinates are calculated as

$$\eta = \operatorname{atanh}(E_z/E_{tot}),$$

$$\phi = \operatorname{atan2}(E_y/E_x),$$

where the function atan2 returns the angle $\phi \in [-\pi, \pi]$. The resulting values for the energy and the ($\eta$, $\phi$)-coordinate are then used to define the grid of elements in a region around the L1 RoI position, as is illustrated in Fig. 6. In the following text, jets reconstructed using this data unpacking method are referred to as "FEB-based jets".

## 5.2 L2 jet finding algorithm

Jets are defined as a cone-shaped object in the $(\eta, \phi)$-space with a given radius $\Delta R = \sqrt{\Delta\eta^2 + \Delta\phi^2}$. The value $\Delta R$ is a parameter of the algorithm that is defined at trigger configuration time. The jet energy and position are found through an iterative procedure with the following steps:

- The L1 jet RoI is used as a seed for the algorithm. A reference jet is created, labeled $j_0$, defined by the L1 jet RoI position with the pre-defined cone radius $\Delta R$ (see left-hand side of Fig. 6). Note that the possible positions of the reference jet $j_0$ are discreet due to the L1 granularity.

- Grid elements that fall within the ($\eta$, $\phi$)-region encompassed by the reference jet $j_0$ are used to recalculate the jet energy and position according to:

$$\eta_{j_1} = \frac{\sum_{i=1}^{k} E_i \eta_i}{\sum_{i=1}^{k} E_i}, \quad (4)$$

$$\phi_{j_1} = \frac{\sum_{i=1}^{k} E_i \phi_i}{\sum_{i=1}^{k} E_i}. \quad (5)$$

  The sum runs over the $k$ grid elements that are contained in the cone defined by the reference jet $j_0$. A grid element is included in the sum if its centre falls within the region spanned by the cone radius $\Delta R$. The total energy and coordinates ($\eta_{j_1}$, $\phi_{j_1}$), computed in Equations (4) and (5), are used to define the new reference jet $j_1$.

- The previous step is repeated with $j_0$ replaced by $j_1$ in Equations (4) and (5), which results in updated coordinates $\eta_{j_2}$ and $\phi_{j_2}$ to define the updated jet $j_2$. This algorithm can be repeated N times to create jet $j_N$.

- A predetermined fixed number of iterations are executed. The energy of the final jet is calculated as the sum of the energy of all the grid elements falling within the cone radius. The position of the jet is obtained as the energy weighted average of the position of each grid elements within the cone, as shown in Fig. 6.

The outcome of this algorithm is a jet defined by its $(\eta, \phi)$ position and total energy. The calorimeter energy scale, at this point, is set at the electromagnetic scale which does not provide an accurate measure of the jet energy. The next Section describes the calibration weights applied to correct the total jet energy for non-electromagnetic shower components.

## 5.3 L2 jet calibration

The ATLAS calorimeter response to the electromagnetic component of a hadron shower is not equal to the response to the non-electromagnetic component. In general, the hadronic response ($h$) is smaller than the electromagnetic response ($e$),

$$e/h > 1.$$

This effect is mainly due to the energy lost in the breakup of nuclei or in nuclear excitation. In order to correct for it, a weight is applied to each element that makes up a jet. Depending on the calibration method used, a jet can be regarded as composed, for example, of individual calorimeter cells or of energy deposits in calorimeter samplings. The calibrated jet energy can then be written as:

$$E_{\mathrm{jet}}^{\mathrm{rec}} = \sum_{\mathrm{i}=1}^{\mathrm{n}} w_{\mathrm{i}} E_{\mathrm{i}} \tag{6}$$

where the sum runs over the n constituents of the jet.

The weights $w_{\mathrm{i}}$ in Equation (6) are extracted using simulated event samples by minimising the function:

$$S = \sum_{\mathrm{m}=1}^{\mathrm{N_{jets}}} \left[ \frac{(E_{\mathrm{m}}^{\mathrm{truth}} - E_{\mathrm{m}}^{\mathrm{rec}})}{\sigma_{\mathrm{m}}} \right]^2 \tag{7}$$

The sum runs over all the jets in the events. The true energy of the m-th jet in the event is labeled $E_{\mathrm{m}}^{\mathrm{truth}}$ and is obtained from the Monte Carlo (MC) truth information, running a cone jet algorithm with radius $\Delta R = 0.4$ over the MC truth particles and finding the truth jet that is closest to the m-th jet. A truth jet is made up of all particles generated, excluding neutrinos and muons which have their own observables, missing transverse energy for neutrinos and reconstructed tracks for muons. By minimising $S$ with respect to the true jet energy, an improvement of the jet energy scale and resolution is obtained.

Different calibration methods can be applied, which differ in the partitioning of the jet energy into calorimeter components (e.g. cells, layers). The calibration method used here is the so-called sampling technique [4]. In this method, individual weights can be applied to the energy deposited in each of the electromagnetic (EM) and hadronic (HAD) calorimeter sampling layers. The energy dependence of the weights is chosen to be:

$$w_{\mathrm{i}} = a + b \log(E). \tag{8}$$

In the implementation used for L2 jets, only two weights are calculated and applied to calibrate the reconstructed jet energy; one weight for the total energy deposited in the EM calorimeter, and one weight for the total energy deposited in the hadronic calorimeter. Furthermore, the $\eta$ range is split in 32 bins with a size of 0.1 in the region $0 < |\eta| < 3.2$. This procedure assumes an azimuthally symmetric response. This assumption will need to be validated with real data. A fit to Equation (8) is performed

Table 2: Parameters used in the L2 jets reconstruction. The value of the cone size radius is defined for each iteration.

| Parameter | Standard Jets | Forward Jets |
|---|---|---|
| Number of iterations | 3 | 3 |
| Cone radius | 0.4/0.4/0.4 | 1/0.7/0.4 |
| $\Delta\eta$ window size | 1.4 | $3 < \lvert\eta\rvert < 5$ |
| $\Delta\phi$ window size | 1.4 | 1.4 |

using the MINUIT2 package [5]. It yields the values of the two parameters of Equation (8) for each $\eta$-bin and for the electromagnetic and hadronic calorimeter sampling. The computed weights are then stored in a configuration file, which serves as input to the L2 jet algorithm that applies the weights to each identified jet.

### 5.4 L2 jet parameters

In the configuration of the L2 jet trigger algorithm, the values of the following parameters need to be set:

- the data unpacking method: cell-based or FEB-based,
- the calorimeter window size $\Delta\eta$ and $\Delta\phi$, where data need to be unpacked,
- the radius of the cone $\Delta R$ used in the jet finding algorithm,
- the number of iterations used in the jet finding algorithm,
- the calibration constants used.

A set of parameter values which yields the optimal balance between short execution times and adequate physics performance must be chosen.

## 6 L2 performance

The performance of the L2 jet algorithm was studied using the simulated dijet event samples described in Table 1.

The parameters used in the reconstruction of the L2 jets are summarised in Table 2. These parameters were found to be a good compromise between the physics performance (e.g., energy resolution and scale) and the algorithm execution time.

The first parameter studied was the number of iterations of the jet finding algorithm. The variation in the jet $(\eta, \phi)$ coordinate and transverse energy after each iteration of the jet reconstruction algorithm is shown in Fig. 7. The largest variation in $E_T, \eta$ and $\phi$ happens after the first iteration and thus it has the largest impact on the measurement precision. This suggests that the number of iterations could possibly even be reduced to 2 without losing too much precision on this measurement.

In order to study the effect of the window size used to unpack calorimeter data, the total time spent by the jet algorithm running with different sizes was studied. Windows with dimensions of $1.0 \times 1.0$ (in $\eta \times \phi$) and $1.4 \times 1.4$ were studied. A window size of $1.0 \times 1.0$ is only slightly larger than the diameter of a jet with $\Delta R = 0.4$. The maximum initial displacement of a jet with respect to a truth jet is approximately 0.2 in $\Delta\eta$ or $\Delta\phi$, as can be seen in Fig. 7. Therefore, a window size of $1.0 \times 1.0$ should be adequate for

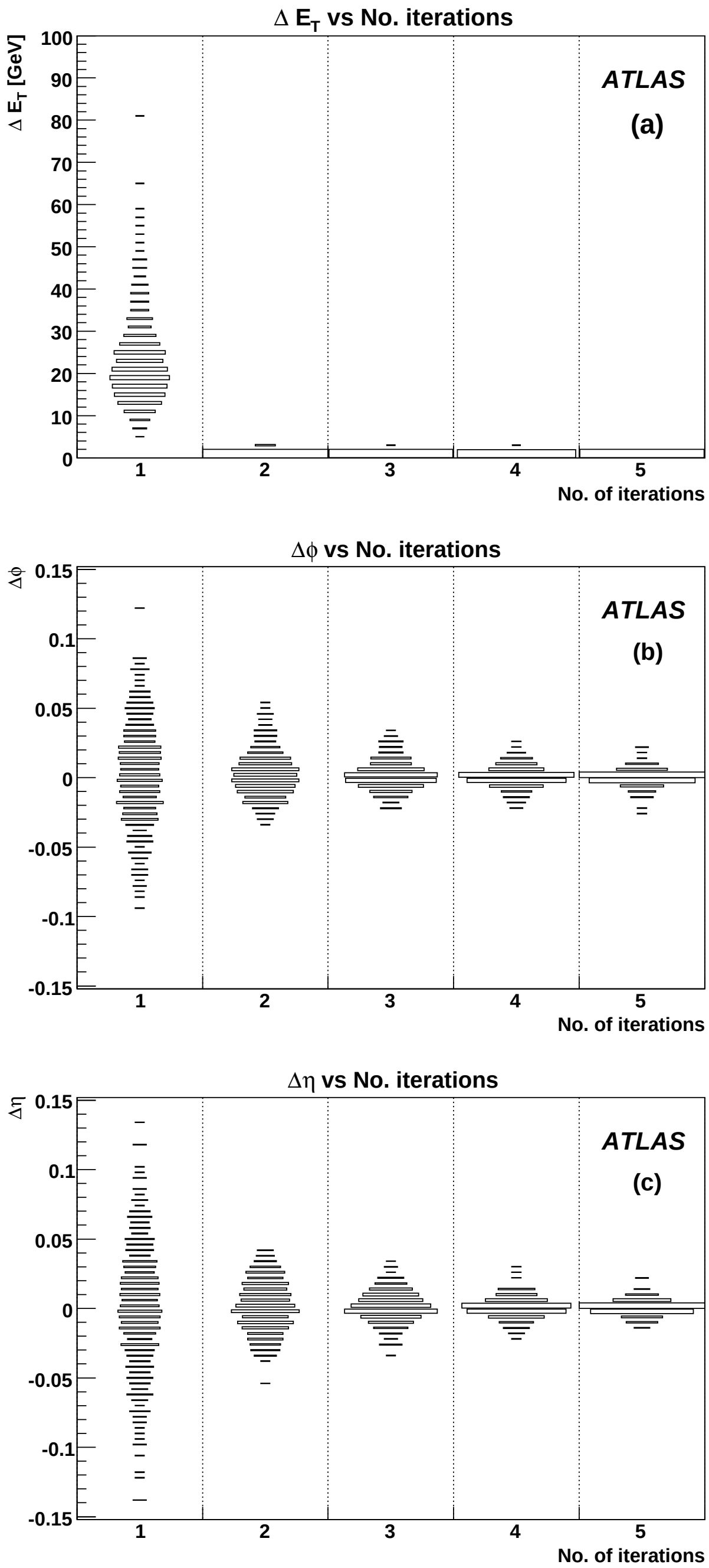


Figure 7: Variation in the (a) transverse energy, (b) $\phi$ position and (c) $\eta$ position of jets as function of number of iterations performed by the L2 jet reconstruction algorithm. The area of the boxes is proportional to the number of entries in each bin.

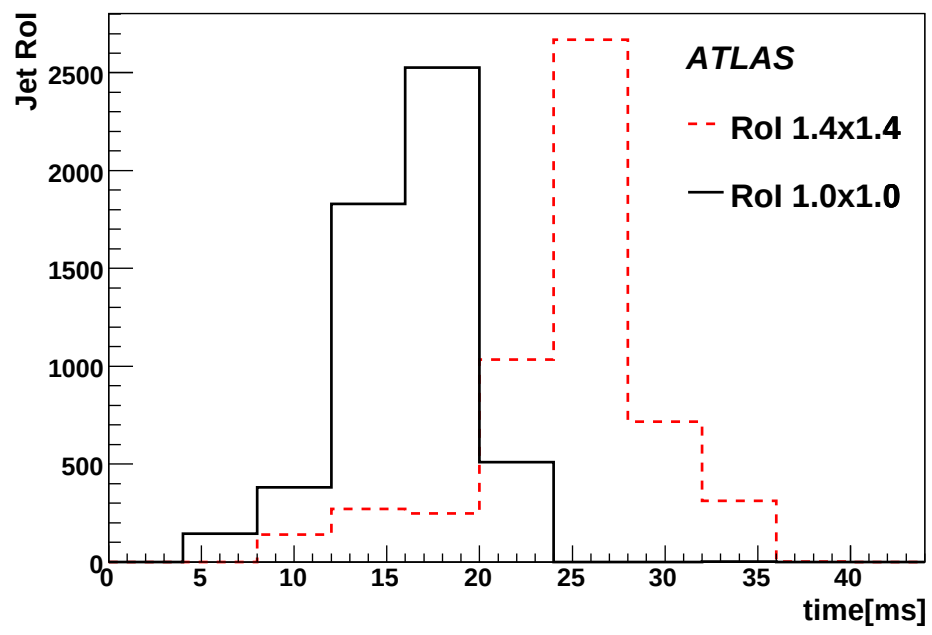


Figure 8: Time spent by the L2 jet reconstruction algorithm for two different window sizes: $1.4 \times 1.4$ (dashed line) and $1.0 \times 1.0$ (solid line).

the jet reconstruction. The total time spent by the L2 jet algorithm for two different window size and using three iterations is shown in Fig. 8. A reduction of the window size from $1.4 \times 1.4$ to $1.0 \times 1.0$ results in a considerable reduction (of order 30%) in the processing time. As shown in the next section, the energy scale and resolution is not significantly affected by this reduction of the size of the window.

## 6.1 Performance for cell-based reconstruction

The calibration constants used to reconstruct cell-based jets in L2 were extracted using dijet events simulated with PYTHIA [6] and following the approach described in Section 5.3. The L2 jet transverse energy scale and resolution for cell-based jets after calibration are presented in Fig. 9. The transverse energy scale is defined as the L2 jet transverse energy divided by the truth jet $E_T$. Truth jets are identified by applying the cone algorithm with $R_{\text{cone}} = 0.4$ on the collection of truth final state particles. The transverse energy scale is close to unity for all the $\eta$ coverage of the L2 jet trigger and all the transverse energies studied, demonstrating that the transverse energy is correctly measured within 2%. The jet transverse energy resolution decreases from 12% for the lowest transverse energies to 4% for transverse energies above 1000 GeV. The resolution curves were fitted with the following expression that includes a stochastic term convoluted with a constant term:

$$\frac{\sigma(E)}{E} = \frac{A}{\sqrt{E}} \oplus B \tag{9}$$

Table 3 shows the result of the fits for all the $\eta$ bins, before and after calibration. A few percent improvement in the resolution is obtained with the current calibration method. A further improvement can be achieved in the future by exploiting the correlation between the fraction of electromagnetic energy and the calibration weights [7].

In another study, two different window sizes were used in order to study the effect on the jet energy scale and resolution. L2 jets were reconstructed using a window size of $0.7 \times 0.7$ (in $\eta \times \phi$) which was chosen to be slightly smaller than the jet cone diameter ($2 \times \Delta R = 2 \times 0.4 = 0.8$) such that some of the energy of the jet may lie outside the window considered. Results were compared with the performance obtained using the window size dimension of $1.4 \times 1.4$. The jet energy calibration constants were calculated independently in both cases and the resulting jet energy scale and resolutions were compared. In both cases, the jet energy scale was found to be within 2% of unity. The resolution of the jets was also found to be similar for both window sizes. This means that the calibration algorithm can adequately correct for a small fraction of the jet energy lost outside the window considered. Therefore, using a window

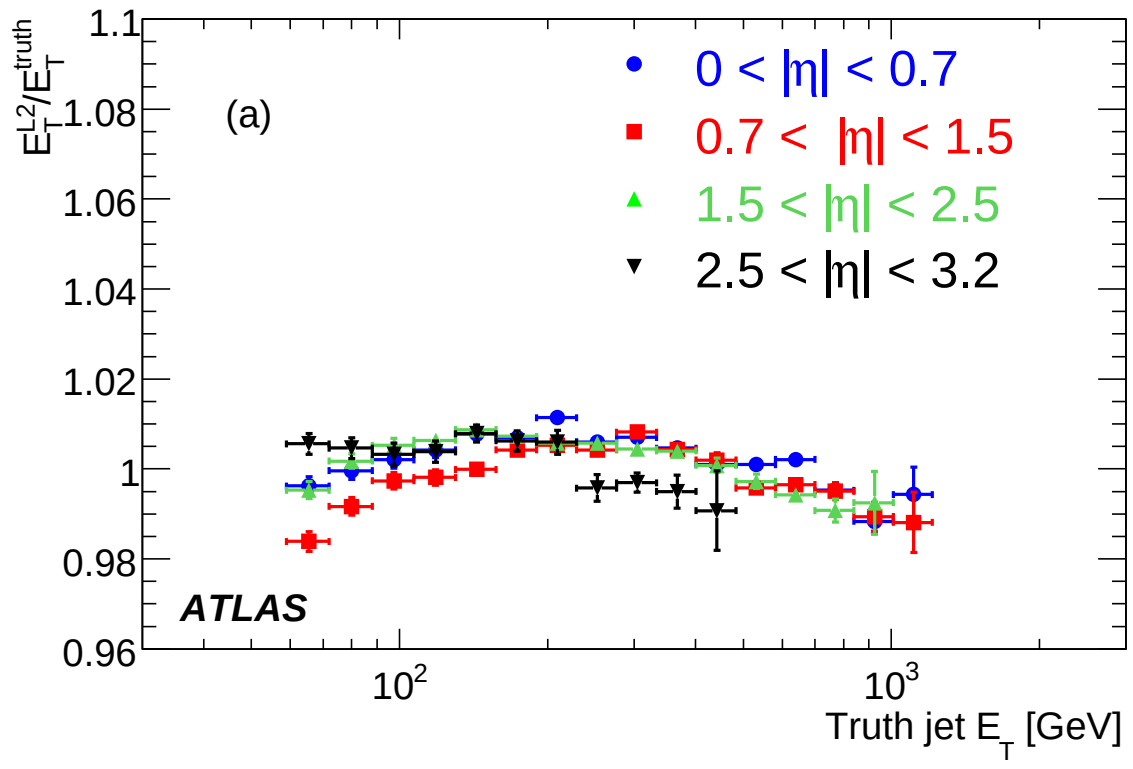


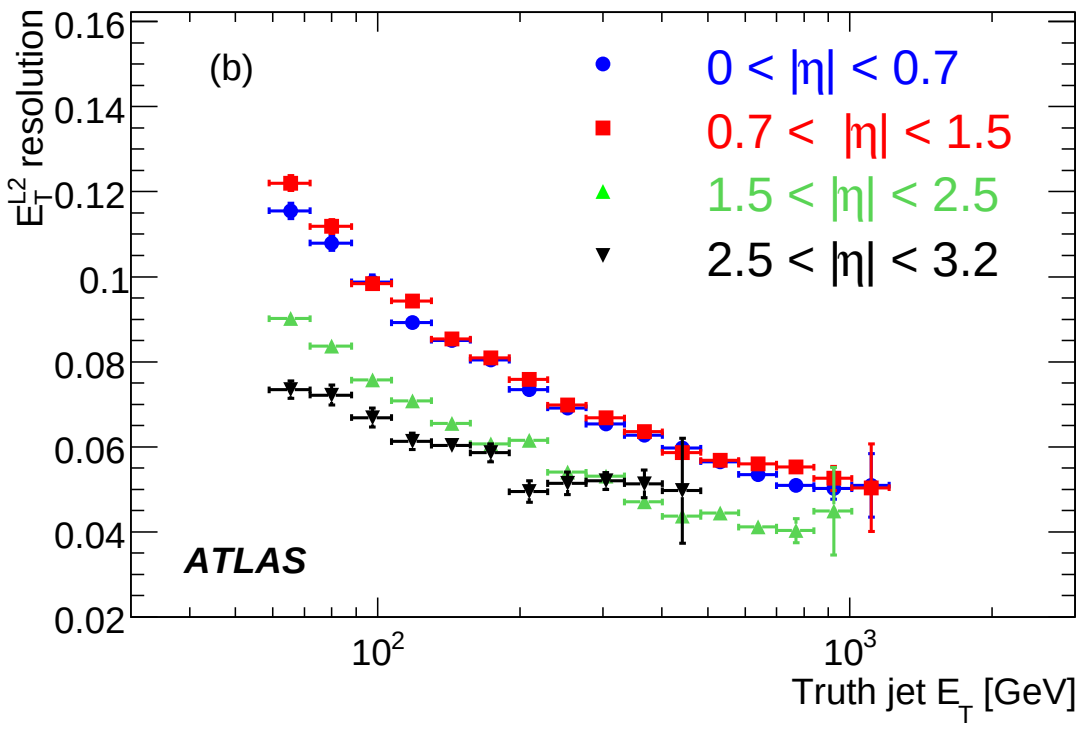


Figure 9: Jet energy scale for the L2 jets as a function of the truth jet $E_T$ (a), for four different bins in $\eta$. Jet energy resolution as a function of the truth energy of the jet (b), for four different bins in $\eta$. These results are obtained after calibration.

Table 3: Results of the jet energy resolution fit as a function of the truth jet energy, before and after applying the calibration. The fit was done using Equation (9).

| $\eta$ region | Before calibration | | After calibration | |
|---|---|---|---|---|
| | A | B | A | B |
| (0.0,0.7) | $1.03 \pm 0.03$ | $0.059 \pm 0.001$ | $0.96 \pm 0.02$ | $0.039 \pm 0.001$ |
| (0.7,1.5) | $1.28 \pm 0.03$ | $0.064 \pm 0.001$ | $1.18 \pm 0.03$ | $0.041 \pm 0.001$ |
| (1.5,2.5) | $1.53 \pm 0.04$ | $0.046 \pm 0.001$ | $1.37 \pm 0.03$ | $0.025 \pm 0.002$ |
| (2.5,3.2) | $1.86 \pm 0.13$ | $0.063 \pm 0.003$ | $1.46 \pm 0.08$ | $0.040 \pm 0.003$ |

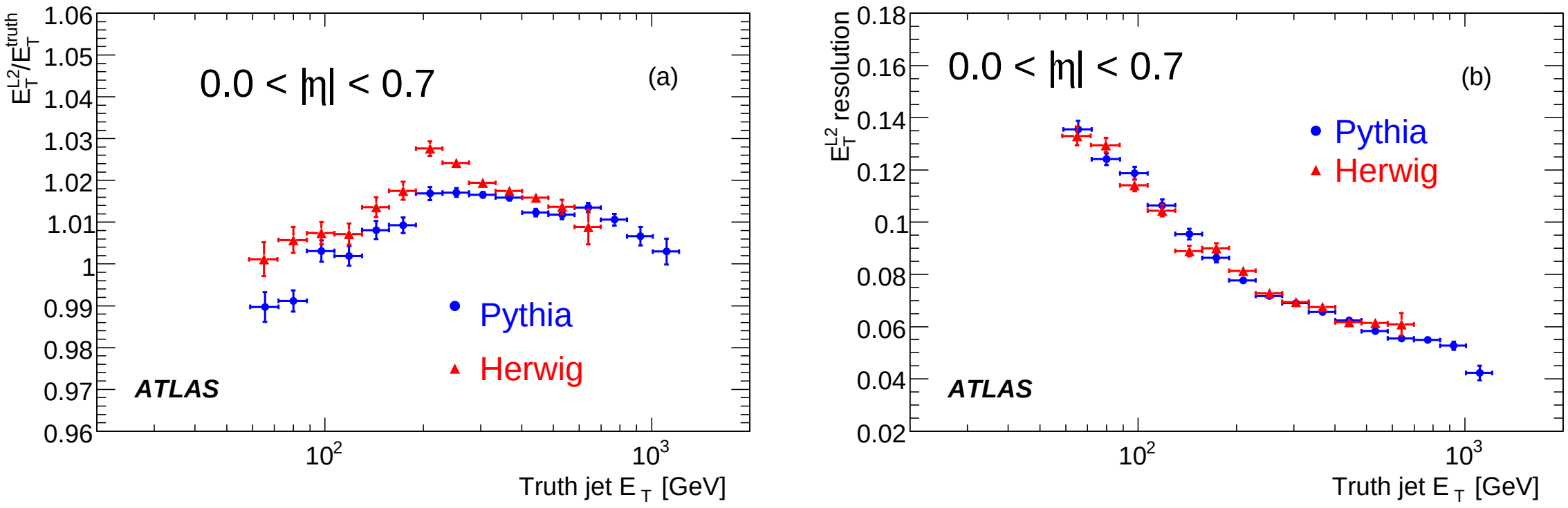


Figure 10: Comparison of the L2 jet energy scale and resolution obtained for two different MC generators, PYTHIA ( blue circles) and HERWIG (red triangles), for jets in the region $0 < \eta < 0.7$.

size of $1.0 \times 1.0$ will reduce processing time while keeping essentially the same physics performance as the larger window ($1.4 \times 1.4$).

The jet energy measured in the calorimeter may be sensitive to the shower development and the hadronisation mechanism that was introduced in the MC simulation. In order to test the sensitivity of the calibration procedure to the simulation, a set of dijet event samples generated with HERWIG [8] was used. Calibration constants were extracted with the PYTHIA dijet samples and used in the reconstruction of the HERWIG dijet events. The jet energy scale and resolution obtained in this way with the HERWIG data sample was compared, in different $\eta$ regions and $E_T$ values, with the one obtained for PYTHIA. Figure 10 shows, as an example, such a comparison in one particular $\eta$ region. Differences between the two generators were found to be smaller than 2%, for all regions of $\eta$ and jet $E_T$, suggesting that the L2 jet energy scale is relatively insensitive to the hadronisation model used in MC generators.

Figure 11 shows the L2 trigger efficiency as function of offline jet $E_T$ after calibration has been applied for four different L2 thresholds. The L1 threshold of 15 GeV was chosen to be significantly smaller than that for L2 in order to avoid a mixture of resolution and jet energy scale effects from the two different trigger levels. The limited sharpness of the curves shown on Figure 11 is, therefore, dominated by the resolution of the L2 jet energy.

Initially, the detector simulation is not expected to provide an exact model of the real detector. The energy measurement will be mainly affected by an incomplete knowledge of the dead material distribution in front of the calorimeters. To study the effect that the limited knowledge of the detector geometry may have on the performance of the trigger and reconstruction algorithms, dedicated MC dijet event samples were produced. They were reconstructed with a geometry where detectors were slightly displaced from their nominal positions and extra dead material was added. This accounted for 7-10% of a radiation length in the inner detector and a few percent of one radiation length in front of the calorimeter. The knowledge of the dead material distribution is assumed to be worst at the interface between different calorimeter subsystems.

The calibration constants obtained assuming a perfect geometry were used to identify L2 jets in the misaligned samples. Hence, the resulting jet transverse energy scale and resolution can serve as an estimation of how much the performance of the L2 jet reconstruction may be degraded due to the limited knowledge of the detector geometry and dead material distribution at the beginning of the data taking period. Figure 12 shows the jet transverse energy scale and resolution obtained for dijet events reconstructed with this mismatch of calibration constants. For most of the pseudo-rapidity region, the linearity with energy is within 3-4%, slightly worse than before. In the region $(1.5 < |\eta| < 2.5)$ the jet

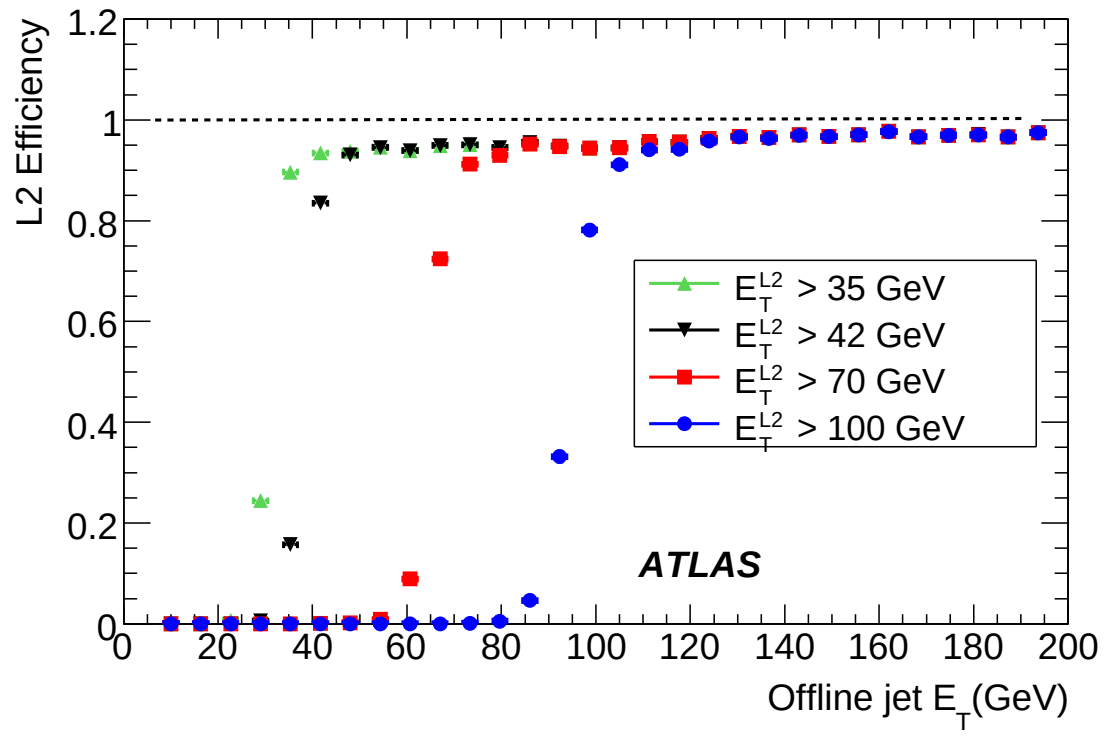


Figure 11: Trigger efficiency as function of offline jet transverse energy for L2 jets after calibration, for four different thresholds (35 GeV, 42 GeV, 70 GeV and 100 GeV). The statistical uncertainty on each point is smaller than the symbols.

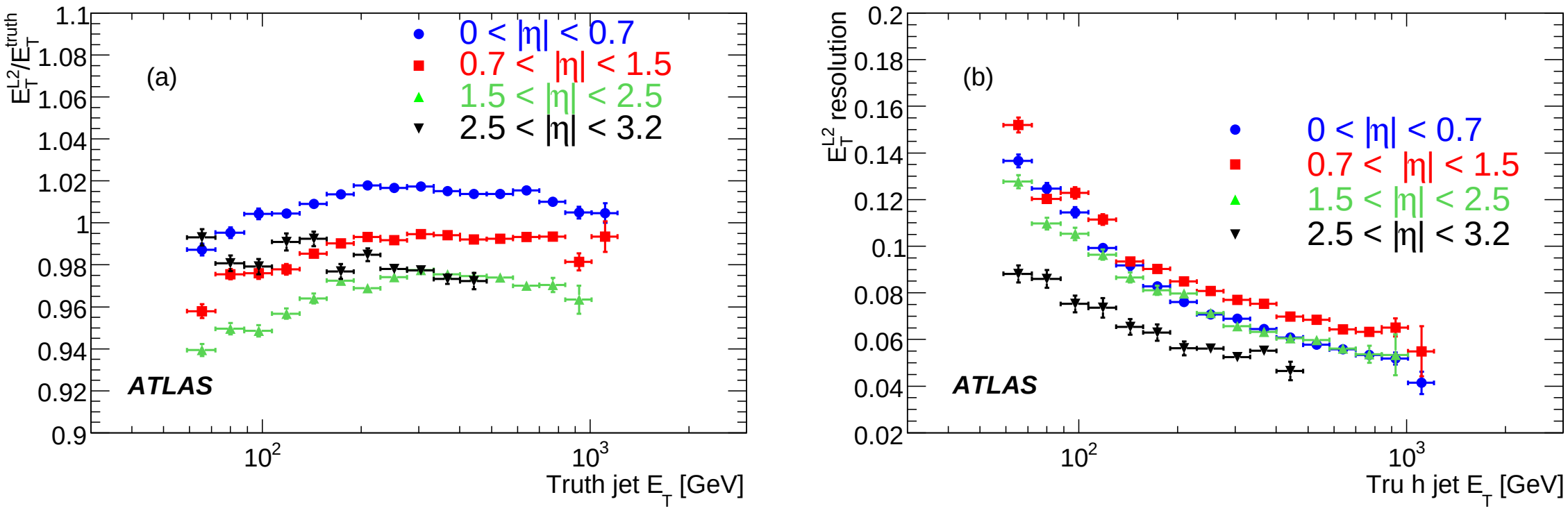


Figure 12: Jet transverse energy scale for L2 jets as a function of the truth jet transverse energy (a), for four different $\eta$ regions. Jet transverse energy resolution as a function of the truth jet transverse energy (b), for four different $\eta$ regions. Both plots were obtained using dijet event samples reconstructed assuming a limited knowledge of the detector's dead material distribution.

energy scale drops to about 94% due to the extra dead material. The transverse energy resolution is also degraded, as shown in Table 4. The performance of the calibration at the beginning of data taking can be improved using in-situ calibration procedures used to extract the calibration constants directly from the data or to correct the detector response in the MC. Several different procedures are currently under study.

The reconstructed jet position is another parameter used to measure the algorithm performance. Currently no selection on jet position is made at the trigger level, but this may prove useful in the future for some physics channels. The position resolution is shown in Fig. 13. This figure is also included for completeness to compare with the FEB unpacking approach that uses a coarser granularity of data.

## 6.2 Performance for FEB-based reconstruction

Studies were done to evaluate the performance of the L2 jet reconstruction using the FEB-based method of data unpacking. For comparison with results presented in Fig. 12, dijet event samples that assume

Table 4: Results of the jet energy resolution fit as a function of the truth jet energy for the event samples where a limited knowledge of the detector's dead material is assumed. The fit was done assuming Equation (9).

| $\eta$ region | After calibration | |
|---|---|---|
| | A | B |
| (0.0,0.7) | $1.04 \pm 0.02$ | $0.038 \pm 0.002$ |
| (0.7,1.5) | $1.24 \pm 0.03$ | $0.055 \pm 0.001$ |
| (1.5,2.5) | $1.95 \pm 0.04$ | $0.018 \pm 0.004$ |
| (2.5,3.2) | $1.66 \pm 0.09$ | $0.039 \pm 0.002$ |

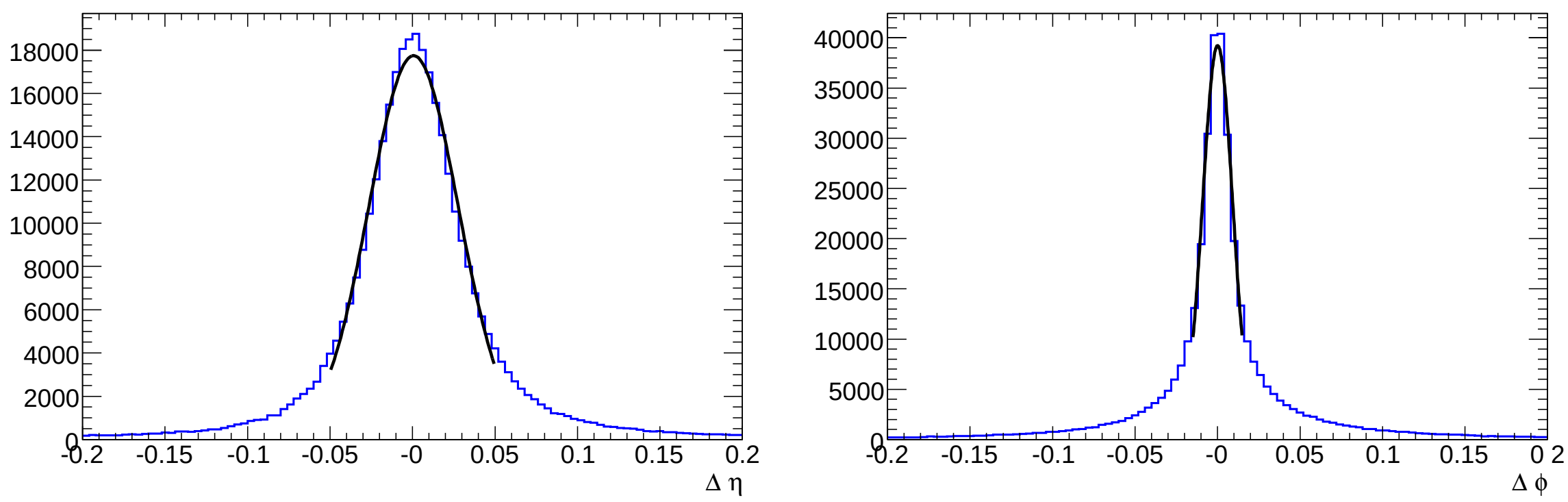


Figure 13: The $\eta$-resolution (a) and $\phi$-resolution (b) of L2 cell-based jets with respect to the truth jet energy. The mean and standard deviation of a Gaussian fit of (a) is 0.0006 and 0.03, while the mean and standard deviation of a Gaussian fit of (b) is 0.00005 and 0.01.

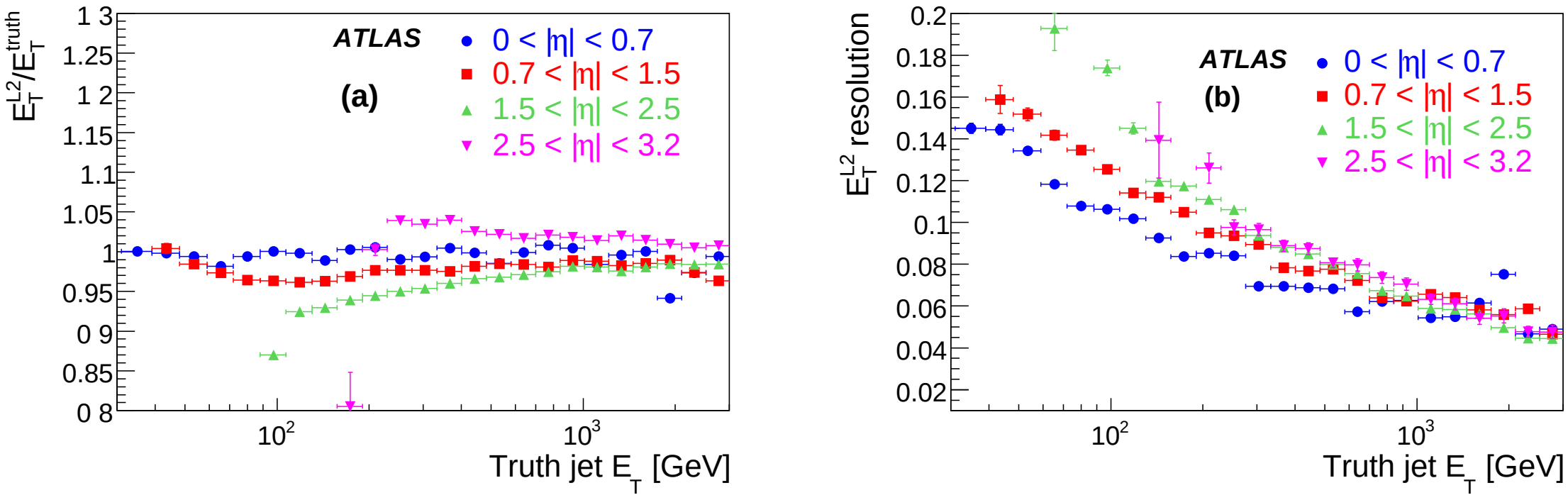


Figure 14: (a) Jet energy scale for the L2 FEB-based jets as a function of the truth jet $E_T$ for four different bins in $\eta$. (b) Jet energy resolution as a function of the truth energy of the jet for four different bins in $\eta$. Both plots were obtained using dijet event samples reconstructed assuming a limited knowledge of the detector's dead material distribution.

a limited knowledge of the detector dead material were used to study the transverse energy scale and resolution of FEB-based jets. The transverse energy of jets reconstructed using the FEB-based data unpacking approach was weighted using the default L2 calibration constants obtained using the cell-based method.

Figure 14 shows the jet transverse energy scale and resolution. The energy scale stays within 5% of unity for most of the pseudorapidity range. The transverse energy resolution distribution was fitted using Equation (9) and the results are presented in Table 5. The transverse energy resolution of FEB-based jets is comparable to that of cell-based jets presented in Table 3. The transverse energy scale and resolution of the FEB-based jets depend strongly on the energy cut-off introduced in Equation (3). It was found that using a cut-off at $2 \cdot \sigma_{\text{noise}}$ gave the best result in terms of transverse energy scale and resolution.

Figure 15 shows the $\eta$ and $\phi$ resolution. These distributions were obtained from the difference between a L2 FEB-based jet and its nearest MC truth jet. These results are comparable with those obtained using the cell-based method as shown in Fig. 13.

The L2 trigger efficiency for different thresholds as function of reconstructed jet transverse energy for FEB-based jets is presented in Fig. 16. The initial slope of these efficiency curves is similar to the results obtained with the cell-based method shown in Fig. 11. This indicates that FEB-based jets have a selection efficiency that is comparable to the cell-based one. The FEB-based jet reconstruction significantly reduces the amount of data to be unpacked at L2 compared to the cell-based approach. The impact of the unpacking choice on the processing time of the jet reconstruction algorithm is presented in Section 6.4.

### 6.3 Forward jets

The implementation of L2 forward jet reconstruction must consider constraints on the precision of the L1 forward jet RoI position. Since each L1 FCAL trigger tower spans the entire $\eta$ range of the FCAL detector, a dedicated optimisation of the jet reconstruction algorithm is required. In order to properly account for the overlap region between the FCAL and endcap calorimeters, a window size of $3.0 < |\eta| < 5$ was chosen. The cone size ($\Delta R$) must also be modified in order to remove any bias from the initial L1 RoI seed position. Similar to the standard jet reconstruction, three iterations of the jet finding algorithm are performed. An initial cone size of $\Delta R = 1$ is used to collect sufficient data to remove the L1 bias. In the second and third iteration of the jet finding algorithm, the cone size is reduced to 0.7

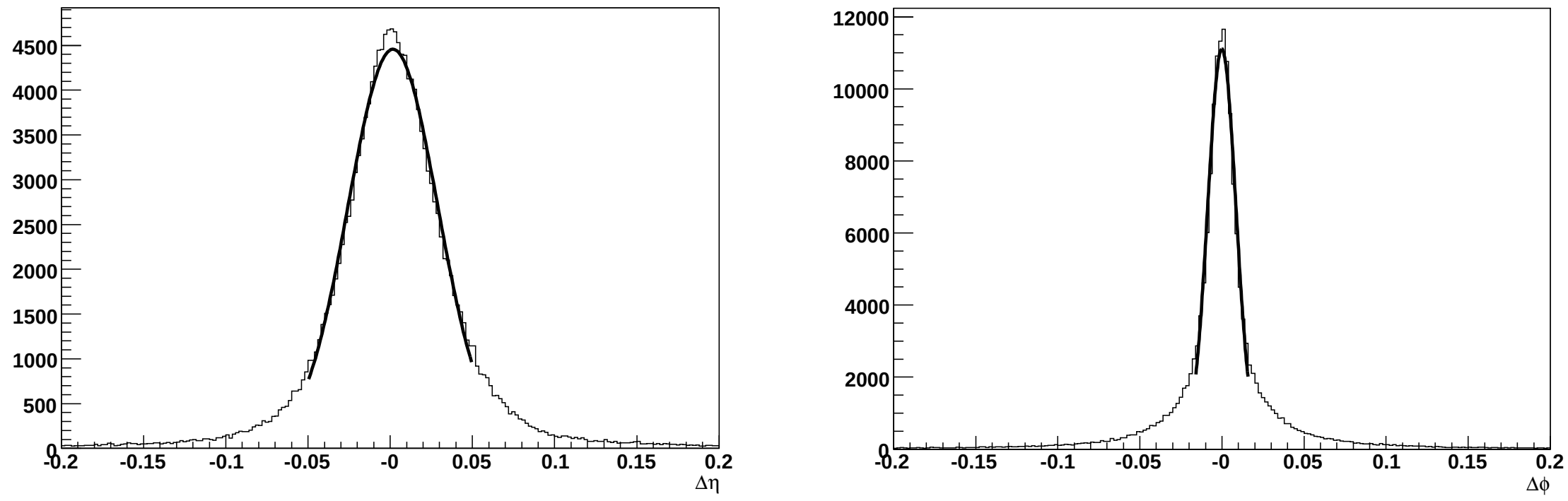


Figure 15: The $\eta$-resolution (a) and $\phi$-resolution (b) of L2 FEB-based jets with respect to the truth jet energy. The mean and standard deviation of a Gaussian fit of (a) is 0.002 and 0.03, while the mean and standard deviation of a Gaussian fit of (b) is -0.00007 and 0.009.

Table 5: Results of the jet energy resolution fit as a function of the truth jet energy for FEB-based jets after applying the default calibration constants obtained using the cell-based data unpacking method. The fit was performed assuming Equation (9).

| $\eta$ region | After calibration | |
|---|---|---|
| | A | B |
| (0.0,0.7) | $0.93 \pm 0.05$ | $0.02\pm 0.01$ |
| (0.7,1.5) | $1.18 \pm 0.05$ | $0.03\pm 0.01$ |
| (1.5,2.5) | $1.56 \pm 0.04$ | $0.05\pm 0.01$ |
| (2.5,3.2) | $1.93 \pm 0.01$ | $0.01\pm 0.01$ |

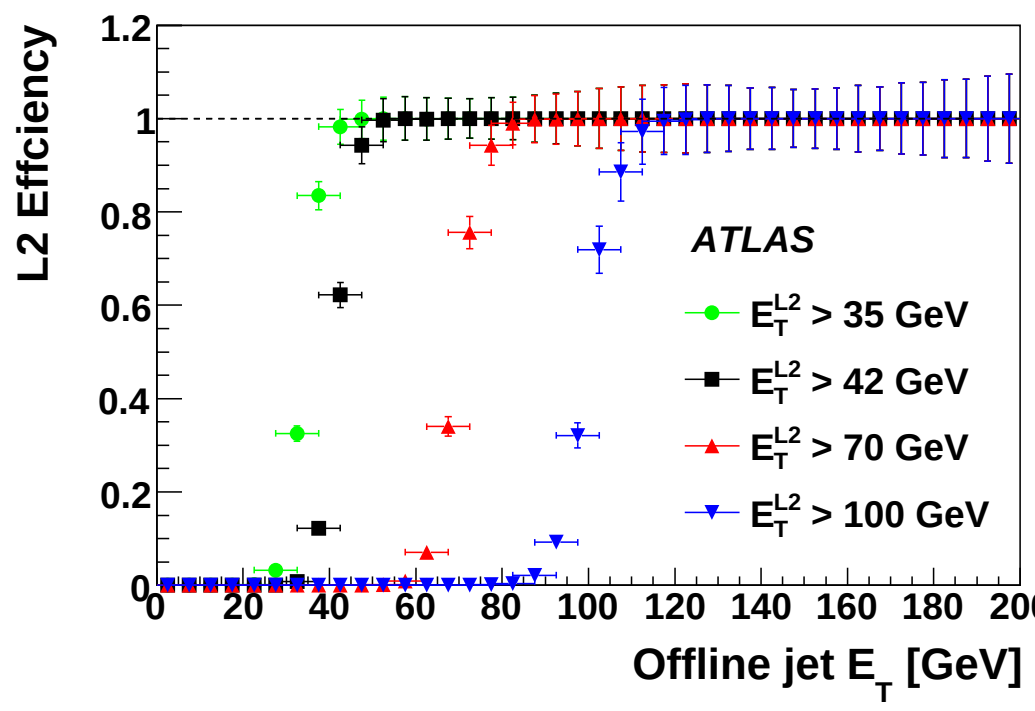


Figure 16: L2 trigger efficiency as function of reconstructed jet transverse energy for FEB-based jets after calibration, for four different thresholds (35 GeV, 42 GeV, 70 GeV and 100 GeV).

Table 6: L2 trigger efficiency for L2 forward jets with respect to truth jets with $E_T > 25$ GeV.

| Tagged Objects | Data unpacking method | |
|---|---|---|
| | cell-based jets (%) | FEB-based jets (%) |
| Highest $p_T^{\text{truth}}$ forward jets | 98±1 | 98±1 |
| All forward jets | 97±1 | 91±1 |

and subsequently to 0.4. Calibration constants for forward jets are derived using the method described in Section 5.3.

The L2 forward jet trigger efficiency was determined with respect to truth jet. Reconstructed jets were matched to truth particle jets using the requirement that $\Delta R = \sqrt{\Delta\eta^2 + \Delta\phi^2} < 0.2$. In addition, truth jets were required to satisfy $p_T$>25GeV. Results from this study are presented in Table 6.

## 6.4 L2 jet timing

The time budget of approximately 40 ms per event and the strong rejection power needed at the L2 trigger impose strong constraints on the L2 algorithms speed and physics performance.

During the commissioning of the ATLAS Trigger/DAQ system, dedicated Technical Runs are performed in order to test HLT algorithm performance. In these runs all the detector Read Out Systems (ROS) and the full Trigger/DAQ infrastructure are dedicated to exercising the data acquisition system and the trigger. Bytestream files[1] containing a mixture of events close to that expected in LHC collisions (mainly QCD jets, mixed with a few $Z$ and $W$ decays to leptons, $t\bar{t}$ events, etc.) are preloaded into the ROS. These events, which contain the RoIs obtained from the L1 simulation, are then processed by the HLT system. The measurements presented here were made during the Technical Run that took place in November 2007.

In order to compare the difference between the cell-based method and the FEB-based method, two data-collection runs were recorded, one with L2 jets reconstructed using the cell-based approach, and one with L2 jets reconstructed using the FEB-based method. The timing distributions obtained from the cell and the FEB-based methods are shown in Fig. 17(a). Both distributions have a similar shape with peaks corresponding to the number of RoIs per event. The FEB-based approach is however almost 50% faster than the cell-based method.

The distribution of the total processing time shown in Fig. 17(a) includes the data unpacking, jet finding algorithm and calibration, as well as the data collection time from the detector Read Out System (ROS) to the L2 processors, which is shown in Fig. 17(b). The data collection time contributes significantly to the total L2 processing time, about 30% for the cell-based method and about 50% for the FEB-based approach. A detailed description of the data preparation methods and performance can be found in [9].

The processing time distribution per RoI for the different steps involved in the L2 jet reconstruction are shown in Fig. 18(a) and (b) for the cell-based and FEB-based method, respectively. These measurements show that the algorithm execution time is dominated by the data unpacking step. The small features observed in the unpacking time distributions outside the main peak come from RoIs that point in regions where less data needs to be unpacked.

Timing measurements of the L2 forward jet algorithm were also performed. A comparison of the reconstruction time between cell-based and FEB-based jets is presented in Table 7. Due to the limited time available during dedicated Technical Runs, these measurements were obtained by running the L2 jet

[1] Files with the same format as the raw data that will come out of the ATLAS detector.

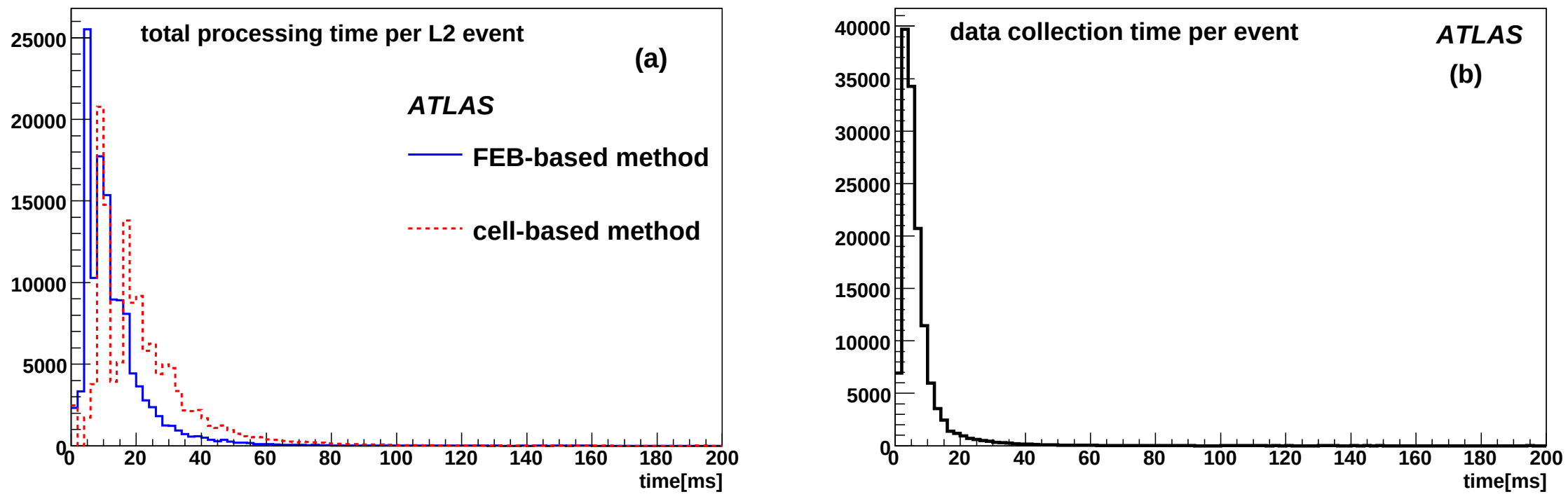


Figure 17: The total processing time per event (a) for the L2 jet algorithm. The solid line is the processing time measured using the FEB-based method (mean 13 ms). The dashed line is the processing time measured using the cell-based method (mean 22 ms). The total data collection time per RoI (b) for both the cell-based and FEB-based data unpacking methods.

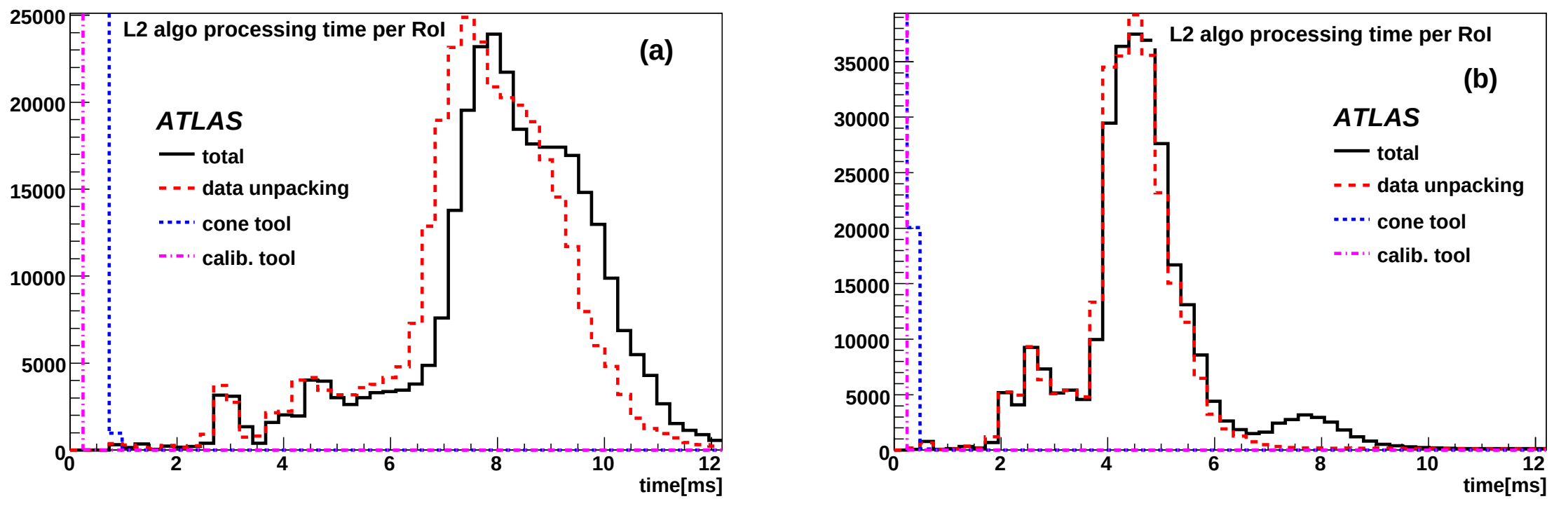


Figure 18: L2 jet algorithm processing time per jet RoI for the cell-based method (a) and FEB-based method (b). The total processing time is shown together with the execution time of the individual steps involved in the L2 jet reconstruction.

Table 7: Comparison of the average L2 jet reconstruction time between the standard and forward jet algorithms in units of ms. Different data unpacking methods are also compared.

| Tagged Objects | Data unpacking method | |
|---|---|---|
| | cell-based jets | FEB-based jets |
| Forward Jets | 0.9 | 0.5 |
| Standard Jets | 6.1 | 2.1 |

algorithms on offline computing resources rather than making use of the full Trigger/DAQ infrastructure. Although the absolute values of reconstruction time should only be taken as an approximate indication of the performance to be expected on the online trigger system, the relative comparison of each measurement is nevertheless informative. For example, although the L2 forward jet algorithm uses a larger $\eta \times \phi$ window than the standard L2 jet algorithm, it requires less than 25% of the total time used to reconstruct standard L2 jets. This is due to the larger granularity of the forward calorimeter, as compared to the liquid Argon barrel and endcap calorimeters, which results in a smaller amount of data being unpacked on average.

# 7 The event filter jet algorithms

The jet reconstruction in the Event Filter (EF) can be divided into two tasks: the input data preparation and the jet finding. These two tasks are carried out by algorithms used in the offline reconstruction but adapted to run in the online trigger environment. Once an EF jet has been identified, it is passed on to a Hypothesis algorithm which validates whether or not the reconstructed jet satisfies a predefined transverse energy threshold.

## 7.1 Input data preparation

There are three calorimeter data preparation algorithms which, respectively, unpack the calorimeter cell information, build trigger towers and construct calorimeter clusters. The implementation of these data preparation algorithms in the EF has been adapted to allow the running of a subset of the algorithms thereby reducing computing time.

The first data preparation algorithm unpacks calorimeter cell information in a predefined window around the position of the jet found at L2. The size of the window where data needs to be unpacked is a parameter that can be adjusted and is configured at trigger initialisation. The energy of each calorimeter cell, at this point, is set at the EM scale [2, 3]. Although not currently used, this unpacking algorithm provides the ability to apply calibration weights to the energy of individual calorimeter cells.

The second data preparation algorithm performs the calorimeter tower reconstruction. A calorimeter tower is an array of calorimeter cells within an $(\Delta\eta, \Delta\phi) = (0.1, 0.1)$ region. The EF calorimeter trigger towers granularity is four times finer than that of the L1. Calorimeter trigger towers are the inputs to the jet finding algorithm described in the next Section.

A third data preparation algorithm, currently being studied, is also available in the EF. This algorithm constructs three-dimensional calorimeter clusters instead of towers. These clusters can alternatively be used as input to the jet finding algorithm instead of calorimeter towers.

| Parameter | Value |
|---|---|
| Window size ($\eta \times \phi$) | $1.6 \times 1.6$ |
| Input objects | towers |
| Proto-jet $E_T$ cut | $E_T > 2$ GeV |
| Jet finding algorithm | cone |
| Jet finding parameter | $R_{\text{cone}} = 0.7$ |
| Calibration scheme | "Energy density"-based cell calibration |
| Final jet $E_T$ cut | $E_T > 10$ GeV |

Table 8: Parameter values for the EF jet reconstruction.

### 7.2 Jet finding algorithm

The EF jet reconstruction uses the offline reconstruction algorithms [10] adapted for the EF. It takes as input any $E_T$ ordered list of calorimeter objects (cells, towers or clusters). The jet reconstruction consists of the following steps:

- Removal of negative energy towers (or clusters) by combining them with adjacent ones. A list of proto-jets is also constructed using a simple pre-clustering algorithm.
- Removal of proto-jets with transverse energy smaller than a given threshold.
- Running of a jet finding algorithm (cone or fast $\text{K}_\text{T}$ algorithm [11]).
- Jet calibration
- Removal of reconstructed jets below a given transverse energy

Many different parameters can be modified as part of the jet reconstruction. The parameters used for the configuration of the EF jet reconstruction are summarized in Table 8. Unless specified explicitly, performance studies described subsequently use this configuration.

All calibration methods available for the offline jet reconstruction can be used by the EF jet algorithm. The default method used in the EF is an "energy density"-based cell calibration described in details in [12].

## 8 Event filter performance

The performance and parameter optimisation of the EF jet reconstruction algorithms were made using the simulated dijet event samples summarised in Table 1.

The window size defining how much data is unpacked was chosen by studying the position difference between offline reconstructed jets and jets found by the L2 system. A distribution of the distance $\Delta R = \sqrt{(\eta_{\text{L2}} - \eta_{\text{offline}})^2 + (\phi_{\text{L2}} - \phi_{\text{offline}})^2}$ is shown in Fig. 19. Most L2 jet RoI are well within $\Delta R < 0.2$ of offline reconstructed jets using a cone algorithm with $\Delta R = 0.7$. A window size of $1.6 \times 1.6$ in $\eta \times \phi$ was therefore chosen to ensure that data is unpacked at the EF in a calorimeter region of sufficient size to reconstruct correctly a jet similar to an offline jet while ensuring complete overlap with the L1 RoI. Additional studies could include the optimization of a variable window size as function of the jet transverse energy as measured by L2.

The EF jet energy scale and resolution were assessed using the same technique as described in Section 6.1. The results are shown in Fig. 20 as a function of the truth jet $E_T$. In general, the energy scale

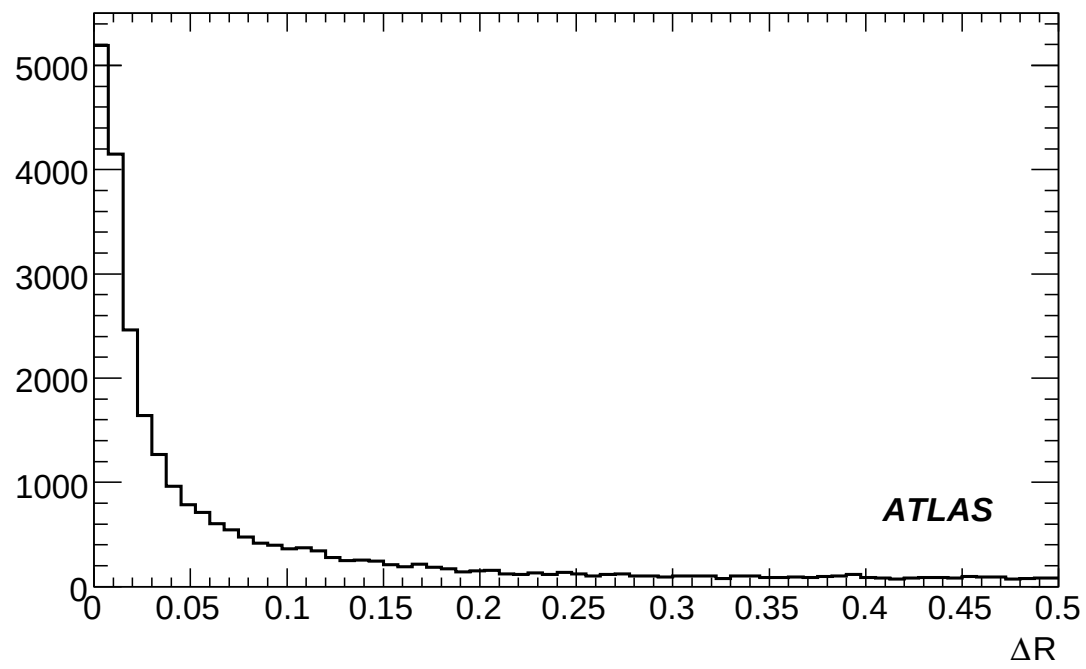


Figure 19: Distance $\Delta R = \sqrt{(\eta_{\mathrm{L2}} - \eta_{\mathrm{offline}})^2 + (\phi_{\mathrm{L2}} - \phi_{\mathrm{offline}})^2}$ between L2 and offline reconstructed jets using a cone algorithm of $R_{\mathrm{cone}} = 0.7$.

improves with increasing truth jet $E_T$. For truth jet energies larger than 200 GeV, the EF jet energy scale is within 2% of unity. The limited window size within which the jet reconstruction takes place in the EF results in some energy leakage that is not corrected for by the offline calibration used in the EF. These results nevertheless suggest that offline calibration are adequate for use in the EF jet reconstruction.

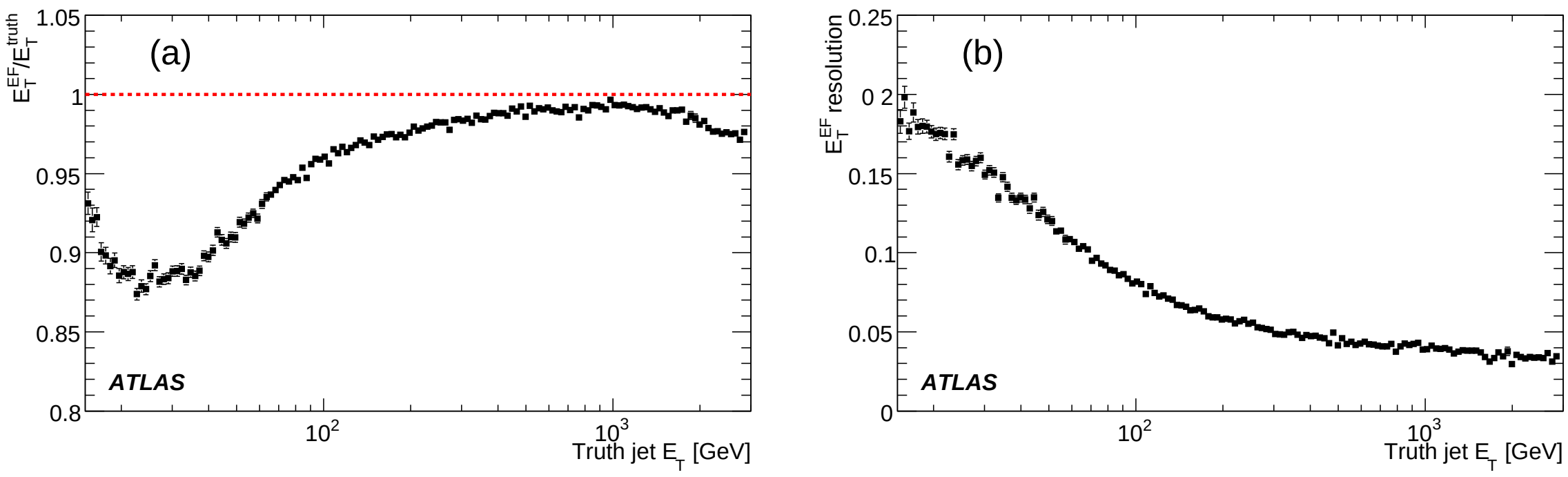


Figure 20: The EF jet transverse energy scale (a) and resolution (b) as function of truth jet $E_T$.

The effect of pile-up corresponding to a luminosity of $10^{33}\mathrm{s}^{-1}\mathrm{cm}^{-2}$ on the jet energy scale and resolution was also studied. Figure 21 shows a comparison of (a) the EF jet transverse energy scale and (b) the EF resolution for simulated dijet event samples with and without pile-up. The effect of pile-up on the overall energy scale is observed to be insignificant, however, a noticeably worse resolution is observed for events with pile-up. The effect is clearly more important for low jet energies where the contribution of pile-up energy can be comparable to the total jet energy.

The EF jet trigger efficiency as function of offline jet energy is shown in Fig. 22 for two different signatures. For an EF trigger threshold close to the L1 threshold, the trigger selection performance is limited by the L1 energy resolution. For EF trigger thresholds significantly higher than the L1 threshold, the excellent EF energy scale improves slightly the performance of the trigger selection as compared to the current L2 capability. These plots emphasize the importance of correctly optimizing the jet trigger thresholds used for each jet signature.

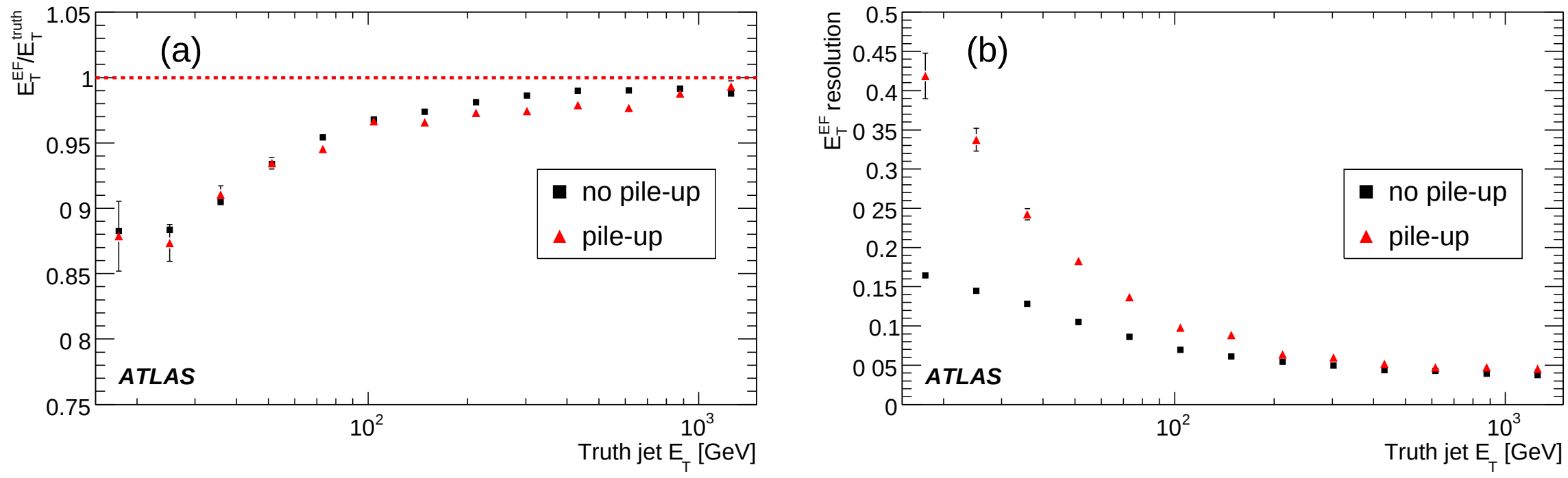


Figure 21: The EF jet transverse (a) energy scale (b) and resolution as a function of truth jet $E_T$ for simulated dijet event samples with and without pile-up.

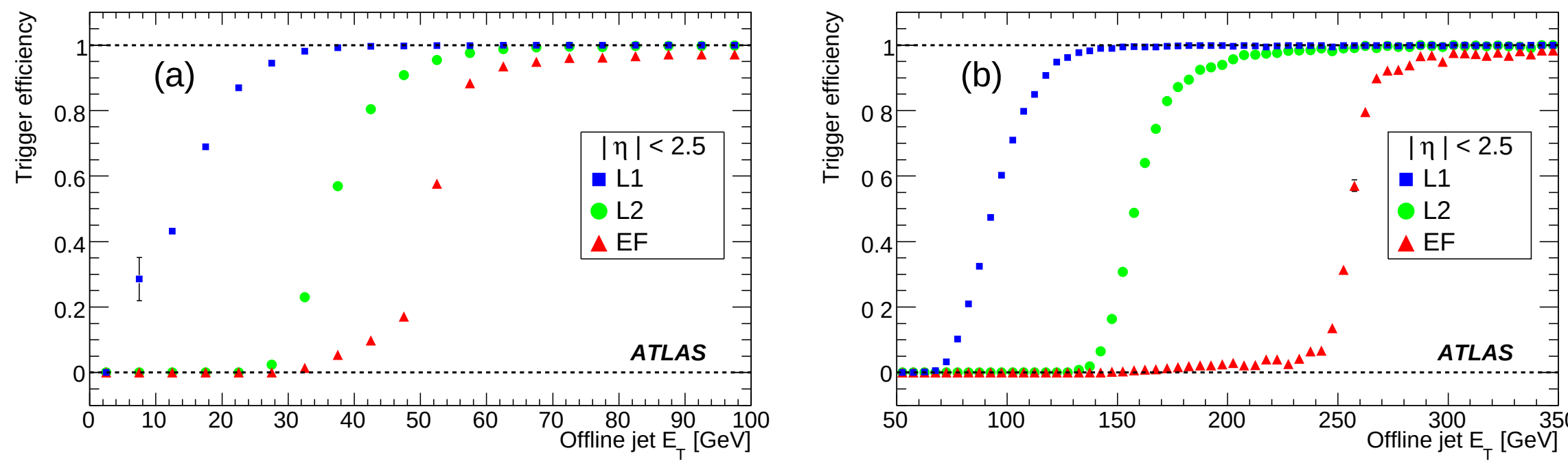


Figure 22: EF trigger efficiency as function of offline jet transverse energy for two different signatures consisting of a set of L1, L2 and EF trigger thresholds: (a) $E_T^{L1} > 10$ GeV, $E_T^{L2} > 30$ GeV, $E_T^{EF} > 50$ GeV; and (b) $E_T^{L1} > 70$ GeV, $E_T^{L2} > 150$ GeV, $E_T^{EF} > 255$ GeV.

### 8.1 Event filter jet algorithm timing

The timing performance of the EF jet reconstruction algorithm has been measured in the November 2007 Technical run described in Section 6.4. Figure 23(a) shows the execution time per RoI for the two EF data preparation steps: the unpacking of calorimeter cell data and the construction of calorimeter towers used by the jet algorithm. Figure 23(b) shows the execution time of the different steps involved in the jet reconstruction described in Section 7.2. The EF jet reconstruction total time per RoI is of order 100 ms. Assuming approximately 4 to 5 jet RoI per event implies a total processing time of order half a second per event which falls within the design budget of approximately one second per event.

## 9 Jet trigger menu

Jet triggers will be used to record many different types of events. Single and multi-jet signatures will identify useful events for various Standard Model QCD measurements. This event sample will also be valuable to study the properties of background events for other analyses and to measure the misidentification efficiency of different offline reconstruction algorithms. Jet trigger requirements are also important when combined with other trigger criteria to identify rare signal events with well-defined topologies.

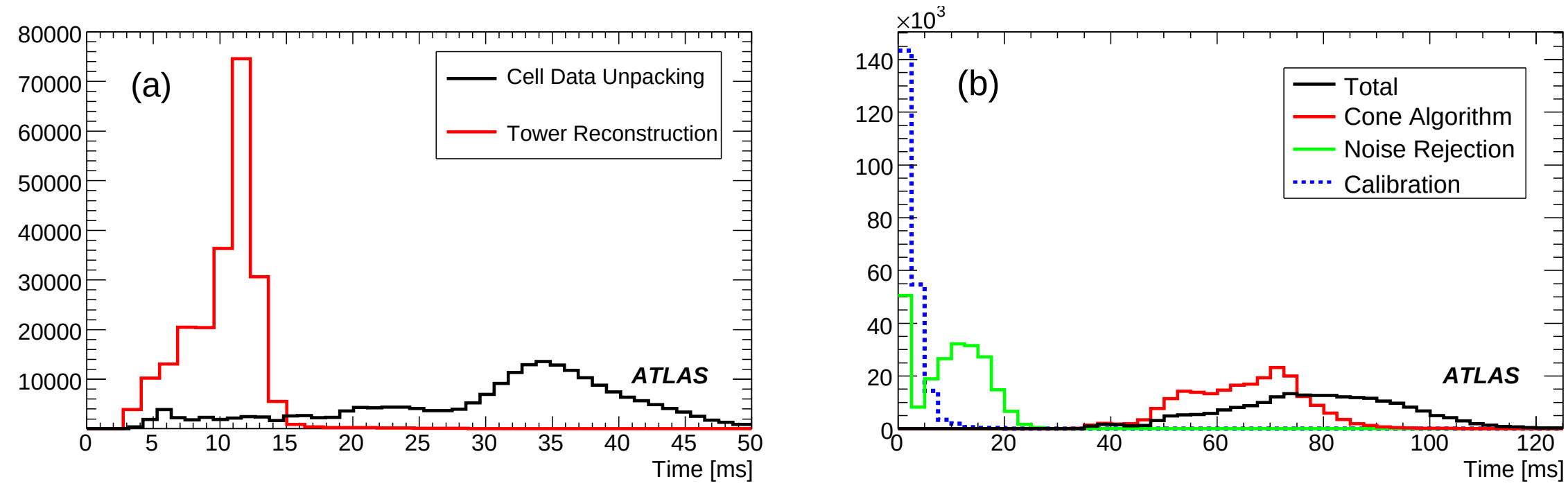


Figure 23: Execution time per RoI for different steps in the EF data preparation (a) and the jet reconstruction (b).

| **Trigger threshold** | J5 | J10 | J18 | J23 | J35 | J42 | J70 | J120 | 3J10 | 3J18 | 4J10 | 4J18 | 4J23 |
|---|---|---|---|---|---|---|---|---|---|---|---|---|---|
| Prescale factor | 300000 | 42000 | 6000 | 2000 | 500 | 100 | 15 | 1 | 150 | 1 | 30 | 1 | 1 |
| Level 1 rate (Hz) | 1 | 4 | 1 | 1 | 1 | 4 | 4 | 8 | 40 | 140 | 40 | 20 | 8 |

Table 9: Summary of the L1 single-jet and multi-jet triggers optimised for the initial data taking period. The L1 prescale factors and expected rate at a luminosity of $10^{31}\text{cm}^{-2}\text{s}^{-1}$ for each threshold are also presented.

The trigger strategy adopted is to define a set of single and multi-jet signatures that, together, will select events with approximately uniform rates over the entire jet energy spectrum. Table 9 shows the set of L1 single and multi-jet signatures and associated L1 prescale factors optimised for this purpose, assuming an initial luminosity of $10^{31}$ cm$^{-2}$s$^{-1}$. The trigger rates were calculated using close to 7 million simulated non-diffractive inelastic events, with an estimated cross-section of 70 mb. These large simulated event samples result in a statistical uncertainty of approximately 0.1 Hz for the rates of unprescaled triggers. Note that a maximum of 8 different jet thresholds can be defined in the L1 system. The L1 thresholds and prescale factors for the single-jet signatures were chosen to significantly limit the L1 output event rate thereby initially avoiding the need to use the HLT at the beginning of data-taking. The L1 jet trigger menu was also designed to be compatible with a luminosity of up to $10^{32}$ cm$^{-2}$s$^{-1}$ without any changes. Figure 24 shows the differential distribution for the number of events selected by the L1 trigger menu as function of offline reconstructed jet transverse energy for 1 fb$^{-1}$ of recorded data.

For commissioning purposes, it is foreseen that the L2 and EF jet algorithms will be initially run for the single-jet signatures in so-called pass-through mode where the result of the HLT selection is recorded with the event data but no events are rejected based on the result of the algorithm selection. The rate of the multi-jet signatures listed in Table 9 will be reduced by using L2 and EF jet requirements. Table 10 summaries the expected rate out of the HLT for all jet signatures foreseen to be run at the beginning of data-taking. In addition to the single and multi-jet signatures discussed above, forward jets signatures are also foreseen. These are also shown in Table 10.

It should be noted that, due to the very large prescales used for the lower threshold jet triggers, the overlap between individual triggers is greatly reduced. This implies that the cumulative rates will rapidly grow, as can be observed in Table 10. The total rate of the entire jet menu has been approximated to be 37 Hz, which represents a little over 18% of the overall trigger output rate budget of 200 Hz.

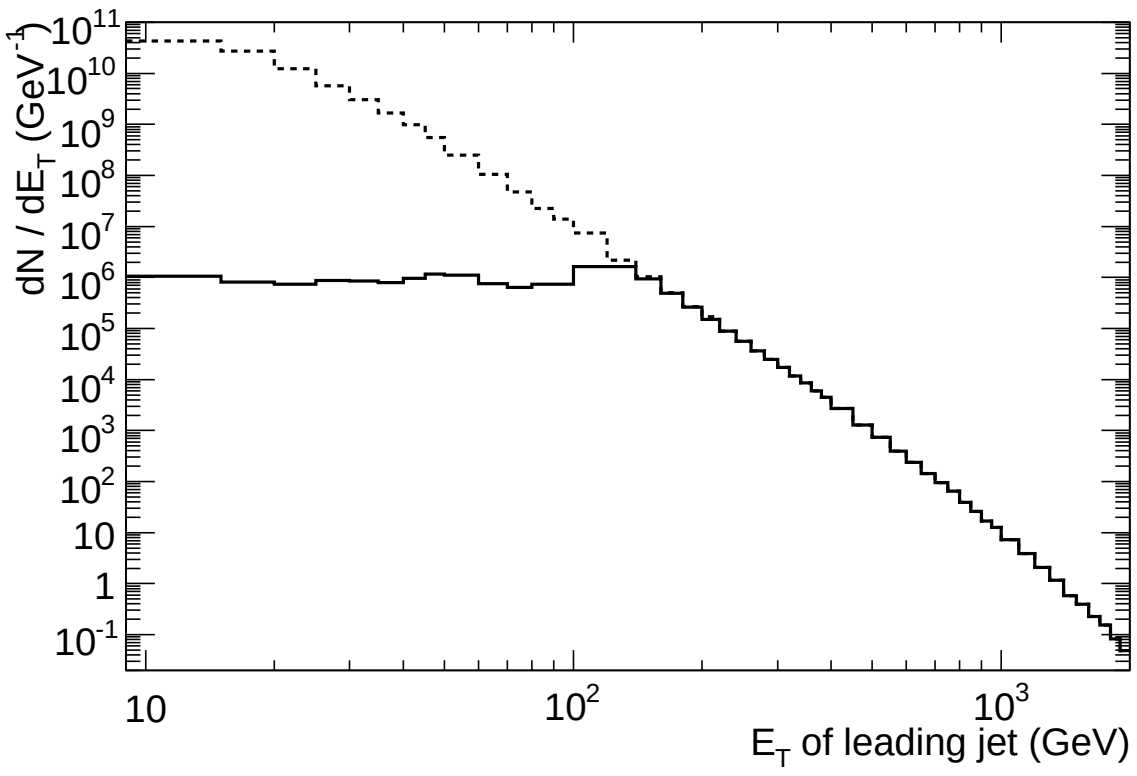


Figure 24: Differential number of events selected as function of the offline reconstructed transverse energy of the leading jet in the event for 1 $fb^{-1}$ of data. The dashed and solid lines show the expected distributions before and after applying the L1 jet trigger menu criteria described in Table 9.

| **Trigger** | **Overall Prescale** | **Rate (Hz)** | **Cumulative Rate (Hz)** |
|---|---|---|---|
| 4j23 | 1 | 6.9 (± 0.8) | 6.9 (± 0.8) |
| 4j18 | 100 | 0.14 (± 0.01) | 7.0 (± 0.6) |
| 4j10 | 300 | 0.045 (± 0.004) | 7.0 (± 0.6) |
| 3j18 | 100 | 0.92 (± 0.03) | 7.9 (± 0.3) |
| 3j10 | 1500 | 0.061 (± 0.002) | 7.9 (± 0.2) |
| Total Multi-Jets | | 7.9 (± 0.2) | |
| j120 | 1 | 8.7 (± 0.9) | 15.3 (± 0.5) |
| j70 | 15 | 4.2 (± 0.2) | 18.7 (± 0.5) |
| j42 | 100 | 3.73 (± 0.06) | 22.3 (± 0.3) |
| j35 | 500 | 1.37 (± 0.02) | 23.6 (± 0.3) |
| j23 | 2000 | 1.37 (± 0.008) | 24.9 (± 0.2) |
| j18 | 6000 | 1.02 (± 0.004) | 26.0 (± 0.1) |
| j10 | 42000 | 3.9 (± 0.003) | 29.9 (± 0.02) |
| j5 | 300000 | 0.9470 (± 0.0004) | 30.8 (± 0.01) |
| Total Single-Jets | | 24.40 (± 0.01) | |
| 2fj70 | 1 | 0 | 30.8 (± 0.01) |
| 2fj35 | 1 | 1.7 (± 0.4) | 32.5 (± 0.01) |
| 2fj18 | 100 | 0.94 (± 0.03) | 33.4 (± 0.01) |
| Total Multi-Fjets | | 2.65 (± 0.09) | |
| fj120 | 1 | 0.9 (± 0.3) | 34.1 (± 0.01) |
| fj70 | 20 | 1.16 (± 0.08) | 35.2 (± 0.01) |
| fj35 | 700 | 0.68 (± 0.01) | 35.8 (± 0.01) |
| fj18 | 7000 | 1.04 (± 0.004) | 36.9 (± 0.02) |
| Total Single-Fjets | | 3.74 (± 0.01) | |

Table 10: Estimated trigger rates for jet signatures foreseen at the beginning of data-taking with a luminosity of $10^{31}$ $cm^{-2}s^{-1}$.

## 10 Summary

A detailed description of the algorithms used to reconstruct jets in the L1, L2 and Event Filter was presented. The performance of each of these algorithms was shown. The physics performance and algorithm timing has been shown to be within the design targets. Nevertheless, optimisation studies to further improve the performance of these algorithms are ongoing. An example trigger menu for jet signatures proposed for the beginning of data-taking was presented. The overall trigger strategy for selecting events based on jet signatures was also described.

# Standard Model

# A Study of Minimum Bias Events

**Abstract**
This note describes the methods developed for the measurement of the properties of minimum bias events during low luminosity running using the ATLAS detector. These methods aim to reconstruct the inclusive pseudorapidity and transverse momentum distributions of charged particles with $p_T > 150$ MeV produced from 14 TeV pp collisions. The triggers used to record minimum bias events are described and their acceptances evaluated. An analysis to measure the inclusive distributions is presented and the systematic uncertainties discussed. Finally, there is an overview of future physics studies with minimum bias samples.

## 1 Introduction

This note focuses on how the ATLAS detector [1] can be used to measure the central pseudorapidity and transverse momentum distributions of charged particles produced in inelastic proton-proton (pp) collisions during early running at the LHC. Measuring the characteristics of these collisions allows an understanding of the physics behind these processes to be developed, particularly their energy dependence. The minimum bias events allow the soft-part of the underlying event in high-$p_T$ collisions to be characterised. Studies of inclusive particle distributions in minimum bias events in pp collisions are important to provide the baseline for measurements in heavy-ion collisions, such as allowing differences in the number of particles to be attributed to QCD effects rather than the simple scaling of the number of nucleons. Finally these interactions will be a major background during low luminosity running ($10^{33}$ cm$^{-2}$s$^{-1}$) and high luminosity running ($10^{34}$ cm$^{-2}$s$^{-1}$), where the average number of such interactions per beam crossing is $\sim 2$ and $\sim 18$, respectively.

The total proton-proton cross-section can be divided into elastic and inelastic components, and the inelastic component can be further divided into: non-diffractive, single diffractive and double diffractive components [2]. The total cross-section ($\sigma_{tot}$) can then be written as:

$$\sigma_{tot} = \sigma_{elas} + \sigma_{sd} + \sigma_{dd} + \sigma_{nd}$$

where these cross-sections are elastic ($\sigma_{elas}$), single diffractive ($\sigma_{sd}$), double diffractive ($\sigma_{dd}$) and non-diffractive ($\sigma_{nd}$), respectively. The cross-sections for the inelastic subprocesses determined using PYTHIA [3], which was used to generate the event samples for this study, are given in Table 1. For comparison, the cross-sections predicted by PHOJET [4] are also shown in Table 1. The PHOJET predictions include central diffraction, however this hard proton-pomeron interaction only contributes to the cross-section at the few per cent level. As this is not simulated in PYTHIA, it was not considered further in this note.

The acceptance of inelastic events is defined by the trigger, which is usually known as a minimum bias trigger. It is designed to avoid bias in the sample, such as selecting high-$p_T$ events by triggering on high-$p_T$ objects. However, some bias is usually introduced due to effects such as the geometrical acceptance of or minimum energy thresholds in the trigger detector. It is therefore not unusual to find different definitions for minimum bias events in the literature. Historically, the minimum bias triggering used in hadron collider experiments [5–11] often used triggers based on forward-backward coincidences that favoured the detection of non-single diffractive inelastic events (NSD), i.e. $\sigma_{nsd} = \sigma_{tot} - \sigma_{elas} - \sigma_{sd}$. Thus, NSD events have often been classified as 'minimum bias events'. In this

note, results from simulated events selected using the minimum bias triggers are presented. These results have been corrected for detector reconstruction and acceptance effects. The results for the non-single diffractive sample are also presented to allow comparison with previous results from hadron-collider experiments. However, this requires correcting for the trigger acceptance for each of inelastic processes, which depends on the physics model used to generate the different processes.

Minimum bias interactions have previously been studied at a range of different energies at the CERN ISR [5], CERN SppS [6–8] and Fermilab's Tevatron [9–11] and RHIC colliders. Based on these results, Monte Carlo models have been tuned to generate predictions for LHC multiplicities [12]. Figure 1 shows a comparison of model predictions for the central charged particle density in NSD $p\bar{p}$ events for a wide range of centre-of-mass energies ($\sqrt{s}$). The data points shown are corrected for detector and trigger effects and efficiencies back to the particle level. Figure 1 compares predictions generated with two different tunings of PYTHIA [3] (ATLAS and CDF tune-A [12]) with the central charged particle density generated with PHOJET [4,13]. It is clear from this figure that there is a large uncertainty in the predicted central particle density of non-single diffractive interactions at the LHC energy even though the models have been tuned to agree with data at lower energies. This uncertainty arises because the energy dependence in a variety of different models for low-$p_T$ hadronic processes is not well understood. Measuring the central particle density at the LHC will thus be crucial to determining the energy dependence of the central particle density and constraining models of inelastic events.

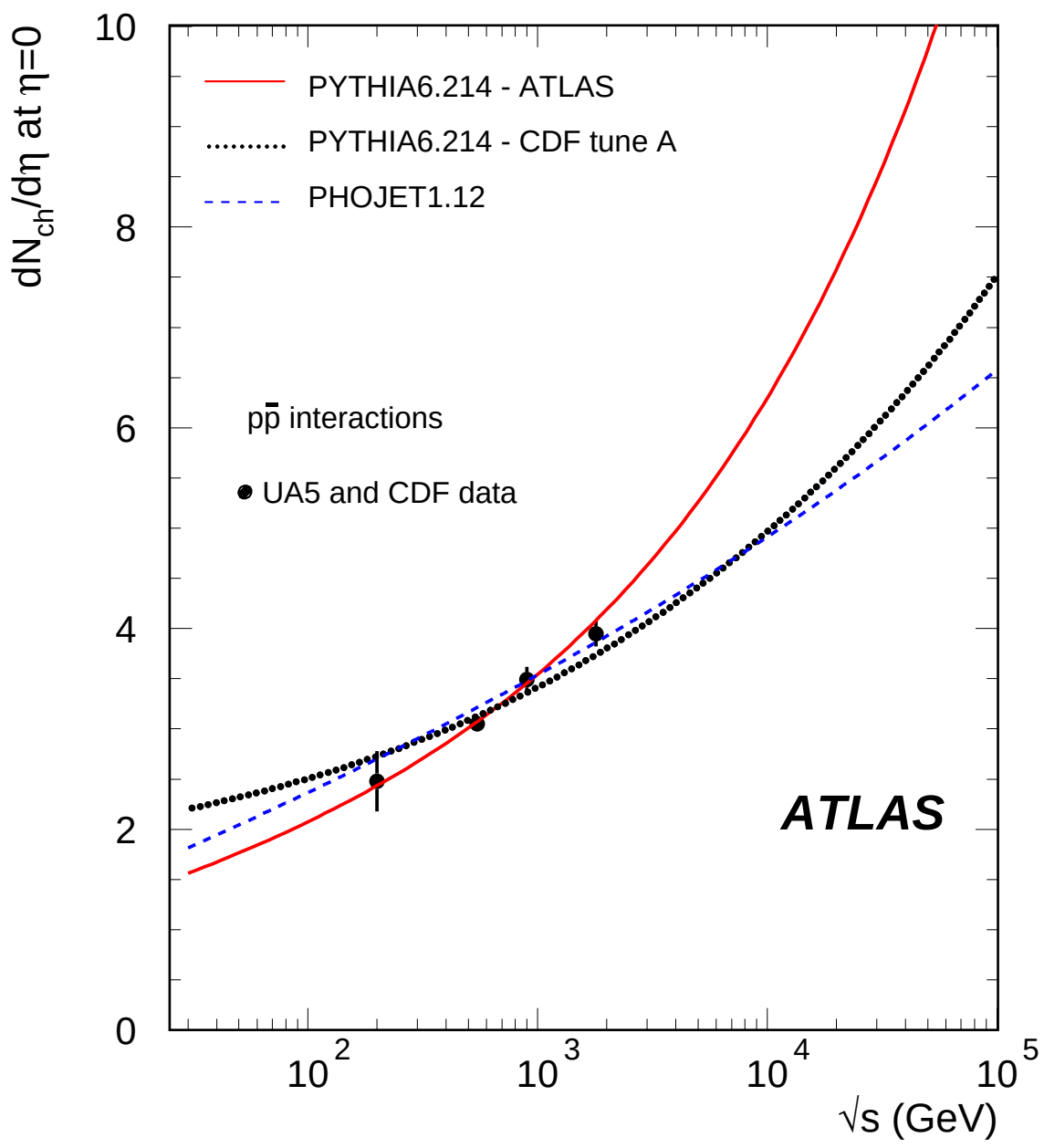


Figure 1: Central charged particle density for non-single diffractive inelastic events as a function of energy. The lines show predictions from PYTHIA using the ATLAS tune and CDF tune-A, and from PHOJET. The data points are from UA5 and CDF $p\bar{p}$ data.

This paper is organized as follows: In section 2 we briefly describe the experimental setup and the trigger strategies that have been developed to accept inelastic collisions with minimum bias. The

analysis procedure, corrections, as well as the systematic effects involved in this study are discussed in section 3. Section 4 presents an overview of future work planned to be done with ATLAS minimum bias measurements. Finally, in section 5 we present our conclusions.

| | **Cross-section (mb)** | |
|---|---|---|
| Process | PHOJET | PYTHIA |
| non-diff. | 69 | 55 |
| single diff. | 11 | 14 |
| double diff. | 4 | 10 |
| central diff. | 1 | - |
| total inelastic | 85 | 79 |
| elastic | 35 | 23 |
| total | 120 | 102 |

Table 1: Cross-section predictions for pp interactions at $\sqrt{s} = 14$ TeV from PYTHIA and PHOJET.

### 1.1 Characteristics of inelastic events

The pseudorapidity ($\eta$) and transverse momentum ($p_T$) distributions of charged particles generated using PYTHIA [3], with the parameter set defined in Ref. [14] and PHOJET [4, 13] with default parameters are shown in Figs. 2(a) and 2(b). These distributions correspond to non-diffractive, single- and double-diffractive inelastic pp interactions at $\sqrt{s} = 14$ TeV, respectively. They clearly show large uncertainties in model predictions for the LHC and that the events are dominated by low-$p_T$ particles with the highest densities found in the central region $|\eta| < 3.0$. Much of this central region is covered by the ATLAS inner detector. The charged particle distributions will be reconstructed from tracks which are measured by the inner detector for $|\eta| < 2.5$, with $p_T$ greater than 150 MeV.

### 1.2 Backgrounds

The main backgrounds in minimum bias events, particularly during early running, will be beam-gas collisions within the beampipe over the length of ATLAS, and beam-halo from interactions in the tertiary collimators in the accelerator. These backgrounds can provide spurious triggers that must be removed from the inelastic event sample to prevent distortion of its characteristics. During early low luminosity running a large fraction of bunch crossings will have no pp interaction. Using a trigger based only on bunch-crossings would result in a large number of empty events, which only contain detector noise, being recorded. Therefore, the trigger must be able to reject such events in order to optimise the use of the trigger bandwidth. In this paper we will present results which include discussions on beam-gas events and empty events but not on beam-halo.

## 2 Simulation and experimental setup

The samples of single-, double-, and non-diffractive inelastic events used in this study were generated with PYTHIA version 6.403 [3]. The PYTHIA event generator was configured according to an underlying event tuning fit [14] from previous experiments and is expected to provide a reasonable description of the non-diffractive part of minimum bias events.

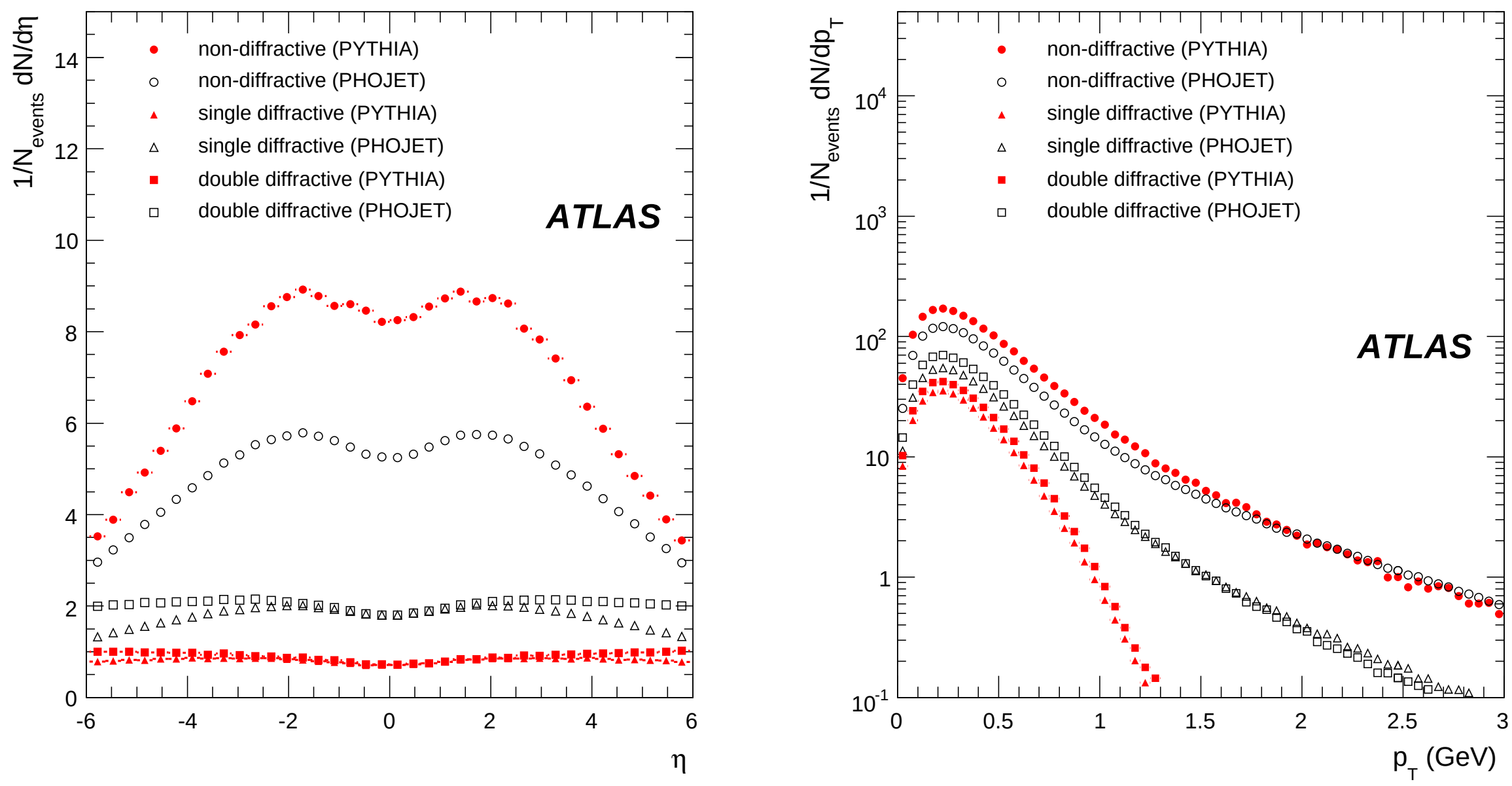


(a) (b)

Figure 2: Pseudorapidity (a) and transverse momentum distribution (b) of stable charged particles from simulated 14 TeV pp inelastic collisions generated using PYTHIA and PHOJET event generators.

### 2.1 The ATLAS inner detector

The pseudorapidity and transverse momentum distributions of charged particles are measured using the ATLAS inner detector. The inner detector is described in detail elsewhere [1]. It consists of, in order of increasing radius, a silicon pixel system, a silicon microstrip system (SCT) and a gas-based transition radiation detector (TRT). The inner detector is mounted inside a solenoid magnet which provides a 2 T magnetic field.

The inner detector sub-detectors are designed as independent but also complementary systems. The pixels cover radii of 50.5 mm-149.6 mm, the SCT covers 299 mm-560 mm and the TRT covers 563 mm-1066 mm. The precision tracking detectors (pixel and microstrip) cover $|\eta| < 2.5$ and are divided into barrel ($|\eta| < 1.4$) and endcaps ($1.4 < |\eta| < 2.5$).

The ATLAS inner detector will provide hermetic and robust pattern recognition, excellent momentum resolution and both primary and secondary vertex measurements for charged tracks above a $p_T$ threshold, which is nominally 500 MeV but can be as low as 100 MeV within $|\eta| < 2.5$. The nominal $p_T$-cut of 500 MeV corresponds to tracks traversing the full inner detector, however as discussed in section 1.1 a measurement of the properties of minimum bias events requires a $p_T$-cut of $\leq$ 200 MeV. A $p_T$-cut of 150 MeV, used in this paper, corresponds to tracks that traverse the precision Si tracker (pixels and SCT) allowing low-$p_T$ tracks to be well reconstructed.

### 2.2 The minimum bias trigger scintillators

The Minimum Bias Trigger Scintillators (MBTS) [15] are mounted on the inner surface of the liquid argon endcap cryostats and cover a pseudorapidity range of $2.12 < |\eta| < 3.85$. The MBTS is constructed from 2 cm polystyrene-based scintillator counters. Each side of the MBTS is made up of 16 counters each of which is split into two regions of equal pseudorapidity ($2.12 < |\eta| < 2.83$, $2.83 < |\eta| < 3.85$) and covering $\frac{\pi}{4}$ in azimuth. The MBTS is read out through the tile calorimeter electronics providing a fast L1 signal, which is discriminated above a voltage threshold, relative to the bunch-crossing signal.

## 3 Minimum bias trigger scenarios

A minimum bias trigger should select inelastic collisions with as little bias as possible, precluding the use of the standard high-$p_T$ triggers. Ideally, the L1 random trigger [16] with beam pickup would be used to accept events with zero bias, and inelastic collisions would be selected offline. However, during early running when the luminosity is expected to be $< 10^{30}$ cm$^{-2}$s$^{-1}$, the random trigger will be very inefficient since the probability of an interaction during a bunch crossing is $< 1\%$. Therefore, validation of the L1 random trigger is required in the high-level-trigger (HLT). The use of tracking in the HLT to validate the random trigger and a dedicated L1 trigger based on the minimum bias trigger scintillators (MBTS) has been studied. A minimum bias trigger stream consisting of the random trigger, track trigger and the MBTS is shown in Figure. 3. The random-based track trigger and the MBTS triggers are discussed in sections 3.1 and 3.2, respectively. In addition, the L1 MBTS could be used in conjunction with the pixel and SCT spacepoints reconstructed at L2 and the Event Filter (EF) track trigger. However, this was not studied in this note.

As the luminosity increases from $\sim 10^{31}$ cm$^{-2}$s$^{-1}$ to $10^{32}$ cm$^{-2}$s$^{-1}$ and higher values, the mean number of interactions per bunch crossing will be around 1. The random trigger will then record interactions for each crossing without requiring further triggers at L2 or EF to reject empty bunch

events. This will allow a zero bias sample to be efficiently accepted by ATLAS using the random trigger.

For the purposes of this note, a luminosity of $10^{31}$ cm$^{-2}$s$^{-1}$ with a bunch spacing of 75 ns has been assumed. The rate of inelastic events under those conditions is 792 kHz, and the mean number of events per bunch crossing is 0.06. Trigger efficiencies were calculated using around $10^5$ events of

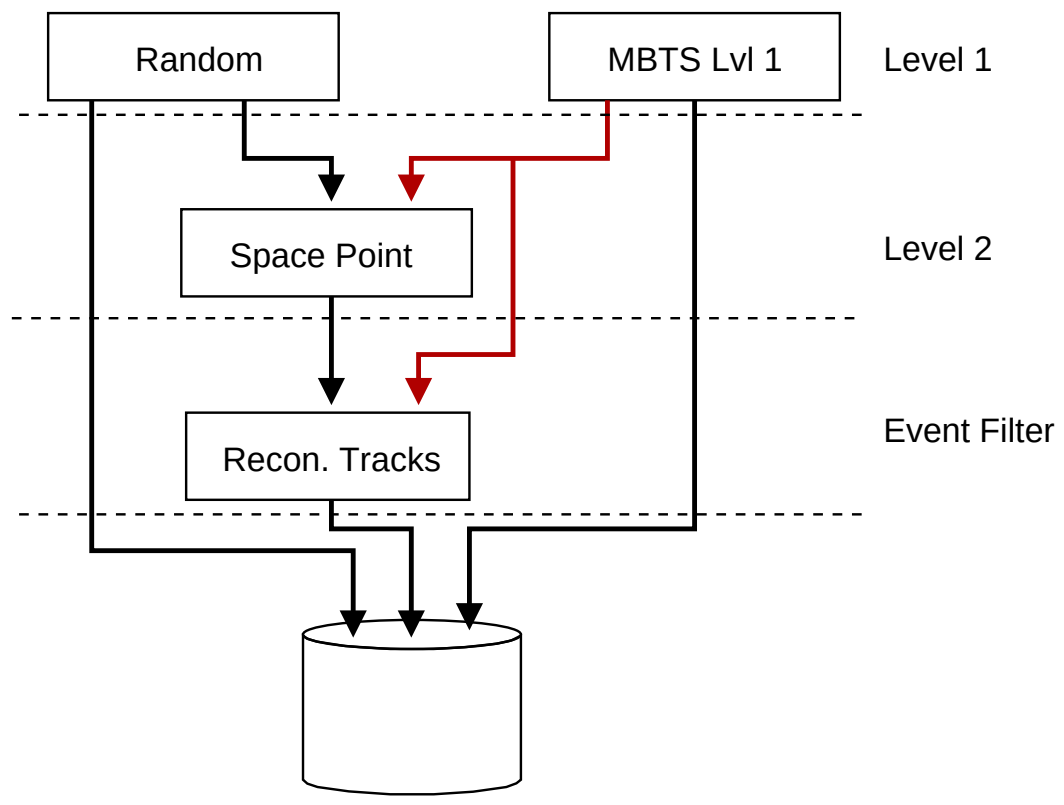


Figure 3: Minimum bias trigger slice.

simulated hydrogen beam-gas, single diffractive, double diffractive, non-diffractive and empty events. Each of these simulated samples were reconstructed and passed through the trigger logic and the trigger efficiency was then calculated from the number of events satisfying the trigger logic.

## 3.1 Random-based track trigger

The selection of minimum bias events by a track trigger is performed by the high-level trigger. The aim is to reject empty bunch and beam-gas events. After the random event selection at L1, the L2 trigger rejects empty events by requiring a minimum number of spacepoints in the pixel and SCT detector. The trigger efficiency curves for different types of events (ddiff: double diffractive, ndiff: non-diffractive and sdiff: single diffractive) are shown in Figure 4. The efficiency of empty bunch events as a function of spacepoints clearly shows that a modest constraint on either the number of pixel or SCT spacepoints rejects events containing only random noise. The thresholds for the number of pixel and SCT spacepoints were determined from the requirement to have a signal to background for non-diffractive to empty bunch events of 100:1. Given that the beam conditions assumed in this paper correspond to a probability of a pp non-diffractive inelastic interaction of 0.05, this corresponds to constraining the efficiency of empty bunch events to be less than $5 \times 10^{-4}$. This requires setting the threshold on the number of spacepoints to be 12 and 3 in the pixel and SCT respectively.

While the spacepoint (SP) trigger reduces the number of beam-gas events accepted, the rate can be further reduced by requiring the presence of reconstructed tracks within a small $z$-region around the nominal interaction point. A full track reconstruction scan with a minimum $p_T$ of 200 MeV is carried out in the EF on the events that pass the SP trigger. These tracks are then used to select events using the cuts defined in Table 2. The trigger efficiency is plotted with respect to the number of reconstructed tracks in Figure. 5. The empty events are rejected by the cut on the number of pixel and SCT spacepoints resulting in an efficiency of zero.

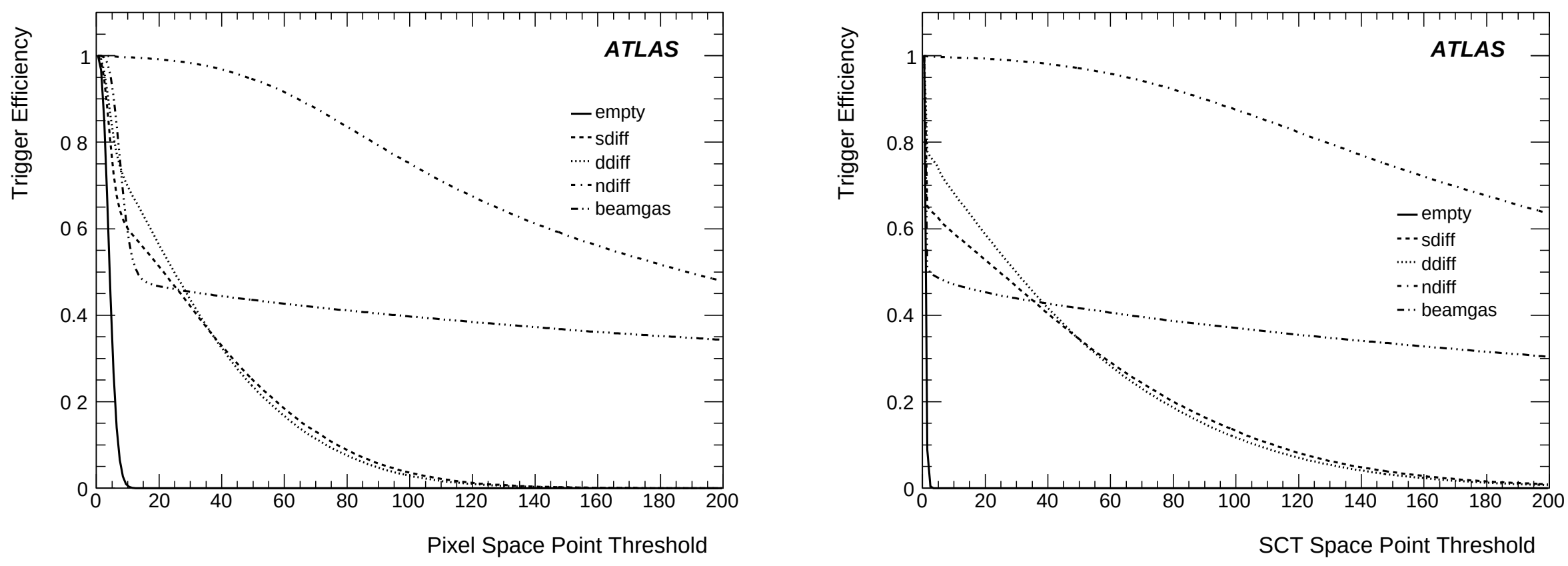


(a) (b)

Figure 4: Trigger efficiency as a function of the number of pixel spacepoints (a) and the number of SCT spacepoints (b).

| Track parameter | 2 Trigger cut |
|---|---|
| Number of tracks | $\geq 2$ |
| Track z0 | $< 200$ mm |

Table 2: Track cuts for EF track trigger.

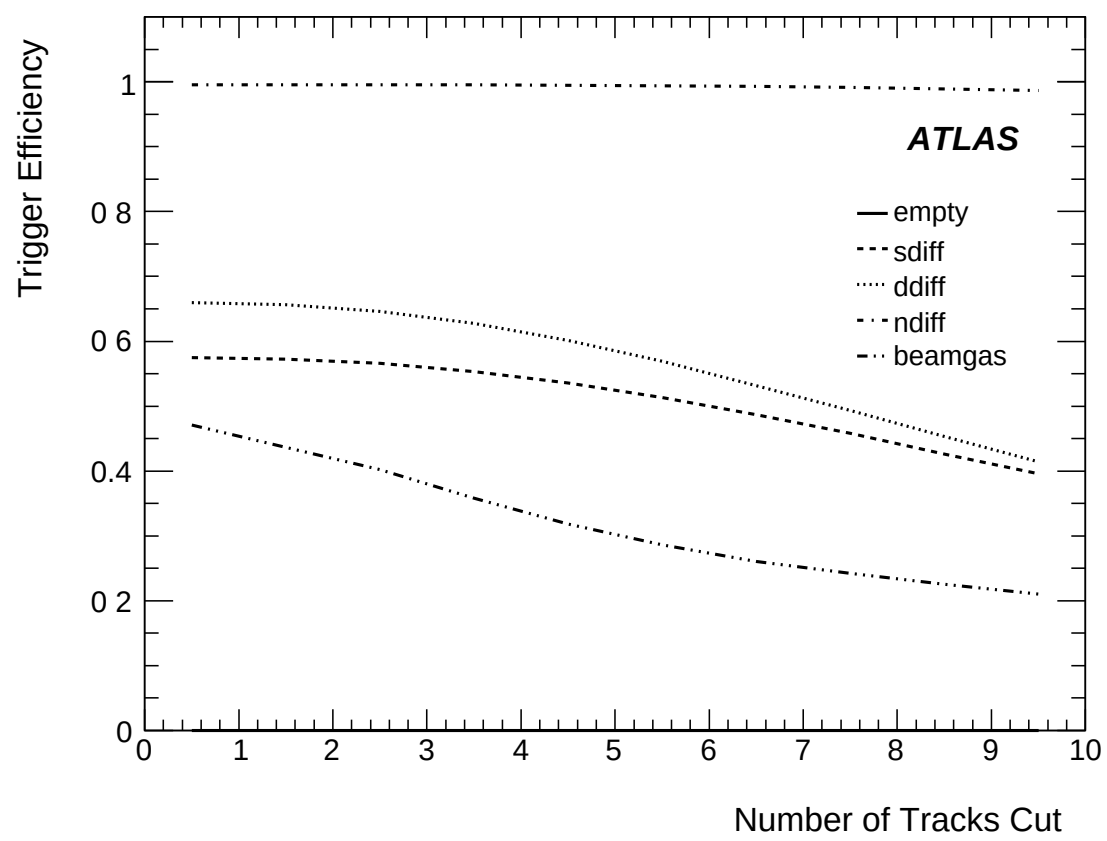


Figure 5: Trigger efficiency as a function of the number of reconstructed tracks for events satisfying the L2 spacepoint requirement.

### 3.2 Minimum bias triggers using the MBTS

Two simple MBTS L1 trigger strategies were considered for minimum bias event selection: MBTS_1_1 and MBTS_2. MBTS_1_1 is defined as at least one MBTS counter above threshold on each side, where the threshold of 40 mV was chosen from measurements of cosmic commissioning data. MBTS_2 is defined as two or more MBTS counters above threshold anywhere in the MBTS system. The trigger efficiencies as a function of MBTS counter threshold for inelastic non-diffractive, double-diffractive, single-diffractive, beam-gas and empty bunch events are given in Fig. 6.

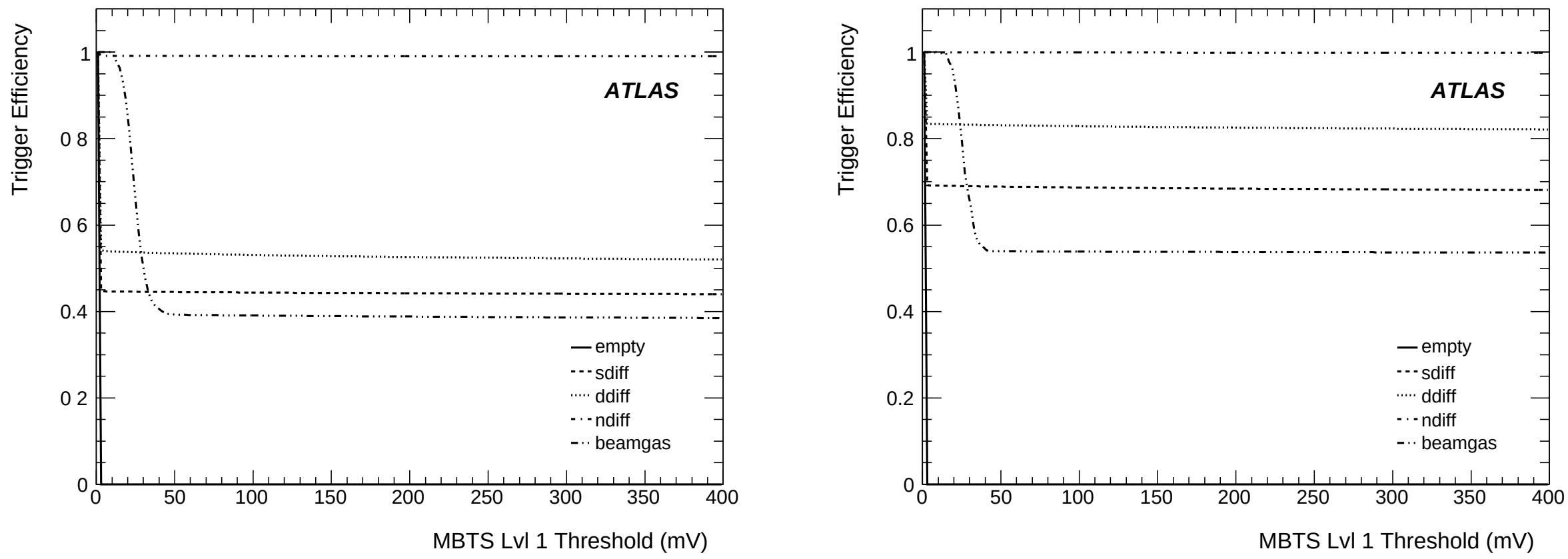


(a) (b)

Figure 6: MBTS L1 trigger threshold scans for the two trigger configurations: (a) MBTS_1_1 (b) and MBTS_2.

### 3.3 Trigger efficiency

A summary of the trigger efficiencies for the track-based and MBTS triggers are given in Table 3. The efficiency of diffractive events is around half of that of non-diffractive events. This is primarily due to the low particle multiplicity in diffractive events and because the triggers cover the central region and are therefore only sensitive to a fraction of the diffractive cross-section corresponding to high mass diffractive states.

The acceptances for the different processes i.e. the efficiencies weighted by the fraction of the inelastic cross-section are given in table 4. This shows that due to lower efficiencies and smaller cross-sections the acceptance of diffractive events is significantly suppressed. The acceptance of the proposed minimum bias triggers is around 85-92% and this will consist of around 80% of non-diffractive events and roughly equal numbers of single and double diffractive events.

## 4 Analysis

The goal of the analysis is to reconstruct the number of primary charged particles per unit of pseudorapidity and per unit of transverse momentum for inelastic pp interactions taken with the minimum bias trigger. Primary particles are defined as particles produced in the pp collision, but excluding secondary

| | MBTS_1_1 | MBTS_2 | SP | SP & EF Tracks |
|---|---|---|---|---|
| Non-diffractive | 99% | 100% | 100% | 100% |
| Double-diffractive | 54% | 83% | 66% | 65% |
| Single-diffractive | 45% | 69% | 57% | 57% |
| Beam-gas | 40% | 54% | 47% | 40% |

Table 3: MBTS and random-track trigger efficiencies.

| | MBTS_1_1 | MBTS_2 | SP | SP & EF Tracks |
|---|---|---|---|---|
| Non-diffractive | 69% | 70% | 70% | 70% |
| Double-diffractive | 7% | 10% | 8% | 8% |
| Single-diffractive | 8% | 12% | 10% | 10% |

Table 4: MBTS and random-track trigger acceptances.

particles from weak decays of strange hadrons or from electromagnetic or hadronic interactions in the detector material.

A sample of ~150,000 inelastic events, with ND, SD and DD events mixed in the proportion given by the PYTHIA cross-sections, was reconstructed and used in this study (MB sample). A typical non-diffractive event contains around 45 reconstructed tracks of which 37 are from primary particles and 8 are from secondary particles. The efficiencies and corrections were derived from half the sample and used in the analysis of the other half of the sample.

## 4.1 Event selection

Two selection criteria are applied to the set of reconstructed events: the event must be triggered by the minimum bias trigger and it must contain at least one reconstructed vertex. The sample of inelastic events selected by the MBTS_2 trigger, described in Section 3.2, was used for this analysis. No beam-gas or pileup events were included in the sample studied here. An additional criterion should be added to exclude events with multiple pp interactions in a bunch-crossing but this was not included in this analysis as the samples of simulated data used for this study contained events with a single pp interaction in a bunch-crossing.

## 4.2 Track selection

Tracks were reconstructed using tracking with the minimum $p_T$ set to 100 MeV rather than the default 500 MeV. To avoid threshold effects, tracks with $p_T > 150$ MeV were used in the analysis. The relative $p_T$-resolution, $\sigma(\frac{1}{p_T})p_T$, was found to be around 1.5% for $|\eta| < 0.8$ and 4.3% for $|\eta| > 1.6$.

Cuts on the measured track parameters were used to select a set of well reconstructed tracks for the analysis. These cuts are listed in Table 5. The requirement of a hit in the inner pixel layer,the b-layer, removes a sizable portion of fake tracks and tracks associated to secondary particles since they frequently do not leave a hit in the first layer of the pixel detector. The cut on the number of pixel and SCT hits is a standard track scoring cut applied during reconstruction and is listed here only for completeness. The values of the cuts were chosen by fitting each of the distributions with a gaussian and cutting at $\sim 3\sigma$.

| | |
|---|---|
| Quality cuts | No. of b-layer hits $\geq 1$<br>No. of Silicon hits $\geq 5$ |
| Resolution cuts | $\lvert\sigma_{d_0}\rvert < 1.6$ mm<br>$\lvert\sigma_{z_0}\rvert < 6.0$ mm<br>$\lvert\sigma_{\phi}\rvert < 0.03$<br>$\lvert\sigma_{\theta}\rvert < 0.015$<br>$\lvert\sigma_{q/p_T}\rvert < 0.0003$ ( GeV)$^{-1}$ |
| Track-to-vertex cut | $N_\sigma < 3$ |

Table 5: Track selection cuts used in this analysis. The resolutions $\sigma^2_{d_0}$, $\sigma^2_{z_0}$, $\sigma^2_{\phi}$, $\sigma^2_{\theta}$ and $\sigma^2_{q/p_T}$ are the five diagonal elements in the track parameter covariance matrix. For more information on the tracking Event Data Model (EDM) see [17].

The track-to-vertex cut is the selection that most effectively cuts away tracks from secondary particles while accepting tracks from primaries. It is made by cutting on the distance of closest approach to the nearest reconstructed vertex, normalized by the error. The normalized distance to the vertex is defined as

$$\Delta R \equiv \sqrt{\left(\frac{\Delta d_0}{\sigma_{d_0}}\right)^2 + \left(\frac{\Delta z_0}{\sigma_{z_0}}\right)^2}, \quad (1)$$

where $\Delta d_0/\sigma_{d_0}$ and $\Delta z_0/\sigma_{z_0}$ are the normalized distances in the transverse and longitudinal directions, respectively. The number of $\sigma$ to the vertex as a function of $\Delta R$ (for a two-dimensional gaussian) is given by

$$N_\sigma = \sqrt{2}\mathrm{erf}^{-1}(1 - e^{-\Delta R^2/2}). \quad (2)$$

Note that the above formula assumes no correlations between the resolutions in the transverse and longitudinal direction.

To determine the effects of the cuts and evaluate efficiencies and acceptances, the reconstructed tracks are matched to the generated primary and secondary particles. The track that has the highest percentage of hits that overlap with the trajectory of a generated particle is matched to that generated particle. A track is well matched to a particle if greater than 50% of its hits overlap with the generated particle's trajectory. A primary track is one matched to a primary particle and similarly a secondary track is one matched to a secondary particle. If a track is not matched to any generated particle, it is classified as a fake.

Table 6 shows the influence of the track cuts on the reconstructed sample. Each row shows the percentage of tracks that did not pass each of the specified cuts. Note that a single track can fail several cuts and therefore give counts in several rows. Three of the rows show the percentage of tracks that fail any of the quality cuts ('Quality'), any of the resolution cuts ('Resolution') and any of the quality or resolution cuts ('Q || R'). The strongest and most effective cut is the track-to-vertex cut, which rejects about 88% of the secondaries. The second last row shows the percentage of reconstructed tracks that are outside the pseudorapidity and $p_T$ range used in the analysis ($|\eta| > 2.5$, $p_T < 150$ MeV).

| Cut | % Cut All | % Cut Primary tracks | % Cut Secondary tracks |
|---|---|---|---|
| b-layer hit | 15.9 | 8.5 | 46.8 |
| $\text{cov}_{d0}$ | 11.5 | 6.0 | 34.2 |
| $\text{cov}_{z0}$ | 9.4 | 5.0 | 27.4 |
| $\text{cov}_{\phi}$ | 8.9 | 5.1 | 24.3 |
| $\text{cov}_{\theta}$ | 4.9 | 4.2 | 8.2 |
| $\text{cov}_{q/p_T}$ | 6.4 | 4.3 | 14.9 |
| Quality | 15.9 | 8.5 | 46.8 |
| Resolution | 16.7 | 10.9 | 40.4 |
| Q \|\| R | 24.6 | 15.6 | 62.1 |
| Track-to-Vtx | 30.7 | 16.9 | 87.8 |
| $\eta$ \|\| $p_T$ | 1.2 | 1.3 | 0.9 |
| Total | 38.6 | 24.6 | 96.5 |

Table 6: Fraction of tracks cut away by the selection cuts.

### 4.3 Procedure & corrections

The $dN_{ch}/d\eta$ and $dN_{ch}/dp_T$ distributions are obtained by starting with the measured number of selected tracks in selected events and applying the following three corrections:

- Track-to-particle correction
- Vertex reconstruction correction
- Trigger bias correction

The first correction accounts for the difference between the number of measured tracks and the number of primary charged particles. This is essentially a correction for the inner detector acceptance and efficiency of the tracking software [18]. The second correction takes into account the vertex reconstruction efficiency and corrects for events that have no reconstructed vertex. These two corrections account for detector dependent effects and produce the minimum bias sample, in which the data have been corrected for detector reconstruction and acceptance but have not been corrected for the trigger acceptance.

The trigger bias correction depends both on the detector simulation and on the physics model used in the event generator to simulate inelastic pp interactions. Different corrections for the trigger can be applied to correct the measured distributions back to different physics processes, i.e. inelastic, non-single-diffractive or non-diffractive interactions.

The track-to-particle correction is applied only at the track level. The vertex correction and the trigger bias correction are applied at the track and event level. The track-level vertex and trigger bias corrections are needed to compensate for any bias on the measured track distribution due to vertex and trigger requirements. All track-level corrections are determined as a function of pseudorapidity ($\eta$), $p_T$ and the $z$-position of the collision vertex, $v_z$. The event level corrections are determined as a function of $v_z$ and the number of reconstructed tracks in the event.

The corrections are 3-D and 2-D histograms but the principal corrections shown in the figures below are plotted as projections of $p_T$ $\eta$ $z$-position or the number of reconstructed tracks as appropriate for clarity.

### 4.3.1 Track-to-particle correction

The number of selected reconstructed tracks differs slightly from the number of primary charged particles due to a number of different effects: the acceptance of the detector, the detector and track reconstruction efficiency, the contribution from secondaries and fakes, and the acceptance of the track selection cuts. The track-to-particle correction takes all these effects into account and is calculated as

$$C_{trk}(\eta, v_z, p_T) \equiv \frac{\text{No. of primary charged particles}}{\text{No. of selected reconstructed tracks}}. \quad (3)$$

The numerator and denominator are calculated for the same events taken through the full detector simulation and reconstruction.

Figure 7 shows the projection of the 3-dimensional track-to-particle correction onto the $\eta$ and $p_T$ axes. The correction is significant and changes rapidly with $p_T$ below 200 MeV where the reconstruction efficiency is lower due to a lower number of hits on a track and the effects of material and multiple scattering. Above 200 MeV the correction is small and has a small dependence on $p_T$.

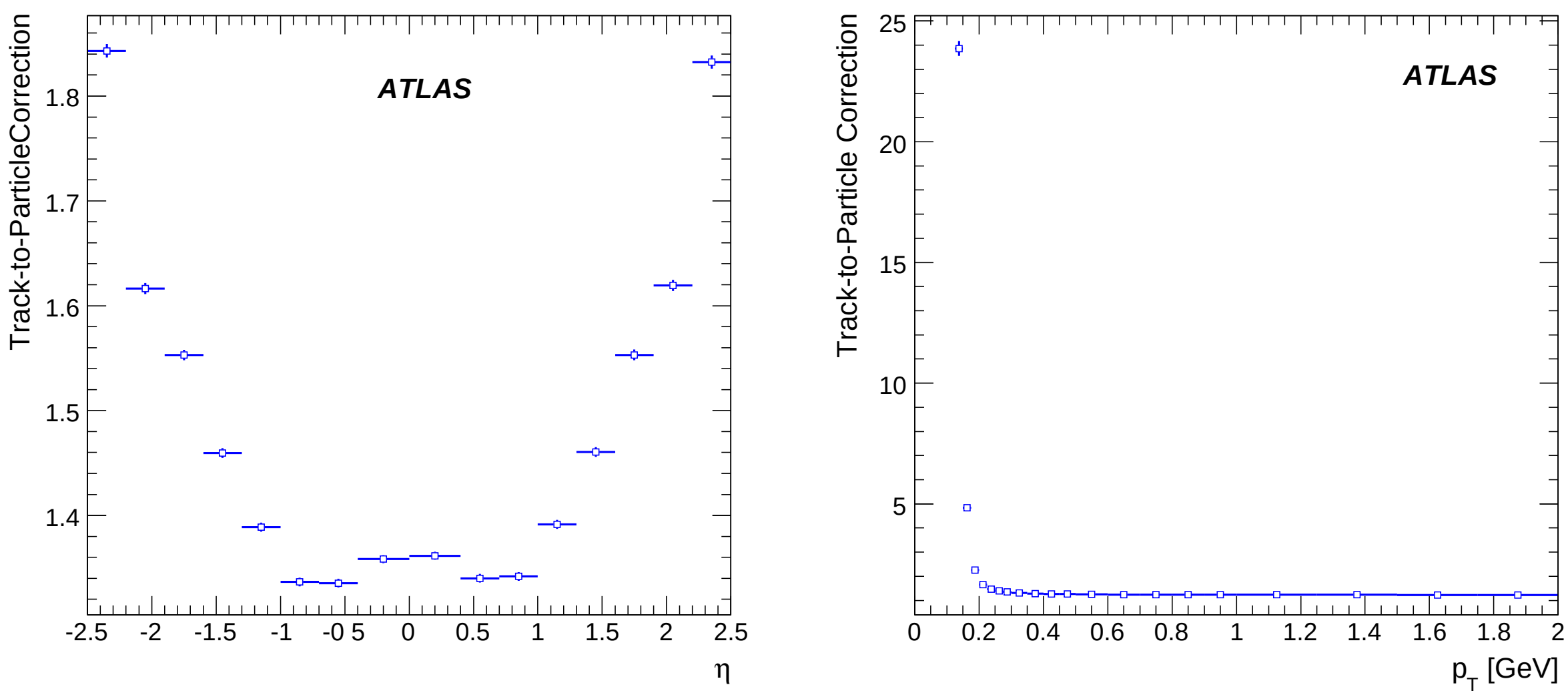


(a) (b)

Figure 7: The track-to-particle correction $C_{trk}$ as a function of (a) $\eta$ and (b) $p_T$.

### 4.3.2 Vertex reconstruction correction

The vertex correction takes into account the bias introduced by events that are not counted because their vertex was not found by the vertex reconstruction algorithm. The event-level correction is calculated as a function of $v_z$ and number of reconstructed tracks in the event:

$$\tilde{C}_{vtx}(v_z, N) \equiv \frac{\text{No. of triggered events}}{\text{No. of triggered events with} \geq 1 \text{ reconstructed vertex}}. \quad (4)$$

The track-level vertex correction is calculated as a function of $\eta$, $v_z$ and $p_T$:

$$C_{vtx}(\eta, v_z, p_T) \equiv \frac{\text{No. of tracks in all triggered events}}{\text{No. of tracks in triggered events with} \geq 1 \text{ reconstructed vertex}}. \quad (5)$$

Although the vertex correction is essentially a detector correction, some model dependence is present since only triggered collisions are taken into account to compute the correction factors. However, the use of only triggered events has the advantage that the correction can be computed directly from the data, without relying on any simulation. Once data are available it will also be possible to compare the properties of the triggered events with no reconstructed vertex to the corresponding events in the simulation. This may help minimize and better estimate any systematic uncertainties.

Figure 8 shows the event-level vertex reconstruction correction as a function of the number of reconstructed tracks, $N$, and vertex $z$-position. A vertex can be defined using a single well reconstructed track. As the number of tracks increases, the correction approaches unity. For events with $N > 10$, a vertex is always found and the correction is no longer required.

Figure 9 shows projections of the 3-dimensional track-level vertex correction onto the $v_z$ and $p_T$ axes. These corrections are found to be at the level of a few percent.

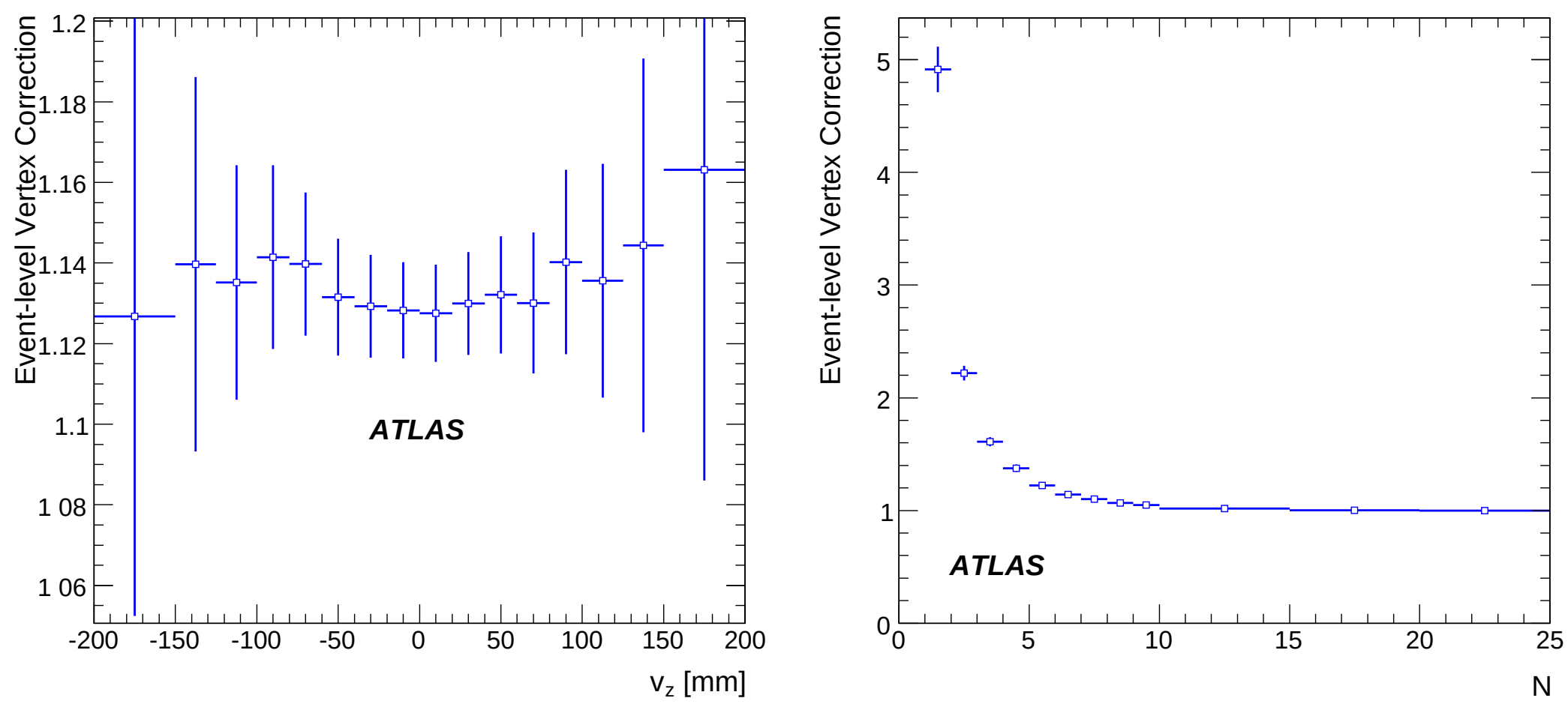


(a) (b)

Figure 8: The event-level vertex correction as a function of (a) $v_z$ and (b) the number of reconstructed tracks ($N$).

### 4.3.3 Trigger bias correction

The trigger bias correction accounts for the difference between the MB sample selected by the minimum bias trigger and the physics process of interest. This correction depends on the trigger detector simulation and on the models of the pp interactions used in the event generator. Therefore, this correction is dependent on the relative cross-sections of the inelastic processes used and on the modelling of particle production for the different inelastic processes.

After an initial model-independent measurement is made, the MB sample, the trigger bias correction can be used to obtain distributions for other samples of interest: non-diffractive (ND), non-single diffractive (NSD) or inelastic (INEL) collisions. Each of these samples corresponds to a different trigger correction and each correction can be applied to yield the final distributions for the different collision samples. In general the correction from MB to NSD collisions is the smallest since these

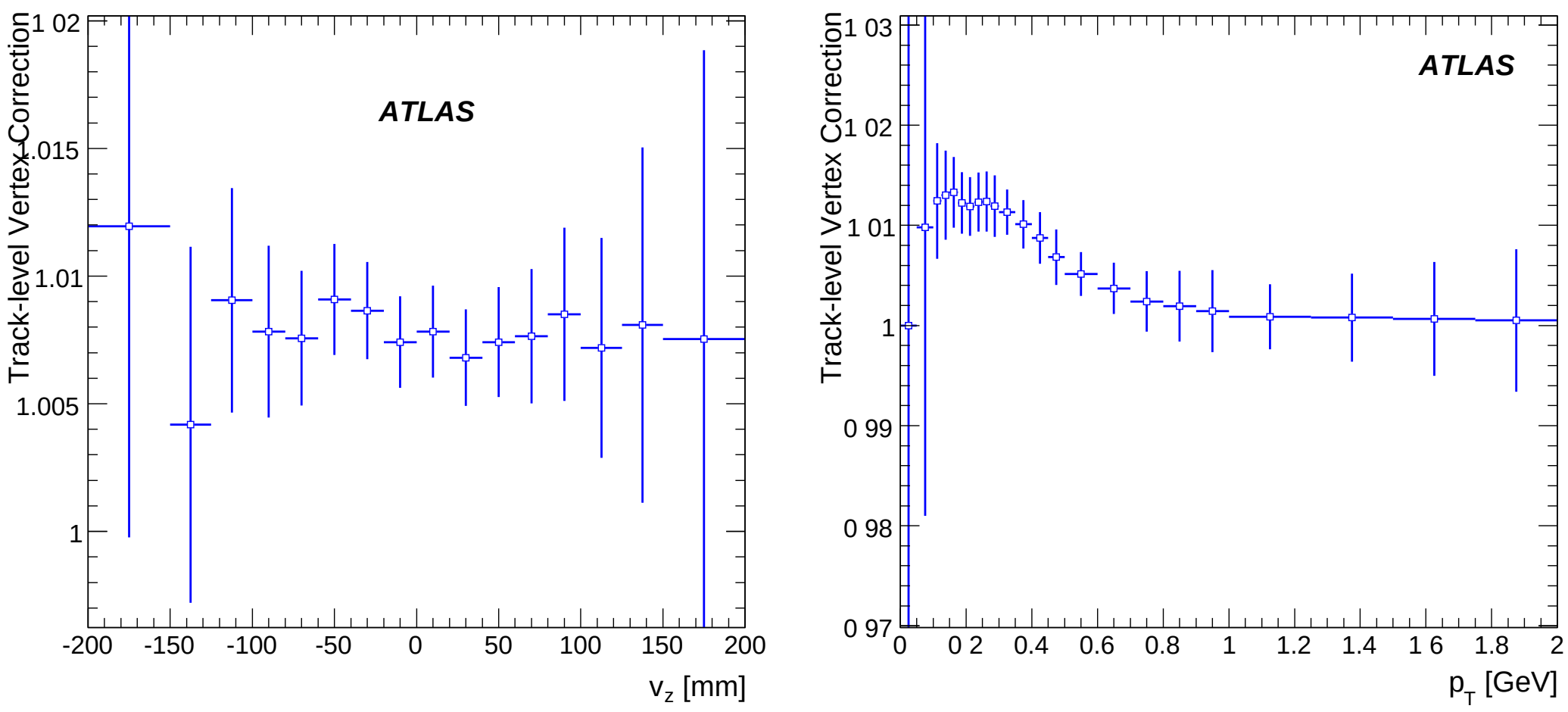


(a) (b)

Figure 9: The track-level vertex correction as a function of (a) $v_z$ and (b) $p_T$.

two event samples are almost identical, while the correction from MB to INEL is on the order of 1.2 (integrated over $N$ and $v_z$).

The trigger bias correction is determined on event level as a function of $v_z$ and multiplicity, defined as in the vertex correction:

$$\tilde{C}_{trig}(v_z, N) \equiv \frac{\text{No. of interactions in sample of interest}}{\text{No. of triggered events}}, \tag{6}$$

and on track-level as a function of $\eta$, $v_z$ and $p_T$:

$$C_{trig}(\eta, v_z, p_T) \equiv \frac{\text{No. of tracks in sample of interest}}{\text{No. of tracks in triggered events}}. \tag{7}$$

In this analysis it is assumed that there is only one interaction per event since the events have been simulated in this way. This will need to be verified in the real data for any events with more than one reconstructed vertex by looking at the multiplicity of each of the vertices and the distance between the vertices. As in the vertex correction, all tracks are counted since the track-to-vertex distance is undefined for events with no reconstructed vertex.

Figure 10 shows the event-level trigger bias correction for the NSD sample as a function of multiplicity and vertex $z$-position. Using the MBTS_2 trigger, the correction is only important for $N < 25$; for higher multiplicities the correction factor is not needed. It is largest for $N < 5$ which are principally diffractive events that have a lower trigger efficiency relative to the non-diffractive events as discussed in section 3.

Figure 11 shows projections of the 3-dimensional track-level trigger bias correction for the NSD sample to two of the three 2-dimensional planes and to the $v_z$ and $p_T$ axes.

### 4.4 Corrected distributions

An $\eta$-$v_z$-$p_T$ (3-D) histogram is filled for each track in each event and weighted with the values of the track-level corrections described above. A $v_z$-$N$ (2-D) histogram is also filled for each event and

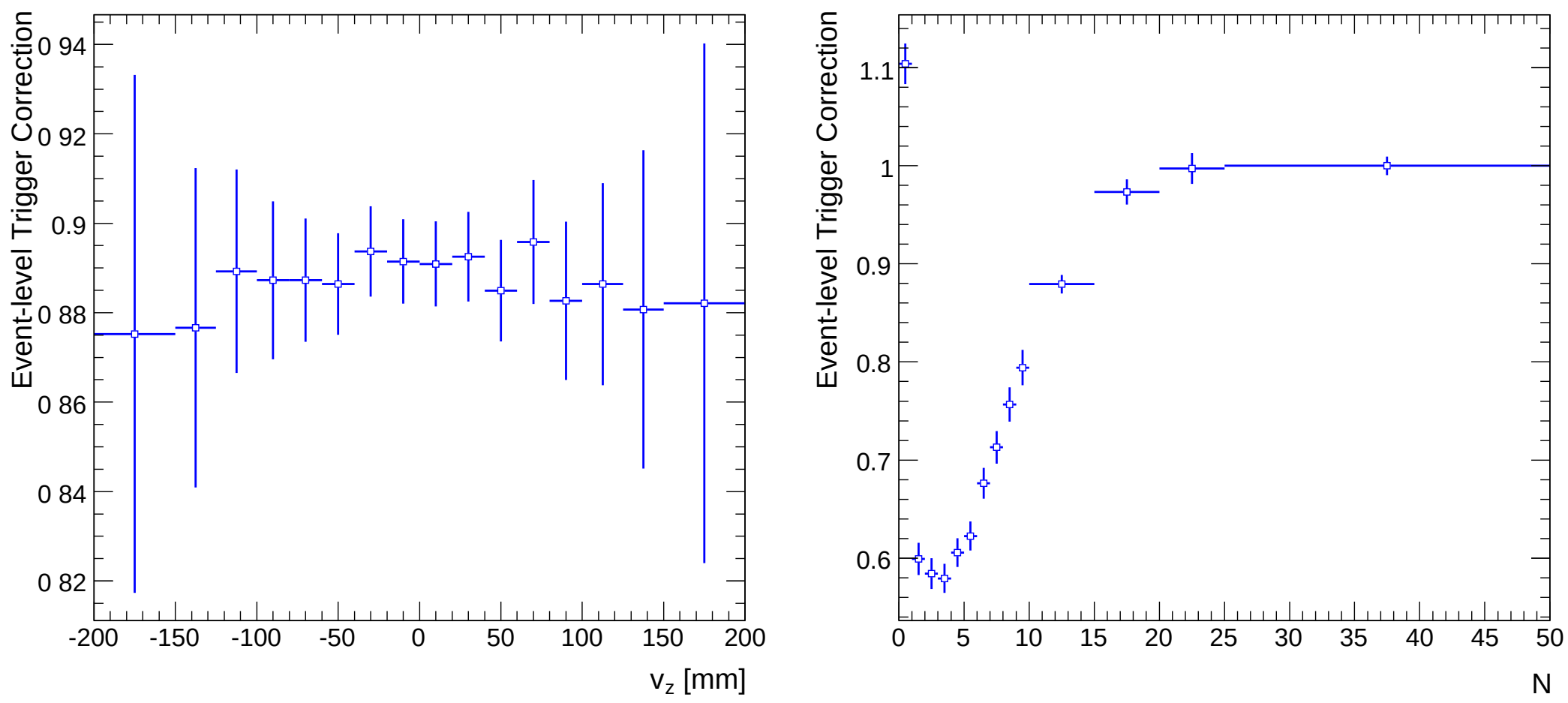


(a) (b)

Figure 10: The event-level trigger bias correction for the NSD sample as a function of (a) $v_z$ and (b) number of reconstructed tracks (N).

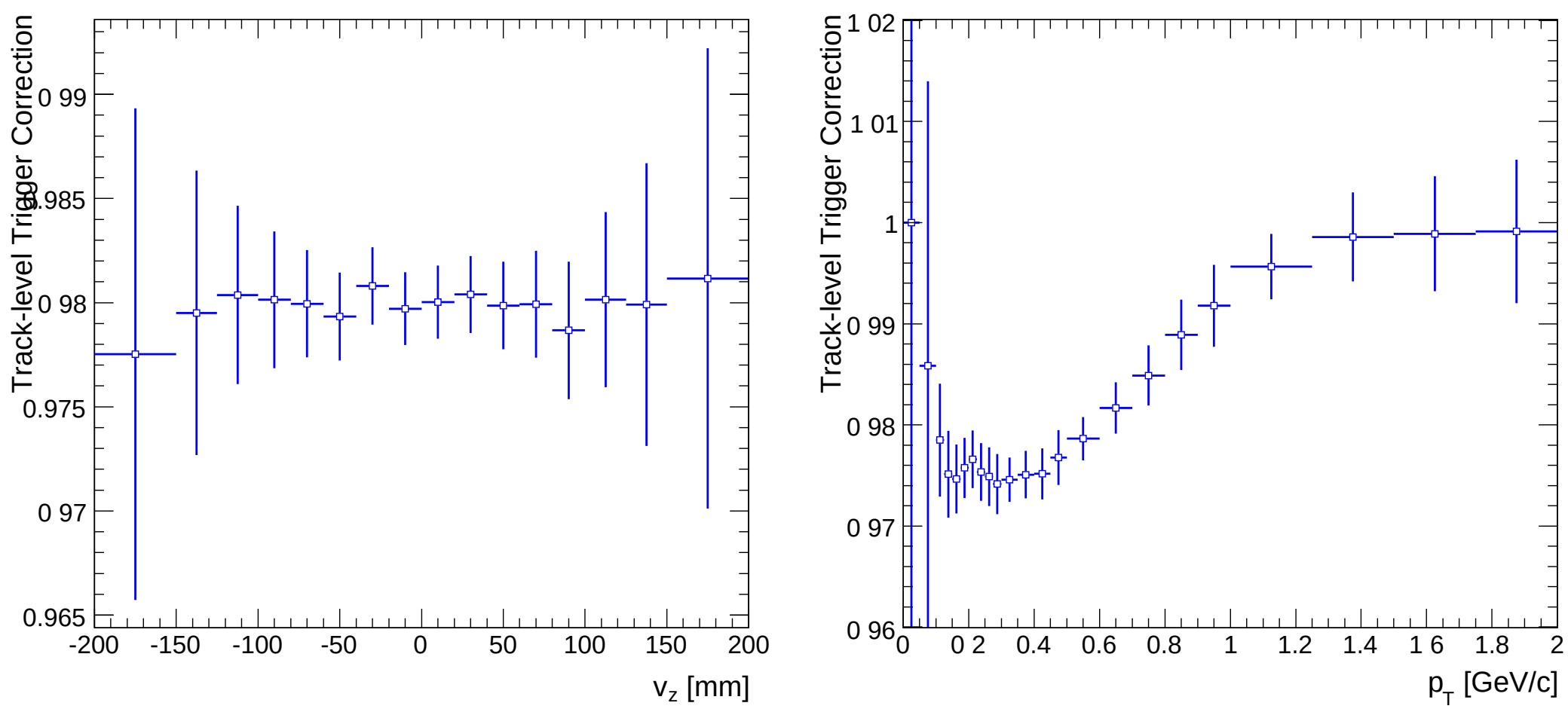


(a) (b)

Figure 11: The track-level trigger bias correction for the NSD sample as a function of (a) $v_z$ and (b) $p_T$.

weighted with the values of the event-level corrections. This histogram is needed to provide the proper normalization factor for the analysis.

After filling the histograms, a vertex range is chosen and the $v_z$ variable is integrated out. In the case of $dN_{ch}/d\eta$, the $p_T$ variable is integrated for $p_T > 150$ MeV; in the case of $dN_{ch}/dp_T$, the $\eta$ variable is integrated over $-2.5 < \eta < 2.5$. Each $\eta$-bin or $p_T$-bin is then divided by the total number of events, calculated by integrating the weighted event distribution histogram over $N$ and $v_z$. Finally, the distribution is normalized by the inverse width of the bins. No correction was made for the effect of the $p_T$ cutoff in the reconstruction. However, the stability of the results will be tested by varying the $p_T$ cutoff in data.

The distributions are presented for the sample events accepted by the minimum bias trigger. For this analysis MBTS_2 is used, and are labelled as MB. The MB distributions have only been corrected for the acceptance and efficiencies associated with reconstructing tracks and vertices. To obtain the NSD distributions to allow comparisons with other results, the MB distributions are corrected for the trigger bias.

The correction procedure can be expressed mathematically for one bin, corresponding to a certain region of phase space, in the following way. The number of particles $P$ and number of interactions $I$ are calculated as:

$$P(\eta, v_z, p_T) = \sum_{events}\sum_{tracks} (C_{trk}(\eta, v_z, p_T) \cdot C_{vtx}(\eta, v_z, p_T) \cdot C_{trig}(\eta, v_z, p_T)), \tag{8}$$

$$I(v_z, N) = \sum_{events} (\tilde{C}_{vtx}(v_z, N) \cdot \tilde{C}_{trig}(v_z, N)). \tag{9}$$

Tracks are weighted by the track-to-particle correction $C_{trk}(\eta, v_z, p_T)$, by the track-level vertex correction $C_{vtx}(\eta, v_z, p_T)$ and the track-level trigger bias correction $C_{trig}(\eta, v_z, p_T)$. Events are weighted by the event-level vertex correction $\tilde{C}_{vtx}(v_z, N)$ and the event-level trigger bias correction $\tilde{C}_{trig}(v_z, N)$.

For a given vertex range $[V_1, V_2]$ the $dN_{ch}/d\eta$ and $dN_{ch}/dp_T$ are then calculated as:

$$\left.\frac{dN_{ch}}{d\eta}\right|_{\eta=\eta'} = \frac{\int_{V_1}^{V_2}\int P(\eta', v_z, p_T)dp_T dv_z}{\int_{V_1}^{V_2}\int I(v_z, N)dN dv_z}, \tag{10}$$

$$\left.\frac{dN_{ch}}{dp_T}\right|_{p_T=p'_T} = \frac{\int_{V_1}^{V_2}\int P(\eta, v_z, p'_T)d\eta dv_z}{\int_{V_1}^{V_2}\int I(v_z, N)dN dv_z}. \tag{11}$$

To exercise the analysis chain, the complete reconstructed sample was divided in half: one part was used to calculate the corrections and the other used as 'data' input to the analysis.

Figures 12 and 13 show the corrected pseudorapidity distribution ($p_T > 150$ MeV) and the corrected transverse momentum spectrum ($|\eta| < 2.5$) for the MB and NSD event samples. Statistical errors on the corrected distributions are shown. The aim was to achieve statistical errors on the corrections of less than 2% per bin. The errors are mostly negligible since sufficient statistics were used to determine the correction factor. Data in regions where small statistical errors could not be achieved (i.e. near the edges of the acceptance) are not included in the analysis.

## 4.5 Systematic uncertainties

The systematic uncertainties have been estimated by changing the parameters in the generation, reconstruction or analysis and then re-evaluating the corrections and applying them to the analysis input sample. Although many of the systematic uncertainties are correlated, the different effects are studied here independently. The systematic uncertainties investigated include:

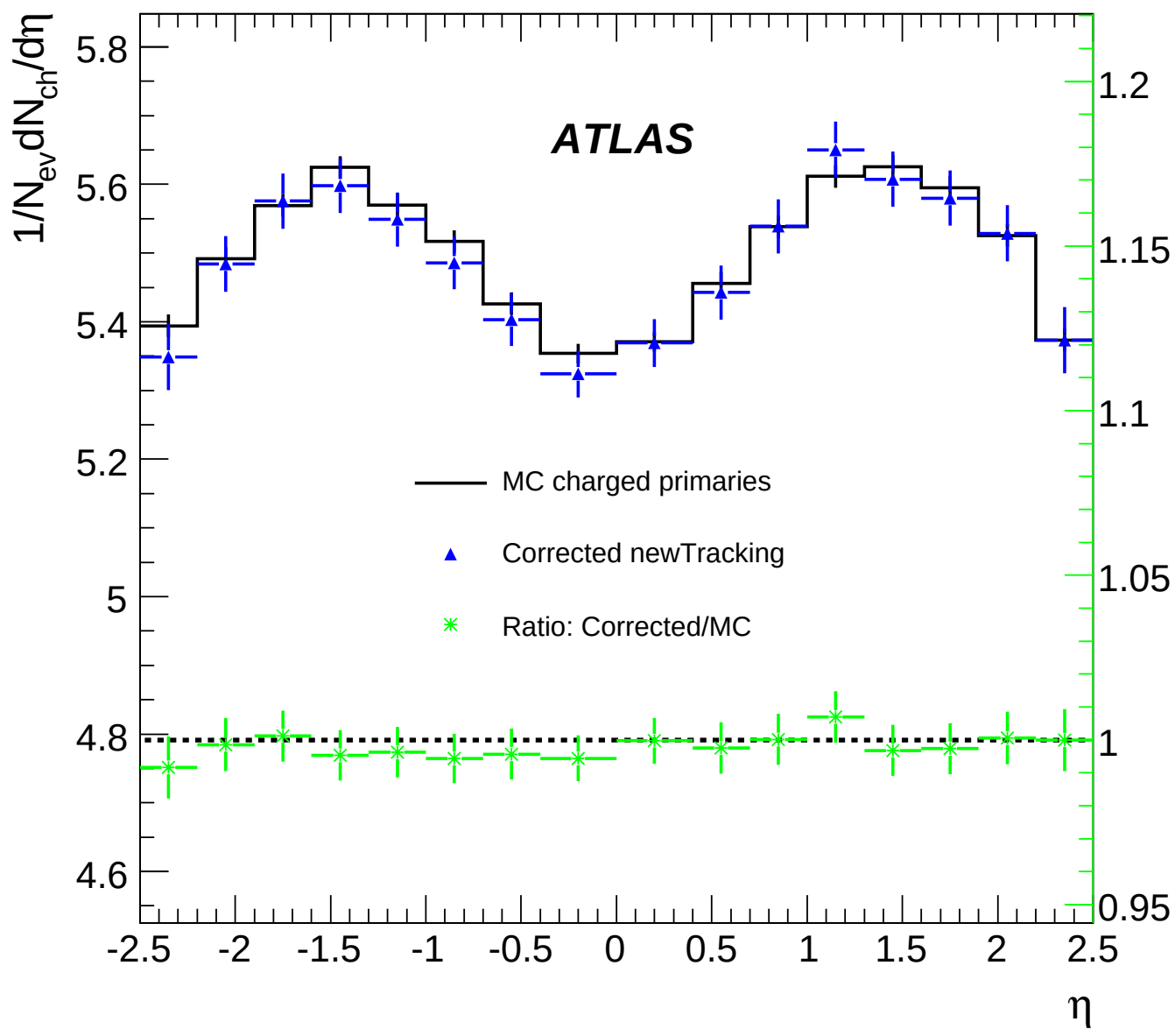


(a)

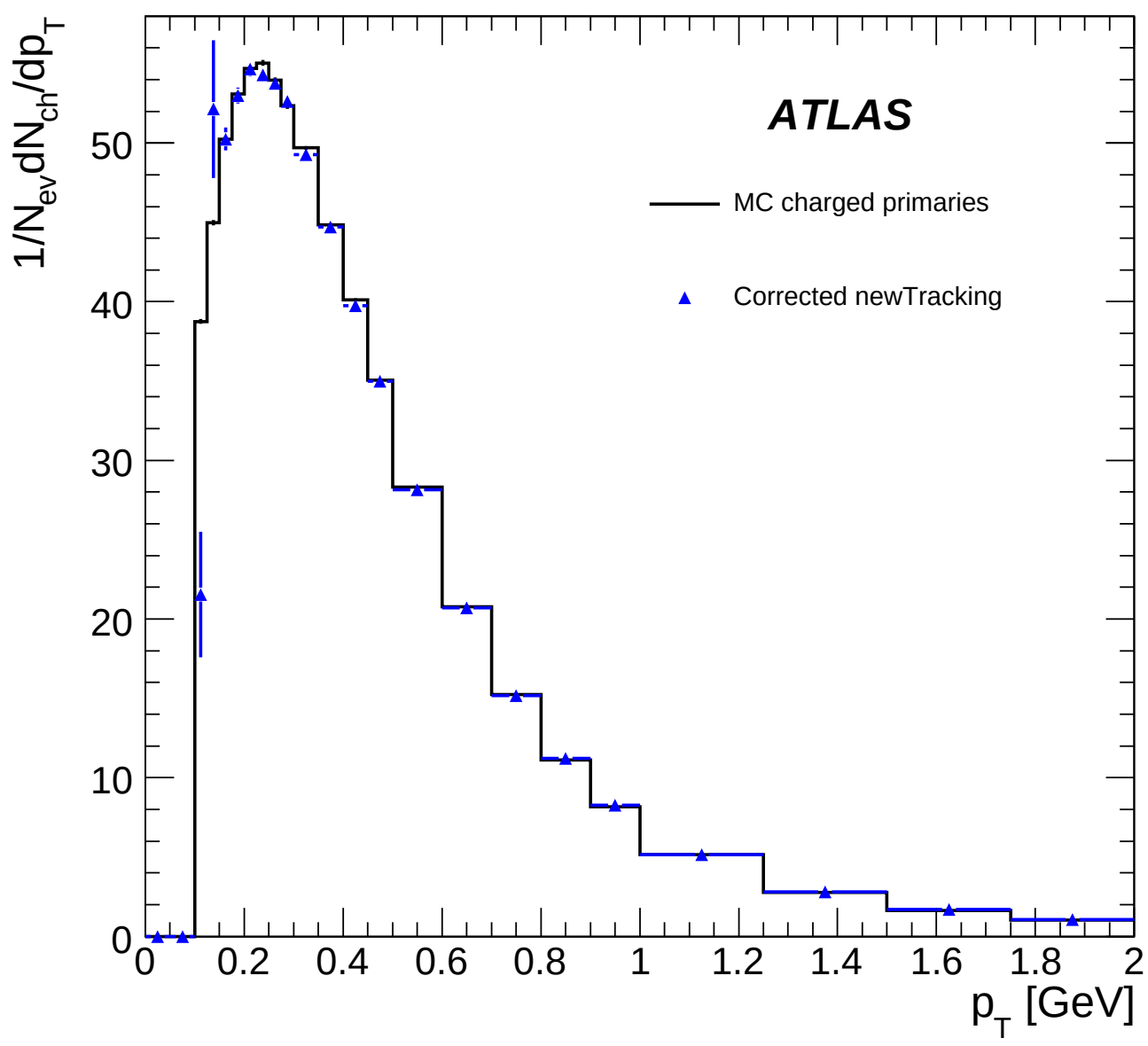


(b)

Figure 12: Corrected normalised pseudorapidity and $p_T$ distributions, together with the input Monte Carlo truth (full line) for the MB sample (blue triangles). (a) $\eta$ distributions for $p_T >$ 150 MeV, where the lower part, the ratio of the analysis result over the Monte Carlo prediction is shown. The corresponding scale is drawn on the right axis. (b) $p_T$ spectra for $|\eta| < 2.5$.

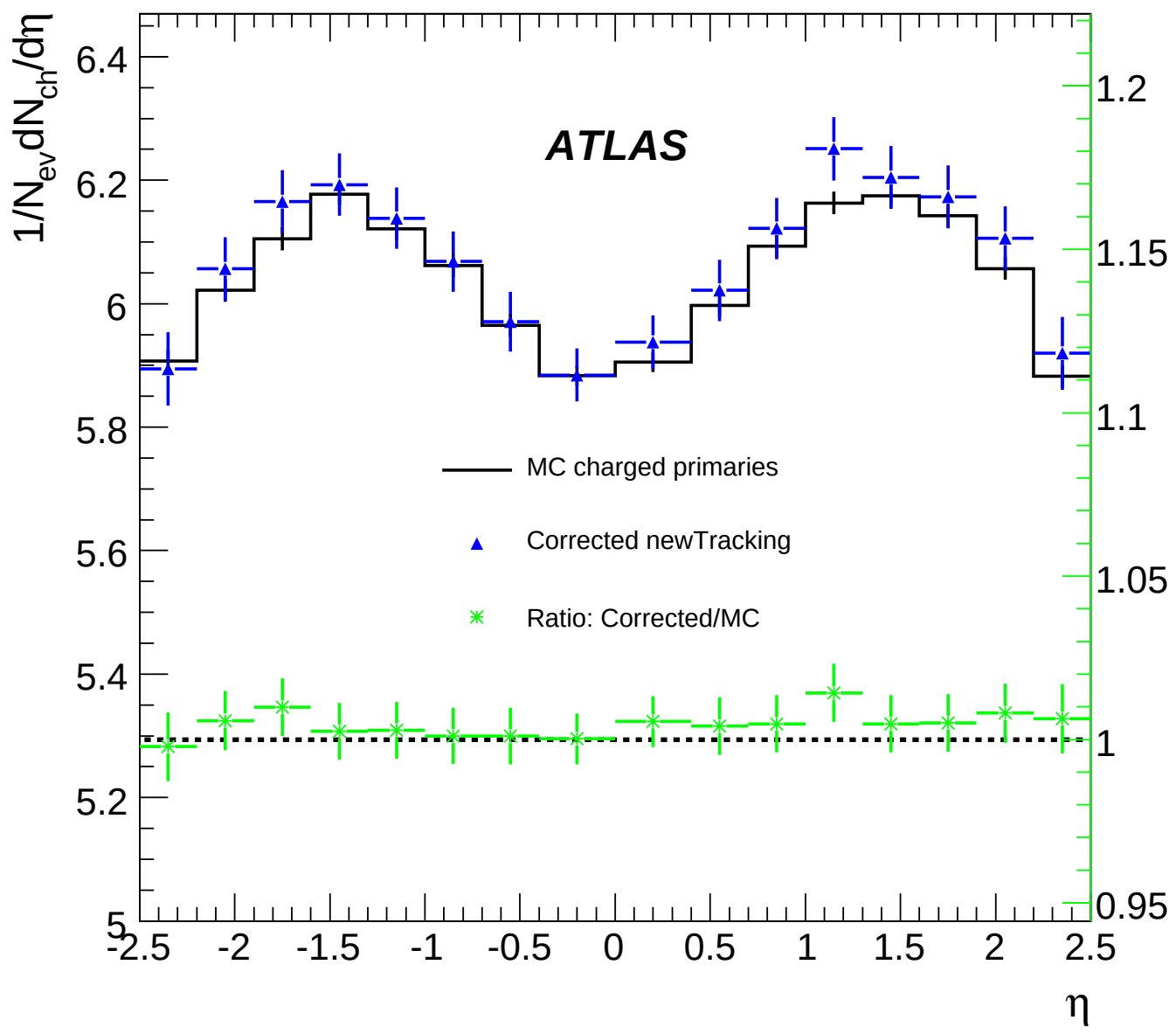


(a)

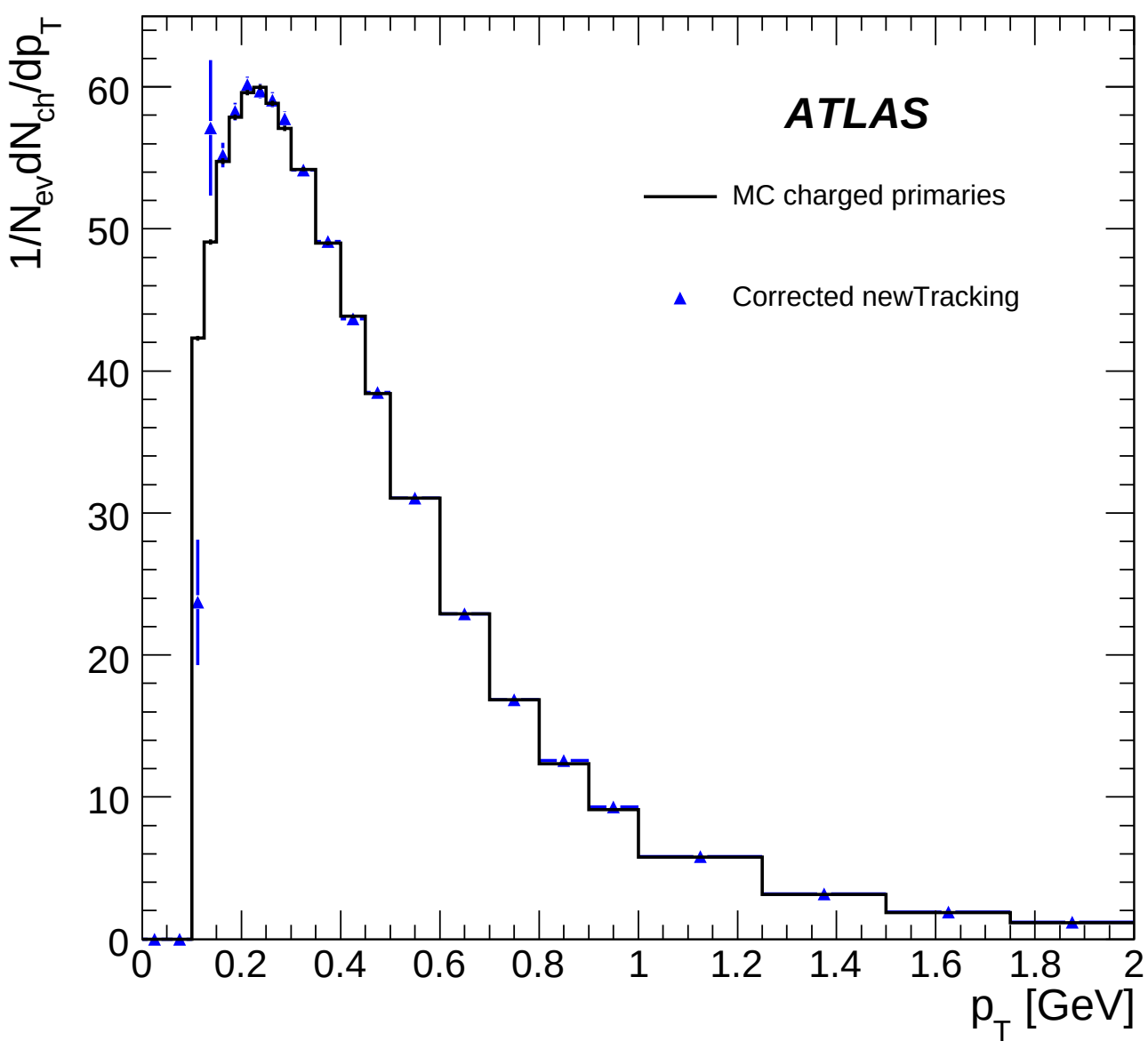


(b)

Figure 13: Corrected normalised pseudorapidity and $p_T$ distributions, together with the input Monte Carlo Truth (full line) for the NSD sample (blue triangles). (a) $\eta$ distributions for $p_T >$ 150 MeV, where the lower part, the ratio of the analysis result over the Monte Carlo prediction is shown. The corresponding scale is drawn on the right axis. (b) $p_T$ spectra for $|\eta| < 2.5$.

- Track selection: Based on the estimated track resolutions ($\sigma$) for their distance from the vertex of the interaction, $d_0$, tracks with $d_0 > 3\sigma$ are removed from the track sample. The error associated with this cut was estimated by varying the cut value by $\pm 0.5\sigma$ and observing the change in the number of accepted tracks. A mis-estimate of $\pm 0.5\sigma$ leads to a change in the number of accepted tracks of less than 2%.

- Secondaries: Before track selection cuts are applied, the number of tracks originating from secondaries is about 25% of the number of tracks originating from primaries. After the track selection the total number of accepted secondary tracks is about 2.2% of the number of the accepted tracks. The error on the number of secondaries was estimated by changing the number of secondaries in the reconstructed sample by $\pm 50\%$. The total number of accepted secondary tracks was found to change by 1.1%. The effect is strongest in the low-$p_T$ region and vanishes only for $p_T$>5 GeV. The systematic uncertainty was therefore estimated to be 1.5%.

- Vertex reconstruction bias: Since the $z$-position of the reconstructed vertex is integrated out, it is possible that a bias from the reconstruction of the vertex could introduce a systematic effect on the measurement. Bias due to the vertex reconstructed was evaluated by repeating the $dN_{ch}/d\eta$ analysis using the generated $v_z$ instead of the reconstructed $v_z$. The observed difference was found to be 0.1%.

- Misalignment: The systematic effect from misalignment was estimated by re-running the reconstruction with a different geometry that corresponded to a misaligned and distorted detector, and comparing the $dN_{ch}/d\eta$ produced with that produced for a detector that is ideally aligned. The misaligned geometry was misaligned both globally and locally. However, the local misalignments had been corrected for using the alignment procedure. The two distributions agreed at the level of 5-6% across the barrel and forward regions. This is the exepected level that can achieved with cosmics and early data. A systematic uncertainty of 6% was attributed to misalignment.

- Beam-gas interactions: The presence of background events coming from beam-gas would result in additional systematics on the measurement.

  The rate of beam-gas interactions is highly dependent on the particular beam conditions during startup such as the number of protons per bunch and the beam current. Studies by the LHC group have estimated the rate of beam-gas collisions within ATLAS to be 100 Hz [19]. Assuming a pp collision rate during early running of $8\times10^5$ Hz this corresponds to approximately 1 beam-gas event per 8000 pp events. Preliminary studies with simulated beam-gas events have also shown that only a very small fraction (<1%) of beam-gas events passed the vertex reconstruction requirement. Therefore the effect of beam-gas events is expected to be small.

  The systematic uncertainty due to beam-gas events was estimated to be $\sim 1\%$.

- Particle composition: The track-to-particle correction is calculated from events generated with PYTHIA [14]. Although $\pi^{\pm}$, $K^{\pm}$, and $p^{\pm}$ compose over 98% of the charged particle multiplicity in these events, the efficiency for detecting each of these is considerably different for $p_T < 500$ MeV. This model dependent effect introduces an additional systematic error on the measurement since relative abundances in PYTHIA could differ from reality.

  Relative abundances of charged pions, kaons and protons and anti-protons were enhanced and reduced by $\pm 50\%$ to estimate the systematic uncertainty due to variations in the particle com-

position. The systematic uncertainty obtained by this method is estimated to be around 2%. This is highly dependent on the low-$p_T$ cut used in track reconstruction.

- Relative process frequency: The trigger bias and vertex reconstruction corrections are calculated from a sample of inelastic events where the relative cross-sections are as predicted by PYTHIA [3]. Since the trigger and vertex reconstruction efficiencies are different for each of the inelastic components (non-diffractive, single diffractive, and double-diffractive), these corrections are dependent on the relative cross sections of the different components.

  The relative cross sections between non-diffractive, single-diffractive and double-diffractive components were enhanced and reduced to estimate the systematic due to the uncertainty from the PYTHIA event generator. The corrections have been calculated by changing the diffractive cross sections by ±50% of the PYTHIA prediction, i.e. 7.1 mb $< \sigma_{SD} <$ 21.4 mb and 5.1 mb $< \sigma_{DD} <$ 15.3 mb, which, as can be seen in Table 1, covers the difference in the predicted relative cross sections between PYTHIA and PHOJET.

  Changing the diffractive cross sections by ±50% changes the result of the analysis by about 4% on the final NSD sample. For the minimum bias sample in which no trigger bias correction is applied, this systematic error is approximately zero.

| Name | Level | Estimated Uncertainty |
|---|---|---|
| Track selection cuts | Analysis | 2% |
| Mis-estimate of secondaries | Analysis | 1.5% |
| Vertex reconstruction bias | Reconstruction | 0.1% |
| Misalignment | Reconstruction | 6% |
| Beam-gas and pileup | Offline Trigger | 1% |
| Particle composition | Generation/Simulation | 2% |
| Diffractive cross sections (NSD sample) | Generation | 4% |
| Total | | 8% |

Table 7: Summary of the various systematic uncertainties and the level at which they are introduced. The total systematic uncertainty assumes each of the individual uncertainties are independent.

A summary of the systematic uncertainties is given in Table 7 along with the step in the full chain at which they are introduced. The uncertainties in the track selection, vertex reconstruction bias and mis-estimate of secondaries are dominated by the low-$p_T$ tracks (<500 MeV). Any uncertainties at the generator level are due to uncertainties in predictions of physics models at LHC-energy and are unavoidable in the final analysis. However, the principal error in the reconstruction arises from alignment of the inner detector. The estimate of 6% is for initial running and is likely to improve.

# 5 Future work

## 5.1 Work required for first data

A number of issues have arisen during the work for this note that need to be studied further. Many of these studies are already in progress and should be completed before the first pp collisions from the LHC are recorded. The main topics currently under investigation are:

- *Studies with other Monte Carlo generators*: The studies of pp interactions in this note have been carried out using PYTHIA [3]. The biases introduced by the trigger will certainly depend on the particular physics models employed by PYTHIA.

  The generator PHOJET [4,13] is being used to simulate samples of LHC inelastic pp collisions (non-diffractive, single- and double-diffractive). The physics studies discussed previously in the analysis section are going to be repeated with PHOJET samples and the results compared with those shown in this note. Comparing the results from analyses in different samples will provide us indications on the systematic uncertainties due to different physics models.

- *Vertexing*: During initial running at low luminosities, there will still be periods where there will be more than one event per bunch crossing. It is therefore important to be confident that an event contains only one interaction by identifying events with a single vertex. Further work is required to look at vertex reconstruction with low levels of additional events to evaluate how well events with several vertices can be reconstructed.

- *MBTS readout*: The MBTS L1 trigger readout is currently being upgraded as part of the tile calorimeter refurbishment. As a part of this refurbishment the 3-in-1 cards connected to the MBTS readout are being switched from low gain to high gain. This modification means that the signal to noise should be 5-6 times better than the values used in Section 3.2. In addition to the L1 improvements, the MBTS will be readout at L2 to allow verification of the L1trigger decision using the precision readout of the Tile Calorimeter electronics. Work is ongoing to implement software to match these requirements. Once the software has been validated, the MBTS L1and L2 trigger performance will be reevaluated.

- *Beam-gas and beam halo estimate from commissioning runs*: Recent efforts to produce more accurate predictions for beam gas rates will allow a re-run of studies on how best to optimize the MBTS and the track-trigger to reduce this background in the measured minimum bias sample. For both cases, strategies on how to estimate the rates of beam-gas and halo events from data during commissioning runs with a single proton beam are beginning to be discussed.

- *ATLAS Beam Conditions Monitor* The Beam Conditions Monitor [20] is designed to distinguish collisions from background through time-of-flight measurement and can improve our understanding of beam-gas and halo rates.

### 5.2 Physics studies beyond first data

Following the first physics studies with minimum bias data a series of applications can be foreseen. Among them we would highlight:

- *Retuning Monte Carlo generators*: LHC predictions for minimum bias events generated with tuned MC generators PYTHIA and PHOJET indicate that although these tuned models give comparable descriptions of lower energy data, they disagree typically by $\sim 30\%$ for minimum bias multiplicity distributions [12].

  Measuring the global properties of minimum bias interactions as presented in this paper will provide enough information to revisit the generator predictions and retune event generators to describe interactions at the LHC energy. This will not only allow us to distinguish which physics model and assumptions better agree with the data, but will also allow the reassessment of systematic uncertainties in various channels.

- *Multiparton interactions*: Charged particle multiplicity distributions [21] have been widely used as important tools for studying multiple particle production in inelastic hadronic events [5, 6, 10, 11]. They are particularly useful when displayed in terms of multiplicity scaled variables known as KNO variables [21]. Plotted as a function of KNO variables, the charged multiplicity distributions provide a clearer display of fluctuations seen for both very low (less than half of the average multiplicity) or very high multiplicity (more than the double the average) events. The rise of the high multiplicity tail in these distributions has been interpreted as an effect caused by multiparton interactions [10, 11] and the LHC measurement of this effect will contribute greatly to a better understanding of the underlying physics in inelastic collisions.

- *Contribution to the measurement of the total cross section*: A measurement of the total inelastic rate by combining the NSD analysis with a measurement of single diffractive cross section by the ALFA detector should be possible. Simultaneously measuring the elastic t-spectrum, which can then be extrapolated to $t \to 0$, will enable us to determine the total cross section ($\sigma_{tot}$) in a luminosity-independent way or to calibrate the absolute luminosity in $\sigma_{tot}$-independent way, using the optical theorem.

- *Baseline for heavy-ion studies*: A measurement of the particle yield from a single minimum bias event can and will be used to cross-check components and assumptions made by models for heavy-ion collisions which will then be tuned according to the LHC data. This procedure is expected to help in generating more reliable predictions for the heavy-ion runs.

## 6 Conclusions

We have investigated methods for measuring the characteristics of inelastic collisions with the ATLAS detector during early LHC running. The main goal of our study was to reconstruct the normalised central pseudorapidity ($\frac{1}{N_{ev}}\frac{dN}{d\eta}$) and transverse momentum ($\frac{1}{N_{ev}}\frac{dN}{dp_T}$) distributions of charged primaries in pp interactions selected by a minimum bias trigger. Triggering strategies for minimum bias event selection have been developed and an analysis to determine the charged particle distributions were presented.

Aiming at selecting inelastic collisions with as little bias as possible, two scenarios were proposed for triggering: a random-based track trigger and using the minimum bias trigger scintillators. The random-based track trigger combines random event selection at L1 with the use of information from the ID in the HLT; spacepoints from the pixel detector and the SCT at L2 and tracks reconstructed in the EF. Minimum bias selection with the MBTS is performed by recording events that pass L1 MBTS hit requirements. It has been shown that both triggers are highly efficient at selecting non-diffractive inelastic samples. However, the acceptance of the diffractive component is highly suppressed relative to the non-diffractive component, and the acceptance of the single-diffractive sample is similar to that of the double-diffractive sample. This means that the ATLAS minimum bias triggers do not select a NSD sample as previous experiments have and therefore model dependent corrections will be required to compare ATLAS measurements to results from previous experiments.

Results were presented for two samples: the MB sample, which has only been corrected for detector effects, and the NSD sample where the MB sample has been further corrected for the trigger acceptance, which is model dependent. The analysis was performed on a sample of $\sim 75,000$ events, corresponding to a luminosity of $10^{-6}$ pb$^{-1}$. The uncertainty on $\frac{1}{N_{ev}}\frac{dN}{d\eta}$ for both samples is dominated by the systematic uncertainty from alignment and there is an additional uncertainty on the NSD sample due to the model dependence of the relative cross-sections. The systematic uncertainty on $\frac{1}{N_{ev}}\frac{dN}{d\eta}$ for

the MB sample was estimated to be 6% and 8% for the NSD sample. This will be sufficient to distinguish between different models of minimum bias events.

# Electroweak Boson Cross-Section Measurements

**Abstract**
This report summarises the ATLAS prospects for the measurement of $W$ and $Z$ production cross-section at the LHC. The electron and muon decay channels are considered. Focusing on the early data taking phase, strategies are presented that allow a fast and robust extraction of the signals. An overall uncertainty of about 5% can be achieved with 50 $pb^{-1}$in the $W$ channels, where the background uncertainty dominates (the luminosity measurement uncertainty is not discussed here). In the $Z$ channels, the expected precision is 3%, the main contribution coming from the lepton selection efficiency uncertainty. Extrapolating to 1 $fb^{-1}$, the uncertainties shrink to incompressible values of 1-2%, depending on the final state. This irreducible uncertainty is essentially driven by strong interaction effects, notably parton distribution uncertainties and non-perturbative effects, affecting the $W$ and $Z$ rapidity and transverse momentum distributions. These effects can be constrained by measuring these distributions. Algorithms allowing the extraction of the $Z$ differential cross-section are presented accordingly.

## 1 W and $Z$ cross-section measurements at the LHC

The study of the production of $W$ and $Z$ events at the LHC is fundamental in several respects. First, the calculation of higher order corrections to these simple, colour singlet final states is very advanced, with a residual theoretical uncertainty smaller than 1% [1]. Such precision makes $W$ and $Z$ production a stringent test of QCD.

Secondly, more specifically for $Z$ production, the clean and fully reconstructed leptonic final states will allow a precise measurement of the transverse momentum and rapidity distributions, respectively $d\sigma/dp_T$ and $d\sigma/dy$. The transverse momentum distribution will provide more constraints on QCD, most significantly on non-perturbative aspects related to the resummation of initial parton emissions, while the rapidity distribution is a direct probe of the parton density functions (PDFs) of the proton. The high expected counting rates will bring significant improvement on these aspects, and this improvement translates to virtually all physics at the LHC, where strong interaction and PDF uncertainties are a common factor.

From the experimental point of view, the precisely measured properties of the $Z$ boson provide strong constraints on the detector performance. Its mass, width and leptonic decays can be exploited to measure the detector energy and momentum scale, its resolution, and lepton identification efficiency very precisely.

Finally, a number of fundamental electroweak parameters can be accessed through $W$ and $Z$ final states ($M_W$, through the $W$ boson decay distributions; $\sin^2\theta_W$, via the $Z$ forward-backward asymmetry; lepton universality, by comparing electron and muon cross-sections). These measurements are long term applications where the understanding of the hadronic environment at the LHC is crucial, and to which the above-mentioned measurements are necessary inputs.

The present note summarises the ATLAS preparations for $W$ and $Z$ cross-section measurements, in the context of the early running of ATLAS and the LHC. The electron and muon decay channels are considered. A baseline integrated luminosity of $\mathscr{L} = 50$ pb$^{-1}$ is assumed for the total cross-section analyses; based on these results, we estimate our prospects for $\mathscr{L} = 1$ fb$^{-1}$. Anticipating that the measurement precision will soon be limited by the above-mentioned theoretical uncertainties, differential cross-section analyses are presented in the second part of this work. We consider the Drell-Yan mass spectrum below the $Z$ peak, and the $y$ and $p_T$ distributions on the $Z$ resonance.

The note is organised as follows. Section 2 gives technical details about cross-section measurements, lists the simulation samples used in the analyses, and reviews the main reconstruction aspects. Sections 3 and 4 describe the selections that allow extraction of the $W$ and $Z$ signals, give the expected statistics, and discuss the uncertainties on the background rates, as these are specific to each channel. Common systematic uncertainties affecting the cross-section determination are discussed in Section 5. Section 6 then presents the expected performance for total cross-section measurement. Differential cross-sections are discussed in Section 7, and Section 8 summarizes our results.

# 2 General discussion

This section describes the general procedure used to extract physical cross-sections, and the simulation samples used to evaluate the expected performance.

## 2.1 Total cross-section measurements

The number of events $N$ passing a given set of selections is expressed as follows:

$$N = \mathscr{L}\, \sigma\, A\, \varepsilon + B \tag{1}$$

where $\mathscr{L}$ is the integrated luminosity; $\sigma$ the signal cross-section; $A$ the acceptance of the signal, defined as the fraction of the signal that passes the kinematic and angular cuts; $\varepsilon$ is the reconstruction efficiency of the signal within the fiducial acceptance; $B$ is the number of background events. In the above, $\varepsilon$ is to be understood as averaged over the phase space accepted by the selections. Conversely, the measured total cross-section is expressed as:

$$\sigma = \frac{N - B}{\mathscr{L}\, A\, \varepsilon} \tag{2}$$

and the overall measurement uncertainty gets contributions from the different terms as below:

$$\frac{\delta\sigma}{\sigma} = \frac{\delta N \oplus \delta B}{N - B} \oplus \frac{\delta\mathscr{L}}{\mathscr{L}} \oplus \frac{\delta A}{A} \oplus \frac{\delta\varepsilon}{\varepsilon} \tag{3}$$

Above, $\delta N \sim \sqrt{N}$ is of purely statistical origin, and the relative uncertainty decreases with increasing luminosity following $\delta N/N \sim 1/\sqrt{\mathscr{L}}$. The terms $\delta B$, $\delta A$ and $\delta\varepsilon$ are of both theoretical and experimental origin. They are considered as systematic uncertainties in the cross-section measurements, but can be constrained *via* auxiliary measurements. We thus expect these terms to improve over time, provided the auxiliary measurements have statistically dominated uncertainties. The machine luminosity $\mathscr{L}$ will be measured with different methods. The uncertainty on this parameter, $\delta\mathscr{L}$, is expected to decrease through improved understanding of the LHC beam parameters and of the ATLAS luminosity detector response [2].

## 2.2 Signal and background samples. Benchmark cross-sections

Our main signals, $W$ and $Z$ events decaying into electrons and muons, are generated using PYTHIA [3]. The analysis described in Section 7, a measurement of the low-mass Drell-Yan cross-section, exploits samples produced using HERWIG [4].

The $W$ and $Z$ samples are filtered at generation-level, requiring at least one lepton within the fiducial acceptance. The electron channels require $|\eta_e| < 2.7$ and $p_T^e > 10$ GeV; the muon channels require $|\eta_\mu| < 2.8$ and $p_T^\mu > 5$ GeV. These filters have an efficiency of about 85% for $Z$ events, and about 65% for $W$ events. In the case of the Z sample, the available energy for the hard process is limited by $\sqrt{\hat{s}} > 60$ GeV. The low-mass Drell-Yan samples have the same fiducial cuts on both leptons, but require $8 < \sqrt{\hat{s}} < 60$ GeV. The W and Z cross-sections are normalised to the NNLO cross-sections as provided by the FEWZ program [1].

The backgrounds considered in the analyses originate from $W$ and $Z$ events decaying to $\tau$-leptons, with subsequent leptonic $\tau$ decays; $t\bar{t}$ events involving at least one semileptonic decay, and from inclusive jet events filtered to favour the presence of real leptons or hadrons misidentified as leptons. Low-mass Drell-Yan events, analysed only in the electron channel ($\gamma^* \to ee$), also account for backgrounds from boson pair production.

The $W \to \tau\nu_\tau$ and $Z \to \tau\tau$ events are produced as the signal samples (generators and filters). The $t\bar{t}$ samples are generated inclusively, using MC@NLO [5] to provide both the final states and the cross-section. They are filtered for the presence of at least one electron or muon, without kinematic constraints. Diboson events, one of the main background in the low-mass Drell-Yan analysis, are generated using MC@NLO. For the $WW$ process only $W \to e\nu$ decays are considered, while the $ZZ$ and the $WZ$ are generated inclusively. No filters are applied.

The jet events are produced using PYTHIA. The transverse momentum defined in the rest frame of the hard interaction is required to be above 15 GeV for jet samples used in W and Z analyses, whilst no such $p_T$ cut is required for the sample used in low-mass Drell-Yan analysis. Jet backgrounds for the muon channels are generated as inclusive jets, then requiring a final state with one (W analysis; $p_T(\mu) > 15$ GeV) or two (Z analysis; $p_T(\mu) > 5$ and 15 GeV) muons from b-hadron decays with $|\eta_\mu| < 2.5$. Background events from hadron punch-through and from decays in flight of long lived particle are found negligible. Background events from cosmic muons can be eliminated in a very efficient way with timing cuts. Relying on the Tevatron results [6], this background is neglected. This appears as a safe approximation for the ATLAS experiment, since the Tevatron experiments are built close to the surface, while the ATLAS detector is $\sim 100$ m underground. In the electron channels, fake electrons are an important issue. Therefore, rather than requiring a true electron in the final state, the events are required to contain at least one narrow cluster of energetic final state particles. In practice, there should exist a tower of size $\Delta\eta \times \Delta\phi = 0.12 \times 0.12$ containing a total transverse energy greater than 17 GeV for use in W and Z analyses and 6 GeV for low-mass Drell-Yan analysis. Events passing this filter are considered likely to produce fake electrons and passed through the simulation step.

All samples are interfaced to the CTEQ6L1 or CTEQ6M parton density sets [7] depending if the generator uses a leading or next to leading order calculation, respectively. The events are processed through full simulation using Geant 6.4 and a special misaligned geometry, as described in Ref. [8]. While summarised here, the cross-sections used are described and justified, together with their uncer-

tainties, in Ref. [9]. Tables 1 and 2 summarise the signal and background samples and their properties. The first column in these Tables indicates explicitely the cases in which a specific leptonic decay have been required at the generation level. For this reason, the second column represents in some case (like for the $t\bar{t}$ dataset) the total production cross section and in other cases the total cross-section multiplied by the leptonic branching ratio(s). The third column indicates the efficiency of the filter which is applied on the generated final states described in the first column.

| Channel | $\sigma(\times B_r)$ | $\varepsilon_{filter}$ | $N_{evt}$ ($\times 10^3$) | $\mathscr{L}$ (pb$^{-1}$) |
|---|---|---|---|---|
| $W \to e\nu$ | 20510 pb | 0.63 | 140 | 11 |
| $\gamma/Z \to ee, \sqrt{\hat{s}} > 60$ GeV | 2015 pb | 0.86 | 399 | 230 |
| $\gamma/Z \to ee, \sqrt{\hat{s}} < 60$ GeV | 9220 pb | 0.022 | 197 | 969 |
| $W \to \tau\nu_\tau$ | 20510 pb | 0.20 | 32 | 8 |
| $Z \to \tau\tau$ | 2015 pb | 0.05 | 13 | 129 |
| $t\bar{t}$ | 833 pb | 0.54 | 382 | 850 |
| Inclusive jets ($p_T$ >6 GeV) | 70 mb | 0.058 | 2480 | 0.0006 |
| Inclusive jets ($p_T$ >17 GeV) | 2333 $\mu$b | 0.09 | 3725 | 0.02 |
| WW$\to (e\nu)(e\nu)$ | 1.275 pb | 1. | 20 | 15608 |
| ZZ | 14.8 pb | 1. | 43 | 2922 |
| WZ | 29.4 pb | 1. | 50 | 1699 |

Table 1: Signals and background samples in the electron channels. $W$ and $Z$ cross-sections are normalised to the NNLO prediction; the $t\bar{t}$ cross-section is computed at NLO; the jet cross-section is the LO result. The filters are described in the text. The number of simulated events and the corresponding integrated luminosity are also indicated.

| Channel | $\sigma(\times B_r)$ | $\varepsilon_{filter}$ | $N_{evt}$ ($\times 10^3$) | $\mathscr{L}$ (pb$^{-1}$) |
|---|---|---|---|---|
| $W \to \mu\nu$ | 20510 pb | 0.69 | 190 | 13 |
| $\gamma/Z \to \mu\mu, \sqrt{\hat{s}} > 60$ GeV | 2015 pb | 0.89 | 446 | 249 |
| $W \to \tau\nu_\tau$ | 20510 pb | 0.20 | 32 | 8 |
| $Z \to \tau\tau$ | 2015 pb | 0.05 | 13 | 129 |
| $t\bar{t}$ | 833 pb | 0.54 | 382 | 850 |
| $b\bar{b} \to \mu + X$ | 766 $\mu$b | $2.1 \times 10^{-4}$ | 110 | 0.67 |
| $b\bar{b} \to \mu\mu + X$ | 25 $\mu$b | $1.6 \times 10^{-4}$ | 140 | 35 |

Table 2: Signals and background samples in the muon channels. $W$ and $Z$ cross-sections are normalised to the NNLO prediction; the $t\bar{t}$ cross-section is computed at NLO; the $b\bar{b}$ cross-section is the LO result. The filters are described in the text. The number of simulated events and the corresponding integrated luminosity are also indicated.

### 2.3 Common selection aspects

As already mentioned, $W$ and $Z$ boson final states are selected through their decays into electrons and muons. The reconstruction of electrons is based on a cluster measured in the electromagnetic

calorimeter, geometrically matching a track reconstructed in the Inner Detector. The identification of isolated high-$p_T$ electrons is then based on the shapes of the electromagnetic showers, and on track reconstruction information. Three sets of identification criteria have been defined. The Loose criterion consists of simple shower-shape cuts; the Medium criterion adds further cuts on shower-shape and on track quality; the Tight criterion tightens the track-matching requirement, adds a cut on the energy-momentum ratio and further selections based on the vertexing-layer hits and on the Transition Radiation Tracker. Electron reconstruction and its performance are described in Ref. [10].

The muon reconstruction is done with the Muon Spectrometer, possibly completed by the Inner Detector. Stand-alone muons are defined as consisting of a reconstructed track in the spectrometer only, and combined muons are the subset of the above that include a matching track in the Inner Detector. Muon reconstruction is documented in Ref. [11].

The measurement of missing energy in the transverse plane ($\not{E}_T$) is an important requirement for $W$ boson cross-section measurements, as significant $\not{E}_T$ reflects the presence of at least one neutrino in the final state. The algorithm exploits the energy deposits in the calorimeter cells, the reconstructed muon tracks, and an estimate of the energy lost in the cryostat. The calorimeter cells are calibrated according to the physical object they represent (electrons or photons, taus, jets and muons). Cells corresponding to electrons, photons and muons are calibrated at the electromagnetic scale, whereas all other cells are calibrated at the hadronic scale. The $\not{E}_T$ value is then computed as the vector sum of the cell transverse energies. If muons are reconstructed in the event, their transverse momentum is added to the calorimetric sum. A complete description of the $\not{E}_T$ reconstruction can be found in Ref. [12].

Jets are reconstructed from calorimeter cells. The Cone algorithm is used, where the jet size parameter, $\Delta R = \sqrt{(\Delta\eta)^2 + (\Delta\phi)^2}$, is set to $\Delta R = 0.7$.

At low luminosity, $\mathscr{L} = 10^{31}$ cm$^{-2}$s$^{-1}$, the relevant trigger items require at least one electron or muon with $p_T > 10$ GeV, at least two electrons with $p_T > 5$ GeV, or two muons with $p_T > 4$ GeV. No isolation criteria are imposed on the leptons. As the LHC luminosity ramps up towards its design value, tighter selections will be needed to control the rates. The thresholds are raised, and isolation criteria are imposed on the electrons. The trigger items relevant for $W$ boson selection require at least one isolated electron with $p_T > 22$ GeV, or one muon with $p_T > 20$ GeV. For $Z$ production, two isolated electrons with $p_T > 12$ GeV or two muons with $p_T > 10$ GeV can be required in addition to the above. The trigger items described above are summarised in Table 3. Many more trigger items exist; a complete description can be found in Ref. [13, 14].

The reconstruction efficiency for electrons and muons, and the resolution of the $\not{E}_T$ reconstruction algorithm are illustrated in Fig. 1. For electrons the Medium identification efficiency is illustrated, and for muons the combined reconstruction efficiency is shown. With looser criteria, higher efficiency and weaker $\eta$ dependence are obtained, at the cost of larger backgrounds. A complete description of the ATLAS detector and its performance can be found in Ref. [15].

| Trigger item | Description |
|---|---|
| `e10, e20` | One electron, $p_T > 10, 20$ GeV |
| `mu10, mu20` | One muon, $p_T > 10, 20$ GeV |
| `2e5` | Two electrons, $p_T > 5$ GeV |
| `2mu4` | Two muons, $p_T > 4$ GeV |
| `e22i` | One isolated electron, $p_T > 22$ GeV |
| `mu20` | One isolated muon, $p_T > 20$ GeV |
| `2e12i` | Two isolated electrons, $p_T > 12$ GeV |
| `2mu10` | Two isolated muons, $p_T > 10$ GeV |

Table 3: Main trigger items relevant for the selection of $W$ and $Z$ boson final states. The first group of trigger items is relevant to the $\mathscr{L} = 10^{31}$ cm$^{-2}$s$^{-1}$ trigger menu; the second group is relevant for the $\mathscr{L} = 10^{33}$ cm$^{-2}$s$^{-1}$ trigger menu.

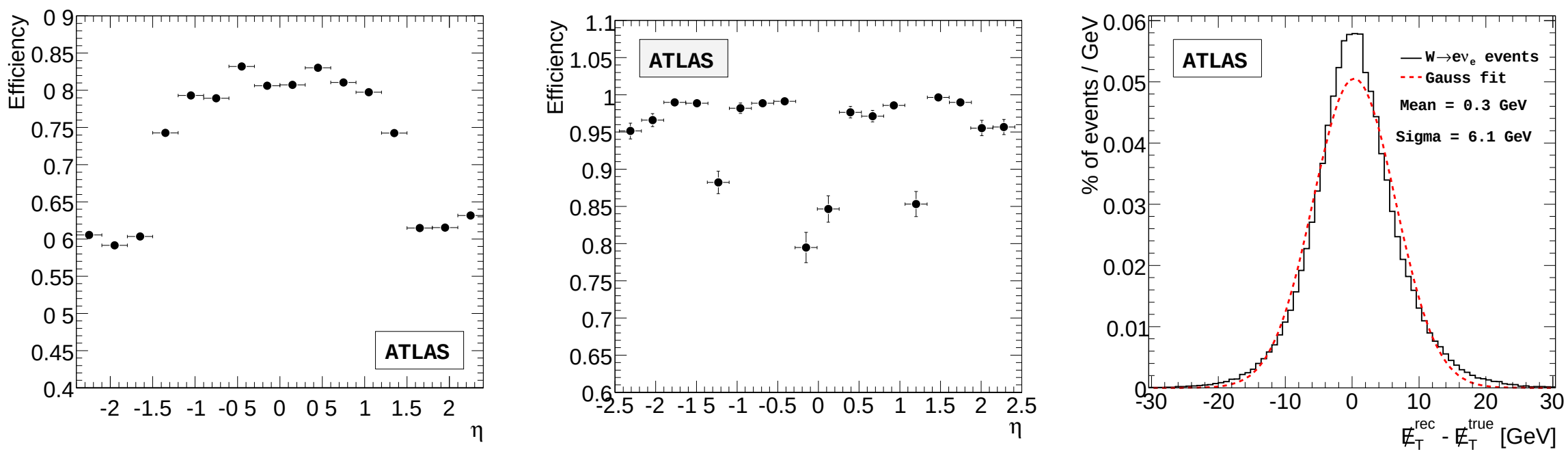


Figure 1: From left to right : Medium identification efficiency for electrons vs. $\eta$; muon combined reconstruction efficiency vs. $\eta$; $\not{E}_T$ resolution. The efficiencies are obtained from $Z$ boson events, and integrated over $p_T > 20$ GeV. The $\not{E}_T$ resolution is obtained from $W \to e\nu$ events.

| Selection | $W \to e\nu$ | jets | $W \to \tau\nu$ | $Z \to ee$ |
|---|---|---|---|---|
| Trigger | $37.01 \pm 0.09$ | $835 \pm 18$ | $1.73 \pm 0.02$ | $6.07 \pm 0.01$ |
| $E_T > 25$ GeV, $\|\eta\| < 2.4$ | $30.84 \pm 0.09$ | $383 \pm 12$ | $1.03 \pm 0.01$ | $3.23 \pm 0.01$ |
| Electron ID | $26.77 \pm 0.09$ | $110 \pm 6$ | $0.91 \pm 0.01$ | $2.95 \pm 0.01$ |
| $\not{E}_T > 25$ GeV | $22.06 \pm 0.09$ | $4.6 \pm 0.7$ | $0.55 \pm 0.01$ | $0.06 \pm 0.01$ |
| $M_T > 40$ GeV | $21.71 \pm 0.08$ | $1.5 \pm 0.4$ | $0.43 \pm 0.01$ | $0.04 \pm 0.01$ |

Table 4: Number of expected signal and background events ($\times 10^4$) in the $W \to e\nu$ channel after all selections, for an integrated luminosity of 50 pb$^{-1}$. The quoted uncertainties refer to the finite Monte-Carlo statistics only; systematic uncertainties are discussed in the text.

# 3 Electron final states

This section describes the event selections in the electron final states, the expected event rates, and estimations of the uncertainties on the remaining backgrounds.

## 3.1 $W \to e\nu$

**Event selection.** The selection of $W \to e\nu$ events proceeds as follows. First, the `e20` trigger item of the $10^{31}$ trigger menu should be passed. Then, exactly one electromagnetic (EM) cluster, matched with a track and such that $E_T > 25$ GeV, $\mid \eta \mid < 1.37$ or $1.52 < \mid \eta \mid < 2.4$, should be present in the event. This object should satisfy the Medium electron identification criterion. Finally, the reconstructed missing transverse energy, reflecting the missing final state neutrino, should satisfy $\not{E}_T > 25$ GeV, and the transverse mass of the $(l, \nu)$ system should satisfy $M_T > 40$ GeV.

Due to the high di-jet cross-section and the high rejection power of the selections, the available Monte Carlo statistics is not sufficient to evaluate this background directly. To overcome this difficulty, the jet background has been estimated by applying the trigger and electron identification selections only, and correcting the result with a factor obtained by computing the rejection power due to the $\not{E}_T$ and $M_T$ cuts only.

The number of signal and background events after the successive cuts are given in Table 4 for an integrated luminosity of 50 pb$^{-1}$. The statistical uncertainty on the expected number of events corresponding to 50 pb$^{-1}$is $\Delta_{N_S} = 0.04 \cdot 10^4$. The resulting transverse mass distribution is shown in Fig. 2.

**Background estimation.** As can be seen from Table 4, jet events constitute the largest background component. In addition, the jet production cross-section and fragmentation properties at the LHC are largely unknown and induce a significant uncertainty on the magnitude of this background; an uncertainty of about a factor 3 is estimated. It is therefore important to develop methods allowing to monitor the jet background using the data. An attempt is presented below.

The principle of the method is to measure the normalisation and shape of the jet background ahead of the $\not{E}_T$ cut, in a sufficiently pure jet sample. It is thus needed to find a jet sub-sample that is free of signal events, but exhibits a transverse mass distribution and jet multiplicity close to that of the jet background to $W \to e\nu$ at this level of the selection. This sub-sample is then used to evaluate the

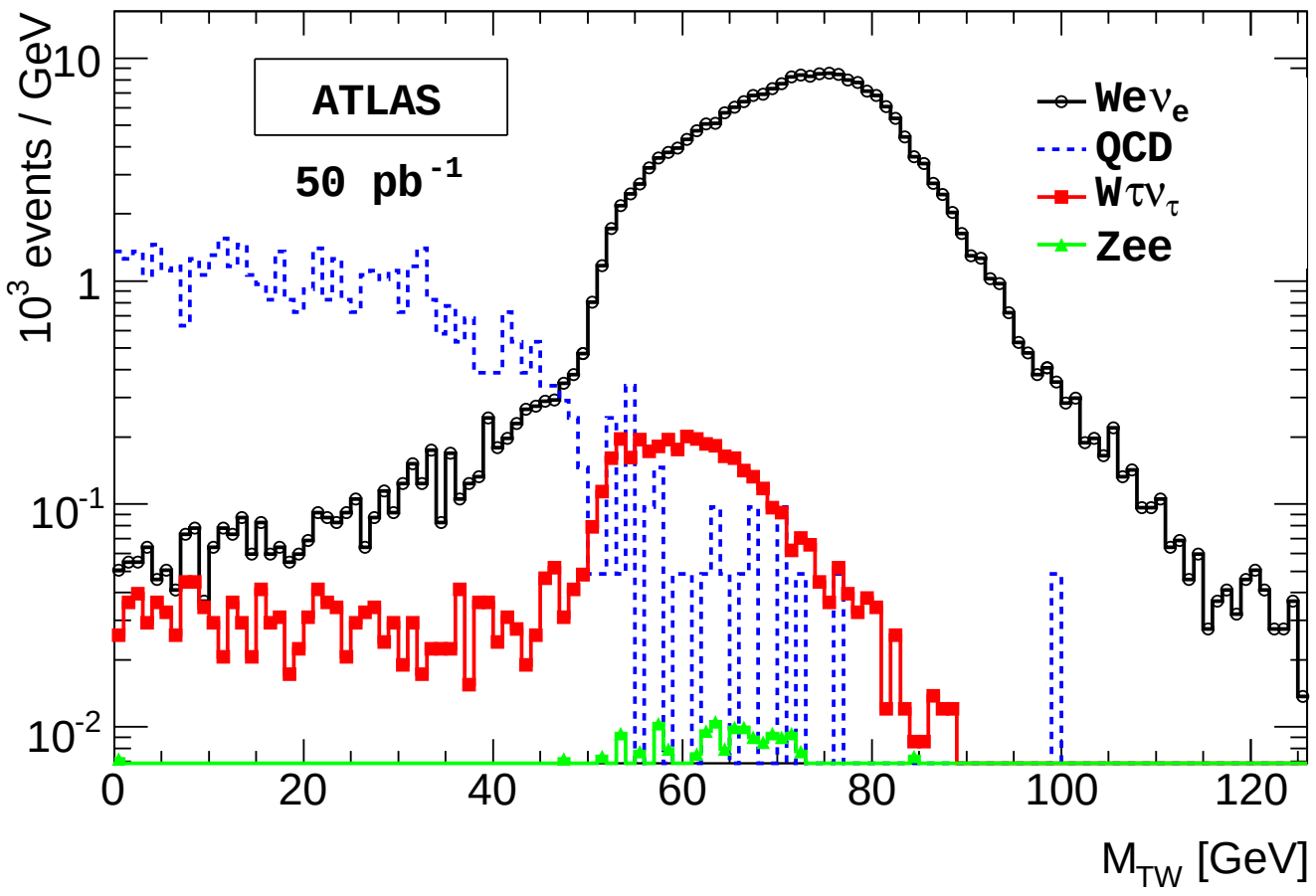


Figure 2: Transverse mass distribution in the $W \to e\nu$ channel, for signal and background after all selections, for $\mathscr{L} = 50$ pb$^{-1}$ after all selections except $M_T$ cut.

rejection of $\not{E}_T$ cut, allowing a realistic estimation of the jet background in the $W \to e\nu$ selection.

In this approach, the signal sample is obtained by applying the same trigger, kinematics and electron identification selection as described before and removing in addition events with a second high-$p_T$ electromagnetic cluster giving an invariant mass, together with first selected electron, close to the $Z$ boson mass ($65 < M_{ll} < 130$ GeV).

The jet background control sample is selected using a single photon trigger with $E_T > 20$ GeV, and subsequent photon identification using the same calorimetric variables as the electron identification. The photon cluster should also satisfy the same kinematics cuts of the electron candidate in the signal sample. There should be no Inner Detector track matching the photon cluster, to reject events with true electrons (e.g. $W$ events) contaminating this photon sample. Simulation studies show that these selections provide a sample essentially composed of jet events, even at high values of $\not{E}_T$, and that the shape of the $\not{E}_T$ distribution is identical, within the statistical precision, to that of the jet background in the $W \to e\nu$ sample (see Fig. 3). Above $\not{E}_T > 10$ GeV, the slope can be described with the convolution of an exponential and a second degree polynomial function.

After the subtraction of the estimated background to the signal sample, the analysis then proceeds applying the $\not{E}_T$ selection mentioned above. This data-driven estimation yields a jet background fraction of $(0+4-0)\%$. The uncertainty corresponds to a number of events, $\delta B = 0.92 \times 10^4$ events. Besides, a relative uncertainty of 3% is assumed on the $W \to \tau\nu$ background, as estimated from the experimental uncertainties on the $W$ and $\tau$ branching fractions. This process thus contributes $\delta B = 0.01 \times 10^4$ events.

### 3.2 $Z \to ee$

**Event selection.** This analysis relies on the `e10` trigger. Events are further preselected by requiring two EM clusters with $E_T > 15$ GeV and $|\eta| < 2.4$. The presence of two electrons in the final state allows application of the Loose electron identification criteria, which we briefly describe below. Three

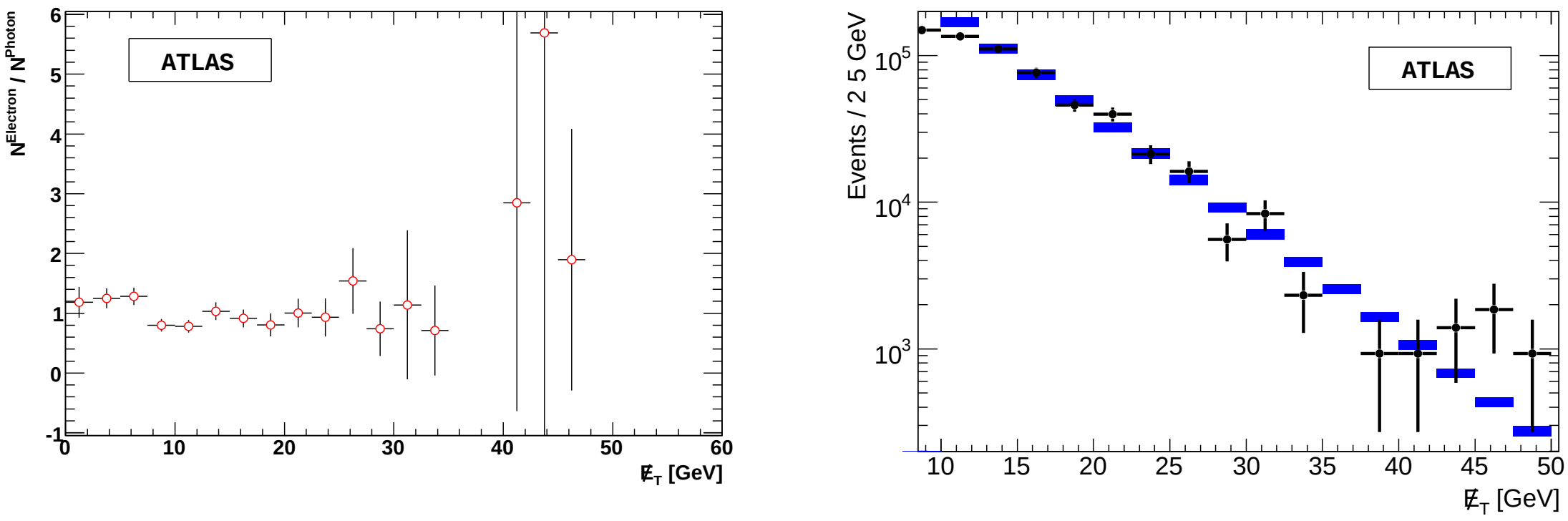


Figure 3: Left: ratio of the $\not{E}_T$ distributions in jet background events and in the control sample. Right: comparison of the jet background (points with error bars) and the fitted background (rectangles), for an integrated luminosity of 50 pb$^{-1}$.

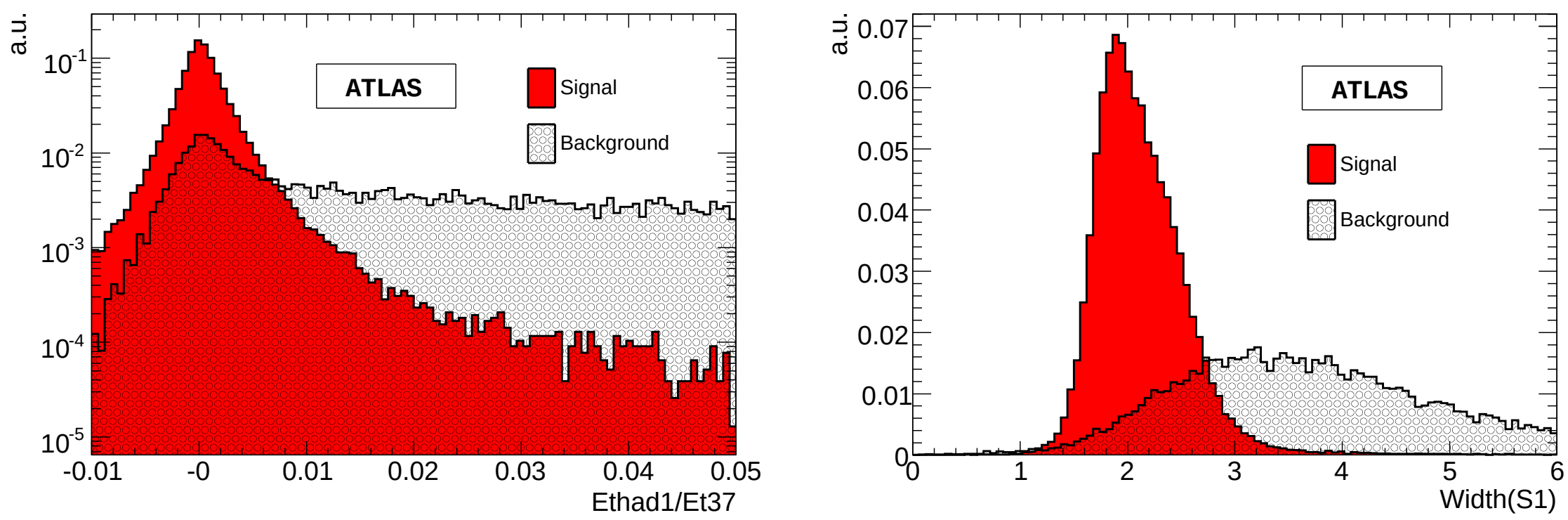


Figure 4: The electron identification criteria described in the text: the cluster hadronic to EM energy ratio (left); the cluster width in the first calorimeter sampling (right). The distributions are normalised to the number of background entries.

discriminant variables are used to separate EM clusters, deposited by electrons, from the hadronic background.

The first one is based on the longitudinal shower shape, and represents the ratio of the transverse energy deposited in the first compartment of the hadronic calorimeter divided by the transverse energy of the EM cluster. This ratio is expected to be small for EM objects, and large for hadronic clusters.

The second and third one are based on the shower width measured in the EM calorimeter. In the second compartment of the EM calorimeter, the width is computed from the ratio of the shower energy deposited in a region of size $\Delta\eta \times \Delta\phi = 0.075 \times 0.175$, divided by the energy deposited within $\Delta\eta \times \Delta\phi = 0.175 \times 0.175$ around the cluster barycenter. In the first compartment, the cluster spread is used, computed as the root-mean-square (RMS) of the cluster energy distribution. These two variables discriminate the narrow EM clusters from the wider hadronic clusters. Distributions of the three discriminators for electrons and hadrons are shown in Fig. 4 and in Fig. 5.

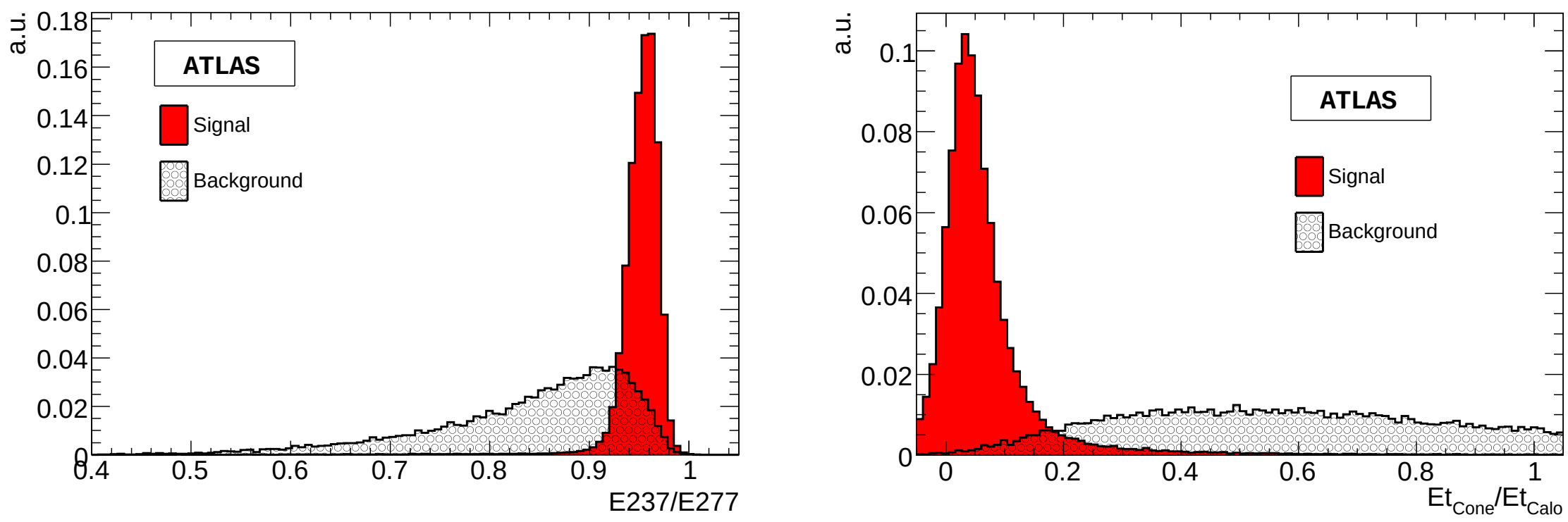


Figure 5: The electron identification criteria described in the text: cluster width in the second calorimeter sampling (left) and electron isolation variable (right). The distributions are normalised to the number of background entries.

| Selection | $Z \to ee$ | jets |
|---|---|---|
| Trigger | $6.70 \pm 0.01$ | $3110 \pm 40$ |
| $E_T > 15$ GeV, $\lvert\eta\rvert < 2.4$, 80 GeV $< M_{ee} <$ 100GeV | $2.76 \pm 0.01$ | $11.1 \pm 0.8$ |
| Electron ID | $2.64 \pm 0.01$ | $0.8 \pm 0.2$ |
| Isolation | $2.48 \pm 0.01$ | $0.2 \pm 0.1$ |

Table 5: Number of expected signal and background events ($\times 10^4$) in the $Z \to ee$ channel after all selections, for an integrated luminosity of 50 pb$^{-1}$. The quoted uncertainties refer to the finite Monte-Carlo statistics only; systematic uncertainties are discussed in the text.

Electrons identified as above are then required to be isolated. The isolation variable is computed from the total measured energy in a cone of size $\Delta R = 0.45$ around and excluding the electron, divided by the electron energy. Electrons are isolated if this ratio is smaller than 0.2. Distributions of this variable for the signal and the background are shown in Fig. 5.

The number of signal and background events after the successive cuts are given in Table 5 for an integrated luminosity of 50 pb$^{-1}$. The expected signal counting rate is $N = (2.48 \pm 0.02) \times 10^4$ events.The uncertainty quoted here is obtained by scaling the statistics of the Monte-Carlo sample to 50 pb$^{-1}$. The resulting di-electron invariant mass distribution is shown in Fig. 6.

**Background estimation.** As in the $W \to e\nu$ analysis, the simulation-based jet background estimate of Table 5 is replaced by a data-driven estimate. In this analysis, the signal and background fractions are estimated simultaneously, *via* a fit to both contributions. The signal is described by the convolution of a Breit-Wigner and a Gaussian resolution function, and the background, completely dominated by jet events, by an exponential function.

At the preselection level (just ahead of the electron identification and without the $M_{ee}$ cut), the background largely dominates the signal and allows to determine the exponential slope. After the identification and isolation cuts, the fit yields a background fraction of $(8.5 \pm 1.5)\%$, or $B = (0.23 \pm 0.04) \times$

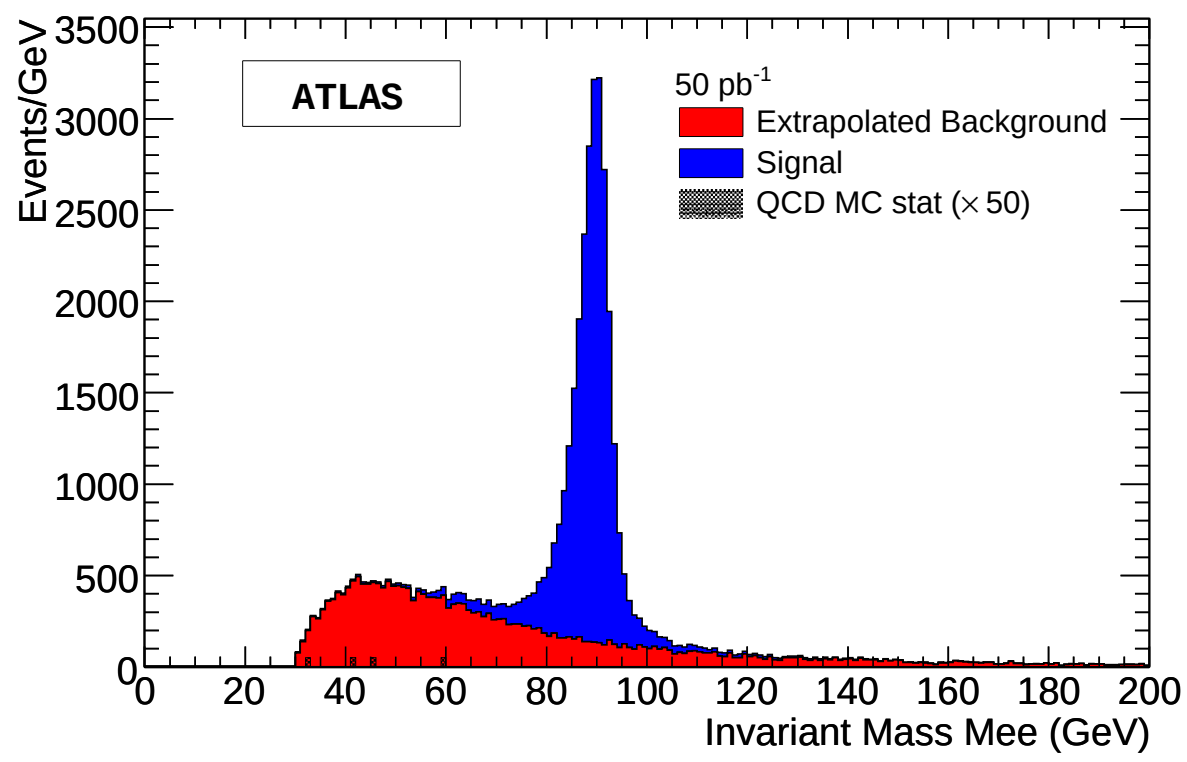


Figure 6: Di-electron invariant mass distribution in the $Z \rightarrow ee$ channel, for signal and background, for 50 pb$^{-1}$, after all selection cuts, except $M_{ee}$ cut.

$10^4$ events. The uncertainty on the background fraction derives from the modelling of the signal and background shapes.

The relatively important background rate is explained by the rather loose identification cuts. The present selections are chosen to illustrate the robustness of the signal extraction, and to exemplify the background extraction method.

# 4 Muon final states

## 4.1 $W \rightarrow \mu\nu$

**Event selection.** The $W \rightarrow \mu\nu$ signal is selected as follows. The events should contain exactly one muon track candidate, passing the mu20 trigger item and satisfying $|\eta| < 2.5$ and $p_T > 25$ GeV. The energy deposited in the calorimeter around the muon track, within a cone of radius $\Delta R = 0.4$, is required to be lower than 5 GeV. The event missing transverse energy should satisfy $\not{E}_T > 25$ GeV, and $M_T > 40$ GeV is required.

For the initial luminosity the $p_T$ cut of the trigger on the muon track is expected to be 20 GeV. Having a higher $p_T$ threshold, however, can further reduce the backgrounds in particular from heavy flavour hadron decays and from decays in flight of long lived particles. The isolation, $\not{E}_T$ and $M_T$ cuts are also effective to reduce those backgrounds.

After all selections, the overall efficiency for the signal is expected to be close to 80%, with very large rejection factors for $b\bar{b}$ and $t\bar{t}$ events. The number of events that are expected to pass the selection criteria for an initial integrated luminosity of 50 pb$^{-1}$ are shown in table 6. The expected background level corresponds to a fraction of $\sim 7\%$. Figure 7 shows the corresponding W transverse mass distribution before the transverse mass cut (last cut of table 6).

**Background estimation.** In contrast to the electron channels, the jet background is less important here and does not dominate the overall background. Muons from heavy flavour decays are rejected

| Selection | $W \to \mu\nu$ | $W \to \tau\nu$ | $Z \to \mu\mu$ | $bb \to \mu X$ | $t\bar{t}$ |
|---|---|---|---|---|---|
| Trigger | $44.44 \pm 0.07$ | $1.53 \pm 0.01$ | $2.03 \pm 0.01$ | $83.34 \pm 0.09$ | $0.53 \pm 0.07$ |
| $p_T > 25\,\text{GeV}, \|\eta\| < 2.5$ | $35.55 \pm 0.06$ | $1.22 \pm 0.01$ | $1.62 \pm 0.01$ | $68.27 \pm 0.08$ | $0.42 \pm 0.06$ |
| Isolation | $34.80 \pm 0.06$ | $1.20 \pm 0.01$ | $1.59 \pm 0.01$ | $9.67 \pm 0.03$ | $0.35 \pm 0.06$ |
| $\not{E}_T > 25$ GeV | $28.59 \pm 0.05$ | $0.72 \pm 0.01$ | $1.10 \pm 0.01$ | $1.00 \pm 0.01$ | $0.30 \pm 0.05$ |
| $M_T > 40$ GeV | $28.03 \pm 0.05$ | $0.57 \pm 0.01$ | $1.10 \pm 0.01$ | $0.10 \pm 0.01$ | $0.24 \pm 0.05$ |

Table 6: Number of expected signal and background events ($\times 10^4$) in the $W \to \mu\nu$ channel, for an integrated luminosity of 50 pb$^{-1}$. The quoted uncertainties refer to the finite Monte-Carlo statistics.

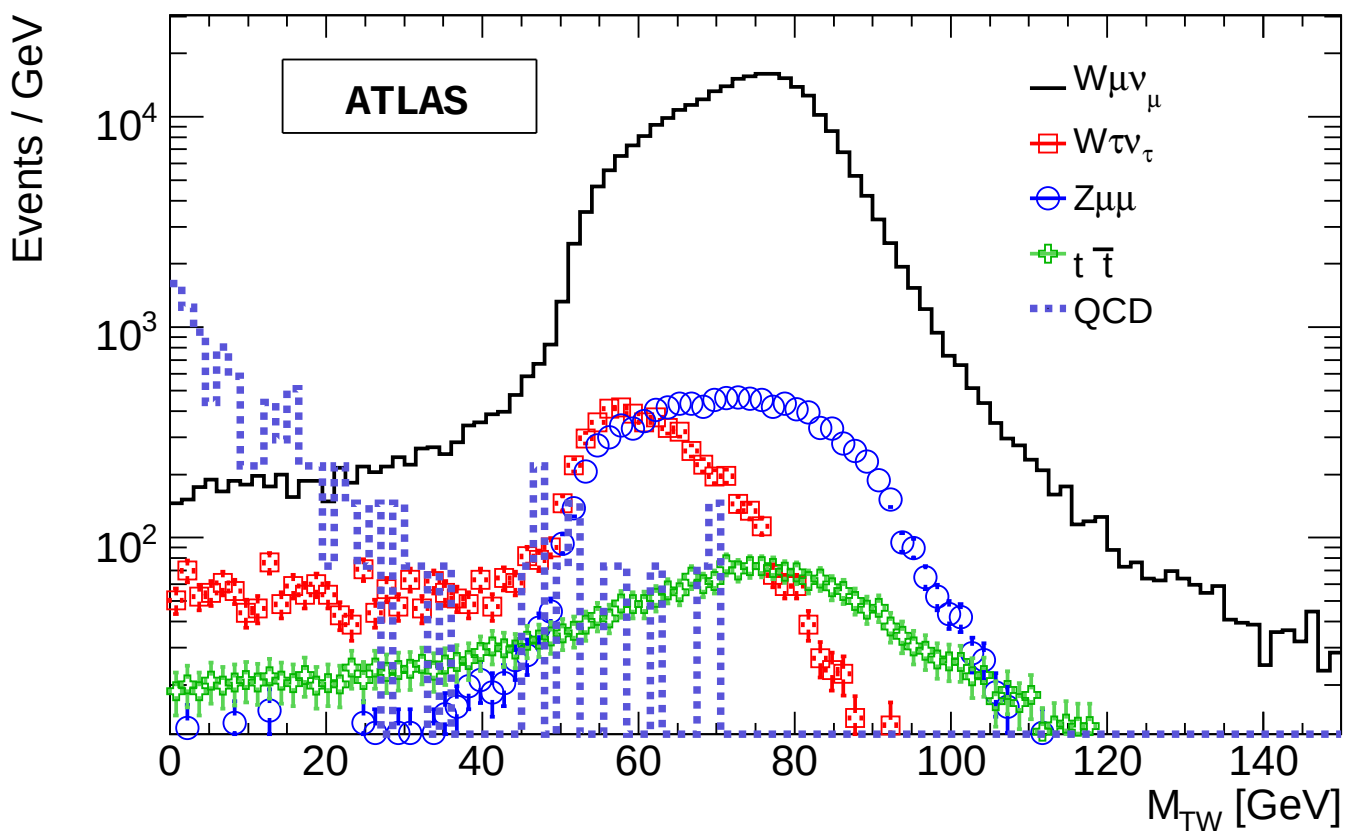


Figure 7: Transverse mass distribution in the $W \to \mu\nu$ channel, for signal and background, for 50 pb$^{-1}$ after all selections except $M_T$ cut.

using the $p_T$ and the isolation cuts, and muons from decays in flight of long lived particles could be further rejected using loose impact parameter cuts. The $t\bar{t}$ background and its uncertainty are small.

As can be seen in Table 6, the dominant backgrounds are expected from $W \to \tau\nu$ and $Z \to \mu\mu$ events. These processes are well understood theoretically, in particular with respect to the $W \to \mu\nu$ signal, and can be safely estimated based on simulation. A relative uncertainty of 3% is assumed on the $W \to \tau\nu$ background, as estimated from the experimental uncertainties on the $W$ and $\tau$ branching fractions. Exploiting the CTEQ6.5 eigenvector sets (cf. Section 5), an uncertainty of 2% is assumed on the $Z$ event rate passing the selections.

The jet background (mostly muons from b-hadron decays) is theoretically not well known. An uncertainty of 100% is assumed on this background component.

A theoretical uncertainty of about 15% on the $t\bar{t}$ cross-section is assumed. In addition, an uncertainty of 10% is considered on the rejection obtained from the isolation cut. This leads to a total uncertainty of about 20% on the $t\bar{t}$ background rate.

### 4.2 $Z \to \mu\mu$

**Event selection.** The $Z \to \mu\mu$ analysis uses the 10 GeV single muon trigger. The triggered data sample is further reduced by requiring at least two reconstructed muon tracks. The present analysis

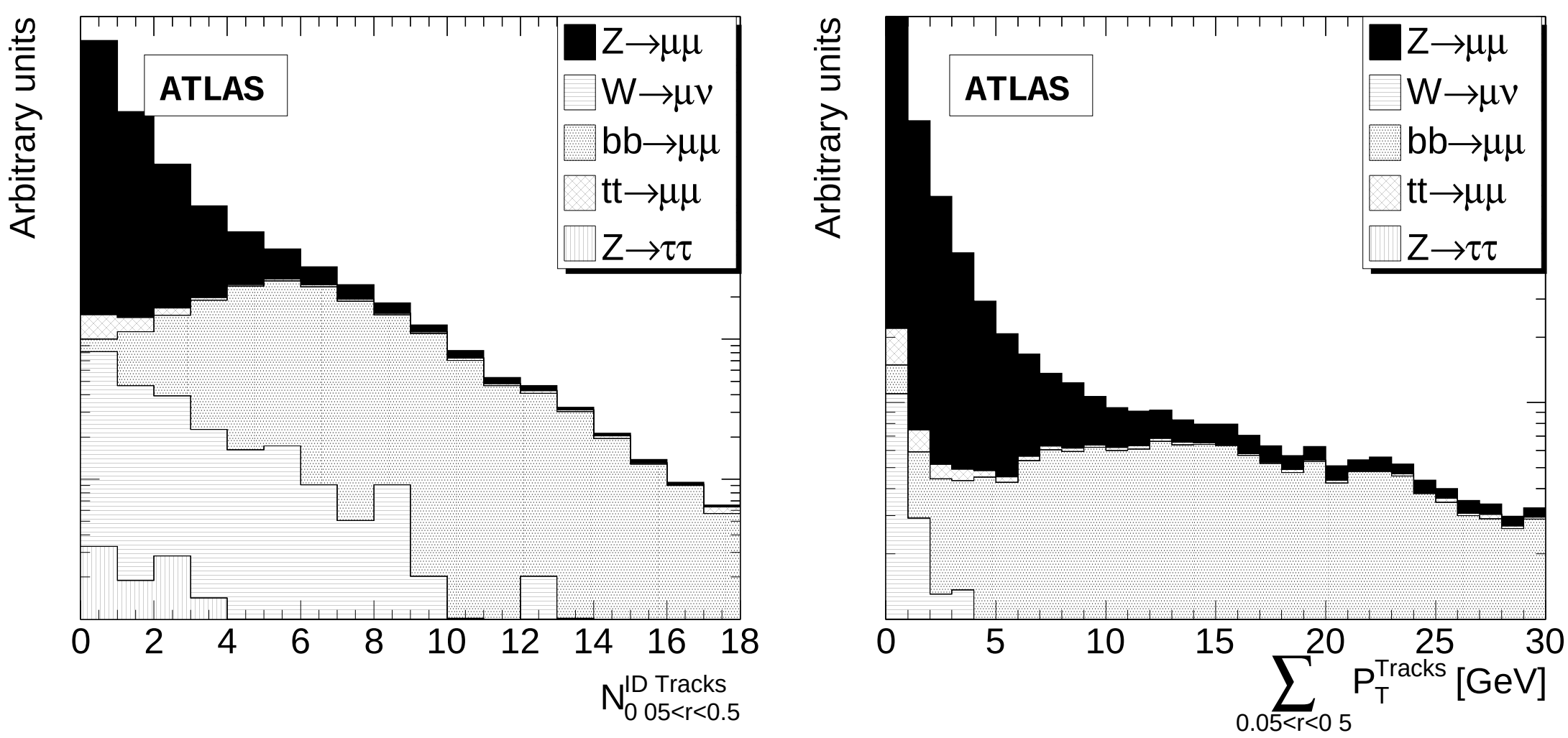


Figure 8: Distributions of the muon isolation variables for signal and backgrounds after all selection cuts. Left : track multiplicity within $\Delta R = 0.5$ around the muon. Right : total transverse momentum of these tracks.

relies on the muon spectrometer only. The reconstructed muon tracks should satisfy $|\eta| < 2.5$ and $p_T > 20$ GeV. The reconstructed charges must be opposite, and the invariant mass of the muon pair is required to fulfil $|91.2$ GeV-$M_{\mu\mu}| <$20 GeV.

Muons in jet events tend to be produced within a decay cascade of further particles, and should therefore not appear isolated in the detector, in contrast to the leptonic decays of $Z$ and $W$ bosons. To quantify the isolation of the muons, the number of Inner Detector tracks within a cone around the candidate muon, as well as the total transverse momentum of these tracks are used. The cone size is $\Delta R = 0.5$, and the muon track itself is excluded from the calculation.

The distributions of the isolation variables for signal and background processes normalised to their cross sections is shown in Fig. 8, after the above-mentioned cuts.

The isolation and $p_T$ cuts are chosen to minimise the statistical uncertainty on the cross-section measurement. The expected number of events after each cut are shown in Table 7. The chosen cuts select about 70% of the $Z \to \mu\mu$ events with muons in the detector acceptance. The residual background fraction of this selection is $0.004 \pm 0.001(stat)$. The corresponding invariant mass distribution is shown in Fig. 9.

**Background uncertainty.** In this channel, the dominant background originates from $t\bar{t}$ events. Besides a theoretical uncertainty of about 15% on the cross-section, an uncertainty of 10% is assumed on the rejection obtained from the isolation cuts. This leads to a total uncertainty of about 20% on the $t\bar{t}$ background rate.

The jet background (mostly muons from b-hadron decays) is expected to be smaller, but is theoreti-

| Selection | $Z \to \mu\mu$ | $bb \to \mu\mu X$ | $W \to \mu\nu$ | $Z \to \tau\tau$ | $t\bar{t}$ |
|---|---|---|---|---|---|
| Trigger | $3.76 \pm 0.01$ | $10.08 \pm 0.04$ | $36.7 \pm 0.1$ | $0.09 \pm 0.01$ | $0.69 \pm 0.01$ |
| 2 muons + opp. charge | $3.33 \pm 0.01$ | $3.00 \pm 0.04$ | $1.14 \pm 0.02$ | $0.04 \pm 0.01$ | $0.35 \pm 0.01$ |
| $M_{\mu\mu}$ cut | $3.04 \pm 0.01$ | $0.26 \pm 0.01$ | $0.04 \pm 0.01$ | $(14 \pm 4) \times 10^{-4}$ | $0.02 \pm 0.01$ |
| $p_T$ cut | $2.76 \pm 0.01$ | $0.125 \pm 0.001$ | $0.004 \pm 0.001$ | $(11 \pm 4) \times 10^{-4}$ | $(134 \pm 8) \times 10^{-4}$ |
| Isolation | $2.56 \pm 0.01$ | $(18 \pm 5) \times 10^{-4}$ | $(9 \pm 5) \times 10^{-4}$ | $(11 \pm 4) \times 10^{-4}$ | $(66 \pm 4) \times 10^{-4}$ |

Table 7: Number of expected signal and background events ($\times 10^4$) in the $Z \to \mu\mu$ channel, for an integrated luminosity of 50 pb$^{-1}$. The quoted uncertainties refer to the finite Monte-Carlo statistics.

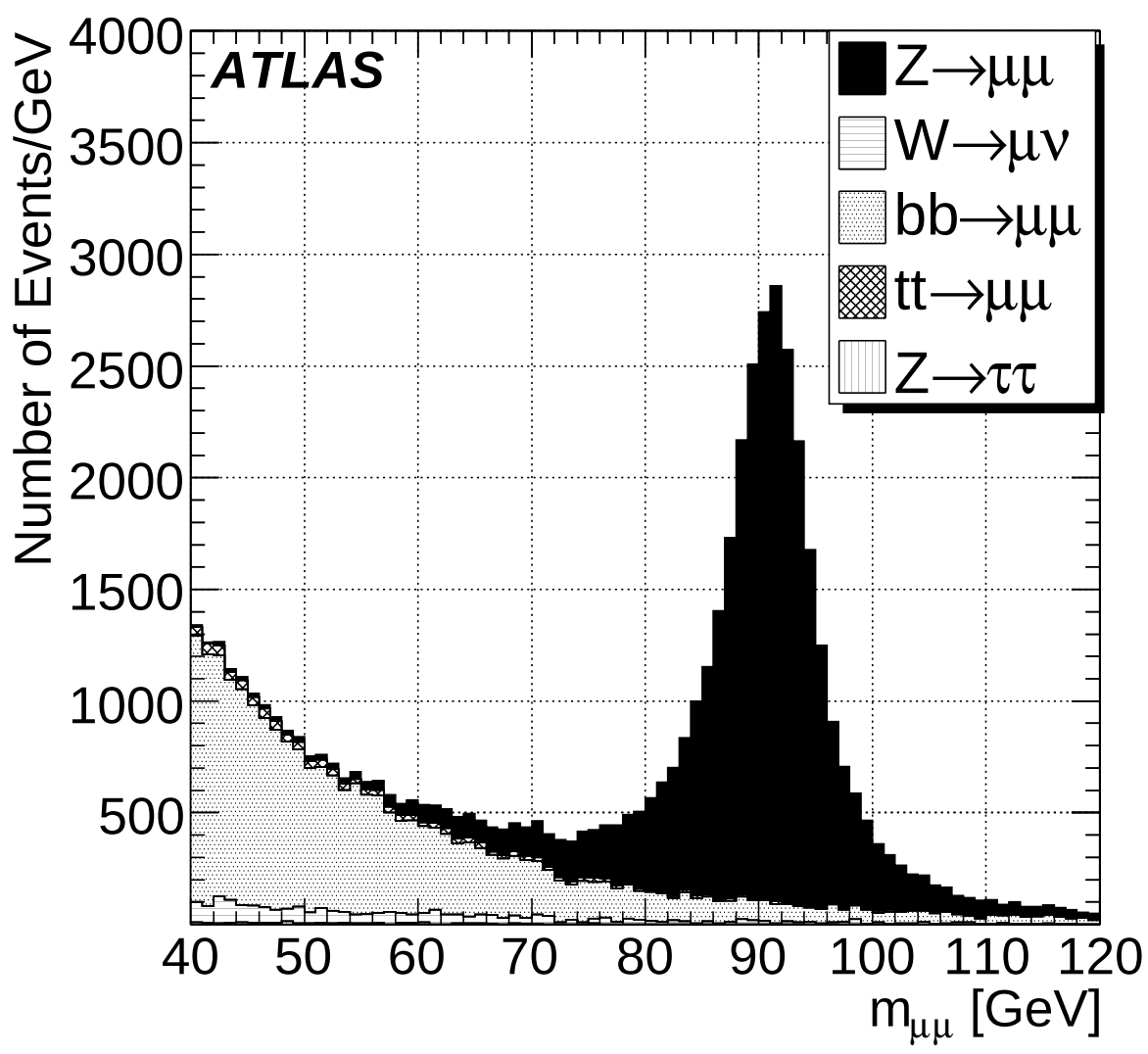


Figure 9: Di-muon invariant mass distribution in the $Z \to \mu\mu$ channel, for signal and background, for 50 pb$^{-1}$, after all cuts, except the isolation and $M_{\mu\mu}$ cuts.

cally not well known. An uncertainty of 100% is assumed on this background component.
The other backgrounds are smaller, theoretically well known in comparison to the above, and contribute negligibly to the overall background uncertainty.

# 5 Common systematic uncertainties

## 5.1 Trigger and reconstruction efficiency

As has been seen in Sections 3 and 4, the selection of leptonic $Z$ boson decays provides clean signals with low backgrounds. This allows determination of the lepton trigger [13, 14] and reconstruction efficiencies [10, 11] using the well-known tag-and-probe method, which is briefly outlined below.

Selecting $Z \to ee$ and $Z \to \mu\mu$ events as in Sections 3 and 4, *i.e.* requiring a single lepton trigger and two reconstructed leptons, the efficiency of a given trigger item is defined as the fraction of the selected events where the second reconstructed lepton passes this trigger item.

The off-line reconstruction efficiency can be determined in a similar way. Requiring one reconstructed lepton satisfying tight identification criteria, and requiring a second isolated, high-$p_T$ object such that the invariant mass of the pair is close to the $Z$ boson mass, provides a sufficiently pure $Z \to ll$ sample; the efficiency of a given identification criterion is then defined as the fraction of events where the second object is indeed identified. Conversely, the efficiency of the isolation cuts can be determined by requiring the second object to be identified, and counting the fraction of events where the isolation cut is passed.

The above methods are exact in the limit where backgrounds vanish. For tight trigger and off-line cuts, this is the case in practice : the background magnitude and uncertainty have a negligible impact on the efficiency determination. Backgrounds are larger when assessing looser identification and isolation cuts, and lower trigger thresholds. In this case, interpreting the observed dilepton mass spectrum as a sum of signal and background contributions (described by the convolution of a Breit-Wigner resonance and a Gaussian resolution function, and by an exponential or polynomial function, respectively) allows to extract the background fraction and correct the computation accordingly. This procedure was performed and shown to provide efficiency estimates that are unbiased within the statistical precision expected for $\mathscr{L} = 50$ pb$^{-1}$.

Figures 10 and 11 illustrate this discussion, on the examples of the `e20` trigger item and Medium electron identification cut, and of the `mu20` trigger item and combined muon reconstruction. For the selections used in the analyses of Section 3 and 4, the overall efficiency can be reconstructed with a precision of $\delta\varepsilon/\varepsilon = 0.02$ for electrons and muons.

The overall reconstruction efficiencies in Fig. 10, 11 reflect the performance of the reconstruction software at the time of writing this note. Improved performance and results are presented in [10, 11, 13, 14].

## 5.2 Theoretical systematic uncertainties

This section presents comparisons on the acceptance for $W \to l\nu$ and $Z \to ll$ events, as obtained from the `Pythia`, `Herwig` and `MC@NLO`. The purpose of this study is to determine the contribution of the

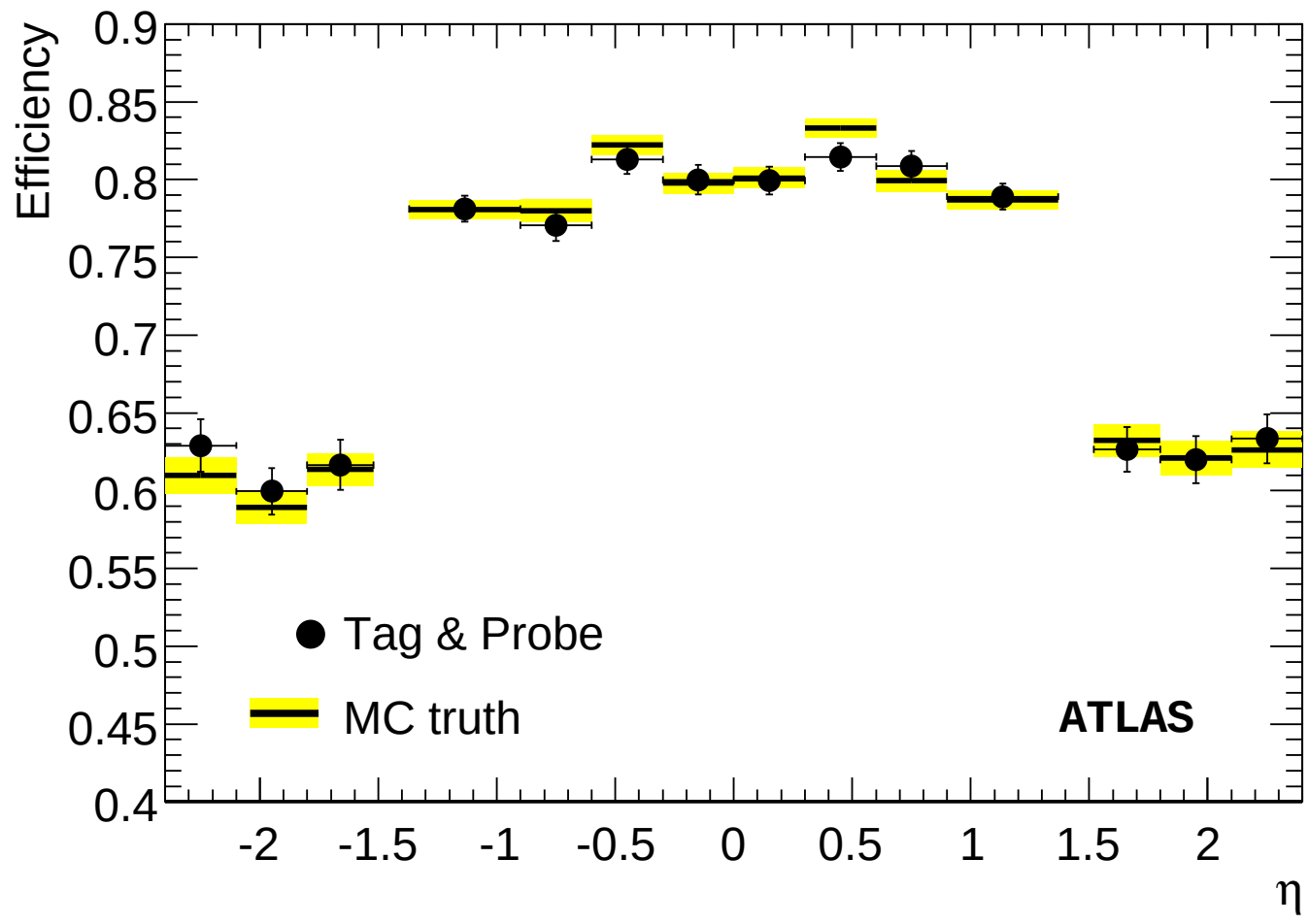


Figure 10: Electron detection efficiency vs. $\eta$, as measured from the tag-and-probe method and compared to the truth, for 50 $pb^{-1}$. The product of the `e20` trigger efficiency, and Medium electron identification efficiency is represented.

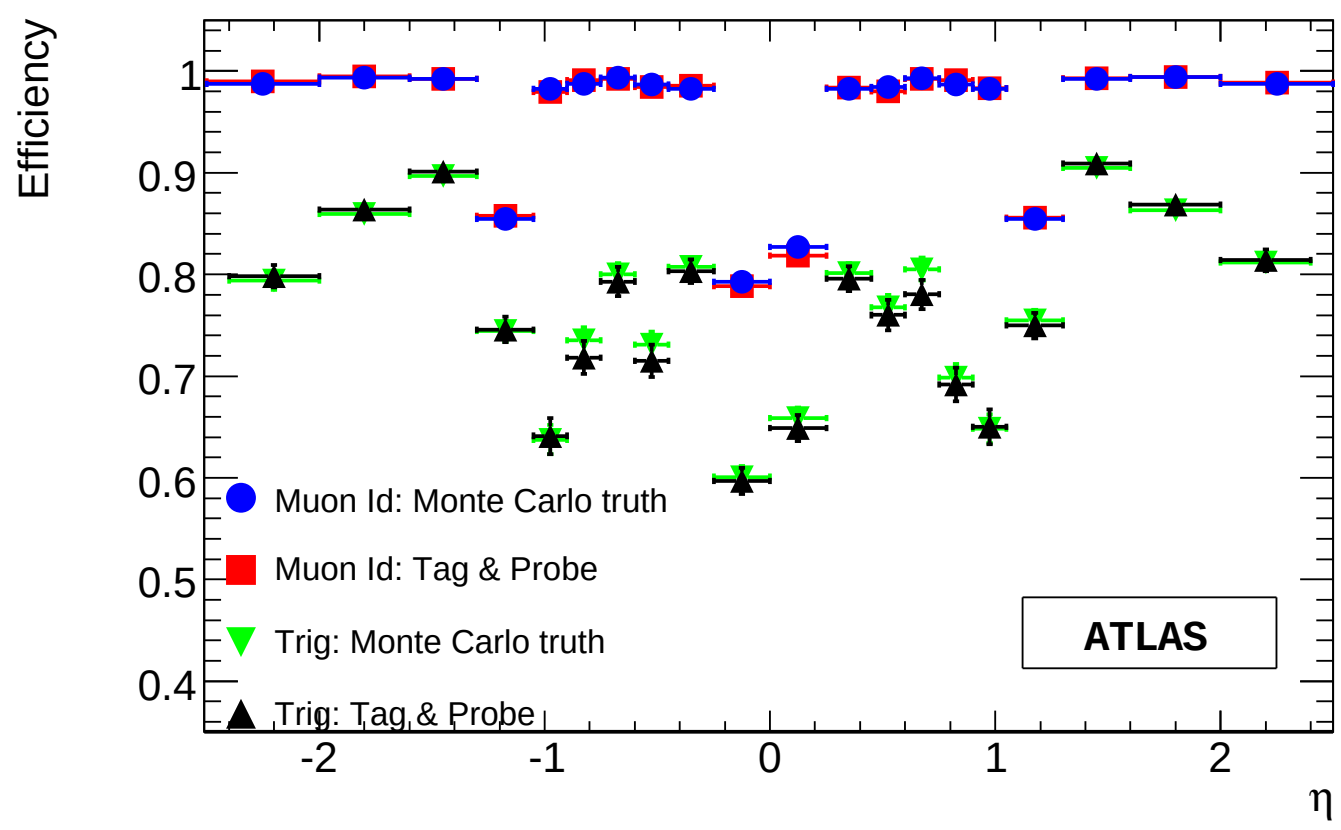


Figure 11: Muon detection efficiency vs. $\eta$, as measured from the tag-and-probe method and compared to the truth, for 50 $pb^{-1}$. The `mu20` trigger efficiency, and the combined muon reconstruction efficiency are represented.

uncertainties on the acceptance to the overall systematic uncertainty on the cross-section.

The kinematic cuts described in Sections 3 and 4 are applied to the generator-level particles for each of the above generators. For $W$ events, the acceptance varies by 2.5% from one program to the other. For $Z$ events the observed variation is of order 3.2%. The sources which could explain the observed differences are the Initial State Radiation (ISR), the intrinsic $k_T$ for the incoming partons, the Underlying Event (UE), final state photon radiation, Parton Density Functions (PDFs) and matrix element corrections applied to the parton shower (ME).

To quantify the impact of the individual sources, samples are generated with ISR, $k_T$, UE and ME all switched off. The impact of each effect is then studied by switching on this effect individually. For the sake of clarity, the discussion is given explicitly for $W \to e\nu$ production only; at the end of the section results are given for both $W$ and $Z$ production.

The effects of electroweak corrections on the acceptance have been studied using PHOTOS. By running alternatively with and without PHOTOS, one obtains an effect of 1.8% for $W$ events.
Switching on ISR, or changing the intrinsic $k_T$ of the incoming partons has an important impact on the lepton $\eta$ and $p_T$ distributions. Specifically, ISR introduces a difference of 10.2% on the acceptance for $W$ events. Similarly, turning on and off the $k_T$, ME and UE, the following differences on the $W$ acceptance are obtained: 1.9% for the intrinsic $k_T$, 1.0% for the UE and no effect for the ME.

For these sources, the systematic uncertainty is estimated as 20% of the above numbers, which amounts to assuming that the models describing the above are correct within 20%. One thus obtains the following uncertainties for $W \to e\nu$: 2.0% (ISR), 0.4% (kT) and 0.2% (UE). In the case of PHOTOS, one has an uncertainty of 0.3%.

The PDFs are an important source of differences in the acceptances. The uncertainty is determined using the CTEQ6.5 PDF uncertainty sets. An uncertainty of 0.9% is found.

To simplify the procedure, we assume no correlations between the different sources and calculate the total uncertainty from the quadratic sum of all numbers, finding $\delta A/A = 2.3\%$ for $W$ events. Repeating the same exercise with $Z$ events gives a very comparable systematic uncertainty. While these uncertainties will be significantly reduced with the analysis of the LHC data, we assume these figures hold for our initial cross-section measurements.

# 6 Total cross-section results

Cross-section results for $\mathscr{L} = 50$ pb$^{-1}$ are presented first. At the end of the section, the performance is extrapolated to higher luminosity.

## 6.1 Results for $\mathscr{L} = 50$ pb$^{-1}$

We gather below the results of the analyses performed in Sections 3 and 4, and of the discussion of systematic uncertainties of Section 5. Table 8 contains our estimations of statistical and systematic uncertainties, the cross-section values and their uncertainties computed according to Equations 2 and 3.

| Process | $N(\times 10^4)$ | $B(\times 10^4)$ | $A\times\varepsilon$ | $\delta A/A$ | $\delta\varepsilon/\varepsilon$ | $\sigma$ (pb) |
|---|---|---|---|---|---|---|
| $W\to e\nu$ | $22.67\pm 0.04$ | $0.61\pm 0.92$ | 0.215 | 0.023 | 0.02 | $20520\pm 40\pm 1060$ |
| $W\to \mu\nu$ | $30.04\pm 0.05$ | $2.01\pm 0.12$ | 0.273 | 0.023 | 0.02 | $20530\pm 40\pm 630$ |
| $Z\to ee$ | $2.71\pm 0.02$ | $0.23\pm 0.04$ | 0.246 | 0.023 | 0.03 | $2016\pm 16\pm 83$ |
| $Z\to \mu\mu$ | $2.57\pm 0.02$ | $0.010\pm 0.002$ | 0.254 | 0.023 | 0.03 | $2016\pm 16\pm 76$ |

Table 8: Measured cross-sections, their uncertainties and overall selection efficiency $A\times\varepsilon$, for an integrated luminosity of 50 pb$^{-1}$. The uncertainty on $N$ is statistical, the other sources are systematic. The quoted cross-section uncertainties include the mentioned statistical and systematic contributions but not an overall luminosity uncertainty.

As can be seen from Table 8, the results are dominated by the systematic error, even for $\mathscr{L} = 50$ pb$^{-1}$. The luminosity uncertainty is common to all cross-sections, and vanishes in cross-section ratios, e.g $\sigma_W/\sigma_Z$. In the $W$ channels, the systematic uncertainty is dominated by the background uncertainty. This can be expected given the important fraction of jet events. This background could be further reduced, notably by requiring the absence of jets, but this would jeopardize the inclusive nature of the cross-section measurement. The efficiency and acceptance uncertainties give a slightly smaller contribution.

The $Z$ channels benefit from smaller backgrounds, due to the presence of two decay leptons. For the same reason, the efficiency uncertainty is also larger than in the $W$ channels. Given the smaller acceptance uncertainty, the efficiency uncertainty is the largest source of uncertainty.

## 6.2 Prospects for $\mathscr{L} = 1$ fb$^{-1}$

For higher integrated luminosity, the statistical uncertainty on the counting ($N$) becomes negligible, and the efficiency uncertainty, which is determined from measurement and also of statistical nature, strongly decreases.

With increased luminosity, a number of modifications will have to be applied to the analyses. Most prominently, in the electron channels, the single electron trigger threshold will be increased to 22 GeV, and the Tight electron identification is expected to be used. In $Z\to\mu\mu$ channel, spectrometer muons are replaced by combined muons, and the muon isolation cuts are refined by exploiting calorimetric information in addition to the track-based isolation used in the low-luminosity analysis. The $W\to\mu\nu$ analysis is unchanged.

Table 9 summarizes the expected signal yields and the cross-section determination in this case. On this timescale, $\mathscr{L}$ might be measured with improved precision, exploiting elastic proton scattering at very small angles [2]. Compared to the low-luminosity analysis, the systematic uncertainties from backgrounds and efficiency are expected to scale with statistics. Without further input, the acceptance uncertainty does not decrease and dominates the result.

| Process | $N(\times 10^5)$ | $B(\times 10^5)$ | $A\times\varepsilon$ | $\delta A/A$ | $\delta\varepsilon/\varepsilon$ | $\sigma$ (pb) |
|---|---|---|---|---|---|---|
| $W\to e\nu$ | $45.34\pm0.02$ | $1.22\pm0.41$ | 0.215 | 0.023 | 0.004 | $20520\pm\ 9\pm516$ |
| $W\to\mu\nu$ | $60.08\pm0.02$ | $4.02\pm0.05$ | 0.273 | 0.023 | 0.004 | $20535\pm\ 7\pm480$ |
| $Z\to ee$ | $5.42\pm0.01$ | $0.46\pm0.02$ | 0.246 | 0.023 | 0.007 | $2016\pm\ 4\pm\ 49$ |
| $Z\to\mu\mu$ | $5.14\pm0.01$ | $0.02\pm0.001$ | 0.254 | 0.023 | 0.007 | $2016\pm\ 4\pm\ 49$ |

Table 9: Measured cross-sections, their uncertainties and overall selection efficiency $A\times\varepsilon$, for an integrated luminosity of 1 fb$^{-1}$. The uncertainty on $N$ is statistical, the other sources are systematic. The quoted cross-section uncertainties include the mentioned statistical and systematic contributions but not an overall luminosity uncertainty.

# 7 Differential cross-sections

As it is clear from the previous sections, total cross-section measurements are dominated by the systematic uncertainty even for modest integrated luminosity. The main cross-section uncertainty is related to the acceptance uncertainty, which in turn comes from our limited knowledge of the underlying physics (notably non-perturbative mechanisms and PDFs). It is therefore important to measure the distributions, which will help to constrain these uncertainties. Three examples are given below, namely the measurement of the Drell-Yan invariant mass spectrum at low mass, and the rapidity and transverse momentum distributions for $Z$ events.

Compared to the inclusive analyses, the differential measurements require larger statistics. For this reason the differential distributions shown in this section refer to an integrated luminosity of 200 pb$^{-1}$, which corresponds to the available statistics of the Monte Carlo signal samples.

## 7.1 Low-mass Drell-Yan production

**Event selection.** This study relies on a low threshold single electron trigger, `e10`, possible at low luminosity. At the reconstruction level, exactly two oppositely charged electrons are required, each satisfying the tight identification criteria, and with $p_T > 10$ GeV. Offline electron reconstruction is limited to $|\eta| < 2.5$. After these preselections, the reconstructed missing transverse energy in the event, $E_T^{miss}$, should be smaller than 30 GeV. Finally, the di-electron invariant mass, $m_{ee}$, is required to be in the range $20 < m_{ee} < 60$ GeV. Table 10 shows the impact of these cuts on signal and background. Only the main backgrounds except the one arising from QCD dijets are considered at this level; the dijet background is discussed separately.

The kinematical acceptance, given by the final state requirement of two electrons satisfying $p_T >$ 10 GeV and $|\eta| < 2.5$, is 17% in the range $20 < m_{ee} < 60$ GeV. For $40 < m_{ee} < 60$ GeV, the acceptance reaches 34%. The `e10` trigger channel provides an efficiency of about 89% for signal events within the kinematical acceptance. The trigger efficiency is about 74% for di-electron pair masses of $m_{ee} = 20$ GeV, and reaches 97% at $m_{ee} = 60$ GeV. The offline electron selection efficiency is 36% [10], and the further kinematic selections have an efficiency of 83%.

**Dijet background estimation.** The background contributions listed in Table 10 are estimated from simulation, by counting the events that pass the selection criteria described above. This method can not be applied to the dijet background which, due to very high rejection factors, can not be simulated

| cut | $\gamma^* \rightarrow ee$ | $\tau\tau$ | $t\bar{t}$ | di-boson |
|---|---|---|---|---|
| Preselections | 2632±12 | 48±3 | 218±3 | 342±3 |
| $E_T^{miss} < 30$ GeV | 2604±12 | 38±3 | 28±1 | 164±2 |
| $20 < m_{ee} < 60$ GeV | 2189±11 | 30±3 | 7±1 | 7±1 |

Table 10: Signal and background event rates, following the selections described in the text. For all samples, the normalization corresponds to 50 pb$^{-1}$. The $\gamma^*/Z \rightarrow ee$ signal with $m_{ee} > 60$ GeV is excluded from these numbers, at all levels of the selection. The dominant dijet background is not displayed here, and discussed separately below.

in sufficient amounts.

An alternative estimation, based on single electron rejection factors, is used. This method relies on the probability for an event to display a single fake electron. On the inclusive jet sample, this probability is found to be $(1.5 \pm 0.1) \times 10^{-4}$. Assuming no correlations, the probability to find a fake di-electron pair with opposite charge is then estimated as one-half of the square of the single fake probability. The invariant mass distribution of fake electron pairs is estimated using Monte Carlo information.

The invariant mass distribution of the signal and backgrounds are displayed in Figure 12 (left). The estimated background represents about one third of the selected sample. It is dominantly composed of inclusive jets (96%), the other backgrounds being negligible.

Two thirds of the dijet events that contain one tight electron candidate are true electrons from heavy quark decays, while the remainder are due to light hadrons mis-identified as electrons and to photon conversions. It may be possible to reduce the jet background further by optimizing the selection criteria for these low $E_T$ electrons. Additional event variables, like hadronic activity estimators, could also be used to reduce or constrain the size of the dijet background. However, the large fraction of electrons from heavy quark decays indicates that the uncertainty on the background composition related to different rejection factors for heavy flavour and light jets under the the Drell-Yan signal is large and requires further study.

**Measurement of the cross-section.** The raw measured differential cross-section may be corrected bin by bin in di-electron mass, as follows:

$$\left(\frac{d\sigma}{dM}\right)_i = \frac{N_i - B_i}{\varepsilon_i A_i \Delta m_i \mathscr{L}} \tag{4}$$

where $N_i$ is the number of signal events in mass bin $i$, $B_i$ is the expected number of background events, $A_i$ is the kinematical and angular acceptance efficiency, $\varepsilon_i$ is the overall efficiency accounting for trigger, identification and further selections. $\Delta m_i$ is the width of bin $i$ and $\mathscr{L}$ represents the integrated luminosity.

In practice, the corrections can be considered to be of two types: those correcting for experimental effects (backgrounds and efficiencies), and those correcting for the kinematic acceptance of the electron $p_T$ and $\eta$ requirements. Figure 12, right, illustrates the two corrections. The agreement between the corrected distribution and the expected distribution from theory provides a technical consistency

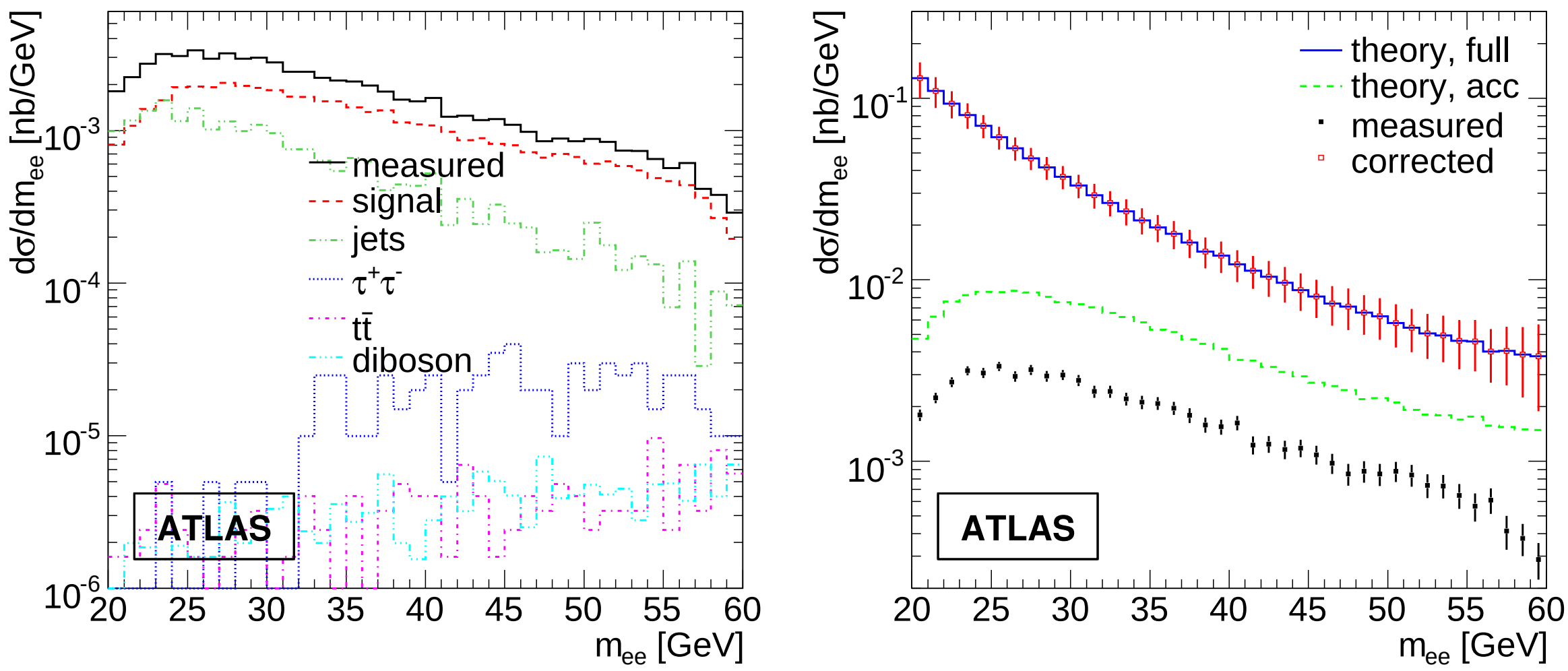


Figure 12: Left: Mass distributions of the measured raw events (solid line), Drell-Yan signal events (dashed line) and background contributions arising from inclusive jets, $\tau\tau$, $t\bar{t}$, di-boson in decreasing numerical importance. Right: invariant mass distributions of the selected sample before any corrections (rectangles with error bars), the theoretical signal within the fiducial acceptance (dashed line), the corrected Drell-Yan signal sample (unshaded rectangles with error bars) and the complete theoretical distribution (solid line).

check of the method.

The integrated cross-section can be calculated by integrating the corrected histogram; the statistical error on the cross-section is estimated by adding the error from each bin in quadrature. The total Drell-Yan cross section in the electron channel, for $20 < m_{ee} < 60$ GeV is $\sigma_{\rm DY} = 1.07$ nb. The corresponding statistical uncertainty is 4% for $\mathscr{L} = 50\ \text{pb}^{-1}$and 1% for $\mathscr{L} = 1\ \text{fb}^{-1}$. This can be compared with the PDF contribution to the cross-section uncertainty, estimated to be 6.9% using `CTEQ6.1M` [7].

The statistical sensitivity provides a natural target for the different contributions to the systematic uncertainty: the knowledge of the background level, and of the efficiency and acceptance corrections should match the statistical sensitivity. The efficiency can be determined with sufficient precision using the methods discussed in Section 5. The acceptance corrections are mostly affected by the transverse momentum distribution of the Drell-Yan pairs in this mass range; this has large theoretical uncertainties, and will have to be measured from data. The invariant mass resolution, about 1 GeV, is found to have no significant effect on the Drell-Yan spectrum. As discussed above, the most serious challenge is posed by the understanding of the jet background, where the current uncertainties are much larger than the expected statistical precision. Further study is needed to assess the precision of techniques to constrain this background directly from data; this is beyond the scope of the current work.

### 7.2 Z differential cross-section : bin by bin correction method

| Category $i$ | Definition | $n_i$ |
|---|---|---|
| 1 | All events | 398 750 |
| 2 | Fiducial and kinematics (generation) | 172 544 |
| 3 | Trigger and off-line (fiducial, kinematics and ID) | 49 754 |
| 4 | Intersection of categories 2 and 3 | 48 436 |

Table 11: Event categories used for the extraction of detector smearing corrections, geometric acceptance and event selection efficiency in the electron channel.

**Electron channel.** In the electron channel, only events that pass the `2e12i` trigger condition are considered. Furthermore, it is required that they contain exactly two, oppositely charged electrons, each of them satisfying $|\eta| < 2.5$ and $P_T > 20$ GeV. Both electrons are required to pass the Tight electron identification criteria.

The background is dominated by hadrons misidentified as electrons in inclusive jet events. Taking into account that with the Tight identification criteria the expected rate of hadrons misidentified as electrons is very low (see [10]), the background to the di-electron signal is neglected in the following.

After all selections, the following event categories are defined. In the following, $n_1$ denotes the total sample size; $n_2$ is the number of events having two generator-level electrons satisfying $|\eta| < 2.5$, $p_T > 20$ GeV, and 75 GeV $< M_{ee} <$ 105 GeV. The number of events satisfying these conditions at the reconstruction level is noted $n_3$; finally, $n_4$ counts the events passing these criteria on both generation and reconstruction levels. Table 11 summarises these definitions and contains values for the $n_i$, directly counted from the simulated signal sample.

**Muon channel.** In the muon channel, only events that pass the `mu20` trigger condition are considered. The events should further contain exactly two oppositely charged muons, each of them satisfying $|\eta| < 2.5$. Both muons should be reconstructed in the Inner Detector and in the Muon Spectrometer. The most energetic muon should satisfy $p_T > 20$ GeV; the second one should have $p_T > 15$ GeV. The muon pair invariant mass should lie between 76 and 106 GeV.

Contamination from $W \to \mu\nu$ and $t\bar{t}$ events effectively disappears after the selection process, but $b\bar{b} \to \mu\mu$ contamination is still about 3.5% of the signal. To minimise the $b\bar{b}$ background, two isolation quantities are studied. The first one is the number of tracks in a cone of size $\Delta R$ =0.45 around the muon track ; the second one is the total calorimeter transverse energy in the same cone. The distributions of these two quantities for the four samples, corresponding to 40 pb$^{-1}$of integrated luminosity, are shown in Fig. 13. A muon track is accepted if the first isolation variable is less than six and the second isolation variable is less than 20 GeV. These cuts are applied to both muons in the event.

The isolation cut efficiency for the signal sample is larger than 98%, and the residual contamination is less than 0.5%. The sample is pure enough at this point that we can neglect the background contamination in the differential cross-section plots.

As in the electron channel, four event categories are defined to allow the computation of the differen-

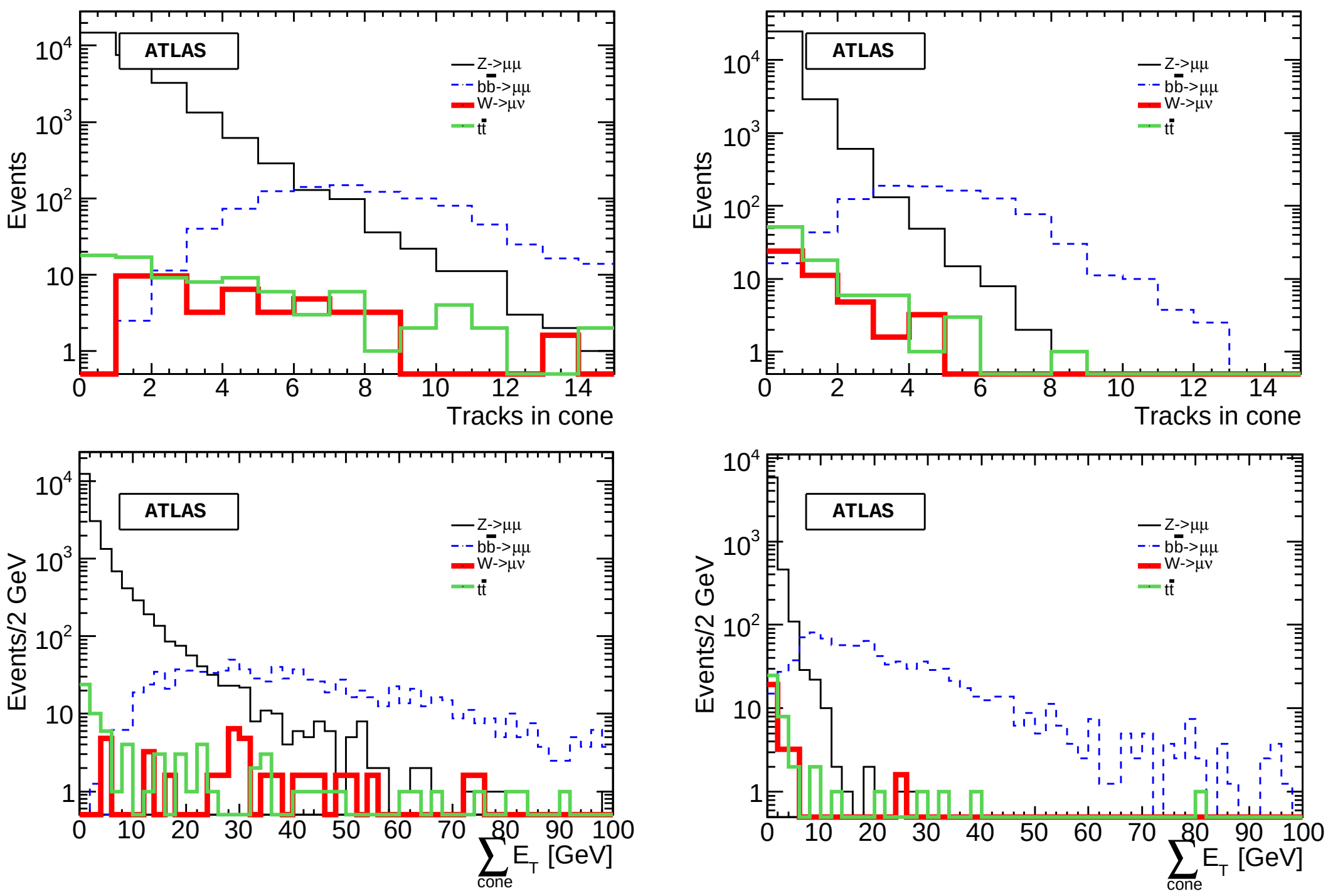


Figure 13: Number of tracks (top) and calorimeter $E_T$ in a cone R = 0.45 (bottom). Distributions for the muon with the higher value of the quantity is shown on the left, and distributions for the muon with the lower value of the quantity on the right. Black line: $Z \to \mu\mu$, red (or dark grey line): $W \to \mu\nu$, dashed line: $b\bar{b} \to \mu\mu$, green (or light grey line): $t\bar{t}$.

| Category $i$ | Definition | $n_i$ |
|---|---|---|
| 1 | All events | 445650 |
| 2 | Fiducial and kinematics (generation) | 234610 |
| 3 | Trigger and off-line (fiducial, kinematics and ID) | 181652 |
| 4 | Intersection of categories 2 and 3 | 180260 |

Table 12: Event categories used for the extraction of detector smearing corrections, geometric acceptance and event selection efficiency in the muon channel.

tial cross-section. The categories and their sizes $n_i$ are given in Table 12.

**Extraction of $d\sigma_Z/dp_T dy$** The $Z$ boson phase space is sliced in rapidity and transverse momentum regions, or bins. In each region, labeled $\alpha$, the differential cross-section is obtained from the raw event count using the usual expression:

$$\sigma_\alpha = \frac{S_\alpha}{\mathscr{L}} \frac{d_\alpha - b_\alpha}{\varepsilon_\alpha A_\alpha}, \tag{5}$$

where $S_\alpha$, $\varepsilon_\alpha$ and $A_\alpha$ respectively represent the detector smearing correction (correcting for event migration to and from bin $\alpha$, due to resolution effects), overall event selection efficiency and geometric acceptance in region $\alpha$; $d_\alpha$ is the observed event count and $b_\alpha$ the estimated background in this region, and $\mathscr{L}$ is the integrated luminosity. In terms of the definitions in Tables 11 and 12, we have:

$$S_\alpha = \frac{n_{3,\alpha}}{n_{4,\alpha}}, \;\; \varepsilon_\alpha = \frac{n_{3,\alpha}}{n_{2,\alpha}}, \;\; A_\alpha = \varepsilon_{filter} \frac{n_{2,\alpha}}{n_{1,\alpha}}, \;\; d_\alpha = n_{3,\alpha}, \tag{6}$$

where the $n_{i,\alpha}$ are computed in each bin $\alpha$ of the $Z$ phase space. The acceptance values account for the generator-level filtering efficiency, as described in Section 2.2.

**Differential cross-section results** In the electron channel, the $Z$ boson phase space was divided in 50 $p_T$ bins and 9 rapidity bins. The $p_T$ bins have a width of 2 GeV, in the range $0 < p_T < 100$ GeV. The rapidity bins have a width of 0.3, in the range $0 < |y| < 2.7$. A good agreement is found between the reconstructed cross-sections and the true distributions, obtained from a statistically independent sample. Figure 14 shows the $Z$ boson differential cross-section in the electron channel in rapidity and transverse momentum bins.

In the muon channel, the $Z$ boson phase space is divided as in the electron channel. A good agreement is again found between the reconstructed cross-sections and the true distributions. Figure 15 shows the $Z$ boson differential cross-section in the dimuon channel in rapidity and transverse momentum bins.

The plots have been normalised to the total NNLO cross-section of 2015 pb times a global acceptance factor of 0.73. Since the correction factors and measurement were both extracted from the same dataset, the good agreement is a check of consistency of the method.

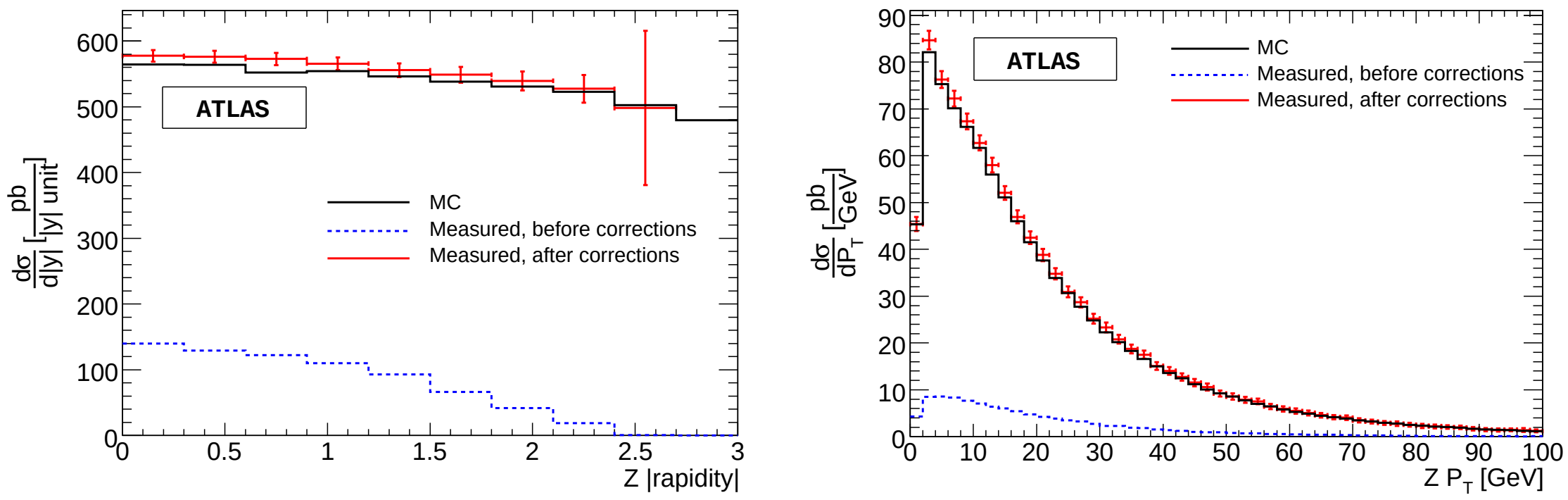


Figure 14: Left: $d\sigma_Z/dy$, integrated over $p_T$. Right: $d\sigma_Z/dp_T$, integrated over $-2.7 < y < 2.7$. Distributions obtained in the electron channel, with a precision corresponding to an integrated luminosity of 200 $pb^{-1}$. The black line histograms correspond to the generated cross-section, the dashed histograms to the measured cross-section before corrections are applied, while the crosses show the measured cross-section after all corrections have been applied.

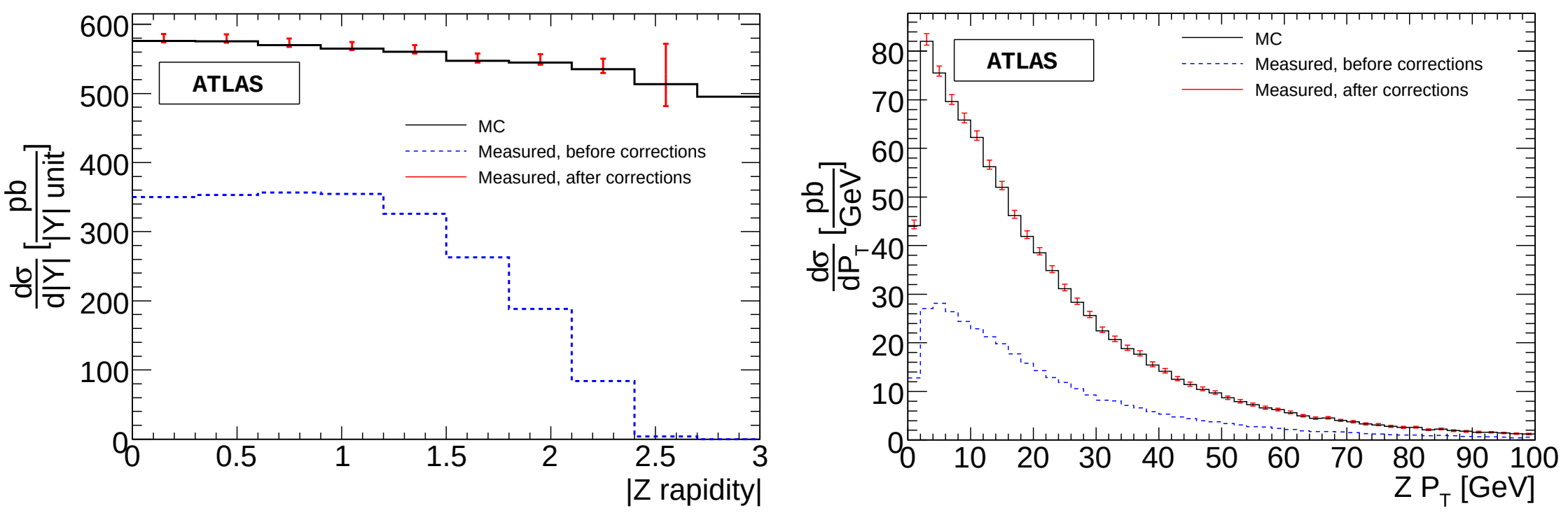


Figure 15: Left: $d\sigma_Z/dy$, integrated over $p_T$. Right: $d\sigma_Z/dp_T$, integrated over $-2.7 < y < 2.7$. Distributions obtained in the muon channel, with a precision corresponding to an integrated luminosity of 200 $pb^{-1}$. The black line histograms correspond to the generated cross-section, the dashed histograms to the measured cross-section before corrections are applied, while the crosses show the measured cross-section after all corrections have been applied. .

### 7.3 Z differential cross-section : alternative method

The method presented here attempts to fully exploit the phase space of the $Z$ boson and its decay products. Writing the cross-section in terms of the complete phase space allows to extract, in addition to the Z boson distributions, possible $p_T$, $\eta$ or $\phi$ dependencies of the lepton selection efficiency.

**Method** The events are classified in bins both for the $Z$ and for the decay leptons. We define $N_{yZ}$ bins along $y^Z$, and $N_{ptZ}$ bins along $p_T^Z$. As before, the $Z$ boson phase space intervals are labelled $\alpha$. In addition, we define $N_{Et}$ intervals in the lepton transverse energy distribution, and $N_\eta$ intervals for the leptons pseudorapidity. The lepton phase space is labelled $i$, $j$ (one index for each lepton).

For each $\alpha$, we measure $N_{ij}^\alpha$, which is the number of lepton pairs reconstructed with one lepton in bin $i$ and one lepton in bin $j$. The following relation holds, in the practical absence of background, as justified in the previous sections:

$$N_{ij}^\alpha = \varepsilon_i \varepsilon_j P_{ij}^\alpha \mathscr{L} \Delta\sigma^\alpha, \tag{7}$$

where $P_{ij}^\alpha$ is the probability, computed on Monte Carlo, that a $Z$ boson produced in bin $\alpha$ decays into two leptons in bins $i$ and $j$; $\varepsilon_i$ is the lepton reconstruction efficiency in bin $i$; $\mathscr{L}$ is the integrated luminosity, and $\Delta\sigma^\alpha$ is the Z production cross-section in bin $\alpha$.

Resolution effects, primarily on the lepton $E_T$, are accounted for as follows. When the $P_{ij}^\alpha$ histograms are filled, the lepton $E_T$ is first smeared according to its expected, $E_T$ and $\eta$ dependent resolution. The smeared quantities are then used to compute the $Z$ variables ($p_t$, $y$). In this way the above equation is unchanged and all detector effects can be incorporated in the $P_{ij}^\alpha$ factors. Writing the above for all $\alpha$, $i$, $j$ provides an over-constrained system whose unknowns are the efficiencies and cross-sections. We can then compute the $\varepsilon_i$ in each bin $\alpha$, up to a factor related to $\mathscr{L}\Delta\sigma^\alpha$.

The system can be solved analytically, using for example the singular value decomposition method, or SVD. A drawback of this method is that it is based on least squares; it is thus not valid in the case of low statistics. In particular, at low luminosity, the statistics are such that several bins contain only a few events. To avoid bias in the efficiency determination, a likelihood using Poisson probabilities is constructed and used to fit the efficiencies numerically. In order to help the fit to converge, we first solve the system using the SVD method, and we use the results as initial parameters of the fit. Since the $\varepsilon_i$ are expected not to depend on $\alpha$, we can compute their weighted average over the bins $\alpha$.

In case of low statistics, the bin sizes should be large enough to integrate a sufficient statistics in each bin. If the lepton reconstruction efficiency is not constant within each bin, the hypothesis that the efficiency does not depend on the $Z$ boson phase space might be violated. To avoid such effects, the lepton binning is chosen such that the efficiency is *a priori* constant within each bin. This results in bins with variable width, which does not affect the method.

In any given $Z$ phase space interval $\alpha$, we can write for each lepton bin $(i, j)$ the following relation:

$$\mathscr{L}\Delta\sigma^\alpha = \frac{N_{ij}^\alpha}{<\varepsilon_i><\varepsilon_j> P_{ij}^\alpha}, \tag{8}$$

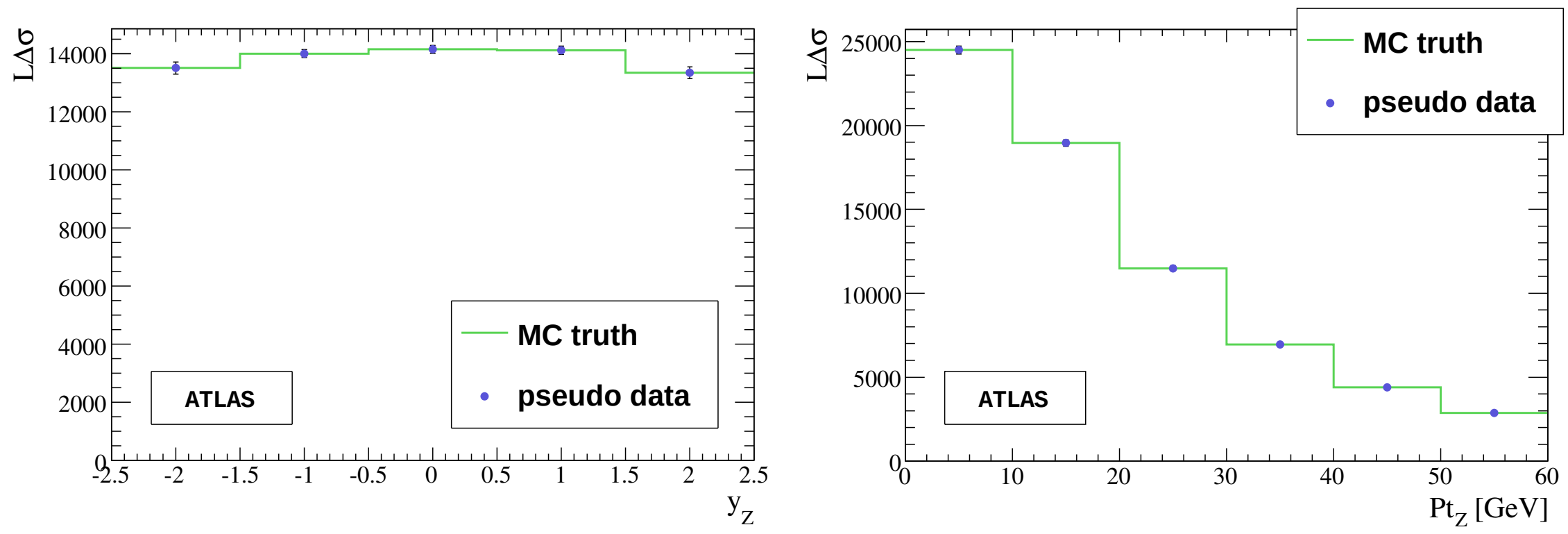


Figure 16: Left: $\mathscr{L}\Delta\sigma$ versus $y^Z$, integrated over $p_T^Z$. Right: $\mathscr{L}\Delta\sigma$ versus $p_T^Z$, integrated over $y^Z$.

where $< \varepsilon_i >$ and $< \varepsilon_j >$ are the average efficiencies computed at the previous step. Finally, $\mathscr{L}\Delta\sigma^\alpha$ is computed by averaging over all $(i, j)$.

**"Classical limit" of the method.** As discussed above, the method proposed here might need significant integrated luminosity to be applied safely. The classical method is reached by simply setting $N_{Et} = N_\eta = 1$, and accordingly computing the acceptance and efficiencies from Monte-Carlo in each $Z$ boson phase space interval $\alpha$.

Resolution effects are taken into account as before, by smearing the Monte Carlo input before determining the acceptance. Once the acceptance and efficiency are determined in each bin $\alpha$, just counting the number of events $N^\alpha$ in the bin and allows to deduce the differential cross-section using the usual cross-section expression in the absence of background:

$$\mathscr{L}\Delta\sigma^\alpha = \frac{N^\alpha}{\varepsilon^\alpha \varepsilon^\alpha A^\alpha}. \tag{9}$$

**Results.** The complete method has been tested on the $Z \to \mu\mu$ samples described in Section 2.2. Due to the limited statistics, the $Z$ phase space was mapped using $N_{ptZ}$ =10 for $0 < p_T^Z < 60$ GeV, $N_{yZ}$ =5 for $-2.5 < y^Z < 2.5$. The muon reconstruction efficiency has no $p_T$-dependence above 10 GeV; this allows to set $N_{Et}$ =1. The definition of the muon $\eta$ intervals is dictated by the detector geometry which affects the efficiency as a function of $\eta$; we set $N_\eta$ =7, with the intervals [-2.7,-1.6], [-1.6,-1.4], [-1.4,-0.1], [-0.1,0.1], [0.1,1.4], [1.4,1.6], [1.6,2.7].

The results are illustrated in Fig. 16. The Z boson rapidity and $p_T$distributions are correctly reconstructed. Measured and true distributions agree within the statistical precision, which varies from 3% in regions where the differential cross-section is high, to about 20% in the tails of the Z boson phase space ($y^Z > 1.5$). In addition, the $\eta$ dependence of the reconstruction efficiency can be measured accurately. Given the interval definition above and the size of our sample, a precision of about 2% is obtained for each point. This is competitive with the tag-and-probe determination described in Section 5, and illustrated in Fig. 17.

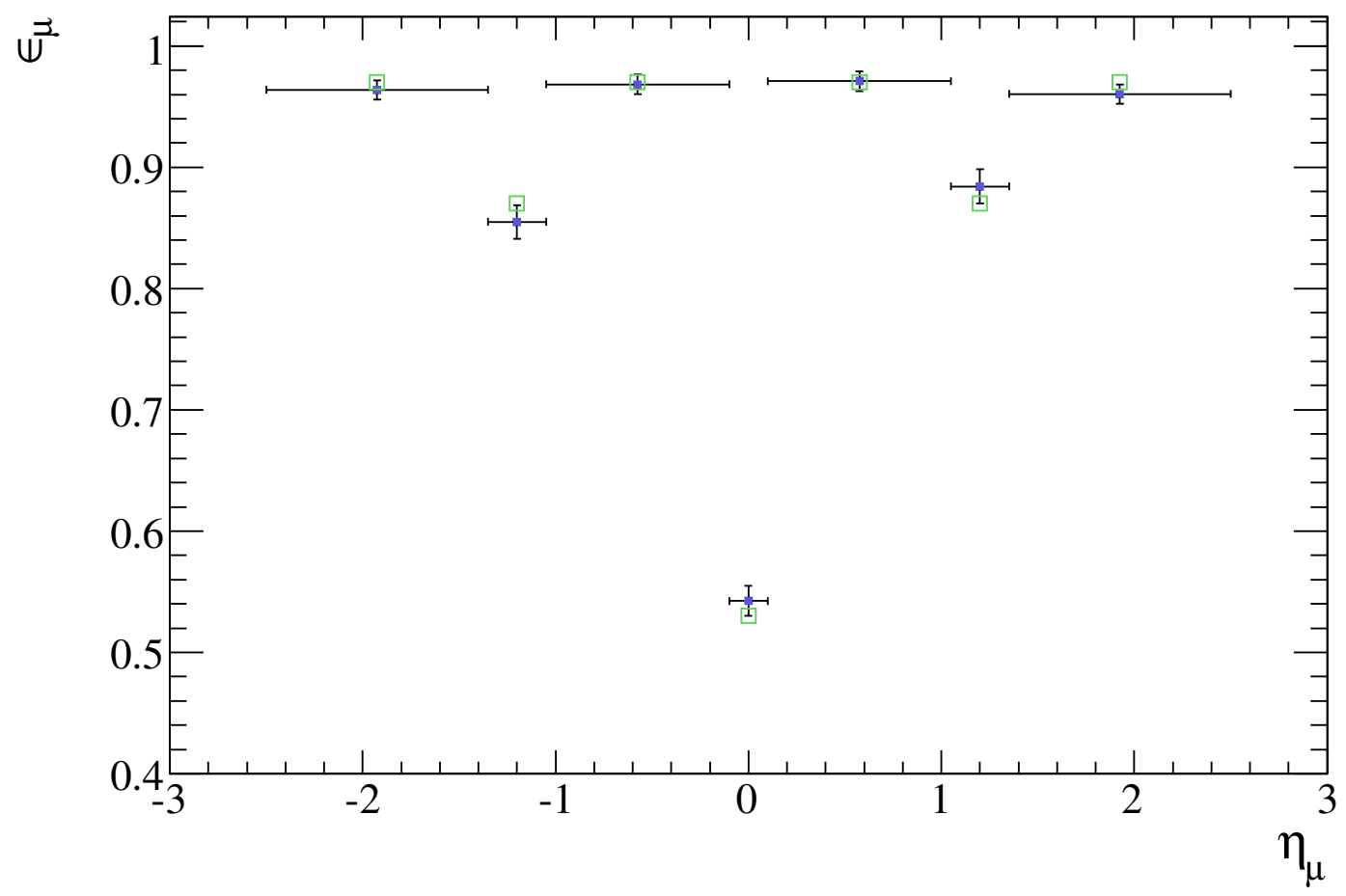


Figure 17: Muon reconstruction efficiency versus $\eta$, measured simultaneously with the differential cross-section.

For cross-checks, the classical limit of the method has been tested using the $Z \to ee$ samples. The intervals are defined as before, except $N_{yZ}$ =10 and $N_{ptZ}$ =20. The following selection criteria are applied: two reconstructed electrons of opposite charge are required in the detector acceptance ($|\eta| \geq 2.5$), with 20 $GeV \leq p_t \leq 80$ $GeV$. Both electrons should pass the Tight identification criterion. $Z$ events are selected around the mass peak (87 $GeV \leq M_Z \leq 95$ $GeV$), with $p_T^Z \leq 60$ $GeV$. The acceptance cuts on the electrons imply $y^Z \leq 2.5$.

The results of the differential cross-section $d\sigma/dy^Z$, integrated over $p_T^Z$, and $d\sigma/dp_T^Z$, integrated over $y^Z$are shown in Fig. 18. The squares represent the raw distribution, the dots represent the measurements, corrected by the acceptance and the efficiency factors. The shows the input value. The measurements are consistent with the input, which proves the consistency of the method.

## 8 Summary and perspectives

This work presents the ATLAS prospects for the measurement of $W$ and $Z$ boson cross-sections at the LHC. In the four considered channels ($W \to e\nu$, $Z \to ee$, $W \to \mu\nu$, $Z \to \mu\mu$), the analyses confirm the high purity of the samples after fairly usual selections (high-$p_T$lepton identification, isolation, and $\not{E}_T$ in the $W$ final states). The jet background is poorly predicted, and dedicated studies are needed to monitor its magnitude using real data. Data-driven methods are presented that seem to have sufficient sensitivity to keep the jet background at a level where it does not prevent a precise cross-section measurement.

With 50 pb$^{-1}$, the background and signal acceptance uncertainties contribute similarly to the measured cross-section uncertainty, at the level of 2-4% depending on the channel. The uncertainty on the integrated luminosity is not included in the present discussion. Extrapolating to 1 fb$^{-1}$, all uncertainties are expected to scale with statistics, except the acceptance uncertainty. This leads to the conclusion (see also, for example, [16, 17]) that the $W$ and $Z$ cross-sections can not be measured to a

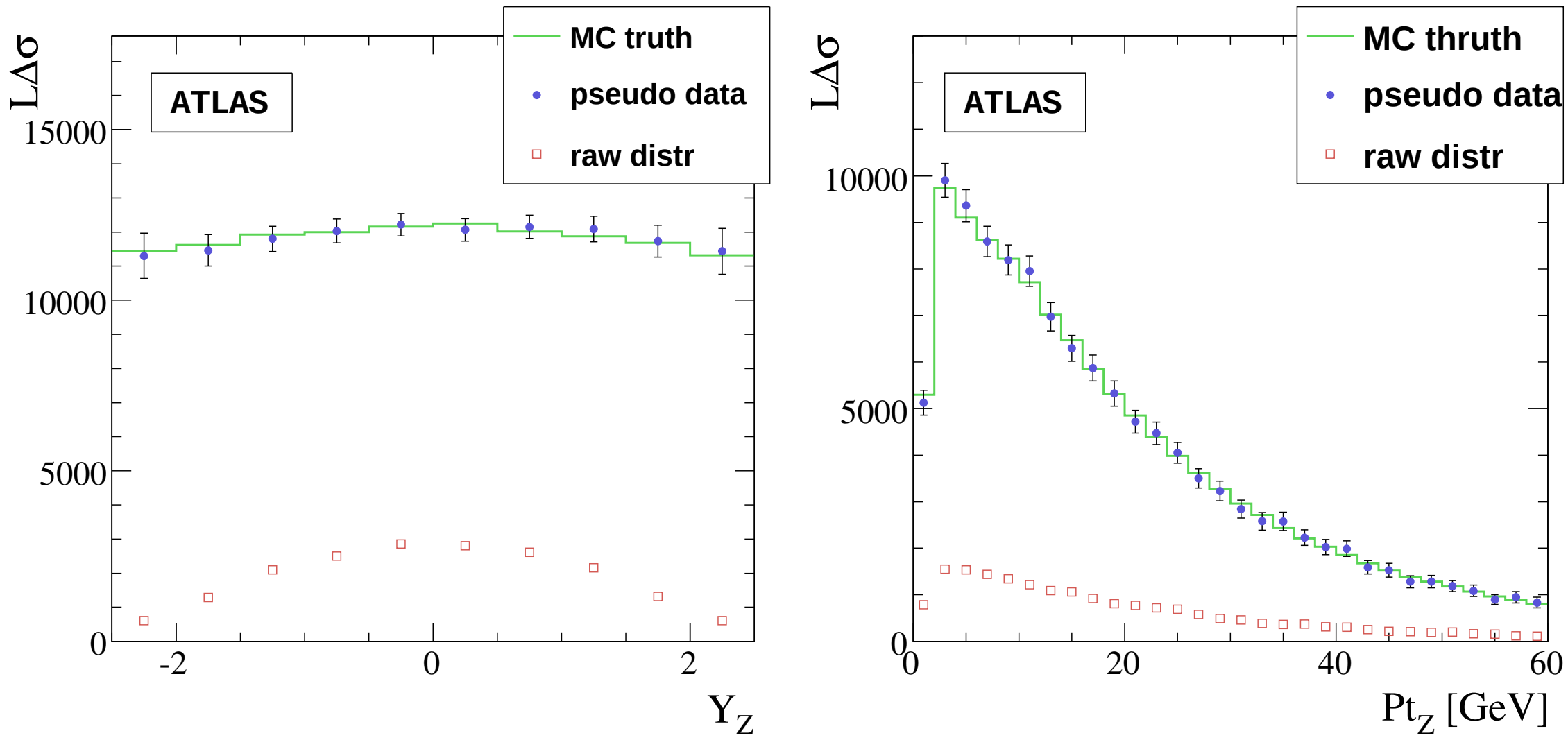


Figure 18: Left: $\mathscr{L}\Delta\sigma$ versus $y^Z$, integrated over $p_T^Z$. Right: $\mathscr{L}\Delta\sigma$ versus $p_T^Z$, integrated over $y^Z$.

precision better than about 2 %.

This argument however ignores the additional input from differential cross-section measurements. In contrast to total cross-sections, the differential ones benefit from small acceptance uncertainties, and have the potential to constrain the uncertainties that affect total cross-sections. The examples of the dilepton mass spectrum below the $Z$ peak, and of the $Z$ boson rapidity and $p_T$ distributions are studied here. The methods presented are shown to provide correct estimations of the differential cross-sections. The natural next step of these analyses, *i.e.* quantify their physical implications, is beyond the scope of this note and reserved for the real data.

# Production of Jets in Association with $Z$ Bosons

**Abstract**
We simulate a measurement of the inclusive $Z(\rightarrow e^+e^-/\mu^+\mu^-)$ + jets cross-section with the ATLAS experiment in pp collisions at 14 TeV for an integrated luminosity of 1 fb$^{-1}$ using fully-simulated signal and background Monte Carlo data sets. The reconstruction of leptons and of missing transverse energy becomes more complex in the presence of a multi-jet final state. We quantify the reconstruction differences with respect to those observed for inclusive $Z$ production. We derive statistical and systematic limitations in terms of probing the perturbative QCD predictions and discriminating between predictions of different event generators.

## 1 Introduction

The production of $W/Z$ + jets in pp collisions at 14 TeV is an important part of the physics program at ATLAS. The processes are interesting in their own right as tests of perturbative QCD at the LHC, as well as forming important backgrounds for both Standard Model and Beyond Standard Model physics processes. The results of the measurements can be compared directly with fixed-order predictions at leading order (LO) and next-to-leading order (NLO) in QCD and the gauge boson mass provides a large scale for the pertubative calculations. In addition, the measurements can be used to test the performance of Monte Carlo event generators that will also be used to simulate the backgrounds for other physics processes.

In this analysis we present a feasibility study of the cross-section measurements for data corresponding to an integrated luminosity of 1 fb$^{-1}$ performed with fully-simulated signal and background Monte Carlo samples. The goal of the analysis is to test the performance of the lepton and jet triggering and reconstruction algorithms in high jet multiplicity events, to develop the necessary analysis techniques (unfolding, background subtraction) and to evaluate the statistical and systematic limitations of the data, in terms of probing the fixed-order QCD predictions and of discriminating between predictions of different Monte Carlo event generators. The primary end-result of the analysis with real ATLAS data will be hadron-level cross-sections. In this note we will concentrate on $Z$ + jets in final states with electrons and muons, but with techniques applicable to the case of $W$ + jets as well. Much of the effort on the triggering and reconstruction of leptons is in common with the inclusive $W/Z$ note [1]; thus, we do not reproduce all of the details from that note, but rather comment on the impact of a multi-jet environment on these issues.

## 2 Reference Cross-Sections and Monte Carlo Datasets

Reference cross-sections are collected in [2], but we briefly discuss here the cross-sections relevant for this note. NLO is the first order at which the $Z$ + jets cross-sections have a realistic normalization (and realistic shape for some kinematic distributions) [3]. The current state of the art for NLO calculations is for $Z$ + 2 jets, although there is ongoing work for the calculation of 3 jet final states. Cross-sections for $Z$ + 0,1 and 2 (3) jet final states can be conveniently calculated at LO and NLO (LO only) using the parton-level MCFM [4] program (version 5.1. interfaced with LHAPDF 5.2.3 [5]), and it is from this program that we determine our reference cross-sections. We use the CTEQ6.1 parton distribution functions (PDFs) [6] and a dynamic renormalization/factorization scale of $m_Z^2 + p_{T,Z}^2$. We apply similar

kinematic cuts on the leptons and jets and the same jet algorithm on the partons as will be described in Section 3. The error on the cross-section stemming from the PDF uncertainty is calculated using the complete set of error PDFs in the CTEQ6.1 set.

## 2.1 Monte Carlo Datasets

The most important Monte Carlo data sets for the signal processes ($Z \rightarrow e^+e^-$ and $Z \rightarrow \mu^+\mu^-$) used in these studies are generated with ALPGEN [7] (v 2.05), interfaced with HERWIG [8] and using the leading order PDF set CTEQ6LL [9]. (Hereafter, when we refer to ALPGEN it is understood that it is interfaced with HERWIG.) The generation is done with a renormalization/factorization scale of $m_Z^2 + p_{T,Z}^2$ and a MLM [10] matching cut at $p_T = 20$ GeV (jets below this cut are generated by the parton shower and not by the matrix element) and $|\eta| < 6$. A discussion of the uncertainty in predictions for $Z$ + jets final states using different matrix element + Monte Carlo calculations and different matching cuts is beyond the scope of this study, but is given in Ref. [10].

The final Monte Carlo data sets are obtained following the standard prescription [10], by merging the samples of $Z$ + n partons (where n=0-5), each sample weighted with the product of the respective sample cross-section, the MLM matching efficiency and the efficiency of the generator-level filter. All but the highest jet multiplicity sample are exclusive, i.e. events are only kept if all jets with $p_T > 20$ GeV and $|\eta| < 6$ are matched to a matrix element parton. The highest multiplicity sample, $Z + 5$ partons, is inclusive; events with additional jets softer than the partons from the matrix element are not discarded. Thus, there can be more than 5 jets in this sample. The di-lepton mass is required to be larger then 40 GeV and lower than 200 GeV. A generator-level filter requires one seeded-cone jet, with a radius of $R = \sqrt{\Delta\eta^2 + \Delta\phi^2} = 0.4$, with $p_T > 20$ GeV and $|\eta| < 5.0$, and two electrons/muons with $p_T > 10$ GeV and $|\eta| < 2.7$ in the event.

For the comparison with the fixed-order theoretical predictions, the merged data sets are normalized to the NLO inclusive $Z \rightarrow e^+e^-$ and $Z \rightarrow \mu^+\mu^-$ cross-sections. Because of the jet filter used in their generation, the fully-simulated data sets can not be used to derive the global normalization factor. For this purpose, we use additional $Z \rightarrow e^+e^-$ and $Z \rightarrow \mu^+\mu^-$ ALPGEN data sets which are produced with the same conditions, but without the generator-level filter applied.

PYTHIA [11] signal and background samples are generated with version 6.323 ($Z \rightarrow e^+e^-$, $Z \rightarrow \mu^+\mu^-$, $Z \rightarrow \tau\tau$, and $W \rightarrow e\nu$) or 6.403 ($t\bar{t}$, filtered QCD multi-jet) using the corresponding ATLAS underlying-event tune [2]. PYTHIA $Z \rightarrow e^+e^-$ and $W \rightarrow e\nu$ events are preselected with a generator-level filter requiring one electron with $p_T > 10$ GeV and $|\eta| < 2.7$. The filter for the corresponding processes with muon final states requires one muon with $p_T > 5$ GeV and $|\eta| < 2.8$. PYTHIA $Z \rightarrow \tau\tau$ events are generated with a filter requiring two electrons/muons with $p_T > 5$ GeV and $|\eta| < 2.8$. For each of the $Z \rightarrow \ell\ell$ samples, the di-electron (di-muon, di-tau) mass is required to be larger than 60 GeV. The jet background for the electron channel is simulated with a PYTHIA QCD multi-jet sample with a minimum hard-scattering transverse momentum of 15 GeV. A generator-level filter requires a jet of $p_T = 17$ GeV clustered in a narrow region of $\Delta\eta/\Delta\phi = 0.06$, a size similar to an electron cluster. The QCD multi-jet background for the muon channel is estimated with a PYTHIA $b\bar{b}$ sample. Two muons with $p_T > 4$ GeV and 6 GeV, respectively, are required in the final state.

## 2.2 Corrections from Parton to Hadron Level

Comparisons of data cross-section measurements and LO/NLO predictions are to be made at the hadron level (particle level). Hence, the MCFM predictions for the observables have to be corrected with respect to the non-perturbative effects resulting from jet fragmentation and from the underlying

event. The impact of the underlying event correction is to add energy to the MCFM jets, while the jet fragmentation correction subtracts energy. Both corrections are expected to decrease with increasing jet $p_T$. The non-perturbative corrections are determined from the current ATLAS PYTHIA tune by comparing the multiplicity and the $p_T$ distribution of jets with a cone radius of 0.4 clustered on the final-state particles in $Z \to \mu^+\mu^-$ Monte Carlo samples generated with PYTHIA 6.403 (a) using the standard ATLAS PYTHIA tune [2] and (b) with fragmentation and multiple-particle interactions switched off. To the extent to which the two partons that can comprise a jet in MCFM mimic the effects of the parton shower in PYTHIA, the corrections derived from the above procedure can be applied to the MCFM output [3]. For jets with cone radius 0.4 with $p_T > 40$ GeV, the effects of fragmentation and underlying event cancel up to a residual correction at the percent level, which is then applied to the MCFM predictions. These corrections are expected to be re-done with the underlying event measurements determined from ATLAS data.

# 3 Particle Identification and Trigger

We adopt as much as possible definitions and cuts in common with the other analyses, and in particular with the inclusive $W/Z$ study [1].

## 3.1 Particle Identification

The electron candidates are required to have $p_T > 25$ GeV, and to lie in the range $|\eta| < 2.4$, excluding the barrel-to-endcap calorimeter crack region ($1.37 < |\eta| < 1.52$). The electrons are required to fulfill the medium electron-identification signature [12], which consists of requirements on the calorimeter shower-shape and the matched track. The $Z$ selection requires two electron candidates with an invariant mass of $81 < m_{ee} < 101$ GeV and $\Delta R > 0.2$ between the electrons. No calorimeter isolation cuts are applied for this analysis, although they will be applied for actual data analysis. There is an implicit isolation cut, however, present in the trigger [13].

A muon candidate requires the combined reconstruction of an inner detector track and a track in the muon spectrometer [14]. Muons are required to have $p_T > 15$ GeV and $|\eta| < 2.4$, with the range $1.2 < |\eta| < 1.3$ being excluded. Isolation is applied by requiring the energy deposition in the calorimeter to be less than 15 GeV in a cone of $\Delta R = 0.2$ around the extrapolation of the muon track. The $Z$ selection requires that there be two muon candidates with an invariant mass of $81 < m_{\mu\mu} < 101$ GeV.

For the analyses in this study, we use jets clustered with the standard ATLAS seeded-cone algorithm with a radius of $R = 0.4$, built from either calorimeter towers ($Z \to e^+e^-$ analysis) or topological clusters [15] ($Z \to \mu^+\mu^-$ analysis), and calibrated to the hadron level. The lepton and jet candidates must be separated by $\Delta R_{lj} > 0.4$. It is required that the jet transverse momentum be larger than 40 GeV and that the jet be in the range $|\eta| < 3.0$.

## 3.2 Trigger Selection

The trigger selection used here is the same as that used in the inclusive analyses [1]. In the electron channel, $Z \to e^+e^-$ + jets events are required to pass the isolated di-electron trigger or the isolated single-electron trigger. In the muon channel, $Z \to \mu^+\mu^-$ + jets events are required to pass the isolated di-muon trigger. The trigger efficiencies at the first, second and event filter levels are evaluated as a function of the jet multiplicity. The trigger efficiency is also studied as a function of the overall hadronic activity, the $p_T$ of the leading jet and the $Z$ transverse momentum. For this purpose, the

generated Monte Carlo information and the data driven tag-and-probe method are compared. Good agreement between the two methods is found. The efficiency for an electron to pass the isolated single-electron trigger is found to decrease with increasing jet multiplicity, $p_T$ of the leading jet and with decreasing distance to the closest jet.

# 4 Measurement of $Z$ + jets Cross-Sections

We study the comparison of theory and measurement for quantities suited to compare with a fixed-order NLO calulation: the inclusive cross-section for $Z \to \ell\ell^{\cdot}$ with at least 1 jet, 2 jets and 3 jets and the differential cross-sections with respect to the $p_T$ of the leading and the next-to-leading jets.

## 4.1 Lepton Reconstruction in a Multi-jet Environment

The presence of additional jets in the event has an impact on the kinematics of both the leptons and jets: the leptons are more boosted (larger $p_T$ and lower $\Delta\phi$ between leptons) in events with jets and the distance between leptons and jets becomes smaller in high-multiplicity events. The average jet $p_T$ increases with the number of jets. Due to the combination of the single electron and di-electron trigger channels used in this analysis and due to the boost of the electrons with large jet activity, the efficiency loss of the isolated electron triggers for large jet multiplicities has only a negligible impact. The total $Z$ reconstruction efficiency (offline+trigger) is stable with respect to both the jet multiplicity and the transverse momentum of the leading jet.

Muon reconstruction efficiencies and rejections for QCD multi-jet background are investigated for different isolation requirements. The isolation requirement for this analysis (see Section 3.1) is chosen such that it presents no significant bias for events with large jet multiplicities and large jet $p_T$ while at the same time providing a sufficiently large rejection for the QCD multi-jet background.

## 4.2 Background Estimation

For the evaluation of backgrounds to the $Z \to e^+e^- + \text{jets}$ signal we consider processes with real electrons ($t\bar{t}$, $W \to e\nu$, $Z \to \tau^+\tau^-$) and QCD multi-jet production. Statistics of the multi-jet background sample are increased by applying a very loose electron selection and then reweighting the events with the rejection from the final electron identification cuts. For the $Z \to \mu^+\mu^-$ analysis the background is dominated by processes with real muons ($t\bar{t}$, $W \to \mu\nu$, and QCD multi-jets). QCD multi-jet backgrounds for isolated highly-energetic muons result mainly from decays of $b\bar{b}$ mesons. We thus use a $b\bar{b}(\to \mu^+\mu^-)$ sample to evaluate this background. For both analyses, all backgrounds are estimated from fully-simulated Monte Carlo samples, generated with PYTHIA. They are compared with the signal distributions, derived from the respective ALPGEN $Z$ + jets data sets.

Table 1 provides an overview of the accepted cross-section and the corresponding signal and background fractions within the selected event sample, for several jet multiplicities and for the electron and the muon channel respectively. The uncertainties displayed in the table are of statistical nature only. The total background is at the level of 5-15 % depending on the jet multplicity. With increasing jet multiplicity, $t\bar{t}$ replaces QCD multi-jet production as the dominant background source in both analysis channels. Due to the larger acceptance and the larger lepton reconstruction efficiency, we obtain more than twice as many signal events in the muon channel. Since the dominant ($t\bar{t}$) background also contains two real leptons, the signal-to-background ratio is comparable in both analyses.

Figures 1(a+b) show the combined distribution of the di-muon mass and the jet multiplicity for events with at least one jet for signal and background events in the $Z \to \mu^+\mu^- + \text{jets}$ channel. Fig-

| | $Z \to \ell\ell + \geq 1$jet | | $Z \to \ell\ell + \geq 2$jets | | $Z \to \ell\ell + \geq 3$jets | |
|---|---|---|---|---|---|---|
| Process | $\sigma$ (fb) | fraction (%) | $\sigma$ (fb) | fraction (%) | $\sigma$ (fb) | fraction (%) |
| $Z \to e^+e^-$ + jets analysis | | | | | | |
| $Z \to e^+e^-$ | 23520±145 | 91.9±0.8 | 4894±45 | 87.9±1.3 | 900±15 | 80.0±2.4 |
| QCD jets | 1545±89 | 6.0±0.4 | 336±42 | 6.0±0.8 | 78±20 | 6.9±1.8 |
| $t\bar{t}$ | 496±28 | 1.9±0.1 | 333±23 | 6.0±0.4 | 146±15 | 13.0±1.4 |
| $W \to e\nu$ | (28±13) | (0.1±0.05) | (5.9±2.6) | (0.1±0.05) | (1.1±0.5) | (0.1±0.05) |
| $Z \to \tau^+\tau^-$ | 3.2±1.2 | 0.01±0.01 | (0.67±0.25) | (0.01±0.01) | (0.1±0.05) | (0.01±0.01) |
| $Z \to \mu^+\mu^-$ + jets analysis | | | | | | |
| $Z \to \mu^+\mu^-$ | 59400±650 | 96.2±1.0 | 12600±300 | 90.1±1.9 | 2450±100 | 89.7±3.7 |
| QCD($b\bar{b}$) | 1230±550 | 2.0±0.9 | 600±300 | 4.3±2.2 | 0±110 | 0.0±4.0 |
| $t\bar{t}$ | 1140±110 | 1.8±1.8 | 790±90 | 5.7±0.2 | 275±50 | 10.2±1.9 |
| $W \to \mu\nu$ | 0±180 | 0.0±0.3 | 0±30 | 0.0±0.2 | 0±5 | 0.0±0.2 |

Table 1: The accepted cross-sections ($\sigma$, in fb) and the corresponding fraction of the total sample (in %) for signal and for the background channels in the $Z \to e^+e^-$ + jets and the $Z \to \mu^+\mu^-$ + jets analyses, after applying the cuts outlined in Section 3. The numbers in brackets are extrapolated from results obtained for a lower jet multiplicity. Errors shown are statistical only.

ures 1(c+d) show the distribution of signal and backgrounds for the $p_T$ of the leading and the next-to-leading jet in the $Z \to e^+e^-$ + jets channel. The jets from the QCD multi-jet background have a similar $p_T$ distribution as the jets from the signal events, while the jets from the $t\bar{t}$ background tend to be harder.

#### 4.2.1 Background Subtraction

The $Z \to \tau^+\tau^-$, $t\bar{t}$ and $W \to e\nu$ backgrounds are subtracted using the Monte Carlo estimates. The systematic uncertainty from the limited background statistics is propagated into the systematic uncertainty of the cross-section measurement. Special care will be needed in validating against data the differential cross-section for the $t\bar{t}$ process, since it is the dominant background for large jet multiplicities. The QCD multi-jet background is expected to be determined with data-driven methods. From the simulations we expect a multi-jet background fraction independent of the jet $p_T$ such that it can be subtracted by applying a global factor. We assume in the following an uncertainty of 20% on the measurement of the QCD multi-jet fraction. The error is propagated into the systematic error on the measured cross-section.

### 4.3 Unfolding of Detector Effects

The reconstructed data have to be unfolded from the detector level to the hadron level, correcting for efficiency, resolution and non-linearities in electron and jet reconstruction. In this study, the individual unfolding corrections are assumed to factorize in leading approximation, and the individual contributions are investigated and corrected for separately. The corrections are detailed in the following for the case of the $Z \to e^+e^-$ channel. Unfolding of the $Z \to \mu^+\mu^-$ final state is done in a similar way. All corrections are derived with fully-simulated ALPGEN Monte Carlo samples.

The dominant correction on the inclusive cross-section for the $Z \to e^+e^-$ channel stems from the

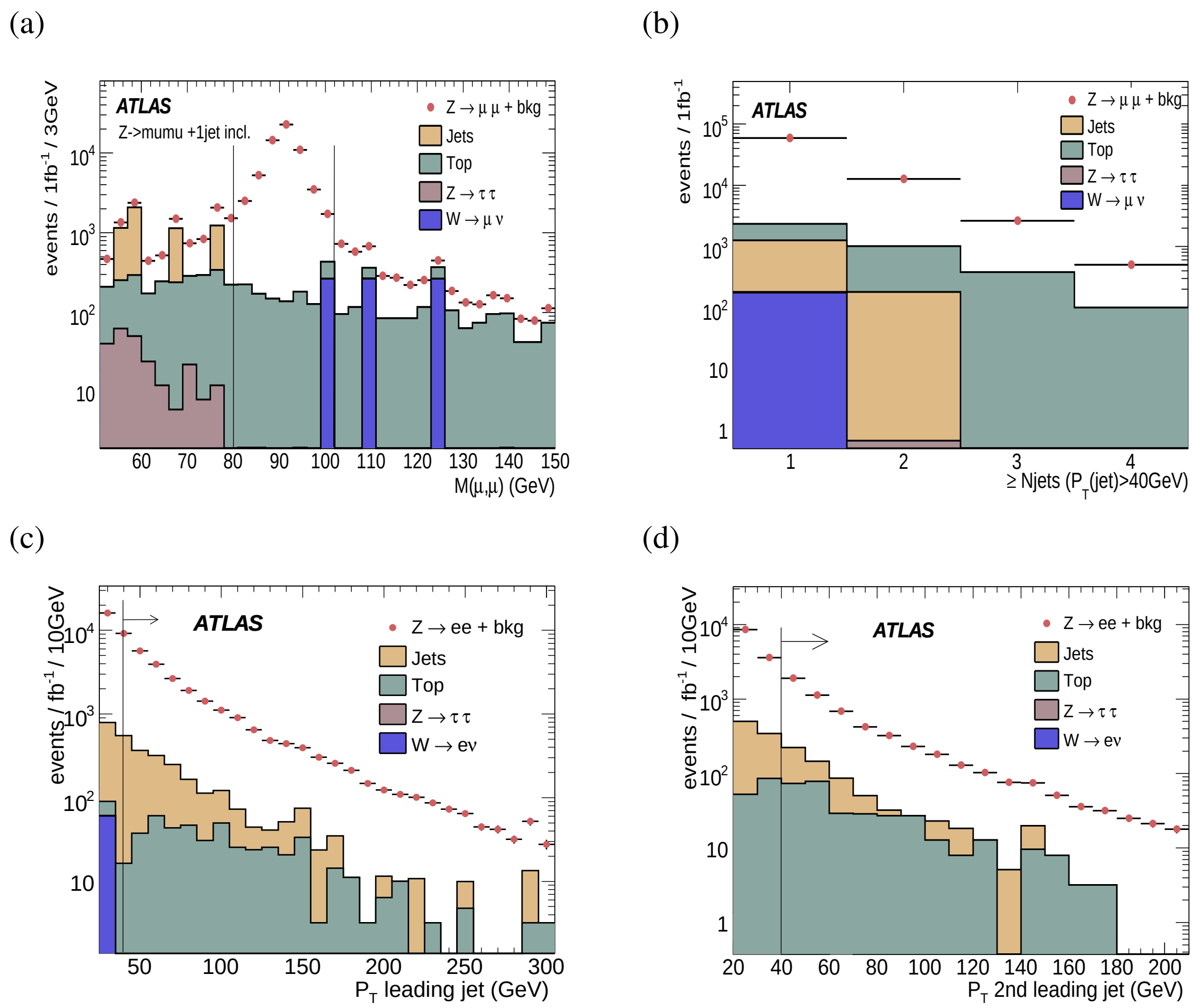


Figure 1: The distribution of the di-muon mass (a) and the inclusive jet multiplicity (b) for signal and backgrounds in the muon channel. In order to provide higher statistics for the background determination, background events in an invariant mass window of 51 – 131 GeV, are scaled down for an invariant mass window of 81 – 101 GeV. Also shown is the distribution of $p_T$ of the leading (c) and the next-to-leading (d) jet in the electron channel for $\int Ldt = 1$ fb$^{-1}$. The vertical lines in (a), (c) and (d) indicate the kinematic cuts applied in the analysis.

electron reconstruction. For each of the two electrons, the cross-section is corrected for the electron reconstruction efficiency, given as a function of the electron pseudo-rapidity and transverse momentum. The cross-section is also corrected for the trigger efficiency with respect to the offline selection. Corrections from jet reconstruction have a comparably small impact on the overall cross-section but bias the jet $p_T$ spectrum since, in general, the detector effects are greater for low $p_T$ jets. The reconstructed jet $p_T$ is corrected for the non-linearity of the jet energy scale, and for each jet in the required selection, the cross-section is corrected for the reconstruction efficiency and for the effect of the jet energy resolution. The uncertainties on deriving these corrections, stemming from the limited Monte Carlo statistics, are taken into account as systematic uncertainties on the cross-section measurement.

Figure 2 compares the distributions of the $p_T$ of the leading jet and the next-to-leading jet, in different unfolding stages, with the $p_T$ distribution of the original (hadron-level) Monte Carlo jets. Within the statistical and systematic errors, the $p_T$ distributions of the Monte Carlo jets and the corrected reconstructed jets are in agreement, thus providing a consistency check for the unfolding corrections.

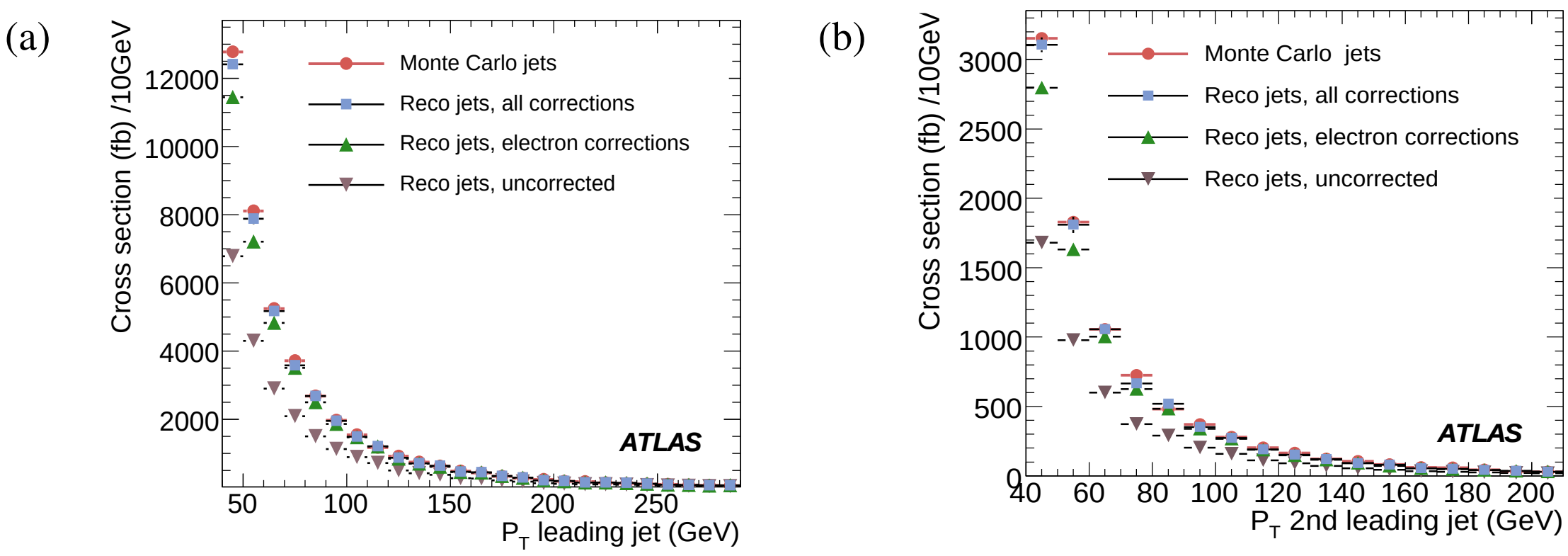


Figure 2: Comparison of the distribution of the $p_T$ of the leading jet (a) and the next-to-leading jet (b) for the generated (hadron-level) Monte Carlo and for the reconstructed quantities without any correction, after the corrections for electron triggering and reconstruction and after applying in addition the jet-releated corrections.

## 4.4 Comparison of Event Generator and MCFM Predictions at the Hadron Level

One of the goals of our study is to evaluate the statistical and systematic precision of the $Z$ + jets cross-section measurement and to compare this precision with the size of the uncertainties that are expected in the first inverse femotbarn of data. Since this study deals with the measurements from the first data, we compare the precision of the measurement with the differences in the predictions of our LO and NLO QCD calculations and with the predictions from matrix element and parton-shower generators.

### 4.4.1 Generator Comparisons

In this section we compare the prediction for the inclusive jet cross-section from the generators PYTHIA and ALPGEN with those from the MCFM partonic level event generator. In order to separate the reconstruction from the generation effects we use only Monte Carlo hadron-level generator information.

Figures 3(a)-(c) show the comparison of the distribution of the jet multiplicities and the $p_T$ of the leading and next-to leading jets for ALPGEN and PYTHIA $Z \rightarrow \mu^+\mu^- +$jets samples with the NLO (LO) calculations from MCFM. The errors on the generator distributions are purely Monte Carlo statistics whereas the errors on the MCFM cross-section correspond to the PDF uncertainties and to the error from the unfolding to the hadron level. MCFM predictions are corrected to the hadron level as specified in section 2.2. The two Monte Carlo samples are normalized to the inclusive NLO $Z$ cross-section, as determined in MCFM.

The NLO MCFM predictions for the $Z$ + 1 jet and $Z$ + 2 jets cross-sections are, in general, greater than the LO predictions by 20 to 30%. PYTHIA predicts a larger $Z$ + 1 jet cross-section than ALPGEN, but also predicts a lower average jet multiplicity. Both Monte Carlo generators predict a lower cross-section than the NLO MCFM calculation for final states with more then one jet. The difference between the predictions of PYTHIA and ALPGEN, and between both generators and MCFM, amounts to 10-60% depending on the jet multiplicity. A comparison of the differential cross-section as a function of the jet $p_T$ indicates that the inclusive cross-sections shown in Figure 3(a) depend very much on the minimum jet $p_T$ required by the selection. PYTHIA predicts larger cross-sections than

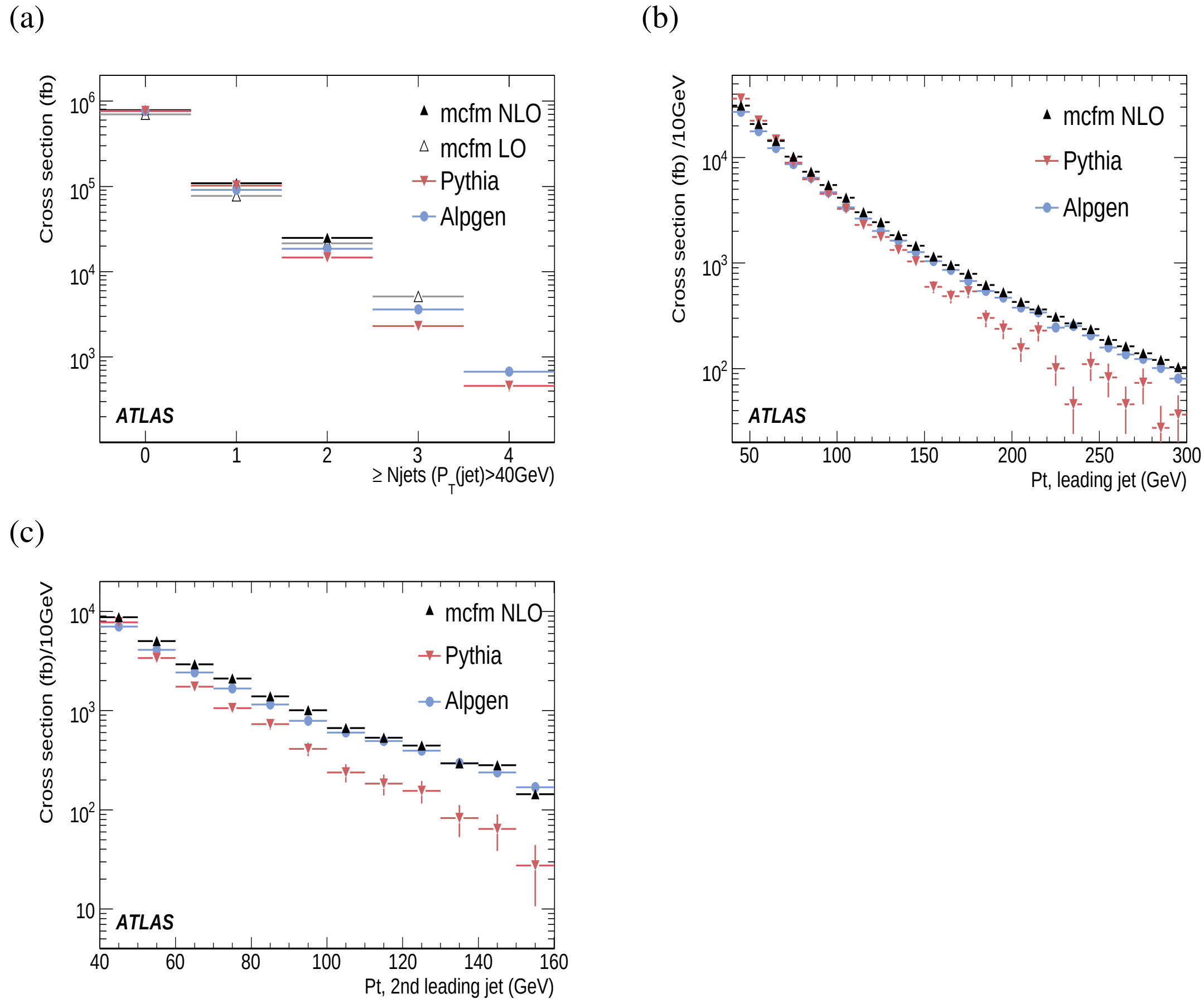


Figure 3: Comparison of the inclusive jet cross-section (a) and the $p_T$ of the leading jet (b) and of the next-to-leading jet (c) for the $Z \to \mu^+\mu^- +$jets channel from PYTHIA and ALPGEN Monte Carlo with NLO (LO) MCFM predictions. The MCFM predictions have been corrected to the hadron level.

even NLO MCFM for low jet $p_T$. But, while the shape of the jet $p_T$ distribution predicted by ALPGEN agrees well with the NLO MCFM prediction, PYTHIA generates a clearly softer $p_T$ spectrum, as expected.

### 4.4.2 Statistical and Systematic Errors

In order to determine the expected precision of the analysis, the cross-section measurement is performed on the fully-simulated ALPGEN $Z$ + jets data sets, which are corrected to the hadron level, following the prescription of Section 4.3. Systematic errors from the corrections are included. In the next step we evaluate the impact of the uncertainties expected for real ATLAS data taking. ATLAS expects a limited precision of the jet energy scale in the first years, starting from uncertainties at the level of 10% and converging eventually towards 1%. We obtain two benchmark scenarios by propagating jet energy scale uncertainties of 5% and 10% into the measured cross-sections. Several backgrounds will be estimated with data-driven methods, introducing additional systematic errors. We account for that in a first approach by adding an error of 20% on the fraction of the multi-jet background for each

jet multiplicity. The statistical uncertainties in the samples are scaled to the number of events expected to be selected for an integrated luminosity of 1 fb$^{-1}$.

Figure 4 compares, for the $Z \to e^+e^- + \text{jets}$ channel, the inclusive jet multiplicity (a) and the $p_T$ of the leading jet (b) for MCFM and the fully-simulated corrected ALPGEN sample. The errors on the MCFM predictions result from the PDF uncertainty and from the errors from the correction for the non-perturbative effects. The errors on the ALPGEN Monte Carlo data include all the statistical and systematic uncertainties described in this section with the jet energy scale uncertainty set to 5%.

An additional systematic uncertainty is introduced on the unfolding correction for the jet resolution due to the uncertainty on the jet resolution measurement and to the uncertainty on the shape of the $p_T$ distribution which we use to derive the corrections. Using corrections from different event generators and varying the jet resolution within its uncertainty results in a systematic error on the cross section at the percent level.

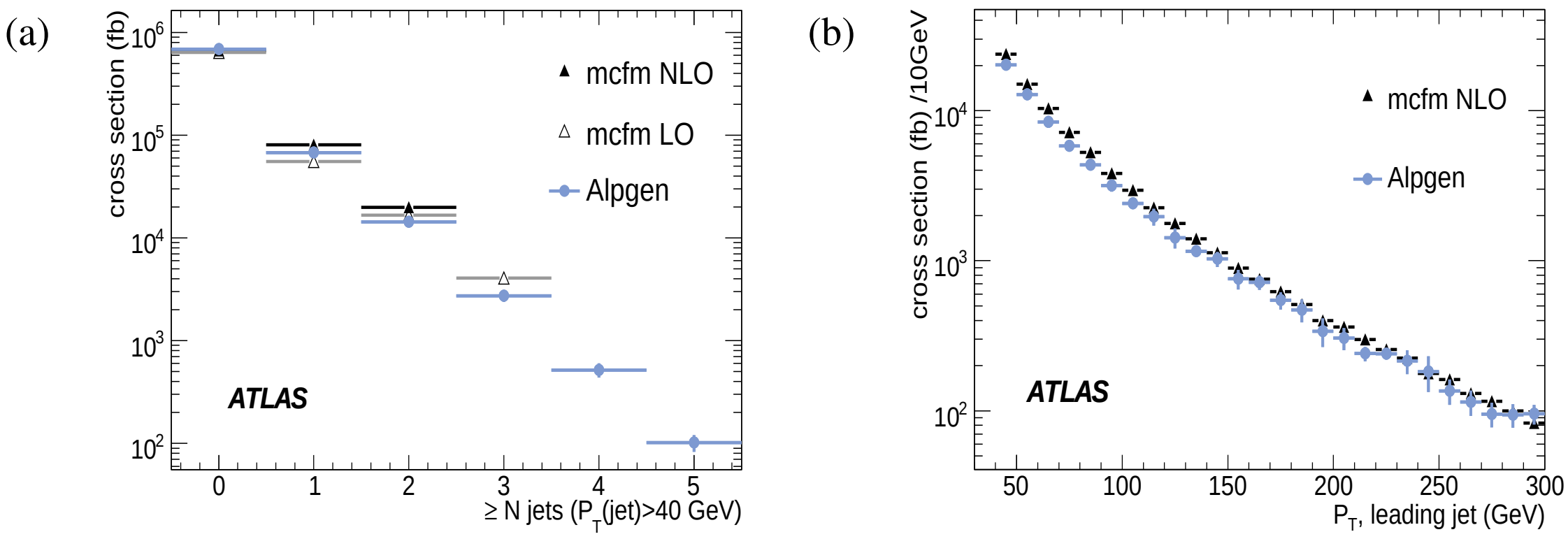


Figure 4: The inclusive jet cross-section (a) and the distribution of the $p_T$ of the leading jet (b), as predicted by NLO (LO) MCFM (corrected to the hadron-level) and by ALPGEN for the $Z \to e^+e^- + \text{jets}$ process.

The uncertainty on the theoretical predictions and on the measured cross section, as shown in Figure 4, are propagated on the data/theory ratio. Figure 5 shows the resulting uncertainty on a ratio of 1 for the inclusive cross-section and for the $p_T$ of the leading jet. The systematic uncertainty on the inclusive cross-section from a jet energy scale uncertainty of 5% is twice as large as as the sum of all the other statistical and systematic uncertainties. In this case, the overall precision on the data/theory ratio expected with the first fb$^{-1}$ of data is at the level of 8-15% for topologies with 1-3 jets. A jet energy scale uncertainty of 10% results in the dominant error on the cross-section. In this case, the total uncertainty on the cross section is at the level of 15-30%, which is at the same order as the typical differences expected between LO and NLO predictions, or between predictions from PYTHIA, ALPGEN and MCFM. Statistical limitations become sizable for large jet $p_T$ ($\gg$ 200 GeV).

## 5 Conclusions

Final states containing $Z$ + jets will serve as one of the Standard Model benchmarks for physics analyses at the LHC. We have simulated cross-section measurements for theoretically well-defined quantities such as the inclusive $Z$ + jets cross-section, and the jet transverse momentum for the leading and next-to-leading jets. An unfolding technique from the detector to the hadron level has been

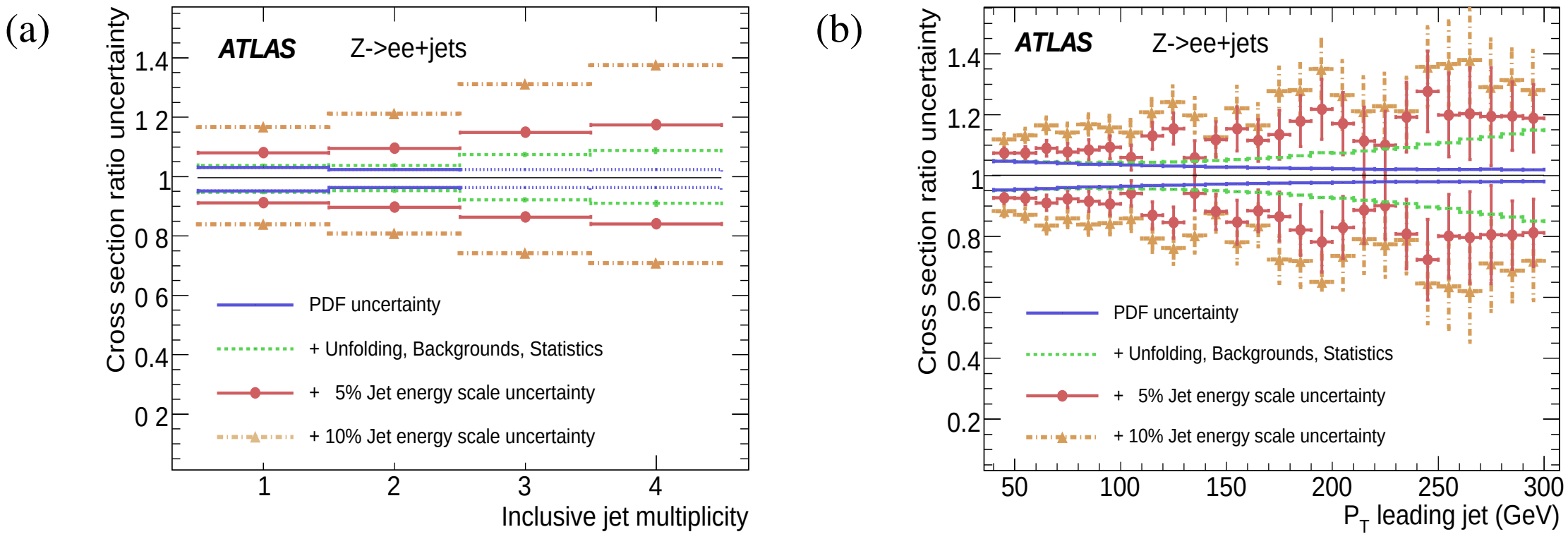


Figure 5: Uncertainty on the ratio of measurement and theory for the inclusive jet cross-section (a) and the $p_T$ of the leading jet (b) for the $Z \rightarrow e^+e^- + \text{jets}$ process.

developed, and results are presented at the hadron level. Theoretical corrections from parton to hadron level, necessary for comparisons of data to parton level predictions, are determined.

The main background sources are found to be QCD multi-jet processes for low jet multiplicities and $t\bar{t}$ for large jet multiplicities and amount to the level of 5-20%, depending on the jet multiplicity. Predictions from the Monte Carlo generators PYTHIA and ALPGEN have been compared with MCFM NLO (LO) calculations. The inclusive cross-section predictions differ by 10-60%, with larger discrepancies for the PYTHIA parton shower prediction (with respect to MCFM and ALPGEN) with increasing jet $p_T$. Statistical and systematic uncertainties on the ratio data/theory have been determined. A jet energy scale uncertainty of 5% would be the dominant systematic uncertainty on the measured cross section, resulting in a total uncertainty of 8-15% for final states with 1-3 jets. A jet energy scale uncertainty of 10% results in an overall precision at the level of 15-30% which is at the same order as the typical differences expected between LO and NLO predictions or between predictions from PYTHIA, ALPGEN and MCFM.

# Measurement of the $W$ Boson Mass with Early Data

**Abstract**
We present new methods for measuring the $W$ mass at ATLAS, and show their performance on simulated data. The experimental systematic uncertaintiess and their impact on the $m_W$ measurement are evaluated, with samples downscaled to 15 pb$^{-1}$. The electron transverse momentum analysis yields a precision of $\delta m_W = 120\,(stat) \oplus 117\,(syst)$ MeV. The systematic uncertainty is dominated by the energy scale. In the muon channel, the transverse mass analysis gives $\delta m_W = 57\,(stat) \oplus 231\,(syst)$ MeV, where the dominant contribution comes from the recoil calibration. PDF uncertainties contributes $\delta m_W = 25$ MeV. Other theoretical uncertainties were not explicitely considered, but expected to be small in comparison to experimental uncertainties in early $m_W$ measurements.

## 1 Introduction to the $W$ mass measurement at the LHC

The mass of the $W$ boson is currently measured to be $m_W = 80.399 \pm 0.025$ GeV [1]. Since the $W$ boson mass and the top quark mass are the largest sources of uncertainty in the indirect determination of the Higgs boson mass, improved precision is desirable. The expected total inclusive $W$ cross section at the LHC is about 20.5 nb for each lepton channel [2]. In 10 fb$^{-1}$ of data, around $30 \times 10^6$ $W$ events will be selected in each leptonic decay channel ($W \to e\nu, \mu\nu$), providing a combined statistical sensitivity of about 2 MeV. Systematic errors will need to be controlled to comparable precision.

The present note is however concerned with the early ATLAS data ($\mathscr{L} \sim 10 - 20$ pb$^{-1}$), with an expected statistical precision of about 100 MeV. In this context, the main sources of systematic uncertainty are of experimental origin (energy and momentum scales, resolution, efficiency). Other sources, related to the theoretical description of $W$ production, play a less significant role.

The aim of the following is to establish an unbiased $W$ mass fit with templates, demonstrate that using $Z$ events for calibration is valid, and show that the experimental sources of systematic uncertainties can be controlled to the level of the statistical sensitivity. In doing so, we test the readiness for the $W$ mass measurement in the first years of data taking.

This note is structured as follows. First a general discussion in Section 2 gives the outline of the analysis, the ingredients involved in a $W$ mass measurement, and the challenges in controlling them. Template based $W$ mass fits are presented in Sections 3 to 5, exploiting the electron and muon channels. In a first step, all ingredients entering the templates are supposed perfectly known. We then evaluate the dependence of the fit results on each of them. Section 6 describes the calibration of the absolute lepton energy scale and resolution using $Z$ decays. This section also discusses the effect of the lepton reconstruction efficiencies. Background uncertainties are treated in Section 7. Section 8 quantifies residual systematic uncertainties after *in situ* calibration. Finally, Section 9 summarizes the analysis and concludes.

# 2 Outline and strategy of the analysis

## 2.1 Simulation and data sets

The simulated $W$ and $Z$ samples on which this note is based are generated using the PYTHIA event generator [3]. Photon radiation is carried out by PHOTOS [4], and $\tau$ decays are handled by TAUOLA [5]. Detector simulation is done using GEANT4 version 4.0 [6], and reconstruction with Athena version 12.0.6. These fully simulated events are reconstructed using the ATLAS software, and used as real data ("pseudo-data"). For the production of templates, we use the generator-level information only, together with estimated detector smearing corrections. Simulated statistics of our main signal samples are shown in Table 1. Our $W$ signal samples correspond to roughly 15 pb$^{-1}$. In the following, all results will be normalized to this luminosity, i.e. downscaling results from the larger $Z$ samples.

| Channel | Nb. of events | Cross section [pb] | $\varepsilon$ of filter | Corresponding $\mathscr{L}$ [pb$^{-1}$] |
|---|---|---|---|---|
| $W \to e\nu$ | 170143 | 20510 | 0.624 | 13.3 |
| $W \to \mu\nu$ | 189903 | 20510 | 0.686 | 13.5 |
| $Z \to ee$ | 377745 | 2015 | 0.857 | 218.7 |
| $Z \to \mu\mu$ | 150650 | 2015 | 0.896 | 83.4 |

Table 1: Number of events, cross sections (at NNLO [2]), and corresponding luminosity of the simulated $W$ and $Z$ signal samples used in the analysis.

## 2.2 Event selection

Since the dijet cross section at hadron colliders is several orders of magnitude larger than the $W$ boson cross section, the hadronic decay modes of $W$ and $Z$ bosons are not usable. Therefore, only the leptonic decay modes $W \to \ell\nu$ and $Z \to \ell\ell$ ($\ell = e, \mu$) are considered. While the $W$ and $Z$ bosons are produced with a small transverse momentum on average (see Figure 2(a)), their longitudinal boost can be large. However, this boost leaves the lepton transverse momenta unchanged, which subsequently is a good distribution to study $W$ boson decays.

$W$ events are required to have one isolated lepton with $p_T$ above 20 GeV and missing transverse energy ($\not{E}_T$) in excess of 20 GeV. $Z$ events are required to have two isolated leptons with $p_T$ above 20 GeV of opposite charge (see Figure 1). The triggers providing these events are an isolated 15 GeV electron trigger and a 20 GeV muon trigger. The electrons are required to pass tight identification criterion [7], and only combined muons (with reconstructed tracks in both the inner detector and the muon spectrometer [8]) are used. Both electrons and muons are required to lie within the tracking range $|\eta| < 2.5$ (see Figure 2(b)). In addition, the calorimeter barrel-endcap transition range $1.3 < |\eta| < 1.6$ is excluded for electrons.

In addition to this basic selection, some other requirements apply. To reject backgrounds from jet and $t\bar{t}$ events, the signal events are required not to have large hadronic transverse activity. A summary of the requirements can be found in Table 2.

The expected numbers of events in 15 pb$^{-1}$ from the above mentioned event selection are summarized in Table 3. Though the expected number of reconstructed $Z$ events is an order of magnitude smaller

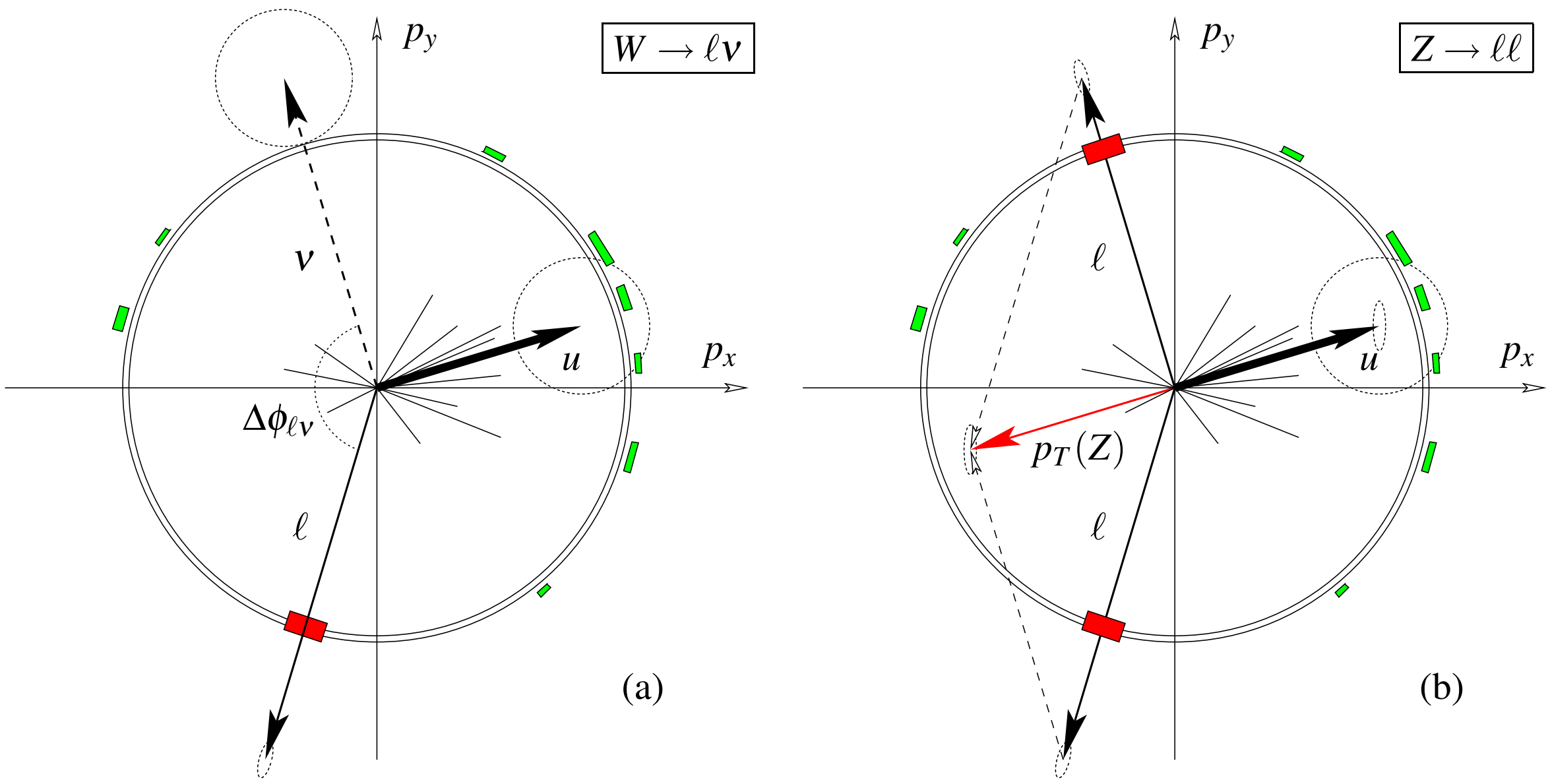


Figure 1: Transverse view of a $W \to \ell\nu$ (a) and a $Z \to \ell\ell$ (b) event. The combined transverse momentum of the recoil $u$, which should match that of the boson, is used to estimate the momentum of the undetected neutrino in the $W \to \ell\nu$ decay. The dotted ellipses represent the uncertainties.

| Requirement | $W \to e\nu$ | $W \to \mu\nu$ |
|---|---|---|
| Reconstructed lepton | $p_T > 20$ GeV, $\lvert\eta\rvert < 2.5$ | $p_T > 20$ GeV, $\lvert\eta\rvert < 2.5$ |
| Isolation | $E_T^{\text{cone}}/E_T < 0.2$ | |
| Missing energy | $\not{E}_T > 20$ GeV | $\not{E}_T > 20$ GeV |
| Crack region | Remove $1.30 < \lvert\eta\rvert < 1.60$ | |
| Recoil momentum | $p_T < 50$ GeV | |
| **Requirement** | $Z \to ee$ | $Z \to \mu\mu$ |
| Reconstructed leptons | $p_T > 20$ GeV, $\lvert\eta\rvert < 2.5$ | $p_T > 20$ GeV, $\lvert\eta\rvert < 2.5$ |
| Isolation | $E_T^{\text{cone}}/E_T < 0.2$ | |
| Crack region | Remove $1.30 < \lvert\eta\rvert < 1.60$ | |
| Recoil momentum | $p_T < 50$ GeV | |

Table 2: Selection criteria for the $W$ and $Z$ decays. See text for details.

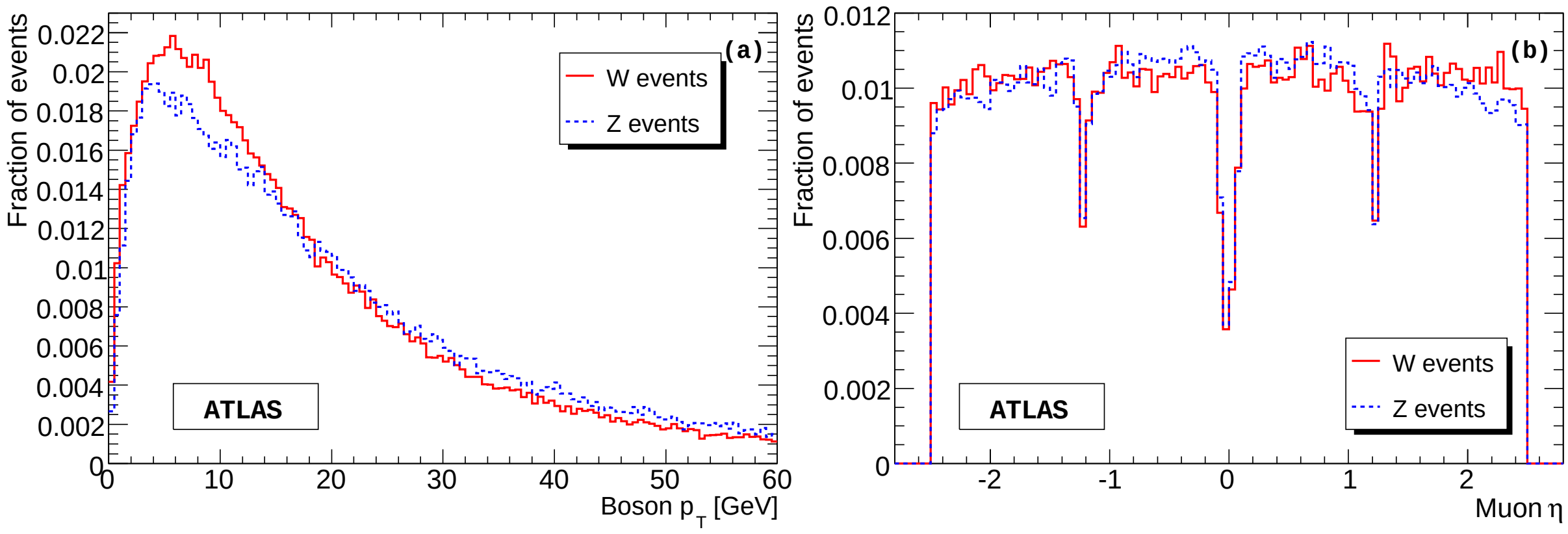


Figure 2: (a) Reconstructed transverse momentum of $W$ and $Z$ bosons. The larger resolution in the $W$ events causes the wider peak at low $p_T$, while at higher $p_T$ the $Z$ spectrum slightly dominates. (b) Distribution in $\eta$ of reconstructed muons from $W$ and $Z$ events.

than that of $W$ events, the fact that $Z$ events are fully reconstructed and thus have much better mass resolution compensates for this deficit.

| Channel | $W \to e\nu$ | $W \to \mu\nu$ | $Z \to ee$ | $Z \to \mu\mu$ |
|---|---|---|---|---|
| Detector acceptance [%] | 44.3 | 45.4 | 42.4 | 39.9 |
| Reconstruction efficiency [%] | 21.7 | 39.1 | 10.4 | 33.4 |
| Nb. of events for 15 $pb^{-1}$ [$10^3$ events] | 66.7 | 120.2 | 3.2 | 10.1 |

Table 3: Acceptances, total reconstruction efficiencies, and resulting statistics for 15 $pb^{-1}$ of data. The $W$ cross sections are inclusive, while the $Z$ cross sections are for invariant masses above 60 GeV. Both are at NNLO and contain the relevant branching fractions. The acceptance is the fraction of events which lies within the detector acceptance, while the total reconstruction efficiency is the overall efficiency for an event to pass all selection criteria including the acceptance.

### 2.3 Input to $W$ mass fit

While the $Z$ decay can be fully reconstructed, and its mass calculated from the invariant mass of the decay leptons, this is not the case for $W$ decays, where the neutrino goes undetected. From the momentum imbalance one can infer the missing energy, but with limited precision and only in the transverse direction. This means that the invariant mass can not be determined, and one is forced to consider other variables sensitive to the $W$ mass. In principle there are three sensitive variables:

- The lepton transverse momentum, $p_T^\ell$.
- The missing transverse momentum, $p_T^\nu \equiv \not{E}_T$.
- The $W$ transverse mass, defined as: $m_T^W \equiv \sqrt{2p_T^\ell p_T^\nu(1-\cos(\phi^\ell - \phi^\nu))}$.

The lepton transverse momentum is measured with an accuracy of about 2% for electrons and muons [7, 8] in the momentum range of interest (see Section 6.2). This is an order of magnitude better com-

pared to the accuracy of the missing transverse energy determination, which has a resolution of about 20-30% [9]. Finally, the $W$ transverse mass combines the two momenta along with the azimuthal angle between them.

All of the above distributions have a Jacobian peak either at $m_W/2$ ($p_T^\ell$ and $p_T^\nu$) or $m_W$ ($m_T^W$), which is sensitive to the $W$ mass. The sharpness of the peak is affected both by the resolution and the boson $p_T$. While the lepton $p_T$ has a very good resolution, the $p_T$ of the boson smears this Jacobian edge. On the contrary, $m_T^W$ is to first order insensitive to the $p_T$ of the boson, but here the edge is smeared by the poor resolution of the missing transverse energy (see Figure 3). Finally, $p_T^\nu$ suffers from both effects, and is therefore the poorest candidate for a fitting variable. Since $m_T^W$ is formed from $p_T^\ell$ and $p_T^\nu$, it is of course statistically correlated with $p_T^\ell$. However, the statistical correlation between $m_T^W$ and $p_T^\ell$ is only about 30%, and since they have different systematic errors, combining the measurements based on these observables could improve the sensitivity.

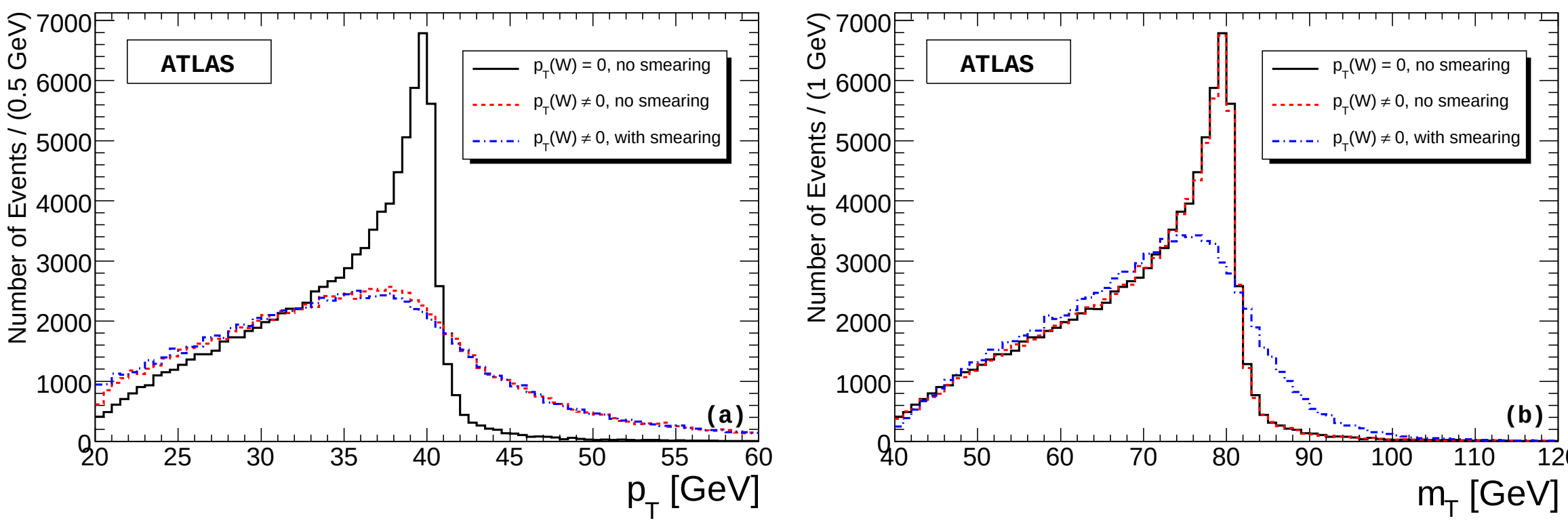


Figure 3: Fitted distributions of $p_T^\ell$ (a) and $m_T^W$ (b), showing the Jacobian peak, and the effects of finite detector resolution (i.e. smearing) and recoil (i.e. $p_T$ of the $W$). While $p_T^\ell$ is more sensitive to the recoil than to the resolution, the converse is true for the $m_T^W$ distribution.

## 2.4 Fitting the $W$ mass with templates

The lepton transverse momentum and $W$ transverse mass distributions, $p_T^\ell$ and $m_T^W$, shown in Figure 3, are the result of several non trivial effects. For this reason no analytical expression describes the distributions in detail, and one is forced to use numerical methods. One method for fitting these distributions is template fitting [10, 11]. Templates of the $p_T^\ell$ and $m_T^W$ distributions are produced with varying $m_W$ values, and compared to the corresponding distribution observed in data (see Figure 6). The comparison is based on a binned $\chi^2$ method.

To estimate the impact of a given effect on the $W$ mass determination, templates unaware of the effect under consideration are produced and subsequently fitted to data, which includes this effect. Assuming an unbiased fit, when the effect is not included in the data (see Sections 3 and 4), the resulting shift in fit value measures the systematic error on the $W$ mass from not including the effect. By gradually changing the size of an effect, the systematic error on the $W$ mass as a function of this effect can be determined. As most effects are small, the dependencies are approximately linear. They are in general

different for the $p_T^\ell$ and $m_T^W$ fits. If an effect can be characterized by one parameter $a$, the systematic errors ($\delta m_W$) can be calculated from the derivative $\partial m_W/\partial a$ times the size of the uncertainty $\delta a$. If more parameters are required, the systematic uncertainty is calculated from all parameters $a_i$ and their covariances $\mathrm{Cov}_{i,j}$.

$$\delta m_W = \frac{\partial m_W}{\partial a}\delta a \quad \text{(Single parameter)} \qquad\qquad \delta m_W{}^2 = \sum_{i,j}\frac{\partial m_W}{\partial a_i}\frac{\partial m_W}{\partial a_j}\mathrm{Cov}_{i,j} \quad \text{(Multi parameter)}$$

In the following, the $\delta a$ are to be understood as relative uncertainties, and the derivatives are with respect to these relative uncertainties. Our results for $\partial m_W/\partial a$ are normalized in MeV/%.

## 2.5 Calibration procedure

The calibration of the absolute energy/momentum[1] lepton scale plays a central role, as it is the largest systematic uncertainty and the starting point of all other calibrations.

### 2.5.1 Average calibration

To first order, a single average lepton scale factor, defined as $\alpha_E = E_{\mathrm{rec}}/E_{\mathrm{true}}$, independently of $\eta$ and $p_T$, can be obtained by demanding that the reconstructed $Z$ peak matches its known mass. While this assures the correct lepton scale and resolution for $Z$ events, the energy scale obtained in this way might not apply to $W$ events. Because of non linearities and non uniformities, the different $p_T$ and $\eta$ distributions in $W$ and $Z$ events can possibly introduce significant bias.

### 2.5.2 Differential calibration

If needed, an upgrade to a differential calibration can be performed, which contrary to the average calibration includes variations of scale and resolution with energy/momentum, $\eta$, and/or $\phi$. The three key ingredients to such a calibration are the precise knowledge of the $Z$ mass, width, and decay kinematics, the non zero transverse momentum of the $Z$ bosons, and the very large sample of $Z$ bosons produced at the LHC.

The calibration uses a large sample of reconstructed $Z \to \ell\ell$ events along with corresponding templates, representing our knowledge of the $Z$ lineshape, which we assume known. Through a comparison of the two in bins of the variables of interest (lepton $p_T$, $\eta$, and $\phi$) one can extract the scale and resolution in each of these bins [12].

Once the absolute scale is set for the leptons, the hadronic recoil scale can be determined by comparing it to the transverse momentum of the leptons from the $Z$. Again, this calibration can be done differentially, measuring the missing momentum response as a function of $p_T(Z)$, the recoil size, and the hadronic transverse activity ($\sum E_T^{\mathrm{hadrons}}$). While this calibration is naturally not as accurate as that of the lepton scale, due to the poorer resolution of the hadronic recoil, it fortunately turns out that the mass fit is less sensitive to the recoil scale (see Section 4.3).

[1] These two terms cover the same aspect, but will generally be used about electrons and muons, respectively. Angular uncertainties are omitted throughout this note, as they play an insignificant role.

#### 2.5.3 Efficiency determination

The efficiency is determined using the "tag and probe" methods [13], again applied to $Z \to \ell\ell$ events. The different spectra in $W$ and $Z$ events is accounted for by determining the efficiency as a function of $p_T$ and $\eta$, which has been found to yield efficiencies compatible between $W$ and $Z$ [7, 8].

The following sections attempts to quantify the above outline.

# 3 Fitting the $W$ mass with templates - electron channel

## 3.1 Modelling templates for $W$ mass fit

As no analytical expression matches the lepton transverse momentum and $W$ transverse mass distributions $p_T^\ell$ and $m_T^W$, we fit the $W$ mass using the template method (see Section 2.4). For the templates of varying $W$ mass to match the measured distribution well, all effects influencing these distributions must be included when producing the templates. The principle influences are scale, resolution, non gaussian tails, efficiency and background effects, which are (with the exception of background) primarily obtained from the similar but more constrained $Z \to \ell\ell$ events, as discussed in Section 6.

Two assumptions have to be validated first, namely the lack of bias of the fit in itself, and the portability of the calibration from the $Z$ to the $W$. The lack of bias is tested by assuming perfectly known physics and detector response. In practice, the detector response is determined at this stage from direct comparisons of the lepton reconstruction to the generator level kinematics, using the $W$ sample that is used in the mass fits. Then the fit is repeated using templates with the detector response estimated from the $Z$ sample, still comparing reconstructed to simulated kinematics. An unbiased result validates the portability, i.e. that detector parameters can indeed be ported from $Z$ to $W$ events, justifying an *in situ* determination of these parameters using $Z$ events.

In addition it has to be tested that the template components can be included without biasing the fit, and thus that a subsequent calibration, which matches the truth, will yield unbiased templates. This is tested in the following. The statistical sensitivity of our $W \to e\nu$ sample, corresponding to 15 pb$^{-1}$ of data, is about 120 MeV, and we provide an estimate of the required precision on the detector response parameters to keep the systematic uncertainty within this limit.

## 3.2 Fits to $m_W$ using templates: Validation of the method

In this section, the detector parameters are determined from fits to $E_{\rm rec}/E_{\rm true}$, the ratio of reconstructed to true energy of the decay electrons. The fits are done using the so-called "Crystal Ball" PDF [14], which aims at describing the result of calorimetric resolution together with upstream energy loss in a single function. It has four parameters and displays a gaussian core, and a power-law tail at low energy. Its expression is, up to normalization factors:

$$CB(x) = \begin{cases} e^{-(\frac{x-\alpha_E}{\sigma_E})^2}, & x > \alpha_E - n\sigma_E \\ (\beta/n - |n| - x)^{-\beta}, & x < \alpha_E - n\sigma_E \end{cases} \quad (1)$$

where $x = E_{\rm rec}/E_{\rm true}$, $\alpha_E$ is the position of the peak, $\sigma_E$ the gaussian width; $n$ gives, in units of $\sigma_E$, the point of transition between the gaussian and power-law descriptions, and $\beta$ is the exponent controlling the tails. The relative normalization of the two components preserves continuity at $\alpha_E - n\sigma_E$, up

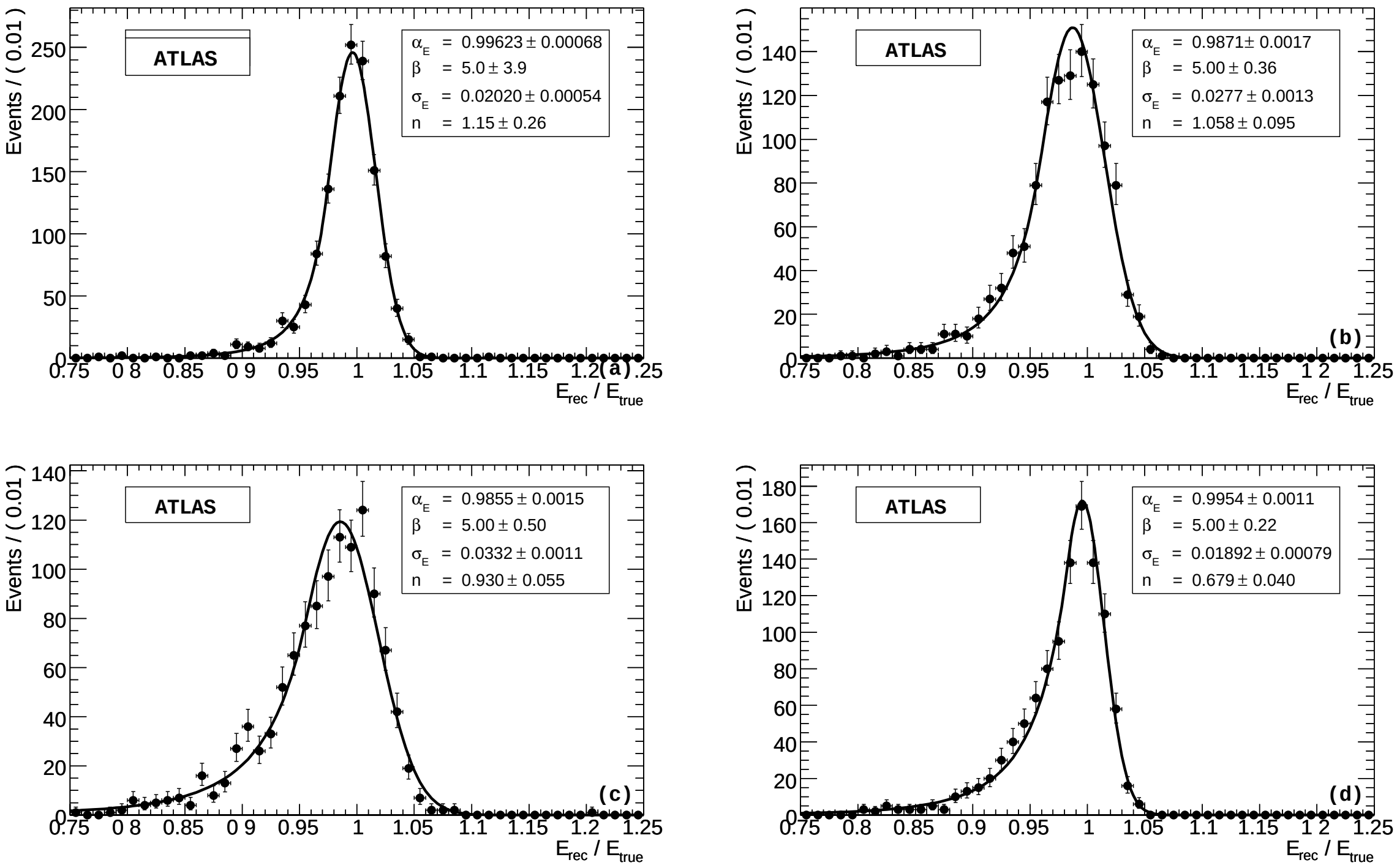


Figure 4: Examples of detector response functions fitted to $E_{\rm rec}/E_{\rm true}$, for $30 < p_T^e < 40$ GeV. From upper left to lower right: $0.4 < |\eta| < 0.5$ (a), $0.8 < |\eta| < 0.9$ (b), $1.3 < |\eta| < 1.4$ (c), and $1.9 < |\eta| < 2.0$ (d).

to the first derivative. While not fully satisfactory from a theoretical point of view (the combination of resolution effects and radiation should in principle be given by a proper convolution), it is very effective in describing the observed response.

The fits are performed vs. $\eta$ and $p_T$. The angular range $0 < |\eta| < 2.5$ is divided in intervals of size $\Delta\eta = 0.1$. In each interval, fits are done for $10 < p_T < 70$ GeV, in intervals $\Delta p_T = 10$ GeV. Figure 4 shows a number of examples, at different values of $\eta$ and $p_T$. The $\eta$ dependence of the parameters, for $30 < p_T < 40$ GeV, are displayed in Figure 5, where the shaded regions correspond to the excluded $|\eta|$ region described in Section 2.2.

In our fits the $\beta$ parameter was constrained to the range $0 < \beta < 5$. As the examples in Figure 4 illustrate, the $\beta$ parameter appears to systematically choose values close to its upper bound, while satisfactory fits are still obtained. Therefore, we fix $\beta = 5$ in the remaining of the analysis, and treat the response functions in terms of $\alpha_E$, $\sigma_E$ and $n$ only. The $p_T$ spectrum templates are produced from generator level $W \to e\nu$ events, where the electrons are smeared using the above function and parameters according to their kinematic variables. Three example template distributions are shown in Figure 6(a), corresponding to three values of $m_W$. The number of events used to produce the templates is increased by repeatedly smearing generator level particles. Although this limits the impact of statistical fluctuations in the templates on the result, template fluctuations are still visible. Distributions on Figures 6 are plotted on a wide range to assure that the entire $p_T$ spectrum is under control. The fitting range

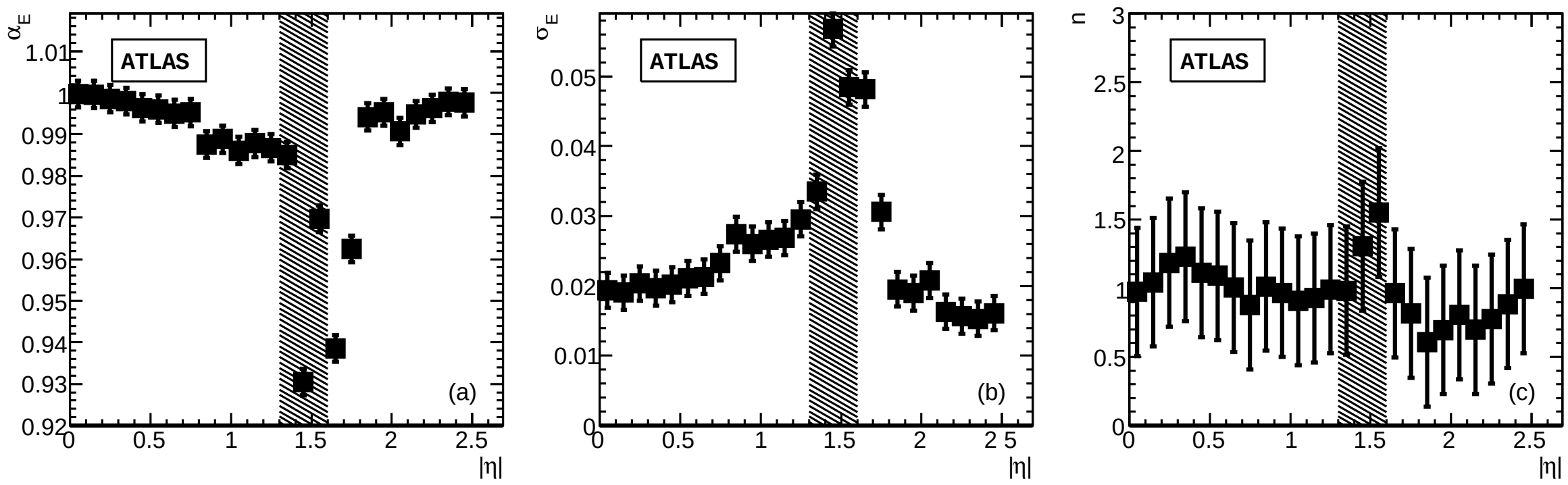


Figure 5: $\eta$ dependence of $\alpha_E$ (a), $\sigma_E$ (b) and $n$ (c), for $30 < p_T^e < 40$ GeV. The shaded regions correspond to the excluded $|\eta|$ region described in the event selection (Section 2.2).

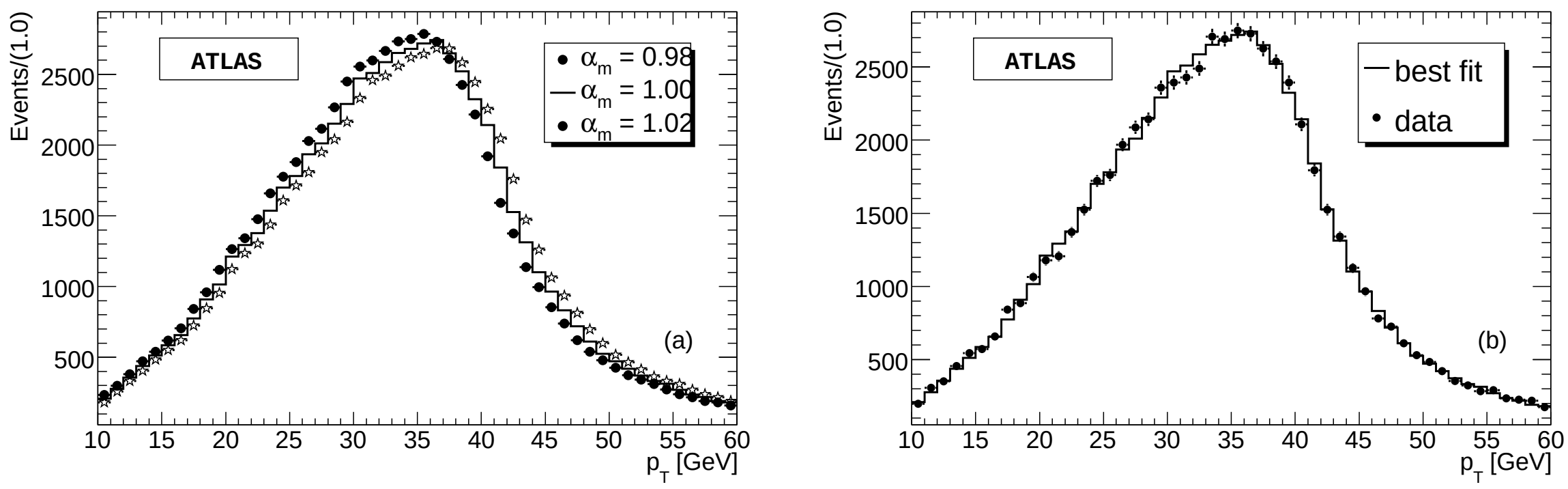


Figure 6: (a) Templates obtained at three example mass points, namely $\alpha_m = m_W/m_W^{true} = 0.98, 1, 1.02$. (b) Best fit template (histogram), compared to the pseudo-data (points).

is chosen to be between 30 and 60 GeV to avoid edge effects and to reject backgrounds (see Section 7).

The mass fit is performed using binned $\chi^2$ comparisons between the pseudo-data and the template histograms. Given that all $p_T$ bins contain at least several hundred events, the $\chi^2$ of a given comparison can be defined as:

$$\chi^2 = \sum_{i=1}^{N} \frac{(n_{i,data} - n_{i,template})^2}{\sigma_{i,data}^2 + \sigma_{i,template}^2}, \tag{2}$$

where the sum is over the histogram bins, and $n$ and $\sigma$ are the bin contents and their errors, respectively. Computed as a function of $\alpha_m = m_W/m_W^{true}$, the $\chi^2$ follows the parabola illustrated in Figure 7, which can be used to determine the $m_W^{fit}$ and its error. The obtained parabola is satisfactory despite the fluctuations of the $\chi^2$ points with respect to the fitted curve, which are due to the finite statistics of the templates. We obtain $m_W^{fit} = 80.468 \pm 0.117$ GeV, to be compared to the input value $m_W^{true} = 80.405$ GeV. The best fit template is shown in Figure 6(b). The stability of this result is verified by repeating this exercise a number of times, with the detector smearing applied independently

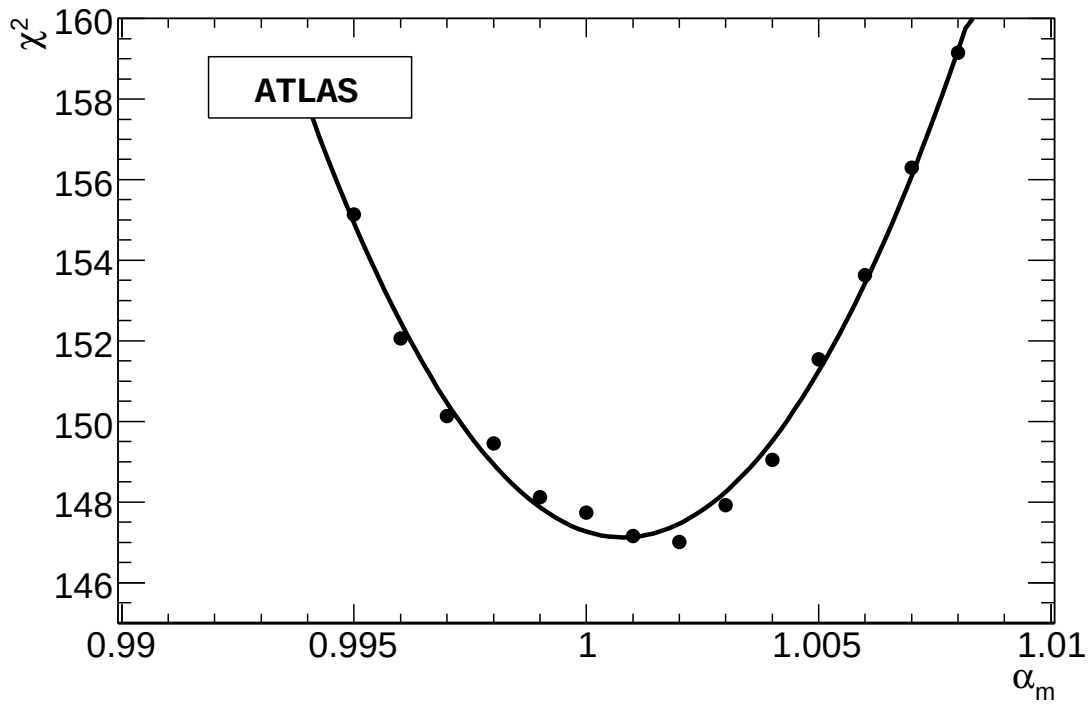


Figure 7: $\chi^2$ vs. $\alpha_m = m_W/m_W^{true}$, for the comparisons of pseudo-data and templates described in the text.

in each exercise (i.e, producing independent sets of templates). The distribution of $m_W^{fit}$ has a spread well compatible with the estimated fit uncertainty.

We thus conclude that within the statistical sensitivity of the $W \to e\nu$ sample, the current procedure provides an unbiased estimate of $m_W$. We now proceed to relax our main assumptions and quantify the dependence of the fit on the detector response parameters.

### 3.3 Sensitivity of $m_W^{fit}$ to the template components

This section quantifies the stability of $m_W^{fit}$ under variations of the assumptions used to produce the templates. As stated earlier, we leave phenomenological considerations aside and concentrate on experimental effects. We study explicitly the effect of non gaussian tails in the detector response function, reconstruction and identification efficiency, and backgrounds. The dependence of $m_W^{fit}$ on the detector scale and resolution was studied in [15], and reviewed here. The bias $\delta m_W = m_W^{fit} - m_W$ as a function of the fractional error on the lepton scale ($\alpha_E$) and resolution ($\sigma_E$) is found to be:

$$\partial m_W/\partial \alpha_E = 800 \text{ MeV}/\%, \quad \partial m_W/\partial \sigma_E = 0.8 \text{ MeV}/\%. \tag{3}$$

The impact of non gaussian tails is studied as follows. Starting from the detector response parametrization decribed in Section 3.2, we suppress the tails of the distribution and assume a pure gaussian response. The parameters describing scale and resolution are kept to their previous value. We then produce templates and perform the fit as above. The procedure is illustrated in Figure 8. The response distribution can be compared to Figure 4 to assess the impact of neglecting the non gaussian part of the distribution. As can be seen, the corresponding templates are biased towards higher $p_T$; we thus anticipate, as otherwise expected, that an underestimation of the tails should imply a negative $\delta m_W$.

We obtain a bias $\delta m_W = -555$ MeV, corresponding to an underestimation of the non gaussian tails by 100%. Denoting $\tau$ the non gaussian fraction of the response function, we thus estimate the bias as a function of the relative error on the tails as:

$$\partial m_W/\partial \tau = -5.5 \text{ MeV}/\%. \tag{4}$$

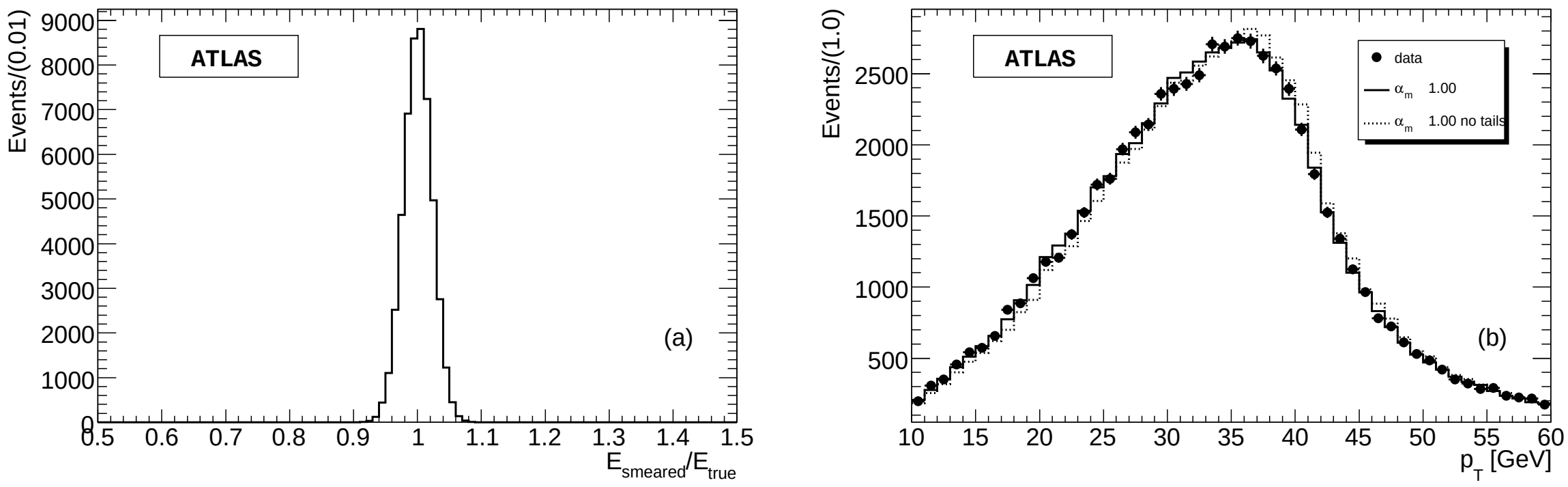


Figure 8: (a) Response function at $0.2 < \eta < 0.3$, $20 < p_T < 30$ GeV, removing the non gaussian part of the distribution. (b) Pseudo-data (points), compared to templates produced assuming $m_W = m_W^{true}$, with non gaussian tails included (full line) or not (dashed line).

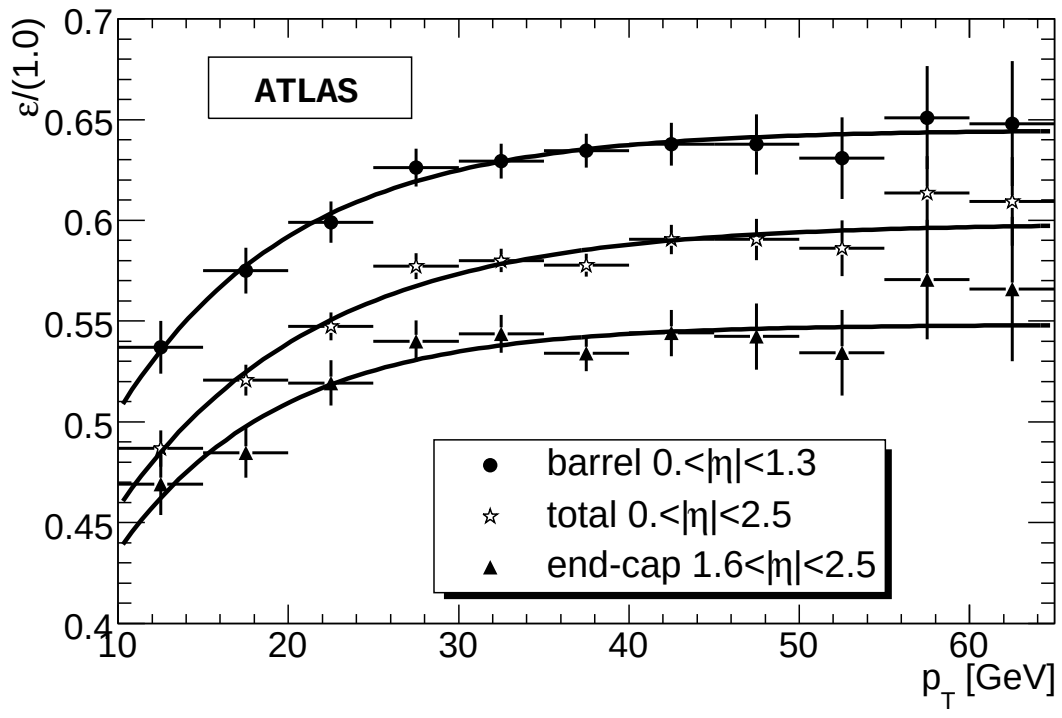


Figure 9: Electron reconstruction efficiency as a function of $p_T$, for different regions in $\eta$.

Distortions in the $p_T$ distribution can also be caused by the lepton reconstruction efficiency, as soon at it has a non trivial $p_T$ dependence, i.e. $\varepsilon_\ell = \varepsilon_\ell(p_T)$. This is the case in the electron channel, as illustrated in Figure 9.

As above, we quantify the impact of this $p_T$ dependence by taking the pseudo-data as they are, but assuming a flat efficiency in the templates. Since $\varepsilon_\ell(p_T)$ is an increasing function of $p_T$, we expect that the templates will be biased towards lower $p_T$ values, inducing a positive shift in $m_W^{fit}$. Performing the mass fit indeed yields $\delta m_W = 360$ MeV (this bias corresponds to a perfectly flat efficiency assumption). We estimate the bias per percent relative error on the $p_T$ dependence of $\varepsilon_\ell$ to be:

$$\partial m_W / \partial \varepsilon_\ell = 3.6 \text{ MeV}/\%. \tag{5}$$

Note that the present analysis only relies on the $p_T$ dependence of $\varepsilon_\ell$ and not on its absolute value.

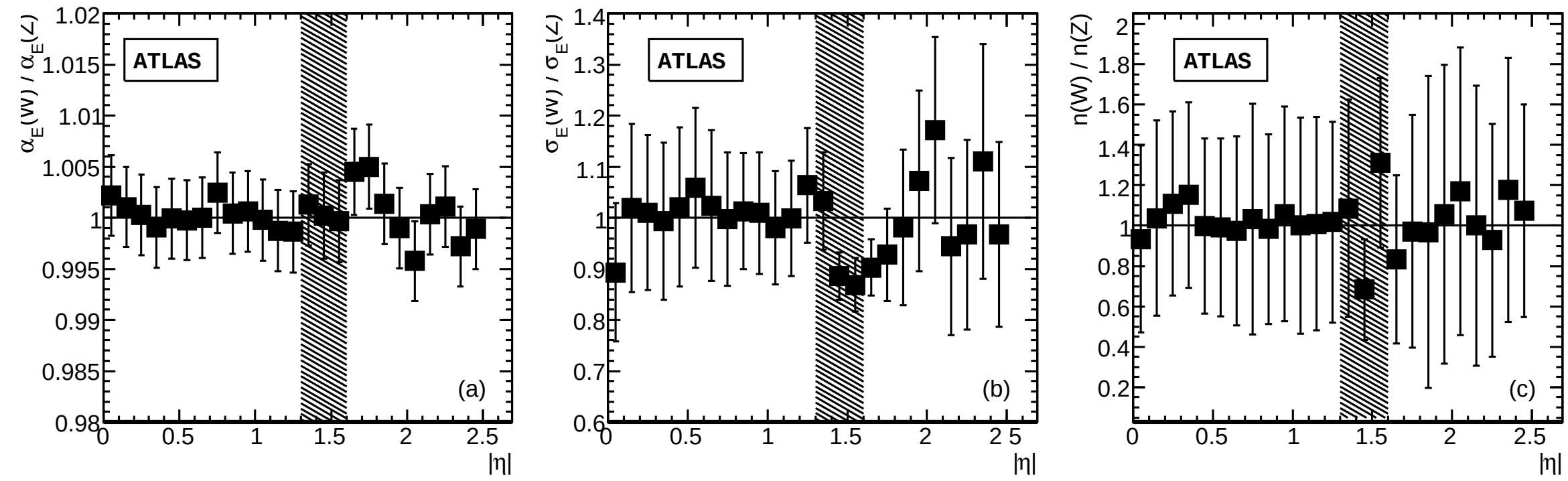


Figure 10: Ratio of the fitted values of $\alpha_E$, $\sigma_E$ and $n$, between $W$ and $Z$ events. Each histogram represents the $\eta$ dependence of the parameter $\alpha_E$ (a), $\sigma_E$ (b) and $n$ (c), for $30 < p_T^e < 40$ GeV.

### 3.4 Comparison of $W$ and $Z$ events

Before explicitly calibrating detector parameters from $Z$ events and applying them in the $m_W$ fit, we verify that this procedure is indeed justified. To this end, we perform the detector response fits as described in Section 3.2 on our $Z \to ee$ sample, obtaining a map of the response parameters $\alpha_E$, $\sigma_E$ and $n$ as a function of $\eta$ and $p_T$ of the electrons.

A first check is to compare the obtained values to those extracted from the $W$ sample. This is illustrated in Figure 10. For all parameters, agreement is found within the statistical sensitivity throughout the analysed electron phase space, except for the resolution parameter in the shaded $\eta$ region, excluded in the study as explained in Section 2.2. We thus expect that templates produced using detector response to $Z$ events will provide an adequate description of $W$ events.

A $m_W$ fit is performed next. Templates are produced from generator level $W \to e\nu$ events, smeared according to detector performance found on $Z$ events. The resulting distributions are shown in Figure 11(a) together with their ratio in Figure 11(b); good agreement is observed. Fitted with a straight line, Figure 11(b) shows a slope of $(3 \pm 2)\ 10^{-4}$, compatible with 0, but which yields a small bias towards higher masses. Accordingly, the result of the fit is $m_W^{fit} = 80.567 \pm 0.118$ GeV, to be compared to $m_W^{fit} = 80.468 \pm 0.117$ GeV obtained with detector performance found on $W$ events. The result is compatible with the input value $m_W^{true} = 80.405$ GeV.

## 4 Fitting the $W$ mass with templates - muon channel

This section repeats the discussion of Section 3 for the muons channel. In addition, a template fit of the $m_T^W$ distribution is described.

### 4.1 Template fits to the muon transverse momentum

Repeating the analysis from Section 3 for the muon channel yields $m_W^{fit} = 80.538 \pm 0.106$ GeV, obtained with detector performance found on $W$ events, and $m_W^{fit} = 80.508 \pm 0.106$ GeV, with detector

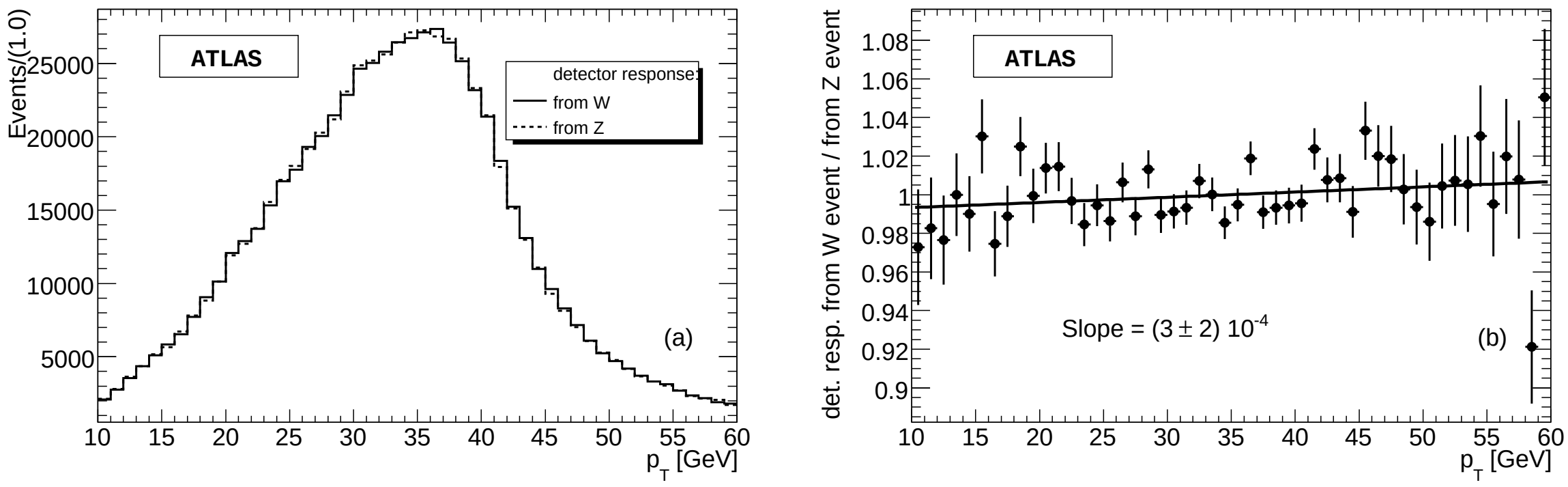


Figure 11: (a) Templates obtained for $\alpha_m = m_W/m_W^{true} = 1$, with detector performance obtained from $W$ events (full line) and from $Z$ events (dashes). (b) Ratio of the previous histograms, fitted with a straight line.

performance found on $Z$ events. The differences with the electron channel are described in the following.

The dependence of the template fit on the scale and resolution is the same in the muon channel as in the electron channel. The muon momentum resolution is generally slightly worse, and whereas the electron resolution improves with $p_T$, the converse is true for the muon resolution. Figure 12 shows four examples of the momentum ratio distributions $p_T^{\rm rec}/p_T^{\rm true}$ for muons – two from $W$ events and two from $Z$ events. As can be seen, the shapes can be modeled well with a core gaussian distribution describing the general muon bias and resolution complemented by an outlier gaussian distribution accounting for the muons, which are encountering parts of the detector with poor muon spectrometer coverage and increased material, resulting in a slightly degraded resolution.

To check the portability, the fitted constants are again compared between the $W$ and the $Z$ events, as can be seen in Figure 13. The correspondance between the fitted parameters is satisfactory. According to Figures 13(b) and 13(d), the resolution parameters are systematically smaller in $W$ events. The difference between the resolutions in $W$ and $Z$ events averaged over $\eta$ is 3.6%, which combined to Eq. 3 leads to a bias of 3 MeV. As will be seen in Section 6, this bias can be neglected given the precision of the *in situ* scale and resolution determination.

Unlike the electron case, the muon reconstruction efficiency does not vary significanly over the momentum range of interest. As can be seen from Figure 14, the efficiency is quite constant above 10 GeV, varying only slightly between barrel ($\varepsilon = 95.8\%$) and endcap ($\varepsilon = 94.3\%$) region, due to increase in material. As the reconstructed muons are required to have a momentum above 20 GeV, the efficiency is essentially constant.

The flatness of the efficiency is not expected to change until at several hundred GeV, where radiative losses grow larger than those due to ionization [1]. Using the "tag-and-probe" method, the hypothesis of a flat muon efficiency will be tested. The uncertainty in the linear fit translates for 15 $\mathrm{pb}^{-1}$ into a systematic error of 16 MeV, which is much less than in the electron channel, as expected.

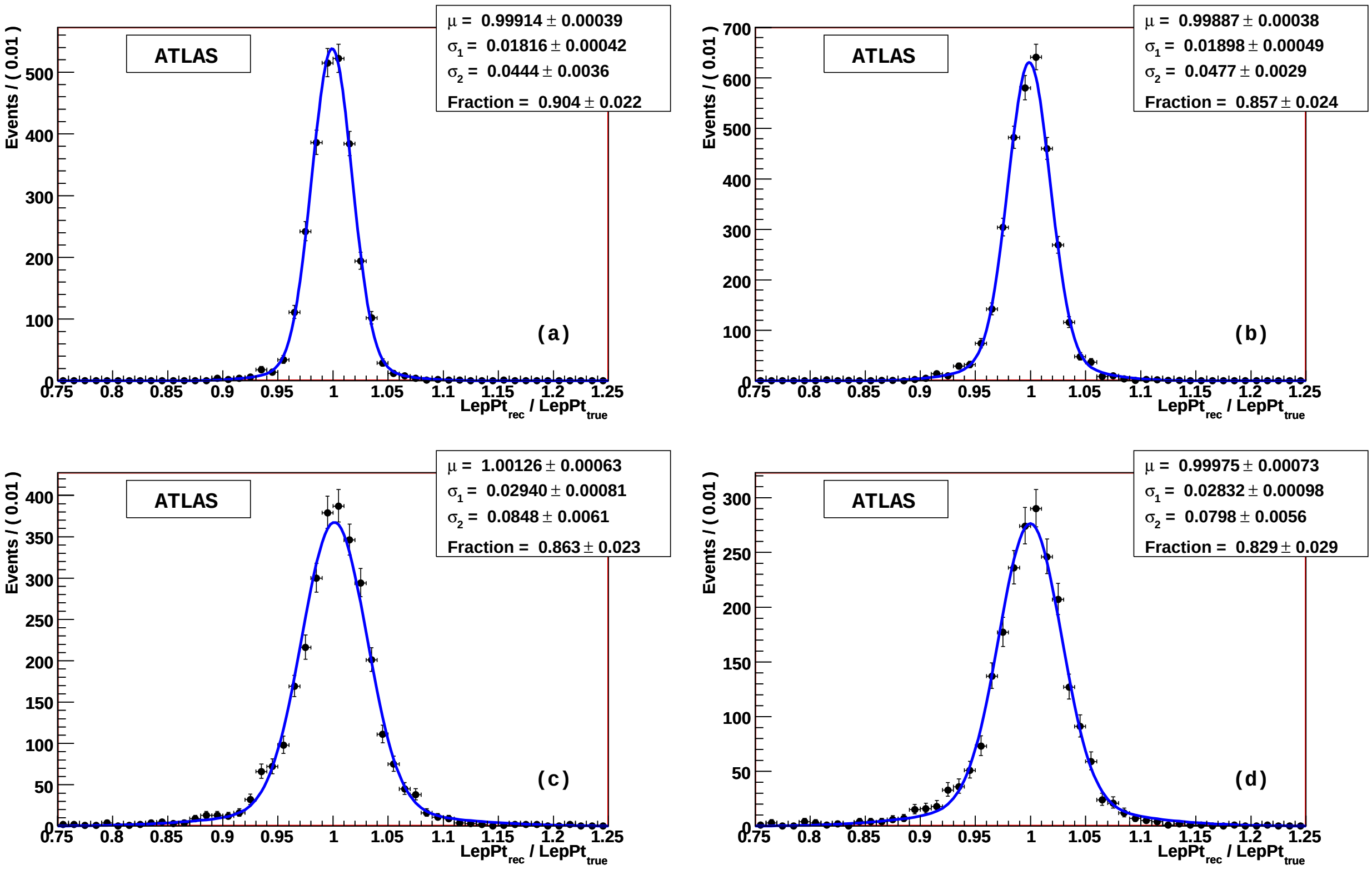


Figure 12: Distributions of transverse momentum ratios $p_T^{\mathrm{rec}}/p_T^{\mathrm{true}}$ for muons from $W$ ((a) and (c)) and $Z$ ((b) and (d)) decays at $|\eta| < 0.36$ ((a) and (b)) and $1.79 < |\eta| < 2.14$ ((c) and (d)) in the momentum range 35-40 GeV.

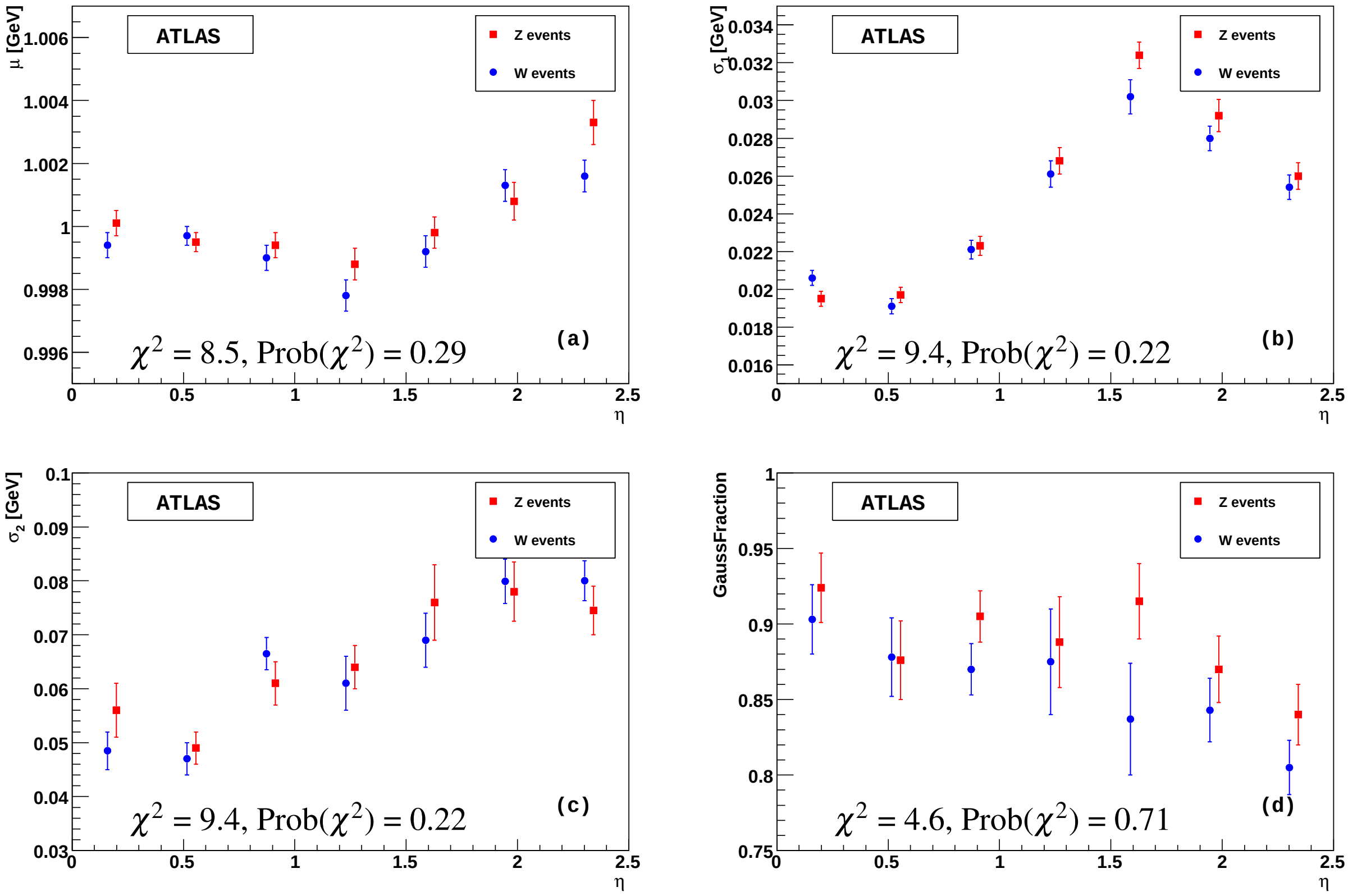


Figure 13: Comparison of fitted mean (a), core resolution (b), outlier resolution (c), and outlier fraction (d) between $W$ (circles) and $Z$ (squares) events for muons in the momentum range $30 < p_T < 35$ GeV in seven bins of $|\eta|$. The markers have been artificially shifted, left ($W$) and right ($Z$), in $|\eta|$ to increase readability.

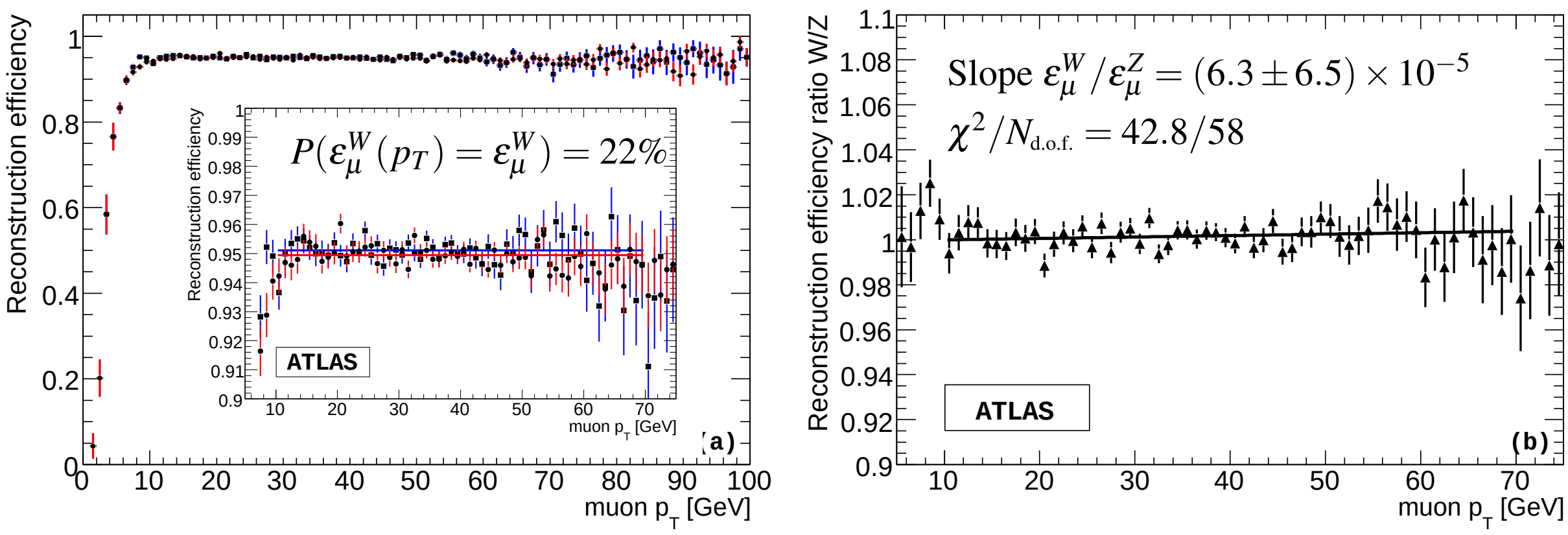


Figure 14: (a) Muon efficiency as a function of $p_T$ for $W$ (circles) and $Z$ (squares) events. Insert shows a zoomed view on the range of interest along with a constant fit to each of the graphs. The hypothesis of a flat probability in the range 10-70 GeV has been tested to be valid. (b) Muon efficiency ratio between $W$ and $Z$ events. No $p_T$ dependence is seen, and a linear fit yields a slope of $(6.3 \pm 6.5) \times 10^{-5}$, consistent with zero.

### 4.2 Fitting the transverse $W$ mass

Having tested the template fitting of the $p_T^\ell$ distribution, we now move to the $m_T^W$ distribution. In addition to the lepton residuals, this fit requires residuals for the missing momentum. Since the missing momentum is a transverse quantity, it does not depend on the $\eta$ of the lepton(s) in the event. However, the detector response depends on the total transverse hadronic activity $\Sigma E_T$ and the recoil momentum perpendicular to the direction of the leptons (cf. Figure 1).

To describe the response, $\Sigma E_T$ is divided in 10 bins in the range [0,200] GeV and recoil momentum into 10 bins in the range [0,40] GeV, each with an additional overflow bin. In each bin, the distributions are well described by two gaussian distributions with a common mean, as the missing momentum residuals are not expected to have any asymmetric tails.

To study the response, it is useful to project the $\not{E}_T$ momentum components onto two axes, defined event by event in the transverse plane. The parallel axis is defined such that the azimuthal angles between the leptons ($\ell$ and $\nu$) and this axis are minimized and equal, and the perpendicular axis is defined as perpendicular to the parallel axis. This reference system adapts better to the individual event topology than fixed axes. Examples of the residual distributions projected on these axes are shown in Figure 15.

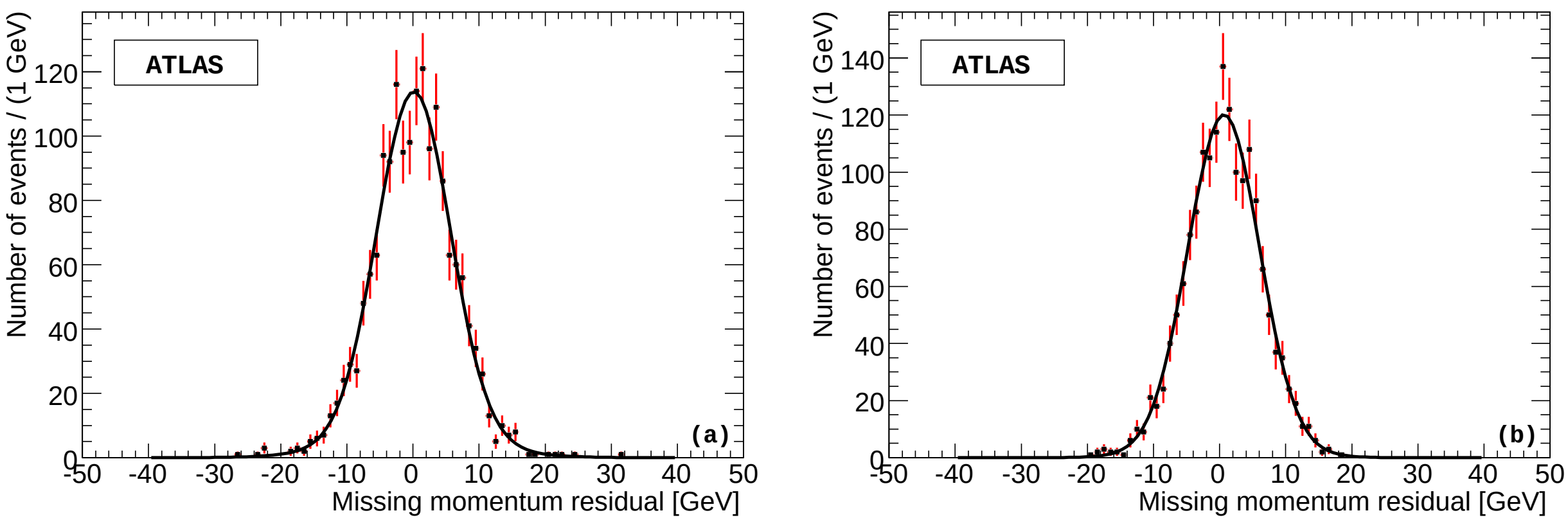


Figure 15: Distribution of $\not{E}_T$ residuals ${\not{E}_T}^{\text{rec}} - {\not{E}_T}^{\text{true}}$ for $W$ decays. The $\not{E}_T$ momentum components along the parallel (a) and perpendicular (b) axes are shown, for a recoil mometum in the range [8,12] GeV, and $\Sigma E_T$ in the range [20,30] GeV. The fitting function describes the distributions well.

Using the above modelling of the missing momentum response, we produce $m_T^W$ templates and test if the fit is unbiased or not. Unlike the lepton case, no additional efficiency curve has to be included, as the missing momentum is calculated for every event. As can be seen from Figure 16, the $m_T^W$ templates match the reconstructed distribution, and the fit is unbiased, giving a fitted value of $80.421 \pm 0.059$ GeV compared to an input value of 80.405 GeV.

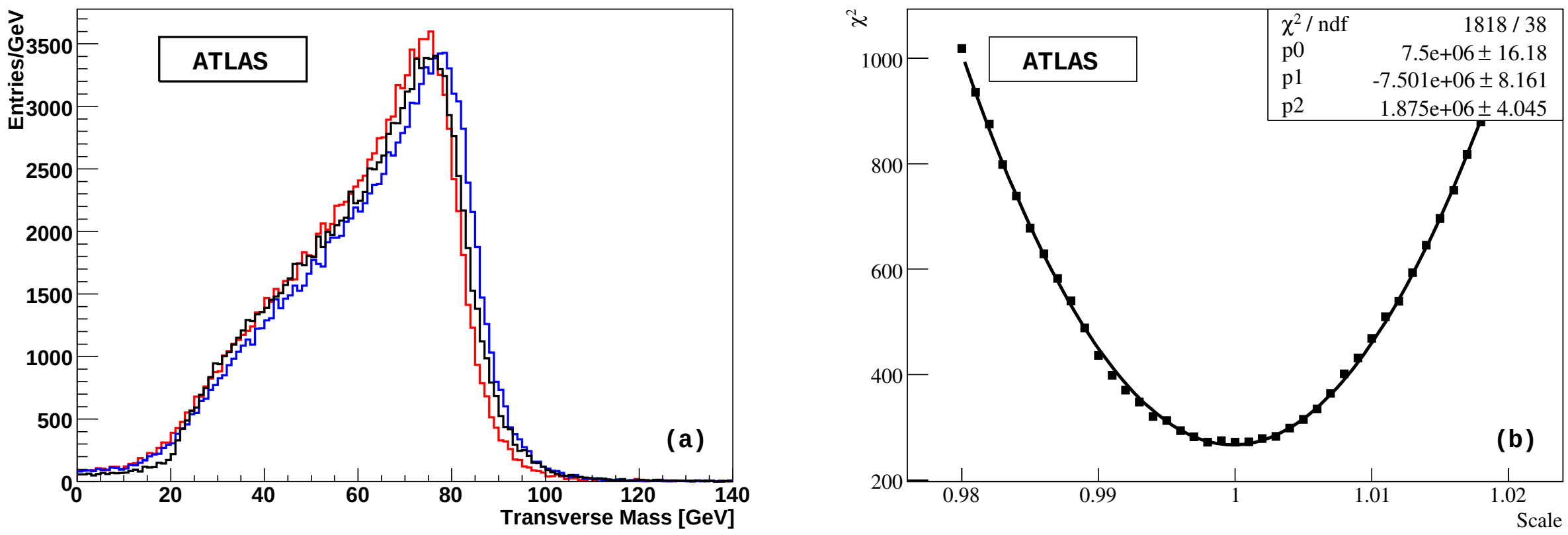


Figure 16: (a) Reconstructed $m_T^W$ distribution (middle curve) along with templates produced with the $W$ mass hypothesis 78.792 GeV (left curve) and 82.008 GeV (right curve), before any kinematic selections. (b) $\chi^2$ value of fitting templates to the reconstructed distribution as a function of the template's (fraction of) $W$ mass hypothesis (compared to the nominal mass). The fit yields $80.421 \pm 0.059$ GeV in agreement with the input value of 80.405 GeV.

### 4.3 Fitting the transverse $W$ mass using the $Z$ events for calibration

The dependence of the $m_W^{fit}$ on the relative recoil scale and resolution uncertainty was determined to be [15]:

$$\partial m_W / \partial \alpha_{recoil} = -200 \text{ MeV}/\%, \quad \partial m_W / \partial \sigma_{recoil} = -25 \text{ MeV}/\%. \tag{6}$$

These parameters can again be measured on $Z$ events. To test the portability from the $W$ to the $Z$ of the $m_T^W$ fit, we model the missing momentum using the $Z$ events and compare this to the one obtained from the $W$ events. However, unlike the lepton case, the detector response is not exactly the same for $W$ and $Z$ events. The difference is caused by the $\not{E}_T$ reconstruction algorithm, which in its current state does not correctly subtract lepton calorimetric signals from the hadronic recoil. This results in a difference between $W$ events where only one lepton is present, and $Z$ events containing two leptons.

To illustrate this point, consider again the parallel and perpendicular axes defined in the previous section. The residuals of the recoil momentum components are projected on both axes, and the response in $W$ and $Z$ events is compared, cf. Figure 17. Along the parallel axis, the average difference between the residuals is $\Delta_{W-Z} = 17 \pm 35$ MeV. Along the perpendicular axis, the difference is larger, $\Delta_{W-Z} = 1964 \pm 35$ MeV. Using a Z-based calibration in the W mass fit is thus expected to be biased; performing this exercise indeed yields $m_W^{fit} = 79.752 \pm 0.062$ GeV compared to the input value of 80.405 GeV. The Z based recoil calibration is thus not exploitable at present.

Instead, we assume that the needed improvements to the $\not{E}_T$ reconstruction algorithm will be done in time for the measurement, providing equal response between $W$ and $Z$ events. The statistical sensitivity based on the Z-based calibration is about 50 MeV for 15 pb$^{-1}$, serving as a lower bound. Given the present uncertainties and the performance of similar analysis [16], we assume that the *in situ* calibration can be performed with a precision of 1 % with an associated uncertainty $\delta m_W = 200$ MeV, according to Equation 6. The effect of pile-up on the missing momentum has not been studied.

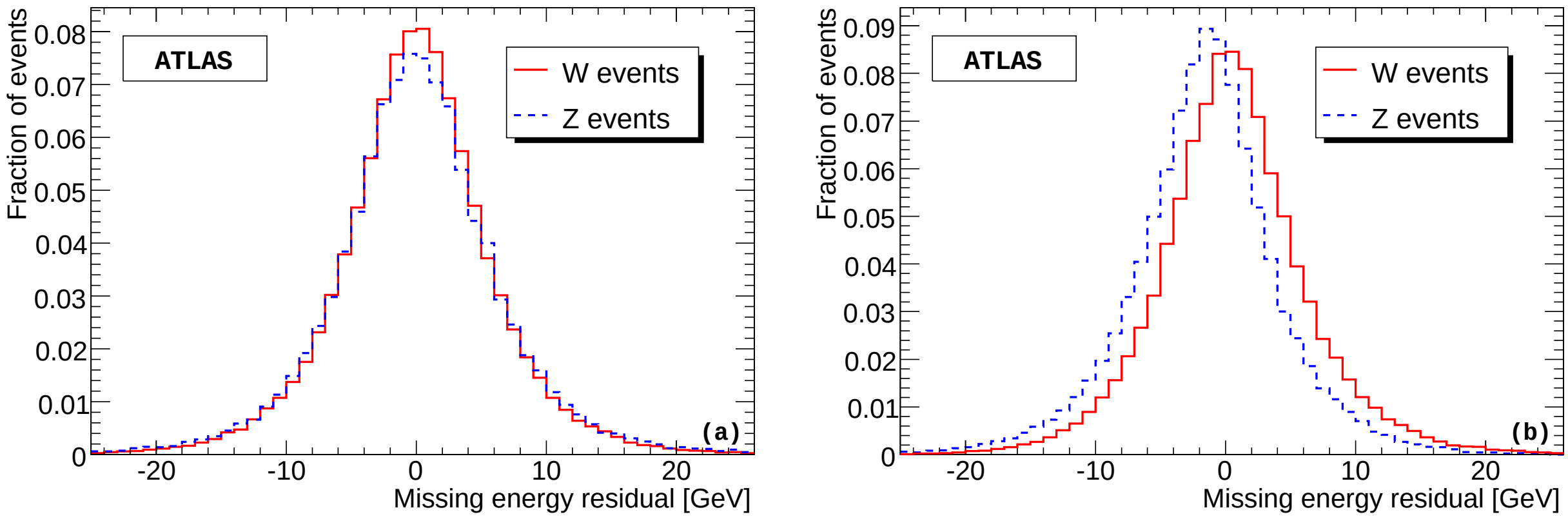


Figure 17: Distribution of $\not{E}_T$ residuals $\not{E}_T{}^{\text{rec}} - \not{E}_T{}^{\text{true}}$ parallel (a) and perpendicular (b) to the lepton axis for $W$ and $Z$ decays.

## 5 Statistical uncertainty as a function of fitting range

As previously stated, the sensitivity to the $W$ mass comes from the Jacobian edge in the fitting distribution. Generally the Jacobian edge is slightly sharper for the $m_T^W$ distribution (see Figure 3), yielding a smaller statistical uncertainty. To test the influence of the fitting range, three different fitting ranges has been tested for the $p_T^\ell$ and $m_T^W$ distributions for the $W \to \mu\nu$ sample. Since the typical $m_T^W$ values are twice as large as the $p_T^\ell$ values, the range size has been chosen accordingly. The result can be seen in Table 4.

| Transverse lepton momentum, $p_T^\ell$ | | Transverse $W$ mass, $m_T^W$ | |
|---|---|---|---|
| Fitting range [GeV] | $\sigma_{\text{stat.}}$ [MeV] | Fitting range [GeV] | $\sigma_{\text{stat.}}$ [MeV] |
| 10-80 | 87 | 20-160 | 54 |
| 20-70 | 93 | 40-140 | 55 |
| 30-60 | 106 | 60-120 | 57 |

Table 4: Statistical uncertainty as a function of fitting range for $p_T^\ell$ and $m_T^W$ fits. The uncertainties in the $p_T^\ell$ fit are larger, because the Jacobian edge is less sharp than in the $m_T^W$ fit (see Figure 3).

As can be seen from Table 4, the $p_T^\ell$ statistical uncertainty changes of about 30% with fitting range, while the $m_T^W$ statistical uncertainty is essentially insensitive to the range, and generally somewhat lower as expected. Considering that most systematic effects (such as backgrounds, electron calibration and efficiency, etc.) are largest at low momenta, a loss in $p_T^\ell$ statistical uncertainty will be countered by a gain in systematic uncertainty. For the $m_T^W$ fit a narrow range is surely preferable. However, it is not possible to quantify this gain until all systematic uncertainties have been calculated.

## 6 Lepton performance determination *in situ*

In this section we review algorithms to calibrate the lepton response using $Z$ events, and feed back the results to the $m_W$ fit.

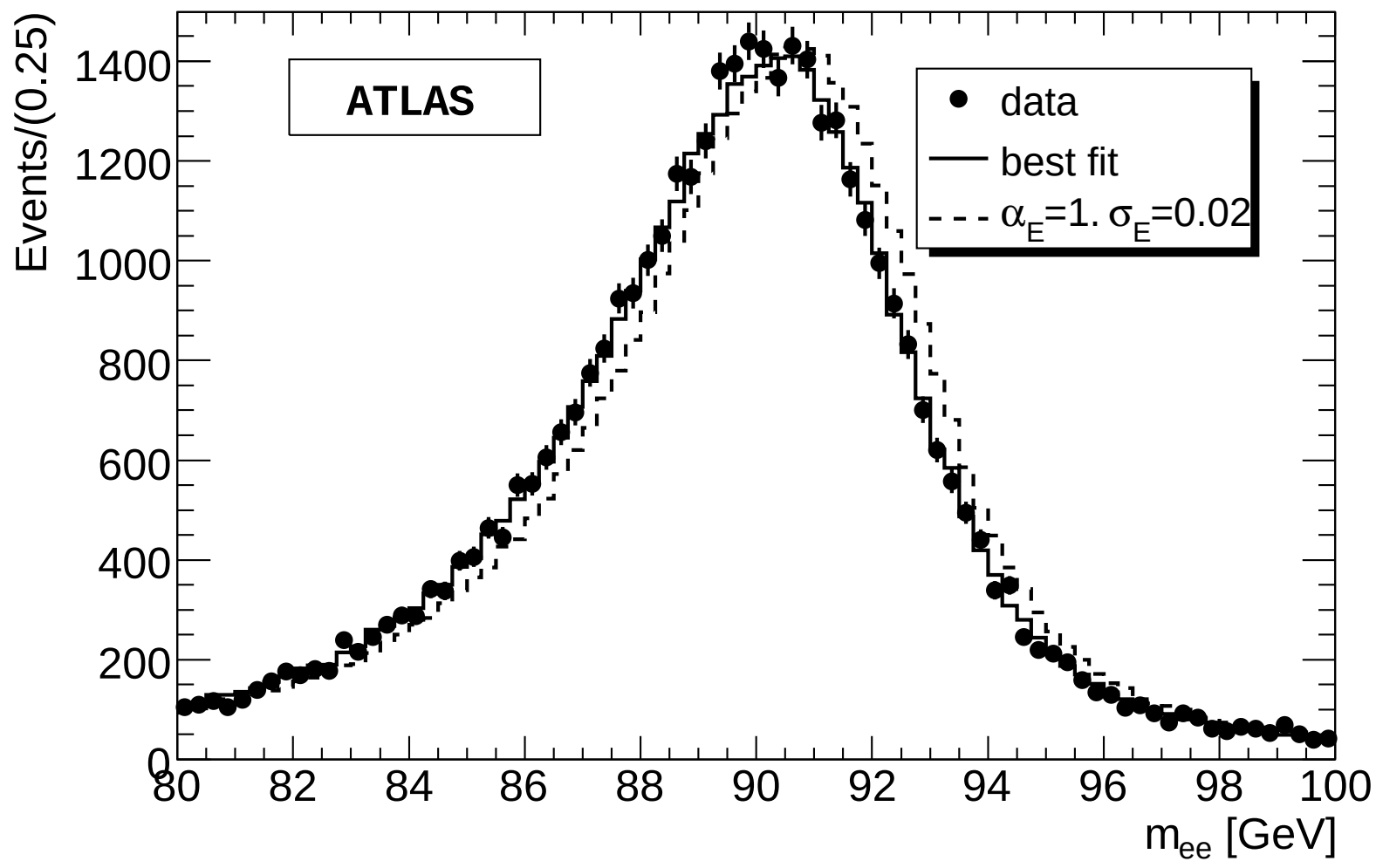


Figure 18: Fully simulated $Z$ pseudo-data (dots with error bars), compared to an example $Z$ resonance template with $\alpha_E = 1$, $\sigma_E = 0.02$, and to the best fit.

## 6.1 Average scale and resolution

We first perform a global scale analysis, to verify whether neglecting possible non linearities in the response can be expected to induce a significant bias. We restrain ourselves to the electron channel. Fixing the non gaussian tail parameters to $n = 0.8$ and $\beta = 5$, as expected from the studies performed in Section 3.3, we produce templates of the $Z$ resonance by varying the electron scale and resolution. These response parameters are applied to generator level electrons as before.

The templates are then fitted to the fully simulated $Z$ peak. A very good fit is obtained, as shown in Figure 18. An average scaling factor of $\alpha_E = 0.9958 \pm 0.0003$, and an average relative resolution of $\sigma_E = 0.0207 \pm 0.0003$ provide a statisfactory description of the resonance. The precision of the fit corresponds to the complete $Z$ event sample, i.e. $\mathscr{L} = 200$ pb$^{-1}$. Scaling to our default luminosity $\mathscr{L} = 15$ pb$^{-1}$, the precision becomes $\delta\alpha_E = 0.0013$ and $\delta\sigma_E = 0.0013$.

## 6.2 Differential scale and resolution

The calibration uses a (large) sample of reconstructed $Z \to \ell\ell$ events and a corresponding simulated sample (representing our knowledge of the $Z$ lineshape). For each event the two leptons are assigned to bins $i$ and $j$ (choosing $i \geq j$) according to energy/momentum, $\eta$, and/or $\phi$. Based on the lepton bins, events are divided into categories $(i, j)$, as shown in Figure 19.

For each category $(i, j)$, the reconstructed sample is compared to the known $Z$ lineshape (obtained from the corresponding simulated sample), and a $Z$ mass resolution function $R_{ij}$ is obtained from requiring that its convolution with the theoretical lineshape matches the reconstructed distribution (see Equation 7). Each of these $Z$ mass resolutions $R_{ij}$ is the direct result of combining two lepton

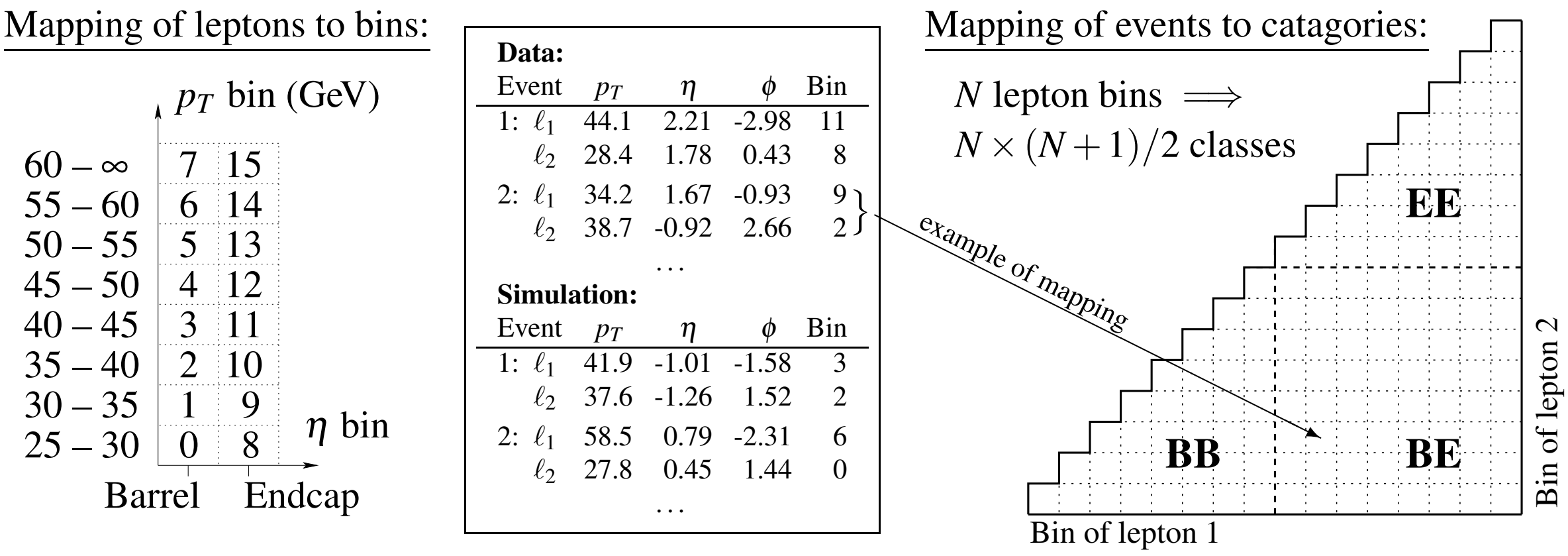


Figure 19: Illustration of scheme to categorize $Z \rightarrow \ell\ell$ events. Each lepton is assigned a bin according to its $p_T$ and $|\eta|$, here eight $p_T$ bins and two $|\eta|$ bins (barrel (**B**) and endcap (**E**)) thus $8 \times 2 = 16$ bins total, as demonstrated (left). $Z \rightarrow \ell\ell$ events from data/simulation (middle box) are then divided into categories (squares right) according to the reconstructed/truth $p_T$ and $|\eta|$ bin of both leptons.

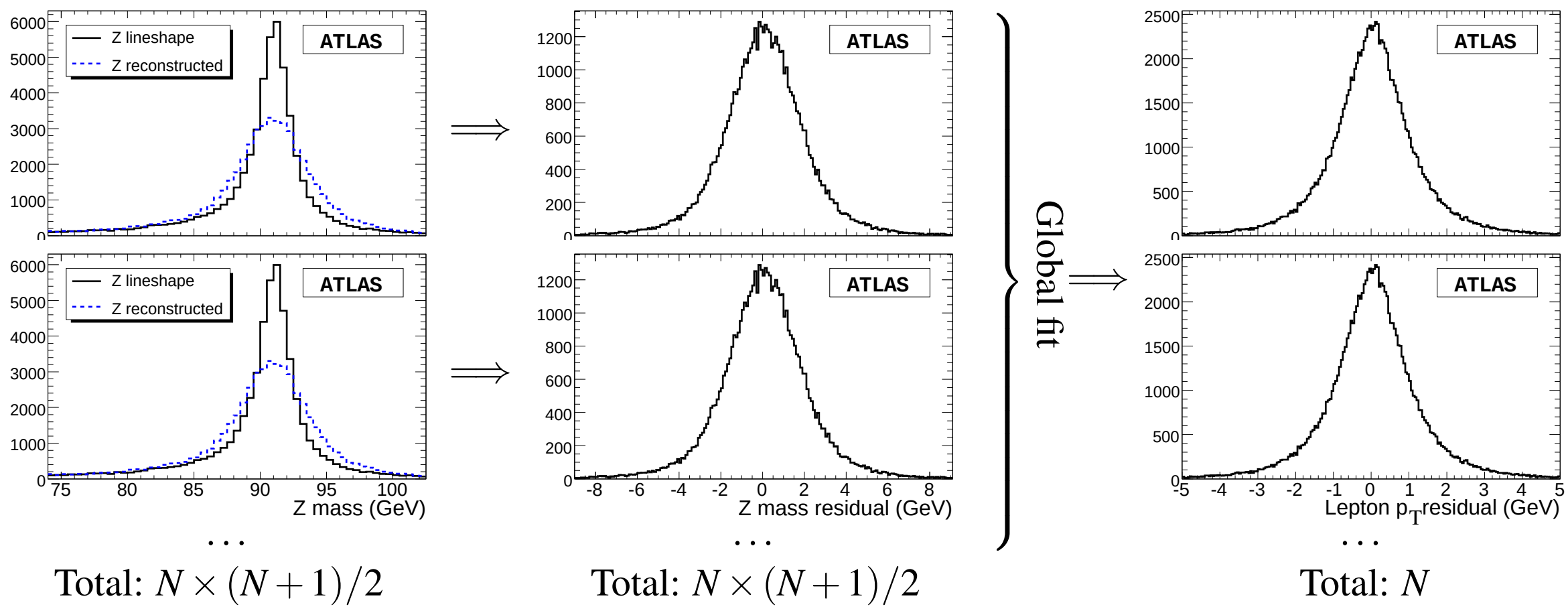


Figure 20: For each category a $Z$ mass resolution function (middle) is determined from folding it with the simulated distribution to match the reconstructed one (left). The lepton bias and resolution parameters are determined for each of the 16 lepton bins, by globally fitting the $16 \times 17/2 = 136$ $Z$ mass resolutions, which each is a result of two individual lepton resolutions (right).

momentum resolutions $R_i$ and $R_j$:

$$f(m_Z)_{ij}^{\text{Reco}} = f(m_Z)_{ij}^{\text{Truth}} \otimes R_{ij}, \qquad R_{ij} = R_i \otimes R_j \tag{7}$$

The complicated lepton scale and resolution calibration can thus be split into two parts, which both saves computing time and allows for intermediate checks and changes.

Given $N$ lepton bins and thus lepton resolution functions to determine, there are $N \times (N+1)/2$ $Z$ mass resolution functions, and thus the overconstrained system can be solved by a global $\chi^2$ fit. This calibration procedure is illustrated in Figure 20, and allows for a determination of the detector response for all combinations of $p_T$ and $\eta$.

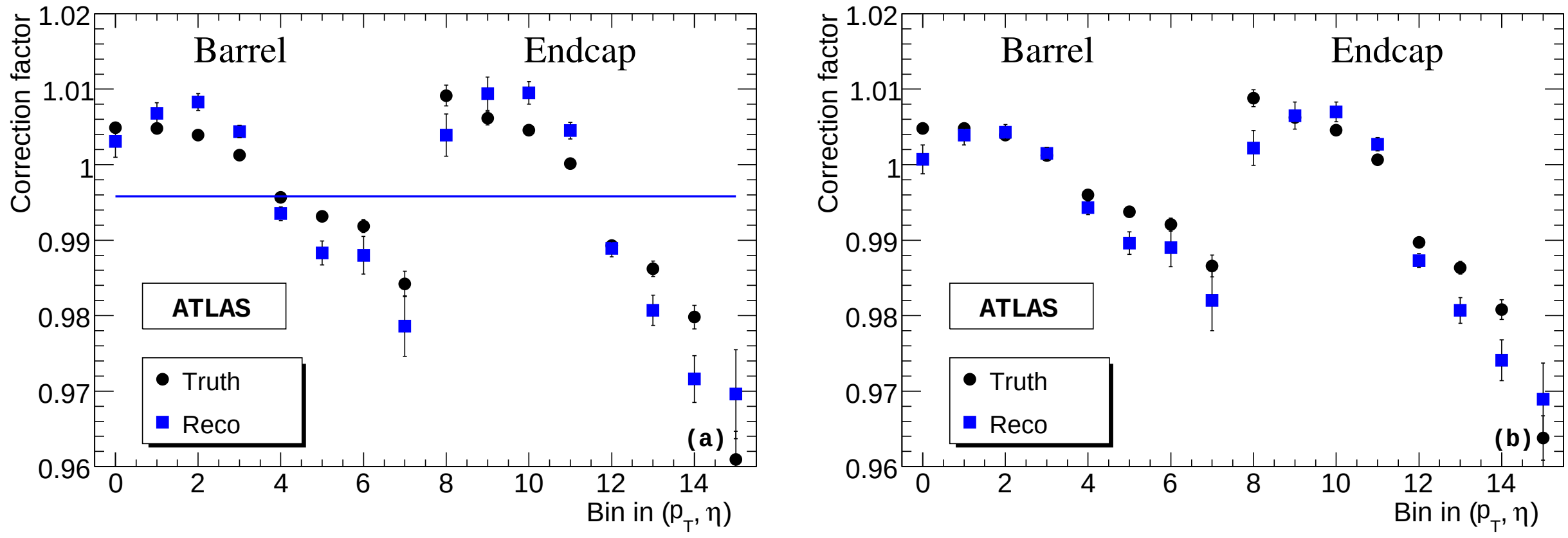


Figure 21: Lepton scale constants for electrons (a) and muons (b) as obtained from simplified calibration to the $Z$ peak (squares) and from truth (circles) using full $Z$ samples (see Table 1). The first eight bins are scale constants for leptons of increasing $p_T$ reconstructed in the barrel, while the last eight are for those in the endcap (see text). The result is in good agreement with average scale of $0.9958 \pm 0.0003$ (indicated by the line in plot (a)) found in Section 6.1.

A simplified version of the above analysis has been performed, with the aim of obtaining only the scales. In this simplified procedure, the resolution functions $R$ in Equation 7 reduces to calibration constants. The result of the calibration using the full $Z$ samples (see Table 1) is shown for both electrons and muons in Figure 21, along with the scales obtained from the truth information.

As can be seen from the figure, the simplified calibration yields the correct behaviour of the scales in general. Some fluctuations around the expected values are observed induced by the above simplifications. Scaled to 15 $\text{pb}^{-1}$, the uncertainties are to be increased by a factor 3.8 and 2.4 for electrons and muons, respectively.

### 6.3 Lepton reconstruction efficiency

The $p_T^\ell$ and $m_T^W$ distributions of $W$ events are also influenced by any $p_T$ and $\eta$ dependence of the lepton reconstruction efficiency. Any difference between the data and the simulation used to produce the templates will induce a difference in the distribution and cause a bias in the $W$ mass fit.

Though the method to obtain the (differential) efficiency is conceptually the same for electrons and muons, the two cases are different in that electron reconstruction is generally more dependent on $p_T$ and $\eta$ than the muon reconstruction. For this reason, we have chosen to consider the electron reconstruction efficiency in the following. The electron reconstruction efficiency can be determined from the data with $Z$ events, using the so-called "tag and probe" method [10], which we briefly summarize here. Events are selected with one well-identified electron, and an additional high $p_T$ object. The invariant mass of these two objects is required to be within 10 GeV of the nominal $Z$ boson mass. Assuming that this selects $Z$ events with enough purity, the identification efficiency is then simply obtained by computing the fraction of events where the second object is indeed identified as an electron. The efficiency of the isolation criterion is obtained in a similar way.

The studies reported in [17] indicate that for an integrated luminosity of 100 pb$^{-1}$, the electron efficiency can be reconstructed with a precision of 1.5% in the range $20 < p_T^e < 70$ GeV. The uncertainty is statistically dominated. Scaling this number down to $\mathscr{L} = 15$ pb$^{-1}$, we anticipate a precision of 3.9%.

## 7 Background uncertainties

The leptonic $W$ channel does not suffer from large backgrounds, due to the high cross section and the cleanness of the signal. The backgrounds are mostly from similar vector boson decays, such as $W \to \tau(\to \ell\nu\nu)\nu$, $Z \to \ell\ell$ (missing one lepton), and $Z \to \tau(\to \ell\nu\nu)\tau$. With the good particle identification capabilities of the ATLAS detector [13], jet events will despite their large cross section not be dominant. The backgrounds from $t\bar{t}$ and $W^+W^-$ events are negligible. The backgrounds and the uncertainties in their sizes are estimated below.

**$W \to \tau(\to \ell\nu\nu)\nu$ events:** A large background (largest in the electron channel) is from $W \to \tau\nu$ events, where the $\tau$ decays into a lepton. This background is irreducible, as the final state is identical to the signal; however, its $p_T^\ell$ and $m_T^W$ distributions are generally below the fitting range, as both the lepton and missing momentum are much reduced, leaving only a tail into the fitting range. Though a quite significant background, its uncertainty is small, as only the $\tau \to \ell X$ branching ratio (1.0%) and the acceptance relative to the signal (2.5%) enter.

**$Z \to \ell\ell$ events:** Another large background (largest in the muon channel) comes from $Z \to ll$ events, where one lepton is either undetected or not identified. A loose lepton identification for the second lepton can reduce this background, and possibly further reduction can be obtained with a $Z$ veto, should the associated uncertainty still be significant. We do not apply this veto in the present analysis. The $Z \to \ell\ell$ background extends significantly into the $p_T^\ell$ and $m_T^W$ distributions, except in the $m_T^W$ electron channel, where the missed electron cluster still reduces the missing momentum, effectively lowering the apparent $m_T^W$. The size of this background has uncertainties from the $W$ to $Z$ cross section ratio $R_{WZ}$ (1.8%), $Z$ veto efficiency (2.0%), and the acceptance (2.5%).

**$Z \to \tau(\to \ell\nu\nu)\tau$ events:** A small background origins from $Z \to \tau\tau$ events, where one $\tau$ decays leptonically, while the other is not identified. While the cross section for such a process is small, it can fake missing momentum. The largest uncertainty in the size of this background comes from the $\tau$ detector response (5.0%), along with cross section ratio $R_{WZ}$ (1.8%), and acceptance (2.5%).

**Jet events:** This background is not studied here, because it can not be evaluated reliably using simulation only. Relying on Tevatron experience [16], we will assume it is small and it will not be discussed in this note.

**Impact of backgrounds:** If the size and shape of the backgrounds were perfectly known, then they would not affect the $W$ mass measurement, as they could be included in the templates. It is thus the uncertainty on the size and shape of the backgrounds, in the fitting range of the $p_T^\ell$ and $m_T^W$ spectra, which gives rise to systematic errors. The uncertainties on the size of backgrounds arise from uncertainties relative to those of the signal events in cross sections, branching ratios and acceptances. These

| Electron channel | | | Muon channel | | |
|---|---|---|---|---|---|
| Process | Evts / 15pb$^{-1}$ | Fraction [%] | Process | Evts / 15pb$^{-1}$ | Fraction [%] |
| $W \to e\nu$ | 45468 | 97.8 | $W \to \mu\nu$ | 83263 | 93.9 |
| $W \to \tau\nu$ | 666 | 1.4 | $W \to \tau\nu$ | 1238 | 1.4 |
| $Z \to ee$ | 305 | 0.7 | $Z \to \mu\mu$ | 3483 | 3.9 |
| $Z \to \tau\tau$ | 30 | 0.1 | $Z \to \tau\tau$ | 153 | 0.2 |

Table 5: Signal and expected backgrounds in 15 pb$^{-1}$ after the event selection described in Section 2.2 and in the $p_T^{\ell}$ range [30,60] GeV.

are obtained from PDG [1] and efficiency studies (see Section 6.3).

The background shapes are determined from simulation. They are essentially unaffected by variations in the production, decay and resolution model. For jet events background, which was ignored in the present study, both normalisation and shape will have to be measured directly from the data.

We first assess the overall impact of the backgrounds. The backgrounds remaining after the selection described in Section 2.2 are given in Table 5, and the $p_T$ spectrum of each process is shown in Figure 22.

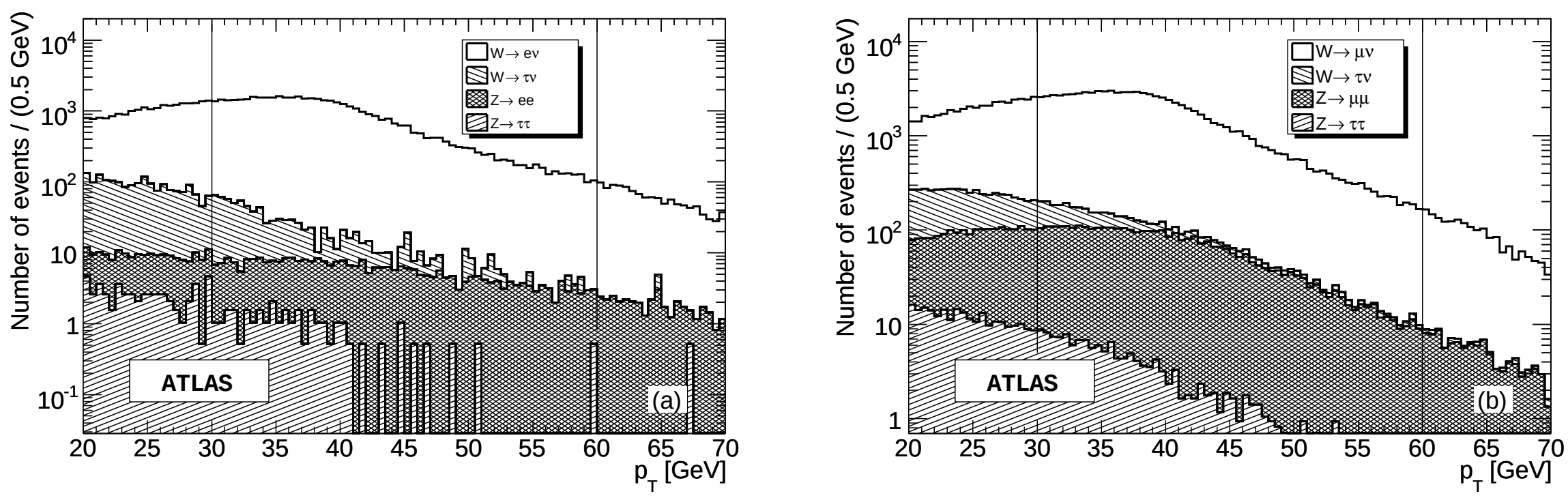


Figure 22: $p_T$ distribution for signal and backgrounds for electrons (a) and muons (b) after the selections described in Section 2.2. The $p_T$ range used in the mass fit is $30 < p_T < 60$ GeV.

Ignoring the background altogether in the templates leads to a bias $\delta m_W = -10$ MeV. This is however the result of a conspiracy : the $W \to \tau\nu$ background alone gives a bias of $-80$ MeV, while the $Z \to \ell\ell$ background gives a bias of $+70$ MeV; both sources of background can vary independently within the uncertainties given above. The other backgrounds have negligible impact. We thus estimate the bias per percent relative error on the background normalization (checked to scale linearly with the size of the background) to be:

$$\partial m_W / \partial N_{\tau\nu-\mathrm{bkg}} = -0.8\ \mathrm{MeV}/\%, \quad (8)$$

$$\partial m_W / \partial N_{\ell\ell-\mathrm{bkg}} = 0.7\ \mathrm{MeV}/\%. \quad (9)$$

We remind that the above does not include the jet events study.

# 8 Summary of uncertainties for $\mathscr{L} = 15\ \mathrm{pb}^{-1}$

The response parameters determined *in situ* using $Z$ events (cf. Section 6.1) are used to produce templates of the $p_T^\ell$ spectrum in $W$ events, as shown in Figure 23, for the electron channel. The resulting fit yields $m_W = 80.466 \pm 0.110$ GeV, with no bias with respect to the true value. This results shows that for $\mathscr{L} = 15\ \mathrm{pb}^{-1}$, propagating a global scale determined on $Z$ events does not induce a significant bias in the analysis. Given that the global scale calibration had a precision of 0.13% and Equation 3, the scale-induced systematic uncertainty is $\delta m_W(\alpha_E) = 110$ MeV. Likewise, the resolution uncertainty contributes $\delta m_W(\sigma_E) = 5$ MeV.

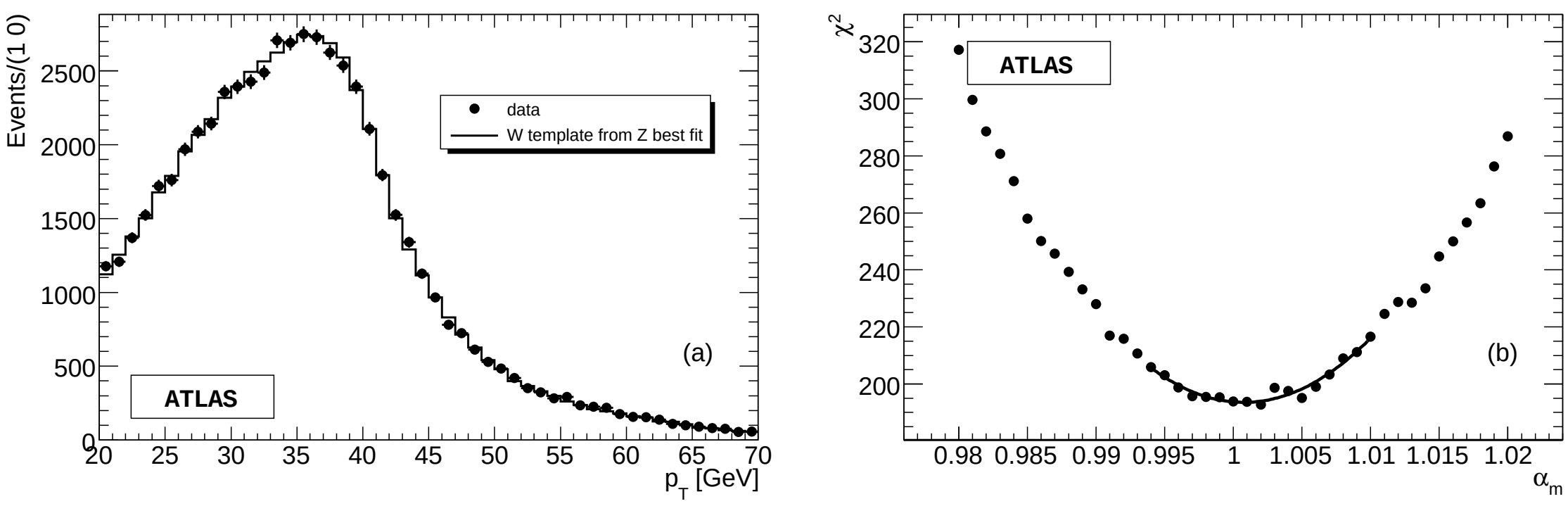


Figure 23: (a) $p_T^\ell$spectrum from fully simulated $W$ decays (dots with error bars), and $p_T^\ell$template obtained assuming the $Z$ based scale and resolution and the true value of $m_W$. (b) Template fit to $m_W$.

We did not attempt to control the non gaussian tails *in situ*. We assume that this contribution can be determined to 5% with 15 $\mathrm{pb}^{-1}$, yielding a contribution of 28 MeV to the systematic uncertainty. As can be seen by comparing Figure 4 and Figure 12, the effect is smaller in the muon channel.

The expected precision of the *in situ* efficiency measurement (Section 6.3) and Equation 5 imply a systematic uncertainty on the fit result of about $\delta m_W = 14$ MeV. This result holds for the electron channel. Given the flatness of the muon reconstruction efficiency, the systematic uncertainty in the muon channel is expected to be much smaller.

It was not attempted to determine the recoil calibration *in situ*. We assume that this calibration can be performed to 1%, yielding a systematic contribution of 200 MeV from Equation 6. This contribution affects the transverse mass analysis only.

The background uncertainty is discussed in the electron channel, and assumed identical in the muon channel. The $Z \to \tau\tau$ background is negligible. Given the current knowledge of the $W$ and $\tau$ branching ratios [1], a realistic estimate of the $\tau$ background uncertainty is 2.5%. Injecting the nominal background in the templates, then varying them within this range, we find a systematic uncertainty $\delta m_W(\mathrm{bkg}) = 2$ MeV. Similarly, the contribution from the $Z \to ee$ background is found to be 2 MeV.

The total background contribution to the systematic uncertainty is thus expected to be 3 MeV.

A detailed discussion of theoretical uncertainties is provided in [15]. At the start-up of LHC, the dominant theoretical contribution comes from the proton parton density functions, which contribute 25 MeV.

In summary, for an integrated luminosity of 15 pb$^{-1}$, we found that the analysis of the $p_T^\ell$ spectrum gives a statistical sensitivity of about $\delta m_W = 110$ MeV per channel, while the transverse mass provides $\delta m_W = 60$ MeV. These numbers have experimental systematic uncertainties of 114 MeV in the $p_T^\ell$ analysis, and 230 MeV in the transverse mass analysis. The uncertainty on the lepton and recoil scales dominate these numbers. Finally, the uncertainty from PDFs is about 25 MeV and comparatively small. Our numbers are summarized in Table 6.

| Method | $p_T(e)$ [MeV] | $p_T(\mu)$ [MeV] | $M_T(e)$ [MeV] | $M_T(\mu)$ [MeV] |
|---|---|---|---|---|
| $\delta m_W$ (stat) | 120 | 106 | 61 | 57 |
| $\delta m_W$ ($\alpha_E$) | 110 | 110 | 110 | 110 |
| $\delta m_W$ ($\sigma_E$) | 5 | 5 | 5 | 5 |
| $\delta m_W$ (tails) | 28 | $< 28$ | 28 | $< 28$ |
| $\delta m_W$ ($\varepsilon$) | 14 | – | 14 | – |
| $\delta m_W$ (recoil) | – | – | 200 | 200 |
| $\delta m_W$ (bkg) | 3 | 3 | 3 | 3 |
| $\delta m_W$ (exp) | 114 | 114 | 230 | 230 |
| $\delta m_W$ (PDF) | 25 | 25 | 25 | 25 |
| Total | 167 | 158 | 239 | 238 |

Table 6: Summary of contributions to $\delta m_W$, for the different fitting methods described in the text. From top to bottom: statistical uncertainty; systematic uncertainties related to the absolute scale, resolution, non gaussian tails, reconstruction efficiency, recoil calibration, and backgrounds; total experimental systematic uncertainty; uncertainty from PDF; total systematic uncertainty.

## 9 Conclusions

This note presents a first attempt towards $m_W$ fits, confronting distributions obtained from fully simulated $W \to e\nu, \mu\nu$ events to templates produced using a simplified detector model.

The simplified detector model used here relies on empirical functions to describe the energy and momentum scale, resolution, and non-gaussian tails, as well as the lepton selection efficiency. These functions are first determined, as a function of $p_T$ and $\eta$, using simulated $W$ events, i.e. the signal itself; the resulting fit, unbiased, provides a technical validation of the method.

In a second step, the model parameters are used as extracted *in situ*, from the analysis of $Z$ events. The $Z$ peak provides the detector scale and resolution; the tag-and-probe method [10], applied to its decay leptons, is used to estimate the efficiency. A fit of the $Z$-based templates to the $W$ data again

shows no significant bias.

Our current approach can thus be considered valid, within a statistical sensitivity of about 60 MeV, for the transverse mass-based fit in the muon channel, and of 120 MeV for the transverse momentum-based fit in the electron channel.

This note has focused on experimental issues. Phenomenological uncertainties, related to QED, QCD and parton density functions are discussed in detail in [15].

# Forward-Backward Asymmetry in $pp \to Z/\gamma^* X \to e^+e^- X$ Events

**Abstract**
This paper describes a study on the measurement of the forward-backward asymmetry in $pp \to Z/\gamma^* X \to e^+e^- X$ events with the ATLAS detector for an integrated luminosity of 100 fb$^{-1}$. Such a measurement can be used to determine the effective weak mixing angle, $\sin^2\theta_{eff}^{lept}$. We will demonstrate that a very high accuracy on the weak mixing angle, $\delta\sin^2\theta_{eff}^{lept} = (1.5(\text{stat}) \pm 0.3(\text{exp}) \pm 2.4(\text{PDF})) \times 10^{-4}$, can be reached. This is possible due to the large cross-section for the production of $Z$ bosons at the LHC and by using electron reconstruction in the forward calorimeters of ATLAS ($2.5 < |\eta| < 4.9$).

## 1 Introduction

The forward-backward asymmetry, $A_{FB}$, measurement is one of the important precision measurements that can be done at the Large Hadron Collider (LHC). It will improve the knowledge of Standard Model parameters and test the existence of physics beyond the Standard Model.

The $Z$ boson events in $pp$ collisions originate from the annihilation of valence quarks with sea antiquarks or from the annihilation of sea quarks with sea antiquarks. Since the valence quarks carry on average a larger momentum fraction than the sea quarks, the boost direction of the dilepton system can indicate the quark direction. However, dilepton events which originate from the annihilation of sea quarks with sea antiquarks do not contribute to the observed asymmetry.

$A_{FB}$ measurements with quarks and leptons at the $Z$ peak provide a precise determination of the weak mixing angle $\sin^2\theta_{eff}^{lept}$. The weak mixing angle is an important parameter in the electroweak theory that describes the mixing between weak and electromagnetic interactions. In the global fit of the Standard Model, the weak mixing angle has an impact on indirect constraints on Higgs mass.

With the experimental capabilities of the ATLAS experiment at the LHC and an expected integrated luminosity of up to 100 fb$^{-1}$ it becomes interesting to perform a study on the measurement of the asymmetry, $A_{FB}$. In order to improve the measurement precision, it will be necessary to detect leptons in the very forward pseudo-rapidity region, which favors electrons over muons at ATLAS. In 100 fb$^{-1}$ data around $1.5\times10^8$ $Z$ events will be produced, of which $\sim 5\times10^6$ decay to an electron-positron pair, providing the measurement of $A_{FB}$ (and $\sin^2\theta_{eff}^{lept}$) with a competitive precision to the current world average value [1], $\delta\sin^2\theta_{eff}^{lept} = 0.00016$.

This paper is organized as follows. In section 2 we present a short introduction of the theoretical aspects of the measurement. An overview of the ATLAS detector is given in section 3. In section 4 the electron measurement in ATLAS is presented, both in the central and forward regions the latter being necessary for a precise measurement of the forward-backward asymmetry. We then present the simulated data samples used in section 5 and the definition of the electron polar angle in section 6. The event selection cuts are discussed in section 7. The charge misidentification is given in section 8. The pile-up effect is discussed in section 9. The systematic uncertainties are summarized in section 10. The expected precision on the asymmetry around the $Z$ mass is shown in section 11.

## 2 Forward-backward asymmetry

In proton-proton collisions, $e^+e^-$ pairs are predominantly produced via the annihilation of a quark $q$ and an antiquark $\bar{q}$. In the Standard Model, quark-antiquark annihilations proceed via an intermediate [2, 3] $\gamma^*$ at low invariant mass $M_{(e^+e^-)}$ or via a $\gamma^*/Z$ interference at $M_{(e^+e^-)}$ around the $Z$ mass. The electroweak neutral current in the Standard Model lagrangian violates parity (due to the presence of vector and axial-vector couplings of the quarks and leptons to the $Z$-boson) and leads to an asymmetry in the polar emission angle of the electron in the rest frame of the electron-positron pair. The asymmetry can only be extracted with respect to the boost direction of the di-electron system since the quark direction is not known (see Sec. 6).

The differential cross-section for the production of an electron-positron pair in $q\bar{q}$ annihilation via the $s$-channel can be written, in the electron-positron rest frame [4, 5], as:

$$\frac{d\sigma}{d\cos\theta} = N_c[(1+\cos^2\theta)F_0(s) + 2\cos\theta F_1(s)] \tag{1}$$

where $\theta$ is the emission angle of the $e^-$ relative to the momentum vector of the quark in the $e^+e^-$ rest frame and s is the center-of-mass energy squared. $F_0(s)$, $F_1(s)$ are form factors and $N_c$ (1/3) is the colour factor:

$$F_0(s) = \frac{\pi\alpha^2}{2s}(q_q^2 q_l^2 + 2Re\chi(s) q_q q_l C_V^q C_V^l + |\chi(s)|^2((C_V^q)^2 + (C_A^q)^2)((C_V^l)^2 + (C_A^l)^2)) \tag{2}$$

$$F_1(s) = \frac{\pi\alpha^2}{2s}(2Re\chi(s) q_q q_l C_A^q C_A^l + |\chi(s)|^2 2C_V^q C_V^l 2C_A^q C_A^l), \tag{3}$$

with

$$\chi(s) = \frac{s}{s - M_Z^2 + is\Gamma_Z/M_Z} \tag{4}$$

where $q_{q,l}$ is the electric charge of the quark or lepton and $C_V$, $C_A$ are the vector and axial-vector coupling to the $Z$.

The angular dependence of the various terms is either $\cos\theta$ or (1+$\cos^2\theta$). Only the $(1+\cos^2\theta)$ term contributes to the total cross-section, as the $\cos\theta$ term integrates to zero. However, it induces a forward-backward asymmetry. Thus, the differential cross-section can be written in a simple expression:

$$\frac{1}{\sigma}\frac{d\sigma}{d\cos\theta} = [\frac{3}{8}(1+\cos^2\theta) + A_{FB}(s)\cos\theta] \tag{5}$$

where $\sigma = \frac{8}{3}F_0(s) \times N_c$ and

$$A_{FB}(s) = \frac{1}{\sigma}[\sigma(\cos\theta > 0) - \sigma(\cos\theta < 0)] \tag{6}$$

$$= \frac{3}{4}\frac{F_1(s)}{F_0(s)} \tag{7}$$

$$A_{FB}(s = M_Z^2) \sim \frac{3}{4}\frac{2C_V^q C_A^q}{(C_V^q)^2 + (C_A^q)^2}\frac{2C_V^l C_A^l}{(C_V^l)^2 + (C_A^l)^2} \tag{8}$$

The forward-backward asymmetry is commonly defined using the number of forward produced ($\cos\theta > 0$) events ($N_F$) and backward-produced ($\cos\theta < 0$) events ($N_B$)

$$A_{FB} = \frac{N_F - N_B}{N_F + N_B} \tag{9}$$

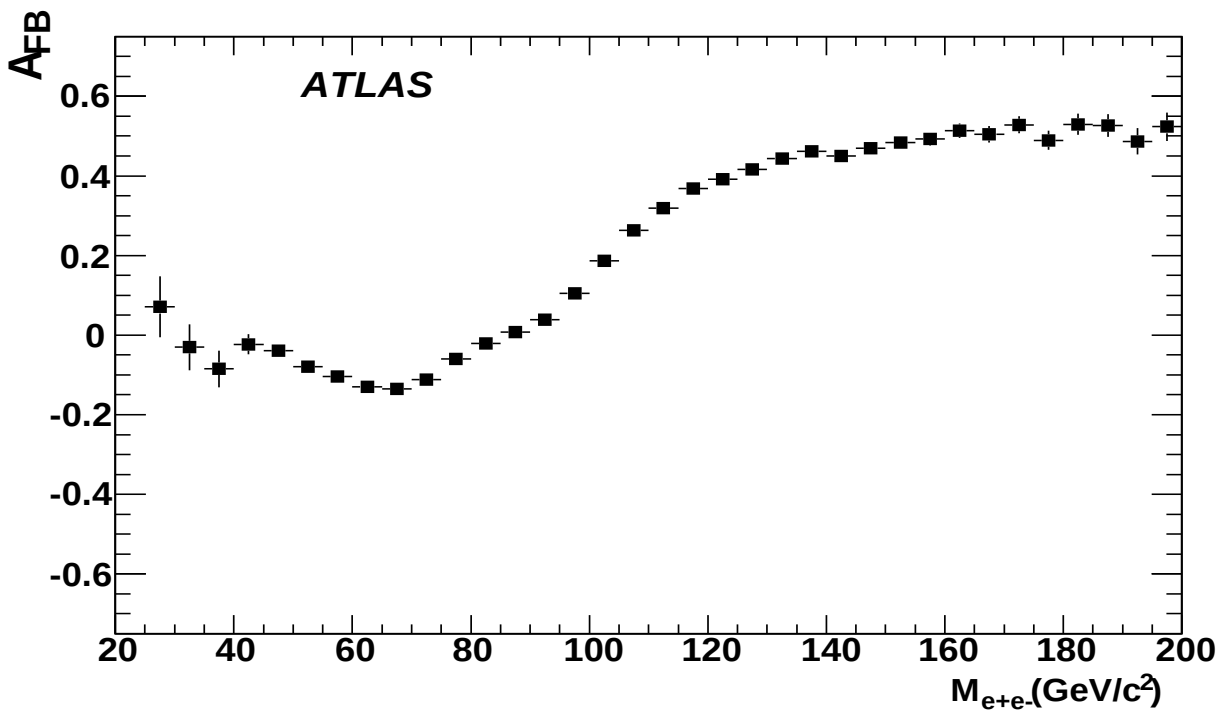


Figure 1: SM prediction of the forward-backward charge asymmetry $A_{FB}$ in the electron pair channel versus the di-electron invariant mass $M_{e^+e^-}$ for $|y_{e^+e^-}| > 1$ and for at least one electron in the central region ($|\eta| < 2.5$).

where

$$F = \int_0^1 \frac{d\sigma}{d\cos\theta} d\cos\theta, \quad B = \int_{-1}^0 \frac{d\sigma}{d\cos\theta} d\cos\theta \tag{10}$$

### 2.1 Dependence of $A_{FB}$ on $M_{e^+e^-}$

In Figure 1 we show the Standard Model prediction, using MRST PDF [6], of the forward-backward charge asymmetry as function of the di-electron invariant mass. At the $Z$-pole we see a small asymmetry (as expected) which is dominated by a small vector coupling in $Z \to e^+e^-$ with the dominant axial coupling. Around the $Z$ mass the asymmetry is linear with the weak mixing angle $\sin^2\theta^{lept}_{eff}$. It was estimated that (in a good approximation) the weak mixing angle can be determined from the measurement of the forward-backward asymmetry (this is the raw forward backward asymmetry measured at detector level) when averaged over the rapidity of the electron pair as follow [7, 8, 9]:

$$A_{FB} = b(a - \sin^2\theta^{lept}_{eff}) \tag{11}$$

the parameters $a$ and $b$ depend on the parton distribution functions (PDFs).

At large invariant mass, $A_{FB}$ is dominated by the properties of the interference between the propagators of the $\gamma^*$ and the $Z$ and is almost constant at a large positive value, close to 0.6, independent of the invariant mass.

## 3 Detector overview

The ATLAS detector has been described in detail in [10]. It consists of an inner tracking system, with pseudo-rapidity coverage of $|\eta| < 2.5$, inside a 2 T solenoidal magnetic field, followed by the calorimeters, and an outer muon spectrometer, with pseudo-rapidity coverage of $|\eta| < 2.7$, installed in a large toroidal magnet system. We briefly describe the parts of the detector relevant to this analysis.

Figure 2 shows a schematic transversal view of the ATLAS calorimeter system. It has been designed to be hermetic to $|\eta| < 4.9$ with a fine lateral and longitudinal segmentation. A liquid argon (LAr) sampling calorimeter in a barrel-endcap geometry provides the electromagnetic and hadronic calorimetry. At $|\eta| < 1.2$ the hadronic calorimetry is completed by a tile (Iron/scintillator) hadronic calorimeter.

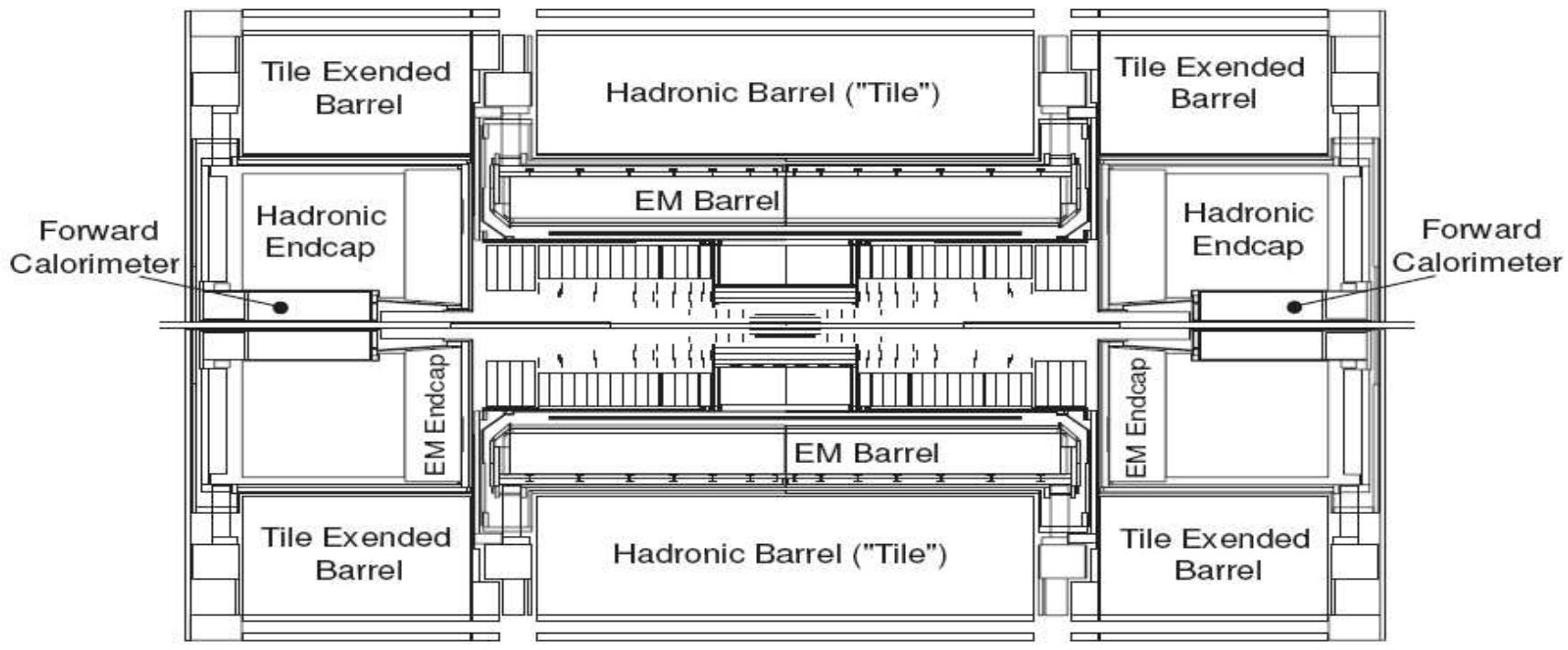


Figure 2: Schematic transversal view (r-z view) of the calorimeters in the ATLAS detector. The cylindrical coordinate system is used with the z axis along the proton-beam direction and r is the transverse coordinate.

The electromagnetic barrel calorimeter, covering a pseudo-rapidity range of $|\eta| < 1.475$, shares its cryostat with the superconducting solenoid, the calorimeter being behind the solenoid. Each electromagnetic calorimeter end-cap (EMEC) covers the pseudo-rapidity range 1.4-3.2. To correct for the energy lost in the material in front of the calorimeters, both end-caps and barrel are preceded with pre-sampler detectors in the region $|\eta| < 1.8$. The performance expected gives for the energy resolution a sampling term of $10\%/\sqrt{E(\text{GeV})}$ and a constant term better than 0.7%. The hadronic end-cap calorimeter (HEC) and the forward calorimeter (FCal) share the same cryostat with the EMEC and cover the pseudo-rapidity range 1.5-3.2 and 3.1-4.9 respectively. The design of the FCal was constrained by the high radiation level in the very forward region. It consists of three consecutive modules along the beam line: one electromagnetic module (FCal1) and two hadronic modules (FCal2 and FCal3). To optimize the resolution, copper was chosen as the absorber for FCal1, while tungsten was used in FCal2 and FCal3, to provide containment and minimize the lateral spread of hadronic showers.

The electron energy can be measured either in the central electromagnetic calorimeter ($|\eta| <2.5$) and/or in the forward calorimeters ($2.5 < |\eta| < 4.9$). The $|\eta| < 2.5$ region corresponds to the EM barrel and the EMEC outer wheel.

# 4 Electron identification

## 4.1 Central electrons

An electron candidate is reconstructed in the central calorimeters if there is a cluster with energy $E$ and a charged track matched to the cluster with the condition of $E/p < 7$ and $|\Delta\eta| < 0.05$ and $|\Delta\phi| < 0.1$. $\Delta\eta$ ($\Delta\phi$) is the difference between the $\eta$ ($\phi$) position of the track and the position of the cluster. The track is then extrapolated to the cluster. The energy of the electron is determined by the total energy it deposits in the EM calorimeter. The transverse momentum ($p_T$) of the electron is determined by the energy measured in the calorimeter and by the track angle at the nominal interaction point. The charge of the electron ($q_e$) is determined from the sign of the curvature of the track.

The development of electromagnetic and hadronic showers is quite different so that shower shape information can be used to differentiate between electrons and hadrons. Electrons deposit almost all their energy in the electromagnetic section of the calorimeter, while hadrons are typically much more

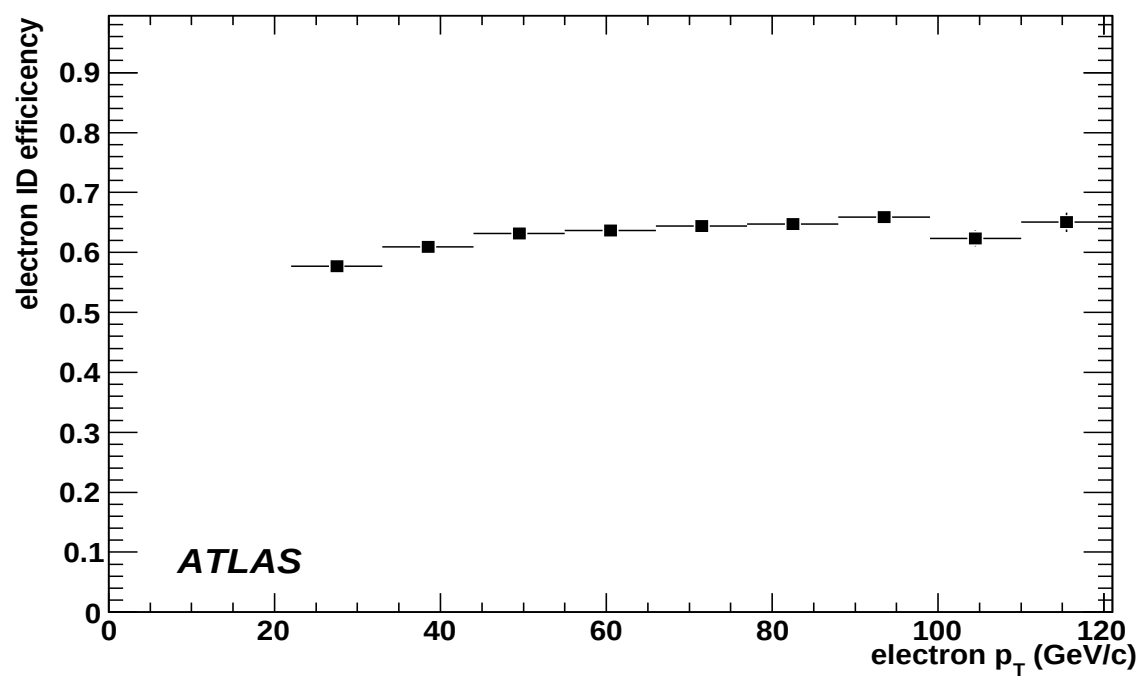


Figure 3: Central Electron ID efficiency vs electron $p_T$ for events with $|\eta| < 1.3$ and $|\eta| > 1.6$.

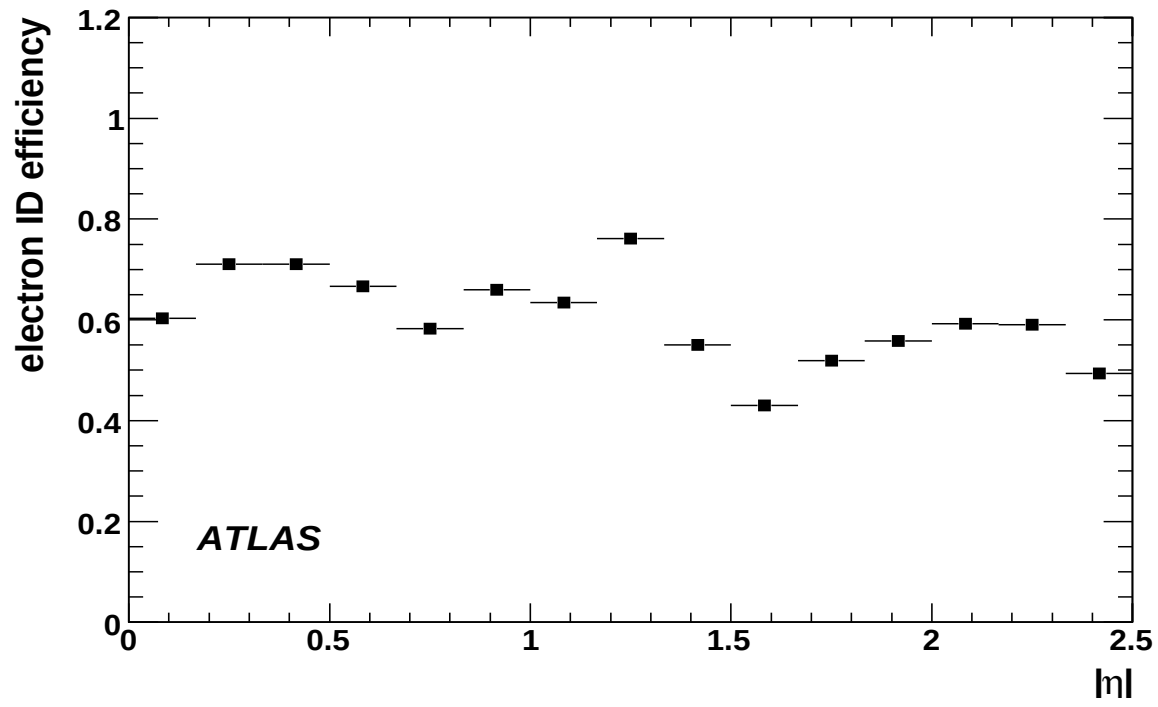


Figure 4: Central Electron ID efficiency vs electron $\eta$. The efficiency is integrated over $p_T > 20$ GeV.

penetrating. To obtain the best discrimination against hadrons, we use both longitudinal and transverse shower shapes. For example, for electrons passing tight identification criteria, the hadronic leakage and the lateral shower shape in the first and second sampling of the calorimeter are used. After the selections based on calorimeter information, we also take into account the inner detector information. We ask for a good track pointing to the calorimeter and fulfilling the consistency of the spatial and energy matching between the calorimeter and the inner detector.

Figure 3 and Figure 4 show the electron identification efficiency, 62.0±1.0 % on average, as a function of the electron transverse momentum and pseudo-rapidity $\eta$ using electrons (and positrons) from $Z \to e^+e^-$ events (the samples used are described in section 5). Tight cuts have been applied to identify the central electrons. We observe an expected increase in the efficiency as function of $p_T$. This dependence (in particular at low $p_T$) is due to the lower performance for low $p_T$ electrons. The drop in the efficiency versus $\eta$ happens at the transition region between barrel and end-cap calorimeters, $1.37 < |\eta| < 1.52$. A more detailed description of the electron identification in the central calorimeters is given elsewhere [11].

## 4.2 Forward electrons

In contrast to the central region, forward electron reconstruction can use only calorimeter information as the tracking system is limited to the central region ($|\eta| < 2.5$). In this case we can not distinguish between an electron, positron or photon. The electron candidate in the forward calorimeter[1] is reconstructed if there is a cluster with $E_T > 20$ GeV. The direction of the electron is defined by the barycenter of the cells belonging to the cluster in the calorimeter.

To discriminate between electron and hadron a multivariate analysis [12] is used. Variables are defined using cluster moments or a combination of them.

The cluster moment of degree $n$ for a variable $x$ is defined as:

$$\langle x^n \rangle = \frac{1}{E_{\text{norm}}} \times \sum_i E_i x_i^n, \tag{12}$$

where $E_{\text{norm}} = \sum_i E_i$, $E_i$ is the cell energy and $i$ is the cell index of the cluster.

[1] the EM forward calorimeters are the EMEC inner wheel ($2.5 < |\eta| < 3.2$) and the front compartment of the FCal. These EM calorimeters are completed by the HEC and the last 2 compartments of the FCal ($3.2 < |\eta| < 4.9$).

We use the following five discriminants in the electron ID:

- $r_i = |\,(\vec{x}_i - \vec{c}\,) \times \vec{u}\,|$:

  Electromagnetic and hadronic showers have different shapes in both the transverse and the longitudinal directions. To evaluate the shower shapes in the transverse direction we use the second moment of the distance $r_i$ of each cell $i$ ($\vec{x}_i$) to the shower axis ($\vec{c}$). $\vec{u}$ is the shower axis.

- $lat_2/(lat_2 + lat_{max})$:

  A modified lateral moment is derived as above. It takes into account the two most energetic cells (shower core). So, $lat_2$ is the second moment of the variable $r_i$ for which we impose that the distance $r = 0$ for the two most energetic cells. For $lat_{max}$ we impose that $r = 4$ cm for the two most energetic cells and $r = 0$ for the remaining cells.

- $long_2/(long_2 + long_{max})$:

  To evaluate the shower shapes in the longitudinal direction we use an equivalent variable to the previous one where the distance of the each cell to the shower centre is used.

- $\frac{1}{E_{\rm norm}} \times \sum_i E_i\ (E_i/V_i)$:

  where $V_i$ is the volume of the cell $i$. The electromagnetic shower is narrow and deposits energy more locally than a hadronic shower.

- $f_{max}$

  Fraction of the energy in the most energetic cell of the cluster. By measuring the energy fractions deposited in the cells of a segmented calorimeter it is usually possible to distinguish incident hadrons from electrons and photons.

In order to combine the electron identification information from the various discriminants into a single quantity that provides optimal discrimination power, we calculate the likelihood for each discriminant based on probability density functions. For each discriminant, the signal likelihood ($p_s$) and the background likelihood ($p_B$) are separately calculated and combined using the likelihood ratio:

$$R_L = \frac{\prod_{j=1}^{N_{var}} p_s(j)}{\prod_{j=1}^{N_{var}} p_s(j) + \prod_{j=1}^{N_{var}} p_B(j)} \tag{13}$$

where $j$ runs over each discriminant. Figure 5 shows the distribution of $R_L$ using all the discriminants described above. Both electrons (signal) and background are shown. As can be seen, good separation is achieved with this variable. The electron identification efficiency determined using $Z \to e^+e^-$ events as function of the electron $p_T$ and $\eta$ as shown in Figures 6 and 7. The drop in the efficiency plot versus $\eta$ happens at the transition region between inner wheel of the electromagnetic end-cap and the FCal.

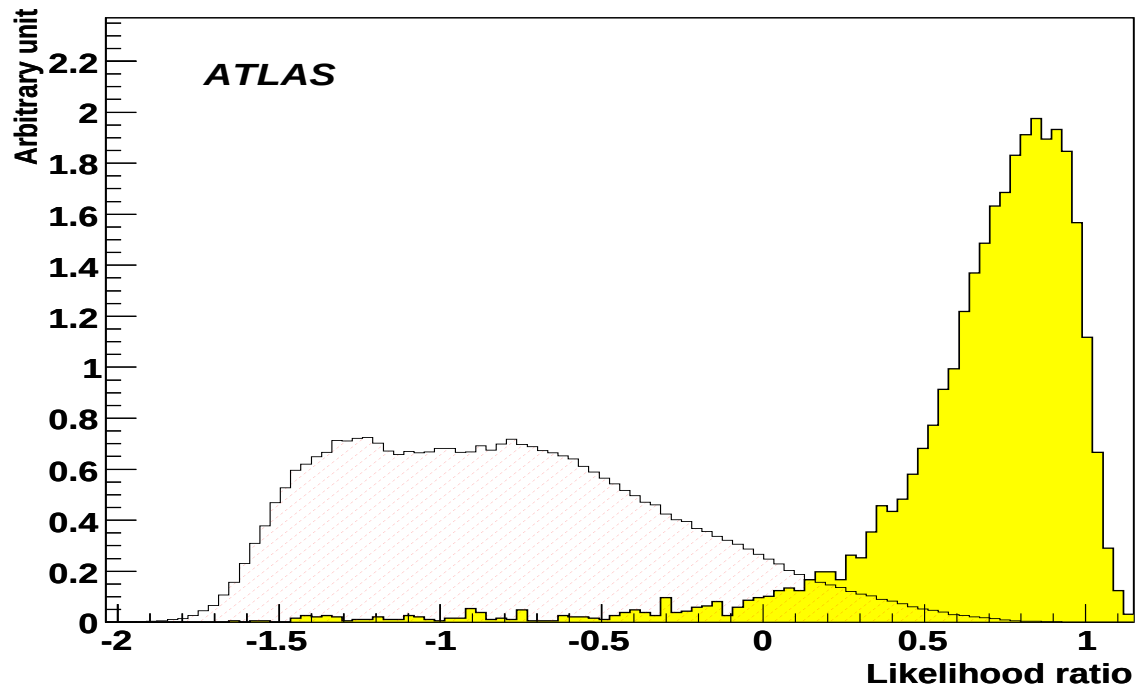


Figure 5: The likelihood ratio $R_L$ for the signal (yellow) and background (solid line). Due to the fact that the likelihood ratio is strongly peaked at 0 and 1, in the plot, a transformation is applied that zooms into the peaks.

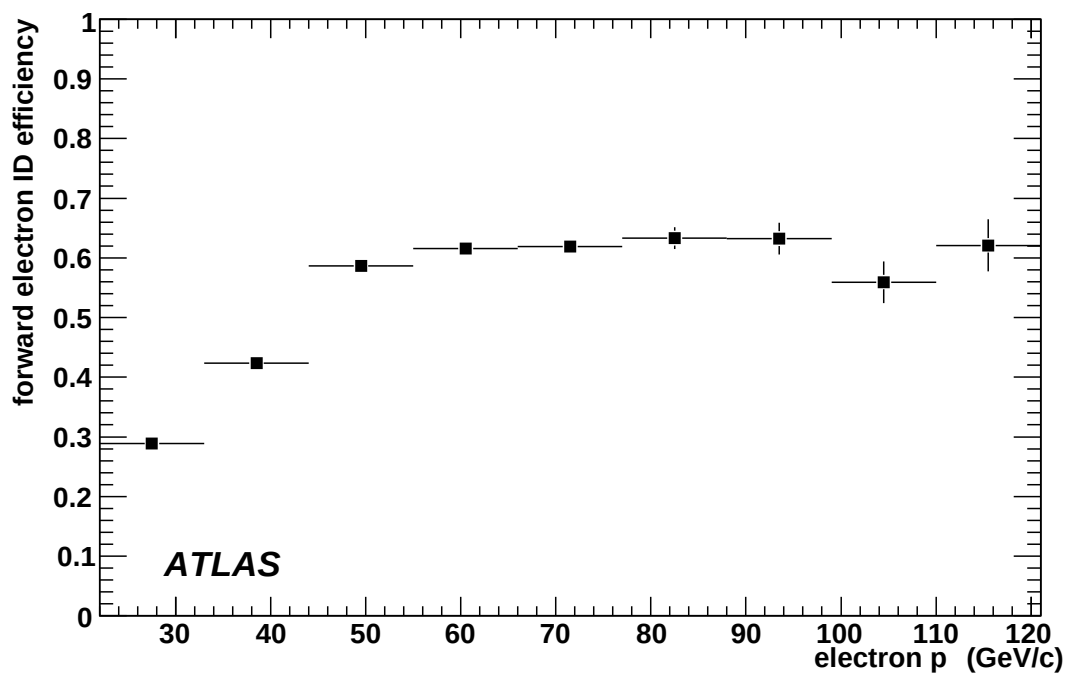


Figure 6: Forward Electron ID efficiency vs electron $p_T$ for events with $2.5 < |\eta| < 4.9$.

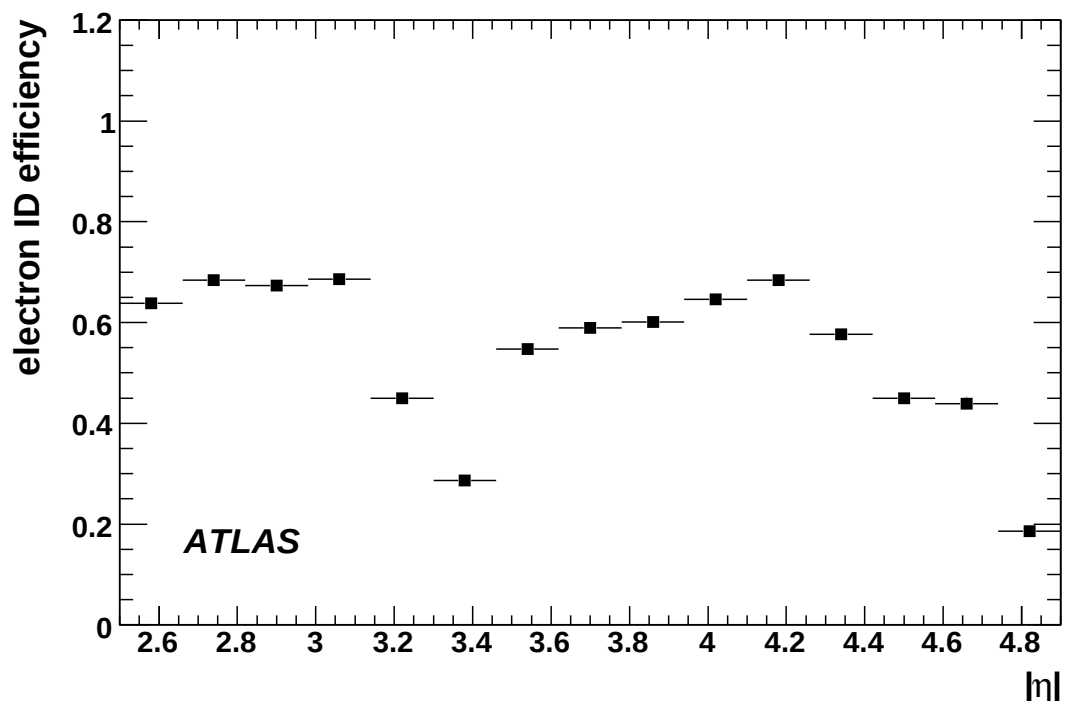


Figure 7: Forward Electron ID efficiency vs electron $\eta$ for events with $p_T > 20$ GeV.

# 5 Monte-Carlo samples

The Monte Carlo events are generated at a centre of mass of $\sqrt{s} = 14$ TeV using PYTHIA [13] (version 6.3) for the signal $pp \to Z/\gamma^* \to e^+e^-$ and the background $pp \to jj$ with the CTEQ6LL [14] parton distribution functions. Table 1 shows the summary of the cross-sections, number of events and the equivalent luminosity for the signal and background simulated events.

## 5.1 Signal

The $pp \to Z/\gamma^* X \to e^+e^- X$ events generated are filtered requiring at least one electron with $|\eta| < 2.7$ and with transverse momentum $p_T > 10$ GeV and a dilepton mass, $\hat{m}$, greater than 60 GeV. Figures 8 and 9 show the $p_T$ distribution of electrons and the invariant mass of the electron pairs in the signal events.

## 5.2 Background

The dominant sources of background to the signal process $pp \to Z/\gamma^* X \to e^+e^- X$ are:

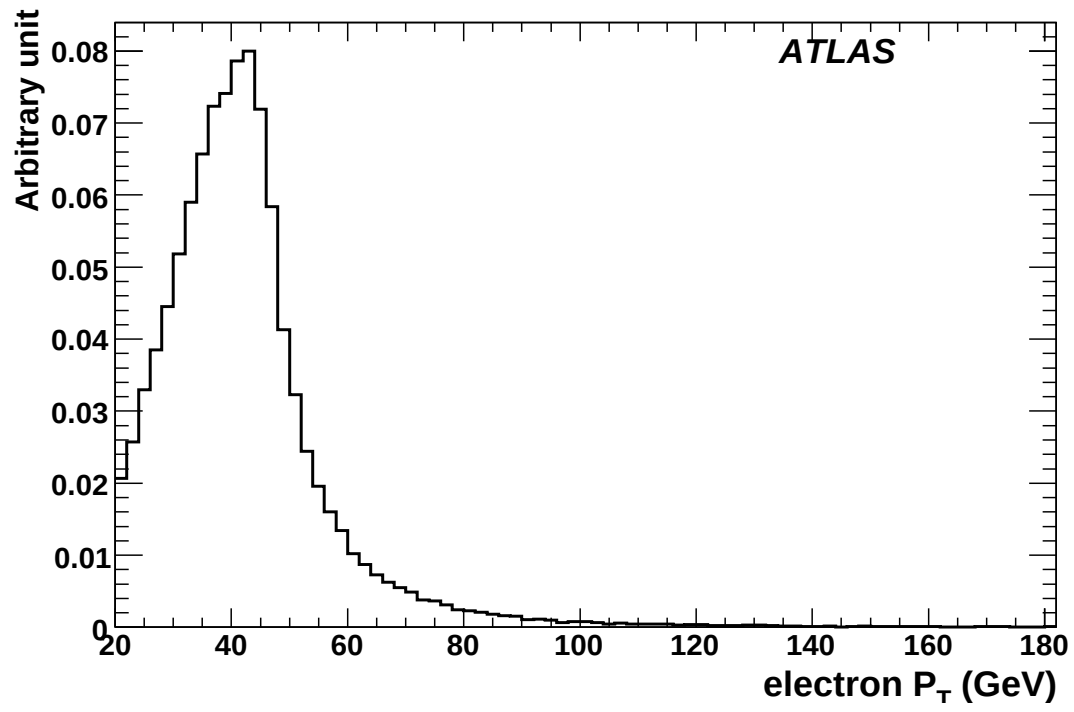


Figure 8: The electron $E_T$ distribution normalized to unit for events with $|\eta| < 1.3$ and $|\eta| > 1.6$.

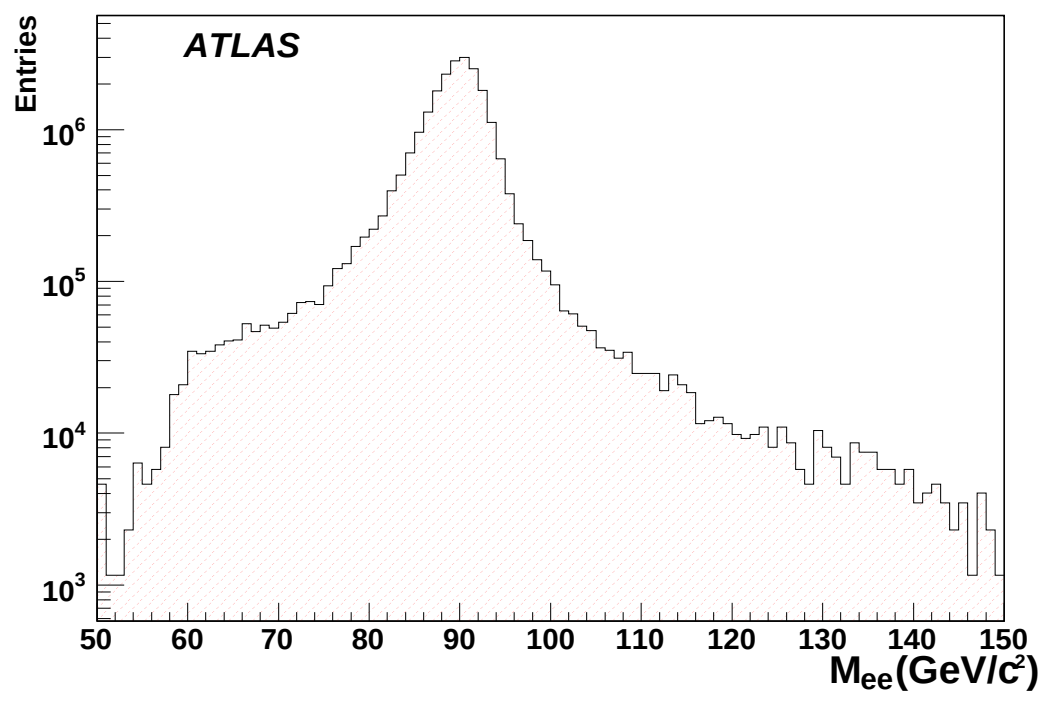


Figure 9: Di-electron invariant mass distribution for events with $|\eta| < 1.3$ and $|\eta| > 1.6$.

- dijet production: this is the largest background when two jets fake an electron. The cross-section of this process is greater by several orders of magnitude than the signal one, and dominates at low transverse momentum.
- $pp \to t\bar{t}X \to e^+e^-X$: The top quark decays into the $W$ boson and a $b$ quark, followed by the $W$ decay into electron and neutrino ($t \to Wb$, $W \to e\nu$). It has the same signature as the signal one as the two electrons of the final state can simulate the two electrons from $Z$.
- $pp \to W + X \to e\nu_e + X$: where $X$ is a photon or a jet misidentified as an electron

These events are passed through a detector simulation program to model detector response. Two detector simulation tools are used: a full detector simulation tool, based on GEANT4 [15], and a fast simulation program, ATLFAST [16] which simulates the detector response using efficiencies and smearing parameters measured from detailed simulation data. ATLFAST was used due to the high di-jet cross-section, the high rejection power of the selections and the available Monte Carlo statistics of fully simulated events which is not sufficient to evaluate the background. The Monte Carlo samples processed with the fast detector simulation were used only for estimation of the background contributions.

| Physics process | Cross section (nb) | $\varepsilon_{filter}$ | Number of events | Equivalent luminosity (pb$^{-1}$) |
|---|---|---|---|---|
| $Z \to e^+e^-$ | 2.015 | 0.126 | $5.10^5$ | 248.14 |
| Inclusive jets | $21.10^4$ | 0.09 | $3.10^7$ | 0.15 |
| $t\bar{t}$ | 0.833 | | $6.10^5$ | 720.3 |
| $W + X$ | 38.2 | | $1.10^6$ | 26.2 |

Table 1: List of signal and background samples.

Ref. [17] discusses expectation for electron triggers at different luminosities. The $Z/\gamma^*$ events used in this measurement are selected by high transverse momentum triggers on one or two electrons. This selection requires at least one isolated electron with $p_T > 20$ GeV or at least two isolated electrons with $p_T > 12$ GeV. At Level 1, electrons are selected by the presence of an energy deposition in an EM calorimeter tower. The Level 2 decision, which is based on the result of the Level 1 trigger, can take into account the information from all ATLAS subdetector systems in the regions of interest (ROIs) and the event filter (EF) performs its task only after the complete event has been assembled in the event builder (EB).

## 6 Angular distribution

The $Z$ production can be either from the annihilation of valence quarks with sea antiquarks or from the annihilation of sea quarks with sea antiquarks. Since the original quark direction is unknown in proton-proton collisions (it can originate with equal probability from either proton), the sign of $\cos\theta$ is not directly measurable.

As the valence quarks have, on average, a much larger momentum than the sea quarks, in valence-sea quark collisions, the longitudinal motion of the dilepton system approximates the quark direction and the angle between the lepton and the quark in the $e^+e^-$ rest frame can be extracted with respect to the beam axis. A lepton asymmetry can thus be expected with respect to the boost direction. For sea-sea quark collisions no such asymmetry is expected, diluting the overall effect.

To minimize the effect of the unknown transverse momenta of the incoming quarks in the measurement of the forward and backward cross-sections, we use the Collins-Soper reference frame [18]. This reference frame reduces the uncertainty in electron polar angle due to the finite transverse momentum of the incoming quarks. The particle four-vectors are transformed to the $e^+e^-$ rest frame and the polar angle $\theta^*$ is measured with respect to the axis, which bisects the two quark momentum vectors.

$$\cos\theta^* = \frac{2}{m(e^+e^-)\sqrt{m^2(e^+e^-)+p_T^2(e^+e^-)}}[p^+(e^-)p^-(e^+) - p^+(e^+)p^-(e^-)] \tag{14}$$

where $p^\pm = \frac{1}{\sqrt{2}}(E \pm p_z)$, $E$ is the energy and $p_z$ is the longitudinal component of the momentum. To take into account our supposition that the $Z$ boost and the quark direction are the same we add the sign of the $Z$ boost to the definition:

$$\cos\theta^* = \frac{|p_z(e^+e^-)|}{p_z(e^+e^-)} \times \cos\theta^* \tag{15}$$

Figure 10 shows the resulting distributions of events for which th $e^+e^-$ invariant mass is close to the $Z$ mass. We should note that the detector itself doesn't introduce false asymmetries with the geometry used in the MC samples.

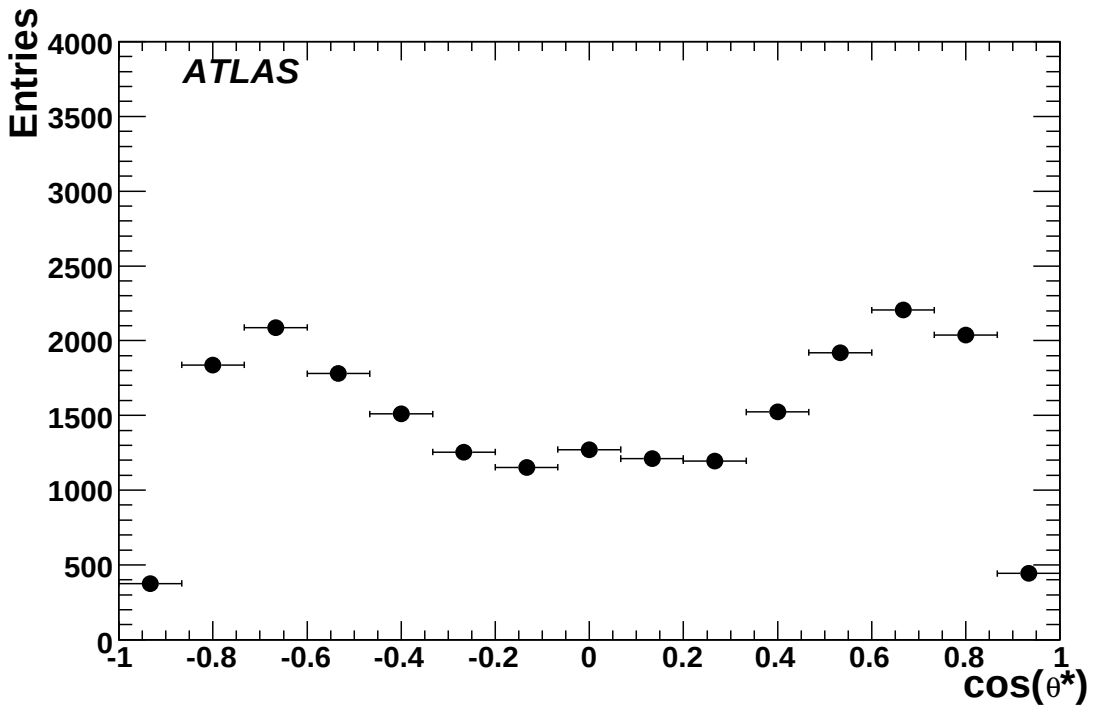


Figure 10: Distribution in $\cos\theta^*$ for reconstructed events in the $Z$ pole region.

# 7 Analysis method

## 7.1 Event selection

In this analysis we search for events with two electrons. An electron must fulfill certain quality and kinematic requirements. It must first satisfy the tight electron criteria outlined in Section 4.1. In addition the following cuts were applied:

- **C1**

  We require an electron transverse momentum higher than 20 GeV ($p_T > 20$ GeV) to emulate the energy threshold of the electron trigger

- **C2**

  We require the $e^+e^-$ invariant mass, $M_{(e^+e^-)}$, to be within a window of 12 GeV around the $Z$ mass, 85.2 GeV $< M_{(e^+e^-)} <$ 97.2 GeV ($Z$ pole)

- **C3** $|y_{e^+e^-}| > 1$

  In contrast to Tevatron $p\bar{p}$ collisions, sea quark effects dominate at the LHC. At central rapidity, $y_{e^+e^-}$, the probability that the valence quark direction and the dielectron boost coincide is lower due the smallness of the valence quark distribution. This reduces the forward backward asymmetry. Since the valence quark dominates at high values of $x$, the events where one parton carries a large fraction of the proton momentum are more sensitive and they give a large rapidity to the dilepton system. In this region the most significant measurements can be performed, as shown in Figure 11. A purer, though smaller, signal sample can thus be obtained by introducing a rapidity cut. For the following studies we will impose a $|y_{e^+e^-}| > 1$ cut.

- **C4** $E_t^{miss} < 20$ GeV

  We require the missing transverse momentum to be less than 20 GeV. This cut rejects efficiently the background coming from $pp \to t\bar{t}X$ channel where both top quarks decay semileptonically.

## 7.2 Analysis cases

A study using fast simulated events [19] showed that the use of forward electrons improves the precision of the forward-backward asymmetry measurement (and the weak mixing angle). As can be seen in Figure 12, a very high electron identification performance in the forward calorimeters is not needed, as already with a rejection of 100 against jets the influence of the remaining QCD background is negligible. Figure 12 shows only the statistical uncertainty on $A_{FB}$ versus the jet rejection in the forward calorimeters (with a fixed electron efficiency of 50%). Thus, it is required that one of the two electrons lies in the central region ($|\eta| < 2.5$), while the other electron may be either in the central region or in the forward region up to $|\eta| = 4.9$:

- $|\eta| < 2.5$ for both electrons (C-C), or
- $|\eta| < 2.5$ for one of the two electrons and $|\eta| < 4.9$ for the other (C-F).

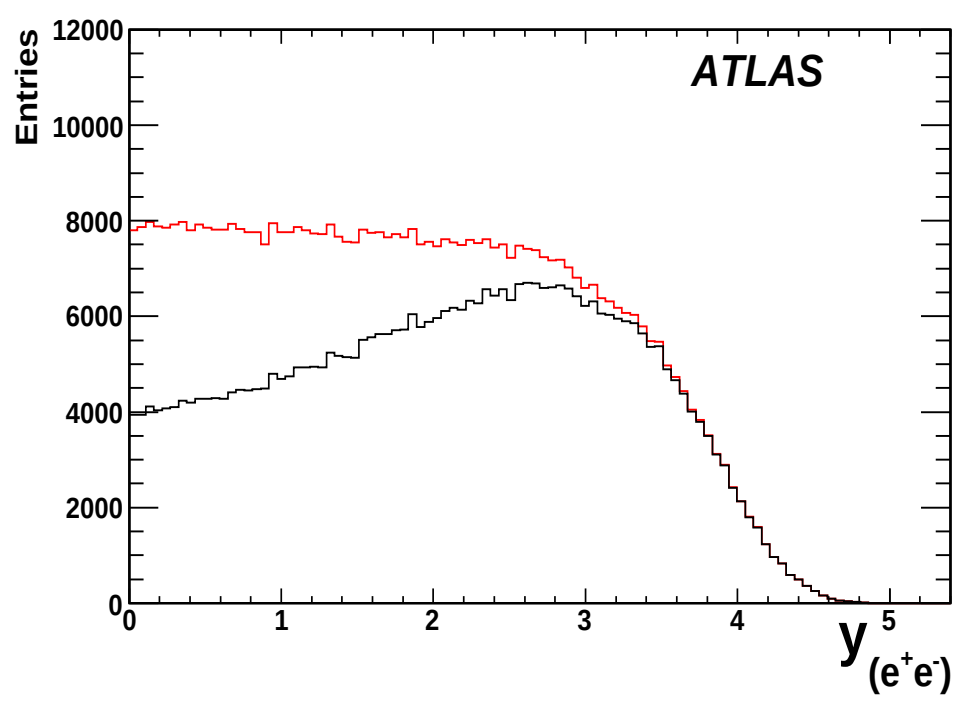


Figure 11: Di-electron rapidity distribution for all events (upper line) and for the events with correct quark direction (lower line).

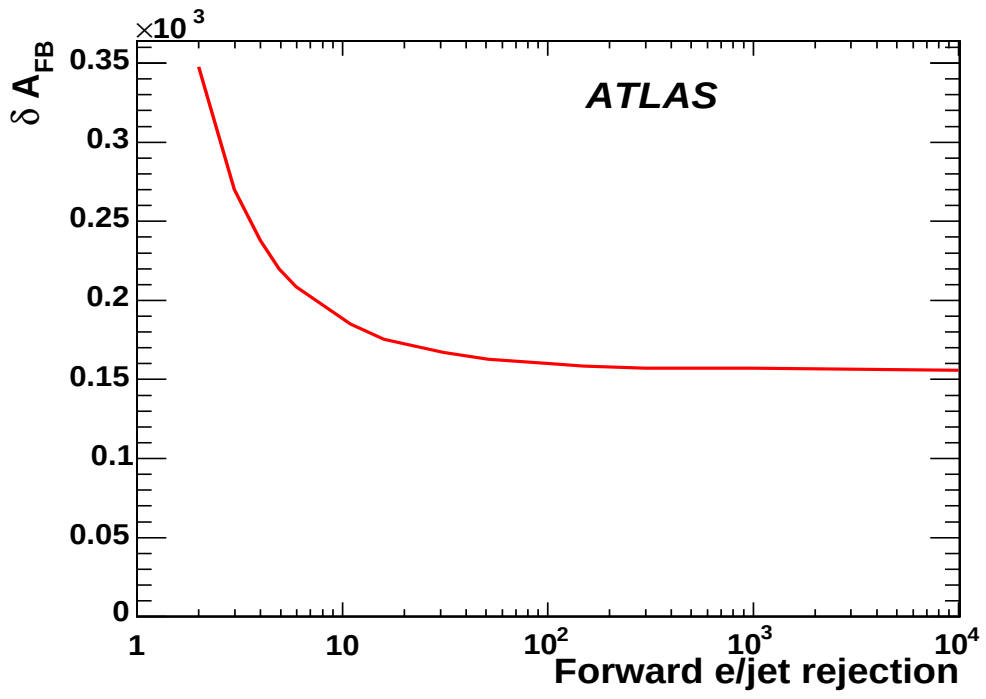


Figure 12: Forward-Backward asymmetry statistical accuracy versus the forward jet rejection in the events where at least one electron is in the central region and keeping the efficiency of the electron in the forward region at 50%.

In the region $2.5 < |\eta| < 3.2$ the calorimeters used are the EMEC and the HEC and for $|\eta| > 3.2$ the forward calorimeter (FCal) is used. Note that we can not reconstruct the electron track in the forward region ($2.5 < |\eta| < 4.9$) as the tracking system of ATLAS is limited to the region $|\eta| < 2.5$. In addition, the forward calorimeters have a coarser granularity than the central ones (a factor 2 in both eta and phi directions). Figure 13 shows the the $E_T$ distribution of electron for events with at least one electron in the central region.

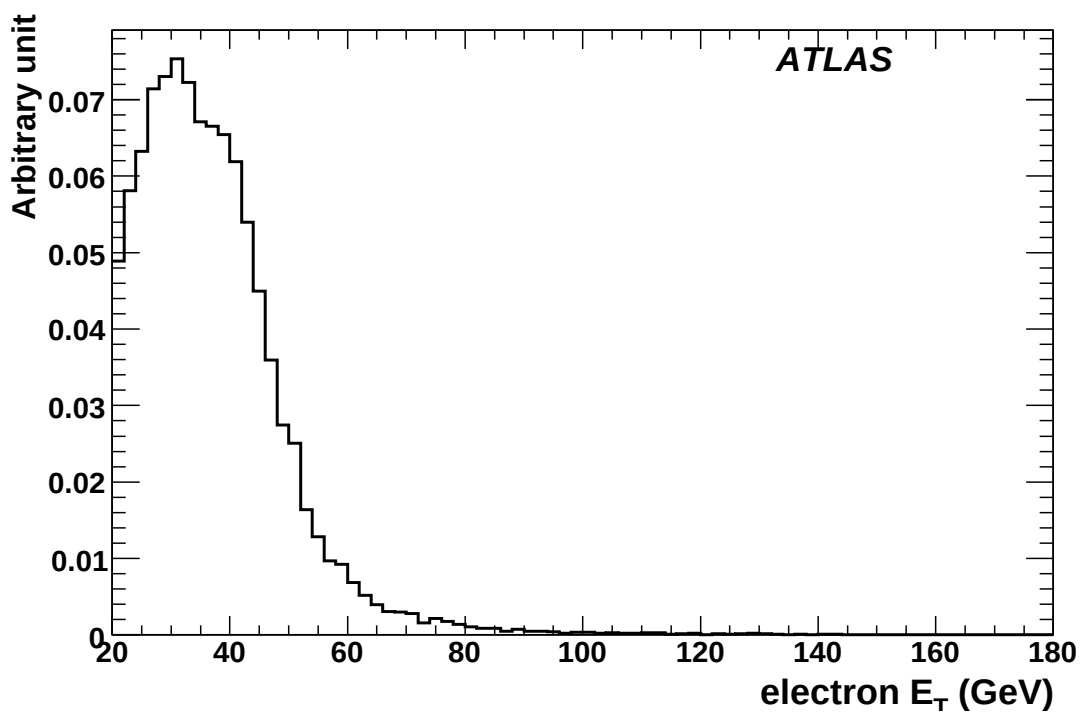


Figure 13: The electron $E_T$ distribution for the central-forward events, normalized to unity.

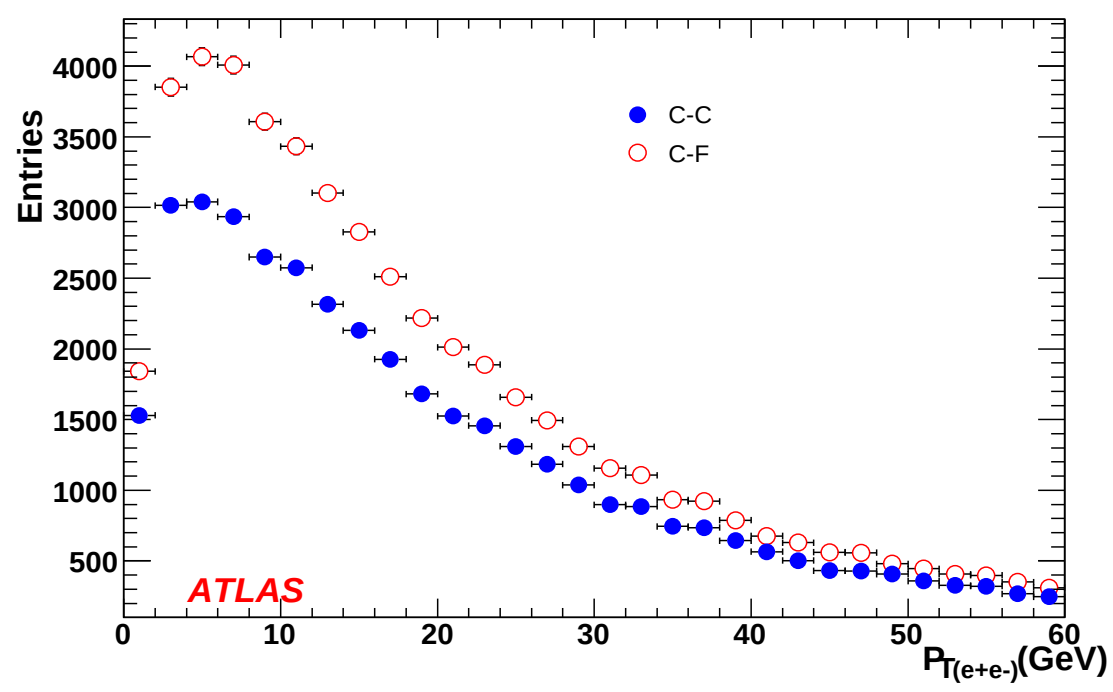


Figure 14: Di-electron $p_T$ distribution for the central-central events (C-C) and the central-forward events (C-F).

Using the forward electrons we gain about 30% in the statistics as shown in figure 14, where we compare the di-electron $p_T$ distribution in the two analysis cases, C-C and C-F. Furthermore the value of the asymmetry is higher in the C-F events. The expected numbers of the signal and background events of each process are shown in table 2 in the two analysis regions, C-C and C-F.

Event statistics corresponding to a 100 fb$^{-1}$ would be required for this analysis to reach a very high precision on the determination of the weak mixing angle from the forward-backward asymmetry measurement. Nevertheless the measurement can be also made at low luminosity in order to test the

consistency with the Standard Model. This can be done as function of $M_{e^+e^-}$ to enhance sensitivity to possible deviations at high mass scales.

### 7.3 A$_{FB}$ calculation

The forward-backward asymmetry is calculated according to Eqs 9 and 10.

If we assume that the distribution of $N_F$ and $N_B$ follows a binomial distribution, then the uncertainty on the two quantities can be written as follows (which is valid only for small asymmetries): $\sigma_{N_F} = \sigma_{N_B} = \sqrt{N_F N_B}/\sqrt{N_F + N_B}$, and the $A_{FB}$ uncertainty is $\sigma_{A_{FB}} = \sqrt{\frac{1-A_{FB}^2}{N}}$.

| Selection cut | Signal | dijet | W+X | $t\bar{t}$ |
|---|---|---|---|---|
| C-C | | | | |
| **C1** | $2.39802\times10^7$ | 4741 | 73312 | 315178 |
| **C2** | $1.87808\times10^7$ | 404 | 6048 | 28204 |
| **C3** | $7.31282\times10^6$ | 144 | 2212 | 9937 |
| **C4** | $7.30496\times10^6$ | 142 | 205 | 741 |
| C-F | | | | |
| **C1** | $3.09742\times10^7$ | 447408 | 2764790 | 336714 |
| **C2** | $2.41535\times10^7$ | 35920 | 172321 | 29824 |
| **C3** | $1.26855\times10^7$ | 35340 | 168486 | 11556 |
| **C4** | $1.25967\times10^7$ | 35128 | 15868 | 864 |

Table 2: Summary of expected number of C-C and C-F signal and background events for 100 fb$^{-1}$ of integrated luminosity. Results are given after the application of cuts **C1**-**C4**.

## 8 Electron charge identification

Any misidentification of the electron charge dilutes the asymmetry. Where possible, the charge of both electrons is measured and it is required that the signs differ. When both electrons are in the central region both charges can be measured and the condition that they have opposite charges removes the events where the charge of one of the two electrons is misidentified. In this case the forward-backward asymmetry is not significantly affected by charge misidentification.

The situation is different when one electron is forward and the other central. Only the charge of the central electron can be measured. A misidentification of the charge of the electron changes the sign of $\cos\theta^*$ and a forward event may be taken as a backward one. This effect can be corrected for if the charge misidentification fraction is known.

To measure the charge misidentification we use two methods:

- MC method

  We use the two electrons coming from $Z$ and we count the number of events where the generated charge ($q_{gen}$) and reconstructed charge ($q_{reco}$) of the individual electrons differ. We define the charge misidentification fraction as follow:

$$r = \frac{N(q_{reco} \neq q_{gen})}{N_{tot}}. \quad (16)$$

- Tag and probe method

  This method is used with the tag electron $e_1$ having the tight electron requirements, $p_T >$20 GeV and $|\eta| < 1.5$. The tag and probe electron pair must have a dielectron mass within ±6 GeV of $M_Z$. The rate at which the second electron $e_2$ has the same charge gives an estimation of the charge misidentification fraction.

The true asymmetry can be deduced from the raw asymmetry using the relation:

$$A_{FB_{true}} = \frac{A_{FB} - r_+ + r_-}{1 - r_+ - r_-}. \tag{17}$$

where $r_-$ is the fraction of true $e^-$ misidentified as $e^+$ and $r_+$ is the fraction of true $e^+$ misidentified as $e^-$.

Figure 15 shows the misidentification fraction as function of the electron pseudo-rapidity obtained with these two methods. The misidentification fraction is small. The difference between the misidentification fraction for $e^+$ and $e^-$ is shown in figure 16.

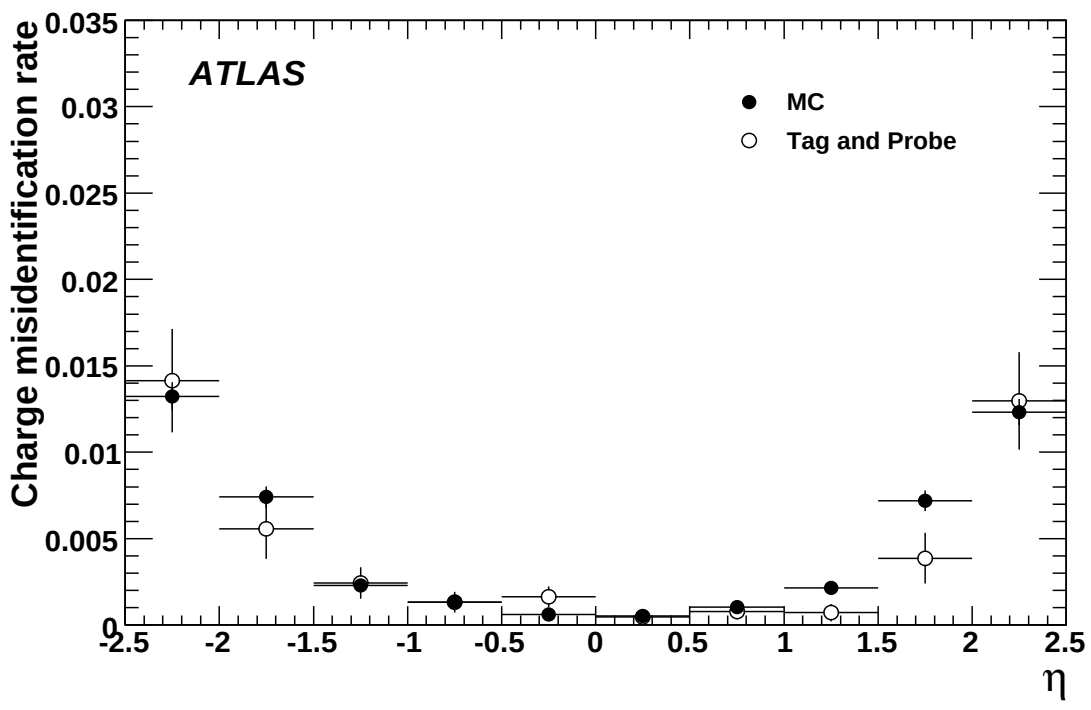


Figure 15: Electron charge misidentification fraction versus the electron pseudo-rapidity: Data method of tag-and-probe measuring the charge misidentification fraction (open points), and Monte Carlo driven charge misidentification fraction (full points).

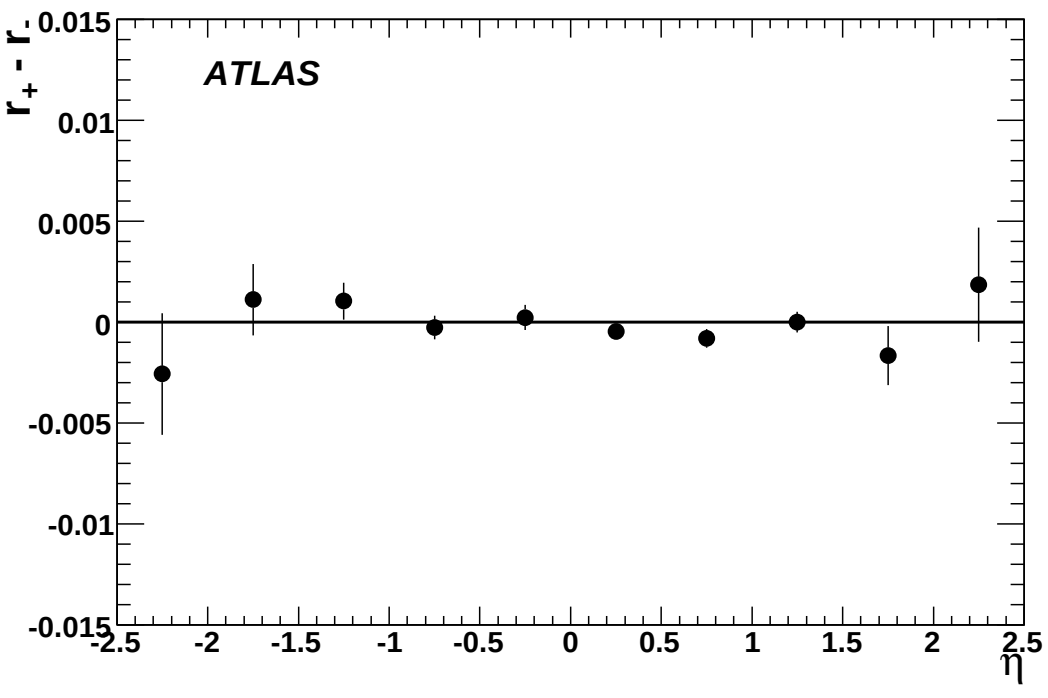


Figure 16: The difference in charge misidentification fraction of positrons and electrons versus the electron pseudo-rapidity.

## 9 Effect of pile-up

Pile-up originates from the fact that several $pp$ collisions can occur during the same bunch crossing. This causes extra activity in the detector and therefore influences the event selection. The effect of pile-up is assessed by superimposing additional events on top of the signal when events are simulated; hits from several bunch crossings are overlaid. In this study, pile-up events corresponding to $10^{33}$ cm$^{-2}$ s$^{-1}$ were investigated.

The pile-up affects the electron selection efficiency due to the presence of additional activity in the events. Figure 17 displays the electron identification efficiency with and without pile-up. We can see also in Table 3 the summary of the results for the C-C and C-F events, after all selection cuts. The results show a loss of about 3% of signal events.

| | C-C | C-F |
|---|---|---|
| No pile-up | 7,304,960 | $1.25967\times10^7$ |
| With pile-up | 7,026,653 | $1.217696\times10^7$ |

Table 3: Number of expected events, at 100 fb$^{-1}$, after all cuts, for the signal, and for C-C and C-F events. The numbers are shown for the two cases: with and without pile-up. The reference pile-up luminosity used is $10^{33}$ cm$^{-2}$ s$^{-1}$.

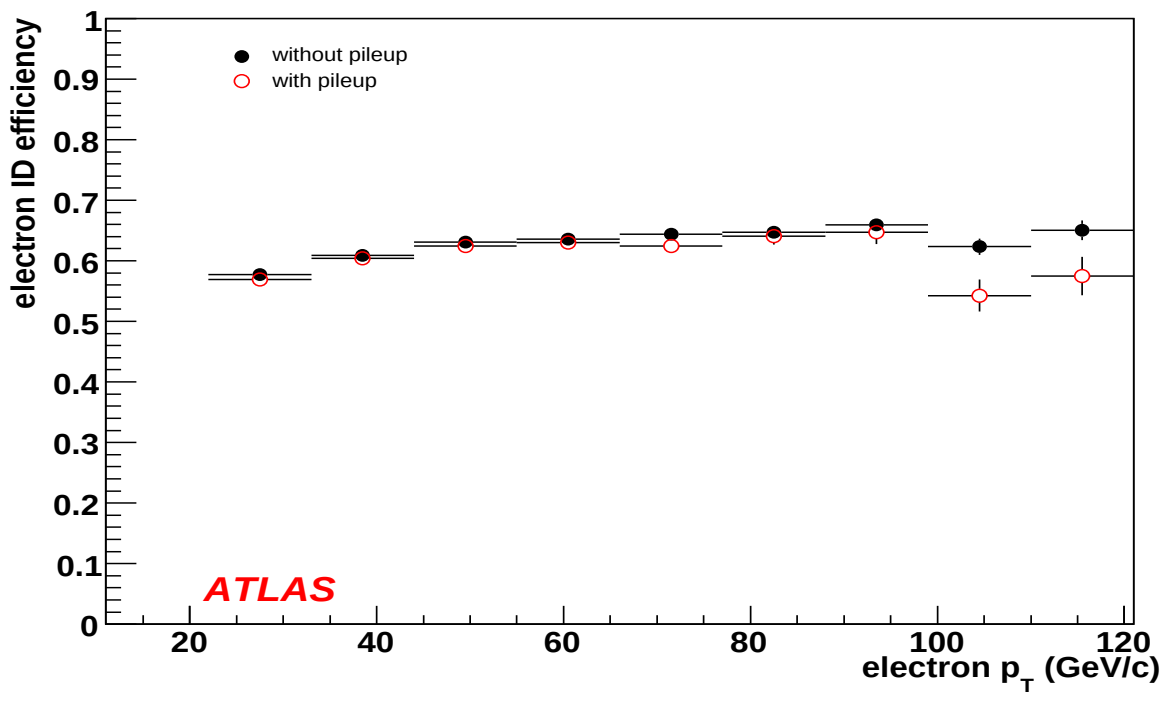


Figure 17: Effect of pile-up on the electron identification efficiency. The reference pile-up luminosity used is $10^{33}$ cm$^{-2}$ s$^{-1}$.

# 10 Systematic Uncertainties

We have considered several sources of systematic uncertainties. There are PDF uncertainties and experimental ones including the energy scale, energy resolution, reconstruction efficiency and the background estimation.

## 10.1 Parton Distribution Functions

Since the vector and axial vector couplings of the $u$ and $d$ quarks to the $Z$ boson are different, the forward-backward asymmetry is expected to depend on the ratio of the $u$ and $d$ quark parton distribution function. Thus, the choice of the parton distribution functions (PDFs) will affect the measured lepton forward-backward asymmetry.

The MRST PDF parametrization [6] is used to assess the uncertainty arising from the PDFs. Thirty eigenvectors are used in this parametrization to indicate the effect of $\pm1\sigma$ variations. Figure 18 shows the $e^+e^-$ asymmetry for each eigenvector. The deviation of A$_{FB}$ from the central value is of the same order of magnitude as the statistical accuracy.

To study the effect of the PDFs on the asymmetry, a PDF reweighting technique is used to reduce the need for Monte Carlo generations and simulation. Thus, we generate the events with the best fit central value ($PDF_0$) and then we reweight to $PDF_i$ of the set $i$ by weighting each event via:

$$\frac{f_{PDF_i}(x_1,Q^2,flav_1).f_{PDF_i}(x_2,Q^2,flav_2)}{f_{PDF_1}(x_1,Q^2,flav_1).f_{PDF_1}(x_2,Q^2,flav_2)} \tag{18}$$

where $xf_{PDF}(x,Q^2,flav)$ is the parton momentum distribution for flavour, $flav$, at scale, $Q^2$, and mo-

mentum fraction, $x$.

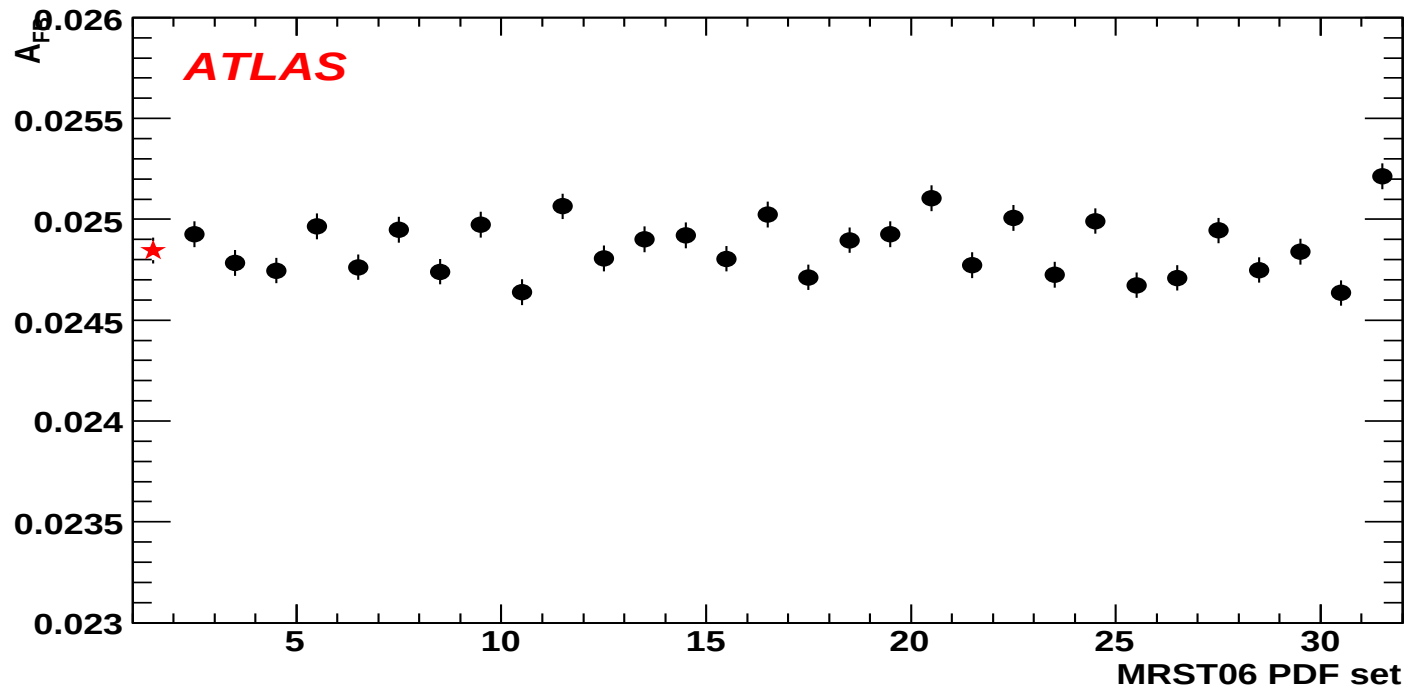


Figure 18: The forward-backward asymmetry for each MRST eigenvector; the red star is the central value.

### 10.2 Background subtraction

The determination of $A_{FB}$ (with real data) requires knowledge of the number of background events and the forward-backward charge asymmetry of the background events. Thus the central value of $A_{FB}$ we showed above can be obtained from the data events by subtracting the background events in the forward and backward regions separately from the *raw* forward-backward asymmetry. The number of background events estimated thus gives rise to a source of systematic uncertainties on $A_{FB}$.

The systematic uncertainty is evaluated by varying the estimated numbers of background events by 30%. The uncertainty value is taken as the shift in $A_{FB}$. The largest shift is less than 0.01%.

### 10.3 Detector performance

Various effects due to uncertainties on the knowledge of the detector performance have to be taken into account:

**Energy scale** The electron energy scale uncertainty, which arises from calorimeter calibration uncertainties, affects the forward-backward asymmetry by causing a shift in the $e^+e^-$ invariant mass over which we integrate $A_{FB}$. The effect is significant in the Z-pole region as can be seen in Figure 1. To take these effects into account, the central calorimeter scale is varied by 0.1% and the forward calorimeter scale is varied by 0.5% to estimate the systematic uncertainties. The positive and negative variations are considered separately and the largest shift is taken as the uncertainty.

**Reconstruction efficiency** The impact of the uncertainties on the electron reconstruction in the $A_{FB}$ measurement can be estimated by removing 0.2% and 0.5% fraction of the reconstructed electrons in the central region and forward region respectively. These values used here are based on our study on the forward calorimeters reconstruction.

**Energy resolution** The forward backward asymmetry is sensitive to the variations in the energy resolution. These variations are mainly due to the dead material distribution, known with an insufficient accuracy. To evaluate the impact of this contribution the central electron energy is degraded by an extra-term 0.01 $E_T$ and the forward calorimeter energy by 0.05 $E_T$.

**Charge identification** A systematic variation in charge misidentification fraction would directly translate into uncertainties of the forward-backward asymmetry. The systematic uncertainty is estimated by comparing the measurement as performed as in the data (tag and probe method), to that using the true Monte Carlo. As it is shown in figure 15 a difference of about 0.1% is observed.

Table 4 reports all the systematic uncertainties at 100 fb$^{-1}$. As expected, the dominant contribution to the overall uncertainty is from the PDF uncertainties.

| Source | $\delta A_{FB}$ (abs) | $\delta \sin^2\theta_{eff}^{lept}$ (abs) |
|---|---|---|
| Energy scale | $2.7\times10^{-5}$ | $1.5\times10^{-5}$ |
| Reco. Eff. | $3.4\times10^{-5}$ | $1.9\times10^{-5}$ |
| Energy resol. | $1.9\times10^{-6}$ | $1.1\times10^{-6}$ |
| Charge ID | $2.6\times10^{-5}$ | $1.4\times10^{-5}$ |
| Background subtraction. | $< 10^{-5}$ | $< 10^{-5}$ |
| PDFs | - | $^{-2.4\times10^{-4}}_{+1.3\times10^{-4}}$ |
| $a$ and $b$ parameters | - | $3\times10^{-5}$ |
| Statistical error | $2.7\times10^{-4}$ | $1.5\times10^{-4}$ |

Table 4: Summary of the systematic and statistical uncertainties on $A_{FB}$ and $\sin^2\theta_{eff}^{lept}$ for events with at least one electron in the central region (C-F). The uncertainty on $\sin^2\theta_{eff}^{lept}$ is determined using Eqs. 11 and $a$ and $b$ parameters from figure 21.

## 11 Results

Figure 19 shows the dielectron invariant mass spectrum expected in terms of the number of events per GeV for an integrated luminosity of 100 fb$^{-1}$ after the application of cuts **C1**, **C3** and **C4**. The yellow histogram corresponds to the signal contribution. The light blue displays the contributions from dijet events. The green and violet histograms display the contribution of $W+X$ and $t\bar{t}$ backgrounds respectively. The background contributions were obtained with MC samples with a fast detector simulation. The signal contribution was obtained with a full detector simulation.

Table 2 shows the expected number of background and signal events for 100 fb$^{-1}$ of integrated luminosity after the application of cuts **C1**-**C4**. It indicates that the contribution from background events is at the level of 0.2%.

In Figure 20 we display the variation of the charge asymmetry versus the rapidity of the two electrons. It is observed that the asymmetry increases when allowing the second electron to be up to $|\eta| = 4.9$. Using $|y_{e^+e^-}| > 1$ and going from C-C to C-F events, the integrated asymmetry increases from 1.3% to 2.7%. The precision on the forward-backward asymmetry improves from $3.7\times10^{-4}$ to $2.7\times10^{-4}$.

Using the parameters $a = 0.23 \pm 0.03$ and $b = 1.832 \pm 0.255$, of Eqs. 11, derived from the linear fit of the figure 21, we can estimate the error expected for $\sin^2\theta_{eff}^{lept}$ from a measurement of the forward-

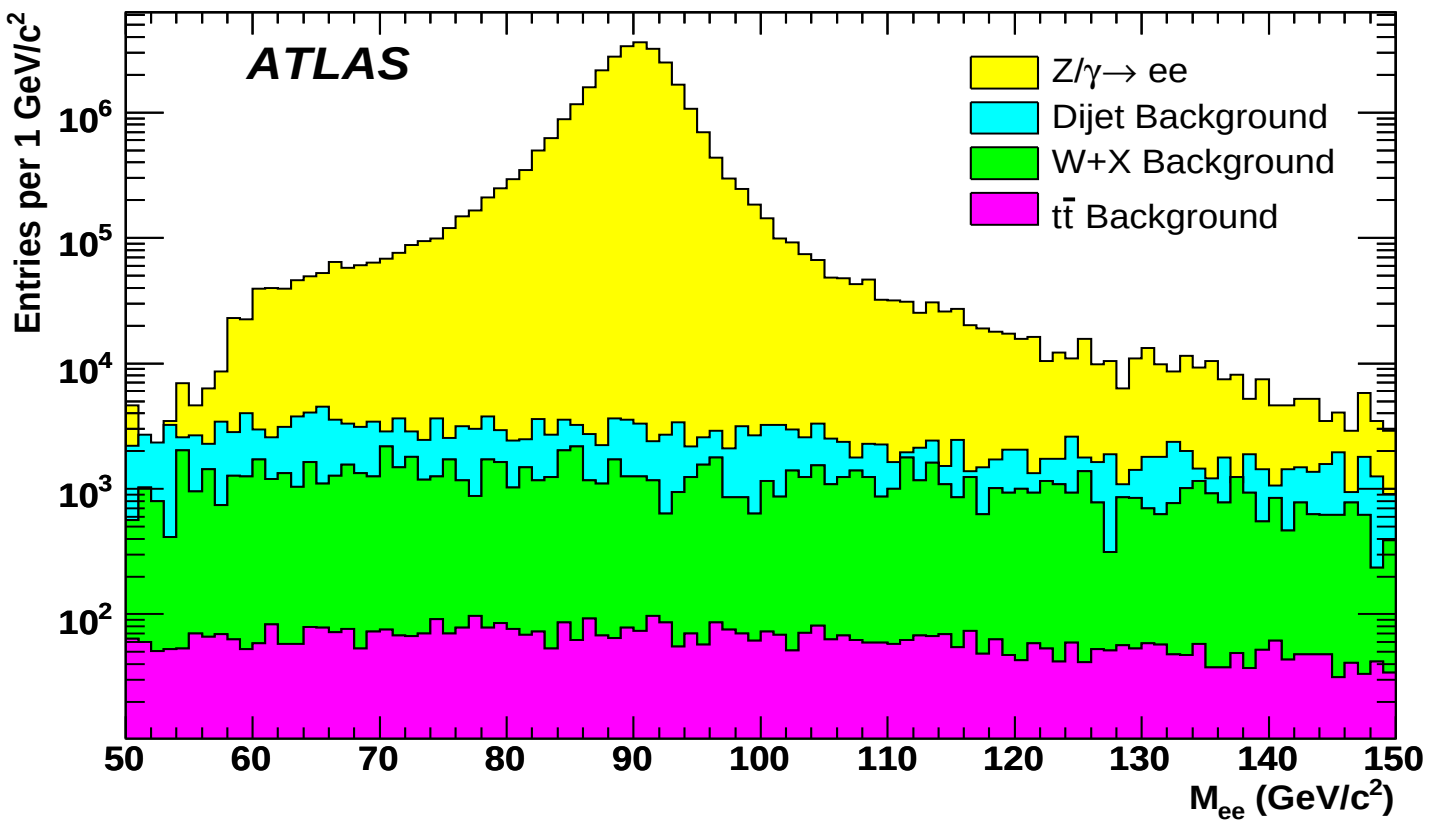


Figure 19: Dielectron invariant mass distribution for the signal and background events normalized to 100 fb$^{-1}$, with the final selection except that the cut **C2** is removed.

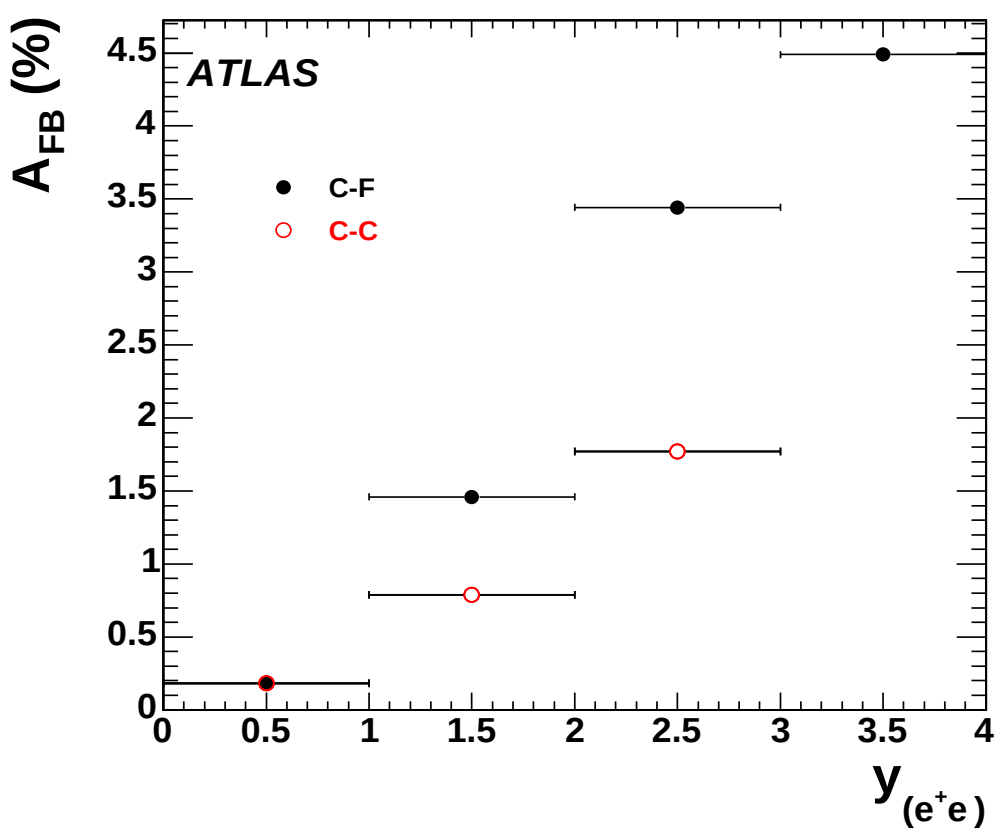


Figure 20: Forward-backward asymmetry versus dielectron rapidity in the C-C events (open points) and in the C-F events (full points).

backward asymmetry at the $Z$ pole, for an integrated luminosity of 100 fb$^{-1}$:

$$\delta \sin^2 \theta_{eff}^{lept} = (1.5(\text{stat}) \pm 0.3(\text{exp}) \pm 2.4(\text{PDF})) \times 10^{-4} \quad (19)$$

## 12 Conclusion

We report on a detailed study of the forward-backward charge asymmetry $A_{FB}$, at LHC with the ATLAS detector, of electron pairs resulting from the process $pp \to Z/\gamma^* X \to e^+e^- X$. This measurement provides a test of the Standard Model. In the vicinity of the $Z$-pole this measurement can be used to determine the effective weak mixing angle $\sin^2 \theta_{eff}^{lept}$. The precision on $\sin^2 \theta_{eff}^{lept}$ obtained is $\delta \sin^2 \theta_{eff}^{lept} = (1.5(\text{stat}) \pm 0.3(\text{exp}) \pm 2.4(\text{PDF})) \times 10^{-4}$ for a 12 GeV mass window around the $Z$ mass, and 100 fb$^{-1}$ of integrated luminosity. Electron identification in the forward region ($2.5 < |\eta| < 4.9$) of

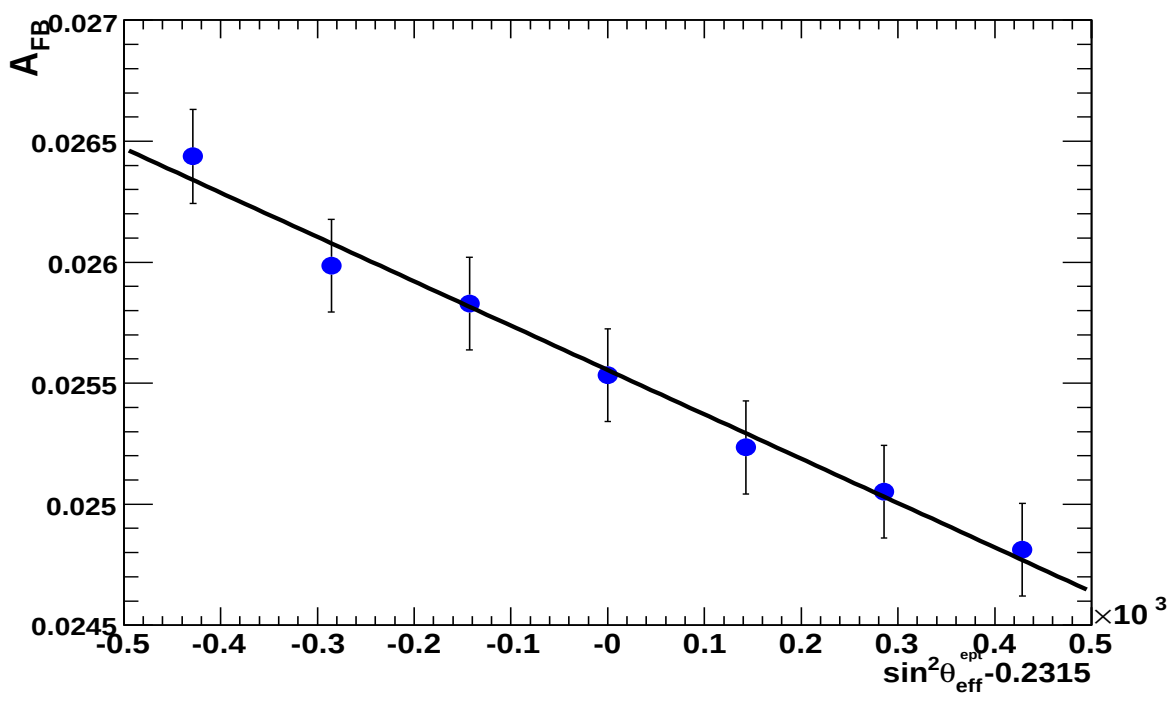


Figure 21: Forward backward asymmetry $A_{FB}$ versus the weak mixing angle $\sin^2\theta_{eff}^{lept}$ at the $Z$ pole. The straight line is a $\chi^2$ fit to the points shown. Fast simulated events are used.

the ATLAS detector is very important for the measurement. In this region an electron ID efficiency of 80% is achieved with less than 3% QCD background.

The main systematic effects relevant for forward-backward asymmetry measurements with 100 fb$^{-1}$ of data are addressed, including systematic uncertainties on detector effects and MRST PDF uncertainty. An advanced technique was developed and used to estimate the error due to the PDF uncertainties. This study showed that the uncertainty in the weak mixing angle determination due to PDF's is of the same order as the measurement statistical error, which means that the precision on the weak mixing angle at LHC can be competitive to the current world average. In addition we expect that in the future the knowledge of the PDF's will improve from the constraints imposed by Tevatron, HERA and first LHC measurements (e.g. using $W$ asymmetry), and the systematic uncertainty due the uncertainty in the PDF's should decrease by the time ATLAS high luminosity data is available. If this is not the case, the asymmetry measurements can be used, conversely, to constrain the parton distribution functions.

# Diboson Physics Studies

**Abstract**
This note presents studies of the sensitivity of the ATLAS experiment to Standard Model diboson ($W^+W^-$, $W^\pm Z$, $ZZ$, $W^\pm\gamma$, and $Z\gamma$) production in pp collisions at $\sqrt{s} = 14$ TeV, using final states containing electrons, muons and photons. The studies use ATLAS simulated data, which include trigger information and detector calibration and alignment corrections. The influence of backgrounds on diboson detection is assessed using large samples of fully simulated background events. The cross-section measurement uncertainties (both statistical and systematic) are estimated as a function of integrated luminosity (from 0.1 to 30 fb$^{-1}$). The studies show that the Standard Model $W^+W^-$, $W^\pm Z$, $W^\pm\gamma$, and $Z\gamma$ signals can be established with significance better than $5\sigma$ for the first 0.1 fb$^{-1}$ of integrated luminosity, and the $ZZ$ signal can be established with 1 fb$^{-1}$ of integrated luminosity. The ATLAS experiment's sensitivity to anomalous triple gauge boson couplings is also estimated. The anomalous triple gauge boson coupling sensitivities can be significantly improved, even with 0.1 fb$^{-1}$ of data, over the results from the Tevatron that use 1 fb$^{-1}$ of data.

## 1 Introduction

This paper presents studies of the ATLAS experiment's sensitivity to diboson ($W^+W^-$, $W^\pm Z$, $ZZ$, $W^\pm\gamma$, $Z\gamma$) production using lepton and photon final states, and the corresponding ability to set limits on anomalous triple gauge boson couplings (TGC). The analysis of diboson production at the LHC provides an important test of the high energy behavior of electroweak interactions. Vector boson self-couplings are fundamental predictions of the Standard Model [1], resulting from the non-Abelian nature of the $SU(2)_L \times U(1)_Y$ gauge symmetry theory, which was demonstrated by precision measurements of $W^+W^-$ and $ZZ$ pair production at LEP II [2].

Any theory predicting physics beyond the Standard Model while maintaining the Standard Model as a low-energy limit may introduce deviations in the gauge couplings at some high energy scale. Precise measurements of the couplings will not only provide stringent tests of the Standard Model, but will also probe for new physics in the bosonic sector. These tests will provide complementary information to other direct searches for new physics at the LHC. Many models predict deviations of vector boson self-couplings from the Standard Model at the $10^{-3} - 10^{-4}$ level [3]. Experiments that can reach this sensitivity could provide powerful constraints on these models. The signature for such anomalous couplings is enhanced diboson production cross-sections, particularly at high transverse momentum ($p_T$) of the bosons. Experimental limits on non-Standard Model TGC's can be obtained by comparing the shape of the measured $p_T$ or mass distributions (or transverse mass, $M_T$, for final states involving $W$) with predictions, provided that the signal is not overwhelmed by background.

The analysis uses over 30 million fully simulated and reconstructed events, with a detector layout and trigger system that reflects the ATLAS experiment as it will operate at LHC turn-on at 14 TeV center of mass energy, thus providing a realistic understanding of the detection of these diboson final states. A *Boosted Decision Tree* [4] technique is applied to selected channels, significantly enhancing measurement sensitivities. These are among the ways this study improves on our understanding and the results of the previous ATLAS diboson studies [5]- [10].

### 1.1 Diboson production cross-sections

Tree-level Feynman diagrams for electroweak diboson production at hadron colliders are shown in Figure 1. The $s$-channel diagram contains the vector-boson self-interaction vertices of interest here. The cross-sections are calculated to next-to-leading-order (NLO) in [11]- [13]. The Standard Model diboson production cross-sections are listed in Table 1.

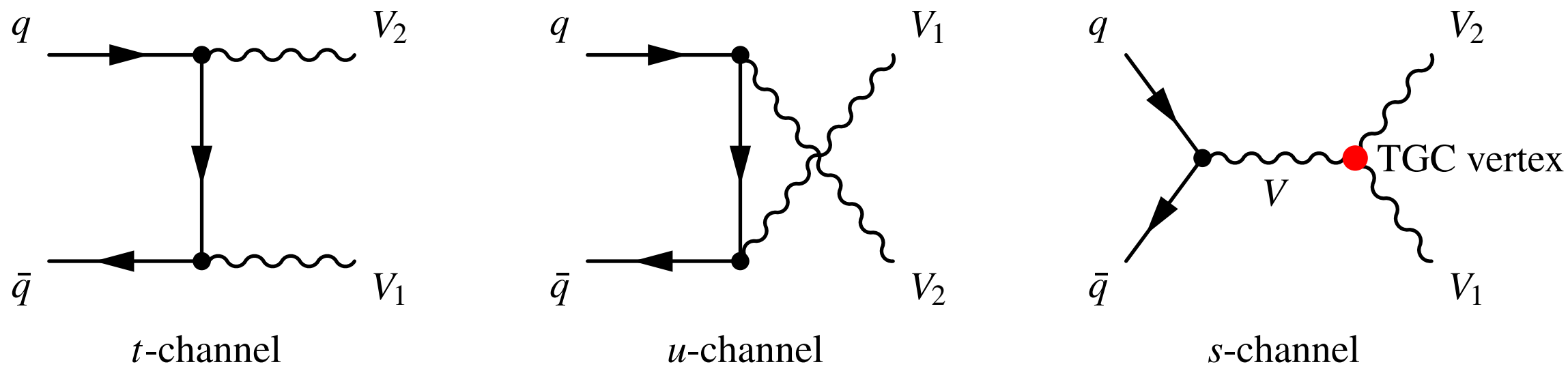


Figure 1: The generic Standard Model tree-level Feynman diagrams for diboson production at hadron colliders; $V, V_1, V_2 = \{W, Z, \gamma\}$. The $s$-channel diagram contains the trilinear gauge boson vertex. In the Standard Model, only $WW\gamma$ and $WWZ$ vertices are allowed.

Table 1: The Standard Model diboson production total cross-sections, calculated to the NLO, at the Tevatron ($\sqrt{s} = 1.96$ TeV) and the LHC ($\sqrt{s} = 14$ TeV). The references in the first column indicate the MC generators used for the calculations, with parton density function (PDF) CTEQ6M and the electroweak parameters [14]. The theoretical uncertainty from the PDF and the QCD scale factor is typically 5%.

| Diboson mode | Conditions | $\sqrt{s} = 1.96$ TeV $\sigma[pb]$ | $\sqrt{s} = 14$ TeV $\sigma[pb]$ |
|---|---|---|---|
| $W^+W^-$ [15] | $W$-boson width included | 12.4 | 111.6 |
| $W^\pm Z$ [15] | $Z$ and $W$ on mass shell | 3.7 | 47.8 |
| $ZZ$ [15] | $Z$'s on mass shell | 1.43 | 14.8 |
| $W^\pm\gamma$ [16] | $E_T^\gamma > 7$ GeV, $\Delta R(\ell,\gamma) > 0.7$ | 19.3 | 451 |
| $Z\gamma$ [17] | $E_T^\gamma > 7$ GeV, $\Delta R(\ell,\gamma) > 0.7$ | 4.74 | 219 |

The LHC diboson production rates will exceed those of the Tevatron by at least a factor 100 (10 times higher in cross-sections and at least 10 times higher in luminosity). Furthermore, because the energy reach at the LHC will be 7 times higher than at the Tevatron, the LHC sensitivity to anomalous TGC's is expected to be improved by orders of magnitude over that which can be reached at the Tevatron or LEP.

### 1.2 Effective Lagrangian for charged TGC's

The most general effective Lagrangian, that conserves $C$ and $P$ separately, for charged triple gauge boson interactions is [19]:

$$L/g_{WWV} = ig_1^V(W^*_{\mu\nu}W^\mu V^\nu - W_{\mu\nu}W^{*\mu}V^\nu) + i\kappa^V W^*_\mu W_\nu V^{\mu\nu} + \frac{\lambda^V}{M_W^2} W^*_{\rho\mu}W^\mu_\nu V^{\nu\rho}$$

where $V$ refers to the neutral vector-bosons, $Z$ or $\gamma$, $X_{\mu\nu} \equiv \partial_\mu X_\nu - \partial_\nu X_\mu$ and the overall coupling constants $g_{WWV}$ are given by $g_{WW\gamma} = -e,\ g_{WWZ} = -e\cot\theta_W$, with $e$ the positive electron charge and $\theta_W$ the weak mixing angle. The Standard Model triple gauge boson vertices are recovered by letting $g_1^V = \kappa^V = 1$ and $\lambda^V = 0$. Experimentally, deviations from the Standard Model couplings is searched for; thus the anomalous coupling parameters are defined as

$$\Delta g_1^Z \equiv g_1^Z - 1, \quad \Delta\kappa_\gamma \equiv \kappa_\gamma - 1, \quad \Delta\kappa_Z \equiv \kappa_Z - 1, \quad \lambda_\gamma, \quad \text{and} \quad \lambda_Z.$$

Note that electromagnetic gauge invariance requires $g_1^\gamma = 1$ or $\Delta g_1^\gamma = 0$.

Studies of three different diboson final states, $W^+W^-$, $W^\pm Z$ and $W^\pm\gamma$ will provide complementary sensitivities to the charged anomalous TGC's [17]. For example, the $\Delta\kappa_V$ terms in $W^+W^-$ production are proportional to $\hat{s}$, defined as the square of invariant mass of the vector-boson pair, whereas these terms are only proportional to $\sqrt{\hat{s}}$ in $W^\pm Z$ and $W^\pm\gamma$ production. $W^+W^-$ production is thus expected to be more sensitive to $\Delta\kappa_V$ than $W^\pm Z$ and $W^\pm\gamma$ production. Conversely, $W^\pm Z$ production is expected to be more sensitive to $\Delta g_1^Z$ than $W^+W^-$ production because terms in $\Delta g_1^Z$ are proportional to $\hat{s}$ in $W^\pm Z$ production. The $\lambda$-type anomalous couplings have an $\hat{s}$ dependence in all three cases, thus the sensitivities will be enhanced at the high center-of-mass energy of the LHC.

With non-Standard Model coupling parameters, the amplitudes for gauge boson pair production grow with energy, eventually violating tree-level unitarity. The unitarity violation is avoided by introducing an effective cutoff scale, $\Lambda$ [18]. The anomalous couplings take a form, for example,

$$\Delta\kappa(\hat{s}) = \frac{\Delta\kappa}{(1+\hat{s}/\Lambda^2)^n},$$

where $\Delta\kappa$ is the coupling value in the low energy limit. The scale $\Lambda$ is physically interpreted as the mass scale where the new phenomenon, which is responsible for the anomalous couplings, would be directly observable. The value of $n = 2$ is used for charged anomalous TGC, and $n = 3$ for neutral anomalous TGC.

### 1.3 Effective Lagrangian for neutral TGC's

In the Standard Model, neutral boson pairs, $ZZ$ and $Z\gamma$, are produced via the $t$-channel diagrams shown in Figure 1. While the Standard Model $ZZZ$ and $ZZ\gamma$ triple gauge boson couplings are zero at tree level, anomalous couplings may contribute. This study considers the effect of anomalous couplings on the production of pairs of on-shell $Z$ bosons only. In this case, the most general form of the $Z^\alpha(q_1)Z^\beta(q_2)V^\mu(P)$ ($V = Z,\ \gamma$) vertex function which respects Lorentz invariance and electromagnetic gauge invariance may be written as [20]

$$g_{ZZV}\Gamma^{\alpha\beta\mu}_{ZZV} = e\frac{P^2 - M_V^2}{M_Z^2}[\, i f_4^V(P^\alpha g^{\mu\beta} + P^\beta g^{\mu\alpha}) + i f_5^V \varepsilon^{\mu\alpha\beta\rho}(q_1 - q_2)_\rho \,]$$

where $M_Z$ is the Z-boson mass and $e$ is the positive electron charge; $q_1, q_2$ and $P$ are the 4-momenta of the two on-shell Z bosons and the $s$-channel propagator, respectively. The effective Lagrangian generating the $g_{ZZV}$ vertex function is

$$L = -\frac{e}{M_Z^2}[f_4^V(\partial_\mu V^{\mu\beta})Z_\alpha(\partial^\alpha Z_\beta) + f_5^V(\partial^\sigma V_{\sigma\mu})\tilde{Z}^{\mu\beta}Z_\beta],$$

where $V_{\mu\nu} = \partial_\mu V_\nu - \partial_\nu V_\mu$ and $\tilde{Z}^{\mu\beta} = \frac{1}{2}\varepsilon_{\mu\nu\rho\sigma}Z^{\rho\sigma}$. The couplings $f_i^V$ ($i = 4,\ 5$) are dimensionless complex functions of $q_1^2$, $q_2^2$ and $P^2$ and are zero at tree level. All couplings are $C$ odd; $CP$ invariance forbids $f_4^V$, while parity conservation requires that $f_5^V$ vanishes. Because $f_4^Z$ and $f_4^\gamma$ are $CP$-odd, contributions

to the helicity amplitudes proportional to these couplings will not interfere with the Standard Model terms, and hence $ZZ$ production is not sensitive to the sign of these couplings. The $CP$ conserving couplings $f_5^V$ contribute to the Standard Model cross-section at the one-loop level, but this contribution is $\mathscr{O}(10^{-4})$ [20].

### 1.4 Current Tevatron results on diboson physics

Diboson production measurements and studies of anomalous TGC's have been performed at the Tevatron with the CDF and D0 experiments, using up to 2 fb$^{-1}$ of integrated $p\bar{p}$ luminosity. Diboson cross-section measurements using $e/\mu$ decay modes, and their event statistics, measurement precision, and background events are summarized in Table 2. The measurments from Tevatron experiments are consistent with the Standard Model predictions based upon NLO matrix element calculations.

Table 2: Summary of Tevatron $p\bar{p} \to$ diboson cross-sections. For the $W^+W^-$, $W^\pm Z$, and $ZZ$ channels total production cross-sections are quoted.

| Process | Source | L fb$^{-1}$ | observed events | background events | $\sigma$(data) [pb] $\pm$ (stat)$\pm$(sys)$\pm$(lum) | $\sigma$(theory) [pb] |
|---|---|---|---|---|---|---|
| $W^+W^-$ | CDF [21] | 0.83 | 95 | 38±5 | 13.6±2.3±1.6 ± 1.2 | 12.4±0.8 |
| ($ee$, $\mu\mu$, $e\mu$) | D0 [22] | 0.25 | 25 | 8.1±.5 | 13.8±4.1±1.1 ± 0.9 | ” |
| $W^\pm Z$ | CDF [23] | 1.1 | 16 | 2.7±0.4 | $5.0^{+1.8}_{-1.4} \pm 0.4$ | 3.7±0.3 |
| ($\ell^\pm \nu \ell^+ \ell^-$) | D0 [24] | 1.0 | 13 | 4.5±0.6 | 2.7 +1.7−1.3 (total) | ” |
| $Z\gamma$ | CDF [25] | 0.2 | 72 | 4.9±1.1 | 4.6 ±0.6 (sta+sys) ± 0.3 | 4.5±0.3 |
| ($\ell^+\ell^-\gamma$) | D0 [26] | 1.0 | 968 | 117±12 | 4.96 ±0.3 (sta+sys) ± 0.3 | 4.7±0.2 |
| $W^\pm\gamma$ | CDF [25] | 0.2 | 323 | 114±21 | 18.1 ±3.1 (sta+sys) ± 1.2 | 19.3±1.4 |
| ($\ell^\pm\nu\gamma$) | D0 [27] | 0.16 | 273 | 132±7 | 14.8 ±1.9 (sta+sys) ± 1.0 | 16.0±0.4 |
| $ZZ$ | CDF [28] | 1.9 | 2 | 0.014 | $1.4^{+0.7}_{-0.6} \pm 0.6$ | 1.5±0.2 |
| ($\ell^+\ell^-\ell^+\ell^-$) | D0 [29] | 1.0 | 1 | 0.13 | $< 4.4$ | ” |

The Tevatron's $p\bar{p}$ collisions produce the charged states of $W^\pm Z$ and $W^\pm\gamma$. These states can be used to study the $W^+W^-Z$ and the $W^+W^-\gamma$ couplings independently, which is in contrast with the anomalous TGC measurements made at LEP [32] from the $W^+W^-$ final state, where certain assumptions relating the $W^+W^-Z$ and the $W^+W^-\gamma$ couplings were made. The Tevatron limits for the $WW\gamma$ and WWZ anomalous TGC's are summarized in Table 3. These limits will improve significantly by combining the constraints from the $W^\pm\gamma$, $W^\pm Z$ and $W^+W^-$ channels, and by increasing the datasets using the expected integrated luminosity of 6 fb$^{-1}$ at the end of the Tevatron running.

## 2 Signal and background modeling and simulated data samples

### 2.1 MC generators used to produce fully simulated events

The diboson physics analyses focus on leptonic decay channels of the boson pairs ($W^+W^-$, $W^\pm Z$, $ZZ$, $W^\pm\gamma$, and $Z\gamma$) produced in the proton-proton collisions at the LHC.

Diboson productions of the $W^+W^-$, $W^\pm Z$, and $ZZ$ final states, as well as the subsequent pure leptonic decays, are modeled by the MC@NLO [15] Monte Carlo generator. The generator incorporates the next-to-leading-order (NLO) QCD matrix elements into the parton shower by interfacing to the HERWIG/Jimmy [33] programs. A branching ratio of 0.0336 for $Z \to \ell^+\ell^-$ and 0.108 for $W^+ \to \ell^+\nu$ is used

Table 3: Anomalous gauge coupling limits (95% C.L.) for $WW\gamma$ and $WWZ$ from the Tevatron experiments, with $\Lambda = 2$ TeV.

| Coupling | Source | L (fb$^{-1}$) | $\lambda_Z$ | $\Delta\kappa_Z$ | $\Delta\kappa_\gamma$ | $\lambda_\gamma$ |
|---|---|---|---|---|---|---|
| $WW\gamma$ from $W^\pm\gamma$ | D0 [27] | 0.16 | | | [-0.88, 0.96] | [-0.2, 0.2] |
| $WWZ$ from $W^\pm Z$ | D0 [24] | 1.0 | [-0.17, 0.21] | [-0.12, 0.29] | | |
| $WWZ$ from $W^\pm Z$ | CDF | 1.9 | [-0.13, 0.14] | [-0.82, 1.27] | | |
| $WWZ = WW\gamma$ | | | | | | |
| from $W^+W^-$ | D0 [30] | 0.25 | [-0.31, 0.33] | [-0.36, 0.33] | | |
| from $W^+W^-$, $W^\pm Z$ | CDF [31] | 0.35 | [-0.18, 0.17] | [-0.46, 0.39] | | |

for each lepton flavor ($e,\ \mu,\ \tau$). The gauge-boson decays into tau leptons are included in the MC event generator and these tau leptons decay to all the possible final states. Hard emission is treated as in NLO computations and soft/collinear emission is treated as in a regular parton shower MC. The matching between these two regions is smooth (no double-counting). $W$-boson width and spin-spin correlations are included in the generator. However, 'zero-width' approximations are used in $W^\pm Z$ and $ZZ$ calculations, and no $Z/\gamma^*$ interference terms are included. MC@NLO does not include anomalous triple gauge boson couplings. The process of $W^+W^-$ production via gluon-gluon fusion and the leptonic decays of the $W$, $gg \to W^+W^- \to \ell\nu\ell'\nu$, is modeled by the MC generator GG2WW [34]. The $W^\pm\gamma$ and the $Z\gamma$ production processes and subsequent leptonic decays of the $W^\pm$ and the $Z$ are modeled by the PYTHIA MC generator [35], which only incorporates the leading-order (LO) QCD matrix elements into the parton shower. To include the off-shell $Z$ and $\gamma^*$ into the $ZZ$ analysis, we have also used PYTHIA to generate the $Z/\gamma^* + Z/\gamma^* \to \ell^+\ell^-\ell'^+\ell'^-$ events. The $Z/\gamma^*$ mass threshold is set to 12 GeV in PYTHIA. Table 4 list all the diboson signal samples used in this paper.

Major physics backgrounds for diboson signal detection come from top pairs and hadronic jets associated with $W$ or $Z$ gauge bosons. We have used MC@NLO to model $t\bar{t} \to \ell + X$ production (700k events). The inclusive $W + X$ and $Z + X$ ($X$ = jets, or $\gamma$) processes are modeled by the PYTHIA (30M events) and ALPGEN [36] (1.1M events).

Whenever LO event generators are used, the cross-sections are corrected to NLO by using k-factors from NLO matrix element calculations to normalize the expected signal and background events.

## 2.2 MC generators for TGC studies

Monte Carlo generators BosoMC [16] and BHO [17] are used for anomalous TGC studies. These MC programs are numerical parton level generators. They are used to calculate both LO and NLO cross-sections for all five diboson final states ($W^+W^-$, $W^\pm Z$, $ZZ$, $W^\pm\gamma$, $Z\gamma$) with anomalous coupling parameters. However, they do not include parton showers automatically. We use the BHO MC to model the $ZZ$, $W^+W^-$, and $Z\gamma$ production cross-sections and kinematics with Standard Model and anomalous couplings. BosoMC is used for $W^\pm Z$ and $W^\pm\gamma$ diboson final-state TGC studies. The calculated diboson production rates from these generators are accurate to NLO. The $W^+W^-$, $W^\pm Z$ and $ZZ$ production calculations with the Standard Model couplings are compared with the MC@NLO calculations using PDF CTEQ6M. We found that both cross-sections and kinematic distributions are in good agreement, as shown in Figure 2. The left plot shows the $W^\pm Z$ production differential cross-section as a function of the transverse mass of the $W^\pm Z$ system. The right plot shows the $W^+W^-$ production differential cross-section as a function of the $W$ transverse momentum. The discrepancies between MC@NLO and the BHO MC are within 2%.

Table 4: Diboson signal production processes, cross-sections including branching ratios of $W/Z$ leptonic decays and fully simulated number of MC events. The MC simulation 'filter' is the event selection at the generator level. The corresponding filter efficiencies are given in the table. We also indicate the MC generators used to produce the MC events and to calculate the cross-sections given in this table.

| Process | cross-section (fb) | $\varepsilon_{\text{filter}}$ | $N_{MC}$ | Generator |
|---|---|---|---|---|
| $q\bar{q}' \to W^+W^- \to \ell^+\nu\ell^-\nu$ | 11718 | 1.0 | 180,000 | MC@NLO |
| $gg \to W^+W^- \to \ell^+\nu\ell^-\nu$ | 540.0 | 0.96 | 180,000 | GG2WW |
| $(\ell = e,\ \mu,\ \tau)$ | | | | |
| $q\bar{q}' \to W^+Z \to \ell^+\nu\ell^+\ell^-$ | 441.7 | 1.0 | 50,000 | MC@NLO |
| $q\bar{q}' \to W^-Z \to \ell^-\nu\ell^+\ell^-$ | 276.4 | 1.0 | 50,000 | MC@NLO |
| $(\ell = e,\ \mu;\ Z$ on mass shell$)$ | | | | |
| $q\bar{q}' \to ZZ \to \ell^+\ell^-\ell^+\ell^-$ | 66.8 | 1.0 | 49,250 | MC@NLO |
| $q\bar{q}' \to ZZ \to \ell^+\ell^-\nu\bar{\nu}$ | 397 | 1.0 | 118,000 | MC@NLO |
| $(\ell = e,\ \mu;\ Z$ on mass shell$)$ | | | | |
| $q\bar{q}' \to ZZ \to \ell^+\ell^-\ell^+\ell^-$ | 159 | 0.219 | 43,000 | PYTHIA |
| $(\ell = e,\ \mu,\ \tau;\ M_{Z/\gamma^*} > 12$ GeV $)$ | | | | |
| (4 leptons $(e,\ \mu)$, $p_T^\ell > 5$ GeV, $\lvert\eta^\ell\rvert < 2.7$) | | | | |
| $q\bar{q}' \to W^+\gamma \to \ell^+\nu\gamma$ | 10220 | 1.0 | 38,400 | PYTHIA |
| $q\bar{q}' \to W^-\gamma \to \ell^-\nu\gamma$ | 6820 | 1.0 | 25,600 | PYTHIA |
| $(\ell = e,\ \mu;\ E_T^\gamma > 10$ GeV $)$ | | | | |
| $q\bar{q}' \to Z\gamma \to \ell^+\ell^-\gamma$ | 5280 | 1.0 | 66,000 | PYTHIA |
| $(\ell = e,\ \mu;\ E_T^\gamma > 10$ GeV $)$ | | | | |

A somewhat different procedure is used in the estimation of neutral triple gauge couplings from ZZ events. In this case, the signal expectation with anomalous couplings was determined from the leading order Monte Carlo calculation of BHO, corrected using a $p_T$ dependent k-factor derived from MC@NLO.

# 3 Diboson event selection

This section discusses features of five diboson signals, ($W^\pm Z$, $W^\pm\gamma$, $W^+W^-$, $Z\gamma$, $ZZ$), the major backgrounds and the analysis cuts required to discriminate between them. The varied event topolology of each diboson signal precludes a common set of universal cuts. Two analysis approaches are employed. The first is based on a sequence of straight cuts on kinematic quantities. The second is a refined, multivariate analysis based on *Boosted Decision Tree* (BDT) which is briefly described in Section 3.2 . Some of the diboson analyses employ both techniques. The major results of these analyses are included here.

## 3.1 Physics objects

The ATLAS detector and its performance is described in detail elsewhere [37]. A brief description of the physics objects used in diboson analysis is given below.

The major physics objects used in diboson physics analysis are electrons, photons, muons, missing $E_T$ ($\not{E}_T$), and hadronic jets. Electrons are identified by their distinctive pattern of energy deposition in the calorimeter and by the presence of a track in the inner tracker that can be extrapolated from the

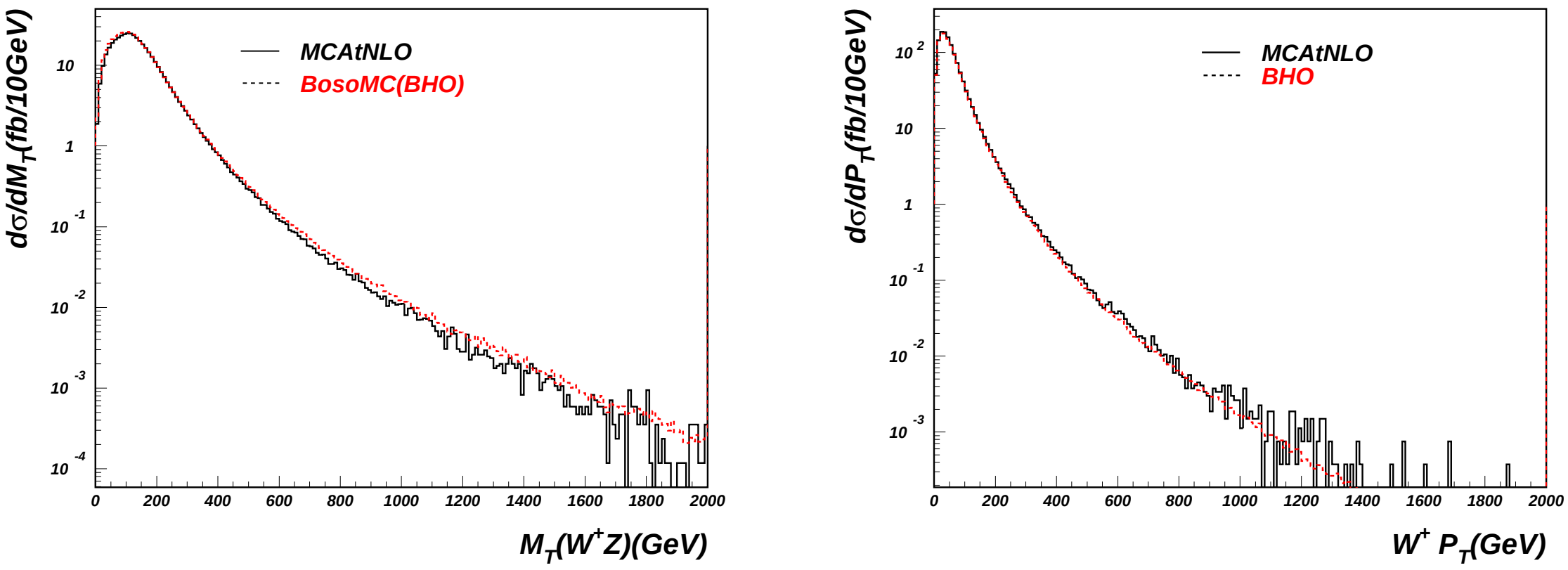


Figure 2: Comparison of MC@NLO MC to BHO MC for $W^{\pm}Z$ and $W^{+}W^{-}$ production. The histograms are normalized to production cross-sections. Left plots: the $W^{\pm}Z$ $M_T$ distribution from $W^{\pm}Z$ production. Right plots: the $W^{+}$ $p_T$ distribution from $W^{+}W^{-}$ production.

interaction vertex to a cluster of energy in the calorimeter. To ensure high trigger efficiency, for events with a single electron, the transverse energy of an single electron must satisfy $E_T > 25$ GeV. For events with dielectrons, both electrons are required to have $E_T > 10$ GeV. The electrons must be isolated from other energy clusters. An electron is required to pass a set of cuts on shower shape, track quality and track to calorimeter cluster matching. Photon identification is similar to an electron in the EM calorimeter, but no charged tracks in the inner tracker should match the EM energy cluster. The average electron identification efficiency in the barrel is about 75% and in the endcaps about 60%.

Muons are reconstructed using information from the outer muon spectrometer (MDT chambers and trigger chambers), the inner tracking detectors and the calorimeters. Muons are identified with a tracking algorithm that associates a track found in the muon spectrometer with the corresponding inner detector track, after the former is corrected for energy loss in the calorimeter. The combined muon detection rapidity coverage is $|\eta| < 2.5$. The minimum $p_T$ of reconstructed muon track is 5 GeV. The candidate muons are required to be isolated in the calorimeter and inner tracker to minimize the contributions of muons originating from hadronic jets. The average muon identification efficiency is about 95%.

The hadronic jets are reconstructed using the fixed-cone jet algorithm. The cone size used in this analysis is 0.7. The jet seed threshold on the transverse energy in a tower is set to $E_s = 1$ GeV, and the final energy cut on a jet is $E_T > 7$ GeV. With this cut the minimum measurable jet $E_T$ should be 20 GeV.

Missing transverse energy, $\not{E}_T$, is calculated from the energy deposited in all calorimeter cells and from muons. A correction is applied for the energy lost in the cryostat. For diboson events with neutrinos in final states, the $\not{E}_T$ resolution we found to be 6.5 GeV based on our studies using $Z \to \ell^{+}\ell^{-}$ MC events.

The ATLAS trigger consists of three levels of event selection: Level-1 (L1), Level-2 (L2) and the event filter (EF). The L2 and EF together form the High-Level Trigger (HLT). According to the present physics trigger menu for initial running, diboson candidate events with multi-lepton final states will be recorded with single muon, single electron, dielectron and dimuon triggers. The trigger efficiencies for diboson events, defined as the fraction of events accepted by the analysis cuts that have satisfied the trigger requirements, are expected to be in a range of 95% - 100%, with the exception of a lower efficiency ($\sim$ 80%) for $W\gamma$ events. The events that are accepted by the trigger are used for the analysis results.

### 3.2 Boosted Decision Trees

A rather new multivariate analysis technique, *Boosted Decision Tree*, has been used in our analysis to improve the detection sensitivity for diboson signals. BDTs have been used in HEP data analysis in recent years [38], and details can be found in the Ref. [4]. A further development, allowing for weighted events, is used in these diboson physics studies [39].

The BDT technique involves a fast 'training' procedure for event pattern recognition. It works with a set of data including both *signal* and *background*. Data are represented by a set of physics variable distributions. A *decision-tree* splits data recursively based on *cuts* on the input variables until a stopping criterion is reached. Every event ends up in a *signal* (score=1) or a *background* (score=-1) *leaf* of the decision-tree. Misclassified events will be given larger weights in the next tree (boosting). This procedure is repeated several hundreds to thousands of times until the performance is optimal. The *discriminator* from the BDT training is the sum of the weighted scores from all the decision-trees. If the total score for a given event is relatively high this event is most likely a signal event, and if the score is low it is likely a background event.

### 3.3 $W^{\pm}Z \rightarrow \ell^{\pm}\nu\ell^{+}\ell^{-}$ selection

The $W^{\pm}Z$ production at hadron colliders uniquely probes the $WWZ$ trilinear gauge boson coupling as shown in Figure 1, the Standard Model tree-level Feynman diagrams.

$W^{\pm}Z$ candidate events have three charged lepton final states, referred to as trileptons, produced when $Z \rightarrow \ell^{+}\ell^{-}$ and $W^{\pm} \rightarrow \ell^{\pm}\nu$, where $\ell^{\pm}$ are $e^{\pm}$ or $\mu^{\pm}$. Standard Model backgrounds can be highly suppressed by requiring three isolated high $p_T$ leptons and large missing transverse energies ($\not{E}_T$). However, the pure leptonic decay mode of the $W^{\pm}Z$ events only has a 1.5% branching ratio [2]. The total cross-sections times the branching ratio are 442 fb and 276 fb for $W^{+}Z$ and $W^{-}Z$, respectively. Event selection with high efficiency is important for early observations of this channel.

Major backgrounds to the $W^{\pm}Z$ trilepton final states come from $ZZ \rightarrow \ell^{+}\ell^{-}\ell^{+}\ell^{-}$ with one lepton undetected; $Z+X \rightarrow \ell^{+}\ell^{-}+X$ (X=jets, or photon) with a jet or a photon faking a lepton; and $t\bar{t} \rightarrow W^{+}W^{-}b\bar{b} \rightarrow \ell+\ell+\ell+X$.

After a trigger, $W^{\pm}Z$ events are selected in two stages: (1) pre-selection with relatively loose cuts, and (2) final selection with tightened cuts or with BDT multivariate discriminator. The overall trigger efficiency for $W^{\pm}Z$ events with trileptons in the final state is (98.9± 1.0)% using a combination of single lepton and dilepton triggers.

The pre-selection of the $W^{\pm}Z$ events is done by identifying three leptons (at least one lepton with $p_T > 25$ GeV) and requiring $\not{E}_T > 15$ GeV in an event with characteristics consistent with $Z$ dilepton decays ($M_{\ell\ell} = (91.18 \pm 20)$ GeV) and $W$ leptonic decays (10 GeV $< M_T(\ell, \not{E}_T) <$ 400 GeV). The overall pre-selection efficiency for $W^{+}Z$ events is 26%, and for $W^{-}Z$ events is 29%. The acceptance difference is due to the differing $\eta$ distributions of the leptons decaying from $W^{+}$ and $W^{-}$. After the preselection, the known background is about 70 times larger than the signal.

To bring the background level below the signal, a rejection power better than 100 in the second stage of event selections is required. To achieve this, events with $\not{E}_T < 25$ GeV and with significant hadronic jet activities are rejected. Events must contain no more than one hadronic jet with $E_T^{jet} > 30$ GeV and $|\eta^{jet}| < 3.0$. The transverse recoil of the $W^{\pm}Z$ system, calculated using the vector sum of the $p_T$ of the charged leptons and $\not{E}_T$, is required to be less than 120 GeV and the sum of the hadronic transverse energy must be less than 200 GeV. These requirements effectively reject the $t\bar{t}$ and hadronic jet events. To reject the $Z+X$ background, the third lepton (not from the $Z$ boson decay) $p_T$ is required to be greater than 20 GeV and 25 GeV for muons and electrons, respectively. Any pair of leptons must satisfy $\Delta R = \sqrt{(\Delta\eta^2 + \Delta\phi^2)} > 0.2$. Leptons must be isolated from other energy clusters based on calorimeter and inner tracker measurements. All three leptons must be associated with isolated tracks that originate

Table 5: Number of expected $W^{\pm}Z$ signal ($N_{WZ}$) and background ($N_B$) for 1 fb$^{-1}$ data with cut-based analysis.

| | $WZ$ | $ZZ$ | $t\bar{t}$ | $Z+$jet | $Z+\gamma$ | Drell-Yan | Total bkg | $N_{WZ}/N_B$ |
|---|---|---|---|---|---|---|---|---|
| $N$ events | **53** | 2.7 | .023 | 1.9 | 0.18 | 2.5 | **7.3** | 7.3 |
| % of background | - | 37 | .32 | 26 | 2.5 | 35 | - | - |

from the same collision point. The dilepton invariant mass best matching the mass of $Z$ must be within the $Z$-mass window of $|M_Z - M_{\mu\mu}| < 12$ GeV or $|M_Z - M_{ee}| < 9$ GeV. These mass windows are set by the mass resolutions. The $M_T$ determined by the third lepton and the $\not{E}_T$ must be within the transverse $W$-mass window: 40 GeV $< M_T <$ 120 GeV.

The total and selected number of the signal and the background events for each trilepton final state, and for 1 fb$^{-1}$ of integrated luminosity, are listed in Table 5. The overall signal efficiency is 8.7% and 7.1% for $W^-Z$ and $W^+Z$, respectively. For 1 fb$^{-1}$ of data, 53 $W^{\pm}Z$ signal events and 7 background events are expected. The dominant background contributions are from $ZZ$, $Z$+jet and Drell-Yan processes, while $Z\gamma$ and $t\bar{t}$ contribute a small fraction of the total background events.

Table 6: Number of expected $W^{\pm}Z$ signal ($N_{WZ}$), and background ($N_B$) for 1 fb$^{-1}$ of data with BDT analysis using the cut BDT $>$ 200.

| | $WZ$ | $ZZ$ | $t\bar{t}$ | $Z+$jet | $Z+\gamma$ | Other | Total bkg | $N_{WZ}/N_B$ |
|---|---|---|---|---|---|---|---|---|
| $N$ events | **128** | 7.7 | 2.8 | 2.5 | 2.0 | 1.1 | **16** | 7.9 |
| % of background | | 48 | 17 | 16 | 12 | 7.0 | - | - |

For 0.1 fb$^{-1}$ of integrated luminosity, only 5 signal events ($N_S$) with 0.7 background event ($N_B$) contamination are expected. The $W^{\pm}Z$ detection significance, defined as the probability from Poisson distribution with mean $N_B$ to observe equal or greater than $N_S + N_B$ events, converted in equivalent number of sigmas (standard deviations) of a Gaussian distribution, will be $3.6\sigma$ only. To improve the detection sensitivity, the BDT analysis technique is employed. This BDT analysis is conducted with a total of 1000 trees with 20 tree-split nodes. Based on the variables used in the cut-based analysis, and the BDT training Gini-index (a measure of a variable effectiveness in separating signal from background), a total of 22 kinematic and topology variables are selected for the BDT training. About 12000 pre-selected signal events and 18000 pre-selected background events are used in this BDT-based analysis, where 50% of the signal and background events are used for the training, and another 50% of statistically independent events allocated to the BDT test sample sets. The BDT output discriminator from the testing sample, used to separate the signals from the background, is shown in Figure 8 (right), see Section 4.1. This spectrum will be used to determine (fit) the $W^{\pm}Z$ production cross-section.

Results of the $W^{\pm}Z$ event selection with the cut BDT $>$ 200 are shown in Table 6. Expected numbers of signal and background events are given for 1 fb$^{-1}$ of integrated luminosity. The estimated background uncertainties are 15-20% due to the limited number of simulated events available. The overall $W^{\pm}Z$ event selection efficiency is about 13.7% for signal, resulting in 12.8 signal and 1.6 background events for 0.1 fb$^{-1}$. The $W^{\pm}Z$ detection significance is expected to be $5.9\sigma$ (including both statistical and 20% systematic uncertainties) for 0.1 fb$^{-1}$ of integrated luminosity.

### 3.4 $W^{\pm}\gamma \to \ell^{\pm}\nu\gamma$ selection

The $W^{\pm}\gamma$ signal events are modeled with PYTHIA, which includes tree-level diagrams as shown in Figure 1 for $W$ production with initial state radiation (ISR) (the $t-$ and $u-$channels) and the $s-$channel production depending on the triple-gauge-coupling $WW\gamma$ vertex. The $WW\gamma$ vertex introduces a destructive interference of *zero amplitude* at $\cos\theta_{\bar{q},\gamma} = \pm 1/3$ for $W^{\pm}$ production, where $\theta_{\bar{q},\gamma}$ is the photon scattering angle to the incoming anti-quarks.

The $W^{\pm}\gamma$ diboson events are selected from $pp$ collisions using pure $W$ leptonic decays. The experimental signature is a final states with one high $p_T$ lepton (electron or muon), one high $p_T$ photon and large $\not{E}_T$. Major backgrounds are from the processes:

- Inclusive $W+X$ production with $W+X \to \ell\gamma\nu + X$ , where the $\gamma$ is from final state radiation (FSR) from the lepton.
- Inclusive $W+X \to \ell\nu + X$ productions, with $X = jets$ and the jets faking a photon.
- Inclusive $Z+X \to \ell\ell + X$ productions, with one lepton escaping detection, and with $X = \gamma$ or jets faking a photon.

A photon isolation cut is effective in suppressing these backgrounds.

To study the $W^{\pm}\gamma$ detection sensitivity about 1.3 million inclusive $W$ events and about 100,000 inclusive $Z$ events were generated with the PYTHIA MC generator. In these datasets, the photon transverse energy threshold is 10 GeV and the lepton and photon separation, $\Delta R(\ell,\gamma)$ was required to be greater than 0.7.

The $W^{\pm}\gamma$ candidates are inclusive $e^{\pm}\gamma$ or $\mu^{\pm}\gamma$ events having one electron or muon observed and the absence of the oppositely charged lepton of the same flavor. The photon selected is the most energetic one in the events. Efficiencies for three trigger types are investigated: isolated muon with $p_T > 20$ GeV, isolated electron with $p_T > 22$ GeV and photon with $E_T(\gamma) > 55$ GeV. The overall $W^{\pm}\gamma$ trigger efficiency is about 80% based on our study.

Background to the $W^{\pm}\gamma$ signal is dominated by inclusive $W^{\pm}$ events with radiated jets faking a photon. Contamination from inclusive $Z$ events with an undetected lepton is also significant. Figure 3 shows the $W^{\pm}\gamma$ signal (first column) scatter plots compared to those for major backgrounds: the inclusive $W$ events with final state radiation (FSR) (the 2nd column), and with fake photon (the third column), and the $Z$ events with one lepton escaping detection and a photon of any type reconstructed (the 4th column).

The BDT method is used to select the $W^{\pm}\gamma$ signal events. Three trainings are done to separate: 1) The $\ell\gamma\nu$ events with FSR photons from other sources, 2) The $W^{\pm}\gamma$ signal photons from fake photons, and 3) The signal photons from the contamination of $Z$ inclusive events. Nineteen variables are used in the BDT analysis. Cuts are applied to the BDT output spectra for the three trainings. The cuts chosen for signal selection correspond to a $W^{\pm}\gamma$ selection efficiency of 65% with a signal to background ratio of 0.95 (0.98) for electron (muon) final states, respectively. The numbers of events selected with the BDT cuts are listed in Table 7. The selected events are further required to pass the trigger requirements as the final accepted events, which are listed in the Table as well.

The BDT discriminates signal photons effectively from FSR and fake photons in the high $E_T(\gamma)$ region. This preserves a high detection efficiency in the region that is sensitive to discovery of phenomenon beyond Standard Model predictions.

### 3.5 $W^{+}W^{-} \to \ell^{+}\nu\ell^{-}\nu$ selection

The production of $W$ pairs has been investigated extensively at LEP and at the Tevatron. The experimental $W^{+}W^{-}$ signature is two high transverse momentum leptons with opposite charge associated with

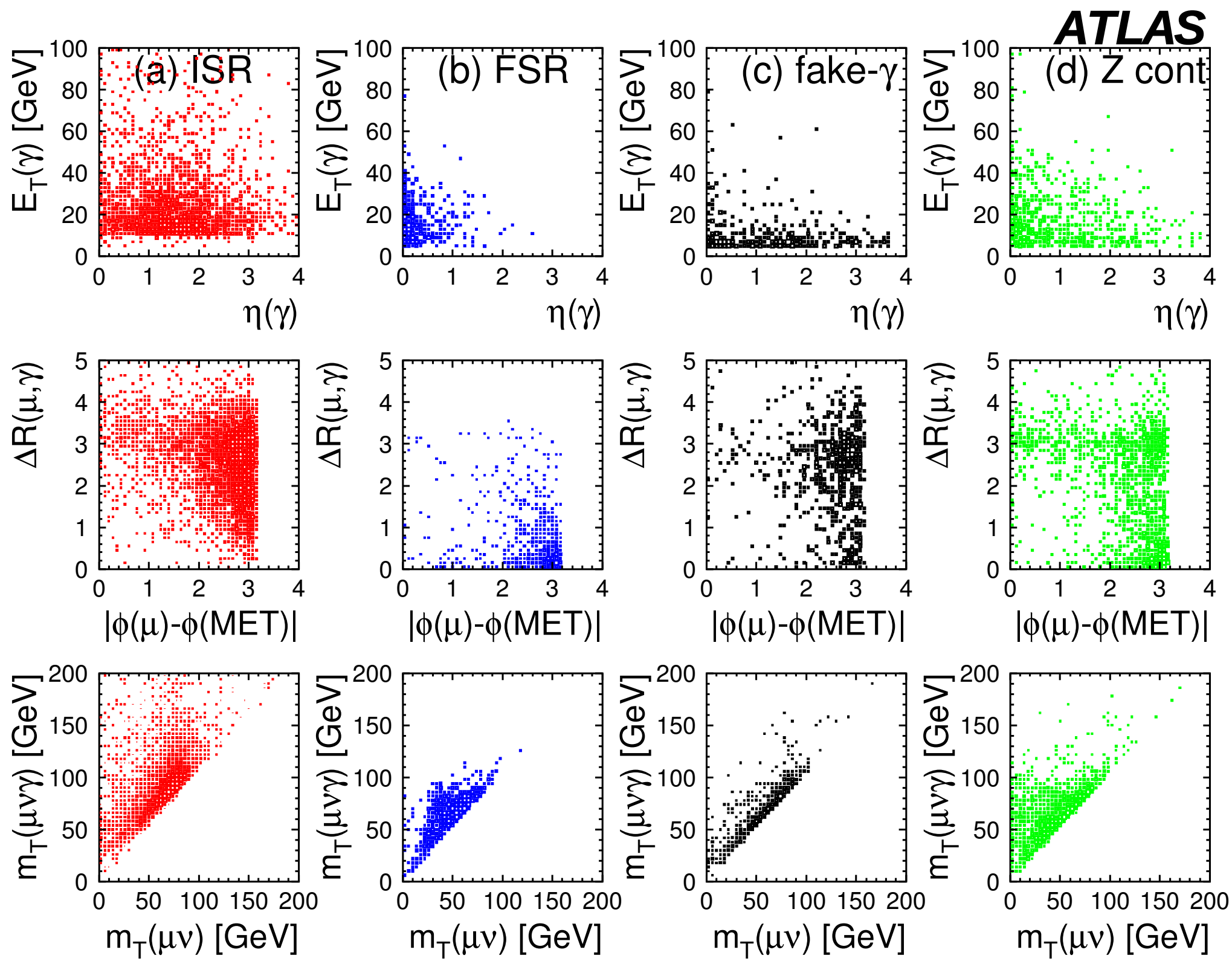


Figure 3: Distributions of the $W^{\pm}(\mu^{\pm}\nu)\gamma$ event variables, with a photon from the $W^{\pm}\gamma$ signal (ISR), and from backgrounds, FSR, fake $\gamma$, and inclusive $Z$ contamination.

large transverse missing energy. $W^+W^-$ production involves both $WWZ$ and $WW\gamma$ triple gauge boson couplings and is most sensitive to the $\Delta\kappa_V$ anomalous coupling parameters. Furthermore, the $W$ pair production provides an important background to Higgs boson searches in the $pp \to H \to WW^{(*)} \to \ell\nu\ell\nu$ channel at the LHC. In dileptonic $W$-decays no Higgs mass peak can be reconstructed, so this background cannot be estimated from the measured data via sideband interpolation. An understanding of the irreducible $W$-pair continuum background is therefore crucial.

At the LHC the major background processes ($t\bar{t}$, inclusive $W$ and $Z$, and Drell-Yan) have much higher cross-sections than $W^+W^-$. It will be necessary to achieve very high background rejection power to suppress the events with mis-identified leptons (from jet, photon and instrumentation effects) and leptonic decays from heavy-flavor quark jets. In the Drell-Yan process the mis-measured $\not\!\!E_T$ also contributes non-negligible background.

Event selection consists of the trigger, pre-selection (two high $p_T$ leptons plus $\not\!\!E_T$), and final selection with a set of conventional cuts or with BDT selection cuts. The $W^+W^-$ events are required to pass one of two high-level trigger paths: a single, isolated electron with $p_t > 25$ GeV or a single muon, with $p_t > 20$ GeV. The trigger efficiencies for $WW \to ee$, $WW \to \mu\mu$ and $WW \to e\mu$ events with two opposite sign isolated leptons ($p_T > 20$ GeV, $|\eta| < 2.5$) are estimated as follows: 98.2 %, 95.9%; and 97.4% respectively.

The cut-based analysis rejects background events with tight lepton identification criteria and isolation requirements as well as with event topology variables which clearly distinguish the signal and significantly suppress the main background processes $t\bar{t}$ and $Z+X$. The $W^+W^-$ leptonic decay events are selected by requiring two well identified isolated leptons with opposite charge, and with $p_T^{\ell} > 20$ GeV and $|\eta| < 2.5$. A jet veto requirement rejects events with any jet ($p_T^{jet} > 20$ GeV) in the rapid-

Table 7: The number of $W^{\pm}\gamma$ signal and background events after pre-selection, BDT selection and trigger requirement, for an integrated luminosity of 1 fb$^{-1}$. The signal and total background are then scaled to NLO cross-sections with the k-factors indicated. For the signal, the k-factor is obtained using BosoMC. For background, the k-factors are obtained by comparing the cross-sections calculated with *MC@NLO* and PYTHIA generators.

| | | Signal | Background | | | |
|---|---|---|---|---|---|---|
| | | $W^{\pm}\gamma$ | W+FSR_$\gamma$ | W+fake_$\gamma$ | $Z(\ell\ell)\gamma$ | Total |
| $\ell = e$ | Pre-selected | 1710 | 11440 | 7890 | 32480 | |
| | BDT selection | 1145 | 242 | 791 | 101 | |
| | Triggered | 966 | 188 | 628 | 93 | |
| | NLO scaled | 1604 (k=1.66) | | | | 1183 (k=1.3) |
| $\ell = \mu$ | Pre-selected | 2680 | 28410 | 10250 | 3950 | |
| | BDT selection | 1793 | 413 | 961 | 409 | |
| | Triggered | 1305 | 177 | 595 | 260 | |
| | NLO scaled | 2166 (k=1.66) | | | | 1342 (k=1.3) |

ity region $|\eta| < 3$. This cut is efficient in $t\bar{t}$ suppression since $t\bar{t}$ contains one or two energetic $b$ jets in addition to the $W^+W^-$ signature. An event with $\not{E}_T < 50$ GeV is rejected to reduce the background arising from the event pileup and from $Z/\gamma*$ events in the Drell-Yan process. To reject the dilepton events from inclusive $Z$ production, a $Z$ mass veto is applied. Finally, angular variable cuts are imposed. For cross-section measurements, events must pass a cut: $\phi_{\ell\ell} < 2$ rad, where $\phi_{\ell\ell}$ is the angle between the transverse momentum of the two leptons. For anomalous TGC studies, this cut is replaced with: $\Phi(\mathbf{p}_T^{\ell^+,\ell^-}, \mathbf{p}_T^{miss}) > 175\,$deg, where $\Phi$ is the angle between the transverse momenta of the lepton pair, and the missing transverse momentum. The first angular cut (selection-A) results in high signal detection efficiency, but it is not optimized for high $p_T(\ell)$ and $p_T(\ell\ell)$ event detection and thus could decrease the sensitivity to anomalous TGC's. The second angular cut (selection-B) results in lower signal detection efficiencies, but has higher efficiency for high $p_T(\ell)$ and $p_T(\ell\ell)$ events. The yields of the $W^+W^-$ selection with the cut-based analysis are summarized in Table 8. Both signal and background events are normalized to 1 fb$^{-1}$ of data. Figure 4 shows the transverse momentum distributions of leptons (left) and lepton pairs (right) after applying the kinematic cuts of Selection-B. The distributions are shown for sum of signal and various backgrounds, and for individual backgrounds, for an integrated luminosity of 1 fb$^{-1}$ of data.

The $W^+W^-$ overall signal detection efficiency is 1.4-3%, depending on the selection cuts and the final lepton states. The signal detection significance for 0.1 fb$^{-1}$ of data is expected to be $4.7\sigma$ (for Selection A), after taking into account a 20% background systematic uncertainty. The detection efficiencies can be improved by using the multivariate BDT technique. In this $W^+W^-$ analysis one thousand decision trees, wherein one tree has 20 *splitting-nodes*, are used to separate signal from background based on the input variables. The input data for the BDT analysis must first pass the pre-selection cuts (two leptons with $p_T^{\ell} > 10$ GeV and $\not{E}_T > 15$ GeV). The pre-selected simulated samples are divided into two equal parts: one sample is used for BDT training and the other to test event selection performance. The BDT output spectra for both signal and background are shown in Figure 8 (left). By varying the location of the cut along the BDT *scores* ($x$-axis), the signal to background ratio can be optimized. Table 9 presents the detection sensitivities with total integrated luminosity of 1 fb$^{-1}$: the selected number of signal events ($N_{WW}$), the corresponding signal efficiency ($\varepsilon_{WW}$), the number of background ($N_{bkg.}$) events, and signal to background ratio ($N_{WW}/N_{bkg}$) are shown. The breakdown of background contributions are also given

Table 8: Yield and total event detection efficiency of the WW selection for 1 fb$^{-1}$ of data. The errors shown are statistical only.

| | efficiency | | $N_{WW}$ | | $N_{\text{background}}$ | $N_{\text{sig.}}/N_{\text{bkg.}}$ |
|---|---|---|---|---|---|---|
| Selection-A | $gg \to WW$ | $q\bar{q} \to WW$ | $gg \to WW$ | $q\bar{q} \to WW$ | | |
| $ee$ | 2.1% | 1.3% | 1.3±0.05 | 17.4 ± 1.1 | 1.4 ±0.3 | 13.3±3.0 |
| $\mu\mu$ | 4.1% | 2.8% | 2.4±0.08 | 36.4 ± 2.2 | 10.7 ± 2.1 | 3.6± 0.8 |
| $e\mu$ | 2.8% | 1.9% | 3.3±0.13 | 50.6± 1.8 | 7.2 ± 1.2 | 7.5 ± 1.3 |
| $ll$ | 3.0% | 2.0% | 7.0±0.16 | 104.4±2.4 | 19.3±2.4 | 5.8 ± 0.8 |
| Selection-B | | | | | | |
| $ee$ | 0.94% | 0.92% | 0.6±0.04 | 12.0 ± 0.9 | 2.8± 1.2 | 4.5 ± 1.9 |
| $\mu\mu$ | 2.1% | 2.0% | 1.1±0.03 | 25.5 ± 1.8 | 4.8±1.0 | 5.5 ± 1.2 |
| $e\mu$ | 1.3% | 1.4% | 1.5±0.09 | 35.3 ± 1.5 | 7.4± 1.3 | 5.0 ± 0.9 |
| $ll$ | 1.4% | 1.4% | 3.2±0.10 | 72.8 ± 2.5 | 15.0 ± 2.0 | 5.1 ± 0.8 |

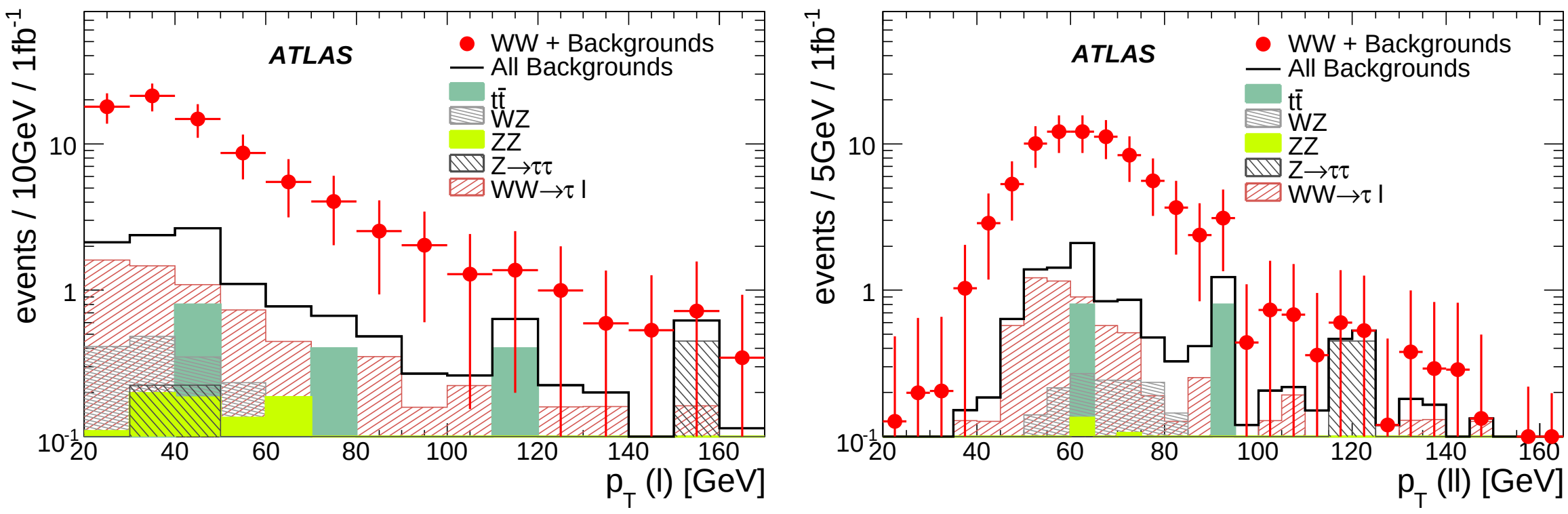


Figure 4: Transverse momentum distributions of leptons (left) and lepton pairs (right) after applying kinematic cuts from Selection-B. The distributions are shown for sum of signal and various backgrounds, and for separated backgrounds for L=1 fb$^{-1}$.

in this table. For initial measurements using early LHC data based on 0.1 fb$^{-1}$ of integrated luminosity, the application of BDT is compelling. As inferred from Table 8, the initial data is expected to yield a total for all decay channels of $\sim$ 10 signal events using conventional cuts, whereas the BDT-based analysis, which gives a similar signal to background ratio as the conventional cuts is expect to yield total 47 signal events. With an estimated background contribution of 9.2 events the $W^+W^-$ detection significance is about 10$\sigma$ (including 20% background systematic uncertainties).

### 3.6 $Z\gamma \to \ell^+\ell^-\gamma$ selection

$Z\gamma$ signals are produced through initial state radiation (ISR) of the photon from the quarks as illustrated by the $t$- and $u$-channel diagrams shown in Figure 1. The $s$-channel $Z\gamma$ production contains the $Z\gamma V$ ($V = Z,\gamma$) vertex, which is forbidden at tree-level in the Standard Model. The cross-section measurement of the $Z\gamma$ production would provide a sensitive probe to anomalous $Z\gamma V$ couplings, which can be investigated through this channel by measuring the $E_T(\gamma)$ distribution, expecially at large values.

The cleanest $Z\gamma$ experimental signature is two high $p_T$ leptons from the decay of the $Z$ boson and an

Table 9: $WW \rightarrow$ leptons detection sensitivities of accepted signal and background events for 1 fb$^{-1}$ of integrated luminosity. Results from the BDT analysis are shown with cuts that give similar signal to background ratio as the cut-based analysis. The quoted efficiencies in the table are the BDT selection efficiencies including the trigger requirements based on pre-selected events.

| Modes | $\varepsilon_{WW}(\%)$ | $N_{WW}$ | $N_{bkg}$ | Background fraction $t\bar{t}$ | $W^{\pm}Z$ | $Z+X$ | $N_{WW}/N_{bkg}$ |
|---|---|---|---|---|---|---|---|
| $e\nu\mu\nu$ | 32.7 | 347± 3 | 64± 5 | 47.7% | 27.8% | 21.8% | 5.4 |
| $\mu\nu\mu\nu$ | 12.1 | 70± 2 | 17± 2 | 54.1% | 34.6% | 11.3% | 4.1 |
| $e\nu e\nu$ | 13.7 | 52± 1 | 11± 1 | 81.4% | 7.2% | 11.4% | 4.7 |

isolated high $p_T$ photon (from ISR). The backgrounds to this $Z\gamma$ signal are: 1) the $Z$ boson production with a FSR photon from the leptons decay from the $Z$, 2) the $Z$ boson production with a fake photon from jets, and 3) a small contamination from $W+X$ production reconstructed as a $\ell^+\ell^-\gamma$ final state. Some of the event variables for different photon sources are shown in Figure 5.

Events are pre-selected with $E_T(\gamma) > 10$ GeV, a value chosen to be above the PYTHIA generator threshold and as low as is reasonably achievable by detector reconstruction. The FSR event rate is almost an order of magnitude higher than the ISR rate in the inclusive $Z$ production process. Backgrounds with a $Z$ boson and a fake photon are comparable to the $Z\gamma$ signal photon rate.

The event selection is conducted with BDTs trained to separate $Z\gamma$ events of different photon types. The training is in two stages: first to identify the FSR photon background events and then to distinguish the signal (ISR) photon from $Z$ events with fake photons. Separate BDT training is done for the electron and muon $Z$ decay channels.

The BDTs are trained with 19 variables in total. As one example, FSR photons can be identified from the opening angle from the nearest lepton. Fake photons originating from high $p_T$ neutral mesons decaying to two photons can not be directly identified within the limits of the spatial resolution provided by the ECAL segmentation. However, they are often accompanied by jet secondaries or underlying remnant particles. By counting the charged tracks in a neighborhood (a cone of 0.45 rad in this case), or summing their energies, parameters useful for differentiating background from isolated ISR photons can be formed.

With the chosen BDT cuts the signal selection efficiency is 67%, and the signal to background ratio is 2.0 and 1.8 for the electron and muon $Z$ decay channels, respectively. The estimated numbers of reconstructed $Z\gamma$ candidates for an integrated luminosity of 1 fb$^{-1}$ are listed in Table 10.

The FSR and fake photons contribute approximately equally to the total background but have important distinctions. The fake photons populate the low $E_T(\gamma)$ region, and are differentiated from the signal photon (ISR) in the $E_T(\gamma) > 20$ GeV region. The shape of the $E_T$ distribution in the low energy region is important for calibration and measurement of the production cross-section with ISR photons. On the other hand, the FSR photons have an $E_T(\gamma)$ distribution similar to the ISR photons which carry signatures of the coupling to the colliding quarks. Event rates in the high $E_T(\gamma)$ region, where the background is primarily FSR, is an important probe of new physics phenomena.

### 3.7 $ZZ \rightarrow \ell^+\ell^-\ell^+\ell^-$ selection

The cleanest experimental signature for $ZZ$ detection is through the four lepton decay channels:

$$pp \rightarrow ZZ \rightarrow e^+e^-e^+e^-,\ \ \mu^+\mu^-\mu^+\mu^-,\ \ e^+e^-\mu^+\mu^-.$$

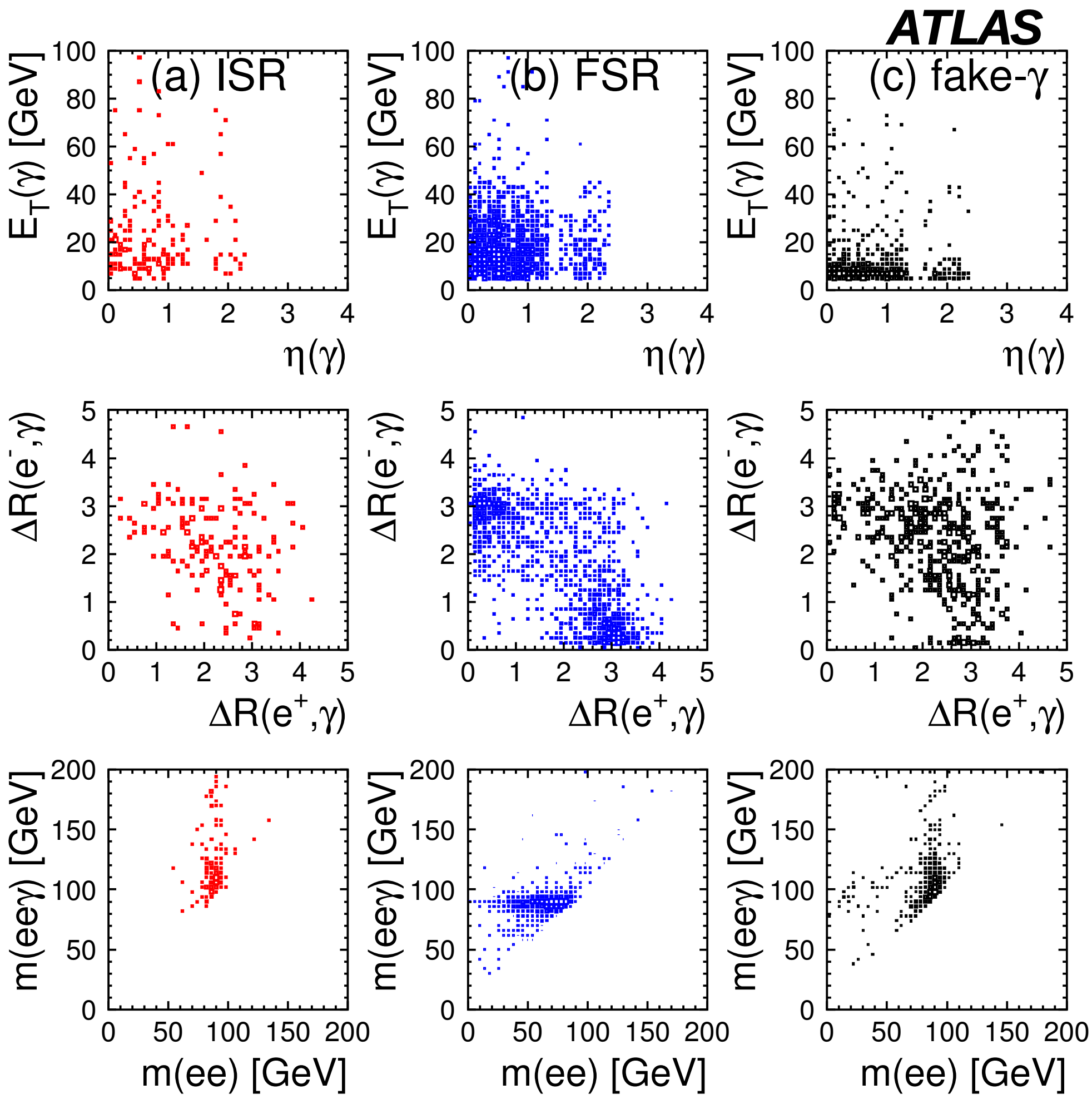


Figure 5: Distributions of $Z(ee)\gamma$ event variables for ISR (left column), FSR (middle column.) and fake photons (right column).

The $ZZ \to 4\ell'$ $(\ell' = e,\ \mu,\ \tau)$ signal is modeled at LO by the PYTHIA event generator using the CTEQ6L PDF. The $Z/\gamma^*$ interference terms are included in the generator. With the 12 GeV mass cut on the dileptons decay from $Z/\gamma^*$, the cross-section times the dilepton decay branching ratio, $\sigma \times BR$ is 159 fb. A filter is applied to the simulated data to pre-select four lepton ($e$ and $\mu$ only) events by requiring that the lepton transverse momenta must be greater than 5 GeV, and the lepton rapidity, $\eta_\ell$, must be in a range $|\ eta\ | < 2.7$. The overall filter efficiency is 0.219. The $\tau$ lepton contribution to the four lepton channels ($4e$, $4\mu$, $2e2\mu$) is less than 4% in the event sample after the filter. The fraction of the on-shell $ZZ$ events (both lepton pair masses are between 70 GeV to 110 GeV) in the sample is about 73%. Next-to-leading-order calculations give higher production cross-sections. The k-factor is about 1.35 when both $Z$'s are on mass shell. However, when the $Z/\gamma^*$ are off the $Z$ mass shell, the k-factor varies from 1.15 to 1.52 for the $(Z/\gamma^*)(Z/\gamma^*)$ mass range from 115 GeV to 405 GeV, which is determined by using the MCFM Monte Carlo calculations [11]. The k-factor is set constant at 1.35 to normalize the $ZZ \to 4\ell$ signal events. The four-lepton events have high trigger efficiencies close to 100%.

Major background processes for four lepton final states are $t\bar{t} \to WWb\bar{b} \to 4\ell + X$ and $Zb\bar{b} \to 4\ell + X$. The $t\bar{t}$ background events, generated using *MC@NLO*, has a total production cross-section of 833 pb. The $Zb\bar{b}$ background events are generated using *AcerMC*, with cross-section scaled to NLO by a k-

Table 10: The number of $Z\gamma$ signal and background events after pre-selection and BDT selection is listed, for an integrated luminosity of 1 fb$^{-1}$. The signal and total background are then scaled to NLO cross-sections with the k-factors indicated. For the signal, the k-factor is obtained using BHO.

| | | Signal | | Background | | |
|---|---|---|---|---|---|---|
| | | $Z\gamma$ | Z+FSR_$\gamma$ | Z+fake_$\gamma$ | $W(l\nu)\gamma$ | Total |
| $\ell = e$ | Pre-selected | 430 | 2760 | 490 | 44 | |
| | BDT selection | 288 | 70 | 74 | 0 | |
| | Triggered | 282 | 65 | 79 | 0 | |
| | NLO scaled | 367 (k=1.3) | | | | 187 (k=1.3) |
| $\ell = \mu$ | Pre-selected | 950 | 7500 | 790 | 930 | |
| | BDT selection | 636 | 173 | 186 | 0 | |
| | Triggered | 578 | 164 | 165 | 0 | |
| | NLO scaled | 751 (k=1.3) | | | | 429 (k=1.3) |

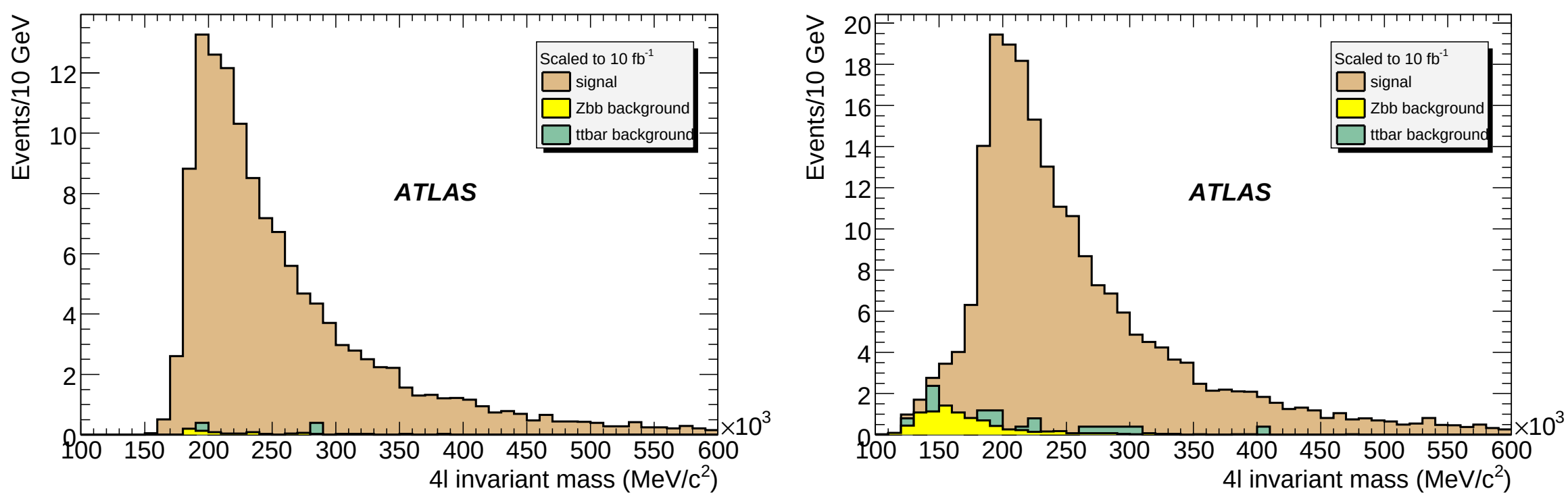


Figure 6: The four-lepton invariant mass distributions of $ZZ$ signal, $Zb\bar{b}$ and $t\bar{t}$ background events with tight (left) and loose (right) $Z$ mass cut on the lepton pairs. The number of events correspond to an integrated luminosity of 10 fb$^{-1}$.

factor. Leptons from b quark decays in these processes are produced in association with hadrons. Their contributions can be highly suppressed by lepton isolation requirements. For muons, the ratio between the transverse energy deposited in a cone around the muon track of radius $\Delta R = 0.4$ and the transverse energy of the muon $E_T^{\mu}$ must be below 0.2. A similar isolation cut is applied to the electron selections. To reject background with leptons not originating from the $Z$ decays, the two opposite sign lepton pairs must have at least one lepton with $p_T$ greater than 20 GeV, and at least one lepton pair must have the invariant mass between 70 GeV- 110 GeV. This is referred to as the loose $Z$ mass cut. A tight $Z$ mass cut requires the second lepton pair also to have invariant mass between 70 GeV- 110 GeV. The separation between the two leptons must be $\Delta R(\ell^+\ell^-) > 0.2$. Table 11 lists the signal selection cut efficiencies for all the four lepton final states. Each cut efficiency value is relative to the previous selection. The quoted uncertainties of the selected numbers of events are statistical only.

The total selection efficiencies for the $Zb\bar{b}$ background using these same tight cuts are $0.13\% \pm 0.06\%$, $0.61\% \pm 0.14\%$, and $0.51\% \pm 0.13\%$ for the $4\mu$, $4e$, and $2\mu 2e$ channels, respectively. For $t\bar{t}$ these efficiencies are $0.07\% \pm 0.07\%$ for all three channels. The expected number of signal and background events for $L = 1$ fb$^{-1}$ requiring tight $Z$ mass cut are given in Table 12. The expected signal and

Table 11: $ZZ \to \ell^+\ell^-\ell^+\ell^-$ signal selection cut efficiencies

| | $4\mu$ [%] | | $4e$ [%] | | $2\mu 2e$ [%] | |
|---|---|---|---|---|---|---|
| Lepton Preselection | 70.7 | | 62.3 | | 65.4 | |
| Pair formation, dR | 99.3 | | 88.0 | | 93.4 | |
| Isolation, $p_T^{max}$ | 81.1 | | 58.6 | | 59.1 | |
| Z Mass | tight | loose | tight | loose | tight | loose |
| | 72.7 | 92.0 | 76.1 | 93.5 | 77.8 | 95.2 |
| Total | $41.4 \pm 0.6$ | $52.4 \pm 0.7$ | $24.4 \pm 0.5$ | $30.0 \pm 0.6$ | $28.1 \pm 0.4$ | $34.3 \pm 0.4$ |

Table 12: Expected number of $ZZ \to \ell^+\ell^-\ell^+\ell^-$ signal and background events at $L = 1$ fb$^{-1}$ using the tight $Z$ mass cut.

| | $4\mu$ events | $4e$ events | $2\mu 2e$ events | Total |
|---|---|---|---|---|
| Signal | 4.5±0.05 | 2.6±0.04 | 6.2±0.06 | 13.3 ±0.09 |
| $Zb\bar{b}$ | 0.01±0.003 | 0.04±0.01 | 0.04±0.01 | 0.08±0.01 |
| $t\bar{t}$ | 0.04±0.04 | 0.04±0.04 | 0.04±0.04 | 0.12±0.07 |
| Total background | 0.05±0.04 | 0.08±0.04 | 0.08±0.04 | 0.20±0.07 |

background events with only one on-shell Z (Loose $Z$ mass cut) are in Table 13. Uncertainties quoted in these tables are statistical only. Figure 6 shows the four-lepton invariant mass distributions for the $ZZ$ signal, $Zb\bar{b}$ and $t\bar{t}$ background with tight and loose $Z$ mass cut on the lepton pairs. Based on these results the ATLAS experiment will establish the $ZZ \to 4\ell$ signal with a significance of $6.8\sigma$ (after taking into account 20% background systematic errors) with the first 1 fb$^{-1}$ of integrated luminosity.

## 3.8 $ZZ \to \ell^+\ell^-\nu\bar{\nu}$ selection

The $ZZ \to \ell^+\ell^-\nu\bar{\nu}$ signature is two high-$p_T$ charged leptons with a large missing transverse energy ($\not{E}_T$) due to the neutrino pair leaving the detector. The main backgrounds will either come from channels with large cross sections, such as $t\bar{t}$ and $Z \to \ell^+\ell^-$, or those with a similar signature to the signal, such as the $W^\pm Z$ diboson channel. Both signal ($ZZ \to \ell^+\ell^-\nu\bar{\nu}$) and background ($t\bar{t}$, $W^\pm Z$, $W^+W^-$, and Drell-Yan dileptons) are modeled by the generator *MC@NLO*, except for high $p_T$ $Z$ ($p_T(Z) > 100$ GeV) events which are modeled by PYTHIA.

To reduce the backgrounds, a set of simple cuts on discriminating parameters is invoked. In general, each cut is used to suppress a particular background channel, as described below.

First, two oppositely charged good quality leptons with $p_T > 20$ GeV are selected. This reduces much

Table 13: Expected $ZZ \to \ell^+\ell^-\ell^+\ell^-$ signal and background events at $L = 1$ fb$^{-1}$ using the loose $Z$ mass cut.

| | $4\mu$ events | $4e$ events | $2\mu 2e$ events | Total |
|---|---|---|---|---|
| Signal | 5.7±0.06 | 3.2±0.04 | 7.6±0.07 | 16.5±0.1 |
| $Zb\bar{b}$ | 0.1±0.01 | 0.5±0.02 | 0.3±0.02 | 0.9±0.1 |
| $t\bar{t}$ | 0.1±0.06 | 0.5±0.14 | 0.4±0.13 | 1.0±0.2 |
| Total background | 0.2±0.06 | 1.0±0.14 | 0.7±0.13 | 1.9±0.2 |

of the $t\bar{t}$ background which contains softer leptons than the signal. This cut also reduces the background from $Z \to \tau^+\tau^- \to \ell^+\ell^-\nu_\ell\bar{\nu}_\ell\nu_\tau\bar{\nu}_\tau$ as the electrons and muons are produced with reduced $p_T$. The leptons must also lie within the inner detector pseudorapidity range, $|\eta| < 2.5$.

The charged lepton pairs are required to have an invariant mass close to the $Z$ mass, specifically $|M_{\ell\ell} - 91.2\text{ GeV}| < 10\text{ GeV}$. This is equivalent to $\sim 5\sigma$ of the signal width, and helps to reduce background combinatorics where the lepton pair does not come directly from a $Z$ decay. A lepton veto is imposed by combining the good quality and loose lepton selection to remove any events with more than two leptons in total. This reduces background from the $W^{\pm}Z$ channel, whose $Z$ has an almost identical signature to the signal, and the neutrino from $W$ decay also appears as $\not\!\!E_T$. The third-lepton veto suppresses the $W^{\pm}Z$ background by $\sim 30\%$. If the lepton from the $W$ is not reconstructed, however, this background channel becomes almost indistinguishable from the signal.

A main characteristic of the signal decay is a large missing transverse energy ($\not\!\!E_T$) from the $Z \to \nu\bar{\nu}$ decay. An important background, due to its large cross section, comes from the $Z \to \ell^+\ell^-$ Drell-Yan process, where jets are produced in addition to the leptons. If these jets are aligned with cracks in the detector, then they will fake $\not\!\!E_T$ as they will not be fully accounted for in the calorimeters. This background can be significantly reduced by applying a 50 GeV $\not\!\!E_T$ cut. The background from $ZZ \to 4\ell$ is also reduced, but this is less significant as it has a much smaller cross-section. The $W^{\pm}Z$ channel is suppressed by this cut as only one neutrino is produced, and hence the $\not\!\!E_T$ distribution is slightly softer.

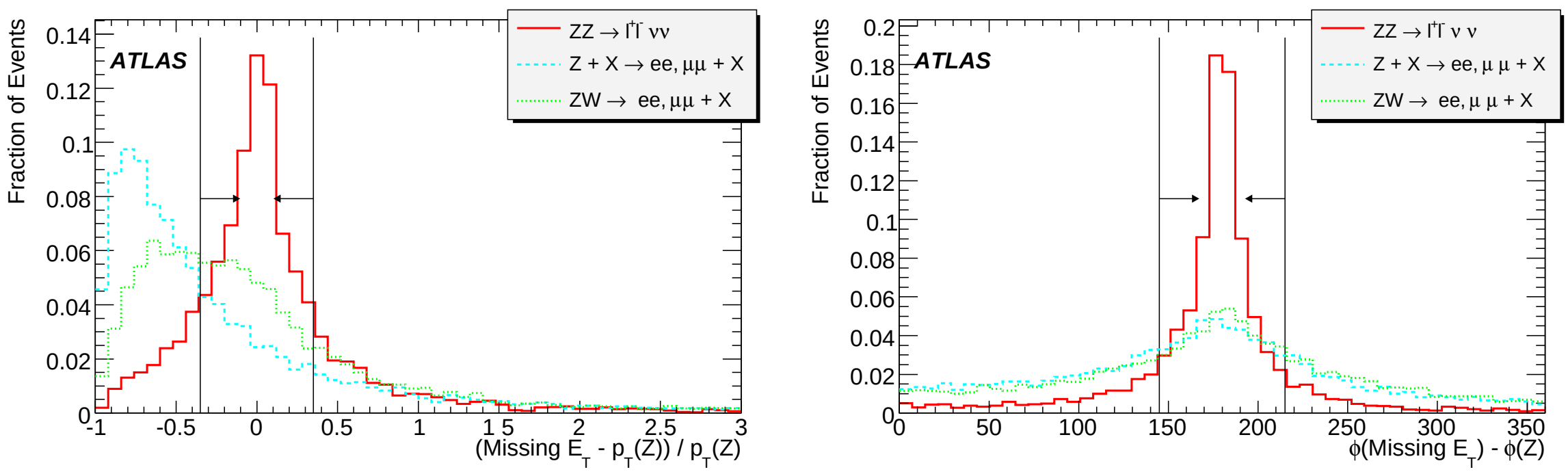


Figure 7: The $\not\!\!E_T - p_T(Z)$ magnitude (left) and angle matching (right) distributions before cuts for $ZZ \to 4\ell$ (solid), $Z \to \ell^+\ell^-$ (dash) and $W^{\pm}Z$ (dot). The plots are normalised to unit area for comparison of distribution shapes.

The signal is expected to have missing $p_T$ equal and opposite to that of the reconstructed $Z$, when the $ZZ$ pair is produced with no initial $p_T$ and they decay back-to-back. Figure 7 (left and right) shows a clear peak in the signal for both magnitude and angle matches. The $W^{\pm}Z$ background shows a worse magnitude match as some of the $W$ momentum is lost to either an electron or muon on decay. This means that the missing $p_T$ will not quite match up with that of the recoiling $Z$. The angular distribution shows a peak in both the $W^{\pm}Z$ and $Z \to ll$ channels. In the case of $W^{\pm}Z$, this is because the $W$ and $Z$ are produced in approximately opposite directions. When the $W$ decays, the neutrino will be deflected and so the peak has a wider distribution. In a similar way, in $Z \to ll$, the $Z$ is likely to be produced with some quarks recoiling against it. These will manifest themselves as jets which can fake $\not\!\!E_T$. Cuts at $(|\not\!\!E_T - p_T(Z)|)/p_T(Z) < 0.35$ and $145^\circ < \phi_{\not E_T} - \phi_Z < 215^\circ$, reduce the $W^{\pm}Z$ background.

A jet veto reduces backgrounds with large hadronic activity. For example, the predominant decay channel for the top quark in $t\bar{t}$ is the $t \to Wb$ final state, resulting in several high $p_T$ jets. Its contribution can be reduced by applying a veto on events containing any jet with $p_T^{jet} > 30$ GeV and $|\eta^{jet}| < 3.0$.

The final cut to be applied is on the $p_T$ of the reconstructed $Z$ boson. This reduces the background

Table 14: Cut flow table for $ZZ \to \ell^+\ell^-\nu\bar{\nu}$ signal and background after cuts for an integrated luminosity of 1 fb$^{-1}$. The values in brackets indicate the percentage of events passing each cut relative to the previous cut. Note that this $Z \to ll$ MC sample already requires $p_T(Z) > 100$ GeV.

| Cut | $ZZ \to \ell\ell\nu\nu$ | $ZZ \to 4l$ | $Z \to ll$ | $t\bar{t}$ | $W^{\pm}Z$ | $W^+W^-$ | $Z \to \tau\tau$ |
|---|---|---|---|---|---|---|---|
| Leptons | 130.1 | 54.3 | 13100 | 4530 | 271.2 | 491.1 | 2170 |
| Third-lepton veto | 101.9 | 3.1 | 1900 | 428.9 | 52.9 | 375.6 | 1690 |
| | (78.3%) | (5.7%) | (14.5%) | (9.5%) | (19.5%) | (76.5%) | (77.9%) |
| Dilepton mass | 100.2 | 2.7 | 1740 | 110.2 | 45.3 | 83.8 | 40.1 |
| | (98.3%) | (87.1%) | (91.6%) | (25.7%) | (85.6%) | (22.3%) | (3.4%) |
| Missing $E_T$ | 38.0 | 0.34 | 3.8 | 17.9 | 9.4 | 18.3 | 0 |
| | (39.9%) | (12.6%) | (0.2%) | (16.2%) | (20.8%) | (21.8%) | (0.0%) |
| Jet veto | 34.4 | 0.30 | 0.44 | 6.0 | 7.6 | 16.7 | 0 |
| | (90.5%) | (88.2%) | (11.6%) | (33.5%) | (80.9%) | (91.3%) | (0.0%) |
| $p_T^Z$ | **10.2** | **0.08** | **0.4** | **3.0** | **1.7** | **0.02** | **0** |
| | (29.7%) | (26.7%) | (90.9%) | (50.0%) | (22.4%) | (0.1%) | (0.0%) |
| Stat. Error [90%$CL$] | 0.2 | 0.01 | 0.2 | 2.1 | 0.1 | 0.22 | [1.6] |

Table 15: The expected $ZZ \to \ell^+\ell^-\nu\bar{\nu}$ signal yields and total signal selection efficiency for 1 fb$^{-1}$ of integrated luminosity. The errors shown are statistical only.

| $N_{\text{signal}}$ | Signal efficiency | $N_{\text{background}}$ | $N_S/N_B$ |
|---|---|---|---|
| $10.2 \pm 0.2$ | 2.6% | $5.2 \pm 2.6$ | $2.0 \pm 0.8$ |

from the single $Z$ channel, whose $p_T(Z)$ distribution drops much faster than the signal. A cut of $p_T(Z) >$ 100 GeV significantly reduces this background, and has a negligible effect on the sensitivity to anomalous couplings, which predominantly manifest at high $p_T$.

Using the single isolated electron trigger (effective $E_T > 22$ GeV) and the single isolated muon trigger (effective $p_T > 20$ GeV), the trigger efficiency for selected $ZZ \to \ell^+\ell^-\nu\bar{\nu}$ events is expected to be 97%.

Table 14 gives a summary of the cuts applied and presents the expected number of events passing the cuts. The final row in each column gives the statistical error. If no events pass cuts, the figure given is the number of expected events at the 90% confidence level. The dominant background is $t\bar{t}$ . Table 15 summarizes the expected yield and sensitivity of the $ZZ \to \ell^+\ell^-\nu\bar{\nu}$ channel.

## 4 Total cross section measurements

A binned likelihood method is used to determine the most likely cross-sections. This likelihood method is also used to extract the sensitivities to the anomalous TGCs which will be described in Section 5. The likelihood is based on Poisson statistics convoluted with Gaussian probabilities to model the signal and background uncertainties. What follows is a more detailed description of the binned likelihood method followed by a description of the statistical and systematic uncertainties for the diboson cross-section measurements using the first 1.0 fb$^{-1}$ of data.

### 4.1 Binned likelihood

In the binned likehood method, expected events are determined from high statistics Monte Carlo simulation, and observed events are also determined from Monte Carlo simulation for this work, but with appropriate statistical fluctuation according to the luminosity. The events are binned by one or more observables. As an example, in the case of the cross-section measurement the BDT output spectrum, an example of which is shown in Figure 8, could be used. In the TGC analysis described in Section 5 the $M_T(VV)$ and $p_T(V)$ spectra are choosen.

For each bin expected signal and background are compared to the observed number of events ($n$ events in the bin) with a likelihood, which is based on Poisson statistics. We assume the systematic uncertainties of the signal and background are Gaussian and uncorrelated for each bin. Thus, two Gaussian distributions are convolved with the Poisson distribution to form the likelihood

$$L = \int_{1-3\sigma_b}^{1+3\sigma_b} \int_{1-3\sigma_s}^{1+3\sigma_s} g_s\, g_b\, \frac{(f_s\nu_s + f_b\nu_b)^n\, e^{-(f_s\nu_s+f_b\nu_b)}}{n!} df_s\, df_b \quad \text{with} \quad g_i = \frac{e^{(1-f_i)^2/2\sigma_i^2}}{\int_0^\infty e^{(1-f_i)^2/2\sigma_i^2}}\ (i=s,b),$$

here the total systematic uncertainty of signal and background appear as $\sigma_s$ and $\sigma_b$, respectively.

From these likelihoods a total log-likelihood is formed from all the bin likelihoods. Some processes may also be separated into multiple channels (such as the three decay combinations of $WW \to ee, e\mu, \mu\mu$). Also, a factor of -2 is included to make this test statistic comparable to a chi-squared distribution. Thus, the negative log-likelihood is

$$-2lnL = -2 \sum_{k=\text{channels}} \sum_{i=\text{bins}} \log(L_i^k).$$

In cross-section measurements, the likelihood is determined as a function of cross-section in each bin of a measured spectrum for each channel (e.g. the BDT output spectrum for the $W^+W^- \to e\nu\mu\nu$ channel as shown on the left in Figure 8). The log-likelihoods are then combined and the minimum of the negative log-likelihood determines the most likely cross-section (or anomalous TGC). The 68% C.L. limits ($\pm 1\sigma$) are taken from the minimum of the negative log-likelihood plus 1.0. To set the 95% confidence-level interval of the anomalous TGC limits, likelihood minimum+1.92 is taken when fitting one parameter, and the minimum+2.99 for a fit of two parameters (e.g. two independent anomalous couplings).

### 4.2 Statistical uncertainties

Based on the diboson event selections described in Section 3, the expected number of signal and background events for 1 fb$^{-1}$, and the expected detection significance of observing the Standard Model signals, are summarized in Table 16, after taking into account the known background contributions and 20% systematic uncertainties of the background estimate. The expected signal statistical uncertainties are also given in the $5^{th}$ column of the table. For 1 fb$^{-1}$ they range from 2.1% to 31% depending on the channel.

For early LHC data with 0.1 fb$^{-1}$ integrated luminosity, the statistical uncertainties are large. However, based on BDT analysis, with an assuption of a 20% systematic uncertainty, the signal detection significances could reach $9.9\sigma$ and $5.9\sigma$ for $W^+W^-$ and $W^\pm Z$, respectively. The overall detection significance is expected to be greater than $10\sigma$ for both $W^\pm\gamma$ and $Z\gamma$ signals with 0.1 fb$^{-1}$ of data.

### 4.3 Systematic uncertainties

The major theoretical uncertainties on the production cross-sections come from the PDF uncertainties and the QCD factorization scaling uncertainties (for NLO calculations). By varying the PDF's and scale values for $W^+W^-$, $W^\pm Z$, and $ZZ$ cross-section calculations, the differences of the calculated cross-sections are found to range from 3.4% to 6.2%.

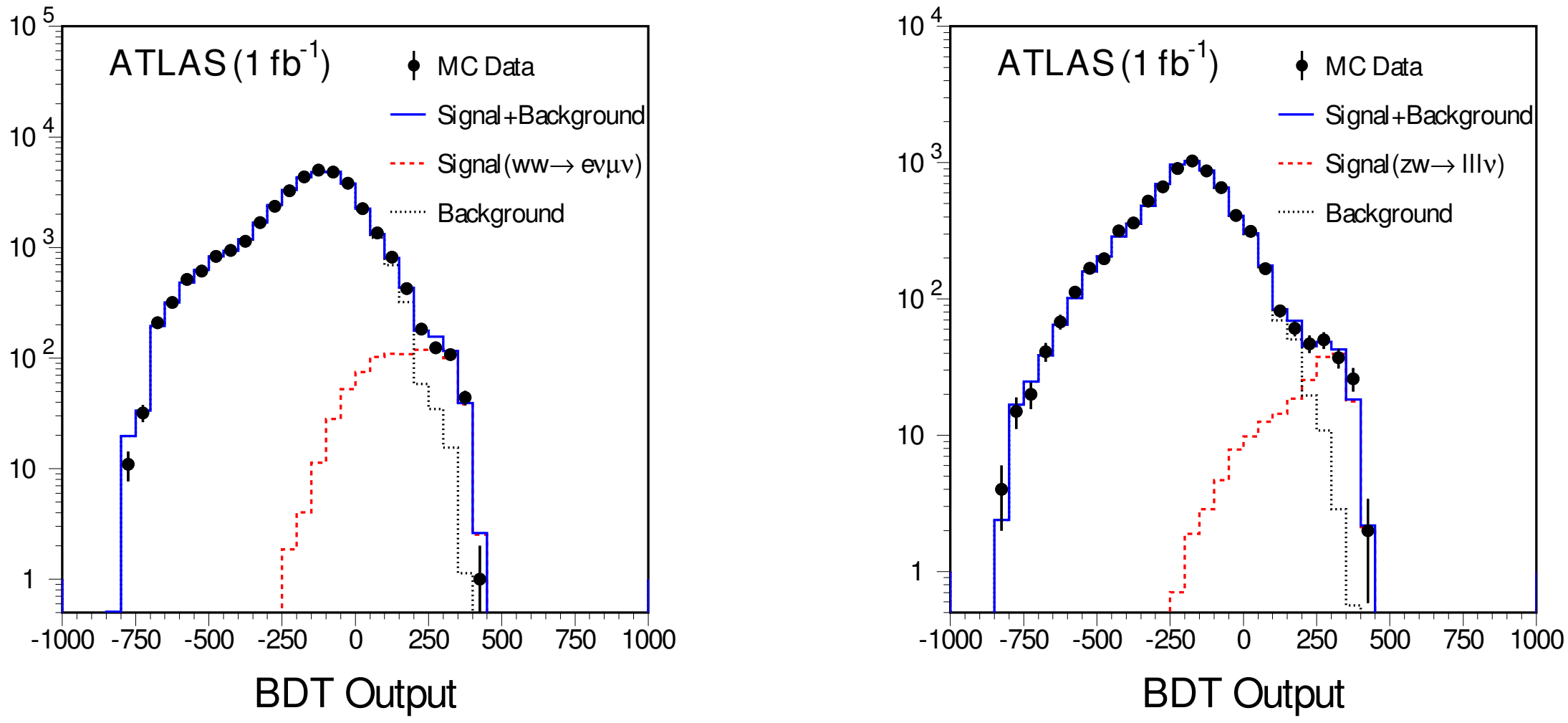


Figure 8: BDT-output spectra from a Monte Carlo experiment for $W^+W^-$ (left) and $W^\pm Z$ (right) detection with 1 fb$^{-1}$. The dots in the plots are Monte Carlo 'mock data'. The dashed histograms represent the signal and the dotted, background.

The major experimental systematic effects in the cross-section measurements arise from the uncertainties of the luminosity determination, the lepton identification efficiencies and energy/momentum resolutions, the jet energy scale and resolutions, and background model and estimate.

A promising possibility for the precise determination of the luminosity is to use the $W$ and $Z$ production and leptonic decays. The estimates show that in this way the luminosity uncertainties could be controlled to $\sim$5% [40]. It should be noted that a 6.5% luminosity uncertainty was quoted in Tevatron Run II physics papers.

The lepton acceptance uncertainty is about 2-3% mainly due to the isolation requirement which involves the hadronic jet energy uncertainties. The lepton trigger efficiency uncertainties also contribute. With large $Z$ samples, this uncertainty could be minimized. $Z$ decays can typically be triggered and identified using only one of the two decay leptons. This leaves the second lepton unbiased from the point of view of trigger and offline identification. The rate at which the unbiased lepton passes the trigger and ID requirements provides a measurement of the respective efficiencies.

In the studies using the ATLAS simulated events the background estimate dominates the systematics with uncertainties of 15-25% for all the diboson channels except for the $ZZ \to \ell^+\ell^-\ell^+\ell^-$ channel, where the background uncertainty should be less than 2%. Even though more than 30 million fully simulated events are used to estimate the background, the analyses are still largely limited by $W + jets$ event sample statistics in the diboson background estimate. Tevatron experiments have used data to estimate the background, and typical uncertainties for diboson physics analyses are around 10% for 1 fb$^{-1}$ of data. With early LHC data (0.1 fb$^{-1}$), the background estimate uncertainty would be comparable to current Tevatron diboson background uncertainties, and with more data the uncertainties of the background estimate should decrease.

The lepton and jet energy resolution uncertainties will contribute to additional background estimate uncertainties which will further propagate to the cross-section measurement uncertainties. A study has been performed in $W^\pm Z$ analysis to estimate the size of such uncertainties. In this study, the $W^\pm Z$ BDTs are first trained with Monte Carlo signal and background events simulated with the 'standard' detector energy resolutions and calibrated energy scale. For independent test samples, 10% and 3% are added to the jet and lepton energy resolutions, respectively, and the reconstructed energy related quantities are

Table 16: Summary of signal and background of all diboson final states ($\ell$ denotes $e$ and $\mu$) for 1 fb$^{-1}$ of integrated luminosity. The 4th column indicates the overall signal selection efficiency and the type of analysis, The 5th column gives the signal statistical uncertainty. The last two columns indicate the $p$-value and the significance (in Gaussian standard deviations) where $p$-value is the probability of the background fluctating to the expected total observation assuming 20% systematic uncertainties.

| Diboson mode | Signal | Background | Signal eff. | $\sigma_{stat}^{signal}$ | $p$-value | Sig. |
|---|---|---|---|---|---|---|
| $W^+W^- \to e^\pm\nu\mu^\mp\nu$ | 347±3 | 64±5 | 12.6% (BDT) | 5.4% | $3.6\times10^{-166}$ | 27.4 |
| $W^+W^- \to \mu^+\nu\mu^-\nu$ | 70±1 | 17±2 | 5.2% (BDT) | 12.0% | $8.8\times10^{-30}$ | 11.3 |
| $W^+W^- \to e^+\nu e^-\nu$ | 52±1 | 11±2 | 4.9% (BDT) | 13.9% | $1.9\times10^{-24}$ | 10.1 |
| $W^+W^- \to \ell^+\nu\ell^-\nu$ | 103±3 | 17±2 | 2.0% (cuts) | 9.9% | $1.4\times10^{-54}$ | 15.5 |
| $W^\pm Z \to \ell^\pm\nu\ell^+\ell^-$ | 128±2 | 16±3 | 15.2% (BDT) | 8.8% | $3.0\times10^{-76}$ | 18.4 |
| | 53±2 | 8±1 | 6.3% (cuts) | 13.7% | $3.1\times10^{-30}$ | 11.4 |
| $ZZ \to 4\ell$ | 17±0.5 | 2±0.2 | 7.7% (cuts) | 24.6% | $6.0\times10^{-12}$ | 6.8 |
| $ZZ \to \ell^+\ell^-\nu\bar{\nu}$ | 10±0.2 | 5±2 | 2.6% (cuts) | 31.3% | $7.7\times10^{-4}$ | 3.2 |
| $W\gamma \to e\nu\gamma$ | 1604±65 | 1180±120 | 5.7% (BDT) | 2.5% | significance > 30 | |
| $W\gamma \to \mu\nu\gamma$ | 2166±88 | 1340±130 | 7.6% (BDT) | 2.1% | significance > 30 | |
| $Z\gamma \to e^+e^-\gamma$ | 367±12 | 187±19 | 5.4% (BDT) | 5.2% | $1.2\times10^{-91}$ | 20.3 |
| $Z\gamma \to \mu^+\mu^-\gamma$ | 751±23 | 429±43 | 11% (BDT) | 3.6% | $5.9\times10^{-171}$ | 27.8 |

Table 17: Change of background acceptance in a test of BDT ($W^\pm Z$ vs. $ZZ$) performed by smearing jet energy, $E_{jet}$, and missing $E_T$, $\not{E}_T$, by an additional 10%; and the lepton energy $E_T^\ell$ by an additional 3%.

| Signal Efficiency | Background Eff. No additional smearing | Background Eff. 10% for $E_{jet}$ & $\not{E}_T$ | Background Eff. 10% for $E_{jet}$& $\not{E}_T$, 3% for $E_T^\ell$ |
|---|---|---|---|
| 40% | 4.0% | 4.2% (+5.7%) | 4.2%(+6.7%) |
| 50% | 8.6% | 8.9% (+3.7%) | 9.0%(+4.8%) |
| 60% | 14.6% | 14.9% (+2.2%) | 15.1%(+3.7%) |
| 70% | 22.3% | 22.7% (+2.0%) | 23.0%(+3.4%) |

'smeared' to reflect the uncertainties of the diboson detection sensitivity (signal to background ratio). In this study the signal efficiencies are fixed and changes to the background acceptance are gauged. The results are summarized in Table 17. As an example, for a BDT signal selection efficiency of 60%, the change of the signal to background ratio is 3.4%. For the $W^{\pm}Z$ cross-section measurement with 1 fb$^{-1}$ of integrated luminosity, the 3.4-6.7% background contribution uncertainty would result in additional cross-section measurement uncertainties of about 2-3%.

### 4.4 Measurement uncertainties vs. selection cuts and luminosities

The cross-section measurement uncertainties are estimated for various event selection cuts and integrated luminosities in the $W^+W^-$ and $W^{\pm}Z$ BDT based analysis. The BDT output spectra are used to build the log-likelihood by using 'mock data', which is a sample of simulated events with appropriate statistics according to the luminosity and the Standard Model. For example, the BDT-output spectra for a Monte Carlo experiment with 1 fb$^{-1}$ of data are shown in Figure 8 for $W^+W^- \to e^{\pm}\nu\mu^{\mp}\nu$ detection (left) and for $W^{\pm}Z \to \ell^{\pm}\nu\ell^+\ell^-$ detection (right). The Standard Model 'mock data' (points) are compared to expected signal (dashed histogram) and background (dotted histogram).

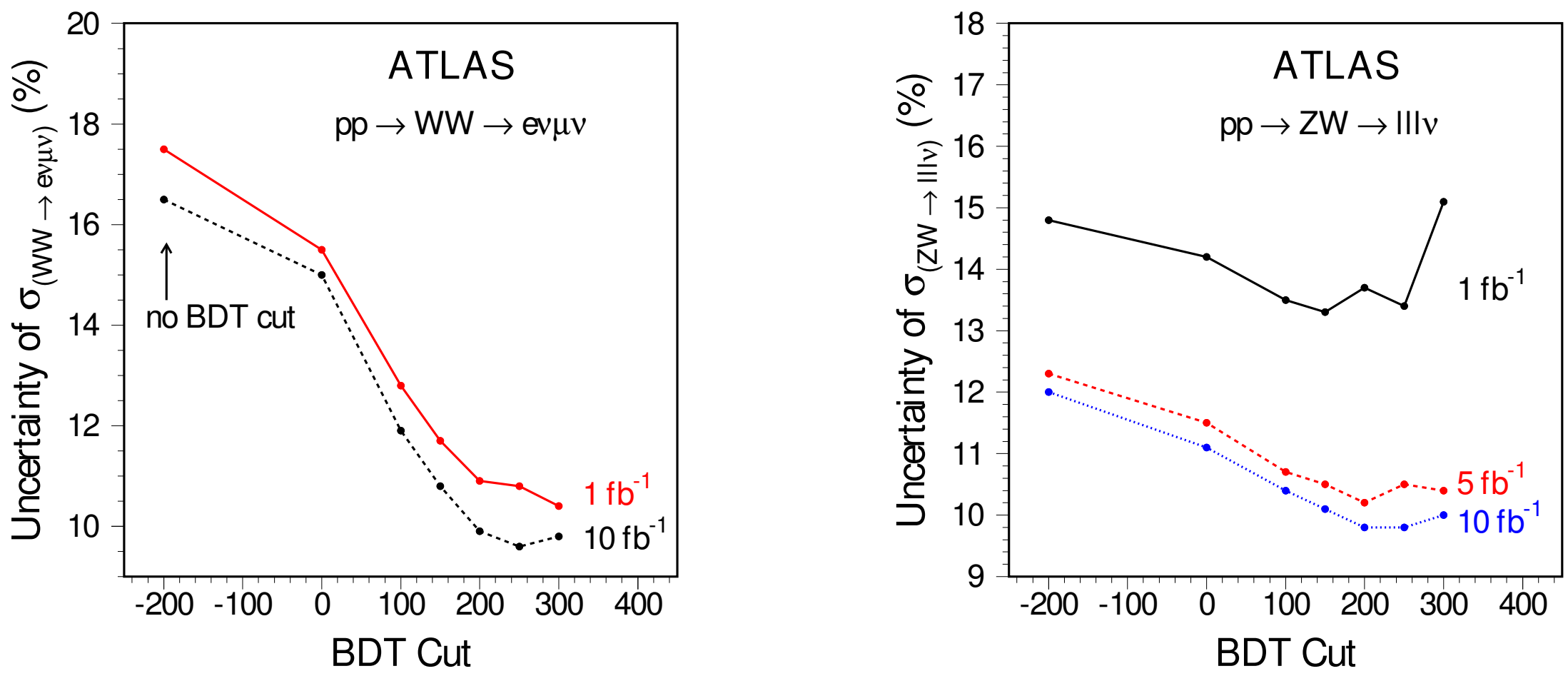


Figure 9: The total relative uncertainties for $W^+W^-$ (left) and for $W^{\pm}Z$ (right) cross section measurements as the BDT cut is varied for different luminosities. The optimal BDT cut is between 200 and 300. An overall 9.2% systematic uncertainty was included in the fitting process.

To understand the optimal cut on the BDT spectra for cross-section measurements, the cuts on the BDT spectra are varied and the cross-section measurements are repeated. A total 9.2% systematic uncertainty is included in the fitting process. Figure 9 shows the cross-section measurement uncertainty as a function of the BDT cut for different integrated luminosities from the $W^+W^-$ and the $W^{\pm}Z$ analysis.

Figure 10 shows the relative cross-section uncertainties as a function of integrated luminosity (with BDT spectrum cut at 200) for $W^+W^-$ (left) and $W^{\pm}Z$ (right) cross-section measurements. From these plots it should be noted that the systematic uncertainty starts to dominate after 5 fb$^{-1}$ of integrated luminosity for $W^+W^-$ cross-section measurements, and after 10 fb$^{-1}$ for $W^{\pm}Z$.

## 5 Sensitivity to anomalous couplings

The signature of anomalous couplings in diboson production is an increase in the cross-section at high values of gauge boson transverse momentum ($p_T$) and diboson transverse mass ($M_T$). The ATLAS

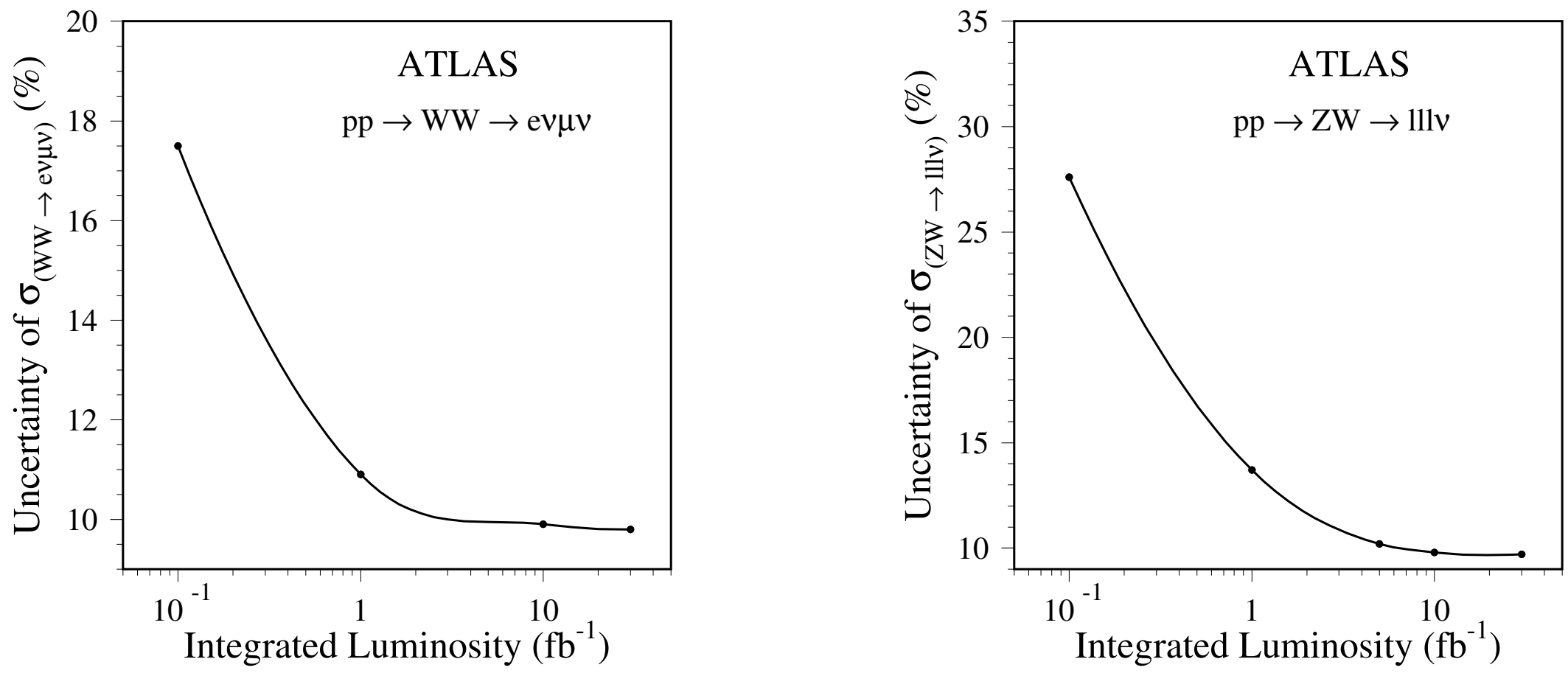


Figure 10: The $W^+W^-$ (left) and the $W^\pm Z$ (right) cross-section measurement uncertainties as a function of integrated luminosity (with BDT spectrum cut at 200). An overall 9.2% systematical uncertainty was included in the fitting process.

sensitivity to anomalous TGC's is investigated by comparing the 'measured' diboson production cross-sections and the vector boson $p_T$ or diboson $M_T$ distributions to models with anomalous TGC's. A binned likelihood fitting procedure using the $M_T$ or $p_T$ spectrum for each channel is followed to extract the 95% C.L. intervals of anomalous coupling parameters. The most dramatic effect is an increase in the high $M_T$ or high $p_T$ cross-sections, so it is important for the binned likelihood calculation to include events up to the highest values of the observables. Details of the binned likelihood method are described in Section 4.1. One- and two-dimensional limits are set on the charged CP-conserving coupling parameters from the $W^+W^-$, $W^\pm Z$, and $W^\pm\gamma$ final states. The $ZZ$ final state is used to probe the neutral anomalous TGC sensitivity.

The values of the form factor scale $\Lambda$ are chosen such that the extracted experimental anomalous coupling limit from data for a certain diboson production process is less than the unitarity limit [41]. For this study, with 0.1-1.0 fb$^{-1}$ of integrated luminosity at early LHC running, $\Lambda$ values of 2-3 TeV are used. The same $\Lambda$ values of 2-3 TeV are also used to estimate the anomalous coupling sensitivities for higher luminosities for simplicity. It should be noted that as the luminosity increases, the $\Lambda$ value should have increased accordingly.

## 5.1 Re-weighting the fully simulated events

To avoid producing an impractically large number of fully simulated events in non-Standard Model anomalous coupling parameter space, a **re-weighting** method was invoked to study the ATLAS detector sensitivities to anomalous coupling parameters. The BHO and the BosoMC calculations are used with different anomalous coupling parameters to re-weight the fully simulated events generated by MC@NLO. As an example, Figure 11 shows the $W^+W^-$ production differential cross-section distributions for Standard Model and some anomalous coupling parameters (left plot) and the corresponding differential cross-section ratio, $\frac{d\sigma(\text{non-SM})/dM_T}{d\sigma(\text{SM})/dM_T}$ (right plot). These ratios have been used as *weights* to re-weight the fully simulated events to probe the anomalous TGC sensitivities. The *weights* are generated in one-dimensional and two-dimensional anomalous coupling space according to parton level kinematics using the BHO and the BosoMC programs. To produce the weights the step size in coupling parameter space ranges from $0.1 \times 10^{-3}$ to $1.0 \times 10^{-3}$. For each point in the coupling parameter space 5 million

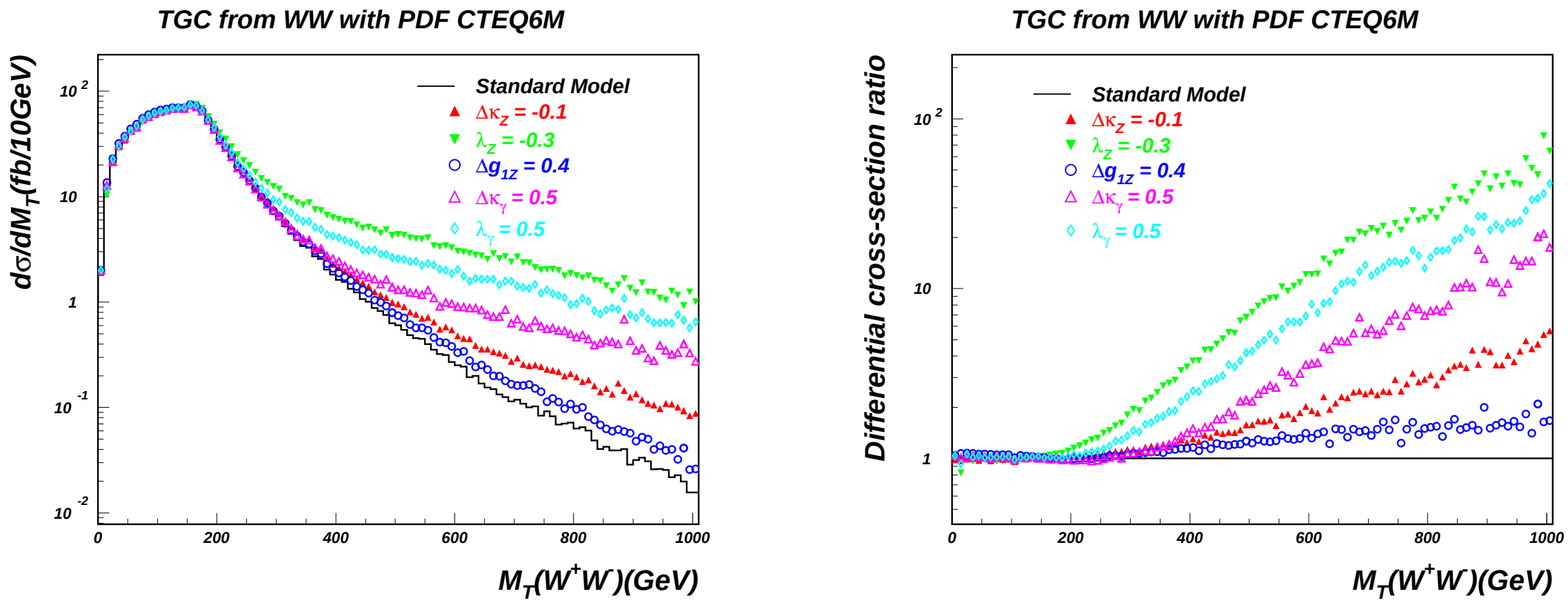


Figure 11: Left: WW transverse mass, $M_T$, distributions. Events are generated with the Standard Model coupling (black line) and anomalous couplings (colored symbols); Right: the corresponding differential cross-section ratio.

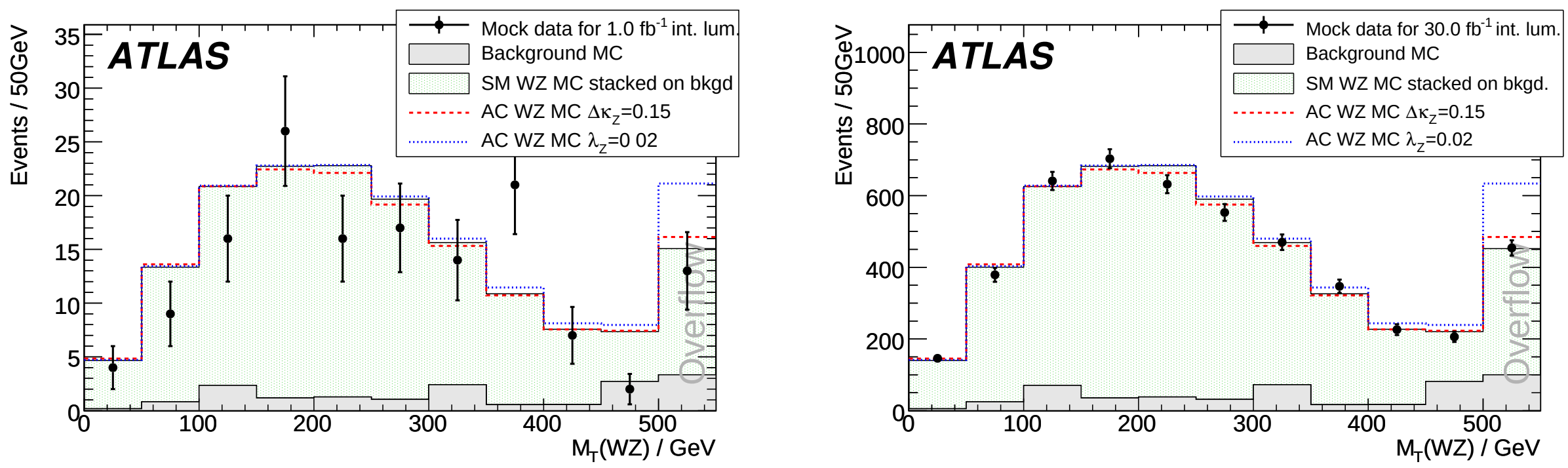


Figure 12: The expected signal+background of the Standard Model, superimposed with 'mock data' (points with error bars showing statistical uncertainty), and the non-Standard Model (anomalous couplings) predicted signal+background histograms (dashed and dotted histograms). The left plot is for 1 fb$^{-1}$ of data and the right plot is for 30 fb$^{-1}$ of data.

events were generated to obtain the theoretical 'reference' distributions. The fully simulated events with the Standard Model couplings are required to pass the event selection cuts, and then reweighted according to the parton level kinematics. The weighted events are equivalent to fully simulated events with the corresponding anomalous couplings. The distributions of variables, sensitive to the anomalous coupling, such as lepton $p_T$, of the data events, can be compared to those of simulated events with anomalous couplings included, to extract the limits on the anomalous couplings. In this study the Standard Model 'mock data' are used to probe the ATLAS detector sensitivities to anomalous triple gauge boson couplings.

## 5.2 $WWZ$ anomalous TGC sensitivity in $W^{\pm}Z$ analysis

The $W^{\pm}Z$ diboson production involves exclusively the $WWZ$ coupling, in contrast to the $W^+W^-$ diboson final state which contains both $WWZ$ and $WW\gamma$ couplings. To extract the 95% C.L. sensitivity intervals of the anomalous parameters, $\Delta\kappa_Z, \Delta g_1^Z$, and $\lambda_Z$, from the $W^{\pm}Z$ diboson final state, both the transverse mass of $W^{\pm}Z$ ($M_T(W^{\pm}Z)$) and the transverse momentum of $Z$ ($p_T(Z)$) spectra are used to fit the anomalous

Table 18: Summary of $WWZ$ one-dimensional anomalous coupling parameter 95% CL sensitivities using the $M_T(W^{\pm}Z)$ fitting for $\Lambda = 2$ TeV and $\Lambda = 3$ TeV for integrated luminosities of 0.1, 1, 10 and 30 fb$^{-1}$.

| Int. Lumi (fb$^{-1}$) | Cutoff $\Lambda$ (TeV) | $\Delta\kappa_Z$ | $\lambda_Z$ | $\Delta g_1^Z$ |
|---|---|---|---|---|
| 0.1 | 2.0 | [-0.440, 0.609] | [-0.062, 0.056] | [-0.063, 0.119] |
| 1.0 | 2.0 | [-0.203, 0.339] | [-0.028, 0.024] | [-0.021, 0.054] |
| 10.0 | 2.0 | [-0.095, 0.222] | [-0.015, 0.013] | [-0.011, 0.034] |
| 30.0 | 2.0 | [-0.080, 0.169] | [-0.012, 0.008] | [-0.005, 0.023] |
| 0.1 | 3.0 | [-0.399, 0.547] | [-0.050, 0.046] | [-0.054, 0.094] |
| 1.0 | 3.0 | [-0.178, 0.281] | [-0.020, 0.018] | [-0.017, 0.038] |
| 10.0 | 3.0 | [-0.135, 0.201] | [-0.015, 0.013] | [-0.013, 0.018] |
| 30.0 | 3.0 | [-0.069, 0.131] | [-0.008, 0.005] | [-0.003, 0.016] |

couplings.

Monte Carlo experiments are performed with 0.1, 1, 10, and 30 fb$^{-1}$ of integrated luminosities to study the anomalous coupling sensitivities. Figure 12 shows the expected signal+background of the Standard Model, superimposed with the 'mock data' (points with error bars), and the non-Standard Model (anomalous couplings) predicted signal+background distributions. Table 18 shows the summary of 1-dimensional 95% C.L. anomalous coupling parameter intervals based on the $M_T(W^{\pm}Z)$ spectra fitting. Results corresponding to 0.1, 1, 10 and 30 fb$^{-1}$ of integrated luminosities for cutoff, $\Lambda = 2$ TeV and $\Lambda = 3$ TeV are listed. It should be noted that even for 0.1 fb$^{-1}$ of integrated luminosity, the ATLAS sensitivity to $WWZ$ anomalous couplings could be much better than the Tevatron limits based on 1 fb$^{-1}$ of $p\bar{p}$ collision data.

To understand the systematic uncertainty effects on the TGC sensitivity, three different systematic uncertainty assumptions are considered: (1) ideal case with no systematic uncertainties: $\sigma_S = 0$, and $\sigma_B = 0$; (2) expected uncertainty of 7.2% for signal, and 12% for background, based on estimate from various contributions; and (3) worse than expected systematic uncertainty of 9.2% for signal, and 18.3% for background. Unless otherwise stated, (3) was used to evaluate coupling limits.

The 95% C.L. 1-dimensional limits for the $WWZ$ anomalous couplings, obtained from the fits to the $p_T(Z)$ assuming $\Lambda = 2$ TeV are shown in Table 19, for different scenarios of systematic uncertainties. From this table it is seen that only when reaching 30 fb$^{-1}$ of integrated luminosity do the systematic uncertainties become significant enough to affect the TGC sensitivities.

The studies on the $WWZ$ anomalous couplings in two-dimensional space are also based on the $p_T(Z)$ fits for different integrated luminosities (0.1, 1, 10 and 30 fb$^{-1}$) and for two cutoff values, $\Lambda = 2$ TeV and 3 TeV. The anomalous coupling limit contours are not very sensitive to these cutoff values. The effects of different systematic uncertainties on the 2-dimensional TGC sensitivity contour are shown in Figure 13. The left plot shows the 95% C.L. anomalous TGC limit contour of $\lambda_Z$ vs. $\Delta\kappa_Z = \Delta g_1^Z$ without systematic uncertainties, and the right plot shows the 95% C.L. TGC limit contour with the systematic uncertainties ($\sigma_S = 9.2\%$, $\sigma_B = 18.3\%$) included. Again, the systematic uncertainties become significant when the integrated luminosity reaches 30 fb$^{-1}$.

## 5.3 $WW\gamma$ anomalous TGC sensitivity in $W^{\pm}\gamma$ analysis

The $W^{\pm}\gamma$ diboson production involves exclusively the $WW\gamma$ triple gauge coupling. To extract the 95% C.L. sensitivity intervals of the anomalous parameters, $\Delta\kappa_\gamma$ , and $\lambda_\gamma$, from the $W^{\pm}\gamma$ diboson final state, the photon transverse energy $E_T(\gamma)$ distribution is used to fit the anomalous couplings, with $\Lambda = 2$ TeV.

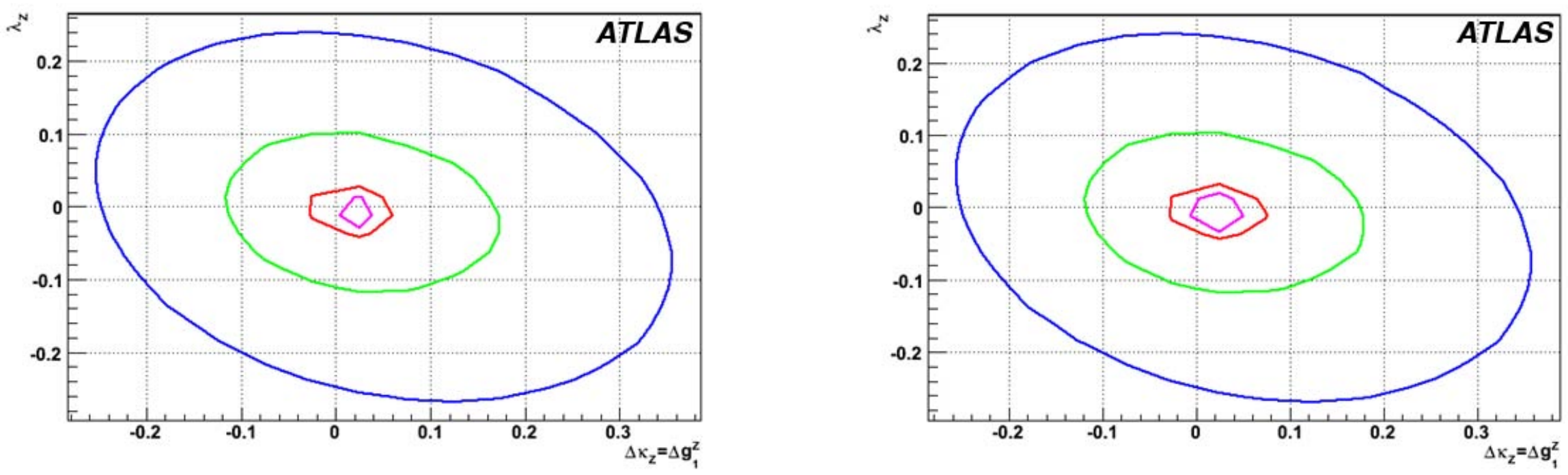


Figure 13: The left plot: 95% C.L. $WWZ$ TGC limit contour of $\lambda_Z$ vs. $\Delta\kappa_Z = \Delta g_1^Z$; without the systematic uncertainties. The right plot: the 95% C.L. $WWZ$ TGC limit contour of $\lambda_Z$ vs. $\Delta\kappa_Z = \Delta g_1^Z$, with systematic uncertainties ($\sigma_S = 9.2\%$, $\sigma_B = 18.3\%$) included. The anomalous coupling limit contours from outer to inner corresponding integrated luminosities of 0.1, 1, 10 and 30 fb$^{-1}$, respectively. The systematic uncertainties become significant when the integrated luminosity reaches 30 fb$^{-1}$.

Table 19: Comparison of $WWZ$ one-dimensional anomalous coupling parameter 95% C.L. sensitivities for different systematic uncertainties. Results obtained in this table are using the $p_T(Z)$ fit for $\Lambda = 2$ TeV for integrated luminosities of 0.1, 1, 10 and 30 fb$^{-1}$.

| Systematic uncertainties | Int. Lumi (fb$^{-1}$) | $\Delta\kappa_Z$ | $\lambda_Z$ | $\Delta g_1^Z$ |
|---|---|---|---|---|
| $\sigma_S = 0$ | 0.1 | [-0.942, 1.130] | [-0.203, 0.193] | [-0.227, 0.324] |
| $\sigma_B = 0$ | 1.0 | [-0.561, 0.664] | [-0.093, 0.082] | [-0.106, 0.154] |
| | 10.0 | [-0.233, 0.231] | [-0.033, 0.024] | [-0.025, 0.061] |
| | 30.0 | [-0.128, 0.136] | [-0.024, 0.013] | [-0.009, 0.047] |
| $\sigma_S = 7.2\%$ | 0.1 | [-0.950, 1.140] | [-0.204, 0.194] | [-0.228, 0.325] |
| $\sigma_B = 12.0\%$ | 1.0 | [-0.574, 0.692] | [-0.093, 0.083] | [-0.106, 0.158] |
| | 10.0 | [-0.228, 0.302] | [-0.033, 0.027] | [-0.022, 0.070] |
| | 30.0 | [-0.164, 0.212] | [-0.026, 0.018] | [-0.009, 0.055] |
| $\sigma_S = 9.2\%$ | 0.1 | [-0.956, 1.150] | [-0.204, 0.194] | [-0.229, 0.326] |
| $\sigma_B = 18.3\%$ | 1.0 | [-0.583, 0.706] | [-0.094, 0.084] | [-0.106, 0.159] |
| | 10.0 | [-0.241, 0.316] | [-0.033, 0.028] | [-0.024, 0.071] |
| | 30.0 | [-0.184, 0.228] | [-0.028, 0.020] | [-0.011, 0.056] |

Table 20: 95% C.L. intervals for the anomalous $WW\gamma$ coupling parameters obtained from fitting the $E_T(\gamma)$ distribution to the NLO expectations using the combined sample of $W(e\nu)\gamma$ and $W(\mu\nu)\gamma$ events, with $\Lambda = 2$ TeV.

| | $W(\ell\nu)\gamma$ | | |
|---|---|---|---|
| | 1 fb$^{-1}$ | 10 fb$^{-1}$ | 30 fb$^{-1}$ |
| $\lambda_\gamma$ | [-0.09, 0.04] | [-0.05, 0.02] | [-0.02,0.01] |
| $\Delta\kappa_\gamma$ | [-0.43, 0.20] | [-0.26, 0.07] | [-0.11,0.05] |

The intervals are calculated for $W^\pm\gamma$ events by combining the electron and the muon decay channels.

Figure 14 shows an $E_T(\gamma)$ distribution from $W^\pm(\ell^\pm\nu)\gamma$ normalized to 1 fb$^{-1}$ of data. The signal expectations at LO and NLO are shown by the dashed and dotted lines on the left in Figure 14. On the right in Figure 14 is shown the 95% confidence contour in the $\lambda_\gamma$-$\Delta\kappa_\gamma$ parameter space for 1 fb$^{-1}$ of data. The 1-dimensional 95% C.L. intervals of $\lambda_\gamma$ and $\Delta\kappa_\gamma$ are listed in Table 20.

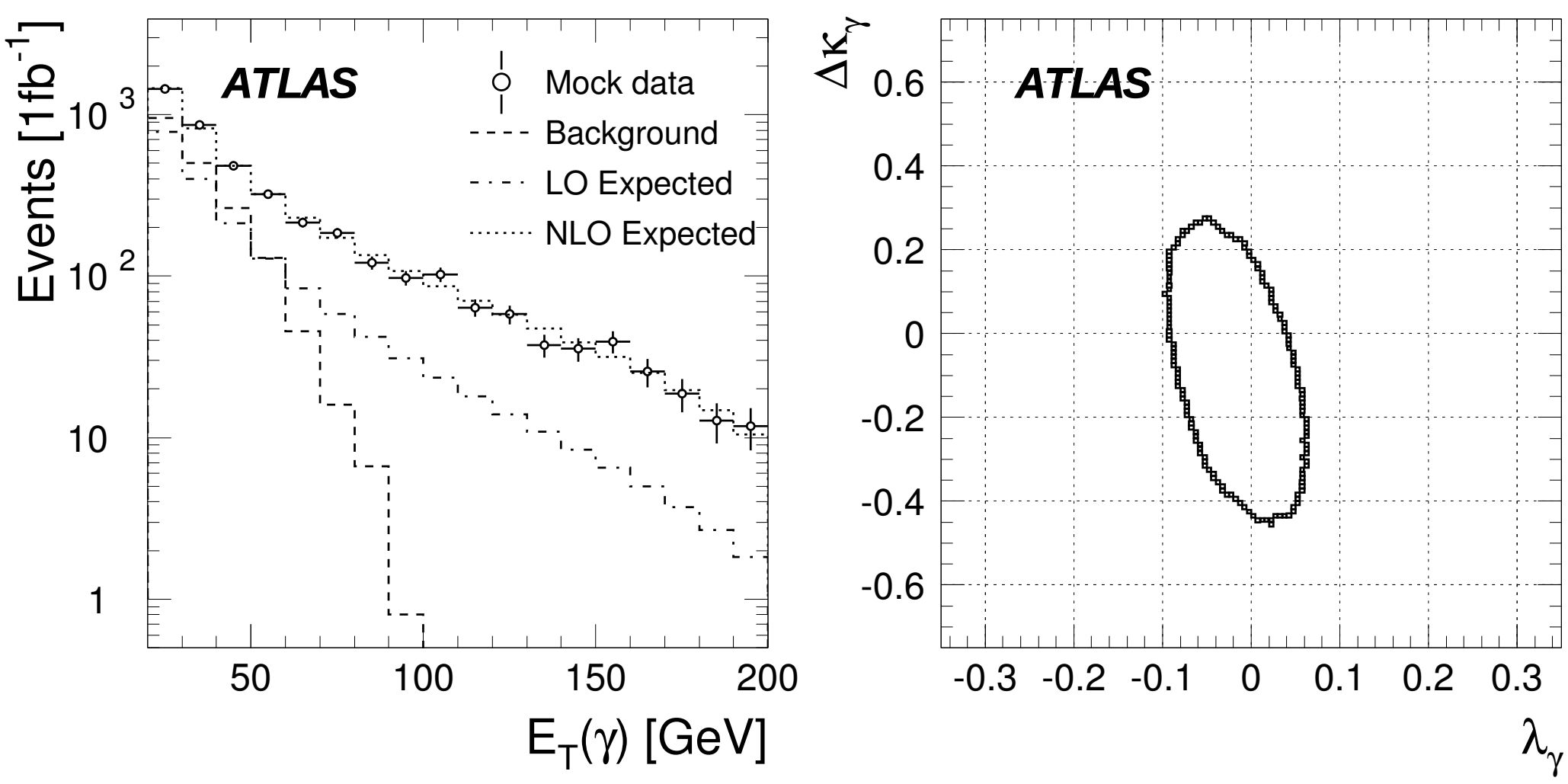


Figure 14: Left: the $E_T(\gamma)$ distributions of $W(\ell\nu)\gamma$ ($\ell = e,\ \mu$) events for 1 fb$^{-1}$ of data. Right: the 95 % confidence contour in the $\lambda_\gamma$-$\Delta\kappa_\gamma$ parameter space ($\Lambda = 2$ TeV) for 1 fb$^{-1}$ of $W^\pm\gamma$ data (with $W \to e\nu,\ \mu\nu$).

## 5.4 $WWZ$ and $WW\gamma$ anomalous TGC sensitivity in $W^+W^-$ analysis

The $M_T$ spectrum of $W^+W^-$ pair is fitted to obtain the $WWZ$ and $WW\gamma$ anomalous TGC sensitivity intervals at 95% confidence level. A comparison of the $M_T(WW)$ distribution of the 'mock data' to that of models with anomalous coupling is shown in Figure 15. Five anomalous coupling parameters, ($\Delta\kappa_Z$, $\lambda_Z$, $\Delta g_1^Z$, $\Delta\kappa_\gamma$, $\lambda_\gamma$), have been studied with only one parameter varied at the time; the remaining parameters are fixed to Standard Model values. One dimensional anomalous coupling sensitivity intervals at 95% C.L. for different integrated luminosities are given in Table 21. The cutoff $\Lambda = 2$ TeV is used in these calculations. The two-dimensional anomalous coupling limits from $W^+W^-$ production with different scenarios relating the anomalous coupling parameters have also been investigated in this study. The two-dimensional contours of the TGC limits at 95% confidence level for 0.1, 1, 10 and 30 fb$^{-1}$ integrated luminosities are shown in Figure 16. The left contours are the limits calculated with the HISZ assumption [43]. The right contours are calculated by assuming $\lambda_Z = \lambda_\gamma$ and $\Delta\kappa_Z = \Delta\kappa_\gamma$.

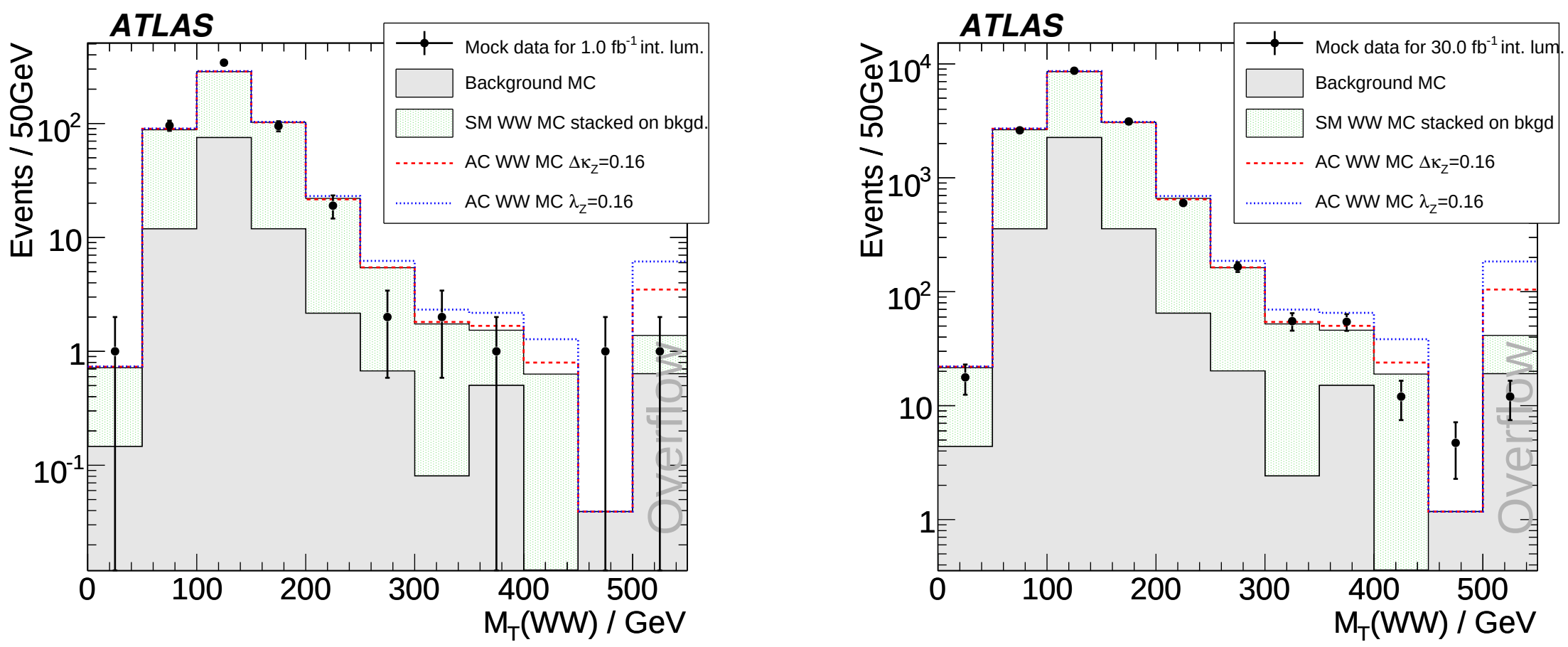


Figure 15: $W^+W^-$ transverse mass distributions for 1 (left) and 30 (right) fb$^{-1}$ of integrated luminosities. The last bins in the plots are 'overflow'-bins.

Table 21: One-dimensional 95% C.L. interval of the $WWZ$ and $WW\gamma$ anomalous coupling sensitivities from the WW final state analysis for 0.1, 1, 10 and 30 fb$^{-1}$ integrated luminosities, with $\Lambda = 2$ TeV.

| Int. Lumi (fb$^{-1}$) | $\Delta\kappa_Z$ | $\lambda_Z$ | $\Delta g_1^Z$ | $\Delta\kappa_\gamma$ | $\lambda_\gamma$ |
|---|---|---|---|---|---|
| 0.1 | [-0.242, 0.356] | [-0.206, 0.225] | [-0.741, 1.177] | [-0.476, 0.512] | [-0.564, 0.775] |
| 1.0 | [-0.117, 0.187] | [-0.108, 0.111] | [-0.355, 0.616] | [-0.240, 0.251] | [-0.259, 0.421] |
| 10.0 | [-0.035, 0.072] | [-0.040, 0.038] | [-0.149, 0.309] | [-0.088, 0.089] | [-0.074, 0.165] |
| 30.0 | [-0.026, 0.048] | [-0.028, 0.027] | [-0.149, 0.251] | [-0.056, 0.054] | [-0.052, 0.100] |

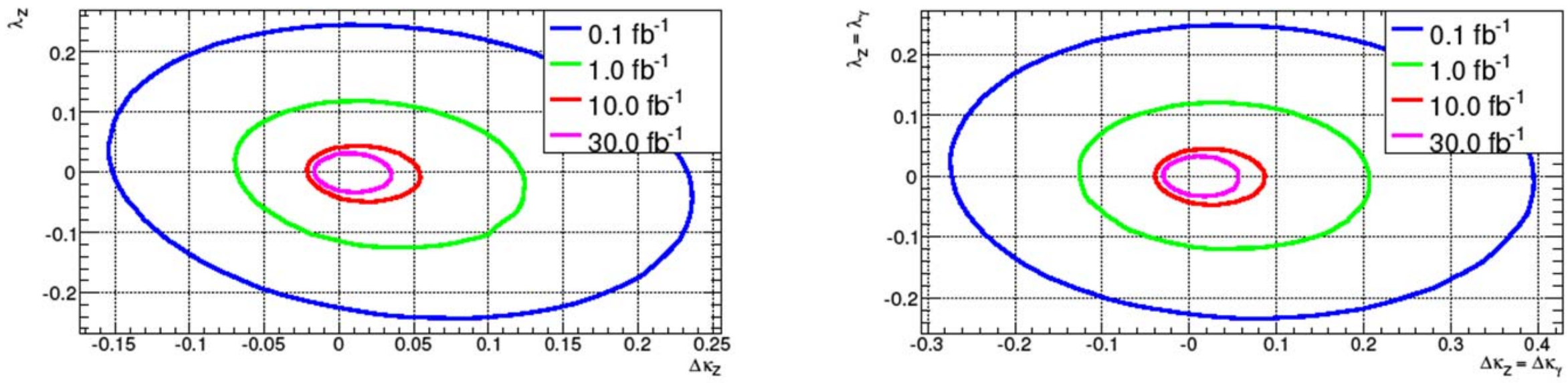


Figure 16: The two-dimensional anomalous TGC limits at 95% C.L. for 0.1, 1, 10 and 30 fb$^{-1}$ integrated luminosities.

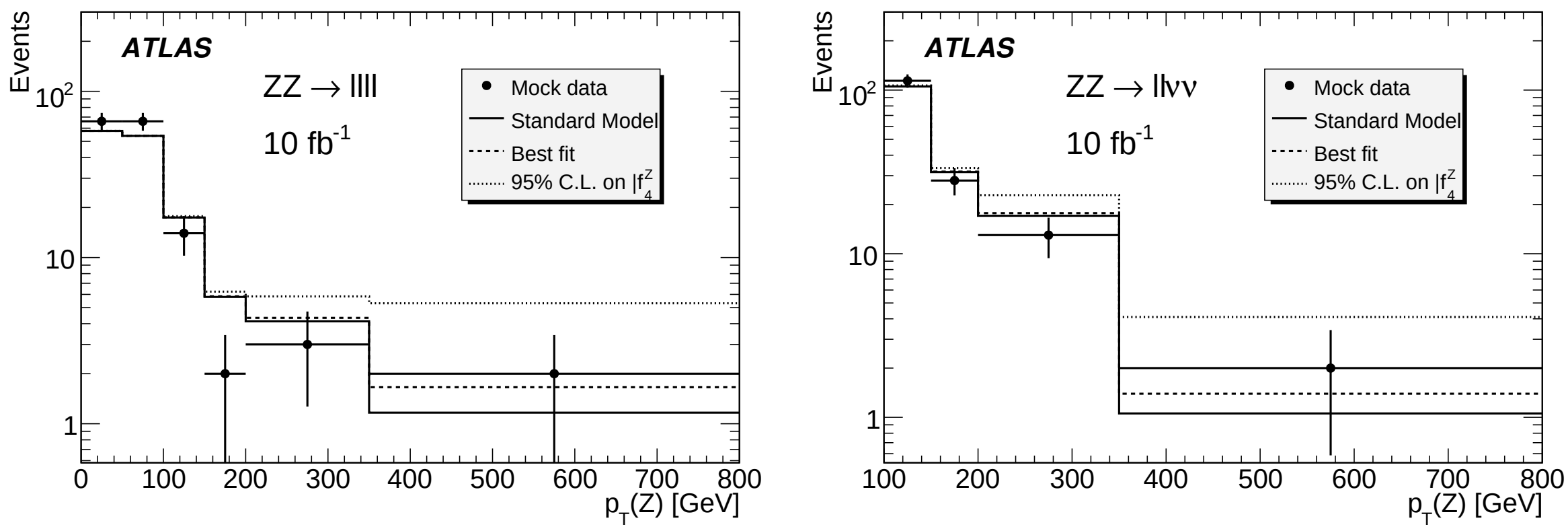


Figure 17: Example of a fit to one 'mock data' sample in each channel. The points show the total number of data events in each bin (not number per unit $p_T$ ). The histograms show the Standard Model prediction (solid), the best fit (dashed) and the 95% C.L. limit on $|f_4^Z|$ (dotted).

## 5.5 $ZZZ$ and $ZZ\gamma$ anomalous TGC sensitivity in $ZZ$ analysis

Measurements of the $pp \to ZZ$ differential cross-section can be used to measure, or set limits on, $ZZZ$ and $ZZ\gamma$ couplings. These couplings are zero at tree level in the Standard Model. Measurements of the couplings provide a sensitive test of the Standard Model, and non-zero values would indicate the presence of new physics beyond the Standard Model.

In order to estimate limits on anomalous couplings which may be obtained from measurements of $ZZ$ production in early ATLAS data, the $p_T$ distribution of the $Z$ boson is considered. In the $ZZ \to \ell\ell\nu\nu$ channel the visible $Z$ boson reconstructed from the charged leptons is used. In the $ZZ \to \ell\ell\ell\ell$ channel one of the two reconstructed $Z$ bosons is chosen in each event at random. Simulated 'mock data' distributions are fitted with the sum of expected signal and background distributions, where the signal distribution depends on the anomalous couplings. A binned maximum likelihood fit is employed, with systematic uncertainties included by convolution with the predictions. Fits are performed to each channel separately, and a combined fit is performed by multiplying together the likelihoods from the two channels assuming no correlated uncertainties.

An example fit for each channel is shown in Figure 17. The results presented here use four $p_T$ bins for the $\ell\ell\nu\nu$ channel and six $p_T$ bins for the 4-lepton channel, as shown in Figure 17. Reasonable modifications to the number or position of $p_T$ bins change the expected limits by up to 15% (12%) in the $\ell\ell\nu\nu$ ($\ell\ell\ell\ell$) channel. Removing the first two $p_T$ bins for the $\ell\ell\ell\ell$ channel, and fitting only the region

Table 22: Expected 95% C.L. intervals on anomalous couplings from fits to the $ZZ \to \ell\ell\ell\ell$ channel, the $ZZ \to \ell\ell\nu\nu$ channel and both channels together for various values of integrated luminosity, with $\Lambda = 2$ TeV. In each case, other anomalous couplings are assumed to be zero.

| | Int. Lumi / fb$^{-1}$ | $f_4^Z$ | $f_5^Z$ | $f_4^\gamma$ | $f_5^\gamma$ |
|---|---|---|---|---|---|
| $ZZ \to \ell\ell\ell\ell$ | 1 | [–0.023, 0.023] | [–0.024, 0.024] | [–0.028, 0.028] | [–0.029, 0.028] |
| | 10 | [–0.010, 0.010] | [–0.010, 0.010] | [–0.012, 0.012] | [–0.013, 0.012] |
| | 30 | [–0.008, 0.008] | [–0.008, 0.008] | [–0.009, 0.009] | [–0.009, 0.009] |
| $ZZ \to \ell\ell\nu\nu$ | 1 | [–0.024, 0.024] | [–0.024, 0.025] | [–0.029, 0.029] | [–0.030, 0.029] |
| | 10 | [–0.012, 0.012] | [–0.012, 0.012] | [–0.014, 0.014] | [–0.015, 0.014] |
| | 30 | [–0.009, 0.009] | [–0.009, 0.009] | [–0.011, 0.011] | [–0.011, 0.011] |
| Combined | 1 | [–0.018, 0.018] | [–0.018, 0.019] | [–0.022, 0.022] | [–0.022, 0.022] |
| | 10 | [–0.009, 0.009] | [–0.009, 0.009] | [–0.010, 0.010] | [–0.011, 0.010] |
| | 30 | [–0.006, 0.006] | [–0.006, 0.007] | [–0.008, 0.008] | [–0.008, 0.008] |

$p_T > 100$ GeV has a negligible effect on the limits.

Table 22 shows the mean expected limits from each channel separately, and from combining the channels, for various values of integrated luminosity. With an integrated luminosity of 1 fb$^{-1}$ the sensitivities of the two channels are very similar. At higher luminosities, the $\ell\ell\ell\ell$ channel becomes somewhat more sensitive, because it has lower background and hence a lower associated systematic uncertainty. With as little as 1 fb$^{-1}$ of data it should be possible to improve the LEP limits [32] on $f_4^Z$, $f_5^Z$ and $f_5^\gamma$ by an order of magnitude using a single channel, while a similar improvement on $f_4^\gamma$ will require both channels.

At an integrated luminosity of 10 fb$^{-1}$, the expected limits have only a low sensitivity to the background level and to the systematic uncertainties. With the same signal efficiency but no background, the limits from the $\ell\ell\nu\nu$ channel improve by 10%, while those from the $\ell\ell\ell\ell$ channel change by only $\sim$0.2%; in the latter case, doubling the background has an effect of only $\sim$0.4%. Reducing all systematic uncertainties to zero improves the limits by 7% (6%) in the $\ell\ell\nu\nu$ ($\ell\ell\ell\ell$) channel. Thus, the background level and systematic uncertainties are unlikely to be important factors in obtaining limits from early data.

As discussed above, the expected limits are affected by the choice of $p_T$ bins. The number of bins is currently limited by the statistics of the fully simulated Monte Carlo events. Future studies would benefit from increased signal Monte Carlo statistics, particularly in the high $p_T$ region. In addition, samples of fully simulated events with anomalous couplings should be used to investigate the dependence of the efficiency at a particular $p_T$ value on the production diagram.

## 6 Summary

This note presents studies of the production of $W^+W^-$, $W^\pm Z$, $ZZ, W^\pm\gamma$ and $Z\gamma$ dibosons from pp collisions at the LHC, using leptonic decays of $W^\pm$ and $Z$ bosons. The simulated measurements are done using the ATLAS detector with full detector simulation and event reconstruction, and the statistics expected in the initial data taking periods. It focuses on the sensitivities that ATLAS can achieve in the early running of LHC, rather than the ultimate sensitivites that ATLAS might reach after running at the design luminosity. The advanced analysis technique BDT is used in analysis of most of the final states, which improves the sensitivities significantly. Table 16 lists the expected numbers of signal and background events using 1 fb$^{-1}$ of data, and the significance of the Standard Model signals after taking into account the known background contributions with 20% systematic uncertainties. It concludes that with

Table 23: 95% C.L. interval of the anomalous coupling sensitivities from $W^+W^-$, $W^\pm Z$, $W^\pm\gamma$ final states with 10.0 fb$^{-1}$ of integrated luminosity and the cutoff $\Lambda = 2$TeV. The table also indicates the variables used in the fit to set the AC sensitivity interval. For reference, some recently published limits from Tevatron and LEP are also listed. These limits caculation assumptions are given in the table as well.

| Diboson, (fit spectra) | $\lambda_Z$ | $\Delta\kappa_Z$ | $\Delta g_1^Z$ | $\Delta\kappa_\gamma$ | $\lambda_\gamma$ |
|---|---|---|---|---|---|
| WZ, ($M_T$) | [-0.015, 0.013] | [-0.095, 0.222] | [-0.011, 0.034] | | |
| $W\gamma$, ($p_T^\gamma$) | | | | [-0.26, 0.07] | [-0.05, 0.02] |
| WW, ($M_T$) | [-0.040, 0.038] | [-0.035, 0.073] | [-0.149, 0.309] | [-0.088, 0.089] | [-0.074, 0.165] |
| $WZ$, (D0) (1.0 fb$^{-1}$) | [-0.17, 0.21] | [-0.12, 0.29] ($\Delta g_1^Z = \Delta\kappa_Z$) | | | |
| $W^\pm\gamma$ (D0), (0.16 fb$^{-1}$) | | | | [-0.88,0.96] | [-0.2,0.2] |
| $WW$, (LEP) ($\lambda_\gamma = \lambda_Z$,$\Delta\kappa_Z = \Delta g_1^Z - \Delta\kappa_\gamma \tan^2\theta_W$) | | | [-0.051,0.034] | [-0.105,0.069] | [-0.059,0.026] |

Table 24: Expected 95% C.L. intervals on anomalous couplings from fits to the $ZZ \to \ell\ell\ell\ell$ channel, the $ZZ \to \ell\ell\nu\nu$ channel and both channels together for 10 fb$^{-1}$ of integrated luminosity, with $\Lambda = 2$ TeV. In each case, other anomalous couplings are assumed to be zero. The 95% C.L. limits on neutral TGC from LEP ZZ detection are also listed.

| | $f_4^Z$ | $f_5^Z$ | $f_4^\gamma$ | $f_5^\gamma$ |
|---|---|---|---|---|
| $ZZ \to \ell\ell\ell\ell$ | [–0.010, 0.010] | [–0.010, 0.010] | [–0.012, 0.012] | [–0.013, 0.012] |
| $ZZ \to \ell\ell\nu\nu$ | [–0.012, 0.012] | [–0.012, 0.012] | [–0.014, 0.014] | [–0.015, 0.014] |
| Combined | [–0.009, 0.009] | [–0.009, 0.009] | [–0.010, 0.010] | [–0.011, 0.010] |
| LEP Limit | [–0.30, 0.30] | [–0.34, 0.38] | [–0.17, 0.19] | [–0.32, 0.36] |

0.1 fb$^{-1}$ of integrated luminosity the Standard Model signals of $W^+W^-$, $W^\pm Z$, $W^\pm\gamma$ and $Z\gamma$ can be established with significance better than $5\sigma$ assuming 20% systematic uncertainties. $ZZ$ production can be established with 1 fb$^{-1}$ of data using the four-lepton decay channels.

Any significant deviation from the Standard Model prediction for these final states can lead to indications of new physics phenomena. In Section 5, the sensitivities to anomalous TGC are presented. The sensitivities are expressed in terms of constraints on the anomalous triple gauge boson couplings in the effective Lagrangian. Table 23 compares the 95% confidence level sensitivity interval for charged anomalous TGC's using observables from different diboson final states with 10 fb$^{-1}$ of integrated luminosity.

The neutral anomalous TGC's can be explored with the $Z\gamma$ and $ZZ$ final states. In this note, only $ZZ$ pairs are used for constraining the anomalous coupling, with the study using $Z\gamma$ still in progress. Both the $ZZ \to \ell^+\ell^-\ell^+\ell^-$ and $ZZ \to \ell^+\ell^-\nu\bar{\nu}$ final states are used to constrain the neutral anomalous TGC parameters ( $f_4^Z$, $f_5^Z$, $f_4^\gamma$, $f_5^\gamma$ ). The 95% C.L. intervals on the anomalous couplings for 10 fb$^{-1}$ of integrated luminosity are listed in Table 24.

The current status of the Monte Carlo generators for diboson is less than satisfactory. MC@NLO is integrated with a parton shower (Herwig), but it does not have matrix elements for the effective Lagrangian beyond the Standard Model with anomalous couplings. The BHO program can generate

parton-level LO and NLO diboson events with anomalous couplings, but can not be correctly integrated with the parton shower programs. In the current analysis, MC@NLO is used to simulate the Standard Model events. The BHO MC with anomalous TGCs is then used to re-weight the events so that the fully simulated events can effectively have the anomalous TGC's, and be used directly to compare with the simulated 'mock data'.

Because of the higher center of mass energy at the LHC, the cross-sections for diboson production are an order of magnitude higher than at the Tevatron. This will allow ATLAS to improve the Tevatron measurements in the early running of the LHC. The large signal statistical significances and signal to background ratios determined from these studies suggest that early observations of these channels will take place at the LHC start up with 0.1 to 1 $fb^{-1}$ of data. Systematic uncertainties will dominate the cross-section measurement errors starting from 5-30 $fb^{-1}$ of data. With increasing luminosity, the constraints on the anomalous couplings will provide important probes of physics beyond the Standard Model.

# Top Quark

# Top Quark Physics

**Abstract**
In the early days of data taking at the LHC, top quark physics will have a role of primary importance for several reasons. At start-up, already with the first few $fb^{-1}$ of integrated luminosity, a top quark signal can be clearly separated from the background even with an imperfectly calibrated detector and the top quark pair production cross-section can be extracted at better than 20% accuracy and with negligible statistical error. The first measurement of the top quark mass will provide feedback on the detector performance and top quark events can be used to understand and calibrate the light jet energy scale and the b-tagging. Additionally in scenarios beyond the Standard Model, new particles may decay into top quarks, therefore a detailed study of the top quark properties may provide a hint of new physics. A good understanding of top quark physics is also essential as top quark events are a background for many new physics searches.

## 1 Introduction

The top quark, discovered at Fermilab in 1995 [1], completed the three generation structure of the Standard Model and opened up the new field of top quark physics. Produced predominantly, in hadron-hadron collisions, through strong interactions, the top quark decays rapidly without forming hadrons, and almost exclusively through the single mode $t \rightarrow Wb$. The W-boson can then decay leptonically or hadronically. The relevant CKM coupling is already determined by the (three-generation) unitarity of the CKM matrix. Yet the top quark is distinguished by its large mass, about 35 times larger than the mass of the next heavy quark, and close to the electroweak symmetry breaking scale. This unique property raises a number of interesting questions. For example if the top quark mass is generated by the Higgs mechanism as the Standard Model predicts and if its mass is related to the top-Higgs-Yukawa coupling, or if it does play an even more fundamental role in the electroweak symmetry breaking mechanism. Non Standard Model physics could first manifest itself in non-standard couplings of the top quark which show up as anomalies in top quark production and decays. By studying the top quark, some of these questions may be answered. Further insight in top quark properties will come from measurements done with the high statistics sample of $t\bar{t}$ pairs such as top quark and W polarization studies sensitive to anomalous Wtb couplings, searches for rare top quark decays indicating the presence of new physics, or for new resonances decaying to $t\bar{t}$ pairs.
The LHC will be a top quark factory, producing millions of $t\bar{t}$ pairs in a sample of 10 $fb^{-1}$, which is expected to be collected during the first years of LHC operation.
The understanding of the experimental signatures for top quark events involves most parts of the ATLAS detector and is essential for claiming potential discoveries of new physics.
Since numerous single (anti-)top quark events are produced via electroweak interactions, the top quark properties, such as the Wtb coupling, can be examined with high precision at the LHC during its first years of running.

## 2 Top quark pair production

In proton-proton collisions top quark pairs are produced through both gluon-gluon and quark-antiquark scattering (Figure 1). The relative importance of both amplitudes depends on the center of mass energy of the collision and nature of the beams: at the LHC the gluon scattering process dominates ($\sim$ 90% of

the cases) while at the Tevatron, production of top quark pairs is kinematically restricted to the quark dominated region. This difference appears also in the values of the cross-section: that is about 100 times larger at the LHC than at the Tevatron. The large top quark mass ($m_t$) ensures that top quark production is a short-distance process, and that the perturbative expansion, given by a series in powers of the small parameter $\alpha_s(m_t)$, converges rapidly. The cross-section used throughout this note for production at the LHC has been calculated up to NLO order including NLL soft gluon resummation, and results in about 833$\pm$100 pb, where the uncertainty reflects the theoretical error obtained from varying the renormalisation scale by a factor of two [2]. The effect of PDF errors accounts for a few percent uncertainty, while varying the top mass by a factor of two, an uncertainty of about 6% on the cross-section is obtained. A calculation that includes NNLL soft-gluon corrections results in a central value for the $t\bar{t}$ cross-section of 872.8 pb [3]. This translates to about 83,000 top quark pairs in a sample of 100 pb$^{-1}$ and of the order of $10^7$ top quark pairs produced per year before any selection or detection criteria are applied.

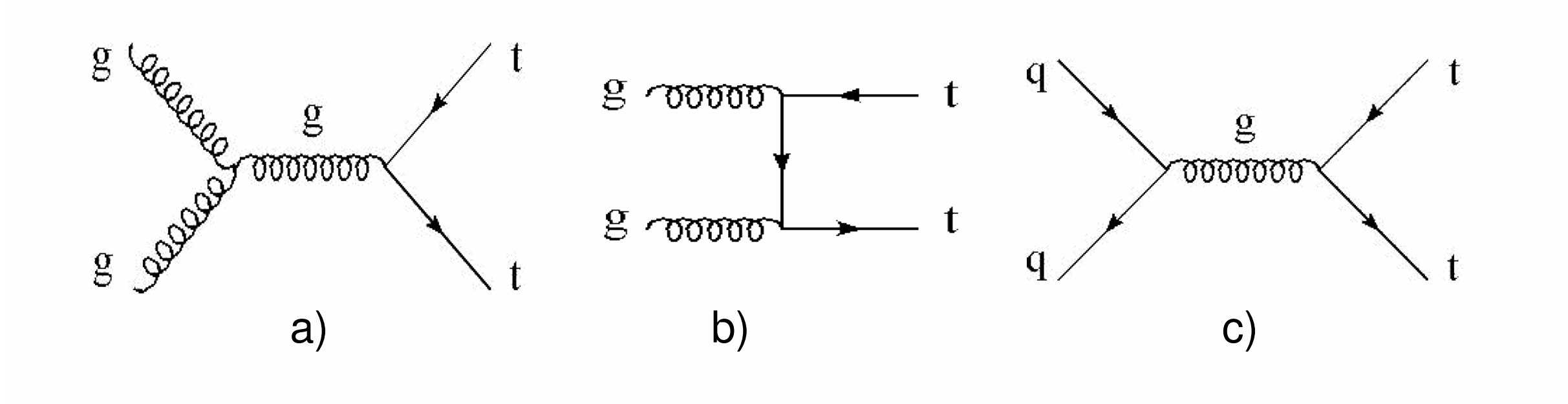


Figure 1: Top production processes at lowest order: gluon-gluon scattering diagrams (a) and b)) and quark-quark scattering diagram c).

## 2.1 Observables and phenomenology

In the Standard Model, the decay of top quarks takes place almost exclusively through the t$\rightarrow$Wb decay mode. A W-boson decays in about 1/3 of the cases into a charged lepton and a neutrino. All three lepton flavors are produced at approximately equal rate. In the remaining 2/3 of the cases, the W-boson decays into a quark-antiquark pair, and the abundance of a given pair is instead determined by the magnitude of the relevant CKM matrix elements. Specifically, the CKM mechanism suppresses the production of b-quarks as $|V_{cb}|^2 \simeq 1.7 \times 10^{-3}$. Thus, the quarks from W-boson decay can be considered as a clean source of light quarks.

From an experimental point of view, one can characterise the top quark decay by the number of W-bosons that decay leptonically. A value of 10.8% and 67.6% has been used for the leptonic and hadronic branching ratio (BR) of the W-boson, respectively [4]. The following signatures can be identified:

- *Fully leptonic:* represents about 1/9 of the $t\bar{t}$ events. Both W-bosons decay into a lepton-neutrino pair, resulting in an event with two charged leptons, two neutrinos and two b-jets. This mode is identified by requiring two high $p_T$ leptons and the presence of missing transverse energy ($\not{E}_T$), and allows a clean sample of top quark events to be obtained. However, this sample has limited use in probing the top quark reconstruction capability of the ATLAS experiment, due to the two neutrinos escaping detection.

- *Fully hadronic:* represents about 4/9 of the $t\bar{t}$ decays. Both W-bosons decay hadronically, which gives at least six jets in the event: two b-jets from the top quark decay and four light jets from the W-boson decay. In this case, there is no high $p_T$ lepton to trigger on, and the signal is not easily

distinguishable from the abundant Standard Model QCD multi-jets production, which is expected to be orders of magnitude bigger than the signal. Another challenging point of this signature is the presence of a high combinatorial background when reconstructing the top quark mass.

- *Semi-leptonic:* represents about 4/9 of the $t\bar{t}$ decays. The presence of a single high $p_T$ lepton allows to suppress the Standard Model W+jets and QCD background. The $p_T$ of the neutrino can be reconstructed as it is the only source of $\not{E}_T$ for signal events.

In this document, top pair production is studied in the semi-leptonic and fully leptonic decay modes.

# 3 Single Top Quark Production

In the Standard Model single-top quark production is due to three different mechanisms: (a) W-boson and gluon fusion mode, which includes the t-channel contribution and is referred to as t-channel or $Wg$ as a whole (b) associated production of a top quark and a W-boson, denoted Wt, and (c) s-channel production. The corresponding diagrams are shown in Fig. 2. We note however that these definitions are valid only at leading order (LO): next to leading order (NLO) calculations may introduce diagrams which cannot be categorised so unambiguously. The total NLO cross-section amounts to about 320 pb at the LHC. Among those channels, the dominant contribution comes from the t-channel processes, which account for about 250 pb; the Wt contribution amounts to about 60 pb while the s-channel mode is expected with a cross-section of about 10 pb [5] [6].

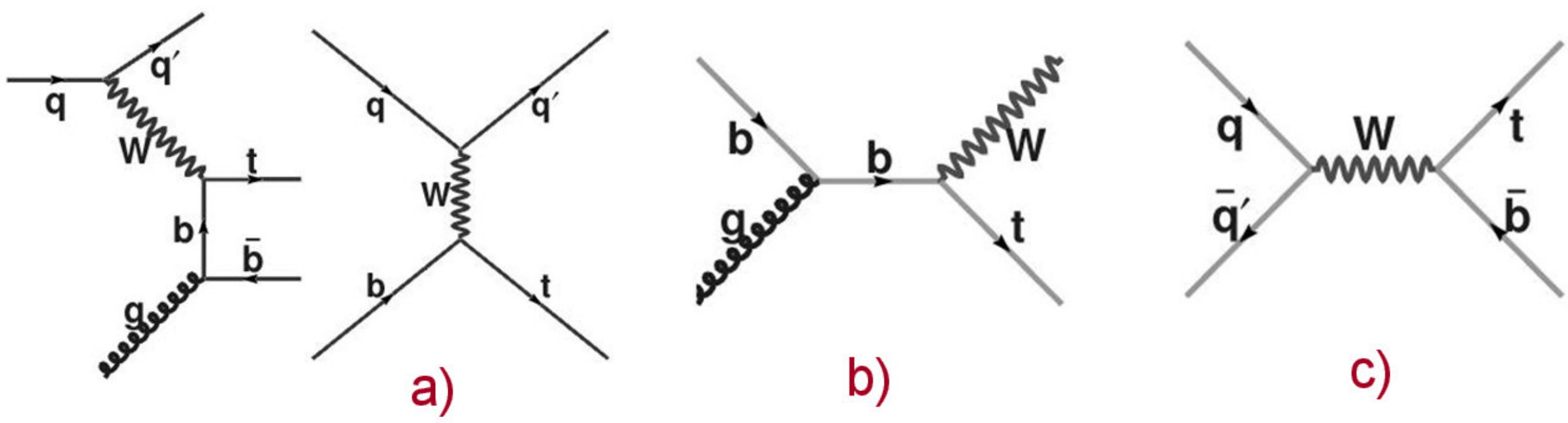


Figure 2: Main graphs corresponding to the three production mechanisms of single-top quark events: (a) t-channel (b) Wt associated production (c) s-channel.

In the following notes, when discussing the analysis strategy in the s- and t-channels, we will use only the leptonic decay of the W-bosons ($l\nu b\bar{b}$ and $l\nu b(\bar{b})q$ final states, respectively)[1]. For the associated Wt production, we will consider events where one of the W-bosons ( either the one produced together with the top quark or the one appearing in the top quark decay) decays leptonically and the other hadronically. The $\tau$ decay modes were included in all relevant simulated event samples, though signal selection is aimed at electron and muon signatures.

We note that in pp collisions, the cross-section for single-top quark is not charge symmetric. The s-channel $t\bar{b}$ final state cross-section is predicted to be a factor 1.6 higher than the one corresponding to the $\bar{t}b$ final state. This ratio is 1.7 if only the t-channel processes are included. This feature is of special interest since it generates a charge asymmetry in the leptonic final state that can be exploited in the analysis to reduce the contamination from the top quark pair production, which constitutes the main

[1]The hadronic decay modes have obvious disadvantages for triggering and the lack of a lepton signature increases the background significantly

background to our signal. On the other hand, the rates for the charge-conjugate processes $W^- + t$ and $W^+ + \bar{t}$ are identical.

Significant sources of uncertainties affect the theoretical predictions of the production cross-sections. The s-channel is known with a precision of 9.1% at NLO [5,7], while the t-channel has an uncertainty of 4.8% [5,7]. An uncertainty of 3% is quoted for the Wt channel [6]. Those uncertainties come from three main sources. The uncertainty in the parton luminosity, depending upon the choice of the parton density functions, is particularly important (2 to 4% for the s- and t-channels), since the b parton or the gluons are involved in the hard processes. The choice for the renormalization and factorization scales accounts for about 2 and 3% uncertainty in s- and t-channel calculations respectively. Finally, a few GeV uncertainty on the top quark mass $m_t$ results in percent level variation of the cross-sections. The uncertainty in $\alpha_s$ enters marginally in the total error at a value below the 1% level.

# 4 Monte Carlo samples

The Monte Carlo samples which have been used for the top quark analyses reported are described in this section. The calculation of many processes benefit from methods such as resummation of next-to-leading log terms and some are calculated at the full NLO accuracy. All the samples used have been normalized using ”K-factors”, to the NLO theoretical cross-section calculations whenever available [8]. The value of $m_t = 175$ GeV has been used for the generation of all samples and all cross-sections correspond to this value. Most samples were processed with the full GEANT4 ATLAS detector simulation and reconstruction code. In some cases, the fast simulation package ATLFAST has been used.
The effect of pile-up in the cavern corresponding to a luminosity of $10^{33}$ cm$^{-2}$s$^{-1}$ has been simulated both for $t\bar{t}$ and single top events.

## 4.1 Simulation of $t\bar{t}$ signal events

Top quark pair production has been simulated using the Monte Carlo generator MC@NLO [9] version 3.1. The hard process of $t\bar{t}$ production is calculated at NLO, so that diagrams that produce one additional parton in the final state are included at matrix element level. The parton density function CTEQ6M [10] is used. Fragmentation and hadronisation is simulated using HERWIG [11] and the underlying event by Jimmy [12]. The $t\bar{t}$ main samples for analysis are a sample of single and double leptonic events and one of fully hadronic events. There are no cuts applied at generation level other than the lepton flavor separation according to W-boson decay type that allowed subdividing the generated events into these two samples. A number of other samples mainly aimed at systematic studies have been produced and are listed in Table 1. Among these AcerMC [13] samples interfaced with PYTHIA for the hadronisation and fragmentation and simulation of the underlying event, aimed at initial and final state radiation (ISR and FSR) studies, that will be discussed later in the text. For the $t\bar{t}$ rare decays samples were produced with TopRex [14] interfaced with PYTHIA for the hadronisation and fragmentation and simulation of the underlying event.

## 4.2 Simulation of single top quark events

For the single top quark signal production, the AcerMC matrix element generator was used in conjunction with PYTHIA, that was used for hadronisation, fragmentation and simulation of the underlying event. The parton density functions CTEQ6M have been used. Compared to TopRex which was previously used in ATLAS, its t-channel generation method is based on a more physically motivated method [15] for combining LO and tree level NLO diagrams. The contribution from NLO diagrams is rather important for the t-channel as the gluon splitting to $b\bar{b}$ tends to be underestimated with the parton shower method.

Table 1: $t\bar{t}$ and single top quark simulated samples used throughout the notes. ($m_t$=175 GeV) is used as default in the generation. Given are a short description of the simulated physics process, the generator used, the production cross-section ($\sigma$), and the K-factor that should be applied to the quoted cross-section.

| **MC@NLO + HERWIG $t\bar{t}$** – fully simulated events – **K-factor = 1.0** | $\sigma$(pb) $\times BR$ |
|---|---|
| Fully leptonic and semi-leptonic $t\bar{t}$ | 450 |
| Fully hadronic $t\bar{t}$ ($m_t$=175 GeV) | 380 |
| Fully leptonic and semi-leptonic $t\bar{t}$ ($m_t$=160 GeV) | 450 |
| Fully leptonic and semi-leptonic $t\bar{t}$ ($m_t$=170 GeV) | 450 |
| Fully leptonic and semi-leptonic $t\bar{t}$ ($m_t$=180 GeV) | 450 |
| Fully leptonic and semi-leptonic $t\bar{t}$ ($m_t$=190 GeV) | 450 |
| Fully leptonic and semi-leptonic $t\bar{t}$ no UE | 450 |
| Inclusive $t\bar{t}$ ($p_T(t) \geq 200$ GeV) | 100 |
| **AcerMC+PYTHIA – $t\bar{t}$** – fully simulated events – **K-factor = 1.0** | $\sigma$(pb) |
| Fully leptonic and semi-leptonic $t\bar{t}$ | 450 |
| Fully leptonic and semi-leptonic $t\bar{t}$, different ISR/FSR (low top mass) | 450 |
| Fully leptonic and semi-leptonic $t\bar{t}$, different ISR/FSR (high top mass) | 450 |
| **AcerMC+PYTHIA single top quark Wt-channel** – full sim. events **K-factor = 1.14** | $\sigma$(pb) |
| Single top quark associated Wt production, semi-leptonic decay | 25.5 |
| **AcerMC+PYTHIA single top quark s-channel** – full sim. events **K-factor = 1.5** | |
| Single top quark s-channel leptonic decay | 2.3 |
| **AcerMC+PYTHIA single top quark t-channel** – full sim. events **K-factor = 0.98** | |
| Single top quark t-channel leptonic decay | 81.3 |
| **TopRex+PYTHIA – $t\bar{t}$ rare decays** – fully simulated events | |
| $t\bar{t} \to bW(\ell\nu)+q\gamma$ | |
| $t\bar{t} \to bW(\ell\nu)+qZ(\ell\ell)$; $\ell = e, \mu$ | |
| $t\bar{t} \to bW(\ell\nu)+qg$ | |

The s-channel and Wt-channel are generated at LO accuracy only. All three channels were generated with W-bosons forced to decay leptonically (e or $\mu$ or $\tau$). In the case of Wt, either the associated W-bosons or the W-bosons from top quark decay is forced to decay leptonically and no dileptonic events are included. The Monte Carlo samples that have been used in the analyses and the corresponding normalisation cross-sections can be found in Table 1.

## 4.3 Simulation of background W + jet events

For the W + jets production, the ALPGEN [16] generator with HERWIG [11] clustering has been used. HERWIG has been used for the simulation of the fragmentation and the hadronisation and Jimmy for the underlying event. The MLM [17] algorithm has been used to match the parton shower and the matrix element calculations. The matching parameters are the minimum $p_T$ of the partons and the minimum $\Delta R$ among two partons, defined as the separation of two objects in the $\eta$-$\phi$ space, $(\Delta R)^2 = (\Delta\eta)^2 + (\Delta\phi)^2$. Here, $\eta$ is the pseudorapidity of an object, defined as $\eta = -\ln(\tan(\theta/2))$ and $\phi$ and $\theta$ are the azimuthal and polar angles, respectively. All cones are defined in $\eta$-$\phi$ space. The values of the matching parameters that are used in this note are $p_T = 20$ GeV and $\Delta R = 0.3$.
A fraction of this background contains heavy quarks. This background is treated separately in ALPGEN

by producing $W + b\bar{b}$ and $W + c\bar{c}$ (plus light jets) samples. The W-boson background samples used throughout the notes are described in Table 2 [2].

### 4.4 Simulation of background $Z \to \ell\ell$+ jets events

The $Z \to \ell\ell$ + n jets background events, where the lepton flavor can be any charged lepton, has been generated with ALPGEN, while HERWIG has been used for the simulation of the fragmentation and the hadronisation and Jimmy for the underlying event. The MLM algorithm has been used to match the parton shower and the matrix element calculations. The matching parameters values are $p_T = 20$ GeV and $\Delta R = 0.3$.
The contribution of those backgrounds to our analyses is non-negligible only when there is at least one jet in addition to the Z. The samples have been generated with up to 5 additional jets. Samples generated with PYTHIA have also been used: the complete list of the samples can be found in Table 3[3].

### 4.5 Simulation of di-boson background

Di-boson events produced with light jets can be a background for the $t\bar{t}$ and single top quark signal. WW, WZ and ZZ processes with all decay modes have been generated with HERWIG: a filter was applied to select those events with an electron or a muon with $p_T > 10$ GeV. WW events with the W-boson decaying leptonically into final states with two electrons, an electron (muon) and a $\tau$-lepton and two $\tau$-leptons have been generated with MC@NLO interfaced with HERWIG for the hadronisation and fragmentation ( see Table 4).

### 4.6 QCD background

QCD multi-jet events are a background for $t\bar{t}$ and single top quark analyses if at least one of the jets in the event is misidentified as an isolated lepton. The level of QCD multi-jet background has large uncertainties with the currently available generation tools, which are based on a leading order description: ALPGEN has been used to generate these events with the same matching parameters as discussed in section 4.3. Given the large cross-section for this process only fast simulated events (ATLFAST) have been produced and used for the present studies. Events with 2 to 5 light jets and $b\bar{b} + 0, 1, 2, 3$ light jets have been generated requiring at least 3 jets in the final state with $p_T > 30$ GeV. Di-jet fully simulated PYTHIA events are available and have been used for comparison with the ATLFAST results (see Table 5). In practice, the level of background will be derived directly from the data and will strongly depend on the lepton fake rate, and can depend on the topology of the event. Different cuts can be applied to strongly reduce this background. In fully hadronic $t\bar{t}$ decays, QCD multi-jet events are the main background and a different strategy has to be developed. These decays are not studied in details in this document.

## 5 Reconstruction of physics objects

We use definitions of high level reconstructed objects (electrons, muons, jets, etc.) that are standard in ATLAS. They are described in the following sections. In the definitions, we often use the distance $\Delta R$ between objects.

[2]The difference in the cross section for the different leptons in the first set of events in Table 2 (fully simulated W boson events with ALPGEN) is due to the truth jet filter. Electrons and $\tau$ leptons can also be reconstructed as a jet: it is more probable for electrons than for $\tau$ leptons that are reconstructed via their visible decay products.

[3]In the truth jet filter applied to the fully simulated Z boson events with ALPGEN in Table 3 jets made from leptons are removed at the filter level.

Table 2: W-boson background samples used throughout the notes. Given are a short description of the simulated physics process, the generator used, the production cross-section ($\sigma$) including filter, matching efficiency for ALPGEN and selection efficiency, and the K-factor that should be applied to the quoted cross-section.

| **ALPGEN + Jimmy – W-boson** – fully simulated events – **K-factor = 1.15** | $\sigma$(pb) |
|---|---|
| W→ $e\nu$ + 2 partons; truth filter 3jets with $p_T^j \geq 30$ GeV | 214 |
| W→ $e\nu$ + 3 partons | 124 |
| W→ $e\nu$ + 4 partons | 54 |
| W→ $e\nu$ + 5 partons | 22 |
| W→ $\mu\nu$ + 2 partons; truth filter 3jets with $p_T^j \geq 30$ GeV | 16 |
| W→ $\mu\nu$ + 3 partons | 65 |
| W→ $\mu\nu$ + 4 partons | 36 |
| W→ $\mu\nu$ + 5 partons | 20 |
| W→ $\tau\nu$ + 2 partons; truth filter 3jets with $p_T^j \geq 30$ GeV | 88 |
| W→ $\tau\nu$ + 3 partons | 87 |
| W→ $\tau\nu$ + 4 partons | 46 |
| W→ $\tau\nu + \geq 5$ partons | 21 |
| **ALPGEN + Jimmy – W boson** – Atlfast sample – **K-factor = 1.15** | $\sigma$(pb) |
| W→ $e\nu$ + 0 parton | 13400 |
| W→ $e\nu$ + 1 parton | 2610 |
| W→ $e\nu$ + 2 partons | 826 |
| W→ $e\nu$ + 3 partons | 239 |
| W→ $e\nu$ + 4 partons | 67.4 |
| W→ $e\nu + \geq 5$ partons | 24.0 |
| W→ $\mu\nu$ + 0 parton | 13400 |
| W→ $\mu\nu$ + 1 parton | 2590 |
| W→ $\mu\nu$ + 2 partons | 826 |
| W→ $\mu\nu$ + 3 partons | 236 |
| W→ $\mu\nu$ + 4 partons | 68.3 |
| W→ $\mu\nu + \geq 5$ partons | 24.3 |
| W→ $\tau\nu$ + 0 parton | 13400 |
| W→ $\tau\nu$ + 1 parton | 2620 |
| W→ $\tau\nu$ + 2 partons | 828 |
| W→ $\tau\nu$ + 3 partons | 239 |
| W→ $\tau\nu$ + 4 partons | 67.7 |
| W→ $\tau\nu + \geq 5$ partons | 24.4 |
| **ALPGEN + Jimmy – W + $\mathbf{b\bar{b}}$** – fully simulated events – **K-factor = 2.57** | $\sigma$(pb) |
| $W + b\bar{b}$ + 0 parton; no filter | 6.26 |
| $W + b\bar{b}$ + 1 parton | 6.97 |
| $W + b\bar{b}$ + 2 partons | 3.92 |
| $W + b\bar{b}$ + 3 partons | 2.77 |
| **ALPGEN + Jimmy – W + $\mathbf{c\bar{c}}$** – fully simulated events – | $\sigma$(pb) |
| $W + c\bar{c}$ + 0 parton; no filter | 6.72 |
| $W + c\bar{c}$ + 1 parton | 7.49 |
| $W + c\bar{c}$ + 2 partons | 4.36 |
| $W + c\bar{c}$ + 3 partons | 2.45 |

Table 3: Z boson background samples used throughout the note. Given are a short description of the simulated physics process, the generator used, the production cross-section ($\sigma$) including filter, selection efficiency and matching efficiency for ALPGEN, and the K-factor that should be applied to the quoted cross-section.

| **ALPGEN + Jimmy – Z boson** – fully simulated events – **K-factor = 1.24** | $\sigma$(pb) |
|---|---|
| $Z \to e^+e^-$+ 1 parton; $p_T^e \geq 10$ GeV, one jet with $p_T^j \geq 20$ GeV | 138 |
| $Z \to e^+e^-$+ 2 partons | 50.5 |
| $Z \to e^+e^-$+ 3 partons | 16.2 |
| $Z \to e^+e^-$+ 4 partons | 4.6 |
| $Z \to e^+e^- + \geq 5$ partons | 1.7 |
| $Z \to \mu^+\mu^-$+ 1parton; $p_T^\mu \geq 10$ GeV, one jet with $p_T^j \geq 20$ GeV | 136 |
| $Z \to \mu^+\mu^-$+ 2 partons | 51.7 |
| $Z \to \mu^+\mu^-$+ 3 partons | 16.3 |
| $Z \to \mu^+\mu^-$+ 4 partons | 4.6 |
| $Z \to \mu^+\mu^- + \geq 5$ partons | 1.7 |
| $Z \to \tau^+\tau^-$+ 1 parton; $p_T^\ell \geq 10$ GeV, one jet with $p_T^j \geq 20$ GeV | 57 |
| $Z \to \tau^+\tau^-$+ 2 partons | 21.3 |
| $Z \to \tau^+\tau^-$+ 3 partons | 7.0 |
| $Z \to \tau^+\tau^-$+ 4 partons | 2.2 |
| $Z \to \tau^+\tau^- + \geq 5$ partons | 0.8 |
| **PYTHIA – Z boson** – fully simulated events – **K-factor = 1.22** | $\sigma$(pb) |
| $Z \to e^+e^-$; $p_T^e \geq 10$ GeV, $m_{\ell\ell} \geq 20$ GeV | 1432 |
| $Z \to \mu^+\mu^-$; $p_T^\mu \geq 10$ GeV, $m_{\ell\ell} \geq 20$ GeV | 1497 |
| $Z \to \tau^+\tau^-$; $p_T^\ell \geq 5$ GeV, $m_{\ell\ell} \geq 20$ GeV | 77 |

### 5.1 Electron definition

Electron candidates are reconstructed and identified by the calorimeters and inner tracker of ATLAS and are reconstructed in the pseudorapidity range $|\eta| < 2.5$. An electron candidate is defined as a medium electron identified by the isEM algorithm [18].

If an electron is found in the calorimeter crack region $1.37 < |\eta| < 1.52$, it is vetoed. The electron has to be isolated based on calorimeter energy: the additional transverse energy $E_T$ in a cone with radius $\Delta R = 0.2$ around the electron axis is required to be less than 6 GeV. The $p_T$ and $|\eta|$ cuts depend on the event signature and are detailed in the various relevant sections.

For electrons above 20 GeV and $|\eta| < 2.5$ and outside the crack region, the average identification efficiency in $t\bar{t}$ events is about 67%, with a purity of about 97%.

### 5.2 Muon definition

Muons are reconstructed by the muon spectrometer and the inner detector. The muon reconstruction is performed using the Staco algorithm [19], and muons are defined from the best match combination of the muon chambers and the tracker information. Muons are reconstructed in the pseudorapidity range $|\eta| < 2.5$ and have to be isolated based on calorimeter energy: the additional transverse energy $E_T$ in a cone with radius $\Delta R = 0.2$ around the muon is required to be less than 6 GeV. The $p_T$ and $|\eta|$ cuts applied to the muons are given in the selection cuts described in the various sections. For muons above

Table 4: Diboson background samples used throughout the notes. Given are a short description of the simulated physics process, the generator used, the production cross-section ($\sigma$) including filter, matching efficiency and selection efficiency for ALPGEN, and the K-factor that should be applied to the quoted cross-section.

| **HERWIG + Jimmy – WW** – fully simulated events – **K-factor = 1.57** | $\sigma$ (pb) |
|---|---|
| 1 e or $\mu$ $p_T^\ell \geq 10$ GeV | 24.5 |
| **HERWIG + Jimmy – ZZ** – fully simulated events – **K-factor = 1.29** | |
| 1 e or $\mu$ $p_T^\ell \geq 10$ GeV | 2.1 |
| **HERWIG + Jimmy – WZ** – fully simulated events – **K-factor = 1.89** | |
| 1 e or $\mu$ $p_T^\ell \geq 10$ GeV | 7.8 |
| **MC@NLO + HERWIG – WW** – fully simulated events – **K-factor = 1.0** | $\sigma$(pb) |
| $W^+W^- \rightarrow e^+\nu\ e^-\nu$; no filter | 1.1 |
| $W^+W^- \rightarrow e^+\nu\ \tau^-\nu$ | 1.1 |
| $W^+W^- \rightarrow \tau^+\nu\ e^-\nu$ | 1.1 |
| $W^+W^- \rightarrow \mu^+\nu\ \tau^-\nu$ | 1.1 |
| $W^+W^- \rightarrow \tau^+\nu\ \mu^-\nu$ | 1.1 |
| $W^+W^- \rightarrow \tau^+\nu\ \tau^-\nu$ | 1.1 |

20 GeV, the average reconstruction efficiency in $t\bar{t}$ events is 88%. The fake rate, defined as the rate at which an object that is not associated with a true muon is mis-identified as a muon, is $0.1 \pm 0.01$ %.

## 5.3 Jet definition

Jets are reconstructed with the standard ATLAS cone algorithm in $\eta - \phi$ space, for $|\eta| < 2.5$ and a cone radius of 0.4, operating on energy depositions in calorimeter towers [20]. In $t\bar{t}$ events, jets coinciding within $\Delta R < 0.2$ with electrons (as defined in section 5.1) are removed.
A jet is identified as originating from a b-quark by determining the probability that it contains a secondary vertex. The three-dimensional impact parameter (IP3D) and the secondary vertex (SV1) [21] algorithms are used and we require that the resulting output weight of the event is larger than 7.05. The requirement on the weight is chosen such that an efficiency of about 60% for jets with $p_T > 30$ GeV in $t\bar{t}$ semi-leptonic events is reached, with a mistag rate of 100.

## 5.4 Missing transverse energy definition

For the missing transverse energy ($\not\!\!E_T$), we determine the sum of five components.

1. the contribution of cells in identified electron or photon clusters;
2. the contribution of cells inside jets;
3. the contribution of cells in topological clusters outside identified objects;
4. the contribution from muons;
5. the cryostat correction.

For the calculation of $\not\!\!E_T$ we use the standard ATLAS variable [22]. The sum of the tranverse energy in semi-leptonic top events is about 500 GeV, which gives a typical $\not\!\!E_T$ resolution of the order of 10 GeV.

Table 5: QCD background samples used throughout the notes. Given are a short description of the simulated physics process, the generator used, the production cross-section ($\sigma$) including filter, matching efficiency and selection efficiency for ALPGEN.

| **PYTHIA – QCD Dijets** – fully simulated events – | $\sigma$(pb) |
|---|---|
| Dijets; $e/\gamma$ filter $p_T \geq 15$ GeV | $1.91 \cdot 10^8$ |
| **Alpgen + HERWIG – jets** – Atlfast – | |
| 2 partons; one jet with $p_T^j \geq 30$ GeV | $1.13 \cdot 10^6$ |
| 3 partons | $2.03 \cdot 10^6$ |
| 4 partons | $1.10 \cdot 10^6$ |
| $\geq 5$ partons | $0.32 \cdot 10^6$ |
| **Alpgen + HERWIG – $b\bar{b}$ + jets** – Atlfast – | $\sigma$(pb) |
| $b\bar{b}$ + 0 parton; one jet with $p_T^j \geq 30$ GeV | $5.3 \cdot 10^3$ |
| $b\bar{b}$ + 1 parton | $33.6 \cdot 10^3$ |
| $b\bar{b}$ + 2 partons | $29.5 \cdot 10^3$ |
| $b\bar{b}+ \geq 3$ partons | $18.9 \cdot 10^3$ |

### 5.5 Overlap removal

In some cases when there is ambiguity in the definition of an object, the overlaps are removed: if an object is identified as an electron it is not counted in the category of photons or taus or jets.

## 6 Systematics

In this section, we list the sources of systematics common to all analysis and describe how they have been consistently treated.

### 6.1 Estimate of the luminosity and its uncertainty

At the LHC start-up only a rough measurement of the machine parameters will be available. The expected uncertainty on the luminosity during this phase will be of the order of 20-30%. A better determination of the beam profiles using special runs of the machine will lead ultimately to a systematic uncertainty of the order of 5%. The proposed ALFA detector will measure elastic scattering in the Coulomb-nuclear interference region using special runs and beam optics, determining the absolute luminosity with an expected uncertainty of the order of 3%. The optical theorem, in conjunction with a precise external measurement of the total cross-section, can achieve a similar 3% precision [23].

### 6.2 Lepton identification efficiency

For the first 100 $pb^{-1}$ of integrated luminosity, the lepton identification efficiency error is expected to be of the order of 1% for electrons and muons, while the error on the fake rate is expected to be 50% and 20%, respectively [18, 19].

### 6.3 Lepton trigger efficiency

The lepton trigger efficiency is measured from data using Z events. We expect the uncertainty to be of the order of 1% for an integrated luminosity of 100 $pb^{-1}$ [24].

### 6.4 Jet energy scale

In the difficult hadron-hadron collision environment, the determination of the jet energy scale is rather challenging [25]. While several methods are proposed such as using $\gamma$+jet events to propagate the electromagnetic scale to the hadronic scale, the jet energy scale depends on a variety of detector and physics effects. This includes non-linearities in the calorimeter response due, for example, to energy losses in "dead" material, and additional energy due to the underlying event. Energy lost outside the jet cone can also affect the measured jet energy. Effects due to the initial and final state radiation (ISR/FSR) modelling could also affect the jet energy scale but they are evaluated separately. The ultimate goal in ATLAS is to arrive at a 1% uncertainty on jet energy scale though such performance is only reachable after several years of study. The jet energy resolution used throughout the note is $60\%/\sqrt{E} \oplus 5\%$.

To estimate the sensitivity of the analyses to the uncertainty on the jet energy scale in early data we have repeated them while artificially rescaling the energies of the jets by $\pm 5\%$. The resulting variation in the measurement from the analyses (e.g. cross-section, mass etc.) gives a good measure of the systematic uncertainty due to the jet energy scale.

Since several analyses are using the missing transverse energy in the event to reduce the backgrounds, the effect of the variation of the jet energy scale on the missing energy has been taken into account by calculating the contribution to the transverse missing energy coming from unscaled jets and subtracting it from the overall missing transverse energy in the event. Finally the contribution from the rescaled jets is calculated and added to the missing transverse energy of the event.

### 6.5 b-tagging uncertainties

The use of b-tagging in $t\bar{t}$ and single top quark events is essential in order to reduce the backgrounds, in particular that from W+jets, and the combinatorial background when reconstructing the top quark. At the beginning of data taking the b-tagging performance will need to be understood and $t\bar{t}$ events will be used as a calibration tool for the determination of the b-tagging efficiency. To avoid having a large dependence on the b-tagging efficiency in the early days of data taking we have studied methods to extract the $t\bar{t}$ cross-section and the top quark mass without applying b-tagging. The uncertainty on the b-jet efficiency is currently estimated to be of the order of $\pm 5\%$ and the uncertainty on the mistag rate is 50% [21].

### 6.6 ISR and FSR systematics

More initial and final state QCD radiation (ISR and FSR) increases the number of jets and affects the transverse momentum of particles in the event. Selection cuts for top quark events include these quantities, therefore ISR and FSR will have some effect on the selection efficiency. In order to evaluate the effect of the ISR and FSR systematics, several studies have been performed using the AcerMC [13] generator interfaced with the PYTHIA parton showering.

Samples of $t\bar{t}$ and single top quark events with separate variations of the PYTHIA ISR and FSR parameters have been generated. The study was limited to parameters which have been shown to have the biggest impact on the reconstructed top mass at the generator level. The choices of the parameters depend on the analysis and include ISR $\Lambda_{QCD}$, the ISR cutoff, FSR $\Lambda_{QCD}$, and the FSR cutoff. The effect of the selection efficiency can be as large as 10%, depending on the analysis cuts.

### 6.7 Parton density uncertainties

The systematic error due to the parton density functions (PDF) uncertainties is evaluated on $t\bar{t}$ signal samples. Both the PDF error sets CTEQ6M and MRST2002 [26] at NLO are used. Both sets have positive and negative error PDFs. In order to evaluate the systematic effect on an observable, the following

formula [27] could be adopted:

$$\Delta X = \frac{1}{2}\sqrt{\Sigma(X_i^+ - X_i^-)^2}$$

where $i$ varies over the set of PDF errors and $X^+$ and $X^-$ is the observable evaluated with the positive and negative error PDFs respectively.

The error PDFs do not guarantee that $X_i^+ > X_0$ and $X_i^- < X_0\ \forall i$ (here $X_0$ is the value of the observable as computed with the default PDF set). An alternative approach has thus been proposed in reference [28], which entails the use of the following asymmetric formulas, that are used here:

$$\Delta X_{max}^+ = \sqrt{\Sigma[max(X_i^+ - X_0, X_i^- - X_0, 0)]^2}$$

$$\Delta X_{max}^- = \sqrt{\Sigma[max(X_0 - X_i^+, X_0 - X_i^-, 0)]^2}.$$

In order to avoid generating a sample for each set of PDF errors, a re-weighting method has been used. Once an event has been generated with a certain PDF, PDF1, it can be re-weighted as if it would have been generated with a different PDF, PDF2, by using the following re-weight factor:

$$event_{(reweights)} = \frac{f_{PDF2}(x_1, flav_1, Q)}{f_{PDF1}(x_1, flav_1, Q)} \frac{f_{PDF2}(x_2, flav_2, Q)}{f_{PDF1}(x_2, flav_2, Q)} \quad (1)$$

The variable $flav_i$ is the flavor of parton $i = 1,2$, which initiates the hard scattering, $x_i$ its momentum fraction, and $Q$ is the mass scale used by the Monte Carlo in the computation of the PDFs.

Eq.1 is an approximation and should be checked by comparing samples obtained with the re-weighting technique and generated ones. Using the re-weighting technique is not the same as generating the samples with the different sets of PDFs, the reason being that in the generator the Sudakov factors of the initial state radiation depend on the PDFs and the dependence does not factorize, i.e. it is not a multiplicative factor. The effect is believed to be generally small, but has been checked for each process by comparing ATLFAST subsamples of our signal events generated with the PDFs and re-weighted samples: the selection efficiencies for those samples have been compared for the $t\bar{t}$ and single top quark analyses and the results agree within the statistical errors.

## 6.8 W+jets normalization

For final states with at least one top decaying leptonically, the ones studied in this document, one of the most important backgrounds is W-boson production in association with jets where the W-boson decays leptonically. The uncertainty on the normalization of this process is particularly relevant in analyses where the background contribution is estimated on the basis of the Monte Carlo expectations rather than fitted from the data.

To evaluate this background we used Monte Carlo samples of W+0,1,2,3,4 jets produced with the ALPGEN Monte Carlo, and we evaluated the selection efficiencies for this background. For the normalisation of the exclusive samples we used the leading order cross-section prediction of the ALPGEN Monte Carlo, with standard settings for the renormalisation and factorisation scales. The inclusive cross-section is then normalised to the NNLO value [8]. However, the normalisation of the exclusive samples has a large theoretical uncertainty due to the matching: we therefore envisage to determine it from data itself.

The exclusive cross-section for W-boson produced in association with jets ($\sigma(\mathrm{W+nj})$) can be extracted from the data by using the following relation:

$$\frac{\sigma(\mathrm{W_{incl}})}{\sigma(\mathrm{W+nj})} = \frac{\sigma(\mathrm{Z_{incl}})}{\sigma(\mathrm{Z+nj})} \quad (2)$$

where the inclusive W-boson and Z-boson cross-sections ($\sigma(\mathrm{W_{incl}})$ and $\sigma(\mathrm{Z_{incl}})$) and the exclusive cross-section for Z produced in association with n jets ($\sigma(\mathrm{Z+nj}), \mathrm{n} = 0-4$) are extracted from the data. Equation (2) is demonstrated to be valid to a few percent level in [29], depending on the selection cuts applied (e.g. $p_{\mathrm{T}}$ cut on jets).

When the data will be available the normalisation of the W+j background will be extracted from relation (2) using Z+nj samples selected in $\mathrm{Z} \rightarrow \mathrm{ee}$ events.

Obtaining the normalisation for this background from the data has shown to be essential, since the uncertainty on the exclusive cross-section of the W+nj background obtained from ALPGEN can be as large as 50% [30]. The largest uncertainties affect the topologies with higher number of jets that are typical of $\mathrm{t\bar{t}}$ events. We have verified that such large variations of the exclusive cross-sections are found when selecting events with at least 4 jets with $p_{\mathrm{T}} > 40$ GeV in ALPGEN W+nj fast simulated samples generated with different sets of matching parameters. While the normalisation has large uncertainties, the shapes of the distributions are found to be basically independent of the choice of the matching parameters. With data driven methods and about 1 $\mathrm{fb}^{-1}$ of luminosity, a 20% uncertainty on the W+nj normalisation should be reachable.

# Triggering Top Quark Events

**Abstract**
Collisions at the LHC occur at a rate of up to 40 MHz, much larger than the 200 Hz storage capacity of the ATLAS experiment. The ATLAS trigger system has the challenging task of rejecting 99.9995 % of the events produced in collisions, while keeping those needed to achieve the physics goals of the experiment. This note evaluates the expected performance of the trigger system in top quark events by investigating the response of the trigger system to single objects such as a muon, an electron or a jet originating from top quark decays. In addition, the methodology needed to efficiently select top quark events in the online trigger system is discussed including methods to determine trigger efficiencies from data.

## 1 Introduction

Triggering at the Large Hadron Collider (LHC) is a challenging task. The selection system is required to reduce the initial bunch-crossing rate of 40 MHz down to a manageable 100-200 Hz while retaining and recording the five in one million most interesting physics events. ATLAS has designed a three-level trigger which aims to select these events, with the highest possible efficiency and lowest possible bias.

The ATLAS trigger is largely based on signatures of high transverse-momentum particles and large missing transverse energy. The first trigger level (L1) is implemented using custom electronics and is based on coarse-resolution data from the calorimeters and dedicated fast muon detectors [1]. The level-2 trigger (L2) and the Event Filter (EF), referred to together as the High Level Trigger (HLT), are implemented in software and run on a commodity computing cluster [2]. The L2 design is based on the concept of Regions of Interest (RoIs). Algorithms request full-resolution data only from the region of the detector corresponding to the L1 candidate object and perform a more refined analysis of its features. The EF works also in a seeded mode although it has access to the full-resolution data of the entire detector. It runs more sophisticated offline reconstruction and selection algorithms.

At the LHC top quarks are mainly produced either in pairs or singly. A large set of different and complex event signatures is expected. For $t\bar{t}$ events, approximately 44% of the decays will be fully hadronic, resulting in a final state with six jets[1]: four light jets from the W-boson decays and two b-jets. Another 44% of the decays will be semi-leptonic, with a final state containing one lepton, one neutrino, two b-jets and two light jets. Roughly 11% of the $t\bar{t}$ decays will be purely leptonic, with two leptons, two neutrinos, and two b-jets in the final state. In single-top events, the W-boson decays hadronically two-thirds of the time and leptonically one-third. More details about the properties and decay modes of top events are discussed in the introduction [3].

The trigger efficiencies for top events of the most relevant single object triggers are studied in Section 2 of this note. The rich topologies of top events results in large overlap between trigger signatures. This feature can be exploited to monitor trigger efficiencies as discussed in Section 3. Finally, in Section 4, the electron trigger efficiency is extracted from $Z \to ee$ events and is compared to the true efficiency in $t\bar{t}$ events.

## 2 Single object trigger performance in top quark events

Results are presented for the basic trigger signatures using electron, muon and jet trigger objects, as well as missing transverse energy ($E_T^{miss}$). The main trigger criteria from the ATLAS trigger tables for the

[1]Neglecting for now initial or final state radiation.

initial data taking phase are evaluated using simulated top quark signal samples. Two types of $t\bar{t}$ signal samples are used in this note. The first is the fully hadronic sample with only jets in the final state and the second is a leptonic sample, consisting of a combination of semi-leptonic and di-leptonic events, and will be referred to as *leptonic* sample in the following. For top quarks produced singly, the s, t, and Wt channels are simulated and only those events which result in an electron or a muon from the W-boson decay are considered here. More details about the data samples are given in [3].

Trigger items are described by a combination of letters and numbers. The trigger object is represented by an abbreviation consisting of one or two letters. It is preceded by a number representing the object multiplicity and followed by a number signifying the transverse momentum ($p_T$) of the trigger threshold. If an isolation requirement is applied, it is indicated by the letter 'i' after the $p_T$ threshold. For example, the item name '2EM18I' represents a trigger on two electromagnetic objects, with a threshold of 18 GeV each, including isolation requirements. The L2 and EF item naming conventions are the same as L1 except that lower-case letters are used. A trigger chain consists of the L1, L2, and EF trigger items an object must satisfy and is referred to using the HLT notation. A *trigger menu* is a list of triggers enabled during a data taking run. For a more complete description, see [4].

The dependence of a trigger on the actual $p_T$ cut will be investigated here by means of *turn-on curves*, which shows the fraction of objects passing a certain trigger as a function of the reconstructed or true simulated $p_T$ of that object. Trigger candidate objects are matched to reconstructed or Monte Carlo simulated objects by cutting on $\Delta R = \sqrt{\Delta\eta^2 + \Delta\phi^2}$ between the two objects.

In the following sections, single object triggers are evaluated in detail first in $t\bar{t}$ events, and then in single-top events.

## 2.1 Electron triggers in $t\bar{t}$ events

### 2.1.1 Introduction

Leptonic $t\bar{t}$ decays will form a statistically large sample of events early on at the LHC. Figure 1 (a) shows the simulated $p_T$ spectrum of the electron from a W-boson decay in leptonic $t\bar{t}$ events. The plot gives an idea of the expected fraction of events the trigger system is confronted with above certain $p_T$ values. The single-electron trigger threshold is expected to be around 20 GeV, indicating that complex multi-object triggers (with correspondingly lowered thresholds) may be unnecessary.

### 2.1.2 Single electron triggers

The L1 electron triggers operate on reduced granularity ($0.1 \times 0.1$ in $\Delta\eta \times \Delta\phi$) calorimeter trigger towers which cover the range $|\eta| < 2.5$. A central cluster of four towers is formed in the electromagnetic and hadronic calorimeters, along with a ring of 12 towers around this central cluster. The ring is used to select candidates using isolation criteria by cutting on the amount of energy deposited around the central cluster. At L2, electromagnetic clusters are formed, tracking is then performed for the first time, and, finally, the reconstructed cluster is matched to a track. In the final stage, the EF, tracking and cluster determination is performed with more accurate algorithms, further refining the trigger decision. More details can be found in [4].

### 2.1.3 Single electron trigger efficiencies

The lowest unprescaled trigger threshold depends on the luminosity. For a luminosity of $10^{33}$ cm$^{-2}$ s$^{-1}$, the e22i trigger is the unprescaled trigger with the lowest $p_T$ threshold. For the start-up luminosity of $10^{31}$ cm$^{-2}$ s$^{-1}$, an e12i trigger is available. The e12i chain has a lower threshold and hence a higher efficiency in leptonic $t\bar{t}$ events.

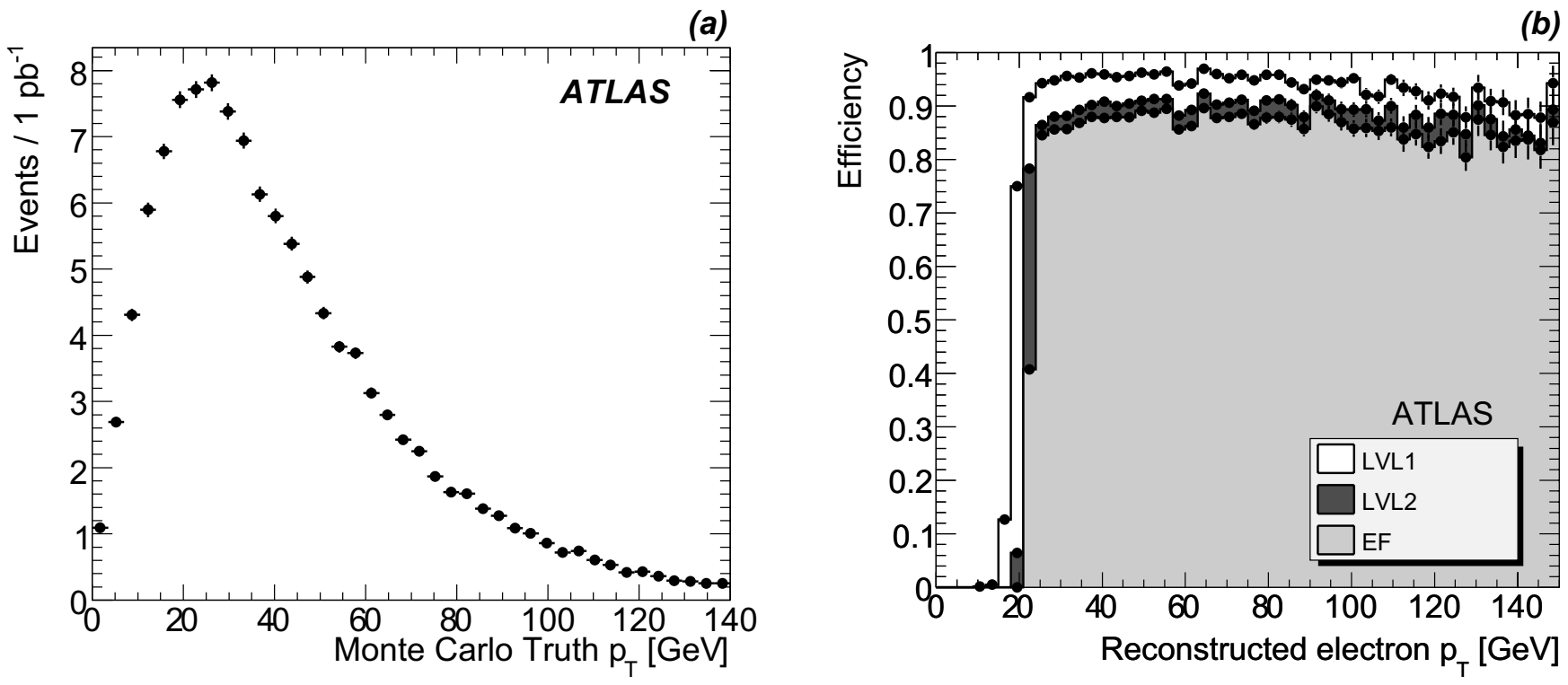


Figure 1: **(a):** $p_T$ spectrum of electrons from the decay W→e$\nu$ in leptonic $t\bar{t}$ events. The number of electrons is scaled to an integrated luminosity of 1 pb$^{-1}$, and is shown as a function of the true simulated electron $p_T$. **(b):** Turn-on curves for the e22i trigger determined with leptonic $t\bar{t}$ events. The efficiencies are shown with respect to an offline selection (excluding the cut on $p_T$) as a function of the offline reconstructed $p_T$ of the electron that fired the trigger.

Table 1: Efficiency of the e22i and e12i trigger for leptonic $t\bar{t}$ events with a W→e$\nu$ decay. The binomial errors $\Delta$ on the efficiencies for the different trigger levels are quoted for an integrated luminosity of 100 pb$^{-1}$, and are calculated as $\Delta^2 = \varepsilon(1-\varepsilon)/N$, where $\varepsilon$ is the efficiency. Note that no matching constraint is imposed between the trigger and the reconstructed or true simulated electrons. Moreover, the efficiencies are determined for $|\eta| < 2.5$ by cutting on the $\eta$ of the trigger as well as the reconstructed or simulated electrons.

| Trigger | Compared to Monte Carlo | Compared to offline selection |
|---|---|---|
| | Eff. [%] | Eff. [%] |
| **e22i:** | | |
| L1 EM18I | $74.7 \pm 0.5$ | $96.0 \pm 0.6$ |
| L2 e22i | $59.6 \pm 0.6$ | $92.7 \pm 0.9$ |
| EF e22i | $52.9 \pm 0.6$ | $89.8 \pm 1.0$ |
| **e12i:** | | |
| L1 EM7I | $83.6 \pm 0.4$ | $98.6 \pm 0.3$ |
| L2 e12i | $66.7 \pm 0.5$ | $92.6 \pm 0.8$ |
| EF e12i | $63.5 \pm 0.5$ | $91.8 \pm 0.8$ |

Table 1 shows the fraction of events passing each trigger level in the e22i chain per 100 pb$^{-1}$. The e22i trigger chain consists of the L1 trigger L1_EM18I, and the e22i trigger at L2 and EF. The efficiency has been calculated for each trigger level with respect to the total number of leptonic $t\bar{t}$ events with a W→e$\nu$ decay and with respect to the number of events reconstructed and selected in the *commissioning* analysis [5], requiring $|\eta| < 2.5$ for the trigger and reconstructed or simulated electrons. The cuts are:

- At least one reconstructed isolated electron with $p_T > 20$ GeV;
- $E_T^{miss} > 20$ GeV;
- At least 3 reconstructed jets with $p_T > 40$ GeV and at least 4 reconstructed jets with $p_T > 20$ GeV.

Also shown in Table 1 are the percentage of events passing each trigger level in the e12i chain.

Compared to e22i, a larger fraction of events passes this looser chain at each level, as expected. Note that the large value of the trigger efficiency with respect to the selection, exceeding 90%, is partly due to jets fulfilling the electron trigger. Eventually this effect has to be accounted for when determining trigger efficiency corrections to be used in an analysis.

Figure 1 (b) shows the turn-on curves for each trigger level in the e22i chain, determined from simulated leptonic $t\bar{t}$ events. The turn-on behaviour of the trigger is sharp at all levels. From the L1 to the HLT the $p_T$ dependence is very similar, the curves reveal a flat plateau once beyond the turn-on region. Note that the slight decrease in efficiencies at large $p_T$ values are due to the isolation requirements of the e22i trigger. This small loss can be recovered using a high-threshold non-isolated trigger, although it has not been considered in this study.

In summary, the single electron trigger efficiency in $t\bar{t}$ events is high for $t\bar{t}$ events with a $W \rightarrow e\nu$ decay in the current physics trigger menus. The electron trigger chain will be, together with the muon chain discussed in the next section, the main trigger for selecting $t\bar{t}$ events in the *golden* leptonic decay channel. Simple and efficient high-$p_T$ isolated single electron triggers exist to select $t\bar{t}$ events, avoiding considerable complications in the determination of trigger efficiencies for complex combined trigger signatures.

## 2.2 Muon triggers in $t\bar{t}$ events

### 2.2.1 Introduction

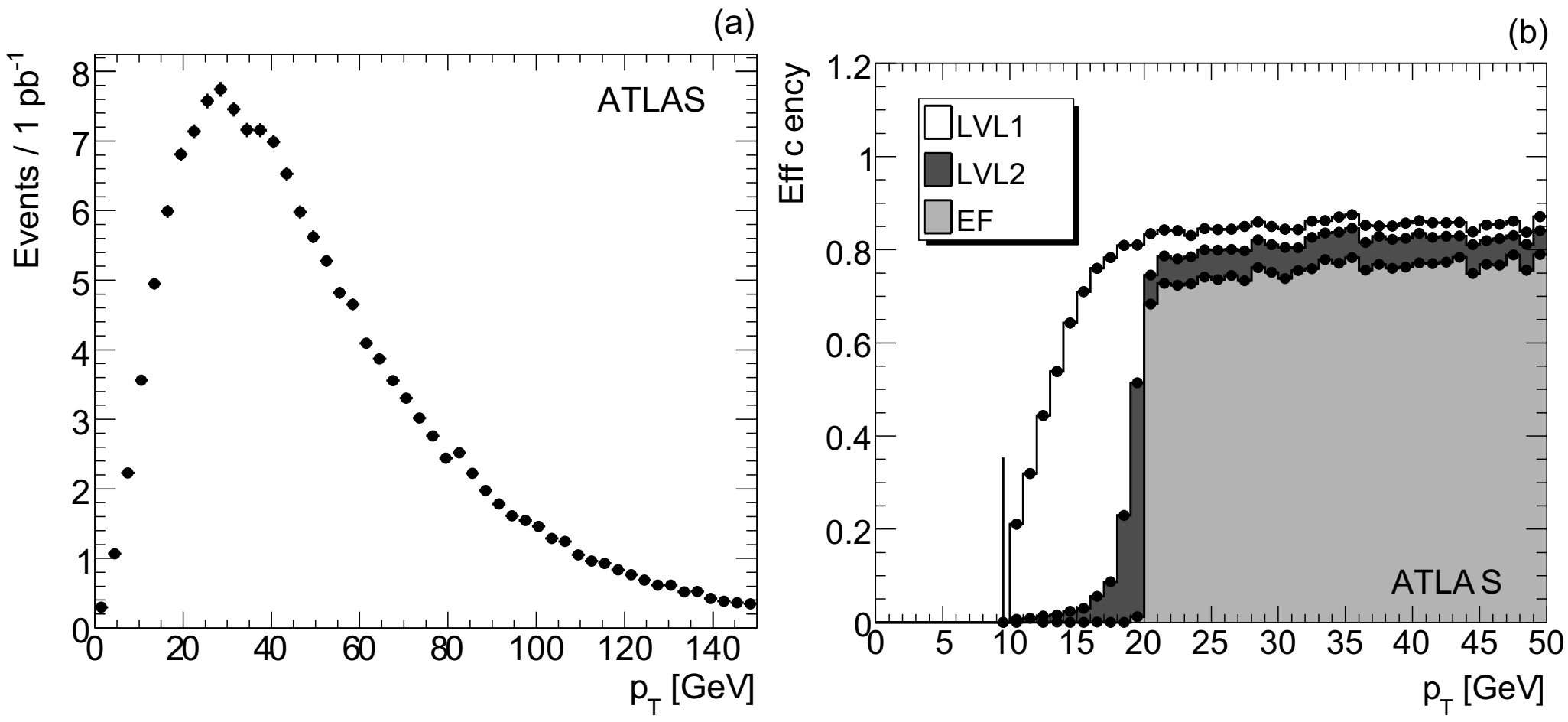


Figure 2: **(a):** $p_T$ spectrum of muons from $W \rightarrow \mu\nu$ decays in leptonic $t\bar{t}$ events. The muon $p_T$ is the true simulated value. The distribution is scaled to correspond to an integrated luminosity of 1 pb$^{-1}$. **(b):** Trigger efficiencies for the mu20 with respect to an offline selection (excluding the cut on $p_T$), given as a function of reconstructed muon $p_T$.

The $p_T$ spectrum of truth muons from the $W \rightarrow \mu\nu$ decay in $t\bar{t}$ events is shown in Fig. 2 (a). As for electrons, rather high-$p_T$ single muon triggers can be used to select $t\bar{t}$ events with at least one W-boson decaying to a $\mu$ and $\nu_\mu$ without much loss of efficiency.

### 2.2.2 Single muon triggers in ATLAS

The L1 muon trigger consists of fast electronics establishing coincidences between hits of different detector layers of the muon system inside programmed geometrical windows. The size of the window defines

the transverse momentum interval corresponding to the deflection of the muon in the toroidal magnetic field. One of six programmable $p_T$ thresholds is assigned to the candidate. The L2 processing consists of three reconstruction steps applied to full granularity data of the region defined by L1. First, the muon candidate is reconstructed in the muon spectrometer. Then inner detector tracks are reconstructed around the muon candidate. Both are combined to form the L2 muons upon which the trigger decision is based. Currently only the $p_T$ of the muon candidate is checked. Isolation requirements or constraints on the $\chi^2$ of the combined inner detector and muon spectrometer track are not imposed here (but might be in future). The muon reconstruction in the Event Filter is done using offline algorithms. The EF muon trigger decision at the moment is also based solely on the $p_T$ of the reconstructed muon candidate. A more complete description of the muon trigger can be found in [4].

### 2.2.3 Single muon trigger efficiencies

Using the standard L1 muon thresholds, six different trigger items have been defined in the simulated samples under study; L1_MU06, L1_MU08, L1_MU10, L1_MU11, L1_MU20 and L1_MU40. The efficiencies for each trigger item are shown in Fig. 3 (a). Each efficiency is calculated with respect to the number of simulated leptonic $t\bar{t}$ events with a W$\rightarrow \mu\nu$ decay. It can be seen that the high-$p_T$ L1 single muon triggers provide a high efficiency for selecting the events. When running at an increased luminosity of $10^{34}$ cm$^{-2}$ s$^{-1}$, the threshold could be raised to 40 GeV without an unacceptable loss in efficiency.

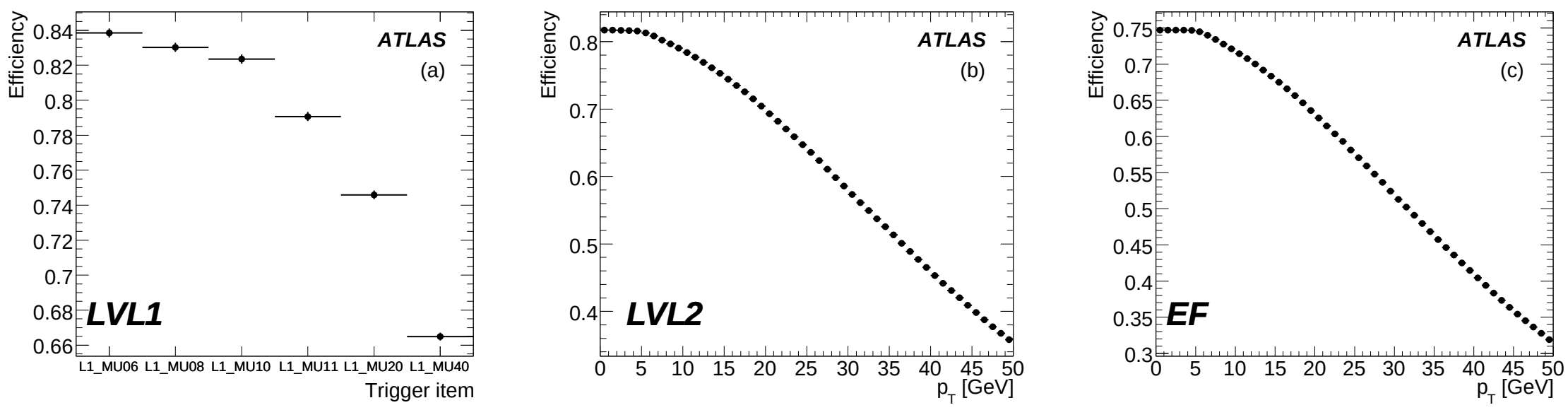


Figure 3: Efficiencies of the single muon triggers for L1 (a), L2 (b), and EF (c), determined using leptonic $t\bar{t}$ events with a W$\rightarrow \mu\nu$ decay. For the L1 items, the efficiencies are calculated for the six available thresholds. The HLT efficiencies are plotted as a function of the $p_T$ cut applied. In Fig. (b) the L2 chain is started by L1 RoIs passing the MU06 threshold and in Fig. (c) the EF chain is started by muon candidates passing the mu06 signature at L2. Note that the efficiencies are calculated with respect to the number of simulated muons within $|\eta| < 2.4$.

Figure 3 (b) shows the absolute efficiency of the L2 single muon trigger in the case of the HLT chain being started by a MU06 RoI. The $p_T$ cut applied by the L2 algorithm is shown on the horizontal axis. Similar efficiencies are shown in Fig. 3 (c) for the EF when the EF chain is started by a L2 muon candidate fulfilling the mu06 signature. The absolute trigger efficiency for the HLT cannot in principle be inferred directly from these plots, as for instance the L2 mu20 processing will only be initiated by L1 RoIs which fulfill the L1_MU20 requirements. On the other hand, the efficiencies for each trigger level, normalized to the number of events with a true simulated muon from a W-boson decay within $|\eta| < 2.4$, have been calculated for a few trigger signatures and are shown in Table 2. Comparing Table 2 with Fig. 3, one can see that the differences are small. Hence these figures can be used to indicate the effects of a change in trigger threshold cuts.

The efficiencies of the single muon trigger signatures have also been evaluated for the events selected by the *commissioning* analysis [5]. The same cuts as described in the electron section 2.1.3 are applied.

Table 2: Efficiencies of the mu06, and mu20 trigger chains for leptonic $t\bar{t}$ events with at least one W$\rightarrow \mu\nu$ decay. The numbers are absolute trigger efficiencies, the reference offline selection is given in [5]. The binomial errors $\Delta$ are calculated as $\Delta^2 = \varepsilon(1-\varepsilon)/N$, where $\varepsilon$ is the efficiency, for an integrated luminosity of 946pb$^{-1}$, which corresponds to the available Monte Carlo statistics. Note that the efficiencies are determined for $\eta < |2.4|$ by cutting on the $\eta$ of the trigger as well as the reconstructed or true simulated muons.

| Trigger | Compared to Monte Carlo Eff. [%] | Compared to offline selection Eff. [%] |
|---|---|---|
| **mu06:** | | |
| L1 | $83.8 \pm 0.3$ | $91.9 \pm 0.4$ |
| L2 | $80.2 \pm 0.3$ | $88.7 \pm 0.4$ |
| EF | $73.1 \pm 0.2$ | $83.1 \pm 0.4$ |
| **mu20:** | | |
| L1 | $74.6 \pm 0.2$ | $86.4 \pm 0.4$ |
| L2 | $66.3 \pm 0.2$ | $82.3 \pm 0.4$ |
| EF | $58.8 \pm 0.2$ | $76.6 \pm 0.4$ |

The results are given in Table 2 for the mu20 trigger chain, which consists of the L1 trigger L1_MU20, and the mu20 trigger at L2 and EF. The table shows that the mu20 signature is effective in selecting events for the commissioning analysis.

One can also determine the turn-on curve of the trigger efficiency for single muons in $t\bar{t}$ events. For that purpose a L1 muon is matched to a reconstructed muon by requiring $\Delta R < 0.15$ while L2 and EF trigger muons are matched by requiring $\Delta R < 0.1$[2]. Figure 2 (b) shows the mu20 trigger chain efficiencies for reconstructed muons as a function of their reconstructed $p_T$. The muons are required to have $|\eta_\mu| < 2.4$, as this is the reach of the muon trigger chambers. The L1 turn-on is less steep than those for the higher trigger levels, due to the coarser $p_T$ threshold assignment at L1 than at the HLT. The thresholds at the different levels are set such as to reach the plateau at the same value of the offline reconstructed $p_T$. At all trigger levels, once the plateau is reached, the efficiencies remain flat versus $p_T$.

In summary, as for the electrons, the single muon trigger efficiency in $t\bar{t}$ events is reasonably large. The mu20 is the most relevant muon trigger for selecting $t\bar{t}$ events with at least one W-boson decaying to a muon.

## 2.3 Jet triggers in $t\bar{t}$ events

Jet triggers are dominated by QCD multi-jet events, which have production cross-sections orders of magnitudes larger than the top signal processes. Therefore it is not obvious that multi-jet triggers giving acceptable rates will correspond to $p_T$ thresholds efficient for selecting top events. In order to characterize jet distributions in $t\bar{t}$ events, Fig. 4 (a) shows the $p_T$ distribution of the six highest-$p_T$ jets (ordered in $p_T$ and defined using a cone jet algorithm with radius 0.4) for hadronic $t\bar{t}$ events as obtained after a full event reconstruction. The characteristics of top events are a large number of high-$p_T$ jets. This is expected to provide some discriminatory power against the background, which has a steeply falling $p_T$ spectrum.

In Fig. 4 (b), turn-on curves are given for different $p_T$ thresholds at L1, for leptonic $t\bar{t}$ events. The plot is obtained by matching a jet triggered at L1 with a corresponding fully reconstructed jet within

[2]The $\Delta R$ cuts were chosen after studying the $\Delta R$ distributions for the different trigger levels.

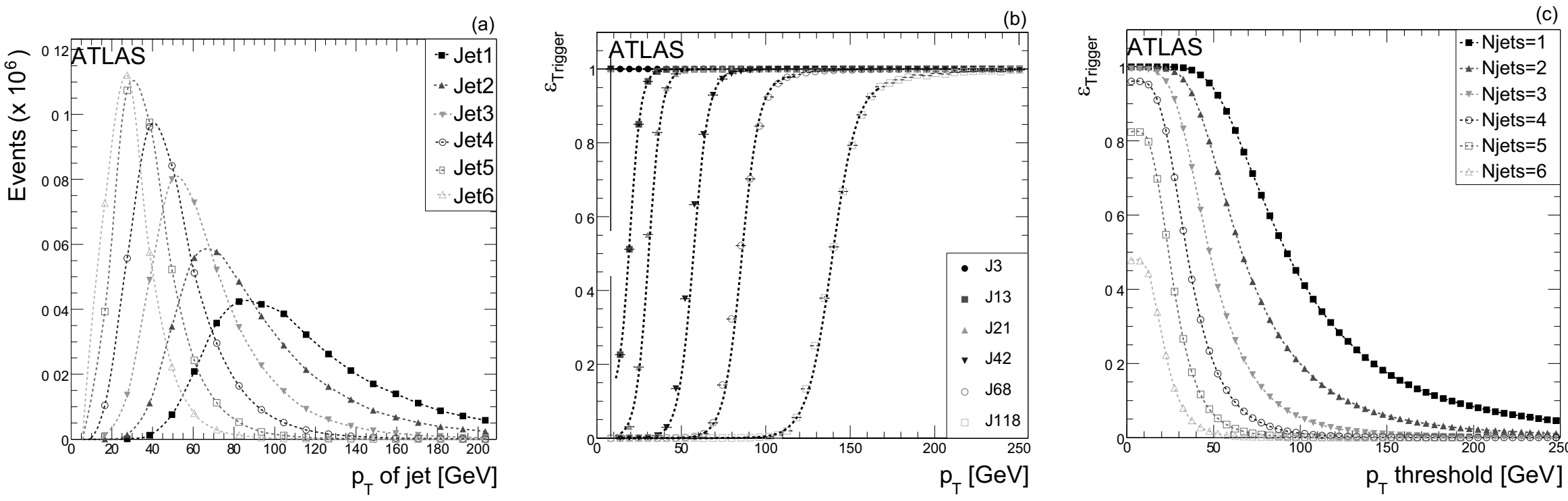


Figure 4: **(a):** Reconstructed $p_T$ distribution of the six leading jets for hadronic $t\bar{t}$ events. **(b):** Turn-on curves relative to the reconstructed jet $p_T$ for jets triggered at L1 at various threshold values for leptonic $t\bar{t}$ events. **(c):** L1 trigger efficiency versus trigger threshold for multi-jet triggers for hadronic $t\bar{t}$ events. The efficiency is calculated with respect to all events in the sample.

$\Delta R < 0.2$. The curves reveal, compared to the lepton triggers, a rather slow turn-on of the L1 jet triggers, resulting from coarse resolution at L1 (cf. [6]).

Table 3: Jet trigger efficiencies for $t\bar{t}$ signal and trigger rates, the latter also for the most important background events (the background samples are described in [3]). The efficiencies are for the whole trigger decision (after the EF), and are calculated with respect to the total number of events. The order of magnitude of the trigger rates are given for a luminosity of $10^{31}$ cm$^{-2}$ s$^{-1}$.

| Process | j20 | | j160 | | 4j50 | |
|---|---|---|---|---|---|---|
| | Eff. [%] | Rate | Eff. [%] | Rate | Eff. [%] | Rate |
| Leptonic $t\bar{t}$ | 99.8 | O($10^{-3}$ Hz) | 11.2 | O($10^{-4}$ Hz) | 9.9 | O($10^{-4}$ Hz) |
| Hadronic $t\bar{t}$ | 100.0 | O($10^{-3}$ Hz) | 12.9 | O($10^{-4}$ Hz) | 21.0 | O($10^{-4}$ Hz) |
| QCD | - | O($10^{3}$ Hz) | - | O(Hz) | - | O($10^{-1}$ Hz) |
| W-boson+Jet | - | O($10^{-3}$ Hz) | - | O($10^{-4}$ Hz) | - | O($10^{-4}$ Hz) |

Figure 4 (c) shows the L1 trigger efficiency for different jet multiplicities as a function of the trigger $p_T$ threshold for hadronic $t\bar{t}$ events. Single jet triggers will be more than 90% efficient for thresholds up to about 60 GeV, and roughly 45% efficient at 100 GeV. Table 3 summarises trigger efficiencies for signal and background rates for a selection of single and multi-jet triggers. Depending on how high the jet trigger thresholds are set, the signal efficiency varies between 10 and 100%. Due to large QCD rates, only high single-jet thresholds or carefully optimised multi-jet triggers are affordable. For example, for the j160, a rate of O(Hz) is to be expected from QCD multi-jets already for an instantaneous luminosity of $10^{31}$cm$^{-2}$s$^{-1}$.

### 2.3.1 Multi-jet triggers for the fully-hadronic $t\bar{t}$ channel

The multi-jet trigger item 4J50 listed in the previous section is intended to be general purpose, applicable to a wide range of physics channels. In order to optimize the signal efficiency of the jet triggers, especially for the fully-hadronic $t\bar{t}$ decays not accessible with lepton triggers, other trigger combinations for multi-jet triggers are studied. The aim is to identify optimum trigger combinations that reduce QCD background rates, while keeping a sizable signal acceptance in the fully-hadronic $t\bar{t}$ decay channel.

Table 4: Set of optimum multi-jet trigger combinations for fully-hadronic $t\bar{t}$ events. The QCD multi-jet background rates are given relative to the background rate of the 4j50 trigger (cf. Table 3). Efficiencies and rates at the EF are calculated with respect to the total number of events. Note that the names of the triggers are inclusive, "4j60_2j100_j170" means 4 jets have to pass the 60 GeV trigger, 2 jets the 100 GeV, and another 1 jet has to pass the 170 GeV trigger.

| Trigger | Signal Efficiency [%] | Relative Background Rate | S/B |
|---|---|---|---|
| 4j60_2j100_j170 | 6 | 0.13 | $2.8 \cdot 10^{-3}$ |
| 5j45_2j60_j100 | 16 | 0.34 | $3.0 \cdot 10^{-3}$ |
| 6j35_5j45_4j50_3j60 | 10 | 0.18 | $3.7 \cdot 10^{-3}$ |

Therefore, trigger combinations requiring at least 4, 5 or 6 jets for a given set of jet thresholds, were tested.

Table 4 summarizes signal efficiencies and increased background suppression for a set of trigger combinations. As figure of merit, the ratio of the trigger efficiency for the signal and QCD events is used. A strong suppression of QCD is observed when tightening the cuts on the jet energies, as expected. The signal-to-background ratio of the 4J50 trigger (cf. Table 3) is improved by a factor of two in the 5J45_2J60_J100 trigger (cf. Table 4) for roughly the same signal efficiency. Overall, the three trigger combinations given in Table 4 are optimal in terms of the figure of merit, and the 5 jet trigger combination leads to the best signal efficiency with the lowest background rate for fully-hadronic $t\bar{t}$ events. It is worth pointing out, that the reliable determination of trigger efficiencies from data is another challenging aspect of multi-jet triggers.

Large uncertainties are inherent in predictions for LHC energies, hence the QCD background trigger rates given here and the jet trigger definitions are preliminary and subject to tuning as soon as data taking starts. This is especially true for the effect of pile-up, which also impacts the signal trigger efficiencies. It is shown nevertheless that it should be possible to trigger on fully-hadronic $t\bar{t}$ decays with reasonable efficiency by optimising the choice of thresholds and multiplicity. Further HLT studies are underway to fully exploit also advanced methods for background suppression, such as multi-variate analysis techniques.

## 2.4 Missing $E_T$ triggers in $t\bar{t}$ events

The L1 $E_T^{miss}$ trigger performance in $t\bar{t}$ events is presented in this section. The HLT performance is not discussed due to ongoing development effort at the time of this writing.

The L1 energy triggers calculate missing transverse energy ($E_T^{miss}$, trigger item name: XE) based on reduced granularity calorimeter data (the trigger towers) without taking muons into account, across an $\eta$ range of $|\eta| < 5.0$ [4].

Figure 5 shows the $E_T^{miss}$ spectra at L1 for $t\bar{t}$ signal and background events normalized to an integrated luminosity of 1 $\text{pb}^{-1}$. Shown is the QCD background for which $E_T^{miss}$ is mainly faked by jet response fluctuations and losses in non-instrumented regions of the calorimeters. In the W-boson+jets background, a neutrino from the W-boson decay is present and produces real $E_T^{miss}$. Single top quark events are also shown, with a combined rate of about one third of $t\bar{t}$. As can be seen from the figure, high QCD rates will not permit inclusive $E_T^{miss}$ triggers with low thresholds, hence L1 trigger efficiencies for $t\bar{t}$ events will be small at larger luminosities. While for example L1_XE30 has an efficiency of 81% for leptonic events (applying no selection cuts in the efficiency calculation), the efficiency decreases to 19% for L1_XE100. For comparison, the lowest un-prescaled $E_T^{miss}$ threshold for the early running at a luminosity of $10^{31}\ \text{cm}^{-2}\ \text{s}^{-1}$ is expected to be L1_XE70 [4].

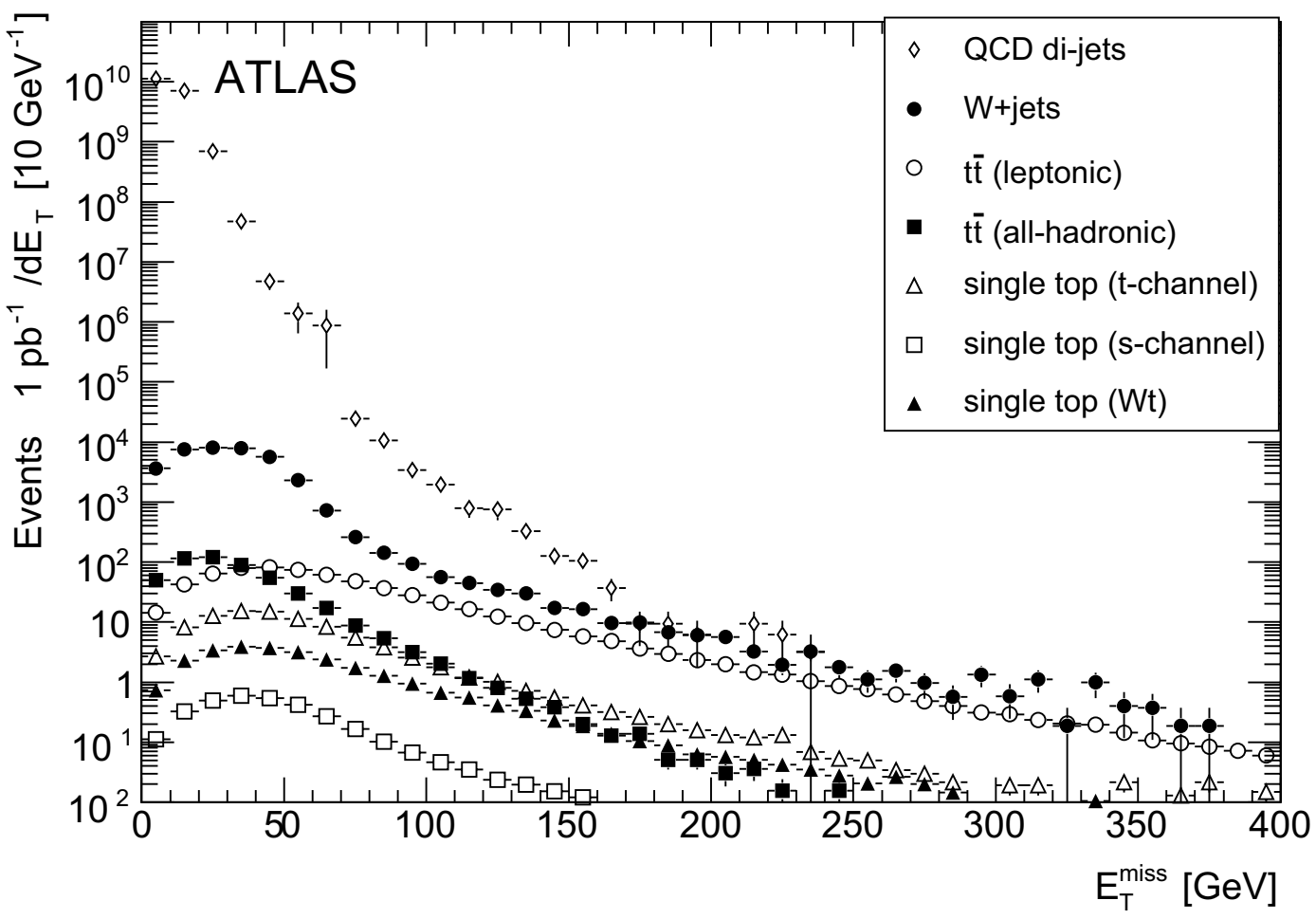


Figure 5: $E_T^{miss}$ spectra as measured by the L1 trigger for leptonic and fully-hadronic $t\bar{t}$ signals and the most significant backgrounds to $t\bar{t}$. The samples are normalized to an integrated luminosity of 1 $pb^{-1}$. Note that the *W + jets* samples used here include W-boson decays to electrons, muons, and taus, as described in Table 2 of [3].

### 2.5 Single-top trigger

The final state of leptonic single-top events in the s, t, and Wt channels are characterized by one high-$p_T$ muon or electron, $E_T^{miss}$, and by one to three jets. Since these leptons are among the high-$p_T$ products of a W-boson decay, a relatively large lepton $p_T$ threshold can be used to select the single-top events at the trigger level. Therefore, the most important triggers for selecting single-top events are the high-$p_T$ muon and electron triggers. Multi-jet and $E_T^{miss}$ triggers can be used in combination with lower threshold electron and muon triggers in order to enhance acceptance.

The turn-on curves for the mu20 and e22i trigger are shown in Fig. 6. The trigger efficiencies are shown with respect to the single-top selection (the cut-based one listed in detail in [7]), and are very similar to the ones found in $t\bar{t}$ events shown in the previous sections. Both triggers exhibit a sharp turn-on behaviour and a flat plateau up to 100 GeV.

The single-top event selection will include the trigger combination e22i OR e55 OR mu20. The efficiency for this combination, as well as the total trigger efficiencies for the individual items in the muon and electron channels, are shown in Table 5. The efficiencies are found to be high, above 80% in the muon channel, and up to 90% in the electron channel.

## 3 Trigger redundancy in top quark events

Due to the rich event topologies, top quark events satisfy many trigger items simultaneously. This redundancy may be exploited to enhance signal over background rates by forming combined trigger items and to monitor the trigger by providing information on trigger efficiencies from data.

Combining suitable trigger items may reduce significantly the background rate. For example, in a combined jet and $E_T^{miss}$ trigger, the $E_T^{miss}$ requirement will suppress the QCD background. For the same allocated bandwidth, the jet $p_T$ threshold of the combined trigger can be lowered compared to the inclusive single jet trigger. Overall this will increase the number of recorded leptonic $t\bar{t}$ events which

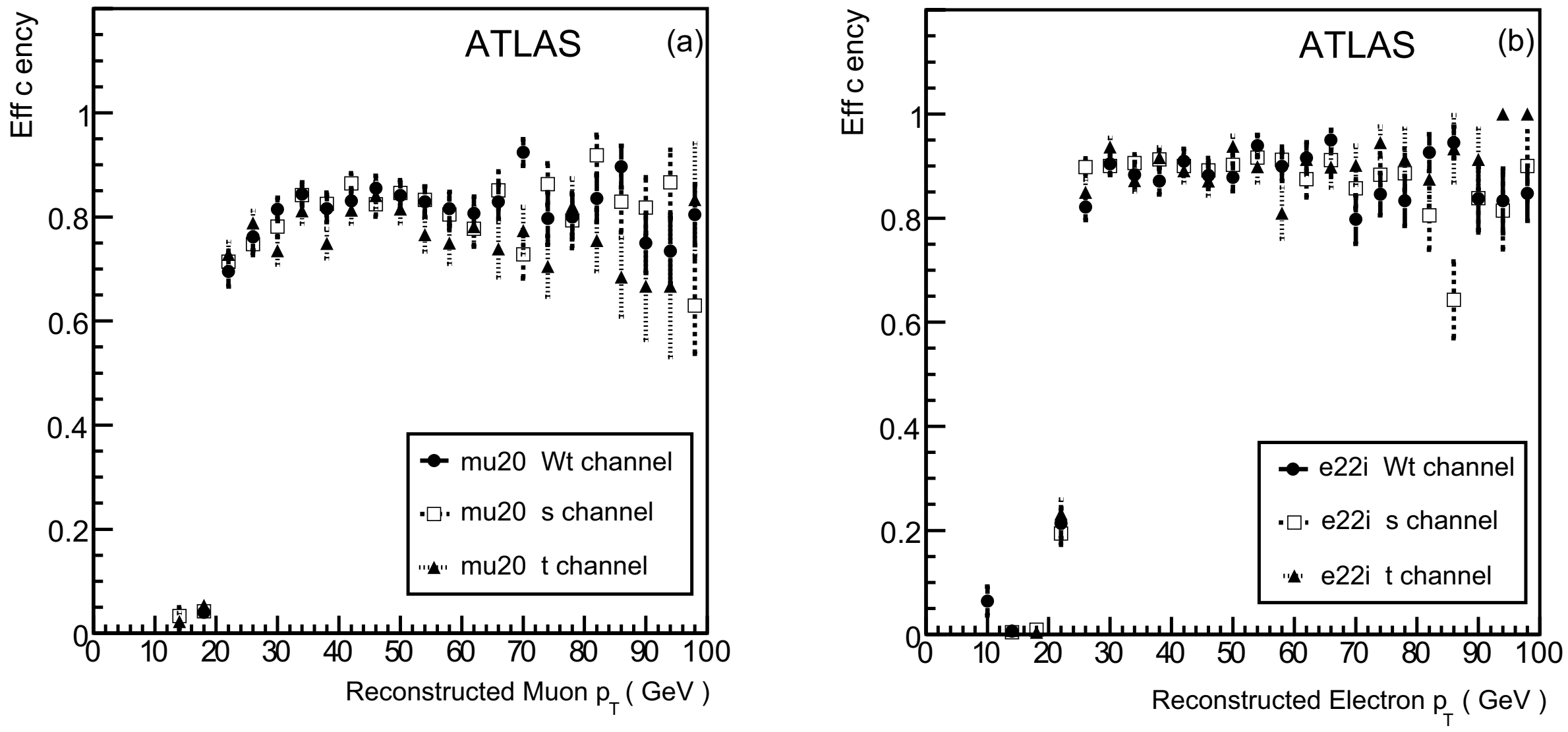


Figure 6: Turn-on curves are shown for the mu20 (a) and the e22i (b) trigger. In both plots, the circles represent Wt-channel single-top, the squares represent s-channel single-top, and the triangles represent t-channel single-top events.

have good acceptance for a moderate threshold on $E_T^{miss}$ .

On the other hand, if there is sufficient overlap between two triggers, the efficiency for one of the triggers can be measured by selecting offline a clean sample of $t\bar{t}$ events which satisfied the other trigger. For example, the lepton triggers used to select leptonic $t\bar{t}$ events could be monitored in events that have been triggered by the combined jet and $E_T^{miss}$ trigger. Possible biases due to correlation should be checked and eventually corrected for.

### 3.1 Trigger item overlap in $t\bar{t}$ events

The correlation among different trigger items is given in Fig. 7, with plot (a) showing overlaps for fully-hadronic $t\bar{t}$ events and plot (b) showing overlaps for leptonic $t\bar{t}$ events. The plots show the percentage of events triggered by the item on the *x*-axis that were also triggered by the item on the *y*-axis. The trigger items on the axes were representative of the ATLAS trigger menu at the time of this writing.

As expected, there is considerable overlap for the leptonic $t\bar{t}$ events, and the event topology of these events is clearly illustrated. The leptonic triggers have very low acceptance in the fully-hadronic $t\bar{t}$ events. Only the very low-$p_T$ muon triggers, which will be prescaled [4], have considerable efficiency due to muons from b-jets. Any combination of trigger items used to enhance the signal-to-background ratio in the fully-hadronic sample will have to rely on the jet triggers (cf. Section 2.3.1) only. For the leptonic channel, trigger combinations involving $E_T^{miss}$ are potentially useful. Studies done at L1 have shown that $E_T^{miss}$ trigger items indeed overlap considerably with the leptonic and jet triggers in leptonic $t\bar{t}$ events.

## 4 Determination of trigger efficiencies from data and application to top quark events

Trigger efficiencies should be determined from data in order to reduce systematic uncertainties, as the shape as a function of $p_T$ and the absolute value of trigger efficiencies are hard to describe precisely in

Table 5: Total trigger efficiencies for the three single-top channels. Efficiencies for the muon (W→ $\mu\nu$) and electron (W→e$\nu$) channels are shown separately. The errors are calculated using the full available Monte Carlo statistics. Note that the reference event selection [7] selects one and only one isolated lepton for the electron and muon channel. Hence the trigger effciency in the Wt channel does not increase when taking the OR of electron and muon triggers as compared to the single electron or muon triggers.

| Sample | Muon Channel | | Electron Channel | |
|---|---|---|---|---|
| | Trigger | Efficiency (%) | Trigger | Efficiency (%) |
| Wt | mu06 | 88.4 ± 0.6 | e22i | 87.1 ± 0.7 |
| | mu20 | 82.5 ± 0.7 | e22i OR e55 | 90.6 ± 0.6 |
| | e22i OR e55 OR mu20 | 82.5 ± 0.7 | e22i OR e55 OR mu20 | 90.8 ± 0.6 |
| s | mu06 | 88.0 ± 0.6 | e22i | 89.2 ± 0.7 |
| | mu20 | 82.6 ± 0.7 | e22i OR e55 | 90.7 ± 0.6 |
| | e22i OR e55 OR mu20 | 82.6 ± 0.7 | e22i OR e55 OR mu20 | 91.0 ± 0.6 |
| t | mu06 | 86.0 ± 0.7 | e22i | 89.5 ± 0.7 |
| | mu20 | 79.6 ± 0.8 | e22i OR e55 | 90.6 ± 0.7 |
| | e22i OR e55 OR mu20 | 79.6 ± 0.8 | e22i OR e55 OR mu20 | 90.9 ± 0.7 |

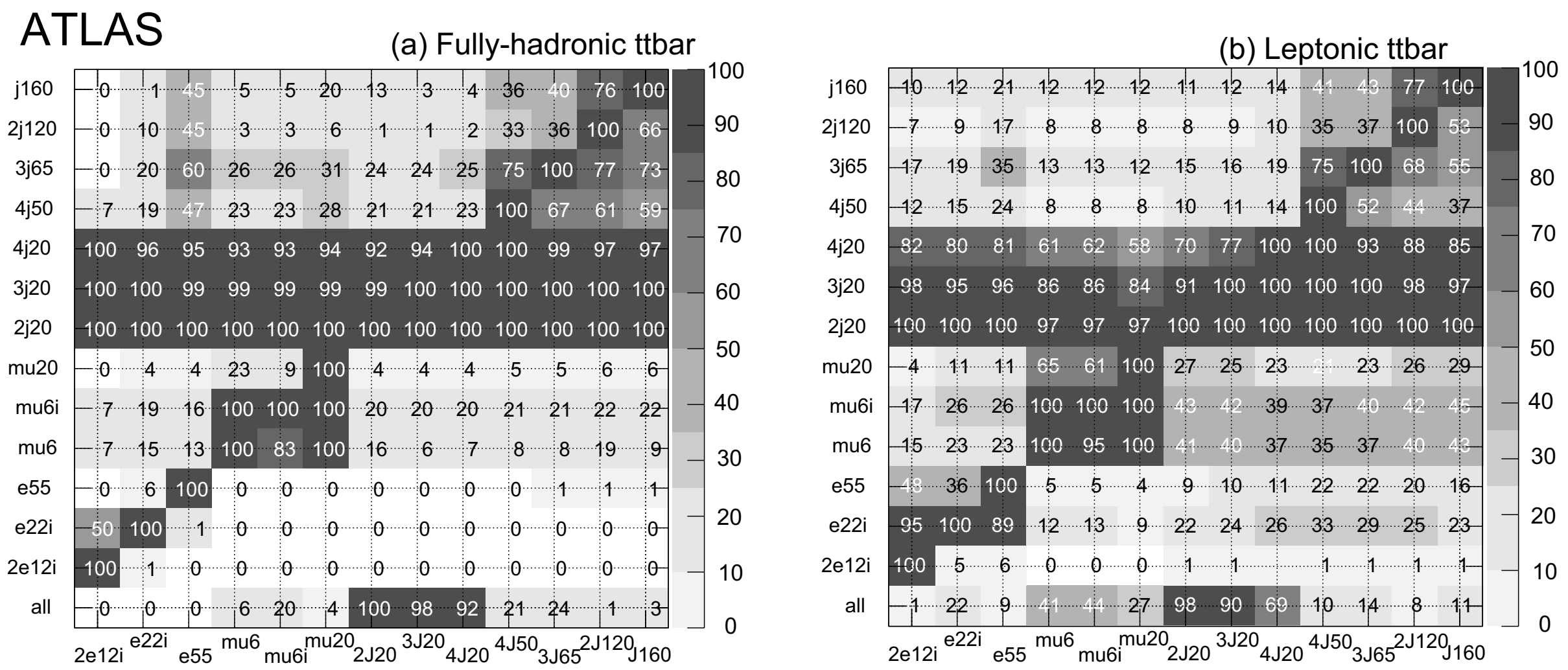


Figure 7: Trigger item overlap at the HLT for fully-hadronic (a) and leptonic (b) $t\bar{t}$ events. The numbers in the plots indicate the percentage of events triggered by the item on the $x$-axis also triggered by the item on the $y$-axis. Note that, by definition, the values on the diagonal have full acceptance. The bottom row in each plot gives the total efficiency of the corresponding item on the $x$-axis.

simulations.

A methodology is tested to determine the electron trigger efficiency for top events in data. Events with Z-boson decays are used to parameterize the electron trigger efficiency as a function of $p_T$ and $\eta$ by means of the *tag and probe* method [8]. These parametrized efficiencies yield correction factors which can then be applied to top events.

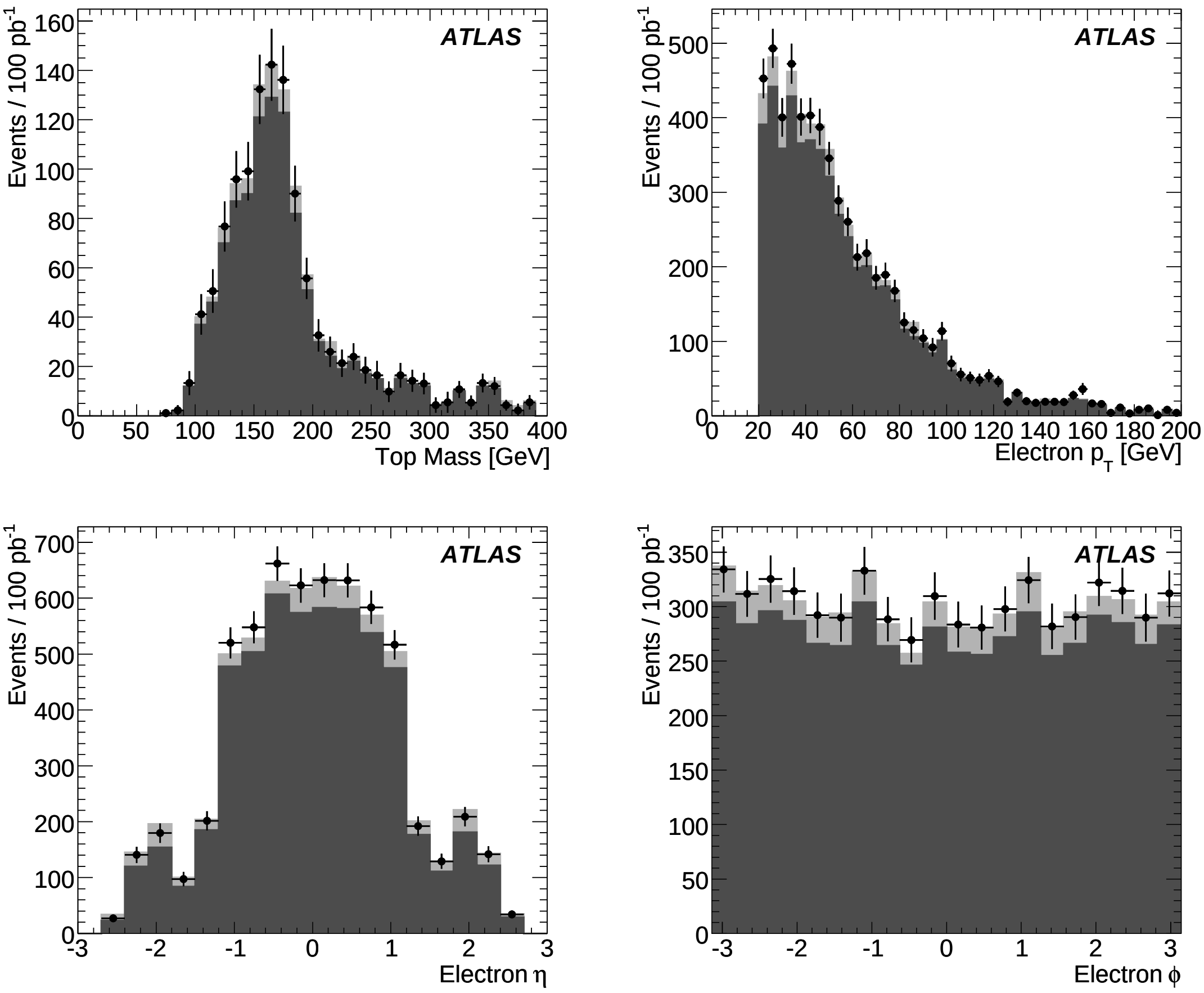


Figure 8: Plots of the reconstructed top mass, reconstructed electron $p_{\mathrm{T}}$, $\eta$ and $\phi$. The darker histogram is the reconstructed value, the black points include the correction from Z→ee. The lighter histogram is from simulation where no trigger selection is applied.

The Standard Model Z-boson decaying to two leptons is relatively free of backgrounds at the LHC, and therefore provides a clean source of events for the *tag and probe* method. This method relies on the ability to obtain a clean sample of Z-boson events using selection cuts after obtaining the initial sample using a single electron trigger. From the di-electron final state one can *tag* the electron that fired the trigger, and *probe* the other in order to measure the trigger efficiency. Once this efficiency is extracted it can be applied to other processes which have been obtained using the same trigger, in this case top events.

Having selected top events using the commissioning selection [5], a correction for the trigger efficiency is applied to recover the number of events that would be measured by an ideal detector. For each event, a weight is applied according to the $p_{\mathrm{T}}$ and the position in $\eta$ of the electron in the event. The weight is taken as the inverse efficiency extracted from look-up tables, which have been obtained with the *tag-and-probe* method.

Figure 8 shows the $p_{\mathrm{T}}$ $\eta$, and $\phi$ distributions of the electron in the $t\bar{t}$ event. The dark histogram represents events that have passed the trigger selection, the black points show the events after the correction is applied. The lighter histogram is from a sample of events run with the trigger disabled. Also shown in the figure is the top quark mass as measured using the above selection, after a correction for the trigger efficiency. It can be seen from the figure that the efficiency correction works well, the effect of the trigger

can be recovered to reproduce the simulated distributions. The effect of additional jet activity in $t\bar{t}$ events as compared to Z→ee events is thus indicated not to have a big effect at the current level of precision.

The final value of the trigger efficiency, used as a correction to the cross section measurement of the leptonic $t\bar{t}$ decay channel with one electron in the final state, is a value $\varepsilon_{trigger}$ of 92.6 ± 0.4 % for an e12i trigger chain and 91.5 ± 0.4 % for an e22i trigger chain for 100pb$^{-1}$ . These values are obtained by taking the ratio of the number of events before and after the trigger correction, and the uncertainty is statistical only. Note that to obtain these numbers, a matching between the trigger and the reconstruction electron candidates has to be done ($\Delta R < 0.2$), in contrast to the numbers in Table 1.

In conclusion, it is shown that the *tag-and-probe* efficiency can be successfully applied to leptonic top decays. The method appears robust, however, the issue of background events passing the Z→ee selection and the possible bias introduced thereby needs to be investigated.

## 5 Conclusions and outlook

The first studies of the expected ATLAS trigger performance for top events have been shown. All levels of the electron and muon triggers are found to be highly efficient for the *golden* leptonic $t\bar{t}$ decays. The jet triggers, especially important for hadronically decaying tops, are much more challenging to use due to the high rate of QCD jets. High jet multiplicities of 5 or 6 jets with optimised $p_{\mathrm{T}}$ thresholds may provide the best acceptance.

Top quark events at the LHC will have especially rich event topologies and will result in many trigger items being fulfilled simultaneously. The overlap of trigger items was investigated to identify combinations of trigger items with improved signal efficiency and trigger rate, as well as potential monitoring triggers. The use of the $E_T^{miss}$ trigger should be further investigated.

The trigger efficiency should be determined from data with the least possible bias. This must be validated with Monte-Carlo simulations, comparing the efficiencies derived with the data driven methods to the true efficiencies. Such a study, albeit not yet complete, was shown for the electron trigger efficiency using simulated $Z \to ee$ samples. By using the trigger efficiency correction to the reconstructed kinematical distributions, it was shown that the simulated top quark distributions are very well recovered.

# Jets from Light Quarks in $t\bar{t}$ Events

**Abstract**
This note describes the properties of jets from light quarks ($u$, $d$, $s$ and $c$) in $t\bar{t}$ events, with the goal of obtaining a good reconstruction of the hadronically decaying W-boson, which will be used for an in-situ jet energy scale measurement and will enter, together with jets from $b$ quarks, into the top quark reconstruction.

## 1 Introduction

The measurement of the top quark mass with a precision of 1 GeV is one of the goals of the LHC experiments. In order to achieve such a precision, several complementary methods can be used [1]. The most straightforward method, namely the measurement of the tri-jet invariant mass from the hadronically decaying top quark in the $t\bar{t} \rightarrow l\nu b\ qqb$ channel, can in principle reach such a precision, provided that the three jets are well reconstructed and calibrated.

This note describes some studies of the reconstruction and calibration of jets from light quarks ($u$, $d$, $s$ and $c$) in $t\bar{t}$ events, with the goal of obtaining the best possible reconstruction of the hadronically decaying W-boson. After a comparison of various jet algorithms, two methods of in-situ jet calibration using the W-boson mass distribution will be presented. Finally, the stability of the W-boson reconstruction with different gluon radiation settings in the Monte Carlo event generator will be shown.

Unless otherwise specified, the studies presented here have been performed using $t\bar{t}$ events generated with MC@NLO + Herwig, with the top quark mass equal to 175 GeV, at least one of the W-bosons decaying leptonically, and a full GEANT simulation of the ATLAS detector. The simulation statistics corresponds to a luminosity of about 0.9 fb$^{-1}$. By default, no pile-up is included in the simulation, except in a few dedicated studies. Triggering was not considered, as $t\bar{t} \rightarrow l\nu b\ qqb$ events will be triggered mainly by the leptonic decay of the second top quark, without biasing the jets on the hadronic side.

## 2 A comparison of jet reconstruction algorithms

The current default jet algorithm used for $t\bar{t}$ event reconstruction in ATLAS is the cone algorithm with a size $\Delta R = 0.4$. A small cone size was indeed found in previous studies to give the best mass resolution and signal over background ratio. This section presents new studies of the optimal jet algorithm, performed by looking at the jet energy, angular resolutions and at the W-boson mass reconstruction. For these studies, the cone and $k_T$ algorithms, with various parameters, were tried. Details on jet algorithms can be found in Ref. [2].

Cone algorithms define jets as the combination of input objects from generated stable particles or calorimeter energy deposits within a cone of radius R around the jet directions $\eta$ and $\phi$. In an iterative procedure, the jets are repeatedly reconstructed until a stable configuration is found. Two different implementations exist: the first one, referred to as seeded cone jet finder, used in this paper, uses high $E_T$ objects in the event as a starting point, whereas the second, seed-less implementation, is much slower but theoretically more accurate. In both scenarios, the jets obtained undergo a split-merge procedure, to define non-overlapping exclusive jets. The resulting jets then need to be calibrated as explained below.

The second class of jet algorithms consists of the $k_T$ algorithm, which reconstructs jets via a clustering procedure. Such jets do not necessarily have a cone-shape with a fixed radius. Instead, the algorithm clusters "nearest" protojets together, depending on their relative transverse momentum. Thus, it also ensures a unique association of input clusters to jets, without the need for split/merge procedures as in cone algorithms.

The $k_T$ algorithm introduces a distance measurement $\Delta R$ and a recombination scheme [3]. The distance measurement decides which pair of protojets should be merged and when to stop merging a protojet any further. Deciding whether to merge the two closest jets or flagging one of the protojets as a final jet is executed recursively in the algorithm, until there are no more mergeable protojets left. The recombination scheme determines how to combine two protojets into a new protojets. Here we use the E recombination scheme.

There are two modes of operation for the $k_T$ algorithm, the inclusive and exclusive mode. The inclusive $k_T$ algorithm is intended for defining inclusive jet cross sections, as they are typically used in hadron-hadron interactions. It has a parameter R, which controls the decision to merge protojets or declare them as final jets, and plays a role comparable to the cone size in cone jet algorithms.

The $k_T$ algorithm in its exclusive mode must account for the proton remnants and the underlying event in the p-p interaction for the calculation of exclusive cross sections. Its parameter $D_{cut}$ controls the scale which separates "beam jets" from the "hard jets" during the clustering. Moreover, the value of $D_{cut}$ can also be determined dynamically if one fixes the desired number N of final jets.

The basic input objects for both cone and $k_T$ jet algorithms are either calorimeter towers, defined as a group of cells in a fixed ($\Delta\eta$, $\Delta\phi$) grid, or topological clusters, defined as a group of cells formed around a seed cell [4]. These calorimeter towers or topological clusters are given at the electromagnetic scale. The jets obtained are then calibrated to the hadronic scale using different weighting schemes such as the H1-style [5], whose major drawback is that the set of corrections is specific for each type of jet finder and each value of its parameters.

The comparisons of jet algorithms were performed on $t\bar{t} \rightarrow l\nu bjjb$ events. In order to consider only jets relevant for the top quark mass measurement, an event selection was performed, requiring:

- a missing transverse energy $> 20\,\text{GeV}$ to account for the unmeasured neutrino.
- one isolated electron or muon, defined as described in [6], with $p_T > 20\,\text{GeV}$ and $|\eta| < 2.5$.
- at least 3 jets with $p_T > 40\,\text{GeV}$, a $4^{th}$ jet with $p_T > 20$ GeV, with $|\eta| < 2.5$.

## 2.1 Jet energy resolution and linearity

This section presents the energy and position measurement performance for various jet algorithms, calibrated with the H1 weighting scheme. Although we investigated, in this section, all the combinations of cone / inclusive $k_T$ algorithm, 2 jets sizes, made from towers / topological clusters, with / without pile-up, for the sake of clarity not all results are shown in the figures.

In the events passing the event selection cuts, the jets considered for the resolution studies are the two jets closest to the two quarks from the W-boson decaying hadronically, when their distance $\Delta R$ to the quark is less than 0.3. As the final goal is to improve the two jet invariant mass reconstruction, the jet energies (and directions) are hereafter compared directly to the ones of the associated quarks.

Figure 1 shows, for four jet algorithms, the distributions of the energy difference between the matched quark and jet, divided by the quark energy, for two different jet energy ranges. Larger jets lead, especially at low energies, to a worse resolution and to larger tails because the jet energies are overestimated when they overlap with other particles in the underlying event.

These energy distributions were fitted with a Gaussian distribution [1]. The width of the distribution, as a function of the energy, is shown on figure 2 and in Table 1. The resolutions obtained here are significantly worse than the detector resolution itself, as shown in Ref. [5], mainly because they include the fluctuations of the energy lost outside the jets, as relevant to determine the best algorithm for the

[1]To make the fit stable, a first Gaussian was fitted between the mean of the histogram $\pm$ 2 times the RMS. A second fit was then performed in the same way using the mean and width of the first fit.

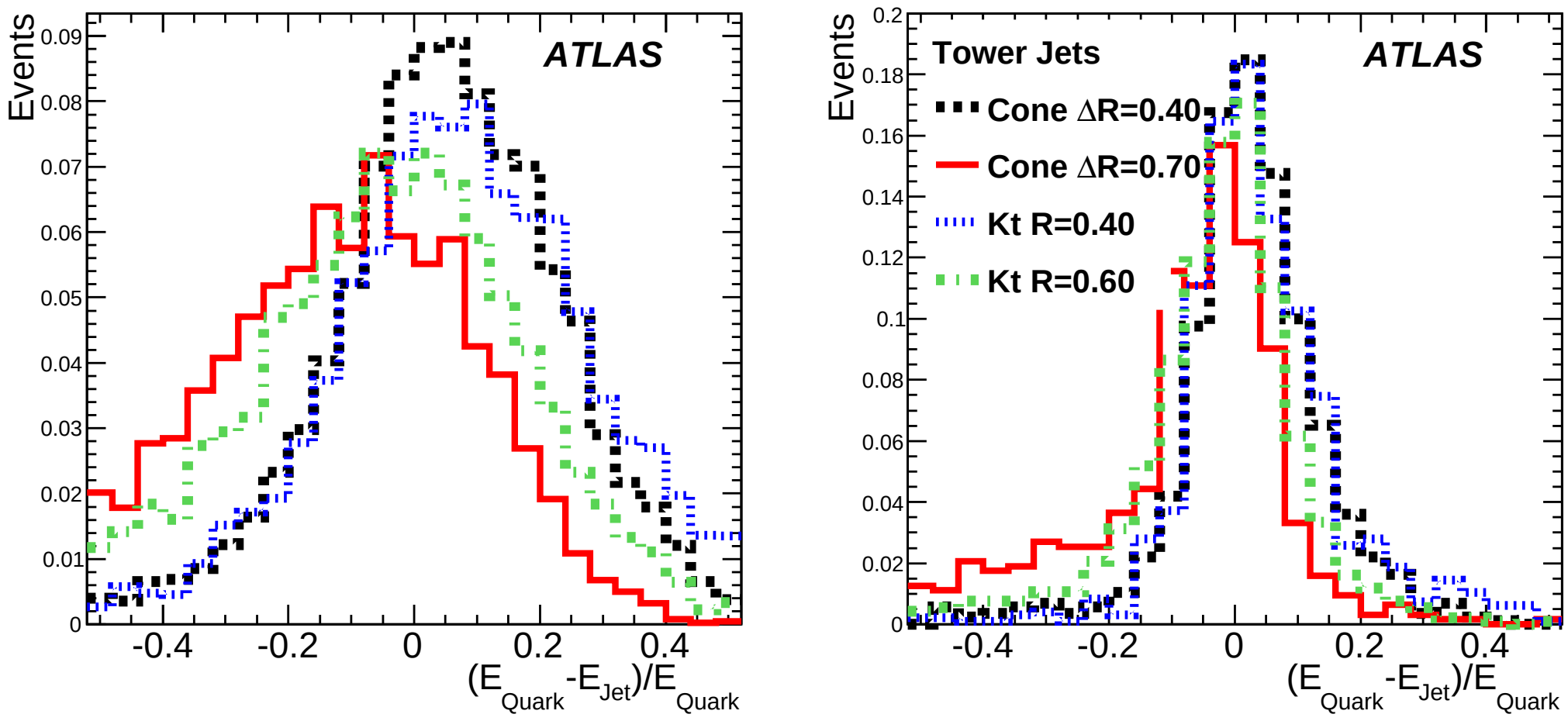


Figure 1: Distributions of ($E_{quark}$ - $E_{jet}$)/$E_{quark}$ for the jets from the W-boson decay, for two different quark energy ranges (left figure: quark energy between 15 and 50 GeV, right figure: quark energy above 350 GeV).

W-boson (and the top quark) mass reconstruction. To quantify this effect, Figure 3 compares these resolutions to the ones obtained by using the jets made from the Monte Carlo hadrons ("Truthjets"). The latter ones are typically 20% better, and comparable to the resolutions shown in [5].

Jet energy resolutions were also studied with events including the pile-up expected at a luminosity of $10^{33}$ cm$^{-2}$s$^{-1}$ (figure 4 and Table 1), showing that:

- When jets are made from towers, small sizes are preferable, especially for the cone algorithm. For example, Figure 4 shows that the energy resolutions of cone 0.4 jets made from towers is degraded by 30% to 50% in presence of pile-up.

- The algorithms using topological clusters seem to behave very well, even in the presence of pile-up, as can also be seen on this figure.

For cone 0.4 algorithms, the reconstructed jet energy is on average 4% lower than the quark energy, for quarks with transverse momenta above 40 GeV. This 4% miscalibration can *a priori* come from the detector calibration itself, from the energy lost outside the jet, or from the underlying event. Figure 5 shows that, in the current simulation, the detector is very well calibrated, to a precision better than 3% for all algorithms, and to better than 1% for the cone 0.4 algorithm.

On the other hand, Figure 6 shows that the out-of-cone energy is important, and accounts for the 4% energy shift observed with cone 0.4.

The difference between the quark and the reconstructed jet energies, for quark energies between 15 and 50 GeV, is shown in Figure 7 with and without pileup, for the cone 0.4 from topological clusters algorithm. For such a small jet size and a topological cluster noise suppression, the average energy is shifted by 2.5 %.

## 2.2 Jet angular resolution

Figure 8 shows, for four jet algorithms, the distributions of the distance $\Delta R$ between the quark and the matched jet, for the lowest and the highest energy bins. It can be seen that smaller jets have a

Table 1: Energy resolution (in %) as a function of the quark energy, for various jet algorithms, with and without pile-up.

| | Energy range [GeV] | | | | | |
|---|---|---|---|---|---|---|
| algorithm | 15 - 50 | 50 - 100 | 100 - 150 | 150 - 250 | 250 - 350 | > 350 |
| Cone ΔR = 0.4, Tower | 18.3 | 14.7 | 13.0 | 11.2 | 9.4 | 8.2 |
| Cone ΔR = 0.4, Topo | 18.5 | 14.8 | 12.6 | 11.0 | 9.5 | 8.3 |
| Cone ΔR = 0.7, Tower | 25.2 | 16.3 | 13.5 | 11.6 | 10.3 | 8.7 |
| Cone ΔR = 0.7, Topo | 20.4 | 14.3 | 12.5 | 11.2 | 9.3 | 8.6 |
| $k_T$ R = 0.4, Tower | 21.4 | 16.7 | 14.4 | 11.7 | 9.6 | 8.0 |
| $k_T$ R = 0.4, Topo | 20.4 | 16.2 | 13.1 | 10.9 | 9.5 | 8.2 |
| $k_T$ R = 0.6, Tower | 22.1 | 16.1 | 13.4 | 11.2 | 10.3 | 8.9 |
| $k_T$ R = 0.6, Topo | 19.3 | 14.8 | 12.6 | 10.6 | 9.8 | 8.2 |
| Cone ΔR = 0.4, Tower, pile-up | 27.8 | 20.0 | 16.1 | 13.6 | 11.1 | 8.8 |
| Cone ΔR = 0.4, Topo, pile-up | 20.9 | 15.8 | 13.1 | 12.2 | 10.6 | 8.5 |
| Cone ΔR = 0.7, Tower, pile-up | 41.7 | 26.8 | 20.4 | 19.1 | 15.8 | 13.4 |
| Cone ΔR = 0.7, Topo, pile-up | 34.2 | 19.4 | 14.6 | 13.2 | 11.8 | 9.6 |
| $k_T$ R = 0.4, Tower, pile-up | 30.4 | 23.7 | 18.3 | 15.5 | 12.8 | 10.3 |
| $k_T$ R = 0.4, Topo, pile-up | 24.4 | 17.3 | 14.6 | 12.9 | 10.9 | 9.3 |
| $k_T$ R = 0.6, Tower, pile-up | 33.3 | 24.0 | 19.6 | 17.0 | 14.8 | 11.6 |
| $k_T$ R = 0.6, Topo, pile-up | 26.1 | 18.3 | 14.4 | 12.8 | 11.6 | 9.3 |

more precise direction determination. This can also be seen from Figure 9 where the full width at half maximum (FWHM) of these distributions are plotted. It was checked that, as for the energy resolution, the same conclusion remains when pile-up is added, and that jets made from topological clusters are less sensitive to pile-up.

It is observed that, for large cone sizes and large transverse momenta, the distributions of the distance between the quark and the matched jet have larger tails (larger fraction of jets with a distance to the associated quark between 0.1 and 0.3), because of a higher overlap probability.

## 2.3 W-boson mass reconstruction

This section compares the W-boson mass reconstruction performances for the cone algorithms with sizes 0.4 and 0.7 and for the $k_T$ jet algorithms with different choices of parameters, both using topological clusters.

The choice of the jet algorithm parameters is crucial for the performance of the W-boson reconstruction. Indeed, the probability, averaged over the W-bosons selected in this section, for the two quarks from the W-boson decay to be at a distance $\Delta R$ smaller than 1.4 is 25%, leading to non-negligible overlap effects for jet sizes of the order of 0.7, as already seen in section 2.1, and eventually to the merging of the two jets coming from the light quarks into a single one. On the contrary, the probability for the 2 quarks from the W-boson decay to be at a distance smaller than 0.8 is only 4%.

The same event selection as the one described in section 2.1 is performed, apart for the requirement that the fourth jet satisfies $p_T > 40$ GeV, which is needed to improve the W-boson purity. On the selected events, the three jets whose four-momentum sum has the highest $p_T$ are chosen to belong to the hadronic top quark. Among these three jets, the di-jet combination whose mass is closest to the known value of the W-boson mass (80.4 GeV) is taken to represent the W-boson. The remaining third jet is assumed to be the *b* quark; no further b-tagging information is used in this study. The distribution for the reconstructed

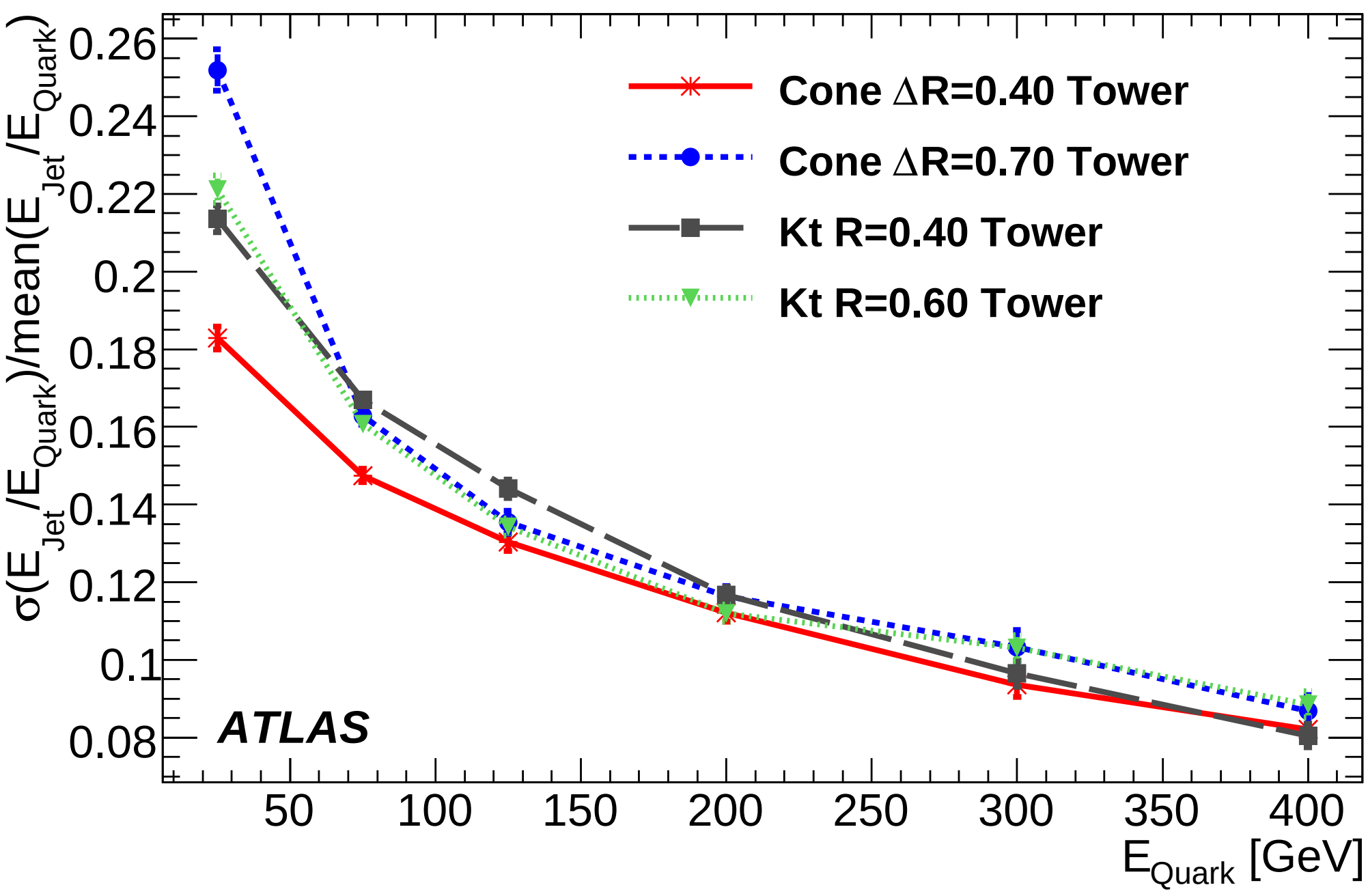


Figure 2: Energy resolution for jets from light quarks (integrated over all $\eta$ values), as a function of the quark energy, for various jet algorithms (events simulated without pileup).

W-boson mass is shown in Figure 10.

To obtain mass values from the invariant mass spectra of Figure 10, the sum of a Gaussian and a $4^{th}$ degree Chebychev polynomial is fitted to the distribution. The polynomial describes the background from wrong jet combinations and from background events, whereas the mean value of the Gaussian and its error are interpreted as the mass ($m_W$) and its statistical error, $\sigma_{stat}$. The width of the Gaussian is referred to as $\sigma_{Gauss}$. Table 2 shows the obtained W-boson mass with the E recombination scheme for various choices of the $k_T$ parameters.

The error resulting from 1% variation on the jet energy scale (JES) is obtained by varying the reconstructed jet energies by $\pm 1\%$ followed by a linear fit across the three W-boson masses (-1%, nominal value, +1%).

With increasing R and $D_{cut}$ parameter values, the reconstructed W-boson (and also top quark) mass rise monotonically. Higher parameter values lead to fewer but more energetic jets and thus their combination has a higher invariant mass. Part of this effect could in principle be absorbed by the in-situ calibration, but on the other hand, the event is more likely to be misreconstructed due to unwanted jet merging, which makes the choice of the three jets maximizing the $p_T$-sum unpredictable.

The efficiencies quoted in Table 2, defined as the number of events in the gaussian part of the mass distribution divided by the initial number of events, which can be as high as 5.8%, drop to less than 4% when the jets become too big.

Table 2 also shows the purity of the W-boson reconstruction, defined as the fraction of events in the gaussian part of the distribution, in the range $[-1\sigma_{gaus}, 1\sigma_{gaus}]$ from the mean value of the gaussian fit. A purity around 64% can be obtained with some choices of the algorithms / parameters, for example cone 0.4 or $k_T$ with R = 0.4, whereas other choices may lead to purities about 15% smaller.

This study indicates that, for the W-boson mass reconstruction in $t\bar{t}$ events, the cone 0.4 algorithm

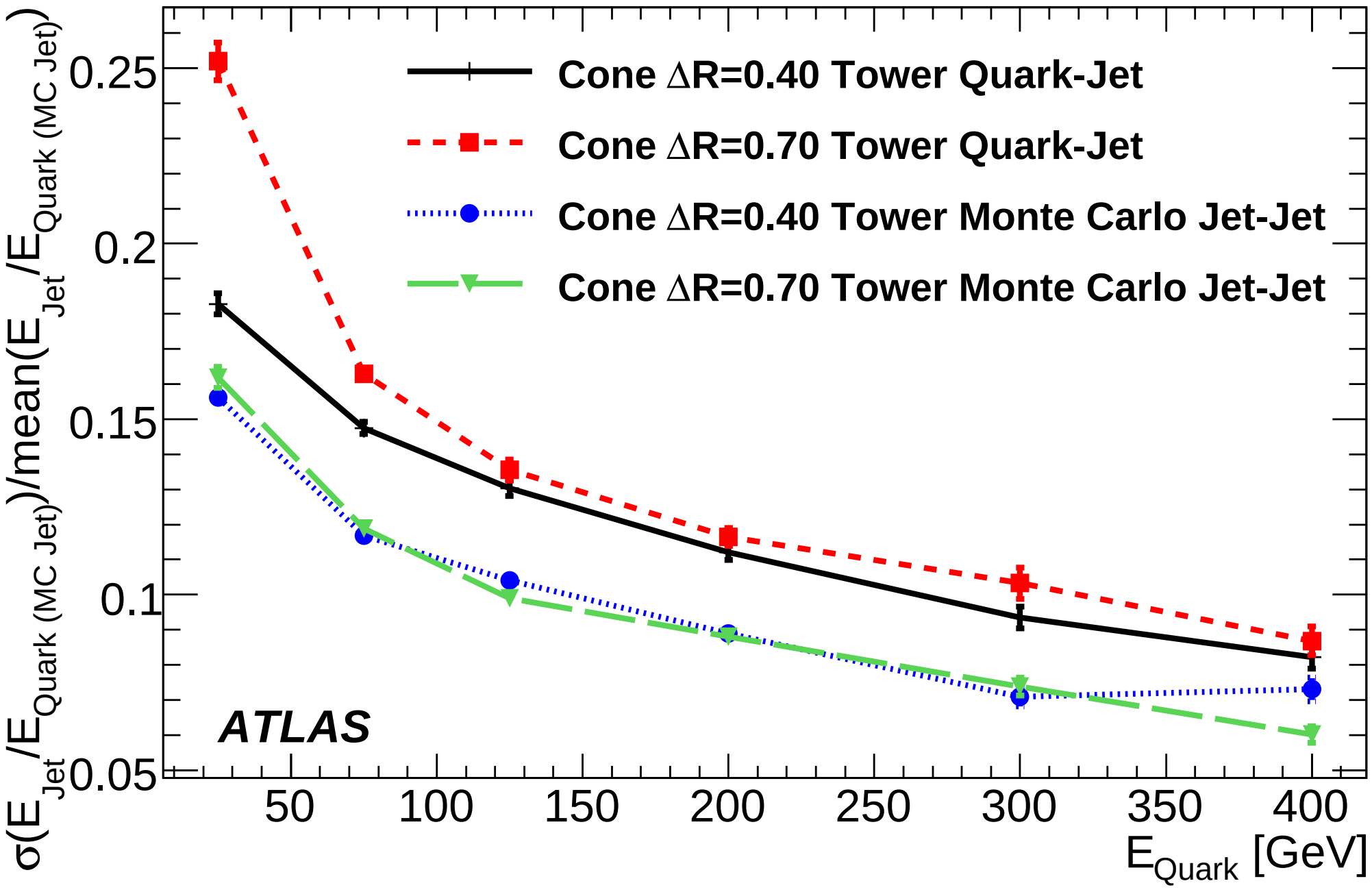


Figure 3: Comparison of the jet energy resolutions with respect to the Monte Carlo hadrons ("Monte Carlo jets") and with respect to the initial quarks from the W-boson decay, for events simulated without pileup.

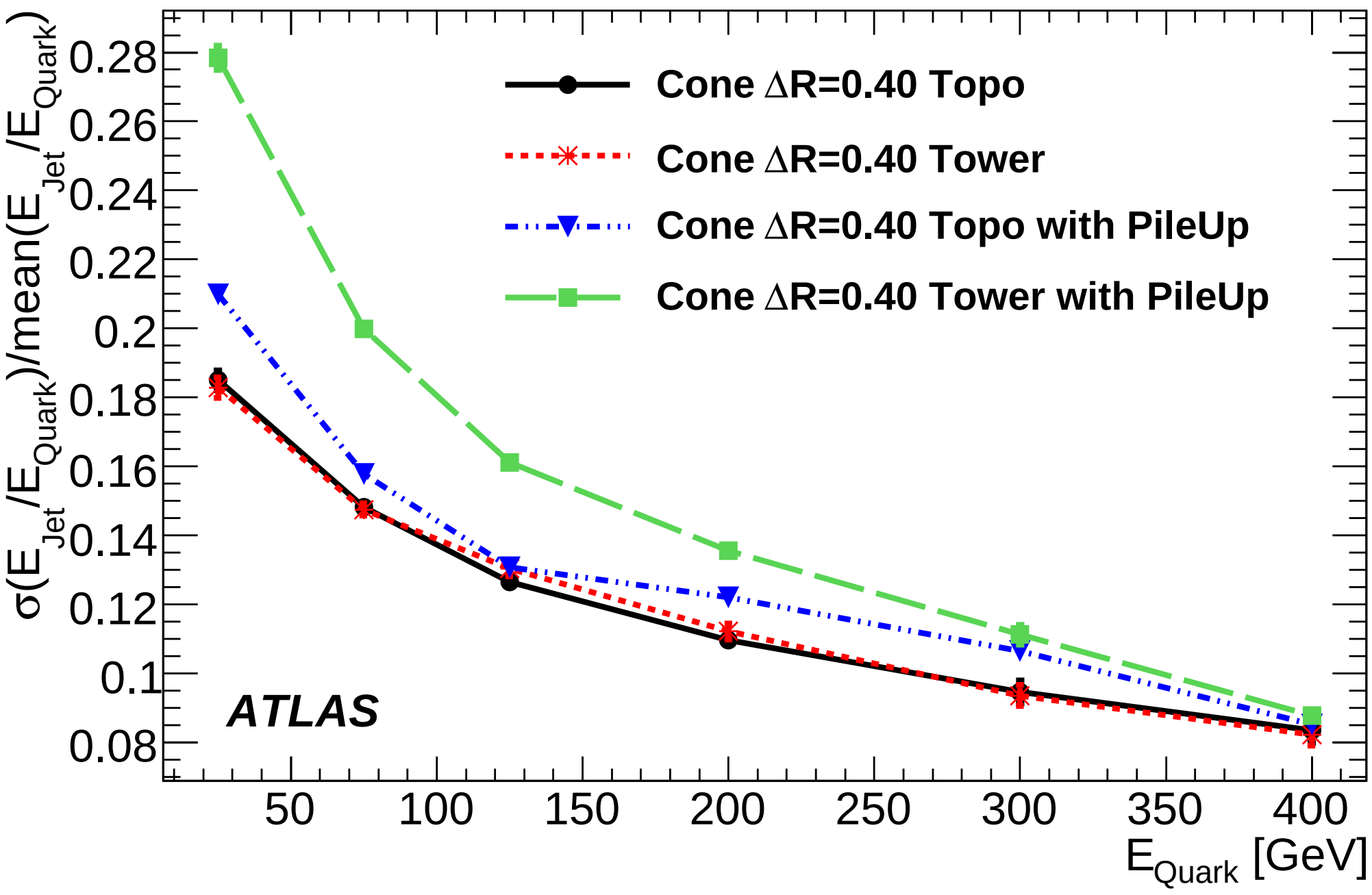


Figure 4: Comparison of the jet energy resolutions for events simulated with and without pile-up.

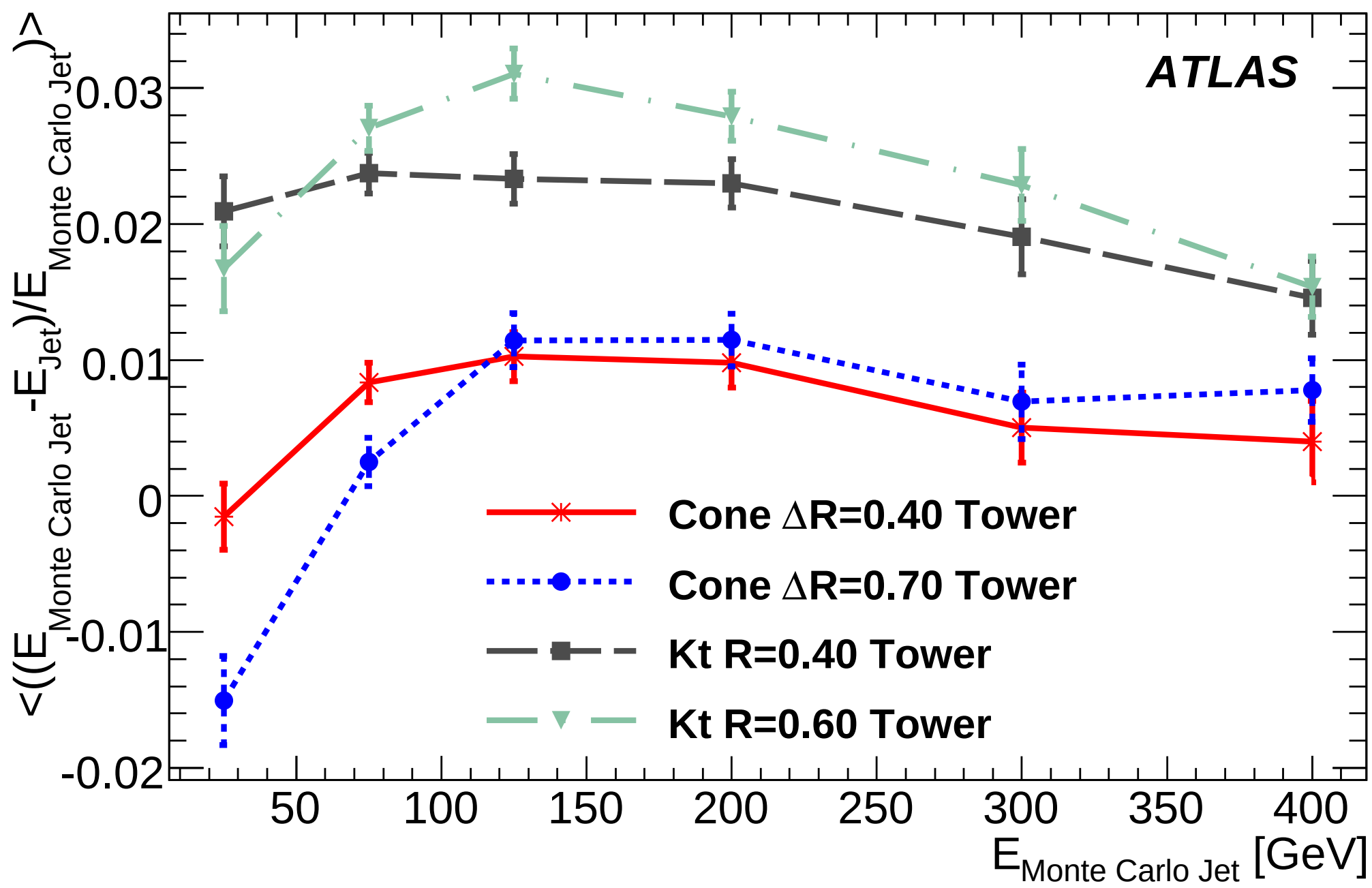


Figure 5: Relative difference between the Monte Carlo jet energy and the reconstructed jet energy, as a function of the Monte Carlo jet energy, for events simulated without pileup.

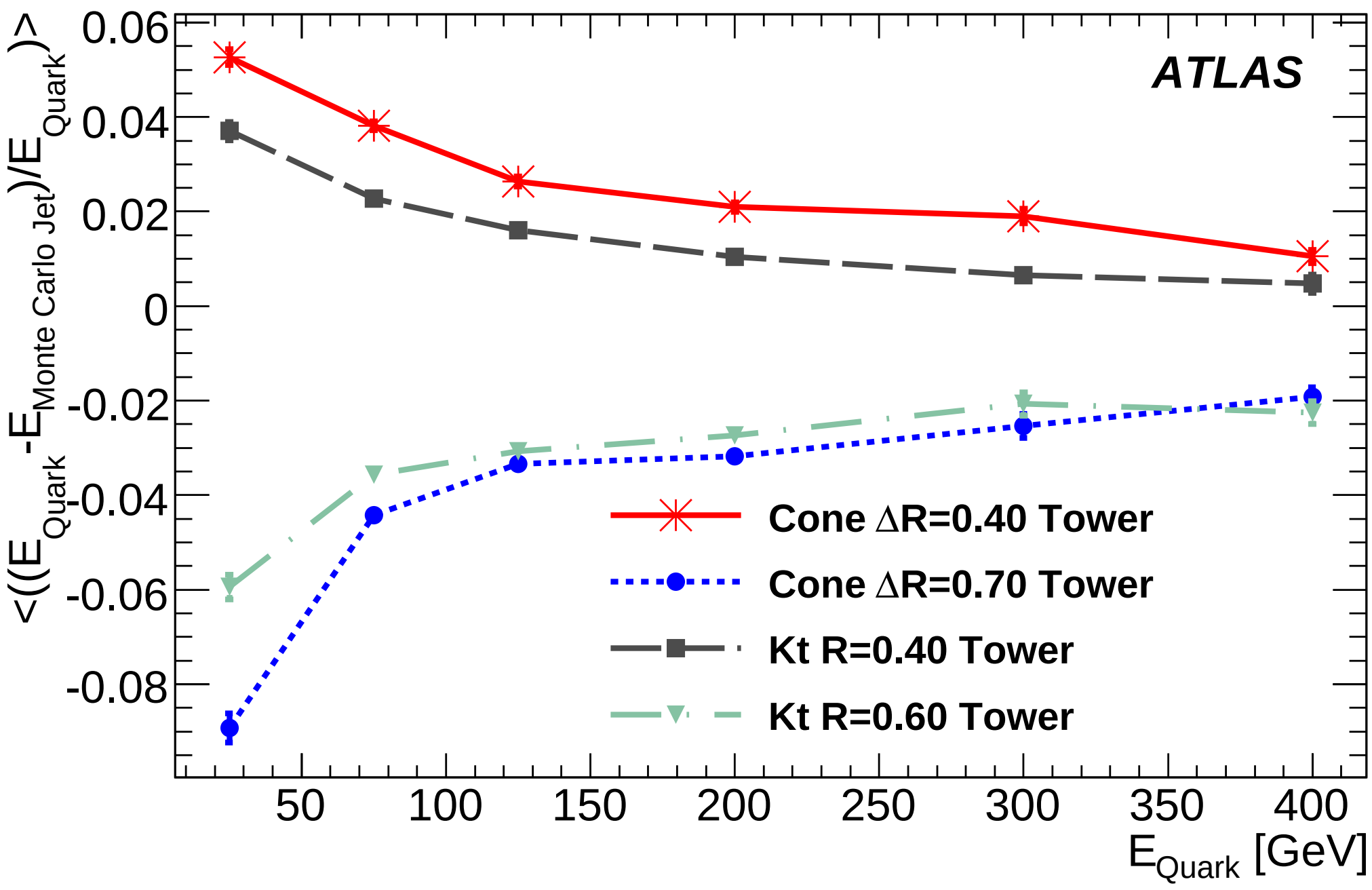


Figure 6: Relative difference between the quark energy and the Monte Carlo jet energy, as a function of the quark energy, for events simulated without pileup.

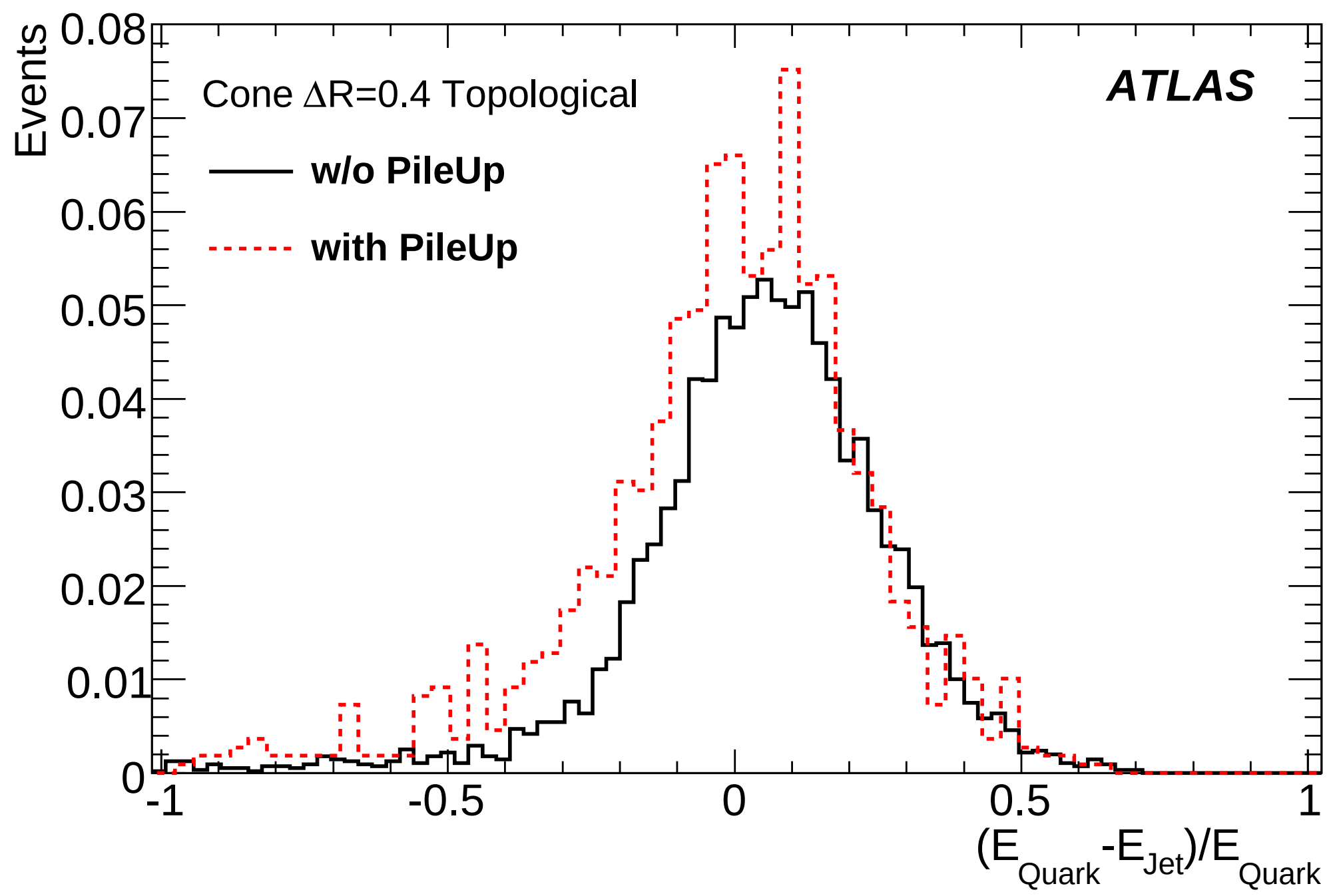


Figure 7: Relative difference between the quark and the jet energy, for quark energies between 15 and 50 GeV, with and without pileup, for the cone 0.4 algorithm from topological clusters.

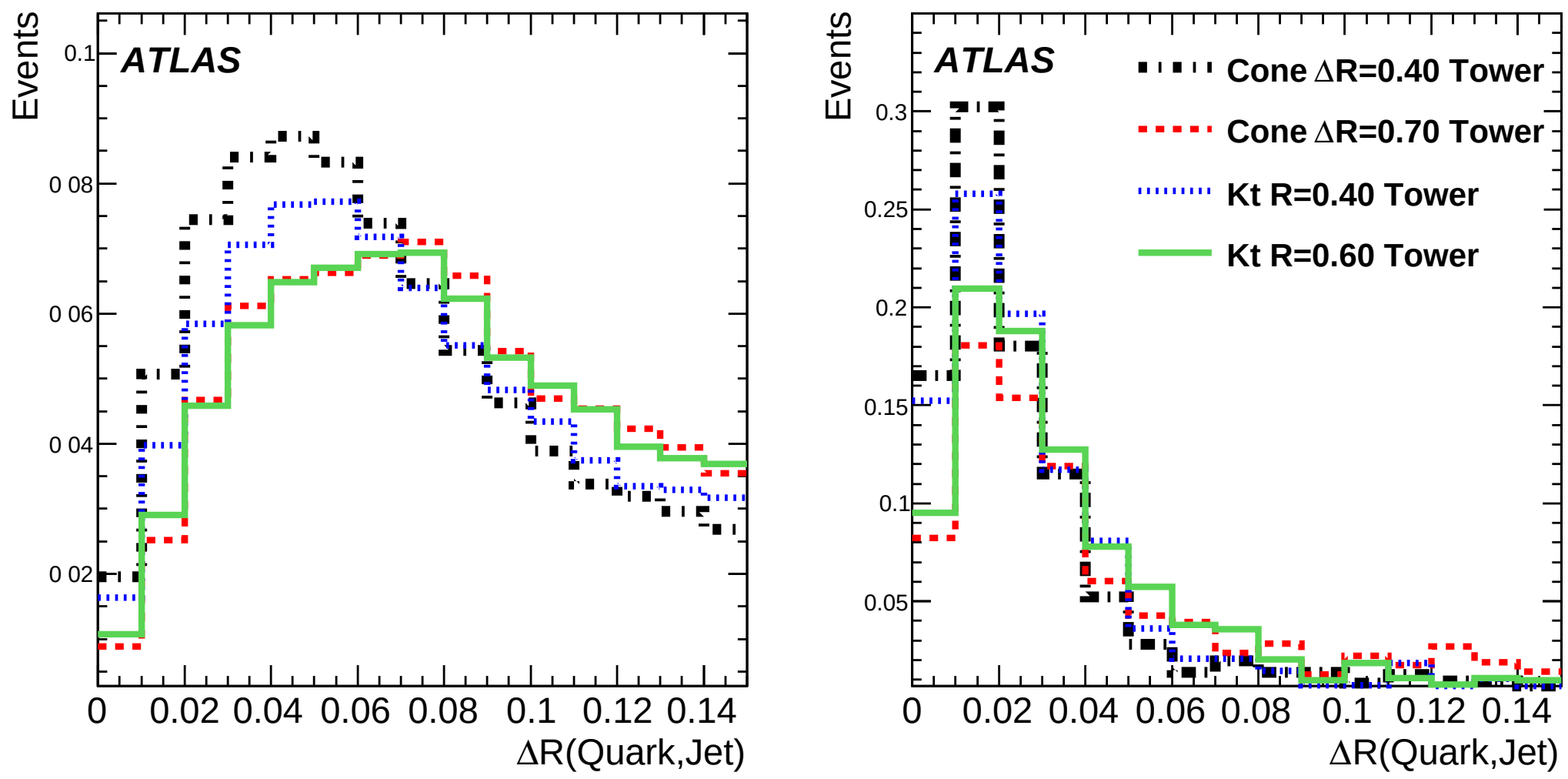


Figure 8: Distance between the jet and the quark, for various jet algorithms / parameters and for two different quark energy ranges (left figure: quark energy between 15 and 50 GeV, right figure: quark energy above 350 GeV)

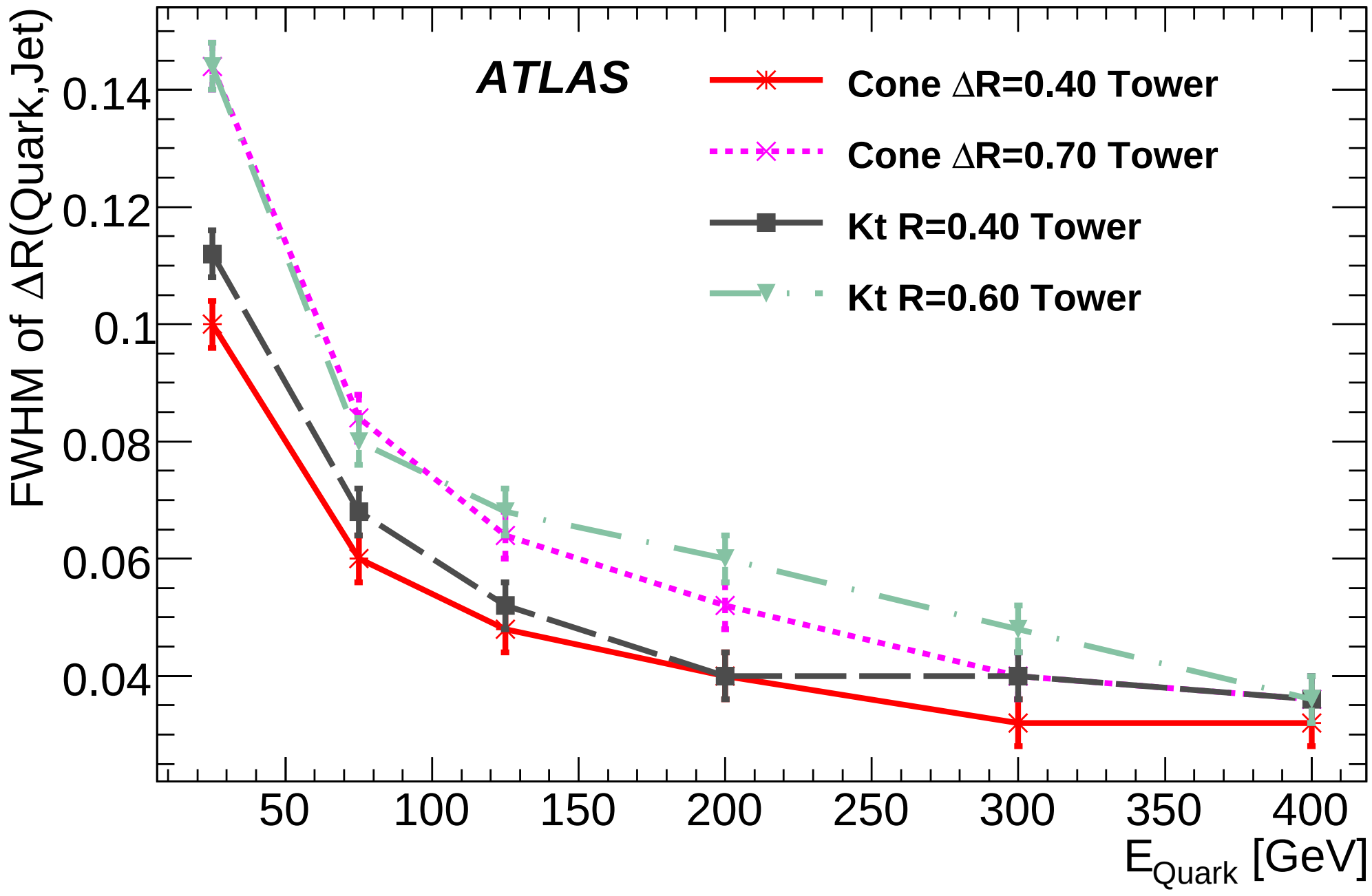


Figure 9: Resolution (defined as the full width at half maximum) on the distance between the jet and the quark, as a function of the quark energy, for various jet algorithms / parameters.

Table 2: Reconstructed hadronic W-boson mass for different $k_T$-parameters with statistical uncertainty, efficiency, purity and the systematic uncertainty from jet energy scale. (*) The cone jet algorithms use uncalibrated topological clusters with H1-style cell weights.

| W mass | | | | | | |
|---|---|---|---|---|---|---|
| $k_T$ parameter | $m_W$ [GeV] | $\sigma_{Gauss}$ [GeV] | $\sigma_{stat}$ [GeV] | efficiency [%] | purity [%] | 1% ΔJES [GeV] |
| R = 0.4 | 78.30 | 7.37 | 0.18 | 4.4 | 63 | ±0.26 |
| R = 0.6 | 79.75 | 7.09 | 0.18 | 4.4 | 63 | ±0.28 |
| R = 0.8 | 80.47 | 6.95 | 0.23 | 3.1 | 58 | ±0.20 |
| $D_{cut} = (10\,\text{GeV})^2$ | 81.25 | 7.03 | 0.17 | 5.4 | 54 | ±0.21 |
| $D_{cut} = (20\,\text{GeV})^2$ | 81.69 | 7.40 | 0.18 | 5.8 | 55 | ±0.27 |
| $D_{cut} = (30\,\text{GeV})^2$ | 81.84 | 7.44 | 0.20 | 5.3 | 57 | ±0.33 |
| $D_{cut} = (40\,\text{GeV})^2$ | 82.00 | 7.12 | 0.22 | 4.2 | 57 | ±0.39 |
| N = 5 | 82.02 | 7.35 | 0.21 | 4.5 | 54 | ±0.34 |
| * Cone R = 0.4 | 81.00 | 6.38 | 0.14 | 5.6 | 64 | ±0.36 |
| * Cone R = 0.7 | 85.15 | 8.03 | 0.26 | 3.9 | 53 | ±0.55 |

leads to the best performance both in terms of resolution and in terms of efficiency / purity.

Finally, in addition to this study which uses jets calibrated with H1-style weights, a study of the W-boson reconstruction using local hadron calibration, described in detail in [5] and in [7], has been started. Although the local hadron calibration algorithm is still being improved, the W-boson mass reconstruction

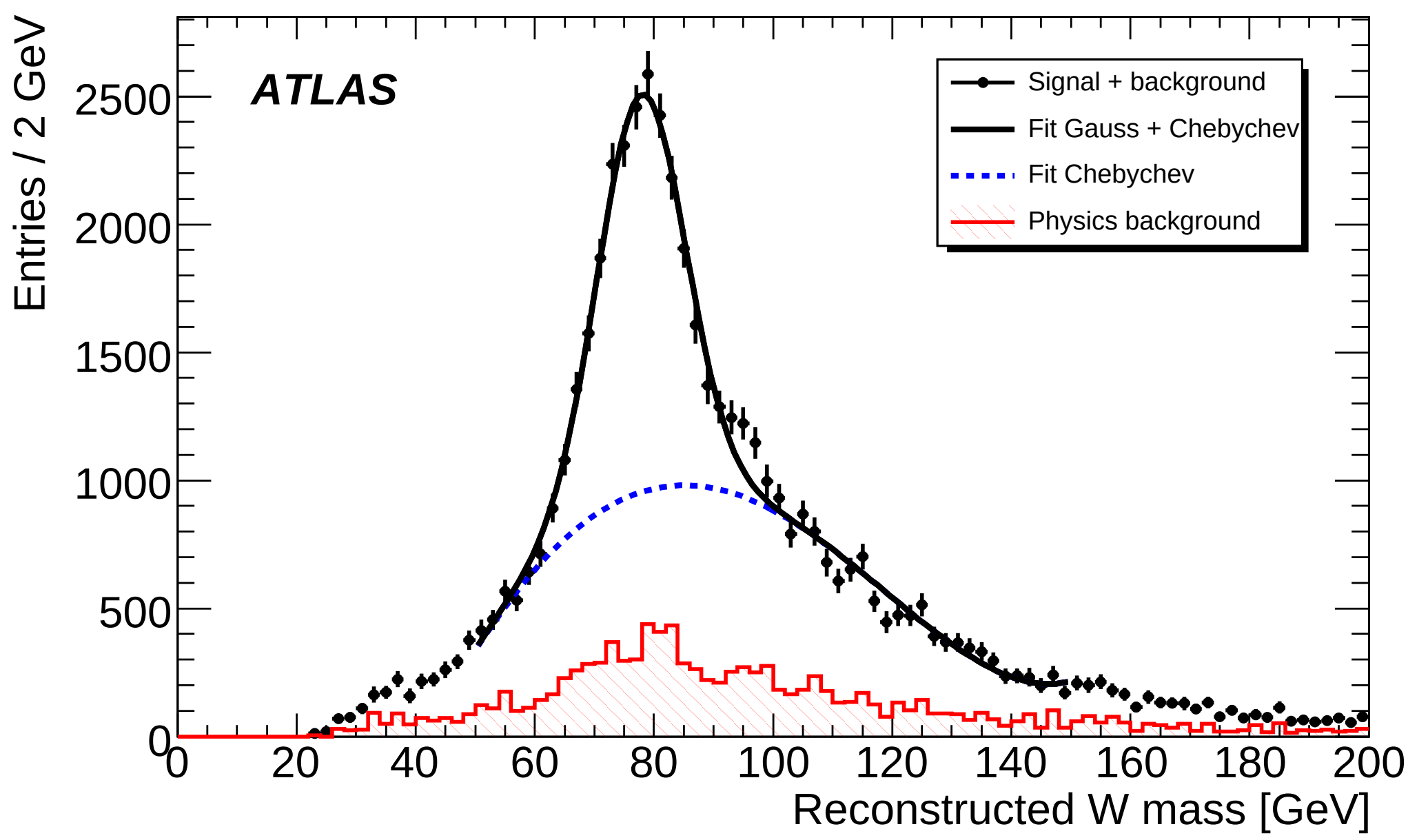


Figure 10: Invariant di-jet mass as an estimate of the hadronically decaying W-boson mass in the selected events. The luminosity used corresponds to 1 $fb^{-1}$ of signal and background events. The inclusive $k_T$ algorithm in the E scheme with R=0.4 is used for jet reconstruction. The data points for signal and background events are shown in black. The hatched histogram shows the distribution for background events (fully hadronic $t\bar{t}$ events, single top events and W + jets events).

capabilities are similar to the ones obtained with H1-weighted jets, both in terms of mass resolution and purity.

### 2.4 Conclusion

The studies shown here indicate that a small jet size should be preferred for a proper reconstruction of the jets in $t\bar{t}$ events, to avoid overlap effects spoiling the energy and angular resolutions and, eventually, leading to the merging of jets. Thus, the cone 0.4 jet algorithm from towers used in the following sections and in most top physics studies is a good choice for looking for top events in the first LHC data, although it could be complemented by an algorithm using topological clusters, which seem to be robust against pileup. The effect of the pile-up on the W-boson mass reconstruction needs however to be assessed.

## 3 Measuring the light jet energy scale

### 3.1 Introduction

The miscalibration of jet energies is one of the main sources of systematic error in the measurement of the top quark mass in the three-jet invariant mass method [1]. Indeed, a 1% error on the jet energy translates into a 0.9 GeV error on the top quark mass, 0.2 GeV coming from the light jet energy scale and 0.7 GeV from the b-jet energy scale. The goal of measuring the top quark mass with a precision of 1 GeV thus puts the limit on the miscalibration at 1% or better.

Several physics processes will be used in ATLAS to achieve such a precision [8]: di-jet events will help check the uniformity of the response as a function of rapidity; $Z$+jet and $\gamma$+jet events will be used to measure the energy scale as a function of energy. However, these processes cannot be applied directly to the jets in $t\bar{t}$ events, for several reasons:

1. These processes contain a mixture of jets from light quarks, jets from $b$ quarks and jets from gluons, with calibration factors probably different by a few % due to their different fragmentation and neutrino content
2. The underlying event may be different in $t\bar{t}$ events
3. Jet selection performed to purify the $t\bar{t}$ sample could lead to a different average jet energy scale in this sample

The use of $W \to jj$ from the $t\bar{t}$ events themselves (in-situ calibration) will thus be very important. After having shown how to select a clean $W \to jj$ signal from $t\bar{t}$ events, two possible methods for extracting the light jet energy scale will be presented. All the studies presented in this section (apart from the check of the stability with different top quark masses shown in section 3.6.3) use the default MC@NLO + HERWIG simulated events (with a top quark mass of 175 GeV), without pileup. The cone 0.4 jet algorithm from towers with the H1-style calibration is used for all studies in this section.

## 3.2 Event and $jj$ pair selection

In order to measure the jet energy scale with a precision of about 1%, a clean $W \to jj$ sample must be selected. In this section the following selection cuts, which are stricter than the ones used previously, are thus used:

- One and only one isolated lepton (electron or muon) with $P_T > 20$ GeV
- Missing $E_T > 20$ GeV
- At least 4 jets with $p_T > p_T^{cut}$
- Among them, 2 and only 2 jets must be tagged as $b$-jets. The other jets are called "light jets" in the following.

In order to increase the purity, the W-boson candidate $jj$ pairs are selected among all light jet pairs which lead to a reconstructed top quark mass between 150 and 200 GeV. In addition, one may want to consider only events with only two light jets.

The estimated number of $jj$ pairs selected for 1 fb$^{-1}$ and the purity of the selection, defined as the fraction of $jj$ pairs with both jets within $\Delta R = 0.25$ of the two W-boson quarks, are shown in table 3 as a function of $p_T^{cut}$, and whether or not one uses only events with two light jets.

Requiring two light jets only reduces the number of $jj$ pairs by about a factor of almost two, but is necessary if one aims to measure the jet energy scale down to 20 GeV, in order to minimize the bias due to the combinatorial background.

Figure 11 shows the $jj$ mass distribution for $p_T^{cut}$ = 40 GeV and exactly two light jets. The combinatorial background (shown in grey) is almost flat. The systematic error due to the knowledge of the combinatorial background will be discussed later. Because of the requirement for 2 b-tagged jets, the background from other processes is of the order of a few percent and is neglected here.

Table 3: Number of selected $W \rightarrow jj$ pairs (for 1 fb$^{-1}$) and purity of the selected pairs (the statistical uncertainty on the purity is about 1%)

| | 2 jets only | | 2 jets or more | |
|---|---|---|---|---|
| $p_T^{cut}$ [GeV] | $jj$ pairs | purity [%] | $jj$ pairs | purity [%] |
| 20 | 3100 | 76.9 | 7100 | 64.3 |
| 30 | 2300 | 77.7 | 4100 | 71.5 |
| 40 | 1200 | 81.0 | 1900 | 79.1 |

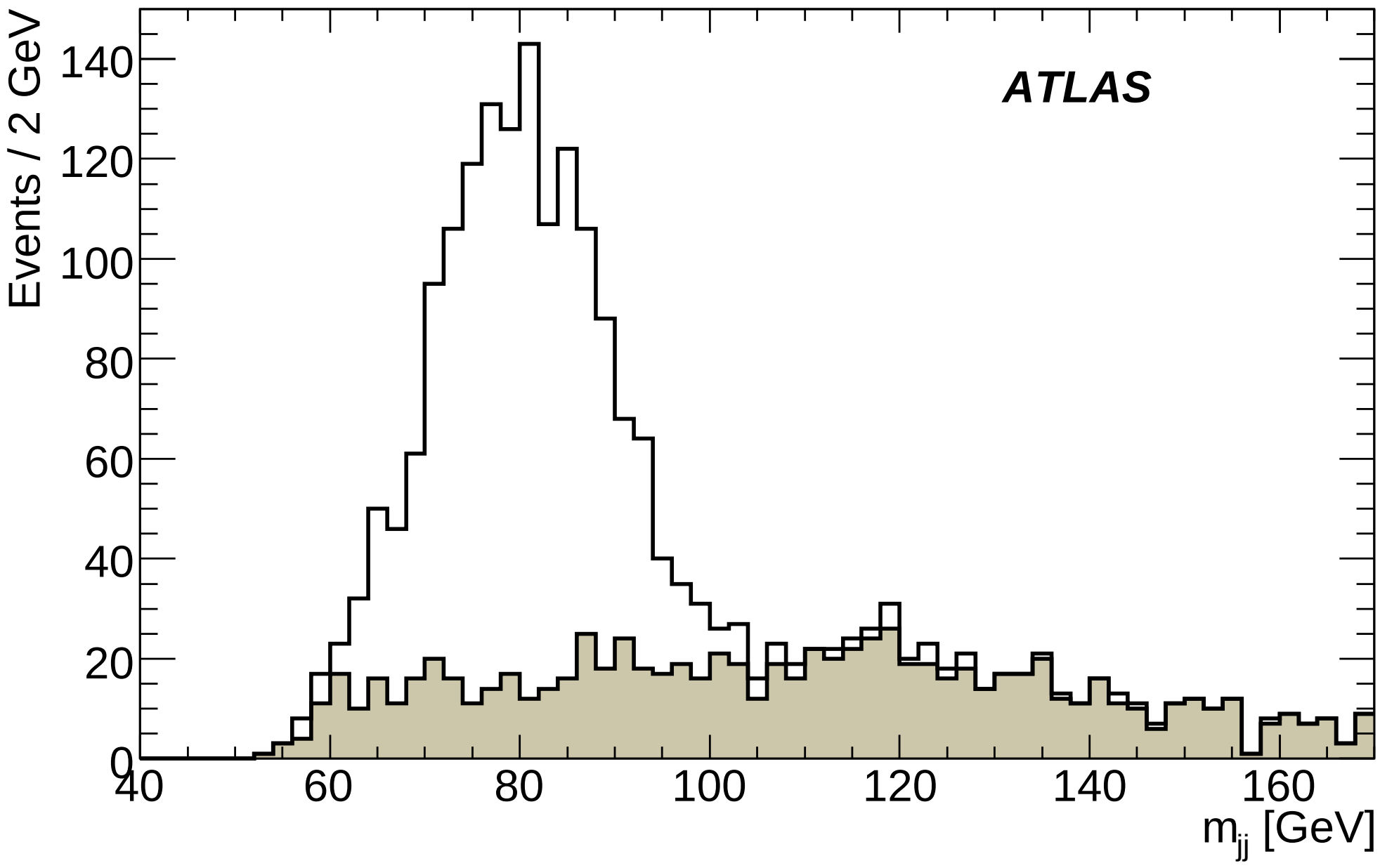


Figure 11: Invariant mass $m_{jj}$ of the selected jet-jet pairs when the two jets are at $\Delta R < 0.25$ from the two quarks from the W-boson decay (white histogram) and when at least one of the jets is not from the W-boson decay (grey histogram).

### 3.3 The $P_T$ cut effect in the Jet Energy Scale measurement

When measuring the jet energy scale, one must be aware of a shift introduced by any cut on the transverse momentum of the jets needed for the event selection. Indeed, when one selects jets with a reconstructed $p_T$ greater than $p_T^{cut}$, jets with a true $p_T$ lower than the cut but a reconstructed one above are selected, whereas jets with a true $p_T$ above the cut but a reconstructed one below are lost. These two effects lead to a ratio (reconstructed energy) / (true energy) higher than one near the cut, even if the calibration is perfect.

This can easily be seen, and the size of the effect measured, with a simple Monte Carlo simulation where the momenta of the quarks from the $W$ boson decays in $t\bar{t}$ are smeared according to the expected jet resolution [2], and a $p_T$ cut on the smeared quark momenta is applied.

Figure 12 shows the apparent calibration as a function of the jet $p_T$ cut, using this simple simulation.

[2]The value $\sigma(E) = 3.8\ \mathrm{GeV} + 0.063 \times \mathrm{E_q}$ was used.

This apparent calibration is greater for jets whose energy is closer to the value of the $p_T$ cut. The size of the effect is fully determined once the $p$, $p_T$ spectra and the detector resolution are known. With the expected resolution of the ATLAS calorimeter, the global apparent calibration is 2% for a $p_T$ cut = 40 GeV.

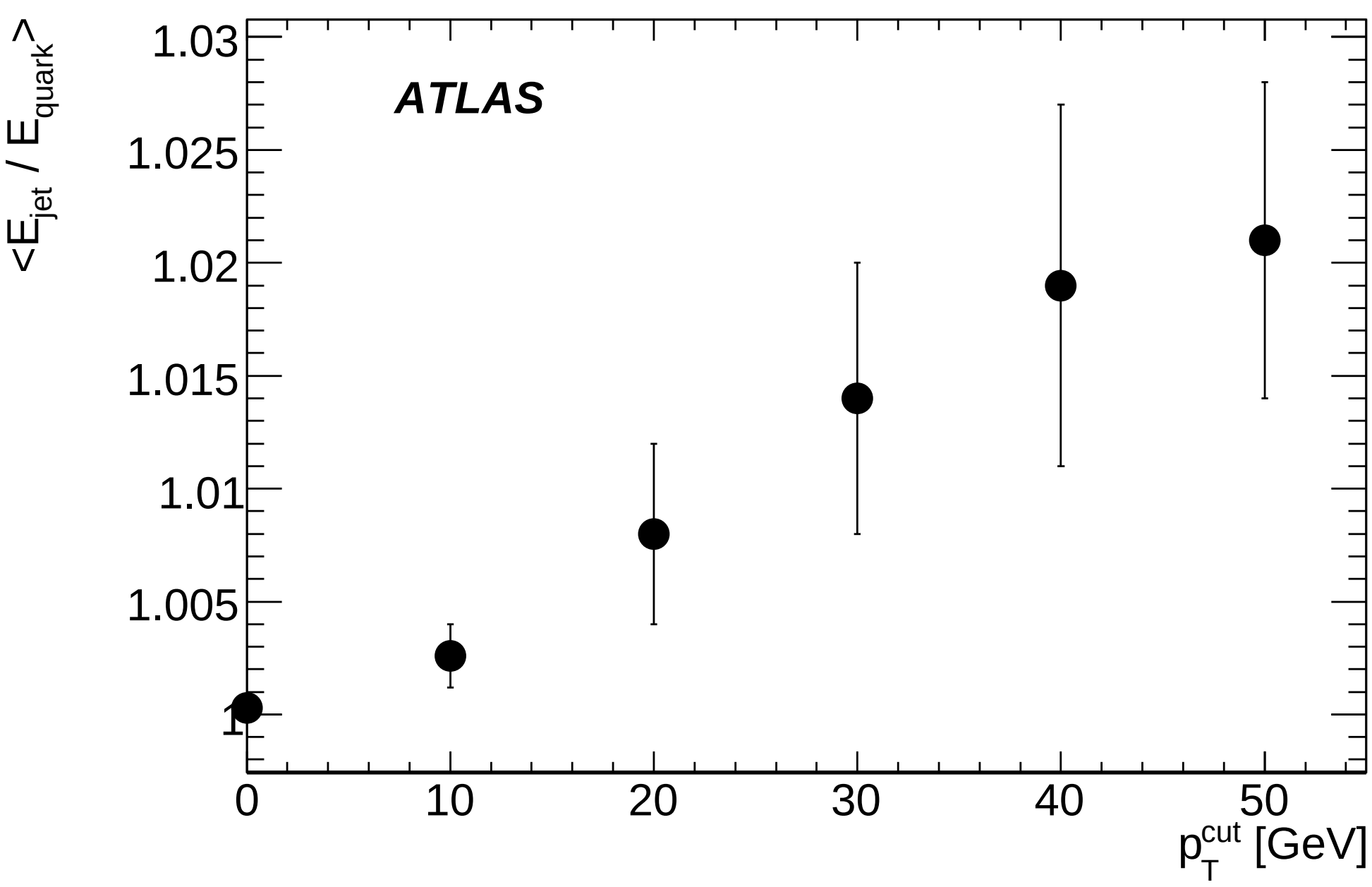


Figure 12: The apparent jet energy scale due to a cut on the transverse momenta of the reconstructed jets. The error bars indicate how the shift varies when changing the jet resolution by ± 20%. This figure was obtained using the simple simulation described in the text.

The shift in the energy scale also leads to a shift on the reconstructed $W$ boson and top quark masses, which are shown in Figures 13 and 14. With a $p_T^{cut}$ at 40 GeV, the bias on the $W$ boson mass is about 1.5 $\pm$ 0.6 GeV, whereas the bias on the top quark mass is about 2.0 $\pm$ 0.4 GeV.

The two jet energy scale determination methods described below handle this effect in a different way:

- The iterative method measures the effective jet energy scale above the $p_T^{cut}$, which will not reflect the bare jet energy scale, but will lead to $W$ boson and top quark masses with the correct values.
- The template method aims at finding the bare jet energy scale (which was by definition 1 in the simple Monte Carlo simulation). The reconstructed $W$ boson and top quark masses will then be shifted, and must be corrected.

### 3.4 Expected Jet Energy Scale

The energy scales fitted by the two methods described below will be compared with the true jet energy scale, obtained by comparing the reconstructed jet energy and the quark energy.

In the simulation version used for this note, the mean value of the ratio of jet to quark energies, when the two jets chosen to form the hadronic W-boson are at $\Delta R < 0.25$ from the two quarks from the W-boson, for quarks with $P_T > 40$ GeV is $0.961 \pm 0.003$, as can be seen in figure 15.

As shown in section 2.1, this 4% miscalibration is mainly due to the out-of-cone energy.

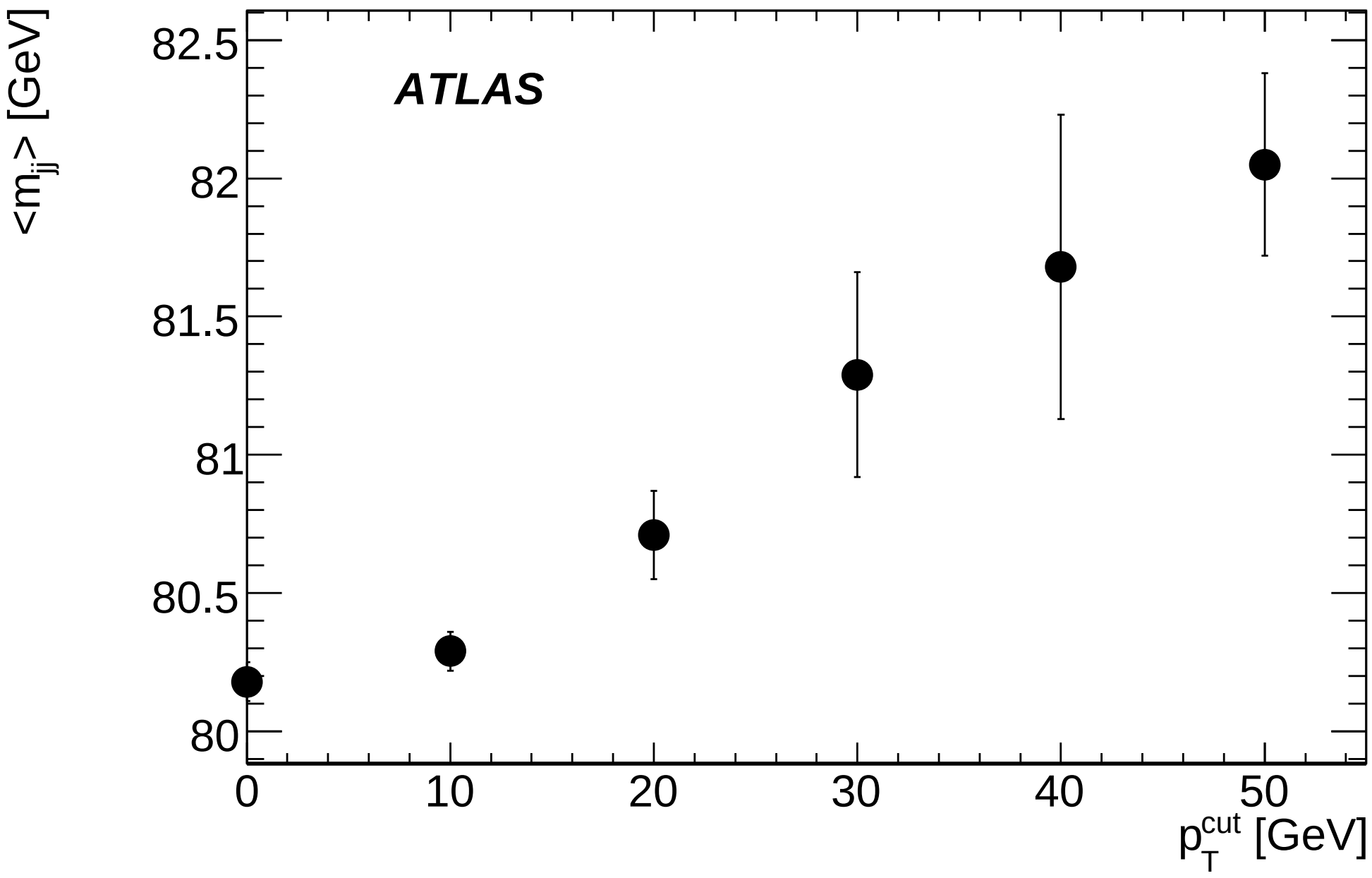


Figure 13: Shift of the reconstructed $W$ boson mass, $< m_{jj} >$, due to the $p_T^{cut}$ on the jet energies. The error bars indicate how the bias varies when changing the jet resolution by $\pm$ 20%. This figure was obtained using the simple simulation described in the text.

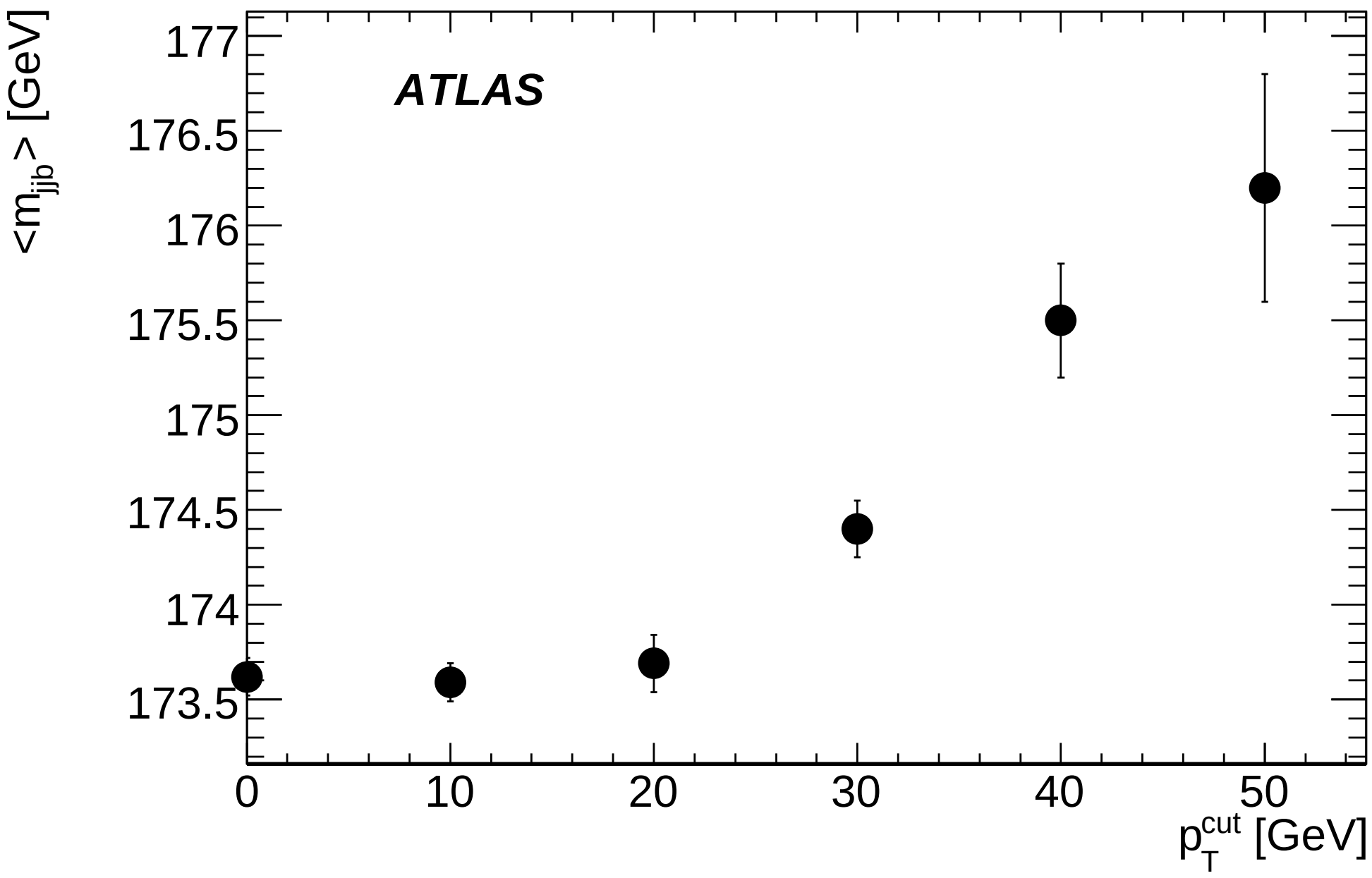


Figure 14: Shift of the reconstructed top mass, $< m_{jjb} >$, due to the $p_T^{cut}$ on the jet energies. The error bars indicate how the bias varies when changing the jet resolution by $\pm$ 20%. This figure was obtained using the simple simulation described in the text.

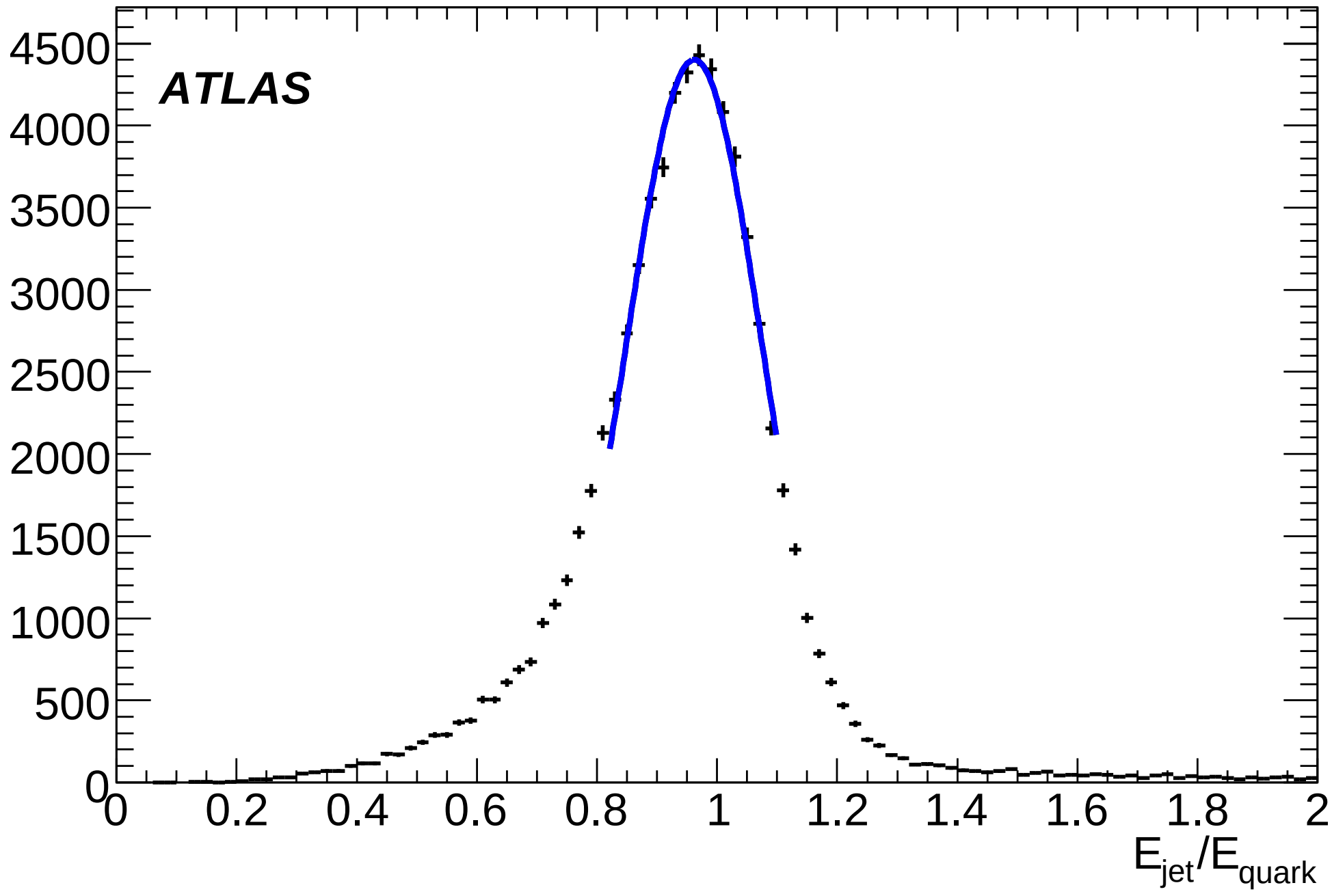


Figure 15: $E_{jet}/E_{quark}$ for jets originating from a W-boson and with $p_T(quark) > 40$ GeV.

## 3.5 The iterative rescaling method

### 3.5.1 Method

This method uses the precisely known W-boson mass as a reference to extract the light jet energy scale. The invariant mass of the two jets with energies $E_1$ and $E_2$ and an opening angle $\theta_{jj}$, originating from the W-boson can be written as :

$$M_{jj} = \sqrt{2E_1E_2(1-\cos(\theta_{jj}))}$$

Figure 16 shows that the angle between the two jets is measured without any significant bias for most jet-jet pairs. Any deviation of $M_{jj}$ therefore comes mainly from the energy miscalibration.

The peak value of this invariant mass matches the PDG W-boson mass value if the appropriate jet energy scale factors $K(E_1)$ and $K(E_2)$ are used:

$$M_W^{PDG} = \sqrt{2(K(E_1)E_1)(K(E_2)E_2)\times(1-\cos(\theta_{jj}))} = \sqrt{K(E_1)K(E_2)}M_{jj}$$

If the jet calibration is independent of the energy ($K = K(E_1) = K(E_2)$), the jet energy scale factor $K$ is simply $K = M_W^{PDG}/M_{jj}$. The simple rescaling gives the effective jet energy scale, taking into account the $p_T$ cut effect from the jet selection.

With 1 fb$^{-1}$ of luminosity, around 4000 jets originating from the W in the $t\bar{t}$ sample are available for calibration. Normalizing the sample to 1 fb$^{-1}$, the truth calibration, for jets with $P_T > 40$ GeV, is $K_{truth} = (E_{jet}/E_{parton})^{-1} = 1.014 \pm 0.002$. The value obtained by the simple rescaling of jet pairs originating from a W boson (no background contribution) is $K = M_W^{PDG}/M_{jj} = 1.014 \pm 0.003$ . The 1% goal on the precision on the global jet energy scale is therefore obtainable with 1 fb$^{-1}$, provided the luminosity dependence is small or can be corrected for.

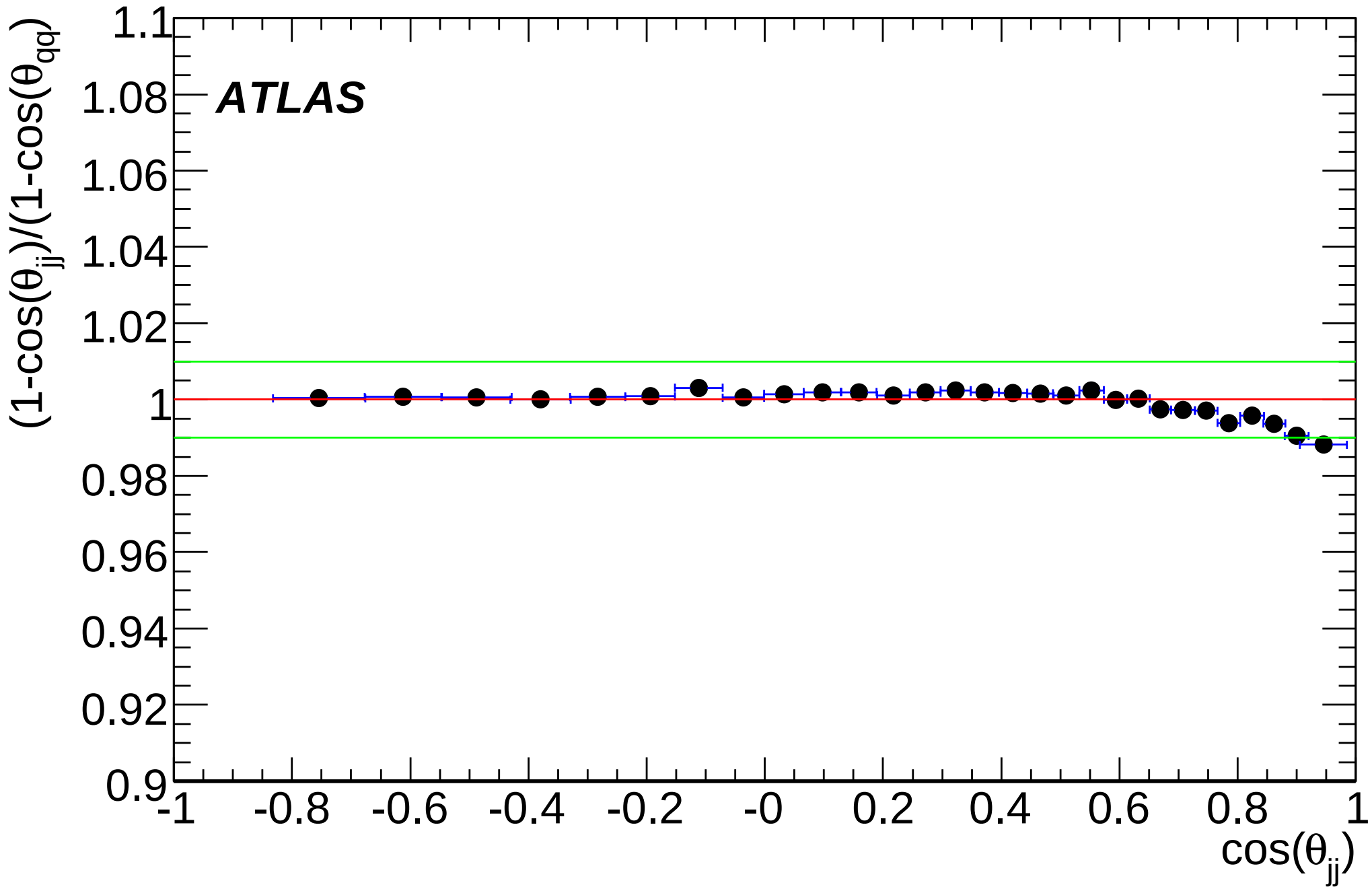


Figure 16: $(1-\cos(\theta_{jjMC}))/(1-\cos(\theta_{jj}))$ as function of the reconstructed $\cos(\theta_{jj})$, where $\theta_{jj}$ is the reconstructed angle between the two jets and $\theta_{jjMC}$ the angle between the two associated quarks.

Since the calibration is not constant over the energy and pseudorapidity, and since the two jets from the W-boson decay have in general different energies, the rescaling method has to be adapted to get the calibration K(var) (var=jet energy, jet pseudorapidity or any other jet variable). Instead of building one jj mass distribution, the first step is to split the studied jet variable in N bins, and to build all associated jj invariant mass distributions. In such a way, the distributions become correlated to the studied variable.

It can be shown that, if the rescaling method is iterated M times, extracting at each iteration j the correction factor for the bin i ($K_i^j$) and recomputing the invariant mass by adding the new $K_i^j$ for the next iteration, the exact correction $K_i$ per bin i can be obtained (thus the function K(var)):

$$K_i = \prod_{j=1,M} K_i^j \quad (1)$$

This method converges in 3-4 iterations, and is stopped when the maximum difference between $K^j$ and $K^{j-1}$ is less than 1%. The number of bins N depends on the available jet statistics.

### 3.5.2 Results

The simple rescaling method allows to determine a global jet energy scale factor. A W-boson mass value $M_W = 79.12 \pm 0.25$ GeV is obtained from a fit with a polynomial+gaussian function to the jet-jet invariant mass spectrum. The resulting jet energy scale factor is thus $K = M_W^{PDG}/M_{jj} = 1.016 \pm 0.003$ [3]. This value is very close to the one obtained previously by a fit on the same sample without combinatorial background included.

Figures 17 and 18 show the effective calibration to be applied on the selected jets after the $t\bar{t}$ selection for 10 fb$^{-1}$ of integrated luminosity [4]. The observed discrepancy between the fit result and the truth $K$ factors is due to the ($\eta$, E) correlations existing on the jet energy scale factors. This is particularly true for high jet $\eta$ and high jet energy. The discrepancy at low energy is a threshold effect which could be removed by lowering the $p_T cut$ on the jets to 30 GeV when measuring the jet energy scale above 40 GeV. Nevertheless the jet energy scale as a function of $\eta$ and E can be extracted with a precision of the order of 1%, without any a priori knowledge of the function shape before the fit. The treatment of the correlation could be resolved by a fit on W mass distributions defined as a function of both jet $\eta$ and E.

For 1 fb$^{-1}$ of integrated luminosity, the number of available jets ($\approx$ 4000) limits the number of bins and therefore the accuracy of the extracted jet energy scale functions ($K(\eta)$ and K(E) ). The resulting plots for 1 fb$^{-1}$ are shown in Figures 19 and 20. A precision better than 2% is obtained in each energy or $\eta$ bin.

### 3.5.3 Systematic uncertainties

The systematic uncertainties associated to each point are of three types:

- the first is related to the method itself. Performing the fit on a large sample shows up to 1% deviation, due to the ($\eta$,E) correlations on the jet energy scale which are not taken into account yet.
- the second is related to the impact of the background on the mass peak measurements. It is rather low since the W-boson sample has a purity above 80%. Further purification cuts can be applied to increase the purity. The associated error depends on the available statistics and is higher for

[3] The W boson mass is set to $M_W = 80.4$ GeV in the simulation.

[4] The calibration factors are large in the barrel-endcap transition region. In principle this dependence is corrected at the level of jet reconstruction algorithm using a $p_T$ balance method in di-jet events. This correction has not been applied on the used sample.

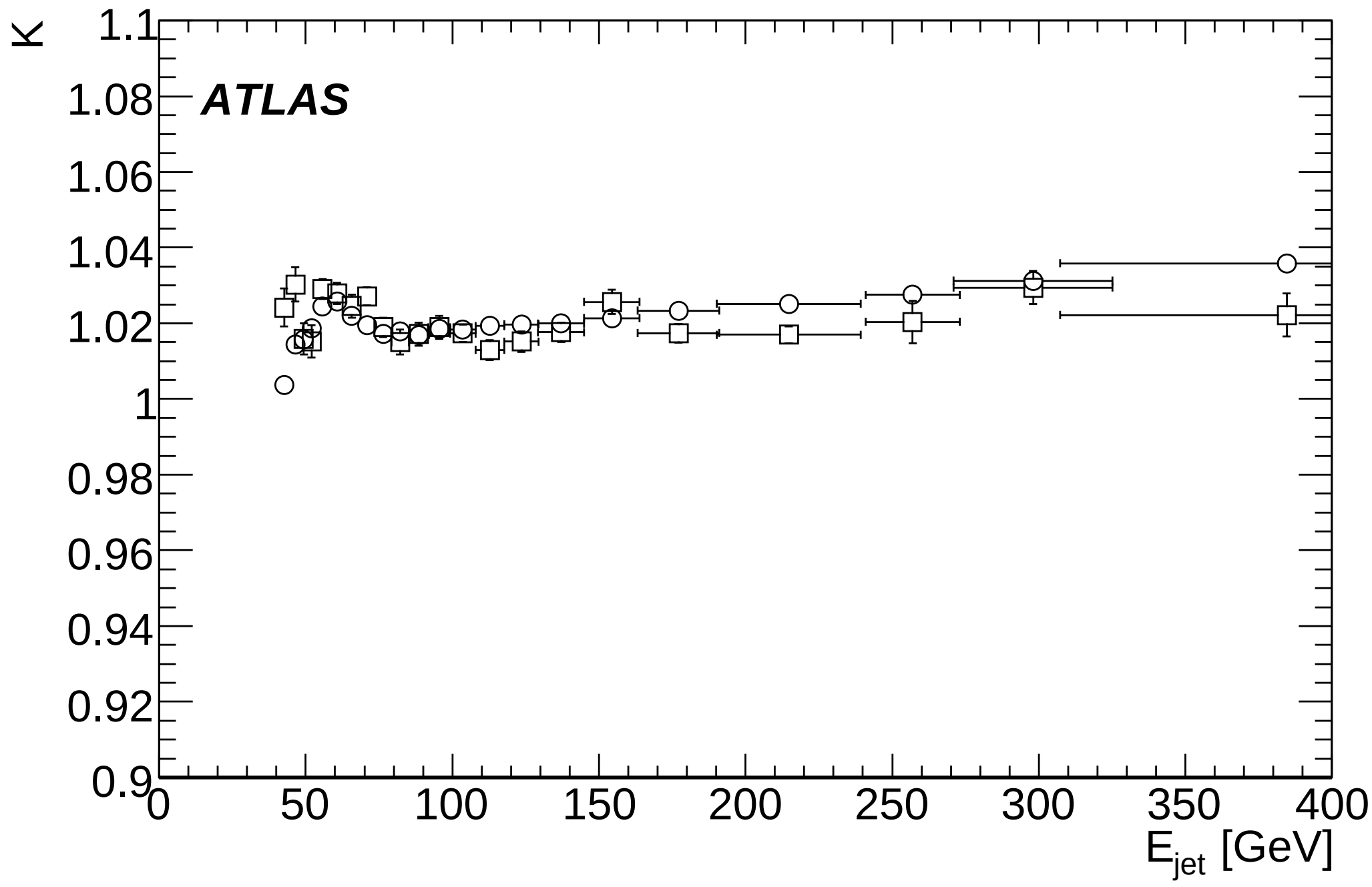


Figure 17: Result of the iterative rescaling fit as a function of $E_{jet}$, with 10 fb$^{-1}$: the expected effective calibration factors are shown as square points, in addition to the fitted calibration factors with circles. The calibration factor obtained from Monte Carlo truth information $K_{truth}(E) = (E_{parton}/E_{jet})$ are shown as square points while the calibration factor $K(E_i)$ obtained in the iterative rescaling procedure as described in section 3.5.1 are described with circles.

low jet energy where the background contribution is higher. It should be noticed that the jet pairs contributing to the background are mainly composed of one jet originating from the W-boson, the others being jets coming from the remaining part of the event or gluon radiation. An event mixing technique, described in [1] paragraph 4.2, will be useful to assess the shape of the background contribution on real data. To evaluate the systematics induced by the background contribution, the background contribution was changed by +20%, separately for the FSR and non-FSR contributions [5]. The fitted $M_W$ values are all within 0.2 GeV, compatible with the statistical uncertainty leading to an uncertainty of $0.003 \pm 0.003$ for the global jet energy scale factor.

- the last comes from the correction factor to be applied on the fitted value to take into account the $p_T$ bias discussed in section 3.3. This correction of the effect is fully determined once the jet p, $p_T$ spectra and energy resolution are known. The size of this correction is $-1\%$ above 100 GeV and up to 2% at 40 GeV. In order to assess the error on these values, the resolution has been rescaled by $\pm 10\%$, leading to variation less than one percent of the correction factors.

[5] The shapes of the background when a jet coming from the W-boson decay is associated with a FSR or a non FSR jet are different. They are taken from a fast simulation Monte Carlo.

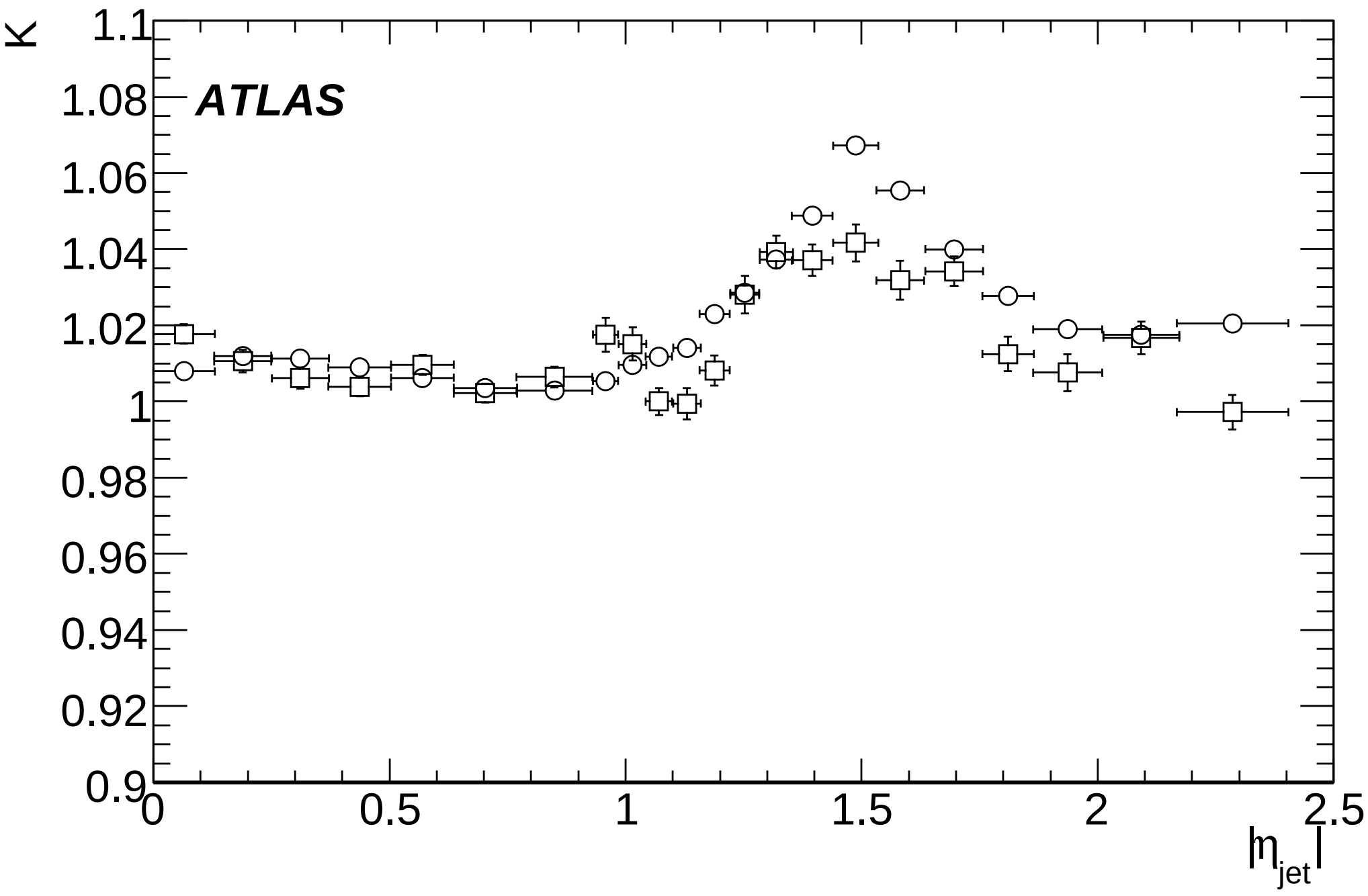


Figure 18: Result of the iterative rescaling fit as a function of $\eta_{jet}$, with 10 fb$^{-1}$: the expected effective calibration factors are shown as square points, in addition to the fitted calibration factors with circles.

## 3.6 The template method

### 3.6.1 Method

The template method for the light jet energy scale determination is similar to the method described in [9] for the electromagnetic energy scale determination using $Z^0 \rightarrow e^+e^-$ events. It uses template histograms with various energy scales $\alpha$ and relative (to the default jet energy resolution) energy resolutions $\beta$. The $\chi^2$ between each template histogram and the "data" is then computed. The minimum of the $\chi^2$ is found in the ($\alpha$, $\beta$) plane. With 1 fb$^{-1}$, this method fits both the average jet energy scale and a relative jet resolution. With enough data, it can be extended to measure these quantities as a function of the jet energy or rapidity.

The template histograms were generated from $W \rightarrow qq$ decays in 1.2 million PYTHIA $t\bar{t}$ events by smearing the quark energies by a gaussian number with width equal to $3.8\ \mathrm{GeV} + 0.063 \times \mathrm{E_q}$, determined on reconstructed jets in an older simulation of $t\bar{t}$ events. As will be shown later, the exact choice of the default energy resolution in the templates is not very important, as it will lead at the end only to a change in the fitted $\beta$ and not in the fitted $\alpha$. However, in order to obtain a $\beta$ value close to 1, the jet angles are also smeared, according to the expected resolution, in the templates and the observed correlation between the two jet energies is added by using correlated gaussian numbers for the energy smearing.

After smearing the quark energies, the same $p_T^{cut}$ as used in the data is applied. This ensures that the fitted energy scale is not affected by the bias described in section 3.3. Figure 21 shows the reconstructed jet-jet mass (for events with two jets or more, with the combinatorial background included), the template histogram for $\alpha$ = 1, $\beta = 1$ and the best fit histogram.

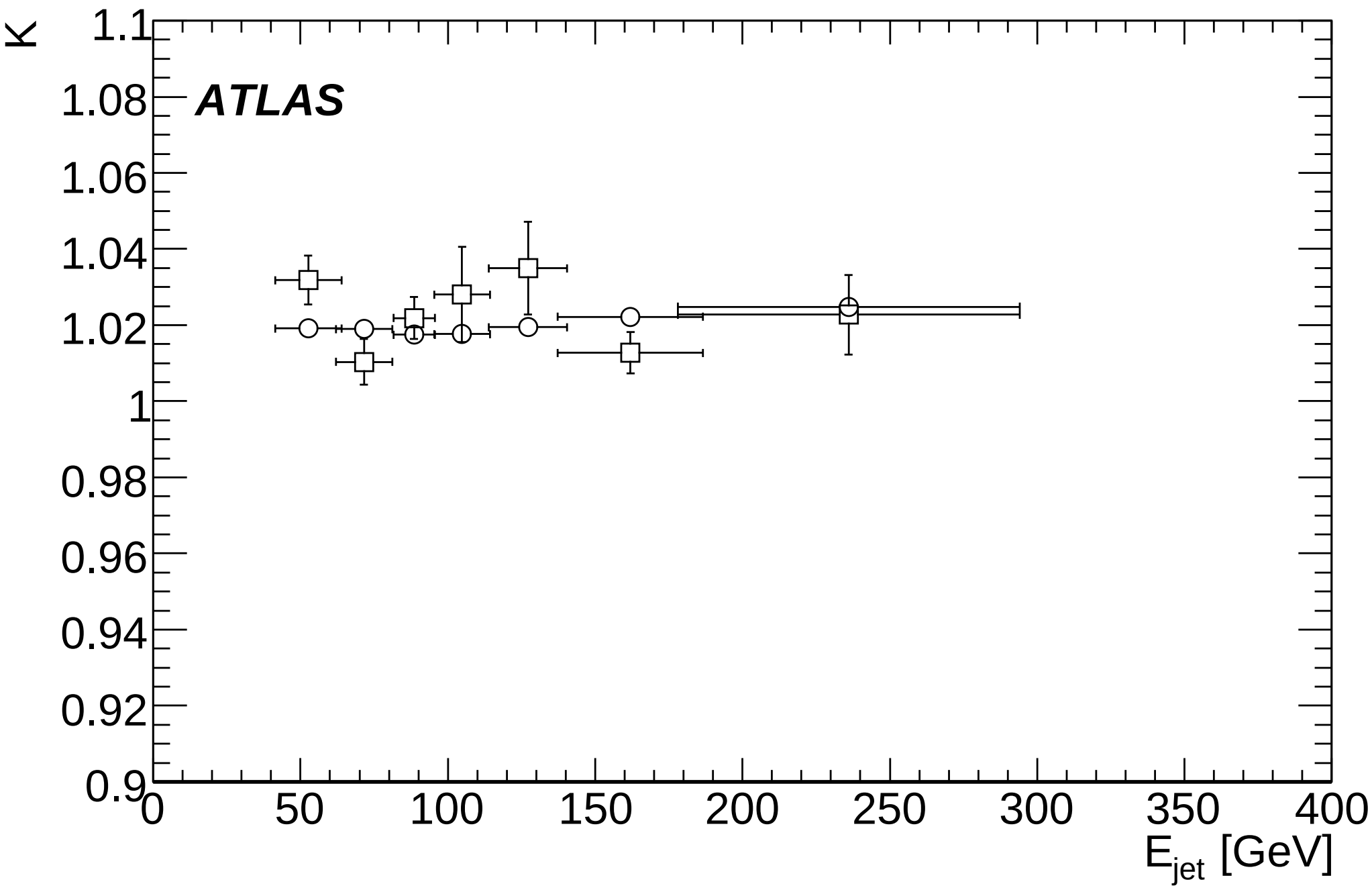


Figure 19: Result of the iterative rescaling fit as a function of $E_{jet}$, with 1 fb$^{-1}$: the expected effective calibration factors are shown as square points, in addition to the fitted calibration factors with circles.

### 3.6.2 Results

Table 4 summarizes the results obtained for jets with $P_T > 40$ GeV. The combinatorial background, which is flat as shown in Figure 11, doesn't affect the fitted jet energy scale (less than 0.6%), but degrades the fitted relative resolution by 30 to 40%. The choice to use or not events with more than two jets is also not changing the fitted jet energy scale.

Finally, these fitted energy scales are in good agreement with the expected value from the Monte-Carlo ($0.961 \pm 0.003$), as obtained in section 3.4.

Table 4: Fitted jet energy scale and relative jet energy resolution using the template method, for jets with transverse momentum greater than 40 GeV. *Good combinations* refers to the cases where the two jets come from the W-boson decay, while the results with *all combinations* include the effect of the combinatorial background.

| | $\alpha$ | $\beta$ |
|---|---|---|
| 2 jets, good combinations | $0.9693 \pm 0.0045$ | $1.145 \pm 0.054$ |
| 2 jets, all combinations | $0.9638 \pm 0.0049$ | $1.434 \pm 0.064$ |
| $\geq 2$ jets, good combinations | $0.9696 \pm 0.0033$ | $1.089 \pm 0.035$ |
| $\geq 2$ jets, all combinations | $0.9660 \pm 0.0036$ | $1.308 \pm 0.041$ |

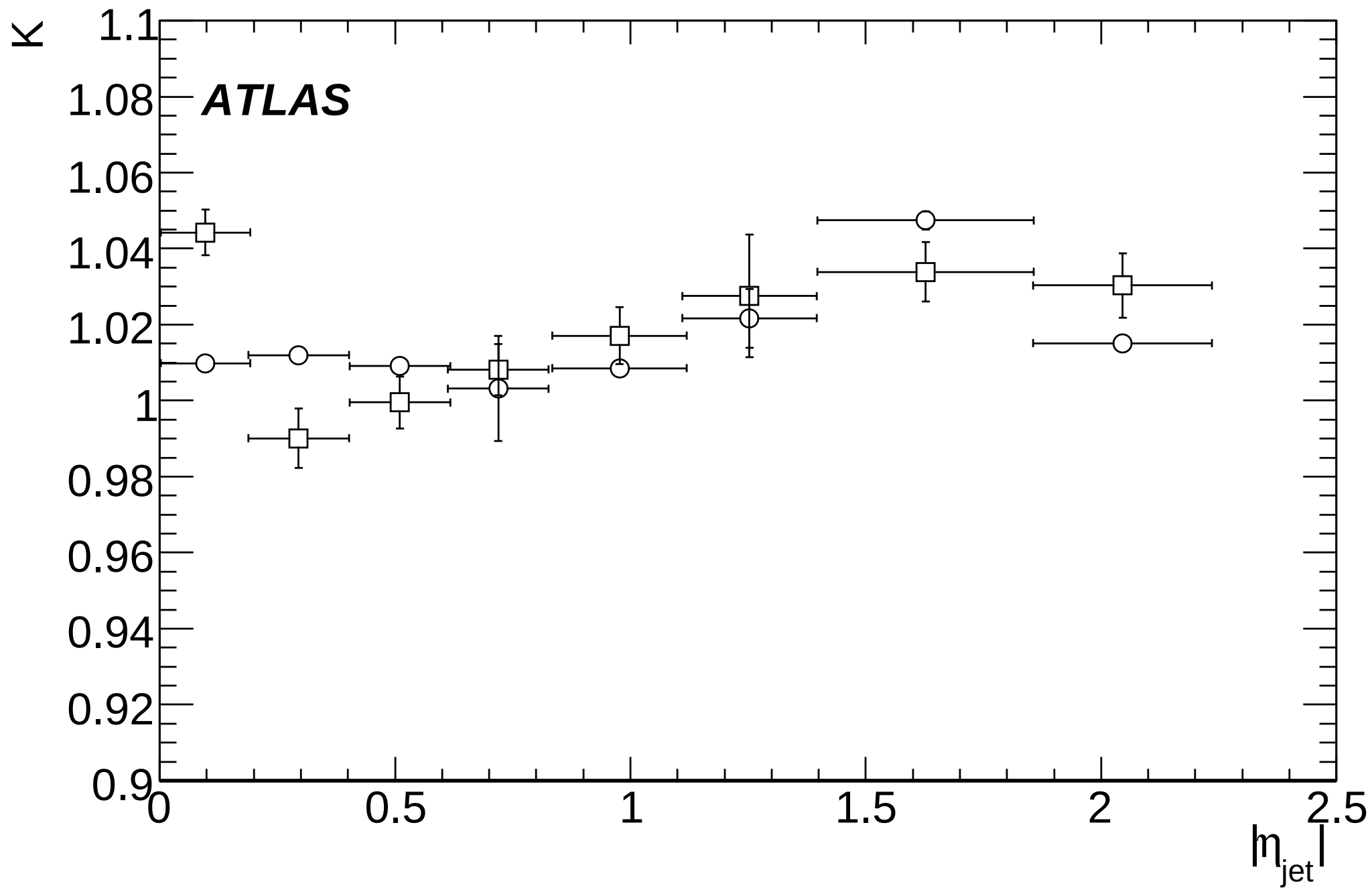


Figure 20: Result of the iterative rescaling fit as a function of $\eta_{jet}$, with 1 fb$^{-1}$: the expected effective calibration factors are shown as square points, in addition to the fitted calibration factors with circles.

#### 3.6.3 Systematic uncertainties

Several checks of the stability of the method have been performed:

- The influence of the combinatorial background, and the choice of using or not events with more than two jets was already described in the previous section. All fitted energy scales are compatible within $\pm$ 0.3 %.
- The stability with respect to the ingredients used in the templates was checked by generating template histograms without the smearing of the angles, or without the correlation between the jet energies, or with a default jet resolution degraded by 20%. The fitted jet energy scales were found to be compatible within $\pm$ 0.3%, and only the fitted relative resolutions were changing. This shows that the method is not very sensitive to the simulation used for the templates.
- The jet energy scales were fitted on $t\bar{t}$ events simulated with various top masses but the same version of the simulation and reconstruction codes. The results are shown in Figure 22. All results are compatible within $\pm$ 0.5%, even when including top masses in a very wide range.

#### 3.6.4 Stability of the template method with smaller integrated luminosities

In order to check the stability of the template method at lower luminosities, the 770 pb$^{-1}$ full dataset was split into sixteen parts, each part thus corresponding to 48 pb$^{-1}$. The fitted jet energy scales in the sixteen pseudo-experiments are shown in Figure 23. The average of the sixteen measurements is 0.9534, in good

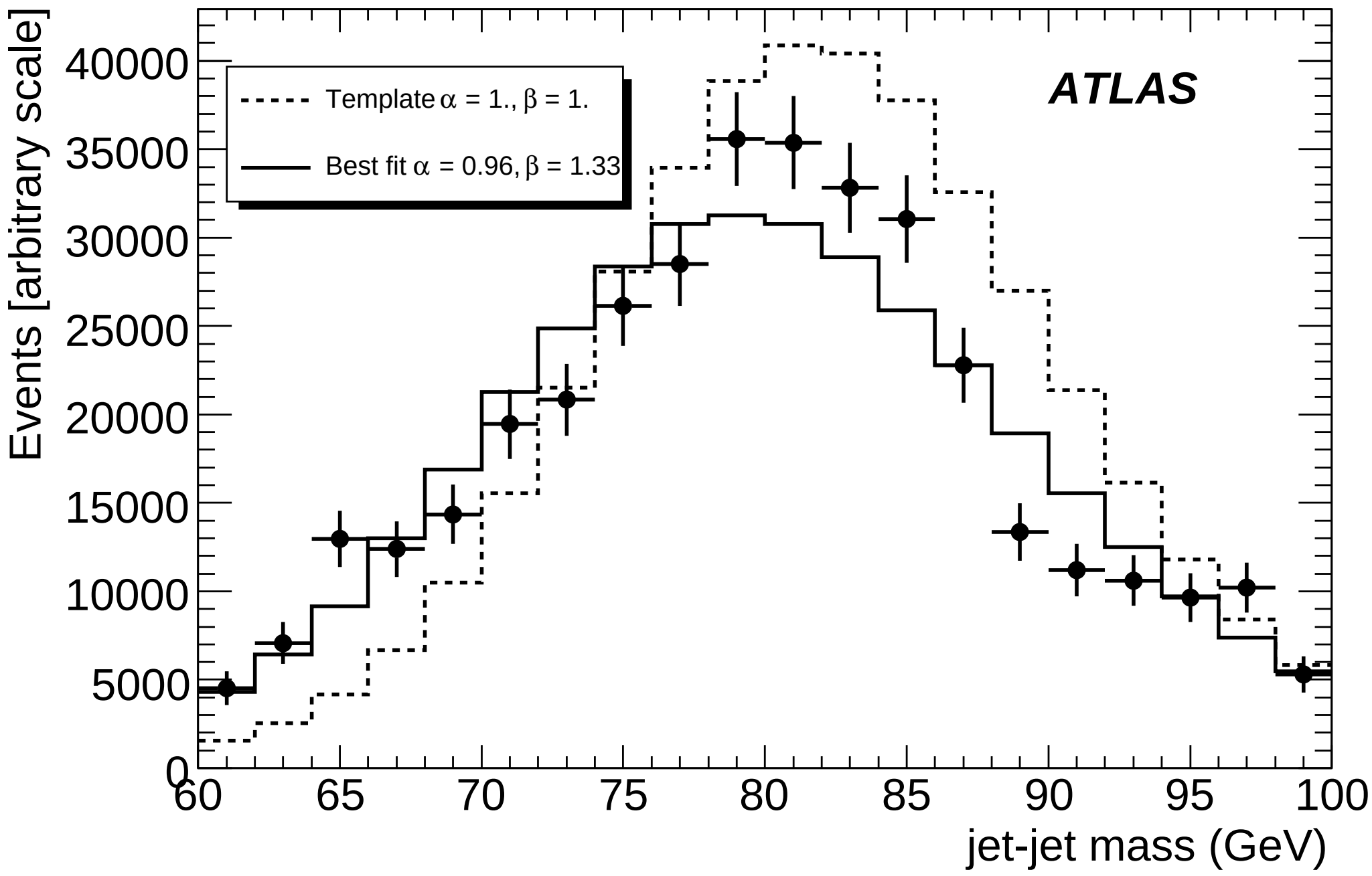


Figure 21: Jet-jet invariant mass in fully simulated events (dots) superimposed on the template histogram with $\alpha = 1$, $\beta = 1$ and the best fit histogram.

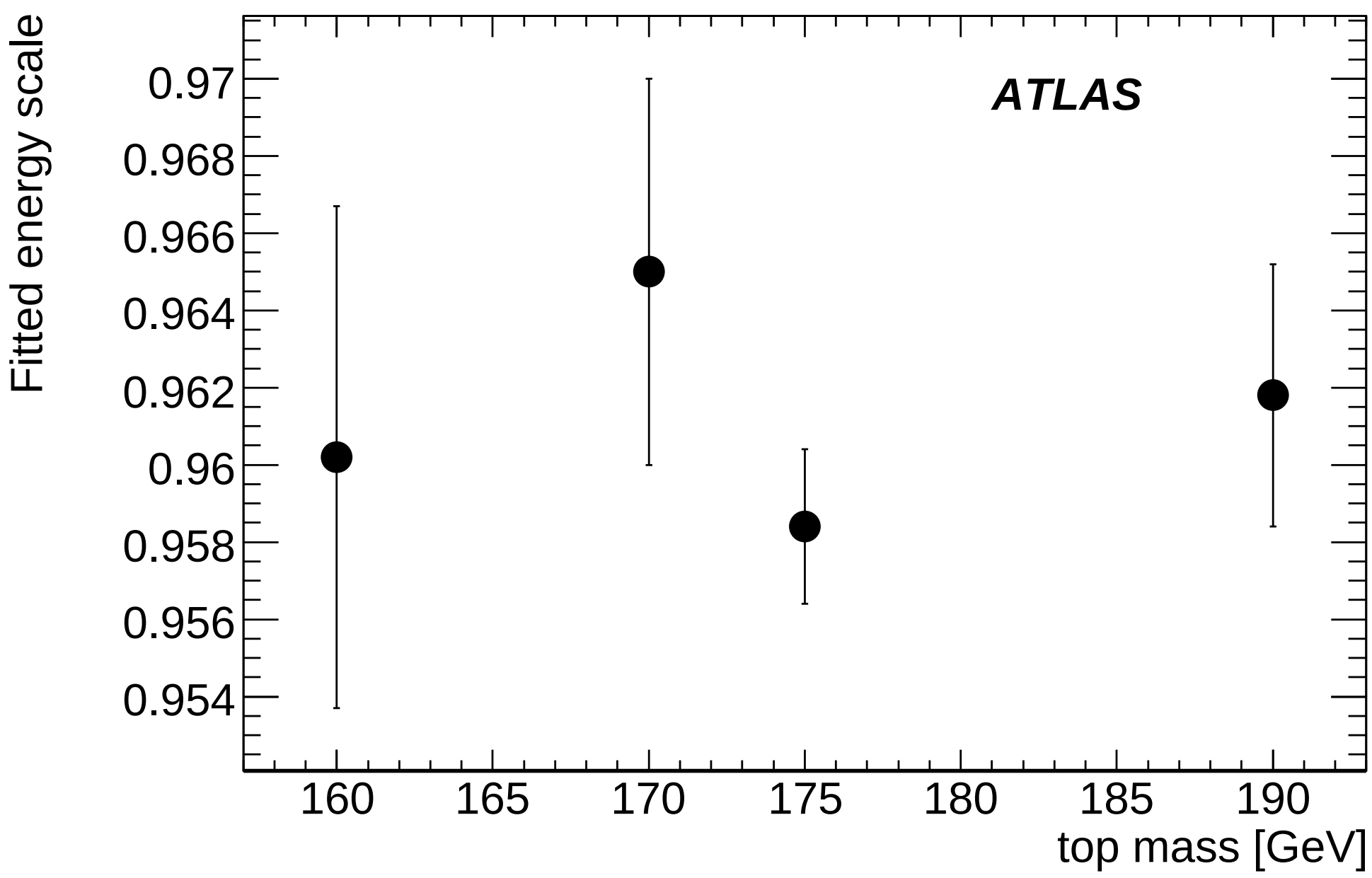


Figure 22: Fitted jet energy scale as a function of the top mass.

agreement with the fit on the full sample ($0.9562 \pm 0.0039$). The RMS of the sixteen measurements is 1.9%, only slightly worse than the value expected by a scaling with the square root of the luminosity (1.6%). In addition, the average of the errors given by the fit is 1.5%, showing that this error is correctly estimated by the fit even at low luminosities.

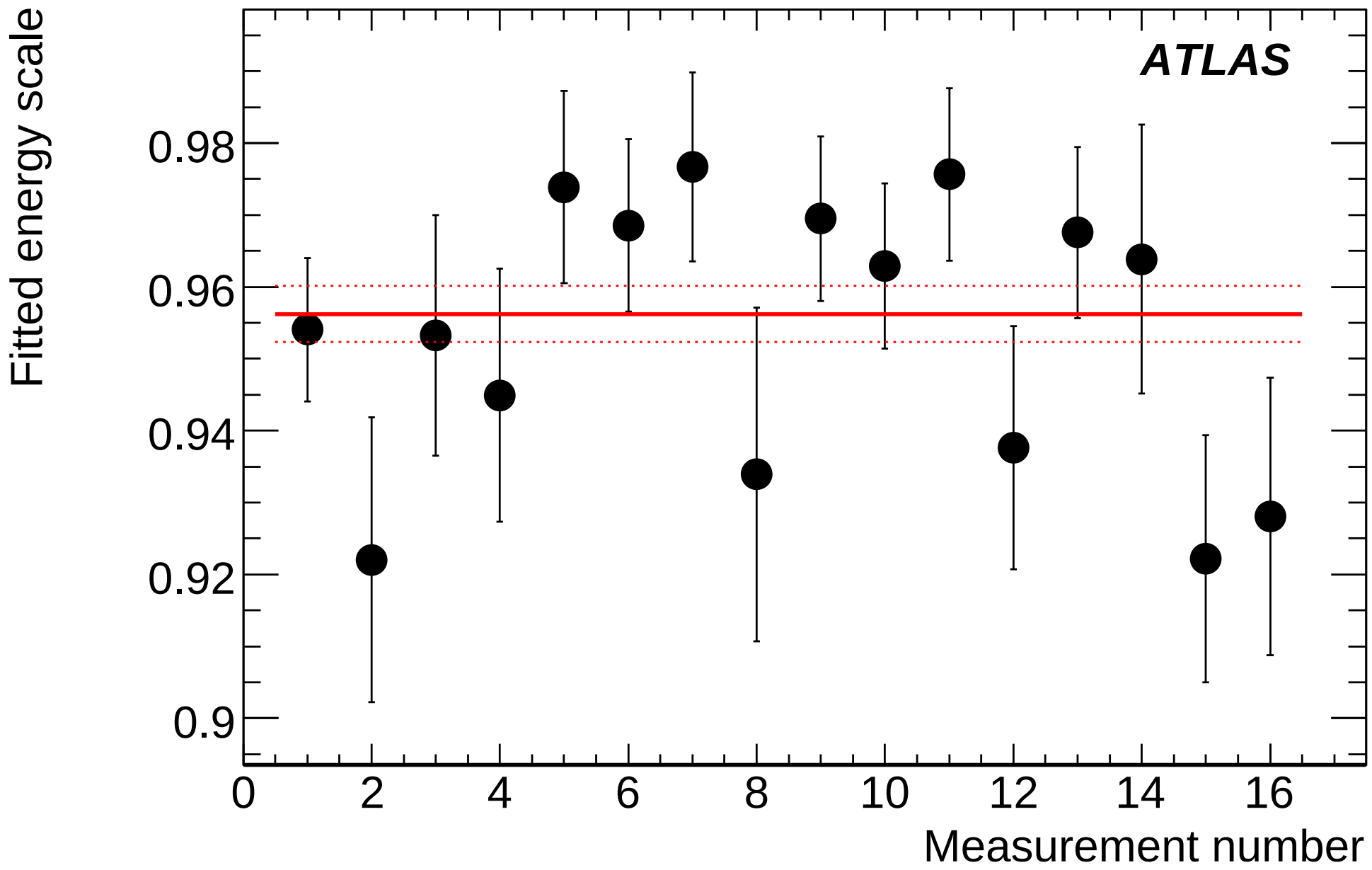


Figure 23: Fitted jet energy scale for sixteen 48 pb$^{-1}$ pseudo-experiments. The horizontal lines show the result of the fit on the full sample with its error bar.

### 3.7 Conclusion on light jet energy scale

Two complementary methods for measuring the light jet energy scale from the $W \to jj$ mass distribution in $t\bar{t}$ events have been studied. The template method is well suited to measure the bare jet energy scale (integrated over some transverse momenta), with a precision which could be around 2% with 50 pb$^{-1}$. Within the statistical precision of the available datasets, no systematic error larger than 0.5% has been identified. The required precision of 1% on the light jet energy scale should thus be achievable. The iterative method, on the other hand, is well suited to measure the jet energy scale of the selected jet sample, possibly as a function of the energy or as a function of rapidity.

In this study. the invariant mass of selected jet pairs present a clear W-boson signal on top of a flat background dominated by combinatorial from top events (Figure 11). This is possible because b jets are tagged with a high efficiency and purity. In the early data-taking phase, b-tagging may be less performant and the level of background will be much higher with contribution from QCD or W+jet events. The performance of the methods for calibration of the light jet energy scale in these conditions need to be evaluated.

The dependence of the jet energy scale with the level of pile-up has to still to be investigated, in order to decide whether a specific correction is needed.

# 4 W-boson mass reconstruction and gluon radiation

The stability of the W-boson reconstruction with different QCD-related jet activity, commonly referred to as initial- and final-state radiation (ISR and FSR) has been evaluated by using the standard HERWIG + MC@NLO simulation and the "low mass" and "high mass" ACERMC + PYTHIA event samples described in [6].

## 4.1 Reconstructed W-boson masses

The same event selection (described in the previous section) and mass reconstructions is applied on the 3 samples, with jets being reconstructed with the $\Delta R = 0.4$ cone algorithm from towers.

Figure 24 shows that the W-boson mass distributions for the 3 samples, normalized to the same number of events after selection, are clearly not compatible before calibration. The results of the mass peak fits are shown in Table 5. The difference between the high mass dataset and the low mass dataset is $1.7 \pm 0.2$ GeV.

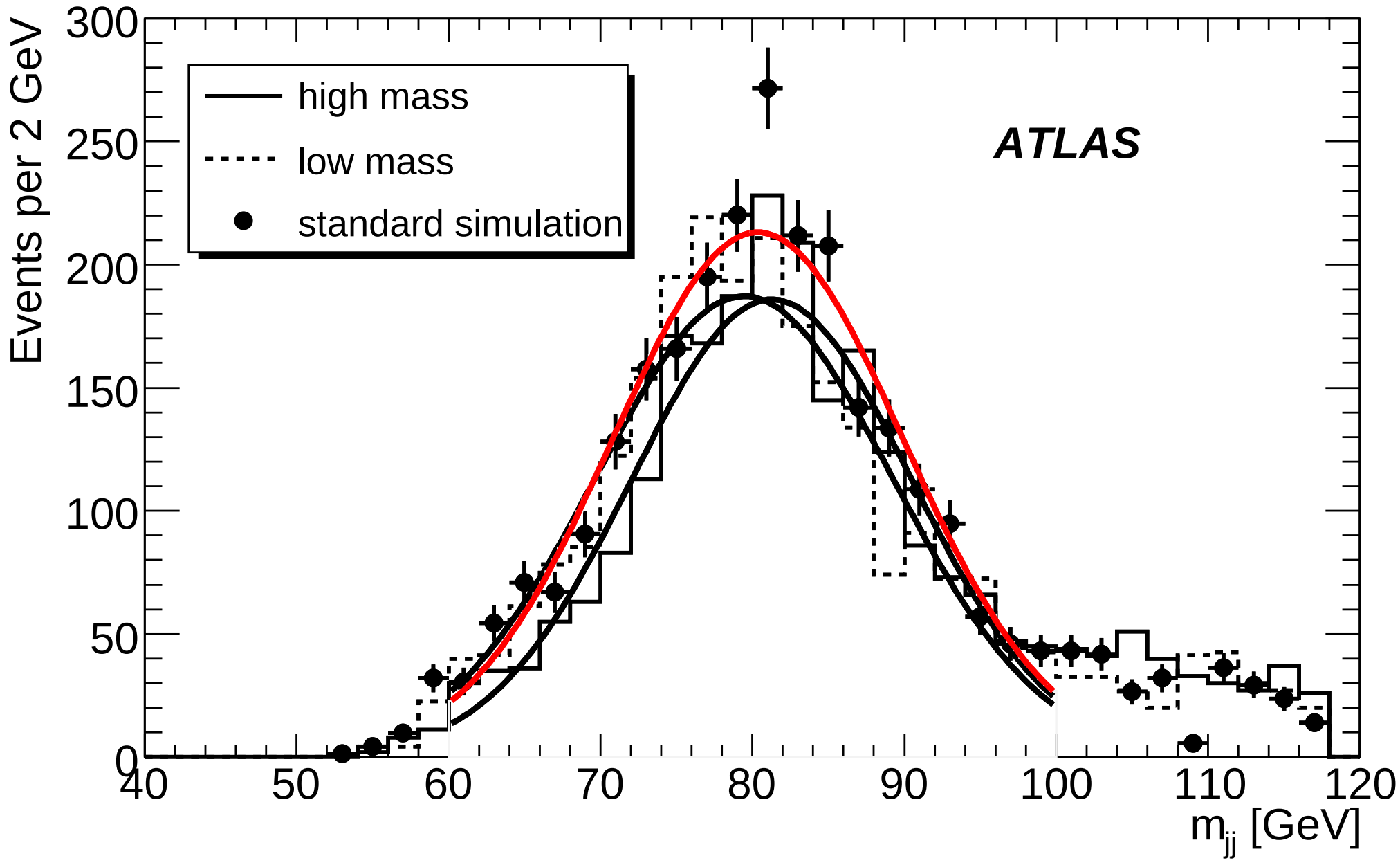


Figure 24: Reconstructed W-boson mass distribution (before calibration) in datasets with different gluon radiation settings. The distributions are normalized to the number of events after selection in the standard simulation and fitted with a gaussian function.

Table 5: Fitted top and W masses for the three datasets with different gluon radiation settings.

| | maximum mass | minimum mass | MC@NLO + Herwig |
|---|---|---|---|
| W mass (GeV) | $81.29 \pm 0.12$ | $79.59 \pm 0.15$ | $80.44 \pm 0.14$ |
| fitted energy scale | $0.9779 \pm 0.0024$ | $0.9489 \pm 0.0026$ | $0.9636 \pm 0.0025$ |
| W mass after calibration (GeV) | $82.46 \pm 0.12$ | $82.38 \pm 0.14$ | $82.63 \pm 0.14$ |

The jet energy scales obtained with the template method described in section 3.6 are also shown in the table, together with the W-boson masses obtained after recalibrating the jet four-momenta. One observes that the calibrated W-boson masses are above the true W-boson mass because of the bias explained in section 3.3, but are all compatible within errors, showing that the calibration method is indeed working well.

## 4.2 Light jet properties

Although the in-situ calibration leads to W-boson mass distributions which are independent of the gluon radiation level, we looked for additional measurable quantities which could help to understand these effects and tune the Monte Carlo simulations.

Figure 25 show the $p_T$ distributions for the light jets coming from the W-boson decay [6]. Only a very small difference, at low energies, is visible, which would probably be very difficult to measure in the data.

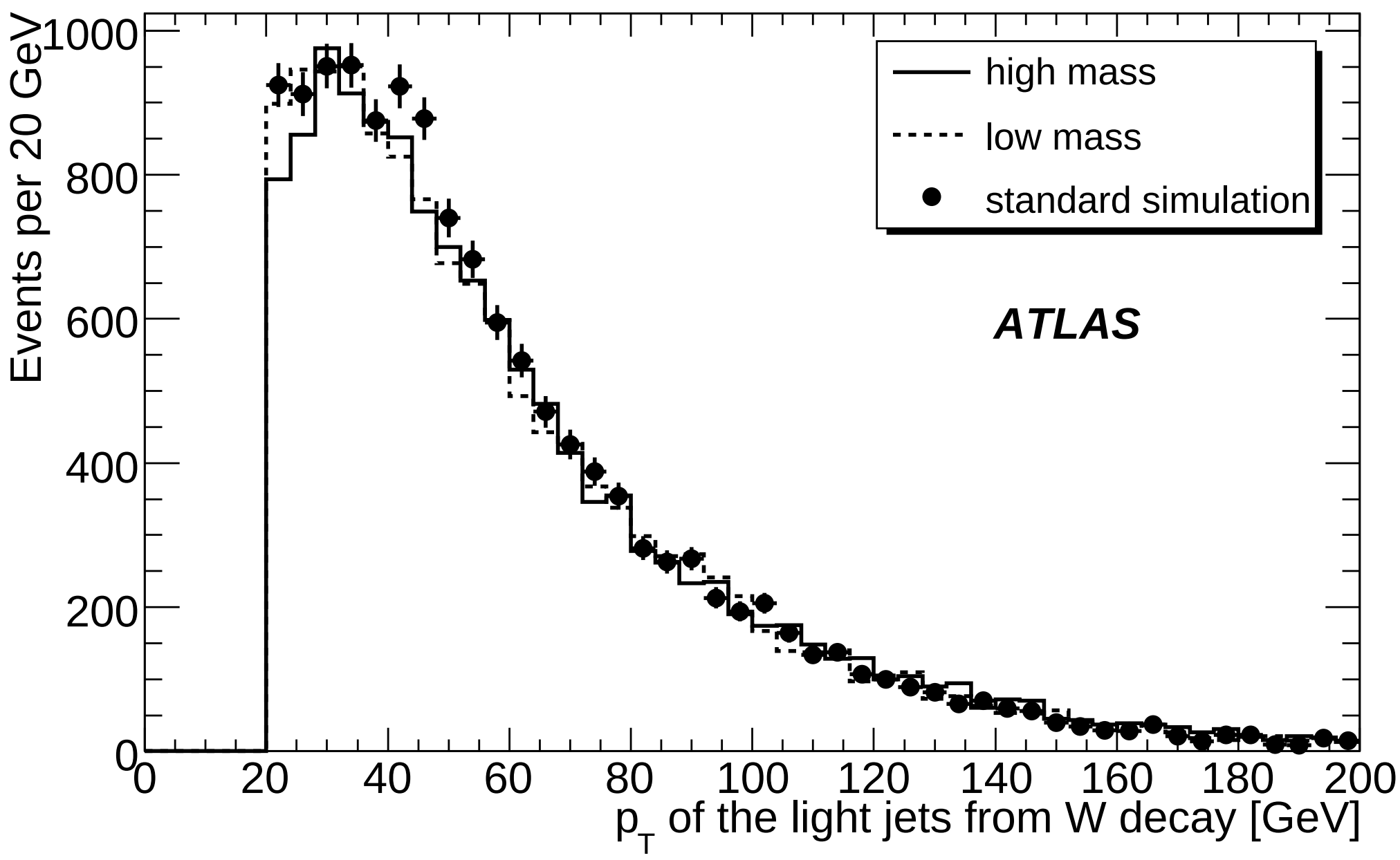


Figure 25: $p_T$ distribution of the jets from the hadronic W-boson decay for datasets with different gluon radiation settings. The distributions are normalized to the number of events after selection in the standard simulation. In order to look for a possible difference at low values, the cut on the $p_T$ value of the jets has been reduced to 20 GeV for this figure.

On the contrary, a very large difference is visible in the number of jets, as can be seen for example in the total number of jets with $p_T > 10$ GeV reconstructed in the events, shown in Figure 26. If the jet reconstruction efficiency is well understood, this distribution in the data could help to tune the simulations, although the number of jets may also depend on the underlying event and may need to be corrected for luminosity.

Finally, no large difference has been seen in the $p_T$ distributions for the jets not assigned to top quark decays, as shown in Figure 27.

[6]A cut $p_T > 20$ GeV is applied in these figures

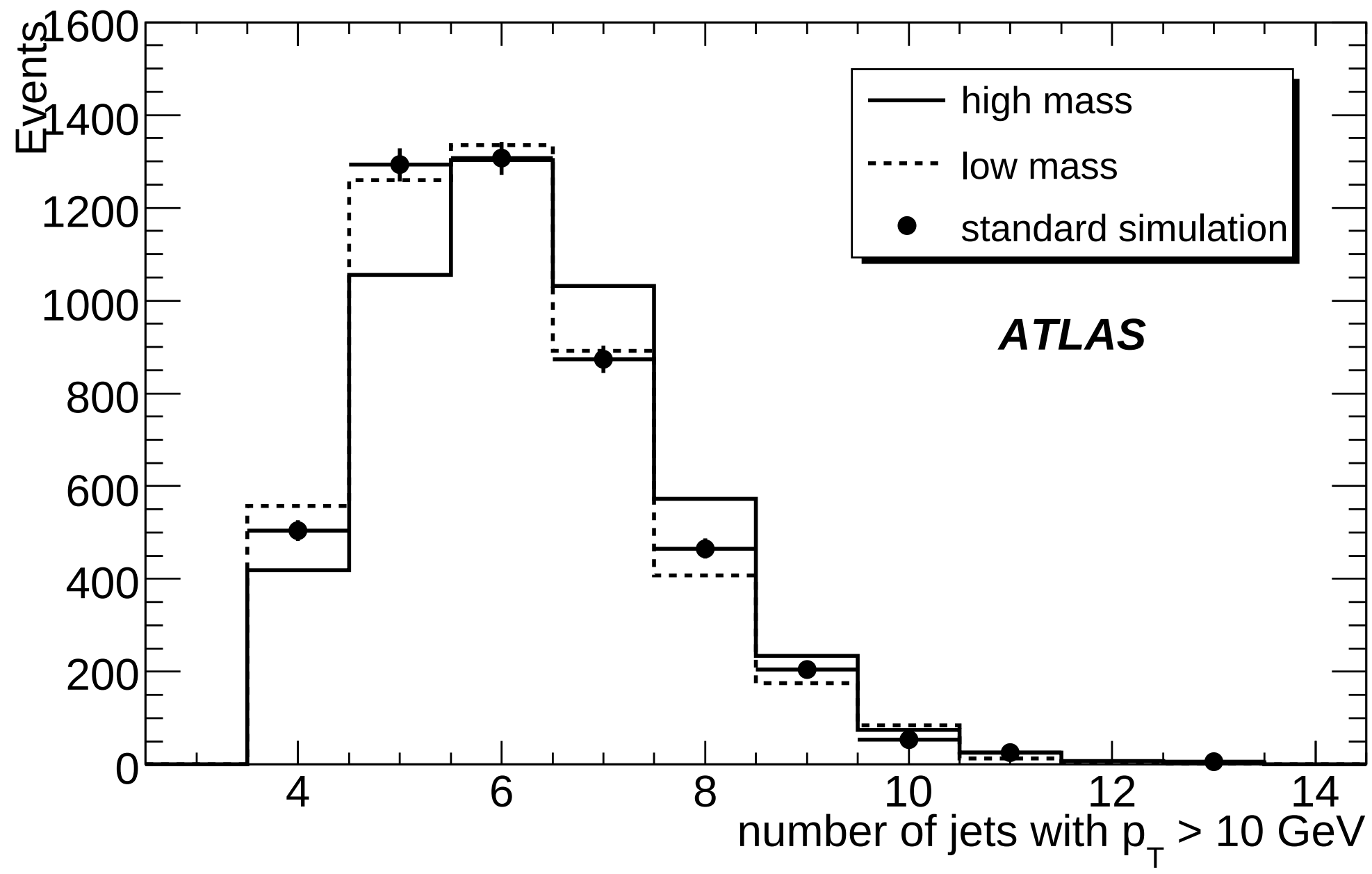


Figure 26: Total number of reconstructed jets with $p_T > 10$ GeV for datasets with different gluon radiation settings. The distributions are normalized to the number of events after selection in the standard simulation.

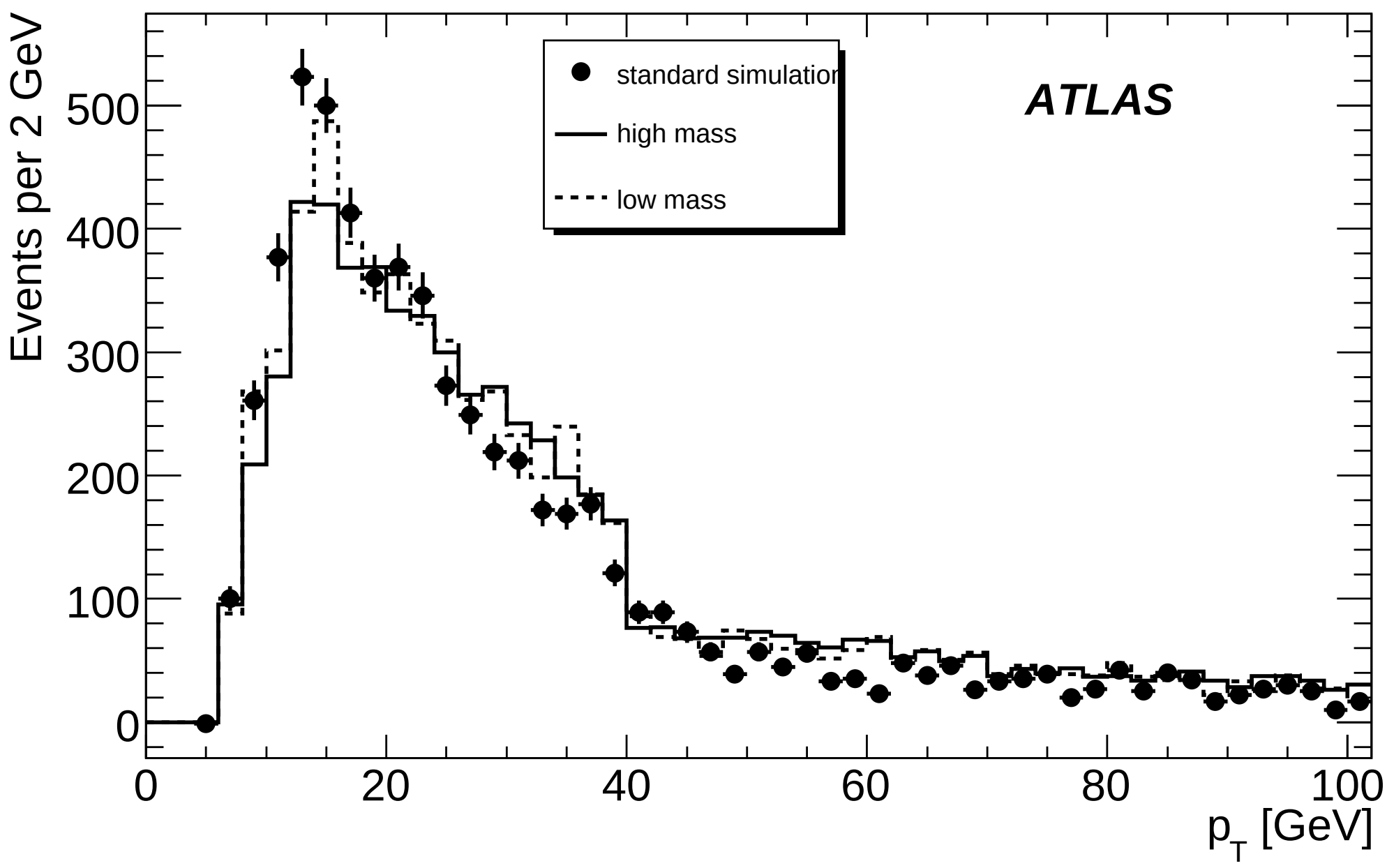


Figure 27: $p_T$ of the additional jets for datasets with different gluon radiation settings. The distributions are normalized to the number of events after selection in the standard simulation.

## 5 Conclusions

This note has presented several aspects of the W-boson reconstruction in $t\bar{t}$ events.

First, it was shown in section 2 that small jet sizes should be prefered to avoid overlap effects. The current default (cone algorithm with $\Delta R = 0.4$) is well suited for the study of the first data, but could be complemented by a reconstruction using topological clusters, which seems robust in the presence of pileup.

Two methods for the in-situ calibration of the light jet energy scale have been studied (section 3). The two methods should be able to provide a light jet energy scale to the 1% level with 1 $\text{fb}^{-1}$, and could perhaps be used to follow the jet energy scale with time and/or luminosity. The two methods use complementary approaches and should be both applied in real data, up to the level of a top quark mass measurement, to allow for cross-checks.

A study of simulations performed with different gluon radiation settings (section 4) has demonstrated that, although the reconstructed W-boson mass can change by a large amount with different settings, the in-situ calibration method is able to correct for it.

More generally, the clean and large W-boson sample which should be available in $t\bar{t}$ events will be a good laboratory to study the performance of jet reconstruction and calibration in real data, and compare them with the performance in Monte Carlo simulated events.

Most studies presented in this note were performed without including pile-up effects. The dependence of the results with luminosity should thus be checked. In addition, a scenario of not yet optimal detector performance, which would lead to higher backgrounds, should also be studied in order to understand the potential of the method for the early phase of data-taking.

Finally, in order to reconstruct properly the top quarks, the knowledge of the *b* jets will be of particular importance. For instance, the *b* jet energy scale is, in our simulation, about 5% lower than the light jet energy scale. This difference should be checked in the data to reach the ultimate precision for the top quark mass measurement.

# Determination of Top Quark Pair Production Cross-Section

**Abstract**
An accurate determination of the top quark pair production cross-section at the LHC provides a valuable check of the Standard Model. Given the high statistics which will be available, corresponding to about one top-quark pair per second, at a luminosity of $10^{33}$ cm$^{-2}$s$^{-1}$, the cross-section measurement can be performed relatively fast after the turnon of the LHC. Prospects for measuring the total top quark pair production cross-section with the ATLAS detector during the initial period of LHC are presented in this note. The cross-section is determined in the semi-leptonic channel, and in the dilepton channel. For the semi-leptonic channel we perform the measurement both with and without relying on the tagging of b-quark initiated jets.

## 1 Introduction

The determination of the top quark pair production cross section is one of the measurements that will be carried out once the first data samples are available at the ATLAS experiment. It casts light on the intrinsic properties of the top quark and its electroweak interactions. Cross section measurements are also an important test of possible new production mechanism, as non Standard Model top quark production can lead to a significant increase of the cross section. New physics may also modify the cross section times branching ratio differently in various decay channels, as for example predicted by Supersymmetric models [1] with charged Higgs particles, $t \rightarrow H^- \bar{b}$, or with super-partners of the top quark, $t \rightarrow \tilde{t}\chi_0$. The selection of top quark events is based on the identification of a jet from a b-quark, assuming a branching ratio $BR(t \rightarrow Wb) = 1$. The consistency of this assumption, performed with kinematic methods, is another important check of the Standard Model prediction but falls outside the scope of this paper.

Last but not least, the top pair production process will be valuable for the in-situ calibration of the ATLAS detector during the commissioning phase. The large cross section and the large signal to background ratio for the semileptonic channel, allows to identify high purity samples with large statistics in a short period of time. Understanding the experimental signatures of top events involves most parts of the ATLAS detector and is essential for claiming discoveries of new physics.

The cross-section values and the Monte Carlo samples which have been used throughout this note, are described in [9].

#### 1.0.1 Cross section measurements at Tevatron

During Tevatron Run I (1992-1996) an integrated luminosity of about 100 pb$^{-1}$at a centre of mass energy of $\sqrt{s} = 1.8$ TeV allowed to measure top pair production cross sections of $6.5^{+1.7}_{-1.4}$ pb and $5.7 \pm 1.6$ pb by the CDF and DØ collaborations respectively [2]. The Tevatron Run II started in 2001 and until spring 2006 about 1 fb$^{-1}$ of $p\bar{p}$ collisions at $\sqrt{s} = 1.96$ TeV have been collected and analysed. At this higher centre of mass energy an increase of about 30% in the cross section is expected. The most recent calculations predict a cross section of $6.7^{+0.7}_{-0.9}$ pb [3] at NLO+NLL or $6.8 \pm 0.6$ pb [4] at NLO+ threshold resummation for a top mass, $m_t$ = 175 GeV. CDF and DØ measured a combined channels cross-section equal to 7.3±0.5(stat) ±0.6(sys)±0.4(lum) [5] and 7.4±0.5(stat)±0.6(sys)±0.4(lum) [6] respectively. All the measurements are in good agreement with the predictions. From a combination of all results, an experimental error of the order of the theoretical error is expected.

# 2 Single lepton channel

In this section the strategy for the determination of the $t\bar{t}$ cross-section in the semi-leptonic decay mode is described. This channel, which has a branching fraction of approximately 45%, has a clear signature, is experimentally easily accessible and is expected not to suffer from large backgrounds.

A robust analysis is presented of the first 100 pb$^{-1}$ ATLAS data, which are expected to be collected during the first few months of the LHC data taking period. In particular it is studied whether a pure $t\bar{t}$ sample can be identified without utilizing the full ATLAS b-tagging capabilities. This is brought about by the fact that efficient tagging of jets originated from the hadronization of b-quarks, called b-tagging from now on, is non trivial and implies a precise alignment of the inner detector, which will probably require several months of data taking. This analysis solely relies on the measurement of jets, leptons and $E_{\mathrm{T}}^{\mathrm{miss}}$ (transverse missing energy), and requires a functioning lepton triggering system. Thanks to the over-constrained kinematics of the $t\bar{t}$ system, with the selected events it will be possible to measure the b-tagging performance and the $E_{\mathrm{T}}^{\mathrm{miss}}$, as well as to calibrate the light jet energy scale.

## 2.1 Event selection

The identification of semi-leptonic $t\bar{t}$ events starts by requiring a highest level (event filter) lepton trigger to have fired. In this study we assume that either the single isolated electron trigger `e22` or the muon trigger `mu20` (for the definition see [7]) has fired. A correct description of the trigger efficiencies is vital for the cross-section determination. The strategy for determining the trigger efficiencies from the Monte Carlo, as well as from the data without relying on Monte Carlo, is not pursued in this note, but included in [7].

Further, we define a candidate $t\bar{t}$ event as having one reconstructed high-$p_T$ isolated lepton (electron or muon), a minimal amount of missing energy and at least four reconstructed jets. The definition of electrons, muons and jets in our analysis has been discussed before.

For our default off-line selection the events are required to fulfil the following:

- One lepton (electron or muon) with $p_T > 20$ GeV.
- $E_{\mathrm{T}}^{\mathrm{miss}} > 20$ GeV.
- At least four jets with $p_T > 20$ GeV.
- Of which at least three jets with $p_T > 40$ GeV.

The fraction of events passing the individual selection requirements and the overall selection efficiency are shown in Table 1 for semi-leptonic events. In this table we split the efficiencies for semi-leptonic $t\bar{t}$ events according to the $W$ decay in the Monte Carlo generator: $t\bar{t}$ (electron) where it decayed to an electron and a neutrino and $t\bar{t}$ (muon) where it decayed to a muon and a neutrino.

We observe a combined efficiency for these requirements which is somewhat larger for the $t\bar{t}$ (muon) events compared to the $t\bar{t}$ (electron) events.

### 2.1.1 Reconstructing $t\bar{t}$ events

Before discussing additional requirements to improve the purity of the $t\bar{t}$ event selection, we present the second step in the event reconstruction. In this step we test the events for compatibility with a $t\bar{t}$ hypothesis. In the $t\bar{t}$ candidates, three of the reconstructed jets are expected to form the hadronic top-quark. In the absence of b-tagging there is an additional ambiguity in choosing the correct three-jet

Table 1: Fraction of events passing the various selection criteria and the combined 'default' selection efficiency for semi-leptonic (electron and muon) analyses respectively. The statistical uncertainties on these numbers are negligible.

| | Trigger eff (%) | Lepton eff (%) | $E_T^{miss}$ eff (%) | Jet req. (I) eff (%) | Jet req. (II) eff (%) | Combined eff (%) |
|---|---|---|---|---|---|---|
| $t\bar{t}$ (electron) | 52.9 | 52.0 | 91.0 | 70.7 | 61.9 | 18.2 |
| $t\bar{t}$ (muon) | 59.9 | 68.7 | 91.6 | 65.5 | 57.3 | 23.6 |

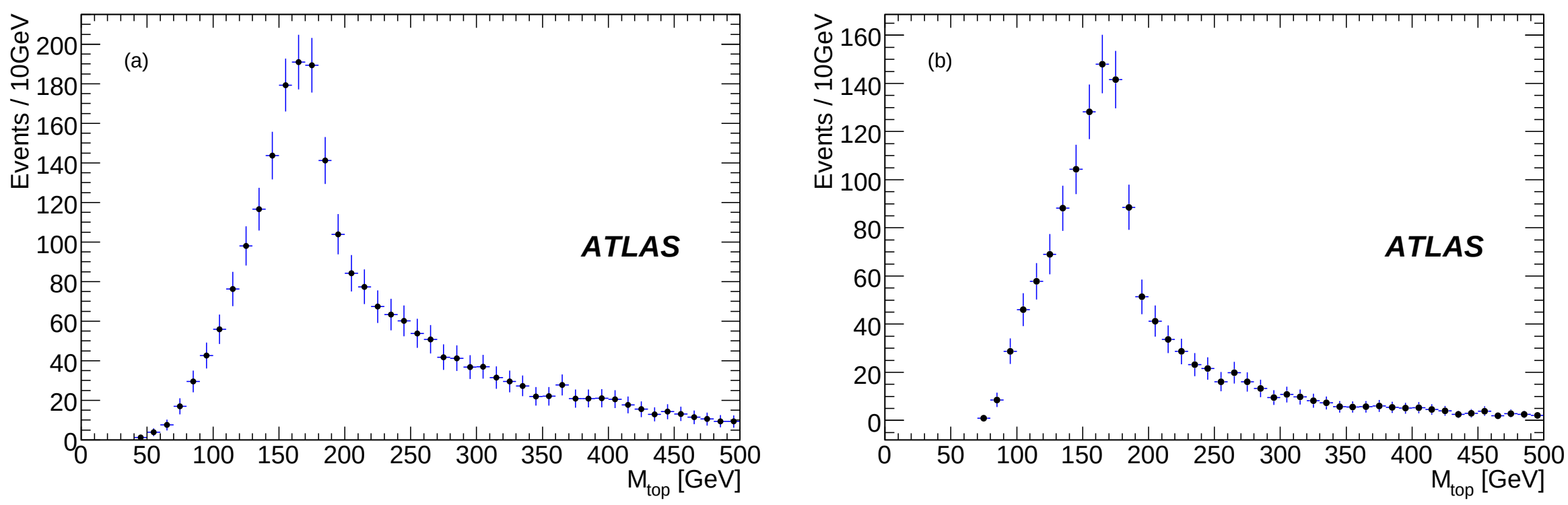


Figure 1: (a): Three-jet invariant mass distribution for the electron analysis default selection, normalised to 100 pb$^{-1}$. The statistical errors in each bin are indicated. (b): The same distribution after the additional $W$-boson mass constraint..

combination among the reconstructed jets. We define our top-quark decay candidate as the three-jet combination of all jets that has the highest transverse momentum sum.

Fig. 1 (a) shows the reconstructed top mass for this selection (from now on referred as default selection) for the $t\bar{t}$ sample. The top mass peak is clearly visible, and the tails of the distributions correspond to the combinatorial background.

### 2.1.2 Selection variations: I

Apart from the default event selection as described above, a number of additional criteria are defined to further increase the purity of the top sample. Here we improve on the simple $t\bar{t}$ analysis by exploiting additional information: every three-jet combination that originates from a top decay also contains a two-jet combination that originates from a $W$-boson decay. To illustrate the presence of the $W$-boson we take the three jets that constitute the top quark, and select from the three combinations of di-jets the one that results in the highest value of the sum of the $p_T$ of the two jets. The $W$-boson mass is then the invariant mass of the two jet system. In Fig. 2 (a) this mass distribution is shown for the electron analysis, and the $W$-boson mass peak around 80 GeV is clearly visible.

However we prefer an unbiased $W$-boson mass distribution in the analysis, for which we choose not to pick/define one particular $W$-boson di-jet pair out of the three combinations, but rather require that at least one of the three di-jet invariant masses is within 10 GeV of the reconstructed mass of the $W$-boson (taken as the peak value of the mass distribution of the $W$-boson candidates). This selection will be referred to as the $W$-boson mass constraint selection. The distribution of all three di-jet combinations in

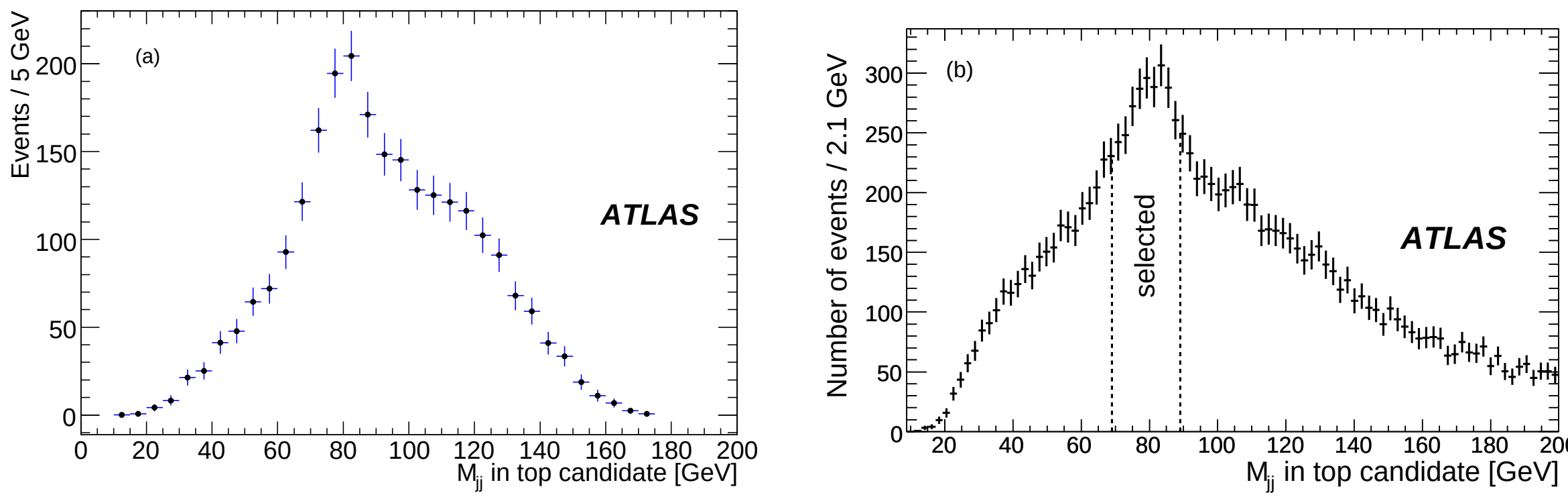


Figure 2: (a): The di-jet combination with highest $p_T$ (left) for the electron analysis. (b): The three di-jet combinations invariant masses among the top-quark candidates in a 100 pb$^{-1}$ event sample for the muon analysis.

the top candidate is shown in Fig. 2 (b). Note that each event enters three times in this distribution. In this figure the background, as discussed in the next section, is already included.

The distribution of the three-jet invariant mass after the additional requirement that at least two jets are compatible with the mass of the $W$-boson is shown in Fig. 1 (b). This requirement shows a substantial reduction in the $t\bar{t}$ combinatorial background compared to Fig. 1 (a). Notice that, compared to the default selection, the top mass peak becomes narrower and the tail of the distribution is reduced. However, the $W$-boson mass constraint also introduces a visible shoulder in the distribution which makes fitting to the data more subtle.

In Table 2 we show the fraction of $t\bar{t}$ events that pass these various selection requirements.

Table 2: Efficiencies at different stages of the electron and muon analyses for several event types: after trigger and event selection (left column), after a cut on the di-jet masses (see text for details) in the top candidate (middle column) and events with, in addition to the di-jet mass cut, a hadronic top mass $141 < m_t < 189$ GeV (right column). The first three rows correspond to the single-lepton final states, the fourth row to the di-lepton final state and the last row to the hadronic final state.

| Event type | **Electron analysis** Trigger+Selection (%) | | | **Muon analysis** Trigger+Selection (%) | | |
|---|---|---|---|---|---|---|
| | | $W$ const. | $m_t$ win | | + $W$ const. | $m_t$ win |
| $t\bar{t}$ (elec) | 18.2 | 9.2 | 4.5 | 0.1 | 0.0 | 0.0 |
| $t\bar{t}$ (muon) | 0.0 | 0.0 | 0.0 | 23.6 | 12.0 | 5.8 |
| $t\bar{t}$ (tau) | 1.4 | 0.7 | 0.3 | 2.0 | 1.0 | 0.5 |
| $t\bar{t}$ (di-lepton) | 2.2 | 1.0 | 0.2 | 3.0 | 1.3 | 0.4 |
| $t\bar{t}$ (hadron) | 0.0 | 0.0 | 0.0 | 0.1 | 0.1 | 0.0 |

### 2.1.3 Background evaluation

We consider a number of background processes. The dominant expected background is $W$-boson+jets, but also single top production, $Z$-boson+jets and $Wb\bar{b}$ are sizeable. Tables 3 and 4 summarise the ex-

Table 3: Number of events which pass the various electron selection criteria for the $t\bar{t}$ signal and for the most relevant backgrounds normalised to 100 pb$^{-1}$.

| **Electron analysis** | | | | | | |
|---|---|---|---|---|---|---|
| Sample | default | $W$ const. | $m_t$ win | $W$ const. + $\|\eta\| < 1$ | $W$ const. + 1 b-tag | $W$ const. + 2 b-tag |
| $t\bar{t}$ | 2555 | 1262 | 561 | 303 | 329 | 208 |
| hadronic $t\bar{t}$ | 11 | 4 | 0.0 | 0.8 | 0.6 | 0.0 |
| $W$+jets | 761 | 241 | 60 | 38 | 7 | 1 |
| single top | 183 | 67 | 23 | 12 | 18 | 7 |
| $Z\rightarrow ll$ +jets | 115 | 35 | 8 | 5 | 2 | 0.4 |
| $W\ b\bar{b}$ | 44 | 15 | 3 | 5 | 5 | 0.7 |
| $W\ c\bar{c}$ | 19 | 6 | 1 | 1 | 0.4 | 0.0 |
| $WW$ | 7 | 4 | 0.4 | 0.0 | 0.0 | 0.0 |
| $WZ$ | 4 | 1 | 0.4 | 0.2 | 0.0 | 0.0 |
| $ZZ$ | 0.5 | 0.2 | 0.1 | 0.0 | 0.0 | 0.0 |
| Signal | 2555 | 1262 | 561 | 303 | 329 | 208 |
| Background | 1144 | 374 | 96 | 63 | 33 | 10 |
| S/B | 2.2 | 3.4 | 5.8 | 4.8 | 10.0 | 20.8 |

pected numbers of signal and background events for the electron and muon analysis respectively. The first column of the two tables shows the event numbers obtained by applying the default selection, whilst the second column gives the corresponding numbers with the $W$-boson mass constraint. All numbers are normalised to 100 pb$^{-1}$. The evaluation of the QCD fake rate deserves a separate discussion. The QCD production of $pp \rightarrow b\bar{b}$ is characterised by a cross-section of about 100 $\mu$b, and can therefore be an important background for our signal. Requiring the presence of a high $p_T$ lepton and missing energy can reduce its contribution, but since the cross-section enhancement relative to the signal is so large, there might be QCD events with a fake lepton and/or poor missing energy reconstruction that pass these requirements as well.

The rate for extra (medium [8]) electrons is studied and found to be roughly $1.0 \times 10^{-3}$ per jet. This number is divided between semi-leptonic $B(D)$ decays and true fakes, i.e. hadronic objects identified as electrons. The origin of extra isolated muons is dominated by semi-leptonic $B$ decays, i.e. by the presence of hard b-quarks. The isolated muon rate per b-parton reaches a few times $10^{-3}$ for b-parton momenta around 40 GeV, while the fake rate is only a few times $10^{-5}$. By studying their origin and dependence on jet/parton kinematics like the $p_T$, $\eta$, jet multiplicity and quark content of the jet, we can get an estimate of the fraction of multi-jet events that will pass the lepton requirement in the event selection. The validity of this approach has been checked using a large sample of di-jet events at various transverse momenta. As a result, the QCD background has been evaluated to be smaller than the $W$-boson+jets background and will not be discussed further.

The distribution of the invariant mass of the three-jet combination that forms the hadronic top-quark candidate with the default selection and with the backgrounds added together, is shown in Fig. 3 (a). The events where the correct jets were selected to reconstruct the hadronically decaying top quark candidate are clearly visible as the mass peak (open histogram) on top of a smooth background distribution. This background is partially composed of events from non-top processes (light shaded histogram), but is dominated by the (combinatorial) background from semi-leptonic $t\bar{t}$ events (dark shaded histogram). The combinatorial background was determined using the matching of the top candidate with the generated

Table 4: Number of events which survive the various muon analysis requirements for the $t\bar{t}$ signal and for the most relevant backgrounds normalised to 100 pb$^{-1}$.

| Muon analysis | | | | | | |
|---|---|---|---|---|---|---|
| Sample | default | $W$ const. | $m_t$ win | $W$ const. + $\|\eta\| < 1$ | $W$ const. + 1 b-tag | $W$ const. + 2 b-tag |
| $t\bar{t}$ | 3274 | 1606 | 755 | 386 | 403 | 280 |
| hadronic $t\bar{t}$ | 35 | 17 | 7 | 6 | 5 | 2 |
| $W$+jets | 1052 | 319 | 98 | 47 | 11 | 0.0 |
| single top | 227 | 99 | 25 | 19 | 19 | 10 |
| $Z \rightarrow ll$ +jets | 84 | 23 | 3 | 2 | 0.5 | 0.0 |
| $W\,b\bar{b}$ | 64 | 19 | 4 | 4 | 5 | 2 |
| $W\,c\bar{c}$ | 26 | 9 | 3 | 0.7 | 0.1 | 0.0 |
| $W\,W$ | 7 | 3 | 0.7 | 0.7 | 0.0 | 0.0 |
| $W\,Z$ | 7 | 3 | 0.8 | 0.5 | 0.0 | 0.0 |
| $Z\,Z$ | 0.7 | 0.3 | 0.1 | 0.0 | 0.0 | 0.0 |
| Signal | 3274 | 1606 | 755 | 386 | 403 | 280 |
| Background | 1497 | 495 | 143 | 84 | 42 | 14 |
| S/B | 2.2 | 3.2 | 5.3 | 4.6 | 9.6 | 20.1 |

top-quark in a cone of size $\Delta R < 0.2$.

In Fig. 3 (b) the reconstructed three-jet mass after the $W$-boson mass constraint is presented. The background is also shown.

Table 3 and 4 show the number of signal and background events in a 100 pb$^{-1}$ data sample. To give an indication of the signal purity in the top mass peak region, in the third column of Tables 3 and 4 we give the number of events in a hadronic top mass region: $141 < m_t < 189$ GeV. Although not all signal events are correctly reconstructed, in both the electron and muon analyses the purity of the signal in the top mass window is close to 80%.

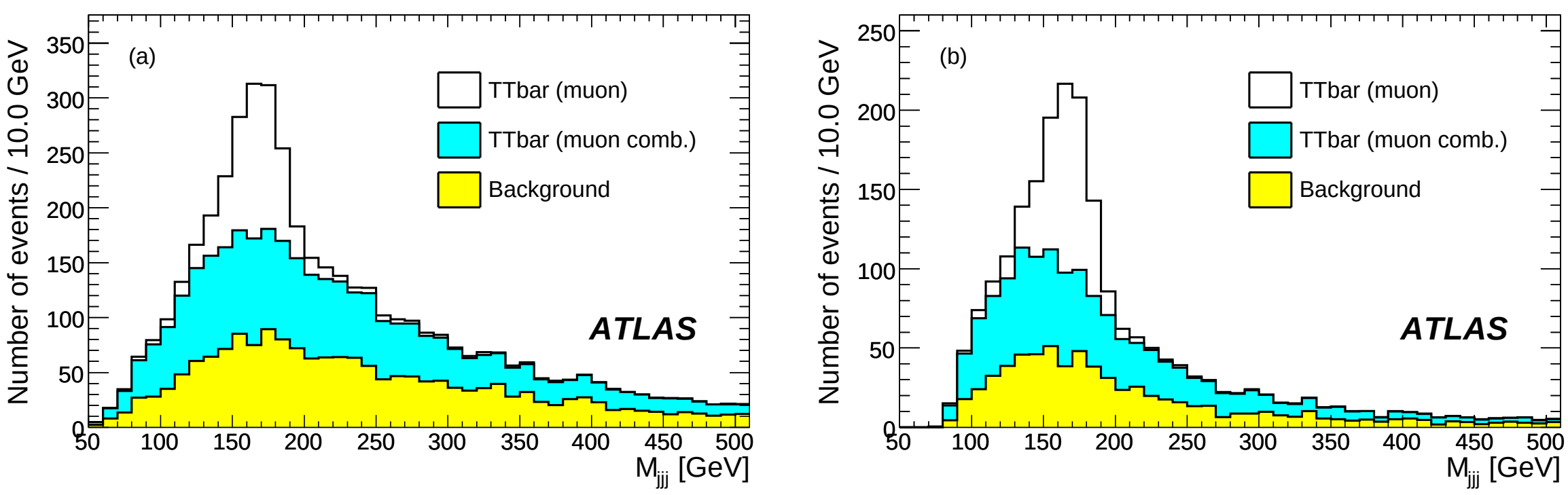


Figure 3: (a): Expected distribution of the three-jet invariant mass after the standard selection. The white area represents the $t\bar{t}$ signal in the muon channel. The dark shaded area is the combinatorial background and the light shaded area represents the background contribution. (b): The same after the $W$-boson mass constraint in a 100 pb$^{-1}$event sample. Both plots are for the muon analysis.

### 2.1.4 Selection variations: II

Additional ways to kinematically select top events other than the $W$-boson mass constraint, or to improve the signal purity after having applied the $W$-boson mass cut itself, were explored. In the commissioning phase, it can happen that the barrel calorimetry will be better calibrated than the forward one. Therefore, it can be useful to apply the additional request that the three highest $p_T$ jets are all at $|\eta| < 1$. The reconstructed top mass in this case is shown in Fig. 4.

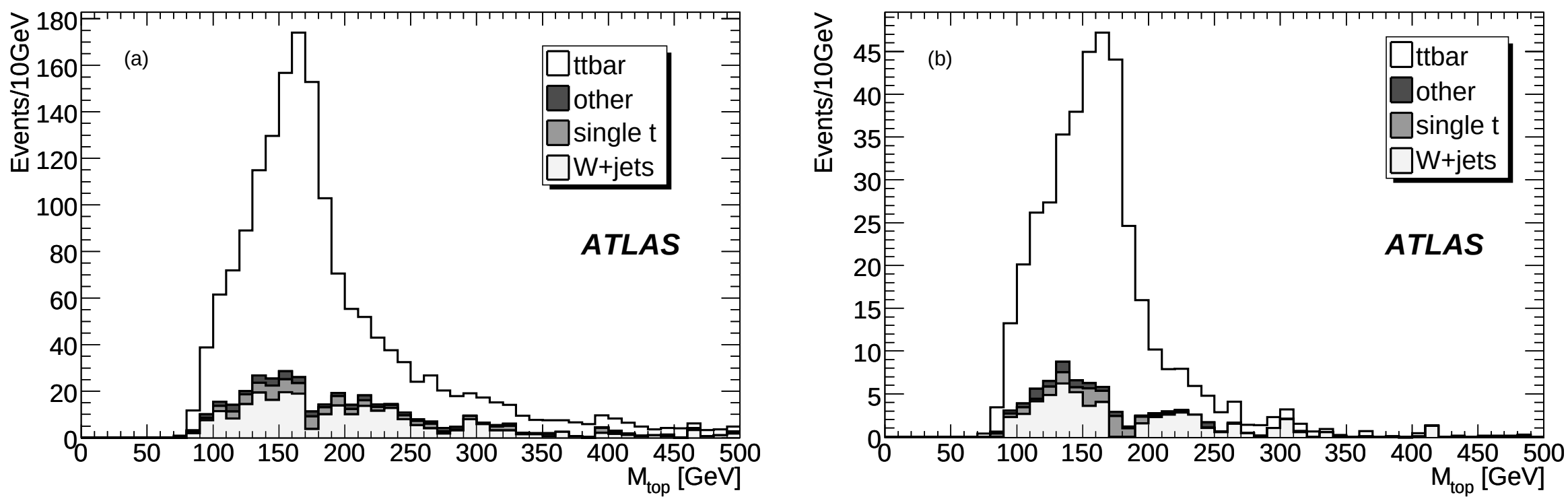


Figure 4: (a): Reconstructed top mass after $W$-boson mass constraint for the electron analysis. The white area represents the $t\bar{t}$ signal in the electron channel, while the three shaded areas corresponds, going from the lighter to the darker, to the background from $W$-boson+jets, single top and all the other background sources considered in the analysis. The distribution is normalised to 100 $pb^{-1}$. (b): The same distribution, but in addition requiring that the three highest $p_T$ jets are at $|\eta|<1$.

The centrality requirement applied after the default selection allows to reach the same signal-over-background that one obtains after applying the $W$-boson constraint. Tables 3 and 4 show the signal-over-background and signal efficiencies for the electron and muon analyses if the centrality requirement is applied in addition to the $W$-boson constraint (fifth column).

Other variables were exploited as well, like the $\cos\theta^*$ [1] and the total invariant mass of the event. In the following no cuts on these variables are used in the analysis.

## 2.2 Determination of the cross-section

In this section two complementary methods to determine the $t\bar{t}$ cross-section of the commissioning analysis are presented. The first method estimates the $t\bar{t}$ signal by performing a maximum likelihood fit on the three-jet invariant mass distribution. The second is based solely on counting the number of top candidate events that pass the selection, and subtracting all backgrounds in order to get the yield of $t\bar{t}$ events in the sample. The two methods are affected by different systematics. Whereas this counting method needs all backgrounds to be properly addressed and normalised, it does not rely on a correct reconstruction of the top quark. The peak-fit method is rather insensitive to background normalisation and efficiencies, but requires a fairly well understood top mass peak.

[1] It is the angle that one jet forms with the direction of the incoming proton in the centre of mass of the event (it is expected that the top decay products are emitted more centrally than the $W$-boson+jets and jets from QCD)

### 2.2.1 Likelihood fit method

To extract the number of completely reconstructed $t\bar{t}$ events (after having applied the default + $W$-boson mass constraint selection) a maximum likelihood fit is performed to the three-jet mass distribution with a Gaussian signal on top of the background described by a Chebychev polynomial, see Fig. 5. It has been verified that the background model correctly describes the combined $t\bar{t}$ combinatorial and background distribution in the signal region by comparing the fitted background to the subset of events that are not fully reconstructed signal as determined from truth matching information.
Using 10000 pseudo-experiments, based on the input from the full simulation Monte Carlo events, one can extract the average fraction of signal events that pass all selection requirements and enter in the peak (i.e., the fraction correctly reconstructed). The average number of events in the peak in the muon analysis, i.e. correctly reconstructed semi-leptonic $t\bar{t}$ (muon) events, in 100 pb$^{-1}$ is 508 events as shown in Fig. 5 (a). This corresponds to an efficiency of $(4.23 \pm 0.57)\%$. For the electron analysis this efficiency is $(2.73 \pm 0.47)\%$.

The signal significance is defined using the likelihood ratio from two hypotheses: the presence of a signal (a peak) and its absence (only the Chebychev polynomial). The amount of data needed to make a statistically significant observation of the $t\bar{t}$ signal depends on the amount of background. For low luminosities, for example for 25 pb$^{-1}$, the sampling fluctuations are too large and there is no typical plot like the one of Fig. 3. To quantify the relation between signal significance, luminosity and the amount of background, 10000 pseudo-experiments based on the full simulation distribution of the three-jet mass as a function of the integrated luminosity, have been modelled and fitted. The expected statistical significance is shown in Fig. 5 (b), where the yellow band is obtained by assuming the nominal level of QCD $W$-boson+jets background, while the red one refers to the case when this background is multiplied by two. .

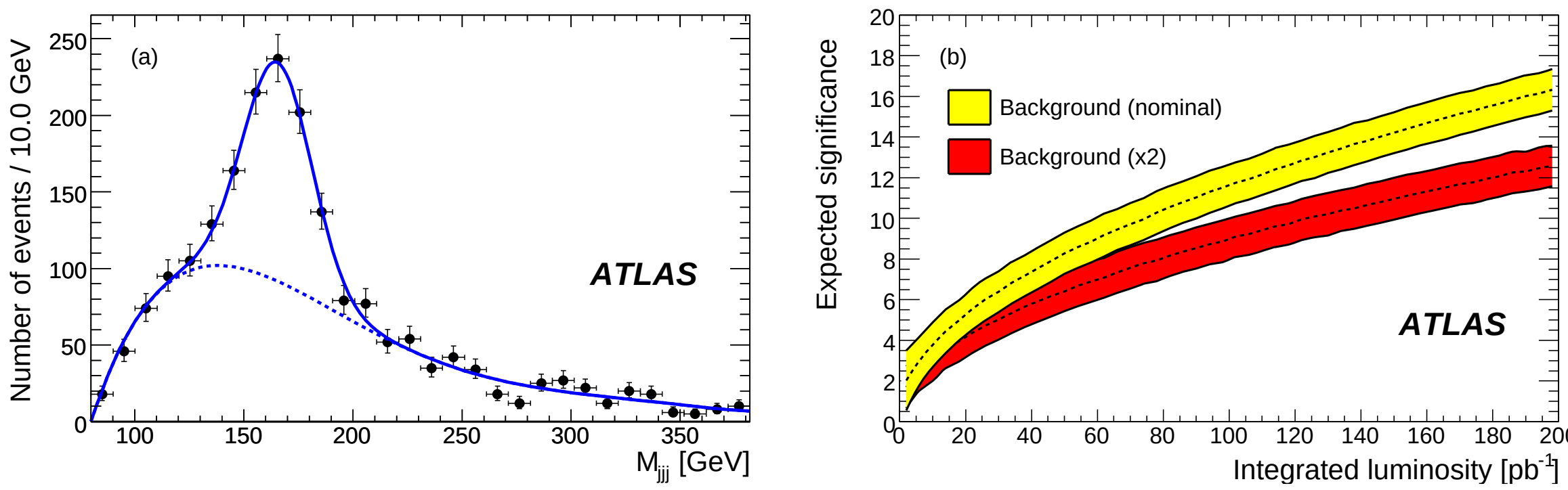


Figure 5: (a): Fit to the top signal. The Chebychev polynomial fit to the background is indicated by the dotted line and the Gaussian fit of the signal events is indicated by the full line. (b): Distribution of the expected statistical significance of the top signal in the peak as a function of the integrated luminosity for two background scenarios. The yellow band is obtained by assuming the nominal level of QCD $W$-boson+jets background, while the red one by assuming that this background is doubled.

To go from a fitted number of properly reconstructed hadronic top quarks to a cross-section, one needs to correct for the event selection efficiency and the hadronic top reconstruction efficiency. The statistical error is estimated from having simulated 100000 pseudo-experiments, applying the fluctuations which are expected in 100 pb$^{-1}$to both signal and background and fitting the peak in both the electron and muon channels.

One of the biggest uncertainties is the correct modelling of the jet multiplicity distribution as it affects

the fraction of $t\bar{t}$ signal events that are correctly reconstructed using the algorithm described earlier. An overview of the systematic uncertainties is given in [9].

#### 2.2.2 Counting method

The $t\bar{t}$ cross-section can be obtained by performing a counting experiment:

$$\sigma = \frac{N_{\text{sig}}}{\mathscr{L} \times \varepsilon} = \frac{N_{\text{obs}} - N_{\text{bkg}}}{\mathscr{L} \times \varepsilon}$$

$N_{\text{bkg}}$, the number of background events estimated from Monte Carlo simulations and/or data samples, is subtracted from $N_{\text{obs}}$, the number of observed events meeting the selection criteria of a top-event signature. This difference is divided by the integrated luminosity $\mathscr{L}$ and the total efficiency $\varepsilon$. The latter includes the geometrical acceptance, the trigger efficiency and the event selection efficiency, and is slightly dependent on $m_t$ . The advantage of using event counts in the commissioning phase is that, early on, the Monte Carlo simulations may not predict the shapes of distributions very well.

In order to perform the counting experiment the Monte Carlo samples were divided into two, statistically independent: one which represents real data, used to obtain $N_{obs}$ and the other one used as a Monte Carlo to obtain both $\varepsilon$ and $N_{bkg}$.

#### 2.2.3 Systematic uncertainties

The main sources of systematic uncertainties are described in [9]. Some relevant points for the analyses presented here are discussed for the case of the default selection plus the $W$-boson mass constraint. The systematic uncertainty on the cross-section due to the luminosity determination, is factorised and mentioned as a separate uncertainty on the overall results. The event selection efficiencies have a nearly linear dependency for jet energy scale variations, which affects the counting method directly. The hadronic top reconstruction efficiency has an inverse dependence on the jet energy scale, caused by the algorithm that picks the three-jet combination that are considered to be the hadronic top. If the jet energy scale is lowered, the jet multiplicity (and therefore the number of three-jet combinations) increases. The probability that the algorithm picks the right combination generally decreases with the number of combinations to choose from.

The effect of ISR and FSR parameter variations (chosen in such a way to maximise the effect on the cross-section measurement) has been evaluated. For the uncertainties related to the PDFs, both the uncertainty coming from CTEQ and MRST, have been considered and the largest one (coming from CTEQ) has been used for the final systematics evaluation.

The main systematics uncertainties for the two analysis are listed in Table 5. For the likelihood method a 5% change in jet energy scale causes a 2.3% (0.9%) change in the combined reconstruction efficiency for the electron (muon) channel. The different jet multiplicity distributions affect not only the event selection efficiencies (jet requirements), but also the overall efficiency of the hadronic top reconstruction algorithm. A comparison between the two generators MC@NLO and ALPGEN has been made to study these systematic effects. ALPGEN predicts 7% and 4% larger selection efficiencies in the electron and muon channels respectively. For the overall efficiencies the values are 10.5% and 4.7% larger in the electron and muon channel. However, these numbers were not added in the final result since there is overlap with the ISR/FSR systematics. Systematics effects on the shape of the fit as well as the normalisation of the peak fit w.r.t. background is estimated with toy Monte Carlo's. Deviations of 14.0 (10.4)% for the electron (muon) channel are found, while the effect of changing the fit-ranges is negligible.
For the counting analysis the uncertainty arising from the Monte Carlo used to generate the signal process has been taken into account as well. This has been done by comparing the cross-section obtained applying the same analysis to $t\bar{t}$ events generated with MC@NLO and with the ACERMC Monte Carlo.

Table 5: Systematic uncertainties on the commissioning likelihood and counting method cross-section measurement, in percent.

| Source | Likelihood fit | | Counting method (elec) | |
|---|---|---|---|---|
| | **Electron** (%) | **Muon** (%) | **Default** (%) | ***W* const.** (%) |
| Statistical | 10.5 | 8.0 | 2.7 | 3.5 |
| Lepton ID efficiency | 1.0 | 1.0 | 1.0 | 1.0 |
| Lepton trigger efficiency | 1.0 | 1.0 | 1.0 | 1.0 |
| 50% more $W$+jets | 1.0 | 0.6 | 14.7 | 9.5 |
| 20% more $W$+jets | 0.3 | 0.3 | 5.9 | 3.8 |
| Jet Energy Scale (5%) | 2.3 | 0.9 | 13.3 | 9.7 |
| PDFs | 2.5 | 2.2 | 2.3 | 2.5 |
| ISR/FSR | 8.9 | 8.9 | 10.6 | 8.9 |
| Shape of fit function | 14.0 | 10.4 | - | - |

The $W$-boson+jets normalization uncertainty has been evaluated using the $Z$-boson+jets sample as discussed in the general introduction of systematic uncertainties, but also varying the level of the expected $W$-boson+jets level by 20%, 50% and even by a factor of two. For the final selection (including the $W$-boson mass constraint) this corresponds to an uncertainty of 4%, 10% and 19% respectively. As reference value to calculate the overall systematic error, the 50% case will be used.

### 2.2.4 Contributions of new physics

Many models of physics beyond the Standard Model contain new particles which couple to top-quarks. For example, in a supersymmetric estension of the Standard Model (SUSY) these new particles are top squarks [10] and in warped extra dimensions they are Kaluza-Klein resonances [11]. Since the new particles are expected around the TeV scale, the typical cross-sections are of order a few pico barns and hence one can expect a few hundreds new physics events in the first 100 pb$^{-1}$ of data. In principle, these new physics events could represent a significant background to the cross-section measurement. For new physics models with much lower masses, e.g. low mass supersymmetry models, there will be many more events: this case will be considered later on. Here the existence of a new particle $V$ which decays only into $t\bar{t}$ p̃airs, $V \rightarrow t\bar{t}$ it is assumed. It is further assumed that this particle has a production cross-section of 5 pb, so that 25/9 pb is the cross-section for the non-fully hadronic decays of $V$. As a model for $V$ a 1 TeV Z' is used. The efficiency for these events with respect to the default selection + the $W$-boson mass constraint is roughly twice the one obtained for Standard Model $t\bar{t}$ events, and the number of events passing the selection will be of the order of 1% or less of the $t\bar{t}$ events. Hence the new particle $V$ will not affect the cross-section determination significantly.

The predictions at several mSUGRA benchmark points [12] have been studied as well. The results are shown in Table 6 and demonstrate that the expected signals are small. At specific parameter points however, like SU4 [12], the cross-section is sizeable and the event topology is similar to that from top quark pairs which results in additional backgrounds as large as the total Standard Model background. In Fig. 6 one can see that the shape of the SU4 supersymmetry signal in the top quark candidate three-jet invariant mass distribution is very similar to that from the Standard Model background.

The separation of $t\bar{t}$ events and these new physics signals is addressed in more detail in [12], dedicated to searches for supersymmetry.

Table 6: Expected number of events in a 100 pb$^{-1}$ data sample at different stages of the analysis for several event types: after trigger and event selection (left column), after a cut on the di-jet masses in the top-quark candidate (middle column) and events with in addition to the di-jet mass cut a hadronic top mass cut $141 < m_t < 189$ GeV(right column).

| Event type | **Electron analysis** | | | **Muon analysis** | | |
|---|---|---|---|---|---|---|
| | Trigger+Selection | $W$ const. | $m_t$ win | Trigger+Selection | $W$ const. | $m_t$ win |
| SU1 | 53 | 9 | 1 | 64 | 12 | 2 |
| SU2 | 10 | 2 | 0.5 | 13 | 3 | 0.7 |
| SU3 | 108 | 22 | 4 | 124 | 26 | 4 |
| SU4 | 1677 | 541 | 155 | 2141 | 700 | 199 |
| SU6 | 29 | 5 | 0.6 | 35 | 6 | 0.6 |
| SU8 | 27 | 5 | 0.6 | 33 | 6 | 0.8 |

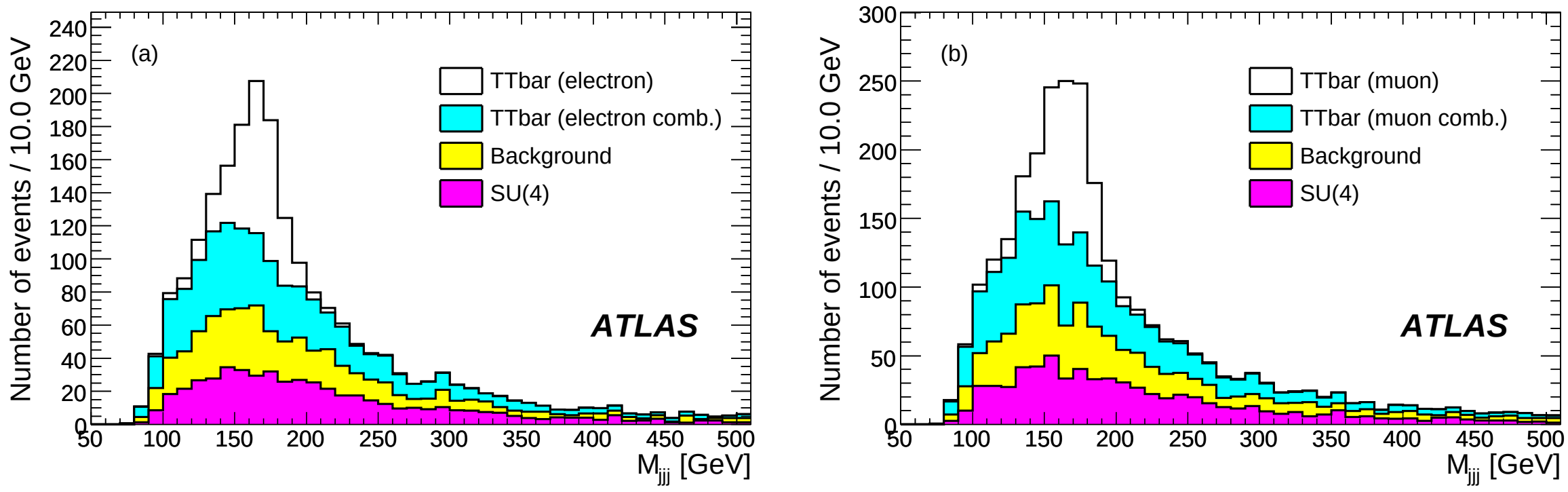


Figure 6: (a): Expected distribution of the three-jet invariant masses among the top-quark candidates after the requirement on the mass of the di-jet system in the electron channel in a 100 pb$^{-1}$ event sample. (b): Same distribution for the muon channel. The light histogram represents the Standard Model background. i.e. the background from mSUGRA point SU4 is shown separately.

## 2.3 Implementation of b-tagging

The possibility to identify b-flavoured jets (b-tagging) will improve the signal to background ratio of the selection. The b-tagging requirements are described in [9]. The number of "tagged" b-jets in the $t\bar{t}$ , single top and $W$-boson+jet events which pass the default selection is shown in Fig. 7.

Tables 3 and 4 list the number of $t\bar{t}$ and background events in the electron and muon channel which survive the default selection plus the $W$-boson mass constraint, and the request of having one and only one, or two and only two b-jets (column six and seven). For all these cases, the corresponding signal to background ratios are given. Requiring one or two b-tagged jets improves the purity of the sample by more than a factor of four, while the signal efficiency is only reduced by a factor of two.

In Fig. 8 the reconstructed three-jet mass is shown when one or two b-tagged jets are required for the default selection (a) and for the default selection + the $W$-boson mass constraint (b). To reconstruct the top mass, we find the three-jet combination with the highest possible $p_T$, obtained by requiring that one and only one of the three jets is a b-jet. The $W$-boson mass constraint can then be applied to the two jets which are not b-tagged (among the three). If the three-jet combination chosen above is such that the two

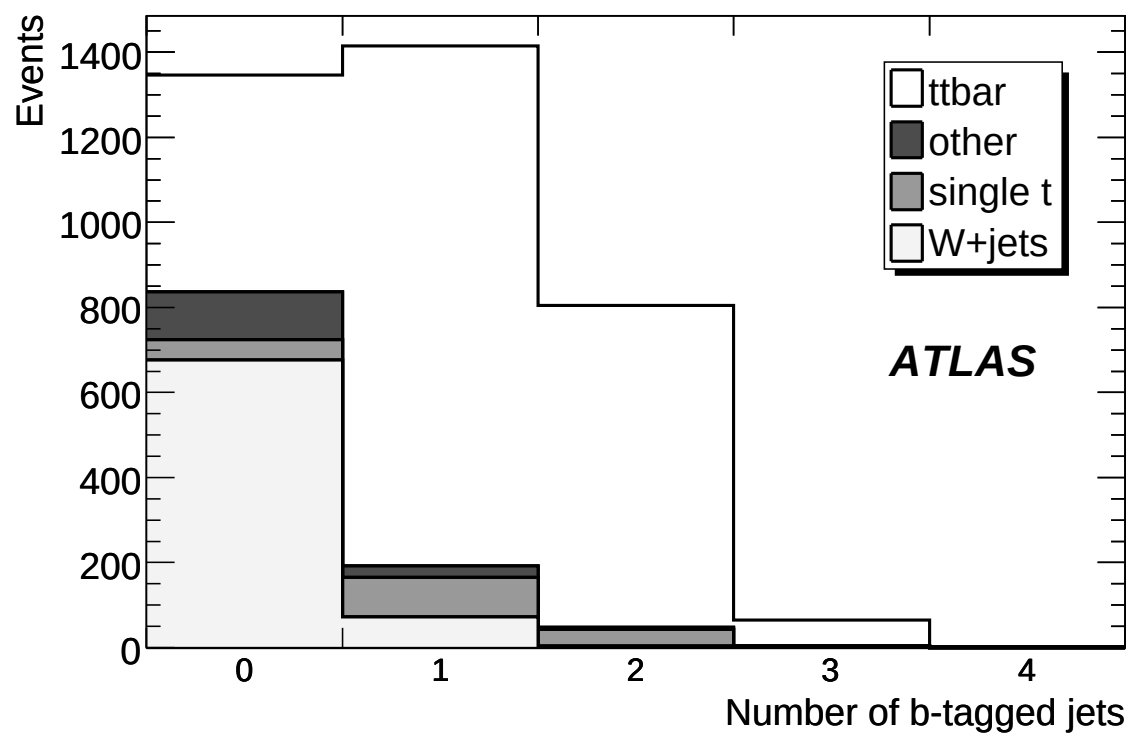


Figure 7: Number of jets tagged as coming from a b-quark in $t\bar{t}$, single top and $W$-boson+jet events after the default electron selection.

non-b-jets don't combine to give a $W$-boson candidate, that event is rejected.

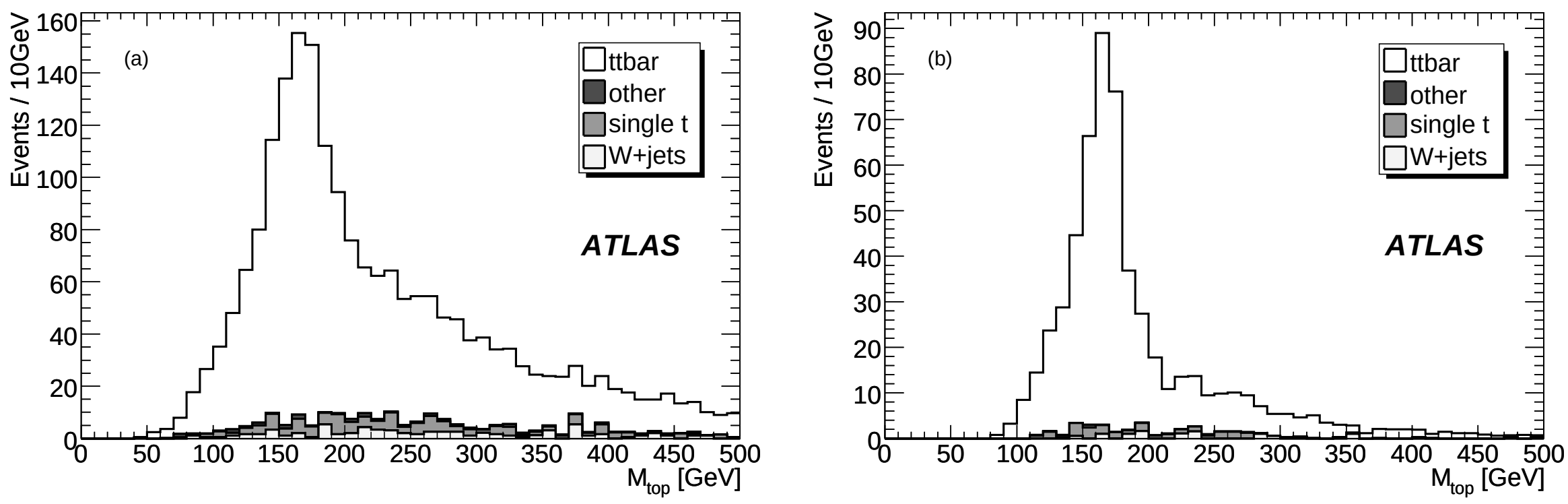


Figure 8: (a): Reconstructed three-jet mass for $t\bar{t}$ , single top and $W$-boson + jet events for the default electron selection, requiring one or two jets tagged as coming from a b-quark. (b): Same distribution for the default selection + the $W$-boson mass constraint and requiring one or two jets tagged as coming from a b-quark.

The statistical error on the cross-section which is obtained by requiring one or two b-tagged jets is 4.5%. The systematic error due to the jet energy scale is in this case of 4.9%, while a wrong normalization of the $W$-boson+jets background by a factor of 20%, 50% or even a factor two, brings a systematic error on the cross-section of 3.4%, 4.7% and 6.9% respectively. A 5% relative error on the b-tagging efficiency is expected from present studies for an efficiency of 50-60% and for a luminosity of 100 pb$^{-1}$. The resulting uncertainty on the cross-section turns out to be negligible. The undertainty on the mistag rate is assumed to be of the order of 50%.

## 2.4 Results

With the first 100 pb$^{-1}$of data, we can observe a $t\bar{t}$ signal and determine its production cross-section. This will be determined with a number of methods and we expect to reach the following accuracies (for

the default selection with the $W$-boson mass constraint, using electron and muons):

$$\text{Likelihood method:} \quad \Delta\sigma/\sigma = \quad (7(\text{stat}) \pm 15(\text{syst}) \pm 3(\text{pdf}) \pm 5(\text{lumi}))\% \tag{1}$$

$$\text{Counting method:} \quad \Delta\sigma/\sigma = \quad (3(\text{stat}) \pm 16(\text{syst}) \pm 3(\text{pdf}) \pm 5(\text{lumi}))\% \tag{2}$$

### 2.5 Differential cross-sections

We studied several differential distributions for $t\bar{t}$ production. First, we present the momentum and rapidity distribution of the hadronically decaying top quarks, after having applied the default selection and the $W$-boson mass constraint. The results are shown in Fig. 9 (a) and (b).

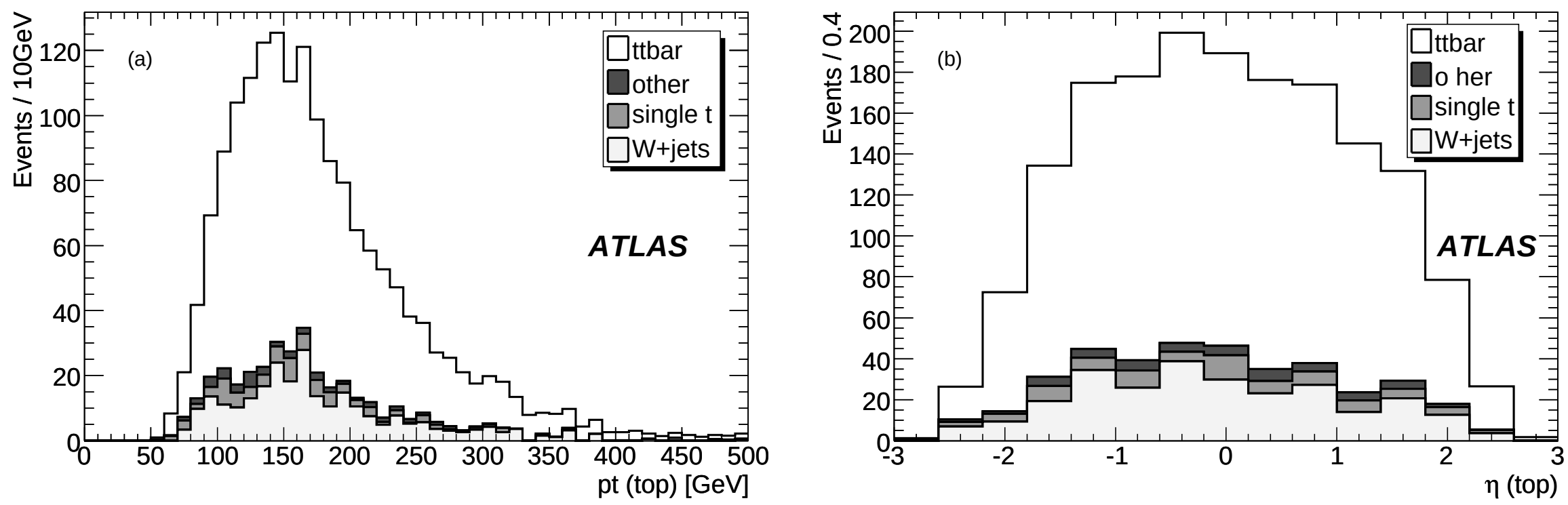


Figure 9: (a): Momentum distribution of the reconstructed three-jet mass for $t\bar{t}$ , single top and $W$-boson+jet events for the electron analysis. Right plot: Rapidity distribution of the hadronically decaying top quark.

A more detailed study has been performed for the differential cross-section as a function of the $t\bar{t}$ system mass, and for several double differential distributions as shown in the following sections.

The differential cross-section for $t\bar{t}$ production can be measured as a function of the invariant mass of the $t\bar{t}$ system in the semi-leptonic channel (with no tau leptons in the final state). Such a measurement provides an important check of the Standard Model and, at the same time, deviations from the $t\bar{t}$ continuum could indicate the presence of new physics, for example new heavy resonances decaying into a $t\bar{t}$ pair [14].

The standard commissioning selection is applied and the momenta of the four jets with highest $p_T$, the lepton and the best estimate of the $E_{\mathrm{T}}^{\mathrm{miss}}$ vector are used as inputs to a least squares fit with the constraints that, in each event, the masses of both the $W$-boson and top-quark are consistent with 80.4 GeV and 175 GeV respectively. The fit procedure is documented in [15]. The goal is to improve the measurement of the reconstructed final state particles' four vectors by incorporating the precise knowledge of the masses of the $W$-boson boson and the top-quark. No b-tagging is used in this analysis and hence there are 12 possible combinations to assign jets to the (anti-) top. All combinations were investigated and and one was chosen: the assignment which returned the smallest weighted sum of squared residuals obtained from the kinematic fit. A simpler reconstruction scheme obtains the $t\bar{t}$ mass by deriving the leptonic $W$-boson momentum from the lepton and missing energy momenta with the $W$-boson mass constraint and then combining it with the momenta of the four highest $p_T$ jets.

The reconstructed mass distributions for the two methods are compared to the true di-top mass distributions in Fig. 10 (a) for the non-hadronic signal. The true di-top mass distribution is in better agreement with the result obtained by making use of the full event fitting technique.

In this case, the expected mass resolution ranges from 5% to 9% between 200 and 850 GeV. A variable bin size of about twice the expected resolution is used to take such variation into account and reduce bin-to-bin migrations. The di-top mass spectrum ($dN/dm_{tt}$), reconstructed with the full event fit, is shown in Fig. 10 (b) for the signal and the backgrounds studied. Backgrounds include: full hadronic top, single top, $W$-boson+jets, $Wb\bar{b}$ , $Wc\bar{c}$ , inclusive $Z$-boson to leptons. The contribution from the di-boson ($WW$,$WZ$ and $ZZ$) backgrounds is negligible.

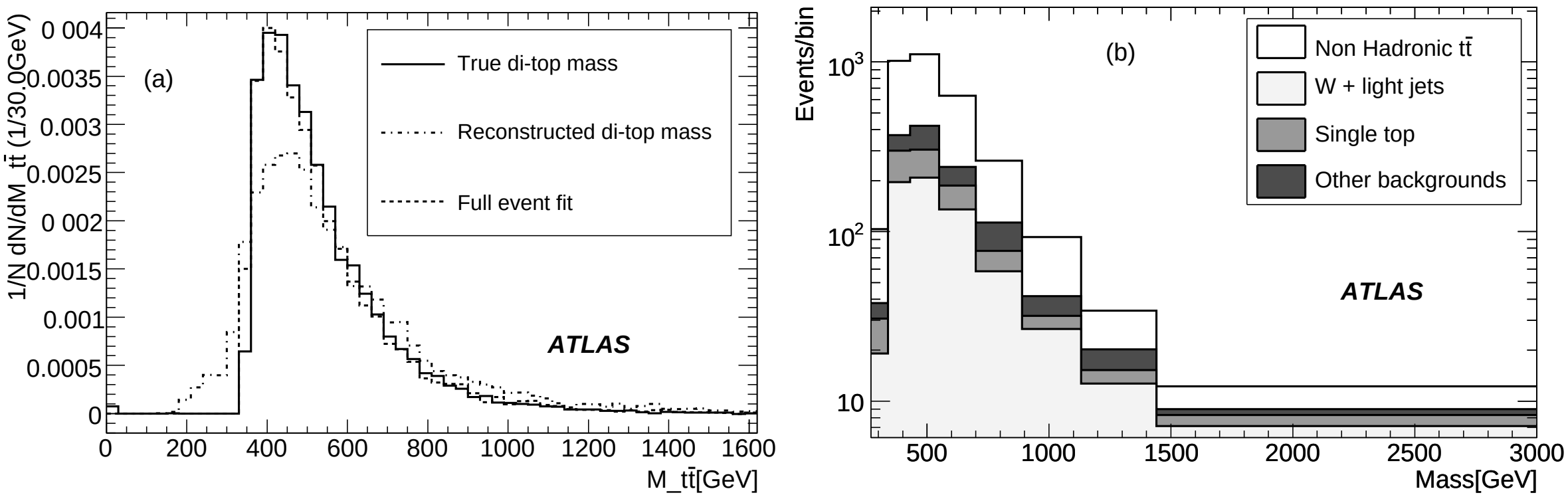


Figure 10: (a): Normalised di-top mass distribution for the more complex (dashed line) and the simple reconstruction (dotted line). The normalised true di-top mass is also shown for reference (solid line). (b): Expected reconstructed di-top mass distribution after all cuts for signal and studied backgrounds, normalised to 100 pb$^{-1}$.

### 2.5.1 Double differential cross-section as function of $p_T$ and $y$

The double differential cross-section for $t\bar{t}$ production is sensitive to possible new physics beyond the Standard Model, e.g. extra dimensions based on studies of the top quark spin correlation [13], which depends on the knowledge of the top quark's momentum. A measurement investigates the decay products of the top quark in its rest frame and therefore good knowledge of its $p_T$ and $y$ as defined in (3), for a top quark of energy $E$ and longitudinal momentum $p_z$, is needed.

$$y = \frac{1}{2}\ln\left(\frac{E+p_z}{E-p_z}\right) \tag{3}$$

Theoretical predictions can be found in [4]. Here we present a feasibility study which, since the neutrino momentum cannot be directly measured, concentrates on the reconstruction of the hadronically decaying top quark in semileptonic $t\bar{t}$ events. Since in this case a high purity is needed, the default event selection is tightened by requiring exactly two b-tagged jets. The reconstruction of the hadronic top quark proceeds as follows: all possible combinations of two non-b-tagged jets with 60 GeV$<$ m$_{jj}$ $<$ 100 GeV are selected as $W$-boson candidates. The nearest b-tagged jet for every $W$-boson candidate is found. The combination with the highest transverse vector sum momentum is then taken as the reconstructed hadronic top quark. This results in a purity of well reconstructed top quarks of 45%. The main background is due to combinatorics.

Figure 11 shows the reconstructed double-differential distribution of the hadronic top scaled to an integrated luminosity of 1 fb$^{-1}$. In (a) the truth distribution of the $t\bar{t}$ signal is presented, while in (b) the distribution of reconstructed hadronic top-quarks is shown. In this distribution the contribution of background (from single top, $W$-boson + jet, $Wb\bar{b}$ and $Wc\bar{c}$), which is very small after the requirement

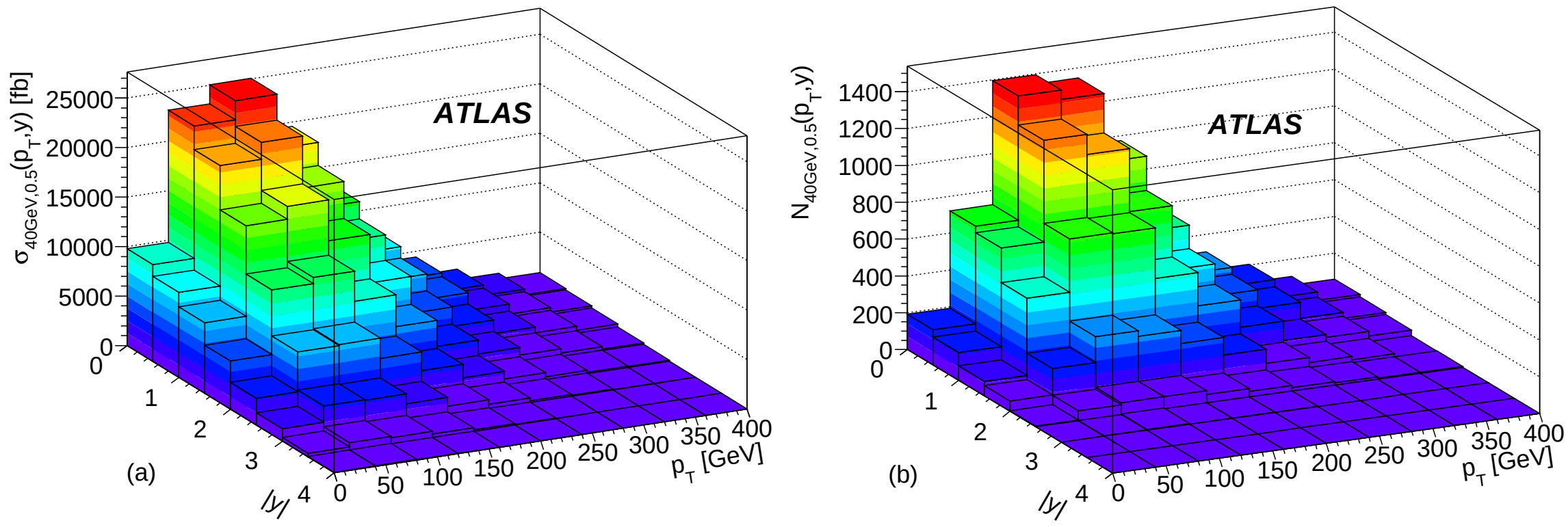


Figure 11: (a): Distribution of $p_T$ and $y$ of hadronically decaying top quarks calculated by MC@NLO. (b): Reconstructed distribution of $p_T$ and $y$ of hadronically decaying top quarks from fully simulated samples scaled to an integrated luminosity of 1 fb$^{-1}$.

of two b-tagged jets, has been added. After such a selection, the number of expected events limits the region of interest to $|y| < 2$ and 50 GeV$<p_T<$ 280 GeV. 1 fb$^{-1}$ is a reasonable statistic which allows to fill a significant number of bins with an appropriate statistical error. A lower integrated luminosity of 100 pb$^{-1}$ would increase the statistical error of the bin contents at the edge of the interesting area from 10% to approximately 30%, which would limit the measurement to a smaller phase-space. The systematic uncertainties of this feasibility study are expected to be small and under control.

The potential to determine the double differential cross-section of $t\bar{t}$ events decaying semileptonically has been estimated. The phase space can be determined with an average efficiency of $3.98 \pm 0.04\%$ and a peak efficiency of 8% near the central rapidity region and for $p_T \leq 140$ GeV.

The main systematic uncertainties for this study will come from the jet energy scale and from the initial and final state radiation: each source contributing with an uncertainty of the order of $\pm$ 15% in the central region.

# 3 Di-Lepton channel

In this section the determination of the $t\bar{t}$ cross-section where both $W$-bosons decay leptonically is presented. This measurement depends crucially on the correct identification of leptons. We proceed by investigating the channel with two electrons, one electron and a muon, and two muons in the final state. Final states with tau leptons are not studied here. The determination of the cross-sections is performed with a simple 'cut and count' method, a template method and a likelihood fit.

## 3.1 Event selection

The di-lepton sample is expected to be triggered with high efficiency using a combination of single-lepton and di-lepton triggers [7]. The overall trigger efficiency for the electron-muon channel is $(97 \pm 1)\%$, for the two electron channel it is $(98 \pm 1)\%$ and for the two muon channel it is $(96 \pm 1)\%$. Due to the trigger OR condition between the channels and the high statistics available for efficiency measurements, the uncertainty is small.

The offline selection of di-lepton events is based on the identification of leptons as described in the introductory section. Two high-$p_T$opposite signed leptons, i.e. two electrons, two muons, or one electron and one muon, are required. These requirements define the preselection sample and additional selection criteria are imposed depending upon the $t\bar{t}$ cross-section extraction method. The expected number of

events produced after the preselection cuts of two isolated opposite charged leptons is shown in Table 7. None of the selections makes use of b-tagging.

Table 7: Simulated Monte Carlo samples and expected events produced for 100 $pb^{-1}$of integrated luminosity. Cross-sections $\sigma$ are at least Next-to-Leading Order total cross-sections, and $\sigma_{\text{eff}}$ are the effective Monte Carlo cross-sections including generator level filter efficiencies (reported in the "Filter(%)" column). The last three columns show the number of preselected events after requiring two opposite signed leptons.

| Sample | $\sigma$(pb) | Filter(%) | $\sigma_{\text{eff}}$(pb) | $e\mu$ | $ee$ | $\mu\mu$ |
|---|---|---|---|---|---|---|
| $t\bar{t}$ (di-lepton) | 833 | 7(2$l$) | 55 | 699 | 312 | 381 |
| $t\bar{t}$ (semi-leptonic) | | 48(1$l$) | 397 | 31 | 20 | 8 |
| $Z\rightarrow e^+e^-$ | 2015 | 86 | 1733 | 5 | 37418 | 0 |
| $Z\rightarrow \mu^+\mu^-$ | 2015 | 89 | 1793 | 153 | 0 | 51139 |
| $Z\rightarrow \tau^+\tau^-$ | 2015 | 5 | 101 | 249 | 101 | 159 |
| $W\rightarrow e\nu$ | 20510 | 63 | 12920 | 42 | 69 | 0 |
| $W\rightarrow \mu\nu$ | 20510 | 69 | 14150 | 152 | 0 | 40 |
| $WW$ | 117 | 35 | 41 | 76 | 32 | 44 |
| $WZ$ | 48 | 29 | 14 | 6 | 41 | 52 |
| $ZZ$ | 15 | 19 | 3 | 1 | 25 | 31 |
| single top | 324 | 31 | 99 | 5 | 3 | 2 |

#### 3.1.1 Lepton identification and isolation

The inclusive di-lepton selection requires at least one of the electrons to be identified with the 'tight-electron' algorithm[2] to improve the fake lepton rejection.

One of the major backgrounds in this channel is given by semi-leptonic $t\bar{t}$ events where the second lepton candidate is faked by a jet. A requirement on the isolation of the electrons lowers this background substantially. Fig. 12 (a) shows the distribution of the variable $\Sigma_{\Delta R<0.2}E_T$, defined to be the energy deposited in a hollow cone with radius $\Delta R$ of 0.2 around the electron candidate. The figure shows this variable for reconstructed electrons in the semi-leptonic $t\bar{t}$ background sample which are close to a monte carlo truth electron ($\Delta R < 0.1$ to a truth electron from the $W$-boson decay) and for reconstructed electrons with $\Delta R > 0.1$. With the requirement $\Sigma_{\Delta R<0.2}E_T < 6$ GeV, the signal is reduced by only $\sim 4\%$ whereas the semi-leptonic $t\bar{t}$ background is reduced by a factor of two.

Two cuts are used to select muons from $W$-boson decays while simultaneously rejecting muons from b-jets. The first cut selects the muon candidate with tracks in the muon spectrometer that match best with tracks in the inner detector. Additionally it is required that the muon is not closer than $\Delta R < 0.2$ to a jet, otherwise the muon is removed.

Fig. 12 (b) shows the distance of muons to the closest jet in the event. Muons that are close ($\Delta R < 0.2$) to a monte carlo truth b-quark are separated from muons that are close to a monte carlo truth muon ($\Delta R < 0.1$). A $\Delta R$ cut of 0.2 removes the muons that do not originate from the $W$-boson decay. This requirement reduces the semi-leptonic $t\bar{t}$ background by $\sim 25\%$, whereas the signal is reduced by only 0.8%.

[2]The tight-electron algorithm makes full use of the TRT information as described in Ref. [16].

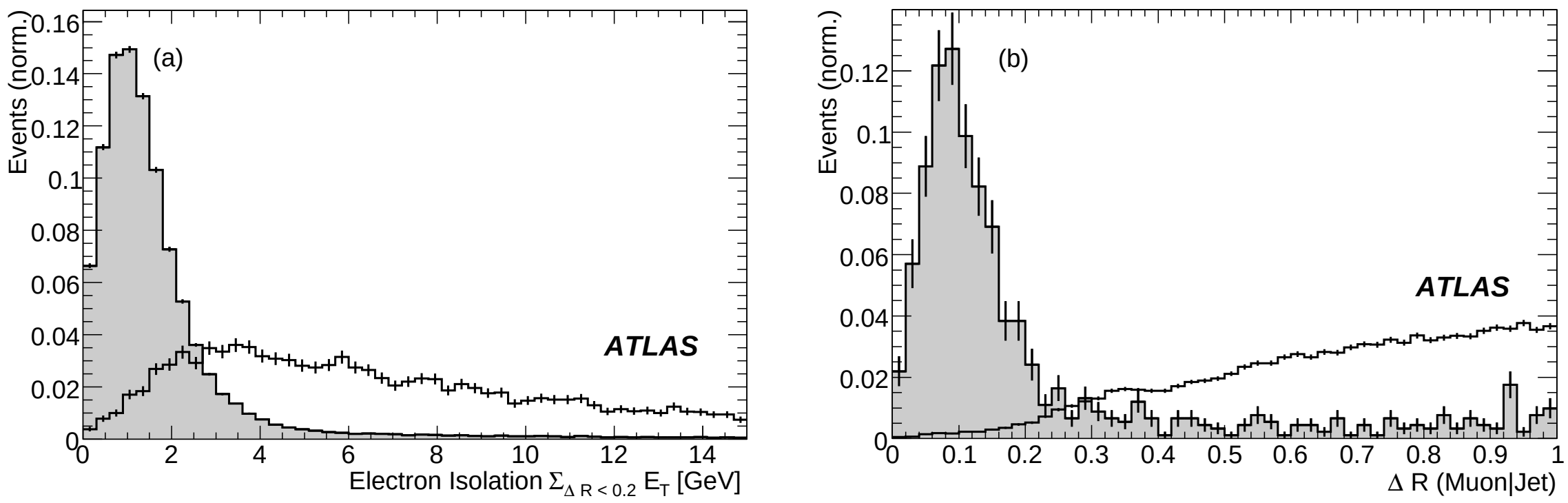


Figure 12: (a) The $\Sigma_{\Delta R<0.2}E_T$ variable for reconstructed electrons of the semi-leptonic $t\bar{t}$ background, in black are electrons matched to truth electrons, in white otherwise. (b) $\Delta R$ from a muon to the closest jet in the semi-leptonic $t\bar{t}$ sample. In white are muons matched to truth muons, in black are muons that are close to a b-quark.

### 3.2 Backgrounds

The backgrounds to the $t\bar{t}$ di-lepton signal can be classified into two main categories: prompt leptons originating from an electroweak decay, or non-isolated non-prompt leptons and jets that are falsely identified as leptons mainly originating from QCD jets.

For the estimation of the electroweak backgrounds we use Monte Carlo samples, as shown in Table 7. For the same-flavour Drell-Yan processes we made an exception, as the size of this contamination depends on the determination of the $E_{\mathrm{T}}^{\mathrm{miss}}$ which is difficult to model. As a default we use the shape of the $E_{\mathrm{T}}^{\mathrm{miss}}$ distribution for events with a di-lepton mass inside a window around the $Z$-boson mass, to correct the shape of the $E_{\mathrm{T}}^{\mathrm{miss}}$ distribution in the Monte Carlo. The same correction factor is used for Drell-Yan processes with di-lepton invariant masses outside the $Z$-boson mass window.

The backgrounds from fake leptons are estimated from the QCD dijet samples. These estimations of fake leptons are subsequently applied to all objects that can lead to fake leptons (e.g. jets) in the inclusive single electron or muon samples.

### 3.3 Cross-section measurement

As mentioned, three methods to determine the cross-section for the di-lepton channel are presented. The di-lepton channel profits from having smaller backgrounds and systematics and the methods are complementary to each other. The robust 'cut and count' method can be replaced with the more sophisticated template and likelihood method with increased accumulated data.

#### 3.3.1 'Cut and count' method

A 'cut and count' analysis is the most straightforward method to determine the cross-section. It can be used as the basis and reference for the more elaborate likelihood methods and act as a cross check between the different approaches.

The selection criteria are defined to maximise the efficiency $\varepsilon$ and the purity $p$ at the same time. The figure of merit is the product

$$\varepsilon \times p \propto \frac{S}{\sqrt{S+B}} = s$$

which is referred to as significance.

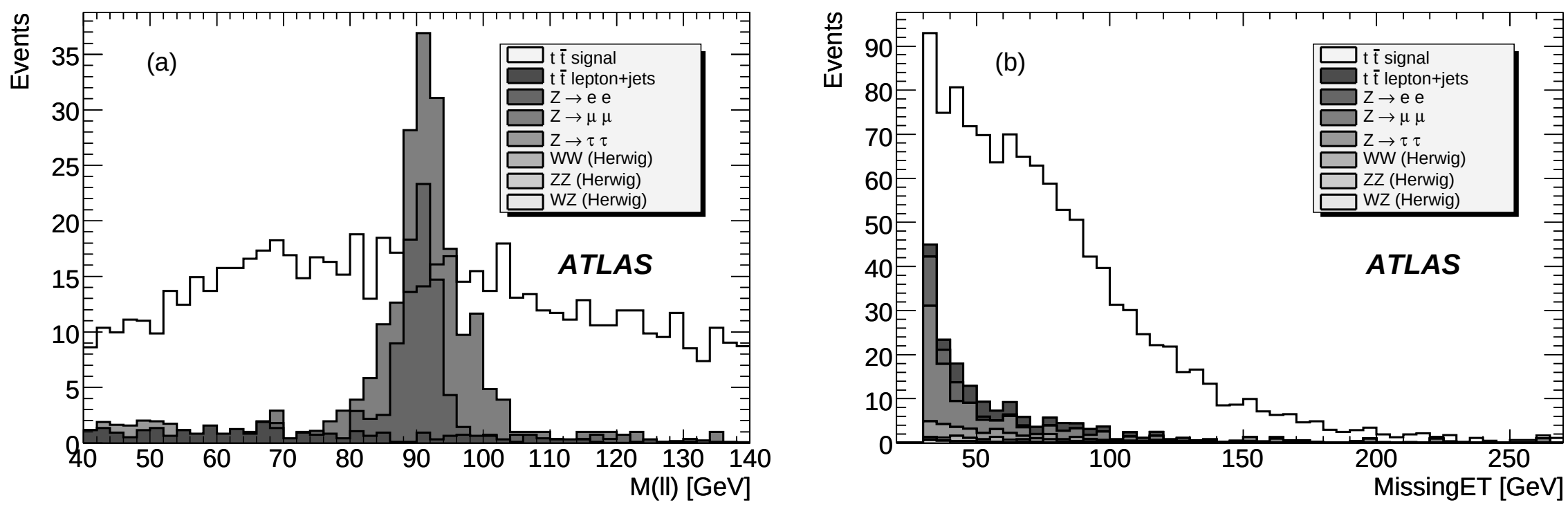


Figure 13: (a) Di-lepton mass in signal, semi-leptonic $t\bar{t}$ and $Z \rightarrow \ell^+\ell^-$ events. (b) Distribution of the $E_{\mathrm{T}}^{\mathrm{miss}}$ for signal and various backgrounds, normalised to 100 pb$^{-1}$

.

The variables that best characterise the signal events are the transverse momenta of the two leptons and the two jets, as well as the missing transverse energy. The jet momenta are high since they originate from the b-quarks, not present in the prominent background processes $Z \rightarrow \ell^+\ell^-$ or dibosons $\rightarrow \ell^+\ell^-$ ($WW$, $WZ$ and $ZZ$). A large amount of missing transverse energy is expected in signal events due to the two escaping neutrinos. In addition a veto on events with a di-lepton invariant mass around the $Z$-boson mass is applied. Fig. 13 (a) shows the di-lepton mass distribution in $Z \rightarrow \ell^+\ell^-$ events. Most of the events are found to have an invariant mass between 85 and 95 GeV. In the case of the $Z \rightarrow \tau^+\tau^-$ events the peak is shifted and broadened, since the visible leptons do not come directly from the $Z$-boson. Also the neutrinos from the $\tau$ decay add to the missing transverse energy. It is therefore expected that this background will be dominant, although the branching ratio for both $\tau$'s decaying leptonically is only $\sim 9\%$. The optimal selection was found from a multidimensional scan of the significance $s$. Exactly two leptons are required and at least two jets. The requirements on the lepton and jet transverse momenta and on the $E_{\mathrm{T}}^{\mathrm{miss}}$ is then varied from 20 to 60 GeV. Finally, the cuts with the maximum significance are used to evaluate the cut and count performance.

As a result, the multidimensional scan indicates that the values of the preselection requirement of 20 GeV for the two leptons and for the two jets with the highest transverse momentum already maximise the significance. One additional cut is imposed: the $E_{\mathrm{T}}^{\mathrm{miss}}$ is required to be at least 30 GeV for a selection of two leptons (all sub-channels together), at least 20 GeV for the $e\mu$ decay channel and at least 35 GeV for the same flavour lepton channels. The distribution of the missing transverse momentum for the signal and the background samples is presented in Fig. 13 (b). The efficiencies, signal over background ratios and the significance for an integrated luminosity of 100 pb$^{-1}$ are shown in Table 8.

The cross-section is derived from:

$$\sigma = \frac{N_{\mathrm{sig}}}{\mathscr{L} \times \varepsilon} = \frac{N_{\mathrm{obs}} - N_{\mathrm{bkg}}}{\mathscr{L} \times \varepsilon}$$

To evaluate the statistical uncertainty on $\sigma$, the error on $N_{\mathrm{obs}}$ is taken to be Gaussian assuming it will be measured in the data. The error on $N_{\mathrm{bkg}}$ is calculated here from Monte Carlo and scaled to the desired luminosity. The relative error on the efficiency $\varepsilon$ (the product of geometrical acceptance and selection efficiency) here is also calculated from the Monte Carlo. The expected statistical error on the cross-section measurements for different integrated luminosities is given in Table 9. In data the efficiencies regarding leptons will be estimated from $Z$ data events using tag and probe. Also the rates of backgrounds containing misidentified leptons can only be measured in data reliably.

Table 8: Number of events which survive the optimised selection criteria for signal and background samples, scaled to a luminosity of 100 $pb^{-1}$.

| dataset | $e\mu$ | $ee$ | $\mu\mu$ | all channels |
|---|---|---|---|---|
| $t\bar{t}$ (di-lepton) | 555 | 202 | 253 | 987 |
| $\varepsilon$ [%] | 6.22 | 2.26 | 2.83 | 11.05 |
| $t\bar{t}$ (semi-leptonic) | 24 | 11 | 4 | 39 |
| $Z \to e^+e^-$ | 0.0 | 9 | 0.0 | 20 |
| $Z \to \mu^+\mu^-$ | 5 | 0 | 51 | 79 |
| $Z \to \tau^+\tau^-$ | 17 | 4 | 6 | 25 |
| $WW$ | 6 | 2 | 2 | 10 |
| $ZZ$ | 0 | 0.2 | 0.4 | 0.9 |
| $WZ$ | 1 | 0.6 | 1 | 3 |
| $W \to e\nu_e$ | 7 | 7 | 0.0 | 14 |
| $W \to \mu\nu_\mu$ | 25 | 0.0 | 7 | 33 |
| single top $Wt$ | 0.7 | 0.5 | 0.0 | 1 |
| single top s-chann. | 0.0 | 0.0 | 0.0 | 0.1 |
| single top t-chann. | 2 | 0.8 | 1 | 4 |
| Total bkg. | 86 | 36 | 73 | 228 |
| $S/B$ | 6.3 | 5.6 | 3.4 | 4.3 |

Table 9: Expected statistical error on the cross-section determination for the cut and count analysis for different luminosities.

| Luminosity [$pb^{-1}$] | | 10 | 100 | 1000 |
|---|---|---|---|---|
| $\Delta\sigma/\sigma$ | $e\mu$ | 14.1 % | 4.5 % | 1.5 % |
| | $ee$ | 23.7 % | 7.6 % | 2.6 % |
| | $\mu\mu$ | 22.5 % | 7.6 % | 3.6 % |
| | All channels | 11.0 % | 3.6 % | 1.5 % |

### 3.3.2 Inclusive template method

The inclusive template method is based on the observation that the three dominant sources of isolated leptons which can be selected in the $e\mu$ channel are $t\bar{t}$, $WW$ and $Z \to \tau\tau$. However, these three processes can be separated looking at the two-dimensional plane spanned by $E_{\mathrm{T}}^{\mathrm{miss}}$ and number of jets, as shown in Fig. 14. Table 7 shows that there might be instrumental effects that introduce non-prompt non-isolated or falsely identified leptons, primarily in single $W$-boson and Drell-Yan decays to muons. These non-prompt non-isolated or falsely identified leptons can weaken the separation of the templates if their contribution is too large. To reduce this effect we add a 'tight-electron' requirement on one of the electrons. Signal sensitivity and robustness against systematic uncertainties are improved by adding also the $ee$ and $\mu\mu$ channels. In the $\mu\mu$ channel we reject events where the $E_{\mathrm{T}}^{\mathrm{miss}}$ is aligned along any of the reconstructed muons. To further reduce the Drell-Yan background in the $ee$ and $\mu\mu$ channels a $E_{\mathrm{T}}^{\mathrm{miss}} > 35$ GeV cut and a Z-veto are added in those channels. The estimated number of events remaining after these additional background rejection cuts are shown in Table 10.

By constructing normalised 2D templates for each channel we determine the relative size of the $t\bar{t}$,

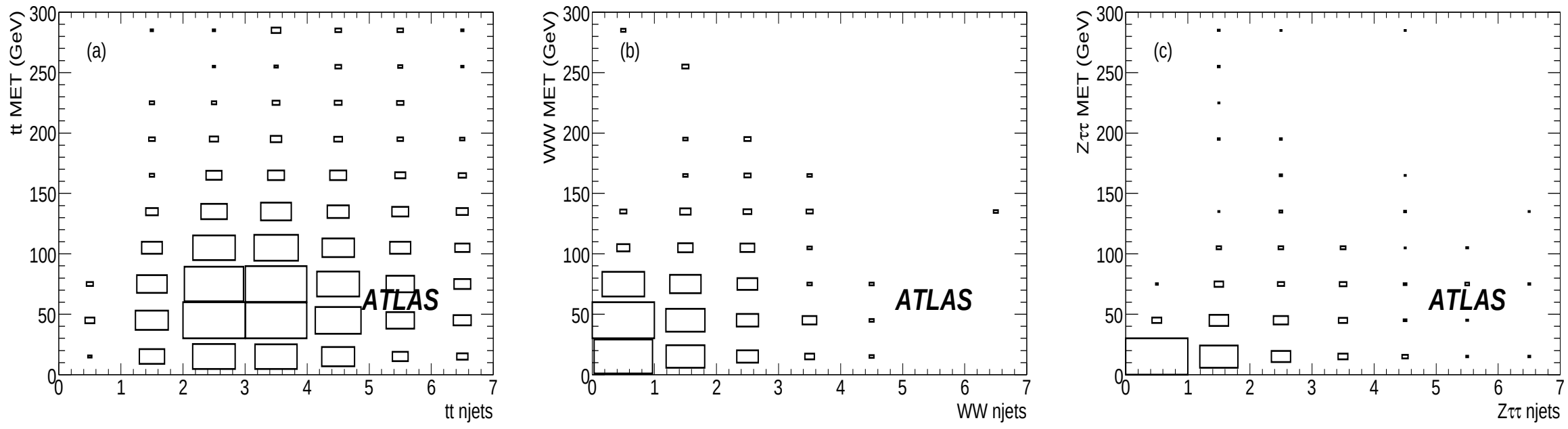


Figure 14: Monte Carlo $e\mu$ templates spanning the plane $E_{\mathrm{T}}^{\mathrm{miss}}$ and number of jets for $t\bar{t}$ (a), $WW$ (b) and $Z \to \tau\tau$ (c).

Table 10: Estimated number of events remaining after the background rejection cuts used in the inclusive template analysis.

| Sample | $e\mu$ | $ee$ | $\mu\mu$ |
|---|---|---|---|
| $t\bar{t}$ (di-lepton) | 516 | 213 | 178 |
| $t\bar{t}$ (semi-leptonic) | 12 | 11 | 2 |
| $Z \to e^+e^-$ | 3 | 9 | 0 |
| $Z \to \mu^+\mu^-$ | 9 | 0 | 18 |
| $Z \to \tau^+\tau^-$ | 151 | 10 | 2 |
| $W \to e\nu$ | 35 | 28 | 0 |
| $W \to \mu\nu$ | 11 | 0 | 11 |
| $WW$ | 57 | 17 | 19 |
| $WZ$ | 5 | 3 | 2 |
| $ZZ$ | 1 | 1 | 1 |
| single top | 3 | 1 | 1 |

$WW$ and $Z \to \tau\tau$ contribution by performing a binned log likelihood fit to pseudo-experiments. To ensure a fit with a pull distribution compatible with a Gaussian with zero mean and unit sigma we relax the template likelihood using nuisance parameters constrained to their expected errors. The nuisance parameters are: common acceptance and common background normalization for each final state. Bias from potential new physics is avoided by adding one free parameter with an associated orthogonal background template. Orthogonal in this context means that an additional background template does not introduce a bias when only SM events are present in the pseudo-experiments. The fit has in total ten free parameters including the $t\bar{t}$, $WW$ and $Z \to \tau\tau$ cross-sections.

Many different configurations of background templates are scanned and the template with the highest probability is selected. For robustness, the fit is first performed in the $e\mu$ channel, and then the fitted parameters are fed as start values into the full fit which includes all channels: $e\mu$, $ee$ and $\mu\mu$. The result from the full fit applied to pseudo-experiments including all systematics is shown in Fig. 15.

Systematic uncertainties enter in two ways: in the acceptance of the two leptons, and in the shapes of the 2D templates exemplified in Fig. 14. The impact of the systematic uncertainties is estimated using pseudo-experiments where all uncertainties and correlations, as discussed in section 3.4, are included, except the luminosity. Note that the acceptance is by definition not affected by the jet energy scale, but has a 2% uncertainty from QCD showering. The shape has a 2% uncertainty from QCD initial

state showering, and 2% from the jet energy scale. QCD final state showering only comes from $t\bar{t}$ and is found not to have any measurable effects on the fit when all channels are included. The projected significance versus integrated luminosity including all systematic uncertainties, except luminosity, is shown in Fig. 15.

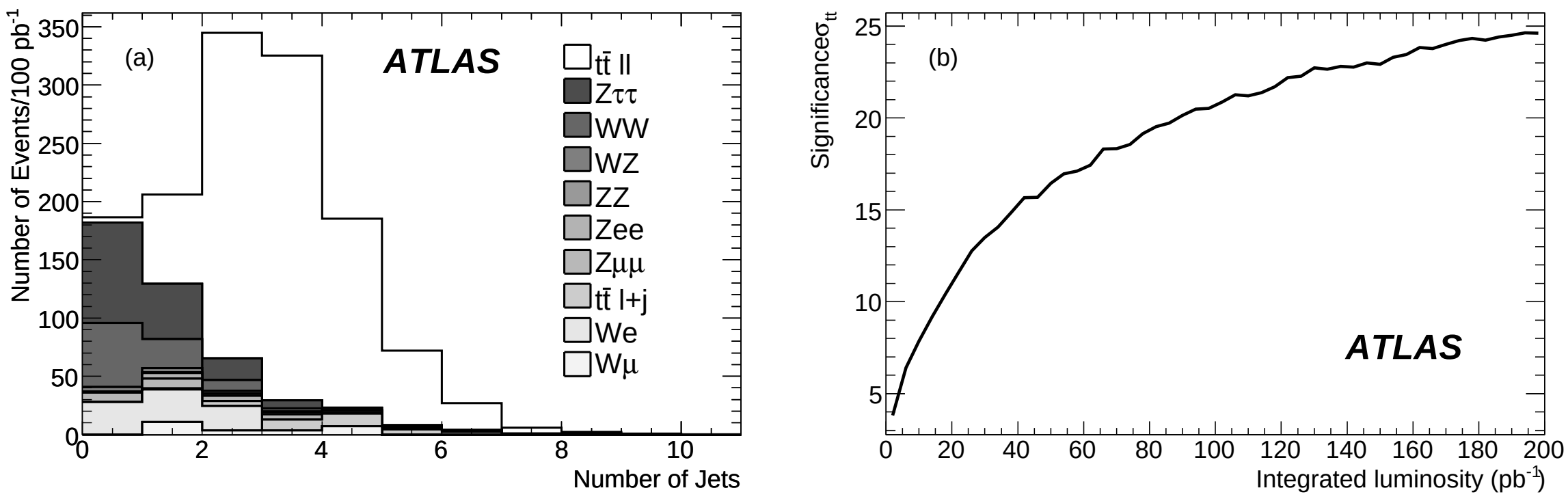


Figure 15: (a) Total composition of the inclusive di-lepton selection versus number of jets. (b) Projected significance from pseudo-experiments for the inclusive template fit versus integrated luminosity including all channels and all systematic uncertainties.

#### 3.3.3 Likelihood method

The likelihood method uses the following log-likelihood function to extract the parameters $N_{\text{sig}}$ and $N_{\text{bkg}}$ given the fixed total number of events $N_{\text{tot}}$ and the measurements $(x_i)$:

$$L = -\sum_{i=1}^{N_{\text{tot}}} \ln(G[x_i|N_{\text{sig}}, N_{\text{bkg}}]) + N_{\text{tot}}$$

with

$$G(x) = N_{\text{sig}} \times S(x) + N_{\text{bkg}} \times B(x)$$

The multidimensional function $G(x)$ is the sum of the functions $S(x)$ which describes the signal distribution and of $B(x)$ that describes the background distribution. The functions are determined by fitting Chebychev polynomials to the signal and background Monte Carlo distributions after the cut and count cuts were applied in the variables $|\Delta\varphi(\text{lepton}_0, E_{\text{T}}^{\text{miss}})|$ ($\Delta\varphi$ between the highest $p_T$ lepton and the missing transverse energy vector) and $|\Delta\varphi|(\text{jet}_0, E_{\text{T}}^{\text{miss}})|$ ($\Delta\varphi$ between the highest $p_T$ jet and the $E_{\text{T}}^{\text{miss}}$ vector). Fig. 16 shows one of the distributions and the solid line shows the fits to the distribution that are used as $S(x)$ and $B(x)$.

The sum of the semi-leptonic $t\bar{t}$ , $Z \to \ell^+\ell^-$ and $WW$ events are considered as background and added up according to their cross-section to produce one single background distribution.

To estimate the error on the cross-section, ensemble tests were performed for different integrated luminosities ranging from 10 pb$^{-1}$to 1 fb$^{-1}$. The relative statistical errors are presented in Table 11.

### 3.4 Systematic uncertainties

The systematic uncertainties have been evaluated according to the standard prescription [9]. In particular, the lepton ID efficiency will be measured from data using $Z$-boson events and the uncertainty is expected to be of the order 1%. The lepton trigger efficiency will be measured from data using $Z$ events with an

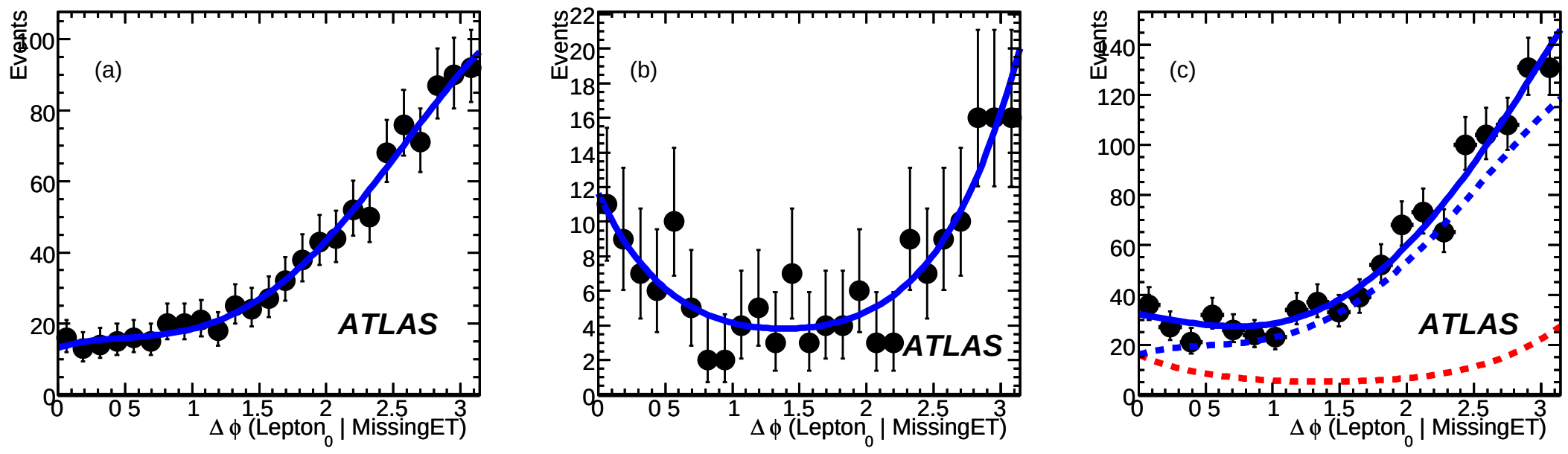


Figure 16: Distribution of $\Delta\varphi$ between the highest $p_T$ lepton and the $E_T^{miss}$ momentum vector for (a) signal, (b) background and (c) signal/background composition according to the ratio obtained from the cut analysis.

Table 11: Expected statistical error evaluated from ensemble tests on the cross-section measurement for the likelihood method for different luminosities. For 10 pb$^{-1}$ the fits in the subchannels do not converge well enough to give results.

| Luminosity [pb$^{-1}$] | | 10 | 100 | 1000 |
|---|---|---|---|---|
| $\Delta\sigma/\sigma$ | $e\mu$ | - | 9.1 % | 2.7 % |
| | $ee$ | - | 16.0 % | 4.6 % |
| | $\mu\mu$ | - | 7.8 % | 2.5 % |
| | All channels | 18.2 % | 5.2 % | 1.7 % |

Table 12: Uncertainties on the cross-section measurement for the cut and count and the likelihood methods.

| | **cut and count method** | | | | **likelihood method** | | | |
|---|---|---|---|---|---|---|---|---|
| $\Delta\sigma/\sigma$ [%] | $e\mu$ | $ee$ | $\mu\mu$ | all | $e\mu$ | $ee$ | $\mu\mu$ | all |
| CTEQ6.1 set | 2.4 | 2.9 | 2.0 | 2.4 | 0.3 | 0.4 | 0.2 | 0.2 |
| MRST2001E set | 0.9 | 1.1 | 0.7 | 0.9 | 0.2 | 0.2 | 0.1 | 0.2 |
| JES-5% | -2.0 | 0.0 | -3.1 | -2.1 | -5.4 | 1.1 | 4.9 | 8.3 |
| JES+5% | 2.4 | 4.1 | 4.7 | 4.6 | 7.8 | 3.9 | -4.6 | -4.4 |
| FSR | 2.0 | 2.0 | 4.0 | 2.0 | 0.2 | 0.4 | 0.0 | 0.3 |
| ISR | 1.1 | 1.1 | 1.2 | 1.1 | 2.5 | 1.8 | 0.0 | 1.7 |
| parameters-1$\sigma$ | | | | | -3.0 | -0.2 | -2.1 | -1.8 |
| parameters+1$\sigma$ | | | | | 3.2 | 0.8 | 2.0 | 2.0 |

uncertainty which should also be of the order 1%. The lepton fake rate uncertainty is estimated from dijet events, and during the initial phase of data taking we expect this uncertainty to be of the order of 50%. The systematic effect due to the uncertainty of the jet energy scale is investigated by scaling the reconstructed jet energies by ±5%. The value of the $E_T^{miss}$ is rescaled accordingly. The initial and final state radiation was investigated and the results are summarised in Table 12 as well.

For the likelihood, the fit parameters of the Chebychev polynomials were varied by 1 $\sigma$ at the same time in the same direction. The results can be seen in Table 12.

The reweighting technique, to assess the systematic uncertainties from parton densities is also described in Section [9]. To cross-check the technique ATLFAST samples with different PDFs were used. For the cut and count method the efficiency using MRST PDFs is larger than the efficiency using CTEQ PDFs. Table 12 summarises the changes in the selection efficiency. For the cut and count analysis this directly translates into an uncertainty on the cross-section determination. For the likelihood method the reweighted events were used to fit the $N_{\rm sig}$ and $N_{\rm bkg}$ to the templates which were generated with the unmodified fully simulated events.

### 3.5 Contribution of new physics

Many models of physics beyond the Standard Model can have a significant branching ratio to the di-lepton final state. In particular the 'cut and count' method is sensitive to these potential contributions. In some SUSY scenarios the contribution can be as large as one-third of the total $t\bar{t}$ signal. This indicates how crucial it is to include all final states in order to verify the global consistency of the $t\bar{t}$ cross-section. The template/likelihood methods are able to consider more distinctive $t\bar{t}$ event properties and can be made much more robust against non-Standard Model sources.

Data-driven methods which consider the full kinematics of the di-lepton $t\bar{t}$ system can further help to disentangle new physics, as described in other notes [12] of this volume.

### 3.6 Results

The final results in percent for the combined di-lepton channels for an integrated luminosity of 100 pb$^{-1}$ are summarised here

$$\text{Cut and Count method:} \quad \Delta\sigma/\sigma = \quad (4(\text{stat})^{+5}_{-2}(\text{syst}) \pm 2(\text{pdf}) \pm 5(\text{lumi}))\% \tag{4}$$

$$\text{Template method:} \quad \Delta\sigma/\sigma = \quad (4(\text{stat}) \pm 4(\text{syst}) \pm 2.(\text{pdf}) \pm 5(\text{lumi}))\% \tag{5}$$

$$\text{Likelihood method method:} \quad \Delta\sigma/\sigma = \quad (5(\text{stat})^{+8}_{-5}(\text{syst}) \pm 0.2(\text{pdf}) \pm 5(\text{lumi}))\% \tag{6}$$

## 4 Discussion and outlook

In this note we have demonstrated that ATLAS will be able to reliably determine the $t\bar{t}$ production cross-section already from the startup period of LHC. We have determined this cross-section for the $t\bar{t}$ system decaying both into a single electron or muon with associated jets, or two electrons or muons with jets. For the single-lepton mode, we have investigated robust selection criteria that do not depend on the $b$-quark tagging. For the di-lepton channel various complementary channels have been investigated.

With only 100 pb$^{-1}$ of accumulated data, we have shown that we can observe the top-quark signal and measure its production cross-section. Various methods have been presented and the corresponding uncertainties studied. Apart from the luminosity uncertainty, the overall uncertainties are of the order of (5-10)% and are dominated by systematics. Consistency between the methods constrain contributions of new physics as they affect the various methods differently.

# Prospect for Single Top Quark Cross-Section Measurements

**Abstract**

At the LHC, the single top quarks will be produced at a third of the rate of top quark pairs. With more than two million single top quark events produced every year during the low luminosity run, a precise determination of all contributions to the total single top quark cross section seems achievable. Comparison between the measured cross section and the theoretical prediction will provide a crucial test of the Standard Model. These measurements will lead to direct measurement of $V_{\mathrm{tb}}$ at the few percent level of precision, and constitute a powerful probe for new physics, via the search for evidence of anomalous couplings to the top quark or the measurements of additional boson contributions to single top quark production.

The single top quark production mechanism proceeds through three different sub-processes resulting in distinct final states, topologies and backgrounds. Given the level of backgrounds affecting the individual selections and the importance of the systematic uncertainties, the use of sophisticated methods appears mandatory to unambigiously observe each process and determine precisely the corresponding cross sections. This report presents the methods developed to optimize the selection of single top quark events in the three channels and establishes the ATLAS potential for the cross section measurements for the early data period and for a 30 fb$^{-1}$ low luminosity run.

## 1 Introduction

The top quark is one of the key particles in the quest for the origin of particle mass. In particular the electroweak interaction of the top quark is sensitive to many types of new physics. The electroweak production of top quark leads to a final state of a single top quark plus other particles. The production cross section is sensitive to contributions from new particles such as new heavy bosons W$'$ or charged Higgs bosons H$^{\pm}$. Other processes such as flavor-changing neutral currents also result in a single top quark final state [1]. Furthermore, single top quark production is an important background to many searches for new physics.

The D0 [2] and CDF [3] collaborations at the Fermilab Tevatron reported evidence for single top quark production and a first direct measurement of the CKM matrix element $V_{\mathrm{tb}}$. This involved advanced analysis methods to extract the small single top quark signal out of the large backgrounds. The Tevatron experiments will collect several fb$^{-1}$ of data, and expect to not only observe single top quark production at the 5 sigma level but also separate two different production modes: the s-channel and the t-channel. However, the single top quark production cross section is small at Tevatron energies and single top quark measurements will be limited by statistics.

At the LHC, the number of signal events is not a problem anymore. The LHC will not only be a strong interaction top quark factory but will also produce several million single top quark events. The cross section for all three modes of single top quark production as well as the CKM matrix element $V_{\mathrm{tb}}$ can be measured with high precision [4]. Once the single top quark signal has been established, detailed measurements of the process will follow, for example of the top quark polarization, ratios of cross sections, and charge asymmetries. Searches in each of the single top quark channels are sensitive to new physics even before the Standard Model single top quark signal is found. The D0 collaboration has published limits on W$'$ boson production and flavor changing neutral currents [5, 6].

This paper describes the cross section measurements for the three single top quark production modes and is organized as follows: the single top quark phenomenology is introduced in Section 2. All studies are based on a common event preselection, presented in Section 3. The prospects for the cross section measurements for the t-channel, s-channel and Wt-channel are then presented in Sections 4, 5, and 6. Finally, we summarize our conclusions in Section 7.

## 2 Phenomenology and strategy for single top quark analyses in ATLAS

In the Standard Model single-top quark production is due to three different mechanisms: (a) W-boson and gluon fusion mode, which includes the t-channel contribution and is referred to as t-channel as a whole (b) associated production of a top quark and a W-boson, indicated as Wt-channel, and (c) s-channel production coming from the quark anti-quark annihilation. Among those channels, the dominant contribution comes from the t-channel processes which account for $246^{+12}_{-12}$ pb [7, 8]. The Wt-channel contribution amounts to $66 \pm 2$ pb [9] while the s-channel mode is expected to have a cross section of $11 \pm 1$ pb [7, 8]. The cross sections are calculated at the next-to-leading Order (NLO) with an input top quark mass of 175 GeV. A 4.3 GeV uncertainty on the top quark mass is included in t-channel and s-channel cross section uncertainties, while the Wt-channel uncertainty includes strong energy scale variations only.

In the Standard Model, the top quark is assumed to decay almost exclusively into a W-boson and a b quark. The W-boson can then decay leptonically or hadronically. In the following, when discussing the analysis strategy in the s- and t-channels, only the leptonic decays of the W-boson are considered ($l\nu b\bar{b}$ and $l\nu b\bar{b}q$ final states, respectively)[1]. For the associated production (Wt-channel), only the modes where a lepton originating either directly from the W-boson produced together with the top quark or from the top quark decay are used. The $\tau$ decay modes are included in all relevant samples and treated as a signal for leptonic $\tau$ decays, since the signal selection is aimed at electron and muon signatures.

Top quark pair production constitutes a dominant background to single-top quark events. The LHC total production cross section computed at NLO with next-to-leading logarithmic resummations is $\sigma(t\bar{t}) = 833 \pm 100$ pb [10] [11], about 3 times larger than the single top quark cross section, and more than 80 times that of the s-channel. Given the fact that single top quark final state topology is characterized by one high $p_T$ lepton, missing energy and jets, $t\bar{t}$ production represents a significant background in its lepton+jets decay mode, i.e. when one of the W's decays leptonically and the other hadronically ($t\bar{t} \to l\nu bjjb$), with a final state containing two jets from the hadronization of b quarks and two jets from light quarks, a high $p_T$ lepton and missing energy. The "dilepton" channel ($t\bar{t} \to l\nu bl\nu b$) where a lepton is lost in acceptance also constitutes a major background. Finally, top quark pairs with one or both W-boson(s) decaying into a $\tau$ lepton where the $\tau$ decays into an electron or a muon, may also survive the selection.

W+jets events constitute another major source of background because of a cross section several orders of magnitude above the one of the single top quark production. The Leading Order (LO) Alpgen [12] generator with the HERWIG [13] parton shower algorithm was used for the generation of W+jets and $Wb\bar{b}$+jets events in this analysis. The corresponding LO cross-sections have also been used. A 20% uncertainty on the W+jets and $Wb\bar{b}$+jets cross-sections is considered in the following. WW and WZ diboson processes were also studied using samples generated with the HERWIG generator. The corresponding cross-sections are reported in [10].

While QCD dijet and multijet processes do not have features of the signal, they contribute to the background due to their overwhelming cross section and small but finite rate of object misidentification.

[1] The hadronic decay modes have obvious disadvantages for triggering and the lack of lepton signature increases the background significantly.

The estimation of this background needs to be done using data-driven methods and has been studied using full simulated dijet events generated with PYTHIA [14].

## 3 Single top quark event preselection

The three production modes of single top quark events are characterized by similar features that motivate a common set of preselection criteria. This preselection, which classifies the events according to exclusive electron and muon selections, is also aimed at reducing the level of three types of processes which constitute the main backgrounds to single top quark analyses: the top quark pair production, the W+jets and the QCD events.

### 3.1 Triggering and event preselection

The triggers for single top quark events rest upon the use of the inclusive isolated electron and muon triggers. Their design and performance for the three levels of trigger are presented extensively elsewhere [15]. Triggering efficiency is 84±1% overall for top pair quark events. Only mariginal differences are seen in single top events. Note that for fast simulation samples like W+jets datasets, where no trigger information is available, a trigger weight derived from turn on curves established on $t\bar{t}$ events is applied to every event.

Selected events must have at least one offline high $p_T$ isolated lepton in the central region with $|\eta| \leq 2.5$. The isolation criterion requires that the energy in a cone of $\Delta R$= 0.2 around the lepton direction be less than 6 GeV and is important for the rejection of QCD background in which a jet can fake an electron or a muon [10]. Muons and electrons are required to have a transverse momentum greater than 30 GeV ensuring that the trigger efficiency is on the plateau and hence less sensitive to trigger uncertainties. Finally, the sign of the highest $p_T$ lepton gives the flavor of the decaying top quark.

The event must then pass a second isolated lepton veto cut, applied to any lepton with a $p_T$ above 10 GeV and in the pseudo-rapidity region $|\eta| \leq 2.5$ in order to reduce contamination from dilepton backgrounds. The rejection of $t\bar{t}$ events in the dilepton channel is increased by a factor 2.5 as this requirement is applied. The impact on the signal efficiency is limited, with a loss of 3.6% in both the s- and t-channel and 2.7% in the Wt-channel.

Events are preselected if at least two jets are reconstructed with a $p_T$ above 30 GeV in the pseudo-rapidity region $|\eta| \leq 5.0$. The $t\bar{t}$ production being the dominant background to single top quark analyses, the use of a jet veto appears mandatory. The event jet multiplicity is thus further required to be lower or equal to four, where the extra jets must be reconstructed with $p_T$ above 15 GeV. It is important to keep this threshold as low as possible, since lowering the veto $p_T$ threshold from 30 to 15 GeV results in a rejection of $t\bar{t}$ events increased from 5 to almost 10. Among those events, the l+jets decay modes are the most sensitive to such a requirement with a rejection rate that is doubled compared to the 30 GeV threshold. Dilepton events are less affected with a rejection rate going from 5 to 6.6. This requirement results in a relative loss of 20 to 25% of the single top quark in the t- and Wt-channel channels and of 7% in the s-channel.

Among the highest $p_T$ jets, one at least must be b-tagged, have a $p_T$ above 30 GeV and be in the central pseudo-rapidity region $|\eta| \leq 2.5$. The b-tagging algorithms are described elsewhere [10]. A cut on the b-tag weight corresponding to a 60% efficiency and a mistag rate of light jets of 100 [16] is used. The mistag rates being small, we use an additional method to estimate backgrounds which do not contain b partons in the final states but are present in the selected sample because of a cross section several orders of magnitude above that of the signal. This is the case for the W+jets production. In those events, each jet is assigned a tag weight based on the parametrization of the mistag rate as function of the jet pseudo-rapidity and transverse momentum. The combination of all jets results in a weight assigned to the event

to be considered as a "1 b-tag inclusive", "2 b-tags" etc... event. Known as the "tagging rate function", this method, which applies on Monte Carlo events, improves the statistics in the distributions of such events, while keeping a global normalization corresponding to the actual mistag rates [17].

Finally the transverse missing energy is required to be larger than 20 GeV. This criterion is consistent with the leptonic decays of a W-boson while reducing the contamination from QCD events.

### 3.2 Preselection efficiency

Preselection efficiencies as well as event yield for an integrated luminosity of 1 $\mathrm{fb}^{-1}$ are reported in Table 1. The single top quark efficiency ranges between 5 and 6% in the electron channel and between 5.5 and 7.5% in the muon channel respectively. These numbers translate into a total of approximately 5,000 and 4,200 single top quark events in the muon and electron channel respectively including $\tau$ decays.

The production of $\mathrm{t\bar{t}}$ events constitutes the dominant source of background to single top quark analyses. The $\mathrm{t\bar{t}}$ events in the $\mathrm{l+jets}$ $(\mathrm{l = e}, \mu)$ modes are selected with an efficiency lower than 5%, resulting in a total of 22,580 events. The $\tau + \mathrm{l}$ (and $\tau\tau$) modes also contribute significantly with about 6,000 expected events. Despite the use of a second lepton veto dilepton events, where a lepton escapes detection, also contribute to the preselected sample significantly with about 4,000 events surviving the selection. The production of W+jets events is also an important source of background, with about 13,000 events in the preselected sample. However these events populate mostly the lower jet multiplicity bin and their selection depends crucially on the mistag rates. The selection of $\mathrm{Wb\bar{b}}$+jets events results in about 1,000 events.

QCD events may also contribute to the selected sample of events. However, the requirements of the presence of at least one isolated identified lepton and one jet tagged as a b-jet reduce very significantly their level as discussed in [11]. Typically QCD background is expected to be smaller than the W+jets background. Further rejection may be achieved by applying more restrictive requirements on the lepton identification, or on the missing transverse energy and its correlation in $\phi$ with reconstructed objects like leptons and jets [18]. The uncertainty on the remaining level of QCD background may be large. The QCD background contamination in selected samples should preferably be monitored, for example by the reconstruction of the transverse W-boson or top quark reconstructed masses [19].

### 3.3 Cross-section determination and systematic uncertainties

Specific selections have been defined for the t-, s- and Wt- channels analyses. A general procedure to determine the cross-section on the selected samples and assess the errors associated to the measurement has been defined and is common to all three single top quark analyses. The following expression is used to calculate the cross section, $\sigma$:

$$\sigma = \frac{N_{sig}}{a \times \mathscr{L}} = \frac{N_{tot} - B}{a \times \mathscr{L}}, \tag{1}$$

where $N_{sig}$ and $N_{tot}$ are the number of signal and all selected events respectively, $B$ is the number of background events, $a$ is the signal acceptance and $\mathscr{L}$ is the luminosity of the data sample. In the following studies, these numbers are estimated using Monte Carlo samples only. Data driven methods are expected be used to estimate specific backgrounds in the forthcoming data analyses.

The propagation of the errors into the cross-section has been done in a consistent way among the three channels. They are combined and propagated to the measured cross section using a Monte Carlo method, which randomly generates $N_{tot}$ according to a Poisson distribution, and varies randomly $B$ and $a$ for every systematic source by an amount chosen around its central value, according to a gaussian distribution. This procedure is performed a few thousand times and the RMS of the resulting distribution is interpreted as the total uncertainty.

Table 1: Preselection efficiency, including trigger, for signal and background events. The uncertainties come from Monte Carlo statistics only. The convention $l = e, \mu$ is used.

| Process | Muon channel | | Electron channel | |
|---|---|---|---|---|
| | $\varepsilon(\%)$ | $N(1fb^{-1})$ | $\varepsilon(\%)$ | $N(1fb^{-1})$ |
| s-channel$\rightarrow$ l | $7.1 \pm 0.1\%$ | 166±3 | $5.8\ \pm 0.\%$ | 136±3 |
| s-channel$\rightarrow \tau$ | $0.7 \pm 0.1\%$ | 8±1 | $0.5 \pm 0.1\%$ | 6±1 |
| t-channel$\rightarrow$ l | $5.9 \pm 0.2\%$ | 3143±80 | $5.2 \pm 0.1\%$ | 2787±76 |
| t-channel$\rightarrow \tau$ | $0.6 \pm 0.1\%$ | 169±19 | $0.3 \pm 0.05\%$ | 92±14 |
| Wt-channel$\rightarrow$ l | $6.8 \pm 0.1\%$ | 1314±27 | $5.6 \pm 0.1\%$ | 1091±24 |
| Wt-channel$\rightarrow \tau$ | $1.0 \pm 0.1\%$ | 93±7 | $0.8 \pm 0.1\%$ | 77±7 |
| $t\bar{t} \rightarrow l + jets$ | $4.9 \pm 0.05\%$ | 11846±130 | $4.4 \pm 0.05\%$ | 10734±124 |
| $t\bar{t} \rightarrow \tau + jets$ | $0.6 \pm 0.05\%$ | 757±34 | $0.5 \pm 0.02\%$ | 625±31 |
| $t\bar{t} \rightarrow ll$ | $5.8 \pm 0.1\%$ | 2257±58 | $4.6 \pm 0.1\%$ | 1762±51 |
| $t\bar{t} \rightarrow l + \tau$ | $7.9 \pm 0.2\%$ | 3055±66 | $6.7 \pm 0.2\%$ | 2595±61 |
| $t\bar{t} \rightarrow \tau\tau$ | $1.6 \pm 0.2\%$ | 158±16 | $1.6 \pm 0.2\%$ | 151±15 |
| Wbb+jets$\rightarrow$ l or $\tau$ | $2.6 \pm 0.05\%$ | 514±21 | $2.1 \pm 0.05\%$ | 424±19 |
| W+jets$\rightarrow$ l or $\tau$ | $0.014 \pm 0.001\%$ | 7437±105 | $0.011 \pm 0.002\%$ | 5449±90 |
| WW | $0.3 \pm 0.04\%$ | 78±11 | $0.3 \pm 0.04\%$ | 65±10 |
| WZ | $0.6 \pm 0.04\%$ | 50±3 | $0.5 \pm 0.04\%$ | 39±3 |

The total uncertainty is calculated using a common procedure as described in Section 4.2.3 in the three individual analyses. The main sources of experimental systematic errors taken into account are the b-tagging efficiency, jet energy scale, luminosity, for which central values are provided in the corresponding sections. Theoretical uncertainties that have been considered are the errors on the background cross-sections, as well as the impact of the uncertainties in the parton distribution functions and the b-quark fragmentation in the selection efficiency. Note that we consider the errors as fully correlated between signal and background for jet energy scale, b-tagging, and luminosity.

# 4 Measurement of the t-channel cross section

The t-channel is the most promising channel for single top quark observation at the LHC. It has the largest theoretical cross section of the three channels and its event features give us reasonable visibility of the signal according to the studies done so far [20]. In addition to the cross section measurement, the t-channel single top quark is one of very few candidate channels for the $V_{tb}$ and the top quark polarization measurements. It is hoped that these measurements will shed light on our understanding of the electroweak symmetry breaking since these quantities have not been studied with high precision in past experiments. Compared to $t\bar{t}$ , the t-channel single top quark analysis suffers from a higher level of background due to its lower jet multiplicity, which makes the selection sensitive to QCD and W+jets backgrounds whose cross sections are several orders of magnitude higher than that of the signal. Thus, while the signal production rate is not statistically limited at the LHC, a good strategy for signal extraction is required and a careful estimation of the background rate is necessary.

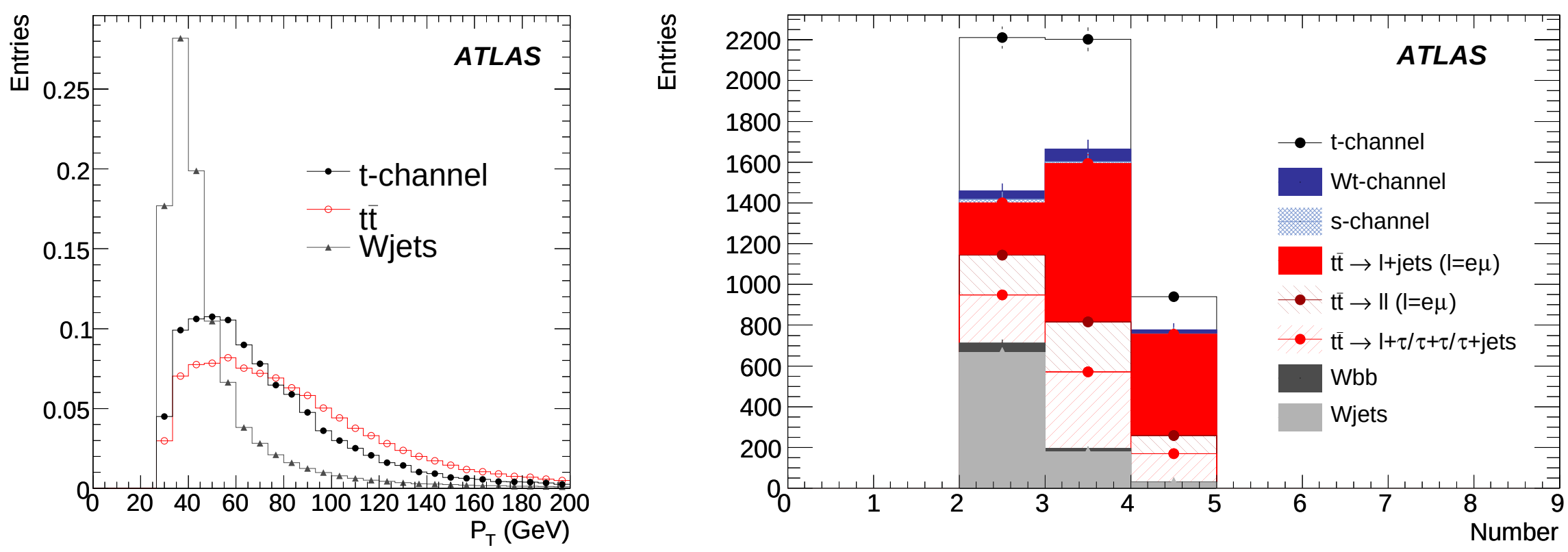


Figure 1: Signal and background distributions of the b-tagged jet $p_T$ and multiplicity of jets with $p_T >$ 30 GeV after the t-channel selection.

## 4.1 Cut based event selection

After the common preselection, the contribution from the background is still high mainly due to W+jets at the lower end of the kinematic distributions (such as in the jet $p_T$ spectrum) and $t\bar{t}$ at the higher end. The b-tagged jets coming from $t\bar{t}$ and the t-channel single top quark tend to have high $p_T$ and to be located more centrally since they mainly originate from top quark decays. On the other hand, b-tagged jets from W-boson events are much softer as they primarily come from mistagged jets originating from extra gluon radiation. Furthermore, the recoiling forward quark in the t-channel single top quark produces a high $p_T$ light jet in the forward direction. This was found to be one of few features useful to reject $t\bar{t}$ background

events because light jets from the $t\bar{t}$ events are typically initiated from hadronic W-boson decays and are therefore more central. A cut on b-tagged jet $p_T > 50$ GeV reduces the W+jets significantly, while a cut on the hardest light jet $|\eta| > 2.5$ can reject $t\bar{t}$. Figure 1 shows the $p_T$ distribution of the hardest b-tagged jet and the composition of the sample in terms of the jet multiplicity after the additional event selection.

Table 2 lists the number of preselected events, the number of events selected after the cut on b-tagged jet $p_T$ and the number after the light jet $\eta$ cut for the signal and the background processes. The signal efficiency is 1.81% and the signal to background ratio after all selection is 37%. Note that the single top quark $\tau$ decays are included in the signal count. By far the largest background contribution comes from the $t\bar{t}$ process; the l+jets top quark pair events contribute the most, although dilepton events have a higher survival probability of 1.36% compared to 0.64% of the l+jets. W+jets is the second largest background while the $Wb\bar{b}$ contribution is relatively small. The s-channel single top quark contribution is almost negligible and the diboson background is even smaller and therefore is not included in the table.

Table 2: Number of events selected after each cut in the t-channel analysis. The last column shows the number of remaining events in the cut-based analysis at the integrated luminosity of 1 fb$^{-1}$. Note that the convention $l = e, \mu$ is used.

| Process | Preselected | | b Jet $p_T$ | | light jet $\eta$ | |
|---|---|---|---|---|---|---|
| | $\varepsilon(\%)$ | N(1fb$^{-1}$) | $\varepsilon(\%)$ | N(1fb$^{-1}$) | $\varepsilon(\%)$ | N(1fb$^{-1}$) |
| t-channel | 7.7% | 6191±112 | 5.5% | 4412±95 | 1.8% | 1460±56 |
| $\mu$ channel | 4.1% | 3312±83 | 2.9% | 2352±71 | 0.9% | 728±40 |
| $e$ channel | 3.6% | 2879±78 | 2.6% | 2060±66 | 0.9% | 732±40 |
| s-channel | 9.0% | 316±5 | 7.0% | 245±4 | 0.8% | 26±1 |
| Wt-channel | 8.9% | 2575±37 | 6.4% | 1854±32 | 0.4% | 122±9 |
| $t\bar{t}$ l+jets | 9.3% | 22580±176 | 7.3% | 17775±158 | 0.6% | 1556±48 |
| $t\bar{t} \to l+\tau/\tau+\tau/\tau$+jets | 4.3% | 7342±104 | 3.4% | 5776±93 | 0.4% | 740±34 |
| $t\bar{t} \to ll$ | 10.4% | 4018±75 | 8.1% | 3143±67 | 1.3% | 520±28 |
| W+jets | 0.025% | 12886±138 | 0.012% | 6082±95 | 0.0017% | 873±36 |
| $Wb\bar{b}$ | 4.7% | 939±27 | 3.0% | 597±22 | 0.4% | 69±8 |
| $S/B$ | | 0.12 | | 0.12 | | 0.37 |
| $S/\sqrt{B}$ ($\sigma$) | | 27.5 | | 23.4 | | 23.4 |
| $\sqrt{S+B}/S$ | | 3.9% | | 4.5% | | 5.0% |

As seen in Table 2, a precision ($\sqrt{S+B}/S$) of 5% can be obtained after the final selection. Note, however, that the selection cut does not optimize the precision nor the significance of the signal observation. From purely statistical arguments, the t-channel cross section can be measured to a few percent accuracy with a few fb$^{-1}$ of data.

In comparison to the previous study in the ATLAS physics TDR [20], which reported S/B a ratio of 3, there is a significant apparent reduction in the event selection performance. The cut-based selection used here is similar to that used in the TDR and we would have expected a similar result. Further investigation revealed several issues with the TDR analysis. Firstly, Monte Carlo (MC) generators changed drastically in recent years. PYTHIA introduced a new parton shower algorithm, which is much more radiative. The matrix-element generator for the signal has also changed to AcerMC, which combines the NLO diagram contribution to the LO ones as opposed to PYTHIA, which only uses LO. For the W+jets background production, Alpgen is used because it is shown that it reproduces better the high $p_T$ tails of jet distributions at the TeVatron [21], while Herwig was used in the TDR. In addition, the TDR analysis did not include some of the crucial background processes, like dilepton $t\bar{t}$ channels and $\tau$ decay modes which are shown to contribute significantly in the present analysis. Therefore, the result of the TDR can

now be seen as too optimistic.

Note that the presence of pile-up affects the reconstruction of the objects selected in the event. In particular, the selection efficiency is directly affected by the presence of extra jets in the events. Specific Monte Carlo samples of single top quark and $t\bar{t}$ events have been produced with or without pile-up events superimposed. The use of events with pile-up results in a decrease of about 25% for signal events, and 34% for $t\bar{t}$ events. No systematic uncertainties will be associated with this value. The use of data will be mandatory for the tuning of the pile-up modeling and the corresponding uncertainty is expected to be brought down to a negligible value with respect to the other sources of error.

## 4.2 Systematic uncertainties

The systematic uncertainties have been evaluated following a procedure common to all three analyses. This procedure is defined here and the results specific to all analyses are reported in the corresponding sections.

### 4.2.1 Experimental systematic uncertainties

Various criteria need to be considered to assess the systematic effects as it depends on a number of detector sub-components and theoretical assumptions. A general procedure has been defined to assess all systematic errors in a consistent way for the three single top quark analyses. The estimate is shown here for the t-channel analysis only.

The effect of the uncertainties on the measured cross section is summarized in Table 4. While the statistical uncertainty is not a constraint for the t-channel cross section analysis even with the early data, systematic uncertainties can only be reduced with detailed understanding of the detector performance and theoretical uncertainties. The measurement in the t-channel is largely limited by such effects.

Table 3: Effect of b-tag and jet energy scale systematics. Numbers are quoted from relative variation of selected event.

| Process | b-tag -5% | b-tag +5% | JES -5% | JES +5% |
|---|---|---|---|---|
| t-channel | -4.0% | +4.8% | -4.8% | +5.0% |
| s-channel | -5.2% | +3.6% | -6.6% | +12.7% |
| Wt-channel | -8.9% | +8.9% | -2.0% | +6.9% |
| $t\bar{t}$ | -6.4% | +6.8% | -3.1% | +5.9% |
| $Wb\bar{b}$ | -4.0% | +8.9% | -12.6% | +12.7% |
| W+jets | -1.6% | +2.5% | -9.9% | +14.6% |
| Total bkgd. | -4.6% | +6.1% | -4.8% | +8.1% |

The uncertainty in the b-tagging performance and jet energy scale can have crucial effects on the measured cross section since we rely on jet kinematics to reduce the background. In particular, the large background contribution from $t\bar{t}$ can cause a large fluctuation on the total event rate with these uncertainties, which leads to a large error on the cross section measured. The jet energy scale uncertainty affects the $p_{\mathrm{T}}$ distribution of jets strongly at the low end of the distribution. Cutting on $p_{\mathrm{T}}$ in this region can lead to a larger uncertainty on the signal acceptance. The effect of 5% variation of the b-tagging performance has been considered together with the corresponding change in the mistag rate of non b-tagged jets. The effect of a 5% change of the jet energy scale determination (JES) has also been evaluated

on the efficiency. The impact of both effects on the selection is summarized in Table 3. It should be noted that the $Wb\bar{b}$ and W+jets backgrounds are more affected by the jet energy scale due to the fact that most of the jets in these sample are in the low-$p_T$ region, which means that they are more likely to enter or leave the acceptance as the scale is changed.

The uncertainty due to the trigger requirements has been estimated by computing the variation in the signal and background rates resulting from a bias of 1% in the inclusive lepton trigger efficiency. This corresponds to the expected level of precision derived from turn on curves determination with the early data. Similarly, a 1% uncertainty in the lepton identification was considered and the corresponding impact on the cross section measurement assessed. These two effects are expected to contribute to less than 2% of the total uncertainty with $1\text{fb}^{-1}$.

Note that for all those results, the limited statistics of the Monte Carlo samples may result in fluctuations in the variable distributions used to discriminate signal from background events. The uncertainty associated to the low statistics is found to be 6-8% with W+jets being the largest cotrigger requirements.

#### 4.2.2 Theoretical and Monte Carlo systematic uncertainties

Since the event selection relies heavily on the jet multiplicity, the number of additional jets from the parton shower can affect the selection efficiency. To evaluate the effect of ISR/FSR uncertainty, two sets of PYTHIA parameter variations were considered for comparison, which control the free parameters in the underlying event, initial- and final-state radiation, showering, and multiple interactions. The variations were combined to maximize the variation in the number of selected events to give a conservative estimate. The effect on the jet multiplicity is shown in Figure 2. It can be seen that the loose jet ($p_T$ > 15 *GeV*) distribution is affected more severely, which affects the selection efficiency due to the jet veto cut. The parameter variation set that gave the largest uncertainty leads to an overall uncertainty in the t-channel selection efficiency of $-11\%$ $+7\%$, and is quoted as a systematic uncertainty.

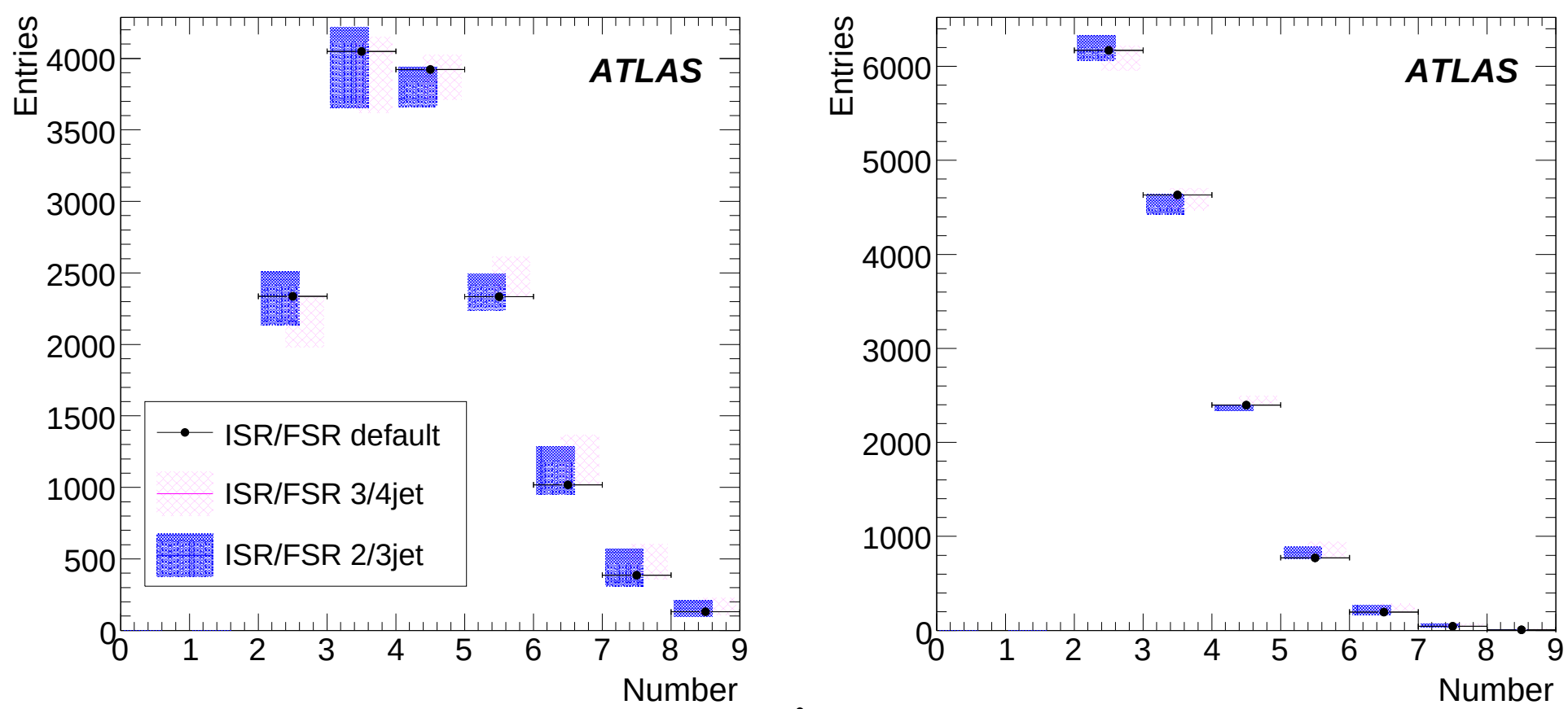


Figure 2: Variation in t-channel multiplicity for the loose ($p_T > 15$ GeV, left) and the tight ($p_T > 30$ GeV, right) jets due to ISR/FSR parameter variation. The distributions are shown without applying the t-channel selection cuts. The jet veto cut ($\leq 4$ loose jets) and the requirement of b-tagged jet were also removed for these plots to show the uncertainty over all multiplicities[2]. Black points are the nominal entries while the mesh band shows variation from "3/4 jet parameter set" and the filled band shows that of "2/3", which are the two sets of parameter variations compared to study different jet multiplicity events.

The uncertainty due to PDF was calculated using a re-weighting method together with error PDF

sets provided by the PDF packages. While the event generation was done only for the central value PDF, weights were calculated in each event according to:

$$w_i^{\pm} = \frac{f_1(x_1, Q; S_i^{\pm}) \cdot f_2(x_2, Q; S_i^{\pm})}{f_1(x_1, Q; S_0) \cdot f_2(x_2, Q; S_0)}, \tag{2}$$

where, $f_1$ and $f_2$ are the PDF values for a given hard scattering process (characterized by flavors $f$ and momentum fractions $x$ of the initial partons, and by $Q$, the event energy scale) evaluated for the $i^{th}$ error PDF pair $S_i^{\pm}$. The variation in efficiency was calculated by applying an event selection similar to the preselection cuts at the generator level. The variations in efficiency were added using the Hessian formalism and the results from CTEQ and MRST PDF error sets [22] [23] were consistent. The effect of the PDF uncertainty on the signal was evaluated to be +1.4% -1.1% and it is a minor contribution to the final systematics. A larger effect of +6.2% -5.5% was seen on the $t\bar{t}$ background. The PDF uncertainty was estimated for the leading contribution from the $t\bar{t}$ process only.

In addition to the above, the t-channel process has a fairly large uncertainty from MC generator predictions. Comparing various combinations of ME and PS generators, it was observed that the Pythia and Herwig parton shower algorithms give significantly different jet multiplicities. While we assume that this difference can be eliminated by tuning the parameters to the observed data[3], the instability of theoretical prediction from matrix element generators is an outstanding issue and a 4.2% variation in signal acceptance was seen by comparing AcerMC+Herwig and MC@NLO+Herwig. We quote this as an estimate for systematic uncertainty from the theoretical prediction. In our present analysis, the background is es-

Table 4: Summary of all uncertainties that affect the measured cross section, shown for the cut-based analysis and the BDT analysis. “Data statistics” represents the Poisson error one would expect from real data

| Source | Analysis of 1 fb$^{-1}$ | | | Analysis of 10 fb$^{-1}$ | | |
|---|---|---|---|---|---|---|
| | Variation | Cut-based | BDT | Variation | Cut-based | BDT |
| Data Statistics | | 5.0% | 5.7 % | | 1.6% | 1.8 % |
| MC Statistics | | 6.5 % | 7.9% | | 2.0 % | 2.5% |
| Luminosity | 5% | 18.3 % | 8.8% | 3% | 10.9 % | 5.2% |
| b-tagging | 5% | 18.1 % | 6.6% | 3% | 10.9% | 3.9% |
| JES | 5% | 21.6% | 9.9% | 1% | 4.4 % | 2.0% |
| Lepton ID | 0.4% | 1.5 % | 0.7% | 0.2% | 0.6 % | 0.3% |
| Trigger | 1.0% | 1.7 % | 1.7% | 1.0% | 3.6 % | 1.7% |
| Bkg x-section | | 22.9% | 8.2% | | 6.9 % | 2.5% |
| ISR/FSR | +7.2 -10.6% | 9.8 % | 9.4% | +2.2 -3.2% | 2.7 % | 2.5% |
| PDF | +1.38 -1.07% | 12.3 % | 3.2% | +1.38 -1.07% | 12.3 % | 3.2% |
| MC Model | 4.2% | 4.2 % | 4.2% | 4.2% | 4.2 % | 4.2% |
| Total | | 45% | 22% | | 22% | 10% |

timated from Monte Carlo and its normalization is currently estimated based on theoretical uncertainties. When data will be available, the $t\bar{t}$ and the W+jets backgrounds will be measured from data as well. It

[3]The current tunings of Pythia and Herwig parton shower weres obtained independently based on extrapolation from the Tevatron data. [24]

is beyond the scope of this paper to discuss methods for the data-driven background estimation and the uncertainty from the background is currently estimated based on theoretical uncertainties.

### 4.2.3 Summary of uncertainties

A large part of the final uncertainty is due to the overwhelming amount of $t\bar{t}$ background. For this reason, the event selection optimized for statistical significance or statistical precision does not give the smallest total uncertainty when systematics are included. Table 4 reports the uncertainties associated the cross section determination. The total uncertainty is dominated by systematic effects related to the background normalization. Reducing the background contamination to increase signal to background ratio would help to minimize the systematic uncertainty. It is therefore desirable to further optimize the selection beyond what can be achieved with simple cuts on the existing variables.

Note that for an analysis based on 10 fb$^{-1}$, the table reports the expected performance assuming a better understanding of the experimental aspects of the detection. Assuming a b-tagging efficiency known with a precision of 3%, the jet energy scale determined at 1%, a luminosity known at 3% and a better understanding of the background to the 3% level and the radiation modeling at 3%, a total systematic uncertainty of 10% seems achievable for the BDT analysis.

## 4.3 Multivariate event selection

In the cut-based analysis no variable was found that was effective to reject the $t\bar{t}$ background. One commonly employed separation technique is a Multivariate Analysis (MVA), which effectively factorizes the process of cut optimization using a set of rules for separating the signal from the background events. In the present analysis, an attempt to eliminate the remaining background contribution from $t\bar{t}$ makes use of the Boosted Decision Tree (BDT) method [25] within the TMVA [26] framework. The sample used is selected with the b-tagged jet $p_{\mathrm{T}}$ cut selection, without applying the forward jet $\eta$ cut. This cut is indeed effective to reduce the $t\bar{t}$ contribution but not very efficient. About 40 object/event level variables were studied using a genetic algorithm, which scanned a large number of variable sets. Extensive studies of this variable sets revealed that a small subset of the variables can achieve signal to background discrimination close to what was achieved using all variables. This set was further refined so that the chosen variables are not too sensitive to JES systematics:

- $p_{\mathrm{T}}$ of the leading b-tagged jet and non-b-tagged jet;
- $\eta$ of the leading non-b-tagged jet and $\cos\theta$ of the leading jet;
- Centrality ($\frac{p_T^{jet0}+p_T^{jet1}}{|p|^{jet0}+|p|^{jet1}}$);
- Scalar sum of the $p_{\mathrm{T}}$ of the two highest energy jets, the transverse missing energy $\not{E}_{\mathrm{T}}$, and the lepton $p_{\mathrm{T}}$;
- $\Delta R$ between the two jets with the highest transverse mometum
- $\Delta R$ between the leading jet and the lepton;
- $\Delta R$ between the leading non-b-tagged jet and the lepton;
- the lepton, and the W-boson transverse mass;
- $\eta$ of the jet with largest $|\eta|$;
- the number of jets with $p_{\mathrm{T}} > 30$ GeV.

Figure 3 (left) shows the BDT output discriminator constructed from the variables discussed above. By cutting on a high value of the discriminator, the $t\bar{t}$ background can be removed more effectively than by cutting on individual input variables. Since the BDT was optimized for the $t\bar{t}$ separation, as expected the output is not effective against W+jets. It can be seen in the right figure that a high level of signal purification is achieved using the BDT discriminator. We optimized the cut on the BDT output by min-

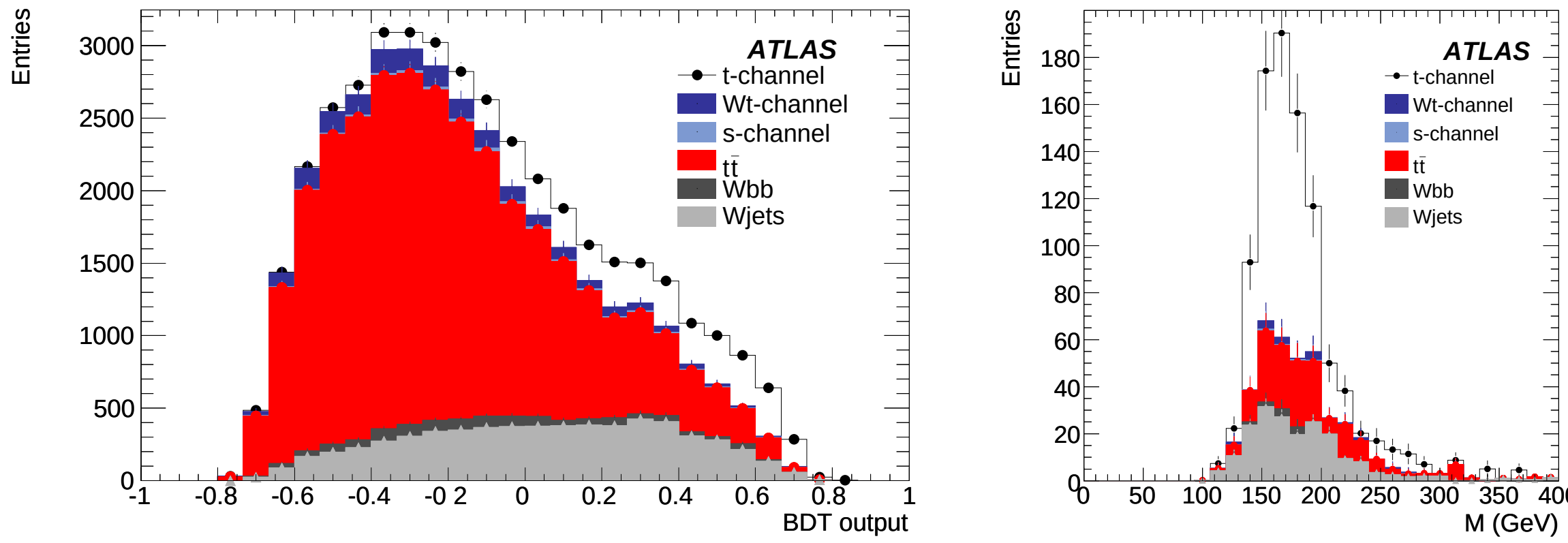


Figure 3: Boosted decision tree output for signal and background after the b-tagged jet $p_{\mathrm{T}}$ cut (left) and leptonic top quark mass distribution using cut on BDT output at 0.6 (right).

imizing the final uncertainty on the measured cross section, including the systematic effects. For each cut, the systematic uncertainties were calculated assuming that the relative systematic uncertainty stays constant for each channel for each source of uncertainty. This is a fairly reasonable assumption considering that the relative uncertainty changes slowly with the cuts and the variables were chosen to avoid larger systematics. For most cut values, the systematic effects are dominant and the total uncertainty is reduced with increasing signal to background ratio. On the other hand, with very tight cuts the statistical error becomes larger than systematics. The overall minimum was found to be at 0.6 where the signal to background ratio is 1.3 and 542 signal events are left as shown in Table 5. The reconstructed top mass distrubution after applying BDT selection is shown in figure 3 (right). The top mass was reconstructed from the lepton, $\not{E}_{\mathrm{T}}$ and the highest $p_{\mathrm{T}}$ b-tagged jet. The W-boson mass was constrained to be 80.4 GeV to obtain two solutions for neutrino kinematics and the one with the smaller $pz$ was selected. The statistical uncertainty is 5.7% while the total uncertainty is 22% as shown in Table 4.

Table 5: Final event yield after the cut on BDT discriminator at $1\mathrm{fb}^{-1}$.

| Process | t-channel | s-channel | Wt-channel | $t\bar{t}$ | W+jets | $Wb\bar{b}$ | $S/B$ |
|---|---|---|---|---|---|---|---|
| N($1\mathrm{fb}^{-1}$) | 542 | 3 | 15 | 184 | 201 | 10 | 1.31 |

## 4.4 Sensitivity at 1 fb$^{-1}$ and the measurement of $|V_{tb}|$

Although the estimated systematic uncertainty of the cut-based analysis is rather large (44.7%), it has been shown that this can be reduced by rejecting the $t\bar{t}$ background using boosted decision trees. The total uncertainty decreases as the $S/B$ ratio increases and the estimated uncertainty at 1 fb$^{-1}$ is

$$\frac{\Delta\sigma}{\sigma} = \pm 5.7\%_{stat} \pm 22\%_{sys} = \pm 23\%. \quad (3)$$

The t-channel cross section is proportional to $|f_L V_{\text{tb}}|^2$, where the parameter $f_L$ is the weak left-handed coupling and $f_L = 1$ in the Standard Model. In the theory predictions, the product $|f_L V_{\text{tb}}|^2$ is always set to unity. Thus, if one measures the cross section, and then divides by the theoretical cross section, one obtains a measurement of $|V_{\text{tb}}|^2$, making the Standard Model assumption that $f_L = 1$ [27].

The relative uncertainty on $V_{\text{tb}}$ is the relative uncertainty on $|V_{\text{tb}}|^2$ divided by two since $\delta|V_{\text{tb}}|/|V_{\text{tb}}| = \delta|V_{\text{tb}}|^2/2|V_{\text{tb}}|^2$. However, there are additional systematic uncertainties in the $V_{\text{tb}}$ measurement due to the presence of the theoretical cross section in the denominator. Here, we quote the uncertainty calculated in [7], in which a theoretical uncertainty of $+3.8\ -4.1\%$ is reported including the contributions due to the strong scale, PDF and top quark mass uncertainties. We use the average of the positive and the negative uncertainties. Therefore, the estimated uncertainty on the measured value of $V_{\text{tb}}$ is

$$\frac{\Delta|V_{\text{tb}}|}{|V_{\text{tb}}|} = \pm 11\%_{stat+sys} \pm 4\%_{theo} = \pm 12\%. \tag{4}$$

### 4.5 Summary

The cross section measurement of the single top quark t-channel was studied in this chapter. The characteristics of the signal and background were investigated in detail and an analysis strategy was developed first using simple cuts and then using boosted decision trees.

While a cut-based event selection can achieve a statistical precision of a few percent at an integrated luminosity of 1 fb$^{-1}$, the $t\bar{t}$ background is difficult to reduce. This results in large systematic uncertainties coming from both experimental and theoretical origins. Uncertainties in the jet energy scale, b-tagging and luminosity all affect the measurement considerably. The uncertainty on the background cross section is also rather large though we expect that it will be constrained at higher accuracy with the data. This is also true for the QCD background, which is not included in our current analysis. The results shown are obtained on samples without any pile-up and no systematic uncertainty was associated to it. Data-driven background estimation methods should be developed once data is available. Among the theoretical issues, the ISR/FSR uncertainty degrades the measurement more significantly than other theoretical effects such as the PDF and the Monte Carlo generator model.

A multivariate background discrimination method is very effective in reducing the background and thus reducing the total uncertainty to nearly a half of the cut-based analysis. We conclude that multivariate analysis tools are highly effective for a t-channel cross section measurement and further studies of these techniques will be very beneficial for the improvement of the analysis in the future. However, to reach a precision at a few percent level, studies of systematic uncertainties and an excellent understanding of the detector response will be necessary.

## 5 Measurement of the s-channel cross section

The measurement of the single top quark s-channel appears the most delicate of the three main single-top quark processes. Suffering from a low cross section compared to the main backgrounds, the event topology makes this channel very sensitive to the presence of both $t\bar{t}$ and W+jets events. Because of the low jet multiplicity of such events, the analysis is also expected to be sensitive to dijet production, despite the tight requirements on the presence of at least two b jets. The s-channel is however one of the most interesting because the production of tb final state events is directly sensitive to contributions from extra W-bosons or charged Higgs bosons as predicted in two Higgs doublet model (2HDM) [28].

The event selection is presented in three steps. A first one makes use of a standard cut-based analysis, and will serve as a reference with the early data. In a second step, likelihood functions designed to improve the discrimination against specific backgrounds are presented together with the sets of discrim-

inant variables that enter their definition. Finally, the selection criteria applied on those likelihoods are defined so that the total uncertainty affecting the cross section determination is minimized.

### 5.1 Sequential cut analysis

After applying the common preselection described in Section 3, only two b-tagged jet events with $p_T$ above 30 GeV are selected, and a jet veto is applied on any other jet with a transverse momentum above 15 GeV. This strong requirement is used to reject $t\bar{t}$ events which represent the dominant background to our signal at this stage. This set of requirements also reduces the W+jets and QCD multijet contamination, since those events feature much softer b jets or no b jets at all. The selection also requires the opening angle $\Delta R$ between the two jets to be between 0.5 and 4.0, the scalar sum of total jet transverse momenta $H_T(\text{jet})$ to be above 80 and below 220 GeV and finally the sum of the transverse missing energy ($\not{E}_T$) and lepton transverse momentum ($p_T$) to be in the range between 60 and 130 GeV.

Selected event yields are reported in Table 6 for the three single top quark processes, and all $t\bar{t}$ and W+jets backgrounds. When adding all contributions, the overall signal efficiency is about 1.1% in the electron channel and 1.6% in the muon channel. This corresponds to a total of about 25 selected candidates for an integrated luminosity of 1 fb$^{-1}$. The signal to background ratio is about 10%. The dominant background is composed of the $t\bar{t}$ events which account for about 60% of the total background yield. Among those, the $t\bar{t}$ in the lepton+jets mode, including $\tau$ decays, contribute about 40%. The remaining backgrounds originate from the $Wb\bar{b}$+jets production, which constitutes about 14% of the background yield, and almost equally from the single top quark t-channel (11%) and W+jets (9%) events. As expected, W+jets events are removed due to the requirement of b-tagged jets, with a final yield depending upon the mistag rate. Diboson contributions (WW and WZ) are found to be negligible.

Table 6: Event yield for signal and background for the cut-based analysis in the 2-jet multiplicity bin for 1 fb$^{-1}$. Uncertainties come from Monte Carlo statistics only. The convention $l = e, \mu$ is used.

| Events in 1fb$^{-1}$ | e channel | $\mu$ channel | $e+\mu$ combined |
|---|---|---|---|
| s-channel | 10.3 ± 0.8 | 14.5 ± 1 | 24.8 ± 1.3 |
| t-channel | 17.0 ± 5.7 | 13.6 ± 5.1 | 30.6 ± 7.9 |
| Wt-channel | 6.5 ± 1.9 | 2.4 ± 1.2 | 8.9 ± 2.3 |
| $t\bar{t} \to l+\text{jets}$ | 18.8 ± 4.3 | 20.5 ± 5.3 | 39.3 ± 5.3 |
| $t\bar{t} \to ll$ | 29.0 ± 5.5 | 15.4 ± 5.0 | 44.4 ± 7.4 |
| $t\bar{t} \to l\tau$ | 20.5 ± 5.6 | 40.9 ± 6.9 | 61.4 ± 8.9 |
| $Wb\bar{b}$+jets | 18.9 ± 2.0 | 19.7 ± 2.0 | 40.6 ± 2.5 |
| W+jets | 14.8 ± 1.4 | 11.0 ± 2.2 | 25.8 ± 2.3 |
| Total Bkg | 125.5± 10.3 | 123.5 ± 10.6 | 251.0 ± 14.2 |
| $S/B$ | 8.2% | 11.7% | 9.8% |
| $S/\sqrt{B}$ | 0.9 | 1.3 | 1.6 |
| $\sqrt{S+B}/S$ | 1.1 | 0.8 | 0.7 |

### 5.2 Likelihood selection

The cut based analysis shows that a simple approach to select single top quark s-channel events is hampered by a high level of background. The use of a likelihood discriminator is aimed at improving the performance of the discrimination against backgrounds in order to purify the selected signal samples. This approach assumes that the distributions entering the definition of likelihood functions are well known and validated on data itself. This result can be achieved by cross-checking at every step the agreement between data and Monte Carlo distributions on selected sub-samples where a high level of $t\bar{t}$ and W+jets background is expected. In the following, only pre-selected events with exactly two jets, both of which are b tagged are considered. A jet veto on any other jet is applied.

#### 5.2.1 Definition of the likelihood functions

The main background processes to our signal show very distinct features in the final state and topology that lead us to define several likelihood functions each devoted to the discrimination of a specific process: three likelihood functions are devoted to $t\bar{t}$ events in the dilepton, the $l+\tau$ and the l+jets decay modes and two likelihood functions have been developed to discriminate against W+jets and t-channel events. Due to the limited Monte Carlo statistics, these likelihood discriminators have been defined by combining both muon and electron channels.

The list of variables entering a likelihood function is derived from a procedure of optimization that selects only the variables that bring a significant discrimination between signal and the considered background. The discriminating power of a given variable is computed using the selection efficiencies for both signal and background in the plane $(\varepsilon_S, \varepsilon_B)$. When the discriminating power of a variable is low, the variation of the background efficiency $\varepsilon_B$ follows that of the signal $\varepsilon_S$. On the contrary, a high discriminating variable results in a larger decrease of $\varepsilon_B$ compared to the variation seen in $\varepsilon_S$. This variation, computed for each $\varepsilon_S$ and integrated over the full range of $\varepsilon_S$ can thus be seen as an estimator of the discriminating power of the variable. In this analysis, the variable is selected if the discriminating power is about a few percent. The use of higher thresholds results in a degradation of the performance due to the loss of discriminating power of the formed likelihoods. The set of discriminant variables is built from 16 relevant kinematical variables:

- the opening angles between the lepton and the jets $\Delta R(l,b_1)$, $\Delta R(l,b_2)$
- the angles $\cos\Delta\Phi(l,b_1)$ and $\cos\Delta\Phi(l,b_2)$;
- the opening angle between the two b tagged jets $\Delta R(b_1,b_2)$;
- the pseudo-rapidity of the b tagged jets $\eta_{b1}$ and $\eta_{b2}$;
- the invariant mass formed by the systems of the two b tagged jets $M_{\mathrm{inv}}(b_1,b_2)$
- the invariant mass formed by the reconstructed $W$ leptonic boson and the b tagged jets $M(W_{\mathrm{lep}},b_1)$ and $M(W_{\mathrm{lep}},b_2)$;
- the transverse momentum of the reconstructed top quark candidates $p_{\mathrm{T}}\ (top_1)$ and $p_{\mathrm{T}}\ (top_2)$
- the sum of the missing transverse energy and lepton transverse momentum $\not{E}_{\mathrm{T}}+\ p_{\mathrm{T}}\ (l)$;
- the transverse mass of the leptonic W-boson candidate $M_{\mathrm{T}}(W_{\mathrm{lep}})$;
- the scalar sum of the jet transverse momenta $H_{\mathrm{T}}(jets)$;
- global event shape variables: sphericity, aplanarity and centrality.

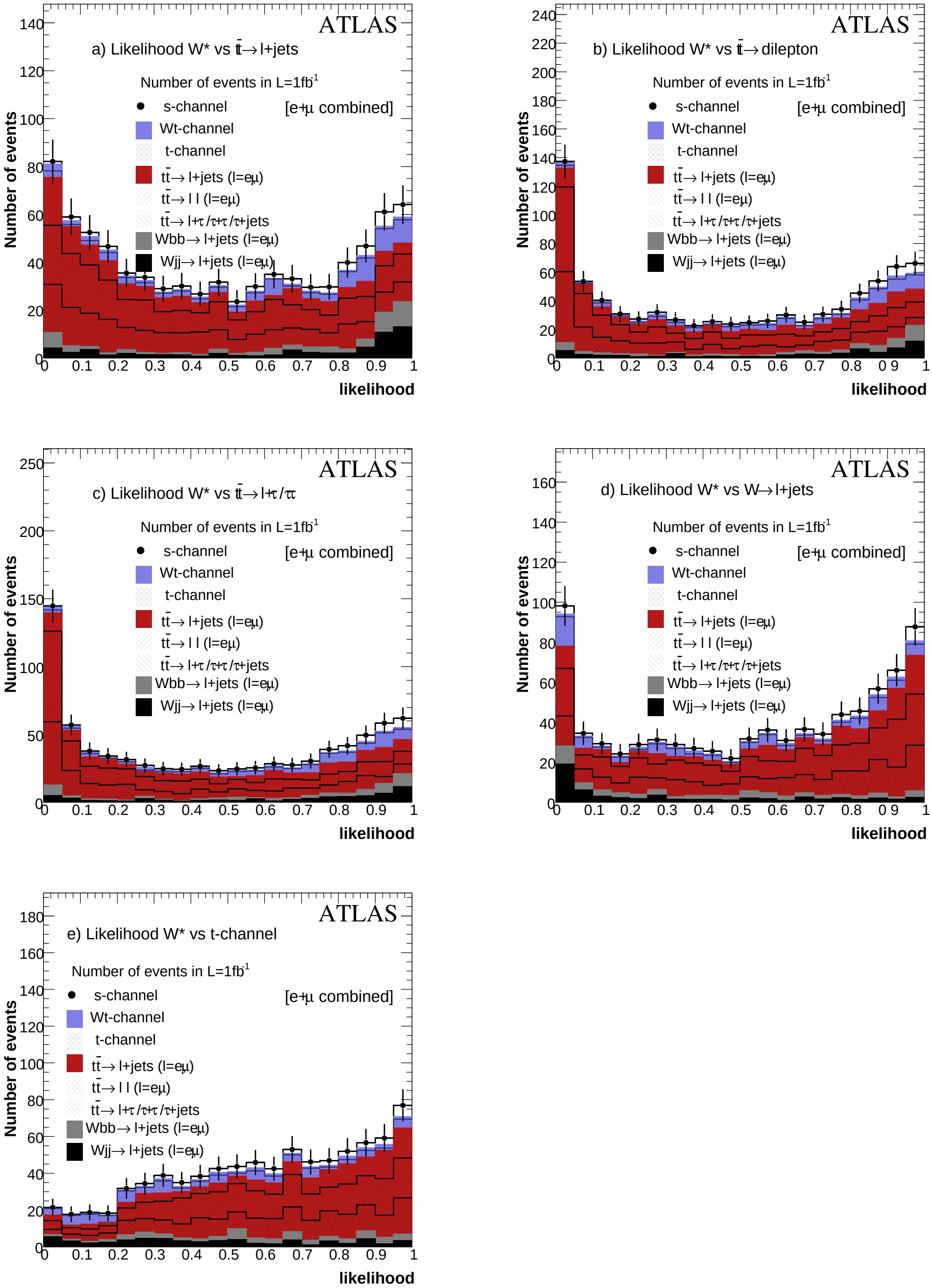


Figure 4: Distributions of the five likelihood functions for an integrated luminosity of $1\text{fb}^{-1}$. a) Likelihood against $t\bar{t}$ in the l+jets channel; b) Likelihood against $t\bar{t}$ in the dilepton channel; c) Likelihood against $t\bar{t}$ in the $1+\tau$ channel; d) Likelihood against W+jets events; e) Likelihood against t-channel

#### 5.2.2 Optimization of the likelihood selection

The likelihood function distributions corresponding to an integrated luminosity of 1 $\text{fb}^{-1}$ are shown in Figure 4. Thresholds have been applied on each of the 5 likelihood values. In the present analysis, the thresholds on the likelihood values have been set so that the total uncertainty affecting the cross section measurement is minimized. The total uncertainty was calculated as described in Section 4.2.3. The main sources of systematic errors taken into account are the b-tagging efficiency, jet energy scale, luminosity and uncertainties on the background level, for which central values are provided in Section 5.3. Note that we consider the errors as fully correlated between signal and background for jet energy scale, b-tagging, and luminosity.

The thresholds set on the five likelihood outputs resulting from the minimization of the total uncertainty are listed below:

$$\mathscr{L}_{\mathrm{t\bar{t}/lepton+jets}} > 0.34,$$
$$\mathscr{L}_{\mathrm{t\bar{t}/dilepton}} > 0.56,$$
$$\mathscr{L}_{\mathrm{t\bar{t}}/\tau+\mathrm{lepton}} > 0.80,$$
$$\mathscr{L}_{\mathrm{W+jets}} > 0.32,$$
$$\mathscr{L}_{\mathrm{t-channel}} > 0.46$$

Table 7 reports the number of events expected for all signal and backgrounds for an integrated luminosity of 1 $\text{fb}^{-1}$. The overall signal to background ratio is improved significantly compared to the sequential cuts analysis with a purity increased from 9.8% to 18.7%. This is an expected outcome of the optimization procedure since the main source of errors comes from the uncertainties in the background.

Table 7: Numbers of expected events for an integrated luminosity of 1 $\text{fb}^{-1}$ expected from the likelihood analysis in the two jet final state events. The results are shown separately for the electron and muon channels. Statistical uncertainties correspond to the Monte Carlo statistics only. The convention $\mathrm{l} = \mathrm{e}, \mu$ is used.

| Events in $1\text{fb}^{-1}$ | e channel | $\mu$ channel | $\mathrm{e}+\mu$ combined |
|---|---|---|---|
| s-channel | $6.3 \pm 0.7$ | $9.1 \pm 0.8$ | $15.4 \pm 1.0$ |
| t-channel | negl. | $1.7 \pm 1.7$ | $1.7 \pm 1.7$ |
| Wt-channel | $1.8 \pm 1.0$ | negl. | $1.8 \pm 1.0$ |
| $\mathrm{t\bar{t}} \to \mathrm{l+jets}$ | $7.6 \pm 2.6$ | $7.7 \pm 3.5$ | $15.3 \pm 4.4$ |
| $\mathrm{t\bar{t}} \to \mathrm{ll}$ | $6.0 \pm 2.6$ | $6.0 \pm 2.6$ | $12.0 \pm 3.8$ |
| $\mathrm{t\bar{t}} \to \mathrm{l}+\tau$ | $6.8 \pm 3.4$ | $14.5 \pm 4.1$ | $21.3 \pm 5.3$ |
| $\mathrm{Wb\bar{b}}$+jets | $10.0 \pm 3.2$ | $7.0 \pm 2.6$ | $17.0 \pm 4.1$ |
| W+jets | $6.2 \pm 1.2$ | $7.3 \pm 2.1$ | $13.5 \pm 2.4$ |
| $\mathrm{WZ+WW}$ | negl. | negl. | negl. |
| Total Bkg | $36.8 \pm 5.8$ | $45.9 \pm 6.6$ | $82.7 \pm 8.6$ |
| $S/B$ | 17.3% | 19.8% | 18.7% |
| $S/\sqrt{B}$ | 1.0 | 1.3 | 1.7 |
| $\sqrt{S+B}/S$ | 1.0 | 0.8 | 0.6 |

The dominant background comes from $t\bar{t}$ production, which contributes about 60% of the total event yield. Among those events, the main contribution originates from the $l+\tau$ decay modes (l=$e,\mu$), which corresponds to 45% of the $t\bar{t}$ yield, followed by the dilepton and the l+jets channels. The production of W$b\bar{b}$+jets represents about 20%, while the W+jets events still contribute 16% of the total background yield. Finally, the contamination originating from the other single top quark channels is smaller than the signal expected event yield with a contribution of 4% from the sum of t- and Wt- channels. The results show a similar statistical sensitivity compared to the standard cut-based analysis, with an improved signal to background ratio.

As described in Section 4.1 the presence of pile-up affects the reconstruction of the objects selected in the event. The use of events with pile-up results in a decrease of about 9% for signal events, and 15% for $t\bar{t}$ events. No systematic uncertainty is associated to this difference, as $t\bar{t}$ events completely dominate the selected sample. Dedicated studies with data will be used to tune all Monte Carlo generators.

## 5.3 Systematic uncertainties

The systematic uncertainties have been evaluated on the likelihood analysis. We follow the procedures defined in Section 4.2.

### 5.3.1 Experimental systematic uncertainties

The uncertainties on the b-tagging efficiency and the mistag rates have been estimated by varying the b weight cut value corresponding to a change of ±5% in the b-tag efficiency with respect to the reference value of 60%. Table 8 shows the impact of such effects on signal and background events. As the b-tag efficiency is varied by +5% and -5%, the signal selection efficiency is shifted by 7.1% and -7.7% respectively. For background events, the impact of such variations result in both cases in an increase of the background level which reaches +15% and +5% respectively because of the change affecting the mistag rates. Table 9 reports the impact on the total cross-section determination. The understanding of the b-tagging performance completely drives the analysis strategy.

Table 8: Effect of the main systematic effects, b-tag efficiency and mistag rate variation and of the jet energy scale variation on the number of expected events for an integrated luminosity of 1 $fb^{-1}$ expected from the likelihood analysis in the two jet final state events. Numbers in parentheses are the relative variations.

| Events in $1fb^{-1}$ | b-tag -5% | b-tag +5% | JES -5% | JES +5% |
|---|---|---|---|---|
| s-channel | 14.2 (-7.7%) | 16.5( +7.1%) | 16.9 (+9.7%) | 15.4 (negl.) |
| t and Wt-channel | 6.9 (+97%) | 3.5 (negl.) | 5.1 (+45%) | 5.1 (+45%) |
| $t\bar{t}$ combined | 47.8 (-1.6%) | 58.1 (+19.5%) | 49.5 (+1.8%) | 46.9 (-3.5%) |
| W$b\bar{b}$+jets | 15.5 (-8.8%) | 16.4 (-3.5%) | 17.0 (negl.) | 17.0 (negl.) |
| W+jets | 17.0 (+26%) | 17.0 (+26%) | 15.4 (+14%) | 15.1 (+11.8%) |
| Total bkg | 87.2 (+5.4%) | 94.9 (+14.8%) | 87.7 (+6%) | 84.9 (+2.6%) |

A variation of -5% and +5% of the jet energy scale has been propagated to the jet reconstruction and the selection efficiencies were re-assessed. Backgrounds change by about 6% and 3% respectively, while the signal acceptance is found to vary by about ±10%. Again, in both cases the background is

increased with any change of the jet energy scale. Indeed, as the jet scale factor is decreased, top quark pair events in the dilepton and l+jets modes tend to have a lower multiplicity, resulting in an increased contamination. As the scale factor goes up, the low multiplicity events are favored while top quark pair production is almost not affected. Table 8 shows the effects of the jet energy scale variation on the signal and background event yield.

The impact of the trigger efficiency and lepton identification uncertainties has been estimated as explained in the t-channel Section. This reflects as an uncertainty of 6% on the total cross-section measurement. Table 9 reports the impact on the total cross-section determination.

Table 9: Summary of all uncertainties that affect the measured cross section. Data statistics is the Poisson error one would expect from real data while MC Statistics is the uncertainty on the estimated quantities due to MC statistics.

| Source of | Analysis for 1 $fb^{-1}$ | | Analysis for 10 $fb^{-1}$ | |
|---|---|---|---|---|
| uncertainty | Variation | $\Delta\sigma/\sigma$ | Variation | $\Delta\sigma/\sigma$ |
| Data Statistics | | 64% | | 20% |
| MC Statistics | | 29% | | |
| Luminosity | 5% | 31% | 3% | 18% |
| b-tagging | 5% | 44% | 3% | 25% |
| JES | 5% | 25% | 1% | 5% |
| Lepton ID | 1% | 6% | 1% | 6% |
| Bkg x-section | 10.3% | 47% | 3% | 16% |
| ISR/FSR | 9% | 52% | 3% | 17% |
| PDF | 2% | 16% | 2% | 16% |
| b-fragmentation | 3.6% | 19% | 3.6% | 19% |
| Total Systematics | | 95% | | 48% |

### 5.3.2 Theoretical and Monte Carlo systematic uncertainties

In the present analysis, uncertainties on the background estimates come from the theoretical uncertainties associated with their cross section. The main background contributions are $t\bar{t}$ (60%), $Wb\bar{b}$+jets (20%) and W+jets (16%), so 10% uncertainties on the $t\bar{t}$ cross-section prediction and 20% on the W+jets and $Wb\bar{b}$+jets events lead to a total of 10.3% uncertainty on the total background. The understanding of the background level is thus a crucial point for a precise cross-section determination. Note that despite very distinct topologies, the selections of the s-, t- and Wt- channels analyses are not orthogonal. However it is believed that correlations can be properly addressed with dedicated study and enough statistics.

The selection of a 2-jet final state is very sensitive to the presence of extra jets originating from gluon radiation. Any uncertainty in the ISR/FSR modeling is thus expected to have a significant impact on the selection efficiencies, in particular for the s-channel and the $t\bar{t}$ events. Specific Monte Carlo samples have been generated with ISR/FSR settings leading to the largest cross section variations. Variations of 5% of the signal selection efficiencies are expected, and are due to the change in the jet multiplicity. The uncertainties however reach large values for the $t\bar{t}$ production, with variations of 17.8% between the two extreme cases. Note that these variations have been assessed with non-calibrated jets, although there is a high expected correlation between jet energy scale and ISR/FSR gluon modeling. An overall

9% uncertainty corresponding to half the variation is quoted. Note that the s-channel being produced via quark-antiquark annihilation, some constraints coming from the use of W-boson events should allow the tuning of the showering interfaces.

Another uncertainty is related to the choice of the PDFs. The procedures used to assess the impact of such choice is presented in Section 4.2.2. The expected bias on the signal is below 3% while this number is below 2% for $t\bar{t}$ events.

Finally the effect of the b-fragmentation parametrization has also been investigated using fast simulation. The b quark fragmentation is performed according to the Peterson parametrization, with one free parameter $\varepsilon_b$. Varying the default value from $\varepsilon_b = -0.006$ by $\pm 0.0025$ [29] and taking the difference as a systematic uncertainty leads to a change of 3.6% in the $t\bar{t}$ and signal selection efficiencies.

Table 9 lists all sources of uncertainties and reports their impact on the cross-section determination. Two cases are considered: one defined by the level of uncertainty in the b-tagging, JES, luminosity that will presumably characterize the early data taking period and a second one assuming a reasonable improvement on those effects with an integrated luminosity of 10 fb$^{-1}$. The assumptions made in the latter case are the same as the ones listed in Section 4.2.3.

### 5.4 Summary

The determination of the s-channel cross section constitutes a challenging measurement, due to the presence of large backgrounds from the $t\bar{t}$ production and from the W+jets channels. With an expected signal to background ratio of 18% the measurement will be hampered not only by a significant statistical uncertainty but also by the systematic effects affecting both signal and background. The measurement with an integrated luminosity of 1 fb$^{-1}$ is thus both statistically and systematically limited, with about 60% of statistical and 90% of systematic uncertainties.

Given the present limitation on the background knowledge, early measurements will have to be devoted to the understanding of the background, in terms of shape and absolute normalization. For this purpose, specific studies performed on enriched background samples can be used. In this area, the knowledge of the effects of the ISR/FSR gluon radiation will need dedicated studies, in particular in $t\bar{t}$ events. The constraint from data itself will thus be very important, and a tuning *à la* CDF [30] will be crucial for the understanding of these radiations. From the detector side, a reliable cross-section measurement requires a good knowledge of the b-tagging tools performance, since double-tag events are considered. A good determination of the jet energy scale is also mandatory for the selection, at the level of better than 5%. PDF and b-fragmentation effects are expected to have a significant impact only at higher luminosity.

In this context, the use of sophisticated statistical methods appears mandatory to discriminate the signal from the background and to establish convincing evidence for a signal. Their use however requires an a priori good understanding of the background normalization and shapes. With an improved situation for the b-tagging and jet energy scale, with a background normalization determined from the data and a better ISR/FSR knowledge, evidence at 3 $\sigma$ should be achievable with 30 fb$^{-1}$.

## 6 Measurement of the Wt-channel cross section

The Wt-channel is characterized by the associated production of a top quark and a W-boson. At the LHC, this single top quark process is the second highest cross-section after t-channel production, with an expected cross section of 66 pb. The final state features two W-bosons and a b jet, making this channel experimentally very close to top quark pair production, from which it differs by the absence of a second b jet in the final state. With a cross section 15 times larger than the Wt-channel processes, top quark pair events will thus constitute the dominant source of backgrounds and drive the definition of the selection

criteria in higher jet multiplicity events. On the other hand, low jet multiplicity events will suffer from W+jets contamination. In this report, we consider only the case where one of the two bosons decays into leptons while the other decays into jets. Two approaches have been used to estimate the sensitivity to the Wt-channel cross section measurement: a sequential cut-based analysis, which will provide reference numbers, and an analysis based on the use of Boosted Decision Trees.

### 6.1 Sequential cuts analysis

The common preselection defined for all three single top quark processes has been extended further to account for the specific topology of Wt-channel events. Selected events must have exactly one high $p_T$ b-tagged jet above 50 GeV. A veto on any other b tagged jet above 35 GeV is applied in order to reject $t\bar{t}$ events. This b-tag veto utilizes a looser b-tag weight cut which has been optimized according to the signal over $t\bar{t}$ background ratio: a ratio of 18% is reached in 2- and 3- jet final states events while 14% is found in 4-jet events. The corresponding efficiency is about 30% on signal events and 10% on $t\bar{t}$ events. For events containing more than three high $p_T$ jets, the selection requires that the invariant mass of the two highest $p_T$ non b-tagged jets to be between 50 and 125 GeV. The number of selected events is reported in Table 10 for signal and backgrounds.

Table 10: Number of expected events for an integrated luminosity of 1 $fb^{-1}$ as function of the jet multiplicity and for the electron and muon channel combined in the sequential cut-based analysis. Errors shown are statistical only. The convention $l = e, \mu$ is used.

| Events in $1fb^{-1}$ | 2 jets (1b1j) | 3 jets (1b2j) | 4 jets (1b3j) |
|---|---|---|---|
| Wt-channel | $435 \pm 16$ | $164 \pm 10$ | $40 \pm 5$ |
| t-channel | $1218 \pm 47$ | $94 \pm 13$ | $58 \pm 11$ |
| s-channel | $42 \pm 2$ | $5 \pm 0.6$ | $0.6 \pm 0.2$ |
| $t\bar{t} \to l+$jets | $1260 \pm 38$ | $664 \pm 27$ | $240 \pm 16$ |
| $t\bar{t} \to$ dilepton | $291 \pm 18$ | $50 \pm 7$ | $17 \pm 4$ |
| $t\bar{t} \to l+\tau$ | $428 \pm 22$ | $55 \pm 8$ | $17 \pm 5$ |
| W+jets | $2983 \pm 71$ | $207 \pm 19$ | $38 \pm 6$ |
| $Wb\bar{b}$+jets | $137 \pm 33$ | $13 \pm 3$ | $6 \pm 2$ |
| TOTAL bkg | $6359 \pm 232$ | $1088 \pm 74$ | $377 \pm 42$ |
| $S/B$ | 6.8% | 15.0% | 10.6% |
| $S/\sqrt{B}$ | 5.4 | 5.0 | 2.1 |
| $\sqrt{S+B}/S$ | 0.19 | 0.21 | 0.51 |

In two jet events (labeled as '1b1j'), the signal yield is about 430 signal events for an integrated luminosity of 1 $fb^{-1}$, with a signal to background ratio of 6.8%, combining both electron and muon channels. The W+jets production is the dominant background with more than 45% of the total background yield. The $t\bar{t}$ contamination in the l+jets channel constitutes about 20% of the background, while the other $t\bar{t}$ modes combined contribute 11%. In this low jet multiplicity bin, the single top quark t-channel contamination is significant and represents 19% of the total.

In the three jet final state events (labeled as '1b2j'), 160 signal events are expected in 1 $fb^{-1}$, with a signal to background ratio close to 15%. In this final state, the use of the hadronic W boson mass

constraint helps improve the rejection of W+jets and the other single top quark events. The background is made up of $t\bar{t}$ events which represent 70% of the total background yield, and of about 20% of W+jets events. Among the $t\bar{t}$ events, the l+jets mode is dominant and constitutes about 85% of the total.

The four jet final state events (labeled as '1b3j') only bring a marginal improvement to the analysis. The number of expected Wt-channel single top quark events is small and expected to be 40 for 1 $fb^{-1}$ with a signal to background ratio of 10.6%. In this high jet multiplicity bin, the main background comes from the $t\bar{t}$ events in the l+jets mode, which constitute 64% of the background. The rest originate mostly from the single top quark t-channel and from W+jets events, each being of the same order as the signal.

## 6.2 Boosted decision tree analysis

As in the s-channel analysis, several multivariate discriminators have been defined to optimize the discrimination against backgrounds. Two Boosted Decision Tree (BDT) functions are defined to separate signal and $t\bar{t}$ events, one devoted to the discrimination from the dominant l+jets channel $\mathscr{D}_{t\bar{t}/l+jets}$, and the other against the dilepton channels $\mathscr{D}_{t\bar{t}/dilepton}$ including $\tau$'s. A function $\mathscr{D}_{W+jets}$ has been formed to separate signal from the W+jets sample, defined as the total contribution from light and heavy flavor jets. Another BDT discriminator is devoted to the separation of the signal and the single top quark t-channel events $\mathscr{D}_{t-channel}$. The analysis being based upon events with jet multiplicity between two and four, specific BDTs have been defined in each jet multiplicity bin for each background. The electron and muon channels are being treated in a combined way, thus leading to the definition of 3 (jet multiplicity) x 4 (background) ie: 12 BDT discriminators in total.

### 6.2.1 Definition of the discriminant variables

The set of discriminant variables is derived from the same procedure of optimization as the one explained for the s-channel analysis. The final set of discriminant variables is built from 25 relevant kinematical variables:

- the opening angles between the lepton and the jets $\Delta R(l,b)$, $\Delta R(l,j_1)$, $\Delta R(l,j_2)$ where $j_1$, $j_2$ and $j_3$ are the non b-tagged $p_T$ ordered jets
- the opening angle between the jets $\Delta R(b,j_1)$, $\Delta R(b,j_2)$, $\Delta R(j_1,j_2)$;
- the angles $\cos\Delta\Phi(j_1,j_2)$ and the pseudo-rapidity of the non b-tagged jets $\eta_{j2}$ and $\eta_{j3}$;
- the invariant mass formed by the sum of all jets $M_{inv}(jets)$ and by the reconstructed W-boson and top quark candidates $M(W+t)$;
- the mass of the hadronic W-boson candidate $M(W_{had})$;
- the sum of the missing transverse energy and lepton transverse momentum $\not{E}_T$+ $p_T(l)$;
- the transverse mass of the leptonic W-boson candidate $M_T(W_{lep})$
- the transverse mass of the systems formed by the b jet and the reconstructed W-bosons $M(b,W_{had})$ and $M(b,W_{lep})$;
- the scalar sum of the jet transverse momenta $H_T(jets)$ and of all objects in the events $H_T(tot)$;
- the transverse jet momenta $p_T$ $(b)$, $p_T(j_1)$, $p_T(j_2)$;
- the longitudinal momentum of the neutrino solution computed with the top quark mass;

- the global event shape variables, sphericity, aplanarity and centrality.

The BDT output distributions associated to each of the four backgrounds are represented in Figure 5 for the 3 jet final state analysis for the electron+muon channels for 1 $fb^{-1}$.

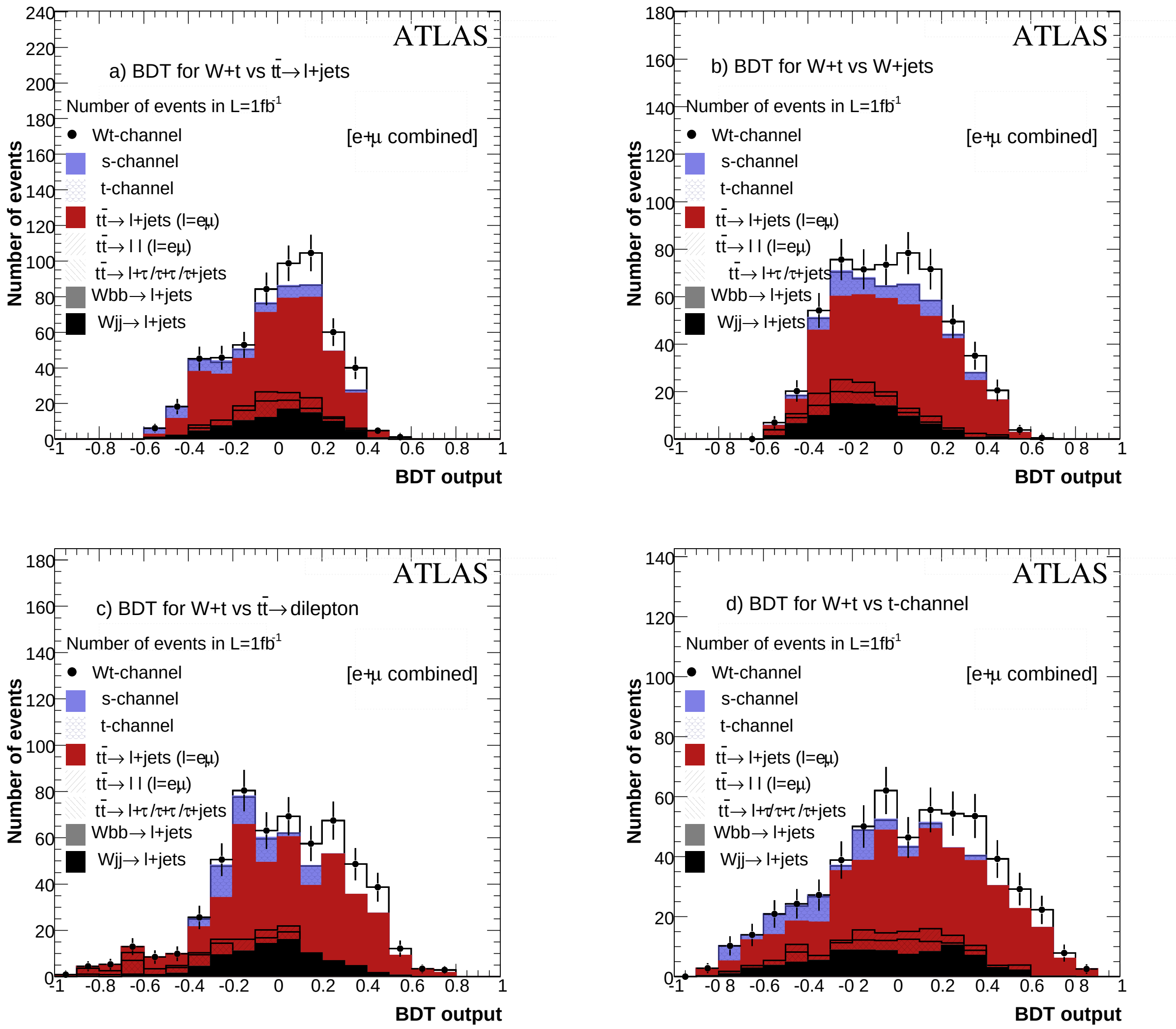


Figure 5: Distributions for the four BDT defined in the 3 jet ('1b2j') final state analysis. a) BDT against $t\bar{t}$ in the 'l+jets' channel; b) BDT against W+jets events; c) BDT against $t\bar{t}$ in the dilepton lepton (+tau) channel; d) BDT against single top quark t-channel events.

#### 6.2.2 Results with the Boosted Decision Trees

Several selections can be designed, allowing for different levels of signal purity. In this analysis as in the s-channel, the cuts on the discriminant BDTs have been set so that the total uncertainty affecting the cross section measurement is minimized.

The thresholds set on the three BDT outputs are:

$$\mathscr{D}_{\mathrm{ttlj}} > 0.6$$
$$\mathscr{D}_{\mathrm{ttl}\tau} > -0.36$$
$$\mathscr{D}_{\mathrm{Wjet}} > 0.30$$
$$\mathscr{D}_{\mathrm{t-chan}} > 0.18$$

The number of events for signal and the different backgrounds is reported in Table 11 for an integrated luminosity of 1 $\mathrm{fb}^{-1}$. In two jet events ('1b1j'), the signal yield is about 60 with a signal to background ratio of 35%, combining electron and muon analyses. This result represents an improvement by a factor 6 compared to the cut-based analysis. This feature is important since one of the main sources of systematic uncertainty originate from the imperfect knowledge of the background levels. Regarding the composition of the background, the W+jets production constitutes the dominant background and contributes 58% of the background yield. $\mathrm{t\bar{t}}$ production in the l+jets mode represents about 40% of the total. The only remaining single top quark events originate from the t-channel which accounts for 6% of the total background yield. The statistical significance for this final state analysis alone is 4.5$\sigma$.

Table 11: Number of expected events for an integrated luminosity of 1 $\mathrm{fb}^{-1}$ as function of the jet multiplicity after the BDT analysis. The convention $\mathrm{l} = \mathrm{e}, \mu$ is used.

| Events in 1$\mathrm{fb}^{-1}$ | 2 jet (1b1j) | 3 jet (1b2j) | 4 jet (1b3j) |
|---|---|---|---|
| Wt-channel | $58.0 \pm 5.8$ | $20.9 \pm 3.5$ | $6.6 \pm 2.0$ |
| t-channel | $10.2 \pm 4.2$ | negl. | $1.7 \pm 1.7$ |
| s-channel | $1.4 \pm 0.3$ | negl. | negl. |
| $t\bar{t} \rightarrow all\,jet$ | negl. | negl. | negl. |
| $t\bar{t} \rightarrow l + jet$ | $56.3 \pm 8.2$ | $41.8 \pm 6.3$ | $13.7 \pm 3.4$ |
| $t\bar{t} \rightarrow dilepton$ | $1.7 \pm 1.2$ | negl. | negl. |
| $t\bar{t} \rightarrow l + \tau$ | negl. | negl. | negl. |
| W+jets | $92.1 \pm 8$ | $3.2 \pm 1.4$ | $0.2 \pm 0.1$ |
| $\mathrm{Wb\bar{b}}$+jets | $3.9 \pm 3.9$ | negl. | negl. |
| Total bkg | $165.6 \pm 9.2$ | $45.1 \pm 6.3$ | $15.6 \pm 3.4$ |
| $S/B$ | 35.0% | 46% | 36.2% |
| $S/\sqrt{B}$ | 4.5 | 3.1 | 1.7 |
| $\sqrt{S+B}/S$ | 0.25 | 0.39 | 0.71 |

In three jet final state events ('1b2j'), about 20 signal events are expected with 1 $\mathrm{fb}^{-1}$, with a signal to background ratio of 46%. Again, one notices a gain of more than a factor 3 compared to the cut-based analysis. The background events are made up almost exclusively (95%) of $\mathrm{t\bar{t}}$ events in the lepton+jets events. The use of the jet-jet invariant mass reduces the W+jets background to a few percent. The selection in this final state alone provides a signal statistical significance of 3.1 $\sigma$.

In the four jet final state channel ('1b3j'), the number of expected Wt-channel single top quark events is 7 with 1 $\mathrm{fb}^{-1}$. In this bin, the signal to background ratio is about 36%, which corresponds to a gain of a

factor 4 compared to the sequential cut analysis. In this jet multiplicity bin, the only competing processes is the $t\bar{t}$ events in the l+jets. The signal significance remains small, around $1.7\sigma$ with a corresponding statistical precision of 72%.

The effect of pile-up has been investigated with signal and $t\bar{t}$ samples produced with a pile-up corresponding to a $10^{32}$ cm$^2$s$^{-1}$ luminosity run. The results show decreases of 18%, 47% and 26% of the signal event yield in the 2, 3 and 4 jet final states. For backgrounds, similar variations of 16%, 39% and 20% in the 2, 3 and 4 jet final states are seen. As expected the increase of the number of light jets seen in the events directly impacts the tight selection of 2- and 3- jet final states. No systematic uncertainty is associated to this effect, as for the two previous analyses, as only dedicated studies using data itself will be used to tune all Monte Carlo signal and background generators.

## 6.3 Systematic uncertainties

The systematic uncertainties have been evaluated on the BDT analysis. We follow the procedures defined in Section 4.2.

### 6.3.1 Experimental systematic uncertainties

The b-tagging uncertainties affect the Wt-channel selection because of the requirement of one b-tagged jet on one side, and the use of a veto for any second b tagged jet on the other side. The variation by 5% of the b tagging efficiency and the corresponding mistag rate results in a 7% change of the signal selection efficiency in the 2 jet final state. The sensitivity is higher in the higher multiplicity bins, with effects of 10% seen in the 3 jet bin. Regarding backgrounds, the impact of this uncertainty increases with the jet multiplicity, with $t\bar{t}$ events being the dominant background. Variations of 3% and 5% are seen respectively in 3 jet and 4 jet final states. Table 12 reports the relative changes in signal and background events. Table 13 reports the impact on the total cross section determination, all final states combined.

Table 12: Effect of the b-tag efficiency and mistag rate variation and of the jet energy scale variation on the number of expected events for an integrated luminosity of 1 fb$^{-1}$ expected from the BDT analysis.

| Process | Events in 1fb$^{-1}$ | b-tag $\pm$ 5% | JES $\pm$ 5% |
|---|---|---|---|
| 2-jet events | | | |
| - signal | 58.0 | $\pm$ 7.0% | $\pm$0.5% |
| - total bkg | 165.6 | $\pm$ 3.0% | $\pm$3.1% |
| 3 jet events | | | |
| - signal | 20.9 | $\pm$10.1% | $\pm$7.0% |
| - total bkg | 45.1 | $\pm$3.2% | $\pm$3.0% |
| 4 jet events | | | |
| - signal | 6.6 | $\pm$3.1% | $\pm$7.9% |
| - total bkg | 15.6 | $\pm$5.1% | $\pm$4.0% |

The precise knowledge of the jet energy scale is important for the Wt-channel analysis because of the requirements made on the mass reconstruction and $p_{\mathrm{T}}$ thresholds used to select jets. A 5% variation

of the jet energy scale has been propagated to the jet reconstruction and the selection efficiencies are re-assessed. Table 12 shows the impact of the jet energy scale variation on signal and background event yields. In two jet events, the effects of the scale variation is about 1% on the signal events and 3% on the sum of all backgrounds. In three jet events a variation of 7% of the signal event yield is observed, with a change of 3% in the backgrounds. In four jet events, variations of 8% and 4% are seen in signal and background respectively. Table 13 reports the impact on the total cross section determination.

An uncertainty of 1% in the lepton identification or in the trigger efficiency would impact the number of selected events. Such an uncertainty would reflect in an uncertainty of 2.6% on the total cross section measurement, which is negligible with respect to the others.

### 6.3.2 Theoretical and Monte Carlo uncertainties

Uncertainties on the background estimates come from the theoretical uncertainties associated to the cross sections. An uncertainty of 10% is quoted for the $t\bar{t}$ events while 20% is associated to the W+jets and $Wb\bar{b}$+jets events. This translates into a total of 12.5% in the 2 jet final state and 10% in the higher jet multiplicity bins where $t\bar{t}$ events completely dominate the background. Note that despite very distinct topologies, the selections of the Wt- and the t- channels analyses are not orthogonal. However it is believed that correlations can be properly addressed.

Table 13: Summary of all uncertainties that affect the measured cross section. Data statistics is the Poisson error one would expect from real data while MC Statistics is the uncertainty on the estimated quantities due to MC statistics. (*) background to 2j (12.5%) and 3j and 4j final states (10%)

| Source of uncertainty | Analysis for 1 $fb^{-1}$ | | Analysis for 10 $fb^{-1}$ | |
|---|---|---|---|---|
| | Variation | $\Delta\sigma/\sigma$ | Variation | $\Delta\sigma/\sigma$ |
| Data Statistics | | 20.6% | | 6.6% |
| MC Statistics | | 15.6% | | |
| Luminosity | 5% | 20% | 3% | 7.9% |
| b-tagging | 5% | 16% | 3% | 6.6% |
| JES | 5% | 11% | 1% | 1.5% |
| Lepton ID | 1% | 2.6% | 1% | 2.6% |
| Bkg x-section | 12.5/10%(*) | 23.4% | 3% | 9.6% |
| ISR/FSR | 9% | 24.0% | 3% | 7.8% |
| PDF | 2% | 5.2% | 2% | 5.2% |
| b-fragmentation | 3.6% | 9.4% | 3.6% | 9.4% |
| Total Systematics | | 48% | | 19.4% |

The selection of a low jet multiplicity final state is very sensitive to the presence of extra jets originating from gluon radiations. Any uncertainty in the ISR/FSR modelling is thus expected to have a significant impact on the selection efficiencies, in particular for the Wt-channel and top quark pair events. We quote an overall 9% uncertainty due to the modelling of gluon radiation in the $t\bar{t}$ events.

The uncertainties in the PDF may affect the topologies as well as the momentum distributions of the final state objects, hence impacting the determination of the selection efficiency for both signal and backgrounds. The procedure to estimate the impact of the choice of the PDF to the selection efficiency is

detailed in Section 4.2.2. Complete studies have been performed in each final state and a 2% uncertainty is quoted for signal and $t\bar{t}$ events.

Finally the effect of the b-fragmentation parametrization has also been investigated following a similar procedure of that defined in Section 5.3.2 from fast simulation. The relative uncertainties of the different channel and final state selection efficiencies due to the b-fragmentation uncertainty is found to be 3.6%.

Table 13 lists all sources of uncertainties and reports their impact on the cross section determination. Two cases are considered: one defined by the level of uncertainty in the b-tagging, jet energy scale and luminosity that will presumably characterize the early data taking period; another assuming reasonable improvements on those effects with an integrated luminosity of 10 $fb^{-1}$. The assumptions made in the latter case are listed in Section 4.2.3.

### 6.4 Summary

The determination of the Wt-channel cross section constitutes a challenging measurement with the early data, due to the presence of important $t\bar{t}$ and W+jets backgrounds. This measurement makes use of the events with a jet multiplicity between two and four jets. With a signal to background ratio of about 30-40%, the analysis requires a good knowledge of the W+jets production in the lower multiplicity bins, and of the $t\bar{t}$ process in higher bins. The estimates of the shapes and normalization of those processes will have to rely upon the use of data. Strategies exist for QCD and W+jets events, but the discrimination against $t\bar{t}$ events remains a challenge.

As for the t-channel single top quark analysis, the cross section determination will very early be dominated by systematic uncertainties. The dominant effect is constituted by the background uncertainties, followed by the modeling of the gluon radiation. From the detector side, the dominant source of uncertainty is the b-tagging because of the imperfect knowledge of the b-tag and b-tag veto efficiencies as well as of the mistag rates. Another source is the determination of the jet energy scale, which affects the reconstruction of the W-boson mass and all the jet energies used in the analysis. Note that the determination of the luminosity to better than 5% is required in order to ensure a good measurement, or the use of ratio of different Wt-channel final states [31] can be used as well with higher luminosity. A 3 $\sigma$ evidence can be reached with a few $fb^{-1}$ of data taking and a precision of 20% on the cross section measurement is achievable with about 10 $fb^{-1}$ provided that improvements are made in both the experimental aspects of the detection (backgrounds from data, b-tagging, jet energy scale and luminosity) and from the theoretical side.

## 7 Conclusion

At the LHC the production of single top quark events accounts for about a third of the $t\bar{t}$ production, which leads to about 2.5 million events per year during a run at $10^{32}cm^2s^{-1}$. Similarly to the situation at the Tevatron, the selection of single top quark events will suffer from the presence of both W+jets and $t\bar{t}$ backgrounds, which are produced at much higher rates. Thus, careful approaches devoted to the understanding of these backgrounds in terms of shape and normalization performed directly from data will have to be defined. Besides, except for the s-channel, single top quark analyses will be very early dominated by the systematic uncertainties, and will require a good control of b-tagging tools and a reliable determination of the jet energy scale.

In a context of low signal over background ratio, the use of sophisticated tools like genetic algorithms, likelihoods and Boosted Decision Trees appears very useful if one wants to establish the signal or to determine its cross section precisely. These techniques, which are now in common use at the Tevatron, will require the use of reliable event samples for modeling signal and backgrounds, that will presumably

be produced from the data. The analyses should also be optimized with respect to the total level of systematic uncertainty, which will be the main limiting factor for 30 $fb^{-1}$ measurements.

Finally, a precise determination of single top quark cross sections can be achieved for a few $fb^{-1}$ in the t-channel and the Wt-channel , while for the s-channel, higher statistics will be required. Their possible interpretation in terms of new physics should thus come at a later stage, once the systematic effects are under control.

# Top Quark Mass Measurements

**Abstract**

This note summarizes studies performed in order to estimate the potential of ATLAS to measure the top quark mass from the first few hundred $pb^{-1}$ of data. The analyses shown here, based on fully simulated events, have been performed for the channel where the top quarks decay into one lepton plus a number of jets, using various methods to extract the top quark mass. The performance of the detector has been evaluated, including triggering, particle identification, and jet reconstruction. For each method, the expected background, statistical and systematic uncertainties, and linearity have been studied. This work shows that the precision on the top quark mass depends mainly on the jet energy scale uncertainty: a precision of the order of 1 to 3.5 GeV should be achievable with 1 $fb^{-1}$, assuming a jet energy scale uncertainty of 1 to 5%.

## 1 Introduction

At the LHC, top quarks will be produced mainly in pairs through the hard processes $gg \to t\bar{t}$ (90%) and $q\bar{q} \to t\bar{t}$ (10%). The corresponding cross section, at next-to-leading order, is 833 pb [1]; therefore, we expect roughly 800 000 $t\bar{t}$ pairs to be produced with an integrated luminosity of 1 $fb^{-1}$, corresponding to about two weeks at a luminosity equal to $10^{33}$ $cm^{-2}s^{-1}$. In contrast to earlier measurements of the top quark mass [2], which concentrated on preserving as much of the signal as possible in order to minimize the statistical uncertainty, the LHC will produce so many top quark pairs that rather stringent requirements can be imposed, to restrict the measurement to regions in which the systematic uncertainties can be well-controlled. With 1 $fb^{-1}$ of data, the measurement will already be completely dominated by systematic uncertainties of the order of 1 GeV. The dominant contribution to the systematic uncertainty is expected to be the jet energy scale. In-situ calibration methods will allow the light jet energy scale to be known to the percent level precision after 1 $fb^{-1}$ of collected data [3]. This high precision can be, at least in part, translated to the $b$-jet energy scale, although differences are expected. Detailed studies with the data will be necessary to reach the desired precision.

The top quark mass measurements described here are based on finding the peak in the invariant mass distribution of the top quark's decay products: a $W$ boson and a $b$-quark jet. This closely corresponds to the pole mass of the top quark. Because of fragmentation effects, it is believed that the top quark mass determination in a hadronic environment is inherently ambiguous by an amount proportional to $\Lambda_{QCD}$ [4]: the intrisic amibiguity is of the order of 100 MeV.

This note describes several ways to measure the top quark mass in the semi-leptonic channel, which corresponds to a $t\bar{t}$ final state where one $W$ boson decays leptonically while the other one decays into two jets. The primary results rely on a full reconstruction of the final state, with the mass estimator taken as the invariant mass of the three jets from the hadronically-decaying top quark. This has been studied for two cases. In the first case, both $b$-jets are identified via displaced vertices, and two different mass measurement methods are presented. In the second case, fewer $b$-jets are identified. This latter case is expected to be important during early running, before the detector is fully calibrated. An analysis which relies on a kinematic fit is also presented. Other $t\bar{t}$ decay channels can also be used, but have not been considered in this note.

Another possibility relies on the determination of the mean distance of travel of b-flavoured hadrons from top quark decays [5]. The top quark mass can be inferred from the averaged lifetime

in the laboratory frame of the b-flavoured hadron, since the bottom quark's boost directly impacts the lifetime of the b-flavoured hadron. Rather than measuring the lifetime, the transverse decay length of the b-flavoured hadron is used in this method. The mean of the transverse decay length distribution is determined for different assumed top quark masses, providing an estimator for the experimental data in the mean decay length as a function of the top quark mass. Since this analysis relies mainly on tracking, the systematic uncertainties are mostly uncorrelated with those of other methods; in particular, the influence of the jet energy scale as a source of uncertainty is negligible. Work is ongoing on this method, which is not shown here.

A last possibility is to estimate the top quark mass from the measured $t\bar{t}$ production cross section. The errors on these quantities are related, in the Standard Model, via $\Delta\sigma_{t\bar{t}}/\sigma_{t\bar{t}} \sim 5\Delta m_{\rm top}/m_{\rm top}$ [6]. This would allow a determination of $m_{\rm top}$ independent of the kinematic reconstruction. In addition, it would be clear that the top quark mass is the one used in the perturbative calculations (i.e. the pole mass). However, even without considering experimental uncertainties, the achievable precision would already be limited to 2 GeV due to the uncertainty of the theoretical calculations. The scale dependence, which is the dominant source of uncertainty, might be reduced to the order of 6 % by performing the computation at higher orders, including the resummation of next-to-leading logarithmic corrections (NLL). This scale dependence can also be reduced by using a different mass definition such as the running mass[1]. This method is not considered further in this note.

## 2 Motivations for a precise top quark mass measurement

Electroweak precision observables in the Standard Model and the Minimal Supersymmetric Standard Model (MSSM) depend on the value of the top quark mass. Therefore, a precise measurement of the top quark mass is important for consistency tests of the Standard Model, to constrain the Higgs boson mass within the Standard Model, and to increase the sensitivity to physics beyond the Standard Model.

The most important dependency of the electroweak observables on the top quark mass arises via the one-loop radiative correction term $\Delta r$ [9], which is related to the $W$ boson mass through the relation $m_W^2 = \frac{\pi\alpha}{\sqrt{2}G_F \sin^2_{\Theta_W}}(1+\Delta r)$ ($\Theta_W$ being the weak mixing angle, $\alpha$ the fine structure constant and $G_F$ the Fermi coupling constant). The top quark mass is present in $\Delta r$ as terms proportional to $m_{\rm top}^2/m_Z^2$, while the Higgs boson mass is present only in terms proportional to $\log(m_H/m_Z)$. Therefore, the dependence on the Higgs boson mass is much weaker than the dependence on the top quark mass.

The precision of the indirect prediction of the Higgs boson mass depends mainly on the uncertainty on the following quantities: the hadronic contribution to the electromagnetic coupling at the scale $m_Z$ $\Delta\alpha_{had}$, $\sin^2_{\Theta_W}$, $W$ boson mass and top quark mass. For the current value of the top quark mass ($m_{\rm top}$ = 172.6 $\pm$ 1.4 GeV) [10], $m_H = 87^{+36}_{-27}$ GeV, implying $m_H < 160$ GeV at 95% C.L [11]. In order to ensure a similar contribution to the indirect prediction of the Higgs mass, the precision on $m_W$ and $m_{\rm top}$ must satisfy $\Delta m_W \simeq 0.07\Delta m_{\rm top}$. At the LHC, we expect to reach an accuracy of 15 MeV on $m_W$ [12] and 1 GeV on $m_{\rm top}$. With these precision measurements, the relative precision on a Higgs boson mass of 115 GeV would be of the order of 18% [13].

[1]The dependency of the perturbative calculations convergence on the quark mass definition has been observed in other observables in which the running mass has led to better results [7, 8]. This would allow as well a determination of the top running mass.

# 3 Top quark mass measurement in the semi-leptonic channel with the standard ATLAS $b$-tagging

In this section we investigate the determination of the top quark mass using the standard ATLAS $b$-tagging.

## 3.1 Physics background

The major sources of background, listed in Table 1, are single top events (Wt and t channels) and $W$ boson production ($W$ boson with $W \to l\nu$ decay, $W$+$b\bar{b}$ and $W$+$c\bar{c}$). It has been shown that backgrounds from $Z+$jet events with $Z \to ll$, $WW$, $WZ$, and $ZZ$ gauge boson pair production have much smaller contributions [14]: therefore, their contribution has not been re-evaluated here. Backgrounds from QCD multi-jets and $b\bar{b}$ production have not been investigated with full simulation. Nevertheless, studies performed on fast simulation have shown that these backgrounds are negligible after leptonic cuts (lepton $p_T$, $\not{E}_T$) [14].

Another source of background comes from $t\bar{t}$ events themselves (fully leptonic or fully hadronic channels). Moreover, $t\bar{t}$ events in which the $W$ boson decays into $\tau\nu_\tau$ are classified in the following way: $\tau$ decaying leptonically belong to signal events, whereas $\tau$ decaying hadronically belong to background (fully hadronic $t\bar{t}$).

Before any selection requirements, the signal to background ratio is of the order of $\frac{800pb}{80mb}$, i.e. $10^{-8}$.

Table 1: Main backgrounds to the semi-leptonic ($\ell$ = e,$\mu$) signal listing the number of events in 1 fb$^{-1}$ before and after the selection cuts (samples used here have comparable weights).

| Process | Number of events | 1 isolated lepton $p_T > 20$ GeV and $\not{E}_T > 20$GeV | >= 4 jets $p_T > 40$ GeV | 2 b-jets $p_T > 40$ GeV |
|---|---|---|---|---|
| Signal | 313200 | 132380 | 43370 | 15780 |
| $W$ boson backgrounds | $9.5 \times 10^5$ | 154100 | 9450 | 200 |
| all-jets (top pairs) | 466480 | 1020 | 560 | 160 |
| di-lepton (top pairs) | 52500 | 16470 | 2050 | 720 |
| single top, t channel | 81500 | 24400 | 1230 | 330 |
| single top, $W$ t channel | 9590 | 8430 | 770 | 170 |
| single top, s channel | 720 | 640 | 11 | 5 |

## 3.2 Combinatorial background

When a $t\bar{t}$ semi-leptonic decay is reconstructed, one must choose which jets in the event to associate with the hadronic $W$ boson decay and also which jets correspond to each of the two $b$-jets. When at least one of these choices is wrong, the event is classified as combinatorial background.

### 3.3 Jets

All particles (electrons, muons, jets) forming the $t\bar{t}$ final state are required to lie within $|\eta| < 2.5$, since $b$-jets cannot be tagged and muons are not reconstructed for higher values of $|\eta|$. See Ref. [15] for a description of the features of jet selection that are common to all top quark analyses; only the points specific to the analyses in this note are discussed below.

#### 3.3.1 Jet calibration

This analysis uses jets defined as in [15]. The jet calibration effects are removed by performing a jet calibration to the parton level using the Monte Carlo information. This allows jet calibration to be disentangled from other effects on the top quark mass measurement from the other effects (selection, reconstruction, measurement methods). This is done by matching jets to partons before radiation (requiring $\Delta R(\text{quark},\text{jet})$ lower than 0.2) and deriving the difference between reconstructed jet energy and the parton energy as a function of the parton energy. This is performed separately for $b$-quark jets with and without muons and for light jets. This procedure leads to a perfect and unbiased jet energy scale.

Uncertainties on the top quark mass measurement arising from the jet energy scale uncertainties will be assesed here by applying miscalibration factors to the calibrated jets.

Jet scales on data will be obtained in a different way. The light jet energy scale determination is explained in detail in a separate note [3]; the $b$-jet energy scale will be measured using $Z$+jets samples. The $Z+$jets yield will be low at the start of LHC running; therefore, the $b$-jet scale will probably be derived from the measured light jet scale, together with a Monte Carlo correction term modelling the difference between the two jet energy scales.

#### 3.3.2 Jet labelling

A jet is called purely electromagnetic if the distance $\Delta R$ to the nearest electromagnetic cluster is lower than 0.2 and the ratio between the cluster and jet energies is greater than 0.8. Only 0.15% of jets produced by hadronisation pass these cuts. Such objects are ignored for the remainder of the analysis.

A jet is tagged as a $b$-jet if its b-weight is larger than 6 (the weight is defined by the three-dimensional space $b$-tagging, described in the introduction of this chapter [15]). The $b$-tagging efficiency for this weight in $t\bar{t}$ events is equal to 62% and the light (i.e. u, d and s flavoured) jet rejection to 130 for isolated ($\Delta R(\text{jet},\text{jet}) > 0.8$) jets with $p_T$ greater than 15 GeV (Fig. 1).

All remaining jets (i.e. non b-tagged jets nor purely electromagnetic jets) are called light jets. Therefore, a b-jet which is not tagged by the b-tag algorithm is considered as a light jet.

#### 3.3.3 Jet multiplicity

- The light jet multiplicity, for signal events, is illustrated in Fig. 2.(a), for all light jets, and for light jets above the $p_T$ cut applied (40 GeV). The multiplicity is often higher than two (47% of events, for jets above the $p_T$ cut), due to the presence of initial and final state radiation. This has of course an impact on the combinatorial background size and shape.
- The $b$-jet multiplicity, illustrated in Fig. 2.(b), reflects the $b$-tagging efficiency.

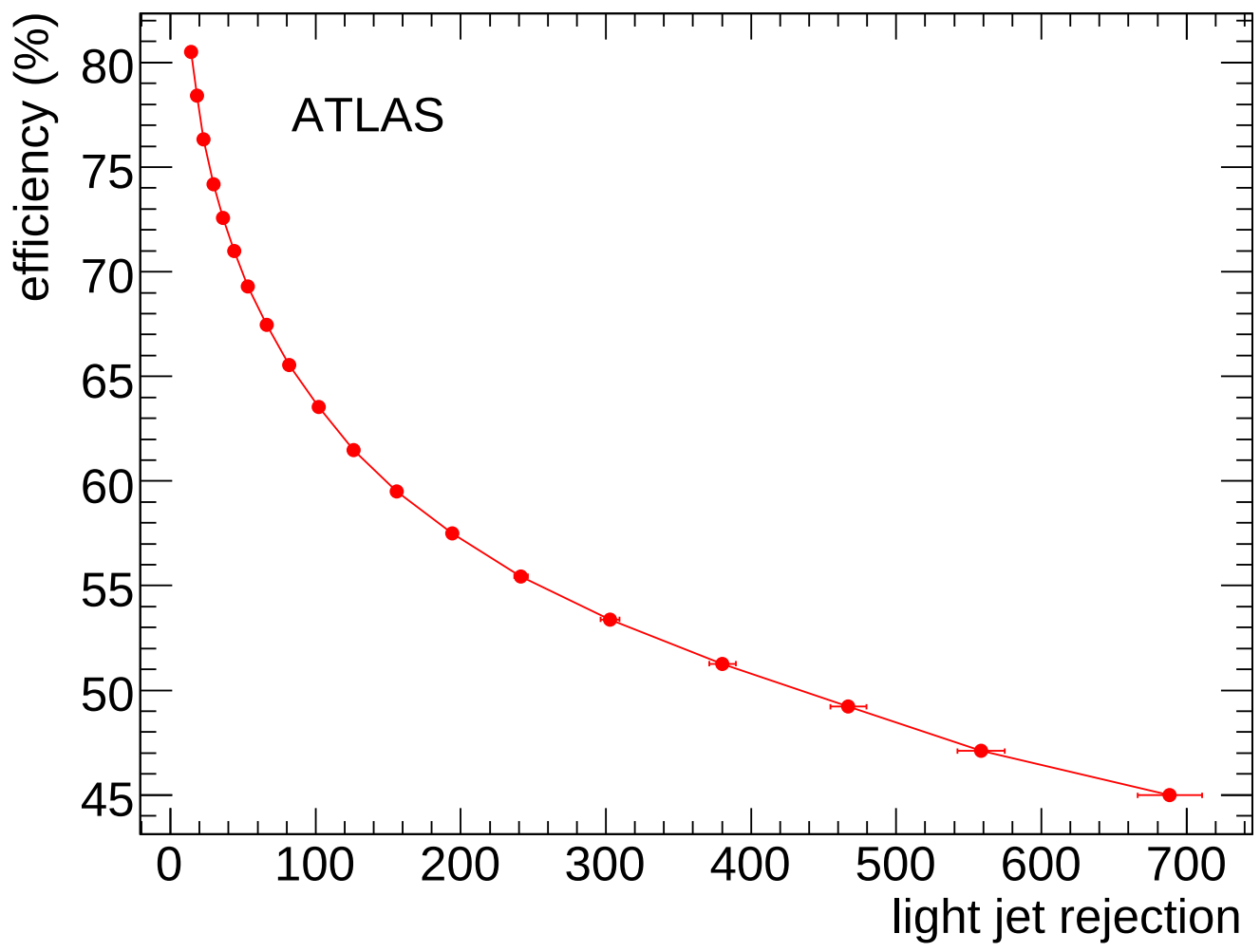


Figure 1: $b$-tagging efficiency as a function of light (i.e. u, d and s flavoured) jet rejection in $t\bar{t}$ events, for isolated ($\Delta R(\text{jet},\text{jet}) > 0.8$) jets with $p_T$ greater than 15 GeV.

### 3.4 Event selection

#### 3.4.1 Trigger

The following Event Filter trigger [16] selections have been applied:

- At least one isolated electron with $p_T$ greater than 25 GeV ("e22i"). This trigger is satisfied by 53% of $t\bar{t}$ $e+$jets events, and 71% of $e+$jets with $p_T$(e) greater than 25 GeV pass.
- At least one isolated muon with $p_T$ greater than 20 GeV ("mu20"). This trigger is satisfied by 59 % of $t\bar{t}$ $\mu+$jets events, and 74% of $\mu+$jets with $p_T(\mu)$ greater than 20 GeV pass.

#### 3.4.2 Standard cuts

A sequence of consecutive cuts is applied in order to reduce the contribution from physics background.

- Exactly one isolated lepton, with $p_T > 20$ (25) GeV for muons (electrons) and $|\eta| < 2.5$. This cut corresponds to the trigger selection. Moreover, the isolation criteria reject a large fraction of the leptonic $b$-decays in the all-jets channel: 99.6% of the $t\bar{t}$ events with both $W$ bosons decaying hadronically are rejected by this first selection step.
- Missing transverse energy cut: $\not{E}_T > 20$ GeV. Together with the lepton requirement, this cut reduces QCD background.
- At least four jets with $p_T > 40$ GeV. Below 40 GeV, jets are known to be less precisely calibrated; the jet energy scale will be discussed later on, as a source of systematic uncertainty. Therefore, they are removed in order to improve the precision of the top quark mass measurement [3]. Only 34% of hadronically-decaying $W$ bosons have both jets passing this requirement. If it is relaxed so that one jet can have $p_T$ down to 20 GeV, 88% of these $W$ bosons pass, but

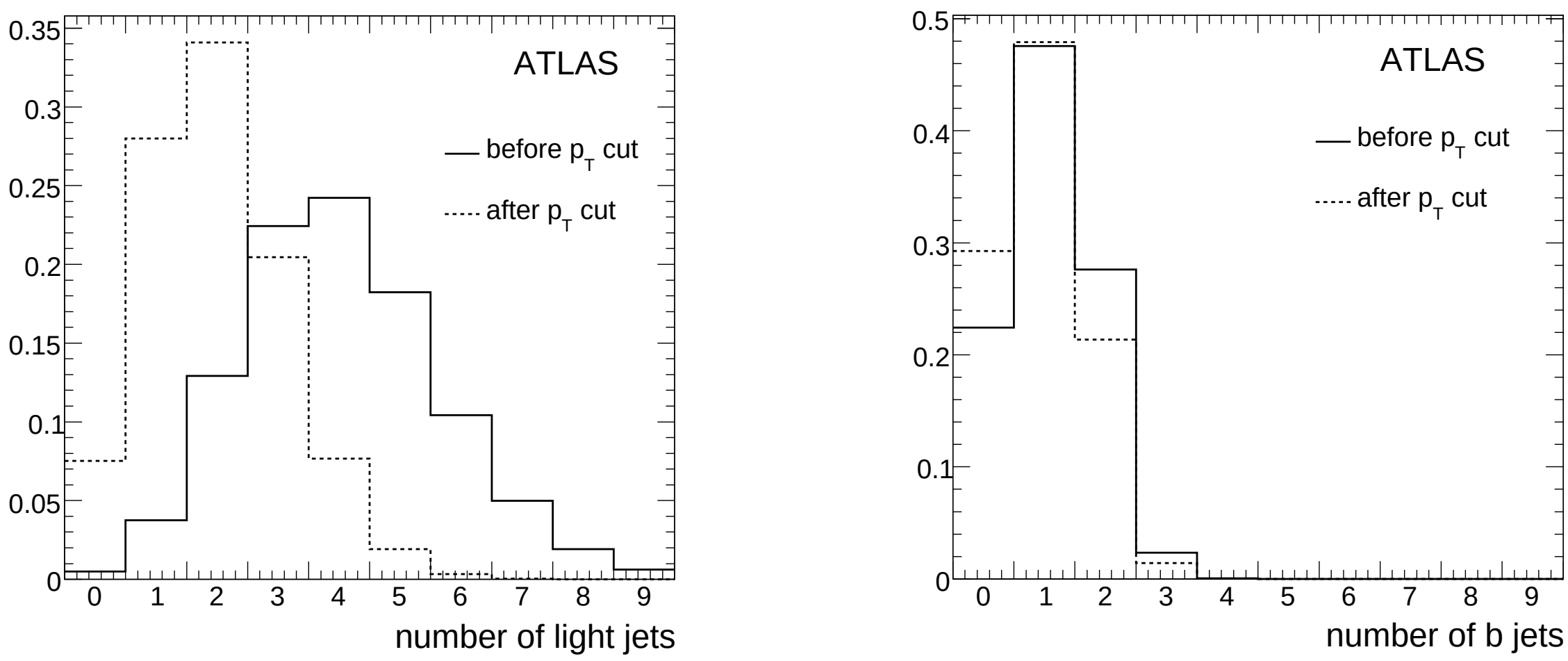


Figure 2: a) Light jet and b) $b$-jet multiplicity in semi-leptonic $t\bar{t}$ events before (solid line) and after (dashed line) requiring that the jet $p_T$ be greater than 40 GeV(plots are normalised to unity).

the combinatorial and physical backgrounds increase. Moreover, due to initial and final state radiation (ISR, FSR), 30% of the signal events have more than two light jets with $p_T > 40$ GeV. Therefore, no requirement is made on the number of light jets.

- Among these jets, exactly two must be $b$-tagged.

### 3.4.3 Purity definition

When a $t\bar{t}$ semi-leptonic decay is reconstructed, one must choose which jets in the event to associate with the hadronic $W$ boson decay and also which jets correspond to each of the two $b$-jets. The success of an algorithm for making such as choice is quantified as its *purity*, the fraction of events in which this choice is correct, based on looking at the Monte Carlo parentage information. For this purpose, jets are matched to the closest Monte Carlo parton with $\Delta R < 0.25$. Purities are defined for identifying the hadronically-decaying $W$ boson (both light jets chosen are within 0.25 from the quark stemming from the $W$ boson), the hadronic b-quark, and the hadronically-decaying top quark. Note that the top quark purity is not the product of the other two due to correlations.

## 3.5 Hadronic $W$ boson mass reconstruction

Several algorithms have been tried to choose the two light jets from the hadronically-decaying $W$ boson. Three have been identified that give the best compromise between efficiency and purity:

- the $\chi^2$ minimization method,
- the geometric method: this method consists in choosing the two closest jets,
- choosing the two light jets that give the mass closest to the known mass of the $W$ boson [14].

The first and last methods are quite similar, but the first one contains in addition an event-by-event rescaling. Therefore, the last method will not be described here.

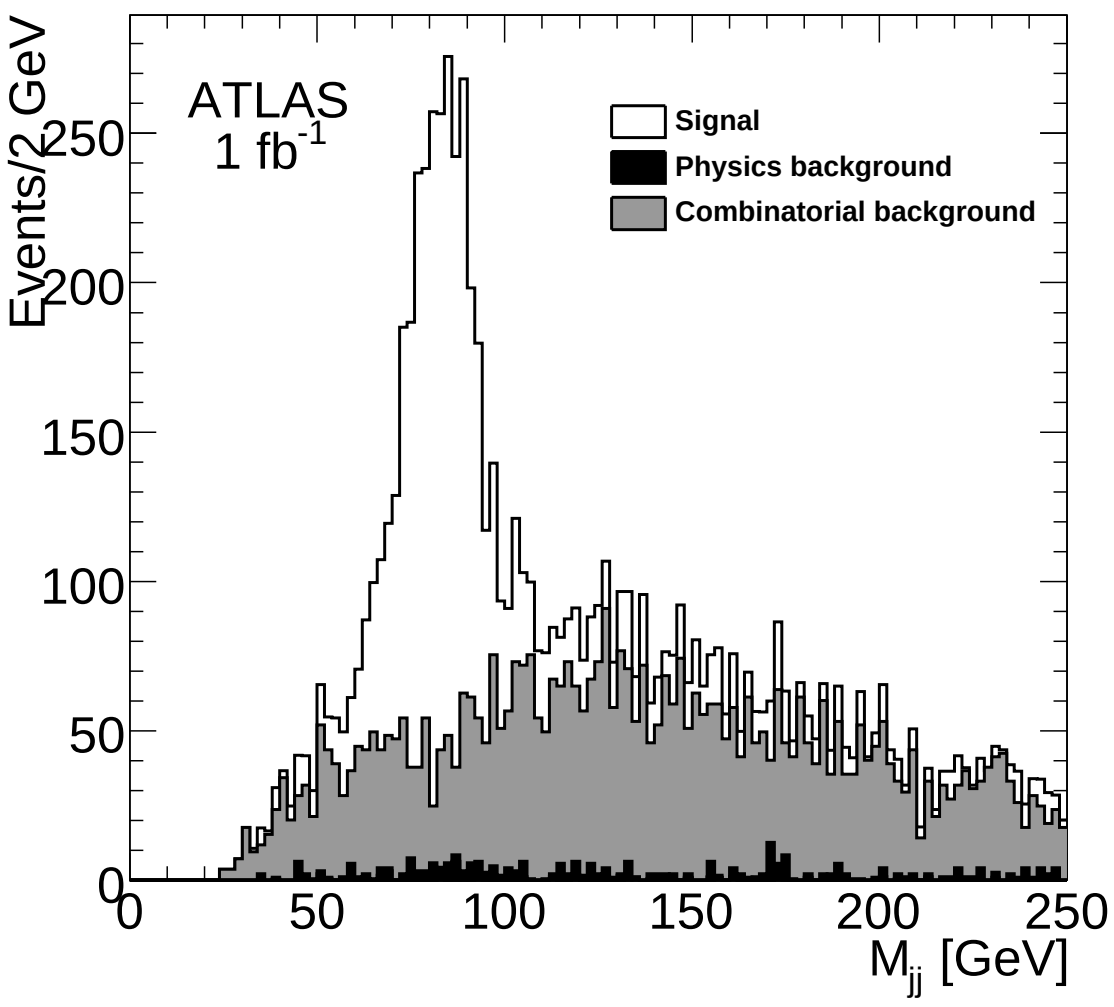


Figure 3: Invariant mass of light jet pairs for events with only two light jets (the contribution of physics background is shown in black, the one from combinatorial background, in grey).

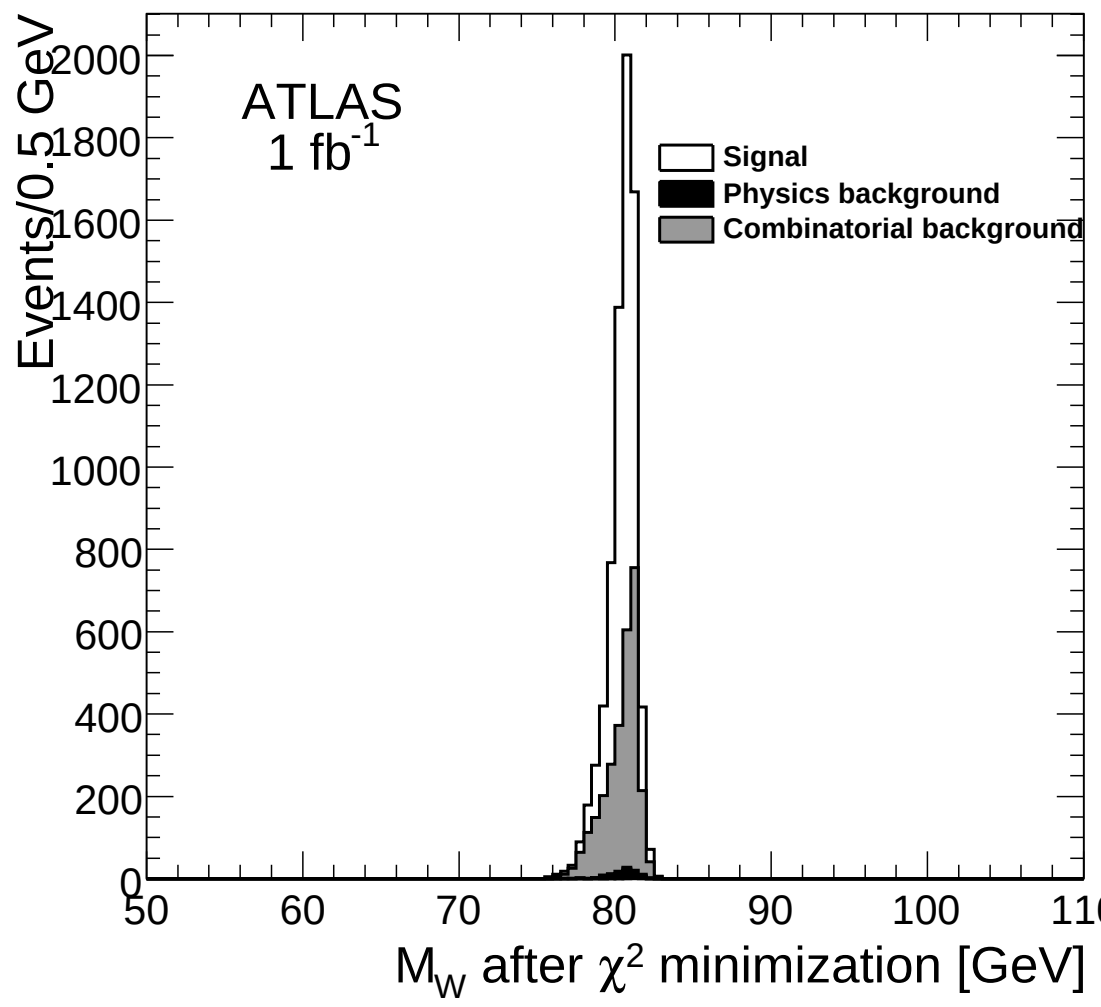


Figure 4: Hadronic $W$ boson mass after $\chi^2$ minimization (the contribution of physics background is shown in black, the one from combinatorial background, in grey).

### 3.5.1 Hadronic $W$ boson mass reconstruction with $\chi^2$ minimization method

Events kept after the selection described above have at least two light jets above the $p_T$ threshold (50% of the events have more than two). Figure 3 shows the distribution of the invariant mass of the light jet pairs, in events with only two light jets. As a first step, we select the hadronic $W$ boson candidates in a mass window of ($\pm$ 30 GeV) around the peak value of this distribution (82 GeV).

The energy scale of the jets may be shifted, due to effects that include the energy lost out of the jet cone and initial and final state radiation effects, not taken into account in the default jet energy calibration [17]. Moreover, the effects of the jet $p_T$ cut applied during event selection (explained in [3], section 3.3) contribute to the jet energy scale shift. To reduce the effect of such shifts on the final measured top quark mass, the jets are rescaled by constraining the pair to the known $W$ boson mass using a $\chi^2$ minimization. The quantity to be minimized is shown in Eq. (1). The first term constrains the jet pair mass $M_{jj}$ to the $W$ boson mass and width from the Particle Data Group [18] ($M_{\mathrm{W}}^{\mathrm{PDG}}$, $\Gamma_{\mathrm{W}}^{\mathrm{PDG}}$). The other two terms are the usual $\chi^2$ terms for scaling the jets with multiplicative constants $\alpha_{E_{\mathrm{j1,j2}}}$; $\sigma_{1,2}$ are the light jet energy resolutions[2].

$$\chi^2 = \frac{(M_{\mathrm{jj}}(\alpha_{E_{\mathrm{j1}}}, \alpha_{E_{\mathrm{j2}}}) - M_{\mathrm{W}}^{\mathrm{PDG}})^2}{(\Gamma_{\mathrm{W}}^{\mathrm{PDG}})^2} + \frac{(E_{\mathrm{j1}}(1-\alpha_{E_{\mathrm{j1}}}))^2}{\sigma_1^2} + \frac{(E_{\mathrm{j2}}(1-\alpha_{E_{\mathrm{j2}}}))^2}{\sigma_2^2}. \quad (1)$$

This $\chi^2$ is minimized, event by event, for each light jet pair. The pair with the smallest $\chi^2$ is kept as the hadronic $W$ boson candidate. This minimization procedure also gives the corresponding energy correction factors $\alpha_{E_{\mathrm{j1}}}$ and $\alpha_{E_{\mathrm{j2}}}$, shown in Fig. 5. Given the jet energy resolution, the first term in Eq. (1) dominates the $\chi^2$. Therefore, the hadronic $W$ boson mass reconstructed with the light jets chosen

[2]The light jet energy resolution has been estimated from the gaussian fit of the difference between a light jet energy and the corresponding quark energy. The expression thus obtained is the following: $\sigma_E = E * \sqrt{(a^2/E) + b^2}$, where a = 0.989 GeV$^{1/2}$ and b = 0.075.

by this $\chi^2$ minimization is very narrow, as shown in Fig. 4. Further on, only the hadronic $W$ boson candidates within a mass window of $\pm$ 2 $\Gamma_{\mathrm{M_W}}$ ($\Gamma^{\mathrm{PDG}}_{\mathrm{M_W}}$ = 2.1 GeV) are kept; this cut is called **C0**.

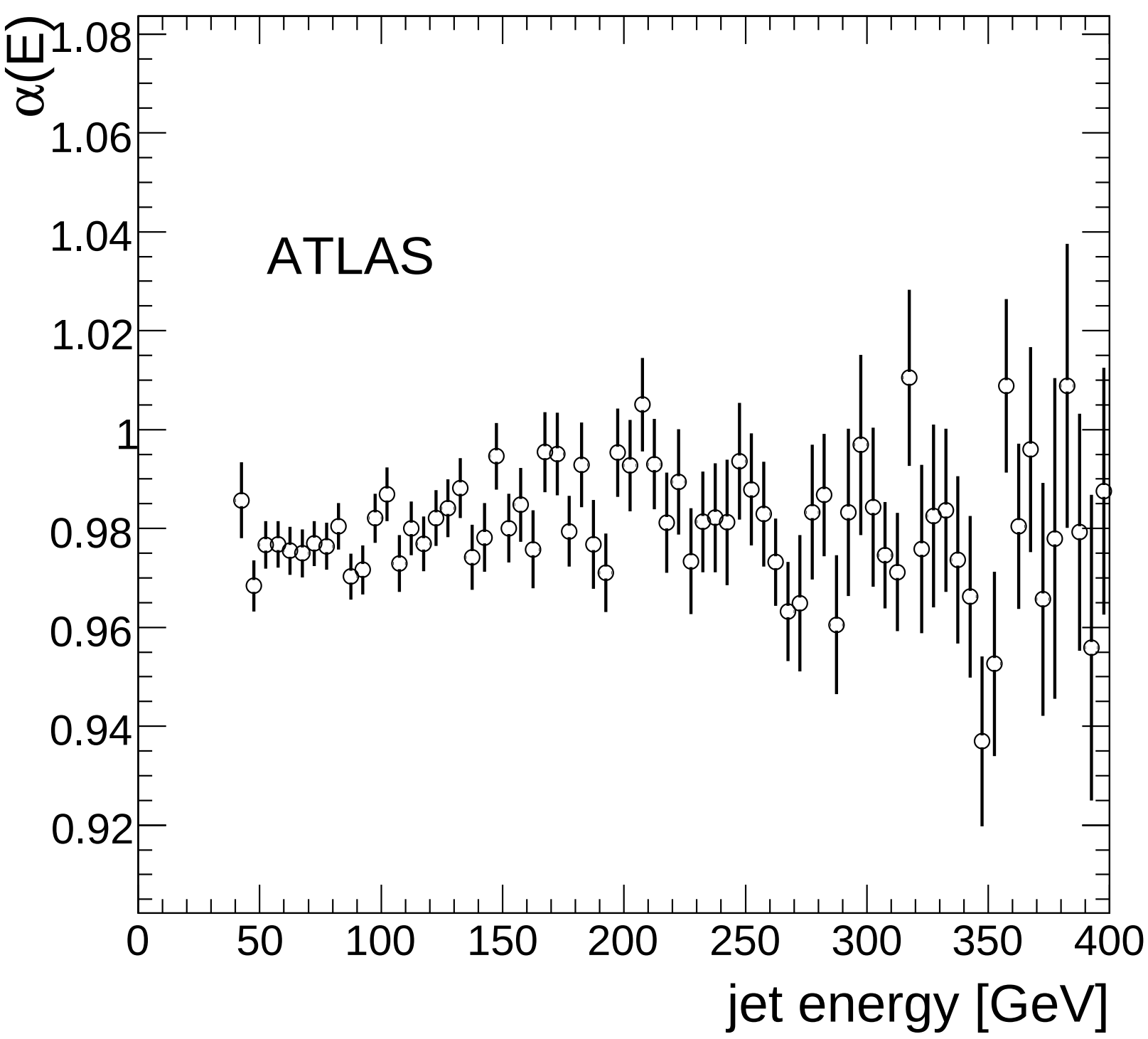


Figure 5: Energy correction factors estimated by the $\chi^2$ minimization.

### 3.5.2 Hadronic $W$ boson mass reconstruction with geometric method

In this method, the light jet pair with the smallest $\Delta R$ distance between the two jets is taken as the hadronic $W$ boson candidate. This method is simple and does not depend on the accuracy of the jet energy scale. The resulting $W$ boson mass distribution is shown in Fig. 6. Only hadronic $W$ boson candidates within a mass window of $\pm$ 2 $\sigma_{\mathrm{M_W}}$ ($\sigma_{\mathrm{M_W}}$ = 10.4 GeV) around the peak value of the invariant mass distribution of all light jet pairs are kept; this cut is called **C1**.

## 3.6 Leptonic $W$ boson mass reconstruction

The main difficulty in reconstructing the leptonic $W$ boson comes from the kinematics of the neutrino. The missing transverse momentum $\not{E}_T$ is used as an estimate of the neutrino transverse momentum. This is, however, only an approximation, illustrated in Fig.7, as there may be other, softer, neutrinos in the event, such as from leptonic $b$-decay. With this hypothesis, $p_T^{\nu}$ is underestimated by more than 2% (miscalibration of $\not{E}_T$ and fake missing $\not{E}_T$ also contributes to this shift).

Using the known $W$ boson mass, four-momentum conservation for the $W \to \ell + \nu$ decay gives a quadratic equation for the longitudinal component of the neutrino momentum $p_z^{\nu}$ ($\ell$ stands for lepton and $\nu$ stands for neutrino):

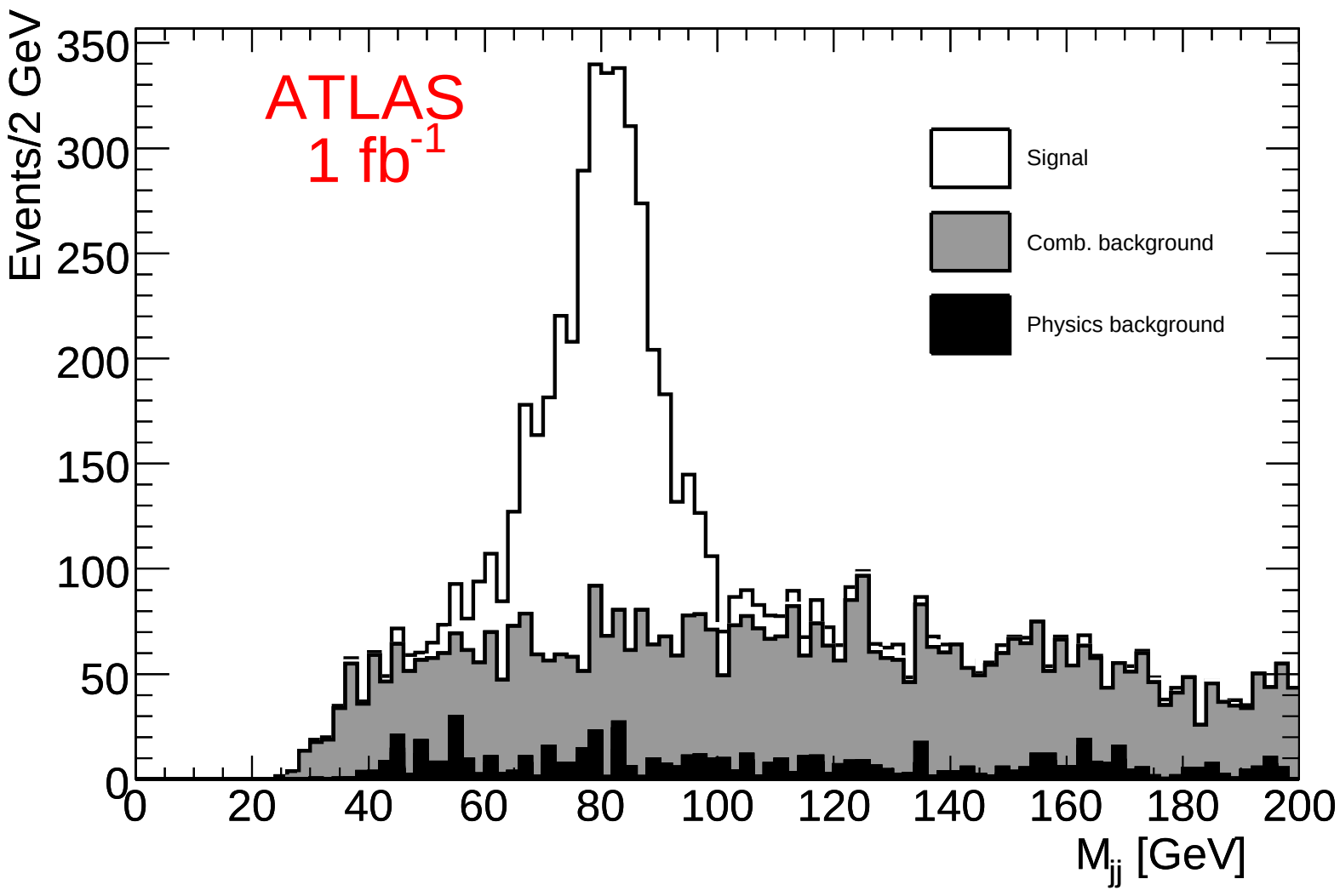


Figure 6: Hadronic $W$ boson mass with the geometric method (the contribution of physics background is shown in black, the one from combinatorial background, in grey).

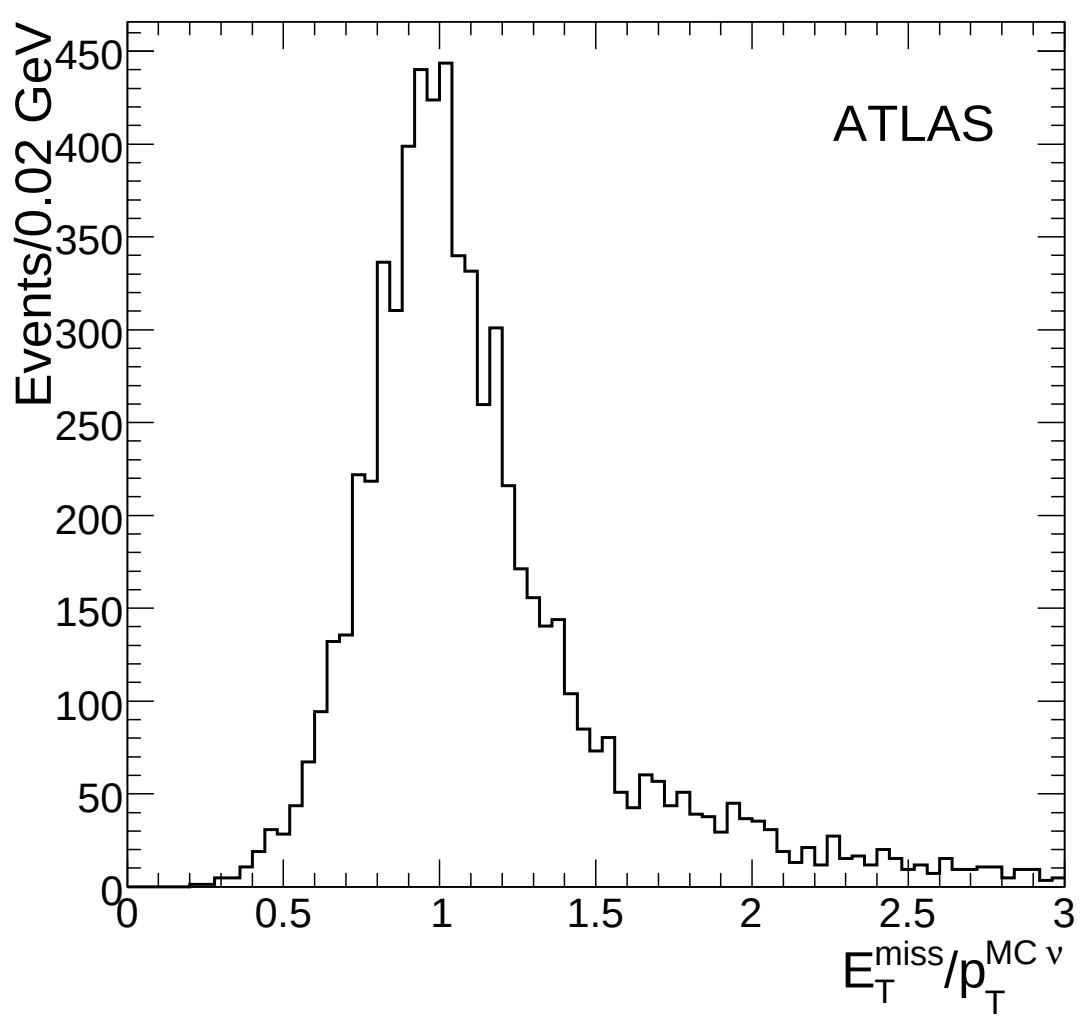


Figure 7: Ratio of $\not{E}_T$ to the generated neutrino transverse momentum, after C0 cut. The neutrino considered here is the one from the leptonic $W$ boson decay.

$$M_{\mathrm{W}}^2 = m_{\mathrm{l}}^2 - 2(p_{\mathrm{x}}^{\mathrm{l}} p_{\mathrm{x}}^{\nu} + p_{\mathrm{y}}^{\mathrm{l}} p_{\mathrm{y}}^{\nu}) + 2E_{\mathrm{l}}\sqrt{\not{E}_T^{\,2} + (p_{\mathrm{z}}^{\nu})^2} - 2(p_{\mathrm{z}}^{\mathrm{l}} p_{\mathrm{z}}^{\nu}).$$

This equation has no solution if the measured $\not{E}_T$ fluctuates such that the neutrino-lepton invariant mass is above the $W$ boson mass; this happens in 30% of the remaining events after the C0 cut. In this case, $p_T^{\nu}$ is reduced until a solution is found, with the restriction that the transverse $W$ boson mass remain below 90 GeV (see Fig. 8). Only 11% of the events still have no solution after this procedure. Otherwise, the equation has two solutions. The choice among the two $p_z^{\nu}$ is performed together with the association of the $b$-jet to the corresponding $W$ boson: the combination giving the smaller difference between the hadronic and leptonic top quark masses is kept.

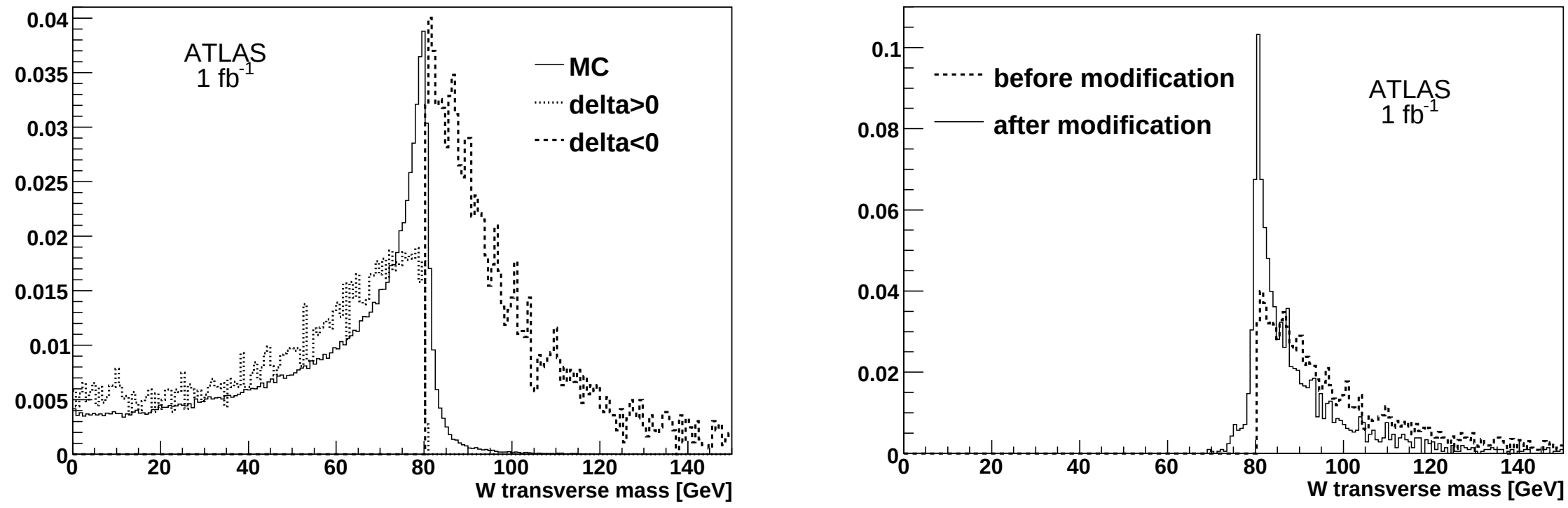


Figure 8: Transverse $W$ boson mass (distributions are normalized to unity). On the left, the Jacobian peak is clearly seen at the generator level (solid line); events for which no $p_z^{\nu}$ solution is found (dashed line) can be distinguished from events for which a solution is found (dotted line). On the right, the effect of the $p_T^{\nu}$ modification for events for which no $p_z^{\nu}$ solution is found, before (dashed line) and after (solid line) modification: a fraction of these events are recovered.

### 3.7 Top quark reconstruction

The two methods used for the hadronic $W$ boson reconstruction lead to two methods for the top quark mass reconstruction. Moreover, additional cuts are applied in order to increase the purity of the selected sample; their relevance is illustrated below.

Once the hadronic $W$ boson is reconstructed, the next step is to choose from the two $b$-jets the one to associate with the hadronic $W$ boson in order to reconstruct the hadronic top quark. Several methods have been investigated:

- choose the $b$-jet that maximizes the hadronic top quark $p_T$.
- Choose the $b$-jet closest to the hadronic $W$ boson.
- Choose the $b$-jet furthest from the leptonic $W$ boson.

All three methods give similar results, but the second method has a slightly higher purity, so that one has been chosen. The remaining $b$-jet and the leptonic $W$ boson then define the leptonic top quark.

The performance of the analyses before any additional cuts is summarized in Tables 2 and 3, in a full top quark mass window and within $\pm$ 3 $\sigma_{m_{\mathrm{top}}}$, where $\sigma_{m_{\mathrm{top}}}$ = 10 GeV.

Table 2: Efficiency of cuts applied and final purities (full top quark mass window and within ± 3 $\sigma_{m_{\text{top}}}$, where $\sigma_{m_{\text{top}}}$ = 10 GeV), with respect to $t\bar{t}$ semi-leptonic sample (e,$\mu$), for both methods.

| Cuts applied | Efficiency (%) | $W$ boson purity (%) | b purity (%) | top purity (%) |
|---|---|---|---|---|
| $\chi^2$ minimization method | | | | |
| Cut C0 | 2.22 ± 0.03 | 59.4 ± 0.8 | 61.6 ± 0.8 | 40.2 ± 0.8 |
| Cut C0 within ± 3 $\sigma_{m_{\text{top}}}$ | 1.40 ± 0.04 | 70.4 ± 0.8 | 77.3 ± 0.8 | 60.7 ± 0.8 |
| Cuts C0, C2 and C3 | 1.25 ± 0.04 | 59.3 ± 0.8 | 82.2 ± 0.8 | 56.5 ± 0.8 |
| Cuts C0, C2 and C3 within ± 3 $\sigma_{m_{\text{top}}}$ | 0.90 ±0.03 | 74.9 ±0.9 | 91.1 ± 0.9 | 73.6± 0.9 |
| Cuts C0, C2, C3, C4 and C5 | 0.91 ± 0.05 | 80.1 ± 0.8 | 92.8 ± 0.8 | 77.1 ±0.8 |
| Geometric method | | | | |
| Cut C1 | 1.26 ± 0.03 | 68.8 ± 0.8 | 69.7 ± 0.8 | 53.8 ± 0.9 |
| Cut C1 within ± 3 $\sigma_{m_{\text{top}}}$ | 1.01 ± 0.04 | 77.6 ± 0.8 | 77.7 ± 0.8 | 65.9 ± 0.8 |
| Cuts C1, C2 and C3 | 0.85 ± 0.03 | 68.7 ±0.9 | 84.7 ±0.8 | 66.1 ±0.9 |
| Cuts C1, C2 and C3 within ± 3 $\sigma_{m_{\text{top}}}$ | 0.70 ± 0.03 | 79.4 ±0.9 | 90.7 ±0.7 | 78.1 ±0.9 |
| Cuts C2, C3, C4 and C5 | 0.57 ± 0.05 | 86.9 ± 0.9 | 94.0 ± 0.6 | 86.4 ± 0.9 |

Table 3: Number of events in 1 fb$^{-1}$ (signal and background) after selection cuts (full top quark mass window and within ± 3 $\sigma_{m_{\text{top}}}$, where $\sigma_{m_{\text{top}}}$ = 10 GeV), for both methods. $\tau$ + jets events correspond to: hadronic $W$ boson → $\tau\nu$ and $\tau$ decays hadronically. Leptonic $\tau$ decays are counted together with the signal.

| Number of events for 1 fb$^{-1}$ | signal | W + jets | $\tau$ →jets | di-lepton | all-jets | single top |
|---|---|---|---|---|---|---|
| $\chi^2$ minimization method | | | | | | |
| Cut C0 | 6946 | 19 | 14 | 191 | 51 | 148 |
| Cut C0 within ± 3 $\sigma_{m_{\text{top}}}$ | 4382 | 12 | 3 | 62 | 20 | 71 |
| Cuts C0, C2 and C3 | 3918 | 7 | 10 | 104 | 28 | 67 |
| Cuts C0, C2 and C3 within ± 3 $\sigma_{m_{\text{top}}}$ | 2863 | 4 | 3 | 24 | 28 | 24 |
| Cuts C0, C2, C3, C4 and C5 | 2850 | 1 | 2 | 10 | 17 | 19 |
| Geometric method | | | | | | |
| Cut C1 | 3949 | 9 | 10 | 19 | 39 | 89 |
| Cut C1 within ± 3 $\sigma_{m_{\text{top}}}$ | 3155 | 4 | 6 | 9 | 7 | 56 |
| Cuts C1, C2 and C3 | 2643 | 4 | 5 | 11 | 33 | 48 |
| Cuts C1, C2 and C3 within ± 3 $\sigma_{m_{\text{top}}}$ | 2198 | 4 | 3 | 5 | 0 | 27 |
| Cuts C2, C3, C4 and C5 | 1785 | 0 | 1 | 2 | 7 | 13 |

### 3.7.1 Additional cuts

Additional cuts can be applied in order to increase the final top purity (combinatorial background rejection).

- Cut **C2**: the invariant mass of the hadronic $W$ boson and the $b$-jet associated to the leptonic $W$ boson must be greater than 200 GeV.

- Cut **C3**: the invariant mass of the lepton and the $b$-jet associated to the leptonic $W$ boson must be lower than 160 GeV.

These cuts are illustrated in Figs. 9 and 10. Their effects on efficiency and purity are shown in Tables 2 and 3.

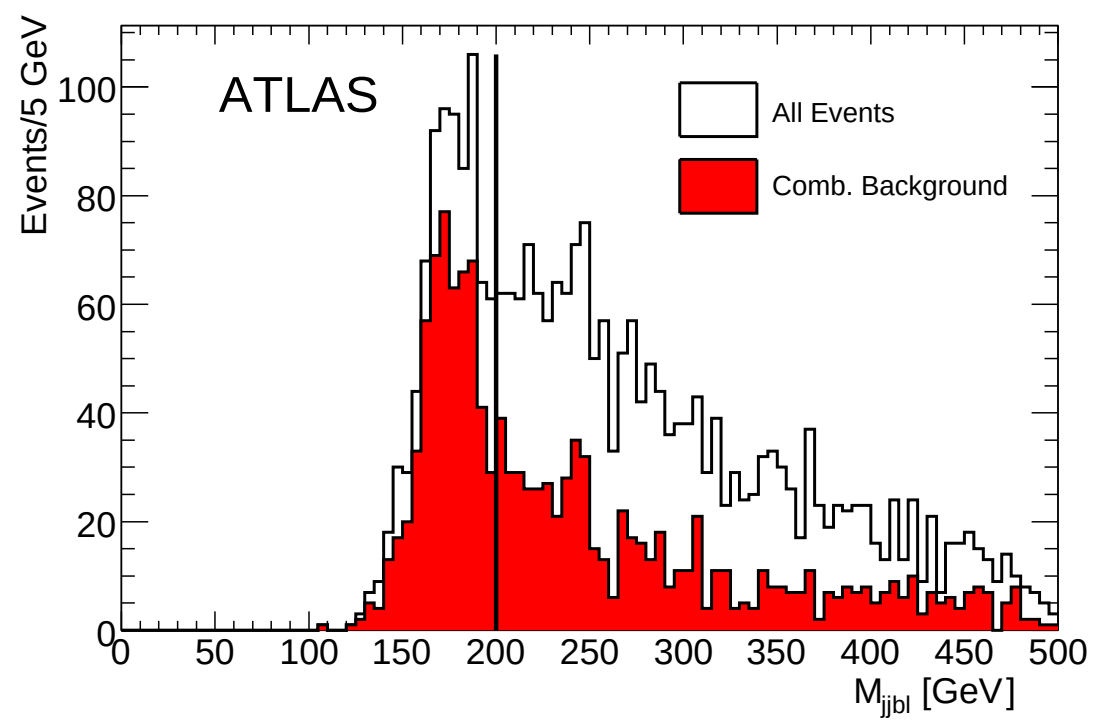


Figure 9: Invariant mass of the hadronic $W$ boson and the leptonic $b$-jet for events satisfying C1. The vertical line corresponds to $M_{jjbl} = M(W_{\rm had}, b_{\rm lep}) > 200$ GeV.

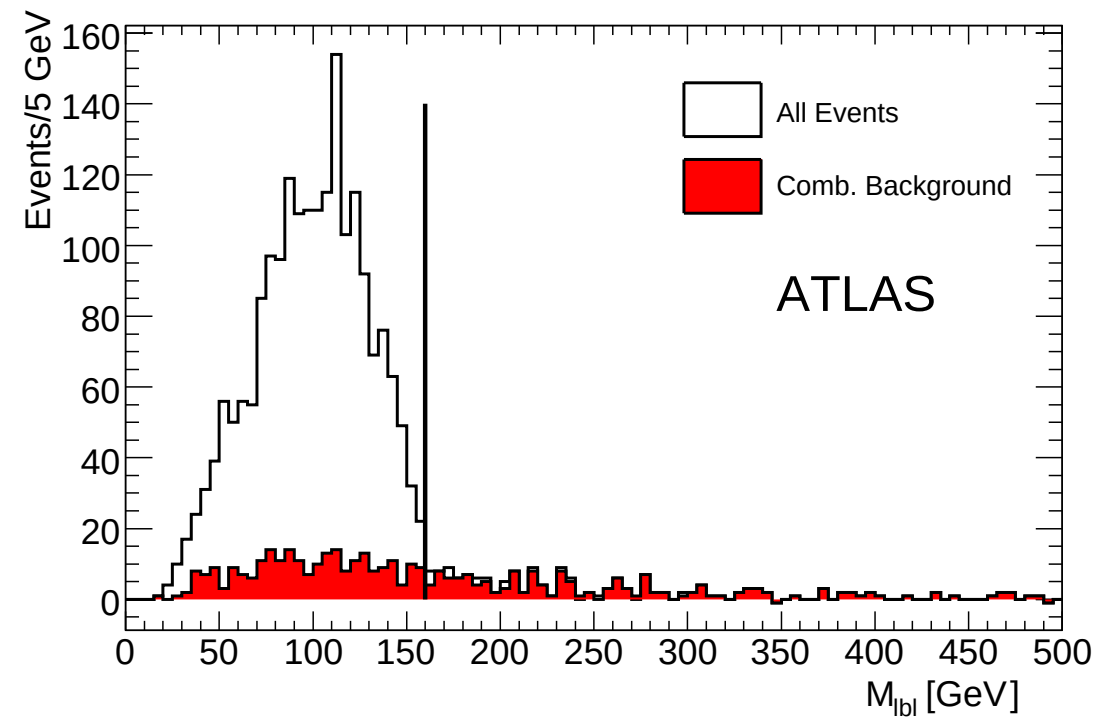


Figure 10: Invariant mass of the lepton and the leptonic $b$-jet for events satisfying C1 and C2. The vertical line corresponds to $M_{lbl} = M(l, b_{\rm lep}) < 160$ GeV.

The combinatorial background can be further suppressed with two more cuts [19]. These are defined based on the following variables, where $E^*$ denotes the energy of a particle in the top quark rest frame:

$$X_1 = E^*_{\rm W} - E^*_{\rm b} = E^*_{\rm j1} + E^*_{\rm j2} - E^*_{\rm b} = \frac{M^2_{\rm W} - M^2_{\rm b}}{M_{\rm top}}, \tag{2}$$

$$X_2 = 2E^*_{\rm b} = \frac{M^2_{\rm top} - M^2_{\rm W} + M^2_{\rm b}}{M_{\rm top}}. \tag{3}$$

We call the peak and width of the $X_{1,2}$ distributions $\mu_{1,2}$ and $\sigma_{1,2}$, as is found from simulated $t\bar{t}$ events with $m_{\rm top} = 175$ GeV that satisfy all previous requirements. Then the two following cuts are defined:

- Cut **C4**: $|X_1 - \mu_1| < 1.5\,\sigma_1$,

- Cut **C5**: $|X_2 - \mu_2| < 2\,\sigma_2$.

These cuts are illustrated in Figs. 11 and 12. With respect to C2 and C3, these cuts reduce the efficiency by 30% but increase the purity to 85%, as shown in Tables 2 and 3. The numbers are identical in the full top quark mass window and within $\pm$ 3 $\sigma_{m_{\rm top}}$, since C4 and C5 restrict the top quark mass to a more stringent window than $\pm$ 3 $\sigma_{m_{\rm top}}$ around the peak value.

Table 4 summarizes the cuts applied in these analyses.

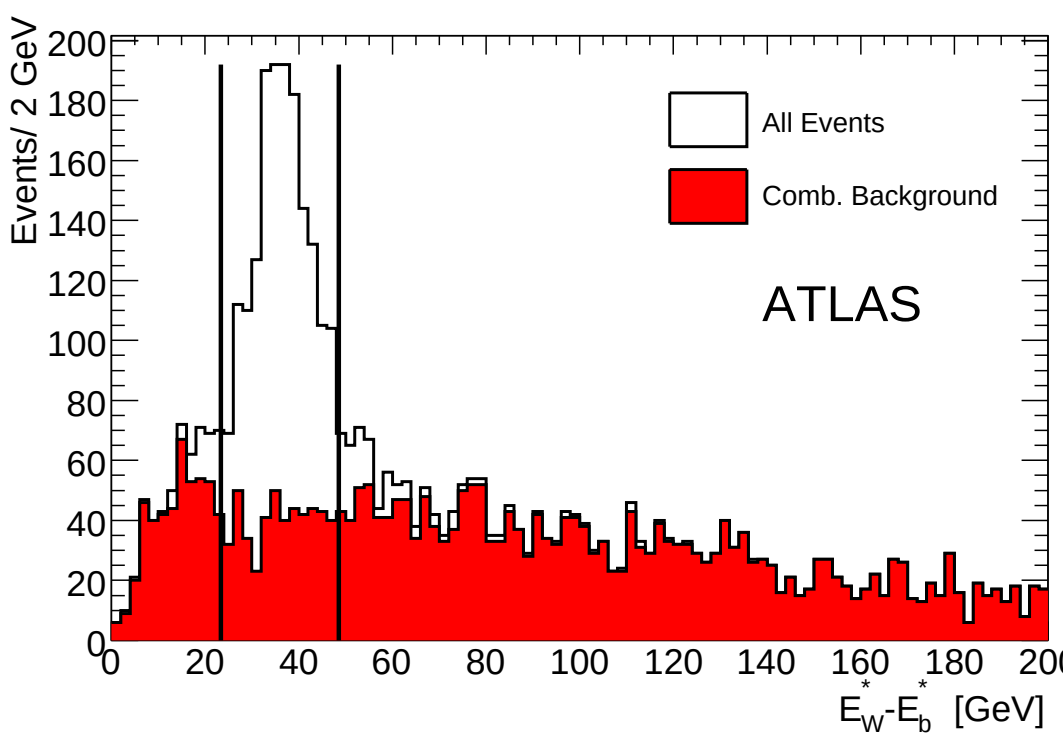


Figure 11: Distribution of $X_1 \equiv E_W^* - E_b^*$ for events passing C2 and C3. Vertical lines correspond to the bounds of the **C4** cut.

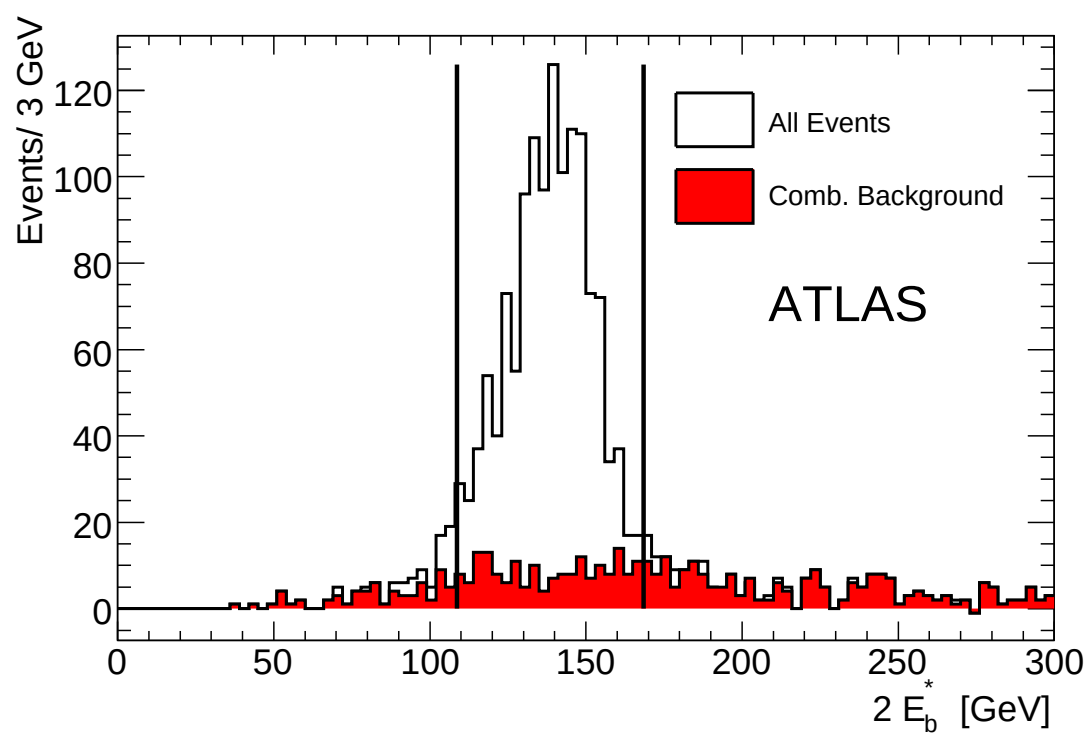


Figure 12: Distribution of $X_2 \equiv 2E_b^*$ for events passing C2, C3, and C4. Vertical lines correspond to the bounds of the **C5** cut.

Table 4: Additional cuts applied, after the event selection, for both methods ($X_i$, $\mu_i$ and $\sigma_i$ are defined in the text of this section).

| Cut label | Description |
|---|---|
| Cut C0 ($\chi^2$ minimization) | $\lvert M_W^{\rm rec} - M_W^{PDG}\rvert < 2\Gamma_{M_W}^{PDG}$ ($M_W^{\rm rec}$ is the reconstructed hadronic W and $\Gamma_{M_W}^{PDG}$ = 2.1 GeV) |
| Cut C1 (geometric method) | $\lvert M_W^{\rm rec} - M_W^{peak}\rvert < 2\,\sigma_{\rm M_W}$ ($\sigma_{\rm M_W}$ = 10.4 GeV) |
| Cut C2 (both methods) | $M(W_{\rm had}, b_{\rm lep}) > 200$ GeV |
| Cut C3 (both methods) | $M({\rm lepton}, b_{\rm lep}) < 160$ GeV |
| Cut C4 (both methods) | $\lvert X_1 - \mu_1\rvert < 1.5\,\sigma_1$ |
| Cut C5 (both methods) | $\lvert X_2 - \mu_2\rvert < 2\,\sigma_2$ |

### 3.8 Top quark mass measurement

In Fig. 13, the hadronic top quark mass reconstructed with the $\chi^2$ minimization method is fit to the sum of a Gaussian and a polynomial (third degree). For 1 fb$^{-1}$, the fit Gaussian has its mean at $175.0 \pm 0.2$ GeV and a width of $11.6 \pm 0.2$ GeV ($\chi^2$/dof = 137/67). It is seen that C2 and C3 do not significantly shift the top quark mass: $m_{\text{top}} = 174.8 \pm 0.3$ GeV with a width equal to $11.7 \pm 0.4$ GeV ($\chi^2$/dof = 82/67: C2 and C3 improve significantly this value).

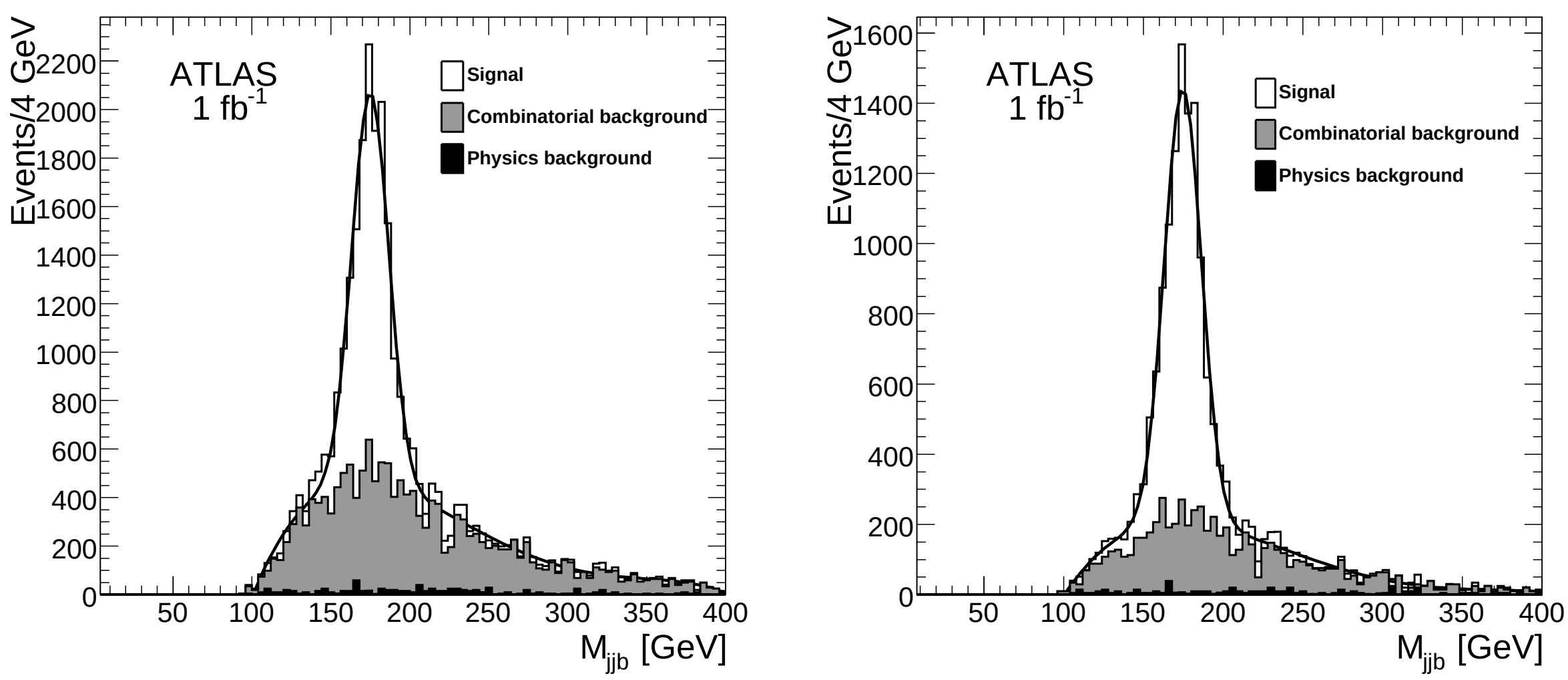


Figure 13: The hadronic top quark mass reconstructed with the $\chi^2$ minimization method, fit with a sum of a Gaussian and a third order polynomial, scaled to 1 fb$^{-1}$. Left, after C0: $m_{\text{top}} = 175.0 \pm 0.2$ GeV, with a width equal to $11.6 \pm 0.2$ GeV. Right, after C2 and C3: $m_{\text{top}} = 174.8 \pm 0.3$ GeV with a width equal to $11.7 \pm 0.4$ GeV

Figure 14 shows the result of the hadronic top mass reconstruction using the geometric method, fit to the sum of a Gaussian and a threshold function[3](left, $\chi^2$/dof = 97/75) and to a pure Gaussian (right, $\chi^2$/dof = 38/24). After all cuts, the Gaussian mean fits to $175.0 \pm 0.4$ GeV with a width of $14.3 \pm 0.3$ GeV. The width is larger than with the $\chi^2$ minimization method since no attempt is made to perform an event-by-event rescaling of the light jets. Nevertheless, the contribution of the light jets to the top quark mass resolution can be removed to first order by computing the top quark mass as $m_{\text{top}} = M_{\text{jjb}} - M_{\text{jj}} + M_{\text{W}}^{\text{peak}}$. The results of this geometric method with rescaling are shown in Fig. 15. The width decreases to 10.6 GeV, consistent with the results from the $\chi^2$ minimization method.

Table 5 summarizes the fit results from all the methods discussed here.

[3]The formula of the threshold function used is the following:

$$A \cdot e^{-\frac{1}{2}\left(\frac{x-M_{top}}{\sigma_{top}}\right)^2} + CstBdf \cdot (x - threshold)^b \cdot e^{-c(x-threshold)} \tag{4}$$

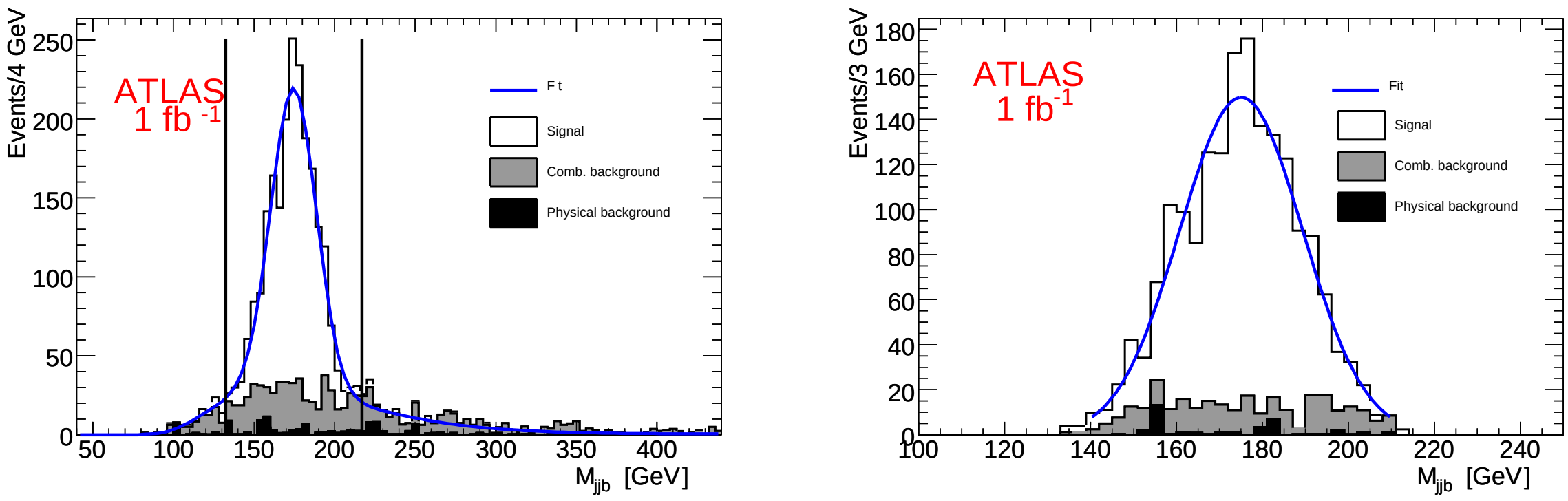


Figure 14: The hadronic top quark mass reconstructed with the geometric method, fit with a sum of a Gaussian and a threshold function (left) and with a pure Gaussian (right), scaled to 1 $\text{fb}^{-1}$. Left, after C1, C2, and C3: $m_{\text{top}} = 174.6 \pm 0.5$ GeV, with a width equal to $11.1 \pm 0.5$ GeV; right, after C2, C3, C4, and C5: $m_{\text{top}} = 175.0 \pm 0.4$ GeV, with a width equal to $14.3 \pm 0.3$ GeV.

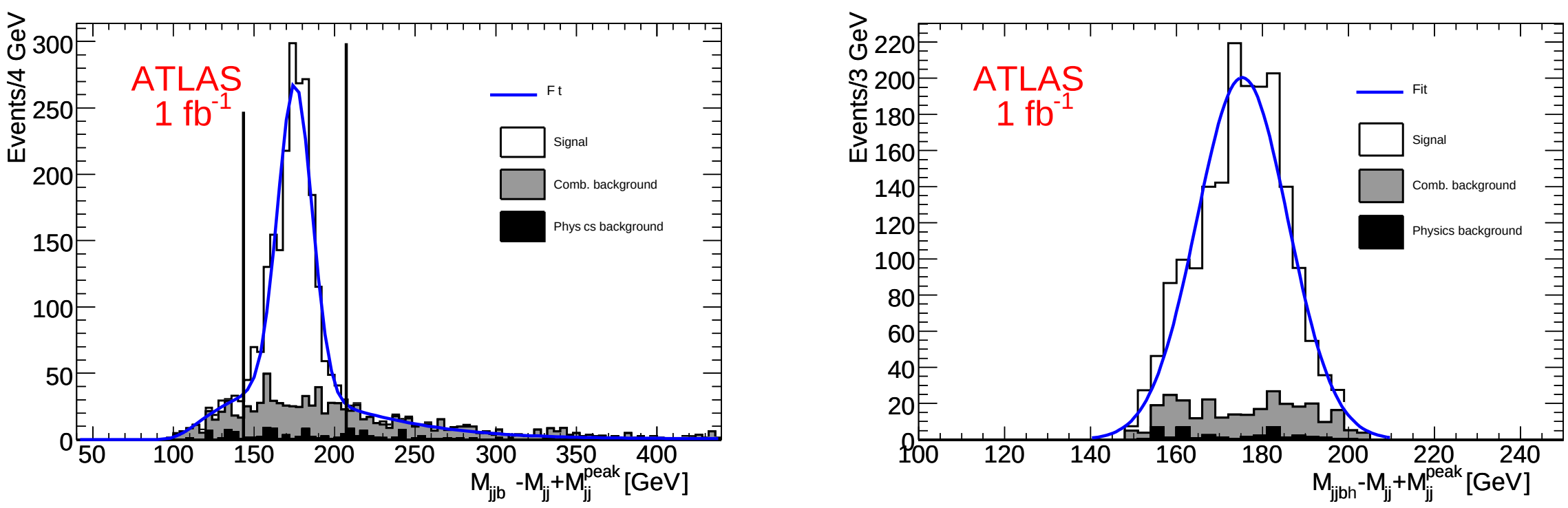


Figure 15: The hadronic top quark mass reconstructed with the geometric method with rescaling, fit with a sum of a Gaussian and a threshold function, scaled to $1\text{fb}^{-1}$. Left, after C1, C2, and C3: $m_{\text{top}} = 175.4 \pm 0.4$ GeV, with a width equal to $10.6 \pm 0.4$ GeV ($\chi^2$/dof = 109/73); right, after C2, C3, C4, and C5: $m_{\text{top}} = 175.3 \pm 0.3$ GeV, with a width equal to $10.6 \pm 0.2$ GeV ($\chi^2$/dof = 43/16).

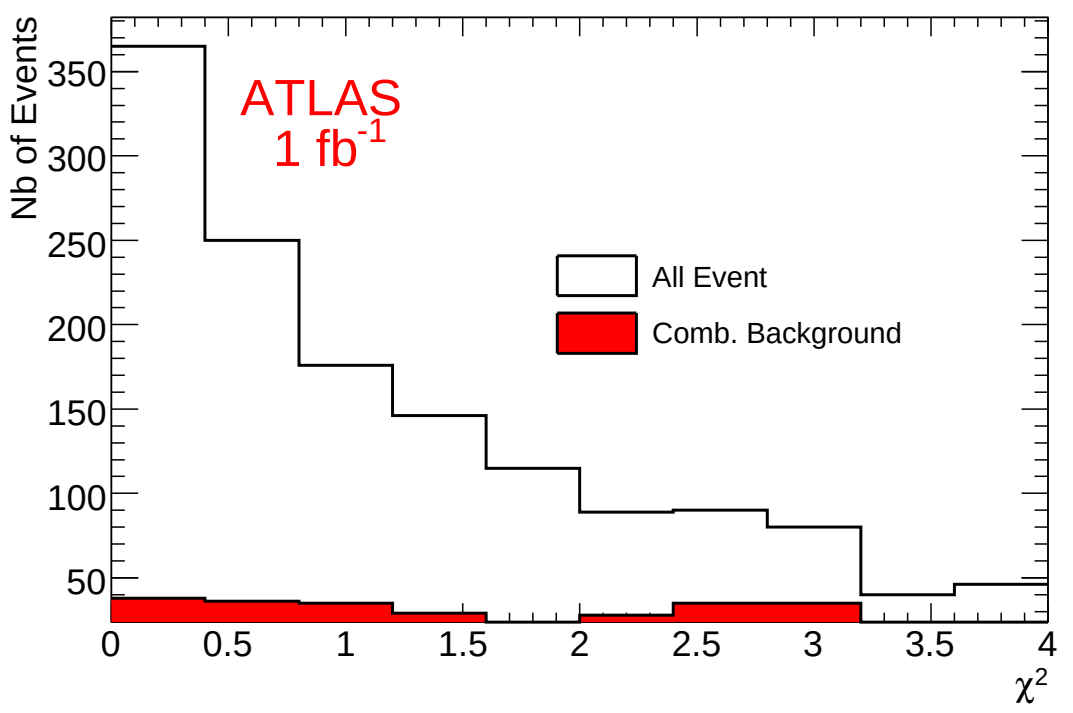


Figure 16: $\chi^2$ distribution (with combinatorial background), with cuts **C2** and **C3** applied.

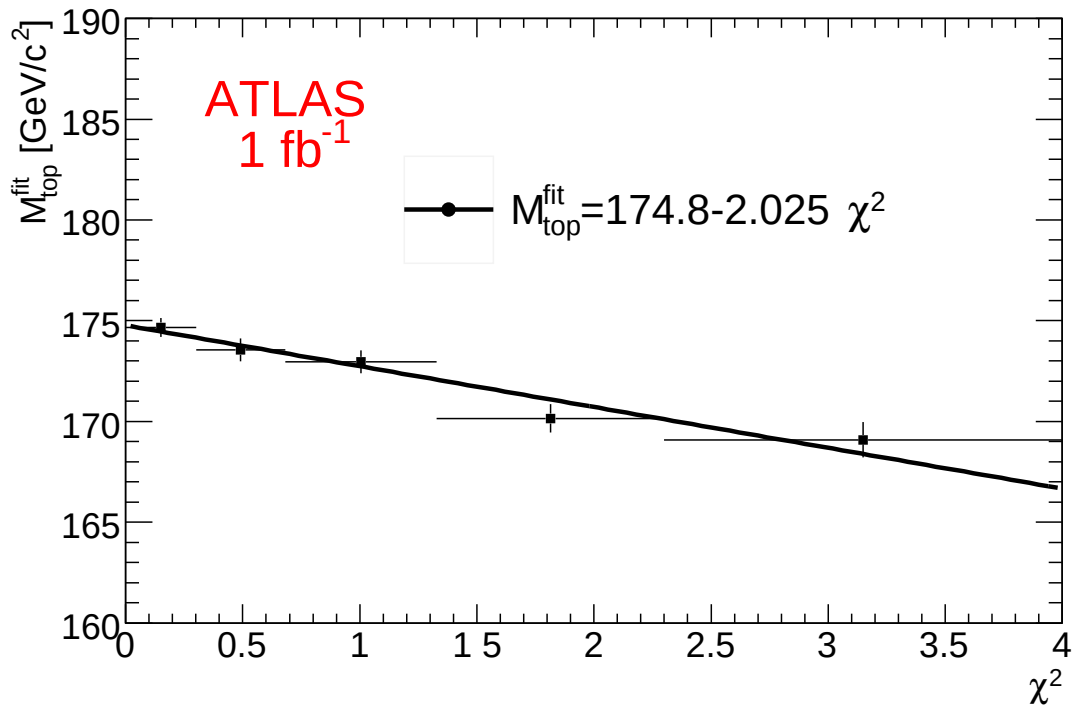


Figure 17: Top quark mass as a function of $\chi^2$. The points are fitted by a linear function to extract $m_{\rm top} = m_{\rm top}^{\rm fit}(\chi^2 = 0)$

.

### 3.9 Kinematic fit

The top quark mass can be extracted from a kinematic fit using a $\chi^2$ based on the entire final state, as defined in equation (5). The terms in the first line consist of the usual $\chi^2$ terms[4], while the last four constrain the object masses. The resolutions are extracted from Monte Carlo. The $\chi^2$ is calculated event by event and the resulting $\chi^2$ distribution, shown in Fig. 16, exhibits a higher purity for lower $\chi^2$ values. The purity of the final sample could be improved by cutting on the $\chi^2$.

$$\begin{aligned}\chi^2 = &\sum_{\rm jets}\Big(\big(\frac{\eta_{\rm i}^{\rm m}-\eta_{\rm i}^{\rm f}}{\sigma_\eta^{\rm i}}\big)^2+\big(\frac{\phi_{\rm i}^{\rm m}-\phi_{\rm i}^{\rm f}}{\sigma_\phi^{\rm i}}\big)^2\Big)+\sum_{\rm jets,lepton}\big(\frac{{\rm E}_{\rm i}^{\rm m}-{\rm E}_{\rm i}^{\rm f}}{\sigma_{\rm E}^{\rm i}}\big)^2+\sum_{\rm x,y,z}\big(\frac{{\rm p}_{\rm i\nu}^{\rm m}-{\rm p}_{\rm i\nu}^{\rm f}}{\sigma_{\rm i\nu}}\big)^2 \\ &+\big(\frac{m_{\rm jj}-M_{\rm W}^{\rm PDG}}{\sigma_W}\big)^2+\big(\frac{m_{l\nu}-M_{\rm W}^{\rm PDG}}{\sigma_W}\big)^2+\big(\frac{m_{\rm jjb_h}-m_{\rm top}^{\rm fit}}{\sigma_t}\big)^2+\big(\frac{m_{l\nu b_l}-m_{\rm top}^{\rm fit}}{\sigma_t}\big)^2.\end{aligned} \tag{5}$$

The $\chi^2$ minimization provides a high constraint on the jets from $W$ boson. In an earlier study [14], it has been shown that the accuracy of the determination of the top quark mass depends on the $\chi^2$. For events in which the b-quark jets are well measured and extra final state effects are small, the $\chi^2$ of the fit is close to 0 and produces a top quark mass value reflecting the Monte Carlo generated top quark mass. For larger $\chi^2$ value, this top quark mass value decreases while the fraction of b-quarks with a large gluon radiation increases. This observation can be used by performing a linear fit to the extracted top quark mass as a function of the $\chi^2$. The fit is shown on Fig. 17 where the top mass distribution is fitted by a gaussian for each $\chi^2$ slice. The most accurate estimate of the top quark mass is obtained by extrapolating this linear fit to $\chi^2 = 0$. This procedure should lead to a lower sensitivity to final state radiation effects of the extracted top quark mass, which is found to be $m_{\rm top} = 174.8 \pm 0.4$ GeV, where the uncertainty is statistical.

This method is more computationally intensive than the others to assess the systematics, and is not considered further in this note.

[4]The $m$ index stands for measured quantities and the $f$ index, for fitted quantities. E, $\eta$ and $\phi$ are respectively the energy, pseudo-rapidity and polar angle of the considered objects.

Table 5: Fitted peaks for the hadronic top quark mass and corresponding widths for several methods. From simulated events with $m_{\mathrm{top}} = 175$ GeV, for 1 fb$^{-1}$.

| Method | Cuts | Top quark mass [GeV] | $\sigma$ [GeV] | Pull bias | Pull width |
|---|---|---|---|---|---|
| $\chi^2$ minimization | **C0** | $175.0 \pm 0.2$ | $11.6 \pm 0.2$ | $-0.3$ | 1.23 |
| $\chi^2$ minimization | **C0**, **C2**, and **C3** | $174.8 \pm 0.3$ | $11.7 \pm 0.4$ | $-0.2$ | 1.11 |
| $\chi^2$ minimization | **C0**, **C2**, **C3**, **C4**, and **C5** | $174.8 \pm 0.3$ | $11.8 \pm 0.4$ | $-0.5$ | 1.00 |
| Geometric | **C1**, **C2**, and **C3** | $174.6 \pm 0.5$ | $14.1 \pm 0.5$ | $-0.39$ | 1.03 |
| Geometric | **C2**, **C3**, **C4** and **C5** | $175.0 \pm 0.4$ | $14.3 \pm 0.3$ | $-0.37$ | 1.11 |
| Geometric and rescaling | **C1**, **C2**, and **C3** | $175.4 \pm 0.4$ | $10.6 \pm 0.4$ | 0.51 | 1.10 |
| Geometric and rescaling | **C2**, **C3**,**C4**, and **C5** | $175.3 \pm 0.3$ | $10.6 \pm 0.2$ | 0.17 | 1.15 |
| Kinematic fit | | $174.8 \pm 0.4$ | | | |

### 3.10 Statistical uncertainties

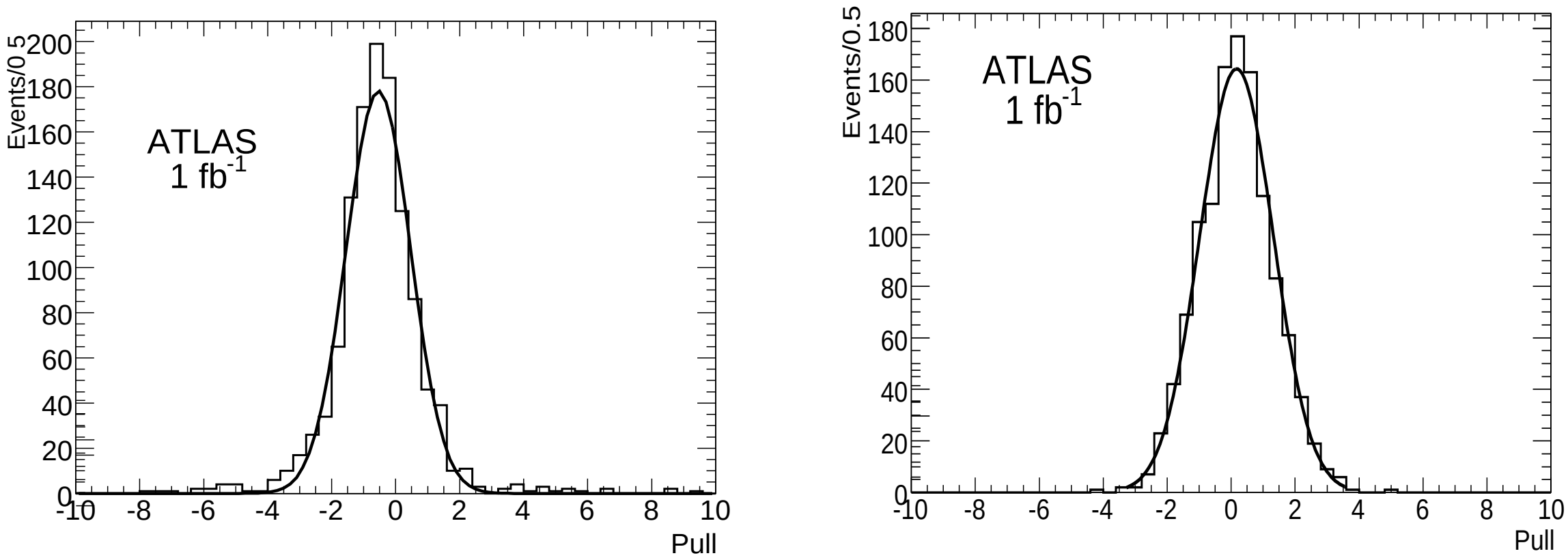


Figure 18: Pull distributions for the top quark mass measurement. Left, $\chi^2$ minimization method and right, geometric method. Both plots after C2, C3, C4, and C5.

The statistical uncertainties quoted have been evaluated using a single simulated experiment with statistics corresponding to 1fb$^{-1}$. To evaluate the reliability of these estimates, a bootstrap resampling technique has been used [20].

The pull distributions for the top quark mass measurement $(M_{\mathrm{i}} - M_{\mathrm{gen}})/\sigma_{M_{\mathrm{i}}}$ have been produced for each method using 1200 pseudo-experiments. The biases and widths are reported in Table 5 for all the analyses shown in this note. Fig. 18 shows the pull distributions for the $\chi^2$ minimization method and for the geometric method, after cuts **C2** through **C5**. All pull mean values are of size 0.5 or less: this indicates that the biases induced by the methods of measuring the top quark mass are of the order of 0.1 to 0.2 GeV depending on the cuts applied. The pull widths are slightly larger than 1; this indicates that the statistical uncertainty of the fit is slightly underestimated.

## 3.11 Systematic uncertainties

The statistical uncertainty on the top quark mass being negligible with a few $fb^{-1}$ of collected data, the total uncertainty will quickly be dominated by the systematic uncertainties. All the contributions are listed below and summarized in Table 6.

Table 6: Systematic uncertainties on the top quark mass measured in the semi-leptonic channel.

| Systematic uncertainty | $\chi^2$ minimization method | geometric method |
|---|---|---|
| Light jet energy scale | 0.2 GeV/% | 0.2 GeV/% |
| b jet energy scale | 0.7 GeV/% | 0.7 GeV/% |
| ISR/FSR | $\simeq$ 0.3 GeV | $\simeq$ 0.4 GeV |
| b quark fragmentation | $\leq$ 0.1 GeV | $\leq$ 0.1 GeV |
| Background | negligible | negligible |
| Method | 0.1 to 0.2 GeV | 0.1 to 0.2 GeV |

### 3.11.1 Jet energy scale (JES)

The effect of the uncertainty of the jet energy scale on the top quark mass measurement has been estimated by multiplying separately the light jet and $b$-jet momenta by several rescaling factors (20 factors, between $-10\%$ and $+10\%$). Neither the event selection nor the $\not{E}_T$ have been changed after this jet energy rescaling.

The resulting top quark mass depends linearly on the rescaling factor. The related systematic uncertainty on the top quark mass can therefore be expressed as a percentage of the light jet and $b$-jet energy scale miscalibration.

- The uncertainty in the $b$-jet energy scale produces an uncertainty in the top quark mass of 0.7 GeV/%. The $b$-jet scale will ultimately be determined with data from $Z+$jets. However, at the start of LHC running, the $Z+$jets statistics will be low, so the $b$-jet scale will be derived from the measured light jet scale togheter with a Monte Carlo correction term modelling the difference between the two jet energy scales. The systematic uncertainty associated with these methods has not been yet evaluated.

- The uncertainty in the light jet energy scale produces an uncertainty in the top quark mass of 0.2 GeV/%. The reduced dependence compared to that of the $b$-jet energy scale is due to the $W$ boson mass constraint used in the rescaling ($\chi^2$ minimization method or kinematical fit) or the definition of the top quark mass estimator (geometric with rescaling method). It has been shown that the light jet energy scale should be known with a precision of 1% in 1 $fb^{-1}$ of data [3]: the corresponding uncertainty on the top quark mass would therefore be 0.2 GeV.

### 3.11.2 Initial and final state radiation (ISR and FSR)

The study of the effect of initial and final state radiation on the top quark mass measurement is still preliminary. Several samples have been simulated for this study, corresponding to different sets of parameters:

Table 7: Top quark mass measured with several ISR/FSR parameters ($\chi^2$ minimization method, after cut C0 and jet calibration).

| Sample | Top quark mass [GeV] | $\sigma$ [(GeV] |
|---|---|---|
| 1 | $176.6 \pm 0.4$ | $13.4 \pm 0.6$ |
| 2 | $176.3 \pm 0.7$ | $13.7 \pm 0.8$ |
| 3 | $176.3 \pm 0.5$ | $12.7 \pm 0.6$ |

- Sample 1: AcerMC $t\bar{t}$ events, with maximum reconstructed top quark mass (half the default value of $\Lambda$(QCD) for FSR; twice the default value of $\Lambda$(QCD) for ISR).
- Sample 2: AcerMC $t\bar{t}$ events, with default ISR and FSR parameters.
- Sample 3: AcerMC $t\bar{t}$ events, with minimum reconstructed top quark mass (twice the default value of $\Lambda$(QCD) for FSR, half the default value of $\Lambda$(QCD) for ISR

For each sample, a specific jet calibration has been applied (separately for $b$-jets and light jets).

The measured top quark masses are summarised in Table 7. A shift on the top quark mass is observed. A first estimate of the systematic uncertainty due to initial and final state radiation can be estimated from the relative difference between these three top quark mass values: it is approximately equal to 0.3 GeV (more statistics would be necessary for a more precise estimate).

This preliminary study shows the impact on the top quark mass measurement of a given change of ISR/FSR parameters, but the estimate of the systematic uncertainty due to ISR or FSR will benefit from a measurement of these effects with ATLAS data. Figure 19 shows, for example, that the jet multiplicity could help to determine the size of the initial and final state radiation contribution. Initial state radiation could be measured with Drell-Yan events, as has been done at the Tevatron.

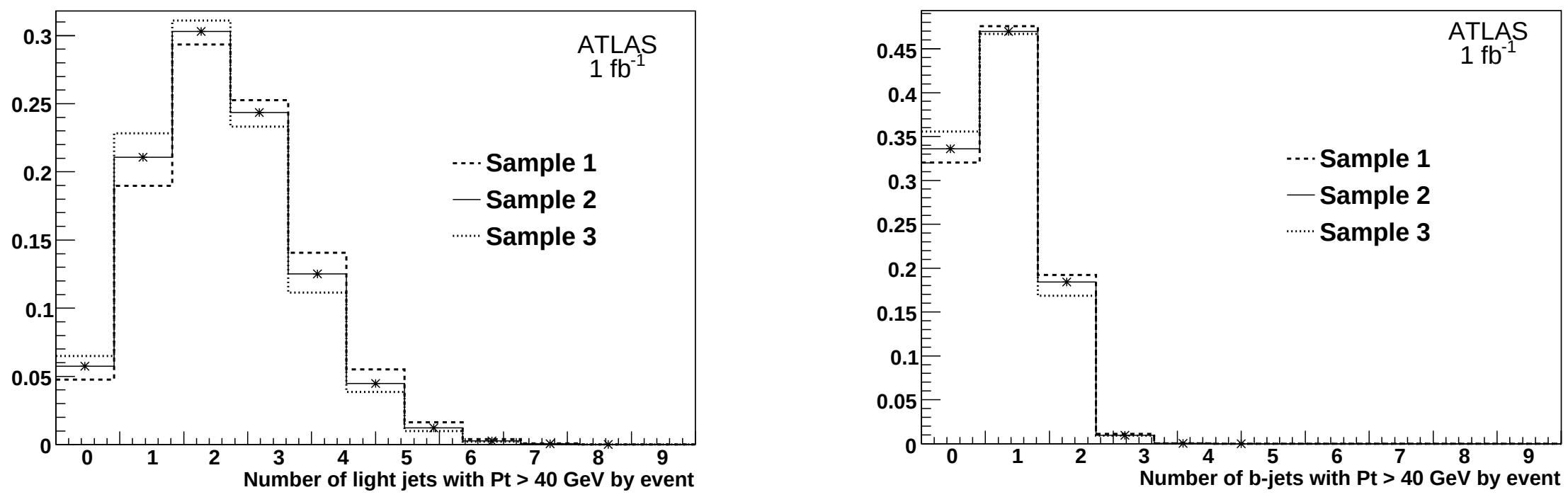


Figure 19: Light and $b$-jet multiplicity for several values of the ISR/FSR parameters ($p_T$(jet) $> 40$ GeV).

### 3.11.3 b-quark fragmentation

The effect of b-quark fragmentation has been estimated by varying the Peterson parameter within its uncertainty (study performed on fast simulation [14]).

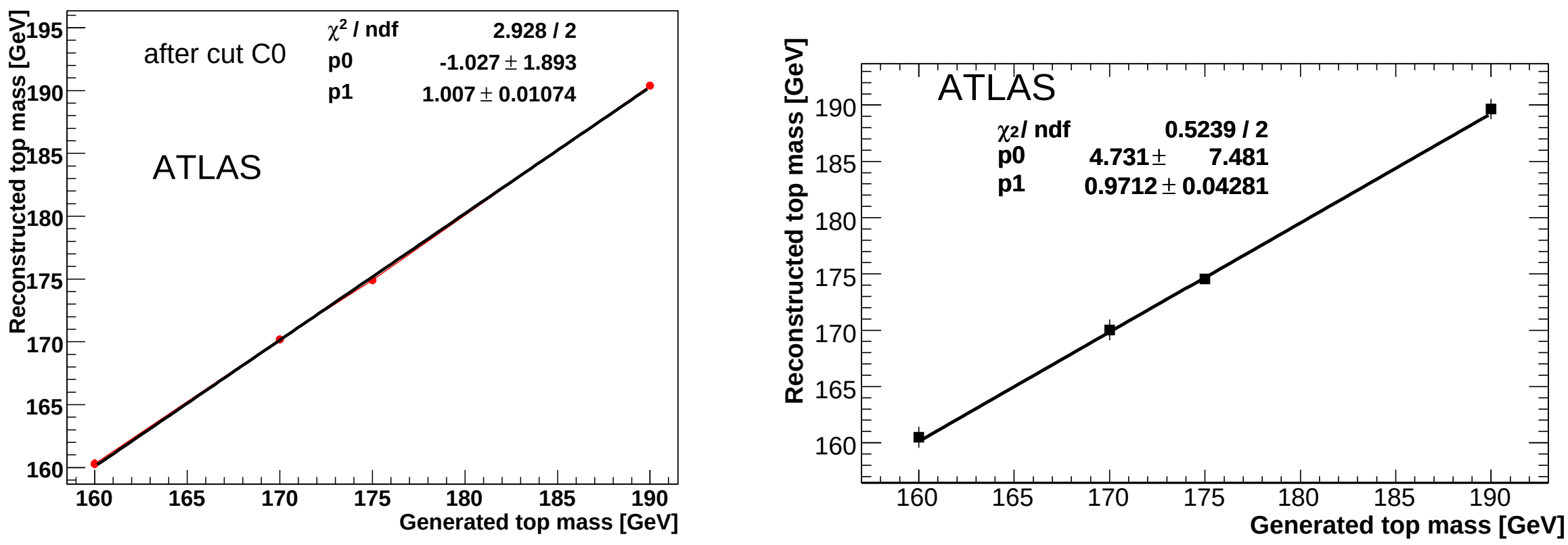


Figure 20: Reconstructed top quark mass with 2 $b$-tagged jets as a function of the generated top quark mass (left: $\chi^2$ minimization method, after C0 cut; right: geometric method after C1 cut). The method has good linearity.

The resulting uncertainty is lower than 0.1 GeV.

#### 3.11.4 Background estimate

Variations in the size of the background have no noticeable effect on the extracted top quark mass. Nevertheless, it is important to extract the shape of the background from data.

### 3.12 Linearity of the method

The analysis leading to the top quark mass measurement has been applied to samples corresponding to several values of generated top quark mass. Figures 20 and 21 show that both methods have good linearity. The reconstructed top quark mass lies above the generated one by an average offset equal to 0.2% of $m_{\text{top}}$. This comes from the $p_T$ jet spectrum which is different for each sample with a different generated top quark mass [3]. The jet energy scale is then a little bit different in each sample due to the $p_T$ cut applied on jets.

## 4 Top quark mass measurement in the semi-leptonic channel with relaxed requirements on the $b$-tagging

At the start of LHC running, the detector will not be optimized and will require a commissioning phase with first data to calibrate its sub-components. This is particularly true for the pixel detector and its capability to tag $b$-jets. Thus, it could be useful to have a top quark mass measurement analysis in which the use of $b$-tagging is reduced. This is addressed in this section by performing a first analysis in which exactly one $b$-jet is $b$-tagged (the assumed $b$-tagging efficiency is the same as before: $\varepsilon_b \simeq 60\%$). This sample has no events in common with the sample used in section 3. In a second analysis, $b$-tagging is not used in the reconstruction of the $t\bar{t}$ events. Even if not used, this sample contains events with 0, 1, or 2 $b$-tagged jets and overlap with the other two samples.

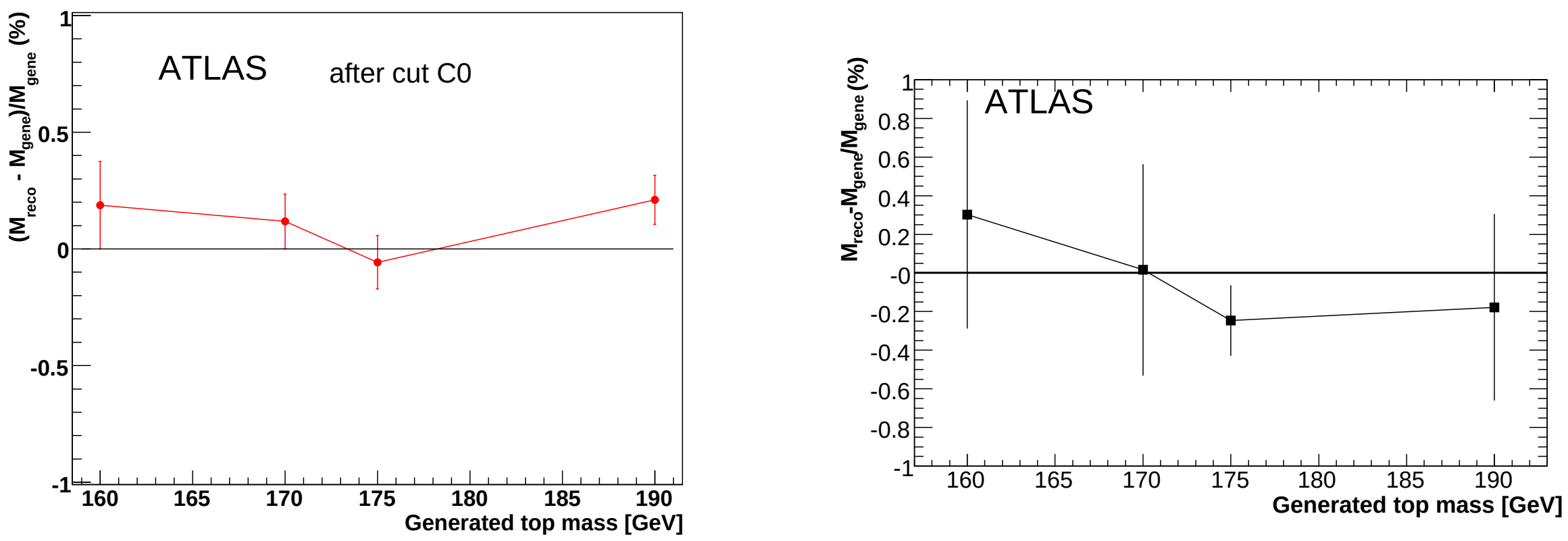


Figure 21: Difference between the reconstructed and generated top quark mass, as a function of the latter (left: $\chi^2$ minimization method, after C0 cut; right: geometric method after C1 cut).

### 4.1 Displaced vertex $b$-tagging, 1 $b$-tagged jet

The sample is here limited to events with no more than five jets. In this case, the best method to select the two light jets from the $W$ boson is the one which minimizes $M_{jj} - M_W^{peak}$. The $b$-jet associated with the hadronic top quark decay is then chosen among the remaining light jets and the tagged $b$-jet. The best choice is the one which minimizes $\sqrt{(X_1 - \mu_1)^2 + (X_2 - \mu_2)^2}$ (see equations (2) and (3)).

A large fraction of the non-hadronic $b$-jets is removed by requiring the $b$-jet to be closer to the reconstructed $W$ boson than the lepton originating from the leptonic top quark decay:

- cut **C6** : $\Delta R(\text{lepton}, b_{\text{had}}) - \Delta R(W, b_{\text{had}}) > 1$

As in previous sections, the purification cuts **C1** and **C3** are applied. Harder purification cuts (**C4** and **C5**) could also be applied. The top quark mass spectra derived after applying the two sets of purification cuts are shown in Fig. 22.

Table 8 summarizes the reconstruction efficiency and purity related to this reconstruction method for both sets of cuts.

Table 8: Single $b$-tagged sample: efficiency of cuts applied and final purities (within $\pm$ 3 $\sigma_{m_{\text{top}}}$), with respect to semi-leptonic events.

| Cuts | Efficiency (%) | $W$ boson purity (%) | b purity (%) | top purity (%) | Number of events |
|---|---|---|---|---|---|
| **C1+C3+C6** | 0.54±0.02 | 70 ±1 | 69±1 | 62±1 | 1063 |
| **C3+C4+C5+C6** | 0.52±0.02 | 71±1 | 70±1 | 63±1 | 1016 |

The systematic uncertainties on the top quark mass measurement, determined in the same way as in the previous section, are summarized in Table 9.

The top quark mass values obtained are shown in Table 10 together with the pull distribution results. The single-tag sample can be added to the double-tagged sample to perform a top quark mass measurement with 50% more signal events.

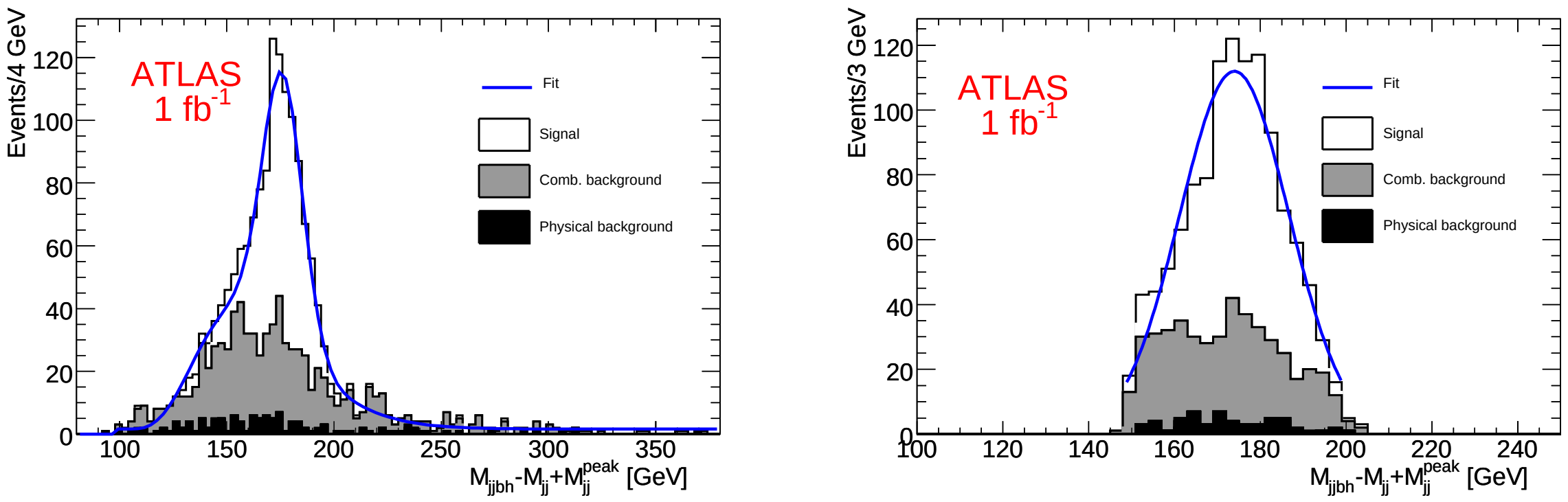


Figure 22: Events with only 1 $b$-tagged jet: $m_{\text{top}} = M_{\text{jjb}} - M_{\text{jj}} + M_W^{\text{peak}}$ with the geometric method and rescaling, after purification cuts (left, **C1**, **C3**, and **C6** cuts: $m_{\text{top}} = 176.0 \pm 0.7$ GeV, with a width equal to $9.4 \pm 0.8$ GeV($\chi^2$/dof = 58/52) ; right, **C3**, **C4**, **C5**, and **C6** cuts: $m_{\text{top}} = 174.0 \pm 0.4$ GeV, with a width equal to $12.7 \pm 0.4$ GeV) ($\chi^2$/dof = 27/13).

Table 9: Systematic uncertainties on the top quark mass with rescaling measured in the semi-leptonic channel, with 1 $b$-tagged jet and no $b$-tagging, using the geometric method.

| Systematic uncertainty | 1 $b$-tagged jet | No $b$-tagging |
|---|---|---|
| Light jet energy scale | 0.3 GeV/% | 0.4 GeV/% |
| b jet energy scale | 0.7 GeV/% | 0.7 GeV/% |
| ISR/FSR | $\simeq 0.4$ GeV | $\simeq 0.4$ GeV |
| b quark fragmentation | $\leq 0.1$ GeV | $\leq 0.1$ GeV |
| Background | $< 1$ GeV | 1 GeV |

Table 10: Fitted top quark mass value and corresponding widths for 1 $b$-tag and no $b$-tag samples, in 1 fb$^{-1}$. Pull bias and width are also given.

| Geometric method with rescaling | Cuts | Top quark mass [GeV] | $\sigma$ [GeV] | pull bias | pull width |
|---|---|---|---|---|---|
| 1 $b$-tagged jet | **C1**, **C3** and **C6** cuts | $176.0 \pm 0.7$ | $9.4 \pm 0.8$ | $-0.48$ | 1.17 |
| 1 $b$-tagged jet | **C3**, **C4**, **C5** and **C6** cuts | $174.0 \pm 0.4$ | $12.7 \pm 0.4$ | $-0.05$ | 1.07 |
| No $b$-tagged jet | **C1** and **C3** cuts | $175.0 \pm 0.4$ | $11.7 \pm 0.5$ | 0.40 | 1.12 |
| No $b$-tagged jet | **C3**, **C4** and **C5** cuts | $175.2 \pm 0.5$ | $12.4 \pm 0.8$ | $-0.15$ | 0.94 |

### 4.2 Top quark mass measurement in the semi-leptonic channel assuming no $b$-tagging

The same analysis is performed with no use of $b$-tagging information. The main difference with the two $b$-tagged analysis comes from the larger contribution of the combinatorial and physics background.

The corresponding top mass spectra are given in Fig. 23 for both sets of cuts. Table 11 summarizes the efficiency and purity related to this method.

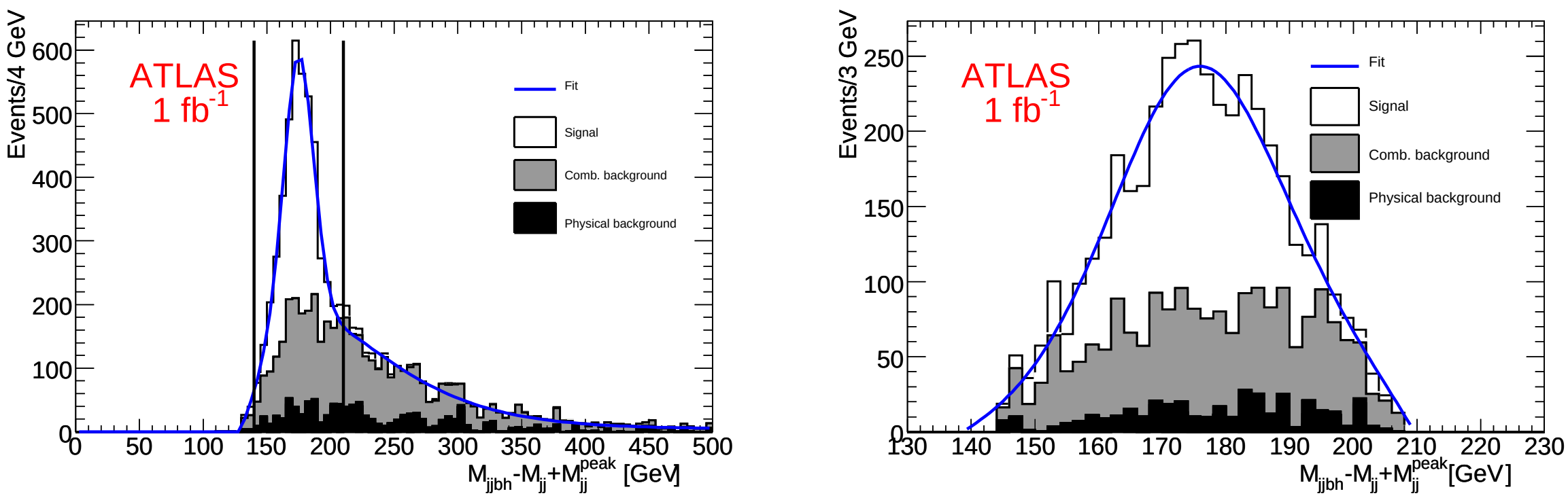


Figure 23: Events without $b$-tagging: $m_{\rm top} = M_{\rm jjb} - M_{\rm jj} + M_W^{\rm peak}$ with the geometric method, after purification cuts (left, **C1** and **C3** cuts: $m_{\rm top}$ = 175.0 ± 0.4 GeV, with a width equal to 11.7 ± 0.5 GeV($\chi^2$/dof = 130/69) ; right, **C3**, **C4** and **C5** cuts: $m_{\rm top}$ = 175.2 ± 0.5 GeV, with a width equal to 12.4 ± 0.8 GeV($\chi^2$/dof = 41/24)).

Table 11: Efficiency of cuts applied and final purities (within ± 3 $\sigma_{m_{\rm top}}$), with respect to semi-leptonic (e,$\mu$) events, assuming no $b$-tagging.

| Cuts | Efficiency (%) | $W$ boson purity (%) | b purity (%) | top purity (%) | Number of events |
|---|---|---|---|---|---|
| **C1+C3** | 1.59 ± 0.03 | 49.5 ± 0.9 | 46.8 ± 0.9 | 58.4 ± 0.9 | 3115 |
| **C3+C4+C5** | 1.48 ±0.03 | 51.1 ± 0.9 | 49.0 ± 0.9 | 58.9 ± 0.9 | 2916 |

The systematic uncertainties on the top quark mass measurement are summarized in Table 9. The obtained top quark mass values together with the pull distribution results are shown in Table 10.

For the second set of cuts (right plot of Fig. 23), the background contribution ($\simeq$ 45 % of the sample) is peaked exactly below the signal contribution, preventing a fit using a parametrization like gaussian function (signal) + polynomial function (background). An event mixing technique has been used to fix the background shape in the parametrization. The event mixing technique consists in replacing one of the two jets momentum associated to the W with a simulated jet whose energy, $\phi$ and $\eta$ distribution is randomly selected according to the global energy, $\phi$ and $\eta$ distributions observed for the jets associated to the W. This technique looks very promising as shown in Fig. 24 comparing data and event mixing samples for the two jet and three jet invariant masses after preselection.

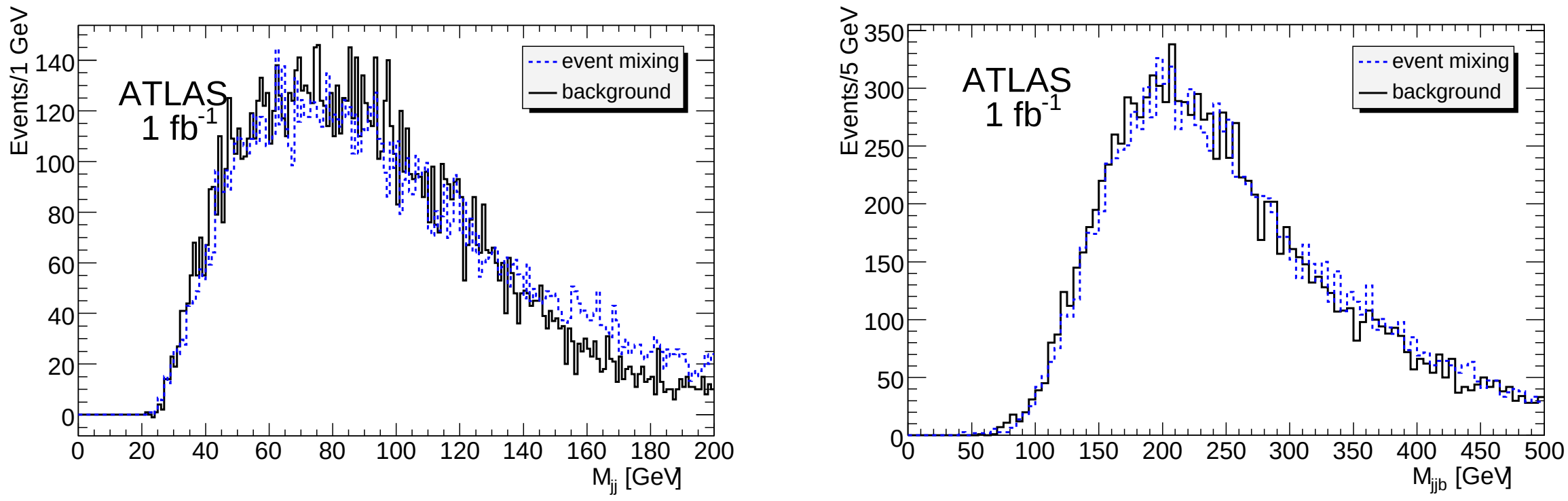


Figure 24: Events without $b$-tagging: comparison between background and event mixing samples for two jets (left) and three jets (right) invariant mass.

# 5 Conclusion

Several methods have been investigated in order to perform an accurate top quark mass measurement with 1 fb$^{-1}$ of collected data, in the $t\bar{t}$ semi-leptonic channel. The best top quark mass determination is achieved with two $b$-tagged events and a top mass estimator taken as the invariant mass of the three jets from hadronically-decaying top quark; the uncertainty on the top quark mass measurement with this analysis will be dominated by systematics, the statistical uncertainty being already small ($\leq$0.4 GeV). The precision on the top quark mass relies mainly on the jet energy scale uncertainty: a precision of the order of 1 to 3.5 GeV should be achievable with 1 fb$^{-1}$, assuming a jet energy scale uncertainty of 1 to 5%. $W$ boson sample can be extracted from the $t\bar{t}$ sample in order to constrain the light jet energy scale. The main uncertainty on the top quark mass measurement will come from the b-jet energy scale.

Events with one or no $b$-tagged jets lead also to an interesting measurement if the background (physical and combinatorial) shape is constrained from data. The estimated precision on the top quark mass value is below 2 GeV (assuming a jet energy scale uncertainty of the order of the percent), with a very good signal over background ratio. These samples are thus very useful for jet energy scale or $b$-tagging studies during the commissioning phase with early data.

# Top Quark Properties

**Abstract**
The ATLAS potential for the study of top quark properties and physics beyond the Standard Model in the top sector, is reviewed in this paper. Measurements of the top quark charge, the spin and spin correlations, the Standard Model decay ($\mathrm{t} \rightarrow \mathrm{bW}$), rare top quark decays associated to flavour changing neutral currents ($\mathrm{t} \rightarrow \mathrm{qX}, \mathrm{X} = \gamma, \mathrm{Z}, \mathrm{g}$) and $\mathrm{t\bar{t}}$ resonances are discussed. The expected sensitivity of the ATLAS experiment is estimated for an integrated luminosity of 1 $\mathrm{fb}^{-1}$ at the LHC. For the Standard Model measurements the expected precision is presented. For the tests of physics beyond the Standard Model , the $5\sigma$ discovery potential (in the presence of a signal) and the 95% confidence level limit (in the absence of a signal) are given.

## 1 Introduction

Several properties of the top quark have already been explored by the Tevatron experiments, such as the mass, charge and lifetime, the rare decays through flavour changing neutral currents (FCNC) and the production cross-sections. The structure of the Wtb vertex and the main top quark decay mode ($\mathrm{t} \rightarrow \mathrm{bW}$) within the Standard Model were also investigated together with the measurements of the W-boson helicity fractions. Many of these studies were performed by reconstructing $\mathrm{t\bar{t}}$ pairs in the semileptonic, dileptonic and fully hadronic decay modes. Given the current Tevatron luminosity, most of these studies are limited by the statistics acquired.

The electric charge of the top quark is one of its fundamental properties and will be probed with high statistics at the LHC. The measurement of the top quark charge can be performed either by identifying the charge of its decay products in the main decay channel $\mathrm{t} \rightarrow \mathrm{bW}$ or by studying radiative top quark processes. At the Tevatron the D0 [1] and CDF [2] collaborations have already initiated the study of the top quark charge and, with the available statistics, they showed that the data gives preference to the Standard Model top quark hypothesis (with a charge of +2/3) over the scenario with an exotic quark (XM) of charge −4/3 and mass ≈ 170 GeV, fully consistent with the present precision electroweak data [3,4]. The D0 and CDF exclude the exotic quark hypothesis with 92 and 87 % confidence, respectively.

As the top quark decays before it can form hadronic bound states, a consequence of its high mass, the spin information of the top quark is propagated to its decay products. This unique behaviour among quarks allows direct top quark spin studies, as spin properties are not washed out by hadronization. Through the measurement of the angular distributions of the decay products the information of the top quark spin can be reconstructed. Top quark spin polarization and correlations in $\mathrm{t\bar{t}}$ events produced at the LHC are precisely predicted by the Standard Model and are sensitive to the fundamental interactions involved in the top quark production and decay. By testing only the top quark decay, the W-boson polarization measurement complements top quark spin studies, helping to disentangle the origin of new physics, if observed. The W-boson polarization states can be measured through the longitudinal ($F_0$), left-handed ($F_\mathrm{L}$) and right-handed ($F_\mathrm{R}$) helicity fractions. At the present, the most stringent limits on the W-boson helicity fractions were obtained at the Run-II of the Tevatron [5–13]. Analysing 1.9 $\mathrm{fb}^{-1}$ of data, the CDF experiment measured [5] $F_0 = 0.62 \pm 0.11$ with $F_\mathrm{R}$ fixed to zero and $F_\mathrm{R} = -0.04 \pm 0.05$ with $F_0$ fixed to the Standard Model expectation for $m_\mathrm{t} = 175$ GeV. For 2.7 $\mathrm{fb}^{-1}$, the D0 experiment measured [10] $F_0 = 0.490 \pm 0.106(\mathrm{stat}) \pm 0.085(\mathrm{syst})$ with $F_\mathrm{R}$ fixed to zero, and $F_\mathrm{R} = 0.110 \pm 0.059(\mathrm{stat}) \pm 0.052(\mathrm{syst})$ with $F_0$ fixed to the Standard Model value.

Within the Standard Model, the Wtb coupling is purely left-handed (at the tree level), and its size is given by the Cabibbo-Kobayashi-Maskawa (CKM) matrix element $V_\mathrm{tb}$. In Standard Model extensions,

Table 1: The values of the branching ratios of the FCNC top quark decays, predicted by the SM, the quark-singlet model (QS), the two-higgs doublet model (2HDM), the minimal supersymmetric model (MSSM) and SUSY with R-parity violation are shown [30–36].

| Process | SM | QS | 2HDM | MSSM | $\not R$ SUSY |
|---|---|---|---|---|---|
| $t \to uZ$ | $8 \times 10^{-17}$ | $1.1 \times 10^{-4}$ | – | $2 \times 10^{-6}$ | $3 \times 10^{-5}$ |
| $t \to u\gamma$ | $3.7 \times 10^{-16}$ | $7.5 \times 10^{-9}$ | – | $2 \times 10^{-6}$ | $1 \times 10^{-6}$ |
| $t \to ug$ | $3.7 \times 10^{-14}$ | $1.5 \times 10^{-7}$ | – | $8 \times 10^{-5}$ | $2 \times 10^{-4}$ |
| $t \to cZ$ | $1 \times 10^{-14}$ | $1.1 \times 10^{-4}$ | $\sim 10^{-7}$ | $2 \times 10^{-6}$ | $3 \times 10^{-5}$ |
| $t \to c\gamma$ | $4.6 \times 10^{-14}$ | $7.5 \times 10^{-9}$ | $\sim 10^{-6}$ | $2 \times 10^{-6}$ | $1 \times 10^{-6}$ |
| $t \to cg$ | $4.6 \times 10^{-12}$ | $1.5 \times 10^{-7}$ | $\sim 10^{-4}$ | $8 \times 10^{-5}$ | $2 \times 10^{-4}$ |

departures from the Standard Model expectation $V_{\rm tb} \simeq 0.999$ [1] are possible [14, 15], as well as new radiative contributions to the Wtb vertex [16, 17]. These deviations might be observed in top quark production and decay processes at LHC. The most general Wtb vertex for on mass shell W-boson, top quark and b quark, containing terms up to dimension five can be written as

$$\mathscr{L} \;\; = \;\; -\frac{g}{\sqrt{2}}\bar{b}\,\gamma^{\mu}\,(V_{\rm L}P_L + V_{\rm R}P_R)\,t\,W_{\mu}^{-} - \frac{g}{\sqrt{2}}\bar{b}\,\frac{i\sigma^{\mu\nu}q_{\nu}}{M_W}\,(g_{\rm L}P_L + g_{\rm R}P_R)\,t\,W_{\mu}^{-} + \text{h.c.}\,, \qquad (1)$$

with $q = p_{\rm t} - p_{\rm b}$ the W-boson momentum and $P_{\rm R(L)}$ the chirality projectors. Additional $\sigma^{\mu\nu}k_{\nu}$ and $k^{\mu}$ terms, where $k = p_{\rm t} + p_{\rm b}$, can be absorbed into this Lagrangian using Gordon identities. If the W-boson is on its mass shell or it couples to massless fermions $q^{\mu}\varepsilon_{\mu} = 0$, and terms proportional to $q^{\mu}$ can be dropped from the effective vertex. The new constants $V_{\rm R}$, $g_{\rm L}$ and $g_{\rm R}$ [18, 19], are vector like ($V_{\rm R}$) and tensor like ($g_{\rm L}$ and $g_{\rm R}$) anomalous couplings, can be related to $f_1^{\rm R}$, $f_2^{\rm L}$ and $f_2^{\rm R}$ in Ref. [20] (and references therein) as $f_1^{\rm R} = V_{\rm R}$, $f_2^{\rm L} = -g_{\rm L}$ and $f_2^{\rm R} = -g_{\rm R}$. If we assume CP is conserved, these couplings can be taken to be real. Within the Standard Model $V_{\rm L} \equiv V_{\rm tb} \simeq 0.999$ and the other couplings ($V_{\rm R}, g_{\rm L}, g_{\rm R}$) vanish at the tree level, while nonzero values are generated at higher orders [21, 22]. Indirect limits on the Wtb vertex anomalous couplings can be inferred from radiative B-meson decays and $B\bar{B}$ mixing [23]. Taking into account the current world average [23], $\text{BR}(\bar{B} \to X_s\gamma) = (3.55 \pm 0.24^{+0.09}_{-0.10} \pm 0.03) \times 10^{-4}$, varying one parameter at a time, the 95% C.L. bounds on $V_{\rm R}$, $g_{\rm L}$ and $g_{\rm R}$ are in the range $[-0.0007, 0.0025]$, $[-0.0015, 0.0004]$ and $[-0.15, 0.57]$, respectively [22, 24].

Flavour Changing Neutral Currents are strongly suppressed in the Standard Model due to the Glashow-Iliopoulos-Maiani (GIM) mechanism [25]. Although absent at tree level, small FCNC contributions are expected at one-loop level, determined by the CKM mixing matrix [26–29]. For the top quark within the framework of the Standard Model, these contributions limit the FCNC decay branching ratios to the gauge bosons, $\text{BR}(t \to qX, X = Z, \gamma, g)$, to below $10^{-12}$. There are however extensions of the SM, like supersymmetry (SUSY) [30], multi-Higgs doublet models [31] and models with exotic (vector-like) quarks [32–34], which predict the presence of FCNC contributions already at tree level and significantly enhance the FCNC decay branching ratios compared to the Standard Model predictions [35, 36]. The branching ratio for the different models are shown in Table 1. FCNC processes associated with the production and decay of top quarks have been studied at colliders and the observed upper limits on the branching ratios at 95% C.L., from the direct searches, are shown in Table 2.

Since the top quark mass is much larger than the other quarks, the top quark may play a privileged role in the electroweak symmetry breaking (EWSB) mechanism. Any new physics connected to the EWSB could be preferentially coupled to the top quark sector. This would lead to deviations from the

[1] Three generations of quarks and unitarity of the CKM matrix are assumed.

Table 2: Experimental observed upper limits on the branching ratios at 95%C.L. for the FCNC top quark decays.

| | LEP | HERA | Tevatron |
|---|---|---|---|
| $BR(t \to qZ)$ | 7.8% [37–41] | 49% [42] | 3.7% [43] |
| $BR(t \to q\gamma)$ | 2.4% [37–41] | 0.75% [42] | 3.2% [44] |
| $BR(t \to qg)$ | 17% [45] | 13% [42,46,47] | 0.1 – 1 % (estimated from [46,48]) |

expected Standard Model $t\bar{t}$ production rate and could distort the top quark kinematics. New resonances and gauge bosons strongly coupled to the top quark are expected in a large variety of models, in particular those with strong EWSB [49–51]. The $t\bar{t}$ final states are also interesting for leptophobic $Z'$ bosons which can appear in Grand Unification Models [52]. These new particles could reveal themselves in the $t\bar{t}$ invariant mass distribution. At the Tevatron experimental upper limits were set at 95 %C.L. for the $\sigma(p\bar{p} \to Z') \times BR(Z' \to t\bar{t})$ with $Z'$ masses between 450 GeV and 900 GeV. A topcolor leptophobic $Z'$ is ruled out below 720 GeV and the cross section of any narrow $Z'$ decaying to a $t\bar{t}$ is less than 0.64 pb at 95%C.L., for $Z'$ masses above 700 GeV [53].

In this note the ATLAS potential for the study of the top quark properties and tests of physics beyond the Standard Model in the top quark sector are reviewed for an expected luminosity of 1 $fb^{-1}$ at the LHC. The note is organized as follows: the basic event selection and the trigger used are reminded in Section 2. In Sections 3, 4, 5 and 6 the studies of the top quark charge, the W-boson and top quark polarisation studies and the Wtb anomalous couplings, the top quark FCNC decays and the production of $t\bar{t}$ resonances are discussed respectively. The summary and conclusions are presented in Section 7. The top quark mass, one of the most important top quark properties, is not investigated here as a separate note is devoted to this issue [54].

# 2 Basic event selection

As most of the studies performed in this paper are related to either the semileptonic ($t\bar{t} \to WWb\bar{b} \to l\nu j_1 j_2 b\bar{b}$ with $l = e, \mu$) or the dileptonic ($t\bar{t} \to WWb\bar{b} \to l\nu l' \nu' b\bar{b}$ with $l, l' = e, \mu$) decays of $t\bar{t}$ events, basic criteria for the event selection were defined for each one of these final state topologies. Changes to the criteria are to be expected depending on the type of top quark property under study. For the background studies several sources were considered, $t\bar{t}$, W+jets, $Wb\bar{b}$+jets, $Wc\bar{c}$+jets, Z+jets, WW, ZZ, WZ and single top events (see the introduction of the Top Chapter for a full list of backgrounds). The backgrounds are also described in more detail separatly for each section of the note. All signal and background events were required to pass the single lepton trigger requirements for electrons and muons. The triggers considered were L1_EM18I (L1_MU20,L1_MU40) for electrons (muons) at L1, e22i (mu20) for electrons (muons) at L2 and e22i (mu20) for electrons (muons) at the Event Filter (EF) [55].

Different selection criteria are applied for the top quark FCNC analyses, as the final state topology is different from those considered above. These criteria are explained in Section 5.

*Semileptonic topology*

In the semileptonic topology, signal events have a final state with one isolated lepton (electron or muon), at least four jets (two of them from the hadronization of b quarks and labelled b-jets) and large transverse missing energy from the undetected neutrino. More information on the signal can be found in the introduction of the Top Chapter. The basic selection criteria were defined by requiring that the events should have:

Table 3: Cumulative efficiencies of the standard top quark selection criteria for the semileptonic and dileptonic type of events with electrons and muons for the pseudorapidity range $|\eta| \leq 2$.

| criterion | $\varepsilon(\%)$ | criterion | $\varepsilon(\%)$ |
|---|---|---|---|
| Semileptonic events | 100 | Dileptonic events | 100 |
| 1 isol.lept. ($p_T$ >25/20 GeV) | 58.9 | 2 isol.lept. ($p_T$ >25/20 GeV) | 35.5 |
| $\geq$ 4 jets ($p_T$ >30 GeV) | 34.2 | $\geq$ 2 jets ($p_T$ >30 GeV) | 31.8 |
| $\geq$ 2 b-tagged | 10.5 | = 2 b-tagged | 8.3 |
| missing $E_T > 20$ GeV | 8.5 | missing $E_T > 30$ GeV | 6.5 |

- exactly one isolated electron (muon) with $|\eta| < 2.5$ and $p_T > 25$ GeV ($p_T > 20$ GeV);
- at least 4 jets with $|\eta| < 2.5$ and $p_T > 30$ GeV;
- at least 2 jets tagged as b-jets;
- missing transverse energy above 20 GeV.

*Dileptonic topology*

In the dileptonic topology, signal events have a final state with two isolated leptons (electrons and/or muons), at least two jets (tagged as b-jets) and large transverse missing energy from the two undetected neutrinos. The basic selection criteria were defined by requiring that the events should have:

- exactly two isolated electrons (muons) with $|\eta| < 2.5$ and $p_T > 25$ GeV ($p_T > 20$ GeV);
- at least 2 jets with $|\eta| < 2.5$ and $p_T > 30$ GeV;
- 2 jets tagged as b-jets;
- missing transverse energy above 30 GeV.

*Selection efficiency*

The efficiency of the basic selection criteria was examined using $\approx$ 450 000 events from the $t\bar{t} \to bWbW \to bqqb\ell\nu, b\ell\nu b\ell\nu$ signal sample. The results are presented in Table 3.

The effect of the trigger efficiency is investigated separately for the individual analysis.

## 3 Top quark charge reconstruction

There are several techniques to determine the electric charge of the top quark at hadron collider experiments [56–58]. The top quark charge measurement presented here is based on the reconstruction of the charges of the top quark decay products. As the dominant decay channel of the top quark is $t \to W^+b(\bar{t} \to W^-\bar{b})$, the top quark charge determination requires the measurement of both the W boson and the b quark charges. While the charge of the W boson can be determined through its leptonic decay, the b quark charge is not directly measurable due to quark confinement in hadrons. In this note two possible ways to determine the b quark charge were investigated:

- The charge weighting technique: this approach is based on finding a correlation between the b quark charge and the charges of the tracks belonging to the b-jet [59, 60].
- The semileptonic b-decay approach: in this case the b quark charge is determined using the semileptonic b-decays ($\mathrm{b} \to \mathrm{c}, \mathrm{u} + \mathrm{W}^-, \mathrm{W}^- \to \ell^- + \bar{\nu}_\ell$), where the sign of the soft lepton indicates the sign of the b quark charge.

Two major issues have to be addressed. The first one is to find the selection criteria to perform the correct pairing of the lepton and the b-jet originated in the same top quark decay. In the Standard Model a b-jet, coming from a b quark, should be associated with a positive lepton ($\ell^+$), while in the exotic case it should be associated with a negative one ($\ell^-$). The second issue is the assignment of a charge to the b-jet selected by the pairing criterion. While the former issue is common for both the approaches, the latter one is tackled in different ways.

## 3.1 Event generation and selection

The standard $\mathrm{t\bar{t}} \longrightarrow \mathrm{W^+bW^-\bar{b}}$ samples were used as signal events in both the semileptonic and the dileptonic channels (only electrons and muons are taken as signal). For the background studies, the $W$+jets sample was used. The $\mathrm{t\bar{t}}$ all jets channels as well as the semileptonic and dileptonic channels of $\tau$ leptons were analysed as they can contribute to background (all jets channel) and to signal (events with the leptonic decays of $\tau$ leptons). In the present analysis the common selection criteria were used, as defined in Section 2. In addition, for each approach, specific criteria were applied to the events.

## 3.2 The lepton and b-jet pairing algorithm

The lepton and b-jet pairing was done using the invariant mass distribution of the lepton and the b-tagged jet, $m(\mathrm{l}, \mathrm{b_{jet}})$. If the assignment is correct, $m(\mathrm{l}, \mathrm{b_{jet}})$ is limited by the top quark mass, otherwise there is no such restriction, as can been seen in Figure 1, where the signal sample with the standard cuts applied was analysed. To find the connection between the b-quarks and reconstructed b-jets and the parton level leptons and reconstructed leptons, the MC truth was used: the matching was treated as successful if the cone difference, $\Delta R$ between b-quark and b-jet was less than 0.4 (in the lepton case $\Delta R < 0.2$). For double b-tagged events, only the b-jets that satisfy

$$m(\mathrm{l}, \mathrm{b}_{\mathrm{jet}}^{(1,2)}) < m_{\mathrm{cr}} \quad \text{and} \quad m(\mathrm{l}, \mathrm{b}_{\mathrm{jet}}^{(2,1)}) > m_{\mathrm{cr}} \tag{2}$$

were accepted. In the di-lepton case both leptons should fulfill the condition (2). The optimal value for the pairing mass cut, $m_{\mathrm{cr}} = 155$ GeV, is a trade-off between the efficiency ($\varepsilon$) and purity ($P$) of the pairing method. The factor $\varepsilon(2P-1)^2$ was maximised to find the optimum working point. As this criterion requires events with two b-tagged jets and one combination for the lepton and b-jet invariant mass must be below $m_{\mathrm{cr}}$ and the other one above $m_{\mathrm{cr}}$, the efficiency of the method is small. On the other hand, this criterion gives a high purity sample as is shown in Section 3.5.1. In the analysis two variants of b-tagged events treatment were considered. In the first one exactly two b-jets were required while in the second one two and more b-jets were allowed (the two with the highest $p_{\mathrm{T}}$ treated as true b-jets). Slightly better results were obtained for the former variant and the results presented here correspond to this case. To suppress the background some additional cuts were tried: W boson mass ($M_{\mathrm{W}}$) window, top quark mass ($m_{\mathrm{top}}$) window, etc. By using the combined W boson and top quark mass window the background can be reduced by factor more than 10 at the expense of a factor 2 loss in signal. The $M_{\mathrm{W}}$ window requires that at least one pair of non b-tagged jets should have an invariant mass within 10 GeV of the W boson mass. The $m_{\mathrm{top}}$ window requires that the reconstructed W boson can be combined with a b-jet (not previously paired with a high-$p_{\mathrm{T}}$ lepton) to give an invariant mass within 40 GeV of the top

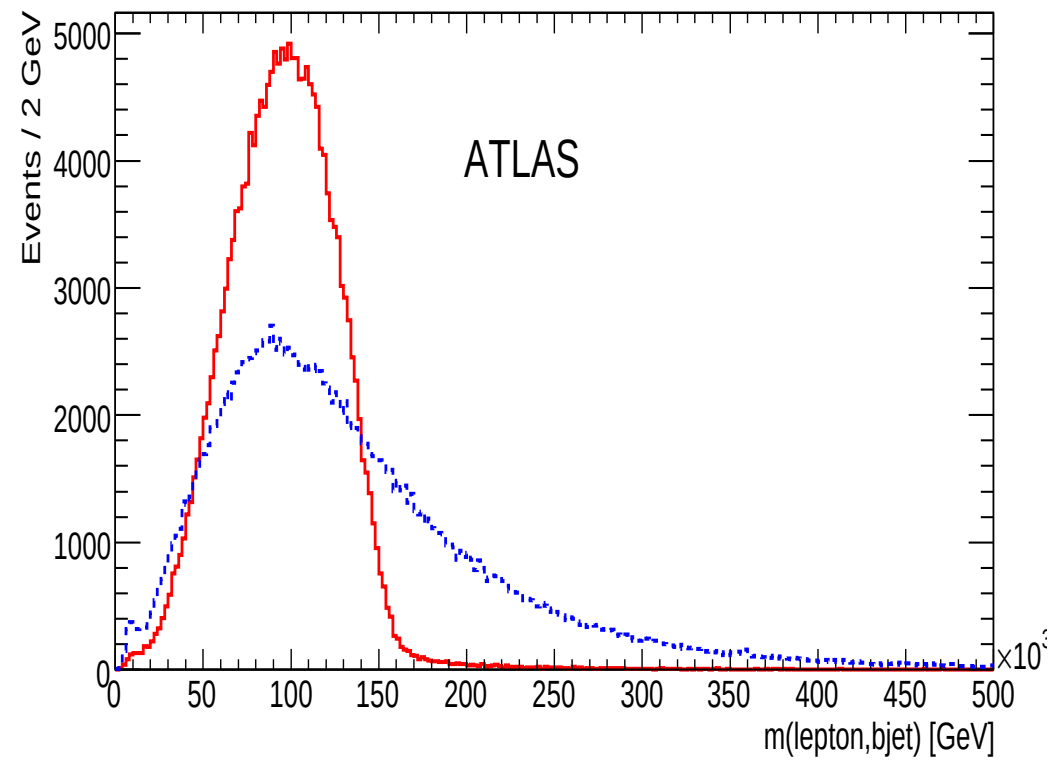


Figure 1: Lepton - b-jet invariant mass spectra for the lepton and b-jet pairs from the same top quark (full line) and from different top quarks (dashed line).

quark mass. This extra mass cut was applied by default in the whole top quark charge analysis using the weighting approach.

### 3.3 The jet track charge weighting approach

The determination of the average b-jet charge was done using a weighting technique in which the b-jet charge is evaluated as the weighted sum of the b-jet track charges:

$$Q_{\rm bjet} = \frac{\sum_i q_i |\vec{j}_i \cdot \vec{p}_i|^\kappa}{\sum_i |\vec{j}_i \cdot \vec{p}_i|^\kappa} \tag{3}$$

where $q_i(p_i)$ is the charge (momentum) of the $i^{\rm th}$ track inside the jet and $\vec{j}$ is the b-jet axis unit vector. The $\kappa$ parameter was optimised for the best separation between b- and $\bar{\rm b}$-jets and the optimum value was found to be $\kappa = 0.5$. In addition, for the charge weighting technique, it was further required, using only tracks with $p_{\rm T} > 1.5$ GeV, that at least two tracks must be found within a cone with $\Delta R < 0.4$ with respect to the jet axis. For b-jets with more than seven such tracks, only the seven with the highest $p_{\rm T}$ are used. The parameters of the weighting procedure are the result of a maximisation of the difference between the mean values of the b- and $\bar{\rm b}$-jet charge distributions - these mean values were found for a set of the parameters values ($\kappa$ and track $p_{\rm T}$) and compared. For the procedure optimisation the signal $t\bar{t}$-sample was used.

### 3.4 Semileptonic b-decay approach

In this approach the pairing procedure described in Section 3.2 is also used. But in this case the b quark charge is determined through its semileptonic decay. The sign of the b-jet charge is determined by the lepton charge within the b-jet,

$$\mathrm{b} \to \mathrm{c},\mathrm{u} + \ell^- + \bar{\nu}\ , \qquad \bar{\mathrm{b}} \to \bar{\mathrm{c}},\bar{\mathrm{u}} + \ell^+ + \nu .$$

The lepton from the b-decay will be identified as a non-isolated lepton inside the corresponding b-jet, and its charge ($Q_{\rm nonIs}$) defines the b quark charge. The non-isolated lepton is searched for among the tracks pointing to the treated b-jet and originating in the corresponding secondary vertex. Several processes can lead to an incorrect b quark charge assignment with this approach. Semileptonic decays of D mesons produced in the B decay chain, and the $\mathrm{B}^0$-$\bar{\mathrm{B}}^0$ mixing are examples of such processes. To

suppress the contribution from D mesons, the non-isolated lepton transverse momentum with respect to the b-jet axis, $p_T^{rel}$, can be used. The fact that the lepton $p_T^{rel}$ from b-decays is, on average, higher than from D meson decays can be used to diminish this contamination. The $p_T^{rel}$ cut was optimised using a sample of $\approx$ 555000 signal $t\bar{t}$ events and the value of 1 GeV has been found as the optimum cut. An additional source of wrong b quark charge assignments is mistagging, i.e. light jets incorrectly tagged as b-jets.

The main drawback of this approach is that, from all the selected lepton plus b-jet pairs, only those with a b-jet containing a non-isolated lepton can be used in the analysis. In addition to that, due to difficulties in selecting a pure sample of electrons from within jets, only muons were taken as the non-isolated leptons.

## 3.5 Results

In both approaches, the Standard Model scenario of top quark production and decay is assumed. The corresponding results are discussed in the following sections.

### 3.5.1 Weighting technique approach results

As a first step, the efficiency and purity of the lepton b-jet pairing was investigated. Using the events which passed the selection criteria, the obtained pairing efficiency is $\varepsilon = 30.5\%$ and the pairing purity is $P = 85.6\%$. The purity of pairing is defined as $P = N_{\rm good}/N_{\rm all}$, where $N_{\rm good}$ ($N_{\rm all}$) is the number of correctly paired lepton – b-jet pairs (all treated pairs) and the Monte Carlo truth is used to find $N_{\rm good}$.

The b-jet charge spectra reconstructed using the Monte Carlo truth and invariant mass pairing procedure for the signal $t\bar{t}$ events are presented in Figure 2, left and right respectively. From Figure 2, the shift of the b-jet charges associated with $\ell^+$, $Q_{\rm bjet}^{(+)}$, and $\ell^-$, $Q_{\rm bjet}^{(-)}$, (or with b and $\bar{b}$ quark in the Monte Carlo case) is clearly seen. The obtained b-jet charge purity, defined as the percentage of b-jets with the correct charge ($Q_{\rm bjet}^{(+)} < 0$ and $Q_{\rm bjet}^{(-)} > 0$), is $P \approx 62\%$. In addition to that, the $Q_{\rm comb}$ b-jet charge spectrum, defined as $Q(\ell) \times Q_{\rm bjet}^{(\ell)}$, has been reconstructed. The influence of trigger was also investigated, namely the lepton level 1 and level 2 triggers as well as the event filter (EF). The results with and without trigger are summarised in Table 4. No significant impact of the trigger is observed. A small asymmetry in favour of the positive b-jet charge, as was revealed by the analysis, is due to the dominance of positive charge in the initial state (two colliding protons).

Note that the peaks at $\pm 1$ in Fig. 2 correspond to the cases when all the tracks pointing to a b-jet have the same charge sign - in this case the weighting procedure (2) gives $Q_{\rm bjet} = \pm 1$.

Table 4: The mean b-jet charge associated with positive ($Q_+$) and negative ($Q_-$) lepton and combined b-jet charge ($Q_{\rm comb}$) without (no) and with (yes) EF trigger; two b-tags required.

| trigger | $Q_+$ | $Q_-$ | $\mid Q_{\rm comb} \mid$ | Nevent | efficiency |
|---|---|---|---|---|---|
| no | -0.092 ± 0.006 | 0.103 ± 0.006 | 0.097 ± 0.004 | 7129 | 100.0 |
| yes | -0.095 ± 0.006 | 0.106 ± 0.006 | 0.101 ± 0.004 | 6130 | 86.0 |

The main background processes for the top quark charge measurement in the semileptonic mode are: $W+$ jets production (the most important background in this mode), QCD multi-jets, di-boson and single top quark production. The single top production is not a genuine background as it gives the same sign of the b-jet charge asymmetry as the signal. For the selection criteria that were used, the ratio of the accepted semileptonic $t\bar{t}$ events to the accepted single lepton ones is more than 15:1. In the dileptonic

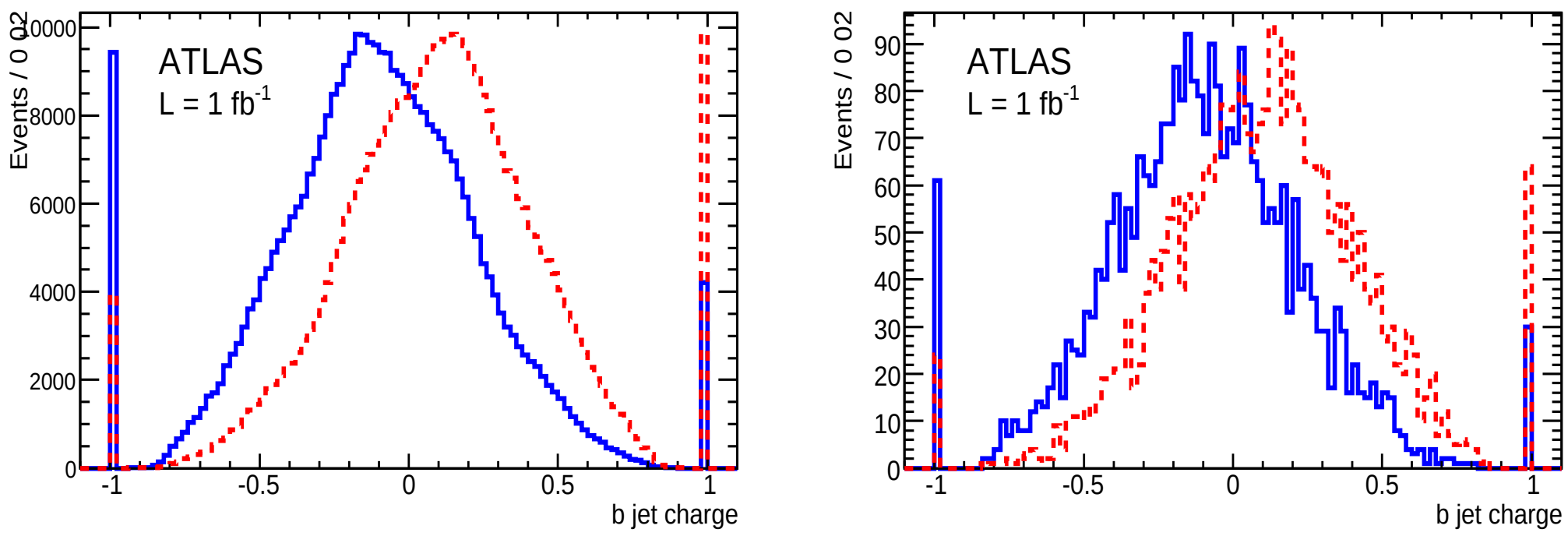


Figure 2: The b-jet charge associated with positive (full line) and negative (dashed line) lepton using the Monte Carlo truth (left) and invariant mass criteria (right) for the $\ell b$-pairing.

mode, the background is composed of Drell-Yan pairs (the most significant background), multi-jet QCD processes and di-boson production [2].

The background studies for this analysis require large Monte Carlo samples (in addition to the standard cuts the invariant mass criterion is highly restrictive) not available at present. The ideal way to determine the basic background parameters, the S:B ratio and background charge asymmetry, is to use the W+jets (dominant background) samples. However, after applying the selection criteria to the available W+jets samples, only a few events remained (20 lepton b-jet pairs). It is clear that due to poor statistics the samples are not suitable for a valuable background analysis. Nonetheless, combining the b-jet charge spectra, obtained for the individual W+jets channels (W+n$\times$jets and W$b\bar{b}$, W$c\bar{c}$ + n $\times$jets) scaled according to their cross sections to 1 fb, a S:B ratio of $\approx$ 38 $\pm$ 8 was obtained. To fix the S:B ratio we need to include other backgrounds and take a regard for the poor statistics. Taking into account only the standard cuts with a loose $M_W$ window ($\pm$30 GeV) a value of 7:1 was obtained for the S:B ratio which is compatible with that of the CDF background studies [2]. As a result, a nominal S:B ratio of 30:1 has been assumed, with 7:1 as a very conservative lower limit for studying systematic uncertainties related to the background.

The poor statistics of the available W+jets samples does not enable the background b-jet charge asymmetry to be determined precisely, the obtained value being $\approx -0.02 \pm 0.05$. On the other hand, as it was shown by CDF [2], no marked background asymmetry is expected. For this reason, as a background, we use the signal events but without the pairing of leptons and b-jets. As a consequence the obtained b-jet charge spectrum is not correlated with the high $p_T$ lepton charge and should not have any charge asymmetry. Assuming the nominal S:B ratio, the spectrum is normalized to 1/30 of the signal statistics. The analysis showed that this background exhibits practically no asymmetry. For the systematics studies, a background corresponding to S:B=7:1 was also considered.

To find a realistic b-jet charge distribution, the signal and background distributions are combined. In Figure 3 (left) the expected b-jet charge ($Q_{\mathrm{comb}}$) distribution combining the signal with the background (full line) and the background itself (dashed line) are shown. From the reconstructed b-jet charge spectra using the two treated backgrounds, the expected mean b-jet charge (assuming the Standard Model) is:

$$Q_{\mathrm{comb}} = -0.094 \pm 0.0042 \text{ (stat)}.$$

The $Q_{\mathrm{comb}}$ value is obtained as the mean value of the signal plus background (S+B) distribution combining signal with the background.

In conclusion, $\approx$ 6000 $\ell$b combinations could be selected for the top quark b-jet charge analysis, using the 1 fb$^{-1}$ sample. The expected combined b-jet charge purity, $N(Q_{\mathrm{bjet}} < 0)/N_{\mathrm{all}}$, is $\approx 0.62 \pm 0.01$ for the Standard Model case.

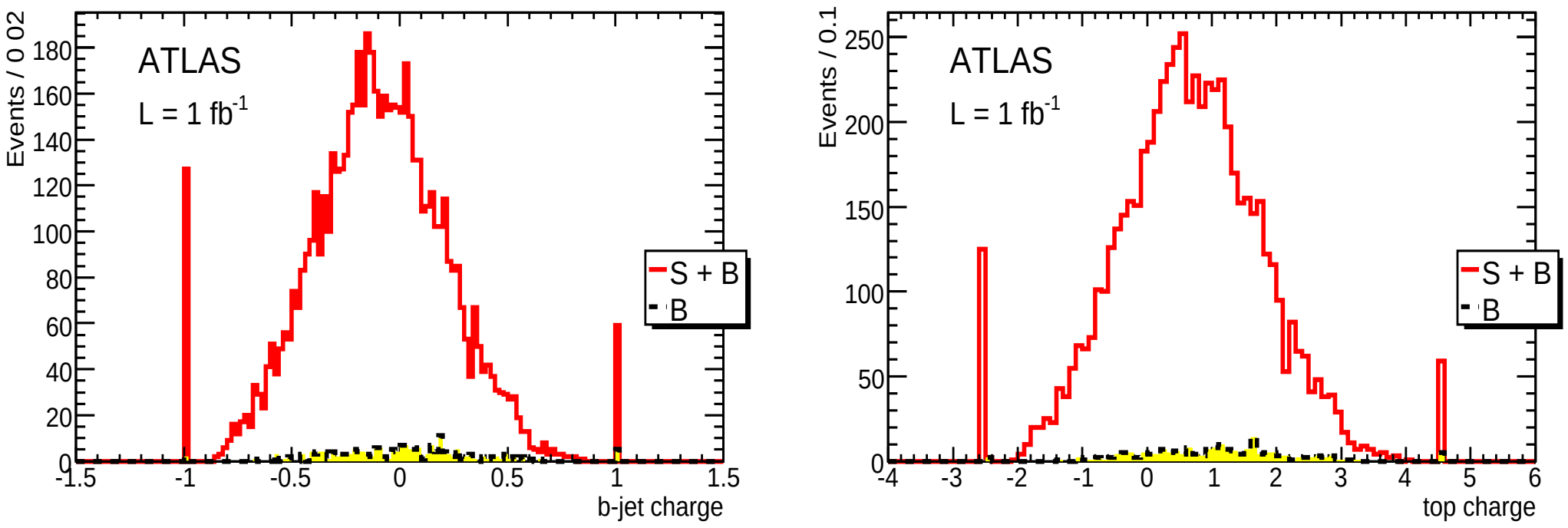


Figure 3: Left: the full S+B b-jet charge ($Q_{\mathrm{comb}}$) distribution (full line) and the background itself (dashed line); right: the reconstructed top quark charge ($Q_{\mathrm{t}}^{\mathrm{comb}}$) (full line) and its background (dashed line).

Taking into account the statistical uncertainty of the obtained mean b-jet charge ($Q_{\mathrm{comb}}$) it can be stated that the obtained value will differ from 0 by more than $20\sigma$. Using a simple statistical treatment it is easy to show that for a reliable determination of $Q_{\mathrm{comb}}$ ($\geq 5\sigma$) a sample of $\approx 0.1$ fb$^{-1}$ should be sufficient. In addition to that the analysis has revealed that the reconstructed b-jet charge is more influenced by the size of the S:B ratio than by the background asymmetry: going from the pure signal b-jet charge spectrum to that of the 7:1 mixture of signal and background, the mean b-jet charge decreased by 14%, while a replacement of the symmetric background by the asymmetric one with an asymmetry 1/4 of the signal one, leads to only 3% change of the charge at the 7:1 S:B ratio.

The direct reconstruction of the top quark charge can be done relying on the obtained value of $Q_{\mathrm{comb}}$ (see above). Using the Standard Model value of the b quark charge ($Q_{\mathrm{b}} = -1/3$) and the mean reconstructed value of the b-jet charge ($Q_{\mathrm{comb}}$), the b-jet charge calibration coefficient $C_{\mathrm{b}} = Q_{\mathrm{b}}/Q_{\mathrm{comb}}$ is $3.54 \pm 0.16$ and the top quark charge then reads:

$$Q_{\mathrm{t}} = Q(\ell^{+}) + Q_{\mathrm{bjet}}^{(+)} \times C_{\mathrm{b}}\,, \qquad Q_{\bar{\mathrm{t}}} = Q(\ell^{-}) + Q_{\mathrm{bjet}}^{(-)} \times C_{\mathrm{b}} \tag{4}$$

where $Q(\ell^{\pm}) = \pm 1$ is the lepton charge and $Q_{\mathrm{bjet}}^{(\pm)}$ is as above.

The reconstructed top quark charge is shown in Figure 3 (right) for the sample of 1 fb$^{-1}$. The absolute value of top quark charge obtained by combining $Q_{\mathrm{t}}$ and $Q_{\bar{\mathrm{t}}}$ for the above mentioned sample is $Q_{\mathrm{t}}^{\mathrm{comb}} = 0.67 \pm 0.06\,(\mathrm{stat}) \pm 0.08\,(\mathrm{syst})$. The statistical error assumes that the relative error of $C_{\mathrm{b}}$ is the same as that of $Q_{\mathrm{comb}}$. The systematic error of $Q_{\mathrm{t}}^{\mathrm{comb}}$ can be studied comprehensively only by using experimental data [2]. In this case the main source of the systematic error is the weighting procedure that influences the coefficient $C_{\mathrm{b}}$, that should be determined independently on the investigated b-jet charge, as well as the mean b-jet charge. In our case only the systematics stemming from determination of the mean b-jet charge were taken into account.

### 3.5.2 Semileptonic b-decay approach results

The charge of the non-isolated lepton found within the b-jet provides discrimination between the Standard Model and the exotic hypotheses on a statistical basis. Figure 4 shows the number of b-jets, which have been paired with positive (left) and negative (right) high $p_{\mathrm{T}}$ lepton and which contain inside a non-isolated lepton, as a function of the charge ($Q_{\mathrm{nonIs}}$) of the contained non-isolated lepton. The mean values

[2] The reconstructed b-jet charge or coefficient $C_{\mathrm{b}}$ for an experimental sample, e.g. dijet $\mathrm{b}\bar{\mathrm{b}}$ data, should be compared with the corresponding Monte Carlo one to look for a possible difference in the b-jet track topology between Monte Carlo and real data.

of the non-isolated lepton charge obtained from these figures are:

$$\bar{Q}^{(+)}_{\rm nonIs} = \frac{N(\ell^+) - N(\ell^-)}{N(\ell^+) + N(\ell^-)} = -0.32 \pm 0.05, \quad \bar{Q}^{(-)}_{\rm nonIs} = \frac{N(\ell^+) - N(\ell^-)}{N(\ell^+) + N(\ell^-)} = 0.30 \pm 0.05,$$

where $\bar{Q}^{(+)}_{\rm nonIs}$ ($\bar{Q}^{(-)}_{\rm nonIs}$) is the mean charge of the non-isolated leptons in the b-jets paired with the positive (negative) high $p_{\rm T}$ lepton. $N(\ell^-)$ ($N(\ell^+)$) is the number of b-jets with a negative (positive) charged lepton. The quantity $Q(\ell) \times Q^{(\ell)}_{\rm nonIs}$, where $Q(\ell)$ is the charge of the lepton paired with the b-jet containing a non-isolated lepton, can be used to combine both histograms. The obtained mean combined charge in the Standard Model is $\bar{Q}^{(\rm comb)}_{\rm nonIs} = -0.31 \pm 0.04$, showing a potential to distinguish between the Standard model and exotic hypothesis even with 1 fb$^{-1}$ of data, as in the case of the exotic scenario the opposite value of $\bar{Q}^{(\rm comb)}_{\rm nonIs}$ is expected.

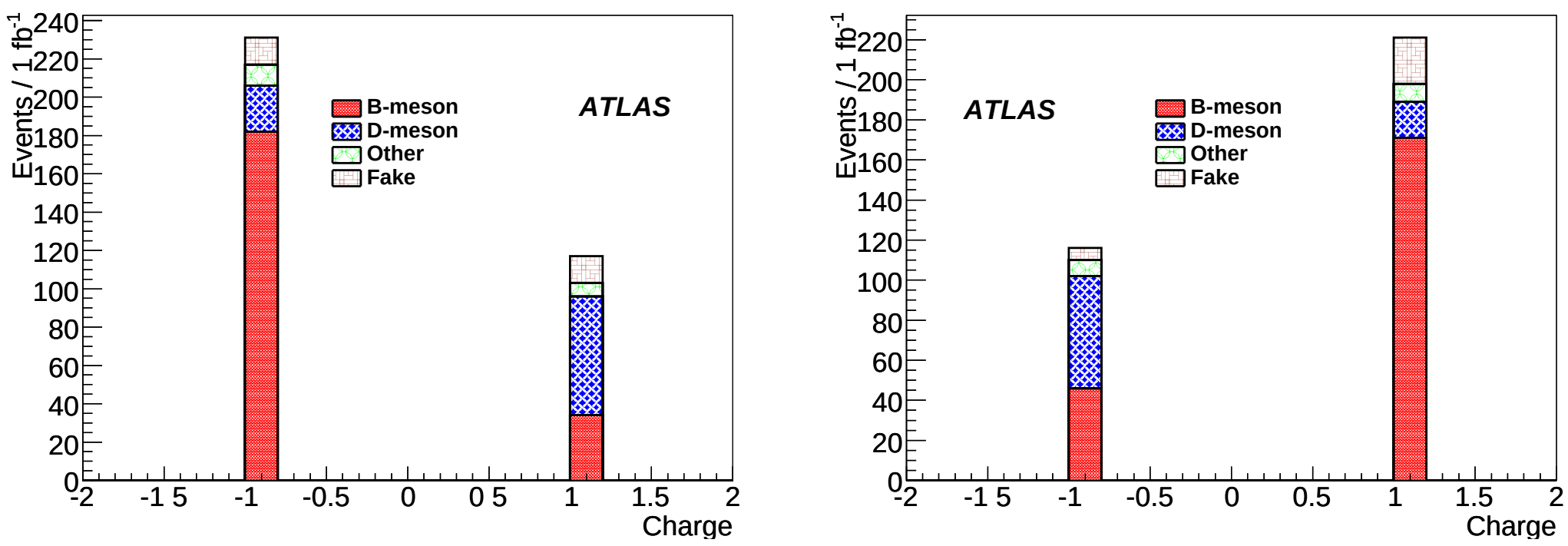


Figure 4: Number of b-jets, containing a non-isolated lepton inside, associated with positive (left) and negative (right) high $p_{\rm T}$ lepton vs the charge of the non-isolated lepton. The contribution of different sources of leptons are marked by different shading styles.

### 3.6 Systematic uncertainties

The systematic studies have been done following the prescription described in Section 6 of Chapter 1. The resulting systematic errors are summarised in Table 5. The systematic uncertainty caused by the top quark mass was estimated from the absolute difference between the $Q_{\rm comb}$ reconstructed at the nominal value (175 GeV) and $m_{\rm t}$=160 and 190 GeV and rescaling to an effective 2 GeV uncertainty. The systematic uncertainty connected with the signal to background ratio was found assuming that this ratio is known with 30% uncertainty. The background asymmetry systematics was estimated assuming that the background charge spectrum asymmetry is at a level of 10% of the signal one.

The systematic uncertainties due to Monte Carlo modeling of $t\bar{t}$ signal were studied by comparing samples with different fragmentation parameters: an AcerMC/Pythia sample and the standard signal MC@NLO/Herwig one, giving values of 18 ± 13% and 16 ± 13% for the weighting and semileptonic approaches, respectively. The large uncertainties are due to a limited Monte Carlo statistics, so these values cannot be considered as a reliable estimate of the corresponding systematic uncertainties and are therefore not included in Table 5.

Pileup background affects the weighting technique procedure, as tracks from additional minimum-bias interactions get included in the jet charges. The associated systematic uncertainty was evaluated by comparing the standard sample to a dedicated $t\bar{t}$ sample including pileup, leading to a shift of 20±18%. Since the Monte Carlo statistical error is large, and it is expected that any pileup effect can be minimised

(at least at moderate LHC luminosities) by applying track impact parameter cuts to eliminate tracks from pileup vertices, this value is also not included Table 5.

Table 5: The systematic uncertainties (%) of the mean reconstructed charge, $Q_{\rm comb}$, the weighting technique and b-decay approaches.

| Source | Weighting (%) | b-decay (%) |
|---|---|---|
| jet scale | 0.7 | 0.3 |
| b-jet scale | 1.9 | 6 |
| $\Delta m_{\rm t}$ | 1.3 | 7 |
| PDF | 0.6 | – |
| ISR | 2.8 | 15 |
| FSR | 7.8 | 8 |
| Pile-up | – | 1.8 |
| Background asymmetry | 1 | – |
| S/B ratio | 9 | – |
| total | 12.5 | 19.3 |

# 4 Polarization studies in $\rm t\bar{t}$ semileptonic events

The measurements of the W-boson and top quark polarizations in $\rm t\bar{t}$ events provide a powerful test of the top quark production and decay mechanisms and are a sensitive probe of new physics. W-boson or top quark spin information can be inferred from the angular distributions of the daughter particles in the W-boson or top quark rest frame, respectively. The W-boson can be produced with right, left or longitudinal polarizations, with corresponding partial widths $\Gamma_{\rm R}$, $\Gamma_{\rm L}$, $\Gamma_0$ that depend on new anomalous couplings [61] ($V_{\rm R}$, $V_{\rm L}$, $g_{\rm L}$ and $g_{\rm R}$) which can appear at the Wtb vertex (see Eq. 1).

## 4.1 W-boson polarization and $\rm t\bar{t}$ spin correlation measurements

The probability for the three helicity states of W-boson produced in top quark decay, $F_0$ (longitudinal), $F_{\rm L}$ (left-handed) and $F_{\rm R}$ (right-handed), can be extracted from the $\Psi$ angular distribution [61] :

$$\frac{1}{N}\frac{dN}{d\cos\Psi} = \frac{3}{2}\left[F_0\left(\frac{\sin\Psi}{\sqrt{2}}\right)^2 + F_{\rm L}\left(\frac{1-\cos\Psi}{2}\right)^2 + F_{\rm R}\left(\frac{1+\cos\Psi}{2}\right)^2\right], \tag{5}$$

where $\Psi$ is the angle between the W-boson direction in the top quark rest frame and the charged lepton direction in the W-boson rest frame obtained by a boost along the W-boson flying direction in the top quark rest frame. The correlation between the parameter couples ($F_0$,$F_{\rm L}$), ($F_0$,$F_{\rm R}$) and ($F_{\rm L}$,$F_{\rm R}$) are -0.9, -0.8 and 0.4, respectively.

Although, in the Standard Model, top quarks are produced unpolarised in $\rm t\bar{t}$ events, their spins are correlated [62]. The production asymmetry $A$ in these events, measures the spin correlation and is defined as

$$A = \frac{\sigma(t_\uparrow\bar{t}_\uparrow)+\sigma(t_\downarrow\bar{t}_\downarrow)-\sigma(t_\uparrow\bar{t}_\downarrow)-\sigma(t_\downarrow\bar{t}_\uparrow)}{\sigma(t_\uparrow\bar{t}_\uparrow)+\sigma(t_\downarrow\bar{t}_\downarrow)+\sigma(t_\uparrow\bar{t}_\downarrow)+\sigma(t_\downarrow\bar{t}_\uparrow)}, \tag{6}$$

where $\sigma(t_{\uparrow/\downarrow}\bar{t}_{\uparrow/\downarrow})$ denotes the production cross-section of a top quark pair with spins up or down with respect to a selected quantisation axis. It can be extracted from the $\theta_1$ and $\theta_2$ angular distributions [62]:

$$\frac{1}{N}\frac{d^2N}{d\cos\theta_1 d\cos\theta_2} = \frac{1}{4}(1 - A|\alpha_1\alpha_2|\cos\theta_1\cos\theta_2), \quad (7)$$

where $\theta_1$ ($\theta_2$) is the angle between the t ($\bar{\mathrm{t}}$) direction, measured in the $\mathrm{t\bar{t}}$ rest frame, and the direction of the t ($\bar{\mathrm{t}}$) decay product in the t ($\bar{\mathrm{t}}$) rest frame obtained by a boost along t ($\bar{\mathrm{t}}$) direction in the $\mathrm{t\bar{t}}$ rest frame, $\alpha_i$ is the spin analysing power of the top quark decay product $i$, which ranges between $-1$ and 1 and measures the degree to which its direction is correlated with the spin of the parent top quark. The parameter $A_{\mathrm{D}}$ defined in [20], used to measure the production asymmetry in another basis, is extracted from the $\Phi$ angular distribution [62] :

$$\frac{1}{N}\frac{dN}{d\cos\Phi} = \frac{1}{2}(1 - A_{\mathrm{D}}|\alpha_1\alpha_2|\cos\Phi), \quad (8)$$

where $\Phi$ is the angle between the direction of flight of the two spin analysers, defined in the t and $\bar{\mathrm{t}}$ rest frames respectively.

Table 6: Standard Model values of W-boson polarization parameters($F_0$, $F_{\mathrm{L}}$, $F_{\mathrm{R}}$) at the next-to-leading order and $\mathrm{t\bar{t}}$ spin correlation parameters ($A$, $A_{\mathrm{D}}$) at leading order for a top quark mass of 175 GeV. For $A$ and $A_{\mathrm{D}}$ the asterisk superscript means that $m_{\mathrm{t\bar{t}}} < 550$ GeV is applied.

| $F_0$ | $F_{\mathrm{L}}$ | $F_{\mathrm{R}}$ | $A$* | $A_{\mathrm{D}}$* |
|---|---|---|---|---|
| 0.695 | 0.304 | 0.001 | 0.422 | -0.290 |

The Standard Model predictions for the W-boson polarization ($F_0$, $F_{\mathrm{L}}$, $F_{\mathrm{R}}$), at next-to-leading order, and $\mathrm{t\bar{t}}$ spin correlation ($A$, $A_{\mathrm{D}}$), at leading order, are given in Table 6. The ATLAS sensitivity to these observables has been evaluated with an ATLFAST simulation [20] and a full simulation of a perfect detector [63]. A precision of 1% to 5% should be achievable with 10 fb$^{-1}$ of data, dominated by systematic uncertainties (in particular from the b-jet energy scale). This measurement requires a complete reconstruction of the $\mathrm{t\bar{t}}$ system and a reliable Monte Carlo description to correct the distortion induced by trigger, cuts and reconstruction. In this section the robustness of the analysis with a realistic detector simulation including triggering is assessed. Only the semileptonic topology of the $\mathrm{t\bar{t}}$ events ($\mathrm{t\bar{t}} \rightarrow \mathrm{WWb\bar{b}} \rightarrow \ell\nu \mathrm{j_1 j_2 b\bar{b}}$ with $\ell = e, \mu$) is used as signal (an analysis using $\mathrm{t\bar{t}}$ dilepton events as signal can improve the results of this study, especially for the $\mathrm{t\bar{t}}$ spin correlation measurements). In this case the most powerful spin analysers of the top quark are the charged lepton ($\alpha_1 = 1$) and the least energetic non b-jet in the top quark rest frame ($\alpha_{\mathrm{jet}} = 0.51$) [64], which are chosen afterwards for the spin correlation measurement.

### 4.1.1 Event simulation and selection

The largest statistics Monte Carlo sample was generated with MC@NLO which does not implement $\mathrm{t\bar{t}}$ spin corelations, so it can only be used to study W-boson polarisation. Spin correlations were studied using the smaller AcerMC sample. In both cases, non-semileptonic $\mathrm{t\bar{t}}$, semileptonic $\mathrm{t\bar{t}}$ which decay to $\tau$, W-boson+jets and single top quark events were considered as background. Semileptonic signal events are characterised by one (and only one) isolated lepton, at least 4 jets, of which at least 2 jets are tagged as b-jets, and missing transverse energy. Following the common criteria for the semileptonic selection, all kinematic cuts are summarised in Table 7 together with the corresponding signal efficiencies.

After kinematic cuts are applied, the event is fully reconstructed, as described below. The angles $\Psi$, $\theta_1$, $\theta_2$ and $\Phi$ are computed and the polarization parameters are extracted.

Table 7: Selection cuts used and corresponding efficiencies for semileptonic $t\bar{t}$ MC@NLO events and $t\bar{t}$ AcerMC events.

| Variables | Cuts | Efficiency (%) | |
|---|---|---|---|
| | | MC@NLO | AcerMC |
| Lepton | exactly 1 identified | 57.8 | 58.7 |
| Jets | at least 4 selected jets | 59.3 | 62.1 |
| b-tagging | at least 2 are tagged as b | 32.0 | 31.0 |
| Missing energy | $p_T^{miss} > 20$ GeV | 92.0 | 92.2 |
| Cumulative efficiency | | 9.4 | 9.8 |

### 4.1.2 W-boson and top quark reconstruction

The energies of all jets are calibrated according to the comparison with the energies of corresponding parton level quarks before selection and reconstruction. Then, in the event reconstruction, the light jet pair with invariant mass $m_{jj}$ closest to the known W-boson mass, $m_W$, is selected to reconstruct the W-boson which decays hadronically. This W-boson is then combined with one of the b-jets to reconstruct the top quark. As there are several possible combinations, the one which gives the mass closest to the top quark mass $m_t$ is assumed to be the correct one. The b-jet which is closest to the lepton in $\Delta R$ among the remaining b-jets, is reserved for reconstructing the other top quark whose daughter W-boson decays leptonically. To reconstruct the W-boson which decays leptonically, the neutrino $p_T$ is taken as the missing transverse energy. Its longitudinal component $p_z$ is determined by constraining $m_{l\nu}$ to $m_W$ [63]. When two solutions for $p_z$ are found, the one giving $m_{\ell\nu b}$ closer to $m_t$ is kept.

Quality cuts $|m_{jjb} - m_t| < 35$ GeV, $|m_{\ell\nu b} - m_t| < 35$ GeV and $|m_{jj} - m_W| < 20$ GeV are applied to reject badly reconstructed events. At this stage, 2.8% of the signal events are kept for MC@NLO events and 2.7% for AcerMC events, corresponding to 7000 signal events for 1 fb$^{-1}$ of data (Table 8). After this event selection, the main background comes from the $t\bar{t} \to \tau + X$ events, where the tau decays to electron or muon. The number of events from W-boson+jets and single top quark channels is less than 3% of the selected number of signal events, so we neglect them in the following sections. Due to this cancellation of the background, the S/B ratio can be affected as much as 20%, which is taken as a systematic uncertainty (see Table 10).

After event reconstruction, the angle $\Psi$ (W-boson polarization) as well as the angles $\theta_1$, $\theta_2$ and $\Phi$ ($t\bar{t}$ spin correlations) are computed using the prescription descibed at the beginning of Section 4. The measured distributions of $\cos\Psi$, $\cos\theta_1 \times \cos\theta_2$ and $\cos\Phi$ are distorted, compared with their distributions at parton level. The detector resolution results in much smaller smearing effect on the final particles than that coming from particle radiation, quark fragmentation-hadronization and final event reconstruction. That is to say, the latter effects dominate the resolution of the reconstructed objects from top quark decay [65]. A correction function, taken from the ATLFAST simulation, is used to recover the distributions at parton level. With ATLFAST data, the correction function is obtained from the ratio between the two normalized distributions of the $\cos\Psi$ (i.e. after reconstruction of the signal and main background events and at parton level of the pure signal). To correct for the distortion, a weight derived from this function is applied on all the reconstructed $\cos\Psi$ of full simulation, on an event by event basis, allowing to recover, as much as possible, the shape of the distribution of the $\cos\Psi$ of the pure signal at parton level. A detailed description of the method can be found in [20].

Table 8: The number of signal (MC@NLO) and the most important background events before (left) and after (right) selection for 1 fb$^{-1}$. $*$: in the single top quark decay W$\rightarrow$ e$/\mu/\tau+\nu$ mode

| | Events for 1 fb$^{-1}$($\times 10^3$) | Selected events full simulation |
|---|---|---|
| **Signal** (t$\bar{\text{t}}$ semileptonic) | 250 | 7000 |
| t$\bar{\text{t}}$ $\rightarrow \tau + X$ | 130 | 710 |
| $W(\rightarrow l\nu)$+ jets | 800 | [10,55] |
| Single t (Wt channel) | 25$^*$ | 90 |
| Single t (t channel) | 80$^*$ | 55 |

### 4.1.3 Impact of the trigger on the analysis

Semileptonic t$\bar{\text{t}}$ events are characterised by a single isolated lepton, which can be used to trigger the events with high efficiency. Applying the trigger selection to events passing the standard selection cuts, 15% of well reconstructed events are lost. The measurement results with and without trigger applied on the data, while keeping all other aspects of the measurement unchanged, were compared. The effect of the trigger on the measurements of the W-boson polarization is almost zero But the effect on the t$\bar{\text{t}}$ spin correlations is not negligible, and is taken as a systematic error, shown in Table 10.

### 4.1.4 Measurement of the W-boson polarization

Figure 5 shows the correction function (left) and the reconstructed $\cos\Psi$ distribution (right) after applying the correction function. This distribution is fitted to Eq. (5) varying $F_0$, $F_{\text{L}}$ and $F_{\text{R}}$, but constrained by $F_0 + F_{\text{L}} + F_{\text{R}} = 1$. The fit is restricted to the region $-0.9 < \cos\Psi < 0.8$, which is the most extended region where the correction is varying slowly. The results are shown in Table 9.

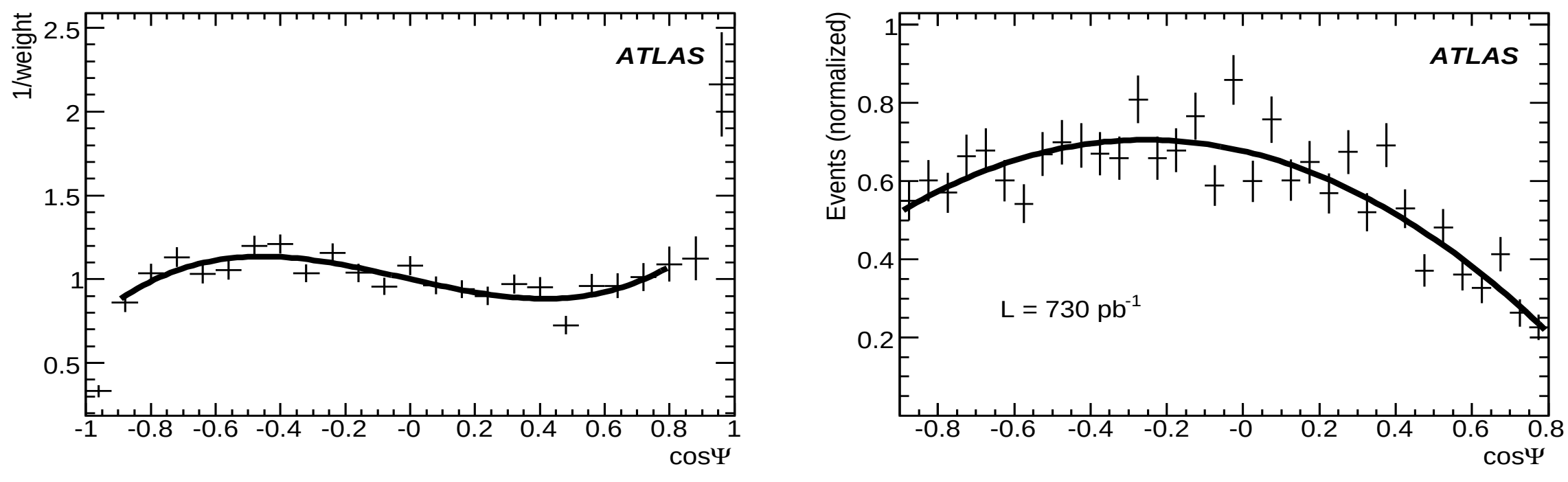


Figure 5: Left: Correction function taken from ATLFAST simulation fitted with a third order polynomial function. Right: Normalised reconstructed and corrected distribution of $\cos\Psi$, the full line corresponds to the fit to Eq. (5). The sample has an integrated luminosity of 730 pb$^{-1}$.

Two complementary methods to extract the W-boson helicity fractions, using the observed angular distribution between the charged lepton direction in the W-boson rest frame and the W-boson direction in the top quark rest frame, are currently under development at ATLAS inspired on Tevatron methods [5,66].

### 4.1.5 Measurement of $t\bar{t}$ spin correlation

At parton level, before any phase space cut, the two estimators $C = -9 \times \cos\theta_1 \cos\theta_2$ and $D = -3 \times \cos\Phi$ are unbiased [20]. Figure 6 shows the reconstructed distributions of $-9 \times \cos\theta_1 \cos\theta_2/0.51$ and $-3 \times \cos\Phi/0.51$ after correction. It should be stressed that, in the evaluation of the $t\bar{t}$ spin correlation parameters, the theoretical value for the spin analyzing power for signal events (0.51) was assumed for both cases. The means of the distributions are unbiased estimators of $A$ and $A_D$, provided corrections for

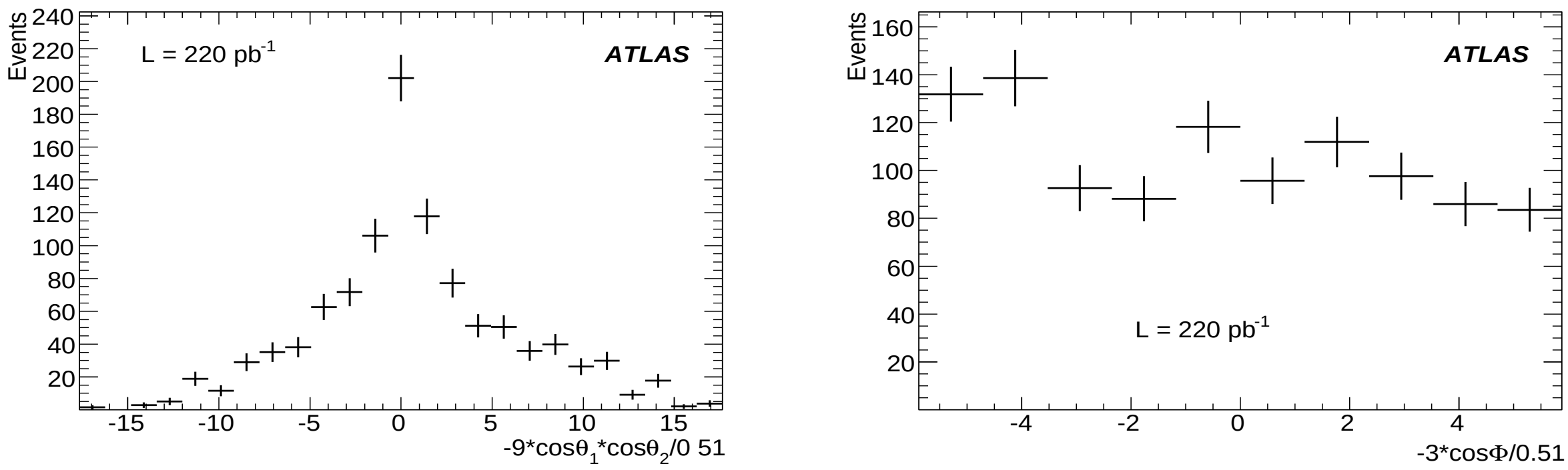


Figure 6: Left: Reconstructed and corrected distribution of $-9 \times \cos\theta_1 \cos\theta_2/0.51$. Right: Reconstructed and corrected distribution of $-3 \times \cos\Phi/0.51$. The integrated luminosity of the sample is 220 $pb^{-1}$ in each case.

physics effects and detector effects are fully carried out.

Due to the fact that the ATLFAST was used to extract the correction function, an additional systematic uncertainty of 0.25 was derived for the $A$ spin correlation parameter. This uncertainty reflects the different parametrizations of the Monte Carlo simulations. Preliminary studies suggest that this shift can be removed by using high statistics full simulation samples to derive the correction functions and for this reason this uncertainty was not included in Table 10.

Table 9: W-boson polarization and top quark spin correlation parameters extracted after triggering. The indicated errors are statistical and systematic, respectively.

| W-boson polarization | $F_L$ | $F_0$ | $F_R$ |
|---|---|---|---|
| | 0.29 ±0.02 ±0.03 | 0.70 ±0.04 ±0.02 | 0.01 ±0.02 ±0.02 |
| $t\bar{t}$ spin correlation | $A$ | $A_D$ | |
| | 0.67 ±0.17±0.18 ±0.25 | -0.40 ±0.11 ±0.09 | |

### 4.1.6 Systematic uncertainties

The systematic uncertainties were estimated with ATLFAST simulation for the factorisation scale, structure function, ISR, FSR, b-fragmentation, top quark mass, hadronization scheme and pile-up effects, and with full simulation for the b-tagging efficiency, b-jet energy scale, light jet energy scale and signal to background ratio (S/B scale) as listed in Table 10, for the measurement of the W-boson helicity fractions and top quark pair spin correlations respectively.

With data, $t\bar{t}$ events can be reconstructed without requiring b-jet tagging on the side of top quark whose daughter W-boson decays leptonically. This can provide a pure sample of b-jets which can be

Table 10: Systematics for W-boson polarization and top quark spin correlation measurement.

| Source of uncertainty | $F_L$ | $F_0$ | $F_R$ | $A$ | $A_D$ |
|---|---|---|---|---|---|
| Factorisation | 0.000 | 0.001 | 0.001 | 0.029 | 0.006 |
| Structure function | 0.003 | 0.003 | 0.004 | 0.033 | 0.012 |
| ISR | 0.001 | 0.002 | 0.001 | 0.002 | 0.001 |
| FSR | 0.009 | 0.007 | 0.002 | 0.023 | 0.016 |
| b-fragmentation | 0.001 | 0.002 | 0.001 | 0.031 | 0.018 |
| Hadronization scheme | 0.010 | 0.016 | 0.006 | 0.006 | 0.008 |
| Pile-up (2.3 events) | 0.005 | 0.002 | 0.006 | 0.001 | 0.005 |
| Input top quark mass (2 GeV) | 0.015 | 0.011 | 0.004 | 0.028 | 0.013 |
| b-tagging efficiency (5%) | 0.007 | 0.002 | 0.005 | 0.027 | 0.07 |
| b-jet energy scale (5%) | 0.02 | 0.002 | 0.02 | 0.07 | 0.015 |
| light-jet energy scale (5%) | - | - | - | 0.11 | 0.017 |
| S/B scale (20%) | 0.004 | 0.002 | 0.001 | 0.000 | 0.004 |
| Trigger | - | - | - | 0.10 | 0.03 |
| TOTAL | 0.03 | 0.02 | 0.02 | 0.18 | 0.09 |

used to check the uncertainty on the b-tagging efficiency [67]. The b-jet energy miscalibration can also be obtained from other control samples, such as Z+b-jet events [68].

## 4.2 Anomalous couplings at the Wtb vertex

The W-boson polarisation is sensitive to new anomalous couplings ($V_L$, $V_R$, $g_L$ and $g_R$) associated with the Wtb vertex. Although the W-boson helicity fractions ($F_0$, $F_L$ and $F_R$) depend on these couplings, the helicity ratios $\rho_{R,L} \equiv \Gamma_{R,L}/\Gamma_0 = F_{R,L}/F_0$, were found to be more sensitive to $V_L$, $V_R$, $g_L$ and $g_R$. The $\rho_R$ and $\rho_L$ observables are independent quantities and take the LO values $\rho_R = 5.1 \times 10^{-4}$, $\rho_L = 0.423$ in the Standard Model. General expressions for $\rho_{R,L}$ in terms of the new anomalous couplings can be found in Ref. [19]. As for the helicity fractions, the measurement of helicity ratios sets bounds on $V_R$, $g_L$ and $g_R$. A third and simpler method to extract information about the Wtb vertex is through angular asymmetries involving the angle of the charged lepton in the W-boson rest frame and the W-boson direction in the top quark rest frame (introduced in the previous section). For any fixed $z$ in the interval $[-1,+1]$, one can define asymmetries as the difference between the number of events above and below $z$, normalised to the total number of events. The most obvious choice is $z = 0$, giving the forward-backward asymmetry $A_{FB}$ [18, 69]. The forward-backward asymmetry is related to the W-boson helicity fractions by $A_{FB} = \frac{3}{4}[F_R - F_L]$. Other convenient choices are $z = \mp(2^{2/3} - 1)$. Defining $\beta = 2^{1/3} - 1$, we have $A_+ = 3\beta[F_0 + (1+\beta)F_R]$ and $A_- = -3\beta[F_0 + (1+\beta)F_L]$. Thus, $A_+$ ($A_-$) only depend on $F_0$ and $F_R$ ($F_L$). The LO Standard Model values of these asymmetries are $A_{FB} = -0.223$, $A_+ = 0.548$, $A_- = -0.840$. They are sensitive to anomalous Wtb interactions, and their measurement allows to probe this vertex without the need of a fit to the distribution of the angle between the charged lepton direction in the W-boson rest frame and the W-boson direction in the top quark rest frame. In the present analysis the ATLAS sensitivity to the $\rho_L$, $\rho_R$, $A_{FB}$, $A_+$ and $A_-$ observables is studied.

### 4.2.1 Event selection

In this section an alternative two-level likelihood analysis is explored for the semileptonic channel, i.e. $t\bar{t} \to W^+bW^-\bar{b}$ where one of the W-bosons decays hadronicaly and the other one decays in the leptonic

channel $W \to \ell\nu_\ell$ (with $\ell = e, \mu$). Any other process constitutes a background to this signal. In particular, it should be noticed that the fully hadronic, dileptonic and semileptonic (with one of the W-bosons decaying into $\tau\nu$) $t\bar{t}$ channels are considered backgrounds to the present analysis. Additionally, the following Standard Model processes were considered as background: single top production, W+jets, $Wb\bar{b}$+jets, $Wc\bar{c}$+jets, $Z \to e^+e^-$, $Z \to \mu^+\mu^-$, $Z \to \tau^+\tau^-$, WW, ZZ and WZ. The likelihood analysis is based on the construction of a discriminant variable which uses distributions of some kinematical properties of the events. In the first analysis level (called the pre-selection), the common selection criteria for the semileptonic topology, with the exception of the b-tagging requirement on jets, was applied to the event. The full event reconstruction was performed using a $\chi^2$, defined by

$$\chi^2 = \frac{(m_{\ell\nu j_a} - m_t)^2}{\sigma_t^2} + \frac{(m_{j_b j_c j_d} - m_t)^2}{\sigma_t^2} + \frac{(m_{\ell\nu} - m_W)^2}{\sigma_W^2} + \frac{(m_{j_c j_d} - m_W)^2}{\sigma_W^2}\,, \tag{9}$$

where $m_t = 175$ GeV, $m_W = 80.4$ GeV, $\sigma_t = 14$ GeV and $\sigma_W = 10$ GeV are the expected top quark and W-boson mass resolutions[3], $\ell$ represents the selected electron or muon, $m_{\ell\nu}$ is the invariant mass of the electron (muon) and the neutrino, and $j_{a,b,c,d}$ corresponds to all the possible combinations among the four jets with highest $p_T$ (with $m_{\ell\nu j_a}$, $m_{j_b j_c j_d}$ and $m_{j_c j_d}$ being the corresponding invariant masses). The neutrino was reconstructed using the missing transverse energy and allowing the $p_z^\nu$ to vary in the range $[-500, +500]$ GeV. The solution corresponding to the minimum $\chi^2$ was chosen. The jets used to reconstruct the hadronic W-boson will be labelled "non-b" jets and the remaining two are labelled "b-jets". It should be stressed that no b-tagging information was used so far. The pre-selection was completed by requiring $\chi^2 < 16$. In the second level (the final selection), signal and background-like probabilities were constructed for each event ($\mathscr{P}_i^{\mathrm{signal}}$ and $\mathscr{P}_i^{\mathrm{back.}}$, respectively) using probability density functions (p.d.f.) built from relevant physical variables: the cosine of the angle between the leptonic top quark and the leptonic "b-jet", the transverse momentum of the hadronic W, the hadronic and leptonic top quark masses, the transverse momentum of the two "b-jets", the transverse momentum of the lepton and the $\sqrt{\chi^2}$ distribution. It should be stressed that the objective is to test the sensitivity for new physics exclusion, under the hypothesis that the Standard Model holds, and the simulation was done assuming no anomalous couplings. Signal ($\mathscr{L}_S = \Pi_{i=1}^n \mathscr{P}_i^{signal}$) and background ($\mathscr{L}_B = \Pi_{i=1}^n \mathscr{P}_i^{back.}$) likelihoods (with $n = 8$, the number of p.d.f.) are used to define a discriminant variable $L_R = \log_{10}(\mathscr{L}_S/\mathscr{L}_B)$. The distribution of this variable is shown in Figure 7(a). The final event selection is done by applying the cut $L_R > 0.1$ on the discriminant variable. The number of signal and background events (normalised to $L = 1$ fb$^{-1}$) after the pre-selection and final selection are shown in Table 11. After the final selection (including the trigger), the dominant backgrounds are W+jets and semileptonic $t\bar{t}$ with taus in the final state (corresponding to 49%, and 29% of the total background, respectively, as shown in Table 12).

The effect of the single lepton trigger on the event selection was studied. The results are summarised in Table 11. In what follows, only events passing the trigger are considered.

Once b-tagging is well understood, additional information can be used. In this case only jets with a positive b-tagging weight were considered as "b-jet" candidates for the $\chi^2$ minimisation method. In addition, the b-tagging weights of these jets were considered as p.d.f.s for the discriminant variable evaluation. In this case, the number of selected signal and background events for $L = 1$ fb$^{-1}$ is expected to be $(6.6 \pm 0.1) \times 10^3$ and $(0.9 \pm 0.1) \times 10^3$, respectively. The $t\bar{t}$ background is expected to be dominant (72% of the total background, mainly due to the semileptonic channel with taus in the final state) and the W+jets and single top processes correspond to 15% and 13% of the total background, respectively. The discriminant variables corresponding to the analysis with and without b-tagging are shown in Figure 7.

[3] These resolutions are taken from the top quark mass measurement analyses [70]. It should be noticed that $\sigma_t$ and $\sigma_W$ can be interpreted as weights for each term of the $\chi^2$ definition. By changing their values by a factor $\sim 2$, the obtained observables ($\rho_L$, $\rho_R$, $A_+$ and $A_-$) are the same within the statistical error.

Table 11: Number of signal $t\bar{t} \rightarrow \ell \nu b\bar{b}q\bar{q}'$ and background events (and corresponding statistical error), normalised to $L = 1$ fb$^{-1}$, after the pre-selection and final selection for the analysis without b-tagging. The effect of the trigger on the event selection is also shown.

| | | $e+\mu$ sample | $e$ sample | $\mu$ sample |
|---|---|---|---|---|
| Total background | presel. | $(15.8 \pm 0.4) \times 10^3$ | $(7.2 \pm 0.3) \times 10^3$ | $(8.6 \pm 0.3) \times 10^3$ |
| | final sel. | $(5.2 \pm 0.2) \times 10^3$ | $(2.5 \pm 0.2) \times 10^3$ | $(2.8 \pm 0.2) \times 10^3$ |
| | trigger | $(4.0 \pm 0.2) \times 10^3$ | $(1.8 \pm 0.2) \times 10^3$ | $(2.1 \pm 0.2) \times 10^3$ |
| Signal | presel. | $(27.4 \pm 0.2) \times 10^3$ | $(12.0 \pm 0.1) \times 10^3$ | $(15.5 \pm 0.1) \times 10^3$ |
| | final sel. | $(15.2 \pm 0.1) \times 10^3$ | $(6.5 \pm 0.1) \times 10^3$ | $(8.8 \pm 0.1) \times 10^3$ |
| | trigger | $(12.6 \pm 0.1) \times 10^3$ | $(5.8 \pm 0.1) \times 10^3$ | $(6.9 \pm 0.1) \times 10^3$ |

Table 12: Background composition and corresponding statistical error, normalised to $L = 1$ fb$^{-1}$, after the final selection, including the effect of the trigger, for the analysis without b-tagging.

| | $e+\mu$ sample | $e$ sample | $\mu$ sample |
|---|---|---|---|
| W+jets, Wbb+jets, W$c\bar{c}$+jets | $(19.6 \pm 1.9) \times 10^2$ | $(8.8 \pm 1.4) \times 10^2$ | $(10.3 \pm 1.4) \times 10^2$ |
| Z+jets | $(1.6 \pm 0.4) \times 10^2$ | $(1.2 \pm 0.4) \times 10^2$ | $(0.5 \pm 0.3) \times 10^2$ |
| WZ, ZZ, WW | $(0.4 \pm 0.2) \times 10^2$ | $(0.3 \pm 0.1) \times 10^2$ | $(0.2 \pm 0.1) \times 10^2$ |
| $t\bar{t}$ (except signal) | $(13.1 \pm 0.6) \times 10^2$ | $(5.4 \pm 0.3) \times 10^2$ | $(7.9 \pm 0.4) \times 10^2$ |
| single top | $(5.3 \pm 0.3) \times 10^2$ | $(2.5 \pm 0.2) \times 10^2$ | $(2.7 \pm 0.2) \times 10^2$ |

#### 4.2.2 Measurement of the angular distribution and asymmetries

The experimentally observed angular distribution, which includes the $t\bar{t}$ signal as well as the Standard Model background, is affected by detector resolution, $t\bar{t}$ reconstruction and selection criteria [71]. In order to recover the Standard Model distribution, it is necessary to subtract the background and correct for the effects of the detector, event selection and reconstruction. For this purpose, two different sets of signal and background event samples were used: one "experimental" set, which simulates a possible experimental result, and one "reference" set, which is used to parametrise the effects mentioned, and correct the previous sample. The procedure is as follows. After subtracting reference background samples, the full "experimental" distribution is multiplied by a correction function $f_c$ in order to recover the Standard Model one. This correction function is determined by assuming that the charged lepton distribution corresponds to the Standard Model. In case that a deviation from Standard Model predictions (corresponding to anomalous couplings) is found, the correction function must be modified accordingly, and the expected distribution recalculated in an iterative process. These issues have been analysed in detail in Ref. [20], where it was shown that this process quickly converges. The correction function is calculated, for each bin of the angular distribution, by dividing the number of generated events by the number of selected events, using the reference sample. In order to avoid non-physical fluctuations due to the limited amount of Monte Carlo statistics, a smoothing procedure was applied to the obtained correction function. The value of $f_c$ is in the range [0.2,1.4]. Other methods of correcting the angular distribution are under investigation.

The procedure of correcting for detector and reconstruction effects in the asymmetries is similar to that used with the full angular distribution, but using only two or three bins. This has the advantage that the asymmetry measurements are less sensitive to the extreme values of the angular distributions,

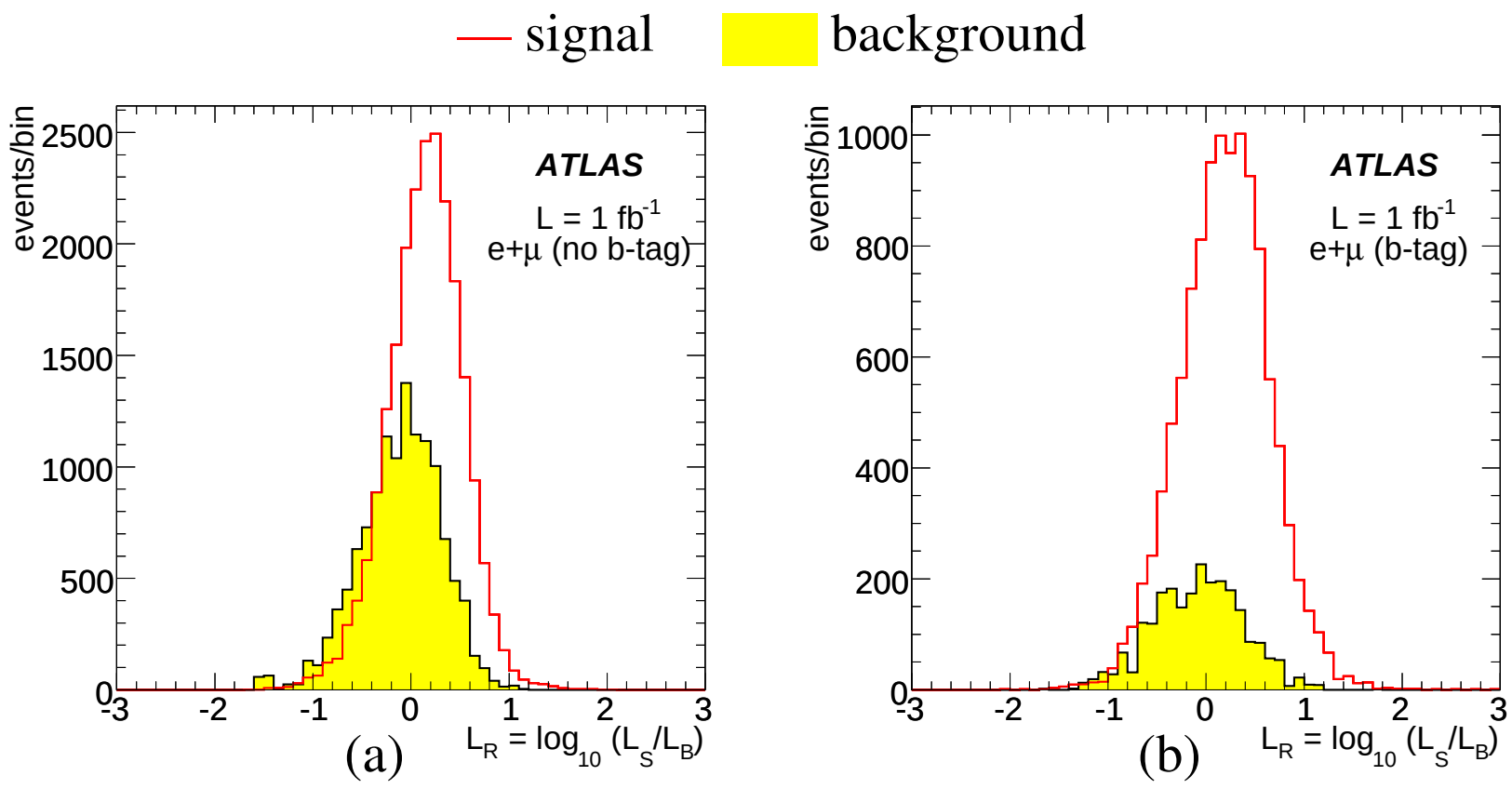


Figure 7: Discriminant variables for the Standard Model background (shaded region) and the $t\bar{t}$ signal (full line), normalised to $L = 1$ fb$^{-1}$ corresponding to the (a) analysis without b-tagging and (b) analysis using the b-tagging weights of the "b-jets" selected by the $\chi^2$ minimisation method.

where correction functions deviate from unity. Moreover, it should be noticed that the extreme bins of the angular distribution have a very significant impact on the measurement of the $\rho_\mathrm{L}$ and $\rho_\mathrm{R}$ helicity ratios. The values obtained from a fit to the corrected distribution, as well as the angular asymmetries $A_\mathrm{FB}, A_\pm$, are collected in Table 13, with their statistical uncertainties. Although the statistical errors of the observables obtained for the analyses without and with b-tagging are similar, the use of this tool allows to improve the signal to background ratio, leading to smaller systematic uncertainties, as discussed below.

Table 13: Expected values and corresponding statistical errors for the helicity ratios and angular asymmetries. The results for an integrated luminosity of 1 fb$^{-1}$ (analyses with and without b-tagging) are shown.

| | $\rho_\mathrm{L}$ | $\rho_\mathrm{R}$ | $A_\mathrm{FB}$ | $A_+$ | $A_-$ |
|---|---|---|---|---|---|
| | | | Analysis without b-tagging | | |
| $e+\mu$ | 0.402 ± 0.050 | -0.008 ± 0.008 | -0.220 ± 0.025 | 0.560 ± 0.024 | -0.845 ± 0.012 |
| | | | Analysis with b-tagging | | |
| $e+\mu$ | 0.453 ± 0.048 | -0.004 ± 0.007 | -0.229 ± 0.026 | 0.542 ± 0.028 | -0.830 ± 0.014 |

### 4.2.3 Systematic uncertainties

The study of the systematic uncertainties considered possible errors from different sources: jet energy scale, luminosity, top quark mass, background level, ISR and FSR, Monte Carlo generator and pile-up. The jet calibration used in the present analyses is described in Ref. [70]. As for the reference analyses, full simulation Monte Carlo samples were used for the study of all the systematic sources of uncertainty. Only the simulated sample used as the "experimental" set (which fakes the data) was changed for each systematic source of uncertainty. The correction function and the Monte Carlo sample used to perform the background subtraction were kept unchanged. The impact on the measurements is summarised in Tables 14 and 15. As the background subtraction is based on Monte Carlo simulation, the background estimation required a luminosity value and therefore the corresponding systematic uncertainty was considered. Once the cross-sections for known backgrounds are measured with data, a data-driven normali-

sation will be possible and the luminosity systematic error should be reduced. Moreover, with data it will be possible to compare the discriminant variable distributions for selected events obtained from data and Monte Carlo. This comparison will allow the correction of systematic uncertainties caused by inaccurate description of data by the Monte Carlo simulation.

Table 14: Sources of systematic uncertainties in the determination of the helicity ratios and angular asymmetries (analysis without b-tagging).

| Source | $\rho_L$ | $\rho_R$ | $A_{FB}$ | $A_+$ | $A_-$ |
|---|---|---|---|---|---|
| Jet energy scale | 0.02 | 0.003 | 0.004 | 0.006 | 0.002 |
| Luminosity | 0.02 | 0.002 | 0.006 | 0.005 | 0.001 |
| Top quark mass | 0.02 | 0.002 | 0.009 | 0.006 | 0.004 |
| Background | 0.01 | 0.002 | 0.005 | 0.003 | 0.002 |
| ISR+FSR | 0.13 | 0.009 | 0.044 | 0.046 | 0.011 |
| MC generator | 0.18 | 0.013 | 0.039 | 0.042 | 0.001 |
| Pile-up | 0.14 | 0.004 | 0.053 | 0.039 | 0.017 |
| Total | 0.27 | 0.017 | 0.080 | 0.074 | 0.021 |

Table 15: Sources of systematic uncertainties in the determination of the helicity ratios and angular asymmetries (analysis with b-tagging).

| Source | $\rho_L$ | $\rho_R$ | $A_{FB}$ | $A_+$ | $A_-$ |
|---|---|---|---|---|---|
| Jet energy scale | 0.04 | 0.001 | 0.010 | 0.004 | 0.002 |
| Luminosity | 0.01 | 0.000 | 0.006 | 0.005 | 0.001 |
| Top quark mass | 0.03 | 0.003 | 0.013 | 0.008 | 0.006 |
| Background | 0.01 | 0.000 | 0.003 | 0.002 | 0.004 |
| ISR+FSR | 0.05 | 0.006 | 0.024 | 0.028 | 0.015 |
| MC generator | 0.01 | 0.008 | 0.009 | 0.011 | 0.000 |
| Pile-up | 0.15 | 0.006 | 0.012 | 0.041 | 0.022 |
| Total | 0.16 | 0.012 | 0.033 | 0.052 | 0.027 |

### 4.2.4 Constraints on the anomalous couplings

Using the parametric dependence of the observables on $V_R$, $g_L$ and $g_R$ (and considering the correlations between them, which are shown in Table 16), constraints can be set on the anomalous couplings. The helicity ratios $\rho_{R,L}$ and the asymmetries $A_\pm$ were used as input for the program `TopFit` [19]. The expected 68% CL allowed regions on the Wtb anomalous couplings for $L = 1$ fb$^{-1}$ (analyses with and without b-tagging) are shown in Figure 8. In addition to the allowed regions of Figure 8, additional solutions can be found at $g_R \sim 0.8$. Such solutions are due to a large cancellation between $\mathcal{O}(g_R)$ and $\mathcal{O}(g_R{}^2)$ terms and can be excluded by the measurement at the LHC of the single top cross-section [72].

Table 16: Correlation matrix for the $A_{\pm}$, $\rho_{R,L}$ observables.

| | $A_+$ | $A_-$ | $\rho_L$ | $\rho_R$ |
|---|---|---|---|---|
| $A_+$ | 1 | 0.16 | -0.73 | -0.14 |
| $A_-$ | 0.16 | 1 | -0.10 | 0.55 |
| $\rho_L$ | -0.73 | -0.10 | 1 | 0.42 |
| $\rho_R$ | -0.14 | 0.55 | 0.42 | 1 |

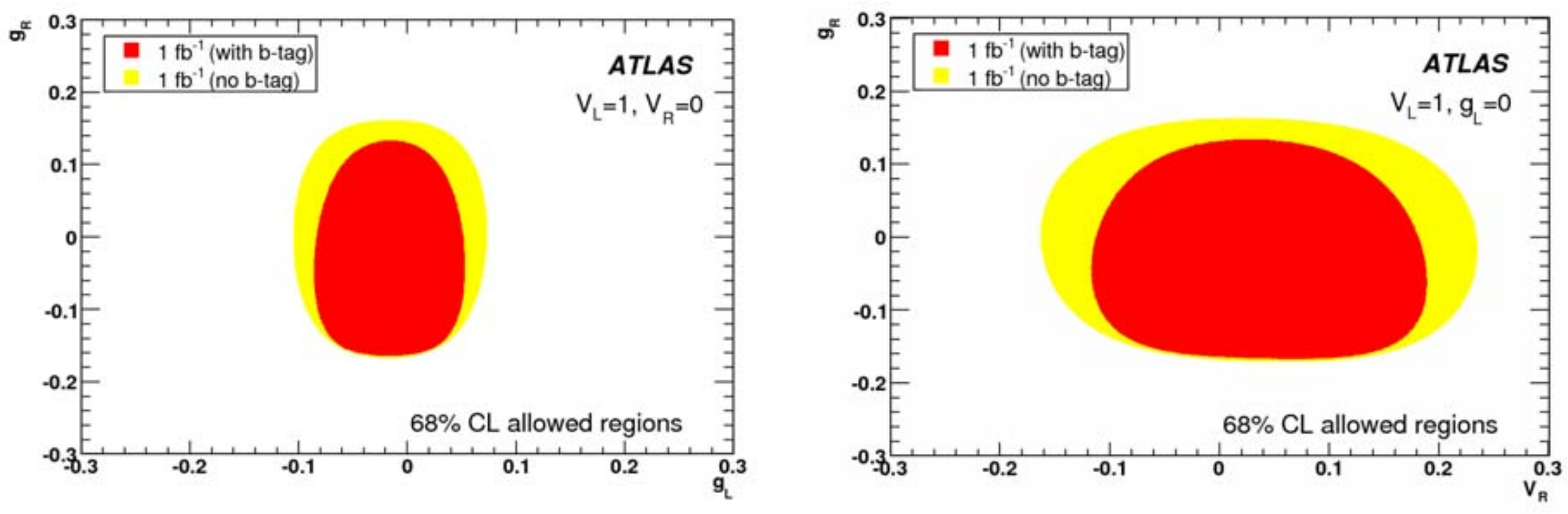


Figure 8: Expected 68% CL allowed regions on the Wtb anomalous couplings for luminosities of 1 fb$^{-1}$ (with and without b-tagging), obtained from the $\rho_{R,L}$ and $A_{\pm}$ observables using `TopFit`.

# 5 Rare Top Quark Decays and FCNC

This section discusses the study of rare top quark decays via FCNC ($t \to qX, X = \gamma, Z, g$) using $t\bar{t}$ events produced at the LHC. These decays are strongly suppressed in the Standard Model at tree level due to the GIM mechanism. In the effective Lagrangian approach [73, 74] the new top quark decay rates to the gauge bosons can be expressed in terms of the $\kappa^g_{tq}$, $\kappa^\gamma_{tq}$, $(|v^Z_{tq}|^2 + |a^Z_{tq}|^2)$ and $\kappa^Z_{tq}$ anomalous couplings to the g, $\gamma$ and Z bosons respectively, and $\Lambda$, the energy scale associated with the new physics.

## 5.1 Event samples

The signal event samples used in this analysis correspond to $t\bar{t} \to b\ell\nu qX$, where $X = \gamma, Z \to \ell\ell, g$ and $\ell = e, \mu$. The following common Standard Model samples were considered as background: fully hadronic, fully leptonic and semi-leptonic $t\bar{t}$, all from MC@NLO; Wt, $s$- and $t$- channels of single top production, all from AcerMC; W+jets, W$b\bar{b}$+jets and W$c\bar{c}$+jets, all from ALPGEN; $Z \to e^+e^-$, $Z \to \mu^+\mu^-$ and $Z \to \tau^+\tau^-$, all from PYTHIA; WW, ZZ and WZ, all from HERWIG.

## 5.2 Event selection

The $t\bar{t}$ final states corresponding to the different FCNC top quark decay modes lead to different topologies according to the number of jets, leptons and photons. There is however a common characteristic of all channels under study, *i.e.* in all of them one of the top quarks is assumed to decay through the dominant Standard Model decay mode $t \to bW$ and the other is forced to decay via one of the FCNC modes $t \to qZ$, $t \to q\gamma$ or $t \to qg$. The QCD backgrounds at hadron colliders make the search for the signal via the fully hadronic channels (when the W and the Z bosons decay to quarks) difficult. For this reason only the

Table 17: Selection cuts applied to the FCNC analyses. For the $t \rightarrow qg$ channel, $E_{vis}$, $p_{T_g}$ and $m_{qg}$ represent the total visible energy, the transverse momentum of the jet associated with the gluon and the reconstructed mass of the top quark with FCNC decay, respectively (see text for details).

| Channel | $t\bar{t} \rightarrow bWq\gamma$ | $t\bar{t} \rightarrow bWqg$ | $t\bar{t} \rightarrow bWqZ$ |
|---|---|---|---|
| Pre-selection | $= 1\ell$ ($p_T > 25$ GeV) | $= 1\ell$ ($p_T > 25$ GeV) | $= 3\ell$ ($p_T > 25, 15, 15$ GeV) |
| | $\geq 2j$ ($p_T > 20$ GeV) | $= 3j$ ($p_T > 40, 20, 20$ GeV) | $\geq 2j$ ($p_T > 30, 20$ GeV) |
| | $= 1\gamma$ ($p_T > 25$ GeV) | $= 0\gamma$ ($p_T > 15$ GeV) | $= 0\gamma$ ($p_T > 15$ GeV) |
| | $\not{p}_T > 20$ GeV | $\not{p}_T > 20$ GeV | $\not{p}_T > 20$ GeV |
| Final selection | $p_{T_\gamma} > 75$ GeV | $E_{vis} > 300$ GeV | 2 $\ell$ same flavour, oppos. charge |
| | | $p_{T_g} > 75$ GeV | |
| | | $m_{qg} > 125$ GeV | |
| | | $m_{qg} < 200$ GeV | |
| Trigger | e22i, mu20 or g55 | e22i or mu20 | e22i or mu20 |

Table 18: The trigger efficiency, in percentage, for the background and signal events after all the other cuts of the FCNC analyses.

| | $t \rightarrow q\gamma$ | | $t \rightarrow qZ$ | | $t \rightarrow qg$ | |
|---|---|---|---|---|---|---|
| | Sig. | Back. | Sig. | Back. | Sig. | Back. |
| Trigger | 99.6 | 99.5 | 99.2 | 95.0 | 83.2 | 82.2 |

leptonic decays of both W and Z to e and $\mu$ were taken into account. Only isolated muons, electrons and photons separated by $\Delta R > 0.4$ from other reconstructed objects, were considered. Specific pre-selection and selection cuts were applied for each FCNC channel, as outlined in Table 17. For the $t \rightarrow q\gamma$ channel, exactly one reconstructed lepton and one reconstructed photon where required in the events. Additionally, at least two jets were required. For the $t \rightarrow qg$ channel, the events had to have exactly one reconstructed lepton, three jets and no reconstructed photon. Finally, for the $t \rightarrow qZ$ channel, only the events with at least two jets, exactly three reconstructed leptons and no reconstructed photons were accepted. Therefore, the selection criteria for these channels are orthogonal. The expected number of background events and signal efficiencies after the final selection are shown in Table 19. The effect of the trigger on the background and signal events, after all the other cuts, is shown in Table 18. It should be noticed that the events of the $t \rightarrow q\gamma$ ($t \rightarrow qZ$) channel can be triggered by the lepton or the photon (one of the three leptons), which results on higher trigger efficiencies, when compared with the $t \rightarrow qg$ channel with just one lepton. For the $t \rightarrow q\gamma$ channel, the dominant backgrounds are $t\bar{t}$, Z+jets and W+jets events, which correspond to 38%, 30% and 29% of the total background. The total background for the $t \rightarrow qZ$ channel is mainly composed of $t\bar{t}$ and Z+jets events (59% and 28% of the total background, respectively), while for the $t \rightarrow qg$ it is mainly composed of W+jets and $t\bar{t}$ events (which correspond, respectively, to 64% and 25% of the total background).

For all the channels, the top quark with Standard Model semileptonic decay ($t \rightarrow b\ell\nu$) cannot be directly reconstructed due to the presence of an undetected neutrino in the final state. The neutrino four-momentum was estimated with a method similar to the one used in Section 4.2, by finding the $p_Z^\nu$ value and the jet combination (and the lepton combination in the case of the qZ topology) which minimizes the following expression:

$$\chi^2 = \frac{\left(m_t^{\rm FCNC} - m_t\right)^2}{\sigma_t^2} + \frac{\left(m_{\ell_a \nu j} - m_t\right)^2}{\sigma_t^2} + \frac{\left(m_{\ell_a \nu} - m_W\right)^2}{\sigma_W^2} + \frac{\left(m_{\ell_b \ell_c} - m_Z\right)^2}{\sigma_Z^2}, \qquad (10)$$

Table 19: The expected number of background events and signal efficiencies after the final selection level (the trigger is included) of the analyses for each FCNC channel. The corresponding statistical errors are also shown. The expected background numbers are normalised to $L = 1$ fb$^{-1}$.

| | e | $\mu$ | $\ell$ |
|---|---|---|---|
| $t\bar{t} \to bWq\gamma$: | | | |
| Total | $(4.4 \pm 0.6) \times 10^2$ | $(2.2 \pm 0.6) \times 10^2$ | $(6.5 \pm 0.7) \times 10^2$ |
| Signal % | $3.6 \pm 0.2$ | $4.1 \pm 0.2$ | $7.6 \pm 0.2$ |
| $t\bar{t} \to bWqZ$: | | | |
| Total | $(0.3 \pm 0.6) \times 10^2$ | $(0.1 \pm 0.6) \times 10^2$ | $(1.3 \pm 0.6) \times 10^2$ |
| Signal % | $1.4 \pm 0.1$ | $2.5 \pm 0.1$ | $7.6 \pm 0.2$ |
| $t\bar{t} \to bWqg$: | | | |
| Total | $(11.0 \pm 0.3) \times 10^3$ | $(8.3 \pm 0.2) \times 10^3$ | $(19.3 \pm 0.4) \times 10^3$ |
| Signal % | $1.3 \pm 0.1$ | $1.5 \pm 0.1$ | $2.9 \pm 0.1$ |

where $m_t^{FCNC}$, $m_{\ell_a \nu j}$, $m_{\ell_a \nu}$ and $m_{\ell_b \ell_c}$ are, for each jet and lepton combination, the reconstructed mass of the top quark decaying via FCNC, the top quark decaying through the Standard Model, the W-boson from the top quark with Standard Model decay and the Z boson from the top quark FCNC decay, respectively. The last term of Eq. 10 was only used in the $t \to qZ$ channel. The following values are used for the constraints: $m_t = 175$ GeV, $m_W = 80.42$ GeV, $m_Z = 91.19$ GeV, $\sigma_t = 14$ GeV, $\sigma_W = 10$ GeV and $\sigma_Z = 3$ GeV[4]. No b-tag information was used to reconstruct the event kinematics. The jet chosen to reconstruct the top quark with Standard Model decay is labeled as b quark. For the $t \to q\gamma$ and the $t \to qZ$ channels, the other jet, which was used to reconstruct the top quark with FCNC decay, is denoted by q quark. For the $t \to qg$ channel, it is assumed that the jet created by the gluon is the most energetic from the two which reconstruct the top quark with FCNC decay and the other is created by the light quark.

Following the selection cuts, a likelihood-based type of analysis was applied, as described in section 4.2. Due to the small statistics of the available full simulation samples, ATLFAST samples were used to obtain the probability density functions (p.d.f.s). The p.d.f.s of the $t \to q\gamma$ channel were built based on the following variables: the mass of the top quark with FCNC decay ($m_t^{FCNC}$); the reconstructed mass of the photon and the b quark ($m_{b\gamma}$) and the transverse momentum of the leading photon ($p_T^\gamma$). For the $t \to qZ$ channel, the p.d.f.s were based on the following physical distributions: the mass of the top quark with FCNC decay ($m_t^{FCNC}$); the minimum invariant mass of the three possible combinations of two leptons ($m_{\ell\ell}^{min}$); the reconstructed mass of the Z and the b quark ($m_{bZ}$); the reconstructed mass of the two quarks ($m_{qb}$); the transverse momentum of the third lepton ($p_T^{\ell_3}$) and the transverse momentum of the light quark ($p_T^q$). The following variables were used to build the p.d.f.s of the $t \to qg$ channel: the mass of the top quark with FCNC decay ($m_t^{FCNC}$); the mass of the top quark with Standard Model decay ($m_{\ell_a \nu j}$); the reconstructed mass of the light and the b quark ($m_{qb}$); the transverse momentum of the b quark ($p_T^b$); the transverse momentum of the light quark ($p_T^q$) and the angle between the lepton and the gluon ($\alpha_{\ell g}$). The distributions of the discriminant variables are presented in Figure 9. It can be seen that ATLFAST describes the fully simulated distributions fairly well, when there are sufficient statistics to tell.

[4]These resolutions are taken from the top quark mass measurement analyses [70]. The effect of changing their values was considered as one of the systematic source of uncertainties. Its impact on the observables is below the statistical error.

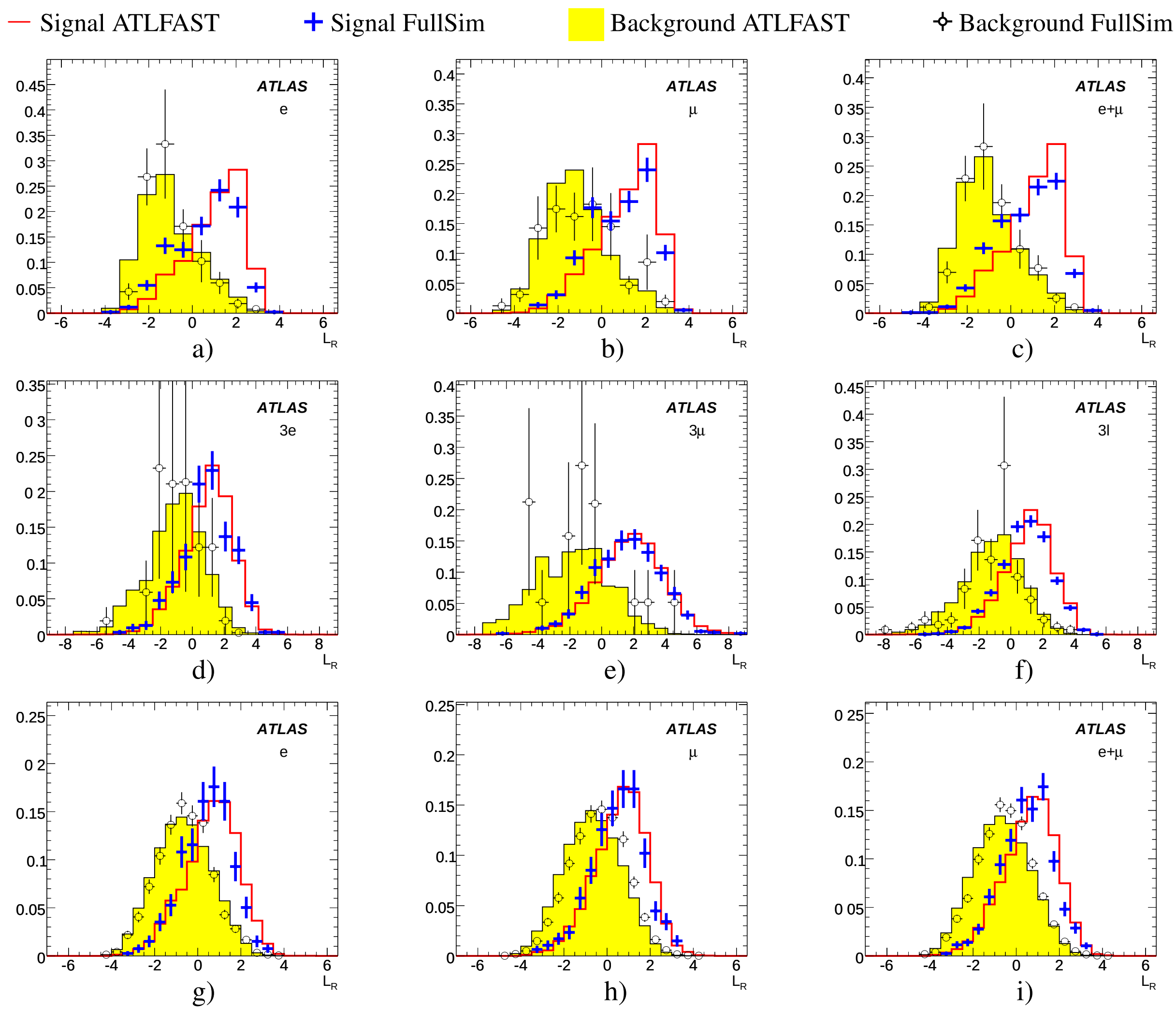


Figure 9: Distributions of the normalised discriminant variables for the expected background and signal after the final selection level of the $t \to q\gamma$ channel when the isolated lepton is identified as (a) an electron, (b) a muon and (c) electron or muon; of the $t \to qZ$ channel when the 3 isolated leptons are identified as (d) electrons, (e) muons and (f) electrons or muons and of the $t \to qg$ channel, when the isolated lepton is identified as (g) an electron, (h) a muon and (i) electron or muon.

Table 20: The expected 95% confidence level limits on the FCNC top quark decay branching ratio, in the absence of signal, are shown for a luminosity of $L = 1$ fb$^{-1}$. The central values are represented together with the $1\sigma$ bands, which include the contribution from the statistical and systematic uncertainties.

| | $-1\sigma$ | Expected | $+1\sigma$ |
|---|---|---|---|
| $t\bar{t} \to$ | bWq$\gamma$: | | |
| e | $4.3 \times 10^{-4}$ | $1.1 \times 10^{-3}$ | $1.9 \times 10^{-3}$ |
| $\mu$ | $4.5 \times 10^{-4}$ | $8.3 \times 10^{-4}$ | $1.3 \times 10^{-3}$ |
| $\ell$ | $3.8 \times 10^{-4}$ | $6.8 \times 10^{-4}$ | $1.0 \times 10^{-3}$ |
| $t\bar{t} \to$ | bWqZ: | | |
| 3e | $5.5 \times 10^{-3}$ | $9.4 \times 10^{-3}$ | $1.4 \times 10^{-2}$ |
| $3\mu$ | $2.4 \times 10^{-3}$ | $4.2 \times 10^{-3}$ | $6.4 \times 10^{-3}$ |
| $3\ell$ | $1.9 \times 10^{-3}$ | $2.8 \times 10^{-3}$ | $4.2 \times 10^{-3}$ |
| $t\bar{t} \to$ | bWqg: | | |
| e | $1.3 \times 10^{-2}$ | $2.1 \times 10^{-2}$ | $3.0 \times 10^{-2}$ |
| $\mu$ | $1.0 \times 10^{-2}$ | $1.7 \times 10^{-2}$ | $2.4 \times 10^{-2}$ |
| $\ell$ | $7.2 \times 10^{-3}$ | $1.2 \times 10^{-2}$ | $1.8 \times 10^{-2}$ |

### 5.3 Results and systematic uncertainties

In the absence of a FCNC top quark decay signal, expected limits at 95% CL were derived using the modified frequentist likelihood method [75] and the discriminant variables obtained with the full simulation samples. No cuts on the discriminant variables were used. Using the Standard Model $t\bar{t}$ production cross-section these limits were converted into limits on the branching ratios. The central values of these limits are shown in Table 20. The branching ratio sensitivity for each FCNC channel, assuming a signal discovery with a $5\sigma$ significance, is on average 3.0 times larger than the central values presented in Table 20.

Several sources of systematic uncertainties were studied following the common criteria used for this note and the results are shown in Table 21. The jet energy scale (the jet calibration used in the present analyses is described in Ref. [70]), the luminosity, the influence of the cross-section values, ISR/FSR, Monte Carlo generator and pile-up effects were studied. For the top quark mass, the full simulation samples were analysed using the values of 170 and 180 GeV (i.e. p.d.f.s, Eq. 10 and limit computation). The systematic error was taken from a linear fit of the values found, for a top quark mass error of $\pm 2$ GeV. To study the effect of the mass resolutions in Eq. 10, and since no cut is applied to the $\chi^2$ distribution, the ratio $\sigma_t/\sigma_W$ was changed by a factor 2. The total systematic uncertainties, computed as the quadratic sum of these individual contributions, are also shown in Table 21. The analysis stability was also cross-checked by varying the kinematic cuts by about 10%, and the maximum relative change on the expected 95% CL limits was 3%, 9% and 5% for the $t \to q\gamma$, $t \to qZ$ and $t \to qg$ channels, respectively. Table 20 shows the obtained central values for the BR limits. The $\pm 1\sigma$ contributions from statistical and systematic uncertainties (added in quadrature) are also shown. The contribution from the luminosity and the absolute value of background level may be reduced with data, by normalising to measured processes.

Figure 10 shows the ATLAS 95% CL expected sensitivity for the first fb$^{-1}$ in the absence of signal, for the $t \to q\gamma$ and $t \to qZ$ channels taking into account the contributions from both the statistical and systematic uncertainties.

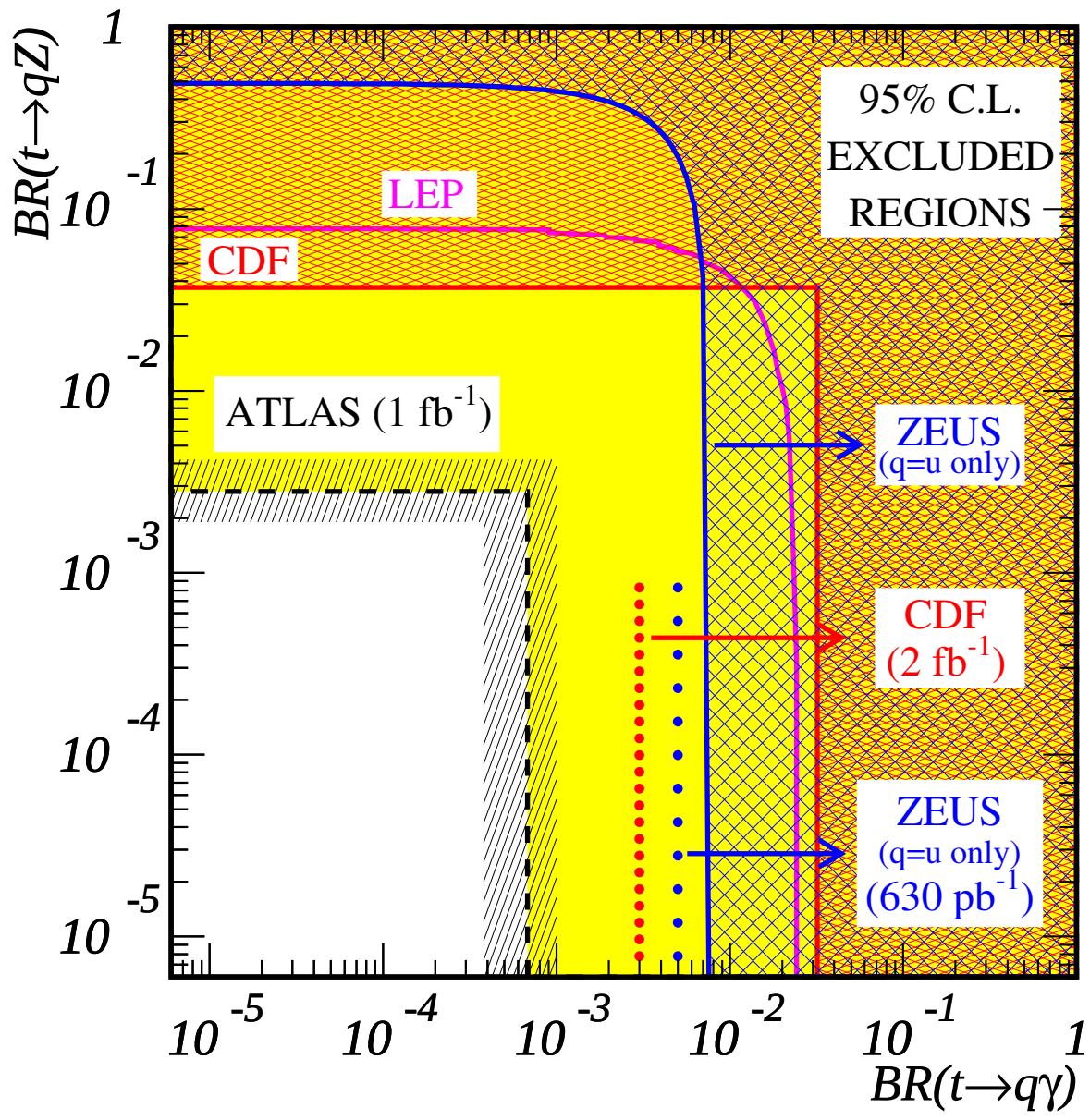


Figure 10: The present 95% CL observed limits on the $BR(\mathrm{t} \to \mathrm{q}\gamma)$ vs. $BR(\mathrm{t} \to \mathrm{qZ})$ plane are shown as full lines for the LEP, ZEUS and CDF collaborations. The expected sensitivity at ZEUS, CDF and ATLAS (together with the statistic plus systematic $1\sigma$ band) is also represented by the dotted and dashed lines.

Table 21: Maximum changes (with respect to the central values of Table 20) of the expected 95% CL limits for each FCNC top quark decay branching ratio for different systematic error sources.

| | $\mathrm{t} \to \mathrm{q}\gamma$ | | | $\mathrm{t} \to \mathrm{qZ}$ | | | $\mathrm{t} \to \mathrm{qg}$ | | |
|---|---|---|---|---|---|---|---|---|---|
| Source | e | $\mu$ | $\ell$ | 3e | $3\mu$ | $3\ell$ | e | $\mu$ | $\ell$ |
| Jet energy calibration | 1% | 2% | 2% | 3% | 2% | 5% | 4% | 4% | 4% |
| Luminosity | 9% | 8% | 10% | 3% | 2% | 6% | 10% | 8% | 10% |
| Top quark mass | 7% | 7% | 6% | 6% | 4% | 12% | 7% | 5% | 5% |
| Backgrounds $\sigma$ | 6% | 10% | 7% | 4% | 7% | 12% | 17% | 16% | 15% |
| ISR/FSR | 21% | 18% | 17% | 6% | 29% | 7% | 3% | 7% | 9% |
| Pile-up | 37% | 21% | 22% | 30% | 14% | 0% | 8% | 10% | 13% |
| Generator | 34% | 18% | 4% | 4% | 14% | 14% | 5% | 0% | 4% |
| $\chi^2$ | 5% | 0% | 4% | 2% | 5% | 7% | 3% | 7% | 9% |
| Total | 56% | 36% | 32% | 32% | 36% | 25% | 24% | 24% | 27% |

# 6 $t\bar{t}$ resonances

New resonances or gauge bosons strongly coupled to the top quark could manifest themselves in the $t\bar{t}$ invariant mass distribution. Due to the large variety of models and their parameters, studies have been done in a model-independent way searching for a "generic" narrow resonance decaying into $t\bar{t}$ (semileptonic channel) [51,76]. The ATLAS discovery potential is assessed using the latest available simulations and detector description, performing the $t\bar{t}$ event reconstruction along the same lines as done for the high precision measurement of the top quark mass in the semileptonic decay channel [54]. Alternative methods are currently under investigation to study the production of high mass resonances ($m > 2$ TeV), where the decay products of the top quark are close together, so requiring a modified selection.

## 6.1 Event generation and selection

In this study, $t\bar{t}$ resonances were produced with PYTHIA. The $Z' \to t\bar{t}$ channel without interference in the $Z'$ production has been chosen [77]. Only the semileptonic decay (electrons and muons) of $t\bar{t}$ pairs is considered. Five samples of $Z'$ resonances with mass 700 GeV (the lower limit from CDF and DØ measurements [78,79]), 1000 GeV, 1500 GeV, 2000 GeV and finally 3000 GeV have been produced.

The common $t\bar{t}$ semileptonic selection criteria were used to accept signal events. Once these criteria are applied, the main source of physical background to be considered in the search for $t\bar{t}$ resonances comes from Standard Model $t\bar{t}$ events. The other sources of background, dominated by W+jets events, are negligible [54].

## 6.2 Data characteristics

### 6.2.1 $t\bar{t}$ reconstruction

Several ways to reconstruct the W-bosons and the top quarks have been investigated [54]. The simplest of them is used here. This method selects, among all jets from light quarks (u,d,c and s), the two jets which are closest in $\Delta R$ to build the hadronic W-boson. The hadronic top quark is reconstructed combining the nearest b-jet to the hadronic W-boson. For the leptonic side, the constraint on the W-boson mass is used to compute the longitudinal momentum of the neutrino, identifying the missing transverse energy as the neutrino transverse momentum. Among the 2 $p_{z^\nu}$ solutions, the one providing the leptonic top quark mass closest to the mean value of the hadronic top quark mass is chosen. To reduce both the physical background and the combinatorial background (which dominates), cuts are applied on the hadronic W-boson mass spectrum, and on both top quark mass spectra. The resulting $t\bar{t}$ mass distribution, given in Figure 11, is used as a starting point for the discovery potential determination.

This method does not need a well understood jet energy scale and it is also one of the most efficient. More sophisticated methods to select the top quarks, such as a kinematic fits of the $t\bar{t}$-pairs (used by the DØ experiment [79]), or a matrix-element motivated method (used by CDF [78]), are not described here. The method used could be considered as a baseline for the $t\bar{t}$ resonance search.

### 6.2.2 Event yield

The reconstruction efficiency of Standard Model $t\bar{t}$ pairs corresponding to the reconstruction scheme and cuts described above is shown in Figure 13. The fact that the produced particles are closer together when the generated $t\bar{t}$ mass is higher explains the drop of efficiency observed as a function of the $t\bar{t}$ mass. Part of the efficiency can be recovered if another jet finding algorithm, better at resolving nearby jets (such as a cone jet algorithm with smaller cone radius), is used.

Figure 14 gives the reconstruction efficiency of $Z'$ resonances as a function of their mass. This reconstruction efficiency is of the same order as for the Standard Model $t\bar{t}$ pairs. Nevertheless, some

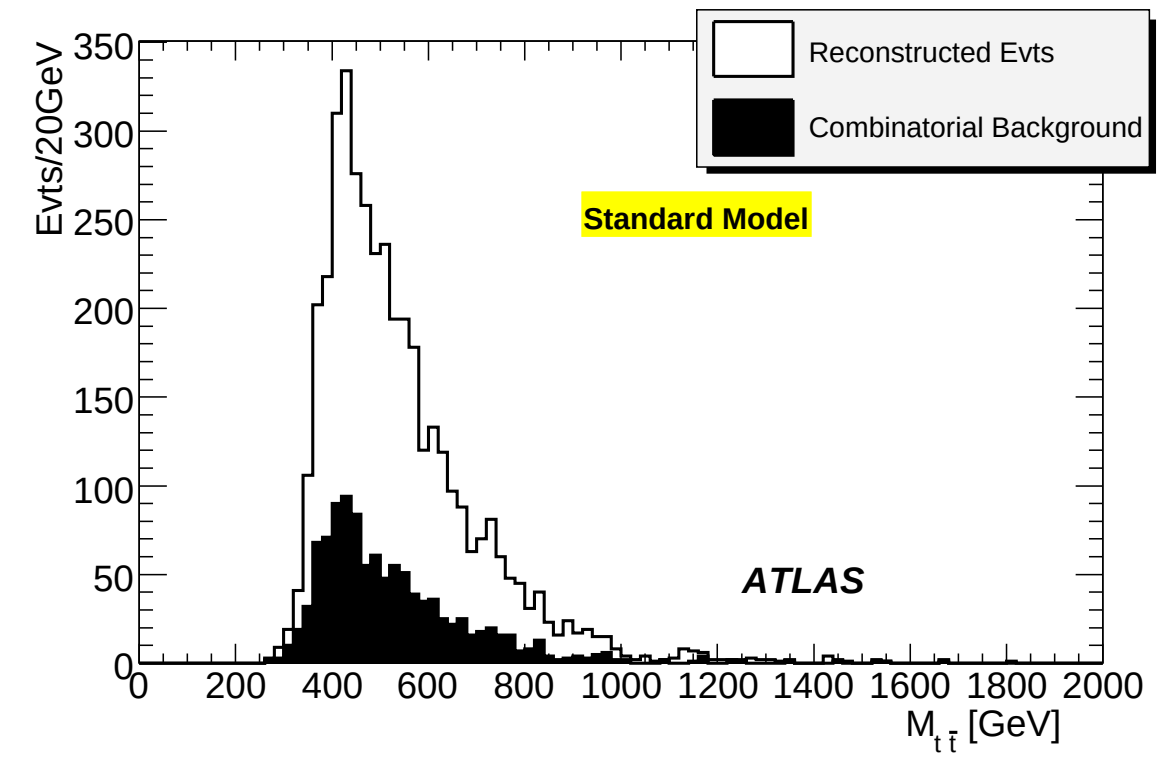


Figure 11: Standard Model $t\bar{t}$ mass spectrum. The black area represents the combinatorial background where at least one of the selected jets or leptons does not match the corresponding parton.

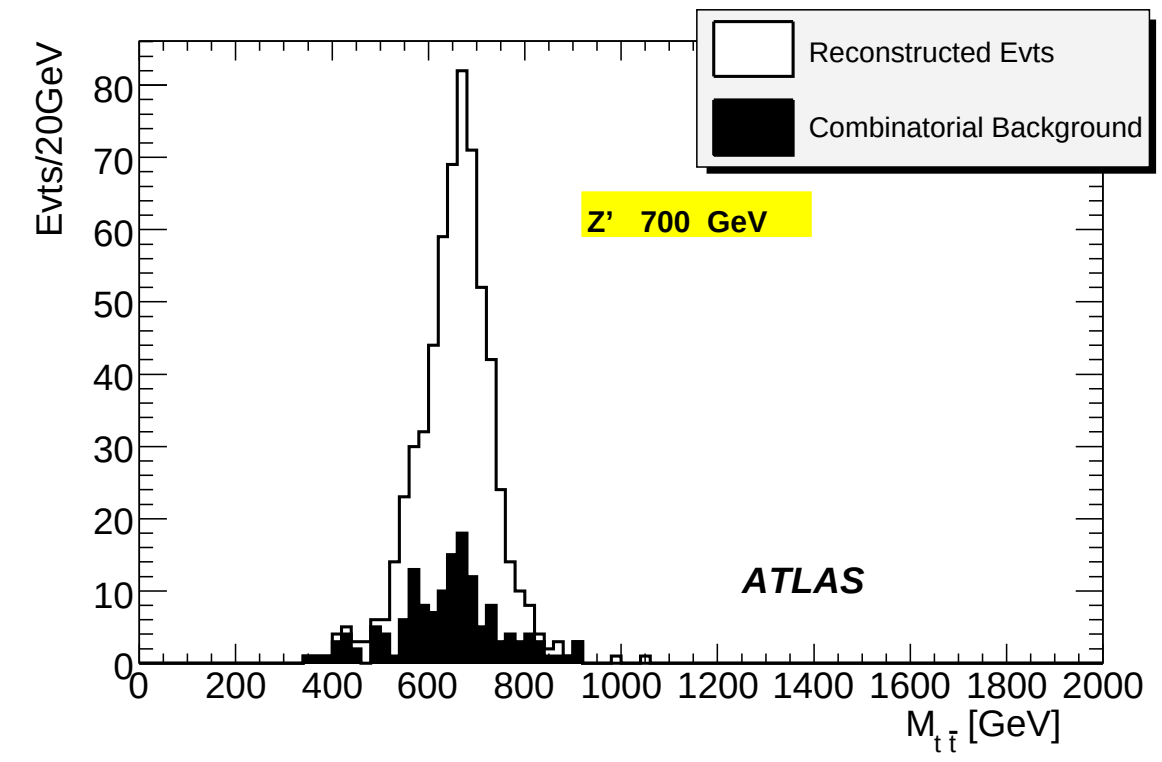


Figure 12: A 700 GeV $Z'$ reconstructed mass spectrum. The black area represents the combinatorial background where at least one of the selected jets or leptons does not match the correspondent parton.

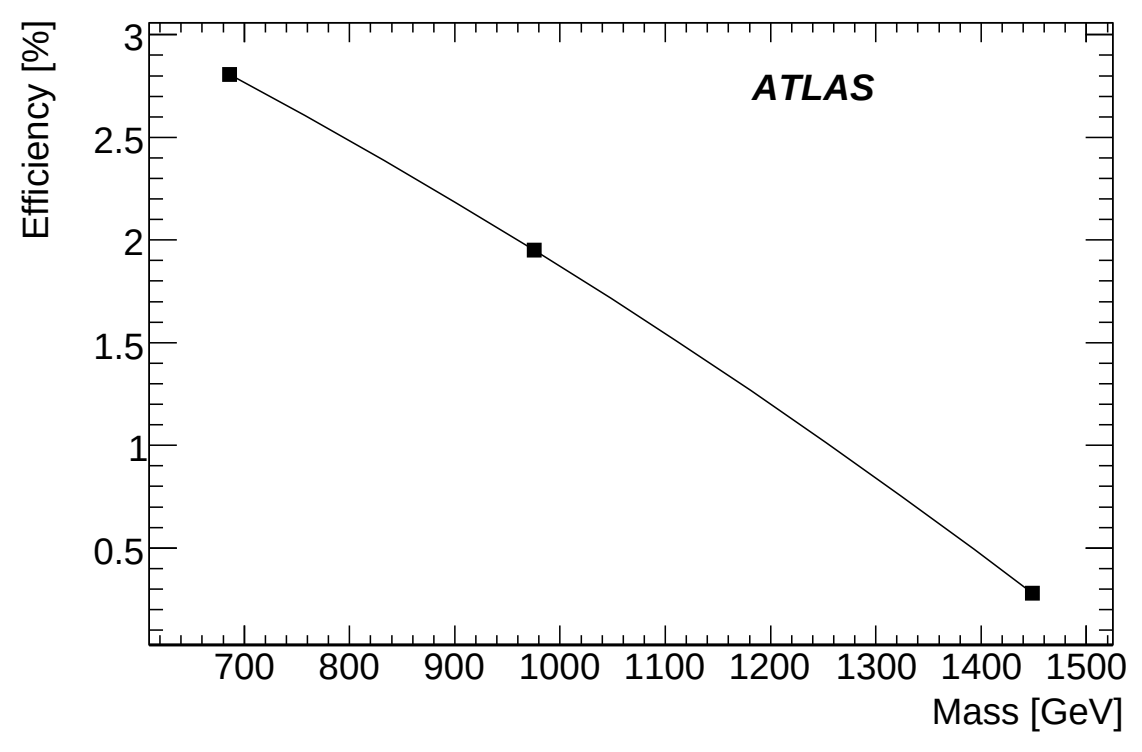


Figure 13: Reconstruction efficiency of Standard Model $t\bar{t}$ pairs as a function of the $t\bar{t}$ mass.

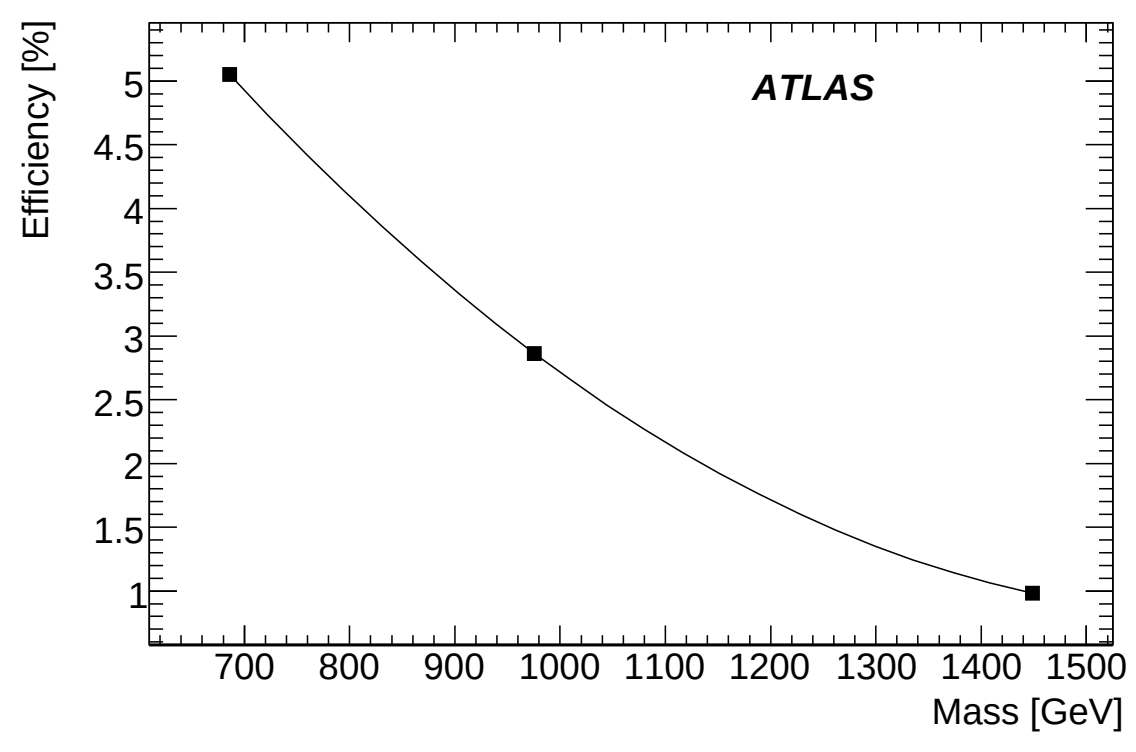


Figure 14: Reconstruction efficiency of $Z' \rightarrow t\bar{t}$ resonances as a function of the $Z'$ mass.

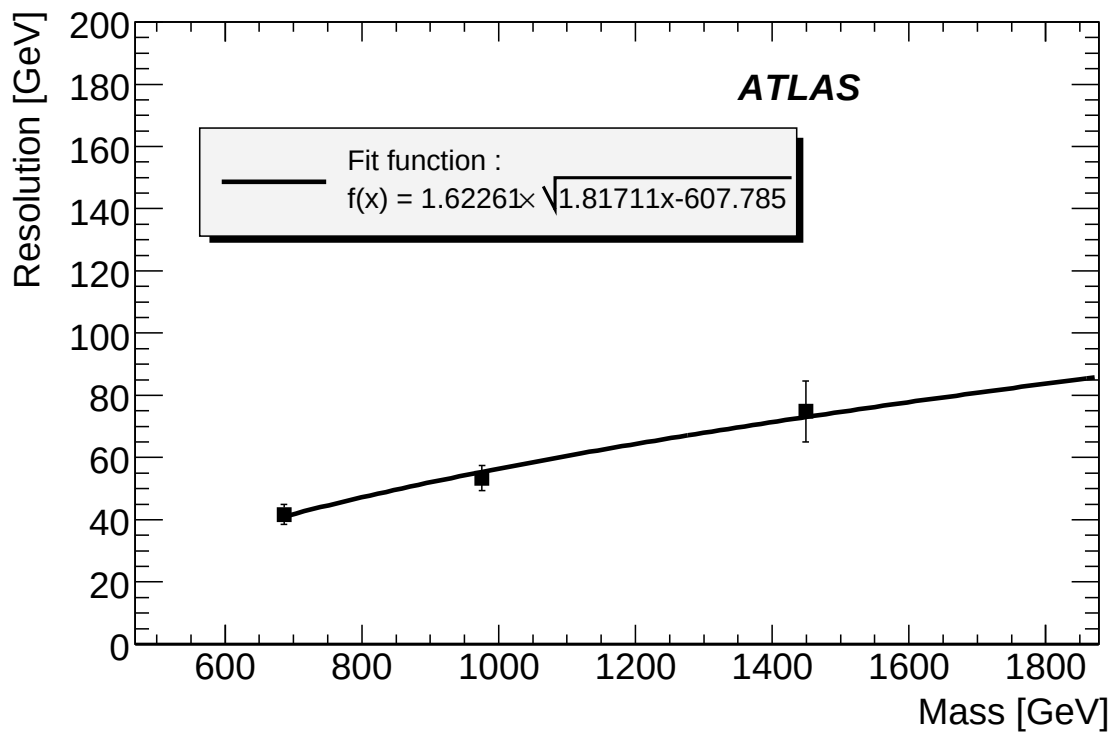


Figure 15: Z′ mass resolution as a function of the Z′ mass.

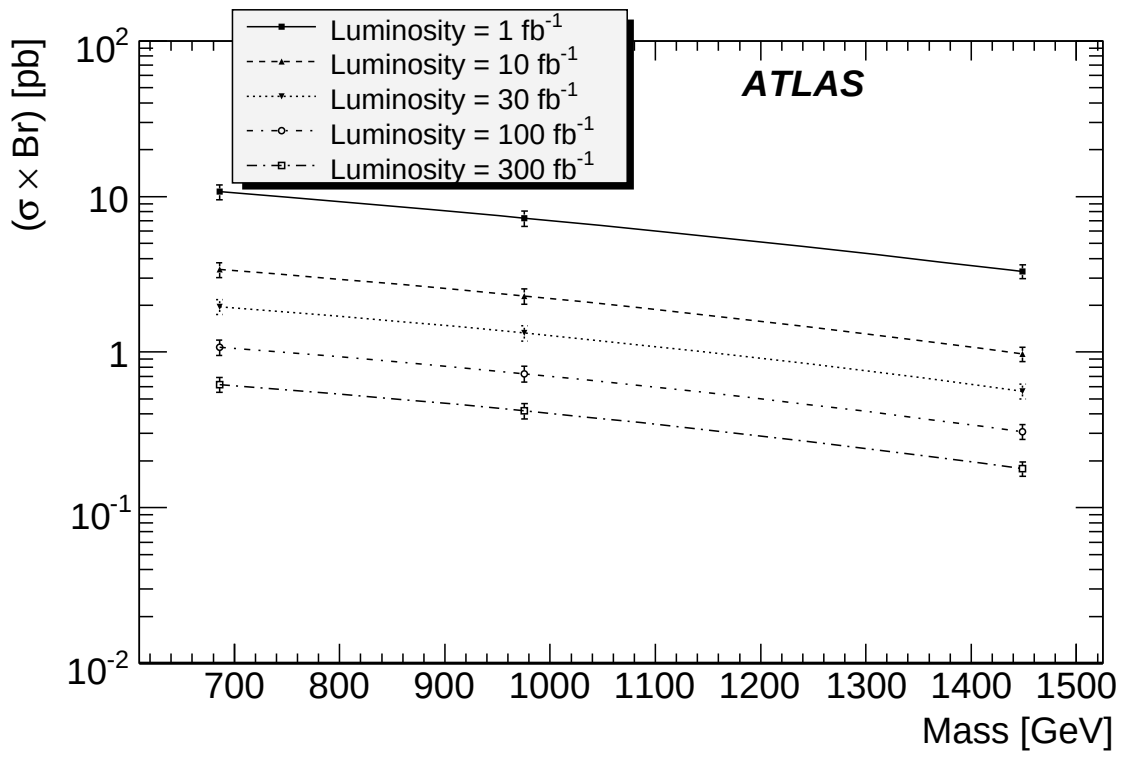


Figure 16: 5$\sigma$ discovery potential of a generic narrow t$\bar{\text{t}}$ resonance as a function of the integrated luminosity.

small differences arise due to the different production mechanism and spin. Thus, the final result is not completely model-independent. The purity (fraction of well reconstructed events among all selected events) in each Z′ sample is of the order of 80-85%.

## 6.3 Discovery potential

### 6.3.1 Method and results

The method used to extract the 5$\sigma$ discovery sensitivity consists of counting the number of Standard Model t$\bar{\text{t}}$ events in a sliding mass window over the invariant mass spectrum. The 5$\sigma$ sensitivity means that an effect is seen over the expected background with a deviation at least 5 times the background fluctuation in the mass window. The width of the window is twice the detector resolution for a given resonance mass. Then, the lowest cross section times branching ratio $\sigma \times Br(\text{Z}' \to \text{t}\bar{\text{t}})$ is computed for the discovery of a resonance at a given mass.

The produced resonances are expected to have a width smaller than the resolution, leading to a gaussian shape for the reconstructed invariant mass. The discovery potential is thus estimated here only for narrow t$\bar{\text{t}}$ resonances.

This method, explained in detail in [76], requires as input the Z′ mass resolution (Figure 15), the reconstruction efficiency of both Standard Model t$\bar{\text{t}}$ events and resonance events, and the purity of the final samples.

The resulting sensitivity is shown on Figure 16. For example, a 700 GeV Z′ resonance produced with a $\sigma \times Br(\text{Z}' \to \text{t}\bar{\text{t}})$ of 11 pb should be discovered with a 5$\sigma$ significance after 1 fb$^{-1}$ of data taking. The t$\bar{\text{t}}$ mass spectrum associated with such a case is shown in Figure 17.

### 6.3.2 Measurement uncertainties

Uncertainties on the sensitivity arise from:

- the reconstruction efficiency for the Z′ signal and t$\bar{\text{t}}$ background. The main contribution arises from the expected error on the b-tag efficiency, which is set to ±5%.
- the background contribution (Standard Model t$\bar{\text{t}}$). The main contribution comes from the t$\bar{\text{t}}$ cross section uncertainty $^{+6.2}_{-4.7}$ %.

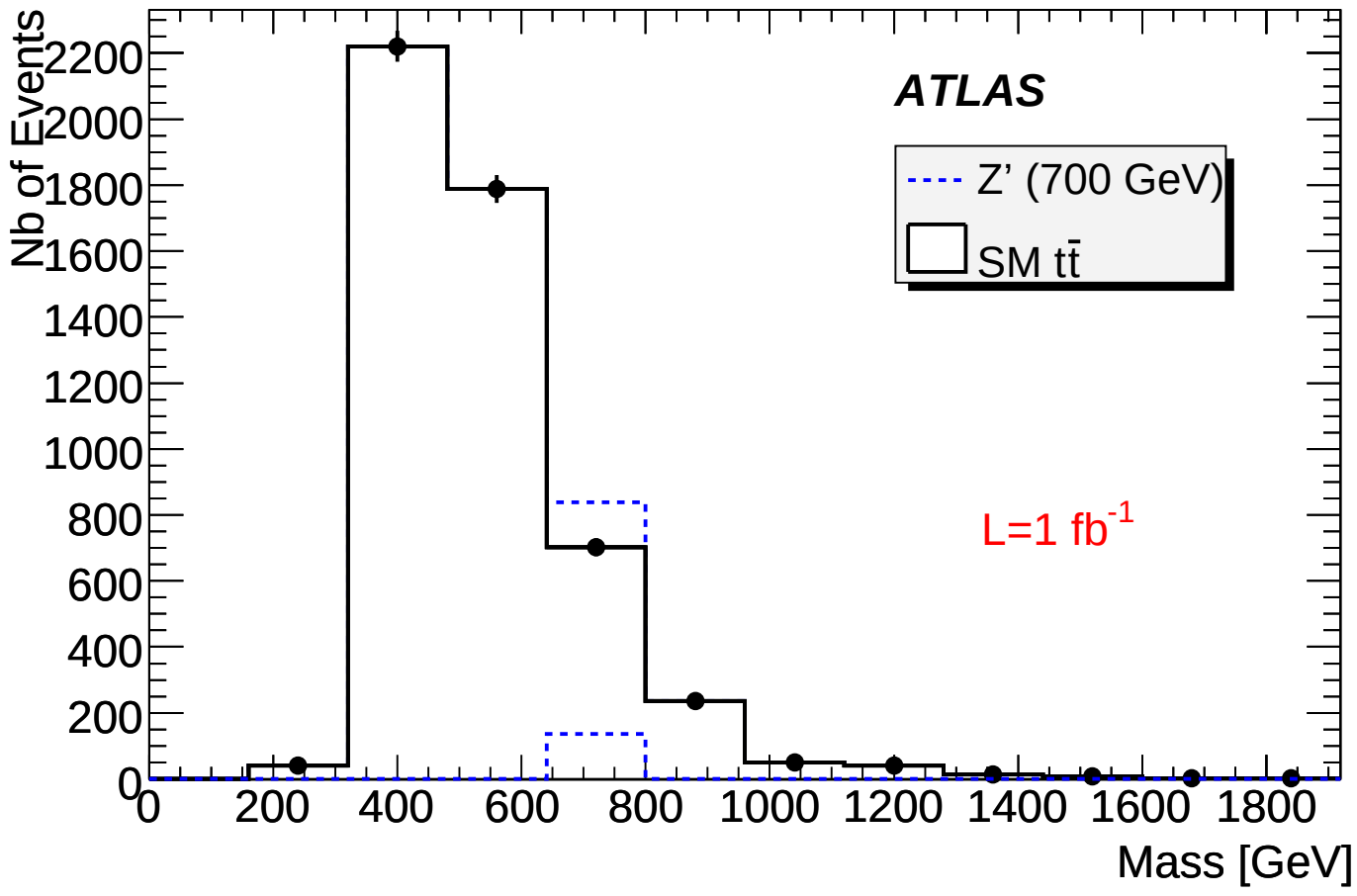


Figure 17: Expected $t\bar{t}$ invariant mass spectrum for the discovery threshold of a 700 GeV $Z'$ resonance $\sigma \times Br(Z' \to t\bar{t}) = 11.07$ pb with 1 fb$^{-1}$ of data.

Table 22: Sources of uncertainties considered on the $5\sigma$ discovery potential computation.

| Source of uncertainty | Error (%) | Effect on the discovery potential (%) |
|---|---|---|
| Reconstruction efficiency | 16.6 | 8.3 |
| Background contribution | $^{+6.2}_{-4.7}$ | 3.1 |
| $t\bar{t}$ mass resolution | $\pm 1\sigma_{\text{resolution}}$ | 2 to 11 |
| Luminosity | 5 | 2.5 |
| Jet energy scale | 5 | - |

- the $t\bar{t}$ mass resolution (Figure 15). The effect on the discovery potential increases with the resonance mass.
- the uncertainty on the integrated luminosity.

The $5\sigma$ discovery potential is given as a function of the $t\bar{t}$ resonance mass. The position of the resonance mass peak depends on the accuracy of the jet energy scale. This will be constrained by the W-boson and top quark events themselves.

All these quantities have been varied within expected errors. The impact on the sensitivity is reported in Table 22.

## 7 Summary and conclusions

The ATLAS sensitivity to the measurement of several top quark properties was reviewed in this paper for an expected luminosity of 1 fb$^{-1}$ at the LHC. The precision of several measurements was estimated, using the full simulation of the ATLAS detector. For the tests of physics beyond the Standard Model associated with the production of top quarks, the 95% CL limit (in the absence of a signal) was presented. In Table 23 a summary of the expected sensitivity on these observables is shown, for a luminosity of 1 fb$^{-1}$. Several sources of systematic errors were considered using an approach common to all studies appearing in this note.

Table 23: Expected ATLAS sensitivity for the top quark properties, for a luminosity of 1 fb$^{-1}$. For the Standard Model measurements the sensitivity is given as the total error divided by the expected Standard Model value, for the searches the absolute value is presented.

| Observables | Expected Precision |
|---|---|
| Top quark charge (2/3 versus -4/3) | $\geq 5\sigma$ |
| Spin Correlations: | |
| $A$ | 50% |
| $A_D$ | 34% |
| W-boson Polarisation: | |
| $F_0$ | 5% |
| $F_L$ | 12% |
| $F_R$ | 0.03 |
| Angular Asymmetries: | |
| $A_{FB}$ | 19% |
| $A_+$ | 11% |
| $A_-$ | 4% |
| Anomalous Couplings: | |
| $V_R$ | 0.15 |
| $g_L$ | 0.07 |
| $g_R$ | 0.15 |
| Top quark FCNC decays (95% C.L.): | |
| $Br(\mathrm{t} \to \mathrm{q}\gamma)$ | $10^{-3}$ |
| $Br(\mathrm{t} \to \mathrm{qZ})$ | $10^{-3}$ |
| $Br(\mathrm{t} \to \mathrm{qg})$ | $10^{-2}$ |
| $t\bar{t}$ Resonances (discovery): | |
| $\sigma \times Br$ ($m_{\mathrm{t\bar{t}}}$=700GeV) | $\geq$ 11 pb |

The sensitivity of the ATLAS experiment to the top quark charge measurement was evaluated. The analysis shows that using the weighting technique, already with 0.1 fb$^{-1}$ it is possible to distinguish with a $5\sigma$ significance, between the b-jet charges associated with leptons of opposite charges, which allows to distinguish the Standard Model from an exotic scenario. For the semileptonic b-decay method the required luminosity is $\simeq$ 1 fb$^{-1}$. The top quark charge itself was reconstructed relying on a Monte Carlo calibration of the b-jet charge. Although the reconstruction of the numeric value of the top quark charge using the weighting technique seems possible with $\simeq$ 1 fb$^{-1}$, it will be necessary to check the performance of the method with real data e.g. di-jet $\mathrm{b\bar{b}}$ events, once available. A more realistic treatment of the background processes will be required for a full understanding of the top quark charge issues.

A study of the W-boson polarisation fractions ($F_0$, $F_L$ and $F_R$) and $\mathrm{t\bar{t}}$ spin correlation parameters ($A$ and $A_D$) has been performed in the semileptonic $\mathrm{t\bar{t}}$ channel. Reconstructed angular distributions were used to set the ATLAS sensitivity to the measurement of the W-boson polarisations and top quark spin correlation parameters.

The W-boson polarisation ratios ($\rho_R$ and $\rho_L$) and the angular asymmetries ($A_+$ and $A_-$) dependence on the anomalous couplings ($V_R$, $g_L$ and $g_R$) was used to find the sensitivity to the Wtb anomalous couplings (for the analyses with and without b-tag).

Top quark rare decays through FCNC processes ($\mathrm{t} \to \mathrm{qZ}, \mathrm{q}\gamma, \mathrm{qg}$) were studied in this note using $\mathrm{t\bar{t}}$ events produced at the LHC. Expected limits on the branching ratios were set at 95% CL in the absence

of signal.

The discovery potential of the ATLAS experiment for narrow $t\bar{t}$ resonances decaying in the semileptonic channel, was studied as a function of the resonance mass.

# $B$-Physics

# Introduction to $B$-Physics

## 1 ATLAS $B$-physics programme

The ATLAS $B$-physics programme covers many aspects of beauty flavour physics. First, by measuring production cross-sections of beauty and charm hadrons and of the heavy-flavour quarkonia, $J/\psi$ and $\Upsilon$, ATLAS will provide sensitive tests of QCD predictions of production in proton-proton collisions at the LHC. Secondly, ATLAS will study the properties of the entire family of $B$ mesons ($B_d^0$, $B^+$, $B_s^0$, $B_c$ and their charge-conjugate states) and $B$ baryons, thereby broadening our knowledge of both the spectroscopic and dynamical aspects of $B$-physics. However, the main emphasis will be on precise measurements of weak $B$ hadron decays. In the Standard Model, all flavour phenomena of weak hadronic decays are described in terms of quark masses and the four independent parameters in the Cabibbo-Kobayashi-Maskawa (CKM) matrix [1]. Enormous quantities of data collected in the past decade by the experiments BaBar, Belle, CDF and D0 allowed very precise measurements of flavour and CP-violating phenomena. Whilst the analysis of the remaining data of these experiments may still push the boundaries, no evidence of physics beyond the Standard Model, nor any evidence for CP violation other than that originating from the CKM mechanism, has yet been found. At the LHC, thanks to the large beauty production cross-section and the high luminosity of the machine, the sensitivity of $B$ decay measurements is expected to substantially improve. Whilst direct detection of new particles in ATLAS will be the main avenue to establish the presence of new physics, indirect constraints from $B$ decays will provide complementary information. In particular, precise measurements and computations in $B$-physics are expected to play a key role in constraining the unknown parameters of any new physics model emerging from direct searches at the LHC.

In ATLAS, the main $B$-physics measurements will be made with an instantaneous luminosity of around $10^{33}$ cm$^{-2}$ s$^{-1}$; however, the $B$-physics potential begins during early data taking at low luminosity ($10^{31}$ cm$^{-2}$ s$^{-1}$). With an integrated luminosity of 10 pb$^{-1}$ ATLAS will already be able to register about $1.3 \cdot 10^5$ events containing $J/\psi \to \mu^+\mu^-$ selected by the low luminosity trigger menu [2]. Recorded events will contain $J/\psi \to \mu^+\mu^-$ produced both directly in proton-proton interactions as well as indirectly from decays of $B$ hadrons. With these statistics, beauty and quarkonia studies will play an important role in the early data-taking period.

In this document we present studies for several periods of beauty measurements in ATLAS. First, there will be a period of integrated luminosity of order of 10 - 100 pb$^{-1}$ when $B$-physics and heavy-flavour quarkonia signatures will serve in helping to understand detector properties and the muon trigger, as well as measuring production cross-sections. The physics analyses in this period will deal with prompt $J/\psi$ and $\Upsilon$ events, along with inclusive $B$ hadron decays to muon pairs via $J/\psi$. Further on, the exclusive decays of $B^+ \to J/\psi K^+$, $B_d^0 \to J/\psi K^{0*}$ and $B_s^0 \to J/\psi\phi$ will be studied.

During the next period, from about 200 pb$^{-1}$ to 1 fb$^{-1}$, we expect to collect the same or higher statistics as are currently available at the Tevatron. During this period we will start to improve upon current measurements of $B$ hadron properties and set new decay rate limits or possibly give evidence for rates above Standard Model predictions for rare decays (e.g. in the channel $B_s^0 \to \mu^+\mu^-$).

In the most important period for ATLAS $B$-physics we expect to achieve about 10 - 30 fb$^{-1}$ at an instantaneous luminosity of mostly around $10^{33}$ cm$^{-2}$ s$^{-1}$. It is expected that ATLAS can achieve this integrated luminosity in about three years. We are preparing to study a large variety of $B$-physics topics covering both the production and decay properties of $B$ hadrons. In this document we give examples of performance studies for this period with polarization measurements of heavy-flavour quarkonia and of the baryon $\Lambda_b$, by the oscillation phenomena of the $B_s^0 - \overline{B_s^0}$ system, and the rare

decay measurement $B_s^0 \to \mu^+\mu^-$. We expect to achieve sensitivities allowing the confirmation of possible contributions of physics beyond the Standard Model.

## 2 Trigger

All $B$-physics studies reported in the current document include trigger reconstruction. The ATLAS trigger comprises three levels and the selection of $B$-physics events is initiated by a di-muon or a single-muon trigger at the first level trigger (L1). At $10^{31}$ cm$^{-2}$ s$^{-1}$ the lowest possible threshold of about $p_T >$ 4 GeV will be used, rising to 6 - 8 GeV at $10^{33}$ cm$^{-2}$ s$^{-1}$ for at least one of the L1 muons. The muon is confirmed at the second trigger level (L2), first using muon-chamber information alone, and then combining muon and inner-detector (ID) information. The use of the more precise muon information available at L2 allows rejection of below-threshold muons that passed the L1 trigger. Combining information from the muon chambers and ID track segments gives rejection of muons from $\pi$ and $K$ decays.

Following the confirmation of the L1 muon(s), cuts on invariant mass and secondary vertex reconstruction of $B$ decay products are used to select specific channels of interest. Channels such as $B^+ \to J/\psi(\mu^+\mu^-)K^+$ and $B_s^0 \to \mu^+\mu^-$ are triggered by requiring two muons fulfilling $J/\psi$ or $B_s^0$ mass cuts. At $10^{31}$ cm$^{-2}$ s$^{-1}$ a single muon is required at L1, with the second muon either originating from an additional L1 trigger or found at L2 in an enlarged Region of Interest (RoI) around the trigger muon. At luminosities above about $10^{33}$ cm$^{-2}$ s$^{-1}$, a L1 dimuon trigger is used, giving an acceptable rate while keeping a low $p_T$ threshold for both of the two muons.

For hadronic final states such as $B_s^0 \to D_s\pi$, ID tracks are combined to reconstruct first the $\phi \to K^+K^-$, then the $D_s \to \phi\pi$ , and finally the $B_s^0$. Two different strategies are used for finding the tracks, depending on luminosity. Full reconstruction of the whole ID can be performed at $10^{31}$ cm$^{-2}$ s$^{-1}$. At higher luminosities, reconstruction will be limited to regions of interest defined by the L1 calorimeter jets.

In all studies reported in the current document the trigger decision was used to accept or reject a given event. The trigger menu used to produce the datasets contained low-$p_T$ L1 muon thresholds of 4, 5 and 6 GeV. These were used to initiate selections at L2 of decays containing $J/\psi \to \mu^+\mu^-$, like $B^+ \to J/\psi(\mu^+\mu^-)K^+$, and $B_s^0 \to \mu^+\mu^-$. The lowest thresholds for $J/\psi$ or $B_s^0$ required two 4 GeV muons. In addition there were selections for $B_s^0 \to D_s\pi$ based either on full ID reconstruction or on a RoI-based reconstruction in L1 Jet RoIs with a 4 GeV threshold. More detailed information on the $B$-physics trigger is presented in three studies in this document. Specific questions on the L1 di-muon trigger performance for $B$-physics are addressed in Section 1 of this chapter. L2 triggering on muons and di-muons is presented in Section 2. Finally, triggers for $B$ decays to purely hadronic final states have been studied in the last part of this chapter, Section 8.

If not stated otherwise, the trigger efficiencies in di-muon events were calculated with respect to Monte Carlo samples generated with cuts of $p_T > 6$ GeV and $p_T > 4$ GeV for the first and second muon respectively and with pseudorapidity cuts of $|\eta| < 2.4$ on both muons.

## 3 Simulation of the events

Some 1 million $B$ hadron events, along with around 400 000 prompt $J/\psi$ and $\Upsilon(1S)$ as well as $c\bar{c}$ events, were produced using PYTHIA 6.4 [3]. The simulations were performed with the CTEQ6L set of parton distribution functions. For quarkonium production, the NRQCD matrix element parameters

in PYTHIA were tuned to fit Tevatron data [4]. In the production of non-resonant $b\bar{b}$ and $c\bar{c}$ flavour-creation, PYTHIA models describing flavour-excitation and gluon-splitting were included. The fragmentation of $b$ quarks and $c$ quarks to hadrons was simulated according to the Peterson fragmentation function with parameter 0.006 and 0.05 respectively. The choice of parameters was motivated by Tevatron measurements [5]. Kinematic selections on final state particles from $B$ decays were applied so that most of the generated $B$ events passed the trigger threshold at the reconstruction stage.

The total $b\bar{b}$ or $c\bar{c}$ cross-section is not well defined in PYTHIA when one includes processes other than those of the lowest order for $b\bar{b}$ ($c\bar{c}$) production, since PYTHIA takes the partons to be massless and therefore the cross-section diverges when the transverse momentum approaches zero. However, only part of the cross section is relevant for our studies - in the phase space of events passing $B$ triggers.

Table 3 summarises the predicted single and di-muon cross-sections from charm, beauty and onia production expected at ATLAS from PYTHIA with cuts on the transverse momenta of the muons of 6 or 4 GeV (as appropriate) and pseudorapidity cuts on both muons of $|\eta| < 2.4$. The cuts reflect the trigger thresholds and were selected to allow most of the simulated events to be accepted by the trigger.

| Process ($\mu$6 threshold) | Cross-section | | Process ($\mu$4 threshold) | Cross-section | |
|---|---|---|---|---|---|
| $bb \to \mu 6X$ | 6.1 | $\mu$b | $bb \to \mu 4X$ | 19.3 | $\mu$b |
| $cc \to \mu 6X$ | 7.9 | $\mu$b | $cc \to \mu 4X$ | 26.3 | $\mu$b |
| $bb \to \mu 6\mu 4X$ | 110.5 | nb | $bb \to \mu 4\mu 4X$ | 212.0 | nb |
| $cc \to \mu 6\mu 4X$ | 248.0 | nb | $cc \to \mu 4\mu 4X$ | 386.0 | nb |
| $pp \to J/\psi(\mu 6\mu 4)X$ | 23.0 | nb | $pp \to J/\psi(\mu 4\mu 4)X$ | 28.0 | nb |
| $pp \to \Upsilon(\mu 6\mu 4)X$ | 4.6 | nb | $pp \to \Upsilon(\mu 4\mu 4)X$ | 43.0 | nb |
| $bb \to J/\psi(\mu 6\mu 4)X$ | 11.1 | nb | $bb \to J/\psi(\mu 4\mu 4)X$ | 12.5 | nb |

Table 1: Predicted PYTHIA cross-sections for various muon and di-muon sources. The numbers following each symbol $\mu$ denote the $p_T$ thresholds that were used in generating the events with PYTHIA.

After PYTHIA simulation, the events were passed through detector simulation (based on GEANT4 [6]) which first modelled the behavior of the particles as they passed through the detector, and simulated the response of the active detector components to the energy deposition by these particles. The layout of the detector used by the simulation code was published in Ref. [7]. The output of the simulation was then reconstructed as if it were real data. Following this, reconstructed muons and hadrons were analysed using the dedicated $B$-physics analysis software.

## 4 Organization of the $B$-physics chapter

The following reports on $B$-physics present our best current understanding of the ATLAS $B$-physics programme. As the $B$ trigger plays a role in all the studies presented here, the chapter starts with two reports, in Sections 1 and 2, dealing with L1 and L2 muon triggers respectively. There follow three physics studies typical for the early data period: physics of heavy-flavour quarkonia (Section 3), measurements of beauty cross-sections (Section 4) and early physics and performance measurements with decays of $B_d^0$ and $B_s^0$ mesons in Section 5. Finally there are Sections 6, 7 and 8, which give typical examples of ATLAS $B$ measurements that can only be achieved during the more advanced data taking period; they are measurements of the polarization of the baryon $\Lambda_b$, the rare decay $B_s^0 \to \mu^+\mu^-$ and

finally the oscillation measurement of the $B_s^0 - \overline{B_s^0}$ system.

# Performance Study of the Level-1 Di-Muon Trigger

**Abstract**
An event with two muons in the final state is a distinctive signal and can be triggered efficiently with the use of the level-1 di-muon trigger. Nevertheless triggering is still an issue if these muon tracks are fairly soft and fake di-muon triggers originating from muons that traverse more than one region of the trigger chambers increase the trigger rate. It is important to provide an acceptable trigger rate, while keeping high trigger efficiency to study low-$p_T$ $B$-physics such as rare $B$ hadron decays or CP violation in the $B$-events, especially in a multi-purpose experiment like ATLAS. In this note, the level-1 di-muon trigger and its expected performance are described.

## 1 Introduction

ATLAS is a multi-purpose experiment and its main focus is the direct search for and study of physics beyond Standard Model. However the indirect search for new physics will also play an important role in revealing the (flavor) structure of a non-trivial Higgs sector, which cannot be obtained by direct searches.

The $B^0_{s,d} \to \mu^+\mu^-$ rare decay is one of the interesting signatures which are sensitive to new physics at the TeV energy scale. The precise measurement of the forward-backward asymmetry [1] or branching ratio in the semi-muonic $B$ decay $B \to \mu^+\mu^- X$ requires good understanding of the possible biases introduced by event selection. Also important are precise measurements of Standard Model parameters, including CP-violation in $B$-events, such as $B_s^0 \to J/\psi\phi$ and $B_d^0 \to J/\psi K_s^0$. The key to the detection of these $B$ signals in ATLAS is to achieve a high trigger efficiency for low-$p_T$ di-muon events, keeping an acceptable trigger rate. It is also essential to understand the acceptance and efficiency using $pp \to J/\psi(\mu^+\mu^-)X$ or $pp \to \Upsilon(\mu^+\mu^-)X$ [2].

First, the level-1 muon trigger is briefly explained in Section 2, then trigger simulation and MC samples used in this note are described in Section 3. The performance of the level-1 single- and di-muon trigger as well as the effect of an algorithm to resolve muons traversing more than one trigger chambers are described in Section 4. Finally, the performance and the impact of various level-1 muon trigger configurations on some example $B$ signal events are discussed in Section 5.

## 2 Level-1 muon trigger

The ATLAS trigger system architecture is organised in three levels; level-1, level-2 and event filter. The level-2 and event filter triggers provide a software-based event selection after the level-1 trigger and events accepted by this trigger chain are finally reconstructed and analyzed offline.

The ATLAS level-1 muon trigger is based on dedicated, fast and finely segmented muon chambers (RPC and TGC) [3] and a trigger logic implemented in hardware. The muon trigger system provides acceptance in pseudo-rapidity up to $|\eta| \sim 2.4$ and in the full azimuthal angle ($\phi$) range. A muon track is triggered by the coincidence of two or three detector stations (consists of chamber doublet or triplet) within a certain road (coincidence window). The transverse momentum of a muon candidate is determined by its deviation from the trajectory of an infinite-momentum track (i.e. a straight line). A three station coincidence is required for any $p_T$ threshold in the endcap and forward regions ($|\eta| > 1.05$) to avoid high trigger rates caused by accidental coincidences due to background.[1] As a result, the acceptance at large $|\eta|$ becomes small for low-$p_T$ muons, as shown in Section 4.

[1]The number of stations used for coincidence is programmable giving the possibility of lowering a $p_T$ threshold at low luminosity if background conditions allow it.

The granularity of the level-1 muon trigger, the size of a Region of Interest (RoI), is $\Delta\eta \times \Delta\phi = 0.1 \times 0.1$ in the barrel region ($|\eta| < 1.05$) and $\sim 0.03 \times 0.03$ in the endcap regions ($1.05 < |\eta| < 2.4$). In short, if two muons leave tracks in the same RoI, they are counted as a single muon candidate by the trigger system. The level-1 muon trigger decisions from different regions are sent to the Muon Central Trigger Processor Interface (MuCTPI) which combines the information and calculates the multiplicity of muon candidates for each $p_T$ threshold over the whole detector. In forming the multiplicities, care is taken to avoid double counting of single-muon tracks, reconstructed separately in adjacent trigger sectors, while retaining a high efficiency for genuine di-muons. The overlap handling is carried out using Lookup Tables (LUT) to give flexibility and programmability. The LUTs are generated based on Monte Carlo simulations of single-muon events, selecting the regions of the level-1 muon system that can be traversed by a single muon.

The overlap handling is mandatory to avoid unacceptably high trigger rates caused by doubly-counted single muons, so called *fake di-muon triggers*. Most of the overlaps are resolved by the MuCTPI, which takes the muon candidate with the higher $p_T$ into account when calculating the muon multiplicity and finding overlapping muon candidates. The MuCTPI consists of independent 16 octant modules and overlap resolving is performed in each module. Therefore, overlapped regions connected to two different octant modules cannot be solved in MuCTPI and should be treated in each subdetector (RPC and TGC) by masking channels or flagging a overlap bit. The parameters used in the overlap handling, such as the combination of RoIs and masked channels are programmable. More details are described in Refs. [3, 4].

## 3 Monte Carlo samples and trigger configuration

Single-muon Monte Carlo samples with various, fixed $p_T$ values are used for performance tests of the level-1 single muon trigger logic. Single-muon events are generated uniformly over the full azimuthal angle range and $|\eta| < 2.7$ with a fixed $p_T$ value. The detector response is simulated using Geant4 [4] with the ATLAS geometry in the Athena framework [5]. Trigger simulation is performed using the trigger configuration for luminosity of $10^{31}$-$10^{33}$cm$^{-2}$s$^{-1}$ running, with a set of $p_T$ thresholds of 4, 5, 6, 11, 20 and 40 GeV. The production vertex is smeared with a Gaussian distribution with $\sigma_x = \sigma_y = 15\ \mu$m and $\sigma_z = 56$ mm. The level-1 single- and di-muon trigger menu items are named as MUx (single-muon trigger) and 2MUx (di-muon trigger with the same threshold for both muons) respectively, where x represents the $p_T$ threshold. The exceptions are MU0 and 2MU0, which correspond to completely opened coincidence windows. However, even in this case, the acceptance of the coincidence windows is limited by connectivity as well as by the dimension of the coincidence matrix ASICs [6].

The trigger efficiencies of various level-1 trigger configurations are also studied using $B$-physics Monte Carlo events. Event generation is done using PYTHIA [7]. Only interesting events associated with at least two muons with $p_T > 6$ GeV (the leading muon) and 4 GeV (the second muon) within $|\eta| < 2.5$ are studied. The $B$-physics samples used in this note are $B_s{}^0 \to \mu^+\mu^-\phi$ (47,650 events), $B_s{}^0 \to \mu^+\mu^-$ (47,450 events) and $B^+ \to J/\psi(\mu^+\mu^-)K^+$ (49,250 events). The last sample is chosen partly as a control sample. The opening angle of the two muons is smaller in this sample than in the other two samples.

In this note, only the level-1 muon trigger is simulated and events triggered by level-1 are studied.

## 4 Level-1 muon trigger efficiency in single-muon events

Detailed performance studies of the single muon trigger were performed in the barrel and endcap regions separately and the results are described in Ref. [8]. The single muon trigger efficiencies as a function of the transverse momentum of the simulated muon, over the whole muon trigger system are shown in

Figure 1. The single muon trigger efficiency is defined by:

$$\varepsilon(\text{MUx}) = \frac{\text{\# of events triggered}}{\text{\# of single muon events with } |\eta| < 2.5}\,. \qquad (1)$$

The range of pseudo-rapidity applied in the efficiency calculation is $|\eta| < 2.5$ and not 2.4 as used in Ref. [8], since $|\eta| < 2.5$ is applied for muons in the $B$-physics samples (Muons with $|\eta| > 2.4$ can be triggered if their momentum is relatively low and the track is bent towards smaller $|\eta|$). The efficiency, $\varepsilon$(MUx), includes all contributions from geometrical acceptance and coincidence window coverage. The trigger efficiency of MU5 and MU6 is lower than MU0 even for high $p_T$ muon tracks. This is because all regions don't have full (100 %) acceptance except MU0 which accepts all muons within station coincidence windows. Figures 2 (a) and (b) show the trigger efficiency as a function of the $|\eta|$ of the simulated

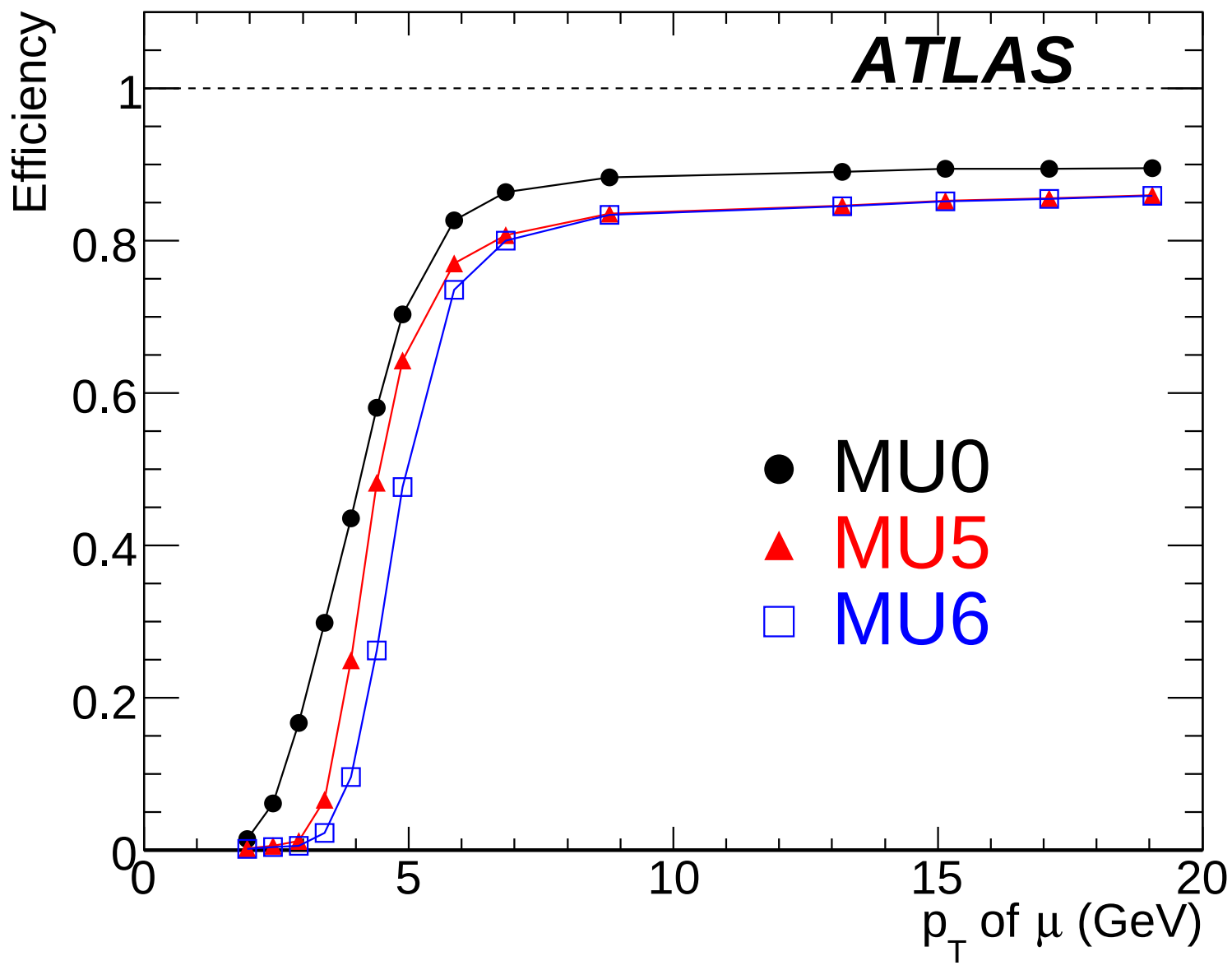


Figure 1: Trigger efficiency as a function of $p_T$ with each low-$p_T$ threshold; MU0 (filled circles), MU5 (filled triangles) and MU6 (open squares) for muons with $|\eta| < 2.5$ at the interaction point.

muon with $p_T$ at threshold and with $p_T$=19 GeV for MU0 and MU6 trigger selections, respectively. The efficiency at the plateau is predominantly determined by geometrical acceptance. Efficiency losses at $|\eta| \sim 0$, 0.4 and 0.7 are caused by cracks and inactive material, like ribs, detector support structures and the elevator hole. Around the transition region between the barrel and the endcap, $|\eta| \sim 1.05$, some efficiency is lost because the station coincidence cannot be satisfied. A small efficiency gap at the boundary between the endcap and forward regions ($|\eta| \sim 2$) is also due to the station coincidence, since the endcap and forward systems are treated separately. However, this effect is smaller than in the transition region between the barrel and the endcap. Poor MU0 efficiency at $p_T$ =4 GeV is seen in the endcap region due to the requirement of three station coincidence, not two as required in the barrel. The trigger was originally designed for $p_T > 6$ GeV as the lowest threshold, but can be applied below this threshold at very low luminosity.

Figures 3 (a) and (b) show the trigger efficiency as a function of $\phi$ ($\phi$=0 and $\pm\pi$ denote directions perpendicular to the beam axis in the horizontal plane) of the simulated muon with $p_T$ at threshold and with $p_T$=19 GeV for MU0 and MU6 trigger selections, respectively. The efficiency loss can be clearly

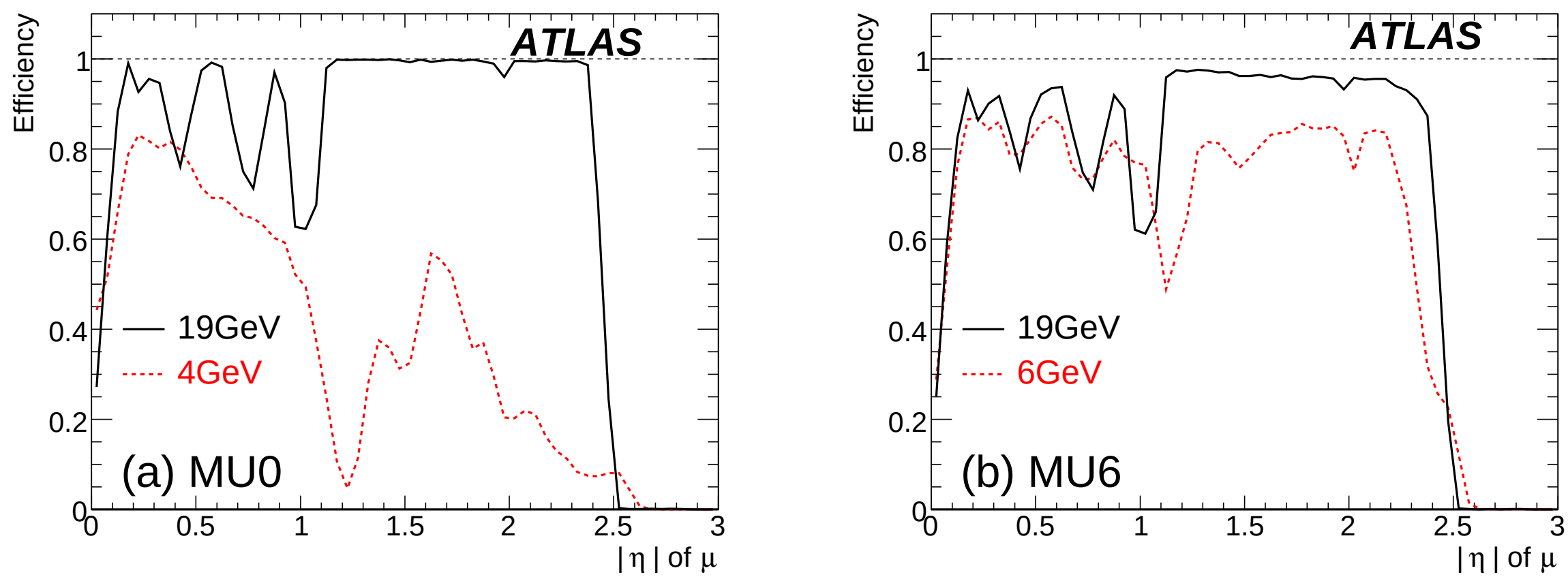


Figure 2: The $|\eta|$ dependency of the trigger efficiency for MU0 (a) and MU6 (b) trigger selections. Solid lines correspond to the simulated muons with $p_T$=19 GeV, dotted lines are for muons with $p_T$ at threshold (in the case of MU0 this threshold is set to 4 GeV).

seen around $\phi \sim -1.2$ and $-2$ at the magnet's feet, especially for high $p_T$ muons. The geometrical structure is in general more visible in efficiency using higher $p_T$ muons, as they produce straighter tracks.

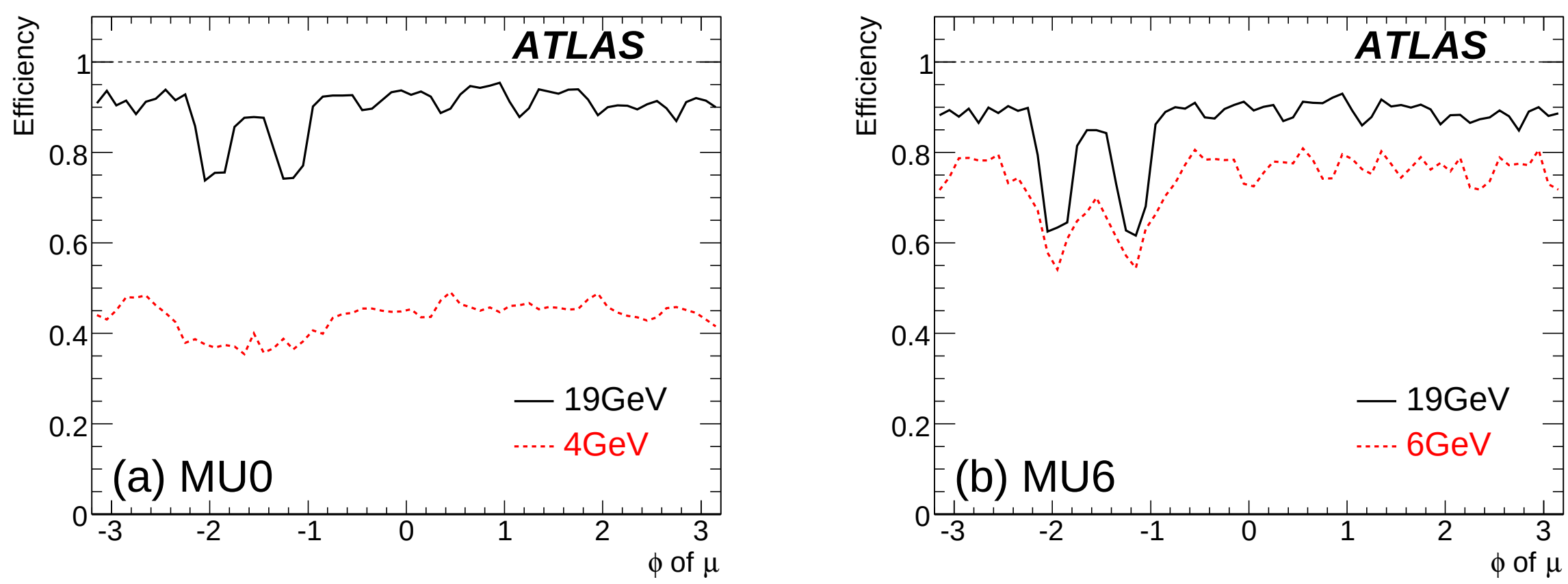


Figure 3: The $\phi$ dependency of the trigger efficiency for MU0 (a) and MU6 (b) trigger selections. Solid lines correspond to the simulated muons with $p_T$=19 GeV, dotted lines are for muons with $p_T$ at threshold (in the case of MU0 this threshold is set to 4 GeV).

## 5 Level-1 muon trigger performance in $B$-physics events

A detailed level-1 trigger simulation is mandatory to find effective trigger menus and thresholds for certain physics processes and to determine the trigger efficiency and rate as well as to study the trigger bias. In this section, di-muon and single-muon trigger efficiencies with different $p_T$ thresholds and the dependency of di-muon trigger efficiencies as a function of the opening angle between simulated muons

are discussed using $B$-physics Monte Carlo samples.

### 5.1 Efficiency with various trigger configurations

The efficiency of the level-1 muon trigger for three different $B$-physics processes, ${B_s}^0 \to \mu^+\mu^-$, ${B_s}^0 \to \mu^+\mu^-\phi$ and $B^+ \to J/\psi(\mu^+\mu^-)K^+$ is studied with various trigger configurations. The efficiencies of the single muon triggers (MU0 and MU6) and the di-muon triggers (2MU0 and 2MU6) are summarized in Table 1. The efficiency of the single muon trigger is high, about 95 %, since multiple muons have a higher probability to be triggered by a single muon trigger. The 2MU0 trigger gives better efficiency than 2MU6 by $\sim$ 16 % for di-muon events with $p_T$ above 6 GeV (4 GeV) for the fastest (second fastest) muon. The efficiency loss due to the MuCTPI overlap handling is seen in all physics processes, although the relative efficiency loss is small, $\sim$ 0.5 %.

| | Efficiency [%] | | |
|---|---|---|---|
| Trigger Menu / Process | ${B_s}^0 \to \mu^+\mu^-$ | $B^+ \to J/\psi(\mu^+\mu^-)K^+$ | ${B_s}^0 \to \mu^+\mu^-\phi$ |
| MU0 | 97.0±0.1 | 96.8±0.1 | 97.0±0.1 |
| MU6 | 93.0±0.1 | 92.9±0.1 | 93.1±0.1 |
| 2MU0 | 67.9±0.2 | 68.8±0.2 | 69.0±0.2 |
| 2MU6 | 51.6±0.2 | 52.9±0.2 | 53.2±0.2 |
| 2MU0 (w/o MuCTPI overlap handling) | 68.2±0.2 | 69.1±0.2 | 69.4±0.2 |
| 2MU6 (w/o MuCTPI overlap handling) | 52.0±0.2 | 53.4±0.2 | 53.7±0.2 |

Table 1: The level-1 trigger efficiency of various configurations. The efficiency is calculated with respect to the number of generated events (not to the number of muons) of three different physics processes. The errors are statistical only.

### 5.2 Opening-angle dependency

In case of measuring some quantity as a function of a certain parameter, one must ensure that the trigger efficiency is independent of that parameter, or correct the measurement to avoid a bias from the trigger. Figure 4 shows the 2MU0 and 2MU6 trigger efficiencies as a function of $\Delta R$ ($=\sqrt{\Delta\eta^2+\Delta\phi^2}$) between the two leading muons in ${B_s}^0 \to \mu^+\mu^-\phi$ events. The $\eta$ and $\phi$ are true parameters of muon flight direction at the production vertex. No offline selection is applied in order to see see the bias from the level-1 trigger. The trigger efficiency clearly depends on the opening angle of the two leading muons. The efficiency is lower at large opening angles because in this case the $B_s$ system is not boosted i.e. momenta of the muons are lower. Since overlap removal between the muon candidates doesn't play a role at large opening angles, this effect is purely due to kinematics. Figure 5 (a) shows the $p_T$ distribution of the leading muons for three different $\Delta R$ ranges: $\Delta R < 0.1$, $0.2 < \Delta R < 0.4$ and $\Delta R > 0.5$. The muons at large opening angles clearly have a softer $p_T$ spectrum. The effect is even more clearly visible for higher threshold di-muon trigger items, as the probability of having muons at large $\Delta R$ with transverse momenta high enough to trigger a high threshold di-muon trigger item is very small, but non-negligible at small opening angles.

The small efficiency loss at very small opening angles ($\Delta R < 0.1$) is however due to the trigger system. In the case that the muons both leave hits in the same RoI, only one muon can be triggered by

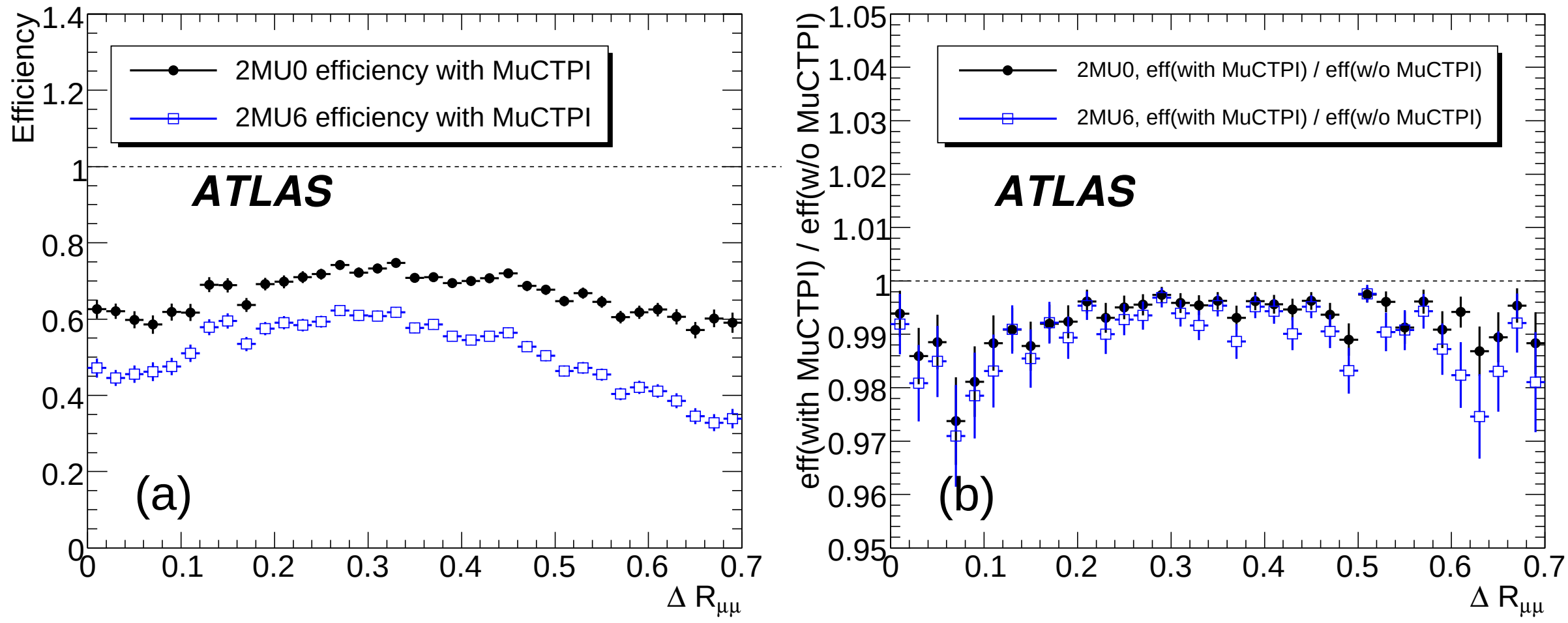


Figure 4: (a) Trigger efficiency as a function of $\Delta R$ in the semi-leptonic rare $B$ decay $B_s^0 \to \mu^+\mu^-\phi$ using the 2MU0 (filled circles) and 2MU6 (open squares) with MuCTPI. (b) Effect of MuCTPI as a function of the opening angle, $\varepsilon$(without MuCTPI)/$\varepsilon$(with MuCTPI) for 2MU0 (filled circles) and 2MU6 (open squares).

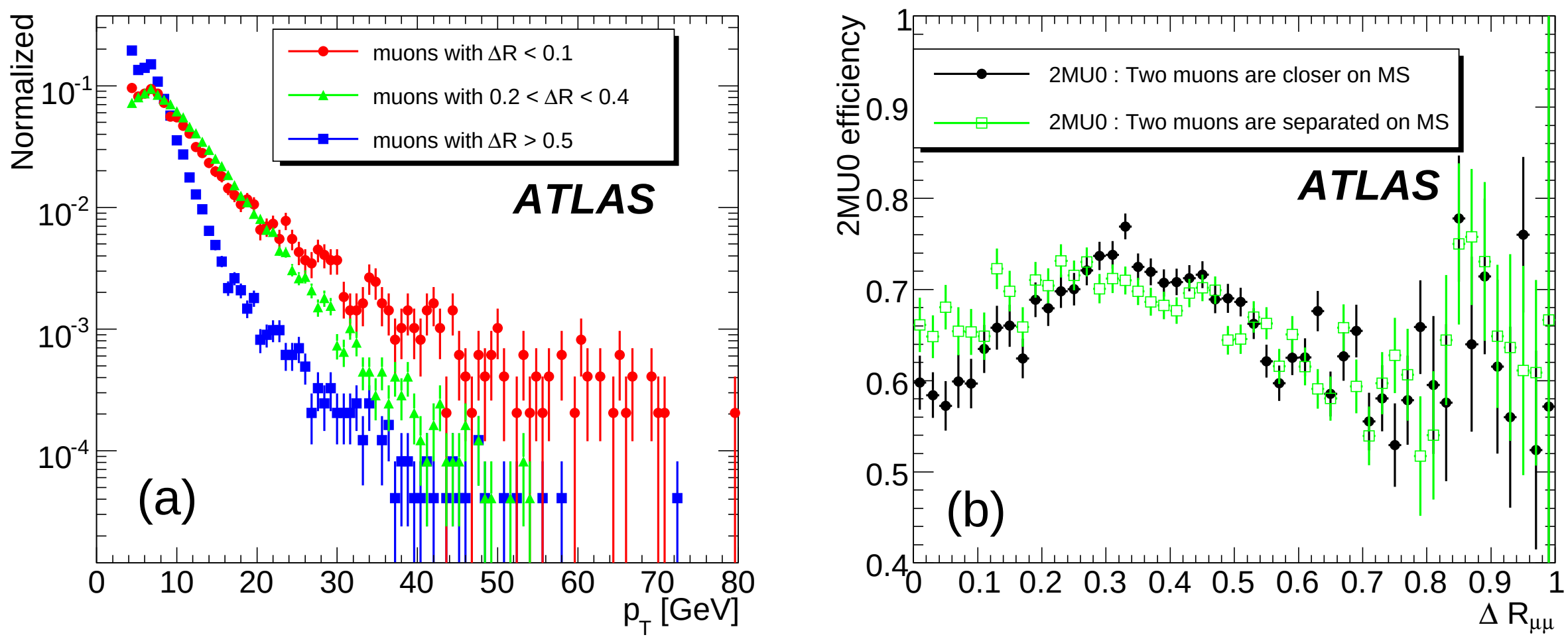


Figure 5: (a) The truth $p_T$ distribution of the two leading muons is different opening angles range in $B_s^0 \to \mu^+\mu^-\phi$ events: $\Delta R < 0.1$ (filled circles), $0.2 < \Delta R < 0.4$ (filled triangles) and $\Delta R > 0.5$ (filled squares). (b) The efficiency of 2MU0 as a function of opening angles for two cases: the $\Delta R$ separation of the two muons is either larger (filled circles) or smaller (open circles) at the muon trigger system compared to that at the primary vertex.

the system. Figure 5(b) shows the 2MU0 efficiency as a function of the opening angle, but events are divided into two classes: In one case we select the muon pairs whose separation is expected to shrink when reaching the muon spectrometer (filled circles), while in the other case (open squares) we select the muon pairs whose separation is expected to grow.

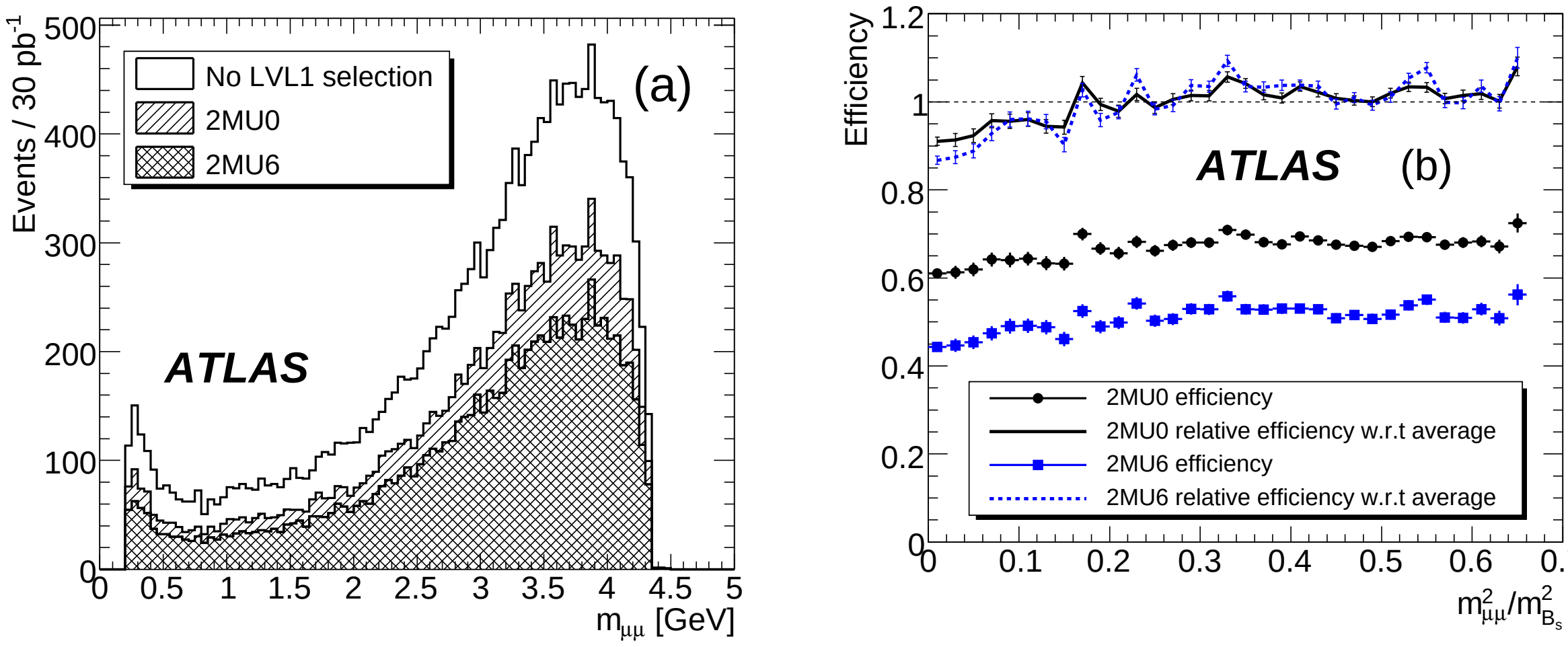


Figure 6: (a) Scaled di-muon invariant mass-squared distribution in ${B_s}^0 \rightarrow \mu^+\mu^-\phi$ events: no level-1 selection (open histogram), triggered by 2MU0 (hatched histogram) and by 2MU6(cross hatched histogram). (b) Efficiencies of 2MU0 (filled circles) and 2MU6 (open squares) as a function of di-muon invariant mass-squared. The two lines show the relative efficiency with respect to the corresponding average value for 2MU0 (solid line) and 2MU6 (dotted line).

One example $B$-physics study is to look at the differential decay rate dBr($B \rightarrow \mu^+\mu^-\phi$) / d$\hat{s}$ or forward-backward asymmetry $A_{FB}$ as a function of di-muon invariant mass to discriminate between Standard Model and new physics contributions. The $\hat{s}$ is $m^2_{\mu\mu}/m^2_B$ and the $A_{FB}(\hat{s})$ is defined as

$$\left(\int_0^1 \frac{d\Gamma}{d\hat{s}\hat{z}} d\hat{z} - \int_{-1}^0 \frac{d\Gamma}{d\hat{s}\hat{z}} d\hat{z}\right) / \int_{-1}^1 \frac{d\Gamma}{d\hat{s}\hat{z}} d\hat{z}, \tag{2}$$

where $\hat{z} = \cos\theta$ ($\theta$ is the angle between the $\mu^+$ and $\phi$ in the $\mu^+\mu^-$ rest frame) and $d\Gamma/d\hat{z}$ is the differential event rate of $B^0_s \rightarrow \mu^+\mu^-\phi$.

Figure 6 (a) shows the $\mu^+\mu^-$ invariant mass distribution with and without level-1 triggers in ${B_s}^0 \rightarrow \mu^+\mu^-\phi$ events. Trigger efficiency with 2MU0 and 2MU6 are shown as a function of $\mu^+\mu^-$ invariant mass in Figure 6 (b). The relative trigger efficiency, normalized to the efficiency averaged over a whole $m^2_{\mu\mu}/m^2_B$ range, is also shown to illustrate the trigger bias. The invariant mass dependency of the level-1 di-muon trigger efficiency is weak compared to the opening angle dependency and almost independent of the trigger threshold. A efficiency drop can be seen at small m$_{\mu\mu}$, which is also because the two muons are triggered as a single muon by the muon trigger system.

Similarly, the forward-backward asymmetry $A_{FB}$ is shown as a function of $\hat{s}$ in Figure7(a). The difference ($A_{FB}$[2MUx] - $A_{FB}$[no selection]) is also shown in Figure 7(b). It should be noted that significant differences between forward and backward samples in terms of momentum, pseudo-rapidity and $\mu^+\mu^-$ opening angle distributions are observed. Nevertheless, as illustrated in Figure 7, the forward-backward asymmetry is more robust to the effects of level-1 di-muon triggering than di-muon mass distributions (see Figure 6).

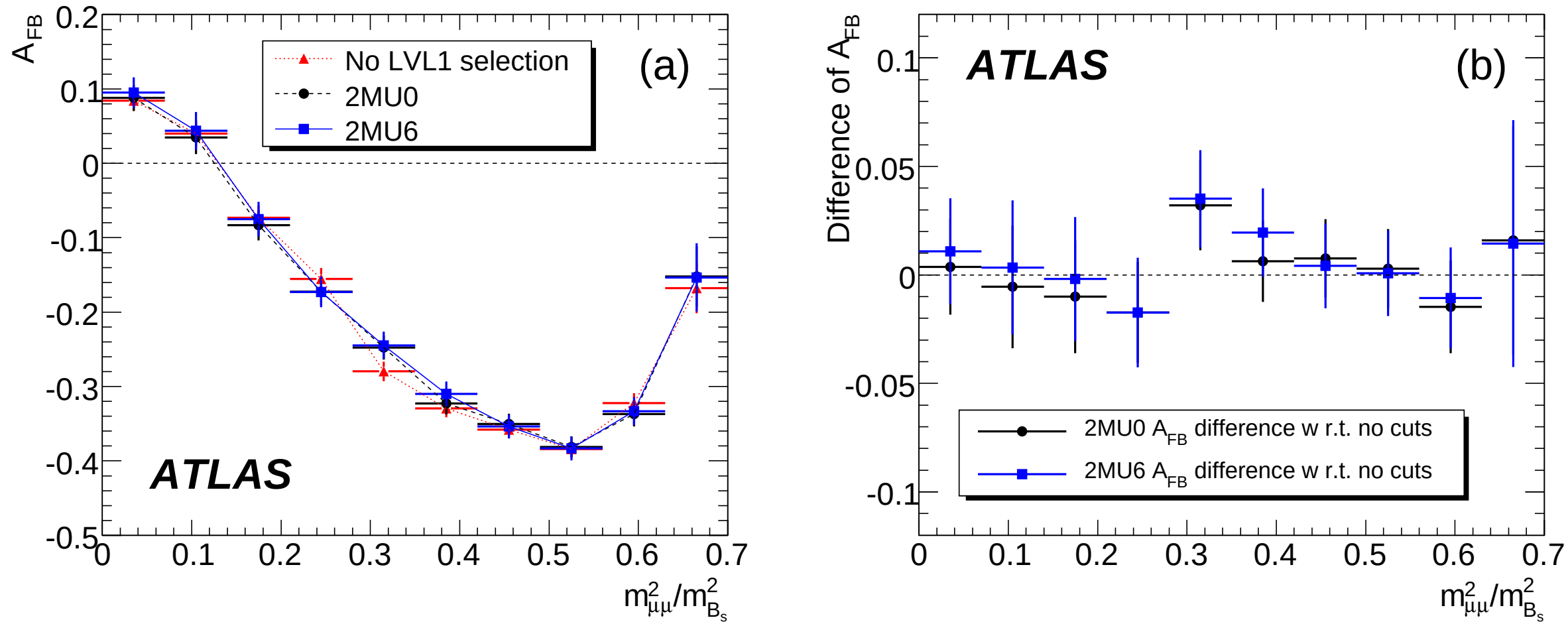


Figure 7: (a) Forward-Backward asymmetry in $B_s^0 \to \mu^+\mu^-\phi$ events: no level-1 selection (filled triangles), triggered by 2MU0 (filled circles) and by 2MU6(open squares). (b) deviation between $A_{FB}$'s with and without triggers ($A_{FB}$[2MUx] - $A_{FB}$[no selection]: 2MU0 (filled circles) and 2MU6 (open squares).

# 6 Summary

The level-1 di-muon trigger is essential for selecting rare $B$ decays that have low-$p_T$ muons in the final state. The rate for a single-muon trigger at the relevant thresholds would be unacceptably high. A detailed level-1 muon trigger simulation is implemented and used for the study of the trigger efficiency and its bias for $B$-physics events. The level-1 di-muon trigger efficiencies of 2MU0 and 2MU6 are high enough (about 70 % and 50 %, respectively) in the interesting $B$-physics events. The fake level-1 di-muon trigger rate is 2.3 kHz [8] while the genuine di-muon event rate is $\sim$ 600 Hz (muons with $p_T > 4$ GeV from $c\bar{c}$ and $b\bar{b}$) at $\mathscr{L} = 10^{33}\mathrm{cm}^{-2}\mathrm{s}^{-1}$. The efficiency loss from the MuCTPI overlap handling is found to be negligible.

The trigger bias in regard to the opening angle and invariant mass of two muons was also studied. An opening angle dependence is clearly seen, of the order of $\pm 10$ % and $\pm 15$ % for 2MU0 and 2MU6, respectively. It is explained by the single-muon efficiency curves and muon kinematics in signal events. This dependency is stronger for higher $p_T$ thresholds. No clear bias is seen due to the presence of the MuCTPI overlap handling. The trigger efficiency is rather flat over the invariant mass and less dependent on the trigger threshold since the invariant mass is not as strongly correlated to muon momenta as the opening angle is.

# Triggering on Low-$p_T$ Muons and Di-Muons for $B$-Physics

**Abstract**
Muon pairs from $J/\psi$ decay are a clear signature of $b$ hadrons. As a large fraction of $b$ hadrons are produced at low-$p_T$, a low rate, efficient di-muon trigger for low-$p_T$ muons and a good understanding of the trigger efficiency are essential. Di-muon final states will also play a key role in calibration, alignment and determination of the trigger efficiencies. The performance of the level-2 dimuon trigger algorithms is discussed, together with a method for reducing backgrounds from decays-in-flight. A strategy for calculating the single muon and di-muon trigger efficiencies at level-1 and level-2 using $J/\psi$ from the data themselves is presented.

## 1 Introduction

$B$-physics is one of the areas of the physics programme of the ATLAS experiment. It includes the study of production cross sections, searches for rare $b$ decays and measurements of CP violation effects. These studies make use of the large $b\bar{b}$ production cross section at the LHC where $b\bar{b}$ pairs are abundant in the low transverse momentum ($p_T$) region. On the other hand, one must extract signals from amongst the large QCD background, mostly composed of light quarks. For this purpose, one of the main channels for $B$ physics study involves decay channels with one or more muons in the final state, especially the channel $J/\psi \to \mu^+\mu^-$.

The output rate of the first level trigger at a luminosity of $10^{33}\mathrm{cm}^{-2}\mathrm{s}^{-1}$ is expected to contain 20 kHz of events where one muon passed the $p_T$ threshold of 6 GeV. Early running is envisioned to include even lower $p_T$ thresholds, down to the lowest threshold achievable in the hardware.

At the second level trigger this rate of events must be reduced to 1-2 kHz, of which 5-10% are available for channels of interest only to $B$-physics. Currently this goal is achieved for level-1 muon triggers by first confirming that a muon over the nominal threshold is reconstructed in the muon spectrometer (MS), and then confirming that there is a matching track in the inner detector (ID). This selection criterion removes many muons from $K$ and $\pi$ decays, but does not by itself produce the required rate reduction. To achieve the required rate $p_T$ thresholds need to be raised and many interesting $b$ events are likely to be filtered out. We therefore focus also on di-muon final states.

We developed an algorithm, TrigDiMuon, which achieves high efficiency at level-2 for the golden CP channels ($J/\psi$), using the identification of relatively low-$p_T$ muons from $J/\psi$ decay. TrigDiMuon searches for di-muon pairs from $J/\psi$ or other resonant sources, when only one of the muons passed the level-1 or level-2 single muon selection. The use of TrigDiMuon can enhance $J/\psi$ efficiency at low-$p_T$ compared to the trigger based on two muons found at level-1, with an acceptable increase in the fake rate.

In this note, we present the performance of the TrigDiMuon algorithm, that looks for a second low-$p_T$ muon partner to a single muon triggered at level-1. We compare it to the performance of an algorithm that requires a di-muon level-1 trigger. The comparison is based on a sample of $J/\psi$ which decayed into muons with low-$p_T$, such that the second muon of each decay may be below the level-1 threshold.

In addition to the foreseen specialised trigger strategies for di-muon signatures, it is important to optimize the rejection of muons from $K$ and $\pi$ decays for the standard single muon selection in order to have the lowest possible threshold on the inclusive single muon trigger. This is achieved

by extrapolating the muon track from the MS back to the interaction vertex and requiring a good match between this track and the associated track from the ID. Muons from light hadron decays do not match accurately the ID track, which in this case is the track produced by the parent hadron (or from a mixture of hits from the parent and the daughter muon) and not by the muon track. This note presents a method implemented at the level-2 trigger for rejecting muons from $K$ and $\pi$ decays. We summarize the rejection power of this method, show that the efficiency loss is minimal for direct muons and estimate the efficiency loss for low-$p_T$ muons from $J/\psi$ decay.

Cross section measurements or searches for rare decays require a good understanding of the efficiency of the event selection. As we are interested in events with rather low-$p_T$, the understanding of the trigger efficiency is crucial and we must have a strategy for measuring it from data with high precision. The tag-and-probe method for measuring the single muon trigger efficiency using $J/\psi$ events is presented, and we demonstrate that the obtained trigger efficiency can be applied to calculate the di-muon trigger efficiency. A calibration trigger is proposed to collect an unbiased sample of single muons with an enhanced $J/\psi$ fraction, and the expected performance of the trigger efficiency measurement in the early days of the data-taking is discussed.

Particles from additional collisions in the same bunch-crossing (pile-up) are not simulated in the samples used for this paper, and therefore this additional background is not taken into account in the analysis.

# 2 Simulated datasets used and production tools

Since the subject of this note is low-$p_T$ muons, we use simulated samples that were produced with especially low-$p_T$ cuts at the event generation level. Samples were generated using the PYTHIA event generator [1]. Except for the minimum bias events, a generator level filter was applied to pre-select efficiently the events in the sample. The di-muons sample passed a filter which required the existence of at least two muons with the appropriate $p_T$ and $\eta$ cuts. For the inclusive muon samples the filter required the existence of a single muon passing the corresponding selection. Events were processed with the full simulation of the ATLAS detector based on the GEANT package [2]. The level-1 simulation, High Level Trigger (HLT) selection and offline event reconstruction were performed.

The following di-muon samples were used:

- An 8000 event sample of the channel $\Lambda_b \to J/\psi\Lambda$ with $J/\psi \to \mu^+\mu^-$, where one of the muons is required at the event generation level to have $p_T > 4$ GeV and the second muon is required to have $p_T > 2.5$ GeV.
- Simulated samples of direct $J/\psi$ production and $b\bar{b} \to J/\psi$ production, with the generator level filter requiring the existence of at least two muons with $p_T > 6$ GeV for the highest $p_T$ muon and $p_T > 4$ GeV for the second highest $p_T$ muon. For both processes, 150000 events were generated.
- Two samples of inclusive muon from $b$ decays were used. A 200000 event sample of $b\bar{b} \to \mu X$ with a generator level filter requiring one muon with $p_T > 4$ GeV, and a sample of 250000 $b\bar{b}$ events with at least one muon with $p_T > 6$ GeV in the final state.
- Since this note deals specifically with methods of rejecting muons from $K$ and $\pi$ decays, a large sample of such decays was required. A minimum bias sample where pions and kaons were

forced to decay inside the inner tracker volume was produced for this purpose. The method developed for producing this "forced-decay" sample is described in detail below.

- Events containing a single muon each were used to study the effect on prompt muons of the method for rejecting muons from $K$ and $\pi$ decays. Some of the single muon samples were simulated with a perfectly aligned detector geometry, while some other samples were simulated with a misaligned detector description.

The trigger and reconstruction software used for some of the samples are of a later version than that used in other notes, because the level-2 and reconstruction software used here have improved significantly in recent version. The software modification will be explained where each package is described. Table 1 summarizes the simulated Monte Carlo samples used in this note.

| Samples | | Generator level filter | Statistics |
|---|---|---|---|
| Signal samples | Direct $J/\psi \to \mu^+\mu^-$ | $p_T^{\mu_{1,2}} > 6,4$ GeV | 150 k |
| | $b\bar{b} \to J/\psi \to \mu^+\mu^-$ | $p_T^{\mu_{1,2}} > 6,4$ GeV | 150 k |
| | $\Lambda_b \to J/\psi\Lambda$ ($J/\psi \to \mu^+\mu^-$) | $p_T^{\mu_{1,2}} > 4,2.5$ GeV | 7.6 k |
| Background samples | $bb \to \mu + X$ | $p_T^{\mu} > 6$ GeV | 250 k |
| | $b\bar{b} \to \mu + X$ | $p_T^{\mu} > 4$ GeV | 185 k |
| | Minimum bias (forced $K$ /$\pi$) | | 114 k |
| | Minimum bias (standard) | | 500 k |

Table 1: Summary of MC samples.

### 2.1 Production tools and samples employed for $K$ and $\pi$ decays

Minimum bias events are the most copious source of pions and kaons. These particles are produced mainly with very low-$p_T$, and typically the muons coming from their in flight decays do not escape the ATLAS hadronic calorimeter. On the other hand pions and kaons with high-$p_T$ have a low probability to decay before the calorimeter because of their high energy. As a consequence it is not efficient to use minimum bias events as a source of decays in flight to muons. Estimates made for this analysis indicated that the simulation of 5000 minimum bias events would be needed to provide one muon from $K$ or $\pi$ decay capable of passing the 6 GeV threshold of the ATLAS level-1 muon trigger.

To increase the statistics of events with charged pions and kaons decaying in flight, a special simulation tool is applied. Since we are interested in decays which happen inside the detector, the decays cannot be made on generator level but must happen in the GEANT simulation. The program to force the $K$ /$\pi$ to decay in the detector is an extension to GEANT which runs before the simulation itself.

In each event a list of all charged pions and kaons with $p_T > 2$ GeV is compiled, and one of the particles in the list is randomly selected. Events with no charged pions or kaons with $p_T > 2$ GeV are dropped at this stage, such that the events are not further simulated and nothing is written to output.

The point of decay is selected by first computing the trajectory length from the production vertex to where the particle would exit the ID (neglecting curvature in the magnetic field). The decay position is then determined by taking a random fraction of this maximum track length. Since particles assigned a late decay have a larger probability to be stopped through hadronic interactions, this procedure may introduce a weak bias towards shorter decay lengths.

For this study a minimum bias sample of 114 k events was simulated with forced decays. The transverse momentum cut of 2 GeV applied to the light hadrons reduces the cross section of the sample to $(43.41 \pm 0.07)\%$ of the minimum bias cross section. In addition to this a minimum bias sample of 500 k events, simulated with the standard GEANT simulator, was also employed to provide a cross-check from a single source for both the muons from $K/\pi$ decay and the prompt muons.

# 3 The muon trigger

The ATLAS trigger system [3] reduces the event rate from the 40 MHz beam crossing rate to ~200 Hz for mass storage, keeping the events that are potentially the most interesting for physics. The first level trigger [4] selection is performed by custom hardware and identifies a detector region for which a trigger element was found. The second level trigger [5] is performed by dedicated software, making its decision based on data acquired from the Region of Interest (RoI) identified at level-1. Eventually, the event filter [5] uses the complete event data, and algorithms adapted from the offline reconstruction, to refine the selection of level-2 and further reduce the trigger rate by about a factor 10. An event must pass all trigger levels to be kept for analysis.

A detailed description of the ATLAS muon trigger and the estimated trigger rates are presented in [6]. Here we give a very brief summary of the principles of the level-1 and level-2 muon triggers.

## 3.1 The level-1 muon trigger

The level-1 muon trigger is based on dedicated fast detectors: the Resistive Plate Chambers (RPC) in the barrel and the Thin Gap Chambers (TGC) in the end-caps [7]. The basic principle of the algorithm is to require a coincidence of hits in the different trigger stations within a predefined angular region, called a ”road”, from the interaction point through the detector. The width of the road is related to the bending of the muon in the magnetic field and thus to the $p_T$ threshold to be applied.

The trigger in both the barrel and the end-cap regions is based on three trigger stations at different distances from the interaction point [4]. The low-$p_T$ triggers (4 to 10 GeV) are derived from a coincidence in two stations, while the high-$p_T$ triggers (over 10 GeV) require an additional coincidence with the third station. In the end-cap, there is an option to use all three stations also for the low-$p_T$ trigger. This option is the one used in the trigger performance studies in Reference [6].

The level-1 trigger provides for each muon candidate the region where it was found, called the region of interest (RoI). For the muon trigger, the size of the level-1 RoI is $\Delta\eta \times \Delta\phi = 0.1 \times 0.1$ in the barrel and $\Delta\eta \times \Delta\phi = 0.03 \times 0.03$ in the end-cap region, respectively.

## 3.2 The level-2 single-muon trigger

The level-2 trigger is a software-based trigger and uses the information of the Region of Interest provided by the level-1 trigger. Level-2 algorithms only process data around the RoI, using the full granularity of the detector readout within the RoI.

The HLT trigger selection proceeds in “trigger chains”. A chain consists of a series of reconstruction and decision (hypothesis) algorithms that process the data in a RoI identified by level-1. The role of the level-2 muon trigger is to confirm muon candidates flagged by the level-1 and to give more precise track parameters for the muon candidate.

The level-2 muon selection is performed in two stages. The first stage is performed by the muFast algorithm [8], which starts from a level-1 muon RoI and reconstructs the muon in the spectrometer, using the more precise Monitored Drift Tubes (MDT) to perform a new $p_T$ estimate for the muon

candidate and creating a new trigger element. The hypothesis algorithm cuts on the estimated $p_T$ and passes the validated trigger elements to the next algorithm.

Track finding in the inner detector is based on the region of the candidate found by muFast. The muFast candidate and ID tracks are passed to the next algorithm, muComb, which matches an ID track with the trigger element from the muon spectrometer and refines the $p_T$ estimate [5].

### 3.2.1 muFast

For level-1 RoI's flagged by the RPC (barrel region) and the TGC (end-cap region), muFast performs global pattern recognition, a local segment fit in each muon station and a fast $p_T$ estimation. The global pattern recognition is designed to select clusters in MDT tubes belonging to a muon track without using the drift time measurements. It is divided into two steps, firstly the pattern recognition in the trigger chambers seeded by the level-1 RoI, and the subsequent MDT pattern recognition seeded by the result of the previous step. In the MDT pattern recognition muon roads are opened in selected MDT chambers, and the loactions of hit tubes are collected. A contiguity algorithm is applied on the selected hits to remove the background.

The track reconstruction approximates a muon track as a series of segments built separately in each MDT chamber. Segments are reconstructed using the drift time measurements and an approximate calibration to obtain hit radii from them. The fitted segments provide a precision measurement of the point where the fitted line crosses the middle of the MDT chamber, called the super-point.

The track bending is measured in a different way in the barrel and the end-cap. In the barrel, the sagitta is computed from the three super-points found in the three stations. In the end-cap, the track bending is measured by the angle $\alpha$ between the track direction measured by the muon chambers in the middle and outer stations of the muon spectrometer, and the direction obtained by connecting the nominal interaction vertex with the mean hit position in the middle station.

The muon transverse momentum is estimated using an inverse linear relationship between the measured sagitta (in the barrel) or $\alpha$ (in the end-cap) and $p_T$. The detector region is divided into bins in $\eta$ and $\phi$ and the parameters of this inverse linear function are estimated in each bin.

### 3.2.2 muComb

The muComb algorithm matches the muon track found by muFast to ID tracks reconstructed at level-2 citeID-CSC. In reconstructing the level-2 ID tracks, only hits from the pixel and SCT detectors were used, for speed.

The matching between muFast and ID track segments proceeds as follows. First, a preselection of ID tracks is made based on the difference in $\eta$ and $\phi$ between muFast and ID track segments. In the barrel, this preselection also makes use of the difference in the $Z$ of the extrapolated track segments at the radius of the barrel calorimeter. A weighted combined $p_T$, and a matching $\chi^2$, are calculated for each ID track passing the preselection in combination with the muon track information, and the ID track giving the lowest $\chi^2$ is selected as the best match to the muFast track.

The version of the algorithm used in this study is improved with respect to that used in [6]. The resolution of muFast tracks assumed in combining with the ID tracks have been retuned, using the correct $p_T$ resolutions for the different end-cap regions, and for the misaligned detector geometry. A specific tuning of the matching windows is used to improve the resolution for very low $p_T$ muons. This algorithm will be referred to as "the baseline muComb selection" in Section 5, where a modified algorithm with better rejection for muons from $K$ and $\pi$ decays is also described, and the performance of the two algorithms is compared.

#### 3.2.3 Level-2 muon hypotheses

The $p_T$ cuts corresponding to each nominal threshold are set so that 90% of the muons at the nominal threshold would pass the selection. The actual cuts used depend on the resolution of the $p_T$ estimation. Therefore the cuts are different for different regions of the detector as well as different between muFast and muComb. Thus for a nominal threshold of 6 GeV the muFast hypothesis cuts at estimated $p_T$ values between 4.5 and 5.4 GeV in the different $\eta$ regions, while the muComb hypothesis cuts at estimated $p_T$ of 5.8 GeV in the barrel and end-cap and 5.6 GeV in the forward region. Since the $p_T$ resolution of muComb is better than that of muFast, the cuts are closer to the nominal thresholds, and reject more muons with $p_T$ below the threshold. A special case is the 4 GeV nominal threshold which is meant to accept lower $p_T$ muons and the cuts are set at 3 GeV in the barrel and 2.5 GeV in the end-cap for both muFast and muComb.

### 3.3 The level-2 di-muon triggers

There are two approaches at level-2 for selecting di-muon events from a resonance such as $J/\psi$ and $\Upsilon$. The first approach is to start from a di-muon trigger at level-1 which produces two muon regions of interest. In this approach, reconstruction of a muon is confirmed separately in each RoI as described above and the two muons are subsequently combined to form a resonance and to apply a mass cut. We will refer to this trigger as the "topological di-muon trigger".

An alternative approach is to start with a level-1 single muon trigger and search for two muons in a wider $\eta$ and $\phi$ region. This approach starts from reconstructing tracks in the inner detector and extrapolating the track to the muon spectrometer to tag muon tracks. Since this method does not explicitly require the second muon at level-1, it has an advantage for reconstructing $J/\psi$ at low-$p_T$. This is implemented in the TrigDiMuon algorithm. The two approaches using either two or one muon RoI are illustrated in Figure 1.

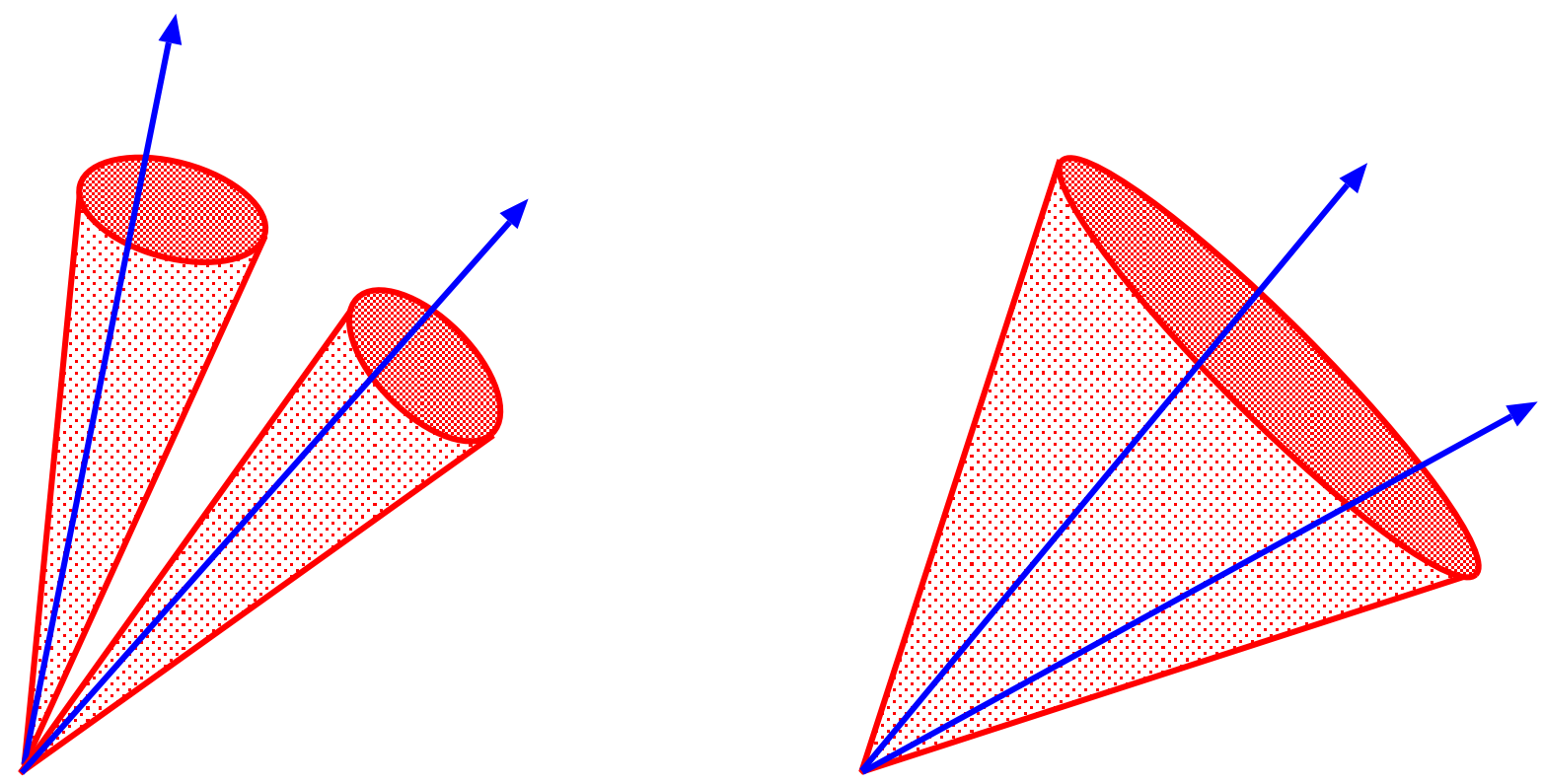

Figure 1: A schematic picture of RoI based di-muon trigger, using two RoI's (left) and seeded by a single muon RoI (right).

### 3.3.1 The TrigDiMuon algorithm

TrigDiMuon is a level-2 trigger algorithm based on associations established between ID tracks and MS hits. Each time a pair of oppositely charged ID tracks, above a minimal invariant mass, is successfully associated with the MS hits, a muon pair object is created with the parameters of these ID tracks. Later, in the hypothesis step, an additional invariant mass selection can be applied, thus selecting interesting physics objects such as $J/\psi$ or $B$.

The motivation for developing this algorithm comes from the fact that while di-muon final states exist in many interesting $B$-physics channels, the cross sections for di-muon final states are orders of magnitude smaller than those for single muons of the same $p_T$. An additional advantage is that resonant final states can also be used to calibrate trigger efficiencies as will be shown in Section 5.

First, the initial muon RoI is extended in order to search for a second muon, which was not triggered by level-1. The input muon may be from a region of interest identified at level-1, but the input rate to the algorithm can be reduced if, prior to the RoI extension, the level-1 RoI would be confirmed in the level-2 trigger by the muFast and (possibly also) the muComb algorithms. The performance of these options is studied in Section 4. The size of the extended RoI is based on the distribution of angular distance in $\eta$ and $\phi$ between two muons from $J/\psi$ decay. Figure 2 shows the probability of including the second muon from $J/\psi$ decays RoI as a function of the extended RoI size for different samples. The current default region size in TrigDiMuon is $\Delta\eta \times \Delta\phi = 0.75 \times 0.75$.

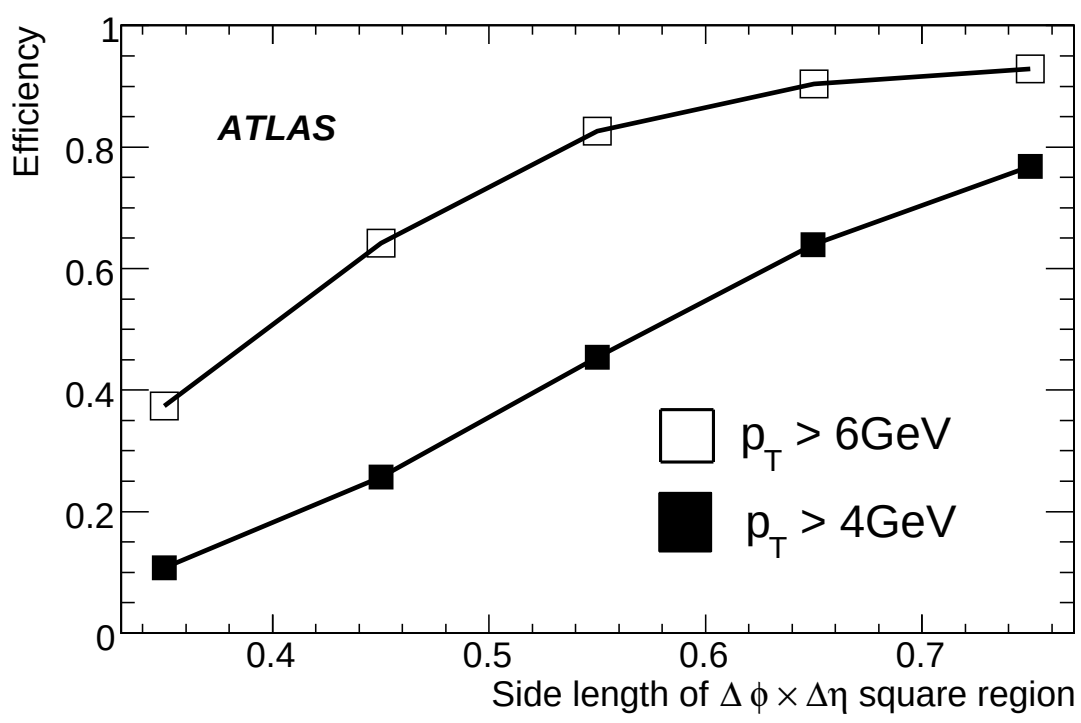


Figure 2: Probability of including the second muon from $J/\psi$ decays RoI as a function of the extended RoI size for different samples. Open squares are from $J/\psi$ decays where one muon has $p_T > 6$ GeV and the other $p_T > 3$ GeV. Full squares are from $J/\psi$ decays where one muon has $p_T > 4$ GeV and the other $p_T > 2.5$ GeV.

The ID tracks in the search region are found using the trigger-tracking program IdScan or SiTrack [9], and are selected if they form a pair of oppositely charged tracks with invariant mass $M > 2.8$ GeV. Each selected track is extrapolated to the different stations of the MS using a formula parameterizing the expected track bending in the magnetic field. The bending parameterization is calculated separately for different regions of the muon spectrometer to account correctly for the inhomogeneous toroidal field in the end-cap region. Figure 3 shows the difference, $\Delta\eta$, between $\eta$ measured in the inner detector and that measured in the middle station of the muon spectrometer. The lines indicate the choice of $\eta$ regions for the parameterization. The parameterization was also subdivided in $\phi$.

The algorithm then searches for muon hits within a road around the extrapolated track. The road size also differs for different $\eta$ regions. If a sufficient number of muon hits are found in the MS, the

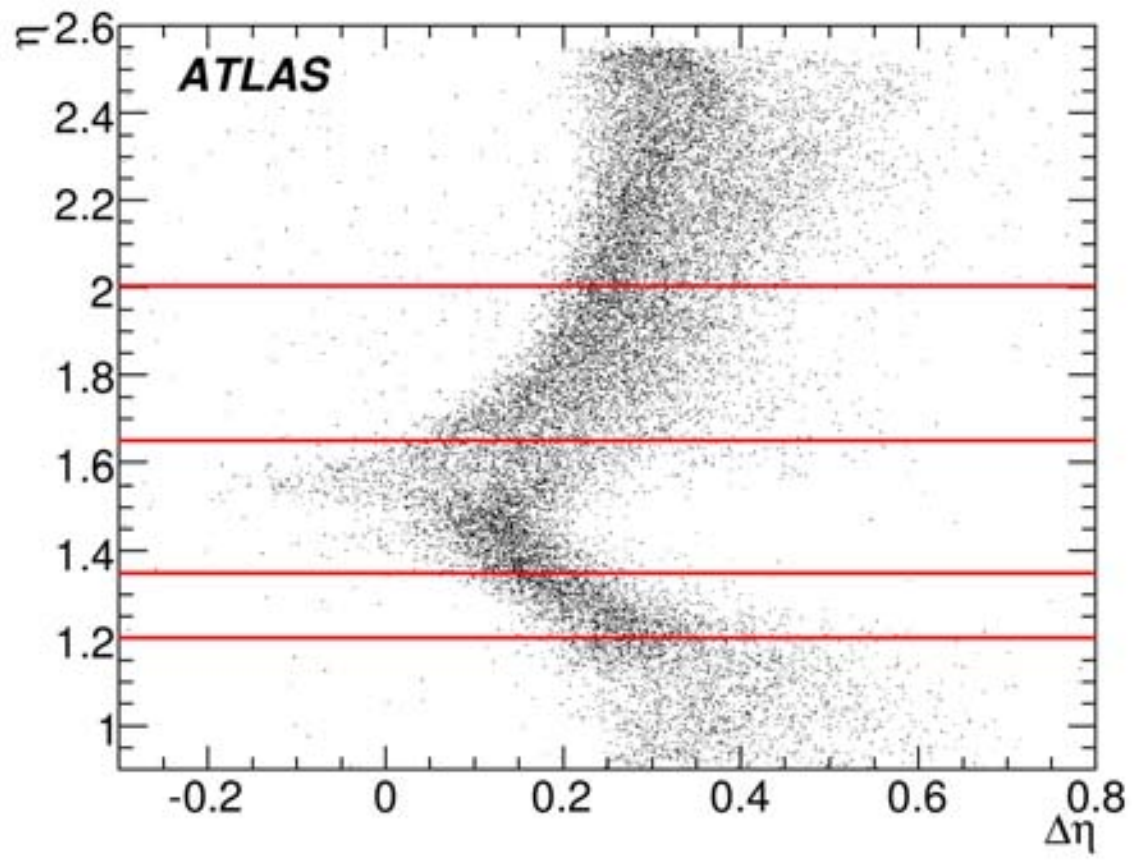


Figure 3: The $\eta$ direction of muons at the intercation point vs. the difference in $\eta$ position between the inner detector and the middle station of the muon spectrometer, for muons with $p_T = 6$ GeV. The lines indicate the choice of $\eta$ regions for the parameterization.

track is identified as a muon. If both tracks from the pair are identified as muons, a muon pair object is created. The two tracks are fit to a common vertex, and the vertex $\chi^2$ is calculated to allow a later selection of only the pairs with a good quality vertex.

## 4 Performance of the level-2 di-muon triggers for $J/\psi$

In this Section the efficiency and fake rates resulting from the two approaches to selecting di-muons at level-2 will be presented and compared. For the TrigDiMuon algorithm we calculate the efficiency and fake rates in three different trigger chains.

In the first configuration, TrigDiMuon runs directly after the level-1 trigger based on the RoI produced by level-1. The performance of this trigger is compared to the efficiency and fake rate of the level-1 di-muon trigger.

The other two trigger chains confirm a single level-2 muon before calling TrigDiMuon. The purpose of these chains is not to reduce the fake di-muon trigger rate, but rather to reduce processing time by reducing the input rate to TrigDiMuon. Since TrigDiMuon starts from ID track reconstruction in an extended region around the muon region of interest, the tracking in the inner detector requires three times longer than if only reconstructing ID tracks in the narrow road used by muComb. Thus, this time consuming process can be avoided for candidates for which the level-1 trigger is not confirmed at level-2.

In the second chain TrigDiMuon runs after muFast. The input to TrigDiMuon in this case is a muon confirmed in the MS, with a cut on the $p_T$ estimated at this stage. The efficiency of this trigger is compared with that of a topological di-muon trigger based on a level-1 di-muon with two oppositely charged muons confirmed in the MS at level-2.

In the third chain TrigDiMuon runs after muComb. The input to TrigDiMuon in this case is a muon confirmed in the MS and the ID, with a cut on the $p_T$ estimated at this stage. The efficiency of this trigger is compared with that of the topological di-muon trigger with two oppositely charged level-2 combined muons, within the same invariant mass window. The invariant mass window can be

applied in this chain, and not the second chain, because only after muComb the momentum resolution is sufficient for a reasonable selection on invariant mass.

The efficiency and fake rates of these trigger sequences were studied for two different trigger thresholds, a 4 GeV threshold that is envisioned to run at initial luminosity, and a 6 GeV threshold that will be the lowest threshold for running at a luminosity of $10^{33}\text{cm}^{-2}\text{s}^{-1}$.

A selection requiring the two muons to be oppositely charged, and the pair to have invariant mass between 2.8 GeV and 3.4 GeV was applied to both TrigDiMuon and the topological di-muon trigger. For the topological algorithm the invariant mass selection was only applied after muComb. The fake rates were calculated as follows: the probability of each of the level-2 strategies to find a di-muon pair in events which contain only a single muon was estimated separately for the $b$ events and the minimum-bias events. This probability was multiplied by rates which were estimated independently from [6], but are consistent with it. The fake probability in $b$ events was taken to be representative of that in all events with prompt muons. For the topological di-muon trigger, the probability was calculated relative to the number of di-muon level-1 triggers, and multiplied by the level-1 fake di-muon trigger rate from [6].

### 4.1 Efficiency relative to events accepted at level-1

Table 2 gives the efficiency, relative to level-1, for the two di-muon trigger algorithms for a trigger threshold of 4 GeV. Table 3 gives the efficiencies, relative to level-1, for a trigger threshold of 6 GeV. The efficiencies in these tables are calculated with respect to the $J/\psi$ events accepted by the corresponding level-1 single muon trigger. In parenthesis we give the efficiency relative to $J/\psi$ events at the starting point of the di-muon algorithm.

One can see that the TrigDiMuon efficiency is significantly higher than the topological di-muon trigger in all cases. As a matter of fact, the topological di-muon trigger, which applies $p_T$ cuts on both muons, can have only a limited acceptance for $J/\psi$ events passing a single muon trigger because the second muon is very frequently below the trigger threshold. To pass the topological di-muon trigger the second, lower $p_T$ muon also has to pass the level-2 selections.

Because TrigDiMuon can reconstruct muons below the level-1 thresholds, the TrigDiMuon efficiency, shown in brackets, remains nearly the same with the different chains. However, the total efficiency is reduced when starting from the single muons accepted by muFast or muComb, because of the $p_T$ cut imposed by those algorithms. In particular, in Table 3 muFast and muComb reject successively more of the muons below the nominal threshold and this explains the big drop from row to row for TrigDimuon and even bigger drop for the topological trigger.

The loss of efficiency in the single muon triggers is smaller for the 4 GeV threshold, because the level-2 cuts for the 4 GeV threshold are quite loose, as mentioned above. When calculating the 4 GeV trigger efficiency using only muons from $J/\psi$ with generated $p_T$ above the 4 GeV threshold, TrigDiMuon has an efficiency of 90% and the topological trigger has an efficiency of 64%.

As discussed earlier, in spite of some loss of efficiency, the trigger chains with TrigDiMuon running after a level-2 confirmed muon might be more suitable to an overall planning of the ATLAS trigger menu due to the reduced input rate they have to sustain.

### 4.2 Fake rates

Table 4 gives the expected fake rates for TrigDiMuon with the different chains described above, for a trigger threshold of 4 GeV at a luminosity of $10^{31}\text{cm}^{-2}\text{s}^{-1}$. Table 5 gives the rates for a trigger threshold of 6 GeV at a luminosity of $10^{33}\text{cm}^{-2}\text{s}^{-1}$.

| Chain starting from | TrigDiMuon (%) | Topological trigger (%) |
|---|---|---|
| level-1 | 73 (73) | 51 |
| muFast | 71 (73) | 43 |
| muComb | 70 (74) | 33 |

Table 2: Efficiency, relative to level-1, of the two di-muon trigger algorithms for a trigger threshold of 4 GeV. In parenthesis is the efficiency calculated relative to $J/\psi$ events that passed the single muon trigger that selects the input to TrigDiMuon. To estimate the efficiency we used a sample of $\Lambda_b \rightarrow J/\psi\Lambda$, where $J/\psi \rightarrow \mu(p_T > 2.5\text{ GeV})\mu(p_T > 4\text{ GeV})$.

| Chain starting from | TrigDiMuon (%) | Topological trigger (%) |
|---|---|---|
| level−1 | 75 (75) | 56 |
| muFast | 67 (77) | 25 |
| muComb | 60 (78) | 15 |

Table 3: Efficiency, relative to level-1, of the two di-muon trigger algorithms for a trigger threshold of 6 GeV. In parenthesis is the efficiency calculated relative to $J/\psi$ events that passed the single muon trigger that selects the input to TrigDiMuon. To estimate the efficiency we used a sample of $\Lambda_b \rightarrow J/\psi\Lambda$, where $J/\psi \rightarrow \mu(p_T > 2.5\text{ GeV})\mu(p_T > 4\text{ GeV})$.

| Source | Chain starting from | Input rate (Hz) | Fake acceptance (%) | Fake rate (Hz) |
|---|---|---|---|---|
| $b+c$ | level−1 | 460 | 0.42 | 1.9 |
| | muFast | 380 | 0.42 | 1.6 |
| | muComb | 340 | 0.43 | 1.5 |
| $K/\pi$ | level−1 | 620 | 0.07 | 0.43 |
| | muFast | 270 | 0.11 | 0.29 |
| | muComb | 170 | 0.09 | 0.15 |
| **Total** | level−1 | 1080 | 0.22 | 2.3 |
| | muFast | 650 | 0.29 | 1.9 |
| | muComb | 510 | 0.32 | 1.6 |

Table 4: Fake rate of the TrigDiMuon algorithm for muons from different sources and total fake rate using a trigger threshold of 4 GeV, at a luminosity of $10^{31}\text{cm}^{-2}\text{s}^{-1}$. The $b$ and $c$ components were estimated from a sample of $b\bar{b} \rightarrow \mu + X$ with $p_T^{\mu} > 4$ GeV and the $K/\pi$ component from the minimum bias sample with forced decays

Fake rates can be further reduced by reconstructing the $J/\psi$ decay vertex from the two muon tracks. Selecting $J/\psi$ with a good quality vertex fit will reduce fake rates from unrelated track combinations. Figure 4 shows the distribution of the vertex $\chi^2$ for true $J/\psi$ decays and for fake di-muon triggers. A cut of $\chi^2 < 30$ reduces the fake rate by 20-30% and only reduces the trigger

| Source | Chain starting from | Input rate (Hz) | Fake acceptance (%) | Fake rate (Hz) |
|---|---|---|---|---|
| $b+c$ | level-1 | 21500 | 0.48 | 103 |
| | muFast | 12500 | 0.60 | 76 |
| | muComb | 10300 | 0.75 | 76 |
| $K/\pi$ | level-1 | 15800 | 0.22 | 34 |
| | muFast | 5000 | 0.30 | 15 |
| | muComb | 3500 | 0.41 | 14 |
| **Total** | level-1 | 37400 | 0.37 | 137 |
| | muFast | 17500 | 0.51 | 91 |
| | muComb | 13700 | 0.66 | 90 |

Table 5: Fake rate of the TrigDiMuon algorithm for muons from different sources and total fake rate using a trigger threshold of 6 GeV, at a luminosity of $10^{33}\text{cm}^{-2}\text{s}^{-1}$. The $b$ and $c$ components were estimated from a sample of $b\bar{b} \rightarrow \mu + X$ with $p_T^{\mu} > 4$ GeV and the $K/\pi$ component from the minimum bias sample with forced decays

efficiency by 1-2%. The vertex position can also be used to reject $J/\psi$ produced at the primary interaction and accept only $J/\psi$ from $b$ hadron decays, but a study of this is outside the scope of this note.

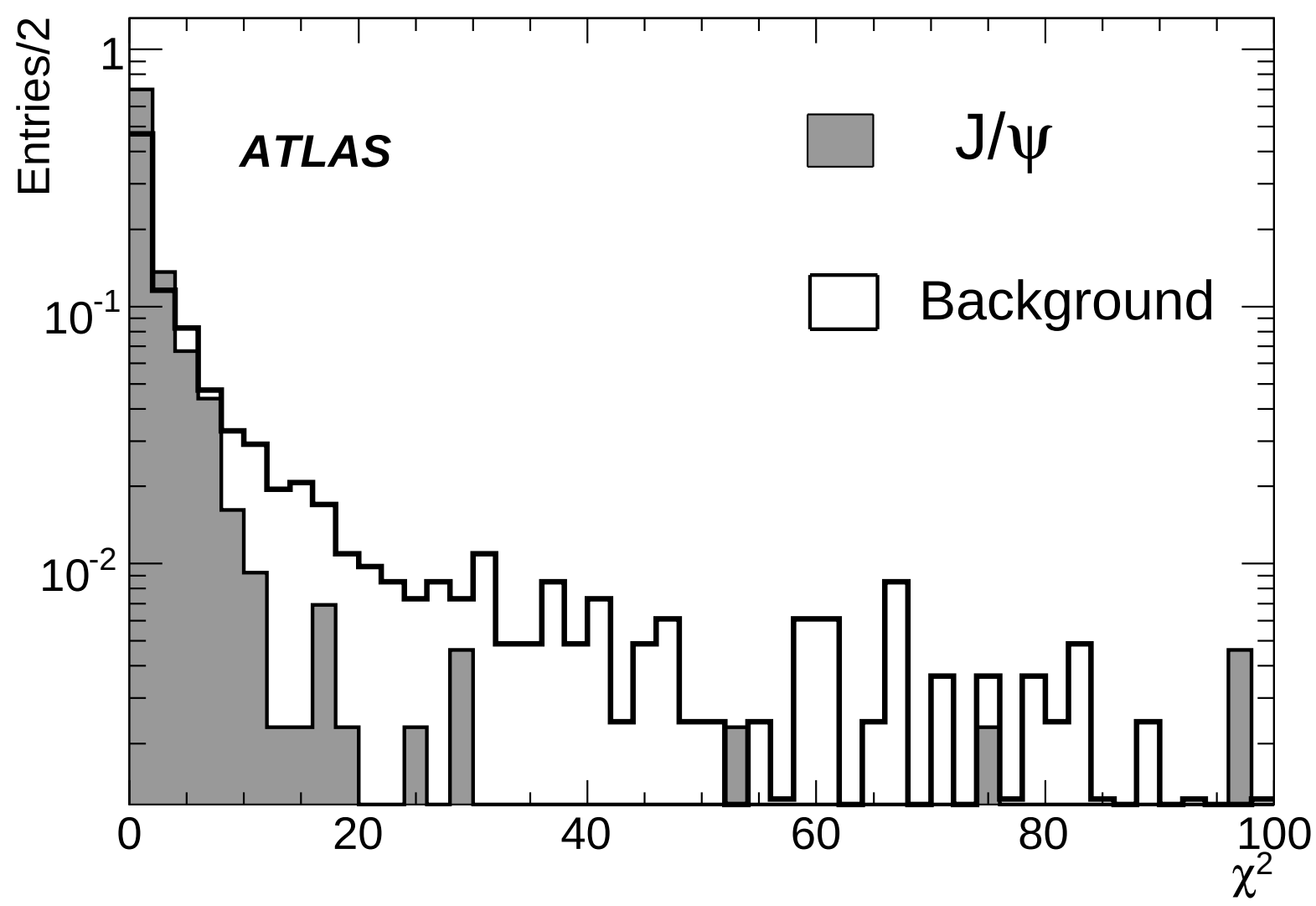


Figure 4: Distribution of the vertex $\chi^2$ for true $J/\psi$ decays (shaded) and for fake di-muon triggers (open histogram).

Finally Table 6 compares the total rates and efficiencies of the two level-2 di-muon algorithms. When TrigDiMuon runs after muComb, the signal to background ratio is worse than when it runs after muFast. Using the input rates from Table 4 and 5, one can calculate that the time needed to reconstruct

the ID tracks is also the smallest when TrigDiMuon runs after muFast, reduced by 1/3 for the 4 GeV threshold and by 1/2 for the 6 GeV threshold with respect to running after L1.

It can be seen from this table that the fake rate for the topological di-muon trigger is small. The fake rate from the TrigDiMuon algorithm is much higher but these output rates from level-2 should be acceptable for the gain in efficiency, at the initial low luminosity.

| Threshold (Luminosity) | Chain starting from | TrigDiMuon | | | Topological | | |
|---|---|---|---|---|---|---|---|
| | | Efficiency (%) | $J/\psi$ rate (Hz) | Total rate (Hz) | Efficiency (%) | $J/\psi$ rate (Hz) | Total rate (Hz) |
| **4 GeV** ($10^{31}$cm$^{-2}$s$^{-1}$) | level$-$1 | 71 | 1.17 | 3.1 | 51 | 0.8 | 24 |
| | muFast | 70 | 1.15 | 2.7 | 43 | 0.7 | - |
| | muComb | 69 | 1.14 | 2.4 | 33 | 0.5 | 0.6 |
| **6 GeV** ($10^{33}$cm$^{-2}$s$^{-1}$) | level$-$1 | 74 | 43 | 151 | 56 | 32.5 | 357.5 |
| | muFast | 66 | 38 | 114 | 25 | 14.5 | - |
| | muComb | 59 | 34 | 109 | 15 | 8.7 | 9.3 |

Table 6: Total rate and efficiency relative to level-1 of the TrigDiMuon algorithm including the vertex cut $\chi^2 < 30$, and of the topological di-muon trigger. The efficiency is estimated from a sample of $\Lambda_b \to J/\psi\Lambda$, where $J/\psi \to \mu(p_T > 2.5\text{ GeV})\mu(p_T > 4\text{ GeV})$.

The $b$-physics trigger rate from TrigDiMuon requires further reduction for $L = 10^{33}$. This can be achieved by introducing an additional trigger with a cut on the $J/\psi$ decay length. Then the trigger without decay length cut will be prescaled to an acceptable rate for calibration and alignment purposes, as for example in Section 6. An event filter algorithm will further reduce the rates for a luminosity of $10^{33}$cm$^{-2}$s$^{-1}$.

### 4.3 Efficiency relative to reconstructed events

Our goal is to maximize trigger efficiency at the level-2 trigger for the muons that can later be identified offline. The efficiencies for the two algorithms and thresholds were re-estimated with respect to the $J/\psi$ events reconstructed with the muon identification program MuGirl [10], which is efficient for low-$p_T$ muons. The resulting efficiencies are given in Table 7 for a trigger thresholds of 4 GeV and 6 GeV respectively. Figure 5 shows the efficiency of TrigDiMuon relative to muons identified by MuGirl for the higher $p_T$ muon (left) and the second muon (right).

| Threshold (Luminosity) | Chain starting from | TrigDiMuon Efficiency (%) | Topological Trigger Efficiency (%) |
|---|---|---|---|
| **4 GeV** ($10^{31}$cm$^{-2}$s$^{-1}$) | level$-$1 | 84 | 58 |
| | muComb | 81 | 42 |
| **6 GeV** ($10^{33}$cm$^{-2}$s$^{-1}$) | level$-$1 | 81 | 45 |
| | muComb | 66 | 17 |

Table 7: Efficiency of the TrigDiMuon and Topological di-muon algorithms for $J/\psi$ reconstructed by MuGirl. To estimate the efficiency we used a sample of $\Lambda_b \to J/\psi\Lambda$, where $J/\psi \to \mu(p_T > 2.5\text{ GeV})\mu(p_T > 4\text{ GeV})$.

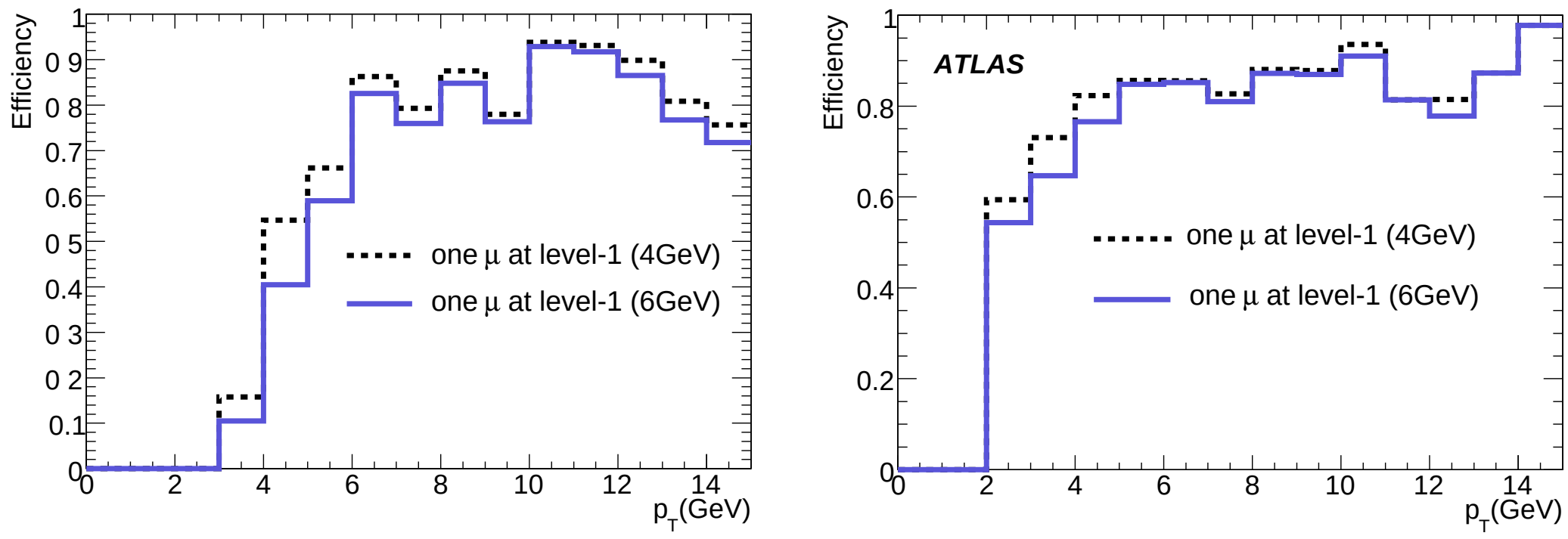


Figure 5: Efficiency of TrigDiMuon relative to muons identified by MuGirl for the higher $p_T$ muon (left) and the second muon (right).

# 5 Rejection of muons from $K$ and $\pi$ decays

The lowest single muon $p_T$ threshold in the original ATLAS HLT design [5] was chosen to be 6 GeV. This is because below this $p_T$ value the rate of muons from $K$ and $\pi$ decays becomes higher than that from $b$ and $c$ decays. Nevertheless, during the low luminosity phase, it is desirable to collect muons with lower $p_T$, both for detector and trigger calibration and for initial physics studies. If thresholds are lowered, $K$ and $\pi$ decays become the dominant source of single muon triggers. These decay muons must be rejected as early as possible so as not to dominate the single muon trigger rate, thus ensuring we can achieve the physics and calibration studies with prompt muons.

## 5.1 Description of the method

The method we describe rejects $K$ and $\pi$ decays based not on their $p_T$ but on the topology of the decay. The track position of a prompt muon, extrapolated from the MS to the interaction vertex, is a gaussian distributed around the ID track position. The corresponding distribution of a decay muon around the light hadron from which it decayed is broader because of the contribution of the decay kink in addition to the multiple scattering effect. Thus if the track seen in the inner detector is that of the $K$ or $\pi$, this discrepancy with prompt muons can be used to reject some of the muons from light hadron decays.

With this method, we can reject the muons from $K$ and $\pi$ decays by using a matching window tuned for prompt muons, with the window width varying according to the track $p_T$. Due to time constraints, a precise propagation of tracks in the magnetic field can not be done at level-2, so instead the muon track from the MS is extrapolated back to the interaction vertex using a parameterization that is a function of measured $\eta$, $\phi$ and $p_T$ (exploiting the linear relationship between the bending and $1/p_T$). Multiple scattering effects can be parameterized in the same way to estimate the corresponding errors of the back extrapolation, which determine the window size.

Three main regions are identified for the field parameterization used to propagate the muon tracks, one in the barrel and two in the end-cap. To account for the relative inhomogeneity of the magnetic field inside each region, the parameters of the back extrapolation are computed as a function of $\eta$, $\phi$, muon charge and spectrometer side ($z$ or $-z$). A different tuning is used for high and low-$p_T$ tracks, to take into account the fluctuations of the energy loss in the calorimeter which are important for the

propagation of the latter. In the end-cap, the innermost station of the MS does not provide a complete geometric coverage, thus for the muons with no hits in the innermost station, a seed from the middle station is used to extrapolate the track back. This is the most difficult case since an extrapolation through the toroids must be performed and the resulting precision is spoiled by a factor of two with respect to the other cases. Therefore the back extrapolation in the end-cap is treated separately for regions with inner station coverage and region without inner station coverage.

The strategy for the level-2 combined muon reconstruction described in [5] is to use only the pixel and the SCT data. With this setup, the decays in flight happening near or after the last SCT layer are reconstructed using mainly the hits of the decaying $K$ or $\pi$. Some rejection of these events can be achieved by checking the $\eta$ and $\phi$ position of the extrapolated MS track relative to the ID track with a matching window whose size is based on the position spread coming out from the muon multiple scattering. Some of the decays between the pixel and the SCT can be rejected by applying a cut on the $\chi^2$ of the Inner Detector fit.

The cuts studied, ordered in terms of increasing rejection are:

- A loose-window cut, using the muon track position from muFast, and the track position from the ID reconsrtruction. The window size was tuned to recover almost 100% of the multiple scattering for a muon $p_T$ equal to the threshold value (4 GeV and 6 GeV);
- A tight-window cut, refining the muon back-extrapolation by exploiting the measurement of the interaction vertex from the ID reconstruction. The window size was tuned to 2.7 $\sigma$ of the multiple scattering spread for that muon $p_T$; the combined $p_T$ estimation is used to tune the window width;
- The normalized $\chi^2$ of the ID track fit is required to be less than 3.2.

## 5.2 Performance of the method

### 5.2.1 Rejection results for $K$ and $\pi$ decays

The 6 GeV threshold is used as a benchmark to estimate the $K$ and $\pi$ rejection achieved by the various cuts. The forced-decay minimum bias sample has been used to study and tune the cuts to reject muons from $K$ and $\pi$ decays. The results shown in Table 8 are expressed in terms of trigger rate, computed using the cross section of these events.

| Cut | $\pi/K$ rate (Hz) |
|---|---|
| Baseline muComb | $3470 \pm 380$ |
| Loose window cut | $2920 \pm 430$ (-16%) |
| Tight window cut | $2800 \pm 440$ (-20%) |
| Tight window + $\chi^2$ cut | $2550 \pm 440$ (-26%) |

Table 8: Expected rate with a 6 GeV single muon threshold from the muComb algorithm for the $\pi$ and $K$ decays. The rejection with respect to the baseline muComb algorithm is shown in parenthesis. These rates were estimated from the forced-decay minimum bias sample

Figure 6 shows the trigger efficiency for muons from $K$ and $\pi$ decays as a function of $p_T$ for the baseline muComb selection, compared to the tight window selection described above.

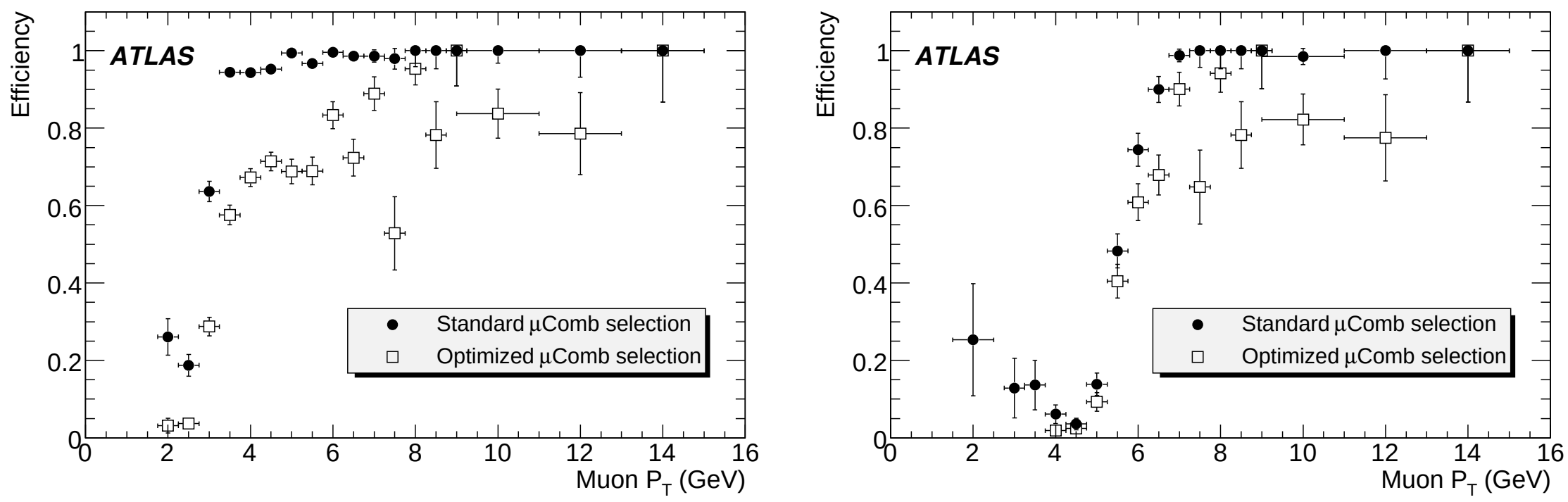


Figure 6: Efficiency for muons from $K$ and $\pi$ decays as a function of $p_T$ for the baseline muComb selection, compared to the optimized muComb selection described in Section 5.1 for the 4 GeV threshold (left) and the 6 GeV threshold (right).

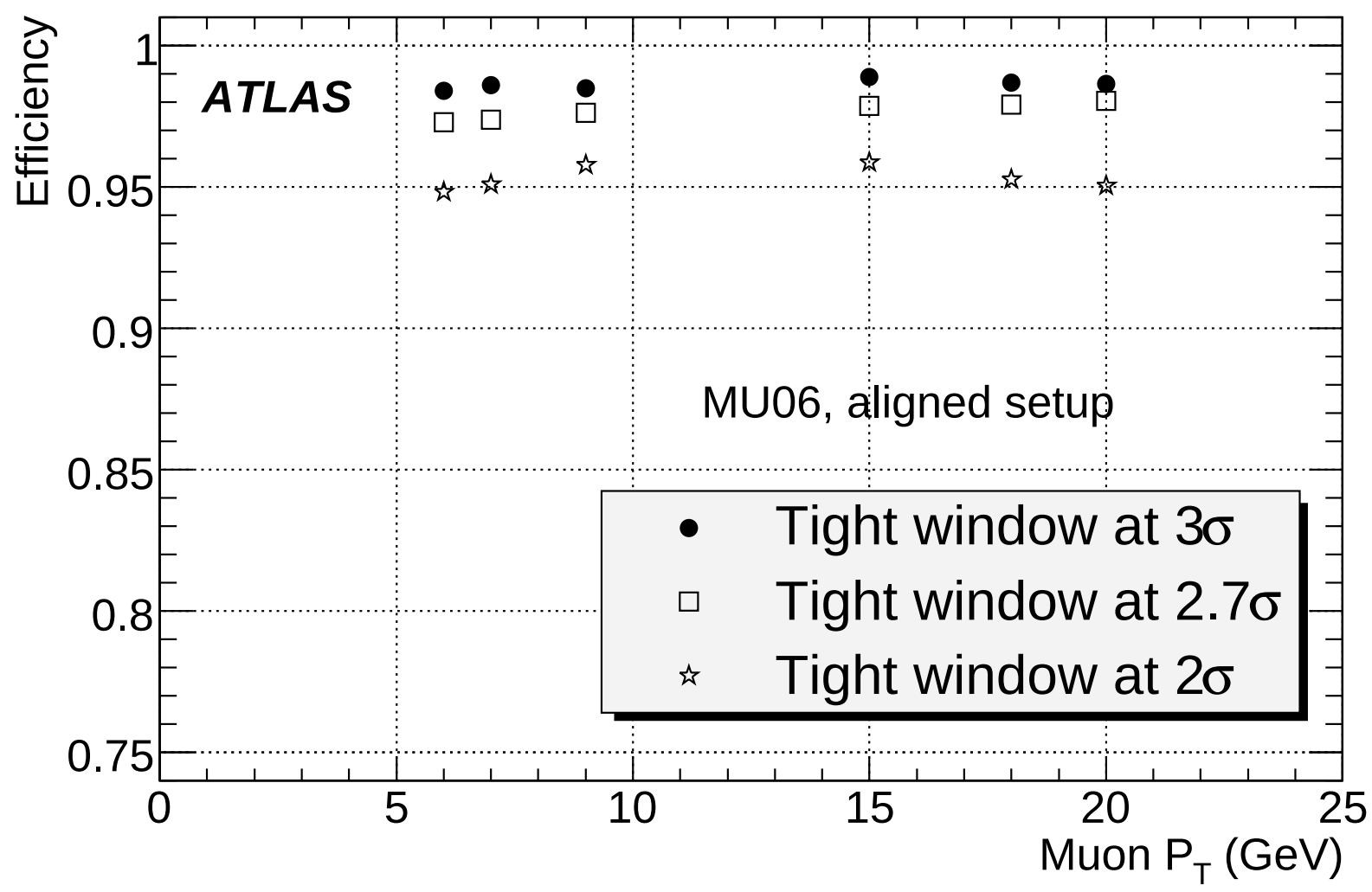


Figure 7: Efficiency of the tight window match on single muons simulated with the aligned detector setup. The efficiency drop for the 6 GeV threshold is shown according to different $\sigma$ cuts.

### 5.2.2 Efficiency loss for prompt muons

The efficiency loss for prompt muons due to the matching window cuts has been estimated with a single muon sample. Figure 7 shows the relative efficiency obtained for the aligned detector setup. The relative efficiency is seen to be almost constant in the $p_T$ range of 4-40 GeV and its value for the cut at 2.7 $\sigma$ is about 98%. The detector misalignment reduces the efficiency plateau to 95% as shown in Figure 8, but the relative efficiency can be recovered by 1% with a specific tuning of the back extrapolator.

The rate reduction for $b$ decays is calculated reliably from the $b \to \mu(4) + X$ sample. The results are shown in Table 9. A good agreement between $b$ events and single muons is found for the efficiency

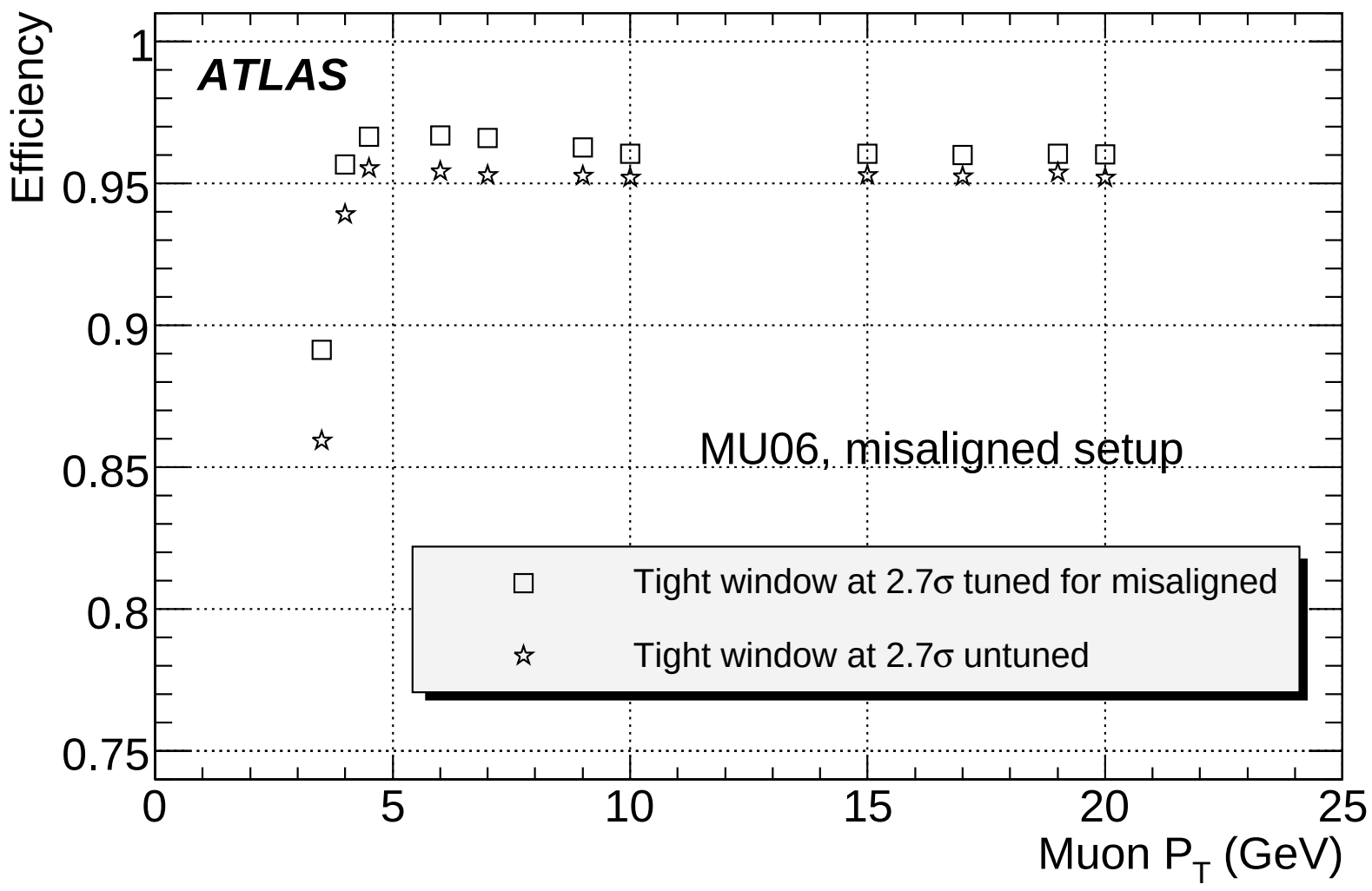


Figure 8: Efficiency of the tight window match at 2.7 $\sigma$ on single muons simulated with the misaligned detector setup. The relative efficiency is shown for the 6 GeV threshold.

loss due to the tight window match.

| Cut | b$\rightarrow \mu(4)+X$ rate (Hz) |
|---|---|
| Baseline muComb | $4850 \pm 20$ |
| Loose window cut | $4780 \pm 20$ (-1.5%) |
| Tight window cut | $4710 \pm 20$ (-3%) |
| Tight window + $\chi^2$ cut | $4560 \pm 20$ (-6%) |

Table 9: Expected rate with a 6 GeV single muon threshold from the muComb algorithm for the $b$ component. In parenthesis is the percentage rejection with respect to the baseline muComb.

### 5.2.3 Resulting muon trigger rates

A coherent description of the full trigger rate after the muComb algorithm is obtained using the standard minimum bias sample and is shown in Table 10 and in Table 11. All the cuts mentioned were applied. A very good agreement is found with both the forced sample for the $K$ /$\pi$ component, and the $b \rightarrow \mu + X$ sample for the $b$ component.

The optimized version of muComb improves the rejection of muons from decays in flight by about 30% at the 4 GeV threshold and of about 20% at the 6 GeV threshold with respect to the baseline muComb algorithm. The rejection of muons from b events is 20% at the 4 GeV threshold and 7% at the 6 GeV threshold. Thus while the total trigger rate is reduced the purity of the sample increases. Given the uncertainties on the estimation of the production cross section for $K$ and $\pi$ this optimization is crucial for the low-$p_T$ single muon trigger.

| Sample | muFast rate (Hz) | muComb rate (Hz) | muComb + $\pi/K$ cuts rate (Hz) |
|---|---|---|---|
| $\pi/K$ | $224\pm15$ | $210\pm9$ | $145\pm9$ |
| $b$ | $145\pm7$ | $140\pm6$ | $114\pm6$ |
| $c$ | $234\pm8$ | $200\pm8$ | $168\pm8$ |

Table 10: Expected output rate of muFast and muComb for a 4 GeV threshold at the $10^{31}\mathrm{cm}^{-2}\mathrm{s}^{-1}$ Luminosity.

| Sample | muFast rate (Hz) | muComb rate (Hz) | muComb + $\pi/K$ cuts rate (Hz) |
|---|---|---|---|
| $\pi/K$ | $5050\pm760$ | $3530\pm380$ | $2860\pm410$ |
| $b$ | $5550\pm600$ | $4900\pm400$ | $4550\pm430$ |
| $c$ | $6900\pm700$ | $5390\pm420$ | $5050\pm450$ |

Table 11: Expected output rate of muFast and muComb for a 6 GeV threshold at the $10^{33}\mathrm{cm}^{-2}\mathrm{s}^{-1}$ Luminosity.

Checking the effect of this method to reject muons from $K$ and $\pi$ decays on the two level-2 di-muon strategies showed that both efficiency and trigger rates are reduced. There is no significant gain in purity from this method for the di-muon selections, because the rejection is achieved by using the $J/\psi$ reconstruction and mass cuts, and most of the fake rate comes from $b$ and $c$ decays.

# 6 Measuring trigger efficiency for low-$p_T$ muons from ATLAS data

## 6.1 Method description

Cross section measurements require a good understanding of the efficiency of the event selections. A precise understanding of the trigger efficiency is crucial and we must have a strategy for measuring it from data with high precision. We study the performance of measuring the trigger efficiency from data with the tag-and-probe method which uses di-muon final states for measuring the single muon trigger efficiency. In this method, a single triggered muon from a reconstructed di-muon decay of a specific particle identified by mass cuts provides the tag that allows us to probe the trigger efficiency of the second muon.

For $B$ physics we are interested in events with rather low-$p_T$ muons, so the tag-and-probe method for measuring the single muon trigger efficiency using $J/\psi$ events is presented. We demonstrate that the obtained trigger efficiency can be applied to calculate the di-muon trigger efficiency. This principle can also be applied to Z decays to calibrate the high-$p_T$ trigger efficiency [11].

### 6.1.1 Measuring single muon efficiency

We use the tag-and-probe method to measure the muon trigger efficiency, using as the calibration sample events collected by a single muon trigger where the $J/\psi$ is found in the offline reconstruction. In this sample, one of the muons forming the $J/\psi$ is triggered, while the other one may or may not be triggered, thus providing an unbiased sample of muons to study the single muon trigger efficiency.

First, the triggered muon is matched to one of the reconstructed muons from an identified $J/\psi$. This muon is called the *tagged*-muon. Once the tagged-muon is identified, the other muon, the *probe*-

muon is used to check whether it is also triggered or not. A matching between the reconstructed muon and the one found at the trigger level is needed here too.

At the high level trigger, the position of the muon found at the trigger level is stored, and can be compared precisely to the reconstructed muon. However, at level-1 the position granularity is that of the RoI. Due to the limited precision of the location of the RoI, care must be taken when matching a muon RoI to the reconstructed muon. The bending of the muon tracks in the magnetic field and the fact that high-$p_T$ decay muons have small opening angles between them also introduces an ambiguity in the matching.

The single muon trigger efficiency, $\varepsilon_{1\mu}$, is calculated as the ratio between the number of probe muons which were triggered ($N_{probe\&triggered}$) to the total number of probe muons ($N_{probe}$),

$$\varepsilon_{1\mu} = \frac{N_{probe\&triggered}}{N_{probe}}. \tag{1}$$

The single muon efficiency can be obtained in detail as a function of kinematic variables ($p_T$, $\eta$, $\phi$) of the muons using as fine a binning as the statistics allows. A fine binning is, in fact, necessary since the efficiency depends on these variables, especially at level-1 where there are sharp changes in the efficiency due to structural features such as the experiment's support structures. Because of this the overall efficiency depends on the distribution of the muons produced. We call the detailed *map* of efficiencies in each region of the phase space a trigger efficiency map.

Our primary goal is to demonstrate that it is possible to obtain the trigger efficiency map from data alone. The di-muon trigger efficiency can then be calculated from it, given the distribution of the parent particles and the decay angular distribution.

### 6.1.2 Calculating di-muon trigger efficiency

The di-muon trigger efficiency can be calculated using the obtained single muon efficiencies, taking into account the dependence on kinematic variables of the muons. For example, the efficiency of $J/\psi$ particles are different depending on the kinematic distribution of the two decay muons. The $J/\psi$ efficiency, $\varepsilon^{J/\psi}$ can be calculated using the single muon trigger efficiency map as

$$\varepsilon_{J/\psi}(p_T^{J/\psi}, \eta^{J/\psi}, \phi^{J/\psi}) = \frac{1}{2\pi} \iint \varepsilon_{1\mu}(p_T^{\mu_1}, \eta^{\mu_1}, \phi^{\mu_1}) \varepsilon_{1\mu}(p_T^{\mu_2}, \eta^{\mu_2}, \phi^{\mu_2}) f(\cos\theta^*) d\cos\theta^* d\phi^*. \tag{2}$$

Here, $f(\cos\theta^*)$ is the angular distribution of the decay muon from the $J/\psi$ where $\theta^*$ represents the decay angle of the muon in the $J/\psi$ rest frame with the $z$-axis taken as the direction of the $J/\psi$ in the laboratory frame. The variable $\phi^*$ is the azimuthal angle of the decay muon in the $J/\psi$ rest frame, normalized as $\int f(\cos\theta^*) d\cos\theta^* = 1$. Kinematic variables of the decay muons ($p_T^{\mu_1}$, $p_T^{\mu_2}$,$\eta^{\mu_1}$, $\eta^{\mu_2}$, $\phi^{\mu_1}$, $\phi^{\mu_2}$) are functions of the $J/\psi$ variables, $\cos\theta^*$ and $\phi^*$. To get the overall efficiency of $J/\psi$ events, the integration of $J/\psi$ variables must be performed in the kinematic region of the cross-section definition.

Note that this formula is universal and can be applied to other resonances such as $B_{s,d} \to \mu\mu X$ using the same single muon efficiency map. The $\cos\theta^*$ distribution depends on the polarization state of the $J/\psi$ which reflects the $J/\psi$ production mechanism. For the unpolarized case, this distribution is flat. In certain analysis, the production mechanism of the parent particle could be of interest, so we cannot assume the distribution to be flat. In such cases, it is necessary to be able to calculate the efficiency as a function of $\cos\theta^*$ as well. In these cases, Equation 2 would become,

$$\varepsilon_{J/\psi}(p_T^{J/\psi}, \eta^{J/\psi}, \phi^{J/\psi}, \cos\theta^*) = \frac{1}{2\pi} \int \varepsilon_{1\mu}(p_T^{\mu_1}, \eta^{\mu_1}, \phi^{\mu_1}) \varepsilon_{1\mu}(p_T^{\mu_2}, \eta^{\mu_2}, \phi^{\mu_2}) d\phi^* \tag{3}$$

without the $\cos\theta^*$ integration. Equation 3 is simply expressing the efficiency of events where $J/\psi$ is produced with a fixed momentum and the decay angle is also fixed, so the decay muon momenta are also fixed. Some variables could be integrated out, but the important thing is that the efficiency with respect to $\cos\theta^*$ can also be obtained from the single muon efficiency map, $\varepsilon_{1\mu}(p_T^\mu, \eta^\mu, \phi^\mu)$.

### 6.2 Performance studies

In order to emulate the efficiency measurement from data, we use $J/\psi$ events with two muons in the offline reconstruction passing the single muon threshold of 6 GeV and a di-muon invariant mass between 2.88 GeV and 3.3 GeV.

#### 6.2.1 Matching of muons at trigger and reconstruction

The first step of the tag-and-probe method is to find out which of the two offline muons was triggered. This is done by finding the best match using $\Delta R = \sqrt{\Delta\eta^2 + \Delta\phi^2}$, where $\Delta\eta$ and $\Delta\phi$ are the difference of $\eta$ and $\phi$ between the offline muon and the triggered muon. For the matching with level-2 muons, track parameters at the perigee are used. On the other hand, for level-1, the position of the triggered muon is taken as the center of the RoI. In this case, the matching must be done carefully as the opening angle of the two muons from the $J/\psi$ is small and the bending, of the rather low-$p_T$ muons, in the magnetic field is non-negligible. The offline muon tracks are extrapolated to the plane of the RPC or TGC chamber which defines the RoI.

Figure 9 shows the $\Delta R$ distrbution between the level-1 RoI and offline track with and without using the extrapolation. The improvement obtained by the extrapolation is significant and as a result a good matching is established by requiring $\Delta R < 0.15$. To further reduce the possibility of having a wrong match, in this study we only use events where the opening angle between the two muons is $\Delta R > 0.4$.

Figure 10 shows the $\Delta R$ distribution between the level-2 muon and the offline track. The resolution of $\Delta R$ becomes an order of magnitude better than at level-1 since the level-2 muon tracks use the measurements from the ID. The condition $\Delta R < 0.005$ is used for the matching between the offline track and the level-2 track.

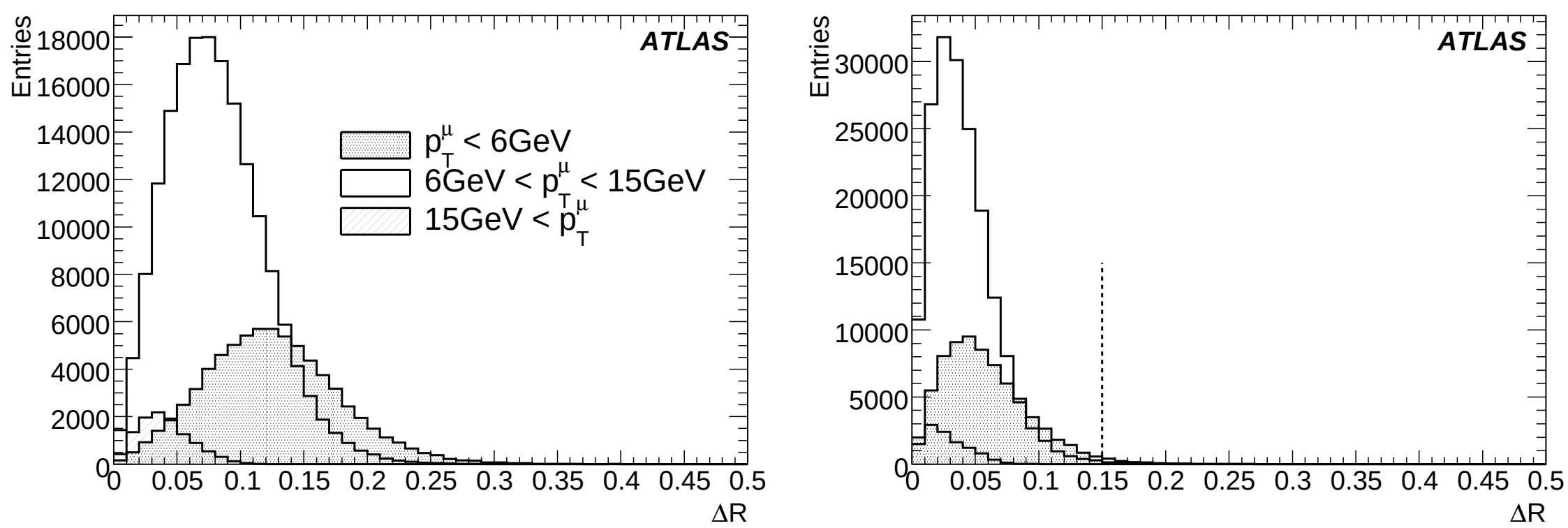


Figure 9: The distribution of $\Delta R$ between the level-1 RoI and the offline muon track, (a) using the offline track parameters at the perigee and (b) by extrapolating the offline track to the RoI position. The dashed line shows the value where the cut was applied for the matching.

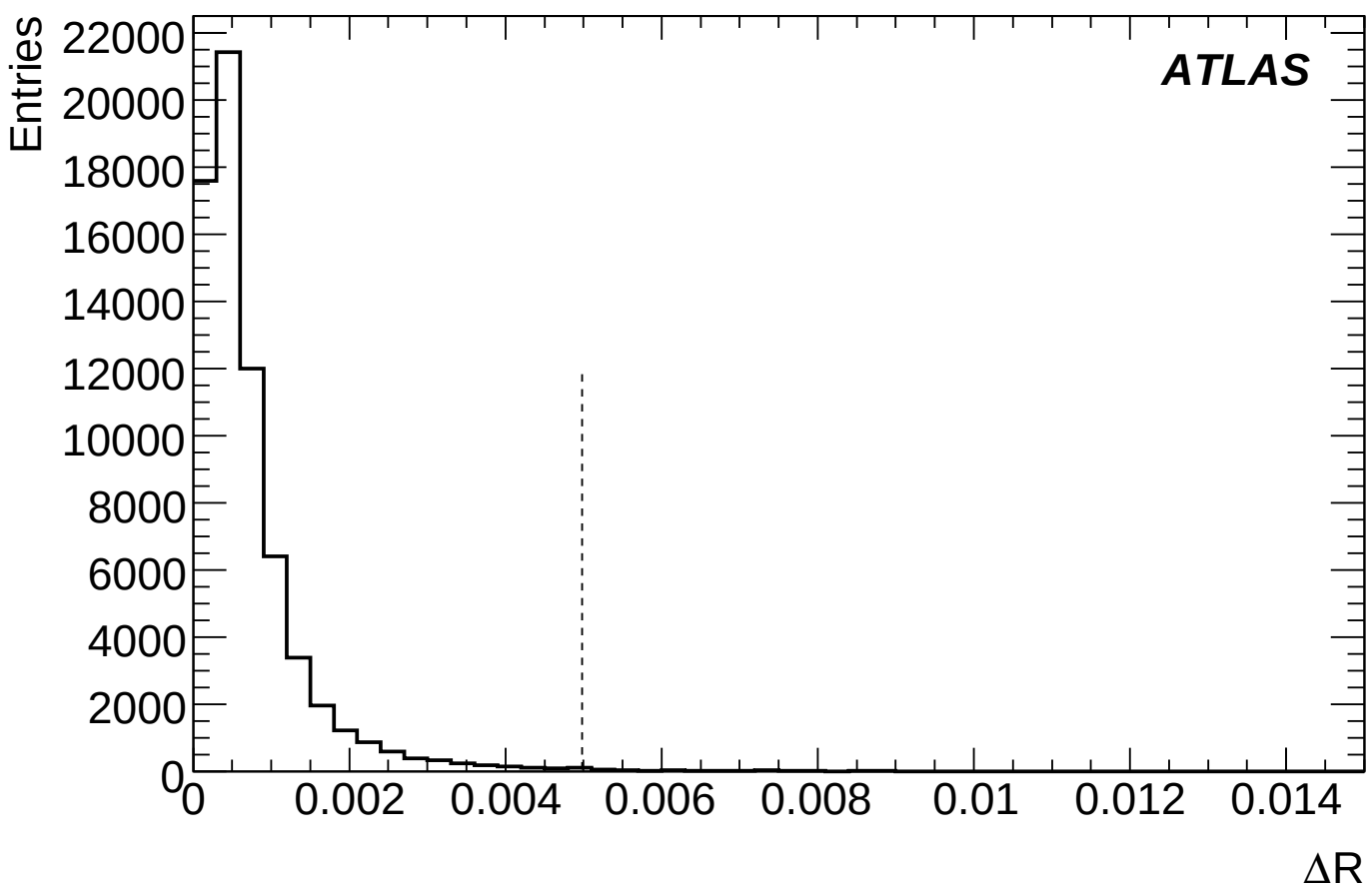


Figure 10: The distribution of $\Delta R$ between the level-2 muon and the offline muon track. The dashed line shows the value where the cut was applied for the matching.

### 6.2.2 Level-1 efficiency

Using the method described in Section 6.1 and the matching criteria, we obtain the single muon efficiency as a function of the $p_T$ of the muon. Figure 11 shows the single muon efficiency as a function of $p_T$ by evaluating how often the probe-muon was triggered. The points are fitted with the function

$$\varepsilon(p_T) = \frac{A}{1+\exp(-a \times (p_T - b)))}. \quad (4)$$

Also shown in Figure 11 is the efficiency measured directly in the single muon Monte Carlo sample which provides an unbiased value of the efficiency. The agreement of the efficiencies obtained by the two methods is around 5% in the turn-on region (4 GeV $< p_T <$ 8 GeV) and becomes smaller as the $p_T$ increases, becoming within a few percent at $p_T >$ 10 GeV.

To calculate the di-muon trigger efficiency using Equation 2, the efficiency curve must be measured in each $\eta$ and $\phi$ region. For this, we divided the detector into $10 \times 10$ regions for the barrel ($-1.05 < \eta < 1.05$). For the end-cap region ($1.05 < |\eta| < 2.4$), we assumed that there is a complete symmetry between octants and divided one octant into $8 \times 6$ regions. Efficiency curves as a function of $p_T$ are obtained in each of the regions. Figure 12 shows the three fit parameters, $A$, $a$ and $b$ in Equation 4 as a function of $\eta$ and $\phi$. The figure shows that the efficiency curve behaves differently for different regions but in a smooth way, except for a small region around $\eta = 0.725$ and $\phi = -1.6$. In this region the efficiency is very low due to the MS layout and as a result the fit is unreliable. These results are used to calculate the di-muon efficiency.

Distributions of offline $J/\psi$ variables, $p_T^{J/\psi}$, $\eta^{J/\psi}$ and $\cos\theta^*$ are shown in Figure 13 after applying the selection criteria of muons ($p_T^\mu >$ 6 GeV and $|\eta^\mu| < 2.4$). The open histograms are for all $J/\psi$ in the MC sample and the filled histograms are for events where the level-1 di-muon trigger, requiring at least two muons with $p_T >$ 6 GeV, has fired. The generator level filter with the $p_T$ cut for the highest (second highest) $p_T$ muon of $p_T > 6(4)$ GeV was applied for the MC sample.

The $\cos\theta^*$ distribution reflects the polarization state of the $J/\psi$ and is flat for the non-polarized case. Since the $\cos\theta^*$ distribution is an interesting quantity to measure in its own right, we do not make any assumptions about this distribution but instead try to measure the efficiency as a function

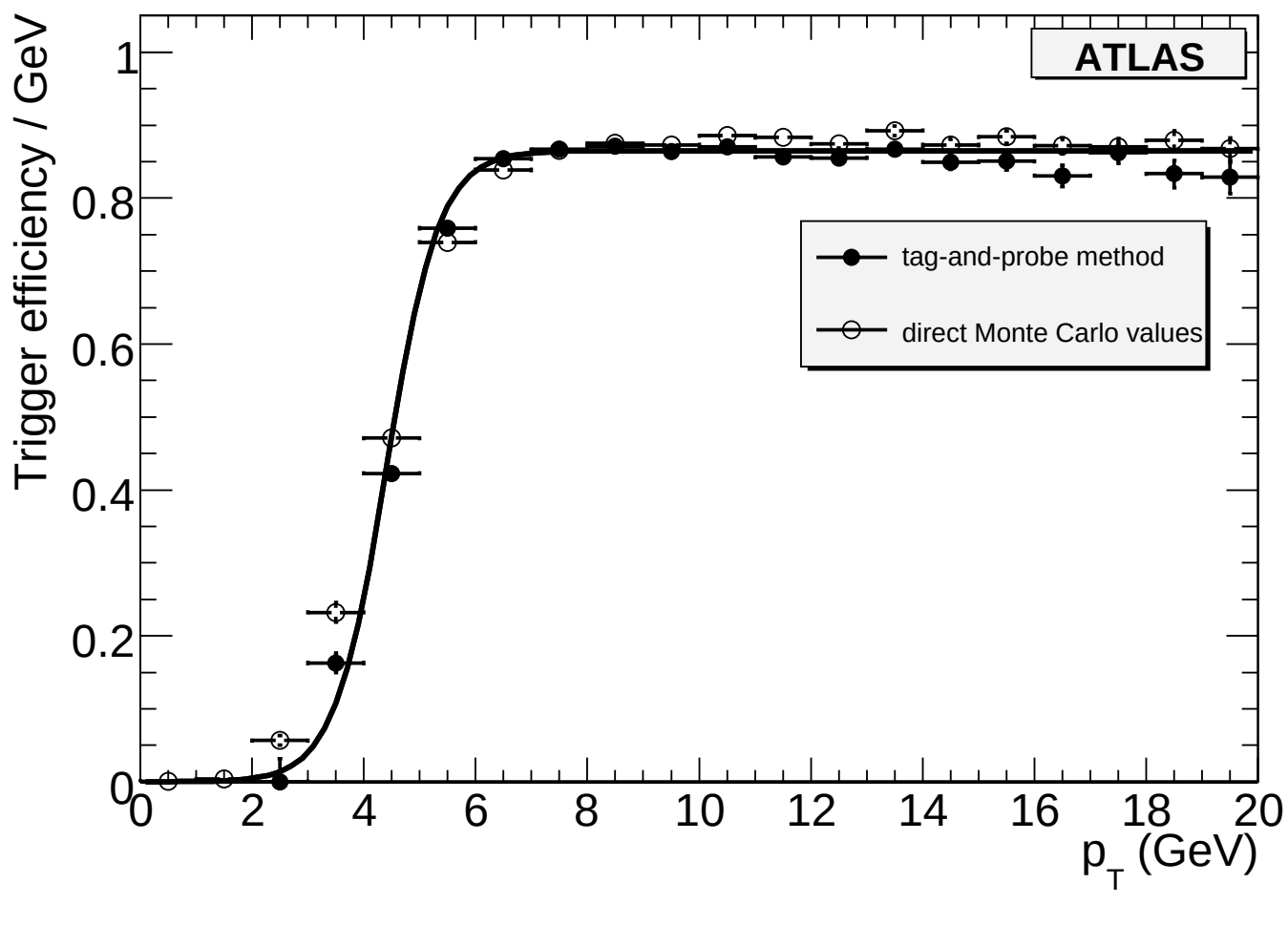


Figure 11: The overall efficiency of the level-1 single muon trigger with respect to the offline selection obtained by the tag-and-probe method (filled circles) and the efficiency estimated from a single muon Monte Carlo sample (open circles). The curve is a fit to the efficiency obtained by the tag-and-probe method from Equation 4.

of $\cos\theta^*$. Although the generated $\cos\theta^*$ distributions were flat, reconstructed distributions will be biased by the $p_T^\mu$ cuts applied at the generator level. At $|\cos\theta^*| = 1$, one of the decay muons flies in the opposite direction to the $J/\psi$ and has low transverse momentum, therefore these events are more likely to be rejected by the $p_T$ cut for the highest (second highest) $p_T$ muon of $p_T > 6(4)$ GeV requirement.

Figure 14 shows the di-muon trigger efficiencies calculated in two methods. The open circles are obtained by checking the decision of the level-1 di-muon trigger for each event. The efficiencies calculated using the parameterization in Figure 12 are shown by the filled circles. They are calculated by assigning an efficiency for each event according to the kinematics of the two muons in the final state using Equation 3. The same technique may be used once data are collected by the experiment. The effect of the systematic uncertainty of the method has been estimated by changing the procedure to obtain the trigger efficiency map, namely by using $16\times16$ bins in $\eta$–$\phi$ in the barrel and $15\times10$ in the endcap. In addition, the fitting function has been changed by adding a linear function $c(p_T-d)$ to the original function for $p_T > d$. This was done to better describe a drop of efficiency for higher $p_T$. $c$ and $d$ are additional fit parameters. Both statistical and systematic errors are a few % in most of the region, but the systematic uncertainty increases where the change of efficiency is rapid. In Figure 14, the statistical and systematic errors are added in quadrature.

Results of the two methods agree within a few % in most regions. Efficiency losses at $\eta^{J/\psi}$ around -1, 0 and +1 are due to the layout of the muon trigger chambers. These plots confirm that the requirement of $p_T^\mu > 6$ GeV on two muons introduces an effective cut of $p_T > 12$ GeV for the $J/\psi$ and the efficiency with respect to $\cos\theta^*$ is flat across the region between $-0.8 < \cos\theta^* < 0.8$. The overall efficiencies calculated in the kinematic region of $p_T^{J/\psi} > 12$ GeV and $|\eta^{J/\psi}| < 2$ are 76.1% and 77.0% using the trigger decision and the trigger efficiency map, respectively, which is in agreement within the statistical and systematic errors. The size of the systematic errors may be improved by creating the efficiency map with finer granularity, which requires more statistics.

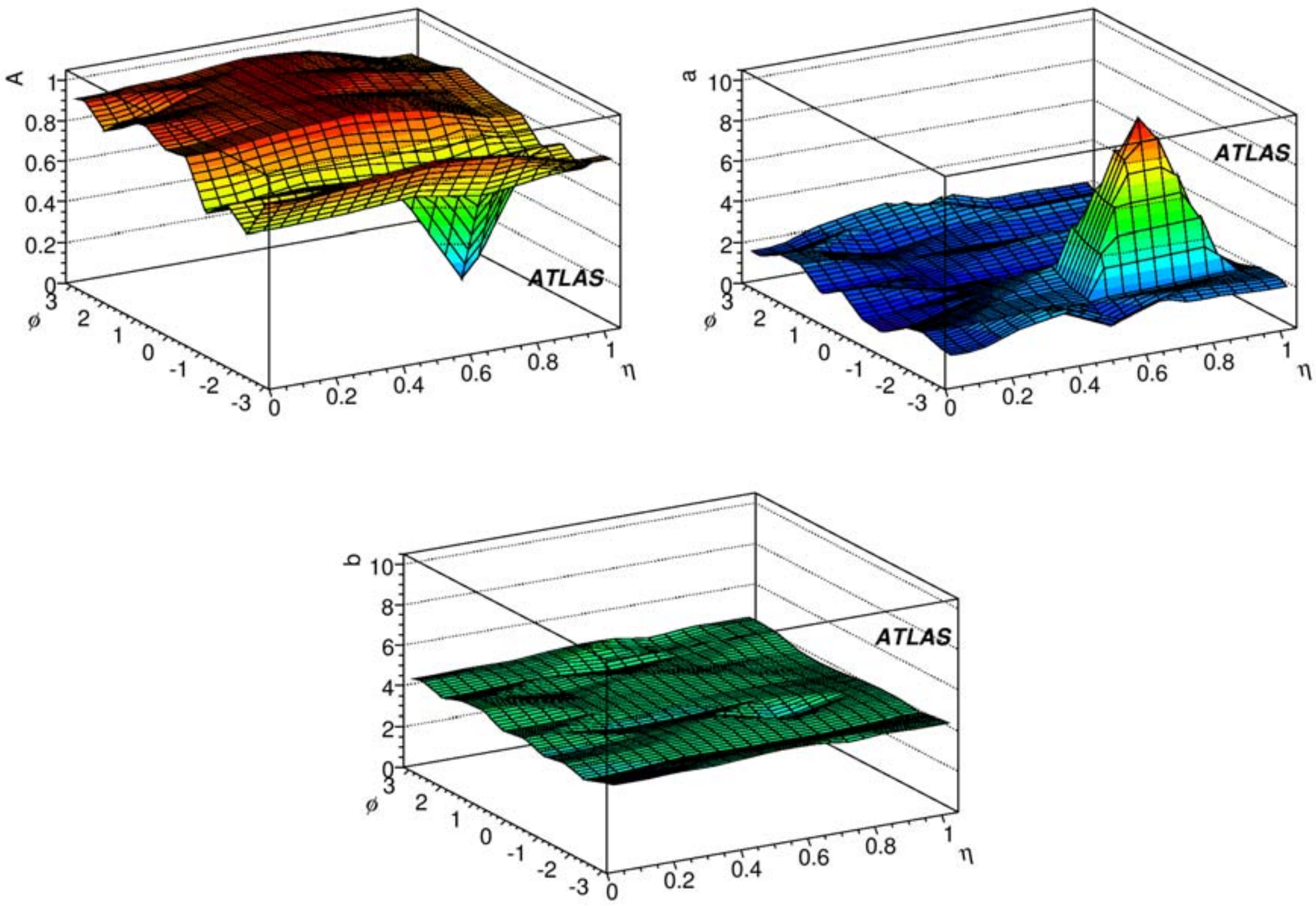


Figure 12: Fit parameters (*A*, *a* and *b* in Equation 4) of the efficiency curve in different $\eta$ and $\phi$ regions.

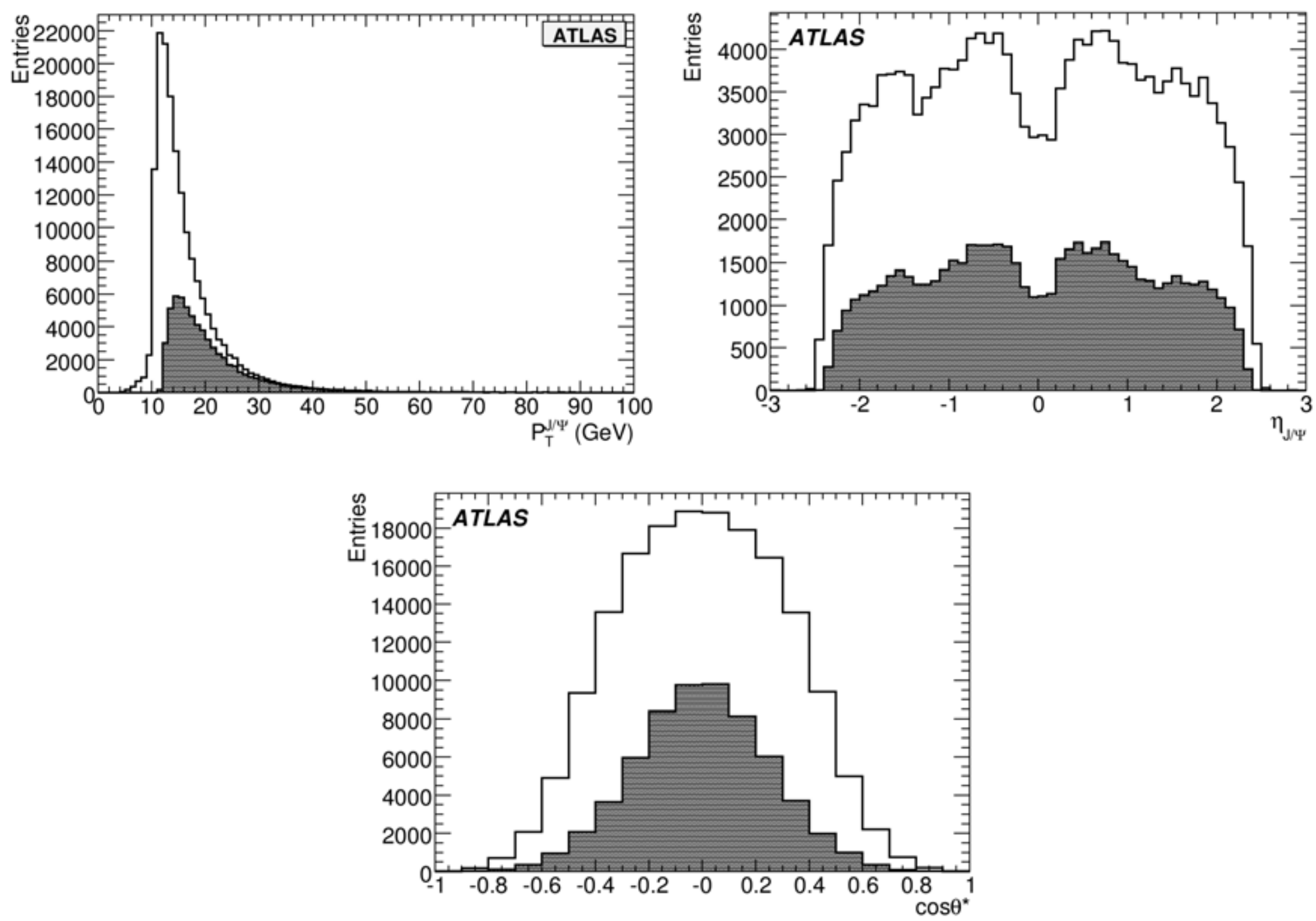


Figure 13: Distributions of $J/\psi$ variables, $p_T$ , $\eta$ and $\cos\theta^*$. The open histograms are for all reconstructed $J/\psi$ s with the generator level cut of $p_T^{\mu_1} > 6$ GeV and $p_T^{\mu_2} > 4$ GeV. Filled histograms are distributions of events passing the level-1 di-muon trigger.

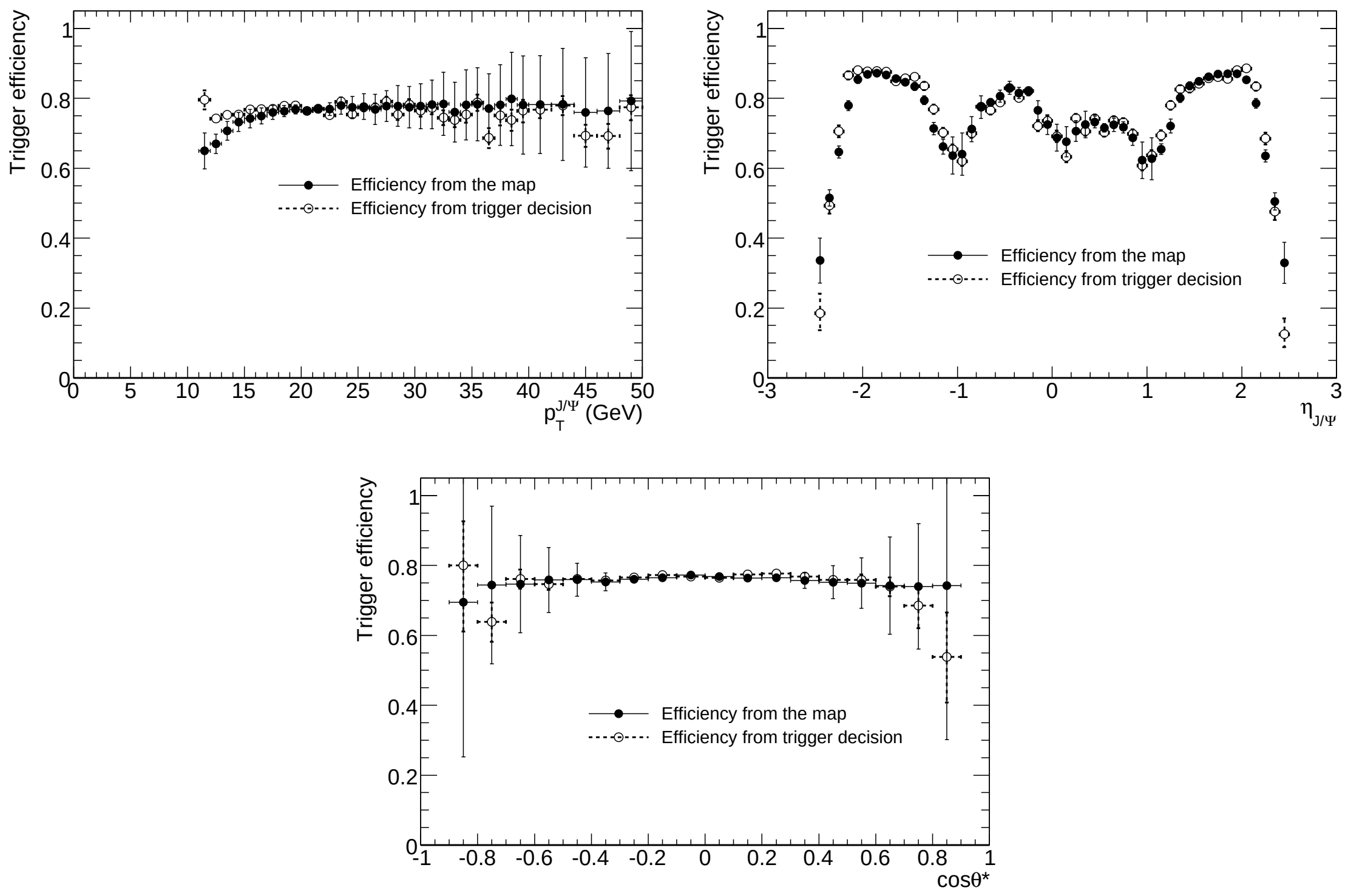


Figure 14: Di-muon trigger efficiency with $p_T^{\mu} > 6$ GeV. Open circles are the result obtained from decision of the level-1 di-muon trigger and filled circles are the efficiency obtained using the parameterization shown in Figure 12.

### 6.2.3 Level-2 efficiency

The single muon level-2 efficiency can be defined in two ways: level-2 efficiency with respect to offline reconstruction (*L2/rec*) and level-2 efficiency with respect to level-1 (*L2/L1*). The *L2/rec* efficiency is obtained in the same way as the efficiency with respect to level-1. The only difference is that the probe muon must have both level-1 and level-2 trigger objects associated to be considered as triggered.

For the calculation of the *L2/L1* efficiency the set of probe muons is restricted to only those that have an associated level-1 RoI. This way the efficiency with respect to level-1 is obtained. Overall efficiencies as a function of $p_T$ are shown in Figure 15. The points were fitted with the functional form of Equation 4. Efficiency curves calculated using all offline muons matched to generated muons are plotted in the same figure to check that the selection of probe muons is unbiased. It is clear that there is a good agreement between the measured efficiency and the direct Monte Carlo efficiency.

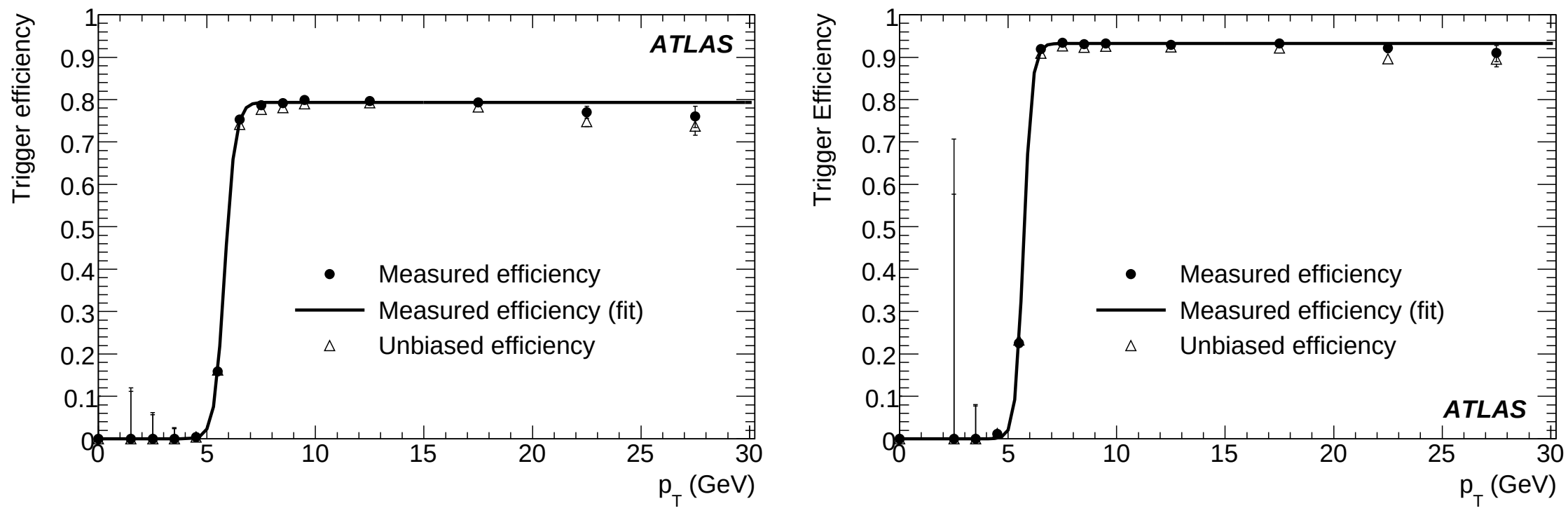


Figure 15: The overall efficiency of the level-2 single muon trigger, (a) with respect to offline reconstruction, (b) with respect to level-1.

To calculate level-2 di-muon efficiency the *trigger efficiency map* was created in the same way as for level-1. The $\eta$–$\phi$ plane was divided into regions as described in Section 6.2.2 and in each region an efficiency curve was constructed and fitted with Equation 4. This was done for *L2/rec* as well as for *L2/L1* efficiencies.

Using the trigger efficiency map for the *L2/rec* efficiency, we can calculate the level-2 trigger efficiency. To check that the map was created correctly, the single-muon efficiency curves were constructed using level-2 muons associated to the offline muons with the $\Delta R$ matching criteria explained in Section 6.2.2. To get an agreement between the two methods one must use the same data sample. In Figure 16, open triangles represent a straightforward calculation of the efficiency using matched trigger objects while the solid circles are the efficiencies calculated using the map. To each offline muon from the sample the probability that it would be triggered was assigned using the map. In each $p_T$ bin the efficiency was calculated as an average of these probabilities. Systematic uncertainties were estimated in the same way as the level-1 efficiency by changing the procedure to calculate the trigger efficiency map. The same method was used for other two variables $\eta$ and $\phi$. The two methods give an agreement within 6% in most regions while the difference gets as large as 15% at some regions ($\phi \simeq -1, -2$) where the efficiency is low and therefore the precision of the fit to create the trigger efficiency map was poor. Also, since the average efficiency is calculated in each bin of the map this causes discrepancies in the regions where efficiency changes rapidly.

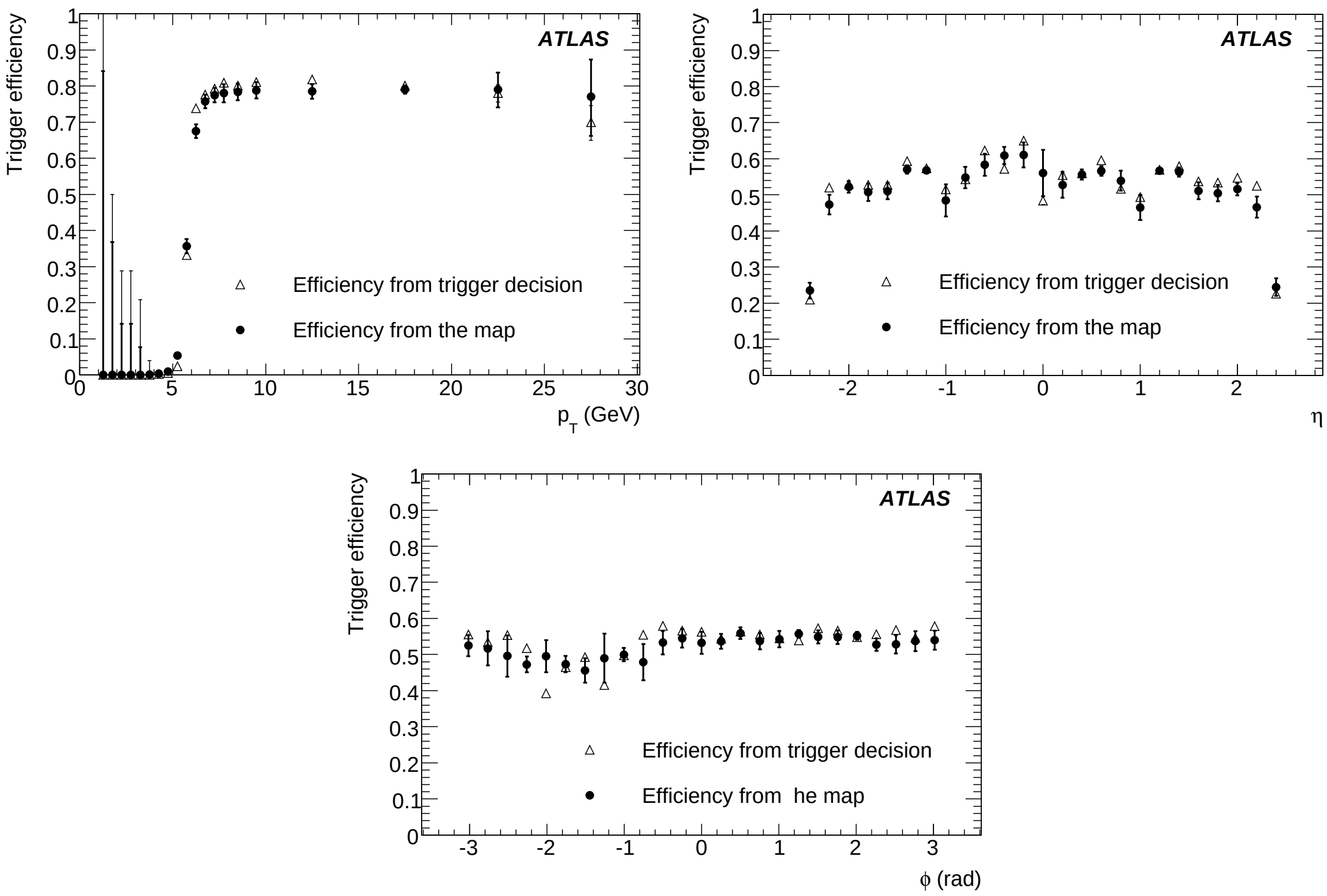


Figure 16: The overall *L2/rec* single-muon trigger efficiency as a function of $p_T$ , $\eta$ and $\phi$. Efficiency is calculated using the *trigger efficiency map* (black circles) and it is compared to the one calculated using matched level-1 and level-2 trigger objects (open triangles).

The efficiency of the level-2 di-muon trigger for $J/\psi$ was calculated in the same way as for level-1. Again two methods were used: firstly direct calculation using the trigger decision for the di-muon trigger and secondly using the efficiency map. Since a simple di-muon trigger without any cut on invariant mass was used, all the muons in the sample must be used in calculation of the efficiency (not just those from $J/\psi$ ). To each offline muon in the event the probability $\varepsilon_i$ that it would be triggered was assigned. The probability that the whole event will be triggered by the di-muon trigger is then

$$\varepsilon^{2\mu} = 1 - \prod_i (1-\varepsilon_i) - \sum_i \varepsilon_i \prod_{j\neq i} (1-\varepsilon_j) \tag{5}$$

where the products and sums run over all decay muons used in the measurement.

The di-muon efficiency is calculated in the same way as the single-muon efficiency, as an average of probabilities $\varepsilon^{2\mu}$ in each bin of a given variable. In our case it is either the $p_T$ , $\eta$, $\phi$ or the $\cos\theta^*$ of the $J/\psi$ reconstructed in the event. Figure 17 shows a comparison of the $J/\psi$ efficiency curves. Like the single muon efficiency results, the agreement between the two methods is within 6% in most regions except for some regions where the available statistics was low. The overall $J/\psi$ efficiency (*L2/rec*) in the kinematic region, $p_T^{J/\psi} > 15$ GeV, $|\eta^{J/\psi}| < 2$ with the cuts on the muons ($p_T^{\mu} > 6$ GeV, $|\eta^{\mu}| < 2.4$) is 69.2% and 69.4% using the trigger decision and the trigger efficiency map, respectively. We have a good agreement on the overall efficiency integrated in the above phase space.

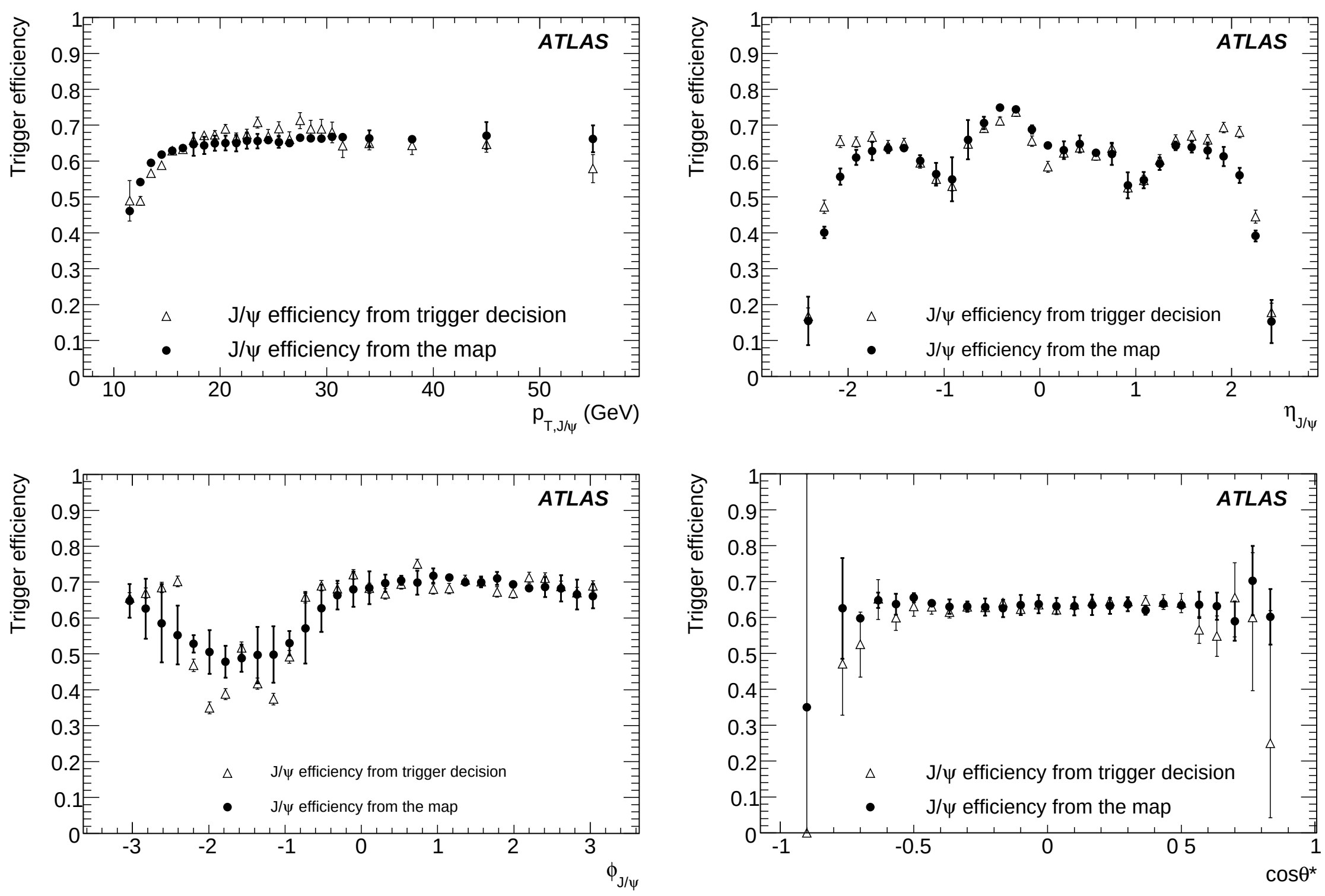


Figure 17: The overall di-muon J/$\psi$ trigger efficiency as a function of $p_T$ , $\eta$, $\phi$ and $\cos\theta^*$. Efficiencies from the trigger decision bit (open triangles) and the ones calculated from the trigger efficiency map (black circles) are shown.

In Figure 18 the efficiency as a function of the distance, $\Delta R$, in the $\eta$–$\phi$ plane between the two

$J/\psi$ muons is shown.

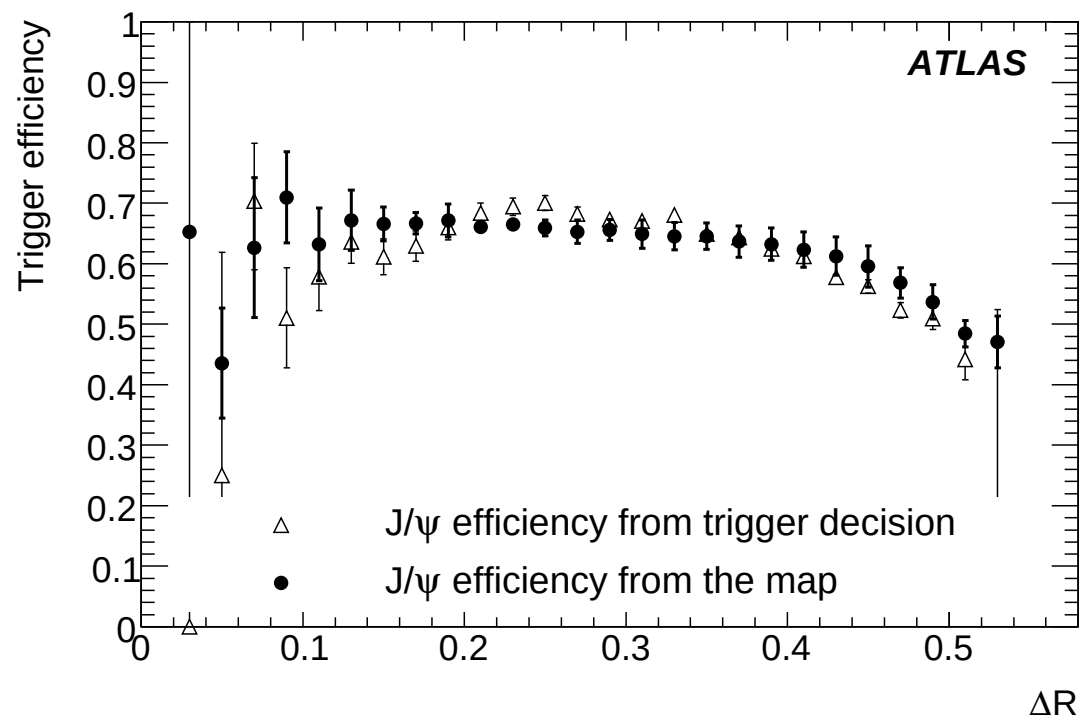


Figure 18: The overall di-muon J/$\psi$ trigger efficiency as a function of $\Delta R$. Efficiencies from the trigger decision bit (open triangles) and the ones calculated from the trigger efficiency map (black circles) are shown.

### 6.3 Requirements on the trigger menu

In order to use the method developed here for real data, we need a large sample of $J/\psi$ events with at least one muon unbiased by the trigger selection. The following selection criteria at each trigger level will satisfy this requirement:

**Level-1** Single muon trigger above a certain threshold;

**Level-2** $J/\psi$ reconstruction within one RoI using the TrigDiMuon algorithm to enhance $J/\psi$ events;

**Event Filter (EF)** No further selection is imposed so as to avoid biasing the sample;

In this Section, we give some estimates of the statistics of the $J/\psi$ sample using this trigger selection and show the precision on the trigger efficiency with a certain luminosity. Taking into account the limit on the EF output rate of 200 Hz it is plausible to use a few Hz of the bandwidth for this calibration trigger. The parameters to optimize are the prescale factor, to reduce the rate when it is too high, and the threshold value.

The level-1 single muon trigger rate has been studied in detail taking into account contributions from different sources. At low-$p_T$ , the main contribution is from $K$ /$\pi$ in-flight decay and muons from $b$ and $c$ quarks. The rejection of these non-$J/\psi$ events by the topological di-muon trigger at level-2 has been studied using a genererated sample of $b\bar{b} \to \mu + X$, where the efficiency of non-$J/\psi$ events to be selected was found to be 0.8%. This factor is used to estimate the rate reduction by the level-2 selection.

Table 12 shows the expected rate of the level-1 single muon trigger and the contribution of the $J/\psi \to \mu^+\mu^-$ to the rate assuming a luminosity of $10^{31}$ cm$^{-2}$s$^{-1}$. The rate of $J/\psi$ events is the sum of $J/\psi$ direct production and from $b\bar{b}$ production. If we could allocate a bandwidth of 1 Hz to this trigger chain with 6 GeV level-1 threshold, we must apply a prescale factor of 3 to reduce the rate down to around 1 Hz. Given the fraction of $J/\psi \to \mu^+\mu^-$ events among this rate is 6%, we get a rate of 0.06 Hz for collecting the calibration sample.

| | rate (Hz) | $J/\psi$ fraction |
|---|---|---|
| level-1 | 380 (0.21) | 0.05% |
| level-2 | 3 (0.19) | 6% |

Table 12: The rates after level-1 and level-2 using the proposed calibration trigger for a luminosity of $10^{31}$ cm$^{-2}$s$^{-1}$ with the threshold of $p_T > 6$ GeV. The contribution of $J/\psi \to \mu^+\mu^-$ process to the rate is also shown in parentheses.

With one year of data, we expect to collect about 300 k $J/\psi$ events according to this triggering strategy, which is comparable to the statistics used in this study (150 k direct $J/\psi$ and 150 k $b\bar{b} \to J/\psi$). Therefore, we expect a similar performance (for $\simeq 1$ Hz calibration trigger rate) to that shown in Section 6.2 after the first year of data-taking. The number of events may increase if we allocate more than 1 Hz for the calibration trigger.

## 7 Conclusions

In this note we presented methods to efficiently select $J/\psi$ events at the second level trigger, reject muons from $K$ and $\pi$ decays and measure the trigger efficiencies from the ATLAS data.

TrigDiMuon is a second-level trigger algorithm that selects efficiently at level-2 events which include $J/\psi$ or other di-muon states, starting from a single level-1 muon trigger. The efficiency of TrigDiMuon for events accepted by level-1 and the level-2 single muon trigger is between 73% for the 4 GeV trigger threshold and 60% for the 6 GeV threshold. This may be compared with the topological di-muon trigger efficiencies of 33% for the 4 GeV threshold and 15% for the 6 GeV threshold. The fake trigger rates of TrigDiMuon are estimated to be 2 Hz for a trigger threshold of 4 GeV at a luminosity of $10^{31}$cm$^{-2}$s$^{-1}$, and 90 Hz for a trigger threshold of 6 GeV at a luminosity of $10^{33}$cm$^{-2}$s$^{-1}$. For the $B$-physics trigger at $L = 10^{33}$cm$^{-2}$s$^{-1}$ the rate will have to be further reduced by means of a decay length cut on the $J/\psi$ decay vertex.

Extrapolating the muon track from MS back to the interaction vertex improves the single muon trigger selection at the level-2 stage and allows increased rejection of muons from $K$ and $\pi$ decays without a significant loss of efficiency for $b$ events. The back extrapolator provides a further reduction factor of about 20% at the trigger threshold of 4 GeV and of about 10% at the trigger threshold of 6 GeV with respect to the output trigger rate of the baseline muComb selection. Despite this good performance, its use for triggering on low $p_T$ di-muon objects is not recommended, since the background to TrigDiMuon is dominated by muons from $b$ and $c$ rather than by muon from $K$ and $\pi$ decays. However, because the rate of muons from $K$ and $\pi$ decays may be even higher than in our simulation, it is important to have it available for the single muon triggers.

The $J/\psi$ trigger efficiency can be measured from ATLAS data using the tag-and-probe method. As the efficiency of the single muon trigger depends on the $\eta$ and $\phi$ regions, it is necessary to measure the efficiency as a function of these variables. With 300k $J/\psi$ events, it is possible to measure the efficiency at level-1 and level-2 with better than 5 % precision for each region. The uncertainty comes mainly from the lack of statistics to measure the efficiencies for each region of $\eta$ and $\phi$ and will improve as more $J/\psi$ events become available. To collect an unbiased $J/\psi$ sample which can be used for the trigger efficiency measurement, we plan to use a level-1 single muon trigger with a $J/\psi$ reconstruction using the inner detector at level-2. With this trigger we can collect around 300k events for an integrated luminosity of 100 pb$^{-1}$ while keeping the rate of this calibration trigger around 1 Hz.

With the methods provided it is possible to collect a large number of $J/\psi$ events at luminosities of around $10^{31}\text{cm}^{-2}\text{s}^{-1}$, without an overly high rate from $K$ and $\pi$ decays. Furthermore, these low-$p_T$ $J/\psi$ events can also be used to calibrate trigger efficiencies from the actual ATLAS data, and provide the trigger efficiencies for analyses that use low-$p_T$ muons.

# Heavy Quarkonium Physics with Early Data


**Abstract**

Results are reported on an analysis on simulated data samples for production of heavy quarkonium states $J/\psi \rightarrow \mu\mu$ and $\Upsilon \rightarrow \mu\mu$, corresponding to an integrated luminosity of 10 pb$^{-1}$. It is shown that the $p_T$ dependence of the cross-section for both $J/\psi$ and $\Upsilon$ should be measured reasonably well in a wide range of transverse momenta, $p_T \simeq 10-50$ GeV. The precision of $J/\psi$ polarisation measurement is expected to reach $0.02-0.06$, while the projected error on $\Upsilon$ polarisation is around 0.2. Observation of radiative decays of $\chi_c$ states, and the feasibility of observing $\chi_b \rightarrow J/\psi J/\psi$ decays are also discussed.


## 1 Introduction and theoretical motivation

The number of $J/\psi \rightarrow \mu^+\mu^-$ and $\Upsilon \rightarrow \mu^+\mu^-$ decays produced at the LHC is expected to be quite large. Their importance for ATLAS is threefold: first, being narrow resonances, they can be used as tools for alignment and calibration of the trigger, tracking and muon systems. Secondly, understanding the details of the prompt onia production is a challenging task and a good testbed for various QCD calculations, spanning both perturbative and non-perturbative regimes. Last, but not the least, heavy quarkonium states are among the decay products of heavier states, serving as good signatures for many processes of interest, some of which are quite rare. These processes have prompt quarkonia as a background and, as such, a good description of the underlying quarkonium production process is crucial to the success of these studies.

This note mainly concentrates on the capabilities of the ATLAS detector to study various aspects of prompt quarkonium production at the LHC. The methods of separating promptly produced $J/\psi$ and $\Upsilon$ mesons from various backgrounds are discussed, and strategies for various measurements are outlined.

### 1.1 Theory overview

Quarkonium production was originally described in a model where the quark pair was assumed to be produced endowed with the quantum numbers of the quarkonium state that it eventually evolved into [1]. This approach, subsequently labelled as the Colour Singlet Model (CSM), enjoyed some success before CDF measured an excess of direct $J/\psi$ production [2], more than an order of magnitude greater than predicted (see Figure 1(a)).

The Colour Octet Model (COM) [5] was proposed as a solution to this quarkonium deficit. COM suggests that the heavy quark pairs produced in the hard process do not necessarily need to be produced with the quantum numbers of physical quarkonium, but could evolve into a particular quarkonium state through radiation of soft gluons later on, during hadronisation. This approach isolates the perturbative hard process from the non-perturbative long-distance matrix elements, which are considered as free parameters of the theory. However, their universality means that their values can be extracted independently from a number of different processes, such as deep inelastic scattering, hadro- and photoproduction.

Hence the good description of the Tevatron data by the Colour Octet Model shown in Figure 1(a) is, at least in part, due to the fact that the values of some parameters were determined from the same data. Tests of other COM predictions have not been so successful: Figure 1(b) shows the polarisation coefficient in $\Upsilon \rightarrow \mu\mu$ decay as a function of its transverse momentum, where the COM prediction disagrees with the data.

A model based on $k_T$ factorisation in QCD showers [6] claims to be able to describe both the lack of transverse polarisation in $J/\psi$ decays [4, 7] and the high cross-section of $J/\psi$ production. Another

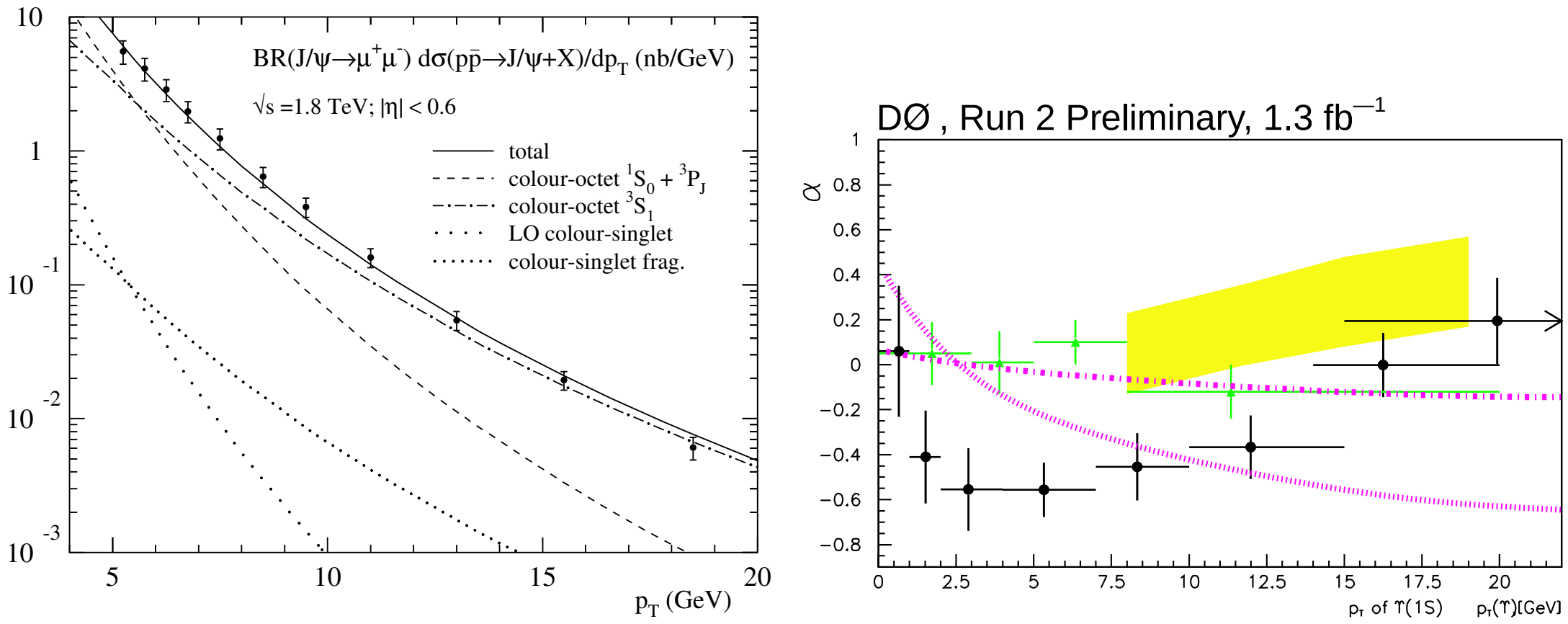


(a) (b)

Figure 1: (a) Differential cross-section of $J/\psi$ production at CDF, with predictions from CSM and COM mechanisms (from [3]). (b) $\Upsilon$ polarisation measured as a function of $p_T$ at DØ (black dots) and CDF (green triangles), compared to the limits of the $k_T$ factorisation model (dashed and dotted curves [6]) and COM predictions [5], depicted by a shadowed band (from [4]).

model [8] argues that the deficit in the cross-section as predicted by CSM can be largely explained by the production of a quarkonium state in association with an additional heavy quark, and also predicts lower levels of polarisation.

In the following, we show that ATLAS is capable of detailed checks of the predictions of various models by measuring not only $p_T$ and $\eta$ distributions of onium states in a wide range of these variables, but also the degree of polarisation and the production of $C$-even states. In the absence of a comprehensive Monte Carlo generator capable of simulating all aspects of all theoretical models, we used the PYTHIA 6.403 generator [9] incorporating the Colour Octet Mechanism, with model parameters fixed through a combination of theoretical and experimental constraints [10]. Inevitably, this simulation is unable to reproduce adequately some features of the data, notably the polarisation angle distributions and hadronic accompaniment of the quarkonium states. However, the simulated samples allowed us to study the acceptance and efficiency of ATLAS to detect all required particles and measure their parameters, across the whole range of the accessible phase space.

## 1.2 Classification of production mechanisms in the simulation

In the following, we will use a simple classification of the quarkonium production mechanisms based on the model implemented in the PYTHIA generator.

A sample diagram describing the leading colour-singlet subprocess $g+g \rightarrow J/\psi + g$ is shown in Figure 2(a). In the accessible range of transverse momenta of $J/\psi$ its contribution is expected to be small. The dominant contribution at the lower $p_T$ comes from the subprocess shown in Figure 2(b), where both singlet and octet $c\bar{c}$ states with various quantum numbers contribute to $J/\psi$ production, through $\chi_{cJ} \rightarrow J/\psi + \gamma$ decays and/or soft gluon emission.

At high $p_T$, the gluon fragmentation subprocess shown in Figure 2(c) becomes increasingly dominant. According to COM, this is unlikely to produce anything other than ${}^3S_1$ quarkonium states. Hence, the fraction of $J/\psi$ mesons produced from $\chi_{cJ}$ decays should decrease with increasing $p_T$. The production mechanisms for the radially excited $\psi'(3686)$ meson follows the same pattern, except for the absence of respective $\chi'_{cJ}$ contributions, thus one should expect different $p_T$ distributions for $J/\psi$ and

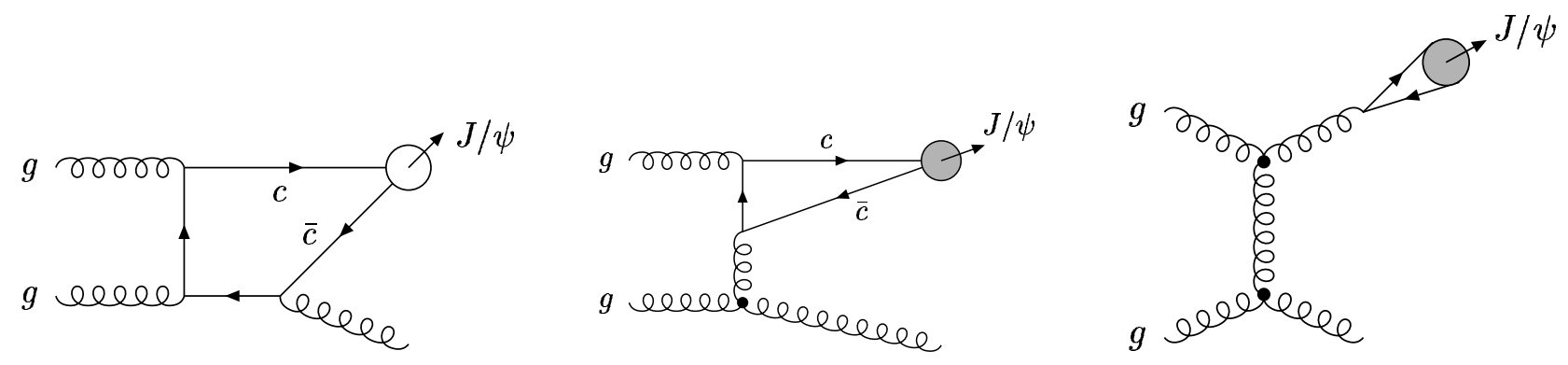


(a) (b) (c)

Figure 2: Some example diagrams for the singlet and octet $J/\psi$ production mechanisms implemented in PYTHIA.

$\psi'$.

The overall picture is expected to be similar for bottomonium production, except the number of radial excitations below the open beauty threshold is now three, and many more radiative transitions are possible between the various $n^3P_J$ and $n^3S_1$ state. However, compared to $J/\psi$, the accessible range of $p_T$ for $\Upsilon$ is significantly extended towards smaller transverse momenta. This opens up the range of $p_T$ dominated by the colour singlet contribution, which may make it directly observable for $\Upsilon$.

### 1.3 $\chi_b \to J/\psi J/\psi$ decay

Despite much higher production cross-sections, $C$-even states of quarkonia are far more difficult to observe than their vector counterparts. The usual way of studying $\chi_{c,b}$ (and $\eta_{c,b}$) states has been so far through radiative decays of or into respective vector states. However, in the high energy hadronic collision environment, observation of the photon in $\chi_b \to \Upsilon + \gamma$ may be problematic (see Section 5.1).

We have performed a feasibility study to assess the capability of ATLAS to observe $\chi_b \to J/\psi J/\psi \to \mu^+\mu^-\mu^+\mu^-$ decay with the standard di-muon trigger. The results are presented in Section 5.2. The observation and measurement of these final states will give a valuable insight into the heavy quark bound state dynamics from several separate viewpoints.

## 2 Trigger considerations

Details of the triggers to be used in ATLAS $B$ physics programme can be found in [11]. This section discusses the trigger signatures relevant for quarkonium production at ATLAS, the implications they have on the measured cross-section, and the expected effects they have on our ability to make various physics measurements.

Two specific types of di-muon triggers dedicated to quarkonium are: the topological di-muon triggers, which require two level-1 regions of interest (RoIs) corresponding to two muon candidates with $p_T$ thresholds of 6 and 4 GeV, and di-muon triggers that only require a single level-1 RoI above a threshold of 4 GeV and searches for the second muon of opposite charge in a wide RoI at level-2. They are discussed in Section 2.1. An additional trigger scenario is based on a single muon trigger with a higher $p_T$ threshold of 10 GeV, discussed in Section 2.2.

### 2.1 Di-muon triggers

Being able to determine the trigger efficiency of measured $J/\psi$ and $\Upsilon$ is crucial to correctly infer the production cross-section of quarkonium at the LHC. Indeed, using $J/\psi$ and $\Upsilon$ (as well as the $Z$ boson) to construct a trigger efficiency map is a necessary step in order to perform cross-section measurements in ATLAS.

Studies are being conducted in ATLAS into developing a calibration method to obtain the low-$p_T$ single muon trigger efficiency, and henceforth the di-muon trigger efficiency of events using real data by virtue of the so-called tag-and-probe method (see Ref. [11] for details). In the absence of data, we have performed our own studies of trigger efficiencies, based on Monte Carlo simulation.

If not stated otherwise, the quoted trigger efficiencies have been calculated with respect to the Monte Carlo samples, generated with the cuts $p_T(\mu_1) > 6$ GeV, $p_T(\mu_2) > 4$ GeV, where $\mu_1$ $(\mu_2)$ is the muon with the largest (second largest) transverse momentum in the event.

The level-1 trigger is a hardware trigger that uses coarse calorimeter and muon spectrometer information to identify interesting signatures to pass to the level-2 and Event Filter stage. Figure 3 shows the various individual level-1 trigger efficiencies as a function of the $p_T$ of the di-muon system. The total trigger efficiency at level-1, running over direct $J/\psi$ events, is 87%.

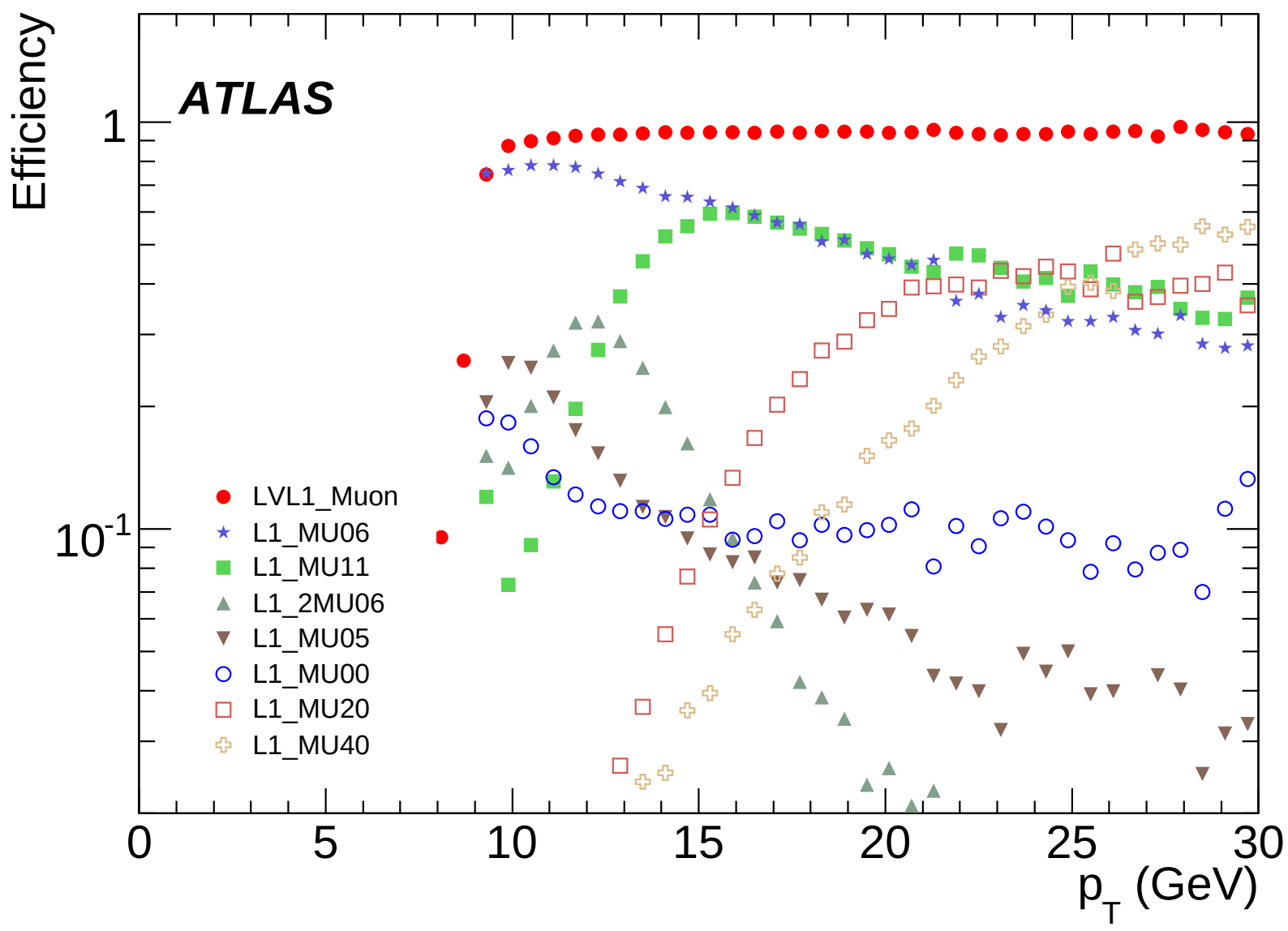


Figure 3: Efficiency of various level-1 triggers for prompt $J/\psi$ events versus $p_T$ of the di-muon system. Only the triggers with efficiencies greater than 2% in some region of $p_T$ are displayed. The relevant triggers are the single muon $p_T$ threshold triggers labelled `L1_MUXX` (where `XX` indicates the $p_T$ threshold in GeV) and the di-muon trigger `L1_2MU06`. Each $p_T$ range of these triggers is exclusive. The efficiency curve labelled `LVL1_Muon` is the sum of all level-1 single muon efficiencies (excluding the di-muon trigger `L1_2MU06`).

The level-2 trigger is software-based and is designed to reduce the output rate of the data, passed to it from level-1, by two orders of magnitude. Within regions of interest defined by the level-1 trigger, full granularity of the detector is accessible. The efficiency of the level-2 triggers for prompt $J/\psi$ events is plotted in Figure 4 as a function of di-muon $p_T$. The total level-2 trigger efficiency in the reconstructed prompt $J/\psi$ events (relative to level-1) is 97%. The di-muon trigger scenario $\mu6\mu4$, considered in the majority of this note, uses all the above trigger signatures.

### 2.1.1 Effect of di-muon trigger cuts on quarkonium rates

Figures 5(a) and 5(b) illustrate the distribution of cross-sections across the values of the $p_T$ of the harder and softer muon from the quarkonium decay without any muon cuts applied at generator level. The lines

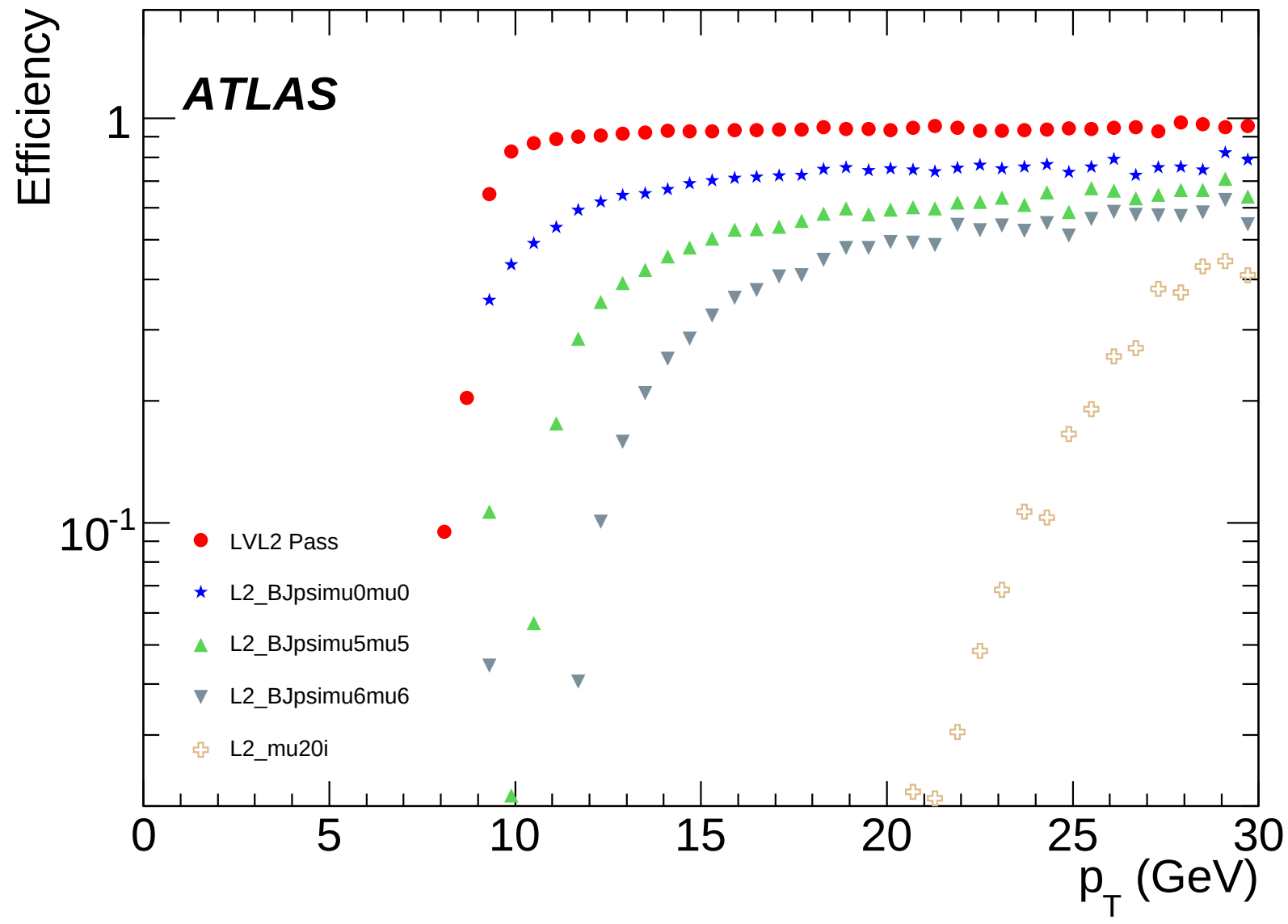


Figure 4: Efficiency of various level-2 triggers for prompt $J/\psi$ events versus $p_T$ of the di-muon system. Only the triggers with efficiencies greater than 2% in some region of $p_T$ are displayed. The level-2 trigger signatures of interest include single muon $p_T$ threshold triggers L2_MUXX and 'TrigDiMuon' triggers L2_BJpsimuXmuY which are specialised for searching for $J/\psi$ [11]. The efficiency curve labelled LVL2_Pass in (b) is the sum of all level-2 efficiencies.

overlaid on the plots represent various nominal muon $p_T$ thresholds: (6 GeV, 4 GeV) and (4 GeV, 4 GeV), as well as the nominal thresholds (10 GeV, 0.5 GeV) corresponding to the single muon trigger $\mu 10$ (see below).

For $J/\psi$ the bulk of the cross-section lies near the (4 GeV, 1 GeV) region, far from the low-$p_T$ muon trigger thresholds proposed for ATLAS, and we see only a small increase in accessible cross-section by lowering the cut on the harder muon from 6 GeV to 4 GeV (although this reduction in the effective $J/\psi$ $p_T$ threshold is useful from a physics standpoint). The situation for $\Upsilon$ is significantly different however, as the relatively large mass of the $\Upsilon$ shifts the bulk of the production to the region near muon $p_T$ thresholds of (5 GeV, 4 GeV). This means that by lowering the di-muon trigger cuts from (6 GeV, 4 GeV), which sits just above the highest density area of $\Upsilon$ production, to (4 GeV, 4 GeV), a much higher fraction of the produced $\Upsilon$ can be recorded, leading to a predicted order-of-magnitude increase in the accessible cross-section. The predicted cross-sections for the processes $pp \to J/\psi(\mu^+\mu^-)X$ and $pp \to \Upsilon(\mu^+\mu^-)X$ (before incorporating trigger and reconstruction efficiencies) for a number of trigger scenarios are presented in Table 1. Although no higher $\psi$ and $\Upsilon$ states have been simulated for this analysis, their expected cross-sections are also shown in Table 1, as estimated using Tevatron results on their relative yields [12]. Numbers include feed-down from $\chi$ states and higher radial excitations to lower ones. Due to the expected ATLAS mass resolution for the $\Upsilon$ states, however, it is unlikely that the higher state resonances will be separable. These predictions have been obtained by extrapolating the Colour Octet Model, tuned to describe the Tevatron results, to the LHC energy. Although every care have been taken to ensure stability of this extrapolation, inevitably there is an uncertainty in the overall scale of the predicted cross-sections (linked to the uncertainties in the parton distribution functions at small $x$), which we estimate at the level of $\pm 50\%$.

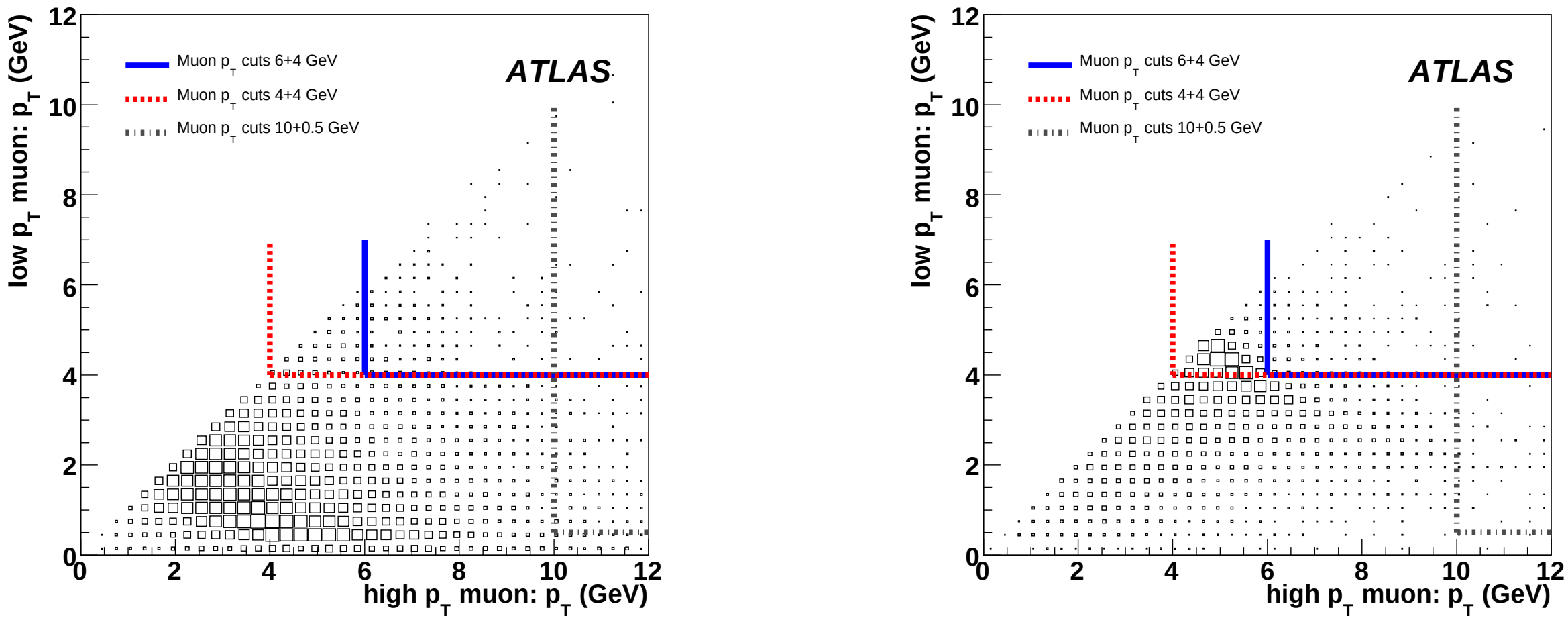


(a) $J/\psi$ (b) $\Upsilon$

Figure 5: Densities of $J/\psi \to \mu\mu$ (a) and $\Upsilon \to \mu\mu$ (b) production cross-section as a function of the two muon transverse momenta. No cut was placed on the generated sample. The overlaid lines represent the nominal thresholds of observed events with various trigger cuts applied: $\mu 6 \mu 4$ (solid line), $\mu 4 \mu 4$ (dashed line) and $\mu 10$+track (dash-dotted line).

It is likely that the cross-section accessible by ATLAS will be higher than the values quoted in Table 1, as during early running the low $p_T$ muon trigger will run with an open coincidence window in $\eta$ at level-1 and no requirement of an additional level-2 di-muon trigger. This trigger item has a turn-on threshold at around 4 GeV, giving the (4 GeV, 4 GeV) trigger scenario described above, but in practice there is a non-zero trigger efficiency below 4 GeV, which, combined with the large rate of low $p_T$ onia, may add a significant extra contribution to the overall observed cross-section. Even including this contribution, the overall rate of signal events from all quarkonium states is likely to remain below the rate of 1 Hz at a luminosity of $10^{31}$ cm$^{-2}$s$^{-1}$, which is a small fraction of the available trigger bandwidth.

| Quarkonium | Cross-section, nb | | | |
|---|---|---|---|---|
| | $\mu 4 \mu 4$ | $\mu 6 \mu 4$ | $\mu 10$ | $\mu 6 \mu 4 \cap \mu 10$ |
| $J/\psi$ | 28 | 23 | 23 | 5 |
| $\psi'$ | 1.0 | 0.8 | 0.8 | 0.2 |
| $\Upsilon$(1S) | 48 | 5.2 | 2.8 | 0.8 |
| $\Upsilon$(2S) | 16 | 1.7 | 0.9 | 0.3 |
| $\Upsilon$(3S) | 9.0 | 1.0 | 0.6 | 0.2 |

Table 1: Predicted cross-sections for various prompt vector quarkonium state production and decay into muons, with di-muon trigger thresholds $\mu 4 \mu 4$ and $\mu 6 \mu 4$ and the single muon trigger threshold $\mu 10$ (before trigger and reconstruction efficiencies). The last column shows the overlap between the di-muon and single muon samples.

#### 2.1.2 Effect of trigger cuts on analysis of octet states

As discussed above, the quarkonium cross-section is composed of three main classes of processes: direct colour singlet production, colour octet production and singlet/octet production of $\chi$ states. Figure 6 illustrates the contributions of these three classes to the overall production rate for $\Upsilon$, once the $p_T$ trigger cuts of 6 and 4 GeV are applied to the muons. Lower $p_T$ trigger cuts will strongly enhance the $\Upsilon$ rate and

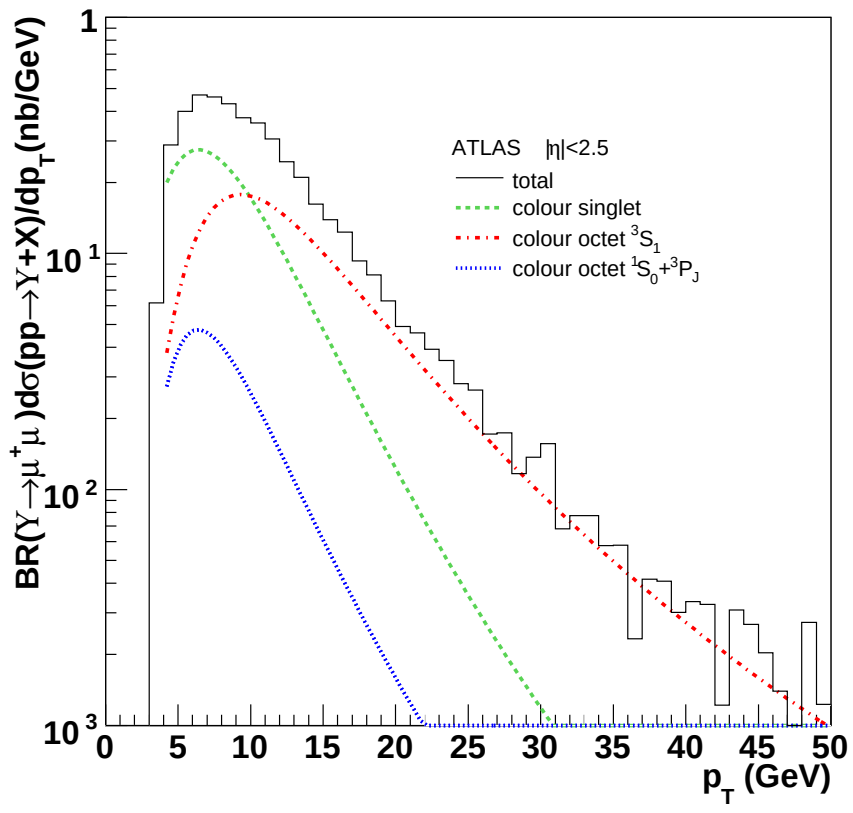


Figure 6: Expected $p_T$-distribution for $\Upsilon$ production, with contributions from direct colour singlet, singlet $\chi$ production and octet production overlaid.

allow for analysis of colour singlet production, which is expected to dominate for $\Upsilon$ with $p_T < 10$ GeV. Lower trigger cuts available during early running, such as the $\mu4\mu4$ trigger described above, will allow the opportunity to extend the low-$p_T$ region down to $p_T \simeq 0$ in the case of $\Upsilon$ and help separate octet and singlet contributions.

#### 2.1.3 Acceptance of $\cos\theta^*$ with di-muon triggers

An important consideration for calculating the di-muon trigger efficiencies of $J/\psi$ and $\Upsilon$ is the angular distribution of the decay angle $\theta^*$, the angle between the direction of the positive muon (by convention) from quarkonium decay in the quarkonium rest frame and the flight direction of the quarkonium itself in the laboratory frame (Figure 7).

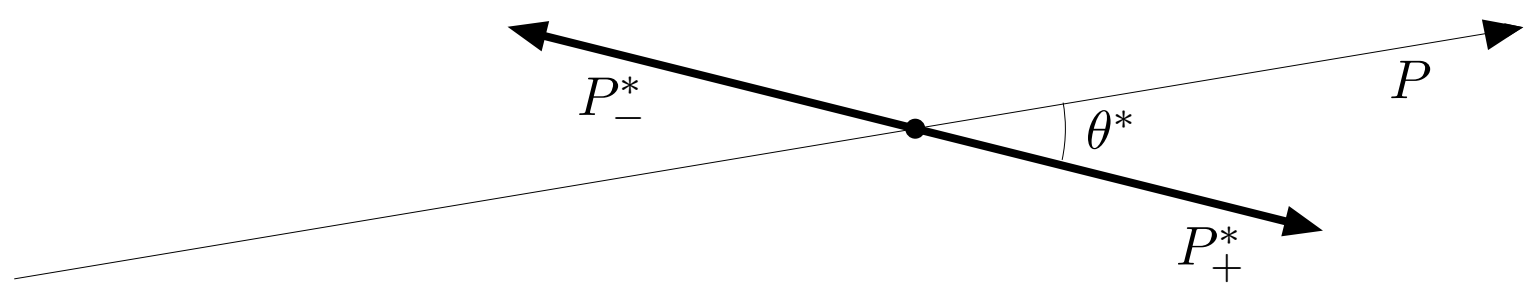


Figure 7: Graphical representation of the $\theta^*$ angle used in the spin alignment analysis. The angle is defined by the direction of the positive muon in the quarkonium decay frame and the quarkonium momentum direction in the laboratory frame.

The distribution in $\cos\theta^*$ may depend on the relative contributions of the various production mechanisms, and is as of yet not fully understood. Crucially, Monte Carlo studies have shown that different production mechanisms (and thus different angular distributions) can have significantly different trigger

acceptances, and without the measurement of the spin-alignment of quarkonium it will be difficult to be sure that the full trigger efficiency has been calculated correctly.

It is clear that $\cos\theta^* \simeq 0$ corresponds to events with both muons having roughly equal transverse momenta, while in order to have $\cos\theta^*$ close to $\pm 1$ one muon's $p_T$ needs to be very high while the other's $p_T$ is very low. In the case of a di-muon trigger, both muons from the $J/\psi$ and $\Upsilon$ decays must have relatively large transverse momenta. Whilst this condition allows both muons to be identified, it also severely restricts acceptance in the polarisation angle $\cos\theta^*$, meaning that for a given $p_T$ of $J/\psi$ or $\Upsilon$ a significant fraction of the total cross-section is lost.

Examples of the polarisation angle distributions for the $\mu 6\mu 4$ trigger are shown by solid lines in Figure 8. Here, the samples for both $J/\psi$ and $\Upsilon$ were generated with zero polarisation, so with full acceptance the corresponding distribution in $\cos\theta^*$ should be flat, spanning from $-1$ to $+1$. Clearly,

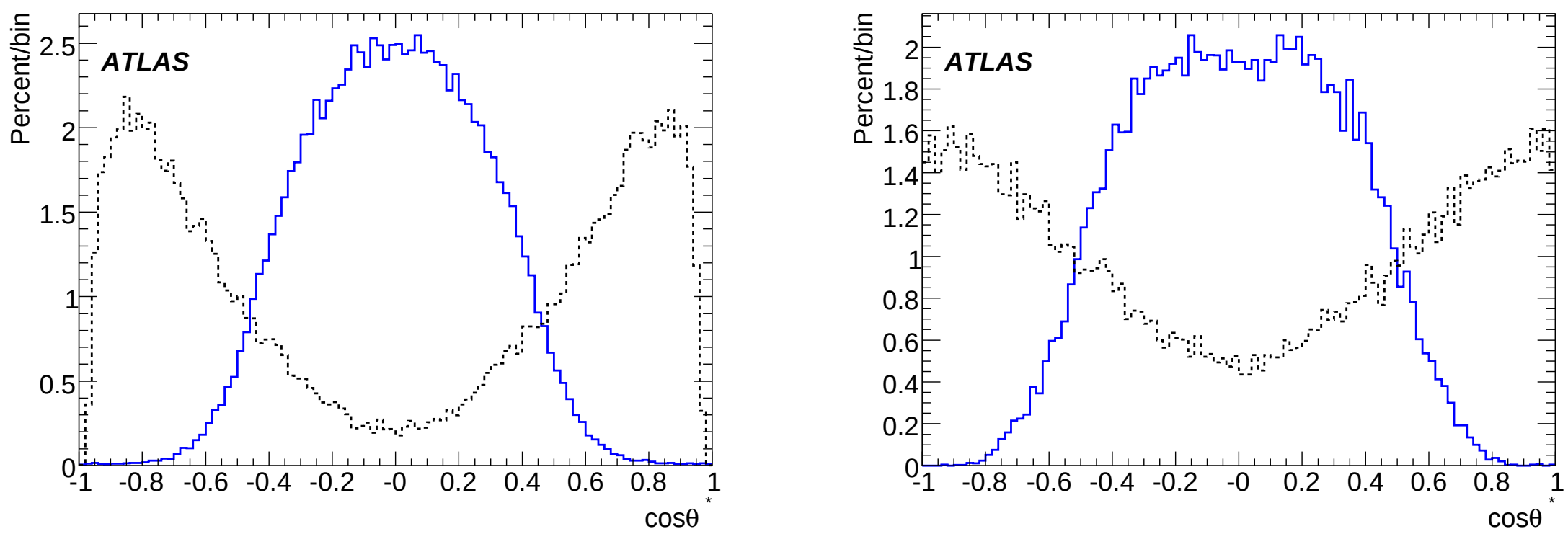


(a) $J/\psi$

(b) $\Upsilon$

Figure 8: Reconstructed polarisation angle distribution for $\mu 6\mu 4$ di-muon triggers (solid line) and a $\mu 10$ single muon trigger (dashed line), for $J/\psi$ (a) and $\Upsilon$ (b). The distributions are normalised to unit area. The generated angular distribution is flat in both cases.

narrow acceptance in $|\cos\theta^*|$ would make polarisation measurements difficult.

## 2.2 Single muon trigger

Another possibility for quarkonium reconstruction is to trigger on a single identified muon. The non-prescaled level-1 single muon trigger `L1_MU10` with a 10 GeV $p_T$ threshold is expected to produce manageable event rates at low luminosities [11]. Once this muon triggers the event, offline analysis can reconstruct the quarkonium by combining the identified muon with an oppositely-charged track in the event. In Figure 5 this trigger corresponds to the dash-dotted lines, with the predicted cross-sections also shown in Table 1. With this trigger (referred to as $\mu 10$ in the following) one removes the need for the other muon to have a large $p_T$, i.e. one has a fast muon, which triggered the event, and one track, whose transverse momentum is only limited by the track reconstruction capabilities of ATLAS, with the threshold around 0.5 GeV.

Thus, the onium events with a single muon trigger typically have much higher values of $|\cos\theta^*|$, as illustrated by the dotted lines in Figure 8, complementing the di-muon trigger sample. So, the single- and di-muon samples may be used together to provide excellent coverage across almost the entire range of $\cos\theta^*$ in the same $p_T$ range of onia.

It's worth noting that the di-muon and single muon samples have comparable cross-sections and similar $p_T$ dependence. They are not entirely independent: at high transverse momenta the two samples

have significant event overlap (see Table 1 for more details), which could be useful for independent calibration of muon trigger and reconstruction efficiencies.

## 3 Reconstruction and background suppression

### 3.1 Quarkonium reconstruction with two muon candidates

In each event which passes the di-muon trigger, all reconstructed muon candidates are combined into oppositely charged pairs, and each of these pairs is analysed in turn. The invariant mass is calculated and, if the mass is above 1 GeV, the two tracks are refitted to a common vertex. If a good vertex fit is achieved, the pair is accepted for further analysis. If the invariant mass of the refitted tracks is within 300 MeV of the nominal mass in the case of $J/\psi$, or 1 GeV in the case of $\Upsilon$, the pair is considered as a quarkonium candidate. The values quoted by the Particle Data Group [13], 3097 MeV and 9460 MeV for $J/\psi$ and $\Upsilon$ respectively, are used throughout this paper, and the widths of the mass windows are chosen to be about six times the expected average mass resolution (see Table 2).

For those pairs for which the vertex fit is successful (more than 99% for both $J/\psi$ and $\Upsilon$), the invariant mass is recalculated. The invariant mass resolution depends on the pseudorapidities of the two muon tracks. To illustrate this effect, all accepted onia candidates are divided into three classes depending on $\eta$ of the muons, and Gaussian fits are performed to determine the resolutions and mass shifts. The results are presented in Table 2. It is found that the mass resolution is the highest when both tracks are

| Quarkonium | $M_{\mathrm{rec}} - M_{\mathrm{PDG}}$, MeV | Resolution $\sigma$, MeV | | | |
|---|---|---|---|---|---|
| | | Average | Barrel | Mixed | Endcap |
| $J/\psi$ | $+4 \pm 1$ | 53 | 42 | 54 | 75 |
| $\Upsilon$ | $+15 \pm 1$ | 161 | 129 | 170 | 225 |

Table 2: Mass shifts and resolutions for di-muon invariant mass distributions after the vertex fit, for $J/\psi$ and $\Upsilon$ candidates.

reconstructed in the barrel area, $|\eta| < 1.05$, degrades somewhat if both tracks are reconstructed in the endcap regions, $|\eta| > 1.05$, and is close to its average value for the mixed $\eta$ events, with one muon in the barrel and the other in the endcap. It should be noted that no significant non-gaussian tails are observed in either of these mass distributions, and the fit quality is good. Also shown in the table are the shifts of the mean reconstructed invariant mass from the respective nominal values. The observed mass shifts are due to a problem with simulation of material effects in the endcap, which has since been understood and corrected.

The reconstructed muon pairs that remain after vertexing cuts are considered to be good quarkonium candidates, and further analysis is done using these pairs only. The transverse momentum distributions of these candidates are shown in Figure 9.

As can be seen from the Figure 9(a), prompt $J/\psi$ are mainly selected with $p_T$ above around 10 GeV, due to the di-muon trigger cuts applied to the events. The decay kinematics of $\Upsilon$ is somewhat different due to its larger mass, thus allowing $\Upsilon$ to be selected with $p_T$ as low as 4 GeV. Even at these, relatively low, statistics one expects to see significant numbers of both types of quarkonia at large $p_T$, which will allow statistically significant high-$p_T$ analyses beyond the reach of the Tevatron.

Figure 10(a) presents the $J/\psi$ acceptance as a function of the $J/\psi$ transverse momentum, relative to the Monte Carlo generated dataset, which requires the two muons to be within $|\eta| < 2.5$ and have transverse momenta greater than 6 and 4 GeV, respectively. Geometric acceptance of the detector and

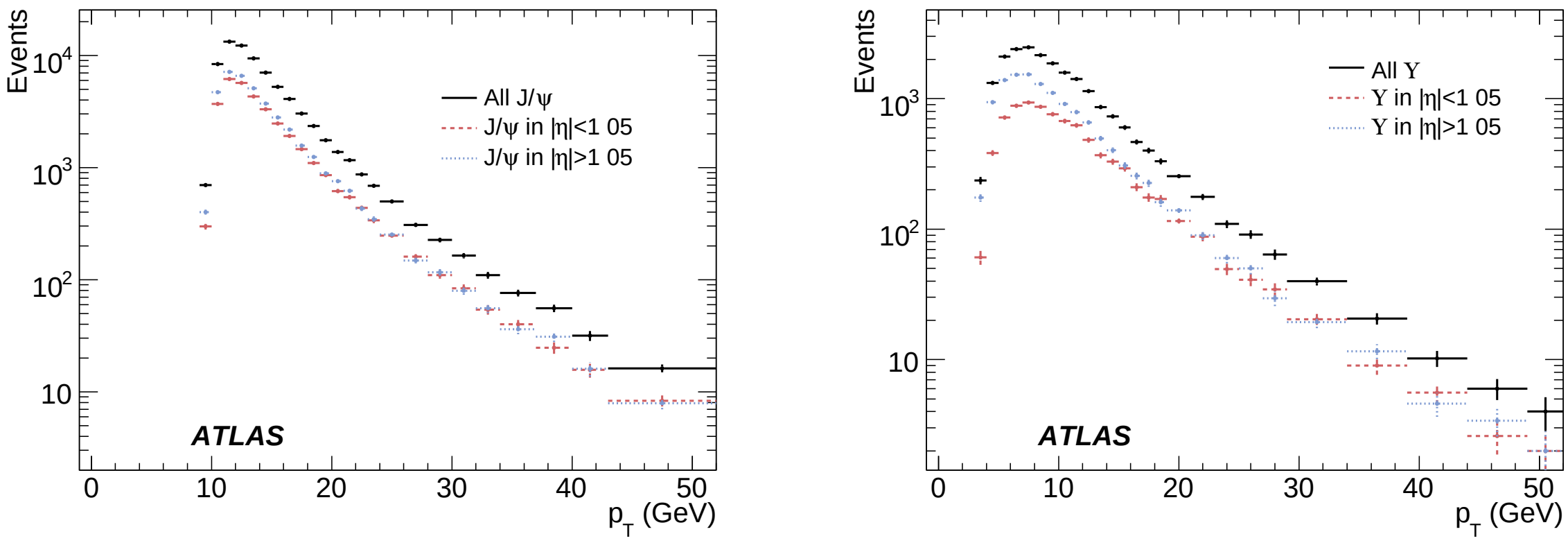


(a) $J/\psi$ (b) $\Upsilon$

Figure 9: Transverse momentum distribution of triggered reconstructed quarkonium candidates, also shown separately for quarkonia found in the barrel and endcap regions of the detector. Statistics shown in the figures correspond to integrated luminosities of about 6 pb$^{-1}$ and 10 pb$^{-1}$ for $J/\psi$ and $\Upsilon$, respectively.

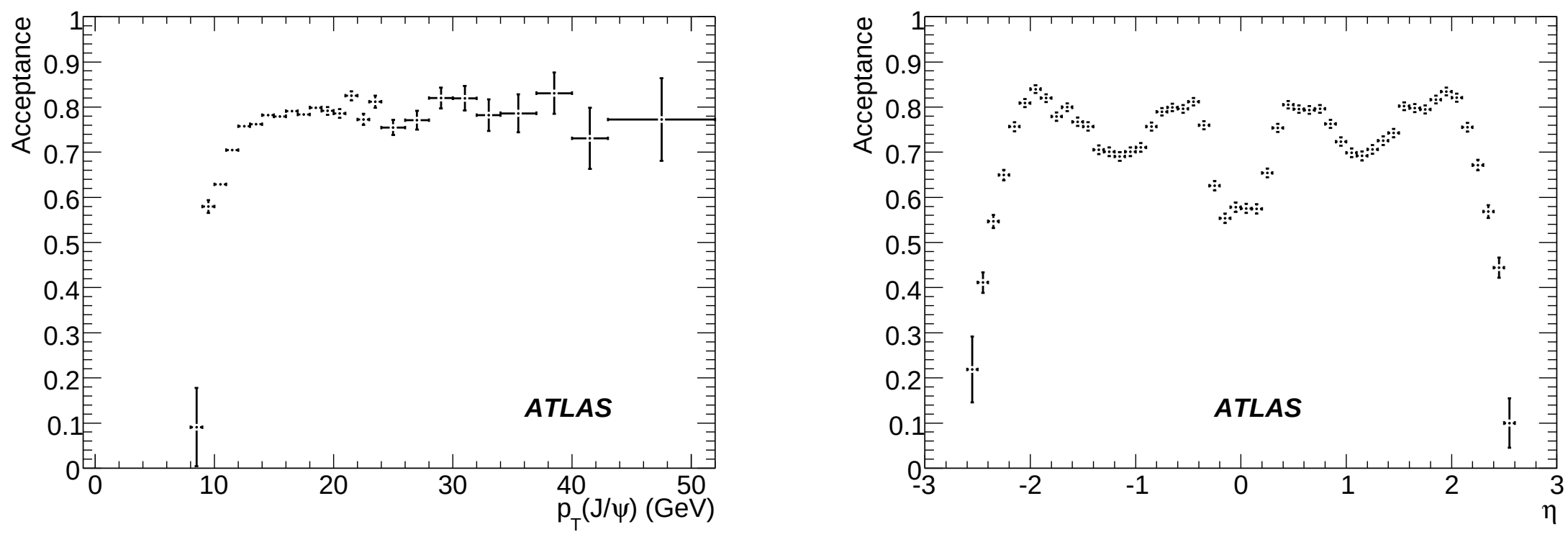


(a) $J/\psi$ acceptance with $p_T$ (b) $J/\psi$ acceptance with $\eta$

Figure 10: Acceptance of reconstructed prompt $J/\psi$ as a function of $J/\psi$ transverse momentum and pseudorapidity (relative to the MC generated dataset with $\mu6\mu4$ cuts).

reconstruction efficiency losses due to vertexing, as well as trigger efficiencies have been taken into account. When $J/\psi$ are produced with a transverse momentum above 10 GeV, we see a sharp rise in the acceptance as $J/\psi$ above this threshold are able to satisfy the muon trigger requirements within a certain kinematic configuration.

The structure in the plot of the $\eta$-dependence of $J/\psi$ reconstruction efficiency, shown in Figure 10(b), highlights the configuration necessary in order for muons from the $J/\psi$ to be able to pass the di-muon trigger, described below. The distribution of reconstructed quarkonium candidates with the angular separation of the two muons, described by the variable $\Delta R = \sqrt{\Delta\phi^2 + \Delta\eta^2}$, is shown in Figure 11. On average, muons from reconstructed $J/\psi$ ($\mu6\mu4$) candidates are separated by $\Delta R \simeq 0.47$, and are restricted from being detected with separations larger than around 0.7.

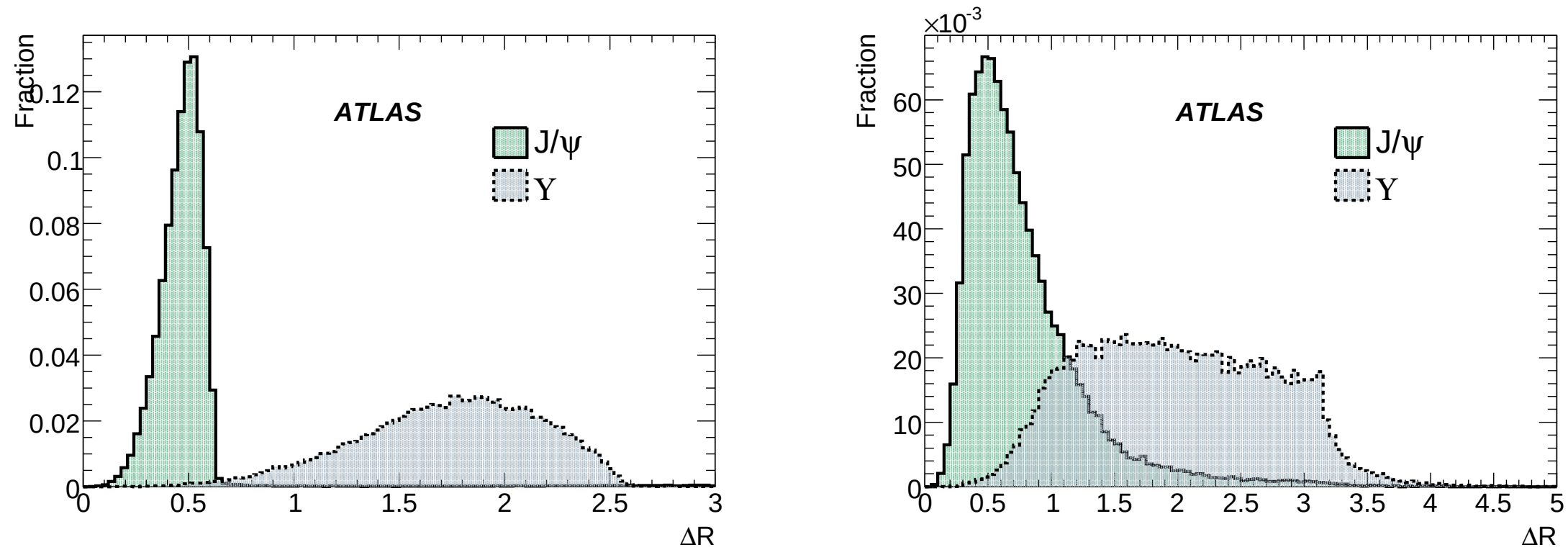


(a) $\Delta R$ with $\mu6\mu4$ cuts

(b) $\Delta R$ with $\mu10$ cuts

Figure 11: Distribution of $\Delta R$ separation of the two muons from $J/\psi$ and $\Upsilon$ candidates with di-muon $\mu6\mu4$ generator-level cuts (left) and single muon $\mu10$ cuts (right) applied.

In comparison, the higher mass of $\Upsilon$ requires the muons in the $\mu6\mu4$ case to have a much larger opening angle, with a broad distribution in $\Delta R$ peaking at around 1.8 and spanning up to 2.6. One can see that for the single $\mu10$ case in Figure 11(b) the distributions are much broader, and generally with smaller separation in $\Delta R$, reflecting the lower $p_T$ constraint on the second muon.

The small separation of muons in $\Delta R$ for the $J/\psi$ ($\mu6\mu4$) case has consequences for the $J/\psi$ reconstruction efficiency as a function of pseudorapidity, shown in Figure 10(b). Significant dips in efficiency are seen near $\eta \pm 1.2$ and $\eta = 0$, due to the muon spectrometer layout [14]. As the muons from $J/\psi$ are on average separated by only $\Delta R = 0.47$, they are subject to similar material and detector effects, and so these effects are carried over into the $J/\psi$ reconstruction with very little smearing. Hence, this distribution has a similar shape to the individual muon reconstruction efficiency distribution in ATLAS.

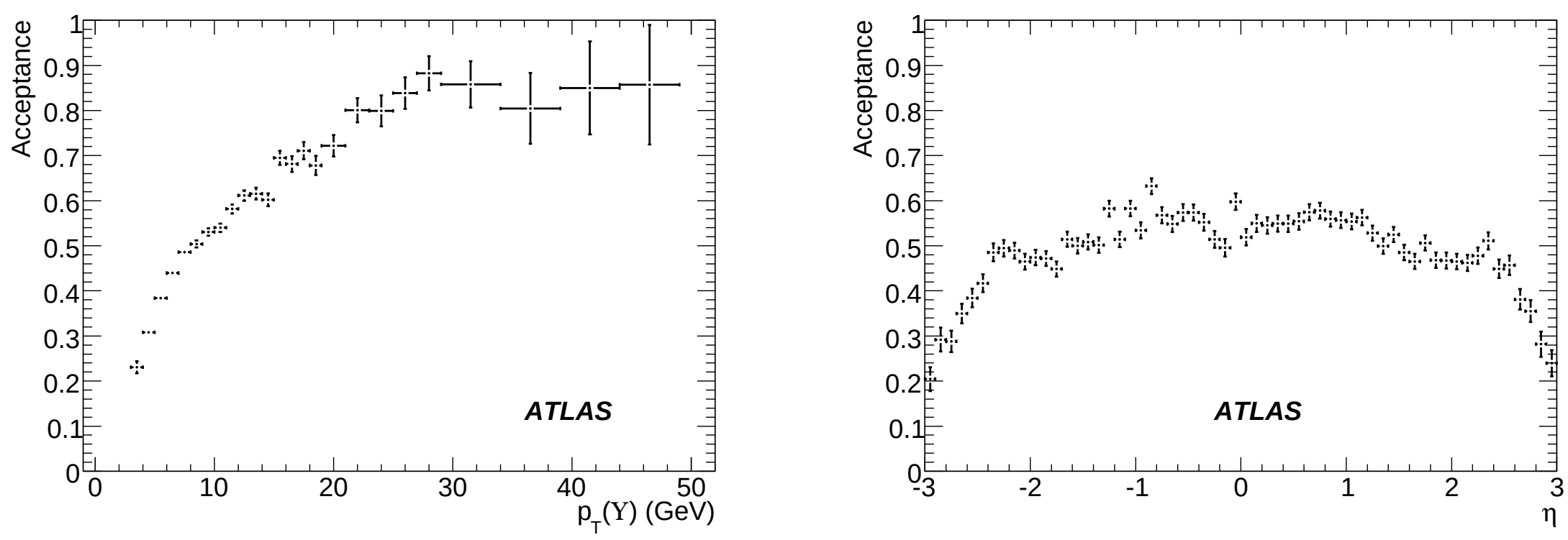


(a) $\Upsilon$ acceptance with $p_T$

(b) $\Upsilon$ acceptance with $\eta$

Figure 12: Acceptance of reconstructed prompt $\Upsilon$ as a function of transverse momentum and pseudorapidity of the quarkonium state (relative to the Monte Carlo generated dataset with $\mu6\mu4$ cuts).

This contrasts with the $\Upsilon$ reconstruction efficiency dependence on pseudorapidity, shown in Fig-

ure 12(b), which is much smoother than in $J/\psi$ case: the two muons have large angular separation and the detector layout effects are smeared over a broader range of $\eta$ values. Figure 12(a) shows the variation of acceptance with the $\Upsilon$ transverse momentum, and reflects the fact that with the $\mu6\mu4$ trigger $\Upsilon$ can be reconstructed with a lower $p_T$ threshold. In the absence of a dedicated topological trigger for $\Upsilon$, trigger efficiency at low $p_T$ suffers due to the differing decay kinematics between $J/\psi$ and $\Upsilon$ as only specialised $J/\psi$ triggers exist in reconstruction software used in this analysis. At larger $p_T$ both acceptances reach a plateau at around 80–85%.

## 3.2 Offline monitoring using quarkonium

The di-muon decays of $J/\psi$ and $\Upsilon$ will be used in both online and offline monitoring at ATLAS. Mass shifts for the reconstructed quarkonium states, plotted versus a number of different variables, have been proposed to monitor detector alignment, material effects, magnetic field scale and its stability, as well as to provide checks of muon reconstruction algorithm performance. The CDF collaboration extensively and successfully used this method, although it took many years at the Tevatron to collect sufficient statistics to allow for the disentanglement of various detector effects [15].

The expected rate of quarkonium production at ATLAS is such that we can expect to be able to perform meaningful monitoring and corrections online. There are many examples of where monitoring of quarkonium mass shifts can be useful in data-taking. Mass shifts in quarkonia as a function of transverse momentum can reveal problems with energy loss corrections and the muon momentum scale. As a function of pseudorapidity this can be a good probe of over- or under-correction of material effects in the simulated detector geometry and of magnetic field uniformity. $J/\psi$ mass shifts in Monte Carlo simulations have already helped to improve muon reconstruction algorithms in ATLAS.

An example of a reconstructed $J/\psi$ mass shift measurement at ATLAS with the statistics corresponding to 6 pb$^{-1}$ is presented in Figure 13. This is the dependence of $\Delta M$ on the difference in curvatures

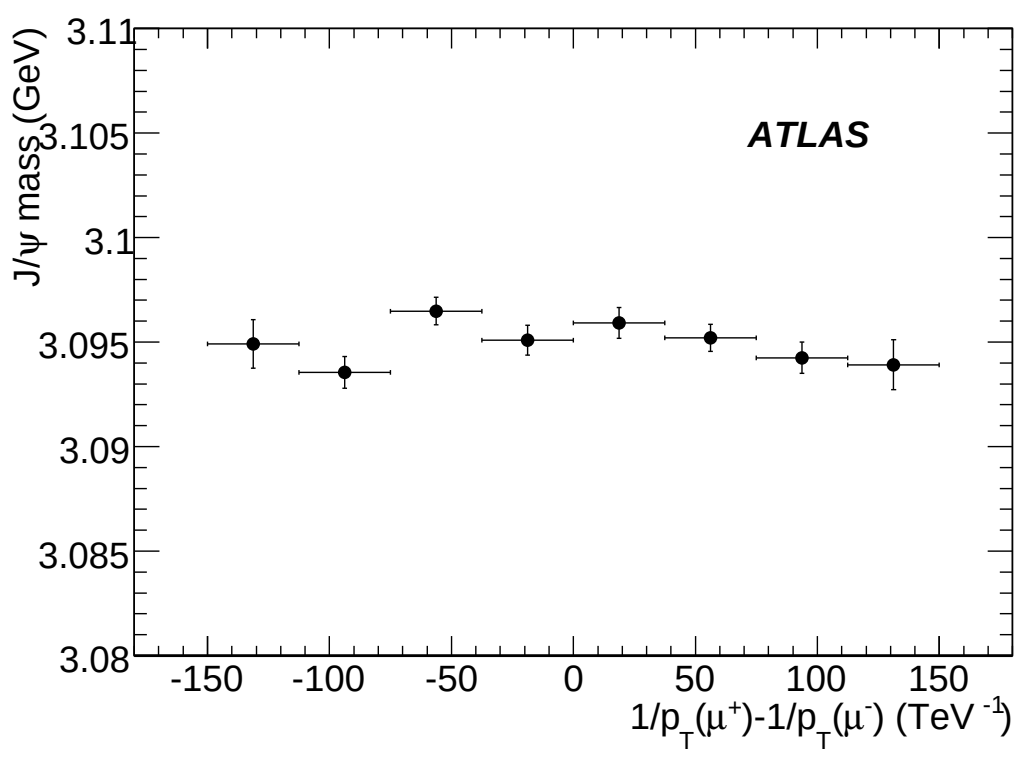


Figure 13: $J/\psi$ mass shift plotted versus the difference of curvature between the positive and negative muons. Statistics corresponds to the integrated luminosity of about 6 pb$^{-1}$.

of positive and negative muons, which allows for checks of a potentially important effect seen at CDF: horizontal misalignments in some detector elements may result in a constant curvature offset that can lead to significant charge-dependent tracking effects. A misalignment may be such that a negative track has a higher assigned curvature (and hence lower momentum) than is truly the case, whilst a positive track would be affected in the opposite way. The sample shown in the figure is simulated with ideal geometry and does not show any significant effects of this kind.

For detector alignment and data monitoring purposes, quarkonium provides a low $p_T$ point for calibration, complementary to the $Z$ boson sample, and allows for the possibility to identify any systematic variations that may develop at higher $p_T$.

In order to be able to analyse mass shifts due to two-variable correlations and disentangle various detector effects, significant statistics of $J/\psi$ and $\Upsilon$ di-muon decays have to be accumulated. A dedicated study is being performed in ATLAS to optimise the strategy of real time and offline monitoring using this method, but these results lie beyond the scope of this note.

### 3.3 Background suppression in di-muon case

The expected sources of background for prompt quarkonium with a di-muon $\mu 6\mu 4$ trigger are:

- indirect $J/\psi$ production from $b\bar{b}$ events;
- continuum of muon pairs from $b\bar{b}$ events;
- continuum of muon pairs from charm decays;
- di-muon production via the Drell-Yan process;
- decays in flight of $\pi^{\pm}$ and $K^{\pm}$ mesons.

The most important background contributions are expected to come from the decays $b \to J/\psi + X$, and the continuum of di-muons from $b\bar{b}$ events. Both of these have been simulated and analysed. The estimated total contribution from charm decays is higher than that from $b\bar{b}$ events. However, this background has not been simulated, as it is not expected to cause problems for prompt quarkonium reconstruction because the transverse momentum spectrum of the muons falls very steeply and the probability of producing a di-muon with an invariant mass within the range of interest is well below the level expected from $b\bar{b}$ events. Only a small fraction of the Drell-Yan pairs survive the di-muon trigger cuts of $\mu 6\mu 4$ in the $J/\psi - \Upsilon$ mass range, which makes this background essentially negligible, as estimated from generator-level simulation. Muons from decays in flight also have a steeply falling muon momentum spectrum, and in addition require random coincidences with muons from other sources in the quarkonium invariant mass range. This is estimated to be at the level of a few percent of the signal rate, spread over a continuum of invariant masses.

All background di-muon sources mentioned above, apart from Drell-Yan pairs, contain muons which originate from secondary vertices, which makes it possible to suppress these backgrounds by removing such di-muons whenever a secondary vertex has been resolved, based on the pseudo-proper time measurement. The pseudo-proper time is defined as

$$\text{Pseudo-proper time} = \frac{L_{xy} \cdot M_{J/\psi}}{p_T(J/\psi) \cdot c}, \tag{1}$$

where $M_{J/\psi}$ and $p_T(J/\psi)$ represent the mass and the transverse momentum of the $J/\psi$ candidate, $c$ is the speed of light in vacuum, and $L_{xy}$ is the measured radial displacement of the two-track vertex from the beamline. Once the two muons forming a $J/\psi$ candidate are reconstructed, the pseudo-proper time is used to distinguish between the prompt $J/\psi$, which have a pseudo-proper time of zero, and $J/\psi$ coming from $B$-hadron decays and hence having an exponentially decaying pseudo-proper time distribution, due to the non-zero lifetime of the parent B-hadrons.

The dependence of the resolution in radial decay length $L_{xy}$ on di-muon pseudorapidity $\eta$ is shown in Figure 14, while the variation of the expected resolution in the pseudo-proper time with di-muon $p_T$ is shown in Table 3. An improvement in the resolution is seen with increasing $p_T$ of the $J/\psi$ and

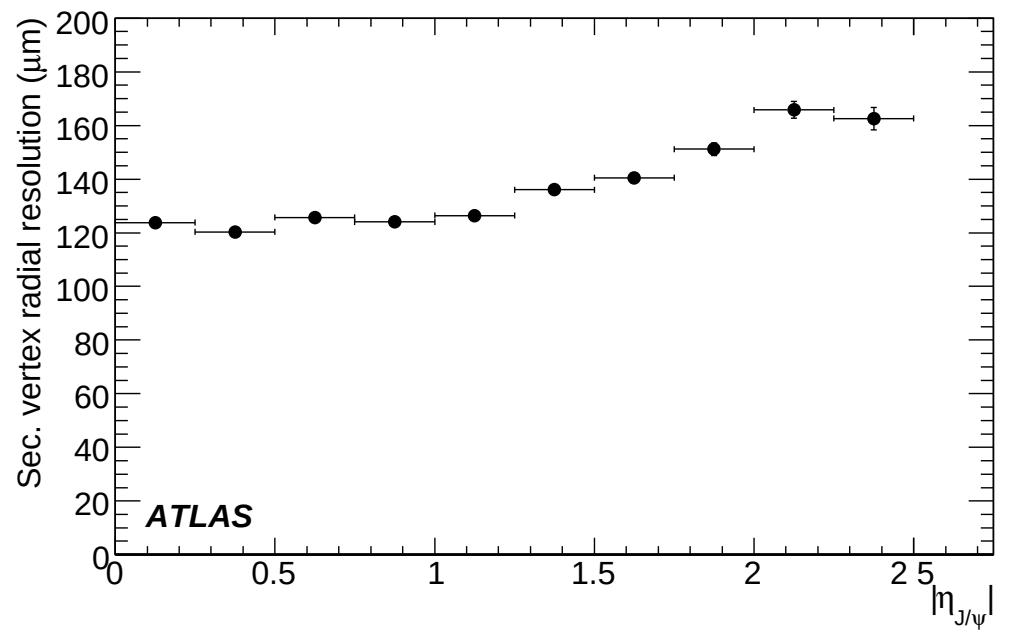


Figure 14: Radial position resolution of secondary vertex for $J/\psi$ decays as a function of the $J/\psi$ pseudorapidity.

| $J/\psi$ transverse momentum (GeV) | $9-12$ | $12-13$ | $13-15$ | $15-17$ | $17-21$ | $>21$ |
|---|---|---|---|---|---|---|
| Pseudo-proper time resolution (ps) | 0.107 | 0.103 | 0.100 | 0.093 | 0.087 | 0.068 |

Table 3: Pseudo-proper time resolution of direct $J/\psi$ events as a function of $J/\psi$ $p_T$.

decreasing $|\eta|$. Here a perfect detector alignment is assumed, with the resulting average resolution estimated at around 0.1 ps.

Figure 15(a) illustrates the pseudo-proper time distribution for both the prompt and indirect $J/\psi$ samples. By making a cut on the pseudo-proper time, one can efficiently separate most of the indirect $J/\psi$ from a prompt $J/\psi$ sample (or vice-versa). The efficiency and purity of the pseudo-proper time cuts for prompt $J/\psi$ are presented in Figure 15(b). A pseudo-proper time cut of less than 0.2 ps allows to retain prompt $J/\psi$ with the efficiency of 93% and the purity of 92%. Note that the distribution shown in Figure 15(a) is, in a sense, self-calibrating: the part to the left of the maximum can be used to determine the resolution $\sigma$, and an appropriate cut of $2\sigma$ can be applied to remove the 'tail' of secondary $J/\psi$ candidates on the right hand side.

The background levels of beauty and Drell-Yan production under the $\Upsilon$ peak are similar to those for the $J/\psi$ , except that here one does not have to contend with sources of non-prompt quarkonia from $B$-decays. However, the $bb \to \mu 6\mu 4$ background continuum under the $\Upsilon$ is more problematic: higher invariant masses around the $\Upsilon$ mean that the two triggered muons will necessarily come from two separate decays, meaning that the pseudo-proper time cut is far less effective.

Fortunately, flags associated to individual reconstructed muon tracks provide further vertexing information, which could be used for suppressing of the $bb \to \mu 6\mu 4$ continuum background. Reconstructed tracks are assigned to either come from the primary vertex, a secondary vertex, or are left undetermined. By requiring that both of the muons combined to make a $J/\psi$ or a $\Upsilon$ candidate are determined to have come from the primary vertex, background from the $bb \to \mu 6\mu 4$ continuum can be reduced by a factor of three or more, whilst reducing the number of signal events by around 5% in both cases.

Figure 16 illustrates the quarkonium signal and main background invariant mass distributions in the mass range $2-12$ GeV, for those events which satisfy the $\mu 6\mu 4$ trigger requirements, with reconstruction efficiencies and background suppression cuts taken into account. Peaks from the $J/\psi$ and $\Upsilon(1S)$ clearly dominate the background. As no higher $\psi$ and $\Upsilon$ states were simulated for this analysis, their peaks are

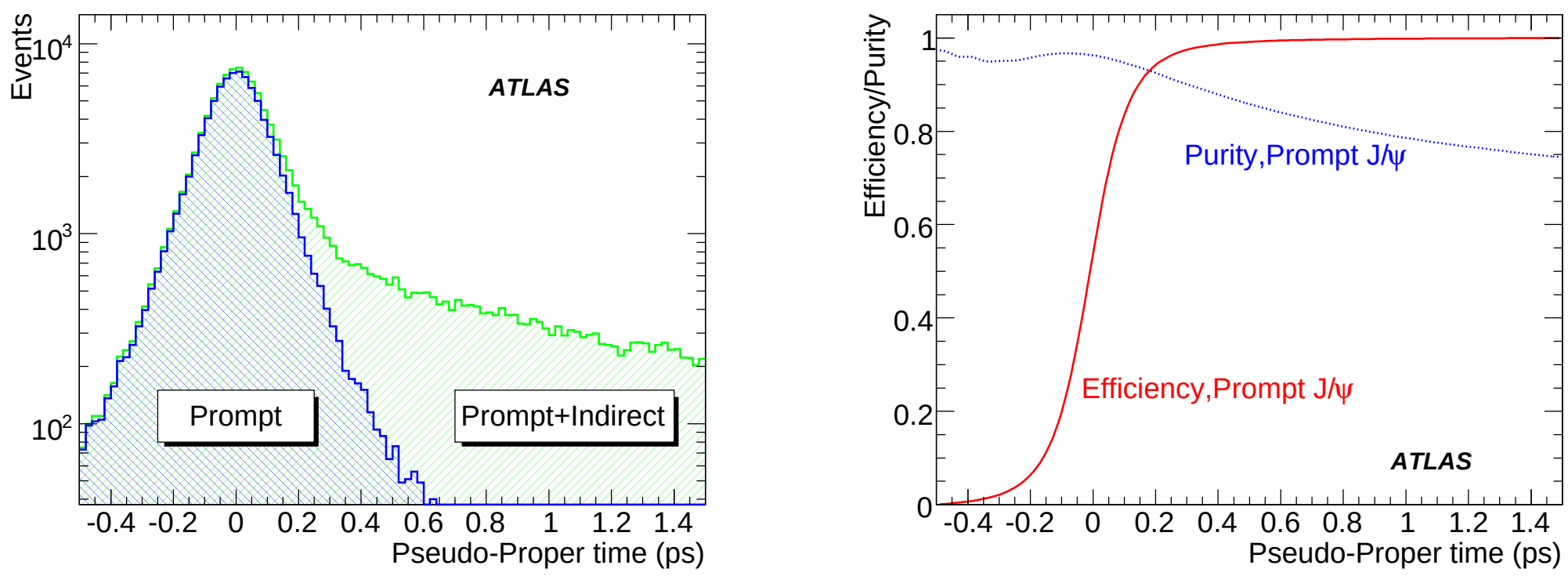


(a) (b)

Figure 15: (a) Pseudo-proper time distribution for reconstructed prompt $J/\psi$ (dark shading) and the sum of prompt and indirect $J/\psi$ candidates (lighter shading). (b) Efficiency (solid line) and purity (dotted line) for prompt $J/\psi$ candidates as a function of the pseudo-proper time cut. Statistics correspond to the integrated luminosity of 6 pb$^{-1}$.

not shown. The dotted line indicates the level of the background continuum before the vertexing cuts.

In conclusion, we find that the level of the backgrounds considered for both $J/\psi$ and $\Upsilon$ do not represent any serious problem for reconstruction and analysis of direct quarkonia with the di-muon $\mu6\mu4$ trigger.

### 3.4 Reconstruction and background suppression with a single muon candidate

By using the $\mu10$ trigger, one selects events with at least one identified muon candidate with $p_T$ above 10 GeV. In this part of the analysis, each reconstructed single muon candidate is combined with oppositely-charged tracks reconstructed in the same event. For both $J/\psi$ and $\Upsilon$ reconstruction, we insist that any other reconstructed track to be combined with the identified trigger muon has an opposite electric charge and is within a cone of $\Delta R = 3.0$ around the muon direction, so as to retain over 99% (91%) of the signal events in the $J/\psi$ ($\Upsilon$) case. As in the di-muon analysis, we require that both the identified muon and the track are flagged as having come from the primary vertex. In addition, we impose a cut on the transverse impact parameter $d_0$, $|d_0| < 0.04$ mm on the muon and $|d_0| < 0.10$ mm on the track, in order to further suppress the number of background pairs from $B$-decays.

The invariant mass distribution for the remaining pairings of a muon and a track is shown in Figure 17(a) for $J/\psi$ with $p_T$ larger than 9 GeV and in Figure 17(b) for $J/\psi$ with $p_T$ larger than 17 GeV. The distributions are fitted using a single gaussian for the signal and a straight line for the background. Clear $J/\psi$ peaks can be seen, with statistically insignificant mass shifts and the resolution close to that in the di-muon sample. It's worth noting that the signal-to-background ratio around the $J/\psi$ peak improves slightly with increasing transverse momentum of $J/\psi$. At higher $p_T$ the $\cos\theta^*$ acceptance also becomes broader, which should help independent polarisation measurements.

For $\Upsilon$ the situation is less favourable, due to the combination of a lower signal cross-section and a higher background. Although the $\Upsilon$ peak can be seen above the smooth background, its statistical significance is rather low. Hence, with this statistics, the use of the single muon sample for $\Upsilon$ cannot be justified, and in the following we will only rely on the di-muon sample.

In conclusion, we expect that the single muon trigger with a 10 GeV threshold can be successfully

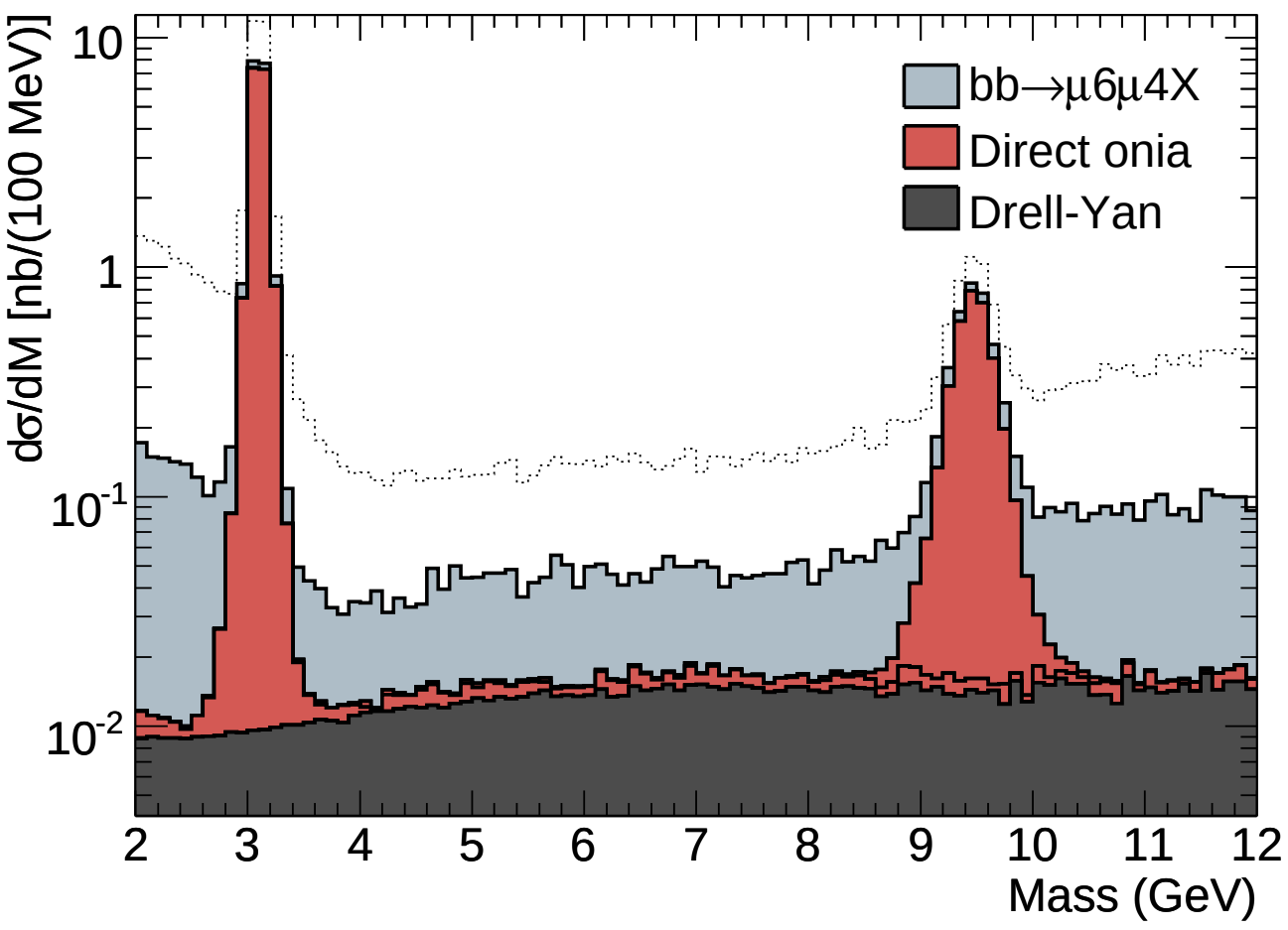


Figure 16: The cumulative plot of the invariant mass of di-muons from various sources, reconstructed with a $\mu6\mu4$ trigger, with the requirement that both muons are identified as coming from the primary vertex and with a pseudo-proper time cut of 0.2 ps. The dotted line shows the cumulative distribution without vertex and pseudo-proper time cuts.

used to select prompt $J/\psi$ events. The expected background here, although much larger than in di-muon case, is well under control. For $\Upsilon$ however, the single muon sample is only likely to be useful at significantly higher statistics and higher transverse momenta.

### 3.5 Summary of cuts and efficiencies

Table 4 summarises the efficiencies of all the selection and background suppression cuts described above, for both the di-muon and single muon trigger samples. Not all cuts are applicable to all samples; those which are not are labelled accordingly. Numbers in italics are estimates in cases where no adequate fully simulated sample was available. The efficiencies for $\mu6\mu4$ samples are calculated relative to the Monte Carlo sample with generator-level cuts on the two highest muon transverse momenta of 6 and 4 GeV. For the $\mu10$ samples, the generator-level cut of 10 GeV was applied to the $p_T$ of the highest-$p_T$ muon. Expected yields $N_S$ of quarkonia for 10 pb$^{-1}$ are given at the bottom of the table, along with background yields $N_B$ within the invariant mass window of $\pm300$ MeV for $J/\psi$ and $\pm1$ GeV for $\Upsilon$, and the signal-to-background ratios at respective $J/\psi$ and $\Upsilon$ peaks for each sample.

For higher, excited quarkonium states with vector quantum numbers the efficiencies are expected to be similar, but not necessarily identical. The biggest differences are expected for $\psi'$, where the production mechanisms as well as decay kinematics are significantly different.

## 4 Polarisation and cross-section measurement

The Colour Octet Model predicts that prompt quarkonia produced in $pp$ collisions are transversely polarised, with the degree of polarisation increasing as a function of the transverse momentum. Other production models predict different $p_T$ dependencies of the polarisation and so this quantity serves as an important measurement for discrimination of these models (see Figure 1(b)).

Quarkonium polarisation can be assessed by measuring the angular distribution of the muons produced in the decay. The relevant decay angle $\theta^*$ is defined in Figure 7. The spin alignment of the parent

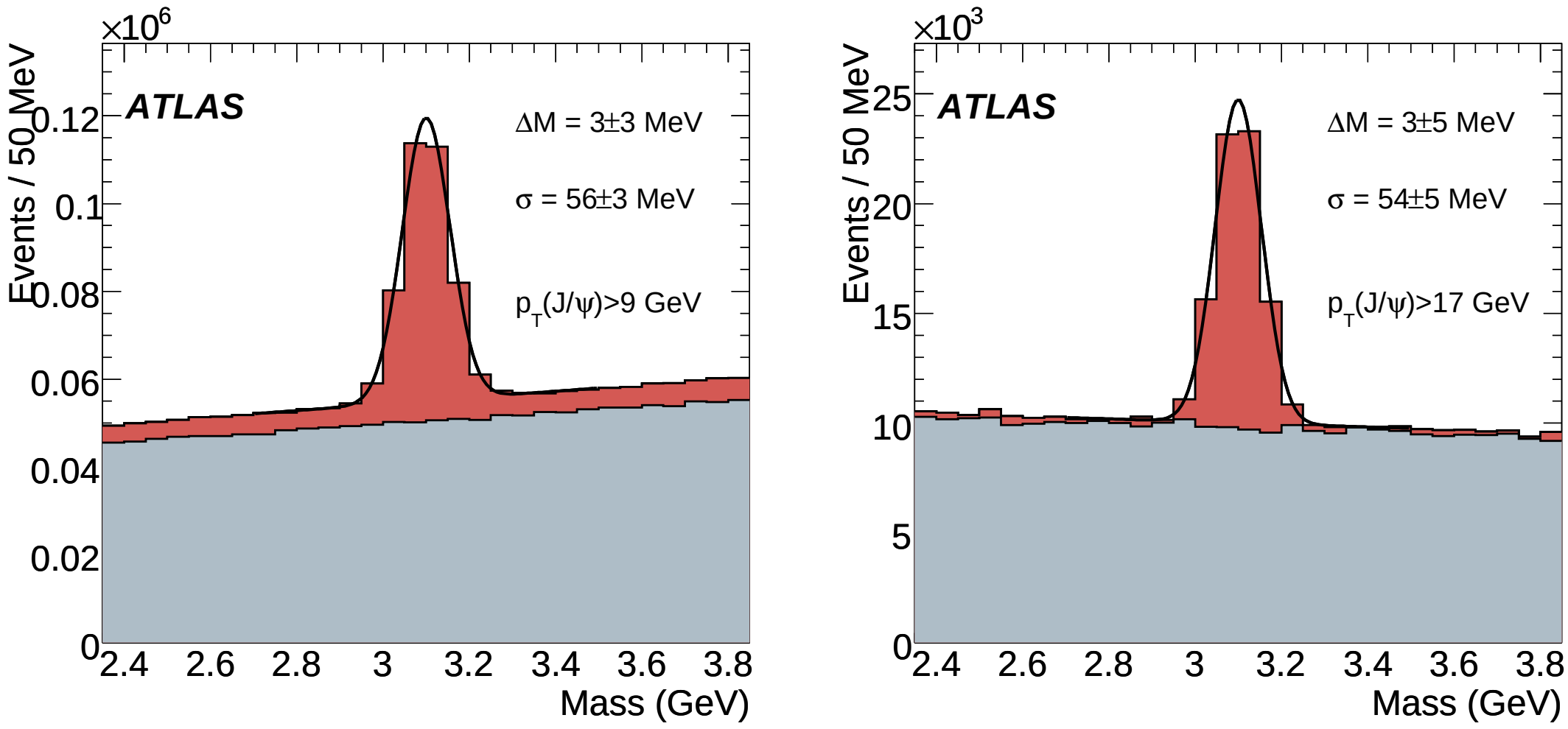


(a) $J/\psi$ with $p_T > 9$ GeV (b) $J/\psi$ with $p_T > 17$ GeV

Figure 17: Prompt quarkonium signal and $bb \to \mu X$ background events selected with the $\mu 10$ trigger, in the mass range around $J/\psi$ with (a) $p_T$ above 9 GeV, and (b) $p_T$ above 17 GeV, corresponding to 10 pb$^{-1}$ of data. The background from $B$ decays is shown in light grey. Cuts described in the text have been applied. The distributions were fitted using the sum of a linear background and a gaussian peak centered at $M = 3097$ MeV $+\Delta M$ with resolution $\sigma$.

vector quarkonium state can be determined by measuring the polarisation parameter $\alpha$ in the distribution

$$\frac{dN}{d\cos\theta^*} = C\,\frac{3}{2\alpha+6}\left(1+\alpha\cos^2\theta^*\right). \tag{2}$$

The choice of parameters in Equation 2 is such that the distribution is normalised to $C$. The parameter $\alpha$, defined as $\alpha = (\sigma_T - 2\sigma_L)/(\sigma_T + 2\sigma_L)$, is equal to $+1$ for transversely polarised production (helicity $= \pm 1$). For a longitudinal polarisation (helicity $= 0$), $\alpha$ is equal to $-1$. Unpolarised production consists of equal fractions of helicity states $+1$, 0 and $-1$, and corresponds to $\alpha = 0$.

The difficulty of quarkonium polarisation measurements is evidenced by the discrepancies between DØ and CDF results shown in Figure 1(b). The problem can be traced to the limited acceptance at high $|\cos\theta^*|$, and hence difficulties in separating acceptance corrections from spin alignment effects (see, e.g., [7]).

Note that the feed-down from $\chi$ state and $b$-hadron decays may lead to a different spin alignment and hence to a possible effective depolarisation which is hard to estimate. In addition, due to the limited statistics, the polarisation measurements at the Tevatron cannot reach the region of high $p_T$, where theoretical uncertainties are expected to be smaller.

At ATLAS we aim to measure the polarisation of prompt vector quarkonium states, in the transverse momentum range up to $\sim 50$ GeV and beyond, with extended coverage in $\cos\theta^*$ which will allow for improved understanding of efficiency measurements and thus reduced systematics. The promptly produced $J/\psi$ mesons and those that originated from $B$-hadron decays can be separated using the displaced decay vertices, as explained above. With a high production rate of quarkonia at LHC, it will be possible to achieve a higher degree of purity of prompt $J/\psi$ in the analysed sample and reduce the depolarising effect from $B$-decays, whilst retaining high statistics.

As explained in Section 2.1.3, with the di-muon trigger signature such as $\mu 6\mu 4$, the acceptance at large values of $|\cos\theta^*|$ (where the difference between various polarisation states is the biggest)

| | Quarkonium<br>Trigger type | $J/\psi$<br>$\mu6\mu4$ | $J/\psi$<br>$\mu10$ | $\Upsilon$<br>$\mu6\mu4$ | $\Upsilon$<br>$\mu10$ |
|---|---|---|---|---|---|
| | MC cross-section | 23 nb | 23 nb | 5.2 nb | 2.8 nb |
| $\varepsilon_{L1}$ | Level-1 trigger | 87% | 96% | 84% | 96% |
| $\varepsilon_{L2}$ | Level-2 trigger | 97% | >99% | 66% | >99% |
| $\varepsilon_{Rec}$ | Reconstruction | 89% | 96% | 93% | 96% |
| $\varepsilon_{Vtx}$ | Vertex fit | 99% | 99% | 99% | 99% |
| $\varepsilon_1$ | $\varepsilon_{L1} \cdot \varepsilon_{L2} \cdot \varepsilon_{Rec} \cdot \varepsilon_{Vtx}$ | 75% | 90% | 51% | 90% |
| $\varepsilon_{t0}$ | Pseudo-proper time cut | 93% | 93% | n/a | n/a |
| $\varepsilon_{Flg}$ | Only primary vertex tracks | 96% | 92% | 95% | 92% |
| $\varepsilon_{\Delta R}$ | Second track inside cone | n/a | 99% | n/a | 91% |
| $\varepsilon_{d0}$ | Impact parameter cut | n/a | 90% | n/a | 90% |
| $\varepsilon_2$ | $\varepsilon_{t0} \cdot \varepsilon_{Flg} \cdot \varepsilon_{\Delta R} \cdot \varepsilon_{d0}$ | 90% | 76% | 95% | 75% |
| $\varepsilon$ | Overall efficiency $\varepsilon_1 \cdot \varepsilon_2$ | 67% | 69% | 49% | 68% |
| | Observed signal cross-section | 15 nb | 16 nb | 2.5 nb | 2.0 nb |
| | $N_S$ for 10 pb$^{-1}$ | 150 000 | 160 000 | 25 000 | 20 000 |
| | $N_B$ in mass window for 10 pb$^{-1}$ | 7000 | 700 000 | 16 000 | 2 000 000 |
| | Signal/Background at peak | 60 | 1.2 | 10 | 0.05 |

Table 4: Predicted and observed cross-sections for prompt vector quarkonia, and efficiencies of various selection and background suppression cuts described in Section 3.

is strongly reduced, especially at low transverse momenta of quarkonium. The kinematic acceptance $\mathscr{A}(p_T, \cos\theta^*)$ of the $\mu6\mu4$ cuts applied at generator level, with respect to the full generator-level sample with no cuts on muon transverse momenta, is shown by the solid lines in Figure 18 for various $p_T$ slices of $J/\psi$. The acceptance is seen to be quite low at $J/\psi$ $p_T$ below 12 GeV, but in higher $p_T$ slices there is an area in the middle of $\cos\theta^*$ range with essentially 100% acceptance, which becomes broader with increasing $p_T$ of the $J/\psi$, but does not go beyond $|\cos\theta^*| \simeq 0.5$.

The acceptance for the single muon trigger sample, shown with the dashed lines in Figure 18, is different: here the areas of 100% acceptance are at high $|\cos\theta^*|$, and the dip in the middle gradually fills up with increasing $p_T$. This sample essentially has a full acceptance at $p_T > 20$ GeV, apart from the drop at $|\cos\theta^*| > 0.95$ due to the cut of 0.5 GeV on the $p_T$ of the track of the second muon.

The plots in Figure 18 were obtained using a dedicated generator-level Monte Carlo sample. The error bars shown in the figure reflect both statistical errors and the uncertainties due to possible dependence on $\eta$ coverage.

The simulated 'raw' measured distributions $dN^{\rm raw}/d\cos\theta^*$, for the same slices of $J/\psi$ transverse momenta, are shown in Figure 19. Again, solid and dashed lines represent the events selected by the di-muon $\mu6\mu4$ and the single muon $\mu10$ triggers, respectively. The sample was generated with zero polarisation. The raw numbers of measured events in the $\mu10$ sample were obtained by fitting the invariant mass distributions with a gaussian peak and a linear background, for each bin of $\cos\theta^*$ in each $p_T$ slice. With the estimated signal-to-background ratios shown in Figure 17(a), this causes an increase in the statistical errors, typically by a factor of 2.

The corrected distributions $dN^{\rm cor}/d\cos\theta^*$ are calculated according to the following formula:

$$\frac{dN^{\rm cor}}{d\cos\theta^*} = \frac{1}{\mathscr{A}(p_T, \cos\theta^*) \cdot \varepsilon_1 \cdot \varepsilon_2} \cdot \frac{dN^{\rm raw}}{d\cos\theta^*} \qquad (3)$$

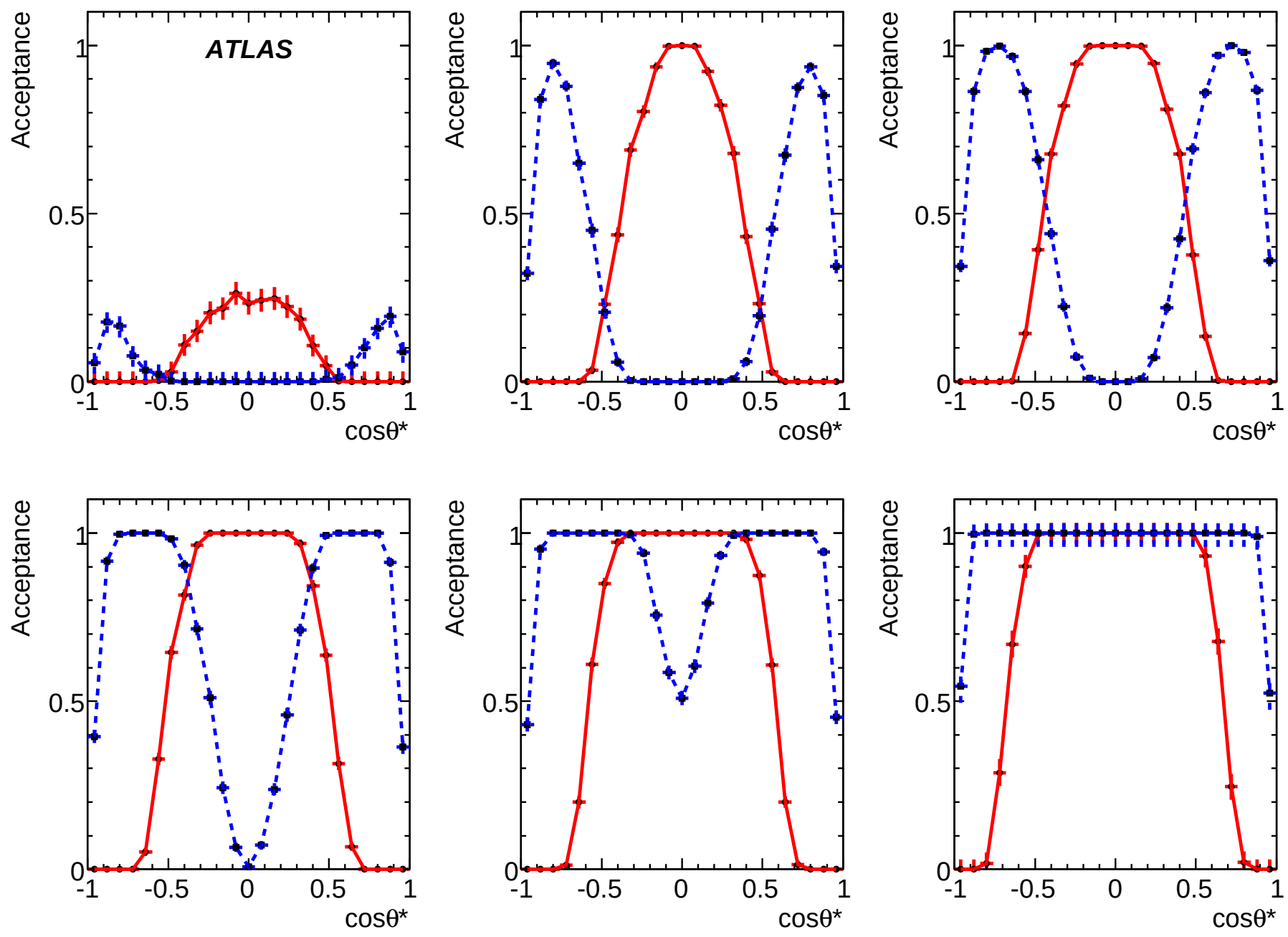


Figure 18: Generator-level kinematic acceptances of the $\mu6\mu4$ (solid lines) and $\mu10\mu0.5$ (dashed lines) cuts, calculated with respect to the sample with no muon $p_T$ cuts, in slices of $J/\psi$ transverse momentum: left to right, top to bottom $9-12$ GeV, $12-13$ GeV, $13-15$ GeV, $15-17$ GeV, $17-21$ GeV, above 21 GeV.

Here $\varepsilon_1$ stands for the trigger and reconstruction efficiency, while $\varepsilon_2$ denotes the efficiency of background suppression cuts for each sample, as defined in Table 4. Their values have been averaged over the accessible phase space within the relevant $p_T$ slice. Studies have shown that while $\varepsilon_1$ depend on $p_T$ (cf. Figure 10(a)), $\varepsilon_2$ remain essentially constant over the phase space of interest. The efficiencies $\varepsilon_1$ and $\varepsilon_2$ for both samples are listed in Table 5, while the acceptances $\mathscr{A}(p_T, \cos\theta^*)$ are shown in Figure 18.

| $p_T$, GeV | $9-12$ | $12-13$ | $13-15$ | $15-17$ | $17-21$ | $>21$ |
|---|---|---|---|---|---|---|
| $\varepsilon_1(\mu6\mu4)$, % | $67\pm1$ | $75\pm1$ | $77\pm1$ | $78\pm1$ | $79\pm1$ | $80\pm1$ |
| $\varepsilon_2(\mu6\mu4)$, % | $90\pm1$ | $90\pm1$ | $90\pm1$ | $90\pm1$ | $90\pm1$ | $90\pm1$ |
| $\varepsilon_1(\mu10)$, % | $86\pm1$ | $89\pm1$ | $90\pm1$ | $90\pm1$ | $90\pm1$ | $90\pm1$ |
| $\varepsilon_2(\mu10)$, % | $76\pm1$ | $76\pm1$ | $76\pm1$ | $76\pm1$ | $76\pm1$ | $76\pm1$ |

Table 5: Efficiencies for the $\mu6\mu4$ and $\mu10$ samples, averaged over each of the six $p_T$ slices.

At high $p_T$ the two samples increasingly overlap, thus allowing for a cross-check of acceptance and efficiency corrections. However, for measurement purposes the $\mu6\mu4$ samples are used whenever possible, complemented by $\mu10$ samples at high $\cos\theta^*$. In order to achieve this, the distributions shown in Figure 19 were appropriately masked and combined. The combined distributions $dN^{\text{cor}}/d\cos\theta^*$, corrected according to Equation 3, are shown in Figure 20. The errors shown in the plots include the statistical errors on the raw data, as well as the uncertainties on the acceptance and efficiencies. These $\cos\theta^*$ distributions are fitted using the Equation 2, with $\alpha$ and $C$ as free parameters for each $p_T$ slice. The fit

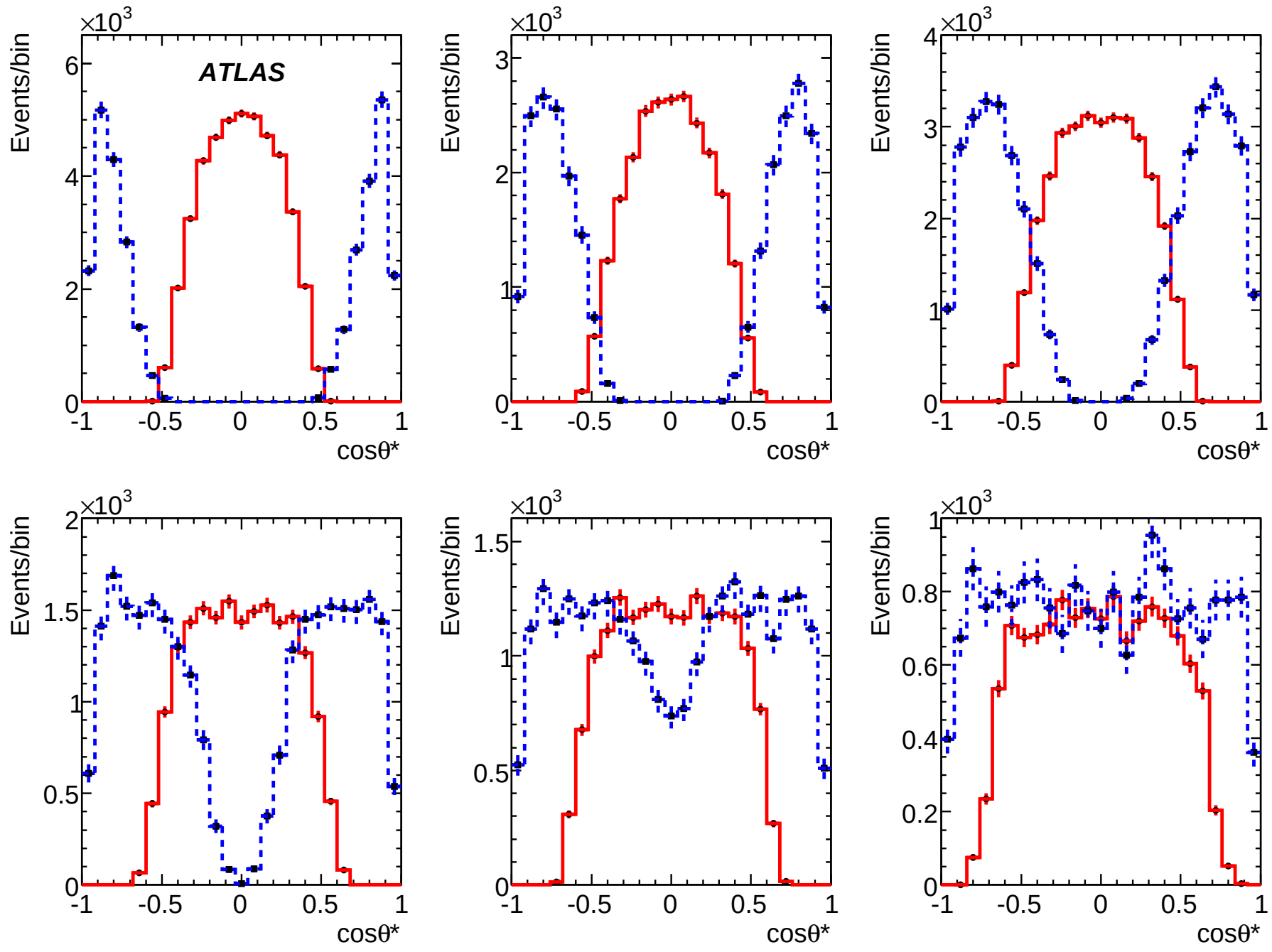


Figure 19: Measured distributions for $\mu6\mu4$- (solid lines) and $\mu10$- (dashed lines) triggered events, in the same $p_T$ slices of the $J/\psi$ candidate as in Figure 18. The simulated data sample is unpolarised. Statistics correspond to 10 pb$^{-1}$.

results are presented in Table 6, with constant $C$ rescaled to the measured cross-section $\sigma$, corresponding to the integrated luminosity of 10 pb$^{-1}$.

To further check our ability to measure the spin alignment of $J/\psi$, the raw distributions shown in Figure 19 were reweighted to emulate transversely polarised ($\alpha_{\text{gen}} = +1$) and longitudinally polarised ($\alpha_{\text{gen}} = -1$) $J/\psi$ samples, and the analysis described above was repeated. The results are shown in Figure 21 and in the middle two sections of Table 6.

A similar analysis can be done for measuring the polarisation and cross-section of $\Upsilon$, but at the integrated luminosity of 10 pb$^{-1}$ these measurements are expected to be far less precise than in $J/\psi$ case. The main reasons are lower $\Upsilon$ cross-sections at high transverse momenta, and higher backgrounds for the $\mu10$ sample. The latter reason, as explained in Section 3.4, means that with these statistics the $\mu10$ sample is essentially unusable, and the limited acceptance of the $\mu6\mu4$ sample at high $|\cos\theta^*|$ makes a precise measurement difficult.

The corrected $|\cos\theta^*|$ distributions for unpolarised $\Upsilon$ from the $\mu6\mu4$ sample are shown in Figure 22. The results of the fit using Equation 2, with normalisation matched to the integrated luminosity of 10 pb$^{-1}$, are shown in the last section of Table 6. With the integrated luminosity increased by an order of magnitude, the $\mu10$ sample should become useful and the estimated errors on $\Upsilon$ polarisation measurement could be reduced by a factor of 5.

The errors shown in Figures 20 —22 and Table 6 include the statistical uncertainties on the measured numbers of events as well as various systematic errors stemming from the uncertainties on acceptances and efficiencies described above.

The overall uncertainty on the integrated luminosity needs to be added to all measured cross-sections, and is expected to be rather large during the initial LHC runs. This uncertainty will not, however, affect the relative magnitudes of the cross-sections measured in separate $p_T$ slices, or the measured values of

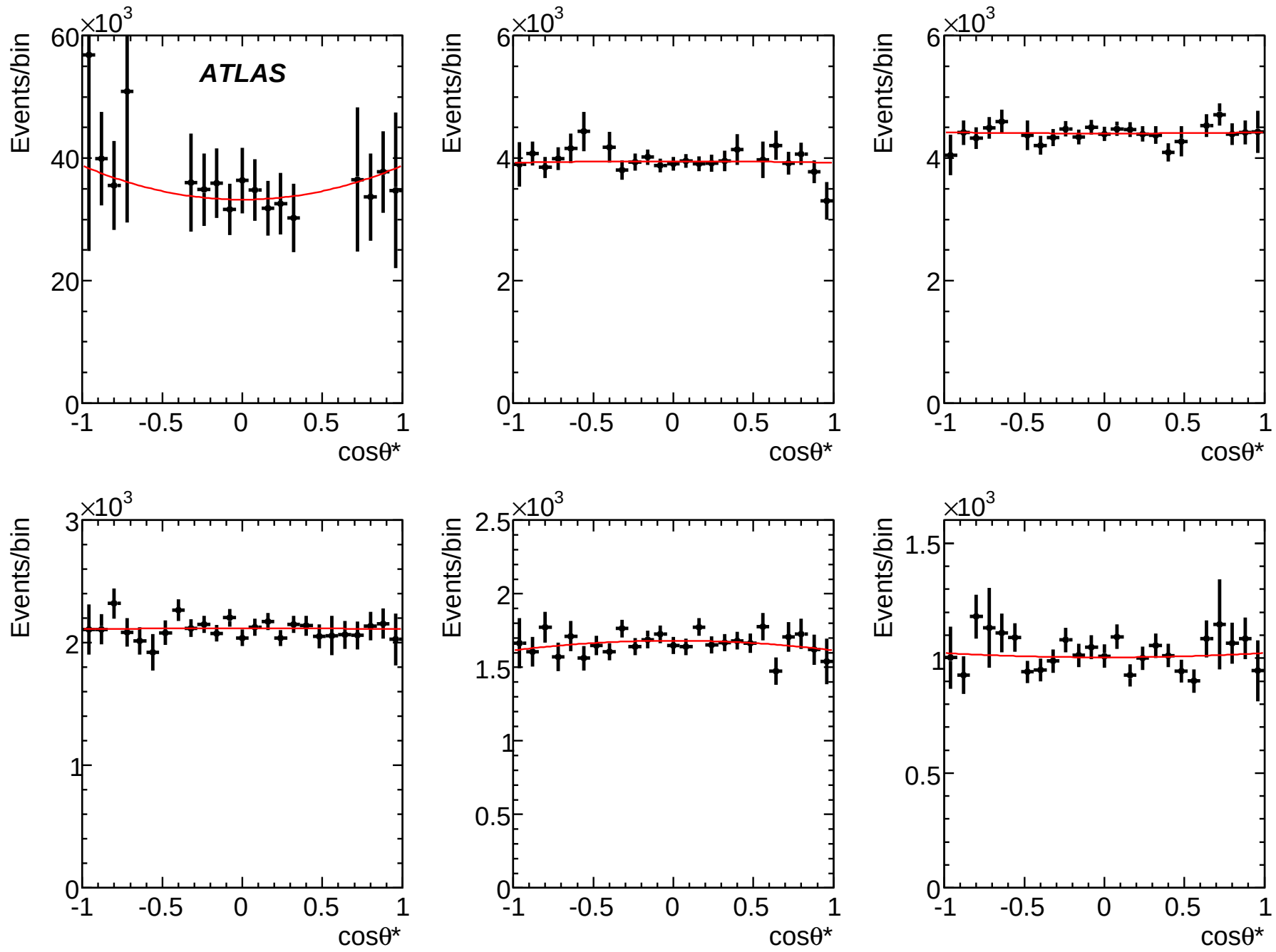


Figure 20: Combined and corrected distributions in $J/\psi$ polarisation angle $\cos\theta^*$, for the same $p_T$ slices as in Figure 18. The data sample is unpolarised ($\alpha_{\text{gen}} = 0$). The lines show the results of the fit using Equation 2, where the fitted values of $\alpha$ are given in Table 6. Statistics correspond to 10 pb$^{-1}$.

| Sample | $p_T$, GeV | $9-12$ | $12-13$ | $13-15$ | $15-17$ | $17-21$ | $>21$ |
|---|---|---|---|---|---|---|---|
| $J/\psi$, $\alpha_{\text{gen}} = 0$ | $\alpha$ | 0.156 $\pm$0.166 | $-0.006$ $\pm$0.032 | 0.004 $\pm$0.029 | $-0.003$ $\pm$0.037 | $-0.039$ $\pm$0.038 | 0.019 $\pm$0.057 |
| | $\sigma$, nb | 87.45 $\pm$4.35 | 9.85 $\pm$0.09 | 11.02 $\pm$0.09 | 5.29 $\pm$0.05 | 4.15 $\pm$0.04 | 2.52 $\pm$0.04 |
| $J/\psi$, $\alpha_{\text{gen}} = +1$ | $\alpha$ | 1.268 $\pm$0.290 | 0.998 $\pm$0.049 | 1.008 $\pm$0.044 | 0.9964 $\pm$0.054 | 0.9320 $\pm$0.056 | 1.0217 $\pm$0.088 |
| | $\sigma$, nb | 117.96 $\pm$6.51 | 13.14 $\pm$0.12 | 14.71 $\pm$0.12 | 7.06 $\pm$0.07 | 5.52 $\pm$0.05 | 3.36 $\pm$0.05 |
| $J/\psi$, $\alpha_{\text{gen}} = -1$ | $\alpha$ | $-0.978$ $\pm$0.027 | $-1.003$ $\pm$0.010 | $-1.000$ $\pm$0.010 | $-1.001$ $\pm$0.013 | $-1.007$ $\pm$0.014 | $-0.996$ $\pm$0.018 |
| | $\sigma$, nb | 56.74 $\pm$2.58 | 6.58 $\pm$0.06 | 7.34 $\pm$0.06 | 3.53 $\pm$0.04 | 2.78 $\pm$0.03 | 1.68 $\pm$0.02 |
| $\Upsilon$, $\alpha_{\text{gen}} = 0$ | $\alpha$ | $-0.42$ $\pm$0.17 | $-0.38$ $\pm$0.22 | $-0.20$ $\pm$0.20 | 0.08 $\pm$0.22 | $-0.15$ $\pm$0.18 | 0.47 $\pm$0.22 |
| | $\sigma$, nb | 2.523 $\pm$0.127 | 0.444 $\pm$0.027 | 0.584 $\pm$0.029 | 0.330 $\pm$0.016 | 0.329 $\pm$0.015 | 0.284 $\pm$0.012 |

Table 6: $J/\psi$ and $\Upsilon$ polarisation and cross-sections measured in slices of $p_T$, for 10 pb$^{-1}$.

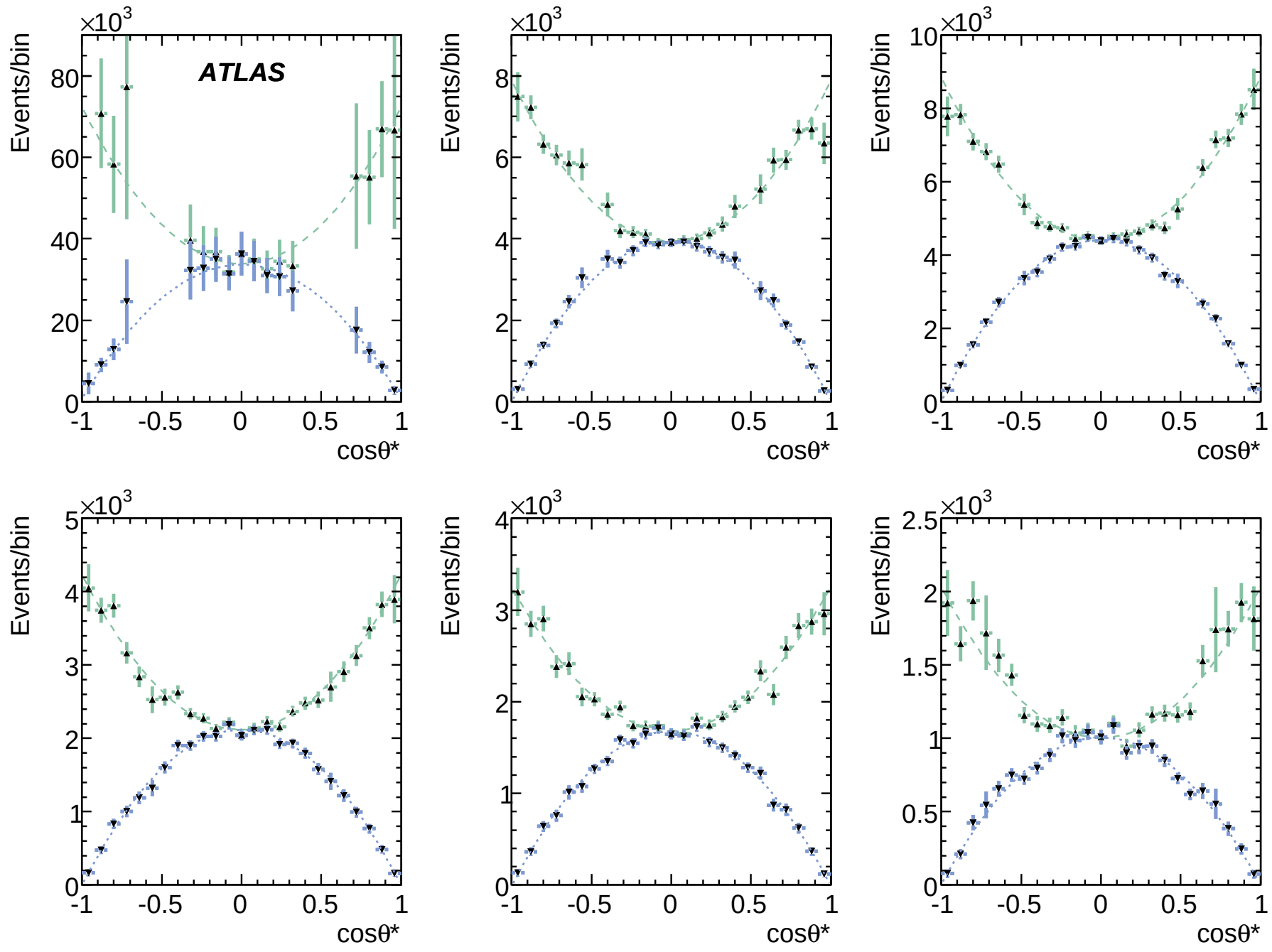


Figure 21: Combined and corrected distributions in polarisation angle $\cos\theta^*$, for longitudinally ($\alpha_{\text{gen}} = -1$, dotted lines) and transversely ($\alpha_{\text{gen}} = 1$, dashed lines) polarised $J/\psi$ mesons, in the same $p_T$ slices as in Figure 18. The lines show the results of the fit using Equation 2, where the fitted values of $\alpha$ are given in Table 6. Statistics correspond to 10 pb$^{-1}$.

the polarisation coefficient $\alpha$. Additional systematic effects have also been studied, such as the influence of finite resolution in $p_T$ and $\cos\theta^*$, changes in binning, details of the functions used for fitting the invariant mass distributions, and variations of cuts used for background suppression. Their respective uncertainties on the measured values of $\alpha$ and $\sigma$ have been found not to exceed a small fraction of the quoted errors, and have thus been deemed negligible.

In conclusion, with the integrated luminosity of 10 pb$^{-1}$ it should be possible to measure the polarisation of $J/\psi$ with the precision of order $0.02-0.06$, depending on the level of polarisation itself, in a wide range of transverse momenta, $p_T \simeq 10-20$ GeV and beyond. In case of $\Upsilon$, the expected precision is somewhat lower, of order 0.20. In both cases, however, the $p_T$ dependence of the cross-section should be measured reasonably well.

## 5 Analysis of $\chi$ production

Quarkonium states with even C parity, such as $\eta_{c,b}$ and $\chi_{c,b}$, have a strong coupling to the colour-singlet two-gluon state, and hence a significantly higher production cross-section than vector quarkonia. Their dominant production mechanism for the phase space area accessible in ATLAS is via the subprocess shown in Figure 2(b) in Section 1. Their detection, however, is rather more difficult due to the absence of purely leptonic decays.

About 30 to 40% of $J/\psi$ and $\Upsilon$ are expected to come from decays $\chi_c \to J/\psi\gamma$ and $\chi_b \to \Upsilon\gamma$. Unfortunately, the energies of the radiated photons tend to be quite small. The ability of ATLAS to detect these photons and resolve various $\chi$ states is analysed in Section 5.1. Another possibility of observing $\chi_b$ and possibly $\eta_b$ states is considered in Section 5.2, where the reconstruction of these states is attempted

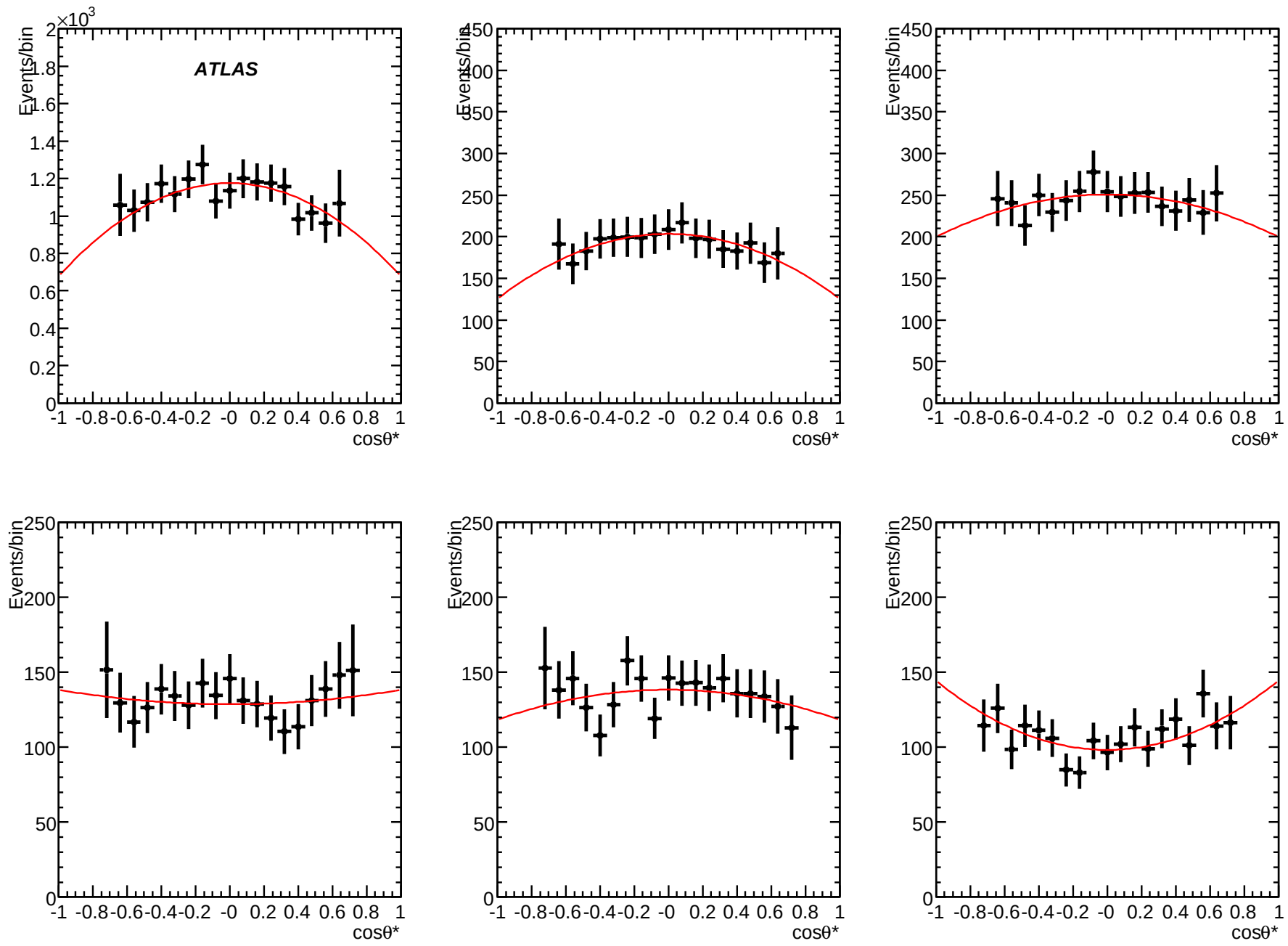


Figure 22: Corrected distributions in polarisation angle $\cos\theta^*$, for unpolarised $\Upsilon$ mesons, in the same slices of $\Upsilon$ transverse momentum as in Figure 18. Only $\mu6\mu4$ sample has been used. Statistics correspond to 10 $\mathrm{pb}^{-1}$.

through their decay into a pair of $J/\psi$, both of which subsequently decay into $\mu^+\mu^-$.

## 5.1 Radiative decays of $\chi_c, \chi_b$ states

Reconstructing $\chi_c$ candidates requires associating a reconstructed $J/\psi$ with the photon emitted from the $\chi_c$ decay. The transverse momentum distribution for all identified photon candidates in events with a prompt $J/\psi$, as measured by the ATLAS electromagnetic calorimeter, is shown in Figure 23(a) (light grey histogram).

For $\chi_c$ reconstruction, each selected quarkonium candidate is combined with every reconstructed and identified photon candidate in the event, and the invariant mass of the $\mu\mu\gamma$ system is calculated. No explicit cut is applied to the $p_T$ of the photon. The $\mu\mu\gamma$ system is considered to be a $\chi$ candidate, if the difference $\Delta M$ between the invariant masses of the $\mu\mu\gamma$ and $\mu\mu$ systems lies between 200 and 700 MeV, and the cosine of the opening angle $\alpha$ between the $J/\psi$ and $\gamma$ momenta is larger than 0.97. The last requirement comes from the observation that for the correct $\mu\mu\gamma$ combinations, the angle $\alpha$ is usually very small (see reconstructed distribution in Figure 23(b)). By analysing Monte Carlo information, it was found that all photons from generated $\chi$ decays were found in the peak near $\cos\alpha = +1$, with the long tail in the reconstructed distribution representing the combinatorial background. The transverse energy distribution for those photon candidates which satisfy the above requirements is presented in Figure 23(a) by the dark histogram. With these cuts, the combinatorial background is strongly reduced.

Figure 24 shows the distribution in $\Delta M$ for the selected $\chi_c$ decay candidates. The expected mean positions of the peaks corresponding to $\chi_0$, $\chi_1$ and $\chi_2$ signals (318, 412 and 460 MeV, respectively) are indicated by arrows. The grey histogram shows the contribution from the background process of $J/\psi$ production from $B$-hadron decays, some of which survive the pseudo-proper time cut.

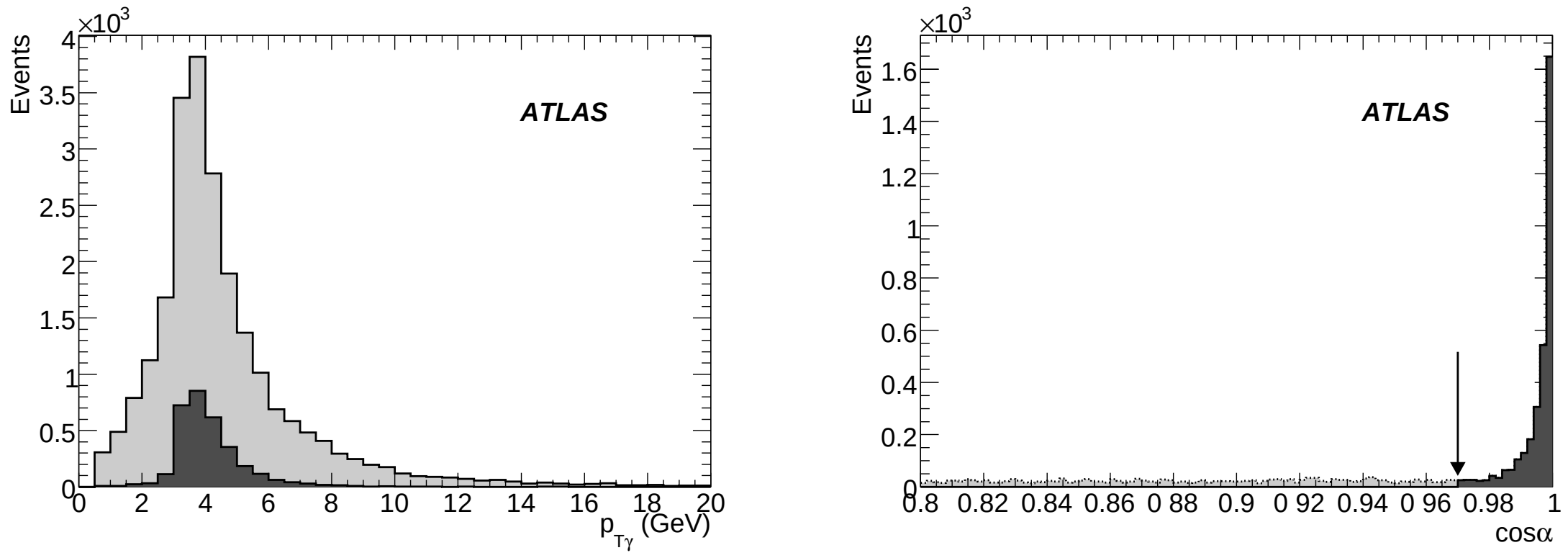


(a) (b)

Figure 23: (a) Transverse momentum distribution of photons reconstructed in prompt $J/\psi$ events. (b) Distribution of $\cos\alpha$ for each reconstructed $\gamma$ in an event. On both plots, the light grey (dark grey) histograms show the distributions before (after) the cut on the opening angle $\alpha$ between the photon and the $J/\psi$ momentum direction. All photons from $\chi \to J/\psi\gamma$ decays have $\cos\alpha > 0.97$, while the vast majority of background combinations fall outside the range shown in plot (b). The sample corresponds to the integrated luminosity of 6 pb$^{-1}$.

The solid line in Figure 24 is the result of a simultaneous fit to the measured distribution, with the three peak positions fixed at their expected values, and the common resolution function $\sigma(\Delta M)$. The resolution in $\sigma(\Delta M)$ is expected to increase with increasing $\Delta M$, and was empirically parameterised as $\sigma(\Delta M) = a \cdot \Delta M + b$. The dashed lines show the shapes of individual peaks and of the background continuum. The fit parameters are the heights of the three gaussian peaks $h_0, h_1, h_2$, the constants $a$ and $b$, and the three parameters describing the smooth polynomial background. The systematic studies include the variation of the background parameterisation and the introduction of a mass shift common for the three resonances. The true amplitudes of the peaks (15, 123 and 87, respectively) are reproduced reasonably well:

$$\begin{aligned} h_0 &= 15 \pm 3(\text{stat.}) \pm 10(\text{syst.}), \\ h_1 &= 101 \pm 4(\text{stat.}) \pm 12(\text{syst.}), \\ h_2 &= 103 \pm 4(\text{stat.}) \pm 9(\text{syst.}), \end{aligned} \tag{4}$$

with a strong negative correlation between the last two. The resolution is found to increase from about 35 MeV at $\chi_0$ to about 48 MeV at $\chi_2$, while the overall reconstruction efficiency of $\chi_c$ states is estimated to be about 4%. It may be possible to significantly improve the resolution by using photon conversions, but this is unlikely to yield a big increase in efficiency.

The procedure of reconstructing $\chi_b$ decays into $\Upsilon + \gamma$ is the same as in the charmonium case, except the di-muon pair is required to be an $\Upsilon$ candidate. However, the higher di-muon mass and hence smaller expected boost makes the photon much softer and hence more difficult to detect. With the available simulated statistics (50 000 events corresponding to 10 pb$^{-1}$), only 20 $\chi_b$ candidates have been found in the appropriate mass window, which gives an efficiency estimate of 0.03%. In order to reliably observe $\chi_b \to \Upsilon + \gamma$ decays, an integrated luminosity of at least 1 fb$^{-1}$ will be needed.

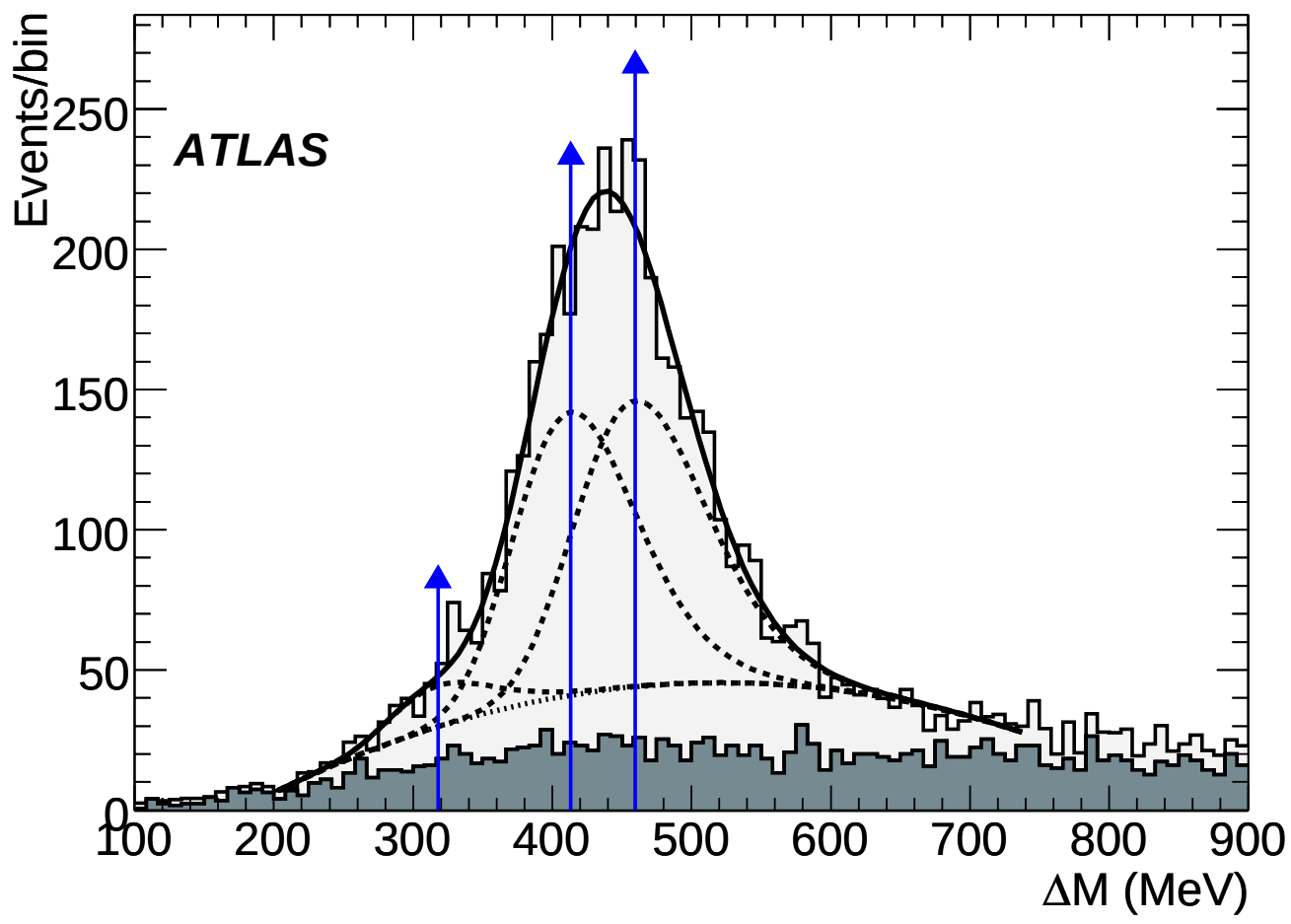


Figure 24: Difference in invariant masses of $\mu\mu\gamma$ and $\mu\mu$ systems in prompt $J/\psi$ events (light grey) with $bb \to \mu6\mu4X$ background surviving cuts (dark grey). The arrows represent the true signal peak positions, and the lines show the results of the fit described in the text. Event yields correspond to an integrated luminosity of 10 pb$^{-1}$.

## 5.2 Analysis of $\chi_b \to J/\psi J/\psi$

Another possibility for measuring $\chi_b$ production is through the decay $\chi_b \to J/\psi J/\psi \to \mu\mu\mu\mu$. The use of this decay for $\chi_b$ detection was proposed in [16], while in [17] the corresponding branching fraction was calculated to be $Br(\chi_{b0} \to J/\psi\, J/\psi) = 2 \times 10^{-4}$.

The predicted total inclusive cross-section of $\chi_{b0}$ production at LHC is estimated at around 1.5 $\mu$b [17], yielding the following theoretical estimate (without any momentum cuts on muons):

$$\sigma(pp \to \chi_{b0} + X) Br(\chi_{b0} \to J/\psi\, J/\psi) = 330\,\text{pb} \tag{5}$$

We use this cross-section in our study. It should, however, be considered as a lower bound, with higher order QCD corrections expected to increase it significantly, especially within the COM approach. This cross-section also does not include other $C$-even states ($\eta_b, \chi_{b2}$ and radial excitations), meaning that the overall combined cross-section of resonant $J/\psi\ J/\psi$ production in the $\Upsilon$ mass region can be at least an order of magnitude higher.

The PYTHIA Monte Carlo generator, used to simulate this process, was modified to include this particular decay. Events for this study are triggered with a di-muon trigger $\mu6\mu4$, as for the $J/\psi$ and $\Upsilon$ di-muon analysis. Out of 50 000 generated $\chi_b \to J/\psi(\mu\mu)J/\psi(\mu\mu)$ events 815, or 1.6%, passed the $\mu6\mu4$ trigger cuts. Taking into account di-muon branching fractions of the two $J/\psi$ mesons, this corresponds to the cross-section $\sigma = 20$ fb after trigger.

The two triggered muons have the highest $p_T$ of the four. The two remaining muons, in many cases, have transverse momenta too low to be identified as muons (i.e. below 2.5 GeV), and sometimes too low to be even reconstructed (below 0.5 GeV).

Two classes of events, remaining after the trigger cuts, have been considered to be useful:

a) events where the two trigger muons came from the same $J/\psi$. Then, the third muon has to be identified by the muon system, while the fourth must at least be reconstructed as a track (124 events);

b) events where each trigger muon came from a different $J/\psi$. The remaining two muons may or may not be identified, but their tracks still need to be reconstructed (330 events).

Thus, taking the trigger, muon identification and track reconstruction efficiencies into account, one expects about 50% of triggered $\chi_b \to J/\psi(\mu\mu)J/\psi(\mu\mu)$ decays to be observed, which amounts to 0.8% of the generated sample, corresponding to the cross-section of 10 fb. Hence, the observed statistics is expected to be around 100 events for the integrated luminosity 10 fb$^{-1}$.

Once the two $J/\psi$ candidates in the event have been reconstructed, a simultaneous fit of the four muon tracks to the common vertex is performed, with $J/\psi$ mass constraints applied to the respective di-muon invariant masses. The resulting distribution is presented in Figure 25(a). The resolution on the $\chi_b$

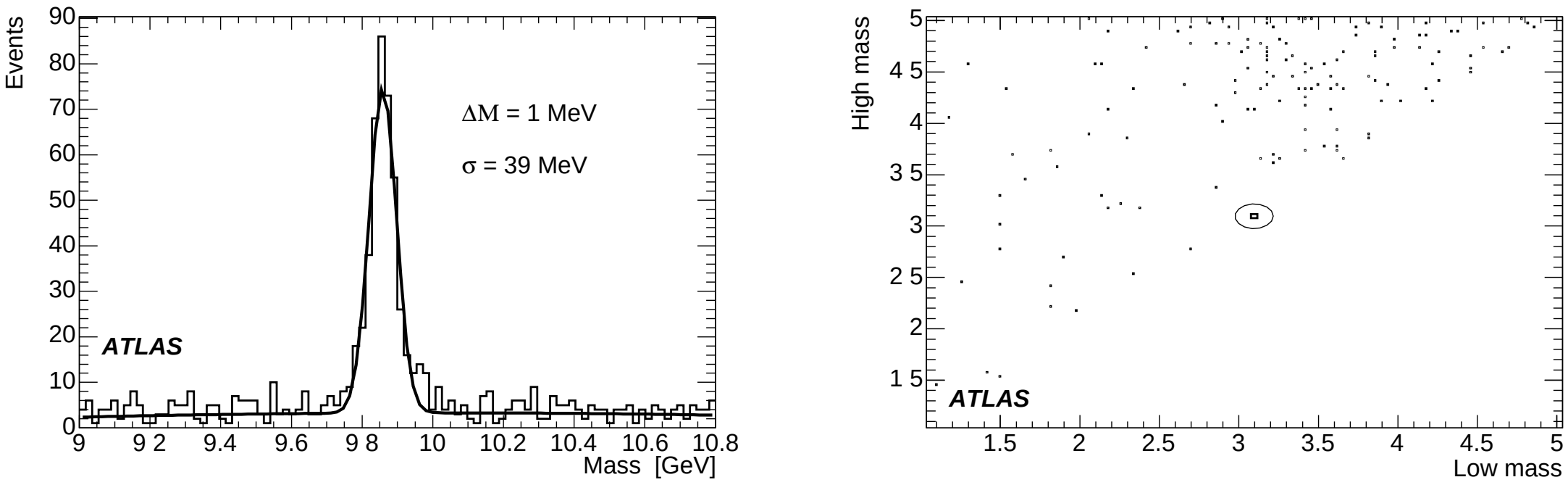


(a) (b)

Figure 25: (a) Reconstructed $\chi_b$ invariant mass, with $J/\psi$ mass constraints applied on the respective di-muon pair masses. (b) Higher di-muon invariant mass plotted versus the lower di-muon invariant mass in $\chi_b \to J/\psi(\mu\mu)J/\psi(\mu\mu)$ events.

mass is estimated to be as good as 40 MeV. Similar resolution should be expected for the reconstructed invariant mass in the decays of other $\chi_{bJ}$ states, while the resolution for $\eta_b \to J/\psi(\mu\mu)J/\psi(\mu\mu)$ should be slightly better.

With two pairs of muons in each signal event, there are two possible pairings of oppositely charged muons. The plot of the invariant mass of one di-muon pair versus the invariant mass of the other is shown in Figure 25(b), using generator-level information. All correct pairings, and none of the incorrect pairings of di-muons fall within the circle of radius 200 MeV (about 3-4 $\sigma$) from the point with coordinates $M_{J/\psi}, M_{J/\psi}$. The incorrect pairings are scattered over the whole area, so by selecting the pairings from the circle defined above, the combinatorial background can be strongly reduced.

The main expected sources of background to $\chi_b \to J/\psi(\mu\mu)J/\psi(\mu\mu)$ decays are the processes of bottom quark production, $pp \to b\bar{b}X$, with each $b$ either decaying into $J/\psi + X$, or into a muon with additional charged tracks. These backgrounds have been analysed with the same Monte Carlo samples used in our study of backgrounds for single $J/\psi$ and $\Upsilon$ production. Within the available statistics, very few background events have survived the signal selection cuts described above, and the background suppression cuts on pseudo-proper time on secondary vertices. Extrapolating these results to the integrated luminosity of 10 fb$^{-1}$ shows that the statistically significant $\chi_b \to J/\psi(\mu\mu)J/\psi(\mu\mu)$ signal peak (or peaks) should be visible on top of the combinatorial continuum, with the expected signal-to-background ratio of 10-20% or above.

In short, so far we have seen no major obstacles in an attempt to search for narrow resonances in the $J/\psi(\mu\mu)J/\psi(\mu\mu)$ invariant mass distributions. However, dedicated high statistics Monte Carlo samples are needed to draw more reliable conclusions.

# 6 Physics reach with early data

During the initial run of the LHC, the integrated luminosity of 1 $pb^{-1}$ with the $\mu6\mu4$ trigger would mean about 15 000 $J/\psi \to \mu\mu$ and 2 500 $\Upsilon \to \mu\mu$ recorded events. If the $\mu4\mu4$ trigger is used, these numbers would increase up to 17 000 and 20 000 respectively, with these additional events mainly concentrated at the lower end of the quarkonium transverse momenta.

Additional, largely independent statistics will be provided by the $\mu10$ trigger: 16 000 $J/\psi$ and 2 000 $\Upsilon$ with transverse momenta above about 10 GeV, with distributions similar to those from the $\mu6\mu4$ samples. Quite separate from these, another 7 000 of $J/\psi \to \mu\mu$ events are expected from $b$-decay events. All these events should be perfectly usable for detector alignment, acceptance and trigger efficiency studies, as well as for understanding tracking and muon system performances.

At the integrated luminosity of about 10 $pb^{-1}$ recorded numbers of $J/\psi \to \mu\mu$ and $\Upsilon \to \mu\mu$ will be roughly equal to the statistics used in this note. With these statistics, the $p_T$ dependence of the cross-section for both $J/\psi$ and $\Upsilon$ should be measured reasonably well, in a wide range of transverse momenta, $p_T \simeq 10-50$ GeV. The precision of $J/\psi$ polarisation measurement can reach $0.02-0.06$ (depending on the level of polarisation itself), while the expected error on $\Upsilon$ polarisation is unlikely to be better than about 0.2. At this stage, first attempts may be made to understand the performance of the electromagnetic calorimetry at low photon energies, and to try and reconstruct $\chi_c$ states from their radiative decays.

With an integrated luminosity of 100 $pb^{-1}$, the transverse momentum spectra are expected to reach about 100 GeV and possibly beyond, for both $J/\psi$ and $\Upsilon$. With several million $J/\psi \to \mu\mu$ and more than 500 000 of $\Upsilon \to \mu\mu$ decays, and a good understanding of the detector, high precision polarisation measurements, at the level of few percent, should become possible for both $J/\psi$ and $\Upsilon$. $\chi_b \to \Upsilon\gamma$ decays could become observable, while other measurements mentioned above will become increasingly precise.

Further increase of the integrated luminosity should make it possible to observe the resonant production of $J/\psi$ meson pairs in the mass range of the $\Upsilon$ system. During the future high luminosity running, the need to keep event rates manageable will mean an increase of thresholds of relevant single- and di-muon triggers, and the prescaling of lower threshold triggers. The higher luminosity will further expand the range of reachable transverse momenta and allow further tests of the production mechanisms, as well as make $\chi_b$ reconstruction easier.

# Production Cross-Section Measurements and Study of the Properties of the Exclusive $B^+ \to J/\psi K^+$ Channel

**Abstract**
In the initial phase of the LHC operation at low luminosity several Standard Model physics analyses will be performed in order to validate the ATLAS detector and trigger system. The $B^+ \to J/\psi K^+$ channel can be observed with the first ATLAS data at LHC and can be used for detector performance studies. This channel will provide a reference in the search for rare $B$ decays. It will also be used to estimate the systematic uncertainties and efficiencies of flavor tagging algorithms, needed for $CP$ violation measurements. The prospects to measure the $B^+$ mass, its total and differential production cross sections and lifetime with the first ATLAS data, are described in this note.

## 1 Introduction

The expected large hadronic cross-section for $b$-quark production and the high luminosity at the LHC leads to copious $b$-quark production, with the presence of a $b\bar{b}$ pair in about one percent of the collisions. Quantitatively, the expected inclusive production cross-section for $pp \to b\bar{b} + X$ at LHC is estimated to be $\sigma_{b\bar{b}} \approx 500$ $\mu$b leading to more than $10^5$ $b\bar{b}$ pairs per second at the LHC design luminosity of $\mathscr{L} \approx 10^{33}$ cm$^{-2}$s$^{-1}$. However, the extrapolation of the $b\bar{b}$ cross-section measurement from the Tevatron energy of 1.8–1.96 TeV [1, 2] to the LHC energy of 14 TeV suffers from large uncertainties. The theoretical predictions are based on NLO QCD calculations with uncertainties smaller than 20 % [3] in the kinematical region of the LHC, originating mainly from scale uncertainties [4], as well as uncertainties due to the parton density functions and the $b$-fragmentation.

A precise measurement of the $b\bar{b}$ inclusive cross-section at the LHC can be used to constrain these theoretical uncertainties. In addition, the large production rate allows for exclusive cross-section measurements shortly after the LHC start up, which have different systematic uncertainties and model dependencies (fragmentation models) from the inclusive ones. Furthermore, the $b\bar{b}$ represents the largest physics background for many processes, therefore its measurement is a prerequisite to any discovery.

In this note the exclusive channel $B^+ \to J/\psi K^+$ is studied extensively and the procedure to measure the differential and total cross-sections with the first 10 pb$^{-1}$ is presented, with event selection based on the identification of the $J/\psi$ decay to two muons.

The exclusive $B^+ \to J/\psi K^+$ decay can be measured during the initial luminosity phase of the LHC, because of the clear event topology and rather large branching ratio. It can serve as a reference channel for rare B decay searches, whose total and differential cross-sections will be measured relative to its cross-section, thus allowing the cancelation of common systematic uncertainties. Furthermore, it can be used to estimate the systematic uncertainties and efficiencies of flavour tagging algorithms, which are needed for CP violation measurements. Finally, the relatively large statistics for this decay allow for initial detector performance studies. In particular, the precise measurement of the well-known mass and lifetime [5] can be used for inner detector calibration and alignment studies.

In Section 2 of the note the Monte Carlo data sets used for this study are described. In Section 3 the $J/\psi$ selection procedure is presented. The $B^+ \to J/\psi K^+$ mass, cross-section and lifetime measurements can be found in Section 4. The expected statistics during the early LHC luminosity phase are discussed in Section 5 together with estimates of the systematic uncertainties.

All the following studies have been done for luminosity of $\mathscr{L} = 10^{32}$ cm$^{-2}$s$^{-1}$. However, since pile-up does not play any role at this luminosity, it is straightforward to rescale the results of these studies to $\mathscr{L} = 10^{31}$ cm$^2$s$^{-1}$, in case this will be the luminosity at startup.

## 2 Monte Carlo Samples

All Monte Carlo (MC) data sets used for the studies presented in this note have been produced using PYTHIA-6.4 [6] without overlaying pileup events.

| Process | $N_{\text{gen}}$ | $\mathscr{L}$ [pb$^{-1}$] | $N_{\text{gen}}(B^+ \to J/\psi K^+)$ |
|---|---|---|---|
| $bb \to J/\psi(\mu 6\mu 4) + X$ | 145 500 | 13.2 | 7 072 |

Table 1: Monte Carlo data set used for the $B^+$ study. The $N_{\text{gen}}(B^+ \to J/\psi K^+)$ events are the ones contained in the whole generated sample.

At the generator level, the ATLAS specific PYTHIA implementation for $B$-physics which provides an interface to PYTHIA-6 [7] was used. The study of the $B^+ \to J/\psi K^+$ channel is done using the inclusive production cross-section of $\sigma(b\bar{b} \to J/\psi(\mu 6\mu 4)X)$, where the numbers in the bracket denote the cuts applied on the muons from the $J/\psi$ decay in order for the generated event to be accepted (one muon with $p_T > 6$ GeV and the other with $p_T > 4$ GeV). The cross-section at the generation level, after implementing these cuts to the muons from the $J/\psi$ decay is 11.1 nb. The total number of generated events and the number of the $B^+ \to J/\psi K^+$ decays found in the sample are given in Table 1. All efficiencies presented in this note, are calculated relative to the generated number of events.

## 3 $J/\psi$ Identification Procedure

A reliable identification of the $J/\psi$ meson in the decay channel $J/\psi \to \mu^+\mu^-$ , as well as the reconstruction of the primary and secondary vertices, are the prerequisites for the $B^+ \to J/\psi K^+$cross-section measurement. For the selection of the $B^+$ candidates a further requirement of a positively charged track ($K^+$ ) originating from the $J/\psi$ secondary vertex is imposed.

The distance $\vec{x}$ between the $pp$ interaction vertex and the secondary vertex of the $B$-decay in the transverse plane is used for the $J/\psi$ identification. In the ATLAS Inner Detector TDR [8], the determination of the position of the primary vertex on an event-by-event basis was demonstrated, and for the $B^+ \to J/\psi K^+$ decay, a vertex resolution of $\sigma_x = 29$ $\mu$m and $\sigma_y = 27$ $\mu$m was estimated. For up-to-date information on the average primary vertex resolution with the staged ATLAS detector see [9].

The vector $\vec{x} = \vec{x}_{\text{prim}} - \vec{x}_B$ from the primary vertex $\vec{x}_{\text{prim}}$ to the secondary $B$-decay vertex $\vec{x}_B$ in the plane normal to the incoming proton beam [10] is used to define the transverse decay length $L_{xy}$, which is actually the projection of $\vec{x}$ onto the direction of the transverse momentum of the $B$ meson:

$$L_{xy} = \frac{\vec{x} \cdot \vec{p}_T}{|p_T|}. \tag{1}$$

The transverse decay length $L_{xy}$ is a signed variable, which is negative if the particle appears to decay before the secondary vertex of its production and positive otherwise. For a zero lifetime sample, a Gaussian distribution peaked at $L_{xy} = 0$ is expected. For exclusive decays, the proper decay length is given by:

$$\lambda = L_{xy} \cdot \frac{m_B}{p_T^B}. \tag{2}$$

For the uncertainty of the transverse decay length $L_{xy}$, only the contribution arising from the uncertainties on the primary and secondary vertex coordinates are taken into account:

$$\sigma_{L_{xy}}^2 = \frac{1}{(p_T^B)^2} \cdot \left( \sigma_x^2 (p_x^B)^2 + 2\sigma_{xy}^2 p_x^B p_y^B + \sigma_y^2 (p_y^B)^2 + \sigma_{x1}^2 (p_x^B)^2 + \sigma_{y1}^2 (p_y^B)^2 \right), \tag{3}$$

where $\sigma_x$, $\sigma_{xy}$, and $\sigma_y$ are the covariance matrix elements of the secondary vertex fit, $\sigma_{x1}$ is the resolution of the primary vertex in $x$, $\sigma_{y1}$ is the resolution of the primary vertex in $y$, $p_T^B$ is the transverse momentum of the $B^+$ meson and finally $p_x$ and $p_y$ are the $x$ and $y$ components of $B^+$ momentum.

Since the $J/\psi$ reconstruction relies on its decay into two muons: $J/\psi \to \mu^+\mu^-$, the first step in the event selection procedure is the identification of the decay muons, which in general have low $p_T$.

If a muon track reconstructed by the muon spectrometer has an inner detector track associated to it, it is considered to be a muon and is used to form the $J/\psi$ candidate. An inner detector track may also be declared a muon candidate and used in the $J/\psi$ mass reconstruction, if it is has hits or track segments in the innermost stations of the muon spectrometer. In either case the $J/\psi$ mass is calculated using the momentum of the muon candidate provided by the inner detector, in order to exploit the better momentum resolution of the inner detector in this $p_T$ region. The main $J/\psi$ selection cuts are as follows:

- All possible di-muons with $p_{T,1} \geq 3.0$ GeV and $p_{T,2} \geq 6.0$ GeV are formed;
- The tracks of each muon pair are then fitted to a common vertex;
- From the vertices found, only the ones with $\chi^2/\text{ndf} < 10$ are retained;
- To select $J/\psi$ mesons originating from the decay of a $B^+$, a cut on the proper decay length, $\lambda > 0.1$ mm, is imposed to reduce combinatorial background from prompt $J/\psi$. If this cut is not imposed, the algorithm identifies all possible combinations consistent with $J/\psi$ decaying to muons in the event;
- $J/\psi$ candidates inside a mass window of 120 MeV around $m_{J/\psi}$ are retained.

The efficiency for all previously mentioned cuts is presented in Table 2, where the efficiency after each cut is computed with respect to the previous.

Given that the sample used does not contain any prompt $J/\psi$, the effect of the cut on the proper decay length $\lambda$ in the table indicates the loss in signal events. As it is explained in the following section, this cut is not applied for the lifetime measurement. The $J/\psi$ reconstruction efficiency is also given for the case of no cut on the $J/\psi$ proper decay length $\lambda$.

The $J/\psi$ invariant mass distribution without a cut on $\lambda$ is shown in Figure 1. The shape can be described by a Gaussian with exponential tails. The $J/\psi$ mass and its resolution, obtained from a Gaussian fit, is 3098 MeV and 57.4 MeV respectively.

| **cut** | with $\lambda$ cut | | no $\lambda$ cut | |
|---|---|---|---|---|
| $J/\psi$ cut | $N_{J/\psi}$ | $\varepsilon_{J/\psi}$ [%] | $N_{J/\psi}$ | $\varepsilon_{J/\psi}$ [%] |
| after vertexing | 123 489 | 84.8 | 123 489 | 84.8 |
| after vtx $\chi^2$ cut | 115 156 | 93.3 | 115 156 | 93.3 |
| after $\lambda$ cut | 84 829 | 73.6 | - | - |
| after mass cut | 81 293 | 95.8 | 105 827 | 91.9 |
| **Total eff** | | 55.8 | | 72.7 |

Table 2: $J/\psi$ reconstruction efficiencies with and without $\lambda$ cut.

In order to study the effect of misalignment, displaced magnetic field and incorrect material map on the $J/\psi$ observed mass position and resolution, a systematic study using different ATLAS geometry configurations was performed, with different combinations of possible misalignments of the calorimeter and the muon spectrometer, as well as a displaced magnetic field map and distorted material in the inner

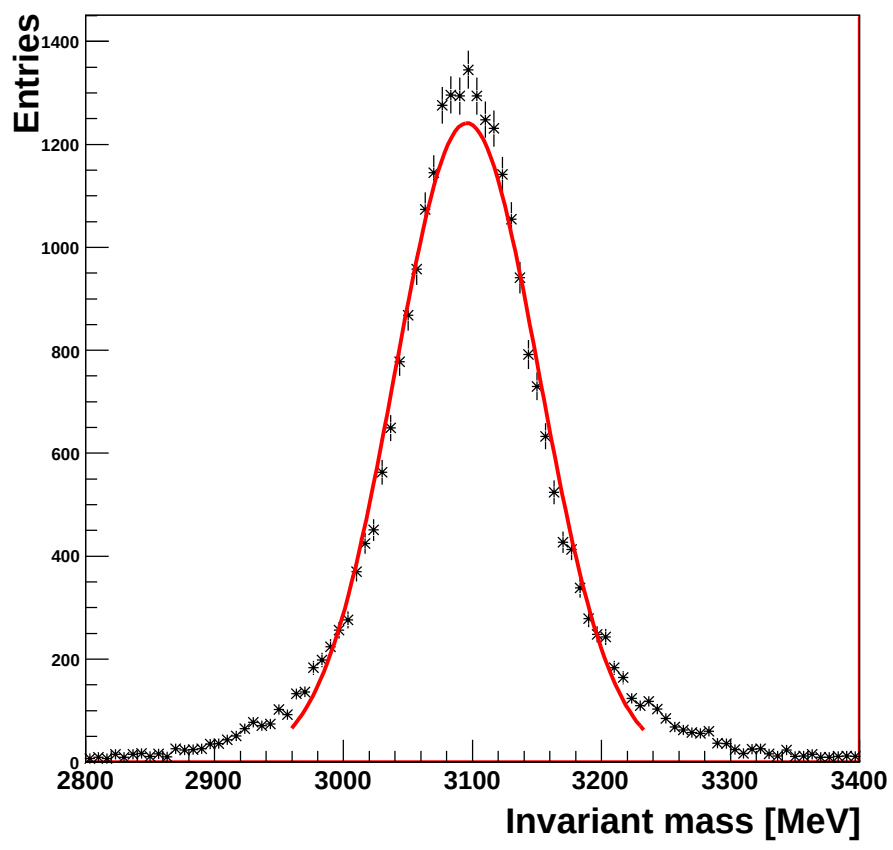


Figure 1: $J/\psi$ invariant mass distribution with a Gaussian fit superimposed.

detector and the calorimeter. These studies were performed using a dedicated $B_s^0 \to J/\psi\phi$ dataset. It was found that the width of the $J/\psi$ mass fit $\sigma(m_{J/\psi})$ is rather stable, varying between 51 and 59 MeV.

The trigger efficiency is about 99 % and was computed from the $J/\psi$ candidates that have at least one muon with $p_T > 6$ GeV in the trigger. This is expected, since at the generation level it is required that both muons from the $J/\psi$ have $p_T > 6$ GeV and $p_T > 4$ GeV.

# 4 Analysis of the $B^+ \to J/\psi K^+$ Channel

The analysis that follows for the selection of $B^+$ events can equally well be applied for the charge conjugate state. Negligible direct $CP$ violation is expected in the $B^\pm \to J/\psi K^\pm$ because for $b \to c + \bar{c}s$ transitions the standard model predicts that the leading and higher order diagrams are characterized by the same weak phase. A measurement of the asymmetry is given in [11]. The main source of asymmetry is the different interaction probabilities for $K^+$ and $K^-$ with the detector material. Other non-CP-violating sources of asymmetry are expected to lead to a lepton energy asymmetry and estimated to be negligible [12].

## 4.1 Event selection

The $B^+$ mesons are reconstructed from a $J/\psi$ and a $K^+$ candidate. The $J/\psi$ selection is described in Section 3 and the $K^+$ candidates are identified using information from the inner detector. Specifically, the procedure comprises the following steps:

- The original collection of tracks is scanned once again (excluding those already denoted as muons) and those with $p_T > 1.5$ GeV and $|\eta| < 2.7$ are retained;
- From this collection, the tracks with positive charge and inconsistent with coming from the primary vertex at one standard deviation level ($|d_0|/\sigma_{d_0} > 1$, where $d_0$ is the impact parameter of the track) are considered to be $K^+$ candidates;
- The $\mu^+\mu^-$ pair considered to be originating from the $J/\psi \to \mu^+\mu^-$ decay and the $K^+$ candidate are fitted to a common vertex. The vector defined by the sum of the $J/\psi$ and $K^+$ momentum vectors is required to point to the primary vertex, and the two muon tracks are constrained to $m_{J/\psi}$;

- Only combinations with vertex $\chi^2/\text{ndf} < 6$ , $p_T(\mu) > 5$ GeV and $\lambda > 0.1$ mm are retained;
- In case that more than two $B^+$ candidates were found in the same event, the one with the smallest vertex $\chi^2/\text{ndf}$ is accepted.

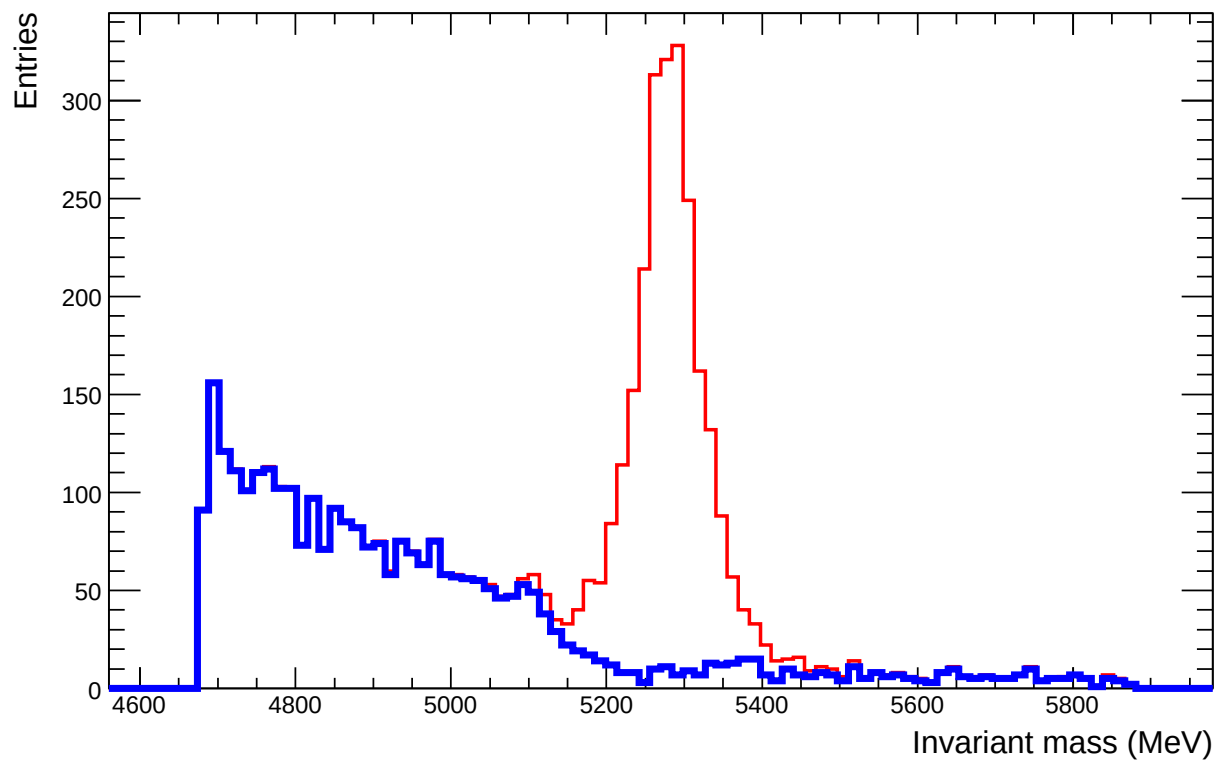


Figure 2: Invariant mass $M(K^+\mu^+\mu^-)$ distribution with the $B^+$ mass peak for signal (red) and combinatorial $b\bar{b}$ -background (blue).

### 4.2 Mass fit

The $B^+$ mass determination has been performed using the sample of $b\bar{b} \to J/\psi X$ decays. The $B^+$ invariant mass distribution $m(K^+\mu^+\mu^-)$ of the candidates fulfilling all cuts is presented in Fig. 2. In the same figure the signal and background events can be seen separately, where the distinction between them is made using the Monte Carlo truth information. The fit to the mass distribution is done by using the maximum-likelihood method, where the probability density function is a Gaussian for the signal region and a linear function for the background:

$$\begin{aligned} L &= \alpha f_{\text{sig}} + (1-\alpha) f_{\text{bkg}} \\ f_{\text{sig}} &= \frac{1}{\sqrt{2\pi}\sigma} e^{-\frac{1}{2}(\frac{m_i - m}{\sigma})^2}, \\ f_{\text{bkg}} &= b(m_i - \frac{w}{2}) + \frac{1}{w}, \end{aligned} \tag{4}$$

where $\alpha$ is the fraction of signal events in the fitted region, $m$ the $B^+$ mass, $b$ is the slope of the background distribution and $w$ defines the range of the fit. The mass range of the fit is taken from 5.15 GeV to 5.8 GeV. This is done in order to reduce contributions from partially reconstructed $B$ meson decays that populate the left side of Fig. 2. The background at the right of the mass peak originates from misidentified $\pi^+$ from $B^+ \to J/\psi\pi^+$ decays.

The result of the $B^+$ mass fit is: $M(B^+) = (5279.3 \pm 1.1)$ MeV with a width of $\sigma(B^+) = (42.2 \pm 1.3)$ MeV. The relative errors scaled properly for an integrated luminosity of $10\ \text{pb}^{-1}$ are about 0.02% and 3.5% respectively. The corresponding fit is presented in Fig. 3. The slight shoulder to the left of the mass distribution is due to the background shape.

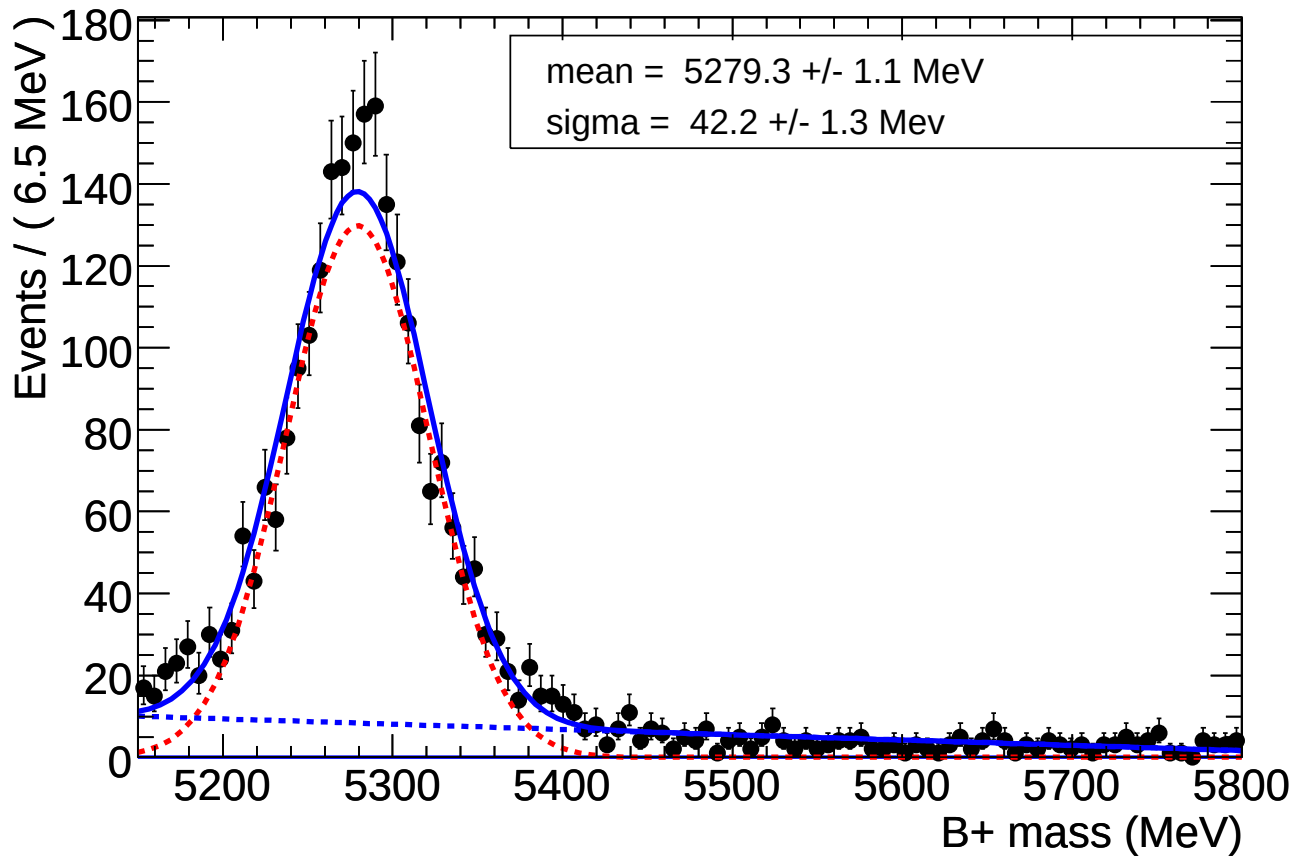


Figure 3: $B^+$ mass fit with the both signal (red) and background (blue) contributions shown separately.

### 4.3 Differential and total production cross-section

The feasibility of the measurement of the $B^+ \to J/\psi K^+$ total and differential production cross-sections at LHC, with the first 10 $\text{pb}^{-1}$, is explored in this section. The dataset used is the $b\bar{b} \to J/\psi X$ and the reconstructed $B^+ \to J/\psi K^+$ candidates were selected based on the event selection described in Section 4.1. The $B^+$ mass fit method described previously was then used to extract the efficiencies in bins of $p_T$ as well as the total efficiency.

The differential cross-section $d\sigma/d\,p_T$ can be obtained from:

$$\frac{d\sigma(B^+)}{d\,p_T} = \frac{N_{\text{sig}}}{\Delta p_T \cdot \mathscr{L} \cdot \mathscr{A} \cdot \text{BR}} \quad (5)$$

where $N_{\text{sig}}$ is the number of reconstructed $B^+$ mesons obtained from the mass fit. The size of the $p_T$ bin is denoted with $\Delta p_T$. Furthermore, $\mathscr{L}$ is the total luminosity and $\mathscr{A}$ the overall efficiency. The branching ratio BR is the product of the world average [5] branching ratios of $\text{BR}(B^+ \to J/\psi K^+) = (10.0 \pm 1.0) \times 10^{-4}$ and $\text{BR}(J/\psi \to \mu^+\mu^-) = (5.88 \pm 0.10) \times 10^{-2}$. The invariant mass spectra of the $B^+$ candidates are fitted in each $p_T$ range using an extended unbinned maximum likelihood fit. The probability density function is a Gaussian for the signal and a linear function for the background region:

$$L = \frac{N_{\text{sig}}}{N_{\text{total}}} \cdot f_{\text{sig}} + \frac{N_{\text{total}} - N_{\text{sig}}}{N_{\text{total}}} \cdot f_{\text{bkg}} \quad (6)$$

where $f_{\text{sig}}$ and $f_{\text{bkg}}$ are the fit functions as described in Equation 4. For this fit, the $B^+$ mass has been fixed to the value obtained from the mass fit in the previous Section 4.2, $m = 5279.3$ MeV. The results for the overall efficiencies and the mass widths of the fits, for the individual $p_T$ bins, are summarized in Table 3. The mass fits in the various $p_T$ regions are presented in Figure 4 whereas the fit over the full $p_T$ range is shown in Fig. 3.

To measure the $B^+$ total cross-section a similar procedure to the one used for the calculation of the differential cross-section is followed, but in this case all $B^+$ with $p_T > 10$ GeV are used to calculate the total efficiency $\mathscr{A}$. The $B^+$ mass distribution is shown in Fig. 3. The results of the total efficiency and the mass width from the fit, for the $B^+$ total cross-section measurement, are presented in Table 4.

| $p_T$ range [GeV] | $p_T \in [10,18]$ | $p_T \in [18,26]$ | $p_T \in [26,34]$ | $p_T \in [34,42]$ |
|---|---|---|---|---|
| $\mathscr{A}$ [%] | 20.1±1.0 | 37.3±1.7 | 45.0±3.1 | 51.6±4.7 |
| $\sigma(B^+)$ [ MeV] | 38.5±2.0 | 42.3±2.1 | 46.1±3.2 | 46.6±4.0 |

Table 3: Efficiency $\mathscr{A}$ and $B^+$ mass width $\sigma(B^+)$ for the various $p_T$ bins.

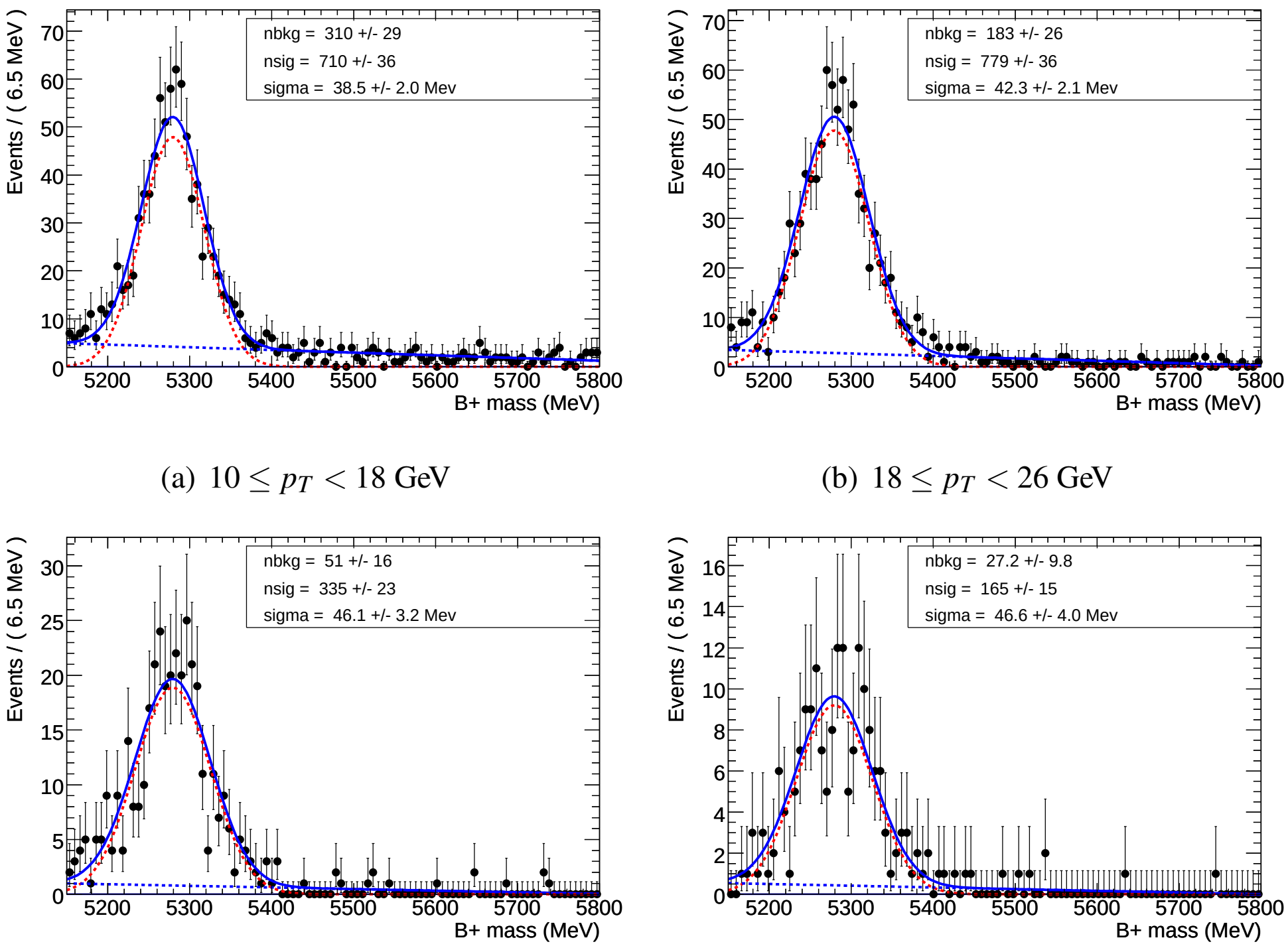


(a) $10 \leq p_T < 18$ GeV

(b) $18 \leq p_T < 26$ GeV

(c) $26 \leq p_T < 34$ GeV

(d) $34 \leq p_T < 42$ GeV

Figure 4: Fit of the $B^+$ mass in various $p_T$ ranges: $p_T \in [10,18]$ GeV (a), $p_T \in [18,26]$ GeV (b), $p_T \in [26,34]$ GeV (c), $p_T \in [34,42]$ GeV (d).

| total cross-section | |
|---|---|
| $\mathscr{A}$ [%] | 29.8±0.8 |
| $\sigma(B^+)$ [ MeV] | 42.2±1.3 |

Table 4: Overall efficiency and $B^+$ mass width for all $B^+$ with $p_T > 10$ GeV.

### 4.4 Lifetime measurement

The measurement of the lifetime $\tau$ of the selected $B^+$ candidates is a sensitive tool to confirm the beauty contents in a sample, in particular the number of the reconstructed $B^+ \to J/\psi K^+$ decays obtained in the $b\bar{b} \to J/\psi X$ dataset. The proper decay time is defined as $t = \lambda/c$. For this analysis, no cut on the proper decay length $\lambda$ (Equation 2) of the $J/\psi$ candidate or the $B^+$ candidate should be applied.

The proper decay time distribution in the signal region $B^+ \to J/\psi K^+$ can be parametrised as a convolution of an exponential function with a Gaussian resolution function, while the background distribution

parametrisation consists of two different exponential functions, where each is convoluted with a Gaussian resolution function. In the $b\bar{b} \to J/\psi X$ no additional zero lifetime events are expected because there are no prompt $J/\psi$ produced. In the realistic case, where zero lifetime events will be present, an extra Gaussian centered at zero is needed in order to properly describe those events. With the model used here, the Gaussian resolution functions depend on the reconstructed uncertainties on an event-by-event basis. In addition, it is assumed, that the distribution of the uncertainties per event are different for the signal and the corresponding background probability density functions (pdf) [13]. The use of conditional pdfs was required in order to take into account the proper decay time error per event. The exponential part of the lifetime distribution has the usual form:

$$F_t(t) = e^{\frac{-t}{\tau}}, \tag{7}$$

where $t$ is the proper decay time and $\tau$ is the lifetime. Accordingly, the convoluted function is then

$$F_c(t) = e^{\frac{-t}{\tau}} \otimes G(t, \mu, s \cdot \sigma_i), \tag{8}$$

where $\mu$ is the mean value of the Gaussian resolution function which parametrises the average bias in each proper decay time measurement. The scale factor of the error is $s$ and $\sigma_i$ is the per event proper decay time error. The conditional pdf on the per event uncertainty is then:

$$F_t(t) = F_c(t|\sigma_i) \cdot P(\sigma_i), \tag{9}$$

where $P(\sigma_i)$ is the distribution of the proper decay time error. The distribution of the proper decay time uncertainty is approximated by a superposition of Gaussian functions. In order to separate between the signal and the background, the proper decay time pdf is multiplied with the $B^+$ mass pdf, described in Section 4.2. A two-dimensional fit to the $B^+$ proper decay time and $B^+$ mass is then performed.

The results of the lifetime fit are presented in Table 5 and shown in Figure 5. The background can be best described with the two lifetime components ($\tau_1$ and $\tau_2$) which are also shown in Table 5. For the events in the mass region of the signal within $M(B^+) \in [5.15, 5.8]$ GeV the proper decay time found from the decay length is compared to the generated $B^+$ lifetime. The result of the comparison is shown in Fig. 5. The differences are well centered at zero with a Gaussian distribution and sigma 0.088 ps. It should be noted that the resolution as well as its $\sigma$ in $\eta$ bins of 0.25 is found to be independent of $\eta$.

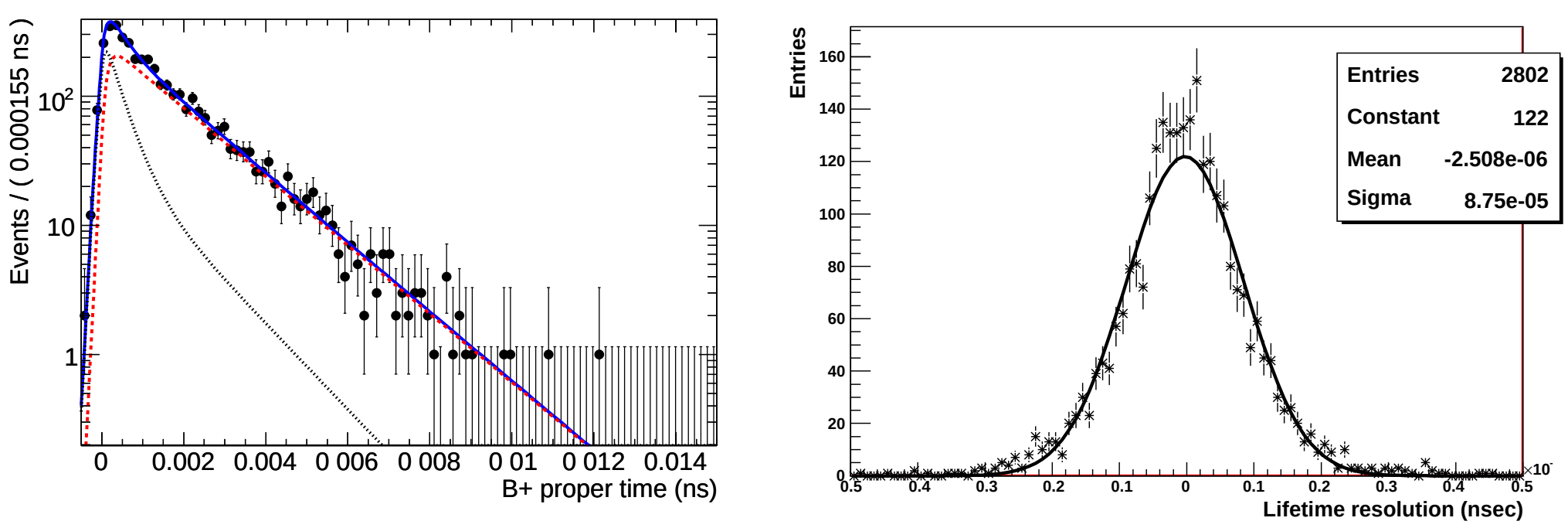


Figure 5: $B^+$ lifetime fit (left) with the signal (dashed red) and the background (dashed black) contributions shown separately and $B^+$ lifetime resolution (right).

| | |
|---|---|
| Signal lifetime $\tau$ [ps] | 1.637±0.036 |
| BG lifetime $\tau_1$ [ps] | 1.320±0.24 |
| BG lifetime $\tau_2$ [ps] | 0.370±0.067 |

Table 5: Results for the lifetime fit.

# 5 Statistical and Systematic Uncertainties

From the analysis presented above, the expected number of reconstructed $B^+$ candidates amounts to 160 per pb$^{-1}$ . This implies that sufficient statistics can be collected for a reliable cross-section measurement, after just a few months of data taking at the initial low luminosity phase of the LHC. This scenario is valid for a luminosity less than $\mathscr{L} = 10^{32}$ cm$^{-2}$s$^{-1}$, since the analysis was performed without pileup events and contains no special trigger requirements or prescaling other than a single muon with $p_T^{\mu} > 6$ GeV at level-1.

For the measurements presented in this note the main sources of systematic uncertainties are the same. The uncertainty from the luminosity in the initial phase is estimated to be 10 % and will be reduced to about 6.5 % after 0.3 fb$^{-1}$ of data. The uncertainty from the PDF's is estimated to be 3 %, while the scale uncertainty of the NLO calculations is about 5 %. Finally, the uncertainty originating from the muon identification is about 3 %. Assuming Gaussian distributions for the above mentioned uncertainties, the total systematic uncertainty of the signal varies from 9.2 % to 12 % and is dominated by the uncertainty in the luminosity.

Given that a statistical precision of $\mathscr{O}(1\ \%)$ will be reached with an integrated luminosity of 0.1 fb$^{-1}$, the contribution of the systematics will dominate the uncertainties of the first measurements. This is the case even for the differential cross-section measurement. Although the statistics in each $p_T$ bin is limited, the total uncertainty is dominated by the systematic uncertainties in the branching ratio of the $B^+ \to J/\psi K^+$ and in the luminosity, which are of the same order. For the exclusive cross-section measurement in the $B^+ \to J/\psi K^+$ channel, the relative uncertainties of the differential and total cross-sections are given in Table 6. Therein, the first row of the table contains the quadratic sum of the statistical uncertainty corresponding to an integrated luminosity of 0.01 fb$^{-1}$ and the uncertainty in the efficiency. The latter is based on the statistics of the Monte Carlo dataset used. The second row is calculated by adding in quadrature the above uncertainty to the systematic uncertainty of the luminosity and the branching ratio for every $p_T$ bin.

For the high statistics $p_T$ bins as well as for the total cross-section, the total relative uncertainty is dominated by systematic errors, originating mainly from the uncertainty in the luminosity, which is assumed to be 10 % for the initial phase, and the 10 % uncertainty in the branching ratio of $B^+ \to J/\psi K^+$. The effect of the assumed background shape on the measurements is estimated to be less than 1 %. Finally, the precision of the lifetime measurement, for the same integrated luminosity is 2.5 %, where no systematic effects are taken into account.

| $p_T$ range [GeV] | $p_T \in [10,18]$ | $p_T \in [18,26]$ | $p_T \in [26,34]$ | $p_T \in [34,42]$ | $p_T \in [10,\inf)$ |
|---|---|---|---|---|---|
| stat. + $\mathscr{A}$ [%] | 7.7 | 6.9 | 10.5 | 13.9 | 4.3 |
| total [%] | 16.1 | 15.8 | 17.6 | 19.8 | 14.8 |

Table 6: Statistical and total uncertainties for the $B^+ \to J/\psi K^+$ differential and total cross-section measurements for an integrated luminosity of 0.01 fb$^{-1}$. Total uncertainties include luminosity and BR systematic uncertainties.

## 6 Summary and Conclusions

In this note, the $B^+ \to J/\psi K^+$ channel using an inclusive $b\bar{b} \to J/\psi(\mu 6 \mu 4)X$ dataset has been studied by developing the $J/\psi$ selection methods and understanding their efficiencies. A method for measuring the $B^+$ mass using a likelihood fit, in order to separate signal and background, is established. The $B^+$ selection efficiency $\mathscr{A}$, both in $p_T$ bins and in the whole $p_T$ region, needed for the calculation of the differential and total cross-sections from real data, is extracted with fit methods similar to those used in the $B^+$ mass measurement case. Finally a likelihood fit, which takes into account the per-event primary vertex error, is performed for the measurement of the $B^+$ lifetime.

The total $B^+ \to J/\psi K^+$ production cross-section can be measured with a statistical precision better than 5% with the first 10pb$^{-1}$ of data. The differential cross-section with precision of the order of 10%. With the same statistics, adequate detector performance studies can be realised using the $B^+$ mass and lifetime measurements.

# Physics and Detector Performance Measurements with the Decays $B_s^0 \to J/\psi\phi$ and $B_d^0 \to J/\psi K^{0*}$ with Early Data

**Abstract**

The decay processes $B_d^0 \to J/\psi K^{0*}$ and $B_s^0 \to J/\psi\phi$ are expected to be observed in large numbers with the ATLAS experiment. During the early data taking period, with an integrated luminosity of around $\sim 10-150$ pb$^{-1}$, it will be possible to measure the masses and proper lifetimes for these decays with sufficient precision to allow them to be used for detector performance checks. Methods for the determination of the mass and lifetime when the performance of the detector and reconstruction software will not be fully understood are presented. A powerful simultaneous fitting technique is used. Understanding the potential for flavour tagging methods will be one of the important goals for $B$-physics with early data. The performance of the jet charge tagger for the self-calibrating decay, $B_d^0 \to J/\psi K^{0*}$, is presented. The implications for the jet charge tag in the $B_s^0 \to J/\psi\phi$ decay are also discussed.

## 1 Introduction

The decay $B_s^0 \to J/\psi\phi$ is one of the most promising channels at the LHC due to its rich physics potential. This will begin with the earliest data taken by ATLAS. Due to the high $b\bar{b}$ cross-section [1] and dedicated $J/\psi$ trigger in ATLAS [2], large statistics data will be quickly accumulated. This will allow the channel to be used for basic measurements of the $B$ mass and lifetime, which will provide a sensitive test of the understanding of the tracking system after only 150 pb$^{-1}$ of data. After collecting only 1 fb$^{-1}$ of data, ATLAS will begin to improve world precisions for these measurements. A similar analysis will also be performed for the channel $B_d^0 \to J/\psi K^{0*}$, where the expected statistics are higher by about a factor of 15.

The analysis methods to be used will evolve with increasing statistics and understanding of the detector and backgrounds. This paper concentrates on the early phase of data taking. More advanced studies of the angular dependence of the decays and CP violation are not covered. In the very early data taking period, the low statistics will not allow an investigation of the full list of theoretical parameters, but rather will concentrate on the mass and lifetime. As the backgrounds will not be well understood either, no hard cuts will be made to reject them, but rather backgrounds topologically similar to the signal will be admitted. This will also reduce the dependence on reconstruction algorithms and trigger behaviour, neither of which will be thoroughly tested when ATLAS starts to take data. In particular, no secondary vertex displacement cuts will be applied, and the dominant background admitted will be from direct $J/\psi$ production.

The simulation of the decays and their backgrounds is described in Section 2, and the reconstruction of the events in Section 3. The methods developed to extract the $B$ hadron mass and lifetime as well as the precision expected to be reached with early data are described in Section 4. A study of the possibilities for flavour-tagging with early data, an important initial step for the CP violation measurements to be done later, is presented in Section 5.

## 2 Monte Carlo production

Table 1 lists the Monte Carlo data samples used in this study. The beauty events were generated by PYTHIA 6.4 [3] using a method described in [4]. For the direct $J/\psi$ decays a special tuning of the

Colour Octet Model was prepared within PYTHIA [5]. In order to make the simulation studies more efficient the initial cuts on the transverse momentum, $p_T$, and the pseudorapidity, $\eta$, were applied at the generator level. To ensure that most of the generated events passed the trigger at the reconstruction stage, only events containing decays of $J/\psi$ into dimuons, with $p_T$ larger than 6 GeV and 4 GeV, both detected within $|\eta| < 2.4$, were retained for detector simulations.

Table 1: Monte Carlo samples used in this study. Cross sections are given by PYTHIA after applying cuts $|\eta| < 2.4$ and $p_T$ larger than 6 GeV and 4 GeV for the first and second muons from $J/\psi$ .

| Process | MC Statistics | Cross section |
|---|---|---|
| $b\bar{b} \to J/\psi X$ | 150 000 | 11.1 nb |
| $pp \to J/\psi X$ | 150 000 | 21.7 nb |
| $B_s^0 \to J/\psi\phi$ | 50 000 | 0.02 nb |
| $B_d^0 \to J/\psi K^{0*}$ | 30 000 | 0.24 nb |

# 3 Analysis of the decays $B_s^0 \to J/\psi\phi$ and $B_d^0 \to J/\psi K^{0*}$

## 3.1 Strategy for analysis of early data

The strategies deployed during the early period of the experiment will differ from those used later. In particular, the low statistics available will not allow a determination of the complete list of physics variables that can in principle be determined from the $B_s^0 \to J/\psi\phi$ and $B_d^0 \to J/\psi K^{0*}$ decays [6]. During the early phase, the compositions of the backgrounds will not be well understood. Furthermore, at this time, the detector and reconstruction software performance will also not be fully understood, and restrictive selection cuts to remove backgrounds may bias the signal in an uncontrolled way. The strategy in these early stages will therefore be to use loose cuts, which will admit more of the background decays. In particular, omitting vertex selections allows a statistically meaningful contribution from prompt $J/\psi$ events. Most of these events fall outside the signal region of the study, and allow a better determination of the vertex resolution, which in turn allows a better overall $B$ lifetime determination. This approach is consistent with the $B$ trigger strategy for early data where no cut on secondary vertex displacement is required.

## 3.2 Reconstruction

Monte Carlo events of signal and background processes, as described in Table 1, were passed through full detector simulation and reconstruction. Trigger algorithms were applied during the reconstruction. Only events accepted by the $J/\psi \to \mu^+\mu^-$ trigger [2] (with thresholds of $p_T > 6$ GeV and $p_T > 4$ GeV for the fastest and second fastest muon) were retained for offline analysis. The reconstructed data objects were then processed as follows.

$J/\psi \to \mu^+\mu^-$ candidates were sought by forming all possible pairs of oppositely charged muon tracks passing the cuts $p_T > 4$ GeV and $|\eta| < 2.4$. Pairs containing at least one muon track with $p_T >$ 6 GeV were fitted to a common vertex. Pairs were assumed to be muons from $J/\psi$ decays if the vertex fit resulted in a fit $\chi^2/n.d.f < 6$ and the invariant mass of the muon pair fell within a 3 $\sigma$ window around the nominal $J/\psi$ mass, with $\sigma = 58$ MeV. This window was chosen by fitting a Gaussian distribution to the invariant mass of the muon pairs in the events $pp \to J/\psi X$ and $bb \to \mu^+\mu^- X$, see Figure 1. The background from non resonant $\mu^+\mu^-$ pairs in the 3 $\sigma$ window is 10%.

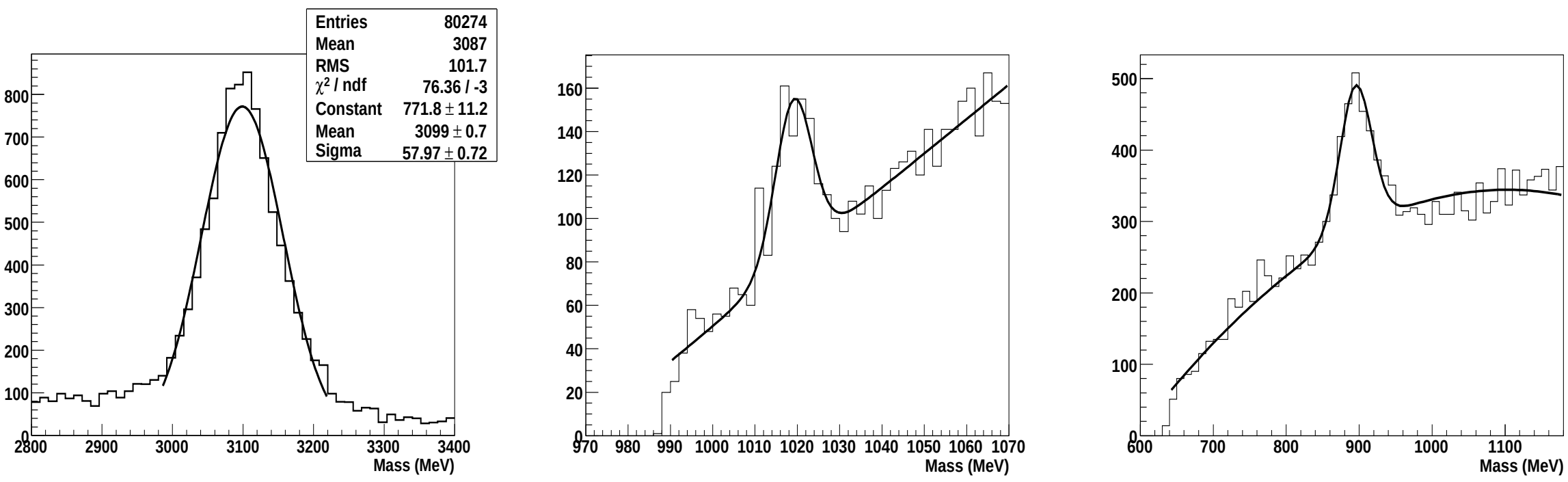


Figure 1: Reconstructed invariant mass distributions of $J/\psi \to \mu^+\mu^-$ (left), $\phi \to K^+K^-$ (middle) and $K^{0*} \to K^\pm\pi^\mp$ (right) candidates.

The $\phi \to K^+K^-$ candidates were reconstructed from all pairs of oppositely charged tracks, not identified as muons, with $p_T > 0.5$ GeV and $|\eta| < 2.5$, which were fitted to a common vertex. These tracks were assumed to be kaons from $\phi$ decays if the vertex fit resulted in a $\chi^2/n.d.f < 6$, and the invariant mass of the track pairs (under the assumption that they were left by kaons) fell within the interval $1009.2 - 1029.6$ MeV. This interval is based on a fit to the invariant mass distribution of the reconstructed $\phi \to K^+K^-$ decay candidates shown in Figure 1. The signal fit used a Breit-Wigner correctly accounting for phase space convoluted with a Gaussian to represent the detector resolution. The background was approximated by a linear function. (Additional terms up to quadratic have no significant influence on the fit.)

The $K^{0*} \to K^\pm\pi^\mp$ candidates were reconstructed by selecting all tracks that had $p_T > 0.5$ GeV and $|\eta| < 2.5$ that had not been previously identified as muons, forming them into oppositely charged pairs and fitting them to a common vertex. These pairs were assumed to be $K^\pm\pi^\mp$ from $K^{0*}$ decays if the fit resulted in a $\chi^2/n.d.f < 6$, the transverse momentum of the $K^{0*}$ candidate was greater than 3 GeV, and the invariant mass of the track pair fell within the interval 790-990 MeV, under the assumption that they were left by $K^\pm\pi^\mp$ hadrons. In Figure 1, the signal has been fitted to a Breit-Wigner function convoluted with a Gaussian and the background has been fitted to a second degree polynomial function.

To find the $B_d^0 \to J/\psi K^{0*}$ candidates, the tracks from each combination of $J/\psi \to \mu^+\mu^-$ and $K^{0*} \to K^\pm\pi^\mp$ candidates were fitted to a common point. The two muon tracks were constrained to the PDG $J/\psi$ mass. These quadruplets of tracks were assumed to be from $B_d^0 \to J/\psi K^{0*}$ decays if the transverse momentum of the $B_d^0$ candidate was greater than 10 GeV and the fit resulted in a $\chi^2/n.d.f < 6$. In the case of more than one candidate per event, the candidate with the lowest $\chi^2/n.d.f$ was retained.

$B_s^0 \to J/\psi\phi$ candidates were sought by fitting the tracks from each combination of $J/\psi \to \mu^+\mu^-$ and $\phi \to K^+K^-$ candidates fitted to a common vertex. The two muon tracks were constrained to the PDG $J/\psi$ mass. These quadruplets of tracks were assumed to be from $B_s^0 \to J/\psi\phi$ decays if the transverse momentum of the $B_s^0$ candidate was greater than 10 GeV and the fit resulted in a $\chi^2/n.d.f < 6$. If there was more than one candidate per event then the candidate with the lowest $\chi^2/n.d.f$ was chosen.

Accepted $B_s^0$ and $B_d^0$ candidates contain a negligible background from non-resonant $\mu^+\mu^-$ pairs, 0.1% and 0.2% respectively, and therefore the background from non-resonant bb $\to \mu^+\mu^-$X events are not considered in this analysis.

Events were accepted in a wide invariant mass window of $\pm 12 \cdot \sigma$ around $B$ hadron mass, where the mass resolution $\sigma$ was determined from recontruction of the $B_d^0$ and $B_s^0$ masses for the two signal channels. The mass resolutions were obtained from fitting a single Gaussian to the Monte Carlo signal. Table 2 shows the number of events that can be expected using the above procedure for an integrated

Table 2: Signal and background statistics of $B_s^0$ and $B_d^0$ candidates expected with 10 pb$^{-1}$.

| | Selected candidates expected with 10 pb$^{-1}$ |
|---|---|
| Signal $B_d^0 \to J/\psi K^{0*}$ | 1024 |
| $pp \to J/\psi X$ background | 1419 |
| $b\bar{b} \to J/\psi X$ background | 3970 |
| Signal $B_s^0 \to J/\psi\phi$ | 76 |
| $pp \to J/\psi X$ background | 2449 |
| $b\bar{b} \to J/\psi X$ background | 1660 |
| All events satifying $B_d^0$ or $B_s^0$ selections | 10323 |

luminosity of 10 pb$^{-1}$.

By the time the LHC reaches a luminosity of $10^{33}$ cm$^{-2}$ s$^{-1}$ and the detector is better understood, it will be safe to apply displaced secondary vertex cuts, which will remove most of the backgrounds. In the studies of exclusive channels of $B$ decays, vertex displacement selections are replaced by cuts on the $B$ hadron decay time. This method avoids any bias on the proper decay time measurements. Table 3 shows the reconstruction efficiences with and without decay time cuts. In particular, by requiring that the proper decay time of the $B_s^0$ candidate is greater than 0.5 ps, additional rejection by a factor of 260 for the $pp \to J/\psi X$ can be achieved while losing 25% of the signal.

Table 3: $B_d^0 \to J/\psi K^{0*}$ ($B_s^0 \to J/\psi\phi$) signal and background reconstruction efficiencies before and after the cut on $B_d^0$ ($B_s^0$) decay time $t$. The applied cut was $t > 0.5$ ps.

| | efficiency [%] | |
|---|---|---|
| | before time cut | after time cut |
| Signal $B_d^0 \to J/\psi K^{0*}$ | 42.0 | 30.4 |
| $pp \to J/\psi X$ background | 0.67 | 0.0064 |
| $b\bar{b} \to J/\psi X$ background | 3.05 | 1.52 |
| Signal $B_s^0 \to J/\psi\phi$ | 40.5 | 30.0 |
| $pp \to J/\psi X$ background | 1.5 | 0.0058 |
| $b\bar{b} \to J/\psi X$ background | 1.1 | 0.8 |

## 4 Simultaneous fit of mass and lifetime of $B_d^0$ and $B_s^0$ with early data

We now turn to methods for extracting physically interesting parameters from the decays of the $B_s^0$ and $B_d^0$ mesons. The first measurements with early data will comprise the mean lifetimes and masses of these mesons.

We perform a simultaneous maximum likelihood fit for each $B_s^0$ and $B_d^0$ mass and proper decay time distributions. The likelihood function $L$ is defined by:

$$L = \prod_{i=1}^{N} \left[ \frac{n_{sig}}{N} \times p_{sig}(t_i, m_i) + \frac{n_{bck1}}{N} \times p_{bkg1}(t_i, m_i) + \frac{N - n_{sig} - n_{bck1}}{N} \times p_{bkg2}(t_i, m_i) \right] \tag{1}$$

where the index $i$ runs over the events, $N = n_{sig} + n_{bck1} + n_{bck2}$ is the total number of reconstructed events in the fit and $n_{sig}$, $n_{bck1}$ and $n_{bck2}$ are the numbers of signal and background events. The terms $p_{sig}$, $p_{bkg1}$ and $p_{bkg2}$ are products of two probability density functions that model the mass $m$ and proper decay time $t$ of the signal and the prompt and non-prompt backgrounds respectively (see Section 2). The number of expected events for the prompt background is $n_{bck1}$ and the corresponding probability density function in formula 1 is $p_{bkg1}$. The probability density function for the non-prompt background is $p_{bkg2}$.

For the signal, the mass distribution is modeled by a Gaussian distribution, whose mean value is the $B$ hadron mass m($B$) and its width $\sigma_m$ is given by the detector mass resolution. Both m($B$) and $\sigma_m$ are determined from the fit. The reconstructed proper decay time distribution for the signal is parameterised by the function:

$$p_{sig}(t_i) = \frac{\int_0^\infty \mathrm{e}^{-\Gamma t}\rho(t-t_i)\,\mathrm{d}t}{\int_{-\infty}^{\infty}(\int_0^\infty \mathrm{e}^{-\Gamma t}\rho(t-t')\,\mathrm{d}t)\,\mathrm{d}t'} \tag{2}$$

where the decay time resolution function $\rho(t-t_i)$ was approximated by a Gaussian of width $\sigma$ which is a free parameter of the fit.

For the background, the mass distribution of the prompt component is assumed to follow a flat distribution as observed in simulated data (see Figure 2). The non-prompt component is modeled with a second order polynomial function where the coefficient of the linear (quadratic) terms, denoted as $c_1$ ($c_2$) in Table 4, are determined from the fit.

The decay time distribution of the prompt background component is parametrised by a Gaussian of width $\sigma$. The non-prompt component was modeled by the sum of two exponential functions, convoluted with the decay time resolution function $\rho$. The two exponential functions are denoted as $\Gamma_1$ and $\Gamma_2$, the constant coefficient between them is $b_1$.

$$p_{bck2}(t_i) = \frac{\int_0^\infty \left(\Gamma_1 \mathrm{e}^{-\Gamma_1 t} + \mathrm{b}_1 \times \Gamma_2 \mathrm{e}^{-\Gamma_2 t}\right)\rho(t-t_i)\,\mathrm{d}t}{\int_{-\infty}^{\infty}(\int_0^\infty (\Gamma_1 \mathrm{e}^{-\Gamma_1 t} + \mathrm{b}_1 \times \Gamma_2 \mathrm{e}^{-\Gamma_2 t})\,\rho(t-t')\,\mathrm{d}t)\,\mathrm{d}t'} \tag{3}$$

### 4.1 $B_d^0 \to J/\psi K^{0*}$ decay

The likelihood function, $-2lnL$ is minimised to extract the $B_d^0$ lifetime $\tau = 1/\Gamma$ and mass m($B$) from the reconstructed events containing a $B_d^0 \to J/\psi K^{0*}$ candidates and backgrounds. This fit corresponds to an integrated luminosity of 10 pb$^{-1}$. The distributions of the reconstructed masses and lifetimes are shown in Figure 2. Table 4 summarises the results of the likelihood fit. The values obtained from the fit agree with the input values used in the simulation (given in the first column) within the statistical errors of the fit. The average lifetime of the $B_d^0$ can be measured with an uncertainty of 10% for 10 $pb^{-1}$.

### 4.2 $B_s^0 \to J/\psi\phi$ decay

The $B_s^0\ \overline{B_s^0}$ system exhibits two mass eigenstates with two lifetimes; the lifetime difference $\Delta\Gamma_s/\Gamma_s$ is expected to be $\mathcal{O}(10^{-1})$. However, with early data (a few hundred pb$^{-1}$), the statistics are insufficient to determine both lifetimes. For the initial period of LHC running, it is assumed that $\Delta\Gamma_s = 0$. The method for the $B_s^0$ fit is the same as for the $B_d^0$ case, the main difference being the smaller fraction of signal events, as shown in the mass and lifetime distributions for the reconstructed events after cuts selecting the $B_s^0$ signal (Figure 3).

Statistics of reconstructed events corresponding to an integrated luminosity of 150 pb$^{-1}$ enables measurements to be made with relative precisions on the $B_s^0$ lifetime of 10% (Table 5). In the fit the background events are weighted by factor of 15, since Monte Carlo statistics were limited to the equivalent of 10 pb$^{-1}$ for the current study.

Table 4: Results of the fit to reconstructed $B_d^0$ candidates corresponding to 10 pb$^{-1}$. The first column shows input values used in simulation.

| Parameter | Simulated value | Fit result with statitical error |
|---|---|---|
| $\Gamma$, ps$^{-1}$ | 0.651 | 0.73 $\pm$ 0.07 |
| m($B$), GeV | 5.279 | 5.284 $\pm$ 0.006 |
| $\sigma$, ps | | 0.132 $\pm$ 0.004 |
| $\sigma_m$, GeV | | 0.054 $\pm$ 0.006 |
| $n_{sig}/N$ | 0.16 | 0.155 $\pm$ 0.015 |
| $n_{bck1}/N$ | 0.062 | 0.595 $\pm$ 0.017 |
| $b_1$ | | 1.08 $\pm$ 0.27 |
| $\Gamma_1, ps^{-1}$ | | 0.67 $\pm$ 0.05 |
| $\Gamma_2, ps^{-1}$ | | 2.4 $\pm$ 0.3 |
| $c_1$ | | -2.75 $\pm$ 0.28 |
| $c_2$ | | 4.7 $\pm$ 1.4 |

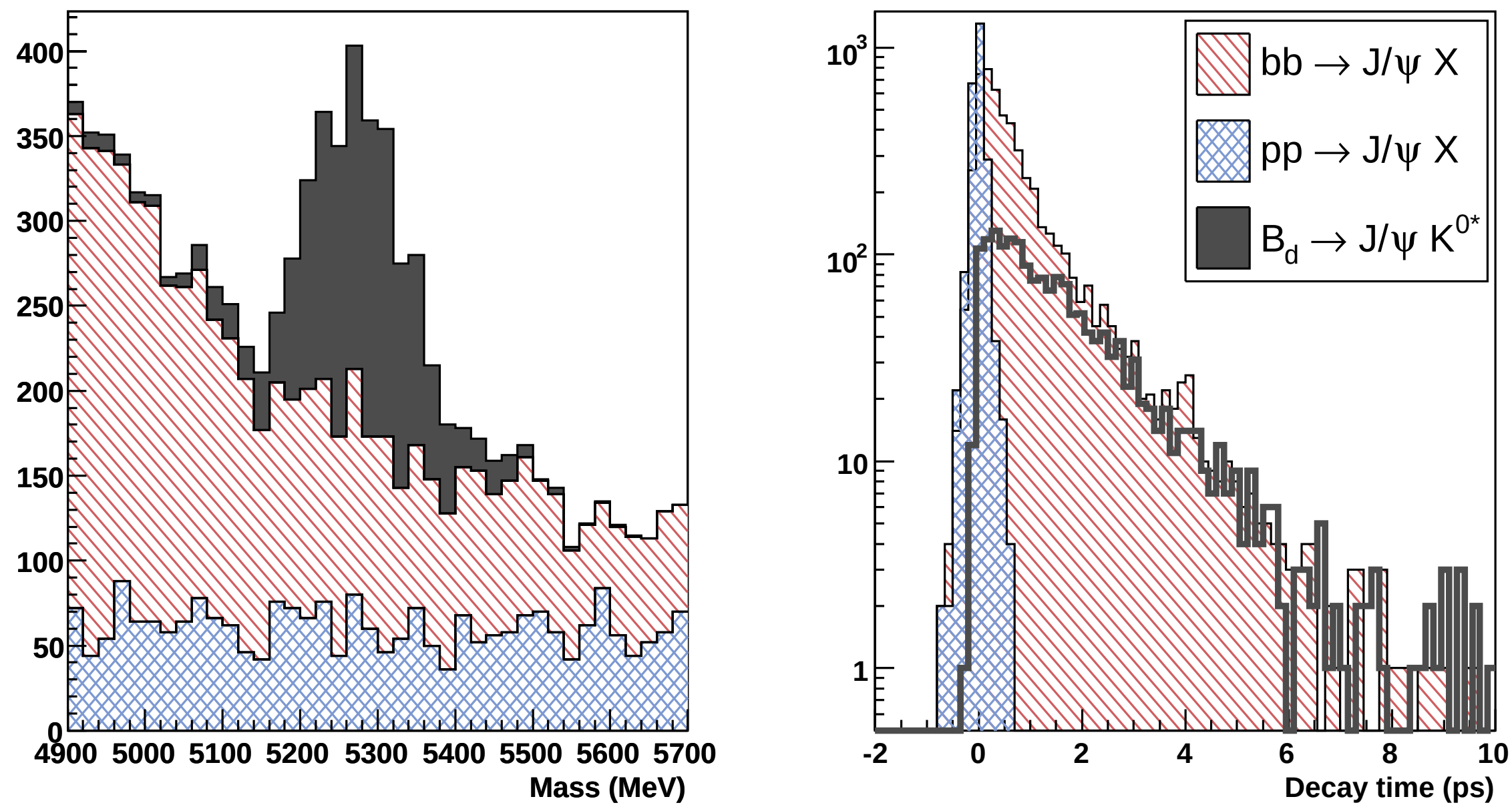


Figure 2: Distributions of the reconstructed $B_d^0$ mass and decay time expected with integrated luminosity of 10 pb$^{-1}$.

## 5 The performance of the jet charge tagger with early data

Most studies of CP-violation and mixing require the identification of the flavour of the neutral $B$ mesons; this is known as *flavour tagging*. Understanding the potential for flavour tagging methods will be one of the important goals with early data. In studies of CP-violation and mixing of neutral $B$ mesons, one must know the flavour of a $B$ meson both at the time of production ($t = 0$) and at the time of decay.

In a small number of cases, the flavour at production can be inferred from the charge of the highest

Table 5: Results from the fit to reconstructed $B_s^0$ candidates corresponding to 150 pb$^{-1}$.

| | Input | Fit result with statistical error |
|---|---|---|
| $\Gamma_s$, ps$^{-1}$ | 0.683 | 0.743 $\pm$ 0.051 |
| m($B$), GeV | 5.343 | 5.359 $\pm$ 0.006 |
| $\sigma$, ps | | 0.152 $\pm$ 0.001 |
| $\sigma_m$, GeV | | 0.061 $\pm$ 0.006 |
| $n_{sig}/N$ | 0.018 | 0.031 $\pm$ 0.005 |
| $n_{bck1}/N$ | 0.397 | 0.379 $\pm$ 0.006 |
| $b_1$ | | 0.023 $\pm$ 0.01 |
| $\Gamma_1, ps^{-1}$ | | 1.35 $\pm$ 0.02 |
| $\Gamma_2, ps^{-1}$ | | 0.44 $\pm$ 0.08 |
| $c_1$ | | -1.44 $\pm$ 0.07 |
| $c_2$ | | 2.14 $\pm$ 0.49 |

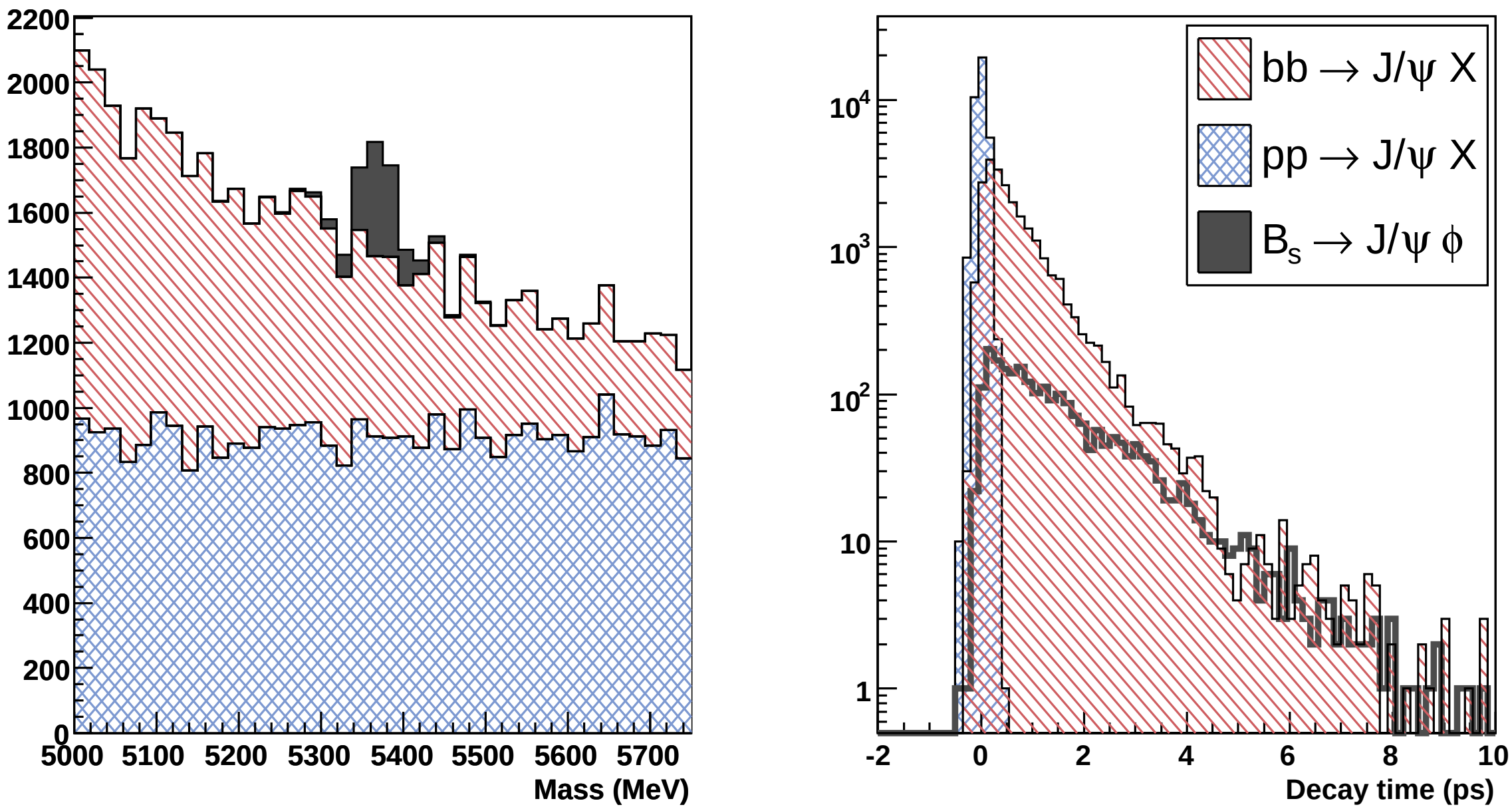


Figure 3: Plots to show the distributions of the reconstructed $B_s^0$ mass and decay time expected with 150 pb$^{-1}$. Background distributions constructed from simulated events corresponding to 10 pb$^{-1}$ were scaled by a factor of 15.

$p_T$ lepton unassociated with the signal decay, with the assumption that this tagging lepton originates from a semi-leptonic decay of the other $B$ hadron in the event. For the majority of the events, one must use the jet charge tagging method. According to fragmentation models, the particles are ordered in the momentum component parrallel to the original quark direction, while charge conservation also imposes charge ordering [7]. These two facts may be used to form a jet charge, which is related to the $b$-quark charge at production. The jet used in this method consists of all tracks that are unassociated with the

signal decay with $p_T > 500$ MeV, $|\eta| < 2.5$, inside a cone of opening angle $\Delta R$ around the $B$ meson in the laboratory frame. The opening angle of the jet cone, $\Delta R$, is defined:

$$\Delta R = \sqrt{\Delta\eta^2 + \Delta\varphi^2} \tag{4}$$

where $\Delta\eta$ and $\Delta\varphi$ are the differences in pseudorapidity and azimuthal angle between the cone wall and the $B$ meson. The jet charge, $Q_{\text{jet}}$, tends to be positive for $\bar{b}$-jets and negative for $b$-jets, thus allowing the $B^0$ meson flavour at production to be inferred. The jet charge is defined as:

$$Q_{\text{jet}} = \frac{\sum_i q_i p_i^\kappa}{\sum_i |p_i|^\kappa} \tag{5}$$

where the $q_i$ is the charge of the $i^{th}$ track in the jet and $p_i$ is a measure of the tracks momentum that can be, for example, the transverse momentum of the track or a projection of the track's momentum along the axis of the $B$ meson's direction. These are referred to as the $p_T$ method and the $p_L$ method respectively. The parameter $\kappa$ controls the relative contribution of the hard and soft tracks in the jet charge. One possible improvement in the algorithm is to remove ambiguous cases such as events with $Q_{\text{jet}}$ close to zero; the smallest allowed value of $|Q_{jet}|$ is called the "exclusion cut". The opening angle of the jet cone, the exclusion cut and $\kappa$ are free parameters and must be tuned to get the best performance from the tagger.

## 5.1 Quantifying the performance of a flavour tagger

The effectiveness of the discrimination between $B^0$ and $\overline{B^0}$ mesons at production time is characterized by two quantities: its *efficiency*, $\varepsilon_{\text{tag}}$, and the *dilution*, $D_{\text{tag}}$. The efficiency is the fraction of $B$ mesons that were tagged either correctly or incorrectly and is described by:

$$\varepsilon_{\text{tag}} = \frac{N_r + N_w}{N_t} \tag{6}$$

where $N_r$ and $N_w$ are the numbers of correctly and incorrectly tagged $B$ mesons respectively, and $N_t$ is the total number of reconstructed $B$ mesons. The *dilution*, also known as the *purity*, is given by:

$$D_{\text{tag}} = \frac{N_r - N_w}{N_r + N_w} = 1 - 2w_{\text{tag}} \tag{7}$$

where $w_{\text{tag}}$ is the wrong tag fraction:

$$w_{\text{tag}} = \frac{N_w}{N_r + N_w} \tag{8}$$

In a typical CP violation study, where the aim is to identify a difference in some property between a particle and its anti-particle, the relationship between the true asymmetry of this property, $A_{true}$, and the asymmetry as measured in the data, $A_{meas}$, will be

$$A_{true} = \frac{1}{D_{\text{tag}}} A_{meas} \tag{9}$$

which is derived in, for instance, [8]. For the small asymmetries expected in the $B$ decays, the statistical uncertainty on $A_{true}$ is, to a good approximation:

$$\sigma_A \approx \frac{1}{\sqrt{\varepsilon_{\text{tag}} D_{\text{tag}}^2 N_t}} \tag{10}$$

The tag algorithm effectiveness is indicated by the *quality factor* or *tagging power*, $Q_{\text{tag}}$:

$$Q_{\text{tag}} = \varepsilon_{\text{tag}} D_{\text{tag}}^2 \tag{11}$$

The quality factor is used as a measure of success when optimising the flavour tagger.

### 5.2 Understanding the jet charge tagger using $B_d^0 \to J/\psi K^{0*}$ decays

During the early data taking phase, there will be too few $B_s^0 \to J/\psi\phi$ decays reconstructed to allow a detailed comparison between the jet charge distribution obtained from the data and that predicted by the Monte Carlo. However, there will be a sufficient number of the analogous $B_d^0 \to J/\psi K^{0*}$ decays to allow such a comparison to be made. Additionally, the final state of the $B_d^0 \to J/\psi K^{0*}$, with a subsequent decay of $K^{0*}$ to charged mesons, allows the initial flavour to be determined in a statistical way, and therefore the decay mode is considered as self-calibrating. The jet charge tagger thus produced will be important for the CP violation studies with $B_d^0 \to J/\psi K_S$, and the study of $B_d^0 \to J/\psi K^{0*}$ will allow us to gain confidence in the tagging performance for $B_s^0 \to J/\psi\phi$.

For this study, the signal decays were reconstructed as described in the Section 3 for both $B_d^0 \to J/\psi K^{0*}$ and $B_s^0 \to J/\psi\phi$ decays. A reconstructed sample of 15000 decays was used for each channel, corresponding to 150 pb$^{-1}$for $B_d^0 \to J/\psi K^{0*}$ and 1.5 fb$^{-1}$for $B_s^0 \to J/\psi\phi$; this defines our working point for the two channels in this study. The quality factor was then maximised by systematically varying the jet charge tagger input parameters $\Delta R$, $\kappa$ and exclusion cut. It was found that optimal results for both $B_s^0$ and $B_d^0$ mesons were obtained using the projection of the track momentum in the direction of the $B$ meson (the $p_L$ method) as the measure of momentum in Equation 5. The other optimal parameters are shown in Table 6. Using these optimised parameters, the jet charge distribution for both $B_d^0 \to J/\psi K^{0*}$ and $B_s^0 \to J/\psi\phi$ are shown in Figure 4.

Table 6: The optimised parameters of the flavour tagging algorithm for both $B_d^0 \to J/\psi K^{0*}$ and $B_s^0 \to J/\psi\phi$.

| Parameter | $B_d^0 \to J/\psi K^{0*}$ | $B_s^0 \to J/\psi\phi$ |
|---|---|---|
| $\kappa$ | 0.9 | 0.8 |
| $\Delta R$ cut | 0.7 | 0.6 |
| Exclusion cut | 0.05 | 0.2 |

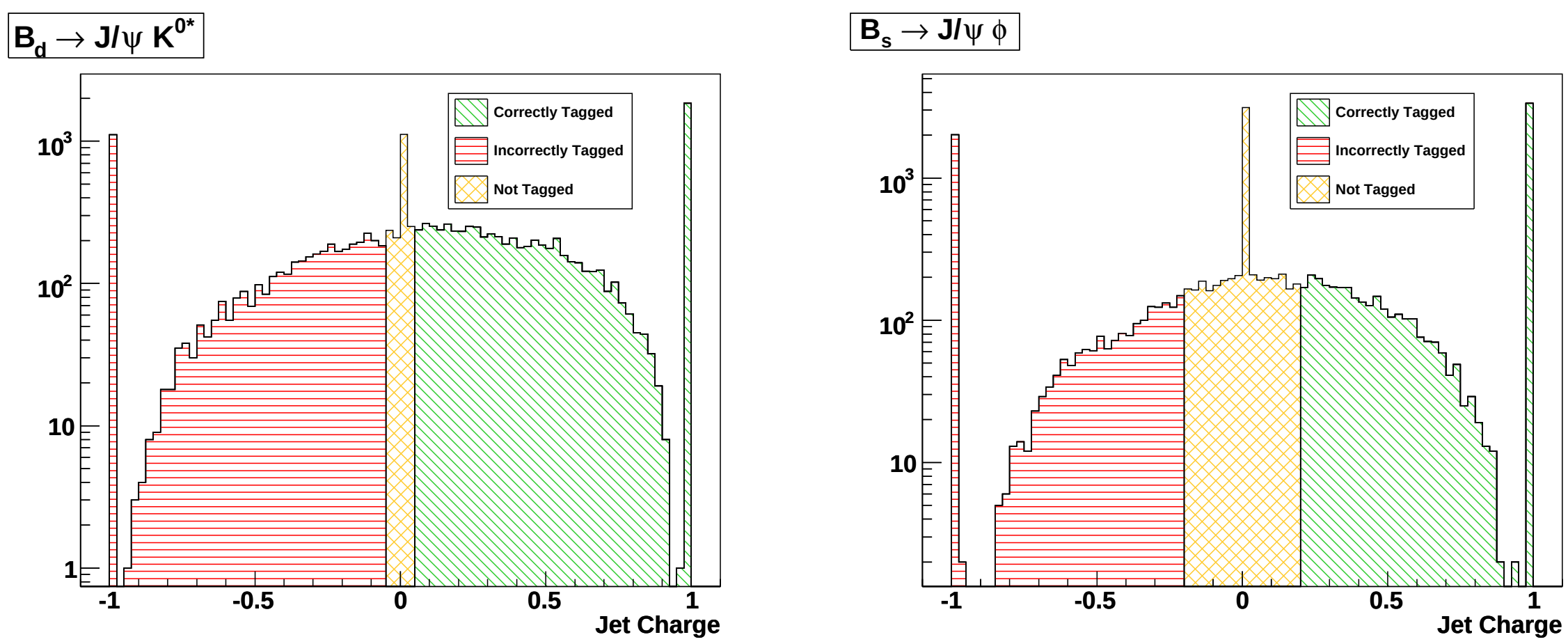


Figure 4: Plots of $Q_{\text{jet}}$ for $B_d^0 \to J/\psi K^{0*}$ (left) and $B_s^0 \to J/\psi\phi$ (right) using their optimised parameters of Table 6 and the equivalent luminosity of Table 7.

One might expect that the different flavour content in the formation of the $B$ mesons will result in a different jet charge behavior for $B_d^0 \to J/\psi K^{0*}$ and $B_s^0 \to J/\psi\phi$. This is indeed what is observed, both

in the optimisation of these jet charges and their distributions. However, one should also note that both the shapes and the optimised parameters, except the exclusion cut, are similar.

The numbers in Table 7 characterise the expected performance of the jet charge tagger. With an integrated luminosty of 150 $\mathrm{pb}^{-1}$, it will be possible to calibrate the jet charge tagger for the $B_d^0$, from the data, with an efficiency of $87.0 \pm 0.3\%$ and a wrong tag fraction of $38.0 \pm 0.4\%$. Calibrating with real data for the $B_s^0$ is more challenging as there is no readily availible and clean self-tagging mode. In this case, the Monte Carlo dependent calibration will be used, but the agreement of the Monte Carlo with real data will be tested indirectly though the $B_d^0 \to J/\psi K^{0*}$ channel.

Table 7: Performance of the flavour tagging algorithm for the optimised values given in Table 6. The errors given in the table are statistical.

| Parameter | $B_d^0 \to J/\psi K^{0*}$ | $B_s^0 \to J/\psi\phi$ |
|---|---|---|
| Equivalent luminosity | 150 $\mathrm{pb}^{-1}$ | 1.5 $\mathrm{fb}^{-1}$ |
| Number of Reconstructed Events | 13948 | 15784 |
| Efficiency, $\varepsilon_{\mathrm{tag}}$ | $0.870 \pm 0.003$ | $0.625 \pm 0.005$ |
| Wrong Tag Fraction, $w_{\mathrm{tag}}$ | $0.380 \pm 0.004$ | $0.374 \pm 0.005$ |
| Dilution, $D_{\mathrm{tag}}$ | $0.240 \pm 0.009$ | $0.251 \pm 0.010$ |
| Quality, $Q_{\mathrm{tag}}$ | $0.050 \pm 0.004$ | $0.039 \pm 0.003$ |

## 6 Summary and conclusion

With the early data, the decays $B_d^0 \to J/\psi K^{0*}$ and $B_s^0 \to J/\psi\phi$ can be used to measure $B$ hadron masses and lifetimes with sufficient precision to permit sensitive tests of the detector performance. In particular, the $B_d^0$ lifetime can be determined with a relative statistical error of 10%, with an integrated luminosity of 10 $\mathrm{pb}^{-1}$, and the same precision will be achieved for the $B_s^0$ lifetime with 150 $\mathrm{pb}^{-1}$. The proposed method of a simultaneous fit of background and signal events allows a sensitive determination of the masses and decay times of $B$ mesons. With early data, the optimal overall precision will be obtained with no cuts on the secondary vertex displacement. This is appropriate for the early data when the performance of the detector and reconstruction algorithms may not be well understood. This strategy is consistent with that of the early $B$-physics triggers, where no displacement cuts on the $J/\psi$ will be applied at the trigger level.

In the early data taking phase, the self-tagging decay $B_d^0 \to J/\psi K^{0*}$ will be used to calibrate the jet charge tag for jets containing a $B_d^0$. This will be of use for physics studies involving $B_d^0$ decays, but also this good understanding for the tagging performance for $B_d^0 \to J/\psi K^{0*}$ will allow the fragmentation modelling for $B_s^0 \to J/\psi\phi$ decays to be improved.

# Plans for the Study of the Spin Properties of the $\Lambda_b$ Baryon Using the Decay Channel $\Lambda_b \to J/\psi(\mu^+\mu^-)\Lambda(p\pi^-)$


**Abstract**

This note summarizes the results of a study of the feasibility of measuring certain spin properties of $\Lambda_b$ baryon in the ATLAS experiment. We present an assessment of approaches for extracting the inclusive $\Lambda_b$ polarization and the parity violating $\alpha_{\Lambda_b}$ parameter for the decay $\Lambda_b \to J/\psi(\mu^+\mu^-)\Lambda(p\pi^-)$ from the reconstructed four final state charged particles. As a key test, we generated Monte Carlo samples of $\Lambda_b$ events of fixed polarization in the ATLAS detector and evaluated our ability to precisely extract the input polarization from the reconstructed events. The physics motivation for the planned measurements in ATLAS include the search for an explanation of the anomalous spin effects in hyperon inclusive production observed at lower energies, tests of various decay models based on HQET, tests of CP in an area not yet directly explored, and the development of $\Lambda_b$ polarimetry as a possible tool for spin analysis in future SUSY and other studies.


## 1 Introduction

We report here plans for the measurement of spin parameters of the $\Lambda_b$ hyperon. We utilize the decay mode $\Lambda_b \to J/\psi(\mu^+\mu^-)\Lambda(p\pi^-)$ to extract the $\Lambda_b$ signal from what is expected to be a low background environment, given that the final state has four charged particles and a displaced secondary vertex. The polarization and parity violating $\alpha_{\Lambda_b}$ parameter will be determined from the relevant angular correlations between the final state particles. We expect to accumulate approximately 13000 $\Lambda_b$ events (and a similar number of $\overline{\Lambda_b}$) with an integrated luminosity of 30 fb$^{-1}$ . This estimation is based on the latest reconstruction software and trigger simulation for the ATLAS experiment.

The $\Lambda_b$ is the lightest baryon containing a $b$ quark, and since its discovery in 1991 by the UA1 Collaboration [1] it has created a great deal of interest. Besides the so−called $\Lambda_b$ lifetime puzzle [2], the $\Lambda_b$ has been the subject of various theoretical studies ranging from proposed tests of CP violation [3], T violation tests and new physics studies [4], measurement of top quark spin correlation functions [5] and the extraction of the weak phase $\gamma$ of the CKM matrix [6]. Specific physics interest in the $\Lambda_b$ parity violating $\alpha_{\Lambda_b}$ parameter studies derives from its ability to serve as a test for various heavy quark factorization models and perturbative QCD (PQCD). $\Lambda_b$ studies are also of interest because of the continuing mystery of why hyperons have consistently displayed large polarizations when produced at energies even up to several hundred GeV and at large $p_T$ where most models predict zero polarization. It is not known if these effects can be explained by some not yet understood effect of existing physics or if they point to new physics altogether. $\Lambda_b$ polarization holds the possibility of illuminating just how polarized $b$ quarks are produced and, indeed, it may have relevance to how fermions are produced in all $pp$ induced processes.

Interest in the studies of the $\Lambda_b$ lifetime parameter derives from the current controversy from Tevatron experiments concerning the question of how much longer the $b$ quark lives in a meson vs. in a hyperon. With an expected increase of a factor of 100 in the statistics at the LHC, we expect to make a definitive statement on this puzzle. Again, this will further constrain the theoretical models which have as their basis PQCD and the Heavy Quark Model. Lifetime measurements will not be examined in this article, since it is not the focus of the current study, though many of the event selection issues, discussed here, might be applicable in the $\Lambda_b$ lifetime studies.

We have examined the primary technical challenges in the measurement of $\Lambda_b$ polarization in ATLAS by generating large samples of $\Lambda_b$ baryons with various known polarizations, allowing them to decay in the detector using model−predicted amplitudes, and then reconstructing these events using standard ATLAS packages. These samples have permitted us to test our ability to reconstruct events and to confirm that we can recover the input polarization and the decay amplitudes. They also have allowed us to compare various polarization extraction methods and to assess the impact of detector corrections and detector resolution effects. We provide here a report on the results of these studies, and on the work we undertook to adapt the EVTGEN [7] decay package to produce polarized $\Lambda_b$ within the ATLAS software framework.

# 2 Theoretical overview

In the quark model the $\Lambda_b$ is a fermion consisting of a $b$ quark accompanied by a di−quark $(ud)$ of total spin zero. In this model the polarization of the $\Lambda_b$ is thus expected to be totally due to the $b$ quark polarization. QCD calculations suggest that the $b$ quark polarization would be small. However, there are models of quark scattering [8], in which spin effects are expected to scale with the mass of the heavy quark, and where the possibility exists for $\Lambda_b$ polarizations to be quite large. We further note that QCD has not been able to predict the very large polarizations that have been observed in the inclusive production of $\Lambda$ hyperons at energies of several hundred GeV. It is hoped that the huge mass difference in the $b$ and $s$ quarks will help elucidate the origin of these unexplained spin effects.

Interest in the $\alpha_{\Lambda_b}$ parameter for the $\Lambda_b$ stems from the fact that HQET models [9] purport to calculate this quantity from rather basic principles of PQCD and factorization. We have an interest in comparing our ultimate measurements of this quantity with these predictions and assessing what constraints they can provide for these models. We provide below a brief overview of the theoretical basis for the polarization and $\alpha_{\Lambda_b}$ measurements.

## 2.1 Heavy quark polarization in QCD

In the Standard Model heavy quark production is dominated by gluon−gluon fusion and $q\bar{q}$ annihilation processes. A non−zero polarization requires an interference between non−flip and spin−flip helicity amplitudes for the $\Lambda_b$ production, with the latter containing an imaginary part. In QCD this complex part can only be generated through loop corrections, so that the relevant diagrams for polarized quarks are $\mathcal{O}(\alpha_s^4)$. The polarization expected from all QCD sub−processes ($g-g$ fusion, $q\bar{q}$ annihilation and $q-q$, $q-g$ scattering) have been calculated [10]. The formulae for the polarization for each one of the four processes is directly proportional to $\alpha_s$, and it depends just on the ratio $x_Q = m_Q/p_Q$ and the scattering angle $\theta_Q$, (all defined in the center of mass frame), and are thus valid for any final−state quark $Q$. The expected polarization in single $b$ quark production by gluon−gluon fusion and $q\bar{q}$ annihilation has been found to be a maximum of 5% for gluon−gluon fusion, and a maximum 10% for $q\bar{q}$ annihilation. When these predictions are compared to the observed $\Lambda$ polarization (due to the s quark polarization) [11], they are found to be an order of magnitude too small. One might not be surprised if the $\Lambda_b$ polarization is, as well, greater than predicted in QCD.

An important result in [10] is the dependence of the polarization on the quark mass. The heaviest quark produced is the most polarized, and the maximum polarization is reached around $x_Q \simeq 0.3$. The $b$ quark polarization is predicted to be an order of magnitude greater than the $s$ quark polarization, which from $\Lambda$ polarization measurements has been found to reach values over 20% at 400 GeV [12].

The measurement of the $\Lambda_b$ polarization in ATLAS in the exclusive channel $\Lambda_b \to J/\psi\Lambda$ proposed here would cover $p_T(\Lambda_b) > 8000$ MeV (because of trigger and reconstruction constraints on the transverse momentum of the final−state particles, see Table 2) and $x_F(\Lambda_b) < 0.1$. It could make a significant

contribution to testing different models of production of polarized baryons in this new kinematic region.

An idea that the heavy quark pre−exists in the incoming proton before scattering and becomes polarized through a direct scattering from an incoming quark provides another pathway for the $\Lambda_b$ to be polarized. This possibility has been discussed by Neal and Burelo [13]. If polarizations are observed in inclusive $\Lambda_b$ production that exceed a few percent, such a mechanism should be given careful attention, since no other existing models can account for such large values.

### 2.2 $\Lambda_b \to J/\psi(\mu^+\mu^-)\Lambda(p\pi^-)$ decay and angular distributions

The proposed study of $\Lambda_b$ polarization would probe not only the production process but also explore the decay of $\Lambda_b$. Decay models predict values for various quantities that can be experimentally observed, thus providing a test of specific HQET/Factorization model [14] assumptions.

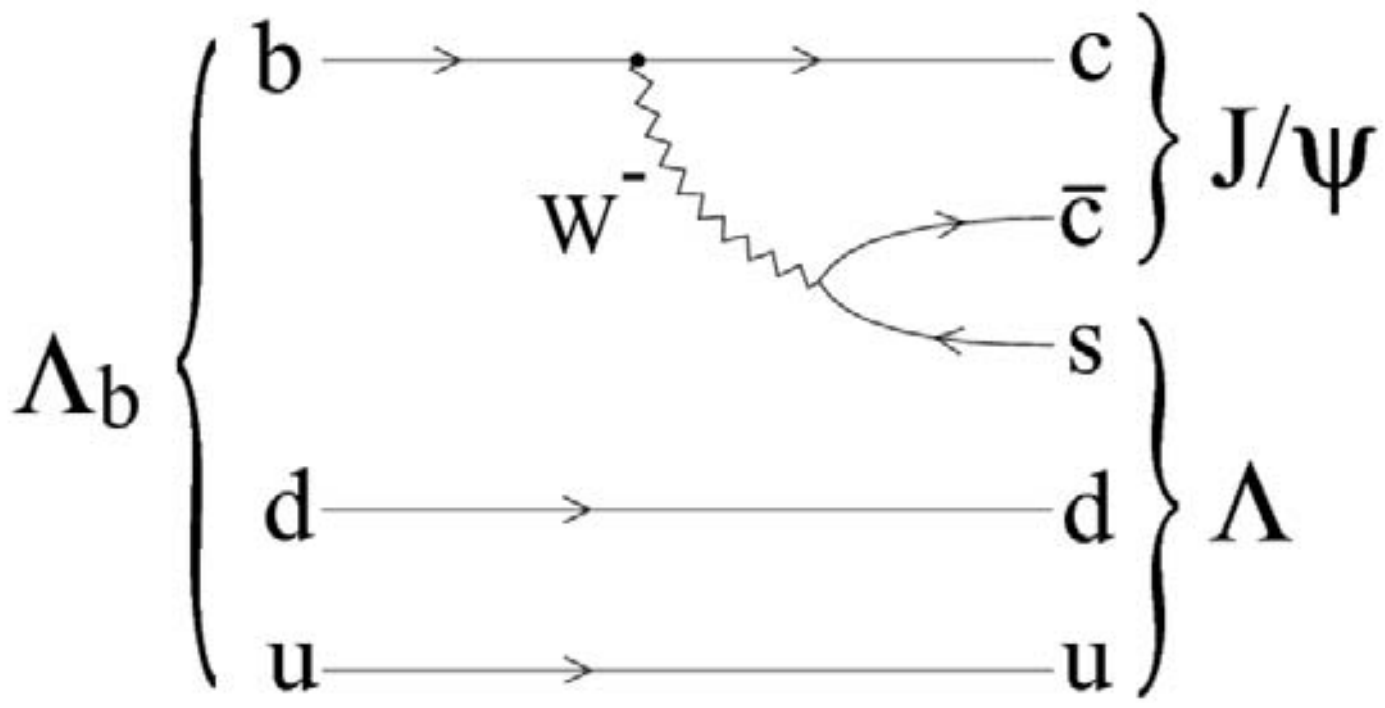


Figure 1: The weak decay of $\Lambda_b$: $\Lambda_b \to J/\psi\Lambda$ .

The fact that $\Lambda_b$ has a significant lifetime suggests that it decays weakly. The dominant decay process would involve the emission of a $W^-$ boson, as illustrated in Figure 1. The spin and parity of the particles involved in the $\Lambda_b \to J/\psi(\mu^+\mu^-)\Lambda(p\pi^-)$ decay are well known. The $\Lambda_b$ with $J^P = \frac{1}{2}^+$ decays to $\Lambda$ with $J^P = \frac{1}{2}^+$ and $J/\psi$ with $J^P = 1^-$. The general amplitudes for the decay of $\Lambda_b(\frac{1}{2}^+) \to \Lambda(\frac{1}{2}^+)J/\psi(1^-)$ is given by:

$$\mathscr{M} = \overline{\Lambda}(p_\Lambda)\, \varepsilon^*_\mu(p_{J/\psi}) \left[ A_1 \gamma^\mu \gamma^5 + A_2 \frac{p^\mu_{\Lambda_b}}{m_{\Lambda_b}} \gamma^5 + B_1 \gamma^\mu + B_2 \frac{p^\mu_{\Lambda_b}}{m_{\Lambda_b}} \right] \Lambda_b(p_{\Lambda_b}), \tag{1}$$

which is parameterized by the four complex decay amplitudes $A_1, A_2, B_1, B_2$ and where $\varepsilon_\mu$ is the polarization vector of the $J/\psi$ .

Given the general amplitude, we may compute the helicity amplitudes. We use helicity amplitudes, because they have a direct physical relationship to the spin parameters we wish to study. Four helicity amplitudes are required to describe the decay completely. We will use the notation $H_{\lambda_\Lambda,\lambda_{J/\psi}}$ for the helicity amplitudes of the decay $\Lambda_b \to J/\psi(\mu^+\mu^-)\Lambda(p\pi^-)$ , where $\lambda_\Lambda = \pm 1/2$ is the helicity of $\Lambda$ and $\lambda_{J/\psi}$ = +1, 0, -1 is the helicity of $J/\psi$ . These four helicity amplitudes: $a_+ = H_{1/2,0}$, $a_- = H_{-1/2,0}$, $b_+ = H_{-1/2,1}$, $b_- = H_{1/2,-1}$ are normalized to unity:

$$|a_+|^2 + |a_-|^2 + |b_+|^2 + |b_-|^2 = 1. \tag{2}$$

In this notation, the $\Lambda_b$ decay asymmetry parameter $\alpha_{\Lambda_b}$ is given by [15]:

$$\alpha_{\Lambda_b} = \frac{|a_+|^2 - |a_-|^2 + |b_+|^2 - |b_-|^2}{|a_+|^2 + |a_-|^2 + |b_+|^2 + |b_-|^2}. \tag{3}$$

The helicity amplitudes $a_+$ , $a_-$ , $b_+$ , $b_-$ are computed directly from the decay amplitudes $A_1$, $A_2$, $B_1$, $B_2$ according to the following equations:

$$\begin{aligned}
a_+ &= \tfrac{1}{m_{J/\psi}}\left\{\sqrt{Q_+}\left[(m_{\Lambda_b}-m_\Lambda)A_1-\tfrac{Q_-}{2m_{\Lambda_b}}A_2\right]+\sqrt{Q_-}\left[(m_{\Lambda_b}+m_\Lambda)B_1+\tfrac{Q_+}{2m_{\Lambda_b}}B_2\right]\right\},\\
a_- &= \tfrac{1}{m_{J/\psi}}\left\{-\sqrt{Q_+}\left[(m_{\Lambda_b}-m_\Lambda)A_1-\tfrac{Q_-}{2m_{\Lambda_b}}A_2\right]+\sqrt{Q_-}\left[(m_{\Lambda_b}+m_\Lambda)B_1+\tfrac{Q_+}{2m_{\Lambda_b}}B_2\right]\right\},\\
b_+ &= \sqrt{2}\left(\sqrt{Q_+}\,A_1\mp\sqrt{Q_-}\,B_1\right),\\
b_- &= -\sqrt{2}\left(\sqrt{Q_+}\,A_1\mp\sqrt{Q_-}\,B_1\right),
\end{aligned} \tag{4}$$

where $Q_\pm = (m_{\Lambda_b} \pm m_\Lambda)^2 - m^2_{J/\psi}$ and $m_{\Lambda_b}$ and $m_\Lambda$ are the $\Lambda_b$ and $\Lambda$ masses respectively [16, 17].

The polarization of the $\Lambda_b$ can be determined from the angular correlations between the $\Lambda_b \to J/\psi\Lambda$ final decay products. The $\Lambda_b$ polarization reveals itself in the asymmetry of the distribution of the angle $\theta$. This angle is defined as the angle between the normal to the beauty baryon production plane and the momentum vector of the $\Lambda$ decay daughter, as seen in the $\Lambda_b$ rest frame. The decay angular distribution can be expressed as:

$$w \sim 1 + \alpha_{\Lambda_b} P \cos(\theta), \tag{5}$$

where $\alpha_{\Lambda_b}$ is the decay asymmetry parameter of $\Lambda_b$ and $P$ is the $\Lambda_b$ polarization [18].

Using the method described in [17], it can be shown that the full decay angular distribution is:

$$w(\vec{\theta},\vec{A},P) = \frac{1}{(4\pi)^3}\sum_{i=0}^{i=19} f_{1i}(\vec{A})\, f_{2i}(P,\alpha_\Lambda)\, F_i(\vec{\theta}) \tag{6}$$

where the $f_{1i}(\vec{A})$ are bilinear combinations of the helicity amplitudes and $\vec{A} = (a_+, a_-, b_+, b_-)$. $f_{2i}$ stands for $P\alpha_\Lambda$, $P$, $\alpha_\Lambda$, or 1, where $\alpha_\Lambda$ is $\Lambda$ decay asymmetry parameter. $F_i$ are orthogonal angular functions defined in Table 1. The $\Lambda_b$ decay asymmetry parameter $\alpha_{\Lambda_b}$ is related to the helicity amplitudes as defined in Equation 3. The five angles $\vec{\theta} = (\theta, \theta_1, \theta_2, \varphi_1, \varphi_2)$ (see Figure 2) in this probability density function (p.d.f.) have the following meanings:

- $\theta$ is the angle between the normal to the production plane and the direction of the $\Lambda$ in the rest frame of the $\Lambda_b$ particle;
- $\theta_1$ and $\phi_1$ are the polar and azimuthal angles that define the direction of the proton in the $\Lambda$ rest frame with respect to the direction of the $\Lambda$ in the $\Lambda_b$ rest frame;
- $\theta_2$ and $\phi_2$, define the direction of $\mu^+$ in the $J/\psi$ rest frame with respect to the direction of the $J/\psi$ in the $\Lambda_b$ rest frame.

There are nine unknown parameters in Equation 6. They are the polarization $P$ and four complex helicity amplitudes: $a_+ = |a_+|e^{i\alpha_+}$, $a_- = |a_-|e^{i\alpha_-}$, $b_+ = |b_+|e^{i\beta_+}$, $b_- = |b_-|e^{i\beta_-}$. Using the normalization condition (see Equation 2) and using the fact that the overall global phase is arbitrary, we can reduce the number of unknown independent parameters to seven.

## 3 Monte Carlo samples

In order to determine if it is feasible to detect polarized $\Lambda_b$'s in the ATLAS experiment and to measure their polarization, Monte Carlo samples of polarized $\Lambda_b$ particles were generated using the standard ATLAS software packages. The generation of polarized $\Lambda_b$ particles and the propagation of their polarization in the decay process required a special treatment, and EVTGEN was adapted for this purpose. The next sections describe how this was implemented in the framework of the ATLAS experiment.

| $i$ | $f_{1i}$ | $f_{2i}$ | $F_i$ |
|---|---|---|---|
| 0 | $a_+a_+^* + a_-a_-^* + b_+b_+^* + b_-b_-^*$ | 1 | 1 |
| 1 | $a_+a_+^* - a_-a_-^* + b_+b_+^* - b_-b_-^*$ | $P$ | $\cos\theta$ |
| 2 | $a_+a_+^* - a_-a_-^* - b_+b_+^* + b_-b_-^*$ | $\alpha_\Lambda$ | $\cos\theta_1$ |
| 3 | $a_+a_+^* + a_-a_-^* - b_+b_+^* - b_-b_-^*$ | $P\alpha_\Lambda$ | $\cos\theta\cos\theta_1$ |
| 4 | $-a_+a_+^* - a_-a_-^* + \frac{1}{2}b_+b_+^* + \frac{1}{2}b_-b_-^*$ | 1 | $1/2\,(3\cos^2\theta_2 - 1)$ |
| 5 | $-a_+a_+^* + a_-a_-^* + \frac{1}{2}b_+b_+^* - \frac{1}{2}b_-b_-^*$ | $P$ | $1/2\,(3\cos^2\theta_2 - 1)\cos\theta$ |
| 6 | $-a_+a_+^* + a_-a_-^* - \frac{1}{2}b_+b_+^* + \frac{1}{2}b_-b_-^*$ | $\alpha_\Lambda$ | $1/2\,(3\cos^2\theta_2 - 1)\cos\theta_1$ |
| 7 | $-a_+a_+^* - a_-a_-^* - \frac{1}{2}b_+b_+^* - \frac{1}{2}b_-b_-^*$ | $P,\alpha_\Lambda$ | $1/2\,(3\cos^2\theta_2 - 1)\cos\theta\cos\theta_1$ |
| 8 | $-3Re(a_+a_-^*)$ | $P,\alpha_\Lambda$ | $\sin\theta\sin\theta_1\sin^2\theta_2\cos\varphi_1$ |
| 9 | $3Im(a_+a_-^*)$ | $P\alpha_\Lambda$ | $\sin\theta\sin\theta_1\sin^2\theta_2\sin\varphi_1$ |
| 10 | $-\frac{3}{2}Re(b_-b_+^*)$ | $P\alpha_\Lambda$ | $\sin\theta\sin\theta_1\sin^2\theta_2\cos(\varphi_1 + 2\varphi_2)$ |
| 11 | $\frac{3}{2}Im(b_-b_+^*)$ | $P\alpha_\Lambda$ | $\sin\theta\sin\theta_1\sin^2\theta_2\sin(\varphi_1 + 2\varphi_2)$ |
| 12 | $-\frac{3}{\sqrt{2}}Re(b_-a_+^* + a_-b_+^*)$ | $P\alpha_\Lambda$ | $\sin\theta\cos\theta_1\sin\theta_2\cos\theta_2\cos\varphi_2$ |
| 13 | $\frac{3}{\sqrt{2}}Im(b_-a_+^* + a_-b_+^*)$ | $P\alpha_\Lambda$ | $\sin\theta\cos\theta_1\sin\theta_2\cos\theta_2\sin\varphi_2$ |
| 14 | $-\frac{3}{\sqrt{2}}Re(b_-a_-^* + a_+b_+^*)$ | $P\alpha_\Lambda$ | $\cos\theta\sin\theta_1\sin\theta_2\cos\theta_2\cos(\varphi_1 + \varphi_2)$ |
| 15 | $\frac{3}{\sqrt{2}}Im(b_-a_-^* + a_+b_+^*)$ | $P\alpha_\Lambda$ | $\cos\theta\sin\theta_1\sin\theta_2\cos\theta_2\sin(\varphi_1 + \varphi_2)$ |
| 16 | $\frac{3}{\sqrt{2}}Re(a_-b_+^* - b_-a_+^*)$ | $P$ | $\sin\theta\sin\theta_2\cos\theta_2\cos\varphi_2$ |
| 17 | $-\frac{3}{\sqrt{2}}Im(a_-b_+^* - b_-a_+^*)$ | $P$ | $\sin\theta\sin\theta_2\cos\theta_2\sin\varphi_2$ |
| 18 | $\frac{3}{\sqrt{2}}Re(b_-a_-^* - a_+b_+^*)$ | $\alpha_\Lambda$ | $\sin\theta_1\sin\theta_2\cos\theta_2\cos(\varphi_1 + \varphi_2)$ |
| 19 | $-\frac{3}{\sqrt{2}}Im(b_-a_-^* - a_+b_+^*)$ | $\alpha_\Lambda$ | $\sin\theta_1\sin\theta_2\cos\theta_2\sin(\varphi_1 + \varphi_2)$ |

Table 1: The coefficients $f_{1i}$, $f_{2i}$ and $F_i$ of the probability density function in Equation 6.

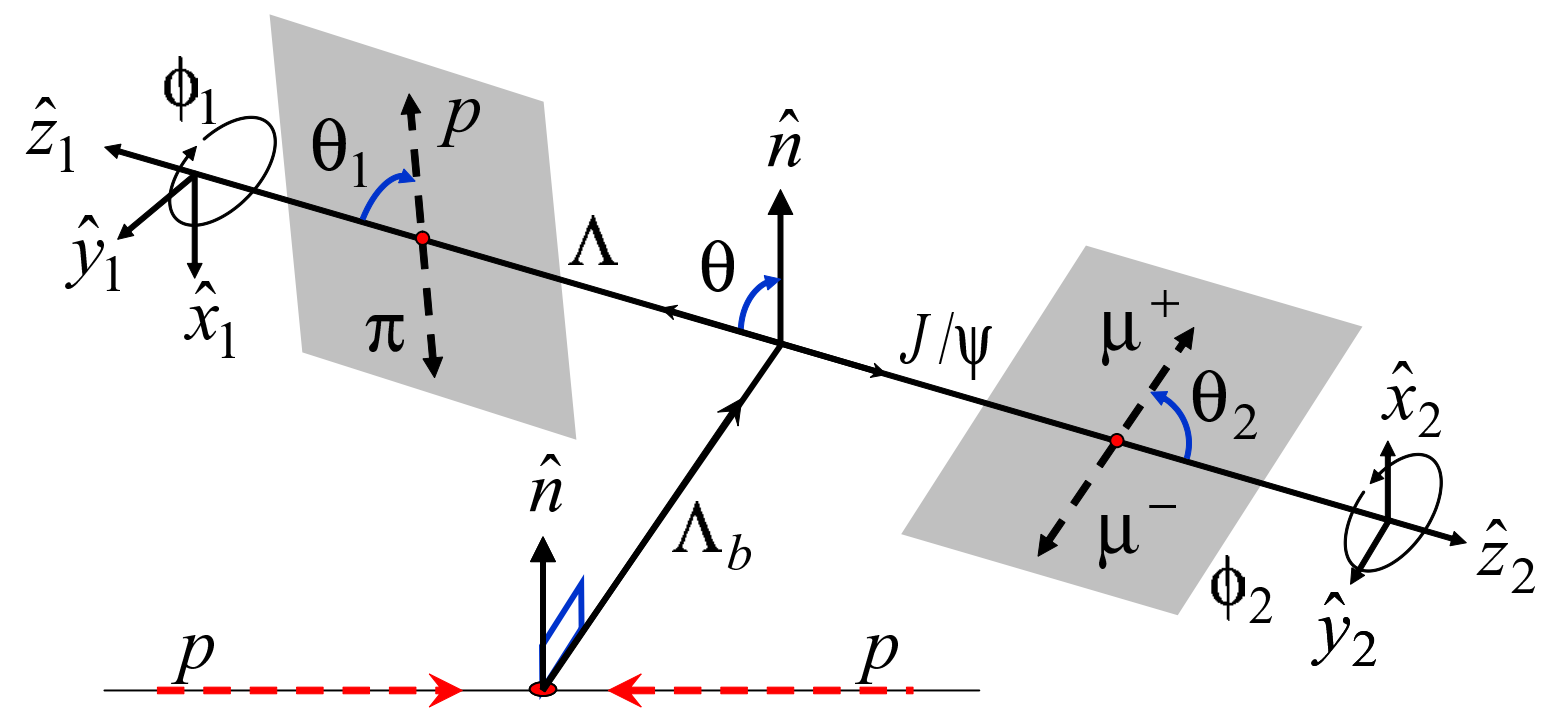


Figure 2: Angles describing the $\Lambda_b \to J/\psi(\mu^+\mu^-)\Lambda(p\pi^-)$ decay.

### 3.1 The generation of polarized $\Lambda_b$ particles

To generate $\Lambda_b$ particles, the PYTHIA 6.4 generator [19] is used. Since PYTHIA does not incorporate polarization information from the decay of $\Lambda_b$ particles, EVTGEN was used to generate the $\Lambda_b$ decay. EVTGEN provides a general framework for implementation of B hadron decays using spinor algebra and decay amplitudes. This framework permits the proper management of spin correlations of very complicated decay processes. EVTGEN is a Monte Carlo generation package itself, but in this case it is used only to decay the $\Lambda_b$ particles produced by PYTHIA.

### 3.1.1 Re−hadronization process and cuts at PYTHIA level

PYTHIA provides mechanisms to produce $b$ quarks, referred to as gluon−gluon fusion, $q$-$\overline{q}$ annihilation, flavor excitation, and gluon splitting. If all these processes are taken into account, beauty quark events would constitute only 1% of the total number of generated events. In addition, the fraction of $b$ quarks hadronizing to $\Lambda_b$ is less than 10%. These make the process of $\Lambda_b$ generation computationally slow. To optimize the generation process, a re−hadronization step of the same event in the $b\bar{b}$ pairs production is used. In order to avoid repetition of $\Lambda_b$ events due to the re-hadronization process, a $\Lambda_b$ pre−selection is implemented at this stage to filter on average only one of the re−hadronized copies of the same event. An additional reason that the $\Lambda_b$ generation process is slow is that around 95% of final state particles (two muons, a proton, and a pion) of the generated $\Lambda_b$ events are outside of the $\eta$ limits ($|\eta| < 2.5$) of the ATLAS detector. In addition, all events must pass the level−1 trigger of the ATLAS trigger system and some pre−reconstruction requirements, such as having a minimum reconstructable transverse momentum. We could not apply these cuts in the PYTHIA step since the kinematics information of the $\Lambda_b$ children is available only at a later stage, when EVTGEN decays the $\Lambda_b$ particles. However, by analyzing the $p_T$ and $\eta$ distributions of $\Lambda_b$ particles before and after cuts (emulating level−1 and level−2 triggers, and requiring $|\eta| < 2.5$) on the final state particles, we estimated $p_T$ and $\eta$ limits, below which the $\Lambda_b$ can not be selected and then applied these cuts in the PYTHIA selection. Figure 3 shows the $p_T$ and $\eta$ distributions from which the $p_T(\Lambda_b) > 6000$ MeV and $|\eta(\Lambda_b)| < 3$ cuts were selected to filter $\Lambda_b$ particles in PYTHIA.

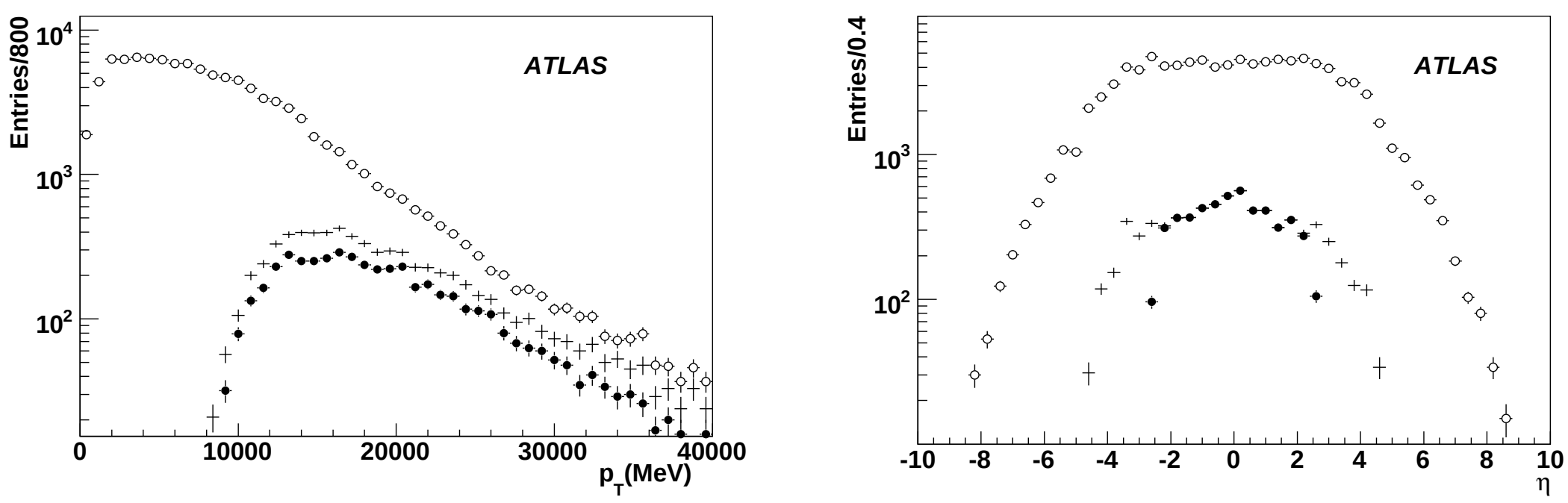


Figure 3: Distributions of $p_T$ (left) and $\eta$ (right) for $\Lambda_b$ particles generated using PYTHIA, without cuts (hollow circle), applying $\eta$ cuts only (cross) and applying all cuts (solid circle) from Table 2.

### 3.1.2 Setting $\Lambda_b$ polarization in EVTGEN

To set the polarization of $\Lambda_b$ particles we used the spin density matrix description of EVTGEN. For the case of spin−1/2 particles like $\Lambda_b$ the density matrix is defined as:

$$\rho = \frac{1}{2}(I + \vec{P} \cdot \vec{\sigma}) \tag{7}$$

where $\vec{P}$ is the polarization vector, and $\vec{\sigma} = (\sigma_1, \sigma_2, \sigma_3)$, where $\sigma_i$ is $i$-th Pauli matrix. In our case $\vec{P}$ is defined as:

$$\vec{P} = P \left( \frac{\hat{z} \times \vec{p}_{lab}(\Lambda_b)}{|\hat{z} \times \vec{p}_{lab}(\Lambda_b)|} \right) \tag{8}$$

where $P$ is the magnitude of the polarization, $\vec{p}_{lab}$ is the momentum of the $\Lambda_b$ in the laboratory frame, and $\hat{z}$ is the $z$ - axis (along the beam direction) in the ATLAS reference system.

To decay polarized $\Lambda_b$ we use the HELAMP model of EVTGEN. This model is capable of simulating a generic two body decay with arbitrary spin configuration, taking as input the helicity amplitudes describing the process. In the case of the decay $\Lambda_b \to J/\psi(\mu^+\mu^-)\Lambda(p\pi^-)$ , as it has been shown in the previous section, there are four complex helicity amplitudes: $a_+$, $a_-$, $b_+$, and $b_-$.

The $\Lambda$ decay into a proton and a pion has been simulated with the same model, using as input parameters the two helicity amplitudes $H_{\lambda_\Lambda,\lambda_p}$ defined in terms of the $\Lambda$ helicity $\lambda_\Lambda$ and the proton $\lambda_p$ helicity as

$$h_- = H_{-\frac{1}{2},-\frac{1}{2}}, \quad h_+ = H_{+\frac{1}{2},+\frac{1}{2}}. \tag{9}$$

The choice of $h_\pm$ is constrained by the experimentally well known $\Lambda \to p\pi^-$ asymmetry parameter [20]

$$\alpha_\Lambda = |h_+|^2 - |h_-|^2 = 0.642 \pm 0.013. \tag{10}$$

Finally, the decay $J/\psi \to \mu^+\mu^-$ has been described with the EVTGEN VLL (Vector into Lepton Lepton) model [7].

#### 3.1.3 Filtering of $\Lambda_b \to J/\psi(\mu^+\mu^-)\Lambda(p\pi^-)$ events

As a last step in the generation process, we apply kinematic cuts on muons, pion and proton to emulate the fiducial acceptance, level−1 trigger, and pre−reconstruction requirements. These cuts are summarized in Table 2.

| Particles | Minimum $p_T$ [MeV ] | Maximum $\lvert\eta\rvert$ |
|---|---|---|
| Protons and $\pi$'s | 500 | 2.7 |
| Most energetic muon | 4000 | 2.7 |
| Other muon | 2500 | 2.7 |

Table 2: Cuts applied at the particle level.

### 3.2 Monte Carlo samples and input model for $\Lambda_b$ decays

As input to the HELAMP class of EVTGEN, the result obtained within the framework of PQCD formalism and the factorization theorem [9] has been used to model the $\Lambda_b$ decay. From the complex amplitudes calculated in this model, $A_1$, $A_2$, $B_1$, $B_2$ in Equation 1, the helicity amplitudes $a_+$ , $a_-$ , $b_+$ , $b_-$ are calculated by using Equation 4. This is summarized in Table 3. In this model, the $\Lambda_b$ decay asymmetry parameter, defined in Equation 3, is $\alpha_{\Lambda_b}$ = -0.457 [1].

| | |
|---|---|
| $A_1 = -18.676 - 185.036i$ | $a_+ = -0.0176 - 0.4229i$ |
| $A_2 = -7.461 - 351.242i$ | $a_- = 0.0867 + 0.2425i$ |
| $B_1 = 15.818 - 162.663i$ | $b_+ = -0.0810 - 0.2837i$ |
| $B_2 = -4.252 + 266.653i$ | $b_- = 0.0296 + 0.8124i$ |

Table 3: PQCD model amplitudes $A_i$ and $B_i$, are given in units of $10^{-10}$ and helicity amplitudes $a_\pm$ and $b_\pm$ are normalized to unity.

By using this decay model as input, two Monte Carlo samples were generated with polarizations of -25% and -75%. These Monte Carlo samples were generated, simulated, and fully reconstructed by using the Athena framework [21].

[1]There is an ERRATA in [9] in the reported value of $\alpha_{\Lambda_b}$ [14].

To show how the angular distributions behave, fast Monte Carlo samples (see section 3.3) were generated using an accepted−rejected method based on the p.d.f. defined in Equation 6. Figure 4 shows the distributions of the five angles for helicity amplitudes from Table 3 and polarizations of 40%, 0%, -40%.

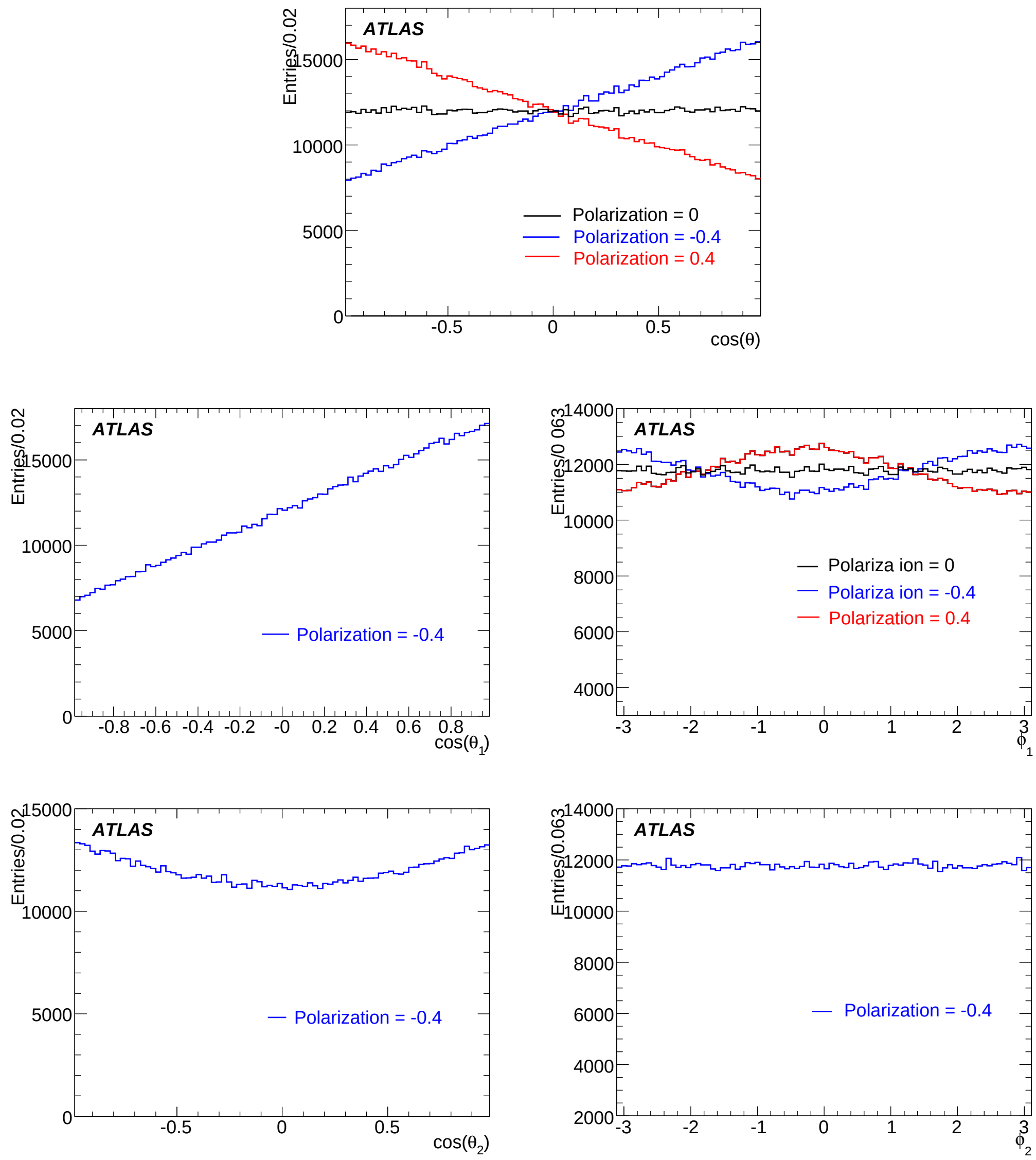


Figure 4: Distributions of the five angles characterizing the decay $\Lambda_b \to J/\psi(\mu^+\mu^-)\Lambda(p\pi^-)$ for different polarization values. For $cos(\theta_1)$, $cos(\theta_2)$, and $\phi_2$, all three distributions for the different polarization values look similar, thus only one polarization case is presented.

### 3.3 Fast Monte Carlo generation

In order to do fast tests of different $\Lambda_b$ decay models and different polarization values, we need to generate large Monte Carlo samples. This represents a problem due to the computer time required to produce a full chain simulated Monte Carlo data. In order to address this problem a fast Monte Carlo generator was

developed. This generator uses Equation 6 to generate angular distributions for the daughters of the $\Lambda_b$ in the $\Lambda_b$ rest frame, and then uses a $(p,\eta)$ distribution derived from phase space of generated events in PYTHIA to compute the kinematic variables of the daughter particles in the laboratory frame. Detector effects are incorporated by using $p_T$ and $\eta$ cuts on final state particles to mimic di-muon triggers and pre−reconstruction requirements. Figure 5 illustrates the strong agreement between angular distributions produced by using PYTHIA and EVTGEN Monte Carlo events and fast Monte Carlo events.

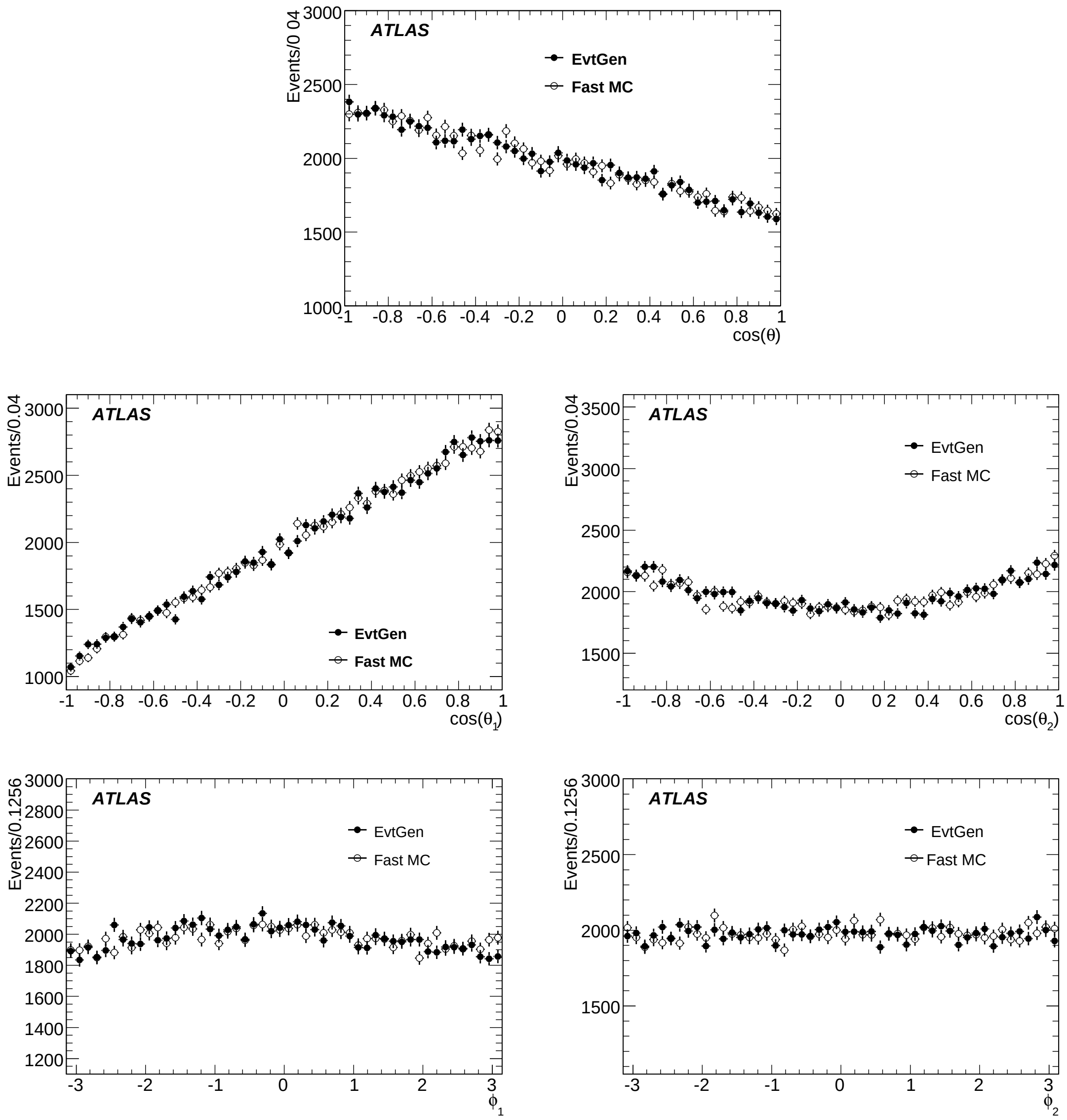


Figure 5: Comparison of Monte Carlo events (PYTHIA + EVTGEN) with fast Monte Carlo generated events. Solid dots represent the Monte Carlo events.

## 4 $\Lambda_b$ reconstruction

The reconstruction of $\Lambda_b$ candidates begins with a search for events with $J/\psi$ candidates. Among these events we search for $\Lambda \to p\pi^-$ candidates, which are then combined with the $J/\psi$ to reconstruct the $\Lambda_b$.

### 4.1 Selection of $J/\psi \to \mu^+\mu^-$ candidates

We search for $J/\psi$ candidates which satisfy the following selection criteria:

- The $\mu^+\mu^-$ candidates must originate at the same reconstructed vertex and the $\chi^2$ of the vertex must be lower than 20;
- The invariant mass of $\mu^+\mu^-$ candidates M($\mu^+\mu^-$) should be within 2800 MeV and 3400 MeV.

The invariant mass distribution of $\mu^+\mu^-$ candidates before applying the invariant mass cuts to select $\Lambda_b$ is shown in Figure 6.

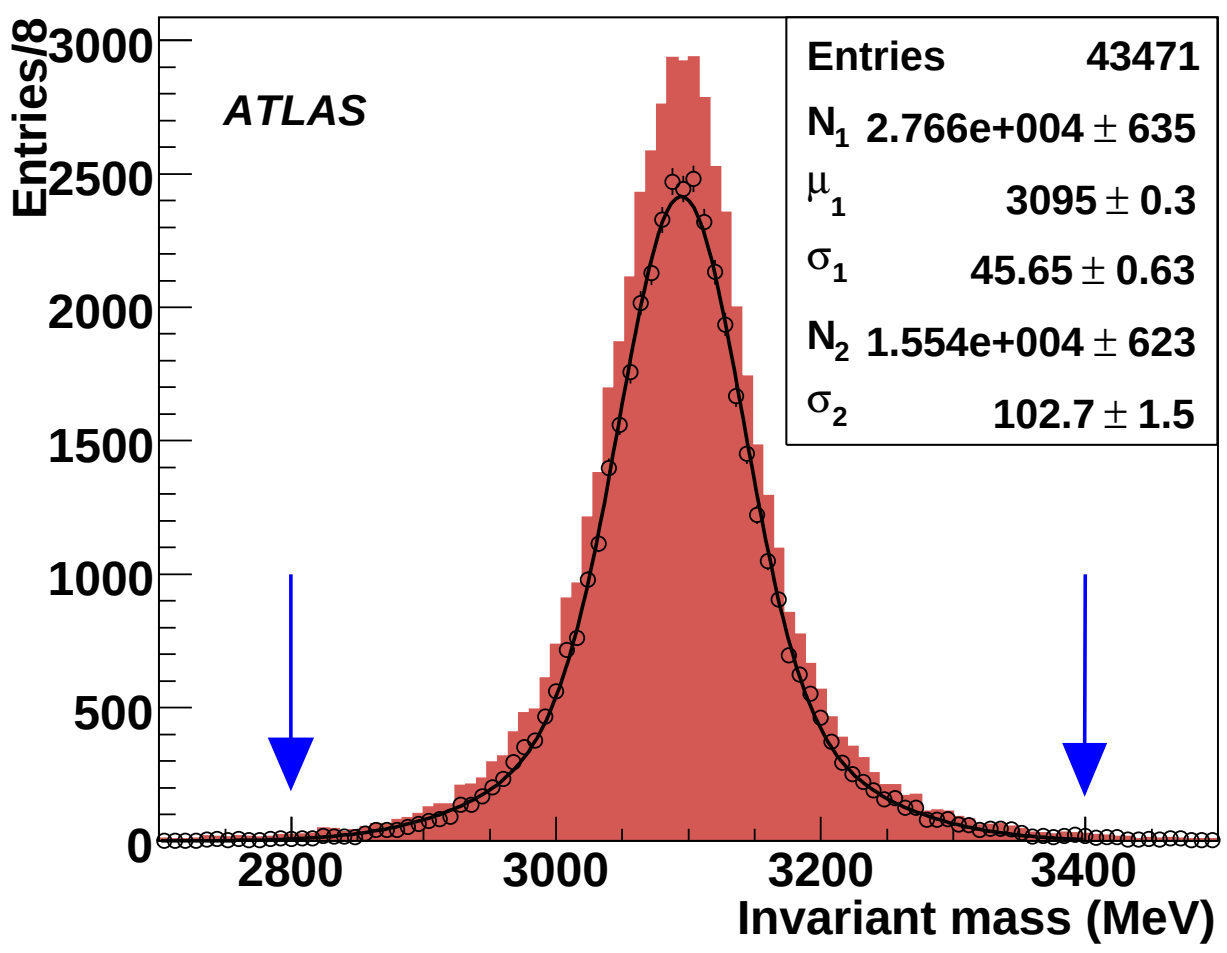


Figure 6: Invariant mass of $\mu^+\mu^-$ candidates. The dark color represents all $J/\psi$ candidates after reconstruction and vertexing requirement. The circles represents $J/\psi$ candidates when level−1 and level−2 trigger signature are required.

### 4.2 Selection of $\Lambda \to p\pi^-$ candidates

From the previously selected events containing a $J/\psi$, $\Lambda$ candidates are selected by applying the following requirements:

- Two opposite charged tracks originating from the same reconstructed vertex.
- The invariant mass of two tracks M($p\pi^-$) should be within 1105 MeV and 1128 MeV range, where for computing M($p\pi^-$), the track with the highest transverse momentum was assumed to be the proton, as observed in 100% of the times in Monte Carlo generations, while the other track was assumed to be a pion.

Many of the $\Lambda$ particles decay outside of the high−precision part of the Inner Detector, which covers a radius of about 40 cm from the beam line, and thus are lost in reconstruction. The decay vertex position of $\Lambda$'s in the $RZ$ plane is presented in Figure 8. If the $\Lambda$ decays outside the 40 cm radius, the number of reconstructed space points (hits in the pixel or silicon layers) is not sufficient for a successful track reconstruction. This effect reduces the fraction of reconstructible $\Lambda$ to around 60%. Figure 7 presents the invariant mass distribution of the $p\pi^-$ candidates before the invariant mass cuts have been applied.

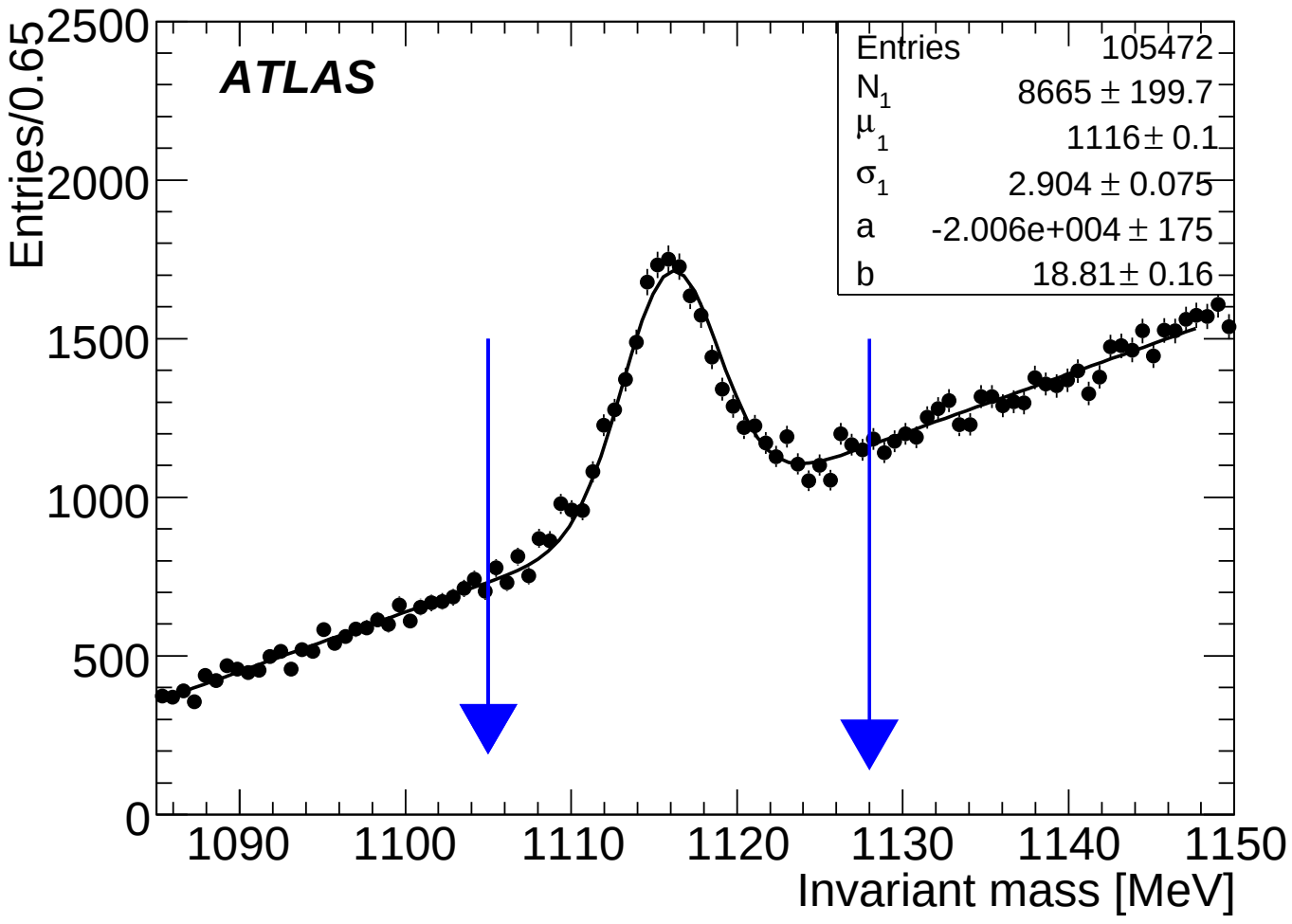


Figure 7: Invariant mass of $p\pi^-$ candidates.

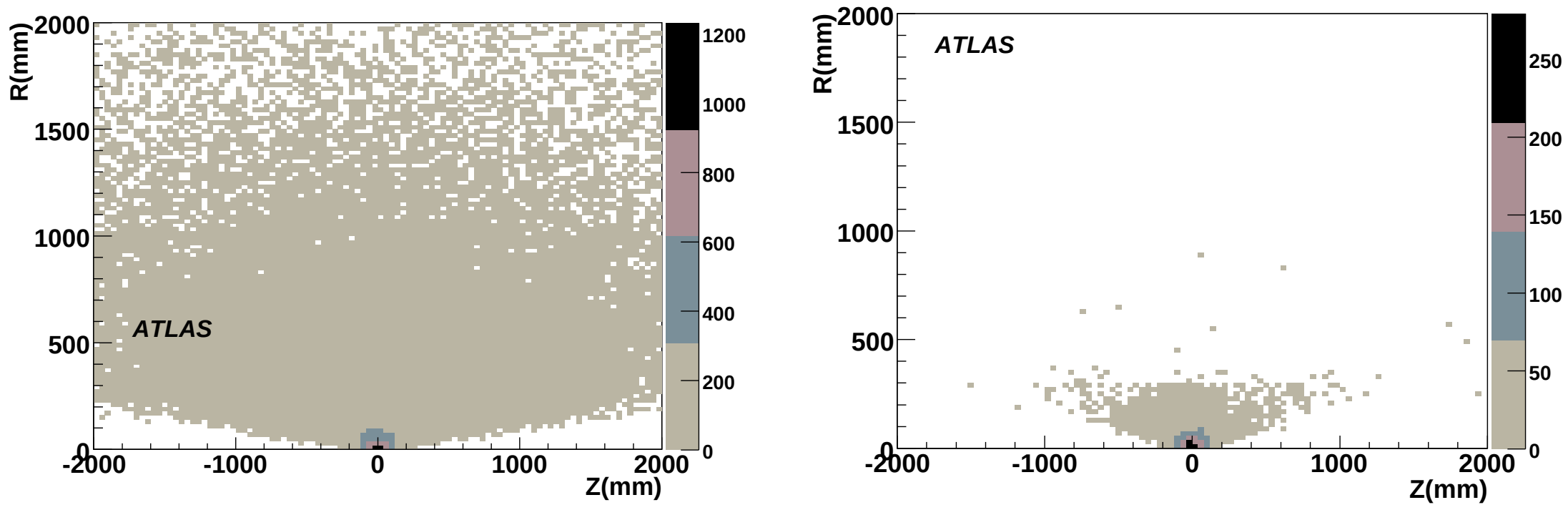


Figure 8: Decay vertex position of $\Lambda$'s in the $RZ$ plane at the generation level (left) and after reconstruction (right)

.

### 4.3 Selection of $\Lambda_b \rightarrow J/\psi(\mu^+\mu^-)\Lambda(p\pi^-)$ candidates

A previous study [22] based on early ATLAS simulation software estimated that the number of $\Lambda_b$ and $\overline{\Lambda_b}$ events which we expect to collect for the integrated luminosity of 30 fb$^{-1}$ is 75000. Using the new fully reconstructed sample we made a new estimation. We used the following expression to calculate the number of events:

$$\mathscr{N} = \mathscr{L}\sigma(\Lambda_b)\mathscr{E}, \tag{11}$$

where $\mathscr{L}$ is the integrated luminosity, $\sigma(\Lambda_b)$ = 7.4 pb is the cross section of $\Lambda_b \rightarrow J/\psi(\mu(p_T > 4000\text{ MeV})\mu(p_T > 2500\text{ MeV}))\Lambda(p(p_T > 500\text{ MeV})\pi(p_T > 500\text{ MeV}))$ , see details of the calculation in Table 4, and $\mathscr{E}$ is an overall $\Lambda_b$ acceptance, which includes the level$-1$ and level$-2$ acceptance for $\Lambda_b \rightarrow J/\psi(\mu^+\mu^-)\Lambda(p\pi^-)$ .

For selecting events with $b$ hadrons at a luminosity below about $10^{33} cm^{-2}s^{-1}$, the first level trigger will require the presence of a muon with $p_T$ > 6000 MeV within the trigger geometric acceptance of $|\eta| < 2.4$. The effect of the level$-1$ trigger threshold on muon $p_T$ is not a sharp cut and a fraction of

| | |
|---|---|
| $\sigma(pp \to \Lambda_b X)$ | 0.00828113 mb |
| BR ( $\Lambda_b \to J/\psi\Lambda$ ) | $(4.7 \pm 2.8) \times 10^{-4}$ [20] |
| BR ( $\Lambda \to p\pi^-$ ) | $(63.9 \pm 0.5) \times 10^{-2}$ [20] |
| BR ( $J/\psi \to \mu^+\mu^-$ ) | $(5.93 \pm 0.06) \times 10^{-2}$ [20] |
| Including cuts | 0.05 |
| **Overall cross-section** | **7.4 pb** |

Table 4: The cross-section calculation of $\Lambda_b \to J/\psi(\mu^+\mu^-)\Lambda(p\pi^-)$ decay.

muons with $p_T$ lower than 6000 MeV will be collected. Figure 9 shows the efficiency of the level$-$1 simulation with nominal $p_T$ threshold of 6000 MeV as function of $p_T$ . Around 69% of events with $J/\psi \to \mu^+\mu^-$ , where one muon has $p_T > 4000$ MeV and the second muon has $p_T > 2500$ MeV, passed the level$-$1 trigger simulation. Therefore a signal dataset with $p_T$ less than 6000 MeV has been chosen to study all possible triggered events with low $p_T$ muons instead of a usual sharp 6000 MeV cut.

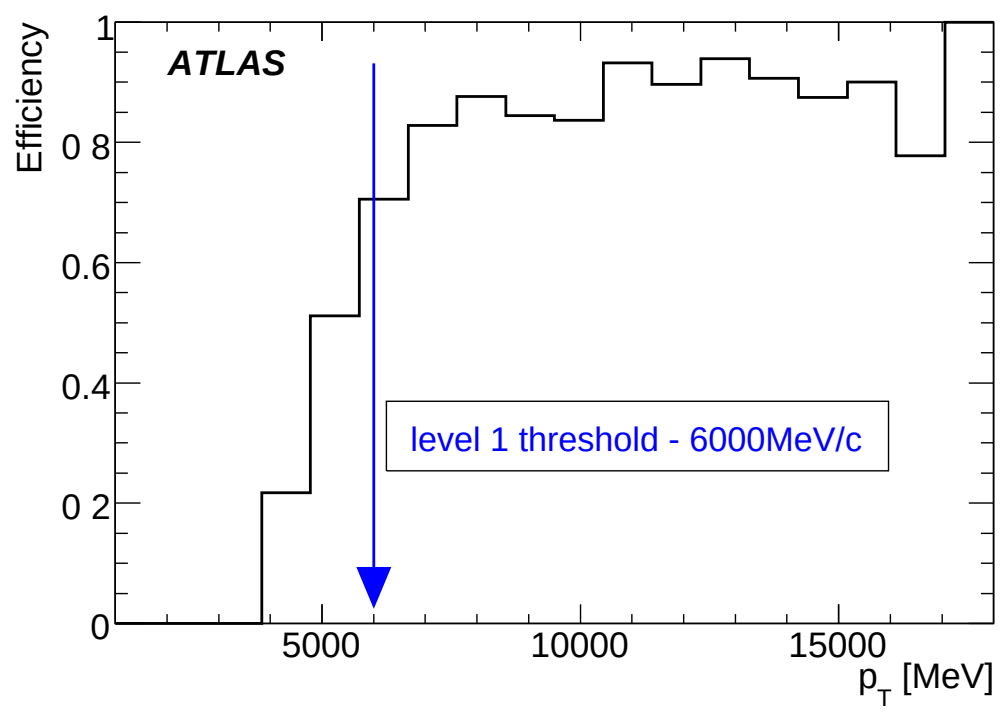


Figure 9: The level$-$1 trigger simulation efficiency as a function of muon $p_T$ , obtained from the $\Lambda_b$ signal sample over the whole detector volume.

Further selections in the high level trigger are based on the Region of Interest (RoI) identified at level$-$1, as follows: a search for a second muon close to the trigger muon is used to select channels containing two final state muons, for example from $J/\psi$ . It is based on expanding the level$-$1 muon RoI to find a second muon which was not triggered by level$-$1. This increases the efficiency of the di$-$muon trigger by extending the $p_T$ acceptance for the second muon down below 6000 MeV. The size of the increased RoI is based on the distribution of angular distance in $\eta$ and $\phi$ between two muons decayed from $J/\psi$ . The Inner Detector tracks which are reconstructed within these RoI, are then extrapolated to the muon system to find the corresponding hits within the window. The Inner Detector tracks associated with the muon spectrometer hits can be identified as muons. The level$-$2 trigger efficiency is found to be around 78% for $\Lambda_b \to J/\psi(\mu(p_T > 4000 \text{ MeV})\mu(p_T > 2500 \text{ MeV}))\Lambda(p(p_T > 500 \text{ MeV})\pi(p_T > 500 \text{ MeV}))$ .

We reconstruct the $\Lambda_b$ by performing a constrained fit to a common vertex for the two muon tracks and $\Lambda$ , with the two muon tracks constrained to the $J/\psi$ mass of 3097 MeV [20]. The reconstruction efficiency depends on the cuts which will be applied on all Inner Detector tracks in the reconstruction stage to reduce the fake rate. The overall efficiency is found to be around 6.1% if the $p_T$ threshold is 500 MeV, see Table 5. Figure 10 shows the invariant mass distribution of $\Lambda_b$ candidates. Simulation of the level$-$1 trigger with level$-$1 $p_T$ thresholds of 6000 MeV and 4000 MeV and level$-$2 trigger, explained above,

included in the analysis. We expect to collect around 13500 (13100) $\Lambda_b \to J/\psi(\mu^+\mu^-)\Lambda(p\pi^-)$ events using 4000 MeV (6000 MeV) level−1 muon threshold for the integrated luminosity of about 30 fb$^{-1}$ .

| level−1 trigger: with $p_T$ threshold | one muon 4 GeV | one muon 6 GeV | two muons 4GeV | two muons 6GeV |
|---|---|---|---|---|
| level−2 trigger: | TrigDiMuon | | Topological trigger | |
| $J/\psi$ reconstruction efficiency including level-1 and level-2 triggers | 42% | 39% | 27.5% | 10% |
| $\Lambda$ reconstruction efficiency | 15% | | | |
| $\Lambda_b$ overall efficiency | 6.1% | 5.9% | 5.4% | 3.5% |

Table 5: The overall $\Lambda_b$ efficiency depending on the trigger strategy.

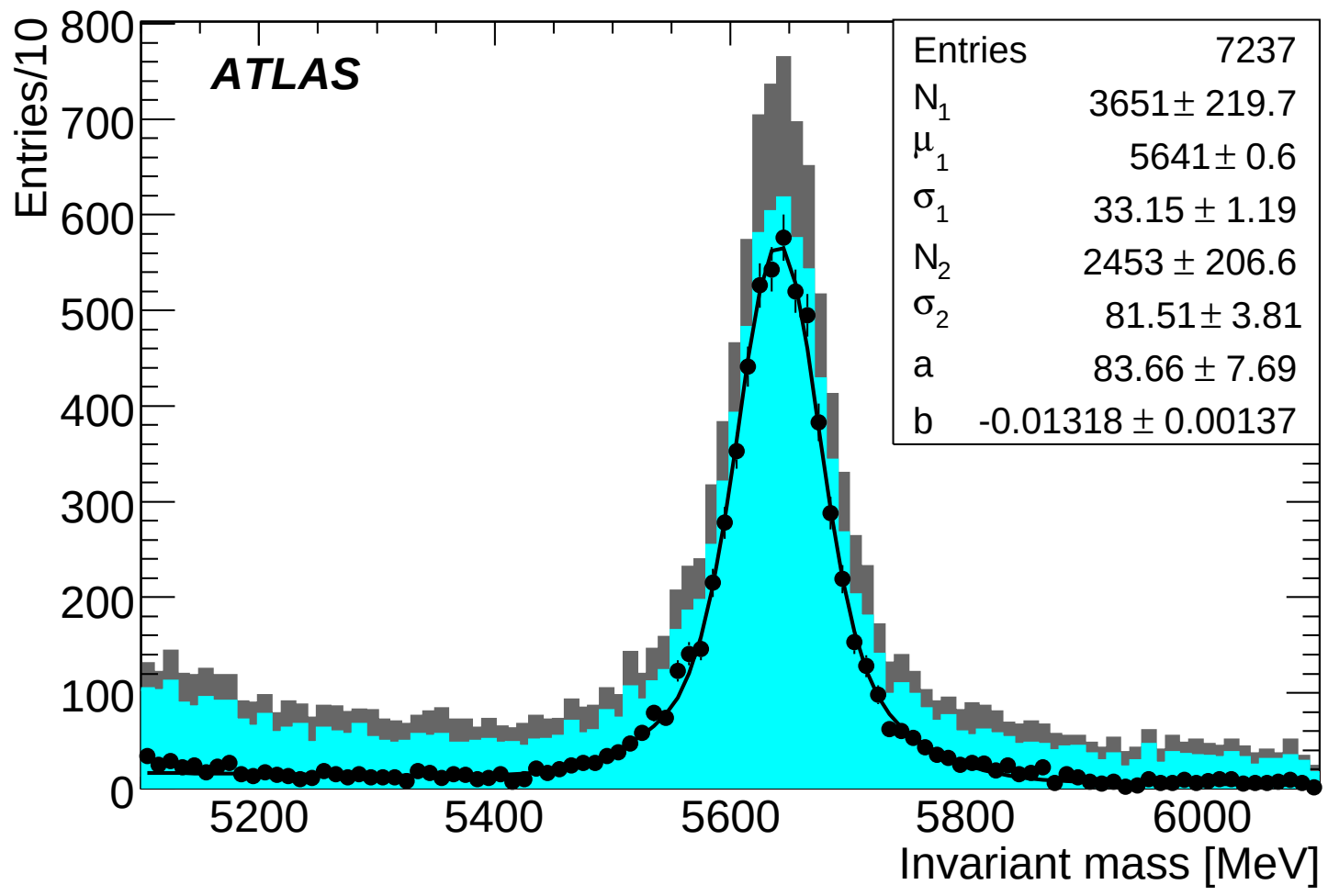


Figure 10: $\mu^+\mu^-$ $\Lambda$ invariant mass distribution. The dark color represents all $\Lambda_b$ candidates after reconstruction and vertexing requirement, and the light color represents the case when a level−1 and level−2 trigger signature is required in addition. Filled circles represents data after all selection cuts. The fit is the result of using double Gaussian and Polynomial functions.

We need to acknowledge that there are other inefficiencies that will appear when we analyze the real data. For example, even if the individual track reconstruction efficiency is as high as 98%, we will have an overall reduction in event rate of about 10%. Even if such reductions occur, we still expect the final sample to be sufficient for a meaningful measurement of the $\Lambda_b$ polarization.

### 4.4 Angular distributions and angular resolutions

The reconstruction efficiency modifies the angular distributions used in the polarization determination. Figure 11 shows how the angular distributions change due to detector acceptance for a Monte Carlo sample with polarization of -75%.

The angular resolution of the five angles is presented in Figure 12. We used this angular resolution in the statistical uncertainty study.

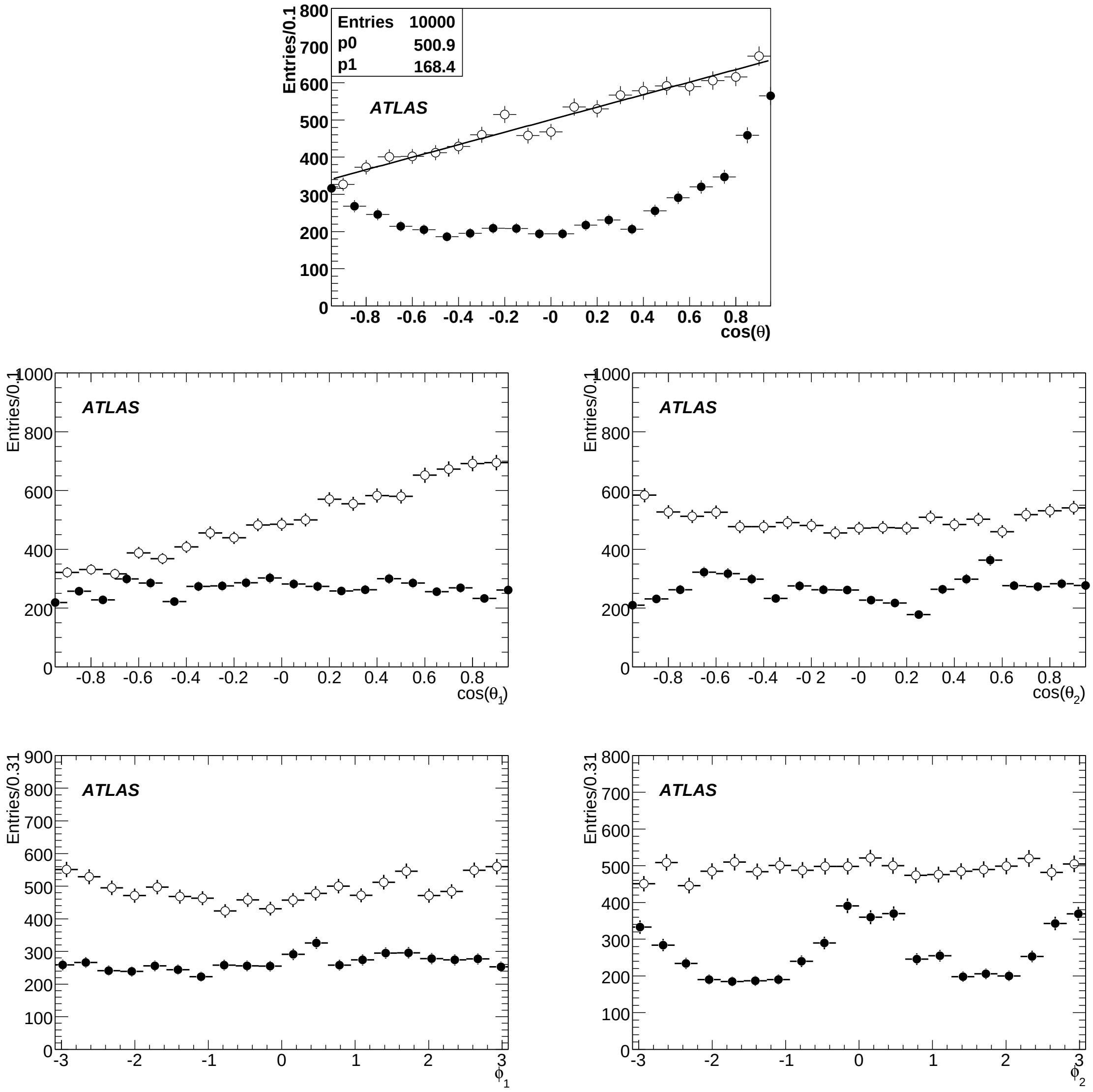


Figure 11: Comparison of fast Monte Carlo events without kinematics and detector acceptance cuts (open circles) and Monte Carlo events after full detector simulation and reconstruction (solid circles).

## 4.5 Background

Due to its production rate the main background source for our $\Lambda_b$ reconstruction will be the prompt production and decay of $J/\psi \to \mu^+\mu^-$ which are then combined with $\Lambda$ candidates in the event. However, the long lifetime of the $\Lambda_b$ allows us to reduce significantly this kind of background by applying a lifetime cut. After a $\Lambda_b$ lifetime cut (a cut of 200 $\mu$m on the proper transverse decay length), this background was found to be negligible and it is not considered in this study.

In order to investigate the different contributions of long−lived background particles not removed by the lifetime cut mentioned above, we used a inclusive $J/\psi$ Monte Carlo sample of $b\bar{b} \to J/\psi X$ requiring in addition to a $J/\psi$ , a $\Lambda$ in each event ($b\bar{b} \to J/\psi \Lambda X$). This $\Lambda$ could be produced along with the $J/\psi$ from a B hadron decay or just be part of the event, and the invariant mass of the $J/\psi + \Lambda$ combination should be within 5100 - 6100 MeV. Figure 13 shows the invariant mass distributions of $\Lambda_b$ candidates

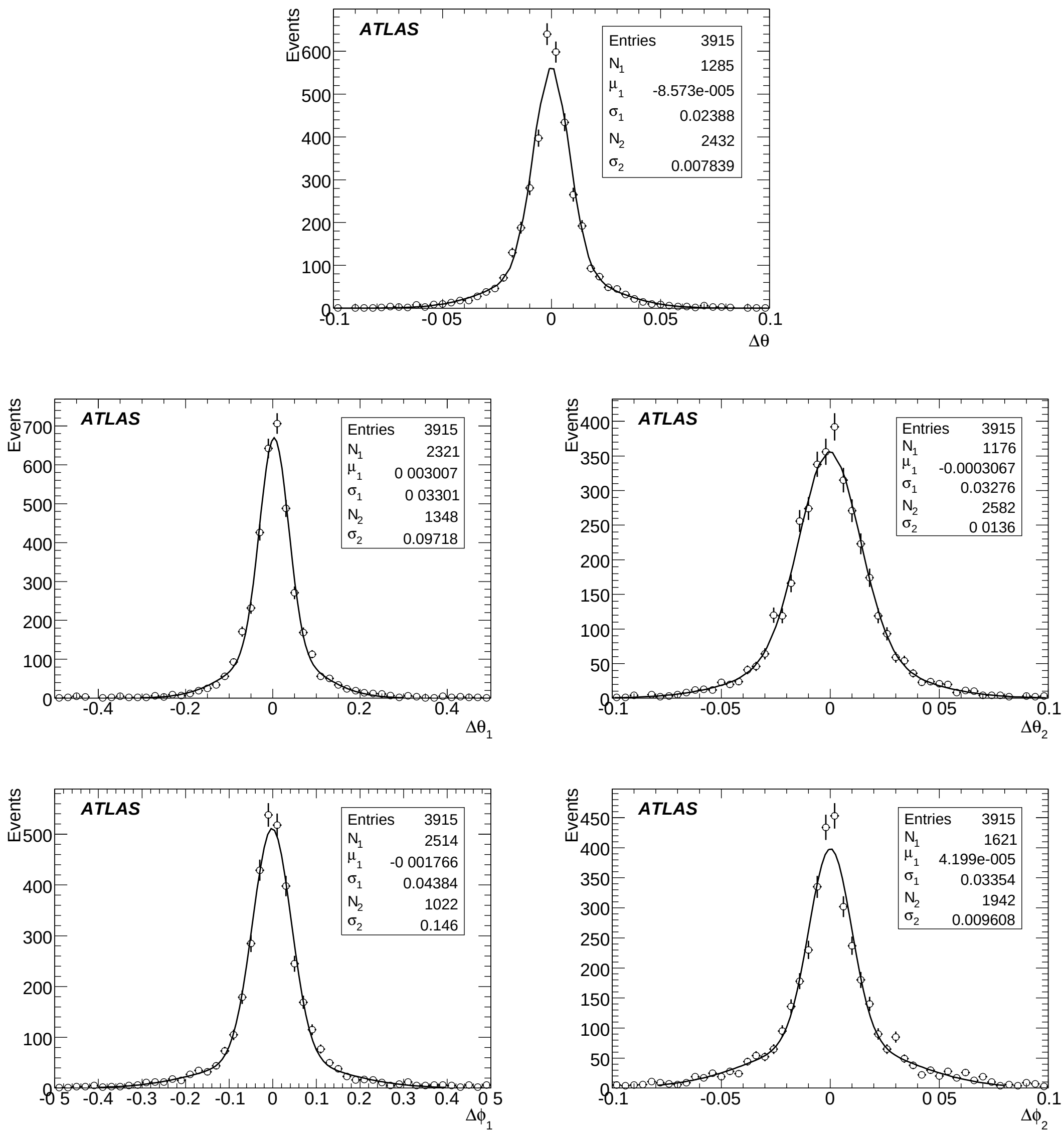


Figure 12: Angular resolution from fully simulated Monte Carlo data. The fit is the result of using double Gaussian distributions.

reconstructed in this Monte Carlo sample. The observed level of background under the $\Lambda_b$ signal is of few percents, and it is considerably reduced after extra cuts like the lifetime cut mentioned above. In Figure 13 another wider distribution due to $\Lambda_b \to J/\psi\Sigma^0(\Lambda\gamma)$ is observed very close to our $\Lambda_b \to J/\psi\Lambda$ signal. This is due to the branching ratios of both decays channels being the same as set by default in PYTHIA. This behavior has not been observed at Tevatron experiments where hundreds of $\Lambda_b \to J/\psi\Lambda$ events are reconstructed. Therefore we expect the branching ratio of the $\Lambda_b \to J/\psi\Sigma^0(\Lambda\gamma)$ decay to be considerably smaller than the branching ratio of the $\Lambda_b \to J/\psi\Lambda$, and that the resulting background will be much smaller than shown.

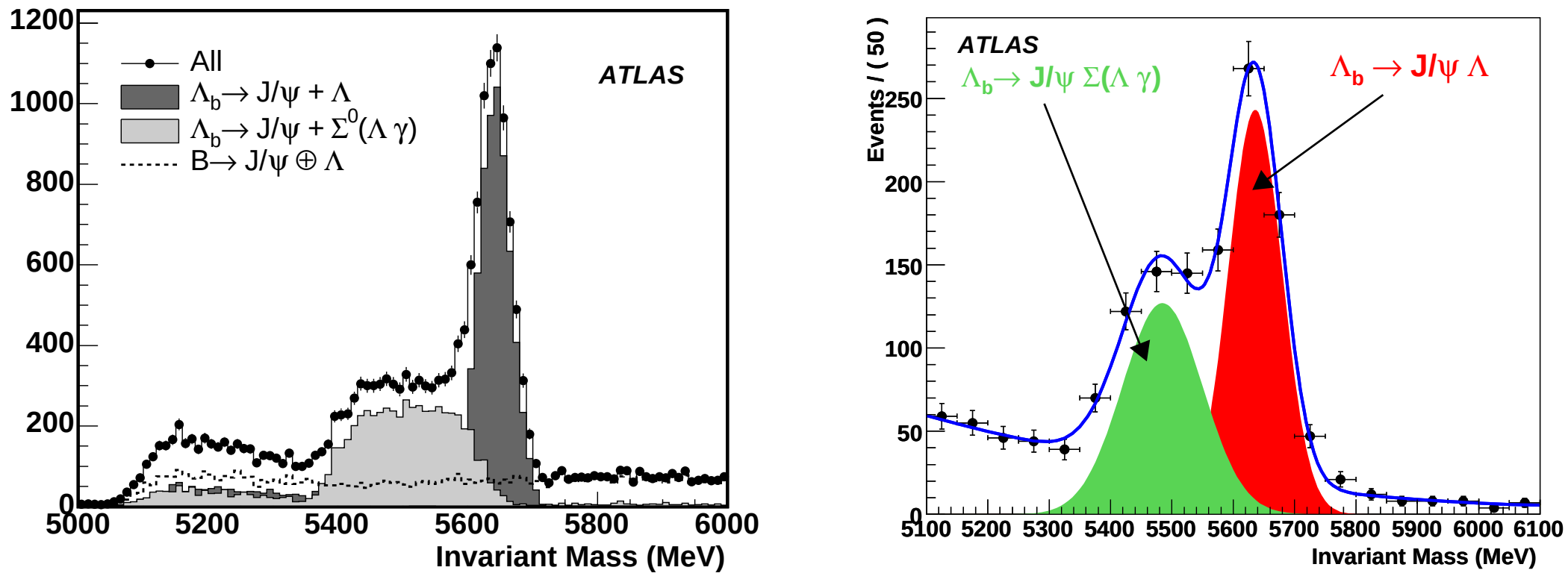


Figure 13: Invariant mass distribution from $\Lambda_b$ candidates identified in $b \to J/\psi \Lambda X$ Monte Carlo sample. Composition at generation level with smearing from reconstruction (left) and fit to the fully reconstructed events (right) after vertexing requirement are shown.

# 5 Extracting $\Lambda_b$ polarization and decay parameters

## 5.1 Fitting method

### 5.1.1 Likelihood function

To extract polarization and decay amplitudes we performed an un−binned maximum likelihood fit to the angular distributions. The log−likelihood function $\mathscr{L}$ is defined by:

$$\mathscr{L} = -2\sum_{j=1}^{N} \log(w_{obs}(\vec{\theta}', \vec{A}, P)), \quad (12)$$

where

$$w_{obs}(\vec{\theta}', \vec{A}, P) = \frac{\int w(\vec{\theta}', \vec{A}, P) T(\vec{\theta}, \vec{\theta}') d\vec{\theta}}{\int \int w(\vec{\theta}', \vec{A}, P) T(\vec{\theta}, \vec{\theta}') d\vec{\theta} d\vec{\theta}'}. \quad (13)$$

$w(\vec{\theta}', \vec{A}, P)$ is the p.d.f defined in Equation 6, $\vec{\theta}'$ are the measured angles, $\vec{\theta}$ are angles without detector effects, and $T(\vec{\theta}, \vec{\theta}')$ is defined as

$$T(\vec{\theta}, \vec{\theta}') = \varepsilon(\vec{\theta}) R(\vec{\theta}, \vec{\theta}'), \quad (14)$$

where $\varepsilon(\vec{\theta})$ is the efficiency function and $R(\vec{\theta}, \vec{\theta}')$ is the resolution function.

In the ideal case the resolution function is:

$$R(\vec{\theta}, \vec{\theta}') = \delta(\vec{\theta} - \vec{\theta}'), \quad (15)$$

then we have

$$w_{obs}(\vec{\theta}', \vec{A}, P) = \frac{w(\vec{\theta}', \vec{A}, P)\varepsilon(\vec{\theta}')}{\sum_{i=0}^{i=19} f_{1i}(\vec{A}) f_{2i}(P\alpha_\Lambda)\mathscr{F}_i}, \quad (16)$$

where $\mathscr{F}_i = \int F_i(\vec{\theta})\varepsilon(\vec{\theta})d\vec{\theta}$ are the acceptance corrections values, which have to be calculated in advance to perform the fit.

The final log−likelihood may be re−written as a sum of two terms:

$$L = -2\sum_{j=1}^{N} [\log(\frac{w(\vec{\theta}', \vec{A}, P)}{\sum_{i=0}^{i=19} f_{1i}(\vec{A}) f_{2i}(P\alpha_\Lambda)\mathscr{F}_i}) + \log(\varepsilon(\vec{\theta}'))]. \quad (17)$$

Since the second term does not depend on the parameters we want to measure, the main challenge is to find the acceptance function.

### 5.1.2 Detector acceptance corrections

The acceptance corrections integral $\mathscr{F}_i = \int F_i(\vec{\theta})\varepsilon(\vec{\theta})d\vec{\theta}$ can be approximated by the following form, using the Monte Carlo integration techniques

$$\mathscr{F}_i \approx \frac{1}{N_{gen}} \sum_{j=0}^{j=N_{acc}} \frac{F_i(\vec{\theta})}{G(\vec{\theta})}, \tag{18}$$

where $N_{gen}$ is the number of generated events, $N_{acc}$ is the number of accepted events after the simulation of the fiducial acceptance and $p_T$ cut and $G$ is the p.d.f which has been used to generate the $\theta$.

If the generation of the events is done using certain p.d.f ($w$), the acceptance can be calculated by the simple expression:

$$\mathscr{F}_i \approx \frac{1}{N_{gen}} \sum_{j=0}^{j=N_{acc}} \frac{F_i(\vec{\theta})}{w(\vec{\theta},\vec{A},P)}. \tag{19}$$

We used this expression to calculate the acceptance in the case when $w$ is the p.d.f from Equation 6.

This method can be used under the assumption that the acceptance does not depend on the measured parameters, and that the angular resolutions are close enough to the ideal resolutions. In order to check the first assumption we plotted the ratio

$$\frac{\int w(\vec{\theta},\vec{A},P)\varepsilon(\vec{\theta},\vec{A},P)d\vec{\theta}}{\int w(\vec{\theta},\vec{A},P)\varepsilon(\vec{\theta},\vec{A},P=0)d\vec{\theta}} \tag{20}$$

for the different polarization values (see Figure 14). No significant dependence of the acceptance on the polarization is observed in this test.

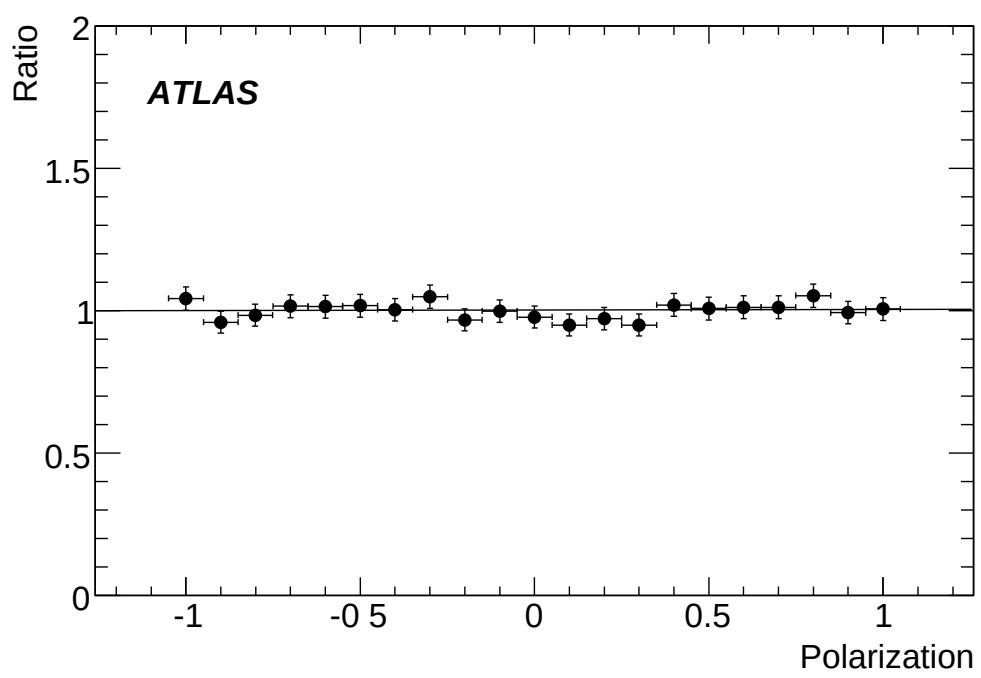


Figure 14: Ratio defined in Equation 20 as a function of polarization.

The angular resolutions are shown in Figure 12. To test the effect of these resolutions, Monte Carlo fits were performed including a smearing of the data based on the Gaussian fits in Figure 12. Fit results with and without smearing are consistent within the statistical uncertainty. Figure 15 shows, as an example, a comparison of fit results for a sample of 2000 events with polarization of -75% when fits are performed on the sample of generated Monte Carlo events.

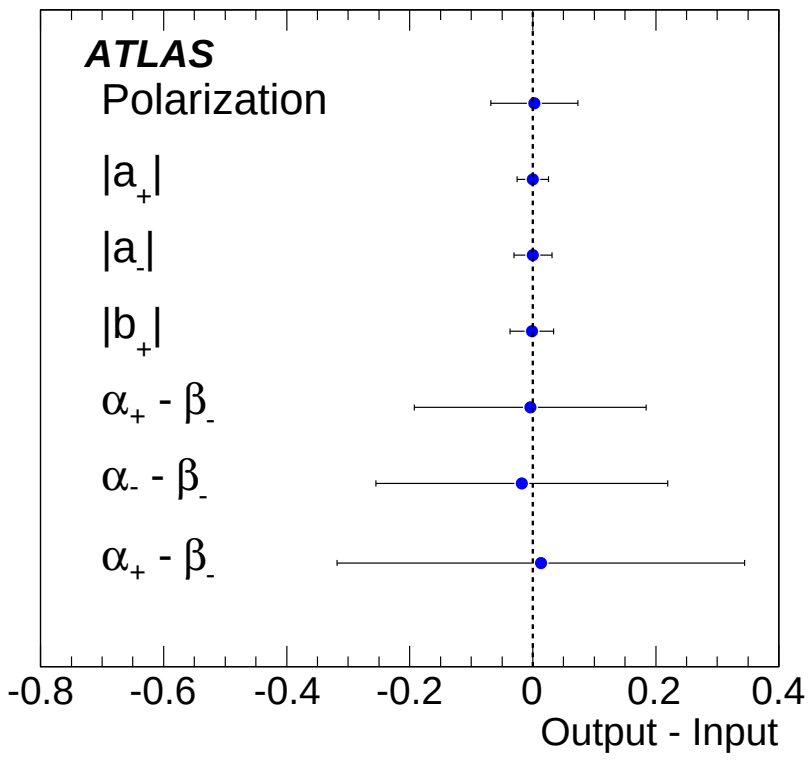


Figure 15: Comparison of fit outputs from generation level Monte Carlo with and without Gaussian smearing due to finite angular resolution. Error bars are statistical uncertainties from the fit with Gaussian smearing included.

### 5.2 Fits to fully simulated Monte Carlo data

In order to extract polarization and decay parameters from the Monte Carlo data samples, final $\Lambda_b$ selection cuts were applied. A proper transverse decay length greater than 200 $\mu m$ is required to remove contamination from prompt produced $J/\psi$ events. The proper transverse decay length for the $\Lambda_b$ candidate is given by:

$$\lambda = \frac{L_{xy}}{(\beta\gamma)_T^{\Lambda_b}} = L_{xy}\frac{cM_{\Lambda_b}}{p_T}, \tag{21}$$

where $(\beta\gamma)_T^{\Lambda_b}$ and $M_{\Lambda_b}$ are the transverse boost and the mass of the $\Lambda_b$, and $L_{xy}$ is a transverse decay length. The transverse decay length is defined as $L_{xy} = \mathbf{L}_{xy} \cdot \mathbf{p}_T / p_T$ where $\mathbf{L}_{xy}$ is the vector that points from the primary vertex to the $\Lambda_b$ decay vertex and $\mathbf{p}_T$ is the transverse momentum vector of the $\Lambda_b$. A minimum $p_T$ of 500 MeV is required for any track used in the $\Lambda_b$ reconstruction. In addition, a $p_T > 4000$ MeV is required for the muon with larger $p_T$, and $p_T > 2500$ MeV for the second muon. These cuts reduce the $\Lambda_b$ sample by 21%, mainly due to the lifetime cut.

Table 6 shows the results of performing a likelihood fit to our fully simulated Monte Carlo data, for a sample of 2000 $\Lambda_b$ events, corresponding to around 5 fb$^{-1}$ of collected data. Figure 16 shows the difference between the input values in Monte Carlo and the extracted values of polarization and decay parameters by the likelihood fit. We used as fitting parameters: $|a_+|$, $|a_-|$, $|b_+|$, $\alpha_+ - \beta_-$, $\alpha_- - \beta_-$, $\beta_+ - \beta_-$, and the polarization $P$.

Detector acceptance corrections in Equation 19 were computed separately from the two Monte Carlo samples with different polarizations which are used in this study. Corrections computed in the Monte Carlo sample of -75% polarization were used in the fit of the Monte Carlo sample of -25% polarization, and vice versa. Due to the limited statistics in the Monte Carlo samples used to calculate the acceptance corrections defined in Equation 19, a bagging (from bootstrap aggregating) technique [23] was used to generate multiple samples in order to avoid the effect of statistical fluctuations. This technique consists of generating replicates of a data set by selecting at random events from the original data set allowing repetition of events. We generated 1000 bootstrap replicates of the fully simulated Monte Carlo data sample. The $\mathscr{F}_i$ factors (Equation 19) were computed from each generated data sample and the average was taken as the value for each of the twenty $\mathscr{F}_i$ correction factors. Systematic uncertainty due to the width of the correction factors distributions in these 1000 generated data sets was estimated by repeating the fit to fully simulated Monte Carlo using the $\mathscr{F}_i$ values from the each of the generated samples, and

| Parameter | Value ± Uncertainty (Polarization = -25%) | Value ± Uncertainty (Polarization = -75%) | Value (Input at generation level) |
|---|---|---|---|
| Polarization | -0.213 ± 0.069 | -0.882 ± 0.064 | -0.25/-0.75 |
| $\lvert a_+\rvert$ | 0.461 ± 0.051 | 0.413 ± 0.023 | 0.429 |
| $\lvert a_-\rvert$ | 0.289 ± 0.058 | 0.161 ± 0.035 | 0.260 |
| $\lvert b_+\rvert$ | 0.259 ± 0.071 | 0.370 ± 0.027 | 0.295 |
| $\alpha_+ - \beta_-$ | -0.991 ± 0.640 | -2.050 ± 0.134 | -1.612 |
| $\alpha_- - \beta_-$ | 0.856 ± 0.364 | 0.681 ± 0.342 | 1.231 |
| $\beta_+ - \beta_-$ | -1.442 ± 0.666 | -2.624 ± 0.187 | -1.849 |

Table 6: Fit results from fully simulated and reconstructed Monte Carlo events with input polarization of -25% and -75%.

assigning the width of the distribution of fitted parameters as a systematic uncertainty. This systematic error (also shown in Figure 16) can be reduced with more Monte Carlo statistics for the $\mathscr{F}_i$ calculation.

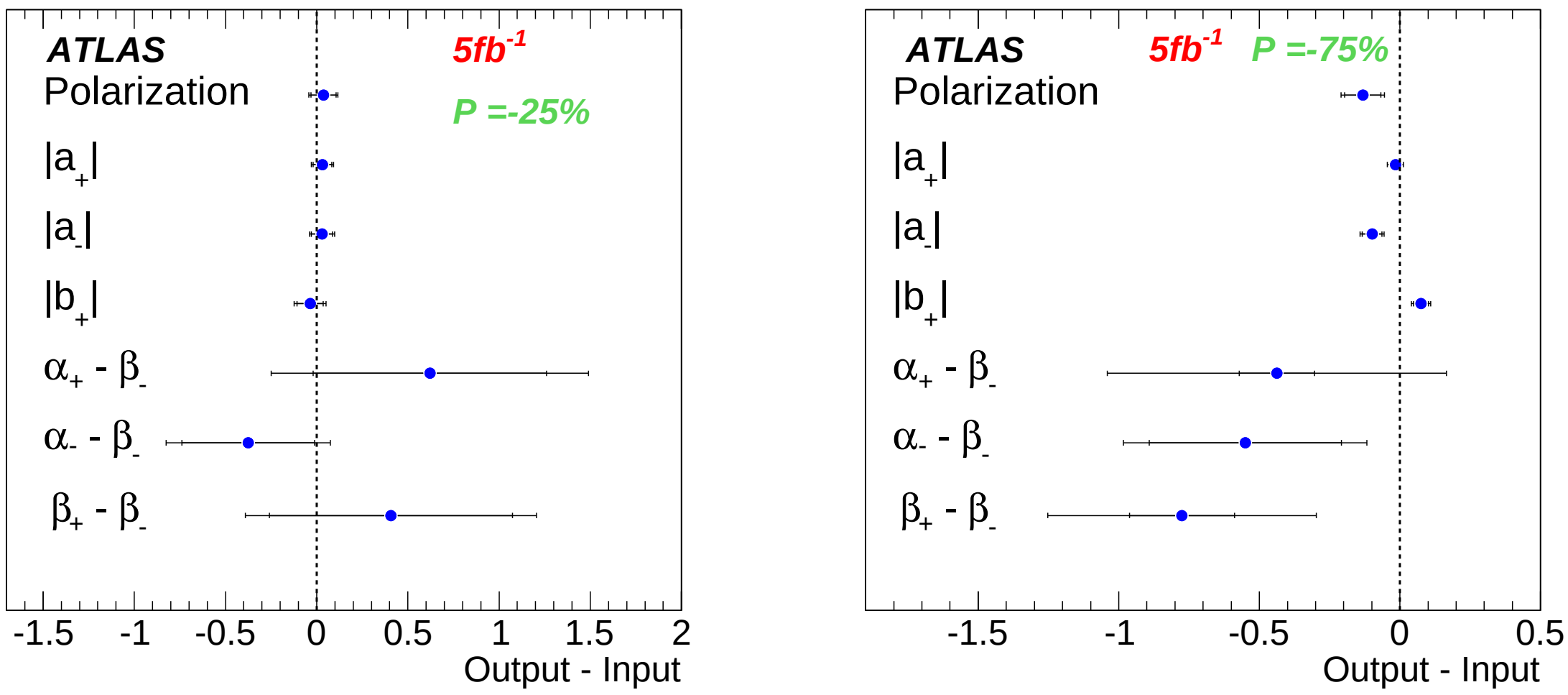


Figure 16: Comparison of fit results for polarization of -25% (left) and -75% (right) with respect to input values from Monte Carlo generation. The statistical and systematic uncertainties are included.

### 5.3 Estimate of statistical uncertainties

To estimate statistical uncertainties as a function of polarization, we used a fast Monte Carlo probabilistic approach to generate polarized $\Lambda_b$ particles. The fast Monte Carlo includes angular resolution from the fully reconstructed samples and detector acceptance simulation. We generated a large number of samples with different values of polarization. A maximum likelihood fit was used to extract the decay parameters and the polarization. Detector acceptance corrections were calculated from high statistics fast Monte Carlo data simulated without polarization. Figure 17 presents the expected statistical uncertainty in the polarization $P$ and in $\alpha_{\Lambda_b}$ as a function of the polarization value for the integrated luminosity of 30 $\text{fb}^{-1}$ . The study was done for $\alpha_{\Lambda_b}$ = -0.457 with the same input model as used in fully simulated Monte Carlo. In Figure 17 also the correlation between $\alpha_{\Lambda_b}$ and $P$ is shown as a function of polarization. The correlation values were extracted from the Maximum Likelihood fit results.

In our study we used specific set of decay amplitudes, presented in Table 3, to demonstrate our ability to extract these parameters. To insure that the success of our analysis techniques did not depend on the amplitudes chosen, we conducted a fast Monte Carlo study using a different model with $\alpha_{\Lambda_b} = 0.1$ [24] to test our procedure in a case of smaller $\alpha_{\Lambda_b}$ value. We found that it is possible to satisfactorily extract $\alpha_{\Lambda_b}$ and the polarization even with such a change in the amplitude values.

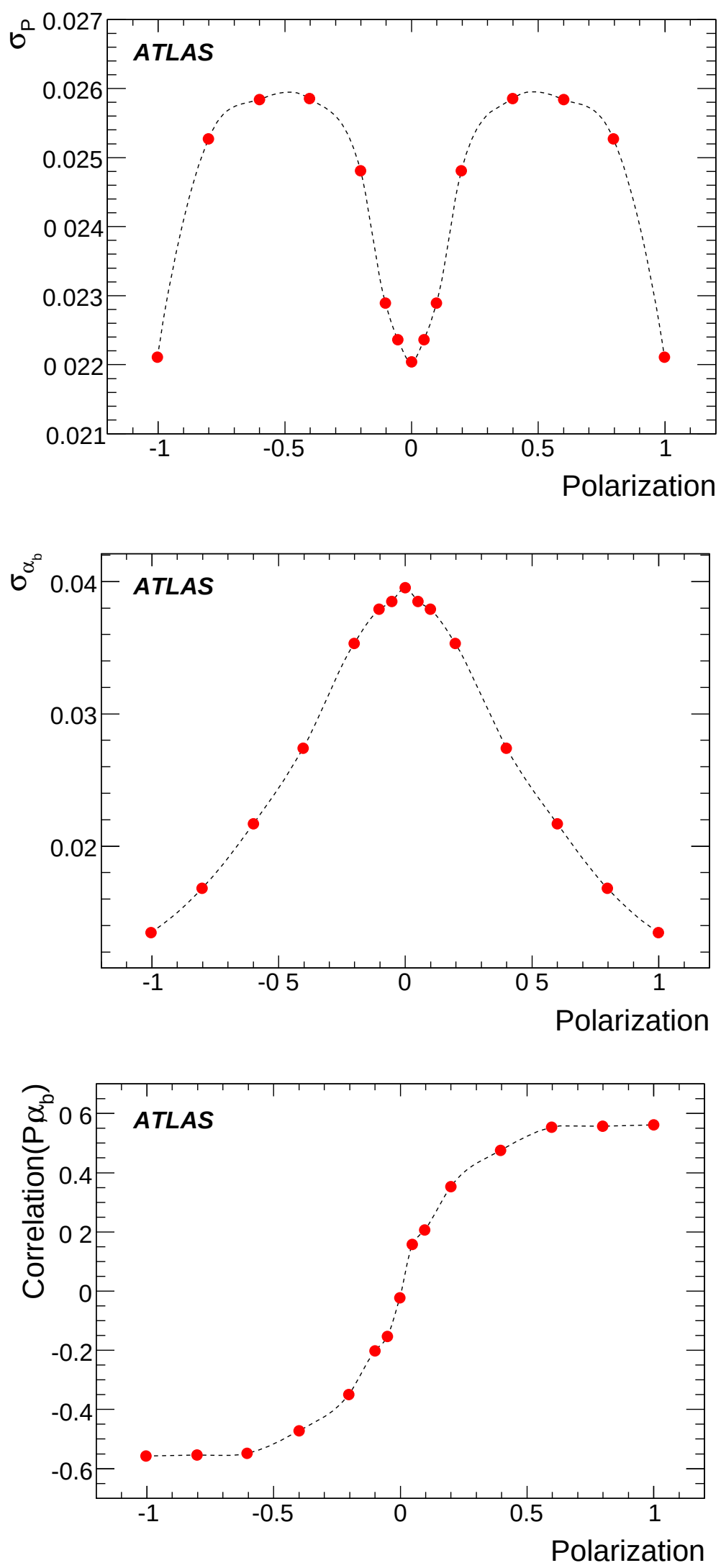


Figure 17: Expected statistical uncertainty on polarization (top) and on $\alpha_{\Lambda_b}$ (center) as a function of the polarization $P$. Bottom plot shows the expected correlation between $\alpha_{\Lambda_b}$ and the polarization $P$. All plots show results from the fast Monte Carlo study, obtained for the expected number of $\Lambda_b$ events in data sample of 30 $fb^{-1}$ .

## 6 Results and conclusions

In this note we have presented the results from a series of studies to determine if polarized $\Lambda_b$ baryons can be reconstructed in ATLAS and have their polarization and $\alpha_{\Lambda_b}$ parameter measured. Our results

indicate that the answer is affirmative. $\Lambda_b$ events should be identifiable through the reconstruction of their four charged final state particles, and the angles between these particles can be measured with sufficient accuracy to determine the parent's polarization. With trigger constraints and detector cuts fully specified, our more complete analysis suggests that the number of events we should expect after 30 $fb^{-1}$ of data will only be 13,000, compared to the 37,500 noted in the ATLAS-TDR [22]. Different additional detector and background effects, which are difficult to model at the current level of detector description, can further reduce the signal sensitivity. These effects could include the detector and trigger inefficiency, misalignment, pile-up events and increased combinatorial background due to e.g. the fake tracks. Nevertheless, even with a reduction of 50%, a polarization measurement with a statistical uncertainty of several percent should be possible in a regime where polarization is larger than 25% as experimentally measured at lower energies. Efforts will continue to develop algorithms to improve the various reconstruction and trigger efficiencies and in consequence providing an enhanced yield of reconstructed particles in data samples.

We note that almost all models predict that the $\Lambda_b$ polarization at the LHC at small Feynman $x$ should be vanishingly small. Measurement of a significant polarization would have to be regarded as a signal of an unexplained effect, either from the domain of existing physics, or of new physics altogether.

We further note that the development of $\Lambda_b$ polarimetry as a tool for studying spin effects at the LHC could be important. For example, members for the SUSY community are quite interested in knowing what fraction of the $b$ quark polarization ends up in the polarization of a $\Lambda_b$, since this could provide a way to test if $b$ quark SUSY partners have the correct handedness. Only a few hundred $\Lambda_b$ decays would be required to, for example, determine if its polarization were 100% or -100%. Challenges clearly exist, however, in determining the polarization transfer fraction, which requires a source of $b$'s such as from $Z \to b\bar{b}$, and in dealing with the fact that only $10^{-5}$ of $b$'s generate decay into the $\Lambda_b$ channel we have described here. Our work on this topic will continue.

Other related studies that should continue include mechanisms for comparing the $\alpha_{\Lambda_b}$ parameters from $\Lambda_b$ and its antiparticle as a test of CP. We will accumulate data on both. If CP is conserved, the two parameters should be equal in magnitude but opposite in sign. While the precision of this test will not be high, and while models predict that any CP violation would be small in this sector, nevertheless, such a test would be unique in this domain and should be made.

Finally, as noted in Section 1, the lifetime of the $\Lambda_b$ remains a topic of significant interest. Such a measurement will be a natural by-product of our efforts to extract the $\Lambda_b$ spin parameters.

# Study of the Rare Decay $B_s^0 \to \mu^+\mu^-$

**Abstract**
We investigate the feasibility of measuring of the rare decay $B_s^0 \to \mu^+\mu^-$ in ATLAS. The contribution of inclusive and of the most important non–combinatorial background is studied.

## 1 Introduction

The rare decays, $B_s^0 \to \ell^+\ell^-$ with $\ell^\pm = e^\pm, \mu^\pm$, or $\tau^\pm$, are mediated by flavour-changing neutral currents that are forbidden in the Standard Model at tree level. The lowest-order contributions in the Standard Model involve weak penguin loops and weak box diagrams that are CKM suppressed. Examples of the lowest-order diagrams are shown in Figure 1. Since the $B_s^0$ meson is a pseudoscalar that has positive C parity and the transition proceeds in an $\ell = 0$ state, the electromagnetic penguin loop is forbidden. The two leptons are either both right-handed or both left-handed leading to additional helicity suppression. Thus, branching fractions expected in the Standard Model are tiny.

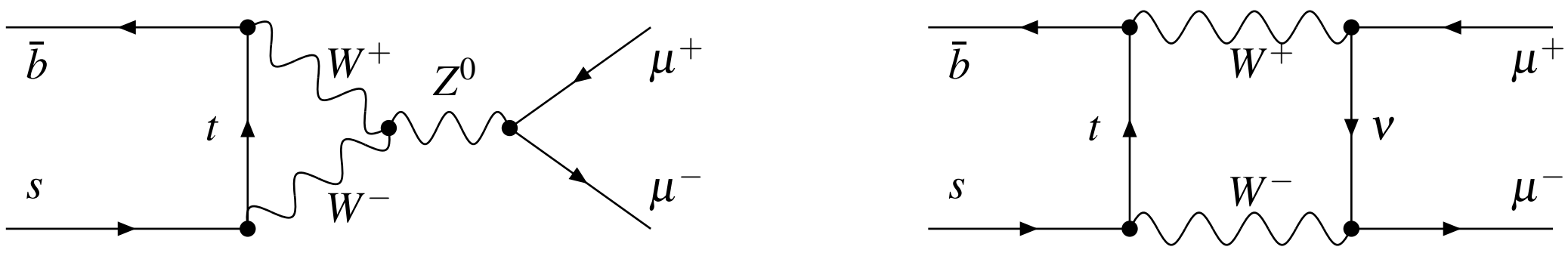


Figure 1: Lowest order Standard Model contributions to $B_s^0 \to \mu^+\mu^-$.

The early searches for rare $B$ meson decays started with radiative penguin decays, first observed by CLEO in 1993, where they presented evidence for the exclusive decay $B \to K^*\gamma$ and for the inclusive decay $B \to X_s\gamma$ a year later [1, 2].

The $B$ factory experiments, BaBar and Belle, have measured these decay modes with more precision. The present world average for the inclusive mode is $\mathscr{B}(B \to X_s\gamma) = (3.55 \pm 0.26) \times 10^{-4}$ [3]. BaBar and Belle also observed the decays $B \to K^{(*)}\ell^+\ell^-$ and $B \to X_s\ell^+\ell^-$ that are two orders of magnitude smaller than $B \to X_s\gamma$ [4, 5]. The decay $B_s^0 \to \mu^+\mu^-$ is expected to be further reduced by three orders of magnitude.

In extensions of the Standard Model, the $B_s^0 \to \mu^+\mu^-$ branching fraction may be enhanced by several orders of magnitude. Thus, several experiments have searched for these decays. The largest $B_s^0$ samples have been collected by CDF and D0 corresponding to a luminosity of 2 fb$^{-1}$ but no signal has been observed. The lowest branching fraction upper limit was set recently by CDF yielding $\mathscr{B}(B_s^0 \to \mu^+\mu^-) < 5.8 \times 10^{-8}$@95% confidence level [6]. This is still about an order of magnitude higher than the Standard Model prediction. As ATLAS has an elaborate muon system extended over a large region of the solid angle, the dimuon final state is expected to be reconstructed with high efficiency and good mass resolution. Thus, there are good prospects for observing this decay in the dimuon channel and measuring its branching fraction with reasonable precision.

## 2 Theoretical description

The Standard Model amplitude for the process $B_{s,d} \to \ell^+\ell^-$ is calculated from the effective Hamiltonian

$$H_{\text{eff}} = -\frac{G_F}{\sqrt{2}}\frac{\alpha}{\pi \sin^2\theta_W} V_{tb}^* V_{tq}(C_{10}(\mu)\mathscr{O}_{10}(\mu) + C_S(\mu)\mathscr{O}_S(\mu) + C_P(\mu)\mathscr{O}_P(\mu)) + h.c., \quad (1)$$

where $C_i(\mu)$ are Wilson coefficients that present the perturbatively calculable short-distance effects and $\mathscr{O}_i(\mu)$ are local operators that describe the non-perturbative long-distance effects of the transition. The scale parameter $\mu$ is of the order of the $b$-quark mass ($\sim 5$ GeV), $\theta_W$ is the weak mixing angle, $\alpha$ is the electromagnetic coupling constant and $V_{tb}^* V_{tq}$ are CKM matrix elements for $t \to b$ and $t \to q = s,d$ transitions, respectively.

The dominant contribution results from the axial-vector operator $\mathscr{O}_{10}$, [7]:

$$\mathscr{O}_{10} = (\bar{b}_L \gamma^\mu q_L)(\bar{\ell}\gamma_\mu \gamma_5 \ell). \tag{2}$$

The Wilson coefficient $C_{10}$ has been determined in the next-to-leading order (NLO) of QCD. The NLO corrections are in the percent range and higher-order corrections are not relevant [8]. In NLO an excellent approximation in terms of the $\overline{MS}$ mass of the top quark, $\bar{m}_t$, is given by:

$$C_{10}(\bar{m}_t) = 0.9636 \left[ \frac{80.4 \text{ GeV}}{M_W} \frac{\bar{m}_t}{164 \text{ GeV}} \right]^{1.52}. \tag{3}$$

The measurements of the top quark mass at the Tevatron, $m_t^{pole} = 171.4 \pm 2.1$ GeV [9], yield an $\overline{MS}$ mass of $\bar{m}_t = 163.8 \pm 2.0$ GeV and the world average of the W-boson mass is $m_W = 80.403 \pm 0.029$ GeV. The accuracy of this approximation is $5 \times 10^{-4}$ for masses of 149 GeV $< \mathrm{\bar{m}_t} <$ 179 GeV.

The other two operators represent scalar and pseudoscalar couplings to the leptons:

$$\mathscr{O}_S = m_b(\bar{b}_R q_L)(\bar{\ell}\ell), \mathscr{O}_P = m_b(\bar{b}_R q_L)(\bar{\ell}\gamma_5 \ell). \tag{4}$$

The Wilson coefficients, $C_S$ and $C_P$, are determined from penguin diagrams that involve the Higgs boson or the neutral Goldstone boson, respectively. Although they are not helicity suppressed, their contributions are tiny in the Standard Model and they may be safely neglected in Standard Model calculations.

The $B_q \to \mu^+\mu^-$ branching fractions including the scalar and pseudoscalar contributions are given by:

$$\begin{aligned} \mathscr{B}(B_q^0 \to \mu^+\mu^-) = {} & \frac{G_F^2 \alpha^2}{64\pi^3 \sin^4 \theta_W} |V_{tb}^* V_{tq}|^2 \tau_{B_q} M_{B_q}^3 f_{B_q}^2 \sqrt{1 - \frac{4m_\mu^2}{M_{B_q}^2}} \\ & \times \left[ \left(1 - \frac{4m_\mu^2}{M_{B_q}^2}\right) M_{B_q}^2 C_S^2 + \left( M_{B_q} C_P - \frac{2m_\mu}{M_{B_q}} C_{10} \right)^2 \right], \end{aligned} \tag{5}$$

where $M_{B_q}$, $\tau_{B_q}$, and $f_{B_q}$ respectively are mass, lifetime and decay constants of the $B_q$ meson. The decay constant is determined in different models, including quark models, QCD sum rules and unquenched lattice theory. The accuracy is presently of the order of $10-15\%$. Evaluating $\alpha$ at the Z-mass scale, $\alpha(M_Z) = 1/128$, the following predictions were made for the $B_q \to \mu^+\mu^-$ branchings fractions in the Standard Model [8]:

$$\begin{aligned} Br(B_s^0 \to \mu^+\mu^-) &= (3.86 \pm 0.15) \times \frac{\tau_{B_s^0}}{1.527\,\text{ps}} \frac{|V_{ts}^* V_{tb}|^2}{1.7 \times 10^{-3}} \frac{f_{B_s}}{240\,\text{MeV}} \times 10^{-9}, \\ Br(B_d^0 \to \mu^+\mu^-) &= (1.06 \pm 0.04) \times \frac{\tau_{B_d^0}}{1.527\,\text{ps}} \frac{|V_{td}^* V_{tb}|^2}{6.7 \times 10^{-5}} \frac{f_{B_d}}{200\,\text{MeV}} \times 10^{-10}. \end{aligned} \tag{6}$$

In extensions of the Standard Model, such as supersymmetry (SUSY), Higgs doublet models or models with extra gauge bosons, scalar-current, pseudoscalar-current or axial-vector current interactions

may arise with new particles in the loop. This yields new contributions in the Wilson coefficients $C_{10}, C_S$, and $C_P$. Since the scalar and pseudoscalar operators are not helicity suppressed, they may give rise to a large enhancement of the branching fraction. Furthermore, the contribution of the pseudoscalar operator may produce destructive or constructive interference with the axial vector operator. Thus, new physics may increase or decrease the branching fraction with respect to the Standard Model value. For example, in the minimal supersymmetric Standard Model (MSSM), the $B_s^0 \to \mu^+\mu^-$ branching fractions are proportional to $\tan^6(\beta)$ [1]. The branching fraction of $B_d \to \mu^+\mu^-$ is expected to be a factor of 40 lower than that for $B_s \to \mu^+\mu^-$, hence, the latter is the focus of this note.

# 3 ATLAS strategy for $B_s^0 \to \mu^+\mu^-$ study

Measurements of the properties of $B$ decays with such extremely low branching fractions in ATLAS is possible namely due to the large beauty cross-section and luminosity of the LHC machine. Thus at luminosity $10^{33}$ cm$^{-2}$ s$^{-1}$ $10^{12}$ $B$ hadron pairs will be produced each year. It is expected that ATLAS will record $10^8$ events with $B$ decays each year by using $B$-physics triggers [10]. Triggers dedicated to rare dimuon $B_s^0 \to \mu^+\mu^-$ decays will be described in the Section 4.2 of this document.

Since the branching fraction is so small in the Standard Model, semileptonic $B$ decays and even some rare $B$ decays may yield substantial backgrounds. The key issue for $B_s^0 \to \mu^+\mu^-$ discovery at the LHC is the suppression of the backgrounds. The ATLAS strategy for observing $B_s^0 \to \mu^+\mu^-$ is as follows.

The first step is to trigger on events containing a $B_s^0 \to \mu^+\mu^-$ candidate using dedicated trigger algorithms which are described in this document. In the offline analysis the selections will be refined to reduce backgrounds. To achieve final separation of signal from background we will employ statistical methods based on several variables. Both parts of the offline selection are described in this paper.

Once recorded data are available, the background in the signal region will be estimated using sidebands in the distribution of the muon pair invariant mass. In the current study the background was estimated using simulated events. Two categories of backgrounds were simulated: the so called combinatorial background from $b\bar{b}$ pairs producing two muons in the final state; and the exclusive backgrounds, coming from two-body hadronic $B$ decays and from the process $B_s^0 \to K^-\mu^+\nu$. The exclusive backgrounds contribute to the signal region and the lower mass sideband only. They do not occur in the higher mass sideband, so their contribution to the signal is estimated separately.

After the number of background events in the signal region has been determined, the number of signal events $N_B$ can be determined from a comparison of the total number of events found in the signal region, and the estimated background. For low statistics an upper limit on $N_B$ corresponding to certain confidence level is determined using appropriate statistical methods. Once $N_B$ is determined the $B_s^0 \to \mu^+\mu^-$ branching fraction, $\mathscr{B}(B_s^0 \to \mu^+\mu^-)$, can be calculated using a relative normalisation to the reference channel $B^+ \to J/\psi(\mu^+\mu^-)K^+$.

This document presents a Monte Carlo simulation study which follows the strategy described above. We start from the trigger level in Section 4.2. This is followed by the offline analysis, optimisation of discriminating variables and finally the determination of background and signal contribution in the signal regions, in Section 4.4. Systematic uncertainties are analysed in Section 5, followed by the start-up strategy in Section 6.

It should be stressed that due to large uncertainty in the predictions of the $b\bar{b}$ production cross-section at the LHC energy this paper cannot derive a precise sensitivity to $\mathscr{B}(B_s^0 \to \mu^+\mu^-)$ at ATLAS but rather to show the ATLAS potential for this study and its discovery capability under some assumptions.

[1] $\tan(\beta)$ is the ratio of vacuum expectation values for charged and neutral Higgs bosons.

# 4 Monte Carlo study

## 4.1 Simulation and event selection

The Monte Carlo simulation samples used in the analysis have been generated as part of the central ATLAS Monte Carlo data production runs, and details of this simulation have been given in the introduction to this chapter.

The list of generated signal and background events is given in Table 1. To ensure that most of the

| Process | # Events |
|---|---|
| $B_s^0 \to \mu^+\mu^-$ | 47.5k |
| $bb \to \mu^+\mu^- X$ | 146.5k |
| $B_s^0 \to K^-\pi^+$ | 50k |
| $B_s^0 \to K^-\mu^+\nu$ | 50k |

Table 1: List of processes and number of events analysed

generated dimuon events passed the trigger, only events containing two muons with $p_T$ larger than 6 GeV and 4 GeV, were retained for detector simulation. For the signal channel $B_s^0 \to \mu^+\mu^-$, multiplying the cross-section reported by PYTHIA with the branching ratio $3.42 \times 10^{-9}$ gave a cross-section of 15 fb. 47.5k events have been generated and passed through the full detector simulation and reconstruction. Simulation of pileup has not been available, thus it was not simulated for either the signal or the background events.

The sample of dominant background process events, $b\bar{b}$ decaying semileptonically giving two muons in the final state, were simulated with the same versions of the software, and with the same kinematic cuts as for the signal. The PYTHIA cross-section for such a sample is estimated to be 110 nb and a total of 146.5k background events that passed the reconstruction stage were used in the physics analysis.

In addition to the combinatorial background, there are several $B$ backgrounds that may contribute to the signal region. These include two and three body decays where two of the final state particles are $K^\pm$, $\pi^\pm$ or $\mu^\pm$. Although the rate for misidentification of kaons or pions as muons, due to punchthrough or decay in flight, is only of the order 0.5%, the small $\mathscr{B}(B_s^0 \to \mu^+\mu^-)$ requires investigation of the other rare $B$ decays. The decay modes which we consider to be most important are summarised in Table 2.

| process | branching fraction | Ref. |
|---|---|---|
| $B^0 \to K^+\pi^-$ | $(1.82 \pm 0.08) \times 10^{-5}$ | [11] |
| $B^0 \to \pi^+\pi^-$ | $(4.6 \pm 0.4) \times 10^{-6}$ | [11] |
| $B^0 \to K^+K^-$ | $< 3.7 \times 10^{-7} @ 90\% CL$ | [11] |
| $B_s^0 \to \pi^+\pi^-$ | $< 1.7 \times 10^{-4} @ 90\% CL$ | [11] |
| $B_s^0 \to \pi^+K^-$ | $< 2.1 \times 10^{-4} @ 90\% CL$ | [11] |
| $B_s^0 \to K^+K^-$ | $< 5.9 \times 10^{-5} @ 90\% CL$ | [11] |
| $B_s^0 \to K^-\mu^+\nu$ | $\sim 1.36 \times 10^{-4}$ | * [2] |
| $B^0 \to \pi^-\mu^+\nu$ | $(1.36 \pm 0.15) \times 10^{-4}$ | [11] |

Table 2: $B$ meson decays contributing to the non-combinatorial background

[2] An estimation based on isospin symmetry and the measurement of $B^0 \to \pi^-\mu^+\nu$.

We have studied one of the two-body and one of the three-body decays with full simulation, namely $B_s^0 \to K^-\pi^+$ and $B_s^0 \to K^-\mu^+\nu$. Contributions from the other channels were estimated to be similar or smaller. To enable the production of a sizable event sample with decay in flight, a special GEANT simulation option which forces kaons and pions from selected $B$ mesons to decay in the inner detector volume was developed [12].The position of the decay is randomly selected from a uniform distribution between the origin and the exit point from the inner detector.

## 4.2 Trigger strategy

We describe trigger methods for selection of the $B_s^0 \to \mu^+\mu^-$ channel developed by the $B$ trigger group as part of the project presented in this document. The full description of the ATLAS B-physics triggers is given in the introduction to the B-physics chapter [10]. The first level trigger (L1) performance for dimuon channels can be found in [13] and details of the second level (L2) dimuon trigger implementation in [12].

At the LHC start-up, the luminosity level is expected to be of order $10^{31}\text{cm}^{-2}\text{s}^{-1}$ and a $p_T$ threshold as low as 4 GeV can be used at L1. The dimuon rate after L1 is expected to be only a few Hz. This admits the possibility of applying L2 track reconstruction in the full volume of the inner detector. This detailed approach allows the study of dimuon background features to understand of their composition. With the subsequent rise of luminosity the L1 $p_T$ threshold will increase to 6 GeV. The dimuon rate after L1 is expected to rise to about 360 Hz at $L = 10^{33}\text{cm}^{-2}\text{s}^{-1}$ and the L2 track reconstruction in the full volume of inner detector will be replaced by the Region of Interest (RoI) guided mode, documented in [12].

The simulation of the trigger in the current study is performed by applying the strategies for luminosity $L = 10^{33}\text{cm}^{-2}\text{s}^{-1}$. At L1 the threshold of $p_T > 6$ GeV has been applied and events containing two L1 muon signatures are analysed further using the L2 topological dimuon trigger algorithm with a threshold of $p_T > 6$ GeV. Following the RoI defined at L1, the muon candidates are reconstructed in the muon spectrometer, then matched to the tracks reconstructed in the inner detector (inside the RoI) and combined into one track. The invariant mass of two opposite sign muons is required to be less than 7 GeV. These muons should also be successfully fitted to a common vertex. Only loose selection criteria are used at this step ($\chi^2 < 10$). Implementation of the third level trigger, the event filter (EF), is not finalised yet and the offline reconstruction efficiency is used to estimate the EF efficiency (the same reconstruction algorithms are supposed to be used at EF).

Results on the L1 and L2 efficiencies as well as an estimated EF efficiency for signal $B_s^0 \to \mu^+\mu^-$ are given in Table 3. The L1 efficiency is defined as the ratio of the $B_s^0 \to \mu^+\mu^-$ events passing the L1 trigger and the input events generated with $p_T > 6$ GeV and $|\eta| < 2.5$ for both muons from the $B_s^0 \to \mu^+\mu^-$ decay. The L2 efficiency is defined as the fraction of events accepted by L1 satisfying the above L2 reconstruction and selection cuts. The efficiency of the event filter is estimated as the fraction of events that both satisfy L2 and also are successfully reconstructed at EF. These values of the efficiency have been used in the subsequent analysis.

Various types of trigger algorithms and trigger thresholds will be used in the real experiment depending on the luminosity achieved, dedicated computing resources available for the online event processing and the actual beauty yield at LHC energies.

Table 3: Trigger efficiency of signal $B_s^0 \to \mu^+\mu^-$. The methods of calculating efficiencies at L1, L2 and EF levels are given in the relevant place in the text.

| L1*L2 efficiency | EF w.r.t L2 | Overall trigger eff. |
|---|---|---|
| 0.52 | 0.88 | 0.46 |

### 4.3 Event reconstruction and analysis.

Muon reconstruction quality is of high importance for the $B_s^0 \to \mu^+\mu^-$ channel. The muon candidates produced by the STACO [14] method were used. This method combines the independently reconstructed inner detector and muon spectrometer tracks. Figure 2 shows the muon reconstruction efficiency as a function of a true muon $p_T$. The efficiency is defined as the number of muon candidates reconstructed and matched to the Monte Carlo particle tracks in the corresponding $p_T$ bin divided by number of generated muons. The $p_T$ spectrum of generated muons is superimposed on the efficiency plot.

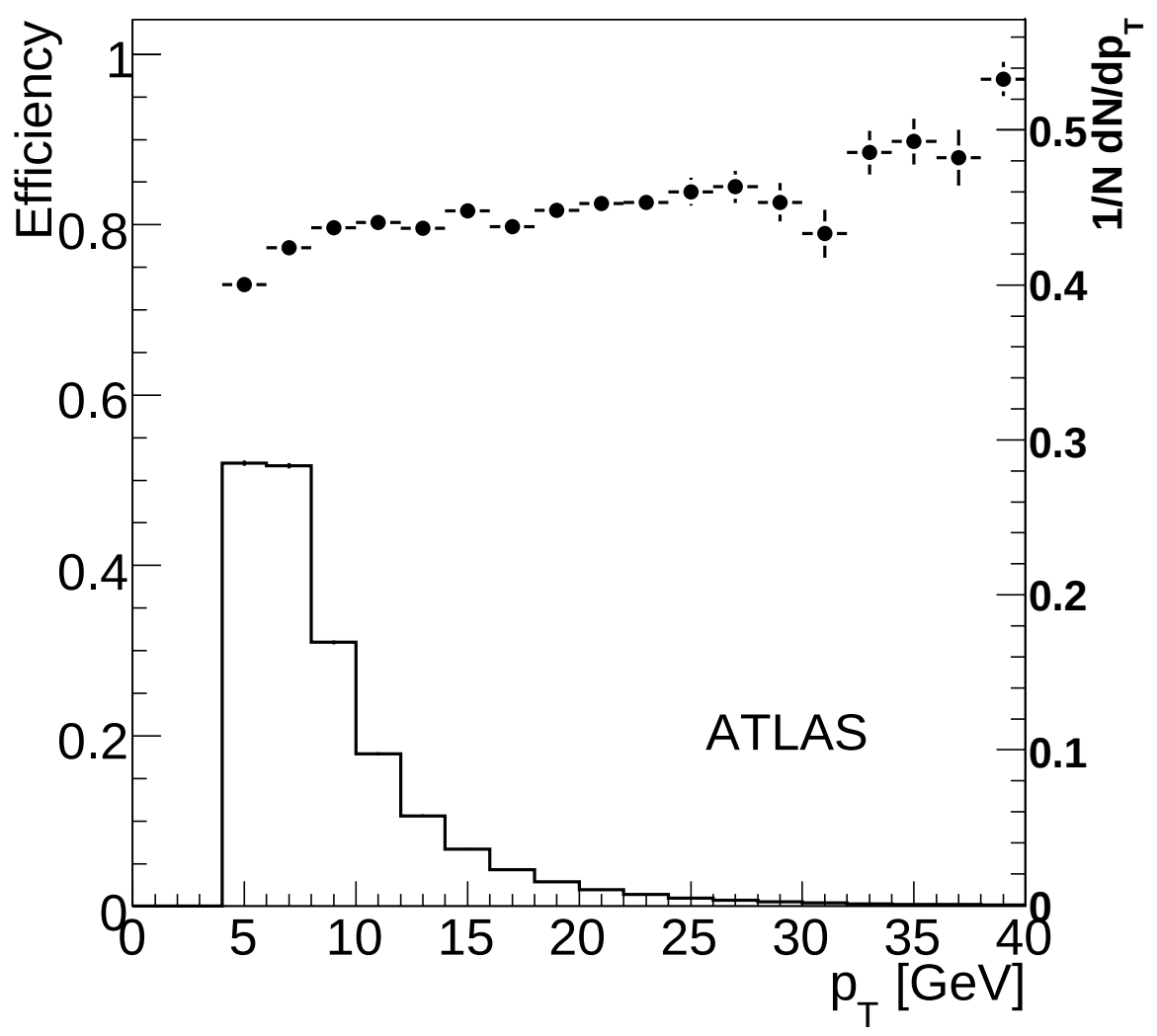


Figure 2: Muon offline reconstruction efficiency as a function of $p_T$. The superimposed histogram - $p_T$ spectrum of muons in the signal events (right scale).

For the physics analysis, we select events containing identified muon pairs with opposite charges. Aside of the kinematic cuts ($p_T^{\mu_1(\mu_2)} > 6(4)$ GeV and $|\eta_{\mu_1,\mu_2}| < 2.5$) no additional cuts have have been applied. These two muons then constitute a $B$ meson candidate. The VKalVrt vertexing package [15] is used to fit tracks into a vertex. We require the vertex quality to have $\chi^2 < 10$. The momentum resolution is important as a narrow mass search window reduces the background contribution. Figure 3 shows the dimuon mass distribution for the cases when both muons are in the barrel region ($|\eta_{\mu_1,\mu_2}| < 1.1$) or in the end-cap ($|\eta_{\mu_1,\mu_2}| > 1.1$). The Gaussian fit (using bins with contents $>$ 10% of maximum) gives $\sigma = 70$ MeV for the barrel and $\sigma = 124$ MeV for the end-cap. We used 90 MeV as an estimate of the invariant mass resolution for the signal events.

In this document we present a cut-based method for signal extraction and background rejection. For the future, we are investigating another method using a boosted decision tree [16].

In the cut based analysis a set of discriminating variables is chosen and using the signal and background simulated events the optimal set of cuts is determined. The signal events are identified by requiring that the dimuon invariant mass is consistent with the mass of $B_s$ meson. To reduce background events where two muons originate from different sources (e.g. independent semileptonic decays of $b$ and $\bar{b}$ quarks), the following discriminating variables were chosen (values used in the final analysis are given in parentheses):

- Transverse decay length of the $B_s$ candidate $L_{xy}$ ($L_{xy}$ >0.5 mm )

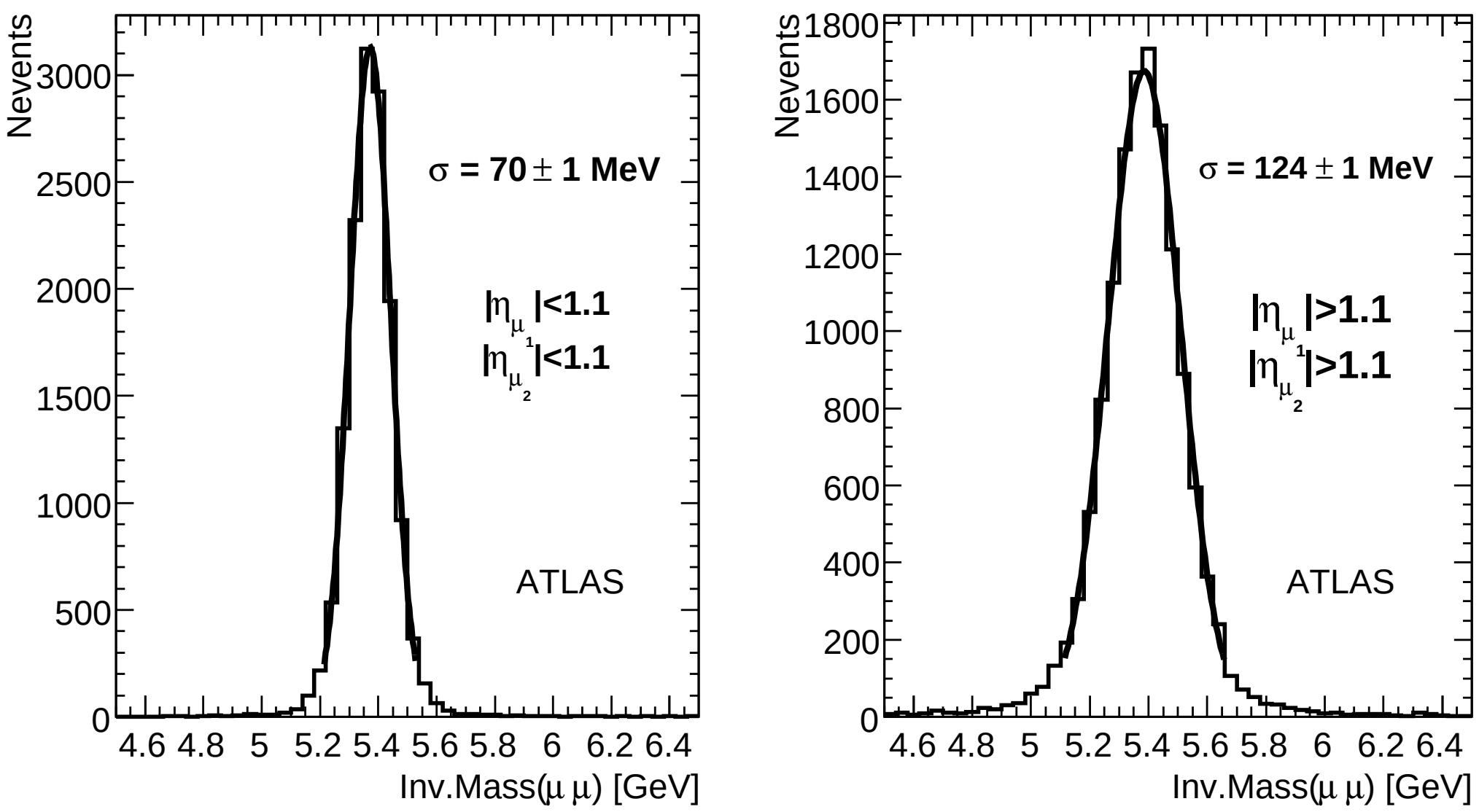


Figure 3: Reconstructed $B_s$ mass in barrel - when both muons have $|\eta| < 1.1$(left) and in the end-cap - both muons have $|\eta| > 1.1$ (right)

- The pointing angle $\alpha$ between the dimuon pair summary momentum and the direction of the decay vertex as seen from the primary vertex ($\alpha$ <0.017 rad )
- Isolation $I_{\mu\mu} = p_T^{\mu\mu}/(p_T^{\mu\mu} + \Sigma_i p_T^i(\Delta R < 1))$ , where the sum is over all tracks with $p_T > 1$ GeV (excluding the muon pairs) within a cone of $\Delta R < 1$, where $\Delta R = \sqrt{(\Delta\eta)^2 + (\Delta\phi)^2}$ and $\Delta\eta$ and $\Delta\phi$ are the pseudorapidity and azimuthal angle of track $i$ with respect to the momentum vector of the muon pair ( $I_{\mu\mu} > 0.9$ ).
- An asymmetric search window for $M_{\mu\mu}, \in [M_{B_s^0} - \sigma, M_{B_s^0} + 2\sigma]$, is used to avoid a possible contribution from $B_d^0 \to \mu^+\mu^-$ decay.

Figures 4, 5 and 6 show distributions of discriminating variables for signal and background events. Due to the low Monte Carlo statistics, it is not feasible to perform a cut analysis as in a real experiment (i.e. applying all cuts simultaneously) since this will leave no events for the analysis. However, some of the discriminating variables show no (or small) correlations between each other. This allows the estimation of the rejection power of such variables separately. Then the product of all efficiencies can provide a reasonable estimate of the total rejection. Table 4 shows a correlation matrix for the variables used in this study. The correlation between the pointing angle $\alpha$ and the transverse decay length $L_{xy}$ is higher than among other variables so the pointing angle and the transverse decay length are examined simultaneously to take their correlation into account. The systematic uncertainty due to this correlation is estimated as +50%.

Figure 7 illustrates the rejection power of each cut. One cut is applied at a time to the combinatorial background events and to the $B_s^0 \to K^-\mu\nu$ and $B_s^0 \to K^-\pi^+$ events where one or two hadrons are misidentified as a muon. The combinatorial background is effectively suppressed by these cuts while the non-combinatorial events are less well rejected.

Table 5 summarizes the output of this cut-based analysis. For the $b\bar{b} \to \mu\mu X$ background in the left column the efficiencies are given separately for each cut, whilst in the right column the combined

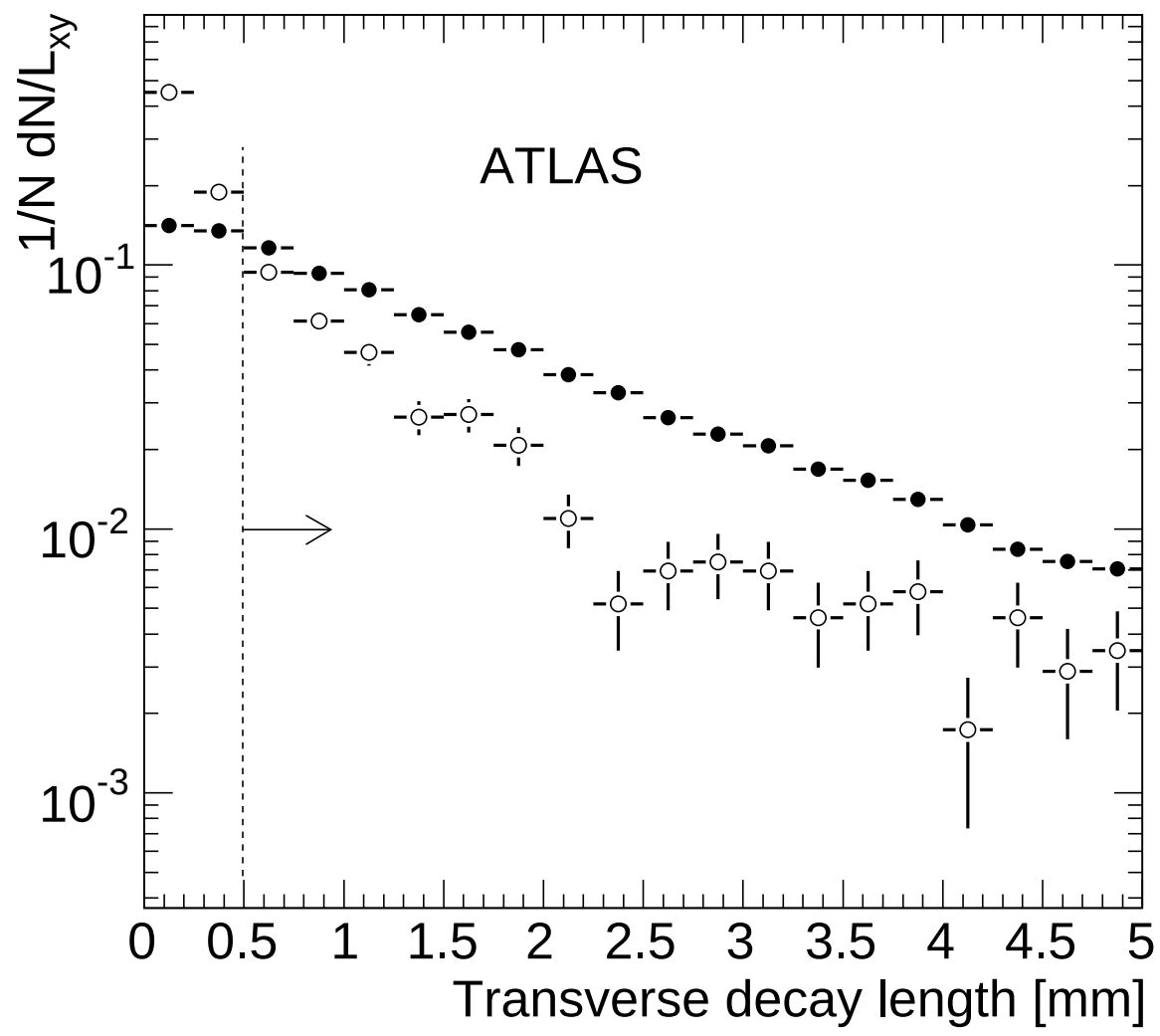


Figure 4: Transverse decay length of reconstructed $B_s^0$ candidates. The signal is shown as closed circles, background as opened circles. Distributions are normalised to 1. The vertical line indicates the lowest transverse decay length allowed for selected events.

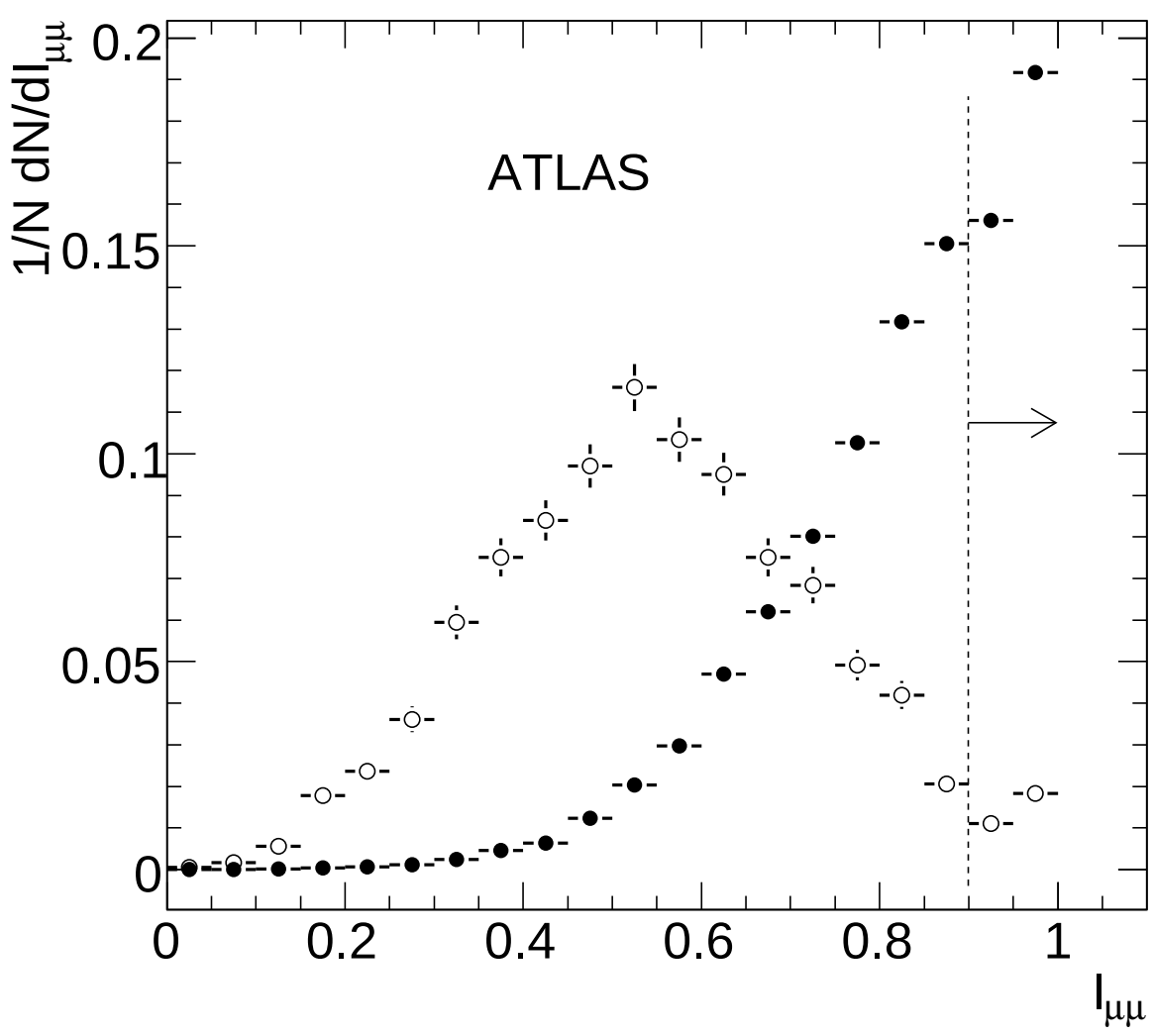


Figure 5: Distribution of isolation variable $I_{\mu\mu}$ for the reconstructed $B_s^0$ candidates. The signal is shown as closed circles, background as opened circles. Distributions are normalised to 1. The vertical line indicates the lowest values of variable $I_{\mu\mu}$ allowed for selected events.

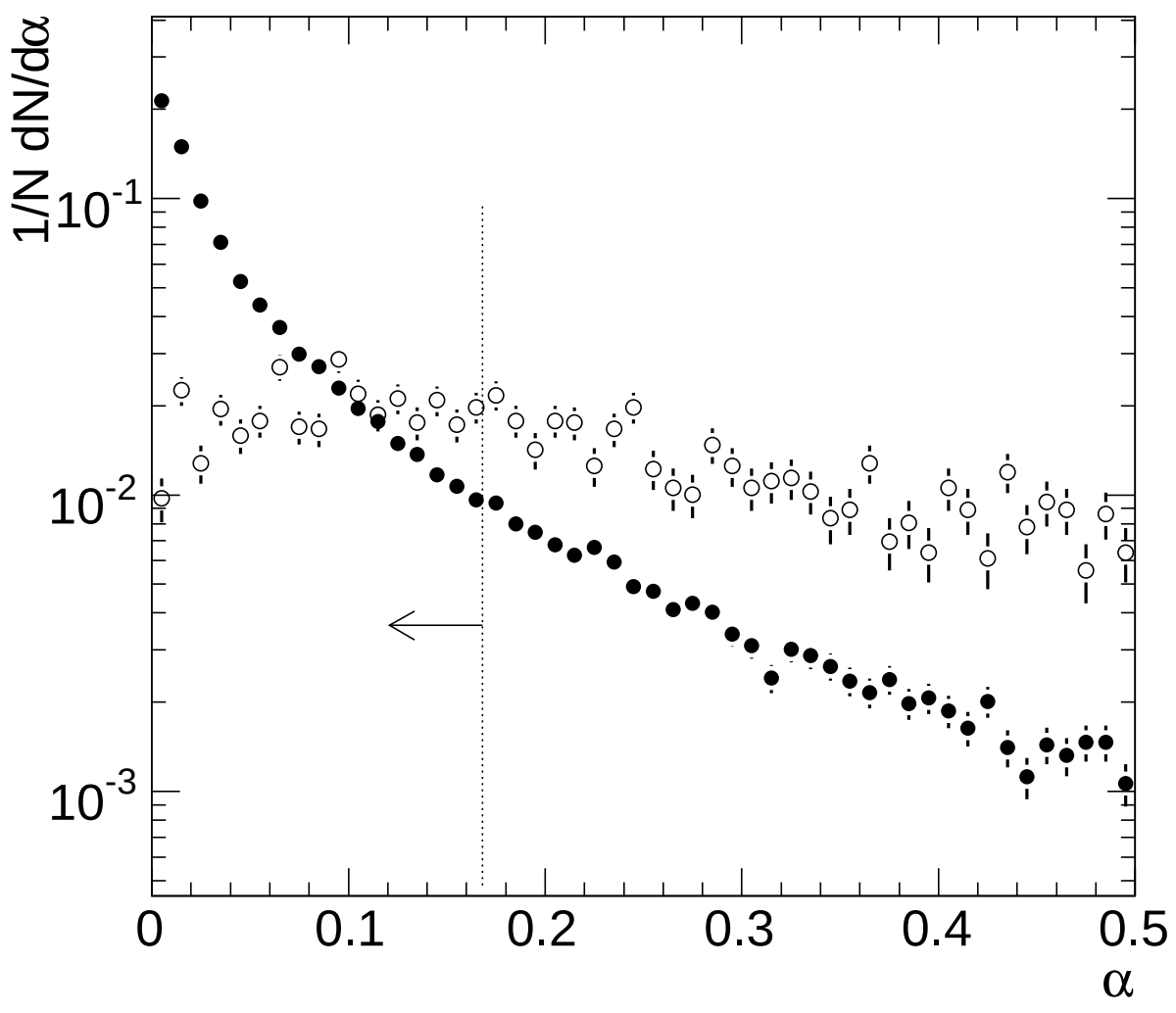


Figure 6: Distribution of pointing angle $\alpha$ for the reconstructed $B_s^0$ candidates. The signal is shown as closed circles, background as opened circles. Distributions is normalised to 1. The vertical line indicates the highest values of $\alpha$ allowed for selected events.

| | $M_{\mu\mu}$ | $I_{\mu\mu}$ | $\alpha$ | $L_{xy}$ |
|---|---|---|---|---|
| $M_{\mu\mu}$ | 1 | -0.09 | 0.04 | -0.03 |
| $I_{\mu\mu}$ | | 1 | -0.07 | -0.03 |
| $\alpha$ | | | 1 | -0.17 |
| $L_{xy}$ | | | | 1 |

Table 4: The linear correlation coefficients among the discriminating variables for background events. The statistical uncertainty is about $\pm 0.05$ for each coefficient.

efficiency is given for the cuts on the pointing angle and on the transverse decay length. As one can see the total rejection is largely overestimated if all cuts are treated separately, so the combined value is used to estimate a total yield of background events. The contribution from $B_s^0 \to K^-\mu\nu$ and $B_s^0 \to K^-\pi^+$ is found to be negligible comparing to the combinatorial background contribution. The errors quoted for the efficiencies are statistical only, so they represent only the size of the available Monte Carlo sample, but not the expected accuracy of the experiment where the initial number of background events will be much higher. More details on uncertainties will be given in Section 5.

Figure 8 shows the dimuon mass distribution for signal and background events after all selection cuts have been applied. For the combinatorial background, the contribution for the left and right side of the signal region is estimated in the same way as for the signal region (Table 5).

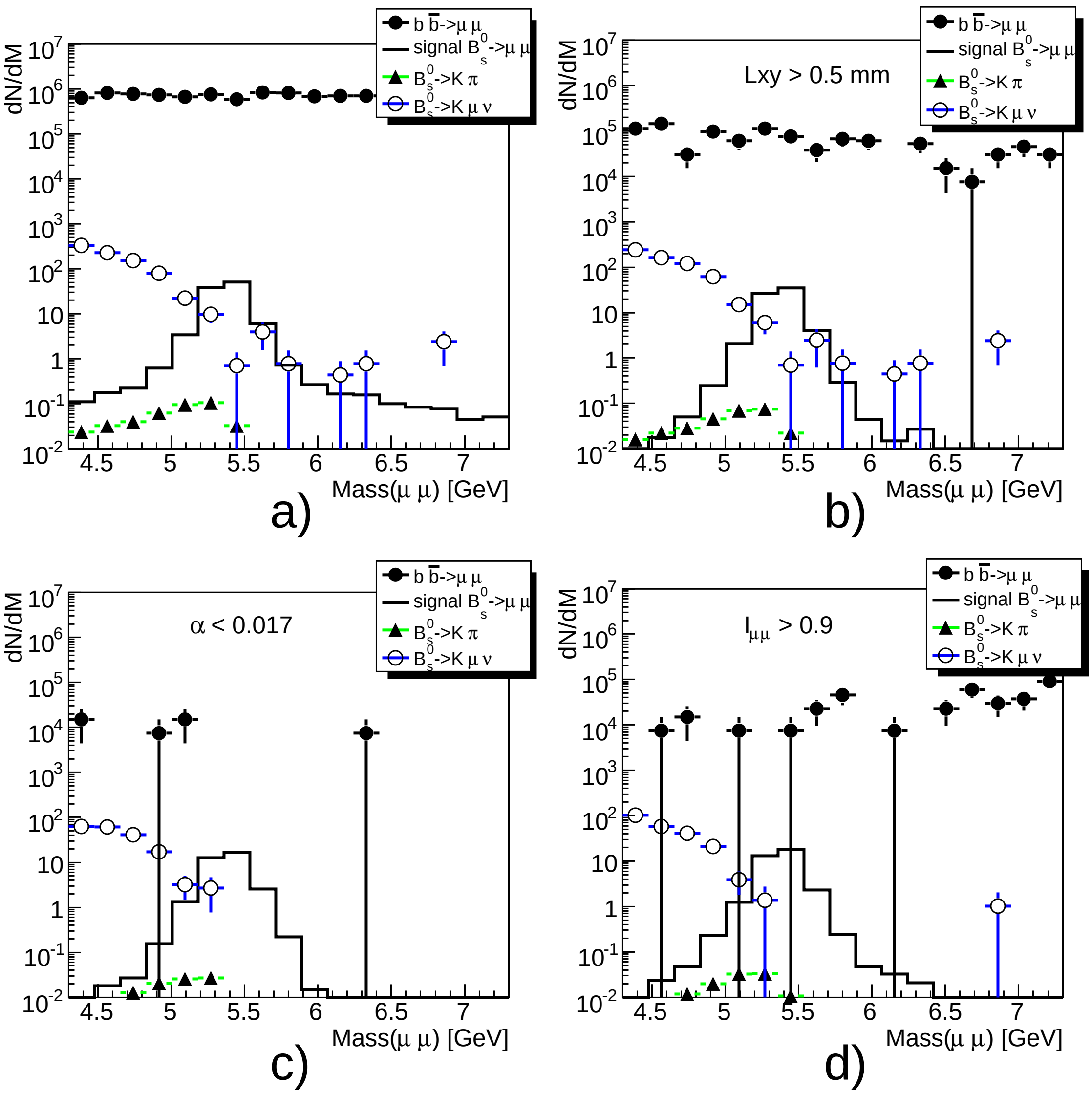


Figure 7: The Monte Carlo di-muon mass distributions for signal $B_s^0 \to \mu^+\mu^-$ (histogram), combinatorial background (closed circles) and non-combinatorial background (open circles and triangles) for (a) preselected events, (b) after cuts on transverse decay length, (c) pointing angle and (d) isolation. The number of events has been scaled to 10 fb$^{-1}$ of integrated luminosity.

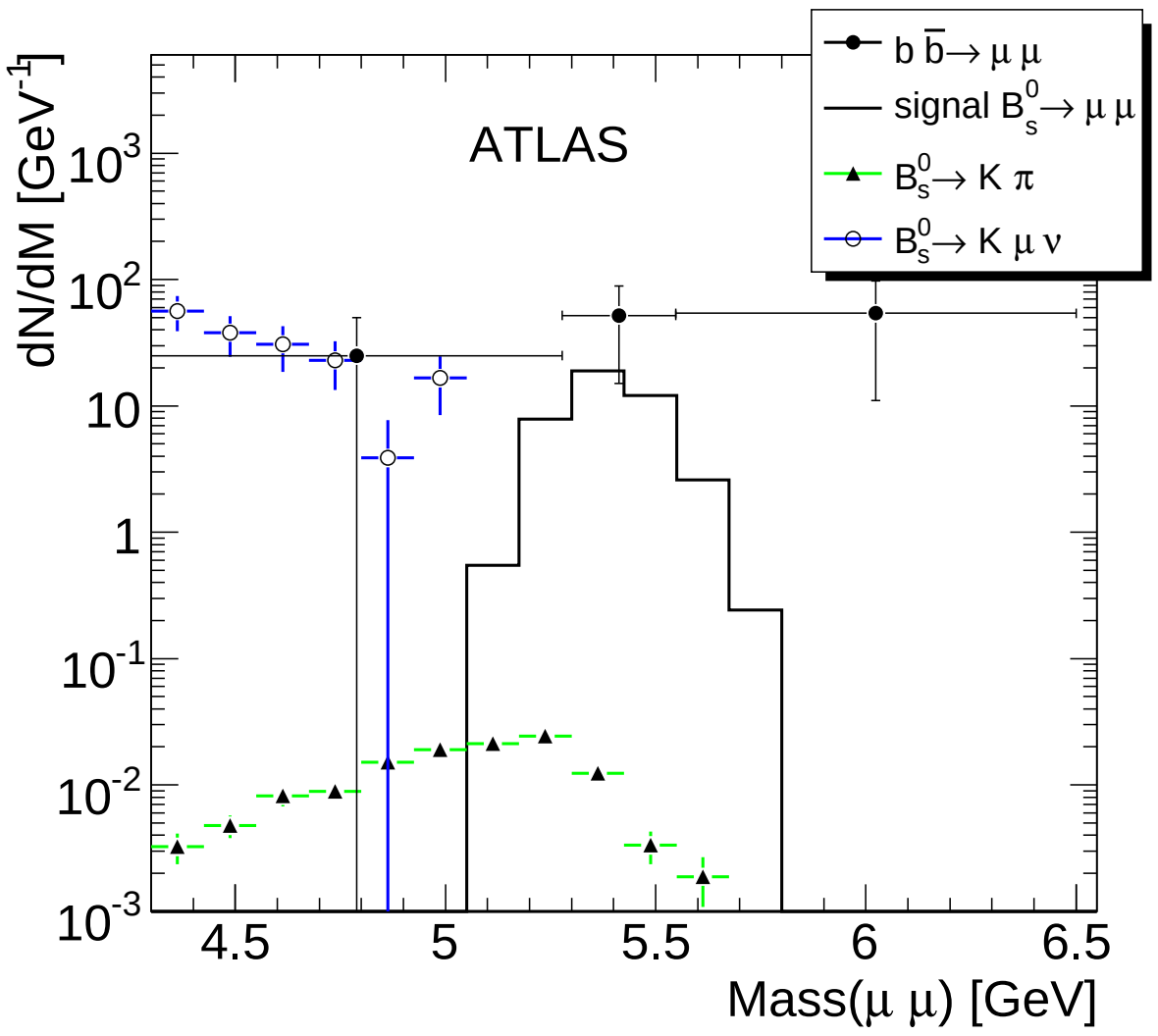


Figure 8: Dimuon mass distribution for surviving events after applying all three cuts. The signal is shown as histogram, combinatorial background by closed circles and non-combinatorial backgrounds by opened circles and triangles. The combinatorial background is estimated assuming a factorisation of applied cuts. Statistics are given for an integrated luminosity 10 fb$^{-1}$.

Table 5: Selection efficiencies and number of signal and background events for integrated luminosity of 10 fb$^{-1}$. Preselection criteria used: 4 GeV $< M(\mu\mu) <$ 7.3 GeV , vertex fit $\chi^2 < 10$ , transverse decay length $L_{xy} <$ 20 mm. Numbers of expected events are computed according to the Standard Model expectation. In the left column for the $b\bar{b} \to \mu\mu X$ background the efficiencies are given separately for each cut, in the right column the combined efficiency is given for the cuts on the pointing angle and on the transverse decay length.

| Selection cut | $B_s^0 \to \mu^+\mu^-$ efficiency | $bb \to \mu^+\mu^- X$ efficiency | |
|---|---|---|---|
| $I_{\mu\mu} > 0.9$ | 0.24 | $(2.6 \pm 0.3) \cdot 10^{-2}$ | |
| $L_{xy} > 0.5$mm | 0.26 | $(1.4 \pm 0.1) \cdot 10^{-2}$ | $(1.0 \pm 0.7) \cdot 10^{-3}$ |
| $\alpha < 0.017$ rad | 0.23 | $(8.5 \pm 0.2) \cdot 10^{-3}$ | |
| Mass in $[-\sigma, 2\sigma]$ | 0.76 | 0.079 | |
| TOTAL | 0.04 | $0.24 \cdot 10^{-6}$ | $(2.0 \pm 1.4) \cdot 10^{-6}$ |
| Events yield | 5.7 | | $14^{+13}_{-10}$ |

## 5 Systematic uncertainties

There are several sources of uncertainty in this analysis. Some of them are relevant only for the Monte Carlo study, whilst others should be taken into account with the real data analysis as well.

In this presented analysis, the expected number of signal and background events is estimated by counting directly instead of the normalisation procedure described in Section 3, which is supposed to be used in a real experiment. Consequently the dimuon trigger and reconstruction efficiencies, and acceptance, are not canceled by those from the reference channel and must explicitly be taken into account. Using the methods developed by ATLAS [12] this systematic uncertainly is estimated to be of about few percent.

There is a theoretical uncertainty of a factor of two in the b-production cross-section at LHC energies,

which clearly affects the Monte Carlo predictions. Consequently all numbers of events from Table 5 scale accordingly.

The difference between the real and simulated kinematic properties of detected particles (e.g., due to not fully accounting for misalignment or material effects) can also introduce a bias in our predictions. However Figure 2 shows that the most of muons from $B_s^0 \to \mu^+\mu^-$ have a $p_T$ in the region of the efficiency plateau so the possible deformation of $p_T$ (as well as $\eta$) spectra could change the resulting efficiency by not more than a few percent. The uncertainty from the cuts factorisation hypothesis is assumed to be approximately 50%.

Some uncertainties will, to a large extent, cancel if we use a normalisation channel $B^+ \to J/\psi K^+$ to estimate the $\mathscr{B}(B_s^0 \to \mu^+\mu^-)$ as the dimuon trigger conditions are similar for both channels. Without experimental data, it is difficult to estimate the uncertainty in the number of background events in a given mass range. In the D0 analysis [17], the sideband extrapolation method is estimated to give an uncertainly of 20-30%. Indeed, this is the main source of uncertainty in the D0 analysis of $B_s^0 \to \mu^+\mu^-$. Corrections should also be made for the contribution of $B_d^0 \to \mu\mu$ decay in the experimental sample.

In total, the uncertainty from the systematic errors discussed in this Section is approximately $\pm 25\%$. In addition, the procedure adopted to estimate backgrounds via cut factorisation has large uncertainties, which are estimated to be of the order of 50%, as discussed above. In addition, a 70% uncertainty arises from Monte Carlo statistics. Overall, we choose to combine these in quadrature to obtain an indicative overall uncertainty on the background of +90%/-75%.

## 6 Start-up strategy

Already at 1 fb$^{-1}$of integrated luminosity, ATLAS can have $O(10^6)$ of dimuon events in a mass window 4 GeV $< M(\mu\mu) <$ 7 GeV (after vertexing and quality cuts). It will allow tuning of the selection procedure either for the cut-based analysis or for the multivariate methods for the background discrimination. Events that survive background suppression will be used to estimate the background contribution to the signal search region. The contribution from combinatorial background will be estimated using the side-bands interpolation procedure. The contribution from exclusive backgrounds due to fake dimuons from hadronic two-body $B$ decays or from $B_s^0 \to K^-\mu^+\nu$, will be determined on the basis of the study of the hadron/muon misidentification probability. The background estimation will be compared with the number of events observed in the signal region. Following this information an upper limit on the number of signal events $N_B$ corresponding to certain confidence level will be determined, using appropriate statistical methods. Finally, the value of $N_B$ will be used to extract the upper limit on the $B_s^0 \to \mu^+\mu^-$ branching fraction, $\mathscr{B}(B_s^0 \to \mu^+\mu^-)$, using a reference channel $B^+ \to J/\psi K^+$. In this procedure a ratio of geometric and kinematical acceptances of the signal and the reference channels will be determined from the Monte Carlo simulation. Trigger and offline reconstruction efficiencies largely cancel for dimuons in these channels. The reference channel $B^+ \to J/\psi K^+$ will also be used to check the Monte Carlo simulation. The efficiency of the final selection cuts on discriminating variables for the signal $B_s^0 \to \mu^+\mu^-$ will be determined using Monte Carlo simulation (validated with the reference channel).

## 7 Conclusions

We have presented the strategy for searching for the rare decay $B_s^0 \to \mu^+\mu^-$ with the ATLAS detector. Whilst we do not expect to observe this decay during the early stages of the LHC, as more luminosity becomes available and our understanding of the backgrounds improves, it should be possible to identify a signal for the process. There are uncertainties due to the relatively unknown beauty production cross-section at the LHC, and also the limited Monte Carlo statistics available for this study. Within these

limitations, assuming the Standard Model, we expect a signal of 5.7 events with a background of $14^{+13}_{-10}$ events for an integrated luminosity of 10 fb$^{-1}$. It is evident that the background uncertainties are large in this study. However, background estimates based on real data will be able to make use of much higher statistics and these will provide reduced uncertainties, as well as allowing the evaluation of more sophisticated methods of analysis.

# Trigger and Analysis Strategies for $B_s^0$ Oscillation Measurements in Hadronic Decay Channels

**Abstract**
The capabilities of measuring $B_s^0$ oscillations in proton-proton interactions with the ATLAS detector at the Large Hadron Collider are evaluated. $B_s^0$ candidates in the $D_s^- \pi^+$ and $D_s^- a_1^+$ decay modes from semileptonic exclusive events are simulated and reconstructed using a detailed detector description and the ATLAS software chain. For the measurement of the oscillation frequency a $\Delta m_s$ sensitivity limit of 29.6 ps$^{-1}$ and a five standard deviation measurement limit of 20.5 ps$^{-1}$ are derived from unbinned maximum likelihood amplitude fits for an integrated luminosity of 10 fb$^{-1}$ . The initial flavour of the $B_s^0$ meson is tagged exclusively with opposite-side leptons. Trigger strategies are proposed for scenarios of different instantaneous luminosities in order to maximise the signal channel trigger efficiencies.

## 1 Introduction

As tests of the Standard Model the CP–violation parameter $\sin(2\beta)$ will be measured with high precision (at the percent level) as well as properties of the $B_s^0$-meson system, like the mass difference of the two mass eigenstates $\Delta m_s$, the lifetime difference $\Delta\Gamma_s/\Gamma_s$ and the weak mixing phase $\phi_s$ induced by CP-violation, with $\phi_s \approx 2\lambda^2\eta$ in the Wolfenstein parametrisation. The different masses of the CP–eigenstates $B_s^L$ (CP–even) and $B_s^H$ (CP–odd) give rise to $B_s$ mixing. The observed $B_s^0$ and $\bar{B}_s^0$ particles are linear combinations of these eigenstates, where transitions are allowed due to non–conservation of flavour in weak–current interactions and will occur with a frequency proportional to $\Delta m_s$. $B_s^0$ oscillations have been observed at the Fermilab Tevatron collider by the CDF collaboration [1] measuring a value of $\Delta m_s = (17.77 \pm 0.10\,(\mathrm{stat}) \pm 0.07\,(\mathrm{sys}))\,\mathrm{ps}^{-1}$ and D0 collaboration [2] reporting a two-sided bound on the $B_s^0$ oscillation frequency with a range of 17 ps$^{-1}$ $< \Delta m_s <$ 21 ps$^{-1}$. Both results are consistent with Standard Model expectations [3]. In ATLAS, the $\Delta m_s$ measurement is an important baseline for the B-physics program and an essential ingredient for a precise determination of the phase $\phi_s$. CP–violation in $B_s^0$-$\bar{B}_s^0$ mixing is a prime candidate for the discovery of non–standard–model physics. For the channel $B_s^0 \to J/\psi\phi$, which has a clean experimental signature, a very small CP-violating asymmetry is predicted in the Standard Model. The measurement of any sizeable effect of the weak–interaction–induced phase $\phi_s$ in the CKM matrix, which lies above the predicted value, would indicate that processes beyond the Standard Model are involved. Furthermore, the determination of important parameters in the $B_s^0$ meson system will be valuable input for flavour dynamics in the Standard Model and its extensions.

In this note an estimation of the sensitivity to measure the $B_s^0$-$\bar{B}_s^0$ oscillation frequency with the ATLAS detector is presented. The signal channels considered are the hadronic decay channels $B_s^0 \to D_s^- \pi^+$ and $B_s^0 \to D_s^- a_1^+$ with $D_s^- \to \phi\pi^-$ followed by $\phi \to K^+K^-$. In the case of $B_s^0 \to D_s^- a_1^+$ the $a_1^+$ decays as $a_1^+ \to \rho\pi^+$ with $\rho \to \pi^+\pi^-$. Including the sub-decay $D_s^- \to K^{*0}K^-$ [4] would increase the event statistics by about 30%. However, for these sub-channels, which require an additional trigger signature, the increase of the overall trigger rate would be unacceptable. Detailed information of the signal and the exclusive background channels is given in Section 2. The high event rate at the Large Hadron Collider (LHC) imposes very selective requirements onto the B-physics trigger strategies, reducing the rate by about six orders of magnitude for recording events. Since an initial "low-luminosity" running period is scheduled with a luminosity starting at $10^{31}$ cm$^{-2}$s$^{-1}$ and rising to $2 \cdot 10^{33}$ cm$^{-2}$s$^{-1}$, followed later on by the design luminosity of the LHC of $10^{34}$ cm$^{-2}$s$^{-1}$, the B-trigger must be flexible enough to cope with the

increasing luminosity conditions. The overall B-trigger strategy as well as the different strategies dealing with the luminosity scenarios in the initial running periods are discussed in Section 3. An important part of the mixing measurement is to identify the flavour at production, i.e., whether the observed $B_s$ meson initially contained a $b$ or a $\bar{b}$ quark. A detailed description of an opposite-side lepton flavour tag and of the various sources of the wrong tag fractions is given in Section 4. The selection of $B_s^0$ candidates with kinematic cuts as well as mass resolutions of the $B_s^0$, are explained and shown in Section 5. A luminosity of $10^{33}$ cm$^{-2}$s$^{-1}$ and no pileup is considered for the detailed analysis of signal and background channels. Strategies for lower and higher luminosities are also discussed in the same section. The results of the signal-candidate selection are used as input to a toy Monte Carlo simulation generating a sample of $B_s^0$ candidates, which is used for the amplitude fit method [5] to obtain the $\Delta m_s$ measurement limits. The construction of the likelihood function, the Monte Carlo sample and the extraction of the $\Delta m_s$ sensitivity are discussed in Section 6.

## 2 Simulated Data Samples

Simulated $b$-quark pairs are generated using PYTHIA [6], with the $\bar{b}$-quark required to decay to one of the specified signal channels. The $b$-quark decays semileptonically producing a muon with $p_T > 6$ GeV within $|\eta| < 2.5$. Details on generation, simulation and reconstruction of the simulated data samples are given in the introduction of the B-chapter [7].

In addition to the simulated signal samples $B_s^0 \to D_s^-(\phi\pi^-)\pi^+$ and $B_s^0 \to D_s^-(\phi\pi^-)a_1^+$, corresponding exclusive background channels that give an irreducible contribution to the selected $B_s^0$ signal were investigated. Two $B_d^0$ decay channels, $B_d^0 \to D^-\pi^+/a_1^+$ and $B_d^0 \to D_s^+\pi^-/a_1^-$, and one $B_s^0$ channel, $B_s^0 \to D_s^{*-}\pi^+/a_1^+$, were simulated for both hadronic decay channels. The dedicated trigger studies described in Section 3 require additional samples, such as the inclusive background channels $b\bar{b} \to \mu 6X$, $b\bar{b} \to \mu 4X$ and $c\bar{c} \to \mu 4X$ containing semileptonic $b$ or $c$ decays requiring one muon with a generated $p_T > 4$ GeV (or 6 GeV) and further decay products ($X$). Also, one particular signal sample (as a choice $B_s^0 \to D_s^- a_1^+$) requiring one muon with a generated $p_T > 4$ GeV (identified by ($\mu 4$)) is used for the trigger studies. A sample of minimum bias events is used for the determination of overall trigger rates. See Table 1 for the number of events generated and the cross-sections calculated from the values given by PYTHIA and the appropriate branching ratios [8]. Errors on the cross-sections include statistical errors and contributions from the uncertainties on the branching ratios.

Effects of pileup and $B$-meson mixing were not included in the simulation of any of the samples.

## 3 Trigger Strategies

The trigger strategy used for the $B_s^0 \to D_s^-\pi^+$ and $B_s^0 \to D_s^- a_1^+$ channels is to identify the $D_s^\pm$ decaying to $\phi(\to K^+K^-)\,\pi$, which is common to both decay channels. At level one (LVL1) a muon is required to enrich the content of the triggered data sample with B-events. The high level trigger (HLT) is split into level two (LVL2) and Event Filter (EF). A search for a $D_s^\pm$ is performed following one of two strategies. The first method, the FullScan approach, performs reconstruction of tracks within the entire Inner Detector. It is an efficient method, but time consuming and its feasibility depends on the background event rate. The second method performs track reconstruction in a limited volume of the Inner Detector only, which is defined by a low-$p_T$ jet region of interest (RoI) identified at LVL1. This RoI-based method is faster but there is a loss in efficiency due to the requirement of a LVL1 jet RoI in the event and the geometrical restriction to the RoI.

The increase of the luminosity after LHC startup affects the trigger in two ways: the trigger rates for the jet and muon trigger will increase, seeding the HLT $D_s^\pm$ algorithm more frequently, and combinatorial

Table 1: Number of events generated and calculated cross-sections for the different signal and exclusive background simulated data samples for the $B_s^0 \to D_s^- \pi^+$ and $B_s^0 \to D_s^- a_1^+$ analysis and particular samples used for dedicated trigger studies. Branching ratios for particle decays into final states are included. $^{*)}$The branching fraction has not been measured yet, only an upper limit exists.

| | Channel | Events | Cross-section [pb] |
|---|---|---|---|
| Signal | $B_s^0 \to D_s^- \pi^+$ | 88 450 | $10.4 \pm 3.5$ |
| | $B_s^0 \to D_s^- a_1^+$ | 98 450 | $5.8 \pm 3.2$ |
| Background | $B_d^0 \to D_s^+ \pi^-$ | 43 000 | $0.2 \pm 0.1$ |
| | $B_d^0 \to D^- \pi^+$ | 41 000 | $6.2 \pm 1.1$ |
| | $B_s^0 \to D_s^{*-} \pi^+$ | 40 500 | $9.1 \pm 2.8$ |
| | $B_d^0 \to D_s^+ a_1^-$ | 50 000 | $< 8.9^{*)}$ |
| | $B_d^0 \to D^- a_1^+$ | 50 000 | $3.7 \pm 2.1$ |
| | $B_s^0 \to D_s^{*-} a_1^+$ | 100 000 | $12.1 \pm 2.7$ |
| Trigger | $B_s^0 \to D_s^- a_1^+ \ (\mu 4)$ | 50 000 | $13.6 \pm 7.6$ |
| | $b\bar{b} \to \mu 6X$ | 242 150 | $(6.14 \pm 0.02) \cdot 10^6$ |
| | $b\bar{b} \to \mu 4X$ | 98 450 | $(19.08 \pm 0.30) \cdot 10^6$ |
| | $c\bar{c} \to \mu 4X$ | 44 750 | $(26.28 \pm 0.09) \cdot 10^6$ |
| | minimum bias | 2 623 060 | $70 \cdot 10^9$ |

background from pileup affects the performance of the selection algorithm. Therefore, trigger menus corresponding to different LHC luminosities are discussed in Sections 3.4 to 3.6.

The trigger efficiencies are presented for the $B_s^0 \to D_s^- a_1^+$ channel. Results for the $B_s^0 \to D_s^- \pi^+$ channel are expected to be similar within a few percent (see Section 3.2).

## 3.1 LVL1 Trigger Selection

The ATLAS hardware allows three LVL1 low-$p_T$ muon trigger thresholds to be defined at once, which can only be adjusted between runs by reconfiguring the lookup tables implemented in the muon trigger firmware. In order to study more than three thresholds we investigated the two available, pre-defined menus (named A and B) with respect to the low-$p_T$ muon trigger thresholds[1].

The three implemented low-$p_T$ thresholds for trigger menu A are: 0 GeV[2] (named `MU00`), 5 GeV (`MU05`) and 6 GeV (`MU06`). For trigger menu B, the three low-$p_T$ thresholds are 6 GeV (`MU06`), 8 GeV (`MU08`), and 10 GeV (`MU10`). Figure 1 shows the efficiencies of the low-$p_T$ LVL1 muon trigger signatures as a function of the true $p_T$ of the muon with the highest $p_T$ in the event.

The LVL1 trigger efficiency depends strongly on the threshold chosen for the transverse momentum of the muon as shown in Table 2. Note that there is a discrepancy between the `MU06` efficiencies from both trigger menus, which will be taken as a systematic uncertainty of the current implementation. Although these dedicated trigger studies have been performed with the $B_s^0 \to D_s^- a_1^+$ sample, the LVL1 efficiencies for $B_s^0 \to D_s^- \pi^+ (\mu 6)$ have been checked and agree well with those in Table 2.

The input to the LVL1 calorimeter trigger is a set of $\sim$7 200 trigger towers with granularity $\Delta\phi \times \Delta\eta \approx 0.1 \times 0.1$ formed by the analogue summation of calorimeter cells. There are separate sets of trigger towers for the EM and hadronic calorimeters. The LVL1 jet algorithm employed here uses a

[1]All presented trigger thresholds are meant to be inclusive, i.e. to include all events fulfilling a trigger signature with a $p_T$ threshold equal to or higher than the indicated one.

[2]This requires a coincidence between the muon chambers without an actual threshold applied. Due to the detector geometry this corresponds to an effective transverse momentum threshold of about 4 GeV.

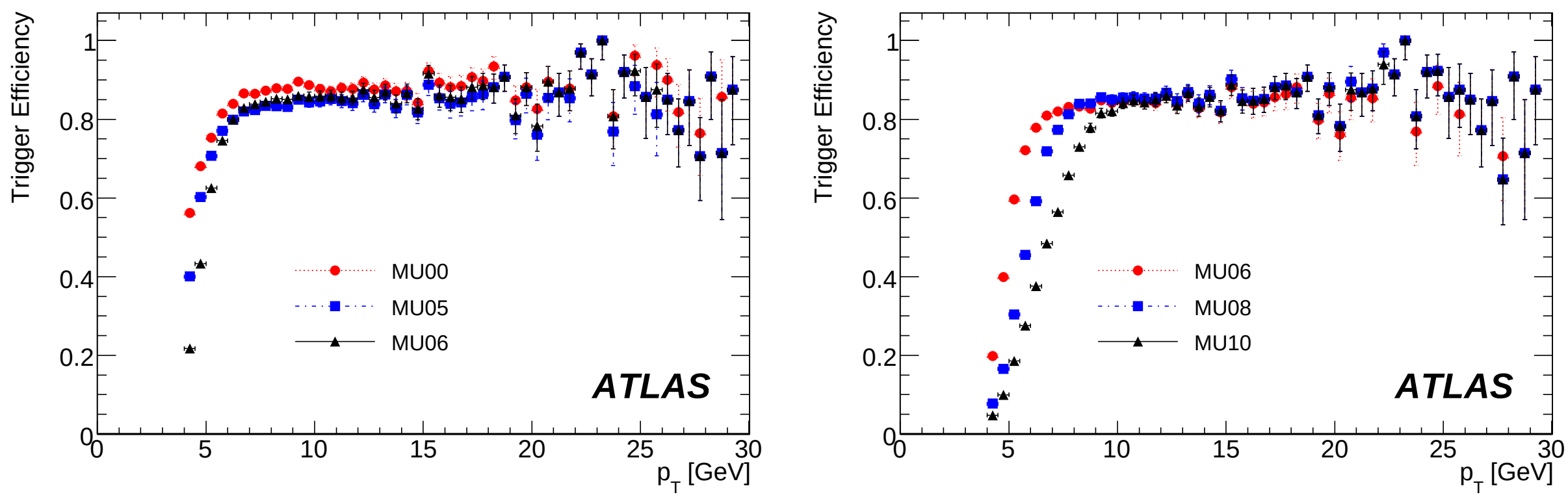


(a) Thresholds for menu A [9]    (b) Thresholds for menu B

Figure 1: Muon trigger efficiency as a function of the true $p_T$ of the muon with the highest $p_T$ in the event for the $B_s^0 \to D_s^- a_1^+(\mu 4)$ for (a) trigger menu A and (b) trigger menu B.

Table 2: LVL1 muon trigger efficiencies for the signal datasets $B_s^0 \to D_s^- a_1^+(\mu 4)$ and $B_s^0 \to D_s^- a_1^+(\mu 6)$ and the exclusive background samples. The first three lines refer to trigger menu A [9], while the last three lines refer to trigger menu B. For $b\bar{b} \to \mu 4X$, $b\bar{b} \to \mu 6X$ and $c\bar{c} \to \mu 4X$, the Monte Carlo data samples are only available using trigger menu A.

| Menu | Threshold | Efficiency [%] | | | | |
|---|---|---|---|---|---|---|
| | | $B_s^0 \to D_s^- a_1^+$ ($\mu 4$) | $B_s^0 \to D_s^- a_1^+$ ($\mu 6$) | $bb \to \mu 4X$ | $bb \to \mu 6X$ | $c\bar{c} \to \mu 4X$ |
| A | MU00 | $75.65 \pm 0.19$ | $86.77 \pm 0.15$ | $71.74 \pm 0.14$ | $86.60 \pm 0.07$ | $70.42 \pm 0.22$ |
| A | MU05 | $68.41 \pm 0.21$ | $82.60 \pm 0.17$ | $63.51 \pm 0.15$ | $81.91 \pm 0.08$ | $62.05 \pm 0.23$ |
| A | MU06 | $58.93 \pm 0.22$ | $81.90 \pm 0.17$ | $52.28 \pm 0.16$ | $81.00 \pm 0.08$ | $50.44 \pm 0.24$ |
| B | MU06 | $61.15 \pm 0.22$ | $83.83 \pm 0.16$ | — | — | — |
| B | MU08 | $44.78 \pm 0.22$ | $77.64 \pm 0.19$ | — | — | — |
| B | MU10 | $34.89 \pm 0.21$ | $65.47 \pm 0.21$ | — | — | — |

cluster of $\Delta\phi \times \Delta\eta$ of approximately $0.4 \times 0.4$ (corresponding to $4 \times 4$ trigger towers). The projections of the vectors of the energy depositions onto the plane perpendicular to the beam axis (transverse energy, $E_T$) are summed over both the electromagnetic and the hadronic layers. The jet algorithm moves the cluster template in steps of 0.2 across the $\phi \times \eta$ plane. An RoI is produced if the $4 \times 4$ cluster is a local $E_T$ maximum (as defined in [10]) and the cluster $E_T$ sum is greater than the required threshold. The jet RoI is usable if the average number of RoIs per event (RoI multiplicity, see Fig. 2 and Table 3) is small, ideally about 1-2. Clearly, a compromise is required as an increased threshold will reduce the multiplicity, but will also give a reduced efficiency for finding the $B$ jet in an event.

For a transverse energy threshold of 4 GeV, which is implemented to initiate the LVL2 $D_s^{\pm}$ trigger in trigger menus A and B, the jet trigger has an acceptance of $(98.36 \pm 0.06)\,\%$ based on all events in the $B_s^0 \to D_s^- a_1^+$ sample.

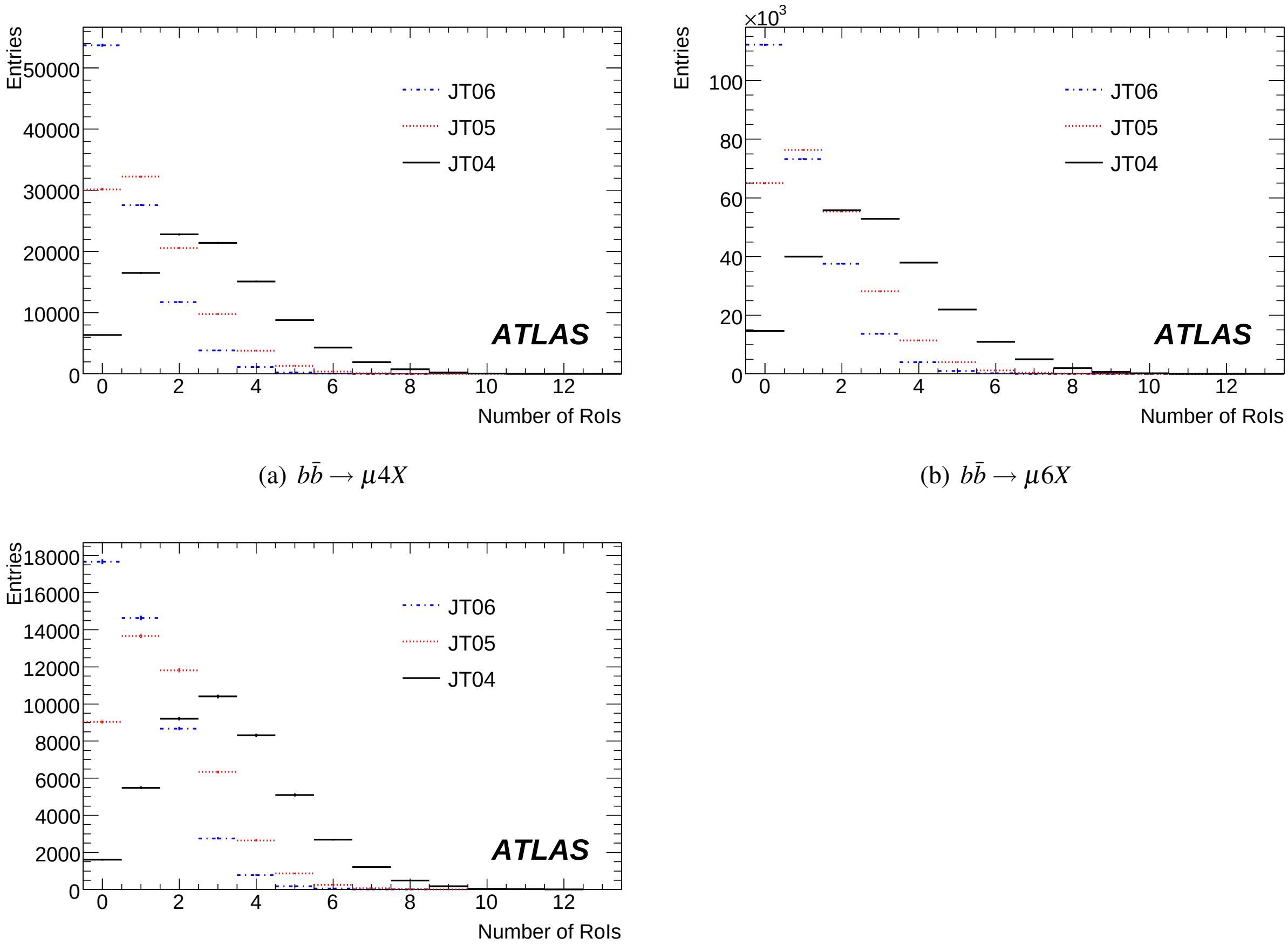


(a) $b\bar{b} \to \mu 4X$

(b) $b\bar{b} \to \mu 6X$

(c) $c\bar{c} \to \mu 4X$

Figure 2: RoI multiplicity distributions for the background samples (a) for $b\bar{b} \to \mu 4X$, (b) for $b\bar{b} \to \mu 6X$ and (c) for $c\bar{c} \to \mu 4X$ as a function of the jet RoI energy threshold [9]. Only RoIs with $\eta < 2.4$ have been taken into account. This corresponds to the requirement that the RoI is to be contained within the solid angle covered by the Inner Detector.

Table 3: Mean and root mean square of the RoI multiplicity distributions (Figure 2) for the background samples as a function of the jet RoI transverse energy ($E_T$) threshold [9]. Only RoIs with $\eta < 2.4$ have been taken into account. A strong anticorrelation between the $E_T$ threshold and the mean RoI multiplicity is observed.

| Threshold | $b\bar{b} \to \mu 4X$ | | $b\bar{b} \to \mu 6X$ | | $c\bar{c} \to \mu 4X$ | |
|---|---|---|---|---|---|---|
| [ GeV] | Mean | RMS | Mean | RMS | mean | RMS |
| 4 | 2.847 | 1.746 | 2.883 | 1.754 | 3.235 | 1.759 |
| 5 | 1.301 | 1.244 | 1.441 | 1.295 | 1.643 | 1.300 |
| 6 | 0.703 | 0.952 | 0.881 | 1.046 | 0.998 | 1.048 |
| 7 | 0.454 | 0.786 | 0.634 | 0.911 | 0.703 | 0.900 |

### 3.2 LVL2 Trigger Selection

The first step of LVL2 is to confirm the LVL1 muon trigger decision using more precise muon momentum measurement. Secondly, information from Inner Detector and muon chambers are combined to give a further improvement in the momentum measurement.

As described above, the LVL2 tracking can be run in either FullScan or RoI-guided modes. In both cases, the same algorithm (named `DsPhiPi`) is used to combine the reconstructed tracks and search first for a $\phi$ and then for a $D_s^\pm$. In the RoI-guided approach tracks are reconstructed in a region $\Delta\phi \times \Delta\eta = 1.5 \times 1.5$ around all jet RoIs with $E_T$ above a certain programmable threshold [11]. A $p_T$ cut of 1.4 GeV is applied to all reconstructed tracks.

Opposite sign track pairs are considered as a $\phi$ candidate if they pass the following cuts: $|\Delta z| < 3$ mm, where $z$ is the distance along the beam line of the track's point of closest approach to the centre of the detector, $|\Delta\phi| < 0.2$ and $|\Delta\eta| < 0.2$.

The tracks are combined using a $K$ mass hypothesis and a cut around the $\phi$ mass $m_\phi(\text{PDG}) = 1019.46$ MeV [8] is applied. Track pairs passing the cut are then combined with all other tracks assuming a $\pi$ mass for the third track. An event is selected if the mass of the track triplet is close to the $D_s^\pm$ mass $m_{D_s}(\text{PDG}) = 1968.2$ MeV. The mass cuts used are 1005 MeV $< m_{KK} <$ 1035 MeV for the $\phi$ candidates and 1908 MeV $< m_{KK\pi} <$ 2028 MeV for the $D_s$ candidates.

The LVL2 track fit masses are shown in Figure 3 and Table 4 for the RoI-guided approach and FullScan. The standard deviations obtained from the Gaussian fits show that the mass cuts used correspond to 3.0 standard deviations for the $\phi$ mass distribution and 2.8 standard deviations for the $D_s^\pm$ mass distribution [9]. The results for the RoI-based approach and those for FullScan agree well.

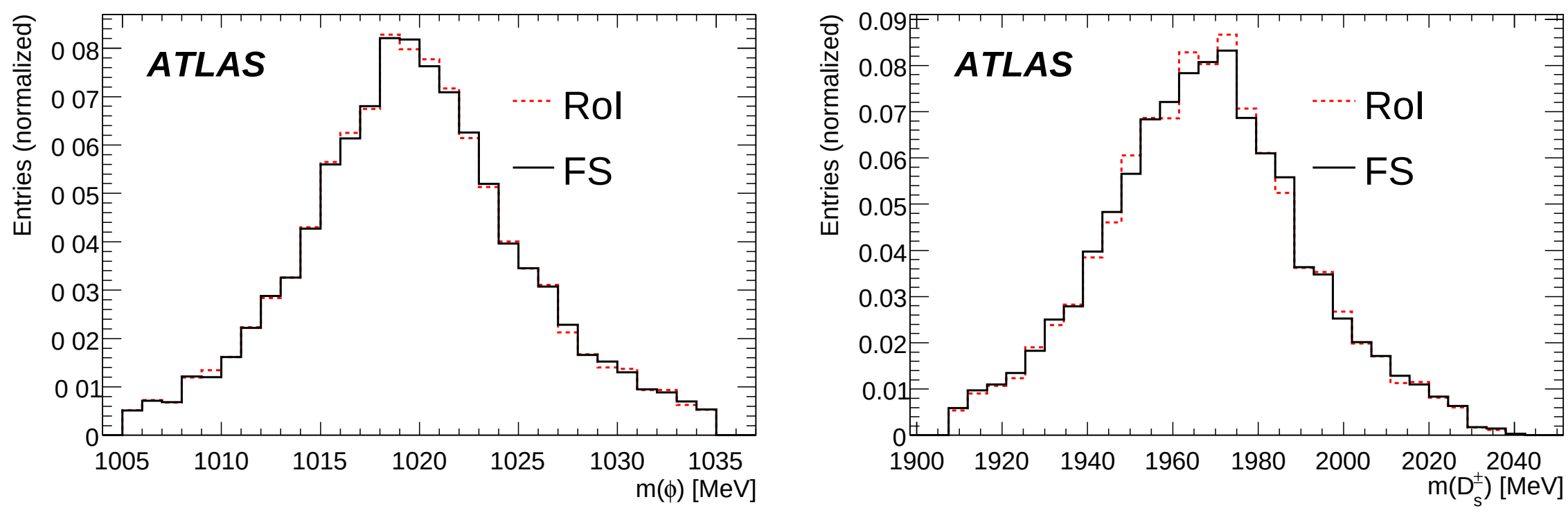


(a) True $\phi$ candidates (b) True $D_s^\pm$ candidates

Figure 3: LVL2 track fit mass distributions of (a) $\phi$ and (b) $D_s^\pm$ candidates (corresponding to a $\phi$ or $D_s^\pm$ particle from the signal decay in the Monte Carlo truth information) for the FullScan- and RoI-based LVL2 trigger signatures from $B_s^0 \to D_s^- a_1^+$ events fulfilling the respective LVL2 $D_s^\pm$ trigger signature and `MU06` [9].

The acceptances of possible trigger strategies up to LVL2 are given in Table 5 for the signal samples and in Table 6 for the background datasets. The LVL2 trigger rates for the $B_s^0 \to D_s^- \pi^+$ channel are expected to be lower by a few percent since the average $p_T$ of the $B_s^0$ candidates and consequently the average $p_T$ of the $D_s$ candidates is smaller for the $B_s^0 \to D_s^- \pi^+$ channel than for the $B_s^0 \to D_s^- a_1^+$ channel due to different track selections (see Fig. 5 and Section 5.1).

Table 4: LVL2 track fit masses for the FullScan- and RoI-based LVL2 trigger signatures (only candidates corresponding to a $\phi$ or $D_s^{\pm}$ particle from the signal decay in the Monte Carlo truth information) from $B_s^0 \to D_s^- a_1^+$ events fulfilling the respective LVL2 $D_s^{\pm}$ trigger and MU06. The table shows the results of Gaussian fits within the trigger mass windows to the mass distributions from Fig. 3. The results for both trigger strategies agree within statistical errors.

| | FullScan-based LVL2 trigger [9] | RoI-based LVL2 trigger |
|---|---|---|
| m($\phi$): mean [MeV ] | 1019.55±0.05 | 1019.52±0.05 |
| m($\phi$): std. dev. [MeV ] | 5.07±0.06 | 5.04±0.05 |
| m($D_s^{\pm}$): mean [MeV ] | 1966.9±0.3 | 1967.0±0.3 |
| m($D_s^{\pm}$): std. dev. [MeV ] | 21.7±0.3 | 21.5±0.3 |

Table 5: Acceptances of LVL2 (RoI and FullScan, FS) for the $B_s^0 \to D_s^- a_1^+$ sample for trigger menus A and B.

| Menu A | | | Menu B | | |
|---|---|---|---|---|---|
| Trigger scenario | Passes (in %) ($\mu$6) | Passes (in %) ($\mu$4) | Trigger scenario | Passes (in %) ($\mu$6) | Passes (in %) ($\mu$4) |
| Events | 50 000 | 50 000 | | 50 000 | 50 000 |
| L2_mu0 | 85.19±0.16 | 72.05±0.20 | L2_mu6 | 77.13±0.19 | 35.03±0.21 |
| L2_mu5 | 79.65±0.18 | 49.39±0.22 | L2_mu8 | 45.54±0.22 | 18.96±0.18 |
| L2_mu6 | 75.66±0.19 | 34.41±0.21 | L2_mu10 | 26.04±0.20 | 10.91±0.14 |
| FS & L2_mu0 | 32.98±0.21 | 22.93±0.19 | FS & L2_mu6 | 29.99±0.21 | 12.71±0.15 |
| FS & L2_mu5 | 30.79±0.21 | 16.75±0.17 | FS & L2_mu8 | 19.14±0.18 | 10.91±0.14 |
| FS & L2_mu6 | 29.38±0.20 | 12.49±0.15 | FS & L2_mu10 | 11.83±0.15 | 4.92±0.10 |
| RoI & L2_mu0 | 28.74±0.20 | 19.14±0.18 | RoI & L2_mu6 | 26.19±0.20 | 11.10±0.14 |
| RoI & L2_mu5 | 26.88±0.20 | 14.26±0.16 | RoI & L2_mu8 | 17.09±0.17 | 7.05±0.12 |
| RoI & L2_mu6 | 25.68±0.20 | 10.91±0.14 | RoI & L2_mu10 | 10.80±0.14 | 4.56±0.09 |

Table 6: Acceptances of LVL2 (RoI and FullScan) for the background samples $b\bar{b} \to \mu 4X$, $b\bar{b} \to \mu 6X$ and $c\bar{c} \to \mu 4X$ (trigger menu A).

| Trigger scenario | passes (in %) ($b\bar{b} \to \mu 4X$) | passes (in %) ($b\bar{b} \to \mu 6X$) | passes (in %) ($c\bar{c} \to \mu 4X$) |
|---|---|---|---|
| events | 98 450 | 242 150 | 44 750 |
| FS & L2_mu0 | 2.00±0.05 | 3.73±0.05 | 2.71±0.08 |
| FS & L2_mu5 | 1.47±0.04 | 3.48±0.05 | 1.93±0.06 |
| FS & L2_mu6 | 1.11±0.03 | 3.32±0.05 | 1.43±0.05 |
| RoI & L2_mu0 | 1.73±0.04 | 3.34±0.05 | 2.34±0.07 |
| RoI & L2_mu5 | 1.30±0.04 | 3.12±0.05 | 1.73±0.06 |
| RoI & L2_mu6 | 1.01±0.03 | 2.99±0.05 | 1.32±0.05 |

### 3.3 Event Filter Selection

The muon confirmation at the Event Filter (EF) employs a muon track reconstruction algorithm using muon detector data only, similar to the algorithm used for offline reconstruction.

The EF $D_s^{\pm}$ selection is very similar to that at LVL2. The track reconstruction can be performed in FullScan or RoI-guided modes, which share a common EF signature. The search for $\phi$ and $D_s^{\pm}$ particles currently uses the same mass cuts as at LVL2, even though better mass resolutions are expected for the EF than for LVL2. In the future the mass cuts might be tightened and additional selection cuts will be added as discussed in section 3.5. EF output rates, which are only available for the minimum bias sample, are discussed in the following subsections.

### 3.4 Trigger Strategies for Early Running

At the lowest luminosities ($10^{31}$ cm$^{-2}$s$^{-1}$ and $10^{32}$ cm$^{-2}$s$^{-1}$), the trigger selection needs to be as efficient as possible, which means running a loose trigger. To estimate rates and perform timing studies a trigger menu with a different set of muon thresholds [12] is applied to the minimum bias sample. Table 7 shows the expected trigger rates for muons at LVL1 and after confirmation at LVL2. The output rates of the `DsPhiPi` trigger at LVL2 and the EF are given in Table 8 for $10^{31}$ cm$^{-2}$s$^{-1}$. For a luminosity of $10^{33}$ cm$^{-2}$s$^{-1}$ and higher, the rates are expected to increase because of event pile-up and cavern background events.

Table 7: Muon rates based on $2.6 \cdot 10^6$ minimum bias events. Rates set in *italics* are based on an interpolation using an exponential approximation of the rate dependence on the muon threshold concerned. Effects caused by event pile-up and cavern-background events are not included.

| Luminosity [cm$^{-2}$s$^{-1}$] | LVL1 | | | | |
|---|---|---|---|---|---|
| | L1_MU00 | | L1_MU06 | | L1_MU10 |
| $10^{31}$ | 1.3 ± 0.02 kHz | | 480 ± 10 Hz | | 266 ± 8 Hz |
| $10^{32}$ | 13.0 ± 0.2 kHz | | 4.8 ± 0.1 kHz | | 2.7 ± 0.1 kHz |
| $1 \cdot 10^{33}$ | 130 ± 2 kHz | | 48 ± 1 kHz | | 27 ± 1 kHz |
| $2 \cdot 10^{33}$ | 260 ± 4 kHz | | 96 ± 2 kHz | | 54 ± 2 kHz |
| | LVL2 | | | | |
| | L2_mu0 | L2_mu5 | L2_mu6 | L2_mu8 | L2_mu10 |
| $10^{31}$ | 450 ± 11 Hz | *213* Hz | 120 ± 6 Hz | *50* Hz | 25 ± 3 Hz |
| $10^{32}$ | 4.5 ± 0.1 kHz | *2.1* kHz | 1.20 ± 0.06 kHz | *500* Hz | 250 ± 30 Hz |
| $1 \cdot 10^{33}$ | 45.0 ± 1.1 kHz | *21* kHz | 12.0 ± 0.6 kHz | *5* kHz | 2.5 ± 0.3 kHz |
| $2 \cdot 10^{33}$ | 90.0 ± 2.2 kHz | *42* kHz | 24.0 ± 1.2 kHz | *10* kHz | 5.0 ± 0.6 kHz |

In addition to the overall allowed output rate, the time constraints of the HLT system are limiting the `DsPhiPi` trigger. The maximum allowed average computing times are 40 ms at LVL2 and 1 s at the EF. Most of the time is taken in the tracking algorithms as can be seen in Table 9 which shows the average CPU time used by the tracking and hypothesis algorithms at LVL2 and EF.

Table 10 summarises the LVL2 efficiencies and the expected numbers for $B_s^0 \to D_s^- a_1^+(\mu 4)$ events before and after the application of event selection cuts in the analysis as well as the estimated LVL2 and EF trigger rates for different luminosity scenarios and different trigger choices. The L2 and EF output rates shown in this table are deduced from the rate information given in Tables 7 and 8. The LVL2 muon trigger efficiency estimates presented in Table 10 are based on the LVL2 results obtained with the $B_s^0 \to D_s^- a_1^+(\mu 4)$ sample shown in Table 5.

Table 8: Output rates for the DsPhiPi trigger based on $2.6 \cdot 10^6$ minimum bias events at $10^{31}$ cm$^{-2}$s$^{-1}$. Rates set in *italics* are based on an interpolation using the results from Table 7.

| | LVL2 | | EF | |
|---|---|---|---|---|
| Muon input trigger | RoI | FullScan | RoI | FullScan |
| L1_MU00 | 23 ± 3 Hz | 31 ± 3 Hz | 14 ± 2 Hz | 19 ± 2 Hz |
| L1_MU06 | 11 ± 2 Hz | 13 ± 2 Hz | 6.4 ± 1.3 Hz | 7.5 ± 1.4 Hz |
| L2_mu0 | 15 ± 2 Hz | 18 ± 2 Hz | 6.1 ± 1.3 Hz | 6.9 ± 1.4 Hz |
| L2_mu5 | *9.5* Hz | *9.9* Hz | *4.5* Hz | *4.5* Hz |
| L2_mu6 | 6.7 ± 1.3 Hz | 6.4 ± 1.3 Hz | 3.5 ± 1.0 Hz | 3.2 ± 0.9 Hz |
| L2_mu8 | *3.9* Hz | *3.3* Hz | *2.1* Hz | *1.7* Hz |
| L2_mu10 | *2.5* Hz | *2.0* Hz | *1.2* Hz | *0.9* Hz |

Table 9: Average CPU times on an HLT computing node (Dual core Intel(R) Xeon(R) CPU 5160 @ 3.00 GHz) using 900 $b\bar{b} \to \mu 6X$ events.

| | | Algorithm | Time/RoI | Time/event |
|---|---|---|---|---|
| LVL2 | tracking | Idscan_RoI | 15 ms | 23 ms |
| | | Idscan_FullScan | | 91 ms |
| | hypothesis | DsPhiPi | | < 1 ms |
| Event Filter | tracking | RoI | 130 ms | 208 ms |
| | | FullScan | | 470 ms |
| | hypothesis | DsPhiPi | | < 1 ms |

At $10^{31}$ cm$^{-2}$s$^{-1}$, once the improvements discussed in Section 3.5 have been applied to the EF algorithms, it should be possible to run the FullScan-based trigger at the LVL1 4 GeV muon rate and to remain within the constraints given by available trigger resources. As the luminosity is increased to $10^{32}$ cm$^{-2}$s$^{-1}$, we will need to raise the muon threshold for the FullScan-based trigger or to move to a RoI-based trigger. However, the muon threshold should be kept as low as possible in order to achieve the highest possible trigger efficiency and to allow for as many $B_s^0 \to D_s^- a_1^+$ events as possible to pass. Compared to earlier publications like [13] the $b\bar{b}$ cross-section as shown in [7] is at the upper limit of what is expected and therefore the muon rates are likely to be overestimated.

### 3.5 Trigger Strategies for Running at $10^{33}$ cm$^{-2}$s$^{-1}$

At $10^{33}$ cm$^{-2}$s$^{-1}$ the trigger needs to remain as efficient as possible while operating within the constraints of the trigger system's resources. The EF output rate is expected to be about 10-20 Hz for $B$-physics.

The muon rates expected at LVL1 and LVL2 for different thresholds and luminosities are included in Table 7. The LVL2 muon rates are the input rates for the LVL2 tracking algorithms. Using the information on jet RoI multiplicities from Figure 2 and Table 3, the computing times from Table 9 and the muon rates in Table 7, trigger strategies are determined for different luminosities.

For a luminosity of $10^{33}$ cm$^{-2}$s$^{-1}$, a LVL1 trigger muon in combination with the RoI-based $D_s$ trigger will be used. It is planned to use thresholds of 6 GeV for the trigger muon and 5 GeV for the jet RoI trigger. It is clear from Table 8 that in order to run such a trigger the LVL2 and EF selections will have to be tightened. This may be achieved by introducing vertex fitting and by reconstructing the $B_s^0$ at the

Table 10: LVL2 and EF output rates for minimum bias events, LVL2 trigger efficiencies and numbers of expected $B_s^0 \to D_s^- a_1^+(\mu 4)$ events without or with selection cuts. Rates set in *italics* are estimates based on the interpolated and extrapolated rates given in Table 7. LVL2 output rates marked by $^\dagger$ are downscaled by a factor two, the estimated rate reduction for a $D_s$ vertex requirement at LVL2. For EF output rates marked by $^\ddagger$, an estimated rate reduction factor of 60 accounting for EF $B_s^0$ reconstruction is applied. (See section 3.5 for details.) The results in the columns "LVL2 eff." and $N_{LVL2\ output}^{B_s^0 \to D_s^- a_1^+}$ (without and with selection cuts applied) are based on the $B_s^0 \to D_s^- a_1^+(\mu 4)$ Monte Carlo data sample. Numbers marked by $^\#$ are corrected for the estimated efficiency loss by a $D_s$ vertex requirement at LVL2. The integrated luminosities and expected event numbers correspond to one year running at the given instantaneous luminosity ($10^7$ seconds).

| $\mathscr{L}$ [cm$^{-2}$s$^{-1}$] | $\int \mathscr{L} dt$ [pb]$^{-1}$ | Trigger set | LVL2 eff. [%] | $N_{LVL2\ output}^{B_s^0 \to D_s^- a_1^+}$ no sel. cuts | $N_{LVL2\ output}^{B_s^0 \to D_s^- a_1^+}$ incl. sel. cuts | LVL2 rate [Hz] | EF rate [Hz] |
|---|---|---|---|---|---|---|---|
| $10^{31}$ | 100 | L2mu0FS | $22.93 \pm 0.19$ | 308 | 63 | $18 \pm 2$ | $6.9 \pm 1.4$ |
| | | L2mu5FS | $16.75 \pm 0.17$ | 225 | 47 | *9.9* | *4.5* |
| | | L2mu6FS | $12.49 \pm 0.15$ | 168 | 35 | $6.4 \pm 1.3$ | $3.2 \pm 0.9$ |
| $10^{32}$ | 1 000 | L2mu6FS | $12.49 \pm 0.15$ | 1 678 | 351 | $64 \pm 13$ | $32 \pm 9$ |
| | | L2mu5RoI | $14.26 \pm 0.16$ | 1 916 | 267 | *95* | *45* |
| | | L2mu6RoI | $10.91 \pm 0.14$ | 1 466 | 322 | $67 \pm 13$ | $35 \pm 10$ |
| $10^{33}$ | 10 000 | L2mu5RoI | $12.01 \pm 0.13^\#$ | $16\,134^\#$ | $3\,582^\#$ | $\mathit{475}^\dagger$ | $\mathit{7.5}^\ddagger$ |
| | | L2mu6RoI | $9.19 \pm 0.12^\#$ | $12\,344^\#$ | $2\,709^\#$ | $335 \pm 93^\dagger$ | $5.8 \pm 2.0^\ddagger$ |
| | | L2mu8RoI | $5.94 \pm 0.10^\#$ | $7\,976^\#$ | $1\,757^\#$ | $\mathit{196}^\dagger$ | $\mathit{3.5}^\ddagger$ |
| | | L2mu10RoI | $3.84 \pm 0.08^\#$ | $5\,159^\#$ | $1\,132^\#$ | $\mathit{126}^\dagger$ | $\mathit{2.0}^\ddagger$ |
| $2 \cdot 10^{33}$ | 20 000 | L2mu6RoI | $9.19 \pm 0.12^\#$ | $24\,687^\#$ | $5\,418^\#$ | $670 \pm 187^\dagger$ | $11.7 \pm 4.0^\ddagger$ |
| | | L2mu8RoI | $5.94 \pm 0.10^\#$ | $15\,953^\#$ | $3\,517^\#$ | $\mathit{392}^\dagger$ | $\mathit{7.1}^\ddagger$ |
| | | L2mu10RoI | $3.84 \pm 0.08^\#$ | $10\,318^\#$ | $2\,264^\#$ | $\mathit{252}^\dagger$ | $\mathit{4.1}^\ddagger$ |

EF level.

Preliminary studies at LVL2 show that a requirement for a vertex fit to the 3 tracks of the $D_s$ candidate can achieve a factor 2 rate reduction for a drop in efficiency from 38% to 32%. This estimate is applied to cells marked by $^\#$ and $^\dagger$ in Table 10. Also, it might be an option to further reduce the rate by tightening the acceptance windows for $m_\phi$ and $m_{D_s}$ on LVL2, but the resulting rate reduction and the expected signal efficiency loss will need to be studied.

A considerable rate reduction at the EF level may be achieved by reconstructing the $B_s^0$. A preliminary study using offline selection cuts (see Section 5.1), which have been relaxed to simulate wider mass window and vertexing requirements for the reconstructed particles, has been performed with the $b\bar{b} \to \mu 4X$ and $c\bar{c} \to \mu 4X$ samples. The resulting rate reduction factor, estimated to be approximately 60, is applied to cells marked by $^\ddagger$ in Table 10. According to this study, the overall trigger and reconstruction efficiency for the $B_s^0 \to D_s^- a_1^+$ signal events will be reduced by about 55%. Although these estimates will need to be confirmed by an implementation of a simplified $B_s^0$ reconstruction at the EF level, reasonable EF output rates are expected to be achievable.

The numbers of expected $B_s^0 \to D_s^- a_1^+(\mu 4)$ events for 10 fb$^{-1}$ of data for a luminosity of $10^{33}$ cm$^{-2}$s$^{-1}$ are given in Table 10. It will be necessary to establish a muon trigger threshold as low as possible to maximise the signal event yield.

### 3.6 Trigger Strategies for Higher Luminosities

As luminosity increases, it is necessary to stay within the limits of the LVL2 trigger processing times and allowable output rates. As Table 10 shows, this will require increasing the muon threshold to 8 GeV or

10 GeV and to add additional trigger elements in the EF as discussed at the end of Section 3.5. Another option, however a less efficient one, is to prescale the 6 GeV rate before running the track reconstruction.

# 4 Flavour Tagging

The measurement of $B_s^0$ oscillations needs the knowledge of the flavour of the $B_s^0$ meson at production time and at decay time in order to classify events as mixed or not mixed. The tagging algorithm tries to determine the flavour at production time, whereas the decay particles of the signal $B_s^0$ determine the flavour at decay time. In this analysis soft muon tagging (see B-physics chapter of [4]) is used and the general application on the simulated data samples is shown in Section 4.2 without applying trigger conditions or any selection cut. Tagging results specific for the hadronic channels under investigation including trigger and selection cuts for $B_s^0$ candidates are given in Section 5.3.

## 4.1 Soft Muon Tagging

In proton-proton collisions $b$ quarks are produced in pairs leaving the signal $B_s^0$ and the opposite side $b$ hadron with the opposite flavour. In the case of a semileptonic decay as shown in Fig. 4, the charge of the produced lepton is correlated with the flavour of the signal $B_s^0$ meson at production time. The charge of the muon with the highest reconstructed $p_T$ is used for the determination of the flavour. Because of the muon trigger, in hadronic $B_s^0$ decay channels soft muon tagging has a high tagging efficiency $\varepsilon_{tag} = N_{tag}/N_{all}$ limited by the muon reconstruction efficiency. Details on the aspects of muon reconstruction and identification in ATLAS can be found in [14].

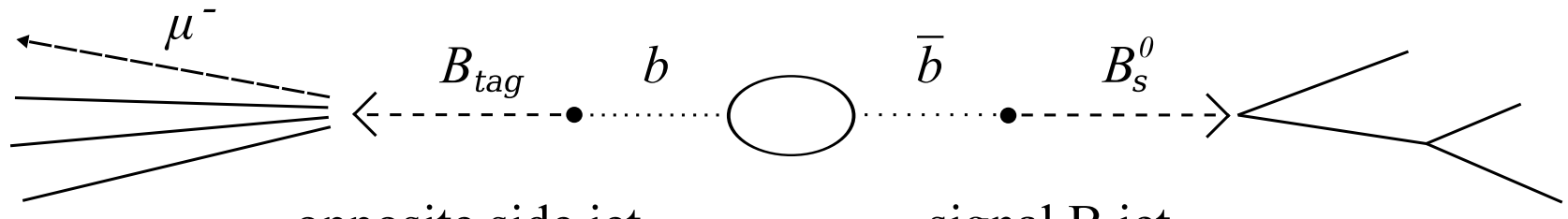


Figure 4: In the case of a signal $B_s^0$, the associated opposite side $b$ hadron decaying semileptonically produces a negatively charged lepton.

The dilution factor is defined as $D_{tag} = \frac{N_c - N_w}{N_c + N_w}$ where $N_c$ is the number of events correctly tagged and $N_w$ is the number of events with a wrong tag. These wrong tags arise from mixing of the tagging $b$ hadron, muons from decays $b \to c \to \mu$, additional $c$ pairs and various particles decaying into muons. The wrong tag fraction $\omega = N_w/(N_c + N_w)$ is the ratio of wrongly tagged events to all tagged events. As the generation of the simulated data does not include $B$ meson oscillations, mixing of the tagging side hadron is introduced artificially using the integrated mixing probabilities $\chi_d$ and $\chi_s$ [8]:

$$\chi_d = \frac{\Gamma(B_d^0 \to \bar{B}_d^0 \to \mu^+ X)}{\Gamma(B_d^0 \to \mu^\pm X)} = 0.188 \pm 0.003 \qquad \chi_s = \frac{\Gamma(B_s^0 \to \bar{B}_s^0 \to \mu^+ X)}{\Gamma(B_s^0 \to \mu^\pm X)} = 0.49924 \pm 0.00003$$

## 4.2 Application to Signal Samples

In Fig. 5 the transverse momentum of the signal $B_s^0$ mesons is compared for the two $B_s^0$ decay channels, the vertical lines show the mean values of the two distributions. This difference arises from the different kinematical configuration due to the condition on all charged final state particles $p_T > 0.5$ GeV at Monte

Carlo generation. As $B_s^0 \to D_s^- a_1^+$ has a total number of six final state particles, the mean transverse momentum of the signal $B_s^0$ is higher compared to $B_s^0 \to D_s^- \pi^+$ with a total number of four final state particles. The difference in the $B_s^0$ transverse momentum spectrum is also expected at offline reconstruction level due to the different $p_T$ selection cuts for the $\pi$ and the $a_1$ combinations (see Section 5.1). This leads in the case of the $B_s^0 \to D_s^- \pi^+$ sample to an overall wrong tag fraction of $\omega = 20.29 \pm 0.14$ % and in the case of the $B_s^0 \to D_s^- a_1^+$ channel to a wrong tag fraction $\omega = 21.05 \pm 0.11$ %, which is higher compared to the $B_s^0 \to D_s^- \pi^+$ channel (see all events in Table 11 in Section 5.3).

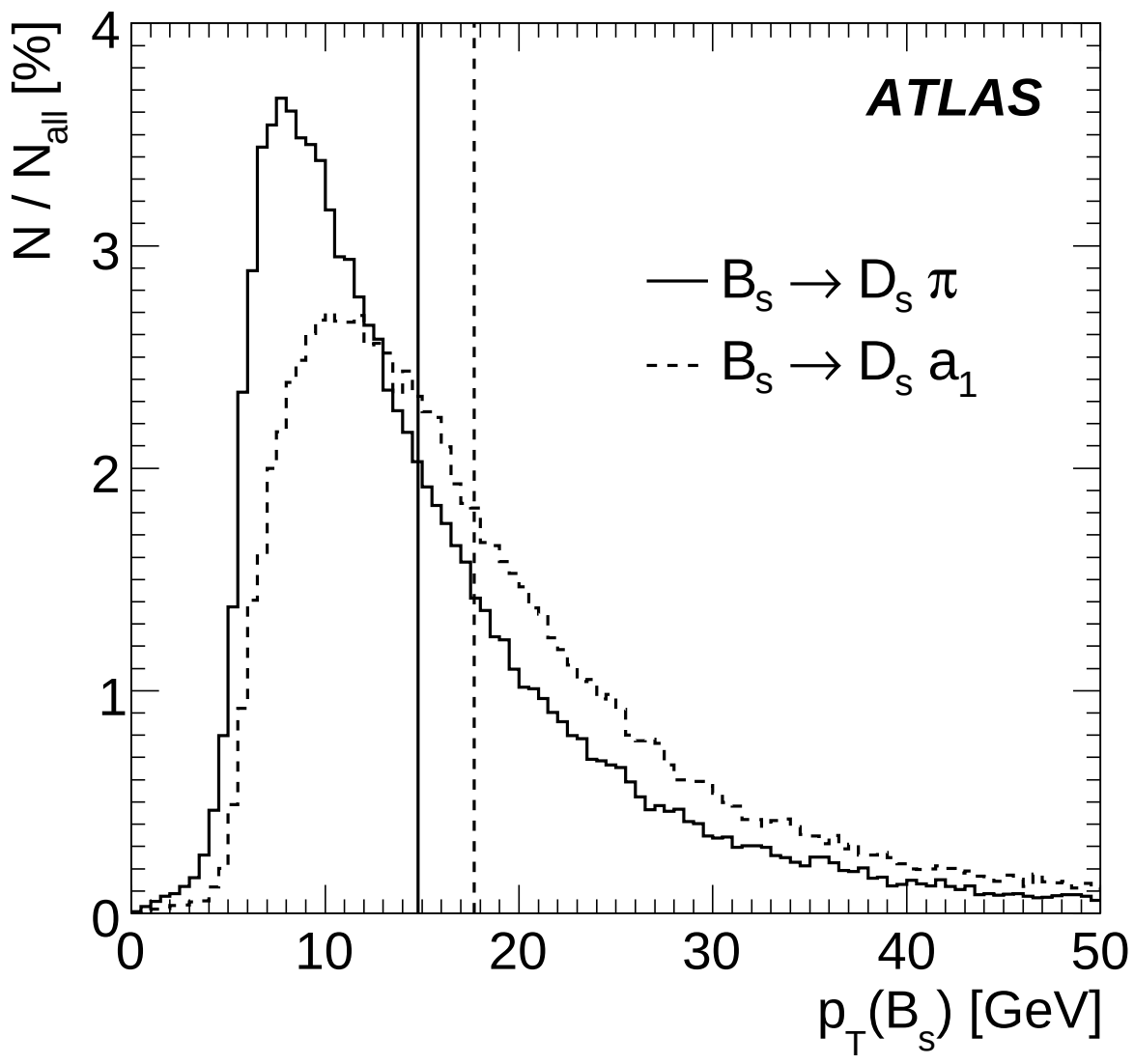


Figure 5: Normalised distributions of signal $B_s^0$ transverse momentum $p_T$ of the two channels $B_s^0 \to D_s^- \pi^+$ and $B_s^0 \to D_s^- a_1^+$. The vertical lines represent the mean values of the distributions, in the case of $B_s^0 \to D_s^- \pi^+$ the mean is $14.80 \pm 0.03$ GeV, in the case of $B_s^0 \to D_s^- a_1^+$ $17.68 \pm 0.03$ GeV. The observed difference is due to the different particle selections.

In Fig. 6 the wrong tag fractions and the sources of these wrong tags are compared for both $B_s^0$ decays channels. The wrong tag fraction is shown as a function of the tagging muon's transverse momentum $p_T(\mu)$, in Fig. 6(a) for $B_s^0 \to D_s^- \pi^+$ and in Fig. 6(b) for $B_s^0 \to D_s^- a_1^+$. In the regime $p_T(\mu) < 11$ GeV the $B_s^0 \to D_s^- a_1^+$ wrong tag fraction is higher. As mentioned above this difference arises from the different track selections of the two decay channels. In both channels the two main sources of wrong tags are mixing of neutral $B$ mesons on the tagging side and muons from cascade decays $b \to c \to \mu$. As the overall wrong tag fraction is decreasing with the muon $p_T$, also the part with a wrong tag due to the cascade $b \to c \to \mu$ is decreasing at the same rate. A further source of mistags are additional $c\bar{c}$-pairs. The wrong tag fraction of this part stays about constant with increasing $p_T(\mu)$. A small part of the wrong tag fraction originates from $J/\psi$, $\phi$, $\rho$, $\eta$ or $\tau$ particles decaying into muons. Additional sources like muons from kaons and pions or hadrons misidentified as muons can be neglected [12].

A $b\bar{b}$ pair produced in proton proton collisions has a transverse momentum equal to zero at first order. Going through fragmentation and hadronisation, the $p_T$ of the signal $B_s^0$ meson and the opposite side $b$ hadron are still correlated, and therefore a muon coming from a semileptonic decay of the $b$ hadron also is correlated with the signal $B_s^0$. Hence a muon from a cascade $b \to c \to \mu$ is more likely to pass the LVL1 muon trigger when $B_s^0$ meson has a higher $p_T$, leading to the increase in wrong tag fraction with $p_T(B_s^0)$. This behaviour is shown in the Fig. 6(c) and 6(d).

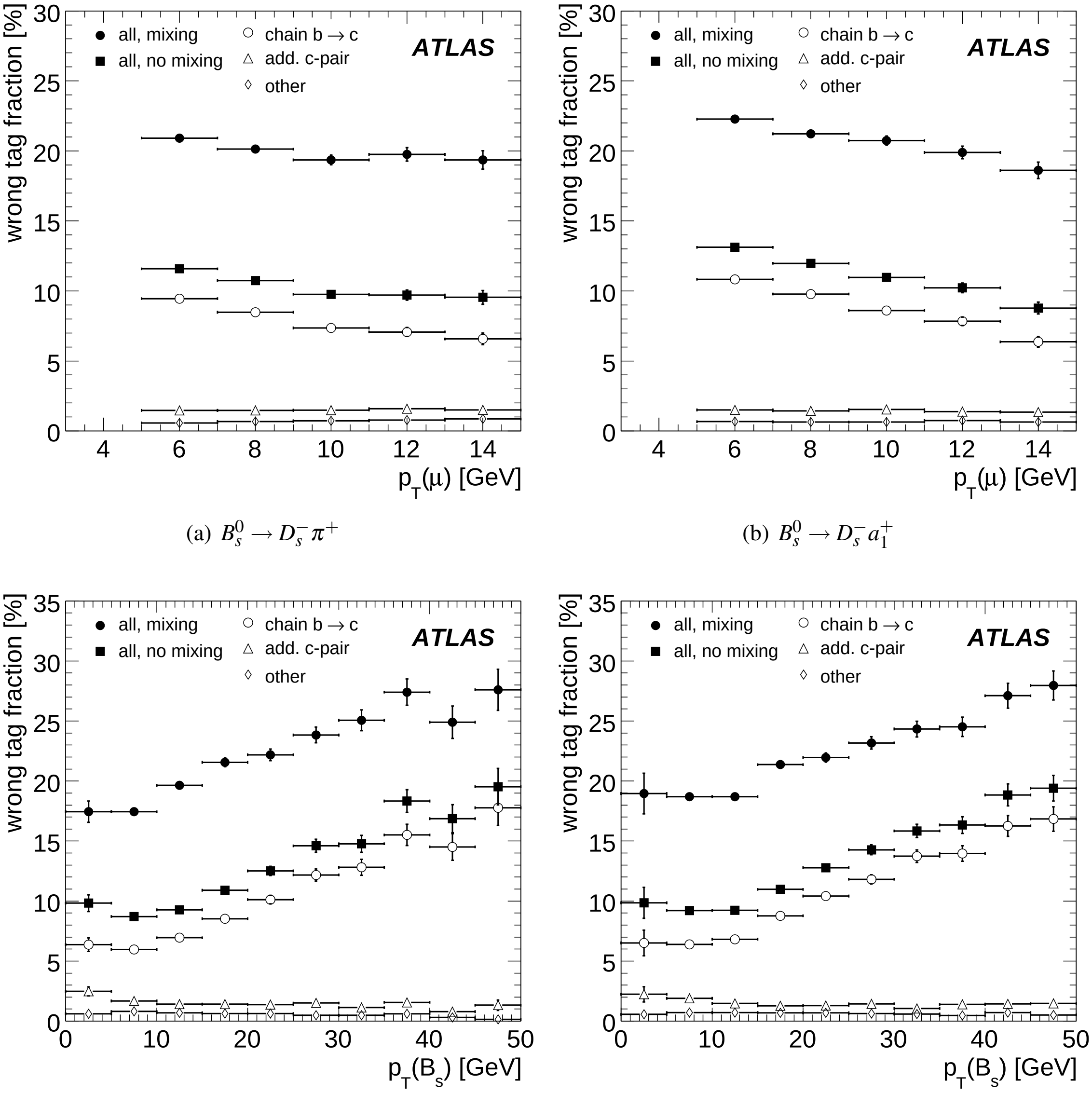


(a) $B_s^0 \to D_s^- \pi^+$

(b) $B_s^0 \to D_s^- a_1^+$

(c) $B_s^0 \to D_s^- \pi^+$

(d) $B_s^0 \to D_s^- a_1^+$

Figure 6: Wrong tag fraction as functions of tagging muons transverse momentum $p_T(\mu)$ in (a) and (b) and wrong tag fractions as functions of signal Monte Carlos $B_s^0$ transverse momentum $p_T(B_s^0)$ in (c) and (d). The wrong tag fraction is shown with mixing of the tagging side $b$ hadron and without mixing. Without mixing, the different sources of wrong tags are shown. The main contribution is coming from $b \to c \to \mu$ followed by additional $c$ pairs. Additional sources shown are muons coming from $J/\psi$, $\phi$, $\rho$, $\eta$ and $\tau$.

### 4.3 Systematic Uncertainties of Soft Muon Tagging

The calibration of the soft muon tagger will be done with events from the exclusive decay channel $B^+ \to J/\psi(\mu^+\mu^-)K^+$ . The high branching ratio and the simple event topology allows the measurement of this channel during the initial luminosity phase at the LHC. Without mixing on the signal side, these events can be used to estimate the systematic uncertainties of soft muon tagging.

For an integrated luminosity of 1 fb$^{-1}$ 160 000 events of the decay channel $B^+ \to J/\psi(\mu^+\mu^-)K^+$ are expected [15] at ATLAS. About 13.5 % events are estimated to have an additional third muon for flavour tagging. Requiring a minimum transverse momentum of 6 GeV for this additional muon, the number of events will be reduced by a factor of three. Assuming that the wrong tag fraction in this channel behaves like in the hadronic $B_s^0$ decay channels, the expected statistical error of the wrong tag fraction would be of the order of 0.1 % for 1 fb$^{-1}$ integrated luminosity.

## 5 Event Selection

For the following analysis selecting $B_s$ candidates the default trigger choices are to require MU06 and JT04 trigger elements at LVL1 and to perform a search for the $D_s \to \phi(K^+K^-)\pi$ decay within a jet RoI at LVL2. Resulting event numbers and plots are given for 10 fb$^{-1}$ unless indicated otherwise.

### 5.1 Signal Event Reconstruction

For the reconstruction of the $B_s$ vertex only tracks with a pseudo-rapidity $|\eta| < 2.5$ are used proceeding via the following steps. The $\phi$ decay vertex is first reconstructed by considering all pairs of oppositely-charged tracks with $p_T > 1.5$ GeV for both tracks. Kinematic cuts on the angles between the two tracks, $\Delta\varphi_{KK} < 10°$ and $\Delta\theta_{KK} < 10°$, are imposed, where $\varphi$ denotes the azimuthal angle and $\theta$ the polar angle. The two-track vertex is then fitted assigning the kaon mass to both tracks. Combinations passing a fit-probability cut [16] of 1% ($\simeq \chi^2$/dof = 7/1) with the invariant mass within three standard deviations of the nominal $\phi$ mass are selected as $\phi$ candidates. The plots in Fig. 7 show the invariant mass distribution for all $m_{KK}$ combinations overlaid with the $\phi$ candidates matching a generated $\phi$ from the signal decay (grey filled area) fitted with a single Gaussian function. For the $B_s^0 \to D_s^- \pi^+$ channel the mass resolution is $\sigma_\phi = (4.30 \pm 0.03)$ MeV and for the $B_s^0 \to D_s^- a_1^+$ channel $\sigma_\phi = (4.28 \pm 0.03)$ MeV. This mass window for accepted $\phi$ candidates is shown by the vertical lines. No trigger selections are applied for the mass plots shown in Fig. 7 to Fig. 9.

From the remaining tracks, a third track with $p_T > 1.5$ GeV is added to all accepted $\phi$ candidates. The pion mass is assigned to the third track and a three-track vertex is fitted. Three-track vertex candidates which have a fit probability greater than 1% ($\simeq \chi^2$/dof = 12/3) and an invariant mass within three standard deviations of the nominal $D_s$ mass are selected as $D_s$ candidates. The plots in Fig. 8 show the invariant mass distribution for all $m_{KK\pi}$ combinations overlaid with the $D_s$ candidates matching a generated $D_s$ from the signal decay (grey filled area) fitted with a single Gaussian function. For the $B_s^0 \to D_s^- \pi^+$ channel the mass resolution is $\sigma_{D_s} = 17.81 \pm 0.13$ MeV and for the $B_s^0 \to D_s^- a_1^+$ channel $\sigma_{D_s} = 17.92 \pm 0.13$ MeV. The $3\sigma_{D_s}$ mass range for accepted $D_s$ candidates is shown by the vertical lines.

For the $B_s^0 \to D_s^- a_1^+$ channel a search is performed for $a_1^\pm$ candidates using three-particle combinations of charged tracks for events with a reconstructed $D_s$ meson. In a first step $\rho^0$ mesons are reconstructed from all combinations of two tracks with opposite charges and with $p_T > 0.5$ GeV, each particle in the combination being assigned a pion mass. A kinematic cut $\Delta R_{\pi\pi} = \sqrt{\Delta\phi_{\pi\pi}^2 + \Delta\eta_{\pi\pi}^2} < 0.650$ is used to reduce the combinatorial background. The two selected tracks are then fitted as originating from the same vertex; from the combinations passing a fit probability cut of 1% ($\simeq \chi^2$/dof = 7/1), those with an invariant mass within 400 MeV of the nominal $\rho^0$ mass are selected as $\rho^0$ candidates. Next a third track with $p_T > 0.5$ GeV from the remaining charged tracks is added to the $\rho^0$ candidate, assuming the

pion hypothesis for the extra track. A kinematic cut $\Delta R_{\rho\pi} < 0.585$ is applied. The three tracks are fitted to a common vertex without any mass constraints. Combinations with a fit probability greater than 1% ($\simeq \chi^2$/dof = 12/3) and with an invariant mass within 325 MeV of the nominal $a_1$ mass are selected as $a_1^\pm$ candidates.

The $B_s^0$ candidates are reconstructed combining the $D_s^\pm$ candidates with $a_1^\pm$ candidates with opposite charge and different tracks. A six-track vertex fit is performed with mass constraints for the tracks from $\phi$ and $D_s$; due to the large $a_1$ natural width the three tracks from the $a_1$ are not constrained to the $a_1$ mass. The total momentum of the $B_s^0$ vertex is required to point to the primary vertex and the momentum of the $D_s$ vertex to the $B_s^0$ vertex. Only six-track combinations with a vertex fit probability greater than 1% ($\simeq \chi^2$/dof = 27/12) are considered as $B_s^0$ candidates.

For the $B_s^0 \to D_s^- \pi^+$ channel for each reconstructed $D_s$ meson a fourth track from the remaining tracks in the event is added. This track is required to have opposite charge with respect to the pion track from the $D_s$ and $p_T > 1$ GeV. The four-track decay vertex is fitted including $\phi$ and $D_s$ mass constraints, and requiring that the total momentum of the $B_s^0$ vertex points to the primary vertex and the momentum of $D_s$ vertex points to the $B_s^0$ vertex. In order to be selected as $B_s^0$ candidates, the four-track combinations are required to have a vertex fit probability greater than 1% ($\simeq \chi^2$/dof = 20/8).

For both channels, $B_s^0 \to D_s^- \pi^+$ and $B_s^0 \to D_s^- a_1^+$, the signed separation between the reconstructed $B_s^0$ vertex and the primary vertex is required to be positive (the momentum should not point backward to the parent vertex). To improve the purity of the sample, further cuts are imposed: the proper decay time of the $B_s^0$ has to be greater than 0.4 ps, the $B_s^0$ impact parameter (shortest distance of the reconstructed $B_s^0$ trajectory from the primary vertex in the transverse plane to the reconstructed $B_s^0$ decay vertex) is required to be smaller than 55 $\mu$m and $p_T$ of the $B_s^0$ must be larger than 10 GeV. The plot in Fig. 9(a) shows the invariant mass distribution for all $m_{KK\pi\pi}$ $B_s$ candidates matching a generated $B_s$ from the signal decay for the $B_s^0 \to D_s^- \pi^+$ channel fitted with a single Gaussian function and giving a mass resolution of $\sigma_{B_s} = 52.80 \pm 0.68$ MeV. For the $B_s^0 \to D_s^- a_1^+$ channel Fig. 9(b) shows the invariant mass distribution for all $m_{KK\pi\pi\pi\pi}$ $B_s$ candidates matching a generated $B_s$ and giving a mass resolution of $\sigma_{B_s} = 40.82 \pm 0.53$ MeV. The difference in the $B_s^0$ mass resolutions is caused by the $p_T$ spectrum of the $\pi$ in the $B_s^0 \to D_s^- \pi^+$ decay being harder than the $p_T$ spectra of the three pions in the $B_s^0 \to D_s^- a_1^+$ decay, as the pion momentum resolution is worse for higher $p_T$. A final mass cut of two standard deviations on the $B_s^0$ candidates is applied for further analysis (see vertical lines in Fig. 9). For some events more than one $B_s^0$ candidate is reconstructed and in that case the candidate with the lowest $\chi^2$/dof from the vertex fit is selected for further analysis.

No relevant effects induced by the trigger selections on fit variables of the mass plots or the kinematic distributions of the $B_s$ candidates are found (discussed in Section 6.3). All differences are within the fit errors.

## 5.2 Background Channels

Two main sources of background are considered: irreducible background coming from a decay channel that closely mimics the $B_s^0$ signal and combinatorial background coming from random combination of tracks.

### 5.2.1 Exclusive Background Channels

The exclusive samples listed in Table 1 are used as irreducible background sources. See Table 12 in Section 6 for the numbers of reconstructed candidates after applying the same selection cuts as for the signal samples. The histograms in Fig. 10 show the invariant mass spectrum of reconstructed $B_s^0$ candidates for the $B_s^0 \to D_s^- \pi^+$ and $B_s^0 \to D_s^- a_1^+$ channel respectively (no trigger selection cut applied). The different contributions are scaled with the cross-section given in Table 1.

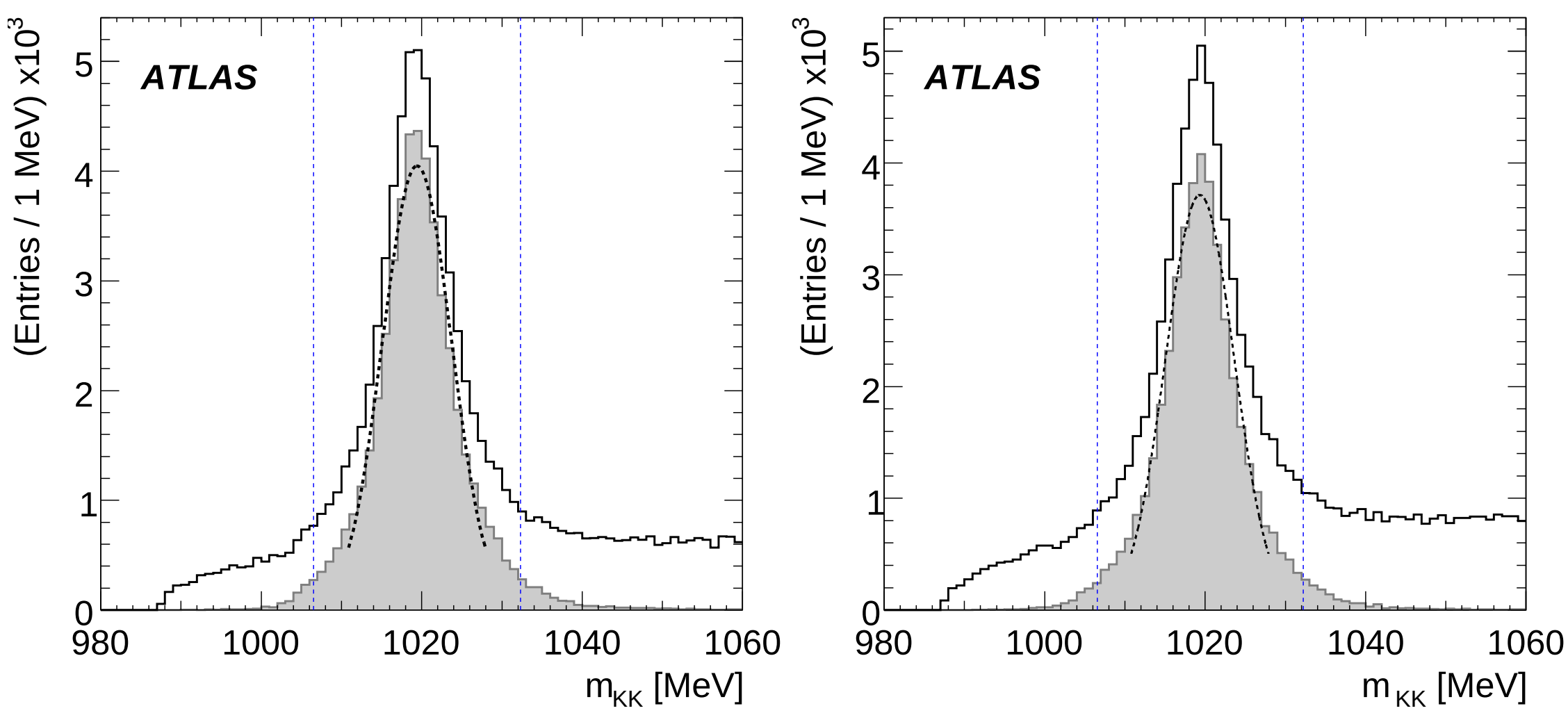


(a) $B_s^0 \to D_s^- \pi^+$ (b) $B_s^0 \to D_s^- a_1^+$

Figure 7: Reconstructed mass $m_{KK}$ for all combinations within the signal sample (black line) and $KK$ candidates corresponding to a $\phi$ particle from the signal decay in the Monte Carlo truth information (grey filled). The standard deviation obtained from a fit within two standard deviations of a Gaussian function (dashed) to the distribution defines the three standard deviation cut range (vertical dashed). No trigger conditions are applied.

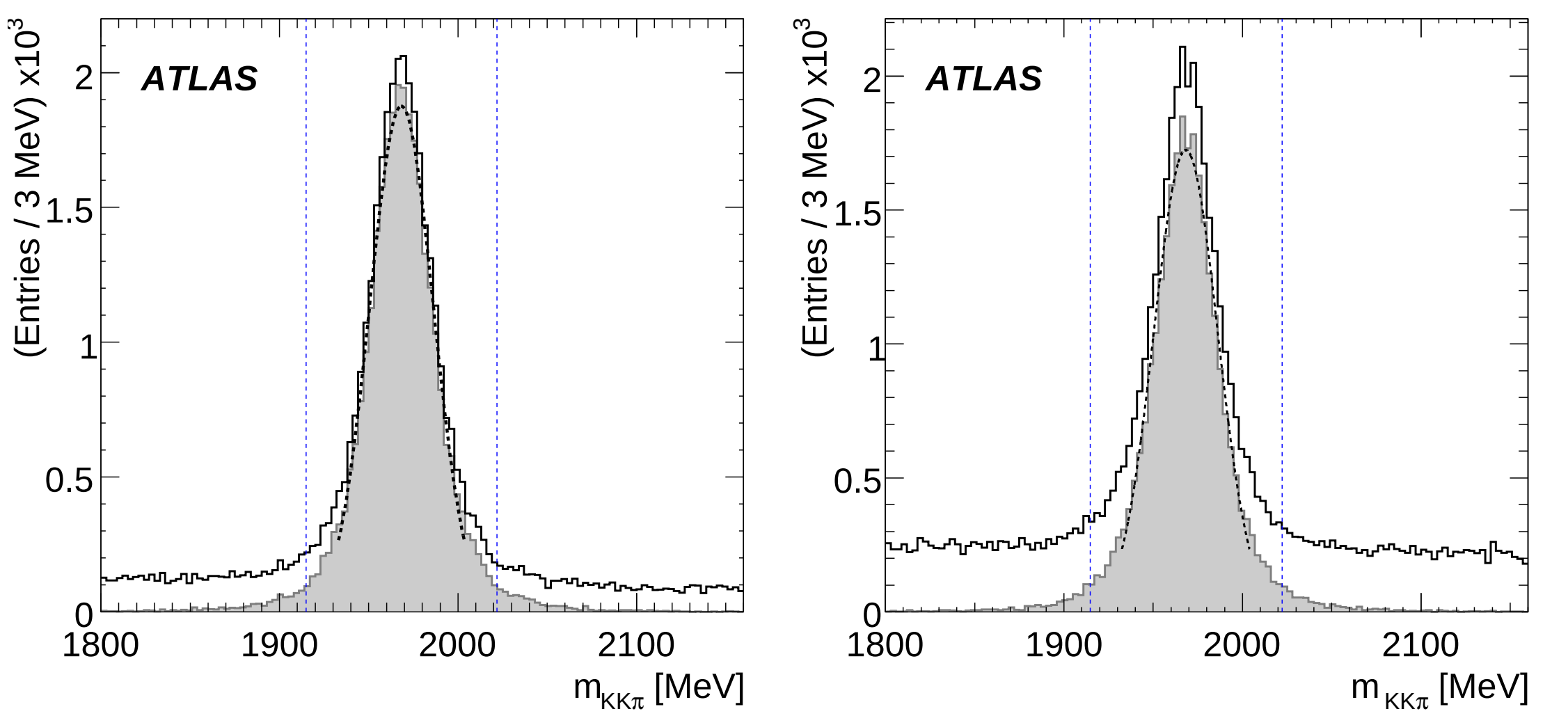


(a) $B_s^0 \to D_s^- \pi^+$ (b) $B_s^0 \to D_s^- a_1^+$

Figure 8: Reconstructed mass $m_{KK\pi}$ for all combinations within the signal sample (black line) and $KK\pi$ candidates corresponding to a $D_s$ particle from the signal decay in the Monte Carlo truth information (grey filled). The standard deviation obtained from a fit within two standard deviations of a Gaussian function (dashed) to the distribution defines the three standard deviation cut range (vertical dashed). No trigger conditions are applied.

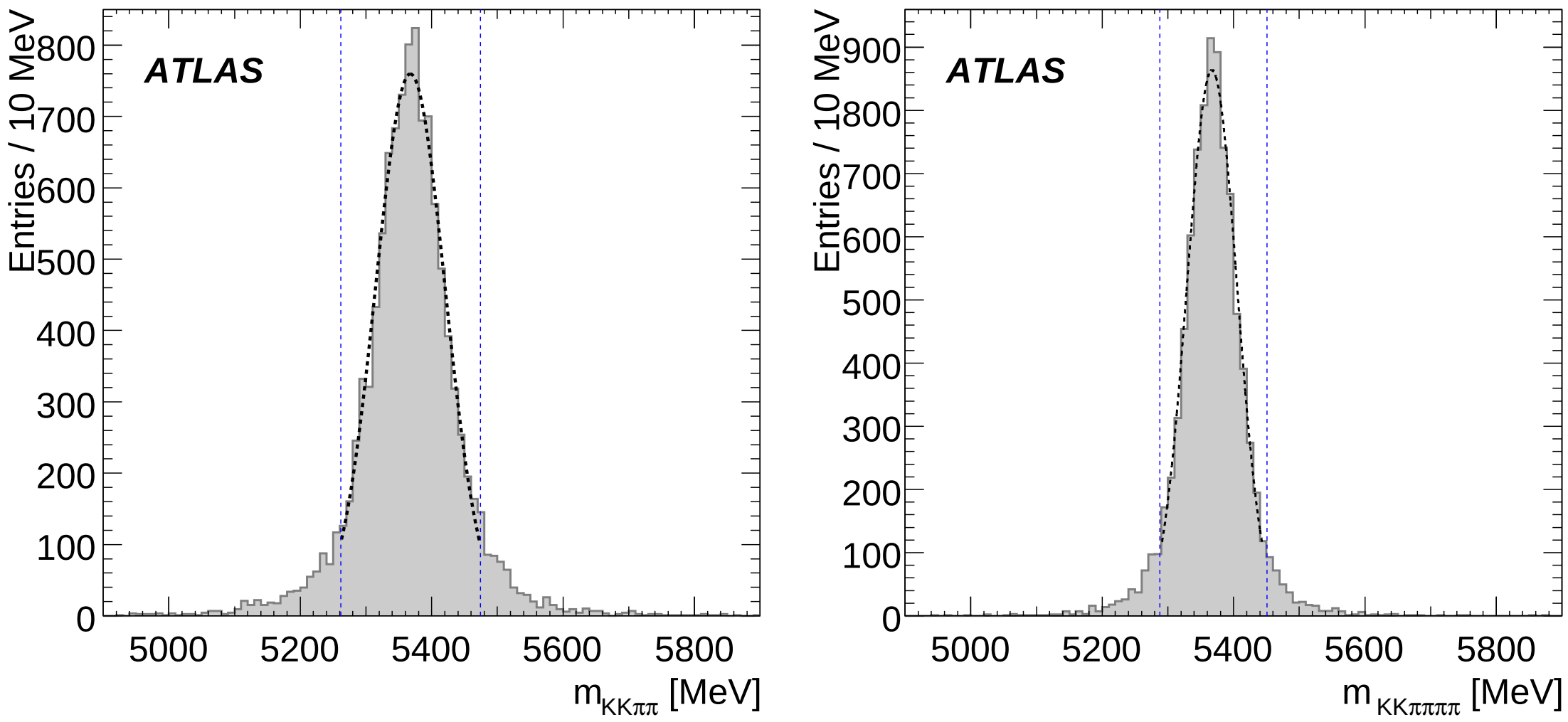


(a) $B_s^0 \to D_s^- \pi^+$ (b) $B_s^0 \to D_s^- a_1^+$

Figure 9: $m_{KK\pi\pi}$ (a) and $m_{KK\pi\pi\pi\pi}$ (b) reconstructed mass and fit of a Gaussian function to the distribution. Each $KK\pi\pi$ ($KK\pi\pi\pi\pi$) candidate displayed corresponds to a $B_s$ particle in the Monte Carlo truth information. The two standard deviation cut range is shown by the vertical dashed lines. No trigger chain applied.

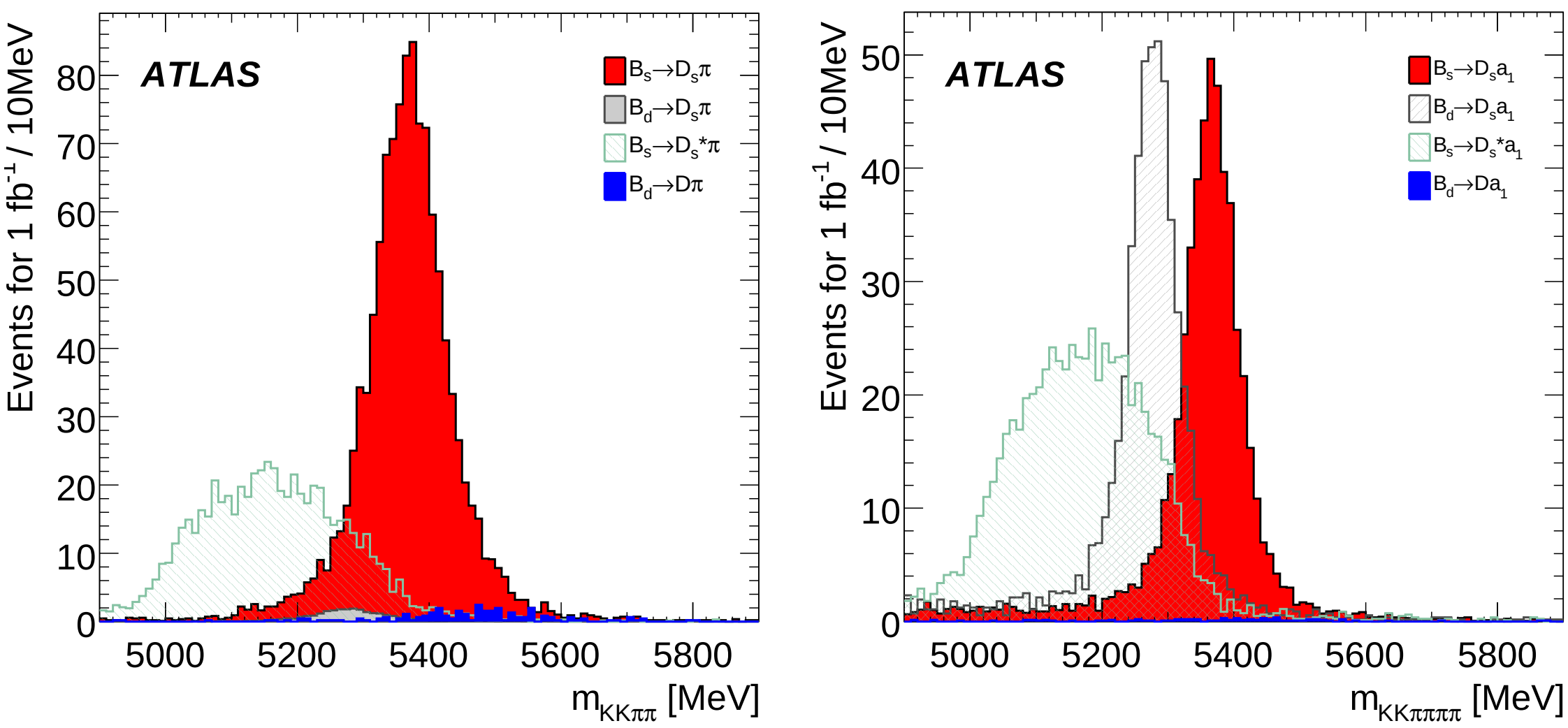


(a) $B_s^0 \to D_s^- \pi^+$ (b) $B_s^0 \to D_s^- a_1^+$

Figure 10: $m_{KK\pi\pi}$ (a) and $m_{KK\pi\pi\pi\pi}$ (b) reconstructed mass for signal and background channels. In (b) the upper limit for the branching fraction of the channel $B_d \to D_s a_1$ is used.

In the case of $B_d^0 \to D_s^+ a_1^-$, there is no measurement of the branching fraction available. The current upper limit is therefore used as a conservative estimate. The similarity to the $B_d^0 \to D_s^+ \pi^-$ channel indicates that the $B_d^0 \to D_s^+ a_1^-$ cross section could be in the same order as the one of the $B_d^0 \to D_s^+ \pi^-$ channel and therefore the $B_d^0 \to D_s^+ a_1^-$ contribution in Fig. 10 (b) could be much smaller.

The $B_s \to D_s^* \pi$ and $B_s \to D_s^* a_1$ channels are treated as a background source, since the momentum and hence the lifetime estimation for this decay is flawed due to the missing photon from the decay of the $D_s^*$.

#### 5.2.2 Combinatorial Background

The limited statistics of the combinatorial background samples (e.g. 242 150 events for $b\bar{b} \to \mu 6X$) do not allow us to give a reasonable estimate for the signal to background ratio. This ratio as well as kinematic properties of the combinatorial background, like the shape of the proper time distribution will be studied once early data are available.

### 5.3 Tagging Results

Soft muon tagging is applied on all available simulated data of the two hadronic signal channels $B_s^0 \to D_s^- \pi^+$ and $B_s^0 \to D_s^- a_1^+$. Table 11 shows the number of events, the tagging efficiency and the wrong tag fractions. The tagging efficiency corresponds to the fraction of events where muon candidates have been successfully reconstructed. The events with at least one muon are separated into events with a good tag and events with a wrong tag resulting in a wrong tag fraction.

Comparing the results for all simulated events between the two signal channels, the difference in the wrong tag fraction is of the order 1 % due to the different kinematical topology. As the RoI trigger applies a $p_T$ cut of 1.4 GeV on all reconstructed tracks, low $p_T$ $B_s^0$ mesons are rejected leading to an increased wrong tag fraction for the triggered events (see Fig. 6(c) and 6(d)). After event reconstruction overall wrong tag fractions of $\omega = 22.30^{+0.56}_{-0.55}$ % for the channel $B_s^0 \to D_s^- \pi^+$ and $\omega = 23.31^{+0.56}_{-0.55}$ % for the channel $B_s^0 \to D_s^- a_1^+$ are observed. Effects of mixing are included as described in Section 4.1.

Table 11: Tagging efficiencies and wrong tag fractions for the signal channels $B_s^0 \to D_s^- \pi^+$ and $B_s^0 \to D_s^- a_1^+$ are shown for three different stages: all simulated events, all triggered events passing the LVL1 muon trigger and the LVL2 RoI trigger, and finally the numbers for the reconstructed events. The errors are statistical only.

| Process | Type of Events | Number of Events | Tagging Efficiency [%] | Wrong Tag Fraction [%] |
|---|---|---|---|---|
| $B_s^0 \to D_s^- \pi^+$ | all events | 88450 | $96.08 \pm 0.07$ | $20.29 \pm 0.14$ |
| | triggered | 21613 | $98.77 \pm 0.07$ | $22.96 \pm 0.29$ |
| | reconstructed | 5687 | $98.79^{+0.14}_{-0.15}$ | $22.30^{+0.56}_{-0.55}$ |
| $B_s^0 \to D_s^- a_1^+$ | all events | 98450 | $95.93 \pm 0.06$ | $21.05 \pm 0.13$ |
| | triggered | 27118 | $98.55 \pm 0.07$ | $23.91 \pm 0.26$ |
| | reconstructed | 5757 | $98.47^{+0.16}_{-0.17}$ | $23.31^{+0.56}_{-0.55}$ |

# 6 Determination of $\Delta m_s$

## 6.1 Methods for the Determination of $\Delta m_s$ and its Measurement Limits

For the determination of the $B_s^0$ oscillation frequency the maximum likelihood method is used. The likelihood $\mathscr{L}$ is a function of the proper time $t$ and the mixing state $\mu$, parametrised by $\Delta m_s$ and $\Delta\Gamma_s$, applied to five classes of events simultaneously: mixed and unmixed $B_s^0$, mixed and unmixed $B_d^0$, and background with lifetime but no mixing. The $B_s^0$ and $B_d^0$ classes have characteristic wrong tag fractions $\omega$, which are determined on event-by-event basis as described previously. By maximising the likelihood $\mathscr{L}$ for a given event sample one can then extract the model parameters.

For obtaining the $5\,\sigma$ discovery and 95% exclusion measurement limits on $\Delta m_s$ the amplitude fit method is used because the maximum likelihood method was found to have some disadvantages in that case [5]. The estimation of the maximum value of $\Delta m_s$ measurable with the ATLAS detector is using $B_s^0$ candidates from the $B_s^0 \to D_s^- \pi^+$ and $B_s^0 \to D_s^- a_1^+$ hadronic channels. The numbers of reconstructed events after applying the trigger selection (L1_MU06 and LVL2 RoI) and the $B_s^0$ selection cuts as well as the expected numbers for an integrated luminosity of 10 fb$^{-1}$are given in Table 12 for all signal and background channels. The effective cross-sections for the various processes can be found in Table 1. Significant background comes from the $\bar{B}_d^0 \to D_s^- \pi^+/a_1^+$ and $B_s^0 \to D_s^{*-} \pi^+/a_1^+$ channels, and from the combinatorial background. Due to limited sample size the estimation of the combinatorial background is very approximate.

The relative fractions of the signal and the background contributions will be determined by a fit of mass shape templates to the reconstructed $B_s^0$ mass distribution employing a wider mass window than used here for the final extraction of $\Delta m_s$, similar to the method used by CDF [1]. The mass shape templates will be determined from Monte Carlo mass distributions of the individual channels. Uncertainties in the knowledge of the shapes will be taken into account as part of the systematical uncertainty.

Table 12: Signal and background samples used for the study of $B_s^0$-$\bar{B}_s^0$ oscillations and number of events as obtained from the analysis as well as expected numbers for an integrated luminosity of 10 fb$^{-1}$.

| Process | Simulated events | Rec. events | Rec. events for 10 fb$^{-1}$ |
|---|---|---|---|
| $B_s^0 \to D_s^- \pi^+$ | 88 450 | 5 687 | 6 657 |
| $B_d^0 \to D_s^+ \pi^-$ | 43 000 | 1 814 | 99 |
| $B_d^0 \to D^- \pi^+$ | 41 000 | 23 | 35 |
| $B_s^0 \to D_s^{*-} \pi^+$ | 40 500 | 495 | 1 116 |
| $B_s^0 \to D_s^- a_1^+$ | 98 450 | 5 757 | 3 368 |
| $B_d^0 \to D_s^+ a_1^-$ | 50 000 | 1 385 | $< 2454$ |
| $B_d^0 \to D^- a_1^+$ | 50 000 | 49 | 36 |
| $B_s^0 \to D_s^{*-} a_1^+$ | 100 000 | 870 | 1 052 |

## 6.2 Construction of the Likelihood Function

The probability density to observe an initial $B_j^0$ meson ($j = d,\ s$) decaying at time $t_0$ after its creation as a $\bar{B}_j^0$ meson is given by

$$p_j(t_0,\ \mu_0) \ = \ \frac{\Gamma_j^2 - (\Delta\Gamma_j/2)^2}{2\Gamma_j}\ \mathrm{e}^{-\Gamma_j t_0} \left( \cosh \frac{\Delta\Gamma_j t_0}{2} + \mu_0 \cos(\Delta m_j t_0) \right) \tag{1}$$

where $\Delta\Gamma_j = \Gamma_{\rm H}^j - \Gamma_{\rm L}^j$, $\Gamma_j = (\Gamma_{\rm H}^j + \Gamma_{\rm L}^j)/2$ and $\mu_0 = -1$.

For the unmixed case (an initial $B_j^0$ meson decaying as a $B_j^0$ meson at time $t_0$), the probability density is obtained by setting $\mu_0 = +1$ in Eq. 1. Here the small effects of CP violation are neglected. Unlike $\Delta\Gamma_d$, which can be safely set to zero, the width difference $\Delta\Gamma_s$ in the $B_s^0$-$\bar{B}_s^0$ system could be as much as 20% of the total width [17].

However, the above probability is modified by experimental effects. The probability as a function of $\mu_0$ and the reconstructed proper time $t$ is obtained as the convolution of $p_j(t_0,\ \mu_0)$ with the proper time resolution $\mathrm{Res}_j(t\,|\,t_0)$:

$$q_j(t,\mu_0) = \frac{1}{N}\int_0^\infty p_j(t_0,\mu_0)\ \mathrm{Res}_j(t\mid t_0)\ \mathrm{d}t_0 \tag{2}$$

with the normalisation factor

$$N = \int_{t_{\rm min}}^\infty \left(\int_0^\infty p_j(t',\mu_0)\ \mathrm{Res}_j(t\mid t')\ \mathrm{d}t'\right)\ \mathrm{d}t\ . \tag{3}$$

Here $t_{\rm min}$ = 0.4 ps is the cut on the $B_s^0$ reconstructed proper decay time. Plots in Fig. 11 show the proper time resolutions, which are parametrised with the sum of two Gaussian functions around the same mean value. The widths from the fit are $\sigma_1$ = (68.4±3.3) fs for the core fraction of 53.2% and $\sigma_2$ = (157.2±5.7) fs for the rest of the tail part of the distribution for the $B_s^0 \to D_s^-\pi^+$ channel. The values for the $B_s^0 \to D_s^- a_1^+$ channel are $\sigma_1$ = (72.5±4.3) fs for the core fraction of (58.0±6.8) % and $\sigma_2$ = (144.7±7.3) fs for the tail part.

Assuming a fraction $\omega_j$ of wrong tags occurring at production and/or decay, the probability becomes

$$\tilde{q}_j(t,\mu) = (1-\omega_j)q_j(t,\mu) + \omega_j q_j(t,-\mu) \tag{4}$$

where $\mu_0$ has been replaced by $\mu$ in order to indicate that now we are talking about the experimental observation of same or opposite flavour tags. For each signal channel, the background is composed of oscillating $B_d^0$ mesons, with probability given by Eq. 4, and of non-oscillating combinatorial background, with probability given by Eq. 5, which results from Eq. 1 and Eq. 4 by setting $\Delta m = 0$ and $\Delta\Gamma = 0$ :

$$\mathrm{p_{cb}}(t,\mu) = \frac{\Gamma_{\rm cb}}{2}\ \mathrm{e}^{-\Gamma_{\rm cb}t}\left[1+\mu\ (1-2\omega_{\rm cb})\right] \tag{5}$$

For a fraction $f_{kj}$ of the $j$ component ($j = s,\ d$, and combinatorial background cb) in the total sample of type $k$, one obtains the probability density function

$$\mathrm{pdf}_k(t,\mu) = \sum_{j=s,d,\mathrm{cb}} f_{kj}\tilde{q}_j(t,\mu)\ . \tag{6}$$

The index $k = 1$ denotes the $B_s^0 \to D_s^-\pi^+$ channel and $k = 2$ the $B_s^0 \to D_s^- a_1^+$ channel. The likelihood of the total event sample is written as

$$\mathscr{L}(\Delta m_s, \Delta\Gamma_s) = \prod_{k=1}^{N_{\rm ch}}\prod_{i=1}^{N_{\rm ev}^k}\mathrm{pdf}_k(t_i,\mu_i) \tag{7}$$

where $N_{\rm ev}^k$ is the total number of events of type $k$, and $N_{\rm ch} = 2$. Each $\mathrm{pdf}_k$ is properly normalised to unity.

Figure 12 shows how the experimental effects, as parametrised in Eqs. 2, 4, and 6, modify the distribution of the proper time $t$ of a Monte Carlo data sample.

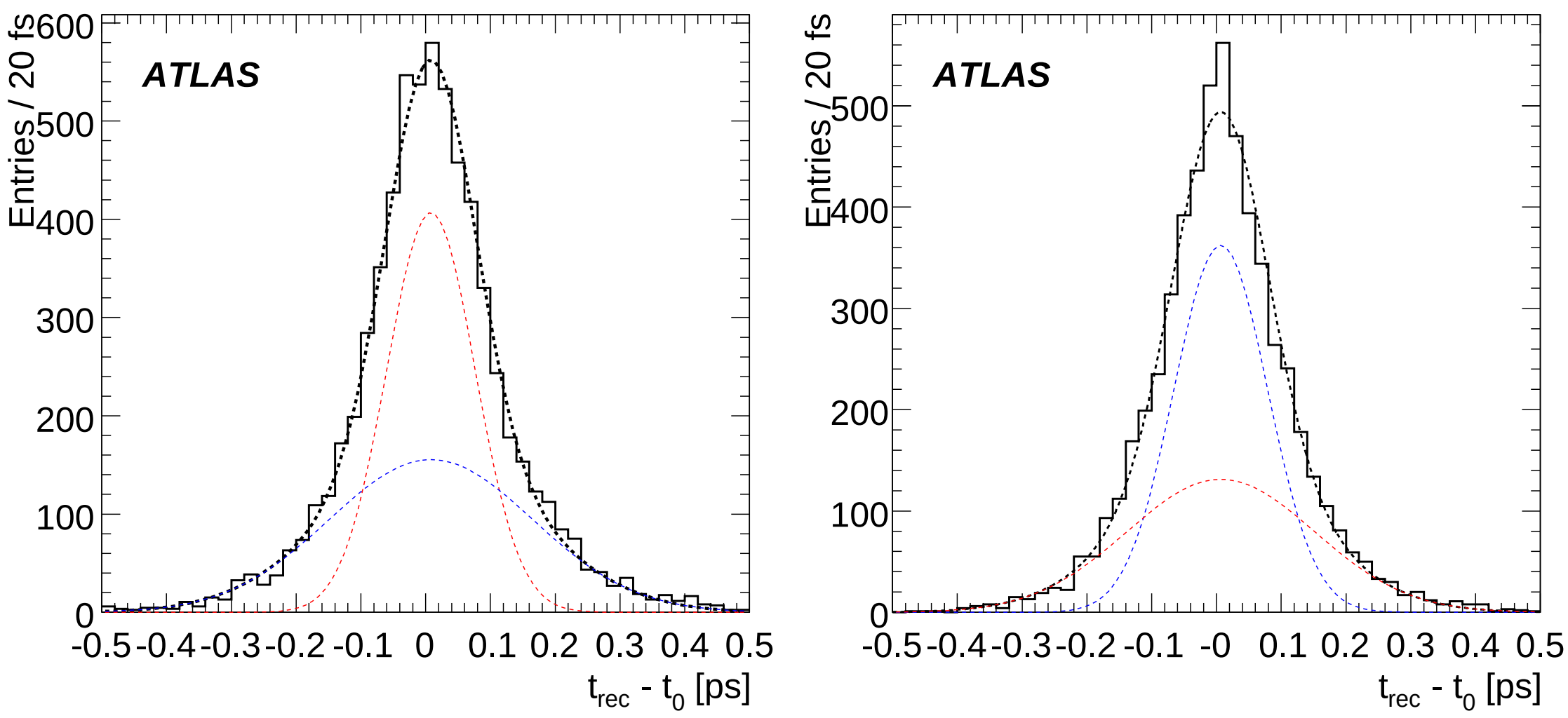


(a) $B_s^0 \to D_s^- \pi^+$

(b) $B_s^0 \to D_s^- a_1^+$

Figure 11: The resolution $\sigma_t$ of the proper time of simulated $B_s^0$ fitted with two Gaussian functions (dashed lines). Both Gaussian functions use a common mean value.

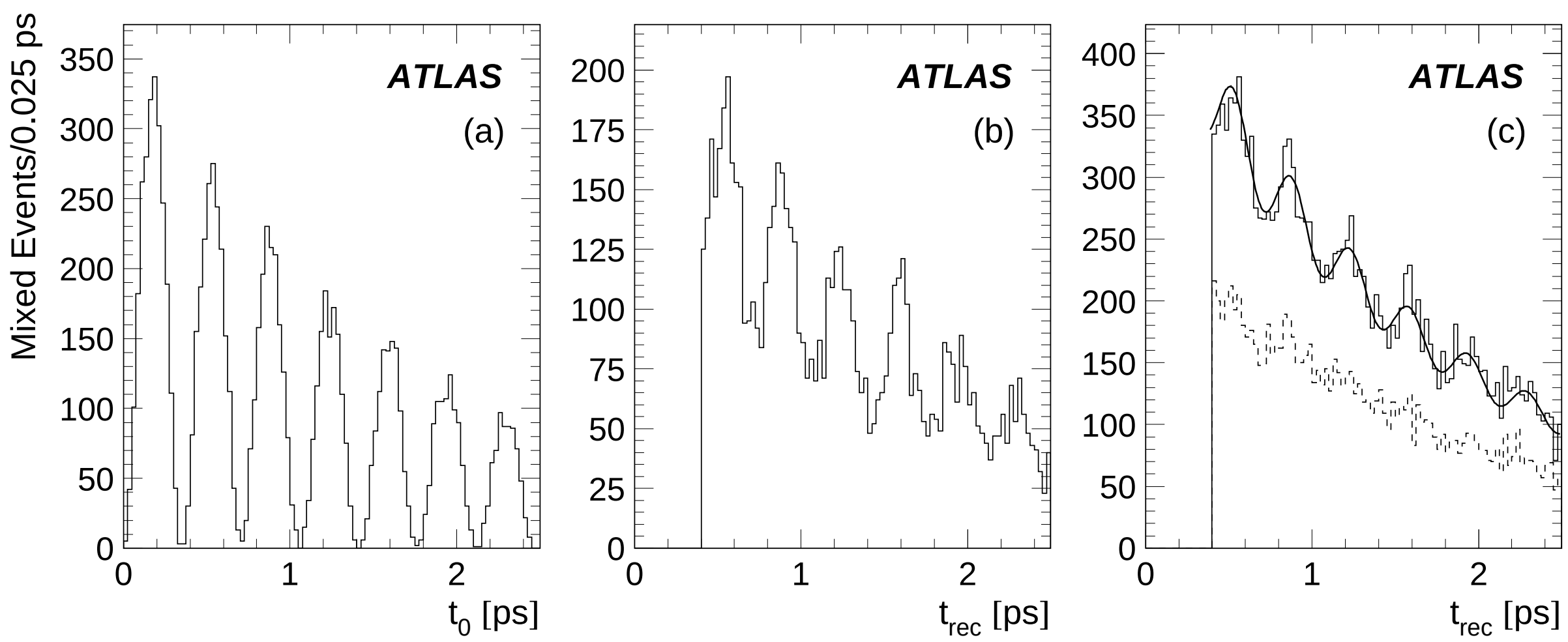


Figure 12: A sequence of plots showing how a true $B_s^0$ oscillation signal with $\Delta m_s^{\text{gen}}$ = 17.77 ps$^{-1}$ (a) is diluted first by the effect of a finite proper time resolution (b) and then by adding background events and including the effect of wrong tags (c). The plots contain samples of events equivalent to 10 fb$^{-1}$ of integrated luminosity. They were generated using the Monte Carlo method described in Section 6.3. Only the case of mixed events is shown. For illustration a $\chi^2$-fit of the function $C\exp(-t/\tau)(1-D\cos(\Delta m_s t))$ is overlaid to the solid histogram in (c), where $D$ can be interpreted as the combined dilution factor: $D \approx 0.1$ here. Note that this is different from the unbinned maximum likelihood fit to the total event sample of mixed and unmixed events which is actually used to derive results in this study. The dashed histogram in (c) describes the contribution from all background sources.

### 6.3 Monte Carlo Data Sample

For the amplitude fit method a simplified Monte Carlo method is applied to generate a $B_s^0$ sample using the following input parameters: for each signal channel $k$ the number of reconstructed signal events $N_{\text{sig}}^{(k)}$ for an integrated luminosity of 10 fb$^{-1}$ and the number of background events $N_{B_{d,s}^0}^{(k)}$ from $B_{d,s}^0$ decays is given in Table 12. For the combinatorial background the ratio $N_{\text{sig}}^{(k)}/N_{\text{cb}}^{(k)}$ is taken to be 1. The wrong tag fraction is assumed to be the same for both $B_s^0$ and $B_d^0$ mesons in a specific signal channel ($\omega_s = \omega_d$), however, the values are slightly different for the two signal channels (see Table 11).

A Monte Carlo sample with $N_{\text{sig}} = N_{\text{sig}}^{(1)} + N_{\text{sig}}^{(2)}$ signal events oscillating with a given frequency $\Delta m_s$ (e.g. $\Delta m_s$ = 100 ps$^{-1}$, which is far off the expected value for $\Delta m_s$), together with $N_{B_{d,s}^0} = N_{B_{d,s}^0}^{(1)} + N_{B_{d,s}^0}^{(2)}$ background events oscillating with frequency $\Delta m_{d,s}$ and $N_{\text{cb}} = N_{\text{cb}}^{(1)} + N_{\text{cb}}^{(2)}$ combinatorial events (no oscillations) is generated according to Eq. 1.

The uncertainty on the measurement of the transverse decay length, $\sigma_{d_{xy}}$ (see Fig. 13), and the true value of the $g$-factor $g_0$ ($g := m/p_T$) as seen in Fig. 14(a), are generated randomly according to the distributions obtained from the simulated samples, fitted with appropriate combinations of Gaussian and exponential functions. For the $B_s^0 \to D_s^- a_1^+$ channel the true $p_T^0$ distribution shown in Fig. 14(b) is fitted with a combination of a parabola function in the low $p_T^0$ region and a sum of two exponential functions in the high $p_T^0$ region. The $g_0$ values are obtained by converting generated $p_T^0$ values at random.

From the computed true decay length, $d_{xy}^0 = t_0/g_0$, the corresponding reconstructed decay length is generated as $d_{xy} = d_{xy}^0 + \sigma_{d_{xy}} \cdot (\mu_{dxy} + S_{d_{xy}}\Omega)$. $t_0$ is the proper time of the generated $B_s$. $S_{d_{xy}}$ is the width and $\mu_{dxy}$ the mean value of the Gaussian shape of the pull of the transverse decay length $\frac{d_{xy}-d_{xy}^0}{\sigma_{d_{xy}}}$ shown in Fig. 15. The fitted values are $S_{d_{xy}} = 1.099 \pm 0.011$ and $\mu_{dxy} = (8.76 \pm 1.47) \cdot 10^{-2}$ for the $B_s^0 \to D_s^- \pi^+$ channel respectively $S_{d_{xy}} = 1.113 \pm 0.011$ and $\mu_{dxy} = (5.40 \pm 1.48) \cdot 10^{-2}$ for the $B_s^0 \to D_s^- a_1^+$ channel. The reconstructed $g$-factor is generated as $g = g_0 + g_0\mu_g + g_0 S_g \Omega'$. The distribution of the fractional $g$-factor $\frac{g-g_0}{g_0}$ as shown in Fig. 16 is fitted with a Gaussian resulting in a width of $S_g = (0.89 \pm 0.01) \cdot 10^{-2}$ and a mean value of $\mu_g = (0.27 \pm 0.12) \cdot 10^{-3}$ for the $B_s^0 \to D_s^- \pi^+$ channel respectively $S_g = (0.82 \pm 0.01) \cdot 10^{-2}$ and $\mu_g = (0.56 \pm 0.11) \cdot 10^{-3}$ for the $B_s^0 \to D_s^- a_1^+$ channel. Both $\Omega$ and $\Omega'$ are random numbers distributed according to the normal distribution. From the transverse decay length and $g$-factor, the reconstructed proper time is then computed as $t = g d_{xy}$. The probability for the event to be mixed or unmixed is determined from the $t_0$ and $\Delta m_s$ (or $\Delta m_d$) values using the expression $(1 - \cos(\Delta m_j t_0)/\cosh(\Delta\Gamma_j t_0/2))/2$ which is left from Eq. 1 after the exponential part has been separated.

For a fraction of the events, selected at random, the state is interchanged between mixed and unmixed, according to the wrong tag fraction $\omega_{\text{tag}}$. Half of the combinatorial events are added to the mixed events and half to the unmixed events.

For the exclusive $B_{d,s}^0$ background channels as well as the combinatorial background, the reconstructed proper time is generated assuming that it has the same distribution as the one for signal $B_s^0$ mesons coming from the $D_s^- \pi^+$ and $D_s^- a_1^+$ sample respectively, no mixing included.

The $\Delta m_s$ measurement limits are obtained applying the amplitude fit method [5] to the sample generated as described in the previous section. According to this method a new parameter, the $B_s^0$ oscillation amplitude $\mathscr{A}$, is introduced in the likelihood function by replacing the term '$\mu_0 \cos \Delta m_s t_0$' with '$\mu_0 \mathscr{A} \cos \Delta m_s t_0$' in the $B_s^0$ probability density function given by Eq. 1. The new likelihood function, similar to Eq. 7, again includes all experimental effects. For each value of $\Delta m_s$, this likelihood function is minimized with respect to $\mathscr{A}$, keeping all other parameters fixed, and a value $\mathscr{A} \pm \sigma_{\mathscr{A}}^{\text{stat}}$ is obtained. One expects, within the estimated uncertainty, $\mathscr{A} = 1$ for $\Delta m_s$ close to its true value, and $\mathscr{A} = 0$ for $\Delta m_s$

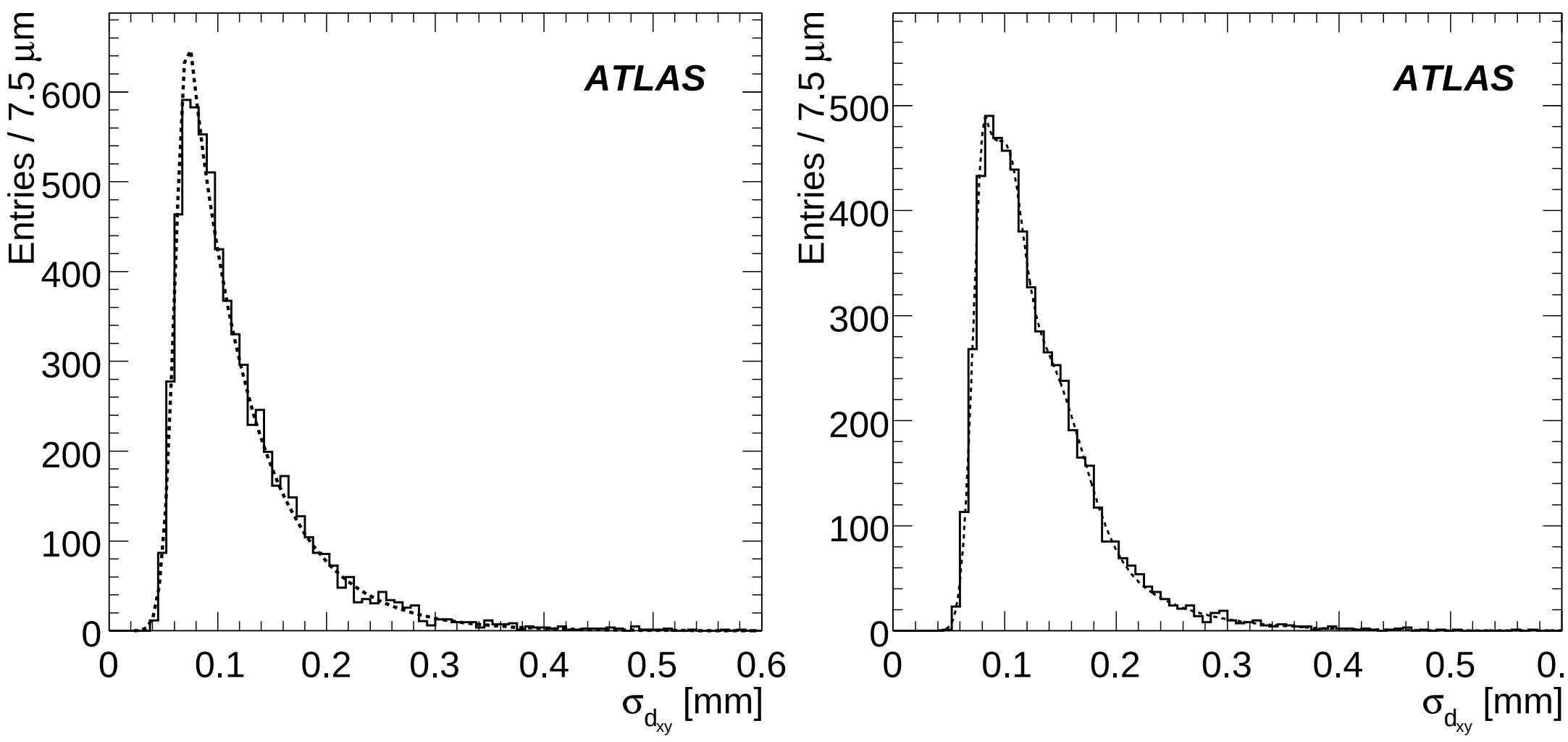


(a) $B_s^0 \to D_s^- \pi^+$ (b) $B_s^0 \to D_s^- a_1^+$

Figure 13: The uncertainty on the measurement of the transverse decay length, $\sigma_{d_{xy}}$ including trigger selection.

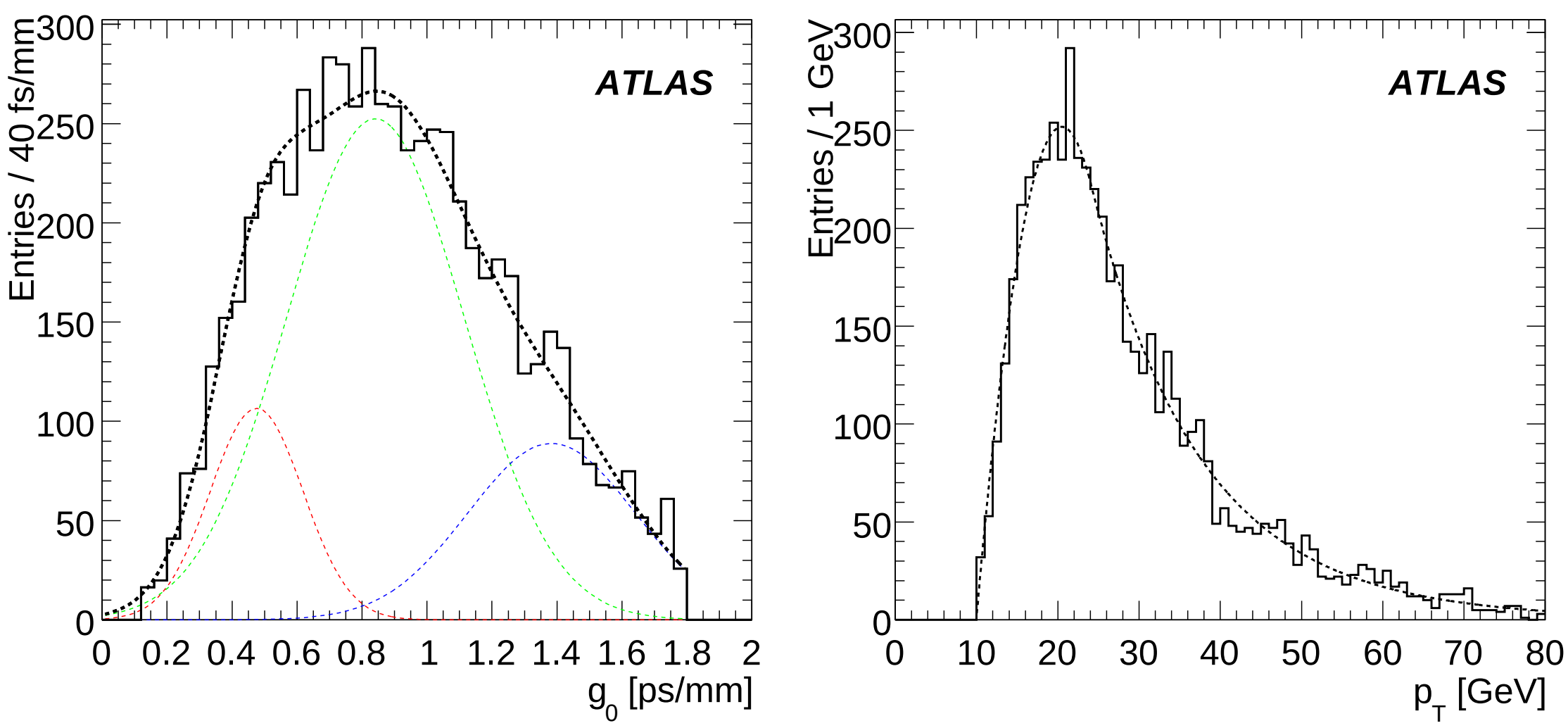


(a) $B_s^0 \to D_s^- \pi^+$ (b) $B_s^0 \to D_s^- a_1^+$

Figure 14: The true value of the $g$-factor $g_0 = t_0/d_{xy}^0$ of simulated $B_s^0$ from the $B_s^0 \to D_s^- \pi^+$ sample (a) fitted with the sum of three Gaussian functions (dashed lines) and the true transverse momentum distribution $p_T^0$ of simulated $B_s^0$ from the $B_s^0 \to D_s^- a_1^+$ sample (b) including trigger selection.

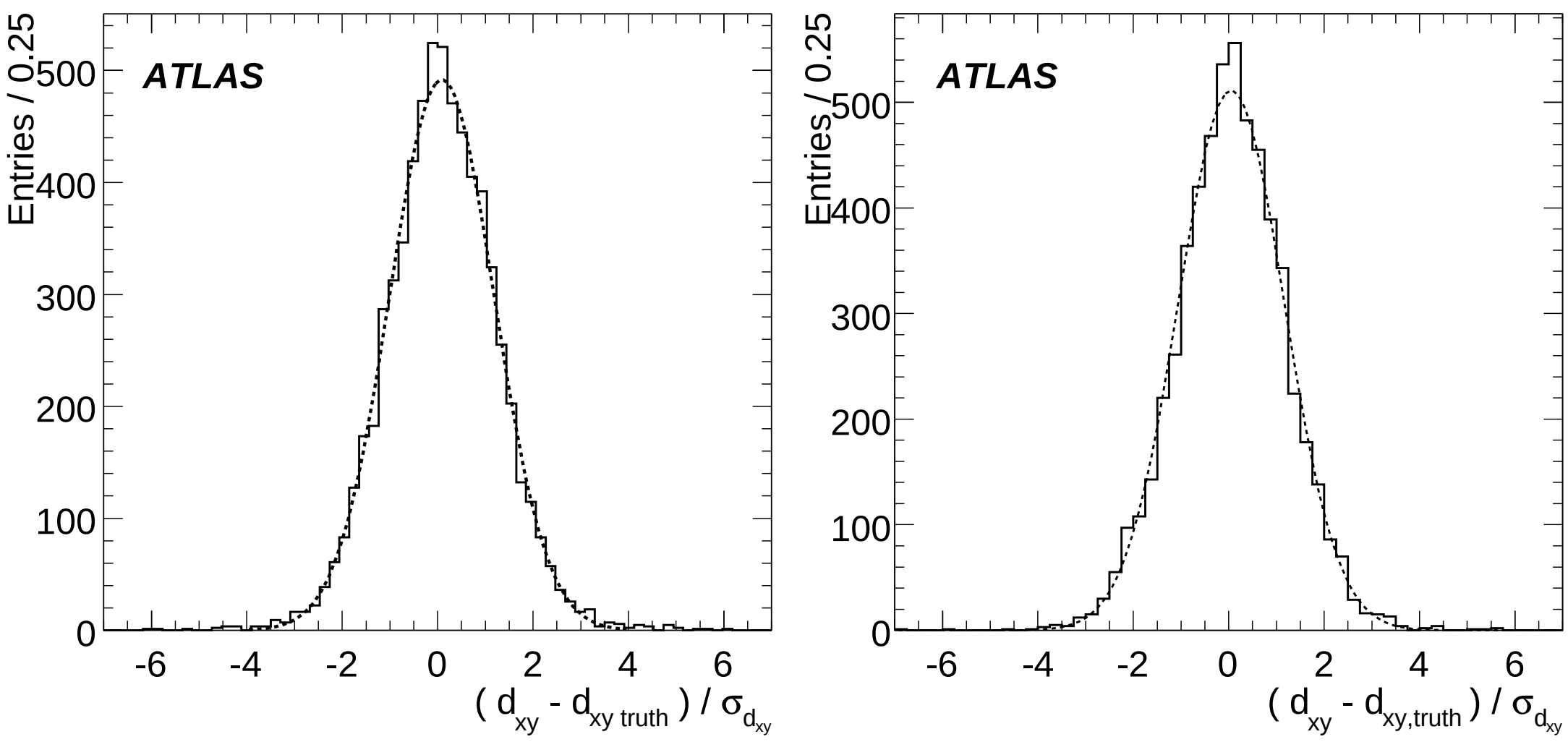


(a) $B_s^0 \to D_s^- \pi^+$ (b) $B_s^0 \to D_s^- a_1^+$

Figure 15: The pull of the measurement of the transverse decay length, $\frac{d_{xy}-d_{xy}^0}{\sigma_{d_{xy}}}$ and fit of a Gaussian function (dashed) to the distribution including trigger selection.

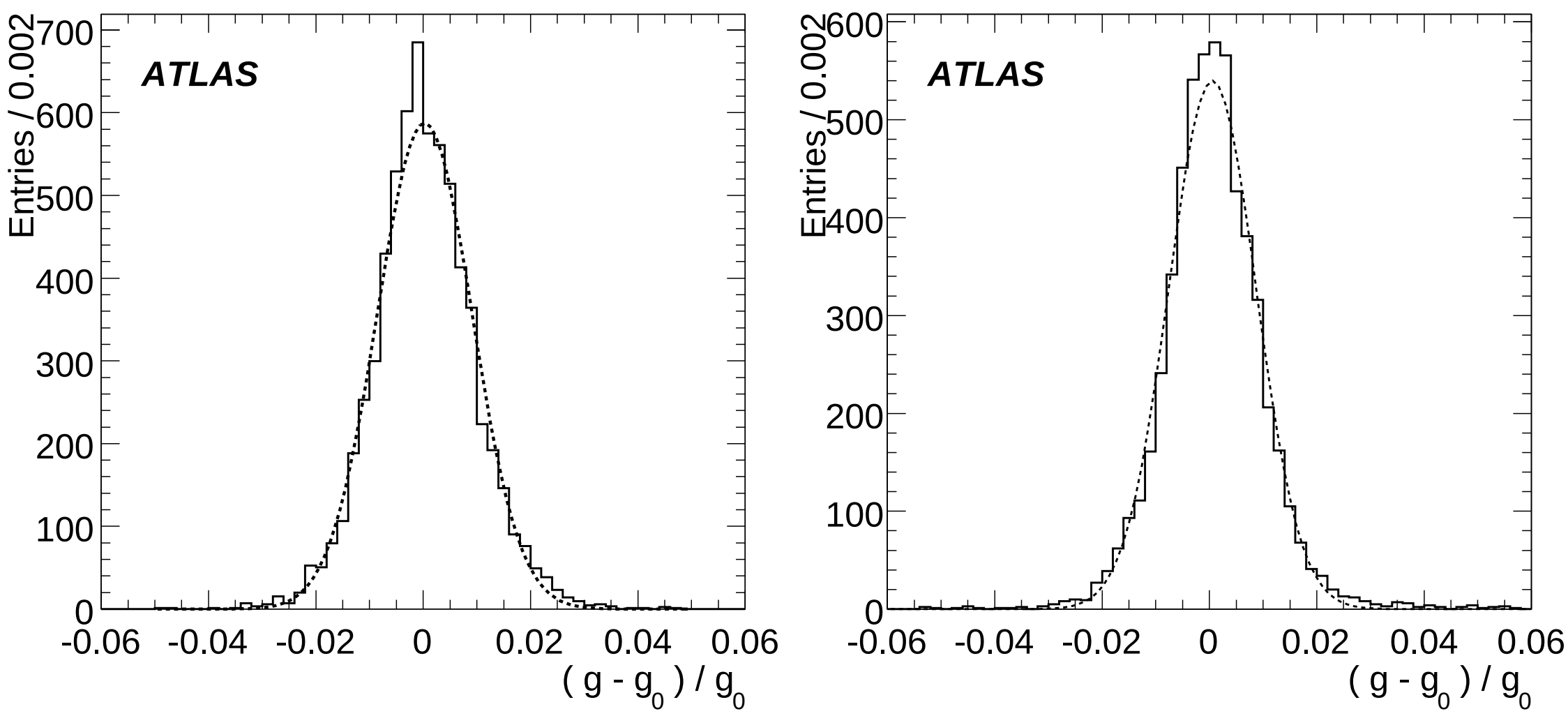


(a) $B_s^0 \to D_s^- \pi^+$ (b) $B_s^0 \to D_s^- a_1^+$

Figure 16: The fractional resolution of the $g$-factor $\frac{g-g_0}{g_0}$ of simulated $B_s^0$ fitted with a single Gaussian function including trigger selection.

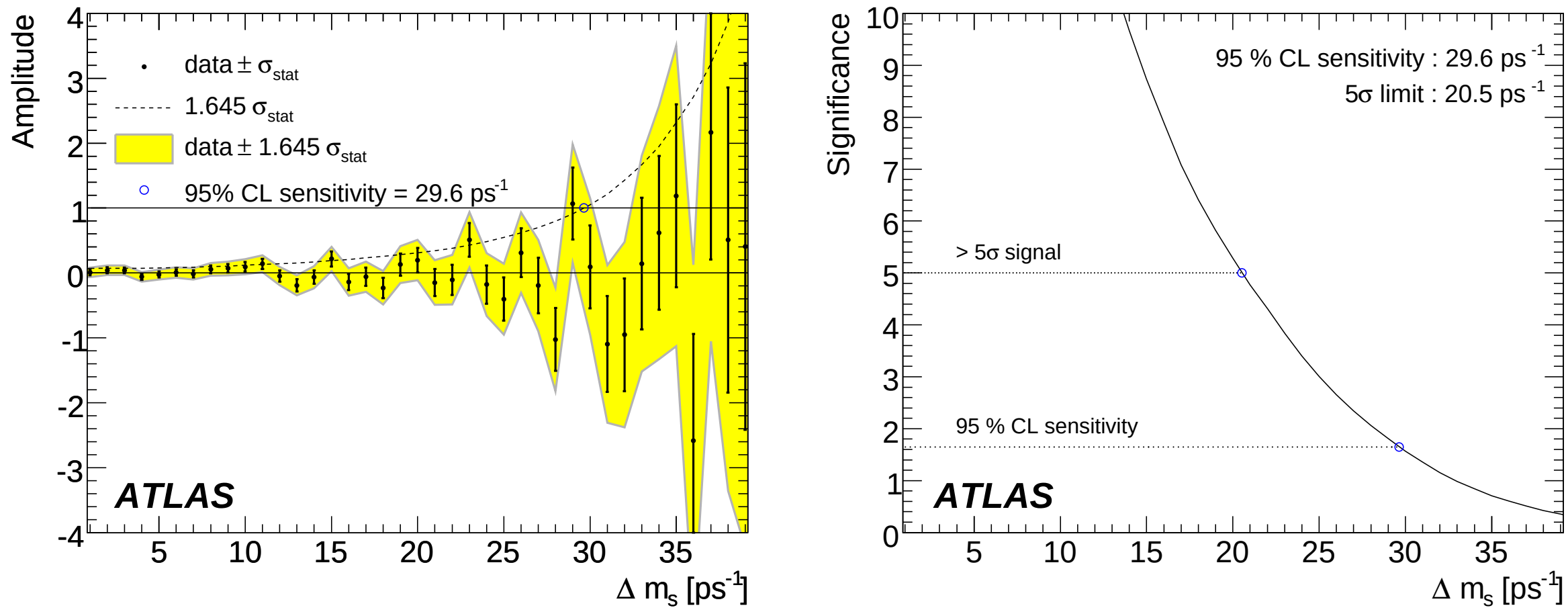


(a) Amplitude vs. $\Delta m_s$ (b) Significance vs. $\Delta m_s$

Figure 17: The $B_s^0$ oscillation amplitude (a) and the measurement significance (b) as a function of $\Delta m_s$ for an integrated luminosity of 10 fb$^{-1}$ for a specific Monte Carlo experiment with $\Delta m_s^{\rm gen}$ = 100 ps$^{-1}$ .

far from the true value. A five standard deviation measurement limit is defined as the value of $\Delta m_s$ for which $1/\sigma_{\mathscr{A}} = 5$, and a sensitivity at 95% C.L. as the value of $\Delta m_s$ for which $1/\sigma_{\mathscr{A}} = 1.645$. Limits are computed with the statistical uncertainty $\sigma_{\mathscr{A}}^{\rm stat}$. A detailed investigation on the systematic uncertainties $\sigma_{\mathscr{A}}^{\rm syst}$, which affects the measurement of the $B_s^0$ oscillation, is presented in [18].

## 6.4 Extraction of the $\Delta m_s$ Sensitivity

For the nominal set of parameters (as defined in the previous sections), $\Delta\Gamma_s = 0$ and an integrated luminosity of 10 fb$^{-1}$, the amplitude $\pm 1\sigma_{\mathscr{A}}^{\rm stat}$ is plotted as a function of $\Delta m_s$ in Fig. 17(a). The 95% C.L. sensitivity to measure $\Delta m_s$ is found to be 29.6 ps$^{-1}$. This value is given by the intersection of the dashed line, corresponding to 1.645 $\sigma_{\mathscr{A}}^{\rm stat}$, with the horizontal line at $\mathscr{A} = 1$.

From Fig. 17(b), which shows the significance of the measurement $S(\Delta m_s) = 1/\sigma_{\mathscr{A}}$ as a function of $\Delta m_s$, the $5\sigma$ measurement limit is found to 20.5 ps$^{-1}$.

The dependence of the $\Delta m_s$ measurement limits on the integrated luminosity is shown in Fig. 18(a), with the numerical values given in Table 13.

Table 13: The dependence of $\Delta m_s$ measurement limits on the integrated luminosity $\mathscr{L}$.

| $\mathscr{L}$ [fb$^{-1}$] | 5 $\sigma$ limit [ps$^{-1}$] | 95% C.L. sensitivity [ps$^{-1}$] |
|---|---|---|
| 3 | 14.5 | 25.0 |
| 5 | 17.0 | 27.0 |
| 10 | 20.5 | 29.6 |
| 20 | 23.7 | 32.0 |
| 30 | 25.3 | 33.2 |
| 40 | 26.4 | 34.1 |

The dependence of the $\Delta m_s$ measurement limits on $\Delta\Gamma_s/\Gamma_s$ is determined for an integrated luminosity

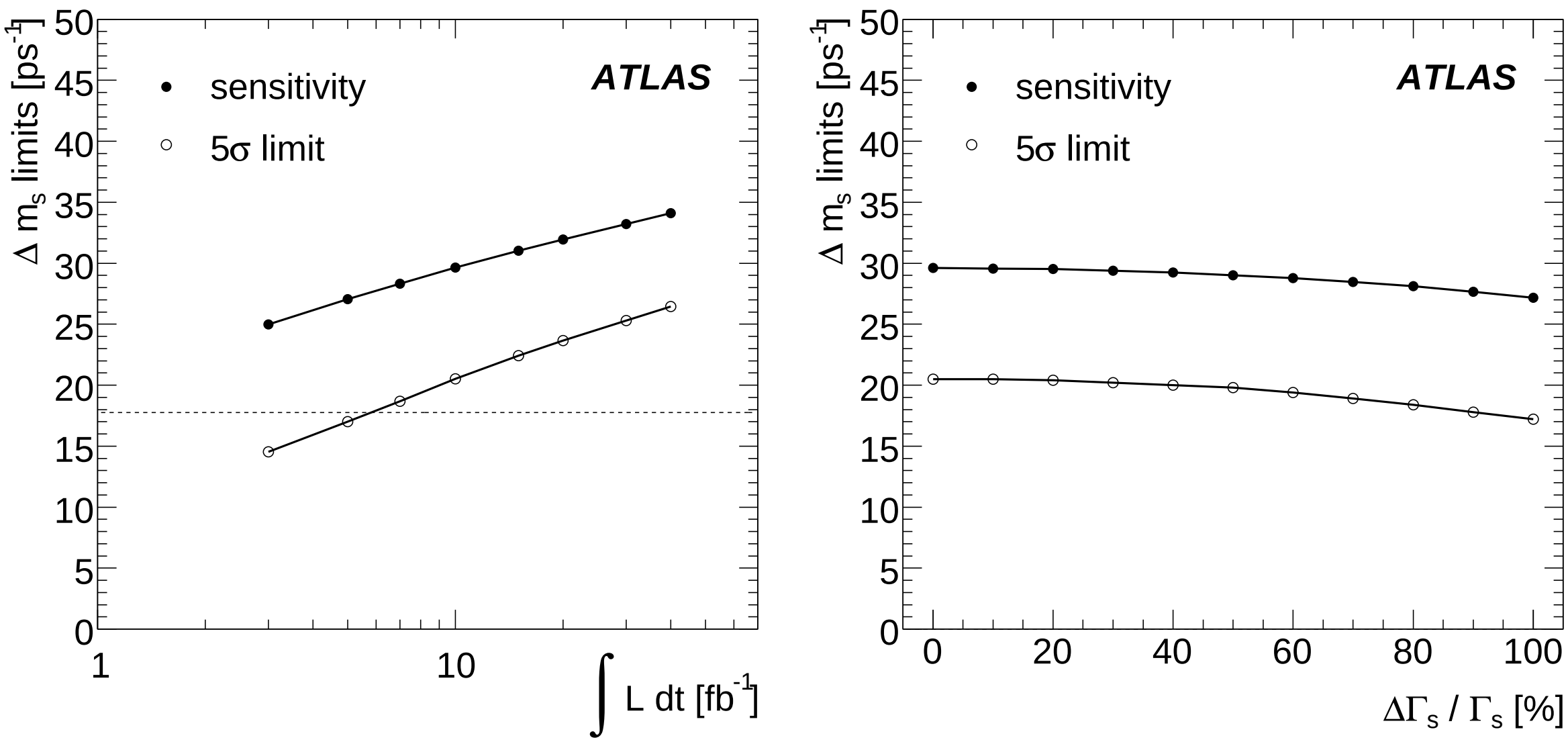


(a) $\Delta m_s$ limits vs. $\int L dt$ (b) $\Delta m_s$ limits vs. $\Delta\Gamma_s/\Gamma_s$

Figure 18: The dependence of $\Delta m_s$ measurement limits (a) on the integrated luminosity and (b) on $\Delta\Gamma_s/\Gamma_s$ for an integrated luminosity of 10 fb$^{-1}$. The dashed horizontal line in (a) denotes the CDF measurement.

of 10 fb$^{-1}$, other parameters having their nominal value. The $\Delta\Gamma_s/\Gamma_s$ is used as a fixed parameter in the amplitude fit method. As shown in Fig. 18(b) no sizeable effect is seen up to $\Delta\Gamma_s/\Gamma_s \sim 30\%$.

### 6.5 Extraction of the $\Delta m_s$ Measurement Precision

Whereas the $\Delta m_s$ measurement limits are obtained by using the amplitude method (see previous section), in case of the presence of an oscillation signal in the data the value of the oscillation frequency $\Delta m_s$ and its precision are determined by minimising the likelihood (given by Eq. 7) with respect to $\Delta m_s$. In this fit $\Delta\Gamma_s$ is fixed to 0, because a study has shown that the systematic uncertainties resulting from varying $\Delta\Gamma_s/\Gamma_s$ in the range 0 to 0.2 (suggested by the present uncertainty) are practically negligible.

An example of the likelihood function is given in Fig. 19(a), in which the $\Delta m_s^{\text{gen}}$ in the Monte Carlo sample has been set to the value measured by CDF for illustration. From this type of graphs the precision of the measurement of $\Delta m_s$ is extracted and plotted in Fig. 19(b) as a function of the integrated luminosity for three values of $\Delta m_s^{\text{gen}}$.

### 6.6 Discussion of Results

In this note it is shown that with an integrated luminosity of 10 fb$^{-1}$ ATLAS is able to verify the CDF measurement of $\Delta m_s = (17.77 \pm 0.10\,(\text{stat}) \pm 0.07\,(\text{sys}))\,\text{ps}^{-1}$ at the five standard deviation level. For these parameters the statistical error on $\Delta m_s$ is calculated to be about 0.065 ps$^{-1}$.

In a preceding study [18] it was found that over a wide range of values for $\Delta m_s$ and integrated luminosity the systematic uncertainty on the measured value of $\Delta m_s$ was smaller by at least a factor of 10 compared to the statistical uncertainty. The list of contributions to that systematic error estimation included the wrong tag fraction with a relative error of 5% compared to 2.5% found in this study. For the reasons mentioned above, the evaluation of systematic effects has not been repeated here. The study of the effect of varying $\Delta\Gamma_s$ (as explained in the previous section) is new, but the contribution to the systematic uncertainty is also very small.

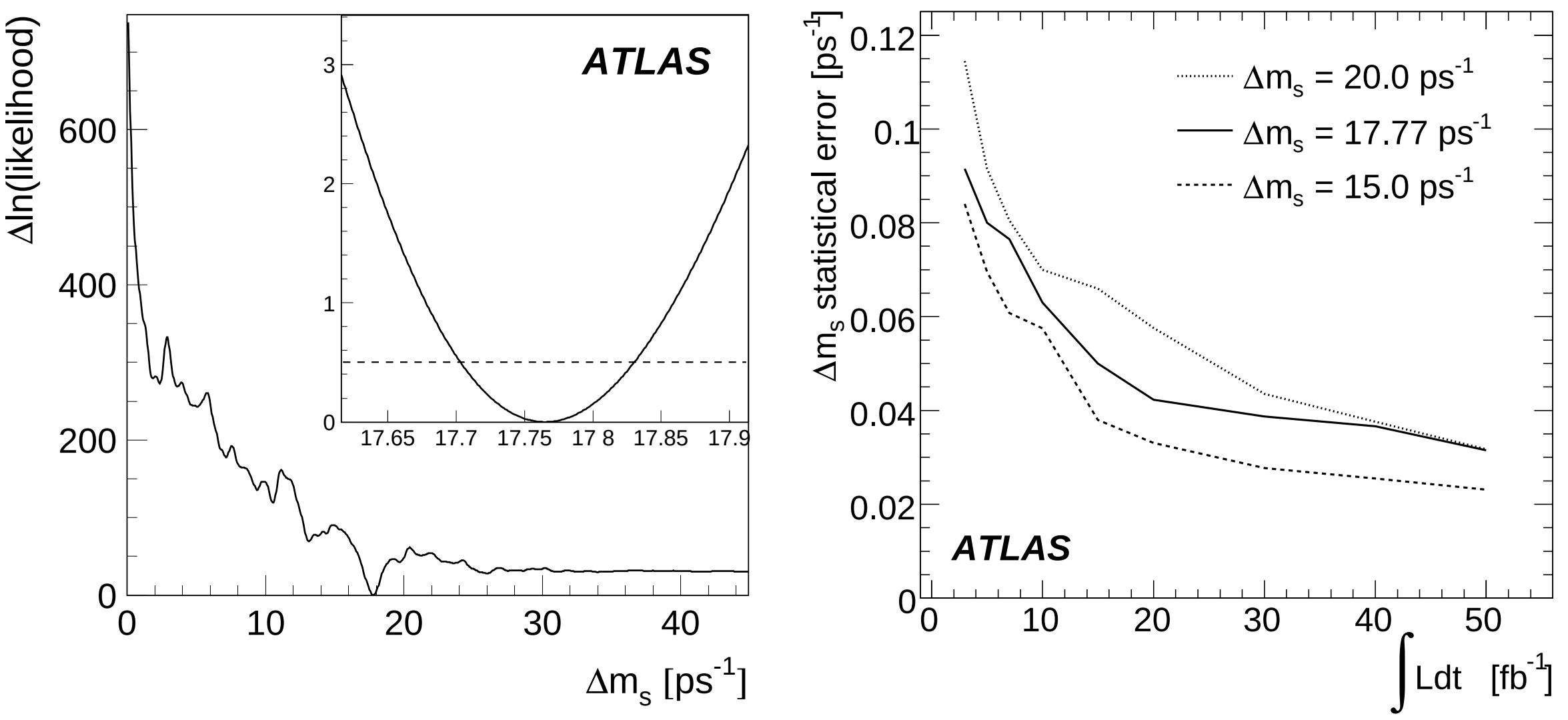


(a) $\Delta\ln$(likelihood) vs. $\Delta m_s$ (b) $\sigma_{\text{stat}}(\Delta m_s)$ vs. $\int Ldt$

Figure 19: (a) The negative natural logarithm of the likelihood for a specific Monte Carlo data sample for an integrated luminosity of 10 fb$^{-1}$ and a true value of $\Delta m_s^{\text{gen}}$ = 17.77 ps$^{-1}$. The inset shows a zoom around the minimum. (b) The statistical error $\sigma_{\text{stat}}(\Delta m_s)$ as a function of the integrated luminosity for values of $\Delta m_s^{\text{gen}}$ of 15, 17.77 and 20 ps$^{-1}$. For comparison: the CDF statistical error on their $\Delta m_s$ measurement is 0.10 ps$^{-1}$.

Systematic uncertainties on the overall trigger efficiencies mainly effect the statistics available for the analysis. However, an important systematic effect for the $\Delta m_s$ measurement would be introduced in case different trigger efficiencies for positively and negatively charged muons are observed. In order to constrain this effect, dimuon events from a calibration channel like $B^+ \to J/\psi K^+$ with $J/\psi \to \mu^+\mu^-$ which are triggered by a single muon trigger could be used.

Clearly LHCb can measure $\Delta m_s$ more precisely than ATLAS ($\sigma_{\text{stat}}(\Delta m_s) \sim 0.01$ ps$^{-1}$ with 2 fb$^{-1}$ of data [19]), but the $\Delta m_s$ measurement with ATLAS is needed for the simultaneous fit of all parameters of the weak sector of the $B_s^0$-$\bar{B}_s^0$ system (weak mixing phase $\phi_s$, $\Delta m_s$, $\Gamma_s$ and $\Delta\Gamma_s$). This will be performed by a combined analysis of the channels described in this note and the $B_s^0 \to J/\psi\,\phi$ channel [20]. The ATLAS measurement is an independent cross-check of the measurements performed by other experiments.

# 7 Summary and Conclusions

We have studied the capabilities of the ATLAS detector to measure $B_s^0$ oscillations in $pp$ collisions at 14 TeV using the purely hadronic decay channels $B_s^0 \to D_s^-(\phi\pi^-)\pi^+$ and $B_s^0 \to D_s^-(\phi\pi^-)a_1^+$. For an integrated luminosity of 10 fb$^{-1}$ a $\Delta m_s$ sensitivity limit of 29.6 ps$^{-1}$ and a five standard deviation measurement limit of 20.5 ps$^{-1}$ is obtained from a likelihood fit employing the amplitude fit method. This result depends only weakly on the lifetime difference $\Delta\Gamma_s$. The trigger is based on a single muon trigger with adjustable muon $p_T$ thresholds between 4 and 10 GeV on all trigger levels and an active search for $D_s \to \phi\pi$ decays by the High Level Trigger. For $10^{31}$ cm$^{-2}$s$^{-1}$ we will be able to afford a muon trigger with the loosest $p_T$ threshold combined with a FullScan $D_s \to \phi\pi$ search. For $10^{32}$ cm$^{-2}$s$^{-1}$, we will need to increase the muon $p_T$ threshold to 6 GeV and possibly employ the RoI-based LVL2 trigger. In both cases, the trigger rates can be kept at an acceptable level. For higher luminosities, we need to implement additional constraints in the HLT in order to reduce the event output rates further.

The offline event reconstruction searches for the hadronic decay of the $B_s^0$ and requires a muon with a minimum $p_T$ of 6 GeV for each event. While the flavour of the $B_s^0$ at decay time is determined from the charge of the $D_s$ particle, the identification of the initial $B_s^0$ flavour at production time is extracted from the charge of the soft muon in the event, taking effects of $B_s^0$ mixing into account. An overall tagging efficiency of $98.8 \pm 0.2\%$ and $98.5 \pm 0.2\%$ as well as average wrong tag fractions of $22.3 \pm 0.6\%$ and $23.3 \pm 0.6\%$ for the $B_s^0 \to D_s^- \pi^+$ and the $B_s^0 \to D_s^- a_1^+$ channels, respectively, are obtained.

About 100 000 Monte Carlo events of each sample have been produced without $B_s^0$ oscillations. Monte Carlo events for several exclusive $B_s^0$ and $B_d^0$ background channels as well as for inclusive background like $b\bar{b} \to \mu X$ and $c\bar{c} \to \mu X$ have been used. The hadronic decay of the signal side $B_s^0$ is reconstructed constraining the masses of intermediate particles in the decay chain. A $B_s^0$ mass resolution of $52.8 \pm 0.7$ MeV and $40.8 \pm 0.5$ MeV is obtained and after the application of all analysis cuts, 6 657 and 3 368 events are expected for an integrated luminosity of 10 fb$^{-1}$ for the $B_s^0 \to D_s^- \pi^+$ and the $B_s^0 \to D_s^- a_1^+$ decay channels, respectively. We have considered several exclusive background channels, contributing to the background inside the $B_s^0$ mass window. The $B_d^0 \to D^- \pi^+ / a_1^+$ channels hardly contribute (percent level of the signal), but the $B_s^0 \to D_s^{*-} \pi^+ / a_1^+$ make a considerable contribution of about 16% ($B_s^0 \to D_s^- \pi^+$) and 31% ($B_s^0 \to D_s^- a_1^+$). While the $B_d^0 \to D_s^- \pi^+$ channel is expected to contribute with about 1.5% relative to the $B_s^0 \to D_s^- \pi^+$ signal, we can only estimate the $B_d^0 \to D_s^- a_1^+$ contribution to be less than about 70% of the $B_s^0 \to D_s^- a_1^+$ signal, given that this decay channel has not yet been observed.

In future, the $B_s^0 \to D_s^{*-} \pi^+ / a_1^+$ channels may be considered signal rather than background. An estimate of the combinatorial background is severely limited by the available Monte Carlo event statistics. We plan to use early data to obtain a realistic estimate. For an integrated luminosity of 100 pb$^{-1}$ at $10^{32}$ cm$^{-2}$s$^{-1}$ we only expect about 90 events in the $B_s^0 \to D_s^- \pi^+$ and $B_s^0 \to D_s^- a_1^+$ channels, while the $B^+ \to J/\psi(\mu\mu)K^+$ decay channel, which has a large branching ratio, may be used to calibrate the soft muon tagging with early data. On LVL2 the processing time for the $D_s \to \phi\pi$ trigger may be reduced by restricting the track reconstruction to Regions of Interest (RoI), seeded by a LVL1 jet energy trigger. This typically leads to a reduction of the trigger efficiency by a few percent. At an instantaneous luminosity of $2 \cdot 10^{33}$ cm$^{-2}$s$^{-1}$ several options will be considered to achieve acceptable Event Filter output rates. Besides further constraining the mass windows of the $D_s \to \phi\pi$ trigger, other improvements are obtained by checking for a good reconstruction quality of the $D_s$ vertex or the implementation of a trigger element in the Event Filter which searches for the full $B_s^0$ decay chain. However, there is an uncertainty of a factor two in the overall $b\bar{b}$ cross-section at the centre-of-mass energy of 14 TeV and therefore the trigger rates may vary.

Due to the achieved sensitivity limit to measure the $B_s^0$ oscillations we expect to be able to verify the CDF measurement of $\Delta m_s = 17.77 \pm 0.10\,(\text{stat}) \pm 0.07\,(\text{sys})$ ps$^{-1}$ at the five standard deviation level with a statistical error on $\Delta m_s$ of about 0.065 ps$^{-1}$. This will provide a reasonable precision which allows us to combine the measurement described in this note with the analysis of the $B_s^0 \to J/\psi\phi$ channel [20] in a simultaneous fit for all parameters of the weak sector of the $B_s^0$-$\bar{B}_s^0$ system.

# Higgs Boson

# Introduction on Higgs Boson Searches

**Abstract**
The investigation of the dynamics responsible for electroweak symmetry breaking is one of the prime tasks of experiments at present and future colliders. Experiments at the CERN Large Hadron Collider (LHC) will be able to discover a Standard Model Higgs boson over the full mass range as well as Higgs bosons in extended models. In this introductory paper the Higgs boson production cross-section and decay branching ratios according to the Standard Model, and its Minimal Supersymmetric extension, are presented and discussed.

## 1 Introduction

The Large Hadron Collider at CERN, almost ready to start colliding proton beams at $\sqrt{s}$=14 TeV, will play an important role in the investigation of fundamental questions of particle physics. While the Standard Model of electroweak [1] and strong [2] interactions is in excellent agreement with the numerous experimental measurements, the dynamics responsible for electroweak symmetry breaking are still unknown. Within the Standard Model, the Higgs mechanism [3] is invoked to break the electroweak symmetry. A doublet of complex scalar fields is introduced, of which a single neutral scalar particle, the Higgs boson, remains after symmetry breaking [4]. Many extensions of this minimal version of the Higgs sector have been proposed, mostly discussed a scenario with two complex Higgs doublets as realized in the Minimal Supersymmetric Standard Model (MSSM) [5].

Within the Standard Model, the Higgs boson is the only particle that has not been discovered so far. The direct search at the $e^+e^-$ collider LEP has led to a lower bound on its mass of 114.4 GeV [6]. Indirectly, high precision electroweak data constrain the mass of the Higgs boson via their sensitivity to loop corrections. Assuming the overall validity of the Standard Model, a global fit [7] to all electroweak data leads to the 95% C.L. $m_H < 144$ GeV. The 95 % C.L. lower limit obtained from LEP is not used in the determination of this limit. Including it increases the limit to 182 GeV [7].

On the basis of the present theoretical knowledge, the Higgs sector in the Standard Model remains largely unconstrained. While there is no direct prediction for the mass of the Higgs boson, an upper limit of $\sim$1 TeV can be inferred from unitarity arguments [8].

Further constraints can be derived under the assumption that the Standard Model is valid only up to a cutoff energy scale $\Lambda$, beyond which new physics becomes relevant. Requiring that the electroweak vacuum is stable and that the Standard Model remains perturbative allows to set upper and lower bounds on the Higgs boson mass [9, 10]. For a cutoff scale of the order of the Planck mass, the Higgs boson mass is required to be in the range $130 < M_H < 180$ GeV. If new physics appears at lower mass scales, the bound becomes weaker, *e.g.*, for $\Lambda = 1$ TeV the Higgs boson mass is constrained to be in the range $50 < M_H < 800$ GeV.

Direct searches for the Standard Model Higgs boson at the Tevatron include looking for its production via gluon fusion and subsequent decay to $WW^{(*)}$ . The observed 95% C.L. limit by CDF (with an integrated luminosity of 2.4 fb$^{-1}$ analysed) is 0.85 pb for $M_H$= 160 GeV, which is about 1.6 times the Standard Model prediction [11]. Similarly the DØ Collaboration excludes at 95% C.L. the production of this boson with a cross-section about 2.4 times the one predicted by the Standard Model ($\sigma_{SM}^{WW}$). Searches at low mass are done studying Higgs bosons produced in association with the $W$ and $Z$, and looking for $H \to b\bar{b}$ with leptonic $W$ an $Z$ decays $(e, \mu)$. CDF sets a 95% C.L. limit to $8.2\times\sigma_{SM}^{b\bar{b}}$, while the one from DØ is $11\times\sigma_{SM}^{b\bar{b}}$, for $M_H$= 115 GeV. Preliminary results on the combination of the results from these two experiments lead to a 95% C.L. limit on the Higgs boson production cross-section to about $5.1\times\sigma_{SM}$ for $M_H$= 115 GeV, and $1.1\times\sigma_{SM}$ for $M_H$= 160 GeV [11].

The Minimal Supersymmetric Standard Model contains two complex Higgs doublets, leading to five physical Higgs bosons after electroweak symmetry breaking: three neutral (two CP-even $h$ and $H$, and one CP-odd $A$) and a pair of charged Higgs bosons $H^{\pm}$. At tree level, the Higgs sector of the MSSM is fully specified by two parameters, generally chosen to be $m_A$, the mass of the CP-odd Higgs boson, and $\tan\beta$, the ratio of the vacuum expectation values of the two Higgs doublets. Radiative corrections modify the tree-level relations significantly. This is of particular interest for the mass of the lightest CP-even Higgs boson, which at tree level is constrained to be below the mass of the $Z$ boson. Loop corrections are sensitive to the mass of the top quark, to the mass of the scalar particles and in particular to mixing in the stop sector. The largest values for the mass of the Higgs boson $h$ are reached for large mixing, characterized by large values of the mixing parameter $X_t := A_t - \mu\cot\beta$, where $A_t$ is the trilinear coupling and $\mu$ is the Higgs mass parameter. If the full one-loop and the dominant two-loop contributions are included [12, 13], the upper bound on the mass of the light Higgs boson $h$ is expected to be around 135 GeV ($\mathrm{m}_h$-max scenario). While the light neutral Higgs boson may be difficult to distinguish from its Standard Model counterpart, the other heavier Higgs bosons are a distinctive signal of physics beyond the Standard Model. The masses of the heavier Higgs bosons $H$, $A$ and $H^{\pm}$ are often almost degenerate.

Direct searches at LEP have given lower bounds of 92.9 (93.3) GeV and 93.4 (93.3) GeV on the masses of the lightest CP-even Higgs boson $h$ and the CP-odd Higgs boson $A$ within the $m_h$-max (no-mixing) scenario. [14] In those scenarios, the mixing parameter in the stop sector is set to values of $X_t$ = 2 TeV and $X_t$ = 0, respectively. Given the LEP results, the $\tan\beta$ regions of $0.9 < \tan\beta < 1.5$ and $0.4 < \tan\beta < 5.6$ are excluded at 95% confidence level for the $m_h$-max and the no-mixing scenarios, respectively [14]. However, it should be noted that the exclusions in $\tan\beta$ depend critically on the exact value of the top-quark mass. In the LEP analysis $m_t$ = 179.3 GeV has been assumed. With decreasing top mass the theoretical upper bound on $m_h$ decreases and hence the exclusions in $\tan\beta$ increase, while for $m_t$ of about 183 GeV , or higher, the exclusions in $\tan\beta$ vanish.

Direct searches at the Tevatron have been performed looking to the tau-pair and b-pair production. With an integrated luminosity of 1.8 $\mathrm{fb}^{-1}$ no excess of events has been observed, and exclusion limits on production cross-section times branching fraction to tau pairs for a Higgs boson mass in the range from 90 to 250 GeV have been set. The expected reach of this search, assuming MSSM Higgs production, extends below $\tan\beta$= 40 for $M_A$ in the mass range $M_A$= 120 to 160 GeV [15].

The charged Higgs boson mass is related to $m_A$ via the tree-level relation $m^2_{H^{\pm}} = m^2_W + m^2_A$ and it is less sensitive to radiative corrections [16]. Direct searches for charged Higgs bosons in the decay modes $H^{\pm} \to \tau\nu$ and $H^{\pm} \to cs$ have been carried out at LEP, yielding a lower bound of 78.6 GeV on $m_{H^{\pm}}$ independent of the $H^{\pm} \to \tau\nu$ branching ratio [17]. At the Tevatron, the CDF and DØ experiments have performed direct and indirect searches for the charged Higgs boson through the process $p\bar{p} \to t\bar{t}$ with at least one top quark decaying via $t \to H^{\pm}b$. These searches have excluded the small and large $\tan\beta$ regions for $H^{\pm}$ masses up to $\sim$160 GeV [18]. Other experimental bounds on the charged Higgs boson mass can be derived using processes where the charged Higgs boson enters as a virtual particle. For example, the measurement of the $b \to s\gamma$ decay rate allows indirect limits to be set on the charged Higgs boson mass [19] which, however, are strongly model dependent [20].

The high collision energy of the LHC will allow the search for Higgs bosons to be extended into unexplored mass regions. The experiments have a large discovery potential for Higgs bosons in both the Standard Model and in the MSSM over the full parameter range. Should the Higgs boson be light, *i.e.* have a mass in the range favoured by the precision electroweak measurements, the experiments at the Tevatron might also get indications of the existence of a Higgs boson.

In this chapter, the potential for Higgs boson searches at the Large Hadron Collider with the ATLAS experiment is reviewed, focussing on the investigation of the Higgs sectors in the Standard Model and in the MSSM.

Table 1: Set of parameters used in the evaluation of the Higgs boson cross-sections (see text).

| | | |
|---|---|---|
| $M_{uds} = 190$ MeV | $M_c = 1.40$ GeV | $M_b = 4.60$ GeV |
| $M_t = 172$ GeV | $M_Z$ =91.187 GeV | $M_W = 80.41$ GeV |
| $G_F = 1.16639 \times 10^{-5}$ GeV$^{-2}$ | $N_F = 5$ | $\Lambda^{LO}_{QCD} = 165$ MeV |
| $\Lambda^{NLO}_{QCD} = 226$ MeV | $\alpha_s^{LO}(M_Z) = 0.130$ | $\alpha_s^{NLO}(M_Z) = 0.118$ |

Table 2: Cross-sections for Higgs boson production via gluon fusion at LO and NLO in the mass range $100 \leq M_H \leq 1000$ GeV. The first column gives the Higgs boson mass, the second the LO order cross-section, the third and fourth column gives the NLO electroweak (EW) and QCD corrections respectively (see text), the last column presents the overall NLO cross-section.

| $M_H$(GeV) | $\sigma_{LO}$ (pb) | $\delta_{QCD}$ | $\delta_{EW}$ | $\sigma_{NLO}$ (pb) | $M_H$(GeV) | $\sigma_{LO}$ (pb) | $\delta_{QCD}$ | $\delta_{EW}$ | $\sigma_{NLO}$ (pb) |
|---|---|---|---|---|---|---|---|---|---|
| 100 | 27.788 | 0.80 | 0.04 | 51.264 | 105 | 25.518 | 0.80 | 0.04 | 47.189 |
| 110 | 23.519 | 0.81 | 0.05 | 43.590 | 120 | 20.170 | 0.81 | 0.05 | 37.579 |
| 130 | 17.491 | 0.82 | 0.06 | 32.809 | 140 | 15.314 | 0.82 | 0.07 | 28.886 |
| 150 | 13.521 | 0.82 | 0.08 | 25.680 | 160 | 12.029 | 0.83 | 0.07 | 22.854 |
| 170 | 10.776 | 0.84 | 0.03 | 20.114 | 180 | 9.715 | 0.84 | 0.02 | 18.080 |
| 190 | 8.812 | 0.85 | -0.01 | 16.212 | 200 | 8.038 | 0.85 | -0.02 | 14.760 |
| 250 | 5.490 | 0.88 | -0.02 | 10.231 | 300 | 4.286 | 0.91 | -0.01 | 8.143 |
| 350 | 4.414 | 0.97 | -0.01 | 8.666 | 400 | 4.124 | 0.93 | 0.00 | 7.923 |
| 450 | 2.945 | 0.91 | 0.00 | 5.622 | 500 | 1.975 | 0.91 | 0.00 | 3.772 |
| 550 | 1.307 | 0.92 | 0.00 | 2.505 | 600 | 0.869 | 0.93 | 0.00 | 1.675 |
| 650 | 0.585 | 0.94 | 0.00 | 1.134 | 700 | 0.398 | 0.95 | 0.00 | 0.777 |
| 750 | 0.275 | 0.97 | 0.00 | 0.541 | 800 | 0.193 | 0.97 | 0.00 | 0.381 |
| 850 | 0.137 | 0.99 | 0.00 | 0.272 | 900 | 0.098 | 1.01 | 0.00 | 0.197 |
| 950 | 0.071 | 1.03 | 0.00 | 0.144 | 1000 | 0.052 | 1.06 | 0.00 | 0.107 |

# 2 Production cross-sections and branching fractions for a Standard Model Higgs boson

## 2.1 Production cross-sections

This section reports on the Standard Model Higgs boson production cross-section via gluon fusion, Vector Boson Fusion (VBF), and the associated production with a Vector Boson ($WH$ and $ZH$), to leading order (LO) and to next to leading order (NLO). The associated production with $t\bar{t}$ is also discussed.

The calculation is performed using the CTEQ6L1 and CTEQ6M Parton Distribution Functions (PDF) at LO and NLO respectively [21]. Single and two-loop calculations of the running strong coupling constant $\alpha_s$ are used for the LO and NLO computation respectively. Table 1 shows the values of the Standard Model parameters used in the computations.

The cross-section of the gluon fusion process is evaluated using the program HIGLU [22]. The calculation is performed with exact NLO matrix element. The renormalization and factorization scales are set to the Higgs boson mass. The values of $\alpha_s$ are calculated according to the values of $\Lambda_{\mathrm{QCD}}$ given in Table 1 with 5 flavors. This approach may overestimate the LO cross-section; however this effect is expected to be negligible compared to the effect from the scale uncertainty.

Table 2 reports the cross-sections for Higgs boson production via gluon fusion at LO and NLO in

Table 3: Cross-sections for Higgs boson production via Vector Boson Fusion at LO and NLO in the mass range $100 \leq M_H \leq 1000$,GeV. The first column gives the Higgs boson mass, the second the LO order cross-section, the third and fourth column gives respectively the NLO electroweak (EW) and QCD corrections (see text), the fifth column presents the overall NLO cross-section.

| $M_H$(GeV) | $\sigma_{LO}$ (pb) | $\delta_{QCD}$ | $\delta_{EW}$ | $\sigma_{NLO}$ (pb) | $M_H$(GeV) | $\sigma_{LO}$ (pb) | $\delta_{QCD}$ | $\delta_{EW}$ | $\sigma_{NLO}$ (pb) |
|---|---|---|---|---|---|---|---|---|---|
| 100 | 5.037 | 0.04 | -0.05 | 5.001 | 105 | 4.835 | 0.04 | -0.05 | 4.802 |
| 110 | 4.633 | 0.04 | -0.05 | 4.608 | 120 | 4.277 | 0.04 | -0.05 | 4.246 |
| 130 | 3.961 | 0.04 | -0.05 | 3.931 | 140 | 3.670 | 0.04 | -0.05 | 3.651 |
| 150 | 3.415 | 0.04 | -0.05 | 3.397 | 160 | 3.173 | 0.05 | -0.05 | 3.154 |
| 170 | 2.956 | 0.05 | -0.04 | 2.976 | 180 | 2.770 | 0.04 | -0.04 | 2.764 |
| 190 | 2.591 | 0.05 | -0.03 | 2.624 | 200 | 2.427 | 0.04 | -0.04 | 2.447 |
| 250 | 1.789 | 0.04 | -0.04 | 1.795 | 300 | 1.355 | 0.05 | -0.04 | 1.358 |
| 350 | 1.053 | 0.04 | -0.06 | 1.037 | 400 | 0.833 | 0.04 | -0.03 | 0.848 |
| 450 | 0.670 | 0.04 | 0.00 | 0.694 | 500 | 0.542 | 0.05 | 0.01 | 0.574 |
| 550 | 0.447 | 0.04 | 0.03 | 0.477 | 600 | 0.371 | 0.04 | 0.04 | 0.402 |
| 650 | 0.311 | 0.04 | 0.06 | 0.341 | 700 | 0.262 | 0.04 | 0.08 | 0.292 |
| 750 | 0.222 | 0.04 | 0.10 | 0.252 | 800 | 0.190 | 0.03 | 0.13 | 0.220 |
| 850 | 0.163 | 0.03 | 0.15 | 0.193 | 900 | 0.140 | 0.03 | 0.19 | 0.170 |
| 950 | 0.121 | 0.03 | 0.23 | 0.152 | 1000 | 0.105 | 0.03 | 0.27 | 0.136 |

Table 4: Cross-sections for the associated Higgs boson production with $W$ bosons at LO and NLO in the mass range $100 \leq M_H \leq 200$ GeV.

| $M_H$(GeV) | $\sigma_{LO}$ (pb) | $\delta_{QCD}$ | $\delta_{EW}$ | $\sigma_{NLO}$ (pb) |
|---|---|---|---|---|
| 100 | 2.476 | 0.22 | -0.06 | 2.877 |
| 110 | 1.855 | 0.22 | -0.06 | 2.154 |
| 120 | 1.414 | 0.23 | -0.07 | 1.641 |
| 130 | 1.095 | 0.23 | -0.07 | 1.269 |
| 140 | 0.860 | 0.23 | -0.08 | 0.995 |
| 150 | 0.684 | 0.24 | -0.09 | 0.787 |
| 160 | 0.550 | 0.24 | -0.12 | 0.615 |
| 170 | 0.447 | 0.24 | -0.10 | 0.511 |
| 180 | 0.366 | 0.24 | -0.11 | 0.417 |
| 190 | 0.303 | 0.25 | -0.09 | 0.349 |
| 200 | 0.252 | 0.25 | -0.09 | 0.292 |

the mass ranges $100 \leq M_H \leq 1000$ GeV. The NLO cross-section is obtained from the LO one as follows: $\sigma_{NLO} = \sigma_{LO} \times (1 + \delta_{QCD} + \delta_{EW})$.

The cross-section of VBF Higgs boson production is estimated with the package VV2H F [22]. The results are reported in Table 3 as a function of the Higgs boson mass in the range 100 GeV$\leq M_H \leq$ 1000 GeV. The renormalization and factorization scales are set to the Higgs boson mass.

The cross-sections for $WH$, $ZH$ and $t\bar{t}H$ production are one to two orders of magnitude below the gluon and vector boson fusion cross-sections. The values are given in Tables 4, 5, and 6 respectively.

The cross-section for the Higgs boson produced in association with $t\bar{t}$ is estimated with the package HQQ [22]. The renormalization and factorization scales are set to $(M_H + 2M_t)/2$. The QCD corrections to this process are known [23,24] and yield a K-factor of about 1.25. However it should be stressed that the main backgrounds in this analysis ($t\bar{t}b\bar{b}$ and $t\bar{t}$jj) are known to LO only. Table 6 reports the cross-section of Higgs boson production associated to $t\bar{t}$ in the mass range 100 GeV$\leq M_H \leq$ 200 GeV.

The package V2HV is used to estimate the Higgs boson production cross-section in association with the $W$ and $Z$ bosons [22]. The renormalization and factorization scales are set to the sum of the invariant masses of the weak boson and of the Higgs boson. Tables 4 and 5 report the cross-sections of these processes in the mass range 100 GeV$\leq M_H \leq$ 200 GeV.

Table 5: cross-sections for the Higgs boson associated production with $Z$ bosons at LO and NLO in the mass range $100 \leq M_H \leq 200$ GeV.

| $M_H$(GeV) | $\sigma_{LO}$ (pb) | $\delta_{QCD}$ | $\delta_{EW}$ | $\sigma_{NLO}$ (pb) |
|---|---|---|---|---|
| 100 | 1.298 | 0.22 | -0.05 | 1.519 |
| 110 | 0.980 | 0.22 | -0.05 | 1.148 |
| 120 | 0.752 | 0.23 | -0.05 | 0.882 |
| 130 | 0.585 | 0.23 | -0.05 | 0.687 |
| 140 | 0.462 | 0.23 | -0.06 | 0.543 |
| 150 | 0.368 | 0.23 | -0.06 | 0.433 |
| 160 | 0.297 | 0.24 | -0.09 | 0.342 |
| 170 | 0.242 | 0.24 | -0.06 | 0.286 |
| 180 | 0.199 | 0.24 | -0.07 | 0.233 |
| 190 | 0.165 | 0.24 | -0.06 | 0.195 |
| 200 | 0.137 | 0.25 | -0.06 | 0.163 |

Table 6: cross-sections for the associated Higgs boson production with $t\bar{t}$ to LO and NLO (courtesy of M. Spira) in the mass range $100 \leq M_H \leq 200$ GeV.

| $M_H$(GeV) | $\sigma_{LO}$ (pb) | $\sigma_{NLO}$ (pb) | K Factor |
|---|---|---|---|
| 100 | 0.873 | 1.088 | 1.25 |
| 110 | 0.680 | 0.848 | 1.25 |
| 120 | 0.537 | 0.669 | 1.25 |
| 130 | 0.428 | 0.534 | 1.25 |
| 140 | 0.345 | 0.431 | 1.25 |
| 150 | 0.282 | 0.352 | 1.25 |
| 160 | 0.232 | 0.291 | 1.26 |
| 170 | 0.193 | 0.243 | 1.26 |
| 180 | 0.162 | 0.204 | 1.26 |
| 190 | 0.137 | 0.174 | 1.27 |
| 200 | 0.117 | 0.149 | 1.27 |

## 2.2 Decays branching ratios

Higgs boson branching ratios are evaluated with the program HDECAY [25]. Here, the default settings defined by the authors are used (see Ref. [25]), except for the parameters specified in Table 1.

Tables 7 and 8 report the most relevant Higgs boson branching ratios for $100 \leq M_H \leq 1000$ GeV: $H \rightarrow b\bar{b}, \tau^+\tau^-, \gamma\gamma, ZZ^{(*)}, WW^{(*)}$ and $t\bar{t}$.

Figure 1 (left) shows the branching fractions and the production cross-section (right) of the Standard Model Higgs boson as a function of its mass (Figures taken from Reference [26]).

## 2.3 Standard Model Higgs boson search in ATLAS

The Standard Model Higgs boson is searched for at the LHC in various decay channels, the choice of which depends by the signal rates and the signal to background ratios in the various mass regions.

The Standard Model Higgs boson channels considered in this volume are:

- $pp \rightarrow H \rightarrow \gamma\gamma$
- $pp \rightarrow H \rightarrow ZZ^{(*)} \rightarrow 4l(l = e, \mu)$
- $pp \rightarrow qqH \rightarrow qq\tau^+\tau^-$
- $pp \rightarrow H \rightarrow W^+W^- \rightarrow l\nu l\nu, l\nu qq$

Table 7: Relevant Higgs boson branching ratios for the mass range 100 $\leq M_H \leq$ 200 GeV.

| $M_H$(GeV) | $\Gamma_H$ (GeV) | $H \to b\bar{b}$ | $H \to \tau^+\tau^-$ | $H \to \gamma\gamma$ | $H \to WW(*)$ | $H \to ZZ^{(*)}$ |
|---|---|---|---|---|---|---|
| 100 | 0.0026 | 0.8117 | 0.08002 | 0.00157 | 0.01019 | 0.00106 |
| 110 | 0.0029 | 0.7694 | 0.07726 | 0.00194 | 0.04454 | 0.00412 |
| 120 | 0.0036 | 0.6773 | 0.06915 | 0.00223 | 0.13310 | 0.01520 |
| 130 | 0.0049 | 0.5249 | 0.05440 | 0.00227 | 0.28880 | 0.03866 |
| 140 | 0.0080 | 0.3414 | 0.03587 | 0.00197 | 0.48540 | 0.06781 |
| 150 | 0.0166 | 0.1742 | 0.01854 | 0.00141 | 0.68310 | 0.08301 |
| 160 | 0.0772 | 0.03960 | 0.00426 | 0.00056 | 0.90150 | 0.04334 |
| 170 | 0.3837 | 0.00837 | 0.00091 | 0.00015 | 0.96540 | 0.02253 |
| 180 | 0.6282 | 0.00536 | 0.00059 | 0.00010 | 0.93460 | 0.05750 |
| 190 | 1.038 | 0.00339 | 0.00038 | 0.00007 | 0.77610 | 0.21870 |
| 200 | 1.426 | 0.00257 | 0.00029 | 0.00005 | 0.73470 | 0.26130 |

Table 8: Relevant Higgs boson branching ratios for the mass range 250 $\leq M_H \leq$ 1000 GeV.

| $M_H$(GeV) | $\Gamma_H$ (GeV) | $H \to WW(*)$ | $H \to ZZ^{(*)}$ | $H \to t\bar{t}$ |
|---|---|---|---|---|
| 250 | 4.046 | 0.7003 | 0.2977 | 0.00000 |
| 300 | 8.505 | 0.6911 | 0.3075 | 0.00007 |
| 350 | 15.60 | 0.6722 | 0.3078 | 0.01878 |
| 400 | 29.30 | 0.5787 | 0.2703 | 0.14990 |
| 450 | 46.55 | 0.5489 | 0.2600 | 0.19030 |
| 500 | 67.56 | 0.5446 | 0.2606 | 0.19420 |
| 550 | 92.55 | 0.5500 | 0.2652 | 0.18430 |
| 600 | 122.3 | 0.5591 | 0.2711 | 0.16940 |
| 650 | 157.7 | 0.5692 | 0.2773 | 0.15310 |
| 700 | 199.7 | 0.5793 | 0.2832 | 0.13710 |
| 750 | 249.2 | 0.5889 | 0.2887 | 0.12210 |
| 800 | 307.7 | 0.5977 | 0.2937 | 0.10840 |
| 850 | 376.5 | 0.6057 | 0.2982 | 0.09586 |
| 900 | 457.4 | 0.6129 | 0.3023 | 0.08458 |
| 950 | 552.5 | 0.6195 | 0.3059 | 0.07445 |
| 1000 | 664.1 | 0.6253 | 0.3092 | 0.06537 |

- $pp \to t\bar{t}H \to t\bar{t}b\bar{b}$
- $pp \to t\bar{t}H \to t\bar{t}W^+W^-$, and $pp \to ZH \to \ell^+\ell^-W^+W^-$

# 3 Higgs Bosons in the Minimal Supersymmetric Extension to the Standard Model

As discussed in the Introduction, in the Minimal Supersymmetric Standard Model two Higgs doublets are required, resulting in three neutral and two charged observable Higgs bosons. The production of neutral Higgs bosons and their decays are different from those in the Standard Model. While decays into ZZ or WW are dominant in the Standard Model for Higgs boson masses above $M_H > 2\ M_W$, for high values of $\tan\beta$ these decay modes are either suppressed in case of the $h$ and $H$ or even absent in the case of the $A$. Instead, the coupling of the Higgs bosons to third generation fermions are strongly enhanced for large regions of the MSSM parameter space.

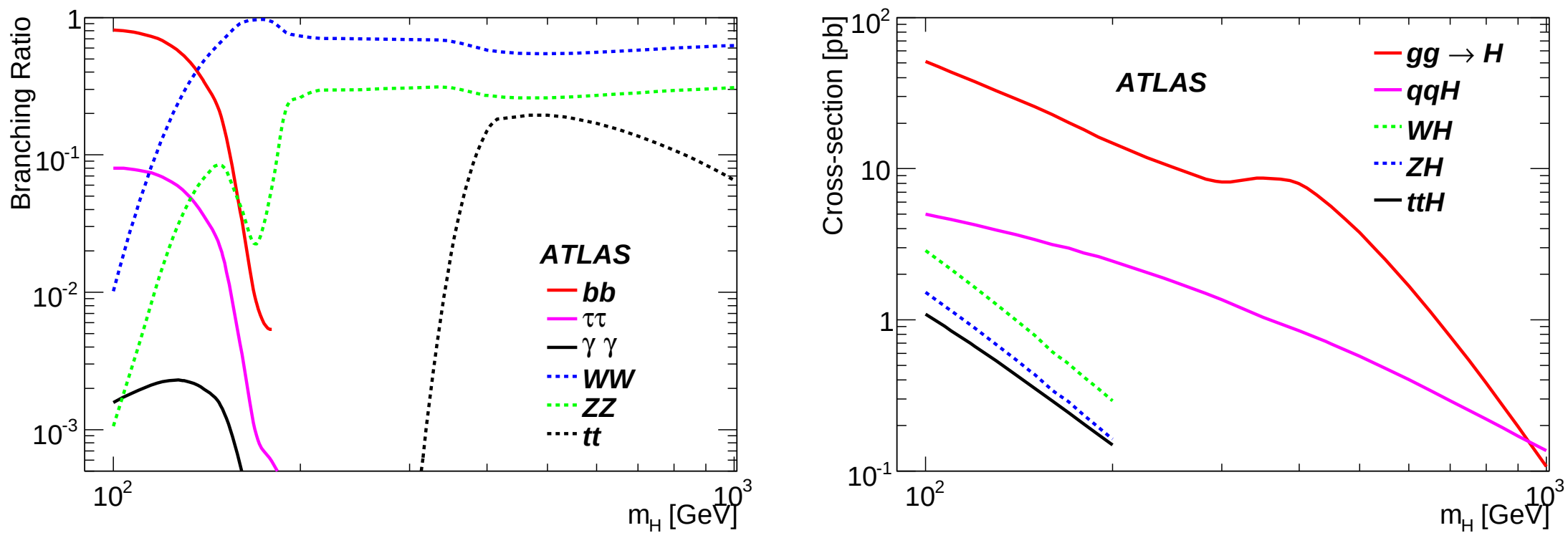


Figure 1: Left: Branching ratios for the relevant decay modes of the Standard Model Higgs boson as a function of its mass. Right: cross-sections for the five production channels of the Standard Model Higgs boson at the LHC at 14 TeV.

### 3.1 Production and decays of neutral Higgs bosons

The Higgs boson production proceeds via two different mechanisms, the direct and the $b$ quark associated production, as described below. In the following $\phi$ stands for either of the three neutral Higgs bosons: $A$, $H$, and $h$. Further details can be found in [27].

**Direct Production:** The diagram for this process is depicted in Fig. 2(a). It dominates in the range of low $\tan\beta$ and its rates are significantly larger than for the Standard Model. For the range of higher $\tan\beta$ it is still dominant for low $m_A$. The cross-section for this process has been calculated at NLO accuracy [22] and the numerical values used here are listed under $\sigma^{\text{direct}}_{h/H/A}$ in Table 9.

**Associated Production:** Different approaches have been followed by theorists to calculate the cross-section for Higgs boson production in association with $b$ quarks, each of them assuming one of the diagrams depicted in Fig. 2(b-d) as their leading order (LO) contribution. The implications connected with this choice are briefly discussed below.

- $gg \to b\bar{b}\phi$:
  The cross-section for this process has been calculated at NLO accuracy for the case of both $b$ quarks at high transverse momentum [28, 29], where this calculation is considered to be reliable.

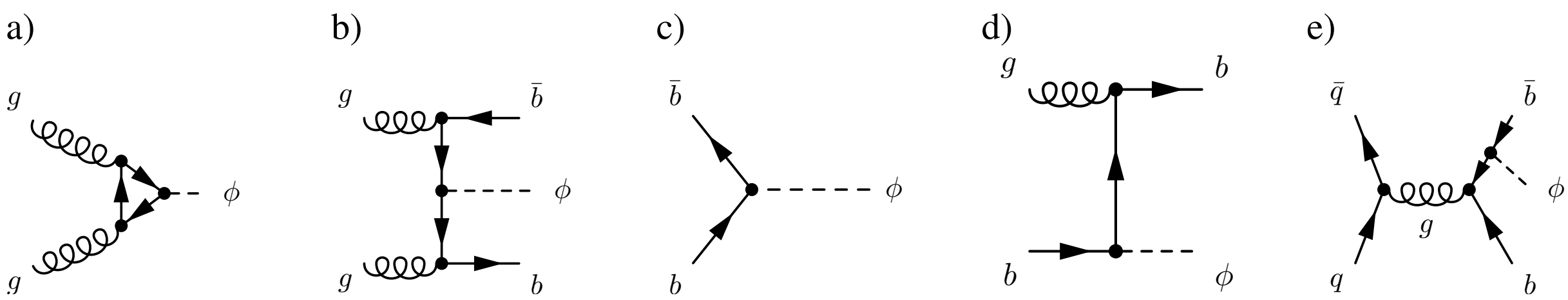


Figure 2: Feynman diagrams contributing to the MSSM Higgs boson production. Diagram a) is called 'direct production', diagrams b) to e) contribute to the $b$ quark associated production. In the above diagrams $\phi$ represents either of the neutral Higgs bosons in the MSSM, $h$, $H$, or $A$.

For the cases where only one or zero $b$ quarks with high $p_T$ are required in the final state, the cross-section has been calculated at NLO accuracy by integrating over the momentum of one or both $b$ quarks, respectively [28, 30, 31]. In these cases the predictions are considered less reliable due to the occurrence of potentially large collinear logarithms of the form $\log(Q/m_b)$ from the gluon splitting process, where $Q$ is the factorization scale which is expected to be much larger than $m_b$.

- $b\bar{b} \to \phi$:
The potentially large collinear logarithms can be dealt with by absorbing them into the parton density function (PDF) of the $b$ quark and thus resumming them to all orders of perturbation theory. The intrinsically still present $b$ quarks from the gluon splitting process are given zero transverse momentum at leading order. At higher order they can acquire transverse momentum. E.g. the process $bg \to b\phi$ is a NLO contribution leading to an observable $b$ quark. The total cross-section for this process has been calculated at NLO [32, 33] and NNLO accuracy [34] and is considered to be reliable when it is not required (but also not vetoed) to observe a $b$ quark.

- $bg \to b\phi$:
This process is a mixture of the two processes discussed above in a sense that one $b$ quark is coming from the matrix element description and the other from the $b$ PDF. This process has been calculated at NLO accuracy [35] and is considered to be reliable if one observes only one $b$ quark with high $p_T$ in the final state. In principle in this case one should also veto any additional $b$ quark which is experimentally challenging due to the limited $b$-tagging efficiency.

- $qq \to b\bar{b}\phi$:
Compared to $gg \to b\bar{b}\phi$ this process only contributes at the 1%-level at LHC energies and is therefore only mentioned for completeness.

A comparison of the predicted cross-sections for both the inclusive and exclusive approach is shown in Figure 3. The blue band corresponds to the NNLO calculation of $b\bar{b} \to \phi$, where the width of the band corresponds to the residual scale uncertainties when varying the default renormalization ($\mu_r = m_H$) and factorization ($\mu_f = m_H/4$) scales. The red band shows the corresponding NLO calculation of $gg \to b\bar{b}\phi$. The scale uncertainty for $gg \to b\bar{b}\phi$ is of the order 20 to 30%, while the scale uncertainty for $b\bar{b} \to \phi$ is much smaller, in particular at high Higgs boson masses. This might be due to the remaining collinear logarithms in the $gg \to b\bar{b}\phi$ calculation. However, contributions from e.g. PDF-uncertainties, in particular for the $b$ PDF have not been included here. Both methods agree within uncertainties. At larger Higgs boson masses, the prediction from $b\bar{b} \to \phi$ is slightly higher than that from $gg \to b\bar{b}\phi$. One of the effects that might explain this is the inclusion of closed-top-loops in the $gg \to b\bar{b}\phi$ calculation, absent in the $b\bar{b} \to \phi$ case. In conclusion, both predictions may be used to normalize an inclusive signal sample, where no cuts have been applied on the momenta of the $b$ quarks at generator level.

Due to the finite integrated luminosity assumed to be available for the analyses discussed in this note, the inclusive normalization is chosen. The cross-section were calculated using FeynHiggs-2.6.2 [36] yielding the cross-section for a Standard Model Higgs boson. The cross-sections in the MSSM were then obtained by scaling them by the ratio of partial widths into $b\bar{b}$:

$$\sigma^{\phi}_{\text{MSSM}}(m_A, \tan\beta) = \sigma_{\text{SM}}(m_\phi) \cdot \frac{\Gamma^{\text{MSSM}}_{b\bar{b}\phi}(m_A, \tan\beta)}{\Gamma^{\text{SM}}_{b\bar{b}\phi}(m_\phi)}. \tag{1}$$

The production cross-sections for all Higgs boson masses considered here and their branching fraction into a $\tau^+\tau^-$ and $\mu^+\mu^-$ final states in the $m_h^{\text{max}}$ scenario are summarized in Table 9 for $\tan\beta = 20$.

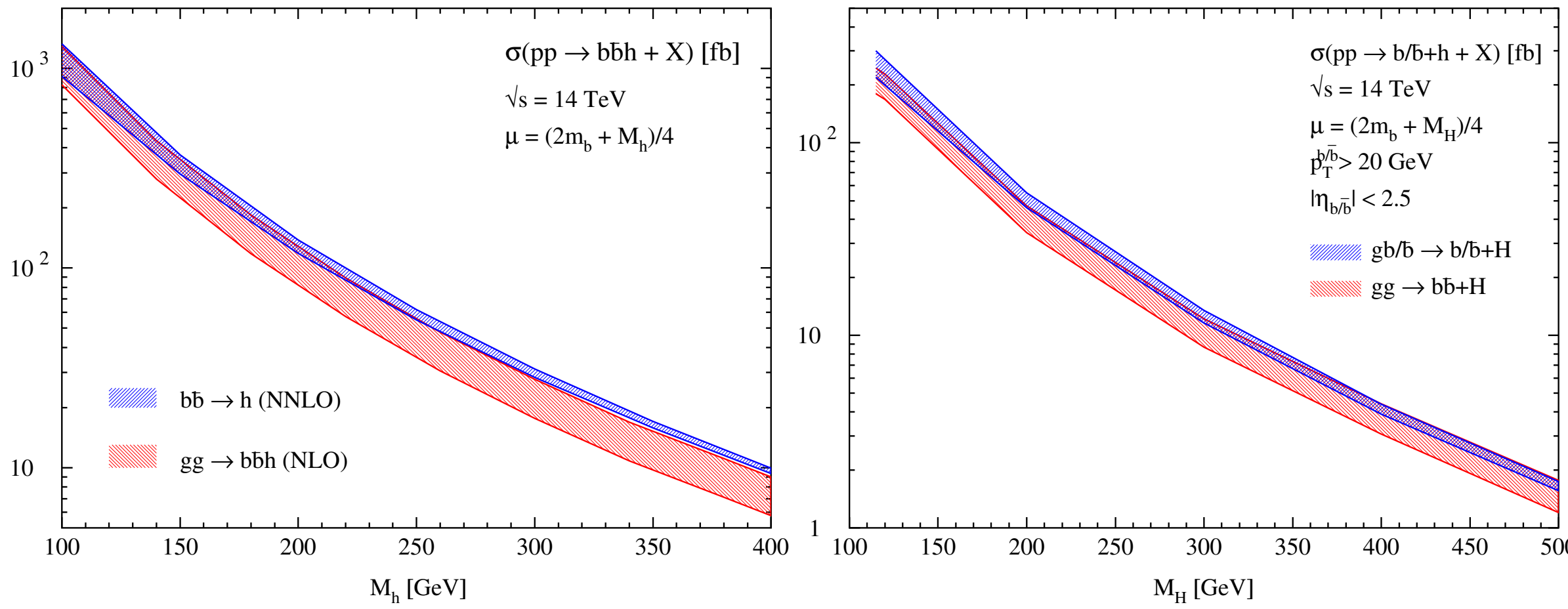


Figure 3: The inclusive cross-sections for the processes $b\bar{b} \to \phi/$ (blue hatched region) and $gg \to b\bar{b}\phi$ (red hatched region) are shown on the right-hand side; the exclusive cross-sections for $bg \to b\phi$(blue hatched region) and $gg \to b\bar{b}\phi$(red hatched region) are shown on the right-hand side. The width of the bands corresponds to the theoretical uncertainty due to the choice of renormalization and factorization scales.

## 3.2 Production and decays of charged Higgs bosons

The search strategies for charged Higgs bosons depend on the charged Higgs boson mass, which dictates both the production and the available decay modes. Below the top quark mass the main production mode is through top quark decays, $t \to H^+b$, and in this range the $H^+ \to \tau^+\nu$ decay mode is dominant. Once above the top quark threshold, production mainly takes place through gb fusion ($g\bar{b} \to \bar{t}H^+$). For such high charged Higgs boson masses, the decay into a top and a $b$ quark dominates, $H^+ \to t\bar{b}$, but $H^+ \to \tau^+\nu$ can still be sizeable and offers a much cleaner signature. The process $gg \to \bar{t}bH^+$ is important for charged Higgs boson production with $m_{H^+}$ around the top mass. Since the LHC will be the first collider $t\bar{t}$ factory, "light" charged Higgs bosons may be copiously produced through the process $q\bar{q}$, $gg \to t\bar{t} \to \bar{t}bH^+$ [37]. While this is the dominant production mode, there are other processes which also contribute to the light charged Higgs boson production, like single top events or diagrams with the same final state as mentioned above ($tbH^+$), but which do not proceed through $t\bar{t}$ production. $H^+$ events through single top production are not considered in this volume.

The charged Higgs boson production cross-section is evaluated for two different MSSM scenarios. They are chosen such that in Scenario A the decay of $H^+$ into SUSY particlces is suppressed, and in Scenario B (also known as the "$m_h$-max" scenario) the mass of the lightest Higgs boson $h^0$ is maximised.

Table 9: Mass, cross-section for direct production, cross section for $b$-associated production, and branching fractions into $\tau^+\tau^-$ and $\mu^+\mu^-$ final states for Higgs bosons in the $m_h^{\max}$ scenario and for $\tan\beta = 20$. All values were obtained using FeynHiggs-2.6.2 and HIGLU.

| Mass / GeV | | | $\sigma^{\text{direct}}_{h/H/A}$/fb | | | $\sigma^{\text{associated}}_{h/H/A}$/fb | | | $\mathscr{B}(h/H/A \to \tau^+\tau^-)$/% | | | $\mathscr{B}(h/H/A \to \mu^+\mu^-)$/% | | |
|---|---|---|---|---|---|---|---|---|---|---|---|---|---|---|
| $A$ | $H$ | $h$ | $A$ | $H$ | $h$ | $A$ | $H$ | $h$ | $A$ | $H$ | $h$ | $A$ | $H$ | $h$ |
| 110 | 129.8 | 109.0 | – | – | – | 314810 | 7579 | 310707 | 8.86 | 9.11 | 8.88 | 0.031 | 0.032 | 0.031 |
| 130 | 134.2 | 124.7 | 92517 | 93941 | 43545 | 189602 | 92897 | 99992 | 9.11 | 9.23 | 9.00 | 0.032 | 0.033 | 0.032 |
| 160 | 160.8 | 128.0 | 32148 | 34706 | 44561 | 97480 | 93102 | 6650 | 9.42 | 9.46 | 8.40 | 0.033 | 0.033 | 0.030 |
| 200 | 200.5 | 128.4 | 9847 | 11377 | 45957 | 45685 | 45095 | 2188 | 9.57 | 9.72 | 7.49 | 0.034 | 0.034 | 0.027 |
| 300 | 300.4 | 128.6 | 955 | 1451 | 46986 | 10312 | 10253 | 979 | 8.22 | 9.51 | 6.27 | 0.029 | 0.034 | 0.022 |
| 450 | 449.8 | 128.6 | – | – | – | 2019 | 2035 | 723 | 6.07 | 6.24 | 5.68 | 0.021 | 0.022 | 0.020 |

For the two scenarios the following set of parameters have been used:

Scenario A:

- $m_t = 175$ GeV
- $M_{\rm SUSY} = 500$ GeV
- $A_t = 1000$ GeV
- $\mu = 200$ GeV
- $M_2 = 1000$ GeV
- $M_3 = 1000$ GeV

Scenario B ("$m_h$-max"):

- $m_t = 170$ GeV
- $M_{\rm SUSY} = 1000$ GeV
- $X_t = 2000$ GeV, where $A_t = X_t + \mu/\tan\beta$
- $\mu = 200$ GeV
- $M_2 = 200$ GeV
- $M_3 = 800$ GeV

$M_{\rm SUSY}$ denotes the soft SUSY-breaking mass parameter in the sfermion sector, $m_t X_t$ is the off-diagonal entry in the stop mass matrix, $\mu$ the Higgsino mixing paramter, and $M_2$ and $M_3$ the soft SUSY-breaking mass parameters in the SU(2) gaugino and the gluino sector, respectively. In the following, numerical NLO cross-sections for the low-mass $H^+$ region are calculated with FeynHiggs (version 2.6.2 [36]). Heavy $H^+$ NLO cross-sections are obtained with Ref. [38] and corrected with the dominant supersymmetry loop corrections (reflecting the altered relation between the bottom quark mass and its Yukawa coupling, $\Delta m_b$) as proposed in Ref. [39].

Figure 4 shows the results for the $tbH^+$ final state as a function of $\tan\beta$ for the MSSM scenarios A and B. The production cross-section has a minimum at $\tan\beta \approx 7$. This is caused by a mimimum in the $H^+tb$ Yukawa coupling and renders the so-called intermediate $\tan\beta$ region ($4 < \tan\beta < 10$) which is experimentally hard to reach.

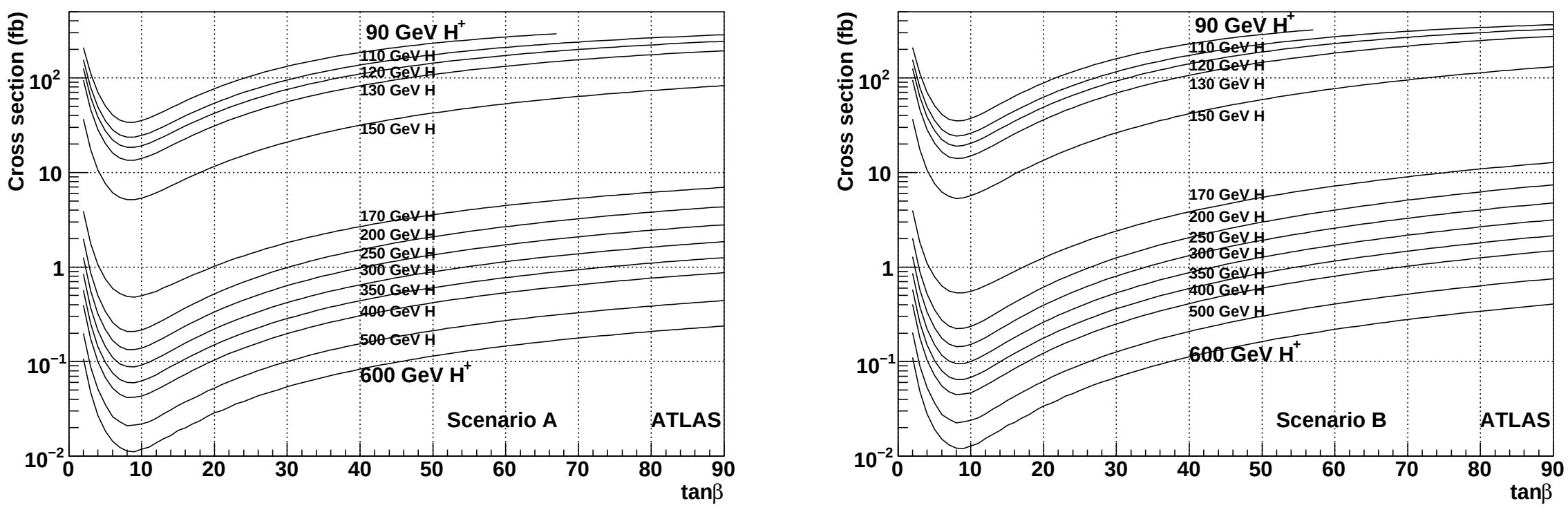


Figure 4: Expected charged Higgs boson production cross-section in the MSSM for scenarios A and B for light [36] and heavy charged Higgs bosons [38].

Below the top quark mass, the charged Higgs boson predominantly decays into a $\tau$ lepton and a neutrino, and for values of $\tan\beta > 5$ this branching ratio is close to 100%, as shown in Figures 5 and 6. Decay modes involving $c\bar{s}$ and $Wh$ are also present, but depending on the value of $\tan\beta$ they are one or two orders of magnitude smaller than $\tau\nu$. The decay of the W boson originating from the associated top quark adds variety to the possible charged Higgs boson signatures, but it also provides handles for signal reconstruction and, even more important, for background rejection.

Once above the top quark mass threshold, the $H^+ \to t\bar{b}$ decay mode shows a rapid growth and soon becomes an important decay mode as shown in Figure 5. Contrary to the light charged Higgs boson,

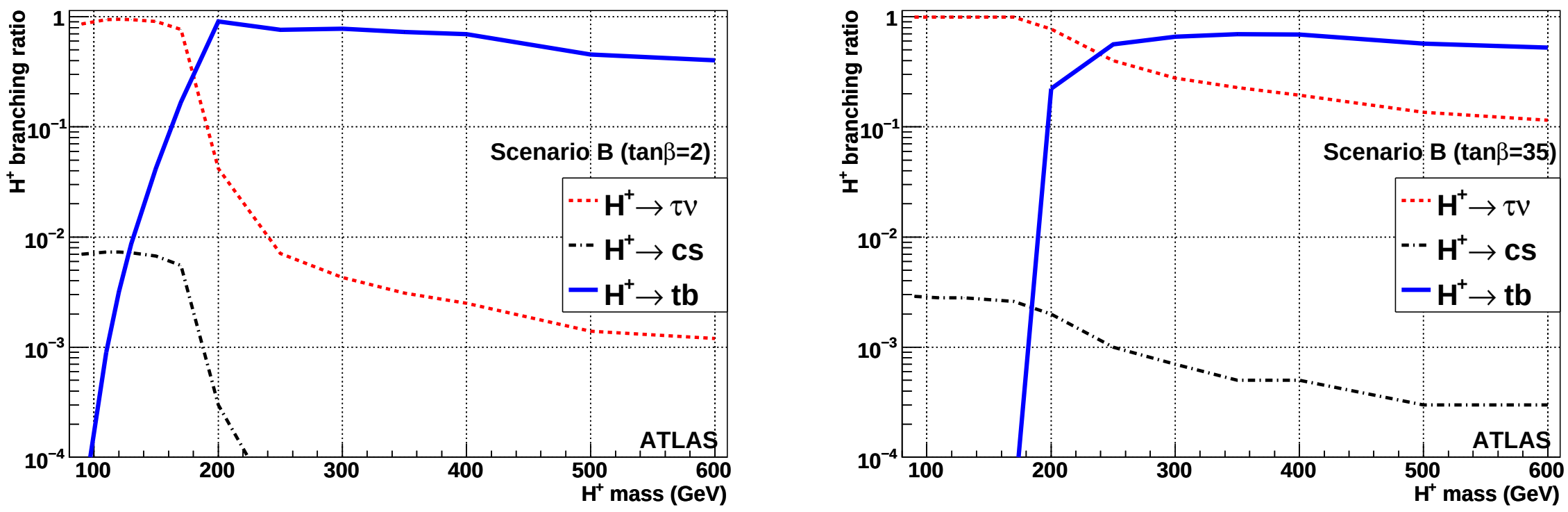


Figure 5: Charged Higgs boson branching ratios as a function of mass for the $m_h$-max scenario for $\tan\beta = 2$ and $\tan\beta = 35$ and three selected decay modes.

for which the $H^+ \to \tau^+\nu$ decay mode is an almost exclusive decay mode, the heavy charged Higgs $H^+$ boson does not solely decay into $t\bar{b}$, but a significant fraction is allowed to decay into other decay modes like $\tau^+\nu$, $W^+h$, $c\bar{s}$ or SUSY particles where kinematically allowed.

The calculations are performed with FeynHiggs [36], as it allows to calculate both the BR($t \to H^+b$) and the H$^+$ decay branching in a consistent way, and includes important corrections to the tree level values. Figure 6 shows the calculated branching ratios for two different charged Higgs boson masses, one light (130 GeV) and one heavy (600 GeV), as a function of $\tan\beta$ .

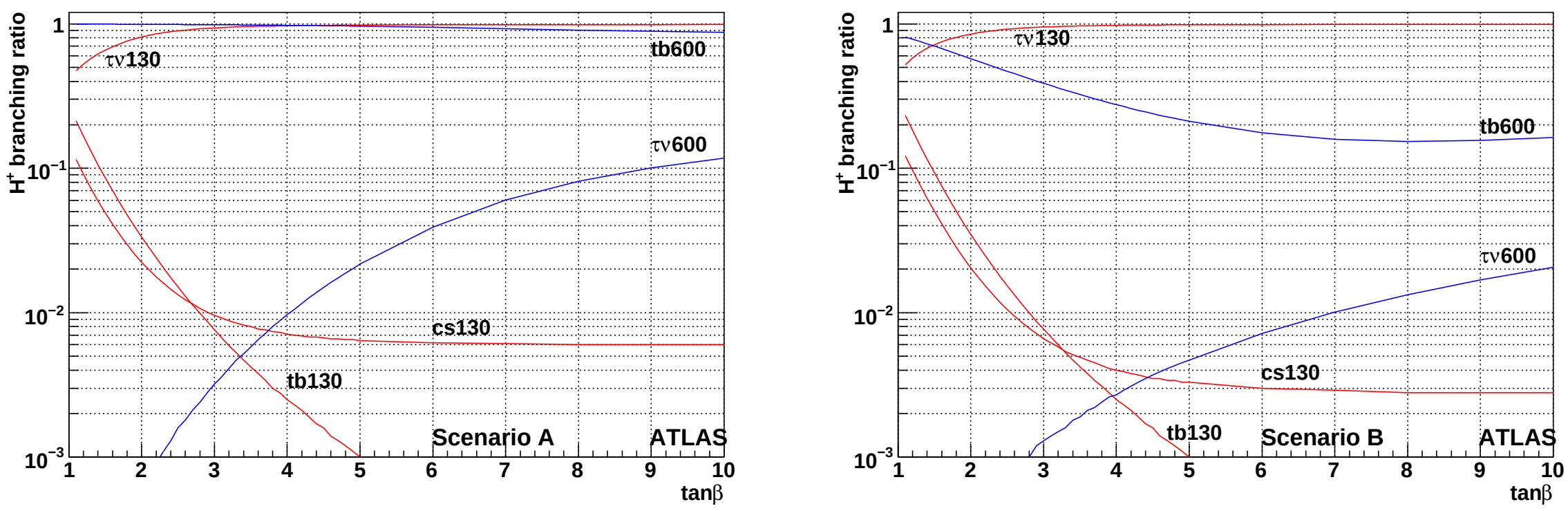


Figure 6: Expected charged Higgs boson branching ratios in the MSSM for scenarios A and B for the example of a light ($m_{H^+} = 130$ GeV) and a heavy ($m_{H^+} = 600$ GeV) charged Higgs boson [36].

# Prospects for the Discovery of the Standard Model Higgs Boson Using the H$\rightarrow \gamma\gamma$ Decay

**Abstract**
The discovery potential for the Standard Model Higgs boson through the $H \rightarrow \gamma\gamma$ decay in the ATLAS detector is reported. Various performance aspects of the Higgs boson mass reconstruction and the photon identification are discussed. Trigger issues are also considered. The potential of an inclusive $H \rightarrow \gamma\gamma$ search and Higgs boson searches in association with one or two high $p_T$ jets are evaluated. Studies of the associated $WH$, $ZH$ and $t\bar{t}H$ production processes are also presented. Finally, the discovery potential is assessed using an unbinned multivariate maximum-likelihood fit. These studies are performed for experimental conditions expected for an instantaneous luminosity of $\approx 10^{33}\,\mathrm{cm}^{-2}\mathrm{s}^{-1}$.

## 1 Introduction

In the mass range $110 < m_H < 140$ GeV the Higgs boson is expected to decay into two photons with a branching fraction large enough to render the search feasible at the LHC [1]. The inclusive search for the Higgs boson in the diphoton decay channel has been studied in ATLAS for many years and constitutes one of the benchmarks for the detector performance [2–4]. In this paper the sensitivity of ATLAS to this channel is re-evaluated with an updated detector description and software. The impact of higher order QCD and electroweak corrections on the discovery potential is also evaluated.

The Higgs boson can be produced in association with hadronic jets of high transverse momentum, $p_{\mathrm{T}}$. Gluons from initial-state radiation in the $gg \rightarrow H$ and qq→qqH Vector Boson Fusion (VBF) processes are the largest contributors to Higgs boson production in association with high $p_{\mathrm{T}}$ hadronic jets. The search for a Higgs boson using the diphoton decay mode in association with one and two jets at the LHC has been suggested [5–7] and a previous analysis reported on feasibility studies for these final states using a fast detector simulation [8, 9]. This paper presents an update of these studies based on a more complete description of the detector simulation.

In addition to the diphoton invariant mass, other discriminating variables are incorporated into the analysis and combined by means of an unbinned maximum-likelihood fit. Photon reconstruction properties and the event topology are used to separate the data sample into categories that are fit simultaneously.

In the search for a Standard Model Higgs boson, the associated production with $W$, $Z$, or $t\bar{t}$ can serve to complement the inclusive Higgs boson and Higgs boson + jets channels and help to determine the Higgs boson coupling to the Standard Model gauge bosons and the Yukawa coupling to the top quark [10]. These measurements would provide consistency checks of the Standard Model. They could also be interpreted in terms of Minimal Supersymmetric Standard Model couplings at high $\tan\beta$ (for $m_H \sim 120$ GeV) and high $m_A$. Deviations from Standard Model rates could imply new physics such as gauge-Higgs unification [11, 12] or the presence of resonances in models of little Higgs [13, 14], Left-Right symmetry [15] or technicolour [16]. The feasibility of diphoton searches in association with weak vector bosons is evaluated using a full detector simulation. This involves searches for diphotons in association with either just missing transverse energy, $E_{\mathrm{T}}^{\mathrm{miss}}$, or $E_{\mathrm{T}}^{\mathrm{miss}}$ and a charged lepton (electron or muon), and updates earlier studies performed with a fast detector simulation [17, 18].

## 2 Monte Carlo event generation

The Monte Carlo (MC) event generation required for this paper is split into two main groups. The first group corresponds to MC samples produced with a full detector simulation based on GEANT4 [19]. These are used for the evaluation of detailed detector effects relevant to the analyses. The second group of MC samples are processed with a fast detector simulation [20] and are used primarily for the evaluation of the analysis sensitivity. For example, the Higgs boson mass resolution, photon identification efficiency and photon-jet rejection are evaluated with a full detector simulation. The resulting photon efficiency and photon-jet rejection are parameterised as a function of the photon $p_{\mathrm{T}}$ and these parameterisations are then applied to the fast detector simulation.

### 2.1 Signal processes

Signal events are generated using PYTHIA [21]: this package implements Leading Order (LO) Matrix Element calculations for all the signal processes considered here. The gluon fusion process is also simulated with the MC@NLO [22, 23] package. This package provides QCD Next-to-Leading-Order (NLO) Matrix Elements [24–27] in addition to a good description of multiple soft-gluon emission at next-to-next-to-leading logarithmic level (NNLL) [28–30]. This is relevant to evaluating the discriminating power of the diphoton $p_{\mathrm{T}}$ and other relevant variables. All signal processes used here are processed through a full detector simulation. Signal events produced via the VBF mechanism are also modeled with HERWIG [31]. All the generated samples for signal processes used here are normalized to the NLO cross-sections [32] taking into account only QCD corrections.

### 2.2 Background processes

Background processes can be split into two main groups: backgrounds coming from the production of two isolated photons, which are usually referred to as irreducible, and reducible backgrounds coming from events with at least one fake photon. Fake photons are mostly due to the presence of a leading $\pi^0$ resulting from the fragmentation of a quark or a gluon. Table 1 gives a summary of the MC packages and the cross-sections (in pb) for the irreducible and reducible backgrounds used here.

In this paper QCD corrections to both signal and background are considered in the inclusive analysis. The irreducible background is mainly due to the $q\bar{q}, qg \to \gamma\gamma$x processes at up to order $\alpha^2\alpha_s$ (namely the Born and Bremsstrahlung contributions) and the $gg \to \gamma\gamma$ up to order $\alpha^2\alpha_s^3$ (referred to as the box contribution). Contributions from photons produced collinear to quarks are also taken into account at NLO. The DIPHOX [33] and ResBos [34–36] programs are used to assess the irreducible background computation. DIPHOX includes all the processes to order $\alpha^2\alpha_s$, including the Bremsstrahlung contribution with the quasi collinear fragmentation of quarks and gluons which are computed at NLO. DIPHOX does not include resummation effects. ResBos includes the Born and box contributions at NLO as well as the bremsstrahlung contribution (but the fragmentation contribution is only at LO). ResBos includes resummation effects to NNLL.

In these computations, a parton level isolation cut of 15 GeV in a cone of $\Delta R = 0.4$ is used. This is checked against the real photon identification cuts using PYTHIA fully simulated events. DIPHOX and ResBos predictions for the total irreducible background agree to better than 10% [37].

Table 1 shows the cross-sections used to normalize the MC generation in the inclusive analysis. The cross-sections for $q\bar{q}, qg \to \gamma\gamma$x and $gg \to \gamma\gamma$ quoted in Table 1 were computed with ResBos for $80 < m_{\gamma\gamma} < 150$ GeV, where $m_{\gamma\gamma}$ is the invariant mass of the diphoton system, and $p_{T\gamma} > 25$ GeV, where $p_{T\gamma}$ is the transverse momentum of the photons. Photons are required to lie in the central region of the detector, $|\eta| < 2.5$. The factorization and renormalization scales are set dynamically as the invariant mass of the diphoton system.

Table 1: Details of the cross-section calculation and event generation for the irreducible and reducible photonic backgrounds used in the $H \to \gamma\gamma$ analysis. The first three columns display the cross-section calculators, kinematic cuts and resulting cross-sections (in pb). The last two columns show the MC packages used for event generation and the number of events generated with a full and fast detector simulations, respectively.

| Process | $\sigma$ calculator | Cuts | $\sigma$(pb) | Full simulation # of events | Fast simulation # of events |
|---|---|---|---|---|---|
| $q\bar{q}, qg \to \gamma\gamma$x | ResBos/ | $80 < m_{\gamma\gamma} < 150$ GeV | 20.9 | PYTHIA/ALPGEN | ALPGEN |
| | DIPHOX | $p_{T\gamma} > 25$ GeV, $\|\eta\| < 2.5$ | | 200000/1300000 | 1670000 |
| $gg \to \gamma\gamma$ | ResBos | $80 < m_{\gamma\gamma} < 150$ GeV | 8.0 | PYTHIA | PYTHIA |
| | | $p_{T\gamma} > 25$ GeV, $\|\eta\| < 2.5$ | | 200000 | 850000 |
| $\gamma j$ | JETPHOX | $p_{T\gamma} > 25$ GeV | $180 \cdot 10^3$ | PYTHIA | ALPGEN |
| | | | | 3000000 | 36700000 |
| $jj$ | NLOJET++ | $p_T > 25$ GeV | $477 \cdot 10^6$ | PYTHIA | ALPGEN |
| | | | | 10000000 | 37000000 |

In order to evaluate irreducible backgrounds for signal-significance computations and to include the effect of high-$p_T$ jets, the ALPGEN MC generator is used [38, 39].[1] This includes $2 \to N$ tree-level Matrix Elements, where $N = 2-5$. The minimum parton $p_T$ and maximum pseudorapidity are set to 20 GeV and $|\eta| < 5$, respectively. A prescription for the merging of matrix elements and parton showers is used [41]. The $p_T$ threshold for the merging of the parton shower and the Matrix Elements is set to 20 GeV and $\Delta R = 0.7$.

The total cross-section of the ALPGEN samples of $\gamma\gamma$+jets events is normalized to the cross-section given in Table 1. The diphoton invariant mass and transverse momentum spectra are re-weighted to the corresponding spectra obtained with ResBos for the Born and Bremsstrahlung contributions. This is motivated by the fact that ResBos provides a NNLL description of the resummation effects that significantly affects the $p_T$ spectrum of the diphoton system up to about 40 or 50 GeV.

The box contribution is simulated with the PYTHIA package using the leading order one loop Matrix Elements with full and fast detector simulations. The cross-section of this sample is normalized to the cross-section given in Table 1. The diphoton invariant mass and momentum spectra are re-weighted to the corresponding spectra obtained with ResBos.

The total inclusive cross-section for the $\gamma j$ process is obtained using the package JETPHOX [42]. This package simulates direct and fragmentation single photon production. The distribution of the photon $p_T$ that is obtained is compared with that for direct production predicted using the PYTHIA package. It is observed that the differential cross-section obtained with JETPHOX is a factor of 2.1 larger than that obtained with PYTHIA, with a weak $p_T$ dependence. Table 1 shows the cross-section for inclusive $\gamma j$ production for $p_{T\gamma} > 25$ GeV.

The inclusive dijet cross-section is computed with the help of the NLOJET++ package [43,44], which takes into account QCD NLO corrections. The dijet cross-section obtained with this program is found to be a factor of 1.3 larger than that obtained with PYTHIA. Table 1 shows the cross-section for inclusive dijet production for $p_T > 25$ GeV.

For purposes of signal significance computation the ALPGEN MC package is used to generate sam-

[1] This generation does not include the box contribution nor the electroweak $\gamma\gamma jj$ process. The latter is generated with the MadGraph package [40] and it is used in the Higgs boson searches in association with hadronic jets only (see Sections 5.2 and 5.3).

Table 2: Event generators and cross-sections for backgrounds used in the analyses of associated production. Photons are required to be in $|\eta| < 2.7$ and to have $p_T > 20$ GeV. For the $W^\pm(\rightarrow e^\pm\nu)\gamma$ process the electron is required to pass the same cuts as the photons. The diphoton invariant mass should be in the interval $90 < m_{\gamma\gamma} < 150$ GeV, when appropriate. For the $b\bar{b}\gamma\gamma$ process the $p_{\mathrm{T}}$ of the b quark is required to be greater than 20 GeV.

| Process | Generator | Cross-section | Number of events |
|---|---|---|---|
| $Z(\rightarrow \ell\ell)\gamma\gamma$ | MadGraph | 2.4 fb | 17500 |
| $Z(\rightarrow \nu\nu)\gamma\gamma$ | MadGraph | 4.9 fb | 17500 |
| $W^+(\rightarrow \ell\nu)\gamma\gamma$ | MadGraph | 3.3 fb | 28200 |
| $W^-(\rightarrow \ell\nu)\gamma\gamma$ | MadGraph | 3.1 fb | 31500 |
| $W^\pm(\rightarrow e^\pm\nu)\gamma$ | PYTHIA | 5.9 pb | 218350 |
| $c\bar{c}\gamma\gamma$ | MadGraph | 257 fb | 5000 |
| $b\bar{b}\gamma\gamma$ | MadGraph | 24.41 fb | 5000 |
| $t\bar{t}\gamma\gamma$ | MadGraph | 1.97 fb | 4900 |

ples of $\gamma$+jets and multi-jets with a fast detector simulation. ALPGEN includes $2 \rightarrow N$ tree-level Matrix Elements, where $N = 2-5$. The minimum parton $p_{\mathrm{T}}$ and pseudorapidity range are set to 20 GeV and $|\eta| < 6$, respectively. The same matching conditions as in the $\gamma\gamma$+jets sample are used. The samples generated with ALPGEN are normalized to cross-sections given in Table 1.

A small background contribution is expected from Drell-Yan $e^+e^-$ faking a photon pair. For this purpose 420k events were generated using the PYTHIA package with a full detector simulation. The generator cross-section after requiring one lepton with $p_{\mathrm{T}} > 10$ GeV and $|\eta| < 2.7$ is 1.23 nb.

Specific backgrounds contributing to the Higgs boson plus lepton and Higgs boson plus missing transverse energy channels are also produced. $W/Z$ + diphoton backgrounds can produce the same topology as the signal. To evaluate the contributions from these backgrounds, the MadGraph [40] generator, which includes the $Z$ and $W$ bosons produced in association with two photons with tree-level diagrams has been used. The cross-sections for these processes, including the leptonic branching ratios, are shown in Table 2, requiring that the photons have $p_T > 20$ GeV and $|\eta| < 2.7$, and that the invariant mass of the two photons be between 90 and 150 GeV.

The $c\bar{c}\gamma\gamma$, $b\bar{b}\gamma\gamma$ and $t\bar{t}\gamma\gamma$ cross-sections shown in Table 2 do not include the branching ratios of leptonic decays. For the $b\bar{b}\gamma\gamma$ process the transverse momentum of the b quark is required to be greater than 20 GeV. In the $t\bar{t}\gamma\gamma$ events, also produced by MadGraph, the photons only come from the top quarks. The contribution from events in which photons arise from the decay products of the $t\bar{t}$ system was roughtly evaluated. Events were generated with PYTHIA using the Matrix Elements for the $t\bar{t}$ process. The production of two additional high $p_{\mathrm{T}}$ photons coming from the final state radiation of the top quark decay products was evaluated. It was found that the $t\bar{t}\gamma\gamma$ background needs to be multiplied by a factor 7.5 to account for this effect: this factor was applied for the analysis of the associated channel. It is probably conservative because the photons radiated by the quarks should not be isolated and would be removed by the analysis cuts.

There is a contribution to the background from events with a $W$ boson and a photon, where the $W$ boson decays into an electron and a neutrino. In some events, the electron can be mis-reconstructed as a photon. The PYTHIA generator is used for this background and the cross-section, including the leptonic branching ratio, is given in Table 2 after requiring $p_T > 20$ GeV and $|\eta| < 2.7$ for both the photon and the electron. For all the processes in Table 2 a full detector simulation was used and QED radiative corrections to the $Z/W$ decays were treated with PHOTOS [45].

# 3 Photon selection and $\gamma\gamma$ reconstruction

## 3.1 Photon reconstruction and calibration

Photons in the calorimeter are reconstructed from clusters of different sizes: in the barrel region ($|\eta| <$ 1.37 ) a $\Delta\eta$x$\Delta\phi = $ 3x7 cluster (in units of middle sampling calorimeter cells) is used for converted photons to recover as much as possible the energy that might be lost due to the opening of the two electrons in the magnetic field. A 3x5 cluster is used for unconverted photons. In the endcap region ($1.52 < |\eta| < 2.37$ ) a 5x5 cluster has been used for both converted and unconverted photons. The chosen cluster sizes are a compromise between maximal energy containment and minimization of the noise. In the simulation, all the effects that contribute to the calorimeter resolution constant term, which by design should be kept below 0.7 %, are taken into account by smearing the reconstructed cells energies. The energy of the photons in the electromagnetic (EM) calorimeter has been reconstructed using appropriate weights for the presampler and back compartments of the calorimeter to correct for the energy lost in the material in front of the calorimeter, for the longitudinal leakage and for the energy losses outside the cluster. Different weights for unconverted photons and electrons are used, as derived from the nominal detector geometry (see Ref. [46] for details). Clusters are then corrected for a series of effects. The variations of the energy with respect to the impact point inside each cell are corrected for. These effects are due to the fact that the shower is not fully contained inside the cluster (the effect is larger for small cluster sizes) and to the calorimeter material structure in $\phi$. The photon candidate position is defined as the barycenter of the associated cluster: the fact that the cells have a finite granularity introduces a bias in the measured position which is corrected for as a function of the particle impact point within the central cell (more details in Ref. [46]).

## 3.2 Photon identification

Powerful photon identification is required to reduce the background from jets faking photons (from $jj$ events and $\gamma j$ events) below the irreducible background. The photon identification relies on the fine segmentation of the electromagnetic calorimeter, especially the first layer, allowing an event-by-event rejection of "isolated" $\pi^0$s, which are the main source of fake photons from jets. The details of the shower-shape variables and the cuts can be found in Ref. [47]. The shower-shape variables include the leakage in the first compartment of the hadronic calorimeter, variables characterizing the lateral size of the shower in the second layer of the EM calorimeter, and variables related to the transverse size in the first layer of the EM calorimeter, together with a search for a second maximum in $\eta$ in the energy deposited in the strips of the first layer. The average efficiency of the calorimeter cuts for photons from Higgs boson decays with $p_{\mathrm{T}} > 25$ GeV has been found to be 83% when pile-up corresponding to $10^{33}\mathrm{cm}^{-2}\mathrm{s}^{-1}$ instantaneous luminosity is added. A track isolation cut is also applied to reduce further the fake background: the sum of the $p_{\mathrm{T}}$ of tracks in a $\Delta R = 0.3$ cone around the cluster position is computed for tracks with $p_{\mathrm{T}} > 1$ GeV and a cut at 4 GeV is applied. For tracks with $\Delta R < 0.1$ of the cluster position additional cuts are applied to remove conversion tracks from the sum. The efficiency of this cut for photons fulfilling all other identification cuts is 98% [47]. The cuts have been optimized using photons from Higgs boson decays as a signal sample and an inclusive jet sample for the fake rate study. The optimisation of the cuts has been done mostly in the $E_T$ range $25-35$ GeV and the same cuts are applied to converted and unconverted photons. Table 3 shows the rejection factors for inclusive jets (normalized to jets found from the simulated particles without detector effects using a cone of size $\Delta R = 0.4$), as measured from the inclusive jet sample. The rejection is also given separately for quark and gluon jets. The difference in rejection comes from the different fragmentation. After all cuts, the dominant background comes from "single" $\pi^0$s. The uncertainty on the rejection (uncertainty on the fragmentation, modeling of the detector response) is close to a factor of two [47].

Table 3: Jet rejections expected for the inclusive jet sample for $E_T > 25$ GeV. The results are shown before and after track isolation cuts for all jets and separately for quark and gluon jets. The errors are statistical only.

| | All | quark-jet | gluon-jet |
|---|---|---|---|
| Rejection (before isolation) | 5070±120 | 1770±50 | 15000±700 |
| Rejection (after isolation) | 8160±250 | 2760±100 | 27500±2000 |

### 3.3 Conversion reconstruction

Considering Higgs boson decays with photons within $|\eta| < 2.5$, about 57% of the selected events have at least one true conversion with a radius smaller than 80 cm. Converted photons may start showering before the beginning of the calorimeter thereby degrading the energy resolution. In addition, the energy deposition in the calorimeter from a converted photon is geometrically broader in $\phi$ than that from an unconverted photon due to the magnetic field in the Inner Detector cavity. On the other hand when a photon from a Higgs boson decay converts, the measurement of the conversion radius can be used to improve the accuracy on the measurement of the photon direction (see Section 3.4).

Conversions are reconstructed by a vertexing algorithm using the reconstructed particle tracks, as explained in more detail in Ref. [48]. An electromagnetic cluster with an associated track which is one of the two tracks of a conversion, is classified as a *double track conversion*. When a photon converts and only one reconstructed track is associated to the corresponding cluster this photon may be mis-identified as an electron: in order to increase the conversion reconstruction efficiency when an electromagnetic cluster has an associated track with no B-layer hit and the associated track does not belong to a reconstructed double track conversion, this object is classified as a *single track conversion*. Including single track conversions in the analysis increases by $\approx 6\%$ the signal efficiency while the overall background is increased by approximately the same factor. The conversion reconstruction efficiency as a function of the conversion radius is shown in Fig. 1: with the reconstruction software version used for this analysis, the overall efficiency is $\approx 66.4\,\%$ for conversions with a radius below 40 cm. The different contributions from single-track and double-track conversion reconstruction are also shown.

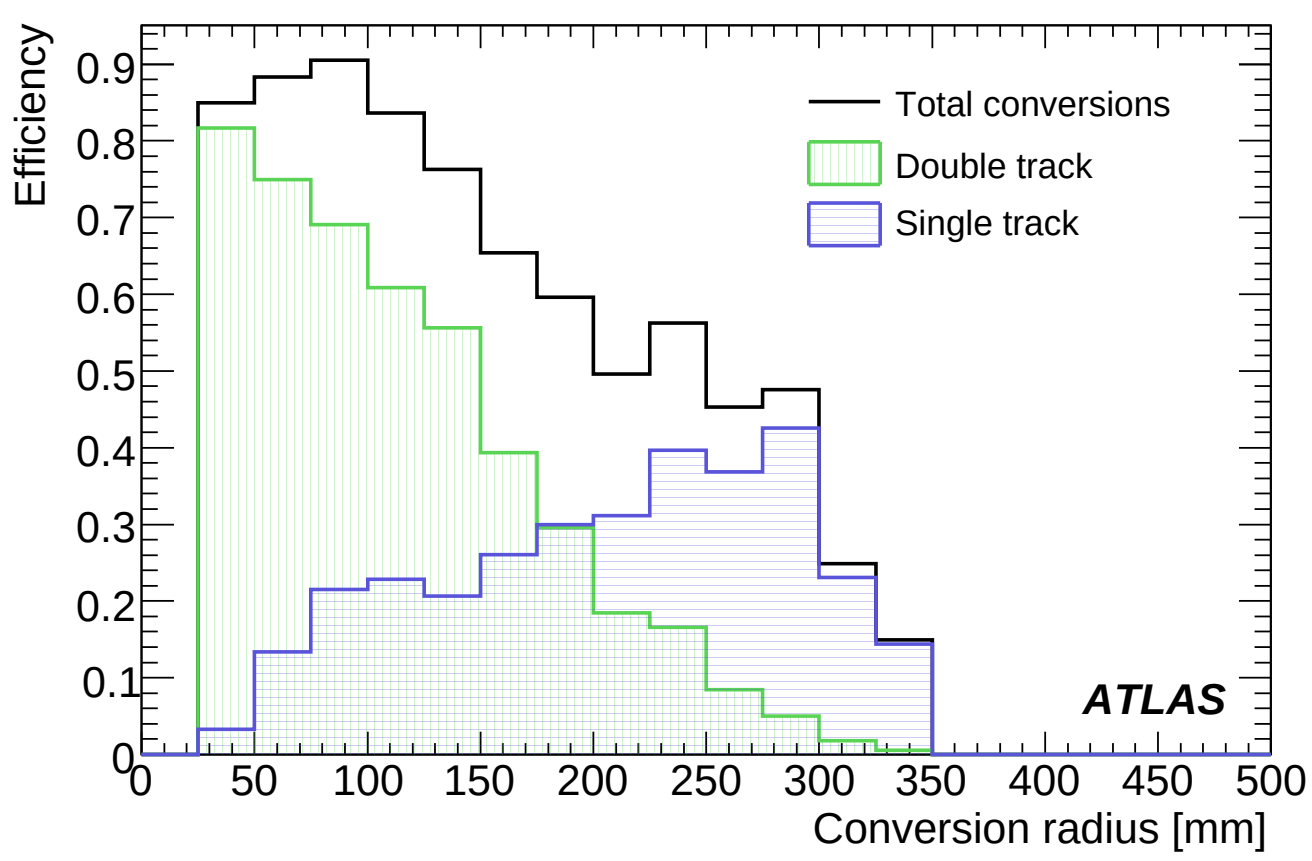


Figure 1: Efficiency of single-track and double-track conversion reconstruction as a function of the conversion radius.

For a given electromagnetic cluster, more than one associated reconstructed conversion can be found:

multiple conversions are due to secondary conversions or fake conversions. Fake conversions arise from two tracks that do not come from a conversion or with only one of the two tracks coming from a conversion. In the case of multiple conversions, the conversion with the smallest radius is selected: the reconstructed conversion is then correctly associated to a converted photon from a Higgs boson decay in 97.6% of simulated signal cases.

The jet contribution to the total background can be evaluated from the analysis of converted photons (see Ref. [48] for more details and updated results). The presence of tracks associated to the conversion provides a measurement of the transverse momentum of the converted photon in the tracker and consequently an evaluation of the ratio $p_{\mathrm{T}}/E_T$, where $E_T$ is the transverse energy of the calorimeter cluster. This variable should be distributed between 0 and 1. As shown in Fig. 2, converted photons have values of $p_{\mathrm{T}}/E_T$ which populate the region around 1 with a large non Gaussian tail to lower values due to bremsstrahlung of the electrons. In contrast, the distribution expected for a converted photon from a $\pi^0$ with the same transverse energy peaks at much lower values since the cluster collects the full energy of both photons from $\pi^0$ decay. It is particularly instructive to note that Fig. 2 also shows that the expected distributions for $p_{\mathrm{T}}/E_T$ for jets passing all the photon identification cuts is very similar to that obtained from single $\pi^0$'s with the same transverse energy: this indicates that the residual background from jets is dominated by single $\pi^0$'s. The shapes in Fig. 2 for converted photons coming from a jet and converted photons from the direct process can be parametrized and used to discriminate between the two components in a data sample of conversion events, allowing an evaluation of the $\gamma j$ and $jj$ percentage in the background.

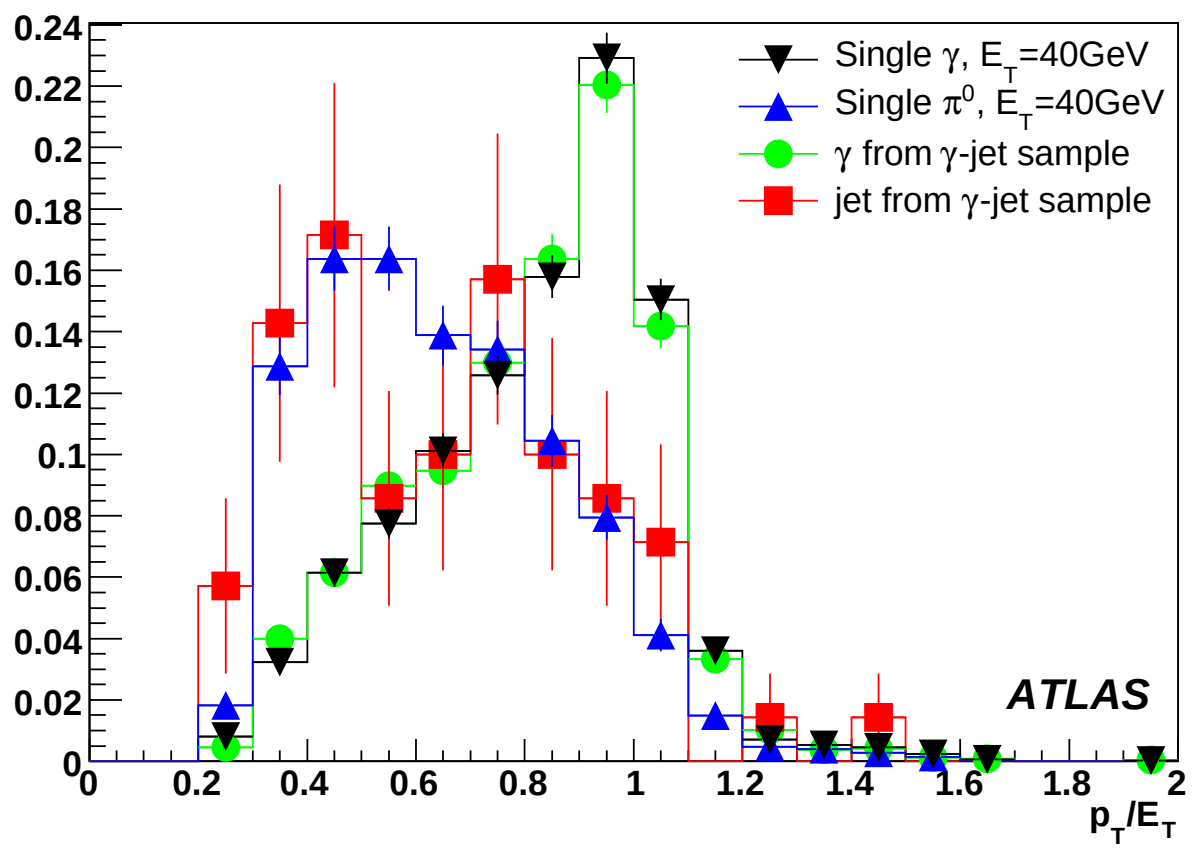


Figure 2: $p_{\mathrm{T}}/E_T$ ratio for conversions where both tracks were reconstructed in different event samples.

### 3.4 Primary vertex reconstruction

Among the reconstructed photons passing the identification cuts, the two with highest $p_{\mathrm{T}}$ are assumed to come from the Higgs boson decay. The azimuthal angle $\phi$ is determined by the cluster barycenter in the second layer of the EM calorimeter. The pseudorapidity $\eta$ relies on the knowledge of the position $z_H$ at which the Higgs boson is produced and decays.

Exploiting the multi-layer structure of the EM calorimeter, an estimate of the direction of each photon can be achieved by fitting a straight line in the $(R,z)$ plane through the cluster barycenters detected in the presampler and in the first and second samplings of the EM calorimeter. The intercept of these lines with the ATLAS beam axis provides two independent measurements of the hard scattering vertex, $z_{\gamma 1}$ and $z_{\gamma 2}$ with their uncertainties. They are combined in a weighted average with the nominal interaction vertex

at $z_0 = 0$ which has a spread of ±56 mm to yield a calorimeter estimate, $z_H^{\text{calo}}$, of the primary vertex. The performance expected for $H \to \gamma\gamma$ events is displayed in Table 4 and Fig. 3 (left plot). The accuracy (RMS) of the primary vertex position reconstruction is 13 mm, 17 mm, 41 mm, when both photons are in the barrel, one in the barrel and one in the endcap, and both in the endcap. In the simulation of the events used for this analysis a 4mm shift of the electromagnetic calorimeter along the z-axis was introduced (see Ref. [49] for more details) with respect to the nominal position. This effect has been taken into account at the reconstruction level, recalculating the pseudorapidity of the cells in each calorimeter layer with respect to the interaction point taking as a radial position reference the geometrical center (in the radial direction) of the cell. To compute the photon direction, the pointing algorithm makes use of the radial position of the barycenter of the shower instead of the geometrical center of the cell biasing the distribution of the reconstructed primary vertex position obtained from calorimeter pointing with respect to the true position as shown in Fig. 3, left plot (see Ref. [50] for more details). When a conversion is reconstructed, its coordinates are added as an extra point to the straight line fit, thus improving the Higgs boson vertex position accuracy; the distribution exhibits large tails with an RMS of 8 mm and a narrow Gaussian core with a width of 0.15 mm.

By adding the reconstructed primary vertex to the linear fit, the best Higgs boson position accuracy is achieved, with a Gaussian width of 0.07 mm (see Table 4 and Fig. 3, plot on the right). In the case of pile-up, more than one primary vertex may be reconstructed by the inner detectors. The discrimination of the hard-scattering vertex from pile-up vertices is done using a likelihood that is a combination of the calorimeter information and the sum of the squares of the $p_T$s of the tracks originating from the vertex (named $P_T^2$ in the following). So

$$\mathscr{L} = \mathscr{L}_{P_T^2} \times \mathscr{L}_{calo} \tag{1}$$

where

$$\mathscr{L}_{P_T^2} = p_H(p_T^2)/p_{MB}(p_T^2) \tag{2}$$

$$\mathscr{L}_{calo} = e^{-\frac{1}{2}\frac{z^2}{56^2}} \times e^{-\frac{1}{2}\frac{(z-z_{calo})^2}{\sigma^2_{z_{calo}}}} / e^{-\frac{1}{2}\frac{z^2}{56^2}} = e^{-\frac{1}{2}\frac{(z-z_{calo})^2}{\sigma^2_{z_{calo}}}} \tag{3}$$

The first component of the likelihood, $\mathscr{L}_{PT2}$, is, for a certain value of $P_T^2$, the probability for the Higgs boson vertex to have this $P_T^2$ divided by the same probability for a minimum bias vertex. The second component is the product of the probabilities that a measured vertex is that of the Higgs boson vertex divided by the probability that the measured vertex position is that of a minimum bias vertex. The vertex misidentification due to pile-up is displayed by the columns labeled 'tail' in Table 4. When the vertex associated to the Higgs boson production is correctly identified among the others, the impact of the direction measurement on the invariant mass resolution becomes negligible with respect to that of the energy measurement.

### 3.5 Invariant mass and signal acceptances

The invariant mass of diphoton pairs has been reconstructed for different signal samples using the tools described in the previous Sections: the `2g17i` trigger selection (see Section 4 for more details) is applied and two identified photons are required to be reconstructed with $p_T > 40$ GeV and $p_T > 25$ GeV within the fiducial region of $0 < |\eta| < 1.37$ and $1.52 < |\eta| < 2.37$. Fig. 4 presents the invariant mass distributions of $\gamma\gamma$ pairs from Higgs boson decay with $m_H = 120$ GeV. The shaded histograms correspond to events with at least one true converted photon with a conversion radius smaller than 80 cm. The converted photons are currently calibrated at the electron scale and the unconverted ones at the photon scale: in the case of the geometry with additional material in front of the calorimeter, the observed difference

Table 4: Performance of the Higgs boson longitudinal vertex position reconstruction, using calorimeter pointing only ($z_H^{\text{calo}}$) and the reconstructed primary vertex ($z_H^{\text{calo+vtx}}$): averages ($\langle\rangle$) and RMS are displayed in mm. 'tail' shows the percentage of events which are outside the histogramme window ($\pm$100 mm and $\pm$1 mm for $z_H^{\text{calo}}$ and $z_H^{\text{calo+vtx}}$ respectively).

| Luminosity | $z_H^{\text{calo}} - z_H^{\text{true}}$ (mm) | | | $z_H^{\text{calo+vtx}} - z_H^{\text{true}}$ (mm) | | |
|---|---|---|---|---|---|---|
| | $\langle\rangle$ | RMS | tail(%) | $\langle\rangle$ | RMS | tail(%) |
| No pileup | 2.3 | 17.3 | 0.09 | -0.008 | 0.10 | 1.2 |
| $10^{33}$ | 3.3 | 17.4 | 0.09 | -0.010 | 0.10 | 13.0 |
| $2 \cdot 10^{33}$ | 2.4 | 17.1 | 0.18 | -0.007 | 0.10 | 18.3 |

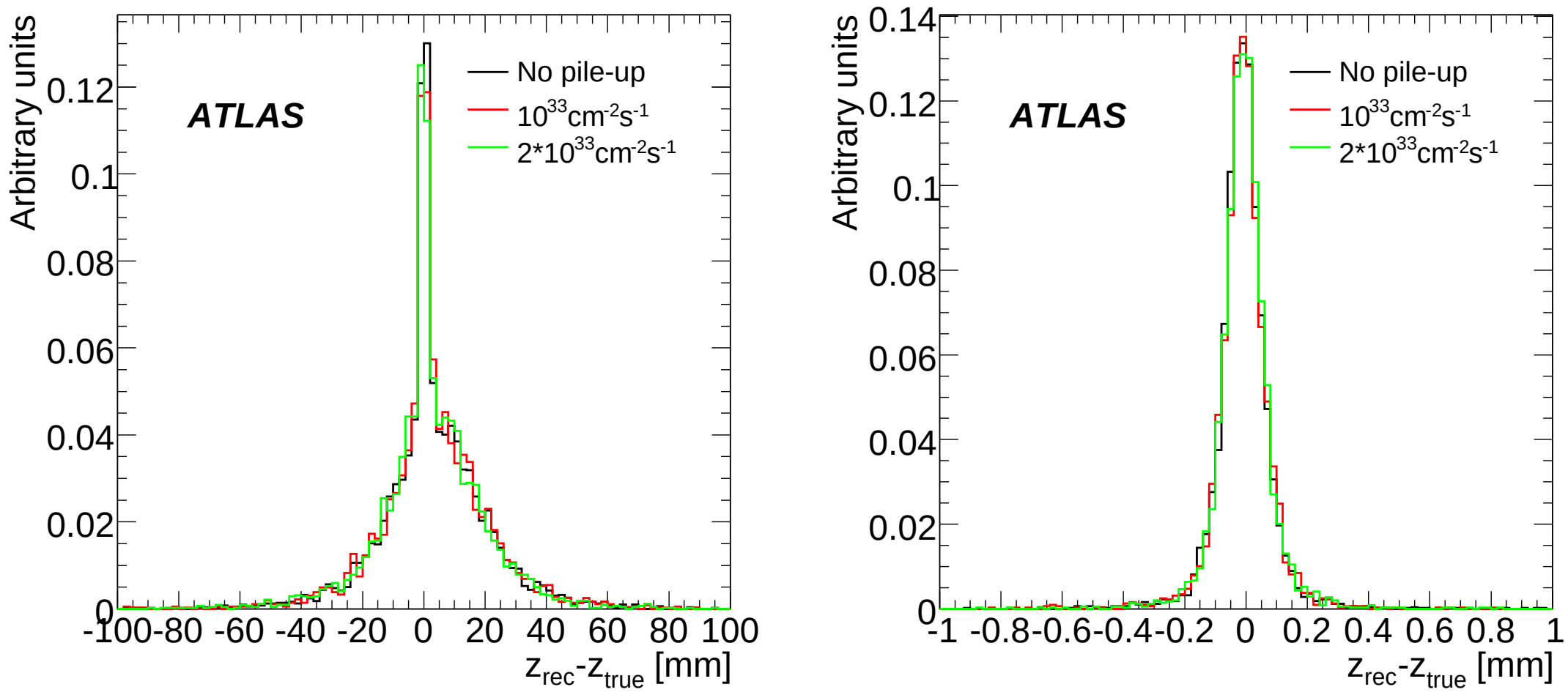


Figure 3: Difference between the reconstructed primary vertex position and the true position obtained from calorimetric pointing and conversion track information (when available) without/with the reconstructed primary vertex (left/right plot), for events without pile-up (black plots) and with pile-up evaluated for $10^{33}$ and $2 \cdot 10^{33}$ cm$^{-2}$s$^{-1}$ (red, green plots). The narrow peak on top of the broader one is due to events in which at least one photon has a reconstructed conversion vertex. In the right plots, the non-Gaussian shape is due to the overlap of barrel-barrel, barrel-endcap and endcap-endcap topologies, which have different resolutions.

between the peak of the distribution of events with no conversions and the one with at least one reconstructed conversion is around 1 %. In the future a specific calibration will be implemented for converted photons using the same prescription as that used for electrons and unconverted photons.

To be consistent with previous studies ( [4], [3]) the mass resolution obtained for the photon pairs has been determined from an asymmetric Gaussian fit ([-2 $\sigma$ , + 3 $\sigma$]) of the invariant mass peak. The asymmetric window is used to reduce the impact of the residual low energy tails in the reported width. Table 5 shows the results for different Higgs boson masses with and without pileup at $10^{33}$ cm$^{-2}$s$^{-1}$ in the case of simulations with additional dead material in front of the calorimeter (see Ref. [49] for more details). Since the energy calibration coefficients have been calculated using the nominal geometry, the presence of additional dead material in front of the calorimeter affects the determination of the value of the invariant mass, moving the peak mean down by a few per mille. In addition it increases the amount of low energy tails: the percentage of events with a reconstructed invariant mass more than 3$\sigma$ from the mean increases from 3.5 % for the nominal geometry to 6.8 % for the distorted geometry.

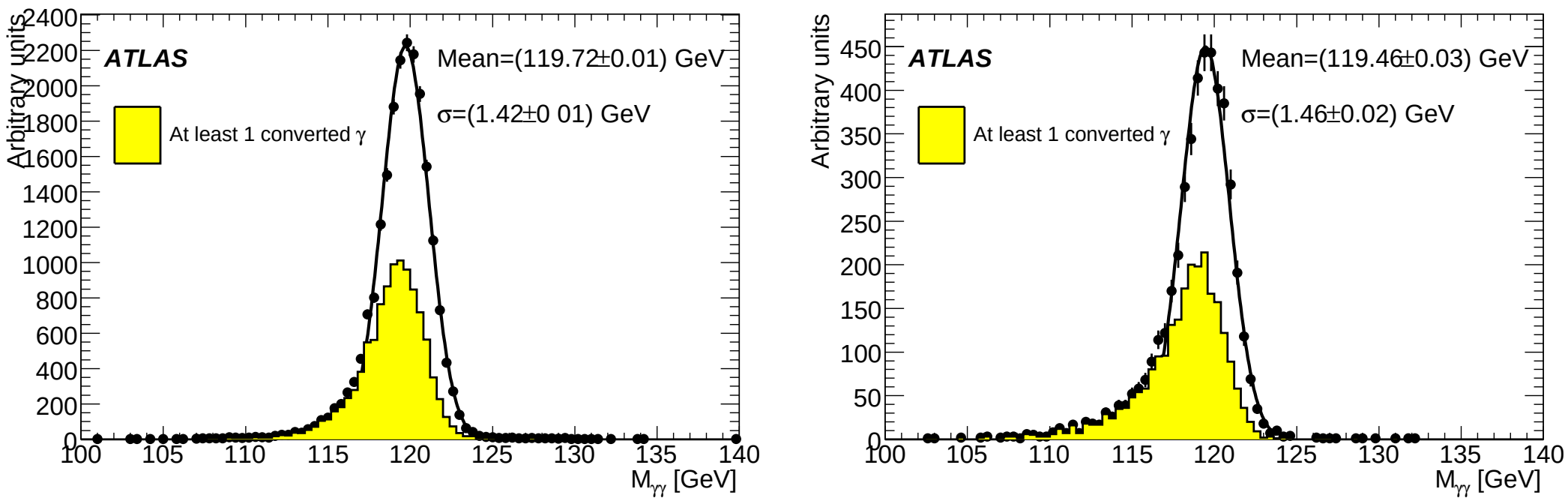


Figure 4: Invariant mass distributions for photons pairs from Higgs boson decays with $m_H = 120$ GeV after trigger and identification cuts; on the left the invariant mass distribution obtained using the nominal geometry simulation is reported while on the right plot the same invariant mass distribution is reported when additional dead material is included in the simulation. The shaded histograms correspond to events with at least one converted photon.

The acceptances after trigger selection (see Section 4 for more details) and analysis cuts together with the mass resolutions for the different values of $m_H$ are reported in Table 5. The mass window for the evaluation of the signal significance (denoted *mass bin*) is defined as $\pm 1.4\sigma$ around the central value. As can be seen in Table 5 the fraction of signal events in the mass bin is 26 % for $m_H = 120$ GeV slightly increasing with the Higgs boson mass. The relative mass resolution $\sigma_m/m$ is close to 1.2 % degrading by a few percent relative when the pileup is added.

Table 5: Efficiencies after trigger, identification and inclusive analysis cuts (see Section 5.1). Reconstructed invariant mass peak positions and resolutions for different Higgs boson masses with and without $10^{33}$ cm$^{-2}$s$^{-1}$ pileup are also reported. Distorted geometry has been used in all cases.

| | $m_H = 120$ GeV | | $m_H = 130$ GeV | | $m_H = 140$ GeV | |
|---|---|---|---|---|---|---|
| | No pileup | Pileup | No pileup | pileup | No pileup | Pileup |
| L1 | 0.66 | 0.64 | 0.69 | 0.65 | 0.68 | 0.66 |
| L2 | 0.54 | 0.52 | 0.55 | 0.52 | 0.56 | 0.53 |
| EF | 0.50 | 0.47 | 0.52 | 0.49 | 0.52 | 0.49 |
| Analysis cuts | 0.36 | 0.32 | 0.38 | 0.35 | 0.39 | 0.36 |
| Mass bin | 0.26 | 0.24 | 0.28 | 0.26 | 0.29 | 0.27 |
| Mass Fitted ($m$) | 119.46 | 119.47 | 129.47 | 129.41 | 139.41 | 139.41 |
| $\sigma_m$, GeV | 1.46 | 1.52 | 1.54 | 1.62 | 1.66 | 1.69 |

### 3.6 Background analysis on fully simulated samples

Large samples of $\gamma\gamma$ and $\gamma j$ events as well as Drell-Yan $Z \to e^+e^-$ were generated with PYTHIA and fully simulated. The differential cross-section at LO as a function of the diphoton invariant mass for $\gamma j$ and $e^+e^-$ events after the analysis procedure described in previous sections is reported in Fig. 5. For the $jj$ background contribution evaluation not enough statistic was available in full simulation and it has been estimated only from fast simulated events using photon efficiency and the jet rejection parametrizations obtained from full simulation.

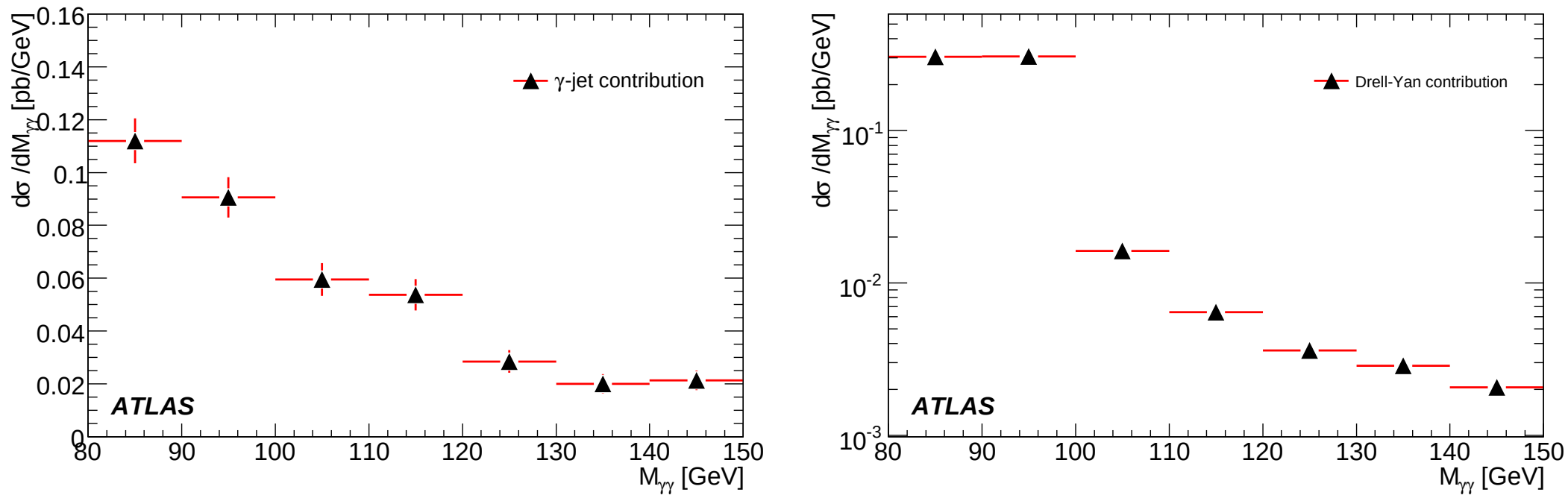


Figure 5: Diphoton candidates invariant mass distribution for $\gamma j$ (left) and Drell-Yan $e^+e^-$(right) events after photon identification and analysis cuts with and without trigger selection.

The goodness of the photon efficiency and jet rejection parametrizations used in Section 5 on fast simulated events to estimate the background contributions has been tested on the $\gamma j$ sample in full simulation: the distribution of the cross-section as a function of the diphoton candidates invariant mass has found to be in agreement within 15 % in the 110-150 GeV mass range with the distribution from the fast simulated sample after photon efficiency and jet rejection parametrization used in Section 5.

### 3.7 Jet tagging in simulation

Section 5 presents two analyses which rely on tagging hadronic jets. Jet tagging is particularly relevant to Higgs boson searches in association with two high $p_T$ jets since such an analysis is intended to isolate Higgs boson production via VBF.

In the VBF production process, it is expected that the two quark-initiated jets are observed in opposite hemispheres and with a large separation in pseudorapidity. Tagging these jets further suppresses the background processes. Furthermore, since there is no colour exchange between the two quarks, the Higgs boson should be observed in a large rapidity gap, where additional activity from QCD jets is small. A central jet veto (CJV) which suppresses the background processes is therefore used in the event selection,

The relative efficiencies for jet tagging are summarized in Table 6. The following four conditions are applied:

- two simulated quarks with $p_T > 40$ GeV and $p_T > 20$ GeV ($|\eta| < 5$) and in opposite hemispheres,
- two reconstructed jets with $p_T > 40$ GeV and $p_T > 20$ GeV ($|\eta| < 5$),
- the two reconstructed jets are in opposite hemispheres,
- differences in $\eta$ between the simulated quarks and the reconstructed jets are smaller than 0.4 (this is referred to as matching).

The present tagging method selects correct tagging jets in 75.7% of the events. Under a pileup condition of $10^{33}\mathrm{cm}^{-2}\mathrm{s}^{-1}$, a degradation of less than 5% is observed. The difference between HERWIG and PYTHIA is about 6%.

## 4 Performance of the photon trigger on $H \to \gamma\gamma$ events.

There are two trigger selections which are foreseen for the $H \to \gamma\gamma$ analyses: `2g17i` and `g55`. The `2g17i` trigger selects events with at least two isolated photons and is efficient for photon $p_T$ above 20 GeV. The

Table 6: Relative efficiencies of the tagging method for VBF $H \to \gamma\gamma$ with $m_H = 120$ GeV (see text).

| Selection | HERWIG (no pileup) | HERWIG (Pileup $10^{33}$) | PYTHIA (no pileup) |
|---|---|---|---|
| step 1 (quark level) | $0.618 \pm 0.002$ | $0.613 \pm 0.003$ | $0.632 \pm 0.002$ |
| step 2 (rec level) | $0.914 \pm 0.002$ | $0.911 \pm 0.002$ | $0.943 \pm 0.001$ |
| step 3 (rec level) | $0.801 \pm 0.002$ | $0.774 \pm 0.003$ | $0.771 \pm 0.002$ |
| step 4 (matching) | $0.757 \pm 0.003$ | $0.726 \pm 0.003$ | $0.713 \pm 0.003$ |

g55 trigger selects events with at least one photon and is efficient for photon $p_{\mathrm{T}}$ above 60 GeV. No isolation is required in the g55 trigger. In the analyses presented in Section 5 only the 2g17i menu has been considered to avoid possible biases in the invariant mass reconstruction: a more detailed description of the photon selection at the High Level Trigger is available [51]. The photon reconstruction at the trigger level relies only on calorimetric information. To calculate $E_T$ and shower shapes variables the High Level Trigger uses algorithms similar to those used for offline analysis using information from the first and second sampling of the calorimeter. For photon trigger menus the isolation is only applied at L1 trigger.

The trigger efficiency has been evaluated for Higgs boson events having two offline reconstructed photons passing the following kinematic cuts: two photons with $p_{T\gamma} > 25$ GeV and $p_{T\gamma} > 40$ GeV respectively both in the region $0 < |\eta| < 1.37$ and $1.52 < |\eta| < 2.37$, passing the identification cuts described in Section 3.2.

Table 7: Efficiency for the 2g17i menu item to trigger on $H \to \gamma\gamma$ events with $m_H = 120$ GeV, normalized with respect to the offline selections.

| Trigger Level | 2g17i Trigger efficiency |
|---|---|
| L1 | 96.3±0.3 |
| L2 Calo | 95.0±0.4 |
| EF Calo | 93.6±0.4 |

Table 7 shows in the first column the efficiency of the 2g17i menu item to select $H \to \gamma\gamma$ events after each trigger level.

The 2g17i trigger menu item is expected to be ∼94% efficient for triggering on Higgs boson decays with two reconstructed photons. The efficiency loss is mainly due to the calorimeter isolation at L1 which is not applied in the offline photon selection.

Trigger rates for 2g17i are studied in the context of the ATLAS trigger development. This trigger is found to be usable (unprescaled) up to luminosities of $10^{33}\,\mathrm{cm^{-2}s^{-1}}$ [51].

# 5 Event selection

In this Section the analysis details of the various event selections are described. The event selection for the inclusive analysis is given in Section 5.1 and for the Higgs boson search in association with jets are given in Sections 5.2 and 5.3. Finally, the event selections for diphoton searches in association with missing transverse momentum and charged leptons are given in Sections 5.4 and 5.5. Here the various event selections are presented as disjoint analyses. In Section 7 the statistical power of these channels is combined and their impact on the Higgs boson discovery potential is evaluated.

### 5.1 Inclusive analysis

The inclusive analysis refers to the search for a resonance in events with two photons that pass certain quality criteria. The analysis reported here follows closely the event selection of past studies [3,4]. The detector performance and optimization studies succinctly presented in Sections 3 and 4 are geared toward maximizing the discovery potential of the inclusive analysis.

The following cuts are applied:

**Ia** At least two photon candidates (see Section 3.2) in the central detector region defined as $|\eta| < 2.37$ excluding the transition region between barrel and endcap calorimeters, $1.37 < |\eta| < 1.52$ (*crack* in the following). At this level it is required that the event passes the trigger selection (see Section 4).

**Ib** Transverse momentum cuts of $40, 25$ GeV on the leading and sub-leading photon candidates, respectively.

The fiducial cuts in **Ia** are motivated by the quality of the off-line photon identification and the fake photon rate (see Section 3.2). The values of the cuts on the transverse momentum of the photon candidates (cut **Ib**) are not varied and are obtained from previous optimization studies [3].

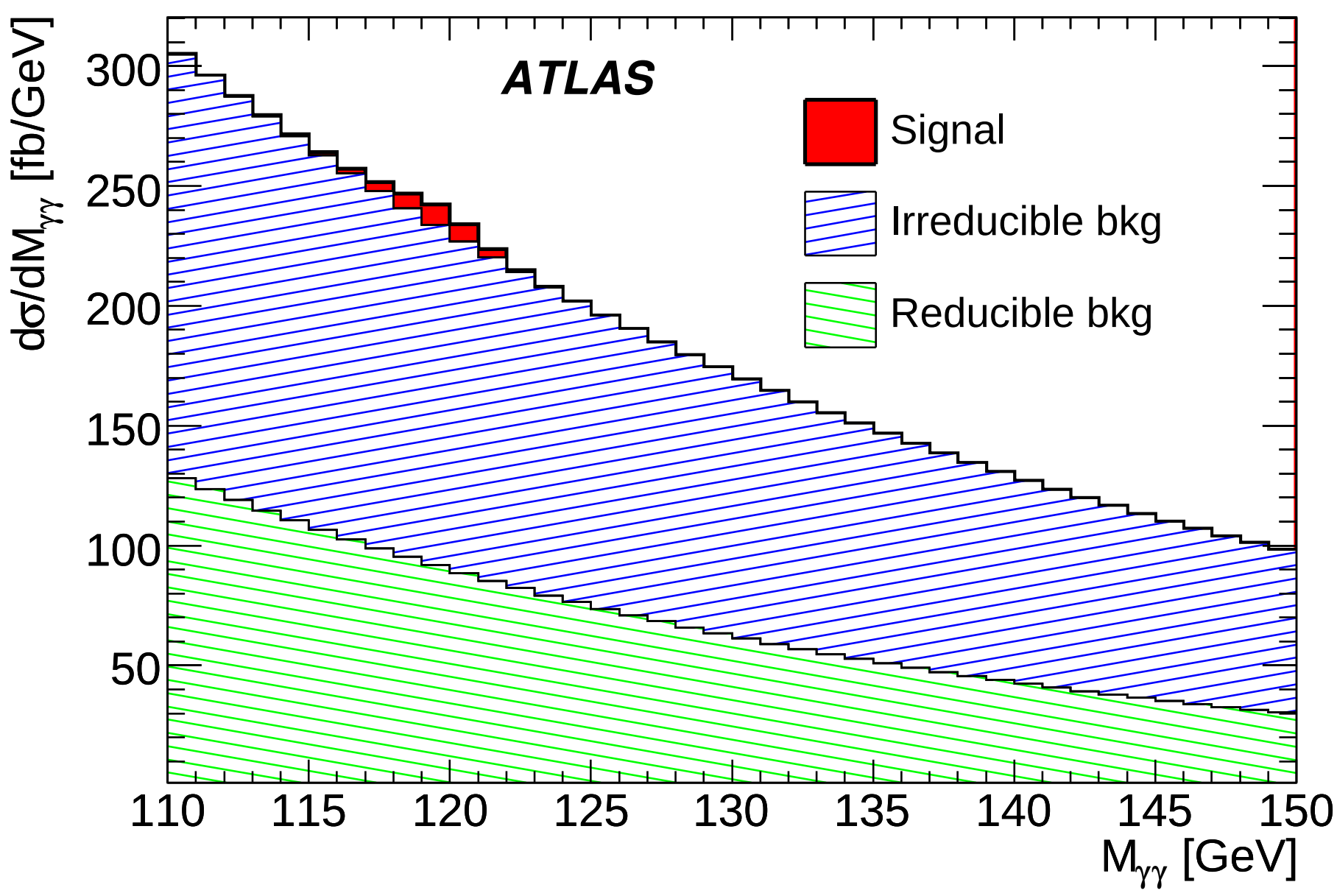


Figure 6: Diphoton invariant mass spectrum after the application of cuts of the inclusive analysis. Results are presented in terms of the cross-sections in fb. The contribution from various signal and background processes are presented in stacked histograms (see text).

Figure 6 shows the expected diphoton mass spectrum after the application of cuts **Ia** and **Ib**. The hashed histogram in the bottom corresponds to the contributions from events with one and two fake photons. The second hashed histogram corresponds to the irreducible backgrounds (see Section 2.2). The background contributions are obtained with MC samples with a fast detector simulation normalized to the cross-sections specified in Section 2.2. The fast detector simulation is corrected in order to reproduce the aspects of the detector performance critical to the analysis, which are obtained with a full detector simulation (see Sections 3 and 4). The expected contribution from a Higgs boson signal for $m_H =$ 120 GeV, obtained with a full detector simulation, is also shown in Fig. 6.

Table 8: Expected cross-sections (in fb) for different signal ($m_H = 120$ GeV) and background processes within a mass window of $m_{\gamma\gamma} \pm 1.4$ of the mass resolution in the no pileup case reported in Table 5. Cuts **Ia** and **Ib** were applied.

| Signal Process | Cross-section (fb) | Background Process | Cross-section (fb) |
|---|---|---|---|
| $gg \to H$ | 21 | $\gamma\gamma$ | 562 |
| VBF $H$ | 2.7 | Reducible $\gamma j$ | 318 |
| $ttH$ | 0.35 | Reducible $jj$ | 49 |
| $VH$ | 1.3 | $Z \to e^+e^-$ | 18 |

Table 9: Summary of the relative systematic uncertainties on the $\gamma\gamma$ and $\gamma j$ processes.

| Potential sources | $\gamma\gamma$ | $\gamma j$ |
|---|---|---|
| Scale dependence | 14% | 20% |
| Fragmentation | 5% | 1% |
| PDF | 6% | 7% |
| Total | 16% | 21% |

Table 8 shows the expected cross-sections (in fb) for background and signal in a mass window of $\pm 1.4$ of the mass resolution in the no pileup case reported in Table 5 around 120 GeV after the application of cuts **Ia** and **Ib**. Table 8 indicates that the relative contribution from events with at least one fake photon constitutes 39% of the total background, about a factor of two larger than evaluated in Ref. [4]. This increase is mostly attributed to three factors. Firstly, a different method for the parametrization of the fake photon background is used here. Secondly, the budget of inactive material in front of the first layer of the calorimeter has increased with respect to the one used in previous studies. Finally, the contribution from fragmentation in the $\gamma j$ process (see Section 2.2) is allowed for here for the first time.

### 5.1.1 Theoretical uncertainties on background prediction

In this Section the theoretical uncertainties of the predictions for prompt single and double photon production used in the inclusive analysis are evaluated.

The irreducible background rate is evaluated using ResBos [34–36], which implements a full matrix element calculation at NLO and the resummed formalism. It thus yields an accurate description of the low $p_{T\gamma\gamma}$ region, as corroborated by recent Tevatron results [52]. In the high diphoton transverse momentum, $p_{T\gamma\gamma}$, domain the precision of its prediction needs to be assessed. To account for the incompleteness of the fixed order calculation two approaches are used: both the renormalisation and factorisation scales ($\mu_{R,F}$) are varied, from $0.5 \times m_{\gamma\gamma}$ to $2 \times m_{\gamma\gamma}$, first assuming $\mu_R = \mu_F$ then independently.

ResBos does not provide the most accurate description of the fragmentation of partons into photons. In particular, it implements the single-photon fragmentation only at LO. An improved estimation of this contribution is given by the DIPHOX program [33] which implements single and double-photon fragmentation at NLO. The predictions of the fragmentation and direct contributions of the two aforementioned calculations agree to within 6%.

The systematic uncertainty related to the parton distribution functions is studied in Ref. [53] and estimated to be of the order of 6%. Systematic uncertainties related to the irreducible background evaluation are summarised in Table 9 they amount to an overall relative uncertainty of 16%.

A similar study has been performed with JETPHOX in order to evaluate theoretical uncertainties of the $\gamma j$ process. The second column of Table 9 reports the results from this study.[2]

### 5.2 Higgs boson plus one jet analysis

In this Section and in Section 5.3 two event selections are presented that take into account the presence and properties of high $p_T$ hadronic jets in association with the photon pair. This analysis follows earlier studies in ATLAS using a fast detector simulation [8].

Parton level studies have indicated that searches for the Higgs boson in association with at least one high $p_T$ jet may have a strong discovery potential [5]. This analysis exploits mainly the fact that the gluon radiation pattern of the two leading Higgs boson production mechanisms differs strongly from the one expected for the reducible and irreducible backgrounds. The leading jet in the $gg \rightarrow Hj$ and $VBF$ mechanisms tends to be harder and be more separated from the diphoton system than in background events. The invariant mass of the two photons and the jet system discriminates well the signal from the background.

The following event selection is chosen after the application of cut **Ia**:

**IIa** Transverse momentum cuts of 45 and 25 GeV on the leading and sub-leading photon candidates, respectively.

**IIb** Presence of at least one hadronic jet with $p_T > 20$ GeV in $|\eta| < 5$.

**IIc** A cut on the invariant mass of the diphoton and the leading jet, $m_{\gamma\gamma j} > 350$ GeV.

The lower bound on the jet $p_T$ is dictated by the ability to calibrate hadronic jets in ATLAS [54]. The large hadronic activity due to the underlying events and multiple proton-proton interactions at the LHC, in conjunction to the significant amount of inactive material before the calorimeter, may make it difficult to lower the $p_T$ threshold. The variable $m_{\gamma\gamma j}$ is the main discriminator used here to improve the signal-to-background ratio. Other discriminating variables could be used to further enhance the analysis sensitivity [5].

Figure 7 displays the resulting diphoton invariant mass spectrum after the application of cuts **Ia** and **IIa-IIc**.

Table 10: Expected cross-sections (in fb) for the Higgs boson plus one jet Analysis. Results are given after the application of cuts **Ia** and **IIa-IIc** (see Section 5.2). In the last row the expected cross-sections within a mass window of $m_{\gamma\gamma}$ of $\pm 2$ GeV around 120 GeV are given.

| Cut | $gg \rightarrow H$ $\sigma$ (fb) | $VBFH$ $\sigma$ (fb) | $VH$ $\sigma$ (fb) | $ttH$ $\sigma$ (fb) | Total $\sigma$ (fb) |
|---|---|---|---|---|---|
| **Ia-IIa** | 28 | 3.6 | 1.7 | 0.49 | 34 |
| **IIb** | 13 | 3.5 | 1.5 | 0.49 | 19 |
| **IIc** | 3.2 | 1.9 | 0.22 | 0.17 | 5.5 |
| Mass Window | 2.3 | 1.4 | 0.17 | 0.13 | 4.0 |

Tables 10 and 11 display the expected cross-sections for signal and background events in the range $110 < m_{\gamma\gamma} < 150$ GeV after the application of cuts **Ia** and **IIa-IIc**. Table 10 illustrates that the leading

[2] It is important to note that the uncertainty of the contribution of the reducible background is dominated by the uncertainty in the determination of the fake photon rejection. This uncertainty may be significantly larger than the uncertainties quoted in Table 9.

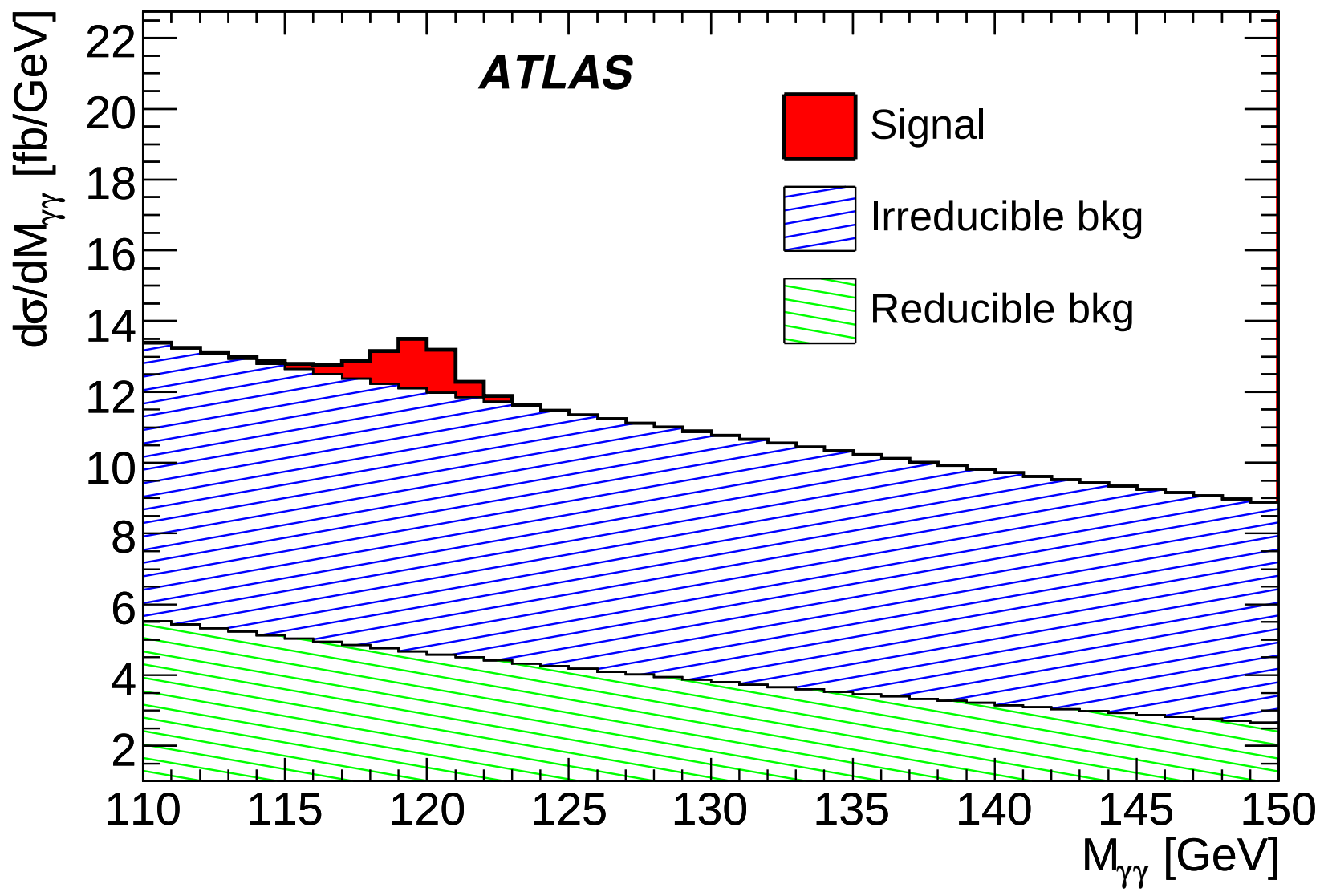


Figure 7: Diphoton invariant mass spectrum in fb obtained with the Higgs boson plus one jet analysis (see Section 5.2). The same procedure as in Fig. 6 in Section 5.1 is used to obtain the histograms in Fig. 7. The same codes for signal and backgrounds are used as in Fig. 6.

Table 11: Expected cross-sections (in fb) of background for the Higgs boson plus one jet Analysis. Results are given after the application of cuts **Ia** and **IIa-IIc** (see Section 5.2). In the last row the expected cross-sections within a mass window of $m_{\gamma\gamma}$ of $\pm 2$ GeV around 120 GeV are given.

| Cut | $\gamma\gamma$ $\sigma$ (fb) | Reducible $\gamma j$ $\sigma$ (fb) | Reducible $jj$ $\sigma$ (fb) | EW $\gamma\gamma jj$ $\sigma$ (fb) | Total $\sigma$ (fb) |
|---|---|---|---|---|---|
| **Ia-IIa** | 9698 | 8498 | 937 | 99 | 19233 |
| **IIb** | 4786 | 4438 | 444 | 99 | 9768 |
| **IIc** | 501 | 824 | 89 | 71 | 1485 |
| Mass Window | 28 | 17 | 2.0 | 1.5 | 49 |

Higgs boson production mechanism after the application of cuts remains the $gg \to Hj$ process, closely followed by the VBF mechanism. It is important to note that the $gg \to Hj$ process has been evaluated at LO ignoring the large QCD NLO corrections.

### 5.3 Higgs boson plus two jets analysis

This Section considers an event selection comprising two photons in association with two high $p_T$ jets, or tagging jets. In this analysis the tagging jets are defined as the two leading jets in the event. The *VBF* Higgs boson process at LO produces two high $p_T$ and relatively forward jets in opposite hemispheres (backward-forward). The pseudorapidity gap and invariant mass of these jets tend to be significantly larger than those expected for background processes. The NLO description of the VBF process does not significantly distort this picture.[3]

[3] About 10% of the VBF events display the feature that a radiated gluon coming from one of the quark lines happens to become a tagging jet. In this class of events the pseudorapidity gap and the invariant mass of the tagging jets appears similar to

A number of variables are chosen that are sensitive to the different kinematics displayed by the signal and background processes [9]. The following is the optimized event selection after the application of cut **Ia**:

**IIIa** Transverse momentum cuts of 50 and 25 GeV on the leading and sub-leading photon candidates, respectively.

**IIIb** Presence of at least two hadronic jets in $|\eta| < 5$ with $p_T > 40, 20$ GeV for the leading and sub-leading jet, respectively. The tagging jets must be in opposite hemispheres, $\eta_{j1} \cdot \eta_{j2} < 0$, where $\eta_{j1}$ and $\eta_{j2}$ correspond to the pseudorapidity of the leading and sub-leading jets, respectively. Finally, it is required that the pseudorapidity gap between the tagging jets be large, $\Delta\eta_{jj} > 3.6$.

**IIIc** Photons are required to have pseudorapidity between those of the tagging jets.

**IIId** Invariant mass of the tagging jets, $m_{jj} > 500$ GeV.

**IIIe** Veto on events with a third jet with $p_T > 20$ GeV and $|\eta| < 3.2$

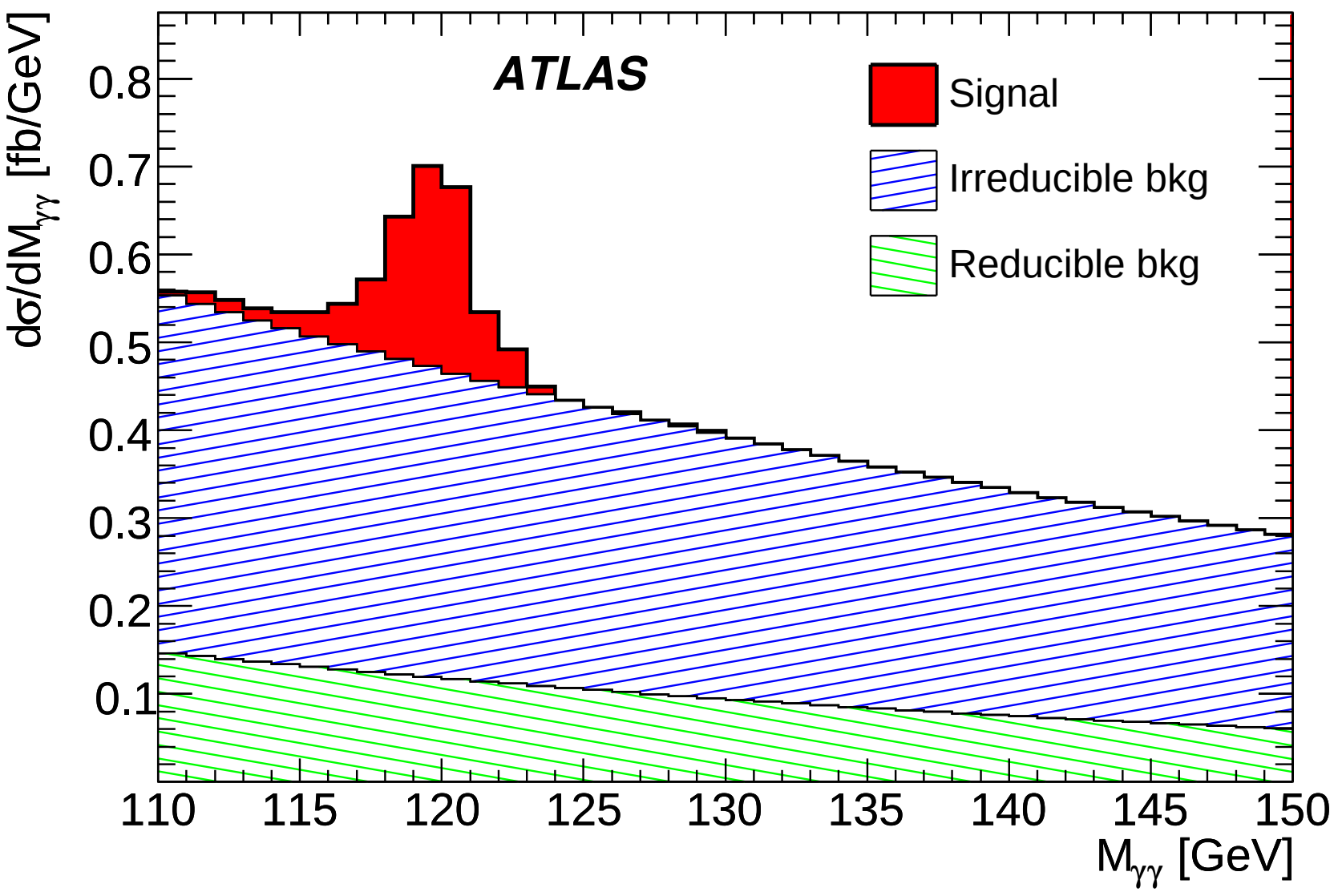


Figure 8: Diphoton invariant mass spectrum obtained with the Higgs boson plus two jet analysis (see Section 5.3).

Figure 8 displays the resulting diphoton invariant mass spectrum after the application of cuts **Ia** and **IIIa-IIIe**.

Tables 12 and 13 display the expected cross-sections for a Higgs boson signal with $m_H = 120$ GeV and background events in the mass range $\pm 2$ GeV around 120 GeV after the application of cuts **Ia** and **IIIa-IIIe**. Table 12 shows that the dominant Higgs boson production mechanism surviving the events selection is the VBF mechanism. Unfortunately, the QCD NLO corrections to the main backgrounds included in Table 13 are not known and therefore these results suffer from large theoretical uncertainties.

The event selections presented in this and the previous Sections have a certain degree of overlap. This is particularly relevant for the VBF Higgs boson production mechanism. In Section 7 the signal significance of a combined analysis is presented that takes into account the event overlap.

that displayed by a typical QCD background process. This effect is well reproduced by the HERWIG generator.

Table 12: Expected cross-sections (in fb) for the Higgs boson plus two jet analysis. Results are given after the application of cuts **Ia** and **IIIa-IIIe** (see Section 5.3). In the last row the expected cross-sections within a mass window of $m_{\gamma\gamma}$ of $\pm 2$ GeV around 120 GeV are given.

| Cut | $gg \to H$ $\sigma$ (fb) | $VBFH$ $\sigma$ (fb) | $VH$ $\sigma$ (fb) | $ttH$ $\sigma$ (fb) | Total $\sigma$ (fb) |
|---|---|---|---|---|---|
| **Ia-IIIa** | 26.40 | 3.53 | 1.68 | 0.50 | 32.11 |
| **IIIb** | 0.63 | 1.44 | 0.02 | 0.01 | 2.10 |
| **IIIc** | 0.55 | 1.39 | 0.01 | 0.01 | 1.96 |
| **IIId** | 0.32 | 1.16 | 0.01 | 0.00 | 1.49 |
| **IIIe** | 0.25 | 1.03 | 0.00 | 0.00 | 1.28 |
| Mass Window | 0.18 | 0.79 | 0.00 | 0.00 | 0.97 |

Table 13: Expected cross-sections (in fb) of background for the Higgs boson plus two jet analysis for $m_H = 120$ GeV. Results are given after the application of cuts **Ia** and **IIIa-IIIe** (see Section 5.3). In the last row the expected cross-sections within a mass window of $m_{\gamma\gamma}$ of $\pm 2$ GeV around 120 GeV are given.

| Cut | $\gamma\gamma$ $\sigma$ (fb) | Reducible $\gamma j$ $\sigma$ (fb) | Reducible $jj$ $\sigma$ (fb) | EW $\gamma\gamma jj$ $\sigma$ (fb) | Total $\sigma$ (fb) |
|---|---|---|---|---|---|
| **Ia-IIIa** | 7417 | 6355 | 710 | 92 | 14574 |
| **IIIb** | 94 | 97 | 13 | 45 | 249 |
| **IIIc** | 70 | 69 | 9.9 | 41 | 189 |
| **IIId** | 33 | 34 | 5.6 | 38 | 111 |
| **IIIe** | 17 | 17 | 2.5 | 26 | 63 |
| Mass Window | 0.86 | 0.42 | 0.06 | 0.59 | 1.95 |

## 5.4 Higgs boson plus missing transverse energy and isolated leptons

The main signal production mechanism contributing to a category of events with two photons, missing transverse energy and isolated leptons will be from $WH \to \ell\nu\gamma\gamma$ and $t\bar{t}H$. The basic selection requires the presence of two energetic photons, missing transverse momentum, and one high energetic lepton. The main backgrounds for this channel are $t\bar{t}\gamma\gamma$, $W\gamma\gamma$ where the $W$ decays to $\ell\nu$ and $W\gamma \to e\nu\gamma$ where the other photon is radiated by the electron or is a fake photon from an additional jet.[4] This latter background is multiplied by a factor 2 to include the $W\gamma \to \mu\nu\gamma$ contribution. Another important background turns out to be $\gamma\gamma$ and $\gamma j$, when fake electrons or muons are reconstructed. ALPGEN was used to produce the diphoton background including a full simulation of the detector. The cross section was multiplied by a factor 1.4 to account approximately for the box diagram (40% of the contribution, see Table 1). For reducible backgrounds (68% contribution, see Table 8), the cross-section of the diphoton background is scaled by a factor 1.68. It is difficult to evaluate the accuracy of this approximation but this diphoton process is not the leading background. A multivariate analysis based on various kinematical variables would require very large Monte Carlo samples for the backgrounds. Here, a simple set of basic event selection criteria is chosen:

**IVa** The first selection criterion sets minima on the transverse momenta of the two reconstructed pho-

[4] Note that there is some double counting when both $W\gamma$ and $W\gamma\gamma$ are included.

tons. The cut requires the $p_{\mathrm{T}}$ of the most energetic photon to be higher than 60 GeV and the $p_{\mathrm{T}}$ of the second more energetic photon to be above 30 GeV. Events where one of the two photons is reconstructed in the crack region are rejected. The cross-sections in the first line of the table 14 are given in the mass range between 110 and 150 GeV.

**IVb** Requiring a cut that the transverse momentum of the most energetic isolated lepton (electron or muon) be higher than 30 GeV suppresses efficiently the diphoton and $W\gamma$ backgrounds. Only the electrons passing a tight electron identification cut are selected.

**IVc** The diphoton background can further be strongly suppressed by requiring missing transverse energy higher than 30 GeV.

**IVd** When an electron is reconstructed, events are removed in with either of the invariant masses of the electron and each of the photons is close to the $Z$ mass (between 80 and 100 GeV): this cut removes the events coming from the $(Z/\gamma^{*})+\gamma \rightarrow e^{+}e^{-}\gamma$ when an electron is reconstructed as a photon.

Table 14: Expected cross-sections (in fb) for the Higgs boson with missing transverse energy and lepton analysis for $m_H = 120$ GeV. Results are given after the application of cuts **IVa-IVc** (see Section 5.4) in the mass range $110 < m_{\gamma\gamma} < 150$ GeV. In the last row the expected cross-sections within a mass window of $m_{\gamma\gamma}$ of $\pm 2$ GeV around 120 GeV are given.

| Cut | $W^{\pm}H \rightarrow \ell\nu\gamma\gamma$ | $t\bar{t}H \rightarrow x\gamma\gamma$ | $W^{\pm}\gamma\gamma \rightarrow \ell\nu\gamma\gamma$ | $t\bar{t}\gamma\gamma$ | $b\bar{b}\gamma\gamma$ | $W^{\pm}\gamma \rightarrow e\nu\gamma$ | $(Z/\gamma^{*})+\gamma \rightarrow e^{+}e^{-}\gamma$ | $\gamma\gamma$ |
|---|---|---|---|---|---|---|---|---|
| | $\sigma$ (fb) | $\sigma$ (fb) | $\sigma$ (fb) | $\sigma$ (fb) | $\sigma$ (fb) | $\sigma$ (fb) | $\sigma$ (fb) | $\sigma$ (fb) |
| **IVa** | 0.328 | 0.467 | 0.509 | 2.53 | 3.78 | 18.0 | 10.28 | 2558 |
| **IVb** | 0.122 | 0.103 | 0.149 | 0.582 | 0.0098 | 0.406 | 2.60 | 0.644 |
| **IVc** | 0.091 | 0.086 | 0.097 | 0.474 | 0 | 0.263 | 0.076 | 0.091 |
| **IVd** | 0.084 | 0.077 | 0.090 | 0.419 | 0 | 0.143 | 0 | 0.091 |
| Mass Win. | 0.064 | 0.062 | 0.0092 | 0.042 | 0 | 0.014 | 0 | 0.010 |

Table 14 summarizes the expected cross-sections for the principal signal and backgrounds after the various cuts used in the analysis. The uncertainty in the background level, due to Monte Carlo statistics only, is estimated to be 10%. However, the background contribution in this analysis can have larger sources of systematic uncertainties. The resulting Higgs boson mass peak and background are shown in Fig. 9 after all cuts have been applied. The reconstructed mass resolution is 1.47 GeV.

## 5.5 Higgs boson plus missing transverse energy

This analysis is intended to select principally the $ZH \rightarrow \nu\nu\gamma\gamma$ events. In order to have an analysis independent from that of Section 5.4, only events which do not have a reconstructed lepton with $p_T >$ 30 GeV have been considered here. It should be noted that cases where the lepton is of low $p_T$ or did not pass the tight selection criterion are included. Thus the $WH$ signal contributes significantly. The main features are the presence of two energetic photons from the Higgs boson and a large transverse missing energy. The dominant backgrounds for this channel are the $t\bar{t}\gamma\gamma$, $Z\gamma\gamma$ and $W\gamma \rightarrow e\nu\gamma$ channels where, in the latter case, the electron can be mis-reconstructed as a photon or the second photon is either radiated by the electron or is a fake photon. Here, the diphoton background is multiplied by the same factors as in Section 5.4 and the $t\bar{t}\gamma\gamma$ by the factor 7.5 as explained in Section 2.2.

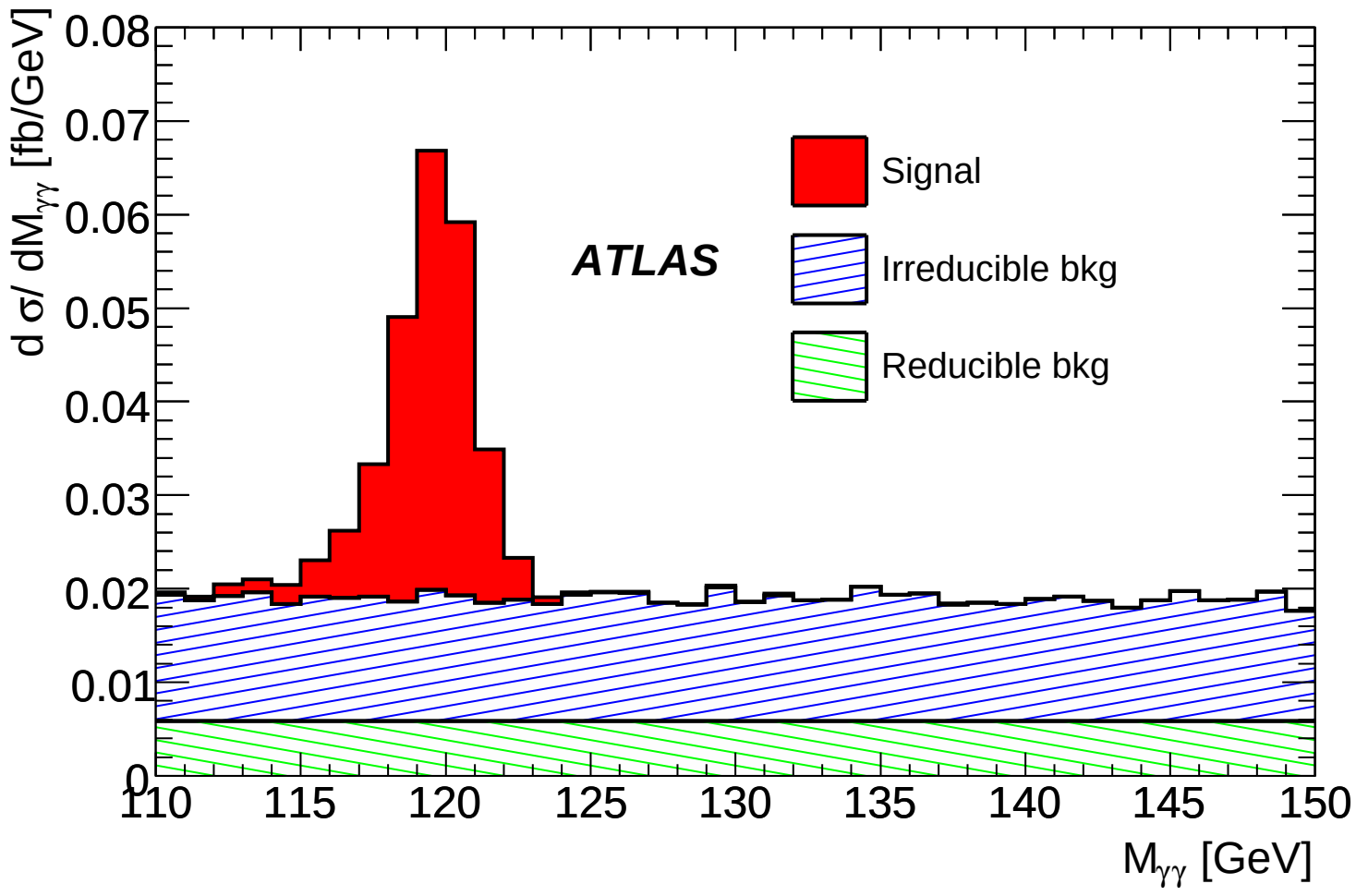


Figure 9: Expected distribution of the invariant mass of the two photons for the signals and main backgrounds after applying the analysis cuts for events having one lepton reconstructed in the final state. Due to a lack of MC statistics for the diphoton and the $W\gamma$ backgrounds, their expected distribution is approximated by showing an average of the number of events passing the analysis cuts in the $m_{\gamma\gamma}$ mass range shown.

**Va** As in Section 5.4, a cut on the transverse momentum of the most energetic photon above 60 GeV and a cut on the second more energetic photon $p_{\mathrm{T}}$ of 30 GeV are applied to suppress the diphoton background. Events where one of the two photons is reconstructed in the crack region are then removed.

**Vb** The selection is then based mostly on the requirement of high missing transverse momentum. A cut of $E_{\mathrm{T}}^{\mathrm{miss}} > 80$ GeV suppresses almost completely the $\gamma\gamma$ background while reducing the $W\gamma$ background by a factor 20 and the $ZH \to \nu\nu\gamma\gamma$ signal by a factor 2.

**Vc** In order to further suppress the $W\gamma$ background, where the electron is often reconstructed as a converted photon, events where either of the photons appears to have converted are rejected.

**Vd** At this point, because of potentially significant background from QCD events, difficult to evaluate, a cut requiring that the scalar sum of the $p_T$ of the jets in the event be larger than 150 GeV is imposed. It suppresses the contribution from the $t\bar{t}\gamma\gamma$ and $b\bar{b}\gamma\gamma$ backgrounds, as well as of the $t\bar{t}H$ signal.

Table 15 summarizes the expected cross-sections after the different cuts applied for this analysis for signal and backgrounds. The expected mass distributions of diphotons from the associated $W/Z$ plus Higgs boson and from the backgrounds are shown in Fig. 10, after the application of all cuts. To account for the $W\gamma \to \mu\nu\gamma$, the $W\gamma \to e\nu\gamma$ background has been multiplied by two in the figure although some double counting is introduced. The uncertainty in the background level, due to Monte Carlo statistics only, is estimated to be 15%. The reconstructed mass resolution is 1.31 GeV. This result is expected to be sensitive to uncertainties in the simulation and reconstruction of $E_{\mathrm{T}}^{\mathrm{miss}}$ tails.

Table 15: Expected cross-sections (in fb) for the Higgs boson with missing transverse energy analysis. Results are given after the application of cuts **Va-Vd** (see Section 5.5) in the mass range $110 < m_{\gamma\gamma} < 150$ GeV. In the last row the expected cross-sections within a mass window of $m_{\gamma\gamma}$ of $\pm 1.8$ GeV around 120 GeV are given.

| Cut | $ZH \to \nu\nu\gamma\gamma$ $\sigma$ (fb) | $WH \to \ell\nu\gamma\gamma$ $\sigma$ (fb) | $t\bar{t}H \to x\gamma\gamma$ $\sigma$ (fb) | $Z\gamma\gamma \to \nu\nu\gamma\gamma$ $\sigma$ (fb) | $W^{\pm}\gamma\gamma \to \ell\nu\gamma\gamma$ $\sigma$ (fb) | $t\bar{t}\gamma\gamma$ $\sigma$ (fb) | $b\bar{b}\gamma\gamma$ $\sigma$ (fb) | $W^{\pm}\gamma \to e\nu\gamma$ $\sigma$ (fb) | $\gamma\gamma$ $\sigma$ (fb) |
|---|---|---|---|---|---|---|---|---|---|
| **Va** | 0.115 | 0.207 | 0.364 | 0.325 | 0.360 | 1.95 | 3.77 | 17.55 | 2558 |
| **Vb** | 0.058 | 0.062 | 0.080 | 0.126 | 0.071 | 0.461 | 0.010 | 0.789 | 0.211 |
| **Vc** | 0.046 | 0.049 | 0.064 | 0.096 | 0.056 | 0.377 | 0.010 | 0.191 | 0.141 |
| **Vd** | 0.042 | 0.042 | 0.006 | 0.093 | 0.050 | 0.021 | 0.005 | 0.120 | 0.073 |
| Mass Win. | 0.034 | 0.033 | 0.0056 | 0.009 | 0.006 | 0.002 | 0.0005 | 0.012 | 0.007 |

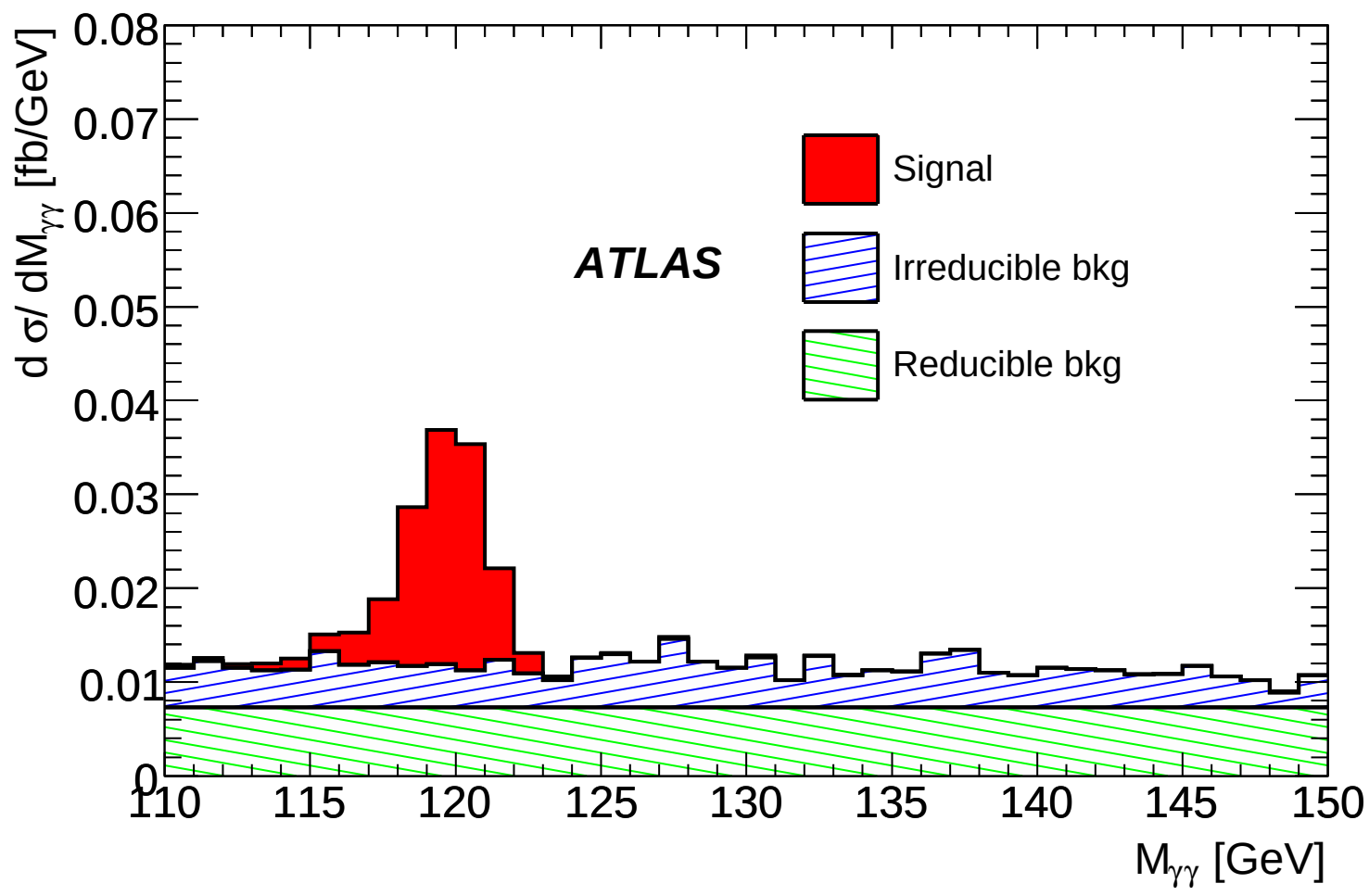


Figure 10: Expected distribution of the invariant mass of the two photons for the signals and main backgrounds after applying all the diphoton and $E_{\mathrm{T}}^{\mathrm{miss}}$ analysis cuts. Due to a lack of statistics for the diphoton and the $W\gamma$ backgrounds, their expected distribution is approximated by showing an average of the number of events passing the analysis cuts in the $m_{\gamma\gamma}$ mass range shown.

# 6 Maximum-likelihood fit

An unbinned extended multivariate maximum-likelihood fit to extract the $H \to \gamma\gamma$ signal and background event yields is performed. With respect to the cut analysis presented in Section 5, the fit takes advantage of further discrimination information from the kinematic and topological properties of $H \to \gamma\gamma$ decays. In addition to the diphoton invariant mass, $m_{\gamma\gamma}$, the transverse momentum of the Higgs boson, $P_{T,H}$, and the magnitude of the photon decay angle in the Higgs boson rest frame with respect to the Higgs boson lab flight direction, $|\cos\theta^{\star}|$, are included. To reduce the model dependence, the most relevant parameters describing the probability density functions (PDF) for the dominant background are freely varied and determined simultaneously with the event yields by the fit. Only the parametric shapes of the PDFs are obtained from Monte Carlo simulation. Sufficient background statistics in the sidebands must be retained to ensure that the floating shape parameters do not increase the statistical errors on the signal

yield. The $m_{\gamma\gamma}$ resolution depends on the diphotons' pseudorapidities, with core Gaussian widths varying between 1.2 and 3 GeV. Disjoint pseudorapidity regions (denoted *categories*), are defined such that the $m_{\gamma\gamma}$ resolution is similar for events belonging to the same category, and maximally different for events of different categories. The $m_{\gamma\gamma}$ resolution also depends on whether or not photons have converted at an early stage in the material in front of the calorimeter (see Section 3.3). $H \to \gamma\gamma$ events can have zero, one or two reconstructed conversions, which are split into corresponding categories. Following Section 5 disjoint categories are also introduced for events accompanied by zero, one and two or more jets, where the last category contains predominantly VBF events. The use of categories separates sub-populations of events with different properties and hence improves the accuracy of the likelihood model.

The signal and background samples used to build the likelihood model are selected with the criteria described in Section 5. The full detector simulation is used for signal processes, while the background processes are generated with the fast simulation. The trigger and photon identification efficiencies, and the misidentification in background events are parametrised using the full detector simulation. The probability of a photon to convert into an $e^+e^-$ pair with one or two tracks being reconstructed, and the energy resolution of the converted photon are parametrised using the full detector simulation.

## 6.1 Likelihood model

The probability density function, $P_i^c$, for an event $i$ in category $c$ is the weighted sum of the probability densities of all components, namely $P_i^c = N_H f_H^c P_{H,i}^c + \sum_{j=1}^{n_{\rm bkg}} N_{B_j}^c P_{B_j,i}^c$, where $N_H$ is the number of $H \to \gamma\gamma$ signal events determined by the fit,[5] $f_H^c$ is the fraction of signal events in category $c$ ($c = 1,\ldots,n_{\rm cat}$, note that $\sum_c f_H^c = 1$) taken from simulation, and $N_{B_j}^c$ is the number of background events of type $j$ ($j = 1,\ldots,n_{\rm bkg}$) found in category $c$, which is determined from the fit. The signal and background PDFs $P_{U,i}^c$, with $U = H, B_j$, are the products $P_{U,i}^c = \prod_{k=1}^{n_{\rm var}} p_U^c(x_{k,i})$ of the PDFs $p_U^c(x_{k,i})$ of the discriminating variables $x_{k,i}$, $k = 1,\ldots,n_{\rm var}$, used in the fit. The extended likelihood over all categories and events is given by $\mathscr{L} = \prod_{c=1}^{n_{\rm cat}} e^{-\overline{N}^c} \prod_{i=1}^{N^c} P_i^c$, where $\overline{N}^c$ ($N^c$) is the total number of events expected (observed) in category $c$ with $c = 1,\ldots,n_{\rm cat}$. The PDFs $P_{U,i}^c$ depend on parameters (coefficients) that may be freely varying in the fit. The parametrisations and hence the adjustable parameters differ between categories in general, but can also span over several categories.

## 6.2 Fit variables

The $H \to \gamma\gamma$ distribution of $m_{\gamma\gamma}$ forms a Gaussian peak with tails to lower values from photon energy losses before the calorimeter. It is well modeled by a *Crystal Ball* (CB) function [55]

$$p_H(m_{\gamma\gamma}) = N \cdot \begin{cases} \exp\left(-t^2/2\right), & \text{for } t > -\alpha, \\ (n/|\alpha|)^n \cdot \exp\left(-|\alpha|^2/2\right) \cdot (n/|\alpha| - |\alpha| - t)^{-n}, & \text{otherwise}, \end{cases} \tag{4}$$

where $t = (m_{\gamma\gamma} - m_H - \delta_{m_H})/\sigma(m_{\gamma\gamma})$, $N$ is a normalisation parameter, $m_H$ is the Higgs boson mass, $\delta_{m_H}$ is a category dependent offset, $\sigma$ represents the diphoton invariant mass resolution, and where $n$ and $\alpha$ parametrise the non-Gaussian tail. To catch outliers an additional broad tail Gaussian is added to Equation 4, which is, however, only relevant for events falling into the "bad" $m_{\gamma\gamma}$ category (see Section 6.3). Within a sufficiently narrow mass window, the background distribution of $m_{\gamma\gamma}$ forms an exponential tail described by a single slope parameter $\xi$.

Because of the large available statistics for simulated signal samples, shape uncertainties due to deficiencies in the functional description are negligible. Systematic effects are due to simulation inaccuracies and are discussed in Section 7. A high-fidelity description of the background PDFs is mandatory because

[5] Throughout this section the subscript $S$ is idenfitied with the $H \to \gamma\gamma$ signal.

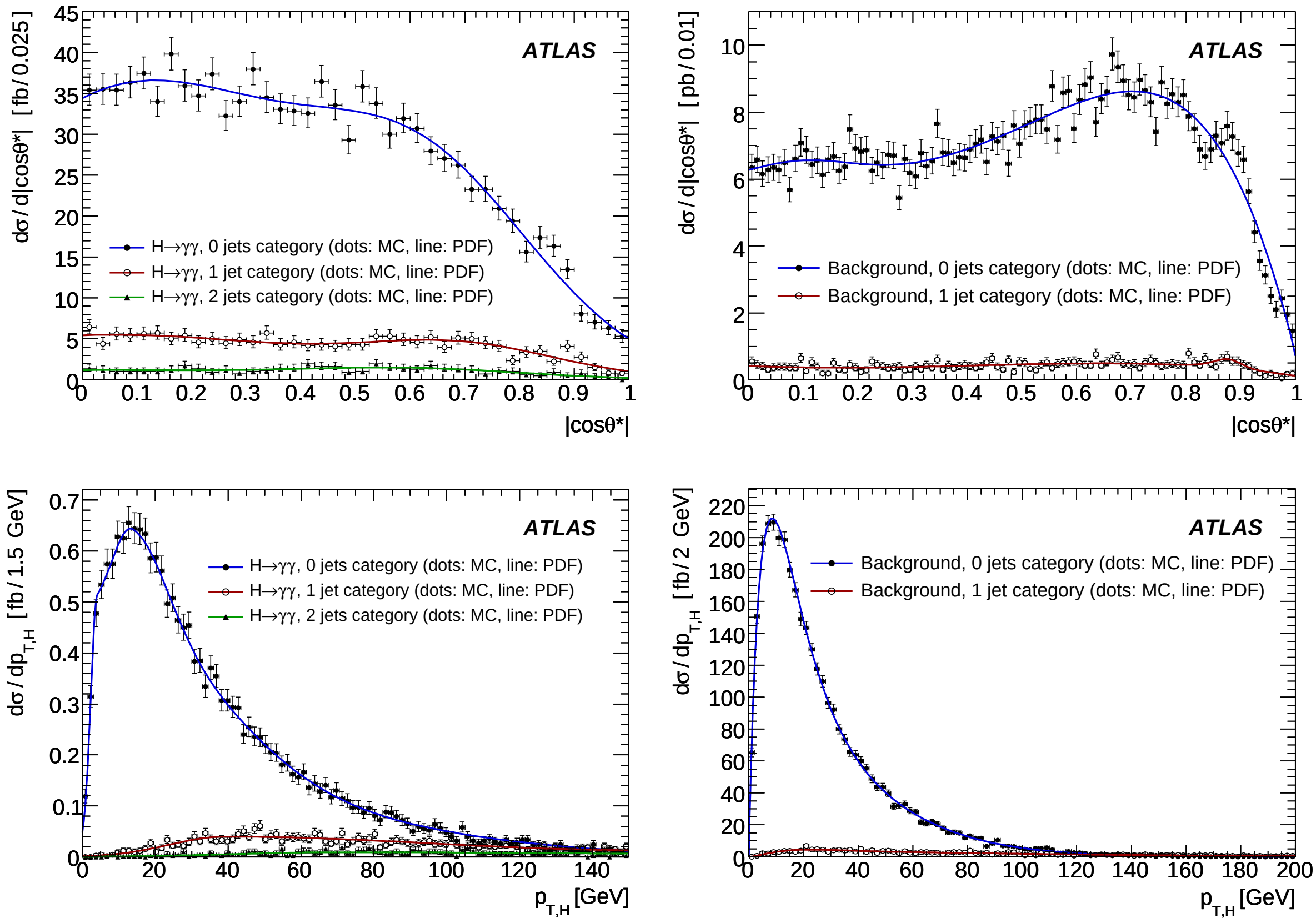


Figure 11: Signal (left) and background (right) distributions of the Higgs boson decay angle, $|\cos\theta^{\star}|$ (top), and the diphoton transverse momentum (bottom) for events with zero jets (full dots), one jet (open circles) and VBF topology (full triangles, not shown for background because of a too low relative cross-section). The corresponding PDF parametrisations are overlaid (see text).

the large majority of the events entering the fit are of that type. It is achieved by flexible parametrisations with a sufficient number of parameters determined by the fit, ensuring a stable fit result with respect to shape redefinitions.

The distribution of $|\cos\theta^{\star}|$ for a scalar Higgs boson is uniform. However, acceptance effects, primarily from the minimum $p_{\mathrm{T}}$ requirements for the photons (See Section 5.1), suppress $|\cos\theta^{\star}|$ values towards one, where the photons are collinear with the Higgs boson lab frame momentum. The empirical signal PDF is interpolated by a double Gaussian function. The phase space for background events from $t$-channel graphs and quark or gluon fragmentation at NLO is enhanced for photons collinear with the diphoton lab momentum, so that the background $|\cos\theta^{\star}|$ distribution exhibits some clustering towards large values. Acceptance suppression competes, however, with this enhancement thus reducing the discrimination power of the variable. It is found that the $|\cos\theta^{\star}|$ distributions differ significantly between the $\gamma\gamma$, $\gamma j$ and $jj$ backgrounds, with stronger enhancements at large $|\cos\theta^{\star}|$ values for the backgrounds originating from jet misidentification. The inclusive shape of these backgrounds is parametrised by the sum of a positively defined third order polynomial and two Gaussian functions. The inclusive signal and background distributions of $|\cos\theta^{\star}|$ for events with and without jets are shown in Fig. 11.

The Higgs boson transverse momentum exhibits a strong rise at low values and a long exponential tail beyond the maximum. The distribution is fitted by a sum of two bifurcated Gaussian functions (distributions where below and above the center half Gaussian distributions with different widths are used) and one symmetric Gaussian. The diphoton transverse momentum distribution for background is

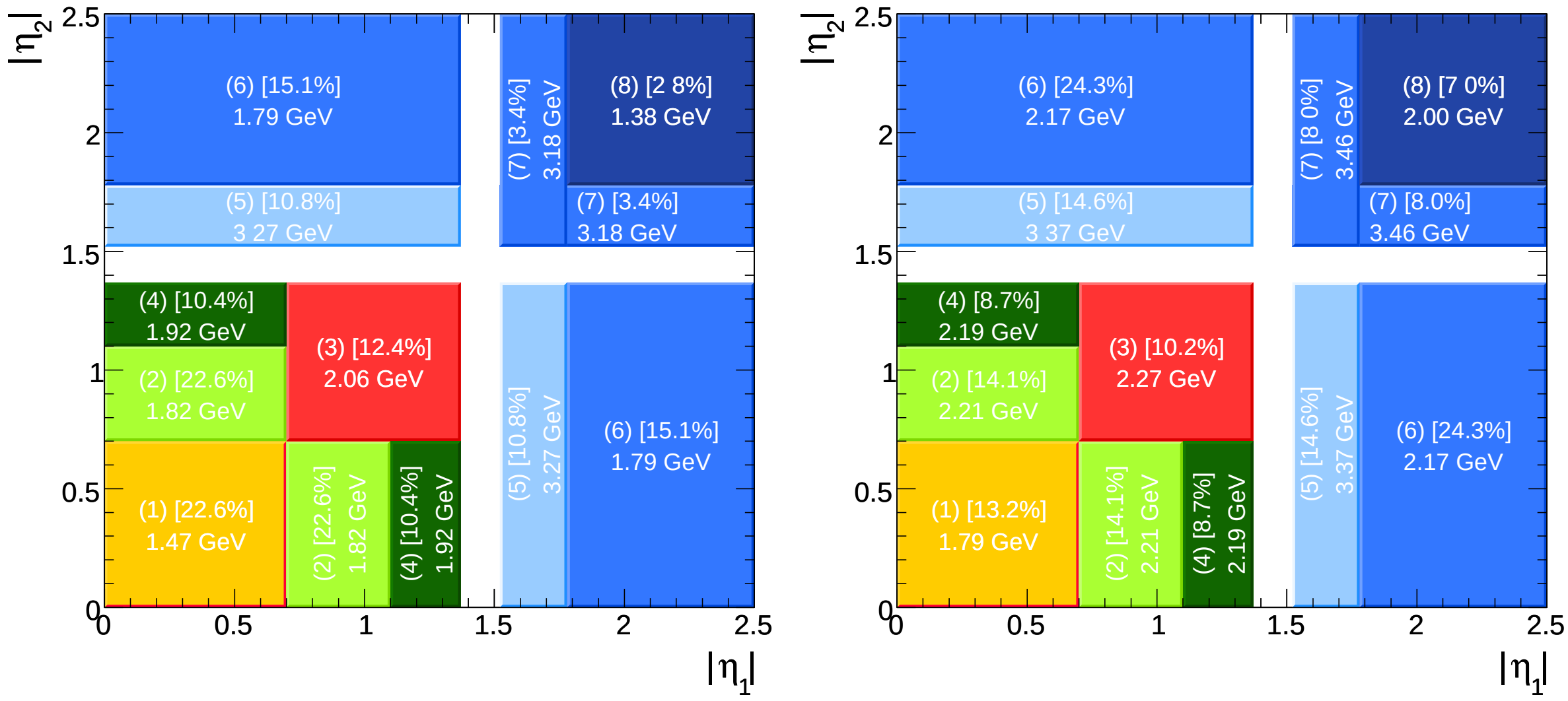


Figure 12: Regions of photon pseudorapidities with different invariant mass resolutions for unconverted photons (left) and at least one converted photon (right). The text per box denotes the region number, the percentage of events occurring in the region and the RMS of the diphoton invariant mass for $H \to \gamma\gamma$ events. To simplify the likelihood model, events with photons in regions (1) and (8) are merged and represent category "good" (signal fraction 24%), events in regions (2), (3), (4) and (6) correspond to category "medium" (60%), and regions (5) and (7) are "bad" (15%).

softer than that for signal and can be described by the sum of three exponential polynomials (see Fig. 11).

### 6.3 Fit categories

Eight photon pseudorapidity regions with different diphoton invariant mass resolution for $H \to \gamma\gamma$ events are identified. They are illustrated in Fig. 12 for unconverted photon pairs (left plot) and events with at least one photon conversion (right plot). The regions are chosen to be symmetric with respect to an interchange of the photons. The crack region is excluded from the photon selection. The $m_{\gamma\gamma}$ RMS values vary between 1.3 GeV in the centre and large-$\eta$ regions to 3.1 GeV in the regions closely beyond the crack. To simplify the likelihood model, the eight regions are merged into three categories that are distinguished in the fit (see Fig. 12).

Two categories are introduced to separate events without and with at least one photon conversion to take account of the worse resolution of the latter events. Photon conversions reconstructed with one or two tracks are not explicitly distinguished. Additional categories are introduced for Higgs boson production with zero, one, and two or more accompanying jets, using the requirements described in Sections 5.2 and 5.3. No separate categories are introduced in the present analysis to distinguish Higgs boson production in association with a $W$ and $Z$ boson or a $t\bar{t}$ quark pair; these are included in the previous selections. Separating them would add some power, but would increase the complexity.

### 6.4 Correlations

The likelihood product used to derive the event PDF $P_U^c$ ignores correlations between the discriminating variables $x_k$. The classification in categories of events with distinct properties improves the accuracy of this assumption. The remaining (positive) linear correlation coefficients is lower than 7% (10%) for signal (background) events among the three fit variables tolerable.

Table 16: $H \to \gamma\gamma$ discovery potential for a Higgs boson mass of 120 GeV, various likelihood setups and a simulated integrated luminosity of $10\,\mathrm{fb}^{-1}$. The left column gives the discriminating variables used in the fit, the second column the categories, the third (fifth) column the average $\Delta\ln\mathscr{L}$ and its statistical uncertainty as derived from the toy MC samples, and the fourth (sixth) column quotes the estimated Gaussian signal significance in units of one standard deviation. Due to the background dominance, the variance of the significance in the toy experiments is approximately one. The fits have been performed with fixed Higgs boson mass (Higgs boson mass floating within $[112, 128]$ GeV). The significance for floating Higgs boson mass has been obtained with toy MC simulation (the $\Delta\ln\mathscr{L}$ cannot be directly interpreted in terms of significance and is given for completeness only). Due to the large number of required toy MC fits it has not been computed for the most involved fit with highest expected significance.

| Fit variables | Categories | Higgs boson mass fixed | | Higgs boson mass floating | |
|---|---|---|---|---|---|
| | | $\langle\Delta\ln\mathscr{L}\rangle$ | Significance $[\sigma]$ | $\langle\Delta\ln\mathscr{L}\rangle$ | Significance $[\sigma]$ |
| $m_{\gamma\gamma}$ | – | $2.67 \pm 0.04$ | $2.31 \pm 0.02$ | $3.54 \pm 0.05$ | $1.44 \pm 0.02$ |
| $m_{\gamma\gamma}$ | $\eta$ | $3.18 \pm 0.05$ | $2.52 \pm 0.02$ | – | – |
| $m_{\gamma\gamma}$ | $\eta$, Conversions | $3.32 \pm 0.05$ | $2.58 \pm 0.02$ | – | – |
| $m_{\gamma\gamma}$ | $\eta$, Conversions, Jets | $5.99 \pm 0.07$ | $3.46 \pm 0.02$ | $6.66 \pm 0.07$ | $2.64 \pm 0.02$ |
| $m_{\gamma\gamma}$, $\lvert\cos\theta^\star\rvert$ | $\eta$, Conversions, Jets | $7.33 \pm 0.08$ | $3.83 \pm 0.02$ | – | – |
| $m_{\gamma\gamma}$, $P_{T,H}$ | $\eta$, Conversions, Jets | $7.03 \pm 0.08$ | $3.75 \pm 0.02$ | – | – |
| $m_{\gamma\gamma}$, $P_{T,H}$, $\lvert\cos\theta^\star\rvert$ | $\eta$, Conversions, Jets | $8.49 \pm 0.08$ | $4.12 \pm 0.02$ | $9.25 \pm 0.09$ | – |

### 6.5 Fit performance

Studies are performed involving large samples of toy MC simulation to assess the discovery potential for fits using likelihood models of increasing complexity.[6] The abundance of signal and background events used in these fits is tuned to the NLO expectation for $10\,\mathrm{fb}^{-1}$ of integrated luminosity, taking into account the trigger and reconstruction acceptance.

The results for fits with fixed and floating Higgs boson mass (the latter only done for the $m_{\gamma\gamma}$-only and the full fits) are summarised in Table 16. For each toy MC sample we perform two fits, one with floating signal yield and another with zero signal to test the background-only hypothesis. The log-likelihood difference, $\Delta\ln\mathscr{L}$, found in these fits, estimates the false discovery probability (p-value). The $\langle\Delta\ln\mathscr{L}\rangle$ values given in Table 16 are obtained from Gaussian fits to well-behaved pull distributions. For fixed Higgs boson mass, the signal significance in terms of $\sigma$ can be approximated by the quantity $\sqrt{-2\Delta\ln\mathscr{L}}$. For floating Higgs boson mass, the extra degree of freedom in the fit yields a higher value of $\langle\Delta\ln\mathscr{L}\rangle$. However in this case the p-value and significance must be evaluated with toy MC simulation of background-only samples. As expected, the obtained significances are lower than in the fixed mass case, in spite of the higher $\langle\Delta\ln\mathscr{L}\rangle$ values.

## 7 Discovery potential

This Section reports on the potential for the discovery of a Higgs boson in the mass range $120 < m_H < 140$ GeV using the event counting computation and the maximum likelihood fit formalism (see Sec-

[6] Although required for real data the evaluation of goodness-of-fit estimators is not discussed here, because the generated toy data is intrinsically consistent with the underlying model.

tion 6). The signal significance is evaluated for each of the analyses presented in Section 5. The discovery reach is given for a combined analysis including the utilization of additional discriminating variables. The last signifies the maximum discovery potential for a Standard Model Higgs boson in the mass range specified above.

Table 17: Expected cross-sections (in fb) of signal (S) and background (B) for the different analyses presented in Section 5 within a mass window of $\pm 1.4\sigma_{\gamma\gamma}$ as a function of the Higgs boson mass (in GeV).

| | Inclusive | | $H+1$jet | | $H+2$jets | | $H$+$E_{\mathrm{T}}^{\mathrm{miss}}$ +$\ell$ | | $H$+$E_{\mathrm{T}}^{\mathrm{miss}}$ | |
|---|---|---|---|---|---|---|---|---|---|---|
| $m_H$ | S | B | S | B | S | B | S | B | S | B |
| 120 | 25.4 | 947 | 4.0 | 49 | 0.97 | 1.95 | 0.134 | 0.077 | 0.075 | 0.037 |
| 130 | 24.1 | 755 | 4.3 | 47 | 0.96 | 1.72 | 0.112 | 0.076 | 0.063 | 0.037 |
| 140 | 19.3 | 610 | 3.9 | 46 | 0.81 | 1.72 | 0.079 | 0.076 | 0.045 | 0.036 |

Table 17 shows the signal and background effective cross-sections for the different analyses presented in Section 5 as a function of the Higgs boson mass.[7] Table 18 displays the corresponding expected signal significances for $10\,\mathrm{fb}^{-1}$ of integrated luminosity. The first sub-column under each analysis shows the signal significance based on event counting, $\sigma(S,B)$, where $S$ and $B$ correspond to the number of signal and background events in a mass window of $\pm 1.4\,\sigma_{\gamma\gamma}$ around $m_H$, respectively, and where $\sigma_{\gamma\gamma}$ is the mass resolution (in the no pileup case reported in Table 5).[8] The values of $\sigma(S,B)$ reported in Table 18 can be compared with earlier studies performed by the ATLAS collaboration [2–4, 8, 9]. The inclusion of pileup decreases the event counting signal significance by at most 10%.

The signal significances reported in the second and third sub-column under each analysis, $\sigma_{1D}^{Fix}$ and $\sigma_{1D}^{Float}$, are obtained by means of a one dimensional fit using the diphoton invariant mass as a discriminating variable by letting the Higgs boson mass fixed and floated, respectively.[9] Each analysis is treated independently from each other. In these fits the Higgs boson mass is let free in the range $110 < m_H < 140$ GeV (except for $m_H = 140$ GeV for which the range was set to $120 < m_H < 150$ GeV) to take into account for the coverage of the analysis.[10] The results in Table 18 do not take into account the event overlap among the three analyses. The last column reports the event counting signal significance of the three analyses combined, taking into account the event overlap.

Table 18 illustrates that the inclusive search for a diphoton resonance is the most sensitive one for a search of a Standard Model Higgs boson. As reported in Section 2, the inclusive analysis is evaluated using QCD corrections for both signal and background processes. This is not the case for the $H+1$jet and other analyses. The discovery potential of $H+1$jet analysis could be further enhanced if QCD NLO corrections were applied on both signal and reducible backgrounds [56, 57].

Table 19 shows two fit-based signal significances compared to $\sigma_{1D}^{Fix}$ and $\sigma_{1D}^{Float}$ reported in Table 18. The values of $\sigma_{C_1 3D}^{Fix}$ and $\sigma_{C_1 3D}^{Float}$ correspond to the signal significance computed by means of a three dimensional fit, including $m_{\gamma\gamma}, P_{T,H}$ and $|\cos\theta^{\star}|$ (see Section 6.2) by means of fixing and floating the mass, respectively. At this stage the event classification according to $|\eta|$ regions is used (see Section 6.3).[11]

[7]The contribution from Drell-Yan is computed for the inclusive analysis only.

[8]The event counting significance for the inclusive and $H+1$jet analyses are approximated by $S/\sqrt{B}$. For the rest of the analyses a Poisson-based computation is used due to the small expected number of background events for $10\,\mathrm{fb}^{-1}$ of integrated luminosity.

[9]For the sake of simplicity, the fit-based signal significance does not include the Drell-Yan contribution.

[10]The range of the fit to the background is set to $110 < m_H < 150$ GeV.

[11]For the final results presented in this section, a simplified fitting model as well as a more conservative classification are used with respect to those considered for Table 16. In particular the diphoton category with two jets is not split into $\eta$ categories

Table 18: Signal significances (expressed in terms of Gaussian sigmas) for a Standard Model Higgs boson as a function of the mass (in GeV) using the different analyses reported in Sections 5.1-5.3 for 10 fb$^{-1}$ of integrated luminosity. Results are reported in terms of the signal significance based on event counting, $\sigma(S,B)$, and a fit-based signal significance, $\sigma_{1D}^{Fix}$ and $\sigma_{1D}^{Float}$ (see text).

| | Inclusive (with K-factors) | | | $H+1$jet (no K-factors) | | | $H+2$jet (no K-factors) | | | Combined |
|---|---|---|---|---|---|---|---|---|---|---|
| $m_H$ | $\sigma(S,B)$ | $\sigma_{1D}^{Fix}$ | $\sigma_{1D}^{Float}$ | $\sigma(S,B)$ | $\sigma_{1D}^{Fix}$ | $\sigma_{1D}^{Float}$ | $\sigma(S,B)$ | $\sigma_{1D}^{Fix}$ | $\sigma_{1D}^{Float}$ | $\sigma(S,B)$ |
| 120 | 2.6 | 2.4 | 1.5 | 1.8 | 1.8 | 1.3 | 1.9 | 2.0 | 1.1 | 3.3 |
| 130 | 2.8 | 2.7 | 1.8 | 2.0 | 2.1 | 1.6 | 2.1 | 2.1 | 1.2 | 3.5 |
| 140 | 2.5 | 2.2 | 1.3 | 1.8 | 1.7 | 1.2 | 1.7 | 2.0 | 1.0 | 3.0 |

Table 19: Signal significances (expressed in terms of Gaussian $\sigma$s) for a Standard Model Higgs boson as a function of the mass for 10 fb$^{-1}$ of integrated luminosity. Different fit-based approaches are used. The significances, $\sigma_{1D}^{Fix}$, $\sigma_{C_13D}^{Fix}$ and $\sigma_{C_23D}^{Fix}$ correspond to a one dimensional fit, to a three dimensional fit using the first type of event classification and to a three dimensional fit (see Section 6.2) using all classifications considered in Section 6.3, respectively using a fix Higgs boson mass (see text). The significances $\sigma_{1D}^{Float}$, $\sigma_{C_13D}^{Float}$ and $\sigma_{C_23D}^{Float}$, correspond to fit based results obtained with a floating Higgs boson mass.

| $m_H$[ GeV] | $\sigma_{1D}^{Fix}$ | $\sigma_{1D}^{Float}$ | $\sigma_{C_13D}^{Fix}$ | $\sigma_{C_13D}^{Float}$ | $\sigma_{C_23D}^{Fix}$ | $\sigma_{C_23D}^{Float}$ |
|---|---|---|---|---|---|---|
| 120 | 2.4 | 1.5 | 3.1 | 2.1 | 3.6 | 2.8 |
| 130 | 2.7 | 1.8 | 3.4 | 2.4 | 4.2 | 3.4 |
| 140 | 2.2 | 1.3 | 3.2 | 2.2 | 4.0 | 3.2 |

The values of $\sigma_{C_23D}^{Fix}$ and $\sigma_{C_23D}^{Float}$ correspond to the maximum achievable sensitivity of all the analyses reported in Section 5 combined. In addition to the procedure followed to obtain $\sigma_{C_13D}^{Fix}$ and $\sigma_{C_13D}^{Float}$, a classification of events according to the presence of hadronic jets is used.

Figure 13 displays a summary of the expected signal significance for the inclusive and final combined analysis for 10 fb$^{-1}$ of integrated luminosity as a function of the Higgs boson mass. The solid circles correspond to the sensitivity of the inclusive analysis reported in Section 5.1 using event counting with background and signal rates assumed. The solid triangles linked with solid and dashed lines correspond to the sensitivity of the inclusive analysis by means of one dimensional fits, with a fixed and floating Higgs boson mass, respectively. The solid squares linked with solid and dashed lines correspond to the values of $\sigma_{C_13D}^{Fix}$ and $\sigma_{C_13D}^{Float}$ given in Table 19, respectively.

The stability of the fits are checked against changes in the composition of the background. The values of $\sigma_{1D}^{Fix}$ for the inclusive analysis are recomputed by increasing and decreasing the reducible background by a factor of two. The fits are redone with the same functional forms as in the nominal analysis and the results are consistent with the expectations obtained by using $S/\sqrt{B}$.

The degradation of the Higgs boson discovery sensitivity due to systematic uncertainties is considered. Various sources of systematic errors are evaluated. The Higgs boson mass resolution has significant impact on the sensitivity. A large degradation of the Higgs boson mass resolution is chosen by the addition of a 1 % constant term in the photon energy resolution. The impact of Higgs boson mass resolution is evaluated for the inclusive analysis by means of one-dimensional fits with a fixed Higgs boson mass.

and the conversion categories are not used.

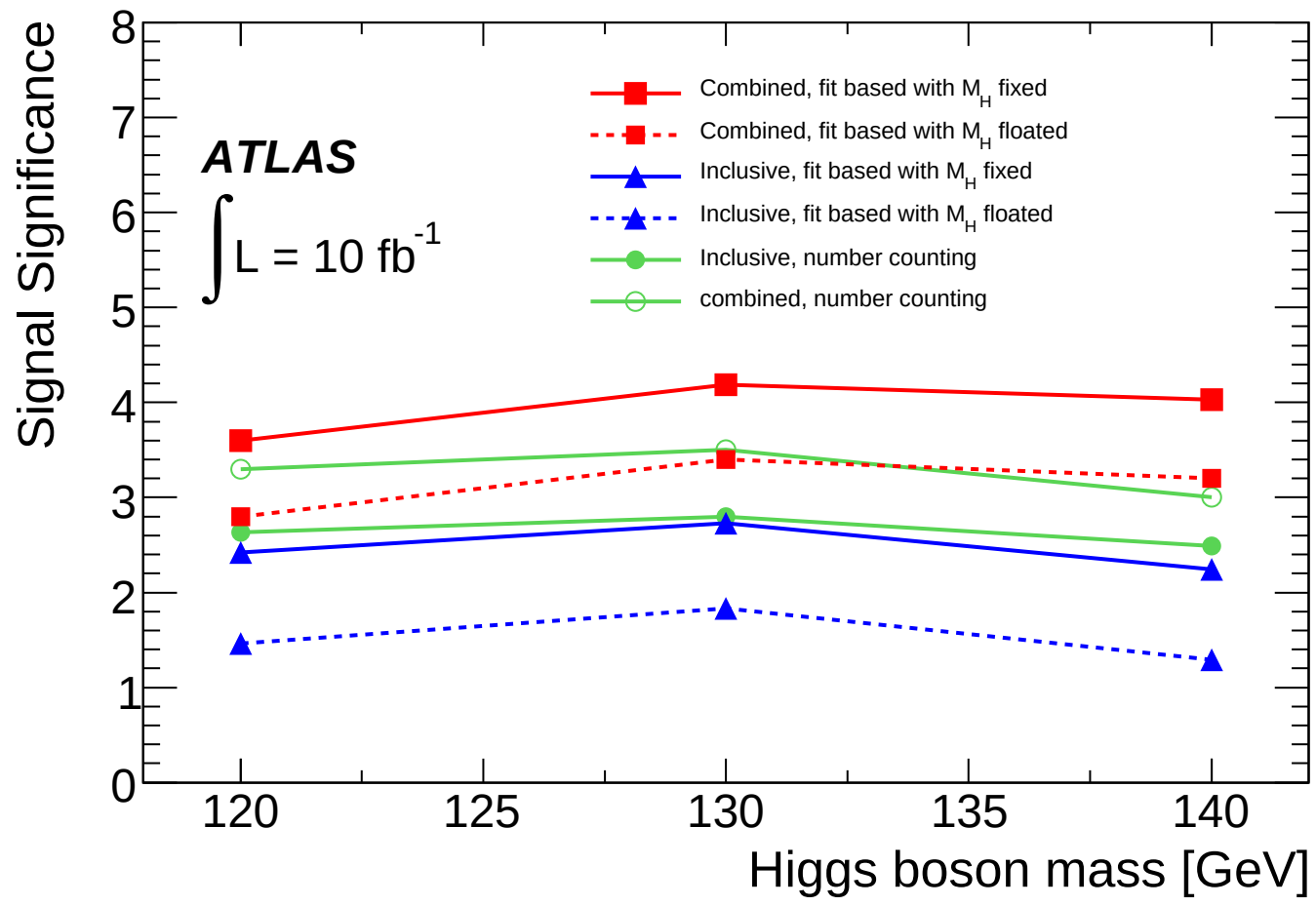


Figure 13: Expected signal significance for a Higgs boson using the $H \rightarrow \gamma\gamma$ decay for $10\,\text{fb}^{-1}$ of integrated luminosity as a function of the mass. The solid circles correspond to the sensitivity of the inclusive analysis reported in Section 5.1 using event counting. The open circles display the event counting significance when the Higgs boson plus jet analyses (see Sections 5.2 and 5.3) are included. The solid triangles linked with solid and dashed lines correspond to the sensitivity of the inclusive analysis by means of one dimensional fits, with a fixed and floating Higgs boson mass, respectively. The solid squares linked with solid and dashed lines correspond to the maximum sensitivity that can be attained with a combined analysis (see text and Table 19).

If the resolution is increased both in toy Monte Carlo experiments and in the fitting function a decrease of 8% in the signal sensitivity is observed. If, however, the resolution is only degraded in toy Monte Carlo experiments, so that the fitting model does not accommodate the Higgs boson mass resolution degradation, then the effect is a 12% reduction in sensitivity.

The systematic uncertainty due to $P_{T,H}$ has been estimated by using PYTHIA signal events, reweighted to NLO using ResBos, and fitted with the nominal model based on MC@NLO event simulation. This inconsistency between generation and fit reduces the significance of the full fit (including all categories and the variables $m_{\gamma\gamma}$, $P_{T,H}$, and $|\cos\theta^{\star}|$, see Table 16) by 5%.

The sensitivity of the associated production channels has been studied separately for the diphoton + $E_{\mathrm{T}}^{\mathrm{miss}}$ + lepton and for the diphoton + $E_{\mathrm{T}}^{\mathrm{miss}}$ analyses (see Sections 5.4-5.5). With 30 $\text{fb}^{-1}$ these channels have the potential to contribute to the overall discovery signal of a SM Higgs boson, but we choose not to report the value of the signal significance because the systematic uncertainties on the background normalization are large given the present status of simulations.

## 8 Conclusions

The feasibility of the search for a Standard Model Higgs boson in the $H \rightarrow \gamma\gamma$ decay with the ATLAS detector at the LHC has been presented. The detector performance issues relevant to the search have been evaluated using a full detector simulation. Triggering effects and the impact of pile-up are also reported. A signal significance based on event counting of 2.6 (4.6) can be obtained with 10 (30) $\text{fb}^{-1}$ of integrated luminosity for $m_H = 120$ GeV in the case of inclusive analysis. Despite the slight degradation with respect to previous studies, the feasibility of the search for a Standard Model Higgs boson in the $H \rightarrow \gamma\gamma$ decay is confirmed. In addition to the inclusive analysis the search for diphotons in association

with jets is considered. The addition of these channels enhances the event counting signal significance to 3.3 (5.7) with 10 (30) fb$^{-1}$ of integrated luminosity for $m_H = 120$ GeV. The expected sensitivity can be enhanced by means of an unbinned maximum-likelihood fit dividing the data sample into categories depending on the event topology and exploiting a number of discriminating variables: the increase in the discovery potential increases up to 3.6 (2.8) in the case of fixed (floating) mass fit for 10 fb$^{-1}$ of integrated luminosity.
The search for $H \to \gamma\gamma$ in association with $Z/W$ or $t\bar{t}$ has been addressed, indicating that a good signal-to-background ratio can be achieved. However, the uncertainty on the background contribution in these analyses is large given the present status of simulations.

# Search for the Standard Model $H \rightarrow ZZ^{*} \rightarrow 4l$

**Abstract**
The Standard Model Higgs boson discovery potential through its observation in the 4-lepton (electron and muon) final state using the ATLAS detector is investigated. Fully simulated signal and background samples were produced with the latest simulation of the ATLAS detector. The samples were subsequently digitized and reconstructed using the ATLAS offline software. The analysis performance dependence on kinematic, lepton reconstruction and isolation cuts is studied for Higgs boson masses ranging from 120 to 600 GeV. The statistical and systematic uncertainties on the background estimation are evaluated and their impact on the Higgs boson discovery potential and exclusion limits is discussed.

## 1 Introduction

The search for the Standard Model Higgs boson is a major goal of the Large Hadron Collider (LHC). The first proton-proton LHC data at 14 TeV center of mass energy are expected in 2009. The Higgs boson mass is a free parameter in the Standard Model, however there is strong expectation motivated by precision electroweak data [1] and direct searches [2] that a low mass Higgs boson (114.4 – 199 GeV, 95% confidence level) should be discovered at the LHC. The experimentally cleanest signature for the discovery of the Higgs boson is its "golden" decay to four leptons (electrons and muons): $H \rightarrow ZZ \rightarrow 4\ell$. The excellent energy resolution and linearity of the reconstructed electrons and muons leads to a narrow 4-lepton invariant mass peak on top of a smooth background. The expected signal to background ratio after all experimental cuts depends on the Higgs boson mass itself. The major component of the background consists of irreducible $ZZ \rightarrow 4\ell$ decays. The most challenging mass region is between 120-150 GeV where one of the $Z$ bosons is off-shell giving low transverse momentum leptons. In this region backgrounds from $Zb\bar{b} \rightarrow 4\ell$ and $t\bar{t} \rightarrow 4\ell$ are important and require tight lepton isolation cuts to keep their contribution well below the $ZZ$ continuum.

## 2 Detector simulation, Monte-Carlo samples, trigger and event reconstruction

In this section a brief summary of the detector simulation is presented and the Monte-Carlo (MC) samples used in the analysis are described. This includes an outline of the various cross-sections and the corresponding theoretical uncertainties. Finally a brief description of the electron and muon trigger and offline reconstruction is presented.

### 2.1 Detector simulation and Monte-Carlo samples

The ATLAS detector is simulated by the GEANT4 [3] software. Simulation, digitization and reconstruction are all performed within the ATLAS software framework ATHENA. The set of $H \rightarrow 4\ell$ samples used in this analysis covers the mass range from 120 to 600 GeV. Simulation of pileup, cavern background and minimum bias events is performed by mixing them with the Higgs boson signal at digitization level [4], [5]. An instantaneous constant luminosity of $10^{33}$ cm$^{-2}$ s$^{-1}$ is assumed. The cavern background consists of thermalized slow neutrons and low energy photons escaping the calorimeters [6]. The expected level of cavern background is increased by an overall "safety factor": in this analysis a

safety factor of 5 is used.
The H$\rightarrow 4\ell$ analysis is sensitive to uncertainties in the knowledge of the material distribution in ATLAS, to distortions of the magnetic fields, and to the accuracy of the Inner Detector (ID) and of the Muon Spectrometer (MS) alignment. Some of these uncertainties have been taken into account by using geometrical layouts including extra material. The layout used in this analysis includes additional material in the ID, and between the barrel presampler and strips and barrel cryostat upstream and downstream the calorimeter. The calibration of the LAr Electromagnetic Calorimeter (LAr EMC) is based on the nominal geometry without this extra material and its inclusion in our study provides a realistic systematic effect in the analysis. A detailed description of the layout used in simulation and reconstruction can be found in [7]. In the analysis presented in this note, uncertainties due to misalignment corrections are not included: the same misaligned layout is used both in simulation and reconstruction.

The Higgs boson signal samples were generated exclusively by PYTHIA [8] (version 6.3 for $m_H =$ 130 GeV and version 6.4 for the rest of the masses), while for the background samples various event generators were used. For the signal, PYTHIA calculates the cross-sections in leading order (LO) taking into account both gluon and vector boson fusion (VBF) diagrams. Next-to-leading order (NLO) effects are considered by scaling the total PYTHIA cross-sections [9]. During generation, a 4-lepton filter was applied to the samples, requiring 4 true leptons with $p_T > 5$ GeV/c within $|\eta| < 2.7$. The filter acceptance and the cross-section as a function of the Higgs boson mass is given in Table 1. The number of available MC events are shown in the last column.

| Process | $\sigma_{LO} \cdot BR\ [fb]$ | $\sigma_{NLO} \cdot BR\ [fb]$ | Filter acc. | Events |
|---|---|---|---|---|
| $H[120] \rightarrow 4l$ | 1.68 | 2.81 | 0.584 | 40K |
| $H[130] \rightarrow 4l$ | 3.76 | 6.25 | 0.633 | 40K |
| $H[140] \rightarrow 4l$ | 5.81 | 9.72 | 0.662 | 40K |
| $H[150] \rightarrow 4l$ | 6.37 | 10.56 | 0.685 | 10K |
| $H[160] \rightarrow 4l$ | 2.99 | 4.94 | 0.704 | 40K |
| $H[165] \rightarrow 4l$ | 1.38 | 2.29 | 0.712 | 40K |
| $H[180] \rightarrow 4l$ | 3.25 | 5.38 | 0.733 | 40K |
| $H[200] \rightarrow 4l$ | 12.39 | 20.53 | 0.753 | 50K |
| $H[300] \rightarrow 4l$ | 7.65 | 13.32 | 0.782 | 10K |
| $H[400] \rightarrow 4l$ | 6.07 | 10.78 | 0.814 | 40K |
| $H[500] \rightarrow 4l$ | 2.98 | 5.12 | 0.842 | 40K |
| $H[600] \rightarrow 4l$ | 1.53 | 2.53 | 0.853 | 40K |

Table 1: Monte Carlo signal data samples, 4-lepton $(e, \mu)$ filter acceptance, LO and NLO cross-sections, and number of events used in the analysis as a function of the Higgs boson mass in GeV (reported in square brackets). The cross-sections in the table include the branching ratio of the Higgs boson to $ZZ^*$ and $Z\rightarrow$ll, l=e,$\mu$.

### 2.1.1 Corrections to leading order cross-sections for background processes

The backgrounds considered in this analysis, together with their cross-sections and K-factors, defined as the ratio between the NLO and the LO cross-sections, are listed in Table 2. For the generation of these samples several MC generators are employed. The $t\bar{t}$ background was generated using MC@NLO [10]. The QCD $ZZ$ was generated with PYTHIA6.3, the $Zb\bar{b}$ background with AcerMC3.1 [11], and the $WZ$ background with HERWIG 6.5 [12] interfaced to Jimmy [13] for simulation of the underlying event. The cross-sections listed in Table 2 are all at LO (except for the $t\bar{t}$ which is at NLO), and they do not include

the lepton filter efficiency. The following corrections are applied to the cross-section of each process to correct for effects of subprocesses not originally included in the generators:

- an additional 30% is applied to the QCD $ZZ$ background cross-section to account for the missing quark box diagram in PYTHIA;
- 8.5 pb are added to the $Zb\bar{b}$ cross-section to account for the $qq \rightarrow Zb\bar{b}$ diagrams that are not included in the generation.

The last column in Table 2 shows the events available for each sample. Each $Z$ boson in the QCD $qq \rightarrow ZZ$ irreducible background is forced to decay into lepton pairs (including all three flavours).

The $Zb\bar{b}$ reducible background was generated using ACERMC3.1 [11] with Parton Density Function (PDF) set CTEQ6L, QCD scales $\mu_R = \mu_F = m_Z$, massive $b$-quark, interfaced to PYTHIA 6.3 for showering and hadronization. The $Z \rightarrow ll$ ($l = e, \mu$) decay is forced at generator level. The full $Z/\gamma^*$ interference is taken into account, with a cut on the resonance mass at 30 GeV. The ACERMC dataset includes only the $gg \rightarrow Zb\bar{b}$ process. The $q\bar{q} \rightarrow (Z/\gamma^*)b\bar{b}$ contribution, where $q$ is a light quark, is not included; its contribution to the total LO $Zb\bar{b}$ cross-section is less than 15%. The AcerMC LO cross-section, including the $Z \rightarrow ll$ branching ratio (BR), evaluated with the above parameter choice, amounts to 52.03 ± 0.03 pb for the $gg \rightarrow (Z/\gamma^*)b\bar{b}$ and 8.64 ± 0.01 pb for the $q\bar{q} \rightarrow (Z/\gamma^*)b\bar{b}$.

| Process | Generator | $\sigma\cdot$ BR [fb] | Corrections | FA | Evts [k] |
|---|---|---|---|---|---|
| $q\bar{q} \rightarrow ZZ \rightarrow 4\ell$ | PYTHIA6.3 | 158.8 | +47.64 | [4$\ell$]0.219 | 100 |
| $gg \rightarrow Zb\bar{b} \rightarrow 2\ell b\bar{b}$ | AcerMC/PYTHIA6.3 | 52030 | +8640 ($q\bar{q} \rightarrow Zb\bar{b}$) | [4$\ell$] 0.00942 | 430 |
| $gg \rightarrow Zb\bar{b} \rightarrow 2\ell b\bar{b}$ | AcerMC/PYTHIA6.3 | 52030 | +8640 ($q\bar{q} \rightarrow Zb\bar{b}$) | [3$\ell$] 0.147 | 200 |
| $gg, q\bar{q} \rightarrow t\bar{t}$ | MC@NLO/Jimmy | 833000 | | [4$\ell$] 0.00728 | 400 |
| $q\bar{q} \rightarrow WZ$ | Jimmy | 26500 | | [3$\ell$] 0.0143 | 70 |
| $q\bar{q} \rightarrow Z$ inclusive | PYTHIA6.3 | $1.5\cdot10^6$ | | [1$\ell$]0.89 | 500 |

Table 2: Background samples, generators used, acceptance of the multi-lepton filter (FA), LO cross-section (except for $t\bar{t}$, which is NLO) and corrections applied. The number of events is given in the last column. For $ZZ$, $l = e, \mu, \tau$ while for the rest $l = e, \mu$. The relative errors on the filter acceptances (FA) are smaller than 0.4%.

#### 2.1.2 Next-to-Leading-Order cross-sections for the background

The $t\bar{t}$ sample, generated by MC@NLO, is the only sample in Table 2 that already includes NLO processes. To evaluate the NLO cross-sections for the other background processes, the program MCFM [14] is used. The overall conditions for all MCFM NLO calculations are the following: CTEQ6M, $\mu_R = \mu_F = m_Z$, $m_b = 0$ full $(Z/\gamma^*)$ interference. Any two final state partons are merged in a single jet if their separation $\Delta R(jj)$ is smaller than 0.7.

In the case of the $Zb\bar{b}$ sample, the following additional selections are applied: $m_Z > 30$ GeV; $p_T(b) > 10\ GeV$, $|\eta(b)| < 2.5$. The NLO cross-section obtained from MCFM is:

$$270.4^{+40.6}_{-35.7}(\mu_R)^{+5.4}_{-8.0}(\mu_F) \pm 12.5(\mathrm{PDF}) \pm 1.0(\mathrm{stat})\ \mathrm{pb}$$

where the first two uncertainties come from the QCD renormalization and factorization scales, varying independently the energy scale of the process from $0.5m_Z$ to $2m_Z$, while the last one quotes the PDF uncertainty, calculated by making use of 40 sets of PDFs for CTEQ6M (20 plus and 20 minus). This

| $q\bar{q} \to Z/\gamma^* Z/\gamma^*$ | |
|---|---|
| $m_{Z/Z^*} > 12$ GeV | |
| $m_{ZZ}$ (GeV) | K-factor |
| [115,125] | 1.15 |
| [125,135] | 1.21 |
| [145,155] | 1.25 |
| [155,165] | 1.34 |
| [175,185] | 1.31 |
| [195,205] | 1.32 |
| [295,305] | 1.40 |
| [395,405] | 1.52 |
| [495,505] | 1.84 |
| [595,605] | 1.81 |

Table 3: K-factors for the ZZ background. The error on the K-factors is dominated by the systematics (from PDF and renormalization and factorization scales) and amounts to 3.3%.

result has proved to be quite stable with respect to the variation of the cuts both on the minimum jet $p_T$ and the minimum jet separation $\Delta R(jj)$. The LO cross-section evaluated with MCFM on the same phase space is $189.9 \pm 0.2$ pb, resulting in a K-factor of 1.42. The effective cross-section can be evaluated as follows

$$\sigma_{eff} = \sigma_{LO} \cdot BR(Z \to ee, \mu\mu) \cdot FA \cdot K = 812.1 \text{ fb},$$

where $\sigma \cdot BR(Z \to ee, \mu\mu)$ is the AcerMC cross-section listed in Table 2 rescaled to include the $q\bar{q}$ contribution, and FA is the acceptance of the generator filter, as reported in Table 2. The statistical error on the filter efficiency is 0.2% for this dataset.

The NLO cross-section for the $q\bar{q} \to ZZ^*$ process is calculated with MCFM, applying the same kinematic selection on the $Z$ boson masses as in PYTHIA ($m_{Z^{(*)}} > 12$ GeV):

$$\sigma_{NLO} = 22.1^{+0.1}_{-0.2}(\mu_R = \mu_F) \pm 0.7(PDF) \text{ pb}$$

where the statistical error is much smaller than the systematic ones. The effective cross-section used in the analysis can be evaluated according to

$$\sigma_{eff} = \sigma_{LO} \cdot [BR(Z \to ll)]^2 \cdot FA \cdot (K + 0.3) = 34.82 \cdot (K(m_{ZZ}) + 0.3) \text{ fb},$$

where $\sigma_{LO} \cdot [BR(Z \to ll)]^2$ is the LO cross-section, that includes the $Z$ branching ratio to leptons from PYTHIA and amounts to $159 \pm 0.05$ (stat) fb. The additional 30% accounts for the correction coming from the $gg \to ZZ^*$ due to the quark box, as listed in Table 2. The relative statistical error on the filter acceptance FA is 0.3% for this process. The mass-dependent $K$-factors used in this analysis are shown in table 3.

The NLO cross-section for the $WZ$ background is evaluated by applying the same cut on the boson masses ($m_{Z^*/W^*} > 20$ GeV) as in the HERWIG-Jimmy generator used for the LO process:

$$W^- : \sigma_{NLO} = 21.7^{+0.5}_{-0.9}(\mu_R = \mu_F) \pm 0.9(\text{PDF}) \text{ pb}$$

$$W^+ : \sigma_{NLO} = 34.8^{+1.2}_{-0.9}(\mu_R = \mu_F) \pm 1.0(\text{PDF}) \text{ pb}$$

where the QCD scale uncertainties are included. The statistical error is negligible. The effective cross-section can then be evaluated as follows:

$$\sigma_{eff} = \sigma_{NLO}(W^+Z + W^-Z) \cdot FA = 807 \text{ fb}$$

where $\sigma_{NLO}(W^+Z + W^-Z)$ refers to the sum of the two contributions. The relative statistical error on the filter efficiency for this process is 0.4%.

## 2.2 Trigger

The simulation of the full ATLAS trigger chain allows to evaluate the impact of the on-line selection on the Higgs boson search. Level-one (LVL1) trigger objects (Region-of-interest, ROI), available in the simulation, correspond to small $\Delta\eta \times \Delta\phi$ regions where a lepton that satisfies the online selection criteria is found. LVL1 muon thresholds are programmable in the $p_T$ range from 4 GeV/c to about 40 GeV/c. Similarly to the LVL1 muon Trigger, electron/photon ROIs can be of eight types, depending on the highest $E_T$ threshold satisfied. The settings in this case are also programmable. In this analysis, two types of electron/photon $E_T$ thresholds are considered, ${E_T}^{thres}$=15 and 22 GeV. For both thresholds, cuts on isolation and on leakage in the hadronic calorimeter are also applied at LVL1. The LVL1 electrons/photons and muons are subsequently confirmed by the High Level Trigger (HLT). The first step is the Level-2 trigger (LVL2) where fast algorithms are used to validate the selected lepton. Events with leptons of a given quality and with energy above certain fixed thresholds are retained for a more accurate reconstruction and selection by the Event Filter (EF), where algorithms based on the offline reconstruction software are used for the final selection. The choice of HLT selection thresholds is optimized for physics performance, assuring that trigger rates satisfy the system latency. The main aspects of the electron and muon trigger systems are described in [15] and [16].

The acceptance of the muon trigger as a function of the generated $p_T$ for the threshold ${p_T}^{thres}$=20 GeV/c, is shown in Fig. 1. The efficiency above threshold is explained by the geometrical coverage of the muon LVL1 trigger detectors; in the barrel, the space occupied by detector feet, supports, services, etc., limits the geometrical acceptance to about 80%, while the efficiency of the trigger algorithm itself is very close to 100%. The trigger efficiencies of electrons for a selection threshold of ${E_T}^{thres}$=22 GeV, including electron identification and isolation cuts, is shown in Fig. 2.

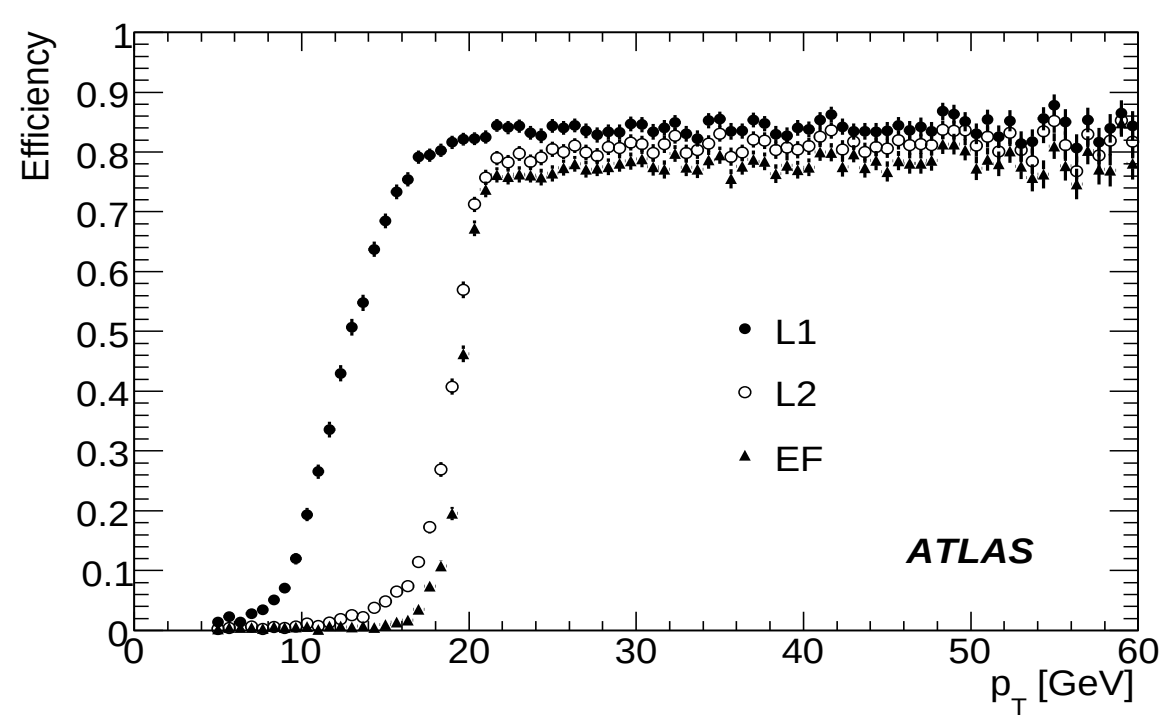


Figure 1: Muon trigger: selection efficiencies of the three trigger levels as a function of the true muon transverse momentum. The selection threshold is ${p_T}^{thres}$=20 GeV/c.

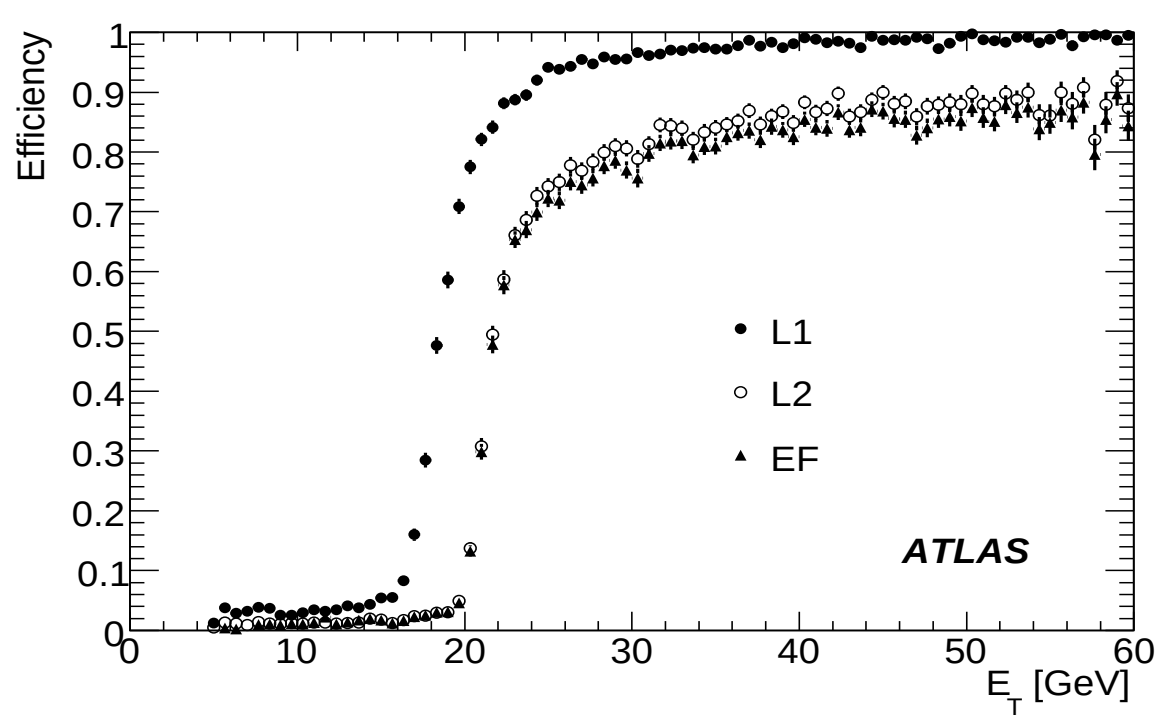


Figure 2: Electron trigger: selection efficiencies of the three trigger levels as a function of the true electron transverse energy. The selection threshold is ${E_T}^{thres}$=22 GeV.

In this paper, several trigger menus foreseen for the LHC running at luminosity L=$10^{33}$ cm$^{-2}$s$^{-1}$, are considered. The trigger efficiency for single and double lepton triggers, for a signal sample with a Higgs

| | Unbiased sample | | | After event selection | | |
|---|---|---|---|---|---|---|
| **Trigger Menu** | $4e$ | $4\mu$ | $2e2\mu$ | $4e$ | $4\mu$ | $2e2\mu$ |
| $1\mu20$ | 0.1 | 95.3 | 71.3 | 0.4 | 98.2 | 72.7 |
| $1e22$i | 94.7 | 0.4 | 68.6 | 99.8 | 0.1 | 78.1 |
| $2e15$i | 76.3 | <0.2 | 33.2 | 98.9 | <0.2 | 60.2 |
| $1\mu20$ or $1e22$i | 94.7 | 95.3 | 95.7 | 99.8 | 98.2 | 98.9 |
| $2\mu10$ or $2e15$i or $1\mu10$ and $1e15$i | 76.4 | 93.3 | 87.8 | 98.9 | 97.6 | 96.9 |

Table 4: Trigger selection efficiencies (in %) for various trigger menus, computed for $H \rightarrow ZZ^* \rightarrow 4l$ ($m_H$=130 GeV), for the full trigger chain LVL1+HLT. The first three columns show the efficiencies computed on the full event sample. The last three columns show the efficiencies for events selected by the $H \rightarrow ZZ^* \rightarrow 4l$ analysis. The absolute errors on the efficiencies are 0.4% for the unbiased samples, and 0.2% for events passing the offline selection.

boson mass of 130 GeV, is shown in Table 4. The results are shown for two classes of events: those filtered at generator level, requiring 4 leptons within $p_T >$ 5 GeV and $|\eta| <2.7$ (unbiased sample), and those passing the signal selection of reconstructed events, described later in this note. A double-lepton trigger with 10 GeV threshold for the muons and 15 GeV threshold for the electrons (isolation required), selects Higgs boson to four lepton decays with an efficiency higher than 97 %. In the following, the single-lepton menu requiring $1\mu20$ or $1e22$i will be applied as trigger selection.

## 2.3 Electron and muon reconstruction

The leptons used for the efficiency studies are required to satisfy the generator level kinematic cuts $|\eta| < 2.5$,and $p_T >$ 5 GeV. The lepton efficiency is defined as the ratio of reconstructed to generated leptons originating from $Z$ decays. The reconstructed leptons include leptons not originating from Z decays (non-prompt leptons) and fakes, so it is important to study these candidates and to estimate the fraction of leptons coming from non-Z decays and fakes. The non-Z lepton fraction is defined as the number of reconstructed leptons matched to true leptons not originating from $Z$ decays, divided by the total number of reconstructed leptons. The fraction of fakes is defined as the number of reconstructed leptons not matched to a true lepton, divided by the total number of reconstructed leptons.

### 2.3.1 Electron reconstruction

The details of electron reconstruction are described in [17] and [18]. Here the electron definitions used in this analysis are briefly summarized. An electron is selected using the offline algorithm requiring:

1. a cluster in the barrel and endcap LAr EMC, reconstructed by the ATLAS offline software;
2. an inner detector track associated with the cluster;
3. cluster containment in the LAr EMC;
4. consistency of the lateral shower shape of the cluster with an electron isolated from hadronic activity;
5. the lateral shower shape to be inconsistent with a $\pi^0 \rightarrow \gamma\gamma$ decay, using the strip section of the LAr EMC;

6. track quality requirements: a certain number of hits on the Pixel and SCT detector is required, and a transverse impact parameter smaller than 0.1 cm.

The containment and isolation requirements (2,3,4) are satisfied using the so-called *LooseElectron* definition. This definition uses shower shape variables calculated with the middle sampling of the LAr EMC. The addition of requirements 5 and 6 corresponds to the *MediumElectron* definition. In this analysis the *MediumElectron* definition with the addition of calorimetric isolation using all cells (EM and hadronic) inside a $\Delta R = \sqrt{\Delta\eta^2 + \Delta\phi^2} = 0.2$ cone (*MediumElectron+CaloIso*) has been used. The isolation cuts are $\eta$-dependent, as described in detail in [17].
True electrons not originating from a $Z$ boson decay represent a background to the $H \rightarrow ZZ^* \rightarrow 4l$ search. Typically they are not isolated and in most cases these electrons originate from heavy quark decays. It should be stressed that leptons from B-mesons are more isolated than those coming from $D$-meson decays, and therefore are more difficult to reject. The isolation also depends on the event topology. The non-Z electrons are a small fraction (less than 1%) of the total number of electrons in the $H \rightarrow ZZ^* \rightarrow 4l$ events. The electron efficiency as a function of the pseudorapidity $\eta$ and the transverse momentum $p_T$ for the *Loose*, *MediumElectron* and *MediumElectron+CaloIso* definitions is shown in Fig. 4. A significant drop in efficiency is observed at low $p_T$. This is due to the loss of discrimination power of the shower shape cuts. A summary of electron efficiencies and fraction of fakes and non-Z electrons is presented in [17]. The total fraction of fakes depends on the isolation cut and is particularly high for lower $p_T$. For low momenta, $p_T < 15$GeV, the fraction of these electrons is about 8% of the total reconstructed electrons in this $p_T$ range. The contribution from electrons which do not come from $Z$ decays is a fraction of the fake electrons dominated by heavy flavour (c,b) decays. The b-originated electrons are a factor of 1.5-2 more than the c-originated electrons.

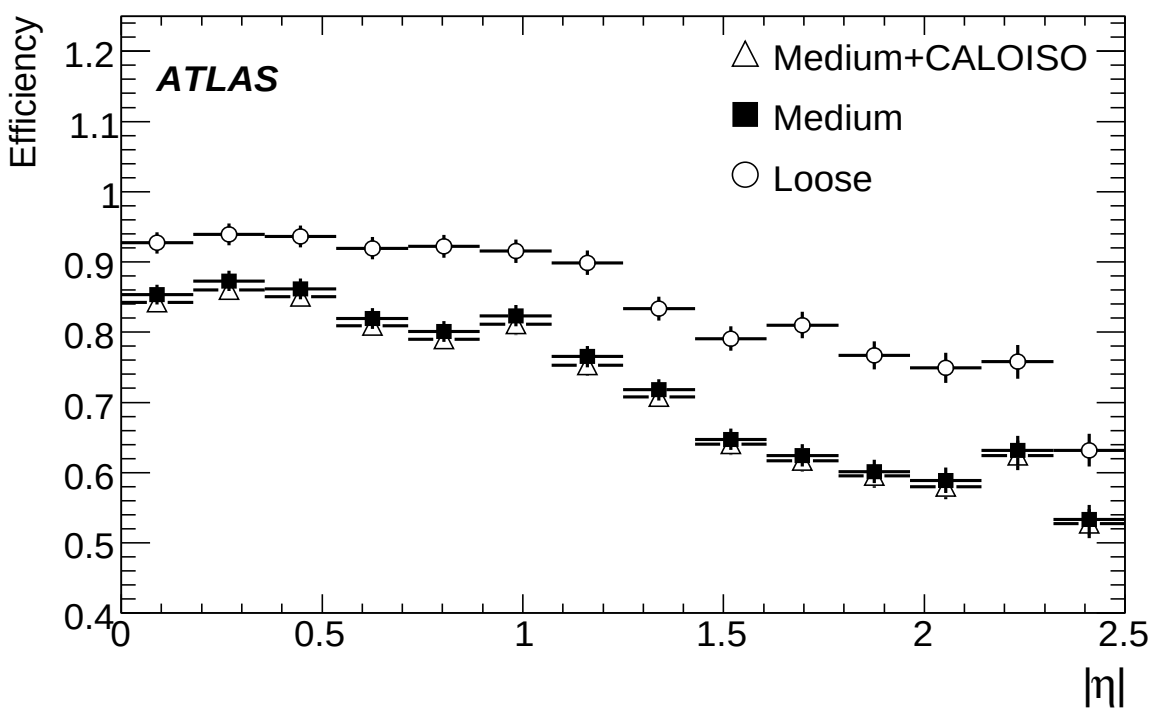


Figure 3: Electron reconstruction efficiency as a function of $\eta$. The electron-id criteria are described in the text.

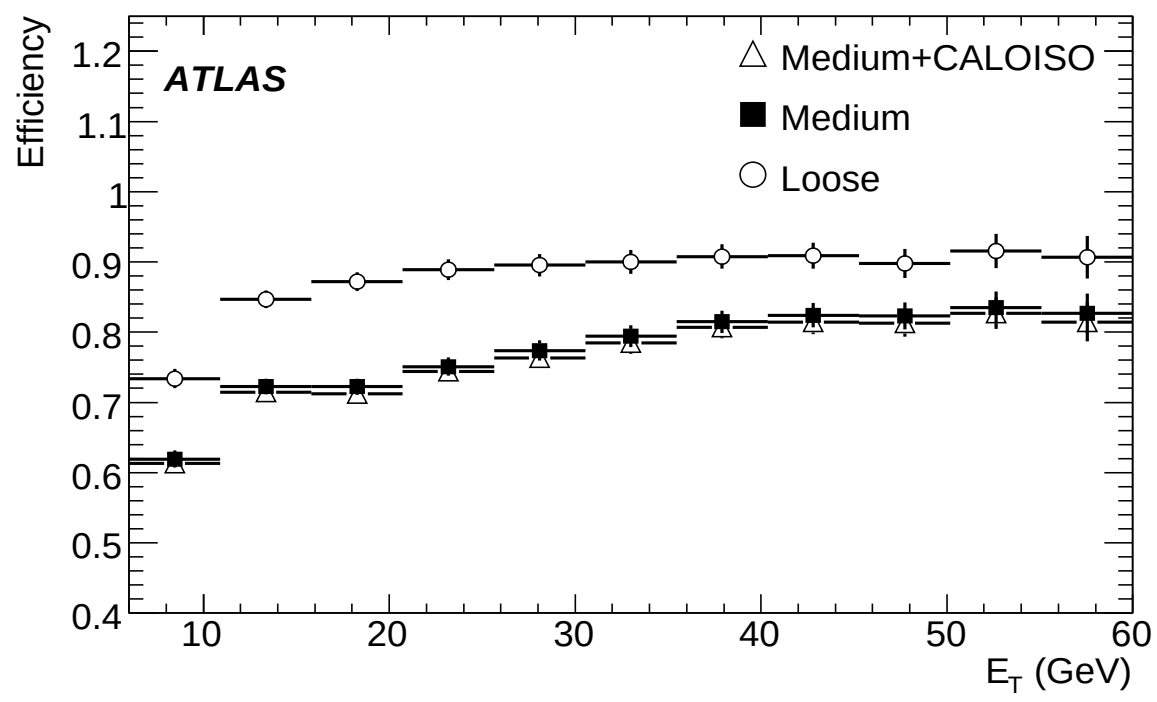


Figure 4: Electron reconstruction efficiency as a function of $p_T$. The electron-id criteria are described in the text.

### 2.3.2 Muon reconstruction

The muon identification in ATLAS relies on the Muon Spectrometer (MS) for standalone reconstruction as well as on the ID and Calorimeters for combined muon reconstruction. In order to combine the muon tracks reconstructed in the ID and the MS, the ATLAS offline muon identification packages have been developed. The purpose of these packages is to associate segments and tracks found in the MS with the corresponding ID track in order to identify muons at their production vertex with optimum parameter resolution. Details on the muon system design and performance can be found in [19]. Details on the algorithms of the muon identification described here can be found in [20].

The standalone muon reconstruction algorithm measures the five parameters of the reconstructed tracks with their associated 5x5 covariance matrix at the entrance of the Muon Spectrometer. The muon momentum is corrected using an energy loss parameterization. Overall, the reconstructed track parameters with their full covariance matrices are provided at three locations: 1) at the entrance to the muon spectrometer, 2) at the entrance to the calorimeters, 3) at the perigee of the track. Standalone tracks reconstructed in the MS are combined with tracks reconstructed in the ID in the region of $\eta$<2.5. This combination improves the momentum resolution for tracks with momenta up to 100 GeV and suppresses the rate of fake muons. A statistical combination of the two sets of track parameters is made, weighted by their corresponding covariance matrices. Given a certain $\Delta R = \sqrt{\Delta\eta^2 + \Delta\phi^2}$ cone, all tracks from the ID are combined with the ones from the MS and the pair giving the best $\chi^2$ is kept. The same procedure is repeated until no more combinations are possible.

An algorithm has been developed in order to recover muons which fail to be reconstructed in the MS (either because they are low $p_T$ muons or because the number of stations is insufficient). The principle of the algorithm is based on the extrapolation of ID tracks to the inner stations of the MS and their matching to a segment reconstructed in these stations that was not yet associated to a combined track. The extrapolation is also performed to medium stations, in the regions of $\eta$ between $1 < \eta < 1.4$, where there is a type of inner stations missing, resulting in a drop of the reconstruction efficiency.

The reconstruction efficiency of muons from a 130 GeV Higgs boson sample as a function of their transverse momentum and $\eta$, is shown in Figs. 5 and 6 respectively.

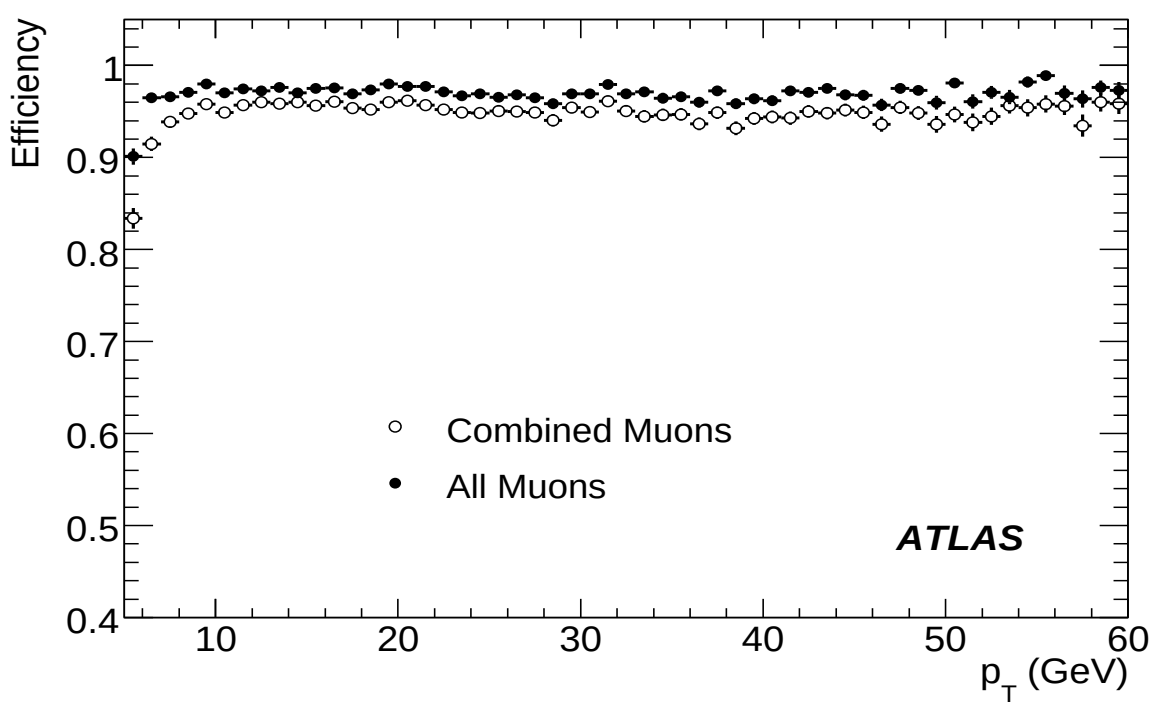


Figure 5: Muon reconstruction efficiency as a function of $p_T$. Empty (filled) markers show the efficiency of the combined (combined+extrapolated from the ID) algorithm. Reconstructed muons of a Higgs boson sample of 130 GeV mass decaying into four muons are used.

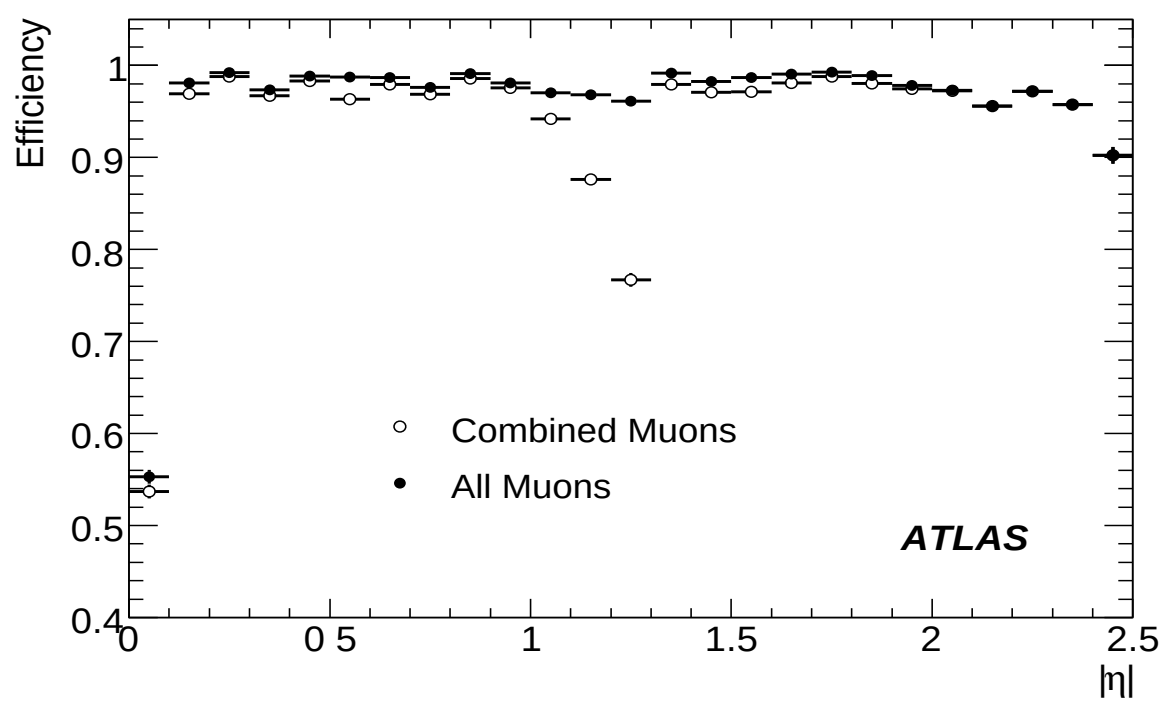


Figure 6: Muon reconstruction efficiency as a function of $\eta$. Empty (filled) markers show the efficiency of the combined (combined+extrapolated from the ID) package. Reconstructed muons of a Higgs boson sample of 130 GeV mass decaying into four muons are used.

## 3 Background Rejection

The large $Zb\bar{b}$ and $t\bar{t}$ background cross-sections compared to the Standard Model Higgs boson cross-section, require further reduction of these processes by applying additional lepton identification criteria. A reduction well below the irreducible $ZZ^*$ background yield is a safeguard against large uncertainties on the production cross-sections of these final states. One can exploit the fact that leptons originating from the $Z$ boson decays are expected to be significantly more isolated than the ones originating from heavy quark leptonic decays. These leptons are also expected to originate from the main interaction point, while $b, c$-originating leptons should come from secondary displaced vertices.

In this section, a set of isolation and impact parameter cuts for background rejection is described. In this analysis the cuts have been chosen so that the expected rate for $Zb\bar{b}$ to 4-leptons ($4e$, $4\mu$ and $2e2\mu$) is no more than one third of the $ZZ$ rate. For Higgs boson masses above 160GeV, the reducible background contribution is expected to be less than 10% of the irreducible $ZZ$.

## 3.1 Muon Isolation

For the muon final state both calorimetric and track isolation criteria have been considered.

- The calorimetric isolation discriminant is defined as the sum of the transverse energy deposited in the calorimeter inside a cone of a given radius $\Delta R = \sqrt{\Delta\eta^2 + \Delta\phi^2}$, around the muon. The energy deposition of the muon itself is subtracted from the isolation energy. The cone energy of the least isolated muon is used as a discriminant.
- The track isolation discriminant is defined as the sum of the transverse momenta of the inner detector tracks in a cone of radius $\Delta R$ around the muon. The inner detector muon track is excluded from the sum. The least isolated track of all muons in the event, is used as discriminant.

In Figs. 7 and 8 the rejection of the $Zb\bar{b}$ background as a function of the $4\mu$ signal efficiency ($m_H$=130 GeV) is presented, for different calorimetric and track isolation cones, respectively. A cone size of 0.2 is adopted for both calorimetric and track isolation as a conservative choice. Moreover, as shown in Figs. 9 and 10, the background rejection improves when the isolation quantities are normalized to the transverse momentum of the muon, thus the normalized isolation discriminants are used in the analysis.

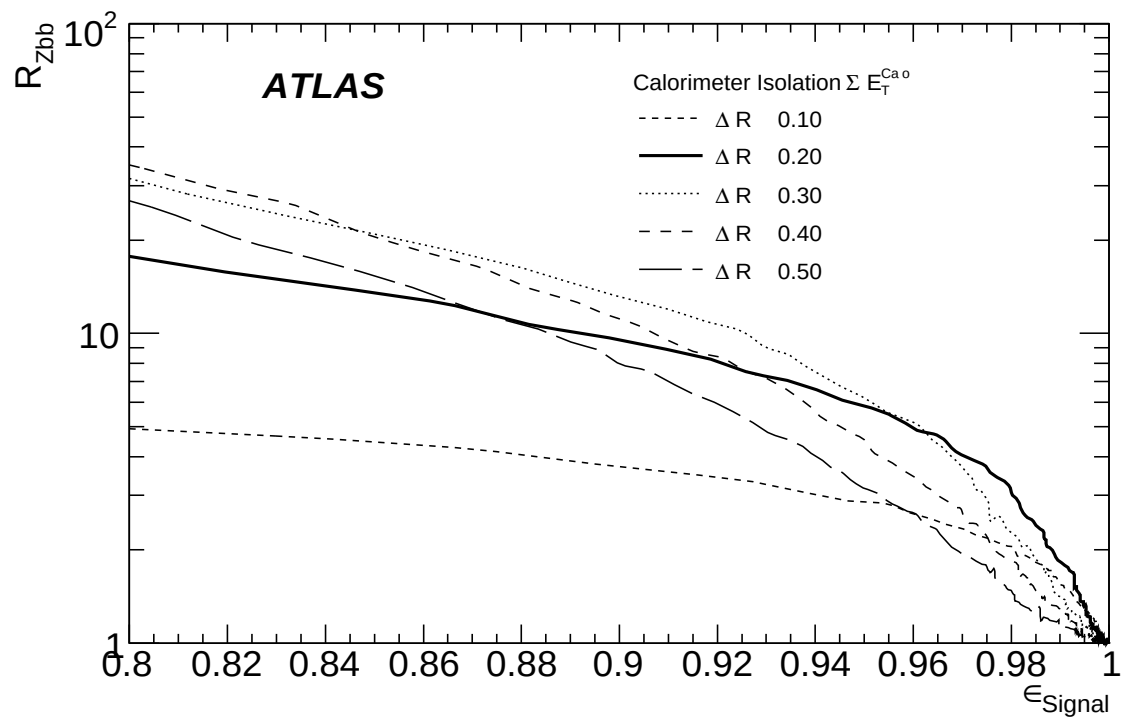


Figure 7: $Zb\bar{b}$ rejection versus $H \rightarrow 4\mu$ efficiency, for $m_H = 130GeV$, for various calorimetric isolation cone sizes.

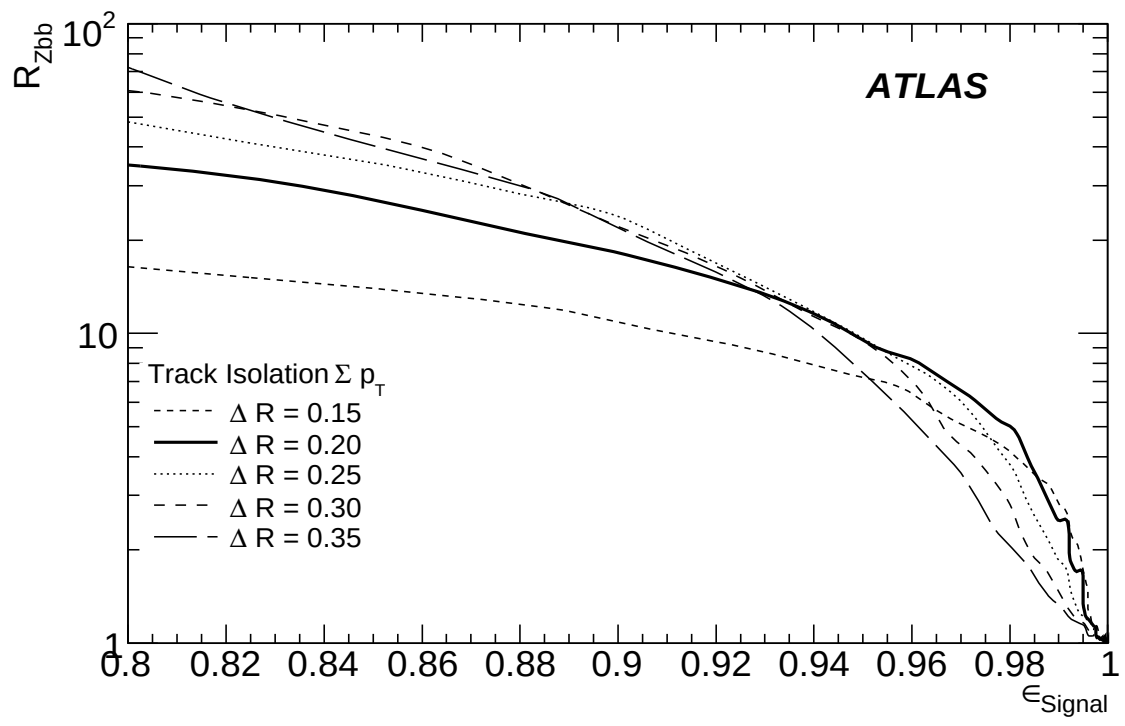


Figure 8: $Zb\bar{b}$ rejection versus $H \rightarrow 4\mu$ efficiency, for $m_H = 130GeV$, for various track isolation cone sizes.

The distributions of the normalized calorimetric and track isolation variables calculated using the cones chosen in this analysis for the signal and the main backgrounds are presented in Figs. 11 and 12, respectively. The cuts on the calorimetric and tracker isolation are chosen so that the efficiency for the signal, after application of the two cuts, is close to 90%. The selection cuts are placed at 0.23 and 0.15 for the calorimetric and the tracker isolation, respectively.

## 3.2 Electron Isolation

For the electron final state both calorimetric and track isolation in the inner detector are considered. Although partial calorimetric isolation along the $\eta$ direction is already part of the electron-id requirements, an extra calorimetric isolation in a cone of $\Delta R = 0.2$ is applied. The definition of the electron

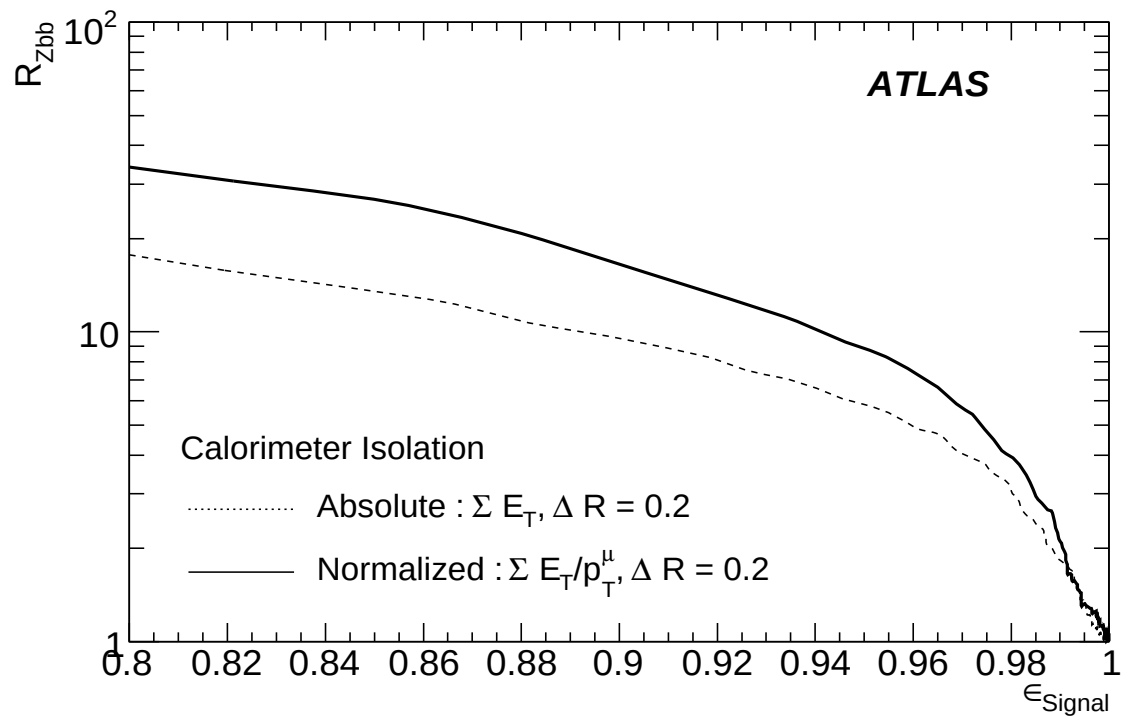


Figure 9: $Zb\bar{b}$ rejection versus $H \to 4\mu$ efficiency, for $m_H = 130GeV$, for standard and normalized calorimetric isolation calculated in a $\Delta R$=0.2 cone around the muon track.

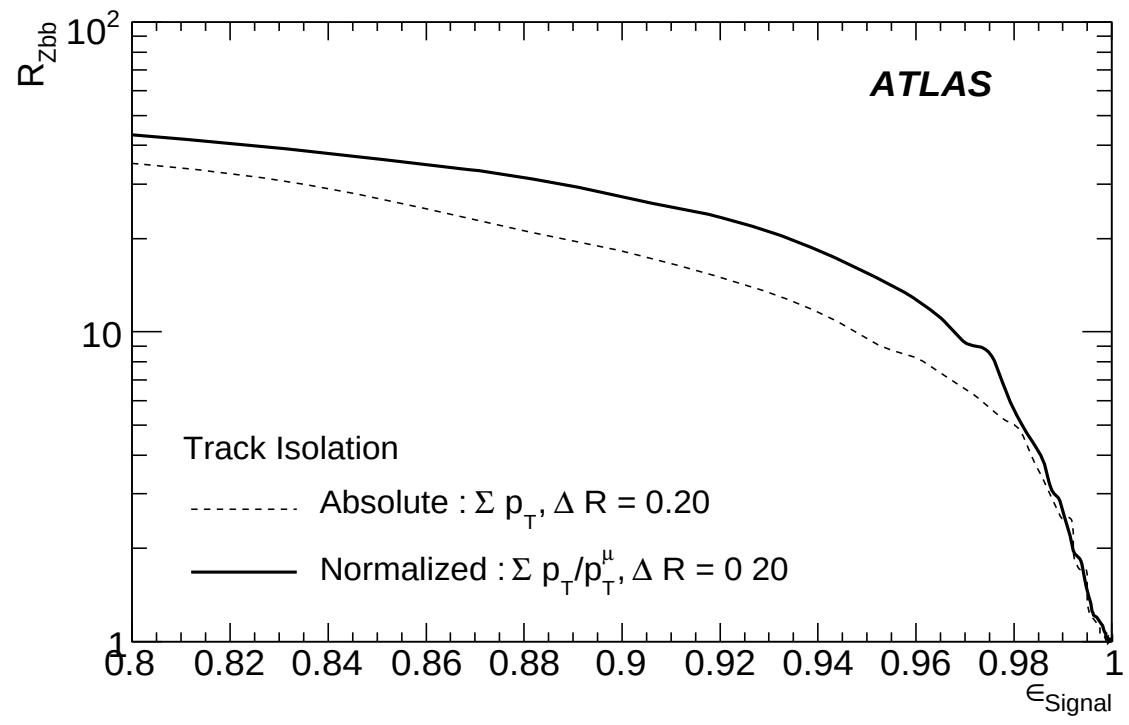


Figure 10: $Zb\bar{b}$ rejection versus $H \to 4\mu$ efficiency, for $m_H = 130GeV$, for standard and normalized track isolation calculated in a $\Delta R$=0.2 cone around the muon track.

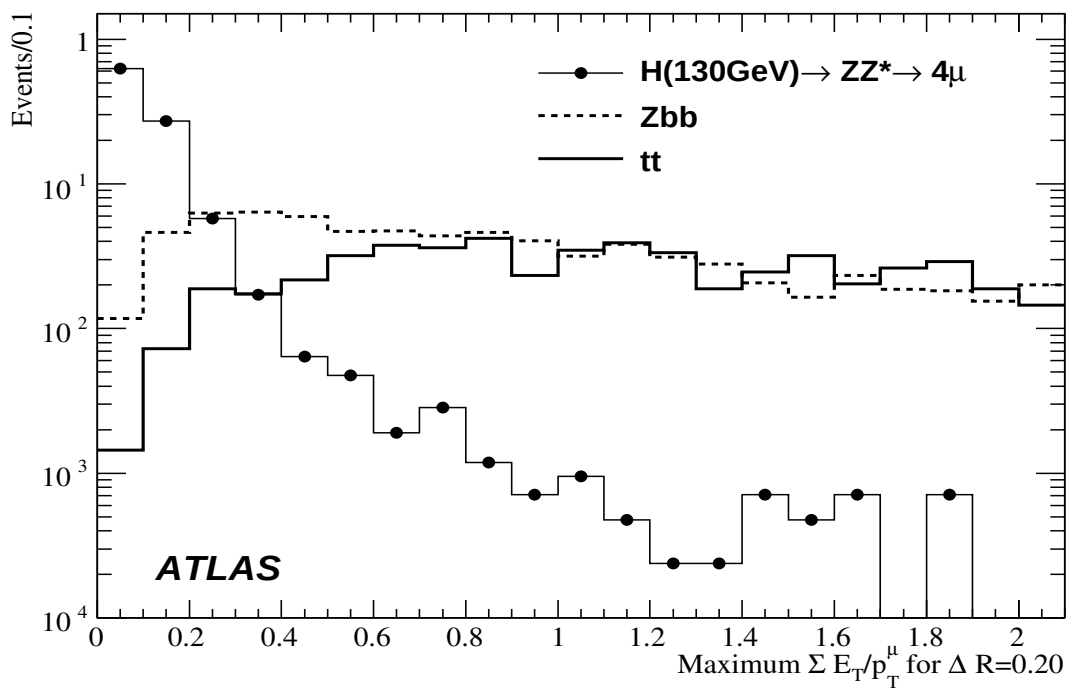


Figure 11: Normalized calorimetric isolation ($\Delta R$=0.2) for the signal ($m_H = 130$), the $Zb\bar{b}$ and $t\bar{t}$ backgrounds for the $4\mu$ channel.

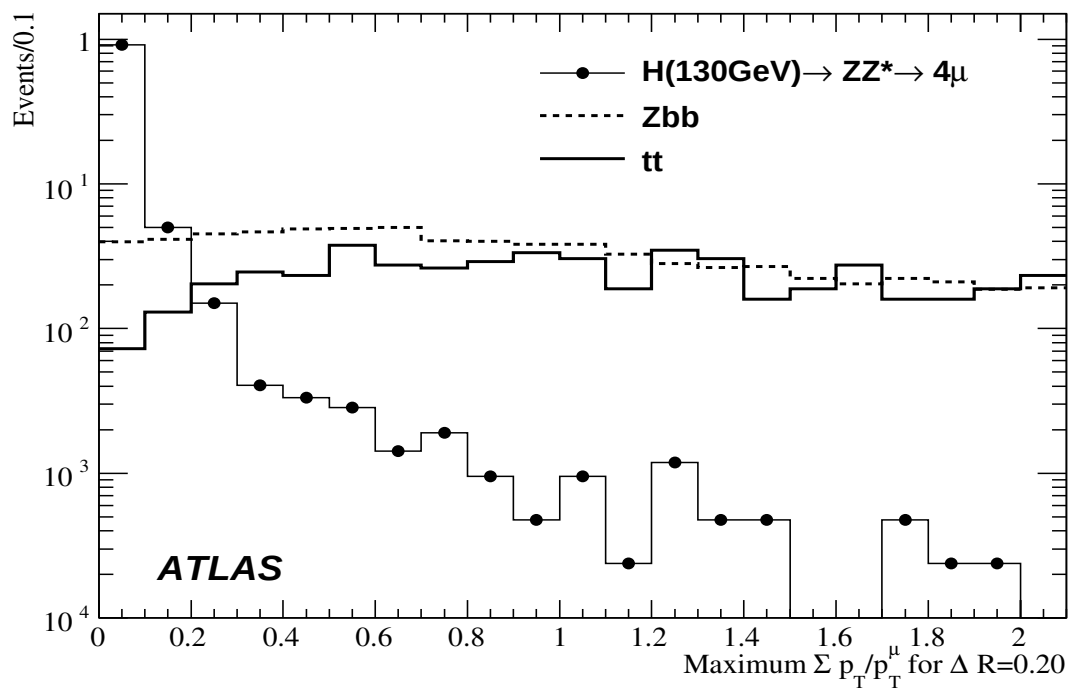


Figure 12: Normalized track isolation ($\Delta R$=0.2) for the signal, the $Zb\bar{b}$ and $t\bar{t}$ backgrounds for the $4\mu$ channel.

track isolation is the same as for the muons: it is the sum of the transverse momenta of inner detector tracks in a cone with radius $\Delta R$ around the direction of the electron track. The $p_T$ of the electron track is excluded from the sum. In an attempt to remove from the sum tracks originating from conversions of bremsstrahlung photons, only tracks which have at least one hit in the B-layer (the innermost layer of the Pixel detector) are considered in the sum. The track isolation is normalized to the electron $p_T$. The distribution of the electron track isolation is shown in Fig. 13, for a cone size $\Delta R$=0.2. For the analysis all leptons must satisfy a $\Sigma p_T/p_T < 0.15$ tracking isolation cut.

## 3.3 Impact parameter analysis

Leptons from $t\bar{t}$ and $Zb\bar{b}$ backgrounds are most likely to originate from displaced vertices. Further rejection of these backgrounds can be achieved by placing a cut on the transverse impact parameter significance (defined as $d0/\sigma_{d0}$, where $d0$ is the distance of closest approach in the transverse plane) of the tracks associated to the leptons. For electrons, bremsstrahlung smears the impact parameter distribution, hence reducing the discriminating power of this cut with respect to muons. The impact parameter is calculated with respect to the event vertex fitted using a set of tracks reconstructed in the ID. This allows to remove the effect of the spread of the vertex position, which at LHC is 15 $\mu$m along each of

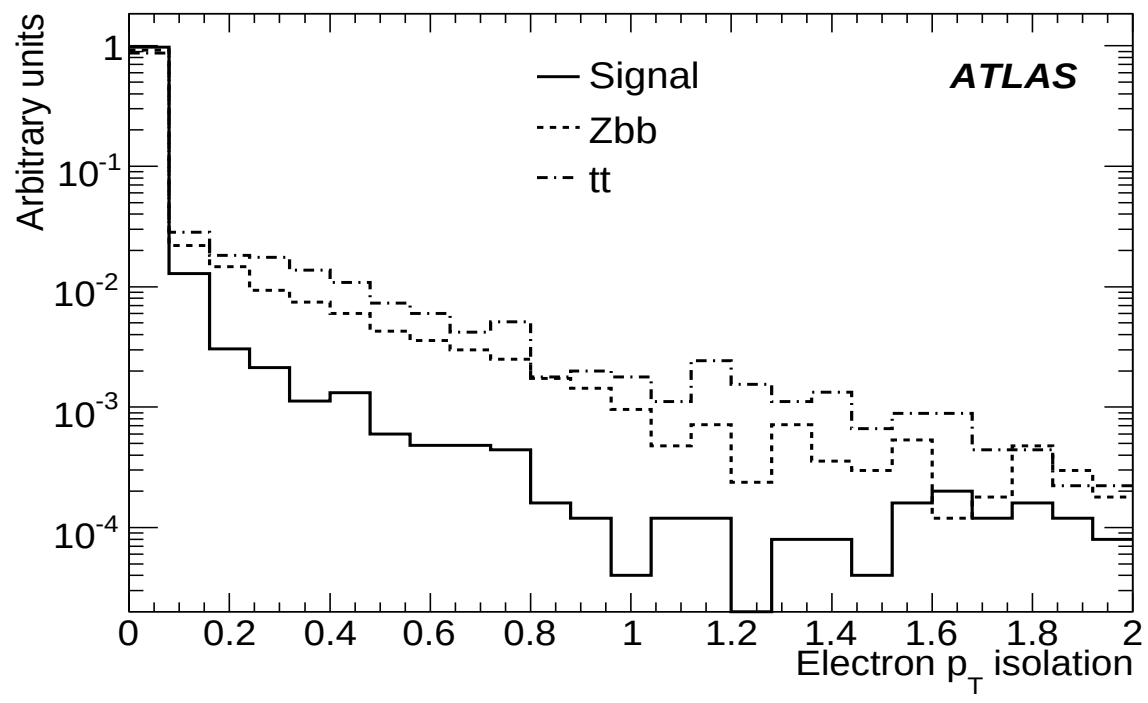


Figure 13: Signal and background distributions for electron track isolation, normalized to the electron track $p_T$, in a cone $\Delta R$ <0.2.

the transverse $x$ and $y$ axes. In Figs. 14 and 15, the transverse impact parameter significance for muons and electrons is shown. Tracks with $d0$ significance greater than 3 are not included in the primary vertex fit, and this causes the shoulder visible in the distributions. The discriminating variable used in the event selection for $Zb\bar{b}$ and $t\bar{t}$ background rejection, is the maximum lepton impact parameter in the event. This is shown in Figs. 16 and 17 for $4\mu$ and $4e$ events, respectively. For electron tracks, the maximum impact parameter normalized to its error is required to be less than 6, while for muons less than 3.5.

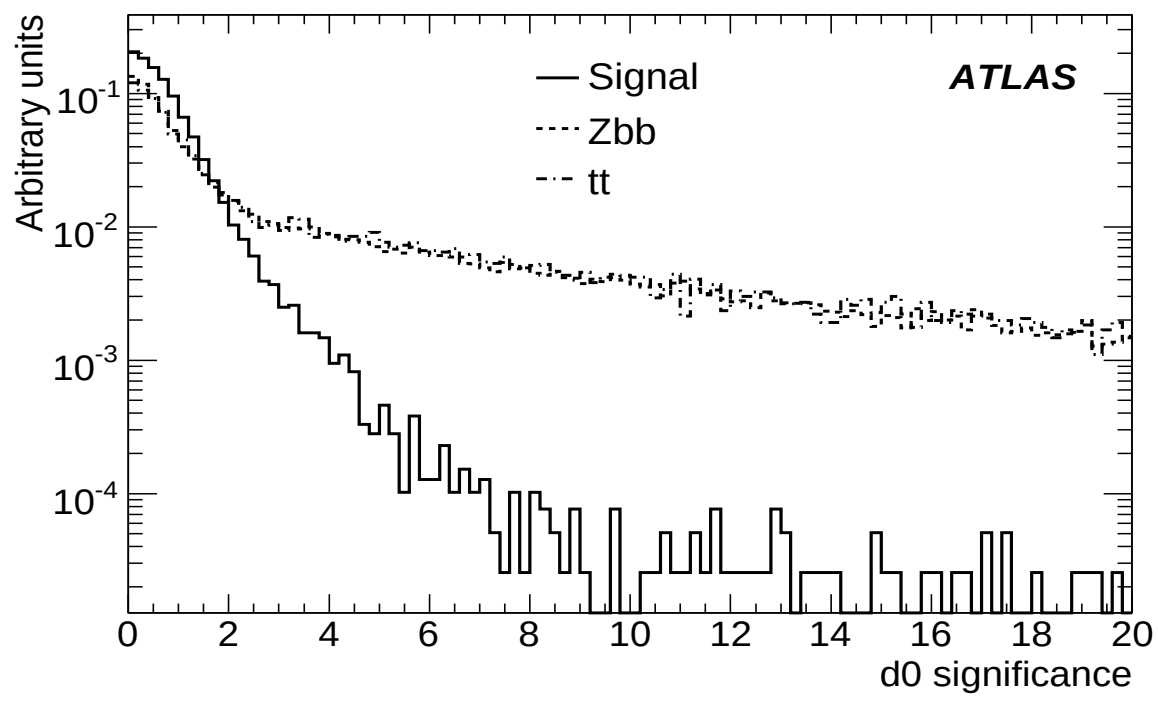


Figure 14: Transverse impact parameter significance for muons from signal and reducible background events.

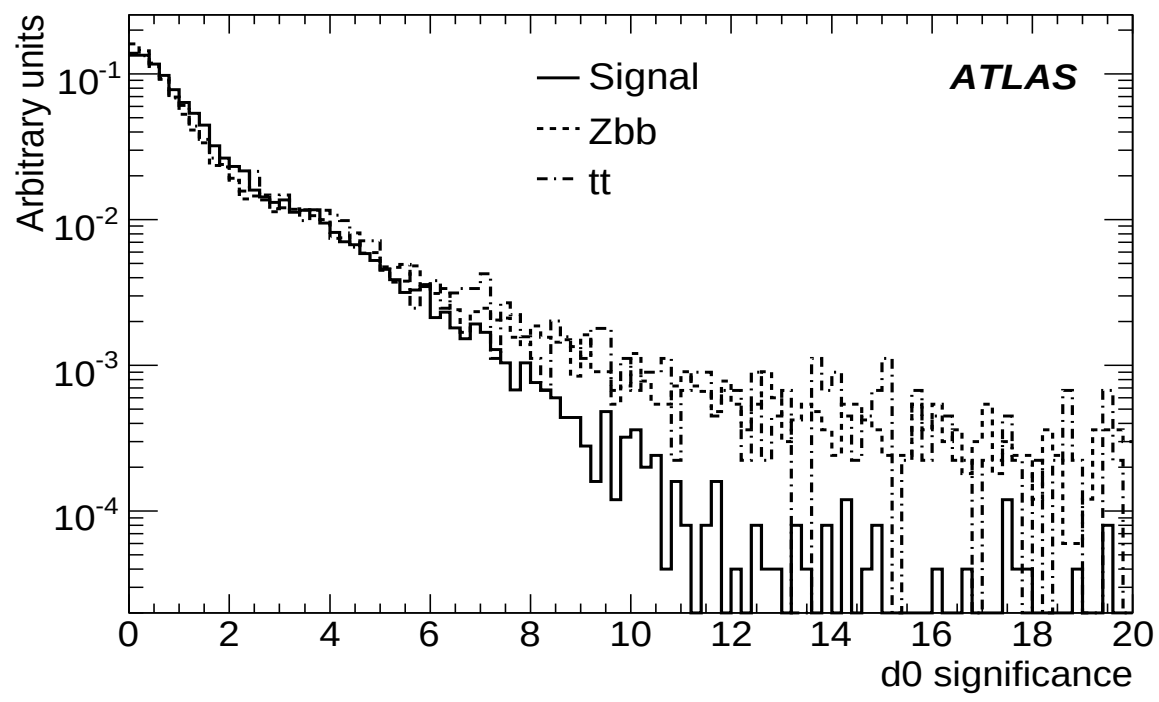


Figure 15: Transverse impact parameter significance for electrons from signal and reducible background events.

# 4 Event selection

In this section the full set of cuts performed in this analysis are summarized and the event kinematic reconstruction is described. These cuts including the isolation and vertexing cuts covered in the previous section are summarized in Table 5.

## 4.1 Event preselection

Events that pass the trigger selection are required to further satisfy certain lepton preselection criteria. An electron must satisfy the *LooseElectron* requirement described in Section 2.3.1, and have an $E_T$>5

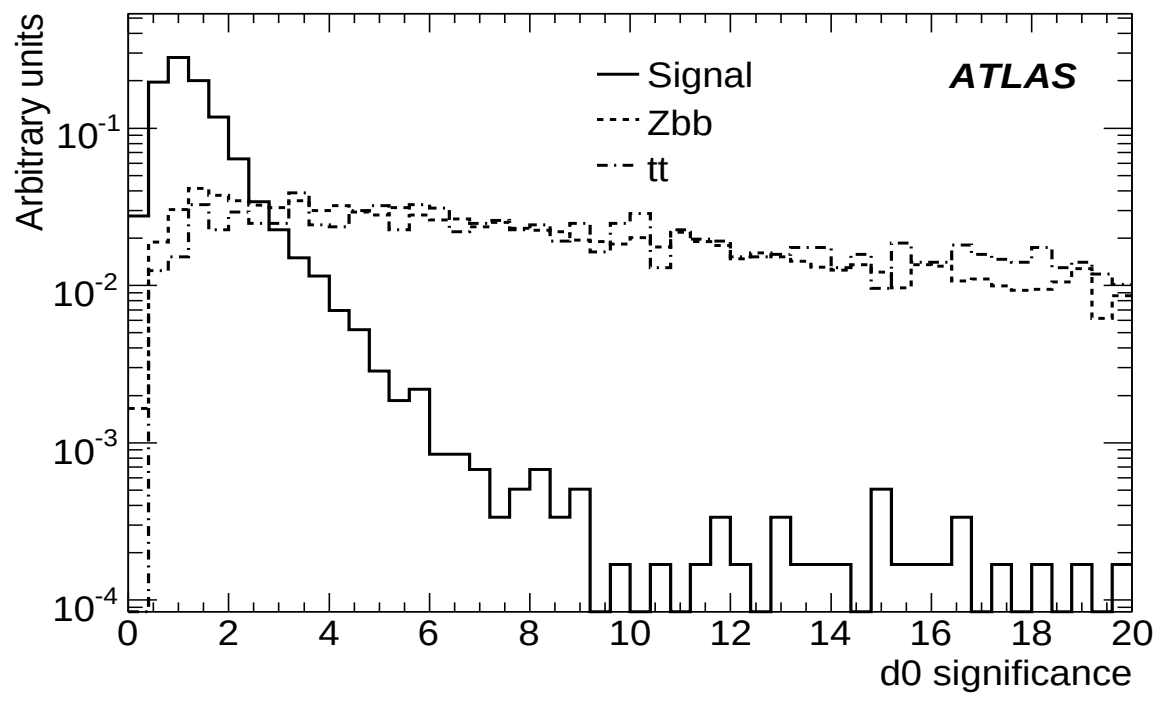


Figure 16: Maximum impact parameter significance in 4-muon events, for signal and reducible backgrounds.

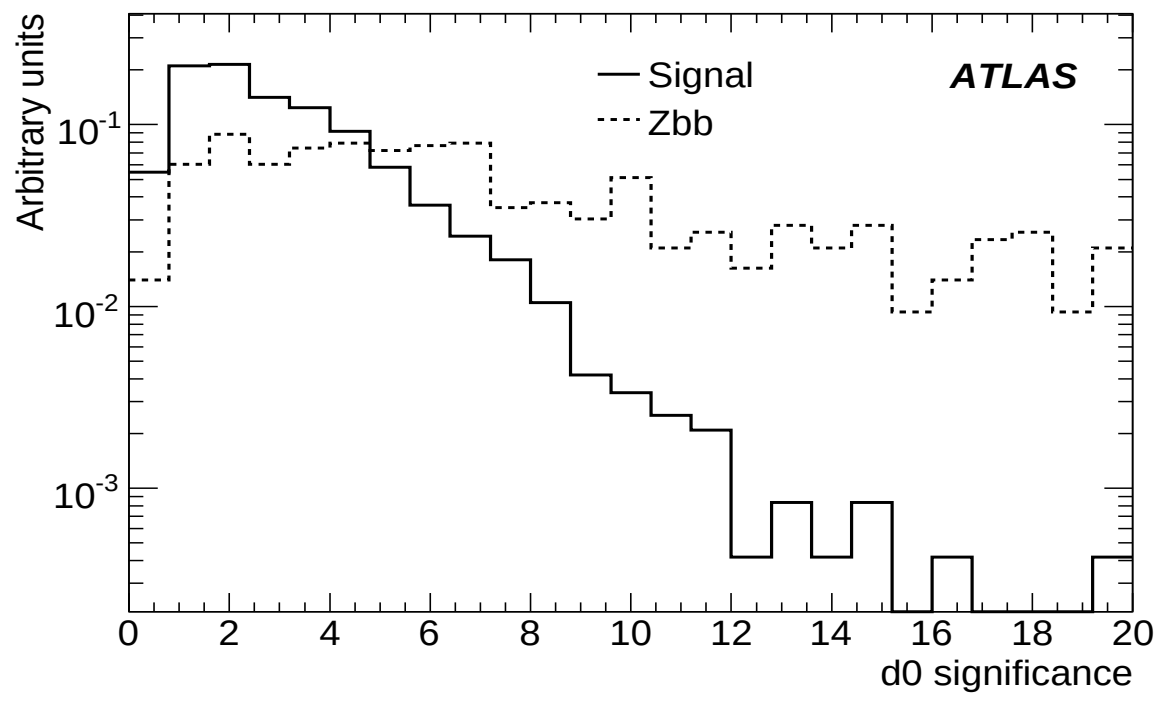


Figure 17: Maximum impact parameter significance in 4-electron events, for signal and reducible backgrounds.

| | |
|---|---|
| Event Preselection | Four leptons: *LooseElectrons* or muons.<br>$p_T$>7 GeVand $\|\eta\| < 2.5$, at least two with $p_T$>20 GeV |
| **Event Selection** | |
| Kinematic Cuts | Lepton quality: 2 pairs of same flavour opposite charge leptons.<br>Electrons must be *MediumElectrons* satisfying the *CaloIso* criterion.<br>For $H$ masses of 200 GeV and higher, four *LooseElectrons* are required instead.<br><br>Z, Z* and Higgs boson reconstruction: single quadruplet with<br>$\|m_{ll1} - m_Z\| < \Delta m_{12}$ GeV, $m_{ll2} > m_{34}$. |
| Isolation and vertexing cuts | Muon Calorimetric isolation ($\Sigma E_T/p_T < 0.23$).<br>Lepton Inner detector track isolation ($\Sigma p_T/p_T < 0.15$).<br>Cut on maximum lepton impact parameter<br>($d_0/\sigma_{d_0} < 3.5$ for muons, $d_0/\sigma_{d_0} < 6.0$ for electrons). |

Table 5: Summary of the analysis cuts for the H→ $4\ell$ analysis. The two lepton pairs are denoted as $m_{ll1}$ and $m_{ll2}$. The values of the mass window $\Delta m_{12}$ and of the cut $m_{34}$ are defined in Table 6.

GeV and $|\eta| < 2.5$. Muons are selected by requiring $p_T$>5 GeV and $|\eta| < 2.5$ (see section 2.3.2). The final stage of event preselection requires at least four leptons with $p_T$>7 GeVand $|\eta| < 2.5$, with at least of two these leptons having $p_T$>20 GeV.

## 4.2 Event selection and kinematic reconstruction

Following Table 5, in this section we review the kinematic cut part of the event selection. These cuts have been optimized separately for each Higgs boson mass considered in this note.

**Lepton quality requirements:**

Events used in this analysis are required to have at least four leptons (e, $\mu$) which can be coupled in pairs of opposite charge and same flavour. These leptons are required to satisfy the requirements:

- Electrons: for Higgs boson masses below 200 GeV, electrons are required to satisfy the *Medium-Electron* quality requirement, and the calorimetric isolation in a $\Delta R = 0.2$ cone (*CaloIsolation*). For masses above or equal to 200 GeV, due to the higher momentum of the decay electrons, the

*MediumElectron* requirement is relaxed and these electrons are required to satisfy the *LooseElectrons* quality (see Section 2.3.1). This is a conservative choice, due to the fact that complete studies on Z+jets background rejection were limited by the number of simulated events available.

- Muons are required to be either reconstructed by the combined reconstruction algorithm, or by the tagging algorithm which is matching tracks in the inner detector to hit patterns in the muon spectrometer (see Section 2.3.2).

**Z, $Z^{(*)}$ and Higgs boson mass reconstruction**

The 4-lepton Higgs boson candidate mass reconstruction proceeds after selecting one single lepton quadruplet in an event. When more than one quadruplet is found, the one with a dilepton mass closest to the nominal $Z$ mass, and with the highest $p_T$ leptons associated to the second $Z$, is chosen. The resolution of the dilepton mass can be improved by applying a $Z$-mass constraint to the pair with a mass closest to the $Z$ invariant mass. When both $Z$'s are on-shell (for Higgs boson masses of 200 GeV and above), the $Z$-mass constraint can be applied to both lepton pairs. The constraint itself is a convolution between the nominal $Z$ Breit-Wigner distribution, and a gaussian distribution centered at the measured $Z$ value with $\sigma$ equal to the experimental resolution. The distribution of the Higgs boson mass reconstructed in the case of a 130 GeV Higgs boson is shown in Figs. 18, 19 and 20 for the $4e$, $4\mu$ and $2e2\mu$ channels respectively. Only the gaussian region of the distribution is considered in the fit in the case of the $4e$ channel: the fraction of events falling within $\pm 2\sigma$ from the mean value is 81.7% in this case. The tail of the distribution is due to electron bremsstrahlung losses upstream the calorimeters.

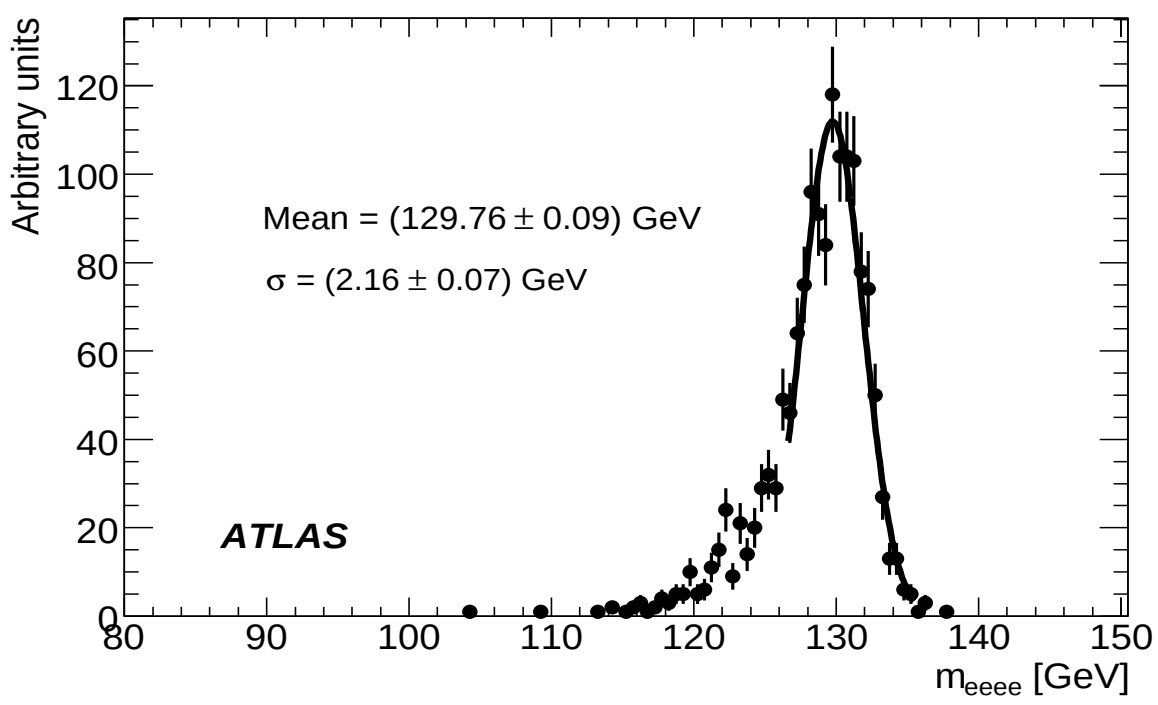


Figure 18: Reconstructed H(130 GeV)$\rightarrow 4e$ mass after application of the Z-mass constraint fit.

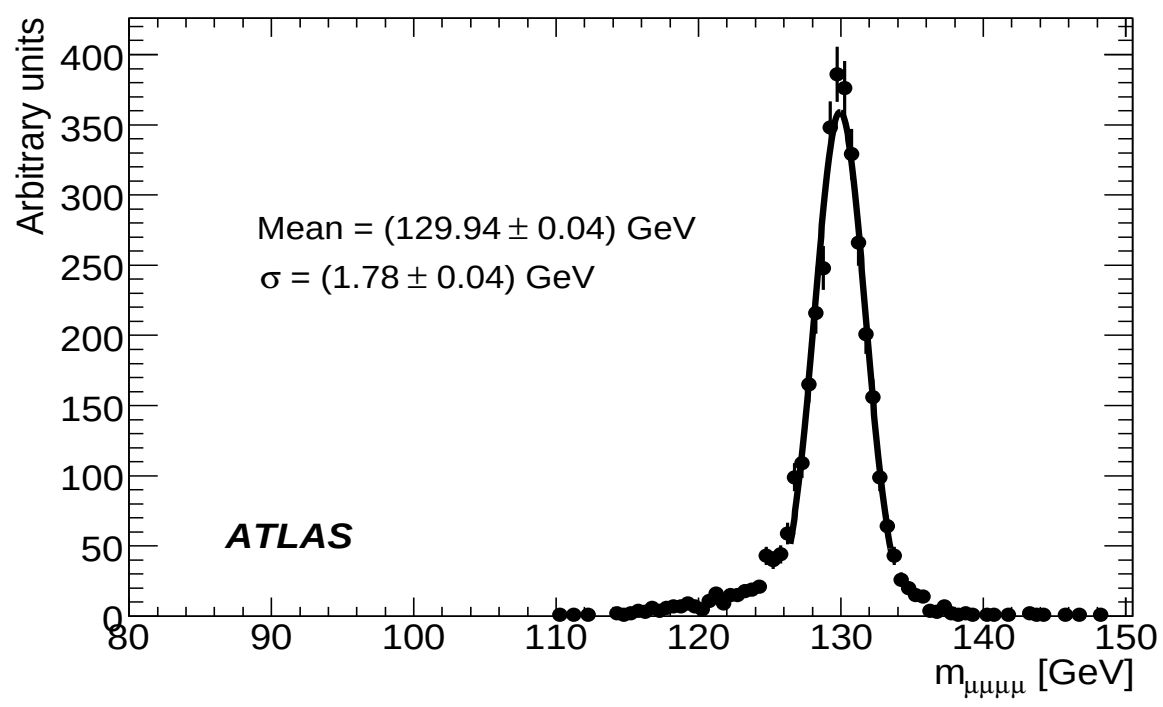


Figure 19: Reconstructed H(130 GeV)$\rightarrow 4\mu$ mass after application of the Z-mass constraint fit.

The Higgs boson mass resolution for masses for which the Higgs boson has a negligible intrinsic width, is shown in Figs. 21 and 22. In these figures it is shown that the Z-mass constraint improves the mass resolution by 10% to 17%. The 4-lepton mass shifts after the Z-mass constraint are shown in Fig. 23 for each of the three decay channels. Independently of the fit, the 4e mass is biased to lower values due to material effects (see Section 2.3.1). For this reason the electron energy has already been corrected by +1% based on the difference of the reconstructed $Z$ mass and the PDG $Z$ mass. The Z-mass constraint is only slightly correcting for these effects. In the $4\mu$ channel, the bias introduced by the Z-mass constraint is negligible.

Finally, the set of kinematic cuts applied to the reconstructed $Z$ invariant masses is shown in Table 6. The cuts have been optimized using the expected distributions for signal and backgrounds, and the expected dilepton resolution. In this table $Z_1$ is the dilepton with a mass closest to the nominal Z mass, while $Z_2$ is the lower mass dilepton pair. To estimate the significance, the events in a $\pm 2\sigma$ mass window are selected.

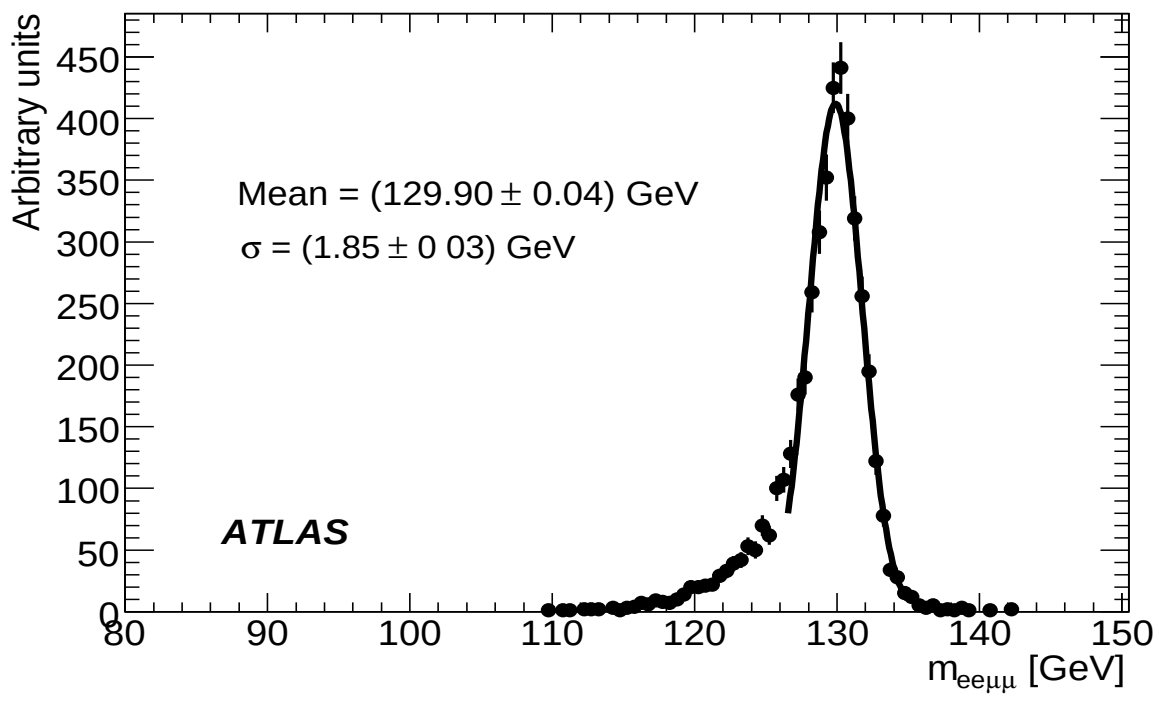


Figure 20: Reconstructed H(130 GeV)$\rightarrow 2e2\mu$ mass after application of the Z-mass constraint fit.

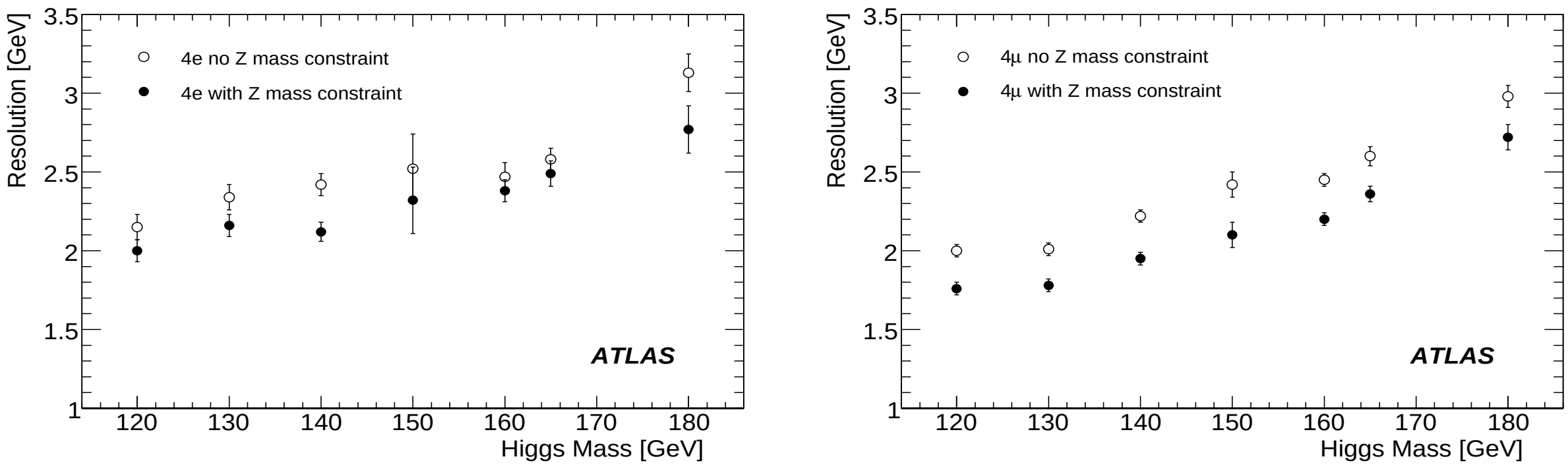


Figure 21: Higgs$\rightarrow 4e$ mass resolution as a function of the Higgs boson mass. Open circles denote the resolution obtained when no Z-mass constraint is applied, while full circles show the resolution in the case of the Z-mass constraint.

Figure 22: Higgs$\rightarrow 4\mu$ mass resolution as a function of the Higgs boson mass. Open circles denote the resolution obtained when no Z-mass constraint is applied, while full circles show the resolution in the case of the Z-mass constraint.

## 4.3 Event selection results

The cut flow for the selection of a 130 GeV Higgs boson is shown in Tables 7 and 8, for signal and backgrounds respectively. The same is shown for the $t\bar{t}$ background in Table 9. In this case, the available MC statistics is not sufficient to determine the number of expected events, and only upper limits at 90% CL are set. The final selection efficiencies are shown in Figs. 24 and 25 The distributions of the reconstructed 4-lepton mass, obtained after all cuts, are shown in Figs. 26, 27, 28 for three of the low-mass values (130, 150, and 180 GeV), and in Figs. 29, 30, and 31 for three of the high-mass ones (300, 400, and 600 GeV).

The number of expected events shown for the three decay channels combined is computed using NLO cross-sections (Section 2.1), for a luminosity of 30 fb$^{-1}$. The NLO cross-sections after the full event selection, are shown in Table 10 for the three decay channels separately and combined. In this table, signal events are selected within a $m_H \pm 2\sigma_{m_H}$ mass window, and systematic errors are not yet taken into account in the significance calculation. Here $\sigma_{m_H}$ is the experimental 4-lepton mass resolution, shown in Table 6. The significances for the each of the three channels, and their combination, are summarized in Fig. 32.

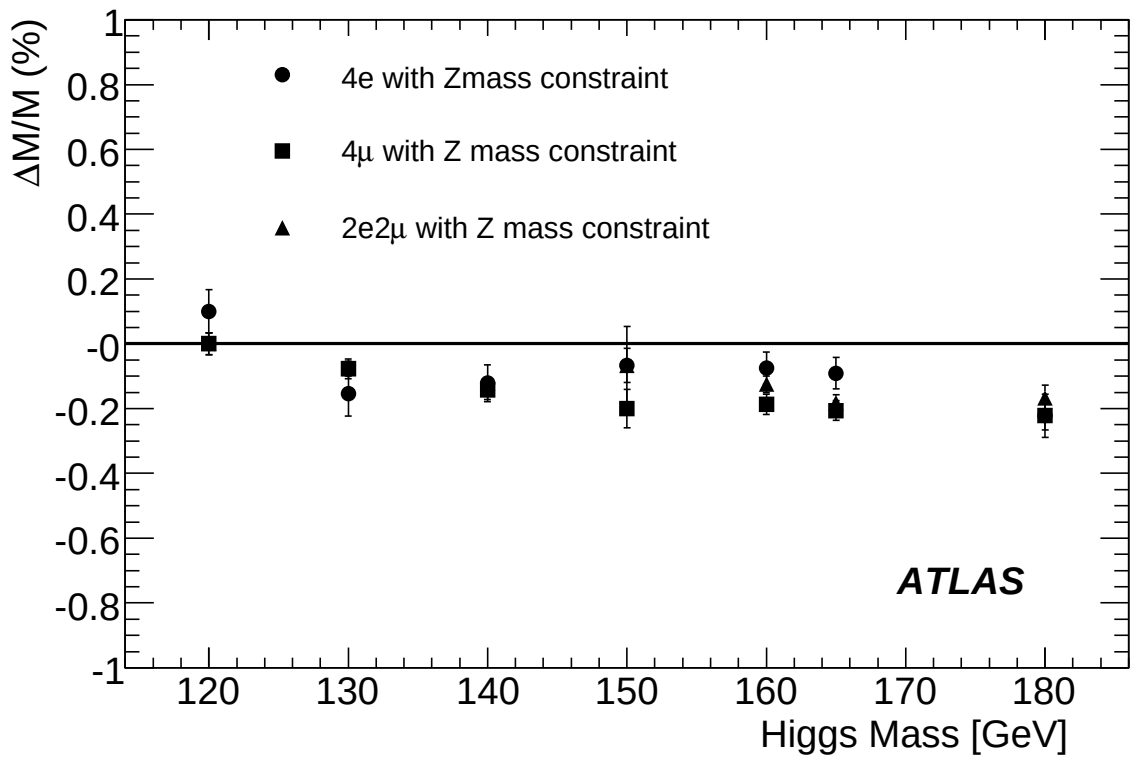


Figure 23: Shift of the mean 4-lepton mass for each of the three decay channels, with and without the Z-mass constraint on the dilepton mass.

| H Mass (GeV) | $Z_1$ mass window (GeV) | $Z_2$ mass cut (GeV) | H mass resolution (GeV) 4e | 4μ | 2e2μ |
|---|---|---|---|---|---|
| 120 | ±15 | >15 | 2.0 | 1.8 | 1.9 |
| 130 | ±15 | >20 | 2.2 | 1.8 | 1.9 |
| 140 | ±15 | >30 | 2.2 | 2.0 | 2.1 |
| 150 | ±15 | >30 | 2.3 | 2.1 | 2.2 |
| 160 | ±15 | >30 | 2.4 | 2.2 | 2.3 |
| 165 | ±15 | >35 | 2.5 | 2.4 | 2.4 |
| 180 | ±12 | >40 | 2.8 | 2.7 | 2.8 |
| 200 | ±12 | >60 | 3.9 | 3.7 | 3.8 |
| 300 | ±12 | ±12 | 8.4 | 8.4 | 8.4 |
| 400 | ±12 | ±12 | 16.5 | 17.3 | 17.2 |
| 500 | ±12 | ±12 | 33.8 | 34.4 | 32.8 |
| 600 | ±12 | ±12 | 52.2 | 57.2 | 53.2 |

Table 6: Cuts applied to the reconstructed leading and sub-leading $Z$ masses, and the Higgs boson mass resolution values used to define the signal region.

## 4.4 Estimates of WZ and Z+Jets backgrounds

The contribution of other potentially dangerous backgrounds have also been estimated. Examples are the WZ$\rightarrow 3\ell$ and the inclusive $Z+X$ where the $Z$ decays leptonically. In the WZ$\rightarrow 3\ell$ background pileup and cavern background are included. WZ is found to give a negligible contribution to the total background: the upper limit on the expected number of events is lower than the limit on $t\bar{t}$ events. The $Z$+jets process provides one of the most serious backgrounds to the 4e-channel at low masses. With the available MC statistics (500K events, see Table 2), no event survives the lepton quality and $p_T$ selection cuts. An estimate based on cut factorization leads to an evaluation of the expected number of events after the full event selection, that would correspond to one event after the initial lepton quality and $p_T$ cuts. This is at the level of about twice the Zbb background, i.e. below the $ZZ^*$continuum. The available MC statistics is therefore not sufficient to set stringent limits on the $Z$+jets background. However, the rejection based on the lepton quality and $p_T$ cuts should guarantee the complete removal of this background.

| Selection cut | Selection step | Signal | | |
|---|---|---|---|---|
| | | 4e | $4\mu$ | $2e2\mu$ |
| Trigger selection | 1 | 94.7 | 95.3 | 95.7 |
| Lepton preselection | 2 | 57.0 | 73.8 | 66.8 |
| Lepton quality and $p_T$ | 3 | 24.7 | 60.5 | 39.7 |
| Z's mass cuts | 4 | 17.1 | 42.9 | 27.6 |
| Calo Isolation | 5 | 17.1 | 39.5 | 25.4 |
| Tracker Isolation | 6 | 16.5 | 38.1 | 24.7 |
| IP cut | 7 | 15.1 | 36.5 | 23.2 |
| H Mass cut | 8 | 12.5±0.3 | 31.4±0.5 | 19.2±0.4 |

Table 7: Fraction of events (in %) selected after each event selection cut, for each of the three decay channels, and for a 130 GeV Higgs boson. The efficiencies of each selection are calculated with respect to the fraction of events in which the Higgs boson decays into the corresponding channel, and that pass the generator filter described in Section 2.1.

| Selection cut | | ZZ | | | Zbb | | |
|---|---|---|---|---|---|---|---|
| | | 4e | $4\mu$ | $2e2\mu$ | 4e | $4\mu$ | $2e2\mu$ |
| Trigger | 1 | 96.6 | 96.6 | 96.6 | 91.4 | 91.4 | 91.4 |
| Preselection | 2 | 13.8 | 17.6 | 31.4 | 2.6 | 9.4 | 12.0 |
| Lepton quality and $p_T$ | 3 | 7.3 | 16.0 | 21.9 | $1.1 \cdot 10^{-1}$ | 2.1 | 1.7 |
| Z mass cuts | 4 | 6.9 | 14.8 | 20.2 | $4.7 \cdot 10^{-2}$ | 1.1 | $8.4 \cdot 10^{-1}$ |
| Calo Isolation | 5 | 6.9 | 13.9 | 19.5 | $4.7 \cdot 10^{-2}$ | $8.5 \cdot 10^{-2}$ | $1.2 \cdot 10^{-1}$ |
| Track Isolation | 6 | 6.8 | 13.6 | 19.2 | $1.3 \cdot 10^{-2}$ | $3.3 \cdot 10^{-2}$ | $4.4 \cdot 10^{-2}$ |
| IP cut | 7 | 6.2 | 13.0 | 17.8 | $5.6 \cdot 10^{-3}$ | $1.1 \cdot 10^{-2}$ | $1.8 \cdot 10^{-2}$ |
| H Mass window | 8 | $5.2 \cdot 10^{-2}$ | $11.3 \cdot 10^{-2}$ | $12.0 \cdot 10^{-2}$ | $1.6 \cdot 10^{-3}$ | $1.2 \cdot 10^{-3}$ | $3.0 \cdot 10^{-3}$ |

Table 8: Fraction of events (in %) selected after each event selection cut for the background processes. The 130 GeV Higgs boson mass selection cuts are applied.

### 4.5 Effects of pile-up and cavern background

The effects of pile-up and cavern background in the analysis are studied for the analysis for the Higgs boson mass of 130 GeV . The signal selection efficiencies are shown in Fig. 33, for the $4\mu$ and $4e$ channel, as a function of the selection cuts listed in Table 7. Table 12 summarizes the results for the three decay channels. The effect of pileup is to decrease the signal selection efficiency by about 10%, for all three decay channels. This decrease is due to a slight decrease in the trigger efficiency, and of the calorimetric and tracker isolation cut efficiencies. Part of the loss can be recovered by reoptimizing these cuts.

## 5 Systematic uncertainties

Central to the H$\rightarrow 4\ell$ analysis is the estimate of the background in a candidate signal region. In this section the systematic uncertainties on quantities associated with the background estimation and the signal efficiency are discussed. First the theoretical uncertainties are presented. Subsequently, the impact of experimental systematic uncertainties on the event selection is discussed. Some of the systematics, like the theoretical uncertainties and systematics on the signal efficiency, affect the estimate of the expected sensitivity. Signal significance extraction from real data is affected by systematics on the knowledge of

| Selection cut | | $t\bar{t}$ | | |
|---|---|---|---|---|
| | | 4e | $4\mu$ | $2e2\mu$ |
| Trigger | 1 | 75.1 | 75.1 | 75.1 |
| Preselection | 2 | 1.0 | 4.7 | 10.1 |
| Lepton quality and $p_T$ | 3 | $6.8{\cdot}10^{-3}$ | $7.3{\cdot}10^{-1}$ | $5.8{\cdot}10^{-1}$ |
| Z mass cuts | 4 | $1.6{\cdot}10^{-3}$ | $2.0{\cdot}10^{-1}$ | $1.0{\cdot}10^{-1}$ |
| Calo Isolation | 5 | $1.6{\cdot}10^{-3}$ | $1.6{\cdot}10^{-3}$ | $5.4{\cdot}10^{-3}$ |
| Track Isolation | 6 | $2.6{\cdot}10^{-4}$ | $2.5{\cdot}10^{-4}$ | $1.0{\cdot}10^{-3}$ |
| IP cut | 7 | $2.6{\cdot}10^{-4}$ | $< 6 \cdot 10^{-4}$ | $2.6{\cdot}10^{-4}$ |
| H Mass window | 8 | $< 6 \cdot 10^{-4}$ | $< 6 \cdot 10^{-4}$ | $< 6 \cdot 10^{-4}$ |

Table 9: Fraction of events (in %) selected after each event selection cut for the $t\bar{t}$ background. For small available statistics 90% CL limits are considered.

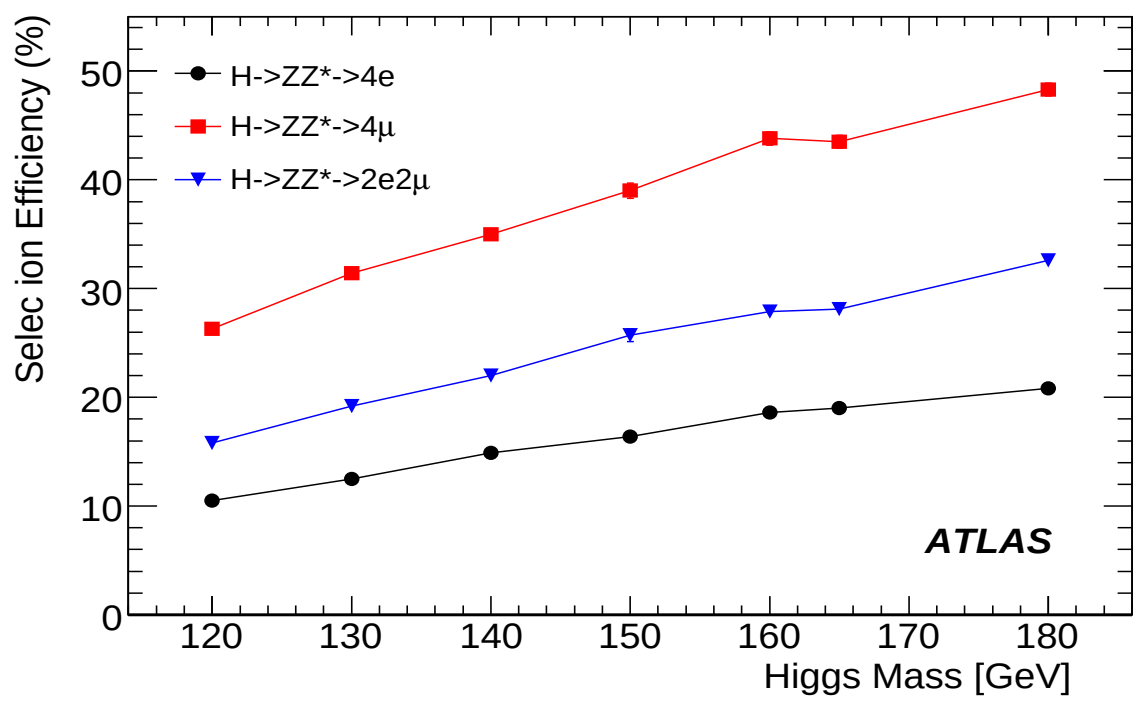


Figure 24: Selection efficiency as a function of the Higgs boson mass, for each of the three decay channels, for the case of only one on-shell Z.

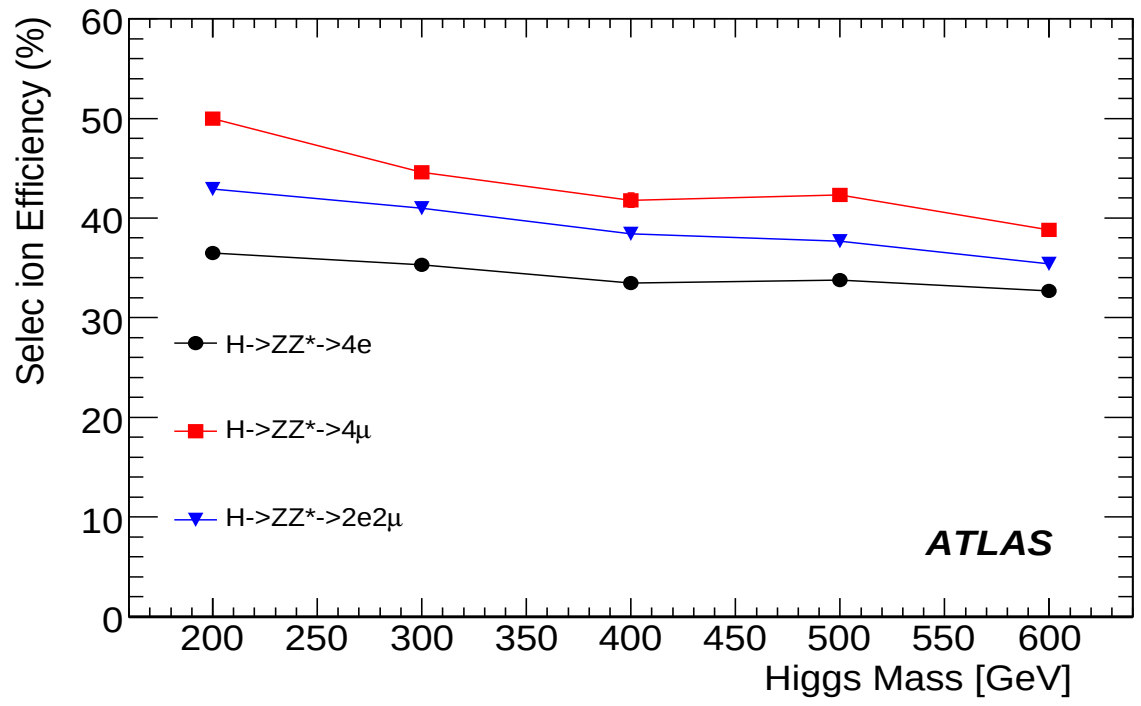


Figure 25: Selection efficiency for as a function of the Higgs boson mass, for each of the three decay channels, for the case of two on-shell Z's.

lepton energy scale and resolution, as determined for example from inclusive $Z$ studies, lepton reconstruction efficiency, and reducible background knowledge from control samples.

## 5.1 Theoretical uncertainties

The major theoretical uncertainties in the prediction of the inclusive background cross-sections are the PDF uncertainties and uncertainties related to the QCD renormalization and factorization scales. Scale uncertainties reflect theoretical uncertainties due to the omission of higher order diagrams. PDF and scale uncertainties have already been discussed and evaluated for the main backgrounds and are only recalled here (see Section 2.1). In summary, for the calculation of the NLO inclusive cross-sections, the QCD scales have been independently varied in the range (0.5-2)$\times$ the energy scale of the process. The PDF uncertainty has been evaluated by making use of 40 sets of PDF's for CTEQ6M (20 plus and 20 minus).

## 5.2 Experimental uncertainties

Systematic effects on the H$\rightarrow 4\ell$ analysis arise from experimental uncertainties related to the lepton reconstruction. The major contributions in the total systematic uncertainty in the H$\rightarrow 4\ell$ yield come from

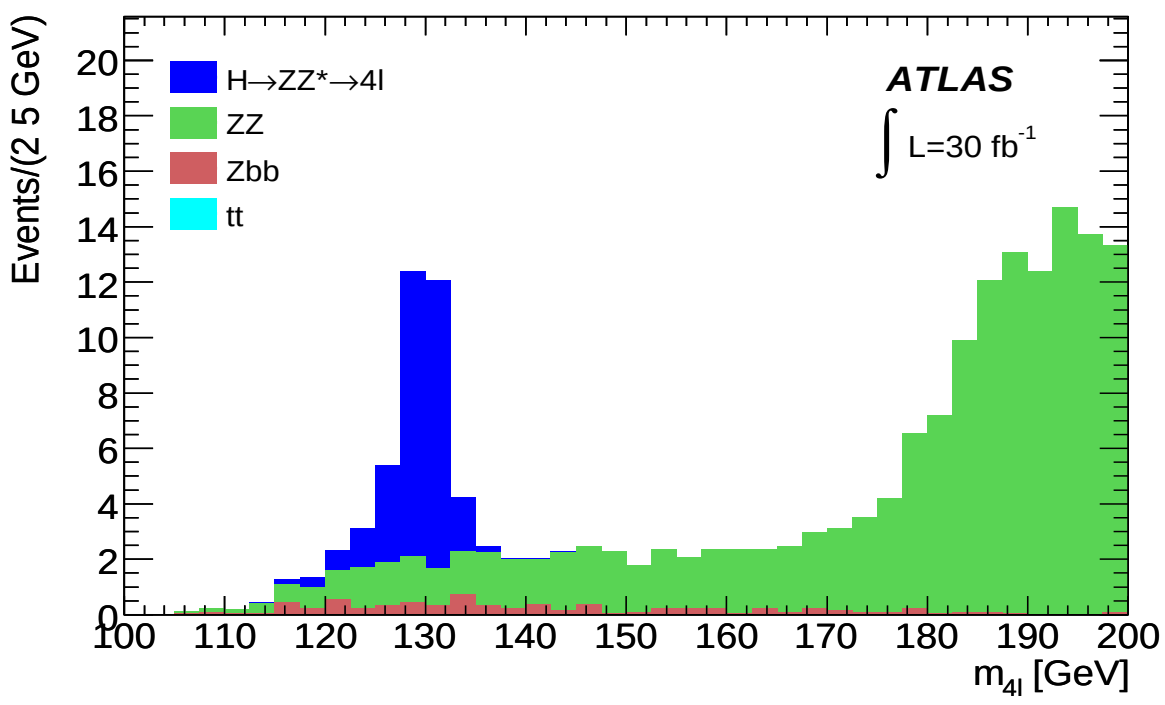


Figure 26: Reconstructed 4-lepton mass for signal and background processes, in the case of a 130 GeV Higgs boson, normalized to a luminosity of 30 fb$^{-1}$.

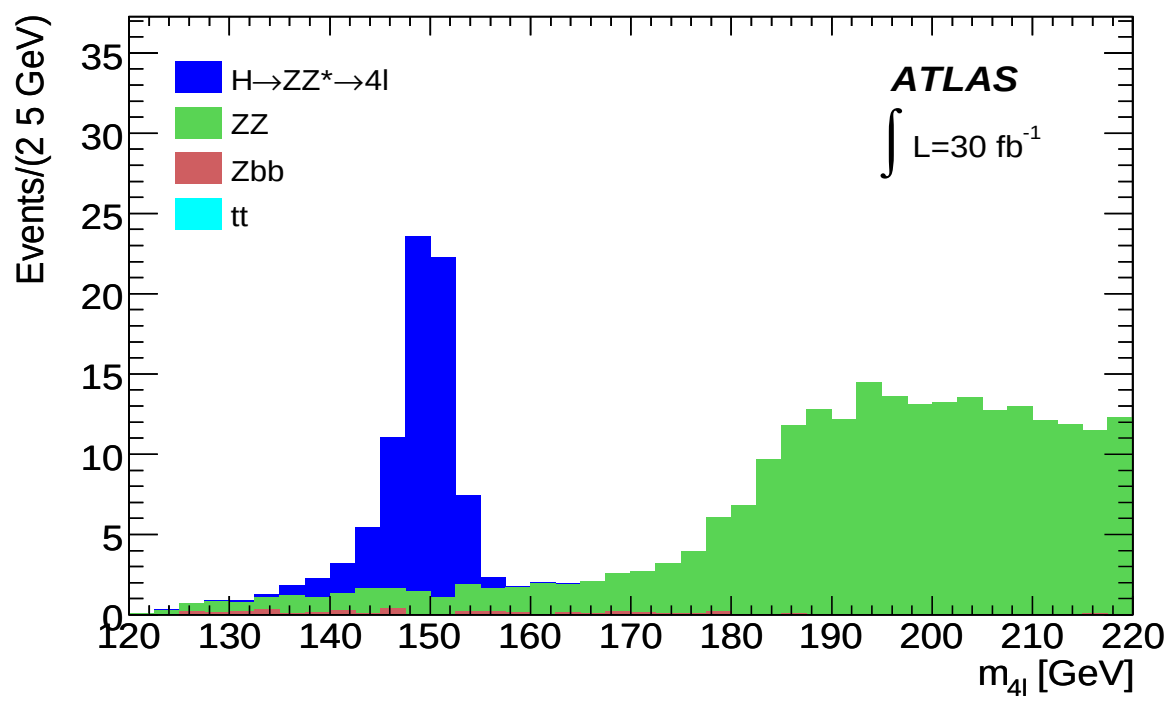


Figure 27: Reconstructed 4-lepton mass for signal and background processes, in the case of a 150 GeV Higgs boson, normalized to a luminosity of 30 fb$^{-1}$.

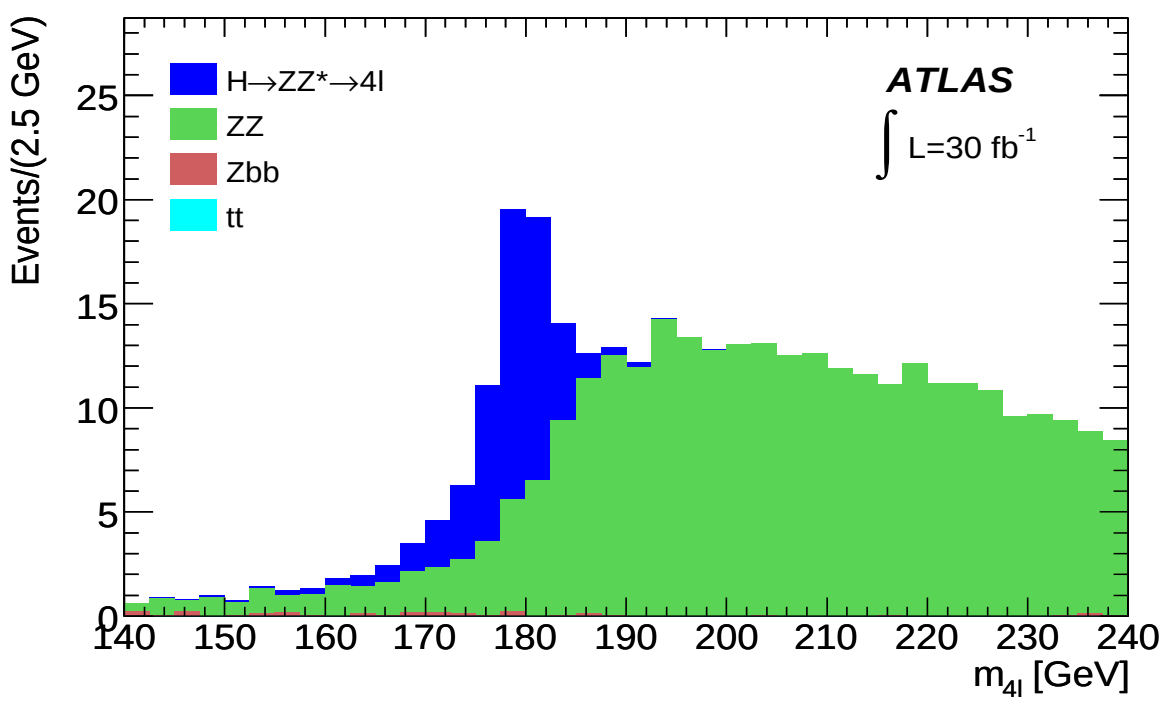


Figure 28: Reconstructed 4-lepton mass for signal and background processes, in the case of a 180 GeV Higgs boson, normalized to a luminosity of 30 fb$^{-1}$.

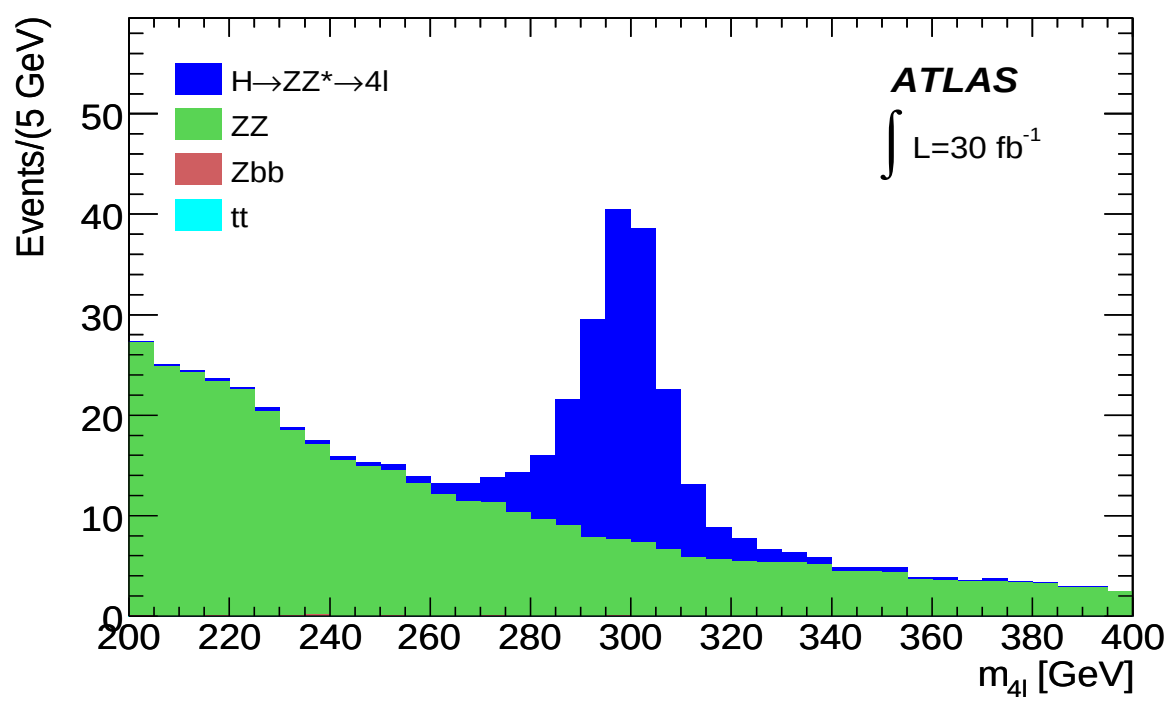


Figure 29: Reconstructed 4-lepton mass for signal and background processes, in the case of a 300 GeV Higgs boson, normalized to a luminosity of 30 fb$^{-1}$.

uncertainties in lepton energy scale, reconstruction and identification efficiency. The impact of these uncertainties on the analysis is studied by applying variations to offline reconstructed variables. The level of these variations has been provided by the performance groups.

**Uncertainties in lepton energy scale**

Uncertainties on the energy scale of electrons arise from the EM calibration. These are considered by varying by $\pm 0.5\%$ the $E_T$ of the reconstructed electrons. Energy scale uncertainties for muons arise due to the imperfect knowledge of the magnetic field. Here the recostructed muon $p_T$ is varied by $\pm 1\%$. These values are assumed on the basis of the foreseen in-situ determination of the detector performance.

**Uncertainties in lepton energy resolution**

The level of knowledge of the material distributions in ATLAS affects the lepton energy reconstruction. To properly evaluate the impact of this contribution on the analysis, the reconstructed electron energies are smeared with a Gauss function using a $\sigma_{E_T} = 0.0073 \cdot E_T$. This extra smearing deteriorates the transverse energy resolution of 50 GeV electrons by a relative 10%. In the muon system, an additional term can be added to this smearing, to take into account misalignment uncertainties. The total muon smearing is $\sigma_{1/p_T} = 0.011/p_T \oplus 0.00017$ (with $p_T$ in GeV). In the $p_T$ range of interest for Higgs boson searches, the second term is negligible. The values of the corrections described above have been chosen so that the

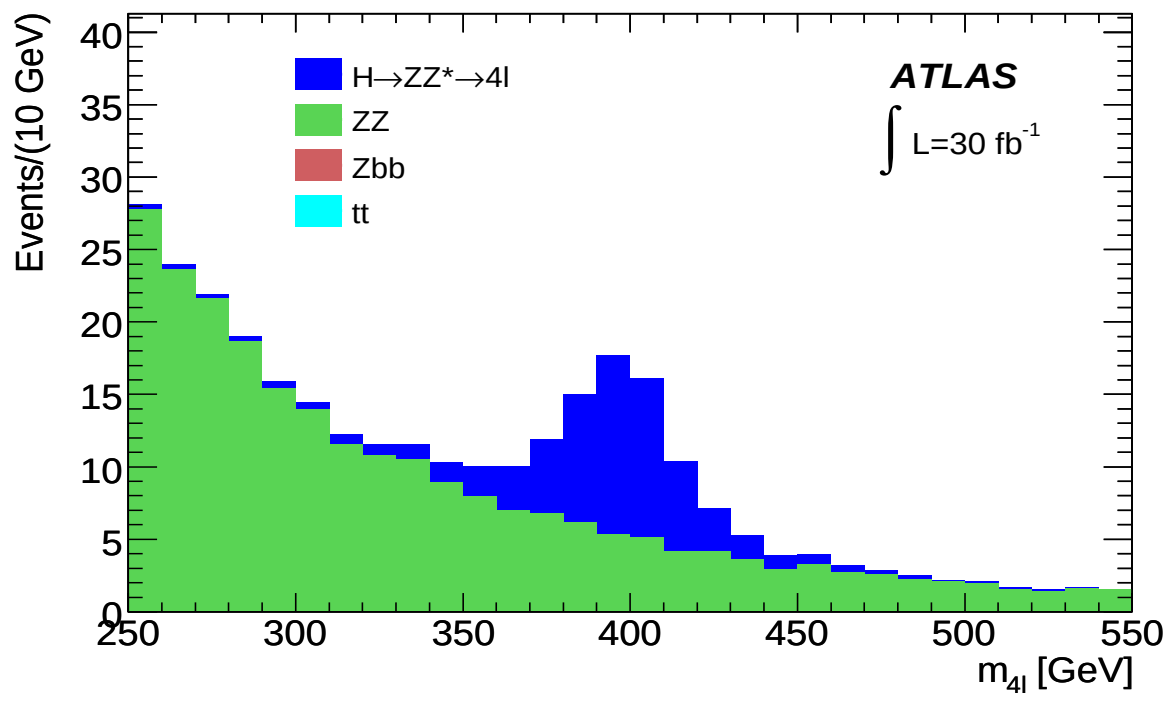


Figure 30: Reconstructed 4-lepton mass for signal and background processes, in the case of a 400 GeV Higgs boson, normalized to a luminosity of 30 fb$^{-1}$.

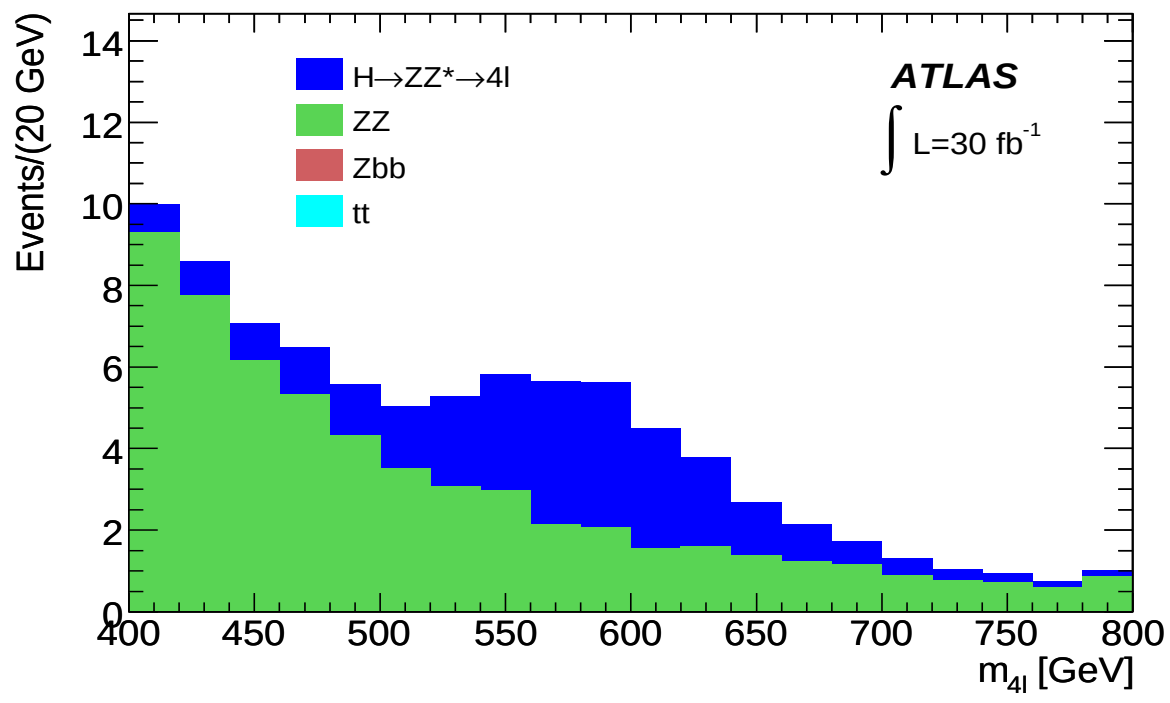


Figure 31: Reconstructed 4-lepton mass for signal and background processes, in the case of a 600 GeV Higgs boson, normalized to a luminosity of 30 fb$^{-1}$.

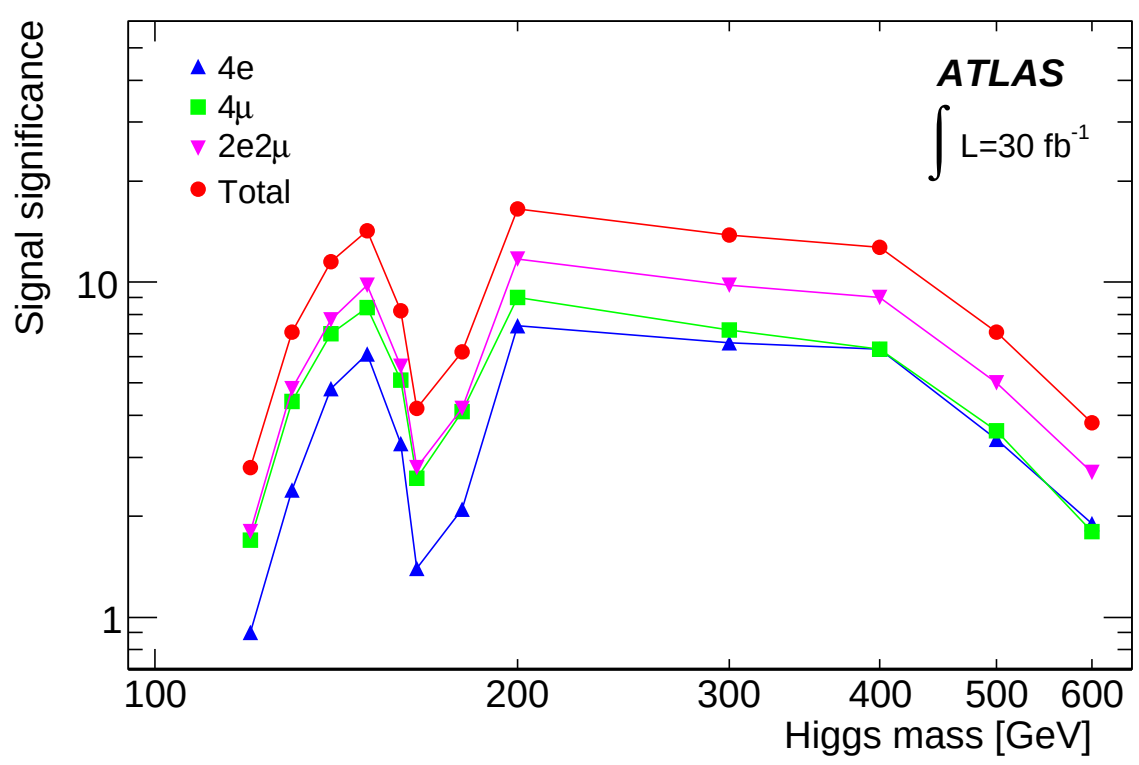


Figure 32: Expected signal significances computed using Poisson statistics, for each of the three decay channels, and their combination.

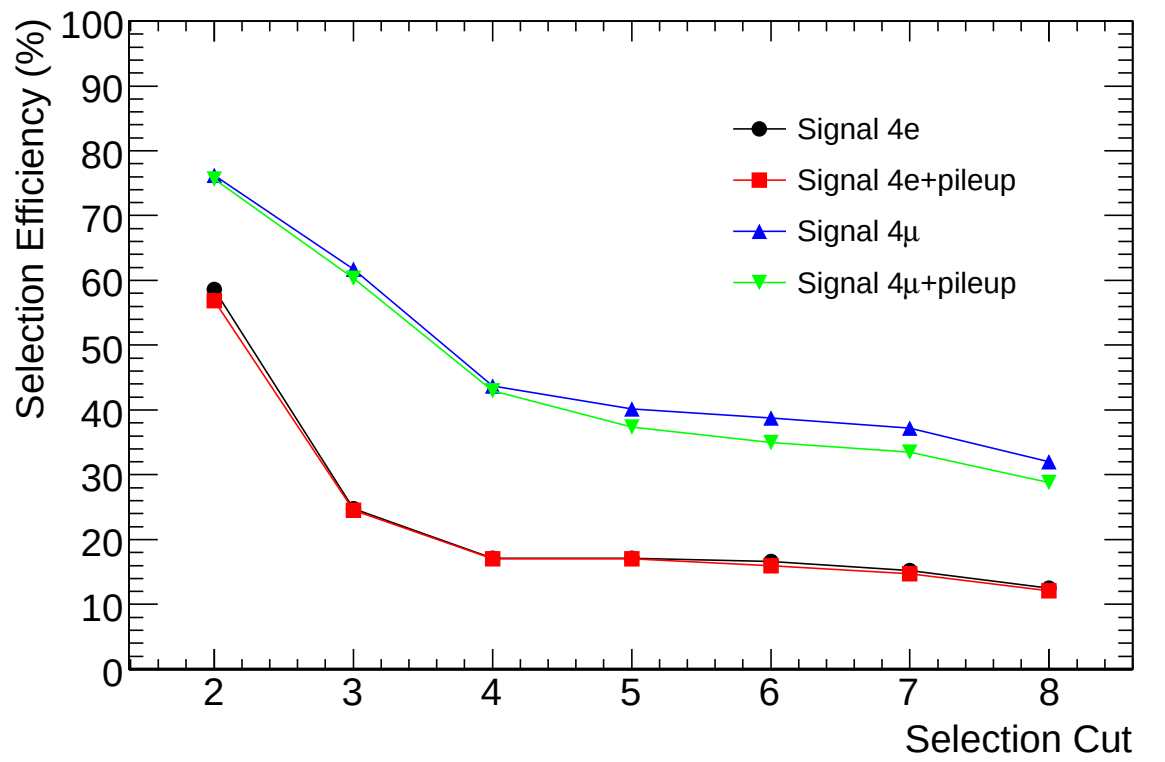


Figure 33: Fraction of selected events with and without pile-up and cavern background, for the cuts in Tables 7 and 8 (130 GeV H→$4e$ and $4\mu$ analyses).

nominal resolution on muon $p_T$ of 3% is increased to 3.3% after applying the extra smearing.

**Uncertainties in lepton reconstruction efficiency**

The impact of uncertainties in the lepton reconstruction efficiency can be estimated by discarding a fixed fraction of leptons before the analysis. The level of uncertainties considered here is 0.2% for electrons and 1% for muons, motivated by performance group studies.

**Material effects in electron efficiency**

Uncertainties in electron efficiency receive a large contribution from uncertainties in the knowledge of material upstream the LAr EMC. Systematic effects influence shower shape discriminants included in the electron identification criteria. Examples of such discriminants are the mean energy fraction in a core of 3×7 middle sampling cells normalized to a window of 7×7 cells, and the mean energy fraction outside a 3-strip core and inside a 7-strip window. As discussed in [17], the presence of extra material shifts and changes the shapes of these distributions, hence reducing the discrimination power of these cuts. The integrated effect in electron efficiency is rather small (less than 2%). However the true systematic uncertainty in the efficiency due to the knowledge of the material depends on how well the material and

| **Mass (GeV)** | | **120** | **130** | **140** | **150** | **160** | **165** | **180** |
|---|---|---|---|---|---|---|---|---|
| **Selection** | | | | | | | | |
| $4e$ | Signal | 0.043 | 0.124 | 0.239 | 0.297 | 0.162 | 0.078 | 0.205 |
| | $ZZ^*/\gamma^*$ | 0.028 | 0.027 | 0.020 | 0.017 | 0.033 | 0.044 | 0.196 |
| | Zbb | 0.006 | 0.013 | 0.004 | 0.004 | $< 0.004$ | 0.006 | 0.002 |
| | $t\bar{t}$ | <0.04 | <0.04 | <0.04 | <0.04 | <0.04 | <0.04 | <0.04 |
| | Significance (30 $fb^{-1}$) | 0.9 | 2.4 | 4.8 | 6.1 | 3.3 | 1.4 | 2.1 |
| $4\mu$ | Signal | 0.108 | 0.311 | 0.563 | 0.707 | 0.381 | 0.177 | 0.476 |
| | $ZZ^*/\gamma^*$ | 0.052 | 0.059 | 0.061 | 0.017 | 0.073 | 0.080 | 0.258 |
| | Zbb | 0.019 | 0.009 | 0.009 | 0.008 | 0.006 | 0.004 | 0.008 |
| | $t\bar{t}$ | <0.04 | <0.04 | <0.04 | <0.04 | <0.04 | <0.04 | <0.04 |
| | Significance (30 $fb^{-1}$) | 1.7 | 4.4 | 7.0 | 8.4 | 5.1 | 2.6 | 4.1 |
| $2e2\mu$ | Signal | 0.130 | 0.381 | 0.709 | 0.932 | 0.485 | 0.229 | 0.642 |
| | $ZZ^*/\gamma^*$ | 0.063 | 0.063 | 0.081 | 0.074 | 0.102 | 0.116 | 0.483 |
| | Zbb | 0.030 | 0.025 | 0.013 | 0.009 | 0.009 | 0.004 | 0.004 |
| | $t\bar{t}$ | <0.04 | <0.04 | <0.04 | <0.04 | <0.04 | <0.04 | <0.04 |
| | Significance (30 $fb^{-1}$) | 1.8 | 4.8 | 7.7 | 9.8 | 5.6 | 2.8 | 4.2 |
| All | Signal | 0.281 | 0.816 | 1.511 | 1.94 | 1.03 | 0.484 | 1.32 |
| | $ZZ^*/\gamma^*$ | 0.143 | 0.150 | 0.163 | 0.151 | 0.208 | 0.240 | 0.938 |
| | Zbb | 0.055 | 0.047 | 0.026 | 0.021 | 0.015 | 0.013 | 0.013 |
| | $t\bar{t}$ | <0.04 | <0.04 | <0.04 | <0.04 | <0.04 | <0.04 | <0.04 |
| | Significance (30 $fb^{-1}$) | 2.8 | 7.1 | 11.5 | 14.2 | 8.2 | 4.2 | 6.2 |

Table 10: Cross-sections (in fb) after the full event selection for the signal and irreducible and reducible backgrounds. The cross-sections are given for each of three channels $4e$ ,$4\mu$ and $2e2\mu$ , and for their combination. When no event is passing the event selection, 90% C.L. limits on the cross-section are set. For each channel and for their combination, the expected significance is given for 30 $fb^{-1}$. It is assumed that the background in the signal region is known with negligible uncertainty. The $t\bar{t}$background is assumed not to contribute to the signficance,

the shower shapes can be measured using data.

### 5.3 Summary of the systematic uncertainties

The impact of the various lepton systematic uncertainties on the Higgs boson signal yield and background rejection is summarized in Table 13 for the various $4\ell$ final states. Variations have been applied to both signal and background samples, and the analysis algorithm ran with the hypothesis of $m_H = 130$ GeV. A 3% uncertainty on the luminosity has been taken into account in the total systematic error calculation. The total systematic error on the signal efficiency has been included in the calculation of the exclusion limits described in the following.

## 6 Background extraction from data and significance estimation

In Section 4 the significance has been obtained assuming that the background is known with a negligible uncertainty. In this section various methods to extract the background from data, evaluate the background uncertainties, and include them in the significance calculation, are presented.

| **Mass (GeV)** | | **200** | **300** | **400** | **500** | **600** |
|---|---|---|---|---|---|---|
| **Selection** | | | | | | |
| $4e$ | Signal | 1.41 | 0.917 | 0.737 | 0.370 | 0.174 |
| | $ZZ^*/\gamma^*$ | 0.689 | 0.342 | 0.225 | 0.228 | 0.179 |
| | Significance (30 fb$^{-1}$) | 7.4 | 6.6 | 6.3 | 3.4 | 1.9 |
| $4\mu$ | Signal | 1.94 | 1.16 | 0.918 | 0.463 | 0.206 |
| | $ZZ^*/\gamma^*$ | 0.867 | 0.468 | 0.379 | 0.346 | 0.296 |
| | Significance (30 fb$^{-1}$) | 9.0 | 7.2 | 6.3 | 3.6 | 1.8 |
| $2e2\mu$ | Signal | 3.33 | 2.13 | 1.69 | 0.825 | 0.377 |
| | $ZZ^*/\gamma^*$ | 1.53 | 0.836 | 0.610 | 0.570 | 0.438 |
| | Significance (30 fb$^{-1}$) | 11.7 | 9.8 | 9.0 | 5.0 | 2.7 |
| All | Signal | 6.68 | 4.21 | 3.34 | 1.66 | 0.76 |
| | $ZZ^*/\gamma^*$ | 3.09 | 1.65 | 1.21 | 1.14 | 0.914 |
| | Significance (30 fb$^{-1}$) | 16.5 | 13.8 | 12.7 | 7.1 | 3.8 |

Table 11: Cross-sections (in fb) after the full event selection for signal and irreducible background. For Higgs boson masses above 180 GeV the contribution of the reducible backgrounds to the total background cross-section is negligible. The cross-sections are given for each of three channels $4e$, $4\mu$ and $2e2\mu$ , and for their combination. When no event is passing the event selection, 90% C.L. limits on the cross-section are set. For each channel and for their combination, the expected significance is given for 30 fb$^{-1}$. It is assumed that the background in the signal region is known with negligible uncertainty.

| | 4e | $4\mu$ | $2e2\mu$ |
|---|---|---|---|
| No Pileup and CB | 12.5±0.3 | 31.4±0.5 | 19.2±0.4 |
| With Pileup and CB | 12.1±0.3 | 28.4±0.5 | 17.0±0.4 |

Table 12: Selection efficiencies in %, after all cuts, for the signal at 130 GeV, and for each of the three decay channels. The efficiencies are shown for the two cases: with and without the addition of minimum bias low-luminosity pileup and cavern background with safety factor 5.

The main challenge in measuring the background for the $4\ell$ channel over a very wide mass range (from $m_{4\ell}$ 120 GeVto $m_{4\ell}$ 600 GeV), comes from the fact that signal and background shapes and cross-sections vary considerably. The dominant background comes from the $ZZ^{(*)}$ continuum. While we expect to measure the $ZZ$ in the high mass region ($M_{4\ell} > 180$ GeV) where the reducible backgrounds become negligible, in the low mass region the significant presence of $Zb\bar{b}$ and $t\bar{t}$ requires their knowledge. Measurements of these backgrounds with early data will provide upper bounds in their expectation after all Higgs boson analysis cuts.

### 6.1 Significance determination

Most of the systematic uncertainties discussed in the previous section do not contribute to the significance determination if the data distributions are fitted, and signal and background are extracted from the fit. The resulting uncertainty in the knowledge of the background reduces the confidence level for claiming a discovery. In this section a fit-based approach for background and significance extraction is presented. A fit of the selected 4-lepton invariant mass, using the signal hypothesis at a fixed mass and applying the profile likelihood ratio method to extract the significance is considered. The fit method, and

| | $Zbb$ | $ZZ$ | $H$ | $Zbb$ | $ZZ$ | $H$ | $Zbb$ | $ZZ$ | $H$ |
|---|---|---|---|---|---|---|---|---|---|
| | $4e$ | | | $4\mu$ | | | $2e2\mu$ | | |
| Scale +0.5% (+1%) | +1.5 | +0.1 | +0.9 | +2.4 | +0.4 | +1.3 | +1.9 | +0.1 | +0.9 |
| Scale -0.5% (-1%) | -1.1 | -0.2 | -0.5 | -2.3 | -0.3 | -2.5 | -1.7 | -0.2 | -1.4 |
| Resolution | -0.5 | -0.1 | -0.4 | +0.1 | -0.1 | -2.6 | -0.2 | -0.1 | -0.5 |
| Rec. efficiency | -1.0 | -0.7 | -0.5 | -3.8 | -4.0 | -3.8 | -2.0 | -2.1 | -1.7 |
| Luminosity | 3 | | | 3 | | | 3 | | |
| Total | 3.6 | 3.1 | 3.2 | 5.4 | 5.0 | 6.0 | 4.1 | 3.7 | 3.8 |

Table 13: Impact, in %, of the systematic uncertainties on the overall selection efficiency, as obtained for a $m_H = 130$ GeVin the $4e$,$4\mu$, and $2e2\mu$ final states.

the results obtained after varying the background shape, the fit range and even including background-only fits without the signal hypothesis, are the main subject of this section. An alternative approach with a global two-dimensional (2D) fit on the $(m_{Z^*}, m_{4\ell})$ plane is also briefly discussed.

The baseline method used as input to the combination of all ATLAS Standard Model Higgs boson searches, is based on a fit of the 4-lepton invariant mass distribution over the full range from 110 to 700 GeV. The method, whose details are described in [21], is summarized below.

The 4-lepton reconstructed invariant mass after the full event selection (except the final cut on the 4-lepton reconstructed mass) is used as a discriminating variable to construct a likelihood function. The likelihood is calculated on the basis of parametric forms of signal and background probability density functions (pdf) determined from the MC. For a given set of data, the likelihood is a function of the pdf parameters $\vec{p}$ and of an additional parameter $\mu$ defined as the ratio of the signal cross-section to the Standard Model expectation (i.e. $\mu = 0$ means no signal, and $\mu = 1$ corresponds to the signal rate expected for the Standard Model). To test a hypothesized value of $\mu$ the following likelihood ratio is constructed:

$$\lambda(\mu) = \frac{L(\mu, \hat{\hat{\vec{p}}})}{L(\hat{\mu}, \hat{\vec{p}})} \quad (1)$$

where $\hat{\hat{\vec{p}}}$ is the set of pdf parameters that maximize the likelihood L for the analysed dataset and for a fixed value of $\mu$ (conditional Maximum Likelihood Estimators), and $(\hat{\mu}, \hat{\vec{p}})$ are the values of $\mu$ and $\vec{p}$ that maximise the likelihood function for the same dataset (Maximum Likelihood Estimators). The profile likelihood ratio is used to reject the background only hypothesis ($\mu = 0$) in the case of discovery, and the signal+background hypothesis in the case of exclusion. The test statistic used is $q_\mu = -2\ln\lambda(\mu)$, and the median discovery significance and limits are approximated using expected signal and background distributions, for different $m_H$, luminosities and signal strength $\mu$. The MC distributions, with the content and error of each bin reweighted to a given luminosity (in the following referred to as "Asimov data", see [21]), are fitted to derive the pdf parameters: in the fit, $m_H$ is fixed to its true value, while $\sigma_H$ is allowed to float in a $\pm 20\%$ range around the value obtained from the signal MC distributions. All parameters describing the background shape are floating within sensible ranges. The irreducible background has been modelled using a combination of Fermi functions which are suitable to describe both the plateau in the low mass region and the broad peak corresponding to the second $Z$ coming *on shell*. The chosen model is described by the following function:

$$f(m_{ZZ}) = \frac{p0}{(1+e^{\frac{p6-m_{ZZ}}{p7}})(1+e^{\frac{m_{ZZ}-p8}{p9}})} + \frac{p1}{(1+e^{\frac{p2-m_{ZZ}}{p3}})(1+e^{\frac{p4-m_{ZZ}}{p5}})} \quad (2)$$

The first plateau, in the region where only one of the two $Z$ bosons is *on shell*, is modelled by the first term, and its suppression, needed for a correct description at higher masses, is controlled by the $p8$ and $p9$ parameters. The second term in the above formula accounts for the shape of the broad peak and the tail at high masses. This function can describe with a negligible bias the $ZZ$ background shape with good accuracy over the full mass range. The $Zb\bar{b}$ contribution is relevant to the background shape only when searching for very light Higgs boson (in this study, only at $m_H = 120$ GeV). In this case, an additional term is added to the $ZZ$ continuum, with a functional form similar to the second part of equation 2. For the signal modelling a simple gaussian shape has been used for $m_H \leq 300$ GeV, while a relativistic Breit-Wigner formula is needed at higher values of the Higgs boson mass. In Figs. 34 and 35 two examples of pseudo-experiments with the resulting fit functions for signal and background are shown.

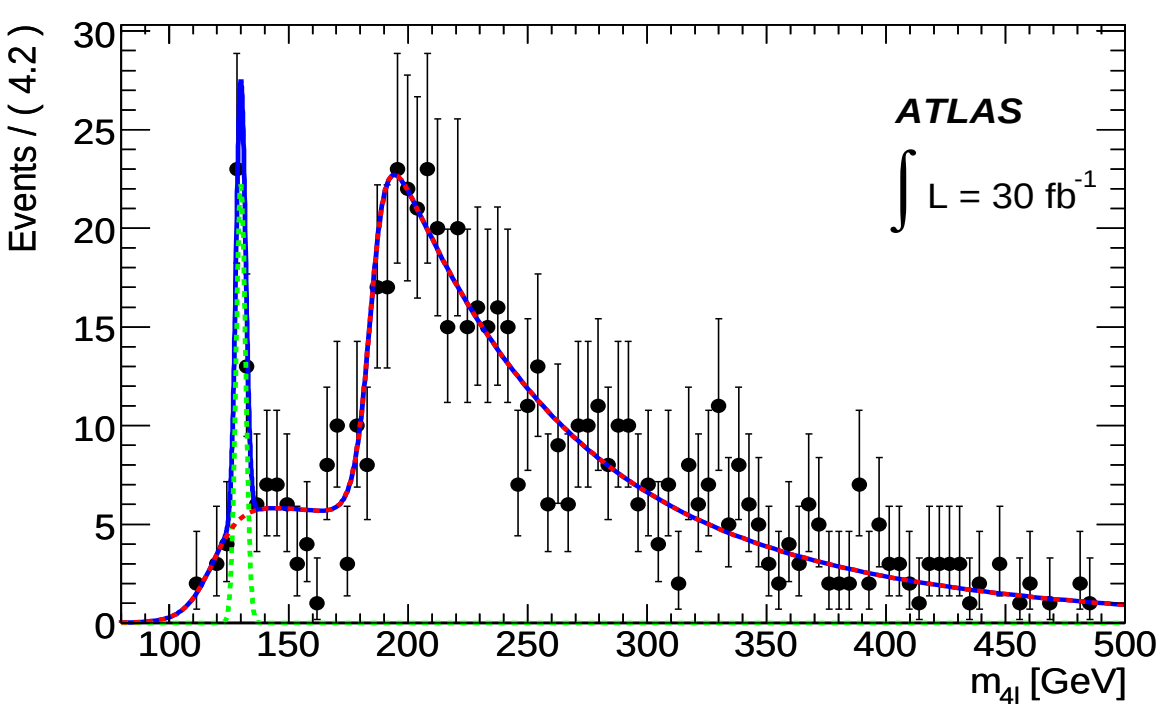


Figure 34: A pseudo-experiment corresponding to 30 fb$^{-1}$ of data for a Higgs boson mass of 130 GeV. The functions fitting the signal and the background are shown.

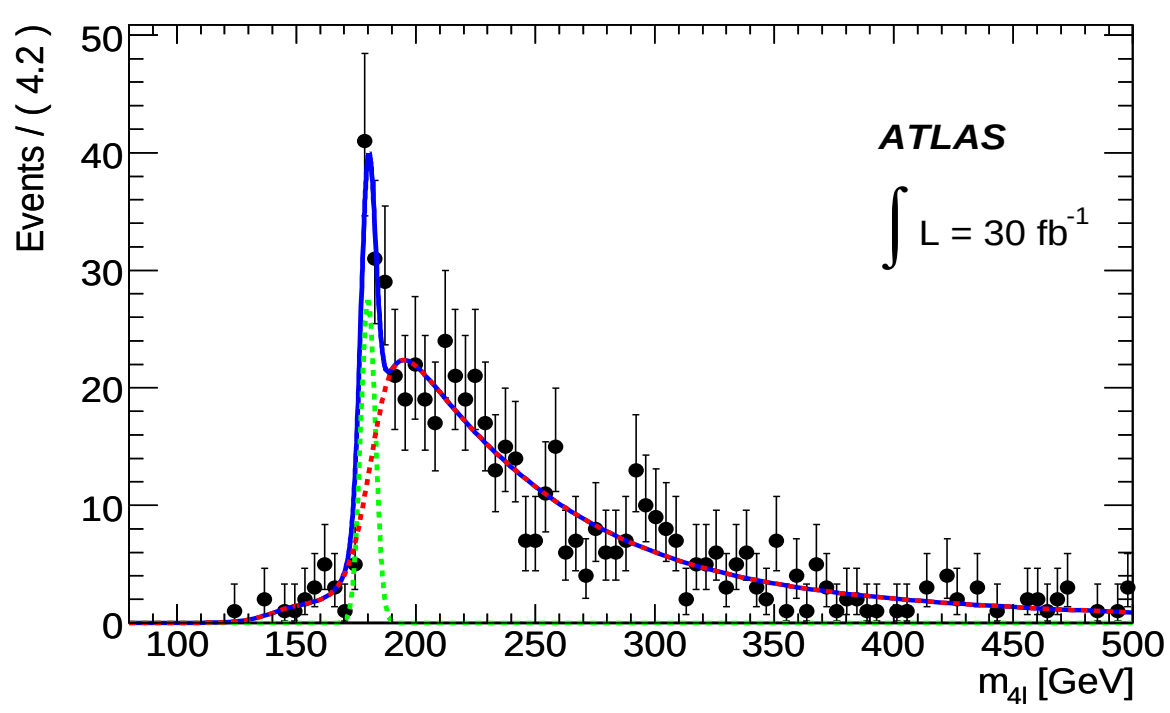


Figure 35: A pseudo-experiment corresponding to 30 fb$^{-1}$ of data for a Higgs boson mass of 180 GeV. The functions fitting the signal and the background are shown.

The results presented in the following approximate the significance from the test statistics as $\sqrt{-2\ln\lambda(\mu)}$. In order for the results of the method to be valid, the test statistic $q_\mu = -2ln\lambda(\mu)$ should be distributed as a $\chi^2$ with one degree of freedom. The results obtained with the strategy described above must thus be validated using toy MC. Such validation tests show a good agreement of the test statistic with the expected $\chi^2$ distribution, as discussed in detail in [21]. This allows to approximate the significance from the test statistic as $\sqrt{-2\ln\lambda(\mu)}$. The significances obtained as the square root of the median profile likelihood ratios for discovery, $-2\ln\lambda(\mu = 0)$ are shown in Table 14 for all $m_H$ values considered in this paper, and for various luminosities. In Fig. 36, the significance obtained from the profile likelihood ratio, after the fit of signal+background is shown. The significance is compared to the Poisson significance shown in Section 4. The slightly reduced discovery potential is due to the fact that several background shape and normalization parameters are derived from the data-like sample.

Concerning exclusion, the median profile likelihood ratios are calculated under the background only hypothesis, and the integrated luminosity needed to exclude the signal at 95% C.L. is the one corresponding to $\sqrt{-2ln\lambda}$=1.64. The integrated luminosity needed for exclusion is shown in Fig. 37.

The median significance estimation with Asimov data can be validated using toy MC pseudo-experiments. For each mass point, 3000 background-only pseudo-experiments are generated. For each experiment, the profile likelihood ratio method is used to find which $\mu$ value can be excluded at 95% CL. The resulting distributions are then analysed to find the median and $\pm1\sigma$ and $\pm2\sigma$ intervals. The outcome of this test is summarized in Fig. 38, where the 95% CL exclusion $\mu$ obtained from single fits on the full MC datasets is plotted as well. As shown, the agreement is good over the full mass range. Fitting the 4-lepton mass distribution with all parameters left free in function 2 allows the fit to absorb

| L | $m_H$ (GeV) | | | | | | | | | | | |
|---|---|---|---|---|---|---|---|---|---|---|---|---|
| ($fb^{-1}$) | 120 | 130 | 140 | 150 | 160 | 165 | 180 | 200 | 300 | 400 | 500 | 600 |
| 1 | 0.5 | 1.1 | 2.0 | 2.3 | 1.3 | 0.7 | 0.9 | 2.6 | 2.3 | 1.9 | 0.9 | 0.6 |
| 2 | 0.7 | 1.6 | 2.8 | 3.3 | 1.8 | 1.0 | 1.3 | 3.7 | 3.2 | 2.8 | 1.3 | 0.8 |
| 5 | 1.0 | 2.4 | 4.5 | 5.2 | 2.9 | 1.6 | 2.1 | 5.9 | 5.1 | 4.2 | 2.1 | 1.2 |
| 10 | 1.5 | 3.5 | 6.3 | 7.3 | 4.1 | 2.2 | 2.9 | 8.3 | 7.2 | 6.0 | 2.9 | 1.8 |
| 30 | 2.6 | 6.0 | 10.9 | 12.7 | 7.0 | 3.8 | 5.1 | 14.4 | 12.7 | 10.4 | 5.3 | 3.3 |

Table 14: The significances obtained from the median profile likelihood ratios for discovery -2ln$\lambda(\mu = 0)$, for all Higgs boson masses considered and for various luminosities.

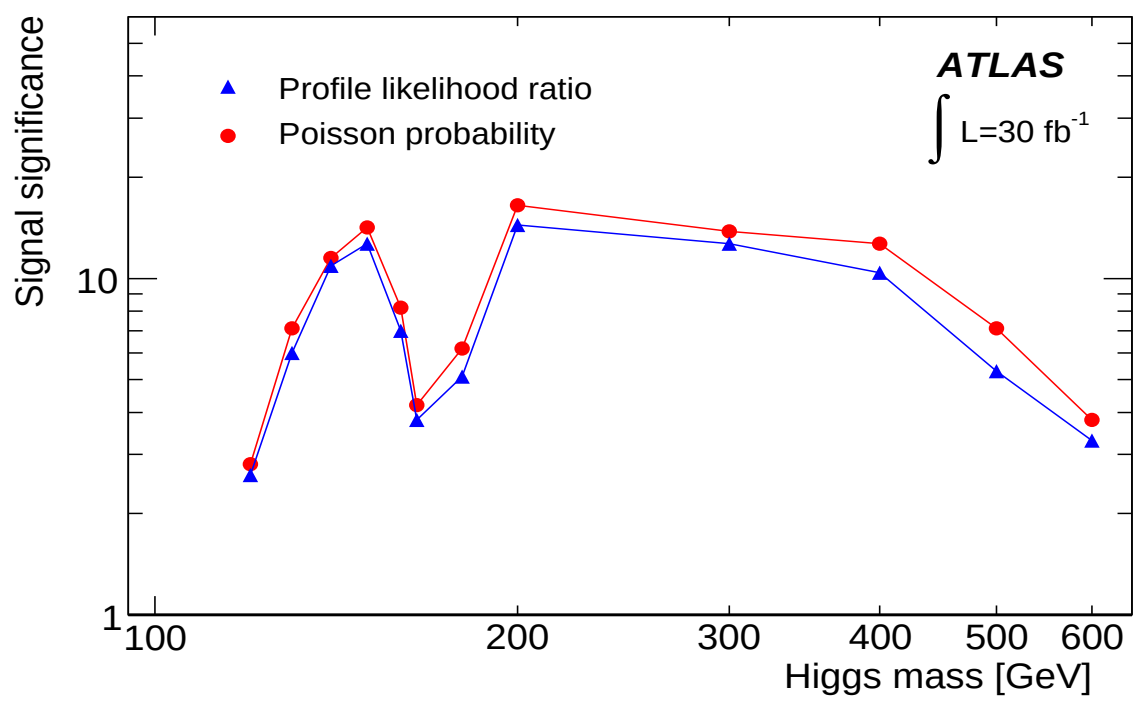


Figure 36: Significance obtained from the profile likelihood ratio, as a function of the Higgs boson mass. The result is compared with the one shown in Section 4 where systematic errors on signal and background have not been included, and the significance has been calculated using Poisson statistics.

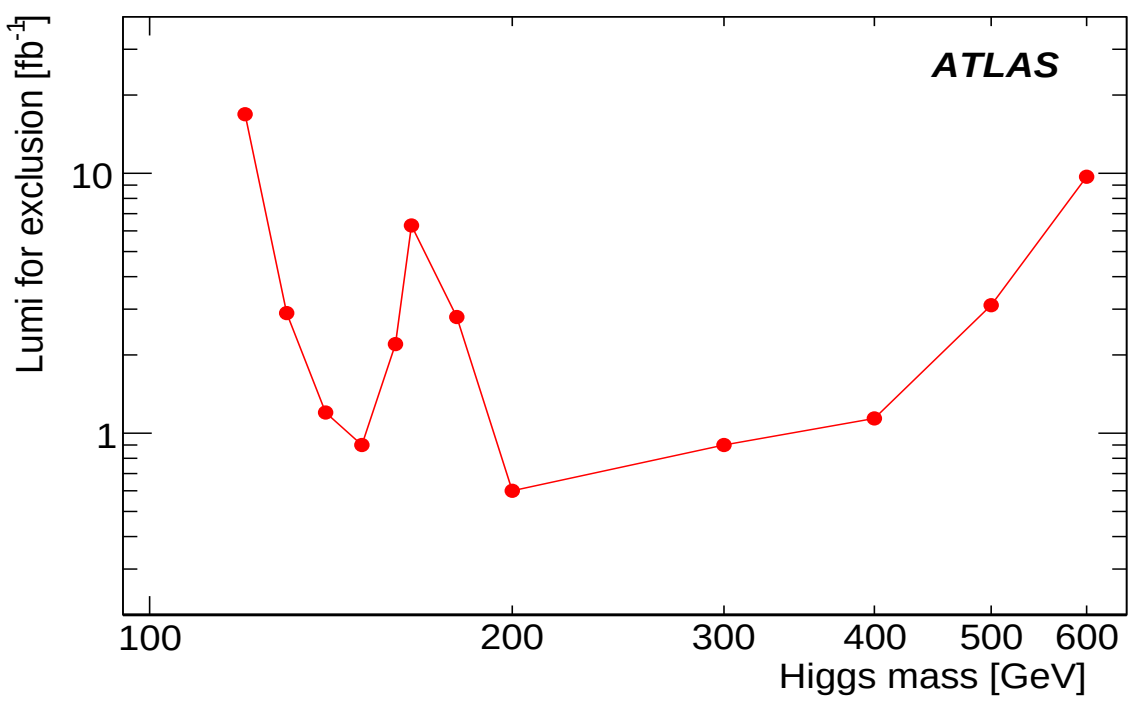


Figure 37: The luminosity needed for exclusion of the Standard Model Higgs boson with the $H \rightarrow ZZ^* \rightarrow 4l$ channel alone, as a function of the Higgs boson mass.

possible systematics. Inclusion of systematic effects on mass scale and resolution (see Section 5), and on the knowledge of the reducible background, have a total effect smaller than 4% on the significance from the fit.

In addition to the baseline method described above, different fit assumptions have been studied and are discussed below for completeness.

A background-only fit can be performed using pseudo-experiments corresponding to 30 $fb^{-1}$ of data, including both signal and backgrounds; the function described in formula 2 is used to describe the background, and the fit is performed over the full range 110-700 GeV, but in this case all the parameters apart from the overall normalization are fixed to the values obtained from a fit of the full MC statistics available. The signal mass window, defined for each value of the Higgs boson mass as described in Section 4, is excluded from the fit (sideband-fit), and the expected background in the signal region is obtained as the integral of the background function in the signal window. The pseudo-experiments fits provide an estimate of background statistical fluctuations and of possible biases introduced by the fit, that must be taken into account in the significance determination. The statistical uncertainty on the background resulting from the fit is no more than 6% over the full mass range.

Another method based on the sidebands consists of calculating, using the full MC statistics, the ratio $\tau = \frac{B_{MW}}{B_{SB}}$ between the background events in the signal mass window ($B_{MW}$) and those in the sidebands ($B_{SB}$). In the pseudo-experiments, the background level in the signal mass window is estimated as: $B_{MW}(i) = \tau \times N^{obs}_{SB}(i)$ and the signal as: $S_{MW}(i) = N^{obs}_{MW}(i) - B_{MW}(i)$. The resulting uncertainty on the

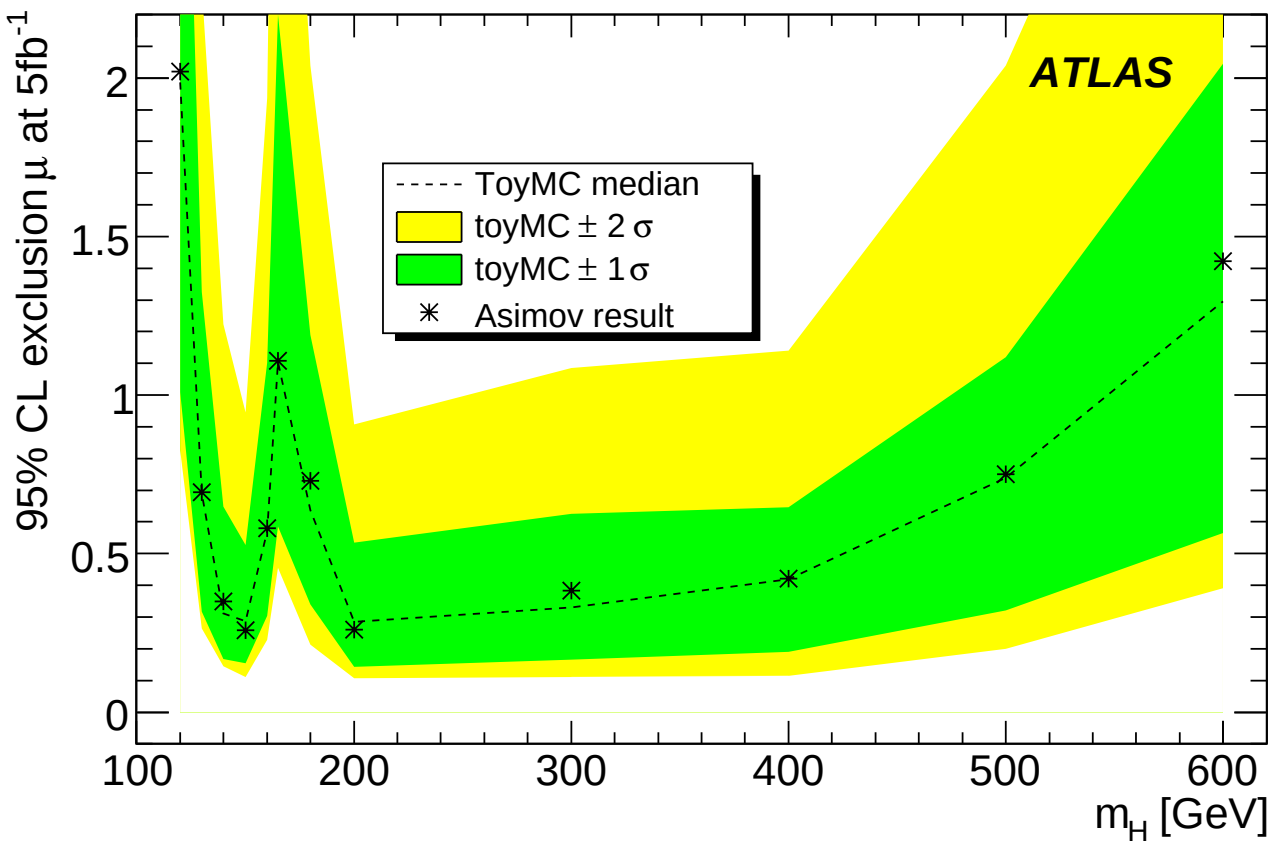


Figure 38: Validation of the median significance estimation with toy Monte-Carlo experiments. The $\mu$ value corresponding to 95% CL exclusion obtained from Asimov data is compared to the one obtained using the median profile likelihood ratio from the toy MC pseudo-experiments.

background is no more than 5% over the full mass range. The significance is then obtained, in both cases, using the profile likelihood ratio method applied to the case of a counting experiment, assuming the background mean, and statistical error on the background as from the fit. The results for 130, 150 and 180 GeV Higgs boson mass are summarized in Table 15. The sideband fit and the $\tau$-ratio method

| Method | Background error | 130 GeV | 150 | 180 |
|---|---|---|---|---|
| Sideband | from fit | 6.6 | 14.0 | 5.9 |
| $\tau$-ratio | from $\tau$-ratio | 6.7 | 14.0 | 5.8 |

Table 15: The signal significance for various Higgs boson masses, obtained including the background uncertainties from the sideband fit and the $\tau$-ratio methods.

can be used to test the impact of some systematic effects, like the mass scale and resolution in the $ZZ$ threshold region at 180 GeV, and the uncertainties on the reducible $Zb\bar{b}$ background. The impact of a $\pm 1\%$ variation of the energy scale on the significance from the fit with all parameters fixed apart from the normalization is of 21% for a Higgs boson mass of 180 GeV, the most critical region. A $\pm 20\%$ variation of the mass resolution has a 7% effect on the significance. Systematics connected to the knowledge of the reducible $Zb\bar{b}$ background have been tested removing this background from the samples used to calculate the fit parameters or the $\tau$-ratio, and have been found to be at the level of 8%.

The presence of uncertainties in the shape of the $m_{4\ell}$ distribution at low masses ($m_{4\ell} < 180$ GeV) and the potential difficulty of predicting the shape in this region using the distribution at higher masses $m_{4\ell} > 180$ GeV, motivate studies where shape information is extracted from restricted fits in this low mass region. Examples are fits in the region from 110 to 170 GeV, where the background is expected to have a simple shape. In this range we perform both background-only fits excluding the signal region, and fits including the signal hypothesis, using the profile likelihood method but a simpler background parametrization (1st order polynomial). For the background-only fits we extract the significance calculating the p-value as defined in [21], using the method described in [22]. The extracted significance has also been calculated throwing large numbers of toy-MC experiments that were used to calculate the p-value, in order to validate the method of [22]. The background-only fits in the 110 to 170 GeV region

give a significance of 5.6$\sigma$ for a 130 GeV Higgs boson. When the signal hypothesis is included using the profile likelihood method, a significance of 5.8$\sigma$ is found. The small increase in the significance is due to the inclusion of the signal hypothesis which renders the statistical test more powerful. These results for the 130 GeV Higgs boson are close to those obtained from the full mass range fit baseline method (Table 14). Independent measurements (e.g. $Zb\bar{b}$ background) are expected to provide constraints in the knowledge of the background thus improving the discovery potential.

The fitting approach discussed so far involves one-dimensional fits of the $m_{4\ell}$ distribution. This approach can be generalized to a global 2D fit on the $(m_{Z^*}, m_{4\ell})$ plane. Such 2D fits, in contrast to the baseline approach, exploit correlations between $m_{Z^*}$ and $m_{4\ell}$, after a single set of cuts independent of the Higgs boson mass. Having a single set of cuts allows the extraction of the signal through both fixed and floating Higgs boson mass fits. The 2D models on the $(m_{Z^*}, m_{4\ell})$ plane, have been obtained for the $ZZ$ and $Zb\bar{b}$ backgrounds. All available signal samples (for masses from 120 to 600 GeV) have been used to develop a one-parameter family of surfaces (the parameter being $m_H$) that adequately models the signal for any intermediate value of the true Higgs boson mass. Table 16 summarizes the expected significance as a function of $m_H$, for both fixed and floating Higgs boson mass window fits, and an integrated luminosity of 30 fb$^{-1}$. These results are obtained using toy MC pseudo-experiments using a median likelihood ratio for the fixed mass case, and a p-value for the floating mass case, as described in [21]. The significances shown in the first row of Table 16 are consistent with the corresponding results of the baseline analysis shown in Table 14.

| Mass (GeV) | 120 | 130 | 140 | 150 | 160 |
|---|---|---|---|---|---|
| 2D fit Fixed Higgs boson Mass | 2.0 | 6.3 | 11.6 | 14.0 | 8.3 |
| 2D fit Floating Higgs boson Mass | 1.1 | 5.6 | 10.9 | 13.2 | 7.5 |

Table 16: Median significance from a 2D global fit on the $(m_{Z^*}, m_{4\ell})$ plane for both fixed and floating Higgs boson mass window and an integrated luminosity of 30 fb$^{-1}$.

# 7 Summary and conclusion

The potential of ATLAS in observing the Higgs boson in its $H \rightarrow ZZ^* \rightarrow 4\ell$ decay mode was presented. For an integrated luminosity of 30 fb$^{-1}$, ATLAS will discover the Higgs boson in the $4\ell$ channel-alone in the mass range from $130-500$ GeV, with the exception of the region around 160 GeV of the WW turn on, where significances of about 4$\sigma$ were found. In the very low mass region of 120 GeV, just above the LEP limit, a significance close to 3$\sigma$ is expected: this is a strong contribution to the combined significance of the various Standard Model decay modes [21]. The $H \rightarrow ZZ \rightarrow 4\ell$ channel is highly sensitive in the high mass region (400 GeV $> m_{4\ell} >$ 200 GeV), and in the 150 GeV region, where the Higgs boson should be discovered with 5 fb$^{-1}$.

The results obtained in this work include studies of the effect of systematic uncertainties on the signal extraction. In parallel to these studies, several attempts to improve the sensitivity of the channel by increasing the signal yield and keeping the same level of background were performed. Examples are the relaxing of the lepton-id criteria for one of the four leptons, the use of calorimeter for muon tagging, and application of multivariate techniques. Although the first results are encouraging, these studies are beyond the scope of this note since they are still in their very early stages.

# Search for the Standard Model Higgs Boson via Vector Boson Fusion Production Process in the Di-Tau Channels

**Abstract**

We outline a search for the Standard Model Higgs boson decaying into a $\tau$-pair in association with two jets, which is produced dominantly by the Vector Boson Fusion (VBF) process. The results indicate significant potential for a discovery in the low mass range. We consider fully leptonic, semi-leptonic, and, for the first time, fully hadronic tau decays. Mass reconstruction, central-jet veto, and jet tagging are discussed, and we present an approach to estimate the background from the data. Additional emphasis has been given to trigger issues and the impact of pileup. The results are based on an improved detector description, including misalignments, the most recent reconstruction software, and modern Monte Carlo event generators, including a revised prediction of the underlying event activity.

## 1 Introduction

The search for the Higgs boson and the source of electroweak symmetry breaking is a primary task of the Large Hadron Collider (LHC). It has been shown [1] that the ATLAS detector is capable of discovering the Standard Model Higgs boson with masses ranging from the LEP limit of 114 GeV [2] to about 1 TeV. The low-mass region is preferred from electroweak precision measurements and in this region ($m_H < 130$ GeV) the searches for Higgs bosons decaying to taus and photons are the most promising for discovery [3–5]. Searches for the Higgs boson produced in Vector Boson Fusion (VBF) tend to have reasonably high signal-to-background ratios, making them more robust to systematic uncertainties.

Within the Standard Model, the ability to observe the Higgs boson in multiple production and decay configurations makes it possible to measure the Higgs boson coupling to fermions and vector bosons [6]. Furthermore, the VBF processes provide a tool for measuring the Higgs boson spin and CP properties [7, 8]. In the context of the Minimal Supersymetric Standard Model, (MSSM), the branching ratio of a Higgs boson decaying to photons is generally suppressed, which makes the search for Higgs boson decaying to taus very important. The complementarity of the coupling of the light and heavy CP-even, neutral Higgs bosons of the MSSM to taus makes it possible to cover most or all of the $m_A - \tan\beta$ plane by reinterpreting the results for a Standard Model Higgs boson decaying into taus in the context of the MSSM [9, 10].

A previous ATLAS analysis outlined the sensitivity to a low mass Higgs boson including the first estimates for the VBF channels [4]. These results were primarily based on a fast simulation that parameterized the results of key detector performance studies performed with a full GEANT simulation. In this note we have considered three decay modes: the lepton-lepton (*ll*-channel), lepton-hadron (*lh*-channel) and the hadron-hadron (*hh*-channel) from VBF $H \to \tau^+\tau^-$ signature. The analysis has been done using state-of-the art Monte Carlo generators, full GEANT-based simulation of the ATLAS detector with realistic misalignments and distortions applied to the expected material in the detector, utilization of our current reconstruction algorithms, and, where possible, incorporation of pileup interactions.

This analysis requires excellent performance from every ATLAS detector subsystem; the presence of $\tau$ decays implies final states with electrons, muons, hadronic tau decays, and missing transverse momentum, while the Vector Boson Fusion production process introduces jets that tend to be quite forward in the detector. Due to the small rate of signal production and large backgrounds, particle identification must be excellent and optimized specifically for this channel. Furthermore, triggering relies

on the lowest energy lepton triggers or exceptionally challenging tau trigger signatures. The detector performance aspects so important to this analysis are described in Refs. [11–19].

### 1.1 Monte Carlo samples

Estimating the sensitivity of ATLAS to this channel requires the state of the art in Monte Carlo tools. The most challenging aspect of the theoretical calculations is the description of jet activity, an area in which the tools have evolved substantially since ATLAS ' first publication on the sensitivity to the VBF processes. Details of the Monte Carlo samples are outlined in Ref. [20]. The signal samples were produced with HERWIG [21] and PYTHIA [22]. The QCD $Z$+jets and $W$+jets samples were produced with ALPGEN [23], which employs the MLM matching [24] between the *hard process* (calculated with a leading-order matrix element for up to 5 jets) and the parton shower of HERWIG. The electroweak (ELWK) $Z$+jets background was simulated with SHERPA [25]. The $t\bar{t}$+jets and diboson background samples were generated with MC@NLO [26]. In all processes with taus, the tau decay was simulated using TAUOLA [27]. Additional photon radiation from charged leptons was simulated with PHOTOS [28]. The production cross-section for the signal is based on the next-to-leading order (NLO) computation and the k-factor (the ratio of the cross-section to that predicted by the lowest order calculation) is around 5% in the target mass range of 100−150 GeV. Note that the k-factor only involves the QCD corrections.

Because the GEANT-based detector simulation is computationally intensive, an event filter was applied to each sample after the parton shower and hadronization. Most processes were required to have at least one lepton in the final state. For background processes a VBF filter was used to remove events that would fail jet-related requirements. The filter bias has been studied and well-validated, but it affects our ability to estimate background rates early in the analysis cut flow. Furthermore, a significant Monte Carlo sample was produced with the ATLAS fast simulation, ATLFAST [29], without any event filter. These ATLFAST samples are used for systematic studies and to aid in the estimation of background rates (see Section 3.4).

The effect of in-time pileup (i.e. other soft $p$-$p$ collisions in the same bunch crossing), out-of-time pileup (i.e. $p$-$p$ collisions in neighboring bunch crossings), and the underlying event (i.e. multi-parton scattering and soft activity in the $p$-$p$ collision of interest) are all important to this analysis. The underlying event has substantial theoretical uncertainty, and different models' predictions for the underlying event activity vary by large factors when extrapolating to the LHC energy range. Fortunately, the underlying event activity will be one of the first measurements at the LHC and will be well measured by the time the analysis described in this Note is performed. The pileup interactions are incorporated early in the simulation chain, at the time when the detector readout is simulated.

## 2 Event selection

### 2.1 Triggering

While the ATLAS trigger system provides several possibilities for triggering that take advantage of the signal's complex final state, we restrict ourselves here to simple robust trigger signatures that are expected to have a low rate and an acceptable selection efficiency [12, 13]. For the $lh$ and $ll$ final states the events are selected by an isolated electron with $p_T \geq 22$ GeV (`e22i`) or an isolated muon with $p_T \geq 20$ GeV (*mu*20). The entire trigger chain has been simulated with the use of our current trigger algorithms and trigger menus; however, the dilepton triggers composed of isolated muons with $p_T \geq 10$ GeV and isolated electrons with $p_T \geq 15$ GeV considered in Ref. [4] were not used in this study. The trigger efficiency for VBF $H \to \tau\tau$ (with $m_H$ = 120 GeV) is 9.0% for events selected by the electron trigger and 9.9% in the case of muons. The trigger efficiencies include detector acceptance and are normalized with

Table 1: The product of efficiency and acceptance for the signal from the `e22i`, `mu20`, and`L1_TAU30_xE40_softHLT` triggers.

| Trigger menu | Efficiency × Acceptance(%) |
|---|---|
| `e22i` | 9.08 ± 0.03 |
| `mu20` | 9.88 ± 0.04 |
| `L1_TAU30_xE40_softHLT` | 3.67±0.02 |

respect to the production cross-section for VBF $H \to \tau\tau$. Additional triggers for the *lh* and *ll* final states, for instance the combined $\tau + e$ or $\tau$+$\mu$ triggers and triggers which take advantage of the tagging jets of the VBF process, are under study.

The all hadronic mode, or *hh*-channel, utilizes a different triggering strategy. Unlike the clean signature of the electron and muon triggers, the single tau trigger is expected to be exposed the large QCD jets background. Therefore only tau trigger in combination with other signatures, like missing $E_T$ or another tau in the event, can be considered. We use `L1_TAU30_xE40_softHLT` as the primary trigger menu for the *hh*-channel in this study. It should be noted that unlike the high $p_T$ single lepton triggers, both the hadronic tau and $E_{\mathrm{T}}^{\mathrm{miss}}$ triggers are based on requirements from the first level of the trigger system with only a loose selection in the high-level trigger [30]. The expected trigger acceptance of `L1_TAU30_xE40_softHLT` is listed in Table 1 as well as those from `e22i` and `mu20` menus. The trigger efficiency for the signal events (for $m_H$ = 120 GeV) is 3.7% for `L1_TAU30_xE40_softHLT`. The disadvantage of the missing $E_T$ trigger is the relatively low efficiency on signal; therefore, alternative menus like double tau menus are now being developed.

## 2.2 Electron and muon reconstruction and identification

Electron candidates are formed from a cluster of cells in the electromagnetic calorimeter together with a matched track. The electron identification includes information from the shape of the shower, tracking information, and the consistency of the track and cluster. ATLAS provides multiple working points that trade electron efficiency for improved rejection of fakes. In this analysis we use the *Medium* class electron as it provides sufficient fake rejection and provides a higher signal efficiency. In addition to the standard electron identification, we require that the energy in an isolation cone of radius $\Delta R = 0.2$ around the electron contains less than 10% of the electron's $E_T$ [1]. The isolation cut is imposed to reject the contamination from hadronic jets.[2] The reconstruction and identification efficiency is fairly flat after $p_T \geq 15$ GeV, where it achieves 69.4 ± 0.2% efficiency while keeping the fake electron contamination at the order of 0.1%. In the VBF $H \to \tau\tau \to \mu\mu + 4\nu$ signal sample, the probability to reconstruct a fake electron was found to be slightly higher, 0.25±0.03%, reflecting some level of process-dependence.

In ATLAS, muon candidates can be seeded from either tracks in the inner detector or in the standalone muon spectrometer. In this analysis we required the highest quality muon candidates, which are formed by extrapolating the track in the muon spectrometer to the interaction point, finding a matching inner detector track, and forming a combined track if the two tracks satisfy various quality requirements [15]. The muon identification is composed of requirements on track quality and hit multiplicity in several muon stations. Similarly to the electrons, we require an isolation condition that the summed $E_T$ within a radius $\Delta R$ of 0.2 is less than 10% of the muon $p_T$ to reject the contamination from jets. The reconstruction and identification efficiency is fairly flat after $p_T \geq 10$ GeV, and it achieves 91.9 ± 0.1% while keeps the fake muon rejection under 0.005%.

[1] Due to a problem with the reconstruction, a correction to the isolation energy in the Tile gap scintillator was required.

[2] A track-based isolation requirement was also studied and shown to have similar performance.

The electron and muon identification criteria are summarized in Table 2. The $p_T$ thresholds for electron and muon identification are chosen to provide stable identification efficiency and sufficient fake rejection. In addition, we require that the $p_T$ of the offline reconstructed lepton must satisfy the $p_T$ thresholds of the corresponding trigger, which is not strictly enforced due to subtle differences between the offline reconstruction and the trigger algorithms.

Table 2: Summary of the identification requirements for electrons and muons.

| Lepton identification |
|---|
| **Electron ID**: Medium |
| Isolation $E_T(\Delta R = 0.2)/p_T \leq 0.1$ |
| $p_T \geq 25$ GeV for trigger electron (`e22i`) |
| $p_T \geq 15$ GeV for other electrons |
| **Muon ID**: Combined muon |
| Isolation $E_T(\Delta R = 0.2)/p_T \leq 0.1$ |
| $p_T \geq 20$ GeV for trigger muon (*mu20i*) |
| $p_T \geq 10$ GeV for other muons |

### 2.3 Hadronic-tau reconstruction and identification

Approximately 65% of tau leptons decays produce hadrons. The majority of hadronic tau decays are composed of *single-prong* candidates with one charged pion, which provides a track and a hadronic shower, and potentially associated neutral pions that provide an additional electromagnetic sub-cluster. In addition, *three-prong* tau decays are also reconstructed, but with a higher rate of fakes from QCD jets. Due to the high momentum of the taus produced in this process, the decay products are collimated into a narrow region. ATLAS currently employs two hadronic tau reconstruction algorithms [31]; both require a calorimeter cluster matching a track; however, one algorithm is seeded by calorimeter clusters and the other is seeded by the track. The two algorithms' efficiencies are complementary in different $p_T$ regimes, and provide rejection strategies for their energy measurement and rejection against jets. The calorimeter-seeded algorithm was used for this analysis.

The calorimeter-seeded algorithm provides a log-likelihood ratio that distills discriminating power from a variety of track quality and shower shape information to discriminate between taus and jets [14]. The discriminating variable is designed to maintain a high tau efficiency while rejecting fake tau candidates from jets, leaving the precise working point to be optimized in the context of a specific analysis. The cuts on the discriminating variable and $p_T$ of the tau candidates were optimized with respect to a simple $s/\sqrt{s+b}$ performance measure. The background sample included $Z$+jets, $W$+jets, and $t\bar{t}$+jets, which comprises a background sample with a representative mixture of real and fake taus. Our modeling of the jet fragmentation indicates that quark-initiated jets are more collimated and have a 6-8 times higher fake rate than gluon-initiated jets. The relative abundance of real and fake tau candidates depends on the kinematic requirements imposed on the sample, thus the optimization should be performed after the final kinematic requirements described in Sections 2.8 and 2.9. However, the limited size of Monte Carlo samples requires that only a subset of the criteria used in the final event selection are applied during the optimization. Several subsets of the final event selection criterion were evaluated, and the final optimization was found to be reasonably stable and nearly independent of $p_T$. After the optimization, the calorimeter-seeded algorithm's log-likelihood ratio was required to be greater than 4, corresponding to an identification efficiency of 50.0±0.2% and a fake jet selection efficiency of $\sim$1% for gluon-initiated jets and $\sim$2.5% for quark-initiated jets.

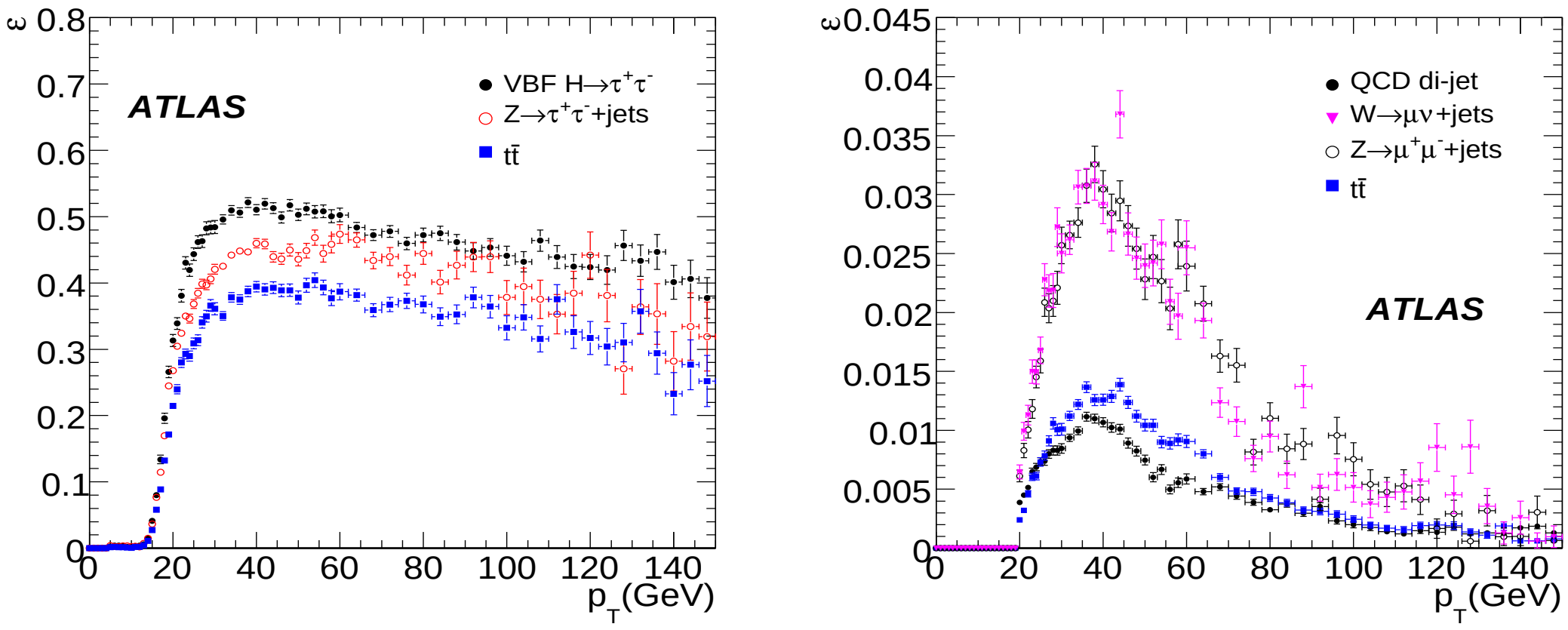


(a) (b)

Figure 1: Reconstruction and identification efficiency of the hadronic tau (a) and the jet-fake rejection efficiency (b) as a function of $p_T$, respectively.

In addition to rejection against jets, an electron-veto was used to reject tau candidates which arise from electrons that have failed the electron identification. This electron-veto was performed by requiring that the tau candidate have at least 0.2% of its energy in the first sampling of the hadronic calorimeter and that the ratio of high-threshold (HT) to low-threshold (LT) hits in the transition radiation tracker (TRT) be less than 20% in the range $|\eta_\tau| < 1.7$. This electron-veto procedure suppresses the electron fake rate by 82.5% while retaining 90% of the hadronic tau candidates selected without the veto.

Finally, we present the hadronic tau reconstruction and identification performance in Fig. 1 (a) and the fake-jet tagging rate (b) as a function of $p_T$, respectively. The selection criteria for the hadronic tau identification is summarized in Table 3.

Table 3: Selection criteria for the hadronic tau identification from the calorimeter-seeded reconstruction algorithm.

| Hadronic tau identification |
|---|
| **Tau ID**: Calorimeter-seeded |
| $p_T \geq 30$ GeV |
| Track multiplicity : 1 or 3 tracks |
| $\|charge\| = 1$ |
| *Log Likelihood Ratio* $\geq 4$ |
| **Electron Veto**: |
| minimum TRT $HT/LT \leq 0.2$ if $\|\eta_\tau\| \leq 1.7$ and $LT \geq 10$ |
| $E_T^{HAD}/p_T \geq 0.002$ in matched electron object |

## 2.4 Jet reconstruction

### 2.4.1 Forward tagging jets

The jet activity of the vector boson fusion process is unique in several ways, providing many handles to suppress backgrounds and isolate a sample of signal events with high purity. The most important features of the VBF process are the presence of two high-$p_T$ quark-initiated "tagging jets", which tend to be relatively forward and well separated in rapidity. Furthermore, due to color coherence in this electroweak process, additional QCD radiation between the tagging jets tends to be suppressed and motivates a Central Jet Veto (CJV) [32]. This section outlines the choice of jet algorithms and their performance, the definition of the tag jets, and several issues related to the CJV.

Figure 2 shows the $\eta$ spectra of the highest and second highest $p_T$ jets in signal and various background samples. Because the VBF jets can be very forward, the jet finding efficiency in this region is important in the analysis. Furthermore, the forward calorimeters (3.1$\leq |\eta| \leq$4.9) do not have a projective geometry, which leads to different challenges for jet reconstruction. ATLAS currently provides collections of jets based on two algorithms (a seeded cone algorithm with split-merge and a $k_T$ algorithm), each with two sets of parameters (the cone size and the $k_T$ cutoff scale), applied to two different input representations of the energy deposits in the calorimeter (towers merged to avoid negative energy fluctuations from electronic noise and clusters based on the ATLAS TopoCluster algorithm) [18]. These different jet algorithms and the different calorimeter pre-clustering result in different performances for jets, especially at low $p_T$ and high $|\eta|$.

Jet identification efficiency and purity are defined to give a quantitative measure of the jet identification. The efficiency and purity were calculated with respect to generator-level jets obtained by running the same jet algorithm on the stable interacting particles after hadronization and before GEANT simulation. To ensure that only hadronic jets are considered, we only use dimuon events, where both taus decay into a muon and neutrinos, or $Z/W$ bosons directly decay into muons. This avoids any bias in the jet reconstruction produced by the presence of electrons. A reconstructed jet is considered to be matched if the corresponding generator-level jet is within $\Delta$R $\leq$ 0.15 for jets with a cone size of 0.4. The matching cone size was chosen to avoid a single generator-level jet being matched to more than one reconstructed jet; with the given parameters this effect is at the order of $10^{-3}$.

The jet reconstruction efficiency in different $|\eta|$ regions and two different clustering algorithms is shown in Fig. 3 as a function of the generator-level jet $p_T$ and $\eta$. The reconstruction efficiency rises over 95% for jets with $p_T$ above 50 GeV. On the other hand, the efficiency drops at $|\eta| \sim 1.5$ and $|\eta| \sim 3.2$ for jets in the range of 20-30 GeV of $p_T$. This drop in efficiency is due to the crack region in the calorimeter or large amounts of dead material in the corresponding $\eta$ region. The jet collections based on calorimeter towers show a drop in efficiencies in the forward region due to a higher seed threshold, while the jet collections based on TopoClusters do not show this loss of efficiency. For this reason, jets based on TopoClusters have been chosen for this analysis.

Correctly identifying the quark-initiated tagging jets from the VBF process is very important for the measurement of Higgs boson spin and CP properties and for making precise correspondence with theoretical calculations [8]. Typically, the tagging jets are found in opposite hemispheres, but there are two approaches to incorporating this requirement in the analysis. One option is to define the tagging jets as the two highest $p_T$ jets in the event, and reject the event from the signal candidates if they are in the same hemisphere (e.g. require $\eta_{j1} \times \eta_{j2} \leq 0$). A second option is to define the first tagging jet to be the highest $p_T$ jet in the event and the second tagging jet to be the highest $p_T$ jet in the opposite hemisphere. In this second approach it is not required that the second tagging jet is the second highest $p_T$ jet in the event. These two strategies were compared, and it was found that the first method more reliably matched the quark-initiated tagging jets from the hard process.

The generator-level jets match the hard-scattered quarks nearly 100% of the time above a certain

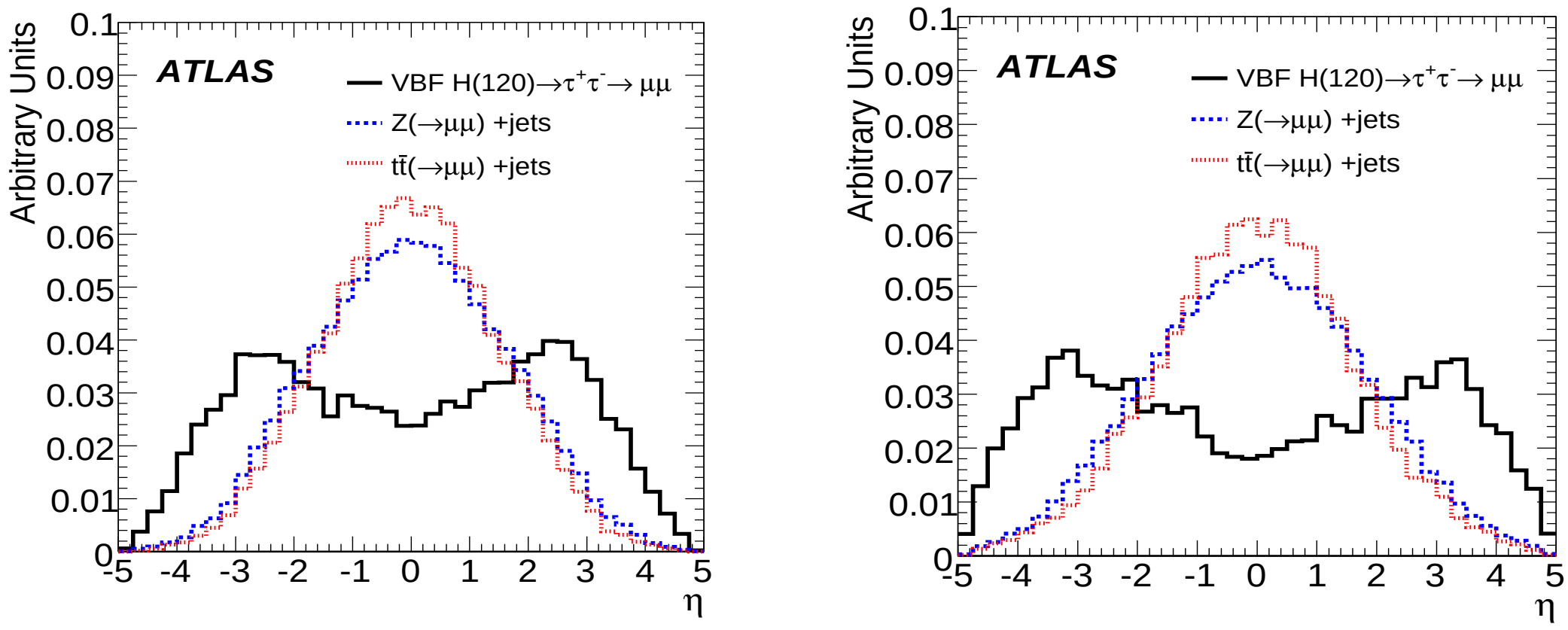


(a) (b)

Figure 2: Pseudorapidity of the highest $p_T$ (a) and the second highest $p_T$ (b) jets for the Cone jet algorithm based on TopoClusters with $R = 0.4$ in VBF $H \rightarrow \tau\tau \rightarrow \mu\mu$ ($m_H$=120 GeV) and background events. Only $p_T$ cuts were applied to jets. Solid (black) histogram is for signal, dashed (red) histogram is for $t\bar{t} \rightarrow WW \rightarrow (\mu\mu)$, and dotted (blue) histogram is for Z$\rightarrow \mu\mu$+n jets.

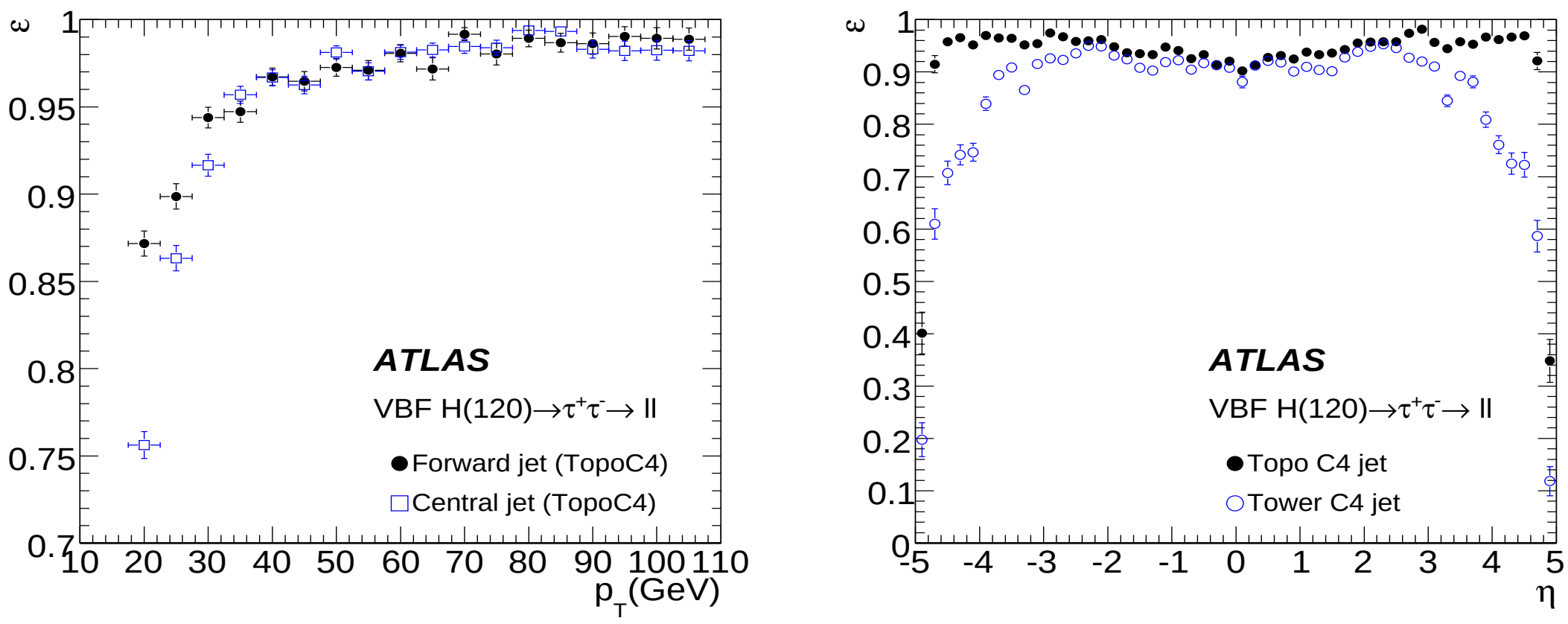


(a) (b)

Figure 3: Jet reconstruction efficiency for the Cone jet algorithm with $R = 0.4$ as a function of the generator-level jet $p_T$ for the jets based on TopoClusters (a) and $\eta$ for Tower- and TopoCluster-based jets (b).

$p_T$ threshold. To estimate the purity of the tagging jets, we define the efficiency with respect to the generator-level jets. The reconstructed tag jets have a high purity over the entire $p_T$ and $\eta$ range and do not show a strong dependence on the jet algorithms. Integrated efficiencies and purities for jets with $p_T \geq 20$ GeV indicate that the TopoCluster-based algorithm has better performance for this analysis. Because additional jets often lie in the central detector region, where we wish to employ a central jet veto, jets with smaller cones are favored for the selection. Furthermore, calorimeter noise (including effects from pileup of minimum-bias events) increases with jet cone radius. Thus, we use the cone jet algorithm with $R = 0.4$ running on TopoClusters as the primary jet algorithm in this analysis.

Having converged on a specific calorimeter pre-clustering and jet algorithm, we now present the kinematic properties of the jets that discriminate between the signal and backgrounds. The $p_T$ cuts on the tagging jets are effective at reducing several backgrounds and Fig. 2 shows that the pseudorapidity distributions are substantially different. Instead of relying directly on the pseudorapidity of the tagging jets, Fig. 4 shows that the pseudorapidity gap (a) and invariant mass of the two tagging jets (b) provide substantial background rejection.

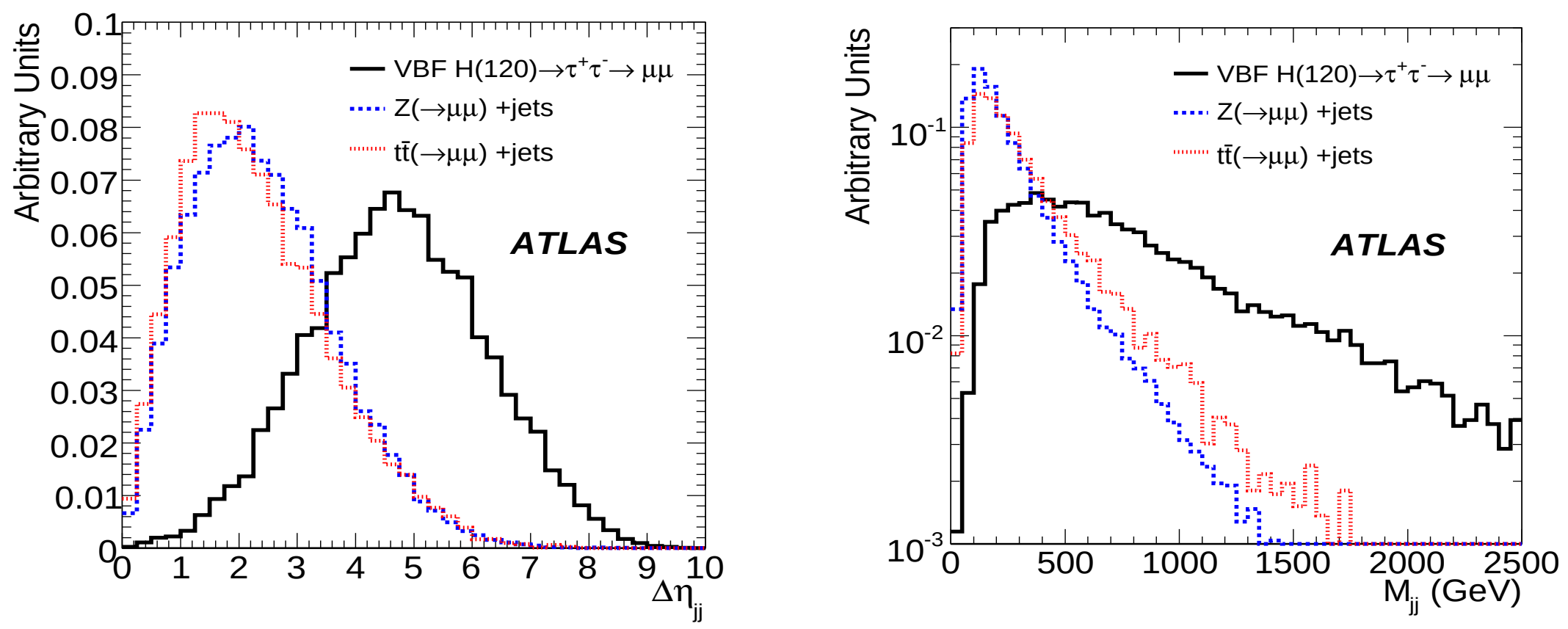


(a) (b)

Figure 4: Pseudorapidity gap between tag jets (a) and invariant-mass distributions of tag jets (b) in VBF $H \to \tau\tau \to \mu\mu$ events ($m_H$=120 GeV). A requirement $\eta_1 \times \eta_2 \leq 0$ is used in addition to the cuts on jet $p_T$. Solid (black) histogram is for signal, dashed (red) histogram is for $t\bar{t} \to WW \to (\mu\mu)$, and dotted (blue) histogram is for Z→ $\mu\mu$+n jets.

### 2.4.2 Central jet veto

As mentioned above, the color coherence in the VBF Higgs boson production leads to a suppression of QCD radiation between the tagging jets. This color coherence is also found in the electroweak $Z$+jets background. In contrast, most of the other backgrounds have a much larger probability for additional QCD radiation in the central region. This is the physical motivation for a central jet veto (CJV). Figure 5 shows the jet multiplicity distribution for the signal and backgrounds after requiring two tagged jets (with $p_T \geq 20$ GeV) in opposite hemispheres. The fraction of signal events with three or more jets is small.

The experimental challenge for the CJV is to provide a cut that is robust against additional minimum bias events (in-time pileup events). The optimization for the central jet veto has been studied in terms of $p_T$ and $\eta$. The probability to have at least one reconstructed jet with $p_T \geq$20 GeV within $|\eta| \leq 3.2$ is 1.6% from a single minimum-bias event. In Fig. 6, we present the trade-off of background rejection versus signal efficiency from varying the $p_T$ threshold on the third highest $p_T$ jet (markers indicate

thresholds of 20 and 30 GeV). A veto based on a fixed $\eta$-window was compared to a dynamic $\eta$-window defined by the $\eta$ of the two tagging jets. We maintain the previous central jet veto requirement: no jets in $|\eta| \leq 3.2$ with $p_T \geq 20$ GeV. Figure 7 shows the efficiency of the central jet veto for signal, irreducible, and reducible backgrounds at varying levels of pileup.

The CJV poses significant theoretical challenges as well. At the parton-level, the CJV efficiency is expected to be known quite well with little theoretical uncertainty. However, the current tools that allow for the full parton-shower and hadronization (prerequisite for an analysis based on a GEANT-based detector simulation) show significant uncertainties. We have observed significant differences between the central jet activity in signal events generated with PYTHIA and those generated with HERWIG. Knowledge of the uncertainty on the CJV is needed for setting limits on the Higgs boson cross-section and for making coupling measurements; however, it is not needed directly in establishing a deviation from the background-only expectation (see Section 5.3).

In future studies we will also include a veto procedure using track information; in particular using vertexing information to reduce the impact of jets from in-time pileup. Furthermore, a track-based veto and the use of timing information in the calorimeter will also be studied to reduce the impact of out-of-time pileup.

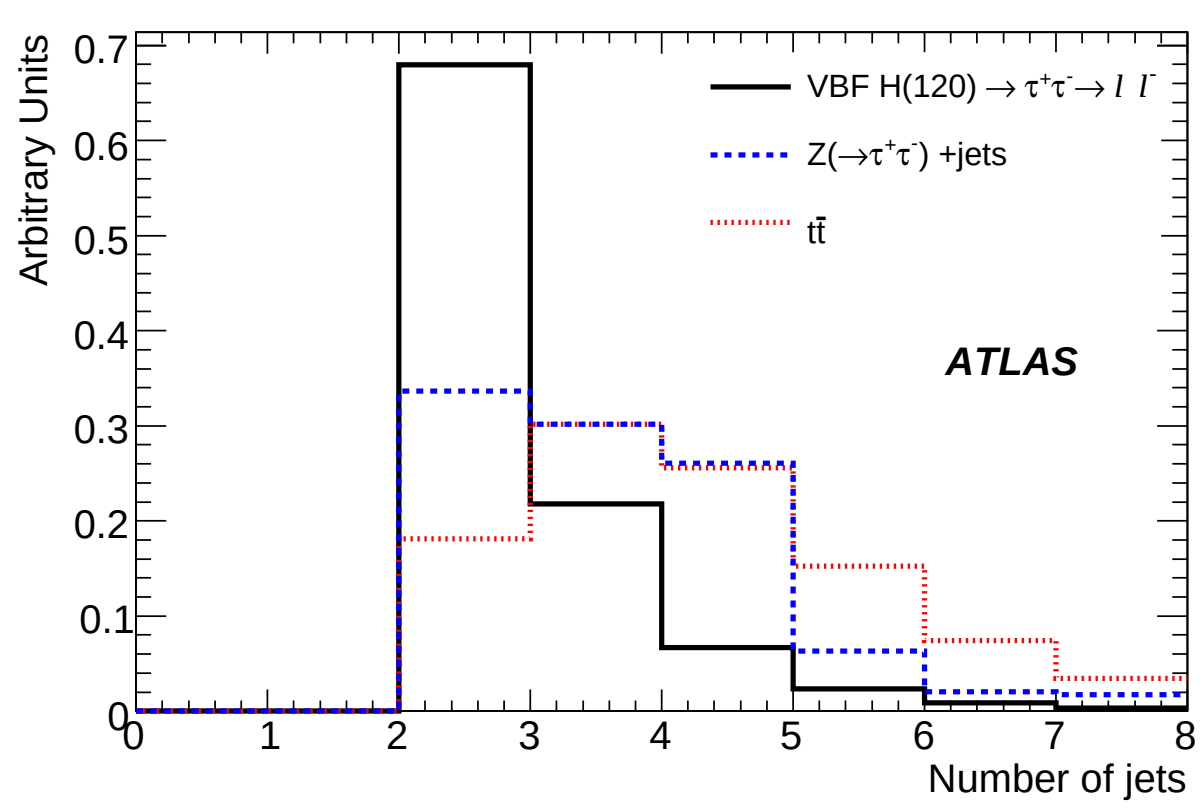


Figure 5: Jet multiplicity distribution for the signal, $Z$+jets, and $t\bar{t}$ background after requiring the cuts up to the N jets $\geq 2$ level in the list of cuts for the $ll$ channel (see Table 5).

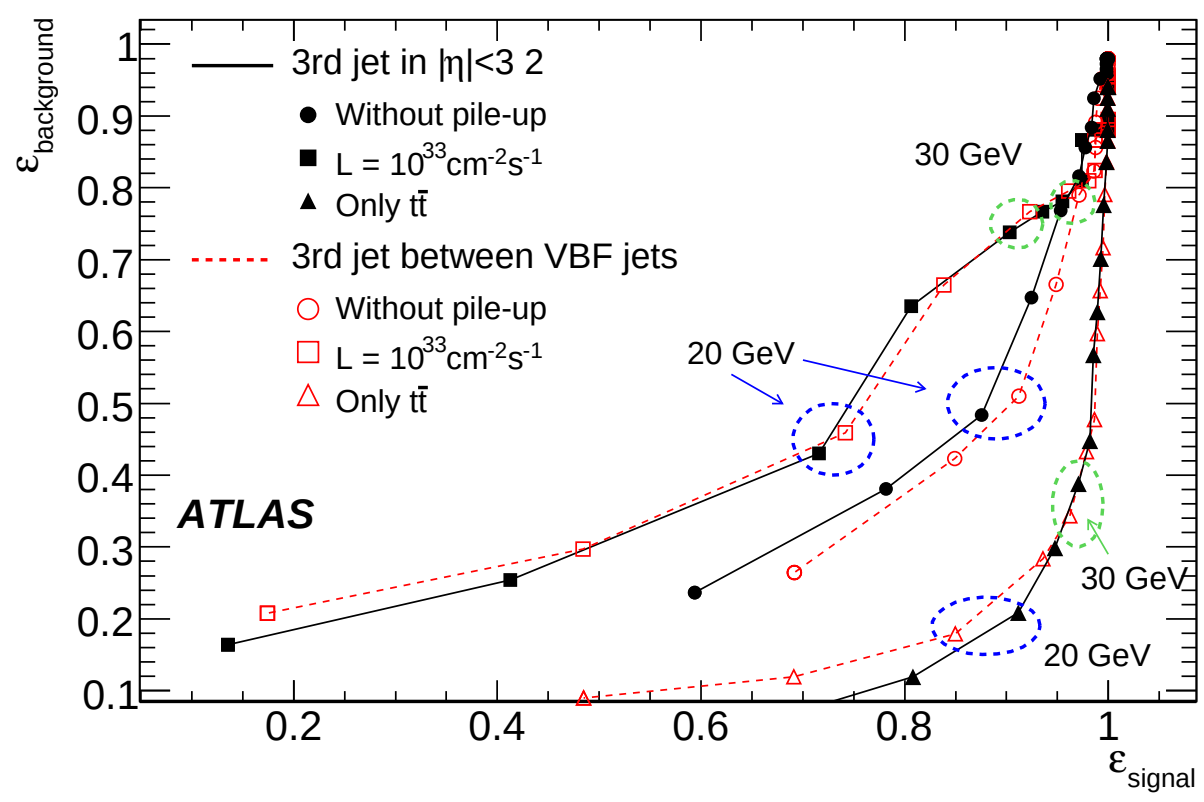


Figure 6: Background rejection versus signal sensitivity for the central jet veto with and without pileup. Also shown is the case for $t\bar{t}$-only background.

### 2.4.3 $b$-jet veto

In the $ll$-channel, the largest background contribution comes from $t\bar{t}(+jets) \to l\nu b l\nu b(+jets)$. By introducing a veto on $b$-tagged forward jets it is possible to reduce this background [19]. Because the tagging jets are fairly forward in the detector, the $b$-tagging requirement is rather loose, i.e. efficient, and the $t\bar{t}$ background can be reduced by a factor 2∼3. Figure 8 demonstrates the efficiency of the $b$-jet veto as a function of the forward jet $p_T$ for the signal and $t\bar{t}$ background. The cut on the $b$-tag weight was optimized to achieve 65.1% reconstruction efficiency for $b$-quark jets, while 9.4% mis-identification efficiency for the light flavor jets is retained. Note that the $b$-jet veto is only used in the $ll$-channel.

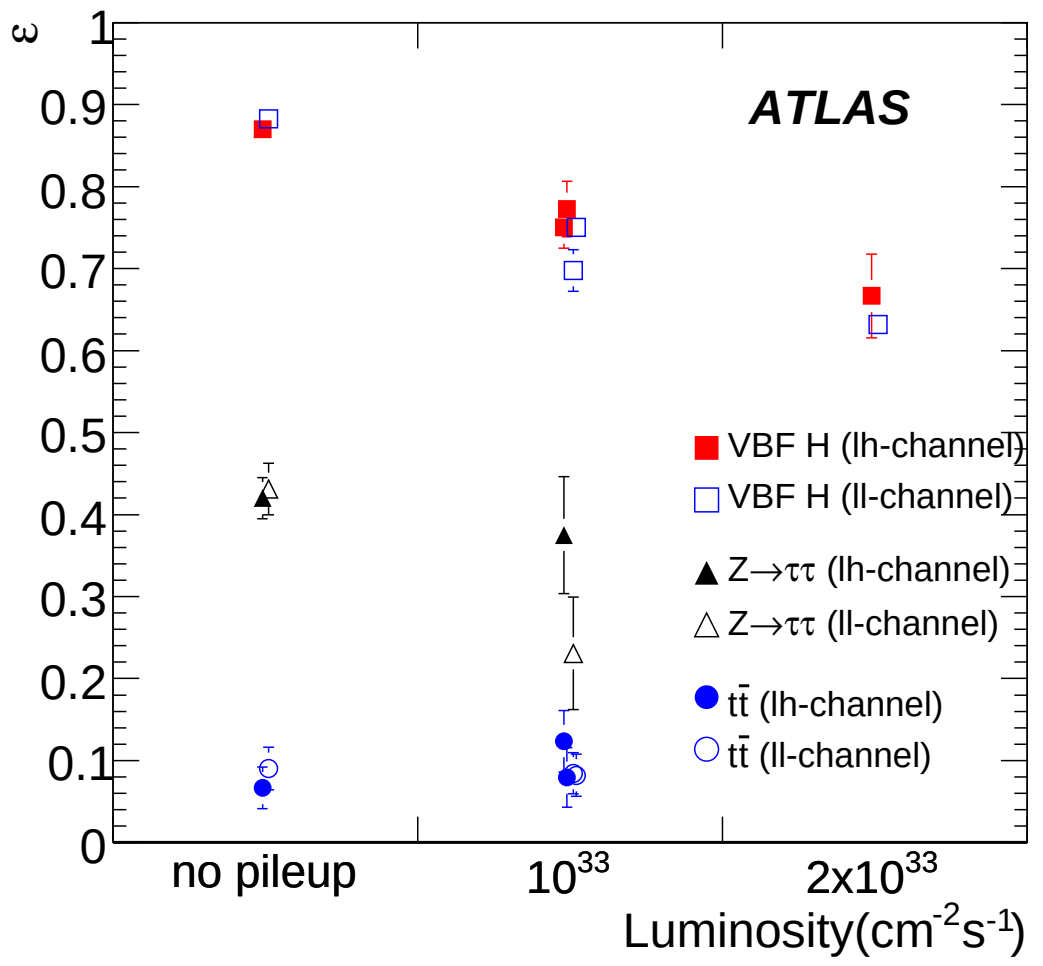


Figure 7: Central jet veto performance in the presence of varying levels of pileup for signal and background samples.

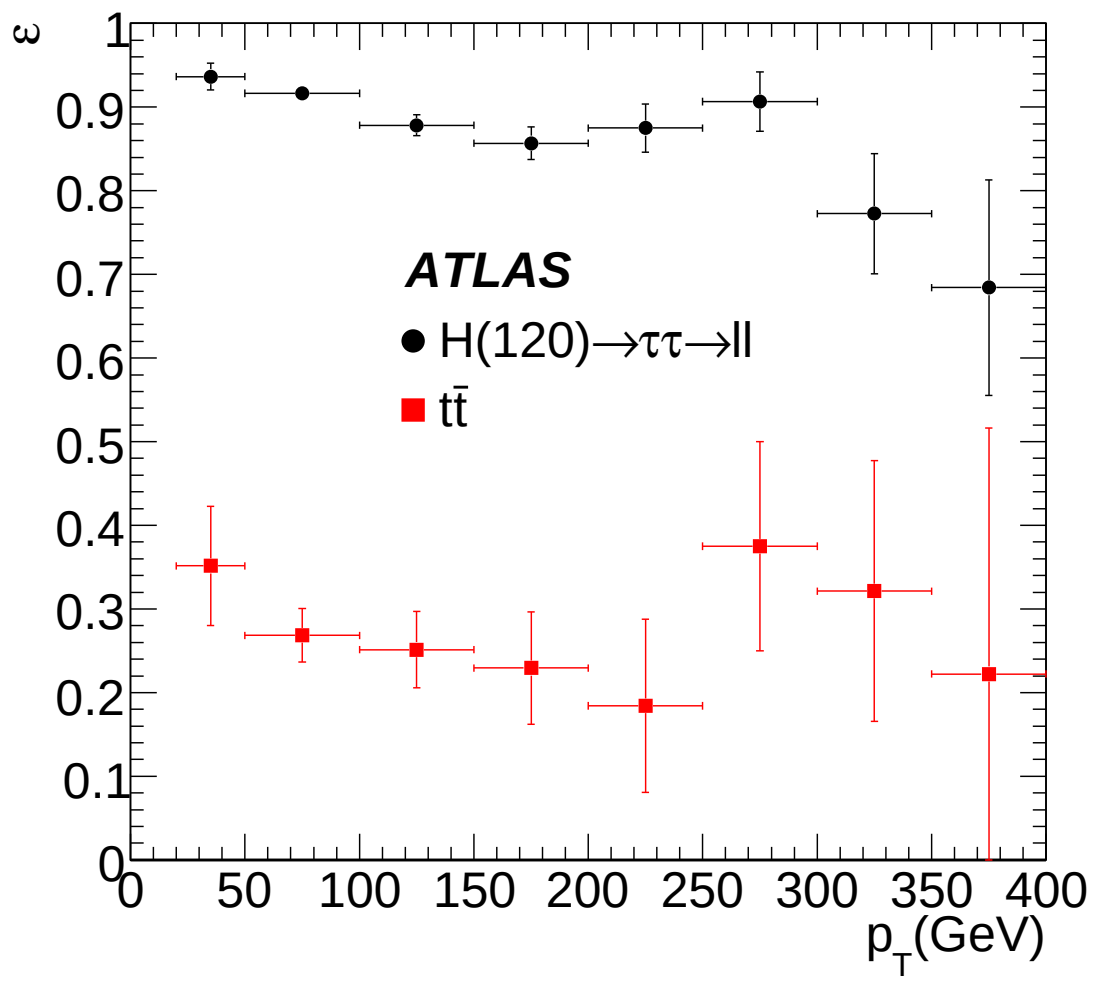


Figure 8: Efficiency of the $b$-jet veto as a function of the forward jet $p_T$ in for the signal and $t\bar{t}$ background.

## 2.5 Missing transverse energy

Significant missing transverse energy ($E_{\mathrm{T}}^{\mathrm{miss}}$ ) is present in $H \to \tau^+\tau^-$ events because neutrinos are always associated with the $\tau$ decays. The performance of the $E_{\mathrm{T}}^{\mathrm{miss}}$ algorithm plays a vital role in this analysis because $E_{\mathrm{T}}^{\mathrm{miss}}$ is used in the mass reconstruction of the tau pair. Ultimately, the $E_{\mathrm{T}}^{\mathrm{miss}}$ resolution is what limits the $m_{\tau\tau}$ resolution. Furthermore, the absolute scale of the $E_{\mathrm{T}}^{\mathrm{miss}}$ must be well calibrated to correctly reconstruct the Higgs boson mass. In addition to the standard $E_{\mathrm{T}}^{\mathrm{miss}}$ algorithm [17], we have made a dedicated correction in the presence of hadronic tau decays. The correction is based on the calibrated tau energy instead of the default treatment of the object that uses a jet calibration. This removes a $\sim 1$ GeV bias in the $E_{\mathrm{T}}^{\mathrm{miss}}$ distribution for the $lh$-channel. The sensitivity of the signal efficiency to the absolute energy scale is presented in Section 5.2. By requiring a large $E_{\mathrm{T}}^{\mathrm{miss}}$ , it is possible to improve the $m_{\tau\tau}$ resolution and reject many backgrounds that do not contain neutrinos (e.g. $Z \to ll$). We require $E_{\mathrm{T}}^{\mathrm{miss}} \geq 30$ GeV for the $lh$-channel and $E_{\mathrm{T}}^{\mathrm{miss}} \geq 40$ GeV for the $ll$- and $hh$-channels.

## 2.6 Mass reconstruction

Although there are several neutrinos in the event, it is possible to reconstruct the $\tau^+\tau^-$ invariant mass by making the approximation that the decay products of the $\tau$ are collinear with the $\tau$ in the laboratory frame. This is a good approximation since $m_H/2 \gg m_\tau$ and hence the taus are highly boosted. This leaves two unknown quantities and two equations: the fraction of each $\tau$'s momentum carried away by neutrinos and the constraints from the two components of $E_{\mathrm{T}}^{\mathrm{miss}}$ . For notational simplicity, consider the $lh$-channel and let $l$ represent the momentum vector for the leptonic visible decay product and $h$ represent the momentum vector for the hadronic visible decay products. By neglecting the $\tau$ rest mass and imposing the collinear approximation, we can write

$$m_{\tau\tau} = \sqrt{2(E_h + E_{\nu h})(E_l + E_{\nu l})(1 - \cos\theta_{lh})} \quad . \tag{1}$$

By introducing the variables $x_h$ and $x_l$, the fraction of the $\tau$'s momentum carried away by the visible decay products, we can re-write the invariant mass as

$$m_{\tau\tau} = \frac{m_{lh}}{\sqrt{x_l x_h}} \qquad \text{for } x_{l,h} \geq 0 \quad . \tag{2}$$

One can easily solve for the $x_\tau$ variables by requiring that the vector sum of the neutrinos coincides with the two measured components of $E_{\mathrm{T}}^{\mathrm{miss}}$ :

$$x_h = \frac{E_h}{E_h + E_{\nu h}} = \frac{h_x l_y - h_y l_x}{h_x l_y + E_{\mathrm{x}}^{\mathrm{miss}} l_y - h_y l_x - E_{\mathrm{y}}^{\mathrm{miss}} l_x} = \frac{N}{D_h} \tag{3}$$

and

$$x_l = \frac{E_l}{E_l + E_{\nu l}} = \frac{h_x l_y - h_y l_x}{h_x l_y - E_{\mathrm{x}}^{\mathrm{miss}} h_y - h_y l_x + E_{\mathrm{y}}^{\mathrm{miss}} h_x} = \frac{N}{D_l}, \tag{4}$$

where we have introduced $N, D_h$, and $D_l$ for convenience. If the two $\tau$s are back-to-back, then these equations are linearly-dependent and one cannot solve for the $x_\tau$s. For this reason, we require that $\cos\Delta\phi_{\tau\tau} \geq -0.9$. Typically the Higgs boson has significant $p_T$ due to the tagging jets. Events that come from the process $X \to \tau\tau$ with no other sources of missing energy should have $0 \leq x_\tau \leq 1$, though resolution effects in $E_{\mathrm{T}}^{\mathrm{miss}}$ may lead to unphysical solutions with either $x_\tau < 0$ or $x_\tau \geq 1$. Equation 2 shows explicitly that cuts on $x_\tau$ will impose constraints on the reconstructed mass for a given event, viz. $m_{\tau\tau} \geq m_{lh}$, which results in an asymmetric distribution for $m_{\tau\tau}$.

The sensitivity of $m_{\tau\tau}$ to a mis-measurement of $E_{\mathrm{T}}^{\mathrm{miss}}$ depends on the orientation of the $\tau$s. This sensitivity can be summarized by a Jacobian factor, $J$. Neglecting correlation between the mis-measurement of the $x$- and $y$-components of $E_{\mathrm{T}}^{\mathrm{miss}}$ , one can define the Jacobian as follows

$$J = \frac{\Delta m_{\tau\tau}}{\Delta E^{miss}_{T\;x/y}} = \sqrt{\left(\frac{\partial m_{\tau\tau}}{\partial E_{\mathrm{x}}^{\mathrm{miss}}}\right)^2 + \left(\frac{\partial m_{\tau\tau}}{\partial E_{\mathrm{y}}^{\mathrm{miss}}}\right)^2} \quad . \tag{5}$$

Thus, we arrive at

$$J = \frac{1}{2}\frac{m_{lh}\sqrt{x_l x_h}}{|N|^3}\sqrt{\left(x_l h_y D_h^2 - x_h l_y D_l^2\right)^2 + \left(x_h l_x D_l^2 - x_l h_x D_h^2\right)^2} \quad . \tag{6}$$

The final mass measurement is a result of a fit to the $m_{\tau\tau}$ distribution, and it is important to incorporate both the asymmetry and the fact that the width of the $m_{\tau\tau}$ distribution is not common for all events. The modeling of the asymmetry and Jacobian scaling of the $m_{\tau\tau}$ distribution is described in Section 4.2.

## 2.7 Summary of the event selection for $ll$-channel

The event selection for $ll$-channel is summarized below, including some kinematic requirements specific to the $ll$-channel.

- Trigger: electron trigger `e22i` or muon trigger `mu20`.
- Trigger lepton: at least one lepton must have a reconstructed $p_T$ greater or equal to the corresponding trigger requirement.
- Dilepton: exactly two identified leptons with opposite charge.
- Missing $E_T$: $E_{\mathrm{T}}^{\mathrm{miss}} \geq 40$ GeV.

- Collinear approximation: $0 \leq x_{l1,l2} \leq 0.75$ and $\cos\Delta\phi_{ll} \geq -0.9$.
The tighter cut on $x_\tau \leq 0.75$ has been found to provide a better background rejection in the $ll$-channel. The shape of the control samples used to estimate the signal sensitivity are obtained after these cuts; additional details of the data-driven background estimation are given in Section 3. In addition to the cuts above, the signal candidates are also required to satisfy the following cuts.

- Jet multiplicity: at least one jet with $p_T \geq 40$ GeV and at least one additional jet with $p_T \geq 20$ GeV.

- Forward jets: in opposite hemispheres $\eta_{j1} \times \eta_{j2} \leq 0$, with tau centrality $\min\{\eta_{j1}, \eta_{j2}\} \leq \eta_{lep_{1,2}} \leq \max\{\eta_{j1}, \eta_{j2}\}$ for the two highest $p_T$ jets.

- $b$-jet veto: the event is rejected if either tag jet has $b$-tag weight greater than 1.

- Jet kinematics: $\Delta\eta_{jj} \geq 4.4$ and dijet mass $m_{jj} \geq 700$ GeV for two forward jets.

- Central jet veto: the event is rejected if there are any additional jets with $p_T \geq 20$ GeV in $|\eta| \leq 3.2$.

- Mass window: $m_H - 15$ GeV $\leq m_{\tau\tau} \leq m_H + 15$ GeV around the test mass $m_H$.

Table 4 summarizes the cross-section for signal events after each of the cuts described above.

Table 4: Signal cross-section (fb) for the $ll$-channel for various Higgs boson masses.

| Mass (GeV) | 105 | 110 | 115 | 120 | 125 | 130 | 135 | 140 |
|---|---|---|---|---|---|---|---|---|
| Cross section (fb) | 394.7 | 372.0 | 341.8 | 309.1 | 266.8 | 225.4 | 180.1 | 135.8 |
| Trigger | 65.6(3) | 65.1(2) | 61.1(2) | 57.2(1) | 51.5(2) | 44.7(1) | 36.5(1) | 28.3(1) |
| Trigger lepton | 56.4(3) | 56.2(2) | 53.2(2) | 49.5(1) | 44.7(2) | 38.9(1) | 31.8(1) | 24.7(1) |
| Dilepton | 5.73(7) | 5.86(6) | 5.80(6) | 5.46(3) | 4.94(5) | 4.30(4) | 3.61(4) | 2.88(4) |
| $E_T^{miss} \geq 40$ GeV | 3.41(5) | 3.49(5) | 3.45(5) | 3.17(3) | 2.94(4) | 2.56(4) | 2.17(3) | 1.78(4) |
| Collinear Approx. | 2.34(5) | 2.38(4) | 2.33(4) | 2.15(2) | 1.95(4) | 1.69(3) | 1.46(2) | 1.16(3) |
| N jets $\geq 2$ | 1.96(4) | 1.97(4) | 1.95(4) | 1.77(2) | 1.61(3) | 1.41(3) | 1.20(2) | 0.95(3) |
| Forward jet | 1.48(4) | 1.49(4) | 1.48(3) | 1.34(2) | 1.21(3) | 1.08(3) | 0.91(2) | 0.73(3) |
| $b$-jet veto | 1.26(3) | 1.30(3) | 1.25(3) | 1.16(2) | 1.04(3) | 0.94(2) | 0.77(2) | 0.64(2) |
| Jet kinematics | 0.70(3) | 0.69(2) | 0.70(2) | 0.63(1) | 0.58(2) | 0.52(2) | 0.43(1) | 0.37(2) |
| Central jet veto | 0.61(2) | 0.60(2) | 0.62(2) | 0.56(1) | 0.50(2) | 0.45(2) | 0.38(1) | 0.32(2) |
| Mass window | 0.52(2) | 0.50(2) | 0.51(2) | 0.45(1) | 0.39(2) | 0.34(1) | 0.29(1) | 0.23(1) |

## 2.8 Summary of the event selection for $lh$-channel

The event selection for $lh$-channel is summarized below, including some kinematic requirements specific to the $lh$-channel.

- Trigger: electron trigger `e22i` or muon trigger `mu20`.

- Trigger lepton: at least one lepton must have a reconstructed $p_T$ greater or equal to the corresponding trigger requirement.

- Dilepton veto: exactly one identified lepton (ensures this sample is disjoint from the $ll$-channel).

- Hadronic $\tau$: exactly one identified hadronic $\tau$ with opposite charge of the lepton.

- Missing $E_T$: $E_T^{miss} \geq 30$ GeV.

- Collinear approximation: $0 \leq x_l \leq 0.75$, $0 \leq x_h \leq 1$, and $\cos\Delta\phi_{lh} \geq -0.9$. The asymmetric treatment of $x_h$ and $x_l$ provides background rejection in the $lh$-channel.
- Transverse mass: in order to further suppress the $W$ + jets and $t\bar{t}$ backgrounds, a cut on the transverse mass of the lepton and $E_{\mathrm{T}}^{\mathrm{miss}}$

$$m_T = \sqrt{2\, p_T^{lep}\, E_{\mathrm{T}}^{\mathrm{miss}} \cdot (1-\cos\Delta\phi)} \leq 30 \text{ GeV} \tag{7}$$

  is required, where $p_T^{lep}$ is the transverse momentum of the lepton in the $lh$-channel and $\Delta\phi$ is the angle between that lepton and $\vec{E}_{\mathrm{T}}^{\mathrm{miss}}$ in the transverse plane. The shape of the control samples used to estimate the signal sensitivity are obtained after these cuts; additional details of the data-driven background estimation are given in Section 3. In addition to the cuts above, the signal candidates are also required to satisfy the following cuts.
- Jet multiplicity: At least one jet with $p_T \geq 40$ GeV and at least one additional jet with $p_T \geq 20$ GeV.
- Forward jets: in opposite hemispheres $\eta_{j1} \times \eta_{j2} \leq 0$, with tau centrality $\min\{\eta_{j1}, \eta_{j2}\} \leq \eta_{lep,\tau} \leq \max\{\eta_{j1}, \eta_{j2}\}$ for the two highest $p_T$ jets.
- Jet kinematics: $\Delta\eta_{jj} \geq 4.4$ and dijet mass $m_{jj} \geq 700$ GeV for two forward jets.
- Central jet veto: the event is rejected if there are any additional jets with $p_T \geq 20$ GeV in $|\eta| \leq 3.2$.
- Mass window: $m_H - 15$ GeV $\leq m_{\tau\tau} \leq m_H + 15$ GeV around the test mass $m_H$.

Table 5 summarizes the cross-section for signal events after each of the cuts described above. With 30 fb$^{-1}$ integrated luminosity, about 20 signal events are expected in the mass window.

Table 5: Signal cross-sections (fb) for the $lh$-channel for various Higgs boson masses.

| Mass (GeV) | 105 | 110 | 115 | 120 | 125 | 130 | 135 | 140 |
|---|---|---|---|---|---|---|---|---|
| Cross section (fb) | 394.7 | 372.0 | 341.8 | 309.1 | 266.8 | 225.4 | 180.1 | 135.8 |
| Trigger | 65.6(3) | 65.1(2) | 61.1(2) | 57.2(1) | 51.5(2) | 44.7(1) | 36.5(1) | 28.3(1) |
| Trigger lepton | 56.4(3) | 56.2(2) | 53.2(2) | 49.5(1) | 44.7(2) | 38.9(1) | 31.8(1) | 24.7(1) |
| Dilepton veto | 50.0(3) | 49.6(2) | 46.7(2) | 43.4(1) | 38.9(2) | 34.0(1) | 27.6(1) | 21.3(1) |
| Hadronic $\tau$ | 7.7(1) | 8.1(1) | 8.1(1) | 8.02(7) | 7.4(1) | 6.68(8) | 5.72(7) | 4.53(9) |
| $E_{\mathrm{T}}^{\mathrm{miss}} \geq 30$ GeV | 4.8(1) | 5.1(1) | 5.08(9) | 4.96(5) | 4.63(8) | 4.16(7) | 3.51(6) | 2.82(8) |
| Collinear Approx. | 3.19(9) | 3.50(8) | 3.51(8) | 3.34(5) | 3.14(7) | 2.77(6) | 2.37(5) | 1.91(6) |
| Transverse mass | 2.53(8) | 2.70(7) | 2.67(7) | 2.46(4) | 2.26(6) | 1.98(5) | 1.64(4) | 1.29(5) |
| N jets $\geq 2$ | 2.12(7) | 2.22(7) | 2.21(6) | 2.02(4) | 1.80(5) | 1.60(4) | 1.32(4) | 1.00(5) |
| Forward jet | 1.61(7) | 1.66(6) | 1.73(5) | 1.52(3) | 1.41(5) | 1.20(4) | 1.03(3) | 0.78(4) |
| Jet kinematics | 0.88(5) | 0.86(4) | 0.92(4) | 0.82(2) | 0.73(3) | 0.65(3) | 0.56(2) | 0.42(3) |
| Central jet veto | 0.77(5) | 0.77(4) | 0.81(4) | 0.72(2) | 0.63(3) | 0.55(2) | 0.50(2) | 0.38(3) |
| Mass window | 0.68(4) | 0.68(4) | 0.70(3) | 0.61(2) | 0.52(3) | 0.44(2) | 0.40(2) | 0.30(3) |

### 2.9 Summary of the event selection for *hh*-channel

The event selection for *hh*-channel is summarized below, including some kinematic requirements specific to the *hh*-channel.

- Trigger: a combination of the hadronic tau and missing $E_T$ trigger `L1_TAU30_xE40_softHLT`.

- Hadronic taus: two identified hadronic taus are required $p_T$ above 35 GeV and 30 GeV with opposite charge.
- Missing $E_T$: $E_{\mathrm{T}}^{\mathrm{miss}} \geq 40$ GeV.
- Collinear approximation: $0.2 \leq x_{h1,h2} \leq 1$, and $\cos\Delta\phi_{hh} \geq -0.9$.
- Di-tau transverse mass: in order to further suppress fake-$\tau$ candidates from $W$ + jets and $t\bar{t}$ backgrounds, a cut on the di-tau transverse mass

$$m_T^{hh} = \sqrt{2\, p_T^{hh}\, E_{\mathrm{T}}^{\mathrm{miss}} \cdot (1-\cos\Delta\phi)} \leq 80 \text{ GeV} \quad (8)$$

  is required, where $p_T^{hh}$ is the transverse momentum of the two hadronic tau system and $\Delta\phi$ represents the azimuthal angle between $p_T^{hh}$ and $\vec{E}_{\mathrm{T}}^{\mathrm{miss}}$. This variable has been identified as potentially useful for the analysis. The optimal value of this cut depends heavily on the relative amount of the $W$+jets, $t\bar{t}$ and QCD backgrounds, therefore, the requirement is kept fairly loose.
- Jet multiplicity, forward jets, angular cuts, and central jet veto in the case of the *lh*-channel.
- Total $p_T$: to reject events with many jets like $t\bar{t}$, a cut on the total $p_T$ is applied:

$$||\vec{p_T}^{h1} + \vec{p_T}^{h2} + \vec{p_T}^{j1} + \vec{p_T}^{j2} + \vec{E}_{\mathrm{T}}^{\mathrm{miss}}|| \leq 60 \text{ GeV} \quad . \quad (9)$$

- Jet kinematics: $\Delta\eta_{jj} \geq 4$ and dijet mass $m_{jj} \geq 700$ GeV for two forward jets.
- Central jet veto: the event is rejected if there are any additional jets with $p_T \geq 20$ GeV in with $|\eta| \leq 3.2$.
- Mass window: $m_H - 15$ GeV $\leq m_{\tau\tau} \leq m_H + 20$ GeV around the test mass $m_H$.

Table 6 summarizes the cross-section for signal events after each of the cuts described above. The events used in the analysis have been generated applying a filter that requires two hadronic taus with $p_T \geq$12 GeV in the final state, produced in $|\eta| \leq$2.7 and with $\Delta\phi \leq 2.9$.

Table 6: Signal cross-sections (fb) for the *hh*-channel for various Higgs boson masses. The events used in the analysis had a filter applied at generation that required two hadronic taus with $p_T \geq$12 GeV in the final state, produced in $|\eta| \leq$2.7 and with $\Delta\phi \leq 2.9$.

| Mass (GeV) | 105 | 110 | 115 | 120 | 125 | 130 | 135 |
|---|---|---|---|---|---|---|---|
| Cross section (fb) | 394.7 | 372.0 | 341.8 | 309.1 | 266.8 | 225.4 | 180.1 |
| Trigger tau & MET | 12.4(2) | 12.1(2) | 12.0(2) | 11.4(1) | 10.4(2) | 9.2(1) | 7.93(1) |
| 2 Hadronic $\tau$s | 1.73(8) | 1.80(8) | 1.93(8) | 1.83(4) | 1.67(7) | 1.52(5) | 1.29(5) |
| $E_{\mathrm{T}}^{\mathrm{miss}} \geq 40$ GeV | 1.34(7) | 1.39(7) | 1.50(7) | 1.43(3) | 1.32(6) | 1.17(5) | 0.99(4) |
| Collinear Approx. | 0.91(6) | 1.02(6) | 1.13(6) | 1.03(3) | 1.00(5) | 0.85(4) | 0.72(3) |
| Di-tau Transverse mass | 0.91(6) | 1.02(6) | 1.13(6) | 1.03(3) | 1.00(5) | 0.85(4) | 0.72(3) |
| N jets $\geq 2$ | 0.77(5) | 0.88(6) | 0.94(5) | 0.86(3) | 0.84(5) | 0.72(4) | 0.61(3) |
| Total $p_T$ | 0.72(5) | 0.84(5) | 0.91(5) | 0.83(3) | 0.80(5) | 0.69(4) | 0.58(3) |
| Forward jet | 0.62(5) | 0.73(5) | 0.75(5) | 0.72(2) | 0.68(4) | 0.58(3) | 0.50(3) |
| Jet kinematics | 0.37(4) | 0.43(4) | 0.41(4) | 0.45(2) | 0.41(3) | 0.36(3) | 0.28(2) |
| Central jet veto | 0.34(3) | 0.38(4) | 0.36(3) | 0.39(2) | 0.35(3) | 0.32(3) | 0.24(2) |
| Mass window | 0.25(3) | 0.35(4) | 0.33(3) | 0.34(2) | 0.29(3) | 0.27(2) | 0.20(2) |

# 3 Background estimation

## 3.1 Overview

Despite the advances in theoretical tools and extraordinarily detailed simulation of the ATLAS detector, it is preferable to estimate backgrounds from data instead of relying entirely on Monte Carlo estimates. Below we describe data-driven background estimation techniques for each of the major backgrounds and estimate the systematic uncertainty of the estimates. Each technique has been developed to address the aspects of the background estimation which are most relevant for the analysis: the shape of the $m_{\tau\tau}$ tail from the irreducible $Z \to \tau\tau$, the fake tau contribution in the *lh*-channel, and the normalization of the QCD backgrounds. Section 4 describes how these techniques are incorporated into the final signal extraction, significance calculation, and mass measurement.

While it is possible to use Monte Carlo to estimate the systematics associated with the data-driven background estimation techniques, we must wait for data until we can employ these methods to produce reliable *estimates* of the backgrounds. Thus, in a feasibility study such as this one we face the additional challenge of *predicting* the expected backgrounds with Monte Carlo. The challenge of predicting our background and the associated uncertainties are distinct from the ones that we will face once we have collected $\sim$30 fb$^{-1}$ of data.

The major challenge in background prediction is related to the limited size of our Monte Carlo samples. Backgrounds from mis-identified leptons are difficult to estimate due to the large rejection factors of the identification algorithms. Even the irreducible $Z \to \tau\tau$ backgrounds are suppressed by several orders of magnitude due to the kinematic requirements. The $t\bar{t}$ background requires particularly large sample sizes because it is suppressed by both identification and kinematic requirements. Table 7 summarizes the size of the Monte Carlo samples used in this study and their corresponding luminosity. Despite the large computing investment and generator-level filters[3], many background samples were not sufficiently large to estimate rates deep in the analysis where rejection is at the order of $10^6$ - $10^8$. Thus, the "full" GEANT simulation was augmented with a fast simulation sample $\sim$100 times larger and a cut factorization method was used to predict the final background rates. While these procedures have large uncertainties, they are only relevant to our ability to estimate our sensitivity and will not plague the analysis once we have data.

## 3.2 Z + jets

While there is some theoretical uncertainty in the $Z \to \tau\tau$+jets background [33], the most serious danger of this background comes from the high-side tail in the $m_{\tau\tau}$ distribution, where we would expect to see the signal. This tail is dominated by instrumental effects; particularly mis-measurement of $E_{\mathrm{T}}^{\mathrm{miss}}$ , which is correlated to instrumental effects related to jet energy mis-measurement. Thus, we have developed a data-driven background estimation technique, which begins with a signal-free $Z \to \mu\mu$+jets sample and transfers the dominant instrumental effects to the $Z \to \tau\tau$+jets sample. This is achieved by replacing the muons with an equivalent tau, and carefully treating the decay of the tau. This technique is justified because the $Z \to \mu\mu$+jets events have identical jet activity and kinematics as $Z \to \tau\tau$+jets (before the tau decays) and because the relevant features of tau decays are well understood. We restrict the technique to the $Z \to \mu\mu$+jets control sample because muons lose only a small fraction of their energy in the calorimeter, and the effect on $E_{\mathrm{T}}^{\mathrm{miss}}$ is easier to estimate. After creating the emulated $Z \to \tau\tau$+jets control sample, the full event selection is applied. The normalization of the $Z \to \tau\tau$ background does not

[3] `VBFCut` : $N_{e/\mu} \geq 1$ or 2, or $N_\tau \geq 1$ with $p_T \geq 10$ GeV, $|\eta| \leq 2.7$ for electron, muon and tau from $W$ and $Z$, respectively. For the hadron level jets with cone size 0.4, $N_{jet} \geq 2$ with $p_T^1 \geq 20$ GeV for the highest $p_T$ jet, $p_T^2 \geq 15$ GeV for the second highest $p_T$ jet, and $|\eta| \leq 5$, $m_{jj} \geq 300$ GeV, $\Delta\eta_{jj} \geq 2$.

Table 7: Details of the background samples used in this study. The VBFCut filter is applied for $Z/W$+jets events at generator level. The corresponding luminosities for $Z/W$+jets and di-boson events are estimated for 2 jets events for $Z$ and 3 jets events for $W$, and $WW$ events for di-boson process respectively. In $t\bar{t}$ samples, the actual physics events are approximately 70% to the total number of simulated events according to the treatment of the negative weighted events.

| Process | cross-section (pb) | Simulated events | Luminosity ($\text{fb}^{-1}$) |
|---|---|---|---|
| $Z \to ll$+jets | 35.1 | 714,500 | 63.1 |
| $W \to l\nu$+jets | 346.3 | 765,000 | 3.43 |
| $t\bar{t}$+jets (full) | 450.0 | 1,012,941 | 1.65 |
| $t\bar{t}$+jets (fast) | 833.0 | 96,250,000 | 84.3 |
| $t\bar{t}$ (2-muon, full) | 32.6 | 904,000 | 20.4 |
| $WW/WZ/ZZ$+jets | 174.2 | 258,094 | 0.57 |
| QCD di-jets (full) | $\sim 1.4 \times 10^9$ | 1,503,250 | $\sim 10^{-6}$ |
| QCD di-jets (fast) | $\sim 1.4 \times 10^9$ | 80,000,000 | $\sim 5 \times 10^{-5}$ |

require this emulation because it can be estimated directly from the height of the $Z$-peak in $m_{\tau\tau}$ spectrum obtained with the signal candidates.

The first task is to obtain a signal-free dimuon control sample. Both loose and tight control samples have been used in this data-driven estimation technique. The loose control sample is used to estimate the ditau background not only from $Z \to \tau\tau$+jets events but also from $t\bar{t}$ or di-boson backgrounds, while the tight control sample can be used for the $Z \to \tau\tau$ background estimation by obtaining relatively pure $Z \to \mu\mu$+jets events. The loose control sample requires only a minimum sets of cuts from the $ll$-channel, hence $\sim 10\%$ of this control sample includes other processes such as $t\bar{t}$, diboson or even $Z \to \tau\tau$. The tight control sample, in contrast, selects pure $Z \to \mu\mu$ events with less than 1% contamination from the other processes. A tighter event selection is used to define this control sample, it is identical to the final event selection in the $ll$-channel but excludes of the $E_{\mathrm{T}}^{\mathrm{miss}}$ , collinear approximation, and mass window cuts. To improve the purity of $Z$ events, a $Z$ mass window cut is used: $m_{\mu\mu} \geq m_Z - 10$ GeV. Due to a strong correlation between the $m_{\tau\tau}$ and $m_{\mu\mu}$ distributions before the $\mu \to \tau$ conversion, the $m_{\tau\tau}$ distribution is biased by a cut on $m_{\mu\mu}$. The lower bound on $m_{\mu\mu} \geq m_Z - 10$ GeV has little influence of the shape in the signal region; however, an upper mass cut causes a large bias. Therefore, no upper bound is placed on $m_{\mu\mu}$. After obtaining the dimuon events, the reconstructed muons are replaced by the Monte Carlo tau, then decayed and simulated in the ATLAS detector simulation and reconstruction software.

Two different techniques for replacing the muon with the tau decay products were evaluated in this study. One uses the TAUOLA decay package [27] and the other uses a simple re-scaling of the momentum and efficiency of taus, since the technical complexity of re-simulation by TAUOLA are rather difficult in the full detector simulation and reconstruction. A comparison of the two different methods provides a quantitative validation of this procedure. Several comparisons of the $Z \to \tau\tau$+jets control sample emulated from $Z \to \mu\mu$ and the true $Z \to \tau\tau$ background were performed after imposing the full event selection criteria for the $ll$- and $lh$-channels. Figure 9 shows the $p_T$ of the leptons and the $E_{\mathrm{T}}^{\mathrm{miss}}$ distributions of the emulated sample compared to the true $Z \to \tau\tau \to ll + 4\nu$ process. Figure 10 (a) depicts the reconstructed visible mass for the true and emulated samples in the $lh$-channel and (b) shows the bin-by-bin ratio of these distributions. Excellent agreement between the true and emulated distributions are also observed in each of the $ll$-, $lh$-, and $hh$-channels in the region of interest. The gray horizontal band in the figure represents $\pm 10\%$ around a ratio of 1, which is used to reflect the uncertainty

in the shape from the tau modeling and sensitivity to the analysis cuts. With this method we are able to accurately model both the shape and normalization of the $Z \to \tau\tau$ backgrounds for all tau decays.

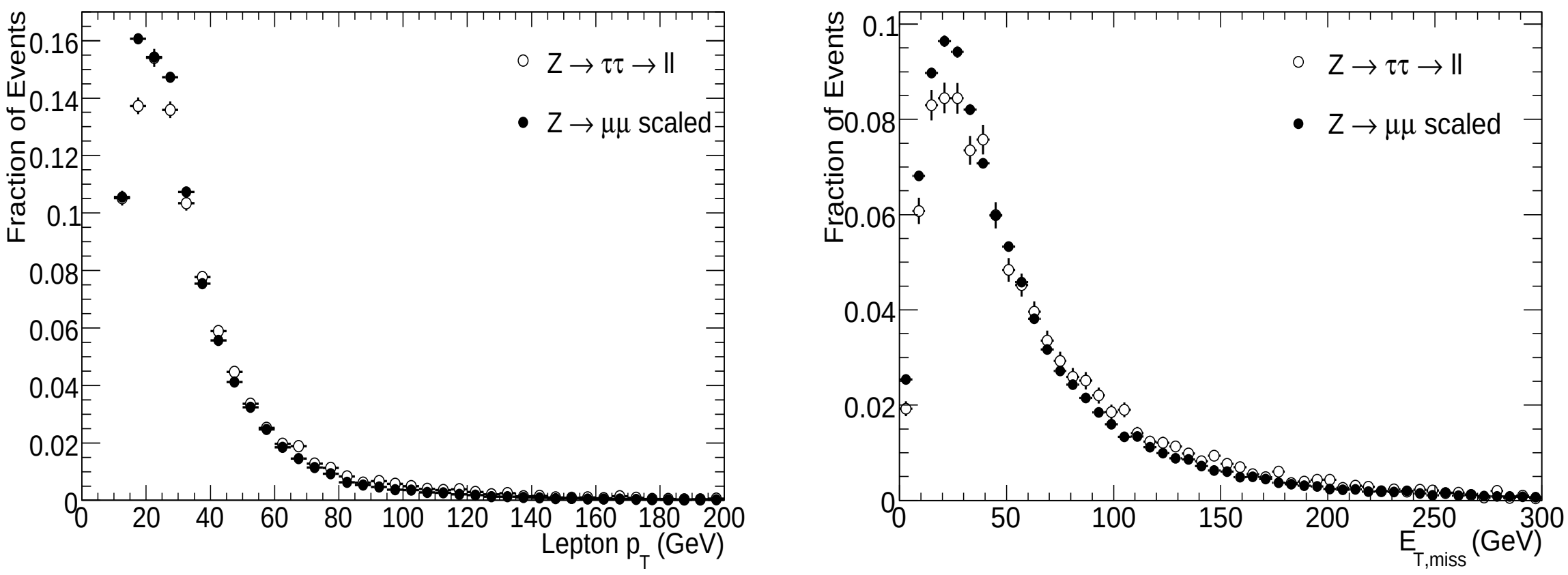


(a) (b)

Figure 9: Transverse momentum of the leptons (a) and missing transverse energy (b) for the two processes rescaled $Z \to \mu\mu$ and true $Z \to \tau\tau \to ll + 4\nu$.

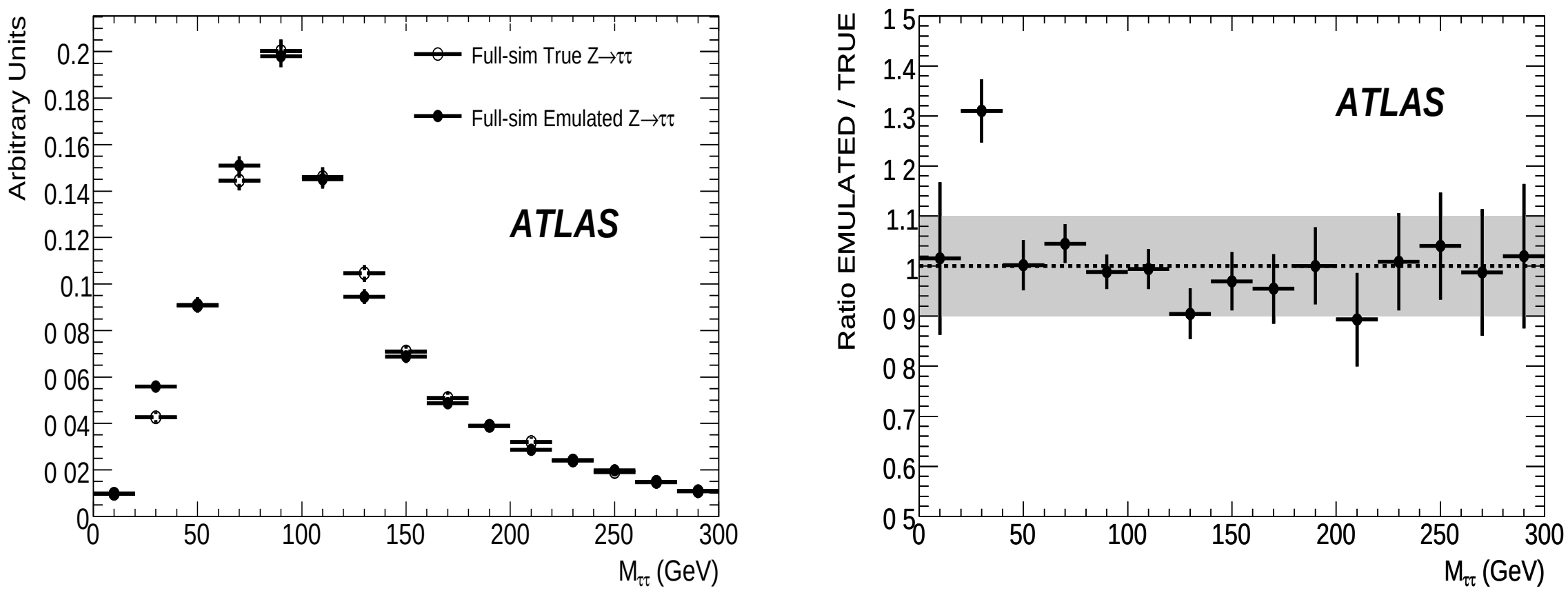


(a) (b)

Figure 10: Reconstructed invariant mass distribution (a) and its bin-by-bin ratio (b) generated from the true and emulated $Z \to \tau\tau \to lh + 3\nu$ events. The gray band represents $\pm 10\%$ around a ratio of 1.

### 3.3 QCD background in $lh$-channel

The method described above estimates the contribution of taus for all processes, including $t\bar{t}$, but does not estimate the contribution of the fake taus. The fake rate of leptons is much smaller as compared to the contribution from real leptons from $W$ and $\tau$ decays in the $ll$-channel. In contrast, in the $lh$-channel roughly half of the $t\bar{t}$ background come from fake hadronic-taus. In addition, the $W$+jets background is comparable to $t\bar{t}$ in the $lh$-channel; therefore, estimating the QCD fake contribution to the $lh$-channel requires a dedicated procedure. Estimation of the QCD fake rate is determined from a data-driven method.

The technique used here exploits the track multiplicity in a cone around the tau candidate. Real taus typically have one or three tracks, with some spread due to tracking efficiency or the presence of spurious tracks. Electrons have dominantly a single track, while jets have a broad distribution with a higher average multiplicity. Figure 11 shows the track multiplicity distribution for taus, electrons, and jets in a cone of radius 0.7 after removing outlying tracks. It is clear that the track multiplicity can be used to constrain the relative abundance of the three components to the distribution. While Fig. 11 was created from Monte Carlo, the electron and jet track multiplicity distributions can easily be obtained from data.

Given a sample of tau candidates, the relative abundance of taus, electrons, and jets can be found by fitting the track multiplicity distribution with the extended likelihood function

$$\begin{aligned} L_{track}(r_{QCD}, r_{tau}) \;=\; & \prod_i^N Pois(n_{exp}^{tot} \times (r_{tau} f_{tau}^i + r_{QCD} f_{jet}^i + (1 - r_{tau} - r_{QCD}) f_{lep}^i) | N_{obs}^i) \\ & \times\, Gaus(N_{obs}^{tot} | n_{exp}^{tot}, \sqrt{n_{exp}^{tot}}) \\ & \times Gaus(N_{lep}^{measured} | n_{exp}^{tot}(1 - r_{tau} - r_{QCD}), \Delta_{lep} n_{exp}^{tot}(1 - r_{tau} - r_{QCD})) \end{aligned} \tag{10}$$

where $n_{exp}^{tot}$ is the total number of events estimated by the fit, $r_{tau}$ ($r_{QCD}$) is the fraction of the tau (jet) contribution with respect to the estimated total number of events, $\Delta_{lep} = 10\%$ is the relative uncertainty on lepton measurement, and $f^i$ is the normalized probability for the $i^{th}$ bin of the track multiplicity distribution. The second term constrains the normalization, and the third term is an additional constraint term for the lepton contribution estimated by an independent analysis. The fit is performed to find the maximum likelihood estimate with MINUIT [34].

The track multiplicity distribution for the QCD jets is modeled from samples of QCD dijets that produce tau candidates with $p_T$ in the range of 17- 280 GeV. No event level selections are applied at this stage. Similarly, the multiplicity distribution of the tau signal and lepton background are modeled with Drell-Yan Monte Carlo samples; however, all analysis requirements up to transverse mass cut are applied. Pseudo-datasets were generated for various luminosities based on the corresponding cross-sections and multiplicity distributions. The highly uncertain QCD multi-jet background was scaled to be five times larger than the estimated rates of $t\bar{t}$ and $W$+jets after event selection. A fit was performed for each of the 2000 pseudo-datasets and the results were used to estimate quantify the performance of the method. The expected error on the fraction $r_{tau}$ is presented in Fig. 12 as a function of luminosity.

The fraction $r_{tau}$ in the signal candidates remaining after the transverse mass cut can be measured to within 5% accuracy with 1 fb$^{-1}$ integrated luminosity. The largest uncertainty in this method comes from the dependence of the track multiplicity on the jet $p_T$. The systematics were estimated by dividing the jet into two samples; those with $p_T \leq 70$ GeV and those with $p_T \geq 70$ GeV. Additionally, the presence of $E_{\mathrm{T}}^{\mathrm{miss}}$ in the pure QCD processes is strongly correlated to the event kinematics. Thus, we assign an additional systematic associated with the variation observed when repeating the method with modified track multiplicity distributions with and without the requirements on $E_{\mathrm{T}}^{\mathrm{miss}}$ . The systematics associated with the QCD shape contribute about 2% to $r_{tau}$ measurement.

### 3.4 Cut factorization method

The analysis cuts described in Section 2 have rejections against backgrounds of the order of $10^8$. Only a few tens of events are expected with 30 fb$^{-1}$ of data, and the background Monte Carlo samples generally correspond to 5 fb$^{-1}$ or less. The lack of sufficiently large Monte Carlo samples requires an approximate procedure to predict the background rate at the end of the analysis. We utilize a cut factorization method in which the analysis cuts are divided into three categories that are roughly uncorrelated so that the rejection can be factorized. The first category are related to the tau decays from the Higgs boson

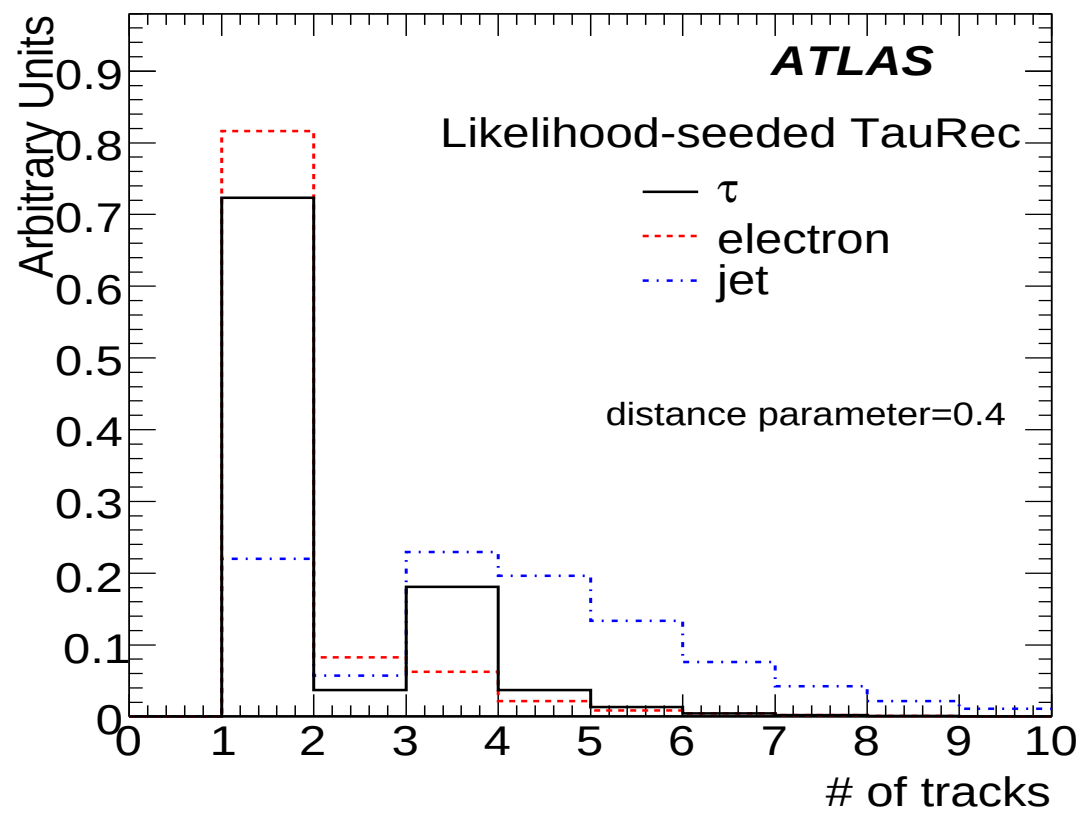


Figure 11: Track multiplicity distribution for QCD fake events and electron-fake events as well as the $\tau$ signal.

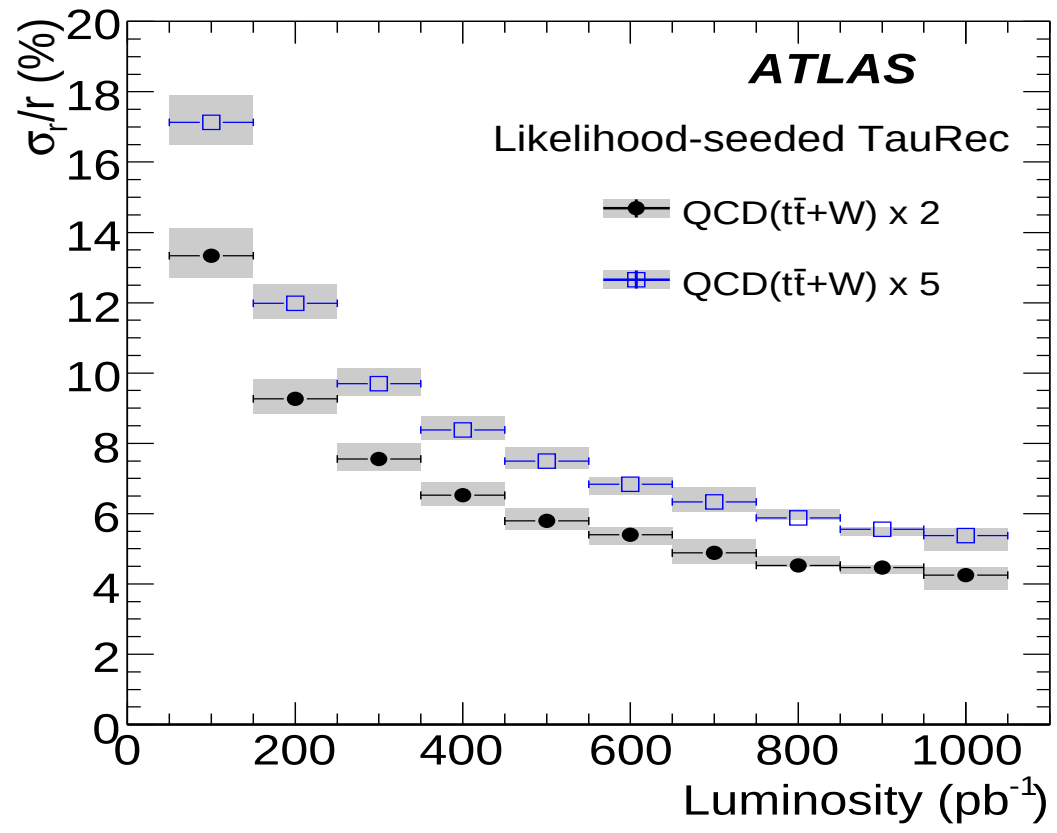


Figure 12: Expected errors of the fraction $r_{tau}$ as a function of luminosity. The QCD events are scaled to $\times 2$ and $\times 5$.

candidate (trigger, lepton ID, hadronic tau ID, $E_{\mathrm{T}}^{\mathrm{miss}}$, the collinear approximation, and the transverse mass) and the rejection is dominated by detector performance issues. The second category of cuts are related to the tagging jets (forward jets, jet separation, and dijet mass) and the rejection is dominated by the kinematic properties of the events. The third category consists of those cuts which are strongly correlated to both the forward tagging jets and the tau decay products (centrality, central jet veto, and mass window cut). The method itself is simple; the background rejection rate is determined for each of the categories individually, and the product of these rejection rates is used to estimate the total rejection rate. Two variations on this cut factorization procedure were considered. In the first approach, the rejection of the jet-related cuts were calculated without any tau-related cuts. The second approach differed in that it included the lepton and tau identification in order to avoid bias effects from the contribution of fake leptons and taus. The residual correlations between the categories of cuts contributes to an uncertainty in this technique. The uncertainties were estimated with the use of larger samples produced with fast simulation and fully simulated samples with generator-level filters, which enrich the backgrounds in the signal-like region. The factorization process was used for all background processes except for the irreducible $Z \to \tau\tau$ background.

For the $W \to l\nu$+jets and $Z \to ll$+jets backgrounds, ALPGEN samples with up to 5-jets corresponding to 4 fb$^{-1}$ integrated luminosity were used. Table 8 shows the average of the jet rejection from the two approaches, which is used as our final prediction. The $Z$ events in $ll$-channel indicate that the jet-related cuts are not strongly correlated with the tau-related cuts, since the rejection rate is the same in the electron and muon modes. The correlation between the categories of cuts was investigated with an ALPGEN $Z \to \tau\tau$ + n jets sample enriched with the `VBFCut` filter. The production kinematics are the same, but the enhanced $E_{\mathrm{T}}^{\mathrm{miss}}$ due to the $\tau$ decays lets a sufficient number of events survive all the analysis cuts. The background predictions with and without cut factorization are consistent within the 20% statistical error. Thus, we assign a 20% systematic uncertainty for this evaluation method on this process.

The $t\bar{t}$ background is the most complicated process in that it includes both irreducible and reducible contributions in all channels. We use $10^6$ $t\bar{t}$ events produced by the MC@NLO generator. The events do not include the process in which both $W$s decay hadronically, thus the sample corresponds to 1.6fb$^{-1}$ of integrated luminosity. While the rejection of the tau-related cuts can be reliably estimated, the rejection of the jet-related cuts suffers from the limited sample sizes. The rejection of the jet-related cuts is expected to be very high ($\times 10^4$ for $t\bar{t}$), resulting in a $\sim$30% statistical error in the final background predictions.

Table 8 summarizes the rejection of the jet-related cuts subdivided into events with a real or fake hadronic tau for the $lh$-channel and by lepton flavor for the $ll$-channel. Again, the average of the two approaches is used for the final background prediction. Table 8 shows that there is a considerable contribution of fake-tau events for the $lh$-channel. In the $ll$-channel, the contribution from semi-leptonic decays of $B$-hadrons was found to be very small. An estimate of the uncertainty in the method was found by comparing the prediction from cut-factorization to the direct estimate from an enriched sample filtered to require at least two muons in the event. Furthermore, the cut-factorization results were compared to a sample produced with the fast simulation that was approximately 100 times larger. Based on these comparisons, we assign 50% uncertainty in the $t\bar{t}$ background prediction.

The diboson background contribution was predicted with the same cut-factorization procedure. Even after using the cut-factorization procedure, the statistical uncertainty in the prediction is very large. Fortunately, the diboson production cross-section is much smaller than the $t\bar{t}$ processes, so the effect of this background is very small. The results of the cut factorization prediction are presented in Table 8, and we assign a large uncertainty of 50% for diboson background rate. While a number of tests were performed to validate the method, it is clear that this is an approximate procedure and that the limited size of the Monte Carlo samples fundamentally limits our ability to predict this background.

Table 8: Rejection rates of the jet-related cuts for $Z/W$+jets, $t\bar{t}$ and diboson events in $ll$- and $lh$-channels, respectively.

| Acceptance | $ll$-channel | | | |
|---|---|---|---|---|
| (%) | $Z \to ee/\mu\mu$+jets | $W \to e\nu/\mu\nu$+jets | $t\bar{t}$ ($ee/\mu\mu/e\mu$) | $WW/WZ/ZZ$ |
| Jet kinematics | 2.32(2) / 2.43(2) | 2.4(3) / 1.5(2) | 0.72(8) / 0.52(5) / 0.60(4) | 0.43(5) / 0.56(3) / 0.33(3) |
| Central jet veto | 1.00(1) / 1.02(1) | 1.1(3) / 0.8(2) | 0.11(3) / 0.04(1) / 0.10(1) | 0.26(5) / 0.26(4) / 0.10(2) |
| Mass window | 0.230(5) / 0.202(4) | 0.1(1) / 0.1(1) | 0.019(7) / 0.002(1) / 0.010(3) | – / – / 0.03(1) |

| Rejection rate | $lh$-channel | | | |
|---|---|---|---|---|
| (%) | $Z \to ee/\mu\mu$ | $W \to e\nu/\mu\nu$ | $t\bar{t}$ (tau / non-tau) | $WW/WZ/ZZ$ |
| Jet kinematics | 2.60(5) / 2.8(1) | 2.7(1) / 2.6(1) | 0.93(6) / 1.11(6) | 0.2(2) / 0.6(1) / 0.50(3) |
| Central jet veto | 1.14(3) / 1.30(9) | 1.7(1) / 1.3(1) | 0.07(3) / 0.12(3) | 0.1(1) / 0.3(1) / – |
| Mass window | 0.22(1) / 0.11(2) | 0.10(3) / 0.05(2) | 0.015(8) / – | – / 0.03(3) / – |

## 3.5 Background for the $hh$-channel

In addition to the $Z$+jets, $W$+jets and $t\bar{t}$ backgrounds, the $hh$-channel also has a background from the pure QCD multi-jet process. The estimation of QCD multi-jets must be made from data. A few handles exist for estimating the pure QCD background. First, one can utilize a sample of same-sign tau candidates to estimate the fake tau contribution since the sign of the tau candidate from QCD is approximately random. Potentially, one can utilize constraints from the track multiplicity distribution as described in Section 3.3. Furthermore, one can loosen the identification requirements on the tau candidates to obtain a sample dominated by QCD fakes, and then extrapolate this background into the signal region using knowledge of the fake tau's likelihood distribution obtained with data. The efficiency of these techniques must also be established with data, but here we assume that the same-sign sample will be able to estimate the QCD background with an uncertainty given by two components. The first component of this uncertainty is the statistical error in the control sample, which scales like $1/\sqrt{N_{SS}}$, where $N_{SS}$ is the size of same sign sample coming from QCD backgrounds. The second component is a systematic uncertainty associated with using the same sign sample to estimate the opposite sign sample. Experience from the Tevatron shows that charge correlations can be of order 13% with an uncertainty of order 3% [35, 36]. Given that the final state requires two additional jets, which can alter the contribution of quark and gluon-initiated

jets, we assume a 10% systematic error associated with the charge correlation. In addition, the same sign sample will also include a contribution from $W \to \tau\nu$+jets, where the Tevatron experiments observed a charge correlation that was much higher with a 40% uncertainty [35]. The experience from the Tevatron provides some insight into this approach, but the results are not directly relevant because the LHC is not a $\bar{p}$-$p$ machine.

As in the $ll$ and $lh$-channels we do not have sufficiently large Monte Carlo samples to predict the background for the $hh$-channel. We employ the same cut factorization method for the $Z$+jets, $W$+jets and $t\bar{t}$ backgrounds as described above. The prediction of the pure QCD multi-jet background from Monte Carlo is hopeless without factorizing the analysis even further. A sample of 80 million QCD dijet events (including $c\bar{c}$ and $b\bar{b}$ processes) were simulated with the fast detector simulation. The tau fake rate was parameterized from full simulation as a function of $\eta$ and $p_T$ and used to re-weight the dijet sample. In order for a pure QCD process to satisfy the event selection, there must be at least four high-$p_T$ jets. Previous studies have shown that the parton shower underestimates the tagging jet requirement by a factor of 2-3 [4]. Therefore, we multiply the prediction from QCD dijets after the forward jet requirement by a factor of 5 to include the underestimate from the parton shower and an additional safety factor. Table 11 shows that the analysis cuts are extremely effective at rejecting QCD backgrounds, but a realistic estimation of the remaining background requires data.

### 3.6 Summary of the background predictions

Tables 9, 10, and 11 summarize the background predictions for the $ll$-, $lh$-, and $hh$-channels, respectively. The tables also indicate the statistical uncertainty on the estimates from the limited size of the Monte Carlo samples. Note that the mass window cut is set with respect to the test Higgs mass of 120 GeV. The effective cross-sections estimated with the cut-factorization method are marked with an asterisk. Furthermore, the predictions from the $t\bar{t}$ sample simulated with the fast simulation are shown, where the effective cross-section was normalized to the fully simulated sample after the collinear approximation for the $ll$-channel and after the transverse mass cut for the $lh$-channel.

Table 9: Summary of the backgrounds for the $ll$-channel. An asterisk is used to indicate cross-sections estimated from the cut factorization method. Note that at least one of taus are decayed leptonically in QCD $Z \to \tau^+\tau^-$+jets process.

| | $Z \to \tau^+\tau^-$+jets($\geq$1) | | $t\bar{t}$ | | $Z \to l^+l^-$+n jets | $W \to l\nu$+n jets | diboson |
|---|---|---|---|---|---|---|---|
| | QCD | ELWK | Full | Fast | (n $\geq$ 1) | (n $\geq$ 1) | $WW/ZZ/WZ$ |
| Cross section (fb) | $168.4\times10^3$ | 1693 | $833\times10^3$ | | $768.6\times10^3$ | $8649\times10^3$ | $174.1\times10^3$ |
| Trigger | $51.5(1)\times10^3$ | 230(1) | $209.8(2)\times10^3$ | | $633.8(4)\times10^3$ | $4411(9)\times10^3$ | $32.0(1)\times10^3$ |
| Trigger lepton | $42.7(1)\times10^3$ | 190(1) | $179.1(2)\times10^3$ | | $588.0(4)\times10^3$ | $3815(9)\times10^3$ | $28.0(1)\times10^3$ |
| Dilepton | $4.25(5)\times10^3$ | 19.2(4) | $21.7(1)\times10^3$ | | $369.9(5)\times10^3$ | $2.5(2)\times10^3$ | $3.95(6)\times10^3$ |
| $E_T^{miss} \geq 40$ GeV | 744(18) | 9.9(3) | 16847( 99) | | 2683( 67) | 1148(176) | 1744( 49) |
| Collinear Approx. | 454(14) | 6.2(2) | 1817( 33) | Atlfast | 104( 12) | 46( 21) | 73( 9) |
| N jets $\geq 2$ | 262( 8) | 5.8(2) | 1722( 32) | 1699(4) | 73( 8) | 14( 6) | 51( 8) |
| Forward jet | 39( 2) | 2.0(1) | 294( 13) | 324(1) | 10( 3) | $\geq$1.2(2)* | 8( 3) |
| $b$-jet veto | 30( 2) | 1.5(1) | 89( 7) | 90.3(9) | 9( 3) | $\geq$1.0(2)* | 5( 2) |
| Jet kinematics | 2.71(5) | 0.57(5) | 11.8(3)* | 26.7(5) | 0.66(3)* | 0.19(4)* | 0.33(5)* |
| Central jet veto | 1.24(3) | 0.43(4) | 1.9(1)* | 2.6(1) | 0.27(1)* | 0.10(2)* | 0.18(4)* |
| Mass window | 0.23(1) | 0.04(1) | 0.10(2)* | 0.06(2) | 0.058(3)* | 0.01(1)* | 0.002(1)* |

Table 10: Summary of the backgrounds for the $lh$-channel. An asterisk is used to indicate cross-sections estimated from the cut factorization method. Note that at least one of taus are decayed leptonically in QCD $Z \to \tau^+\tau^-$+jets process.

| | $Z \to \tau^+\tau^-$+jets($\geq$1) | | $t\bar{t}$ | | $Z \to l^+l^-$+n jets | $W \to l\nu$+n jets | diboson |
|---|---|---|---|---|---|---|---|
| | QCD | ELWK | Full | Fast | (n $\geq$ 1) | (n $\geq$ 1) | $WW/ZZ/WZ$ |
| Cross section (fb) | 168.4×10³ | 1693 | 833×10³ | | 768.6×10³ | 8649×10³ | 174.1×10³ |
| Trigger | 51.5(1)×10³ | 230(1) | 209.8(2)×10³ | | 633.8(4)×10³ | 4411(9)×10³ | 32.0(1)×10³ |
| Trigger lepton | 42.7(1)×10³ | 190(1) | 179.1(2)×10³ | | 588.0(4)×10³ | 3815(9)×10³ | 28.0(1)×10³ |
| Dilepton veto | 38.4(1)×10³ | 171(1) | 156.4(2)×10³ | | 216.5(4)×10³ | 3811(9)×10³ | 23.7(1)×10³ |
| Hadronic $\tau$ | 3062( 42) | 19.3(4) | 5224( 56) | | 20250(156) | 32537(1012) | 704( 30) |
| $E_T^{miss}$ $\geq$ 30 GeV | 850( 20) | 12.1(3) | 4251( 50) | | 468(26) | 21001( 801) | 474( 26) |
| Collinear Approx. | 514( 15) | 7.8(2) | 606( 19) | | 17( 3) | 324( 46) | 32( 6) |
| Transverse mass | 415( 13) | 6.5(2) | 176( 10) | Atlfast | 11( 2) | 67( 18) | 14( 3) |
| N jets $\geq$ 2 | 235( 7) | 6.0(2) | 162(9) | 167(1) | 8( 1) | 49( 11) | 7( 1) |
| Forward jet | 40( 3) | 2.3(1) | 32(4) | 26.1(4) | 1.3(6) | $\geq$2.9(3)* | 3( 1) |
| Jet kinematics | 2.7(1) | 0.72(6) | 1.8(1)* | 3.6(1) | 0.10(1)* | 0.7(1)* | 0.06(1)* |
| Central jet veto | 1.2(1) | 0.49(5) | 0.25(4)* | 0.43(5) | 0.047(6)* | 0.43(6)* | 0.02(1)* |
| Mass window | 0.11(2) | 0.04(1) | 0.012(5)* | 0.03(1) | 0.008(1)* | 0.020(6)* | 0.001(1)* |

Table 11: Summary of the backgrounds for the $hh$-channel. An asterisk is used to indicate cross-sections estimated from the cut factorization method and/or an additional safety factor. Note that both taus are decayed hadronically in QCD $Z \to \tau^+\tau^-$+jets and $W \to \tau\nu$+jets processes. The $t\bar{t}$ sample is required to have at least one lepton in the top decay.

| | $Z \to \tau^+\tau^- + jets(\geq 1)$ | | $t\bar{t}$ | $W \to \tau\nu + n\,jets$ | QCD di-jet |
|---|---|---|---|---|---|
| | QCD | ELWK | | $(n \geq 1)$ | (× 5) |
| Cross section (fb) | 40.3×10³ | 1693 | 833 ×10³ | 922×10³ | 19.1 10¹² |
| Trigger tau & MET | 1756(15) | 126(1) | 78177(232) | 39600(400) | |
| 2 Hadronic $\tau$s | 161(4) | 4.9(2) | 373(16) | 317(33) | 2.756(3) 10⁶* |
| $E_T^{miss}$ $\geq$ 40 GeV | 108(4) | 3.7 (2) | 335(15) | 243(29) | 0.97(3) 10³* |
| Collinear Approx. | 72(3) | 2.3 (1) | 43(5) | 20(7) | 1.7(2) 10²* |
| Di-tau Transverse mass | 72(3) | 2.3(1) | 39(5) | 18(7) | 1.6(2) 10²* |
| N jets $\geq$ 2 | 46(2)* | 2.1(1) | 34(5)* | 8(3)* | 0.86(4) 10²* |
| Total $p_T$ | 40(2)* | 1.9(1) | 24(4)* | 8(3)* | 0.75(3) 10²* |
| Forward jet | 17(1)* | 1.1(1) | 9(2)* | 3(1)* | 23(3)* |
| Jet kinematics | 1.4(1)* | 0.43(6) | 0.6(2)* | 0.5(4)* | 8(3)* |
| Central jet veto | 0.7(1)* | 0.36(6) | 0.16(9)* | 0.3(3)* | 4(1)* |
| Mass window | 0.08(3)* | 0.03(1) | 0.03(3)* | 0.1(1)* | 1(1)* |

# 4 Signal sensitivity

## 4.1 Overview

In this section we outline the method for extracting the signal significance from the data and measuring the Higgs boson mass. In addition to a simple method based on counting the number of events in a mass window, we present a method based on fitting the $m_{\tau\tau}$ spectrum. Particular care has been given to the incorporation of uncertainty in both the rate and shape for the signal and backgrounds. The fitting strategy constrains the shape of the $Z \to \tau\tau$ background from the data-driven techniques described in Section 3.2. The data-driven estimates for the $t\bar{t}$ and $W$+jets background are not finalized, thus we rely on Monte Carlo to constrain the shape of those backgrounds and allow for large variation in the shape. In the $lh$-channel, the normalization of the fake-tau contribution is constrained from the track multiplicity measurement outlined in Section 3.3. Sections 4.2 and 4.3 describe the parameterization of the signal and backgrounds, while Sections 4.4 and 4.5 describe how the control samples are incorporated to constrain the normalization and shape uncertainties. The expected signal sensitivity is estimated by considering a hypothetical data set given by the median of the signal-plus-background estimate. The normalization of the background predictions are described in Section 3.6. Due to the limited size of the Monte Carlo samples, the shape of the $m_{\tau\tau}$ spectrum for the $t\bar{t}$ and $W$+jets background was obtained from an earlier point in the event selection, just after the collinear approximation requirement in the $ll$-channel and after the transverse mass cut in the $lh$-channel (see Sections 2.7 and 2.8). Similarly, the shape of the $Z \to \tau\tau$ background in the control sample, which can be estimated by the data-driven techniques described in Section 3.2, was obtained from a $Z \to \tau\tau$ Monte Carlo sample just after the same points in the event selection. Depending on how the analysis evolves, the shape of the $m_{\tau\tau}$ spectrum from an earlier point in the event selection – which is dominated by backgrounds – may also provide a useful control sample to validate the data-driven background estimation techniques since it is fairly stable in the later stages of the event selection.

For the first time ATLAS has investigated the $hh$-channel. While the signal efficiency and $m_{\tau\tau}$ mass resolution are roughly comparable to the $ll$- and $lh$-channels, a reliable estimate of the QCD background can only be provided with data. Therefore, we do not report on an estimated sensitivity for the $hh$-channel below.

## 4.2 Shape parameterization for $H \to \tau\tau$ and $Z \to \tau\tau$

The shape of the $m_{\tau\tau}$ distribution for signal and $Z \to \tau\tau$ events is dictated by the resolution of $E_{\mathrm{T}}^{\mathrm{miss}}$ and the kinematics of the collinear approximation. As discussed in Section 2.6, the width of the $m_{\tau\tau}$ distributions is given by the $E_{\mathrm{T}}^{\mathrm{miss}}$ resolution scaled by a Jacobian factor and the cuts on the $x_\tau$ variables introduce an asymmetry in the $m_{\tau\tau}$ distribution. A full solution would include an event-by-event Jacobian scaling of the width and truncation at $m_{\tau\tau} \geq m_{vis}/\sqrt{x_1^{cut} x_2^{cut}}$. An approximate parameterization can be constructed with these features in mind. First, we consider three sub-samples of events with $J \leq 2$, $2 < J \leq 5$, and $J > 5$. Let, $\langle J \rangle_i$ and $N_i$ denote the mean of the Jacobian and the number of events in the $i^{th}$ subsample, respectively. We account for the Jacobian scaling of the width by using a triple Gaussian with identical mean, normalizations according to the ratios of the $N_i$ and widths according to the ratios of $\langle J \rangle_i$. To account for the asymmetry introduced by the $x_\tau$ cuts, we modulate the triple Gaussian by an efficiency envelope derived from the $m_{vis}$ spectrum. The efficiency envelope reflects the probability that $m_{\tau\tau}$ is greater than $m_{\mathrm{vis}}$ and is parameterized as: $1/2 + 1/2\,\mathrm{erf}\{[m_{\tau\tau} - \langle m_{vis} \rangle]/\sqrt{2}\sigma_{vis}\}$, where $\langle m_{vis} \rangle$ and $\sigma_{vis}$ are the mean and standard deviation of the spectrum for those events which fail the $x_\tau$ cuts. The parameter $\langle m_{vis} \rangle$ depends on the Higgs boson mass and was linearly parameterized by

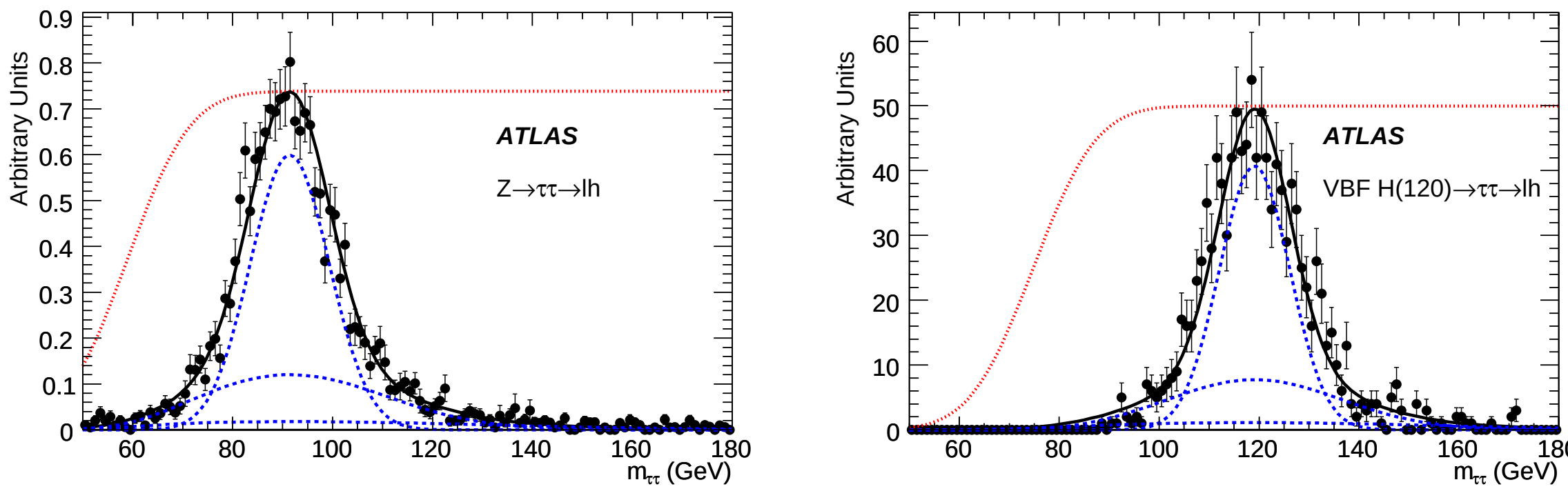


(a) (b)

Figure 13: Figures (a) and (b) show the result of a fit to a pure Monte Carlo samples of $Z \to \tau\tau$ and signal ($m_H = 120$ GeV) in the $lh$-channel, respectively. The dashed lines represent the three components of the model and the dotted curve represents the erf() efficiency envelope. These samples do not include pileup.

$0.576\,m_H + 60$ GeV; the width $\sigma_{vis}$ was fixed at 10 GeV. We parametrize the $m_{\tau\tau}$ distribution as

$$L_{H/Z}(m_{\tau\tau}|m,\sigma_{H/Z}) = \mathscr{N}\left[\frac{1}{2}+\frac{1}{2}\,\mathrm{erf}\left(\frac{m_{\tau\tau}-\langle m_{vis}\rangle}{\sqrt{2}\sigma_{vis}}\right)\right] \times \sum_{i=1}^{3} N_i\, Gaus(m_{\tau\tau}|m,\sigma_{H/Z}\langle J\rangle_i), \qquad (11)$$

where the resulting function's normalization constant $\mathscr{N}$ was found by numerical integration with the ROOFIT package [37].

The $Z \to \tau\tau$ control sample described in Section 3.2 is used to constrain the mean, $m_Z$, and the overall width of the distribution, $\sigma_Z$, which are the only free parameters in the $Z \to \tau\tau$ background model. The error bars in the control sample were scaled by 10% to account for the 10% shape uncertainty in the $\mu \to \tau$ rescaling method and the extrapolation from the control region to the final signal region. Figure 13 shows the result of the fit to the $Z \to \tau\tau$ and Higgs boson signal in the $lh$-channel.

An alternate parameterization was also considered. This parameterization was also composed of three components, but the asymmetry in the shape was modeled with a "bifurcated" or "dimidated" Gaussian [38], in which the width of the lower half was smaller than the width of the upper half of the distribution. This alternate parameterization did not fix the ratio of the widths and normalizations for the three components based on the distribution of the Jacobian; instead, these parameters were themselves parameterized as a piece-wise linear function of $m_H$. Each of the three components required four parameters to model the widths, together with two parameters for normalization constants, and two parameters for the linearity in the Higgs boson mass. In total the signal model was represented by 16 parameters. Results from these two parameterizations were in good agreement.

## 4.3 Shape parameterization for $t\bar{t}$ and $W$+jets

In contrast to the irreducible $Z \to \tau\tau$+jets background, the $W$+jets background is dominated by situations in which one of the tau decay products comes from a $W$ decay and the second tau decay product is a fake from a jet. The $t\bar{t}$ background is even more complicated because the decay products from top contain a real tau contribution as well as a fake tau contribution. It is difficult to estimate this background using Monte Carlo because one must understand in detail both the jet kinematics for as well as the lepton or $\tau$ fake rate as a function of $p_T$ and $\eta$. Instead, it is desirable to estimate this background with data. Since the

data-driven strategy for the $W$+jets background is under development, we simply rely on Monte Carlo to provide a control sample for this background. Figure 14(a) shows the shape of the fully simulated $W$+jets background and the fast simulation $t\bar{t}$ background sample after the transverse mass cut. The shapes are consistent within statistical errors. While the shapes from those backgrounds remain stable through the final stages of the event selection, a conservative 50% error is applied to each bin in the combined $t\bar{t}$ and $W$+jets control sample to reflect uncertainty in how this shape changes as the remainder of the analysis cuts are applied.

The shape of the QCD background was parameterized with the following equation

$$L_{QCD}(m_{\tau\tau}|a_1,a_2,a_3) = \mathcal{N}\left(\frac{1}{m_{\tau\tau}+a_1}\right)^{a_2} m_{\tau\tau}^{a_3} \quad . \tag{12}$$

The form is motivated by a competition between the parton distribution functions and the matrix element. In the $lh$-channel, the normalization of the backgrounds with fake taus can be constrained by using the track multiplicity constraint described in Section 3.3; however, there is an additional uncertainty associated with how well the fake fraction can be extrapolated from the control sample to the signal like region. We apply a conservative 50% systematic on this fraction associated with the extrapolation. Figure 14(b) and (c) show the result of the simultaneous fit to the fake tau background described in the Section 4.5 with (solid) and without (dashed) the signal contribution for the $ll$- and $lh$-channel, respectively. The variation reflects the magnitude of the shape uncertainty.

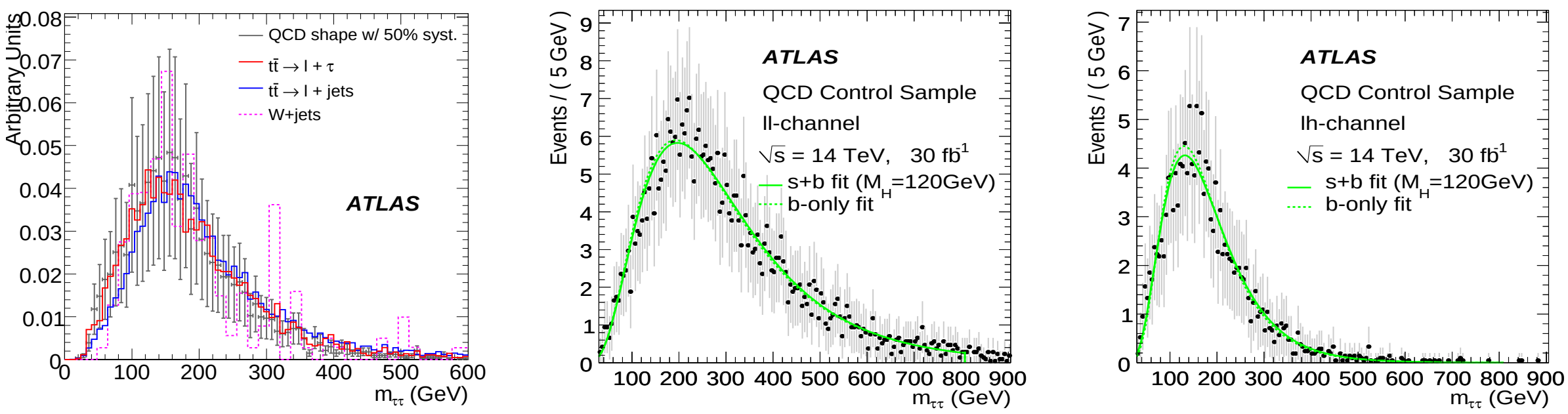


(a) (b) (c)

Figure 14: Figure (a) shows that the shapes are similar for these backgrounds and that the shape is stable in the final stages of the cut flow. The $m_{\tau\tau}$ spectrum for $t\bar{t}$ and $W$+jets backgrounds after all cuts for the $ll$-channel (b) and $lh$-channel (c) with a fit to the spectrum. The solid and dashed curves show the result of the simultaneous fit to the control sample and signal candidates with and without the signal contribution, respectively.

## 4.4 Signal significance neglecting shape uncertainty

For a given hypothesized Higgs boson mass, $m_H$, the mass window has been defined as $m_H - 15$ GeV $\leq m_{\tau\tau} \leq m_H + 15$ GeV. A simple approach to estimating the expected significance of the signal is to count events in this range and calculate the probability for at least this many events from the background-only prediction. An alternate approach is to fit the $m_{\tau\tau}$ spectrum and use the resulting signal yield as a test statistic. Table 12 shows the significance obtained from number counting assuming a 10% background uncertainty as was done in Ref. [4] and the result from the fitted signal yield. The next subsection provides a final result indicating a more realistic treatment of both normalization and shape uncertanties.

Table 12: Expected signal significance for several masses based on number counting in a mass window with 30 fb$^{-1}$ of data. Results are shown neglecting uncertainty in the background rate and incorporating it with two methods (see text). These results do not include the impact of pileup, which is discussed in Section 4.7.

| | $ll$-channel | | $lh$-channel | | combined | |
|---|---|---|---|---|---|---|
| $m_H$ | Counting | Fitted Yield | Counting | Fitted Yield | Counting | Fitted Yield |
| 105 | 2.20 | 2.43 | 2.85 | 3.46 | 3.80 | 4.17 |
| 110 | 2.46 | 2.88 | 3.45 | 4.19 | 4.46 | 5.06 |
| 115 | 2.86 | 3.26 | 4.18 | 4.96 | 5.32 | 5.96 |
| 120 | 2.80 | 3.17 | 4.23 | 4.73 | 5.36 | 5.72 |
| 125 | 2.67 | 2.96 | 3.97 | 4.32 | 5.08 | 5.28 |
| 130 | 2.42 | 2.73 | 3.54 | 3.88 | 4.62 | 4.77 |
| 135 | 2.17 | 2.37 | 3.38 | 3.60 | 4.35 | 4.25 |
| 140 | 1.74 | 2.00 | 2.66 | 2.83 | 3.55 | 3.35 |

### 4.5 Incorporating control samples and shape uncertainty

By fitting the $m_{\tau\tau}$ spectrum to a model that accurately describes the signal and various backgrounds it is possible to directly incorporate uncertainty in the background shape and take advantage of the shape of the signal within the mass window. In order to constrain the background rate and shape, we simultaneously fit the signal candidates and the background control samples outlined in Section 3. The fit is performed twice, once letting the signal parameters float (the maximum likelihood estimates denoted with a single $\hat{}$ ) and once constraining the signal normalization to be zero (the conditional maximum likelihood estimates denoted with a double $\hat{\hat{}}$). The ratio of these likelihoods is referred to as the profile likelihood ratio, $\lambda$,

$$\lambda(\mu=0) = \frac{L(data|\mu,\hat{\hat{b}}(\mu),\hat{\hat{\nu}}(\mu))}{L(data|\hat{\mu},\hat{b},\hat{\nu})} \quad , \tag{13}$$

where $\mu$ represents the signal strength in units of the Standard Model expectation and $\nu$ represents the nuisance parameters needed to describe the shape. If the Higgs boson mass is specified, the distribution of $-2\log\lambda$ ratio asymptotically approaches a $\chi^2$ distribution with the number of degrees of freedom given by the number of parameters of interest[4]. The motivation for $\mu$ is that it enforces the relationship of the Standard Model branching ratios when combining the individual channels, maintaining the property that the distribution of $-2\log\lambda$ is $\chi^2$ with one degree of freedom. This improves the power compared to a method which lets the signal in each channel vary independently. If the Higgs boson mass is not fixed, then one must take into account the "look-elsewhere" effect, which is discussed in more detail in Section 6. The likelihood function used in the simultaneous fit is simply a product of the likelihoods from the individual measurements:

$$\begin{aligned} L(data|\mu,m_H,\nu) &= L_{track}(\text{track multiplicity}|r_{QCD}) \\ &\times L_Z(\text{Z+jets control}|m_Z,\sigma_Z) \\ &\times L_{QCD}(\text{QCD control}|a_1,a_2,a_3) \\ &\times L_{s+b}(\text{signal candidates}|\mu,m_H,\sigma_H,m_Z,\sigma_Z,r_{QCD},a_1,a_2,a_3), \end{aligned} \tag{14}$$

where the $a_i$ are the parameters used to parameterize the fake-tau background and $\nu$ represents all nuisance parameters of the model: $\sigma_H, m_Z, \sigma_Z, r_{QCD}, a_1, a_2, a_3$. When using the alternate parameterization

[4]When constraining $\mu \geq 0$, the distribution for the background-only hypothesis is modified such that $-2\log\lambda(\mu=0) \sim 1/2\delta(0) + 1/2\chi^2_1$, and this is taken into account in computing the p-value.

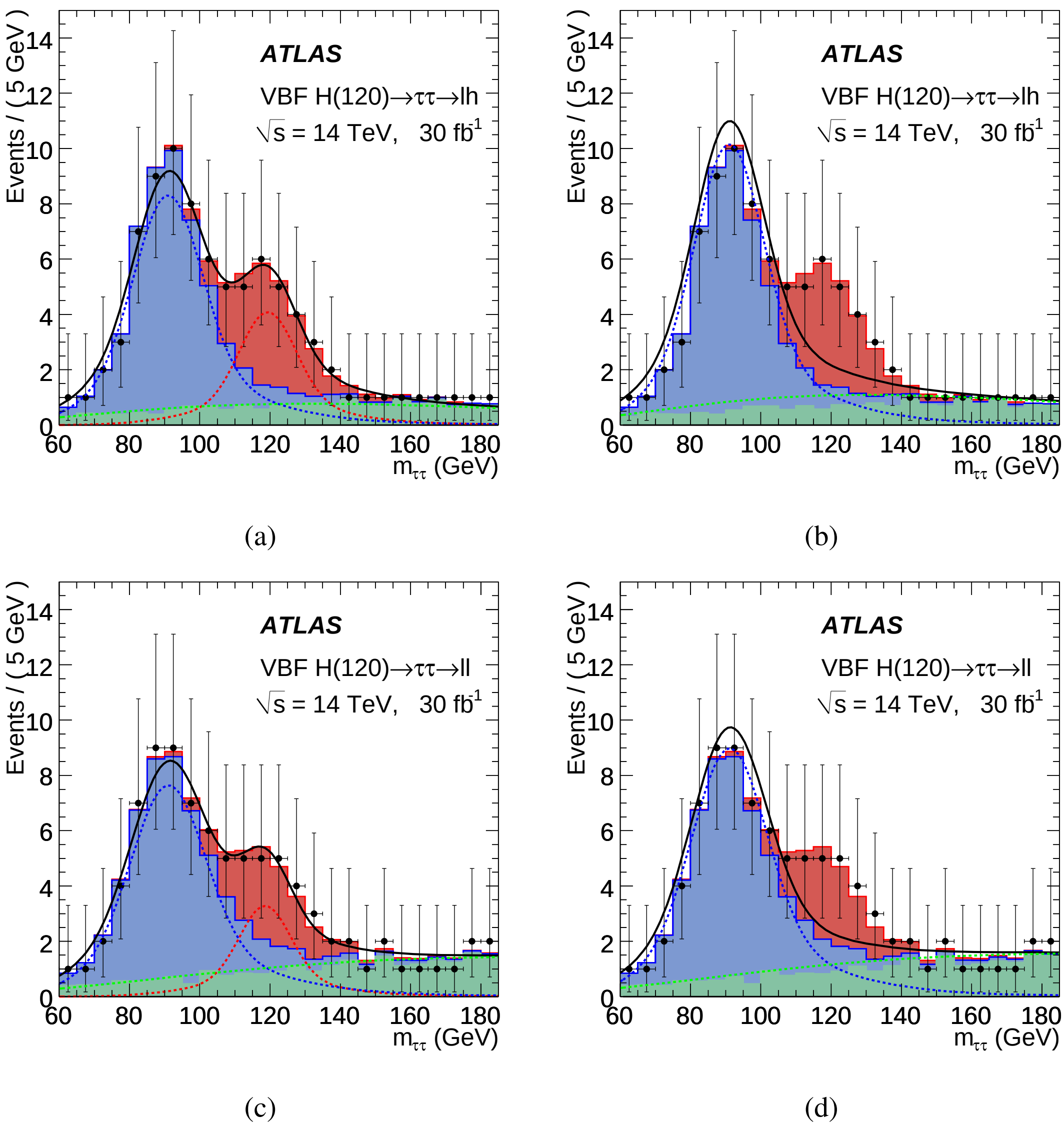


Figure 15: Example fits to a data sample with the signal-plus-background (a,c) and background only (b,d) models for the *lh*- and *ll*-channels at $m_H = 120$ GeV with 30 fb$^{-1}$ of data. Not shown are the control samples that were fit simultaneously to constrain the background shape. The fits are performed to the signal and background expectation (histograms), while the overlaid data with error bars are only indicative of a possible data set. These samples do not include pileup.

of the signal, the exact form of Equation 14 is modified to coincide with parameters of that model.

Figure 15 shows the fit to the signal candidates for $m_H = 120$ GeV with (a,c) and without (b,d) the signal contribution. It can be seen that the background shapes and normalizations are trying to accommodate the excess near $m_{\tau\tau} = 120$ GeV, but the control samples are constraining the variation. Table 13 shows the significance calculated from the profile likelihood ratio for the $ll$-channel, the $lh$-channel, and the combined fit for various Higgs boson masses with 30 fb$^{-1}$ of data. Finally, we present the expected significance as a function of Higgs boson mass in Fig.16.

Table 13: Expected signal significance for several masses based on fitting the $m_{\tau\tau}$ spectrum with 30 fb$^{-1}$ of data. Background uncertainties are incorporated by utilizing the profile likelihood ratio. These results do not include the impact of pileup, which is discussed in Section 4.7.

| $m_H$ | $ll$-channel | $lh$-channel | combined |
|---|---|---|---|
| 105 | 1.95 | 2.41 | 3.10 |
| 110 | 2.44 | 3.35 | 4.15 |
| 115 | 2.98 | 4.07 | 5.04 |
| 120 | 2.92 | 3.87 | 4.85 |
| 125 | 2.75 | 3.75 | 4.65 |
| 130 | 2.46 | 3.38 | 4.18 |
| 135 | 2.21 | 3.32 | 3.99 |
| 140 | 1.80 | 2.70 | 3.24 |

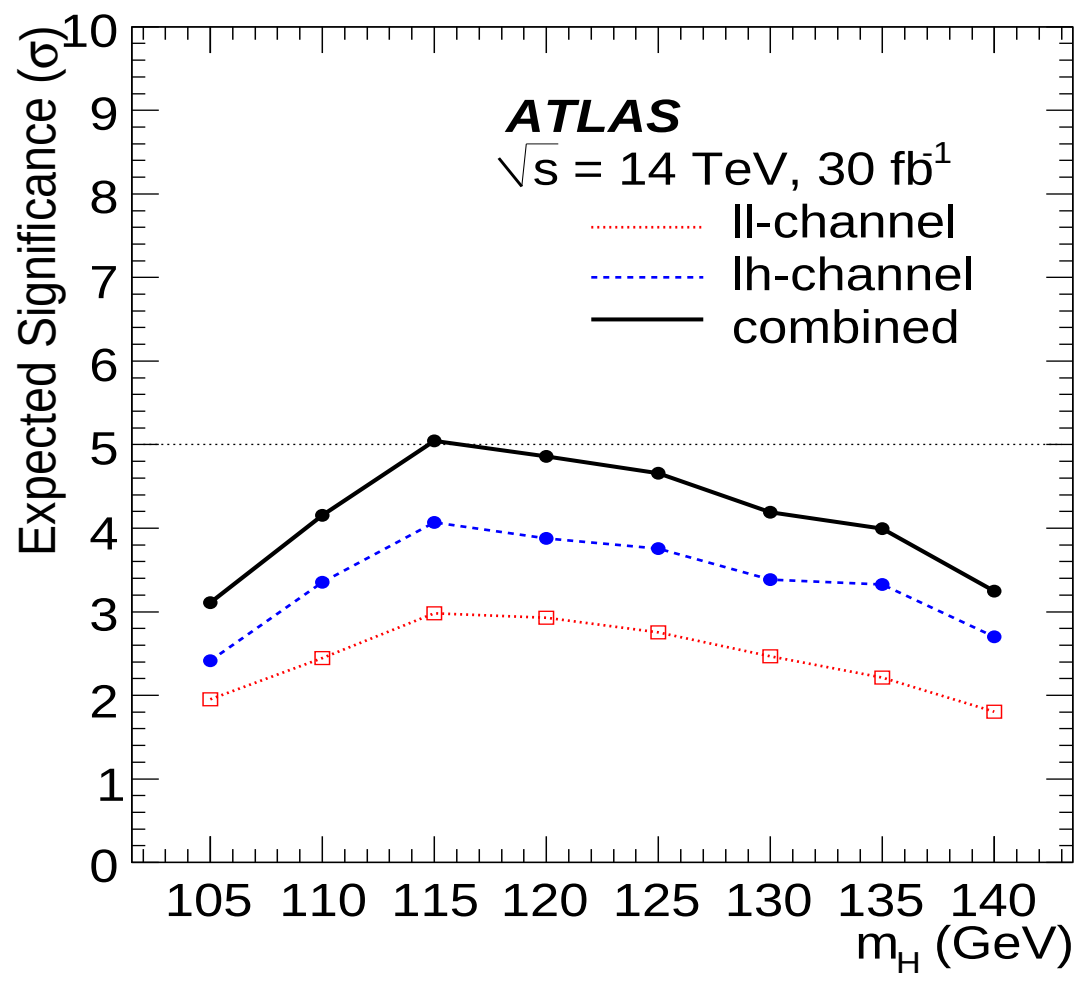


Figure 16: Expected signal significance for several masses based on fitting the $m_{\tau\tau}$ spectrum. Background uncertainties are incorporated by utilizing the profile likelihood ratio. These results do not include the impact of pileup.

### 4.6 Mass determination

The mass parameter $m_H$ and its error can be determined from the fits described above; however, the parameter in the model may not be the best estimate of the physical Higgs boson mass. Similarly, the error on the mass parameter from the fit should be validated with a large number of pseudo-experiments. Figure 17 (a) shows the relationship of the input Higgs boson mass and the reconstructed Higgs boson mass (i.e. the parameter $m_H$ in Equation 11) obtained with 2000 pseudo-experiments per input mass point. Figure 17 (b) shows the $m_{\tau\tau}$ resolution for the signal as a function of the input Higgs boson mass. When scaling the deviation by the MINOS errors, the pull distribution of $m_H$ was found to be consistent with the normal distribution $N(0,1)$. The mass resolution is found to be in the range of 8∼10 GeV. A similar mass resolution was found in the $hh$-channel when analyzing signal Monte Carlo samples.

### 4.7 Influence of pileup

The presence of pileup has three major effects on the analysis. First, additional $p$-$p$ interactions can produce hadronic activity in the central region which causes events to fail the central jet veto. Secondly, the presence of pileup generally degrades the $E_T^{\text{miss}}$ resolution, which, in turn, reduces the efficiency of the collinear approximation cuts and degrades the $m_{\tau\tau}$ resolution. Thirdly, pileup degrades the hadronic

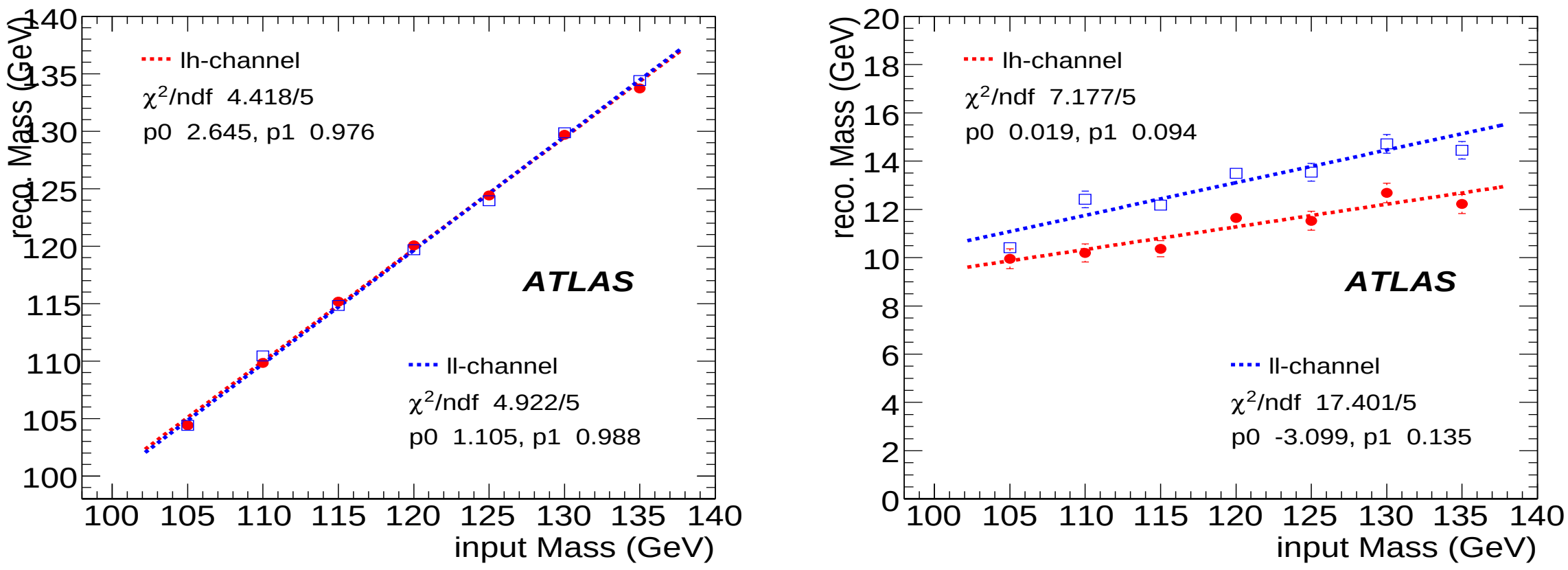


(a) (b)

Figure 17: The linearity of the fitted mass versus the input mass (a) and the mass resolution versus the input mass (b). These results do not include the impact of pileup, which is discussed in Section 4.7.

tau identification. Fortunately, the electron and muon identification have been shown to be quite robust against pileup [15, 16]. While the jet performance is affected by pileup, the analysis is fairly robust against those effects.

The simulation of pileup is technically very challenging since it is performed at a very low level in the detector simulation. Limited samples with pileup were available at the time of writing, and the $E_{\mathrm{T}}^{\mathrm{miss}}$ and hadronic tau identification algorithms were not re-tuned in this context. The distribution of the log likelihood ratio discriminant for the calorimeter-based hadronic tau identification algorithm is shifted to lower values for both real taus and jet fakes. By simply adjusting the cut on the log likelihood ratio to 0, the same signal efficiency can be maintained with approximately a 50% drop in jet rejection. By re-tuning the discriminant in the context of pileup, improved jet rejection should be possible. In all three channels, the mass resolution is degraded from $\sim$9.5 to $\sim$ 11.5 GeV for $m_H = 120$ GeV due to the degradation of the $E_{\mathrm{T}}^{\mathrm{miss}}$ resolution. Figure 8 shows that the central jet veto survival probability drops from $\sim$88% to $\sim$75% at $10^{33}$ cm$^{-2}$ s$^{-1}$ and $\sim$65% at $2 \times 10^{33}$ cm$^{-2}$ s$^{-1}$. Studies indicate that the use of tracking and calorimeter timing information can be used to mitigate this loss in signal efficiency.

Given the lack of background samples simulated with pileup and the need to re-optimize the reconstruction and analysis in that context, we do not report signal significance estimates.

# 5 Systematic uncertainties

## 5.1 Overview

The data-driven background estimation methods described above have been developed so that uncertainty in the background shape and normalization are included directly into the significance calculation. Because the discovery criterion is simply testing the presence or absence of the signal, it is not sensitive to some of the sources of systematic uncertainty. In contrast, measurement of the Higgs boson mass is sensitive to the energy scale of electrons, muons, hadronic taus and $E_{\mathrm{T}}^{\mathrm{miss}}$ . Furthermore, measurement and exclusion of $\sigma(pp \to qqH) \times BR(H \to \tau\tau)$ are sensitive to the uncertainty on the signal selection efficiency. Below we discuss the impact of these systematics on the analysis.

### 5.2 Systematic mis-measurement of the signal

First, we consider the purely experimental sources of systematics. The approach here is to assume that once we have data, the estimates of the energy scales, resolutions, and efficiencies might be systematically biased. Estimates of the systematic uncertainty from various sources are given in Table 14. The uncertainty estimates are common for all this series of notes, except for the uncertainties on the central jet veto and forward jet tagging efficiency. There is not yet a dedicated study of the expected uncertainty on these efficiencies from data, thus we assume uncertainty on the reconstruction efficiency to be half the tau identification efficiency and point out that most effects relevant to the jets have already been included in the jet energy scale uncertainty. We use the nominal detector performance as the central value and then manipulate the Monte Carlo signal to reflect these changes. For instance, in the case of the electron energy scale uncertainty, we coherently change all electrons to have 0.5% higher $E_T$, modify the $E_{\mathrm{T}}^{\mathrm{miss}}$ vector accordingly, and recalculate the signal efficiency. This is done individually for each source of systematic and upward and downward fluctuations are treated separately. In the case of the jet energy scale, only some elements of the uncertainty are relevant for $E_{\mathrm{T}}^{\mathrm{miss}}$ . A study of $E_{\mathrm{T}}^{\mathrm{miss}}$ projected onto the direction of the reconstructed $Z$ in $Z \to ll$+jets with a subset of the analysis cuts indicated that the $E_{\mathrm{T}}^{\mathrm{miss}}$ scale can be measured within 5%; thus we only manipulate the $E_{\mathrm{T}}^{\mathrm{miss}}$ vector according to a 5% jet energy scale shift. In the case of systematic uncertainty on resolution, we only considered a degradation in the resolution by the tabulated amount. Finally, for systematics on reconstruction and identification efficiency we assume a 1-to-1 transfer to the uncertainty on the signal efficiency and include a factor of two when the signal efficiency scales as the square (e.g. the electron efficiency in the $ll$-channel). Table 14 summarizes the effect of systematic mis-measurement on the signal efficiency.

Table 14: Estimated scale of systematic mis-measurements and their effect on the signal efficiency. † When varying the jet energy scale, only a 5% mis-measurement of the jet energy was used in manipulating the $E_{\mathrm{T}}^{\mathrm{miss}}$ vector. See text for details.

| Source | Relative uncertainty | Effect on signal efficiency |
|---|---|---|
| luminosity | ±3% | ± 3% |
| muon energy scale | ± 1% | ± 1% |
| muon energy resolution | $\sigma(p_T) \oplus 0.011 p_T \oplus 1.7\ 10^{-4} p_T^2$ | ± 0.5% |
| muon ID efficiency | ±1 % | ± 2% |
| electron energy scale | ± 0.5% | ± 0.4 % |
| electron energy resolution | $\sigma(E_T) \oplus 7.3\ 10^{-3} E_T$ | ± 0.3 % |
| electron ID efficiency | ± 0.2% | ± 0.4% |
| tau energy scale | ± 5% | ± 4.9% |
| tau energy resolution | $\sigma(E) \oplus 0.45\sqrt{E}$ | ± 1.5% |
| tau ID efficiency | ± 5% | ± 5% |
| jet energy scale† | ± 7% ($\lvert\eta\rvert \leq 3.2$)<br>± 15% ($\lvert\eta\rvert \geq 3.2$)<br>± 5% (on $E_{\mathrm{T}}^{\mathrm{miss}}$ ) | $^{+16\%}/_{-20\%}$ |
| jet energy resolution | $\sigma(E) \oplus 0.45\sqrt{E}$ ($\lvert\eta\rvert \leq 3.2$)<br>$\sigma(E) \oplus 0.67\sqrt{E}$ ($\lvert\eta\rvert \geq 3.2$) | ± 1% |
| b-tagging efficiency | ± 5% | ± 5% |
| forward tagging efficiency | ± 2 % | ± 2% |
| central jet reconstruction efficiency | ± 2 % | ± 2% |
| total summed in quadrature | | ±20% |

### 5.3 Theoretical uncertainty

In addition to the effect of systematic mis-measurement on the signal efficiency, theoretical uncertainties also limit our ability to estimate the signal efficiency. Next-to-leading order QCD calculations are now available for the vector boson fusion process. A dedicated study [39] investigated the overall renormalization and factorization scale dependence (2%) as well as the parton distribution function (PDF) uncertainties (3.5%). Next-to-leading order electroweak corrections are also quite large for the vector boson fusion process, giving a 3% uncertainty for the full next-to-leading order calculation [40]. Recently, the dominant next-to-leading order QCD corrections to the Higgs boson plus three jets have been calculated for vector boson fusion, providing a scale uncertainty on the parton-level central jet veto survival probability of 1% [41].

While the parton-level theoretical uncertainties are under very good control and below the level of both the statistical error and measurement-related systematics, the same is not true for the theoretical uncertainty related to the parton-shower and underlying-event. We rely on Monte Carlo simulations that model the parton-shower, hadronization, and underlying event to simulate the detector response. The uncertainty in these calculations is not comparable to the accuracy of the parton-level predictions. The central jet veto efficiency was studied with the signal process generated with PYTHIA (with various tunings), HERWIG and SHERPA and the fast detector simulation. After the analysis cuts, the different generators differ by 41%. Studies focusing specifically on the matrix element–parton shower matching indicate a substantially smaller uncertainty [33, 42]. We will measure the underlying event [43, 44] and tune the parton shower and hadronization with data, but it is likely that this contribution of the uncertainty will remain significant. Currently there is no estimate of the expected uncertainty related to the parton-shower, hadronization, and underlying event tuning. Clearly, this is an area that deserves attention as such a large uncertainty will hinder exclusions if a Higgs boson does not exist in this mass range and cross-section and coupling measurements if one does. After discussions with the authors of PYTHIA, HERWIG and SHERPA we feel that the residual uncertainty in the parton shower after tuning to the data will be less than the 18% uncertainty quoted for the jet energy scale. Thus, the uncertainty in the signal efficiency will be dominated by the jet energy / $E_{\mathrm{T}}^{\mathrm{miss}}$ scale uncertainty and the precise uncertainty in the parton shower is not relevant. Table 15 summarizes the theoretical uncertainties for the signal production.

Table 15: Theoretical uncertainties which affect the estimation of the signal efficiency.

| Source | Relative uncertainty | Effect on signal efficiency |
|---|---|---|
| PDF uncertanties | $\pm 3.5\%$ | $\pm 3.5\%$ |
| scale dependence on cross-section | $\pm 3\%$ | $\pm$ 3% |
| scale dependence CJV efficiency | $\pm$ 1% | $\pm$ 1% |
| parton-shower and underlying event | $\pm \leq 10\%$ | $\pm < 10\%$ |
| total summed in quadrature | | $\pm < 10\%$ |

## 6 Discussion

The expected signal significances in Table 13 are qualitatively consistent with the results found in Ref. [4]; however, the predicted cross-sections in Tables 9 and 10 are significantly different. In particular, the initial cross-section of the $Z$+jets background is smaller by nearly a factor of four and the $t\bar{t}$ background in the $ll$-channel is larger by nearly a factor of 2. Much of this difference reflects the evolution

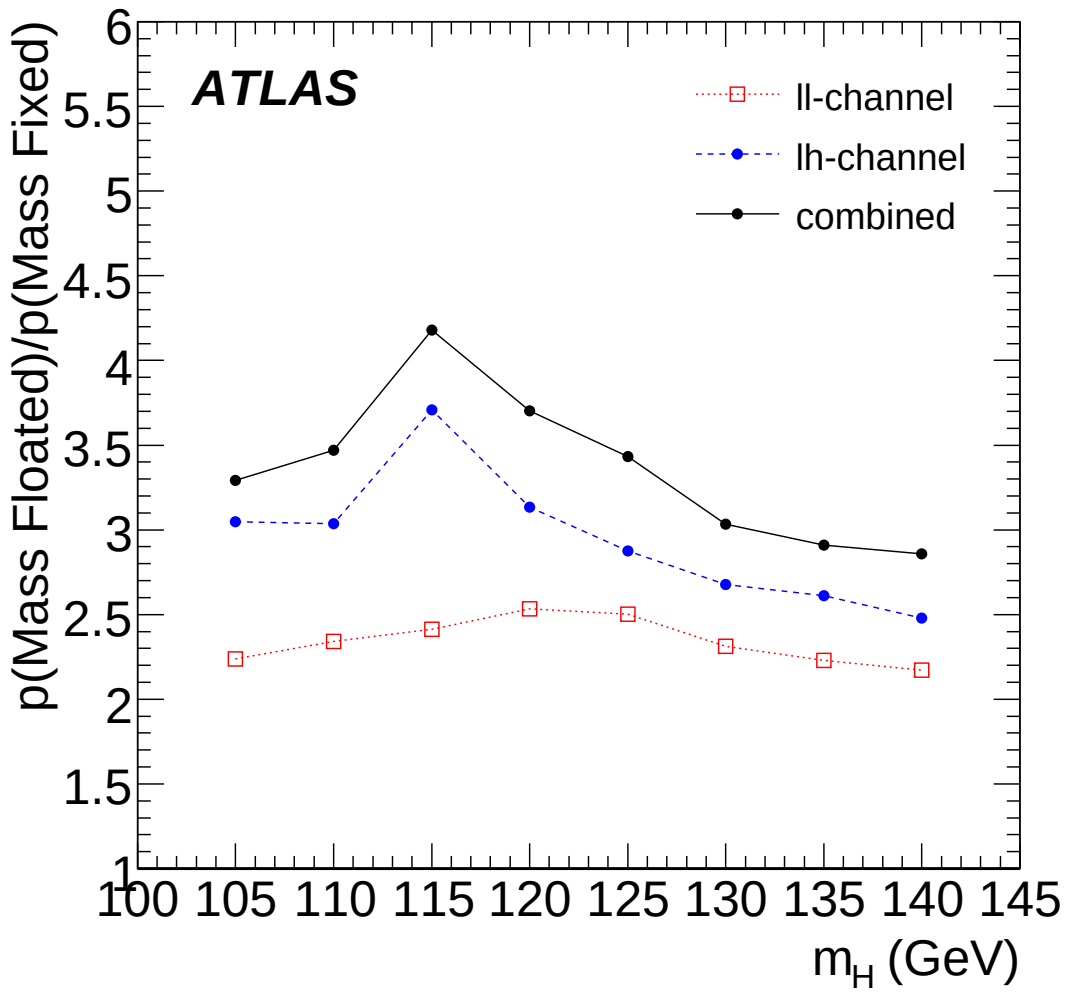


Figure 18: The ratio of expected p-values for the floating and fixed mass fits as a function of the Higgs boson mass. This plot summarizes the impact of the "look-elsewhere" effect in this analysis.

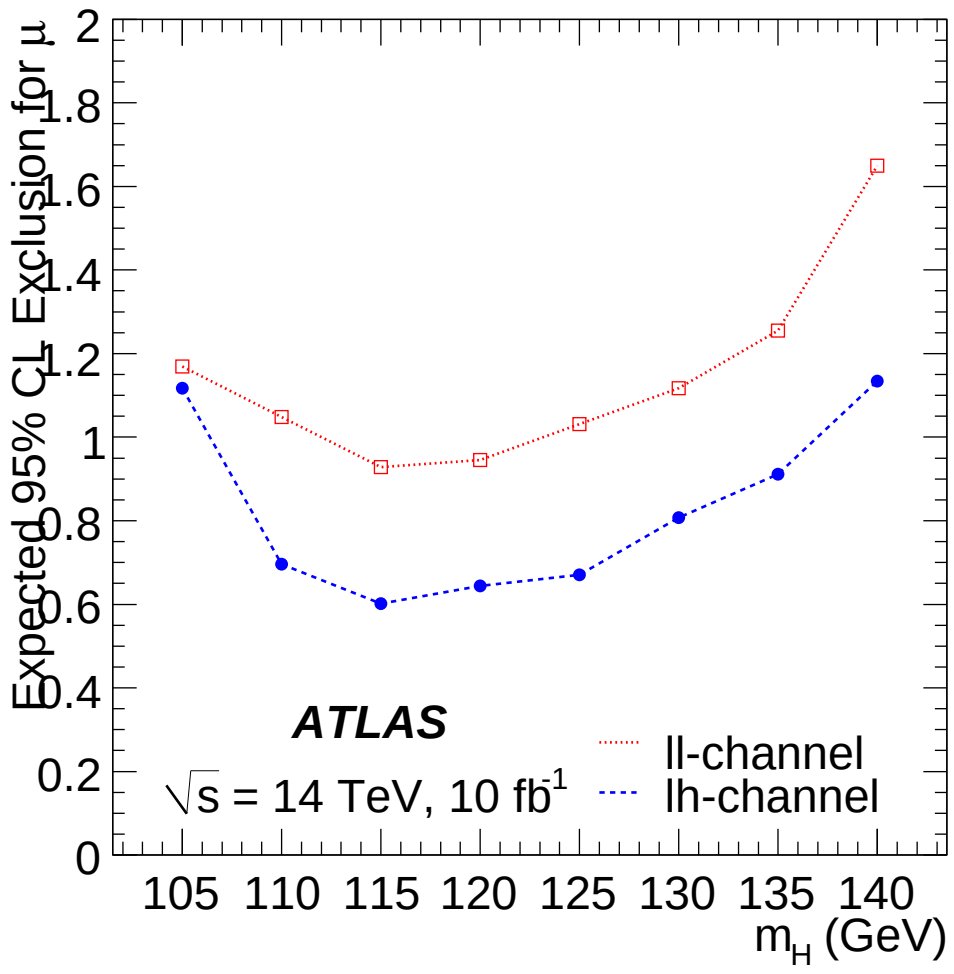


Figure 19: Expected 95% exclusion of the signal rate in units of the Standard Model expectation, $\mu$, as a function of the Higgs boson mass for the $ll$ and $lh$-channels with 10 fb$^{-1}$ of data. The exclusion takes into account the uncertainty on the signal efficiency described in Section 5.

in the Monte Carlo generators for these challenging backgrounds. In the case of the $Z$+jets background, differences in the choice of renormalization and factorization scales and parton density functions are the source for part of the discrepancy; however, approximately a factor of two comes from the treatment of soft and collinear divergences. After substantial investigation, we concluded that the $Z$+jets background estimate used in Ref. [4] was conservative. In the case of $t\bar{t}$+jets, the necessary matching between matrix elements and parton showers had not been developed when Ref. [4] was written. The recipe used to merge the $t\bar{t}$+0,1,2 jet samples at that time and the more realistic $b$-tagging performance limits our ability to understand in detail the source of the differences. We are confident that MC@NLO and our current detector simulation and offline reconstruction provide a superior prediction of this background.

For the first time, ATLAS has investigated the potential of the $hh$-channel. Much of this work has been devoted to the development and study of tau and missing $E_T$ triggers. It now appears that the trigger is feasible, and the reconstruction of the signal maintains an efficiency and $m_{\tau\tau}$ mass resolution comparable to the $ll$- and $lh$-channels. The open question for this channel is the size of the QCD background – a question that can only be answered with data.

The results shown in Section 4 are based on a fixed mass hypothesis. If one leaves the Higgs boson mass a free parameter in the likelihood fits, then one must take into account the "look-elsewhere" effect. Naively, one would expect the magnitude of the effect to be rather small for this channel given the $\sim$10 GeV mass resolution and the $\sim$ 30 GeV mass range of interest. Detailed study shows that it is a mixture of two effects. First, the distribution of $-2\log\lambda(\mu=0)$ is not $\chi^2$-distributed under the background-only hypothesis; it has a longer tail which raises the p-value. Secondly, the median of $-2\log\lambda(\mu=0)$ under the signal-plus-background hypothesis is systematically larger in the floating mass case because the model can adapt to fluctuations in the signal mean. Figure 18 summarizes the impact of these two effects.

Future work for this channel includes an extensive study of performance in the context of pileup, including optimization of the hadronic tau identification and $E_{\mathrm{T}}^{\mathrm{miss}}$ reconstruction. The data-driven background estimation for the $t\bar{t}$ and $W$+jets backgrounds must be further developed. In order to measure $\sigma(pp \to qqH) \times BR(H \to \tau\tau)$ at the desired level, the systematics associated with the signal efficiency must be reduced. This requires a technique to measure the central jet veto and forward jet tagging efficiency in data. Additionally, the use of the $E_{\mathrm{T}}^{\mathrm{miss}}$ projection method should be further developed to mitigate the impact of the jet energy scale uncertainty.

While the results shown in Section 4 only include the contribution of Higgs bosons produced via vector boson fusion, an additional 10% contribution from the gluon-fusion production process could be expected at $m_H = 120$ GeV. Finally, the theoretical uncertainty associated with the parton shower and underlying event, which currently gives the largest uncertainty on the signal efficiency, must be addressed. The tuning of the parton shower, hadronization, and underlying event model will be among the earliest measurements at the LHC. If the residual theoretical uncertainty is not reduced sufficiently, a different strategy may need to be found for the central jet veto. The expected exclusion power based on the uncertainty in the signal efficiency is shown in Fig. 19.

## 7 Summary and conclusion

The sensitivity of the ATLAS detector to a Standard Model Higgs boson produced via vector boson fusion with subsequent decay into taus has been investigated with state-of-the-art Monte Carlo generators, a full GEANT-based simulation of the ATLAS detector and our current trigger and reconstruction algorithms. Particular emphasis has been placed on data-driven background estimation strategies and the estimation of the associated uncertainties in normalization and shape. The impact of pileup has not been fully addressed; however, results without pileup indicate that a $\sim 5\sigma$ significance can be achieved for a Higgs boson mass in the range 115 – 125 GeV after collecting 30 fb$^{-1}$ of data and combining the *ll*- and *lh*-channels. The Higgs boson mass resolution is approximately 10 GeV, leading to approximately 3.5% mass measurement. The *hh*-channel has also been investigated and gives similar results for signal and non-QCD backgrounds as the other channels; however due to the challenge of predicting the QCD background, we do not report on an estimated sensitivity for that channel. Currently, measurement-related systematics and large theoretical uncertainties limit the ability for a measurement of the product of cross-section and branching ratio. Future work is needed to constrain these uncertainties in order to measure the Higgs boson couplings, spin and CP properties.

# Higgs Boson Searches in Gluon Fusion and Vector Boson Fusion using the $H \to WW$ Decay Mode

**Abstract**

The prospects for Higgs searches in the $H+0j$ ($H \to WW \to e\nu\mu\nu$), $H+2j$ ($H \to WW \to e\nu\mu\nu$), and $H+2j$ ($H \to WW \to \ell\nu qq$) channels at ATLAS are presented, including realistic effects such as trigger efficiencies and detector misalignment, with an emphasis on practical methods to estimate the backgrounds using control samples in data. With 10 fb$^{-1}$ of integrated luminosity, one would expect to be able to discover a Standard Model Higgs boson in the mass range $135 < m_H < 190$ GeV and to be able to measure its mass with a precision of about 7 GeV if its true mass is 130 GeV or about 2 GeV if its true mass is 160 GeV.

## 1 Introduction

The nature of the Electroweak symmetry breaking sector is arguably the most important unknown in particle physics today. An introduction to the Higgs mechanism and an overview of the possible search channels for a (standard model) Higgs boson at the LHC can be found in Ref. [1]. This note studies the sensitivity of Higgs boson searches in the decay mode $H \to WW$ to a Standard Model Higgs boson. Special emphasis is placed on in-situ control samples that can be used to estimate the background contributions using data and on the systematic uncertainties associated with these background determination methods.

The note is organised as follows: Section 2 discusses the production mechanisms and final states under study, the relevant backgrounds for each process and the Monte Carlo generators used to model them. In Section 3, the performance of the reconstruction for the various final-state particles is described. In Section 4, an overview of statistical issues common to most of the analyses is presented. Section 5 describes the event selection and control samples for a search for $H \to WW \to e\nu\mu\nu$ in events where no hard jets are reconstructed in the detector. In Section 6, the analysis of the $H+2j$, $H \to WW \to e\nu\mu\nu$ channel is presented, and in Section 7, an analysis of events where one of the $W$ bosons decays to two jets is discussed. Finally, in Section 8, the combined sensitivity of the ATLAS detector to the presence of neutral scalar resonances in the $W$ pair final state is presented.

## 2 Physics Processes and Monte Carlo

In the Standard Model, there are two dominant production modes of the Higgs boson in the kinematic region of interest for the $H \to WW$ decay mode: gluon fusion and Vector Boson Fusion (VBF). In studies of the former production mode, one must consider background processes that yield two leptons and significant missing transverse energy in the final state; in the latter, the final state also includes two hard jets (from the struck quarks) which tend to be well-separated in pseudorapidity. This note considers the following signal and background processes:

- Higgs boson production via gluon fusion: This is the dominant production mechanism for the signal considered in the analysis of Section 5; it is modeled with 48500 events generated by MC@NLO [2, 3]. A Higgs boson decay matrix element similar to the one used in Ref. [4] was used to reweight this Monte Carlo sample to include the complete spin correlations for the Higgs decay.

- Higgs boson production via Vector Boson Fusion: This is the dominant production mechanism for the Higgs boson searches of Sections 6 and 7. This process is modeled with the generators provided in PYTHIA [5] (14500 events), HERWIG [6] (18050 events), and SHERPA [7] (19950 events).

- $pp \to WW$ production. This is the dominant background contribution to the analysis studied in Section 5; in that analysis, it is modeled with 178450 events generated by MC@NLO. It is important to note that MC@NLO only calculates the $O(\alpha_s^0)$ and $O(\alpha_s^1)$ contributions to this process; it does not compute the $O(\alpha_s^2)$ contributions. Because of the large gluon luminosity expected at LHC it is important to include at least the gluon-initiated component (where the $W$ pair production proceeds via a quark box) for the analysis of the fully jet-vetoed final state. This contribution is modeled with 176300 events produced using the generator in Ref. [8].

  In the Vector Boson Fusion searches, the contributions from processes like $qq \to WWqq$ (mediated by the exchange of a weak boson or a gluon) become important. Therefore, a sample of 171250 events generated with ALPGEN [9, 10] is used instead of MC@NLO to model the $pp \to WWjj$ background in the analysis of Section 6.

- $t\bar{t}$ production. The two top quarks decay into a pair of $W$ bosons and two b jets. The cross-section is expected to be dominated by doubly resonant $t\bar{t}$ production in the Vector Boson Fusion searches of Sections 6 and 7; however, for the search of Section 5, one must also take care to handle the single-top background correctly. The absolute cross-section for the contribution from single top production is presently only known with leading-order uncertainties, but since the search presented in Section 5 normalises the top background using data, it is only necessary to consider its potential impact on the extrapolation from the b-tagged to the b-vetoed region. In the estimation of the sensitivity of all channels under study, a sample of 538300 events generated with MC@NLO is used to model this background. Two MadEvent [11–13] samples are used to estimate the importance of the contribution from single-top production in the $H+0j$ channel: one contains all matrix elements for $pp \to WWbb$; the other contains only the doubly-resonant contribution.

- $Z \to ll$ production. In channels with two electrons or two muons, the direct decays of $Z \to ee$ and $Z \to \mu\mu$ dominate, but this background can also contribute to $e\mu$ final states when the $Z$ decays to a pair of $\tau$ leptons which both decay leptonically. This background is modeled with a sample of 163200 events generated with PYTHIA in the $H+0j$ analysis of Section 5 and with a sample of 38150 events produced using ALPGEN in the $H+2j$ analysis of Section 6.

- W+$n$ jets production, with $n \leq 5$. This is the dominant background for the lepton-hadron channel; however, it can also play a role in the dilepton channels as a source of fakes. This background is modeled using a sample of 411424 events generated with ALPGEN.

  A special subset of this background is W+c+$n$ jets, with $n \leq 4$. The charm production in association with a W boson through processes like $gs \to Wc$ is enhanced, due to the large parton density function of strange sea quarks and the large CKM matrix element $|V_{cs}|$. Decays of charm mesons are a potential source of leptons. Lepton isolation cuts are tuned on this kind of background, to reach the necessary rejection against leptons from semileptonic charm- or b- decays. [14].

- $b\bar{b}$, $c\bar{c}$ and QCD multi-jet. Due to the large cross section of these processes they could be a further source of background. The large amount of CPU power needed makes a Monte Carlo simulation of these backgrounds currently unrealistic. For the dilepton channels the requirement of two leptons and missing transverse energy should reduce these backgrounds to a negligible level.

Table 1 gives an overview of the cross sections of the signal and background samples used in this note. Detailed tables of the Higgs boson cross sections and branching ratios can be found in Ref. [1].

| Process | Generator | Cross-section(pb) |
|---|---|---|
| $gg \to H \to WW$ ($M_H = 170$ GeV) | MC@NLO | 19.418 |
| VBF $H \to WW$ ($M_H = 170$ GeV) | PYTHIA/Sherpa | 2.853 |
| VBF $H \to WW$ ($M_H = 300$ GeV) | HERWIG | 0.936 |
| $qq/qg \to WW$ | MC@NLO/Alpgen | 111.6 |
| $gg \to WW$ | GG2WW | 5.26 |
| $pp \to t\bar{t}$ | MC@NLO | 833 |
| $Z \to \tau\tau$+jets | PYTHIA/ALPGEN | 2015 |
| $W$+jets | ALPGEN | 20510 |

Table 1: Overview of the Monte Carlo generators and cross sections for the Higgs boson signal and background processes used in this note. The $W$+jets cross-section listed is the cross-section per lepton flavor.

# 3 Reconstruction

The final states considered here all include leptons. Section 3.1 briefly describes the lepton selection criteria and other central aspects of the reconstruction used in this note. Then, in Section 3.2, a few important details about jets, $E_T^{miss}$, and pile-up are briefly discussed. Note that the results presented in this note were made using version 12 of the ATLAS software, whereas the results on detector performance in Refs. [15] and [16] are based on release 13.

## 3.1 Leptons

The electron selection consists of the following criteria :

- Electrons are identified using the standard ATLAS criteria described in Ref. [15]. In the $H+0j$ ($H \to WW \to e\nu\mu\nu$) and $H+2j$ ($H \to WW \to \ell\nu qq$) channels, tight electrons are required. In the case of the $H+2j$, $H \to WW \to e\nu\mu\nu$ analysis of Section 6, a sufficiently strong rejection against the $W$+jets background can be achieved by applying cuts on the VBF tagging jets that it is possible to use a medium electrons instead.

- A track match is required. The transverse impact parameter significance of the matched track, $d_0/\sigma_{d_0}$, is required to be less than 10.

- Candidates are required to be well isolated. Calorimeter isolation in a $\Delta R$ cone of 0.2 is required to be less than 5 GeV, and $\Sigma p_T$ of tracks in a cone of $\Delta R < 0.4$, excluding tracks from other lepton candidates and tracks with $p_T < 1$ GeV, is required to be less than 5 GeV.

- Kinematic acceptance: $p_T > 15$ GeV and $|\eta| < 2.5$ (The $p_T$ threshold is larger for the lepton-hadron channel). Electron candidates in the crack region, $1.37 < |\eta| < 1.52$, are excluded.

It has been shown that similar isolation criteria can effectively suppress the background from W+c+jet events, where a second lepton comes from charm decay, to negligible levels.

$W$+jets is one of the main sources of fake backgrounds for the dilepton channels; it is crucial to achieve a good rejection against this background. The probability for a jet to be misreconstructed as an electron was derived from a W($\to \mu\nu$)+jets sample which contains no true high $p_T$ electrons. The average efficiency to reconstruct an electron candidate without isolation cuts in the $gg \to H \to WW$ signal events is 60.3±0.5%; the additional isolation criteria reduce this to 50.0±0.5% efficiency. With all cuts,

the average fake rate as determined from $W \to \mu\nu$+jets events is $(1.7 \pm 0.2) \times 10^{-4}$ before isolation and $(6.7^{+1.5}_{-1.3}) \times 10^{-5}$ after isolation.

The muon selection consists of a similar set of cuts :

- Combined muons and best match to an inner detector track [16].
- Calorimeter Isolation based on the $etcone20$ variable [16] ($etcone20 < 2.5$ GeV) and track isolation (in a cone of $\Delta R < 0.4$) as for electrons, but with a different $p_T$ threshold, $\Sigma p_T < 3$ GeV.
- Transverse impact parameter significance less than 10.
- $p_T > 15$ GeV and $|\eta| < 2.5$ (The $p_T$ threshold is larger for the lepton-hadron channel)

As for the electrons, similar isolation criteria were shown to be effective against the W+c+jet background. The performance for muons is better than for electrons. The average efficiency to reconstruct a muon candidate in $gg \to H \to WW$ signal events is $94.2 \pm 0.1\%$ before isolation cuts and $77.1 \pm 0.2\%$ after applying cuts. The corresponding fake rate in $W \to e\nu$+jets events is $(1.7^{+0.6}_{-0.5}) \times 10^{-5}$ after isolation cuts are applied.

### 3.2 Jets, Missing $p_T$, and Pile-up

In the $H + 2j$ searches of Sections 6 and 7, the presence of at least two jets with $p_T > 20$ GeV and a large difference in pseudorapidity are required. It is frequently the case that at least one of these jets will land in the forward region of the detector. To minimise the effect of calorimeter noise and maximise efficiency for the jets of interest to this study, jets reconstructed from topological clusters are used, rather than jets reconstructed from calorimeter towers [17]. For VBF Higgs boson signal processes the efficiency for jets with a cone size of $\Delta R = 0.4$ from topological clusters is 94.3% in the central detector region and 95.3% in the forward region. For jets from calorimeter towers the efficiency is significantly lower with 92.4% in the central detector region and 84.1% in the forward region.

All the searches considered in this note make use of analysis techniques that either select or give a large weight to events without additional jets; in either case, events are effectively rejected if they contain extra jets beyond those jets (if any) that are explicitly required by the analysis. Even in the first years of "low-luminosity" data taking at the LHC, analyses will be affected by the pile-up of multiple collisions per beam crossing. Low $p_T$ QCD interactions (minimum-bias) are present simultaneously to a hard interaction. Such pile-up events will give rise to jet activity that can sometimes cause interesting signal events to be erroneously rejected; in selecting a jet algorithm, it is necessary to consider this effect. Single minimum-bias events have a three times lower probability to be discarded by the central jet veto cut when using jets reconstructed with a cone size of $\Delta R = 0.4$ instead of jets with a cone size of $\Delta R = 0.7$. Furthermore, in the presence of pile-up the jet calibration has to correct for the additional energy that is not from the hard interaction. Due to the well defined jet area, this is much easier done with a cone jet algorithm than with a $k_t$ jet algorithm.

For the reasons mentioned above, the analyses in this note have used jets reconstructed from topological clusters with a cone algorithm of $\Delta R = 0.4$ [17]. Jet efficiencies increase with $E_T$ and are rather stable with $|\eta|$. However it should be noted that the efficiency drops at $|\eta| \geq 4.8$ near the boundary of the forward calorimeter, as expected. Therefore this note uses only jets with $|\eta| \leq 4.8$. Jets within a cone of $\Delta R < 0.4$ of any electron candidate are ignored.

There are other ways to minimise the impact of pile-up; the tracking system of the ATLAS detector is designed to separate the vertices of the different inelastic collisions in an event. Jets built of tracks emerging from the same vertex as the leptons from the Higgs boson decay should be insensitive to pile-up if the vertex reconstruction is working as expected.

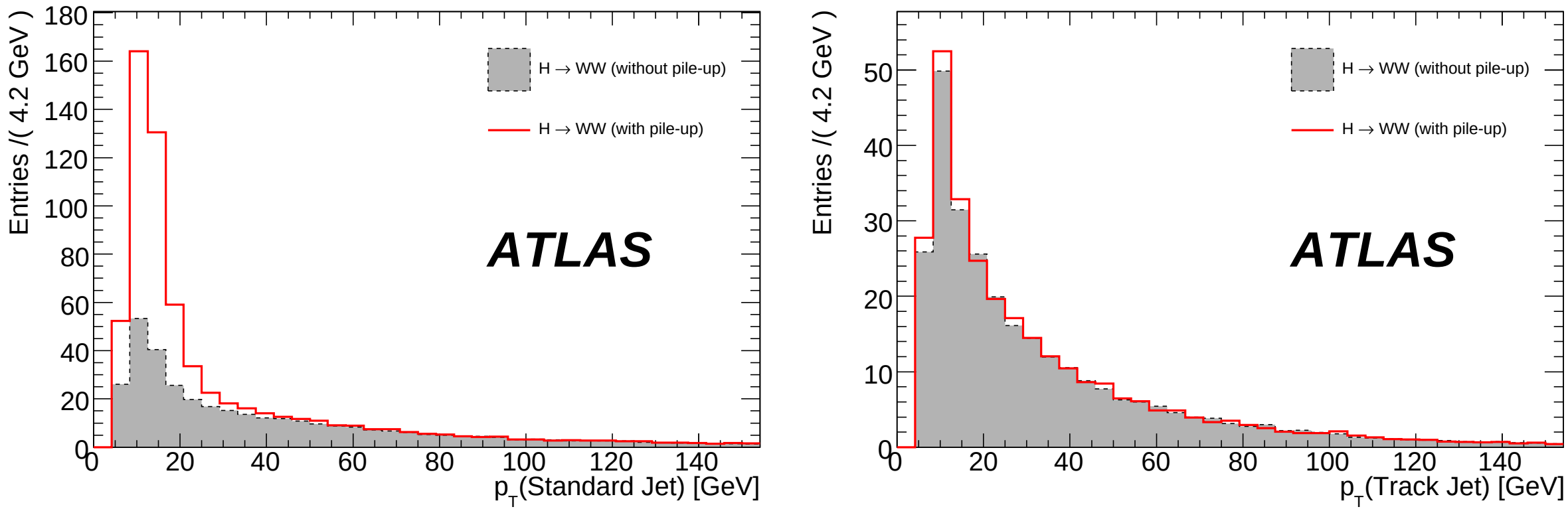


Figure 1: Jet $p_T$ distributions ($|\eta| < 2.5$) for the $H \rightarrow WW$ signal sample with (red line) and without (gray filled area) pile-up. Left: Standard jets. Right: Track jets.

While the $H \rightarrow WW$ searches presented in this note use standard calorimeter jets for the jet veto, it is important to illustrate the potential of track jets. The $H+2j$, $H \rightarrow WW \rightarrow e\nu\mu\nu$ analysis has been studied as an example.

Track jets are reconstructed from only those tracks that have $p_T > 0.5$ GeV and with at least 7 hits in the pixel and semiconductor tracker. The Higgs boson production vertex is taken to be the vertex for which the sum of the transverse energies of the emerging tracks is largest. All tracks that emerge from this vertex and are not associated with an isolated lepton are fed into the standard ATLAS cone $\Delta R = 0.4$ jet algorithm. Leptons are considered isolated if the sum of the transverse energies of all tracks from the same vertex as the lepton in a cone of $0.01 < \Delta R < 0.2$ around the lepton is less than 5 GeV.

The independence of the track jets to activity from pile-up is reflected in the jet $p_T$ distributions in the $H \rightarrow WW$ signal sample without and with pile-up (pile-up was generated with an average of 2.3 additional events) in Figure 1. The additional $pp$ collisions in presence of pile-up created additional standard jets at low transverse energies while the multiplicities of track jets are unaffected by pile-up.

A central jet veto cut is constructed using track jets in the acceptance region of the inner detector of $|\eta| < 2.5$ and standard jets at larger pseudorapidities (up to $|\eta| < 3.2$). The cut on the transverse momentum of the track jets is chosen such that the same rejection efficiency is obtained for $H \rightarrow WW$ events without pile-up as for the jet veto based on standard jets. This requires a lowering of the $p_T$ cut from 20 GeV for standard jets to 12.3 GeV for track jets. Table 2 shows the fraction of events passing

| | $H \rightarrow WW$ | | $t\bar{t}$ | |
|---|---|---|---|---|
| | no pile-up | with pile-up | no pile-up | with pile-up |
| std jets ($|\eta| < 2.5$) | $72.0 \pm 1.0$ | $63.0 \pm 1.2$ | $28.6 \pm 3.4$ | $19.7 \pm 3.3$ |
| track jets | $72.0 \pm 1.0$ | $73.5 \pm 1.1$ | $28.6 \pm 3.4$ | $25.9 \pm 3.6$ |
| std jets ($|\eta| < 3.2$) | $65.4 \pm 1.0$ | $57.0 \pm 1.2$ | $24.0 \pm 3.2$ | $16.3 \pm 3.0$ |
| combination | $65.8 \pm 1.0$ | $65.9 \pm 1.1$ | $24.0 \pm 3.2$ | $23.1 \pm 3.5$ |

Table 2: Fraction of events (%) passing the central jet veto

the different options for the central jet veto. The central jet veto based on standard jets is sensitive to pile-up while the track jets show only small sensitivity to pile-up. The combination of the track jet veto in $|\eta| < 2.5$ and the standard jet veto in $|\eta| > 2.5$ shows robustness against pile-up.

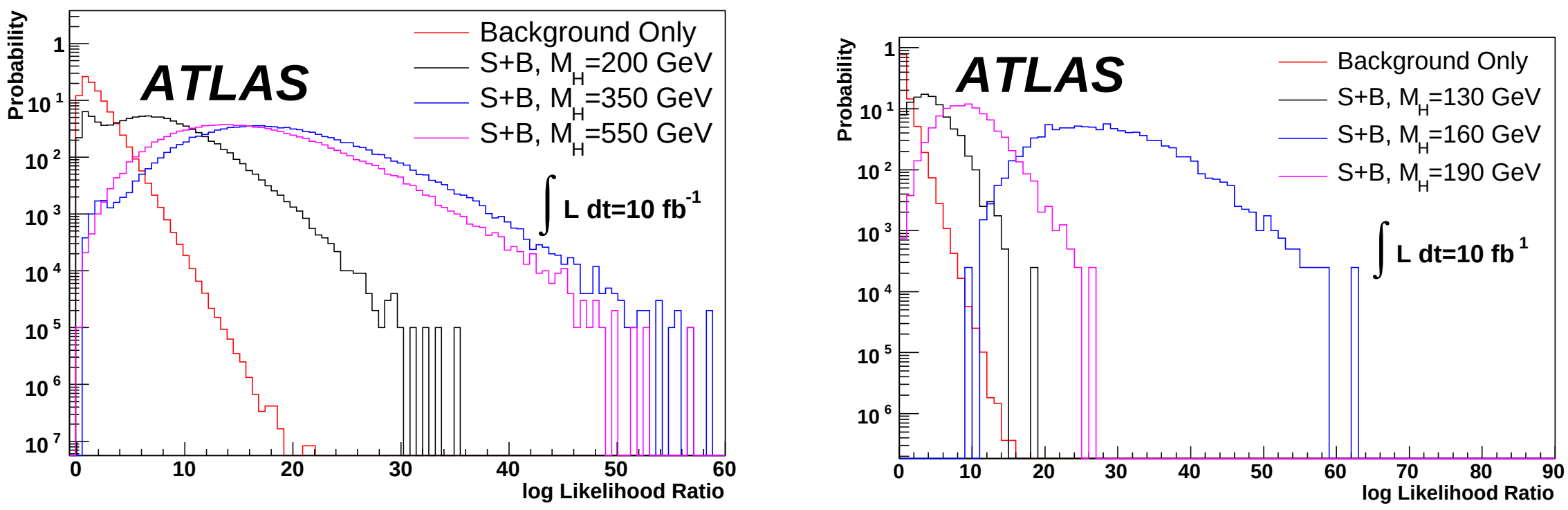


Figure 2: Left: The log Likelihood Ratio distributions for toy Monte Carlo corresponding to the $H \to WW \to \ell\nu qq$ analysis at 10 fb$^{-1}$ for background-only outcomes (red) and signal-plus background outcomes with several values of the Higgs boson mass. Right: The same plot, but for the $H+2j$, $H \to WW \to e\nu\mu\nu$ analysis.

The performance of the missing transverse momentum reconstruction is strongly affected by the presence of pile-up. In this study, the cell based reconstruction algorithm described in Ref. [18] has been used.

Fortunately, the $H \to WW \to \ell\nu\ell\nu$ searches are not very sensitive to the resolution of $E_T^{miss}$. Since the Higgs boson mass cannot be directly reconstructed in the $2\ell 2\nu$ final state, a transverse mass is used; although this does have a peak centred around the true value of the Higgs boson mass, the distribution is very broad (several tens of GeV). Therefore a degraded $E_T^{miss}$ resolution is unlikely to significantly alter the transverse mass shape unless the degradation is rather severe. The $H+2j$ ($H \to WW \to \ell\nu\ell\nu$) analysis of Section 6 makes use only of this transverse mass and observables related to the leptons and the jet activity in the event; the $E_T^{miss}$ cut in that analysis serves mostly to reject backgrounds like $Z \to ee/\mu\mu$ which have no intrinsic $E_T^{miss}$. The $H+0j$ analysis of Section 5 does make use of the transverse momentum of the Higgs boson candidate, which is obviously very sensitive to the $E_T^{miss}$ resolution; however, that analysis involves a fitting algorithm developed with exactly this concern in mind and includes checks that demonstrate that the algorithm still works reliably in the presence of a degraded $E_T^{miss}$ measurement.

In the case of the $H \to WW \to \ell\nu qq$ analysis of Section 7, the $E_T^{miss}$ is used in the reconstruction of the Higgs boson candidate's invariant mass. In principle, then, a degraded $E_T^{miss}$ resolution does propagate into the Higgs boson mass resolution, and so problems with the $E_T^{miss}$ reconstruction are more harmful to the $H \to WW \to \ell\nu qq$ channel than to the dilepton channels. However, for a Standard Model-like Higgs boson with a mass above $\approx 300$ GeV, the natural width of the Higgs boson tends to become rather large, and the $E_T^{miss}$ resolution is again not the primary concern of the analysis.

Mismeasurements of jets and detector inefficiencies can lead to instrumental $E_T^{miss}$. It is very hard to predict such effects without real data and no cuts to reduce instrumental $E_T^{miss}$ are currently foreseen.

# 4 Statistical Formalism

The sensitivity estimates presented in this note use a fit-based hypothesis testing procedure. Unless stated otherwise, selection cuts are always independent of the Higgs boson mass hypothesis, no matter how broad the range of masses where the search is sensitive. To take advantage of the discriminating power contained in variables for which the distribution depends significantly on the Higgs boson mass

| Selection | Selection cuts | $gg \to H$ | $t\bar{t}$ | $WW$ | $Z \to \tau\tau$ | $W + jets$ |
|---|---|---|---|---|---|---|
| pre-selection | Lepton Selection+$M_{ll}$ | 166.4 | 6501 | 718.12 | 4171 | 209.1 |
| | $p_T^{miss} > 30$ GeV | 147.7 | 5617 | 505.25 | 526.3 | 181.6 |
| | $Z \to \tau\tau$ Rej. | 145.8 | 5215 | 485.12 | 164.2 | 150.4 |
| | Jet Veto | 61.80 | 14.84 | 238.35 | 31.91 | 76.12 |
| | b-veto | 61.56 | 6.85 | 237.87 | 30.76 | 76.12 |
| signal region | $\Delta\phi_{ll} < 1.575$, $M_T < 600$ GeV | 50.6±2.5 | 2.3±1.6 | 85.4±2.7 | <1.7 | 38±38 |
| control region | $\Delta\phi_{ll} > 1.575$, $M_T < 600$ GeV | 10.9±1.1 | 4.6±2.3 | 151.9±3.6 | 30.8±4.2 | 38±38 |
| b-tagged signal region | $\Delta\phi_{ll} < 1.575$ | - | 1.14±1.14 | - | - | - |
| b-tagged control region | $\Delta\phi_{ll} > 1.575$ | - | 5.71±2.55 | - | - | - |

Table 3: Cut flows (in fb) for $M_H = 170$ GeV in the $H + 0j$, $H \to WW \to e\nu\mu\nu$ channel. A '-' indicates that the corresponding contribution is ignored in the fit. The WW background contains the two processes $q\bar{q} \to WW$ and $gg \to WW$.

hypothesis or is not precisely predicted by theory, a maximum-Likelihood fit to those variables is used. To perform the hypothesis test itself, a Likelihood Ratio is used, $\lambda = L_{s+b}/L_{bg-only}$. The numerator and denominator of this ratio are obtained from separate fits to the data, with the number of signal floating as a free parameter (subject to the constraint $N_s \geq 0$) in the fit for $L_{s+b}$, and with the number of signal fixed to zero in the fit for $L_{bg-only}$. The sampling distributions of $\lambda$ at a given luminosity (in the presence of signal and in its absence) are studied by generating pseudo-experimental outcomes and performing the full fit to each outcome.

Throughout this note, when a fit is performed, the Higgs boson mass $m_H$ is allowed to float as a free parameter in the fit. The resulting significance estimates therefore represent the probability that, in the absence of signal, a background fluctuation anywhere in the allowed mass range would produce an excess at least as significant as the observed excess. This is slightly different from the convention adopted in some other studies, where the significance represents the probability that a background fluctuation would produce an excess that is consistent with a specified mass and at least as significant as the observed excess.

Figure 2 shows two examples of the Likelihood Ratio probability distributions. The left plot shows the sampling distribution of the Likelihood Ratio for the $H \to WW \to \ell\nu qq$ fit discussed in Section 7 (assuming a negligible background from fakes), and the right plot shows the corresponding distributions for the analysis of Section 6.2. A fit to the real data would yield one value of $\lambda$; to obtain a p-value, one would integrate the red distribution in the acceptance region, i.e. from the observed value of $\lambda$ to infinity. In the absence of data, the expected significance is computed by taking the "observed" value of $\lambda$ to be the median of the signal-plus-background distribution for a given mass. The expected significance $Z$ reported here is one-sided, i.e., $Z = \sqrt{2}\text{erfc}^{-1}(2p)$. When the number of pseudo-experimental outcomes in the acceptance region is too small, the background-only Likelihood Ratio distribution is extrapolated with an exponential decay.

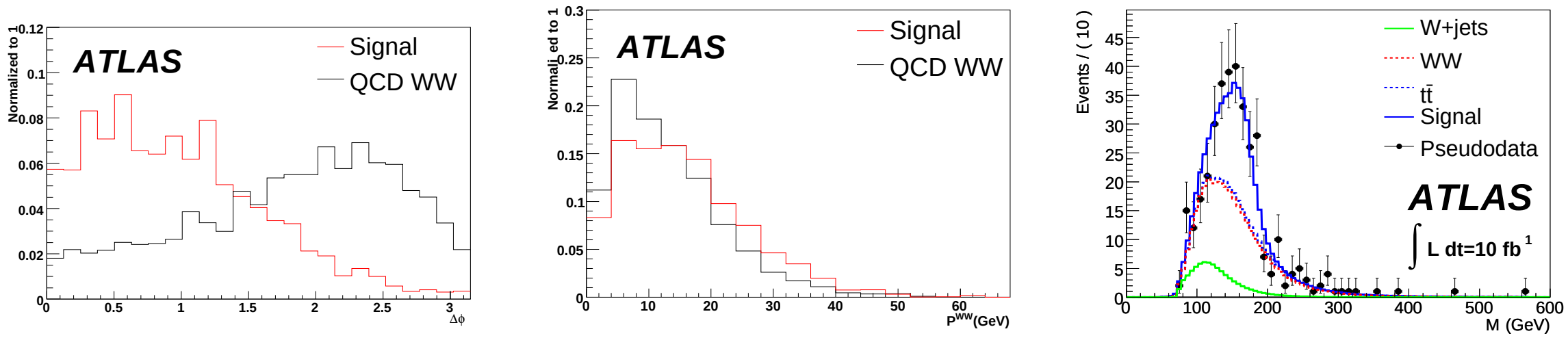


Figure 3: Left: transverse opening angle $\Delta\phi_{ll}$ of the two leptons after preselection cuts. Middle : transverse momentum $p_T^{WW}$ of the WW system after preselection cuts. Right: transverse mass $M_T$ for events with $\Delta\phi_{ll} < 1.575$ and $p_T^{WW} > 20$ GeV, in a fitted toy Monte Carlo outcome containing a Standard Model Higgs boson with $M_H = 170$ GeV, after 10 fb$^{-1}$ of integrated luminosity. The Likelihood Ratio in this outcome is 30.69, which is typical for this value of $M_H$ and this luminosity.

# 5 Leptonic $W$ Pair Production with No Hard Jets

The $H+0j$, $H \to WW \to \ell\nu\ell\nu$ channel has been shown to have a strong discovery potential. [19–22] The basic event selection consists of only a few simple cuts:

- Require that the event has exactly two isolated, opposite-sign leptons (electron or muon) with $p_T > 15$ GeV.
- To suppress backgrounds from single-top production, backgrounds from dileptonic decays of $b\bar{b}$ and $c\bar{c}$ resonances, and lepton pairs from $b \to c$ cascade decays, require that the invariant mass $m_{ll}$ of the leptons is between 12 GeV and 300 GeV.
- Require that the event has $E_T^{miss} > 30$ GeV
- To suppress backgrounds from $Z \to \tau\tau$, reconstruct the invariant mass of a hypothetical $\tau$ pair using the collinear approximation [23]. If the energy fractions $x_\tau^1$ and $x_\tau^2$ are both positive and the invariant mass of the $\tau$ pair $M_{\tau\tau}$ is in the range $|M_{\tau\tau} - M_Z| < 25$ GeV, the event is rejected.
- To suppress backgrounds from top quark decays, reject events that contain any hard jets with $p_T > 20$ GeV and $|\eta_j| < 4.8$.
- To further suppress the top background, reject events with any jets with $p_T > 15$ GeV and a b-tagging weight greater than 4.

Table 3 shows the cross-sections in the $e\mu$ channel for signal and background after these cuts.

The trigger efficiency for signal is sufficiently large for this analysis to be viable. Events are required to pass at least one of the ATLAS single-lepton or double-lepton triggers. The level-1 trigger menus used here are 2EM15I, EM25I, EM60I, MU20, and MU40. For the Level 2 trigger and the event filter, events are required to pass the e25i, 2e15i, e60, or mu20i triggers. Details on the trigger menus can be found in Ref. [24]. For the $e\mu$ channel considered here the trigger efficiency for L1 is 99.0%; for L2 it is 96.7%, and for the EF it is 95.2% The efficiency is quite high, and the trigger efficiency does not distort the shapes of the kinematic variables of interest in the signal in a significant way. Table 3 does not include the trigger efficiency, but it is taken into account as an overall scale factor for the signal and background when generating toy Monte Carlo to test the fitting algorithm described in the next section and in the calculation of the expected statistical significance for this channel.

After the basic event selection (all cuts from Table 3 including the cut on $\Delta\phi_{ll}$), the signal to background ratio is only ≈1:4, but there is still a good deal of discriminating power in the final state. This analysis focuses on three variables:

- the transverse opening angle $\Delta\phi_{ll}$; a cut on this variable exploits differences in the spin correlations in the WW system in the Higgs signal and the WW background,
- the transverse mass $M_T$, as defined in Ref. [25],
- the transverse momentum of the $WW$ system, $p_T^{WW}$, which tends to be slightly larger for signal than for WW background because gluon-initiated processes tend to have more initial-state radiation than quark-initiated processes.

The distributions of these variables are shown in Figure 3. To make use of the discriminating power of these variables, a maximum-likelihood fit is performed.

## 5.1 Fitting Algorithm

The fit is a 2-dimensional fit of transverse mass and $p_T^{WW}$ in two bins of the dilepton opening angle $\Delta\phi_{ll}$ in the transverse plane. [26] The present study considers a fit to only those events with one electron and one muon, since an in-situ background extraction procedure for the $Z \to ee/\mu\mu$ background has not yet been studied. After the preselection cuts and the additional requirement that $M_T < 600$ GeV is applied, the remaining events are separated into two subsamples, one with $\Delta\phi_{ll} < 1.575$ and the other with $\Delta\phi_{ll} > 1.575$. Table 3 shows the cross-sections for the signal and various backgrounds in these two regions. The region with large $\Delta\phi_{ll}$ (control region) is enriched in background and the region with small $\Delta\phi_{ll}$ (signal region) is enriched in signal.

The top background is estimated with the help of b-tagged control samples with the same kinematic cuts as the signal-enriched and background-enriched regions. Table 3 includes estimates of the top cross-section in the b-tagged control regions. In the present study, the b-tagging efficiency is assumed to be well-known (i.e., the ratio of cross-sections in the b-tagged and b-vetoed regions is taken from Monte Carlo), and the contamination from processes with only light jets in the b-tagged regions is ignored. The $t\bar{t}$ cross-section estimates in Table 3 are based on Monte Carlo samples that use MC@NLO to model the top background; even if the b-tagging is well-understood, the ratio of the top cross-section in the b-tagged and b-vetoed regions is sensitive to the treatment of the single-top contribution to the background. Moreover, studies based on fast simulation samples have shown that there are also differences in the shape of the distributions of $M_T$ and $p_T^{WW}$ between top background models based on doubly-resonant top production and models that include the singly-resonant and nonresonant processes. (For example, the single-top contribution leads to a larger number of events with large $M_T$ and small $p_T^{WW}$.) Nevertheless, in the present study, standalone fits are performed on the b-tagged control samples before the fit to the b-vetoed regions begins, and the top background in the b-vetoed regions is estimated by extrapolating both the shape and the normalization from the b-tagged region to the b-vetoed region based on ratios obtained from MC@NLO.

The $Z \to \tau\tau$ background is normalized and its shape is determined by studying a sample of $Z \to \mu\mu$ events taken from real data, where the reconstructed muons are replaced by simulated taus [23]. Two-muon events with a dimuon invariant mass between 82 and 98 GeV are selected, and the same jet veto that is used in the rest of this analysis is applied to the selected events. The effective cross-section after these cuts is roughly 360 pb. The reconstructed muons in the event are replaced with simulated tau leptons, and the remaining event selection cuts are applied. The efficiency of those cuts is about 0.07%, leaving an effective cross-section of roughly 250 fb. To be conservative, the present study assumes that efficency factors will lower this figure to about 200 fb. A standalone fit to these "data-Monte-Carlo"

events is performed to determine the shape and normalization of the $Z \to \tau\tau$ background; in the final fit, the shape parameters are fixed and the normalization is rescaled by a factor that is assumed in this study to be well-predicted. The parameters obtained from the fit to the $Z \to \tau\tau$ data-Monte-Carlo control sample are not allowed to float in the main fit.

At the time of this writing, an in-situ determination of fake backgrounds from processes such as $W$+jets has not yet been studied. For this study, a fixed probability distribution has been used to represent the fake backgrounds, assuming for the moment that the shape and normalization of fake backgrounds are well-predicted. Because of the limited size of the available $W$+jets Monte Carlo sample, there are not enough Monte Carlo events remaining after cuts to provide a prediction of the shape of the transverse mass vs. $p_T^{WW}$ distribution for $W$+jets; the shape of this distribution is therefore taken from a set of events with loosened isolation and shower shape cuts. This treatment ignores any potential systematic uncertainty on the $W$+jets background; future studies will attempt to address this in more detail.

Once the fits to the control samples are completed, a simultaneous fit of the two $\Delta\phi_{ll}$ bins in the b-vetoed region is performed. A few of the parameters that describe the shape of the transverse mass and $p_T^{WW}$ distributions of the $WW$ background are allowed to float in the fit. The $p_T^{WW}$ distributions for the $WW$ background in the two regions are taken to be the same up to a parabolic distortion factor. The normalizations of the $WW$ background are free to float independently; however, we add a penalty term of the form $(R_{fit} - R_{true})^2/\sigma_R^2$, where $R_{fit}$ is the ratio of the best-fit number of $WW$ background events in the small-$\Delta\phi_{ll}$ region over the number in the large-$\Delta\phi_{ll}$ region, $R_{true}$ is the Monte Carlo prediction of the ratio taken from the central-value calculation, and $\sigma_R$ is the uncertainty in the prediction of $R_{true}$, taken to be 10%. This value of $\sigma_R$ is chosen to be larger than the actual variation of R due to changes in the $Q^2$ scale definition (about 5%) so that the constraint term does not cause a large bias in the observed value of $R$; such a bias could be expected to lead to a degradation in the sensitivity of the hypothesis test. The value of $\sigma_R$ has not been optimized; such a study may be performed in the future.

In order to demonstrate the robustness of the fit against systematic uncertainties, toy Monte Carlo has been used to compute the sampling distributions for the Likelihood Ratio in several scenarios where the "true" probability distribution has been distorted to model various sources of systematic error. Seven distorted scenarios are considered: four altered $Q^2$ scale choices (factorization and renormalization scales raised and lowered by factors of 8), two alternative top background models (based on leading-order $pp \to WWbb$ and leading-order $pp \to tt \to WWbb$), and one alternative model of all irreducible backgrounds where the $x$ and $y$ components of $E_T^{miss}$ have been independently smeared by 5 GeV each. These alternative models have been derived using fast simulation because large-statistics Monte Carlo samples are needed in order to be sure any change in the shape of the observables is actually due to the systematic uncertainty under study and not merely a statistical fluctuation. In these systematic scenarios, the background contribution from fakes is ignored. Figure 4 shows the Likelihood Ratio distribution for background-only outcomes (upper left) and the distribution of pulls of the fitted Higgs boson mass (upper right) for signal-plus-background outcomes with a true Higgs boson mass of 170 GeV. Both distributions are nearly independent of the systematic distortions.

The lower left plot in Figure 4 shows the linearity of the mass determination as a function of the true Higgs mass. The line shows the mean of a Gaussian fit to the region around the peak of the distribution of best-fit Higgs masses in the toy Monte Carlo sample for the case of nominal detector performance. The green band shows the width of the Gaussian and is a direct measure of the variability of the mass estimate on repetition of the experiment; the error bars show the median fit error. The typical variability of the mass determination at 10 fb$^{-1}$ ranges from 5.2 GeV at $M_H$ =130 GeV to 1.6 GeV at $M_H$ =160 GeV to 4.2 GeV at $M_H$ =190 GeV.

Figure 4 also shows the expected significance for an integrated luminosity of 10 fb$^{-1}$. This channel is most promising for Higgs boson masses near the $WW$ threshold.

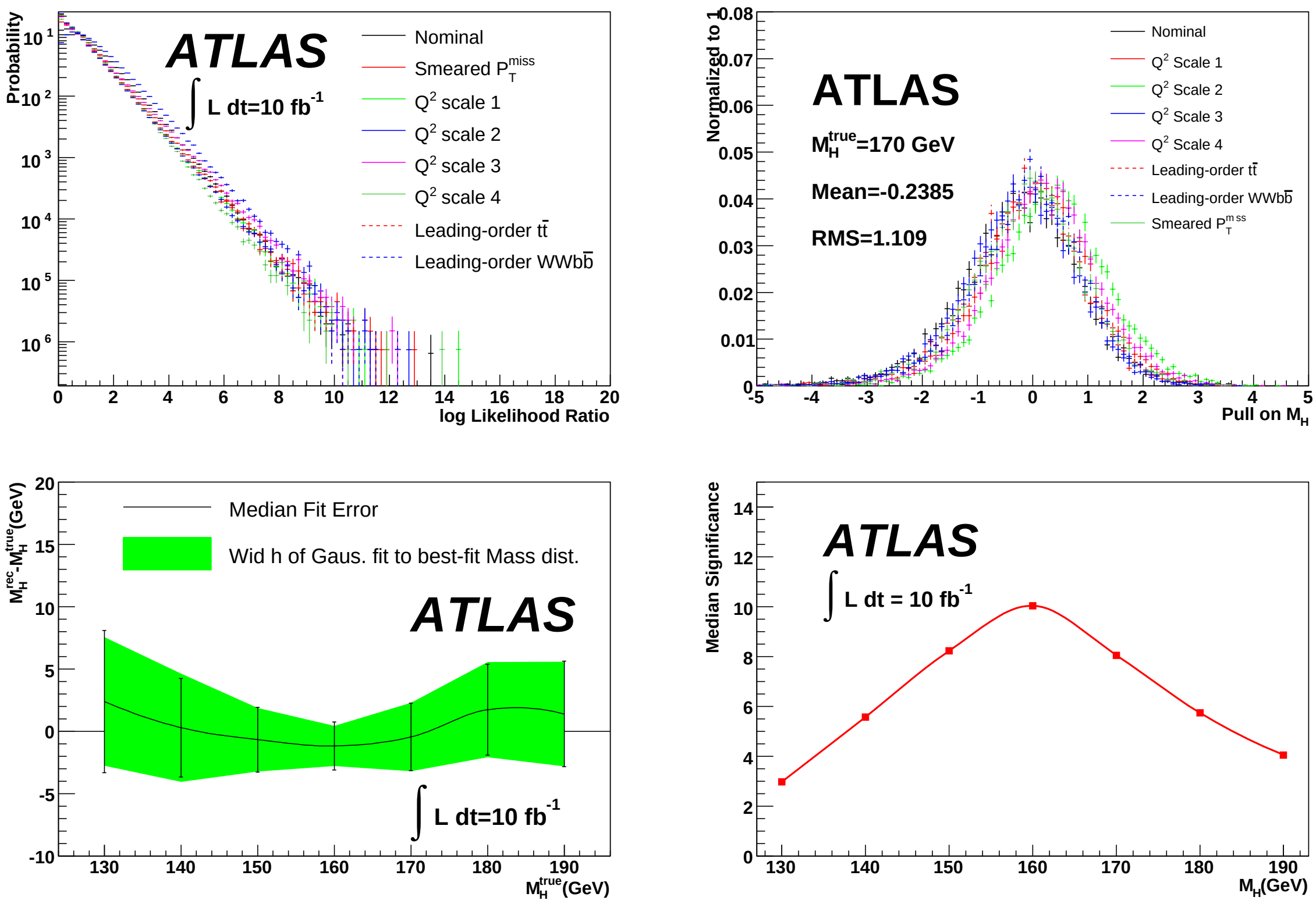


Figure 4: Upper Left: The log Likelihood Ratio distributions for background-only toy Monte Carlo outcomes corresponding to 10 fb$^{-1}$ in the $H+0j$, $H \to WW \to l\nu l\nu$ analysis. Upper Right: The corresponding pull distributions for $M_H$=170 GeV. Lower Left: The linearity of the mass determination. Lower Right: the expected significance for an integrated luminosity of 10 fb$^{-1}$.

| Region | Signal, $M_H = 170$ GeV (fb) | $t\bar{t}$ | $WW$ | $Z \to \tau\tau$ | $W$+jets |
|---|---|---|---|---|---|
| Signal-like | 28.65±0.80 | 1.14±1.14 | 29.35±1.59 | <1.74 | 38±38 |
| Control | 1.47±0.27 | 5.71±2.55 | 61.13±2.33 | 4.06±1.53 | <114 |
| b-tagged | 0 | 6.85±2.80 | 0.11±0.09 | 1.16±0.82 | <114 |

Table 4: Cross-sections (in fb) after all cuts for a number-counting analysis with a test mass of $M_H =$ 170 GeV in the $H+0j$, $H \to WW \to e\nu\mu\nu$ channel.

## 5.2 Cross-check with a Number-Counting Analysis

A number counting analysis of this channel has been performed as a cross-check. The signal region is defined by the basic event selection in addition to the following few additional cuts: $p_T^{WW} > 10$ GeV, $M_{ll} < 64$ GeV, $\Delta\phi_{ll} < 1.5$, and $50 < M_T < 180$ GeV. A control region is defined by the requirements $p_T^{WW} > 10$ GeV, $80 < M_{ll} < 300$ GeV, and $\Delta\phi_{ll} > 1.5$. In order to normalize the top background, it is also useful to define a b-tagged region with similar kinematic cuts: $p_T^{WW} > 10$ GeV and $\Delta\phi_{ll} > 1.5$. Table 4 shows the cross-sections for signal and backgrounds in these regions. Only one of the $W$+jets events passes all the selection cuts for the signal-like region; it corresponds to a cross-section of 38 fb. None of the $W$+jets events survives in either control region. Because this background is modeled with Alpgen $W+n$ jets, with $n \leq 5$, and because the Monte Carlo was generated with different effective

luminosities for different jet multiplicities, it is not straightforward to compute a meaningful upper bound on the $W$+jets cross-section in the control regions. The quoted cross-section is simply three times the cross-section observed in the signal region (which is based on one Monte Carlo event).

The ratios of background cross-sections in the various regions have to come from a theoretical prediction; in a number-counting approach there is no simple way to constrain them with data. A similar background normalization strategy was studied in Refs. [27] and [28]; in Ref. [27], the theoretical error on the ratio of $WW$ cross-sections in the signal-like region and the control region was 5%, the systematic error on the ratio of the top background cross-sections in the signal-like and b-tagged regions was 9%, and the uncertainty on the ratio of the top background cross-sections in the control region and the b-tagged regions was also 9%. These systematic errors have not been re-evaluated in the context of this analysis, nor have the relevant instrumental systematic errors been computed; this cross-check simply uses these values as the systematic uncertainties on the extrapolations.

Using the cross-sections for $M_H = 170$ GeV in Table 4 and taking into account the additional 95.2% trigger efficiency on signal and background, the luminosity-dependent errors on the normalizations of the control samples (ignoring $W$+jets in the Control and b-tagged regions), and the systematic uncertainties described above, the expected significance of the number-counting analysis (in the Gaussian approximation, for $\int L\,dt = 10$ fb$^{-1}$) is $8.5\sigma$. This figure ignores the error on the $W$+jets background prediction; in a future study, it will be important to normalize the $W$+jets background using data. In the interest of performing a meaningful comparison with the fit analysis, it is helpful to compute the significance ignoring the $W$+jets background altogether: $10.4\sigma$. This is comparable to the result from the fit; if the contribution from the $W$+jets background is ignored there, the significance is $8.8\sigma$ at the same luminosity.

# 6 Leptonic W Pair Production with Two Hard Jets

This Section presents the analysis of events where a $W$ pair is produced in association with two hard jets and both $W$ bosons decay leptonically. [29–32] In Section 6.1, the main discriminators in this channel are reviewed. For background extraction and significance estimation, two complementary approaches are considered; both of them are fit-based. In Section 6.2, an analysis that uses a Neural Network to combine several jet variables into a single discriminator and fits the Neural Network output and $M_T$ is presented. This method has the advantage that the correlations among variables related to jets (described in Section 6.1) are naturally taken into account by the Neural Network, and the weak correlations between the jet variables and the lepton variables are largely taken into account by the way the control sample is used in the fit. However, it has the disadvantage that the Neural Network must be trained ahead of time; in principle, this may introduce some model-dependence to the procedure. Ideally, one would like to directly model all kinematic variables in a maximum Likelihood fit; Section 6.3 describes an analysis that takes this approach with a fit of five kinematic variables. This method has the advantage that there is no training stage and therefore has the potential to be more model independent than the analysis of Section 6.2. However, because of the difficulty in constructing a fully correlated probability distribution for a 5-dimensional space, the probability distributions used in the fit of Section 6.3 include only the largest correlations among the kinematic variables and ignore some smaller correlations. It is possible to make some corrections to the shapes of the distributions to model correlations not inherently present in the fit model; Section 6.3 explores the prospects for extracting such corrections from b-tagged control samples. The combined significance estimate at the end of this note will be based on the analysis of Section 6.2, but further studies of both approaches are needed in order to converge on a final strategy for the analysis of real data.

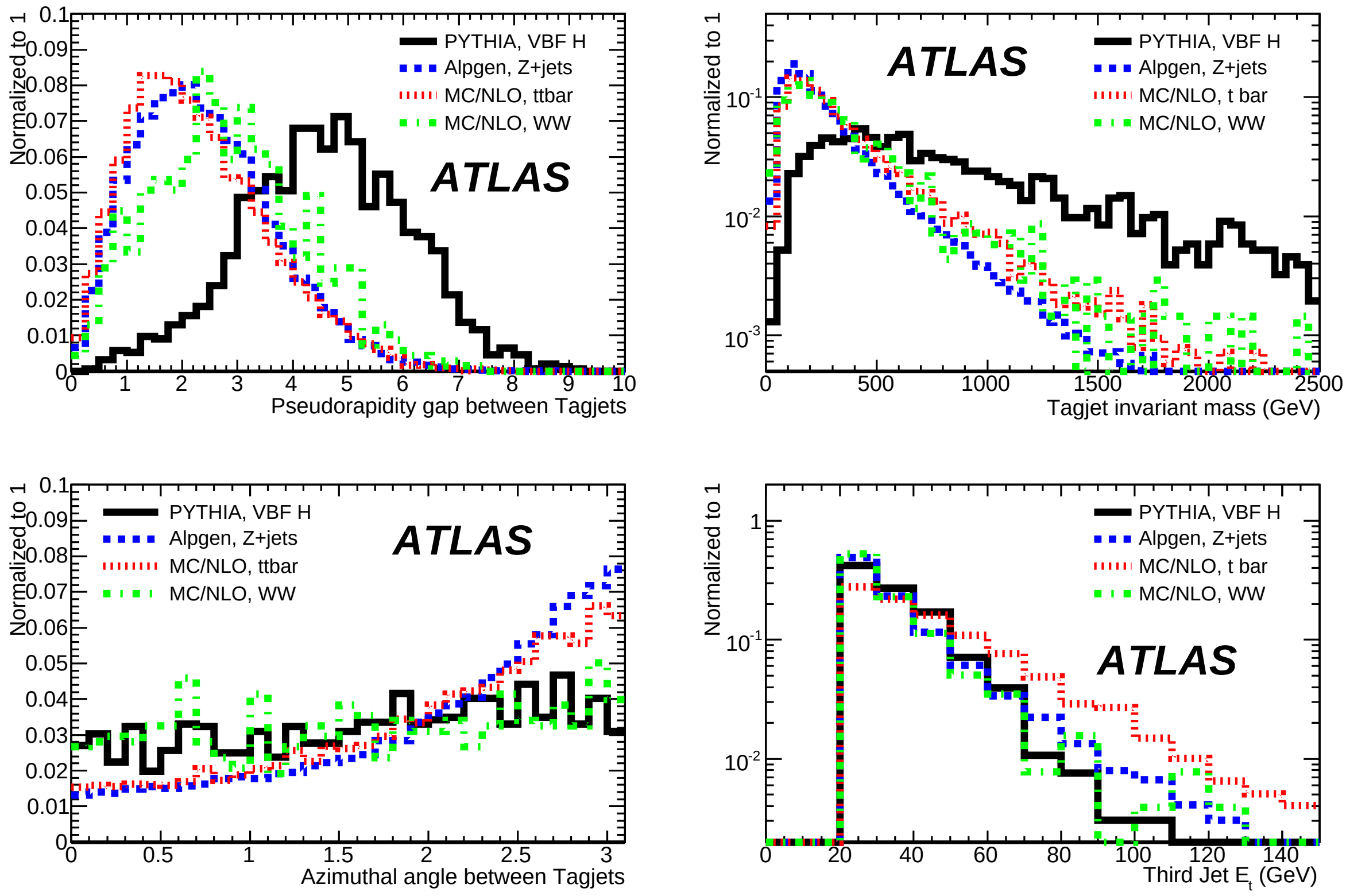


Figure 5: Pseudorapidity gap between tag jets (left top plot), invariant-mass distributions of tag jets (right top plot), azimuthal angle gap between tag jets (left bottom plot) and $E_t$ of the third jet in VBF $H \to WW \to \mu\nu\ \mu\nu$ Pythia events (m(H)=170 GeV). A requirement $\eta_1\eta_2 \leq 0$ is used in addition to the requirement jet $E_t > 20$ GeV.

### 6.1 Handles for Suppressing the Top Background

In both of the fit-based approaches under study, $t\bar{t}$ is the dominant background after the preselection cuts, and in both approaches, similar discriminators are used to suppress it. Two discriminators warrant special attention: jet kinematics, which is discussed in Section 6.1.1, and b-tagging, which is discussed in Section 6.1.2.

#### 6.1.1 Jet Kinematics

One of the main advantages of any search for a Higgs in association with two jets is the possibility to suppress QCD backgrounds with cuts on jet kinematics. The Vector Boson Fusion Higgs signal features several distinctive characteristics:

- The two jets arising from struck quarks (often referred to as “tag” jets) tend to be the highest-$p_T$ jets in the event, and they tend to be well-separated in pseudorapidity;
- they tend to have a large invariant mass;
- there is very little jet activity in the region between the two tag jets.

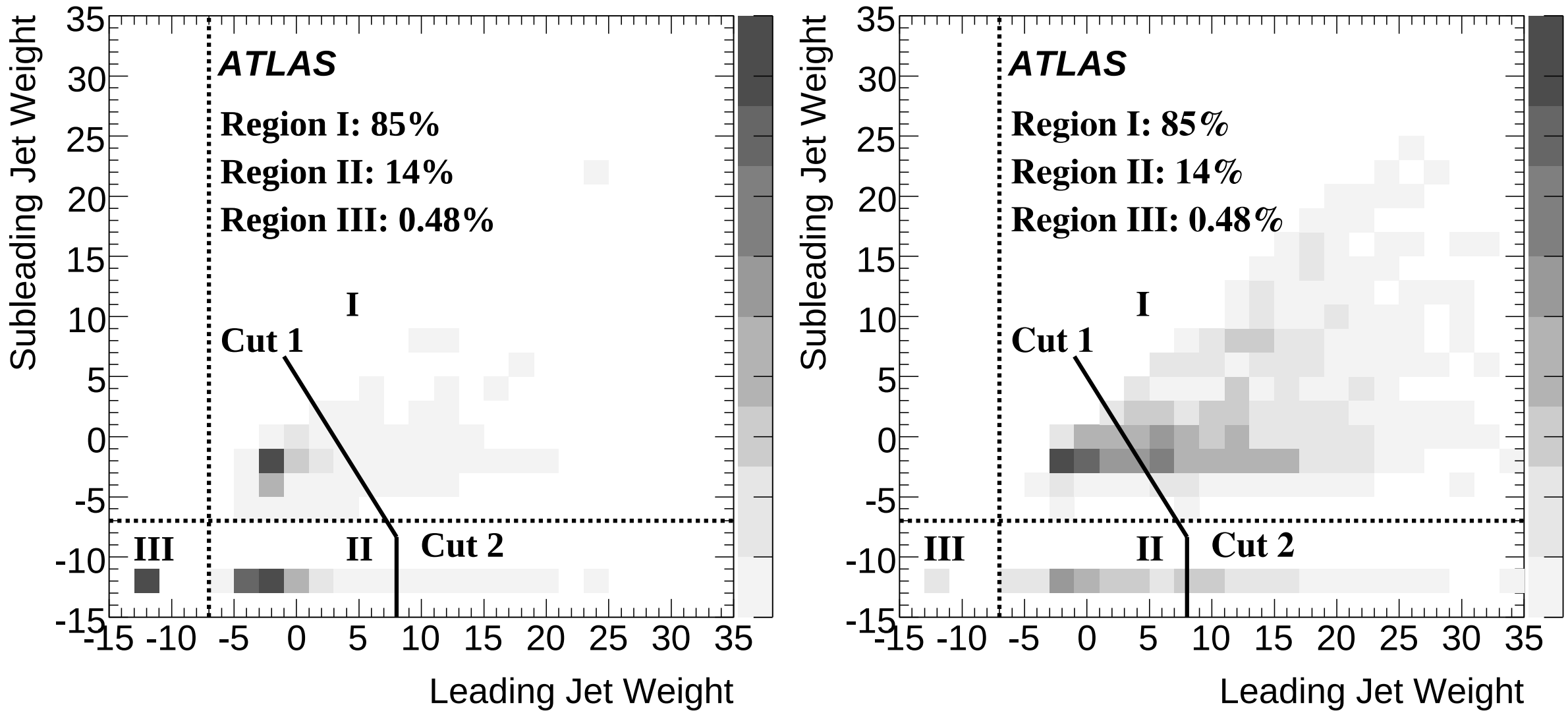


Figure 6: The distribution of leading versus sub-leading jet weights in the events for signal (left plot) and the $t\bar{t}$ background (right plot). The plots are divided in three regions: (I) where there is a non-default $b$-tagging weight for more than one jet, (II) where there is only $b$-tagging information for one jet in the event and (III) where there are no jets with $b$-tagging information in the event.

Figure 5 shows the distributions of the pseudorapidity gap between the tag jets, their invariant mass, and the $p_T$ of the third-highest $p_T$ jet in the event. The figure also shows the distribution of the azimuthal angle between the tagjets, $\Delta\phi_{jj}$. It is possible to enhance the discrimination against backgrounds with a cut on $\Delta\phi_{jj}$, but in this study no such cut is applied on the grounds that this angle is needed in a measurement of the spin and CP properties of the Higgs boson [33].

Only one Higgs boson mass is shown for the signal process in Figure 5, but it is worth noting that the dependence of these variables on the Higgs boson mass is rather weak. It is also worth noting that Pythia 6.4 predicts a somewhat harder $p_T$ spectrum for the third jet than Herwig and Sherpa do. This would amount to a difference as large as a few tens of percent in the survival probability if a jet veto cut were applied.

It has been shown in previous studies that hard cuts on the kinematic variables discussed here can provide a strong rejection against the top background [25,34,35]. As part of the present study, it has been checked that the high signal-to-background ratio that can be achieved in this channel is not sensitive to degradations arising from detector alignment effects.

#### 6.1.2 Rejecting Top with B-tagging

Another effective way to reject a significant fraction of the $t\bar{t}$ events while retaining a high efficiency for the signal is to apply a veto on events containing jets with high $b$-tagging weights. The $t\bar{t}$ events naturally contain two jets originating from $b$-quarks whereas the heavy flavor content in the signal events is limited at tree level to having $c$ quarks in the tagging jets, and then typically in only one of the two tagging jets. This makes vetoing against events with more than one jet with high $b$-tagging weight especially efficient. Figure 6 shows the distribution of leading versus sub-leading jet weights in the events for signal (left plot) and the $t\bar{t}$ background (right plot). The plots are divided in three regions: I where there is a non-default $b$-tagging weight for more than one jet, II where there is only $b$-tagging information for one jet in

| Cut | Signal (170 GeV) | $t\bar{t}$ | $WW$+jets | $Z \to \tau\tau$ | W+jets |
|---|---|---|---|---|---|
| Lepton Selection | 30.20 | 8317 | 838.96 | (2096) | 1323 |
| Forward Jet Tagging | 17.27 | 946.6 | 32.77 | 79.30 | 31.83 |
| Leptons Between Jets | 16.47 | 617.8 | 22.92 | 55.13 | 27.91 |
| $Z \to \tau\tau$ Rejection | 15.68 | 561.8 | 21.20 | 39.03 | 27.91 |
| $p_T^{miss}$, $M_T$, $m_T^{ll\nu}$ | 12.78 | 425.9 | 15.28 | 0 | 13.96 |
| b-veto | 12.67 | 206.72 | - | - | - |
| signal box, b-jet Veto | 9.28±0.27 | 28.5±5.7 | 4.75±0.30 | - | 4.3±4.3 |
| signal box, no b-jet Veto | 9.65 | 114.2 | 4.99 | - | 6.07 |
| Control, b-jet Veto | 3.02±0.15 | 89±10 | 9.78±0.43 | - | 7.9±5.0 |
| Control, no b-jet Veto | 3.13 | 311.7 | 10.28 | - | 7.89 |

Table 5: Cut flows (in fb) for $M_H = 170$ GeV in the $H+2j$, $H \to WW \to e\nu\mu\nu$ channel. Numbers in parentheses are affected by generator-level filter cuts.

the event and III where there are no jets with $b$-tagging information in the event[1]. The fraction of events in each of the three regions is:

- 38% of the events in region I, 48% of the events in region II and 14% of the events in region III for the Higgs signal.
- 85% of the events in region I, 14% of the events in region II and 0.5% of the events in region III for the $t\bar{t}$ background.

The selection in the leading versus sub-leading jet weights plane was optimized for the highest increase in the number-counting significance after the following cuts have been applied:

- Two leptons, $p_T > 15$ GeV
- Missing transverse energy $E_T^{miss} > 30$ GeV
- At least two jets with $p_T > 20$ GeV and $|\eta| < 4.8$
- The two jets with highest transverse momentum are required to be in opposite hemispheres, with $\Delta\eta(jet1, jet2) > 3$
- Require that both leptons are between the two leading jets in pseudorapidity.

Two cuts are applied on the b-tagging weights:

- weight(jet1)+0.6×weight(jet2)<3 for events in region I
- weight(jet1)<8 for events in region II

These cuts provide a strong rejection against the top background and a large acceptance for the Higgs signal.

## 6.2 Two-dimensional Fit

To select the events that are used in the two-dimensional fit, the same preselection cuts that were described in the previous section are applied, plus two additional cuts:

[1] Jets can lack $b$-tagging information if they fall outside the acceptance of the inner tracker, or because there are no tracks with high impact parameter significance in the jet.

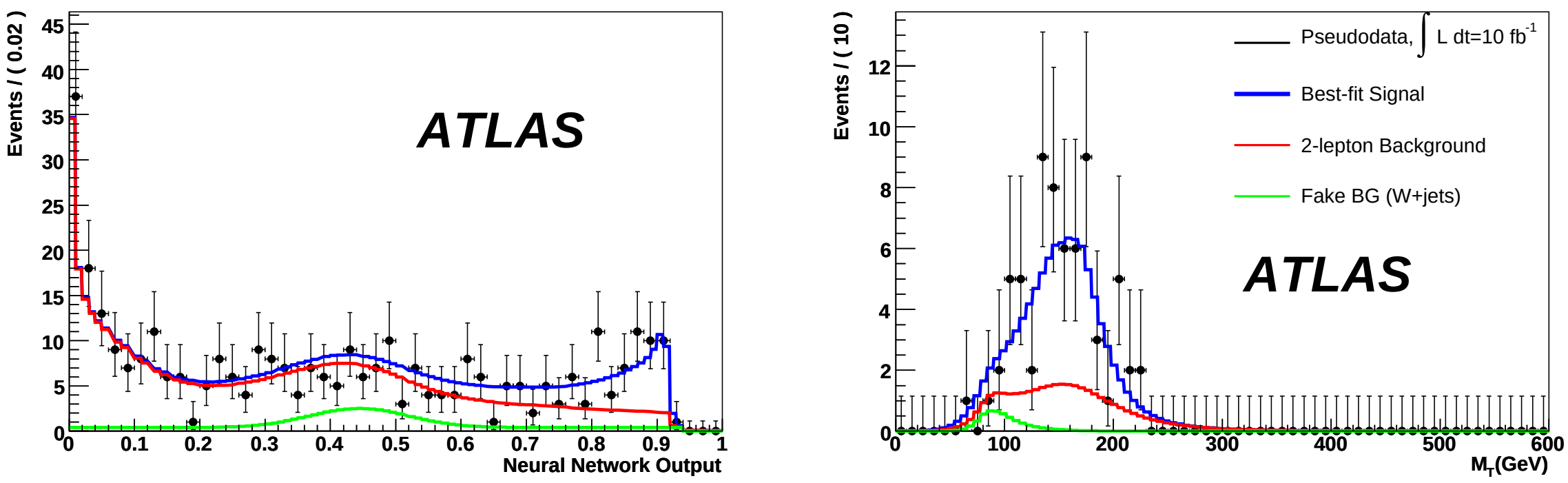


Figure 7: An example fit to a toy Monte Carlo outcome corresponding to 10 $\text{fb}^{-1}$ of integrated luminosity in the $H+2j$, $H \to WW \to e\nu\mu\nu$ channel. The pseudodata contains a Standard Model Higgs boson with a true mass of 170 GeV. Left: The Neural Network output distribution in the signal box, for events with $50 < M_T < 180$ GeV. Right: The transverse mass distribution for events in the signal box with Neural Network output larger than 0.8. The Likelihood Ratio for this outcome was 22.62, which is typical if the signal model used as truth is derived from a Monte Carlo sample generated with SHERPA.

- To reject backgrounds from $Z \to \tau\tau$, assume that the leptons are coming from $\tau$ decays and compute $x_\tau^1$ and $x_\tau^2$, the fractions of the tau energies carried by the visible leptons in the collinear approximation. If $x_\tau^1$ and $x_\tau^2$ are positive and $M_{\tau\tau}$ is close to the $Z$ mass ($|M_{\tau\tau} - M_Z| < 25$ GeV), the event is rejected.
- Require that the transverse mass $M_T$ is between 50 GeV and 600 GeV, and that the transverse mass $m_T^{ll\nu}$ (as defined in Ref. [25]) is at least 30 GeV.

In this analysis, electrons are selected using the standard ATLAS medium electron selection. The definition of muons, and the isolation cuts applied to both, are the same as in the $H+0j$, $H \to WW \to e\nu\mu\nu$ analysis in Section 5.

Events surviving the cuts above are used in the fit; they are partitioned into a signal box and a control region defined by additional cuts on the pseudorapidity gap between the leptons and the dilepton opening angle in the transverse plane. An event lies in the signal box if it has $\Delta\phi_{ll} < 1.5$ and $\Delta\eta_{ll} < 1.4$; otherwise, it lies in the control region. Table 5 shows the cross-sections for signal and background in these two regions.

The trigger efficiency has been studied in the context of this analysis. The same trigger menus as in the $H+0j$, $H \to WW \to e\nu\mu\nu$ analysis are used, and the results are similar: a trigger efficiency of 99.0% after Level 1, 96.8% after Level 2, and 94.5% after the Event Filter. These efficiencies are quite high, and as for the $H+0j$ analysis, it has been checked that the trigger efficiency does not significantly change the shape of the most important kinematic variables. Table 5 does not include the trigger efficiency, but it is taken into account as an overall scale factor for signal and background when generating toy Monte Carlo to test the 2-dimensional fit described in the next section.

After the preselection described above, a four-variable Neural Network is used to further enhance the separation between the signal and the background. The inputs to the Neural Network are:

- $\Delta\eta_{jj}$, the pseudorapidity gap between the tagjets
- $M_{jj}$, the invariant mass of the tagjets
- $p_T^{veto}$, the transverse momentum of the leading non-tag jet in the region $|\eta| < 3.2$, and

- $\eta^* = \eta_3 - (\eta_1 + \eta_2)/2$, the pseudorapidity gap between the tag-tag system and the third jet. It is set to -9 if no third jet is present.

The Neural Network classifies events as signal-like if they have large $\Delta\eta_{jj}$, large $M_{jj}$, low $p_T^{veto}$, and large $|\eta^*|$. Figure 7 shows the Neural Network output and transverse mass distributions in the signal box. The fit is a two-dimensional fit to these two quantities; both the signal model and the background model are uncorrelated product probability density functions (PDFs). The fit does not distinguish among the various types of background.

An in-situ normalization of fake backgrounds has not yet been implemented for this channel. For the present study, a static model has been used to approximate the fake background from $W$+jets. The same functional form used for the irreducible backgrounds was fitted to a set of $W$+jets Monte Carlo events that pass a loosened lepton selection, and its normalization was scaled to the cross-section obtained with the tighter selection cuts actually used in the analysis. In the present study, none of the parameters governing the shape or normalization of the fake background are allowed to float in the fit; systematic errors on the shape and normalization of the $W$+jets background are ignored.

The Neural Network output distribution for the background in the signal box is taken to be the same as the distribution in the control region, but it is multiplied by a linear extrapolation factor. Apart from the slope of this extrapolation factor, all parameters governing the shape of the Neural Network output distribution in the two regions are required to be the same.

A check on the impact of altered jet energy scales on the Neural Network output distributions for signal and background has been performed; changing the jet energy scale by 5% in the region with tracking and 10% elsewhere does not change the Neural Network output shape in a meaningful way. This is not surprising, since the Neural Network inputs are slowly varying functions of the jet energy. Similarly, raising the jet $p_T$ thresholds from 20 GeV to 30 or 40 GeV to mimic a degraded efficiency for low-$p_T$ jets does not have a large impact on the ratio of the Neural Network output distributions in the signal box and the control regions.

Given the fact that the Neural Network output shape is insensitive to the jet energy scale, one would expect that the ratio of distributions in the signal box and control region would be insensitive to the jet energy scale as well. Other possible sources of uncertainty in the ratio of Neural Network output distributions are underlying event activity and the $Q^2$ scale uncertainty; these sources of uncertainty have not been checked explicitly at this time. However, it is reasonable to think they might be small because the cuts on jets in the signal box are the same as in the control region and because the distributions of variables related to jets are not strongly correlated to the lepton angular variables. If the uncertainty in the ratio of distributions is sufficiently small, then one can fix the parameter governing the extrapolation factor; if it is too large, one must allow it to float in the fit. In the latter case, it may be possible to constrain it somewhat by looking into a b-tagged control sample; this possibility has been studied in the context of the five-dimensional fit discussed in the next section. All of these possibilities require further study in the context of the two-dimensional fit; this note will simply consider both scenarios: allowing the slope of the extrapolation factor to float in the fit and fixing it to a predetermined value taken from Monte Carlo. In both cases, it is assumed that the transverse mass distributions are well-predicted, and that the Neural Network output distribution is not; in practice this means that the $M_T$ parameters are set to fixed values while most of the parameters describing the Neural Network output shape are allowed to float in the fit.

The expected significance is shown as a function of the true Higgs mass in Figure 8 for the two treatments of the extrapolation factor considered in this study. The sensitivity is comparable to what was quoted in previous studies of this channel, at least for the case where the extrapolation parameter is constrained to a fixed value.

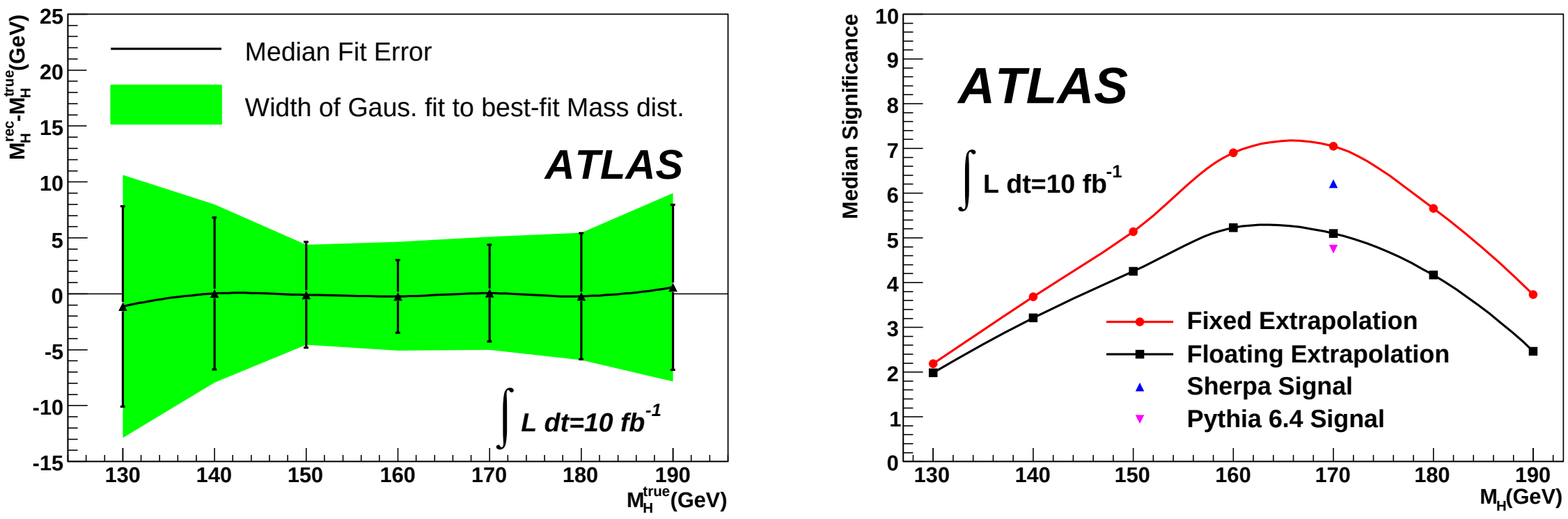


Figure 8: Left: The Neural Network output distribution for the $H+2j$, $H \to WW \to e\nu\mu\nu$ channel in the signal box. Right: The expected significance at 10 fb$^{-1}$. The blue and magenta triangles were computed using a fixed extrapolation.

### 6.3 Five-dimensional Fit

A conceptually simpler but more technically challenging strategy is to directly fit the most important kinematic variables. This section describes such an approach to the $H+2j$, $H \to WW \to \ell\nu\ell\nu$ channel.

The selection cuts used in this analysis are similar to those used in section 6.2, with a few slight differences:

- Either two or three jets with transverse momentum $p_T > 20$ GeV and pseudorapidity $|\eta| < 4.9$ are allowed.
- The missing transverse momentum $E_T^{miss}$ is greater than 20 GeV.
- The pseudorapidity gap between the tagjets is required to satisfy $|\Delta\eta_{jj}| > 2.5$, and the invariant mass of the tagjets is required to lie in the range $m_{jj} \in [600, 3000]$ GeV.
- A $b$-tagged jet is defined as having a displaced vertex significance greater than 4.5.

The values of the Higgs boson mass, $m_H$, and the cross section, $\sigma(VBF\ H \to WW)$, are determined with an unbinned maximum likelihood fit to the distributions of $x = (M_T, \Delta\phi_{ll}, \Delta\eta_{ll}, \Delta\eta_{jj}, m_{jj})$, as obtained from the selected data sample. A multidimensional kernel estimation technique [36] has been used to help model the kinematic distributions in the control samples.

Events in the selected data sample are divided into categories based on four properties: the flavors of the reconstructed lepton pair ($\mu\mu$, $\mu e$, and $ee$); the number of reconstructed good jets (`2jet` and `3jet`); events with and without a $b$-tagged jet (`btag` and `bveto`, where a $b$-tagged jet is defined as having a displaced vertex significance greater than 4.5); and events that fall into the signal region, '`sigbox`', having $|\Delta\phi_{ll}| < 1.5$ and $|\Delta\eta_{ll}| < 1.4$, and those that fall outside, into the '`sideband`' region. The b-vetoed `sigbox` region is denoted as region 1, the b-vetoed `sideband` as region 2, the b-tagged `sigbox` region as region 3, and the b-tagged `sideband` region as region 4.

Most $t\bar{t}$ background events contain a $b$-tagged jet, whereas most $WW$ background events fall into the `bveto` category. The majority of Higgs boson events is expected to be found in physics region 1. Events outside of this region are not considered as signal candidates. These regions are defined to be the background control samples.

For $t\bar{t}$, $WW$, and $H \to WW$ events, the selected samples are mostly dominated by di-leptonic $W$ decays. For this reason, the relative fractions of $\mu\mu$, $ee$, and $\mu e$ events ($f_{\mu\mu}$, $f_{ee}$, $1-f_{\mu\mu}-f_{ee}$) are taken to be identical for background and signal events.

The signal model is obtained from fully simulated Monte Carlo events, and is factorized into four shapes, modelling $M_T$, $\Delta\phi_{ll}$, $\Delta\eta_{ll}$ and $(\Delta\eta_{jj}, m_{jj})$ respectively. The signal shape is assumed to be independent of the lepton pair category and the number of good jets per event.

The Higgs boson transverse mass distribution, $M_T$, is described with a double-sided exponential, having two different lifetimes, convoluted with a Gaussian. The mean of the Gaussian is interpreted as the Higgs boson mass, $m_H$, and is left free in the fit. The left-handed lifetime, $\tau_L$, comes from the missing $z$-component in the transverse mass calculation. The parameter $\tau_L$ is a linear function of $m_H$, and equals 25 GeV for a Higgs boson mass of 130 GeV, and 45 GeV for a Higgs boson mass of 170 GeV. The right-handed component, $\tau_R$, is the result of the non-validity of the approximation $m_{l\bar{l}} = m_{\nu\bar{\nu}}$ in the definition of $M_T$, and is fixed to 20 GeV for all Higgs boson masses. The width of the Gaussian, $\sigma_{MET}$, is interpreted as the resolution on the missing transverse energy, and is fixed to 15 GeV.

The distributions of $\Delta\phi_{ll}$ and $\Delta\eta_{ll}$ in signal are described with single Gaussians, with means fixed to zero, and widths of $\sigma_{\Delta\phi} = 1.09$ and $\sigma_{\Delta\eta} = 0.62$ respectively.

The tagging jet observables $\Delta\eta_{jj}$ and $m_{jj}$ in signal are strongly correlated, and are modeled with a 2-dimensional (2D) kernel estimation function.

For these observables, the largest unmodelled correlation is between $\Delta\phi_{ll}$ and $\Delta\eta_{ll}$, and is found to be 14% from the available signal Monte Carlo sample.

The collection of background events in regions 2–4 serves as a data control sample for the determination of: a) the probability density functions, and b) the number of background candidates in physics region 1. The background shape is assumed to be independent of the lepton pair category and number of good jets per event. The shape is factorized into two pieces, describing $(M_T, \Delta\phi_{ll}, \Delta\eta_{ll})$ and $(\Delta\eta_{jj}, m_{jj})$ respectively.

The background distribution of $(M_T, \Delta\phi_{ll}, \Delta\eta_{ll})$ – observables mostly independent of the jet characteristics – is modelled with a 3D kernel estimation function, using the events in physics region 3. A small efficiency correction is applied to the observable $M_T$, calculated as the ratio the distributions of $M_T$ in physics regions 2 and 4, both determined using 1D kernel estimation functions.

Accounting for the hard $p_T$ spectrum of the $b$-tagged jet sample, the distribution of $(\Delta\eta_{jj}, m_{jj})$ is modelled with a 2D kernel estimation function using the background events in physics region 2. A correction function is applied to $\Delta\eta_{jj}$, determined as the ratio the $\Delta\eta_{jj}$ distributions in physics regions 3 and 4, again determined using 1D kernel estimation functions.

The unmodelled correlations between these observables are found to be smaller than 10%, as obtained from the $t\bar{t}$ and $WW$ background Monte Carlo samples.

The number of background events in physics region 1 is partially estimated from the number of background events in categories 2–4. For this we assume the ratio of events in the `sideband` to `sigbox` regions ($f_{\text{sigbox}-\text{sideband}}$) to be identical for the `btag` and `bveto` categories. Given the ratio of events of physics regions 2 and 4 ($f_{\text{bveto}-\text{btag}}$), determined independently for the `2jet` and `3jet` samples, estimates for the number of background events in physics region 1 can be obtained. These estimates are expressed as the total number of background events, $n_{\text{bkg}}$, and the relative fraction of background event in the `3jet` category, $f_{\text{bkg}-3\text{j}}$.

Finally, the fitted number of Higgs boson signal events, $n_H$, is determined seperately for the `2jet` and `3jet` categories.

### 6.3.1 Toy Monte Carlo Studies

A fit example to approximately 1 fb$^{-1}$ of fully simulated Monte Carlo events, including Higgs boson signal events with a mass of 170 GeV, is demonstrated in the left side of Fig. 9. The plot shows the events

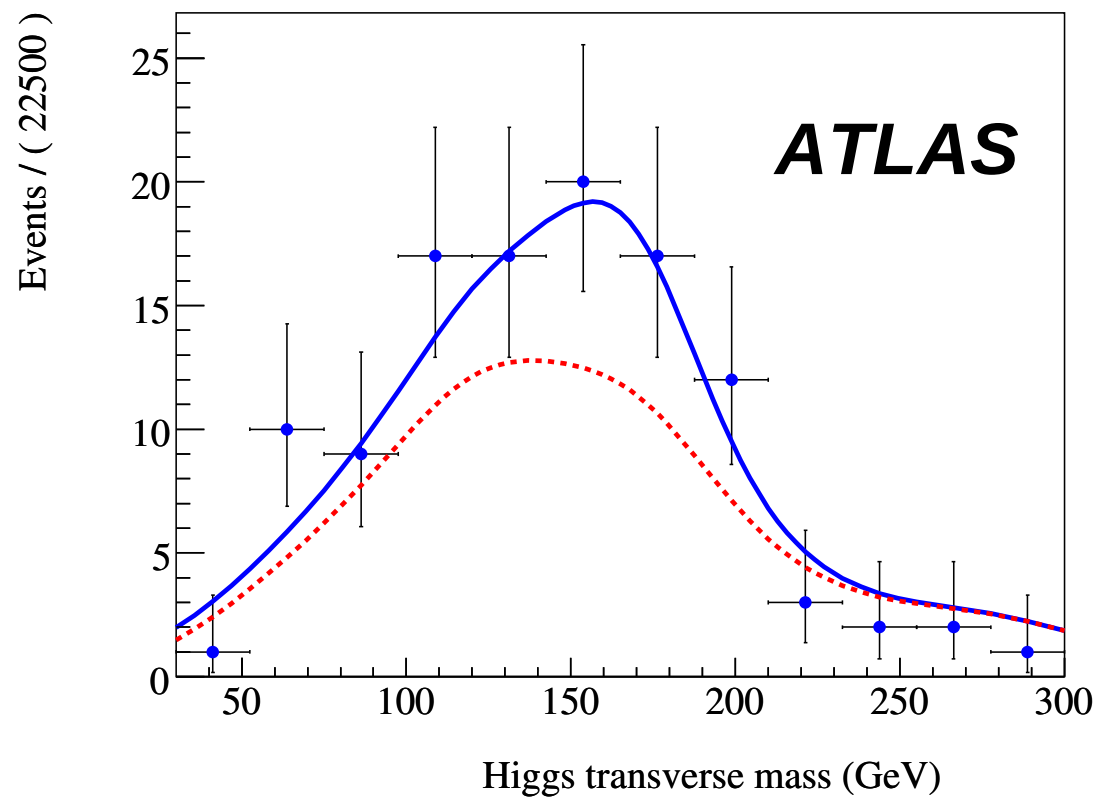


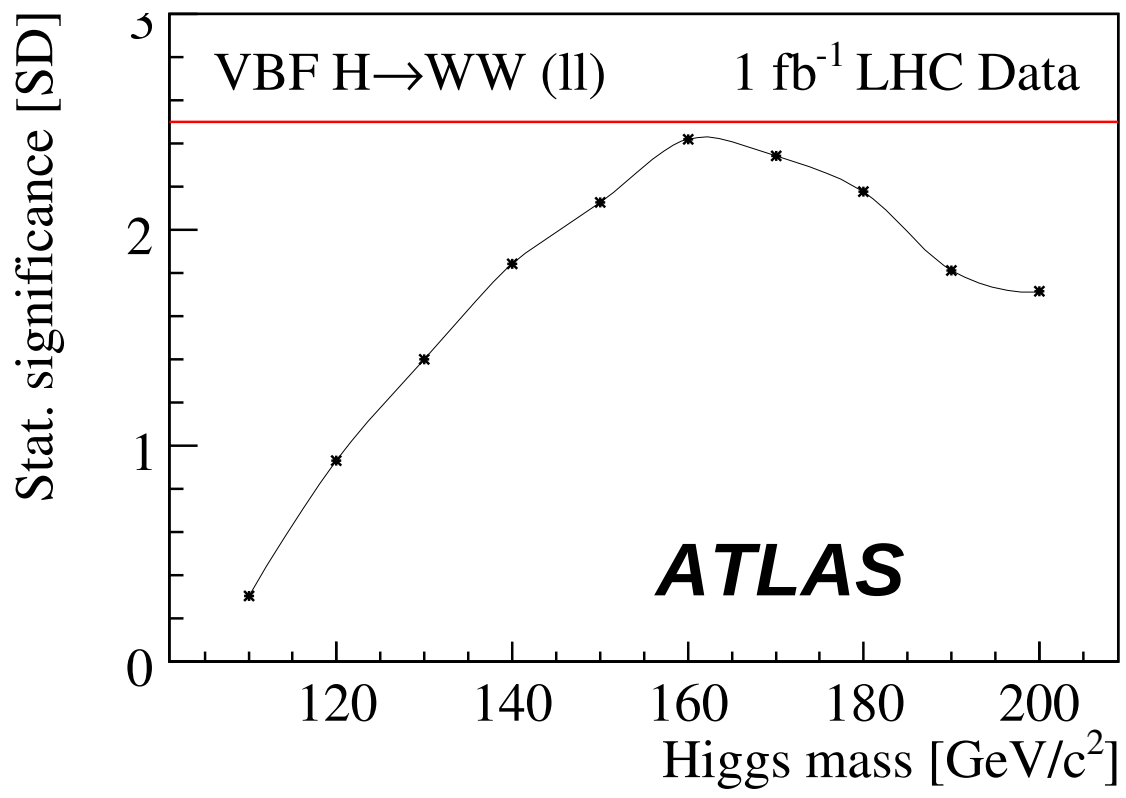


Figure 9: Left: The Higgs boson transverse mass distribution for events in the defined physics signal region, corresponding to approximately 1 fb$^{-1}$ of data, as obtained from fully simulated Monte Carlo events. The solid (blue) curve is the total fit projection. The background contribution is represented by the dashed (red) curve. Right: The expected statistical sensitivity to Standard Model $VBF\ H \to W^+W^- \to l^+\nu l^-\bar{\nu}$ decays for 1 fb$^{-1}$ of ATLAS data, using the five-dimensional fit of Section 6.3.

| Fit parameter | Value | Global corr. | Parameter | Value | Global corr. |
|---|---|---|---|---|---|
| $f_{bveto-btag;\mathrm{2jet}}$ | $0.89 \pm 0.14$ | 72% | $n_B$ | $90.5 \pm 7.4$ | 93% |
| $f_{bveto-btag;\mathrm{3jet}}$ | $3.02 \pm 0.24$ | 88% | $n_{H;\mathrm{2jet}}$ | $18.6 \pm 5.5$ | 25% |
| $f_{bkg-3jet}$ | $0.72 \pm 0.03$ | 80% | $n_{H;\mathrm{3jet}}$ | $9.6 \pm 7.9$ | 38% |
| $f_{sigbox-sideband}$ | $2.64 \pm 0.19$ | 86% | $\sigma_{MET}$ | 15000 | – |
| $f_{ee}$ | $0.11 \pm 0.01$ | 34% | $\sigma_{\Delta\phi_{ll}}$ | $0.62 \pm 0.03$ (fixed) | – |
| $f_{\mu\mu}$ | $0.47 \pm 0.01$ | 34% | $\sigma_{\Delta\eta_{ll}}$ | $1.09 \pm 0.06$ (fixed) | – |
| $m_H$ | $168 \pm 8$ | 12% | $\tau_R$ | 20000 (fixed) | – |

Table 6: The free parameters for the five-dimensional fit described in Section 6.3, and their best-fit values for the example toy Monte Carlo outcome in the left side of Figure 9.

in the defined physics region 1, with the total and background fit projections overlaid. Here the signal shape has been obtained from fully simulated Monte Carlo events. The background probability density function has been obtained using kernel estimation from the (Monte Carlo) events in the background control samples.

The corresponding fit results are given in Table 6. The Higgs boson mass found is consistent with the generated value. The total numbers of Higgs boson and background events found are consistent with the input values of 27 and 86 respectively. The fit is most sensitive to Higgs boson events in the `2jet` signal region – where most Higgs boson events are expected, and contributions from $t\bar{t}$ background events are minimal. For Higgs boson masses in the range $[160, 170]$ GeV, the signal to background ratio is this category is approximately 1 : 1. The error on the fitted number of background events in the signal region is found to be 7.4. When not including the estimated number of background event from the background control samples in the fit, the error on the number fitted number of background events doubles. The error on the fitted number of Higgs boson events rises by 25%, and the correlation of the fitted number of Higgs boson events with all other fit parameters – mostly correlated to the number of background events – rises from 25% to 45%. The correlation of the Higgs boson mass with all other fit parameters is 12%.

The statistical properties of this fitting algorithm have been studied with about 15k toy Monte Carlo

outcomes like the one discussed in the previous paragraph. The toy outcomes have been generated with the same number of events seen in, and the shapes obtained from, the fully simulated Monte Carlo signal and background samples remnant after the event selection, corresponding to 1 $\text{fb}^{-1}$ of data. The mass of the generated Higgs boson sample is varied between 110 and 200 GeV, in steps of 10 GeV. The Higgs boson cross-section is adjusted accordingly [1]. The fit algorithm described above has been performed on each generated pseudo-experiment.

The fitted number of Higgs boson events in the `2jet` signal region – most sensitive to the Higgs signal – shows an average bias of less than 0.8 event over the entire generated mass range. Roughly 12 Higgs boson events are needed in this region before a clear signal peak can be picked up by the fitter. The error on the fitted number of `2jet` Higgs boson events varies between 3.1 and 5.6 events for no generated Higgs boson events to the maximum cross-section between 160 and 170 GeV respectively.

The average fit error on the Higgs boson mass parameter is about 14.0 GeV in background-only outcomes, or about 9.4 GeV in outcomes containing both background and a 160 GeV Higgs signal with cross-section as given by the Standard Model. In Fig. 9, note that the background distribution, dominated by $t\bar{t}$ events, peaks at around 160 GeV. Pseudo-experiments with no generated signal events tend to peak at this mass, where statistical fluctuations are largest. For 1 $\text{fb}^{-1}$ samples, starting from the generated Higgs boson mass of 110 GeV with a value of $+1\sigma$, the fit show a decreasing bias in the mass pull distribution, which is consistent with zero at 160 GeV, and becomes $-1\sigma$ at 200 GeV. The width of the pull is consistent with one in the generated mass range of $[150, 190]$ GeV, and grows for lower cross-section, upto $1.7\sigma$ at 120 GeV.

The average statistical sensitivity derived from the toy Monte Carlo study is shown in the right side of Figure 9. At 1 $\text{fb}^{-1}$ of data, the sensitivity to *VBF* $H \to W^+W^- \to l^+\nu l^-\bar{\nu}$ decays is greater than one for Higgs boson masses greater than 120 GeV. (Below 120 GeV the $H \to W^+W^-$ branching ratio becomes too small for this measurement to be effective.) It reaches a maximum of $2.5\sigma$ at the Higgs boson mass of 160 GeV.

# 7 W Pair Production with Two Hard Jets in the Lepton-Hadron Channel

This section describes the analysis of events where the $W$ pair is produced in association with two hard jets and one $W$ boson decays leptonically (with the other $W$ decaying to jets). [25, 29, 37–42] The dominant backgrounds to this final state are $W$+jets and $t\bar{t}$ production; it is possible that QCD multijets will also play an important role. Because of the large jet multiplicity, there are large theoretical uncertainties in the predicted normalization of the backgrounds, especially for $W$+jets and QCD multijets. Therefore, this section will emphasize studies of the signal, the $W$+jets background, and the $t\bar{t}$ background, with discussion of how to normalize the backgrounds given data. Quantitative predictions of the discovery sensitivity will not be included here.

Because there is only one neutrino in the final state, this channel permits a better estimate of the Higgs candidate invariant mass than the dilepton channels. Taken together with the large sensitivity of the dilepton channels for Higgs masses near 160 GeV, this means that the $H \to WW \to l\nu qq$ channel is most interesting for the study of Higgs bosons with masses in excess of roughly 250-300 GeV.

The most obvious way to estimate the invariant mass of a $H \to WW \to \ell\nu qq$ candidate is to assume that both $W$ bosons are on-shell. One can then use the $W$ mass constraint for the $W \to \ell\nu$ system to estimate the $z$ momentum of the neutrino.[2]

The analysis of this channel uses reconstructed jets to measure the invariant masses of Higgs boson candidates; it is therefore necessary to consider out-of-cone corrections to the jet energies that were not important in the dilepton channels. Eventually, these corrections should be taken from real data;

[2] The $W$ mass constraint yields a quadratic equation which can have 0, 1, or 2 real solutions. If there are two solutions, the one with smaller $|p_z^\nu|$ is used; if there is no real solution, the imaginary term is ignored and only the real part is used.

| Cut | $W$+jets | ttbar | Signal ($M_H = 300$ GeV) |
|---|---|---|---|
| Leptonic $W$ Selection | 2353291* | 128654 | 174.27 |
| Hadronic $W$ Selection | 134483 | 70872 | 73.26 |
| Forward Jet Tagging | 1076.8 | 1929 | 23.16 |
| Lepton Between Jets | 867.0 | 1679 | 22.93 |
| $M_{jj}$ | 131.0 | 367.7 | 9.16 |
| Central Jet Veto | 57.98 | 58.24 | 8.43 |
| $\Delta\eta_{j1,l}$ | 16.07 | 47.96 | 6.93 |
| b-jet Veto | 16.07 | 14.84 | 6.06 |
| Trigger Selection | 13.06 | 12.40 | 5.08 |
| $167 < M_{l\nu qq} < 1000$ GeV | 13.1±4.7 | 12.4±3.4 | 5.08±0.29 |
| Control Regions | | | |
| b-tagged | 0 | 26±5 | 0.75±0.12 |
| Small-$\Delta\eta_{jj}$ Control Region | 500.16 | 778.76 | 20.79 |
| Small-$\Delta\eta_{jj}$ Control Region(b-veto) | 441.64 | 186.13 | 19.95 |
| Small-$\Delta\eta_{jj}$ (b-veto and Trigger) | 369.38 | 157.47 | 15.06 |
| Small-$\Delta\eta_{jj}$ ($167 < M_{l\nu qq} < 1000$ GeV) | 358±29 | 154±12 | 15.1±0.5 |
| Small-$\Delta\eta_{jj}$ Control Region(b-tag) | 37±14 | 493±22 | 2.3±0.2 |

Table 7: Cross-sections (in fb) for the signal and various backgrounds after successive cuts in the $H \to WW \to l\nu qq$ channel. A '*' indicates a number that is biased by generator-level cuts.

however, such a study has not yet been performed. For the present study, the out-of-cone correction from ATLFAST-B [43] has been applied to all jets. After applying this correction to the reconstructed jet momenta, the following cuts are applied:

- Leptonic $W$ selection. The event must contain exactly one hard lepton, with transverse momentum $p_T > 25$ GeV for electrons and $p_T > 20$ GeV for muons. It is also required that the missing transverse momentum, $p_T^{miss}$, be larger than 30 GeV.
- Hadronic $W$ selection. Out of all jets with large transverse momentum, $p_T > 30$ GeV, the two whose invariant mass is closest to the known value of the $W$ mass are selected. It is required that the reconstructed invariant mass of these two jets be between 64 GeV and 90 GeV.
- Forward Jet Tagging. Select the two jets (excluding those from the $W$ decay) that have the highest transverse momentum. These are labelled as the "tagging" jets; by construction, they are disjoint from the jets that form the $W \to qq$ candidate. Require that they be high-$p_T$ ($p_T^{j1} > 50$ GeV, $p_T^{j2} > 30$ GeV), that they lie in opposite hemispheres ($\eta_{j1} \cdot \eta_{j2} < 0$), and that they be well-separated in pseudorapidity ($|\eta_{j1} - \eta_{j2}| > 4.4$).
- Require that the lepton be between the tagjets in pseudorapidity.
- $M_{jj}$: require that the invariant mass of the two tagging jets be greater than 1500 GeV.
- Central Jet Veto: reject the event if it contains any extra jets (in addition to the two jets from the decay of the $W$ and the two tagging jets) with $p_T > 30$ GeV and $|\eta| < 3.2$.
- b-jet veto: apply the b-jet veto cuts described in Section 6.1.2.

| Trigger Level | Electron Channel | Muon Channel |
|---|---|---|
| L1 | 97.1% | 85.3% |
| L2 | 94.3% | 81.2% |
| EF | 92.4% | 79.6% |

Table 8: Trigger efficiencies with respect to offline cuts for the H+2j lepton-hadron channel, for a Higgs boson mass of 170 GeV.

- $\Delta\eta_{j1,l}$: require that the pseudorapidity gap between the leading jet from the decay of the $W$ and the lepton be no larger than 1.5.
- Reconstruct the mass of the Higgs boson candidate, $M_{l\nu qq}$, by applying the $W$ mass constraint to the lepton-$p_T^{miss}$ system. The resolution is enhanced by performing a kinematic fit to the mass of the hadronically decaying $W$, minimizing the function $\chi^2 = (M_W^{rec} - M_W^{true})/\Gamma_W + (\Delta E_{j1})^2/\sigma_{j1}^2 + (\Delta E_{j2})^2/\sigma_{j2}$, where $\sigma$, the energy resolution for the jets, is given for now by the parameterization used in ATLFAST, $\sigma/E_j = 0.03 \oplus 0.5/\sqrt{E_j/\text{ GeV}}$. The left plot in Figure 10 shows the distribution of the Higgs boson candidate invariant mass after all cuts for a representative Higgs boson mass of 300 GeV. The figure includes the mass peak obtained in a few distorted scenarios.

Table 7 shows the cut flow for $W$+jets, for $t\bar{t}$, and for signal at a representative Higgs boson mass of 300 GeV. At the time of this writing, a detailed estimate of the background from QCD multijet events is not available.

The trigger efficiency for this channel has been estimated, considering only the single-lepton triggers: EM25I, EM60, MU20, and MU40 at Level 1; e25i, e60, and mu20i for Level 2 and the Event Filter. Table 8 shows the trigger efficiency for signal ($M_H = 170$ GeV) events that pass the offline cuts. Because the present analysis relies only on single-lepton triggers, the trigger efficiency for this channel is not as high as for the two-lepton channels. However, it has been checked that the trigger efficiency does not distort the $M_{l\nu qq}$ shape for signal in a meaningful way. At the time of this writing, the trigger efficiency for the backgrounds has not been explicitly computed; in Table 7, the trigger efficiency for background is assumed to be the same as the trigger efficiency for signal.

The control sample for this channel is a region with loosened cuts on $\Delta\eta_{jj}$ and $M_{jj}$. In particular, the event selection is the same as in the signal-like region, except that the pseudorapidity gap between the tagjets is required to be less than 4 instead of larger than 4.4, and the lower bound on the invariant mass of the tagjets is lowered to 500 GeV. Table 7 shows the contributions from the various subprocesses after these cuts are applied. Since both the signal-like region and the control region have a nontrivial contribution from top events, it is necessary to define b-tagged control samples to normalize the top background. These samples have the same kinematic cuts as the signal-like and control regions, but the b-veto cut is reversed. The cross-sections in these two regions are also shown in the table.

The shape of the $W$+jets background in the signal-like region is the same as the $W$+jets background in the control region to within the available Monte Carlo statistics. Likewise, the top background in the b-tagged region with the same kinematic cuts as the signal-like region has the same shape as the top background in the b-vetoed signal-like region. In fast simulation, there is a small difference between the top background shapes in the control region and the b-tagged region with the same kinematic cuts as the control region, but that distortion is small compared to the statistical errors expected at the luminosities considered here. These statements are robust against a variety of systematic uncertainties, namely changes in the $Q^2$ scale (a factor of 4 up and down for both top and $W$+jets) as well as altered jet energy scales ($\pm 5\%$ in the region with tracking, $\pm 10\%$ elsewhere), degraded jet energy resolution (Gaussian smearing to roughly double the resolution), degraded $E_T^{miss}$ resolution (modeled by smearing the $x$ and

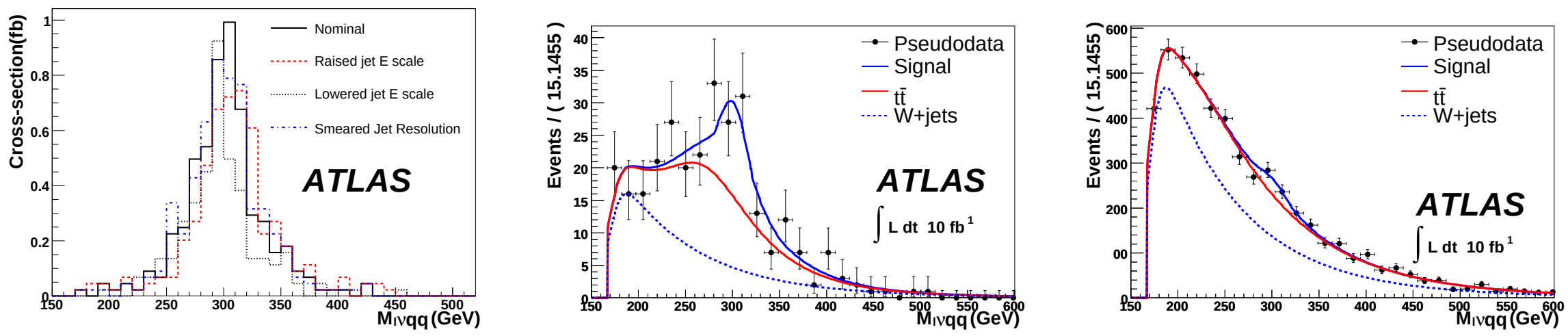


Figure 10: Left: the reconstructed invariant mass of the $M_{l\nu qq}$ system for fully simulated signal Monte Carlo events with a true Higgs boson mass of 300 GeV. The red dashed and black dotted curves show the mass peaks obtained when the jet energy scale is raised and lowered by 7% in the region with $|\eta| < 2.5$ and 15% elsewhere. The blue dot-dashed curve shows the result obtained when the jet energy resolution is smeared by $45\%/\sqrt{E}$ in the region with $|\eta| < 2.5$ and $63\%/\sqrt{E}$ elsewhere. Middle: A toy Monte Carlo outcome corresponding to 10 fb$^{-1}$ of integrated luminosity for the signal-like region of the $H+2j$, $H \to l\nu qq$ analysis described in Section 7. Here, the background from QCD multijets is assumed to be negligible. Right: the corresponding distribution in the control region.

$y$ components of $E_T^{miss}$ by 5 GeV each), and $E_T^{miss}$ reconstruction that has been degraded by artificially inducing a shift of 3% in the reconstructed $E_T^{miss}$.

The background is normalized and the signal extracted with the help of a fitting algorithm which implements a simultaneous binned fit in the signal-like region and the main control region in the range $167 < M_{l\nu qq} < 1000$ GeV. The fit proceeds in two steps: first, standalone fits of the top background model to the b-tagged regions are performed. All parameters governing the shape of the top background distributions are allowed to float in this fit. After the fits to the b-tagged regions, these parameters are fixed, and the histograms are rescaled by the ratio of cross-sections in the b-tagged and b-vetoed regions to obtain estimates of the top background normalization and shape in both b-vetoed regions.

The main fit is a simultaneous fit of the $M_{l\nu qq}$ distribution in the signal-like and control regions. All shape parameters for the $W$+jets background are free to float in the fit, but the parameters governing the top background shape and normalization remain fixed. The $W$+jets background shape is assumed to be the same in the signal-like region and the control region, but the normalizations in the two regions are independent. Likewise, the signal shape is assumed to be the same in both regions, but the ratio of signal cross-sections is parameterized as a function of Higgs boson mass based on Monte Carlo. Figure 10 shows the result of an example fit to a toy Monte Carlo outcome with a true Higgs boson mass of 300 GeV.

The performance of the fitting algorithm has been evaluated by generating pseudo-experimental outcomes corresponding to a luminosity of 10 fb$^{-1}$ with and without true signal events. However, because of the large theoretical uncertainty in the prediction of the $W$+jets background and the lack of a precise estimate of the QCD multijet background, an estimate of the expected significance is omitted here.

# 8 Conclusions

The prospects for a search for a Standard Model Higgs boson in the $WW$ decay mode have been studied, using a realistic model of the ATLAS detector, including effects such as trigger efficiencies, backgrounds from fakes, and realistic misalignments. Three channels have been considered: $H+0j$ with $H \to WW \to e\nu\mu\nu$, $H+2j$ with $H \to WW \to e\nu\mu\nu$, and $H+2j$ with $H \to \ell\nu qq$. In-situ background normalization techniques for all three channels have been proposed, and their effectiveness and robustness has been

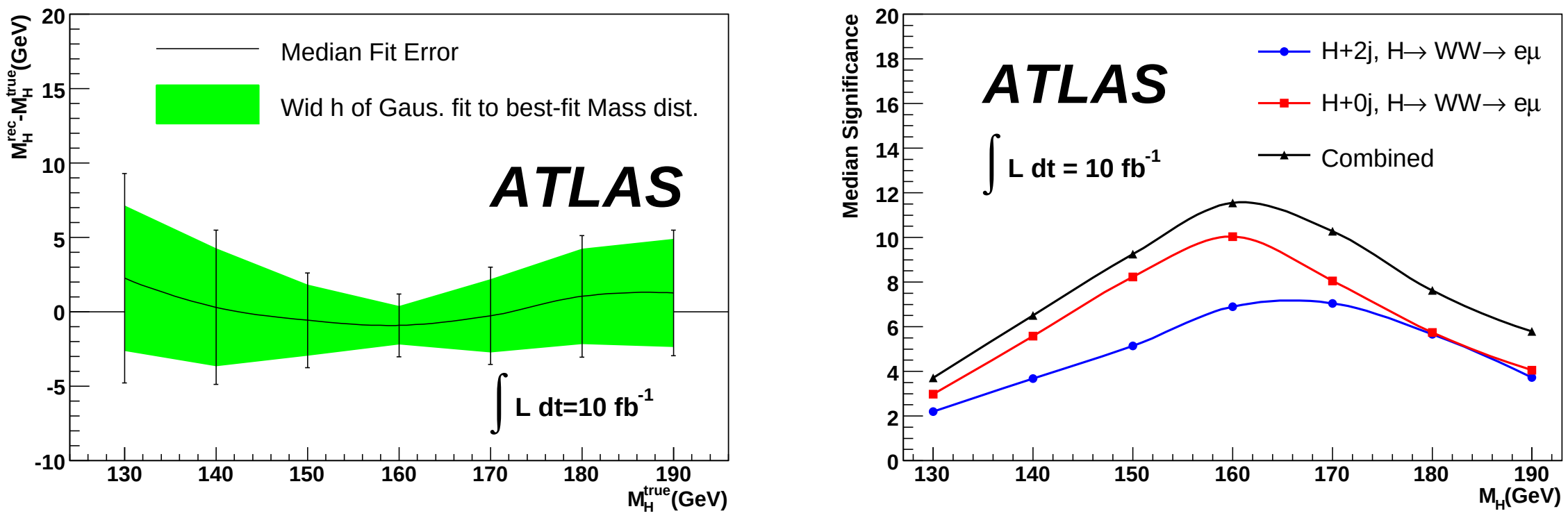


Figure 11: Left: The linearity of the mass determination for the combined fit of $H+0/2j$, $H \to WW \to e\nu\mu\nu$. Right: The expected significance at 10 fb$^{-1}$.

demonstrated with detailed toy Monte Carlo studies.

The $H+0j$, $H \to e\nu\mu\nu$ channel is very promising for Higgs boson masses in the region around the $WW$ threshold. For other masses, this channel is still very promising, but the analysis is more difficult because of large systematic uncertainties in the background prediction. It has been shown that these uncertainties can be controlled, and that the background can be normalized, with a two-dimensional fit in the transverse mass and the $p_T$ of the $WW$ system. With 10 fb$^{-1}$ of integrated luminosity, one would expect to be able to reach a $5\sigma$ discovery with the $H \to WW \to e\nu\mu\nu$ channel alone if there is a Standard Model Higgs with a mass betwen $\sim$ 140 GeV and $\sim$ 185 GeV. A measurement of the mass of the Higgs boson at this luminosity would have a precision of less than 2 GeV for a Standard Model Higgs boson with a mass of 160 GeV, or a precision of less than 4 GeV for a Standard Model Higgs boson with a mass of 140 GeV.

The $H+2j$, $H \to e\nu\mu\nu$ channel has a smaller event rate than the $H+0j$ channel but a similar significance. With 10 fb$^{-1}$ of integrated luminosity, one would expect to be able to reach a $5\sigma$ discovery in the $H \to WW \to e\nu\mu\nu$ channel alone if there is a Standard Model Higgs boson with a mass between 150 GeV and 180 GeV. A measurement of the Standard Model Higgs boson mass at this luminosity would typically return a precision less than $\sim$4-5 GeV if the Higgs boson mass is 160 GeV, or less than 8 GeV if the Higgs boson mass is 140 GeV.

In the range below $M_H = 200$ GeV, both of the dilepton channels are important; it is therefore interesting to consider a combined fit of the two. In the combined fit, a shared mass parameter has been used for the two channels, but the signal normalizations have been allowed to float independently to preserve model-independence. Figure 11 shows the linearity of the mass determination and the expected significance of a combined fit of the two dilepton channels as a function of the true Higgs boson mass. As in the other linearity plots in this note, the green band represents the width of a gaussian fit to the region around the peak of the best-fit mass distribution and the error bars show the median fit error. The combined significance for a Standard Model Higgs is above the $5\sigma$ level for $M_H$ larger than about 140 GeV.

The $H+2j$, $H \to \ell\nu qq$ channel is most interesting for Higgs masses above $\sim$ 250 GeV. It has been shown that the background normalization can be estimated from data, but it is difficult to make a strong statement about the discovery potential of this analysis until a measurement of the $W$+4 jets background can be obtained from first data.

# Search for $t\bar{t}H(H \to b\bar{b})$

**Abstract**

For a light Higgs boson, with $m_H \leq 135$ GeV, the largest decay mode is $H \to b\bar{b}$. Events where the Higgs boson is produced in association with a $t\bar{t}$ pair manifest a distinct signature due to the presence of two $W$ bosons and four $b$ quarks. Topological and kinematical quantities are used to reconstruct the $t\bar{t}$ system. The identification of an additional $b\bar{b}$ pair from the Higgs boson decay is used to further reduce the background.

In this analysis we focus on the sensitivity to a light Standard Model Higgs boson with the ATLAS detector in the channel $t\bar{t}H(H \to b\bar{b})$ using the semi-leptonic final state with 30 fb$^{-1}$ of integrated luminosity. The relevant backgrounds to the channel are investigated and the impact of their associated systematic uncertainties is explored.

## 1 Signal

At the LHC $t\bar{t}H$ production is dominated (90%) by gluon fusion, as illustrated in Fig. 1. The remaining 10% arises from quark-antiquark interactions. For a Higgs boson mass between 115 GeV and 130 GeV the production cross-section times branching ratio to $b\bar{b}$ varies between roughly 0.4 and 0.2 pb at leading order. The top quarks decay almost exclusively to $bW$, and therefore the various final states can be classified according to the decays of the $W$ bosons.

The all-hadronic channel is the one with the highest branching fraction, with a value of 43%. Unfortunately, the large QCD multijet cross-section does not allow easy triggering with jets. Only tight requirements on the jet $p_T$ and on the jet multiplicity could lead to reasonable rates in the first level of the trigger, but these requirements come at the expense of signal efficiency. This is being studied for this final state together with the use of $b$-tagging at the second level of the trigger to reduce the jet $p_T$ threshold.

The fully-leptonic final state analysis is probably the least feasible, despite presenting a simpler signature to trigger on, given the presence of two isolated leptons. The branching fraction (5%) is low and the two neutrinos prevent the reconstruction of the top quarks.

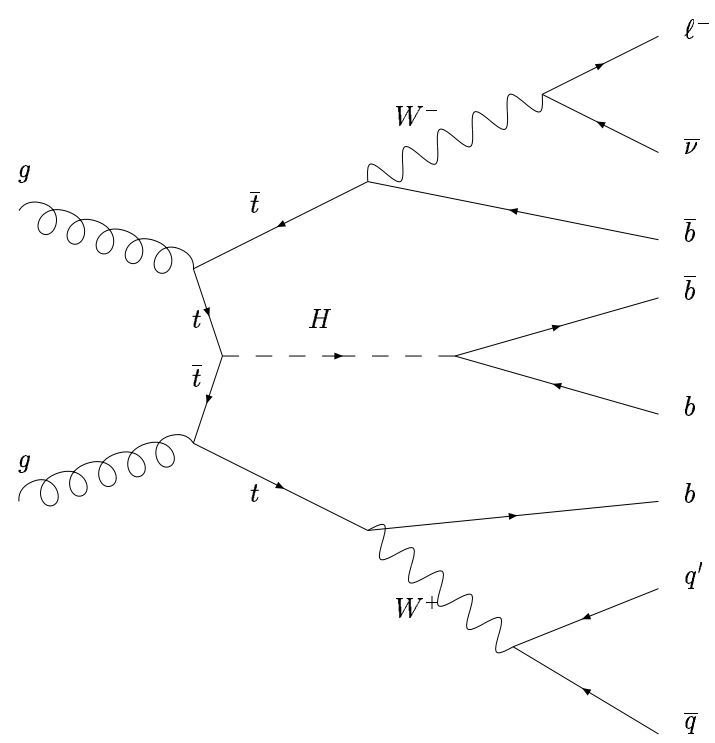


Figure 1: One of the Feynman diagrams for $t\bar{t}H$ production in the semi-leptonic final state.

The semi-leptonic final state is a good compromise with a branching fraction of about 28% excluding tau leptons. The experimental signature consists of one energetic isolated lepton, a high jet multiplicity with multiple $b$-tags, and missing transverse energy from the escaping neutrino, as shown in Fig. 1. The

trigger relies on the presence of the high-$p_T$ lepton. The non-$b$-tagged jets can be used for reconstructing the hadronically decaying $W$ boson, decreasing the possible combinatorial permutations.

## 2 Physics backgrounds

The production of $t\bar{t}$ events is the main background for the $t\bar{t}H$ process. Given the high jet multiplicity in the signal process ($\geq$ 6 jets), only $t\bar{t}$ events produced together with at least two extra jets contribute to the preselected data sample. Since most of these extra jets come from the hadronisation of light quarks, this contribution is greatly reduced by asking for four jets to be identified as $b$-jets.

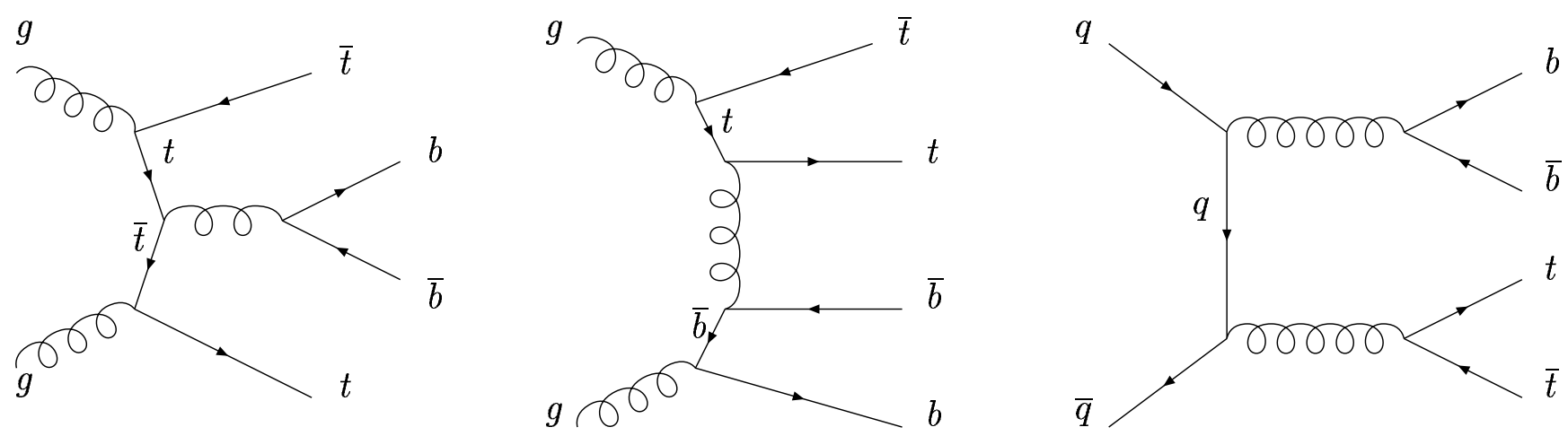


Figure 2: Example of Feynman diagrams for the $t\bar{t}b\bar{b}$ QCD production.

The irreducible background comes from $t\bar{t}b\bar{b}$ production. This can proceed via QCD or electroweak, (EW), interactions with a total cross-section of the order of 9 pb. Some of the Feynman diagrams involved in the two production mechanisms are shown in Fig. 2 and Fig. 3. While the QCD production cross-section is ten times larger than the EW production, the latter is also important. The two $b$-jets not coming from the $t\bar{t}$ decay have large momenta and also have a total invariant mass which is typically close to the $Z$ boson mass, and can therefore contaminate the signal region.

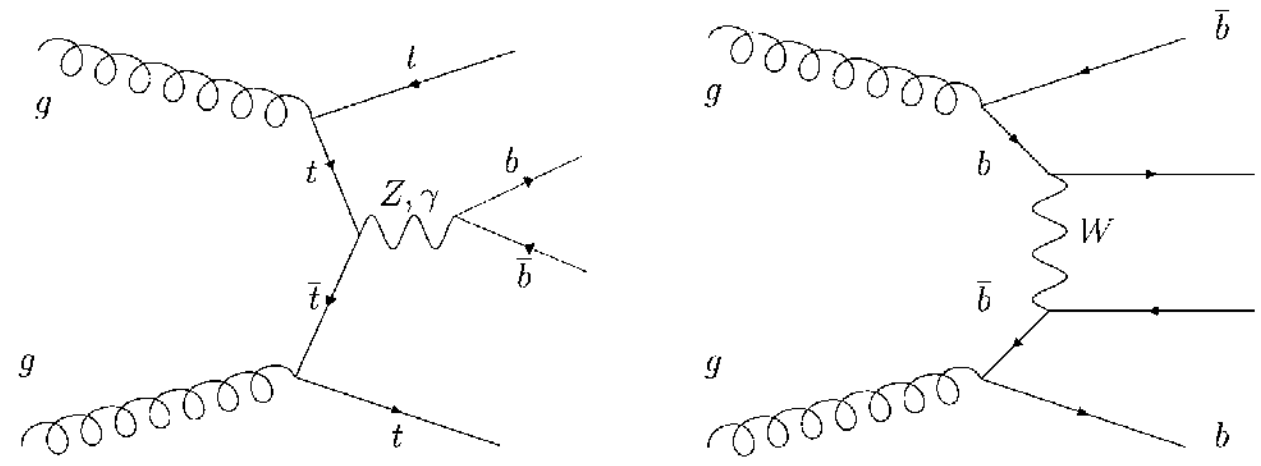


Figure 3: Example of Feynman diagrams for the $t\bar{t}b\bar{b}$ EW production.

The $t\bar{t}c\bar{c}$ background cross-section is 60% higher than $t\bar{t}b\bar{b}$ [1], so it could also play an important role. However, it is found upon investigation that due to the $c$-jet rejection factor, the $t\bar{t}c\bar{c}$ background plays a negligible part in comparison with the $t\bar{t}b\bar{b}$ background. No dedicated sample is therefore simulated for this study. Some $t\bar{t}c\bar{c}$ events are however present in the inclusive $t\bar{t}$ +jets sample used.

Several other backgrounds, such as $W$+jets, $tW$ production and QCD multijet production, could also have a non-negligible impact on the analysis. Even though the $W$ plus two jets inclusive cross-section is about 1200 pb per lepton flavor [2], it has been shown [3] and confirmed in this analysis that the contribution can be reduced to a negligible level if the four $b$-tags requirement is applied. This is also true for the less abundant $tW$ background, which has a cross-section of 9.5 pb [4]. Even when four $b$-jets are requested in the event, contamination via QCD $b\bar{b}b\bar{b}$ production, which has a cross-section of a few hundred nb [5], is still possible. The reconstruction of the $t\bar{t}$ system allows a certain degree of safety against non-top background. None of these samples are presented in what follows.

## 3 Monte Carlo samples and cross-sections

This study uses the leading order cross-sections for the signal and $t\bar{t}b\bar{b}$ samples. No calculation has yet been performed for $t\bar{t}b\bar{b}$ at next-to-leading order, (NLO). This study also uses a $t\bar{t}$ background, simulated at next-to-leading order. This is the only NLO Monte Carlo sample used, and the only $t\bar{t}$ large sample available in ATLAS at the time this analysis was performed. No K-factors are applied to the samples simulated at leading order in the significance estimates.

The signal sample is generated for a Higgs boson mass of $m_H$ = 120 GeV with PYTHIA [6] 6.403. The exact generated process was $pp \rightarrow t\bar{t}H\mathrm{X} \rightarrow \ell\nu bq\bar{q}'bb\bar{b}\mathrm{X}$, with $\ell$ = e or $\mu$. The factorization and renormalization scales used are identical and are listed in Table 1. The signal and the $t\bar{t}b\bar{b}$ events are generated with a lepton filter requiring at least one electron or one muon with pseudorapidity $|\eta| < 2.7$ and transverse momentum above 10 GeV. The leading order production cross-section used is $\sigma(t\bar{t}H) =$ 537 fb [7]. The branching ratios $H \rightarrow b\bar{b}$ of 67.5% at 120 GeV [7], $W \rightarrow \ell\nu$ of 10.66% [8], and $W \rightarrow$ hadrons of 67.6% [8] is applied. Finally the lepton filter efficiency of $\varepsilon = 0.953$ is also applied. The resulting cross-section is 100 fb.

For both $t\bar{t}b\bar{b}$ QCD and EW samples, the exact process generated is gg$\rightarrow t\bar{t}b\bar{b}\mathrm{X} \rightarrow \ell\nu bq\bar{q}'bb\bar{b}\mathrm{X}$, with $\ell$ = e or $\mu$. Both processes can be initiated by a $q\bar{q}$ pair, but only the dominant gluon fusion is simulated, with the cross-sections being increased to allow for the $q\bar{q}$ pair production. For the $t\bar{t}b\bar{b}$ QCD sample, AcerMC 3.4 [9] is used and interfaced to PYTHIA 6.403 for the simulation of the initial and final state radiation, hadronisation and decay. The $t\bar{t}b\bar{b}$ EW sample is generated using AcerMC 3.3 and PYTHIA 6.403. The leading order $t\bar{t}b\bar{b}$ QCD cross-section is $\sigma(pp \rightarrow t\bar{t}b\bar{b}) = 8.2(gg)(+0.5(q\bar{q}))$ pb and the lepton filter efficiency is $\varepsilon = 0.946$. For the $t\bar{t}b\bar{b}$ EW sample, the leading order cross-section is $\sigma(pp \rightarrow t\bar{t}b\bar{b}) = 0.90(gg)(+0.04(q\bar{q}))$ pb and the lepton filter efficiency is $\varepsilon = 0.943$.

The reducible $t\bar{t}$ background events are generated with the MC@NLO [10] program, interfaced to HERWIG [11] and Jimmy [12]. The events in this sample correspond to the processes $pp \rightarrow t\bar{t} \rightarrow (\ell\nu, q\bar{q}')b\ell\nu b$ with $\ell = \mathrm{e}, \mu, \tau$. The generator versions used are MC@NLO 3.1 and HERWIG 6.510. For the inclusive $t\bar{t}$ cross-section we use the NLO+NLL calculation of $\sigma(pp \rightarrow t\bar{t}) = 833$ pb. The $t\bar{t}$ sample is also produced using a filter requiring one electron or one muon with pseudorapidity $|\eta| < 2.7$ and transverse momentum above 14 GeV. The $t\bar{t}$ filter also applies requirements on the jets in the generated events which are reconstructed using a seeded fixed-cone algorithm with a cone size of $\Delta R$ = 0.4 [13], by requiring at least:

- six jets with $p_T >$ 14 GeV and $|\eta| < 5.2$
- four jets with $p_T >$ 14 GeV and $|\eta| < 2.7$

The efficiency of this generator filter on inclusive $t\bar{t}$ events is 0.146.

For the $t\bar{t}$ sample, about 10% of events are $t\bar{t}b\bar{b}$ and are removed following the overlap treatment explained in Ref. [14] together with their associated cross-section.

Table 1 summarizes the cross-sections, calculated using the Monte Carlo generators, of the different processes considered for this analysis, together with the corresponding numbers of generated events and the equivalent integrated luminosity. All branching fractions and filter efficiencies are included.

## 4 Analysis overview

The analysis consists of an initial preselection requirement which is applied to the events to ensure that the fundamental physics objects associated with $t\bar{t}H$ are reconstructed. Following preselection, three different analysis techniques are implemented in order to reconstruct the top quark pairs and the Higgs boson through the identification of their decay products.

Table 1: Summary of the different samples used for the analysis. The cross-sections are taken from the generators and include all branching fractions and filter efficiencies. The fourth column shows the equivalent integrated luminosity, taking into account all corrections (see text). For the scale calculations, $m_H = 120$ GeV and $m_t = 175$ GeV are used. $\max({p_T}^2{}_t, {p_T}^2{}_{\bar{t}})$ corresponds to the higher of the two values of ${p_T}^2$ when both the top and anti-top quarks are considered.

| Process | $\sigma$ (fb) | Events | L (fb$^{-1}$) | Fact. & Renorm Scale | PDF set |
|---|---|---|---|---|---|
| $t\bar{t}H$ (LO) | 100 | 92750 | 931 | $Q^2 = m_t^2 + \max({p_T}^2{}_t, {p_T}^2{}_{\bar{t}})$ | CTEQ6L1 |
| $t\bar{t}b\bar{b}$ QCD (LO) | 2371 | 98350 | 42 | $Q = m_H/2 + m_t = 235$ GeV | CTEQ6L1 |
| $t\bar{t}b\bar{b}$ EW (LO) | 255 | 24750 | 97 | $Q = m_H/2 + m_t = 235$ GeV | CTEQ6L1 |
| $t\bar{t}$ filtered (NLO) | 109487 | 710321 | 6.5 | $Q^2 = m_t^2 + \frac{1}{2}({p_T}^2{}_t + {p_T}^2{}_{\bar{t}})$ | CTEQ6M |

The identification and association of decay products is directly related to the quality of the reconstructed Higgs boson signal. It mainly suffers from the misassociation of the four $b$-tagged jets to the original partons. For this reason, the initial cut-based approach is complemented by two multivariate algorithms, called the pairing likelihood and constrained mass fit.

# 5 Preselection

At the preselection level we require that the event passes the trigger requirement to identify at least one high-$p_T$ lepton (muon or electron) coming from the decay of one of the $W$ bosons. We then require that the event reconstruction identifies exactly one isolated high-$p_T$ lepton (muon or electron). Vetoing the presence of a second isolated lepton is intended to remove additional sources of background. After the lepton requirements are met, we require at least six calorimeter jets, of which at least four must be loosely $b$-tagged jets from the decay of the top quarks and the Higgs boson.

## 5.1 Trigger requirements

The presence of one high-$p_T$ lepton, together with missing transverse momentum, is a distinct signature of $W$ boson production. These leptons can generally be used to trigger on $W$ production with high efficiency. A logical OR of the single isolated electron (`e22i`) [15], high-$p_T$ electron (`e55`) [15] and single muon (`mu20`) [16] triggers is used. The inclusion of the `e55` trigger is found to improve the efficiency for high-$p_T$ electrons where the `e22i` trigger efficiency was reduced due to the isolation requirement. Missing energy triggers were not available at the time of writing, but could be used in future analysis.

The trigger efficiency is approximately 82% for the semileptonic top decays for those events which would otherwise pass the offline analysis. This is included consistently in the following sections.

## 5.2 Reconstructed high $p_T$ lepton selection

In this Section we explain the selection criteria used for reconstructed electrons and muons produced in the semi-leptonic decay of the $t\bar{t}$ system. As previously mentioned, exactly one high-$p_T$ isolated electron or muon must be reconstructed for the event to pass the preselection.

To be considered for the analysis, reconstructed electrons must have transverse momentum $p_T >$ 25 GeV and pseudorapidity $|\eta| < 2.5$. Further calorimeter-based cuts are applied to the loose electron [17] definition. An isolation cut is also applied to the candidate electrons in the form of an upper limit of 0.15 on the ratio of the $p_T$ of the additional tracks inside a cone of size 0.2 in $\Delta R$, ($\sqrt{\Delta\eta^2 + \Delta\phi^2}$), around the electron track to the electron $p_T$.

Muon candidates are reconstructed using a combination of the Inner Detector and Muon Spectrometer [18]. They must pass the acceptance cuts $p_T > 20$ GeV and $|\eta| < 2.5$. In order to remove poorly reconstructed muons, cuts are applied to the muon track fit quality and its transverse impact parameter, which helps to discriminate against muons generated by the decay of long-lived mesons.

An isolation cut of 0.30 on the ratio between the transverse energy deposited inside a cone of size 0.2 in $\Delta R$ around the muon track and the muon $p_T$ is applied.

## 5.3 Jets

To reconstruct the energy of the partons produced in the original collision, calorimeter jets are reconstructed using a seeded fixed-cone algorithm with a cone size of $\Delta R$ = 0.4 [13]. Cuts on $p_T > 20$ GeV and $|\eta| < 5.0$ are initially applied. Only events with at least 6 jets are kept for the analysis. All electrons reconstructed as jets are identified and removed from the jet collection according to the electron overlap removal procedure described in Section 5.3.1. Reconstructed muons which are not isolated are combined into jets where applicable (see Section 5.3.3), and only after this step are the jet energies calibrated for residual effects. The jet multiplicity is shown in Fig. 4, calculated after electron overlap removal and the jet $\eta$ and $p_T$ cuts.

In the following analyses the concept of 'correct' jets is important. This is defined by finding the closest reconstructed jet to each parton, after final state radiation. This match is in $\Delta R$ space and must be closer than 0.4. A $W$ or $H$ boson is correctly matched if both the jets being used are associated to the partons from its decay, while for a top quark the matching refers to the $b$ quark jet only. Normally 'correct' is applied in this note only to one quark or boson at a time.

### 5.3.1 Treatment of overlaps between jets and electrons

Since most electrons are also reconstructed by the jet algorithm it becomes necessary to identify them in the jet collection in order to avoid double counting. The criteria for the jet-electron overlap removal is the following: each jet matching a well-reconstructed electron (i.e. fulfilling the cuts defined in Section 5.2) within a $\Delta R$ of 0.2, and for which the ratio of the electron to the jet transverse momenta is greater than 0.75, is discarded from the jet collection. About 4% of the jets in the signal sample are removed by this selection, 99% of them being actual true electrons.

### 5.3.2 *b*-tagging

*b*-jets are identified using the IP3D+SV1 tagger [19], which exploits both the impact parameter of tracks and the properties of an inclusive secondary vertex, using a likelihood approach which leads to a single discriminating variable: the *b*-tagging weight. In order to allow for a projected decrease in light jet rejection of approximately 30%, the *b*-tagging weight for this study is increased by 0.9 for central jets ($|\eta| < 2.5$) having no associated ($\Delta R < 0.3$) heavy quark or lepton ($b$,$c$,$\tau$) in the Monte Carlo simulation history. The *b*-tag weight spectrum for *b*-, *c*-, and light jets in the signal sample is shown in Fig. 4. A cut on the weight defines which jets will be eventually identified as *b*-jets in the analysis. The rejection of *c*- and light jets versus the *b*-jet efficiency, obtained by varying the weight cut, is also shown in Fig. 5. The different samples exhibit very similar behaviour. The rejection for "purified" jets is shown in the bottom row of Fig. 5. Purified jets are those which have no heavy flavor ($b$, $c$, $\tau$) quark or lepton within 0.8 in $\Delta R$.

In the preselection, a loose set of criteria are initially used to define a sample of jets as a first step to identifying $b$ jets for the analyses. The requirements are that the jet is in the central region of the detector $|\eta| < 2.5$, and has *b*-tag weight $\geq 0$. If there are fewer than four of these jets, the event is discarded.

The cut-based analysis and the pairing likelihood (Sections 8 and 9) require that there are at least four $b$-jets having $b$-tag weight $\geq 5.5$. The $b$-tag weight $\geq 0$ working point implies a $b$-tagging efficiency of about 85% and a rejection of light ($c$-) jets of about 8.6 (2.4), whereas the working point at $b$-tag weight $\geq 5.5$ implies a $b$-tagging efficiency of about 65% and a rejection of light ($c$-) jets of about 60 (6).

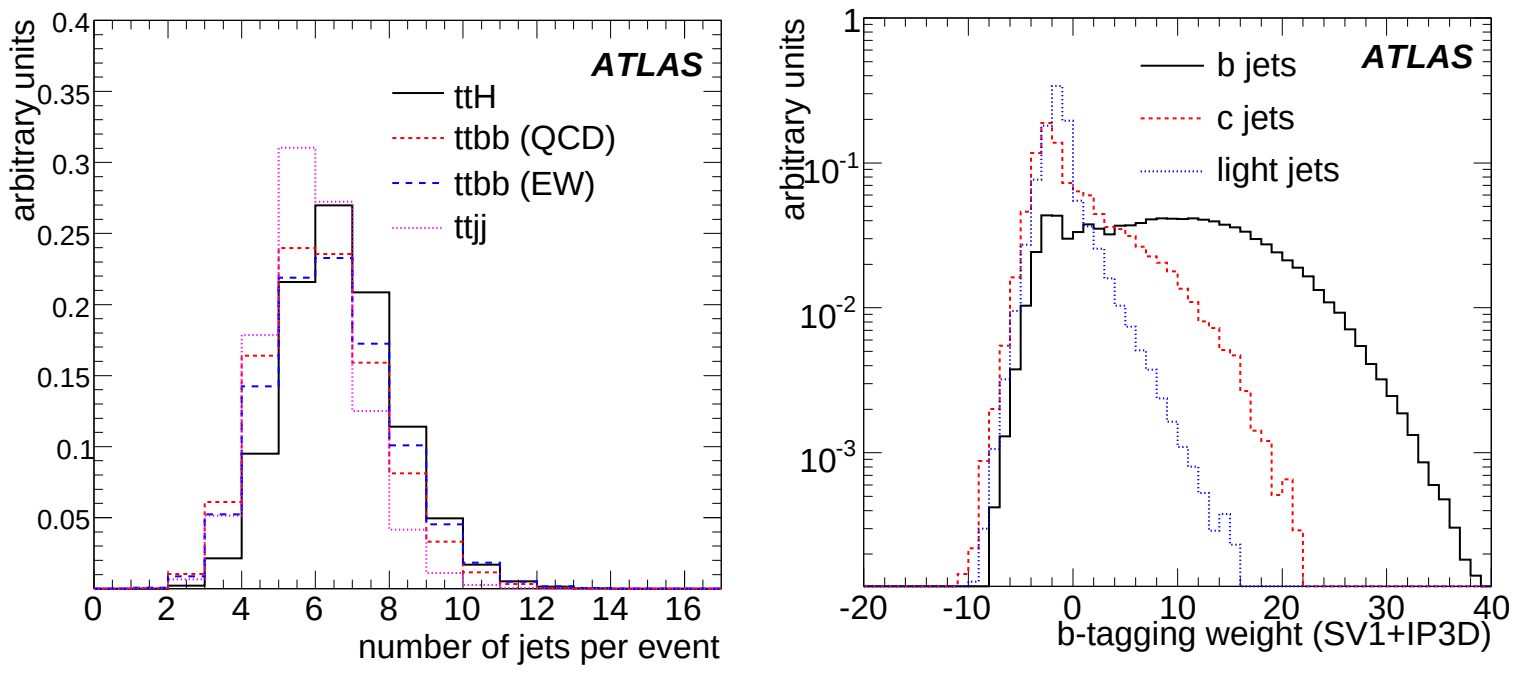


Figure 4: Left: Multiplicity of jets, inside $p_T$ and $\eta$ acceptance. Right: Distribution of $b$-tagging weight for $b$-, $c$- and light jets in $t\bar{t}H$ events, using the IP3D+SV1 tagger.

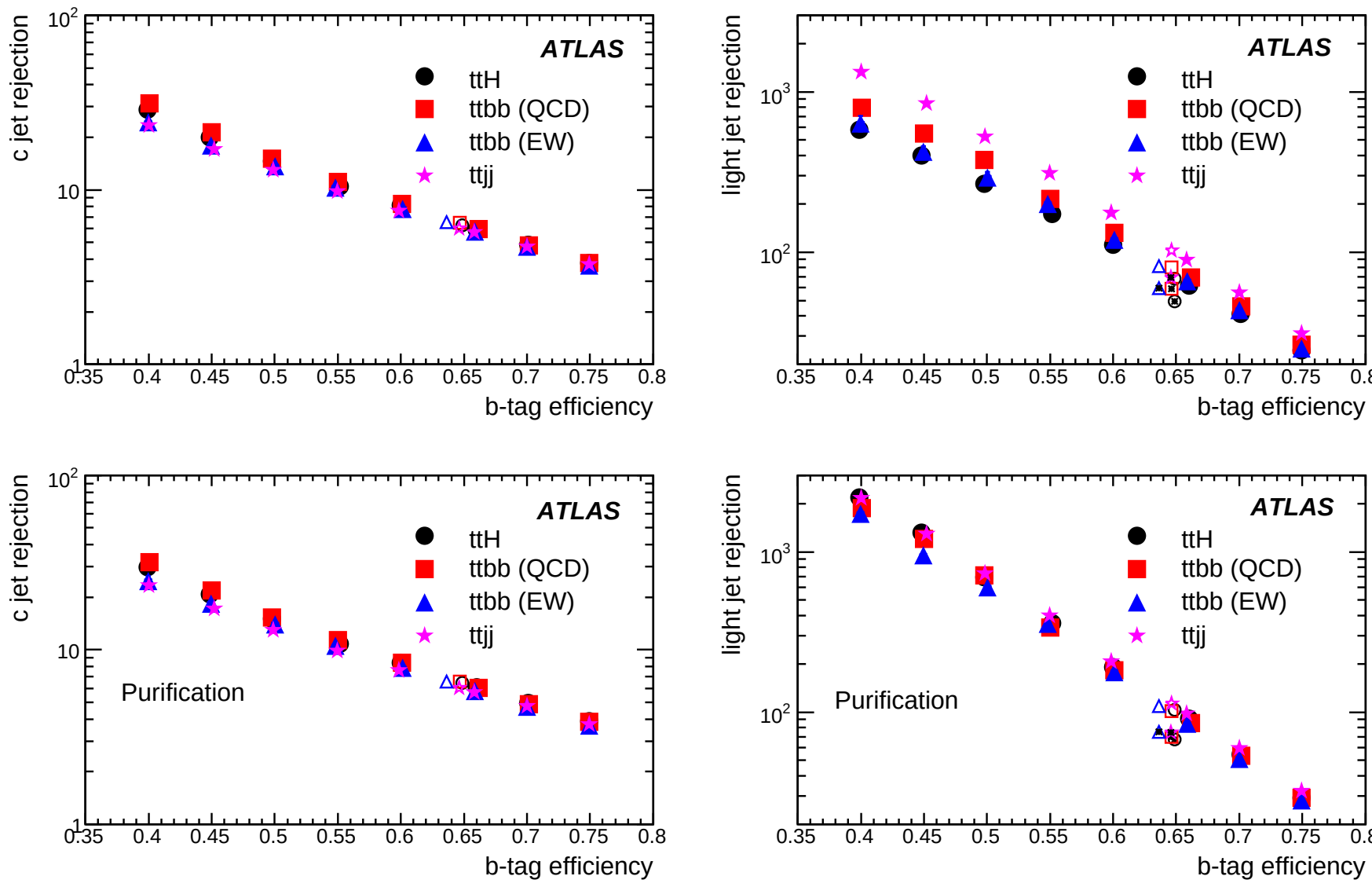


Figure 5: Rejection of light and $c$-jets versus $b$-tagging efficiency. Red open markers indicate the working point ($b$-tag weight $\geq 5.5$) used for the cut-based and paring likelihood analyses, open markers with a central dot (right plots) represent the performance when the 30% performance degradation is applied. The lower plots show results for purified jets, where no heavier quark existed within a wide cone of $\Delta R < 0.8$.

### 5.3.3 Treatment of low $p_T$ muons

About 20% of the time a $B$-meson decay cascade gives rise to a muon. With a four $b$-jet signature in this channel, these muons, also called *soft muons* are present in almost every event. In order to improve the estimate of the momentum of the original $b$ quark, these muons must be used to correct the jet four-momenta by adding the muon four-momentum to a jet. Two different algorithms (high and low-$p_T$ [20])

identify the muons, which are required to be within $\Delta R < 0.4$ of the jet axis. Among the candidates from the high-$p_T$ algorithm which are separated from the selected hard lepton by $\Delta R > 0.1$, only the one with the best track quality is considered for addition. All the neighbouring candidates from the low-$p_T$ algorithm are considered for addition, provided they fulfil $p_T > 4$ GeV and $p_T < 100$ GeV, $|\eta| < 2.5$ and chi-squared per degree of freedom $\chi^2/n < 30$ for the combined fit. In addition, a loose anti-isolation cut is applied, requiring that the energy reconstructed in the calorimeter within a cone of size 0.2 in $\Delta R$ around the muon track divided by the muon $p_T$ is higher than 0.1. Table 2 shows that adding low $p_T$ muons to jets improves both the mean jet $p_T$ and resolution. Fig. 6 shows that there is also an improvement in the Higgs boson mass and resolution, when the correct jet combination is chosen.

### 5.3.4 Calibration

A Monte Carlo based jet correction has been derived to take into account residual calibrations, *e.g.* out-of-cone effects and neutrinos. The parametrization was derived from full simulation, so that the jet four-momentum is corrected by a flavor dependent rescaling factor which scales all components of the four-momentum. Table 2 shows that for $b$-jets, the residual calibration brings the jet and associated parton $p_T$ into agreement from an offset of 5.4 to $-0.5$ GeV. Each of the analyses uses the light jet correction for those jets which are assigned to the W boson and corrects the other four jets as $b$ quark originated. The impact of the calibration on the Higgs boson mass peak where the correct jet combination is chosen is shown in Fig. 6. Both the mass peak location and the resolution are improved.

Table 2: The true parton $p_T$ minus the measured jet $p_T$ with and without adding muons and making the out-of-cone correction. The quoted values are the results of a Gaussian fit in the region ±20 GeV.

| Treatment | Value | No Calibration | Added muons | Calibrated | Both corrections |
|---|---|---|---|---|---|
| All $b$ jets | Mean, GeV | 5.7±0.05 | 5.4±0.05 | -0.1±0.04 | -0.5±0.04 |
| | Sigma, GeV | 10.0±0.05 | 9.8±0.05 | 10.2±0.05 | 10.0±0.04 |
| $b$ jets with muons | Mean, GeV | 26±2.6 | 7.6±0.2 | 11.4±0.5 | 2.3±0.2 |
| | Sigma, GeV | 18.2±1.1 | 11.8±0.2 | 13.6±0.4 | 12.0±0.2 |
| All light jets | Mean, GeV | 2.4±0.04 | 2.3±0.04 | -1.4±0.04 | -1.5±0.04 |
| | Sigma, GeV | 8.4±0.04 | 8.4±0.04 | 8.9±0.04 | 8.3±0.04 |
| Light jets with muons | Mean, GeV | 10.5±0.9 | 2.4±0.4 | 7.3±0.7 | -1.0±0.4 |
| | Sigma, GeV | 11.3±0.8 | 10.9±0.5 | 11.5±0.6 | 10.7±0.4 |

## 5.4 Results of preselection on signal and background

The effect of the event preselection on signal and background samples is illustrated in Table 3. The efficiency for the four $b$-tag cut is different in the signal and in the irreducible background because the two additional $b$-jets have different $p_T$ and $|\eta|$ spectra. The preselection, at a level of four loose $b$-tags, removes a large fraction of the signal, but also reduces the backgrounds to a level where they can be handled more easily; for example 98% of the $t\bar{t}$X background is removed. It also ensures that the selections are a subset of the requirements placed at generator level. The selected sample has approximately a 0.6% signal component, with a little more than 8% of the background being irreducible $t\bar{t}b\bar{b}$. Further tightening the $b$-tagging requirement to four jets with weights of at least 5.5 reduces the samples with four $b$ quarks by a further factor of four while removing 90% of the remaining $t\bar{t}$X background.

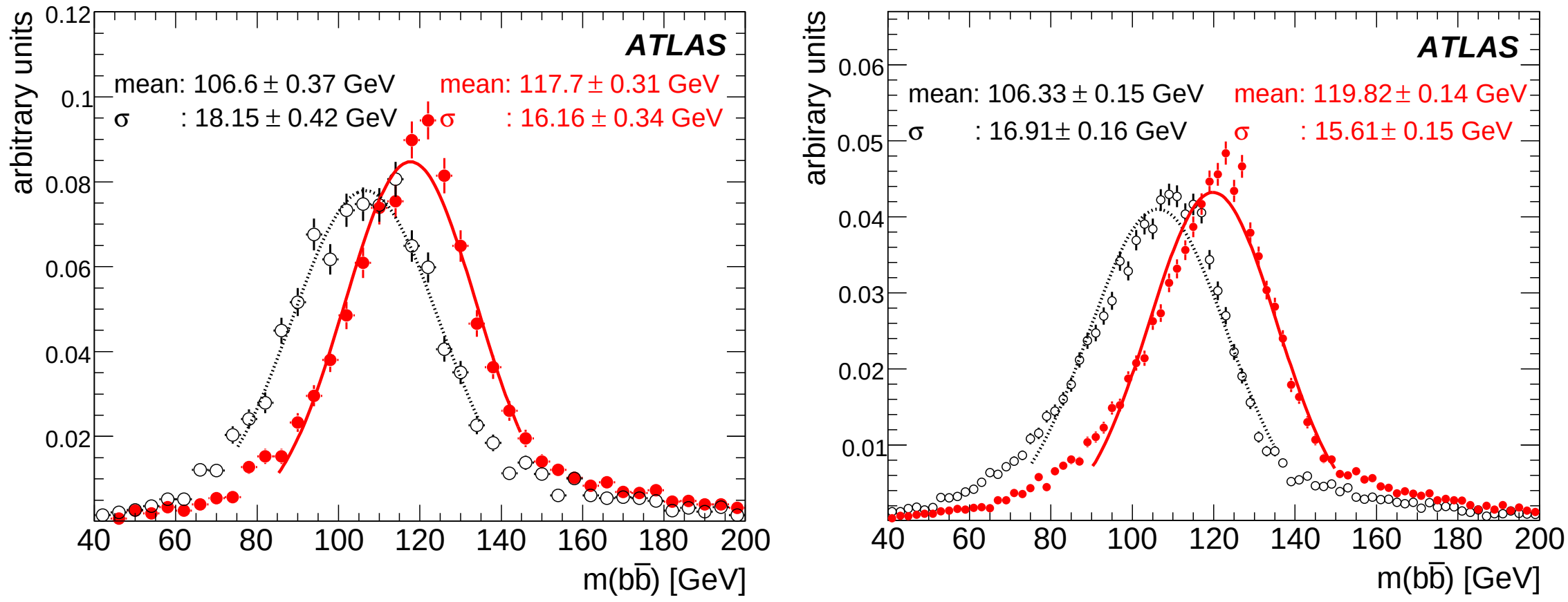


Figure 6: Effect of adding a reconstructed low $p_T$ muon to at least one of the $b$ jets (left), and the effect of jet calibration (right) shown for the Higgs boson mass where the correct jet combination is chosen. Red solid (black open) markers and solid (dotted) line show the mass distribution and fitted values for jets after (before) correction. The effect of the accidental wrong muon matches can be seen in the shape distortion. All distributions are normalized to unity.

Table 3: Cross-sections after each preselection cut for signal and background. The last row shows the further effect of tightening the $b$-tag requirements to the level of the final selection in the cut based and pairing likelihood analyses. In the last column the contribution of $t\bar{t}b\bar{b}$ has been removed. The errors are statistical only.

| Preselection cut | $t\bar{t}H$(fb) | $t\bar{t}b\bar{b}$(EW) (fb) | $t\bar{t}b\bar{b}$(QCD) (fb) | $t\bar{t}$X (fb) |
|---|---|---|---|---|
| lepton | $57. \pm 0.2$ | $141 \pm 1.0$ | $1356 \pm 6$ | $63710 \pm 99$ |
| $+ \geq 6$ jets | $36 \pm 0.2$ | $77 \pm 0.9$ | $665 \pm 4$ | $26214 \pm 64$ |
| $+ \geq 4$ loose $b$-tags | $16.2 \pm 0.2$ | $23 \pm 0.7$ | $198 \pm 3$ | $2589 \pm 25$ |
| $+ \geq 4$ tight $b$-tags | $3.8 \pm 0.06$ | $4.2 \pm 0.2$ | $30 \pm 0.8$ | $51 \pm 2$ |

## 6 Reconstruction of the hadronically decaying $W$ bosons

Reconstructing the $W$ boson four-momenta is necessary to reconstruct the top quarks. The reconstruction of the hadronically decaying $W$ boson is done in different ways by the three analyses.

For the cut-based and pairing likelihood analyses (see Sections 8 and 9), the highest four $b$-tagged jets (with $b$-jet weight $\geq 5.5$) are excluded from the hadronically decaying $W$ reconstruction, however all other jets are paired to form $W$ candidates. Figure 7 shows the mass distribution and multiplicity of all $W$ candidates for the cut-based analysis. Only candidates within 25 GeV of the true $W$ mass are kept. Even with these cuts the hadronically decaying $W$ candidate multiplicity is still very high. All jets used to form these $W$ candidates are calibrated with the light-jet calibration. The likelihood analyses do not require an explicit cut upon this mass as the definition of the likelihood imposes it automatically, but the constrained fit likelihood (Section 10) imposes a requirement that the mass be between 30 and 150 GeV to reduce the number of combinations to be evaluated.

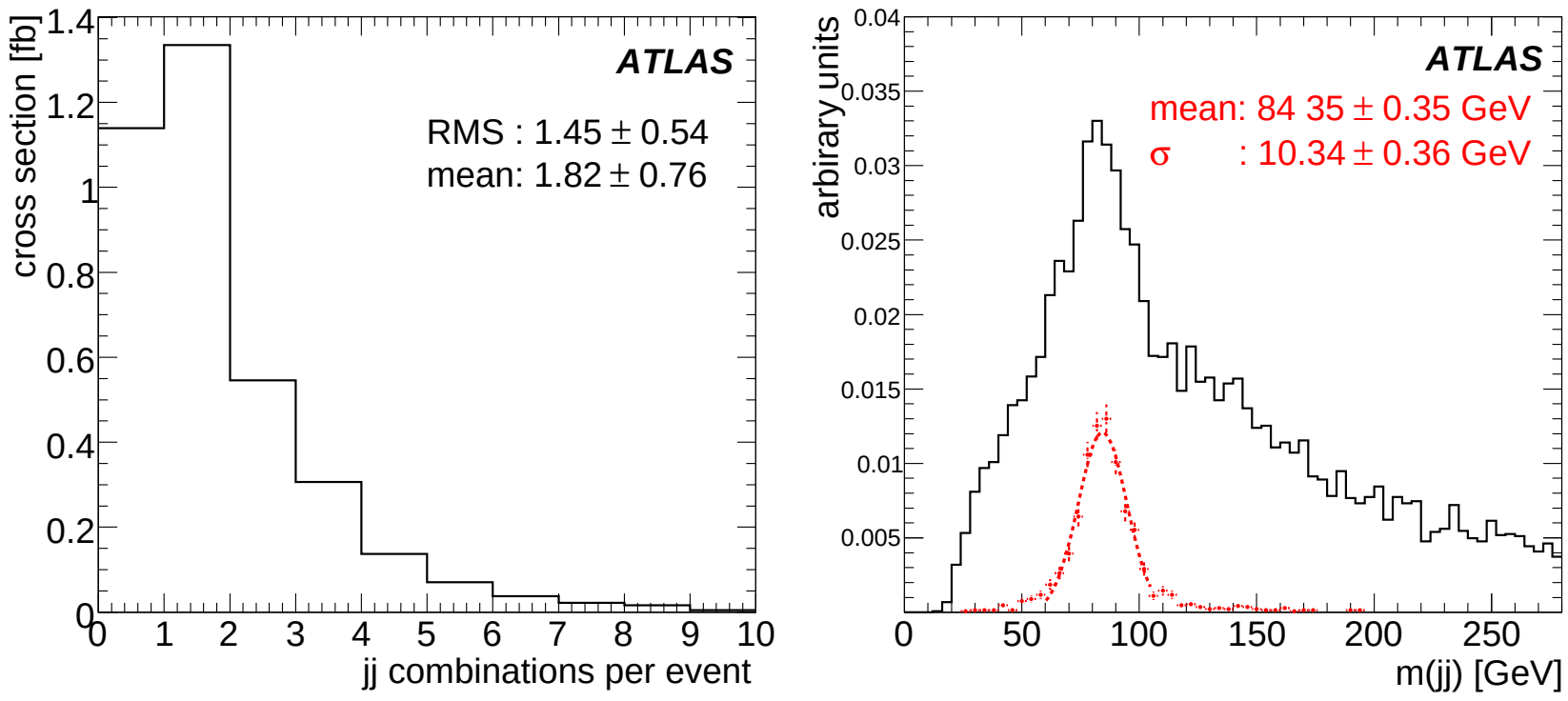


Figure 7: Cross-sections of hadronically decaying $W$ combinations per event giving candidates within 25 GeV of the true $W$ mass in the signal sample. (left). Right: Invariant mass spectrum for hadronically decaying $W$ candidates normalized to unity. The dotted line shows combinations where the jets from the $W$ are correctly matched.

# 7 Reconstruction of the leptonically decaying $W$ bosons

When reconstructing the leptonically decaying $W$, we use the lepton four-momentum as measured in the detector. The neutrino, transverse momentum can be inferred by measuring the imbalance of the transverse energy in the event. This measured quantity is referred to as missing transverse energy.

## 7.1 Neutrino $p_z$ estimation

Once the missing transverse energy is identified with $p_{T\nu}$, the invariant mass of the sum of the lepton and neutrino four-momenta can be constrained to the $W$ boson mass [21]. Because of the limited measurement resolution on the transverse missing energy, for a significant fraction of events the quadratic constraint equation does not have a real solution. In this case the "$\Delta = 0$ approximation" can be made by dropping the imaginary part of those solutions with complex roots. Another method (the "collinear approximation") assumes that the $W$ boson decay products are produced preferentially in the same direction (due to the large top quark mass boosting the $W$ boson). For the collinear approximation, one can assume that $p_{z\ell} = p_{z\nu}$.

Considering $t\bar{t}H$ events where one lepton is reconstructed, 72% of the time $p_{z\nu}$ has real solutions. In this case, both solutions are carried forward into the analyses, and the best performing final state solution is used. For these events the $p_{z\nu}$ resolution is 19.5 GeV. In the other 28% of cases where there are no real $p_{z\nu}$ solutions, the $\Delta = 0$ approximation is used and the $p_{z\nu}$ resolution is 40 GeV. This performs better than the collinear approximation where the $p_{z\nu}$ resolution is 54 GeV. The quality of the $W$-boson reconstruction can be seen in Fig. 8. For the events where there is no real solution for $p_{z\nu}$, the direction and the mass of the $W$ boson is better represented by the $\Delta = 0$ approximation.

Since the mass constraint is lost when using the $\Delta = 0$ approximation (the same would be true for the collinear approximation) there is an actual cut for the reconstructed $W$ boson mass. The cut-based and pairing likelihood analyses only consider $W$ candidates having a mass less than 140 GeV, while the constrained fit analysis does not make an explicit requirement on this but the poor $\chi^2$ will remove extreme cases.

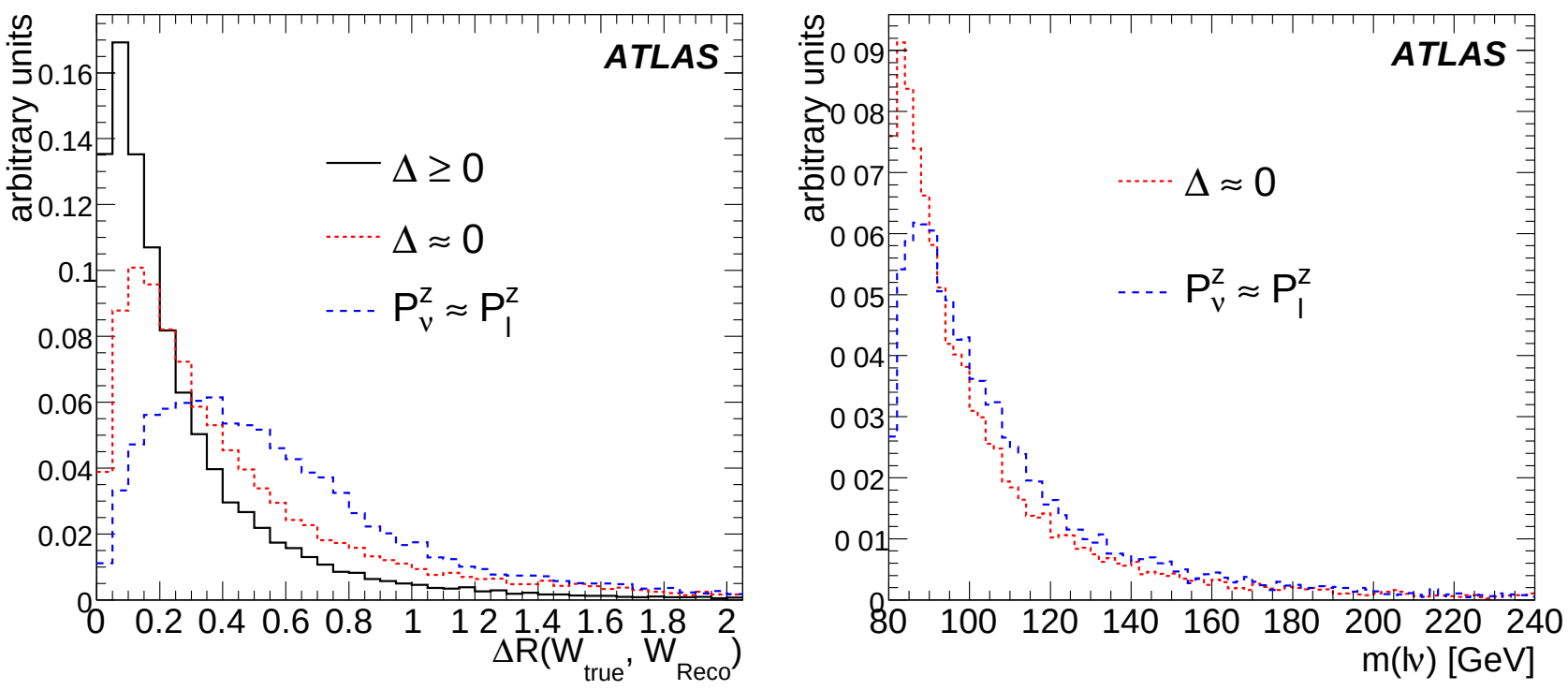


Figure 8: Distributions of $\Delta R$ between the true and the reconstructed $W$ boson (left) and of the reconstructed leptonically decaying $W$ mass (right) for events where a solution for $p_z$ is found (solid black line) and events where an approximation is used (dotted red for $\Delta = 0$, dashed blue for the collinear approximation). All distributions are normalized to unity.

# 8 Cut-based analysis

This section describes the algorithms used to reconstruct the $t\bar{t}$ system and the Higgs boson. The $b$-jets are associated with the leptonically and hadronically decaying $W$ boson candidates, to build a list of top quark candidates. The combination of $b$-jets resulting in the best reconstruction for the top quark candidates is taken as the final choice. It is important to note that the $b$-jets themselves satisfy the cut on $b$-tag weight $\geq 5.5$ and that if there are more than four of these jets, then the four with the highest weight are treated as $b$-jets. Events where there is no combination giving a satisfactory top quark mass reconstruction are discarded. The two remaining $b$-jets are used to form the Higgs boson candidate.

## 8.1 Top-antitop quark system and combinatorial background

In each event, top quarks are reconstructed by pairing two $b$-jets with the $W$ boson candidates in the way which minimizes the $\chi^2$ expressed as:

$$\chi^2 = \left(\frac{m_{jjb} - m_{top}}{\sigma_{m_{jjb}}}\right)^2 + \left(\frac{m_{l\nu b} - m_{top}}{\sigma_{m_{l\nu b}}}\right)^2, \qquad (1)$$

where $\sigma_{m_{jjb}}$ and $\sigma_{m_{l\nu b}}$ are the reconstructed mass resolutions estimated in simulated signal events and are 13 and 19 GeV respectively. Only combinations fulfilling $|m_{jjb} - m_{top}| < 25$ GeV and $|m_{l\nu b} - m_{top}| < 25$ GeV are considered for the $\chi^2$ calculation. The top quark mass distributions for the chosen combination in the signal sample is shown in Fig. 9. The two remaining $b$-tagged jets are used to form the Higgs boson candidates. The mass distribution for all Higgs boson candidates in the signal sample is shown in Fig. 10, and in the same plot the signal and physics background cross-sections are adjacent. Here the difficulty of the analysis is clearly shown, requiring dedicated studies to measure the background normalization and its shape in data.

As a final cut, to discriminate against $t\bar{t}$ events where no Higgs boson is produced, only events in a mass window of 30 GeV from the nominal Higgs boson mass are used for the final estimation of the cut-based analysis significance.

The effect of the final selection for the cut-based analysis on signal and background samples is shown in Table 4. The selections have reduced the signal by a factor of sixteen from the preselection, but the signal to background has increased from 0.006 to 0.11. The irreducible $t\bar{t}b\bar{b}$ background is 46% of the total.

Table 4: Accepted cross-section after each successive mass-window selection cut for signal and background in the cut-based analysis. In the last column the contribution of $t\bar{t}b\bar{b}$ is removed. Errors are statistical only.

| cut | $t\bar{t}H$(fb) | $t\bar{t}bb$(EW) (fb) | $t\bar{t}bb$(QCD) (fb) | $t\bar{t}$X (fb) |
|---|---|---|---|---|
| $W_{\text{had}} + W_{\text{lep}}$ | $2.49 \pm 0.05$ | $2.9 \pm 0.2$ | $18.2 \pm 0.7$ | $22.5 \pm 1.9$ |
| + $t\bar{t}$+Higgs | $2.04 \pm 0.05$ | $2.2 \pm 0.2$ | $14.7 \pm 0.6$ | $14.3 \pm 1.5$ |
| + Higgs boson mass window | $1.00 \pm 0.03$ | $0.52 \pm 0.07$ | $3.6 \pm 0.3$ | $4.9 \pm 0.9$ |

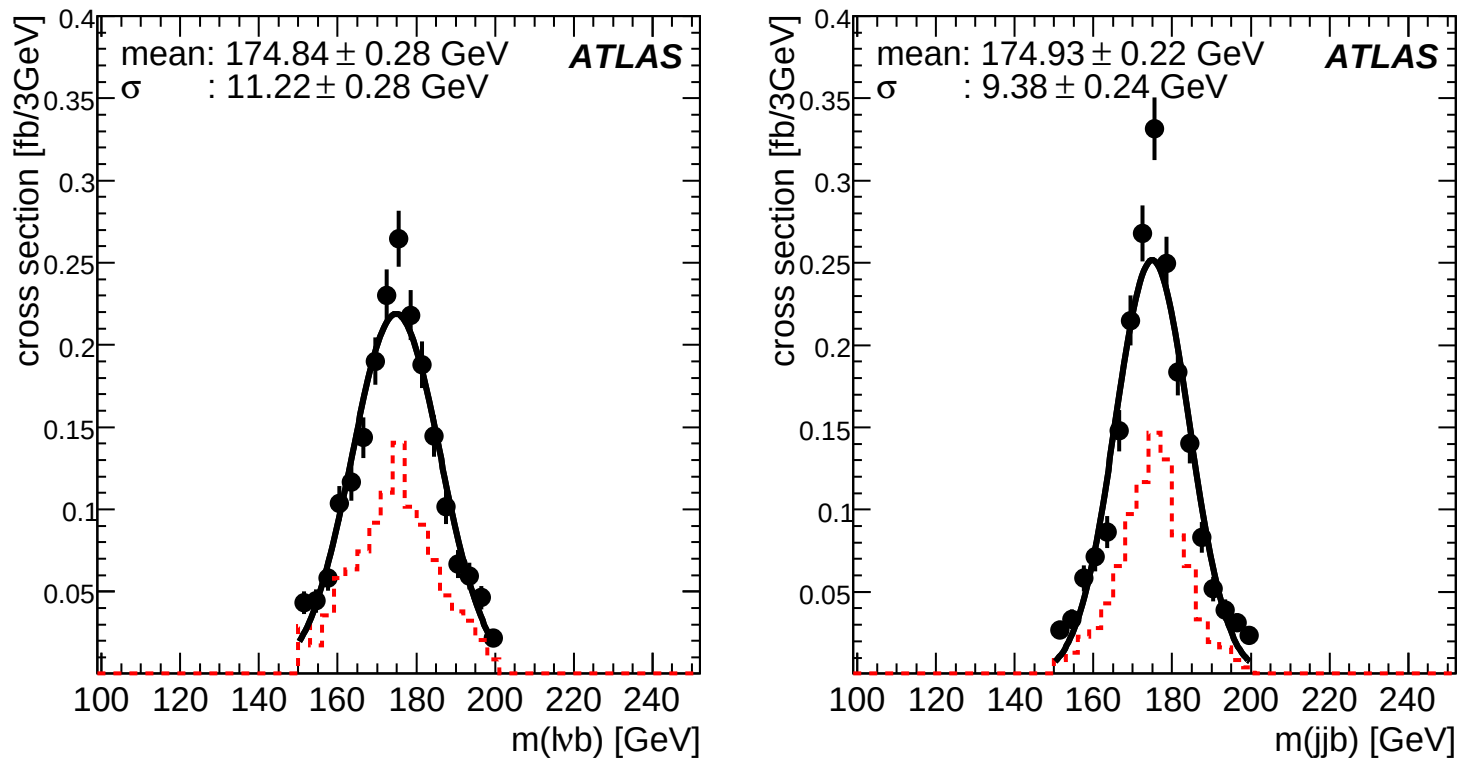


Figure 9: Reconstructed invariant mass spectrum for selected leptonic (left) and hadronic (right) top quark candidates in the signal sample. The dotted red line indicates the candidates formed by assigning the correct $b$-jet to the top quark being considered. All distributions are given in cross-section.

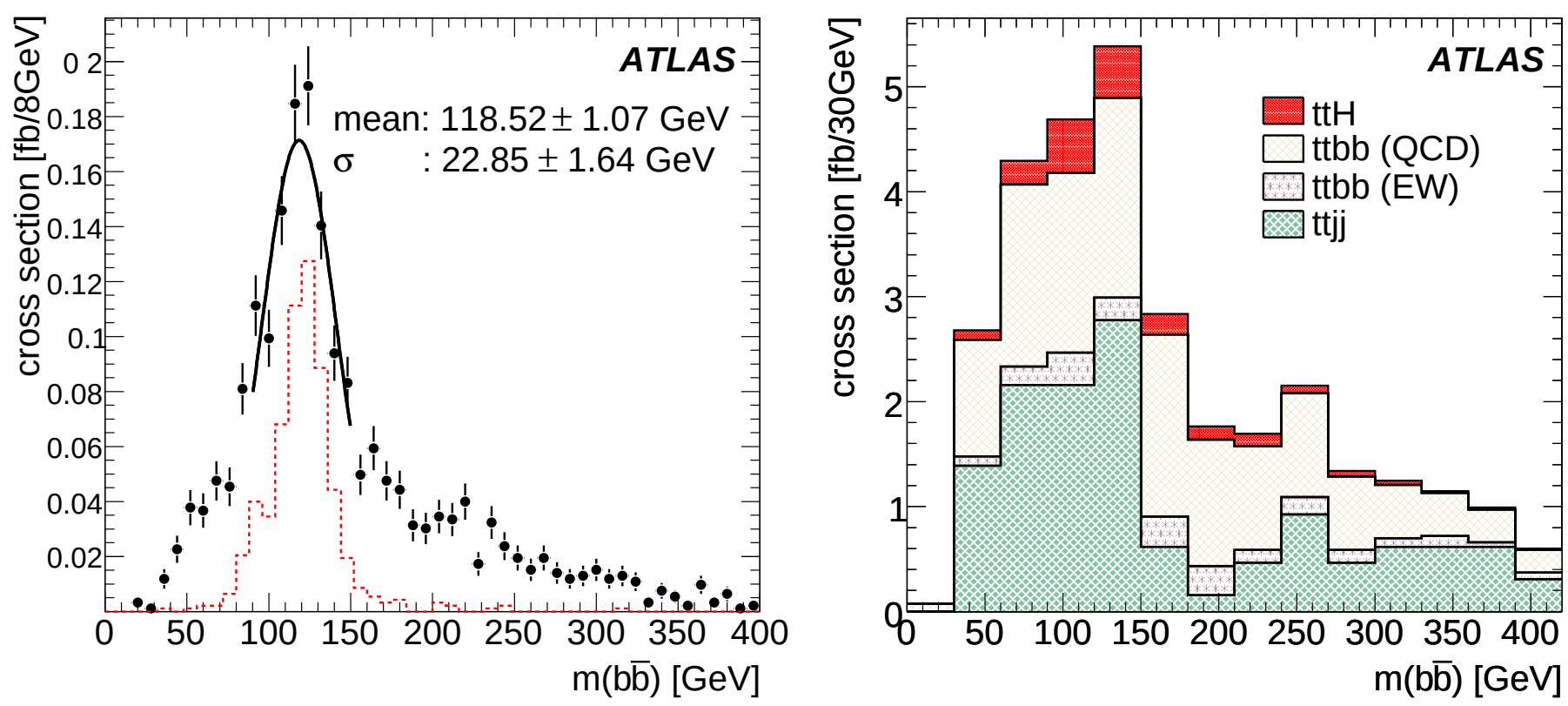


Figure 10: Left: Reconstructed invariant mass spectrum for Higgs boson candidates in the signal sample. The dotted red line indicates the candidates formed by assigning the correct $b$-jets. Right: Reconstructed invariant mass spectrum for signal and backgrounds after the cut-based selection. All distributions are given in cross-section.

## 9 Pairing likelihood analysis

In the previous Section we used a cut-based approach to identify the top quark decay products. A straightforward improvement to such an approach is to use several discriminating topological distributions combined together in order to build a pairing likelihood. As a first step the analysis considers only top quark properties as likelihood templates. Even though Higgs boson properties could help associating $b$-jets, if used, those could lead to bias in the background distributions. A correct combination is obtained when the objects used for the reconstructed variables match the Monte Carlo partons, regardless of whether other objects are correctly associated. On the other hand all the wrongly reconstructed objects are used to form the wrong combination templates. The variables used, shown in Fig. 11, are:

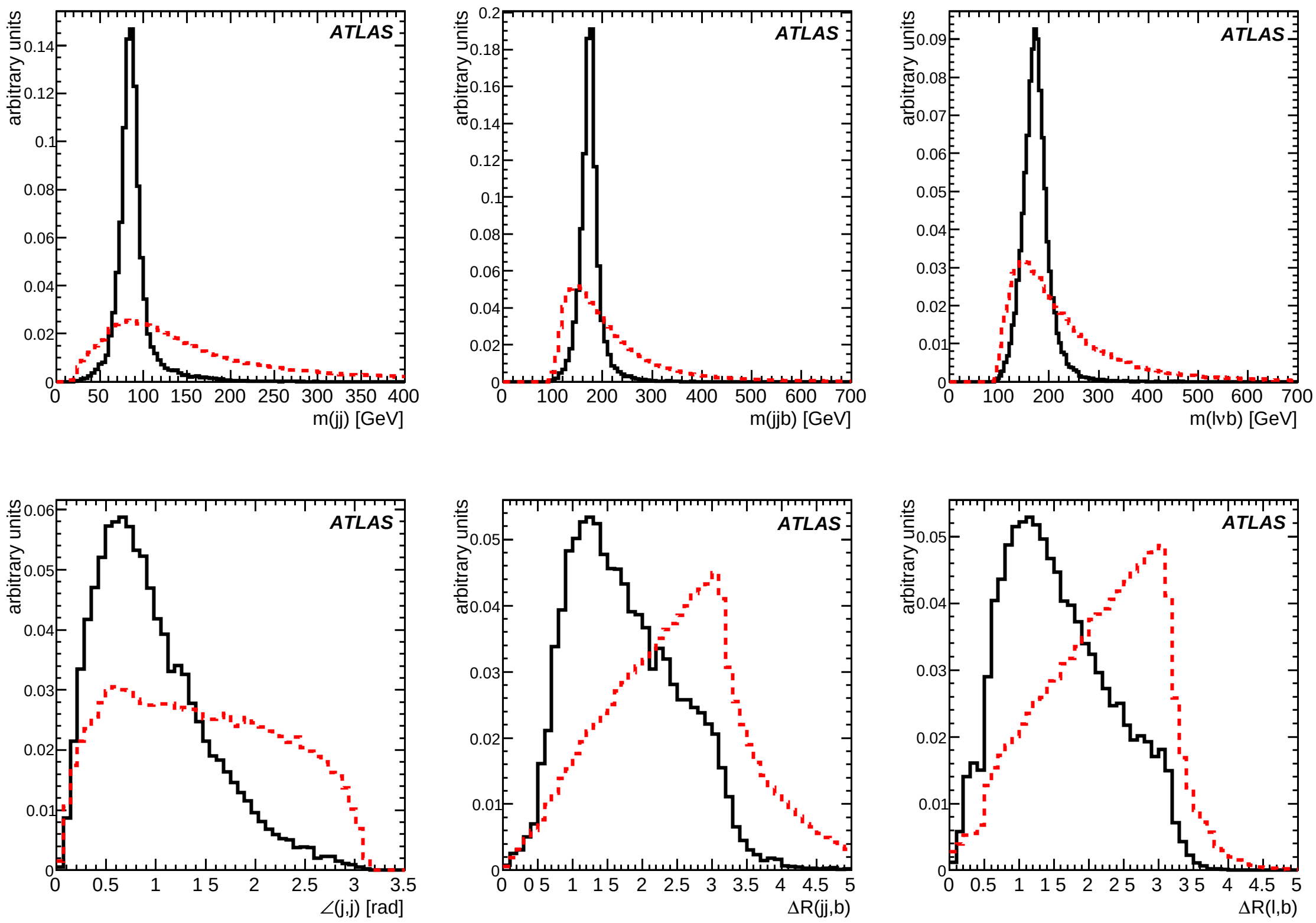


Figure 11: Pairing likelihood templates for top quark topological distributions, derived from the $t\bar{t}H$ signal sample. Solid lines represent the correct combination while the dotted lines show the combinatorial background in the signal itself. See text for a description of the variables.

- $m_{jj}$: The invariant mass of the light jets from the hadronic $W$ decay.
- $m_{jjb}$: The invariant mass of the hadronic top decay products.
- $m_{l\nu b}$: The invariant mass of the leptonic top decay products.
- $\angle(j, j)$: The angle between the light jets from the hadronic $W$ decay.
- $\Delta R(jj, b)$: The distance in $R$ between the hadronic $W$ and $b$ jet from the hadronic top decay.
- $\Delta R(l, b)$: The distance in $R$ between the lepton and the $b$ jet from the leptonic top decay.

The output of the pairing likelihood for the correct and wrong $b$-jet combinations is shown in Fig. 12. As shown in this plot, even though the correct distributions are peaked at 1, the wrong combinations still have a large probability of being selected. The only combination used is the one which maximizes the likelihood output. In order to avoid the presence of a large combinatorial contribution a cut on the likelihood output of 0.9 is used to select well-reconstructed events. After this cut, $b$-jets are associated to reconstruct the Higgs boson, as shown in Fig. 12, and the two top quarks, shown in Fig. 13. The invariant mass distribution for the selected Higgs boson candidates for signal and backgrounds is shown in Fig. 14.

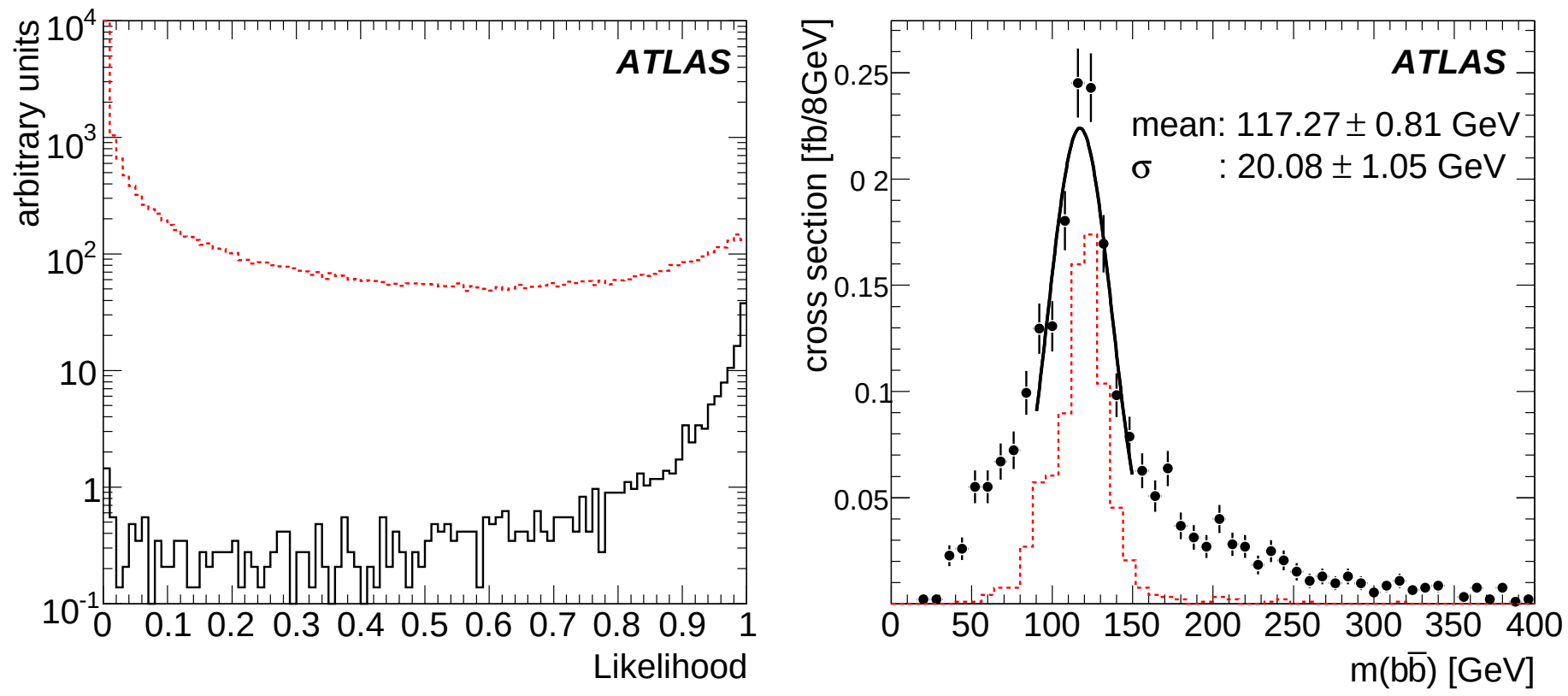


Figure 12: Left hand side: combinatorial likelihood output for $t\bar{t}H$ events. Black solid (red dotted) histogram indicates the correct (wrong) combinations. Right hand side: invariant mass for the Higgs boson candidates reconstructed using the maximum likelihood configuration, after applying a cut on the likelihood. Dotted histogram indicates the correct combinations. The differential cross-section is shown in fb.

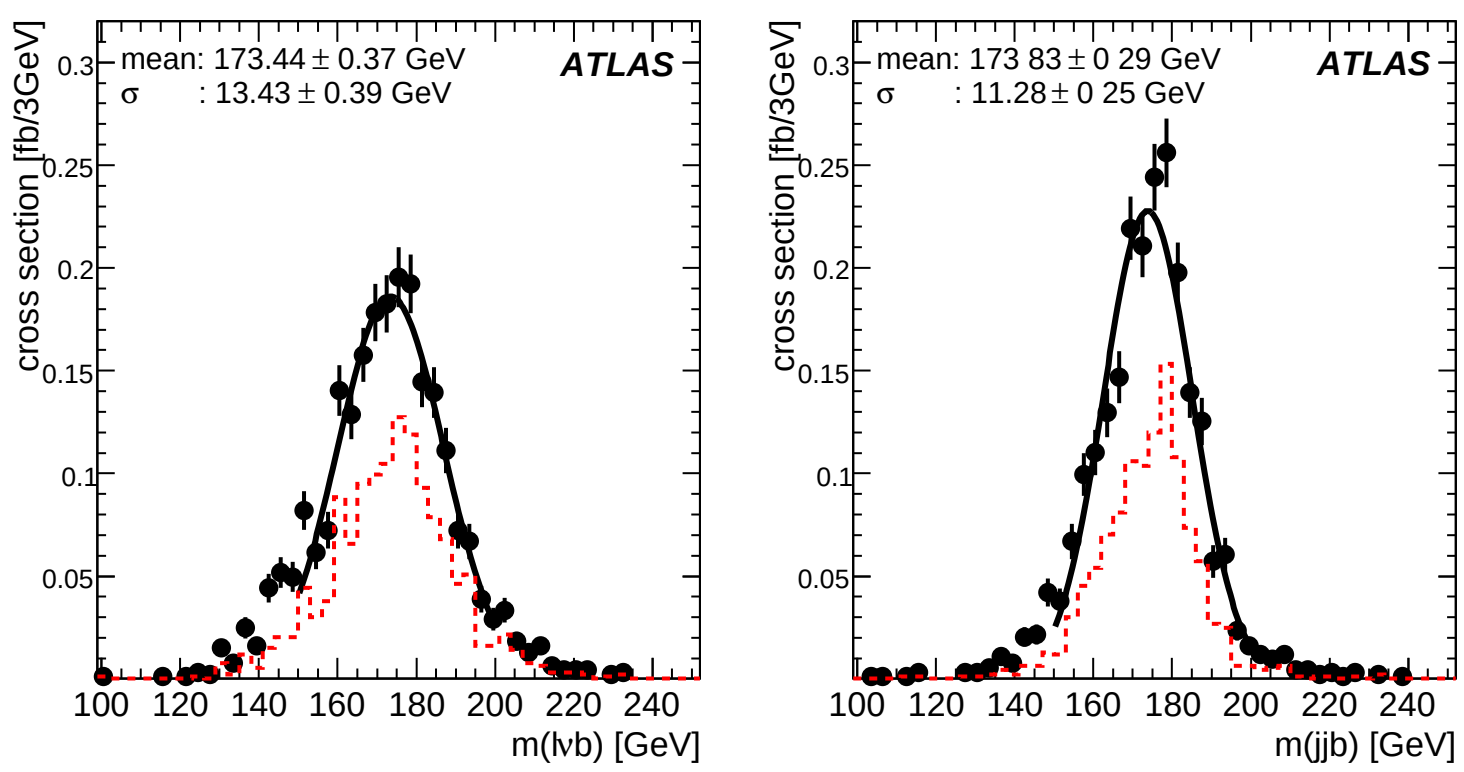


Figure 13: On the left (right) hand side is shown the leptonic (hadronic) top quark candidates reconstructed invariant mass using the maximum likelihood configuration, after applying a cut on the likelihood output. The dotted histogram indicates the correct $b$ quark jet for the top quark being considered. The differential cross-section is shown in fb.

A final cut on the reconstructed Higgs boson mass, requiring it to be within 30 GeV of the Higgs boson nominal mass is applied. The event yield for the whole analysis using the pairing likelihood is shown in Table 5. This analysis reduces the signal by a factor thirteen, and produces a sample which has a signal to background ratio of 0.1. The irreducible $t\bar{t}b\bar{b}$ background increases to 45% of the total.

Table 5: Cross-sections after each selection cut for signal and backgrounds for the pairing likelihood analysis. In the last column the contribution of $t\bar{t}b\bar{b}$ has been removed. Errors are statistical only.

| applied cuts | $t\bar{t}H$(fb) | $t\bar{t}b\bar{b}$(EW) (fb) | $t\bar{t}b\bar{b}$(QCD) (fb) | $t\bar{t}$X (fb) |
|---|---|---|---|---|
| Leptonic $W$ | $3.6 \pm 0.06$ | $4.1 \pm 0.2$ | $29 \pm 0.8$ | $48 \pm 2.7$ |
| + Best likelihood $> 0.9$ | $2.3 \pm 0.05$ | $2.5 \pm 0.2$ | $16 \pm 0.6$ | $19 \pm 1.7$ |
| + Higgs boson mass window | $1.2 \pm 0.04$ | $0.68 \pm 0.08$ | $4.6 \pm 0.3$ | $6.5 \pm 1.0$ |

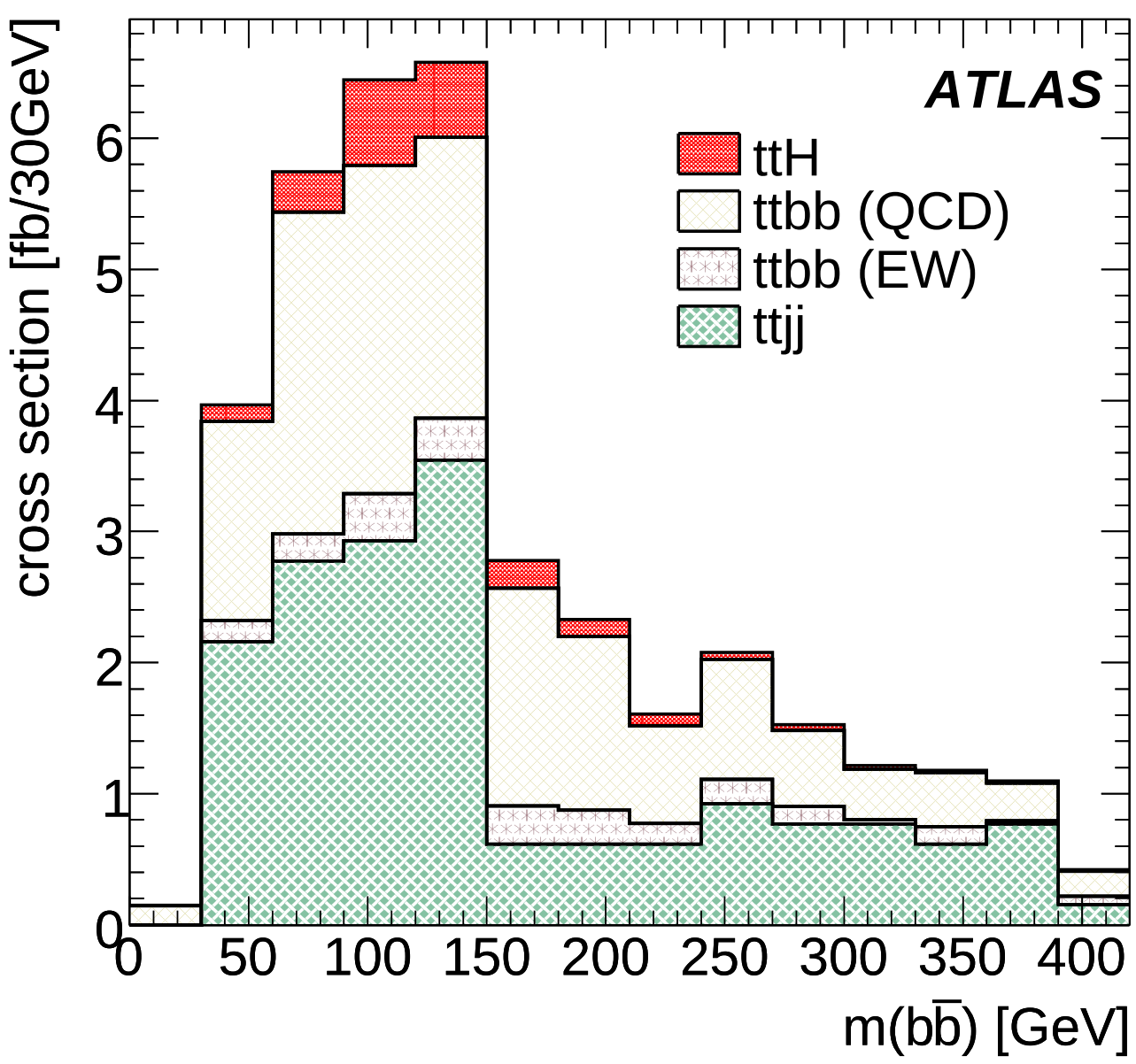


Figure 14: Reconstructed invariant mass spectrum for Higgs boson candidates for signal and backgrounds after pairing likelihood selection. The differential cross-section is shown in fb.

# 10 Constrained fit analysis

An alternative analysis uses a mass-constrained fit to the measured missing energy and jet and lepton four-momenta to help with the jet combinatorics. There are six quarks produced from the top quark and Higgs boson decays, and these are matched to the reconstructed jets. The $\chi^2$ from this fit is used in a likelihood technique together with kinematic variables, $b$-tagging and jet charge. Then, all jet combinations passing loose criteria are tested and the one with the best likelihood is chosen. The analysis starts from the preselection as described in Section 5. The signal events are then separated from background in a second likelihood step.

## 10.1 Mass-Constrained fit technique

The fit varies a scale factor for the four-momenta of the jets and the z component of the neutrino momentum. Adjustments to the jet momenta and masses through the scale parameters, $f^i$, produce accompanying changes in the missing energy and hence in the parameters used in the reconstruction of the leptonically decaying $W$ as the transverse components of its neutrino are taken to be the missing energy. The longitudinal component of the momentum of decaying $W$ boson's neutrino $p_{z\nu}$ is the last fit parameter. The parameters are constrained by the estimated jet errors and by the masses of the top quarks and $W$ bosons. These later are included as approximate Gaussian $\chi^2$ contributions calculated using the masses inferred from the current parameters as indicated in the following equation:

$$\chi^2 = \sum_{i=1}^{6} \left( \frac{f^i_{jet} - 1}{\sigma^i_{jet}/P^{i,initial}_{jet}} \right)^2 + \frac{(m_W^{lep} - 80.425)^2}{\sigma_W^2} + \frac{(m_t^{lep} - 175)^2}{\sigma_t^2} \tag{2}$$

where the $W$ and top quark widths $\sigma_W$ and $\sigma_t$ are 2.1 and 1.5 GeV respectively.

To simplify the fit, the hadronic top quark and $W$ are forced to be exactly on mass shell. The scale factor of the higher $p_T$ jet from the $W$ is externally varied, while the other two scale factors are calculated to give the correct masses. The momenta of all six jets are varied, but these three are linked. There are therefore five free parameters and not seven. This implies that these two particles are fixed to their nominal masses, while the leptonic top quark and $W$ are given widths.

The calculation of the momentum of the neutrino from the leptonic $W$ decay normally has two solutions, as discussed in Section 7. Both of these are used as starting points for the fit, to ensure that it does not find only one local minimum, and the fit with the larger $\chi^2$ is discarded. If the $W$ neutrino solutions had complex roots then the real part of these is used as an initial value.

The jet momenta are calibrated as discussed in Section 5.3.4, and the following errors are used in the fit:

$$\sigma_{P_{light}}/P_{light} = 0.988/\sqrt{p_T} \oplus 0.035 \tag{3}$$

$$\sigma_{P_b}/P_b = 0.888/\sqrt{p_T} \oplus 0.125 \tag{4}$$

where the momenta are measured in GeV. This form comes from comparing reconstructed jet $p_T$ with simulated initial quark $p_T$; in other words it includes not only detector effects but also fragmentation. Both light and $b$-jet momentum errors are treated as Gaussian distributed; for $b$ quark jets in particular this is not a good description as the frequent presence of a neutrino gives tails to the measured energy response.

The fit adjusts the momenta of all the jets, including those from the Higgs boson, but the fitted Higgs boson mass is not used in the analysis, as it offers no improvement.

### 10.2 *b*-tag information

In order to use the *b*-tagger output, as described in Section 5.3.2, for a likelihood, the distributions of weights expected for *b*-jets and non *b*-jets in this environment are required. This is done using jets taken from the signal simulation, where only jets with a parent quark within 0.4 in $\Delta R$ and no second quark within 0.6 in $\Delta R$ are used. The ratio of smoothed weight distributions with *b*-jet over light flavored jets $(u,d,s)$ is shown in Fig. 15.

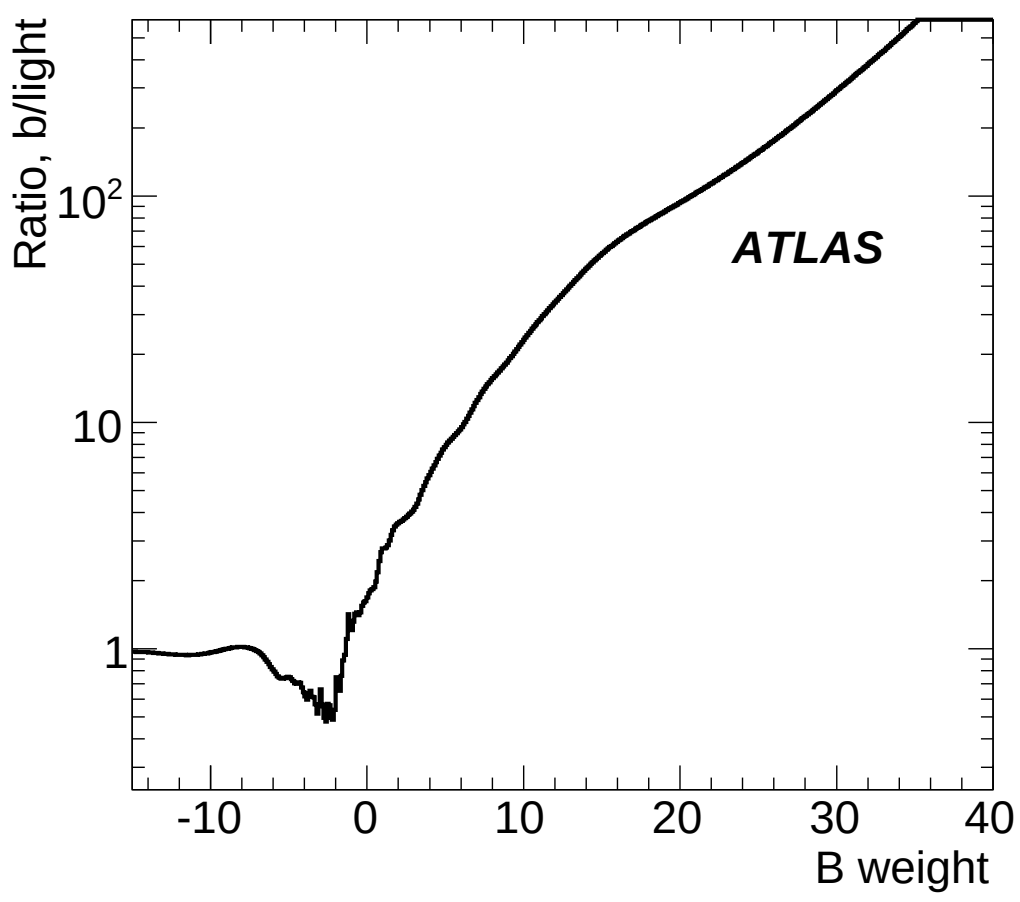


Figure 15: The *b*-tagging likelihood ratio extracted from signal simulation as the ratio of *b*-weight distributions for *b*-jets and light jets. The degree of smoothing reflects the statistical precision at each point.

In the analysis, four jets are taken to be from *b* quarks. For each jet we compute $\mathscr{L}_b^i$, the ratio of $b/light$ for jet $i$ as shown in Fig. 15. In selecting combinations the sum $\Sigma_i \log_{10} \mathscr{L}_b^i$ is taken over the four jets. There is no requirement made on the jets from the $W$ boson.

### 10.3 Jet charge

The assignment of jets to quarks can benefit from the jet charge measurement as we know the expected charges of the quarks involved. The jet charge is the momentum weighted sum of the charged particle charges within the jet, and it shows some correlation with the initial quark charge. The $\bar{b}(b)$ quark has only a charge of (-)1/3, and furthermore, after hadronisation there are oscillations which reduce the sensitivity, but there is some information.

The analysis requires exactly one high $p_T$ lepton which has charge $Q_l$, and therefore we know the expected charge of both of the $b$ quarks associated to the top quarks via the relationship $sign(Q_l) = sign(Qt_{lept}) = -1 \times sign(Qt_{had})$, and the sum of the charges of the jets from the hadronically decaying $W$ boson $QW_{had} = -1 \times Q_l$. The measured values of these are then compared with the expectations using a likelihood. The sum of the Higgs boson jet charges is much less sensitive because the expected value is zero, but it is also used.

The jet charge plots in Fig. 16 are calculated using jets from the signal sample. The $W$ plots are made using only true light jets which were tagged as light jets, and the $b$ plots only from tagged $b$-jets associated with a $b$ quark. This is to ensure that the jet charge is independent of $b$-tag information. The W wrong combinations distribution has spikes at integral values which generally involve at least one jet outside the tracker acceptance contributing a charge of zero. These are less frequent in the correct combinations which tend to be central. The other distributions reflect $b$-tagged jets which must therefore have charged

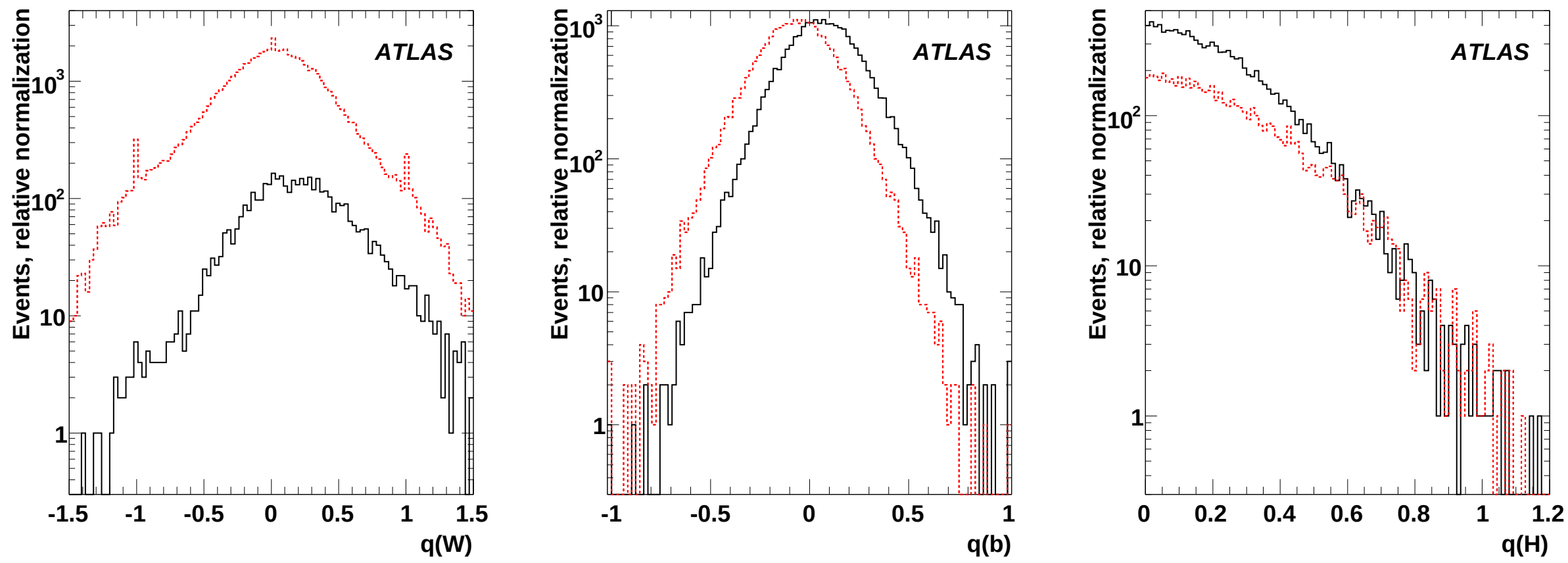


Figure 16: Left: The jet charge of $W$ boson candidates, based on the sum of the jet-charges of the two jets signed by the high-$p_T$ lepton. Center: Jet charges of individual $b$-jets from top quarks; signed so that those where quark charge agreeing with the lepton charge from $W$ decay are positive. Right: The magnitude of the sum of the jet charges of the jets assigned to the Higgs boson. Wrong $b$-jet combinations have a somewhat flatter distribution. Correct combinations are solid (black) and wrong combinations are dashed (red).

particle tracks. It is assumed that the jet charge information can be calibrated from the plentiful top quark pair events. The normalizations are arbitrary, as they offset every combination equally. The jet-charge is used as $\mathscr{L}_{jet-charge} = \mathscr{L}^{hadronic\ top}_{q(b)} \times \mathscr{L}^{leptonic\ top}_{q(b)} \times \mathscr{L}_{q(W)} \times \mathscr{L}_{q(H)}$.

### 10.4 Likelihood analysis for jet assignment

All possible assignments of jets to quarks are evaluated in turn. Those combinations which fail a loose quality requirement are discarded. This quality requirement is that:

- Both of the jets assigned to the Higgs boson must have a $b$-likelihood greater than zero. From Fig. 15 that corresponds to a cut of about -2 on the $b$-tagging variable.
- Selections on the mass of the $W$ and top quark which decay to jets. The $W$ mass calculated from the jets without fitting must lie between 30 and 150 GeV, and the top quark mass between 100 and 250 GeV. Note that the jet energy correction factor applied depends upon whether or not the jet is considered to originate from a $b$ quark in this hypothesis.
- Total $b$-likelihood greater than 8. This is the sum of the log-likelihoods of the four jets which are assigned to $b$-jets; this roughly corresponds to the mean $b$-weight of the jets being 4 or greater.

In the preselected signal sample there are a mean of 5811 combinations to be tested per event, but the above quality requirement reduces this to 233; a considerable saving in time. 90% of the preselected signal events have at least one combination passing the above requirements.

Events which pass these selections are processed by the $t\bar{t}H$ fitting code. Correct or wrong combinations are then used to define a likelihood ratio.

The elements of that likelihood ratio are as follows:

- The $\log_{10}$ of the $\chi^2$ from the fit.

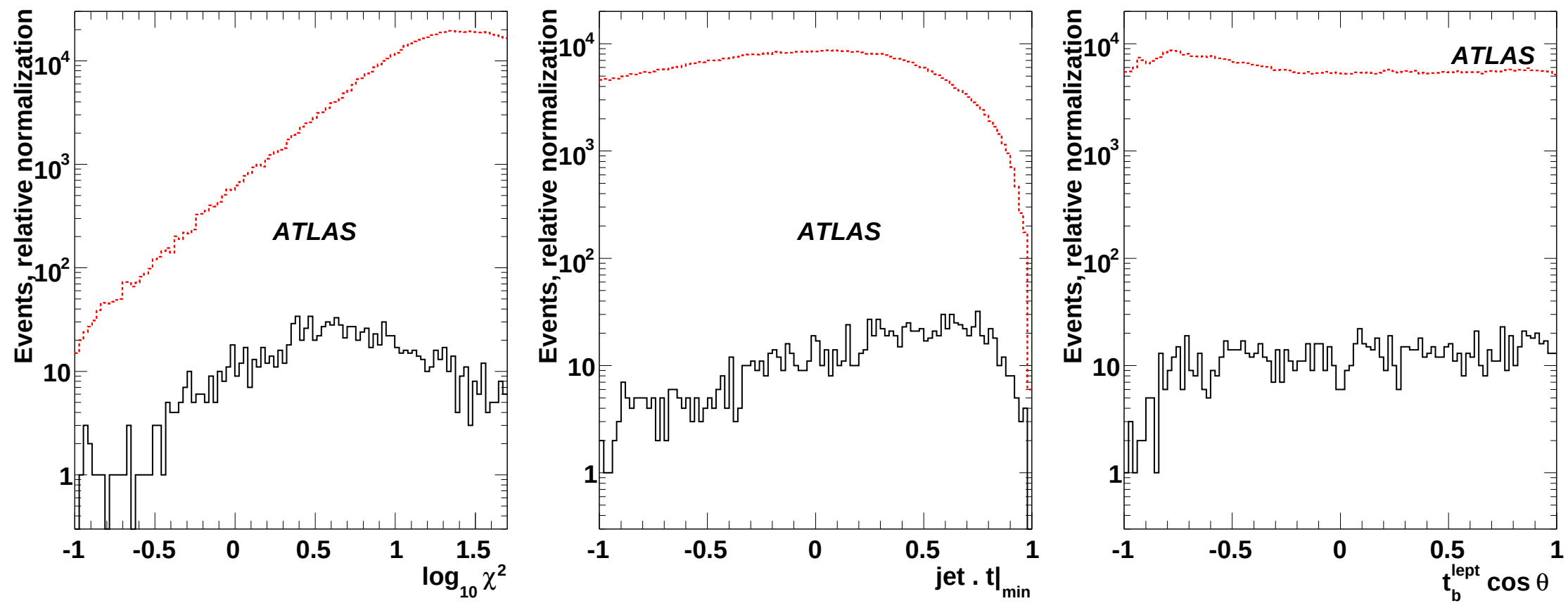


Figure 17: Left: The $\chi^2$ of the fit. Center: jet $\cdot$ top$|_{\text{min}}$. The cosine of the minimum angle between top quarks and their daughter jets. Correct combinations have more collimated tops. Right: The $\mathrm{t}_\mathrm{b}^{\mathrm{lept}} \cos\theta^*$. The decay angle in the rest frame of the leptonically decaying top quark of the $b$-jet relative to the top quark direction in the lab frame. Correct combinations are solid (black) and wrong combinations are dashed (red).

- jet $\cdot$ top$|_{\text{min}}$: The minimum cosine of the angle between the top quarks and any of their four jets in the $t\bar{t}H$ center of mass frame.
- $\mathrm{t}_\mathrm{b}^{\mathrm{lept}} \cos\theta^*$: The decay angle in the rest frame of the leptonically decaying top quark of the $b$-jet relative to the top quark direction in the lab frame.
- $\mathrm{t}_\mathrm{b}^{\mathrm{lept}} \Delta\mathrm{R}$: the distance in $\Delta R$ space between the lepton and the $b$ quark assigned to the same top quark.
- $|\eta|^{\max}$: The maximum $|\eta|$ of the considered jets. Jets from the $t\bar{t}H$ system tend to be more central than those from the underlying event.
- $\cos\theta^*_{\mathrm{H-jet}}$: Measured in the Higgs boson rest frame, this is the cosine of the angle between the higher $p_T$ of the two jets from the Higgs boson and the boost applied to shift from the lab frame to the Higgs boson rest frame.
- $m_t$: The hadronic top quark mass before the fit is performed.

The variables are displayed in Figs. 17 and 18 for correct and wrong combinations. In this case 'correct' implies that all six quarks are correctly assigned. Fig. 19 shows how each variable would perform if used individually to separate correct and incorrect pairings. For each variable combinations are selected by the likelihood ratio found using that variable alone. The Fig. shows the fraction of wrong pairings which would be accepted as a function of the fraction of correct ones. The fit $\chi^2$ is the most powerful single variable over much of Fig. 19, but the masses of the hadronic top quark and $W$ work well at high efficiency while jet $\cdot$ top$|_{\text{min}}$ is also rather powerful. Clearly the variables have correlations, and these are taken into account by evaluating the likelihood in a 3D space defined by the three variables under study. This explicitly includes the correlations, but is limited by the simulation statistics required to populate the space. The combination of $\chi^2$, jet $\cdot$ top$|_{\text{min}}$ and $\mathrm{t}_\mathrm{b}^{\mathrm{lept}} \cos\theta^*$ was adopted as the most powerful set of three variables found.

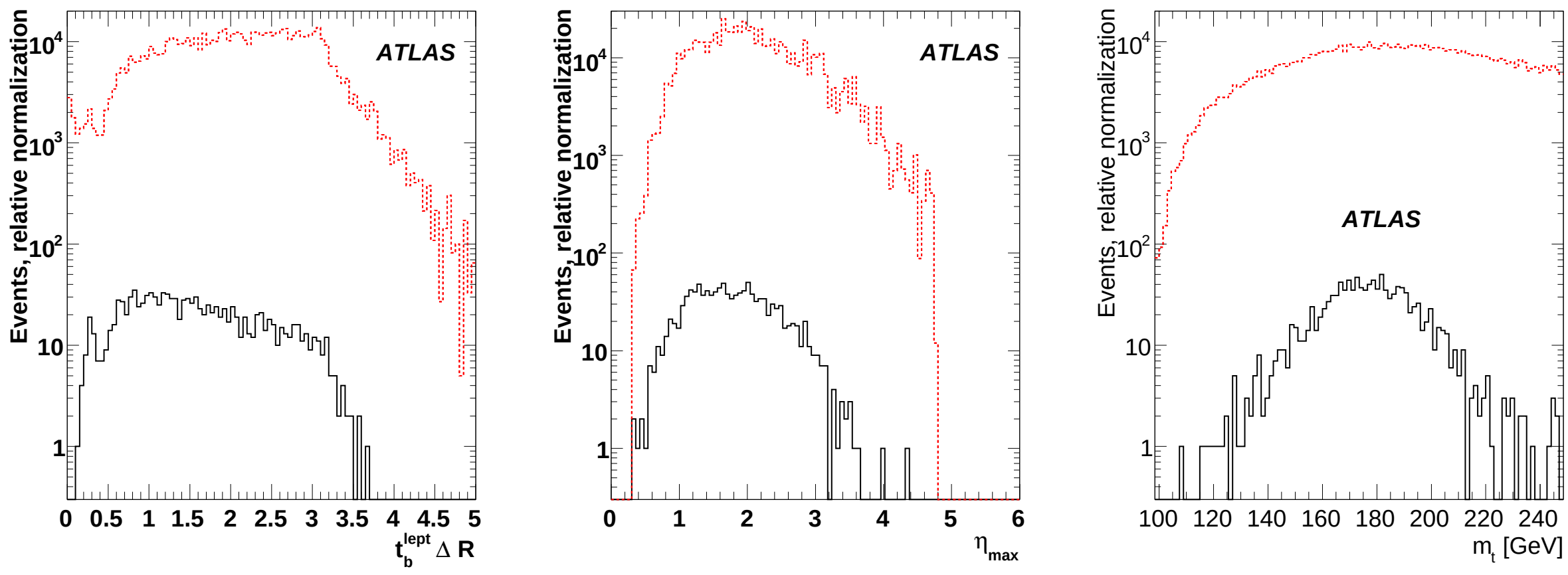


Figure 18: Left: The distribution of distances, $\Delta R$, between the lepton and the $b$ quark from the leptonic top quark. Center: The maximum $|\eta|$ of any jet in the combination being tested. Correct combinations are more central. Right: The reconstructed mass of the hadronically decaying top quark, before any fit is performed. Correct combinations are solid (black) and wrong combinations are dashed (red).

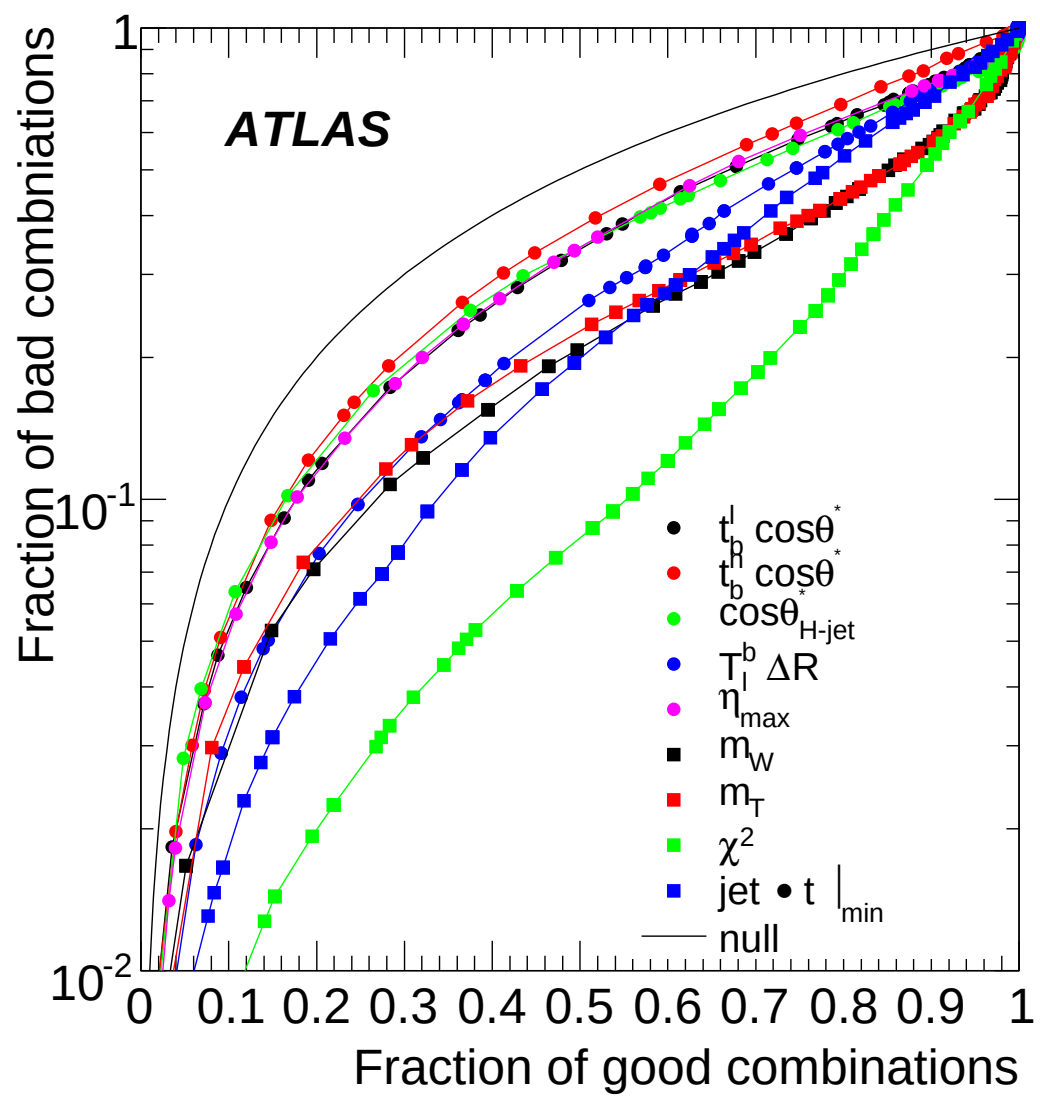


Figure 19: The performance of a range of possible variables if they are used individually to find the correct quark-jet pairing in a $t\bar{t}H$ event. For a given efficiency for selecting the correct pairing (x axis), what fraction of the incorrect pairings will also be chosen (y axis). The line labelled null shows the effect of selecting combinations at random.

The remaining variables were tested to see whether their addition as uncorrelated likelihood contributions made a significant improvement, and those that did were included. The final likelihood is defined as follows:

$$\mathscr{L}_{pairing} = \log \mathscr{L}^{3D}_{\log\chi^2,\mathrm{jet}\cdot\mathrm{top}|_{\mathrm{min}},\mathrm{t_b^{lept}}\cos\theta^*} + \log \mathscr{L}_{\mathrm{t_b^{lept}}\Delta\mathrm{R}} + \log \mathscr{L}_{|\eta|^{\mathrm{max}}} + \log \mathscr{L}_{\cos\theta^*_{\mathrm{H-jet}}} + \log \mathscr{L}_{m_t} + \log \mathscr{L}_{b-tag} + \log \mathscr{L}_{jet-charge} \quad (5)$$

The combination which produces the largest likelihood for each event is adopted. The quality of the chosen combination is examined in Section 11.

### 10.5 Signal and background separation

The separation of signal from background is again done using the likelihood technique. There are two rather different backgrounds considered: the $t\bar{t}j\bar{j}$ component for which $b$-tagging is the primary tool and the 'irreducible' $t\bar{t}b\bar{b}$ background which differs from the signal only in kinematic ways, which can be exploited to give some separation. The variables used to separate signal and background are:

- $\mathscr{L}_{pairing}$: From combinatorics.
- $\Sigma_i \log_{10} \mathscr{L}_b^i$: Sum of the log-likelihoods of the four jets used as b's in the combination chosen.
- $\Sigma^H_{b-tag}$: The sum of the $b$-tagging weight of the two jets from the Higgs boson.
- $\Delta\eta(\mathrm{H,top})^{\mathrm{min}}$: The difference in $\eta$ between the Higgs boson and the closer top quark.
- $\cos(tH)^{max}$: The higher of the two angles between top quarks and Higgs boson in the center of mass of the $t\bar{t}H$ system.
- $H_p$ in C.o.M: The Higgs boson momentum in the center of mass frame of the $t\bar{t}H$ system.
- $\cos\theta^*_{\mathrm{H-jet}}$: As defined in Section 10.4.

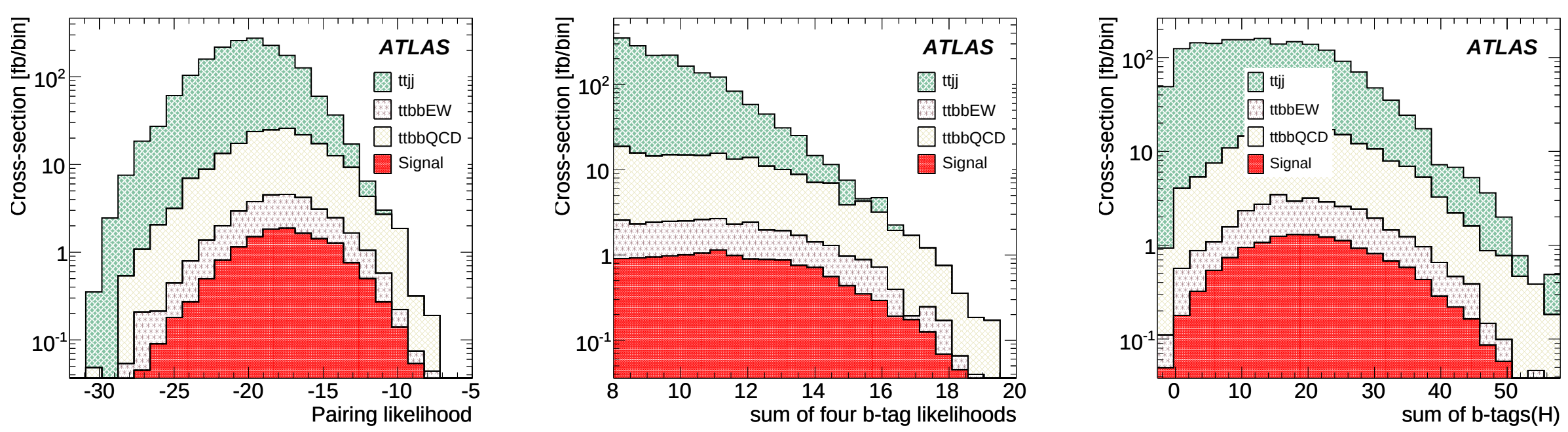


Figure 20: Left: Distributions for signal and backgrounds of the likelihood used to find the combinatorics solution. Center: The sum of the $b$-tag likelihoods of the four jets used as $b$'s in the chosen combination. There was cut at 8 in the combinatorics preselection. Right: The sum of the $b$-tags of the jets associated to the Higgs boson. This variable removes $t\bar{t}j\bar{j}$ more than $t\bar{t}b\bar{b}$. Histograms are filled for every event where there is a successful fit. In all histograms, the sum of the individual histograms is shown. They are stacked to indicate relative contributions.

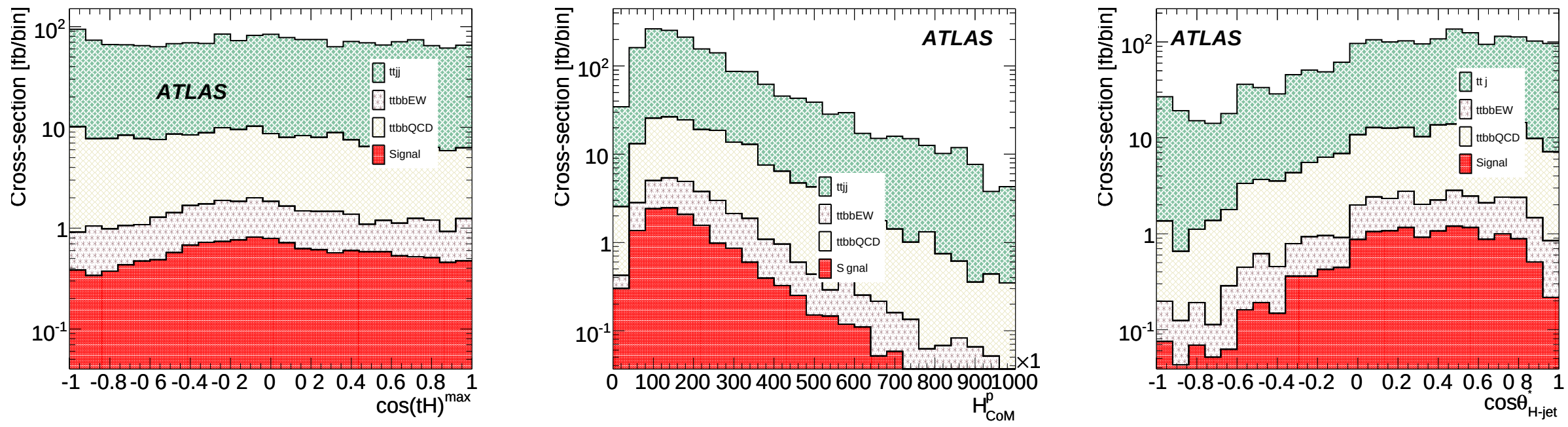


Figure 21: Left: The maximum cosine of the angle between either top quark and the Higgs boson when boosted into the $t\bar{t}H$ center of mass frame. Center: The momentum of the Higgs boson candidate in the center of mass frame. Note that the true Higgs bosons have a lower momentum than the background. Right: The $\cos\theta^*$ of the higher $p_T$ of the jets from the Higgs boson. This tends to be more central for signal events. In all histograms, the sum of the individual histograms is shown. They are stacked to indicate relative contributions.

Figures 20 and 21 show the data used to produce the likelihood ratio for most of the variables. The backgrounds fall into two basic classes - the $t\bar{t}j\bar{j}$ and the $t\bar{t}b\bar{b}$, and to deal with these two 3D likelihoods are defined. The first includes the three quantities which contain $b$-tagging information: $\mathscr{L}_{pairing}, \Sigma_i \log_{10} \mathscr{L}^i_b, \Sigma^H_{b-tag}$. These are all powerful but are highly correlated and therefore benefit from a correct treatment of those correlations. The second is $\cos(tH)^{max}, H_p, \cos\theta^*_{\mathrm{H-jet}}$, which carries discrimination based on event kinematics. It too has important correlations. These likelihoods are combined as if independent, with one further likelihood, derived from $\Delta\eta(\mathrm{H,top})^{\mathrm{min}}$, added as well. All the likelihoods have been smoothed so that the expected fluctuations are below 10%, and there is therefore little over-training, as separate test and training samples are maintained.

The distributions used to define the likelihood are constructed using all events for which a constrained fit was made, and all the components are normalized to the cross-sections at that stage. The signal separation likelihood is:

$$\mathscr{L}_{s/b} = 1/3\left(\log \mathscr{L}^{3D}_{\mathscr{L}_{pairing},\Sigma_i \log_{10}\mathscr{L}^i_b,\Sigma^h_{b-tag}} + \log \mathscr{L}_{\Delta\eta(\mathrm{H,top})^{\mathrm{min}}} + \log \mathscr{L}^{3D}_{\cos(tH)^{max},H_p,\cos\theta^*_{H-jet}}\right) \quad (6)$$

The factor of 3 makes this an average, rather than a sum, and is there purely for convenience. Note that there is nothing in this definition to prefer correctly paired signal events.

Figure 22 shows the distribution of the final likelihood. It is generally dominated by $t\bar{t}j\bar{j}$ events, but at the largest likelihood values the $t\bar{t}b\bar{b}$ and signal events are more prevalent. It can be seen that any $t\bar{t}H$ analysis will be selecting a tail of the signal, and controlling this will be important. The signal to background ratio, within the mass window, rises to about 25%, and any rise above that is in a region affected by lack of simulation statistics.

The final choice of working point will depend upon the details of the systematic error evaluation. The tighter the selection on the likelihood the higher the signal to background ratio but the smaller the samples in data and simulation; the latter is an important consideration.

The maximum significance which might be expected in a measurement (evaluated as s/$\sqrt{b}$) ignoring all systematic uncertainties, is obtained by cutting at a log likelihood of -4.44. This would yield a significance of 2.78$\sigma$, but at low purity. Reducing the mass range by requiring that the candidate has a mass within the range 90 to 150 GeV does not appear to improve the results when systematic errors are not considered.

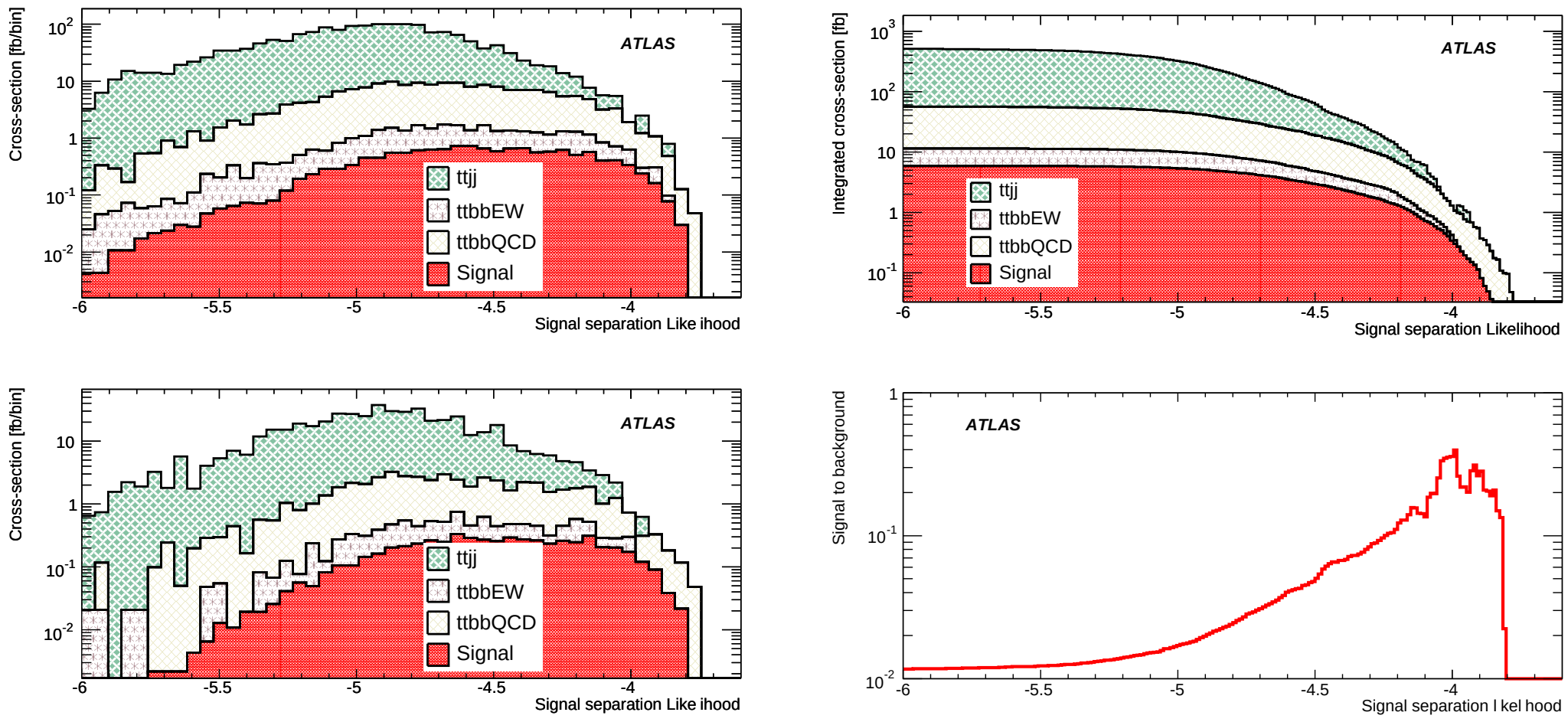


Figure 22: Left: The total signal separation likelihood. The top figure shows all events while the bottom shows only those within a Higgs boson candidate mass window of 90 to 150 GeV. Right: The integrated version of the lower left plot, so the total event rates passing any cut can be seen. The bottom half of this plot is the signal to background ratio implied.

If an arbitrary ten per cent error on the background level is assumed then the significance for this cut, evaluated as s/$\sqrt{b+(\delta b)^2}$, decreases to below 0.5$\sigma$, while the highest significance is around 1.8$\sigma$ for a cut at -4.05. This is shown numerically in Table 6, where the expected event rates are shown for three different cut values. No final choice is really possible without complete evaluation of systematic errors, but -4.2 with the mass window cut applied does seem to be a plausible working point. The statistical significance for this selection is 2.18$\sigma$. At this point the signal is reduced by a factor of twelve from the preselection, but signal to background ratio has become 0.125±0.01. The irreducible $t\bar{t}b\bar{b}$ background has increased to 50% of the total.

Table 6: The accepted cross-sections for signal and the main backgrounds at various stages of the analysis. The $t\bar{t}j\bar{j}$ cross-section suffers from limited statistics.

| Selection | $t\bar{t}H$(fb) | $t\bar{t}b\bar{b}$(EW) (fb) | $t\bar{t}b\bar{b}$(QCD) (fb) | $t\bar{t}$X (fb) |
|---|---|---|---|---|
| Initial Sample | 100 | 255 | 2371 | 109487 |
| Pass preselection | 16 | 23 | 198 | 2589 |
| Fit quality requirements | 14 | 20 | 165 | 1584 |
| $\mathscr{L}_{s/b}$ > -4.40 | 4.9 | 5.1 | 35 | 58 |
| $\mathscr{L}_{s/b}$ > -4.20 | 2.5 | 2.3 | 13.9 | 11.9 |
| $\mathscr{L}_{s/b}$ > -4.10 | 1.4 | 0.96 | 7.11 | 4.5 |
| Mass window 90 to 150 GeV. | | | | |
| $\mathscr{L}_{s/b}$ > -4.40 | 2.3±0.07 | 1.4±0.17 | 10.8±0.7 | 22±3.1 |
| $\mathscr{L}_{s/b}$ > -4.20 | 1.3±0.05 | 0.62±0.12 | 4.6±0.5 | 5.3±1.5 |
| $\mathscr{L}_{s/b}$ > -4.10 | 0.71±0.04 | 0.23±0.07 | 2.5±0.35 | 2.2±1.0 |

The distribution of the masses of the candidates can be seen in Fig. 23, at a cut of $\mathscr{L}_{s/b} > -4.2$. The right hand side of Fig. 23 shows details of the mass distribution for signal only.

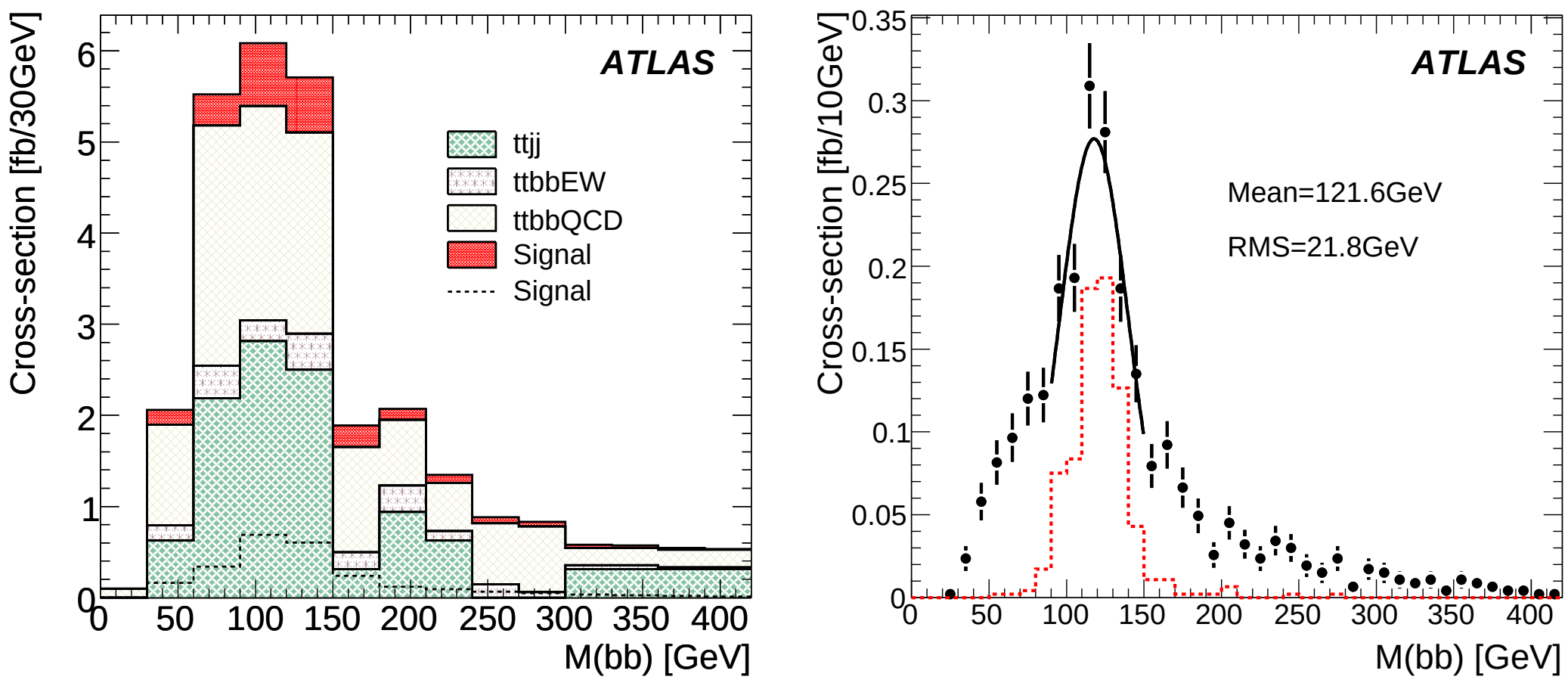


Figure 23: The mass distribution after cutting at -4.2 in $\mathscr{L}_{s/b}$. Left: All samples, showing the contributions stacked. The signal distribution is also shown separately at the bottom. Right: Signal only, the dashed (red) line equals events where the correct jets from the Higgs boson are selected.

## 11 Comparison between the three analysis techniques

The performance of the cut-based, pairing likelihood and the constrained fit analyses in terms of purity versus selection efficiency can be seen in Fig. 24. For the likelihood analyses, the different working points are obtained by varying the final cut on the likelihood discriminant. In the case of the cut-based analysis, the same variation is achieved by loosening or tightening the mass-window cuts on the hadronically decaying $W$ and the reconstructed top quarks. For this section the efficiency is defined as the selection efficiency relative to the total events simulated. The purity is defined in terms of the correctness of the assignment of $b$-jets used to reconstruct the final objects. For instance, one has a *pure* hadronic top quark when the $b$-jet matches the true $b$ parton from the top quark decay, regardless of whether the same happens for the hadronic $W$ boson decay products.

Fig. 24 clearly shows the increase of performance when using more information (likelihood) than just the mass of the reconstructed particles (cut-based).

The chosen working points are indicated with solid markers on Fig. 24. Those points have not been optimized in terms of statistical significance, because of a lack of statistics for the $t\bar{t}$X background and because due consideration of the systematic errors should also influence the decision. However, the significance does not change much with the choice of the cut on the pairing likelihood output since this likelihood is not designed to discriminate signal from physics background events.

The ability of the three analyses to correctly identify objects in the event is compared in Table 7. The likelihood-based assignments perform noticeably better than the cut-based analysis. The signal efficiency and statistical significance are also improved.

## 12 Background Shapes

The success of this analysis relies on the accurate knowledge of the background level and shape. Monte Carlo predictions are affected by large systematic uncertainties, as the background rejection depends critically also on the jet flavor composition. For this reason it is mandatory to develop methods to measure background directly from real data.

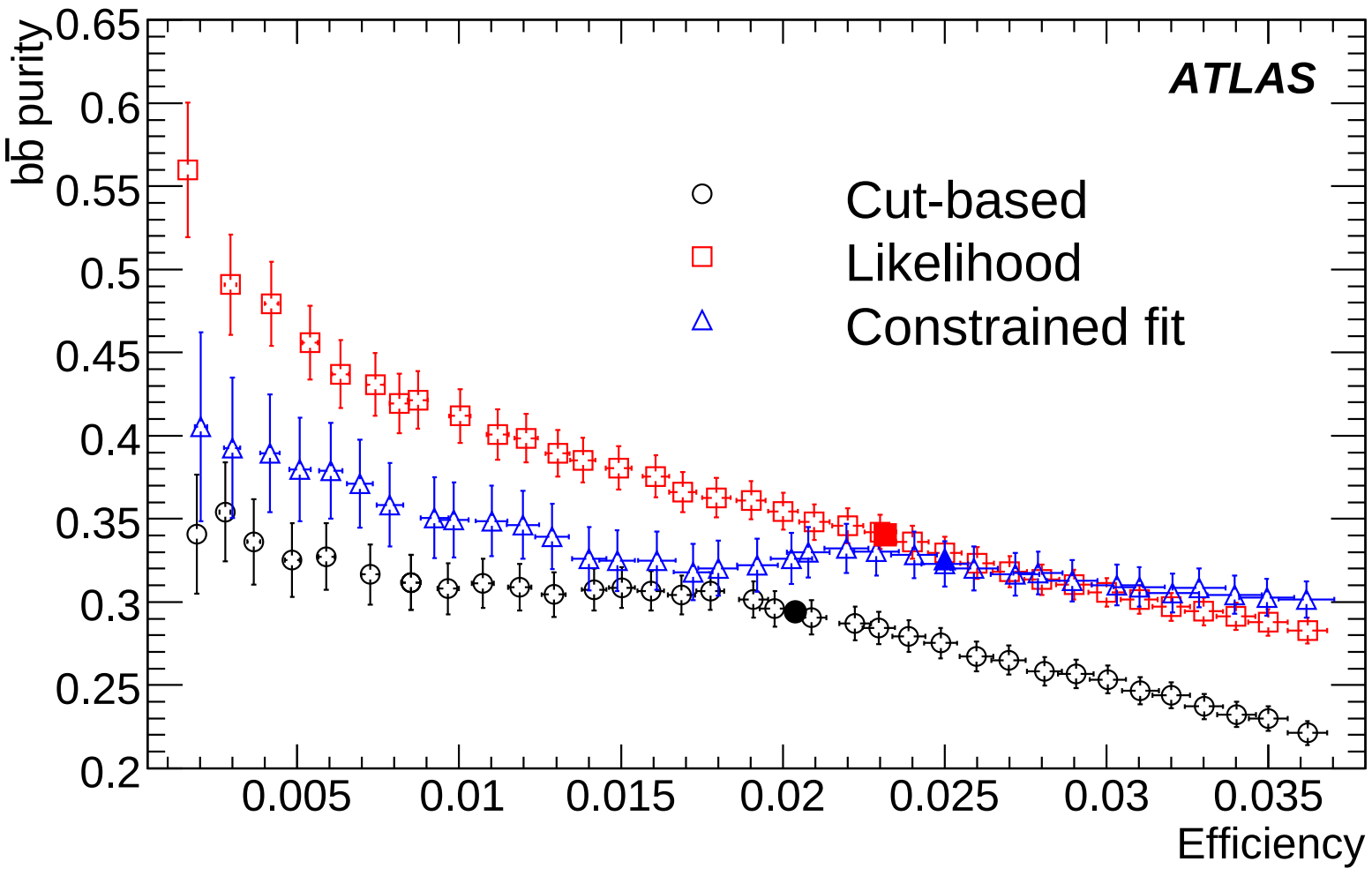


Figure 24: Comparison of the purity of reconstructed $b\bar{b}$ invariant mass before the final mass window cut versus selection efficiency for the cut-based, the pairing likelihood and the constrained fit analysis. The solid markers show the selected working points.

Table 7: A comparison of quality criteria for the three analyses at their working points. The mass window of 90 to 150 GeV is only applied for the last two rows.

| | Cut Based | Pairing likelihood | Constrained fit |
|---|---|---|---|
| $b$ jet from Hadronic top correct | 44.4±1.1% | 49.2±1.1% | 51.0±1.5% |
| $b$ jet from Leptonic top correct | 50.5±1.2% | 57.4±1.1% | 56.2±1.5% |
| Higgs boson jets correctly chosen | 29.4±1.0% | 34.0±1.0% | 32.0±1.4% |
| Four $b$ quarks correct | 23.3±1.0% | 27.5±1.0% | 27.1±1.3% |
| Higgs boson mass peak resolution, GeV | 22.8±1.6 | 20.1±1.1 | 22.3±2.1 |
| Signal Efficiency | 2.04±0.05% | 2.32±0.05% | 2.49±0.07% |
| Signal to background | 0.110±0.014 | 0.103±0.014 | 0.123±0.019 |
| $s/\sqrt{b}$, 30fb$^{-1}$ | 1.82 | 1.95 | 2.18 |

One important result of the present study is that the Higgs boson candidate mass spectrum depends weakly upon the $b$-tagging working point. This is shown in Fig. 25, which reports the difference in $b\bar{b}$ invariant mass shape for the $t\bar{t}b\bar{b}$ and $t\bar{t}$+jets processes after applying the pairing likelihood analysis with the loose and tight $b$-tagging requirement as defined in Section 5.3.2

The complete determination of the background shape from data depends crucially on the relative contributions of the $t\bar{t}b\bar{b}$ and $t\bar{t}$+jets distributions, which in turn depends on the strength of the $b$-tagging cut applied. The $b\bar{b}$ invariant mass can be studied for a $b$-tag requirement, the "medium $b$-tag", between the loose and tight, such that the possible presence of signal can be still neglected. We choose a medium working point corresponding to a $b$-tagging weight cut of 3, such that the ratio of the contribution of $t\bar{t}b\bar{b}$ with respect to $t\bar{t}$+jets goes from 11% to 30%, with a signal contamination of less than 3%.

One strategy contemplated is to use the $t\bar{t}b\bar{b}/t\bar{t}$+jets fraction coming from the Monte Carlo prediction and the total number of events from the data to normalize the Monte Carlo at the loose working point where the signal level is less than 1%. Using the Monte Carlo jet flavor composition and the ratio of the $b$-, $c$- and light jet efficiencies $(\varepsilon^{medium}_{b,\,c,\,light}(p_T,\eta)/\varepsilon^{loose}_{b,\,c,\,light}(p_T,\eta))$ at the loose and medium working

points, it is then possible to predict the shape and normalization at the medium working point.

The data reduction when moving the $b$-tag quality from "loose" to "medium" is explained by the ratio of the $b$-tag efficiencies at these two working points applied to the $t\bar{t}b\bar{b}$ and $t\bar{t}$+jets data. With a 50 pb$^{-1}$ data sample, the $b$-tagging efficiency of $b$- and $c$-jets will be known with an accuracy of 5% [19], while the rejection of light quark jets will be measured with a 10% uncertainty. We expect a significantly more accurate knowledge of the $b$-tagging performance with a data sample of approximately 30 fb$^{-1}$. This will allow the measurement of the background level of $t\bar{t}b\bar{b}$ and $t\bar{t}$+jets as a function of the $b\bar{b}$ invariant mass for the loose and medium $b$-tagging working points. These measurements can be used to verify and tune with data the background prediction given by the Monte Carlo simulation, which will be used to extrapolate the event yeld expectation of known processes when the $b$-tag quality is moved from "medium" to "tight". This extrapolation can be monitored, and eventually further corrected, by looking at the comparison with the measured data outside the mass window, where the number of signal events expected is small (about 4%). If necessary, this procedure can be extended by asking for three $b$-tagged jets to further constrain the background composition and its shape and absolute normalization, to achieve the 5% systematic uncertainty necessary for the analysis of this processes.

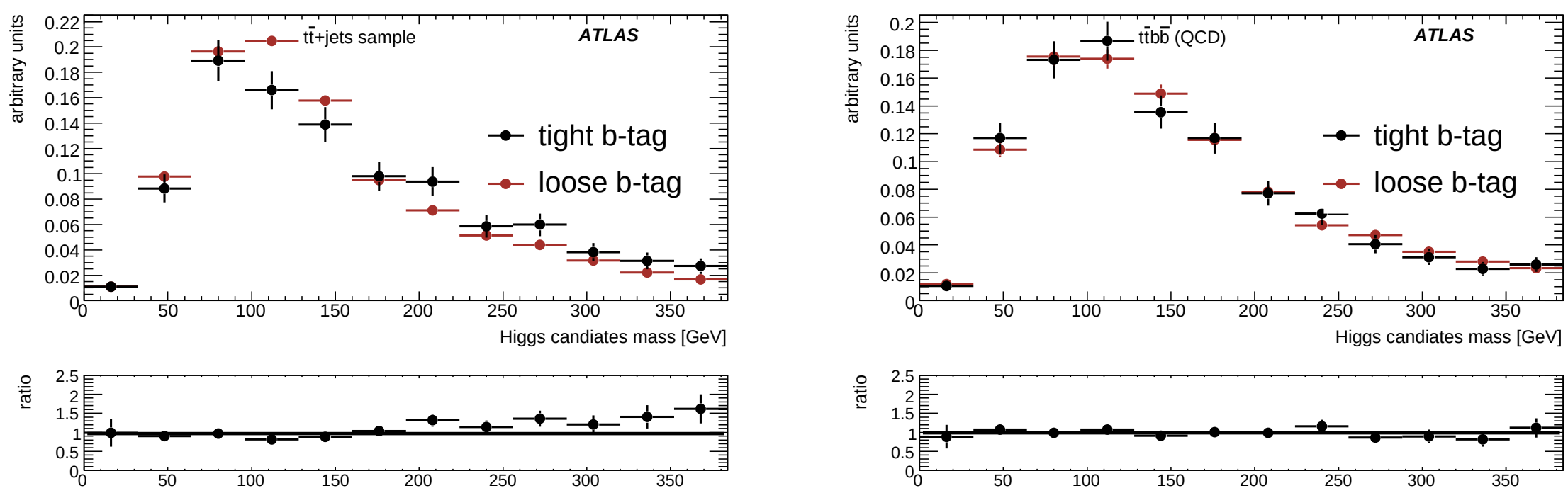


Figure 25: Ratio of the invariant mass spectrum for Higgs boson candidates after combinatorial likelihood analysis and using a loose and tight cut on the $b$-tag weight. Left hand side: $t\bar{t}$+jets, right hand side: $t\bar{t}b\bar{b}$. The signal region shows very consistent behaviour.

## 13 Systematic uncertainties

The evaluation of systematic uncertainties, especially in the background level, is of vital importance in this analysis. Unfortunately it has not yet been brought to a satisfactory level and a robust method to infer background shapes and normalization from data, vital for this channel, still needs to be developed. Following the estimation of systematic uncertainties due to the standard detector effects, Table 8 shows the various contributions for all three analyses. It is noticeable how important the jet uncertainties are for both signal and background. Indeed the knowledge of the jet energy and of the $b$-tagging performance have a crucial impact on the kinematic quantities used for the reconstruction of the $t\bar{t}$ system and for the correct identification of the $b$-jets used for the analysis. Large fluctuations on the background estimations arise due to the lack of statistics for the $t\bar{t}$X sample, giving rise to a relative statistical error up to 20%.

While the theoretical uncertainties for the signal and background normalization are quite large, their impact can be reduced by making direct measurements. This is certainly the case for the $t\bar{t}$ cross-section, where the theoretical uncertainties associated with the NLO+NLL calculation are around 12% [14] while with only 100 pb$^{-1}$ of data, a direct measurement of the cross-section for the semileptonic final state using $b$-tagging could be performed with a much smaller error [22]. The $t\bar{t}b\bar{b}$ background is only calculated at LO, the cross-section calculation has a strong scale dependence, a factor 4 when changing

from $Q^2_{QCD} = \hat{s}$ to $Q^2_{QCD} =< p_T^2 >$ [9]. Even though the signal cross-sections used for this work are LO, NLO calculations are already available with a theoretical uncertainty including errors coming from parton distribution functions of the order of 15-20% [23] (compared to the 100-200% uncertainty of the LO cross-section).

Table 8: Effect of the various systematic uncertainties on the signal and background efficiencies.

| Source | | Cut-based | | Likelihood | | Constrained fit | |
|---|---|---|---|---|---|---|---|
| | | signal | background | signal | background | signal | background |
| Electron | energy scale | ± 0.5% | ± 2% | ± 0.3% | ± 3% | ± 1% | ± 3% |
| | resolution | ± 0.5% | ± 0.6% | ± 0% | ± 1% | ± 0.2% | ± 4% |
| | efficiency | ± 0.2% | ± 2% | ± 0.2% | ± 1% | ± 0.5% | ± 0.2% |
| Muon | energy scale | ± 0.7% | ± 3% | ± 0.6% | ± 0.2% | ± 0.4 % | ± 4% |
| | resolution | ± 0.8% | ± 0.6% | ± 0.3% | ± 0.4% | ± 1% | ± 3% |
| | efficiency | ± 0.3% | ± 0.1% | ± 0.8% | ± 0.1% | ± 0.4% | ± 0.1% |
| Jet | energy scale | ± 9% | ± 5% | ± 9% | ± 14% | ± 9% | ± 8% |
| | resolution | ± 0.3% | ± 7% | ± 1% | ± 5.5% | ± 5% | ± 14% |
| | *b*-tag | ± 16% | ± 20 % | ± 18% | ± 20% | ± 16% | ± 20% |
| | *b* mis-tag | ± 0.8% | ± 5% | ± 1.1% | ± 3% | ± 3% | ± 10% |
| summed in quadrature | | ± 18% | ± 22% | ± 20% | ± 25% | ± 19% | ± 28% |

### 13.1 Effect of pile-up on signal

The portion of the semi-leptonic $t\bar{t}H$ signal sample used here is simulated a second time, but with the anticipated effects of pile-up and cavern background included. It is important to stress that the same generated events are used as input for both pile-up and non pile-up samples. The pile-up actually applied to the events is that expected for running at instantaneous luminosity $\mathscr{L}$ of $10^{33} cm^{-1} s^{-1}$.

The effect of pile-up on the preselection of events (applicable to all three analyses) is shown in Table 9, and as can be seen, the effect of the trigger requirement is the most significant.

Table 9: The effect of pile-up on the samples at successive stages of preselection with relative efficiencies.

| Quantity \ Sample | $t\bar{t}H$ $\sigma$ (fb) No pile-up | $t\bar{t}H$ $\sigma$ (fb) pile-up |
|---|---|---|
| Starting Sample Generated | 100 | 100 |
| Pass Trigger (`e22i`, `e55`, `mu20`) | 65 (65%) | 62 (62%) |
| One high-$p_T$ Lepton | 56 (87%) | 53 (86%) |
| $\geq$ 6 jets ($p_T >$ 20 GeV, $\lvert\eta\rvert < 5$) | 36 (64%) | 34 (64%) |
| $\geq$ 4 central *b*-jet candidates, ($\lvert\eta\rvert < 2.5$ & *b*-jet weight $> 0$) | 16 (45%) | 15 (44%) |
| **Preselected** | **16 (45%)** | **15 (44%)** |

The distribution of the number of high $p_T$ leptons in the events before preselection is shown in Fig. 26 for the pile-up and non pile-up samples. They are very similar, suggesting that the electron and muon reconstruction are not significantly affected by pile-up; however the trigger requirement reduces the number of events with pile-up available to the rest of the preselection.

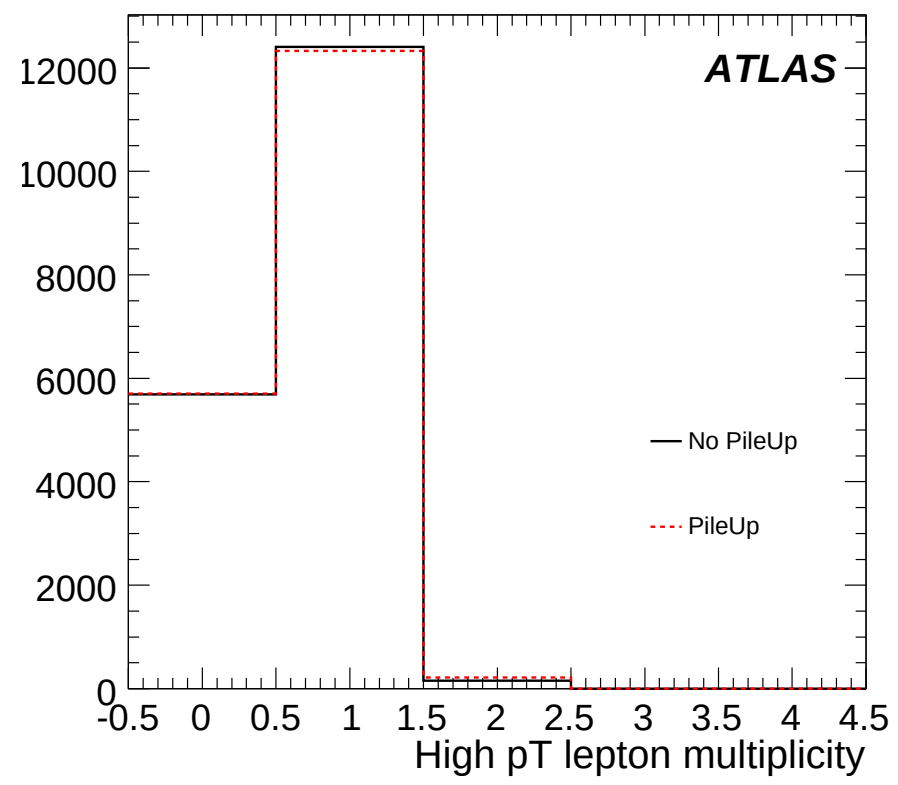


Figure 26: High $p_T$ lepton multiplicity ($e^{\pm}, \mu^{\pm}$) for pile-up and non pile-up samples (before preselection).

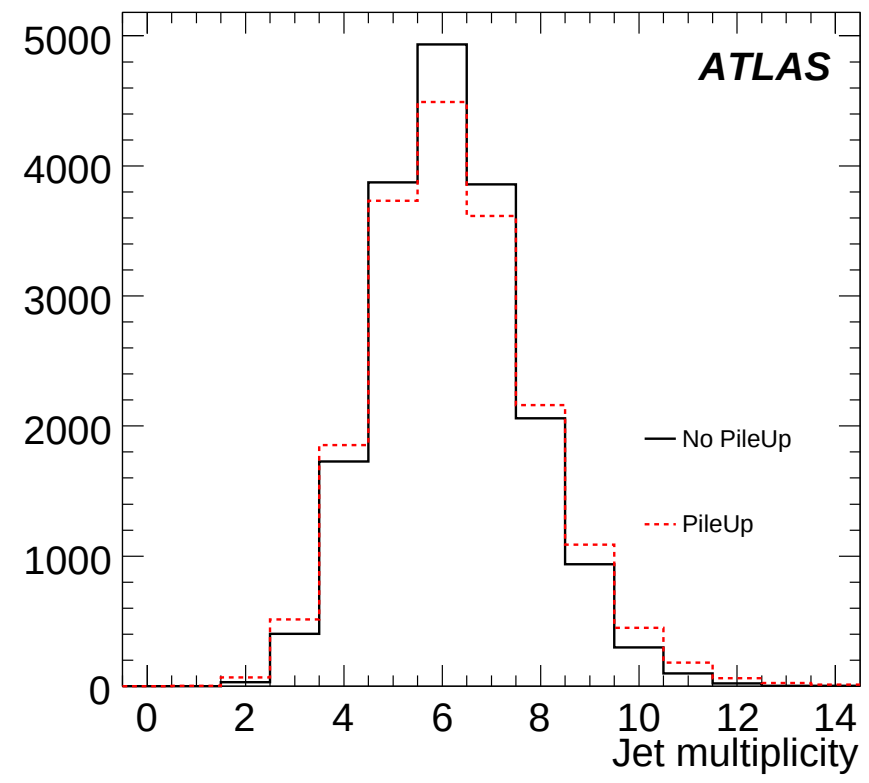


Figure 27: Jet multiplicity (before preselection) for pile-up and non pile-up samples. The cuts $p_T < 20$ GeV and $|\eta| < 5$ are applied to the individual jets.

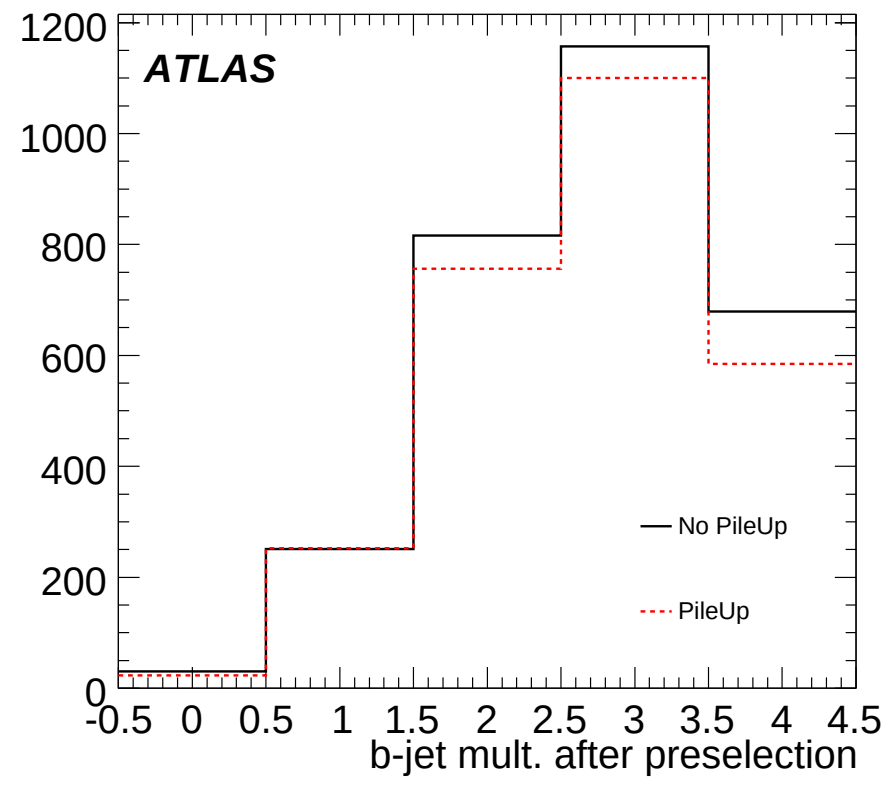


Figure 28: $b$ jet multiplicity (after preselection) for pile-up and non pile-up samples. The cuts $p_T < 20$ GeV, $|\eta| < 2.5$ and $b$jet weight $> 5.5$ are applied to the individual jets.

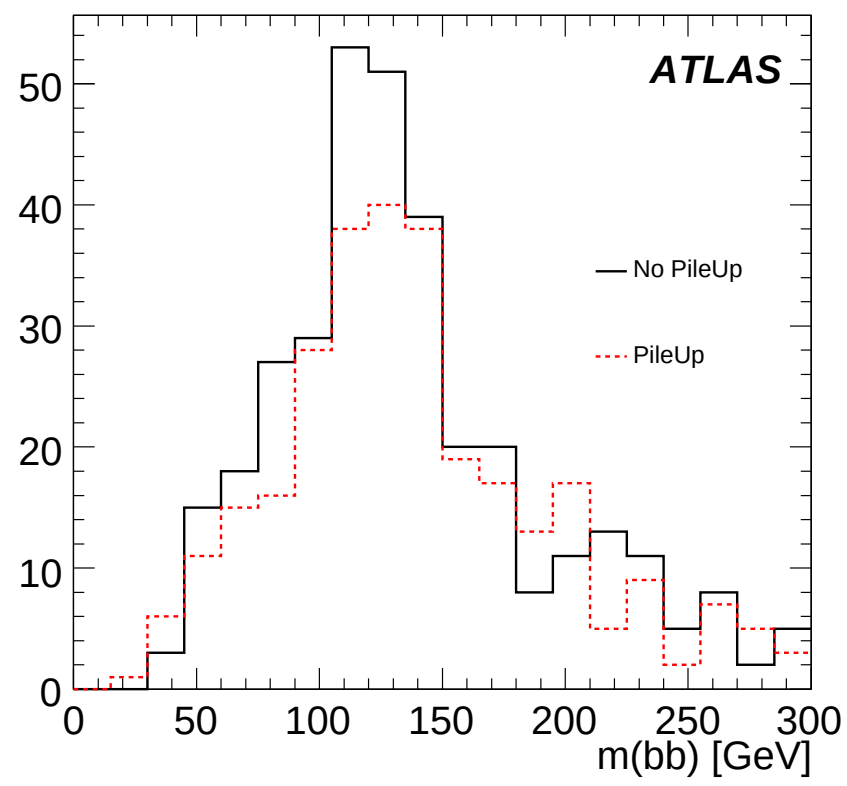


Figure 29: Reconstructed Higgs boson mass peak ($m_{bb}$) for pile-up and non pile-up samples.

Figure 27 shows the jet multiplicity before preselection. It can be seen that the number of events having exactly 6 jets is reduced by approximately 10%, and the number of events having more than 7 jets is increased. The net effect will be an increase in the combinatorial background, though the extra jets will typically have a low $p_T$.

For the cut-based and pairing likelihood analyses, candidate $b$-jets are designated as those jets lying in the central region of the detector ($|\eta| < 2.5$), with $p_T > 20$ GeV and $b$-jet weight $> 5.5$, however, if there are more than four of these then the jets with the highest $b$-jet weights are used.

The number of $b$-jets in the events both with and without pile-up after the other preselection cuts are applied can be seen in Fig. 28, where there is a reduction in the number of events having the requisite four $b$-jets.

The extent of the reduction in events at the various stages of the cut-based analysis is shown in Table 10. The most pronounced difference comes from the reduction in the number of $b$-jets, and the net

effect of pile-up is a ∼12% reduction in the number of events where it is possible to reconstruct a Higgs particle, as shown in Fig. 29. This effect will also manifest itself in the backgrounds, since they all have either two or four $b$-quarks. The most interesting background pile-up study would now be with the $t\bar{t}j\bar{j}$ sample, since this relies on mis-tagging of light jets to create a physics background, however this is beyond the scope of this study which only examines the signal.

It should be noted that in the course of this study, the $b$-jet efficiency and light-jet rejections were studied, however no visible differences were observed. This can be explained by the fact that even a 1% drop in efficiency from 50% to 49% causes almost an 8% drop in events having four $b$-jets.

Table 10: The extent of the reduction in events for pile-up and non pile-up samples for the cut-based analysis with relative efficiencies in parentheses. The harshest reduction comes from the four $b$-jet requirement.

| | $t\bar{t}H$ $\sigma$ (fb) | |
|---|---|---|
| Quantity \ Sample | No pile-up | pile-up |
| Preselected events | 16.0 | 14.8 |
| $\geq 4$ $b$-jets ($b$-jet weight $> 5.5$) | 3.7 (23.1%) | 3.2 (21.5%) |
| Had & Lep $W$ inside mass-window | 2.5 (66.4%) | 2.1 (66.8%) |
| $t$, $\bar{t}$-quarks rec. in mass-window | 2.0 (81.6%) | 1.8 (83.1%) |

## 14 Significance estimates

The number of remaining events in the Higgs boson mass window (30 GeV around the nominal Higgs boson mass) have been used to compute a crude estimate of the statistical significance for this channel with 30 fb$^{-1}$. For such a channel in which the signal and backgrounds are very alike, this naive estimate is not the most relevant figure of merit, but it is still useful to compare analyses. For the cut-based analysis, a significance of 1.8 is achieved with signal to background ratio of approximately 0.11. It is worth noting that the addition of the low $p_T$ muons to jets and the residual jet calibrations performed in Sections 5.3.3 and 5.3.4 improved the cut-based analysis significance by 0.3. With the pairing likelihood approach the significance is 1.95 for a signal to background ratio of 0.1. Finally the constrained fit likelihood gives 2.2 (1.7) for a signal over background value of 0.12 (0.14), obtained with a cut on $\mathscr{L}_{s/b}$ of -4.2 (-4.1). Figure 30 shows the total significance $S/\sqrt{B+(\Delta B)^2}$ as a function of the systematic error on the background ($\Delta B$) for the different analyses. As is shown in the Fig., only a background uncertanity level below 10% allows exploitation of the statistical power of the mass constrained fit analysis with respect to the cut-based analysis, and even less for the case the pairing likelihood. Even for a robust analysis such as the cut-based approach, the large systematic uncertainties estimated in Table 8 provide a clear indication that a data driven background estimation is necessary.

## 15 Conclusion

We performed a baseline sensitivity study for the detection of a Standard Model Higgs boson decaying to $b\bar{b}$ when produced together with a $t\bar{t}$ pair. After the definition of a common preselection, three different techniques are used, all aimed at the reconstruction of the $t\bar{t}$ system. The first one is based on the reconstruction of the top quark and $W$ candidate masses (cut-based analysis). The second one (pairing likelihood analysis) uses a more complete description of the kinematic properties of the $t\bar{t}$ system to build

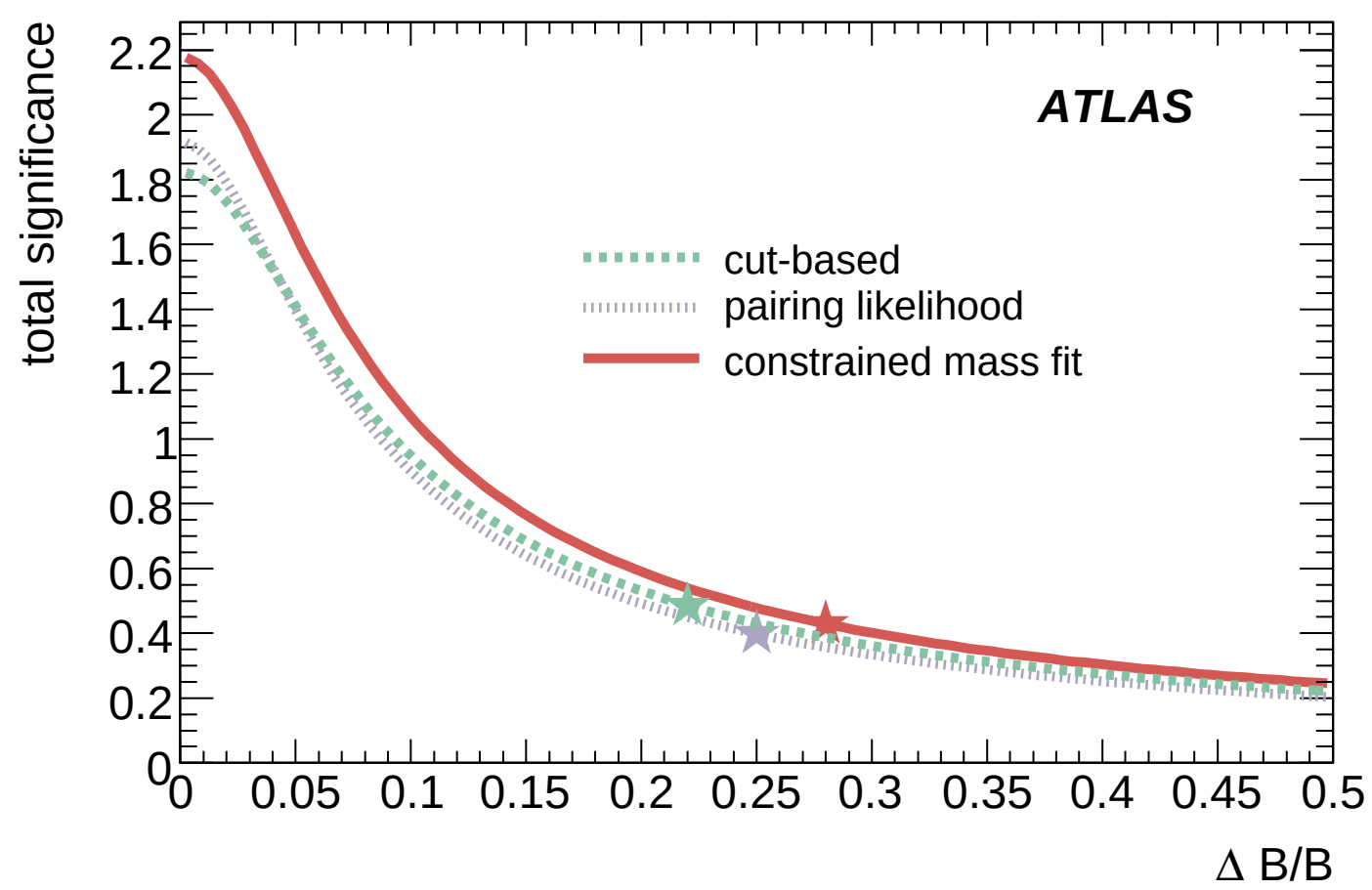


Figure 30: Comparison of the total significance as function of systematic uncertainties ($\Delta B$), for the cut-based, the pairing likelihood and the constrained fit analysis. Markers indicate the significance corresponding to the background uncertainty estimated in Table 8.

a likelihood discriminant and isolate the jets coming from the Higgs boson decay. The third approach (constrained fit) uses the known masses and jet errors as constraints to produce a combinatoric likelihood, and a second likelihood to separate signal from background. While the cut-based analysis is certainly the most stable one, relying only on the reconstructed invariant masses of the top quark candidates, it also performs worse with respect to the other two likelihood based analyses. On the other hand, these likelihood based analyses can be used successfully only after all kinematical variables are well understood together with their correlations. Although beyond the scope of this work, the use of more advanced multivariate techniques is foreseen to reduce both the combinatorial and physics background.

The statistical significance obtained for the three approaches was 1.82 for the cut-based, 1.95 for the pairing likelihood and 2.18 for the constrained mass fit at signal-to-background ratios of 0.11, 0.10 and 0.12 respectively. All the analyses suffer drastic reduction in significance as the overall systematic uncertainty increases. The most important individual uncertainties are those for the jet energy scale and $b$-tagging efficiency.

From this study emerges the necessity of a strong $b$-tagging algorithm which is important not only to suppress the $t\bar{t}$+jets physics background but also to help reduce the combinatorial background by improving the hadronically decaying $W$ reconstruction. It is also clear that the combinatorial background, responsible for the dilution of the Higgs boson mass peak, needs to be further reduced, possibly using multivariate techniques, in order to improve the statistical significance of the channel. Improvements in the mass peak resolution would also enhance the ability of a shape analysis from two perspectives; firstly it would be easier to select a signal-depleted region for any shape fits, and secondly the mass peak itself would become more pronounced.

The results presented in this work can be compared with a previous ATLAS study [3] performed using fast simulation with a parametrized $b$-tagging efficiency which had a higher performance than the one used here and also used PYTHIA in order to simulate the $t\bar{t}$+X background. It resulted in a significance of 1.9 and 2.6 respectively for the cut-based and likelihood analyses. The results presented in this note can also be compared with a recent CMS study [24] reporting a significance of 1.8 for the electron channel and 1.6 for the muon channel, in both cases for an integrated luminosity of 60 fb$^{-1}$. While a detailed comparison between the two experiments is not attempted in this work, it is noteworthy that the jet energy resolution quoted the CMS paper could be a key factor in explaining the improved

sensitivity seen by ATLAS for this channel.

The measurement of the background normalization from data is vital for this channel. Subsequent studies must be performed in this regard. Further methods of extracting shape information from data must also be developed, in particular, the extraction of the signal in the presence of a *quasi-signal-like* background as is exhibited in the invariant mass plots at the ends of the analyses. The shape information and any estimate of the significance obtained from it could be used in conjunction with the counting experiment information to improve the overall significance.

# Study of Signal and Background Conditions in $t\bar{t}H, H \rightarrow WW^{(*)}$ and $WH, H \rightarrow WW^{(*)}$

**Abstract**

In this note we present Monte Carlo studies of the associated Standard Model Higgs boson production in the $t\bar{t}H$ and $WH$ channels with the decay $H \rightarrow WW^{(*)}$. These channels are intended to provide information on the Higgs boson's couplings. We study the two- and three lepton final states in $t\bar{t}H$ and three lepton final states in $WH$, based on the full ATLAS detector simulation.

## 1 Introduction

The discovery and subsequent study of the Higgs boson is one of the main aims of the Large Hadron Collider (LHC) at CERN. The possible mass range of the Standard Model Higgs boson is bounded by the lower limit set at LEP of 114 GeV and reaches to about 1000 GeV [1]. The ATLAS experiment will use all possible channels to extract information on it, because comparing the rates in the different channels will allow information on the couplings to be extracted.

The sensitivity of ATLAS to a Higgs boson produced in gluon fusion or via vector boson fusion and decaying to $W$ quark pairs has been discussed elsewhere in this volume [2]. This note contains the results of studies of the Higgs boson in the same decay mode but produced in association with either top quarks, ($t\bar{t}H, H \rightarrow WW^{(*)}$), or a $W$ boson ($WH, H \rightarrow WW^{(*)}$). The cross-sections for these processes are significantly lower than for inclusive Higgs production, and the additional activity makes them more complex to reconstruct, but the presence of extra signatures gives more possibilities for the reduction of the background.

This note explores techniques to exploit these signatures, and the signal and background conditions are studied in both channels. A full simulation of the ATLAS experiment is employed to estimate these, which represents an improvement over the fast simulation used in previous studies of $t\bar{t}H$ [3] and $WH$ [4,5]. The marginal production rates and numerous background sources, many with large cross-sections, make this analysis difficult, and both the background and signal need to be established in some detail. Nevertheless, if the background can be well estimated, then for integrated luminosities of several tens of fb$^{-1}$ measurements should be possible.

The backgrounds considered in detail here arise from the inclusive $t\bar{t}$ process, from $t\bar{t}$ produced in association with gauge bosons, and from gauge bosons produced inclusively or in pairs. Unfortunately, it has not been possible to model all the relevant backgrounds with a complete simulation at the statistical level required; this is true for example of inclusive QCD multijet events. Section 2 describes the considered signal and background processes. Sections 3 and 4 give the details of $t\bar{t}H$ and $WH$ analysis accordingly. Section 5 discusses the results, including the signal-to-background ratio that can be achieved in these two channels.

## 2 Signal and background Monte Carlo samples

Signal and background were produced with various generators, through a realistic ATLAS detector simulation based on the GEANT 4 package [6].

### 2.1 Signal generation

Events with a Higgs boson decaying to a $W$ pair produced in association with a $t\bar{t}$ pair or with a $W$ boson can be searched for at hadron colliders by requiring the presence of lepton pairs ($\ell = e, \mu$).

In particular, for the two-lepton final states, like-sign leptons are selected; this allows a strong reduction of the large background produced by the $Z$ or $t\bar{t}$ leptonic decays. In order to improve the efficiency of the Monte Carlo data sample production, generated events were filtered before their processing through the ATLAS detector simulation.

For the $WH$ channel, only events with three leptons in the final state were selected. These leptons had to pass loose $\eta$ and $p_{\mathrm{T}}$ cuts.

Samples of $t\bar{t}H$ with at least two leptons were generated and filtered for different Higgs boson masses between 120 and 200 GeV using the PYTHIA 6.4 generator [7]. Results obtained with these samples were normalized to the Next-to-Leading Order (NLO) cross-sections and branching ratios reported in Ref. [1]. Only the $m_H$ = 170 GeV mass point was studied for the $WH$ channel, where signal events were generated with the MC@NLO program [8].

Table 1 summarizes the most important characteristics of the signal samples used for this note.

Table 1: Signal samples generated for the $t\bar{t}H$ and $WH, H \rightarrow WW^{(*)}$ analyses.

| Process | $m_H$ [GeV] | $\sigma_{tot}$(NLO) [fb] | Final states | Generator | $\sigma \times BR \times \varepsilon_{filter}$ [fb] | $N$(events) |
|---|---|---|---|---|---|---|
| $t\bar{t}H$ | 120, 130, 140 | 669, 534, 431 | $t\bar{t}H \rightarrow 4W$ (2L) | PYTHIA 6.4 | 3.60, 6.25, 8.51 | ~40k |
| | 150, 160, 170 | 352, 291, 243 | | | 9.68, 10.49,9.31 | per $m_H$ |
| | 180, 190, 200 | 204, 174, 149 | | | 7.62, 5.50, 4.42 | |
| $t\bar{t}H$ | 120, 130, 140 | 669, 534, 431 | $t\bar{t}H \rightarrow 4W$ (3L) | PYTHIA 6.4 | 2.34, 4.05, 5.49 | ~40k |
| | 150, 160, 170 | 352, 291, 243 | | | 6.31, 6.91, 6.15 | per $m_H$ |
| | 180, 190, 200 | 204, 174, 149 | | | 5.00, 3.54, 2.86 | |
| $WH$ | 170 | 511 | $WH \rightarrow WWW$ (3L) | MC@NLO | 3.42 | 80k |

### 2.2 Background samples for $t\bar{t}H, H \rightarrow WW^{(*)}$

The main backgrounds for the $t\bar{t}H, H \rightarrow WW^{(*)}$ final states are $t\bar{t}$, $t\bar{t}W$, $t\bar{t}Z$, $t\bar{t}t\bar{t}$ and $t\bar{t}b\bar{b}$. Single top events have been neglected. Jets from QCD production and $WZ$ production processes are also sources of background. However, lepton identification with isolation and a jet multiplicity requirement are expected to reject a large fraction of these. The background from QCD multijet production has not been properly estimated so far and it is hoped that the selection requirements reduce it to an acceptable level.

A special MC@NLO sample is filtered for a pair of like-sign or more than two leptons with a $p_{\mathrm{T}} > 13$ GeV and $|\eta| < 2.6$ at the generator level. It results in a filter acceptance of 0.0384. In addition, when there are three or more generated leptons, events with oppositely charged leptons from $W$ bosons falling into a special domain ( $p_{\mathrm{T}} \geq 30$ GeV and $||\eta| - 1.5| \leq 0.2$ for electron, $p_{\mathrm{T}} \geq 15$ GeV and $||\eta| - 1.25| \leq 0.2$ for muon ) were rejected. This results in a small bias, analysis dependent.

The $Wb\bar{b}$ sample was produced by the ALPGEN generator with only leptonic $W$ boson, a generator level filter led to an additional 0.02 acceptance, and a 2.57 K-factor [6] was also included. Leading order $t\bar{t}W + jets$ samples were produced with ALPGEN [9]. The minimum $p_{\mathrm{T}}$ for the additional jets was 15 GeV, while the maximum $|\eta|$ was 6.0. The generated jets were also required to be separated by a distance $\Delta R = \sqrt{\Delta\eta^2 + \Delta\phi^2}$ larger than 0.4. MLM matching [9] was performed to avoid double counting of additional jets.

Samples of $t\bar{t}Z$, $t\bar{t}t\bar{t}$, $t\bar{t}b\bar{b}$ and $t\bar{t}b\bar{b}$(EW) were produced with the leading order generator ACERMC [10]. The $t\bar{t}Z$ events are normalized to the total cross-section recently calculated at NLO [11], while other ACERMC samples are normalized to LO. In the $t\bar{t}Z$ sample, the decay $Z \rightarrow \ell\ell$ was forced. The $t\bar{t}b\bar{b}$(EW) sample contains the electroweak contribution to the production of $t\bar{t}b\bar{b}$. For both $t\bar{t}b\bar{b}$ samples, the final

states containing four $b$-jets, two light jets and a lepton (muon or electron) were generated. Table 2 summarizes the characteristics of all background samples relevant for the $t\bar{t}H$ analysis.

Table 2: The samples used to estimate the background contribution in the $t\bar{t}H, H \to WW^{(*)}$ analysis. $\mathscr{L}$ denotes the effective integrated luminosity available from Monte Carlo statistics.

| Process | Generator | $\sigma_{tot}$ [fb] | $\sigma \times BR \times \varepsilon_{filter}$ [fb] | $N$(events) | $\mathscr{L}$ [fb$^{-1}$] |
|---|---|---|---|---|---|
| $t\bar{t}$ | MC@NLO | 833000 | 450000 | 440k | 0.98 |
| $t\bar{t}$ pre-filtered | MC@NLO | 833000 | 32000 | 350k | 10.9 |
| $t\bar{t}b\bar{b}$(EW) | ACERMC 3.3 | 900 | 244 | 6.5k | 26.6 |
| $t\bar{t}b\bar{b}$ | ACERMC 3.3 | 8200 | 2244 | 44k | 19.6 |
| $Wb\bar{b}$ | ALPGEN | $2.1 \times 10^5$ | 1387.8 | 20k | 14.4 |
| $t\bar{t}W$ + 0 jets | ALPGEN | 189 | 25.3 | 20k | 790 |
| $t\bar{t}W$ + 1 jets | ALPGEN | 156 | 20.7 | 20k | 966 |
| $t\bar{t}W$ + $\geq 2$ jets | ALPGEN | 237 | 34.0 | 18k | 529 |
| $t\bar{t}Z$ | ACERMC 3.4 | 1090 | 87.0 | 19k | 218 |
| $gg \to t\bar{t}t\bar{t}$ | ACERMC 3.4 | 2.2 | 1.44 | 21k | 14583 |
| $qq \to t\bar{t}t\bar{t}$ | ACERMC 3.4 | 0.48 | 0.31 | 7k | 22580 |

### 2.3 Background samples for $WH, H \to WW^{(*)}$

The $t\bar{t}$ and $Wb\bar{b}$ samples as given in Table 2 are used also in this analysis. For the irreducible diboson $WZ/ZZ$ backgrounds only the fully leptonic decays were considered; this was done with the MC@NLO generator. The ALPGEN $t\bar{t}W$+0 jet sample described in Section 2.2 was analyzed to account for the $t\bar{t}W$ background and as it gives a negligible accepted cross-section the samples with additional jets were not considered. The huge $W$+jet background was generated with HERWIG [12] and was normalized to the NLO production cross-section [6] with a filter applied, requiring at least one electron (muon) with $p_T \geq 10$ GeV and $|\eta| \leq 2.7$ ( $p_T \geq 5$ GeV and $|\eta| \leq 2.8$ ).

An overview of all background samples used for the $WH$ analysis is given in Table 3.

Table 3: List of background samples for the $WH$ analysis. $\mathscr{L}$ denotes the effective integrated luminosity available from Monte Carlo statistics.

| Process | Generator | $\sigma_{tot}$ [fb] | $\sigma \times BR \times \varepsilon_{filter}$ [fb] | $N$(events) | $\mathscr{L}$ [fb$^{-1}$] |
|---|---|---|---|---|---|
| $t\bar{t}$ no all-hadronic | MC@NLO | 833000 | 450000 | 440k | 0.98 |
| $t\bar{t}$ pre-filtered | MC@NLO | 833000 | 32000 | 350k | 10.9 |
| $WZ$ | MC@NLO 3.10 | 47760 | 750 | 36k | 48 |
| $ZZ$ | MC@NLO 3.10 | 14750 | 72.5 | 50k | 690 |
| $W$+jets | HERWIG | $1.91 \times 10^8$ | $2.8 \times 10^7$ | 60k | 0.0214 |
| $Wb\bar{b}$ | ALPGEN | $2.1 \times 10^5$ | 1387.8 | 20k | 14.4 |
| $t\bar{t}W$ + 0 jet | ALPGEN | 189 | 25.3 | 20k | 790 |

## 3 Selection of the $t\bar{t}H$, $H \to WW^{(*)}$ two and three-lepton final states

In this study, the high $p_T$ single lepton trigger is used for the $t\bar{t}H$ two-lepton ($2L$) events and three-lepton ($3L$) analyses, with a trigger efficiency larger than 96% for both channels at offline selection level. Cut-based analyses were performed, based on the standard ATLAS reconstruction of a medium quality electron [13], combined muons [14], and cone size of $\Delta R = \sqrt{\Delta\eta^2 + \Delta\phi^2} = 0.4$ (cone-0.4) tower jets [15].

Signal data sets for nine Higgs boson masses in the range between 120 and 200 GeV were analysed. In the following, numbers will be given mainly for the most promising Higgs boson mass of 160 GeV including cut flow information for the 120 and 200 GeV mass points.

### 3.1 Event selection in $t\bar{t}H$, $H \rightarrow WW^{(*)}$

The event selection is based on the analysis of final states with at least two reconstructed leptons and jets; from now on we refer to this analysis as "basic selection". Each lepton or jet is required to have transverse momentum $p_T > 15$ GeV and to lie in the pseudorapidity region $|\eta| < 2.5$. Finally, two-lepton ($2L$) events are required to have at least six reconstructed jets, while events with three leptons ($3L$) must have at least four jets.

The $2L(3L)$ selection retains 36.1% (35%) of the Higgs boson events with $m_H = 160$ GeV, while reducing the various backgrounds (see Tables 4 and 5).

For both selections, further suppression of the main background sources can be done by isolation. The isolation criteria require that the transverse energy deposited in the calorimeter around the lepton in a cone size $\Delta R = 0.2$ be below 10 GeV (calorimeter isolation), the maximum $p_T$ of extra tracks reconstructed in the Inner Detector around the lepton track in a cone size $\Delta R = 0.2$ be below 2 GeV (tracker isolation) and the angular separation $\Delta R_{lep-clj}$ between the lepton and the closest jet be greater than 0.2 for an electron or 0.25 for a muon (cone isolation). This is referred to as "standard isolation" and it allows the reduction of the $t\bar{t}$ background by more than a factor of 10 (170) in the $2L$ ($3L$) analysis. The $t\bar{t}Z/t\bar{t}W$ backgrounds are suppressed by a factor 2 ($t\bar{t}Z$, $2L$) to 5 ($t\bar{t}W + 2$jets, $3L$).

Further reduction of the $t\bar{t}$ background in the dilepton final state can be achieved by requiring exactly two like-sign isolated leptons. This requirement suppresses the large $t\bar{t}$ processes with two leptonic $W$-decays, as well as the contribution from the $t\bar{t}Z$ process.

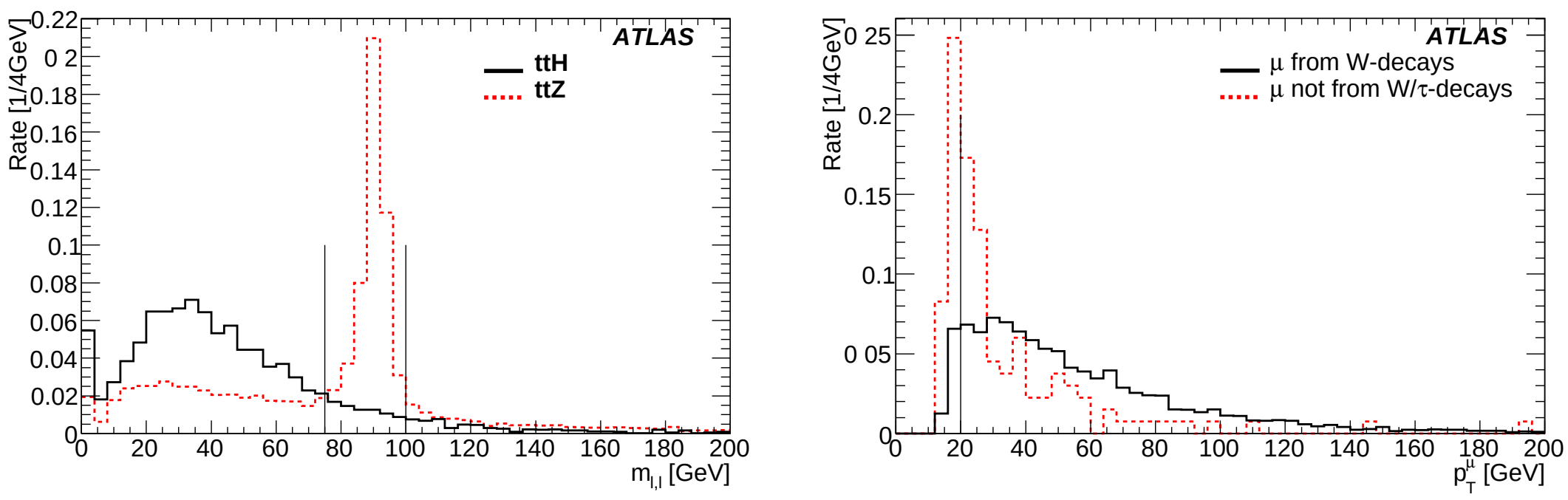


(a) Dilepton invariant mass distributions (b) $p_T$-distributions of isolated and non-isolated muons

Figure 1: (a) Dilepton invariant mass distributions in $t\bar{t}H$ ($2L$) and $t\bar{t}Z$ and (b) $p_T$ -distributions of muons passing the loose isolation criteria. The solid distribution shows electrons from $W$ decays in the 160 GeV signal sample, the dotted distribution shows muons in $t\bar{t}$, which could not be matched to a generator-level muon from a $W$- or $\tau$-decay. All distributions are normalized to unity.

In both final states, $t\bar{t}Z$ can be suppressed further by an explicit $Z$-veto: events that contain a lepton pair of opposite charge and same flavour with an invariant mass between 75 GeV $< m_{\ell\ell} <$ 100 GeV are rejected. This veto includes all leptons passing the selection criteria and loose $p_T$-cut, here set to 6 GeV. The dilepton invariant mass distributions in $t\bar{t}H$ and $t\bar{t}Z$ are shown in Figure 1(a). The $Z$-veto decreases the $t\bar{t}Z$-contribution roughly by 75%, while 98% of the signal survive in the $2L$ analysis. In the $3L$ case, 83% of the signal events pass the $Z$-veto, while 80% of the $t\bar{t}Z$ contribution is suppressed.

At this stage of the $2L$ selection, 73% of the remaining $t\bar{t}$ events have at least one muon from a semi-leptonic heavy quark decay, while the fraction of events with electrons of this origin is only 20% (identification of electrons embedded in jets is more difficult than that of muons).

Further rejection of these muons from $t\bar{t}$ events is achieved by requiring the reconstructed muon $p_T$

to be larger than 20 GeV, as leptons from heavy quark decays tend to be softer than leptons from $W$ decays (see Figure 1(b)). After this cut, the fraction of events with at least one muon from semileptonic decay versus the one with at least one electron from semileptonic decay is respectively 46% and 41% .

The detailed cut flows, with the corresponding accepted cross-sections, are listed in Tables 4 and 5. From now on the "basic selection" is quoted with the filter efficiency allowed for. In the $2L$ analysis we have used the special MC@NLO$t\bar{t}$ sample described in Section 2.2 and a small bias has been computed at various stages of the cut flow. The different values have been found to be compatible and a correction of $1.15 \pm 0.10$ has been applied in the $t\bar{t}$ line in Table 4. In the $3L$ analysis the standard MC@NLO$t\bar{t}$ sample has been used, and therefore no correction has been applied.

In both cases, the largest background contribution is expected from $t\bar{t}$ events. The accuracy of the MC prediction of the total background expectation is limited by the available statistics and by the intrinsic accuracy of the simulation tools. In the case of $t\bar{t}$ the basic cross-section error is large, but the ALPGEN predictions for higher additional jet multiplicities suffer from even larger uncertainties. There may also be background contributions from $W$ bosons with multijets which have not been reliably estimated or QCD multijet production or other sources which it has not been possible to simulate.

Table 4: Cut flow and expected cross-sections [fb] for the $t\bar{t}H$ ($2L$) analysis. The errors presented are statistical only. Some backgrounds, such as $W$+jets, $b\bar{b}$ and $t\bar{t}jj$ have not been included.

| Sample | $\sigma_{Total} \cdot BR$ | Basic sel. | Calo iso. | Track iso. | Cone iso. | Like-sign | $Z$-veto | $p_T^\mu$ |
|---|---|---|---|---|---|---|---|---|
| $t\bar{t}H$ ($2L^{truth}$, 120 GeV) | 3.9 | 1.05 | 0.80 | 0.65 | 0.52 | 0.52 | 0.51 | 0.45±0.01 |
| $t\bar{t}H$ ($2L^{truth}$, 160 GeV) | 11.1 | 4.01 | 3.02 | 2.57 | 2.09 | 2.09 | 2.04 | 1.85±0.03 |
| $t\bar{t}H$ ($2L^{truth}$, 200 GeV) | 4.7 | 1.83 | 1.43 | 1.24 | 1.05 | 1.04 | 1.02 | 0.95±0.01 |
| $t\bar{t}bb$ (EW) | 259.0 | 15.8 | 4.1 | 0.9 | 0.3 | 0.2 | 0.2 | 0.11±0.07 |
| $t\bar{t}b\bar{b}$ | 2360. | 177. | 31.7 | 6.3 | 1.8 | 0.9 | 0.9 | 0.5±0.2 |
| $t\bar{t}$ | 833000. | 6170. | 1970. | 870. | 500. | 16.0 | 16.0 | 7.4±1.1 |
| $t\bar{t}t\bar{t}$ | 2.68 | 0.65 | 0.33 | 0.26 | 0.20 | 0.07 | 0.07 | 0.06±0.00 |
| $t\bar{t}W$+0j | 61.1 | 1.17 | 0.46 | 0.30 | 0.19 | 0.10 | 0.10 | 0.09±0.01 |
| $t\bar{t}W$+1j | 50.5 | 2.09 | 0.93 | 0.66 | 0.48 | 0.23 | 0.23 | 0.21±0.02 |
| $t\bar{t}W$+≥2j | 76.9 | 8.6 | 4.9 | 4.1 | 3.3 | 1.58 | 1.54 | 1.40±0.05 |
| $t\bar{t}Z$ | 110. | 25.7 | 20.5 | 18.1 | 13.7 | 1.6 | 1.2 | 1.14±0.07 |
| $Wb\bar{b}$ | 66721. | 1.6 | 0.14 | - | - | - | - | - |
| **Total background** | | | | | | | | 10.3±1.1 |

Table 5: Cut flow and expected cross-sections [fb] for the $t\bar{t}H$ ($3L$) analysis. The errors presented are statistical only; systematic uncertainties are also important. Some backgrounds, such as $W$+jets, $b\bar{b}$ and $t\bar{t}jj$ have not been included.

| Sample | $\sigma_{Total} \cdot BR$ | Basic sel. | Calo iso. | Track iso. | Cone iso. | $Z$-veto | $p_T^\mu$ |
|---|---|---|---|---|---|---|---|
| $t\bar{t}H$ ($3L^{truth}$, 120 GeV) | 2.5 | 0.66 | 0.46 | 0.38 | 0.29 | 0.24 | 0.20±0.00 |
| $t\bar{t}H$ ($3L^{truth}$, 160 GeV) | 7.1 | 2.53 | 1.78 | 1.47 | 1.14 | 0.95 | 0.82±0.02 |
| $t\bar{t}H$ ($3L^{truth}$, 200 GeV) | 3.1 | 1.16 | 0.82 | 0.70 | 0.55 | 0.43 | 0.39±0.01 |
| $t\bar{t}$ | 833000. | 1600. | 230. | 50.0 | 9.3 | 7.2 | 2.1±2.1 |
| $t\bar{t}W$+0j | 61.1 | 0.78 | 0.17 | 0.08 | 0.04 | 0.03 | 0.03±0.01 |
| $t\bar{t}W$+1j | 50.5 | 1.07 | 0.28 | 0.14 | 0.08 | 0.08 | 0.06±0.01 |
| $t\bar{t}W$+≥2j | 76.9 | 2.77 | 0.85 | 0.60 | 0.50 | 0.42 | 0.38±0.03 |
| $t\bar{t}Z$ | 110. | 15.0 | 8.6 | 6.8 | 5.3 | 1.05 | 0.86±0.06 |
| **Total background** | | | | | | | 3.4±2.1 |

## 3.2 Projective likelihood estimator for electron isolation

A projective likelihood estimator, called IsolationLikelihood [13], was developed in the course of the $t\bar{t}H, H \to WW^{(*)}$ ($2L$) analysis. Alternative to the standard isolation, this tool is meant to combine the separation power of several isolation variables into a single, more powerful one. It uses the likelihood

ratio method to reject electrons from semi-leptonic heavy quark decays using a different set of isolation variables than those described in Ref. [13]:

- The additional transverse energy deposited in a cone of size $\Delta R = 0.2$ around the electron cluster.
- The sum of the $p_{\mathrm{T}}^2$ of all additional tracks measured in a $\Delta R = 0.2$ cone around the electron cluster.
- The transverse impact parameter significance $|Ip|/\delta(Ip)$ of the electron.

In addition, the “cone isolation” cut is also used as in the standard isolation analysis.

When tuned to give the same electron isolation efficiency obtained with standard isolation in the $t\bar{t}H$ analysis, the IsolationLikelihood allows a higher rejection of non isolated electron background by a factor 1.5 to 4, as shown in Figure 2. Using this projective likelihood estimator could suppress the $t\bar{t}$ background from $7.4 \pm 1.1$ pb to $5.7 \pm 1.0$ pb, while keep the same signal and other backgrounds selection efficiencies. It shows a potential improvement of this analysis which could be adopted at a small increase in complexity, but it has not been used in this document.

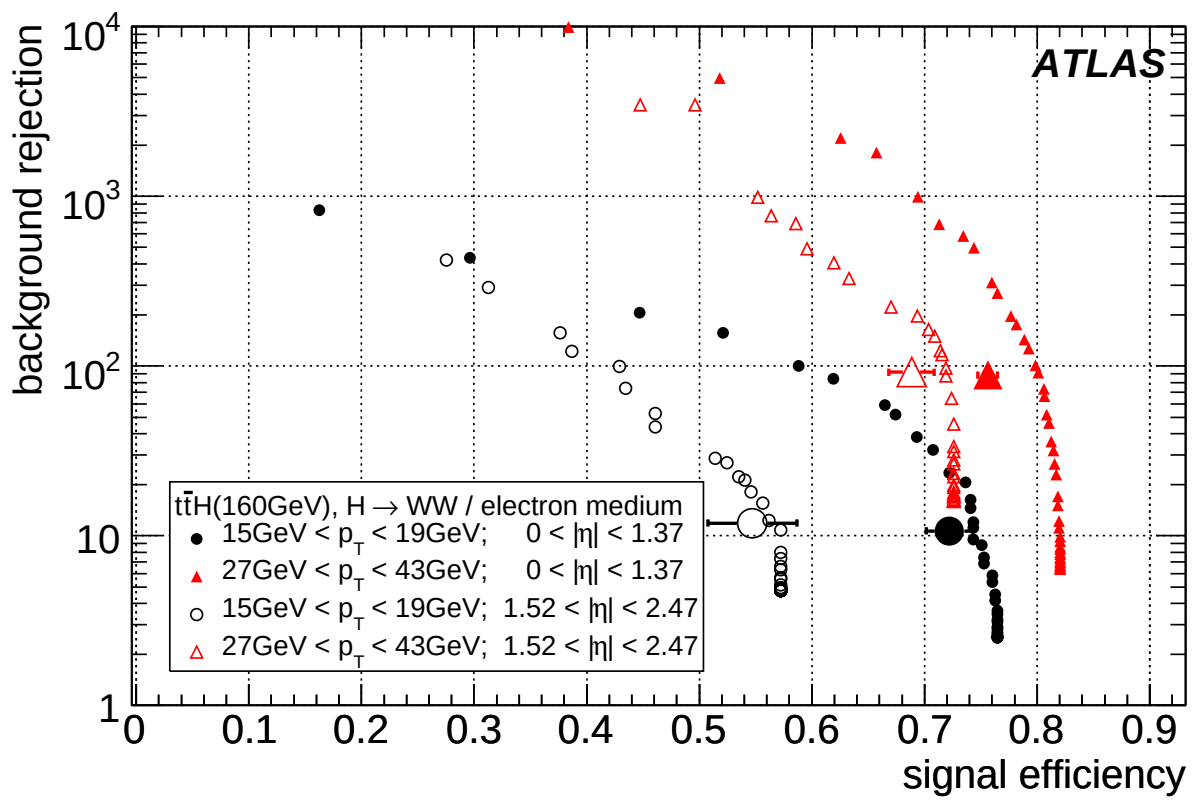


Figure 2: Non-isolated electron rejections vs. signal efficiencies obtained by the IsolationLikelihood estimator for four different $p_{\mathrm{T}}$ and $\eta$ intervals. The large points mark the working point of the standard isolation cuts for comparison and indicate the size of the error bars, which are not shown for the curves.

## 4 *WH* analysis

Only the three lepton final state, $W(H \to WW^{(*)}) \to 3\,(l\nu)$ is described below. The analysis of the larger cross-section dilepton final state, which has an important $W + jet$ background is currently ongoing.

The basic selection requires three leptons that satisfy the lepton identification criteria, i.e. medium electron [13] and standard muon [14]. The lepton $p_{\mathrm{T}}$-thresholds were set to 35 GeV for the leading and 15 GeV for the other lepton. As seen from Fig. 3(a) the cut on the leading lepton reduces the backgrounds much more than the signal. The presence of these leptons also ensures that any signal is efficiently recorded by the ATLAS trigger system

In addition to a 6 GeV calorimeter isolation and a 0.25 cone isolation as described in Section 3.1, $p_{\mathrm{T}}^{track}/p_{\mathrm{T}}^{lepton} \leq 0.05$ were forced, where $p_{\mathrm{T}}^{track}$ is of the track with maximal $p_{\mathrm{T}}$ in a cone of $\Delta R$=0.2 (0.3) around the muon (electron). Furthermore, a cut on the lepton three-dimensional impact parameter $I_{3D}/\sigma_I \leq 2.5$ was employed to reject leptons from bottom quarks.

In order to reduce the $WZ$ background, a $Z$ veto is applied by requiring that no opposite sign and same flavor lepton pair has an invariant mass between mass 65 and 105 GeV (Fig. 3(b)). In addition only events with $E_T^{\text{miss}} \geq 30$ GeV were kept. To further reduce the backgrounds, we ask the sum of the $p_T$ of all the jets (which were preselected above 20 GeV from cone-0.7 tower jets [15]) to be smaller than 120 GeV, as seen in Figure 3(c).

For additional rejection of $t\bar{t}$ and $t\bar{t}W$, events having at least one jet fulfilling a loose $b$-tag [16] are removed. In order to exploit the spin correlations in the $H \to WW^{(*)}$ signal, the minimum angular separation ( $\Delta R$ ) between lepton pairs is required to be in the range of [ 0.1 $\sim$ 1.5 ] (so called "H-S cut").

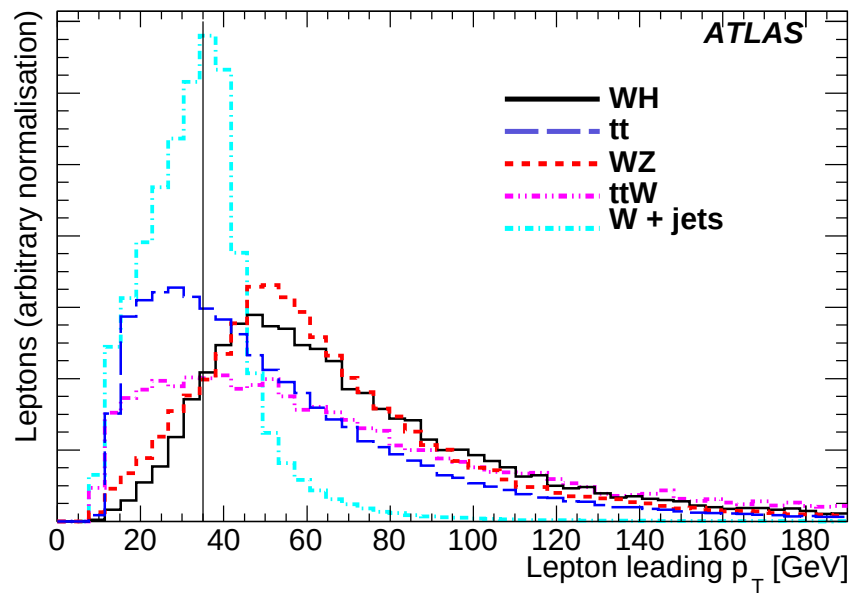


(a) Lepton $p_T$

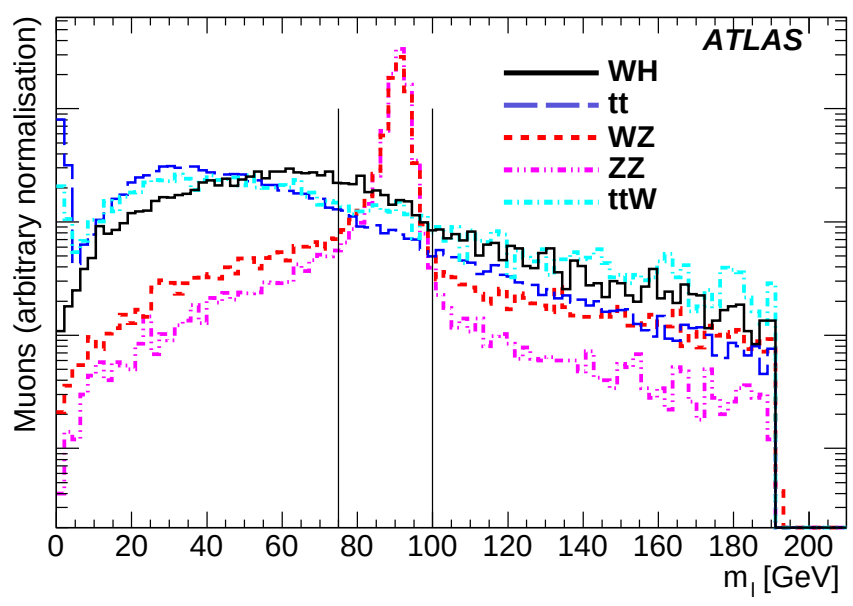


(b) lepton-lepton invariant mass

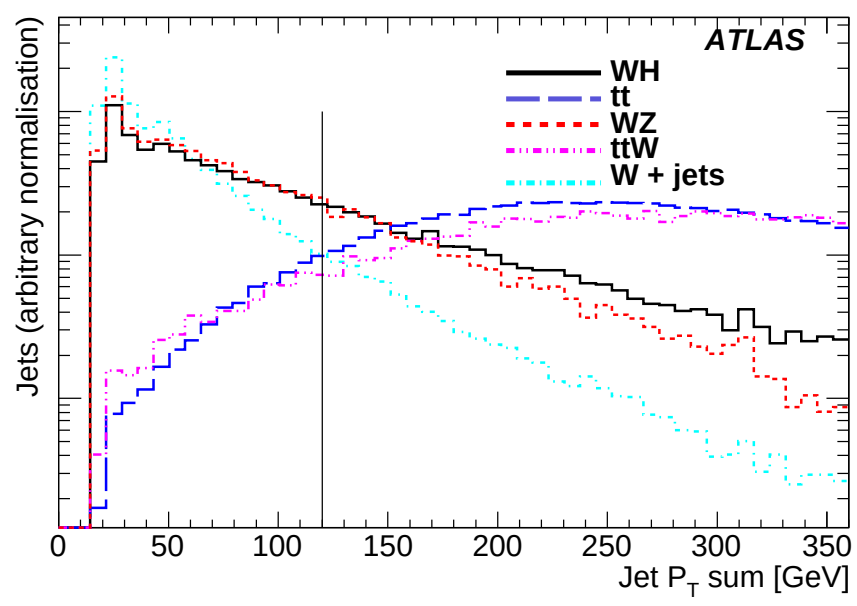


(c) Jet $p_T$-sum

Figure 3: $p_T$-distribution for the leading leptons in the $WH(3L)$ signal, $t\bar{t}$, $WZ$, $t\bar{t}W$ and $W$+jet-production(a), invariant mass of all the lepton pairs (b) and sum of the $p_T$ of all jets (c) for the $WH$ (3L) signal and the relevant backgrounds. All these plots are done after loose cut.

Table 6 summarises the cross-sections after the cut flow described above. The filtered MC@NLO$t\bar{t}$ sample described in Section. 2.2 has been used here. In order to take into account the bias introduced by this sample, a correction of $2.36 \pm 0.6$ has been applied. The background rate from $W$ bosons with multijets, QCD multijet production or other sources, which have not yet been possible to simulate, have their contribution still under study. The errors on the background are, at this stage, much larger than the size of the expected signal.

# 5 Discussion

## 5.1 Uncertainties in the analyses

Several systematic uncertainties affect the results presented in this paper. There are theory uncertainties associated to the the choice of the Parton Distribution Functions (PDFs), to the choice of the renormal-

Table 6: $WH$ (3L) cut flow and corresponding cross-sections. The errors presented are statistical only; systematic uncertainties are also important. Some backgrounds, such as $WWW$, single top and $t\bar{t}Z$ have not been included.

| | Input [fb] | Basic sel. | Isolation | $Z$-veto | $E_T^{miss}$ | H-S | ($b$-) jet veto |
|---|---|---|---|---|---|---|---|
| $WH$ (3L) | 5.04 | 1.18 | 0.62 | 0.53 | 0.47 | 0.36 | $0.31 \pm 0.02$ |
| $WZ$ | 750. | 165.5 | 1.41 | 0.74 | 0.63 | 0.21 | $0.10^{+0.08}_{-0.06}$ |
| $t\bar{t}$ | 833000. | 3564.3 | 6.45 | 6.11 | 5.10 | 1.02 | $0.34^{+0.70}_{-0.3}$ |
| $ZZ$ | 72.5 | 34.5 | 0.13 | 0.06 | 0.013 | 0.008 | $0.005 \pm 0.001$ |
| $t\bar{t}W$ | 61.1 | 1.35 | 0.22 | 0.21 | 0.19 | 0.07 | $0.003^{+0.005}_{-0.003}$ |
| $Wb\bar{b}$ | 66721. | 3.1 | - | - | - | - | - |
| $W \to e\nu$+jets | $2.05 \cdot 10^7$ | 17.6 | - | - | - | - | - |
| $W \to \mu\nu$+jets | $2.05 \cdot 10^7$ | 27.6 | - | - | - | - | - |
| **Total background** | | | | | | | $0.45 \pm 0.70$ |

ization and factorization scales, to the description of the initial and final state radiation and to the model used to simulate the heavy quark fragmentation. In order to evaluate the size of these uncertainties, the theory parameters above mentioned have been varied within intervals corresponding to sensible choices. Concerning the PDFs, the MRST2000-LO set was used at the place of the CTEQ6L1.

For the $t\bar{t}H$ analysis, the theory uncertainties have been found to induce a 9% change of the signal cross-section, dominated by the PDF choice. The impact to the $t\bar{t}$ process, which is the most important source of background to this signal, has been found to be 12% in Ref. [6]. An additional 5%, found in study of the signal process, associated to the uncertainty of the initial and final state radiation, has been included in quadrature, giving an overall 13% uncertainty on the total cross-section. However, the background sample is dominated by $t\bar{t}$ with extra jets, and the uncertainty on this rate is of order a factor two. For $WH$, the PDF uncertainty was found to be less than 5%, and energy scale uncertainty even smaller [17]. Including these effects and others (ISR,FSR) we get a total theoretical uncertainty of 9%.

The effect of experimental systematic uncertainties has been also investigated. The main sources of these uncertainties are represented by the knowledge of the integrated luminosity, the energy scale and the energy resolution of electrons, muons and jets, as well as the tag efficiency of $b$-jets and the rejection of light quarks. The level of these uncertainties and the impact on the overall event selection is presented in Tables 7 and 8. Pile-up events will decrease the detector performance and the impact needs to be properly addressed in future studies. However, the relatively low jet transverse momentum threshold of 15 GeV in the $t\bar{t}H$ analyses may be sensitive to this. The overall systematic uncertainty expected in the $t\bar{t}H$ analysis is 10% (10%) for the 2L (3L) signal and 15% (18%) for those backgrounds which have been quantified. In the case of the $WH$ analysis the overall systematic uncertainty is about 10% for the signal, and about 20% for those background systematics which have been estimated. In each case the total background uncertainty is much larger than this at present.

## 5.2 Conclusion

The $t\bar{t}H, H \to WW^{(*)}$ and $WH, H \to WW^{(*)}$ processes have been studied using two- and three-lepton final states. The signal and main backgrounds have been estimated using a full GEANT based simulation of the detector. The estimated accepted cross-sections in fb of signal and background for these processes are 1.9:10 ($t\bar{t}H$ 2L), 0.8:3.4 ($t\bar{t}H$ 3L) and 0.3:0.4 ($WH$ 3L) respectively. The signal is small and clear distinguishing features such as resonance peaks have not been established. The backgrounds are larger and their uncertainties have not been fully controlled. The analysis is therefore very challenging.

Accurate estimations of the background level using large simulation samples (made with more efficient simulation packages) as well as direct measurements using control samples from real LHC data are essential if a good signal significance is to be reached. For example the production of $W$ bosons with

Table 7: Overview of the experimental systematic uncertainties on the signal and background predictions related to the $t\bar{t}H$ channel in those channels studied. All numbers are in %.

| Source of the uncertainty | | $t\bar{t}H$ (2L) | | $t\bar{t}H$ (3L) | |
|---|---|---|---|---|---|
| | | Δ signal (%) | Δ background (%) | Δ signal (%) | Δ background (%) |
| Luminosity | 3 | 3 | 3 | 3 | 3 |
| Electron ID efficiency | 0.2 | 0.2 | 0.2 | 0.3 | 0.3 |
| Muon ID efficiency | 1 | 1.0 | 1.0 | 1.5 | 1.5 |
| Electron $E_T$ scale | 0.5 | 0.1 | 0.1 | 0.2 | 0.3 |
| Muon $E_T$ scale | 1 | 0.5 | 0.2 | 0.7 | 1.0 |
| Electron $E_T$ resolution | | 0.1 | 0.1 | 0.1 | 0.2 |
| Muon $p_T$ resolution | | 0.6 | 2.2 | 0.3 | 0.9 |
| Jet energy scale | 7 | 1.2 | 4.9 | 2.7 | 10 |
| Jet energy resolution | | 1.0 | 1.4 | 1.9 | 5.7 |
| Electron isolation efficiency | 1 | 1 | 1 | 1.5 | 1.5 |
| Muon isolation efficiency | 1 | 1 | 1 | 1.5 | 1.5 |
| Experimental uncertainty | | ±3.9 | ±6.6 | ±5.2 | ±12.3 |

Table 8: Overview of the experimental uncertainties on the signal and background predictions related to the $WH$ channels in those channels studied. All numbers are in %.

| Source of the uncertainty | | $WH$ 3L selection | | | | |
|---|---|---|---|---|---|---|
| | | Δ $WH$ (3L) (%) | Δ $WZ$ (%) | Δ $t\bar{t}$ (%) | Δ $ZZ$ (%) | Δ $t\bar{t}W$ (%) |
| Luminosity | 3 | 3 | 3 | 3 | 3 | 3 |
| Electron ID efficiency | 0.2 | 0.3 | 0.3 | 0.2 | 0.9 | 1.1 |
| Muon ID efficiency | 1 | 1.5 | 1.7 | 1.9 | 1.0 | 1.7 |
| Electron energy scale | 0.5 | 0.06 | 0.06 | 0.2 | 0.02 | 0.07 |
| Muon energy scale | 1 | 0.2 | 0.1 | 1.0 | 0.08 | 0.7 |
| Muon $p_T$ resolution | | 0.1 | 0.03 | 0.2 | 0.02 | 0.4 |
| Jet energy scale | 7 | 2.5 | 2.6 | 17.4 | 2.3 | 13.6 |
| Jet energy resolution | | 0.005 | 0.03 | 1.9 | 0.5 | 0.7 |
| $b$-tag eff. / light jet rej. | 5 / 32 | 1.0 | 1.0 | 2.7 | 0.8 | 3.2 |
| Experimental uncertainty | | ±4.3 | ±14.5 | | | |

large numbers of jets need to be measured, as does the fake contribution from $b$-jets. These two channels should then contribute to the measurement of the Standard Model Higgs boson properties, in particular the couplings of this boson to top and to the $W$.

# Discovery Potential of $h/A/H \to \tau^+\tau^- \to \ell^+\ell^- 4\nu$

**Abstract**
This note describes a study of the discovery potential for the supersymmetric Higgs bosons $h/H/A$ in proton-proton collisions at a center-of-mass energy of 14 TeV in final states with $\tau$ lepton pairs with the ATLAS detector at the LHC. The Higgs bosons are produced in association with $b$ quarks and decay into a $\tau\tau$ final state where both $\tau$ leptons decay leptonically. The signature of Higgs bosons with masses between 110 and 450 GeV is analyzed and the discovery potential is assessed. The analysis is based on an integrated luminosity of $30\,\text{fb}^{-1}$.

All results are obtained using full simulation of the ATLAS detector. No pile-up or cavern background has been considered in this analysis. In addition a procedure for estimating the shape and the normalization of the irreducible $Z \to \tau^+\tau^-$ background from data is investigated. The discovery potential as a function of $m_A$ and $\tan\beta$ is shown for the $m_h^{\text{max}}$ MSSM benchmark scenario.

## 1 Introduction

In the Minimal Supersymmetric Standard Model (MSSM), the minimal extension of the Standard Model, two Higgs doublets are required, resulting in five observable Higgs bosons. Three of them are electrically neutral ($h$, $H$, and $A$) while two of them are charged ($H^{\pm}$). At tree level their properties like masses, widths and branching fractions can be predicted in terms of only two parameters, typically chosen to be the mass of the CP-odd Higgs boson, $m_A$, and the tangent of the ratio of the vacuum expectation values of the two Higgs doublets, $\tan\beta$.

In the MSSM the couplings of the Higgs bosons to fermions and bosons are different from those in the Standard Model resulting in different production cross-sections and decay rates. While decays into $ZZ$ or $WW$ are dominant in the Standard Model for Higgs boson masses above $m_H \sim 160$ GeV, in the MSSM these decay modes are either suppressed like $\cos(\beta-\alpha)$ in the case of the $H$ (where $\alpha$ is the mixing angle of the two CP-even Higgs bosons) or even absent in case of the $A$. Instead, the coupling of the Higgs bosons to third generation fermions is strongly enhanced for large regions of the parameter space. The decay of the neutral Higgs bosons into a pair of $\tau$ leptons therefore constitutes an important discovery channel at the LHC. The production of the Higgs bosons can proceed via two different processes: gluon-fusion or production in association with $b$ quarks.

In this note, the discovery potential of neutral MSSM Higgs bosons, produced via associated production with $b$ quarks and decaying into a pair of $\tau$ leptons in ATLAS at the LHC is discussed. Only $\tau$ lepton decays into electrons and muons are considered here. Higgs bosons in the mass range between 110 and 450 GeV are analyzed for an integrated luminosity of $30\,\text{fb}^{-1}$. Both the shape and the normalization of the $Z \to \tau\tau$ background which is dominant for low Higgs boson masses are estimated from $Z \to \mu\mu$ and $Z \to ee$ events in data. The results are interpreted in the $m_h^{\text{max}}$ scenario as a function of the two parameter of the model, $m_A$ and $\tan\beta$ [1]. Studies concerning the semileptonic and the fully hadronic final state are not included in this note. These studies are ongoing and will be published separately.

This note is organized as follows. In Section 2 the signal and background processes are introduced and their cross-sections are discussed. In Section 3 the analysis is discussed. After a description of the selection, a procedure to estimate the shape and the normalization of the irreducible $Z \to \tau^+\tau^-$ background from data is detailed before the discovery potential in the $m_A - \tan\beta$ plane is assessed. In Section 4 the results are summarized.

# 2 Signal and background processes

## 2.1 Higgs boson production

The production mechanism of Higgs bosons in the MSSM is discussed in the introductory section of this chapter. The Higgs boson masses, their production cross-section and their branching fraction into a pair of $\tau$ leptons are summarized in Table 1.

Table 1: Masses, cross-sections for $b$-associated production, and branching fractions into the $\tau^+\tau^-$ final state for Higgs bosons in the $m_h^{\max}$ scenario and for $\tan\beta = 20$.

| Mass / GeV | | | $\sigma_{h/H/A}^{\text{associated}}$/fb | | | $\mathscr{B}(h/H/A \to \tau^+\tau^-)/\%$ | | |
|---|---|---|---|---|---|---|---|---|
| A | H | h | A | H | h | A | H | h |
| 110 | 129.8 | 109.0 | 314810 | 7579 | 310707 | 8.9 | 9.1 | 8.9 |
| 130 | 134.2 | 124.7 | 189602 | 92897 | 99992 | 9.1 | 9.2 | 9.0 |
| 160 | 160.8 | 128.0 | 97480 | 93102 | 6650 | 9.4 | 9.5 | 8.4 |
| 200 | 200.5 | 128.4 | 45685 | 45095 | 2188 | 9.6 | 9.7 | 7.5 |
| 300 | 300.4 | 128.6 | 10312 | 10253 | 979 | 8.2 | 9.5 | 6.3 |
| 450 | 449.8 | 128.6 | 2019 | 2035 | 723 | 6.1 | 6.2 | 5.7 |

The theoretical uncertainty on the inclusive production cross-section, i.e. without imposing any requirements on the $p_T$ of the $b$ jets at generator level, is estimated taking into account contributions from the scale uncertainty and from the uncertainty on the parton distribution functions. The scale uncertainty is obtained from Ref. [2] as a function of the mass of the Higgs boson. The contribution from the Parton Density Functions (PDFs) is estimated by exchanging MRST2002 for MRST2004 parton distribution functions. Since the cross-sections obtained with MRST2004 are smaller than that with MRST2002 they are considered conservative. Therefore, half of the difference observed with this variation is taken as a systematic uncertainty.

The total uncertainty on the cross-section, obtained by adding the PDF and scale uncertainties in quadrature, is displayed in Fig. 1 as a function of the Higgs boson mass $m_A$. For Higgs boson masses as low as $m_A = 100$ GeV the total theoretical uncertainty is of the order of 20%. This uncertainty decreases to values below 10% for $m_A = 400$ GeV. For low Higgs boson masses the contribution from the scale uncertainty dominates over that from the parton distribution functions while for high Higgs boson masses the situation is reversed.

## 2.2 Background processes

The following background processes are relevant and have been considered in this analysis (for details see Ref. [3]).

- $Z \to \ell\ell$: The Drell-Yan production of $Z$ bosons and their subsequent decay into a pair of leptons constitutes an important source of background. The production cross-section has been calculated to NNLO accuracy and was found to be $\sigma_{Z\to\ell\ell} = (2015 \pm 60)$pb[1] .

  Events with the $Z$ boson decaying to a pair of $\tau$ leptons constitute an irreducible background. In particular for low Higgs boson masses, due to the limited invariant mass resolution in the $\tau^+\tau^-$ final state this background is problematic and needs therefore to be estimated directly from data.

- $t\bar{t}$ production: The cross-section for this process has been calculated at NLO+NLL accuracy and was found to be $\sigma_{t\bar{t}} = (833 \pm 100)$ pb. This background is dominant for Higgs boson masses beyond $m_A = 200$ GeV.

[1]The cross-section is given for a cut on the invariant mass of the lepton pair of $m_{\ell\ell} > 60$ GeV.

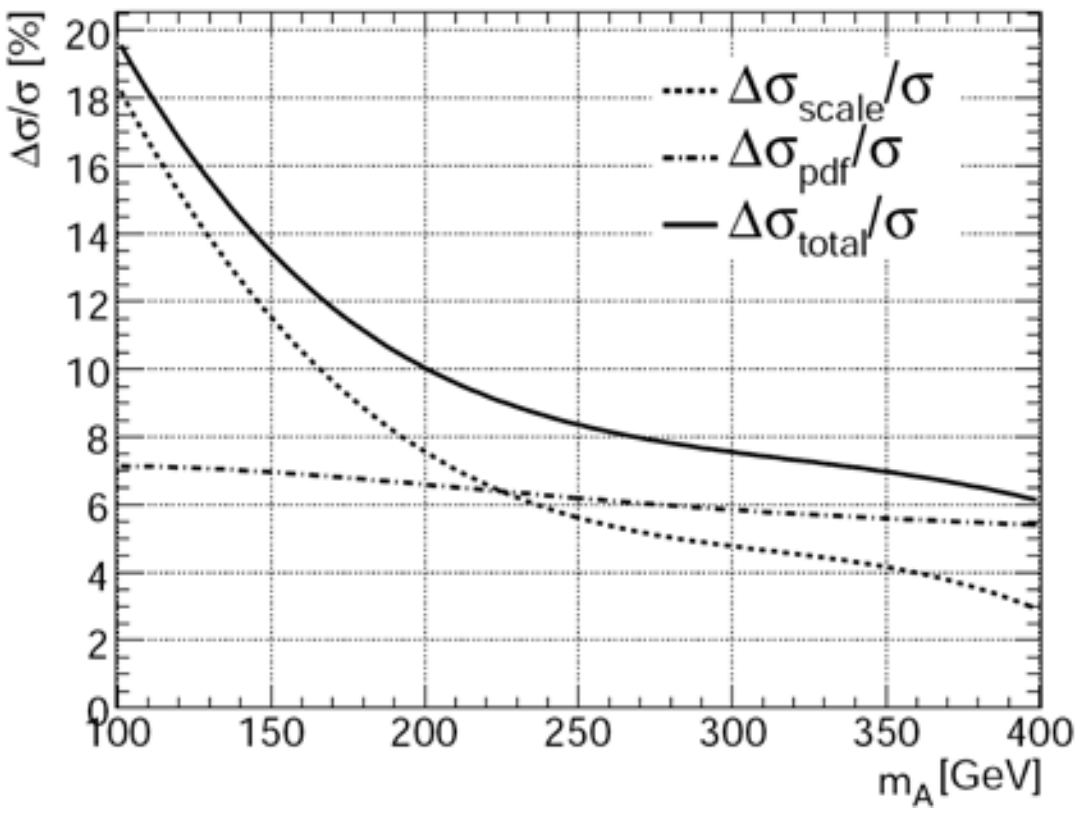


Figure 1: Systematic uncertainty on the signal cross-section for associated Higgs boson production as a function of $m_A$. The dashed line corresponds to the contribution from the scale uncertainty, the dash-dotted line to that introduced by the uncertainty on the parton distribution functions, and the solid line is the sum of both contributions added in quadrature.

- $W$+jets production: The production cross-section for this process has been calculated at NNLO accuracy and was found to be $\sigma_{W+\text{Jets}} = 20510\,\text{pb}$. This dataset was complemented by a $Wb\bar{b}$ sample whose cross-section has been calculated to NLO accuracy ($\sigma_{Wb\bar{b}} = 176.9\,\text{pb}$).

### 2.3 Event generation

The Monte Carlo samples have been generated using the SHERPA [4], PYTHIA [5], HERWIG [6], ALPGEN [7], and MC@NLO [8] Monte Carlo generators. Except for SHERPA, all external matrix element generators are interfaced to HERWIG to produce the parton shower. The $\tau$ leptons are decayed using either SHERPA or TAUOLA [9]. Initial and final state radiation of photons is simulated using PHOTOS [10]. Event filters have been applied for all processes in order to increase the event generation efficiency. Details on Monte Carlo simulation are given in Ref. [3].

## 3 Analysis of the exclusive lepton lepton final state

The experimental signature consists of two leptons from the $\tau$ decays and missing transverse energy, $\not{E}_T$, due to the neutrinos from the $\tau$ decays. At least one jet tagged as coming from a $b$ quark is required in the event and therefore the $b$ quark associated production is dominant here.

### 3.1 Preselection

The preselection cuts are grouped into 'Trigger Selection', '$b$-Tagging', 'Lepton Selection', and cuts related to the reconstruction of the invariant mass of the Higgs boson.

**Trigger selection:** An isolated muon (electron) with transverse momentum of at least 20 GeV (25 GeV) or two isolated electrons with $p_T \geq 15$ GeV, or one electron with $p_T \geq 15$ GeV and a muon with $p_T \geq 10$ GeV are required.

**b-Tagging:** Since the Higgs boson is produced in association with $b$ quarks, at least one jet has to be identified as coming from a $b$ quark in order to suppress backgrounds from processes involving light

quarks. Jets are reconstructed using a cone algorithm with radius $\Delta R = 0.4$[2], and a $b$-tagging weight of $\geq 3$ is required in order for the jet to be labeled as coming from a $b$ quark [11].

**Lepton selection:** Electrons are required to have $p_T \geq 10$ GeV, be within $|\eta| < 2.5$ and pass the medium electron selection [12]. No isolation criteria are applied. Muons are reconstructed using the algorithm described in Ref. [13]. A minimum $p_T$ of 10 GeV and $|\eta| < 2.5$ are required. An isolation cone of opening angle $\Delta R = 0.2$ around the muon track is used with a maximum $E_T$ of 6 GeV deposited in the calorimeters. Two leptons of opposite charge are required if the event is to be considered for further analysis. In case more than two leptons fulfill the above requirements the lepton pair with the highest scalar sum of $p_T$ is selected.

**Collinear approximation:** The invariant mass of the Higgs boson candidate is reconstructed using the collinear approximation [14]. In this approximation the masses of the particles involved in the decay of the $\tau$ lepton are small compared to their momenta, so that the direction of the $\tau$ lepton can be approximated by the direction of its observed visible decay products. The method assumes furthermore that the missing energy observed in the event is entirely due to neutrinos from the $\tau$ lepton decays. In addition the Higgs boson is required to have some amount of transverse momentum. If that is not the case the two $\tau$ leptons from the Higgs boson decay are back-to-back. An accurate reconstruction of the transverse momenta of the $\tau$ leptons is not possible in that case and the resolution of the invariant mass of the $\tau$ pair will be poor. Neglecting the masses of all leptons, the invariant mass of the Higgs boson candidate can be reconstructed via

$$m_{\tau^+\tau^-} = \frac{m_{\ell\ell}}{\sqrt{x_1 x_2}}. \tag{1}$$

The quantity $x_i = p_{T\,\ell_i}/p_{T\,\tau_i}$ is the fraction of the $\tau$ lepton momentum carried by its visible decay products. They are calculated from $\not{E}_T$ in the event and the transverse momentum of the visible leptons. For this calculation $\not{E}_T$ is decomposed into two components, each of them pointing along the direction of the charged decay products of the $\tau$ lepton. This fraction is required to be within physical bounds ($0 < x_i < 1$). In order for the solution to be numerically stable a cut on the angle between the visible decay products of the $\tau$ lepton of $\Delta\phi_{\ell\ell} < 3$ is imposed. This also improves the invariant $m_{\tau\tau}$ resolution.

The accepted cross-section for the above preselection is detailed in Table 2 for the signal and the dominant background contributions.

Table 2: Cross-section in fb passing the preselection criteria as described in the text. The numbers for the signal samples are given assuming $\tan\beta = 20$.

| Process | Trigger | Lepton Selection | 1 or 2 jets | $> 0$ b-tags | Coll. Approx |
|---|---|---|---|---|---|
| $m_A = 110$ GeV | 1837.6 | 1154.8 | 628.6 | 175.7 | 118.4 |
| $m_A = 130$ GeV | 1511.8 | 971.6 | 544.6 | 172.1 | 115.7 |
| $m_A = 160$ GeV | 987.4 | 656.4 | 374.9 | 119.1 | 80.8 |
| $m_A = 200$ GeV | 497.9 | 340 | 199.3 | 63.9 | 44.9 |
| $m_A = 300$ GeV | 139.4 | 98.8 | 60.2 | 20.4 | 13.9 |
| $m_A = 450$ GeV | 25.3 | 18.3 | 11.2 | 3.7 | 2.4 |
| $t\bar{t}$ | 255114 | 48045.9 | 7804.8 | 5479 | 1096.2 |
| $Z \to \tau\tau$ | 47026.8 | 27654.4 | 14053.2 | 665.1 | 440.6 |
| $Z \to ee$ | 1.4 E6 | 797747 | 421393 | 16197.8 | 2848.4[3] |
| $Z \to \mu\mu$ | 1.3 E6 | 704275 | 345491 | 16811.5 | 3223.4[3] |
| $W$+Jets | 17.2 E6 | 91042.8 | 44612.4 | 1537.5 | 122.4[3] |

[2] The radius $\Delta R$ of the cone is defined as $\Delta R = \sqrt{\Delta\phi^2 + \Delta\eta^2}$

[3] Results were obtained using cut factorization.

### 3.2 Event selection

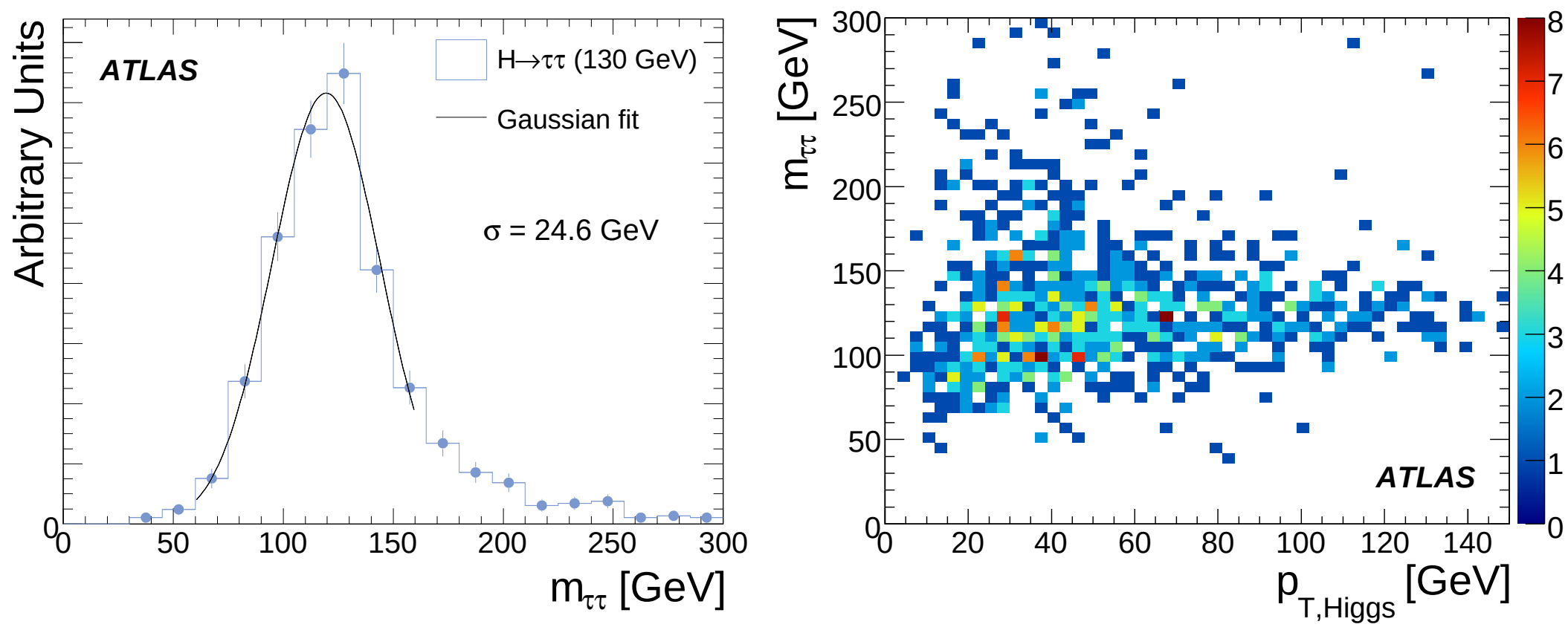


Figure 2: Invariant $m_{\tau^+\tau^-}$ distribution for a Higgs boson of mass $m_A = 130$ GeV (left) after preselection cuts. The width has been determined by a fit of a single Gaussian to the peak region of the distribution. The right-hand plot shows the $m_{\tau^+\tau^-}$ distribution as a function of the $p_T$ of the Higgs boson.

Cuts on kinematic variables have been optimized in an iterative procedure in order to maximize the statistical significance $S/\sqrt{B}$ for a potential Higgs boson signal. Since the composition of the background depends on the signal mass hypothesis, this has been done separately for each mass point. In addition, the optimization has been done separately for the $ee$, $\mu\mu$, and for the mixed $e\mu$ final states.

The following variables are considered: To suppress background from $t\bar{t}$ production only events with less than three jets are selected. The invariant dilepton mass $m_{\ell\ell}$ has to be well below $m_Z$ in order to suppress $Z \to ee$ and $Z \to \mu\mu$ events. Since there are neutrinos from the $\tau$ decays in the event, missing transverse energy is required in the event. The transverse momentum of the jet tagged as coming from a $b$ quark has to be above a certain value which depends on the Higgs boson mass under study. Requirements on the maximum $p_T$ of the leading lepton as well as on that of the lepton-lepton system $p_{T,\ell\ell}$ are imposed. In addition, the angle $\Delta\phi$ between the two leptons is restricted. The cut values and the accepted cross-section for a Higgs boson mass of $m_A = 130$ GeV are given in Table 3.

The resolution of the $m_{\tau^+\tau^-}$ distribution using the collinear approximation for a Higgs boson of mass $m_A = 130$ GeV is illustrated in Fig. 2 (left) after applying preselection cuts. The width as extracted from a fit of a single Gaussian to the peak region is $\sigma = 25$ GeV compared to the natural width of a Higgs boson of that mass between less than 100 MeV up to a few GeV depending on $\tan\beta$. The $m_{\tau^+\tau^-}$ distribution versus the $p_T$ of the Higgs boson is illustrated on the right-hand side of Fig. 2, showing that the invariant mass resolution improves with $p_T$ of the Higgs boson.

### 3.3 Selection results

The accepted cross-sections after all cuts for different Higgs boson masses and the various backgrounds are summarized in Table 4 for $\tan\beta = 20$. The reconstructed $\tau\tau$ invariant mass distributions are displayed in Fig. 3 for all masses considered. The vertical solid lines mark the mass window defined as $m - 1.65\sigma < m_{\tau^+\tau^-} < m + 2\sigma$ where $\sigma$ denotes the invariant $m_{\tau^+\tau^-}$ resolution for a given mass hypothesis as determined from Monte Carlo simulations. All candidate events falling inside this mass window are used to calculate the significances for a Higgs boson signal.

In the low mass range, the invariant mass resolution is of the order of 25 GeV. A potential signal for a Higgs boson is nearly indistinguishable from the irreducible $Z \to \tau^+\tau^-$ background which dominates in this mass range over the contribution from $t\bar{t}$ processes, which contribute at the $(10-20)\%$ level. This

makes it necessary to estimate the shape and the normalization of the $Z \rightarrow \tau^+\tau^-$ background directly from data. A procedure has been developed and will be discussed in detail in Section 3.5.

Table 3: Accepted cross-section in fb for optimized cuts for $m_A = 130$ GeV and $\tan\beta = 20$. The values of the cuts applied are stated for the $ee/\mu\mu$ (upper row) and $e\mu$ (lower row) subchannels. In case only one number is given it applies to all leptonic subchannels.

| Variable | Selection | $H \rightarrow \tau\tau$ | $t\bar{t}$ | $Z \rightarrow \tau\tau$ | $Z \rightarrow ee$ | $Z \rightarrow \mu\mu$ | $W$+Jets |
|---|---|---|---|---|---|---|---|
| Precuts | | 115.7±5.1 | 1096±35 | 441±16 | 3223±123 | 2848±108 | 122±40 |
| $p_T$ $b$-jet | $(15-66)$ GeV | 90.0±4.5 | 443±22 | 337±14 | 2756±113 | 2481±101 | 91±35 |
| $m_{\ell\ell}$ | $(27-70)$ GeV<br>$(0-70)$ GeV | 72.6±4.1 | 138±12 | 326±14 | 134±25 | 92±19 | 60±28 |
| $x_1 \cdot x_2$ | $(0.04-0.4)$<br>$(0.0-0.5)$ | 64.1±3.8 | 108±11 | 251±12 | 47±15 | 36±12 | 40±23 |
| $p_{\mathrm{T}}^{\mathrm{miss}}$ | $(20-\infty)$ GeV<br>$(15-\infty)$ GeV | 52.2±3.5 | 102±11 | 171±10 | 4.3±4.5 | 5.1±4.6 | 33±21 |
| $p_{\mathrm{T}}^{H}$ | $(0-\infty)$ GeV<br>$(0-70)$ GeV | 47.7±3.3 | 57.9±7.9 | 159.4±9.8 | 4.3±4.5 | 4.6±4.3 | 28±19 |
| $p_{\mathrm{T},\ell\ell}$ | $(0-45)$ GeV<br>$(0-60)$ GeV | 46.5±3.3 | 38.2±6.5 | 155.8±9.6 | 3.9±4.3 | 2.5±3.2 | 23±17 |
| $\Delta\Phi_{\ell\ell}$ | $(2.24-3)$<br>$(2-3)$ | 43.3±3.1 | 32.8±6.0 | 107.5±8.0 | 3.6±4.1 | 3.8±4.0 | 21±17 |
| $p_T$ leading $\ell$ | $(10-80)$ GeV | 43.3±3.1 | 32.8±6.0 | 107.5±8.0 | 3.6±4.1 | 3.8±4.0 | 21±17 |
| Mass Window | $(111-198)$ GeV | 28.4±2.6 | 19.7±4.6 | 22.1±3.6 | 1.8±2.9 | 2.0±2.8 | 12±12 |

Table 4: Accepted cross-section for all Higgs boson mass hypotheses analyzed. The cross-section in fb for signal and background after all selection cuts is given (except for the cut on the mass window) for $\tan\beta = 20$.

| | $H \rightarrow \tau^+\tau^-$ | $t\bar{t}$ | $Z \rightarrow \tau^+\tau^-$ | $Z \rightarrow e+e^-$ | $Z \rightarrow \mu^+\mu^-$ | $W$+jets |
|---|---|---|---|---|---|---|
| $m_A = 110$ GeV | 34.4 ± 2.9 | 24.0 ± 5.1 | 62.1 ± 6.1 | 1.9 ± 3.0 | 2.7 ± 3.3 | 8 ± 11 |
| $m_A = 130$ GeV | 28.4 ± 2.6 | 19.7 ± 4.6 | 22.1 ± 3.6 | 1.8 ± 2.9 | 2.0 ± 2.8 | 12 ± 12 |
| $m_A = 160$ GeV | 18.7 ± 1.2 | 39.3 ± 6.6 | 8.4 ± 2.2 | 1.4 ± 2.5 | 2.0 ± 2.9 | 1.8 ± 4.9 |
| $m_A = 200$ GeV | 10.9 ± 0.6 | 28.4 ± 5.6 | 5.4 ± 1.8 | 2.0 ± 3.1 | 2.1 ± 2.9 | 3.7 ± 7.0 |
| $m_A = 300$ GeV | 2.7 ± 0.1 | 32.8 ± 6.0 | 3.0 ± 1.3 | 0.4 ± 1.4 | 1.7 ± 2.6 | 5.8 ± 8.8 |
| $m_A = 450$ GeV | 0.50 ± 0.03 | 50.2 ± 7.4 | 1.8 ± 1.0 | 0.4 ± 1.4 | 0.3 ± 1.1 | 4.1 ± 7.3 |

In the medium mass range, the contributions from $Z \rightarrow \tau^+\tau^-$ events and from $t\bar{t}$ processes become equally important. The mass resolution for signal events is now of the order of $(30-40)$ GeV leading to a broad structure which is indistinguishable from that of background events.

In the high mass range ($m_A = 300$ to 450 GeV), the cross-section for the signal process decreases rapidly. The invariant mass resolution for signal events is between $(50-80)$ GeV so that a discovery with an integrated luminosity of $\mathscr{L} = 30\,\mathrm{fb}^{-1}$ in this channel will not be possible.

### 3.4 Systematic uncertainties

In order to assess the impact of systematic uncertainties, each of these has been applied in turn and the impact on the result of the analysis is evaluated. The uncertainties assumed on the energy and momentum resolution of muons, electrons, photons, and jets are conservative estimates assuming non-optimal performance of the corresponding algorithms at the beginning of data taking.

1. For muons the uncertainty on the reconstructed $p_T$ is $\sigma(1/p_T) = (0.011/p_T \otimes 0.00017)$ with $p_T$

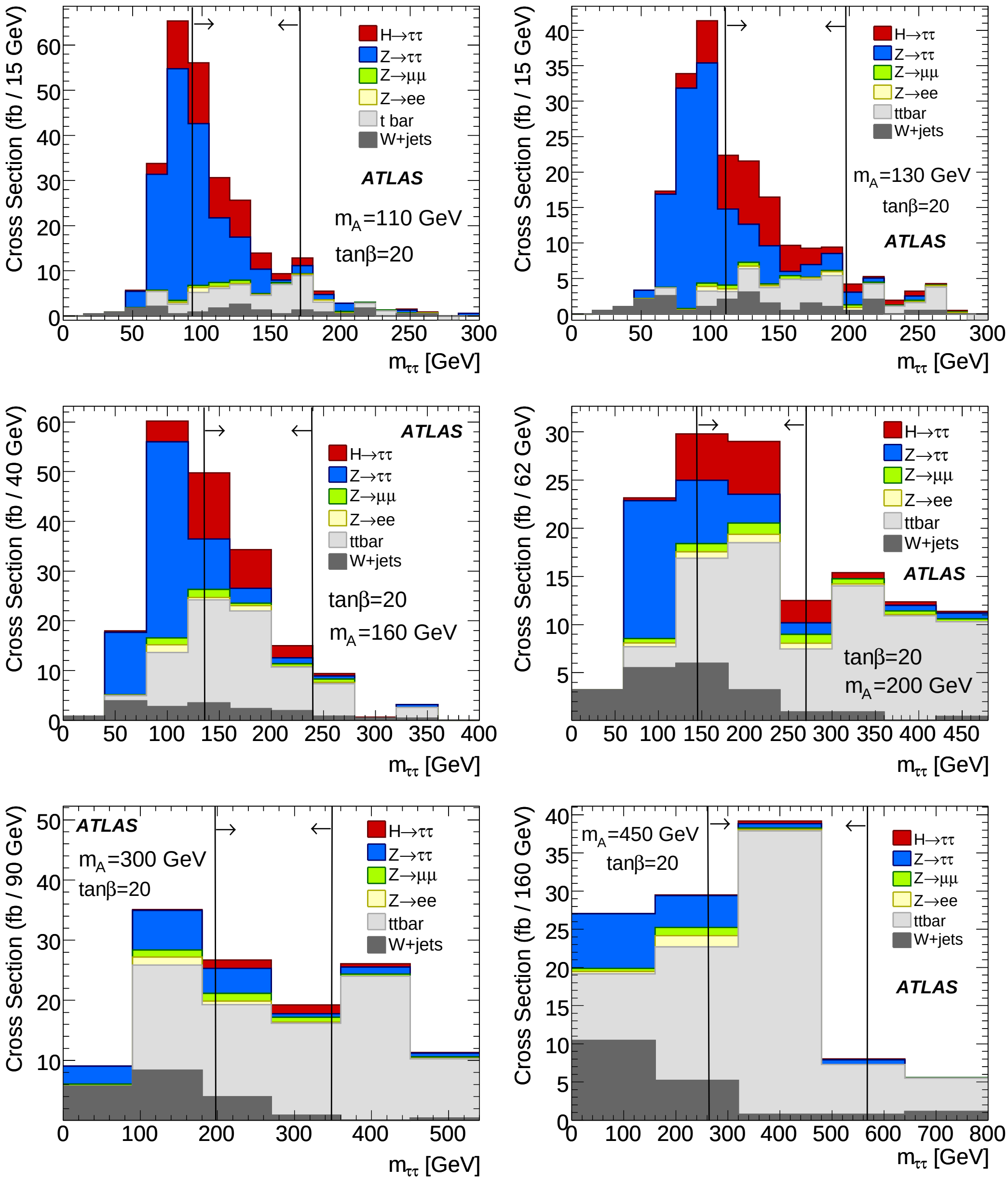


Figure 3: Invariant $m_{\tau^+\tau^-}$ distribution for signal and background events. The distributions are shown after all selection cuts with the nominal masses and $\tan\beta$ values as indicated in the plots. The vertical lines indicate the mass window used for calculating the signal significance.

Table 5: Effects of systematic uncertainties as described in the text for different Higgs boson masses for $\tan\beta = 20$. The effects of the systematic uncertainties considered are listed in percent. The total uncertainty is obtained by adding the individual contributions in quadrature.

| | $m_A = 110$ GeV | | | $m_A = 130$ GeV | | | $m_A = 160$ GeV | | |
|---|---|---|---|---|---|---|---|---|---|
| Uncertainty / % | Signal | $t\bar{t}$ Bkg | $W$+jets | Signal | $t\bar{t}$ Bkg | $W$+jets | Signal | $t\bar{t}$ Bkg | $W$+jets |
| $b$-tagging efficiency | 2.8 | 3.9 | 0.8 | 4.0 | 4.1 | 0.7 | 6.6 | 3.3 | 0.5 |
| Jet energy scale | 0.3 | 6.0 | 0.9 | $< 0.1$ | 5.3 | 1.6 | 1.1 | 4.0 | 1.5 |
| Jet resolution | 8.3 | 0.8 | 0.3 | $< 0.1$ | 0.2 | 1.2 | 6.1 | 0.4 | 1.2 |
| Electron energy scale | 0.7 | 0.5 | 1.0 | 0.4 | 0.1 | 1.0 | 0.4 | 0.5 | 0.9 |
| Electron resolution | 0.7 | 0.6 | 1.0 | 1.6 | 0.2 | 1.0 | 0.4 | 0.2 | 0.4 |
| Muon energy scale | 0.7 | 0.9 | 0.7 | 1.2 | 0.6 | 0.7 | 0.4 | 0.4 | 0.7 |
| Muon resolution | 1.4 | 0.6 | 2.5 | 0.8 | 1.1 | 2.4 | 1.7 | 0.4 | 2.4 |
| Electron efficiency | $< 0.1$ | 0.5 | 0.4 | $< 0.1$ | 0.5 | 0.4 | $< 0.1$ | 0.4 | 0.4 |
| Muon efficiency | $< 0.1$ | 0.8 | 0.7 | 0.8 | 0.2 | 0.7 | $< 0.1$ | 0.9 | 0.7 |
| Light jet rejection | $< 0.1$ | $< 0.1$ | 3.4 | 0.4 | $< 0.1$ | 3.4 | 0.7 | $< 0.1$ | 3.4 |
| Total exp. uncertainty | 9 | 7.4 | 4.7 | 4.6 | 6.8 | 4.9 | 9.2 | 5.3 | 4.8 |

| | $m_A = 200$ GeV | | | $m_A = 300$ GeV | | | $m_A = 450$ GeV | | |
|---|---|---|---|---|---|---|---|---|---|
| Uncertainty / % | Signal | $t\bar{t}$ Bkg | $W$+jets | Signal | $t\bar{t}$ Bkg | $W$+jets | Signal | $t\bar{t}$ Bkg | $W$+jets |
| $b$-tagging efficiency | 4.8 | 3.2 | 0.5 | 4.2 | 3.1 | 0.3 | 4.8 | 3.8 | 0.1 |
| Jet energy scale | 0.6 | 3.6 | 1.5 | 0.9 | 3.1 | 0.6 | 0.4 | 2.1 | 0.7 |
| Jet resolution | 8.0 | 1.6 | 2.7 | 0.8 | 0.4 | 2.7 | 0.7 | 0.5 | 2.4 |
| Electron energy scale | 0.5 | 0.5 | 0.9 | $< 0.1$ | 0.2 | 0.8 | 0.4 | 0.2 | 1.2 |
| Electron resolution | 0.3 | 0.1 | 0.4 | $< 0.1$ | $< 0.1$ | 0.3 | $< 0.1$ | 0.4 | 0.3 |
| Muon energy scale | 0.5 | 0.4 | 0.7 | 0.5 | 0.4 | 0.7 | 0.9 | 0.3 | 0.9 |
| Muon resolution | 0.3 | 0.1 | 2.4 | 0.8 | 0.3 | 2.3 | 0.4 | 0.7 | 2.2 |
| Electron efficiency | 0.3 | 0.3 | 0.4 | 0.8 | $< 0.1$ | 0.4 | 0.4 | 0.1 | 0.4 |
| Muon efficiency | 0.3 | 1.0 | 0.7 | 0.8 | 1.3 | 0.7 | 1.1 | 1.0 | 0.7 |
| Light jet rejection | 0.8 | $< 0.1$ | 3.4 | 0.4 | $< 0.1$ | 3.4 | 0.4 | $< 0.1$ | 3.4 |
| Total exp. uncertainty | 9.4 | 5.3 | 5.4 | 4.6 | 4.7 | 5.1 | 5.1 | 4.6 | 5.1 |

given in GeV. The uncertainty on the energy scale is estimated to be ±1%, and that on the reconstruction efficiency is assumed to be 1% and flat in $p_T$.

2. For electrons and photons the uncertainty on the reconstructed $E_T$ is $\sigma(E_T) = 0.0073 \cdot E_T$. The uncertainty on the energy scale is estimated to be ±0.5%, and that on the reconstruction efficiency is assumed to be 0.2% and being flat in $E_T$.

3. For jets with $|\eta| < 3.2$ ($|\eta| > 3.2$) the uncertainty on the jet energy scale is taken to be ±3% (±10%) and the jet energy resolution is assumed to be $45\%\sqrt{E}$ ($63\%\sqrt{E}$).

4. For the $b$-tagging a degradation of the tagging efficiencies of 5% is taken as systematic uncertainty. For the $Z$+light jets background an uncertainty on the rejection rate of ±10% is assumed.

A detailed description of the sources of systematic uncertainties can be found in [3]. The impact of the systematic uncertainties on the number of signal and $t\bar{t}$ background events[4] inside the mass window is summarized in Table 5 for Higgs boson masses between $m_A = 110$ and 450 GeV. A sample of $48M$ $t\bar{t}$ events from fast simulation has been used for these studies. The $Z \to \tau\tau$ background in this analysis is estimated from sidebands in data as described in the next Section. The number of events from $Z \to \mu\mu$ and $Z \to ee$ processes after all selection cuts is small compared to that from $t\bar{t}$. Their contribution to the total systematic uncertainty is small compared to that from $t\bar{t}$ processes.

### 3.5 Estimation of $Z \to \tau^+\tau^-$ shape and normalization from data

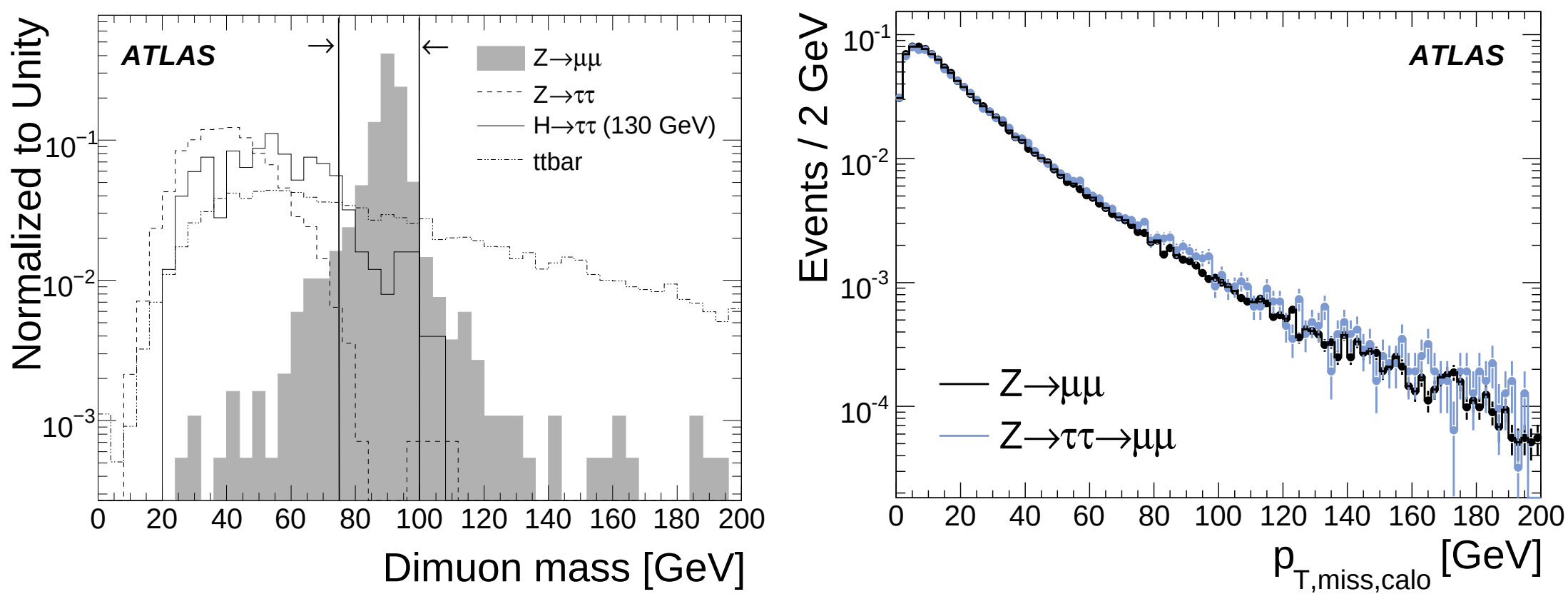


Figure 4: Distribution of the invariant dilepton mass $m_{\ell\ell}$ (left). The shaded area illustrates the distribution from $Z \to \mu\mu$ events, and the solid line that from signal events. The contribution from $Z \to \tau^+\tau^-$ and $t\bar{t}$ events is illustrated by the dashed and dotted lines, respectively. The cut on the invariant $\ell\ell$ mass is indicated by the solid vertical lines. The distribution of $p_T^{\text{miss}}$ in the calorimeter for $Z \to \mu\mu$ and $Z \to \tau\tau \to \mu\mu + X$ events (right). All plots shown are after preselection cuts.

As discussed above the estimation of the shape and the normalization of the irreducible $Z \to \tau^+\tau^-$ events from data is of great importance in particular for low Higgs boson masses where this background is dominant. Procedures to estimate both the shape and the normalization of $Z \to \tau^+\tau^-$ events from data are needed. This procedure is based on $Z \to \mu\mu$ and $Z \to ee$ events selected from a sideband region which is free of Higgs boson events, in contrast to the signal region. The method proceeds in three steps:

[4]The impact on the number of $Z \to \tau^+\tau^-$ events is not listed here since this background is estimated from data.

1. As a first step, a pure sample of $Z \to \mu\mu$ or $Z \to ee$ events is selected from a sideband region as described below.
2. The shape of the $m_{\tau\tau}$ spectrum is estimated by adapting the four-momenta of the muons such that they appear like coming from $Z \to \tau\tau$ events.
3. The normalization of the $Z \to \tau\tau$ background is estimated from a double-ratio comparing the number events found in data and in Monte Carlo in the sideband region as well as in the signal region.

This procedure is described in detail below.

### 3.5.1 Definition of signal and sideband regions

Events of the type $Z \to ee$ and $Z \to \mu\mu$ are selected from a sideband region (called region B) with very high purity and with an event topology similar to that from $Z \to \tau^+\tau^-$ events in the signal region (called region A). The following cuts are applied in order to select $Z \to ee$ and $Z \to \mu\mu$ events in region B:

- The invariant mass of the lepton-lepton system is required to be within 75 GeV $< m_{\ell\ell} <$ 100 GeV, where $\ell\ell$ is either a $ee$ or a $\mu\mu$ final state.
- At least one jet identified as coming from a $b$ quark has to be found in the event.
- The number of jets allowed in the event has to be less than three.

The main cut defining region B is that on the invariant lepton-lepton mass. A distribution illustrating the cut on $m_{\ell\ell}$ is displayed in Fig. 4 (left) showing the large amount of $Z \to \mu\mu$ events selected by the cuts above that only have a small contamination of events coming from $t\bar{t}$ processes and events containing Higgs boson decays. The number of events from $Z \to ee$ and $Z \to \mu\mu$ is around 500 000 for an integrated luminosity of $30\,\text{fb}^{-1}$ and thus much larger than the number of $Z \to \tau\tau$ events expected in the signal region. The purity of the $Z \to \mu\mu$ ($Z \to ee$) control sample in region B after all cuts mentioned above is 99.1% (97.9%) with a contribution of 0.04% (0.05%) of events from the signal process[5] with the remainder coming from $t\bar{t}$ processes.

### 3.5.2 Estimation of the $Z \to \tau^+\tau^-$ shape from data

The method of estimating the shape of $Z \to \tau^+\tau^-$ events from data is based on the assumption that in the calorimeter this type of events is indistinguishable from that of type $Z \to \mu\mu$. This method has been proven to work in a vector boson fusion $H \to \tau^+\tau^-$ analysis [15].

The muons are minimum ionizing particles and their energy deposit in the calorimeter only weakly depends on their momentum. Therefore, the missing energy signatures of both types of events in this detector component are very similar as illustrated on the right-hand side of Fig. 4. Altering the energy of muons in $Z \to \mu\mu$ events so that they correspond to those from $Z \to \tau^+\tau^- \to \mu\mu + 4\nu$ events leads to identical distributions of $p_{T,\mu}$, $p_{T,\text{miss}}$ and $m_{\tau^+\tau^-}$ for both classes of events. The following procedure is adopted:

- Three dimensional reference histograms from $Z \to \tau^+\tau^- \to \mu\mu + 4\nu$ events from Monte Carlo in region A are created. The following variables calculated in the $Z$ rest frame are used:

[5] A Higgs boson mass of $m_A = 130$ GeV and $\tan\beta = 20$ is assumed here. The contamination of Higgs boson events is even smaller for the other signal masses considered in this analysis.

1. The absolute value of the Gottfried-Jackson angle $\xi$ between the $Z$ boson and the negatively charged muon.
2. The energy of the muon with $\cos\xi > 0$.
3. The energy of the muon with $\cos\xi < 0$.

- The components of the momentum vector of the muons in $Z \to \mu\mu$ events from region B are altered in a way that they match those from $Z \to \tau^+\tau^- \to \mu\mu + 4\nu$ events using the above reference histogram:

$$p_{i,\text{altered}} = \frac{p_i}{|\vec{p}|} \cdot E_{\mu,\text{altered}}. \quad (2)$$

  For each event, the angle $\xi$ is calculated and then new energies for the muons are chosen randomly from the reference histogram. After applying this procedure, the muon momenta are boosted back into the lab frame.

- The missing energy in the event is re-calculated according to the new muon momenta:

$$\vec{p}_{T,\text{miss, altered}} = \vec{p}_{T,\text{miss}} - \sum_{\mu} p_{T,\text{altered}} + \sum_{\mu} p_{T,\text{old}}. \quad (3)$$

A comparison of $p_{\text{T,miss}}$, of $x_1 \cdot x_2$ from the collinear approximation and of the invariant $m_{\tau^+\tau^-}$ mass from altered $Z \to \mu\mu$ in comparison with $Z \to \tau^+\tau^- \to \mu\mu + 4\nu$ events is shown on the left-hand side, middle and right-hand side of Fig. 5, respectively. Good agreement within the statistical uncertainties of the samples used is observed.

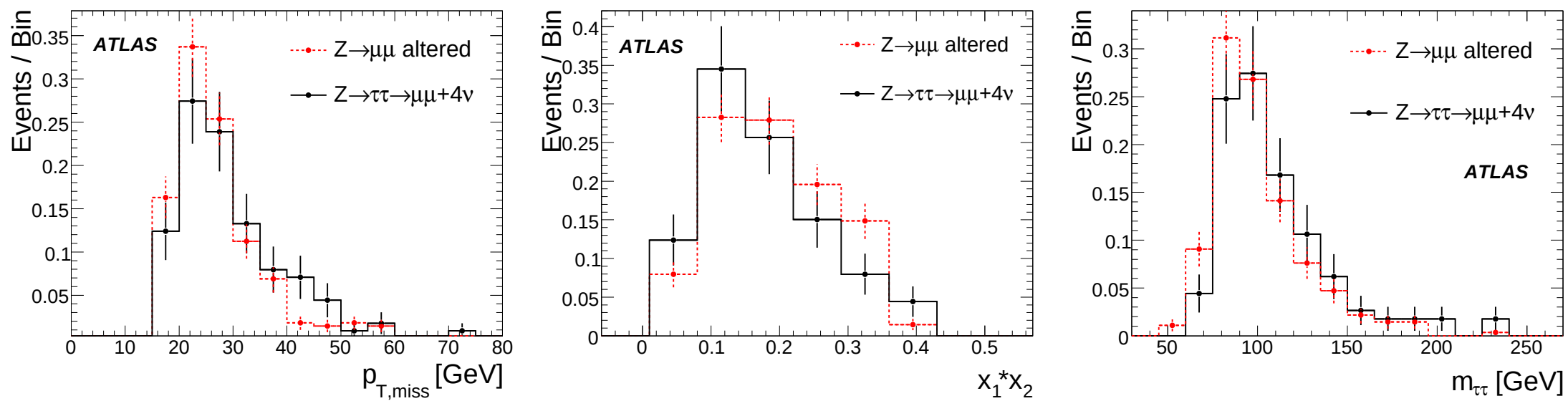


Figure 5: Comparison of the $p_{\text{T,miss}}$, $x_1 \cdot x_1$, and $m_{\tau\tau}$ spectra for events of type $Z \to \mu\mu$ and $Z \to \tau\tau$. Good agreement within the statistical uncertainties is observed.

The same shape as extracted from region B using $Z \to \mu\mu$ events to estimate $Z \to \tau\tau \to \mu\mu + X$ events in region A is also used for $Z \to \tau\tau \to ee + X$ and $Z \to \tau\tau \to \mu e + X$ events. As shown in Figure 6, the $m_{\tau\tau}$ shapes are identical within statistical uncertainties justifying this procedure.

### 3.5.3 Estimation of the background normalization from data

In order to estimate the number of background events from the $Z \to \tau^+\tau^-$ background process for this analysis, the same definition of the sideband region has been used as described above.

The number of $Z \to \tau^+\tau^-$ background events in the signal region A can then be obtained by re-weighting the number of events found in data in region B by the predicted ratio of the number of events found in Monte Carlo in region A relative to that found in region B. In order for this method to be valid, the following two conditions have to hold:

$$\frac{(Z \to \ell\ell)^{\text{B}}_{\text{Data}}}{(Z \to \ell\ell)^{\text{B}}_{\text{MC}}} = \frac{(Z \to \tau^+\tau^- \to \ell\ell + 4\nu)^{\text{B}}_{\text{Data}}}{(Z \to \tau^+\tau^- \to \ell\ell + 4\nu)^{\text{B}}_{\text{MC}}} \quad (4)$$

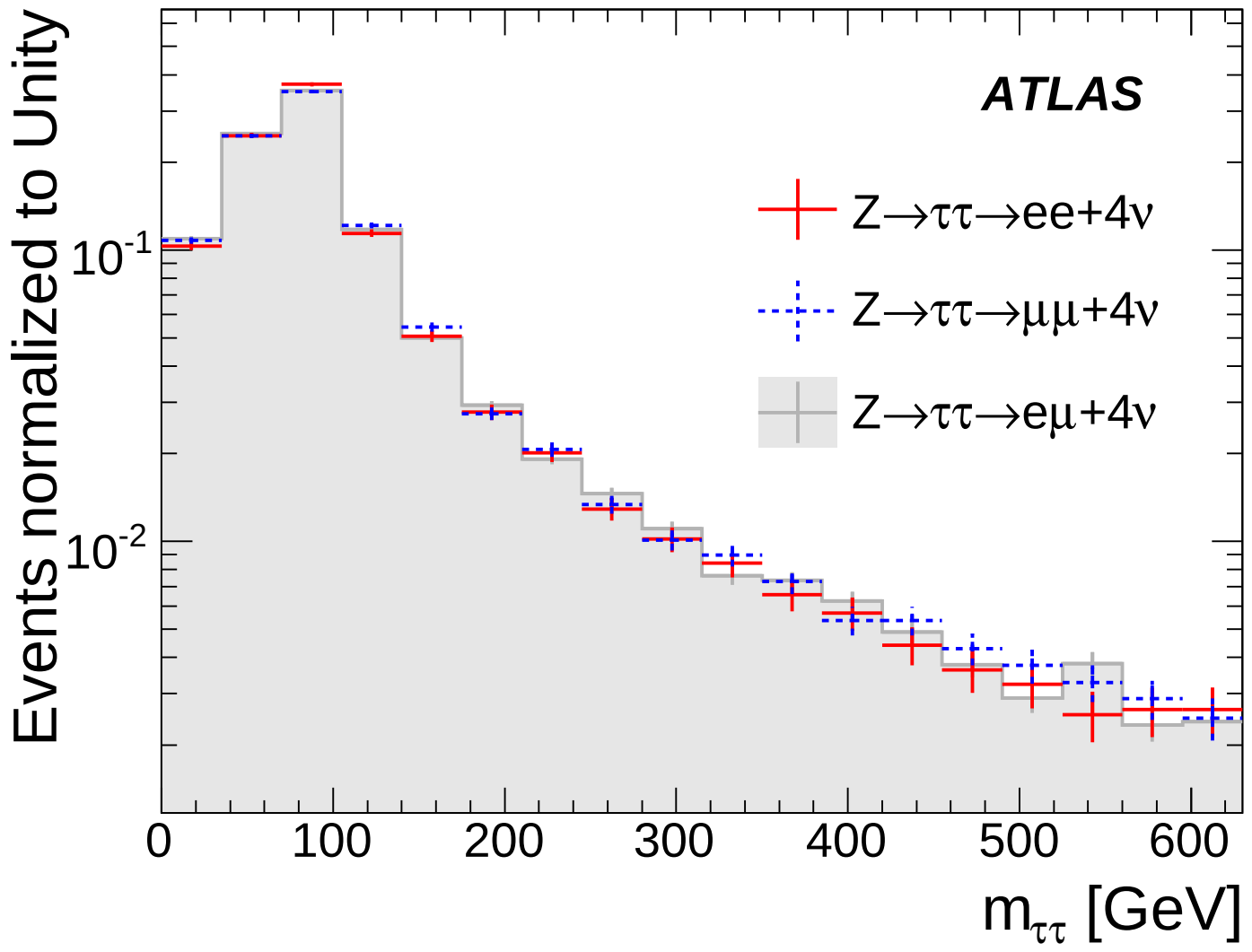


Figure 6: Comparison of the $m_{\tau\tau}$ shape for $Z \to \tau\tau \to \mu\mu + X$, $Z \to \tau\tau \to ee + X$, and $Z \to \tau\tau \to \mu e + X$ events. Good agreement is observed within the statistical uncertainties of the samples used.

$$\frac{(Z \to \ell\ell)^{\mathrm{B}}_{\mathrm{Data}}}{(Z \to \ell\ell)^{\mathrm{B}}_{\mathrm{MC}}} = \frac{(Z \to \tau^+\tau^- \to \ell\ell + 4\nu)^{\mathrm{A}}_{\mathrm{Data}}}{(Z \to \tau^+\tau^- \to \ell\ell + 4\nu)^{\mathrm{A}}_{\mathrm{MC}}}, \tag{5}$$

where the first condition means that $Z \to \ell\ell$ events behave like $Z \to \tau^+\tau^- \to \ell\ell + 4\nu$ events, and when combined with the second it implies that $Z \to \tau^+\tau^- \to \ell\ell + 4\nu$ events in region A and in region B behave identically.

The calculation of the number of events from the $Z \to \tau^+\tau^-$ process in data is performed in bins of $p_T$ of the leading lepton vs. the $p_T$ of the subleading lepton with a bin size of $2 \times 2$ GeV$^2$ which was found to give unbiased results in previous Monte Carlo based studies [16]. This ensures that the method is less dependent on the differences in $p_T$ in the signal and the sideband region. Once events from real data are available the method has to be validated and the influence of a possible difference between data and the Monte Carlo prediction on the results has to be checked.

This procedure also allows to easily take into account differences in acceptance and trigger efficiencies by simply applying the appropriate factors to the reweighting procedure. The number of $Z \to \tau^+\tau^- \to \ell\ell + 4\nu$ events in region A is then given by

$$(Z \to \tau^+\tau^- \to \ell\ell + 4\nu)^{\mathrm{A}}_{\mathrm{Data}} = \sum_{i,j} \frac{\left((Z \to \tau^+\tau^- \to \ell\ell + 4\nu)^{\mathrm{A}}_{\mathrm{MC}}\right)_{ij}}{\left((Z \to \ell\ell)^{\mathrm{B}}_{\mathrm{MC}}\right)_{ij}} \cdot \left((Z \to \ell\ell)^{\mathrm{B}}_{\mathrm{Data}}\right)_{ij}, \tag{6}$$

where $i$ and $j$ indicate the corresponding bin in $p_T$. The statistical uncertainty on this method is calculated according to Gaussian error propagation. The method has been tested using two independent Monte Carlo samples, one to fill the reference histogram and one to test the reweighting procedure. Good agreement within statistical uncertainties between the expected number of events in region A and that actually found was observed.

The application of this method is straight forward for the $ee$ and the $\mu\mu$ final state. For the $e\mu$ final state the procedure has to be adapted since there are no $Z \to e\mu$ decays. In order to estimate the number of background events in that case, events from $Z \to ee$ and $Z \to \mu\mu$ processes are used. Additional correction factors are applied to account for the differences in trigger efficiencies and selection efficiencies in that case.

### 3.5.4 Systematic uncertainties of the background estimation procedure

Table 6: The effects of systematic uncertainties on the calculation of the normalization.

| Uncertainty Effect | Value | Systematic Uncertainty |
|---|---|---|
| Muon resolution | $\sigma\left(\frac{1}{p_T}\right) = \sqrt{\left(\frac{0.011}{p_T}\right)^2 + 0.00017^2}$ | 0.16 % |
| Muon energy scale | $\pm 1\%$ | 1.35 % |
| Muon efficiency | $\pm 1\%$ | 0.7 % |
| Electron resolution | $\sigma(E_T) = 0.0073 \cdot E_T$ | 1.9 % |
| Electron energy scale | $\pm 0.5\%$ | 0.85 % |
| Electron efficiency | $\pm 0.2\%$ | 0.1 % |

The impact of the systematic uncertainties on the estimation of the number of background events from $Z \to \tau^+\tau^-$ processes as described in Section 3.4 has been evaluated. The uncertainties on jet energy scale and jet resolution, as well as on the $b$ tagging efficiency are expected to be negligible since the same effects would apply to the signal region as well as to the sideband region. The remaining contributions to the systematic uncertainty on the background estimation procedure are coming from the energy scale, efficiency and resolution connected with the electrons and muons in the final state. These contributions as expected in $30\,\text{fb}^{-1}$ of data are summarized in Table 6.

In addition, systematic uncertainties due to the different acceptances and trigger efficiencies for the $Z \to ee/\mu\mu$ samples from the sideband region compared to $Z \to \tau\tau \to ee/\mu\mu/e\mu + X$ samples in the signal region need to be taken into account. Since the reweighting of events is done as a function of the $p_T$ of the leading and the subleading lepton, these are easy to apply. Since this analysis is aimed at an integrated luminosity of $30\,\text{fb}^{-1}$, it is assumed that by then these uncertainties are evaluated to a very high precision.

The overall systematic uncertainty has been evaluated to be 2.6% by dividing the available MC events into two independent samples. This uncertainty has been taken into account for the final results.

### 3.5.5 Summary of the background estimate from data

Both methods described above are now combined to estimate the $Z \to \tau\tau$ background and thereby the significance of a possible Higgs boson signal in the low mass region. First, the number of events of type $Z \to \tau^+\tau^-$ is estimated using the method described in Section 3.5.3. Then, the shape of this background is determined as described in Section 3.5.2. The background is then subtracted from the $m_{\tau^+\tau^-}$ distribution.

## 3.6 Results and discovery potential

The discovery potential for the $h/H/A \to \tau^+\tau^- \to \ell\ell 4\nu$ channel is now assessed. All sub-channels, i.e. $ee$, $e\mu$ and $\mu\mu$ are combined to calculate the significance of a Higgs boson signal. A mass window $m - 1.65\sigma < m_{\tau^+\tau^-} < m + 2\sigma$ is applied to calculate the final significance where $\sigma$ denotes the invariant mass resolution of the Higgs boson signal of the corresponding mass.

The shape and the normalization of the $Z \to \tau^+\tau^-$ background is estimated from the sideband region in data as described in Section 3.5. There is no corresponding procedure available yet for the $t\bar{t}$ background so that all experimental systematic uncertainties are taken into account for all calculations below. Theoretical uncertainties are treated separately. The significance of a potential Higgs boson signal in the

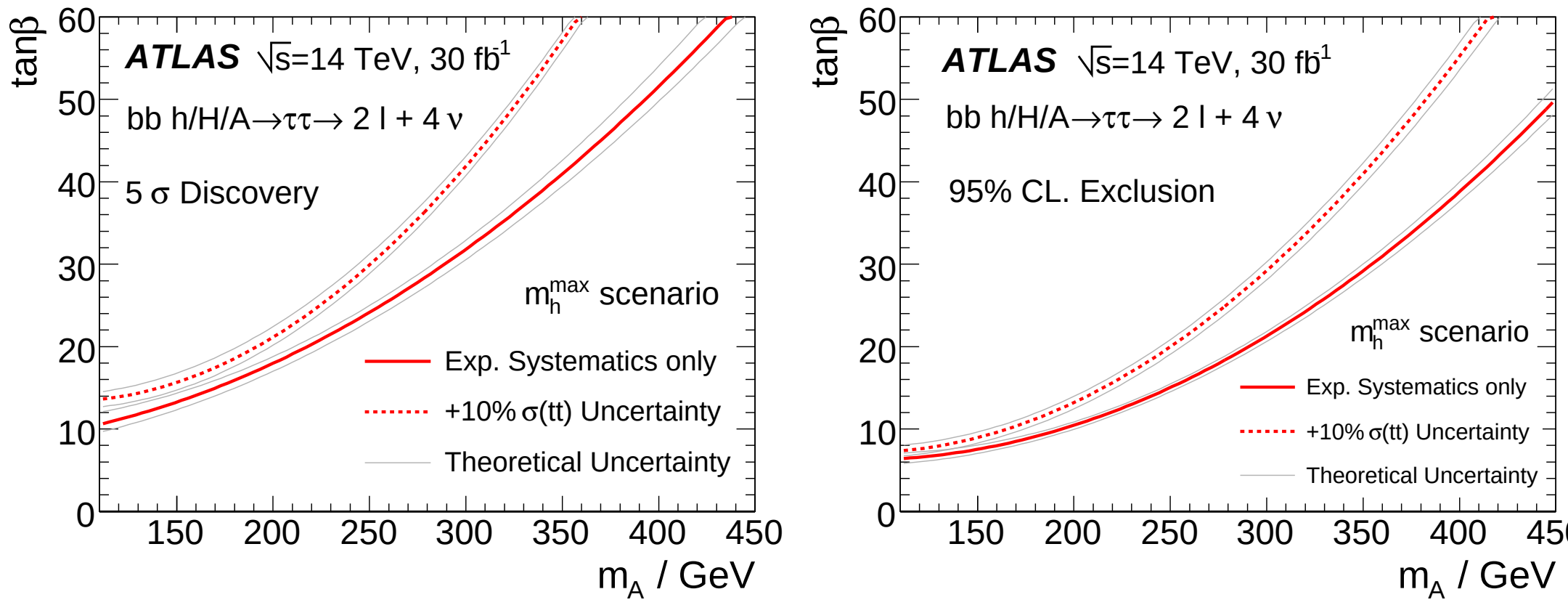


Figure 7: The five $\sigma$ discovery potential (left) and the 95% exclusion limit (right) as a function of $m_A$ and $\tan\beta$. The solid line represents the main result of the analysis. The dashed lines indicate the discovery potential and exclusion limit including an addition 10% uncertainty on the $t\bar{t}$ cross-section. The bands represent the influence of the systematic uncertainty on the signal cross-section.

given mass window is calculated as

$$\text{Sign.} = \frac{S}{\sqrt{N_{t\bar{t}} + (\Delta_{\text{sys}}^{t\bar{t}})^2 + N_{\text{Z}\to\tau\tau} + (\Delta_{\text{sys}}^{Z\to\tau\tau})^2 + N_{\text{W}\to\ell\nu} + (\Delta_{\text{sys}}^{W\to\ell\nu})^2 + N_{\text{Z}\to\text{ee}} + N_{\text{Z}\to\mu\mu}}}, \tag{7}$$

where $S$ is the number of signal events, $N_{t\bar{t}}$ and $\Delta_{\text{sys}}(t\bar{t})$ are the statistical and systematic uncertainties on the $t\bar{t}$ background. The quantities $N_{Z\to\tau^+\tau^-}$ and $\Delta_{\text{sys}}(Z \to \tau^+\tau^-)$ are the statistical and systematic uncertainties on the $Z \to \tau^+\tau^-$ background, respectively. The term in the denominator is dominated by the contribution from $t\bar{t}$ events; the contributions from $N_{\text{Z}\to\text{ee}}$ and $N_{\text{Z}\to\mu\mu}$ and their corresponding systematic uncertainties are negligible.

The discovery potential and the 95% exclusion limit in the $m_h^{\text{max}}$ scenario as a function of $\tan\beta$ and $m_A$ and for an integrated luminosity of $30\,\text{fb}^{-1}$ is displayed in Fig. 7 on the left-hand side and right-hand side, respectively. The uncertainty on the signal cross-section is indicated as bands in the plots. The calculation of the significance includes both statistical and experimental systematic uncertainties on the background. The uncertainties on the background cross-sections are not taken into account in the main results since they are assumed to be measured with high precision at the time of the analysis. However, since the cross-section for the $t\bar{t}$ background might not be measured at high precision in the region of phase space relevant to this analysis, the discovery potential including a 10% uncertainty on that background is displayed in addition (dashed line).

The discovery potential for Higgs bosons is shown in Fig. 8 as a function of $\tan\beta$ for various $m_A$ values, and in Fig. 9 as a function of $m_A$ for various $\tan\beta$ values.

## 4 Conclusion

In this note a study of the discovery potential for the supersymmetric Higgs bosons $h/H/A$ in proton-proton collisions at a center-of-mass energy of 14 TeV with the ATLAS detector at the LHC has been presented. The final state $h/H/A \to \tau^+\tau^-$ in $b$ quark associated production has been investigated with at least one jet identified as coming from a $b$ quark and with both $\tau$ leptons decaying leptonically.

A significant improvement for the discovery potential can be achieved if this channel is combined with the $\ell$-had and had-had channel, where one $\tau$ lepton decays leptonically (either electron or muon)

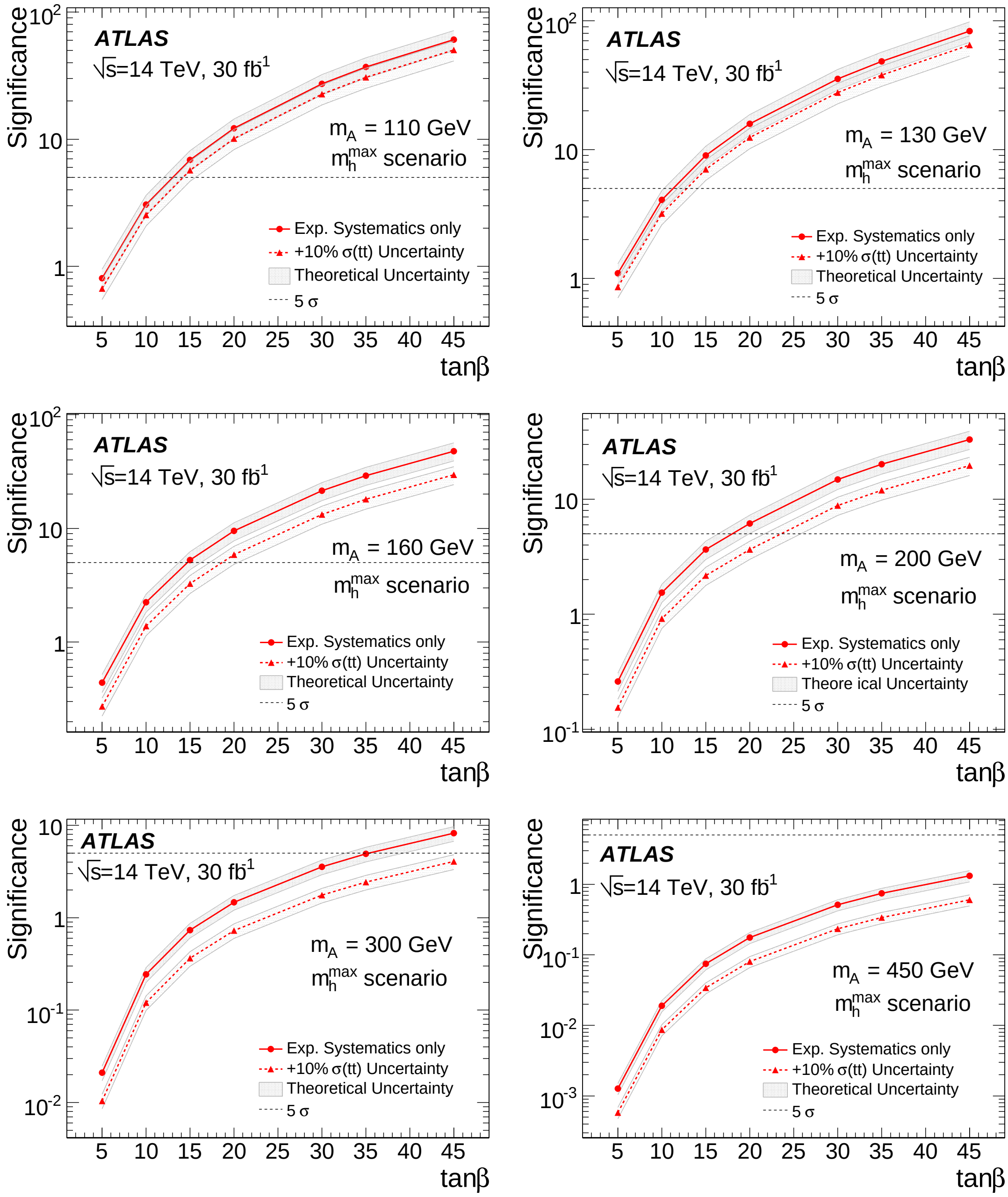


Figure 8: The discovery potential as a function of $\tan\beta$ and for Higgs boson masses as indicated in the plots. The solid lines indicate the discovery potential including experimental systematic uncertainties. The dashed lines indicate the discovery potential including an additional systematic uncertainty on the $t\bar{t}$ cross-section. The bands indicate the impact of the systematic uncertainty on the signal cross-section.

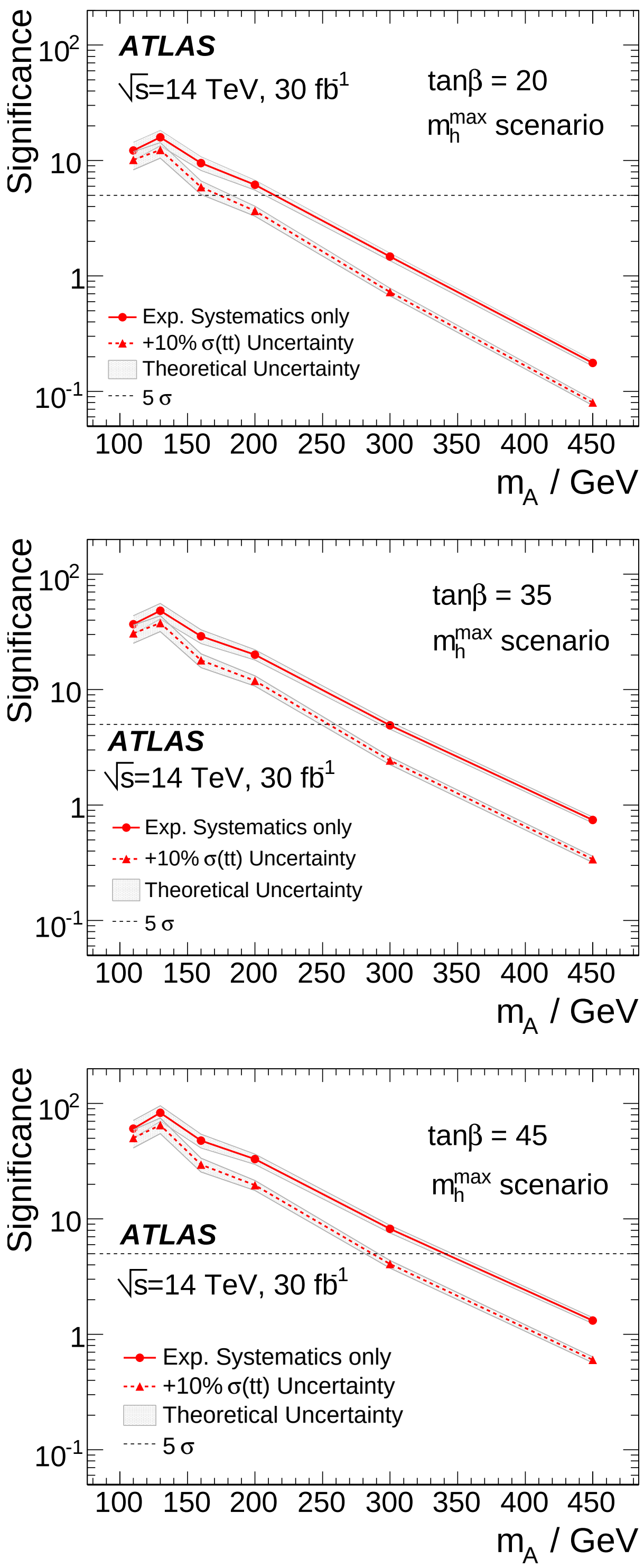


Figure 9: The discovery potential as a function of $m_A$. The solid lines indicate the discovery potential including experimental systematic uncertainties for $\tan\beta$ values as indicated in the plots. The dashed lines indicate the discovery potential including an additional systematic uncertainty on the $t\bar{t}$ cross-section of 10%. The bands indicate the impact of the systematic uncertainty on the signal cross-section.

and one hadronically, or both $\tau$ leptons decay hadronically [17].

# Search for the Neutral MSSM Higgs Bosons in the Decay Channel $A/H/h \rightarrow \mu^+\mu^-$

**Abstract**
Motivated by the high muon momentum resolution and identification efficiency achievable with the ATLAS detector, the observability of $A/H/h \rightarrow \mu^+\mu^-$ channel is explored. The high experimental resolution in this decay mode compensates to some extent for the suppression of the branching ratio, with respect to the $A/H/h \rightarrow \tau^+\tau^-$ decays. The analyses are performed in the Higgs mass range from 100 to 500 GeV. Two main analysis strategies are applied - the search for the dimuon final states resulting from the direct $A/H/h$ production is combined with the search for the associated $b\bar{b}A/H/h$ production mode.

The studies are optimized for the early stage of data taking up to an integrated luminosity of 30 fb$^{-1}$. All results are obtained combining the fast and the full simulation of the ATLAS detector with the nominal detector layout and the trigger efficiencies included. In addition, dedicated data samples are produced to study the impact of the pile-up and cavern background on the analysis performance. The estimation of the background contribution from the experimental data and the contribution of theoretical and experimental uncertainties are also addressed. The discovery potential is shown in the $m_A$-$\tan\beta$ plane in context of the $m_h^{max}$ MSSM benchmark scenario.

## 1 Introduction

In the framework of the Standard Model, the observability of the Higgs boson in the decay channel $H \rightarrow \mu^+\mu^-$ is very unlikely, since the branching ratio for the Higgs decay into muons is very small and the backgrounds from several Standard Model processes are large. As opposed to the Standard Model predictions, the decay of neutral MSSM Higgs bosons $A$, $H$ and $h$ into two muons is strongly enhanced in the MSSM for large values of $\tan\beta$ and can be used either as a discovery channel or for the exclusion of a large region of the $m_A$-$\tan\beta$ parameter space (see Ref [1]).

Compared to the dimuon channel, the $A/H/h \rightarrow \tau^+\tau^-$ decays have a substantially larger branching ratio which scales as $(m_\tau/m_\mu)^2$ and thus provide a promising discovery signature, as discussed in Ref [2]. Nevertheless, the $\tau$ identification represents an experimental challenge. The $\mu^+\mu^-$ final state, on the other hand, has the advantage of a very clear signature in the detector. Furthermore, a full reconstruction of the Higgs boson final state is possible, which allows for a direct mass measurement. The dimuon channel provides for the most accurate Higgs boson mass measurement.

In this note, the potential for the discovery of the neutral MSSM Higgs bosons is evaluated in the dimuon decay channel. The study concentrates on the region of $(m_A - \tan\beta)$ plane with $m_A$ >110 GeV and intermediate $\tan\beta$ values between 10 and 60, which is still uncovered by the current exprimental limits [3, 4]. A detailed study of the ATLAS discovery potential for this channel has been recently performed in the low mass region below 130 GeV [5]. The study includes also higher Higgs boson masses up to 400 GeV.

In Section 2, the relevant production and decay rates of the MSSM Higgs boson are briefly discussed in the context of the $m_h^{max}$ LHC scenario [6], as well as the production mechanisms of the major background processes. In Section 3, the Monte Carlo simulation and data samples used for the analysis are described. Section 4 provides a short description of the detector performance obtained from the simu-

lation, related to the reconstructed particles which will be present in the final state. The event selection criteria, the resulting efficiency of the signal selection and the corresponding background rejection shall be described in Section 5. Discussion of the systematic uncertainties is presented in Section 6. Section 7 describes the methods for the estimation of different background contributions from the the real data. The obtained results are finally represented by the discovery contours in the ($m_A$-$\tan\beta$) plane (see Section 9). The note concludes with Section 10.

## 2 Signal and background processes

In this Section, the properties of the signal shall be briefly described, as well as the background processes relevant for the MSSM Higgs boson searches in the dimuon final state at the LHC.

### 2.1 Signal production and decays

The characteristic production and decay properties of all MSSM Higgs bosons are determined at tree-level by the values of the two free parameters $\tan\beta$ and the mass $m_A$ of the $A$ boson. These properties have been calculated at NNLO with the Feynhiggs 2.6.2 package [7] in the $m_h^{max}$ scenario, as summarized in Ref [1].

The direct $gg \to A/H/h$ production via the gluon-gluon fusion is an analogue to the Standard Model Higgs boson production. This process is important in the region of low $\tan\beta$ values (below 10), where the Higgs bosons couple most strongly to up-type quarks. For larger values of $\tan\beta$, the rate of the $b\bar{b}A/H/h$ Higgs production in association with b-quarks becomes dominant, due to the enhanced couplings to the b-quarks.

There are two approaches to calculate the signal rates for the associated $b\bar{b}A/H/h$ production mode. In the first approach, the production cross-section for the $gg \to b\bar{b}H$ process has been calculated at NLO in Ref [8, 9]. This calculation is most reliable in the case where both outgoing b-quarks have a high transverse momentum (above ∼15 GeV). The inclusive cross-section without any cut on the transverse momenta of the b-quarks is less accurate, due to additional collinear logarithms which appear in the calculation due to the presence of low-momentum b-quarks. An alternative approach is the calculation of the inclusive cross-section for the $b\bar{b} \to H$ process, for which the collinear logarithms can be absorbed in a parton density function for the b-quarks and resummed to all orders of perturbation theory [10, 11]. This later calculation has been implemented in a parametrized way into the Feynhiggs package, which was finally used for the evaluation of the signal cross-sections, as mentioned previously.

The different higher-order calculations mentioned above have been extensively compared in Ref [12]. The two approaches agree within uncertainties. For the $gg \to b\bar{b}H$ calculation, the uncertainty related to the variation of the renormalization and the factorization scale amounts to 20-30%. For the $b\bar{b} \to H$ calculation, which is used for the analysis, the scale uncertainty is much smaller, less than 10%. However, it should be noted that the uncertainty on the b parton density function has not been included here. In order to estimate this pdf-uncertainty, a calculation of the $b\bar{b} \to H$ cross-section is performed with two different parton density functions, MRST2002 and MRST2004. The observed difference of ∼14% is taken as the estimate of the pdf-uncertainty. Adding the 10% scale uncertainty to this in quadrature, the total theory uncertainty for the signal is estimated to be ∼17% for $H$ boson masses up to 500 GeV.

The production cross-section of $H$ and $A$ bosons increases approximately quadratically with increasing $\tan\beta$, while the $h$ boson production is $\tan\beta$-dependent only for $m_A$ <130 GeV. Also the branching ratio of $H$ and $A$ boson decays into $\mu^+\mu^-$ pairs become enhanced with increasing values of $\tan\beta$. The $h$ boson decay is rather insensitive to the two mentioned parameters. The increase of the cross-sections and branching ratios with $\tan\beta$ make the $A/H/h \to \mu^+\mu^-$ decay channel a promising Higgs signature in MSSM.

Additionally, the signal is enhanced due to the mass degeneracy of the neutral Higgs bosons. For $A$ boson masses $m_A$ <130 GeV, the $h$ and $A$ bosons are degenerate in mass, while the heavy $H$ boson mass is rather constant (∼130 GeV). In case of $m_A$ ≈130 GeV, all three bosons have very similar masses. For $m_A$ >130 GeV, the $h$ boson mass reaches its maximum value of ∼130 GeV, independent of the $A$ boson mass, while the $A$ and $H$ bosons become degenerate in mass. Thus the signal can be observed as the sum of all two or three degenerate mass states.

## 2.2 Background processes

The processes with two muons in the final state, which give a major background contribution in the searches for the $A/H/h$ signal are depicted in Figure 1.

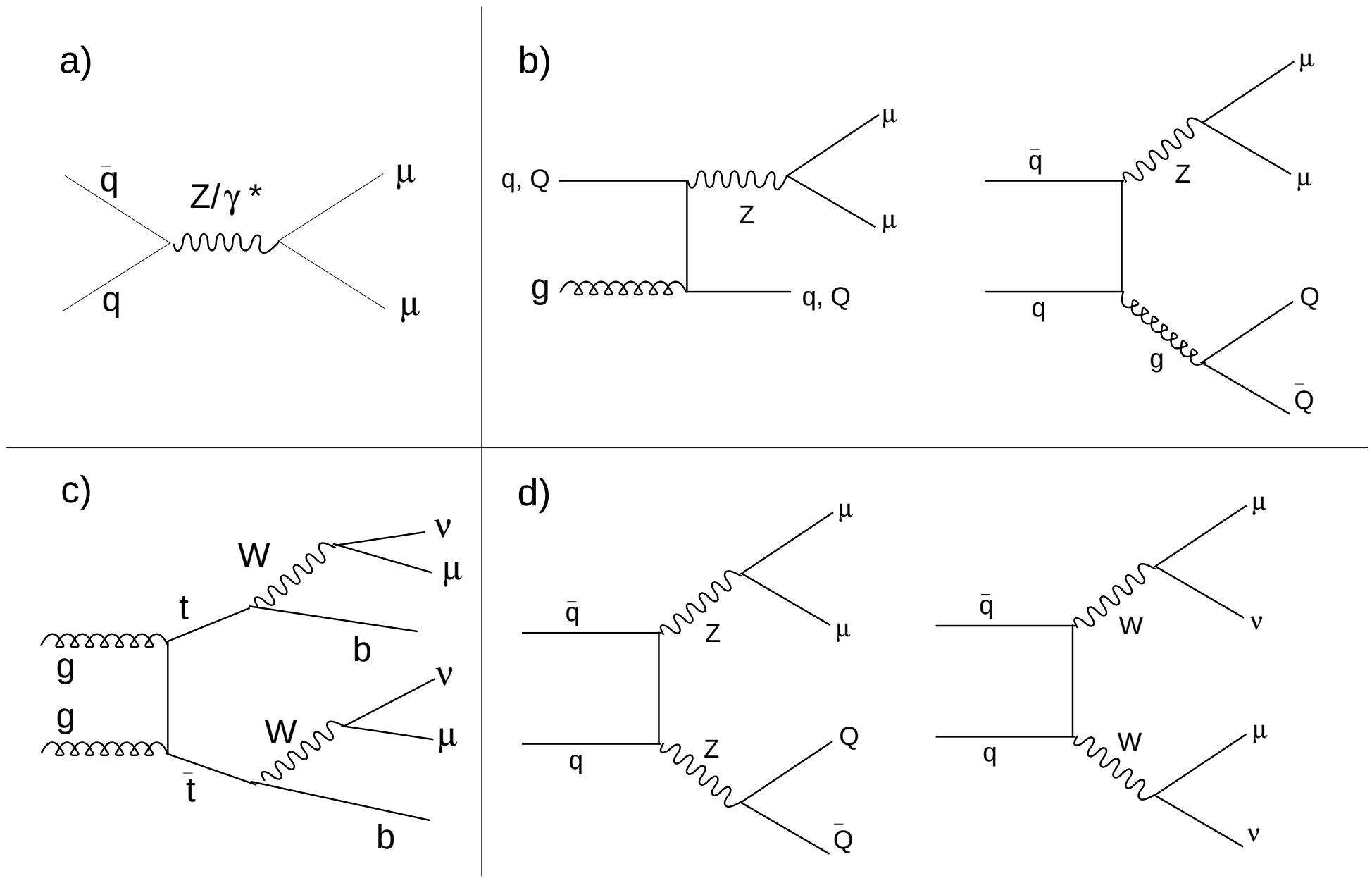


Figure 1: Tree-level Feynman diagrams of the dominant background processes with two isolated muons in the final state: a) Drell-Yan $Z$ boson production, b) $Z$ boson production in association with jets, c) $t\bar{t}$ production and d) $ZZ$ and $WW$ production. $q$ is a general symbol for $u$ and $d$ quarks, while $Q$ stands for the $b$ and $c$ quarks.

The dominant background process with a very large production rate of ∼1 nb is the Drell-Yan $Z$ boson production, with subsequent $Z$ decay into two muons. The invariant dimuon mass peaks at the $Z$ resonance, such that the search for the $A/H/h$ becomes unfeasible for the Higgs masses below 100 GeV. Even for the higher Higgs boson masses, the tail of the $Z$ resonance still provides an overwhelming background. The Drell-Yan background can be suppressed by requiring the presence of one or more additional b-jets, originating from the associated $b\bar{b}A$ production. The major backgrounds remaining after this requirement are the $Z$ boson production in association with the light jets or b-jets and the $t\bar{t} \rightarrow (W^+b)(W^-\bar{b}) \rightarrow (\mu^+\nu b)(\mu^-\nu\bar{b})$ background. The $t\bar{t}$ background can be distinguished from the signal by a higher jet activity and a large missing energy caused by the neutrinos from $W$ decays. Additional background from $WW$ and $ZZ$ diboson productions is expected to be small, due to the much lower production rates.

# 3 Data samples

Two different Monte Carlo generators are used for the signal production. PYTHIA 6.4 [13] has been used to generate the direct $gg \to \phi \to \mu\mu$ and associated $gg \to b\bar{b}\phi \to b\bar{b}\mu\mu$ processes (where $\phi = A,\ H,\ h$). The SHERPA [14] event generator (version 1.0.9) combines all three associated production mechanisms, $gg \to b\bar{b}\phi$, $bg \to b\phi$ and $b\bar{b} \to \phi$, in a coherent way without double-counting. This is accomplished by the CKKW algorithm [15] for the matching of the parton showers to the quark emission from the matrix elements. Both generators provide leading-order cross-sections, which have been rescaled to the Feynhiggs NNLO values. The studies at the D0 experiment [16] have shown that the differential SHERPA distributions are in a good agreement with the real data and can simply be normalized to the previously described inclusive higher-order cross-sections. The comparison of the samples produced by the two generators will be described in Section 6.1. A generator filter requiring at least two muons with $p_T$ >5 GeV and $|\eta|$ <2.7 is applied to each event for all signal data samples, after the showering and before writing out the events into permanent storage. The background samples are listed in Table 1, together with the corresponding NLO cross-sections. The details of the cross-section computation can be found in Ref [17]. In addition to already mentioned generators, the MC@NLO 3.1 [18] and AcerMC 3.4 packages [19] have been used for the event generation.

Full simulation of the detector response has been performed for all signal and background event topologies, within the ATHENA software framework which uses the GEANT4 [20] package for the description of the detector response. In addition, due to the large background production rates, it is necessary to increase the number of simulated events by means of the parametrized fast detector simulation (Atlfast [21]). The presented analyses are based on the combination of both simulation types. The samples obtained with the detailed simulation have been used for the tuning of the parametrized description of the detector performance in the fast simulation. The tuning procedure provides a very good agreement with the full detector simulation. Any remaining differences are treated as systematic uncertainty.

| Process | Generator | $\sigma \times BR$ [pb] | Filter efficiency | Number of events | Simulation type |
|---|---|---|---|---|---|
| $t\bar{t}$; $2\mu$-filter | MC@NLO | 833 | 0.072 | 500 000 | full sim. |
| $t\bar{t}$; $1\ell$-filter | MC@NLO | 833 | 0.556 | 600 000 | full sim. |
| $(Z \to \mu\mu)$+0-3 light jets | SHERPA | 2036 | 0.490 | 5 000 | full sim. |
| $(Z \to \mu\mu)$+1-3 b-jets | SHERPA | 52.3 | 0.914 | 5 000 | full sim. |
| $b\bar{b}(Z \to \mu\mu)$ | AcerMC/PYTHIA | 45 | 0.788 | 280 000 | full sim. |
| $ZZ \to b\bar{b}\mu\mu$ | PYTHIA | 0.151 | 0.724 | 10 000 | full sim. |
| $WW$ | PYTHIA | 116.8 | 0.35 | 50 000 | full sim. |
| $t\bar{t}$, no filter | MC@NLO | 833.0 | 1.0 | 100 000 000 | Atlfast |
| $(Z \to \mu\mu)$+0-3 light jets | SHERPA | 1165.9 | 0.855 | 30 000 000 | Atlfast |
| $(Z \to \mu\mu)$+0-3 b-jets | SHERPA | 52.3 | 0.914 | 1 000 000 | Atlfast |

Table 1: Background data samples with corresponding NLO cross-sections.

All mentioned data samples have been simulated assuming there are no additional $pp$-interactions per event. However, at luminosities of $10^{33}$ cm$^{-2}$s$^{-1}$ one expects to have 2-3 such pile-up interactions superimposed to the hard scattering. In addition, the neutron and photon background of the muon spectrometer (so called cavern background) may increase the muon trigger rate and degrade the muon reconstruction performance [22]. In order to study the impact of the pile-up and cavern background on the analysis performance, dedicated $b\bar{b}A$, $t\bar{t}$ and $Zb\bar{b}$ data samples have been simulated with the realistic pile-up and cavern background contribution. The simulated cavern background is assumed to be five times higher

than the prediction of GCALOR [23] and FLUKA [24] simulations, to account for the uncertainty of the calculation.

# 4 Detector performance

A detailed description of the ATLAS detector and its performance is given in Ref [25]. The details of the detector layout, the software framework used for the Monte Carlo production, as well as the details of the reconstruction of fully simulated events can be found in Ref [26]. In this Section, the performance of the reconstruction algorithms is shortly described, concentrating on the key objects for the analyses: muon identification and momentum measurement, jet reconstruction, b-tagging and the measurement of the missing transverse energy ($E_T^{miss}$). First, the results obtained in absence of pile-up and cavern background in the detector are shown. These are subsequently compared to the results obtained when both pile-up and cavern background are taken into account.

## 4.1 Reconstruction performance without pile-up and cavern background

In ATLAS, the muon reconstruction is performed by combining the information of the muon spectrometer and the inner detector. Staco and MuTag [22] reconstruction packages are used for the study. The average muon reconstruction efficiency is (97.15±0.04)%. This is reduced to (95.44±0.05)% if a match between the muon spectrometer track and the inner detector track is required. The momentum resolution of low-$p_T$ muons is mostly dominated by the inner detector performance, while the high-$p_T$ muon reconstruction is more sensitive to the muon spectrometer performance. The average muon momentum resolution is better than 3%, which allows for an excellent dimuon mass resolution, as shown in Figure 2 for the A-boson ($m_A$=200 GeV) produced via the associated $b\bar{b}A$ and the direct $gg \rightarrow A$ production mode. As expected, the experimental dimuon mass resolution does not depend on the Higgs production mode. Table 2 summarizes the dimuon mass resolutions obtained for different $A$ boson masses.

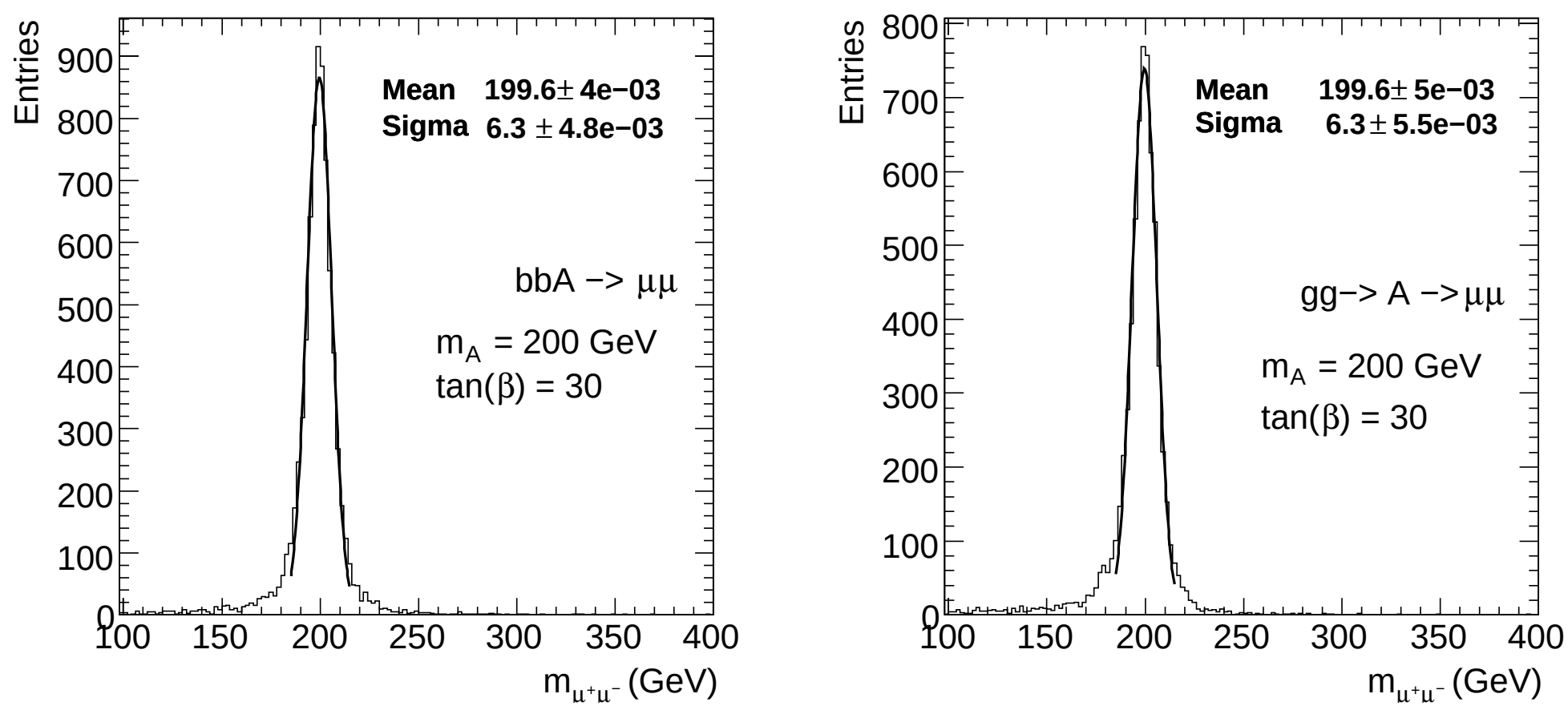


Figure 2: Dimuon mass distribution for the $b\bar{b}A$ and $gg \rightarrow A$ signal samples with an $A$ boson mass of 200 GeV and $\tan\beta$=30. The distributions are fitted by the Gauss function.

Characteristic of the $b\bar{b}A$ signal are the b-jets with generally rather low transverse momenta, as shown in Figure 3(a). Since the efficiency of the b-jet reconstruction decreases with the $p_T$, the number of reconstructed b-jets will in general be smaller for the signal than for the $t\bar{t}$ or $ZZ \rightarrow b\bar{b}\mu\mu$ backgrounds, where the b-jets are more energetic (see Figure 3(b)). A detailed study was performed to identify the

| | *A* boson mass (GeV) | | | | | |
|---|---|---|---|---|---|---|
| (GeV) | 110 | 130 | 150 | 200 | 300 | 400 |
| Natural width | 2.16 | 2.48 | 2.80 | 3.60 | 5.61 | 8.46 |
| Reconstructed $\sigma$ | $2.59 \pm 0.02$ | $3.83 \pm 0.03$ | $4.11 \pm 0.04$ | $6.29 \pm 0.05$ | $10.2 \pm 0.2$ | $15.0 \pm 0.3$ |
| Reconstructed mass | 109.818 $\pm$ 0.006 | 129.738 $\pm$ 0.005 | 149.796 $\pm$ 0.006 | 199.589 $\pm$ 0.005 | 298.82 $\pm$ 0.04 | 399.37 $\pm$ 0.04 |

Table 2: The natural width of the $A$ boson and the expected width of the dimuon resonance based on Monte Carlo simulated data are shown for the $b\bar{b}A$ signal at different mass points and with $\tan\beta$=30.

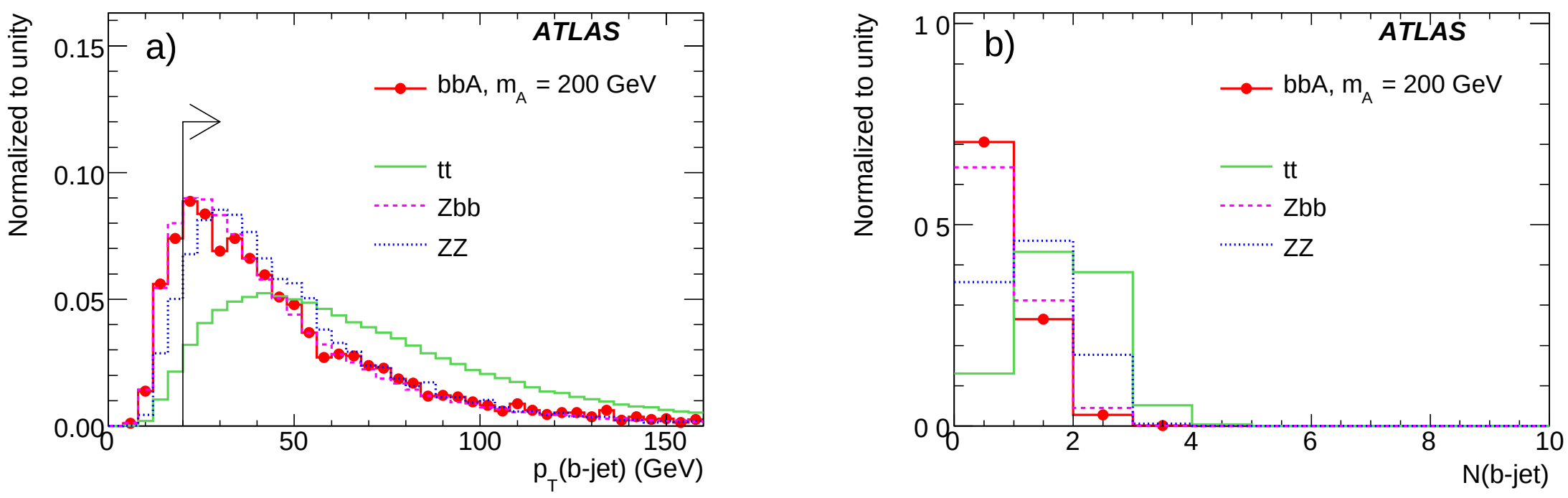


Figure 3: a) Transverse momentum $p_T$ of the b-jets in the $b\bar{b}A$ signal and the dominant background processes and b) the number of reconstructed b-jets per event. The selection criteria for the b-jets are described in the text.

optimum b-jet selection criteria. The best jet reconstruction performance is observed for the jet cone algorithms with the cone size of $\Delta R = \sqrt{\Delta\eta^2 + \Delta\phi^2} = 0.4$, compatible with the performance of the $k_T$ algorithm for the same cone size. After a jet is selected as described, the b-tagging algorithm is performed to determine whether the jet originates from the b-quark. The minimum $p_T$ value of 20 GeV is required for each b-jet in order to reduce the contribution of the calorimeter noise and of the mistagged light- or c-jets. The rejection of light- and c-jets is essential for the suppression of the $Z + jet$ background. One could extend the lower $p_T$-bound down to $\sim$15 GeV without a large change of the rejection rate. However, the impact on the final signal significance will be rather small, while the agreement between the full and the fast simulation is shown to decrease.

Several b-tagging algorithms have been studied in order to define the optimum selection of the low-$p_T$ b-jets coming from the signal. The best rejection is obtained by IP3DSV1 [27], which is based on the information obtained from the transverse and longitudinal impact parameter significances of the tracks and from the reconstructed secondary vertex. The distribution of the b-tagging weight obtained by this algorithm is shown in Figure 4 for the b-jets and the light jets in the $b\bar{b}A$ signal sample at 200 GeV and in the $t\bar{t}$ background sample. The arrow indicates the optimum cut value of 4. Figure 5 shows the obtained b-tagging efficiency for the $b\bar{b}A$ signal sample, in dependence on the b-jet $E_{\mathrm{T}}$ (Figure 5a)) and $\eta$ (Figure 5b)). The kinematic cuts of $p_T >$20 GeV and $|\eta| < 2.5$, as well as the IP3DSV1 weight-cut of 4 have been applied for the b-jet selection. The resulting b-tagging efficiency is (64.1$\pm$0.81)% for the $b\bar{b}A$ signal sample, with a light-jet rejection of (80$\pm$1) for the $Z + jet$ sample. Systematic detector-related

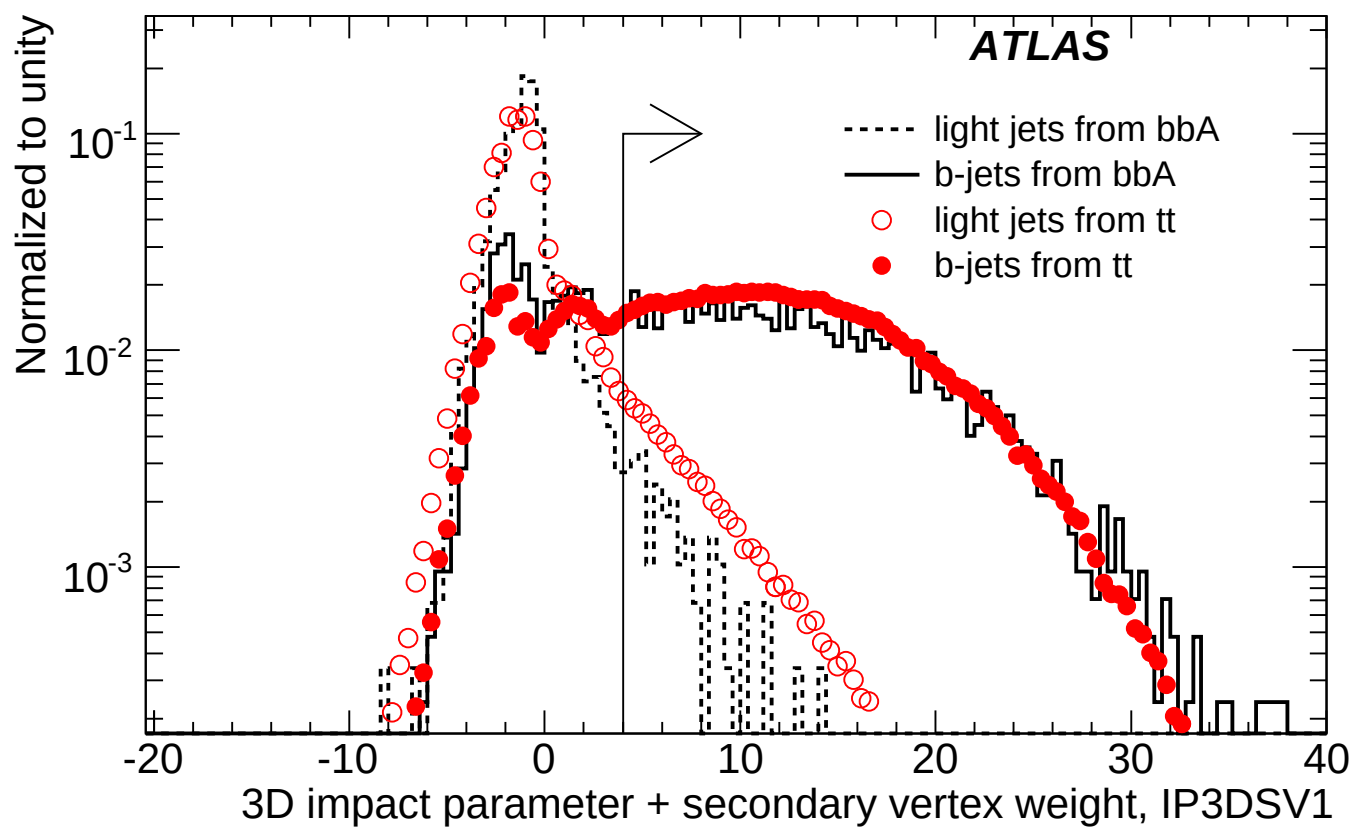


Figure 4: Distribution of the b-tagging weight for the b-jets and the light jets in the $b\bar{b}A$ signal and $t\bar{t}$ background sample, obtained by the IP3DSV1 b-tagging algorithm. The arrow indicates the optimum cut value of 4, which is used for the selection of the b-jets in the analyses.

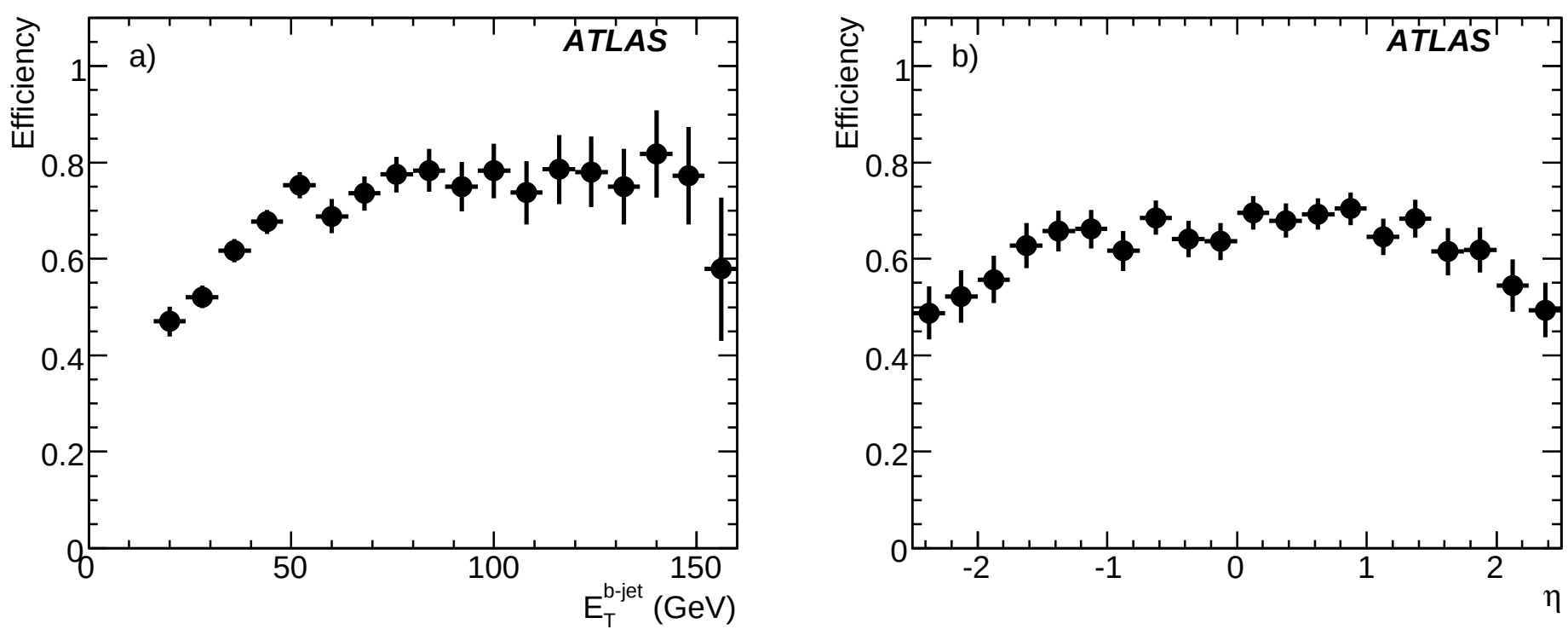


Figure 5: B-tagging efficiency in dependence of b-jet transverse energy $E_{\mathrm{T}}$ (a) and pseudorapidity $\eta$ (b), evaluated for the $b\bar{b}A$ signal sample at $m_A$ =200 GeV and $\tan\beta$ =30. IP3DSV1 b-tagging weight cut of >4 has been applied.

uncertainties are not included here.

The final important reconstruction object is the missing transverse energy ($E_T^{miss}$), which allows for the suppression of the $t\bar{t}$ background. In the signal processes, there is no neutrino contribution, such that the measured $E_T^{miss}$ value is dominated by the experimental resolution. The reconstruction algorithm for the calculation of the missing transverse energy is described in detail in Ref [28]. The distributions of $E_{T(x,y)}^{miss}$ components in the signal samples have a Gaussian part with a width $\sigma$=(7.8±0.1) GeV, while the non-Gaussian tails (above 5$\sigma$) are found to contribute less than 1.5% to the overall distribution. $E_T^{miss}$ is sensitive to pile-up effects, as will be described in the next subsection.

## 4.2 Reconstruction performance under influence of pile-up and cavern background

In the following, the detector performance related to the analysis is evaluated in dependence on the pile-up at luminosities of $10^{33}$ cm$^{-2}$s$^{-1}$ and cavern background (five times higher than the expectation). In

Figure 6(left), the efficiency and the fake rate of the muon reconstruction is shown for the $b\bar{b}A$ signal sample simulated without and with the pile-up contribution as a function of the pseudorapidity. The corresponding momentum resolution is shown in Figure 6(middle). Similar results are obtained also for the background samples. As can be seen from the plots, muon reconstruction is only marginally

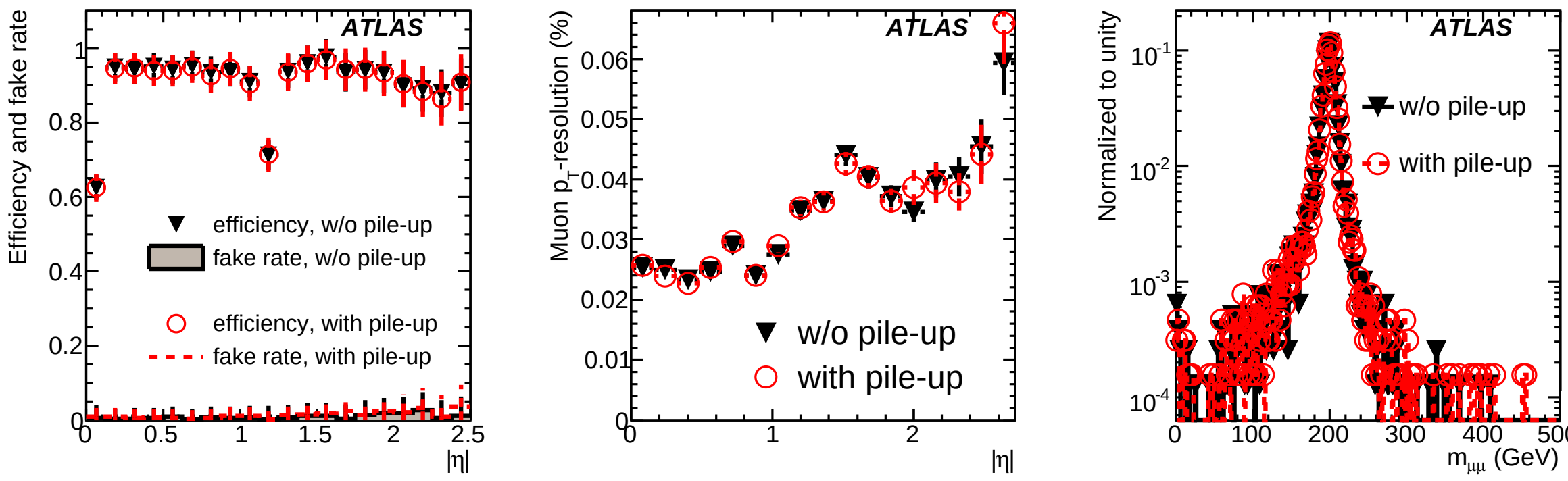


Figure 6: Efficiency and the fake rate (left) and resolution (middle) of the muon reconstruction as a function of the $|\eta|$ (for $p_T$ >20 GeV), with and without the pile-up contribution in the $b\bar{b}A$ signal sample with $m_A$=200 /GeV. The right plot shows the corresponding dimuon mass distribution.

influenced by pile-up. Consequently, the dimuon invariant mass also remains unaffected, as shown in Figure 6(right) for $m_A$=200 GeV.

On the contrary, the reconstruction of the missing transverse energy is substantially affected by pile-up in the calorimeter. The degradation of the $E_T^{miss}$-resolution mainly affects the selection of events with a small true missing energy (signal and the $Z$ background) as shown in Figure 7(left); $t\bar{t}$ events, characterized by a large missing energy, are rather insensitive to pile-up (see Figure 7(right)). This

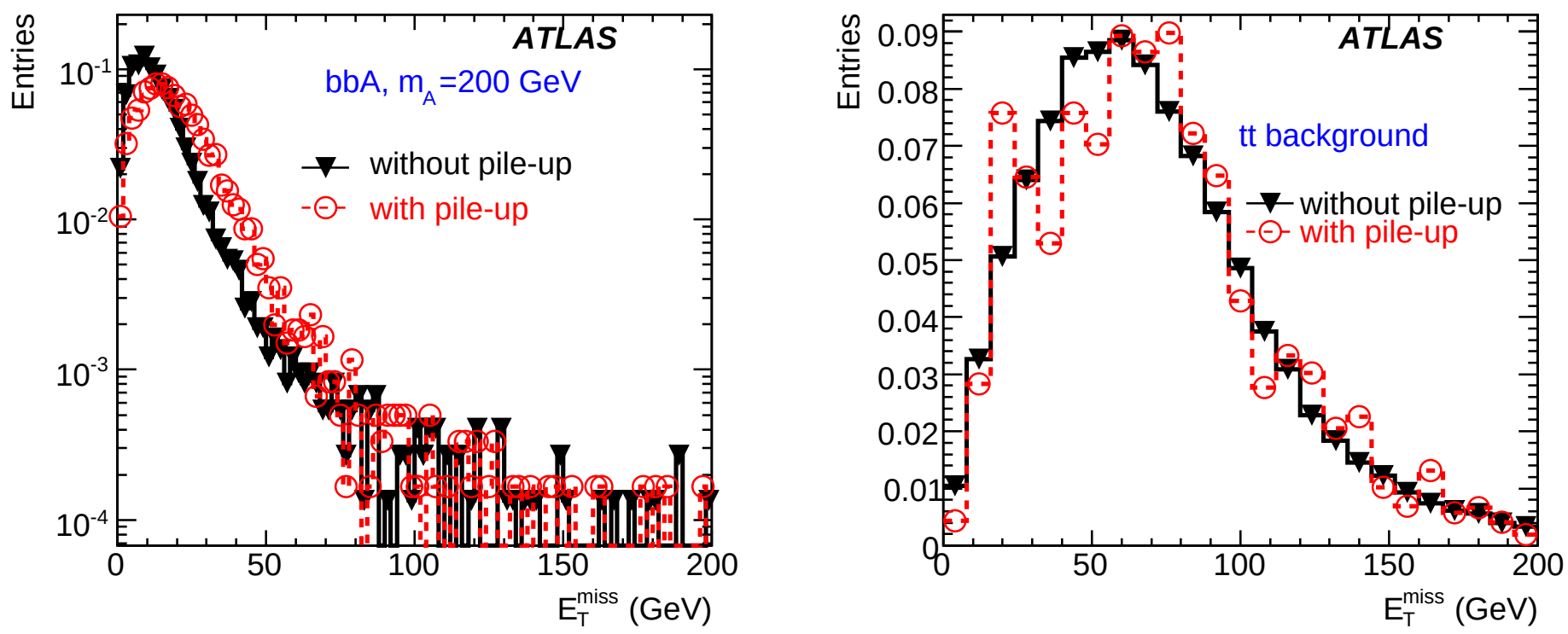


Figure 7: Missing transverse energy distribution for the $b\bar{b}A$ signal at 200 GeV (left) and the $t\bar{t}$ background (right),with and without pile-up.

effect must be taken into account during the optimization of the event selection criteria. For instance, an event selection cut at $E_T^{miss}$ <30 GeV, which is reasonable without pile-up, would reject too many signal events, once the pile-up contribution is included. Therefore, this analysis cut should rather be set to at least 40 GeV in the realistic LHC environment.

The change of the calorimeter response under the influence of pile-up affects also the jet reconstruction. Due to a higher calorimeter activity, one expects an increase in the number of reconstructed jets. This can be observed in Figure 8(left), showing the jet multiplicity in the $Zb\bar{b}$ background events.

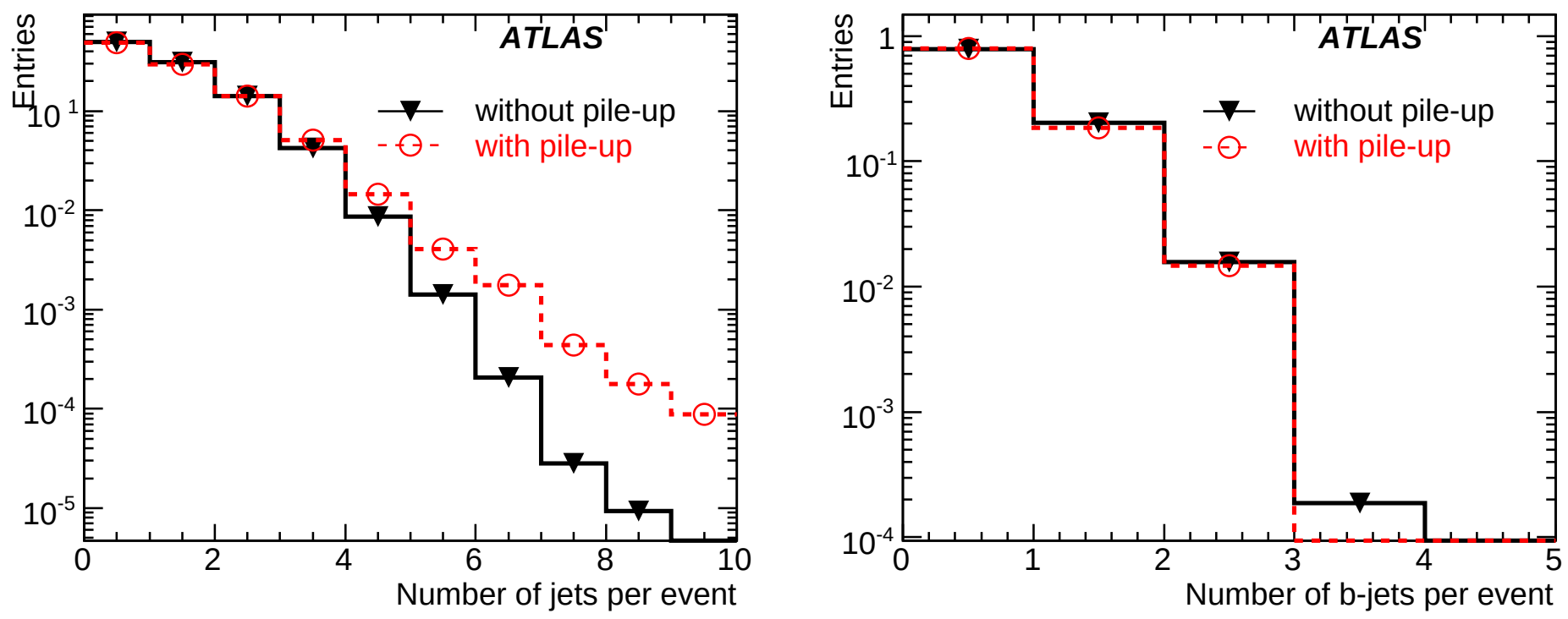


Figure 8: Total number of jets per event (left) and the number of b-jets (right) in the $Zb\bar{b}$ background sample, with and without the pile-up contribution.

No significant impact of the pile-up is observed on the b-tagging (see Figure 8(right)), due to the additional tracking and vertex information.

## 5 Event selection

The search for the MSSM Higgs bosons can be performed by several different approaches, related to the number of jets one requires to be present in the final state. As mentioned before, due to a large signal production rate in the associated production mode, the presence of the b-jets in the final state can help to suppress the Drell-Yan background. On the contrary, the remaining events with 0 b-jets provide for a high signal rate on top of the smoothly distributed background, even at low integrated luminosity.

The event selection methods are optimized separately for the two cases:

- Signatures with 0 b-jets in the final state.
- Signatures with at least one b-jet in the final state.

The two mentioned final states are uncorrelated and therefore complementary. In the case of 0 b-jets in the final state, the dominant background is the Drell-Yan $Z$ boson production, while in the case of at least one b-jet the $t\bar{t}$ background has the biggest contribution, especially for Higgs masses above 130 GeV, which are further away from the $Z$ resonance.

Before describing the selection criteria, the preselection of the events is discussed, common to the two signatures above. The preselection is defined by the kinematic cuts on muon $p_T$ and $|\eta|$, together with the muon isolation criteria.

### 5.1 Preselection

The main characteristics of the signal signatures is the presence of two isolated muons of opposite charge in the final state. The $p_T$-distribution of the muons is shown in Figure 9 for the signal and background processes. The signal is characterized by the relatively high-$p_T$ muons, while the background has muons

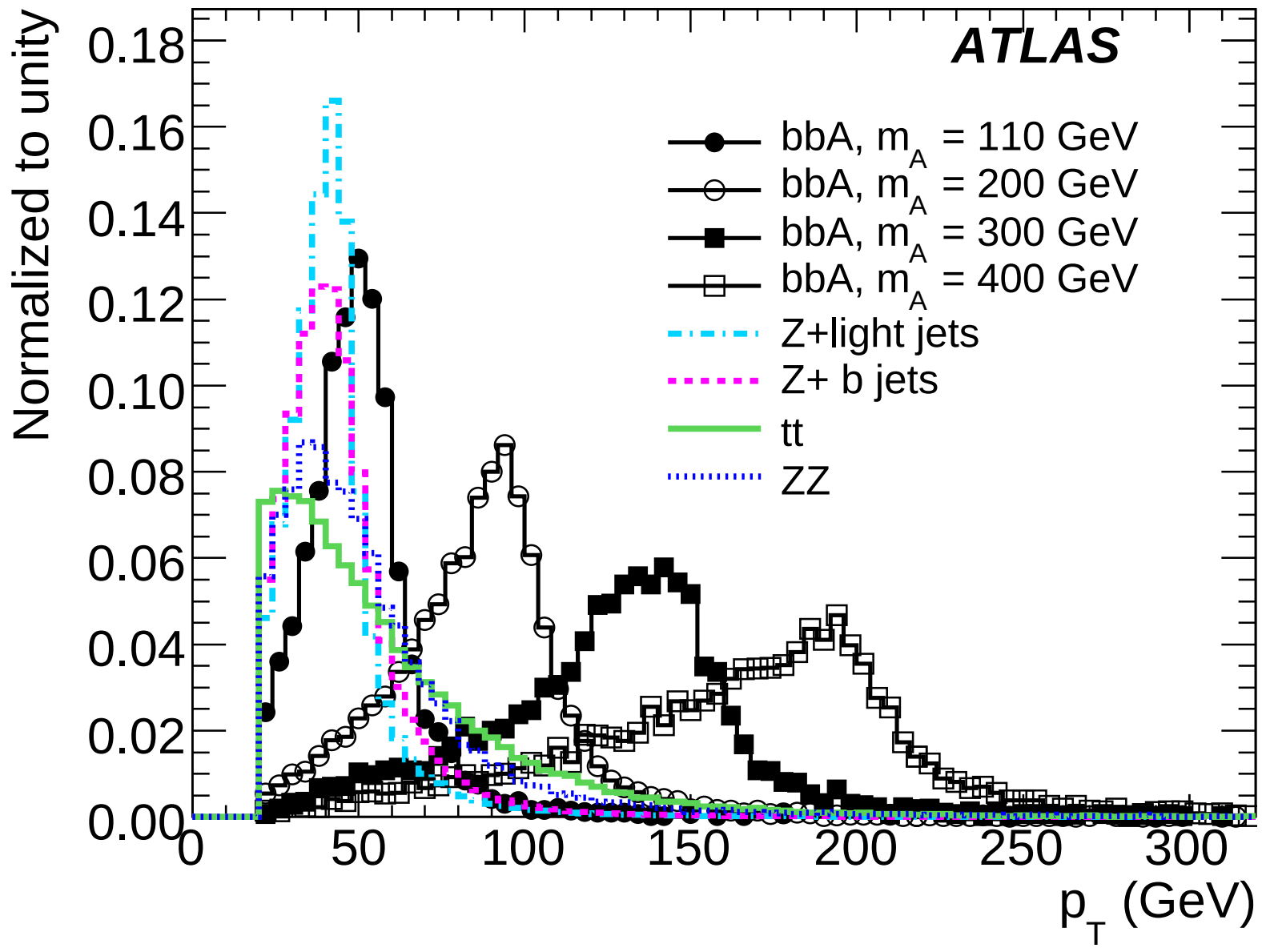


Figure 9: Distribution of the muon transverse momentum $p_T$ for the muons in the signal and background events.

of lower momenta. The muon $p_T$ distribution in the signal is highly correlated to the Higgs boson mass. Therefore, the lower bound on the muon $p_T$ is kept at a relatively low value, in order to allow for a general search in a broad Higgs mass range. At preselection level, both muons are required to have a $p_T$ >20 GeV and to be in the pseudorapidity range $|\eta|$ <2.7.

The selected high-$p_T$ muons are required to be isolated, in order to reject the processes in which the muons originate from the hadronic decays. The applied isolation criteria require the calorimeter energy $E_T$ deposited in a cone of size $\Delta R$ = 0.4 around a given muon, divided by the muon $p_T$ to be lower than 0.2. The distribution of this isolation variable is shown in Figure 10(left) for different signal and background processes.

The isolation criteria significantly decrease the $t\bar{t}$ background, where one of the muons comes from the b-decays. The power of rejection of the non-isolated muons originating from the b-quarks in the $t\bar{t}$ background is shown in Figure 10(right) for the standard calorimeter isolation ($E_T^{cone0.4}$), and for the isolation normalized by the muon $p_T$.

Due to the high muon momenta, the signal can be efficiently triggered by the single high-$p_T$ muon trigger. The efficiency of the trigger selection for the dimuon signal events is shown to be around 95% for all studied mass points. The detailed study of the trigger selection efficiency for events which pass all offline event selection criteria will be presented in Section 5.3. In the following, the results of the event selection without the trigger requirement are presented at first.

## 5.2 Signatures with 0 b-jets and with at least one b-jet in the final state

The large $Z$ boson background contribution can be reduced by requiring that the jets which are present in the final state are tagged as b-jets. Therefore, assuming a fully performing b-tagging algorithm, the following set of selection criteria can be applied:

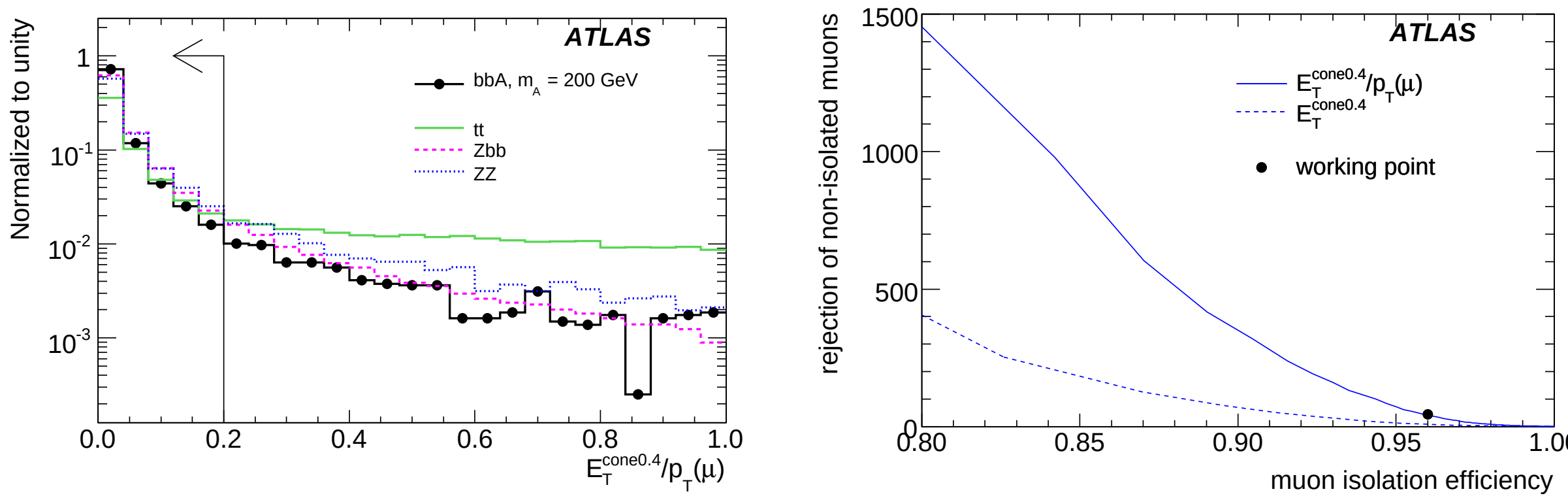


Figure 10: (left) Muon isolation variable $E_T^{cone0.4}/p_T(\mu)$, shown for different signal and background processes. Here, the $E_T^{cone0.4}$ is the energy measured in the calorimeters in cone $\Delta R = 0.4$ around a given muon. (right) Rejection of the non-isolated muons originating from the b-quarks in $t\bar{t}$ events as a function of the selection efficiency for isolated muons, shown for the two isolation variables described in the text. The filled circle indicates the working point with the isolation cut at $E_T^{cone0.4}/p_T(\mu) <0.2$ .

- Events are required to pass the preselection criteria and to have a missing transverse energy $E_T^{miss}$ <40 GeV. This cut is particularly effective in rejecting the $t\bar{t}$ and $WW$ background, which are characterized by a high missing energy due to the presence of neutrinos in the final state (see Figure 11(left)).

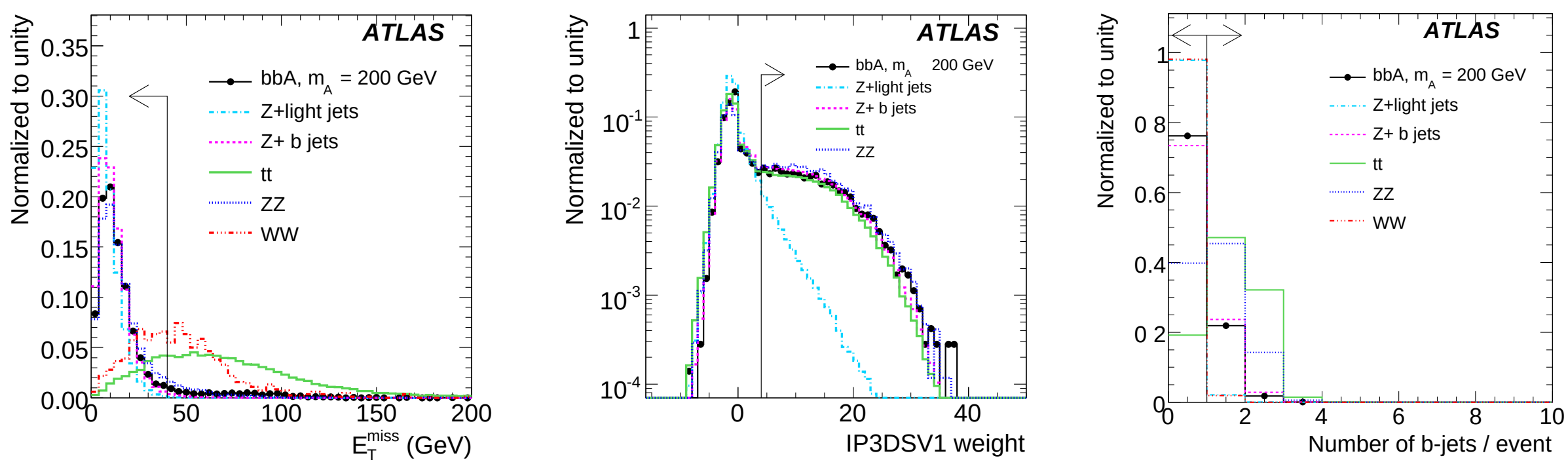


Figure 11: (left) Missing transverse energy, (middle) distribution of the b-tagging IP3DSV1 weight after requiring $p_T$ >20 GeV and $|\eta|$ <2.5 and (right) the multiplicity of reconstructed b-jets per event, after applying the $p_T$- and $\eta$-cuts and requiring the b-tagging IP3DSV1 weight greater than 4. Distributions for the $b\bar{b}A$ signal ($m_A$=200 GeV) and for the major background processes are shown. Arrows indicate the cuts applied in the analysis.

- Subsequently, the number of b-jets is counted in each event, requiring $p_T$ >20 GeV, $|\eta|$ <2.5 and the b-tagging IP3DSV1 weight greater than 4. The distribution of the b-tagging weights before

and the b-jet multiplicity after applying the weight-cut are shown in Figure 11. As shown in the middle plot, the b-tagging weight is effective in reducing the background from $Z + light\ jets$, at the expense of a significant loss of the signal. This cut is scarcely effective against the $t\bar{t}$ background, which, however, can be reduced by the additional cuts discussed below. Therefore, the analysis is divided into a channel with 0 b-jets (in which the $Z$ background is dominant) and the channel with at least one b-jet (in which the $t\bar{t}$ background plays an important role and can be further suppressed).

- Further $t\bar{t}$ rejection criteria have been studied for the channel with at least one b-jet.
  - Two muons originating from the decay of the same particle (Higgs boson) tend to be emitted back-to-back, especially if this particle has a low transverse momentum. As opposed to that, the muons originating from the two different particles (as in the $t\bar{t}$ events) are not correlated and can be separated by any angle. Therefore, the cut is applied on the angle $\Delta\phi_{\mu\mu}$ between the two muons by requiring $|\sin\Delta\phi_{\mu\mu}|$ <0.75. The $|\sin\Delta\phi_{\mu\mu}|$ distributions for the signal and background processes are shown in Figure 12(a).
  - In addition, several discriminating variables related to the hadronic activity in the events have been studied: the $p_T$ distribution of the b-jets, the number of jets per event, or a sum of the transverse momenta of all jets in the event ($\sum p_T^{jet}$). The distributions of two of these variables are shown in Figure 12b) and c). The latter is shown to provide the highest rejection against the $t\bar{t}$ background, while at the same time remaining relatively robust under the the influence of pile-up. A cut at $\sum p_T^{jet}$ <90 GeV is required.

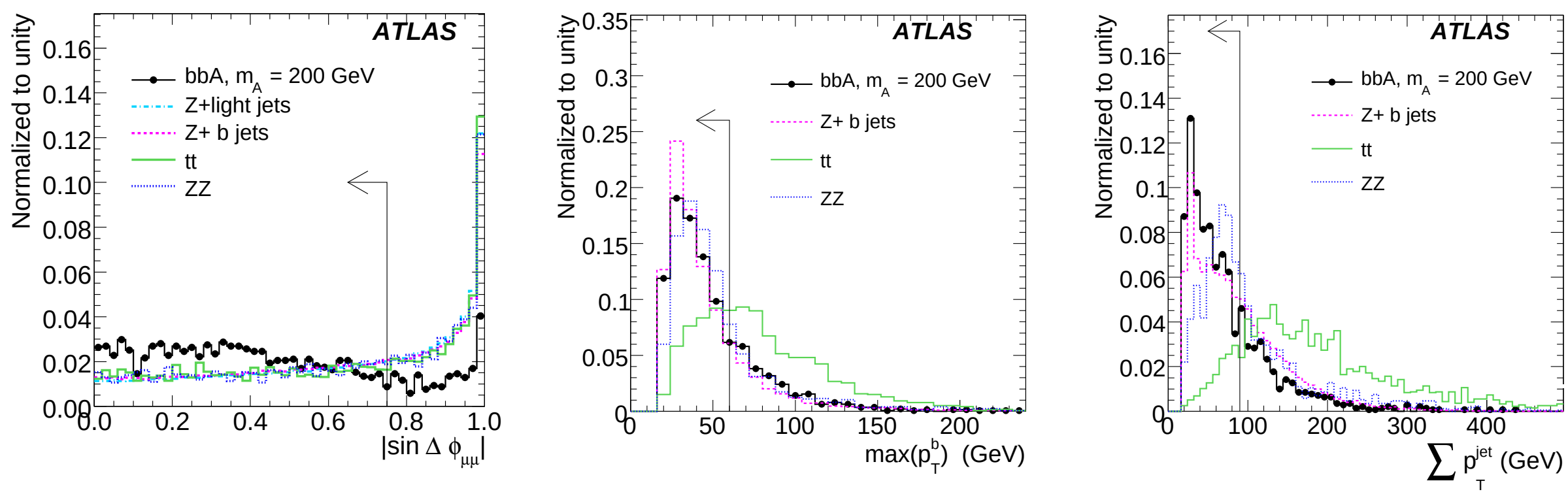


Figure 12: Discriminating variables against the $t\bar{t}$ background: a) $|\sin\Delta\phi|$ shown for the $b\bar{b}A$ signal ($m_A$=200 GeV) and for the major background processes, b) the maximum $p_T$ of the b-jet, and c) $\sum p_T^{jets}$-distribution for the signal and background processes. Arrows indicate the cuts applied in the analysis.

- The final number of events which is used for the calculation of the signal significances is evaluated in a mass window $\Delta m$=$m_A \pm 2\sigma_{\mu\mu}$ around the $A$ boson mass, where $\sigma_{\mu\mu}$ is the expected $m_A$-dependent width of the dimuon resonance (see Table 2).

The signal and background event rates after each of the cuts described above are shown in Tables 3 and 4.

A signal selection efficiency of 7-10% is reached for the channel with at least one b-jet in the final state. The dominant background processes are almost equally the $Z+jet$ and the $t\bar{t}$ events. The signal

| **Cut** | *bbA* **(fb)** 130 GeV | *bbA* **(fb)** 150 GeV | *bbA* **(fb)** 200 GeV | *bbA* **(fb)** 300 GeV | *bbA* **(fb)** 400 GeV | $gg \to A$ **(fb)** 200 GeV |
|---|---|---|---|---|---|---|
| All events | $13.4 \cdot 10^{1}$ | $8.7 \cdot 10^{1}$ | $31.5 \cdot 10^{0}$ | $6.9 \cdot 10^{0}$ | $19.3 \cdot 10^{-1}$ | $32.3 \cdot 10^{-1}$ |
| muon preselection | $8.8(2) \cdot 10^{1}$ | $5.9(1) \cdot 10^{1}$ | $22.3(3) \cdot 10^{0}$ | $5.0(1) \cdot 10^{0}$ | $14.4(3) \cdot 10^{-1}$ | $28.6(3) \cdot 10^{-1}$ |
| $E_T^{miss}$ <40 GeV | $8.3(2) \cdot 10^{1}$ | $5.4(1) \cdot 10^{1}$ | $20.3(2) \cdot 10^{0}$ | $4.4(1) \cdot 10^{0}$ | $11.7(2) \cdot 10^{-1}$ | $25.9(3) \cdot 10^{-1}$ |
| nr. of b-jets=0 | $6.6(1) \cdot 10^{1}$ | $4.3(1) \cdot 10^{1}$ | $15.5(2) \cdot 10^{0}$ | $31.6(8) \cdot 10^{-1}$ | $8.2(2) \cdot 10^{-1}$ | $25.1(3) \cdot 10^{-1}$ |
| $\Delta m$ | $\mathbf{5.6(1) \cdot 10^{1}}$ | $\mathbf{34.1(8) \cdot 10^{0}}$ | $\mathbf{128.0(1) \cdot 10^{-1}}$ | $\mathbf{26.4(8) \cdot 10^{-1}}$ | $\mathbf{6.9(2) \cdot 10^{-1}}$ | $\mathbf{204.2(1) \cdot 10^{-2}}$ |
| nr. of b-jets≥1 | $16.2(7) \cdot 10^{0}$ | $11.2(5) \cdot 10^{0}$ | $4.8(1) \cdot 10^{0}$ | $12.1(5) \cdot 10^{-1}$ | $3.5(1) \cdot 10^{-1}$ | $8.2(5) \cdot 10^{-2}$ |
| $\lvert\sin\Delta\phi_{\mu\mu}\rvert$ <0.75 | $11.7(6) \cdot 10^{0}$ | $8.3(4) \cdot 10^{0}$ | $4.0(1) \cdot 10^{0}$ | $10.7(5) \cdot 10^{-1}$ | $3.2(1) \cdot 10^{-1}$ | $4.8(4) \cdot 10^{-2}$ |
| $\sum p_T^{jets}$ < 90 GeV | $9.3(5) \cdot 10^{0}$ | $6.5(3) \cdot 10^{0}$ | $2.9(1) \cdot 10^{0}$ | $7.1(4) \cdot 10^{-1}$ | $1.7(1) \cdot 10^{-1}$ | $2.0(3) \cdot 10^{-2}$ |
| $\Delta m$ | $\mathbf{8.3(5) \cdot 10^{0}}$ | $\mathbf{5.3(3) \cdot 10^{0}}$ | $\mathbf{23.9(8) \cdot 10^{-1}}$ | $\mathbf{5.9(4) \cdot 10^{-1}}$ | $\mathbf{14.8(8) \cdot 10^{-2}}$ | $\mathbf{1.7(3) \cdot 10^{-2}}$ |

Table 3: $\sigma \times BR$ for the signal processes at $\tan\beta$=30 after each selection cut. Numbers in brackets represent the statistical error on the last digit.

| **Cut** | $Z + light\ jet$ **(fb)** | $Z + b\ jet$ **(fb)** | $t\bar{t}$ **(fb)** | $ZZ \to bb\mu\mu$ **(fb)** | $WW$ **(fb)** | Total |
|---|---|---|---|---|---|---|
| All events | $2036.0 \cdot 10^{3}$ | $52.3 \cdot 10^{3}$ | $833.0 \cdot 10^{3}$ | 151.0 | $116.8 \cdot 10^{3}$ | |
| muon preselection | $727.7(2) \cdot 10^{3}$ | $333.8(4) \cdot 10^{2}$ | $57.6(1) \cdot 10^{2}$ | $61.3(8) \cdot 10^{0}$ | $6.7(2) \cdot 10^{2}$ | |
| $E_T^{miss}$ <40 GeV | $726.2(2) \cdot 10^{3}$ | $330.1(4) \cdot 10^{2}$ | $132.7(6) \cdot 10^{1}$ | $56.1(8) \cdot 10^{0}$ | $3.0(2) \cdot 10^{2}$ | |
| nr. of b-jets=0 | $710.7(2) \cdot 10^{3}$ | $242.3 \cdot 10^{2}$ | $25.6(3) \cdot 10^{1}$ | $22.3(5) \cdot 10^{0}$ | $2.9(2) \cdot 10^{2}$ | |
| $\Delta m$ (130 GeV) | $\mathbf{35.4(1) \cdot 10^{2}}$ | $\mathbf{8.8(2) \cdot 10^{1}}$ | $\mathbf{20.2(8) \cdot 10^{0}}$ | $\mathbf{1.5(4) \cdot 10^{-1}}$ | $\mathbf{1.8(4) \cdot 10^{1}}$ | $\mathbf{3.7(1) \cdot 10^{3}}$ |
| $\Delta m$ (150 GeV) | $\mathbf{152.5(8) \cdot 10^{1}}$ | $\mathbf{3.4(1) \cdot 10^{1}}$ | $\mathbf{16.0(7) \cdot 10^{0}}$ | $\mathbf{0.7(3) \cdot 10^{-1}}$ | $\mathbf{2.6(5) \cdot 10^{1}}$ | $\mathbf{15.9(1) \cdot 10^{2}}$ |
| $\Delta m$ (200 GeV) | $\mathbf{58.9(5) \cdot 10^{1}}$ | $\mathbf{8.4(7) \cdot 10^{0}}$ | $\mathbf{11.6(6) \cdot 10^{0}}$ | $\mathbf{0.6(1) \cdot 10^{-1}}$ | $\mathbf{1.6(3) \cdot 10^{1}}$ | $\mathbf{62.5(6) \cdot 10^{1}}$ |
| $\Delta m$ (300 GeV) | $\mathbf{18.1(3) \cdot 10^{1}}$ | $\mathbf{1.8(3) \cdot 10^{0}}$ | $\mathbf{5.1(4) \cdot 10^{0}}$ | $\mathbf{0.1(1) \cdot 10^{-1}}$ | $\mathbf{0.2(2) \cdot 10^{1}}$ | $\mathbf{19.0(4) \cdot 10^{1}}$ |
| $\Delta m$ (400 GeV) | $\mathbf{7.8(2) \cdot 10^{1}}$ | $\mathbf{0.8(2) \cdot 10^{0}}$ | $\mathbf{2.0(2) \cdot 10^{0}}$ | $\mathbf{0.1(1) \cdot 10^{-1}}$ | $\mathbf{0.2(2) \cdot 10^{1}}$ | $\mathbf{8.4(3) \cdot 10^{1}}$ |
| nr. of b-jets≥1 | $154.9(3) \cdot 10^{2}$ | $87.7(2) \cdot 10^{2}$ | $107.1(6) \cdot 10^{1}$ | $33.8(6) \cdot 10^{0}$ | $0.6(2) \cdot 10^{1}$ | |
| $\lvert\sin\Delta\phi_{\mu\mu}\rvert$ <0.75 | $84.3(2) \cdot 10^{2}$ | $49.8(2) \cdot 10^{2}$ | $61.7(4) \cdot 10^{1}$ | $19.0(5) \cdot 10^{0}$ | $0.3(2) \cdot 10^{1}$ | |
| $\sum p_T^{jets}$ <90 GeV | $44.3(1) \cdot 10^{2}$ | $33.0(1) \cdot 10^{2}$ | $8.9(2) \cdot 10^{1}$ | $10.7(3) \cdot 10^{0}$ | $0.3(2) \cdot 10^{1}$ | |
| $\Delta m$ (130 GeV) | $\mathbf{3.1(1) \cdot 10^{1}}$ | $\mathbf{15.1(9) \cdot 10^{0}}$ | $\mathbf{7.7(5) \cdot 10^{0}}$ | $\mathbf{0.7(3) \cdot 10^{0}}$ | $\mathbf{<0.7 \cdot 10^{0}}$ | $\mathbf{5.5(2) \cdot 10^{1}}$ |
| $\Delta m$ (150 GeV) | $\mathbf{14.8(8) \cdot 10^{0}}$ | $\mathbf{6.0(6) \cdot 10^{0}}$ | $\mathbf{6.2(4) \cdot 10^{0}}$ | $\mathbf{0.2(1) \cdot 10^{-1}}$ | $\mathbf{<0.7 \cdot 10^{0}}$ | $\mathbf{2.8(1) \cdot 10^{1}}$ |
| $\Delta m$ (200 GeV) | $\mathbf{6.7(5) \cdot 10^{0}}$ | $\mathbf{2.0(3) \cdot 10^{0}}$ | $\mathbf{5.8(4) \cdot 10^{0}}$ | $\mathbf{<0.1 \cdot 10^{-1}}$ | $\mathbf{<0.7 \cdot 10^{0}}$ | $\mathbf{1.5(1) \cdot 10^{1}}$ |
| $\Delta m$ (300 GeV) | $\mathbf{2.2(3) \cdot 10^{0}}$ | $\mathbf{0.6(2) \cdot 10^{0}}$ | $\mathbf{2.3(3) \cdot 10^{0}}$ | $\mathbf{<0.1 \cdot 10^{-1}}$ | $\mathbf{<0.7 \cdot 10^{0}}$ | $\mathbf{5.8(8) \cdot 10^{0}}$ |
| $\Delta m$ (400 GeV) | $\mathbf{1.1(2) \cdot 10^{0}}$ | $\mathbf{0.1(1) \cdot 10^{0}}$ | $\mathbf{0.4(1) \cdot 10^{0}}$ | $\mathbf{0.1(1) \cdot 10^{-1}}$ | $\mathbf{<0.7 \cdot 10^{0}}$ | $\mathbf{2.3(7) \cdot 10^{0}}$ |

Table 4: $\sigma \times BR$ for the background processes obtained after each event selection cut. The upper limits are evaluated at 90% CL. Numbers in brackets represent the statistical error on the last digit.

events are mainly lost due to the limited b-jet reconstruction efficiency, as discussed previously. This is in agreement with a much larger signal selection efficiency of 40-50% in the case of the channel with 0 b-jets. Here, the $Z$ background is observed to have a dominant contribution. The invariant dimuon mass distributions obtained for the two channels are shown in Figure 13.

In the initial phase of the detector running, the b-tagging algorithms still may not have the optimum performance. Therefore, the discovery potential has also been evaluated for the case where no b-tagging requirement is imposed on the reconstructed jets. Due to the large $Z$ background, the signatures with no b-tagging requirement will provide a similar discovery potential as the analysis with 0 b-jets in the final state.

### 5.3 Trigger selection

Previously calculated event selection efficiencies have been obtained assuming that each analysed event can be triggered. In this Section, a realistic trigger description is included to evaluate the effect of the

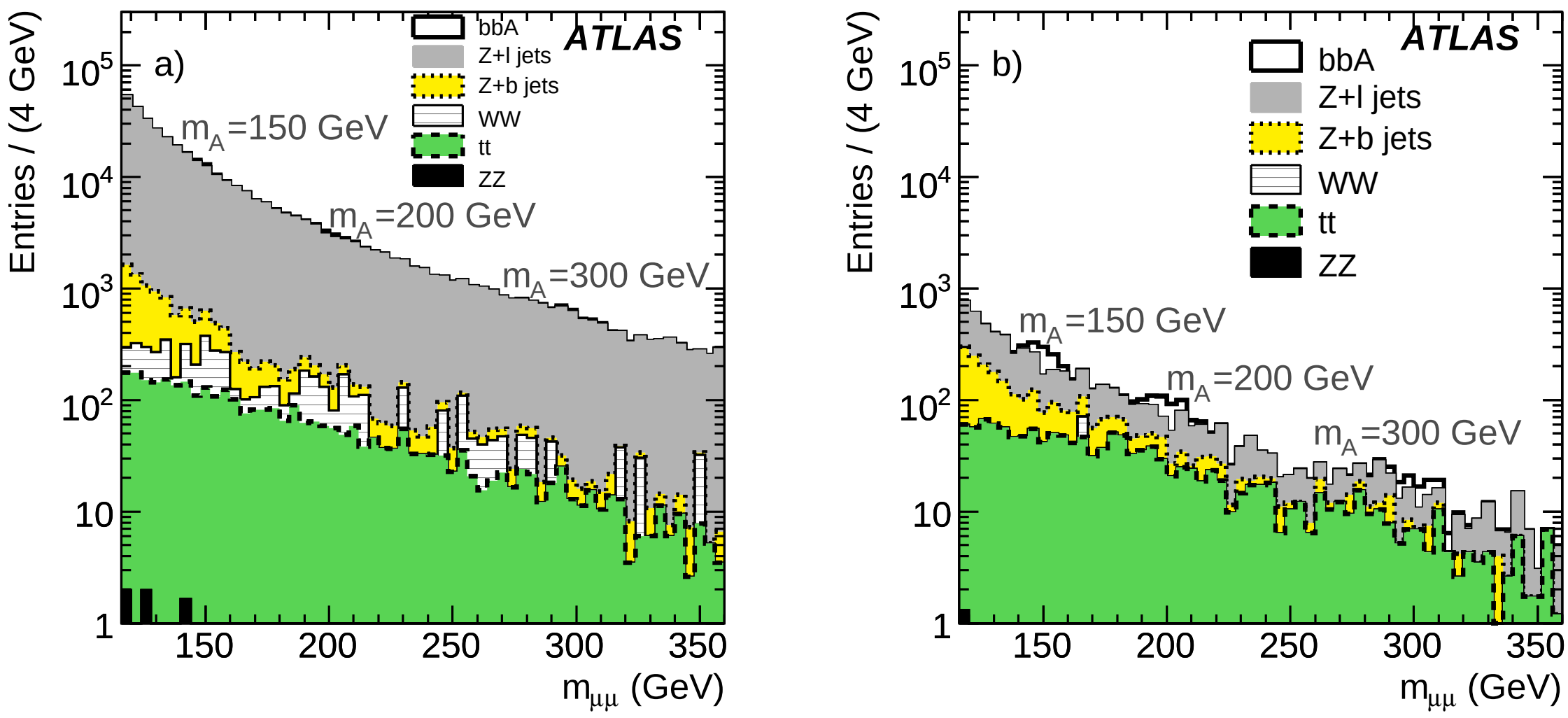


Figure 13: Invariant dimuon mass distributions of the main backgrounds and the $A$ boson signal at masses $m_A$=150, 200 and 300 GeV and $\tan\beta = 30$, obtained for the integrated luminosity of 30 fb$^{-1}$ . B-tagging has been applied for the event selection. The production rates of $H$ and $A$ bosons have been added together. a) for the 0 b-jet final state and b) for the final state with at least 1 b-jet.

limited trigger efficiency on the final event selection.

Since the signal processes are characterized by the two high-$p_T$ muons in the final state (see Figure 9), the most reliable trigger item is a single high-$p_T$ muon with $p_T$ >20 GeV. The single-muon trigger efficiency is mostly limited by the geometrical acceptance of the trigger chambers in the muon spectrometer, as discussed in detail in [29]. The efficiency of the trigger selection for events passing all previously described event selection criteria is shown in Table 5 for the signal at 200 GeV and for the background samples. A very similar trigger selection efficiency can be observed for all signal and background samples. The relatively high final trigger selection efficiency of ∼95% corresponds to the decrease of the signal significance by 2-3% compared to the previously shown results. It has been shown that the observed efficiencies are effectively stable to a level of about ±0.5% at any level of the event selection, since the selection criteria are not affecting the muon kinematics. Furthermore, the trigger selection does not induce any bias to the dimuon invariant mass distribution.

| Dataset | L1 | High level trigger |
|---|---|---|
| $bbA$, 200 GeV | 97.2 | 95.0 |
| $t\bar{t}$ | 97.1 | 95.1 |
| $b\bar{b}Z$ | 97.1 | 94.8 |

Table 5: Trigger selection efficiency (%) of events which pass all event selection criteria described in Section 5.2. The results are listed for the signal and the two background processes.

The studies described above have been performed under the assumption that the trigger threshold for

a single high-$p_T$ muon is 20 GeV. If this threshold should be higher, some of the signal events might be lost by the trigger selection. In order to evaluate the sensitivity of the signal selection on the varying trigger thresholds, a dedicated study has been performed. New trigger thresholds of 20, 26 and 30 GeV are emulated by the requirement that at least one reconstructed muon with a $p_T$ above the given threshold exists in the event. This muon is also required to have a matching L1 region-of-interest in the muon spectrometer, in order to take into account the holes in the geometrical trigger acceptance. The trigger selection efficiencies for the new trigger thresholds are shown in Figure 14 for events remaining after each of the analysis cuts applied on the $b\bar{b}A$ signal with mass $m_A$=200 GeV. No visible degradation of the

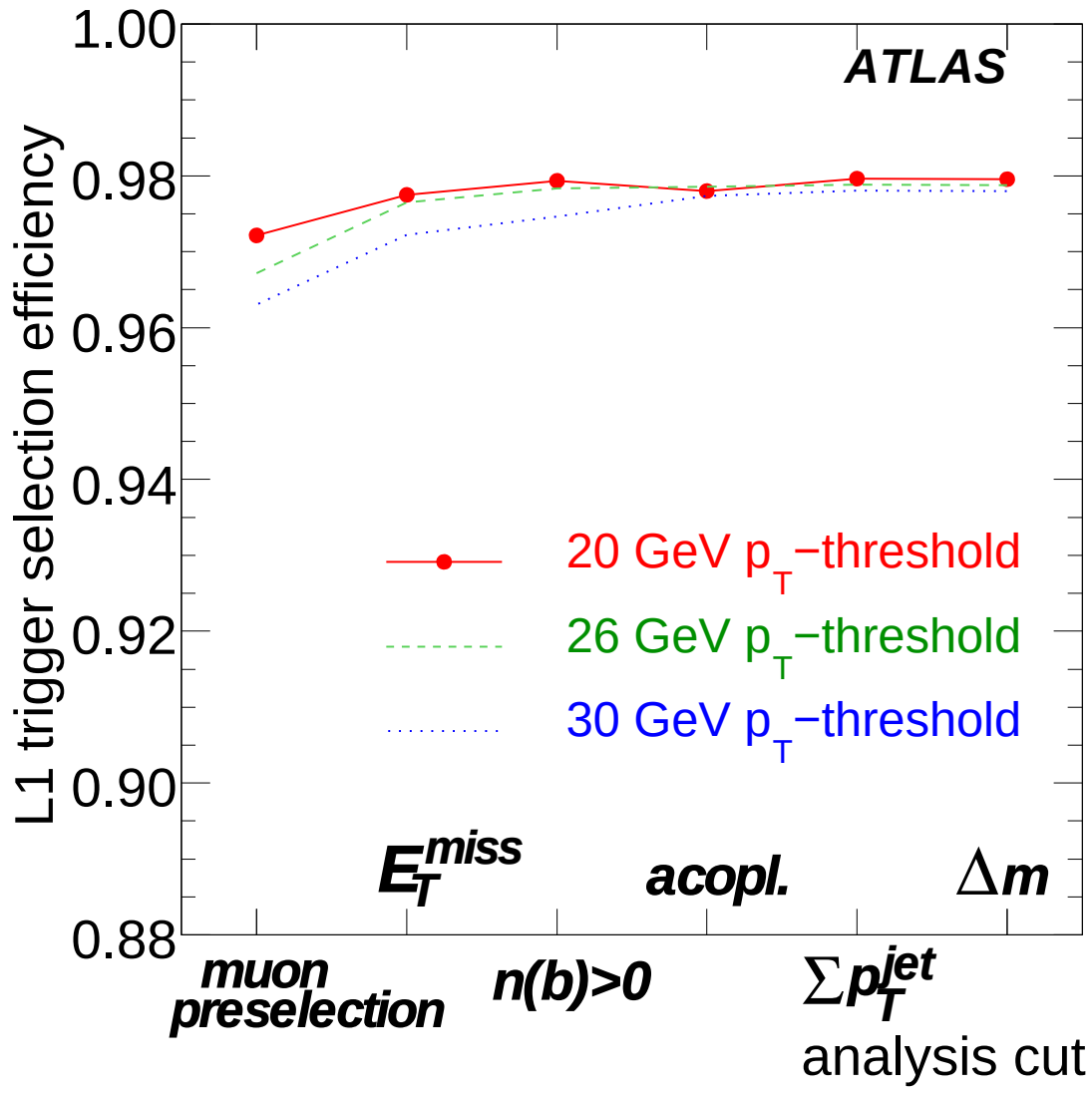


Figure 14: Trigger selection efficiency of the $b\bar{b}A$ signal events ($m_A$=200 GeV) in dependence on the offline analysis selection. Different curves show the results obtained for different trigger thresholds.

final signal selection is observed for the trigger threshold variations of up to 30 GeV. Similar behavior is observed also for the signal at other mass points. This result can be explained by a rather high muon momenta in the signal samples, such that there is only a small fraction of signal events in which both muons have a $p_T$ below 30 GeV. The dependence of the $t\bar{t}$ and $Zb\bar{b}$ background selection on the trigger threshold of up to 30 GeV is shown to be smaller than 2% after all analysis cuts.

# 6 Systematic uncertainties

Several sources of systematic uncertainties can affect the total yields of both the signal and the background events after applying the selection criteria specified in the previous Section. In this Section, the influence of the theoretical and the detector-related systematic uncertainties is evaluated.

## 6.1 Theoretical uncertainties

As mentioned previously, the Higgs boson production in association with the b-quarks has been simulated with PYTHIA and SHERPA Monte Carlo generators and the obtained differential distributions are scaled to the Feynhiggs NNLO cross-sections. It is important to remark that no cuts have been applied on the transverse momenta of the generated b-quarks, since also the event topologies with less than two

observed b-jets are studied in the presented analyses. (An outgoing b-quark can be considered as observable if its momentum is above ∼15 GeV). Therefore, the corresponding NNLO cross-sections also have to be calculated in an inclusive way, with no constraints on the kinematics of the b-quarks.

As discussed in Section 2, the total theoretical uncertainty on the inclusive cross-section amounts to ∼17%, depending on the Higgs boson mass. In addition, since the experimental search distinguishes between the final states with 0 and those with 1 or more b-jets (with $p_T$ >20 GeV), it is important to show that the proportion between the mentioned different event topologies which is obtained from the Monte Carlo generators also agrees with theory predictions. For this purpose, the fraction of generated events is calculated which have *exactly* one b-quark with $p_T$ >15 GeV and $|\eta|$ <2.5 in the final state (before hadronization). This number is then compared to the ratio of the MCFM cross-section calculated for the $bg \to bH$ process with the same cuts on the b-quarks, divided by the inclusive cross-section. The result of the comparison is shown in Figure 15, being comparable to the 20% uncertainty. The SHERPA

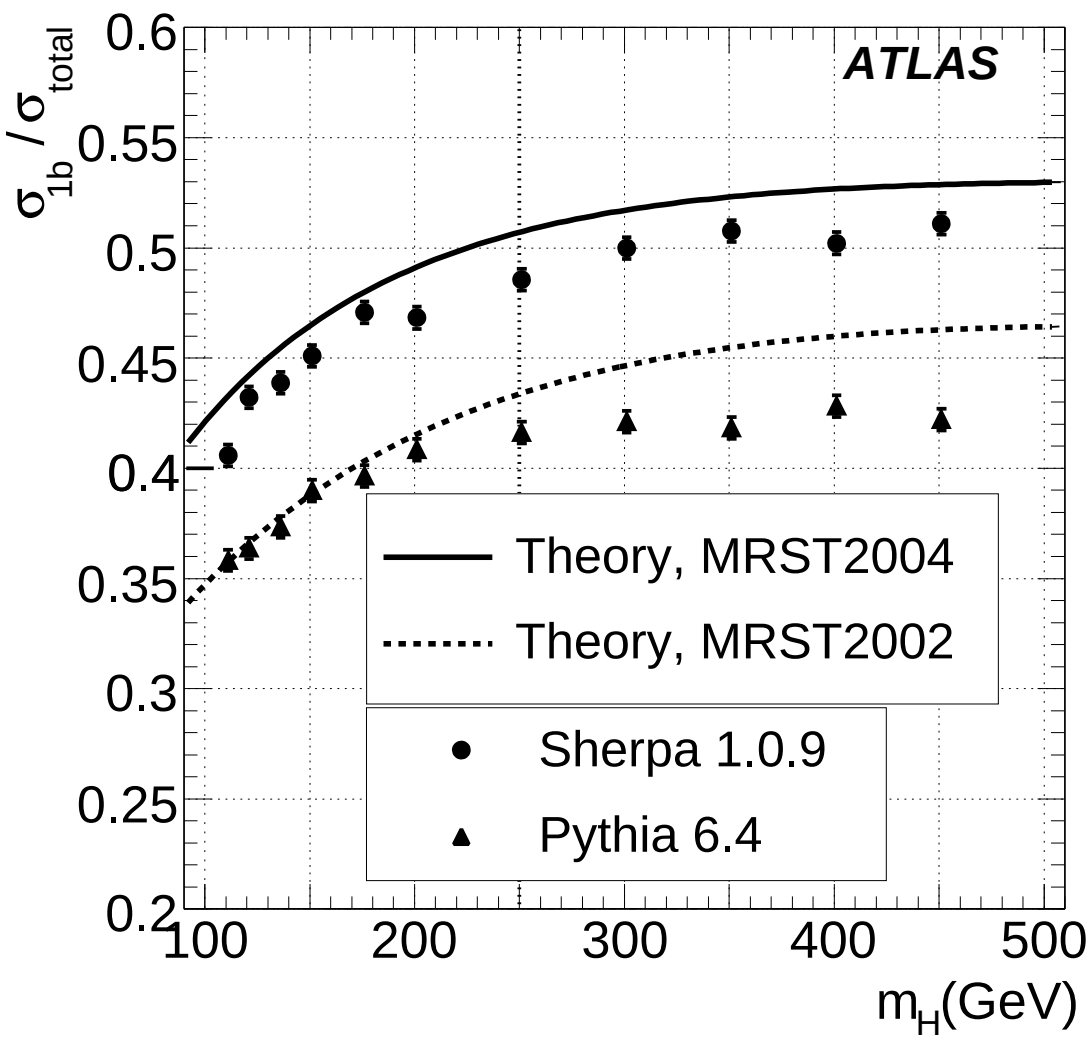


Figure 15: Fraction of events $\sigma_{1b}$ with exactly one b-quark with $p_T$ >15 GeV and $|\eta|$ <2.5, relative to the total inclusive cross-section $\sigma_{total}$, shown in dependence on the Higgs boson mass $M_H$. The lines show the theory prediction as the ratio between the MCFM $bg \to Hb$ and $b\bar{b} \to H$ cross-sections. Solid line: MRST2004 pdf set, dashed line: MRST2002 pdf set. Circles: result from SHERPA 1.0.9. Triangles: PYTHIA 6.4 $gg \to b\bar{b}H$ process.

prediction is about 20% higher than PYTHIA and agrees better with the theory calculation with the MRST2004 pdf set. The MRST2002 pdf set gives a better compatibility with PYTHIA. Taking into account the theory uncertainties mentioned above, one can conclude that the samples produced with both generators can reproduce the NLO predictions from a single b-quark rate.

The observed differences between SHERPA and PYTHIA are visible also in the differential b-quark distributions. The differential $p_T$ and $\eta$-distributions of the leading b-quark, as obtained by the PYTHIA and SHERPA generators are compared in Figure 16 for the signal at the mass of 200 GeV. The comparison is performed on a parton level, before hadronization. A slightly harder transverse momentum spectrum and a more central pseudorapidity of the b-quarks is observed in the SHERPA events.

## 6.2 Detector-related systematic uncertainties

Systematic uncertainties related to the muon and the jet reconstruction, as well as the uncertainties on the b-tagging performance have been evaluated with rather conservative estimates on the level of under-

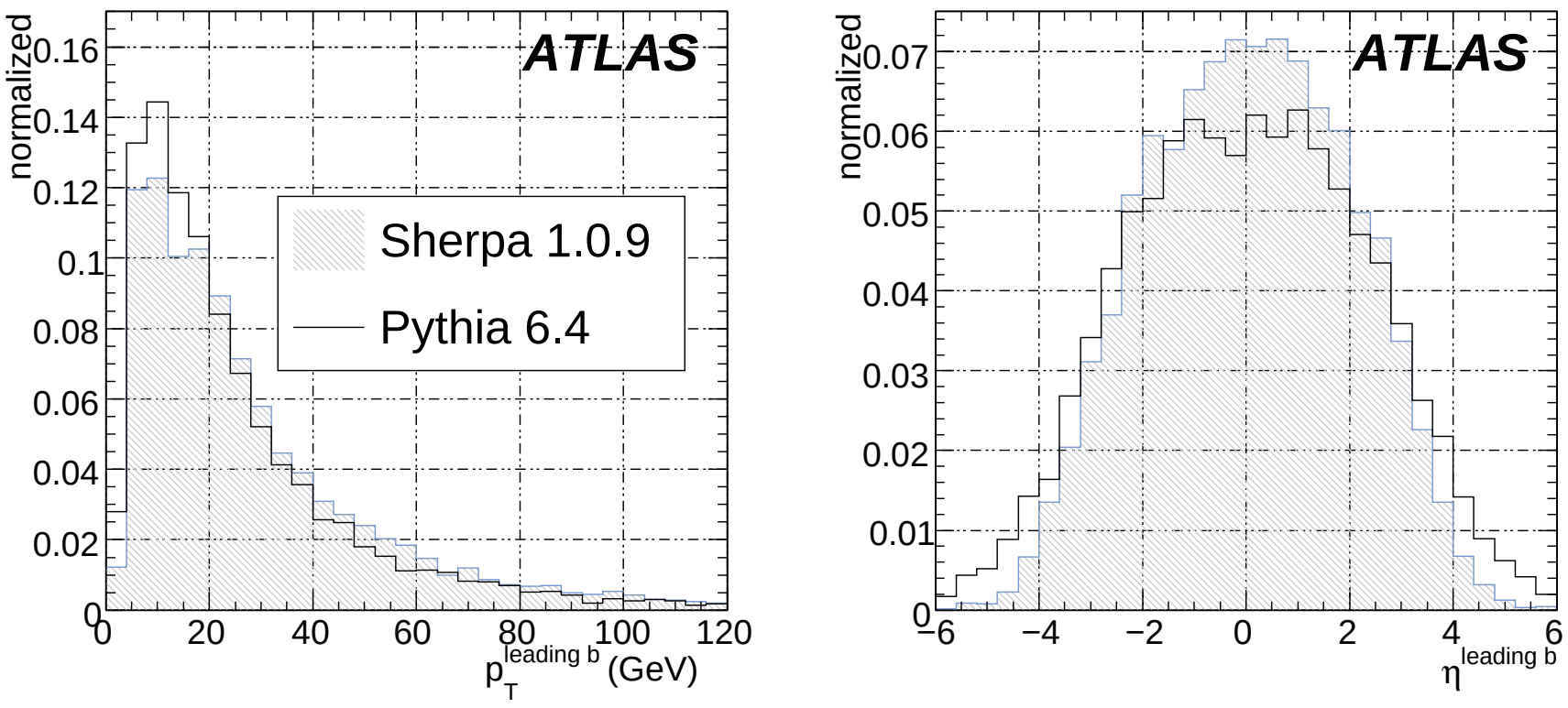


Figure 16: Differential $p_T$ and $\eta$ distribution of the leading b-quark, obtained by SHERPA (gray histogram) and by the PYTHIA $gg \to b\bar{b}H$ calculation (solid line). All histograms have been normalized to unity.

standing of the detector performance. This level is assumed to correspond approximately to an integrated luminosity of 1 fb$^{-1}$. For each of the above effects, the corresponding change of the reconstructed missing transverse energy is taken into account. Further more, systematic effects due to the differences between the fast and the full simulation are taken into account by means of the comparison between the full and fast simulation.

**Muon reconstruction** uncertainties are treated separately for the reconstruction efficiency, muon momentum resolution and the muon momentum scale. The efficiency of the muon identification is assumed to be known with an accuracy of $\pm 1\%$, based on the results of the tag-and-probe method for the muon efficiency measurement from the real data [30]. Systematic errors of the muon momentum scale are taken to be $\pm 1\%$, arising for instance from the non-perfect knowledge of the magnetic field. The incomplete understanding of the material distributions inside the detector, as well as possible residual detector misalignment can lead to an additional smearing of the muon momentum resolution. Based on early detector calibration, the additional smearing is expected to be $\sigma(\frac{1}{p_T}) = \frac{0.011}{p_T} \oplus 0.00017$ (in 1/(GeV)). The first term enhances the effect of the Coulomb scattering, while the second enhances the contribution from the misalignment.

**Jet reconstruction** uncertainties are estimated by the jet performance group [25]. In the pseudorapidity region below $|\eta|$=3.2, the jet energy scale uncertainty of $\pm 3\%$ and a resolution uncertainty of $\sigma(E) = 0.45 \cdot \sqrt{E}$ are assumed. For $|\eta| >$3.2, the corresponding values are $\pm 10\%$ and $\sigma(E) = 0.63 \cdot \sqrt{E}$ respectively.

**b-tagging** efficiency and the fake rate are crucial for the described analysis. Conservative relative uncertainty of $\pm 5\%$ on the b-tagging efficiency and $\pm 10\%$ uncertainty on the rejection of the light jets have been assumed.

The results obtained after implementing each of the above systematic uncertainties separately into the analysis are shown in Table 6 for the signal at $m_A$=150 GeV and for the backgrounds within the corresponding mass window. Signals at different mass points are affected by a similar amount. The background uncertainties are also rather independent of the dimuon mass region, but from one exception. The muon momentum scale mostly affects the background in the lower mass region, close to the $Z$ resonance, while the deviations become smaller for the higher signal masses, i.e. one observes $\pm 11\%$, $\pm 5\%$ and $\pm 3\%$ for the Higgs masses of 110, 130 and 200 GeV respectively.

| Systematic uncertainty [%] | Signal | $t\bar{t}$ | Z+ b jet | Z+light jet |
|---|---|---|---|---|
| Muon efficiency | ±2 | ±1 | ±2 | ±2 |
| Muon $p_T$ scale | ±2 | ±3 | ±3 | ±5 |
| Muon resolution | -2 | -4 | 3 | 3 |
| Jet energy scale | ±1 | ±5 | ±2 | ±2 |
| Jet energy resolution | -1 | -3 | -1 | -1 |
| b-tagging efficiency | ±4 | ±3 | ±4 | ±2 |
| b-tagging fake rate | ±1 | ±0 | ±0 | ±6 |
| Full-Atlfast corrections | 0 | +8 | -10 | +5 |
| Total | ±5 | ±12 | ±12 | ±11 |

Table 6: Relative deviation of the selection efficiency in % for the signal at $m_A$=150 GeV and for the background events after imposing each of the systematic uncertainties separately, as described in the text. The total deviation is given as quadratic sum of the separate contributions, including the one-sided corrections.

# 7 Background estimation based on the measured data

As previously discussed, theoretical and experimental uncertainties can lead to systematic errors in determination of the background rates. Additional information on the shape and the size of the background contributions can be collected from the real data. As shown before, two major backgrounds surviving all event selection criteria are the $Z+jet$ and the $t\bar{t}$ processes. Two strategies to estimate their contributions from the measured data shall now be described.

The first method makes use of the fact that the branching ratio for $A/H$ boson decays into two electrons is negligible compared to the dimuon decay channel. Therefore, since one doesn't expect any signal in the dielectron final state, one can use this signature to determine the total background contribution. Additionally, the signatures with one electron and one muon in the final state provide the contribution of the $t\bar{t}$ background alone, since the $Z+jet$ processes do not contribute to this final state. Thus, one can separately measure the two background contributions. The background estimation based on the $e^+e^-$-channel has been discussed in detail in [31]. Good agreement has been demonstrated between the dilepton invariant mass distributions for the $Z\to\mu\mu$ and the $Z\to ee$ processes. In this paper, the emphasis is given to a similar procedure for the determination of the $t\bar{t}$ background.

The goal of the second method is to define a set of event selection criteria which allow for the higher selection efficiency for the particular background process, while simultaneously rejecting all other signal and background contributions. Such background-enriched control sample can be used to better understand the shape of the invariant dimuon mass distribution.

## 7.1 Background estimation based on the $e^+e^-$ and $\mu^\pm e^\mp$ signatures

The estimation of the $t\bar{t}$ background is important for the analysis channel with at least one b-jet in the final state. A study is performed using fast simulation of the $t\bar{t}$ background, in order to obtain a reliable statistical accuracy. The detector performance given by the fast simulation has been adjusted such to reproduce the performance obtained with the full simulation, as mentioned previously. Based on the studies in [31], the shape of the dilepton invariant mass distribution obtained for the $\mu^+\mu^-$, $e^+e^-$ and $\mu^\pm e^\mp$ final state in the $t\bar{t}$ process are expected to be very similar. The total number of background events selected in each of the three final states will be different due to different reconstruction efficiencies for muons and electrons. However, since these efficiencies can be experimentally measured with an accuracy

of better than 1%, this effect can be corrected for rather precisely. Additional differences may occur due to an additional calorimeter activity in presence of electrons. This can be taken into account by rejecting all reconstructed jets which overlap with reconstructed electrons.

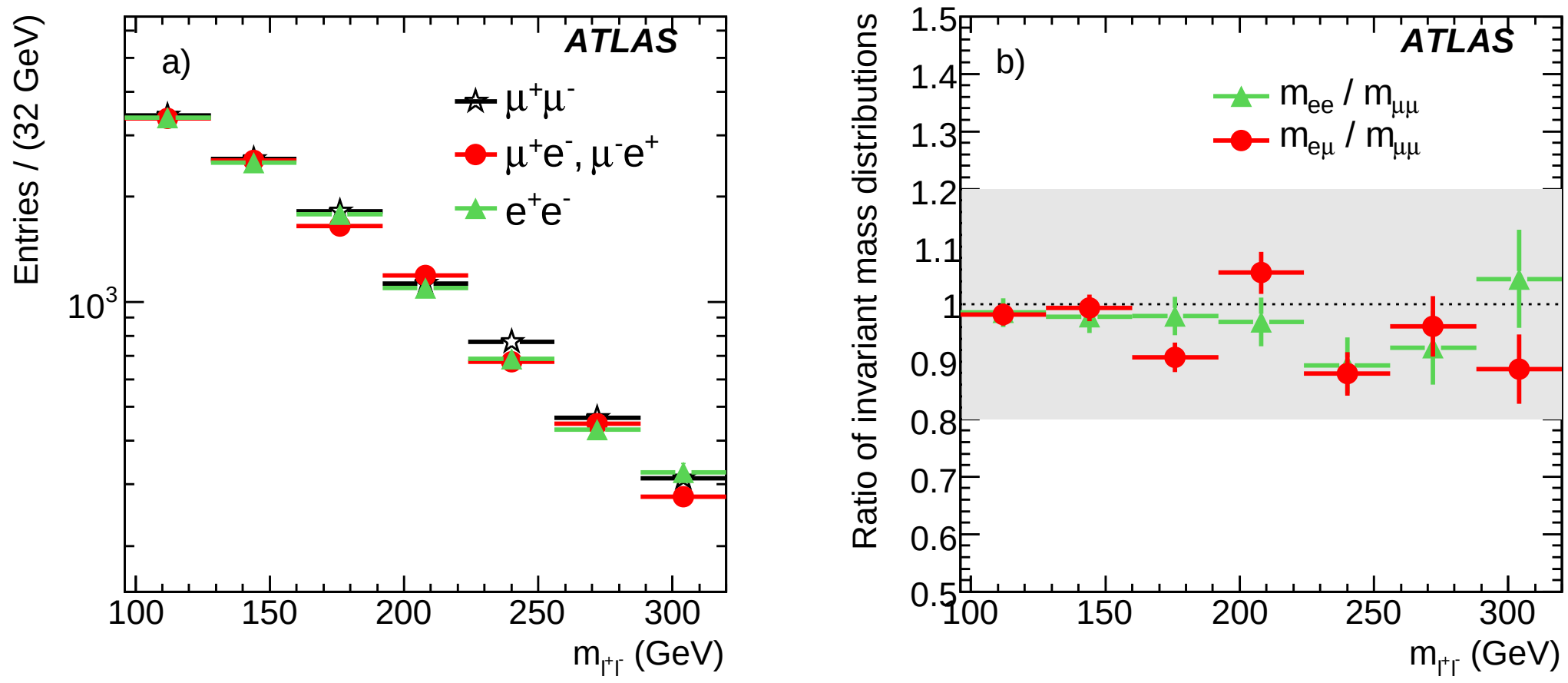


Figure 17: (a) The $\mu^+\mu^-$ invariant mass distribution (stars) of the $t\bar{t}$ background, and its estimates obtained from the $e^+e^-$ (triangles) and $\mu^\pm e^\mp$ (full circles) final states. (b) Corresponding ratios of estimated and actual $\mu^+\mu^-$ invariant mass distributions.

Figure 17(a) shows the invariant mass distributions obtained for the $\mu^+\mu^-$, $e^+e^-$ and $\mu^\pm e^\mp$ final states in $t\bar{t}$ events. The dielectron distribution has been scaled down by a factor $0.84^2$, to account for the difference between the electron and muon reconstruction efficiency. Similarly, the $\mu^\pm e^\mp$-distribution has been scaled down by $0.5\times0.84$. Figure 17(b) shows the corresponding ratios of estimated and actual $\mu^+\mu^-$ invariant mass distribution. The subtraction of the $\mu^\pm e^\mp$ sample from the total background is illustrated in Figure 18 for $\mathscr{L}$ =30 fb$^{-1}$, for the analysis requiring identification of at least one b-jet.

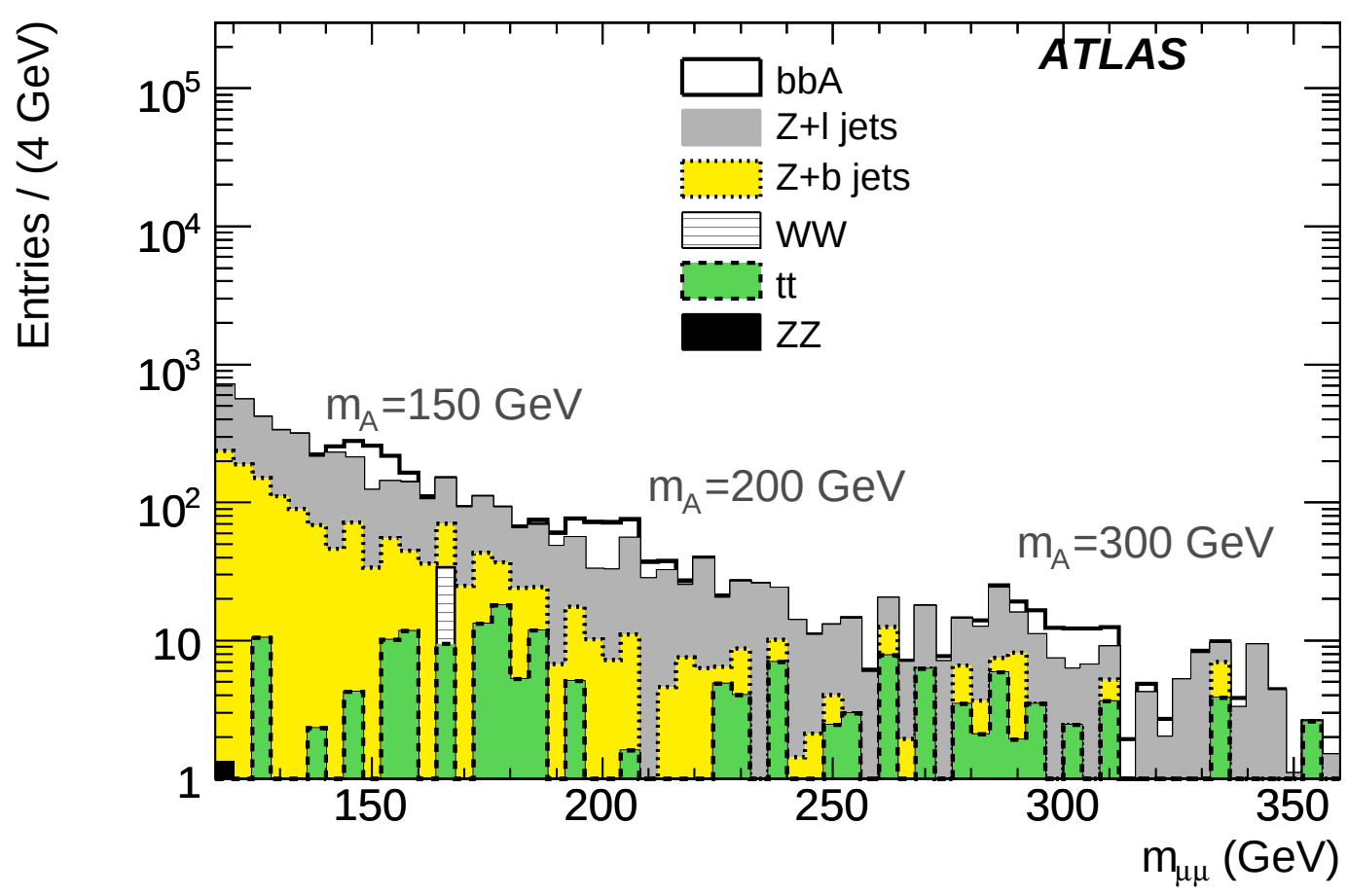


Figure 18: Invariant dimuon mass as in Figure 13(b), after the subtraction of the $t\bar{t}$ background estimated in the $\mu^\pm e^\mp$ final state. The distributions correspond to an integrated luminosity of $\mathscr{L}$ =30 fb$^{-1}$.

### 7.2 Background estimation from the control samples

In analyses with 0 b-jets in the final state the dominant background is the $Z+jet$ process. Since the topology of this process is very similar to the signal topology, it is rather difficult to define a set of selection criteria which would allow for the extraction of the $Z$ background and simultaneous rejection of the signal.

Contrary to that, the $t\bar{t}$ process is characterized by a relatively high missing transverse energy and a high jet activity. Since this background is important for the analysis with at least one jet in the final state, the event selection criteria of this analysis are modified by selecting only events with a missing transverse energy above 60 GeV and by removing the cut on the $\sum p_T^{jet}$-variable. All other selection criteria remain the same. The purity of the $t\bar{t}$ control sample obtained after the described event selection is shown in Figure 19(a). All remaining processes are suppressed to a negligible amount. Figures 19(b) and (c)

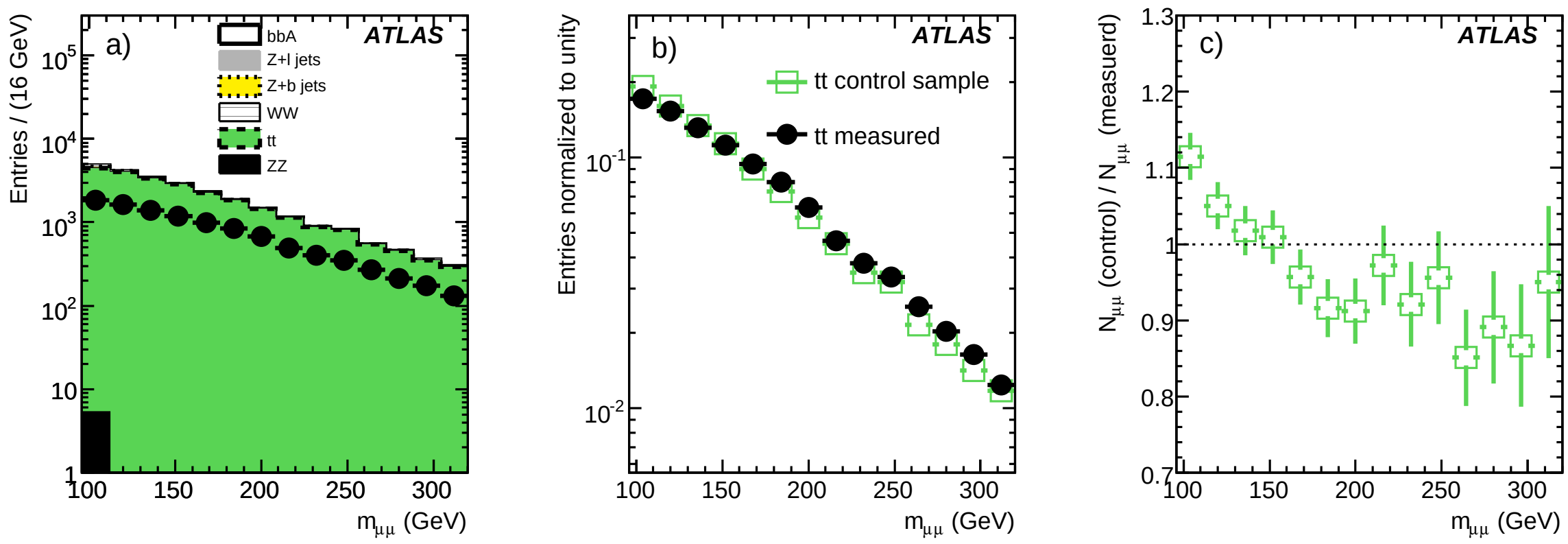


Figure 19: (a) $t\bar{t}$ control sample (stacked histogram) obtained as described in text. Full circles indicate the actually measured $\mu^+\mu^-$ invariant mass distribution for the $t\bar{t}$ background, after applying the standard analysis selection cuts. (b) The measured (full circles) and the control $t\bar{t}$ distributions (open squares), normalized to the same number of events. (c) The ratio of normalized distributions from (b).

show the $t\bar{t}$ background obtained with the selection criteria described in Section 5.2 for at least one b-jet in the final state ($t\bar{t}$ measured) compared to the distribution obtained from the $t\bar{t}$ control sample. Both distributions are normalized to the same number of events. The shape of the $t\bar{t}$ background can be estimated by the described procedure with an accuracy of 10-20%.

### 7.3 Fit function for the parametrization of the background

The background can be parametrized by the function $f_B$ consisting of a Breit-Wigner and an exponential contribution,

$$f_B(x) = \frac{a_1}{x} \cdot \left[ \frac{1}{(x^2 - M_Z^2) + M_Z^2\Gamma_Z^2} + a_2 \cdot exp(-a_3 \cdot x) \right], \tag{1}$$

where $x$ is the running dimuon mass, while $a_1$, $a_2$ and $a_3$ are the free parametrization parameters. The mean $M_Z$ and the width $\Gamma_Z$ of the Breit-Wigner distribution describe the $Z$ resonance. The parameters $a_{2,3}$ describing the exponential part can be determined by the fit on the background estimated from data, as described in the previous subsection. The overall normalization factor $a_1$ is determined by the fit on the side bands of the dimuon mass distribution. Figure 20 shows the result of the fit on the background distribution obtained after all analysis cuts for the case with at least one b-jet in the final state. The

two dashed lines correspond to the errors on the $a_{2,3}$ parameters, as obtained from the $e^+e^-$ data at an integrated luminosity of 10 fb$^{-1}$.

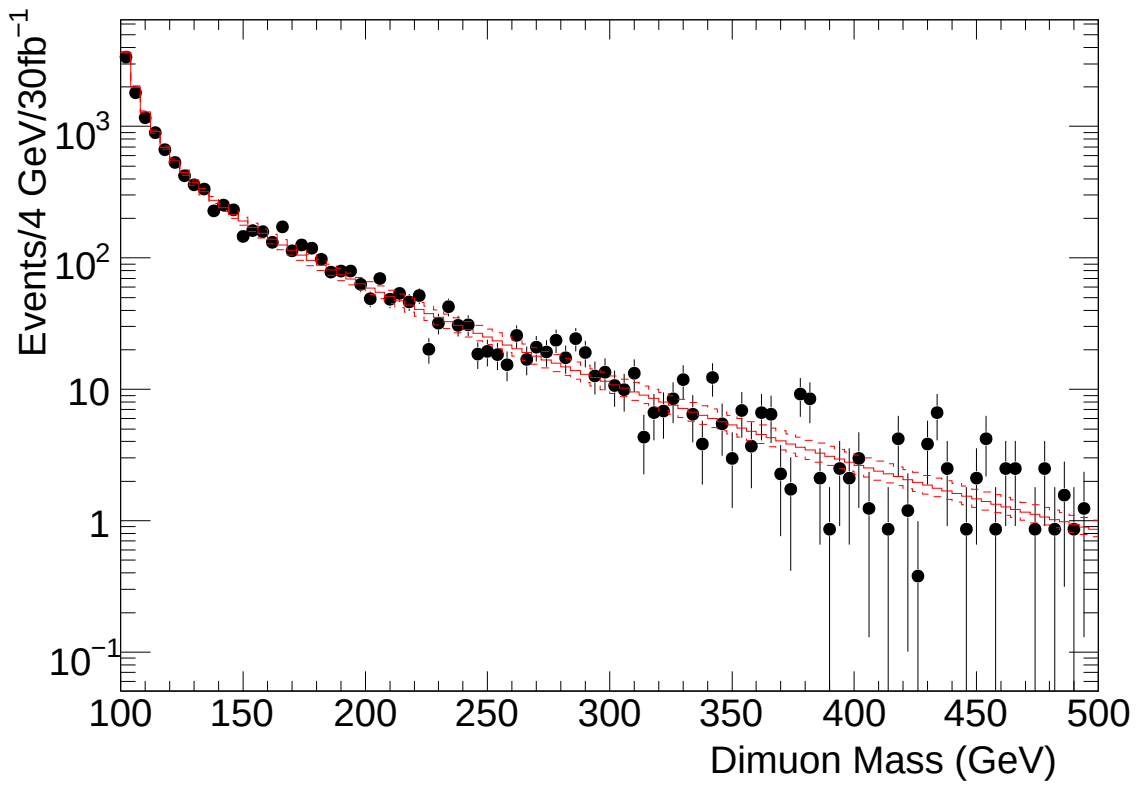


Figure 20: Fit (full line) of the background function $f_{bkg}$ (see Eq. 1) on the background distribution (full circles) resulting from the analysis with at least one b-jet in the final state. The dashed lines represent the shape variations given by the errors on the $a_2$ and $a_3$ parameters, from the fit on the $e^+e^-$ data.

The accuracy of the background parametrization by the described method has been tested by means of large number of toy Monte Carlo experiments at different integrated luminosities $\mathscr{L}$. Typical results of the background extraction in the mass window from 188-212 GeV are shown in Figure 21 for an integrated luminosity of 15 fb$^{-1}$ (left) and 3 fb$^{-1}$ (right). The empirically evaluated expected uncertainty of

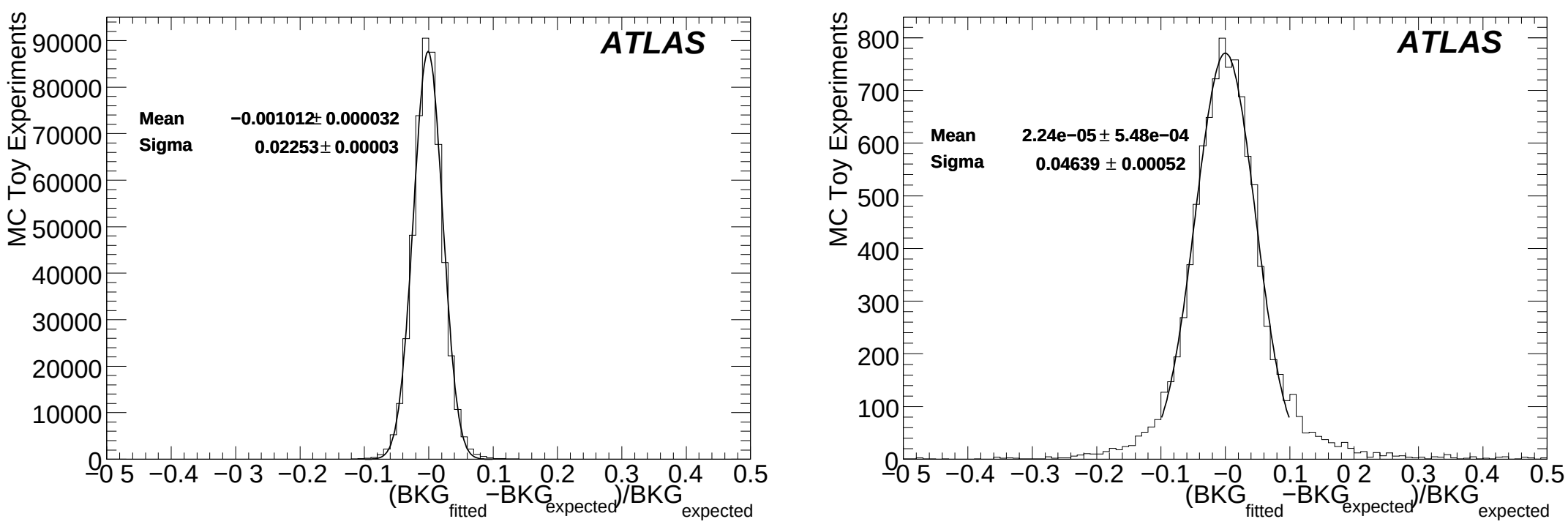


Figure 21: Normalized difference of the expected number of background events ($BKG_{expected}$) in the mass window from 188-212 GeV, and the number ($BKG_{fitted}$) obtained from the fit method described in the text (left) for an integrated luminosity of =15 fb$^{-1}$ and (b) =3 fb$^{-1}$.

the background determination from the fit to data is $\sim 8\%/\sqrt{\mathscr{L}[\mathrm{fb}^{-1}]}$. The variation of the exponential shape, due to the errors on the fit parameters $a_2$ and $a_3$, plays a non-marginal role only for dimuon masses above 300 GeV, decreasing the relative uncertainty to $\sim 10\%/\sqrt{\mathscr{L}[\mathrm{fb}^{-1}]}$. Conservatively the latter expression will be used in the analyses discussed below. Similar fit procedure has been performed also for the analysis with 0 b-jets in the final state. Due to the larger amount of background, a smaller background uncertainty of $\sim 2\%/\sqrt{\mathscr{L}[\mathrm{fb}^{-1}]}$ can be obtained in this case, including the systematic uncertainty on the shape parameters $a_2$ and $a_3$.

# 8 Evaluation of the signal significance

Two approaches have been applied to evaluate the statistical significance of the observed signal. The ”fixed mass” approach provides the significance based on the number of signal and background events which are expected in a given range of the dimuon invariant mass. However, the location of the signal mass peak can be determined only if the signal rates are sufficiently high, or if the signal is already discovered in the $A/H/h \to \tau\tau$ decay channel. In a more general ”floating mass” approach, no constraints are applied on the dimuon invariant mass.

The number of signal and background events can be determined from the fit of the function $f_{SB}$ to the data,

$$f_{SB}(x) = p_0 \cdot f_B + p_1 \cdot f_S = p_0 \cdot f_B + p_1 \cdot \frac{1}{\sigma_A\sqrt{2\pi}} \cdot exp\left(-\frac{(x-p_2)^2}{2\sigma_A^2}\right), \tag{2}$$

where $f_S$ is the Gaussian distribution describing the signal. $p_2$ is the mass of the signal (which can be fixed or left as a free, ”floating” parameter) and $\sigma_A$ is the width of the Higgs resonance. $p_0$ is the background scale and $p_1$ the total number of signal events. The number of signal ($N_S$) and background ($N_B$) events used for the calculation of the signal significance is extracted from the fit, by integration in a window of $\pm 2\sigma_A$. The signal significance is evaluated by means of the profile likelihood method [32], using the obtained number of signal and background events as an input and taking into account the background uncertainty ($10\%/\sqrt{\mathscr{L}[\text{fb}^{-1}]}$) as discussed in Section 7.3 above.

The results obtained for the signal significance have been cross-checked by the large number of Monte Carlo pseudoexperiments for several different integrated luminosities. The pseudoexperiments are based on the Monte Carlo distributions and are divided into the ”background-only” experiments (*BO*), containing only background contributions, and the ”signal-plus-background” experiments (*SpB*) having both the signal and the background contribution. In addition to the profile likelihood calculation in each of these experiments, the signal significance can be estimated also from the log-likelihood ratios. The log-likelihood ratio $\ln Q_{BO}$ and $lnQ_{SpB}$, both defined as $(N_S + N_B)\ln\frac{N_S+N_B}{N_B} - N_S$, are evaluated for each *BO*- and *SpB*-pseudoexperiment. The signal significance is obtained from the probability of a Type-II error, defined by the fraction of *BO* pseudoexperiments which have a log-likelihood ratio $\ln Q_{BO}$ larger than the median of the $\ln Q_{SpB}$-distribution. The probability of a Type-I error, i.e. the number of *SpB*-pseudoexperiments which fall below the median of the $\ln Q_{BO}$-distribution, was used to determine the 95% CL limits.

Figure 22(left) shows the comparison of the signal significances obtained in the fixed-mass approach. The solid line shows the results of the profile likelihood method, while the dots are the results given by the Type-II error probabilities obtained in the pseudoexperiments. At very low luminosities, the profile-likelihood estimation based on average values seems to slightly overestimate the significance. However, at luminosities close to those needed for the $5\sigma$-significance, the two calculations give equivalent results. Figure 22(right) shows the degradation of the signal significance observed once the floating-mass approach is applied in which the Higgs mass is left as a free parameter of the fit (usually reffered to as a look-elsewhere effect). In general, the ratio of the Type-II error probabilities obtained from the fixed-mass and the floating-mass approaches is constant and approximatelly equal to the explored mass range divided by the signal mass width. Correspondingly, the ratio of signal significances is lower at low luminosities, while the difference between the two approaches is reduced to 5% or less at the luminosities close to those needed for a $5\sigma$-discovery.

## 8.1 Influence of the systematic uncertainties on the signal significance

In order to include the systematic uncertainties in the calculation of the signal significance, one should perform additional large number of pseudoexperiments for each of the systematic effects. However, the

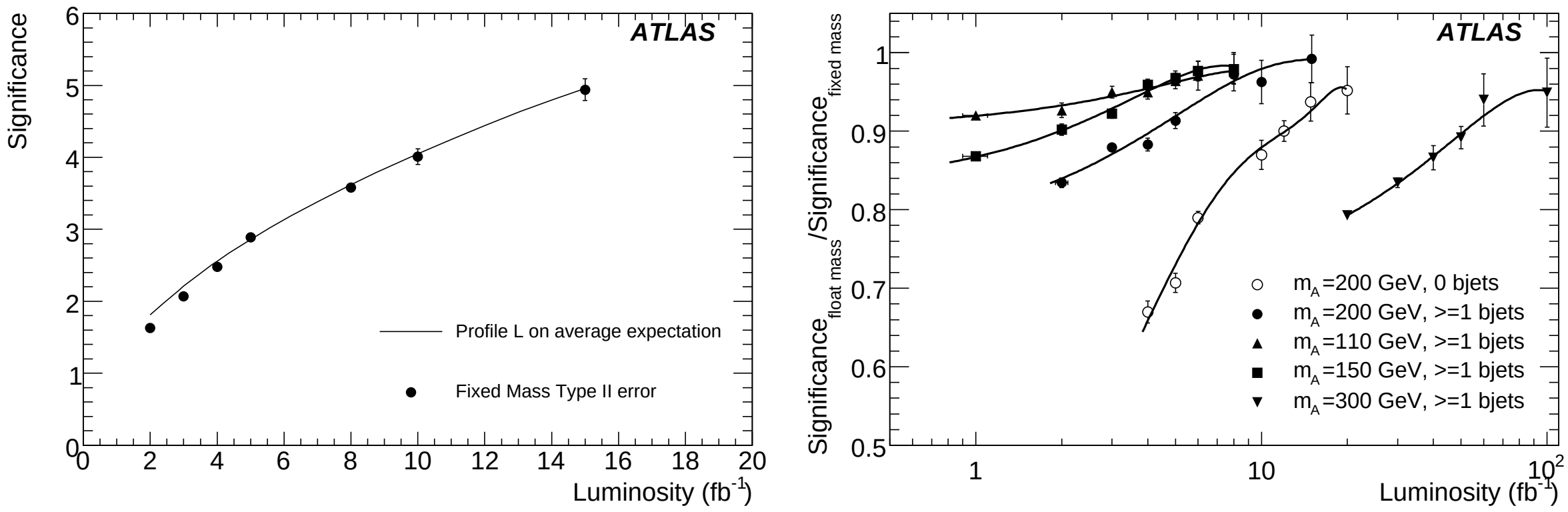


Figure 22: (left) Signal significance obtained in the fixed-mass approach for the signal at $m_A$=200 GeV and $\tan\beta$=30, as a function of integrated luminosity. The full line corresponds to the results obtained by the profile likelihood method using the average number of expected signal and background events, while the dots show the results obtained from the Type-II error probability from a large amount of pseudoexperiments. Both estimations include the background uncertainty from the $f_{SB}$-fit. (right) Ratio of signal significances obtained with the floating- and the fixed-mass approach, as a function of the integrated luminosity.

difference between the signal significance in the floating-mass and in the fixed-mass approach is small compared to the effect of the additional systematic uncertainties. Thus, in the following the fixed-mass approach is used and the full treatment of the look-elsewhere effect is left for the future studies.

By means of the background estimation from the data, the amount of the background events underneath the signal can be determined with an accuracy of $\sim$10%/$\sqrt{\mathscr{L}[\text{fb}^{-1}]}$, as described in Section 7. This can be achieved independently of the systematic uncertainties discussed in Section 6.

Therefore, the influence of the systematic uncertainties on the signal significance is given by the corresponding changes in the expected number of signal and background events. The numbers entering the significance calculation are therefore changed accordingly, taking into account that the systematic uncertainty is different for different background processes. The signal significance is calculated separately for each systematic effect and the deviations from the original signal significance are added in quadrature. Figure 23 shows the signal significance obtained for two different masses $m_A$, as a function of $\tan\beta$. The contributions of the uncertainty in the background determination from the fit to data and of the experimental systematic uncertainties to the expected significance are shown separately.

## 9 Discovery potential and the exclusion limits

The previously described methods for the evaluation of the signal significance and of the exclusion limits have been applied to all mass points studied. Table 7 summarizes the signal significances obtained for different signal mass points at $\tan\beta$=30 and $\mathscr{L}$=10 fb$^{-1}$, for the analysis with 0 b-jets and with at least one b-jet in the final state. The luminosities needed to reach the 5$\sigma$ signal significance and the 95% CL exclusion limit are given in Table 8, for different signal masses at $\tan\beta$=30. The results shown so far do not include the possible degradation due to the influence of pile-up and cavern background. The degraded resolution of the missing transverse energy is expected to cause a $\sim$15% change in the final selection of the signal and the $Z$ background. The $t\bar{t}$ background is characterized by a large $E_T^{miss}$ and is therefore somewhat less sensitive to the $E_T^{miss}$ performance under pile-up ($\sim$10% change in the selection

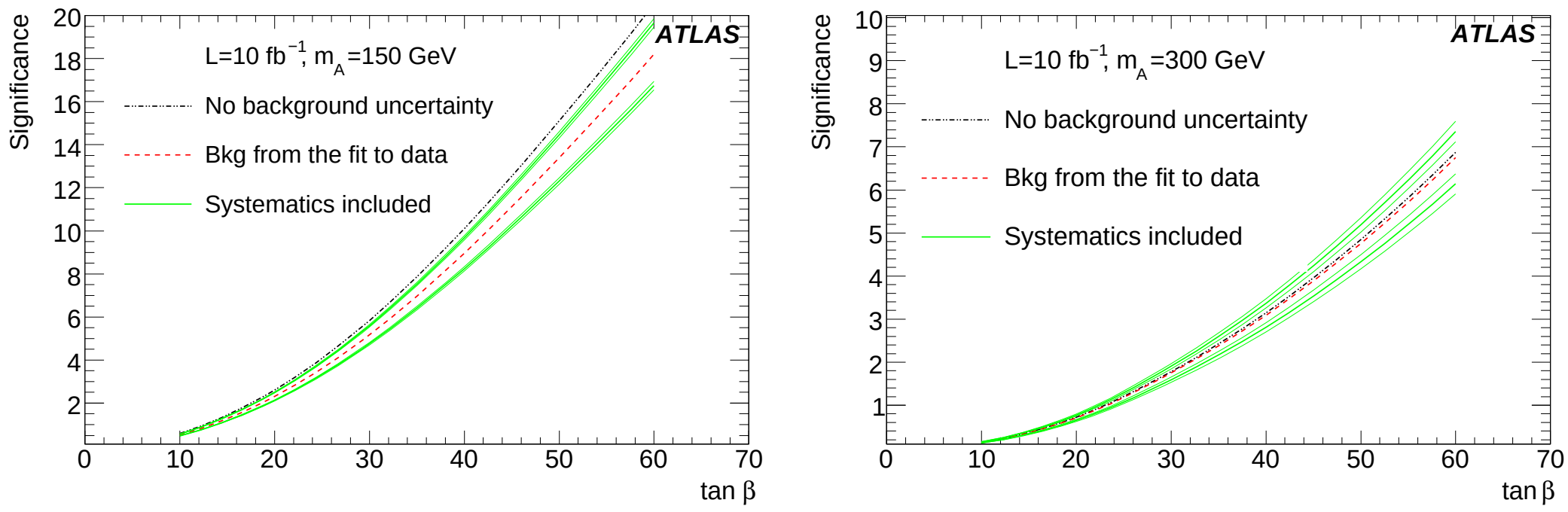


Figure 23: Signal significance as a function of $\tan\beta$ for the integrated luminosity of 10 fb$^{-1}$ and $m_A$=150 GeV (left) and $m_A$=300 GeV (right). The dotted curves are obtained assuming a negligible error of the background determination. The dashed curves show the result once the background uncertainty is taken into account, as obtained from the fit to the data. The solid curves additionally include the systematic uncertainties. The width of the solid curves indicates the errors in the background shape parametrization.

| $m_A$ (GeV) | Signal significance | | | | | | | |
|---|---|---|---|---|---|---|---|---|
| | No background uncertainty | | Backgr. uncertainty of 10%/$\sqrt{\mathscr{L}[\text{fb}^{-1}]}$ | | Experimental syst. uncertainty | | Theoretical uncertainty | |
| | 0b | ≥1b | 0b | ≥1b | 0b | ≥1b | 0b | ≥1b |
| 110 | 5.5 | 6.4 | 2.7 | 4.5 | 2.4 - 3.1 | 3.8 - 5.3 | 2.2 - 3.2 | 3.7 - 5.3 |
| 130 | 5.6 | 6.6 | 3.6 | 5.3 | 3.4 - 3.8 | 4.8 - 5.9 | 3.1 - 4.1 | 4.5 - 6.1 |
| 150 | 5.2 | 5.8 | 4.1 | 5.2 | 3.9 - 4.3 | 4.8 - 5.6 | 3.9 - 4.7 | 4.5 - 5.9 |
| 200 | 3.1 | 3.8 | 2.8 | 3.6 | 2.7 - 2.9 | 3.3 - 3.9 | 2.5 - 3.1 | 3.2 - 4.0 |
| 300 | 1.2 | 1.8 | 1.1 | 1.8 | 1.0 - 1.2 | 1.6 - 2.0 | 1.0 - 1.2 | 1.7 - 1.9 |
| 400 | 0.5 | 0.8 | 0.5 | 0.8 | 0.4 - 0.5 | 0.7 - 0.9 | 0.4 - 0.5 | 0.7 - 0.9 |

Table 7: Signal significance for signal at different mass points, with $\tan\beta$=30 and $\mathscr{L}$=10 fb$^{-1}$. The numbers are shown for different levels of the background uncertainty. For degenerate $A$, $H$ and $h$ boson states with the same mass, the production rates have been summed.

efficiency). Table 9 summarizes the results obtained with pile-up effects taken into account. Only small changes are observed compared to the previous results.

The results shown in Tables 7 and 8 can be extrapolated to other $\tan\beta$ values. The only difference which occurs for the a given signal mass point when changing the $\tan\beta$ value is the change of the production rates and of the natural width of the $A/H/h$ bosons. In the $(m_A - \tan\beta)$ plane which is of interest for this analysis, the resolution of the mass measurement is mostly dominated by the experimental resolution. Nevertheless, the variation of the natural Higgs width is taken into account in the calculation by increasing the background contribution correspondingly to the expected change of the mass window.

Using the previously described production cross-sections and branching ratios, the $5\sigma$-discovery curves are obtained as shown in Figure 24(left) separately for the analysis with 0 and with at least one b-jet in the final state. The luminosity needed for the exclusion of the signal hypothesis at a 95% confidence level is shown in Figure 24(right).

The signatures with the b-jets in the final state allow for the highest discovery potential. The 0 b-jet final state plays nevertheless an important role. Since the search in this final state is uncorrelated to the

| $m_A$ (GeV) | $\mathscr{L}$[fb$^{-1}$] for a 5$\sigma$ discovery (No systematics) | | | $\mathscr{L}$[fb$^{-1}$] for 95% C.L. exclusion | | | | | |
|---|---|---|---|---|---|---|---|---|---|
| | | | | No systematics | | | With systematics | | |
| | 0b | $\geq$1b | comb. | 0b | $\geq$1b | comb. | 0b | $\geq$1b | comb. |
| 110 | 20.9 | 12.2 | 7.7 | 1.8 | 2.0 | 0.9 | 2.7 | 2.8 | 1.4 |
| 130 | 12.0 | 8.9 | 5.1 | 1.9 | 1.4 | 0.8 | 3.6 | 1.9 | 1.2 |
| 150 | 10.5 | 9.3 | 4.9 | 1.6 | 1.5 | 0.8 | 2.2 | 2.0 | 1.0 |
| 200 | 25.9 | 19.1 | 11.0 | 3.7 | 3.1 | 1.7 | 4.9 | 3.8 | 2.1 |
| 300 | 174.6 | 81.6 | 55.6 | 38.6 | 13.8 | 10.2 | 43.8 | 16.4 | 12.0 |
| 400 | 1124.0 | 444.8 | 318.7 | 320.0 | 75.1 | 60.8 | 361.0 | 86.5 | 69.8 |

Table 8: Integrated luminosity (in fb$^{-1}$) needed for a 5$\sigma$ signal significance and 95% CL exclusion of the signal hypothesis, shown for the signal at $\tan\beta$=30. For degenerate $A$, $H$ and $h$ boson states with the same mass, the production rates have been summed.

| $m_A$ $(GeV)$ | $\mathscr{L}$[fb$^{-1}$] for 5$\sigma$ discovery | $\mathscr{L}$[fb$^{-1}$] for 95% CL exclusion |
|---|---|---|
| 110 | 8.7 | 1.9 |
| 130 | 5.8 | 1.6 |
| 150 | 5.6 | 1.3 |
| 200 | 12.5 | 2.6 |
| 300 | 62.9 | 13.6 |
| 400 | 359.5 | 87.5 |

Table 9: Luminosity needed for the 5$\sigma$ signal significance (no systematics) and the 95% CL exclusion of the signal hypothesis (with systematics), obtained for the combined analyses, with the pile-up effects taken into account. The results are shown for $\tan\beta$=30.

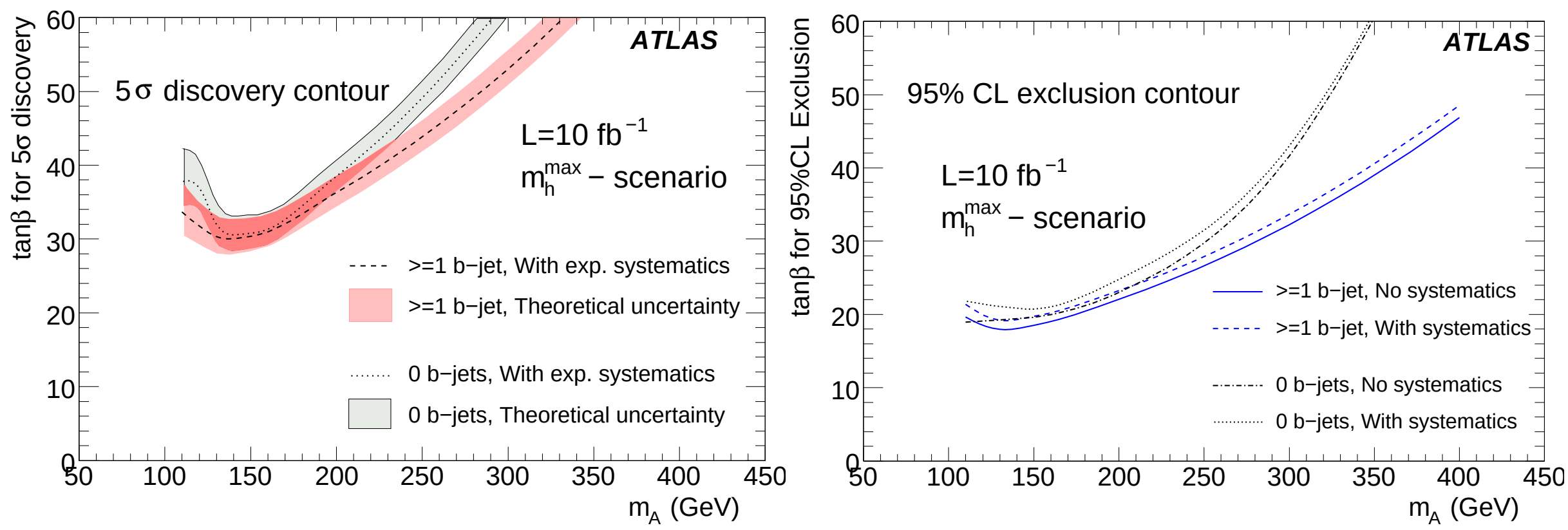


Figure 24: $\tan\beta$ values needed for the 5$\sigma$-discovery (left) and for the 95% CL exclusion of the signal hypothesis (right), shown in dependence on the $A$ boson mass.

previous one, one can quadratically add the signal significances obtained from the two analyses. The 5$\sigma$-discovery curves obtained from the combination of both analyses, as well as the combined 95% CL exclusion limits are shown in Figure 25. At an integrated luminosity of 10 fb$^{-1}$, the discovery can be reached for $m_A$ masses up to 350 GeV with $\tan\beta$ values between 25 and 60. For $m_A$ masses below 110 GeV the sensitivity drops rapidly as shown in Ref [5], due to the increasing Drell-Yan background

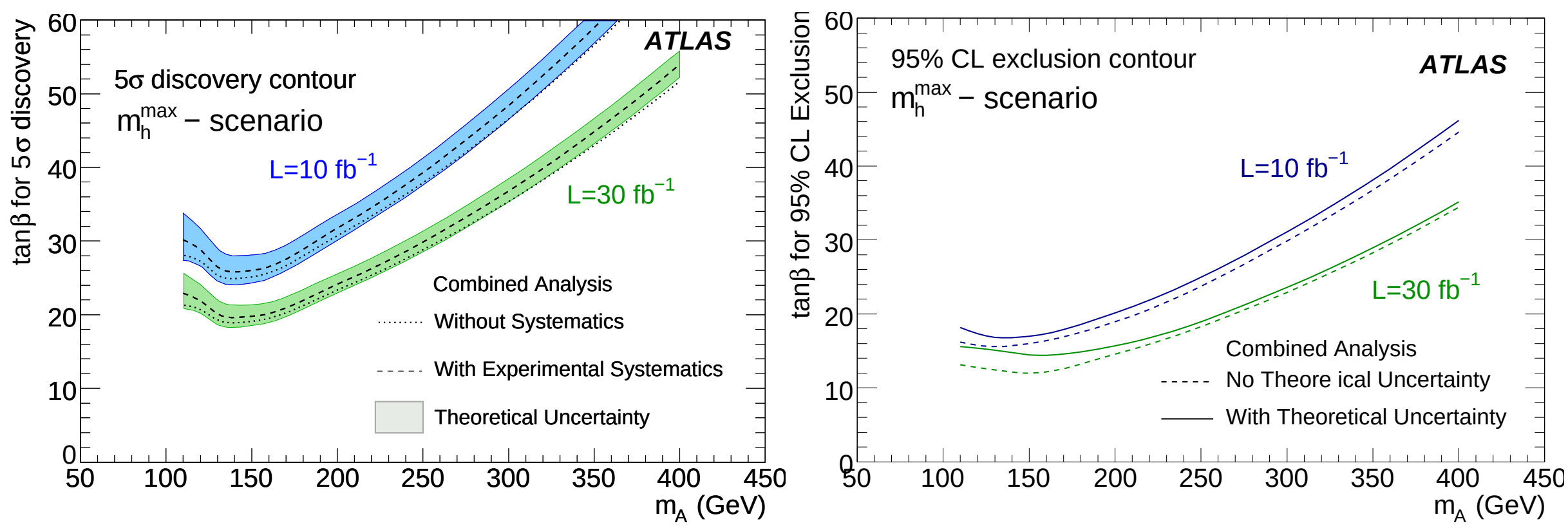


Figure 25: Combined analyses results: (left) $\tan\beta$ values needed for the $5\sigma$-discovery at $\mathscr{L}$=10 fb$^{-1}$ and $\mathscr{L}$=30 fb$^{-1}$, shown in dependence on the $A$ boson mass and (right) combined 95% CL exlucion limits.

close to the pole mass of the $Z$ boson. The $\tan\beta$ values above $\sim$16 can be excluded already with 10 fb$^{-1}$ of integrated luminosity in case of the Higgs boson masses up to 200 GeV.

# 10 Conclusions

In this note, the potential for the discovery of the neutral MSSM Higgs boson is evaluated in the dimuon decay channel. As opposed to the Standard Model predictions, the decay of neutral MSSM Higgs bosons $A$, $H$ and $h$ into two muons is strongly enhanced in the MSSM. In addition, the $\mu^+\mu^-$ final state provides a very clean signature in the detector.

The event selection criteria are optimized in the signal mass range from 100 to 500 GeV, separately for the signatures with 0 b-jets and with at least one b-jet in the final state. The obtained combined result shows that an integrated luminosity of 10 fb$^{-1}$ allows for the discovery for $m_A$ masses up to 350 GeV with $\tan\beta$ values between 30 and 60. Three times higher luminosity allows for an increased sensitivity down to $\tan\beta$=20. The theoretical and detector-related systematic uncertainties are shown to degrade the signal significance by up to 20%. This takes into account that the background contribution can be estimated from the data with an accuracy of $\sim$2-10%/$\mathscr{L}$[fb$^{-1}$].

# Sensitivity to an Invisibly Decaying Higgs Boson

**Abstract**
Many extensions of the Standard Model include Higgs bosons decaying predominantly or partially to non-interacting particles such as the SUSY Lightest Supersymmetric Particle (LSP). To set limits on the production cross-section times the branching fraction to invisible decay products of such Higgs bosons with the ATLAS detector requires an examination of specific production modes such as the associated production ($ZH$) or the vector boson fusion (VBF) process. The predominant Standard Model backgrounds for these processes are $ZZ \rightarrow \ell\ell\nu\nu$ for the $ZH$ channel and jets from QCD processes and $W^{\pm}$ or $Z$ bosons produced in association with jets for the VBF channel. The sensitivity to an invisibly decaying Higgs boson is investigated in this paper using fully simulated ATLAS data for both signal and background. The ATLAS potential for triggering these events is also discussed.

## 1 Introduction

Some extensions of the Standard Model predict that Higgs bosons could decay into stable neutral weakly interacting particles, leading to invisible Higgs boson decays. The Higgs boson decay products could be for example neutralinos, gravitinos, gravitons or Majorons [1-3]. In the case of the Minimal Supersymmetric Standard Model (MSSM), if R-parity is conserved, Higgs bosons decaying into a pair of neutralinos may in some cases even dominate [1]. Being the lightest supersymmetric particles, neutralinos would be stable and would leave the detector without decaying, remaining invisible. If R-parity is violated, then Higgs bosons could decay into Majorons, which would interact too weakly to allow detection [2]. Some theories with extra dimensions also predict invisible Higgs boson decays, and have the added advantage of generating neutrino masses [3]. This search is sensitive to any boson coupling to Z or W and decaying invisibly. The combined LEP Higgs boson mass limit in this channel is 114.4 GeV [4].

At the Large Hadron Collider, Higgs boson production could occur through several mechanisms. To select and identify events with an invisibly decaying Higgs boson one must be able to trigger on a signature that is visible in the event. This is possible for channels such as Vector Boson Fusion (VBF) $qqH$ [5], $t\bar{t}H$ [6] and the associated production processes, $ZH$ and $W^{\pm}H$ [7, 8]. Although gluon fusion has a much higher Higgs boson production cross-section than these modes [9], it is not possible to trigger on these events when the Higgs boson decays invisibly.

In this paper, the ATLAS sensitivity to an invisibly decaying Higgs boson is determined in a way that does not depend on a specific extension of the Standard Model. The analysis uses the variable $\xi^2$ which is defined as,

$$\xi^2 = BR(H \rightarrow inv.)\frac{\sigma_{BSM}}{\sigma_{SM}} \quad (1)$$

where $\sigma_{BSM}$ represents the "Beyond the Standard Model" cross-section and $\sigma_{SM}$ represents the Standard Model cross-section. In the case for which the Higgs boson decays entirely to the invisible mode, $\xi^2$ is the ratio between the non-Standard Model cross-section and the Standard Model cross-section. Only two of the three possible production modes are considered in this paper, VBF and associated production. In addition, for associated production, only the $ZH$ mode is considered as the background to the $W^{\pm}H$ signal is overwhelming [10].

In this paper, the Monte Carlo samples, for both the VBF and the $ZH$ channels, the trigger, the event selection, the systematic uncertainties and the results are discussed. We conclude by summarizing the limits on $\xi^2$ for an invisibly decaying Higgs boson.

This analysis compares signals and backgrounds that have been generated using Standard Model processes. The Higgs boson signal events are simulated to be invisible by changing the properties of the Higgs boson decay chain. In reality an invisibly decaying Higgs boson would be expected to result from a process not contained within the Standard Model, and in this case backgrounds associated with this new physics would be important. However, consideration of "Beyond the Standard Model" backgrounds is beyond the scope of this paper as they would have to be considered in the context of each specific model. This analysis assumes Standard Model backgrounds and serves as a limiting case for the indication of a particle that behaves like a Higgs boson that is not consistent with the Standard Model.

# 2 The Vector Boson Fusion $qqH$ production channel

The vector boson fusion (VBF) channel has the second largest production mode after gluon fusion and has the largest production cross-section for observable invisible Higgs boson decays. The VBF invisible Higgs boson production mode, Figure 1, is characterized by two outgoing jets resulting from the interacting quarks, and large missing transverse energy from the Higgs boson. The topology of the jets is particularly useful in selecting the events as the jets are preferentially separated in pseudo-rapidity ($\eta$) and are correlated in the azimuthal angle $\phi$. In addition, the lack of colour flow between the two jets leads to minimal jet activity between the two tagging jets which is potentially useful for selecting events. At high luminosity however, central jet activity resulting from overlapping events may become problematic for cuts based on this event characteristic.

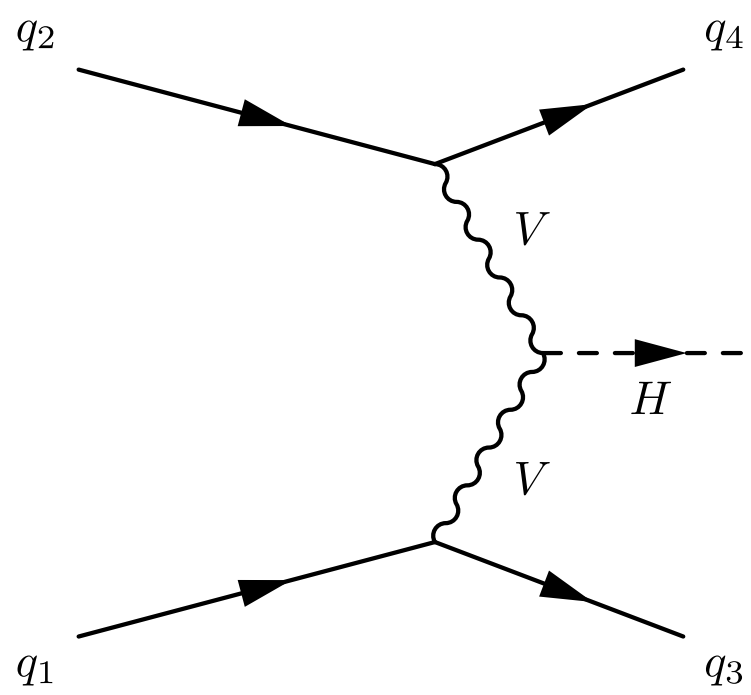


Figure 1: Feynman diagram of the VBF process. The V represents either a Z or W boson.

The study of the VBF channel for this paper includes a mass scan with an estimate of the sensitivity to Higgs boson masses between 110 to 250 GeV. This is based on fully reconstructed signal and background events. In addition to the sensitivity, the trigger acceptance for this channel has been investigated.

## 2.1 Monte Carlo generation for the VBF analysis

Signal and background samples were generated using the standard version of ATLAS software used for this set of papers. The samples were used to determine the sensitivity of ATLAS to an invisible Higgs boson, taking into account the trigger and analysis efficiencies. A number of backgrounds with signatures similar to the signal have been studied and are listed here. In all cases, $\ell$ represents $e$ or $\mu$.

1. Dijet production from QCD processes form a major background due to the very large cross-section for these processes. Fake missing energy measurements can arise from poorly instrumented regions or inefficiencies in the detector.

2. W+jet processes with $W \to \ell\nu$ mimic the signal when the lepton is outside the detector acceptance. The neutrino provides the missing energy signature.

| Higgs boson Mass [GeV ] | Cross-section [pb] | # Events | Generator |
|---|---|---|---|
| 110 | 4.63 | 10k | JIMMY |
| 130 | 3.96 | 30k | JIMMY |
| 130 | 3.93 | 10k | PYTHIA6.4 |
| 200 | 2.41 | 10k | JIMMY |
| 250 | 1.79 | 10k | JIMMY |

Table 1: Invisible Higgs boson samples generated for this analysis. Leading-order cross-sections for the Higgs boson produced via Vector Boson Fusion were evaluated by the ATLAS Higgs Working group [15].

3. Z+jet processes with $Z \to \nu\nu$ constitute an irreducible background.
4. Z+jet with $Z \to \ell\ell$ forms a background to the signal when the leptons are not within the acceptance of the detector.

Event generation for the VBF channel has proved challenging given that the predicted $\eta$ distributions of tagging jets differ greatly according to the event simulation model used. The HERWIG-JIMMY package [11], [12], [13] represents an average response of the available models and has been used to generate data for the Higgs boson mass scan. Signal events have been generated with both HERWIG and PYTHIA [14] at a Higgs boson mass of 130 GeV to estimate the contribution of this effect to the systematic uncertainty. To generate an invisible Higgs boson sample, the Higgs boson is forced to decay into two Z bosons which are subsequently forced to decay into neutrinos. The set of data samples produced for this analysis is summarized in Table 1. The VBF signal Monte-Carlo was produced to leading order. The difference between LO and NLO cross-section is negligible, $\sim 1\%$[9], therefore the LO cross-sections were been used for the signal in this note. Table 1 includes the Standard Model VBF Higgs boson production cross-section that were used [9].

For an ideal detector, event selection cuts efficiently remove the QCD dijet background. However, this background is considered because of the large cross-section for the process and the presence of poorly instrumented regions and dead regions generating false missing transverse energy ($E_T^{miss}$) signals. In order to provide enough statistics throughout the full transverse momentum ($p_T$) spectrum, the QCD background was divided into several $p_T$ ranges to produce several sub-samples, of approximately equal number of events, as shown in Table 2 [16]. The surviving background comes from the high $p_T$ bins allowing reasonable statistics in the final sample. Thus the binning is used to generate the QCD background within a reasonable amount of computer time and allow for the very high rejection factor for this background in this analysis.

In previous ATLAS studies of the invisible Higgs boson produced via the VBF process, the PYTHIA package has been used to generate both the W+jet and Z+jet backgrounds. However, the PYTHIA implementation for these backgrounds only includes the matrix element term for the $q\bar{q} \to gV$ and $qg \to qV$ processes. The PYTHIA implementation tends to underestimate the Z+2jets process because it does not include a complete matrix element calculation. In contrast, ALPGEN [17] provides an exact matrix calculation at tree level for up to 3 partons. For this reason, ALPGEN was used to produce both the W+jet and Z+jet backgrounds. Within ALPGEN, there are two different Z+jet implementations, one which only includes QCD matrix element terms and the second which includes the QCD and EW matrix element terms. In the second case, only on-shell bosons are created without $Z/\gamma^*$ interference, in contrast, the first case does include these effects. By comparing events generated with the two different implementations it was found that the QCD-only process underestimates the background by $\sim$25%. The effect of a non-zero Z boson width was checked by varying the Z boson mass. This showed that the result was insensitive to

| Jet Sample | $\sigma(pb)$ | $p_T$range [GeV ] |
|---|---|---|
| J0 | $1.76\times10^{10}$ | 8-17 |
| J1 | $1.38\times10^{9}$ | 17-35 |
| J2 | $9.33\times10^{7}$ | 35-70 |
| J3 | $5.88\times10^{6}$ | 70-140 |
| J4 | $3.08\times10^{5}$ | 140-280 |
| J5 | $1.25\times10^{4}$ | 280-560 |
| J6 | $3.60\times10^{2}$ | 560-1120 |
| J7 | $5.71\times10^{0}$ | 1120-2240 |

Table 2: Cross-section and $p_T$ range for the QCD dijet background samples generated with PYTHIA. The cross-sections given in the second column are for the specific $p_T$ range with no other cuts.

the use of on-shell bosons. Therefore, all ALPGEN samples used in this study were generated using the option that included both QCD+EW terms.

For the Z+jet background, three samples were produced for each of the two decay modes, $Z \rightarrow \nu\nu$ and $Z \rightarrow \ell\ell$. Two exclusive samples were produced for the one and two parton final states and one inclusive sample was produced for three or more partons. Default ALPGEN settings were used to generate events except for a cut to remove very low $E_T^{miss}$ events by setting $E_T^{miss} > 10$ GeV and a change in acceptance to ensure complete $\eta$ coverage by opening the phase space setting for both jets and leptons ($|\eta_j| < 6$ and $|\eta_\ell| < 6$).

In this study, only the leptonic decay of the W boson from the W+jet background was considered. The $E_T^{miss}$ arises from a combination of the $E_T^{miss}$ associated with the neutrino and the lepton energy in the case where the lepton escapes detection. The W+jet background was generated in the same manner as for the Z+jet background.

## 2.2 Trigger

The major challenge for triggering candidate events for the VBF invisible Higgs boson analysis is to retain signal events whilst reducing the very large QCD background to an acceptable level. These problems are particularly acute with the first level trigger (L1) which can easily be overwhelmed by the QCD background. A trigger for these signal events is possible using a relatively high $E_T^{miss}$ cut while selecting one or two jets of moderate transverse energy. For triggers of this type, QCD backgrounds dominate. In order to produce an acceptable rate for the High Level Trigger (HLT), the trigger menu items used to select invisible Higgs boson events should add no more than a few Hz of trigger rate, even at the highest luminosities.

The trigger study of this note is based on the standard full ATLAS simulation of the L1 trigger. The HLT has not been considered, as at the time of the study HLT algorithms for $E_T^{miss}$ had not been fully implemented and there had been no simulation of forward jets. Jets are classified into central jets $|\eta| < 3.2$ and forward jets $3.2 < |\eta| < 5$.

Data for the trigger study consisted of the sample of VBF Invisible Higgs boson events with a Higgs boson mass of 130 GeV produced using HERWIG and a sample of QCD dijet produced using PYTHIA as described in Section 2.1. The results of this study are shown in Table 3 for $10^{31}$ cm$^{-2}$ s$^{-1}$ luminosity. The acceptances shown in Table 3 give the effect of the trigger on the VBF Higgs boson samples used in the VBF analysis. As such the acceptance is defined as the number of signal events that survive both the trigger and the data selection cuts described in Section 2.3.1, divided by the number of events that survive the selection cuts alone. The trigger rates in Table 3 are the expected raw rates for the specific

| Trigger Menu | Acceptance[%] normalized to offline cuts | Rates[Hz] for $\mathscr{L}= 10^{31}$ cm$^{-2}$ s$^{-1}$ |
|---|---|---|
| L1_XE60 | 99 | $5.0 \pm 1.3$ |
| L1_XE70 | 98 | $1.5 \pm 0.6$ |
| L1_XE80 | 96 | $0.6 \pm 0.1$ |
| L1_XE100 | 84 | $0.2 \pm 0.1$ |
| L1_XE120 | 70 | $0.1 \pm 0.1$ |
| L1_FJ23+XE70 | 78 | $0.9 \pm 0.6$ |
| L1_J23+L1_XE70 | 83 | $1.4 \pm 0.6$ |
| L1_J23+L1_XE100 | 73 | $0.2 \pm 0.1$ |
| L1_FJ23+L1_XE100 | 66 | $0.0 \pm 0.0$ |
| L1_FJ23+L1_J23+L1_XE70 | 62 | $0.9 \pm 0.6$ |
| L1_FJ23+L1_J23+L1_XE100 | 55 | $0.0 \pm 0.0$ |

Table 3: Signal acceptance and level one trigger rates for the VBF invisible Higgs boson channel based on full ATLAS simulations and with $m_H = 130$ GeV . The L1 trigger menu items are $E_T^{miss}$ (L1_XE) central jet (L1_J) and forward jet (L1_FJ), see text. The number following the menu object indicates the trigger threshold given in GeV. Both single and combined triggers are shown in this Table. The values are given for a luminosity of $10^{31}$ cm$^{-2}$ s$^{-1}$ and do not account for pile-up effects.

trigger. One expects an overlap with other trigger signatures such that the additional rate produced by these signatures will be less than the calculated raw rates.

The numbers given in Table 3 are the best estimation we currently have of the trigger rates. In reality beam conditions and detector effects could lead to much higher values. It is clear that the trigger strategy will depend on these background effects and on the luminosity. At low luminosities, ($10^{31}$ cm$^{-2}$ s$^{-1}$) it is likely that a simple trigger based on $E_T^{miss}$ alone such as $E_T^{miss} > 70$ GeV (L1_XE70) will be sufficient. However if the trigger rate for this item is higher than expected, VBF invisible Higgs boson events could still be triggered using a higher $E_T^{miss}$ trigger such as $E_T^{miss} > 80$ GeV (L1_XE80) or $E_T^{miss} > 100$ GeV (L1_XE100) whichever one can be used without pre-scaling. As the luminosity increases or if backgrounds are worse than expected, it will be necessary to use a combined trigger for this channel. The numbers in Table 3 suggest that triggers based on $E_T^{miss}$ and either a forward or central jet would be sufficient. However the addition of a single jet to an $E_T^{miss}$ trigger provides a relatively small reduction in rate due to correlations that can occur when high energy jets are mis-measured. There is concern that this could be amplified by pile-up effects. Requiring a forward jet plus a central jet plus $E_T^{miss}$ is expected to solve these problems albeit with a reduction of signal acceptance. For a luminosity of $10^{33}$ cm$^{-2}$ s$^{-1}$, the most conservative trigger option is a combined trigger with $E_T^{miss} > 100$ GeV and a forward and a central jet each with a $p_T > 23$ GeV. This trigger will have an acceptance rate for the signal of 55% as shown in Table 3. Note that the uncertainties shown in Table 3 are statistical only. A further major uncertainty in trigger rates are pile-up effects which have not been considered here. In practice, adjustments will be required to select the optimum trigger based on the experimental rates observed at the LHC.

Invisibly decaying Higgs boson events produced via vector boson fusion can be selected using a combination of $E_T^{miss}$ and jet triggers with a small impact on the overall L1 trigger rate. For low luminosities ($10^{31}$ cm$^{-2}$ s$^{-1}$) it is expected that a $E_T^{miss}$ trigger of 70 GeV or greater will be sufficient. For higher luminosities such as $10^{33}$ cm$^{-2}$ s$^{-1}$, a trigger with $E_T^{miss} > 100$ GeV and a forward and central jet each with $p_T > 23$ GeV will be required.

### 2.3 Event selection for the VBF channel

Two separate methods are used to extract the signal: the first method is called the cut-based analysis, the second is the shape analysis. Both analyses are conducted by first applying the selection cuts described in the next sub-section. For the cut-based analysis, the signal is extracted using all the cuts including a cut on $\phi_{jj}$, the angle between the two tagged jets in the transverse plane. This is described in Section 2.5. For the shape analysis, this last cut is not applied. Instead, the shape of the $\phi_{jj}$ distribution is used to extract the fraction of signal events. This is described in Section 2.6. Results are derived separately with these two methods.

The selection cuts described in this section were developed using a signal sample with $m_H = 130$ GeV.

#### 2.3.1 Selection cuts

The event selection is based on standard ATLAS definitions for jets, leptons and missing transverse energy [18]. A primary characteristic of a signal event is the presence of two jets from the VBF process. Events are selected based on each of the two highest $p_T$ jets in the event which are referred to as the "tagging jets". These tagging jets are required to have a $p_T > 40$ GeV and be in the rapidity range $|\eta_{j1,2}| < 5$. Cuts on the product and difference of the pseudorapidity of the two jets are used, $\eta_{j1} \cdot \eta_{j2} < 0$ and $\Delta\eta > 4.4$, respectively. Kinematic distributions of the tagged jets for the signal and background are shown in Figure 2. In the upper two plots of this Figure, it can be seen that signal events and the W+jet and Z+jet backgrounds have very similar $p_T$ distributions. When the W+jet and Z+jet events are generated with only one parton the $p_T$ distributions are much softer. When the two parton and three parton components are added the $p_T$ distribution becomes harder and similar to the signal.

The second major event characteristic used to select events is a large $E_T^{miss}$from the invisible decay of the Higgs boson. A cut on this variable significantly reduces the QCD background as no real $E_T^{miss}$is expected for QCD events. In this analysis there is a requirement that $E_T^{miss} > 100$ GeV. The $E_T^{miss}$ distributions for the signal and backgrounds are shown in Figure 3.

The majority of QCD dijet background events will produce soft jets resulting in the tagging jets having a low invariant mass. This feature can be used to reject QCD events by requiring a minimum invariant mass of 1200 GeV for the tagging jets. The invariant mass distribution of the tagging jets is shown in Figure 3. The QCD dijet background can be further reduced by requiring that the direction of the measured $E_T^{miss}$ is not correlated with the tagging jets. A missing transverse energy isolation variable, I, is defined for this purpose as $I = min[\phi(E_T^{miss}) - \phi(j_{1,2})]$. Events with a small value of $I$ are expected to result from mis-measured jets caused by dead material and cracks in the detector. This is illustrated in Figure 4 which shows that QCD dijet events preferentially have a small value of $I$. A selection of $I < 1$ rad has been used which is compatible with previous analyses. The W+jet and Z+jet backgrounds can be reduced by rejecting events with any identified lepton. For this reason, events with electrons or muons with a $p_T > 20$ GeV are rejected, as are events containing $\tau$-jets with a $p_T > 30$ GeV. These cuts are based on an earlier ATLAS fast simulation study.

A key aspect of the VBF Higgs boson search is the electroweak nature of the signal,and this can be used to suppress backgrounds by using the fact that the signal has no color flow between the interacting quarks at tree level. Although the W+jet and Z+jet backgrounds include both electroweak and QCD terms, the cross-section is dominated by the QCD contribution. Therefore, unlike the signal, the majority of background events have QCD activity in the central region. The presence of this extra QCD radiation between the two tagging jets provides, in principle, a powerful tool to suppress this background. In practice the difference is diluted both by the underlying event and pile-up. The Underlying Event (UE) arises from interactions of the spectator partons and is not consistently modeled by the available event generators. For example, the ratio of the average jet multiplicity from the UE between HERWIG/PYTHIA is

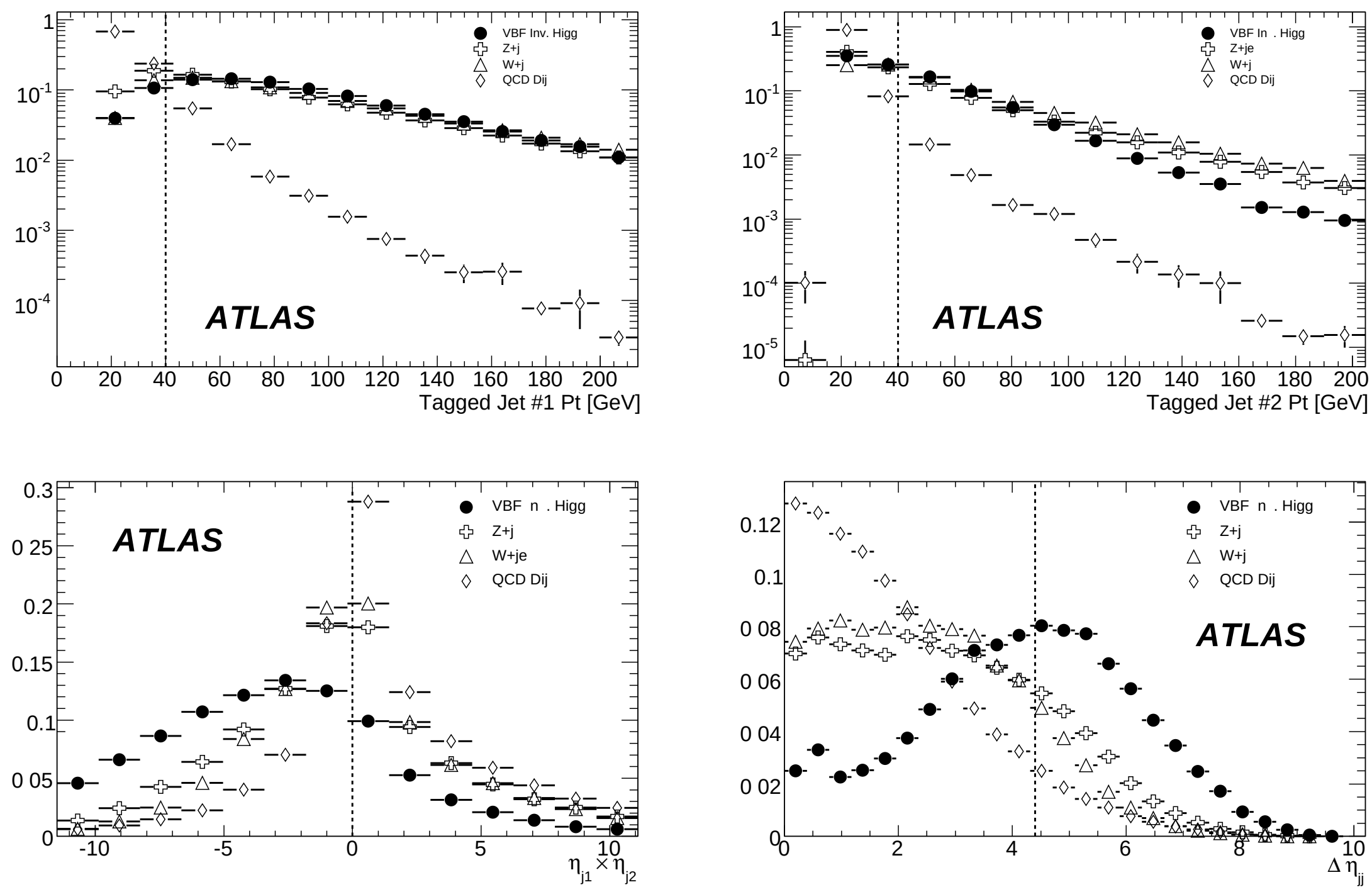


Figure 2: Comparison of tagged jet properties for signal events ($m_H = 130$ GeV) and the three major backgrounds. The upper left plot shows the $p_T$ of the leading tagged jet, the upper right plot shows the $p_T$ of the jet with the second highest $p_T$. The lower left plot shows the product of the directions of the two tagged jets in pseudo-rapidity ($\eta_{j1} \times \eta_{j2}$), and the lower right plot shows the difference in $\eta$ between the two tagged jets ($\Delta\eta$). The enhancement at $\Delta\eta$ of 0.5 in the VBF signal results from a single high $p_T$ jet being reconstructed as two jets. The filter cut described in Section 2.1 has been applied to the W+jet and Z+jet Monte-Carlo data, but no trigger cuts have been applied. The distributions are normalized to unity. The vertical dotted lines show the cut values used in the analysis.

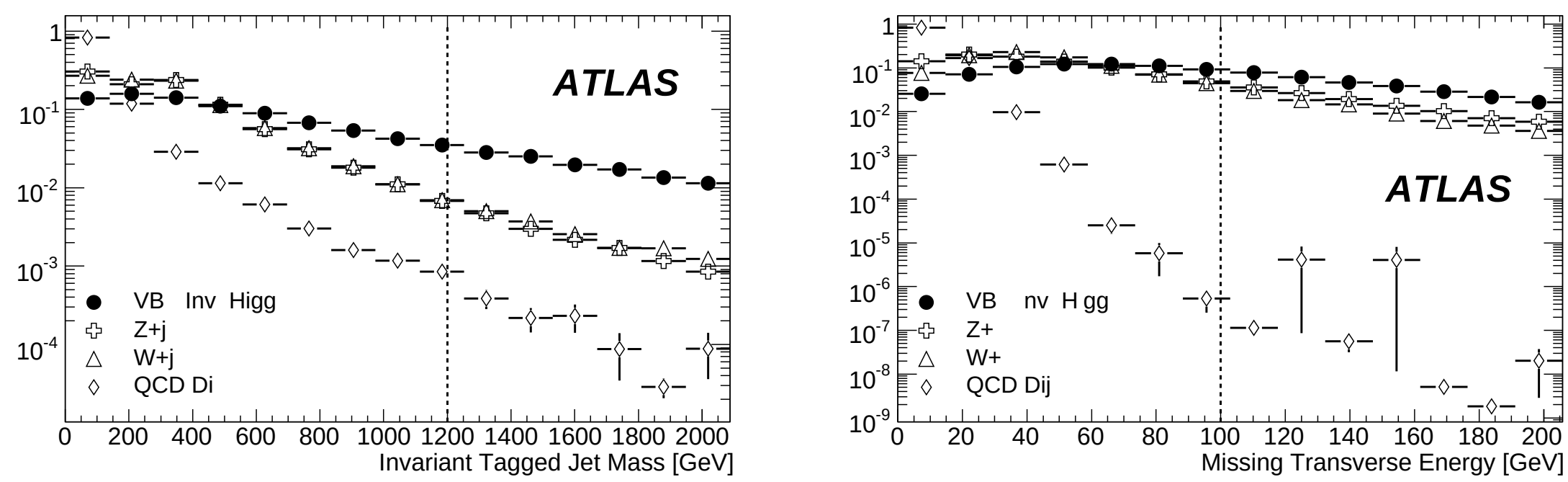


Figure 3: The reconstructed invariant mass of the tagging jets (left) and the $E_T^{miss}$ (right) for the invisible Higgs boson signal ($m_H = 130$ GeV) and the three main backgrounds. Single events in the high $E_T^{miss}$ tail of each individual sample (J0, J1 etc) can result in a spike with a large error. The filter cut described in Section 2.1 has been applied to the W+jet and Z+jet Monte-Carlo data, but no trigger cuts have been applied. The distributions are normalized to unity.

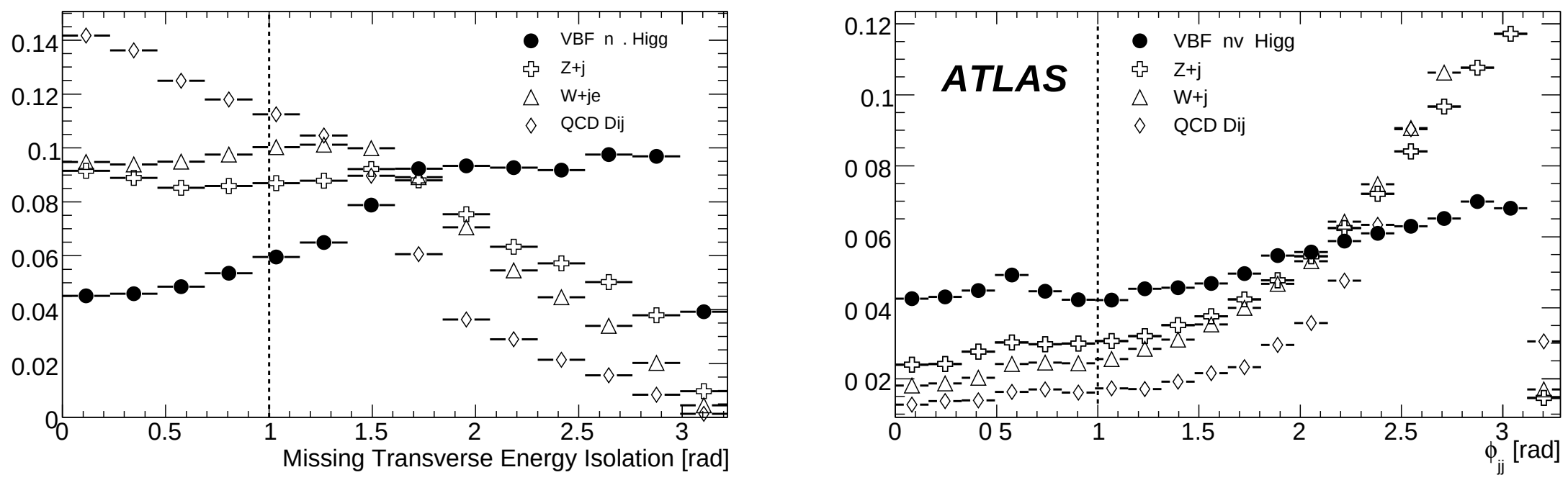


Figure 4: The distribution of the reconstructed $E_T^{miss}$ isolation variable (I) is shown in the right hand plot and the azimuthal angle between the tagging jets ($\phi_{jj}$) is shown in the right hand plot for the invisible Higgs boson signal ($m_H = 130$ GeV) and the three main backgrounds. The filter cut described in Section 2.1 has been applied to the W+jet and Z+jet Monte-Carlo data, but no trigger cuts have been applied. The distributions are normalized to unity.

between 1.38 and 1.85. Therefore PYTHIA generates events with fewer jets from the UE, but these jets have on average a higher $p_T$ . If a cut is applied to remove events that have a central jet that exceed a specific $p_T$ value, the so called Central Jet Veto (CJV) cut, fewer PYTHIA events will survive than HERWIG events. Although there is a clear difference in the topology between the signal and background, the added contribution from the UE has a large effect on the efficiency of this cut. In the same way pile-up, which results from central activity unrelated to the event of interest can also reduce the effectiveness of this CJV cut. The effect of pile-up has not been studied, as suitable data samples were not available. For this analysis, a central jet veto is used requiring that there are no additional jets with $p_T > 30$ GeV for $|\eta| < 3.2$. It should be stressed that this cut is applied after the selection of the two tagging jets which can be located anywhere within the full $\eta$ range including $|\eta| < 3.2$. So this cut does not bias the selection of the tagging jets, nor does it introduce a bias with respect to the trigger which has elements that allow jets to be located within an $|\eta| < 3.2$.

Unlike the signal which is uniquely produced via Vector Boson Fusion, the W+jet and Z+jet backgrounds can be produced by the $q\bar{q} \to gV$ and $qg \to qV$ processes in which the second jet comes from a radiative process. Therefore, the difference in $\phi$ between the two tagged jets is different for the signal and the radiative background as can be seen in Figure 4. This difference provides additional discriminating power and is used in the analysis presented in Section 2.5 requiring $\phi_{jj} < 1$ rad. Moreover, the $\phi_{jj}$ variable motivates the shape analysis presented in Section 2.6.

The selection cuts along with the surviving cross-sections after each cut are shown in Table 4 for a Higgs boson mass $m_H = 130$ GeV and the three main backgrounds. Table 5 shows the effect of the cuts for the four Higgs boson mass values considered in this study. The cross-sections for W+jet and Z+jet processes were calculated to LO but have been normalized to the results calculated with the generator FEWZ at NNLO which results in a value for the total cross-section which is known to within $\sim 10\%$[1] [16].

The first cut applied to the data simulates the effect of the L1 trigger with the most conservative menu option given in Table 3 and discussed in the previous section namely, a $E_T^{miss} > 100$ GeV, a central jet with $p_T > 23$ GeV and a forward jet with $p_T > 23$ GeV. This cut reduces the QCD dijet background rate by approximately 7 orders of magnitude. The effect of the trigger on the W+jet and Z+jet backgrounds is smaller with a reduction of two orders of magnitude, by contrast the signal is reduced by about 50%.

[1]This includes the PDF and QCD scale uncertainties.

| Selection Cuts | Higgs boson $m_H = 130$ GeV | W+jet | Z+jet | QCD |
|---|---|---|---|---|
| Initial $\sigma$ (fb) | $3.93\times10^3$ | $1.24\times10^6$ | $4.08\times10^5$ | $1.91\times10^{13}$ |
| L1 Trigger | $2.71\times10^2$ (0.07) | $1.42\times10^4$ (0.01) | $6.31\times10^3$ (0.02) | $1.52\times10^6$ (<0.01) |
| + Tagged jets | $1.47\times10^2$ (0.54) | $1.35\times10^3$ (0.10) | $6.16\times10^2$ (0.10) | $1.81\times10^5$ (0.12) |
| + $M_{jj}$ | $1.11\times10^2$ (0.76) | $6.64\times10^2$ (0.49) | $3.76\times10^2$ (0.61) | $1.28\times10^5$ (0.71) |
| + $E_T^{miss}$> 100 GeV | $1.08\times10^2$ (0.97) | $4.70\times10^2$ (0.71) | $2.69\times10^2$ (0.72) | $2.84\times10^3$ (0.02) |
| + Lepton veto | $1.07\times10^2$ (1.00) | $3.01\times10^2$ (0.64) | $2.62\times10^2$ (0.97) | $2.76\times10^3$ (0.97) |
| + $I > 1$ rad | $9.60\times10^1$ (0.89) | $1.49\times10^2$ (0.49) | $2.11\times10^2$ (0.81) | 3.61 (<0.01) |
| + Central jet veto | $8.93\times10^1$ (0.93) | $1.10\times10^2$ (0.74) | $1.32\times10^2$ (0.63) | 0.07 (0.02) |
| + $\phi_{jj} < 1$ rad | $4.50\times10^1$ (0.50) | $1.94\times10^1$ (0.18) | $4.21\times10^1$ (0.32) | 0.07 (1.00) |

Table 4: Cross-section in fb for a Higgs boson ($m_H = 130$ GeV) and background samples at each step of the selection process. Initial cross-section for W/Z+jet are quoted after VBF filter and NNLO corrections. The first cut is the effect of the L1 trigger simulation with a $E_T^{miss}$ of 100 GeV, a central jet with $p_T > 23$ GeV and a forward jet with $p_T > 23$ GeV (Table 3). The central jet veto is applied to jets other than the two tagging jets and does not bias any of the other cuts. Numbers in parentheses are the efficiencies for each cut.

| Higgs boson mass [GeV] | 100 | 130 | 200 | 250 |
|---|---|---|---|---|
| Initial $\sigma(fb)$ | 4630 | 3930 | 2410 | 1780 |
| L1 Trigger | 322 | 271 | 240 | 168 |
| + Tagged jets | 166 | 147 | 134 | 93 |
| + $M_{jj}$ | 126 | 111 | 100 | 73 |
| + $E_T^{miss}$> 100 GeV | 121 | 108 | 98 | 70 |
| + Lepton veto | 121 | 107 | 98 | 70 |
| + $I > 1$ *rad* | 108 | 96 | 86 | 63 |
| + Central jet veto | 94 | 89 | 79 | 59 |
| + $\phi_{jj} < 1$ *rad* | 43 | 45 | 39 | 30 |

Table 5: Cross-section in fb for each signal mass at each step in the selection cuts. The first cut is the effect of the L1 trigger simulation with a $E_T^{miss}$of 100 GeV, a central jet with $p_T > 23$ GeV and a forward jet with $p_T > 23$ GeV.

The jet tagging cuts reduces all three backgrounds by a factor of 10. Although a L1 $E_T^{miss}$is applied a large fraction of events still survive because of the the L1 $E_T^{miss}$resolution. The other cuts that have a large impact on the QCD rate are the $E_T^{miss}$ cut and the $E_T^{miss}$ isolation cut. Together they reduce this background to a negligible level. The effect of these selection cuts on the Z+jet and W+jet backgrounds are less dramatic. The lepton veto reduces the W+jet and Z+jet by $\sim 36\%$ and $\sim 3\%$, respectively. The lepton veto cut removes few events in the Z+jet channel as the $E_T^{miss}$ cut removes most of the Z$\rightarrow \ell\ell$ decay mode. The remaining Z+jet events are dominated by the Z$\rightarrow \nu\nu$ mode. Leptons are only identified for $|\eta| < 2.5$, so the lepton veto cut does not remove all the W+jet background events due to this limited $\eta$ range. Therefore, electrons and $\tau$-jet in the forward region ($|\eta| > 2.5$) are mis-identified as jets most of the time. In a similar manner, muons in the forward direction are generally not identified and result in fake $E_T^{miss}$.

## 2.4 Systematic uncertainties

Three major types of systematic uncertainties are considered. One arises from the implemention of the Monte Carlo generators, the second from the experimental systematic uncertainties and the third from

| Selection Cuts | HERWIG 130 GeV | PYTHIA 130 GeV |
|---|---|---|
| Initial $\sigma(fb)$ | $3.93\times10^3$ (1.000) | $3.93\times10^3$ (1.000) |
| Pre-Cut ($E_T^{miss}$> 80 GeV) | $1.76\times10^3$ (0.448) | $1.78\times10^3$ (0.453) |
| + Tagged Jets | $4.07\times10^2$ (0.231) | $4.10\times10^3$ (0.230) |
| + $M_{jj}$ | $2.45\times10^2$ (0.602) | $2.45\times10^3$ (0.598) |
| + $E_T^{miss}$> 100 GeV | $2.05\times10^2$ (0.837) | $2.14\times10^3$ (0.873) |
| + Lepton Veto | $2.05\times10^2$ (1.000) | $2.12\times10^2$ (0.991) |
| + $I > 1$ *rad* | $1.84\times10^2$ (0.898) | $1.80\times10^2$ (0.849) |
| + Central Jet Veto | $1.59\times10^2$ (0.864) | $1.07\times10^1$ (0.594) |
| + $\phi_{jj} < 1$ *rad* | $7.43\times10^1$ (0.467) | $4.93\times10^1$ (0.461) |

Table 6: Comparison between HERWIG and PYTHIA generated samples on the selection cuts for the 130 GeV Higgs boson mass. The cross-section results are quoted in fb and the cut efficiency is given in parenthesis. The major difference occurs in the last two rows of this table.

the theoretical knowledge of the production cross-sections.

Two event generator effects are discussed, the first is the treatment of the Underlying Event (UE) and the second is the effect of using a fixed Z boson mass. To illustrate the effect of the UE, signal events have been generated with two different event generators, HERWIG and PYTHIA that treat the UE in different ways. Table 6 shows the efficiency of each selection cut used in the VBF analyses for the two generators. The difference in cross-sections between the two samples are within $\sim 2\%$ of each other for cuts up to the $E_T^{miss}$ isolation cut. However, once the central jet veto is applied, fewer Pythia events survive resulting in a large difference of $\sim 49\%$. This is believed to be the result of the difference in the modeling of the UE in the two generators; Pythia tends to produce fewer but harder jets, resulting in more events being removed by the central jet veto cut, see Section 2.3. If the number of background events is underestimated due to a combination of this cut and the choice of generator the sensisitivity to the signal will be artificially enhanced. A systematic study of the effect of generator choice on the central jet veto cut would require a large number of data and background samples to be produced with a variety of generators and this is beyond the scope of this paper. It is not clear which generator represents reality best. For consistency both the background and signal samples were generated using HERWIG. When real data becomes available it will be possible to measure the magnitude of central jet activity directly and use this to tune the generators.

The use of ALPGEN requires the use of a fixed Z mass. The effect of the missing off-shell terms from the background samples was checked using the $Z \to \nu\nu$ analysis by adding a $E_T^{miss}$ contribution randomly generated by a Breit-Wigner distribution using the Z mass and width parameters. This study indicated that the effect of using a fixed mass Z boson was negligible.

Two methods are considered in this paper to extract the signal significance. The first, a cut-based analysis, relies on the number of signal and background events after all cuts have been made. The second, a shape analysis relies on the ratio of the number of background events contained in two regions of the $\phi_{jj}$variable distribution; namely the number of events for $\phi_{jj} < 1$ divided by all events. The systematic uncertainties of interest are the ones related to these three quantities; the number of signal and background events and the background shape ratio. They are shown in Table 7.

The event reconstruction variables that result in the largest systematic uncertainties are the jet resolution and the jet energy scale. For the jet energy resolution, the systematic uncertainty was estimated by smearing the momentum of the jets using a Gaussian distribution with a width given by $\sigma(E) = 0.45\sqrt{E}$ for $|\eta| < 3.2$ and $\sigma(E) = 0.63\sqrt{E}$ for $|\eta| > 3.2$. Changing the jet energy magnitude also affects the

| Systematics | Higgs boson 130 GeV | Background | |
|---|---|---|---|
| | | Cut-Based | Shape |
| Luminosity | 3 % | | $\sim 0\%$ |
| Jet energy resolution:<br>$\sigma(E) = 0.45\sqrt{E}$ for $\|\eta\| < 3.2$<br>$\sigma(E) = 0.63\sqrt{E}$ for $\|\eta\| > 3.2$ | 0.8 % | 5.3 % | 4.5 % |
| Jets energy scale:<br>$+7\%$ for $\|\eta\| < 3.2$<br>$+15\%$ for $\|\eta\| > 3.2$ | 4.0 % | 3.2 % | 0.2 % |
| Jets energy scale:<br>$-7\%$ for $\|\eta\| < 3.2$<br>$-15\%$ for $\|\eta\| > 3.2$ | 10.0 % | 19.5 % | 2.8 % |
| Total | 10.5 % | 20.4 % | 5.3 % |

Table 7: Experimental contributions to the systematic uncertainty. The systematic uncertainty on the Jet Energy Scale (JES) is asymmetric. Only the largest (negative) JES systematic uncertainty is included in the total experimental uncertainty shown in the last row of the table.

$E_T^{miss}$, so for each event, $E_T^{miss}$ was recalculated for the x- and y-components. The analysis was then repeated to determine the change in the number of signal and background events and the shape ratio. In the same way the effect of changes in the jet energy scale was investigated by shifting the overall scale by $\pm 7\%$ for $|\eta| < 3.2$ and $\pm 15\%$ for $|\eta| > 3.2$. Again this affects the $E_T^{miss}$, and this change was taken into account. As can be seen in Table 7, there is an asymmetric dependence on the Jet Energy Scale, (JES), so positive and negative deviations are considered separately. Lepton reconstruction was analyzed and found not to contribute to the final systematic uncertainty. The JES uncertainties used here are conservative for 30 fb$^{-1}$, but the impact of this choice on the sensitivity limits presented in this paper is small. Finally, 3% was assigned to the uncertainty in the luminosity. To get the experimental systematic uncertainty for each analysis the terms were added in quadrature giving an overall systematic uncertainty of 20% on the number of background events which applies to the cut-based analysis and an uncertainty of 5.3% on the background shape ratio that applies to the shape analysis.

In addition to the uncertainty in the UE and the reconstruction algorithms, there is a systematic uncertainty which arises from the uncertainty in the absolute cross-section of the backgrounds. The main backgrounds to the invisible Higgs boson channel are Z+jet and W+jet. The total cross-sections for these processes have been corrected to NNLO and are known to $\sim 10\%$, (see Section 2.3). However the cuts used to select the VBF process, result in a very restricted phase space which makes it difficult to determine the systematic uncertainty on the cross-section for the final data samples. This means that the systematic uncertainty on the number of background events due to the cross-section is currently unknown, could be very large and is likely to dominate other uncertainties. The shape of the $\phi_{jj}$ distribution on the other hand is quite well constrained by theory and based on previous studies has a systematic uncertainty of 10% [5]. At NLO it is expected to be 5%. In this analysis a conservative value of 10% is assumed for the uncertainty due to the cross-section which when combined with the much smaller experimental effect leads to an overall systematic uncertainty of 11.3% on the background shape.

### 2.5 Cut-based analysis

This analysis uses the selection cuts summarized in Section 2.3.1. The signal significance is calculated based on the number of signal and background events that remain after all cuts. The limitation of this analysis is that the systematic uncertainty on the background cross-sections described in the previous section could be very large. One way of dealing with this uncertainty is to use experimental data of the Z+jet channel where the Z decays to two leptons. After correction for the detector acceptance, these events can be used to infer the value of the cross-section for the irreducible background in which the Z decays to two neutrinos. A similar analysis can be done for the W+jet background. These so called "data driven" corrections will be the subject of a future publication. In the next section we report on an alternative method which uses the shape of the azimuthal angle distribution between the jets to reduce the dependence of the systematic error on the cross-section. In the current section the sensitivity to an invisible Higgs boson without systematic errors are calculated to provide baseline numbers for the sensitivity to an invisible Higgs boson produced via the VBF process.

The number of signal and background events after the $\phi_{jj}$ cut is shown in Tables 4 and 5. These numbers can be used to calculate the 95 % CL sensitivity of $\xi^2$ for the invisible Higgs boson, given the assumed backgrounds. This is done by calculating the number of signal events required to increase the total event count by a factor 1.64 times the uncertainty on the number of background events as shown in Equation 2.

$$1.64\sigma_B = N_S\xi^2 \tag{2}$$

Here $N_S$ is the number of signal events after the selection cuts and $\sigma_B = \sqrt{N_B}$ . The results of this analysis gives a $\xi^2$ for an integrated luminosity of 30 fb$^{-1}$ at 95 % C.L. for Higgs mass between 110 GeV and 250 GeV of $\sim 5-8\%$ in the case when systematic uncertainties are not included. An additional 6% statistical uncertainty[2] arises from the limited number of events in the data samples.

### 2.6 Shape analysis

The shape analysis is motivated by a marked difference in the $\phi_{jj}$ distribution between the signal and the W/Z+jet background as shown in Figure 5, which is taken from reference [5]. This plot shows that the backgrounds peak above a $\phi_{jj}$ of 1 while the signal is higher at low $\phi_{jj}$ values. To characterize the shape of the $\phi_{jj}$ distribution the ratio R has been defined as the number of events with $\phi_{jj} < 1$ divided by the total number of events as shown in Equation 3. As the proportion of signal in the sample increases the value of R increases.

$$R = \frac{\int_0^1 \frac{d\sigma}{d\phi_{jj}}}{\int_0^\pi \frac{d\sigma}{d\phi_{jj}}} \tag{3}$$

The advantage of the shape analysis described in this section over other analyses is that it does not require a knowledge of the absolute cross-section but rather the ratio of the number of events for ($\phi_{jj} < 1$) to all events. As such, the systematic error associated with the absolute cross-section is reduced to a negligible amount. However, as discussed in Section 2.4, there is a systematic uncertainty associated with the knowledge of the $\phi_{jj}$ distribution which is known to $\sim$10% or better. In addition, the systematic uncertainties due to detector effects are much smaller for this ratio than they are for the number of background events, which is the relevant variable for a pure cut-based analysis (Table 7). The overall systematic uncertainty on the ratio R has been calculated to be 11.3%.

Equation 3 can be re-written in the context of this analysis and expanded to provide a background-

[2] Based on a binomial error calculation.

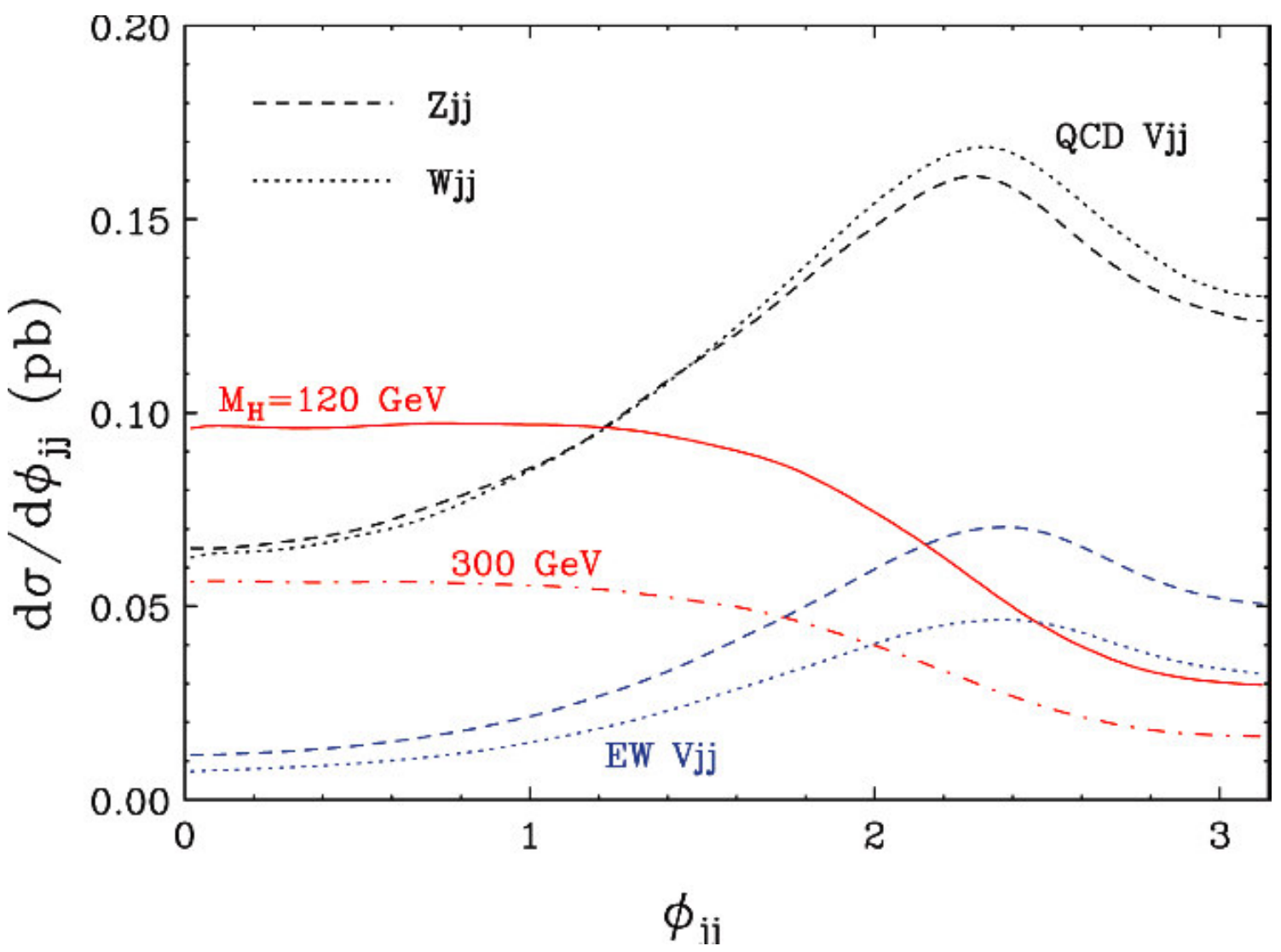


Figure 5: The $\phi_{jj}$ distribution for the signal and background in radians [5]. The solid and dash-dotted lines represent the expected distribution for the Higgs boson signal with Higgs boson masses of $m_H = 120$ and $m_H = 300$ GeV respectively. The dotted and dashed lines represent the distributions expected from the backgrounds. The plot shows the distributions after VBF selection cuts have been made. Note that for the $\phi_{jj}$ plot shown earlier did not have these cuts applied, (Figure 4).

only term, as shown in Equation 4.

$$R = \frac{N_B^1}{N_B^\pi}\left[1 + \xi^2\left(\frac{N_S^1}{N_B^1} - \frac{N_S^\pi}{N_B^\pi}\right) + \cdots\right] \tag{4}$$

Here $N_B^1$ and $N_S^1$ are the number of events within $\phi_{jj} < 1$ and $N_B^\pi$ and $N_S^\pi$ are the number of events within the entire $\phi_{jj}$ range. The first term of Equation 4 provides the expected ratio for the background contribution. However, since the ratio between the signal and background are not the same in the presence of a signal, a non-zero value is expected in the second term. The variation from the 'background only' ratio dictates the sensitivity to new physics. The ratio $N_B^1/N_B^\pi$ can be determined using the Stndard Model theoretical prediction or by a data driven technique.

The shape analysis applies the selection cuts discussed in Section 2.3.1 but not the $\phi_{jj}$ cut. Therefore, the results from Table 4 before this last cut are used. The first term of Equation 4 is calculated to be $0.254 \pm 0.007$. To determined the 95% C.L. sensitivity limit a variation of $1.64\sigma_R$ is required, where $\sigma_R$ is the uncertainty on the ratio R from Equation 4. Therefore, the first order $\xi^2$ terms from Equation 4 is set to the required 95% CL sensitivity limit, that is $1.64\sigma_R$, as shown in Equation 5.

$$1.64\sigma_R = \xi^2\left(\frac{N_S^1}{N_B^1} - \frac{N_S^\pi}{N_B^\pi}\right)\left(\frac{N_B^1}{N_B^\pi}\right) \tag{5}$$

Solving for $\xi^2$ provides the 95 % CL sensitivity limit for the invisible Higgs boson. The results of this analysis are shown in Figure 6. Without systematic errors, the shape analysis gives a value of $\xi^2$ that ranges from 11 to 19%. This can be compared with the simple cut-based analysis which gave a $\xi^2$ value that ranged from 5 to 8%. So although the shape analysis method removes the dependence on the absolute cross-section of the backgrounds there is a reduction in the sensitivity to the signal. To include

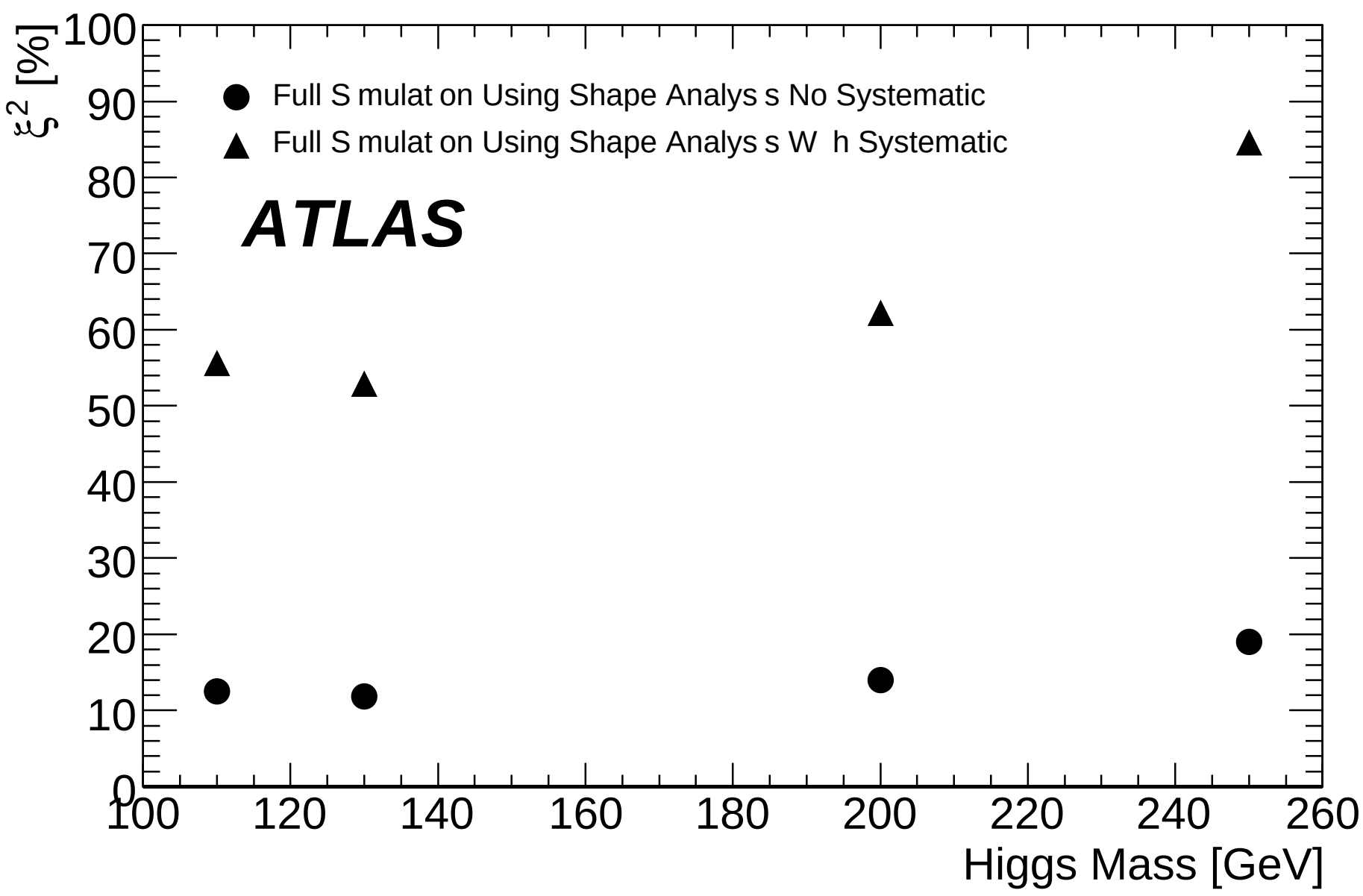


Figure 6: Sensitivity for an invisible Higgs boson at 95% C.L. via the VBF channel using shape analysis for an integrated luminosity of $30 fb^{-1}$ with and without systematic uncertainties. The black triangles (circles) are the results from this analysis with (without) systematic uncertainties.

the systematic uncertainties, the uncertainty on the background becomes $\sigma_R = \sqrt{\sigma_R{}^2 + \alpha^2 R^2}$, where $\alpha$ is the fractional systematic uncertainty given in Table 7. The result obtained that includes systematic uncertainties gives value of $\xi^2$ of around 60% for $m_H$ between 100 and 200 GeV. This sensitivity is dominated by the systematic uncertainty that arises from the theoretical knowledge of the shape of the $\phi_{jj}$ distribution. Using calculations at NLO could reduce this uncertainty by a factor of 2 greatly enhancing the sensitivity of this method of analysis.

### 2.7 Summary for the VBF invisible Higgs boson channel

The study described above has investigated the sensitivity of the ATLAS detector to a Higgs boson particle produced by the VBF process that has an invisible decay mode. It should be stressed that these results do not include pile-up which can reduce the sensitivity. It has been shown that with 30 fb$^{-1}$ of data it is possible to detect this process over a wide range of masses if the Beyond Standard Model cross-section is more than 60% of the Standard Model cross-section for a Higgs mass range of up to 200 GeV and 100% of the Higgs boson decays are invisible. Triggering for this channel is possible using a trigger requiring large $E_T^{miss}$plus a forward and a central jet of moderate $p_T$. Triggers of this kind would be useful up to luminosities of at least $10^{33}$ cm$^{-2}$ s$^{-1}$.

## 3 The associated *ZH* production channel

The Feynman diagram for associated production in the ZH channel is shown in Figure 7. The signal of an invisibly decaying Higgs boson in the *ZH* channel can be detected when the *Z* boson decays into two leptons, which can be used for triggering the event. The presence of an invisibly decaying Higgs boson is detected from the missing transverse energy.

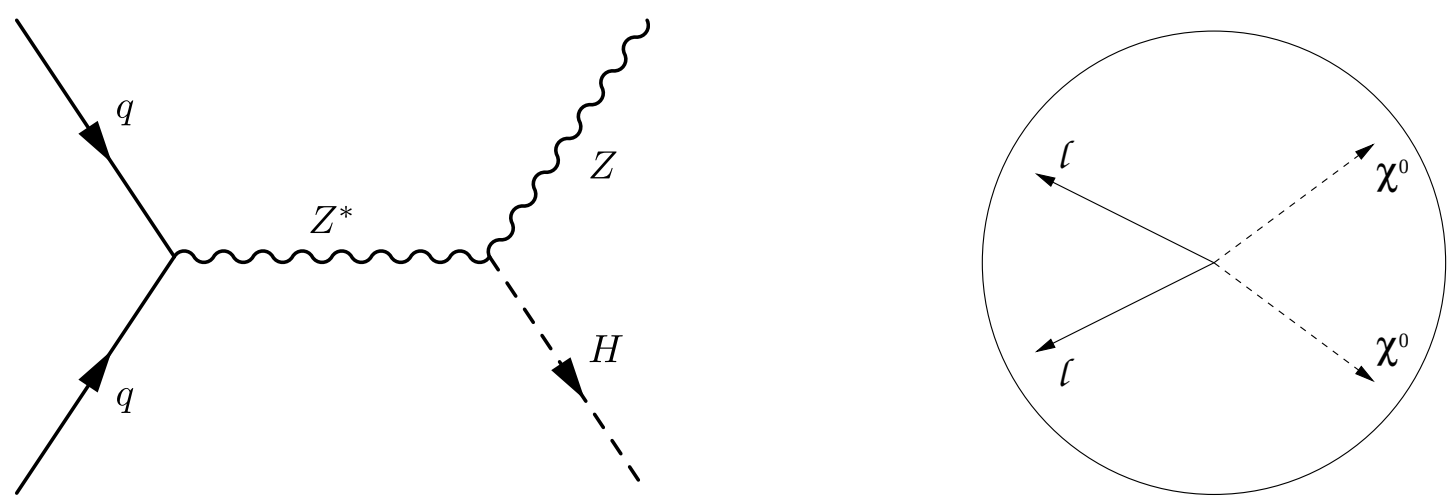


Figure 7: (Left): The Feynman diagram for Higgs boson associated production with a Z boson. (Right): A representation of the decay of a Higgs boson into two invisible neutralinos represented by $\chi^0$ recoiling against the two leptons coming from the $Z$ decay.

Various backgrounds with signatures similar to the signal have been studied and are listed here. In all cases, unless otherwise specified, $\ell$ represents $e$ or $\mu$.

1. The $ZZ \rightarrow \ell\ell\nu\nu$ final state gives the same signature as the signal (irreducible background) and is the main background;
2. The $t\bar{t} \rightarrow b\ell^+\nu\bar{b}\ell^-\nu$ process mimics the signal when the two $b$-jets are not reconstructed, or when a second lepton results from a b quark decay;
3. The $W^+W^- \rightarrow \ell\nu\ell\nu$ process mimics the signal but can be greatly reduced by cutting on the $Z$ mass;
4. The $ZZ \rightarrow \nu\bar{\nu}\tau\bar{\tau}$ and $\tau \rightarrow \ell\nu\bar{\nu}$ can also mimic the signal.
5. The $ZZ \rightarrow \ell\bar{\ell}\tau\bar{\tau}$ and $\tau \rightarrow \ell\nu\bar{\nu}$ can pass the selection criteria if some particles are missed;
6. The $ZW \rightarrow \ell\ell\ell\nu$ decay mode also simulates the signal when one lepton is not detected;
7. The $Z$ plus jets background, with $Z \rightarrow \ell\bar{\ell}$ (Drell-Yan process) final state can be mistaken for the signal when poor jet reconstruction leads to missing transverse energy.

## 3.1 Monte Carlo generation for the ZH channel

The signal and the background events have been generated using different particle generators chosen according to which process they simulate best. The events are fully simulated then reconstructed using ATHENA. The diboson production cross-sections are taken from Ref. [16]. All events were passed through a filter immediately after generation. The two filters used are described in detail in Section 3.2.2. Only the few samples generated with the simpler lepton filter will be mentioned here. All other samples were generated with the filter containing $m_Z$ and $E_T^{miss}$ cuts.

Details on the generated events and pre-defined parameters are given below. All generators use the CTEQ6M structure functions to generate the processes.

- Seven signal samples were generated using PYTHIA, with $m_H = 110$, 120, 130, 140, 150, 200 and 250 GeV. Only the samples with $m_H = 130$ and 140 GeV used the lepton filter. To generate the invisibility of the Higgs boson, the $H$ is produced as a stable particle, which goes undetected.
- $ZZ \rightarrow \ell^+\ell^-\nu\bar{\nu}$ is generated using PYTHIA. This is the main and irreducible background. One $Z$ decays to a lepton pair while the other is allowed to decay to any flavor of neutrino pairs.

- $t\bar{t}$ is produced with the MC@NLO generator with all top quark decay modes allowed. The filter described in detail in Section 3.2.2 retains mostly $t\bar{t} \to b\ell^+\nu\bar{b}\ell^-\bar{\nu}$ events with $\ell = e$ or $\mu$ only. Hadronic top quark decays with a subsequent $b \to c\ell\nu$ decay are also included.
- $W^+W^- \to \ell^+\nu\ell^-\nu$ is generated with MC@NLO.
- $ZZ \to \tau^+\tau^-\nu\nu$ and $ZZ \to \ell\ell\tau^+\tau^-$ events are generated with PYTHIA. The $\tau$ was allowed to decay hadronically and leptonically in both samples.
- $ZW^\pm \to \ell^+\ell^-\ell^\pm\nu$ samples are produced with the MC@NLO generator. Since the cross-sections for $ZW^+$ and $ZW^-$ production are different, two different datasets were generated. The ZW samples were not filtered since MC@NLO does not include the Z width at generation, leading to problems when trying to apply a cut on the Z mass at the generator level. For these datasets, we use samples generated with the lepton filter which selected events containing at least two leptons. These events need to be reweighted after full reconstruction to account for the distorted Z mass distribution.
- $Z \to \ell^+\ell^- + jet$ events are produced using the SHERPA generator.

## 3.2 Event selection for ZH channel

The event selection is made in three stages:

1. For simulated samples, a filter is applied immediately after the event Monte Carlo generation to avoid unnecessary simulation of a large fraction of the background which would otherwise be readily rejected at the preselection level. The event filter cuts are loose preselection cuts applied on true Monte Carlo quantities.
2. The preselection cuts use fully reconstructed variables and aim at rejecting most backgrounds, retaining only the most likely events to be used at the final selection level.
3. The final selection uses a multivariate analysis (Boosted Decision Tree) to refine the selection cuts while retaining a high signal efficiency.

### 3.2.1 Analysis framework for the ZH channel

The standard ATLAS selection criteria are used to identify these objects [18].

### 3.2.2 Filter for the ZH channel

The filter decision is based on true Monte Carlo quantities. The filter must not reject events that would have passed the preselection cuts to avoid introducing biases, and must have a large rejection efficiency against background. To do so, the filter uses cuts looser than, but similar to, the preselection cuts, namely:

- The events must contain at least two leptons with $p_T$ >4.5 GeV of same flavor but opposite charge within $\eta < 2.7$.
- The reconstructed mass of these two leptons must be within ±25 GeV of the $Z$ mass.
- The events must satisfy $E_T^{miss} > 50$ GeV. The $E_T^{miss}$ is computed from a vectorial sum over all invisible, stable particles such as Higgs bosons and neutrinos, and all lost particles falling outside the calorimeter fiducial region of $p_T > 5.0$ GeV and $\eta > 5.0$.

| channel | $ZH \to \ell\ell$ inv. $m_H$=130 GeV | $ZZ$ $\ell\bar{\ell}\nu\nu$ | $t\bar{t}$ | $WW$ $\ell\nu\ell\nu$ | ZZ $\tau\tau\nu\nu$ | ZZ $\ell\ell\tau\tau$ | ZW $\ell\ell\ell\nu$ | Z+jet $\ell\ell + jet$ |
|---|---|---|---|---|---|---|---|---|
| # of generated events | 10 K | 50 K | 104 K | 70.4 K | 50 K | 50 K | 250 K | 27.25K |
| filter efficiency | 75.4% | 29.2% | 1.13% | 4.5% | 5.16% | 71.4% | 100% | 0.0154% |
| $\sigma$*BR (fb) | 46.2 | 728.0 | 833000.0 | 5245.6 | 364.0 | 123.0 | 820.0 | 3105063 |
| filter cuts | 34.9 | 212.6 | 9412.9 | 236.5 | 18.8 | 87.8 | 820.0 | 478.4 |
| after trigger | 32.5 | 198.5 | 8620.6 | 217.1 | 14.2 | 77.8 | 735.3 | 460.6 |
| $E_T^{miss} > 90$ GeV cut | 14.0 | 83.8 | 3254.5 | 46.6 | 1.9 | 4.1 | 85.9 | 105.8 |
| $p_T$ lepton cut | 10.1 | 61.6 | 1596.2 | 30.5 | 0.4 | 1.5 | 24.0 | 46.2 |
| $m_Z \pm 20$ GeV cut | 9.6 | 60.2 | 1187.4 | 19.9 | 0.0 | 1.2 | 19.7 | 43.2 |
| b-tag cut | 9.3 | 58.6 | 358.0 | 18.7 | 0.0 | 1.1 | 19.0 | 17.7 |

Table 8: Monte Carlo estimates of the cross-section times branching ratio in (fb) for the signal with $m_H = 130$ GeV and background processes following successive preselection cuts described in Section 3.2.4 for the ZH analysis. The number of generated events refers to filtered events. The cross-sections are given at NLO as calculated in Ref. [16] for Standard Model processes and from Ref. [9] for the ZH production cross-sections.

| | $ZH \to \ell\ell$ inv. | | | | | | |
|---|---|---|---|---|---|---|---|
| $m_H$ (GeV) | 110 | 120 | 130 | 140 | 150 | 200 | 250 |
| # of generated events | 10 K | 50 K | 10 K | 10 K | 10 K | 10 K | 10 K |
| filter efficiency | 47.0% | 49.6% | 75.4% | 75.9% | 56.4% | 64.6% | 70.1% |
| $\sigma$·BR (fb) | 77.3 | 59.4 | 46.2 | 36.6 | 29.1 | 11.0 | 5.2 |
| after filter cuts | 36.3 | 29.4 | 34.9 | 27.7 | 16.4 | 7.1 | 3.6 |
| after trigger cuts | 34.0 | 27.8 | 32.5 | 26.0 | 15.6 | 6.8 | 3.5 |
| after $E_T^{miss} > 90$ cut | 18.3 | 15.6 | 14.0 | 12.5 | 10.0 | 4.9 | 2.7 |
| after $p_T$ lepton cut | 13.5 | 11.6 | 10.1 | 9.3 | 7.4 | 3.7 | 2.0 |
| after $m_Z \pm 20$ GeV cut | 13.2 | 11.4 | 9.6 | 8.7 | 7.3 | 3.6 | 2.0 |
| after b-tag cut | 13.0 | 11.1 | 9.3 | 8.5 | 7.1 | 3.5 | 2.0 |

Table 9: Monte Carlo estimates of the cross-section (fb) for the signal with seven different Higgs mass hypotheses following successive preselection cuts described in Section 3.2.4 for the ZH channel analysis. The number of generated events refers to filtered events. Two samples were generated using a simpler filter that retained events containing at least two leptons, ($m_H = 130$ and 140 GeV) whereas all other samples used a filter that required finding two leptons forming a Z boson and large $E_T^{miss}$, as described in Section 3.2.2. The efficiency for this filter increases with the Higgs mass hypothesis.

A simpler lepton filter with only the first selection cut was used for the $ZW \to \ell\ell\ell\nu$ samples, for two signal samples with Higgs mass hypothesis $m_H = 130$ and 140 GeV, and for the $ZZ \to \ell\ell\ell\ell$ sample used for a normalization study.

The filter reduces the total CPU time needed for full reconstruction by more than a factor of 1000. The results are summarized for a signal with a Higgs boson mass hypothesis of 130 GeV and the background samples in Table 8. In Table 9, the effect of the filter, trigger and preselection cuts are shown for the signal at other Higgs boson masses.

### 3.2.3 Trigger for ZH channel

An invisibly decaying Higgs boson in the $ZH$ channel can be detected when the $Z$ decays into two leptons. We trigger on such events using the full simulation of the trigger and by requiring either one or

two isolated, high $p_T$ leptons satisfying any of the following trigger signatures:

- single electron trigger: one isolated lepton with $p_T > 22$ GeV.
- single muon trigger: one muon with $p_T > 20$ GeV.
- di-electron trigger: two isolated electrons with $p_T > 15$ GeV.
- missing transverse energy trigger: $E_T^{miss} > 100$ GeV.

The di-muon trigger with a lower $p_T$ momentum cut was not implemented in this analysis but will be used in the future. The overall trigger efficiency of 92.8% compares well with what is retained when applying cuts on fully reconstructed variables, selecting events containing either one electron with $p_T > 25$ GeV, two electrons with $p_T > 15$ GeV or one muon with $p_T > 20$ GeV. The effect of the trigger on all samples studied is given in Table 8.

#### 3.2.4 Event preselection

A first preselection is applied to reject most backgrounds, in particular the $t\bar{t}$ and (Z+jet) backgrounds. The following cuts are applied:

- the event must satisfy one of the trigger signatures described in the previous section;
- large missing transverse energy, i.e. $E_T^{miss} > 90$ GeV.
- the event must contain exactly two leptons of the same flavor but opposite charge with $p_T > 15$ GeV;
- an anti-b-tag is applied to further suppress the $t\bar{t}$ background.
- a loose cut on the invariant mass of the two leptons, namely $|m_{\ell\ell} - m_Z| < 20$ GeV, is applied to reject some $t\bar{t}$ background without reducing the signal efficiency.

The $\sigma \cdot BR$ for the signal and the background processes listed in Section 3 are shown in Table 8. The effects of the filter and preselection cuts are also shown in this table.

#### 3.2.5 Final event selection

In order to improve the sensitivity of this channel, the most discriminative variables are used to form a multivariate analysis (Boosted Decision Tree or BDT). Sixteen different variables are used as inputs to the BDT, namely:

- the missing transverse energy $E_T^{miss}$,
- the transverse mass $m_T = \sqrt{2p_T^{\ell\ell} \cdot E_T^{miss}(1-\cos\Delta\phi)}$ where $\Delta\phi$ is the azimuthal angle between the dilepton system and $\vec{p_T}^{miss}$,
- the cosine of the angle between $\vec{p_T}^{miss}$ and the most energetic lepton,
- the reconstructed Z mass,
- the transverse momentum of each lepton,
- the cosine of the angle between the two leptons in the transverse plane and in 3-dimensions,

- the cosine of the angle between ${\vec{p_T}}^{miss}$ and ${\vec{p_T}}^{Z}$,
- the cosine of the angle between the most energetic jet and ${\vec{p_T}}^{miss}$,
- the energy found in a cone ΔR = 0.1 rad around each lepton; the lepton isolation variable is particularly useful for muons where very little energy is found around isolated muons but not so much for electrons where the energy deposit is much wider,
- the energy of each of the three most energetic jets, and
- the total number of jets.

These variables are shown in Figures 8 to 11 for the signal and the background after applying the preselection cuts listed in Section 3.2.4. Each distribution has been normalized to unity. The various backgrounds have been regrouped: the irreducible background, $ZZ \to \ell\ell\nu\nu$, the non-resonant background: $t\bar{t}$ and $WW$, and finally all other backgrounds containing at least one Z boson: $ZZ \to \ell\ell\tau\tau$, $ZZ \to \tau\tau\nu\nu$, $ZW$ and $(Z + jet)$.

Each of the main backgrounds after the preselection cuts, namely $t\bar{t} \to b\ell\nu\ b\ell\nu$, $WW \to \ell\nu\ell\nu$, $ZZ \to \ell\ell\tau\tau$, $ZW \to \ell\ell\ell\nu$ and (Z+jet) are compared to the signal to train a Boosted Decision Tree (BDT) [19]. Each one of these backgrounds is trained separately. Nothing is gained from training a BDT against the irreducible background, $ZZ \to \ell\ell\nu\nu$, so this background is not used. An ensemble (forest) of decision trees is successively generated from the training sample, where each new tree is trained by giving increased weights to the events that have been misclassified in the previous tree. The classifier response is obtained as the sum of the classification results for each tree, weighted by the purity obtained for all training events in that tree. The large number of decision trees in the forest increases the performance of the classifier and stabilizes the response with respect to statistical fluctuations in the training sample.

For each background type, a set of weights is established. Half the Monte Carlo events contained in each file is used for the training and the other half for the analysis. The file containing the signal and all the backgrounds, including the less important ones, is then analyzed using these weights. Each event is assigned a weight corresponding to the likelihood of being identified as signal or background. The BDT weight distributions are shown in Figure 12. Each plot shows the output variable distribution for Boosted Decision Trees trained against different backgrounds, namely, from top plot to bottom plot, the $t\bar{t} \to b\ell\nu b\ell\nu$, $WW \to \ell\nu\ell\nu$, $ZZ \to \ell\ell\tau\tau$, $ZW \to \ell\ell\ell\nu$ and Z+jet. The $ZZ \to \tau\tau\nu\nu$ background is not used to train a specific BDT since too few events survive the preselection cuts. The cuts on the five BDT outputs are adjusted to minimize the value of $\xi^2$, defined in Equation 2.

The same procedure is repeated using different Higgs boson mass hypotheses ranging from $m_H = 110$ GeV to $m_H = 250$ GeV, optimizing the cuts each time. The numbers of events surviving all Boosted Decision Trees cuts for each Monte Carlo sample and these seven Higgs boson mass hypotheses are given in Table 11. The BDT inputs variables are also ranked during each training against a particular background. Each time, the input variables are assigned a weight proportional to their importance in separating power. The sum of these weights are given in Table 10, showing which variables offer the best separation power. The order of importance varies for each of the trees but all variables are useful in at least one tree. The Z mass is the overall most discriminative variable, mostly due to its very high ranking in the BDT trained against $t\bar{t}$ and $WW$ backgrounds.

## 3.3 Systematic uncertainties

### 3.3.1 Background cross-section

Since this is a counting experiment, one is looking for events in excess of what is expected by the Standard Model. However, exactly what is expected from Standard Model backgrounds is not well

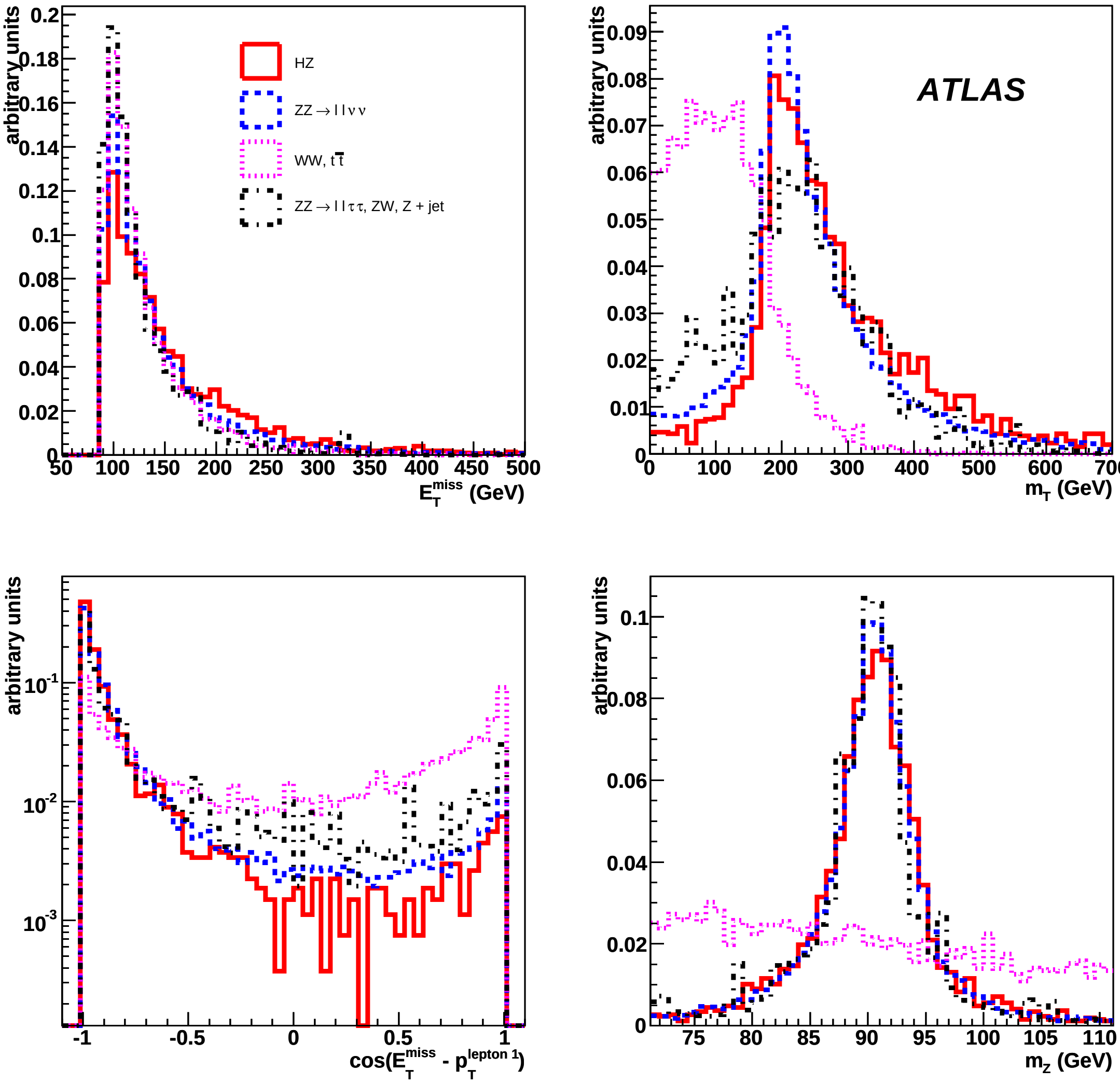


Figure 8: Input variables used by the Boosted Decision Tree for the signal with $m_H = 130$ GeV and the main backgrounds. Top left: Missing $E_T$. Top right: transverse mass, defined as the reconstructed mass in the transverse plane, namely $m_T^2 = E_T^2 - p_T^2$. Bottom left: cosine of the angle between the missing $E_T$ and highest momentum lepton in the transverse plane. Bottom right: reconstructed Z mass. Each plot has been normalized to unity. The combined samples had first been scaled to the same luminosity.

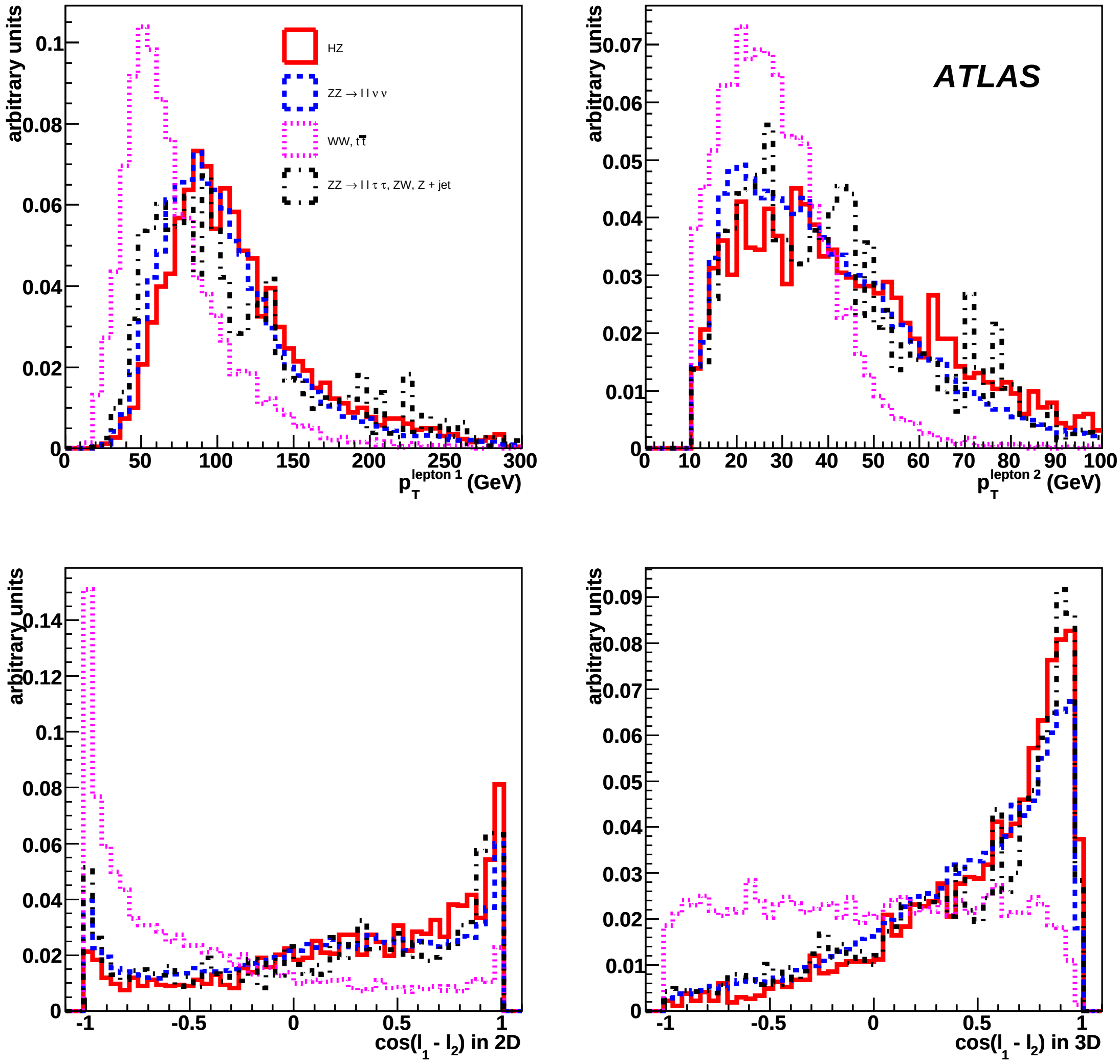


Figure 9: Input variables used by the Boosted Decision Tree for the signal with $m_H = 130$ GeV and the main backgrounds. Top left: Transverse momentum of the most energetic lepton. Top right: Transverse momentum of the second lepton. Bottom left: Cosine of the angle between the two leptons in the transverse plane and, Bottom right: in 3-dimensions. Each plot has been normalized to unity. The combined samples had first been scaled to the same luminosity.

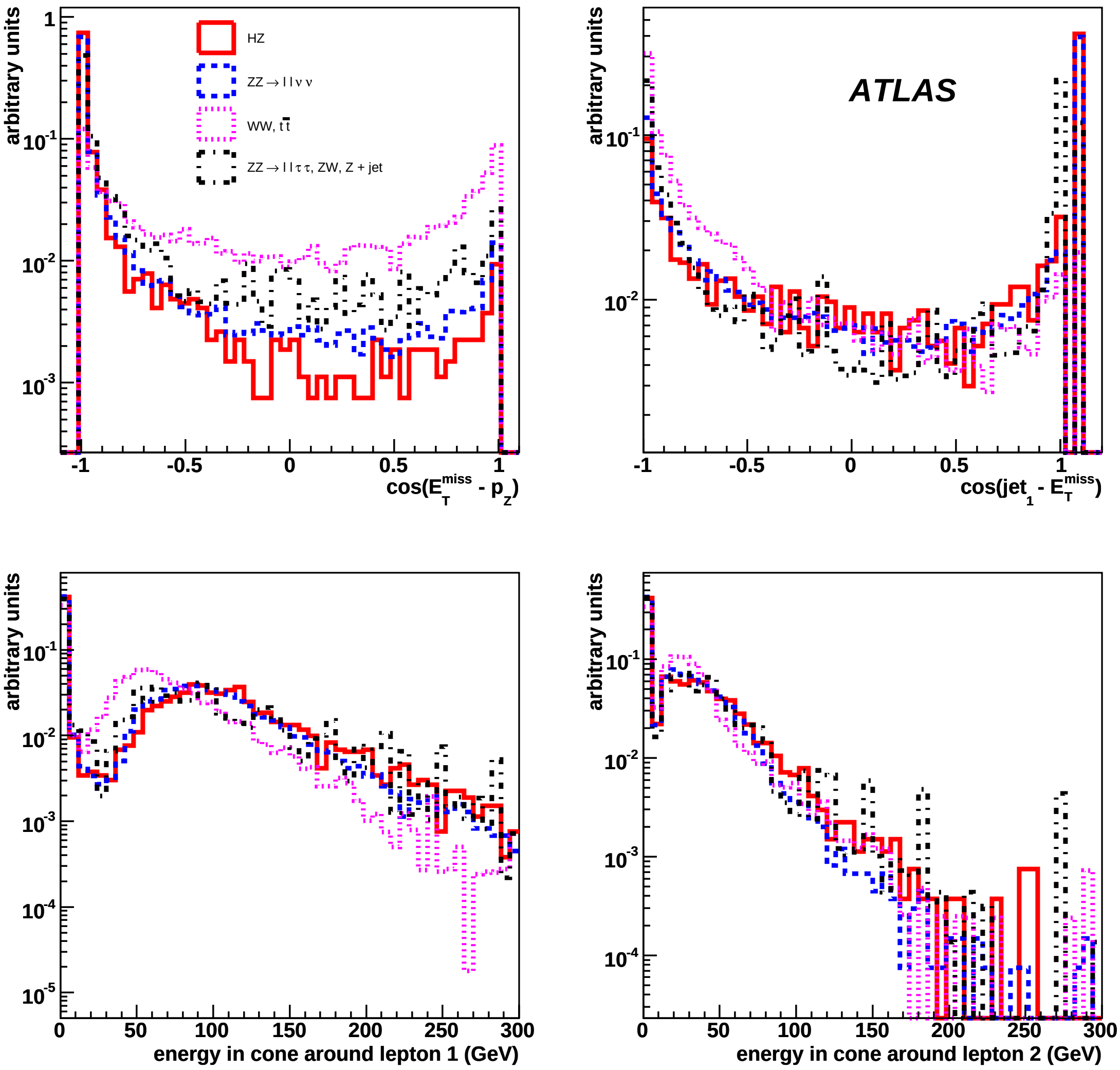


Figure 10: Input variables used by the Boosted Decision Tree for the signal with $m_H = 130$ GeV and the main backgrounds. Top left: The cosine of the angle between the direction of missing $E_T$ and the reconstructed Z transverse momentum. Top right: The cosine of the angle between the most energetic jet and the direction of missing $E_T$. Bottom left: The energy contained in a cone of 0.10 rad around the most energetic lepton. Bottom right: The energy contained in a cone of 0.10 rad around the second lepton. Each plot has been normalized to unity. The combined samples had first been scaled to the same luminosity.

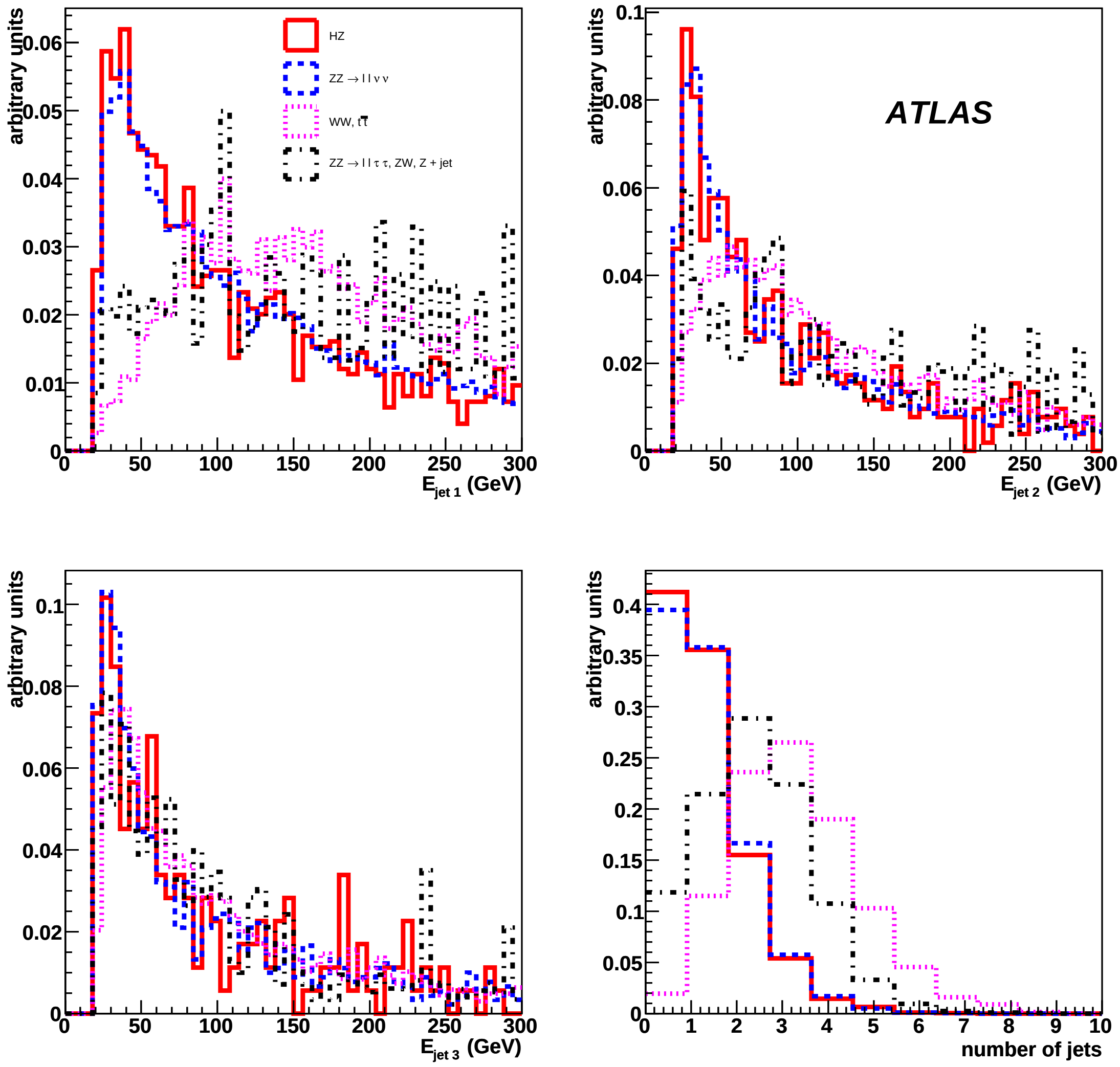


Figure 11: Input variables used by the Boosted Decision Tree for the signal with $m_H = 130$ GeV and the main backgrounds. Top left: The energy distribution for the most energetic jet; Top right: for the second and, Bottom left: third most energetic jets. Bottom right: the number of jets in the event. Each plot has been normalized to unity. The combined samples had first been scaled to the same luminosity.

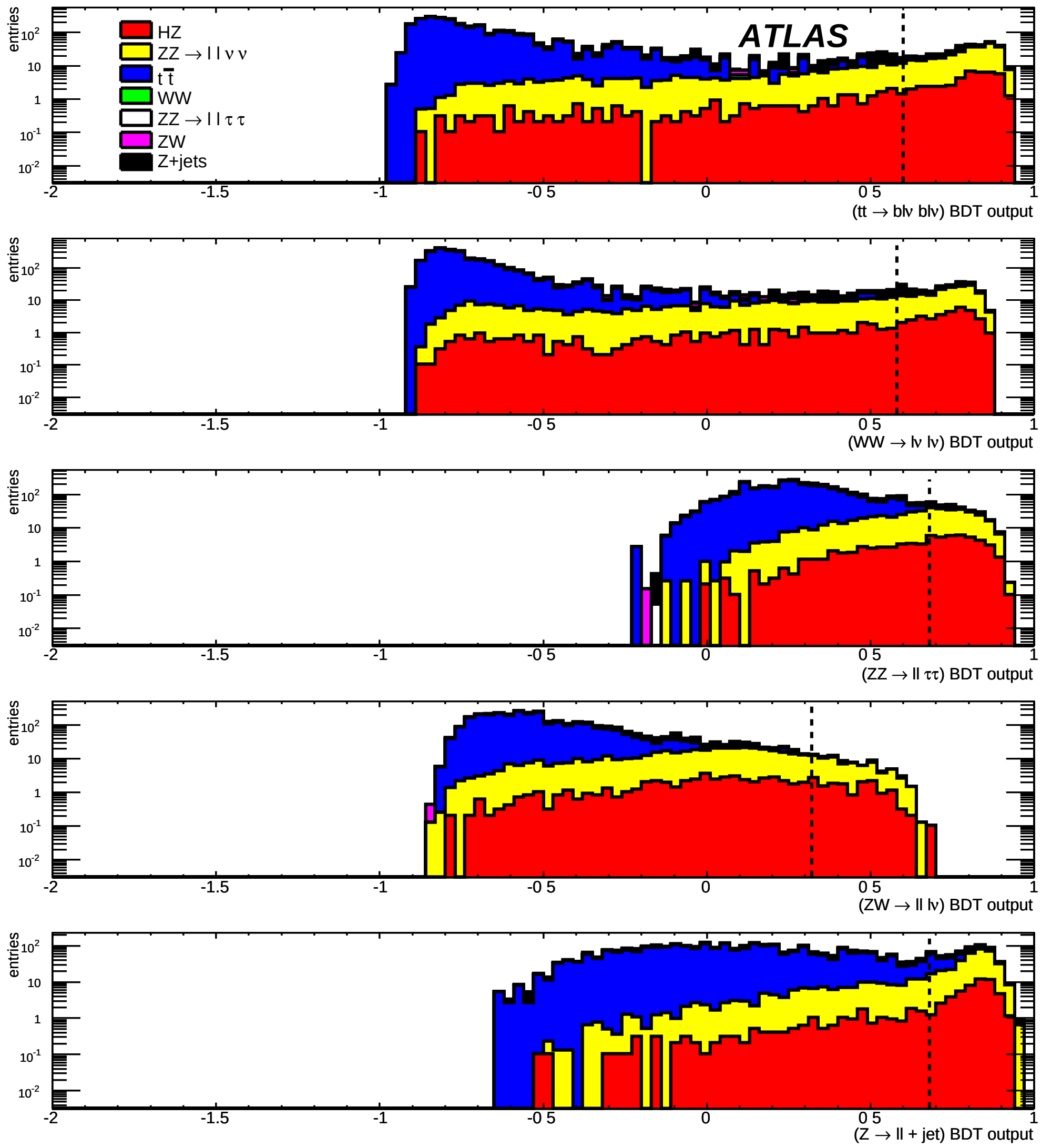


Figure 12: The Boosted Decision Tree (BDT) output variables obtained after comparing half the signal events to five different backgrounds separately, namely, from top to bottom: $t\bar{t} \to b\ell\nu b\ell\nu$, $WW \to \ell\nu\ell\nu$, $ZZ \to \ell\ell\tau\tau$, $ZW \to \ell\ell\ell\nu$ and $Z \to \ell\ell +$ jets. The BDT assigns values close to +1 for a signal-like event and -1 for background-like events. The distributions are shown for the signal and all types of background when using the other half of the events for the analysis. The Boosted Decision Trees trained against the $ZW$ background offers the best separation power. The $\xi^2$ decreases further once additional cuts on the other BDT output variables are applied, namely the WW BDT output, then the $ZZ \to \ell\ell\tau\tau$ BDT output, the (Z+jet) BDT output and finally the $t\bar{t}$ BDT output. All BDT output cut values are indicated by a vertical dashed line. Nothing is gained from training a BDT against the irreducible background, $ZZ \to \ell\ell\nu\nu$ for all Higgs boson mass hypotheses, so it is not used.

| input variable to BDT | sum of weights |
|---|---|
| Z mass | 0.57 |
| $\cos \ell\ell$ (in 2D) | 0.46 |
| $\cos \ell\ell$ (in 3D) | 0.42 |
| $\cos(E_T^{miss} - \vec{p}_Z)$ | 0.41 |
| # of jets | 0.40 |
| transverse mass | 0.39 |
| $\cos(jet - E_T^{miss})$ | 0.32 |
| $\cos E_T^{miss} - p_T^{lepton\#1}$ | 0.31 |
| $E_T^{miss}$ | 0.28 |
| $p_T^{lepton\#1}$ | 0.27 |
| $E_{jet\#1}$ | 0.26 |
| $E_{jet\#2}$ | 0.23 |
| $p_T^{lepton\#2}$ | 0.21 |
| energy in a cone around lepton # 1 | 0.17 |
| energy in a cone around lepton # 2 | 0.16 |
| $E_{jet\#3}$ | 0.13 |

Table 10: Order of importance for the 16 input variables used to train the separate Boosted Decision Trees used for the analysis at $m_H = 130$ GeV. The second column gives the sum of the weights given to each input variable by the five separate BDT used for the analysis. These weights are not used for the analysis per se but give an idea of the relative importance of each input variable. In particular, the first four variables are used to mostly reject the non-resonant background.

| channel | $ZH$ $\ell\bar{\ell}$ inv. | $ZZ$ $\ell\bar{\ell}\nu\nu$ | $t\bar{t}$ | WW $\ell\nu\ell\nu$ | ZZ $\tau\tau\nu\nu$ | ZZ $\ell\ell\tau\tau$ | ZW $\ell\ell\ell\nu$ | Z+jet $\ell\ell$+jet |
|---|---|---|---|---|---|---|---|---|
| | expressed as cross-sections given in fb | | | | | | | |
| $m_H = 110$ GeV | 1.31 | 3.44 | 0.00 | 0.01 | 0.00 | 0.02 | 0.37 | 0.03 |
| $m_H = 120$ GeV | 2.99 | 13.64 | 0.18 | 0.28 | 0.00 | 0.16 | 2.16 | 0.10 |
| $m_H = 130$ GeV | 0.89 | 2.93 | 0.00 | 0.00 | 0.00 | 0.02 | 0.33 | 0.00 |
| $m_H = 140$ GeV | 0.98 | 3.51 | 0.00 | 0.00 | 0.00 | 0.03 | 0.48 | 0.00 |
| $m_H = 150$ GeV | 1.32 | 6.37 | 0.18 | 0.00 | 0.00 | 0.06 | 0.62 | 0.00 |
| $m_H = 200$ GeV | 0.62 | 4.83 | 0.18 | 0.00 | 0.00 | 0.04 | 0.42 | 0.00 |
| $m_H = 250$ GeV | 0.31 | 2.50 | 0.00 | 0.00 | 0.00 | 0.03 | 0.24 | 0.02 |
| | # of events corresponding to 30 fb$^{-1}$ | | | | | | | |
| $m_H = 110$ GeV | 39.2 | 103.1 | <2.7 | 0.2 | <0.01 | 0.8 | 11.0 | 0.8 |
| $m_H = 120$ GeV | 89.6 | 409.3 | 5.4 | 8.5 | 0.1 | 4.7 | 64.7 | 3.0 |
| $m_H = 130$ GeV | 26.8 | 87.9 | <2.7 | <0.06 | <0.01 | 0.8 | 9.9 | <0.36 |
| $m_H = 140$ GeV | 39.5 | 191.1 | 5.4 | <0.06 | <0.01 | 1.7 | 18.7 | <0.36 |
| $m_H = 150$ GeV | 36.5 | 173.0 | 5.4 | 0.2 | <0.01 | 1.8 | 19.0 | <0.36 |
| $m_H = 200$ GeV | 18.5 | 145.0 | 5.4 | <0.06 | <0.01 | 1.3 | 12.7 | <0.36 |
| $m_H = 250$ GeV | 9.3 | 74.9 | <2.7 | <0.06 | <0.01 | 1.0 | 7.2 | 0.7 |

Table 11: Monte Carlo estimates of the cross-sections in fb surviving the final Boosted Decision Tree selection cuts for each background process and seven mass hypotheses for the ZH channel. The corresponding numbers of events for 30 $fb^{-1}$ of total integrated luminosity are also given. The final cuts on the Boosted Decision Tree output variables were set separately for each BDT output variable and for each mass hypothesis, each time optimizing the sensitivity $\xi^2$.

known, given the theoretical uncertainties on the Standard Model production cross-sections and this leads to the main source of systematic uncertainty. The current best estimates for each of these cross-sections from the next-to-leading order calculation is about 6% for $ZZ$ and 5% for $ZW$ [20]. The uncertainty on the (Z+jet) cross-section has no impact on the final results since this background is negligible. Several control samples can be used to constrain the ZZ and ZW cross-sections. For ZZ, one can use the four lepton final state (even including $\tau$) but this will require a large data sample (of the order of at least 30 $fb^{-1}$) to reduce the statistical uncertainty. For the ZW cross-section, one can use a ZW control sample with events containing three identified leptons. Both these cross-sections will be measured in ATLAS data. Uncertainties associated with kinematic distributions have not been taken into account at this point. A combined theoretical uncertainty of 5.8% obtained from a weighted average is assigned to the background production cross-section.

One could in principle use $ZZ \rightarrow \ell\ell\ell\ell$ events from data to calibrate the number of events coming from $ZZ \rightarrow \ell\ell\nu\nu$ decays. Such an approach was proposed in [21] where one would first select a pure sample of $ZZ \rightarrow \ell\ell\ell\ell$ events by finding two Z bosons, then declaring one Z to decay invisibly. This would work in the absence of other backgrounds but it is not possible to completely eliminate the $ZW \rightarrow \ell\ell\ell\nu$ background. More importantly, such a technique has a very low efficiency: about 1.8% of all $ZZ \rightarrow \ell\ell\ell\ell$ survive the preselection cuts, with $\ell$ here being $e, \mu$ or $\tau$. Only a dozen of events would survive all of the BDT selection cuts for 30 fb$^{-1}$ of data. Hence, it is deemed impossible to calibrate quantitatively the $ZZ \rightarrow \ell\ell\ell\ell$ cross-section using this technique. However, one could still check the effect of the preselection cuts on $ZZ \rightarrow \ell\ell\ell\ell$ events with two leptons declared invisible as described above to ensure that the main and irreducible background, $ZZ \rightarrow \ell\ell\nu\nu$, behaves as expected under these cuts. About 85 $ZZ \rightarrow \ell\ell\ell\ell$ events are expected to pass the preselection cuts, as opposed to 163 $ZZ \rightarrow \ell\ell\nu\nu$ events for 30 fb$^{-1}$ of integrated luminosity. After the preselection cuts, the $ZZ \rightarrow \ell\ell\nu\nu$ background corresponds to about 36% of the total number of selected events in the absence of non Standard Model contributions, as seen from Table 8. This method would provide a normalization of the cross-section using data at about 11% uncertainty level.

### 3.3.2 Effect related to the training of the Boosted Decision Tree

Since half the events are used for training the BDT, and the other half for testing, this arbitrary choice has a slight effect on the outcome. For the central value of this analysis, we used every other event for the training. To estimate the effect of this choice, the analysis was redone using the first half of the events for training, and the second half for testing. Since we are only using Monte Carlo events, this second choice does not introduce additional time-dependent effects that one would expect with real data. The difference in the results, namely $+0.2\%$ signal events and $+0.7\%$ background events, is ascribed as a contribution to the systematic uncertainty.

### 3.3.3 Lepton momentum resolution effect and energy scale effect

Different tests are done to assess the contributions to the systematic uncertainty from the lepton momentum resolution and the uncertainty on the lepton energy scale. Each time, new modified input variables are used to retrain the BDT and assess the overall effect by comparing the new number of selected signal and background events to the original numbers of events selected. All contributions to the systematic uncertainty are summarized in Table 12.

The tests performed are:

- The lepton momenta are smeared using a Gaussian distribution. A constant sigma of 0.73% is used for electrons. For muons, the sigma is calculated using the following formula: $\sigma(p_T) = [(0.011 \cdot p_T)^2 + (0.00017 \cdot p_T^2)^2]^{1/2}/p_T$ with $p_T$ in GeV. The smearing is applied to one type of leptons at a time.

| | signal | background |
|---|---|---|
| electron reconstruction efficiency | ±0.2% | ±0.2% |
| electron $p_T$ resolution (±0.73%) | +0.5% | +1.7% |
| electron energy scale (±0.5%) | +1.1% | +2.1% |
| sub-total for electrons (43% of events) | +1.2% -0.2% | +2.7% - 0.2% |
| muon reconstruction efficiency | ±1.0% | ±1.0% |
| muon $p_T$ resolution (see formula in text) | +1.1% | +1.9% |
| muon energy scale (±1%) | +1.0% | +2.2% |
| sub-total for muons (57% of events) | +1.8% - 1.0% | +3.1% - 1.0% |
| combined contributions for leptons | +1.5% - 0.7% | +2.9% - 0.7% |
| jet energy scale (±7% or ±15%) | +0.8% | +0.2% - 2.2% |
| jet energy resolution effect on $E_T^{miss}$ | -2.2% | -0.4% |
| luminosity | - | ±3.0% |
| cross-section | - | ±5.8% |
| filter effects | ±1.4% | ±1.4% |
| Boosted Decision Tree training effects | ±0.2% | ±0.7% |
| total | +2.2% - 2.6 % | +7.3% - 7.1% |

Table 12: Contributions to the systematic uncertainties. The Higgs boson mass was set to 130 GeV to assess these uncertainties. The final background uncertainty is rounded-off to ±7.2%.

- For each type of lepton, a multiplicative scaling factor is applied to simulate an energy scale uncertainty of ±0.5% for electrons and ±1.0% for muons.

### 3.3.4 Jet momentum resolution effect and energy scale effect

Three different modifications are done in turn to the jet energy to evaluate the contributions from the jet energy scale and jet energy resolution to the missing energy evaluation. After each modification, the missing $E_T$ is recalculatedEach contribution is shown in Table 12. The three modifications made to the jet energy are:

- Jet energy scale: the jet energy is increased by ±7% for jets within $\eta \leq 3.2$ and ±15% for jets within $\eta > 3.2$.
- Jet energy resolution: the jet energy is smeared using a Gaussian by $0.45 \cdot \sqrt{E}$ for jets within $\eta \leq 3.2$ and $0.63 \cdot \sqrt{E}$ for $\eta > 3.2$.

## 3.4 Results for the ZH channel

The sensitivity with 30 fb$^{-1}$ of data is evaluated in terms of $\xi^2$ with $\xi^2 = 1.64\sigma_B/N_S$ for a 95% CL as for the VBF analysis where $\sigma_B$ is the combined statistical and systematic uncertainty as detailed in Section 2.5. The values of $\xi^2$ are summarized in Table 13.

## 3.5 Cross-checks with a cut-based analysis

To ensure that the Boosted Decision Tree performed as expected, we duplicated a previous ATLAS analysis performed using a cuts-based approach [21]. The same cuts as were applied to our current

| $m_H$ | # signal | # background | $\sigma_B$ | $\xi^2$ |
|---|---|---|---|---|
| 110 GeV | 39.2 | 115.9 | 13.6 | 56.7% |
| 120 GeV | 82.5 | 449.4 | 38.3 | 76.2% |
| 130 GeV | 26.8 | 98.6 | 12.2 | 74.4% |
| 140 GeV | 39.5 | 217.0 | 21.3 | 88.5% |
| 150 GeV | 36.5 | 199.4 | 20.0 | 89.9% |
| 200 GeV | 18.5 | 164.4 | 17.3 | 153.4% |
| 250 GeV | 9.3 | 83.8 | 10.9 | 191.6% |

Table 13: The sensitivity with 30 $fb^{-1}$ at 95% confidence level calculated in terms of $\xi^2$ for seven different mass hypotheses for the ZH channel.

Monte Carlo samples after the filter and trigger cuts of this analysis, and using the signal generated with $m_H = 130$ GeV. These cuts are:

1. Filter cuts as in this analysis
2. Trigger cuts as in this analysis
3. Lepton cuts: select events containing no more than two leptons with $p_T > 7$ GeV. Electrons must have $p_T > 15$ GeV within $|\eta| < 2.5$ and muons are selected if $p_T > 10$ GeV and $|\eta| < 2.4$. Two leptons of the same flavor but opposite charge are required.
4. $Z$ mass: the recontructed $Z$ mass must be within 10 GeV from the pole mass.
5. $E_T^{miss} > 100$ GeV.
6. Jet veto: all events containing a jet having $p_T > 30$ GeV within $|\eta| < 4.9$ are rejected.
7. b-jet veto: all events containing a b-tagged jet having at $p_T > 15$ GeV within $|\eta| < 4.9$ are rejected.
8. Transverse mass: $m_T > 200$ GeV.

The two analyses can be compared after the MET cut. The sensitivity with 30 $fb^{-1}$ at 95% confidence level calculated in terms of $\xi^2$ for this cut-based analysis is 87.9% for $m_H = 130$ GeV. This compares well with what was obtained with the BDT technique ($\xi^2$ = 74.4% for the same mass value with the BDT approach). The difference in sensitivity increases further for higher Higgs mass hypotheses. The results are given in Table 14.

# 4 Comparison of results and summary

The sensitivity of ATLAS to an invisibly decaying Higgs boson produced via the VBF and $ZH$ channel has been examined. A comparison between the sensitivities of the two channels can be seen in Figure 13. This plot shows that the channels have a similar sensitivity for low Higgs boson masses. It is possible to look at combined statistics for the ZH analysis and the VBF shape analyses although the analysis techniques are different. Clearly the improvement in sensitivity by combining statistics is not large. Of far greater significance in the analysis of real ATLAS data would be the observation of a significant excess of events in two different and distinct channels. An observation of this kind would give credibility to the hypothesis that a particle is being generated that behaves like a Higgs boson and decays invisibly.

| channel | $ZH$ $\ell\bar{\ell}$ inv. | $ZZ$ $\ell\bar{\ell}\nu\nu$ | $t\bar{t}$ | WW $\ell\nu\ell\nu$ | ZZ $\tau\tau\nu\nu$ | ZZ $\ell\ell\tau\tau$ | ZW $\ell\ell\ell\nu$ | Z+jet $\ell\ell$+jet |
|---|---|---|---|---|---|---|---|---|
| $\sigma$*BR in fb | 46.2 | 728.0 | 833000.0 | 5245.6 | 364.0 | 123.0 | 820.0 | 3105062.8 |
| after filter | 34.9 | 212.6 | 9412.9 | 236.5 | 18.8 | 87.8 | 820.0 | 478.4 |
| after trigger | 32.5 | 198.5 | 8620.6 | 217.1 | 14.2 | 77.8 | 735.3 | 460.6 |
| $p_T$ lepton +ID + charge cut | 23.8 | 148.1 | 4451.2 | 158.4 | 2.7 | 37.6 | 177.5 | 221.6 |
| after $m_Z \pm 10$ GeV cut | 20.7 | 133.0 | 1654.7 | 51.1 | 0.0 | 28.8 | 125.4 | 192.9 |
| after $E_T^{miss} > 100$ cut | 7.4 | 44.9 | 460.7 | 7.3 | 0.0 | 0.8 | 13.7 | 27.9 |
| no jet with $p_T > 30$ GeV | 4.2 | 23.7 | 5.8 | 0.8 | 0.0 | 0.3 | 4.2 | 0.1 |
| b-tag cut for jet with $p_T > 15$ GeV | 4.2 | 23.6 | 4.8 | 0.8 | 0.0 | 0.3 | 4.1 | 0.1 |
| after $m_T > 200$ GeV cut | 3.9 | 21.3 | 0.5 | 0.4 | 0.0 | 0.3 | 3.4 | 0.1 |

Table 14: Monte Carlo estimates of the cross-sections in fb after applying simple cuts for each background process and one mass hypothesis of $m_H = 130$ GeV for the ZH channel. The corresponding $\xi^2$ would be 87.9%.

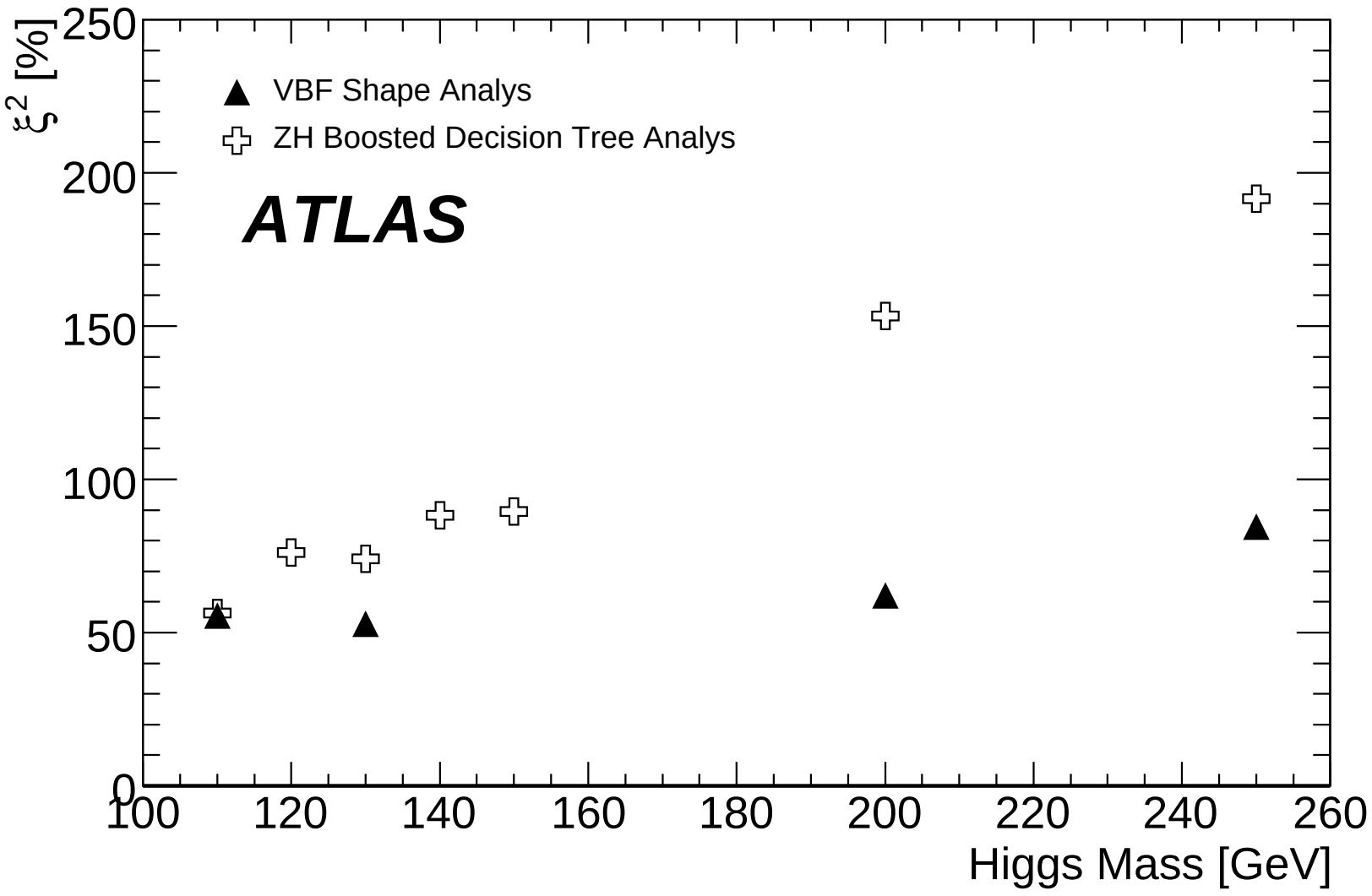


Figure 13: Sensitivity to an invisible Higgs boson with ATLAS for both the VBF and $ZH$ channels with 30 fb$^{-1}$of data assuming only Standard Model backgrounds. The open crosses show the sensitivity for the $ZH$ analysis and the solid triangles show the sensitivity for the VBF shape analysis for 95 % CL. Both these results include systematic uncertainties.

In summary, a study using fully simulated ATLAS data has shown that the ATLAS experiment will be sensitive to an invisibly decaying Higgs boson in both the VBF and *ZH* production channels assuming only Standard Model backgrounds. It is clear that the analysis will require a good understanding of the experimental systematic uncertainties. If the decay of a Higgs boson was entirely in the invisible mode, this analysis has shown that with 30 $fb^{-1}$of data, ATLAS will be sensitive to a situation in which the beyond the Standard Model cross-section is of the order of 80% the Standard Model Higgs boson cross-sections for a Higgs Boson mass of less than 150 GeV. The VBF analysis has a sensitivity of better than 90% up to a Higgs Boson mass of 250 GeV.

# Charged Higgs Boson Searches

**Abstract**
The discovery of a charged Higgs boson would be tangible proof of physics beyond the Standard Model. This note presents the ATLAS potential for discovering a charged Higgs boson, utilizing five different final states of the signal arising from the three dominating fermionic decay modes of the charged Higgs boson. The search covers the region below the top quark mass, taking into account the present experimental constraints, the transition region with a charged Higgs boson mass of the order of the top quark mass, and the high-mass region with a charged Higgs boson mass up to 600 GeV. All studies are performed with a realistic simulation of the detector response including all three trigger levels and taking into account all dominant systematic uncertainties. Results are given in terms of discovery and exclusion contours for each channel individually and for all channels combined, showing that the ATLAS experiment is capable of detecting the charged Higgs boson in a significant fraction of the ($\tan\beta$, $m_{H^\pm}$) parameter space with its first 10 fb$^{-1}$ of data. The so-called intermediate $\tan\beta$ region (around $\tan\beta = 7$) is experimentally hard to reach but exclusion sensitivity is given in this area.

## 1 Introduction

Charged Higgs bosons ($H^\pm$)[1] are naturally predicted in many non-minimal Higgs scenarios, such as Two Higgs Doublet Models (2HDM), and models with Higgs triplets including Little Higgs models. Their discovery would be a definite signal for the existence of New Physics beyond the Standard Model, possibly the first experimental evidence for the Minimal Supersymmetric Standard Model (MSSM) if it is realised in nature, and the supersymmetry mass scale is high enough that sparticles escape discovery. The following analyses will only consider the 2HDM, in particular the so-called type II-2HDM, which is the Higgs sector of the MSSM.

The search strategies for charged Higgs bosons depend on their hypothesized mass, which dictates both the production rate and the available decay modes. Below the top quark mass, the main production mode is through top quark decays, $t \to H^+b$, and in this range the $H^+ \to \tau\nu$ decay mode is dominant. Above the top quark threshold, production mainly takes place through gb fusion ($gb \to tH^+$), and for such high charged Higgs boson masses the decay into a top quark and a $b$ quark dominates, $H^+ \to tb$ . A more detailed discussion of the different signal final states used for this study including backgrounds, cross-sections and branching ratios can be found in Reference [1].

Charged Higgs boson searches involve several higher level reconstructed physics objects such as electrons, muons, jets, jets tagged as $b$ jets and finally jets identified as $\tau$ jets. These objects are reconstructed by dedicated ATLAS algorithms and details about their performance can be found in References [2–6]. In the following, only efficiencies will be quoted for the different objects.

The note is organized as follows: Section 2 summarizes the trigger menus used for the signatures under investigation. The analysis sections are divided into light (Section 3) and heavy (Section 4) $H^+$ studies, which are in turn divided according to the signal final state: First, $t\bar{t} \to bH^+bW \to b\tau(had)\nu bqq$ is discussed in Section 3.1, then $t\bar{t} \to bH^+bW \to b\tau(lep)\nu bqq$ in Section 3.2. Section 3.3 addresses $t\bar{t} \to bH^+bW \to b\tau(had)\nu b\ell\nu$. Two studies for heavy $H^+$ are described, namely for $gg/gb \to t[b]H^+ \to bqq[b]\tau(had)\nu$ (Section 4.1) and $gg/gb \to t[b]H^+ \to t[b]tb \to bW[b]bWb \to b\ell\nu[b]bqqb$ (Section 4.2). In Section 5, systematic uncertainties and their impact are discussed, followed by a description of the

[1] In the following, the charged Higgs boson will be denoted $H^+$ , but $H^-$ is always implicitly included.

data-driven $t\bar{t}$ control sample method which is needed for the extraction of the background shape and normalization in all $H^+$ studies. Section 6 provides combined results from the different search topologies. The charged Higgs boson results are summarized in Section 7.

## 2 Trigger

The production and decay modes of the charged Higgs boson studied here lead to final states containing the following: two to four $b$ jets, light jets from hadronic decays of $W$ bosons, one or more neutrinos from $W$ or $H^+$ decays, and for most channels a tau lepton decaying either hadronically or into an electron or muon plus neutrinos. These event characteristics suggest the following ATLAS Trigger [7] menus, one for an instantaneous luminosity of $L = 10^{31}\text{cm}^{-2}\text{s}^{-1}$, and two alternatives for $L = 10^{33}\text{cm}^{-2}\text{s}^{-1}$, based on these trigger signatures:

- $L = 10^{31}\text{cm}^{-2}\text{s}^{-1}$: `xE70`, `e22i`, `mu20`, `xE30_L1_TAU13`, `xE20_3j20_L1_TAU13`
- $L = 10^{33}\text{cm}^{-2}\text{s}^{-1}$: `xE80`, `e55`, `mu40`, `xE50_L1_TAU30`, `xE40_3j20_L1_TAU30` or
  `xE80`, `e22i_xE30`, `mu20_xE30`, `xE50_L1_TAU30`, `xE40_3j20_L1_TAU30`

where xE represents a missing transverse energy trigger selection, and e, mu, tau and j correspond to electron, muon, tau, and jet triggers, respectively. A number before each symbol indicates the required trigger multiplicity, as in 3j for three jets. Values after these symbols are the approximate trigger thresholds in transverse momentum, and a final i means that isolation requirements are applied. `L1_` indicates that the stated threshold for the following trigger selection is applied at the first trigger level while the high level trigger selection is softer. Combined triggers are indicated by the juxtaposition of two or more selections. Each analysis uses a subset of items of the two $10^{33}$ menus.

The menus were chosen based on a careful study of the signal event characteristics and the most realistic trigger signature rates available. The chosen signatures meet the requirements of the trigger bandwidth budget [7] at all trigger levels. For each of the channels studied in the following sections, the trigger efficiencies will be evaluated.

## 3 Light Charged Higgs Boson Searches

In this section, charged Higgs boson searches in the three light $H^+$ channels ($m_{H^+} < m_t$) selected for investigation are presented:

- $t\bar{t} \to bH^+bW \to b\tau(had)\nu bqq$
- $t\bar{t} \to bH^+bW \to b\tau(lep)\nu bqq$
- $t\bar{t} \to bH^+bW \to b\tau(had)\nu b\ell\nu$

where for a tau lepton, $\tau(had)$ indicates its hadronic decay and $\tau(lep)$ its decay to an electron or muon plus neutrinos. If the charged Higgs boson is light, then BR($t \to bW$) might not be close to unity, as it is in the Standard Model prediction. This means that the expected background from Standard Model $t\bar{t}$ decays (the dominant background to all $H^+$ searches) is reduced. Since the number of events after the event selection is $N_{t\bar{t}}^{\text{SM}}$ for the background-only hypothesis, and $N_{H^+} + N_{t\bar{t}}^{\text{MSSM}}$ in the signal+background case, the number of observable excess events is thus not simply the number of $H^+$ events ($N_{H^+}$) — it is decreased by the difference in the SM and the MSSM $t\bar{t}$ prediction: $N_{excess} = N_{H^+} - (N_{t\bar{t}}^{\text{SM}} - N_{t\bar{t}}^{\text{MSSM}})$. This is taken into account consistently.

All plots, tables and results in this section are based on the trigger menu for an instantaneous luminosity of L=$10^{33}\text{cm}^{-2}\text{s}^{-1}$ presented in Section 2, on the signal cross-sections for the $m_h$-max MSSM scenario (see Reference [1]) and the background cross-sections given in Reference [8].

## 3.1 $t\bar{t} \to bH^+bW \to b\tau(had)\nu bqq$

The channel in which both the $\tau$ and the $W$ boson decay hadronically has a relatively high cross-section, making it a priori one of the most promising in searches for $H^+$ lighter than the top quark. On the other hand, the absence of leptons, as well as the high hadronic activity, is a challenge, particularly for the trigger. The events are characterised by a $\tau$ jet, four other jets, two of which are initiated by $b$ quarks, and missing energy. The main expected backgrounds for this channel are $t\bar{t}$ events where one top quark decays hadronically and the other to a hadronically decaying $\tau$. For this study all $t\bar{t}$ decay modes have been considered as background, as well as single top, $W$+jets and QCD dijet events. For ATLAS, this channel has previously only been studied using fast simulation [9].

### 3.1.1 Preselection

**Trigger** Either the `xE50_L1_TAU30` or the `xE40_3j20_L1_TAU30` trigger signature (see Section 2) is required for this study.

**Cuts I** Following the trigger, a set of cuts was used to preselect the signal and suppress the background. This first cut set relates primarily to the multiplicity of the offline analysis objects (which are different from the trigger objects, leading in some cases to lower optimum cut values for the offline objects). These are given in Table 3.1.1.

**Cuts II** After the first set of cuts, the $W$ is reconstructed from the pair of light jets with invariant mass $m_{jj}$ closest to the nominal $W$ mass, $m_W$. The parent top quark is then found by pairing the reconstructed $W$ with the $b$ quark which leads to an apparent top quark mass $m_{jjb}$ closest to the nominal value $m_t$. Cuts are made both on the $W$ and the top quark reconstructed masses. The top quark on the $H^+$ side cannot be fully reconstructed due to the presence of the neutrino, but information on its azimuthal angle, $\phi$, and transverse momentum, $p_T$, can still be extracted.

At this stage a large number of background events still remains, mainly from Standard Model $t\bar{t}$ as well as from QCD and single-top production. In order to discriminate against the latter two, two cuts (items 8 and 9 in Table 3.1.1) are applied aiming to enhance the topology of $t\bar{t}$ decays.

Table 1: $t\bar{t} \to bH^+bW \to b\tau(had)\nu bqq$: Applied precuts. A $b$-tagging cut is applied to all jets, yielding $b$ jet reconstruction efficiency of about 70%. For $\tau$ jets, a cut on the $\tau$-tagging variable is applied such that a reconstruction efficiency of about 32% is obtained for a pseudorapidity $|\eta| < 1.5$, and a tighter cut (due to the high QCD jet fake-rate for large $\eta$) for $|\eta| > 1.5$ leads to an efficiency of about 20% in this region. The $E_T^{miss}$ cut is increased to 50 GeV if only the signature xE50_L1_TAU30 has triggered the event.

| Cuts I | Cuts II |
|---|---|
| 1. exactly 1 $\tau$-tagged jet with $p_T > 35$ GeV | 6. $\lvert m_W^{rec} - m_W\rvert < 30$ GeV |
| 2. exactly 2 $b$-tagged jets with $p_T > 15$ GeV | 7. $\lvert m_t^{rec} - m_t\rvert < 40$ GeV |
| 3. at least 2 non-tagged jets with $p_T > 15$ GeV | 8. $\Delta\phi(p_T^{\text{hardest top}}, p_T^{\text{softest top}}) > 2.5$ |
| 4. veto on isolated leptons with $p_T > 5$ GeV | 9. $p_T^{\text{hardest top}} / p_T^{\text{softest top}} < 2$ |
| 5. $E_T^{miss} > 40/50$ GeV | |

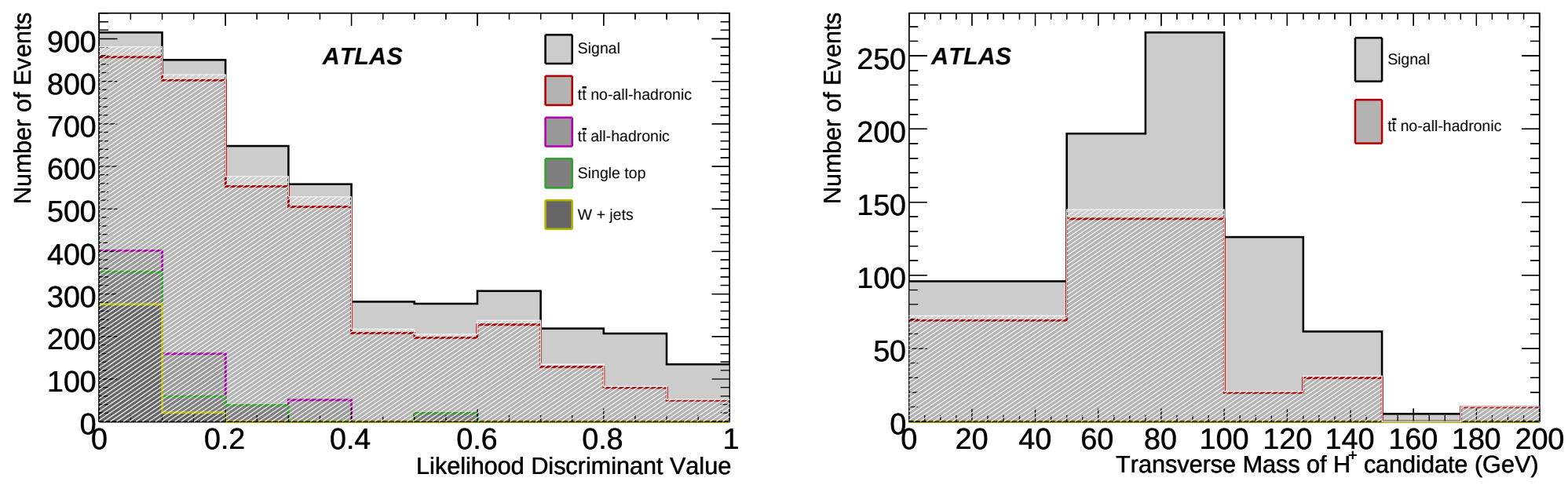


(a) (b)

Figure 1: $t\bar{t} \to bH^+bW \to b\tau(had)\nu bqq$: Likelihood distribution (a) and transverse mass (b) for an $H^+$ mass of 130 GeV and the corresponding background, stacked and normalized to the cross-section with $\tan\beta = 20$. The hatched area shows what would be expected from a Standard Model only scenario. A cut at 0.6 is made on the likelihood and at 65 GeV on the transverse mass.

### 3.1.2 Likelihood Discriminant

Following these cuts, the remaining background is dominated by $t\bar{t}$ events, particularly those in which one $W$ decays hadronically and the other to a $\tau$ and a neutrino, and all reducible backgrounds are effectively suppressed. In order to discriminate between this background and the signal, a Likelihood Discriminant method has been implemented. The variables used (listed in Table 2) have been selected to reflect the two characteristics which distinguish the signal from the background: the heavier mass of the $H^+$ compared to the $W$, with its effects on the kinematics of the event; and the difference in $\tau$ polarization depending on the parent. This difference, exploited in variable I, will cause the leading track within the $\tau$ jet to be harder [10]. Also selected combinations of variables have been included, as correlations are not included in the definition of the likelihood used. Likelihood discriminants are constructed for the five different studied $H^+$ mass points (90, 110,120,130 and 150 GeV). Figure 1a shows the resulting likelihood distribution for an $H^+$ mass of 130 GeV and the background.

Table 2: $t\bar{t} \to bH^+bW \to b\tau(had)\nu bqq$: Variables used for the Likelihood Discriminant. $\tau$ denotes the $\tau$ jet in the event, $b_{H^+}$ is the $b$ jet coming from the same top as the presumed $H^+$, and $M(\tau, b_{H^+})$ is the invariant mass of the $\tau$ jet and the $b_{H^+}$ jet. $\Delta R = \sqrt{(\Delta\phi)^2 + (\Delta\eta)^2}$ represents a distance.

| | | | |
|---|---|---|---|
| I. | $p_T^{\text{leading }\tau\text{ track}}/p_T^{\tau}$ | V. | $p_T^{\tau}/p_T^{b_{H^+}}$ |
| II. | $1-\cos(\Delta\phi(\tau, p_T^{miss}))$ | VI. | $M(\tau, b_{H^+})$ |
| III. | $1-\cos(\Delta\phi(H^+, b_{H^+}))$ | VII. | $\Delta R(\tau, b_{H^+})$ |
| IV. | $M(\tau, b_{H^+}) \cdot \Delta R(\tau, b_{H^+})$ | | |

A cut is applied on the obtained likelihood, requiring a value higher than 0.6 for an event to be retained (0.8 for a charged Higgs boson mass $m_{H^+} = 150$ GeV, due to the better signal-background separation). The transverse mass of the $H^+$ is then calculated for the retained events (shown in Fig. 1b). A final cut is then applied to this reconstructed transverse mass of the $H^+$ candidate (at 50 GeV for $m_{H^+} = 90$ GeV, at 60 GeV for $m_{H^+} = 110$ and 120 GeV, at 65 GeV for $m_{H^+} = 130$ GeV and at 75 GeV for $m_{H^+} = 150$ GeV). These cut values have been selected to optimize the significance considering both systematic and statistical uncertainties.

### 3.1.3 Results

Table 3: $t\bar{t} \to bH^{+}bW \to b\tau(had)\nu bqq$: Selection cut flow. For each sample, the cross-sections after cuts are given in fb and for $\tan\beta = 20$ in the first line and the relative cut efficiencies in the second line (in italics). Standard Model cross-sections are given for all the backgrounds. There are insufficient Monte Carlo statistics for the QCD events – see discussion in the text for a cross-section estimate.

| Channel | | | All events | Trigger | Cuts I | Cuts II |
|---|---|---|---|---|---|---|
| $H^+$ | 90 GeV | [fb] | 38388 | 3384 | 404 | 136 |
| | | [/] | | *0.088* | *0.119* | *0.337* |
| $H^+$ | 110 GeV | [fb] | 27147 | 2871 | 286 | 112 |
| | | [/] | | *0.106* | *0.100* | *0.391* |
| $H^+$ | 120 GeV | [fb] | 21363 | 2563 | 255 | 94 |
| | | [/] | | *0.120* | *0.099* | *0.368* |
| $H^+$ | 130 GeV | [fb] | 15666 | 2136 | 217 | 79 |
| | | [/] | | *0.136* | *0.102* | *0.364* |
| $H^+$ | 150 GeV | [fb] | 5869 | 982 | 88 | 30 |
| | | [/] | | *0.167* | *0.089* | *0.342* |
| $t\bar{t} \geq 1$ lepton | | [fb] | $4.52{\cdot}10^5$ | 56287 | 852 | 307 |
| | | [/] | | *0.125* | *0.015* | *0.360* |
| $t\bar{t}$ hadronic | | [fb] | $3.81{\cdot}10^5$ | 1746 | 37 | 21 |
| | | [/] | | *0.005* | *0.021* | *0.571* |
| single top | | [fb] | 112500 | 7700 | 63 | 17 |
| | | [/] | | *0.068* | *0.008* | *0.277* |
| W + jets | | [fb] | 277800 | 15489 | 73 | 30 |
| | | [/] | | *0.056* | *0.005* | *0.409* |
| QCD dijets | | [fb] | $3.2{\cdot}10^8$ | $3.18{\cdot}10^5$ | 5 | – |
| | | [/] | | *0.001* | $1.7{\cdot}10^{-4}$ | – |

Table 3 shows the selection cut flow for the five different signal mass points and all backgrounds studied in this note. The expected cross-section after the trigger and after the first and second set of cuts is presented, as well as the relative efficiency of each step.

It is difficult to draw conclusions about the suppression of the QCD background, due to its very high cross-section at the LHC. This leads to large statistical uncertainties since no simulated event survives the entire analysis chain. In fact already the expected cross-section after Cuts I (see Table 3) is governed by large statistical uncertainties. A very conservative upper limit for the expected number of QCD events can be obtained by assuming that the expected efficiency of QCD events is limited from above by the efficiency of the hadronic $t\bar{t}$ events for all cuts for which the number of simulated QCD events is not sufficient to draw definitive conclusions[2]. This assumption is very conservative since the second set of cuts have been specifically designed to identify the $t\bar{t}$ topology in order to reject QCD. Nevertheless, using this assumption, the cross-section of surviving QCD events is limited by 55 fb after the likelihood cut (for $m_{H^+} = 130$ GeV), while after the cut on the transverse mass this estimate is down to 50 fb. Given the conservative nature of this estimate, the assumption that the background from QCD events is negligible compared to the one from $t\bar{t}$ events is justified.

Table 4 shows the final expected cross-section following the cut on the likelihood discriminant and the further cut on the transverse mass of the $H^+$ candidate. The shape of the likelihood discriminant

[2] After Cuts II, even more conservatively, this limit is taken from the efficiency for the $t\bar{t} \geq 1$ $e/\mu/\tau$ events, due to the limited number of surviving hadronic $t\bar{t}$ events at this point.

Table 4: $t\bar{t} \to bH^+bW \to b\tau(had)\nu bqq$: Final event selection results. The cross-sections after all cuts are given in fb and for $\tan\beta = 20$ as well as the relative cut efficiencies. Standard Model cross-sections are given for $t\bar{t}$ . Backgrounds not tabulated have been found to be negligible.

| Channel | | Cut | Signal [fb] | Signal [/] | $t\bar{t} \geq 1\ e/\mu/\tau$ [fb] | $t\bar{t} \geq 1\ e/\mu/\tau$ [/] |
|---|---|---|---|---|---|---|
| $H^+$ | 90 GeV | LH > 0.6 | 56.2 | *0.413* | 55.8 | *0.182* |
| | | $mT$ >50 GeV | 35.3 | *0.628* | 32.1 | *0.574* |
| $H^+$ | 110 GeV | LH > 0.6 | 53.6 | *0.478* | 52.7 | *0.172* |
| | | $mT$ >60 GeV | 35.1 | *0.655* | 27.9 | *0.529* |
| $H^+$ | 120 GeV | LH > 0.6 | 42.6 | *0.455* | 45.5 | *0.148* |
| | | $mT$ >60 GeV | 32.5 | *0.764* | 29.0 | *0.636* |
| $H^+$ | 130 GeV | LH > 0.6 | 38.3 | *0.483* | 50.7 | *0.165* |
| | | $mT$ >65 GeV | 31.4 | *0.819* | 25.9 | *0.510* |
| $H^+$ | 150 GeV | LH > 0.8 | 14.0 | *0.467* | 26.9 | *0.088* |
| | | $mT$ >75 GeV | 9.3 | *0.662* | 10.3 | *0.385* |

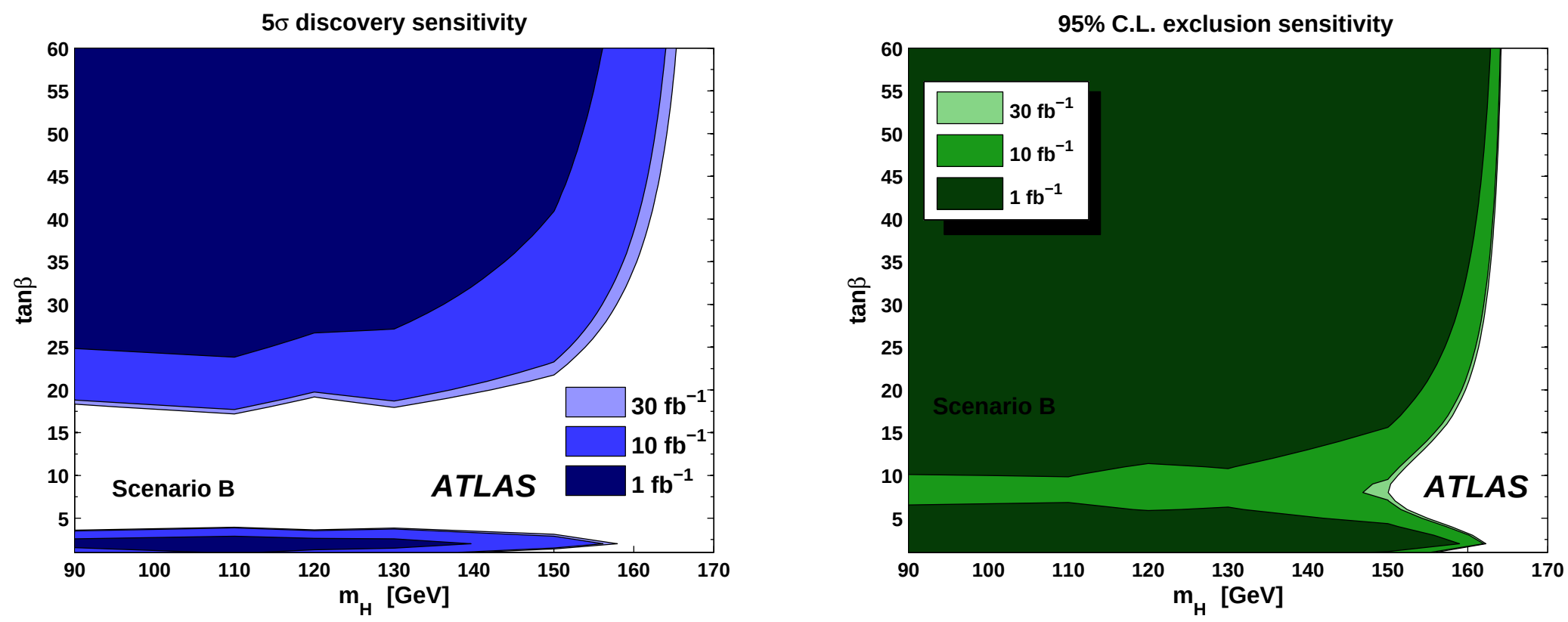


Figure 2: $t\bar{t} \to bH^+bW \to b\tau(had)\nu bqq$: Discovery (left) and exclusion contour (right) for Scenario B ($m_h$-max) [1]. Systematic and statistical uncertainties are included. The systematic uncertainty is assumed to be 10% for the background, and 24% for the signal (see Sections 5.2 and 5.1). The lines indicate a $5\sigma$ significance for the discovery and a 95% CL for the exclusion contour.

depends on the mass point for which the analysis is performed, i.e. one will need to run a separate analysis using a different likelihood discriminant for each mass point. Therefore the background rejection will naturally depend on the mass point for which it is evaluated.

The shape-based Profile Likelihood method (see Section 6) is applied on the entire transverse mass histogram for each masspoint, in order to extract the significance of the signal hypothesis. Fig. 2 shows the discovery contour in the ($\tan\beta$, $m_{H^+}$) plane for an integrated luminosity of $\mathscr{L} = 10$ fb$^{-1}$ as well as the exclusion reach for $\mathscr{L} = 1$ fb$^{-1}$.

### 3.2 $t\bar{t} \to bH^+bW \to b\tau(lep)\nu bqq$

The events of the leptonic $\tau$ channel are characterized by a single isolated lepton, and large missing energy due to three neutrinos in the final state. A full reconstruction of the event is therefore impossible. Instead, kinematic properties of the event are used to discriminate between the signal and the main

background, which is the Standard Model semi-leptonic $t\bar{t}$ process.

The branching ratio to $\tau$ leptons of a light charged Higgs boson is expected to be almost 100%, while the Standard Model $W$ decay is universal with respect to lepton flavor. This means that the most prominent signature of the signal is an excess of kinematically $\tau$-like events.

Furthermore, it is possible to construct a quantity that is bound by the charged Higgs boson mass (in analogy to the transverse mass in a $W$ decay). This 'charged Higgs boson transverse mass' can be used to provide further discrimination power against the background, as well as a direct indication for the charged Higgs boson mass.

### 3.2.1 Preselection

**Trigger** The trigger requirement is based on either an isolated lepton or missing transverse energy (xE). Events are required to pass one of the following three trigger signatures (see Section 2): `e22i_xE30`, `mu20_xE30`, or `xE80`.

**Offline** Events that have passed the trigger are selected based on the following cuts:

- Exactly one isolated lepton with $p_T > 5$ GeV. If the event has not been triggered by the `xe80` signature then the requirement is raised to 20 GeV or 25 GeV (for the `e25i+xe30` and `mu20i+xe30` requirement, respectively).
- Missing transverse energy $E_T^{miss} > 120$ GeV
- At least 4 jets with $p_T > 40$ GeV and $|\eta| < 2.5$
- Exactly two out of the four leading jets are tagged as $b$ jets

Events surviving these selection cuts are expected to be dominantly $t\bar{t}$ events.

### 3.2.2 Method for final state reconstruction

First, the hadronic $W$ is reconstructed from the light (non $b$-tagged) jets. All possible pairs of light jets are considered and the one which gives an invariant mass closest to $m_W$ is selected.

The assignment of the two $b$ jets to the hadronic & leptonic sides is done using the angular correlation between the $b$ jets and their associated particles, i.e. the lepton and the reconstructed $W$. The charge correlation between the lepton and its associated $b$ jet is also used. The charge of the $b$ jet is defined in the following way:

$$Q_{jet} = \frac{\sum (p_L^i)^\alpha q^i}{\sum (p_L^i)^\alpha} \tag{1}$$

where the sum is over tracks $i$ belonging to the jet, $q^i$ is the charge associated to the track, and $p_L^i$ is the longitudinal component of the track momentum with respect to the jet direction. The parameter $\alpha$ is optimized to give maximal separation between $b$ and $\bar{b}$ jets. The value obtained from optimization with MC data is $\alpha = 0.5$.

A likelihood ratio combining the angular and charge correlations is used. This likelihood ratio is defined between two hypotheses corresponding to the two possibilities of assigning the $b$ jets. The hypothesis which achieves the higher likelihood ratio score is adopted. This algorithm selects the correct $b$ jets assignment for about 70% of the events, when no cut is applied to the likelihood score of the selected hypothesis. Higher purity can be achieved at the cost of lower efficiency by placing such a cut, but this is not done for this analysis.

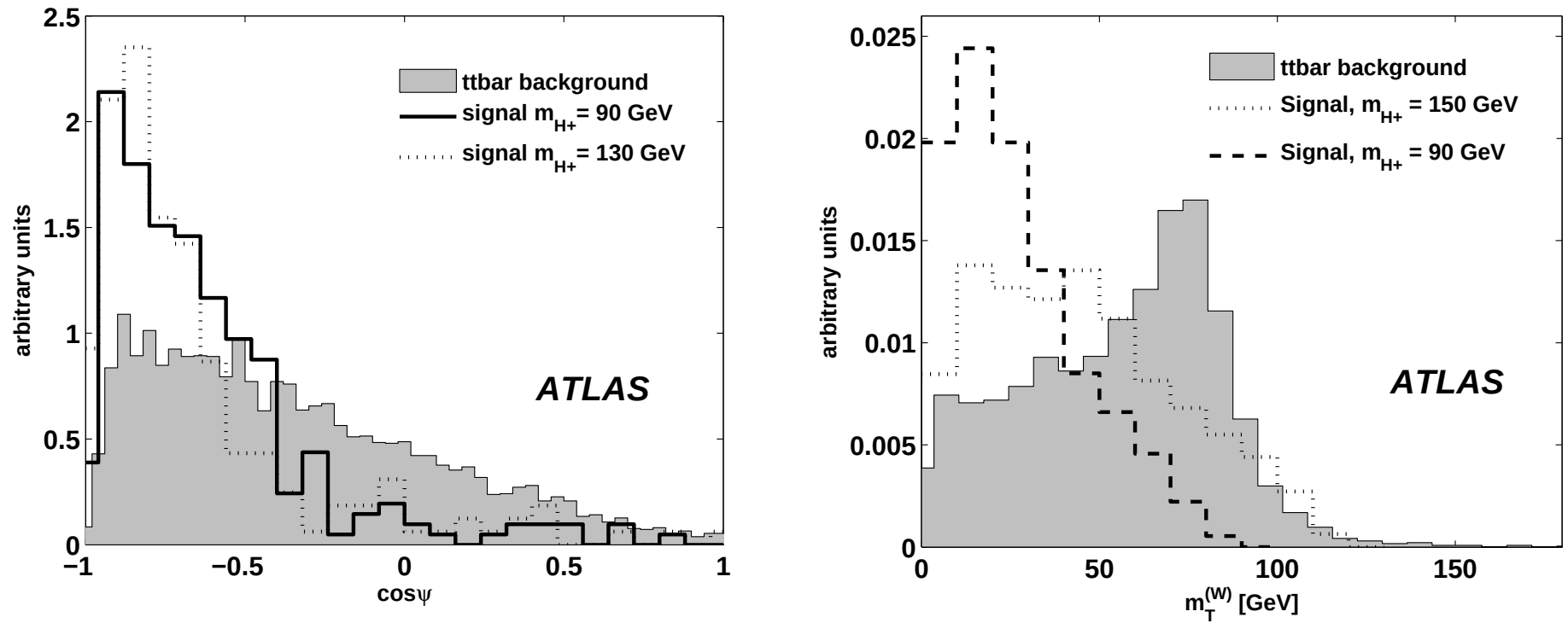


(a) (b)

Figure 3: $t\bar{t} \rightarrow bH^+bW \rightarrow b\tau(lep)\nu bqq$: (a) $\cos\psi$ distribution, (b) W transverse mass distribution, for signal and background, after selection cuts, for charged Higgs boson masses of 90 and 150 GeV.

Once the $b$ jets are assigned, the reconstructed top quark mass $m_{top}^{rec}$ on the hadronic side of the event can be evaluated, and only those events which satisfy 100 GeV $< m_{top}^{rec} <$ 300 GeV are retained for further analysis.

### 3.2.3 Further reduction of $t\bar{t}$ background

In order to reduce the number of background events in which the lepton comes directly from a $W$ decay, the decay angle $\cos\psi$ is used

$$\cos\psi = \frac{2m_{\ell b}^2}{m_{top}^2 - m_W^2} - 1. \tag{2}$$

In such events, $\psi$ is the angle between the lepton and top quark directions in the $W$ rest frame (In the limit that the $b$ quark is massless). The top quark, due to its large mass, couples mostly to the longitudinal polarization component of the $W$ boson. As a result the $\cos\psi$ distribution would have a large ($\approx 70\%$) contribution that is symmetric around $\cos\psi = 0$ [11–13]. However, both the left-handed component of the $W$ boson and the indirect leptons coming from $\tau$ decays contribute to the lower $\cos\psi$ region, and this region is further enhanced by the selection cuts that favor events with large missing energy and therefore leptonic $\tau$ decays. The signal contribution is similar to that of the indirect leptons, and the less energetic $b$ jets, due to the higher $H^+$ mass, push $\cos\psi$ to even lower values. This can be seen in Fig. 3 (a). As a further selection cut, events are required to have $\cos\psi < -0.8$.

The transverse mass of the $W$ (in the hypothesis of a leptonic W decay), $m_T^W$, calculated using the missing transverse momentum and the lepton transverse momentum, provides further discrimination against the $t\bar{t}$ background with a direct lepton from a $W$. The separation is most distinct for low charged Higgs boson masses and is shown for two different $H^+$ masses in Fig. 3 (b).

**Charged Higgs Boson Transverse Mass** A generalized transverse mass for the charged Higgs boson (in the hypothesis of a $H^+ \rightarrow \tau\nu$ decay) in this channel can be defined. The derivation and properties of this variable are described in Reference [14].

$$(m_T^{H^+})^2 = \left(\sqrt{m_{top}^2 + (\vec{p}_T^{\,lep} + \vec{p}_T^{\,b} + \vec{p}_T^{\,miss})^2} - p_T^b\right)^2 - (\vec{p}_T^{\,miss} + \vec{p}_T^{\,lep})^2 \tag{3}$$

This transverse mass satisfies $m_{H+} < m_T^{H^+} < m_{top}$. Figure 4 shows the distributions of $m_T^{H^+}$ for several charged Higgs boson masses.

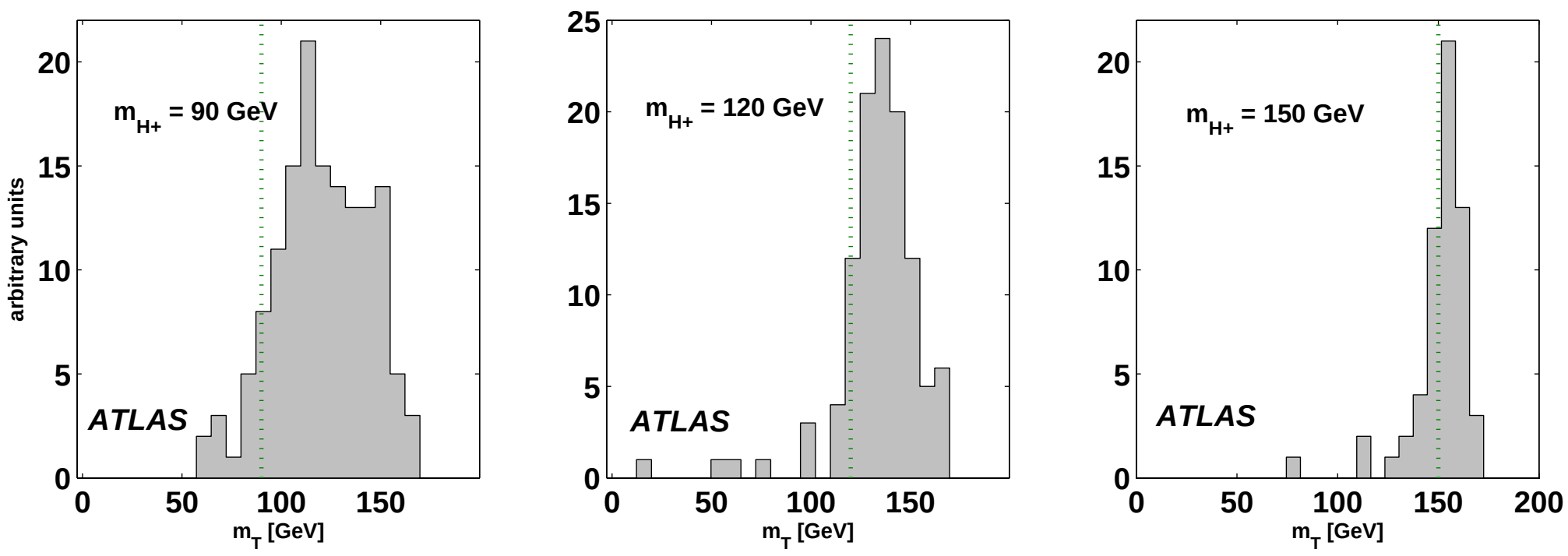


Figure 4: $t\bar{t} \to bH^+bW \to b\tau(lep)\nu bqq$: $m_T^{H^+}$ distribution, after selection cuts, for $m_{H^+} = 90$, 120, and 150 GeV. The dotted line represents the nominal simulated mass.

### 3.2.4 Results

The cross-sections of signal and background events surviving selection cuts are shown in Table 5.

Table 5: $t\bar{t} \to bH^+bW \to b\tau(lep)\nu bqq$: Selection cut flow. For each sample, the cross-sections after cuts (in fb) are given in the first line and the relative cut efficiencies in the second line (in italics). The signal cross-sections correspond to $\tan\beta = 20$.

| Channel | | | All events | Trigger | pre-selection | reco. cuts | $cos\psi$ | $m_T^W$ |
|---|---|---|---|---|---|---|---|---|
| $H^+$ | 90 GeV | [fb] | 20760 | 8408 | 239 | 190 | 74 | 70 |
| | | [/] | | *0.405* | *0.028* | *0.79* | *0.39* | *0.94* |
| | 110 GeV | [fb] | 14710 | 6401 | 138 | 104 | 42 | 37 |
| | | [/] | | *0.43* | *0.021* | *0.75* | *0.40* | *0.89* |
| | 120 GeV | [fb] | 11560 | 5305 | 125 | 82 | 32 | 23 |
| | | [/] | | *0.46* | *0.023* | *0.66* | *0.39* | *0.71* |
| | 130 GeV | [fb] | 8510 | 4103 | 75 | 49 | 23 | 21 |
| | | [/] | | *0.48* | *0.018* | *0.65* | *0.47* | *0.88* |
| | 150 GeV | [fb] | 3180 | 1747 | 33 | 23 | 12 | 10 |
| | | [/] | | *0.55* | *0.019* | *0.70* | *0.54* | *0.79* |
| $t\bar{t} \geq 1\ e/\mu/\tau$ | | [fb] | 452000 | 209339 | 1963 | 1317 | 257 | 144 |
| | | [/] | | *0.46* | *0.009* | *0.67* | *0.19* | *0.56* |
| QCD dijet $p_T$=280-1120 GeV | | [fb] | $12.9 \cdot 10^6$ | 213000 | $< 50$ | $< 50$ | $< 50$ | $< 50$ |
| | | [/] | | *0.017* | $< 2.5 \cdot 10^{-4}$ | - | - | - |
| W+jets | | [fb] | $31.2 \cdot 10^6$ | $7.69 \cdot 10^6$ | 173 | 86.4 | $< 80$ | $< 80$ |
| | | [/] | | *0.25* | $2.2 \cdot 10^{-5}$ | *0.50* | - | - |

Figure 5 (a) shows the $m_T^W$ differential cross-section for $\tan\beta = 20$. Figure 5 (b) shows the corresponding distributions for $m_T^{H^+}$.

The statistical significance of the signal is calculated from both the $W$ transverse mass and the $H^+$ transverse mass distributions, after event reconstruction and all selection cuts. The final significance is taken as the maximum of these two. In Fig. 6, the $5\sigma$ discovery contour is plotted in the ($m_{H^\pm}$,$\tan\beta$) plane.

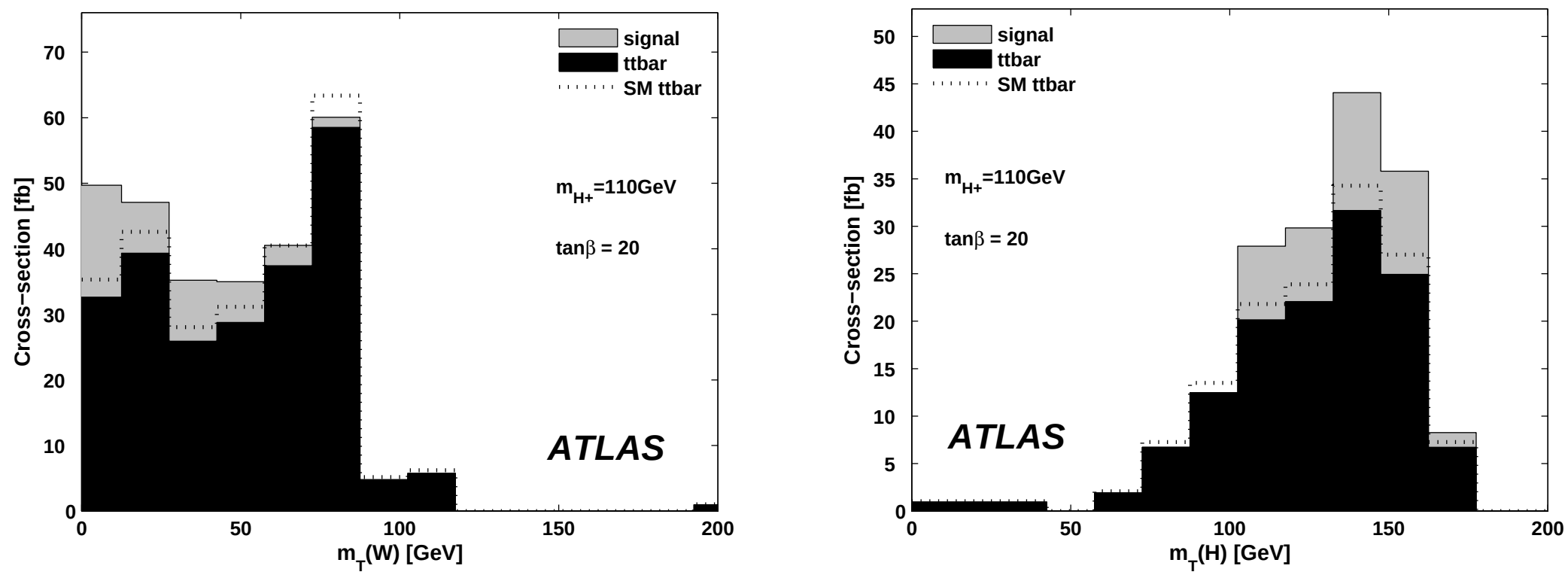


(a) (b)

Figure 5: $t\bar{t} \to bH^+bW \to b\tau(lep)\nu bqq$: Transverse mass differential cross-section for signal and background, for $\tan\beta = 20$, and for the hypothesis of (a) $W$, and (b) $H^+$ .

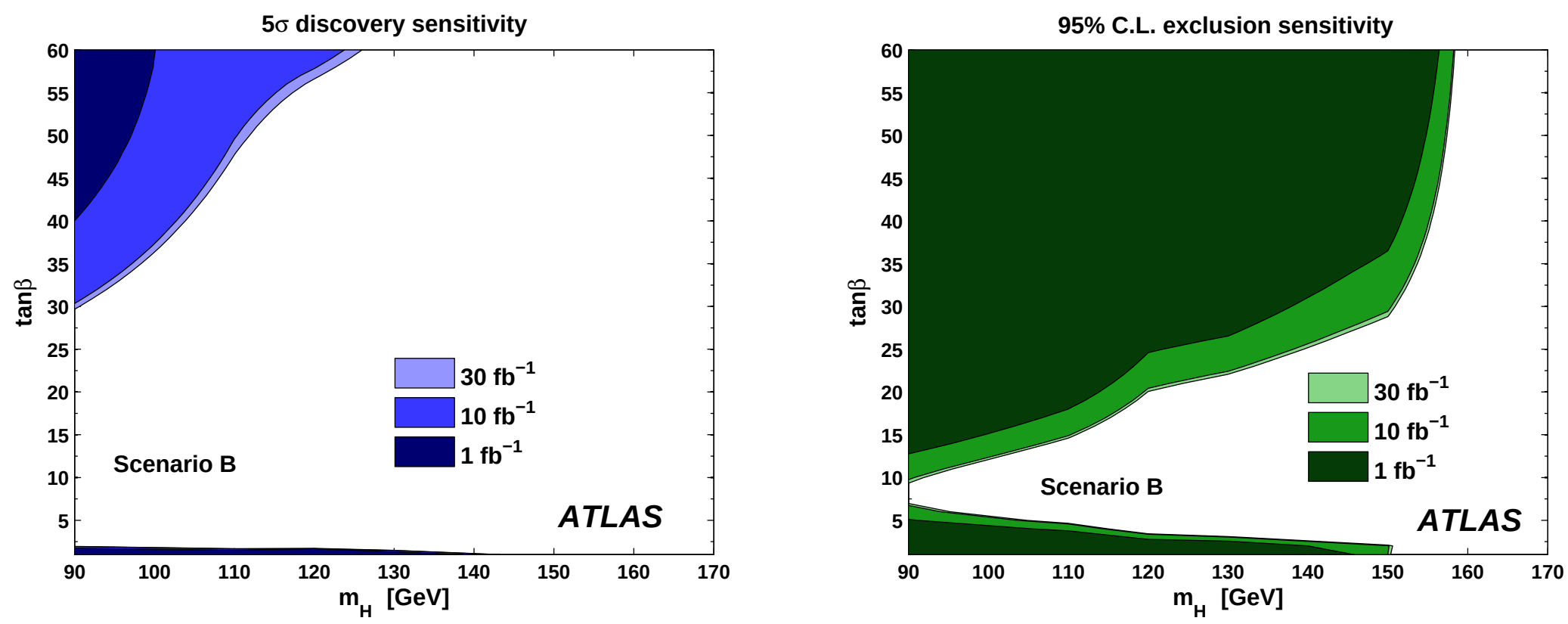


Figure 6: $t\bar{t} \to bH^+bW \to b\tau(lep)\nu bqq$: Discovery (left) and exclusion contour (right) for Scenario B ($m_h$-max) [1]. Systematic and statistical uncertainties are included. The systematic uncertainty is assumed to be 10% for the background, and 35% for the signal (see Sections 5.2 and 5.1). The lines indicate a $5\sigma$ significance for the discovery and a 95% CL for the exclusion contour.

### 3.3 $t\bar{t} \to bH^+bW \to b\tau(had)\nu b\ell\nu$

In this channel a light charged Higgs boson is produced in $t\bar{t}$ decay. The $W$ boson from one top quark decays into leptons, while the $\tau$ lepton originating from the $H^+$ decay results in a jet. At least three neutrinos are present in the event and thus the reconstruction of the complete event is impossible. Due to the high branching ratio of this charged Higgs boson decay (BR($H^+ \to \tau\nu$ ) ≈ 100% for low $m_{H^+}$), the signal can be observed as an excess of tau leptons in the final state over the main background of Standard Model $t\bar{t}$ production.

Due to the production mechanism, the Standard Model $t\bar{t}$ events will be the most important background to this channel. However, significant contributions from other backgrounds are still possible, and thus important to study: In the inclusive process $pp \to W$+jets, the final state is similar to the signal signature if one of the jets is mis-tagged as a $\tau$ jet or a lepton and the $W$ boson decays to the appropriate other object. Furthermore, the backgrounds constituted by single top quark events are considered. Single

top quarks can be produced with an associated $W$ boson or b jet, and thus produce final states similar to the signal. Their contribution to the total background is however expected to be small due to the smaller cross-sections. Finally, the overwhelming background of QCD dijet is expected to be completely suppressed by demanding a high quantity of missing transverse energy and an isolated lepton.

### 3.3.1 Event Selection

**Trigger:** Events are required to pass one of the following three trigger items (see Section 2): `e22i_xE30`, `mu20_xE30`, or `xE40_3j20_L1_TAU30`.

**Selection:** To select signal events and suppress the Standard Model $t\bar{t}$ background several small differences between the processes are exploited. The following cuts are applied:

1. $N_{e,\mu} \geq 1$: At least one isolated lepton with $p_T > 10$ GeV; muons with $|\eta| < 2.7$ and electrons with $|\eta| < 2.5$ are considered.
2. $N_{\text{jets}}=N_{\text{light jets}}+N_{\text{b jets}}+N_{\tau\text{ jets}} \geq 3$: At least three jets with $p_T > 20$ GeV.
3. $N_{\tau\text{ jet}} \geq 1$: At least one of the jets is required to be $\tau$-tagged. A $\tau$ jet quality cut is applied, leading to a $\tau$ jet reconstruction efficiency of about 30%.
4. $N_{\text{b jet}} \geq 1$: At least one $b$ jet with $p_T > 20$ GeV is required, and a $b$ jet quality cut is applied, leading to a $b$ jet efficiency of about 60%.
5. $p_T^\tau > 40$ GeV
6. $p_T^e > 25$ GeV or $p_T^\mu > 20$ GeV: This cut is only applied if the event was triggered by the appropriate lepton signature. If both lepton trigger requirements are met then only the cut on $p_T^e$ is applied.
7. $q_\tau + q_l = 0$: The $\tau$ jet and the lepton are required to have opposite charge.
8. $E_T^{miss} > 175$ GeV

Figure 7 (a) shows the jet multiplicity of the signal and the main $t\bar{t}$ background. While there are more jets for the $t\bar{t}$ background than the signal a cut on jet multiplicity suppresses other backgrounds. Figure 7 (b) displays the $\tau$ transverse momentum ($p_T^\tau$). Cut 6 is only applied, if the event was triggered by a lepton signature. For such events, a $p_T$ cut is applied to the reconstructed lepton according to the trigger threshold.

### 3.3.2 Results

In Table 6 the cross-sections for signal and background after each cut are shown. The first two cuts select events which include a lepton and three jets, assumed to be the two $b$ jets and a $\tau$ jet. These cuts are motivated by the final state under study, but do not reduce the most important leptonic $t\bar{t}$ background. Requiring one jet to be a $\tau$ jet removes more than 95% of the $t\bar{t}$ background while at least one quarter of the signal remains, depending on the charged Higgs boson mass. Only one jet is required to be $b$-tagged, which is motivated by the fact that $m_{H^+} > m_W$ and thus the $b$ quark produced in $t \to bH^+$ is in average softer (the most probable value for $p_T^b$ is about 55 GeV for $t \to bW$, and between 15 GeV for $m_{H^+} = 150$ GeV and 45 GeV for $m_{H^+} = 90$ GeV for $t \to bH^+$). The $b$-tagging efficiency decreases quickly for low $p_T^b$ [5] and thus the number of reconstructed $b$ jets decreases with increasing $m_{H^+}$.

Due to the mass difference mentioned above and the $\tau$ polarizations being different depending on whether the $\tau$ jet originated from an $H^+$ or $W$, the $\tau$-jet is expected to be harder for the signal than the background, motivating a cut on $p_T^\tau$. A harder $\tau$ jet in turn leads to more missing energy, thus making

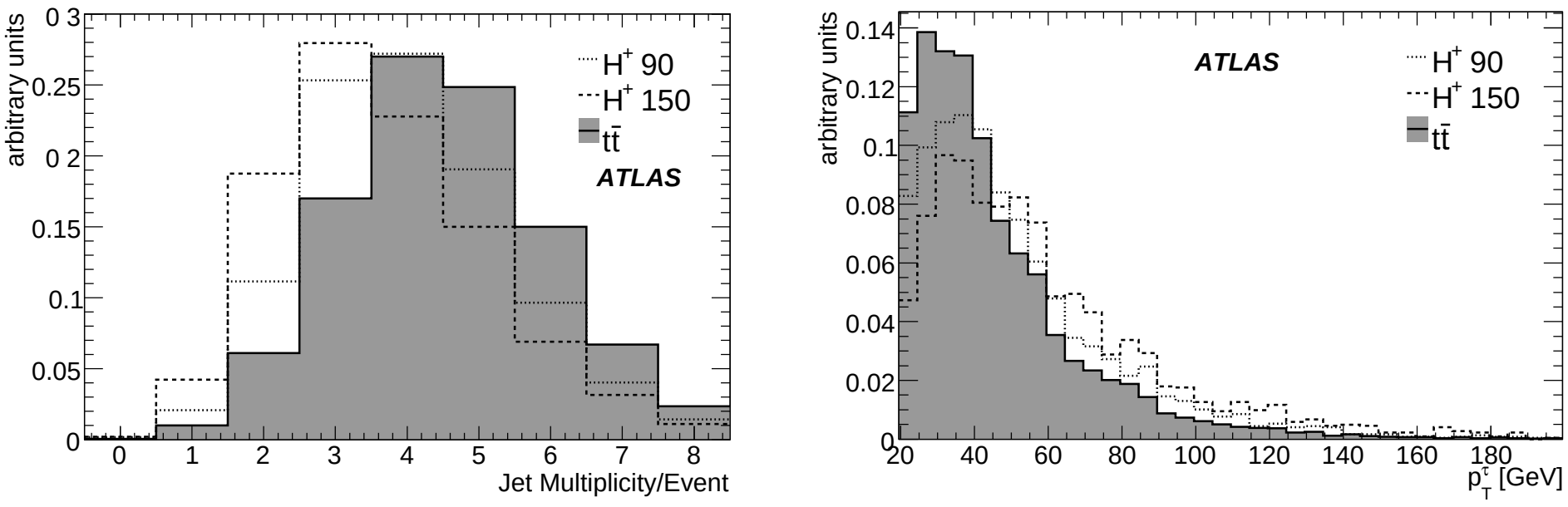


(a) (b)

Figure 7: $t\bar{t} \rightarrow bH^{+}bW \rightarrow b\tau(had)\nu b\ell\nu$: (a) Multiplicity of jets and (b) $\tau$ jet transverse momentum ($p_T^\tau$).

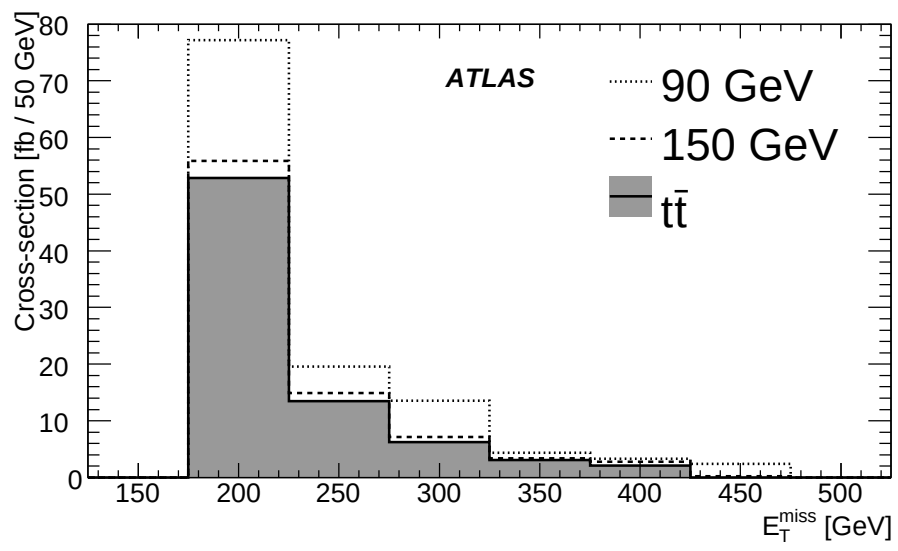


Figure 8: $t\bar{t} \rightarrow bH^{+}bW \rightarrow b\tau(had)\nu b\ell\nu$: $E_T^{miss}$ differential cross-section after all cuts for $\tan\beta = 20$. The signal contributions for the different mass hypothesis are individually stacked on top of the background distribution.

the $E_T^{miss}$ an attractive variable to cut on. A high value is required for $E_T^{miss}$ . This is mainly needed to optimize the significance with respect to the systematic error. Multidimensional optimization attempts have shown that above respective thresholds, the significance is not very sensitive to changes in the values of other continuous selection cuts (such as $p_T^\tau$).

The normalized number of events as a function of the missing transverse energy is shown in Fig. 8. A clear excess is observable for low charged Higgs boson masses, while close to the top quark mass there is only sensitivity for very high values of $\tan\beta$. Figure 9 shows the final discovery and exclusion contours for the channel under investigation. A significant region of the parameter space above $\tan\beta = 30$ is covered by this decay mode, while the sensitivity for low and intermediate $\tan\beta$ is limited.

# 4 Heavy Charged Higgs Boson Searches

In this chapter, charged Higgs boson searches in the two heavy $H^{+}$ ($m_{H^{+}} \gtrsim m_t$) channels selected for investigation are presented:

- $gg/gb \rightarrow t[b]H^{+} \rightarrow bqq[b]\tau(had)\nu$
- $gg/gb \rightarrow t[b]H^{+} \rightarrow t[b]tb \rightarrow bW[b]bWb \rightarrow b\ell\nu[b]bqqb$

The notation $[b]$ implies the additional $b$ given in the production mode $gg \rightarrow tbH^{+}$, which is not produced in the mode $gb \rightarrow tH^{+}$. All plots, tables and results in this section are based on the trigger menu for an

Table 6: $t\bar{t} \to bH^+bW \to b\tau(had)\nu b\ell\nu$: Event selection cut flow. For each sample, the cross-sections after cuts are given in fb and for $\tan\beta = 20$ in the first line and the relative cut efficiencies in the second line (in italics).

| Channel | | | All events | Trigger | $\geq$1 e,$\mu$ | $\geq$3 jets | $\geq 1\tau$ | $\geq$1 b | $\tau$ $p_T$ | $\sum$ q | $E_T^{miss}$ |
|---|---|---|---|---|---|---|---|---|---|---|---|
| $H^+$ | 90 GeV | [fb] | 12098 | 6219 | 4972 | 4248 | 1092 | 929 | 586 | 582 | 44 |
| | | [/] | | *0.51* | *0.80* | *0.85* | *0.26* | *0.85* | *0.63* | *0.99* | *0.08* |
| | 110 GeV | [fb] | 8570 | 4510 | 3534 | 2986 | 772 | 650 | 439 | 431 | 30 |
| | | [/] | | *0.53* | *0.78* | *0.84* | *0.26* | *0.84* | *0.67* | *0.98* | *0.07* |
| | 120 GeV | [fb] | 6737 | 3611 | 2868 | 2440 | 654 | 535 | 360 | 354 | 23 |
| | | [/] | | *0.54* | *0.79* | *0.85* | *0.27* | *0.82* | *0.67* | *0.98* | *0.06* |
| | 130 GeV | [fb] | 4954 | 2670 | 2112 | 1730 | 512 | 399 | 270 | 265 | 20 |
| | | [/] | | *0.54* | *0.79* | *0.82* | *0.30* | *0.78* | *0.67* | *0.98* | *0.07* |
| | 150 GeV | [fb] | 1853 | 1048 | 836 | 626 | 177 | 130 | 94 | 94 | 7 |
| | | [/] | | *0.57* | *0.80* | *0.75* | *0.28* | *0.74* | *0.72* | *1.00* | *0.07* |
| $t\bar{t} \geq 1$ $e/\mu/\tau$ | | [fb] | 452000 | 169612 | 137928 | 122547 | 4760 | 4006 | 1915 | 1730 | 78 |
| | | [/] | | *0.37* | *0.81* | *0.89* | *0.04* | *0.84* | *0.48* | *0.90* | *0.04* |
| single top | | [fb] | 112500 | 30180 | 25065 | 18081 | 271 | 168 | 47 | 38 | 0 |
| | | [/] | | *0.27* | *0.83* | *0.72* | *0.02* | *0.61* | *0.28* | *0.81* | *0.0* |
| W$\to e\nu$+jets | | [fb] | 476012 | 144997 | 114152 | 53060 | 780 | 90 | 40 | 29 | 0 |
| | | [/] | | *0.30* | *0.79* | *0.46* | *0.01* | *0.12* | *0.44* | *0.74* | *0.0* |
| W$\to \mu\nu$+jets | | [fb] | 157800 | 48372 | 43003 | 41493 | 582 | 70 | 40 | 26 | 0 |
| | | [/] | | *0.31* | *0.89* | *0.96* | *0.01* | *0.12* | *0.57* | *0.64* | *0.0* |
| W$\to \tau\nu$+jets | | [fb] | 277755 | 23187 | 9443 | 6920 | 187 | 20 | 12 | 3 | 0 |
| | | [/] | | *0.08* | *0.41* | *0.73* | *0.03* | *0.10* | *0.61* | *0.22* | *0.0* |

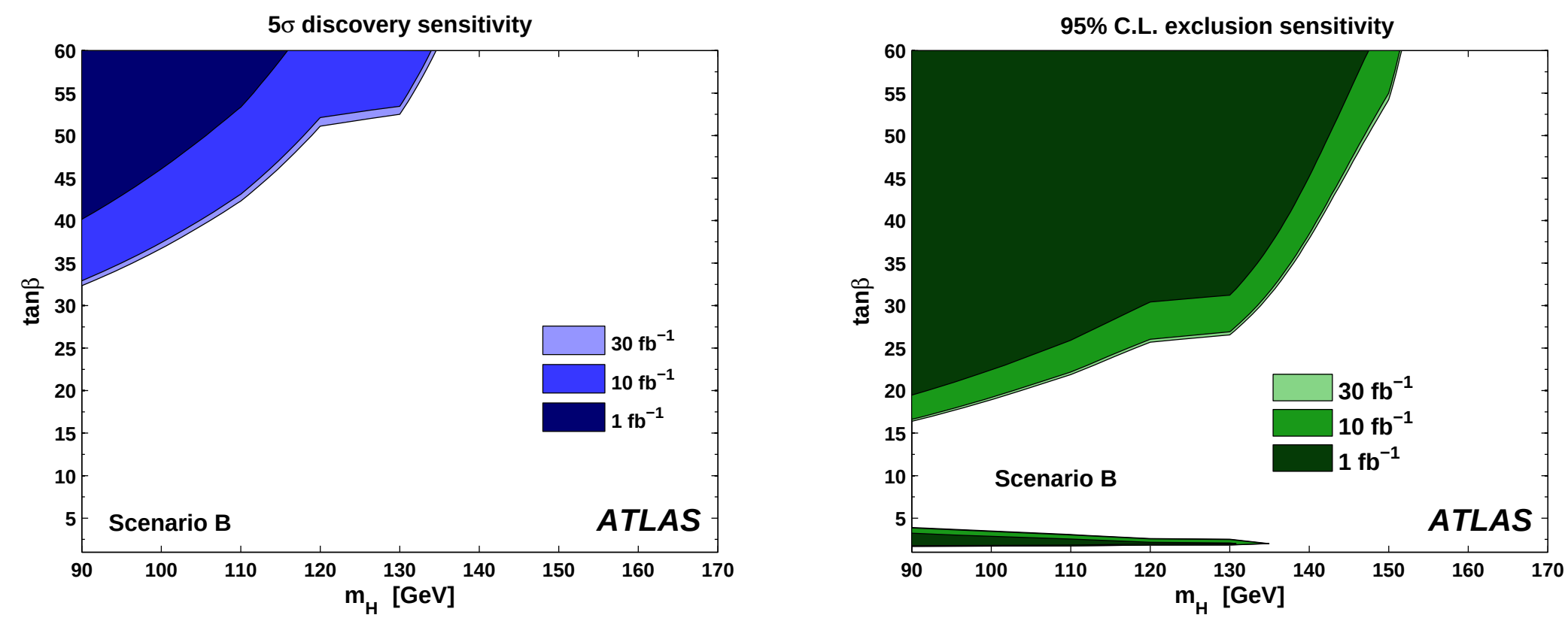


Figure 9: $t\bar{t} \to bH^+bW \to b\tau(had)\nu b\ell\nu$: Discovery (left) and exclusion contour (right) for Scenario B ($m_h$-max) [1]. Systematic and statistical uncertainties are included. The systematic uncertainty is assumed to be 10% for the background, and 41% for the signal (see Sections 5.2 and 5.1). The lines indicate a $5\sigma$ significance for the discovery and a 95% CL for the exclusion contour.

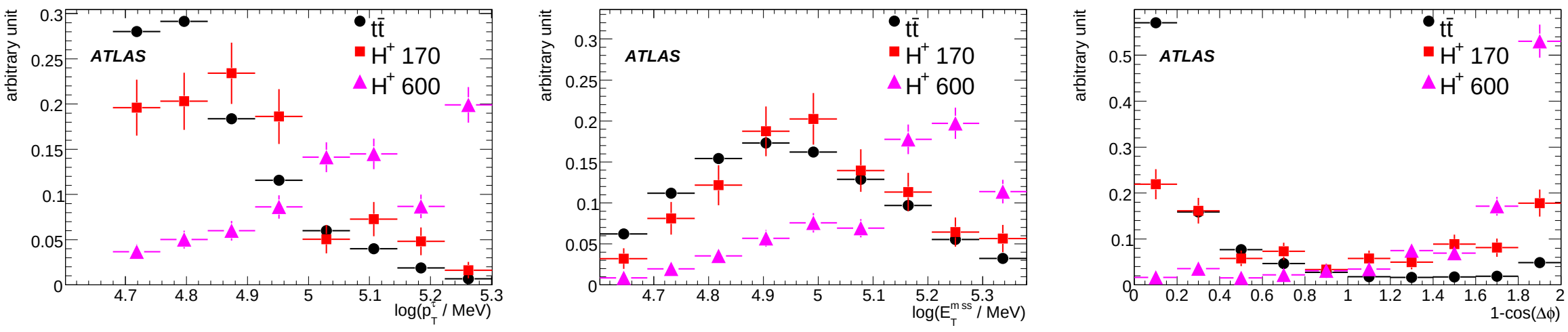


Figure 10: $gg/gb \rightarrow t[b]H^+ \rightarrow bqq[b]\tau(had)\nu$: Probability density functions for $t\bar{t}$ and two signal mass points. The plots show, from the upper left to the lower right, $\log(p_T^\tau)$, $\log(E_T^{miss})$, and $1-\cos(\Delta\phi)$.

instantaneous luminosity of L=$10^{33}$cm$^{-2}$s$^{-1}$ presented in Section 2, on the signal cross-sections for the $m_h$-max MSSM scenario (see Reference [1]) and the background cross-sections given in Reference [8].

## 4.1 $gg/gb \rightarrow t[b]H^+ \rightarrow bqq[b]\tau(had)\nu$

The following section describes the event selection for the $H^+ \rightarrow \tau\nu$ channel for the case $m_{H^+} \gtrsim m_t$. The signal final state is characterised by a hard $\tau$ jet, large missing transverse momentum (due to the neutrino), one or two $b$ jets, two light jets, a W boson and a top quark (the mass of which can be reconstructed) and has previously been investigated in Reference [15].

The main background to this signal channel are $t\bar{t}$ decays, in particular when one of the top quarks decays to a $\tau$ jet, $t \rightarrow b\tau(had)\nu$, and the other one hadronically, $t \rightarrow bqq$. However, other $t\bar{t}$ modes can also contribute when some of the objects in the event are not correctly reconstructed, e.g. a light jet as a $\tau$ jet. Other backgrounds to be considered are single top, W+jets and QCD multi-jet events. A parametrized detector simulation has been used to evaluate the leptonic $(e,\mu,\tau)$ $t\bar{t}$ background. The parametrization has been adjusted (using a smaller full detector simulation $t\bar{t}$ sample) such that the distributions of all the quantities in the events agree with the full detector simulation.

### 4.1.1 Preselection

**Trigger:** The trigger selection employs items of the $H^+$ trigger menu (see Section 2). Events are required to pass at least one of the two following trigger item combinations: `xE40_3j20_L1_TAU30` or `xE50_L1_TAU30`.

**$\tau$ jet reconstruction:** Only $\tau$ jet candidates with transverse momenta greater than 15 GeV with a pseudorapidity outside of the crack region $1.4 < \eta < 1.6$ are considered, and a high cut on the $\tau$ quality variable is placed yielding a $\tau$ reconstruction efficiency of about 20% and leading to a high rejection of parton jets. Exactly one $\tau$ jet is required in the event, followed by a cut of $p_T^\tau > 50$ GeV in order to reduce the QCD background at an early stage.

**Jet reconstruction:** At least three more jets with $p_T > 15$ GeV are required, and exactly one of them has to be $b$-tagged. For this purpose, a $b$-tagging cut is applied to all jets with a $b$-tagging efficiency of about 70%.

**Missing Transverse Energy:** A soft cut of 40 GeV on the transverse missing energy $E_T^{miss}$ is applied to remove most of the QCD background already in the preselection, while it affects the $t\bar{t}$ background only slightly (about 30% of the $t\bar{t}$ events with leptons are removed).

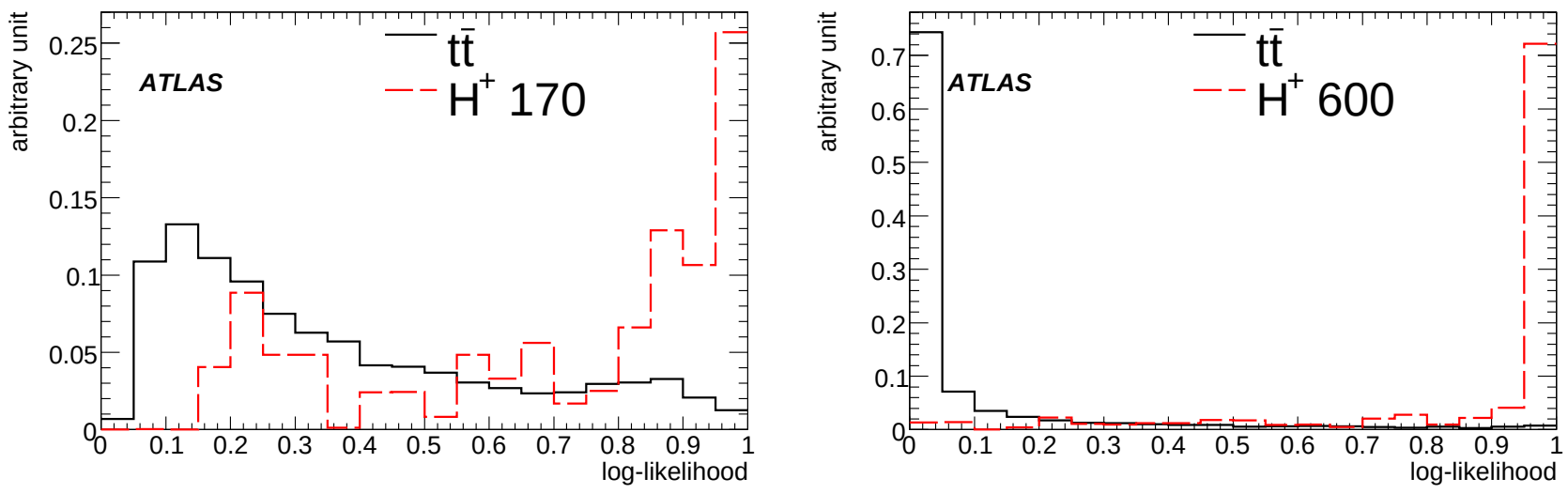


Figure 11: $gg/gb \to t[b]H^+ \to bqq[b]\tau(had)\nu$: Likelihood distributions. The areas are normalized to unity. The background likelihood for 170 GeV does not peak at 0 because of the trigger selection (and to a smaller degree the less stringent preselection cuts for the PDF determination) which already removes most of the events in the first bins.

**W boson and top quark reconstruction:** For further discrimination of all backgrounds without hadronically decaying top quarks, an attempt is made to reconstruct a W boson and a top quark. For all combinations of the $b$ jet with two other jets, the minimum value of

$$\chi^2 = \frac{(m_{jj} - m_W)^2}{\sigma^2_{m_W}} + \frac{(m_{b(jj)^r} - m_t)^2}{\sigma^2_{m_t}} \tag{4}$$

is calculated for each event ($m_W$ and $m_t$ are the nominal $W$ and $t$ masses, $(jj)^r$ are the two jets used in the $W$ reconstruction, rescaled to the $W$ mass, and $\sigma_{m_W}$ and $\sigma_{m_t}$ represent the resolution of the $W$ and $t$ mass reconstruction, 10 GeV and 15 GeV). The resulting $\chi^2$ values are required to be smaller than 3.

**Lepton veto:** To eliminate events with leptons (in particular leptonic $t\bar{t}$ modes), events with at least one isolated lepton ($e$, $\mu$) with $p_T^{\ell} > 7$ GeV are rejected.

The preselection aims at suppressing the reducible background such that only $t\bar{t}$ events involving one hadronic and one semileptonic top quark decay survive. $W$+jets events are successfully removed by requiring a $b$-tagged jet and a reconstructed top quark in the events, single top events by requiring a hard $\tau$ jet and high $E_T^{miss}$ together with a reconstructed hadronically decaying top quark, and QCD events by requiring high $E_T^{miss}$ (and additionally by requiring $b$- and $\tau$-tags and a reconstructed top quark).

### 4.1.2 Likelihood for further reduction of the $t\bar{t}$ background

After the preselection cuts, the background is dominated by $t\bar{t}$ events with one $W$ decaying hadronically, and the other one to a hadronically decaying $\tau$ lepton and a neutrino (75% of the remaining background). An uncorrelated likelihood approach has been chosen to reduce this background, employing the following discriminant variables: (i) $p_T^{\tau}$, (ii) $E_T^{\rm miss}$, (iii) $\Delta\varphi$ (azimuthal angle between the $\tau$ jet and the missing momentum), (iv) $H_{\rm T}$ (scalar sum of the $p_T$ of all jets in the event (excluding $\tau$ jets), and (v) $p_T^{\rm ratio}$ (ratio between the transverse momenta of the $\tau$ jet and the hardest jet not used for the top quark reconstruction).

The probability density functions (PDFs) have been created for each signal mass point and for the $t\bar{t}$ sample (with at least one leptonic W decay) using full detector simulation. Slightly less stringent preselection cuts ($p_T^{\tau}$ >40 GeV instead of 50 GeV, at least 1 b jet instead of exactly 1, and a maximum W/top reconstruction $\chi^2$ of 6 instead of 3) and no trigger selection are used to obtain the PDFs in order to keep a sufficiently high number of simulated events. The PDFs for two mass points and the $t\bar{t}$ background are shown in Fig. 10.

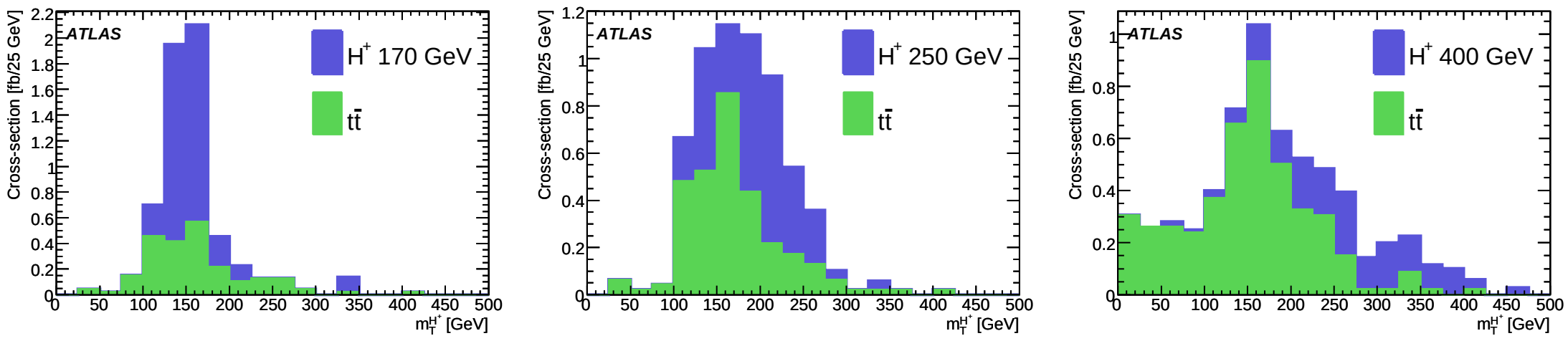


Figure 12: $gg/gb \to t[b]H^+ \to bqq[b]\tau(had)\nu$: $H^+$ transverse mass distributions for three signal mass points and the corresponding background for $\tan\beta = 35$. The signal events are stacked on top of the background events.

The likelihood distributions for two example mass points are shown in Fig. 11. A cut requiring the likelihood to be higher than 0.9 ($m_{H^+} = 400$ and 600 GeV) or 0.95 (other mass cases) is applied and the charged Higgs boson transverse mass of the remaining events is plotted as a final distribution to extract the significance of the $H^+$ signal. These transverse mass distributions are shown in Fig. 12. A clear excess can be observed for lower $H^+$ masses, but the shapes are similar for signal and background. Varying cuts to change the signal-to-background ratio would be a quick way to establish the excess if it was statistically unambiguous, followed by data-driven background estimation methods as described in Section 5.2. For higher $H^+$ masses, sensitivity is only given for higher values of $\tan\beta$, but the signal and background shapes are clearly distinguishable.

**Event selection summary:** The following list summarizes the event selection:

- Trigger: Trigger `xE40_3j20_L1_TAU30` or `xE50_L1_TAU30`
- Cut A: exactly one $\tau$ jet with $p_T^\tau > 50$ GeV, and $E_T^{miss} > 40$ GeV
- Cut B: at least three additional jets
- Cut C: exactly one of the additional jets $b$-tagged
- Cut D: veto on a lepton ($e$, $\mu$) with $p_T^\ell > 7$ GeV
- Cut E: W boson and top quark reconstructed with $\chi^2 < 3$
- Cut F: likelihood value greater than 0.95 (0.9 for $m_{H^+}$ 400 and 600 GeV)
- Cut G: $m_{H^+}$ in a certain mass window

### 4.1.3 Results

The cross-section of signal and background events surviving preselection cuts are shown in Table 7. All backgrounds except for $t\bar{t}$ with at least one leptonic W decay mode are efficiently suppressed at this stage. The QCD dijet selection efficiency is too small to draw definite conclusions due to its high cross-section, but assuming that dijet events in $p_{\mathrm{T}}$ -bins lower than 140 GeV cannot produce a hard $\tau$ jet and three jets with the top quark-invariant mass (plus large missing $E_{\mathrm{T}}$ ), that the numbers presented for Cut B are of the right order of magnitude and that the relative efficiencies for the remaining cuts is smaller than for $t\bar{t}$ events the conclusion can be drawn that the background from dijet events is negligible.

The efficiencies of the remaining Cuts F (likelihood) and G ($m_{H^+}$ ) are shown separately in Table 8; as for these cuts the background efficiencies depend on the $H^+$ mass hypothesis. At this point, all other backgrounds except for $t\bar{t}$ events with $W$ decays to $e$, $\mu$ or $\tau$ are negligible.

A signal-to-background ratio of the order of 1 has been achieved, resulting in robustness with respect to systematic uncertainties. The cross-section for very heavy $H^+ \to \tau\nu$ production is small but a very

Table 7: $gg/gb \to t[b]H^+ \to bqq[b]\tau(had)\nu$: Preselection cut flow. For each sample, the cross-sections after cuts are given in fb and for $\tan\beta = 35$ in the first line and the relative cut efficiencies in the second line (in italics). The size of the QCD dijet sample is too small to draw accurate conclusions.

| Channel | | | All events | Trigger | Cut A | Cut B | Cut C | Cut D | Cut E |
|---|---|---|---|---|---|---|---|---|---|
| $H^+$ | 170 GeV | [fb] | 1346 | 280 | 76 | 67 | 36 | 35 | 14.7 |
| | | [/] | | *0.21* | *0.27* | *0.88* | *0.54* | *0.96* | *0.42* |
| | 200 GeV | [fb] | 551 | 139 | 40 | 33 | 19 | 18 | 7.4 |
| | | [/] | | *0.25* | *0.28* | *0.83* | *0.57* | *0.97* | *0.41* |
| | 250 GeV | [fb] | 184 | 58 | 17 | 15 | 7.8 | 7.6 | 2.9 |
| | | [/] | | *0.32* | *0.30* | *0.84* | *0.54* | *0.97* | *0.38* |
| | 400 GeV | [fb] | 28 | 11 | 3.3 | 2.8 | 1.5 | 1.4 | 0.58 |
| | | [/] | | *0.39* | *0.31* | *0.84* | *0.52* | *0.98* | *0.41* |
| | 600 GeV | [fb] | 4.5 | 1.7 | 0.52 | 0.46 | 0.24 | 0.23 | 0.10 |
| | | [/] | | *0.39* | *0.30* | *0.87* | *0.52* | *0.97* | *0.43* |
| $t\bar{t} \geq 1\ e/\mu/\tau$ | | [fb] | 452000 | 56300 | 1669 | 1532 | 697 | 518 | 188 |
| | | [/] | | *0.12* | *0.03* | *0.92* | *0.46* | *0.74* | *0.36* |
| $t\bar{t}$ hadronic | | [fb] | 381000 | 1746 | 37 | 37 | 16 | 16 | 5 |
| | | [/] | | *0.005* | *0.02* | *1.00* | *0.43* | *1.00* | *0.33* |
| QCD dijet $p_{\mathrm{T}}$ =140-1120 GeV | | [fb] | $3.2 \cdot 10^8$ | $3.2 \cdot 10^5$ | 1285 | 1234 | 106 | 106 | - |
| | | [/] | | *0.001* | *0.004* | *0.96* | *0.09* | *1.00* | - |
| W+jets | | [fb] | 341200 | 19170 | 2518 | 1892 | 314 | 314 | 13 |
| | | [/] | | *0.06* | *0.13* | *0.75* | *0.17* | *1.00* | *0.04* |
| single top | | [fb] | 112500 | 7570 | 161 | 132 | 39 | 37 | 12 |
| | | [/] | | *0.07* | *0.02* | *0.81* | *0.30* | *0.93* | *0.32* |

good discrimination against the background can be obtained with the likelihood method. The resulting discovery contours are presented in Fig. 13 and show that a sizable region of the MSSM space which has not been explored experimentally before can be covered. Sensitivity is given for large $\tan\beta$ and $m_{H^+} < 500$ GeV.

## 4.2 $gg/gb \to t[b]H^+ \to t[b]tb \to bW[b]bWb \to b\ell\nu[b]bqqb$

In this section the analysis strategy for the $H^+ \to tb$ channel is outlined. This is a particularly challenging signal final state which includes 3 (or 4) $b$ quarks, 2 light quarks, 1 high $p_T$ lepton and one neutrino. The channel has previously been studied for ATLAS using fast detector simulation [16], and several aspects of the previous analysis have been adopted. However, the use of full simulation for the present study showed the need to add several new cuts to improve the signal to background ratio.

### 4.2.1 Trigger

Events are required to pass one of the following trigger signatures (see Section 2): `e22i_xE30`, `mu20_xE30`, or `xE40_3j20_L1_TAU30`.

### 4.2.2 Preselection

Following the trigger, the events are required to pass a set of preselection criteria which define the minimum requirements needed for the event reconstruction:

- exactly 1 isolated lepton ($e$ or $\mu$) with $p_T^e > 25$ GeV, $p_T^\mu > 20$ GeV and $|\eta| < 2.5$.

Table 8: $gg/gb \to t[b]H^+ \to bqq[b]\tau(had)\nu$: Event selection results. The cross-sections after cuts are given in fb and for $\tan\beta = 35$ as well as the relative cut efficiencies. The remaining background consists only of $t\bar{t}$ events with at least one W decay to $e$, $\mu$ or $\tau$ plus $\nu$.

| Channel | | Cut | Signal | | Background | |
|---|---|---|---|---|---|---|
| | | | [fb] | [/] | [fb] | [/] |
| $H^+$ | 170 GeV | LH>0.95 | 3.8 | *0.26* | 2.3 | *0.012* |
| | | $m_T^{H^+} > 100$ GeV | 3.8 | *0.99* | 2.1 | *0.91* |
| | 200 GeV | LH>0.95 | 3.1 | *0.42* | 3.2 | *0.017* |
| | | $m_T^{H^+} > 120$ GeV | 2.9 | *0.92* | 2.4 | *0.74* |
| | 250 GeV | LH>0.95 | 1.4 | *0.47* | 3.1 | *0.017* |
| | | $m_T^{H^+} > 150$ GeV | 1.1 | *0.77* | 2.0 | *0.63* |
| | 400 GeV | LH>0.9 | 0.47 | *0.80* | 4.5 | *0.024* |
| | | $m_T^{H^+} > 250$ GeV | 0.26 | *0.56* | 0.33 | *0.074* |
| | 600 GeV | LH>0.9 | 0.075 | *0.76* | 2.5 | *0.013* |
| | | $m_T^{H^+} > 300$ GeV | 0.044 | *0.58* | 0.15 | *0.062* |

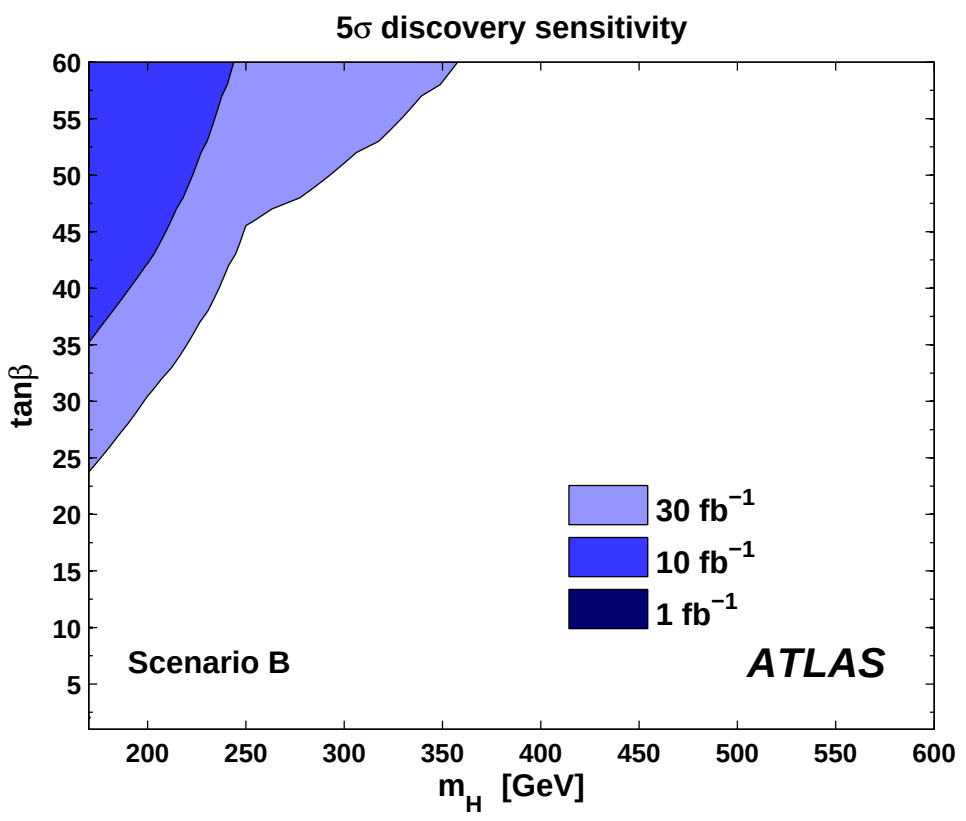


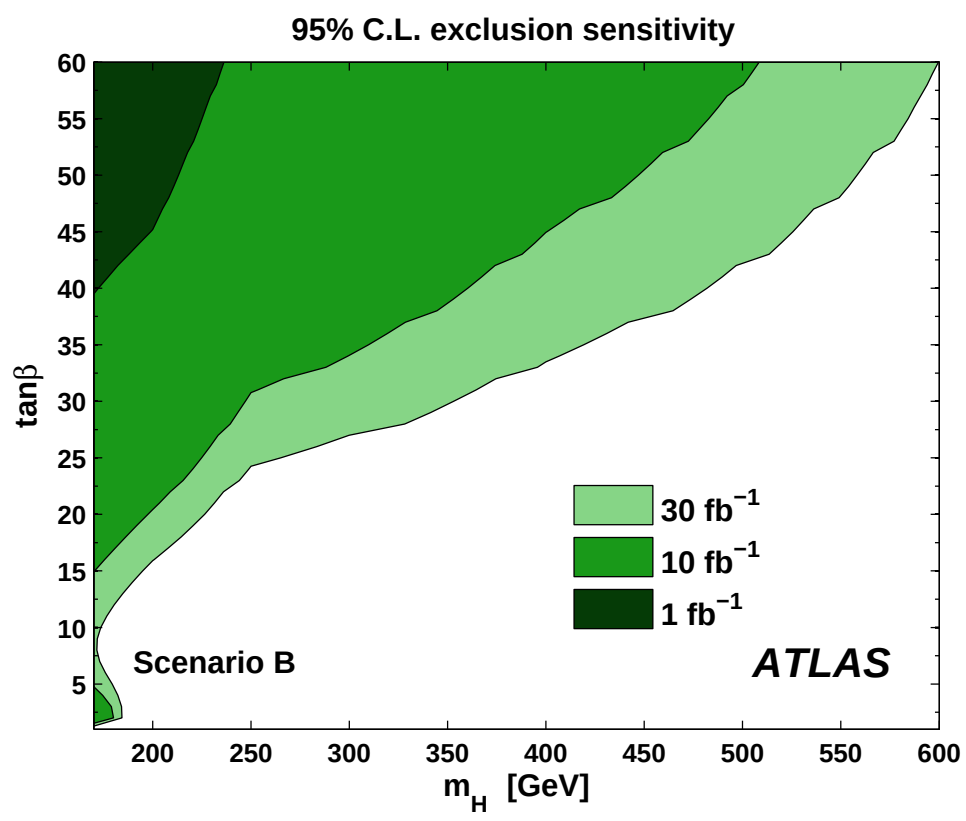


Figure 13: $gg/gb \to t[b]H^+ \to bqq[b]\tau(had)\nu$: Discovery (left) and exclusion contour (right) for Scenario B ($m_h$-max) [1]. Systematic and statistical uncertainties are included. The systematic uncertainty is assumed to be 10% for the background, and 44% for the signal (see Sections 5.2 and 5.1). The lines indicate a $5\sigma$ significance for the discovery and a 95% CL for the exclusion contour.

- at least 5 jets with $p_T > 20$ GeV and $|\eta| < 5$.
- at least 3 $b$-tagged jets with $|\eta| < 2.5$.

**Lepton selection:** Electron candidates that pass the cuts on transverse momentum and pseudorapidity mentioned above are required to pass further identification and isolation criteria based on the shower shape in the electromagnetic calorimeter and the quality of the track in the inner detector. Concerning isolation, the energy in a cone of $\Delta R = 0.20$ around the electron is required to be less than 20% of the electron transverse energy. The same isolation criteria is also applied to muon candidates.

**Jet selection and *b*-tagging criteria:** Jet multiplicity is one of the main sources of combinatorial background and the $b$-tagging efficiency is the major factor reducing the overall efficiency of the preselection. From jets passing the preselection cuts mentioned above, those within the range $|\eta| < 2.5$ are considered

$b$-tagged if passing a $b$-tagging quality cut leading to a $b$ jet reconstruction efficiency of about 60%. Only events with at least 3 $b$-tagged jets are accepted for the rest of the analysis.

#### 4.2.3 Reconstruction

Apart from the objects expected in the event final state, jets from the underlying event are also present leading to an increase of the jet multiplicity.

**Reconstruction of the leptonic $W$:** After the preselection is performed, all physics objects necessary for the complete reconstruction of the event are present except for the neutrino coming from the leptonically decaying $W$. In order to reconstruct the four-momentum of the neutrino, its transverse component is identified with the missing transverse momentum, and the longitudinal component is computed using the $W$ mass constraint, leading to 0, 1 or 2 real solutions. In about 25% of the cases no real solution can be found. In order to recover those events, the approximation of neglecting the imaginary part of the solution is applied and has been found to have only a small effect on the top quark mass resolution.

**The Combinatorial Likelihood:** The next step is to associate the reconstructed physics objects with objects of the event, assuming a signal event topology. The number of possible combinations is very large and depends strongly on the number of light jets in the event. In order to overcome this combinatorial background, a likelihood function is defined. The likelihood formalism used in this analysis is based on $m$ variables and should discriminate between $n$ classes of events. For each of the $m$ variables, first the probability density functions $f_i^j(x_i)$ for each of the $n$ classes are determined: Then the probability for an event to be of class $j$ when the value $x_i$ is measured for variable $i$ is given by:

$$p_i^j(x_i) = \frac{f_i^j(x_i)}{\sum_{k=1}^{n} f_i^k(x_i)} \tag{5}$$

The information about all $m$ variables are combined, ignoring correlations, to define the likelihood $\mathscr{L}_j$ that an event belongs to class $j$ when measuring the values $x_i$ for variables $i = 1, ..., m$:

$$\mathscr{L}_j = \frac{\prod_{l=1}^{m} p_l^j(x_l)}{\sum_{k=1}^{n} \prod_{l=1}^{m} p_l^k(x_l)} \tag{6}$$

A combinatorial likelihood is used to discriminate between the two classes of events: the correct and the wrong combinations. The likelihood is based on 8 variables:

- $m_{jj}$: The invariant mass of two light jets.
- $m_{jjb}$: The invariant mass of two light jets and one $b$ jet.
- $m_{\ell\nu b}$: The invariant mass of the lepton, one of the two solutions of the neutrino and one $b$ jet.
- $p_T(b_H)$: The transverse momentum of the $b$ jet associated to the charged Higgs boson decay.
- $\Delta R(j,j)$: The distance in the azimuthal-pseudorapidity plane ($\Delta R = \sqrt{\Delta\phi^2 + \Delta\eta^2}$) between two light jets.
- $\Delta R(jj,b)$: $\Delta R$ between the sum of two light jets and one $b$ jet.
- $\Delta R(\ell,b)$: $\Delta R$ between the isolated lepton and one $b$ jet.
- $\Delta R(b_H,t_H)$: $\Delta R$ between the $b$ jet and the top quark associated to the charged Higgs boson decay.

The likelihood is computed for each combination, and the combination with the highest likelihood in the event is chosen. If the maximum likelihood in the event is found to be less than 0.7, the event is rejected.

#### 4.2.4 Final Event Selection

After the event reconstruction is complete the physical backgrounds must be suppressed, of which the largest is $t\bar{t} + jets$. In order to reduce this background requiring 4 $b$ jets in the event is found to be crucial. The events passing the 4 $b$ jets requirements are then passed to a likelihood cut. The likelihood function used is based on 5 variables:

- $\eta_{b_H}$: The pseudorapidity of the $b$ jet associated to the charged Higgs boson decay.
- $\sum_b w_b$: The sum of the $b$-tagging weights of the 3 $b$ jets associated to the top quark and charged Higgs boson decay.
- $< \mathscr{L} >$: The average combinatorial likelihood in the event.
- $\Delta R(b_H, b_t b_t)$: The distance in the $(\phi, \eta)$ plane between the $b$ jet associated to the charged Higgs boson decay and the system of the two $b$ jets associated to the top quark decays.
- $p_T^{b_1}/p_T^{b_2}$: The $p_T$ ratio of the two $b$ jets not associated to the top quark decay, $b_1$ being the one with the lowest $p_T$.

A cut on the output of this likelihood is applied. Its value is optimized to maximize the charged Higgs boson signal significance. This final step improves the significance by only 10 to 15%. The limited number of simulated events at this stage of the analysis, especially for the backgrounds, made it very difficult to optimize the choice of variables to include in the likelihood. Larger background samples will be needed to be able to define a more performant likelihood as it was done in Reference [16] with parametrized detector simulation.

#### 4.2.5 Results

Table 9 shows the selection cut flow after the different steps of the analysis for all simulated signal masses, and for the background for one $H^+$ mass hypothesis. Table 10 presents the results for all simulated $H^+$ hypotheses. In Fig. 14, the reconstructed charged Higgs boson mass for signal and physics background is shown. Since the limited number of simulated events does not allow the construction of a performing final selection likelihood, currently no $H^+$ discovery or exclusion power can be extracted from this channel on its own and thus no contours are shown. It, however, contributes to the combined $H^+$ sensitivity.

# 5 Systematic Uncertainties and Background Extraction From Data

The observation of a charged Higgs boson signal will be subject to statistical and systematic uncertainties. The systematic uncertainties stem from two sources: theoretical and experimental, and both are discussed in Section 5.1. Section 5.2 addresses how to extract the dominating $t\bar{t}$ background from real data using a novel technique with so-called control samples.

## 5.1 Systematic Uncertainties

#### 5.1.1 Theoretical Systematic Uncertainties

Uncertainties in the expected production cross-sections for background and signal processes affect the discovery/exclusion potential of the channels under investigation. For all channels the uncertainty of the $t\bar{t}$ background is particularly interesting since this is the dominant background. A 12% uncertainty on

Table 9: $gg/gb \to t[b]H^+ \to t[b]tb \to bW[b]bWb \to b\ell\nu[b]bqqb$: Selection cut flow. The cross-sections in fb are given after each cut and for $\tan\beta = 35$ as well as the relative cut efficiencies. The backgrounds are shown for a charged Higgs mass boson hypothesis of $m_{H^+} = 250$ GeV.

| Channel | | All events | Trigger | Preselection | Reconstruction | 4 b tags | Selection |
|---|---|---|---|---|---|---|---|
| 200 GeV | [fb] | 105 | 64 | 2.3 | 2.2 | 0.18 | 0.16 |
| | [/] | | *0.61* | *0.036* | *0.93* | *0.08* | *0.89* |
| 250 GeV | [fb] | 170 | 108 | 8.1 | 7.3 | 1.06 | 0.65 |
| | [/] | | *0.63* | *0.075* | *0.88* | *0.11* | *0.61* |
| 400 GeV | [fb] | 65 | 45 | 4.5 | 3.9 | 0.43 | 0.26 |
| | [/] | | *0.69* | *0.10* | *0.88* | *0.11* | *0.59* |
| 600 GeV | [fb] | 22 | 16 | 1.8 | 1.7 | 0.27 | 0.18 |
| | [/] | | *0.75* | *0.12* | *0.92* | *0.16* | *0.67* |
| $t\bar{t}$ + jets | [fb] | 112000 | 74400 | 1040 | 875 | 35.0 | 11.7 |
| | [/] | | *0.66* | *0.014* | *0.84* | *0.04* | *0.33* |
| $t\bar{t}\ b\bar{b}$ (QCD) | [fb] | 2240 | 1575 | 130 | 117.8 | 17.6 | 6.8 |
| | [/] | | *0.70* | *0.083* | *0.90* | *0.15* | *0.38* |
| $t\bar{t}\ b\bar{b}$ (EW) | [fb] | 244 | 155 | 14.4 | 13.0 | 2.0 | 0.39 |
| | [/] | | *0.63* | *0.09* | *0.90* | *0.15* | *0.19* |

Table 10: $gg/gb \to t[b]H^+ \to t[b]tb \to bW[b]bWb \to b\ell\nu[b]bqqb$: Event selection results. The cross-sections after all cuts are given in fb and for $\tan\beta = 35$ as well as the global selection efficiencies.

| Channel | Signal | | $t\bar{t}$ + jets | | $t\bar{t}\ b\bar{b}$ (QCD) | | $t\bar{t}\ b\bar{b}$ (EW) | |
|---|---|---|---|---|---|---|---|---|
| | [fb] | [/] | [fb] | [/] | [fb] | [/] | [fb] | [/] |
| 200 GeV | 0.16 | *0.0015* | 21.0 | *0.00019* | 12.2 | *0.0054* | 1.05 | *0.0043* |
| 250 GeV | 0.65 | *0.0038* | 11.7 | *0.00010* | 6.76 | *0.0030* | 0.74 | *0.0030* |
| 400 GeV | 0.26 | *0.0039* | 15.9 | *0.00014* | 8.91 | *0.0040* | 1.05 | *0.0043* |
| 600 GeV | 0.18 | *0.0084* | 19.8 | *0.00018* | 10.5 | *0.0047* | 1.21 | *0.0049* |

the NLL calculations is expected[3] leading to $\sigma_{t\bar{t}} = 833 \pm 100$ pb [17]. Other backgrounds considered have similar or smaller uncertainties.

The branching ratios BR($t \to H^+b$) and BR($H^+ \to \tau\nu, cs, tb$) have been determined with the Feyn-Higgs package, and similar systematic uncertainties apply [18]:

- $\Delta$BR$(t \to H^+b)/$BR $< 10\%$
- $\Delta$BR$(H^+ \to \tau\nu)/$BR $< 5\%$
- $\Delta$BR$(H^+ \to cs, tb)/$BR $< 10\%$

In the high-mass region, the dominant systematic uncertainties on the charged Higgs boson production cross-section stem from the renormalization scale and factorization scale dependence and are calculated to be smaller than 20% in the whole MSSM space. The decrease of the cross-section due to supersymmetry loop corrections has been taken into account by adjusting the cross-sections with an additional

[3]The $t\bar{t}$ cross-section will be measured in early LHC studies and transform this theoretical uncertainty into a much smaller experimental uncertainty, despite possible $H^+$ effects in the measurement.

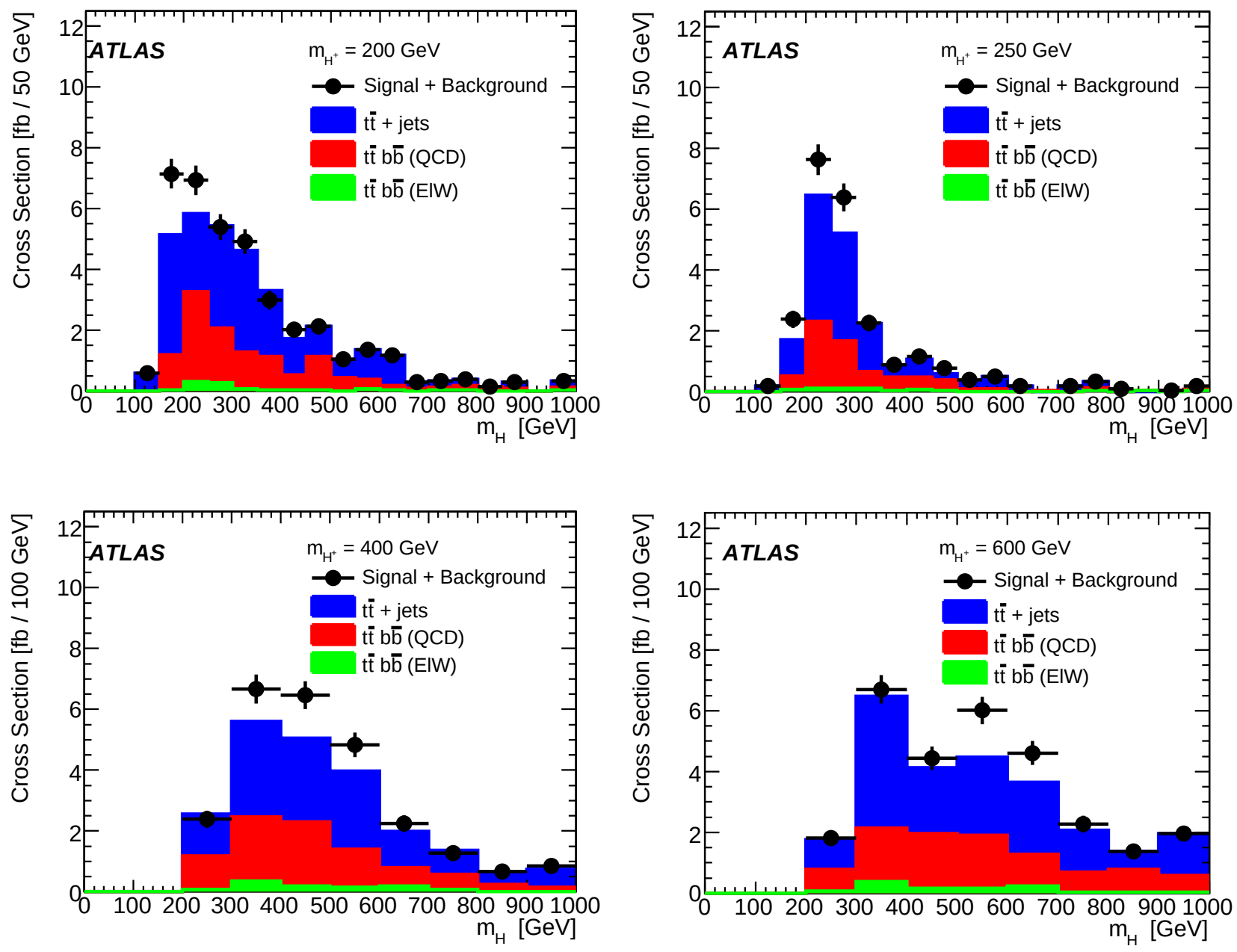


Figure 14: $gg/gb \to t[b]H^+ \to t[b]tb \to bW[b]bWb \to b\ell\nu[b]bqqb$: Reconstructed $H^+$ mass. The value of $\tan\beta$ has been chosen such that the pure statistical significance results in a value of 5.

factor as proposed in Reference [19], reflecting the altered relation between the bottom quark mass and its Yukawa coupling, $\Delta m_b$. Remaining supersymmetry loop corrections are shown to be negligible.

### 5.1.2 Experimental Systematic Uncertainties

Several quantities are subject to experimental systematic uncertainties: Tagging and reconstruction efficiencies, energy scales, energy resolutions and the luminosity determination. A detailed list of the systematical uncertainties considered, including their numerical value, is given in Table 11. Each systematic effect has been evaluated individually using the given uncertainty on an event-by-event basis.

The systematic uncertainty of the missing transverse energy is indirectly considered by taking the effects of the other systematic effects in the missing transverse momentum calculation into account. The systematic uncertainty estimates are generally conservative, in particular for the results assuming an integrated luminosity of 30 fb$^{-1}$. However, due to the usage of control samples to estimate the $t\bar{t}$ background, the uncertainty values have a negligible impact on the $H^+$ discovery sensitivity. Furthermore, tests have shown that even for the exclusion sensitivity the effect is very small: Assuming that the total experimental systematic uncertainty on the results could be halved, the change in sensitivity would only be of the order of 0.1 in $\tan\beta$ for fixed values of $m_{H^+}$.

The dominant systematic uncertainty for all $H^+$ channels is the jet energy scale, with values between 10% and 30%. Similarly, channels with hadronic $\tau$ decays are strongly affected by the $\tau$ jet energy scale. The channel $H^+ \to tb$, requiring 4 $b$-tags, is strongly affected by uncertainties in $b$-tagging efficiency and rejection of light jets. The total experimental systematic uncertainty for the different $H^+$ channels is between about 15% and 40% for the signal, affecting mainly the exclusion sensitivity. Similar values apply for the main background, $t\bar{t}$, which would remove most of the discovery potential. Thus a technique for a data-driven estimation of the background has been developed, greatly reducing the systematic uncertainty on the background. These "$t\bar{t}$ control samples" are discussed in the following section.

Table 11: Effects of systematic uncertainties for all channels under investigation. The numbers are given in terms of percentage changes in cross-section. The channels are: 1: $t\bar{t} \to bH^+bW \to b\tau(had)\nu bqq$ (see Section 3.1), 2: $t\bar{t} \to bH^+bW \to b\tau(lep)\nu bqq$ (see Section 3.2), 3: $t\bar{t} \to bH^+bW \to b\tau(had)\nu b\ell\nu$ (see Section 3.3), 4: $gg/gb \to t[b]H^+ \to bqq[b]\tau(had)\nu$ (see Section 4.1) and 5: $gg/gb \to t[b]H^+ \to t[b]tb \to bW[b]bWb \to b\ell\nu[b]bqqb$ (see Section 4.2).

| Uncertainty | Value | 1 | | 2 | | 3 | | 4 | | 5 | |
|---|---|---|---|---|---|---|---|---|---|---|---|
| | | S | B | S | B | S | B | S | B | S | B |
| $\tau$ E Resolution | $0.45 \times \sqrt{E}$ | -2 | +3 | - | - | +8 | -3 | -4 | -1 | - | - |
| $\tau$ E Scale | $-5\%$ | -2 | +5 | - | - | 0 | -9 | -15 | -21 | - | - |
| | $+5\%$ | -5 | -5 | - | - | +8 | +1 | +4 | +28 | - | - |
| $\tau$-tag Efficiency | $\pm 5\%$ | -5 | -2 | - | - | -8 | -1 | -8 | -5 | - | - |
| Jet E Resolution | $0.45\sqrt{E}, \lvert\eta\rvert < 3.2$<br>$0.63\sqrt{E}, \lvert\eta\rvert > 3.2$ | -2 | -3 | -8 | +5 | +8 | +3 | -12 | -3 | -2 | -4 |
| Jet E Scale | +7(15)%,$\lvert\eta\rvert < (>)3.2$ | -9 | +12 | +29 | +22 | +35 | +19 | +4 | -18 | +9 | +8 |
| | -7(15)%,$\lvert\eta\rvert < (>)3.2$ | -5 | -5 | -21 | -12 | -19 | -17 | -31 | +15 | -8 | -6 |
| $b$-tag Efficiency | $\pm 5\% \varepsilon_{btag}$ | 0 | -14 | +4 | -6 | 0 | -3 | -7 | +3 | -8 | -10 |
| $b$-tag Rejection | -10% | -7 | +10 | 0 | +1 | 0 | 0 | -2 | -3 | -4 | +6 |
| | +10% | +7 | -2 | 0 | 0 | 0 | -1 | -3 | -1 | 0 | -5 |
| $\mu$ E Resolution | $0.011/P_T \oplus 0.00017$ | 0 | 0 | -4 | +1 | 0 | +1 | 0 | 0 | -4 | -5 |
| $\mu$ E Scale | -1% | 0 | 0 | 0 | +1 | +4 | -1 | 0 | 0 | -4 | -6 |
| | +1% | 0 | 0 | -4 | -1 | 0 | 0 | 0 | 0 | +4 | +7 |
| $\mu$ Efficiency | $\pm 1\%$ | 0 | 0 | 0 | -1 | 0 | 0 | 0 | -2 | -2 | -1 |
| $e$ E Resolution | $0.0073 \times E_T$ | 0 | 0 | 0 | 0 | 0 | -1 | 0 | 0 | -4 | -4 |
| $e$ E Scale | -0.5% | 0 | 0 | 0 | +1 | 0 | -1 | 0 | 0 | -4 | -5 |
| | +0.5% | 0 | 0 | 0 | -1 | +4 | -1 | 0 | 0 | +4 | +6 |
| $e$ Efficiency | $\pm 0.2\%$ | 0 | 0 | 0 | 0 | 0 | 0 | 0 | 0 | 0 | -1 |
| Luminosity | -3% | -3 | -3 | -3 | -3 | -3 | -3 | -3 | -3 | -3 | -3 |
| | +3% | +3 | +3 | +3 | +3 | +3 | +3 | +3 | +3 | +3 | +3 |

## 5.2 $t\bar{t}$ Control Samples

The $t\bar{t}$ process, in particular with one or more $\tau$ leptons in the final state, is the dominant background to all analyses presented in this note. As the relative contributions from this background in different jet multiplicities are not known, the subtraction of these backgrounds using a data-driven method is necessary.

The method is data-driven in the sense that it uses $t\bar{t} \to WbWb \to \mu\nu b\mu\nu b$ and $t\bar{t} \to WbWb \to \mu\nu bqqb$ events collected by ATLAS to model the $t\bar{t}$ backgrounds with one or more taus in the final state. After applying a minimal set of event selection criteria on the data (optimized for both efficiency and purity), one or two leptons from the events are removed and the 4 momenta of the removed objects are scaled into tau leptons with corrections for the mass. The tau leptons are fed into TAUOLA [20] for decay and the decay products are passed to the ATLAS detector simulation and reconstruction software. Finally the result is merged with the original event from which the leptons were removed to constitute a control sample.

The result is a data-driven control sample for each of the $t\bar{t}$ final states which potentially constitute a background to one of the $H^+$ analyses, using events from data that can be easily and efficiently triggered. With this method both the shape and the normalization of these backgrounds for all $\tau$ decays (i.e., leptonic and hadronic decays, or in the case of analyses requiring two taus the lepton-hadron and hadron-hadron modes) can be modelled.

### 5.2.1 Obtaining $t\bar{t}$ Control Samples from Data

To extract the $t\bar{t}$ control samples from data a set of selection criteria is applied. These are designed to optimize the efficiency and purity of the samples.

**Dimuonic channel** In order to extract the $t\bar{t} \to WbWb \to \mu\nu b\mu\nu b$ sample, events with at least two isolated muons with $p_{\mathrm{T}} > 20$ GeV are selected. To reject muons coming from Z decays, events with a dimuonic invariant mass in the range 70-110 GeV are rejected. A further requirement is that the missing transverse energy in the event is larger than 40 GeV.

Table 12 summarizes the result of the above selection. The efficiency of the signal (dimuonic $t\bar{t}$ events) to survive this selection is 28% and the sample purity is estimated to be 71%.

Table 12: Efficiency and purity for collecting $t\bar{t}$ dimuonic events. Each $mu20$ indicates a generator level cut requiring one muon with $p_T > 20$ GeV.

| Process | cross-section [fb] | efficiency | events [fb-1] |
|---|---|---|---|
| $t\bar{t}$ signal | 9310 | 0.284 | 2641 |
| $t\bar{t}$ background | 823690 | $4.96 \cdot 10^{-4}$ | 407 |
| W+Jets | 202400 | $1.61 \cdot 10^{-4}$ | 33 |
| Z+Jets | 210290 | $1.45 \cdot 10^{-3}$ | 305 |
| $b\bar{b}(mu20mu20)$ | 261000 | $1.23 \cdot 10^{-3}$ | 322 |
| Total background | - | - | 1067 |

**$\mu$+jet channel** The selection criteria for the $t\bar{t} \to WbWb \to \mu\nu bqqb$ sample are designed to reject $b\bar{b} \to 1\mu + X$ events which have a large cross-section. Events are accepted if an isolated muon is found and two jets in the event with transverse momenta above 40 GeV have an invariant mass within 20 GeV of the nominal W mass. Events with high $p_T$ muons in the jets are rejected. A missing transverse energy cut of 40 GeV is applied, as well as a requirements of least two more jets with transverse momenta above 40 GeV. At least one of these jets is required to be a $b$-tagged. The overall transverse energy of the event is required to be larger than 250 GeV and events with a high $p_T$ isolated electron are rejected.

The results of the above selection are summarized in Table 13. The selection efficiency for signal events is 8.6% and the signal purity is 74%.

Table 13: Efficiency and purity for collecting $t\bar{t}$ muon+jets events.

| Process | cross-section [fb] | efficiency | events [fb-1] |
|---|---|---|---|
| $t\bar{t}$ signal | 119040 | $8.62 \cdot 10^{-2}$ | 10263 |
| $t\bar{t}$ background | 713960 | $1.80 \cdot 10^{-3}$ | 1287 |
| W+Jets | 202400 | $5.61 \cdot 10^{-3}$ | 1134 |
| Z+Jets | 210290 | $2.84 \cdot 10^{-4}$ | 74 |
| $b\bar{b}(mu20)$ | 13600000 | $8.40 \cdot 10^{-5}$ | 1147 |
| Total background | - | - | 3642 |

### 5.2.2 Method Validation

To thoroughly test the $t\bar{t}$ control sample method, the produced events have to be run through the different analyses and the obtained shapes and normalizations have to be compared to the ones obtained in these analyses. However, a global check can be done by comparing various distributions from the two cases (a) "real" $t\bar{t}$ events and (b) "scaled" $t\bar{t}$ control sample events where muons have been replaced by $\tau$ leptons. This has been done separately for three final states of interest.

Basic preselection cuts on quantities like transverse momenta have been applied, lower than or at most equal to the preselection cuts applied in the analyses which use the quantities plotted in the following. The MC@NLO event weights have not been taken into account in order to increase the available statistics since this has been shown not to bias the comparison.

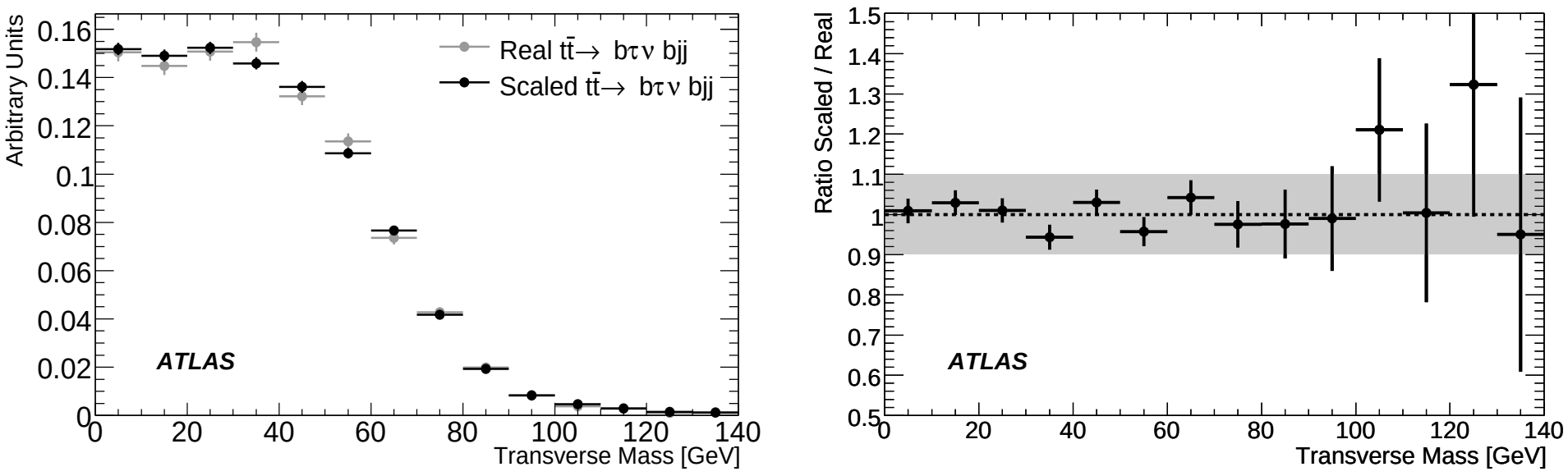


Figure 15: $t\bar{t} \to b\tau(lep)\nu bqq$: Left: $W \to \tau(lep)\nu$ transverse mass, both for the real and the scaled $t\bar{t}$ events. Right: The corresponding bin-by-bin ratio. The gray band represents $\pm 10\%$ around a ratio of 1.

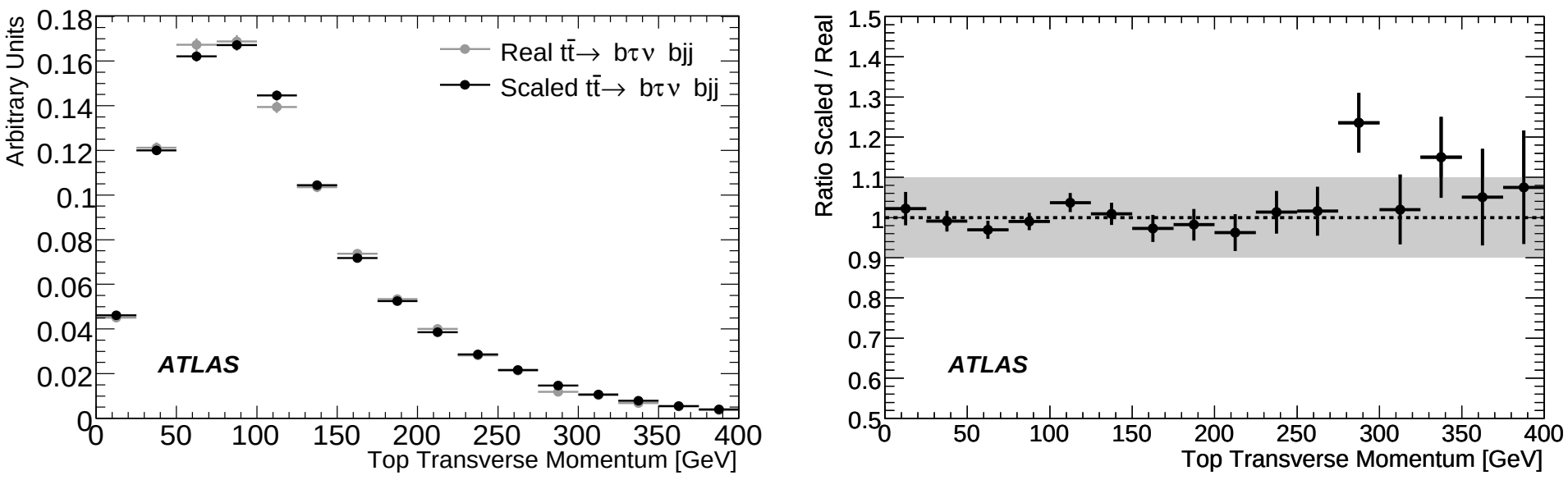


Figure 16: $t\bar{t} \to b\tau(had)\nu bqq$: Left: $t \to b\tau(had)\nu$ momentum, both for the real and the scaled $t\bar{t}$ events. Right: The corresponding bin-by-bin ratio. The gray band represents $\pm 10\%$ around a ratio of 1.

In Fig. 15, the $W \to \tau(lep)\nu$ transverse mass is shown for the $t\bar{t} \to b\tau(lep)\nu bqq$ mode. Here, for the scaled events, a muon has been replaced by a leptonically decaying $\tau$. For the modes $t\bar{t} \to b\tau(had)\nu bqq$ and $t\bar{t} \to b\tau(had)\nu b\ell\nu$, in the scaled events a muon has been replaced by a hadronically decaying $\tau$. The $t \to b\tau(had)\nu$ momentum is shown in Fig. 16, demonstrating the success of this replacement.

The presented plots demonstrate that the dominant background of all $H^+$ studies, $t\bar{t}$, can be modelled with the $t\bar{t}$ control sample method. Even without the intended further refinement of the method, in the regions of interest quantities of the $H^+$ analyses can be modelled within a 10% error margin. This is remarkable in particular for complex quantities, i.e. variables extracted from the combination of several objects (like the top quark mass), and gives confidence that the $t\bar{t}$ control sample method allows to reproduce the relevant correlations in the event. Thus for all results a systematic $t\bar{t}$ background uncertainty of 10% is assumed (while the signal systematic uncertainty is extracted from Monte Carlo events, see Section 5.1).

# 6 Combined Results

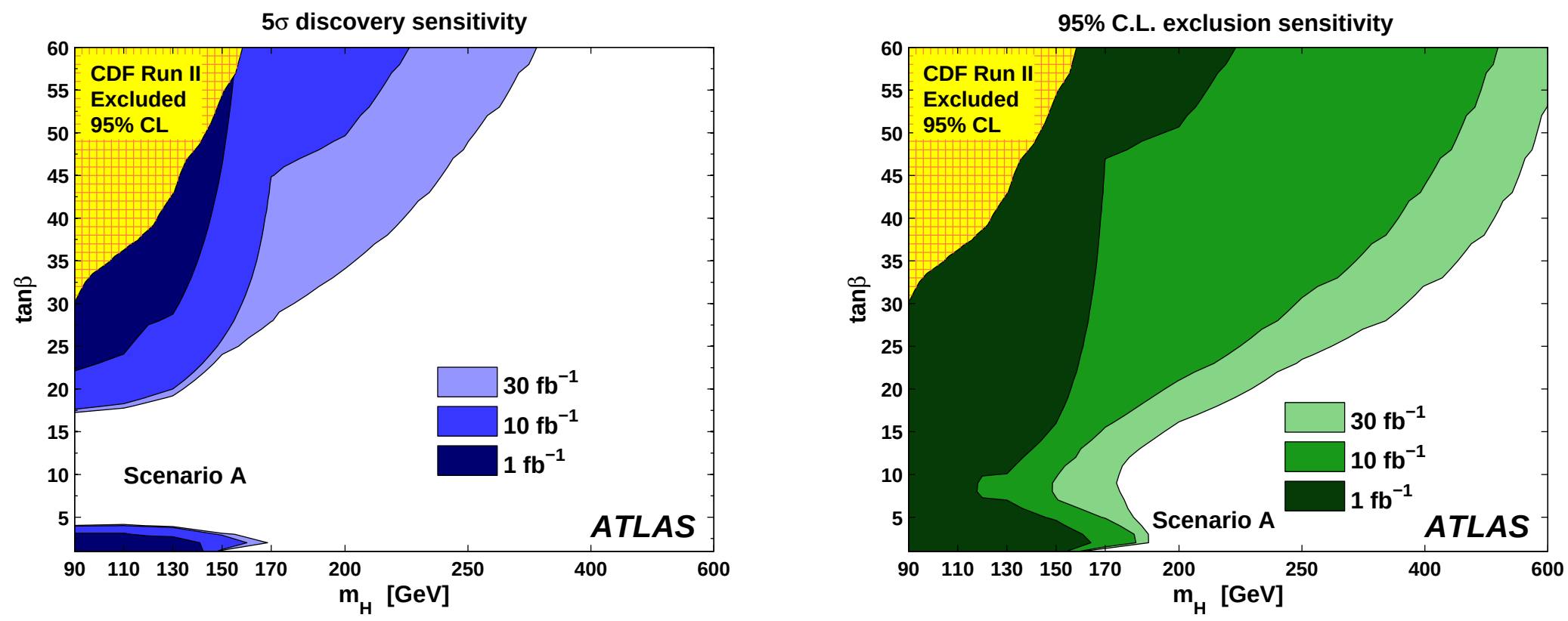


Figure 17: Scenario A: Combined Results. Left: Discovery contour, Right: Exclusion contour. Systematic and statistical uncertainties are included.

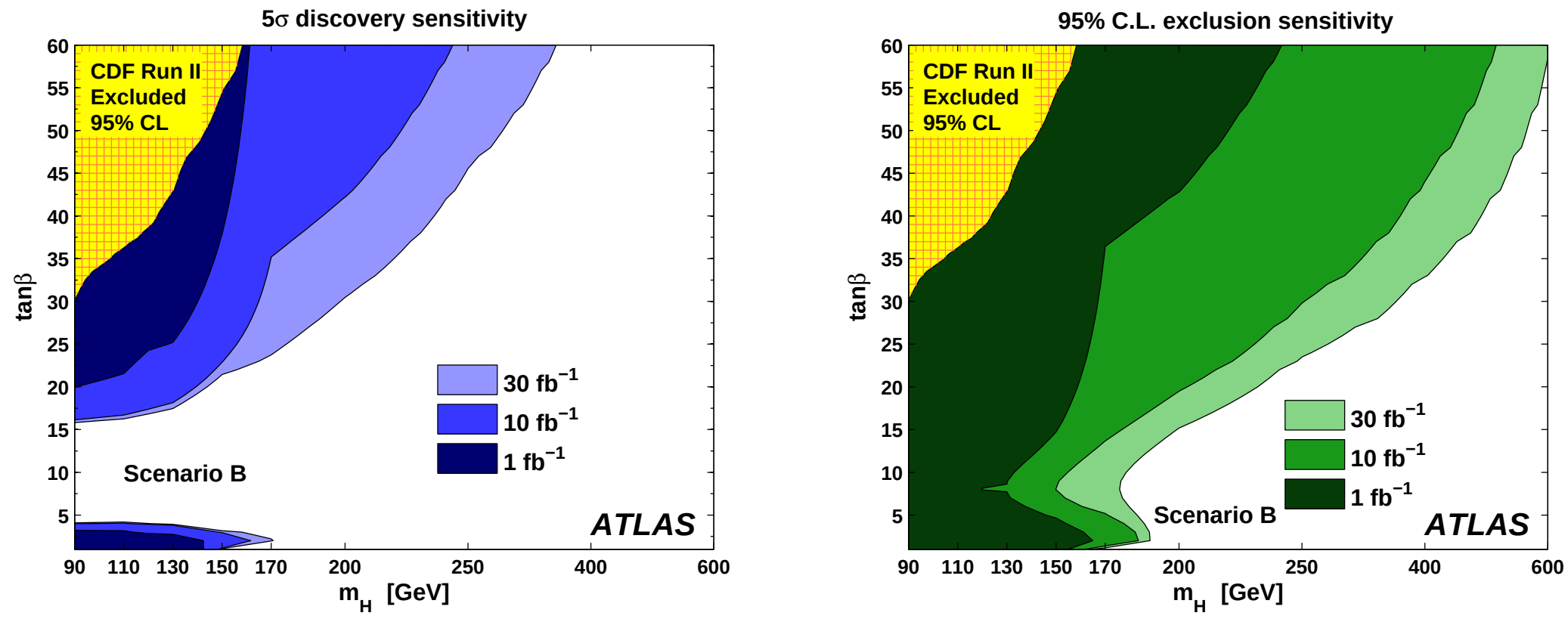


Figure 18: Scenario B ($m_h$-max): Combined Results. Left: Discovery contour, Right: Exclusion contour. Systematic and statistical uncertainties are included.

The sensitivities for discovery and exclusion are calculated with the Profile Likelihood method [21–23], which includes statistical and systematic uncertainties. The results are summarized in combined discovery and exclusion contours for all $H^+$ channels for two MSSM Scenarios A and B [1], and three different integrated luminosities (1, 10, and 30 fb$^{-1}$). Figure 17 shows the result for the MSSM Scenario A. 5$\sigma$ discovery contours and 95% CL exclusion contours are shown. Figure 18 shows the same results for the MSSM Scenario B ($m_h$-max). Previous studies have shown that the dependence of the $H^+$ discovery sensitivity on the specific choice of the MSSM parameter values is generally very small, with the exception of the Higgsino mixing parameter $\mu$ [24, 25]. The discovery significance is calculated for both cases assuming a systematic background uncertainty of 10% following the study of $t\bar{t}$ control samples (Section 5.2). The signal systematic uncertainties are discussed in Section 5.1. The statistical uncertainties arising from the use of Monte Carlo samples with a finite number of events have been consistently taken into account.

A discovery sensitivity is given for a large part of the $m_{H^+}$ -tan$\beta$ space for both scenarios, but the difficult intermediate tan$\beta$ region, where the $H^+$ cross-section has its minimum, is not covered.

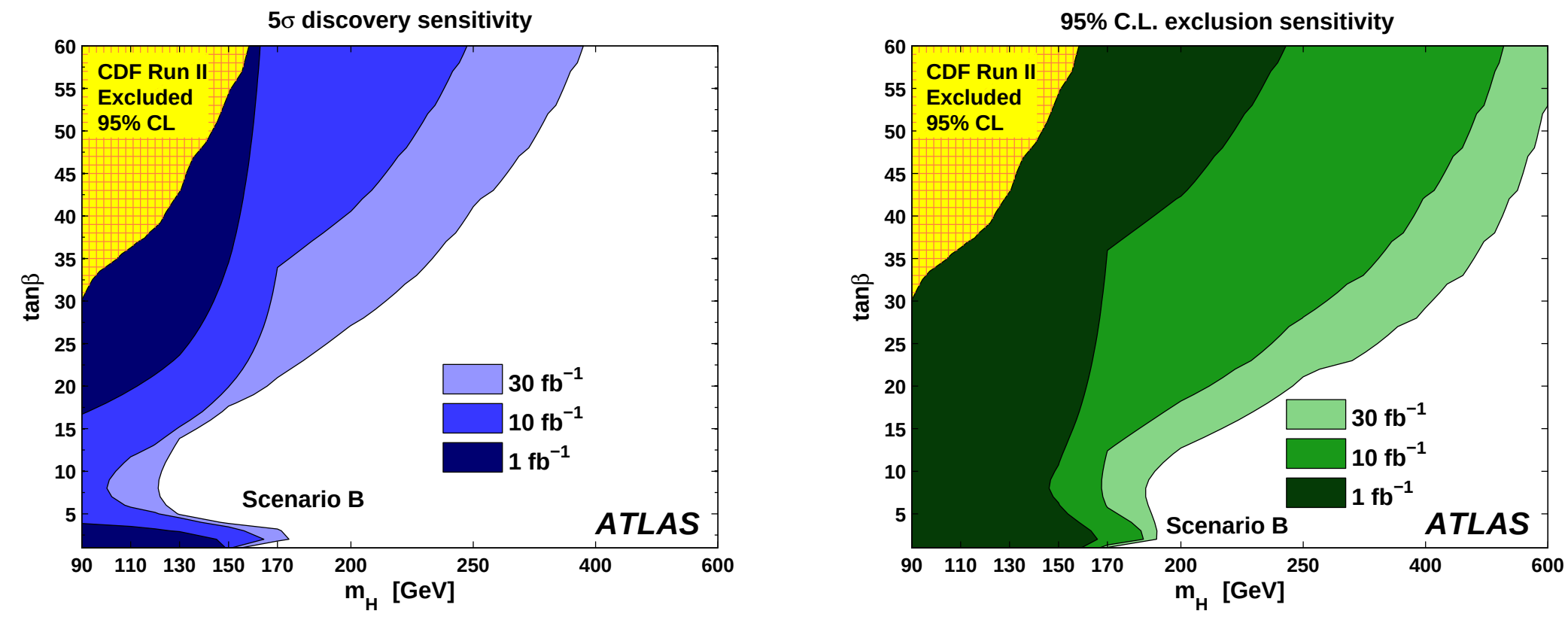


Figure 19: Scenario B ($m_h$-max): Combined Results. Left: Discovery contour, Right: Exclusion contour. Statistical errors arising from simulation statistics are neglected.

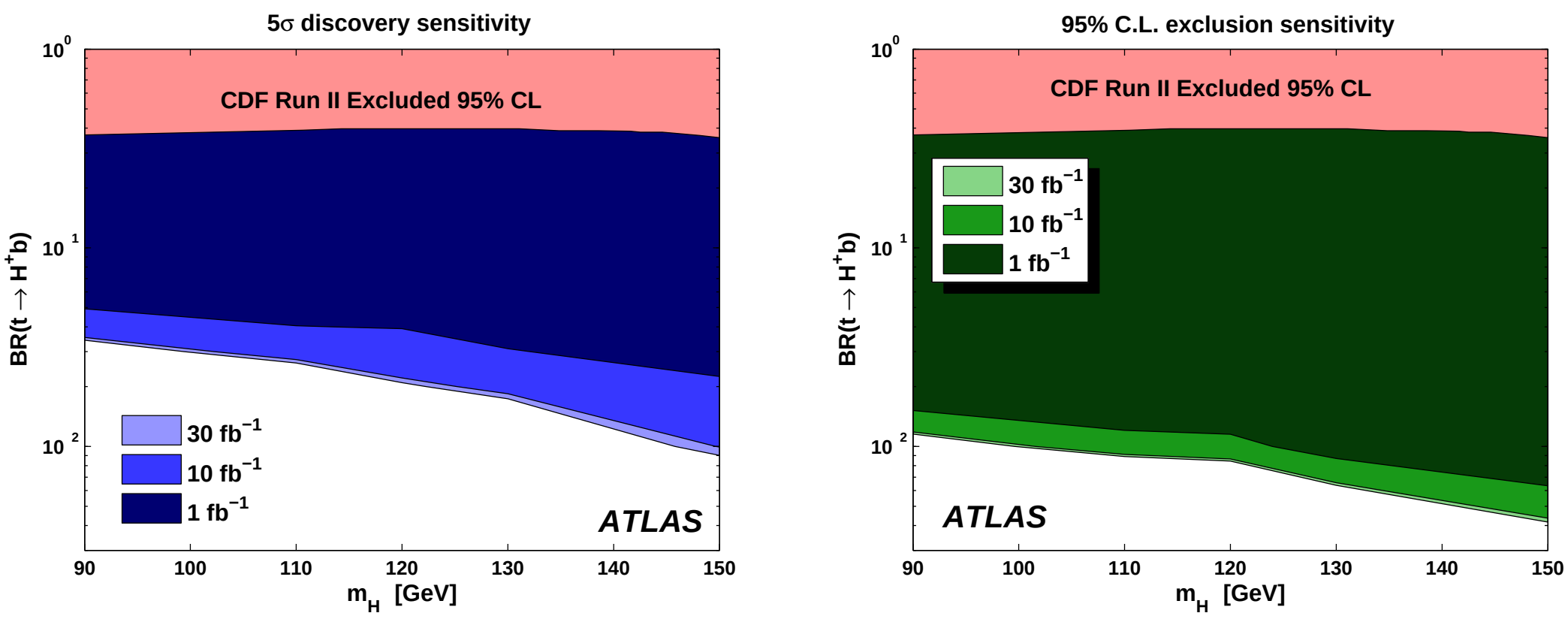


Figure 20: Model-independent: Combined light $H^+$ results in the ($m_{H^+}$, BR($t \to H^+b$))-plane. Left: Discovery contour, Right: Exclusion contour. Systematic uncertainties and statistical uncertainties are included. BR($H^+ \to \tau\nu$) = 1 is assumed.

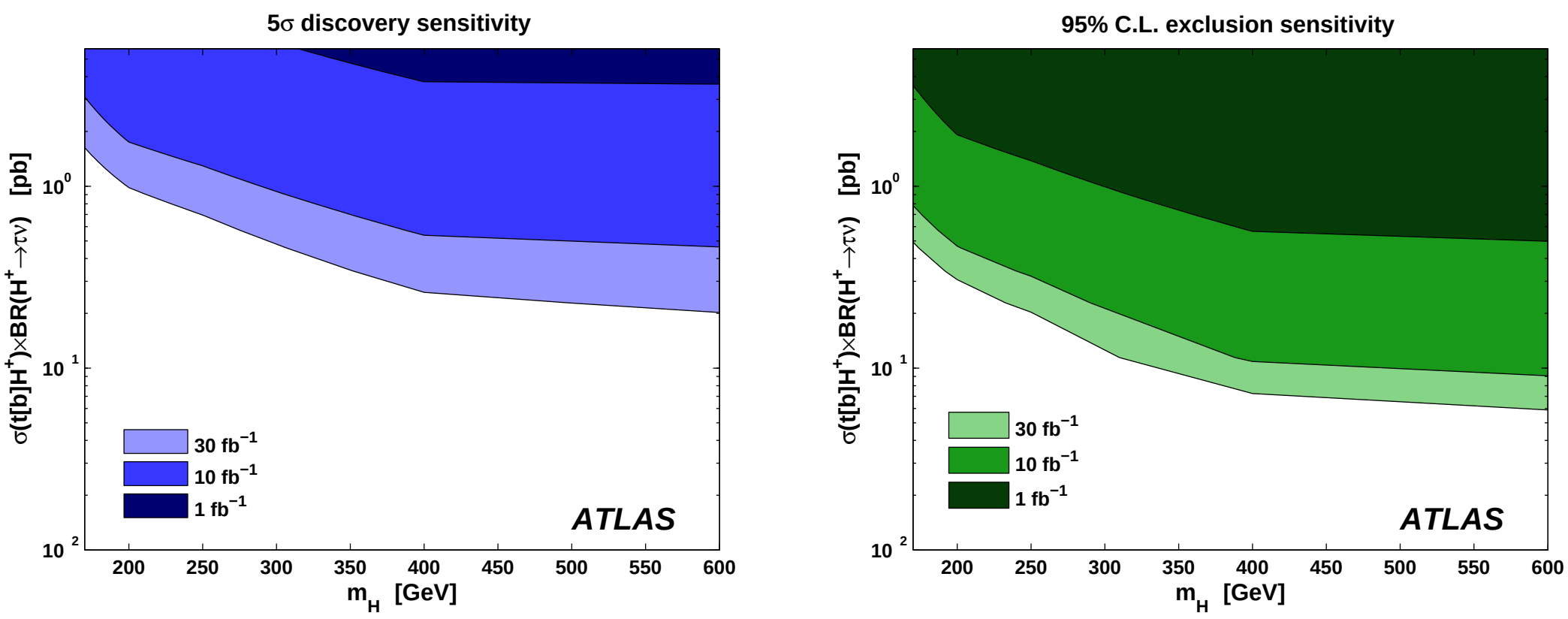


Figure 21: Model-independent: Heavy $H^+$ results in the ($m_{H^+}$, $\sigma$)-plane. Left: Discovery contour, Right: Exclusion contour. Systematic uncertainties and statistical uncertainties are included.

However, a charged Higgs boson in this region could be excluded for values up to the top quark mass. Additional integrated luminosity does not give any further $H^+$ sensitivity for a low-mass $H^+$ after a few fb$^{-1}$ have been recorded. The reason is the statistical error from the small $t\bar{t}$ Monte Carlo sample which is equivalent to about 1 fb$^{-1}$ at the LHC and makes extrapolation to higher luminosities difficult. Repeating the low-mass $H^+$ studies with larger Monte Carlo samples is thus expected to lead to a significantly larger discovery and exclusion reach, as can be seen in Fig. 19 in which the statistical uncertainty arising from simulation statistics is neglected (i.e. it is assumed that the number of simulated events is much larger than the number of expected real-data events).

Figure 20 shows the discovery and exclusion contour in terms of the branching ratio $t \to H^+b$ as a function of $m_{H^+}$. With one year of low luminosity data, it will be possible to discover the charged Higgs boson if BR($t \to H^+b$) is larger than about 1-3%, and to exclude it even if this branching ratio is well below the percent level. Similar contours for a heavy $H^+$ are presented in Fig. 21; here the y-axis shows the cross-section for the process $gg/gb \to t[b]H^+ \to t[b]\tau\nu$. Sensitivity is given for a cross-section of the order of 0.1pb. Both figures are model-independent in the sense that they can be interpreted in the context of any MSSM, other SUSY, or even non-SUSY scenario.

# 7 Conclusions

The ATLAS potential for discovering or excluding the existence of a charged Higgs boson in two different MSSM scenarios has been evaluated for five different final states of the $H^+$ signal. Significant improvements of present day constraints can already be achieved with limited data (less than 1 fb$^{-1}$, about one month at low luminosity at the LHC) although it may not qualify as early physics due to its dependency on higher level reconstruction objects.

Below the top quark mass charged Higgs bosons are predominantly produced in top quark decays and the main decay mode is $H^+ \to \tau\nu$ . Three different signal final states have been studied and analyzed separately, each of them separately outperforming the present sensitivity from the Tevatron experiments already with 1 fb$^{-1}$ of data. The combined performance of the three channels yields a discovery reach for 10 fb$^{-1}$ which covers $\tan\beta$ values down to 20 and up to 4 for all charged Higgs boson masses up to about 150 GeV. For intermediate $\tan\beta$ region (around $\tan\beta = 7$), no discovery sensitivity is present, but a charged Higgs boson could be excluded in this region. The current sensitivity is primarily limited by simulation statistical uncertainties, it is thus expected that a larger production of simulated events will greatly improve the situation and give access to the intermediate $\tan\beta$ region.

Two analyses have been conducted in the search for a heavy charged Higgs boson ($m_{H^+} > m_t$), a region presently uncovered in direct searches. Here, the main production mode is through $gb$ fusion ($gb \to tH^+$) and the decay into a top and a $b$ quark dominates. However, the dominant $tb$ decay mode suffers from large irreducible backgrounds, and the combinatorial background. Consequently the discovery potential for a heavy charged Higgs boson is dominated by the $\tau\nu$ decay mode, which despite its significantly smaller branching ratio allows for more efficient background suppression. The discovery reach in the context of the MSSM ($m_h$-max) strongly depends on the charged Higgs boson mass and reaches from ($m_{H^+} = 200$ GeV, $\tan\beta = 28$) to ($m_{H^+} = 350$ GeV, $\tan\beta = 58$) for an integrated luminosity of 30 fb$^{-1}$. Additionally, the model-independent discovery reach for a charged Higgs boson as a function of its production cross-section has been evaluated. A light $H^+$ sensitivity for a BR($t \to H^+b$) down to the percent level is given for a discovery, and well below that level for an exclusion. For a heavy $H^+$ decaying to $\tau\nu$, sensitivity is given for cross-sections of the order of 0.1pb.

The results presented in this note give confidence that the LHC and the ATLAS detector will be able to probe an extended Higgs sector over a sizable region of the MSSM parameter space. For a high SUSY mass scale, the charged Higgs boson could be the first signal of New Physics (and indication for Supersymmetry) discovered.

# Statistical Combination of Several Important Standard Model Higgs Boson Search Channels


**Abstract**

In this note we describe statistical procedures for combination of results from independent searches for the Higgs boson. Here only the Standard Model Higgs is considered, although the methods can easily be extended to non-standard Higgs models as well as to other searches. The methods are applied to Monte Carlo studies of four important search channels: $H \to \tau^+\tau^-$, $H \to W^+W^- \to e\nu\mu\nu$, $H \to \gamma\gamma$ and $H \to ZZ^{(*)} \to 4$ leptons. The statistical treatment relies on a large sample approximation that is expected to be valid for an integrated luminosity of at least 2 fb$^{-1}$. Results are presented for the expected statistical significance of discovery and expected exclusion limits.


## 1 Introduction

Higgs searches will exploit a number of statistically independent decay channels. One wishes to combine all of the information from them to provide a single measure of the significance of a discovery or limits on Higgs production. The approach taken in this paper is based on frequentist statistical methods, where effects of systematic uncertainties are incorporated by use of the profile likelihood ratio.

The statistical procedures used for establishing discovery and setting limits are described in Section 2. These methods are very general and can be applied to the combination of results of essentially any search that will be carried out at the LHC. Section 3 summarizes the four search channels for the Standard Model Higgs boson considered in this note: $H \to \tau^+\tau^-$, $H \to W^+W^- \to e\nu\mu\nu$, $H \to \gamma\gamma$ and $H \to ZZ^{(*)} \to$ 4 leptons.

The statistical treatment requires knowledge of the distribution of a test statistic based on the profile likelihood ratio. To determine these distributions by Monte Carlo so as to establish discovery at a high level of significance would require an enormous amount of simulated data, which is not practical at present. Therefore the distributions have been estimated using the functional form expected to hold in the large sample limit. Investigations shown in Section 3 indicate that this approximation should be reliable for an integrated luminosity above 2 fb$^{-1}$.

In Section 4 we show the result of the combination. For different values of the integrated luminosity and hypothesized Higgs mass, we present the signal significance expected assuming the Standard Model Higgs production rate, as well as expected upper limits on the Higgs production cross section, under the hypothesis of no Higgs signal.

The channels considered here focus on the search for a Higgs boson in the low-mass range. It is planned to include other channels in the future, e.g., further final states from the $W^+W^-$ and $ZZ$ modes. This will improve sensitivity especially at higher Higgs mass values.

## 2 Statistical methods

In this section we describe the general statistical model and likelihood function, first for a single channel and then generalized to multiple channels. In Section 2.2 we give the procedure used to establish discovery based on a frequentist significance test, where the effects of systematic uncertainties are incorporated by use of the *profile likelihood ratio*. Section 2.3 covers the corresponding methods for setting limits. For both discovery and exclusion one requires the sampling distribution of the statistic used in the test; this

is described in Section 2.4. Section 2.5 discusses a series of approximations used to determine expected values of the discovery significance and exclusion limits.

The approach taken in this note is to carry out tests for discovery and exclusion for fixed values of the Higgs mass $m_H$. In principle the entire procedure is then repeated for all masses, resulting in limits on or a measurement of $m_H$. In practice, an interpolation is made between finite steps in $m_H$.

## 2.1 The statistical model and likelihood function

First we consider the case of a single search channel. The measurement results in a set of numbers of events found in kinematic regions where signal could be present. These typically correspond to a histogram of a variable such as the mass of the reconstructed Higgs candidate, with the numbers of entries denoted by $\mathbf{n} = (n_1, \ldots, n_N)$. In some cases one may consider a histogram with only one bin, i.e., the measured outcome is simply a number of candidate events found. The number of entries in bin $i$, $n_i$, is modeled as a Poisson variable with mean value

$$E[n_i] = \mu L \varepsilon_i \sigma_i \mathscr{B} + b_i \equiv \mu s_i + b_i \,, \tag{1}$$

where $L$ is the integrated luminosity, $\varepsilon_i$, $\sigma_i$ and $\mathscr{B}$ are the signal efficiency, Higgs cross section, and branching ratio, and $b_i$ is the expected number of background events. Here $\mu$ is a signal strength parameter defined such that $\mu = 0$ corresponds to the absence of a signal; $\mu = 1$ gives the signal rate $s_i$ expected from the Standard Model. If we consider a fixed Higgs mass $m_H$, the only parameter of interest is $\mu$. All other adjustable parameters needed to specify the model are called *nuisance parameters*.

In principle the expected background values $b_i$ can be predicted using Monte Carlo models for Standard Model processes. In the measurements considered here, however, the systematic uncertainty in the Standard Model prediction is in many cases quite large, and this would severely limit the sensitivity of the search. Therefore data regions where one expects only a very small amount of signal (*control regions*) are used to constrain the background in the signal region (see also below).

For the $i$th bin of a histogram of a discriminating variable $x$, the expected signal and background can be written

$$s_i = s_{\text{tot}} \int_{\text{bin}\, i} f_s(x;\theta_s)\, dx \,, \tag{2}$$

$$b_i = b_{\text{tot}} \int_{\text{bin}\, i} f_b(x;\theta_b)\, dx \,, \tag{3}$$

where $s_{\text{tot}}$ and $b_{\text{tot}}$ are the total expected numbers of events in the histograms, $f_s(x;\theta_s)$ and $f_b(x;\theta_b)$ are the probability density functions (pdfs) of $x$ for signal and background, and $\theta_s$ and $\theta_b$ represent sets of *shape parameters*.

The parametric forms of the pdfs $f_s(x;\theta_s)$ and $f_b(x;\theta_b)$ are determined from Monte Carlo simulations or data control samples. In the following we will use $\theta = (\theta_s, \theta_b, b_{\text{tot}})$ to refer to all of the nuisance parameters. The signal normalization $s_{\text{tot}}$ here is not an adjustable parameter, but rather is fixed equal to the Standard Model prediction.

In addition to the measured histogram $\mathbf{n}$, some search channels also make use of a set of subsidiary measurements $\mathbf{m} = (m_1, \ldots, m_M)$ in control regions where one expects mainly background events. These can be modeled as being Poisson distributed with mean values

$$E[m_i] = u_i(\theta) \,, \tag{4}$$

where the $u_i$ are calculable quantities depending on a set of parameters, at least some of which are the same as those entering into the predictions for $s_i$ and $b_i$ above. In practice the subsidiary measurements

are constructed so as to provide information on the background normalization $b_{\rm tot}$ and sometimes also on its shape.

If the measurement is based on counting events in a given kinematic region, i.e., without using the shape of a distribution, in the formalism above the histograms have a single bin. The value $s = s_{\rm tot}$ is then the Standard Model prediction for the signal and $b = b_{\rm tot}$ is the (unknown) expected background. There are then no shape parameters, and $b$ itself plays the role of $\theta$ as the single nuisance parameter. In this case the subsidiary measurement $m$ is made in a control region where signal is absent (or can to good approximation be neglected), and has an expectation value

$$E[m] = u = \tau b \;, \tag{5}$$

where $\tau$ is a scaling constant whose value can be estimated from a Monte Carlo simulation.

The likelihood function is the product of Poisson probabilities for all bins:

$$L(\mu,\theta) = \prod_{j=1}^{N} \frac{(\mu s_j + b_j)^{n_j}}{n_j!} e^{-(\mu s_j + b_j)} \;\; \prod_{k=1}^{M} \frac{u_k^{m_k}}{m_k!} e^{-u_k} \;. \tag{6}$$

Equivalently the log-likelihood is

$$\ln L(\mu,\theta) = \sum_{j=1}^{N} \left(n_j \ln(\mu s_j + b_j) - (\mu s_j + b_j)\right) \; + \; \sum_{k=1}^{M} (m_k \ln u_k - u_k) + C \;, \tag{7}$$

where $C$ represents terms that do not depend on the parameters and thus can be dropped. Here and in (6) the parameters $\theta$ enter through Eqs. (2), (3), and (4).

In the case where the presence of signal in the histogram $\mathbf{n}$ gives a peak sitting on a smooth background, one does not need a subsidiary measurement $\mathbf{m}$. Rather, as long as the number of parameters in the models for the signal and background distributions is smaller than the total number of bins measured, one can determine the strength parameter $\mu$ from the histogram $\mathbf{n}$ alone. Here the regions away from the peak (the sidebands) play the role of the subsidiary measurement by providing information on the background level. Of course if an additional subsidiary measurement is available, this will improve the accuracy of the background determination, which will increase the sensitivity of the analysis.

In the case of several independent search channels, the method described above is generalized in a straightforward manner. For each channel $i$ there is a likelihood function $L_i(\mu,\theta_i)$. Its general form is given by Eq. (6), except that all quantities carry an additional index $i$ to label the channel except the global strength parameter $\mu$, which is assumed to be the same for all channels. Since the channels are statistically independent, the full likelihood function is given by the product

$$L(\mu,\theta) = \prod_i L_i(\mu,\theta_i) \;, \tag{8}$$

where $\theta$ here represents all of the nuisance parameters.

Systematic uncertainties are effectively included in the analysis through the nuisance parameters $\theta$. The model must be sufficiently flexible, i.e., it must contain enough parameters, so that for at least some point in its parameter space it can be regarded as representing the truth. One must exercise some restraint in achieving this, however, as an increasing number of nuisance parameters leads to a decrease in sensitivity to the parameters of interest. Some of the components of $\theta$ may be common among different channels, e.g., parameters relating to uncertainty in the integrated luminosity. These then represent a common (correlated) systematic uncertainty.

As an example, consider the signal efficiency $\varepsilon$ that enters in the relation between the cross section and expected number of signal events. Suppose the efficiency has been estimated to have a value $\hat{\varepsilon}$ and systematic uncertainty $\sigma_{\hat{\varepsilon}}$. To incorporate this uncertainty into the model, we can regard the measured

value $\hat{\varepsilon}$ as a random variable whose true value $\varepsilon$ is treated as a nuisance parameter. For the pdf $f_\varepsilon(\hat{\varepsilon};\varepsilon,\sigma_{\hat{\varepsilon}})$ one could use, e.g., a Gaussian distribution centred about $\varepsilon$, or for a quantity such as the efficiency which must lie in the range $0 \leq \varepsilon \leq 1$ one could use a pdf that automatically satisfies this constraint (e.g., a beta distribution). For whatever choice is deemed appropriate, the likelihood (6) is multiplied by $f_\varepsilon(\hat{\varepsilon};\varepsilon,\sigma_{\hat{\varepsilon}})$, evaluated with the best estimate $\hat{\varepsilon}$, and the parameter $\varepsilon$ is included in the set of nuisance parameters $\theta$.

To test a hypothesized value of $\mu$ we construct the profile likelihood ratio,

$$\lambda(\mu) = \frac{L(\mu,\hat{\hat{\theta}})}{L(\hat{\mu},\hat{\theta})} \,. \tag{9}$$

Here $\hat{\hat{\theta}}$ in the numerator denotes the value of $\theta$ that maximizes $L$ for the specified $\mu$, i.e., it is the conditional maximum-likelihood estimator (MLE) of $\theta$ (and thus is a function of $\mu$). The denominator is the maximized (full) likelihood function, i.e., $\hat{\mu}$ and $\hat{\theta}$ are the MLEs. The presence of the nuisance parameters broadens the profile likelihood ratio as a function of $\mu$ relative to what one would have if their values were fixed. This reflects the loss of information about $\mu$ due to the systematic uncertainties.

The likelihood ratio (9) and procedures for incorporating systematic uncertainties applied here differ somewhat from those used for the searches carried out at LEP. Some of these differences are discussed further in Appendix A.

From the definition of the profile likelihood ratio one can see that $0 \leq \lambda \leq 1$, with $\lambda(\mu) = 1$ implying good agreement between the data and the hypothesized value of $\mu$. Equivalently it is convenient to work with the quantity

$$q_\mu = -2\ln\lambda(\mu) \,, \tag{10}$$

so that high values of $q_\mu$ correspond to poor agreement between the data and the hypothesized $\mu$. The statistic $q_\mu$ will have a sampling distribution $f(q_\mu|\mu')$. Here $\mu$ refers to the strength parameter used to define the statistic $q_\mu$, entering in the numerator of the likelihood ratio, and $\mu'$ is the value used to define the data generated to obtain the distribution (i.e., the 'true' value). For the special case $\mu' = \mu$ and for a sufficiently large data sample, the pdf $f(q_\mu|\mu)$ approaches a limiting form related to the chi-square distribution, discussed further in Section 2.4. For $\mu' \neq \mu$, the distribution of $q_\mu$ is shifted to higher values, reflecting the decreased agreement between the data generated with $\mu'$ and the hypothesis tested by $q_\mu$, as indicated in Fig. 1. The two cases of particular interest are $\mu = 0$, the background-only hypothesis, and $\mu = 1$, the hypothesis of background plus signal present at the Standard Model rate.

The level of compatibility between data that give an observed value $q_{\mu,\mathrm{obs}}$ for $q_\mu$ and a hypothesized value of $\mu$ is quantified by giving the $p$-value

$$p_\mu = \int_{q_{\mu,\mathrm{obs}}}^{\infty} f(q_\mu|\mu)\, dq_\mu \,. \tag{11}$$

This is the probability, under the assumption of $\mu$, of seeing data with equal or greater incompatibility, as measured by $q_\mu$, relative to the data actually obtained. This is illustrated in Fig. 1, where the shaded area indicates the $p$-value of the hypothesized $\mu$. The figure also indicates the median value of $q_\mu$ under the assumption of a different value of the strength parameter $\mu'$ used to generate the data. For $\mu$ and $\mu'$ values that are increasingly different, the median $\mathrm{med}[q_\mu|\mu']$ moves further to the right. An observed value of $q_\mu$ at this median would give a correspondingly small $p$-value for $\mu$.

## 2.2 Establishing discovery

To establish discovery we try to reject the $\mu = 0$ (background-only) hypothesis, i.e., that there is no Higgs signal present. To do this we use the statistic $q_0 = -2\ln\lambda(0)$. One expects to find a low value

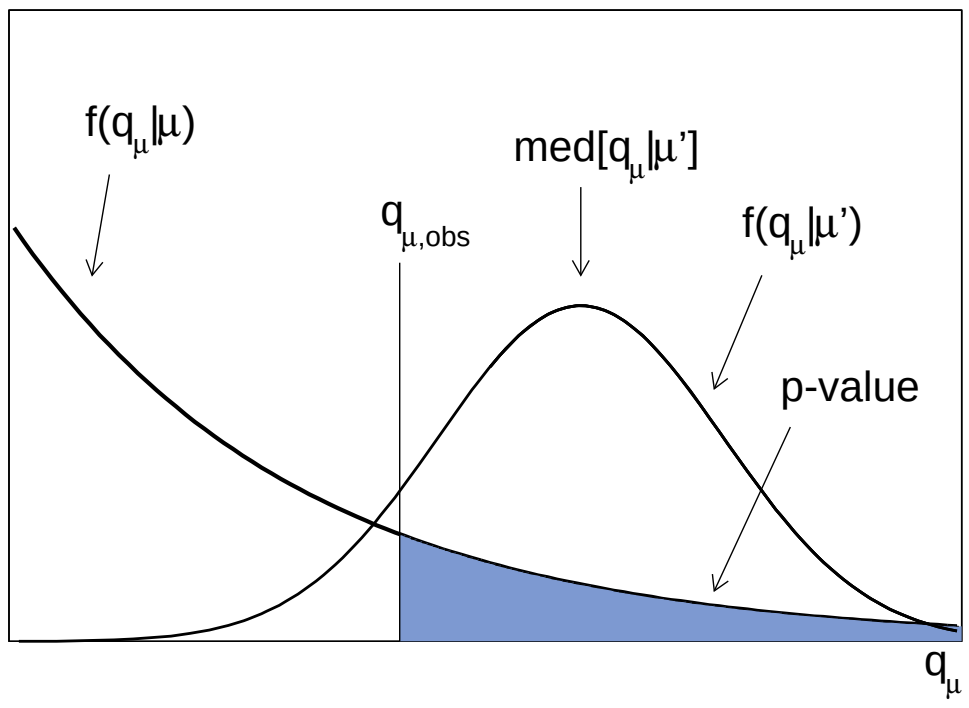


Figure 1: Illustration of the determination of the $p$-value of a hypothesized value of $\mu$. The left-hand curve indicates the pdf of $q_\mu$ for data generated with the same value of $\mu$ as was used to define the statistic $q_\mu$; this is used to determine the $p$-value of $\mu$, shown as the shaded region. The right-hand curve indicates the pdf of $q_\mu$ for data generated with a different value of the strength parameter, $\mu'$.

of $\lambda(0)$ (high $q_0$) if the data include signal. Here even though one is testing the hypothesis that the Higgs does not exist, the definition of $q_0$ depends on the hypothesized Higgs mass $m_H$. It enters through the denominator of the likelihood ratio (9), which contains the maximum-likelihood estimator $\hat{\mu}$ for the strength of a Higgs signal at the mass $m_H$. By defining the test statistic in this way one maximizes the probability of rejecting the $\mu = 0$ hypothesis if the Higgs boson exists at the specified mass. This search procedure is then carried out for all values of $m_H$ (in practice an interpolation is carried out between finite steps in $m_H$).

A given data set will result in an observed value $q_{0,\mathrm{obs}}$ of $q_0$. The level of compatibility between the data and the no-Higgs hypothesis is quantified by giving the $p$-value

$$p_0 = \int_{q_{0,\mathrm{obs}}}^{\infty} f(q_0|0)\, dq_0 \; . \tag{12}$$

This is the probability, under the assumption of $\mu = 0$ (background only), of seeing data as signal-like or more so relative to the data actually obtained. A small value is interpreted as evidence against $\mu = 0$, i.e., a discovery of the signal.

One can define the *significance* corresponding to a given $p$-value as the number of standard deviations $Z$ at which a Gaussian random variable of zero mean would give a one-sided tail area equal to $p$. That is, the significance $Z$ is related to the $p$-value by

$$p = \int_{Z}^{\infty} \frac{1}{\sqrt{2\pi}} e^{-x^2/2}\, dx = 1 - \Phi(Z) \; , \tag{13}$$

where $\Phi$ is the cumulative distribution for the standard (zero mean, unit variance) Gaussian. Equivalently one has

$$Z = \Phi^{-1}(1-p) \; , \tag{14}$$

where $\Phi^{-1}$ is the quantile of the standard Gaussian (inverse of the cumulative distribution). In (13) and (14) the subscript 0 was dropped as these relations hold for all $p$-values, not only those of the $\mu = 0$ hypothesis. The relation between $Z$ and $p$ is illustrated in Fig. 2.

A significance of $Z = 5$ corresponds to $p = 2.87 \times 10^{-7}$. For a sufficiently large data sample, one would obtain a $p$-value of 0.5 for data in perfect agreement with the expected background. With the definition of $Z$ given above, this gives $Z = 0$. If the data fluctuate below the expected background, $Z$ becomes negative.

Note that according to the definition (14), a $p$-value of 0.05 corresponds to $Z = 1.64$. This should not be confused with a $1.96\sigma$ fluctuation of a Gaussian variable that gives 0.05 for the two-sided tail area.

The significance of a discovery $Z$ depends on the data obtained. To quantify our ability to discover a hypothesized signal in advance of seeing the data, we report the *median* significance under the assumption that the signal is present at the Standard Model rate, $\mu = 1$. Since $Z$ is a monotonic function of $p_0$, and $p_0$ is also a monotonic function of $q_0$, we have for the median significance,

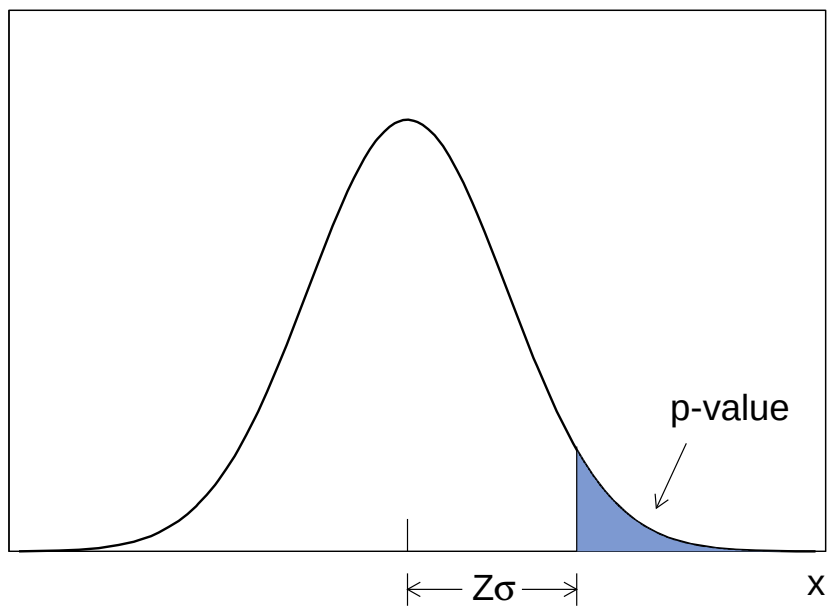


Figure 2: Illustration of the correspondence between the significance $Z$ and a $p$-value.

$$Z_{\rm med} = \Phi^{-1}(1 - p_{0_{\rm med}}) = \Phi^{-1}(1 - p_0(q_{0_{\rm med}})) \; . \tag{15}$$

This can be obtained from the median value of $q_0$ found using data generated under the assumption of $\mu = 1$.

A complete evaluation of the median significance is computationally difficult, as it requires a large number of repeated simulations of the full set of experimental outputs and determination of $Z(q_0)$ from the combination of all channels. Therefore in this note we have used the approximate methods described in Section 2.5, which allow one to estimate quickly the median significance.

### 2.3 Setting limits

In addition to establishing discovery by rejecting the $\mu = 0$ hypothesis, we can consider the alternative hypothesis of some non-zero $\mu$ and try to reject it. A $p$-value is computed for each $\mu$, and the set of $\mu$ values for which the $p$-value is greater than or equal to a fixed value $1 - \mathrm{CL}$ form a confidence interval for $\mu$, where typically one takes a *confidence level* $\mathrm{CL} = 95\%$ . The upper end of this interval $\mu_{\rm up}$ is the upper limit (i.e., $\mu \leq \mu_{\rm up}$ at 95% CL).

To compute the $p$-value for a hypothesized $\mu$ we first consider again the test statistic $q_\mu = -2\ln\lambda(\mu)$ as initially defined in (9) and (10). For purposes of computing limits, we introduce a modification to this definition as described below.

If the data are incompatible with the hypothesized $\mu$, one expects a large value of $-2\ln\lambda(\mu)$, i.e., $\lambda(\mu)$ close to zero. If a data set generated according to the hypothesis $\mu$ gives a large value of $-2\ln\lambda(\mu)$, this can be the result of either an upward or downward fluctuation in $\hat{\mu}$ relative to $\mu$. This is illustrated in the scatterplot of $\hat{\mu}$ versus $-2\ln\lambda(\mu)$ shown in Fig. 3(a), which is from a toy Monte Carlo study with $\mu = 0.8$. The projection of the points on the $\hat{\mu}$ axis is shown in Fig. 3(b). Note that $\hat{\mu} \geq 0$ is imposed; the reasons for and consequences of this requirement are discussed in Section 2.4.

For purposes of setting an upper limit, however, we want to determine the smallest $\mu$ such that there is a fixed small probability (one minus the confidence level) to find data as compatible with that value of $\mu$ or less, relative to the degree of compatibility found with the real data. Therefore the data with upward fluctuations in $\hat{\mu}$ are not counted when computing the $p$-value, because they would be compatible with some larger $\mu$. Therefore for purposes of computing limits we redefine $q_\mu$ to be[1]

$$q_\mu = \begin{cases} -2\ln\lambda(\mu) & \hat{\mu} \leq \mu \; , \\ 0 & \text{otherwise.} \end{cases} \tag{16}$$

The distribution of the new $q_\mu$ thus corresponds to the lower branch only of the U-shaped scatterplot shown in Fig. 3(a).

[1]Equivalently, one could retain the definition $q_\mu = -2\ln(L(\mu,\hat{\hat{\theta}})/L(\hat{\mu},\hat{\theta}))$ by placing an upper bound on $\hat{\mu}$ equal to $\mu$, i.e., by imposing $0 \leq \hat{\mu} \leq \mu$. In this way, when $\hat{\mu} = \mu$ then one has $q_\mu = 0$ just as in the case of discovery when testing $\mu = 0$.

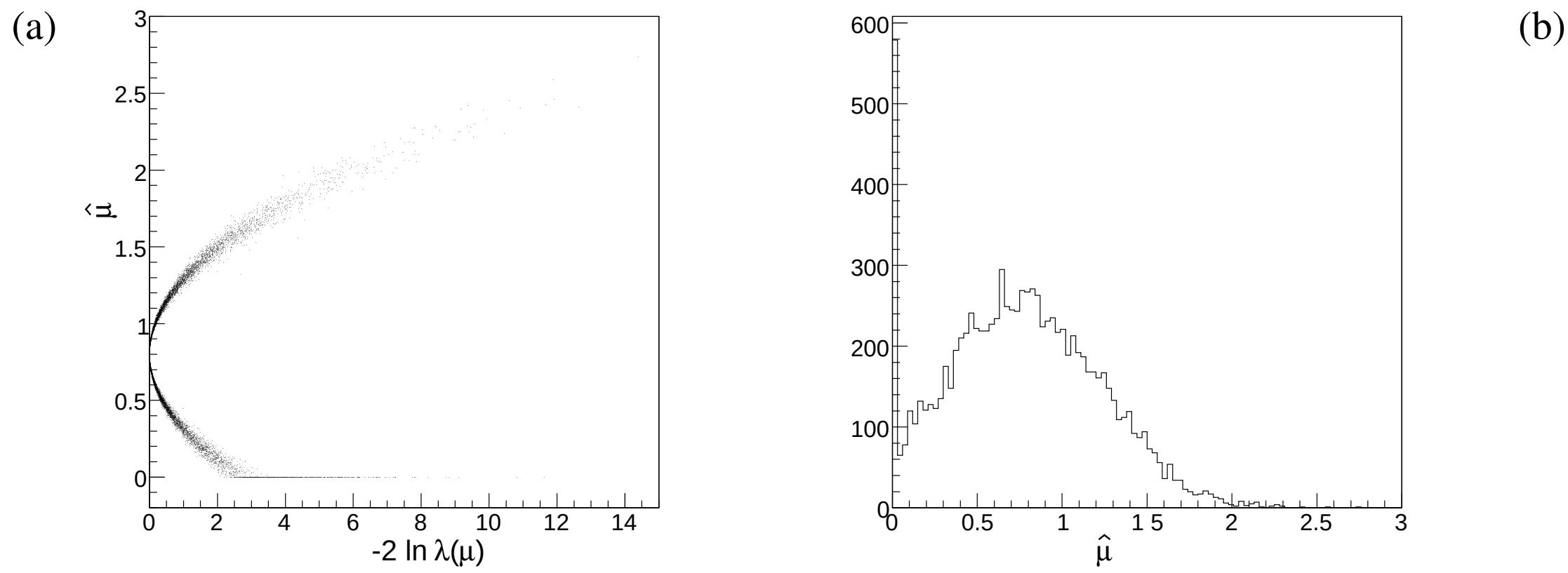


Figure 3: (a) Scatterplot of $\hat{\mu}$ versus $-2\ln\lambda(\mu)$; (b) the distribution of $\hat{\mu}$ (see text).

Using the new definition (16), the $p$-value is given by the integral of $f(q_\mu|\mu)$ from the observed value $q_{\mu,\mathrm{obs}}$ to infinity as in Eq. (11) and as illustrated in Fig. 1. The $p$-value is computed in this manner for all values of $\mu$, and the upper limit $\mu_{\mathrm{up}}$ at 95% confidence level is the largest value of $\mu$ for which the $p$-value is at least 0.05.

The result can be summarized by giving the upper limit on $\mu$ as a function of the Higgs mass $m_H$. Specifically, if we can reject the hypothesis $\mu = 1$ at a certain confidence level, then the corresponding value of $m_H$ is regarded as excluded for a Standard Model Higgs. The lowest mass value not excluded is the lower limit $m_{\mathrm{lo}}$.

One is also interested in the median limit under the assumption that there is no Higgs. As in the case of the discovery significance, a full calculation of the median limit is difficult as it requires a large number of repeated simulations based on the full profile likelihood ratio. For purposes of this note, therefore, we use the approximation techniques described in Section 2.5.

## 2.4 Sampling distribution of the likelihood ratio

To determine the $p$-values required for both discovery and exclusion we need the sampling distribution, assuming data generated according to a given value of $\mu$, of the statistic $q_\mu$, i.e., $f(q_\mu|\mu)$. For the case of discovery significance we use $q_0 = -2\ln\lambda(0)$, and for setting limits we use $q_\mu = -2\ln\lambda(\mu)$ for $\hat{\mu} \leq \mu$ and $q_\mu = 0$, otherwise.

To claim discovery we require $p$-values for $\mu = 0$ down to around $10^{-7}$, and therefore to do this with a Monte Carlo simulation requires an extremely large number of simulated measurements. In practice this is only carried out for simple test cases. Even for setting limits at 95% confidence level, it is often not practical to use Monte Carlo.

Under a set of regularity conditions and for a sufficiently large data sample, *Wilks' theorem* says that for a hypothesized value of $\mu$, the pdf of the statistic $-2\ln\lambda(\mu)$ approaches the chi-square pdf for one degree of freedom [2]. More generally, if there are $n$ parameters of interest, i.e., those parameters that do not get a double hat in the numerator of the likelihood ratio (9), then $-2\ln\lambda(\mu)$ asymptotically follows a chi-square distribution for $n$ degrees of freedom. A proof and details of the regularity conditions can be found in standard texts such as [3].

In the searches considered here, the data samples are generally large enough to ensure the validity of the asymptotic formulae for the likelihood-ratio distributions. In our case, however, the distributions are modified because of constraints imposed on the expected number of events.

Usually when searching for a new type of particle reaction one regards the mean number of events contributed to any bin from any source, signal or background, to be greater than or equal to zero. In

some analyses it could be meaningful to consider a new effect that suppresses the expected number of events, e.g., the presence of a new decay channel could mean that the number of decays to known channels is reduced. Here, however, we will regard any contribution to an expected number of events as non-negative.

Assuming only non-negative event rates, the maximum-likelihood estimators for the parameters are constrained, e.g., $\hat{\mu} \geq 0$. As a consequence, if the observed number of events is below the level predicted by the background alone, then the maximum of the likelihood occurs for $\mu = 0$, i.e., negative $\mu$ is not allowed. We can consider the effect of having $\hat{\mu} = 0$ on the distribution of $q_\mu$ for two cases: $\mu = 0$ and $\mu > 0$.

For $\mu = 0$, i.e., when computing the discovery significance, if $\hat{\mu} = 0$ one has (see (9)),

$$\lambda(0) = \frac{L(0, \hat{\hat{\theta}})}{L(\hat{\mu}, \hat{\theta})} = \frac{L(0, \hat{\hat{\theta}})}{L(0, \hat{\theta})} = 1 \; , \tag{17}$$

since $\hat{\mu} = 0$ and therefore $\hat{\hat{\theta}} = \hat{\theta}$. The statistic $q_0 = -2\ln\lambda(0)$ is therefore equal to zero. This can be seen in the scatterplot of $q_0$ versus $\hat{\mu}$ in Fig. 4(a). Figure 4(b) shows the corresponding $q_0$ distribution with the peak visible at $q_0 = 0$. The superimposed curve is a chi-square distribution multiplied by one half, corresponding to the half of the events with $\hat{\mu} > 0$.

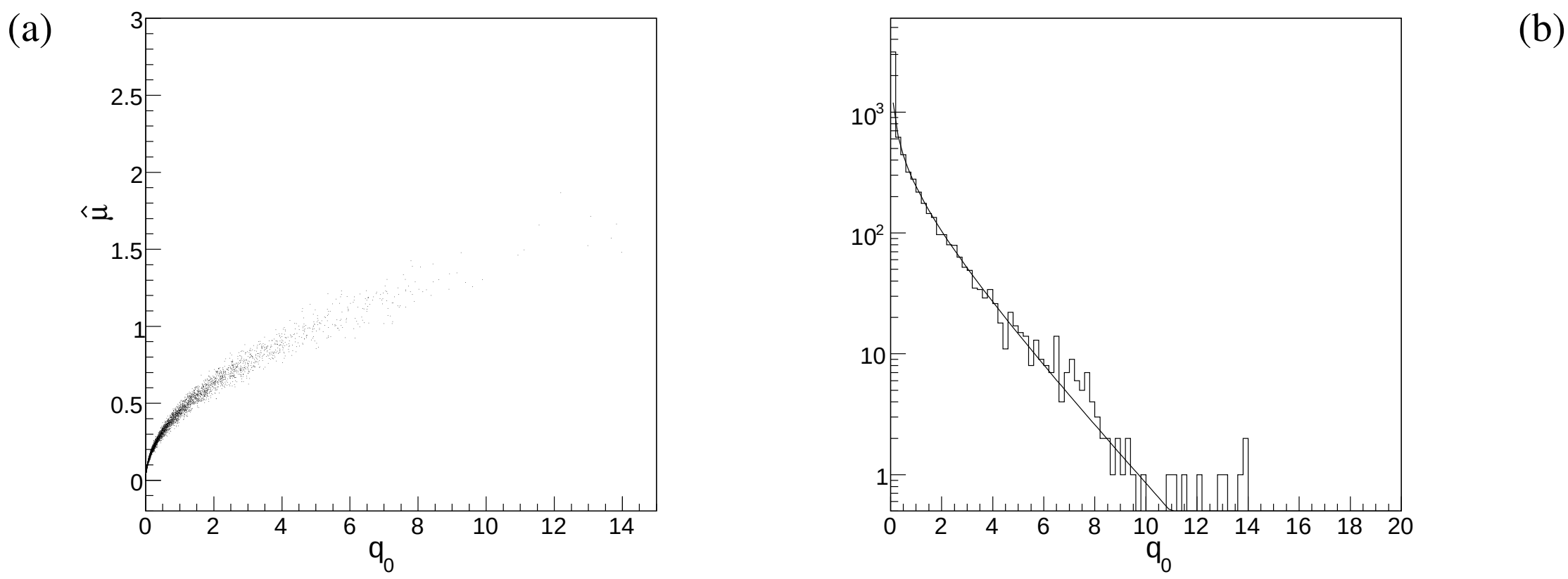


Figure 4: (a) Scatterplot of $\hat{\mu}$ versus $q_\mu$ from a Monte Carlo study with $\mu = 0$; (b) the distribution of $q_0$ (see text).

From Fig. 4(b) one can see that except for the spike at $q_0 = 0$ (when $\hat{\mu} = 0$), the pdf of $q_0$ can be well approximated by the chi-square pdf. Assuming a fraction $w$ for the cases with $\hat{\mu} > 0$ one has the pdf

$$f(q_0|0) = w f_{\chi^2_1}(q_0) + (1-w)\delta(q_0) \; . \tag{18}$$

In the usual case where upward and downward fluctuations of $\hat{\mu}$ are equally likely we have $w = 1/2$. The $p$-value of the background-only hypothesis given an observation $q_{0,\mathrm{obs}}$ greater than zero is therefore

$$p = \int_{q_{0,\mathrm{obs}}}^{\infty} w f_{\chi^2_1}(q_0)\, dq_0 = w(1 - F_{\chi^2_1}(q_{0,\mathrm{obs}})) \; , \tag{19}$$

where $F_{\chi^2_1}$ is the cumulative chi-square distribution for one degree of freedom.

The second case to consider is $\mu > 0$, e.g., when one wants to set an upper limit on $\mu$. Under the hypothesis $\mu$, one obtains $\hat{\mu} > \mu$ and $\hat{\mu} \leq \mu$ with approximately equal probability. Figure 5 shows the distributions of $q_\mu$ for both cases $\hat{\mu} > \mu$ and $\hat{\mu} \leq \mu$ obtained from the scatterplot Fig. 3(a), from a Monte Carlo study with $\mu = 0.8$.

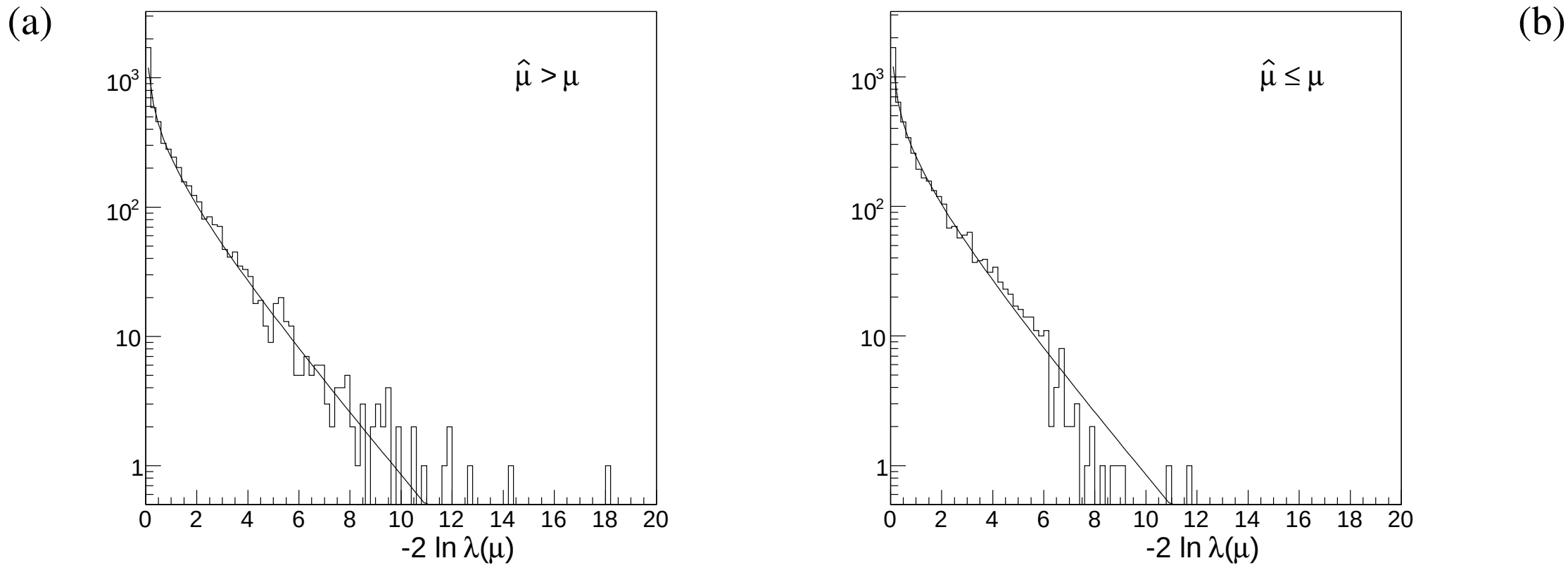


Figure 5: Distributions of $-2\ln\lambda(\mu)$ for (a) $\hat{\mu} > \mu$ and (b) $\hat{\mu} \leq \mu$. The superimposed curves are chi-square distributions for one degree of freedom normalized to half the number of entries in the original distribution (see Fig. 3(a)).

From Fig. 5(a) one can see that for $\hat{\mu} > \mu$, the data follow the chi-square pdf quite accurately. This portion of the distribution is ignored, however, when setting upper limits on $\mu$, because of the modified definition of $q_\mu$ (20) used for limits,

$$q_\mu = \begin{cases} -2\ln\lambda(\mu) & \hat{\mu} \leq \mu\,, \\ 0 & \text{otherwise}\,. \end{cases} \tag{20}$$

Suppose now $\hat{\mu} \leq \mu$ with a probability $w$; in practice this is close to one half. (Note for the case of $q_0$, $w$ is the probability of $\hat{\mu} > \mu$. The different definitions of $w$ are used so as to give similar forms for $f(q_0|0)$ and $f(q_\mu|\mu)$.) Thus for $\hat{\mu} > \mu$ one has from (20) $q_\mu = 0$, and therefore the distribution has a delta function at $q_\mu = 0$ with weight $1-w$. The pdf of $f(q_\mu|\mu)$ can therefore be written

$$f(q_\mu|\mu) = wf(q_\mu|\mu, \hat{\mu} \leq \mu) + (1-w)\delta(q_\mu)\,. \tag{21}$$

where $f(q_\mu|\mu, \hat{\mu} \leq \mu)$ is the conditional pdf for $q_\mu$ given $\hat{\mu} \leq \mu$.

For $\hat{\mu} \leq \mu$, one may sometimes find $\hat{\mu}$ equal to zero, i.e., the lower edge of the allowed range, as can be seen in the scatterplot of $\hat{\mu}$ versus $q_\mu$ shown in Fig. 3(a). Although for the case $\mu = 0$ this gave a peak at $q_0 = 0$, here it gives

$$\lambda(\mu) = \frac{L(\mu, \hat{\hat{\theta}})}{L(\hat{\mu}, \hat{\theta})} = \frac{L(\mu, \hat{\hat{\theta}})}{L(0, \hat{\theta})}\,, \tag{22}$$

which in contrast to (17) is not equal to unity. The effect of having $\hat{\mu} = 0$ on the distribution of $q_\mu$ is therefore more complicated than was the case for $q_0$.

In general for $\hat{\mu} \leq \mu$, the distribution of $q_\mu$ falls off more steeply than the chi-square distribution. This is seen in Fig. 5(b). Therefore a $p$-value based on the chi-square formula will be larger than the true $p$-value, and the corresponding significance $Z$ will be smaller. The upper limits obtained for $\mu$ are therefore larger, i.e., a smaller set of $\mu$ values is excluded.

If $\mu$ is sufficiently large, then $\hat{\mu}$ is very rarely pushed to zero and $f(q_\mu|\mu, \hat{\mu} \leq \mu)$ approaches a chi-square distribution for one degree of freedom. For purposes of the present study, the chi-square approximation is adequate, but gives somewhat conservative limits. That is, we take the distribution of $q_\mu$ to be

$$f(q_\mu|\mu) = wf_{\chi^2_1}(q_\mu) + (1-w)\delta(q_\mu) \tag{23}$$

and use $w = 0.5$. One has therefore the same pdf for $q_\mu$ using the modified definition (20) as was found in (18) for $q_0$ based on the original definition, $q_0 = -2\ln\lambda(0)$.

To summarize the result above, the pdf of $q_\mu$ can be approximated by a mixture of a chi-square pdf for one degree of freedom with weight $w$ and a delta function at zero with weight $1-w$. This holds both for discovery ($\mu = 0$) and setting limits ($\mu > 0$).

Consider now the variable

$$u = \sqrt{q_\mu} = \sqrt{-2\ln\lambda(\mu)} \,, \tag{24}$$

which has the pdf

$$f(u) = \Theta(u) w \sqrt{\frac{2}{\pi}} e^{-u^2/2} + (1-w)\delta(u) \,, \tag{25}$$

where $\Theta(u) = 1$ for $u \geq 0$ and is zero otherwise. The second term in (25) follows from the fact that the values $q_0 = 0$ and $u = 0$ occur with equal probability, $1-w$. Furthermore if a variable $x$ follows the standard Gaussian distribution, then one can show $x^2$ follows a chi-square distribution for one degree of freedom. Therefore if $x^2$ follows a $\chi^2$ distribution, then $\sqrt{x^2}$ follows a Gaussian scaled up by a factor of two for $x > 0$ so as to have a total area of unity.

The $p$-value of the hypothesis $\mu$ for a non-zero observation $q_{\mu,\mathrm{obs}}$ is therefore

$$p = P(q_\mu \geq q_{\mu,\mathrm{obs}}) = P(u \geq \sqrt{q_{\mu,\mathrm{obs}}}) = 2w \int_{\sqrt{q_{\mu,\mathrm{obs}}}}^{\infty} \frac{1}{\sqrt{2\pi}} e^{-u^2/2}\, du = 2w(1 - \Phi(\sqrt{q_{\mu,\mathrm{obs}}})) \,. \tag{26}$$

Combining this with Eq. (14) for the significance $Z$ gives

$$Z = \Phi^{-1}(1 - 2w(1 - \Phi(\sqrt{q_{\mu,\mathrm{obs}}}))) \,. \tag{27}$$

In the usual case where the weights of the chi-square and delta-function terms are equal, i.e., $w = 1/2$, Eq. (27) reduces to to the simple formula

$$Z = \sqrt{q_{\mu,\mathrm{obs}}} \,. \tag{28}$$

### 2.5 Approximate methods

To determine the discovery significance or to set limits using a given data set, one must carry out the global fit described above. For this one needs first to combine the likelihood functions for the individual channels into the full likelihood function containing a single strength parameter $\mu$, and use this to find the profile likelihood ratio. It is possible, however, to find approximate values for the *median* discovery significance and limits in a way that only requires as input the separate profile likelihood ratio values from each of the channels. This is very useful especially in the planning phase of a search that combines multiple channels.

The procedure relies on two separate approximations. First, we estimate the median value of the profile likelihood ratio $\lambda(\mu)$ by evaluating the likelihood function with a single, artificial data set in which all statistical fluctuations are suppressed, as described in Section 2.5.1. Second, to determine the significance values from the likelihood ratios, we use the asymptotic form of the distribution of $-2\ln\lambda(\mu)$ valid for sufficiently large data samples. This is described in Section 2.5.2, and its validity is checked for the individual channels in Section 3. Here the limitations of the approximation are investigated and for one case where it is found to be insufficiently accurate (the discovery significance for the channel $H \to W^+W^-$ plus no jets), an alternate procedure is followed.

### 2.5.1 Approximation for the median likelihood ratio

To find the median discovery significance and limits, the median likelihood ratios $\lambda_i(\mu)$ are first found for each channel separately, and then combined to give the full median likelihood ratio. One can estimate the median $\lambda_i(\mu)$ by the value of the likelihood ratio evaluated with a single artificially constructed data set, in which all statistical fluctuations are suppressed and the data values **n** and **m** are replaced by their expectation values for a given integrated luminosity and a hypothesized strength parameter $\mu_{\mathrm{A}}$. We refer to this as an 'Asimov' data set.[2] It replaces having to simulate a large number of experiments from which one would determine the median.

As before, $\mu_A = 0$ is the background only hypothesis and $\mu_{\mathrm{A}} = 1$ corresponds to background plus signal present at the Standard Model rate. The median referred to thus pertains to what one would obtain with a large number of experiments generated under the assumption of $\mu_{\mathrm{A}}$. The approximation is in fact more accurate if one uses noninteger values for numbers of events in the log-likelihood (the factorial terms are in any case absent) so that the Asimov likelihood $L_A$ is found by substituting

$$n_j = \mu_A s_j + b_j \tag{29}$$

$$m_k = u_k \, , \tag{30}$$

into the likelihood function (6) for each channel. Here for $s_j$, $b_j$ and $u_k$, one needs in principle the expectation values, i.e., these quantities should have no statistical errors. In practice they are estimated using a Monte Carlo sample corresponding to an integrated luminosity substantially larger than what is considered for the data. The numbers of signal and background events are then scaled to the desired luminosity. The other nuisance parameters such as shape parameters are estimated as would be done with any other data set; we refer below to the resulting values as $\theta_{\mathrm{A}}$. Because the Asimov data set has no statistical fluctuations, the $\theta_{\mathrm{A}}$ are simply the values one would derive from a very large Monte Carlo data sample.

The estimate of the median likelihood ratio used for the *i*th channel is therefore

$$\lambda_{A,i}(\mu) = \frac{L_{A,i}(\mu, \hat{\hat{\theta}})}{L_{A,i}(\hat{\mu}, \hat{\theta})} \approx \frac{L_{A,i}(\mu, \hat{\hat{\theta}})}{L_{A,i}(\mu_{\mathrm{A}}, \theta_{\mathrm{A}})} \, , \tag{31}$$

where $L_{A,i}$ denotes the likelihood function (6) evaluated with the Asimov data values (29) and (30). The approximation used for the final step in (31) exploits the fact that ML estimate of $\hat{\mu}$ is very close to the input value $\mu_A$ when the likelihood function is constructed using the Asimov data set.

Note that if the likelihood functions for the individual channels were to be constructed with data containing statistical fluctuations rather than with the artificial Asimov data, then the ML estimate of the strength parameter, $\hat{\mu}$, would in general be different for each channel. The full likelihood function (8) used for the combination, however, contains a single global $\mu$. We can now exploit the fact that for the Asimov data one has $\hat{\mu} \approx \mu_A$ for all of the channels and thus obtain the median likelihood ratio for the combination as the product of the individual $\lambda_{A,i}(\mu)$,

$$\lambda_A(\mu) = \prod_i \lambda_{A,i}(\mu) \, . \tag{32}$$

Monte Carlo studies show that Eq. (32) provides an excellent approximation to the median value one would find from data generated with $\mu_{\mathrm{A}}$ as the strength parameter.

[2] The name of the Asimov data set is inspired by the short story *Franchise*, by Isaac Asimov [1]. In it, elections are held by selecting a single voter to represent the entire electorate.

For purposes of quantifying how likely we are to discover the Higgs if it exists, we report the significance obtained from the $p$-value of $\mu = 0$ with an Asimov data set that corresponds to $\mu_A = 1$,

$$\lambda_{s+b}(0) = \prod_i \frac{L_{s+b,i}(0, \hat{\hat{\theta}})}{L_{s+b,i}(1, \theta_\mathrm{A})} . \tag{33}$$

Here the subscript $s+b$ refers to the Asimov data set; the argument 0 denotes the value of $\mu$ being tested. That is, Eq. (33) approximates what one would obtain with data generated with signal and background for the median value of $\lambda(0)$, which is used to test the background-only ($\mu = 0$) hypothesis. Equation (33) provides the median $q_{0_\mathrm{med}} = -2\ln\lambda_{s+b}(0)$, and from this the $p$-value and significance $Z$ are found using equation (15).

To determine the limits on $\mu$ that we expect to set if the Higgs does not exist (or is beyond our reach), we find the $p$-value of a hypothesized $\mu$ using the likelihood ratio $\lambda(\mu)$ based on Asimov data for background only ($\mu_A = 0$),

$$\lambda_b(\mu) = \prod_i \frac{L_{b,i}(\mu, \hat{\hat{\theta}})}{L_{b,i}(0, \theta_\mathrm{A})} . \tag{34}$$

That is, $\lambda_b(\mu)$ approximates the median value of $\lambda(\mu)$ one would obtain from data generated according to the background-only hypothesis. The value of $\lambda_b(\mu)$ is used to determine the median $q_{\mu_\mathrm{med}}$, which is used to find the median $p$-value, $p_{\mu_\mathrm{med}}$. This is computed for all $\mu$ and the point where $p_{\mu_\mathrm{med}} = 0.05$ gives the 95% CL upper limit.

Because $\hat{\mu} \approx \mu_\mathrm{A}$ holds for each channel individually when using Asimov data, it is possible to determine the values of the likelihood ratio entering into (32) separately for each channel, which simplifies greatly the task of estimating the median significance that would result from the full combination. It should be emphasized, however, that the discovery significance or exclusion limits determined from real data require one to construct the full likelihood function containing a single parameter $\mu$, and this must be used in a global fit to find the profile likelihood ratio.

Furthermore, some systematic errors, e.g., the uncertainty in the integrated luminosity, are common to all channels and correspond to a common nuisance parameter. When using Asimov data, the values of such parameters will be fitted to the same values in all channels. Thus the correlations between common systematics are taken into account just as they would be in a global fit of all channels.

A limitation of the procedure with Asimov data is that it only provides an estimate of the median likelihood ratio. To obtain an uncertainty band on the expected (median) discovery and exclusion sensitivities as a function of $m_H$ one would have to simulate a large number of experiments.

### 2.5.2 Approximate relation between likelihood ratio and significance

To compute the $p$-values we need the distribution $f(q_\mu|\mu)$ of $q_\mu = -2\ln\lambda(\mu)$. For a sufficiently large data sample the pdf of $q_\mu$ takes on a well defined limiting form related to the chi-square distribution, as discussed in Section 2.4. Assuming this form, the $p$-value of the hypothesis $\mu$ is found to be

$$p_\mu \approx 1 - \Phi(\sqrt{q_\mu}) , \tag{35}$$

and the significance $Z$ is given by the formula

$$Z = \Phi^{-1}(1 - p_\mu) \approx \sqrt{-2\ln\lambda(\mu)} . \tag{36}$$

For estimating the median discovery significance we use equations (35) and (36) together with the equation (33), the likelihood ratio based on Asimov data containing signal and background. To find the

median limit on $\mu$, we use again equations (35) and (36) but with the likelihood ratio based on Asimov background data only, equation (34).

The validity of this approximation is investigated for each channel by generating distributions of $q_\mu$ for $\mu = 0, 1$ using a fast Monte Carlo simulation and comparing the resulting histograms with the expected asymptotic form. These comparisons are shown in Section 3.

### 2.6 Consequences of testing many Higgs mass values

The statistical significance of a potential discovery is quantified by giving the *p*-value of the no-Higgs hypothesis, i.e., the probability, under the assumption of background only, that that one would see data with equal or less compatibility with this hypothesis relative to the data obtained. Finding this probability below a specified threshold (e.g., the $5\sigma$ threshold, or $p < 2.87 \times 10^{-7}$) corresponds to claiming discovery of a Higgs boson.

The approach taken in this analysis is to compute the *p*-value of the no-Higgs hypothesis separately as a function of the Higgs mass. The threshold *p*-value is thus the false discovery rate for Higgs boson of a given mass. Further one should also estimate the probability, under the assumption of background only, that this *p*-value will fall below the discovery threshold for *any* mass within the range considered. By searching for the Higgs within a broad range of hypothetical masses, one increases the probability of observing what appears to be a signal at some mass, and so the effective significance of the discovery is reduced. In HEP this is sometimes referred to as the "look-elsewhere effect".

To first approximation the effective increase in the false-discovery rate is given by the number of statistically separate mass ranges explored. If a certain data set would give a *p*-value of $m_\mathrm{H}$ below the discovery threshold, then the same data would in general also indicate discovery for other masses very close by. Roughly speaking, the mass range in which a given data set would indicate discovery is set by the mass resolution for the Higgs candidate. So the factor by which the *p*-value is inflated is given by the mass range explored divided by the average mass resolution. Monte Carlo studies can be used to validate and refine this approximation; this approach is planned for future analyses.

An alternative to considering fixed Higgs masses is to treat both the strength parameter $\mu$ and the Higgs mass $m_\mathrm{H}$ as free parameters in the likelihood ratio. For example, to establish discovery one computes the *p*-value of the no-Higgs hypothesis. As before, this is the probability, under the assumption of no Higgs, of finding data with equal or lesser compatibility with $\mu = 0$ relative to the data obtained. In contrast to the fixed-mass case, however, "less compatible" here means having a lower likelihood ratio for any allowed value of the Higgs mass; the lowest value comes when the denominator contains the fitted maximum-likelihood estimator $\hat{m}_\mathrm{H}$. In practice the fitted value of the Higgs mass is restricted to lie within a stated range. This has been done for the Higgs searches using the $\gamma\gamma$ [4] and $\mathrm{W}^+\mathrm{W}^-$ [5] channels, with the aim of extending this method to a combination of all channels.

## 3 Combination of Higgs search channels

In this section a brief description of each of the four search channels is given. For each channel, the method used to obtain the likelihood ratio is described, and values of the test variable $q_\mu$ as defined by Eq. (10) for discovery and by Eq. (16) for limits are tabulated for several values of the integrated luminosity $L$ and Higgs mass $m_\mathrm{H}$. For the discovery sensitivity where one tests $\mu = 0$, the median value of $q_0$ is given under the assumption of $\mu = 1$; for exclusion sensitivity, the median of $q_1$ is given under the assumption of $\mu = 0$.

In addition, for each channel we show distributions of $q_\mu$ under the assumption of $\mu$ for the two cases $\mu = 0$ and $\mu = 1$. For the approximations used in this note to be valid, these should be close to the asymptotic form described in Section 2.4. This limiting form for the distribution $f(q_\mu|\mu)$ is a mixture of

a delta function at $q_\mu = 0$ and a chi-square distribution for one degree of freedom, where each component has equal weight. We refer to this as a $\frac{1}{2}\chi_1^2$ distribution. For the case of $\mu = 0$ we compare directly the Monte Carlo distribution of $q_0$ with the $\frac{1}{2}\chi_1^2$ distribution and the delta-function term at $q_0 = 0$ is clearly visible. For $\mu = 1$ we show the equivalent comparison but for reasons of convenience only the events with $\hat{\mu} \leq \mu$ are shown, i.e., the events with $\hat{\mu} > \mu$ that contribute to the delta function at $q_1 = 0$ are left off. That is, for exclusion it is the conditional pdf $f(q_1|\mu = 1, \hat{\mu} \leq 1)$ that is compared to a chi-square distribution for one degree of freedom; there is no delta function term.

All channels use data driven background estimation methods. This way, the uncertainties in the background shape and normalization are treated within the framework of the profile likelihood as nuisance parameters. Using control samples the effect of many uncertainties like energy scales and fake rates on the background estimate can be constrained by the control samples. Uncertainties on the signal efficiency do not affect the discovery sensitivity which is testing the presence or absence of a signal; however, this is not the case for exclusion sensitivity. As one would expect, uncertainty in the signal efficiency does reduce the exclusion sensitivity. This uncertainty was incorporated into the profile likelihood calculation by adding an extra term to the likelihood function for every channel as described in Section 2.1: a Gaussian relating the nominal efficiency estimated in an auxiliary measurement, the true efficiency, and the uncertainty of that auxiliary measurement.

### 3.1 $H \to \gamma\gamma$

Details on the $H \to \gamma\gamma$ channel are given in Ref. [4]. We perform an unbinned maximum-likelihood fit to extract the signal and background event yields by using the diphoton invariant mass, $m_{\gamma\gamma}$, as a discriminating variable. The $H \to \gamma\gamma$ distribution of $m_{\gamma\gamma}$ forms a Gaussian peak with tails to lower values from photon energy losses before the calorimeter. It is well modelled by a *Crystal Ball* function. The signal probability density function, $p_H(m_{\gamma\gamma})$, is given by

$$p_H(m_{\gamma\gamma}) = N \cdot \begin{cases} \exp\left(-t^2/2\right), & \text{for } t > -\alpha\,, \\ (n/|\alpha|)^n \cdot \exp\left(-|\alpha|^2/2\right) \cdot (n/|\alpha| - |\alpha| - t)^{-n}, & \text{otherwise}, \end{cases} \tag{37}$$

where $t = (m_{\gamma\gamma} - m_H - \delta m_H)/\sigma(m_{\gamma\gamma})$, $N$ is a normalisation parameter, $m_H$ is the Higgs boson mass, $\delta m_H$ is an offset and $\sigma$ represents the diphoton invariant mass resolution. The non-Gaussian tail is parametrised by $n$ and $\alpha$. We include an additional, broader Gaussian term in Eq. 37 to improve the description of the tails of the distribution. Within a sufficiently narrow mass window, the background of $m_{\gamma\gamma}$ is modeled by an exponential distribution with a single slope parameter $\xi$.

The resulting median profile likelihood ratios for discovery, $\lambda(\mu = 0)$ (using toy $s+b$ Monte Carlo experiments and taking the median of the $\lambda(\mu = 0)$ distribution ) are given in Table 1 for a few Higgs masses at some given luminosities.

The distribution of the test statistic $q_0$ under the null background only hypothesis, for $m_H = 120$ GeV with an integrated luminosity of 2 and 10 fb$^{-1}$, is shown in Fig. 6. A $\frac{1}{2}\chi_1^2$ distribution is superimposed, showing the validity of the asymptotic approximation.

The median profile likelihood ratio for exclusion, $-2\ln\lambda(\mu)$ (using toy background-only Monte Carlo experiments and taking the median of the $\lambda(\mu)$ distribution) is given in Table 2 for a few Higgs masses at several integrated luminosities and for a signal strength $\mu = 1$, corresponding to a Standard Model Higgs Boson.

The distribution of the test statistic $q_1$ for $\hat{\mu} \leq 1$ under the $s+b$ hypotheses for $m_H = 150$ GeV with an integrated luminosity of 2 and 10 fb$^{-1}$ is shown in Figures 7. A $\chi_1^2$ distribution is superimposed, showing the validity of the asymptotic approximation.

Table 1: Median values of $-2\ln\lambda(\mu)$ (evaluated at $\mu = 0$) obtained from fits to simulated data generated with $H \to \gamma\gamma$ signal plus background ($\mu = 1$) for several values of the Higgs mass and integrated luminosity.

| L | $m_H$ (GeV) | | | |
|---|---|---|---|---|
| ($fb^{-1}$) | 115 | 120 | 130 | 140 |
| 1 | 0.35 | 0.55 | 0.75 | 0.67 |
| 2 | 0.75 | 1.07 | 1.45 | 0.95 |
| 5 | 1.95 | 2.95 | 3.65 | 2.55 |
| 10 | 3.95 | 5.86 | 7.35 | 5.05 |
| 30 | 11.85 | 17.72 | 21.99 | 15.05 |

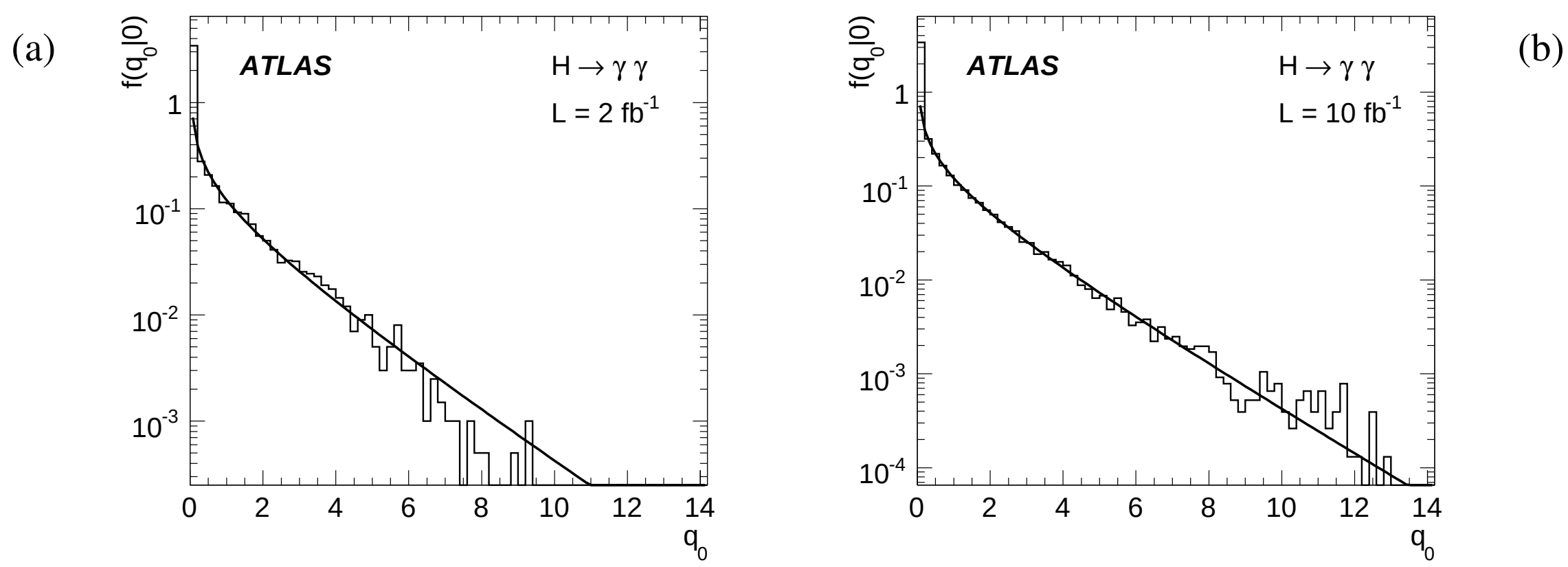


Figure 6: The distribution of the test statistic $q_0$ (for $H \to \gamma\gamma$), under the null background only hypothesis, for $m_H = 120$ GeV with an integrated luminosity of 2 (a) and 10 (b) $fb^{-1}$. A $\frac{1}{2}\chi_1^2$ distribution is superimposed.

Table 2: Median values of $-2\ln\lambda(\mu)$ (evaluated at $\mu = 1$) obtained from $H \to \gamma\gamma$ background-only ($\mu = 0$) simulated data for several values of the Higgs mass and integrated luminosities.

| L | $m_H$ (GeV) | | | |
|---|---|---|---|---|
| ($fb^{-1}$) | 115 | 120 | 130 | 140 |
| 1 | 0.72 | 0.73 | 0.91 | 0.67 |
| 2 | 0.97 | 1.21 | 1.39 | 0.97 |
| 5 | 2.11 | 2.59 | 3.13 | 2.23 |
| 10 | 3.55 | 4.87 | 5.71 | 4.04 |
| 30 | 8.47 | 10.50 | 11.63 | 9.00 |

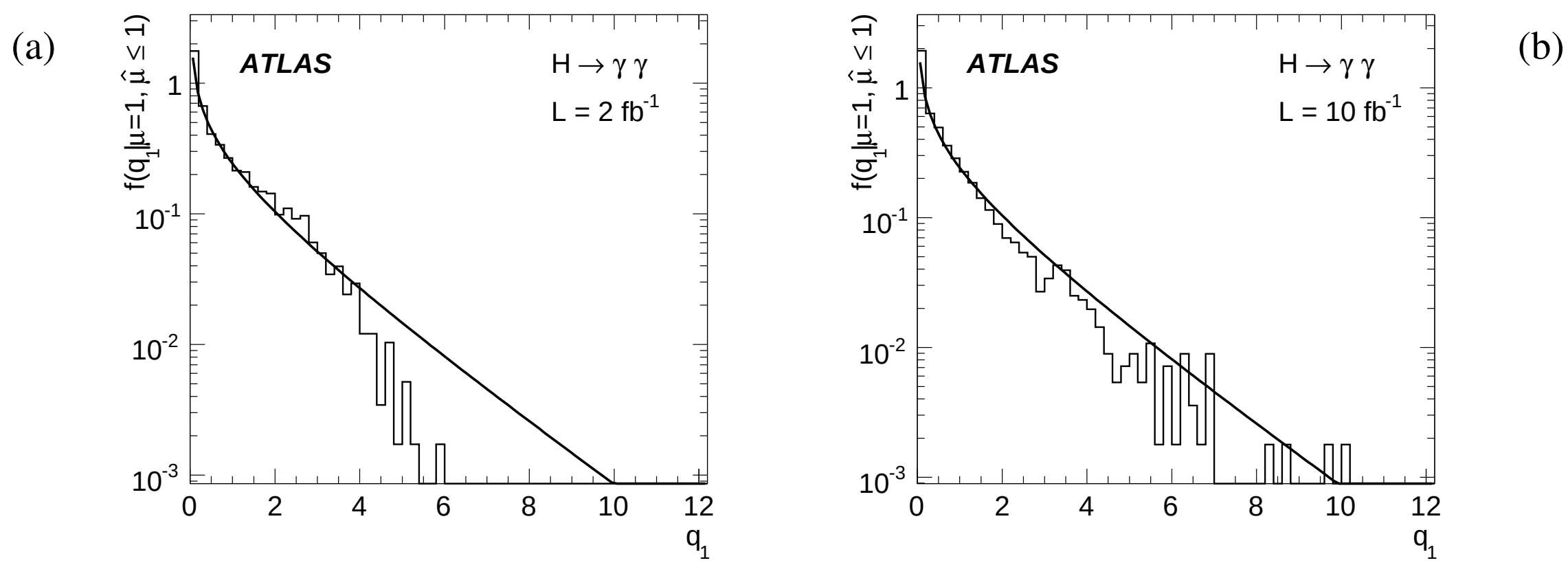


Figure 7: The distribution of the test statistic $q_1$ for $\hat{\mu} \leq 1$ under the $s+b$ hypothesis (for $H \rightarrow \gamma\gamma$), for $m_H = 120$ GeV with an integrated luminosity of (a) 2 fb$^{-1}$ and (b) 10 fb$^{-1}$. A $\chi^2_1$ distribution is superimposed.

### 3.2 $H \rightarrow W^+W^-$

The $H \rightarrow W^+W^-$ search is divided into two topologies, production of a Higgs with no jets ($H+0j$) and with two additional jets ($H+2j$), using in both cases the decay mode $H \rightarrow WW \rightarrow e\nu\mu\nu$. The present study does not yet consider the final states $e\nu e\nu$ or $\mu\nu\mu\nu$, nor those with hadronic W decays. Future inclusion of these channels is expected to improve the search sensitivity particularly for the high Higgs mass region. The search is described in detail in Ref. [5].

#### 3.2.1 $H+0j$

The analysis of the $H+0j$ channel uses a two dimensional maximum-likelihood fit of the transverse mass and the transverse momentum of the WW system in two bins of the dilepton opening angle in the transverse plane. The fit includes control samples to measure the backgrounds from $t\bar{t}$ and Z $\rightarrow \tau\tau$.

The QCD WW background requires particular attention. Its distributions of Higgs-candidate transverse mass and $p_{\mathrm{T}}$ are described with functions containing several adjustable (nuisance) parameters, and several others whose values are determined from a full Monte Carlo simulation and thereafter treated as fixed. The distribution of the test statistic $q_0$ under the background-only ($\mu = 0$) hypothesis is shown in Fig. 8(a) for $m_H = 150$ GeV for an integrated luminosity of 10 fb$^{-1}$. The same fixed QCD WW shape parameters are used both to generate the data and for calculating the likelihood ratio. A $\frac{1}{2}\chi^2_1$ distribution is superimposed, showing the level of agreement of the asymptotic approximation.

For this channel, further investigation of the systematic uncertainties was carried out. For the fixed shape parameters related to $p_T$ and transverse mass distributions for the QCD WW background, the values used to generate the data were varied relative to what was used when determining the likelihood ratio. This was done in a manner that minimized the sensitivity of the resulting $q_0$ distribution to variations in other fixed parameters such as the QCD $Q^2$ scale. The resulting distributions of $q_0$ are thus no longer expected to follow the $\frac{1}{2}\chi^2_1$ form, as can be seen in Fig. 8(b).

Because the chi-square approximation is not valid in this case, the $p$-values are calculated using the $q_0$ distribution obtained directly from the Monte Carlo. An exponential is fitted to the tail region in order to extrapolate to large $q_0$ values, and the median value of $q_0$ under the hypothesis of signal plus background is determined using the same variation of the background parameters. It was found that the median $p$-value of the background-only hypothesis, with the median computed under assumption of the $s+b$ hypothesis, is very similar to the original case where the QCD shape parameters are not varied and the $\frac{1}{2}\chi^2_1$ distribution is used.

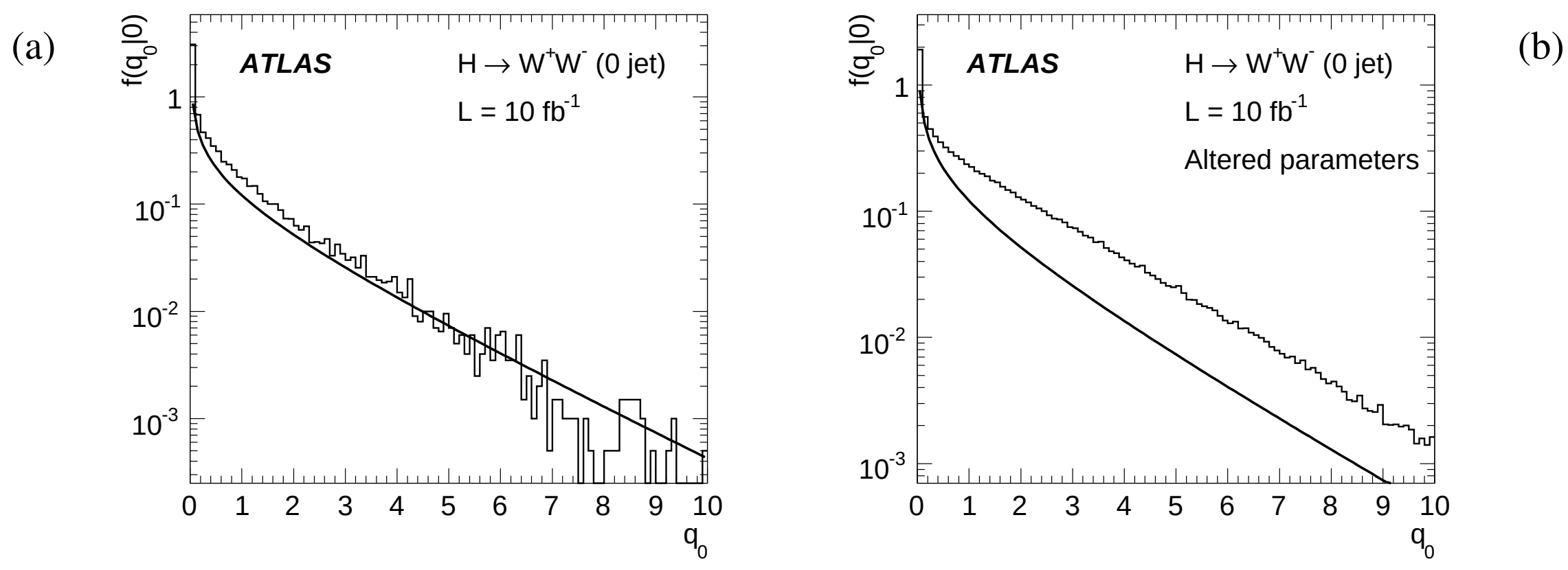


Figure 8: The distribution of the test statistic $q_0$ for $H+0j \rightarrow WW+0j$, under the background-only hypothesis, with the same fixed QCD WW shape parameters used at both the generator and the fit level, for $m_H = 150$ GeV and for an integrated luminosity of 10 fb$^{-1}$ (a) with the same shape parameters for event generation and fitting; (b) with altered shape parameters. A $\frac{1}{2}\chi_1^2$ distribution is superimposed.

For the combination of results for discovery (i.e., testing $\mu = 0$), we have used the $p$-values as described above for the case of the $H+0j$ channel. The same variation of QCD WW parameters was also investigated for the case of exclusion (i.e., testing $\mu > 0$, and in particular $\mu = 1$), and it was done as well for the $H+2j$ channel. In those studies, however, the distributions, under the assumption of $\mu$, of the test statistic $q_\mu$ were found to agree quite well with the expected $\frac{1}{2}\chi_1^2$ distribution, even after the parameter variation. Therefore for these cases we have based the combination of results on the asymptotic approximations for the $q_\mu$ distributions (as done in this paper for the other Higgs channels).

To simplify the comparison with the other channels, the median $p$-values of the background-only ($\mu = 0$) hypothesis for the $H+0j$ channel were converted into effective values of the variable $q_0 = -2\ln\lambda(0)$ according to $q_0 = Z^2 = \left(\Phi^{-1}(1-p)\right)^2$. These are given in Table 3 for several Higgs masses and integrated luminosities.

Table 3: Median values of $-2\ln\lambda(\mu = 0)$ obtained from $H+0j$ fits using data simulated under the assumption of signal plus background ($\mu = 1$) for several values of the Higgs mass and integrated luminosity.

| $L$ | $m_H$ (GeV) | | | | | | |
|---|---|---|---|---|---|---|---|
| (fb$^{-1}$) | 130 | 140 | 150 | 160 | 170 | 180 | 190 |
| 1 | 1.64 | 4.30 | 9.56 | 16.52 | 15.39 | 7.37 | 3.17 |
| 2 | 2.87 | 8.60 | 19.36 | 34.67 | 30.58 | 13.73 | 5.63 |
| 5 | 6.55 | 20.15 | 42.77 | 74.39 | 63.58 | 31.54 | 13.54 |
| 10 | 11.52 | 33.27 | 70.67 | 113.33 | 103.44 | 51.06 | 22.78 |

The median profile likelihood ratio for exclusion, $\lambda(\mu)$ (using background-only MC experiments and taking the median of the $\lambda(\mu)$ distribution), is given in Table 4 for several Higgs masses and integrated luminosities for the signal strength $\mu = 1.0$, corresponding to a SM Higgs Boson.

The distribution of the statistic $q_1$ for $\hat{\mu} \leq 1$ under the $s+b$ hypothesis is shown in Fig. 9 for $m_H = 150$ GeV and for an integrated luminosity of 2 and 10 fb$^{-1}$. A $\chi_1^2$ distribution is superimposed, showing the validity of the asymptotic approximation.

Table 4: Median values of $-2\ln\lambda(\mu = 1)$ obtained from $H+0j$ fits using simulated background-only ($\mu = 0$) data for several values of the Higgs mass and integrated luminosity.

| $L$ | $m_H$ (GeV) | | | | | | |
|---|---|---|---|---|---|---|---|
| (fb$^{-1}$) | 130 | 140 | 150 | 160 | 170 | 180 | 190 |
| 1 | 1.37 | 3.93 | 8.69 | 14.83 | 14.23 | 7.26 | 3.26 |
| 2 | 2.57 | 7.47 | 15.59 | 25.20 | 23.65 | 12.91 | 5.84 |
| 5 | 5.85 | 15.26 | 30.05 | 45.60 | 41.13 | 25.41 | 12.02 |
| 10 | 10.01 | 24.69 | 45.24 | 62.13 | 57.69 | 37.42 | 20.03 |

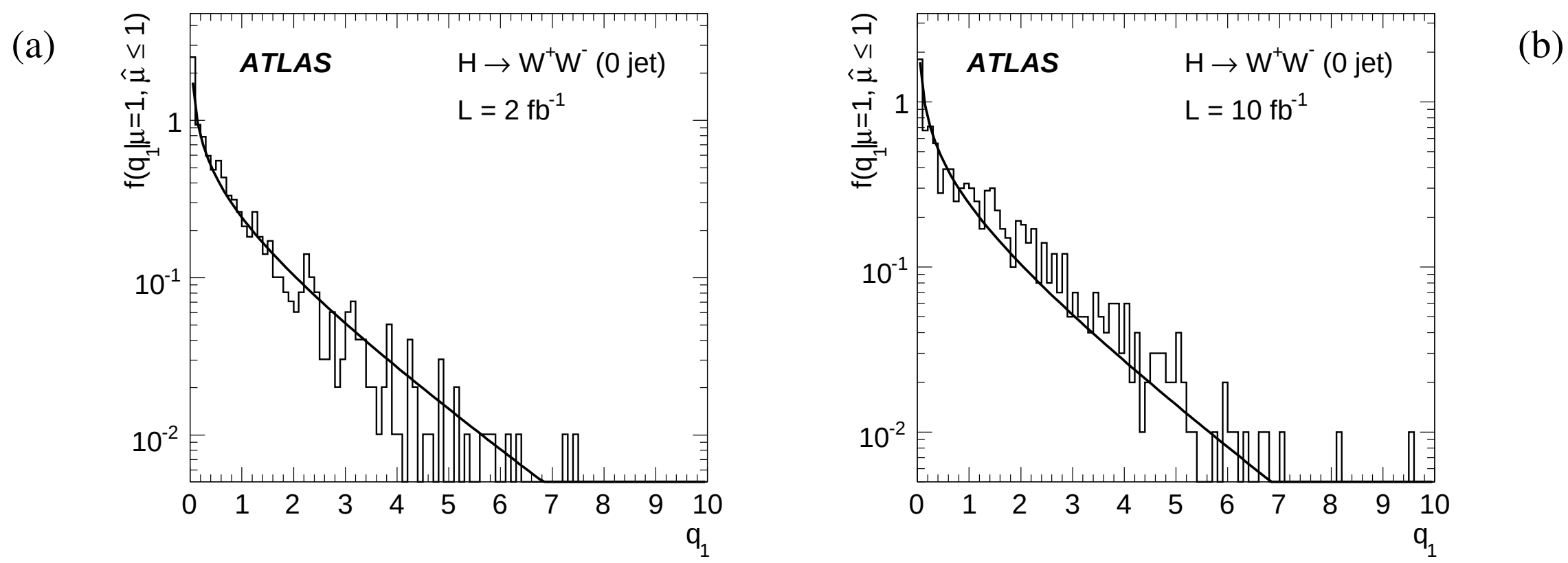


Figure 9: The distribution of the test statistic $q_1$ for $\hat{\mu} \leq 1$ under the $s+b$ hypothesis for $H+0j \to WW+0j$, for $m_H = 150$ GeV with an integrated luminosity of (a) $2\,\text{fb}^{-1}$ and (b) $10\,\text{fb}^{-1}$. A $\chi^2_1$ distribution is superimposed.

### 3.2.2 $H+2j$

The $H+2j$ analysis uses a two-dimensional fit based on the transverse mass and the output of a Neural Network, which takes as input several kinematic variables related to the jet activity in the event. The fit is performed simultaneously in signal-enriched and background-enriched regions distinguished by lepton angular variables, which are nearly uncorrelated to the jet variables used in the Neural Network.

The median profile likelihood ratios for discovery, $-2\ln\lambda(\mu=0)$ (using toy $s+b$ MC experiments and taking the median of the $\lambda(\mu=0)$ distribution), are given in Table 5 for several Higgs masses and integrated luminosities.

Table 5: Median values of $-2\ln\lambda(\mu)$ for $\mu=0$ obtained from $H+2j$ fits to simulated data with signal plus background ($\mu=1$) for several values of the Higgs mass and integrated luminosity.

| $L$ | $m_H$ (GeV) | | | | | | |
|---|---|---|---|---|---|---|---|
| (fb$^{-1}$) | 130 | 140 | 150 | 160 | 170 | 180 | 190 |
| 1 | 0.39 | 0.89 | 1.61 | 2.73 | 2.89 | 1.85 | 1.23 |
| 2 | 0.70 | 1.75 | 3.22 | 5.34 | 5.67 | 3.66 | 2.07 |
| 5 | 2.01 | 5.29 | 8.87 | 14.20 | 14.58 | 9.22 | 5.71 |
| 10 | 3.82 | 9.14 | 16.56 | 26.35 | 26.05 | 16.68 | 9.93 |

The distribution of the statistic $q_0$ under the background-only hypothesis is shown in Fig. 10 for $m_H = 150$ GeV and for an integrated luminosity of 2 and 10 fb$^{-1}$. A $\frac{1}{2}\chi_1^2$ distribution is superimposed, showing the validity of the asymptotic approximation.

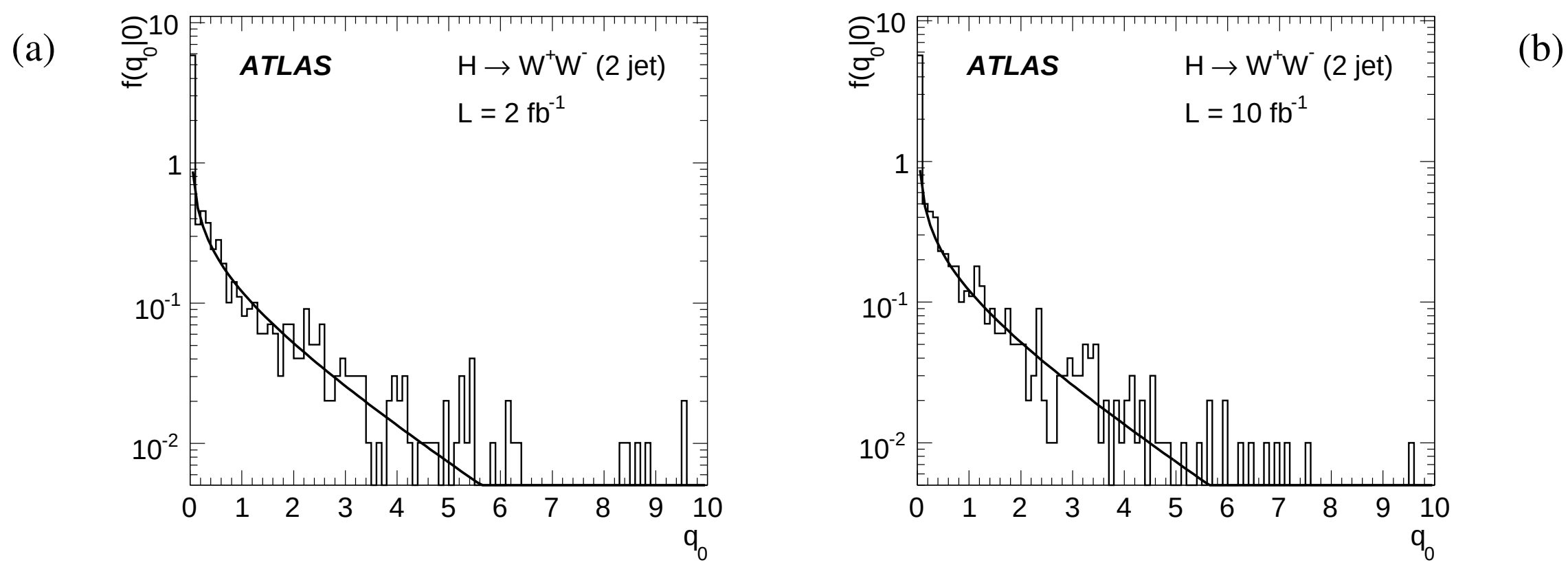


Figure 10: The distribution of the test statistic $q_0$ (for $H+2j \to WW+2j$), under the null background only hypothesis, for $m_H = 150$ GeV and for an integrated luminosity of 2 (a) and 10 (b) $fb^{-1}$. A $\frac{1}{2}\chi_1^2$ distribution is superimposed.

The median profile likelihood ratio for exclusion, $-2\ln\lambda(\mu)$ with $\mu=1$, where the median is computed using background-only MC data, is given in Table 6 for several Higgs masses and integrated luminosities.

The distribution of the test statistic $q_1$ for $\hat{\mu} \leq 1$ under the $s+b$ hypothesis is shown in Figures 11 for $m_H = 150$ GeV with an integrated luminosity of 2 and 10 fb$^{-1}$. A $\chi_1^2$ distribution is superimposed, showing the validity of the asymptotic approximation.

Table 6: Median values of $-2\ln\lambda(\mu)$ for $\mu = 1$ obtained from $H+2j$ fits to simulated background-only ($\mu = 0$) data for several values of the Higgs mass and integrated luminosity.

| $L$ | $m_H$ (GeV) | | | | | | |
|---|---|---|---|---|---|---|---|
| (fb$^{-1}$) | 130 | 140 | 150 | 160 | 170 | 180 | 190 |
| 1 | 0.37 | 0.75 | 1.31 | 2.29 | 2.44 | 1.99 | 0.95 |
| 2 | 0.87 | 2.49 | 5.06 | 8.18 | 7.74 | 3.85 | 1.90 |
| 5 | 1.40 | 3.13 | 5.86 | 9.58 | 9.90 | 7.89 | 4.59 |
| 10 | 2.80 | 6.56 | 10.72 | 16.14 | 16.62 | 13.94 | 8.74 |

(a) (b)

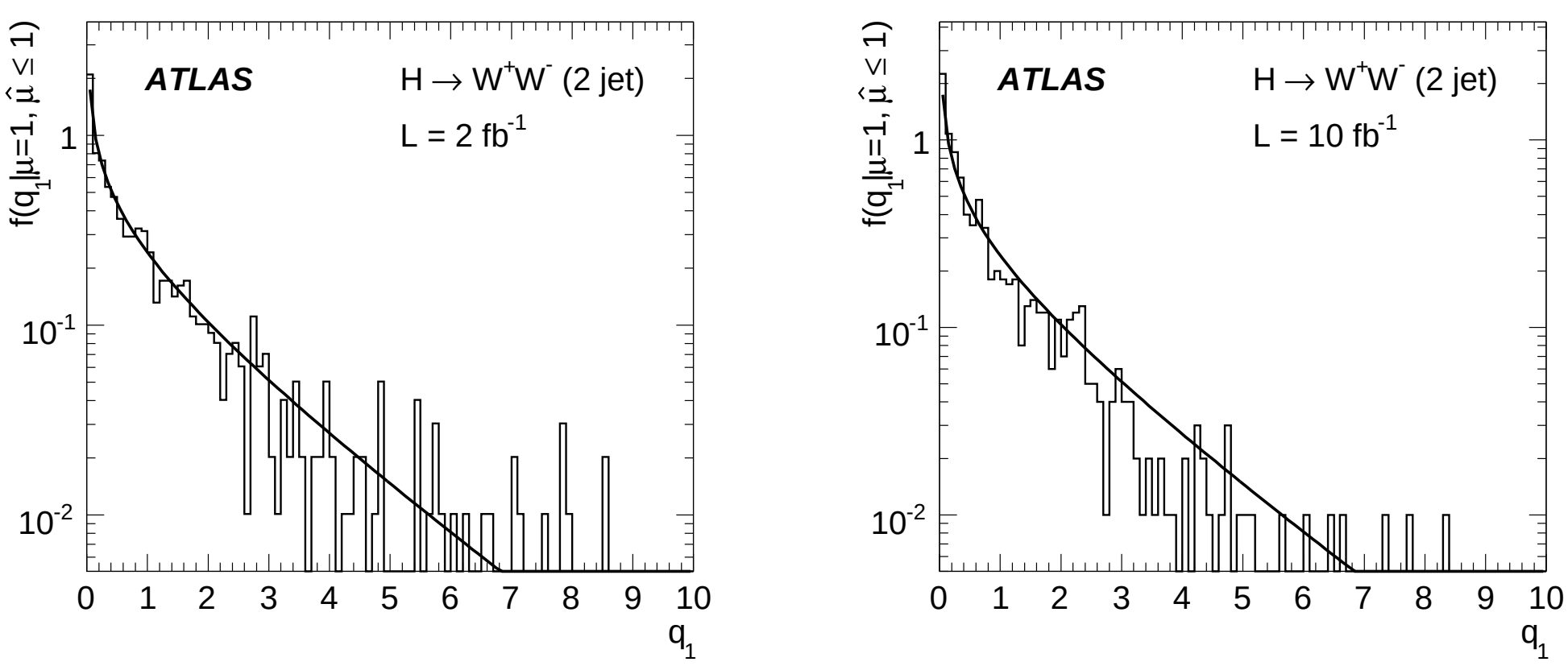


Figure 11: The distribution of the test statistic $q_1$ for $\hat{\mu} \leq 1$ under the $s+b$ ($\mu = 1$) hypothesis (for $H+2j \rightarrow WW+2j$), for $m_H = 150$ GeV with an integrated luminosity of (a) $2\,\mathrm{fb}^{-1}$ and (b) $10\,\mathrm{fb}^{-1}$. A $\chi^2_1$ distribution is superimposed.

### 3.3 $H \to \tau^+\tau^-$

The sensitivity of the ATLAS detector to a Higgs boson produced via Vector Boson Fusion and decaying to tau leptons has been investigated [6].

Two tau decay channels are considered: $ll$ and $lh$. Although there are several neutrinos in the event, it is possible to reconstruct the $\tau^+\tau^-$ invariant mass, $m_{\tau\tau}$, by making the collinear approximation, in which the decay products of the $\tau$ are assumed to be collinear with the $\tau$ direction in the laboratory frame. After other event selection criteria are imposed, the spectrum of $m_{\tau\tau}$ is used to extract the signal. Particular care has been given to the incorporation of uncertainty in both the rate and shape for the signal and backgrounds.

A data-driven background estimation technique has been established for the major backgrounds. Each technique has been developed to address the aspects of the background estimation which are most relevant for the analysis: the shape of the $m_{\tau\tau}$ tail from the irreducible $Z \to \tau\tau$, the fake tau contribution in the $lh$-channel, and the normalization of the QCD backgrounds.

In addition to the $m_{\tau\tau}$ spectrum from the $Z \to \tau\tau$ and QCD control samples, a track multiplicity distribution is used to constrain the fraction of QCD events in the $lh$-channel. This likelihood term is denoted $L_{track}(r_{QCD}, r_{tau})$, where $r_{tau}$ ($r_{QCD}$) denotes the fraction of real taus (fakes from jets) in the sample. The track multiplicity distribution for the QCD jets is modelled from samples of QCD di-jets that produce tau candidates.

The shape of the $m_{\tau\tau}$ distribution for signal and $Z \to \tau\tau$ events is dictated by the resolution of $\not{E}_T$ and the kinematics of the collinear approximation. The parameterization of $m_{\tau\tau}$ for the signal and $Z \to \tau\tau$ background are based on the kinematics of the collinear approximation and reproduces an asymmetric distribution with non-Gaussian tails. This distribution is dependent on the overall width width, $\sigma_{H/Z}$, and a mean, $m_{H/Z}$.

A $Z \to \tau\tau$ control sample is used to constrain the mean, $m_Z$, and the overall width of the distribution, $\sigma_Z$, which are the only free parameters in the $Z \to \tau\tau$ background model. The error bars in the control sample were scaled to 10% to account for the 10% shape uncertainty in the $\mu \to \tau$ rescaling method.

The shapes for $W$+jets and $t\bar{t}$ are very similar and are modelled with a single distribution. A conservative 50% error is applied to each bin in the combined QCD (i.e., $t\bar{t}$ and $W$+jets) control sample to reflect uncertainty in how this shape changes as the remainder of the analysis cuts are applied.

The shape of the QCD background was parametrized with the following equation:

$$L_{QCD}(m_{\tau\tau}|a_1, a_2, a_3) = \mathcal{N} \left( \frac{1}{m_{\tau\tau} + a_1} \right)^{a_2} m_{\tau\tau}^{a_3} \,. \tag{38}$$

The form is motivated by a competition between the parton distribution functions and the matrix element. In the $lh$-channel, the normalization of the backgrounds with fake taus can be constrained by using the track multiplicity method described above. We apply a conservative 50% systematic on this fraction.

By fitting the $m_{\tau\tau}$ spectrum to a model that accurately describes the signal and various backgrounds it is possible to directly incorporate uncertainty in the background shape and take advantage of the shape of the signal within the mass window. We utilize the profile likelihood ratio as our test statistic. The likelihood function corresponding to the simultaneous fit is simply a product of the likelihoods from the individual measurements:

$$\begin{aligned} L(data|\mu, m_H, \nu) &= L_{track}(\text{track multiplicity}|r_{QCD}) \times L_Z(\text{Z+jets control}|m_Z, \sigma_Z) \\ &\times L_{QCD}(\text{QCD control}|a_1, a_2, a_3) \\ &\times L_{s+b}(\text{signal candidates}|\mu, m_H, \sigma_H, m_Z, \sigma_Z, r_{QCD}, a_1, a_2, a_3), \end{aligned} \tag{39}$$

where the $a_i$ are the parameters used to parametrize the QCD background and $\nu$ represents all nuisance parameters of the model: $\sigma_H, m_Z, \sigma_Z, r_{QCD}, a_1, a_2, a_3$.

The data-driven background estimation methods described above have been developed so that uncertainty in the background shape and normalization are included directly into the significance calculation. Because the discovery criterion is simply testing the presence or absence of the signal, it is not sensitive to some of the sources of systematic uncertainty. In contrast, measurement and exclusion of $\sigma(pp \to qqH) \times BR(H \to \tau\tau)$ are sensitive to the uncertainty on the signal selection efficiency. Both experimental and theoretical sources of uncertainty on the signal efficiency have been evaluated. The jet energy scale uncertainty dominates in this channel, and a signal efficiency uncertainty of 18% was used when estimating the exclusion sensitivity.

The median profile likelihood ratios for discovery, $\lambda(\mu = 0)$ (using the Asimov data sets with $\mu_A =$ 1), are given in Table 7 for a few Higgs masses at some given luminosities. The distribution of the test statistic $q_0$ under the null background only hypothesis, for $m_H = 130$ GeV with an integrated luminosity of 2 and $10\,\text{fb}^{-1}$, is shown in Figure 12. A $\frac{1}{2}\chi^2_1$ distribution is superimposed, showing the validity of the asymptotic approximation.

Table 7: Median values of $-2\ln\lambda(\mu = 0)$ obtained from $H \to \tau^+\tau^-$ simulated data generated with signal plus background ($\mu = 1$) for several values of the Higgs mass and integrated luminosity.

| L | $m_H$ (GeV) | | | |
|---|---|---|---|---|
| ($\text{fb}^{-1}$) | 110 | 120 | 130 | 140 |
| 1 | 0.59 | 0.88 | 0.72 | 0.46 |
| 2 | 1.18 | 1.74 | 1.40 | 0.89 |
| 5 | 2.91 | 4.43 | 3.34 | 2.08 |
| 10 | 5.81 | 8.23 | 6.37 | 3.91 |
| 30 | 17.2 | 23.6 | 17.6 | 10.6 |

The resulting median profile likelihood ratio for exclusion, $\lambda(\mu)$ (using the Asimov data sets, $\mu_A =$ 0), is given in Table 8 for a few Higgs masses at some given luminosities and signal strength $\mu = 1.0$ corresponding to a Standard Model Higgs Boson.

The distribution of the test statistic $q_1$ for $\hat{\mu} \leq 1$ under the $s+b$ hypotheses for $m_H = 130\,\text{GeV}$ with an integrated luminosity of 2 and 10 $\text{fb}^{-1}$ is shown in Figures 13. A $\chi^2_1$ distribution is superimposed, showing the validity of the asymptotic approximation.

Table 8: Median values of $-2\ln\lambda(\mu = 1)$ obtained from $H \to \tau^+\tau^-$ background-only ($\mu = 0$) simulated data for several values of the Higgs mass and integrated luminosities.

| L | $m_H$ (GeV) | | | |
|---|---|---|---|---|
| ($\text{fb}^{-1}$) | 110 | 120 | 130 | 140 |
| 1 | 0.64 | 0.71 | 0.52 | 0.32 |
| 2 | 1.31 | 1.41 | 1.03 | 0.64 |
| 5 | 3.30 | 3.42 | 2.52 | 1.56 |
| 10 | 6.93 | 7.79 | 7.18 | 3.48 |
| 30 | 15.3 | 16.6 | 12.8 | 8.48 |

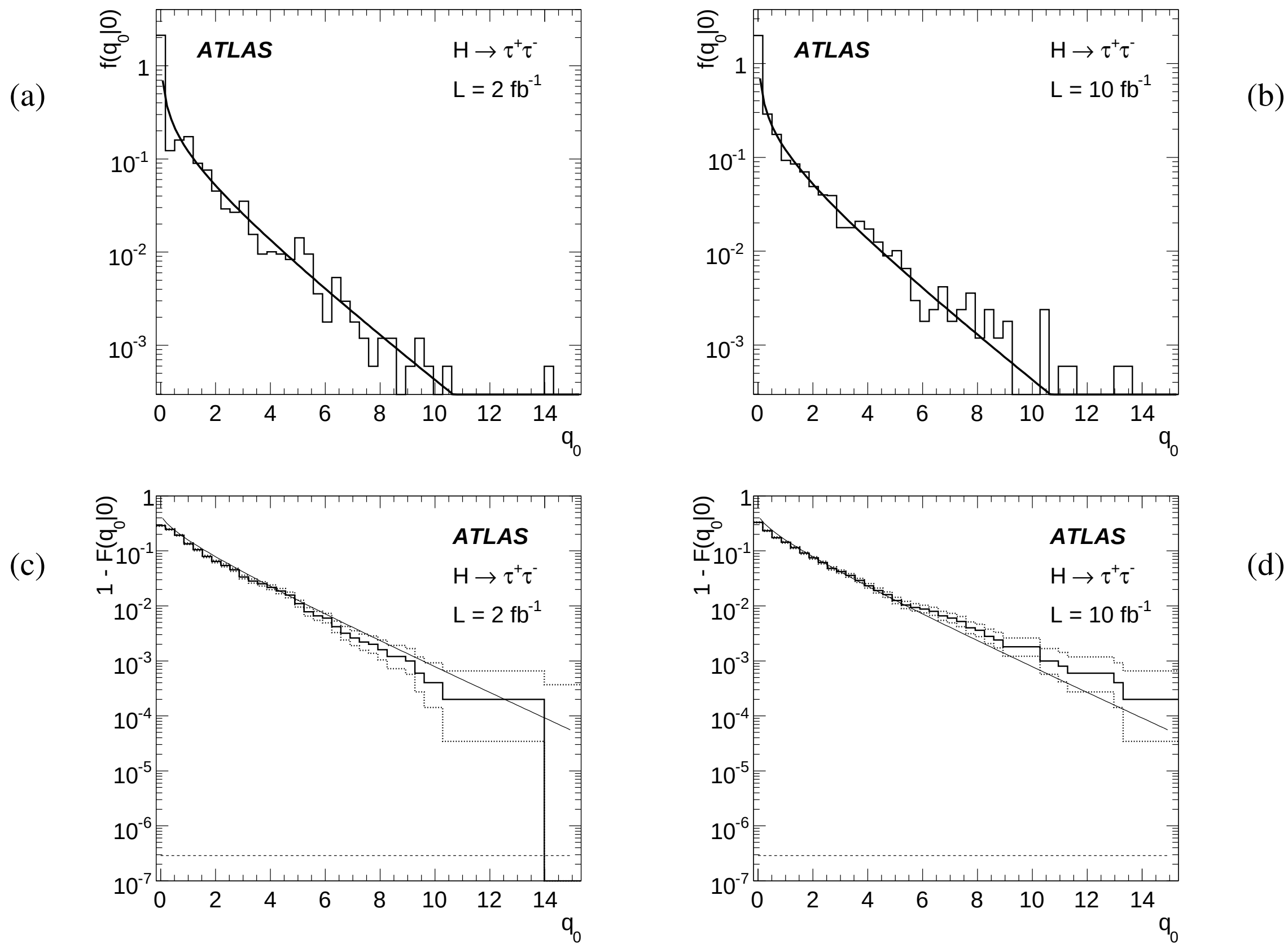


Figure 12: The distribution of the test statistic $q_0$ for $H \to \tau^+\tau^-$ under the null background-only hypothesis, for $m_H = 130\,\mathrm{GeV}$ with an integrated luminosity of 2 (a) and 10 (b) $\mathrm{fb}^{-1}$. A $\frac{1}{2}\chi_1^2$ distribution is superimposed. Figures (c) and (d) show $1 - F(q_0)$ where $F(q_0)$ is the corresponding cumulative distribution. The small excess of events at high $q_0$ is statistically compatible with the expected curves, as can be seen by comparison with the dotted histograms that show the 68.3% central confidence intervals for $p = 1 - F(q_0|0)$. The lower dotted line at $2.87 \times 10^{-7}$ shows the $5\sigma$ discovery threshold.

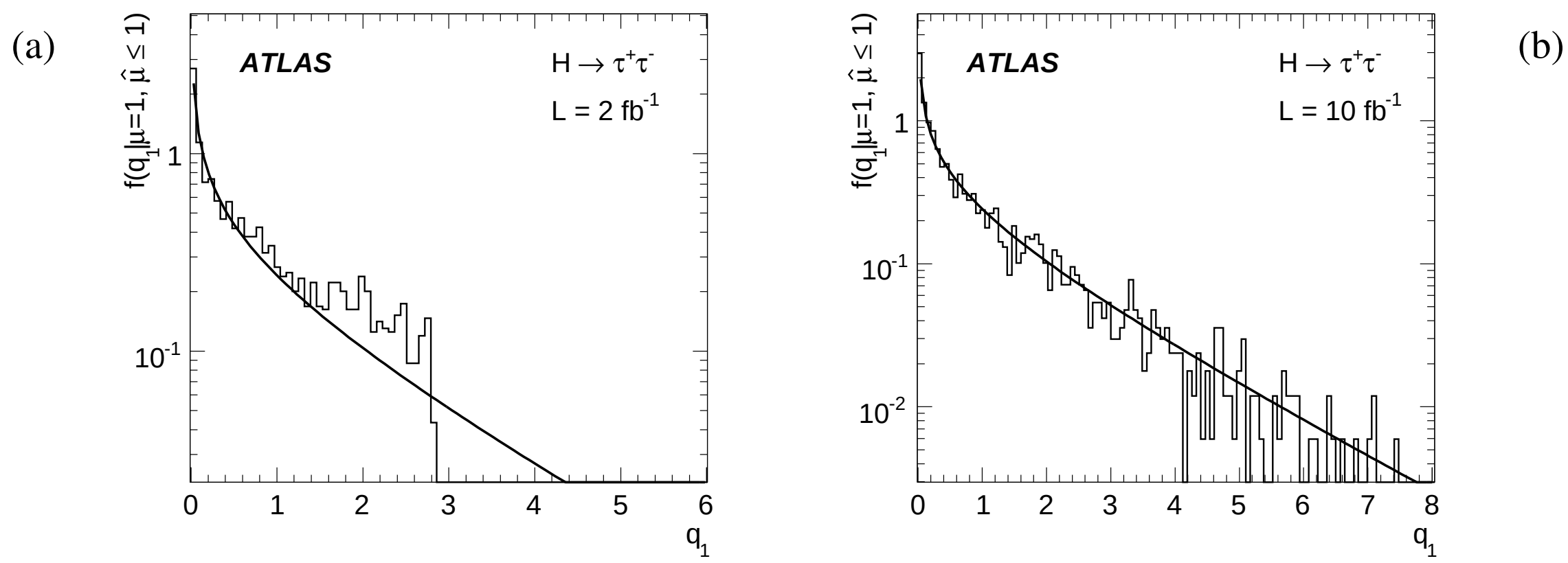


Figure 13: The distribution of the test statistic $q_1$ for $\hat{\mu} \leq 1$ under the $s+b$ hypothesis for $H \to \tau^+\tau^-$, for $m_H = 130\,\mathrm{GeV}$ with an integrated luminosity of (a) 2 $\mathrm{fb}^{-1}$ (b) and 10 $\mathrm{fb}^{-1}$. A $\chi_1^2$ distribution is superimposed.

### 3.4 $H \to ZZ^{(*)} \to 4l$

Details of the $H \to ZZ^{(*)} \to 4l$ channel can be found in Ref. [7]. The main challenge of this channel for what concerns statistical analysis is its relevance over a very wide mass range (from $m_H$ around 120 GeV up to 700 GeV), over which the shapes and cross sections of both signal and background show considerable variations.

While in principle it would be possible to divide the mass range in different regions and use separate models for each of them, using a unique background model for the whole phase space is a better approach, since it allows to estimate discovery and exclusion significance at any mass, without having to worry about boundaries between different models. This will be needed of course when the analysis is performed using real data, where $m_H$ is unknown.

The main background after event filtering in this channel is the irreducible $ZZ \to 4l$ process. Reducible backgrounds such as $Zb\bar{b} \to 4l + X$ or $t\bar{t}$ give a negligible contribution to the overall shape, with the only exception of the $m_H = 120\,\mathrm{GeV}$ case, where $Zb\bar{b} \to 4l + X$ modifies the background shape in the low mass region, and it must therefore be taken into account.
The irreducible background has been modelled using a combination of Fermi functions which are suitable to describe both the plateau in the low mass region and the broad peak corresponding to the second Z coming *on shell*. The chosen model is described by the following function:

$$\frac{p0}{(1+e^{\frac{p6-M_{ZZ}}{p7}})(1+e^{\frac{M_{ZZ}-p8}{p9}})} + \frac{p1}{(1+e^{\frac{p2-M_{ZZ}}{p3}})(1+e^{\frac{p4-M_{ZZ}}{p5}})} \,. \tag{40}$$

The first plateau, in the region where only one of the two Z bosons is *on shell*, is modelled by the first term, and its suppression, needed for a correct description at higher masses, is controlled by the $p8$ and $p9$ parameters. The second term in the above formula accounts for the shape of the broad peak and the tail at high masses. This function can describe with a negligible bias the ZZ background shape with good accuracy over the full mass range.

As already mentioned, the $Zb\bar{b}$ contribution is relevant only when searching for very light Higgs bosons (in this study, only $m_H = 120\,\mathrm{GeV}$). In this case, an additional term is added to the ZZ continuum, with a functional form similar to the second part of equation 40. For what concerns signal modelling, a simple Gaussian shape has been used for $m_H \leq 300\,\mathrm{GeV}$, while a relativistic Breit-Wigner formula was needed to properly describe the big tails arising at higher values of the Higgs mass.

In the fits to determine the profile likelihood ratio, $m_H$ is fixed to the hypothesized value, while $\sigma_H$ is allowed to float in a $\pm 20\%$ range around the value obtained from the signal Monte Carlo distributions. All the parameters describing the background shape are floating within sensible ranges. Given the complexity of the model involved, the fit can from time to time get trapped into local minima. While there is no easy way to avoid this problem, the fake measurements obtained in this case are easy to distinguish from the correct ones, and a repetition of the fit from a different starting point is enough to solve the problem.

The resulting median profile likelihood ratios for discovery, $-2\ln\lambda(\mu = 0)$, with the median computed using toy $s+b$ MC data (i.e., with $\mu = 1$), are given in Table 9 for several values of the Higgs mass and integrated luminosity.

The distribution of the test statistic $q_0$ under the null background-only hypothesis, for $m_H = 200$ GeV with an integrated luminosity of 2 and $10\,\mathrm{fb}^{-1}$, is shown in Fig. 14. A $\frac{1}{2}\chi^2_1$ distribution is superimposed, showing the validity of the asymptotic approximation.

The resulting median profile likelihood ratio for exclusion, $\lambda(\mu)$ (using toy background-only MC experiments and taking the median of the $\lambda(\mu)$ distribution), is given in Table 10 for several Higgs masses and luminosities using a signal strength $\mu = 1.0$ corresponding to a Standard Model Higgs boson.

The distribution of the test statistic $q_1$ for $\hat{\mu} \leq 1$ under the $s+b$ hypothesis for $m_H = 200$ GeV with

Table 9: Median values of $-2\ln\lambda(\mu = 0)$ obtained from $H \to ZZ^{(*)} \to 4l$ simulated data generated with signal plus background ($\mu = 1$) for several values of the Higgs mass and integrated luminosity.

| L | $m_H$ (GeV) | | | | | | | | | | | |
|---|---|---|---|---|---|---|---|---|---|---|---|---|
| (fb$^{-1}$) | 120 | 130 | 140 | 150 | 160 | 165 | 180 | 200 | 300 | 400 | 500 | 600 |
| 1 | 0.22 | 1.20 | 3.98 | 5.35 | 1.65 | 0.49 | 0.86 | 6.86 | 5.20 | 3.55 | 0.88 | 0.31 |
| 2 | 0.44 | 2.40 | 7.96 | 10.7 | 3.30 | 0.98 | 1.71 | 13.7 | 10.4 | 7.68 | 1.74 | 0.62 |
| 5 | 1.05 | 5.97 | 19.9 | 26.7 | 8.26 | 2.46 | 4.27 | 34.3 | 25.8 | 17.8 | 4.34 | 1.54 |
| 10 | 2.18 | 12.0 | 39.8 | 53.5 | 16.5 | 4.9 | 8.55 | 68.7 | 51.6 | 35.5 | 8.47 | 3.11 |
| 30 | 6.56 | 35.8 | 120 | 160 | 48.9 | 14.8 | 25.6 | 206 | 162 | 108 | 27.9 | 10.9 |
| 60 | 13.1 | 71.6 | 239 | 321 | 99.1 | 29.5 | 51.3 | 407 | 310 | 213 | 52.6 | 18.6 |

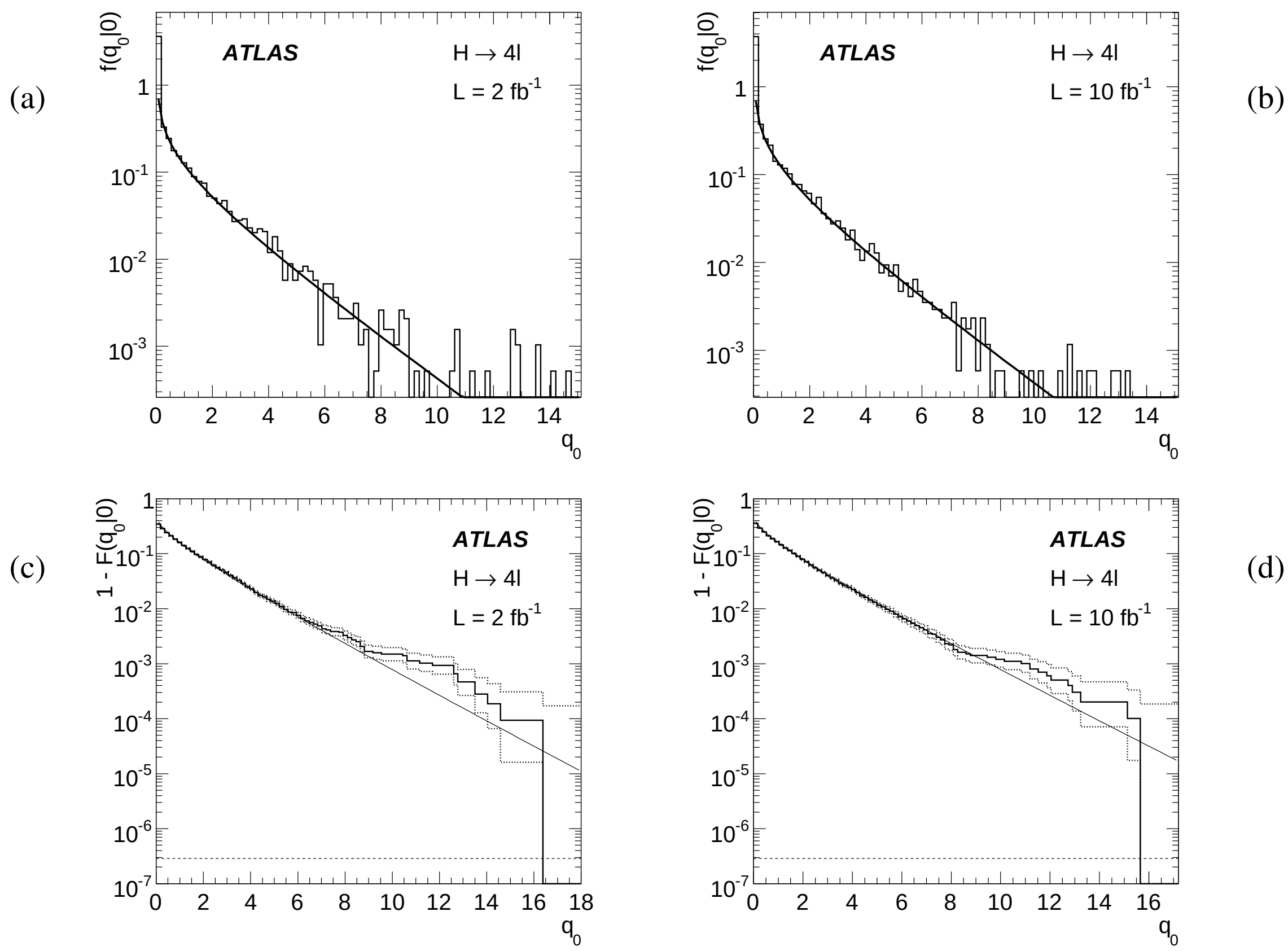


Figure 14: The distribution of the test statistic $q_0$ (for $H \to 4l$), under the null background only hypothesis, for $m_H = 200$ GeV with an integrated luminosity of 2 (a) and 10 (b) fb$^{-1}$. A $\frac{1}{2}\chi_1^2$ distribution is superimposed. Figures (c) and (d) show $1 - F(q_0)$ where $F(q_0)$ is the corresponding cumulative distribution. The small excess of events at high $q_0$ is statistically compatible with the expected curves, as can be seen by comparison with the dotted histograms showing the 68.3% central confidence intervals for $p = 1 - F(q_0|0)$. The lower dotted line at $2.87 \times 10^{-7}$ shows the $5\sigma$ discovery threshold.

Table 10: Median values of $-2\ln\lambda(\mu = 1)$ obtained from $H \to ZZ^{(*)} \to 4l$ simulated data generated with background only ($\mu = 0$) for several values of the Higgs mass and integrated luminosity.

| L | $m_H$ (GeV) | | | | | | | | | | | |
|---|---|---|---|---|---|---|---|---|---|---|---|---|
| ($fb^{-1}$) | 120 | 130 | 140 | 150 | 160 | 165 | 180 | 200 | 300 | 400 | 500 | 600 |
| 1 | 0.16 | 0.93 | 2.25 | 3.12 | 1.23 | 0.39 | 0.51 | 4.86 | 2.86 | 2.36 | 0.87 | 0.28 |
| 2 | 0.33 | 1.85 | 4.50 | 6.22 | 2.4 | 0.75 | 1.90 | 9.64 | 5.68 | 4.69 | 1.74 | 0.56 |
| 5 | 0.83 | 4.60 | 11.2 | 15.4 | 6.09 | 2.28 | 4.74 | 23.5 | 14.0 | 11.6 | 4.32 | 1.39 |
| 10 | 1.60 | 9.14 | 22.0 | 30.3 | 12.1 | 3.86 | 5.05 | 45.2 | 27.3 | 22.7 | 8.57 | 2.77 |
| 30 | 4.78 | 26.8 | 63.0 | 85.3 | 35.1 | 11.4 | 14.7 | 105 | 74.1 | 63.0 | 24.9 | 8.22 |
| 60 | 8.90 | 51.7 | 117 | 155 | 66.9 | 22.3 | 28.0 | 174 | 129 | 113 | 47.6 | 16.1 |

an integrated luminosity of 2 and 10 $fb^{-1}$ is shown in Figures 15. A $\chi^2_1$ distribution is superimposed, showing the validity of the asymptotic approximation.

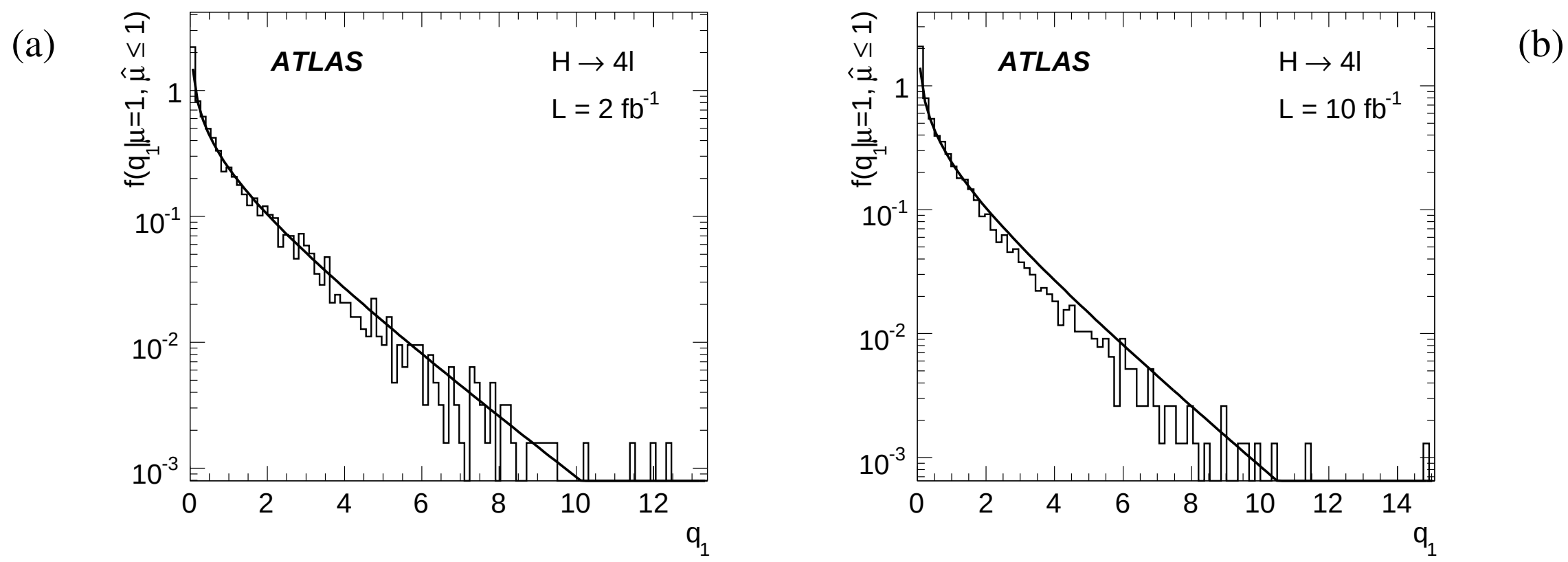


Figure 15: The distribution of the test statistic $q_1$ for $\hat{\mu} \leq 1$ under the $s+b$ hypothesis (for $H \to 4l$), for $m_H = 200$ GeV with an integrated luminosity of (a) 2 $fb^{-1}$ and (b) 10 $fb^{-1}$. A $\chi^2_1$ distribution is superimposed.

### 3.5 Limitations of the approximations used

The distributions shown in Sections 3.1 through 3.4 show varying levels of agreement between the asymptotic chi-square form and the results of Monte Carlo simulations. For the $WW$ (0 jet) channel (Fig. 8), the discrepancy in the distribution of $q_0$ is very large, and this is understood to arise from the special manner in which the systematic uncertainties for this channel were treated. The distribution of $q_0$ for the $WW$ (0 jet) channel therefore does not use the asymptotic formula. This is the only channel for which the approximation was not applied.

For other cases such as the distribution of $q_1$ for the $H \to \gamma\gamma$ channel shown in Fig. 7, the Monte Carlo distribution falls off significantly faster than the chi-square curve. This means that the significance with which one excludes the tested hypothesis will be less when estimated from the chi-square curve, leading to conservative limits. As the integrated luminosity increases, one expects to the asymptotic formula to become more accurate.

In some of the distributions such as that of $q_0$ for the $H \to ZZ^{(*)} \to 4l$ channel shown in Fig. 14, the Monte Carlo simulation indicates a slight excess over the chi-square curve in the tail region. The level of the excess is not statistically significant in the part of the distribution that can be meaningfully assessed

given the amount of Monte Carlo data available (out to $q_0$ between around 9 to 16, i.e., to the level of a 3 to $4\sigma$ discovery). At present it is not practical to verify directly that the chi-square formula remains valid to the $5\sigma$ level (i.e., out to $q_0 = 25$). Thus the results on discovery significance presented here rest on the assumption that the asymptotic distribution is a valid approximation to at least the $5\sigma$ level.

The validation exercises carried here out indicate that the methods used should be valid, or in some cases conservative, for an integrated luminosity of at least 2 fb$^{-1}$. At earlier stages of the data taking, one will be interested primarily in exclusion limits at the 95% confidence level. For this the distributions of the test statistic $q_\mu$ at different values of $\mu$ can be determined with a manageably small number of events. It is therefore anticipated that we will rely on Monte Carlo methods for the initial phase of the experiment.

# 4 Results of the combination

## 4.1 Combined discovery sensitivity

The full discovery likelihood ratio for all channels combined, $\lambda_{s+b}(0)$, is calculated using Eq. 33. This uses the median likelihood ratio of each channel, $\lambda_{s+b,i}(0)$, found either by generating toy experiments under the $s+b$ hypothesis and calculating the median of the $\lambda_{s+b,i}$ distribution or by approximating the median likelihood ratio using the Asimov data sets with $\mu_{A,i} = 1$. Both approaches were validated to agree with each other. The discovery significance is calculated using Eq. 36, i.e., $Z \approx \sqrt{-2\ln\lambda(0)}$, where $\lambda(0)$ is the combined median likelihood ratio.

The resulting significances per channel and the combined one are shown in Fig. 16 for an integrated luminosity of 10 fb$^{-1}$.

(a) (b)

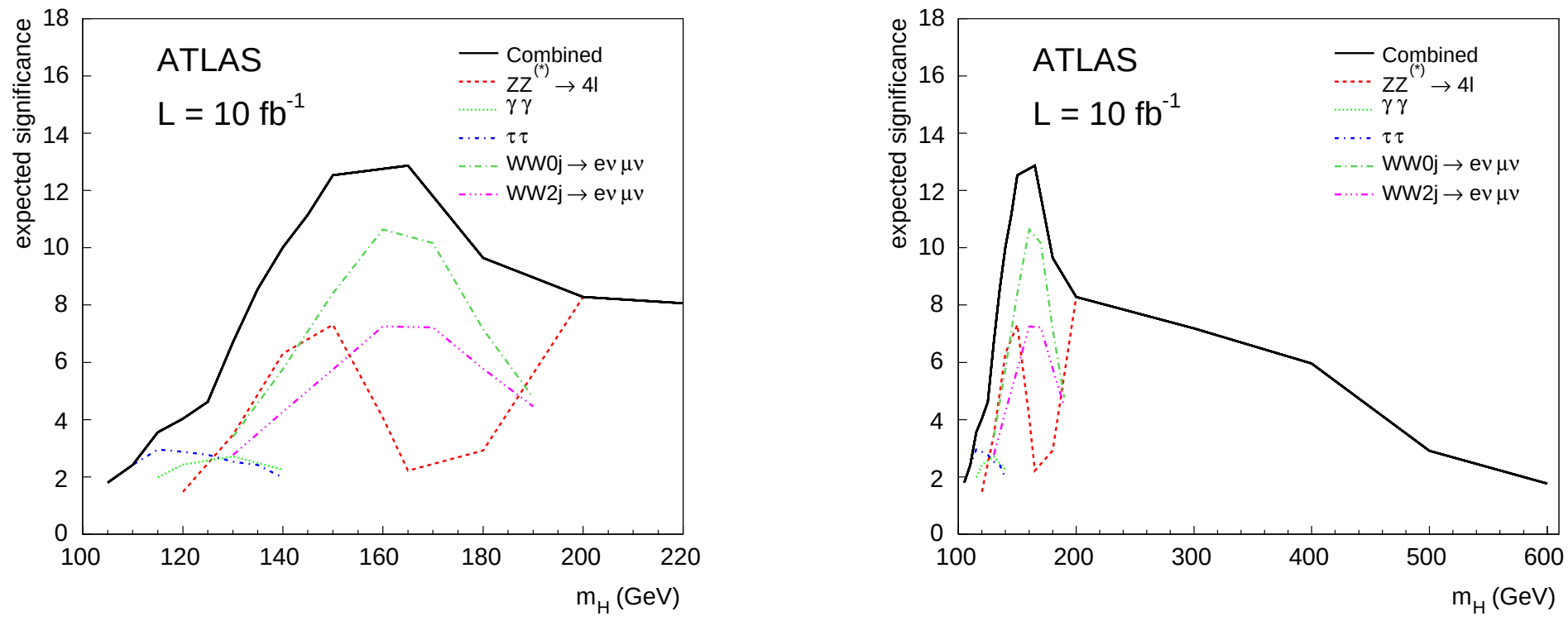


Figure 16: The median discovery significance for the various channels and the combination with an integrated luminosity of 10 fb$^{-1}$ for (a) the lower mass range (b) for masses up to 600 GeV.

The median discovery significance as a function of the integrated luminosity and Higgs mass is shown colour coded in Fig. 17. The full line indicates the $5\sigma$ contour. Note that the approximations used do not hold for very low luminosities (where the expected number of events is low) and therefore the results below about 2 fb$^{-1}$ should be taken as indications only. In most cases, however, the approximations tend to underestimate the true median significance.

## 4.2 Combined exclusion sensitivity

The full likelihood ratio of all channels used for exclusion for a signal strength $\mu$, $\lambda_b(\mu)$, is calculated using Eq. 34 with the median likelihood ratios of each channel, $\lambda_{b,i}(\mu)$, calculated, either by generating

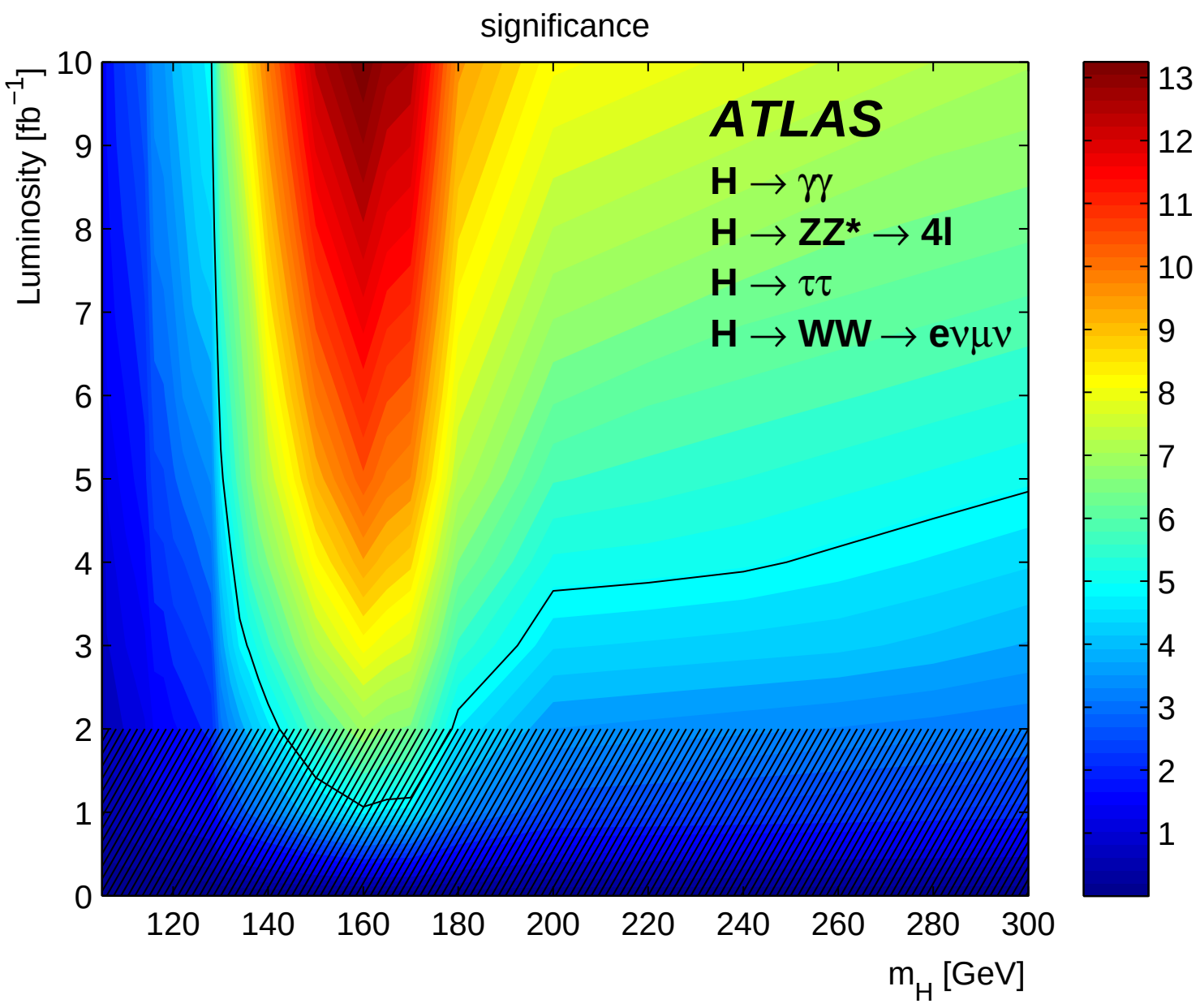


Figure 17: Significance contours for different Standard Model Higgs masses and integrated luminosities. The thick curve represents the $5\sigma$ discovery contour. The median significance is shown with a colour according to the legend. The hatched area below 2 fb$^{-1}$ indicates the region where the approximations used in the combination are not accurate, although they are expected to be conservative.

toy experiments under the $b$-only hypothesis and calculating the median of the $\lambda_{b,i}$ distribution or approximating the median likelihood ratio using the Asimov data sets with $\mu_{A,i} = 0$. Both approaches were checked to agree with each other. A signal strength $\mu = 1$ corresponds to the Standard Model Higgs boson.

Any exclusion of $\mu(m_H)$ smaller than 1 corresponds to an exclusion of a Standard Model Higgs boson with a mass $m_H$. To probe the median sensitivity for excluding a Standard Model Higgs boson we follow Eq. 35 and calculate the corresponding $p$-value for $\mu = 1$, $p_1$ for a given luminosity at a given Higgs mass. A $p$-value of 0.05 corresponds to a significance (Eq. 36) of 1.64. The resulting $p_1$ for the various channels as well as for the combination, for a luminosity of 2 fb$^{-1}$, are shown in Fig. 18. Note that any $p$-value below 0.05 indicates an exclusion. We therefore conclude that with a luminosity of 2 fb$^{-1}$ ATLAS has the median sensitivity to exclude a Standard Model Higgs boson heavier than 115 GeV at the 95% Confidence Level. This can also be seen from Fig. 19, which shows the luminosity required to exclude a Higgs boson with a mass $m_H$ at a given confidence level from the combination of the four channels explored in this note.

The sharp increase in the required luminosity for lower $m_H$ seen in Fig. 17 reflects the decrease in sensitivity to the Higgs when using only the set of channels considered here. Further developments will increase the sensitivity in this region. For example, improved analysis methods for the $H \rightarrow \gamma\gamma$ channel are described in Ref. [4], including a separation of the events into those with zero or two accompanying jets. Additional final states such as $t\bar{t}H$ with $H \rightarrow b\bar{b}$ will help somewhat, although the contribution to the sensitivity will be small because of the large uncertainties in the background.

For the $WW$ channel, the present study includes only the $e\nu\mu\nu$ decay mode, but it is planned to include $e\nu e\nu$, $\mu\nu\mu\nu$ and $qql\nu$ as well. The $ZZ^{(*)}$ channel here only includes $Z$ decays to $ee$ and $\mu\mu$, but in future analyses $qq\nu\nu$ will be included. The additional $WW$ and $ZZ^{(*)}$ modes have been found to have sensitivity for a high-mass Higgs. Finally, combination with the results from ATLAS with those of CMS will of course result in an overall increase in sensitivity.

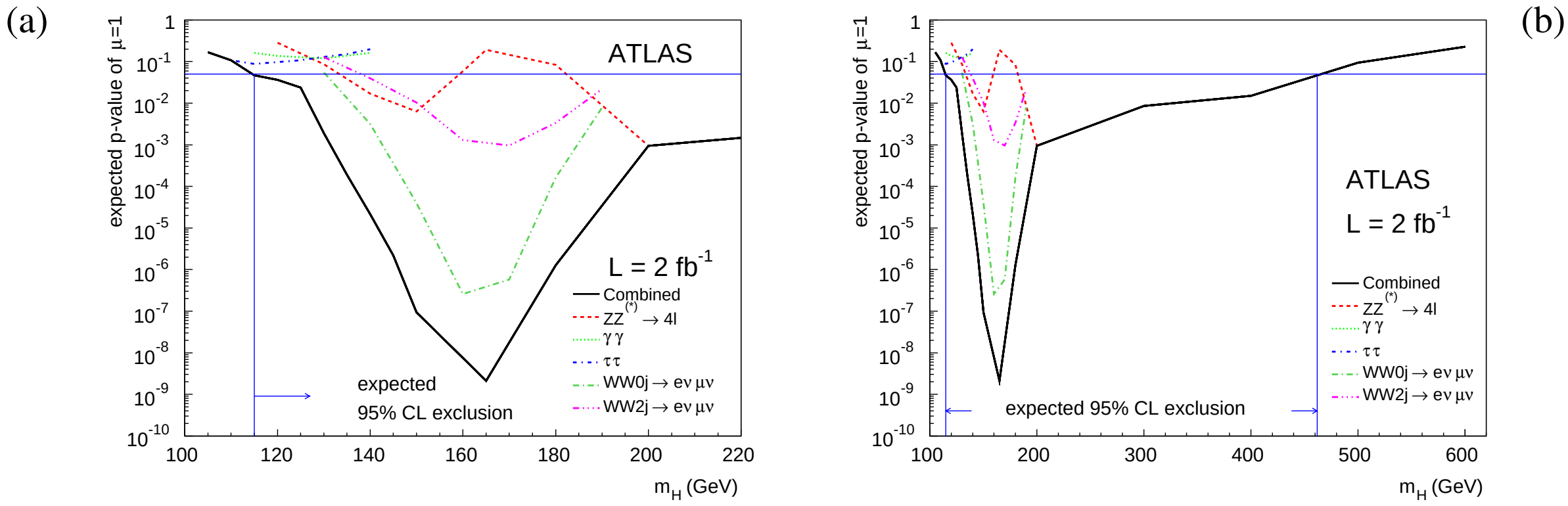


Figure 18: The median $p$-value obtained for excluding a Standard Model Higgs Boson for the various channels as well as the combination for (a) the lower mass range (b) for masses up to 600 GeV.

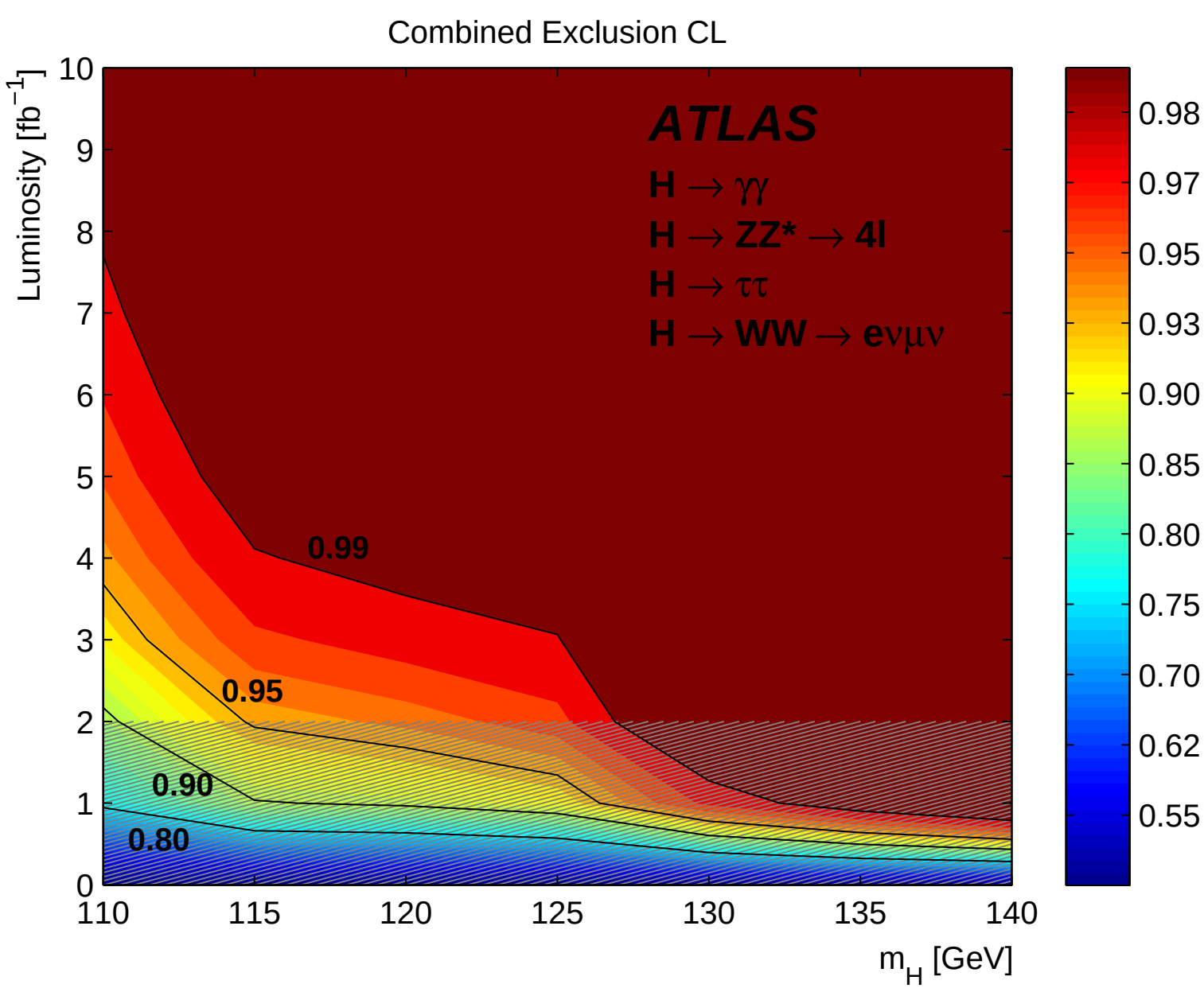


Figure 19: The expected luminosity required to exclude a Higgs boson with a mass $m_H$ at a confidence level given by the corresponding colour. The hatched area below 2 fb$^{-1}$ indicates the region where the approximations used in the combination are not accurate, although they are expected to be conservative.

# 5 Conclusions

The procedure for combination of search results based on the profile likelihood ratio has been applied to a study of the search for the Standard Model Higgs boson using four search channels: $H \to \tau^+\tau^-$, $H \to W^+W^- \to e\nu\mu\nu$, $H \to \gamma\gamma$ and $H \to ZZ^{(*)} \to 4$ leptons. The combination method is very general and can be applied to essentially any search that will be carried out at the LHC.

The study here has not exploited all of the search channels that will be investigated and therefore the current estimates of the sensitivity can be regarded as conservative. For example, using further decay modes in the $ZZ$ and $WW$ channels will provide additional sensitivity especially for a Higgs boson in the higher mass range.

The studies have exploited a series of useful approximations that allow one to determine the median discovery and exclusion sensitivities from a combined fit in a manner that only requires separate input ingredients from the individual channels. The determination of the significance for a given (e.g., real) data set, however, will require a simultaneous fit of all of the channels.

It is not practical at present to generate enough Monte Carlo data to verify directly that the tail of the profile likelihood distribution is well described to the level required for discovery at the $5\sigma$ level, corresponding to an upper tail area of $2.87 \times 10^{-7}$. The estimates of discovery significance presented here therefore rely on the assumption that the large-sample approximation used remains valid out to this level.

The validation studies shown in Section 3 indicate that the approximations used should be reasonably accurate or lead to conservative limits for an integrated luminosity of at least 2 $\text{fb}^{-1}$. For the earlier stages of the experiment it is expected that one will need to rely on Monte Carlo methods, which should be feasible for exclusion limits at the 95% confidence level.

The profile likelihood ratio treats systematic errors by associating the uncertainties with adjustable (nuisance) parameters. Other methods for treating systematic uncertainties can also be considered. Using Bayesian methods, for example, one would associate a prior probability density with the nuisance parameters. We plan to develop and use this and other approaches in parallel with the profile likelihood method for searches at the LHC.

The study presented in this paper provides the discovery significance for a Higgs boson of a specific mass. That is, the traditional discovery threshold $p$-value of $2.87 \times 10^{-7}$ corresponding to a $5\sigma$ effect for a given hypothesized Higgs mass refers to the false discovery rate for a Higgs of that mass. The false discovery rate for a Higgs of *any* mass is higher, and several approaches are being pursued to quantify this (the so-called 'look-elsewhere effect'). The most mature of these methods involves using a simultaneous fit of the Higgs mass $m_{\text{H}}$ and the strength parameter $\mu$ (or equivalently the Higgs production rate), as has been discussed in the studies of $H \to \gamma\gamma$ [4] and $H \to W^+W^-$ [5].

To summarize, the studies based on the four channels considered in this note confirm the good discovery and exclusion sensitivities already shown in the ATLAS Technical Design Report (TDR) [10]. Furthermore the results here are based on better knowledge and a more realistic simulation of the detector than what is described in the TDR. Because of the approximations used, the present studies are valid only for luminosities above $2\,\text{fb}^{-1}$. With a luminosity of $2\,\text{fb}^{-1}$ the expected (median) sensitivity is at the $5\sigma$ level or greater for discovery of a Higgs boson in the mass range between 143 and 179 GeV, and the expected upper limit at 95% confidence level on the Higgs mass is 115 GeV.

# A Comparison with procedures used at LEP

In this appendix we compare the procedures described in the present analysis with those used in searches carried out at LEP. More details on these methods be found in [8]. The important differences involve the definition of the test statistic used and the treatment of systematic uncertainties. In addition, the LEP

analyses adopted a special procedure to prevent spurious exclusion due to a downward fluctuation of the number of events (the $\mathrm{CL}_s$ method, see below).

In the LEP Higgs searches, a hypothesized value of the Higgs mass $m_H$ was tested by constructing the statistic

$$Q = \frac{L_{s+b}}{L_b} = \frac{L(\mu = 1)}{L(\mu = 0)} , \tag{41}$$

where as before $b$ ($\mu = 0$) represents the background-only hypothesis and $s + b$ ($\mu = 1$) refers to background plus signal at the rate predicted by the Standard Model. For convenience the equivalent logarithmic variable $q = -2\ln Q$ was used.

The sampling distribution of $Q$ was determined by Monte Carlo simulation. The Monte Carlo was also used to incorporate systematic errors by sampling values of the corresponding nuisance parameters from pdfs that reflected their uncertainties. That is, one effectively integrated the product of the likelihood and prior pdfs for the nuisance parameters.

For a give observed value $q_{\mathrm{obs}} = -2\ln Q_{\mathrm{obs}}$, the $p$-values for the $s$ and $s + b$ hypotheses were determined as

$$p_{s+b} = \int_{q_{\mathrm{obs}}}^{\infty} f(q|s+b)\, dq \equiv \mathrm{CL}_{s+b} , \tag{42}$$

$$p_b = \int_{-\infty}^{q_{\mathrm{obs}}} f(q|b)\, dq \equiv 1 - \mathrm{CL}_b . \tag{43}$$

Having determine the $p$-values, the LEP analyses then based exclusion of the $s + b$ hypothesis not on the $p$-value of $s + b$ but rather on the ratio $\mathrm{CL_s}$, defined as

$$\mathrm{CL_s} = \frac{\mathrm{CL_{s+b}}}{\mathrm{CL_b}} = \frac{p_{s+b}}{1 - p_b} . \tag{44}$$

The signal-plus-background hypothesis was said to be excluded at confidence level $\mathrm{CL} = 1 - \alpha = 0.95$ if one finds

$$\mathrm{CL}_s < \alpha . \tag{45}$$

Since $\mathrm{CL}_b \leq 1$, one has $\mathrm{CL}_s \geq \mathrm{CL}_{s+b}$. Therefore the $\mathrm{CL}_s$ method will not exclude as large a region of parameter space as that based on the signal-plus-background $p$-value ($\mathrm{CL}_{s+b}$ method). As the $\mathrm{CL}_{s+b}$ method was designed to provide an interval that brackets the true value of the parameter with a probability of at least $1 - \alpha$, the $\mathrm{CL}_s$ limit must cover the true parameter with a greater probability; it is in this sense conservative. The $\mathrm{CL}_s$ method was devised so as to avoid the problem where a downward fluctuation in the number of background events can lead to exclusion of the Higgs mass considered, even for hypothesized mass values where one does not expect to be sensitive to Higgs production [9].

In contrast, in the present analysis we test a hypothesized value of the strength parameter $\mu$ using

$$q_\mu = -2\ln \frac{L(\mu, \hat{\hat{\theta}})}{L(\hat{\mu}, \hat{\theta})} , \tag{46}$$

as described in Section 2. With this definition, the sampling distribution of the test statistic $f(q_\mu|\mu)$ approaches a well defined form related to the chi-squared distribution for a sufficiently large data sample. The ability to exploit this approximate form is very useful as the relevant $p$-value for a $5\sigma$ discovery is $2.87 \times 10^{-7}$, and therefore to determine this from Monte Carlo would require an extremely large

number of simulated experiments. At LEP this was not a crucial issue as the statistical treatment focused primarily on exclusion limits at 95% CL, not on discovery at the $5\sigma$ level.

Systematic uncertainties here have been incorporated using the profile likelihood, rather than with the integrated likelihoods used at LEP. For the uncertainties most relevant to the analyses at the LHC, the broadening of the likelihood function obtained by both procedures is similar. It is the profile likelihood ratio and not the ratio of integrated likelihoods, however, that approaches the chi-squared form in the large sample limit in accordance with Wilks' theorem.

The $\mathrm{CL}_s$ method has not been applied in the present analysis but studies of its application to searches in ATLAS are ongoing.

A final difference with the LEP procedures concerns the definition of significance $Z$. Here we have defined its relation to the $p$-value as the number of standard deviations of a Gaussian variable that would give a one-sided tail area of $p$, as described in Section 2.1. A significance of $Z = 5$ corresponds to $p = 2.87 \times 10^{-7}$. The LEP Higgs group defined this relation using a two-sided fluctuation of a Gaussian variable, i.e., a $5\sigma$ significance corresponded to $p = 5.7 \times 10^{-7}$.

# Supersymmetry

# Supersymmetry Searches

**Abstract**

This chapter serves as an introduction to a collection of six articles that detail the strategy foreseen to search for Supersymmetry with the ATLAS detector at the Large Hadron Collider, concentrating on the initial data taking period with an expected integrated luminosity of about 1 $\text{fb}^{-1}$ . We review here the phenomenology of Supersymmetry with ATLAS and discuss how events are simulated and reconstructed, concentrating on aspects related to Supersymmetry searches. We also introduce many of the experimental variables that are used throughout this collection.

## 1 Introduction

Supersymmetry (SUSY) is one of the theoretically favoured candidates for physics beyond the Standard Model. The main motivation is to protect the Higgs boson mass from quadratically diverging radiative corrections, in a theory where the Standard Model is valid only up to a high scale $\Lambda$. The proposed solution postulates the invariance of the theory under a symmetry which transforms fermions into bosons and vice-versa.

The basic prediction of SUSY is thus the existence, for each Standard Model particle degree of freedom of a corresponding sparticle, with spin different by half a unit. The SUSY generators commute with the $SU(2)\times U(1)\times SU(3)$ symmetries of the Standard Model, and with the Poincaré group. It follows that with unbroken SUSY the partner particles would have the same quantum numbers and masses as the Standard Model particles. Since no superpartner has been observed to date, SUSY must be broken. A common approach to the phenomenological study of SUSY is to assume the minimal possible particle content, and to parametrise the SUSY-breaking Lagrangian as the sum of all the terms which do not reintroduce quadratic divergences into the theory. The model thus obtained is called Minimal Supersymmetric Standard Model (MSSM) and is characterised by a large number of parameters ($\sim 100$). In order to warrant the conservation of baryonic and leptonic quantum numbers, a new multiplicative quantum number, $R$-parity, is introduced, which is 1 for particles and -1 for the SUSY partners. Models where $R$-parity is violated can be formulated, but in the current volume we concentrate on models with $R$-parity conservation.

The consequences of $R$-parity conservation are that sparticles must be produced in pairs, and that each will decay to the lightest SUSY particle (LSP) which must be stable. Cosmological arguments suggest that stable LSPs should be weakly interacting and so would escape direct detection at ATLAS, resulting in the characteristic feature expected for SUSY events – an imbalance of the transverse energy measured in the detector, abbreviated here as $E_{\mathrm{T}}^{\mathrm{miss}}$ . The associated signatures will provide sensitivity to a large class of models. The aim of the simulation studies is to ensure that the ATLAS experiment will have rapid sensitivity to a large ensemble of SUSY models, and to develop a general search strategy. It is not possible to explore in full the 100-dimensional parameter space of the MSSM. It is therefore necessary to adopt some specific assumptions for the SUSY breaking, resulting in models defined by a small number of parameters at the SUSY breaking scale. Two models will be studied in detail:

- mSUGRA, where SUSY breaking is mediated by gravitational interaction;
- GMSB, where SUSY breaking is mediated by a gauge interaction through messenger gauge fields.

These two models give quite different topologies, due to the different nature of the lightest SUSY particle, which is the lightest neutralino for the mSUGRA case and the gravitino for the GMSB case. For each

of these models a set of benchmark points has been defined, on which full simulation studies have been performed.

The analysis work presented in the different SUSY notes is a coherent body of work based on an agreement among the many groups performing the analyses on many subjects both theoretical and experimental. All of the analyses are based on a common implementation of the SUSY model, common background datasets have been used, and all of the analyses apply a common definition of the physics objects in the detectors, optimised for the specific environment of the SUSY searches. In the following sections a quick overview of these common issues will be given, which should be read before approaching the detailed studies documented in the different notes.

# 2 Signal and Background generation

The simulated data used for the studies documented here have been produced by Monte Carlo simulation inside the official ATLAS software and production frameworks, and for all of the samples a detailed simulation of the detector has been performed.

## 2.1 Signal samples

For the detailed SUSY analysis a set of benchmark points in the mSUGRA and GMSB frameworks were chosen, with the aim of exploring sensitivity to a wide class of of final-state signatures.

For the mSUGRA points, the principle that the predicted cosmological relic density of neutralinos should be consistent with the observed density of cold dark matter was used for guidance. In order to reproduce the observed relic density, the model parameters must result in a spectrum which ensures efficient annihilation of the neutralinos in the early universe. In the mSUGRA scenario this is possible only in restricted regions of the parameter space where annihilation is enhanced either by a significant higgsino components in the lightest neutralino or through mass relationships. The points chosen are defined in terms of the mSUGRA parameters at the unification scale:

- SU1 $m_0 = 70$ GeV, $m_{1/2} = 350$ GeV, $A_0 = 0$, $\tan\beta = 10$, $\mu > 0$. Coannihilation region where $\tilde{\chi}_1^0$ annihilate with near-degenerate $\tilde{\ell}$.
- SU2 $m_0 = 3550$ GeV, $m_{1/2} = 300$ GeV, $A_0 = 0$, $\tan\beta = 10$, $\mu > 0$. Focus point region near the boundary where $\mu^2 < 0$. This is the only region in mSUGRA where the $\tilde{\chi}_1^0$ has a high higgsino component, thereby enhancing the annihilation cross-section for processes such as $\tilde{\chi}_1^0\tilde{\chi}_1^0 \to WW$.
- SU3 $m_0 = 100$ GeV, $m_{1/2} = 300$ GeV, $A_0 = -300$ GeV, $\tan\beta = 6$, $\mu > 0$. Bulk region: LSP annihilation happens through the exchange of light sleptons.
- SU4 $m_0 = 200$ GeV, $m_{1/2} = 160$ GeV, $A_0 = -400$ GeV, $\tan\beta = 10$, $\mu > 0$. Low mass point close to Tevatron bound.
- SU6 $m_0 = 320$ GeV, $m_{1/2} = 375$ GeV, $A_0 = 0$, $\tan\beta = 50$, $\mu > 0$. The funnel region where $2m_{\tilde{\chi}_1^0} \approx m_A$. Since $\tan\beta \gg 1$, the width of the pseudoscalar Higgs boson $A$ is large and $\tau$ decays dominate.
- SU8.1 $m_0 = 210$ GeV, $m_{1/2} = 360$ GeV, $A_0 = 0$, $\tan\beta = 40$, $\mu > 0$. Variant of coannihilation region with $\tan\beta \gg 1$, so that only $m_{\tilde{\tau}_1} - m_{\tilde{\chi}_1^0}$ is small.
- SU9 $m_0 = 300$ GeV, $m_{1/2} = 425$ GeV, $A_0 = 20$, $\tan\beta = 20$, $\mu > 0$. Point in the bulk region with enhanced Higgs production

The SUSY particle mass spectra for each of these points are listed in Table 2.

The leading-order and next-to-leading-order cross-sections at 14 TeV center of mass for the chosen points are given in Table 1. These have been calculated with the program PROSPINO 2.0.6 [1–3], using the default settings, and the parton distribution set CTEQ6M [4].

Table 1: Production cross-sections at leading order ($\sigma^{\mathrm{LO}}$) and next-to-leading order ($\sigma^{\mathrm{NLO}}$), numbers of Monte Carlo events simulated (N) and corresponding integrated luminosity for the SUSY benchmark points used by ATLAS.

| Label | $\sigma^{\mathrm{LO}}$ (pb) | $\sigma^{\mathrm{NLO}}$ (pb) | N | L ($\mathrm{fb}^{-1}$) |
|---|---|---|---|---|
| SU1 | 8.15 | 10.86 | 200 K | 18.4 |
| SU2 | 5.17 | 7.18 | 50 K | 7.0 |
| SU3 | 20.85 | 27.68 | 500 K | 18.1 |
| SU4 | 294.46 | 402.19 | 200 K | 0.50 |
| SU6 | 4.47 | 6.07 | 30 K | 4.9 |
| SU8.1 | 6.48 | 8.70 | 50 K | 5.7 |
| SU9 | 2.46 | 3.28 | 40 K | 12.2 |

Table 2: Particle mass spectrum (in GeV) for the SUSY benchmark points.

| Particle | SU1 | SU2 | SU3 | SU4 | SU6 | SU8.1 | SU9 |
|---|---|---|---|---|---|---|---|
| $\tilde{d}_L$ | 764.90 | 3564.13 | 636.27 | 419.84 | 870.79 | 801.16 | 956.07 |
| $\tilde{u}_L$ | 760.42 | 3563.24 | 631.51 | 412.25 | 866.84 | 797.09 | 952.47 |
| $\tilde{b}_1$ | 697.90 | 2924.80 | 575.23 | 358.49 | 716.83 | 690.31 | 868.06 |
| $\tilde{t}_1$ | 572.96 | 2131.11 | 424.12 | 206.04 | 641.61 | 603.65 | 725.03 |
| $\tilde{d}_R$ | 733.53 | 3576.13 | 610.69 | 406.22 | 840.21 | 771.91 | 920.83 |
| $\tilde{u}_R$ | 735.41 | 3574.18 | 611.81 | 404.92 | 842.16 | 773.69 | 923.49 |
| $\tilde{b}_2$ | 722.87 | 3500.55 | 610.73 | 399.18 | 779.42 | 743.09 | 910.76 |
| $\tilde{t}_2$ | 749.46 | 2935.36 | 650.50 | 445.00 | 797.99 | 766.21 | 911.20 |
| $\tilde{e}_L$ | 255.13 | 3547.50 | 230.45 | 231.94 | 411.89 | 325.44 | 417.21 |
| $\tilde{\nu}_e$ | 238.31 | 3546.32 | 216.96 | 217.92 | 401.89 | 315.29 | 407.91 |
| $\tilde{\tau}_1$ | 146.50 | 3519.62 | 149.99 | 200.50 | 181.31 | 151.90 | 320.22 |
| $\tilde{\nu}_\tau$ | 237.56 | 3532.27 | 216.29 | 215.53 | 358.26 | 296.98 | 401.08 |
| $\tilde{e}_R$ | 154.06 | 3547.46 | 155.45 | 212.88 | 351.10 | 253.35 | 340.86 |
| $\tilde{\tau}_2$ | 256.98 | 3533.69 | 232.17 | 236.04 | 392.58 | 331.34 | 416.43 |
| $\tilde{g}$ | 832.33 | 856.59 | 717.46 | 413.37 | 894.70 | 856.45 | 999.30 |
| $\tilde{\chi}_1^0$ | 136.98 | 103.35 | 117.91 | 59.84 | 149.57 | 142.45 | 173.31 |
| $\tilde{\chi}_2^0$ | 263.64 | 160.37 | 218.60 | 113.48 | 287.97 | 273.95 | 325.39 |
| $\tilde{\chi}_3^0$ | 466.44 | 179.76 | 463.99 | 308.94 | 477.23 | 463.55 | 520.62 |
| $\tilde{\chi}_4^0$ | 483.30 | 294.90 | 480.59 | 327.76 | 492.23 | 479.01 | 536.89 |
| $\tilde{\chi}_1^+$ | 262.06 | 149.42 | 218.33 | 113.22 | 288.29 | 274.30 | 326.00 |
| $\tilde{\chi}_2^+$ | 483.62 | 286.81 | 480.16 | 326.59 | 492.42 | 479.22 | 536.81 |
| $h^0$ | 115.81 | 119.01 | 114.83 | 113.98 | 116.85 | 116.69 | 114.45 |
| $H^0$ | 515.99 | 3529.74 | 512.86 | 370.47 | 388.92 | 430.49 | 632.77 |
| $A^0$ | 512.39 | 3506.62 | 511.53 | 368.18 | 386.47 | 427.74 | 628.60 |
| $H^+$ | 521.90 | 3530.61 | 518.15 | 378.90 | 401.15 | 440.23 | 638.88 |
| $t$ | 175.00 | 175.00 | 175.00 | 175.00 | 175.00 | 175.00 | 175.00 |

Even though the SUSY benchmark points are all based on the mSUGRA scenario, they provide a rather wide range of possible decay topologies. They share some common features, for example for all these points the gluino mass is less than 1 TeV, and the ratio $m_{\tilde{g}}/m_{\tilde{\chi}_1^0} = 6$–8. For all points except SU2, the squark and gluino masses are comparable. Hence gluinos and squarks would be copiously produced and would decay giving relatively high $p_T$ jets, possibly leptons, and $E_{\mathrm{T}}^{\mathrm{miss}}$ . These features are relatively general – although not guaranteed – but the cold dark matter density predictions are very specific to mSUGRA.

The signatures from GMSB models are often very different because the $\tilde{\chi}_1^0$ is no longer the lightest supersymmetric particle, so it would be expected to decay, and (depending on its lifetime) may do so within the detector. Given its particular experimental signature, a dedicated article of this chapter is devoted to the relevant signatures [5], where the GMSB benchmarks are discussed in detail.

## 2.2 Backgrounds

The Standard Model background processes most relevant to SUSY searches are $t\bar{t}$, $W+$jets, $Z+$jets, jet production from QCD processes and diboson production.

Different Monte Carlo generators were used for different processes, in the attempt of optimising the reliability of the estimate for the Standard Model backgrounds. See Ref. [6] for a more detailed description.

For $t\bar{t}$ production, which is the dominant background for many of the signatures, the MC@NLO [7,8] generator was used. It includes full next-to-leading order QCD corrections, affording a quite stable absolute cross-section prediction and a good description of the final state kinematics for events with up to one additional QCD jet. QCD showering and fragmentation are performed using the HERWIG [9,10] program. Single top production is expected to add a very small contribution [11], but has also been investigated.

Typically SUSY analyses require large jet multiplicities. It is very important to simulate correctly the kinematics of the additional jets for processes like $W+$jets and $Z+$jets. For these processes we therefore used the ALPGEN [12] generator, which at leading order in QCD and electroweak interactions, calculates the exact matrix elements for multiparton hard processes in hadronic collisions. Showering and hadronisation are provided through the HERWIG program. In order to achieve a correct description of the jet multiplicities, it is necessary to perform a correct match between the jets produced by the matrix-element generator and the ones produced by parton showering. This topic is the subject of active theoretical work. Background samples of $W$ and $Z$ containing at least four jets were produced with ALPGEN, using MLM matching [13] and with the jet matching cut set at 40 GeV. Contributions from processes with matrix-element parton multiplicities between one and five were summed to produce the multi-jet sample. A filter was applied at the generator level requiring at least four jets with transverse momentum above 40 GeV, with at least one of these having a transverse momentum above 80 GeV, and with missing transverse energy above 80 GeV.

The leading-order cross-sections were normalised to the results from next-to-next-leading-order calculations by applying a $k$ factor of 1.15 (1.27) for $W$ ($Z$) production respectively. The $k$ factor was calculated by comparing the inclusive leading-order cross-section to the NNLO calculation of [14]. For topologies involving fewer than four jets, the PYTHIA generator [15] was used for the generation of the $W$ and $Z$ boson backgrounds.

For the simulation of QCD multi-jet samples, ALPGEN would also be an appropriate choice. For practical reasons, however, it was impossible to generate ALPGEN samples with sufficiently large numbers of events to simulate the backgrounds for these studies. As a backup solution we have used a shower Monte Carlo, PYTHIA, for which adequate statistics could be generated by producing samples in slices the of $p_{\mathrm{T}}$ of the hard scattering. A filter at generation level is applied to the PYTHIA events, requiring for the hardest jet a transverse momentum above 80 GeV, for the second jet transverse momentum above

Table 3: Kinematic boundaries, number of events and integrated luminosity of the PYTHIA QCD background samples.

| Label | $p_T$ range (GeV) | N | Int. Lumi (pb$^{-1}$ ) |
|---|---|---|---|
| J4 | 140-280 | 18 k | 19.6 |
| J5 | 280-560 | 60 k | 168.4 |
| J6 | 560-1120 | 20 k | 296.6 |
| J7 | 1129-2240 | 4 k | 754.7 |
| J8 | >2240 | 4 k | 181000 |

40 GeV and missing transverse energy above 100 GeV. The numbers of events which were used for detailed simulation, and the corresponding integrated luminosity for each $p_{\mathrm{T}}$ slice are shown in Table 3.

The contributions of the diboson processes $WW$, $ZZ$ and $WZ$ are almost negligible for multi-jet analyses as they are strongly suppressed by typical SUSY selections requiring a large number of jets with high transverse momenta and large missing transverse energy. However they are very important when searching for direct gaugino production [16]. The corresponding data samples were generated at leading order with the HERWIG Monte Carlo, including the full off-shell structure for $Z/\gamma$. The cross-sections were then normalised to the next-to-leading-order cross-sections calculated with the MCFM code [17].

For all the samples except QCD jet production a sample corresponding to at least 1 fb$^{-1}$ of integrated luminnosity was simulated.

Pile-up and cavern background simulations were generally not included in the signal and background samples except in a few cases where this is specifically indicated.

# 3 Object Identification for SUSY analysis

Supersymmetric events are expected to be characterised by several high-momentum jets and missing transverse energy. Leptons[1] and taus are also present in a large fraction of the events for the benchmark points considered. The analyses documented in this Chapter use common particle identification criteria, which are briefly described in this section.

## 3.1 Jets

Because of the relatively large multiplicity of jets in SUSY events, a narrow cone is preferable in the reconstruction of jets. The algorithm used to reconstruct jets in the analysis documented here is the cone algorithm [18] with a cone size of 0.4. Most analyses presented do not rely on secondary-vertex tagging ("$b$ tagging") but when it is applied, standard ATLAS algorithms [19] are used.

## 3.2 Missing transverse energy

The measurement of the transverse energy imbalance in the detector plays a crucial role in the searches for Supersymmetry with R-parity conservation, and the requirement of a large value of $E_{\mathrm{T}}^{\mathrm{miss}}$ is a common feature of all the analyses presented in this chapter.

[1] Except where explicitly indicated, in this Chapter the word "lepton" and the corresponding symbol $\ell$ should be understood to mean $\ell \in \{e, \mu\}$, i.e. they do *not* include taus.

For the studies presented here, the $E_T^{miss}$ is calculated from the calorimeter cells, with calibration weights derived separately for cells associated to different objects (jets, electrons, photons, taus, and non-associated clusters due to the soft part of the event) [20]. Sources of fake missing energy, such as dead or noisy parts of the calorimeter, fake muons, beam-gas and beam-halo events, cosmic rays and electronics problems are not considered here. A detailed discussion of the strategies to remove sources of fake $E_T^{miss}$ in early data and to measure the $E_T^{miss}$ resolution and scale can be found in [20].

The contribution from non-gaussian tails in the $E_T^{miss}$ measurement can be strongly suppressed by requiring a minimum angular separation between the $E_T^{miss}$ vector and the jets in the event. This cut also suppresses the contributions from jets containing hard neutrinos from the leptonic decays of charmed and beauty mesons. This issue is discussed in detail in Ref. [21].

## 3.3 Electrons

In SUSY searches, the requirement of a high $p_T$ electron is normally associated with additional requirements on jets and $E_T^{miss}$ . For typical SUSY analyses the background from the production of QCD jets can be reduced by cuts other than those requiring any lepton(s). Therefore stringent rejection against jets is not needed in SUSY studies, and relatively mild electron identification cuts can be applied, leading to a significant gain in efficiency especially for searches involving many leptons. Jets reconstructed within a cone[2] of $\Delta R = 0.2$ of an identified electron are discarded from the jet list. This procedure prevents the same object being reconstructed both as a jet and as an electron.

A standard algorithm called "eGamma" [22] was used for the electron identification and reconstruction, using the "medium" purity cuts.

The transverse isolation energy in a cone of $\Delta R < 0.2$ around the electron, computed using the calorimetric information, is used to select isolated electrons. This quantity is required to be smaller than 10 GeV. In the available data-sets this variable was incorrectly calculated, but a significant bias is introduced by this problem only in the crack region $1.37 < |\eta| < 1.52$. In this region the electron identification and measurement are also degraded because of the large amount of material in front of the calorimeter and the crack between the barrel and extended barrel of the calorimeters [23]. Events with an electron reconstructed in this region are therefore rejected.

Finally, an electron is rejected if it is found within a distance $0.2 < \Delta R < 0.4$ of a jet, since such candidates are likely to be associated with the decay of a particle within that jet.

## 3.4 Muons

Muons were reconstructed using an algorithm (STACO), which performs a statistical combination of a track reconstructed in the muon spectrometer with its corresponding track in the inner detector [24]. A reasonable quality of combination was guaranteed with a loose requirement that the tracks should match with $\chi^2 < 100$. If more than one Inner Detector track matched a track from the Muon Spectrometer, only the one with best match (smallest distance $\Delta R$) was kept. The total calorimeter energy deposited in a cone of $\Delta R < 0.2$ around the muon was required to be less than 10 GeV. Finally, muons found within a distance $\Delta R < 0.4$ of a jet were discarded.

## 3.5 Taus

As will be described in subsequent analyses, final states containing taus are one of the possible signatures to be expected within the SUSY rich phenomenology. Taus are challenging objects to identify. When taus undergo leptonic decays, their products can be detected in electron- or muon-specific analyses. However, in approximately 65% of cases, a tau decays hadronically and the resultant jet needs to be disentangled

[2]The cone is defined by $\Delta R^2 = \Delta\eta^2 + \Delta\phi^2$, where $\eta$ and $\phi$ are the pseudorapidity and azimuthal angle respectively.

from the overwhelming jet backgrounds expected in ATLAS. In addition, tau signatures always contain an important source of missing energy in the form of neutrinos which usually make it impossible to fully reconstruct the tau-jet energy.

Elsewhere in this volume [25] two tau reconstruction algorithms, one calorimeter-based and the other one track-based, are described. Unless otherwise stated, taus being used in this Chapter follow the recommendations from [25], with a likelihood discriminant of 4, within the fiducial region $|\eta| < 2.5$ and $p_{\mathrm{T}} > 15$ GeV. In addition, although not crucial for the results, an overlap removal procedure has been applied. Since the electron identification efficiency is higher than the tau identification efficiency, a tau object found in a vicinity ($\Delta R < 0.4$) of an electron object (following the prescriptions defined in the electron performance section of the present note) is considered to be a fake and is removed. Also, since hadronically decaying taus are also usually reconstructed as jets, when a calorimeter jet is found within $\Delta R < 0.4$ of a reconstructed tau-jet, then the non-tau jet is disregarded.

## 3.6 Treatment of systematic uncertainties

### 3.6.1 Detector uncertainties

Supersymmetry searches and measurements will inevitably be subject to uncertainties due to the imperfect understanding of the behaviour of the detector. Many of the performance characteristics of the detector will be constrained from the data themselves, with an accuracy which will depend on the size of the data-set available. In this chapter, we have used estimations of these uncertainties based on the precision which is likely to be achievable with 1 fb$^{-1}$ of integrated luminosity. The detector systematics – energy scale, resolution, or efficiency variations – are applied both to the "true" backgrounds and to the control samples; data-driven methods will therefore be less sensitive to detector systematics than purely Monte Carlo-driven estimates.

**Jets and $E_{\mathrm{T}}^{\mathrm{miss}}$** We assume an overall uncertainty on the jet energy scale of 5%. We apply this as a global uncertainty, independently of jet $p_T$ and $\eta$, and identically for light-quark jets and jets from $b$ quarks.

The jet energy resolution will be studied on di-jet samples and jets from $W$ boson decay in $t\bar{t}$ events. After such measurements we expect a residual uncertainty of 10% on the resolution.

The missing transverse energy of the event, $E_{\mathrm{T}}^{\mathrm{miss}}$ , is calculated from the transverse vector sum of high $p_T$ objects like leptons and jets with a further component from unclustered energy. Part of the uncertainty in $E_{\mathrm{T}}^{\mathrm{miss}}$ is thus correlated with jet energy scale uncertainties, but also a wrong calibration of unclustered energy can affect $E_{\mathrm{T}}^{\mathrm{miss}}$ . When jets are rescaled or smeared, the $E_{\mathrm{T}}^{\mathrm{miss}}$ is recalculated with the new jets. This takes into account the part of the $E_{\mathrm{T}}^{\mathrm{miss}}$ uncertainty that is correlated with the jets. For the low $p_T$ part related to the unclustered energy, we first subtract leptons and jets with $p_T > 20$ GeV from the $E_{\mathrm{T}}^{\mathrm{miss}}$ , apply a 10% uncertainty on the remainder, and then add the leptons and jets back in.

**Electrons** For electrons we estimate an uncertainty on the identification efficiency (including the trigger) of 0.5%. Furthermore, we assume an uncertainty on the electron energy scale of 0.2%, and on the energy resolution of 1%. All these uncertainties are assumed to be independent of $p_T$ and $\eta$.

**Muons** For muons we estimate an uncertainty on the identification efficiency of 1% for muons with $p_T < 100$ GeV, plus a 3% extrapolation uncertainty to a $p_T$ of 1 TeV. Furthermore, we assume an uncertainty on the muon $p_T$ scale of 0.2%, and on the $p_T$ resolution of 4% below 100 GeV, whereas the $p_T$ resolution at 1 TeV is assumed to be $10 \pm 1$%. All these uncertainties are assumed to be independent of $\eta$.

**Heavy-flavour tagging** Where $b$ tagging is used, we assume a relative uncertainty of 5% on the $b$-tagging efficiency, and 10% on the light-quark rejection.

### 3.6.2 Monte Carlo uncertainties

Although data-driven background estimation methods try to use Monte Carlo simulations as little as possible by construction, some dependence remains for example when trying to understand the composition of control samples when Monte Carlo-based corrections, distribution shapes or redecay methods are used.

We estimate uncertainties from the use of Monte Carlo samples by comparing different event generators, particularly Alpgen and MC@NLO, and by variation of generator parameters. For the vector boson ($V \in W,Z$) + jets Alpgen samples used in this note, the standard value for the renormalization and factorization scales was $Q^2 = M_{\ell\ell}^2 + p_T^2(V)$ where the two leptons are the vector boson decay products. The $p_T$ threshold of the partons in the matrix-element calculation was 40 GeV, the minimum distance $\Delta R_{jj}$ between partons was 0.7, and the parton distribution function used was CTEQ6L.

For systematic studies we vary:

- The renormalization and factorization scales between $0.5Q$ and $Q$ as defined above. We have also used a sample with a different scale definition $Q^2 = \sum p_T^2(\text{partons})$.
- The $p_T$ threshold of partons in the matrix-element calculation, between 15 and 40 GeV.
- The distance in $\Delta R$ between partons in the matrix-element calculation, between 0.35 and 0.7.
- The parton distribution function, comparing CTEQ6L to MRST2001J.

# 4 Global Variables

The SUSY analyses use some global event variables, built out the momenta of jets, leptons and $\mathbf{p}_T^{\text{miss}}$, which have a good signal discriminating power. We give here a detailed definition of these variables as used in the work described in the SUSY notes.

## 4.1 Effective mass

The **effective mass** ($M_{\text{eff}}$ ) is a measure of the total activity in the event. It is defined as:

$$M_{\text{eff}} \equiv \sum_{i=1}^{4} p_T^{\text{jet},i} + \sum_{i=1} p_T^{\text{lep},i} + E_{\text{T}}^{\text{miss}}$$

where the sums run respectively over the four highest $p_{\text{T}}$ jets within $|\eta| < 2.5$, and over all of the identified leptons. This variable is useful in discriminating SUSY from Standard Model events. It has also the interesting properties that for SUSY events it the $M_{\text{eff}}$ distribution peaks at a value which is strongly correlated with the mass of the pair of SUSY particles produced in the proton-proton interaction. It can therefore be used to quantify the mass-scale of SUSY events [26].

## 4.2 Transverse sphericity

The **transverse sphericity** ($S_T$) is defined as:

$$S_T \equiv \frac{2\lambda_2}{(\lambda_1 + \lambda_2)} \tag{1}$$

where $\lambda_1$ and $\lambda_2$ are the eigenvalues of the $2\times 2$ sphericity tensor $S_{ij} = \sum_k p_{ki}p^{kj}$ . The tensor is computed using all jets with $|\eta| < 2.5$ and $p_T > 20$ GeV, and all selected leptons.

SUSY events tend to be relatively spherical ($S_T \sim 1$) since the initial heavy particles are usually produced approximately at rest in the detector and their cascade decays emit particles in many different directions. QCD events are dominated by back-to-back configurations ($S_T \sim 0$).

### 4.3 Transverse mass

The **transverse mass** $M_T$ is defined by:

$$M_T^2(\mathbf{p}_T^\alpha, \mathbf{p}_T^{\text{miss}}, m_\alpha, m_\chi) \equiv m_\alpha^2 + m_\chi^2 + 2\left(E_T^\alpha E_T^{\text{miss}} - \mathbf{p}_T^\alpha \cdot \mathbf{p}_T^{\text{miss}}\right) \tag{2}$$

where

$$E_T^\alpha \equiv \sqrt{(\mathbf{p}_T^\alpha)^2 + m_\alpha^2}\,, \qquad E_T^{\text{miss}} \equiv \sqrt{(\mathbf{p}_T^{\text{miss}})^2 + m_\chi^2}\,, \tag{3}$$

$m_\alpha$ and $\mathbf{p}_T^\alpha$ are the mass and transverse momentum of some visible particle and $\mathbf{p}_T^{\text{miss}}$ is the missing-transverse-energy two-vector. The parameter $m_\chi$ is the mass of the invisible particle, which is usually assumed to be zero.

This variable is useful when one parent particle decays to one visible and one invisible daughter particle, for example $W \to e\nu$ where it is clear that the mass of the invisible particle (neutrino) can indeed be safely neglected.

### 4.4 Stransverse mass

The **stranverse mass** $m_{\text{T2}}$ variable can be defined in terms of the transverse mass (Eq. (2)) by:

$$m_{T2}^2(\mathbf{p}_T^\alpha, \mathbf{p}_T^\beta, \mathbf{p}_T^{\text{miss}}, m_\alpha, m_\beta, m_\chi) \equiv \min_{\not{\mathbf{q}}_T^{(1)} + \not{\mathbf{q}}_T^{(2)} = \mathbf{p}_T^{\text{miss}}} \left[\max\left\{M_T^2(\mathbf{p}_T^\alpha, \not{\mathbf{q}}_T^{(1)}; m_\alpha, m_\chi),\; M_T^2(\mathbf{p}_T^\beta, \not{\mathbf{q}}_T^{(2)}; m_\beta, m_\chi)\right\}\right] \tag{4}$$

where $m_\chi$ is the trial mass for the lightest SUSY particle and $\mathbf{p}_T^{\alpha,\beta}$ are the transverse momenta of two visible particles (each of which is a canididate decay product of one of the two SUSY parent particles). The vector sum of the dummy variables $\mathbf{q}_T^{(1)}$ and $\mathbf{q}_T^{(2)}$ is constrained to equal the total $\mathbf{p}_T^{\text{miss}}$ 2-vector, so the missing transverse momentum is required as an input to the $m_{\text{T2}}$ calculation. One can consider $m_{\text{T2}}$ to be a variable formed from dividing $\mathbf{p}_T^{\text{miss}}$ into two parts in all possible combinations that satisfy the kinematics of the event (for some $m_\chi$, which is here taken to be zero) and calculating the transverse mass for each decay branch. The resulting value is the best lower limit on the mass of a pair-produced SUSY particle that could have decayed to the observed final state with the given $\mathbf{p}_T^\alpha$, $\mathbf{p}_T^\beta$ and $\mathbf{p}_T^{\text{miss}}$.

The original purpose for $m_{\text{T2}}$ was to provide information on the masses of pair-produced SUSY particles, decaying semi-invisibly [27, 28]. The variable was first proposed to determine SUSY particle masses in “simple” two-body decays, such as two-jet or two-lepton final states, but it can also be used in more complicated cases, especially if it is possible to unambiguously determine which particles came from which branch of the decay.

## 5 SUSY Studies in ATLAS

In this collection of papers we review the techniques that will be used to search for SUSY with ATLAS at the LHC turn-on. Most of the studies presented are based on a total integrated luminosity of $1\text{fb}^{-1}$ that may be collected during the first year of operation of the LHC. Larger datasets are assumed for studies that investigate SUSY models with a lower cross-section as an illustration of studies that can be carried out once the LHC is in full operation.

The first two articles in this collection concentrate on the background estimation for SUSY searches:

- "Data-driven determinations of $W$, $Z$, and top backgrounds to Supersymmetry" [11],
- "Estimation of QCD backgrounds to Searches for Supersymmetry" [21].

The following two articles discuss SUSY searches and measurements:

- "Prospects for Supersymmetry Discovery Based on Inclusive Searches" [29],
- "Measurements from Supersymmetric events" [30].

The last two papers focus on searches for specific signatures:

- "Multi-lepton Supersymmetry searches" [16],
- "Supersymmetry signatures with high-$p_T$ photons or long-lived heavy particles" [5].

# Data-Driven Determinations of $W$, $Z$ and Top Backgrounds to Supersymmetry

**Abstract**
The Standard Model processes of $W$ boson, $Z$ boson and top quark production each in association with jets constitute major backgrounds to searches for Supersymmetry at the LHC. In this note, we estimate the contribution of these backgrounds for a basic SUSY selection, and discuss methods to derive them from the initial 1 $fb^{-1}$ of integrated luminosity at ATLAS.

## 1 Introduction

### 1.1 Motivation

The Large Hadron Collider (LHC) will provide excellent opportunities to search for new physics beyond the Standard Model, and the ATLAS detector [1] is a general purpose experiment to explore such new physics. Supersymmetry (SUSY) is a theoretically attractive model for new physics beyond the Standard Model, and searching for Supersymmetry is one of the main objectives of ATLAS. The actual search strategy is described elsewhere in this volume [2].

It is clear, however, that any discovery of new physics can only be claimed when the Standard Model backgrounds are understood and are under control. It is expected that at the LHC, Monte Carlo predictions will not be sufficient to achieve this: the backgrounds will have to be derived from the data themselves, possibly helped by Monte Carlo. The development and description of such data-driven background estimation is the topic of this note. We note that for a complete understanding of the backgrounds, multiple, independent methods are desired. Each of these may be sensitive to a specific background source, and affected by specific systematic effects. Only their consistency in combination allows for sufficient confidence in the control of the background to claim a discovery when a signal appears to be present.

### 1.2 Data-driven methods: scope of this note

The general aim of data-driven methods is to estimate from the data the Standard Model backgrounds and their uncertainties in a "signal" region, in which new physics may be present. Such a signal region is typically obtained after applying selection cuts, or multivariate methods, and the new physics is searched for as an excess in the number of selected events over background, or as an excess in certain regions of certain distributions.

The background estimation is performed by selection of "control samples", from which predictions in the signal region are derived. Good control samples should be as close as possible to the signal region, yet free of SUSY signal, give an unbiased estimate of background in the signal region, have sufficient statistics, and small theoretical uncertainties. This note intends to describe a number of ideas on selection of such control samples for SUSY searches. Good control of the composition of control samples is important for a correct extrapolation into the signal region.

The methods described in this note should not be regarded as the final word on these procedures, but rather present a number of ideas. Each of these ideas will have to be pursued further, and the effect of other systematic uncertainties will need to be studied. Furthermore, SUSY selection cuts will evolve, and so the methods will need to evolve too. We do believe, however, that a first indication of the uncertainties that can be expected can be given.

This note deals with top, W and Z backgrounds to SUSY searches with primary squark or gluino production, and assuming $R$-parity conservation. The initial priority is to simulate results for 1 $\mathrm{fb}^{-1}$ of integrated luminosity, and for the understanding of the detector expected. The other important QCD background of quark (other than top) and gluon jet production is treated elsewhere in this volume [3]. Backgrounds to alternative production models are also described elsewhere: direct gaugino production [4], and photonic and long-lived particle (such as $R$ hadron) signatures [5].

### 1.3 SUSY contamination

If SUSY is discoverable, it is likely that SUSY events will creep into the control samples, thereby affecting the background estimates. In general, SUSY events, mistakenly regarded as Standard Model physics, will lead to an overestimation of the background, and thus to a reduced SUSY event excess. The extent to which this happens will be analysis- and SUSY model-dependent.

Since we do not know whether SUSY exists, we quote the SUSY contamination effects separately from the other systematics. We will do so by running each data-driven estimation method not only over background samples, but also on a number of SUSY signal samples [6]. The samples represent various regions of mSUGRA parameter space, and together give an impression of the effects. The SU1 sample is a point in the stau coannihilation region, the SU2 sample in the focus point region, and the SU3 sample in the bulk region. The SU4 point is a low-mass point, just above the Tevatron limits. It has a very large cross-section, and kinematic distributions that are typically only slightly harder than the Standard Model background. As will be shown, this model has the largest SUSY contamination effect on the background estimates.

There are a number of ways that the data-driven methods can take the presence of SUSY into account:

1. Iteration. The Standard Model background is evaluated under the assumption that there is no SUSY. This will overestimate the background if there is SUSY, and reduce any excess. Nevertheless, if an excess is seen, the underlying assumption in the background estimation has been proven wrong, and a correction can be applied. This correction can be derived from the properties of the observed excess, and will lead to a new background estimate. An example of such a procedure is the "new MT method" described in section 3.3.3. However, other implementations are possible, and perhaps necessary, as well.

2. A combined fit determining the composition of the control sample, allowing for a possible SUSY contribution.

Both methods are investigated in this note. Nevertheless it is clear that these are preliminary ideas that require further investigation. Most likely, some form of iteration on the background determinations will be necessary.

### 1.4 Layout of this note

A number of important prerequisites for the studies presented here are described in an introductory note [6]:

- the physics processes that form the background to SUSY searches and how they are simulated, as well as a few SUSY event samples (SU1–SU8) that serve to estimate the effect of SUSY on our background estimates;
- the definition of objects like electrons, muon, taus, jets and missing transverse energy, and common variables like the effective mass $M_{\mathrm{eff}}$ ;

- the origin and common treatment of various systematic uncertainties, both from the simulation and from the performance of the detector.

Furthermore, the trigger menu that was used is described elsewhere [7]. In this note we then discuss the W, Z, and top-quark backgrounds and their data-driven estimation for two different SUSY search modes:

1. the mode with one isolated electron or muon (section 2);
2. the no-lepton mode, with a veto against isolated leptons (section 3).

# 2 One-lepton search mode

## 2.1 Selection

The one-lepton search mode is expected to play a major role in the SUSY search, since the requirement of an isolated lepton will be effective in suppressing QCD background. In this search mode, we require one isolated electron or muon, with a $p_T$ of more than 20 GeV. We veto events with a second identified lepton with a $p_T$ of more than 10 GeV, so that we have no overlap with the di-lepton search mode.

We demand at least four jets with $|\eta| < 2.5$ and $p_T > 50$ GeV, at least one of which must have $p_T > 100$ GeV. The transverse sphericity $S_T$ should be larger than 0.2, and the missing transverse energy $E_{\mathrm{T}}^{\mathrm{miss}}$ should be larger than 100 GeV and larger than $0.2M_{\mathrm{eff}}$, where $M_{\mathrm{eff}}$ is the effective mass[1]. The transverse mass $M_T$ reconstructed from the lepton and $E_{\mathrm{T}}^{\mathrm{miss}}$ should be larger than 100 GeV.

## 2.2 Backgrounds in Monte Carlo

In many SUSY models after the selection cuts have been applied clear excesses will be observed in the high $E_{\mathrm{T}}^{\mathrm{miss}}$ and high effective mass regions, as shown in Figure 1. The dominant background process for the one-lepton mode is $t\bar{t}$ (90%), with $W^{\pm}+$jets (10%) being the subdominant process. The neutrino emitted from the $W^{\pm}$ decays produces the $E_{\mathrm{T}}^{\mathrm{miss}}$ in the both processes. Smaller contributions come from $Z+$jets, diboson and single top events and from QCD processes. It is interesting to note that the major $t\bar{t}$ background does not come from the semileptonic ($t\bar{t} \to b\bar{b}\ell\nu q\bar{q}'$) top pair events which are reduced by the $M_T$ and $E_{\mathrm{T}}^{\mathrm{miss}}$ cuts, but rather from the double leptonic ($t\bar{t} \to b\bar{b}\ell\nu\ell\nu$) top decay where one lepton is not identified.

## 2.3 Data-driven estimation strategies

We discuss a variety of different methods to estimate the background from data. These methods differ in their approach and therefore are influenced by different systematic uncertainties, and they focus on different aspects of the background:

1. estimation of W and $t\bar{t}$ background from a control sample formed by reversing one of the selection cuts (on $M_T$) (section 2.3.1);
2. estimation of the semileptonic $t\bar{t}$ background by explicit kinematic reconstruction and selection on top mass (“top box”) (section 2.3.2);
3. estimation of the double leptonic $t\bar{t}$ background, where one lepton is missed, by explicit kinematic reconstruction of a control sample of the same process with both leptons identified (section 2.3.3);

[1] The variables $S_T$, $M_{\mathrm{eff}}$ and $M_T$ are defined elsewhere in this volume [6].

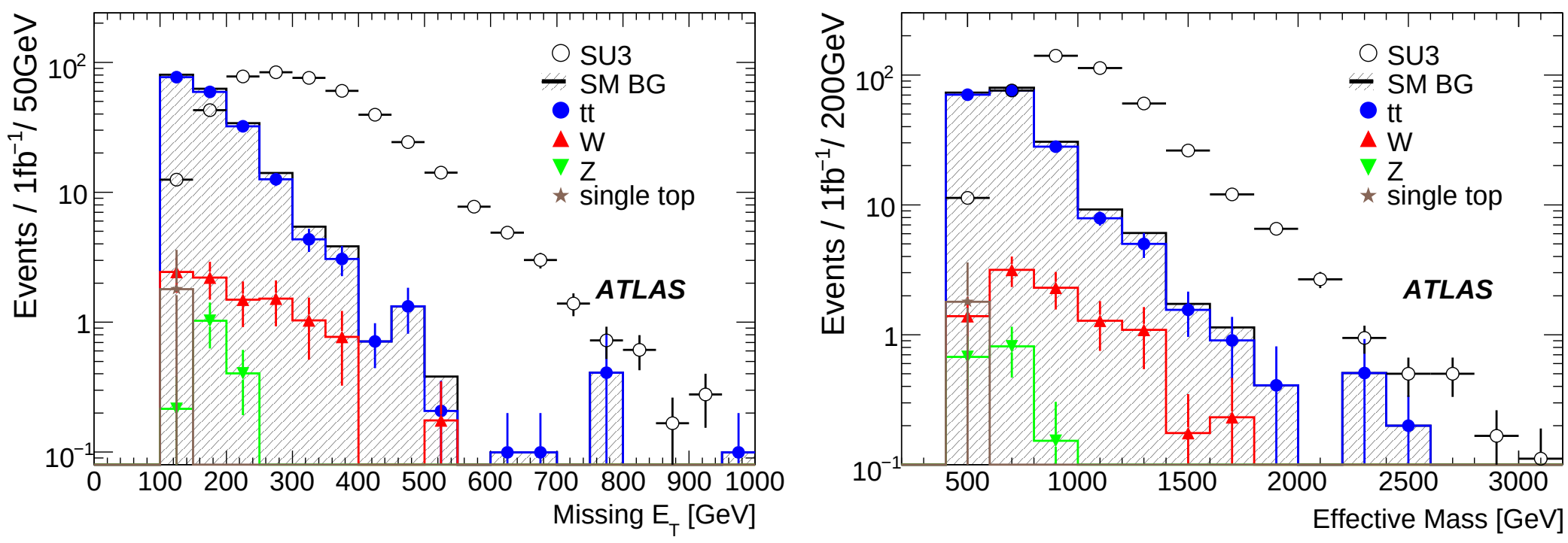


Figure 1: The $E_T^{miss}$ and effective mass distributions for the background processes and for an example SUSY benchmark point (SU3) in the one-lepton mode for an integrated luminosity of 1 fb$^{-1}$. The black circles show the SUSY signal. The hatched histogram show the sum of all Standard Model backgrounds; also shown in different colours are the various components of the background.

4. estimation of that same double leptonic $t\bar{t}$ background from a control sample derived by a cut on a new variable HT2 (section 2.3.4);
5. estimation of $t\bar{t}$ background by Monte Carlo redecay methods (section 2.3.5);
6. estimation of W and $t\bar{t}$ background using a combined fit to control samples (section 2.3.6).

### 2.3.1 Creating a control sample by reversing the $M_T$ cut

The transverse mass $M_T$ is constructed from the identified lepton and the missing transverse energy. In the narrow-width limit $M_T$ is constrained to be less than $m_W$ for the semileptonic $t\bar{t}$ and the $W^{\pm}$ processes. Figure 2 shows that $M_T$ is only weakly dependent on $E_T^{miss}$. This variable is therefore suitable for the estimation of the background distribution itself. Events with small $M_T$ ($< 100$ GeV) are selected as the control sample, in which the $t\bar{t}$ ($\sim 84\%$) and $W^{\pm}$ ($\sim 16\%$) processes are enhanced over the SUSY and the other background processes. The large $M_T$ ($> 100$ GeV) region is referred to as the signal region. Since, for the control sample, the other selection criteria are identical to those for events in the signal region, the same kinematic distributions including $E_T^{miss}$ can be obtained. The number of events for the various processes in signal region and control sample is summarized in the Table 1.

Table 1: Number of background events and estimated numbers for $t\bar{t}$, $W^{\pm}$ and QCD processes without SUSY signal, normalized to 1 fb$^{-1}$.

| | Signal Region | Control Sample |
|---|---|---|
| $t\bar{t}(\ell\nu q\bar{q})$ | 51 (25%) | 1505 (77%) |
| $t\bar{t}(\ell\nu\ell\nu)$ | 140 (70%) | 132 (7%) |
| $W^{\pm}(\ell\nu)$ | 10 (5%) | 305 (16%) |
| SUSY(SU3) | 450 | 317 |

The normalization factor is obtained from the event numbers of the signal region and the control sample ($100 < E_T^{miss} < 200$ GeV), in which the SUSY signal contribution is expected to be relatively

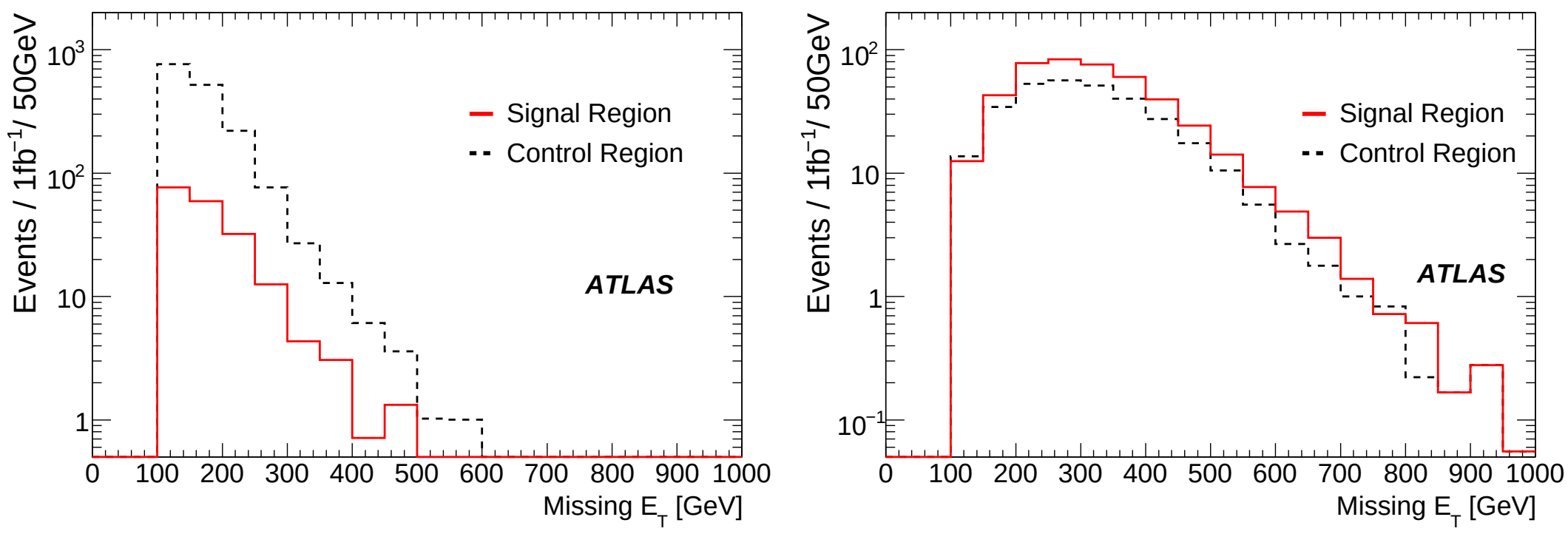


Figure 2: The $E_T^{miss}$ distribution for $t\bar{t}$ (left) and SUSY (SU3, right) signal. In both figures, the solid and dashed histograms show the $E_T^{miss}$ distribution for $M_T > 100$ GeV and $< 100$ GeV, respectively. The numbers are normalized to 1 fb$^{-1}$.

small. Figure 3 shows the $E_T^{miss}$ and $M_{eff}$ distributions which are obtained using this method to estimate the size of these backgrounds, and, for comparison, the true background distributions. The numbers of events with $E_T^{miss} > 100$ GeV and $> 300$ GeV are listed in Table 2. The prediction and the true values agree within the uncertainties, although somewhat less well for high $E_T^{miss}$.

The $t\bar{t}$ event composition of the control sample differs from that of the signal sample, since the $M_T$ cut removes a much larger proportion of the semileptonic $t\bar{t}$ events. The control sample is still able to predict the background in the signal sample within statistical uncertainties. Nevertheless, the resulting systematic shift needs to be investigated, and would be desirable to obtain independent estimates of the fully-leptonic and semileptonic $t\bar{t}$ backgrounds separately.

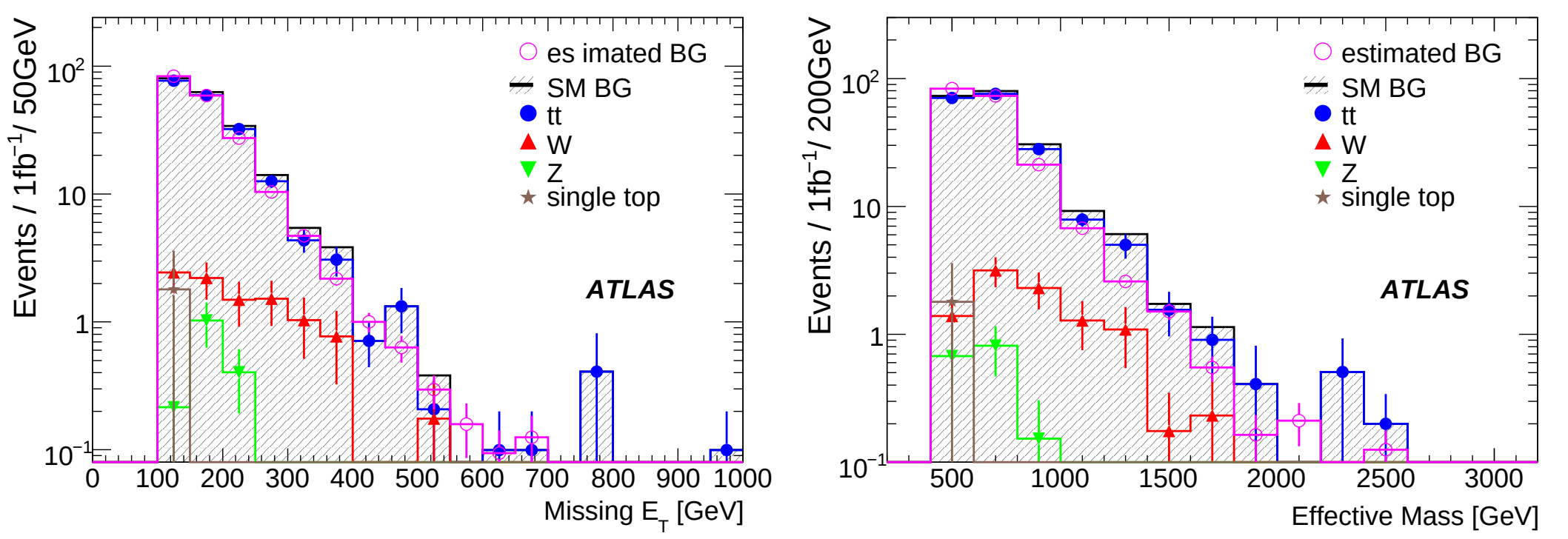


Figure 3: The $E_T^{miss}$ and effective mass distributions of the background processes for the one-lepton mode with an integrated luminosity of 1 fb$^{-1}$. The open circles show the estimated distributions with the $M_T$ method. The hatched histogram shows the true sum of all Standard Model backgrounds; different symbols show the various contributions to the background.

**SUSY signal contamination** If supersymmetric particles are produced they are also likely to contribute to the control samples. The estimated $E_T^{miss}$ distribution with the presence of a SUSY signal (SU3 point) is shown in Figure 4 (left), and the numbers are listed in Table 3. The background is overestimated due to the SUSY contamination, and the inferred $E_T^{miss}$ distribution is biased towards larger values. However,

Table 2: Numbers of background events and estimated numbers for the sum of all background processes without SUSY signal, normalized to 1 fb$^{-1}$

| | $E_T^{miss} > 100$ GeV | $E_T^{miss} > 300$ GeV |
|---|---|---|
| True BG | $203 \pm 6$ | $12.4 \pm 1.6$ |
| Estimated BG | $190 \pm 8$ | $9.4 \pm 0.7$ |
| Ratio(Est./True) | $0.93 \pm 0.05$ | $0.76 \pm 0.11$ |

the amount of the over-estimation is smaller than the SUSY signal itself, and a clear excess can still be observed, as shown in the figure. The same exercise was repeated for other SUSY signal points, as also shown in Table 3.

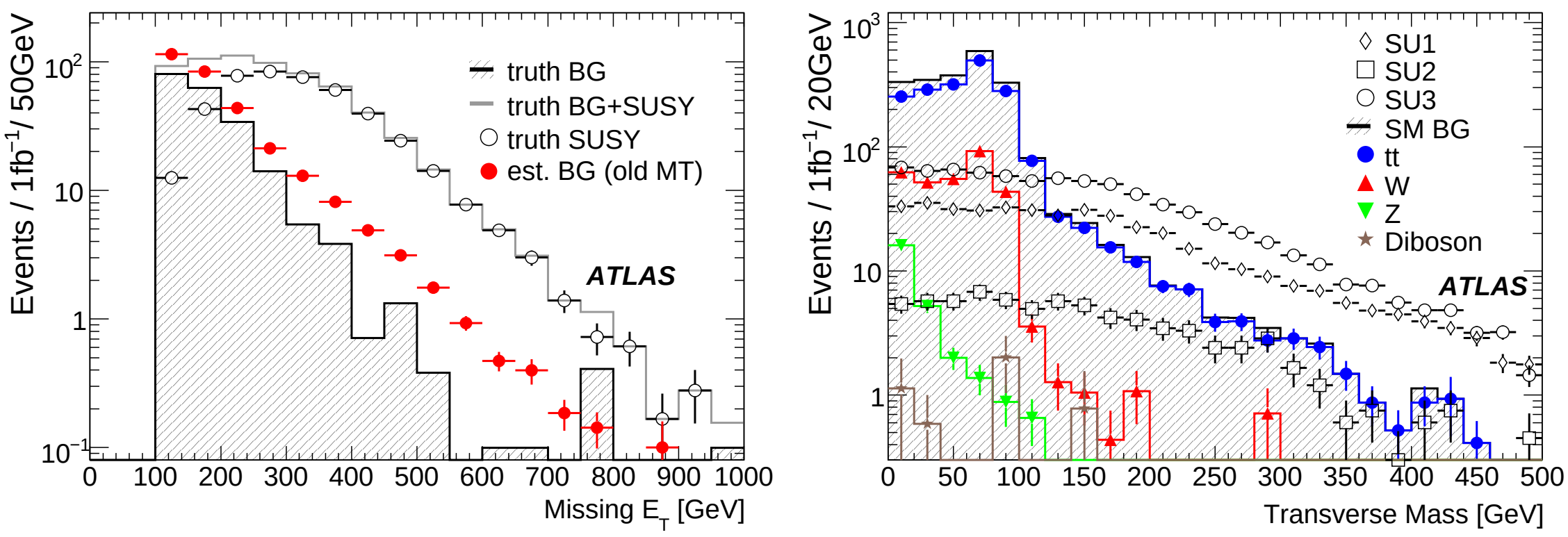


Figure 4: Left: the $E_T^{miss}$ distribution of the background processes for the one-lepton mode with an integrated luminosity of 1 fb$^{-1}$. The red dots show the estimated distributions with the $M_T$ method, with SUSY signal (SU3) present. The hatched histogram shows the sum of all Standard Model backgrounds, and the OPEN histogram shows the SUSY signal (SU3). Right: the transverse mass distributions of the various SUSY signals (SU1, SU2 and SU3) with an integrated luminosity of 1 fb$^{-1}$. Background processes are superimposed for comparison. The hatched histogram shows the sum of all Standard Model backgrounds.

**Correcting for SUSY signal: "New MT method"** If, even for overestimated backgrounds, the presence of a concrete SUSY excess is observed in data, we can try to correct the background estimates.

One possible procedure is described here, referred to as the "new MT method". More advanced implementations of such a correction procedure are possible and should be studied.

The new MT method makes use of the observation that in the one-lepton search mode, the $M_T$ distribution of backgrounds falls off steeply beyond $\sim 100$ GeV, whereas for many SUSY signal models this distribution falls only slowly. This is illustrated in Figure 4 (right). By making a general ansatz for the shape of the SUSY $M_T$ distribution, and neglecting to first order the Standard Model background at high $M_T$, the SUSY contamination can be subtracted from the control sample. Obviously, remaining Standard Model background in the high $M_T$ region and variations in the $M_T$ shape for various SUSY signals are to be treated as systematic uncertainties on the method. Nevertheless, the data itself will tell what the $M_T$ shape is.

Table 3: Number of background events and estimated numbers for all background processes with SUSY signal, normalized to 1 fb$^{-1}$. Also the total number of events (SUSY + background) is shown.

| | $E_T^{miss} > 100$ GeV | $E_T^{miss} > 300$ GeV | $E_T^{miss} > 100$ GeV | $E_T^{miss} > 300$ GeV |
|---|---|---|---|---|
| True BG | $203 \pm 6$ | $12.4 \pm 1.6$ | $203 \pm 6$ | $12.4 \pm 1.6$ |
| | SU1 | | SU4 | |
| Estimated BG | $225 \pm 9$ | $21.6 \pm 1.1$ | $2366 \pm 102$ | $165 \pm 12.7$ |
| True BG+SUSY | $463 \pm 7$ | $194 \pm 4$ | $3177 \pm 79$ | $415 \pm 29$ |
| | SU2 | | SU6 | |
| Estimated BG | $200 \pm 9$ | $10.9 \pm 0.7$ | $213 \pm 9$ | $16.3 \pm 0.9$ |
| True BG+SUSY | $249 \pm 7$ | $34 \pm 2$ | $365 \pm 9$ | $129 \pm 5$ |
| | SU3 | | SU8 | |
| Estimated BG | $296 \pm 10$ | $33.3 \pm 1.4$ | $206 \pm 9$ | $13.7 \pm 0.8$ |
| True BG+SUSY | $653 \pm 8$ | $245 \pm 4$ | $354 \pm 8$ | $115 \pm 5$ |

In the simplest ansatz used here, the ratio of SUSY signal between the control sample $M_T < 100$ GeV and signal region $M_T > 100$ GeV is assumed to be constant for all SUSY signal samples. The normalization factor is obtained from the number of events in the signal region and the corrected control sample in the interval 100 GeV $< E_T^{miss} < 150$ GeV (instead of 100 – 200 GeV) to suppress the SUSY contribution in the normalization region. The statistical error becomes relatively larger when the narrow band is used for normalization, but the over-estimation of the normalization factor due to the SUSY signal can be suppressed. A lower $E_T^{miss}$ region, such as $E_T^{miss} = 70 - 100$ GeV, could be used for the normalization in future studies.

Figure 5 shows the $E_T^{miss}$ and the effective mass distributions of the estimated background processes. The true distributions of the background processes are also superimposed. The numbers in regions of $E_T^{miss} > 100$ GeV and 300 GeV are listed in Table 4. A reasonable agreement between the prediction and the true values is observed. For high values of $E_T^{miss}$, the method tends to subtract too much SUSY contamination and underestimates the background. More study is needed. The SU4 benchmark point is a special case because it has a particularly light SUSY particle spectrum.

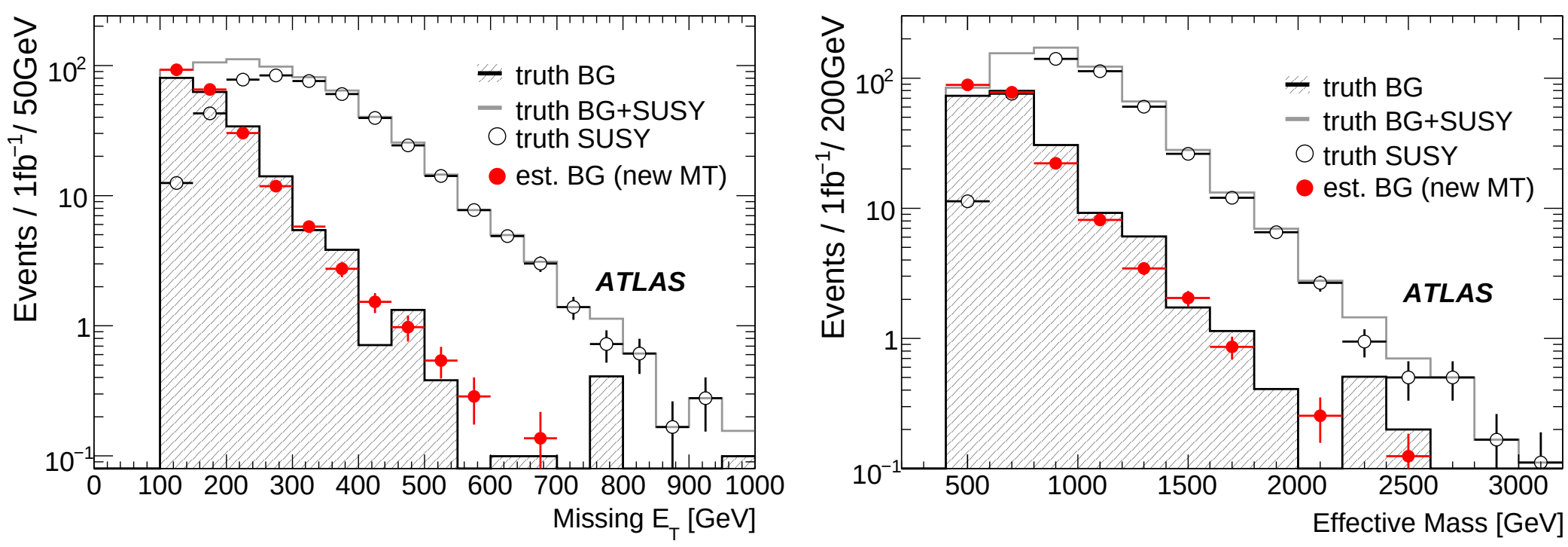


Figure 5: The $E_T^{miss}$ and effective mass distributions of the background processes for one lepton mode with an integrated luminosity of 1 fb$^{-1}$. The red dots show the estimated distributions with the “new $M_T$” method. The hatched histogram show the sum of all Standard Model backgrounds. The open circles indicate the SUSY (SU3) signal.

Table 4: Numbers of background events and estimated numbers for all background processes in the presence of various SUSY signals, using the new MT method. The numbers are normalized to 1 fb$^{-1}$.

| | $E_{\mathrm{T}}^{\mathrm{miss}} > 100$ GeV | $E_{\mathrm{T}}^{\mathrm{miss}} > 300$ GeV | $E_{\mathrm{T}}^{\mathrm{miss}} > 100$ GeV | $E_{\mathrm{T}}^{\mathrm{miss}} > 300$ GeV |
|---|---|---|---|---|
| True BG | $203 \pm 6$ | $12.4 \pm 1.6$ | $203 \pm 6$ | $12.4 \pm 1.6$ |
| | SU1 | | SU4 | |
| Estimated BG | $186 \pm 11$ | $8.9 \pm 0.8$ | $1382 \pm 98$ | $48.3 \pm 12.7$ |
| True BG+SUSY | $463 \pm 7$ | $194 \pm 4$ | $3177 \pm 79$ | $415 \pm 29$ |
| | SU2 | | SU6 | |
| Estimated BG | $183 \pm 11$ | $8.8 \pm 0.8$ | $185 \pm 11$ | $8.1 \pm 0.9$ |
| True BG+SUSY | $249 \pm 7$ | $34 \pm 2$ | $365 \pm 9$ | $129 \pm 5$ |
| | SU3 | | SU8 | |
| Estimated BG | $212 \pm 11$ | $12.3 \pm 1.0$ | $180 \pm 11$ | $6.6 \pm 0.8$ |
| True BG+SUSY | $653 \pm 8$ | $245 \pm 4$ | $354 \pm 8$ | $115 \pm 5$ |

The systematic uncertainties[2] for the MT method are summarized in Table 5. As well as variation of jet energy scale and lepton identification efficiency, the ALPGEN Monte Carlo was compared to MC@NLO, and parameters in ALPGEN (minimum $p_T$ of partons and minimum $\Delta R$ between partons) were varied. This method is stable against these systematic uncertainties at the $\sim 15\%$ level. More work is needed to estimate the SUSY contamination effects.

Table 5: Systematic uncertainties of the one-lepton background estimations with the MT method, excluding those related to SUSY signal contamination. Numbers are normalized to 1 fb$^{-1}$

| | Syst. error |
|---|---|
| Jet energy scale | $< 5\%$ |
| Lepton ID efficiency | 7% |
| MC@NLO vs ALPGEN | 8% |
| Monte Carlo parameter variation (ALPGEN) | $< 5\%$ |

### 2.3.2 Topbox: a control sample for semileptonic top-pair background

**Top mass reconstruction and "topbox" cuts** This section describes a data-driven method, denoted the "topbox method", for estimating the $t\bar{t}$ background where one top decays leptonically, and the other hadronically.

For semileptonic $t\bar{t}$ events, the invariant mass of the leptonically decaying $W$ boson can usually be reconstructed by assuming that the neutrino from the $W$ decay is responsible for all missing energy. This is a fair assumption; after removal of fake $E_{\mathrm{T}}^{\mathrm{miss}}$ (noisy/dead calorimeter cells etc.) in the event-cleaning procedure, the resolution on $E_{\mathrm{T}}^{\mathrm{miss}}$ is expected to be approximately equal to $0.55\sqrt{\sum E_T}$ [1], which is much smaller in a typical $t\bar{t}$ event than the $E_{\mathrm{T}}^{\mathrm{miss}}$ from the escaping neutrino. The fact that the mass of the leptonically decaying top can be reconstructed satisfactorily (see below) further justifies the assumption.

The core of the method is to construct both the semileptonic and the hadronic top decays in a $t\bar{t}$ event following the procedure below:

[2]Throughout this note systematic uncertainties have been calculated according to the procedures outlined in the introduction to this chapter [6].

- The leptonic $W$ is assumed to decay into the observed lepton and a neutrino which is responsible for all missing energy. The $p_x$ and $p_y$ components of the neutrino momentum are hence taken to be the $x$ and $y$ components of $E_{\mathrm{T}}^{\mathrm{miss}}$ . The $p_z$ component of the neutrino can be calculated using a $W$ mass ($m_W$ ) constraint. The four-vector of the leptonic $W$ is the sum of the four-vectors of the lepton and the reconstructed neutrino. For events with transverse mass $M_T$ less than $m_W$ , two solutions can be found. In the case of $M_T > m_W$ no real solution is possible and, in such cases, the momentum of the leptonic $W$ is taken from the transverse components of the lepton and $E_{\mathrm{T}}^{\mathrm{miss}}$ .
- The leptonic top is then reconstructed by taking the solution with the best reconstructed top mass ($m_{\text{top-lep}}$ ) from combinations of a jet and one of the above leptonic $W$ solutions. The jet is taken from the pool of the four highest-$p_T$ jets in the event. The best reconstructed top mass is defined to the one that is closest to the nominal top mass $m_t$ .
- The hadronic $W$ is then taken to be formed from the best reconstructed $W$ mass ($m_{\text{W-had}}$ ) among the two-jet combinations from the remaining three jets in the pool. The best reconstructed $W$ mass is defined to be the invariant closest to $m_W$ .
- Finally, the hadronic top is taken to be the one with the best reconstructed top mass ($m_{\text{top-had}}$ ) among combinations of the hadronic $W$ and one of the remaining jets.

The plots in Figure 6 show the distributions of the reconstructed masses $m_{\text{top-lep}}$ , $m_{\text{W-had}}$ , and $m_{\text{top-had}}$ after the mass reconstruction procedure described above. The distributions are made for $t\bar{t}$ , SU3 and $W+$jets event samples with standard one-lepton cuts, except for a modified $M_T$ requirement (see below in the control sample section). As expected, the topbox mass reconstruction procedure offers a very good separating power between $t\bar{t}$ and other processes.

The topbox cuts are then defined as follows: $|m_{\text{top-lep}} - m_t| < 25$ GeV, $|m_{\text{W-had}} - m_W| < 15$ GeV, and $|m_{\text{top-had}} - m_t| < 25$ GeV.

**Topbox control sample** To make the topbox control sample, events are selected with the standard SUSY search cuts in the one-lepton mode, with the exception that $M_T > 100$ GeV is replaced by $M_T < m_W$. In addition, the above topbox cuts are applied.

Table 6 shows the number of events of various processes in the topbox control sample. The $t\bar{t}+$jets process makes up more than 95% of the topbox control sample if no SUSY signal is present.

Table 6: Composition of the topbox control sample. Numbers shown correspond to an integrated luminosity of 1 fb$^{-1}$ . The last five columns show the numer of SUSY events which would enter into the topbox control sample.

| Process | $t\bar{t}+$jets | $W+$Jets | SU1 | SU2 | SU3 | SU4 | SU6 |
|---|---|---|---|---|---|---|---|
| Events | 340.9 | 6.8 | 1.8 | 0.4 | 4.9 | 243.6 | 0.4 |

**SUSY signal contamination** Table 6 also shows the number of SUSY events, for various signal samples, in the topbox control sample, for 1 fb$^{-1}$. In this method, SUSY contamination is in general small. This fact makes the topbox method a good supplement to the other methods (e.g. the MT method). The exception is the SU4 benchmark point, which has a larger contribution because its light spectrum makes it rather similar to the $t\bar{t}$ background.

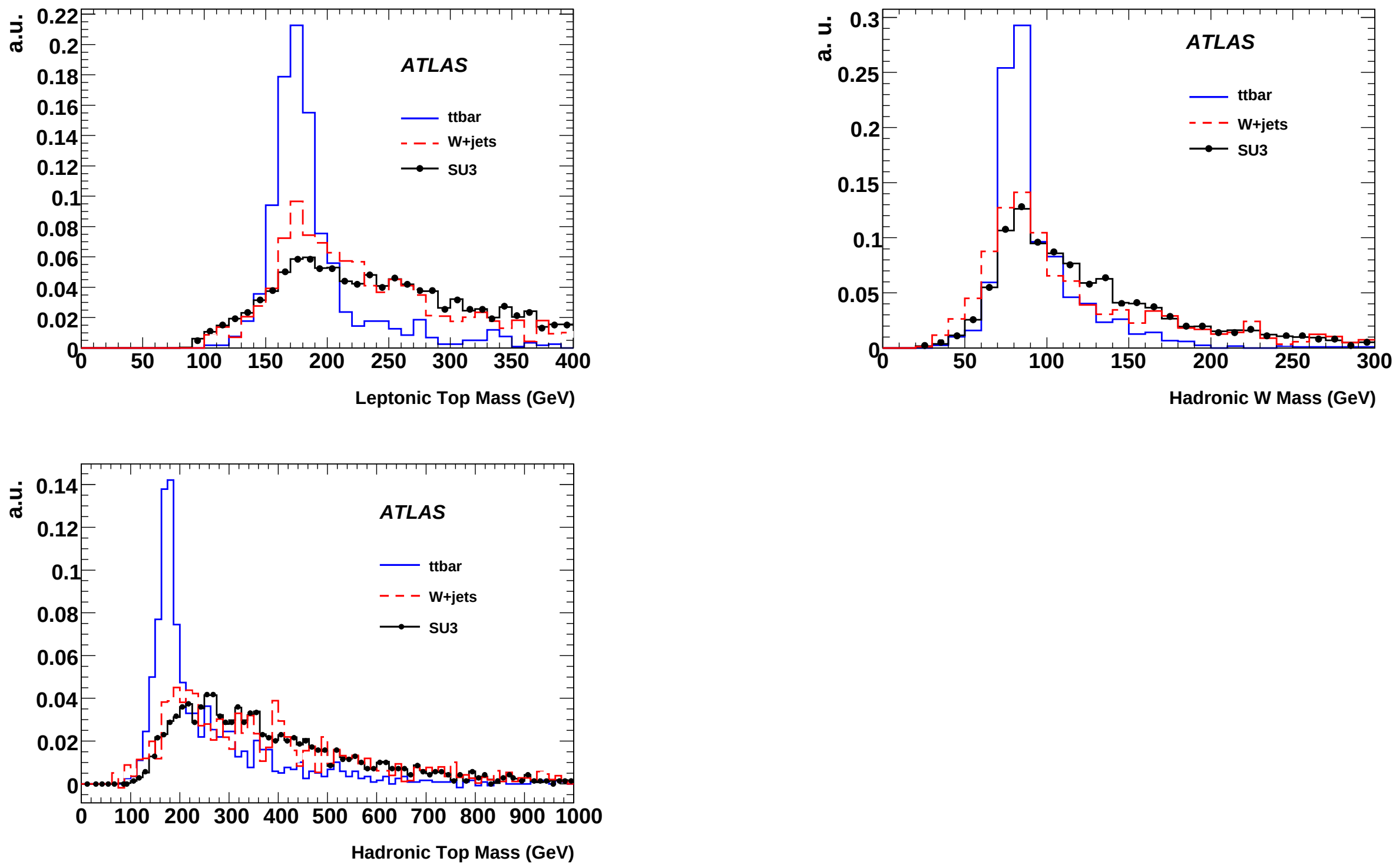


Figure 6: Normalized distributions for reconstructed $m_{\text{top-lep}}$ , $m_{\text{W-had}}$ , and $m_{\text{top-had}}$ for $t\bar{t}$ , $W$ + jets, and SU3 SUSY events, using the "topbox" method.

**Estimation of the $t\bar{t}$ background in the signal region** The $t\bar{t}$ contamination in the signal region is estimated by multiplying the number of events in the data topbox by a scaling factor $R_{tt}$ . $R_{tt}$ is defined as the ratio of the number of Monte Carlo $t\bar{t}$ events in the signal region (those that pass the one-lepton cuts) to that in the topbox control sample. The procedure is summarized by the following equations:

$$N_{t\bar{t}}^{\text{signal-region}}(\text{data}) = N_{t\bar{t}}^{\text{topbox}}(\text{data}) \cdot R_{tt} \quad (1)$$

$$R_{tt} \equiv N_{t\bar{t}}^{\text{signal-region}}(\text{MC}) / N_{t\bar{t}}^{\text{topbox}}(\text{MC}) \quad (2)$$

With fully simulated Monte Carlo samples, $R_{tt}$ is determined to be 0.386 . The model dependence (variation of Monte Carlo generator and generator parameters) of this number is treated as a systematic uncertainty.

**Systematics** The systematic uncertainties of the topbox method are summarized in Table 7. The largest source of uncertainty is from the jet energy scale uncertainty; this is expected since the method relies heavily on the reconstruction of top and $W$ masses. The Monte Carlo model dependency of $R_{tt}$ is estimated by comparing MC@NLO and ALPGEN, and by variation of the ALPGEN parameters, and amounts to 8%. Finally, it is expected that extra jets due to event pile-up may affect the mass reconstruction resolution. However, this is relevant only in high luminosity scenarios, beyond the scope of this note. The statistical uncertainty on the topbox control sample normalization is estimated to be 5% for 1 fb$^{-1}$ given that the effective cross-section of $t\bar{t}$ in the topbox is about 400 fb.

Table 7: Systematic uncertainties of the topbox method for 1 fb$^{-1}$.

| Source | Contribution [%] |
|---|---|
| Jet energy scale | 20 |
| $E_{\mathrm{T}}^{\mathrm{miss}}$ scale | 2 |
| Monte Carlo Model dependence of $R_{tt}$ | 8 |
| Total | 22 |

#### 2.3.3 Di-leptonic top with one lepton missed: kinematic reconstruction

**Introduction** Fully leptonic $t\bar{t}$ events may contribute to the one-lepton SUSY search sample if one of the two leptons originating from the $W$ decay is not identified. Such events can be classified as: (1) events with one tau (51%); (2) events where one lepton is misidentified due to inefficiency of the lepton identification algorithms (20%); (3) events where one lepton is lost inside a jet (17%); (4) events where one lepton is not in the $p_T$ or $\eta$ acceptance (9%); and (5) events with two tau leptons (3%).

The method discussed here is based on the selection of a sample enhanced in $t\bar{t} \to b\bar{b}\ell\nu\ell\nu$ events by requiring that the events satisfy a set of kinematic constraints particular to the $t\bar{t} \to b\bar{b}\ell\nu\ell\nu$ process. This sample, denoted as the control sample, with two isolated identified leptons, is used to estimate the contribution from the first two categories of events listed above. The contribution from category (1) is estimated by replacing one of the leptons in the control sample with a tau, and category (2) is estimated by removing one of the two leptons. The contribution from the categories (3)–(5) is not estimated from the control sample. Events were required to fire either the 4j50 multi-jet trigger or the j80_xE50 jet plus $E_{\mathrm{T}}^{\mathrm{miss}}$ trigger [7].

**Selection of the control sample** The following requirements are imposed to select events in the control sample: two isolated oppositely-charged leptons (electron or muon), with $p_{\mathrm{T}} > 10$ GeV and at least one with $p_{\mathrm{T}} > 20$ GeV; at least three jets with $|\eta| < 2.5$ and $p_{\mathrm{T}} > 50$ GeV at least one of which must have $p_{\mathrm{T}} > 100$ GeV. Note that in contrast to the SUSY one-lepton search selection given in Sec. 2.1 only three jets are required, since the misidentified lepton or tau can produce the fourth jet.

For $t\bar{t} \to b\bar{b}\ell\nu\ell\nu$ events the two leptons, two $b$ jets and the $x$- and $y$-components of the $E_{\mathrm{T}}^{\mathrm{miss}}$ -vector satisfy the following kinematic constraints:

$$\begin{aligned}
(p_\nu + p_{\ell^+})^2 &= m_W^2, \\
(p_{\bar{\nu}} + p_{\ell^-})^2 &= m_W^2, \\
(p_\nu + p_{\ell^+} + p_b)^2 &= m_t^2, \\
(p_{\bar{\nu}} + p_{\ell^-} + p_{\bar{b}})^2 &= m_t^2, \\
p_{\nu_x} + p_{\bar{\nu}_x} &= E_{\mathrm{T,x}}^{\mathrm{miss}}, \\
p_{\nu_y} + p_{\bar{\nu}_y} &= E_{\mathrm{T,y}}^{\mathrm{miss}},
\end{aligned} \tag{3}$$

where $p_{\ell^\pm}$, $p_{\nu/\bar{\nu}}$, $p_{b/\bar{b}}$ are the lepton, neutrino and $b$-quark momenta respectively and $m_W$ and $m_t$ are the $W$ boson and top quark masses. We assume that the only source of $E_{\mathrm{T}}^{\mathrm{miss}}$ is a pair of neutrinos, which is a fair assumption as shown in the previous section.

The final state contains two unknown neutrino momenta and the above system of equations has a two- or four-fold ambiguity, as the solution is given by a quartic equation which can be solved with standard analytical techniques [8]. Since there are at least three jets in each event, all possible combinations of

jet pairs made from the three highest $p_T$ jets are considered. Jet pairs for which the above system of equations has real solutions are denoted as $b$-jet pairs[3]. Figure 7 (left) shows the number of $b$-jet pairs for the various processes contributing to the control sample.

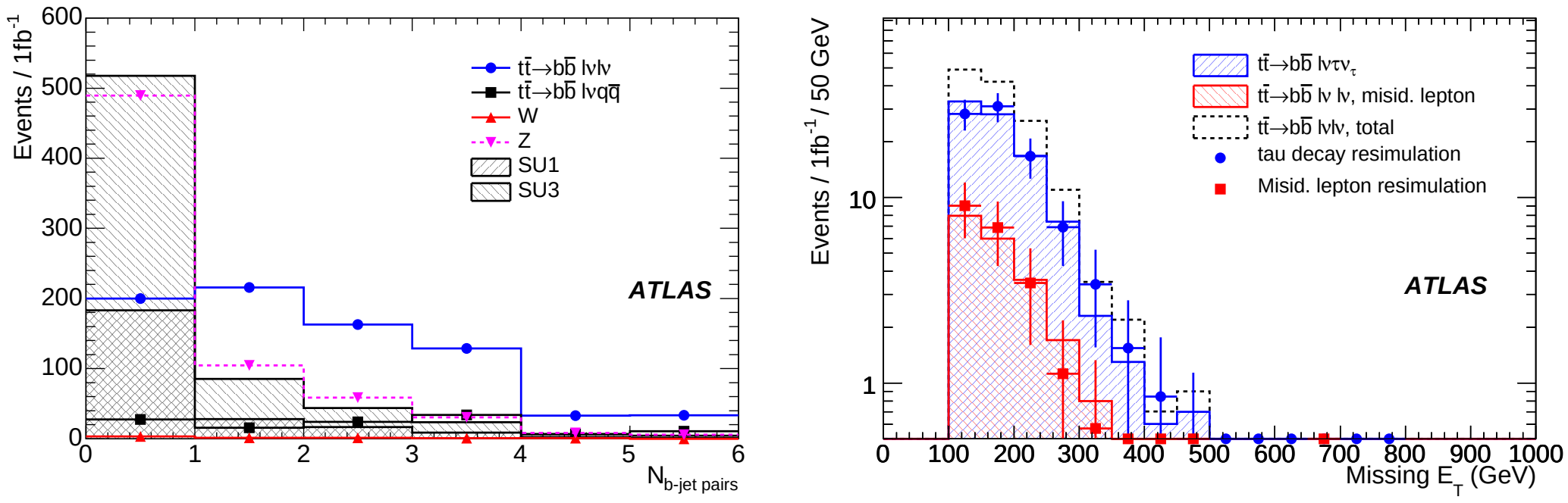


Figure 7: Left: distribution of number of $b$-jet pairs for events passing the control sample requirements in the kinematic reconstruction method. The fraction of $t\bar{t} \to b\bar{b}\ell\nu\ell\nu$ events with no $b$-jet pairs is dominated by events with at least one $b$ jet which is not among the three highest-$p_T$ jets. Right: distribution of $E_T^{miss}$ for $t\bar{t} \to b\bar{b}\ell\nu\ell\nu$ events with one tau lepton and events with a misidentified lepton compared to the estimation from resimulated events with an integrated luminosity of 1 fb$^{-1}$. The requirement on the number of $b$-jet pairs is not applied to the resimulated events. The distribution of all $t\bar{t} \to b\bar{b}\ell\nu\ell\nu$ events is also shown.

**Replacement procedure** Each event in the control sample is used as a seed for producing a series of resimulated events. One of the two identified leptons in the seed event is replaced by tau lepton and a set of 1000 tau decays are simulated using the TAUOLA package [9]. The same procedure is repeated for the second lepton in the seed event, yielding a total of 2000 events for every seed event. Each resimulated event is weighted by a factor of $1/\varepsilon$, where $\varepsilon$, the identification efficiency for the replaced lepton, is estimated from simulations.

The contribution of events where one lepton evades identification is estimated as follows. If the replaced lepton is an electron then a jet with the same momentum is substituted instead of it. If the lepton is a muon it is replaced by a so-called stand-alone muon (defined as a track in the muon spectrometer with no match to a track in the inner detector) justified by the fact that most muons not passing the muon definition are stand-alone muons. This procedure is applied to each of the two leptons in the seed events, resulting in two resimulated events for each seed event. The resimulated events are re-weighted with $\frac{1-\varepsilon}{\varepsilon}$.

For both kinds of resimulated events, the SUSY one-lepton search selection are subsequently applied.

As a closure test of the replacement procedures described above, the $E_T^{miss}$ distribution for resimulated $t\bar{t} \to b\bar{b}\ell\nu\ell\nu$ events passing the control sample selection apart from the requirement of $b$-jet pairs, is compared to the Monte Carlo prediction. The result is shown in Fig. 7 (right) and shows good agreement.

**Normalization** The number of $t\bar{t} \to b\bar{b}\ell\nu\ell\nu$ events in the signal region is estimated by scaling of the sum of described above contributions with two scaling factors. The first factor takes into account the other categories of $t\bar{t} \to b\bar{b}\ell\nu\ell\nu$ events that are not estimated by this method. This first factor is

[3]Note that within this section only kinematical conditions have used to identify these $b$-jet pairs – no secondary-vertex requirement is used.

Table 8: Estimated background corresponding to an integrated luminosity of 1 fb$^{-1}$ for different mSUGRA benchmark points. The second column shows the relative increase of the estimated background with respect to the estimation without contamination from the SUSY signal. The third column shows the number of SUSY events. The Monte Carlo prediction of $t\bar{t} \rightarrow b\bar{b}\ell\nu\ell\nu$ background in the one lepton search mode is 136 events. The errors in the first column are statistical only.

| SUSY point | Estimated Background | Relative change [%] | True Signal Events |
|---|---|---|---|
| No signal | $120 \pm 14$ | | |
| SU1 | $137 \pm 15$ | 15 | 260 |
| SU2 | $127 \pm 15$ | 5.9 | 45 |
| SU3 | $176 \pm 18$ | 47 | 454 |
| SU4 | $604 \pm 38$ | 405 | 2960 |
| SU6 | $129 \pm 16$ | 7.8 | 162 |
| SU8 | $124 \pm 14$ | 3.8 | 100 |

estimated from Monte Carlo to be $R^{\mathrm{MC}} = 1.4 \pm 0.1$. The second normalization factor, $R^{b\text{-jet pair}}$, takes into account the efficiency of $t\bar{t} \rightarrow b\bar{b}\ell\nu\ell\nu$ events to pass the requirement on the number of $b$-jet pairs; it is defined as the ratio of resimulated events before and after the $b$-jet pair selection in a normalization region, $80 \leq E_{\mathrm{T}}^{\mathrm{miss}} \leq 120$ GeV, and found to have the value $R^{b\text{-jet pair}} = 1.4 \pm 0.1(\mathrm{stat}) \pm 0.1(\mathrm{syst})$.

**Presence of SUSY** A possible SUSY signal could have an effect on the background estimation in two ways: 1) by satisfying the kinematic constraints in Eq. 3 and therefore enter the control sample and 2) by entering the normalization region giving a systematic contribution to the scale factor $R^{b\text{-jet pair}}$. In Fig. 8 the estimated $t\bar{t} \rightarrow b\bar{b}\ell\nu\ell\nu$ background is shown with and without the contamination of a SUSY signal (SU3) while Tab. 8 gives the estimated number of $t\bar{t} \rightarrow b\bar{b}\ell\nu\ell\nu$ events in the presence of different SUSY signals.

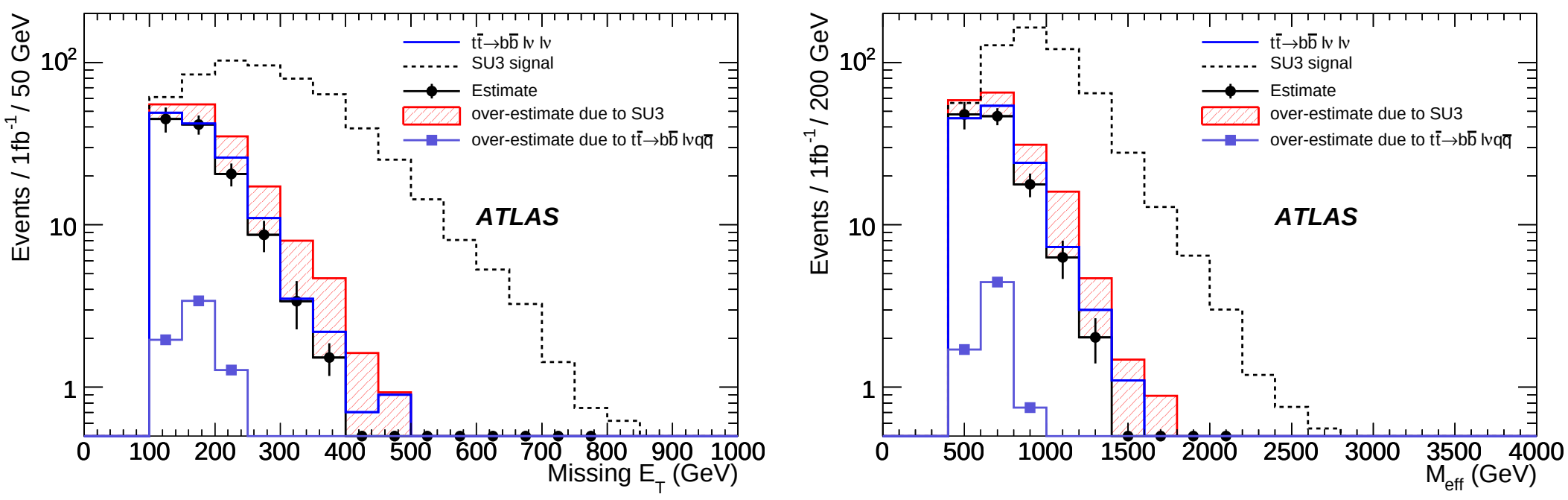


Figure 8: The $E_{\mathrm{T}}^{\mathrm{miss}}$ (left) and $M_{\mathrm{eff}}$ (right) distributions for the estimated and true $t\bar{t} \rightarrow b\bar{b}\ell\nu\ell\nu$ contribution for the one-lepton SUSY search. Black points (red area) represent the estimation without (with) the presence of a signal from SUSY (SU3).

**Systematic Uncertainties** The systematic uncertainties for this method are summarized in Tab. 9. The uncertainty from the replacement procedure is estimated by comparing number of resimulated events to the Monte Carlo prediction, see Fig. 7(right). The uncertainty of $R^{\mathrm{MC}}$ is estimated by comparing

Table 9: Breakdown of systematic uncertainties in the kinematic reconstruction method.

| Source | Contribution [%] |
|---|---|
| Replacement | 10 |
| $R^{\mathrm{MC}}$ | 10 |
| Jet Energy Scale | 9 |
| $R^{b\text{-jet pair}}$ stat. | 9 |
| $R^{b\text{-jet pair}}$ syst. | 8 |
| Background Subtraction | 3 |
| Jet Energy Resolution | 1 |
| $E_{\mathrm{T}}^{\mathrm{miss}}$ scale | 1 |
| Total | 21 |

MC@NLO and ALPGEN. The statistical uncertainty of $R^{b\text{-jet pair}}$ is calculated using binomial errors. The systematic uncertainty of this factor takes into account the difference in the shapes between $E_{\mathrm{T}}^{\mathrm{miss}}$ distribution of the resimulated samples with and without applying the kinematical constraints in Eq. 3. The uncertainty due to background subtraction is dominated by the presence of $t\bar{t} \to b\bar{b}q\bar{q}\ell\nu$ events in the control sample. The systematic effects resulting from uncertainties in the lepton identification efficiency, the trigger efficiencies and the energy scale and resolution are expected to be much smaller.

#### 2.3.4 Dileptonic top with one lepton missed: HT2

**Introduction** In this section we describe a method, denoted the "HT2 method", to estimate background from dileptonic $t\bar{t}$ production where one of the leptons is not identified. It relies on the (near) independence of $E_{\mathrm{T}}^{\mathrm{miss}}$ and the variable HT2. This variable is defined as:

$$\mathrm{HT2} \equiv \sum_{i=2}^{4} p_T^{\mathrm{jet}\,i} + p_T^{\mathrm{lepton}}. \qquad (4)$$

In the HT2 method, the shape of the $E_{\mathrm{T}}^{\mathrm{miss}}$ distribution is estimated from dileptonic $t\bar{t}$ events with low HT2. This distribution is then normalized to the number of events at large HT2, but with low missing $E_{\mathrm{T}}$ , and can then be used to estimate the remaining backgrounds in the signal region of large HT2 and large $E_{\mathrm{T}}^{\mathrm{miss}}$ .

For this method to work, the shape of the $E_{\mathrm{T}}^{\mathrm{miss}}$ distribution needs to be independent of HT2. Note that in Equation 4, the leading jet $p_{\mathrm{T}}$ was excluded from the sum in order to reduce the correlation with $E_{\mathrm{T}}^{\mathrm{miss}}$ . The correlation between the hightest-$p_{\mathrm{T}}$ jet and $E_{\mathrm{T}}^{\mathrm{miss}}$ is likely to be due to simple kinematics, i.e. to first approximation, the rest of the event recoils against this leading jet. This is illustrated in Figure 9 which shows the $E_{\mathrm{T}}^{\mathrm{miss}}$ distribution (at Monte Carlo "truth level") in slices of leading and sub-leading jet $p_{\mathrm{T}}$ . The reduced dependence of the $E_{\mathrm{T}}^{\mathrm{miss}}$ shape on the jet $p_{\mathrm{T}}$ in the second-leading jet case is apparent, and will be further diminished by detector resolution effects.

To further reduce the correlation between HT2 and $E_{\mathrm{T}}^{\mathrm{miss}}$ , the $E_{\mathrm{T}}^{\mathrm{miss}}$ significance was used. This is to remove the correlation which arises from the fact that the $E_{\mathrm{T}}^{\mathrm{miss}}$ resolution depends on $\sum E_{\mathrm{T}}$, where $\sum E_{\mathrm{T}}$ is clearly related to HT2. A simple form of $E_{\mathrm{T}}^{\mathrm{miss}}$ significance was used here, defined as $E_{\mathrm{T}}^{\mathrm{miss}}$ significance = $E_{\mathrm{T}}^{\mathrm{miss}}$ /$[0.49 \cdot \sqrt{\sum E_{\mathrm{T}}}]$.

The results shown here are from a data sample consisting of the sum of $t\bar{t}$ (semi-leptonic and dileptonic decay modes) plus W($l\nu$)+jets (where $l = e, \mu, \tau$). The trigger used in this analysis was the logical OR of the 4j50 multi-jet, the e22i single electron and the mu20 single muon triggers [7].

A control sample defined by HT2 $< 300$ GeV was used to estimate the shape of the $E_{\mathrm{T}}^{\mathrm{miss}}$ significance. The assumption is that this shape is independent of HT2 so it can be used to predict the shape of the $E_{\mathrm{T}}^{\mathrm{miss}}$

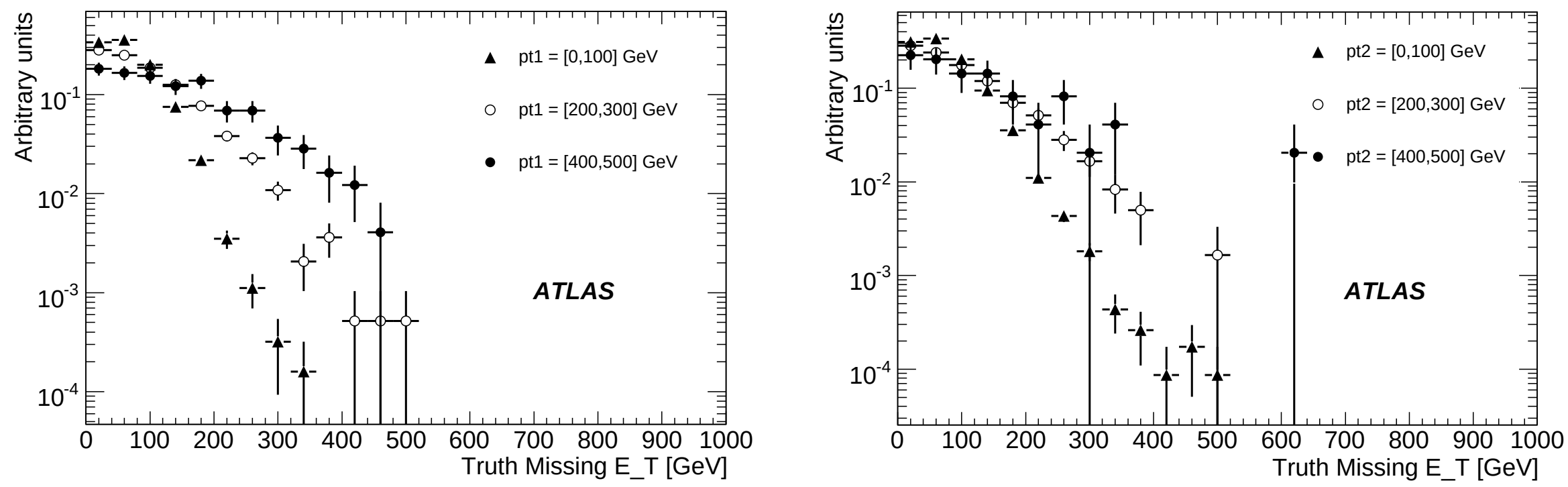


Figure 9: Missing $E_{\mathrm{T}}$ distribution in "lepton+jet" $t\bar{t}$ events with $M_T > 100$ GeV at Monte Carlo "truth" level. Left: as a function of truth leading-jet $p_{\mathrm{T}}$. Right: as a function of truth second-leading jet $p_{\mathrm{T}}$.

Table 10: Predicted and actual background levels as a function of $E_{\mathrm{T}}^{\mathrm{miss}}$ significance cut for an integrated luminosity of 1 fb$^{-1}$ in the HT2 analysis. A rough equivalent $E_{\mathrm{T}}^{\mathrm{miss}}$ cut is listed, but the $E_{\mathrm{T}}^{\mathrm{miss}}$ cut is not sharp.

| $E_{\mathrm{T}}^{\mathrm{miss}}$ sig. cut | Rough equivalent $E_{\mathrm{T}}^{\mathrm{miss}}$ cut [GeV] | Predicted BG | Actual BG |
|---|---|---|---|
| 14 | 180 | $57.3 \pm 5.5$ | $60.6 \pm 3.2$ |
| 16 | 200 | $34.8 \pm 4.5$ | $39.2 \pm 2.6$ |
| 18 | 220 | $19.1 \pm 3.1$ | $23.6 \pm 2.0$ |
| 20 | 240 | $10.1 \pm 2.1$ | $15.1 \pm 1.5$ |
| 22 | 260 | $6.2 \pm 1.8$ | $9.8 \pm 1.2$ |
| 24 | 280 | $3.8 \pm 1.5$ | $6.2 \pm 0.9$ |
| 26 | 300 | $1.3 \pm 0.7$ | $3.5 \pm 0.6$ |

significance in the signal "band" defined by HT2 $>$ 300 GeV. The normalization of the prediction in the signal band was obtainined by the number of events with HT2 $>$ 300 GeV, but at low $E_{\mathrm{T}}^{\mathrm{miss}}$, specifically $8 < E_{\mathrm{T}}^{\mathrm{miss}}$ significance $< 14$. A comparison of this predicted background with the correct background is shown in Figure 10 (left). The agreement between the predicted background and the actual background in Fig. 10 is reasonable, indicating that the correlation between HT2 and $E_{\mathrm{T}}^{\mathrm{miss}}$ significance is small. A numerical comparison of predicted and actual background levels can be seen in Table 10. For each value of the $E_{\mathrm{T}}^{\mathrm{miss}}$ significance cut, a rough equivalent in $E_{\mathrm{T}}^{\mathrm{miss}}$ is listed as a guide, but it should be emphasized that the cut in $E_{\mathrm{T}}^{\mathrm{miss}}$ is not sharp. The number of events is for HT2 $>$ 300 GeV, which corresponds approximately to a cut on the effective mass of $M_{\mathrm{eff}} > 600$ GeV.

The ratio of observed to predicted backgrounds for a $E_{\mathrm{T}}^{\mathrm{miss}}$ significance cut of 14 is $1.06 \pm 0.12$; while the ratio is consistent with unity, we take the uncertainty on the ratio (12%) as a systematic uncertainty due to possible correlations between HT2 and $E_{\mathrm{T}}^{\mathrm{miss}}$ significance. Monte Carlo samples with larger numbers of events would provide one possible way to further study the potential for correlations.

The distribution of the "orthogonal" variable, namely HT2, was predicted in a similar way. The HT2 distribution was measured in a control region defined by $8 < E_{\mathrm{T}}^{\mathrm{miss}}$ significance $< 14$. This distribution was then normalized to the number of events at large $E_{\mathrm{T}}^{\mathrm{miss}}$ significance and low HT2, specifically, $E_{\mathrm{T}}^{\mathrm{miss}}$ significance $> 14$, and 150 GeV $<$ HT2 $<$ 300 GeV. The results are shown in Fig. 10 (right).

The near independence of HT2 and $E_{\mathrm{T}}^{\mathrm{miss}}$ significance should provide an important tool in understanding jet energy and $E_{\mathrm{T}}^{\mathrm{miss}}$ performance in the complex events that make up the background to SUSY searches. After all the SUSY selection cuts have been applied, the jet energy performance can be studied

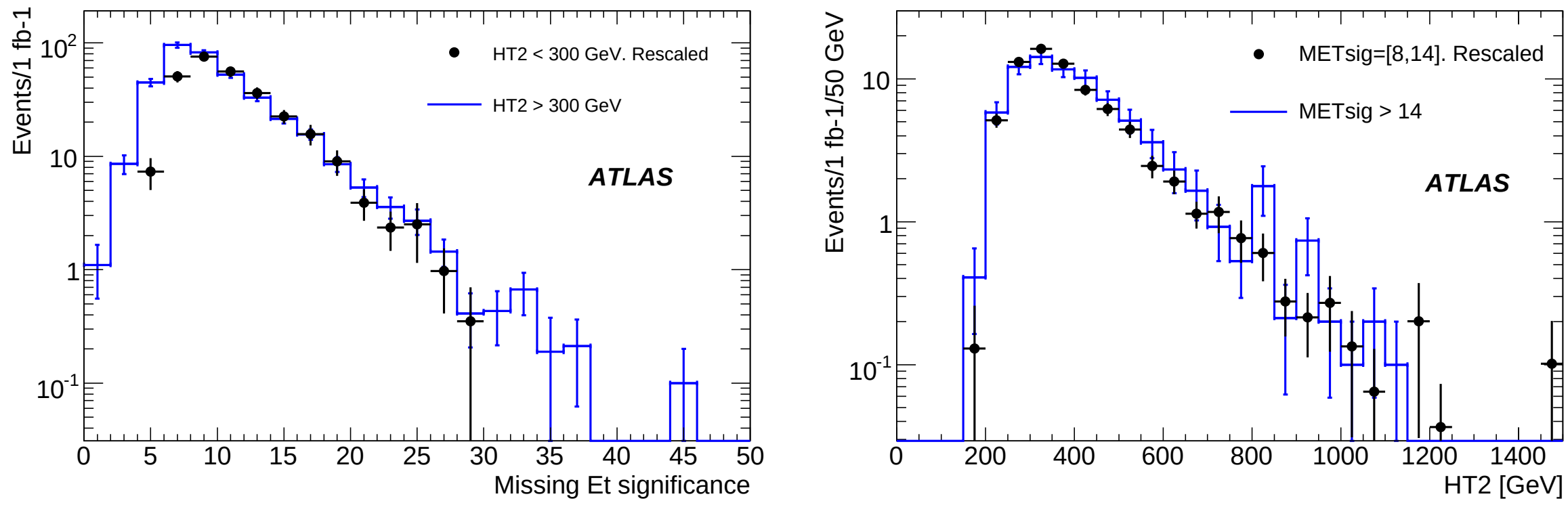


Figure 10: Left: Points: Predicted $E_{\mathrm{T}}^{\mathrm{miss}}$ significance distribution in a $t\bar{t}$ plus $W+$jets sample. Histogram: actual $E_{\mathrm{T}}^{\mathrm{miss}}$ significance distribution. Right: Predicted HT2 distribution in the same sample. Histogram: actual HT2 distribution.

Table 11: Predicted and actual background levels (for 1 fb$^{-1}$, HT2 method) for $E_{\mathrm{T}}^{\mathrm{miss}}$ significance $> 14$ as a function of systematic effects applied to the reconstructed objects.

| Modification | Predicted BG | Actual BG | Actual/predicted |
|---|---|---|---|
| Baseline | $57.3 \pm 5.5$ | $60.6 \pm 3.2$ | $1.05 \pm 0.12$ |
| Energy scaled up | $64.1 \pm 5.5$ | $79.3 \pm 3.7$ | $1.24 \pm 0.12$ |
| Energy scaled down | $45.5 \pm 4.5$ | $47.3 \pm 2.7$ | $1.04 \pm 0.12$ |
| Jet resolution smearing | $55.5 \pm 5.1$ | $65.3 \pm 3.4$ | $1.18 \pm 0.12$ |

by looking at the HT2 distribution for low $E_{\mathrm{T}}^{\mathrm{miss}}$ events; conversely, the $E_{\mathrm{T}}^{\mathrm{miss}}$ distribution can be studied by selecting events with low HT2. Events in the tails of these distributions can be examined for signs of detector problems.

**Systematic uncertainties due to detector miscalibrations** The results of systematic uncertainties due to detector performance are summarized in Table 11. The energy scale variations change the background level by about 30% while the worsening jet energy resolution results in about a 10% increase in background. However the predictions tend to change in the same direction as the actual backgrounds, and generally continue to provide reasonable determinations. We assign a 20% systematic uncertainty due to detector effects.

**Systematic uncertainties due to event generation parameters** The systematic uncertainties in the method due to changes in Monte Carlo event generation parameters were studied with ALPGEN. The parton $p_T$ cut in ALPGEN was changed from 40 to 15 GeV and the renormalization scale was reduced by a factor of 2. The results of the studies are summarized in Table 12. We assign a 20% systematic uncertainty due to event generation uncertainties.

**Background estimation in the presence of SUSY** In this section, we repeat the background estimation in the presence of SUSY signal. Figure 11 (left) shows the $E_{\mathrm{T}}^{\mathrm{miss}}$ significance distributions for the true background, true signal, and the estimated background, as well as the observed distribution of signal plus

Table 12: Predicted and actual background levels (for 1 fb$^{-1}$, HT2 method) for $E_{\mathrm{T}}^{\mathrm{miss}}$ significance $> 14$ as a function of changes in the Monte Carlo generation parameters.

| Modification | | | | |
|---|---|---|---|---|
| $t\bar{t}$ | $W+$jets | Predicted BG | Actual BG | Actual/predicted |
| PT40, scale 1.0 | PT40, scale 0.5 | $73.3 \pm 5.8$ | $63.9 \pm 3.2$ | $0.87 \pm 0.11$ |
| PT40, scale 0.5 | PT40, scale 0.5 | $133.8 \pm 7.2$ | $109.2 \pm 3.6$ | $0.82 \pm 0.05$ |
| PT15, scale 1.0 | PT40, scale 0.5 | $91.1 \pm 12.6$ | $72.5 \pm 6.0$ | $0.80 \pm 0.13$ |

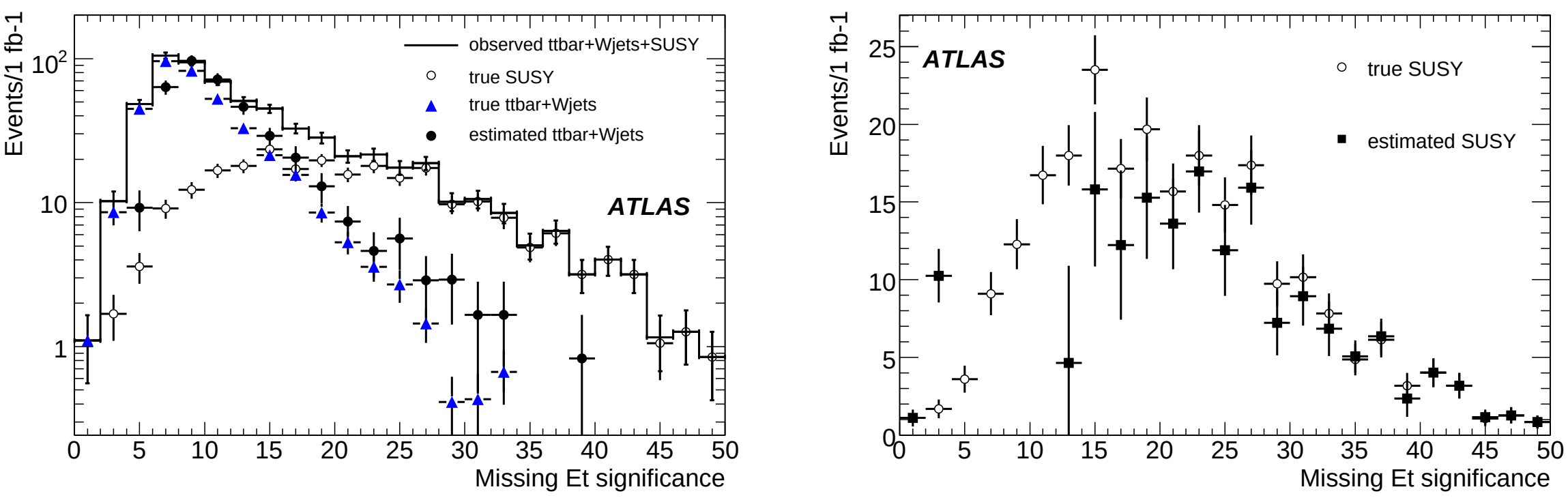


Figure 11: Left: Histogram: observed $E_{\mathrm{T}}^{\mathrm{miss}}$ significance distribution for the sum of $t\bar{t}$ plus $W+$jets background plus SUSY signal. Open circles: SUSY signal. Blue triangles: true $t\bar{t}$ plus $W+$jets background. Black filled circles: estimated background. The SUSY signal shown here is the "1 TeV SUSY" point (see text). Right: Open circles: true SUSY signal as a function of $E_{\mathrm{T}}^{\mathrm{miss}}$ significance. Black: estimated SUSY yield, obtained from the difference of the observed $E_{\mathrm{T}}^{\mathrm{miss}}$ significance distribution minus the estimated background distribution.

background. The SUSY signal here is the so-called "1 TeV SUSY" point ($m_0 = m_{\frac{1}{2}} = 400$ GeV, $\tan\beta = 10$, A=0, $\mu > 0$).

Because of the signal contamination in the control region, the background level is overestimated, leading to an underestimation in the excess of signal over background. Nevertheless, it is clear that by cutting harder on $E_{\mathrm{T}}^{\mathrm{miss}}$ significance, for example, the signal can still be clearly seen over the estimated background. A comparison of the estimated signal yield to the true signal is shown in Figure 11 (right).

The results for all the tested SUSY points are summarized in Table 13.

### 2.3.5 Top background estimation with top redecay simulation

**Introduction** It is possible to isolate a pure biased sample of fully-leptonic $t\bar{t}$ events by selecting low $E_{\mathrm{T}}^{\mathrm{miss}}$ (to reduce SUSY signal) opposite sign dilepton events where one and only one pair of invariant mass combinations $m_{\ell j}$ between the two leptons and two hardest jets ($b$ jets if tagging available) gives values below the expected endpoint from $t \to Wb \to \ell\nu b$ decays: $m_{\ell j}^{\max} = \sqrt{m_t^2 - m_W^2}$ (neglecting $m_b$).

A possible use of such a sample is to estimate the background of fully-leptonic $t\bar{t}$ events to SUSY searches. One can reconstruct the kinematics of the decaying particles ($W$'s or top quarks), remove their inferred decay products from the reconstructed event (including the event $E_{\mathrm{T}}^{\mathrm{miss}}$ ), redecay the reconstructed $W$'s or top quarks using an event generator (e.g. PYTHIA) and then merge the simulated re-decay products back into the parent ('seed') event. By redecaying particles earlier in the decay chain (i.e. the top rather than the $W$) the kinematic bias obtained from the event selection can be minimised.

Table 13: True and estimated background and signal, using the HT2 method, when the background estimation is performed in the presence of SUSY signal. The numbers are for an integrated luminosity of 1 fb$^{-1}$ except for the SU4 point where 100 pb$^{-1}$ is used.

| SUSY point | $E_T^{miss}$ sig. cut | True BG | Est. BG | True signal | Est. signal | True $S/\sqrt{B}$ | Estimated $S/\sqrt{B}$ |
|---|---|---|---|---|---|---|---|
| SU1 | 16 | $39.2 \pm 2.6$ | $100.5 \pm 10.4$ | $219.7 \pm 8.7$ | $158.4 \pm 13.8$ | 35.1 | 15.8 |
| | 20 | $15.1 \pm 1.5$ | $53.1 \pm 7.8$ | $167.0 \pm 7.6$ | $128.9 \pm 11.0$ | 43.0 | 17.7 |
| | 24 | $6.2 \pm 0.92$ | $33.1 \pm 6.5$ | $120.8 \pm 6.4$ | $93.8 \pm 9.2$ | 48.6 | 16.3 |
| SU2 | 14 | $60.6 \pm 3.2$ | $69.1 \pm 6.4$ | $30.4 \pm 2.3$ | $21.9 \pm 7.5$ | 3.9 | 2.6 |
| | 16 | $39.2 \pm 2.6$ | $43.1 \pm 5.3$ | $24.0 \pm 2.1$ | $20.2 \pm 6.2$ | 3.8 | 3.1 |
| | 18 | $23.6 \pm 2.0$ | $24.1 \pm 3.7$ | $18.3 \pm 1.8$ | $17.9 \pm 4.6$ | 3.8 | 3.6 |
| | 20 | $15.1 \pm 1.5$ | $13.9 \pm 2.7$ | $13.5 \pm 1.6$ | $14.7 \pm 3.5$ | 3.5 | 3.9 |
| SU3 | 16 | $39.2 \pm 2.6$ | $198.1 \pm 22.5$ | $328.1 \pm 14.9$ | $169.2 \pm 27.2$ | 52.4 | 12.0 |
| | 20 | $15.1 \pm 1.5$ | $119.9 \pm 18.5$ | $228.9 \pm 12.5$ | $124.1 \pm 22.4$ | 59.0 | 11.3 |
| | 24 | $6.2 \pm 0.92$ | $62.9 \pm 13.7$ | $144.7 \pm 9.9$ | $88.0 \pm 16.9$ | 58.3 | 11.1 |
| SU4 | 16 | $3.92 \pm 0.26$ | $120.7 \pm 8.7$ | $76.4 \pm 4.0$ | $-40.4 \pm 9.6$ | 38.6 | -3.7 |
| | 20 | $1.51 \pm 0.15$ | $47.4 \pm 5.5$ | $37.4 \pm 2.8$ | $-8.5 \pm 6.1$ | 30.4 | -1.2 |
| | 24 | $0.62 \pm 0.09$ | $17.8 \pm 3.3$ | $18.8 \pm 2.0$ | $1.6 \pm 3.9$ | 23.9 | 0.4 |
| SU6 | 16 | $39.2 \pm 2.6$ | $71.5 \pm 7.2$ | $140.5 \pm 5.3$ | $108.2 \pm 9.3$ | 22.4 | 12.8 |
| | 20 | $15.1 \pm 1.5$ | $36.5 \pm 5.0$ | $108.8 \pm 4.7$ | $87.4 \pm 7.0$ | 28.0 | 14.5 |
| | 24 | $6.2 \pm 0.92$ | $25.1 \pm 4.3$ | $79.3 \pm 4.0$ | $60.3 \pm 6.0$ | 31.9 | 12.0 |
| 1 TeV | 16 | $39.2 \pm 2.6$ | $61.1 \pm 6.8$ | $155.0 \pm 5.7$ | $133.1 \pm 9.2$ | 24.7 | 17.0 |
| | 20 | $15.1 \pm 1.5$ | $27.6 \pm 4.4$ | $118.1 \pm 5.0$ | $105.6 \pm 6.8$ | 30.4 | 20.1 |
| | 24 | $6.2 \pm 0.92$ | $15.6 \pm 3.5$ | $84.5 \pm 4.2$ | $75.1 \pm 5.6$ | 34.0 | 19.0 |

This technique has a number of advantages over conventional Monte Carlo techniques. In particular the event generator is used purely for modelling relatively well-understood decay and hadronisation processes – initially poorly understood aspects of process generation, such as parton distributions and the underlying event model, are effectively obtained from the data. In principle this technique is applicable also to other background processes such as $Z \to \tau^+\tau^-$, which could be modelled by replacing identified electrons or muons in $Z \to \ell^+\ell^-$ control sample events with redecayed taus.

It should be noted that this technique is at best an approximation, assuming as it does the factorisation of each $t\bar{t}$ event into two independent tops, and hence neglecting effects such as colour connection and spin correlations between the tops and other partons in the event. It is therefore unlikely to be competitive with a detailed Monte Carlo study using a fully tuned generator and validated parton distribution functions. In the early days of data-taking however it potentially provides a route to a rapid direct estimate of $t\bar{t}$ background from data complementary to, and independent from, more conventional estimates.

**Seed event selection** Seed events were selected from the ‘data’ with cuts designed to maximise the number of fully leptonic $t\bar{t}$ (‘$2\ell$-$t\bar{t}$’) events while minimising the number of Standard Model backgrounds or SUSY signal events. Events were required to pass the j45_xE50 jet + $E_T^{miss}$ trigger [7]. Single and dilepton triggers were not included in this study, but are planned to be added in future analyses. Subsequently, the following criteria were applied: $N_{jet} \geq 2$, $p_T(\text{jet}_2) > 20$ GeV, two Opposite Sign (OS) isolated leptons should be present, $p_T(\ell_2) > 10$ GeV, if the two leptons are of the same flavour $|m_{\ell\ell} - m_Z|$ $> 15$ GeV and $m_{\ell\ell} > 10$ GeV is required, $|m_{\tau\tau} - m_Z| > 15$ GeV, where $m_{\tau\tau}$ is calculated assuming the neutrinos travel parallel to their parents, and $E_T^{miss} < \frac{1}{2}(p_T(\ell_1) + p_T(\ell_2))$. The upper limit on $E_T^{miss}$ as a function of lepton $p_T$ rejects SUSY signal events. For the purposes of this early-data study $b$-tagging was assumed to be either not available or not well-understood.

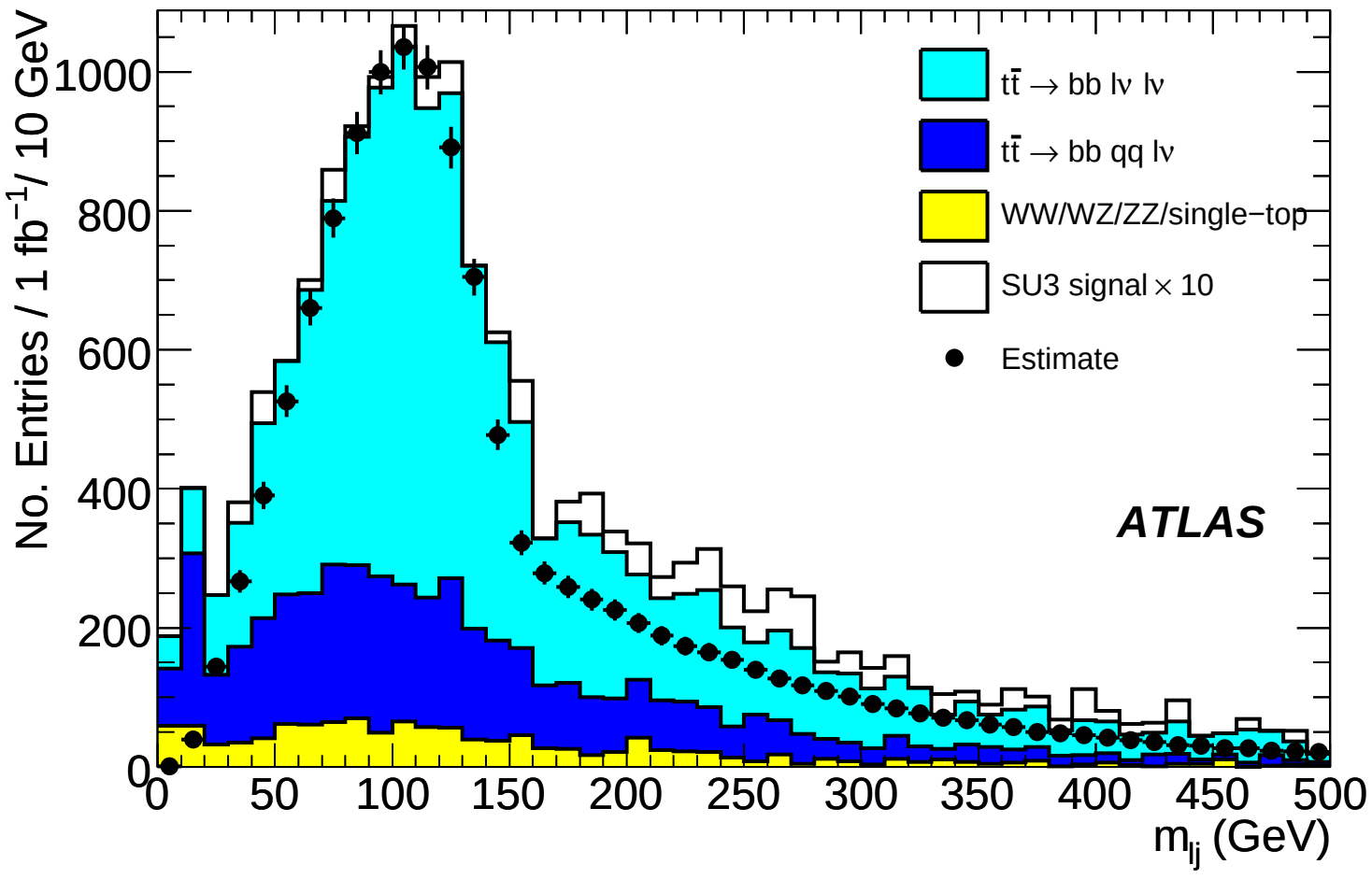


Figure 12: Distributions of $m_{\ell j}$ values for various different Standard Model backgrounds and SUSY signal. The histograms show distributions of 'data' events selected with the $2\ell$-$t\bar{t}$ selection. The data-points show the equivalent distribution estimated with redecay simulation, normalised to the peak.

**Kinematic Reconstruction and Redecay Simulation** The $m_{\ell j}$ distribution of selected events (the histogram in Figure 12) contains a prominent edge at the expected position $m_{\ell j}^{\max}$ = 155.4 GeV ($m_t$ = 175 GeV). Events were further selected in which one and only one of the two possible pairs of $\ell j$ combinations obtained from the two leptons and two hardest jets gave $m_{\ell j}$ values which were both less than $m_{\ell j}^{\max}$.

For $2\ell$-$t\bar{t}$ signal events the two ($b$) jets, two leptons and $E_{\mathrm{T}}^{\mathrm{miss}}$ components satisfy the constraints of Eq. 3 given in section 2.3.3. These constraints, assuming massless neutrinos, leptons and jets, may be solved for the 8 unknown 4-momentum components of the neutrinos. The constraints together give a quartic equation which can be solved with standard analytical techniques. If no solution was obtained then to maximise statistics the real part of the least imaginary solution was taken. If multiple solutions were obtained the solution with the smallest mean $|p_z|$ of the reconstructed top was used.

This selection results in 2207 dileptonic $t\bar{t}$ events for 1 fb$^{-1}$. The contamination by other Standard Model processes and SUSY signal events was 912 events, dominated by semileptonic $t\bar{t}$ events, as shown in Figure 12. The SUSY contamination in the sample is small, due to the tight selection cuts.

Four-vectors of the two reconstructed top quarks from each event were passed to a modified version of PYTHIA 6.4 [10]. 1000 redecayed tops were produced from each reconstructed seed top, with each $W$ forced to decay to $e$, $\mu$ or $\tau$. This 'recycling' of seed events increases the statistics of decay resimulated events for the final $E_{\mathrm{T}}^{\mathrm{miss}}$ estimation process but leads to correlations between resimulated events derived from the same seed event. These correlations were taken into account in the final uncertainties quoted below. Decay products were passed to the ATLAS fast simulation program, and then merged back into their parent seed events.

As a cross-check of the estimation procedure redecayed events were passed through the same selection as their parent seed events and the distribution of $m_{\ell j}$ constructed and normalised to the seed $m_{\ell j}$ distribution. This is shown in Fig. 12 and indicates good agreement below $m_{\ell j}^{\max}$.

**Use in one-lepton search background estimate** Decay resimulated events were subjected to the standard one-lepton SUSY search selection described in section 2.1, with one modification: $M_T(\ell, E_{\mathrm{T}}^{\mathrm{miss}}) >$ 150 GeV.

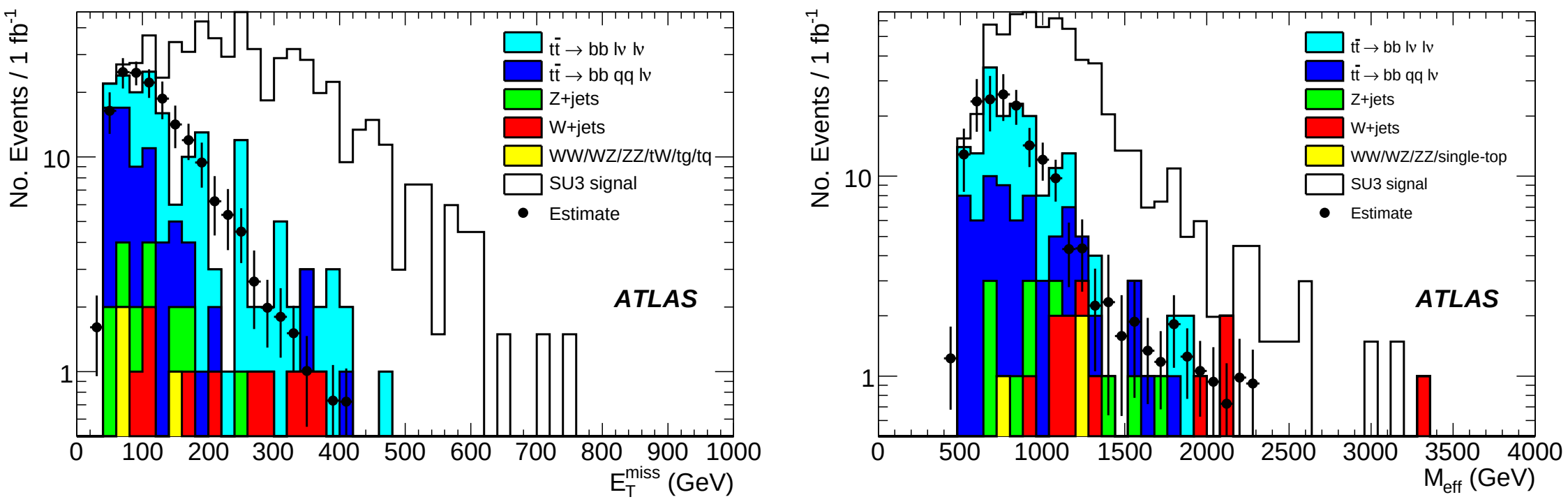


Figure 13: $E_{\mathrm{T}}^{\mathrm{miss}}$ (left) and $M_{\mathrm{eff}}$ (right) distributions of events passing the basic one-lepton SUSY selection cuts described in the text. Note that $Z+$jet and $W+$jet backgrounds are under-represented in these plots for $E_{\mathrm{T}}^{\mathrm{miss}} < 80$ GeV or $M_{\mathrm{eff}} < 350$ GeV due to filter requirements applied to the respective Monte Carlo samples.

The remaining background at high $E_{\mathrm{T}}^{\mathrm{miss}}$ following such cuts is dominated by semi-leptonic $t\bar{t}$ events, and to a lesser extent leptonic $W+$jets, $Z+$jets, single-top and di-boson events. For all these backgrounds one expects primarily a Jacobian peak in the event $M_T(\ell, E_{\mathrm{T}}^{\mathrm{miss}})$ distribution near $M_W$ ($M_Z$ for $Z+$jets). This $1\ell$-Jacobian background was estimated with the $E_{\mathrm{T}}^{\mathrm{miss}}$ distribution of events selected with the same cuts, with the exception of the $M_T(\ell, E_{\mathrm{T}}^{\mathrm{miss}})$ cut, which was reversed to require $M_T(\ell, E_{\mathrm{T}}^{\mathrm{miss}}) < 100$ GeV.

The $1\ell$-Jacobian and the $2\ell$-$t\bar{t}$ background estimates were compared to ‘data’ events subjected to the one-lepton selection criteria described above. The two estimates were simultaneously normalised to the ‘data’ $E_{\mathrm{T}}^{\mathrm{miss}}$ distribution in two bins: $40 < E_{\mathrm{T}}^{\mathrm{miss}} < 100$ GeV and $100 < E_{\mathrm{T}}^{\mathrm{miss}} < 140$ GeV. The total ($2\ell$-$t\bar{t}$ + $1\ell$-Jacobian) normalised estimate is plotted in Fig. 13(left) together with the ‘data’ $E_{\mathrm{T}}^{\mathrm{miss}}$ distribution. For $E_{\mathrm{T}}^{\mathrm{miss}} > 200$ GeV the agreement between the estimate and the ‘data’ is good: $30.7 \pm 9.8$ (30%) estimated, versus 39 ‘observed’.

In Fig. 13(right), the $M_{\mathrm{eff}}$ distribution is shown, with the estimate normalised with the same factors as used in Fig. 13(left). The semi-leptonic $t\bar{t}$ forms a larger fraction of the background at large $M_{\mathrm{eff}}$ compared to at large $E_{\mathrm{T}}^{\mathrm{miss}}$ because it produces a larger number of jets than fully-leptonic $t\bar{t}$.

**SUSY contamination** The shape of the estimated distribution is effectively insensitive to the presence of SUSY signal, primarily due to the low $E_{\mathrm{T}}^{\mathrm{miss}}$ requirement in the $2\ell$-$t\bar{t}$ selection.

However, there may be SUSY signal in the normalisation region. The bias in the estimate will be proportional to the amount of SUSY in the normalization region, which is largest for the samples with the highest SUSY cross-section: SU3 and SU4. The effect of admixture of SU3 or SU4 signal events is shown in Table 14. In case of SU3, the background estimate is 60% higher, in case of SU4 as large as a factor 15. For both samples, however, the excess of signal events is still significant.

In principle the contamination effect could be reduced by normalization to more signal-free regions. The $2\ell$-$t\bar{t}$ and $1\ell$-Jacobian estimates could be normalized for example to the tail and peak of the $M_T(\ell, E_{\mathrm{T}}^{\mathrm{miss}})$ distribution at low $E_{\mathrm{T}}^{\mathrm{miss}}$ .

### 2.3.6 Combined fit method

The fit-based method for measuring the background, as described in this section, aims to improve upon the plain $M_T$ sideband subtraction method for the one-lepton SUSY search mode. By analysing data in a L-shaped region at both low-$E_{\mathrm{T}}^{\mathrm{miss}}$ in the full $M_T$ range and at low $M_T$ in the full $E_{\mathrm{T}}^{\mathrm{miss}}$ range, and

Table 14: Estimated and true background with the redecay method, in case of no SUSY signal, or presence of SU3 or SU4 SUSY signals. For the latter, also the amount of observed signal plus background events is shown, proving that although the background is overestimated, the excess is still present.

| Sample | Estimated BG | True BG | Observed signal + BG |
|---|---|---|---|
| No SUSY | $30.7 \pm 9.8$ | 39 | 39 |
| SU3 | $50.8 \pm 13.6$ | 39 | 392 |
| SU4 | $456 \pm 102$ | 39 | 1230 |

performing a two-dimensional extrapolation into the SUSY signal region we hope to enhance the background estimation. In a fit, correlations between $E_{\mathrm{T}}^{\mathrm{miss}}$ and $M_T$ can be taken into account. Furthermore, an explicit assumption can be put in that there is a finite SUSY contamination in the control sample.

For our purposes, we define in this analysis a sideband (SB) region and a SUSY signal (SIG) region. For both regions, we apply the standard SUSY one-lepton selection cuts defined in section 2.1, with the exception of the cut on $M_T$. The SB region is defined by the following additional cut: $E_{\mathrm{T}}^{\mathrm{miss}} < 200$ GeV *or* $M_T < 150$ GeV. In this analysis, the SIG region is defined by: $E_{\mathrm{T}}^{\mathrm{miss}} > 200$ GeV *and* $M_T > 150$ GeV. This classification assumes that the mass scale of SUSY is much higher than 200 GeV, which is true for all SUSY points considered in this note except SU4.

The analysis described here has been applied to electron + jets events only, but we expect the muon + jets analysis to be completely analogous.

The main Standard Model backgrounds we expect are single-leptonic $t\bar{t}$ decays, double-leptonic $t\bar{t}$ decays, and $W+$jets events. Other backgrounds, such as $Z+$jets, diboson production or single top are negligible for 1 fb$^{-1}$. The trigger used was an OR of the e22i single electron trigger, and the 4j50 multi-jet trigger [7].

**Shape of the backgrounds** We will try to fit the contributions of the backgrounds in three observables: $M_T$, $E_{\mathrm{T}}^{\mathrm{miss}}$ and $m_{top}$. Here $m_{top}$ is the invariant mass of the three jets in the event with the largest vector-summed $p_T$ [11].

We construct probability density functions (p.d.f.'s) that model the major contributing processes in the three observables after the event selection. We have done this both with and without explicitly taking into account correlations between the three observables. Most of these correlations are in any case consistent with zero.

Figure 14 shows the distribution of the three observables for each type of background we consider, as well as for one SUSY signal sample (SU3). The empirical model that we use to describe each background type is overlaid on the data. This comparison of shapes demonstrates that there is sufficient information in these three observables to be able to measure each of them in a combined fit.

We perform the procedure of constructing an empirical model from Monte Carlo for multiple SUSY data points. A striking feature of a comparison of $E_{\mathrm{T}}^{\mathrm{miss}}$ and $M_T$ distributions of these SUSY points *in the SB region* is that, with the exception of lower mass SUSY point SU4, they are all quite similar in shape. We can thus construct a model-independent 'Ansatz' shape to describe the SUSY contamination at low energy.

To validate this procedure, we perform fits to a "data" sample consisting of either background only, or background plus SU3 SUSY signal. The yields we find in a fit to 1 fb$^{-1}$ are listed in Table 15 and are in agreement within errors with the truth values of the fitted event mix. If, in contrast, a SU3 SUSY signal would be present in data, but the fit would not allow for SUSY (see table), the fit would overestimate dileptonic $t\bar{t}$ and $W+$jets, and underestimate semileptonic $t\bar{t}$. This is as expected, as the SUSY signal has a long tail in $M_T$, but no substantial peak in the top mass.

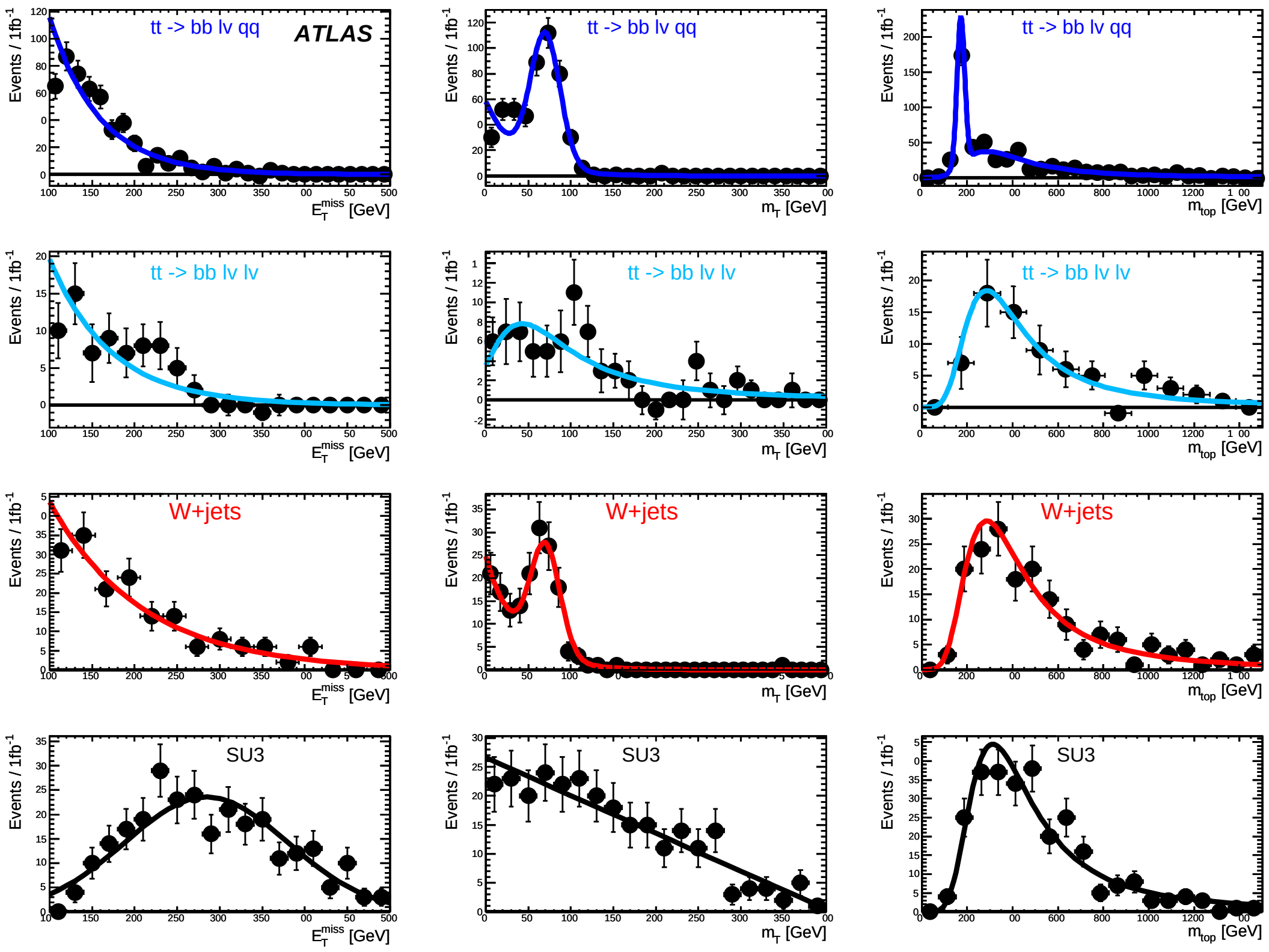


Figure 14: Distributions in missing $E_T$ (left), $M_T$ (middle) and $m_{top}$ (right) of single lepton $t\bar{t}$ (top row), double lepton $t\bar{t}$ (second row), W + jet events (third row) and SUSY SU3 (last row). Each distribution is overlaid with a projection of the three-dimensional model that is fitted to that sample.

The same table also lists the yields obtained from a fit in which all components are taken as a simple uncorrelated product of three one-dimensional p.d.f.'s. The yields and their uncertainties are very similar to those from the fit with models that include correlations and indicates that the effect of correlations in the description of the background components is minor.

**Fitting the data** The fit with fixed shapes relies on simulated events to determine the shapes of the various background components, while all yields are fitted from the data. The next step is to release as many of the shape parameters in the fit to the data; in total there are 15 of these parameters. In the limit that all parameters can be floated, with the exception of the SUSY ansatz shape, the method becomes almost independent on simulation input and fully data driven. It turns out that on a 1 fb$^{-1}$ sample we can float all but two of the $W+$jets and 1-lepton $t\bar{t}$ shape parameters. These two are the fraction of events that give the correct top quark mass in $m_{top}$ for single lepton $t\bar{t}$ and the fraction of events with a correctly constructed $W$ boson in $M_T$ in single lepton $t\bar{t}$ .

Floating the 2-lepton $t\bar{t}$ shape parameters in addition causes the fit to become unstable because the shapes of the dilepton component and that of the SUSY ansatz model are very similar. We are currently investigating the possibility of introducing additional constraints on the shape of the di-lepton $t\bar{t}$ events in order to solve this.

The result of the 1 fb$^{-1}$ fit with floating parameters is shown in Figure 15.

The final step in the analysis is to extrapolate the yields of the standard model background components from the sideband region to the signal region. Table 16 shows the yields from the combined fit with

Table 15: Yields from the combined fit with fixed shapes for the various modeled physics processes. The leftmost column lists the yield fitted with models that include correlations in the model structure. The middle column shows these yields for models in which correlation terms have been disabled. The rightmost column gives the true yields in the fitted event mix.

| Component | Fitted Yield | Fitted Yield w/o correlations | True Yield |
|---|---|---|---|
| no SUSY in data, no SUSY component in fit | | | |
| $W+$jets | $186\pm33$ | $200\pm27$ | 173 |
| 1-lepton $t\bar{t}$ | $507\pm32$ | $482\pm32$ | 502 |
| 2-lepton $t\bar{t}$ | $53\pm13$ | $55\pm16$ | 70 |
| SUSY | - | - | 0 |
| no SUSY in data, SUSY component in fit | | | |
| $W+$jets | $185\pm33$ | $201\pm27$ | 173 |
| 1-lepton $t\bar{t}$ | $507\pm32$ | $482\pm32$ | 502 |
| 2-lepton $t\bar{t}$ | $52\pm15$ | $56\pm16$ | 70 |
| SUSY | $4.4\pm3.0$ | $-7.1\pm16.0$ | 0 |
| SU3 in data, no SUSY component in fit | | | |
| $W+$jets | $292\pm35$ | $261\pm34$ | 173 |
| 1-lepton $t\bar{t}$ | $386\pm30$ | $356\pm31$ | 502 |
| 2-lepton $t\bar{t}$ | $338\pm25$ | $389\pm34$ | 70 |
| SUSY | - | - | 271 |
| SU3 in data, SUSY component in fit | | | |
| $W+$jets | $181\pm40$ | $194\pm35$ | 173 |
| 1-lepton $t\bar{t}$ | $521\pm36$ | $509\pm35$ | 502 |
| 2-lepton $t\bar{t}$ | $35\pm23$ | $15\pm30$ | 70 |
| SUSY | $280\pm22$ | $293\pm24$ | 271 |

floating shapes extrapolated to the signal region while propagating all (correlated) parameter uncertainties; for comparison the same table also shows the results with shapes kept fixed. The fits describe the data within the statistical uncertainties.

Table 16: Yields from the combined fit with either fixed or floating shapes in the sideband region extrapolated to the full parameter space, the truth yields in full parameter space, the extrapolated yields into the signal region and the truth yields in the signal region.

| Component | Extrap. Yield in FULL | | True | Extrap. Yield in SIG | | True |
|---|---|---|---|---|---|---|
| | Shape Fixed | Shape Floating | FULL | Shape Fixed | Shape Floating | SIG |
| $W+$jets | $205\pm45$ | $227\pm68$ | 173 | $0.5\pm0.4$ | $-1.2\pm2.7$ | 2 |
| 1-lepton $t\bar{t}$ | $476\pm35$ | $485\pm59$ | 502 | $0.4\pm0.2$ | $-1.1\pm3.9$ | 0 |
| 2-lepton $t\bar{t}$ | $62\pm38$ | $17\pm54$ | 70 | $4.5\pm2.9$ | $4.7\pm7.9$ | 5 |
| SUSY SU3 | $273\pm33$ | $287\pm38$ | 271 | $92.7\pm2.8$ | $95.6\pm4.0$ | 91 |

**Effect of SUSY signal** While table 16 shows the results of the fit for the various backgrounds and the SUSY signal in the case of the SU3 sample, Table 17 shows results, for fits with floating shapes, for other SUSY signal samples.

As has been noted earlier, SU4 is a special case. It is too close to the background for the Ansatz of the shape to be valid, and it can not be fitted very well.

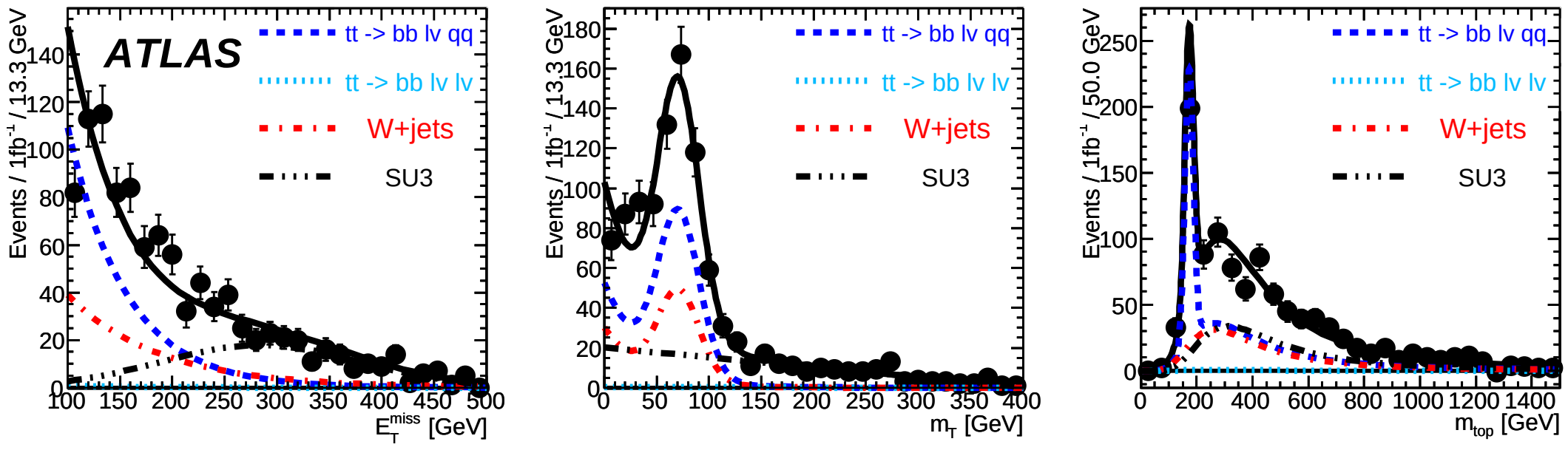


Figure 15: Distribution of missing $E_T$ (left), $M_T$ (center) and $m_{top}$ (right) of a 1 fb$^{-1}$ mix of $t\bar{t}$ , $W+$ jets standard model events and SUSY SU3 events overlaid with projections of the combined model on these observables that was fitted to this mix of events with floating yield parameters and floating shaped parameters. For each projection the contributions of the 1-lepton $t\bar{t}$ ttbar contribution (dark blue), 2-lepton $t\bar{t}$ ttbar contribution (light blue), $W+$jet contribution (red) and ansatz SUSY constribution (black) are shown.

Table 17: Yields from the combined fit with floating shapes in the sideband region extrapolated to the full parameter space, the truth yields in full parameter space, the extrapolated yields into the signal region and the truth yields in the signal region, for various SUSY samples.

| Component | Extrap. Yield in FULL | True in FULL | Extrapolated Yield in SIG | True in SIG |
|---|---|---|---|---|
| | | SU1 | | |
| $W+$jets | $215 \pm 56$ | 173 | $1.8 \pm 2.0$ | 2 |
| 1-lepton $t\bar{t}$ | $486 \pm 54$ | 502 | $0.6 \pm 0.6$ | 0 |
| 2-lepton $t\bar{t}$ | $19 \pm 35$ | 70 | $0.5 \pm 1.7$ | 5 |
| SUSY SU1 | $154 \pm 30$ | 129 | $50.1 \pm 2.6$ | 46 |
| | | SU2 | | |
| $W+$jets | $226 \pm 51$ | 173 | $1.3 \pm 1.0$ | 2 |
| 1-lepton $t\bar{t}$ | $452 \pm 49$ | 502 | $0.7 \pm 0.5$ | 0 |
| 2-lepton $t\bar{t}$ | $81 \pm 32$ | 70 | $8.0 \pm 5.0$ | 5 |
| SUSY SU2 | $0 \pm 10$ | 14 | $1.0 \pm 6.1$ | 4 |
| | | SU6 | | |
| $W+$jets | $215 \pm 54$ | 173 | $0.5 \pm 0.8$ | 2 |
| 1-lepton $t\bar{t}$ | $469 \pm 52$ | 502 | $0.2 \pm 0.6$ | 0 |
| 2-lepton $t\bar{t}$ | $29 \pm 34$ | 70 | $2.6 \pm 2.7$ | 5 |
| SUSY SU6 | $117 \pm 29$ | 86 | $38.8 \pm 2.5$ | 35 |
| | | SU8 | | |
| $W+$jets | $239 \pm 53$ | 173 | $2.5 \pm 2.4$ | 2 |
| 1-lepton $t\bar{t}$ | $485 \pm 51$ | 502 | $1.0 \pm 1.2$ | 0 |
| 2-lepton $t\bar{t}$ | $66 \pm 45$ | 70 | $15.1 \pm 12.0$ | 5 |
| SUSY SU8 | $34 \pm 26$ | 79 | $34.5 \pm 13.0$ | 46 |

**Systematics** Table 18 summarizes the results of a series of systematic studies. These studies are quantified in terms of relative variations of the measured SUSY cross-section, which we define as the counted number of events in the data in the SIG region minus the fitted and extrapolated number of SM model events expected in that same region, and divided by the efficiency for the selection of SUSY events, including the SIG region cuts, as measured from a pure sample of simulated SU3 SUSY events.

The second column of Table 18 lists the relative variation under the influence of each systematic variation if no extrapolation is applied in the fit and the analysis is performed without the SIG region cut. This column thus quantifies the effect on the shape fit stability only. The third columns shows the effect of each systematic study on the fitted yield in the signal region using extrapolation. Most effects are here of the order of 5-7%. The uncertainties due to Monte Carlo statistics are explicitly listed in the table.

Table 18: List of studied systematic uncertainties for the combined-fit method.

| Systematic variation | w/o extrapolation ± MC-error [%] | in SIG ± MC-error [%] |
|---|---|---|
| Jet energy scale | $1.9 \pm 0.7$ | $3.1 \pm 1.1$ |
| Jet energy resolution | $3.7 \pm 0.5$ | $0.5 \pm 0.1$ |
| Electron energy scale | $4.8 \pm 0.6$ | $5.6 \pm 1.4$ |
| Electron energy resolution | $9.0 \pm 1.2$ | $7.2 \pm 1.4$ |
| Electron identification efficiency | $0.5 \pm 0.2$ | $3.6 \pm 2.3$ |
| Soft $E_{\mathrm{T}}^{\mathrm{miss}}$ scale | $8.1 \pm 1.8$ | $7.4 \pm 3.4$ |

# 3 No-lepton search mode

## 3.1 Selection

In the no-lepton search mode, we veto all events with an identified electron or muon with a $p_T$ of more than 20 GeV. We demand at least 4 jets with $|\eta| < 2.5$ and $p_T > 50$ GeV, one of which must have $p_T > 100$ GeV. The transverse sphericity $S_T$ should be larger than 0.2, and the missing transverse energy $E_{\mathrm{T}}^{\mathrm{miss}}$ should be larger than 100 GeV and larger than 0.2 $M_{\mathrm{eff}}$, where $M_{\mathrm{eff}}$ is the effective mass. We add one more cut against the QCD background: the minimum value of the difference in azimuthal angle between the $E_{\mathrm{T}}^{\mathrm{miss}}$ vector and the three highest-$p_T$ jets should be larger than 0.2. This cut is futher discussed in a dedicated note on QCD background estimation [3].

## 3.2 Backgrounds in Monte Carlo

Figure 16 shows the distributions of $E_{\mathrm{T}}^{\mathrm{miss}}$ and $M_{\mathrm{eff}}$ after all the selections are applied.

## 3.3 Data-driven estimation strategies

In this section we discuss data-driven estimation strategies for the no-lepton search mode. The strategies we have studied are:

1. estimation of $Z$ ($\rightarrow \nu\bar{\nu}$) plus jets from $Z$ ($\rightarrow \ell^+\ell^-$) plus jets, purely from data (replace method, section 3.3.1);

2. estimation of $Z$ and $W$ in an analogous way, but also helped by Monte Carlo (section 3.3.2);

3. estimation of $W$ and $t\bar{t}$ background from a one-lepton control sample derived by reversing one of the selection cuts (on $M_T$) (section 3.3.3);

4. estimation of the cross-section of the $t\bar{t} \rightarrow b\bar{b}q\bar{q}'\tau\nu$ process with hadronic tau decay (section 3.3.4).

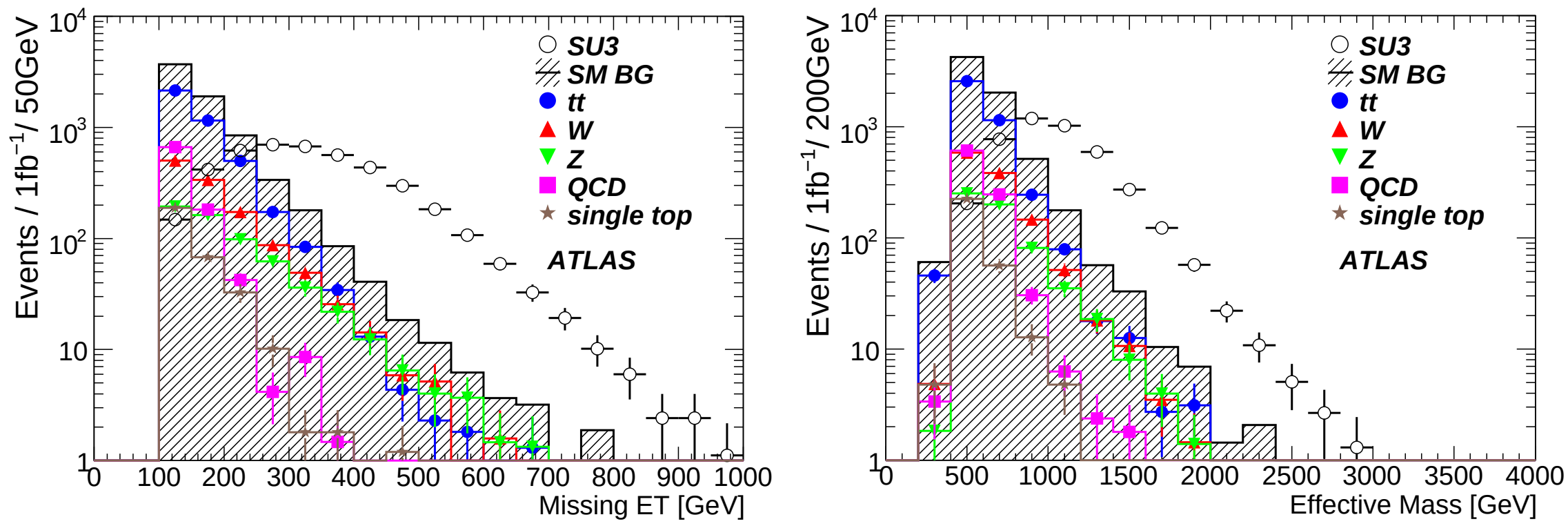


Figure 16: The $E_{\mathrm{T}}^{\mathrm{miss}}$ and effective mass distributions of the SUSY signal and background processes for the no-lepton mode with an integrated luminosity of 1 fb$^{-1}$. The open circles show the SUSY signal (SU3 point). The shaded histogram shows the sum of all Standard Model backgrounds; different symbols show the various components.

### 3.3.1 Replace method: $Z \to \nu\bar{\nu}$ from $Z \to \ell^+\ell^-$

**Introduction** The $Z \to \nu\bar{\nu}$ background is one of the main background process in the no-lepton channel. In order to estimate and reproduce the number of expected background events, as well as the shape of the $E_{\mathrm{T}}^{\mathrm{miss}}$ and $M_{\mathrm{eff}}$ distributions, $Z \to \ell^+\ell^-$ events are selected, and the charged leptons are replaced by neutrinos. However, as the ratio of branching-ratios $\mathrm{Br}(Z \to \ell^+\ell^-)/\mathrm{Br}(Z \to \nu\bar{\nu})$ is small, statistical uncertainties will tend to be relatively large. Two solutions are proposed :

1. Taking the distribution shape from $Z \to \ell^+\ell^-$ data but constraining it via a fit plus the assumption of a smooth evolution of the fitting parameters when relaxing the cuts. This is the method described in this section.

2. Taking the distribution shape from Monte Carlo simulation as described in the next section ( Section 3.3.2).

The Monte Carlo method is more sensitive to generator-level and detector systematic uncertainties, but does not suffer from the larger statistical uncertainties, whereas the replace method precision is limited by the number of events in the control sample, but less sensitive to systematic uncertainties from the detector. Both methods have to account for the fact that the detected charged lepton pairs will not cover the full phase space of the neutrinos.

**Control Sample Selection** The control sample selection is identical to the no-lepton SUSY search selection, except that two electrons or two muons are required, and that the missing $E_T$ ($E_{\mathrm{T}}^{\mathrm{miss}}$) is replaced by $p_T(\ell^+\ell^-) \simeq p_T(Z)$. Thus it is assumed that neutrinos are the main contribution to $E_{\mathrm{T}}^{\mathrm{miss}}$ when the $Z$ boson decays into two neutrinos, such that $E_{\mathrm{T}}^{\mathrm{miss}}$ is roughly equivalent to $p_T(Z)$ for this physics process. The $E_{\mathrm{T}}^{\mathrm{miss}}$ resolution of ATLAS is sufficient for this to be a good approximation. In addition to pairs of isolated charged leptons, a sample composed of $Z \to e^{\pm}X$ is added, where $X$ is a non-isolated electron or an electron-like object with very loose cuts. This additional sample is used to increase the statistics and measure the electron identification efficiency via the "tag-and-probe" method. The goal of the tag-and-probe method is to select on one side a good electron (tag) and look at the other side to the nature of the object (the probe) which matches the constraint on the $Z$ mass. Two cuts are added to reject the

remaining backgrounds : $81 < M_Z(\ell^+\ell^-) < 101$ GeV, and Missing $E_T < 30$ GeV. After all cuts the number of selected events is summarized in Table 19. In particular, Table 19 shows the effect of an upper cut on $E_{\mathrm{T}}^{\mathrm{miss}}$ in order to reject $t\bar{t}$ background in the $Z \to e^{\pm}X$ channel. It has been verified that the efficiencies measured with the tag-and-probe method agree with efficiencies obtained directly from Monte Carlo, confirming that $X$ is indeed dominated by electrons rather than hadronic jets.

Table 19: Number of selected $Z \to \ell^+\ell^- + \geq 4$ jets events in 1 fb$^{-1}$ in the replace-method analysis.

| Process | No $E_{\mathrm{T}}^{\mathrm{miss}}$ cut | $E_{\mathrm{T}}^{\mathrm{miss}} < 30$ GeV |
|---|---|---|
| $Z \to e^+e^- +$n jets | 18.2 | 14.1 |
| $Z \to e^{\pm}X +$n jets | 25.6 | 19.6 |
| $Z \to \mu^+\mu^- +$n jets | 33.2 | 26.1 |
| $Z \to \tau^+\tau^- +$n jets | 1.6 | 0. |
| $t\bar{t} \to bb\ell\nu\ell\nu +$n jets | 56.2 | 0.3 |
| $t\bar{t} \to bb\ell\nu qq +$n jets | 506.6 | 2.5 |

**Lepton identification and acceptance corrections** A number of correlations must be applied in order to derive $Z \to \nu\bar{\nu}$ distributions from $Z \to \ell^+\ell^-$: (1) a fiducial correction, since we cannot detect $e$ and $\mu$ leptons beyond $|\eta| = 2.5$; (2) a kinematics correction for the additional cuts used to select $Z \to \ell^+\ell^-$, including the $Z$ invariant-mass window, the $p_T$ cut on the leptons, and the $E_{\mathrm{T}}^{\mathrm{miss}}$ cut; and (3) a correction for the lepton identification efficiency. The first two effects have to be computed from simulation whereas the lepton identification efficiency can be measured from collision data using the tag-and-probe method.

After all corrections, the distribution can be summarized by the following formula:

$$N_{Z\to\nu\bar{\nu}}(E_{\mathrm{T}}^{\mathrm{miss}}) = N_{Z\to\ell^+\ell^-}(p_T(\ell^+\ell^-)) \times c_{\mathrm{Kin}}(p_T(Z)) \times c_{\mathrm{Fidu}}(p_T(Z)) \times \frac{\mathrm{Br}(Z \to \nu\bar{\nu})}{\mathrm{Br}(Z \to \ell^+\ell^-)}, \quad (5)$$

where $N_{Z\to\nu\bar{\nu}}(E_{\mathrm{T}}^{\mathrm{miss}})$ is the corrected number of events per bin of missing $E_T$, $N_{Z\to\ell^+\ell^-}(p_T(\ell^+\ell^-))$ is the raw number of control sample events as a function of $p_T(Z)$, $c_{\mathrm{Kin}}$ and $c_{\mathrm{Fidu}}$ are the kinematic and fiducial corrections. The $E_{\mathrm{T}}^{\mathrm{miss}}$ and $M_{\mathrm{eff}}$ distributions of $Z \to e^+e^- + e^{\pm}X$ and $Z \to \mu^+\mu^-$ events after all corrections are compared to $Z \to \nu\bar{\nu}$ distributions in Figure 17.

For very high values of $E_{\mathrm{T}}^{\mathrm{miss}}$ and $M_{\mathrm{eff}}$ , statistics is low. In order to present a smooth prediction of the background, for example as a function of $M_{\mathrm{eff}}$ , a fit of the shape has been performed. Of course, the fit is also affected by the low statistics in the tail, but by relaxing the jet $p_T$ and $E_{\mathrm{T}}^{\mathrm{miss}}$ cuts and observing how the fit parameters evolve with the cuts, a smooth prediction can be made.

**Systematic uncertainties** The effects of various Monte Carlo generator and detector systematic uncertainties are summarized in Table 20. The main generator systematic uncertainty is due to the variation of the renormalization scale and affects the acceptance correction, whereas the principal detector systematic uncertainty is related to the soft part of the missing transverse energy which is not taken into account when replacing neutrinos by charged leptons.

**SUSY contamination in the control sample** Due to the tight control sample selection cuts, in particular $E_{\mathrm{T}}^{\mathrm{miss}} < 30$ GeV, the SUSY contamination in the control sample in 1 fb$^{-1}$ is negligible: 0.1 events for SU1, 0.4 events for SU2, 0.9 events for SU3, and $< 0.07$ events for SU6.

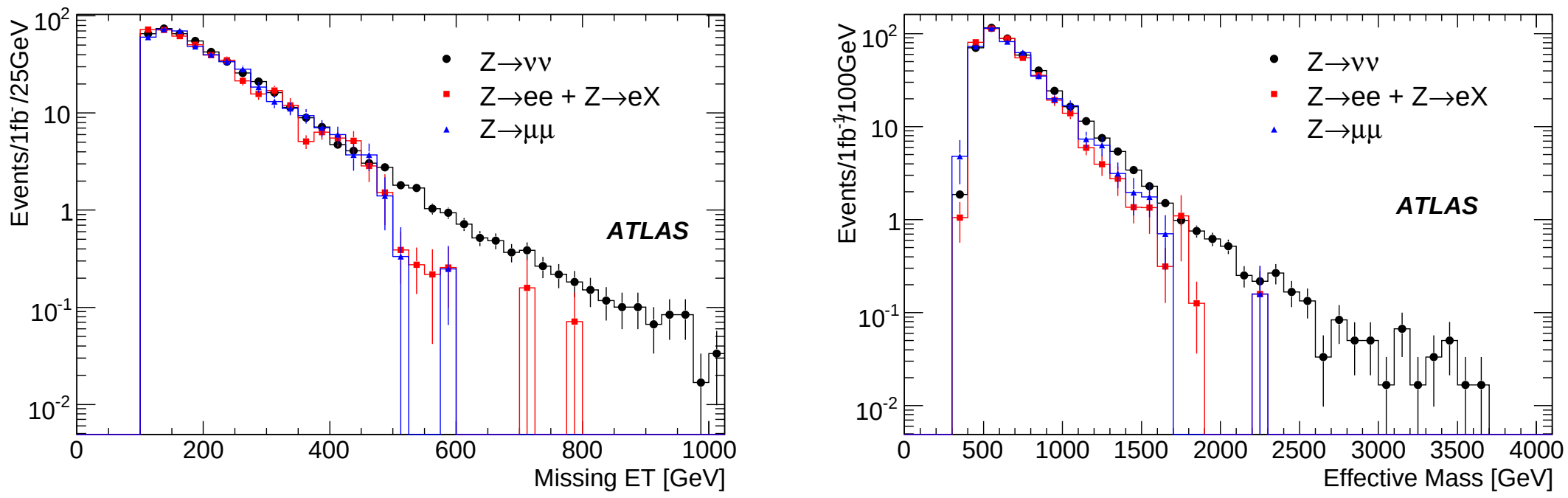


Figure 17: (Left) $E_{\rm T}^{\rm miss}$ distribution after all corrections for $Z \to \nu\bar{\nu}$, $Z \to e^+e^- + e^{\pm}X$ and $Z \to \mu^+\mu^-$ processes. The number of events corresponds to an integrated luminosity of 1 fb$^{-1}$. (Right) $M_{\rm eff}$ distribution for the same physics processes.

Table 20: Summary of systematic uncertainties for the replace method.

| Description | Relative uncertainty $\Delta N/N$ [%] |
|---|---|
| MC generator systematics | |
| ALPGEN parameter variation | 6.3 |
| Detector systematics | |
| Electron energy scale | 0.05 |
| Electron energy resolution | 0.03 |
| Electron id efficiency | 0.50 |
| Muon energy scale | 0.30 |
| Muon energy resolution | 0.39 |
| Muon id efficiency | 1.00 |
| $E_{\rm T}^{\rm miss}$ scale (soft part) | 4.5 |
| Total systematics (quadratic sum) | $\sim 8$ |
| Total statistics | $\sim 13$ |
| Total uncertainties | $\sim 15$ |

### 3.3.2 $Z$ and $W$ background estimates from $Z \to \ell^+\ell^-$ plus Monte Carlo shape

A modification of the method of the previous section is described here. Denoted the "MC method", it uses only the number of events in the $Z(\to \ell^+\ell^-)+$jets control sample for normalization, but otherwise relies on Monte Carlo simulation of kinematical distributions of events. The same normalization factor is used for both the $Z(\to \nu\bar{\nu})$ background and the $W(\to \ell\nu)$ background since the production mechanism is very simliar.

**Control sample** The event selection of the control sample demands: two opposite-sign same-flavour leptons with $p_T > 20$ GeV; $E_{\rm T}^{\rm miss} < 40$ GeV; a di-lepton mass $M_{\ell\ell}$ within $\pm$ 10 GeV of $m_Z$. Subsequently, the standard no-lepton SUSY cuts are applied after replacing $\ell^+\ell^-$ with $\nu\bar{\nu}$.

The number of events selected with 1 fb$^{-1}$ is summarized in Table 21. The contamination from $t\bar{t}$ events is about $10^{-2}$ and therefore not significant. The number of events estimated by a full simulation sample (with event filter $p_T^Z > 80$ GeV) is $72 \pm 3$ for 1 fb$^{-1}$, which leads to a statistical uncertainty on the estimation of 12%.

The method has been tested with a pseudo data sample prepared with Monte Carlo parameters set

Table 21: Number of events in the MC method control sample (for 1 fb$^{-1}$).

| process | $\mu$ mode | e mode | $\mu$ + e (sum) |
|---|---|---|---|
| $Z(\rightarrow \ell^+\ell^-)$ | $51 \pm 1$ | $34 \pm 1$ | $85 \pm 1$ |
| $Z(\rightarrow \tau^+\tau^-)$ | $< 0.1$ | $< 0.1$ | $< 0.1$ |
| $t\bar{t}$ | $1.0 \pm 0.3$ | $0.5 \pm 0.2$ | $1.5 \pm 0.4$ |

Table 22: Systematic uncertainty from variation of Monte Carlo generator parameters (ALPGEN) for the MC method. The pseudo data sample is discussed in the text.

| $Z(\rightarrow \nu\bar{\nu})$ | $E_{\mathrm{T}}^{\mathrm{miss}} > 300$ GeV | $M_{\mathrm{eff}} > 800$ GeV |
|---|---|---|
| pseudo data sample | 5% | 3% |
| half renormalization scale | 2% | 0% |
| lower parton $p_T$ | 9% | 4% |
| $W(\rightarrow \ell\nu)$ | $E_{\mathrm{T}}^{\mathrm{miss}} > 300$ GeV | $M_{\mathrm{eff}} > 800$ GeV |
| pseudo data sample | 16% | 7% |
| half renormalization scale | 4% | 2% |
| lower parton $p_T$ | 12% | 15% |

Table 23: Systematic uncertainty from detector performance in the MC method.

| | $E_{\mathrm{T}}^{\mathrm{miss}} > 300$ GeV | $M_{\mathrm{eff}} > 800$ GeV |
|---|---|---|
| Jet energy scale | 6% | 6% |
| Jet energy resolution | 1% | 1% |
| $E_{\mathrm{T}}^{\mathrm{miss}}$ soft component scale | 1% | $< 1\%$ |
| Lepton energy scale | $< 1\%$ | $< 1\%$ |
| Lepton identification efficiency | 2% | 2% |

differently from the standard Monte Carlo sample[4]. In this pseudo data sample, the shape of $E_{\mathrm{T}}^{\mathrm{miss}}$ and $M_{\mathrm{eff}}$ distributions is not affected; only the normalization changes significantly. The MC method is able to recover such a change and predict the background correctly.

**Systematic uncertainty** The MC method relies on Monte Carlo for the shapes of the background distributions. The systematic uncertainty from the Monte Carlo is estimated by using samples with different Monte Carlo parameters (pseudo real data, half renormalization scale sample, lower parton threshold sample). The results are summarized in Table 22; the deviation of the samples is 16% at most.

Other potential sources of systematic error in the background estimation from uncertainties in detector performance are summarized in Table 23. Since Monte Carlo is used, an uncertainty in the experimental jet energy scale affects the MC method prediction significantly more than for the replace method (Section 3.3.1).

**SUSY contamination of the control sample** The tight selection cuts of the control sample makes any SUSY contamination negligible (as was also the case for the replace method discussed in Section 3.3.1).

[4]renormalisation and factorisation scale 0.8 times the default, a minimum parton $p_T$ of 30 GeV (rather than 40 GeV), and separation between partons $\Delta R_{jj} = 0.6$ (rather than 0.7).

### 3.3.3 Control sample constructed from an $M_T$-selected one-lepton sample

**$W$ and top backgrounds** The $t\bar{t}$ and $W^{\pm}$ + jets processes contribute to the background in the no-lepton mode, when the lepton emitted in the $W^{\pm} \to \ell\nu$ process is not identified. The reasons why the emitted lepton is missed are summarized in Table 24. Hadronic decay of $\tau$'s and the lepton being out of acceptance (the $p_T$ of the lepton is required to be larger than 20 GeV) are the dominant reasons, and similar reasons/ratio's are observed in both $t\bar{t}$ and $W^{\pm}$ + jets processes.

Table 24: Numbers of events in which leptons are missed for various reasons in the semi-leptonic and leptonic decays of top and $W^{\pm}$. The numbers are normalized to 1 fb$^{-1}$ and listed after the SUSY no-lepton selection is applied.

| | $t\bar{t}$ | $W^{\pm}$ + jets |
|---|---|---|
| $W^{\pm} \to \tau \to$ hadron | 1993 (43%) | 773 (42%) |
| Out of acceptance | 1805 (39%) | 762 (41%) |
| Isolation (close to jet) | 807 (18%) | 322 (17%) |

The production processes $t\bar{t}$ and $W^{\pm}$ with $W^{\pm} \to \ell\nu$ where the lepton is identified constitute good control samples from which to estimate these background processes in the no-lepton mode, since similar kinematic distributions are expected except for the presence of the lepton. For the control sample, the same kinematic selections as for the signal in the no-lepton mode are applied. In addition, exactly one isolated lepton (e or $\mu$ with $p_T$ larger than 20 GeV) is required and the transverse mass between this lepton and the $E_{\mathrm{T}}^{\mathrm{miss}}$ is required to be smaller than 100 GeV to enhance the $t\bar{t}$ and $W^{\pm}$ processes. After these selections, this identified lepton is treated as if had been missed and all kinematic variables are recalculated.

The distributions of $t\bar{t}$ and $W^{\pm}$ + jets events differ after the SUSY selections are applied. The $E_{\mathrm{T}}^{\mathrm{miss}}$ for $W^{\pm}$ tends to be larger than for $t\bar{t}$ events, since the boost factor of the $W^{\pm}$ is larger for $W^{\pm}$ + jets after several high $p_T$ jets are required. Therefore the reproduced distributions are sensitive to the mixture of the $t\bar{t}$ and $W^{\pm}$ + jets events in the control sample. The fractions of the $t\bar{t}$ and $W^{\pm}$ + jets processes in the control sample are 81% and 19%, respectively, close to the actual backgrounds in the signal region, which are 73% and 27% respectively. The systematic errors due to the uncertainties in the cross-section of $t\bar{t}$ and $W^{\pm}$ + jets will be discussed later.

The estimated distribution is normalized with the data at 100 GeV $< E_{\mathrm{T}}^{\mathrm{miss}} <$ 200 GeV, where the contribution of the SUSY signal is expected to be small. The estimated distributions are slightly harder than the true distributions of the background processes, but similar distributions are obtained. The number of estimated background events is summarized in the two top rows in Table 25. Reasonable agreement is observed at high $E_{\mathrm{T}}^{\mathrm{miss}}$ .

**QCD, W and top background without and with SUSY signal** In a dedicated note within this volume [3] various methods for the estimation of QCD background from data are discussed.

In this section, we include QCD background, and estimate it as follows. It has been shown that after the removal of events with mismeasured $E_{\mathrm{T}}^{\mathrm{miss}}$ (e.g. noisy or dead calorimeter cells), as discussed elsewhere in this volume [12], semi-leptonic heavy quark ($b$,$c$) decays are the dominant contribution to large $E_{\mathrm{T}}^{\mathrm{miss}}$ in the QCD background. A function is derived to represent the momentum fraction taken by the neutrino in $b$ and $c$ quark decays. This function is then applied to a control sample taken from data (at least four jets with $p_T$ larger than 50 GeV, $p_T$ of the leading jet larger than 100 GeV, $E_{\mathrm{T}}^{\mathrm{miss}}$ smaller than 100 GeV) dominated by light-quark QCD events. The resulting distributions are normalized

to data in the region of $\Delta\phi_{\min} < 0.2$, where $\Delta\phi_{\min}$ is the minimum value of the difference in azimuthal angle between the $E_T^{miss}$ vector and any of the three highest $p_T$ jets, as discussed in section 3.1 and the dedicated note [3].

The $W^\pm$ and $t\bar{t}$ background processes can also be estimated from the data, as discussed in the previous paragraph. Since all these processes are present in the data simultaneously, the background estimations should be done together. The same holds true for the presence of SUSY signal, which would contaminate the control samples.

Figure 18 (top row) shows the estimated and true distributions of $E_T^{miss}$ and the effective mass for the combined background processes, without SUSY signal. The background distributions are reproduced well and the correct normalizations are obtained for all the background processes. The numbers of events are also summarized in Table 25.

Table 25: Number of true and estimated (MT method) background events in the no-lepton mode, for $t\bar{t}$, $W^\pm$ and QCD processes without SUSY signal, normalized to 1 fb$^{-1}$.

| | $E_T^{miss} > 100$ GeV | $E_T^{miss} > 300$ GeV |
|---|---|---|
| $t\bar{t}$ and $W^\pm$ only | | |
| True BG (top and $W^\pm$) | $6894 \pm 83$ | $276 \pm 17$ |
| estimated BG (top and $W^\pm$) | $7018 \pm 269$ | $311 \pm 28$ |
| QCD, $t\bar{t}$ and $W^\pm$ | | |
| True BG (QCD, top and $W^\pm$) | $8077 \pm 90$ | $300 \pm 17$ |
| Estimated BG | $8158 \pm 273$ | $327 \pm 28$ |
| Ratio(Est./True) | $1.01 \pm 0.04$ | $1.09 \pm 0.11$ |

If SUSY exists, SUSY signal can contaminate the background estimations. This is illustrated in Figure 18 (middle row) and Table 26. In the figure, the SUSY SU3 signal point is used; the table shows the variation over a number of SUSY samples. The figure and the table show that SUSY contamination causes a decrease of SUSY event excess by typically 30%, but that the considered points, with the exception of SU2, are still observable with 1 fb$^{-1}$.

Table 26: Number of background events and estimated (MT method) numbers for $t\bar{t}$, $W^\pm$ and QCD processes, as well as various SUSY signals, in the no-lepton mode, normalized to 1 fb$^{-1}$.

| | $E_T^{miss} > 100$ GeV | $E_T^{miss} > 300$ GeV | $E_T^{miss} > 100$ GeV | $E_T^{miss} > 300$ GeV |
|---|---|---|---|---|
| True BG (QCD,top and $W^\pm$) | $8077 \pm 90$ | $300 \pm 17$ | $8077 \pm 90$ | $300 \pm 17$ |
| | SU1 | | SU4 | |
| Estimated BG | $8493 \pm 283$ | $510 \pm 39$ | $27527 \pm 588$ | $1409 \pm 83$ |
| True BG + SUSY signal | $9152 \pm 96$ | $1078 \pm 33$ | $34209 \pm 185$ | $3535 \pm 59$ |
| | SU2 | | SU6 | |
| Estimated BG | $8198 \pm 274$ | $329 \pm 30$ | $8362 \pm 279$ | $431 \pm 35$ |
| True BG + SUSY signal | $8193 \pm 91$ | $351 \pm 19$ | $8930 \pm 95$ | $924 \pm 30$ |
| | SU3 | | | |
| Estimated BG | $9188 \pm 299$ | $633 \pm 44$ | | |
| True BG + SUSY signal | $11333 \pm 106$ | $2113 \pm 46$ | | |

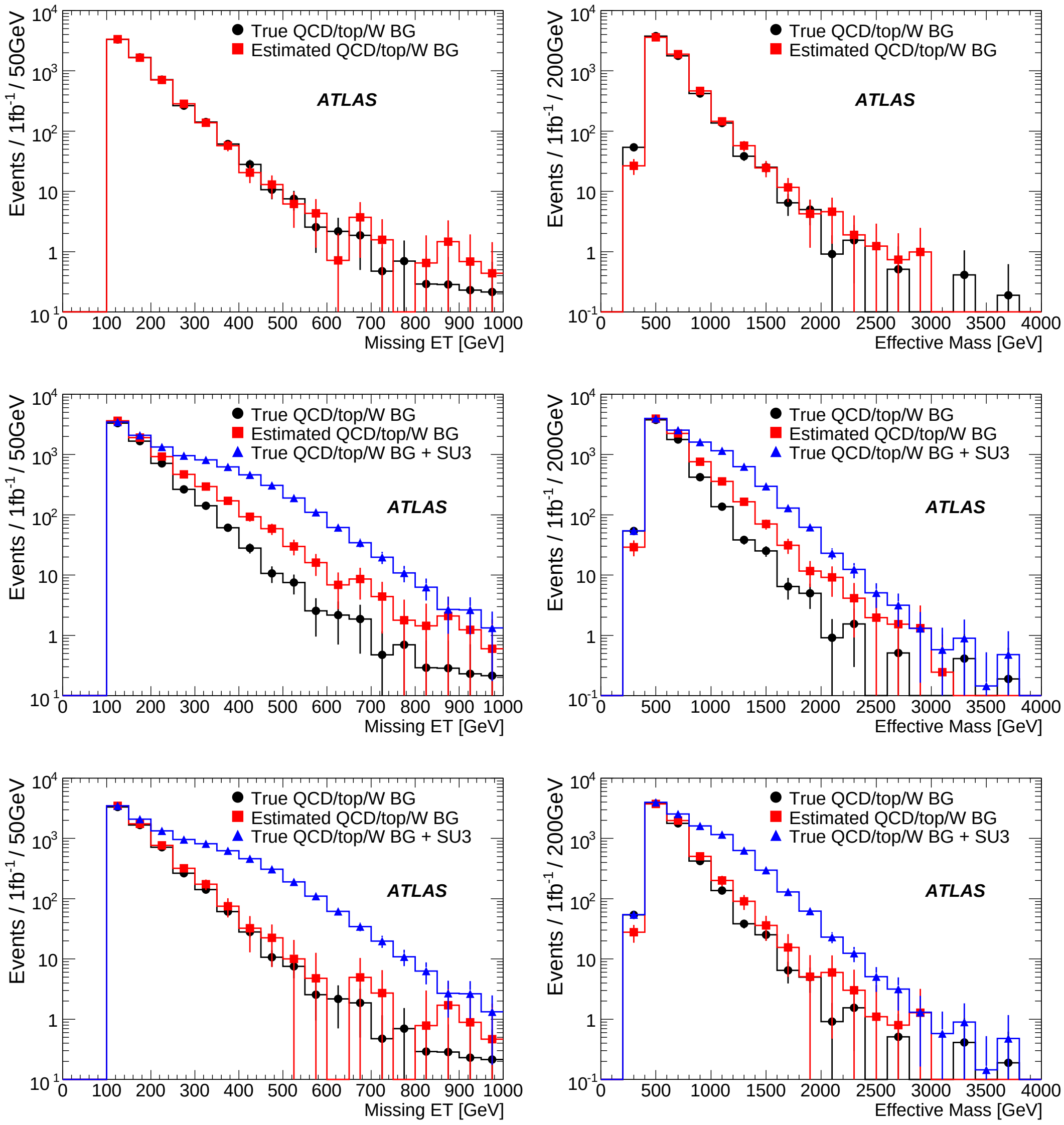


Figure 18: The estimated distributions of $E_{\mathrm{T}}^{\mathrm{miss}}$ and the effective mass of the $t\bar{t}$, $W^{\pm}$ and QCD backgrounds in the no-lepton mode with a luminosity of 1 fb$^{-1}$. In the plots in the top row, no SUSY signal is present in the data, and black/red histograms show the true/estimated (MT method) background distributions. In the plots in the middle and bottom rows, a SUSY SU3 signal is present in the data, and the blue histograms show the background plus the SUSY signal. In the two plots in the middle row, no correction was applied. In the two bottom plots, a correction with the “new MT method” was performed.

**New MT method** The “new MT method”, discussed in section 2.3.1, provides a first rough method to correct for the presence of SUSY signal in the control sample, once a SUSY excess has been observed in data. With this method, the distributions shown in the bottom row of Figure 18 are obtained. The figure shows the estimated background distributions with the new MT method, compared to the true background distributions, when SU3 SUSY signal is present in data.

**Systematic uncertainties** The systematic errors of this method are summarized in Table 27. The ALPGEN parameter variation includes a variation of the relative fraction of $t\bar{t}$ and $W+$jets in the sample. Although the actual number of background events varies with jet energy scale and Monte Carlo generator parameters by some 25%, the data-driven MT method is able to predict the background in the signal region to within an assigned systematic error of 15%.

Table 27: Systematic uncertainties of the MT-method background estimations in the no-lepton mode without SUSY signal. Also changes of the absolute background numbers are listed. Numbers are normalized to 1 fb$^{-1}$.

| | Syst. error | Change in background level |
|---|---|---|
| Jet energy scale | $< 5\%$ | 25% |
| Lepton energy scale | $< 5\%$ | 0% |
| Lepton Efficiency | $< 5\%$ | $< 1\%$ |
| MC@NLO vs ALPGEN | $< 5\%$ | 16% |
| ALPGEN parameter variation | $< 5\%$ | 8% |

### 3.3.4 Top pairs with $\tau$ decay

**Introduction** The precise determination of the cross-section for the process of top-pair production with one tau that decays hadronically, $t\bar{t} \to W(q\bar{q}')W(\tau_{\rm had}\nu_\tau)b\bar{b}$, is relevant because it constitutes an important background to SUSY searches with no leptons and significant $E_{\rm T}^{\rm miss}$ . In fact, if no tau veto is applied, about 65% of the total $t\bar{t}$ background in the no-lepton mode corresponds to events containing one tau.

**Event reconstruction and selection** The topology of $t\bar{t} \to W(qq')W(\tau_{\rm had}\nu)b\bar{b}$ events consists of two light–quark jets, one tau, and two $b$ jets.

The control of the tau fake rate is very important in a busy environment like $t\bar{t}$ where the purity of the reconstructed tau sample is low due to the large jet multiplicity. A high tau purity is needed in order to reduce the internal combinatorial background and the background from the semileptonic ($t\bar{t} \to W(\ell\nu_\ell)W(q\bar{q}')b\bar{b}$ where $\ell \in \{e,\mu\}$) decays of $t\bar{t}$. In this analysis, we use a calorimeter-based tau reconstruction algorithm [1], and require a minimum visible $p_T$ of the identified tau of 25 GeV.

The event is built independently on the hadronic side and the leptonic side, and topology variables useful to identify $t\bar{t}$ events and reject QCD and $W+$jet background are extracted. On the hadronic side, a hadronic $W$ invariant mass is built choosing, among all the combinatorial possibilities of di-jets, the pair of jets with closest invariant mass to its PDG value. The hadronic top is built combining this hadronic $W$ with the closest identified $b$ jet in $\Delta R$. The $b$-jet identification is loose, with a 75% efficiency for $b$ jets from top decay.

The $E_{\rm T}^{\rm miss}$ is combined with the identified tau in order to build a leptonic $W$ transverse mass. We assume a collinear approximation for the decay of the tau (the visible products of the hadronic decay of the tau and the associated $\bar{\nu}_\tau$ are collinear), and determine the invariant transverse mass of the leptonic $W$. The resulting leptonic $W$ is then combined with the closest (in $\Delta R$) $b$ jet to constitute the leptonic top.

For each event, a reconstructed tau will build a leptonic $W$ (and top) in combination with $E_{\rm T}^{\rm miss}$ , and will have an associated hadronic $W$ (and top). If there is more than one reconstructed tau we select the one that is associated with the jet pair that gives the hadronic $W$ invariant mass closest to its PDG value.

Once the event is built, topology variables suitable for $t\bar{t}$ selection are computed and selection cuts are applied: $E_{\rm T}^{\rm miss} > 35$ GeV, no identified electron or muon with $p_T > 15$ GeV should be present in the event, the angle $\Delta\phi$ between the two reconstructed top quarks should be larger than 2.5, the ratio between the

transverse momentum of the two reconstructed top quarks should be smaller than 2, the angular distance (ΔR) between the reconstructed $b$ jets should be larger than 1, and the angle $\Delta\phi$ between the missing momentum vector and the hadronic $b$ jet should be larger than 0.5.

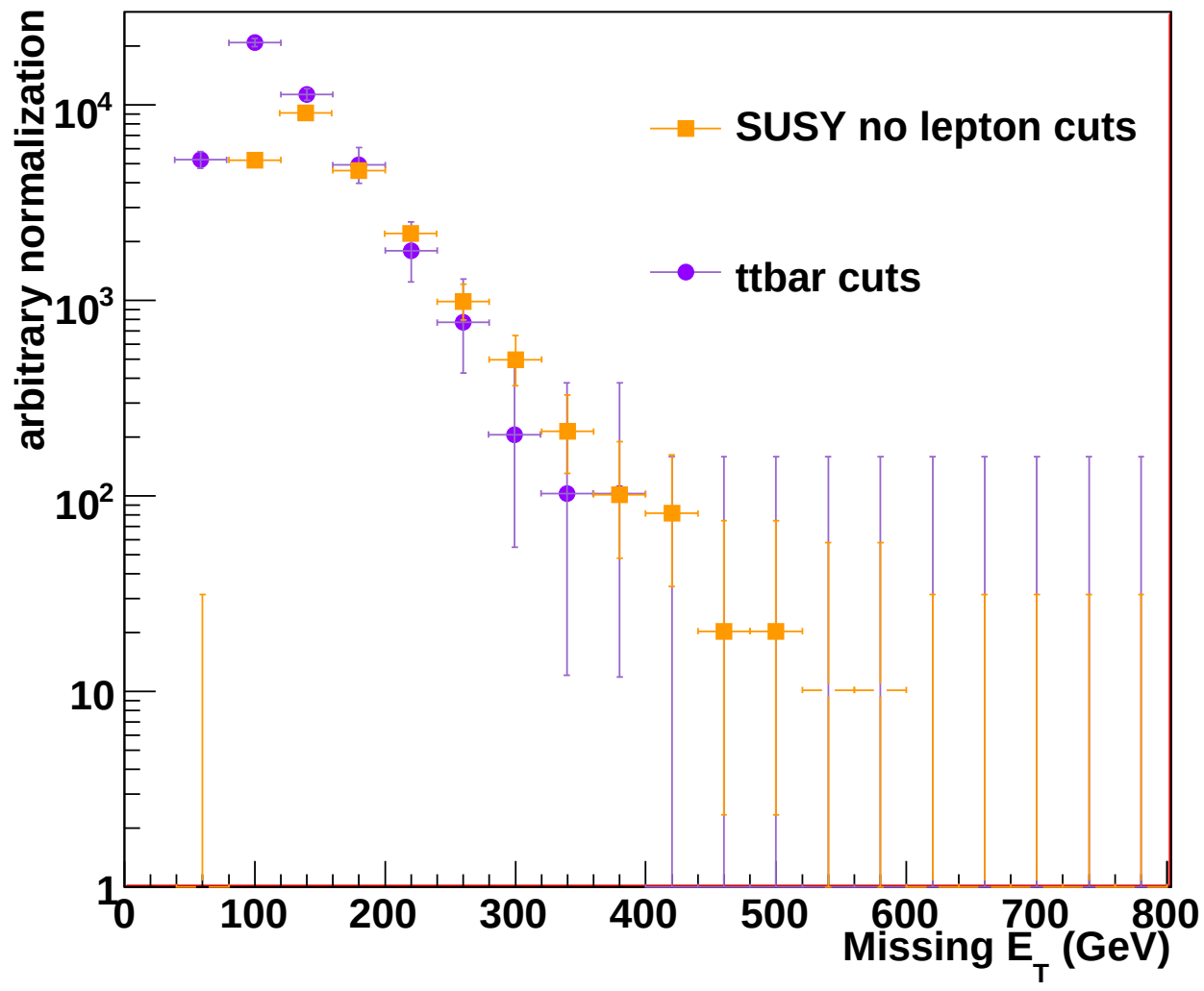


Figure 19: Missing $E_T$ of $t\bar{t}$ events with $(q\bar{q}', \tau_{\rm had})$ for (circles) $t\bar{t}$ selection as in this analysis (control sample) and (squares) for the SUSY no-lepton mode selection.

With the loose $b$ tagging used here, 2910 $t\bar{t}(q\bar{q}', \tau_{\rm had})$ events are selected for 1 fb$^{-1}$. The background consists of QCD and $W+$jets, for which 110 and 100 events respectively are selected. Therefore, with loose $b$ tagging a good signal-to-background ratio can already be reached, although the uncertainty on the QCD background number is large. If one were to use tighter $b$ tagging (60% efficiency), 1650 $t\bar{t}(q\bar{q}', \tau_{\rm had})$ events would be selected, against 2 QCD events and no $W+$jet events.

In the presence of SUSY, the numbers of SUSY signal events that would pass the event selection with the loose $b$ tag for 1 fb$^{-1}$ are given in Table 28. They are generally small, with the exception of the SU4 point.

Table 28: The number of SUSY events remaining after $t\bar{t}(q\bar{q}', \tau_{\rm had})$ selection for different SUSY points, for 1 fb$^{-1}$.

| | SU1 | SU2 | SU3 | SU4 | SU6 | SU8 |
|---|---|---|---|---|---|---|
| $N_{evt}$ | $22 \pm 1$ | $5 \pm 1$ | $155 \pm 5$ | $1700 \pm 60$ | $70 \pm 4$ | $45 \pm 3$ |

**Estimation of the $t\bar{t}$ with $(q\bar{q}', \tau_{\rm had})$ background to SUSY.** Figure 19 shows that the selection applied in order to identify the $t\bar{t}$ events with $(q\bar{q}', \tau_{\rm had})$ introduces little bias in the $E_{\rm T}^{\rm miss}$ distribution, as compared to the SUSY no-lepton mode selection.

Applying the SUSY no-lepton mode cuts, we estimate 210 $t\bar{t}$ $(q\bar{q}', \tau_{\rm had})$ events as remaining background to the SUSY no-lepton mode for 1 fb$^{-1}$.

**Systematic uncertainties** In the data-driven analysis of $t\bar{t}$ decays in the $(q\bar{q}', \tau_{\rm had})$ final state, the most relevant contributions to the detector uncertainties are the jet and $E_{\rm T}^{\rm miss}$ energy scale, $b$-tagging efficiency and $\tau$ identification efficiency.

Table 29: Systematic variation of the $t\bar{t}(q\bar{q}', \tau_{\rm had})$ cross-section due to detector-related uncertainties.

| Systematic variation | Cross-section variation [%] |
|---|---|
| Jet energy scale | 2.5 |
| $b$-tagging efficiency | 7.5 |
| Light quark rejection in b-tag | 1.3 |
| $\tau$-identification efficiency | 3.4 |
| Light quark rejection in $\tau$-identification | 4.5 |

The $b$-tagging efficiency plays an important role in the $t\bar{t}(q\bar{q}', \tau_{\rm had})$ reconstruction, since one of the selection criteria is that the two $b$ jets expected in the final state should be reconstructed and correctly identified[5].

The systematic contribution to the measurement of the $t\bar{t}(q\bar{q}', \tau_{\rm had})$ cross-section due to the $\tau$ identification efficiency has been estimated by varying the $\tau$ identification efficiency and the light quark rejection factor by 10%.

The uncertainty on the QCD background in the $t\bar{t}(q\bar{q}, \tau_{\rm had})$ sample is large due to the limited number of Monte Carlo events which could be generated. Probably tight $b$ tagging is required in this analysis. Further study is needed on this topic.

## 4 Multi-lepton and tau search modes

As well as the one-lepton and no-lepton search modes described earlier, there is considerable SUSY discovery potential in the multi-lepton and tau search modes, as discussed elsewhere in this volume [2,4].

The data-driven estimation of backgrounds in these modes, particularly the opposite-sign dilepton and the tau modes, can use the methods that have been described earlier in the context of the one-lepton search mode have also proven to be useful. These include the MT method, the HT2 method, the kinematic reconstruction method and the redecay method. Furthermore, for the same-sign dilepton mode, a technique based on lepton isolation, as described in the note on QCD backgrounds [3] could be further developed.

## 5 Discussion

The methods presented in this note represent a number of ideas on how top, $W$ and $Z$ backgrounds to SUSY searches can be extracted from the data, with appropriately chosen control samples. The results indicate that we expect, with 1 fb$^{-1}$, to be able to measure in the no-lepton mode:

- the $Z \to \nu\bar{\nu}$ background with two different methods to 8–13% stat. error, 10–15% syst. error;
- the $t\bar{t}$ background with hadronic tau decay to $< 6\%$ stat. error, 10–15% syst. error (but with a caveat for the QCD background);
- and the sum of top, $W$ and QCD backgrounds with the MT method to 4–8% stat. error, and 15% syst. error.

[5]This is in fact the only analysis in this note where $b$ tagging is used.

In the one-lepton mode:

- the MT method gives the sum of $t\bar{t}$ and W background to 4–8% stat. error, 15% syst. error;
- the semileptonic $t\bar{t}$ background can be estimated to 5% stat. error, 22% syst. error;
- we can determine the fully leptonic $t\bar{t}$ background to 10% stat. error, 20% syst. error in at least three independent ways;
- and we have a combined fit method to extract all components.

These methods can also be applied to the multi-lepton and tau search modes.

The results obtained with the MT method in this note are stable with respect to systematic variations in detector performance and Monte Carlo parameters and cross-sections to the 15% level. However, the MT method measures a sum of semileptonic and fully leptonic $t\bar{t}$ and $W/Z+$jets background; it relies on a control sample with different composition than the signal sample, and there is a subtle interplay between the $t\bar{t}$ and $W/Z+$jets components of the background. More work is needed to understand possible systematic effects. It is desirable to understand the individual components of the backgrounds as well, and the various other methods discussed in this note appear to succeed in this.

The presence of SUSY signal would affect the background estimates, at a level that depends on the SUSY signal properties, as well as on the method. Methods with very tight control samples (replace method, topbox method) see almost no effect. For the other methods, the background is overestimated by typically 20–30% for samples like SU1, SU2, SU3 and SU6. If a SUSY excess is nevertheless observed (which is possible with 1 fb$^{-1}$), a correction for the background overestimation can be applied. First ideas have been presented in this note, using the MT method, and the combined fit method. More work is needed in this area. The SU4 benchmark point is a special case because of its light spectrum. It produces events with kinematics which are similar to the Standard Model backgrounds and its cross-section is high, so many methods would struggle to provide background predictions. It would, however, not be missed [11].

# Estimation of QCD Backgrounds to Searches for Supersymmetry

**Abstract**

This note deals with techniques for estimating QCD multijet backgrounds to inclusive jets + $E_{\mathrm{T}}^{\mathrm{miss}}$ + leptons searches for Supersymmetry. The note documents how the backgrounds may be estimated using a data set with an integrated luminosity of 1 fb$^{-1}$ or less. The likely systematic and statistical uncertainties in such estimates are also discussed. Data-driven approaches discussed include jet smearing with functions derived from data and techniques for background estimation using control samples. Monte Carlo based techniques using a variety of generators and simulation tools are also discussed with a view to establishing their likely precisions and hence the optimum simulation strategy.

## 1 Introduction

In searches for supersymmetry in events containing jets and missing transverse energy ($E_{\mathrm{T}}^{\mathrm{miss}}$) the biggest background-determination challenge will be in understanding QCD jet events. Although fast simulation studies [1] indicate that such backgrounds should be sub-dominant in the high $E_{\mathrm{T}}^{\mathrm{miss}}$ and large jet multiplicity region, the situation with real data is still uncertain, and likely to be less favourable for SUSY searches. In particular non-Gaussian tails to the detector jet response function, which can be caused by dead material, jet punch-through, pile-up of machine backgrounds and other effects, can dramatically increase the cross-section of high $E_{\mathrm{T}}^{\mathrm{miss}}$ QCD background. QCD jet events in which such processes have occurred shall be referred to in this note as “fake” $E_{\mathrm{T}}^{\mathrm{miss}}$ events in what follows, since in such events the net $p_T$ of non-interacting particles measured in a perfect detector is inherently small. Events in which the $E_{\mathrm{T}}^{\mathrm{miss}}$ vector is dominated by contributions from non-interacting particles such as neutrinos or the Lightest Supersymmetric Particle (in the case of SUSY signal) shall be referred to as “real” $E_{\mathrm{T}}^{\mathrm{miss}}$ events.

Accurate estimation of QCD jet backgrounds is difficult for a number of reasons. Processes generating fake $E_{\mathrm{T}}^{\mathrm{miss}}$ from event mis-measurement are expected to be poorly modeled in current `GEANT4` [2] simulations – this situation will improve once validation has been performed with real data. Furthermore theoretical and experimental uncertainties will conspire to decrease further the systematic precision of Monte Carlo estimates while the large QCD cross-section will limit statistical precision by rendering production of unbiased `GEANT4` samples corresponding to more than a few pb$^{-1}$ unfeasible. This note addresses these problems, assessing the relative magnitudes of some of the uncertainties, identifying techniques for minimising the contribution of such backgrounds in the SUSY signal region and studying novel Monte Carlo- and data-driven approaches to accurate background estimation. It should be understood that the background estimates presented here are not intended to be definitive; rather the focus is on developing and evaluating the performance of tools for future use. This note should be read in conjunction with the notes outlining ATLAS $E_{\mathrm{T}}^{\mathrm{miss}}$ performance [3], data-driven estimation strategies for $W$, $Z$ and top backgrounds [4] and inclusive SUSY searches [5]. The main variables sensitive to the presence of SUSY signal which have been considered in this note are $E_{\mathrm{T}}^{\mathrm{miss}}$ and the ‘effective mass’ defined by $M_{\mathrm{eff}} = \sum_{i=1}^{4} |p_T(j_i)| + E_{\mathrm{T}}^{\mathrm{miss}}$ [5], where the sum runs over the four leading jets satisfying the conditions described below.

Wherever possible events appearing in $E_{\mathrm{T}}^{\mathrm{miss}}$ or $M_{\mathrm{eff}}$ distributions have been selected with a standard jet selection requiring at least four jets with $p_T > 50$ GeV and $|\eta| < 2.5$, at least one of which must have

$p_T > 100$ GeV. In some sections describing studies of new simulation techniques with limited available simulation statistics these cuts have been relaxed – these cases are highlighted in the text. In addition the performance of the different background estimation techniques has been compared by assessing the uncertainties of the background estimates for events passing a standard baseline set of cuts for the jets + $E_{\mathrm{T}}^{\mathrm{miss}}$ + 0-lepton (1-lepton for Section 5.2) channel described elsewhere in this volume [5].

## 2 Fake $E_{\mathrm{T}}^{\mathrm{miss}}$ rejection

### 2.1 Jet fiducialisation in $\eta$

In QCD multijet events the generation of large fake $E_{\mathrm{T}}^{\mathrm{miss}}$ from jets falling in poorly instrumented regions of the detector is a significant mechanism by which such events mimic SUSY signal. Conversely the large number of such events means they can provide a useful in-situ probe of the the ability of the detector to measure $E_{\mathrm{T}}^{\mathrm{miss}}$. Here we consider the use of such events for defining non-fiducial regions where jets may be expected to be poorly measured and hence capable of generating significant fake $E_{\mathrm{T}}^{\mathrm{miss}}$.

The $E_{\mathrm{T}}^{\mathrm{miss}}$ resolution of the ATLAS detector is known to scale with $\sqrt{\Sigma E_{\mathrm{T}}}$ [6], where $\Sigma E_{\mathrm{T}}$ is the scalar sum over the transverse energies of all calorimeter objects. For this reason the $E_{\mathrm{T}}^{\mathrm{miss}}$-significance defined by $S = E_{\mathrm{T}}^{\mathrm{miss}}/\sqrt{\sum E_{\mathrm{T}}}$ is frequently used to distinguish real $E_{\mathrm{T}}^{\mathrm{miss}}$ from fake $E_{\mathrm{T}}^{\mathrm{miss}}$. Because QCD multijet events typically have little energy in invisible particles the value of $\langle S\rangle$ is determined by the $E_{\mathrm{T}}^{\mathrm{miss}}$ resolution of the detector. Much of the variance in $S$ is due to intrinsic shower fluctuations, but because the reconstructed $E_{\mathrm{T}}^{\mathrm{miss}}$ is dominated by the transverse energy near the highest-$p_T$ jets, some is due to the non-uniformity in energy resolution of the detector. A measurement of $\langle S\rangle$ in a sample of events in which one of the highest-$p_T$ jets points into a particular region of the calorimeter can therefore be used as a measure of the relative performance of that region.

In this study, the $E_{\mathrm{T}}^{\mathrm{miss}}$-significance of QCD jet events was used to calculate $\langle S\rangle$ in $\eta$ bins across the calorimeter. Each event $S$ value was used in two $\eta$ bins – one for each of the two highest-$p_T$ jets. The same techniques could be used generate a full $(\eta,\phi)$ map of the calorimeter. A sample of PYTHIA [7] simulated QCD jet events with $p_T > 280$ GeV and corresponding to an integrated luminosity of 23.8 pb$^{-1}$ was used for the study [8]. Events were required to pass the 160 GeV high-level single-jet trigger [9] and to possess at least two high-$p_T$ jets with $p_T(j_1) > 100$ GeV and $p_T(j_2) > 50$ GeV. These two jets were required to be back-to-back in the transverse plane such that $\pi - |\Delta\phi(j_1, j_2)| < 0.4$. Events were vetoed if they contained an isolated lepton with $p_T > 10$ GeV and $|\eta| < 2.5$ or if $E_{\mathrm{T}}^{\mathrm{miss}}$ was greater than 80 GeV. The $\Delta\phi$ and $E_{\mathrm{T}}^{\mathrm{miss}}$ cuts eliminate contamination from events with significant true $E_{\mathrm{T}}^{\mathrm{miss}}$, eg. $Z$ ($\rightarrow \nu\bar{\nu}$) + jets and heavy quark jets. While the $E_{\mathrm{T}}^{\mathrm{miss}}$ cut does, in principle, prohibit the identification of catastrophically unresponsive regions with this method, regions with consistently poor response would be readily identifiable via the relative deficiency of events with jets pointing into them.

The resulting calorimeter map is shown in Fig. 1. The degraded response in the tile barrel and tile barrel-extended barrel gaps ($\eta \sim 0$ and $0.6 < |\eta| < 0.8$) and the LAr barrel-endcap and hadronic LAr-forward calorimeter transition regions ($1.3 < |\eta| < 1.5$ and $3.1 < |\eta| < 3.3$) are all clearly visible. The principal application of these data is to define non-fiducial regions of the detector, such that events with jets pointing into these regions can be rejected. Defining a cut value $\langle S\rangle_{\min}$ on $\langle S\rangle$, a region is defined to be non-fiducial if $\langle S\rangle > \langle S\rangle_{\min}$ in all of the corresponding bins in the calorimeter map (Fig. 1) and if $\langle S\rangle - \langle S\rangle_{\min} > 3\varepsilon$ in at least one of those bins, where $\varepsilon$ is the uncertainty on the mean $\langle S\rangle$. Fig.1 also shows the non-fiducial regions obtained in this way with $\langle S\rangle_{\min} = 0.95$.

The jet fiducialisation process has been tested on the QCD jet sample which was used to define the regions and also on SUSY events from model point SU3 [10]. Events were required to satisfy the jet cuts listed in Section 1 and events with a jet with $p_T > 40$ GeV pointing into one of the non-fiducial regions depicted in Fig.1 were rejected. The efficiency of these cuts for the QCD and SUSY samples was

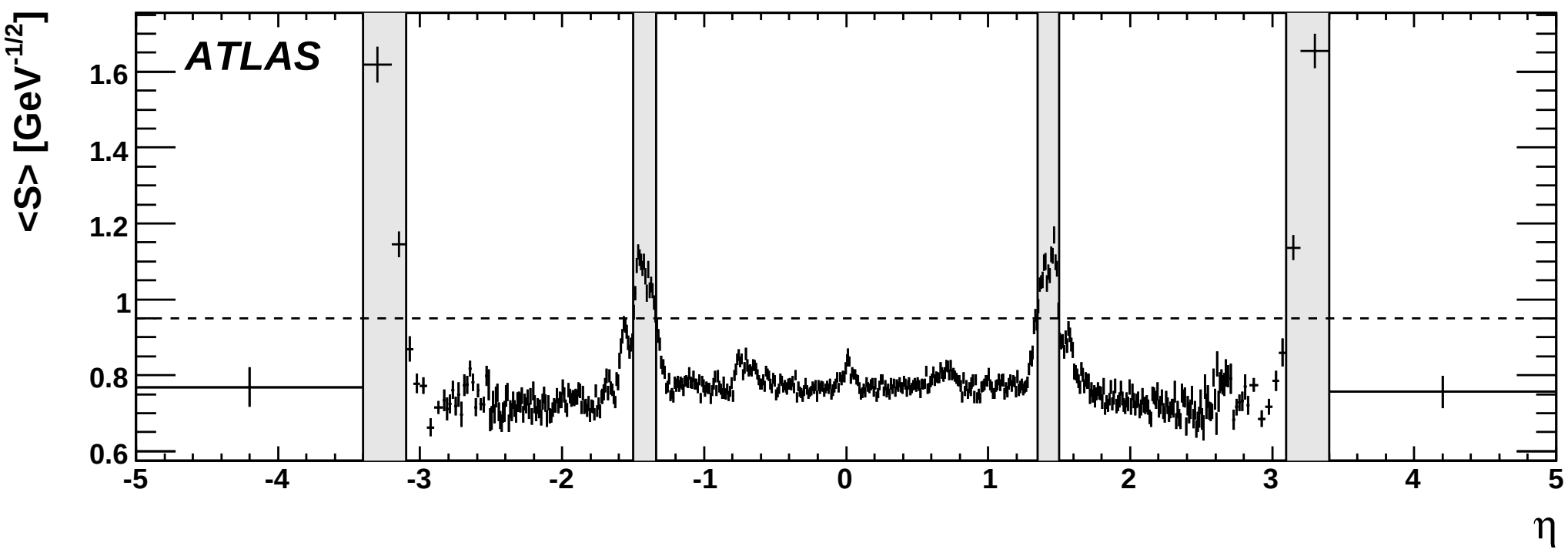


Figure 1: $\langle E_{\mathrm{T}}^{\mathrm{miss}}/\sqrt{\Sigma E_T}\rangle$ vs. $\eta_j$, where $j$ is the first or second highest-$p_T$ jet in each event (points). The non-fiducial regions selected with a cut at 0.95 (dashed line) are also shown (shaded areas; see text). These regions correspond to the LAr barrel-endcap and HEC-forward transition regions.

$(74.8\pm0.4)\%$ and $(75.4\pm0.6)\%$ respectively for $E_{\mathrm{T}}^{\mathrm{miss}} > 0$ GeV, and $(67.5\pm3.7)\%$ and $(75.5\pm0.7)\%$ for $E_{\mathrm{T}}^{\mathrm{miss}} > 100$ GeV. The effect of these cuts on the simulated data-samples is clearly small however they may be of more use with real data acquired with an imperfect detector, and the technique could be used as an on-line monitor of calorimeter performance.

## 2.2 Jet–$E_{\mathrm{T}}^{\mathrm{miss}}$ $\phi$ correlations

One of the main methods used to reduce the QCD multijet background in $E_{\mathrm{T}}^{\mathrm{miss}}$ + jets inclusive SUSY searches at the Tevatron has been the elimination of events in which the $E_{\mathrm{T}}^{\mathrm{miss}}$ is closely associated with one of the leading jets in the transverse plane [11, 12]. This section provides a brief exploration of this method of background elimination in ATLAS.

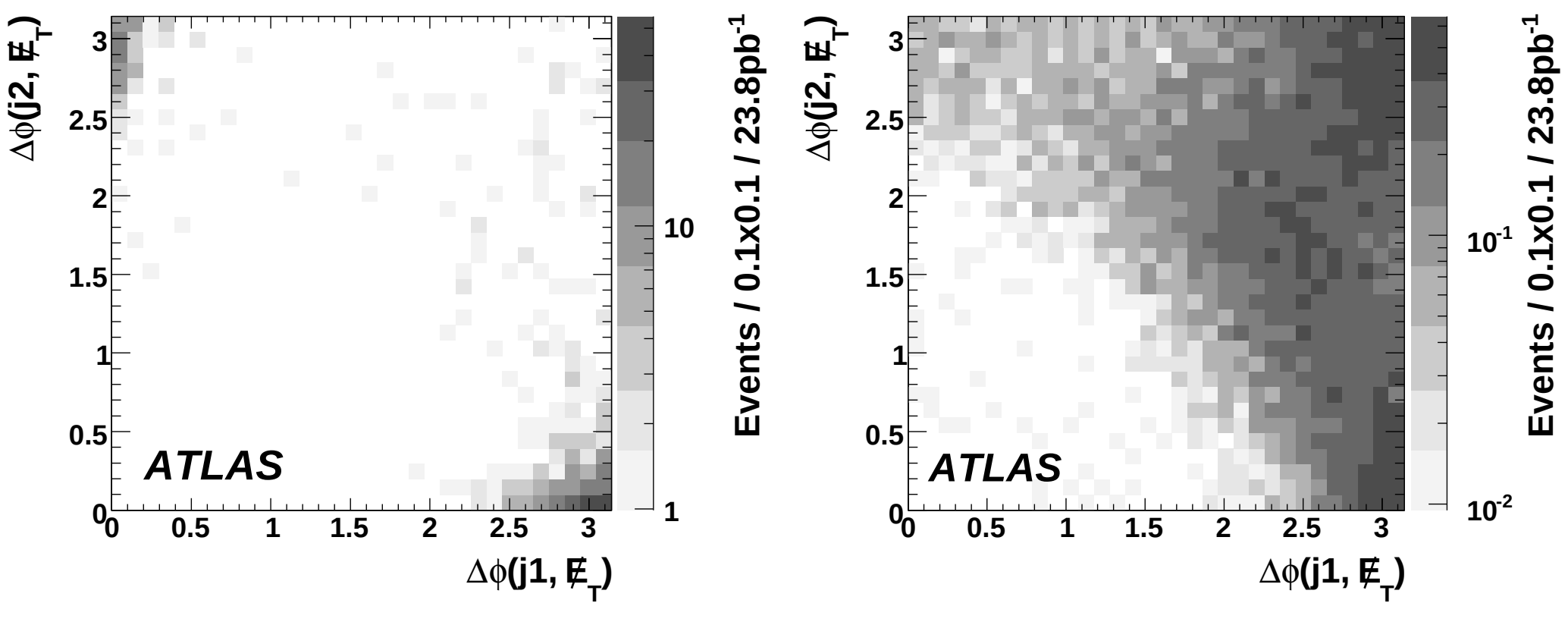


Figure 2: The $|\Delta\phi(j_1,E_{\mathrm{T}}^{\mathrm{miss}})| - |\Delta\phi(j_2,E_{\mathrm{T}}^{\mathrm{miss}})|$ plane for QCD multijet (left) and SUSY SU3 events (right) passing SUSY jet cuts and $E_{\mathrm{T}}^{\mathrm{miss}} > 100$ GeV.

Fig. 2 shows the $|\Delta\phi(j_1,E_{\mathrm{T}}^{\mathrm{miss}})| - |\Delta\phi(j_2,E_{\mathrm{T}}^{\mathrm{miss}})|$ plane for events from the QCD multijet sample described in Section 2.1 which pass the jet cuts described in Section 1 and possess $E_{\mathrm{T}}^{\mathrm{miss}} > 100$ GeV (note that this sample has at least one jet with $p_{\mathrm{T}} > 280$ GeV – significantly higher than the SUSY leading-jet $p_{\mathrm{T}}$ requirement of 100 GeV). Almost all events are confined to small regions around $(\pi,0)$ and $(0,\pi)$, with a small number of events in which the $E_{\mathrm{T}}^{\mathrm{miss}}$ is not associated with one of the two highest-$p_T$ (reconstructed) jets lying outside these regions. For comparison, the $|\Delta\phi(j_1,E_{\mathrm{T}}^{\mathrm{miss}})| - |\Delta\phi(j_2,E_{\mathrm{T}}^{\mathrm{miss}})|$ plane for events from a SUSY (SU3 model) sample [8] passing the above cuts is shown in Fig. 2. No such correlation with $E_{\mathrm{T}}^{\mathrm{miss}}$ is observed for this SUSY benchmark point.

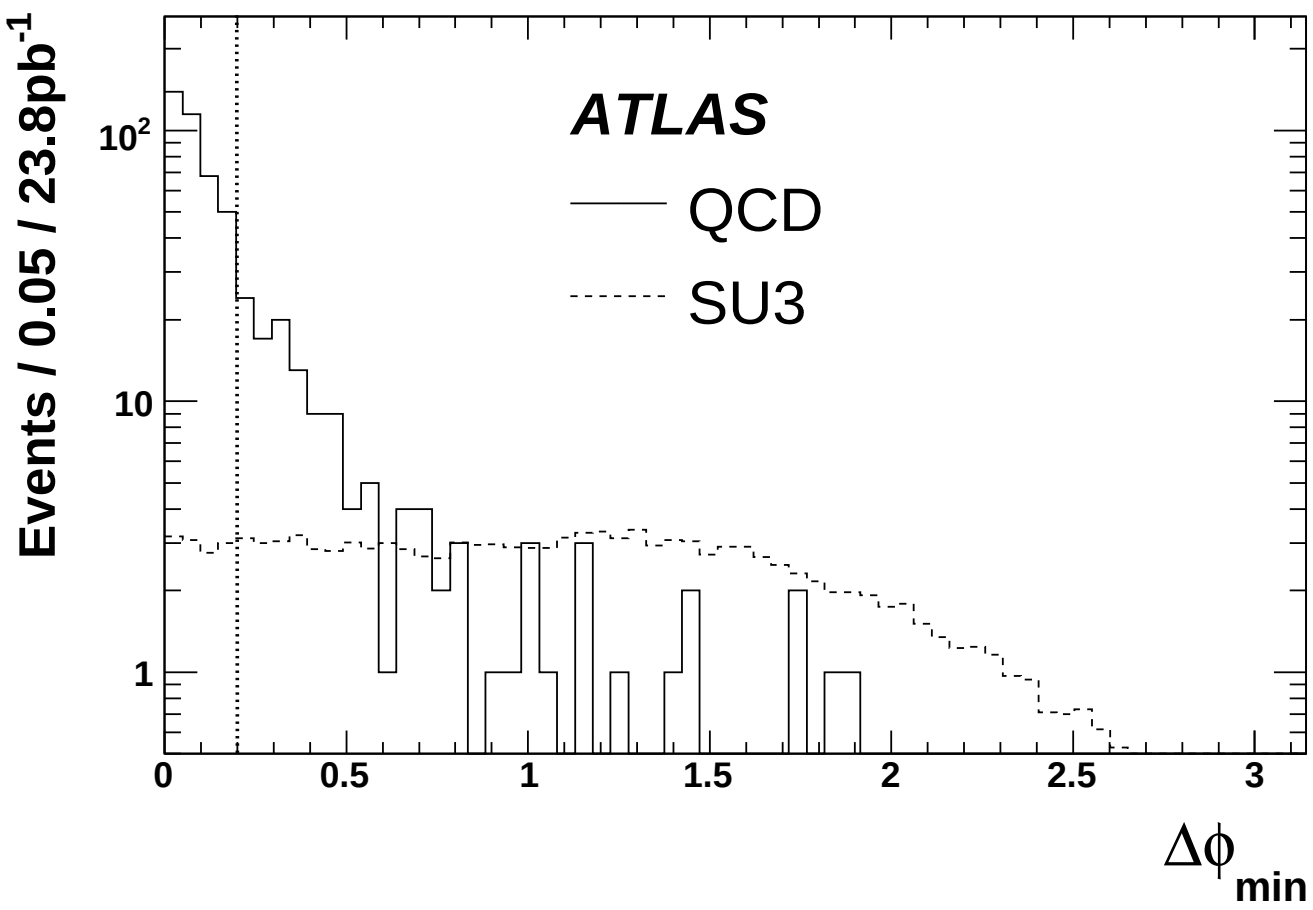


Figure 3: The $\Delta\phi_{\rm min}$ distribution for QCD multijet (solid line) and SU3 SUSY (dashed line) events passing the SUSY jet cuts and with $E_{\rm T}^{\rm miss} > 100$ GeV, as described in the text. Good discrimination is achieved with a cut at $\Delta\phi_{\rm min} = 0.2$ (vertical dotted line).

For the purposes of background reduction, this technique can be generalised by considering three correlations between the $E_{\rm T}^{\rm miss}$ vector and the leading three jets by defining $\Delta\phi_{\rm min}$ as:

$$\Delta\phi_{\rm min} = \min(\Delta\phi(j_1, E_{\rm T}^{\rm miss}), \Delta\phi(j_2, E_{\rm T}^{\rm miss}), \Delta\phi(j_3, E_{\rm T}^{\rm miss})). \tag{1}$$

The $\Delta\phi_{\rm min}$ distribution for the QCD and SUSY samples passing the SUSY jet cuts and with $E_{\rm T}^{\rm miss} > 100$ GeV are shown in Fig. 3. Also shown is the position of a cut at $\Delta\phi_{\rm min} > 0.2$ used in Ref. [5].

## 2.3 Calorimeter and tracking cuts

Events in which jet reconstruction problems lead to fake $E_{\rm T}^{\rm miss}$ can be identified with several variables [3]. Cuts on the fraction of the total energy of the jet ($E_{\rm Total}$) deposited in the outermost layers of the tile calorimeter $E_{\rm Tile2}$ and in the hadronic endcap calorimeter $E_{\rm HEC}$ can be defined to veto events with likely problems in jet containment. Additional cuts on the fraction of the jet energy deposited in the cryostat between the tile and the liquid argon ($E_{\rm Cryo}$ – estimated from the energy in the closest calorimeter layers) and in the gap and crack scintillators ($E_{\rm Gap}$) can be used to reject events with large depositions in dead material. It is also possible to identify events in which there have been problems reconstructing the missing transverse momentum by comparing the value calculated from charged tracks ($E_{\rm T,Trk}^{\rm miss}$) with the usual calorimetric $E_{\rm T}^{\rm miss}$.

Table 1: Cross-section of SUSY SU3 signal and QCD jet backgrounds before and after application of the selection and cleaning cuts proposed in the text. Quoted statistical uncertainties in cross-sections after cuts derive from statistics of the Monte Carlo samples used.

| Sample | No cuts [pb] | Selection cuts [pb] | Cleaning cuts [pb] |
|---|---|---|---|
| SUSY SU3 | 27.51 | $9.654 \pm 0.019$ | $6.792 \pm 0.016$ |
| QCD (280 GeV $< p_T <$ 560 GeV) | $1.25\times10^4$ | $17.2 \pm 1.3$ | $9.7 \pm 1.0$ |
| QCD (560 GeV $< p_T <$ 1120 GeV) | 360.0 | $4.22 \pm 0.13$ | $1.805 \pm 0.082$ |
| QCD (1120 GeV $< p_T <$ 2240 GeV) | 5.71 | $0.434 \pm 0.003$ | $0.130 \pm 0.002$ |

Table 1 shows the cross-section of SU3 signal and multijet background following application of the SUSY jet cuts described in Section 1 together with a cut requiring $E_{\rm T}^{\rm miss}$ >100 GeV, and in addition

applying the fake $E_{\mathrm{T}}^{\mathrm{miss}}$ cleaning cuts discussed above. The latter cuts were applied to jets with $p_T > 150$ GeV with the following acceptance criteria : $E_{\mathrm{Tile2}}/E_{\mathrm{Total}} < 0.1$, $E_{\mathrm{Cryo}}/E_{\mathrm{Total}} < 0.2$, $E_{\mathrm{Gap}}/E_{\mathrm{Total}} < 0.2$, $E_{\mathrm{HEC}}/E_{\mathrm{Total}} < 0.5$ and $E_{\mathrm{T,Trk}}^{\mathrm{miss}} - E_{\mathrm{T}}^{\mathrm{miss}} < 50$ GeV. Note that the specific cut values are analysis dependent and should be adapted for other cases. The table shows that the cleaning cuts are effective at removing between 40% (low $p_T$) and 70% (high $p_T$) of the multijet background, while rejecting $\sim$30% of the remaining signal after selection cuts.

### 2.4 Cosmic backgrounds and rejection cuts

Searches for supersymmetry must contend with backgrounds in which fake $E_{\mathrm{T}}^{\mathrm{miss}}$ arises from a variety of sources. Examples in which the fake $E_{\mathrm{T}}^{\mathrm{miss}}$ is generated independently from a hard scatter include dead or noisy calorimeter channels, accelerator-related background and cosmic rays in which a muon undergoes a hard bremsstrahlung. A range of data-cleaning tools is being prepared to reject fake $E_{\mathrm{T}}^{\mathrm{miss}}$ in ATLAS; here we focus on timing information from the hadronic tile calorimeter (TileCal). As one gains a better understanding of the first data, this will be combined with information from other subdetectors.

Calorimeter timing can be a powerful tool to remove fake $E_{\mathrm{T}}^{\mathrm{miss}}$ backgrounds from cosmic rays [13]. Timing in ATLAS is defined such that particles from the nominal interaction point ($x = 0$, $y = 0$, $z = 0$) will arrive in the calorimeter cells at time $t = 0$. Cosmic ray events, however, will arrive at random times with no correlation to the LHC beam structure. One of the methods to reject cosmic rays is to calculate the "up-minus-down" time, in which one divides the TileCal into two segments, an upper segment, with $\phi > 0$ and a lower segment with $\phi < 0$, where $\phi$ is the angle in the transverse plane. One then calculates the average time of cells in each segment, weighted by the energy of each cell $i$:

$$t_{\mathrm{up(down)}} = \sum_i (E_i^{\mathrm{up(down)}} \times t_i) / \sum_i E_i^{\mathrm{up(down)}}. \quad (2)$$

The difference between the two quantities $t_{\mathrm{up}}$ and $t_{\mathrm{down}}$, should reflect the average time-of-flight. The resulting "up-minus-down" time should be centered at $t = 0$ for particles from the interaction point, as demonstrated in Fig. 4. In this figure, the "up-minus-down" time is calculated for a simulated Monte Carlo QCD dijet sample. As expected, the signal distribution peaks near zero. Note that the simulation includes electronic noise, and so the timing distribution has a spread of several ns.

Depending on their trajectory, cosmic ray muons traveling from the top to the bottom of ATLAS at near the speed of light will have a time-of-flight of typically 18-20 ns over their travel distance of approximately 6-7 m [14] in the TileCal. This is demonstrated in Fig. 5 for a simulated Monte Carlo sample of cosmic ray muons. In this case, the distribution peaks near -18 ns (the time of flight is negative as expected for a particle traveling from the top to the bottom). The width of the simulated distribution is 3 ns, allowing for a good separation of the background from the signal peaking at $t = 0$. Further details of calorimeter timing studies with cosmic rays in ATLAS can be found in Ref. [14].

## 3 Monte Carlo systematics

### 3.1 Generator systematics: proton PDF and underlying event

In this section, uncertainties on the Monte Carlo generation of QCD events arising from proton parton distribution functions (PDFs) and modeling of the underlying event are briefly investigated. These uncertainties provide a limit to the precision with which the novel detector simulation strategies described in Sections 4.1 and 4.2 can model QCD backgrounds to SUSY. Due to the large number of Monte Carlo samples required for this study, each with different PDF or underlying event parameters, it was unfeasible to work with GEANT4 detector simulation samples. Therefore dedicated samples of events were generated in the same $p_T$ ranges as the GEANT4 samples used elsewhere in this note [8] and passed through

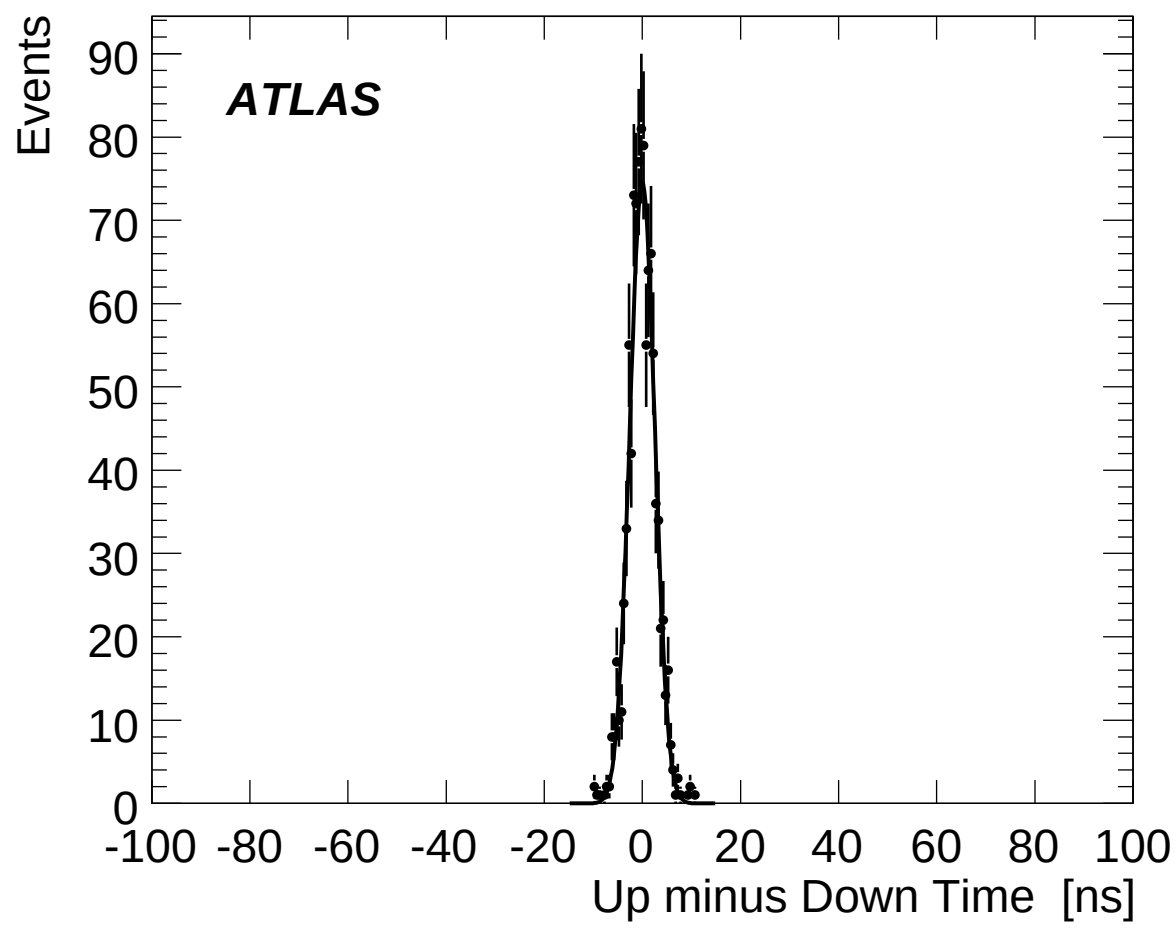


Figure 4: "Up-minus-down" timing distribution in TileCal for a simulated Monte Carlo di-jet sample. As expected, the distribution peaks near 0.

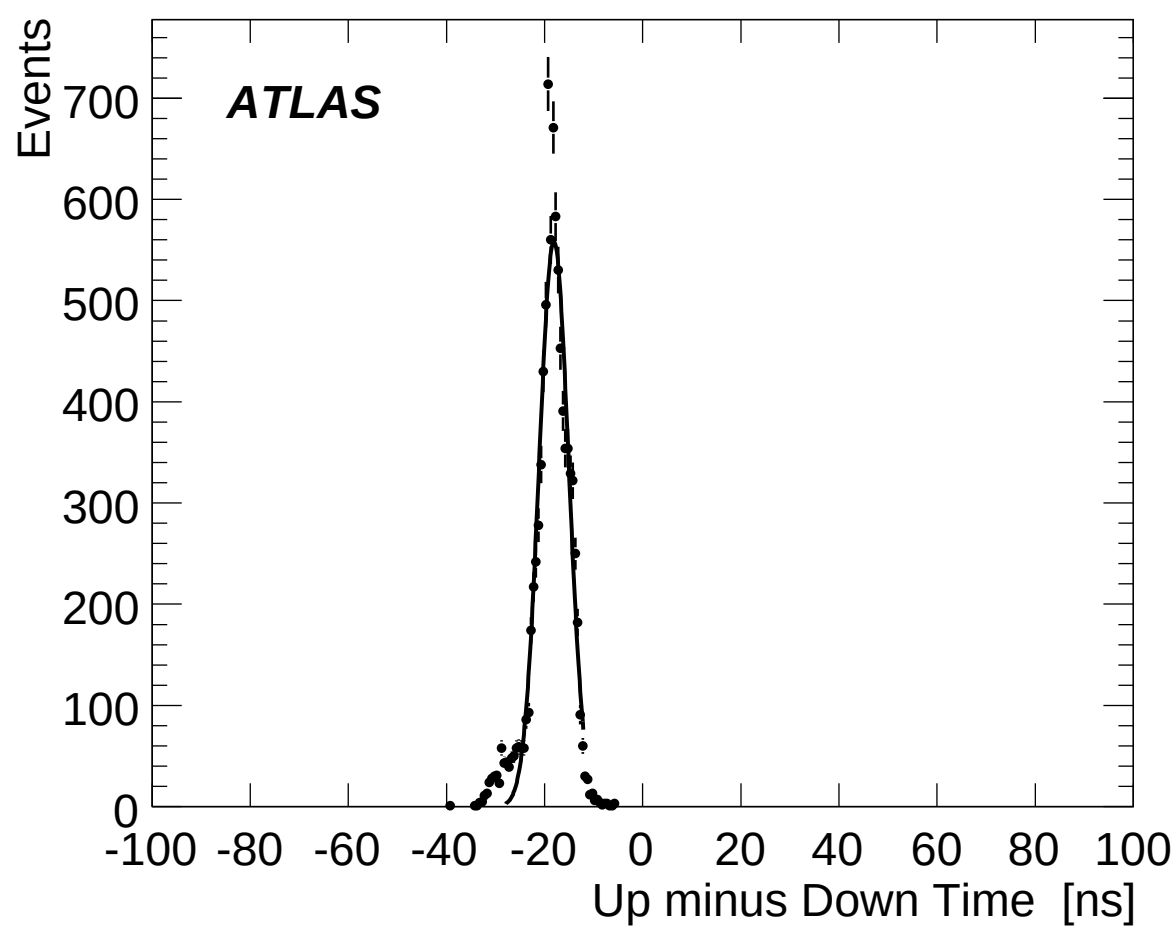


Figure 5: "Up-minus-down" timing distribution in TileCal for a simulated Monte Carlo cosmic ray sample. As expected, the distribution peaks near -18 ns.

the ATLAS fast detector simulation [15]. For all distributions shown in this section, QCD events were selected with the common jet cuts described in Section 1.

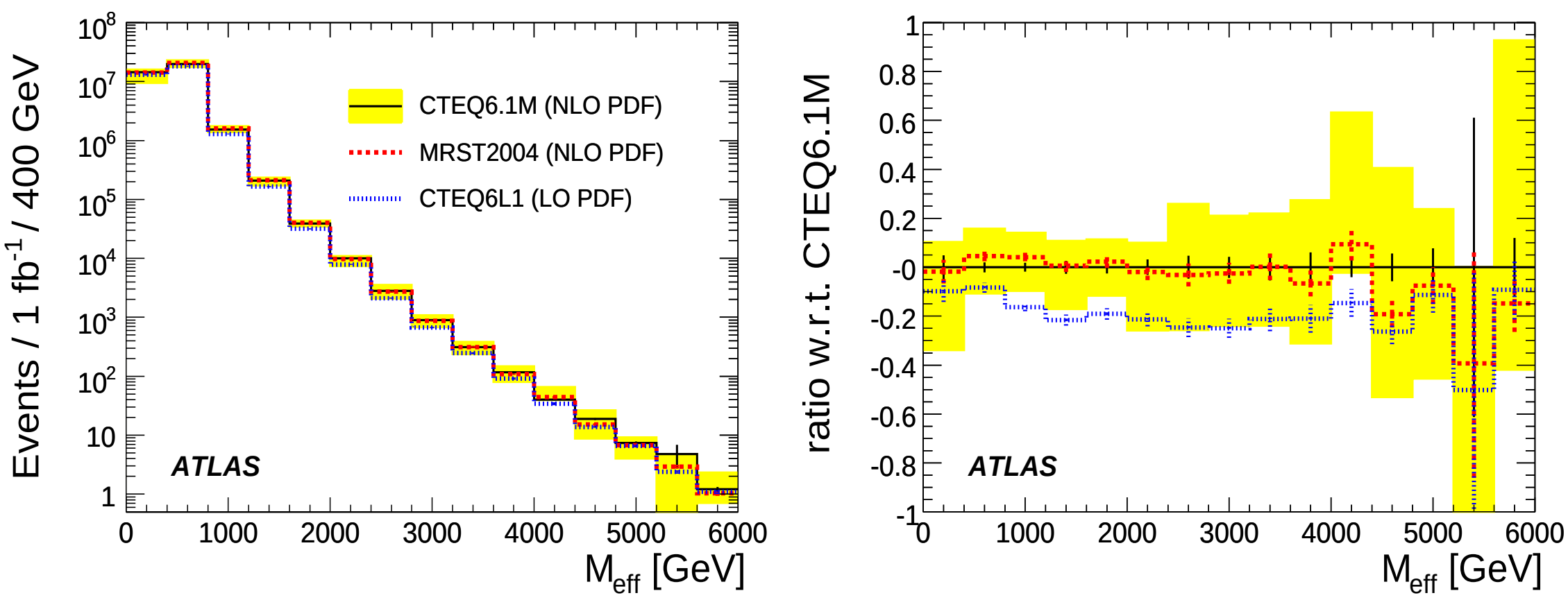


Figure 6: The $M_{\mathrm{eff}}$ distribution for simulated QCD events satisfying SUSY jet cuts described in Section 1. The left plot shows the number of events per bin, scaled to $1\mathrm{fb}^{-1}$ and the right plot shows the fractional difference with respect to CTEQ6.1M. In each case, the solid histogram and shaded band shows the results from the CTEQ6.1M proton PDF. The dashed and dotted histograms show the results from the MRST2004 and CTEQ6L1 PDFs, respectively. The error bars reflect the Monte Carlo statistics.

**Proton Parton Distribution Functions** All calculations of cross-sections at the LHC rely on a knowledge of the proton parton distribution functions (PDFs). High-$p_T$ QCD events will be particularly sensitive to the gluon PDF at large $x$ which, in current global fits [16–19], is relatively poorly known. In this section, the impact of the PDF uncertainties on distributions of observables sensitive to the presence of SUSY signal is estimated.

Events were generated using `PYTHIA 6.403` [7], with the following PDFs: CTEQ6L1 [20], CTEQ6.1M [17]

and MRST2004nlo [21]. Note that CTEQ6L1 is a leading order (LO) PDF, while CTEQ6.1M and MRST2004nlo sets are both next-to-leading-order (NLO) sets. CTEQ6.1M provides a set of error PDFs ($S_i^{\pm}$), corresponding to the $N$ eigenvector directions. Note that the $S_i^{\pm}$ only reflect the experimental uncertainties on the data, and do not include any information on theoretical assumptions used in the fit. The total PDF uncertainty arising from experimental sources, on a given observable, $\Sigma$, can then be constructed according to:

$$\Delta\Sigma^{\pm} = \sqrt{\sum_{i=1}^{N} \left(\max\left(\pm\Sigma(S_i^{+}) \mp \Sigma(S_0), \Sigma(\pm S_i^{-}) \mp \Sigma(S_0), 0\right)\right)^2}; \tag{3}$$

where $\Delta\Sigma^{+}(\Delta\Sigma^{-})$ are the upper (lower) uncertainties on the observable and $S_0$ represents the central (best fit) PDF value. For CTEQ6.1M, there are 40 error PDF sets, $S_i^{\pm}$, corresponding to $N = 20$ eigenvector directions. Since the uncertainty band obtained from CTEQ6.1M only reflects the uncertainties of the experimental data, the predictions of this PDF are also compared here to those from MRST2004nlo (an alternative, independent NLO PDF) and to CTEQ6L1[1].

The distribution of $M_{\mathrm{eff}}$ for the different PDFs considered, and the fractional uncertainty with respect to the central prediction of CTEQ6.1M, is shown in Fig. 6. The CTEQ6.1M uncertainty band ranges from $\sim 20-50\%$, as $M_{\mathrm{eff}}$ increases. The largest contribution to this band comes from eigenvector 15, which is mainly sensitive to the gluon at large $x$. The results from MRST2004nlo lie well within the CTEQ6.1M uncertainty band across the whole range of the distribution. The prediction from CTEQ6L1, a LO PDF, lies below those of the NLO PDFs. For $M_{\mathrm{eff}} \gtrsim 1$ TeV, the difference is approximately constant and at the level of $\sim 20\%$. The validity of using NLO versus LO PDFs, with LO matrix-element generators, is discussed in a recent publication [22]. However, these results indicate that the difference between the results from LO and NLO PDFs is smaller than the CTEQ6.1M uncertainty band.

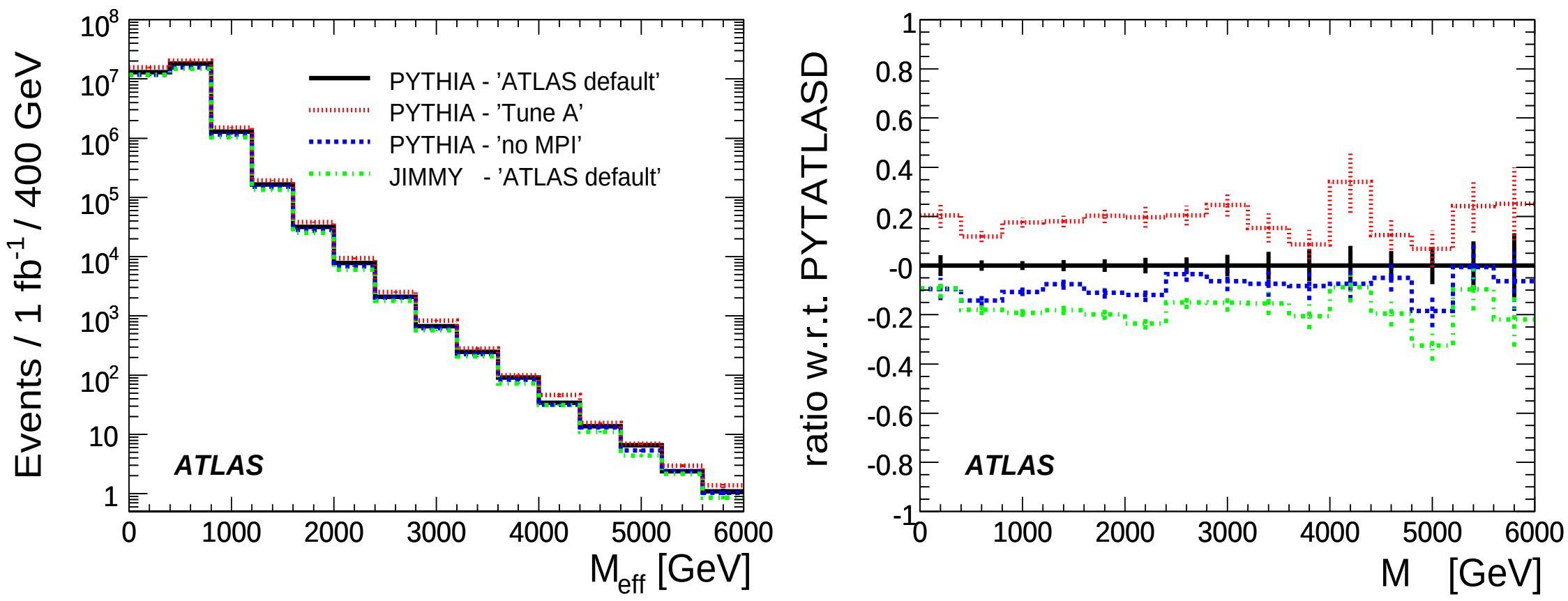


Figure 7: The $M_{\mathrm{eff}}$ distribution for simulated QCD events satisfying SUSY jet cuts described in Section 1. The left plot shows the number of events per bin and the right plot shows the fractional uncertainty with respect to the "PYTHIA – ATLAS default" model for the underlying event. In each plot, the solid histogram shows the results from the "PYTHIA – ATLAS default", the dotted histogram shows "PYTHIA – Tune A", the dashed histogram shows "PYTHIA – no MPI" and the dot-dashed histogram shows "JIMMY – ATLAS default". The error bars reflect the Monte Carlo statistics.

[1] Note that the PYTHIA parameters used by default within ATLAS are tuned to the CTEQ6L1 proton PDF. Strictly, if a different PDF is used, a retuning of the Monte Carlo parameters to data should be performed. This has not been done in the current study, and this should be bourne in mind when interpreting the results.

**Underlying Event** Monte Carlo generation of QCD events is also sensitive to the underlying event. Multi-parton interactions are modelled in several Monte Carlo generators. For this study, the PYTHIA 6.403 [7], and HERWIG 6.507 [23] (used in conjunction with JIMMY v4 [24]) are used. Further details of the underlying-event models compared are summarised below. Note that, unless otherwise stated, CTEQ6L1 is used as the proton PDF.

The underlying event models considered are briefly summarised below:

- "PYTHIA – ATLAS default" : the ATLAS default underlying-event tune for PYTHIA 6.4 [8].
- "PYTHIA – Tune A" : a model based on the "PYTHIA-TuneA" [25] to Tevatron data. Note that since underlying-event tunes also depend on the PDFs, CTEQ5L [26], appropriate to Tune A, has been used here.
- "PYTHIA – no MPI" : a model in which multi-parton interactions have been switched off. All other parameters were left at the PYTHIA 6.403 defaults. This model does not describe the Tevatron data, but is included for comparison.
- "JIMMY – ATLAS default" - the ATLAS default underlying event tune for HERWIG 6.5 + JIMMY 4 [8].

The distribution of $M_{\mathrm{eff}}$ is shown in Fig. 7. The left plot shows the number of events per bin (scaled to 1 $\mathrm{fb}^{-1}$) while the right plot shows the fractional ratio with respect to "PYTHIA – ATLAS default". As expected, the model with no underlying event contribution, "PYTHIA – no MPI", lies below the predictions of all other PYTHIA models. The highest model prediction is given by the "PYTHIA – Tune A" model, based on "PYTHIA-TuneA" to Tevatron data. The ATLAS default tunes, "PYTHIA – ATLAS default" and "JIMMY – ATLAS default", show differences of $\sim 20\%$. However, some of this difference may come from the different normalisations of PYTHIA and HERWIG and not totally due to the underlying event model. The overall spread in predictions from these models, is $\sim 40\%$ and is approximately constant across the range of $M_{\mathrm{eff}}$ shown.

### 3.2 Jet energy scale uncertainty

A significant source of irreducible experimental uncertainty in Monte Carlo QCD background estimates will arise from the uncertainty in the jet-energy-scale (JES). In this section, a simple estimate of the impact of the JES uncertainty is performed. For each event, the energy and momentum of each jet has been scaled by a constant factor, corresponding to a 10%, 5%, 3% and 1% uncertainty on the JES. The $E_{\mathrm{T}}^{\mathrm{miss}}$ has also been adjusted accordingly, by an amount corresponding to the change in the sum of the jet momenta in the $x$ and $y$ directions. All such scaling is performed prior to any selection cuts. For this study, the QCD dijet events generated with PYTHIA have been used, with a fake $E_{\mathrm{T}}^{\mathrm{miss}}$ filter applied to enhance the rate of potentially mismeasured events [3, 8].

Figure 8 shows the resulting $M_{\mathrm{eff}}$ distribution. The shaded bands show the uncertainties on the distribution for the four different values of the JES uncertainty considered. For a 10% value, the uncertainty on the $M_{\mathrm{eff}}$ distribution ranges from $\sim 50-150\%$. An improved understanding of the JES to a level of 5% reduces the uncertainty on $M_{\mathrm{eff}}$ by more than a factor of two, while a 3% value shows an improved uncertainty on $M_{\mathrm{eff}}$ of between $\sim 10-30\%$. If the challenging goal of a 1% JES uncertainty is achievable, the results indicate that the uncertainty on high-$p_{\mathrm{T}}$ events could be much reduced, to a level of $\sim 5-10\%$.

### 3.3 Generator comparison: PS vs. ALPGEN

The recent development of event generators in which regions of phase-space populated by a multi-parton matrix-element calculation are matched to those populated by a parton shower algorithm offer

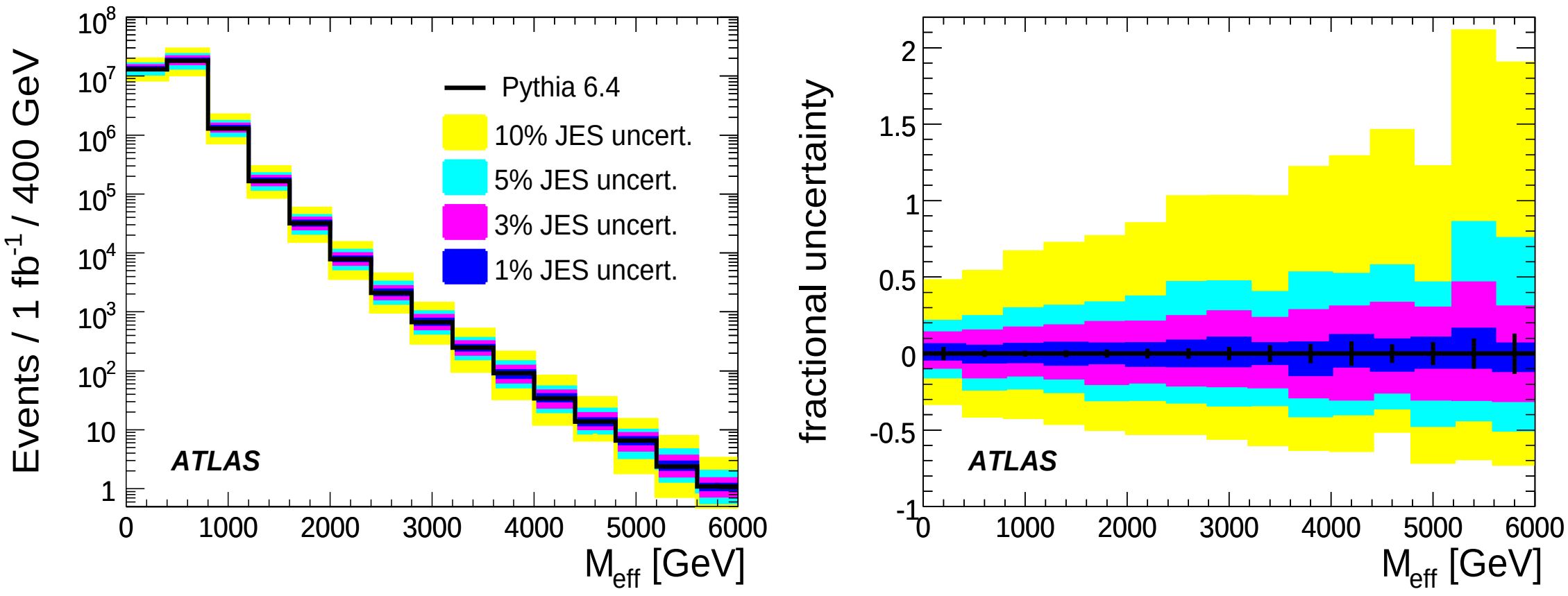


Figure 8: The $M_{\text{eff}}$ distribution for simulated QCD events satisfying SUSY jet cuts described in Section 1. The left plot shows the number of events per bin and the right plot shows the fractional uncertainty with respect to the central prediction. The shaded bands show the estimated uncertainty on the observable for assumed JES uncertainties of 10% (light), 5% (medium-light), 3% (medium) and 1% (dark).

the prospect of improved simulation of high $p_T$ QCD jet events. In this section we compare the physics performance of such a generator, namely the leading-order ALPGEN [27] code, with that of a conventional stand-alone parton-shower generator (PYTHIA 6.403 [7]).

Multijet QCD events were generated with ALPGEN 2.05+JIMMY and with PYTHIA 6.403 and the resulting distributions of observables sensitive to SUSY signal events compared using fast simulation [15]. PYTHIA events were generated in different $p_T$ ranges of the two leading partons to study the high energy tails with sufficient statistics (see Ref. [8]). Such sliced-sample production is also possible in ALPGEN. Only events satisfying the standard SUSY jet cuts described in Section 1 were studied. The number of PYTHIA events passing the jet selection cuts is 2.1 times larger than the number of ALPGEN events for the same integrated luminosity[2]. To facilitate comparison of the two samples, they were both normalized to 1 fb$^{-1}$ and ALPGEN samples were further multiplied by a factor of 2.1. Error bars on each histogram below are based on the Monte Carlo statistics used in this study.

Figure 9 shows the $p_T$ distribution of the leading four jets for ALPGEN and PYTHIA events. The new parton shower scheme available in PYTHIA 6.403 can produce hard subleading jets with similar $p_T$ to those produced by ALPGEN while the two leading jets are somewhat softer. Even though PYTHIA also generates $2 \rightarrow 2$ scattering using matrix-element calculations, the two highest-$p_T$ jets are softer because additional partons are emitted from the leading partons using splitting functions.

The use of fast detector simulation for the comparison of event generators described in this section prevents accurate study of the impact of generator differences on 'fake' $E_{\text{T}}^{\text{miss}}$ distributions. Nevertheless, 'real' $E_{\text{T}}^{\text{miss}}$ events in which heavy quark decays generate neutrinos can potentially dominate in the tail of the $E_{\text{T}}^{\text{miss}}$ distribution following application of the cuts described in Section 2 and Ref. [3]. Fast detector simulation can legitimately be used to study this background permitting comparison of PYTHIA and ALPGEN predictions of $E_{\text{T}}^{\text{miss}}$ and $M_{\text{eff}}$ distributions. Figure 10(a) shows the $E_{\text{T}}^{\text{miss}}$ distributions obtained from PYTHIA and ALPGEN while Figure 10(a) shows the relative difference between the distributions of the $E_{\text{T}}^{\text{miss}}$ .

[2]No $b\bar{b}$ or $c\bar{c}$ pair production events where the heavy quark pair is produced in the hard scatter rather than through gluon splitting were generated in PYTHIA. This contribution is 10 % of the total $b\bar{b}$ cross-section and negligible compared to the total multi-jet cross-section.

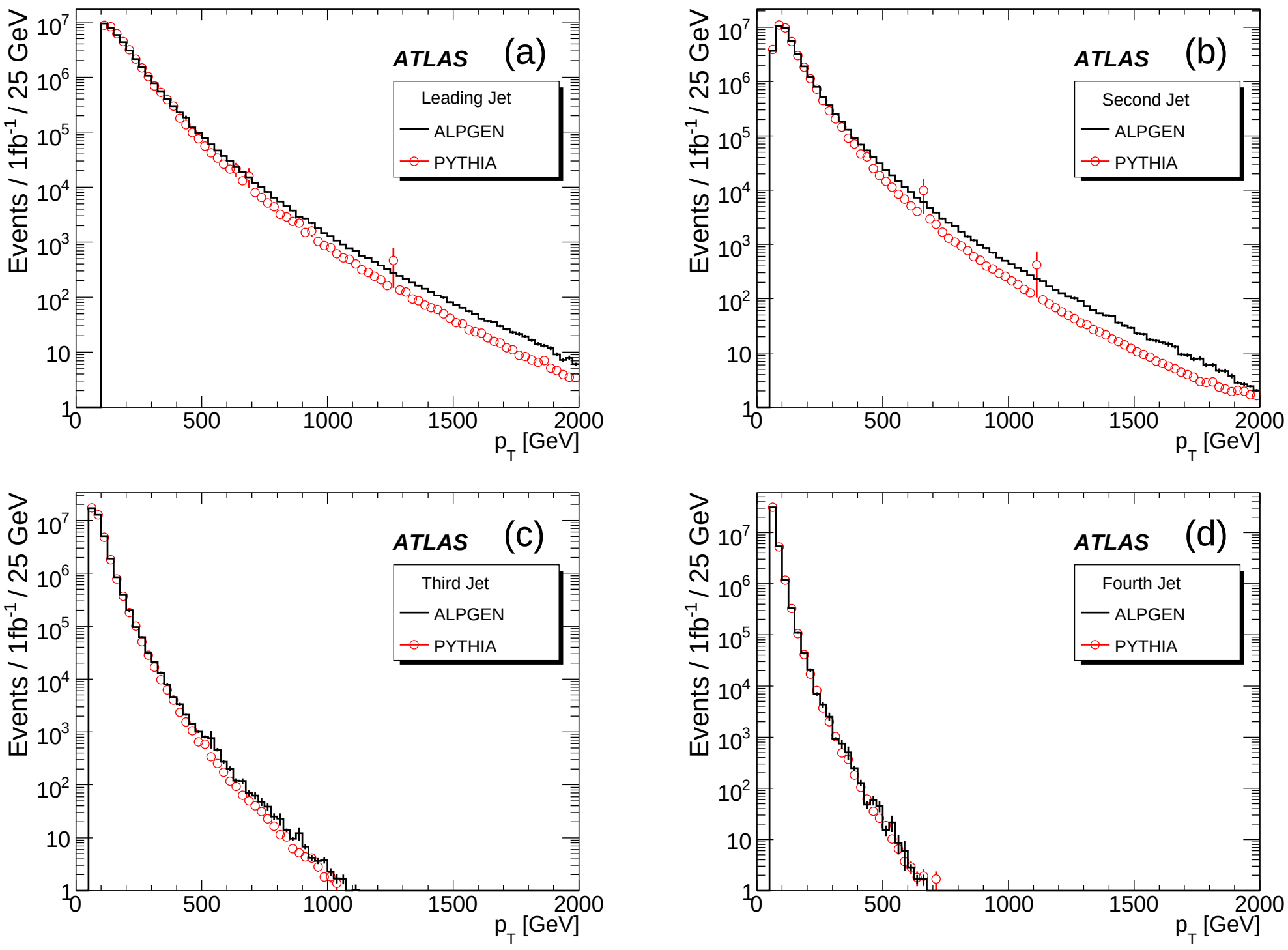


Figure 9: The $p_T$ distributions of the four highest-$p_T$ jets; (a) the leading, (b) the second (c) the third and (d) the fourth jet for PYTHIA (open circles) and ALPGEN (open histogram) events.

PYTHIA predicts somewhat larger $E_{\mathrm{T}}^{\mathrm{miss}}$ than ALPGEN, although limited Monte Carlo statistics prevent firm conclusions from being drawn at large $E_{\mathrm{T}}^{\mathrm{miss}}$ . This effect likely originates from the fact that multi-jet events containing gluon splitting are not fully included in ALPGEN because of the requirements of matrix element – parton shower matching. [3] The equivalent $M_{\mathrm{eff}}$ distributions and the relative differences between generators are shown in Figure 11. The $M_{\mathrm{eff}}$ distribution of ALPGEN events is harder than that of PYTHIA events, as expected from Figure 9(a) and (b).

# 4 Monte Carlo estimates

## 4.1 Jet smearing with Transfer Function

**Introduction** In this section, a Monte Carlo based detector simulation technique is described which models the response function of the ATLAS calorimeter to jets as a function of energy and $\eta$ using Monte Carlo "truth" information, and then uses this response function to smear the energies of jets and other objects in Monte Carlo QCD jet events. The truth-derived response function used here is referred to as the particle-jet transfer function (PJTF) and in principle may be measured also from data using techniques such as those described in Section 5.1.

[3] To avoid double counting of partons generated by the matrix element calculation and parton shower algorithm, the MLM matching scheme employed by ALPGEN requires that matrix element partons satisfy cuts on $p_T$ (>40 GeV) and $\Delta R$ between partons ($>0.7$). As a result of these requirements, processes such as $ggbb$ generating multiple partons with $\Delta R_{bb} < 0.7$ are rejected during matching. Events with gluon splitting generated in the parton shower algorithm are however kept.

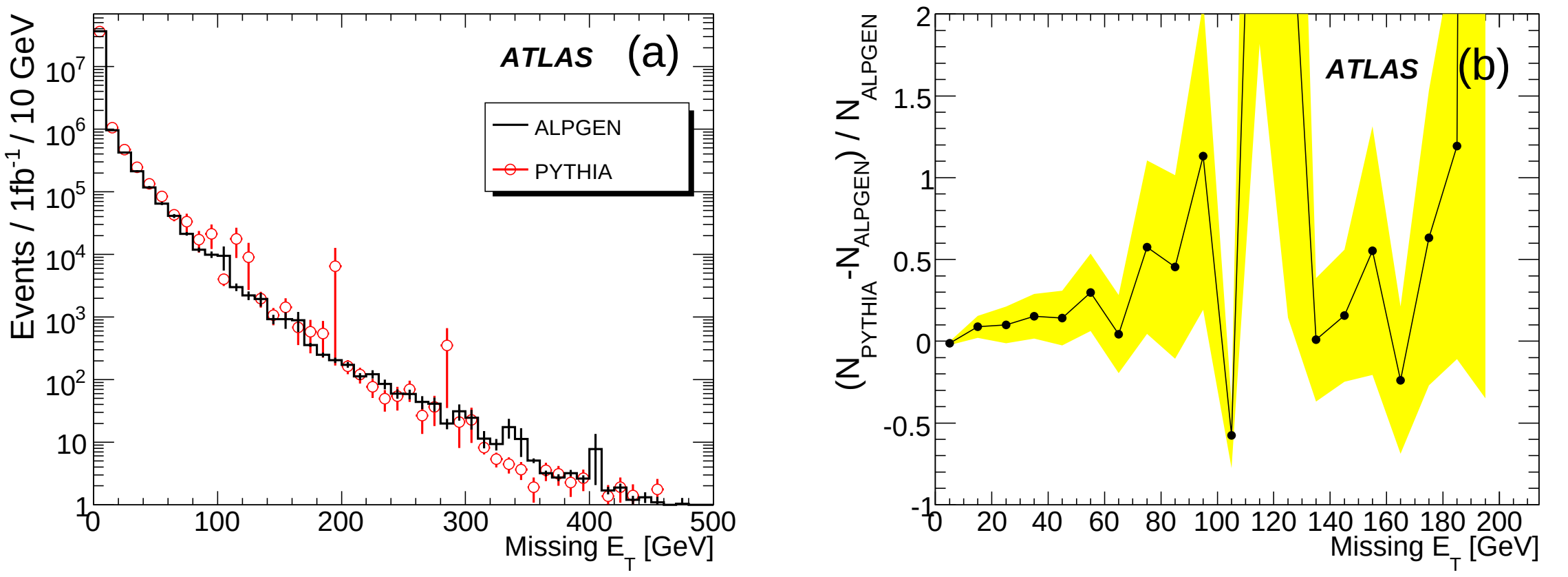


Figure 10: (a) ATLFAST $E_{\mathrm{T}}^{\mathrm{miss}}$ distribution for PYTHIA(circles) and ALPGEN (histogram) and (b) the relative difference ($\mathrm{N}_{\mathtt{PYTHIA}}$-$\mathrm{N}_{\mathtt{ALPGEN}}$)/$\mathrm{N}_{\mathtt{ALPGEN}}$ of the $E_{\mathrm{T}}^{\mathrm{miss}}$ distributions. The shaded band is the Monte Carlo statistical error.

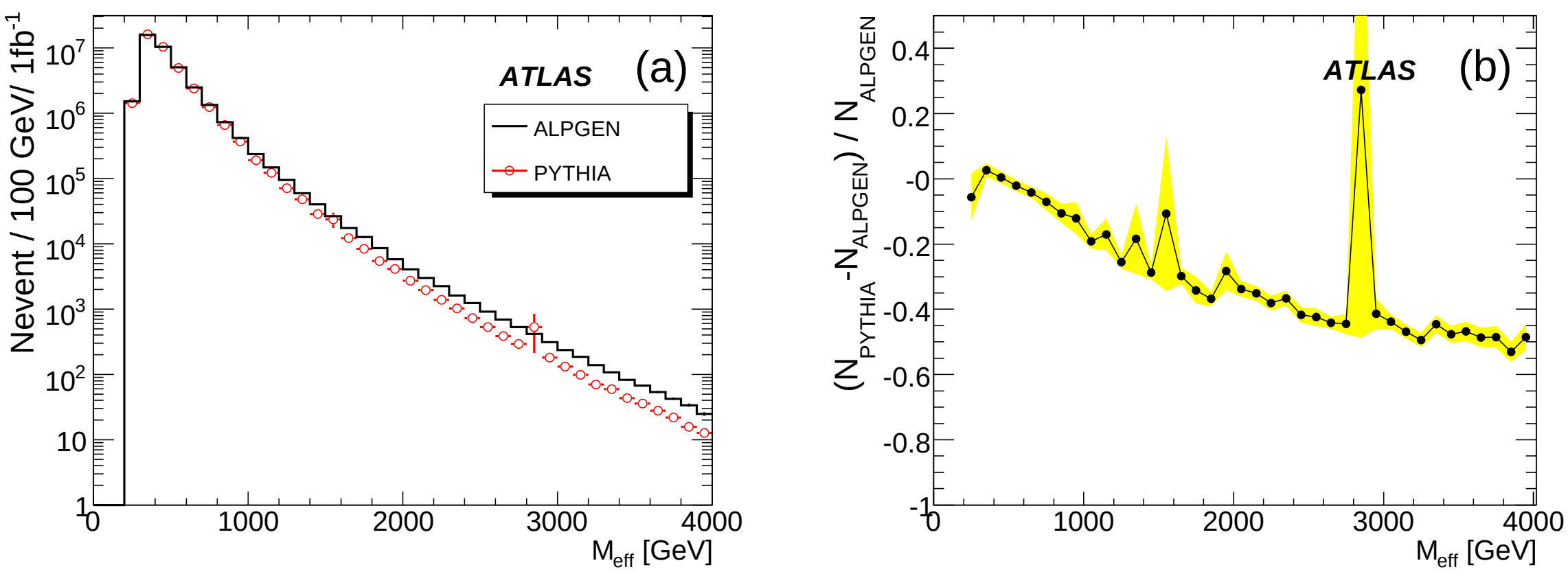


Figure 11: (a) ATLFAST $M_{\mathrm{eff}}$ distribution for PYTHIA(circles) and ALPGEN (histogram) and (b) the relative difference ($\mathrm{N}_{\mathtt{PYTHIA}}$-$\mathrm{N}_{\mathtt{ALPGEN}}$)/$\mathrm{N}_{\mathtt{ALPGEN}}$ of the $M_{\mathrm{eff}}$ distributions. The shaded band is the Monte Carlo statistical error.

**Modeling jets – reconstruction of the PJTF** In this section we use the following terminology for two different types of jets formed from Monte Carlo events – “particle jets”, made from all the generated particles in the event except muons and neutrinos, and “calorimeter jets” formed from topological clusters in the calorimeter.

In general there is a high efficiency for reconstructing jets in the calorimeter, even though their energy can potentially be mis-measured by a large amount. Jet fragmentation to particles of different types and the response of the detector material to these particles in different $\eta$ regions can produce tails in the PJTF. It is possible for there not to be a one-to-one match between generator-level jets and calorimeter jets even though the same jet clustering algorithms are used.

To reconstruct the PJTF it is necessary to measure the response of the calorimeter to generator-level jets. This is accomplished by matching these to calorimeter jets, and requiring a good jet isolation in both jets (separation from nearest jet $\Delta R > 0.8$), to ensure that accidental jet overlaps do not contribute

to the tails in the PJTF. Jet energy responses are modeled with double Gaussians. The ratio of the 1st and 2nd Gaussian components is empirically known to follow the ratio of energy deposits in the electromagnetic and hadron calorimeter and hence the 1st/2nd Gaussian ratio is fixed to the mean value of $E_{\mathrm{had}}/(E_{\mathrm{em}}+E_{\mathrm{had}})$ with a fitting procedure. GEANT4 simulated PYTHIA dijet samples with $p_T > 17$ GeV [8] are used to estimate the PJTF. Reconstructed jet energies are corrected with an energy-density based weighting scheme [28]. The fits are performed before the jet energy corrections have been applied, i.e. the PJTF is obtained at the calorimeter cell energy calibration level. To use the PJTF on reconstructed jets, cell-to-jet energy corrections are applied after the PJTF. For the $E_{\mathrm{T}}^{\mathrm{miss}}$ calculation the cell-level correction is used directly to retain consistency with full GEANT4 simulation $E_{\mathrm{T}}^{\mathrm{miss}}$ .

**Modeling $E_{\mathrm{T}}^{\mathrm{miss}}$** All truth jets (formed from stable truth particles expected to shower in the calorimeter) with energy greater than 10 GeV are smeared with the PJTF. This avoids double counting of interacting particles (electron, photon, tau) due to the potentially different treatment of merging and splitting of objects, for example when the photon is close to a jet. This simplification of reconstructing jets from all stable showering particles mis-measures the energy scale of isolated electrons, photons and taus by a small amount, however the ATLAS procedure of identifying and calibrating electromagnetic and hadronic showers separately ensures that the resulting bias is minimised. This study is additionally focussed on QCD jet events and hence any such bias is still less important. Muons in the event are simulated with the ATLFAST fast simulation code [15] due to the prohibitive time overhead associated with full simulation of the ATLAS tracking systems. Finally in order to account for the potentially large fluctuation in the underlying event, we vectorially sum all the particles that are not part of jets above 10 GeV to form a "soft jet". This soft jet is added to the above jet sum to compensate for the contribution from soft particles outside of the jet cones. Once all the objects in the event are defined, the $E_{\mathrm{T}}^{\mathrm{miss}}$ in the event is calculated by summing their four-vectors, enabling comparison of the performance of the PJTF technique with full GEANT4 simulation.

**Performance** The performance of the PJTF technique was assessed by comparing PJTF predictions with those obtained from full GEANT4 simulation using the $p_T > 17$ GeV dijet samples described above. The event-by-event comparison was performed using both ATLFAST and fully reconstructed objects in the GEANT4 simulation samples.

The jet multiplicity difference between full GEANT4 simulation and PJTF, and between full GEANT4 simulation and ATLFAST are shown in Fig. 12. Only jets with $p_T > 50$ GeV and $|\eta| < 2.5$ were used, but no event selection based on the jet multiplicity was performed.

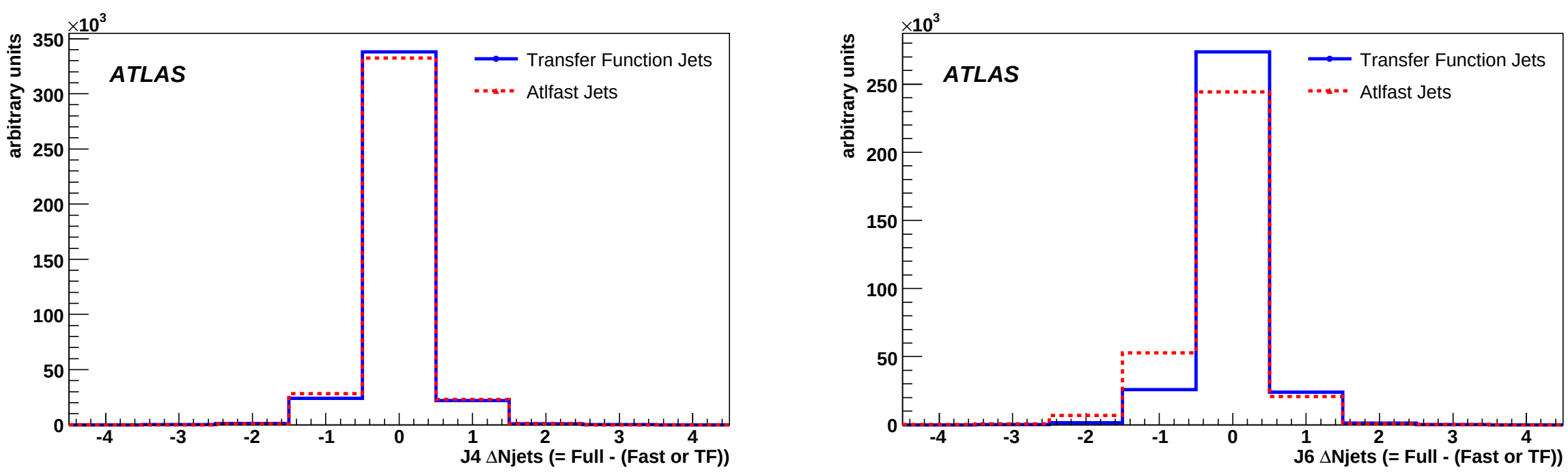


Figure 12: Event by event difference in number of jets between full GEANT4 simulation and PJTF (blue,solid) and between full GEANT4 simulation and ATLFAST (red,dashed). The left plot is for 140 GeV $< p_T <$ 280 GeV events, and the right one is for 560 GeV $< p_T <$ 1120 GeV events.

There is good agreement in jet multiplicity between the full GEANT4 simulation and the PJTF events. The ATLFAST events have a somewhat higher multiplicity due to the simplified calorimeter map used for jet reconstruction. For the PJTF technique to be effective, it is important that the multiplicity at reconstruction level is not biased. This is clearly demonstrated in Fig. 12. Good agreement was found for jet $p_T > 40$ GeV, which can be explained as follows. The topological clustering jet algorithm has a $\sim$100% efficiency so if the transfer function is correctly scaling and smearing the energy the jet multiplicity could only be biased by jets splitting and/or merging.

Another performance check is the comparison of the scalar $p_T$ sum of the four leading jets obtained from full GEANT4 simulation, PJTF, and ATLFAST events. Events passing the typical SUSY jet cuts listed in Section 1 are used. Fig. 13 shows the comparisons for 140 GeV $< p_T <$ 280 GeV and 560 GeV $< p_T <$ 1120 GeV events. Good agreement is observed between the full GEANT4 simulation and PJTF events, while ATLFAST events have significant deviations for the same reasons as for Fig. 12.

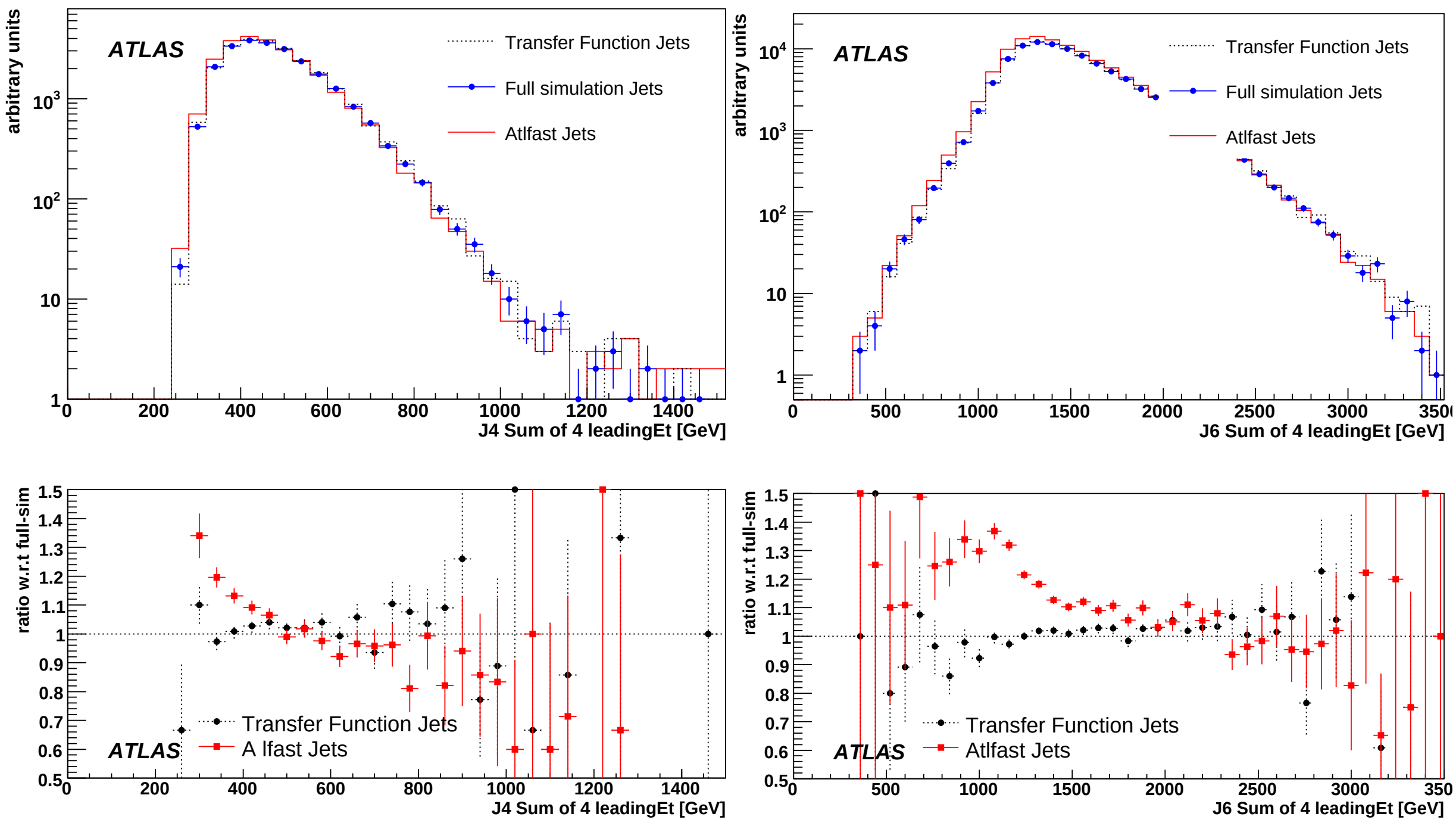


Figure 13: Top: distributions of the scalar $p_T$ sum of the four leading jets in PJTF (black, solid), ATLFAST (red, solid) and fully GEANT4 simulated jets (blue, points). The left plot is for 140 GeV $< p_T <$ 280 GeV events, and the right plot is for 560 GeV $< p_T <$ 1120 GeV. Bottom: the ratio with respect to the fully GEANT4 simulated jets.

In order to check the performance of $E_{\mathrm{T}}^{\mathrm{miss}}$ modeling the typical SUSY jet cuts described in Section 1 were applied, with no $E_{\mathrm{T}}^{\mathrm{miss}}$ cut applied initially. We define fake $E_{\mathrm{T}}^{\mathrm{miss}}$ as follows:

$$E_T^{\mathrm{miss,fake}} = |\mathbf{p}_T^{\mathrm{miss,reco}}| - |\mathbf{p}_T^{\mathrm{miss,true}}|$$

where $\mathbf{p}_T^{\mathrm{miss,reco}}$ is the missing transverse momentum vector reconstructed using either the PJTF technique, GEANT4 simulation, or ATLFAST, and $\mathbf{p}_T^{\mathrm{miss,true}}$ is the transverse vector of the sum of the neutrino momenta.

Fig. 14 shows the fake $E_{\mathrm{T}}^{\mathrm{miss}}$ distribution obtained from the PJTF technique and from full GEANT4 simulation. Also shown are the ATLFAST results, and the PJTF results without the soft-jet corrections.

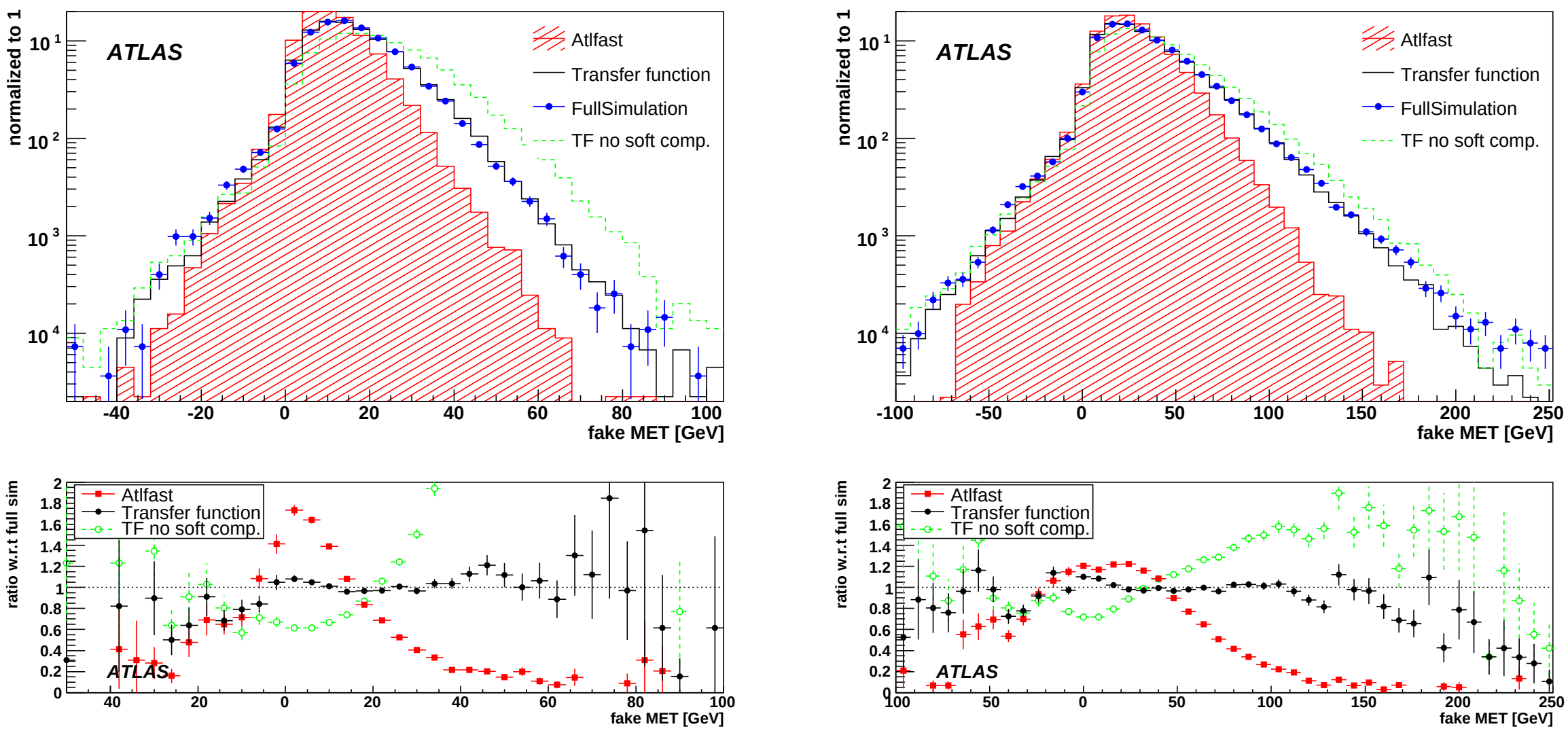


Figure 14: Fake $E_{\mathrm{T}}^{\mathrm{miss}}$ distributions from ATLFAST (red hatched), PJTF technique (black, solid) and full GEANT4 simulation (blue, points). Also shown with the dashed green line is the PJTF distribution without soft-jet correction. The left plot is for 140 GeV $< p_T <$ 280 GeV events, and right plot is for 560 GeV $< p_T <$ 1120 GeV. Bottom plots are the ratio of each distribution with respect to the full GEANT4 simulation.

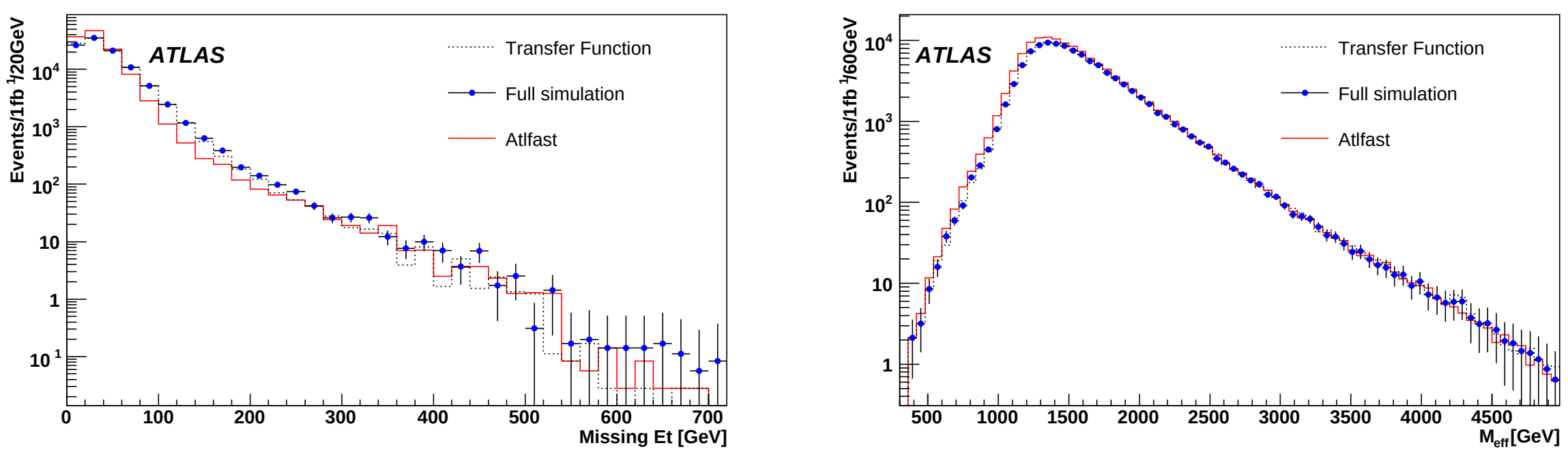


Figure 15: $E_{\mathrm{T}}^{\mathrm{miss}}$ (left) and $M_{\mathrm{eff}}$ (right) distributions of 560 $< p_{\mathrm{T}} <$ 2240 GeV PYTHIA QCD jet events simulated with ATLFAST (light/red solid), the PJTF technique (dark/black, solid) and full GEANT4 simulation (points). Events were required to pass the SUSY jet cuts described in Section 1.

Table 2: Numbers of $560 < p_T < 2240$ GeV PYTHIA QCD jet events passing cuts for 1 fb$^{-1}$ data, starting from an original total of $3.5 \times 10^5$ events. Errors are also normalized to 1 fb$^{-1}$. The cut definitions are described in the text.

| Method | Cut 1 | Cut 2 | Cut 3 |
|---|---|---|---|
| Full simulation | $(103.8 \pm 0.3) \times 10^3$ | $(65.7 \pm 0.3) \times 10^3$ | $4.3 \pm 2.1$ |
| PJTF technique | $(104.9 \pm 0.3) \times 10^3$ | $(62.3 \pm 0.2) \times 10^3$ | $2.2 \pm 1.5$ |
| ATLFAST | $(120.0 \pm 0.3) \times 10^3$ | $(86.1 \pm 0.3) \times 10^3$ | $2.1 \pm 1.5$ |

It can be seen from the figure that $E_T^{miss}$ estimated with the PJTF technique and with full GEANT4 simulation agree at the 20% level for fake $E_T^{miss} < 80$ GeV (200 GeV) for 140–280 GeV (560–1120 GeV) dijet events. The figures highlight the improvement in performance relative to the naive ATLFAST estimate of $E_T^{miss}$, with the agreement between PJTF and full GEANT4 simulation persisting down into the tail region a factor $10^{-3}$ below the peak.

Fig. 15 shows the performance of the PJTF technique for estimating QCD backgrounds to the $E_T^{miss}$ and $M_{eff}$ distributions obtained following application of the SUSY jet cuts described in Section 1. In these plots the event-weights and errors are normalized to 1 fb$^{-1}$. Good agreement is seen between the PJTF distributions and the full simulation distributions over the full $p_T$ range, while the naive ATLFAST distributions show larger deviations from the full simulation results.

Additional cuts equivalent to the full SUSY cuts described in Section 1 were applied to assess further the consistency between the PJTF results and those obtained with full GEANT4 simulation. In Table 2 the remaining numbers of events after each stage in the event selection are shown. Cut 1 is the standard SUSY jet cut applied above. Cut 2 adds the requirement $\Delta\phi_{min} > 0.2$ (see Section 2.2) while Cut 3 requires additionally $E_T^{miss} > \max(100\text{ GeV}, 0.2M_{eff})$ and $M_{eff} > 800$ GeV. Despite limited statistics the results show that the PJTF approach provides reduced event selection bias in comparison with the naive ATLFAST simulation.

## 4.2 Fast GEANT4 simulation

**Introduction** The simulation of the ATLAS detector is currently based on GEANT4 v.4.8.3 [2]. GEANT4 simulation is very time consuming however and so its use for QCD background simulation at moderate or low $p_T$ is not viable. The fully GEANT4-based approach (refered to as full simulation in the following) is especially slow in the simulation of electromagnetic cascades in the calorimeters due to their complicated geometry. A fast simulation tool (ATLFAST [15]) based on Gaussian smearing is also available, and is used for studies of systematic uncertainties elsewhere in this note, but it does not reproduce with sufficient accuracy the tails of distributions interesting for SUSY analysis, i.e. $E_T^{miss}$ or $M_{eff}$. For this reason, an intermediate approach to the simulation of the ATLAS detector, based on the parameterisation of the calorimeters' response, is under study [29]. In the so-called 'fast' GEANT4 simulation, electrons entering the calorimeters may receive different treatments according to their energy. For high energy particles a shower parameterisation is used while for medium energy particles (below 1 GeV) the detector response is taken from a shower library ("frozen showers"), and low energy particles (below 10 MeV) are "killed" by depositing their energy in a single spot in the calorimeter.

**Performance** We tested fast GEANT4 simulation algorithms by simulating PYTHIA [7] dijet events with moderate transverse momentum (280 GeV $< p_T <$ 560 GeV). The fast GEANT4 simulation can be performed using different options, which differ in the treatment applied to electromagnetic particles (shower parameterization, frozen showers or "killing") in different parts of the calorimeters (for more details,

see [29]). In the following, the fast GEANT4 options are numbered from 1 to 3, the first one being the most conservative option, i.e., the closest to full simulation, and the third one the fastest option currently available. The measured fast GEANT4 simulation time is ∼60% lower than that required for full simulation. The three fast GEANT4 options differ among themselves by only a few percent of the full simulation time. To study the effect of the fast GEANT4 simulation algorithms in SUSY searches, we compared the distributions of typical quantities, i.e. the jet multiplicity and transverse momentum, $E_T^{miss}$ , transverse sphericity, and effective mass in 5000 fast GEANT4 simulated events with the ones obtained from full simulation (see Figs. 16, 17 and 18). We reconstructed jets with $p_T > 50$ GeV and $|\eta| < 2.5$. Due to the limited number of Monte Carlo events, no requirement was applied to the jet multiplicity, except in the $E_T^{miss}$ and $M_{eff}$ distributions, where the standard SUSY jet cuts described in Section 1 were applied. Transverse sphericity and effective mass quantities were calculated only for events with at least 2 jets.

Whilst the largest jet transverse momentum is well reproduced in the fast simulated samples, the average jet multiplicity is lower than that of the fully simulated sample by 1-2%. The most conservative sample (option 1) is in agreement with the full simulation distribution inside the statistical uncertainty. The width of the jet $p_T$ resolution is consistent across the samples, however the jet energy scale is too low by 1-2% (see Fig. 17). Due to the limited number of events available only a qualitative comparison is possible in regions of interest for SUSY searches for $E_T^{miss}$ (higher than 100 GeV) and transverse sphericity (larger than 0.2). Within the statistical precision possible the fast simulated distributions agree well with those obtained from full GEANT4 simulation. The size of the available samples preclude a full assessment of performance for the baseline SUSY cuts described in the introduction, since no events pass all cuts when using any of the three options. Similarly, from an equivalent number of GEANT4-simulated events, none pass all the cuts.

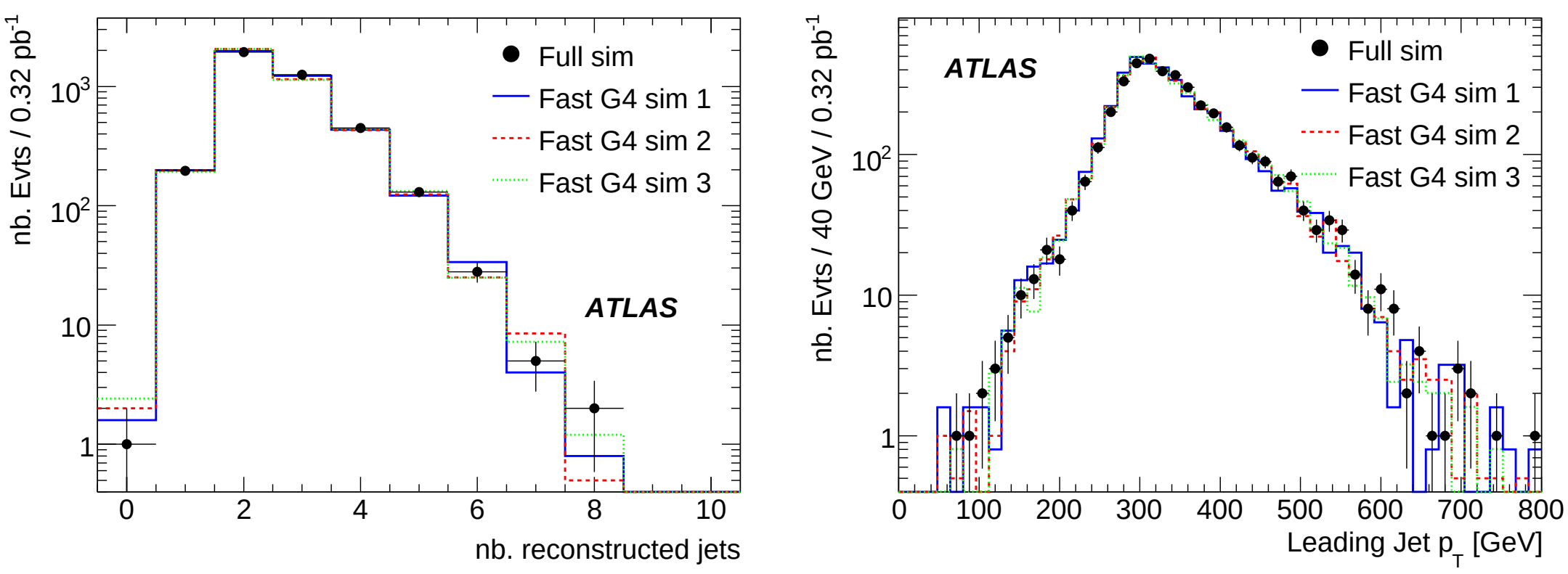


Figure 16: Reconstructed jet multiplicity (left) and highest jet $p_T$ (right) for PYTHIA dijet samples simulated with different options. The statistical uncertainty only is shown for the full simulation sample.

Overall, the fast GEANT4 simulation has been shown to be a promising faster alternative to the full GEANT4 based simulation. The reconstructed quantities in fast GEANT4 simulated samples are in good agreement with full simulation. Work is on-going to further reduce the simulation time, and to improve the modeling of the jet energy scale.

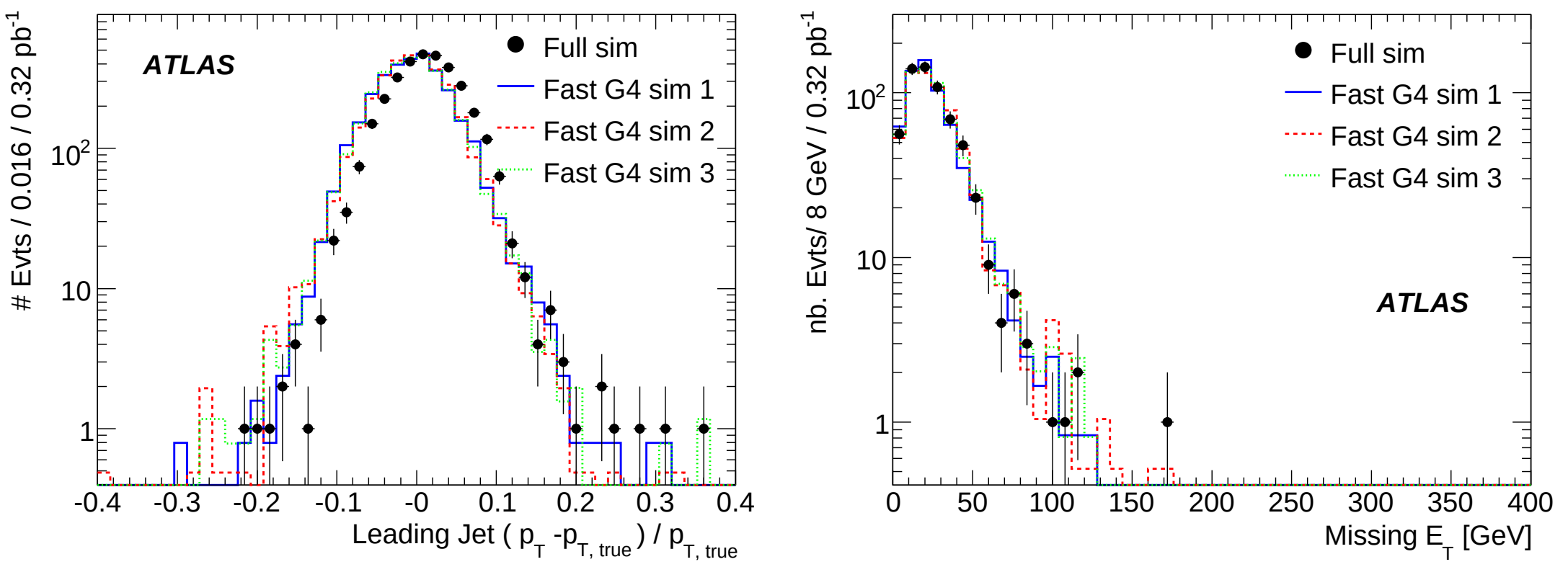


Figure 17: Leading jet $p_T$ resolution (left) and $E_{\mathrm{T}}^{\mathrm{miss}}$ (right) for PYTHIA dijet samples simulated with different options. The statistical uncertainty only is shown for the full simulation sample.

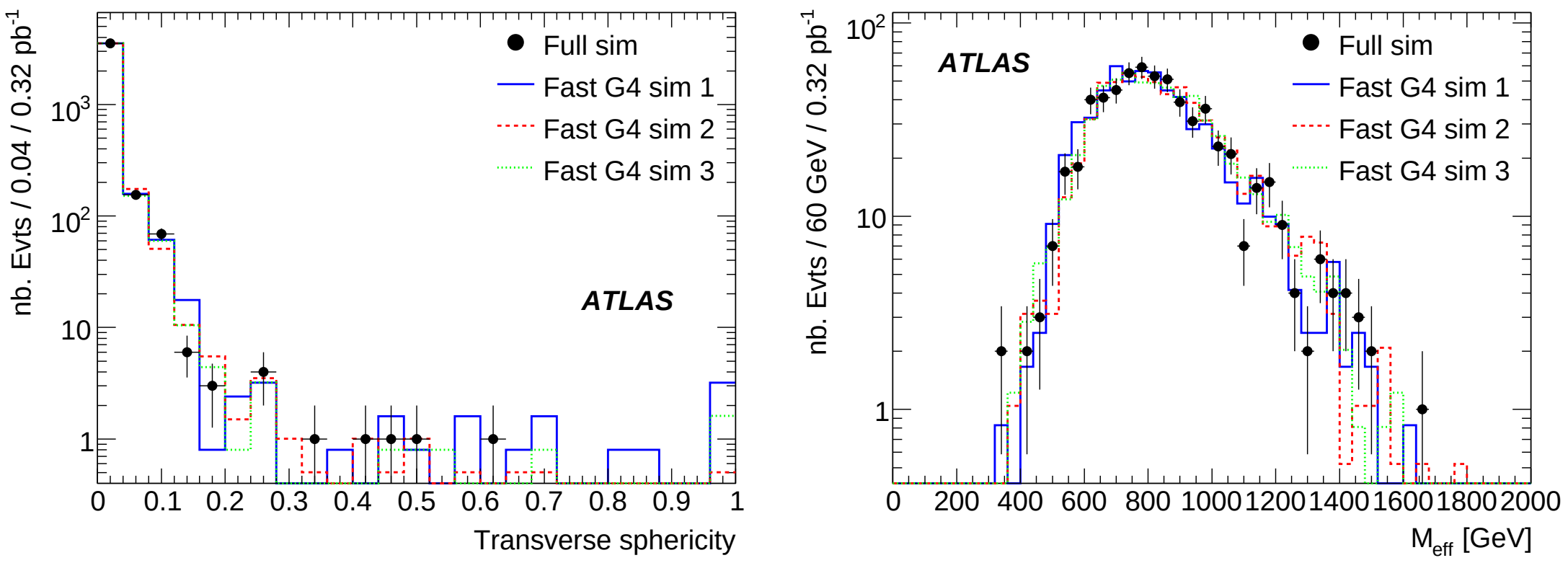


Figure 18: Transverse sphericity (left) and $M_{\mathrm{eff}}$ (right) for PYTHIA dijet samples simulated with different options. The statistical uncertainty only is shown for the full simulation sample.

# 5 Data-driven estimates

## 5.1 Jet smearing in the zero-lepton mode

**Introduction** The inherent systematic and statistical uncertainties in Monte Carlo based QCD background estimates will limit their use until sufficient data have been acquired to understand both the ATLAS detector and the underlying physics of QCD processes at 14 TeV. For this reason data-driven background estimates with minimal reliance on Monte Carlo simulation will be a priority for the early phase of data-taking.

The method described in this section reduces dependence on Monte Carlo simulation by smearing jet transverse momenta in low $E_{\mathrm{T}}^{\mathrm{miss}}$ QCD multijet data with a data-measured jet response function $R$ defined as the distribution of event-by-event ratios of measured jet $p_T$ to true jet $p_T$. This response function could potentially be used inter-changeably with the Monte Carlo-truth derived Particle Jet Transfer Function (PJTF) discussed in Section 4.1. Consequently the method described below provides a route to practical realisation of the PJTF technique for use with real data. The technique can be broken down into three distinct parts which are outlined below.

**Step 1: Gaussian response function measurement** The "$E_{\mathrm{T}}^{\mathrm{miss}}$ projection method" [30] applied to $\gamma$ + jet events allows a measurement of the Gaussian response $R^G$ of the ATLAS calorimeters to jets. The limited number of events prevent the measurement of the non-Gaussian tail with this technique (see Step 2). The method uses transverse momentum conservation to constrain the $p_T$ of all activity associated with a reconstructed jet with the $p_T$ of a hard photon. $R^G$ is obtained from the distribution of the photon-jet $p_T$ balance $R_1$:

$$R_1 = 1 + \frac{\mathbf{p}_T^{\mathrm{miss}} \cdot \mathbf{p}_T(\gamma)}{|\mathbf{p}_T(\gamma)|^2}, \tag{4}$$

where $\mathbf{p}_T(\gamma)$ is the photon transverse momentum, and $\mathbf{p}_T^{\mathrm{miss}}$ is the missing-transverse-momentum two-vector.

A GEANT4-simulated PYTHIA [7] sample equivalent to $23.8\,\mathrm{pb}^{-1}$ of $\gamma$ + jets events was used, with events required to pass a 60 GeV single-photon trigger [9]. With the additional requirement that events should contain one and only one reconstructed jet, the jet response was measured with the distribution of $R_1$ values defined above. The $R_1$ distribution was measured in a number of photon $p_T$ bins, and each was fitted with a Gaussian function. The standard deviations from these fits to the response distributions are shown in Fig. 19 with the statistical error bars derived from the fit errors. The $p_T$ dependence of the widths of the $R_1$ distributions was fitted with the parametric form

$$\sigma_R = A + \frac{B}{\sqrt{p_T}} + \frac{C}{p_T}\,, \tag{5}$$

shown in the figure to describe well the GEANT4 simulated data. The sampling ($A$) and stochastic ($B$) terms in the above formula were used to calculate the width of $R^G$ as a function of jet $p_T$, while the constant term ($C$) was used to estimate the additional smearing of $E_{\mathrm{T}}^{\mathrm{miss}}$ not associated with jets (see Step 3 below).

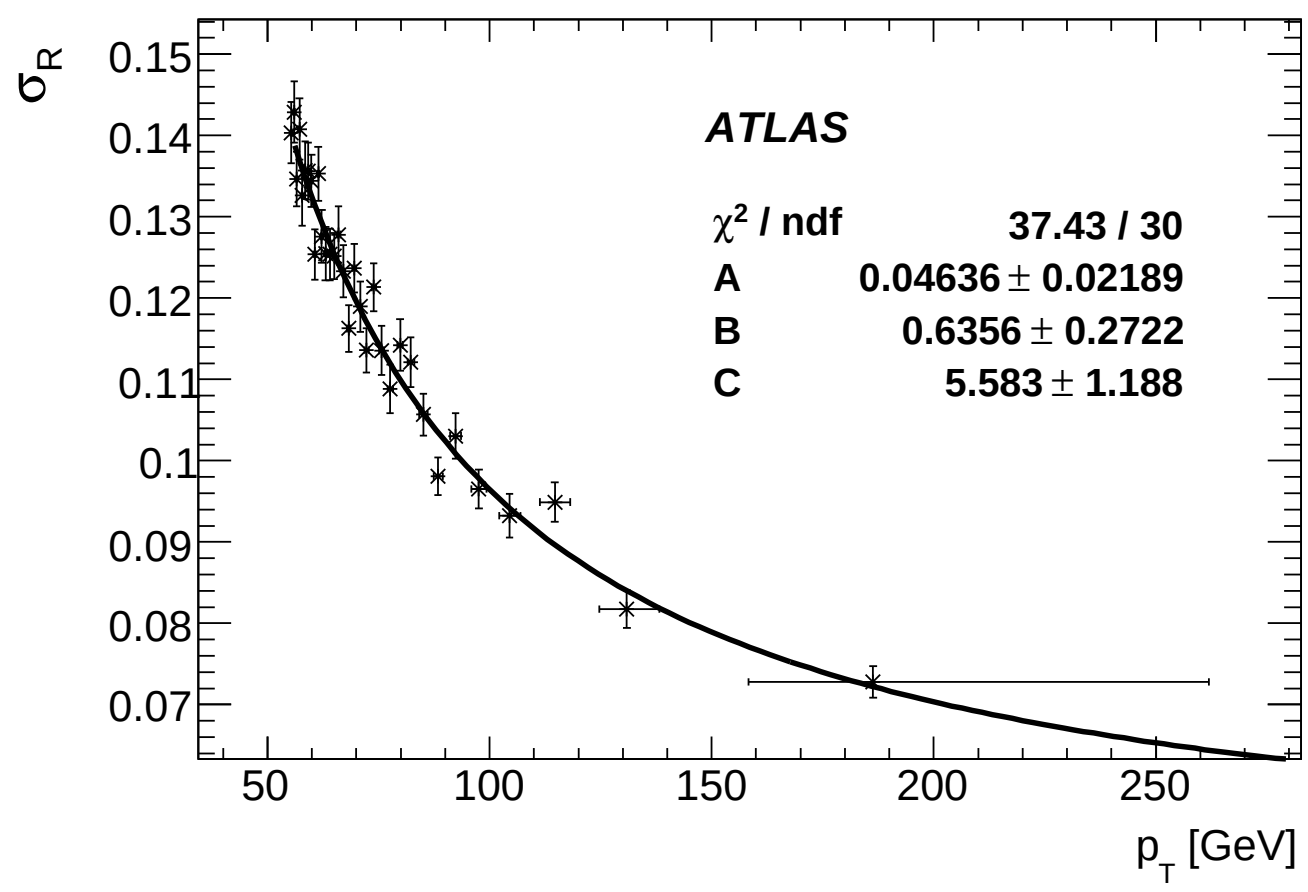


Figure 19: Standard deviations of Gaussian fits to measured response distributions vs. $p_T^{\gamma}$. The fit is of the functional form $\sigma_R(p_T) = A + B{p_T}^{-1/2} + C{p_T}^{-1}$.

**Step 2: Full (Gaussian + non-Gaussian) response function measurement** In order to reproduce the tail of the QCD multijet $E_{\mathrm{T}}^{\mathrm{miss}}$ distribution it is necessary to characterise the non-Gaussian response of the calorimeters to jets. The next stage in the technique therefore uses events in which the $E_{\mathrm{T}}^{\mathrm{miss}}$ vector can be unambiguously associated in $\phi$ with a single jet $J$ to measure that response. The response of the

ATLAS calorimeters to jet $J$ if its $p_T$ lies in the non-Gaussian tail can be obtained from

$$R_2 = \frac{\mathbf{p}_T(J) \cdot \mathbf{p}_T(J,\mathrm{true})}{|\mathbf{p}_T(J,\mathrm{true})|^2} , \tag{6}$$

where $\mathbf{p}_T(J)$ denotes the reconstructed transverse momentum of jet $J$. If it is assumed that the $E_{\mathrm{T}}^{\mathrm{miss}}$ in these events is dominated by fluctuations in $\mathbf{p}_T(J)$ then $\mathbf{p}_T(J,\mathrm{true})$ can be approximated by

$$\mathbf{p}_T(J,\mathrm{true}) \simeq \mathbf{p}_T(J) + \mathbf{p}_T^{\mathrm{miss}} . \tag{7}$$

The non-Gaussian response of ATLAS, $R^{NG}$, is measured with the distribution of event $R_2$ values.

QCD jet events from the sample described in Section 2.1 were required to pass one of the high-$p_T$ multijet or $E_{\mathrm{T}}^{\mathrm{miss}}$ triggers [9], and required to contain at least three jets with $p_T > 250$, 50 and 25 GeV, respectively. Events were also required to possess $E_{\mathrm{T}}^{\mathrm{miss}} > 60$ GeV. Events were further required to contain a jet unambiguously associated with the $E_{\mathrm{T}}^{\mathrm{miss}}$ vector - i.e. events with one and only one jet parallel or anti-parallel to the $E_{\mathrm{T}}^{\mathrm{miss}}$ vector. Of the remaining jets in each event, the two with the highest $p_T$ were required to have $p_T > 250$, 50 GeV. In this way, the selected events predominantly had a 'Mercedes' type configuration, with $E_{\mathrm{T}}^{\mathrm{miss}}$ parallel or anti-parallel to the $p_T$ of one of the jets. The $p_T$ of this jet was used in Eqn. 6 to measure $R_2$. The estimate of the jet response non-Gaussian tail $R^{NG}$ obtained from the distribution of $R_2$ values is plotted as the histogram in Fig. 20(right) with a normalisation obtained with the procedure described below. Although not considered in detail here, the $W(\rightarrow \tau\nu) + 2$ jet background could have been removed with a cut on the multiplicity of inner detector tracks per jet.

Next we combine the Gaussian and non-Gaussian components of the jet response $R$ measured previously. The relative normalisation of the Gaussian ($R^G$) and non-Gaussian ($R^{NG}$) components can be measured using the balance of transverse momenta of jets in dijet events. The relative normalisation can be obtained from the ratio of the numbers of dijet events with respectively one and zero jets with $p_T$ lying in the non-Gaussian tail of the jet response function. We therefore need to define a variable which can classify dijet events according to the number of jets with $p_T$ lying in the non-Gaussian tail. A suitable variable is provided by $R_3(j)$, defined as the projection of the transverse momentum of each jet $j'$ in the event onto the event $E_{\mathrm{T}}^{\mathrm{miss}}$ :

$$R_3(j) = 1 + \frac{\mathbf{p}_T^{\mathrm{miss}} \cdot \mathbf{p}_T(j')}{|\mathbf{p}_T(j')|^2}, \tag{8}$$

which measures the response to the other jet $j$ in the event (cf. Eqn. 4). If $R_3(j)$ lies below some threshold then jet $j$ can be considered to lie in the non-Gaussian tail.

Events were selected from the QCD jet event sample used above with the requirement that they passed a 160 GeV single-jet trigger requirement [9] and contained two and only two jets, back-to-back in the transverse plane. The distribution of $R_3(j)$ for these events is plotted as data-points in Fig. 20(right). The ratio of the integral of the low tail of the $R_3(j)$ distribution to the integral of a Gaussian function fitted to the peak is used to obtain the relative normalisation of the non-Gaussian and Gaussian components of the jet response function. The full jet response function including normalised Gaussian and non-Gaussian components is plotted in Fig. 20 (left).

As a 'closure test' of the reconstruction of the full response function, jets in a subset of the selected dijet events with low $E_{\mathrm{T}}^{\mathrm{miss}}$ -significance (see Section 2.1) were smeared to reproduce the $R_3(j)$ distribution. The resulting distribution is shown in Fig. 20 (right) for comparison with 'data'.

**Step 3: Seed event selection and jet $p_T$ smearing** In order to estimate the $E_{\mathrm{T}}^{\mathrm{miss}}$ and $M_{\mathrm{eff}}$ distributions of QCD multijet events the jet response function $R$ measured in Step 2 was used to smear jet transverse momenta in multijet events with low $E_{\mathrm{T}}^{\mathrm{miss}}$ (referred to below as 'seed events'). Seed events were

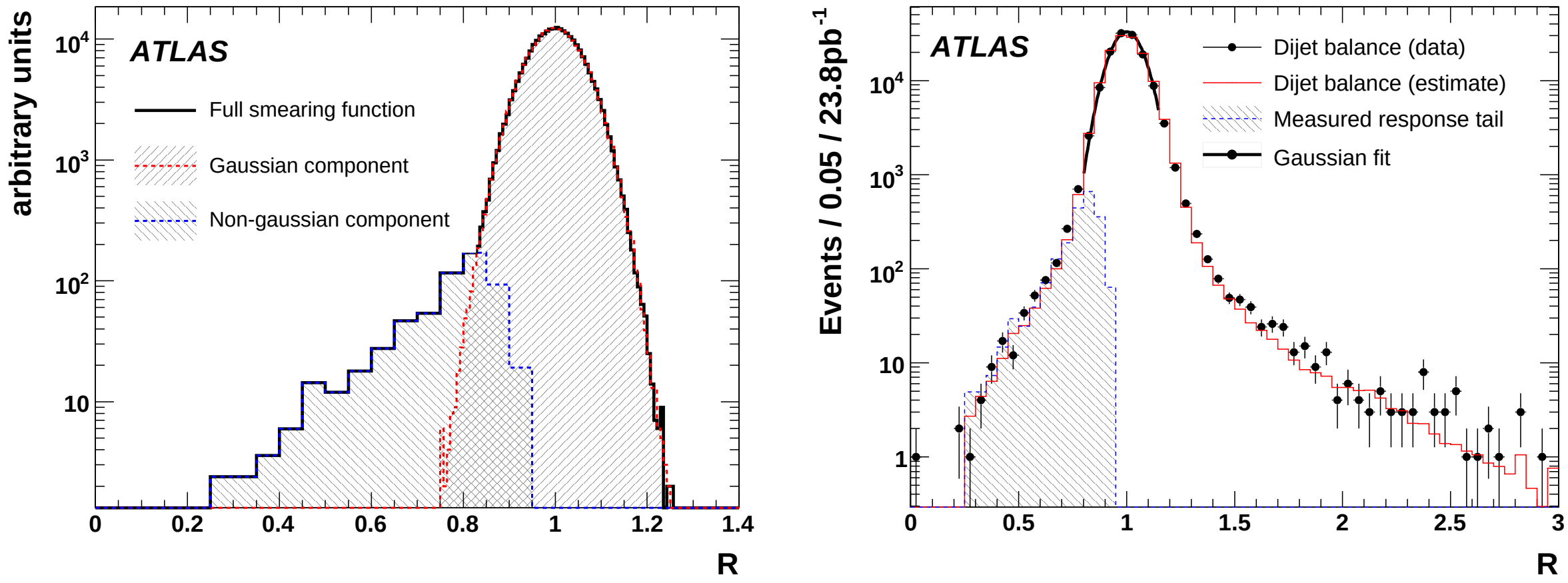


Figure 20: Left – smearing function for a jet of 250 GeV (thick line), with Gaussian and non-Gaussian components (right and left facing hatches respectively) shown separately. Right – dijet balance distribution (points) compared with the equivalent estimated distribution obtained from the jet response function to provide a 'closure test' of the technique. Also shown are a Gaussian fit to the region $0.8 < R_3(j) < 1.15$ (thick line), and the non-Gaussian tail distribution (dashed histogram) measured with 'Mercedes' events normalised to the tail of the dijet balance distribution.

selected from the same PYTHIA jet sample used in Step 2. Seed event candidates were required to pass one of the ATLAS high-$p_T$ jet triggers [9] together with the standard SUSY jet selection cuts described in Section 1 with a reduced jet $p_T$ threshold ($p_T(j_i) > 45$ GeV for $i = 2,3,4$) in order to avoid biasing the final estimate after smearing. Both the $E_{\mathrm{T}}^{\mathrm{miss}}$ significance (see Section 2.1) and the equivalent quantity constructed from $E_{\mathrm{T}}^{\mathrm{miss}}$ derived only from in-cone jet energy were required to be less than 0.5 GeV$^{1/2}$ and 0.7 GeV$^{1/2}$, respectively. The latter of these two cuts ensures that the hadronic activity in the selected seed events is well contained in the jets that will be smeared.

Smeared events were constructed from each selected seed event by smearing the transverse momenta of their constituent jets with the jet response function $R$ determined in Step 2. The $E_{\mathrm{T}}^{\mathrm{miss}}$ of smeared events was calculated by replacing the contribution to the $E_{\mathrm{T}}^{\mathrm{miss}}$ from the $p_T$ of seed jets with a contribution from the $p_T$ of the equivalent smeared jets. An additional contribution $\mathbf{p}_{T,\mathrm{C}}^{\mathrm{miss}}$ to $\mathbf{p}_T^{\mathrm{miss}}$ generated by the constant term measured in Step 1 was also taken into account. The smeared $E_{\mathrm{T}}^{\mathrm{miss}}$ was therefore given by the magnitude of

$$\mathbf{p}_T^{\mathrm{miss}\prime} = \mathbf{p}_{T,\mathrm{C}}^{\mathrm{miss}} - \sum_i \mathbf{p}'_T(j_i) + \sum_i \mathbf{p}_T(j_i), \tag{9}$$

where primed quantities are smeared.

Fig. 21 shows the $E_{\mathrm{T}}^{\mathrm{miss}}$ and $M_{\mathrm{eff}}$ distributions for 23.8pb$^{-1}$ of GEANT4 simulated 'data' and smeared seed events passing the SUSY jet cuts listed in Section 1. The estimate was normalised to GEANT4 simulated PYTHIA QCD jet events ('data') with $E_{\mathrm{T}}^{\mathrm{miss}} < 50$ GeV and somewhat tighter jet cuts ($p_T(j_4) >$ 60 GeV) imposed to ensure a reasonable sample of 'data' events obtained with the multi-jet triggers [9].

The dominant sources of systematic uncertainty in the estimate were $p_T$ bias in the selection of events and finite statistics in the non-Gaussian tail measurement. Uncertainties on the $A$, $B$, and $C$ parameters in Eqn. 5 and the relative normalisation of the Gaussian and non-Gaussian parts of the jet response function were also considered.

Good agreement can be seen between the estimated and GEANT4 'data' $E_{\mathrm{T}}^{\mathrm{miss}}$ and $M_{\mathrm{eff}}$ distributions. Applying the full SUSY cuts described in Section 1, $2.36 \pm 0.09\,(\mathrm{stat}) \pm 1.44\,(\mathrm{syst})$ events are estimated versus 1 'observed'. The uncertainty in the estimate is therefore $\sim 60\%$ for 23.8 pb$^{-1}$. If these figures

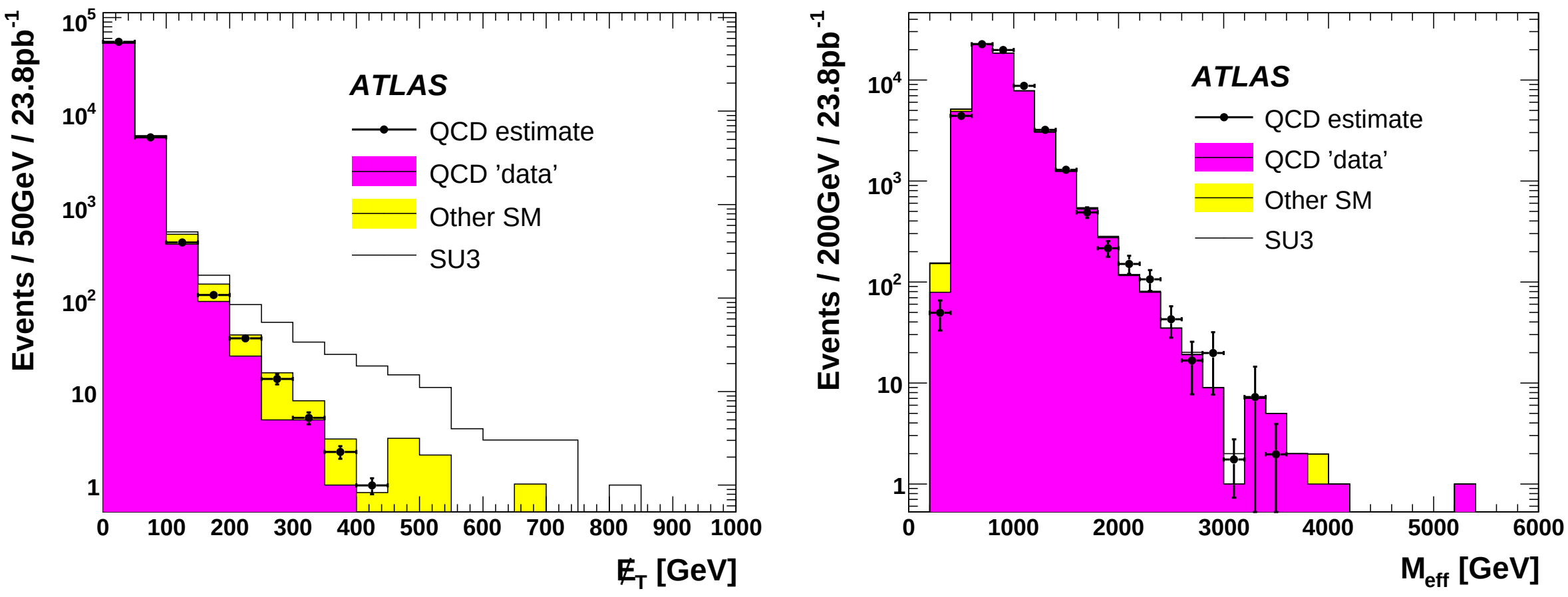


Figure 21: $E_T^{\rm miss}$ (left) and $M_{\rm eff}$ (right) distributions for smeared events and GEANT4 'data' passing 0-lepton SUSY jet cuts. Also included for comparison are 23.8 pb$^{-1}$ of SUSY (SU3) and the summed contribution from $Z \rightarrow \nu\bar{\nu}$ + jets, $W \rightarrow \ell\nu$ + jets and $t\bar{t}$ + jets.

are extrapolated to 1 fb$^{-1}$ then the equivalent calculated uncertainties are $\sim$ 0.6% (stat.) and 12.6% (syst.), with the systematic uncertainty reduced through access to increased statistics at step 2. However, residual Standard Model (e.g. $W(\rightarrow \tau\nu)+2$ jets) contamination of the 'Mercedes' control sample has not been included in these uncertainty estimates and therefore we conservatively assume the same 60% systematic uncertainty for 1 fb$^{-1}$ of data. Systematic bias in the background estimate caused by the presence of SUSY signal is expected to be small – in particular the fractional contribution of SUSY events to the $E_T^{\rm miss}$ normalisation region is $\lesssim 10^{-4}$.

## 5.2 Lepton isolation in the one-lepton mode

**Introduction** Although QCD multijet background is not expected to be significant after application of SUSY signal cuts in the 1-lepton mode [5], it still must be estimated. This will be particularly important when preparing control samples for data-driven background estimates of other backgrounds; it is important to ensure that such samples (which will have looser cuts than the final SUSY selection) are not significantly contaminated by QCD background. The technique studied here is also relevant for QCD backgrounds in other lepton+jets channels, such as $W(\rightarrow l\nu)$+jets or $t\bar{t}$ in the semileptonic channel, although the kinematic cuts used in searches for Supersymmetry (e.g. lepton + $E_T^{\rm miss}$ transverse mass, $M_T$ > 100 GeV) might select different background mechanisms.

The method, which has been used at the Tevatron, relies on the (near) independence of $E_T^{\rm miss}$ and lepton isolation. The shape of the $E_T^{\rm miss}$ distribution is estimated from multijet events containing a non-isolated lepton. This distribution is then normalized to the number of events containing an isolated lepton, but with low $E_T^{\rm miss}$. The method, if valid, would be complementary to the jet-smearing technique described in Section 5.1 applied to one-lepton events, thereby allowing an important crosscheck of the systematic uncertainties of the two methods.

In evaluating this method, it is important to understand potential correlations between isolation and $E_T^{\rm miss}$. One possible mechanism can arise when there are multiple sources of leptons in QCD multijet events; if the sources each have a different $E_T^{\rm miss}$ shape, and the relative mixture of the sources changes with lepton isolation, this will appear as a correlation between $E_T^{\rm miss}$ and isolation. A possible example involves electrons arising from jet mis-identification and electrons from heavy-flavour decay. Another example might involve leptons from charm versus leptons from bottom quark decays. Another possible

source of correlation arises from the kinematics of the heavy quark itself; the $p_T$ of the lepton, the associated neutrino, and the remainder of the heavy quark fragmentation (which will be related to the isolation) are all coupled.

Preliminary indications from studies of GEANT4 simulated QCD dijet samples suggest that the primary source of electrons is a jet faking an electron, while the muon case is dominated by the decay of $B$ hadrons. Thus the issue of multiple background sources alluded to above is not a serious concern according to these Monte Carlo studies, but should be kept in mind for real data.

We first study the applicability of the lepton isolation technique in a pure sample of $b\bar{b}$ + jets. We then move to a more realistic sample containing a mixture of $t\bar{t}$ and $W+$jets events in addition to the $b\bar{b}$ + jets sample, focussing on the muon channel. The dominant systematic is found to be contamination of control samples by $t\bar{t}$ events.

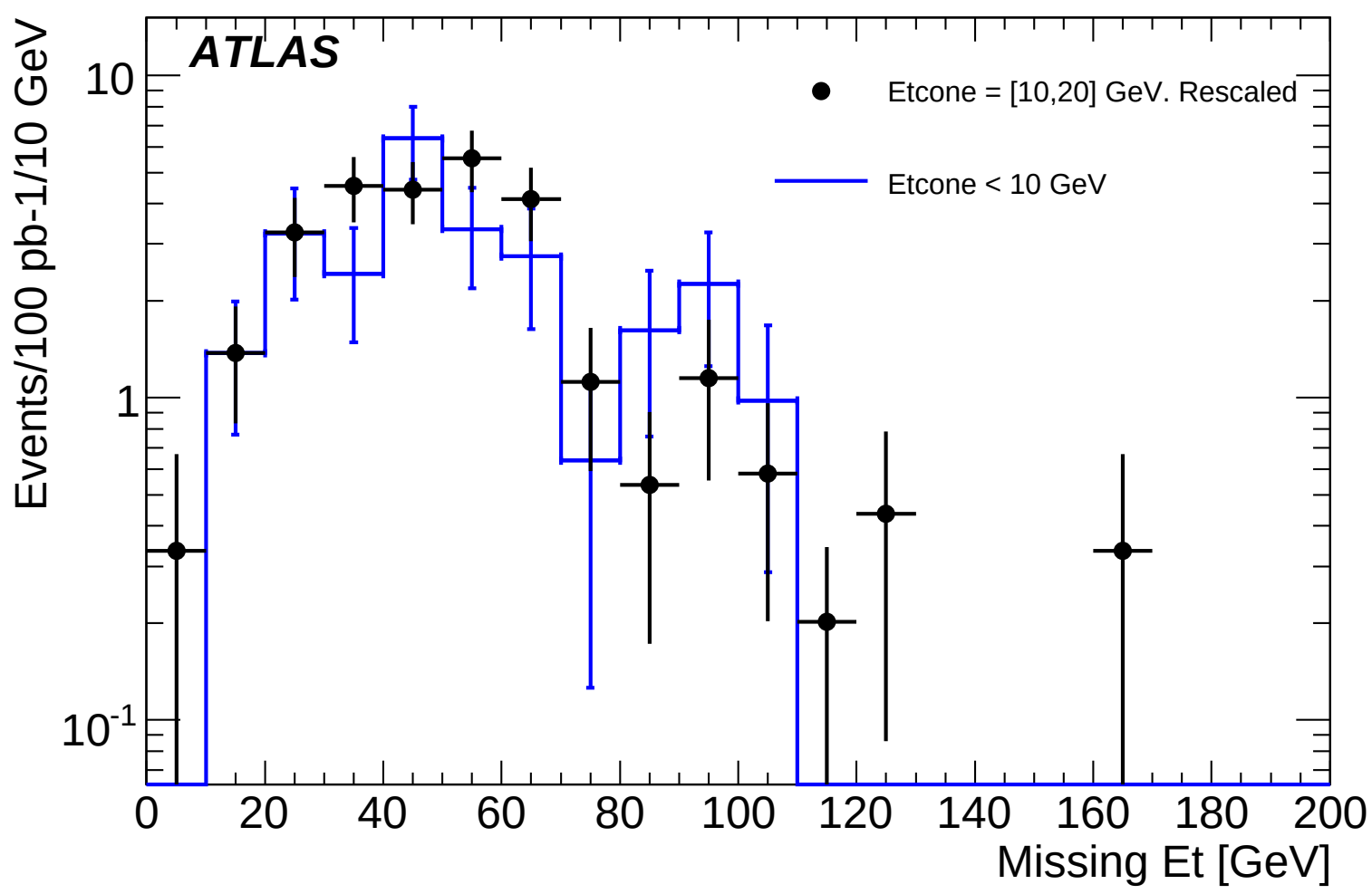


Figure 22: Points: $E_{\mathrm{T}}^{\mathrm{miss}}$ distribution for the non-isolated lepton sample. Histogram: $E_{\mathrm{T}}^{\mathrm{miss}}$ distribution for the isolated lepton sample. The two distributions have been normalized to the same area in the region $E_{\mathrm{T}}^{\mathrm{miss}} = [10,20]$ GeV.

**Lepton isolation versus $E_{\mathrm{T}}^{\mathrm{miss}}$ for $b\bar{b}$ + jets** The dependence of $E_{\mathrm{T}}^{\mathrm{miss}}$ on lepton (e,$\mu$) isolation was studied with a sample of $b\bar{b}+$jets events generated with ALPGEN [27], corresponding to an integrated luminosity of about 200 pb$^{-1}$. At the event generator level, one of the $b$ quarks was forced to undergo semileptonic decay to $e$ or $\mu$; also applied at the event generator level were filters requiring true $E_{\mathrm{T}}^{\mathrm{miss}}$ $> 30$ GeV, four or more truth jets with $p_{\mathrm{T}} > 40$ GeV, a leading truth jet with $p_{\mathrm{T}} > 80$ GeV and truth lepton with $p_{\mathrm{T}} > 10$ GeV. In the offline analysis, the events were required to satisfy the SUSY jet cuts described in Section 1. The events were required to have one and only one lepton with $p_T$ greater than 20 GeV, where the lepton could be either an electron or muon. Events with additional leptons (with $p_T$>10 GeV) were rejected.

A (non-isolated) control sample was defined by the requirement $E_T^{\mathrm{cone}} = [10,20]$ GeV, where $E_T^{\mathrm{cone}}$ is the $E_T$ inside an $\eta-\phi$ cone of radius 0.2 centered around the lepton, excluding the lepton $p_T$. The (isolated) signal sample was defined as $E_T^{\mathrm{cone}} < 10$ GeV. The $E_{\mathrm{T}}^{\mathrm{miss}}$ distribution of the control sample was normalized in the region $E_{\mathrm{T}}^{\mathrm{miss}} = [10,20]$ GeV to the signal sample. This normalized $E_{\mathrm{T}}^{\mathrm{miss}}$ distribution was used to estimate the $E_{\mathrm{T}}^{\mathrm{miss}}$ distribution for isolated leptons. Fig. 22 shows the $E_{\mathrm{T}}^{\mathrm{miss}}$ distribution for the isolated and non-isolated samples. The agreement is good, suggesting that the correlation between $E_{\mathrm{T}}^{\mathrm{miss}}$ and lepton isolation in $b\bar{b}+$jets events is small.

It is also apparent from Fig. 22 that the absolute level of background is low, even though typical SUSY selection cuts on $E_{\mathrm{T}}^{\mathrm{miss}}$ , transverse mass and effective mass (typically $E_{\mathrm{T}}^{\mathrm{miss}} > 100$ GeV, $M_T > 100$ GeV, and $M_{\mathrm{eff}} > 800$ GeV) have not yet been applied. We will return to this point later. The $M_{\mathrm{eff}}$ distribution after the jet and lepton selection cuts is shown in Fig. 23(left). Ideally, one would like to demonstrate that the lack of correlation between lepton isolation and $E_{\mathrm{T}}^{\mathrm{miss}}$ still holds even after the other SUSY selection cuts, such as those on $M_T$ and $M_{\mathrm{eff}}$, have been applied. This was not possible in this study with fully simulated data due to the limited number of events available.

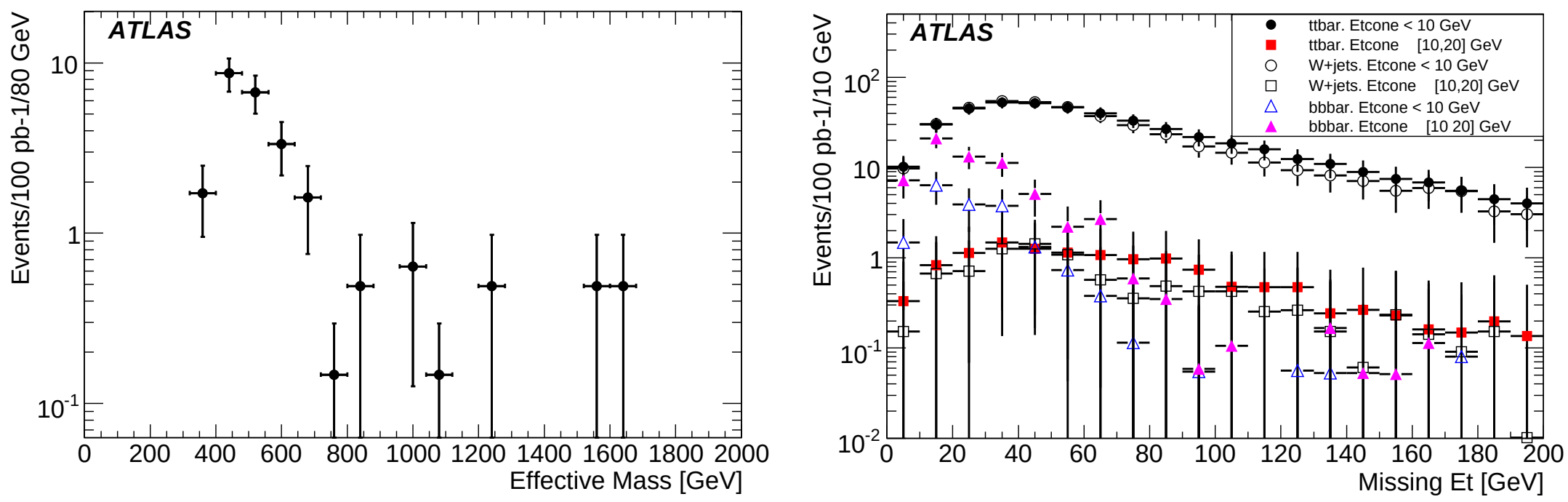


Figure 23: Left: distribution of the effective mass in the $b\bar{b}$ +jets sample after the jet and lepton selection cuts. Right: $E_{\mathrm{T}}^{\mathrm{miss}}$ distributions for $t\bar{t}$ isolated muon sample (black filled circle), $t\bar{t}$ non-isolated muon sample (red filled square), $W+$jets isolated muon sample (black open circle), $W+$jets non-isolated muon sample (black open square), $b\bar{b}$ isolated muon sample (blue open triangle), $b\bar{b}$ non-isolated muon sample (magenta filled triangle).

**Contamination from $W$ + jets and $t\bar{t}$** An important systematic effect comes from the contamination of the control samples by $W+$jets and $t\bar{t}$ events. This was studied with fast simulation samples which had no generator-level cuts. The sample sizes corresponded to integrated luminosities of approximately 200 pb$^{-1}$, 6 fb$^{-1}$ and 3 fb$^{-1}$ for $b\bar{b}+$jets, $t\bar{t}$ and $W+$jets, respectively. The $E_{\mathrm{T}}^{\mathrm{miss}}$ distribution, after the event selection cuts described above is shown in Fig. 23(right) for the $b\bar{b}$, $t\bar{t}$ and $W+$jets samples for isolated and non-isolated muons defined as above. Looking first just at the $b\bar{b}$ events, there is good agreement in the shape of the $E_{\mathrm{T}}^{\mathrm{miss}}$ distribution for the isolated and non-isolated samples; this reinforces the findings from the `GEANT4`-simulated $b\bar{b}+$jets samples on the independence of muon isolation and $E_{\mathrm{T}}^{\mathrm{miss}}$ . It is also clear that the non-isolated sample is dominated by $b\bar{b}$ at low $E_{\mathrm{T}}^{\mathrm{miss}}$ . However, the $E_{\mathrm{T}}^{\mathrm{miss}}$ tail is dominated by $t\bar{t}$ and $W+$jets events, even in the non-isolated sample.[4] This implies that the shape of the $E_{\mathrm{T}}^{\mathrm{miss}}$ distribution in the control sample will be distorted by the high $E_{\mathrm{T}}^{\mathrm{miss}}$ tail. It is possible to increase the number of $b\bar{b}$ events in the control sample by further loosening the isolation requirement, however this has the disadvantage of extrapolating into the isolated signal region over a large range, which could be susceptible to correlations between isolation and $E_{\mathrm{T}}^{\mathrm{miss}}$ . Furthermore, even if the isolation requirement in the control sample were successfully loosened, there is the problem of normalization. In the isolated sample, $t\bar{t}$ and $W+$jets completely dominate, even at very low $E_{\mathrm{T}}^{\mathrm{miss}}$ ; this means that the normalization of the $E_{\mathrm{T}}^{\mathrm{miss}}$ distribution will be distorted when extrapolating from the control to the signal region.

One might consider correcting the event yields in the control regions for the presence of $t\bar{t}$ and $W+$jets events. An early CDF $t\bar{t}$ cross-section analysis from Run2 [31] made corrections based on an assumed $t\bar{t}$ cross-section. D0 [32] used the so-called "matrix method" where additional input on the

[4] As an added complication the shape of this tail will change after the application of the $M_T$ cut which selects primarily dileptonic $t\bar{t}$ decays.

efficiency for $t\bar{t}$ and $b\bar{b}$ events to pass the isolation cuts was used to extract the multijet background. The utility of these methods for the SUSY search could be further studied.

Despite the limitations of the technique in the presence of $t\bar{t}$ and $W+$jets events background it should be noted that this study also indicates that QCD backgrounds are unlikely to be dominant in the 1-lepton channel. Even with application of just the standard SUSY 4-jet and lepton selection cuts, the QCD multijet background in the 1-lepton channel is significantly smaller than the $t\bar{t}$ and $W+$jets backgrounds and applying the remaining SUSY selection cuts further suppresses the QCD multijet background. Consequently in the baseline analysis the accuracy of QCD multijet background estimates is less important than that of the estimates other backgrounds described in Ref. [4]. It should be kept in mind however that for some background studies the jet or lepton selection cuts may be further loosened, in which case the the lepton isolation technique might be applicable without the problem of contamination from other backgrounds.

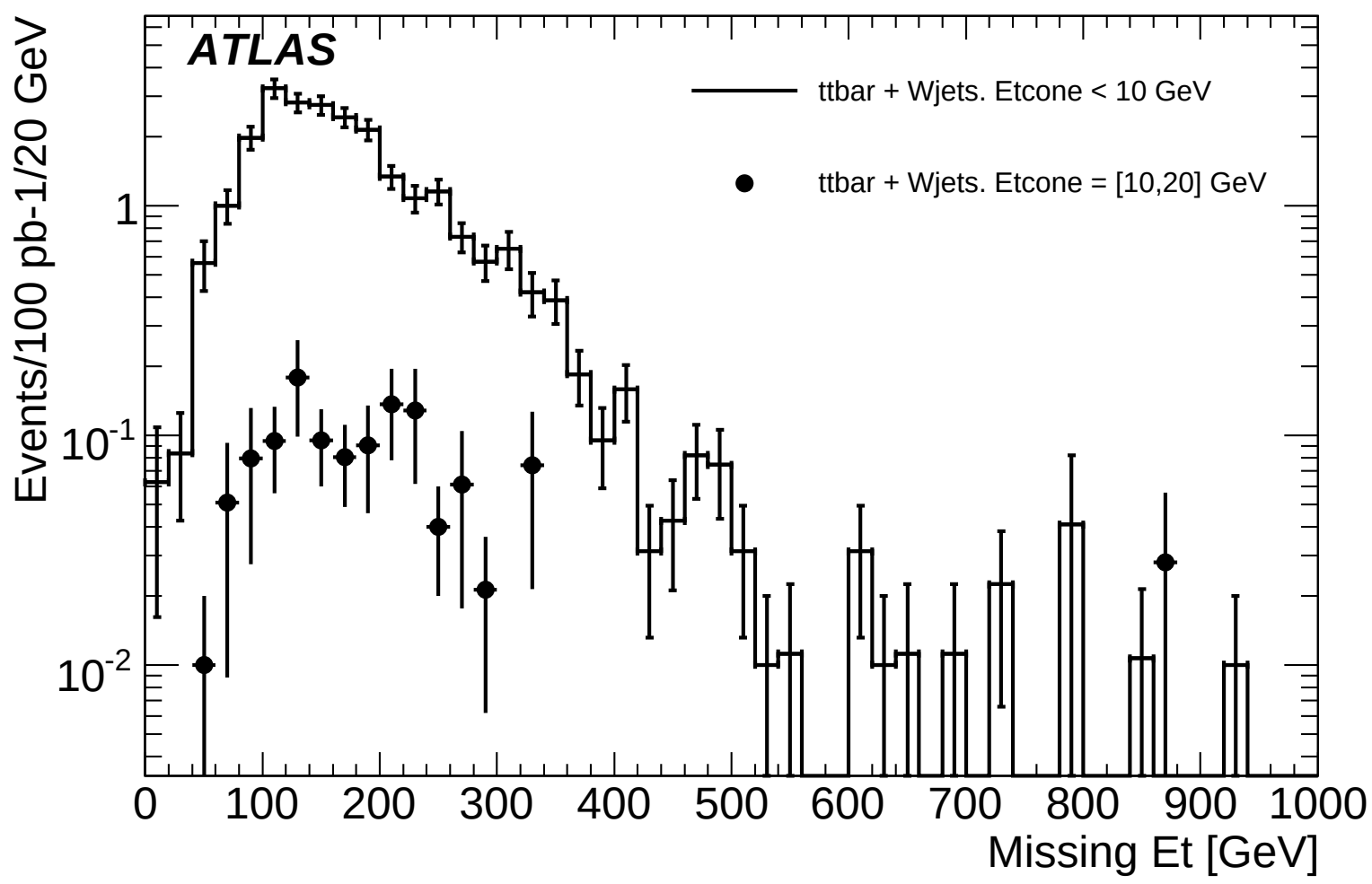


Figure 24: Black circles: $E_{\mathrm{T}}^{\mathrm{miss}}$ distribution for $t\bar{t}$ plus $W+$jets samples where the lepton is non-isolated ($E_T^{\mathrm{cone}} = [10,20]$ GeV). Histogram: $E_{\mathrm{T}}^{\mathrm{miss}}$ distribution for $t\bar{t}$ plus $W+$jets samples with isolated leptons ($E_T^{\mathrm{cone}} < 10$ GeV).

**Upper limit on $E_{\mathrm{T}}^{\mathrm{miss}}$ distribution for $b\bar{b}$ + jets** If the multijet background is large compared to the background from $t\bar{t}$ (either because the existing Monte Carlo simulation happens to underestimate the contribution or because the selection cuts were further loosened), then issues of $t\bar{t}$ contamination will become negligible and the methods described in the previous sections should work well to estimate the multijet background. On the other hand, if the multijet background is small compared to $t\bar{t}$ as expected, one can obtain the $E_{\mathrm{T}}^{\mathrm{miss}}$ distribution from the control region ($E_T^{\mathrm{cone}} = [10,20]$ GeV) as an upper limit on the multijet background; the (reasonable) assumption here (confirmed in Monte Carlo) is that the $E_T^{\mathrm{cone}}$ distribution for multijet background is flat (or falling) as $E_T^{\mathrm{cone}}$ approaches zero. After the $M_T$ cut, this control region is expected to be dominated by $t\bar{t}$ and $W+$jets events containing non-isolated leptons with the QCD multijet contribution playing only a very small role. Thus this will be an upper limit on the multijet background, but the number of events so obtained will in any case be many fewer than from the main $t\bar{t}$ and $W+$jets backgrounds (with isolated leptons) and so are likely to be negligible.

The contribution from $b\bar{b}+$jets has been compared to `GEANT4`-simulated `ALPGEN` $t\bar{t}$ and $W+$jets samples (corresponding to integrated luminosities of about 3 fb$^{-1}$). In addition to the event selection cuts listed above (jet and lepton selection), the events were required to have $M_T > 100$ GeV and $M_{\mathrm{eff}} >$

800 GeV. The $E_{\mathrm{T}}^{\mathrm{miss}}$ distribution for the sum of $t\bar{t}$ and $W+$jets samples for events with non-isolated ($E_T^{\mathrm{cone}} = [10,20]$ GeV) leptons is shown in Fig. 24; the contribution of $b\bar{b}+$jets cannot be seen compared to that from $t\bar{t}$ plus $W+$jets. Furthermore, all contributions are negligible compared to the $t\bar{t}$ plus $W+$jets background with isolated leptons. Counting the number of non-isolated $t\bar{t}$ plus $W+$jets with $E_{\mathrm{T}}^{\mathrm{miss}} >$ 100 (200) GeV, one would find an upper limit on the multijet background in the lepton channel of $10 \pm 2$ $(5 \pm 1)$ events for an integrated luminosity of 1 fb$^{-1}$. The true number of multijet background events would be estimated to be roughly an order of magnitude lower.

## 6 Summary

This paper has examined a wide range of techniques for estimating QCD backgrounds in searches for Supersymmetry at ATLAS. The focus of the paper has been on outlining the strategies which could be used and assessing the likely uncertainties associated with them. It is important to note that only by comparing uncorrelated results from a number of such independent techniques can a robust QCD background estimate be obtained.

Given the difficulty of obtaining accurate estimates of QCD jet backgrounds generating fake $E_{\mathrm{T}}^{\mathrm{miss}}$ it will be essential to reduce this source of background using the techniques described in Section 2 and also the more general cuts described in Ref. [3]. The remaining backgrounds can be estimated using Monte Carlo (Section 3 and 4) or data-driven (Section 5) techniques.

All Monte Carlo based estimates will be subject to systematic effects arising from parton distribution and underlying event uncertainties (Section 3.1), likely to be of order 20% in each case for events satisfying the baseline SUSY cuts described in Ref. [5]. A jet energy scale uncertainty of 5% (Section 3.2) will contribute a further $\sim$ 30%. While the precision with which Monte Carlo generators model QCD jet physics at 14 TeV is difficult to assess prior to data-taking, the difference (Section 3.3) between a 'traditional' parton shower dijet estimate and that obtained from one of the newer matched matrix-element + parton-shower generators for the baseline SUSY cut set is of order 50% if an accurate normalisation to data can be obtained. To these effects must be added luminosity uncertainties ranging from 20–30% at start-up (from machine parameters) reducing to <3–5% (from total cross-section measurements and $W/Z$ counting).

In addition to the above systematic effects, Monte Carlo simulation-based estimates will be subject to detector simulation uncertainties – due to the imperfect description of the response of ATLAS to QCD jets, and statistical uncertainties caused by the large QCD cross-section and access to finite computing resources. It is impossible to assess the accuracy of current full (GEANT4) or fast (ATLFAST, 'fast G4'–Section 4.2 or PJTF–Section 4.1) detector simulations without recourse to data, however the results of Section 4.1 suggest that uncertainties $\sim$ 100% in the understanding of the response of the ATLAS calorimeters to jets may lead to similar uncertainties in the background estimate.

Data-driven background estimates have the advantage that they are less prone to input systematics, and can in some cases benefit from large statistics in control channels used to measure detector performance. However, additional systematic uncertainties have to be considered due to the potential contamination of control samples with non-QCD events, and from relying on Monte Carlo simulation to extrapolate from the control into the signal region. The data-driven estimate for the jets + $E_{\mathrm{T}}^{\mathrm{miss}}$ + 0-leptons channel described in Section 5.1 could potentially give a combined uncertainty of $\lesssim$ 60% for 1 fb$^{-1}$ of integrated luminosity and the baseline SUSY cut set, although a precise figure is difficult to obtain without further Monte Carlo data. In the jets + $E_{\mathrm{T}}^{\mathrm{miss}}$ + 1-lepton channel the QCD jet background is potentially far less significant, in which case the technique described in Section 5.2 is likely to generate a very conservative upper limit on the background a factor $\sim$10 above the true background for 1 fb$^{-1}$.

# Prospects for Supersymmetry Discovery Based on Inclusive Searches

**Abstract**
This note describes searches for generic SUSY models with $R$-parity conservation in the ATLAS detector at the CERN Large Hadron Collider. SUSY particles would be produced in pairs and decay to the lightest SUSY particle, $\tilde{\chi}_1^0$, which escapes the detector, giving signatures involving jets, possible leptons, and $E_{\mathrm{T}}^{\mathrm{miss}}$. The integrated luminosity simulated is 1 fb$^{-1}$ . This article relies on work published elsewhere in this collection, where the Standard Model backgrounds for SUSY are discussed.

## 1 Introduction

This note describes the search for generic SUSY with $R$ parity, so that SUSY particles are produced in pairs and decay to the lightest SUSY particle, $\tilde{\chi}_1^0$, which escapes the detector, giving signatures involving jets, possible leptons and $E_{\mathrm{T}}^{\mathrm{miss}}$, an imbalance in the transverse energy measured in the detector. Most of the introductory information necessary to the understanding of this document is given in the introductory SUSY note [1], which should be read before this one. These include a brief description of the theoretical framework, a definition of the SUSY benchmark models SUn studied in the detailed analyses, a description of the Monte Carlo samples used for signal and background. Common identification criteria for jets, taus and leptons have been adopted throughout the analyses in this note, and are also described in [1] as well as the definition of a few global variables relevant for the analysis, such as effective mass ($M_{\mathrm{eff}}$), stransverse mass ($m_{\mathrm{T2}}$) and transverse sphericity ($S_T$). The background uncertainties used throughout this work are based on Standard Model background studies documented in [2, 3]. Special signatures associated, e.g., with Gauge Mediated SUSY Breaking are treated elsewhere [4].

Two different approaches have been used to develop the inclusive search strategy described here. Firstly, detailed studies have been carried out for various signatures (jets + $E_{\mathrm{T}}^{\mathrm{miss}}$ + 0 leptons, jets + $E_{\mathrm{T}}^{\mathrm{miss}}$ + 1 lepton, ...) using data-sets fully simulated with Geant 4 for specific SUSY signal parameters and for the relevant Standard Model backgrounds. These detailed studies are used to develop deeper understanding of how best to reconstruct these relatively complex events and to define strategies for separating the signal from the Standard Model backgrounds. In order to simplify the procedure of combining the results from the different analyses, the various leptonic signatures have been defined so that they are exclusive. For example the 1-lepton signature rejects all events in which more than one lepton is present. However, no attempt is made to combine the different analyses in the present document.

Secondly, the insight gained from studying specific points has been applied to several scans over subsets of the SUSY parameter space, Since large numbers of signal points must be studied, these scans are of necessity based on fast, parameterized simulation. The goal is to verify that the different sets of basic cuts studied on benchmark points provide sensitivity to a broad range of SUSY models. The results shown in this document will be used as a basis for the development of a strategy for SUSY discovery with early ATLAS data.

### 1.1 Trigger

The trigger efficiency for the inclusive SUSY signals at the benchmark points has been studied based on the complete simulation of all three trigger levels of ATLAS. For all the analyses we adopted the trigger thresholds defined for $2 \times 10^{33}$ cm$^{-2}$s$^{-1}$ in the High Level Trigger TDR [5]. These triggers are discussed

futher elsewhere [6] in this volume, where a detailed explanation of the naming convention for the trigger menu items appearing in Table 1 can be found.

The jet triggers, denoted by "JETS", consist of the logical "or" of the following triggers:

- j400: 1 jet with $p_T > 400$ GeV;
- 3j165: 3 jets with $p_T > 165$ GeV;
- 4j110: 4 jets with $p_T > 110$ GeV.

The $E_{\mathrm{T}}^{\mathrm{miss}}$ trigger "j70_xE70", requires $E_{\mathrm{T}}^{\mathrm{miss}} > 70$ GeV accompanied by a jet with $p_T > 70$ GeV. The lepton triggers are "e22i": an isolated electron efficient for $p_T > 25$ GeV; "2e12i": two isolated electrons efficient for $p_T > 15$ GeV, "mu20": a muon with $p_T > 20$ GeV and "2mu10": two muons with $p_T > 10$ GeV.

Since the goal of this note is to develop a generic SUSY search, only the basic trigger building blocks have been considered. More complex triggers combining different objects can be easily implemented in the trigger menus. Also, only triggers which are not prescaled have been used.

The trigger efficiencies for the signal events passing the the 0, 1, 2, and 3-lepton selections defined in the following sections are listed in Table 1 for different requirements on jet multiplicity. In general the j70_xE70 is highly efficient, although there is some loss for SU4, the low-mass point with a very large cross-section. The j70_xE70 trigger is also very efficient for the $\tau$ and $b$ modes, described in Sections 6 and 7 below.

The basic performance of the leptonic and j70_xE70 triggers will be determined from Standard Model events such as $Z$ and $\bar{t}t$ using the methods described in [6]. It may be useful to check that performance by comparing (Monte Carlo) samples of SUSY events selected with multiple triggers. For the 0-lepton selection, the efficiency for JETS trigger alone is in the range 30-70%, except for the very low mass point SU4. This provides a useful redundancy in the early phases, as the $E_{\mathrm{T}}^{\mathrm{miss}}$ trigger may require a longer time than the other triggers to be completely understood, but only j70_xE70 has an efficiency close to one. For the topologies involving leptons, both the single lepton triggers and j70_xE70 have typical efficiencies in excess of 80%, so comparing them should be quite effective.

### 1.2 Systematic uncertainties and statistical procedure

To assess the discovery potential of the different analyses it is necessary to take into account systematic uncertainties. SUSY searches will address very complex topologies, typically with many jets in the final state. The prediction of the Standard Model backgrounds to these topologies will require a complex interplay of Monte Carlo and data-driven methods. The development of these methods and the estimate of the corresponding uncertainties are described in detail in [2] and [3]. The approximate uncertainties for an integrated luminosity of $1\,\mathrm{fb}^{-1}$ are estimated to be:

- 50% for the background from QCD multijet events,
- 20% for the background from $t\bar{t}$, $W+$jets, $Z+$jets, and $W/Z$ pairs.

The limited Monte Carlo statistics is also taken into account and all systematic uncertainties are added in quadrature. The background can never be known exactly. Uncertainties on the background are incorporated in the significance by convoluting the Poisson probability that the background fluctuates to the observed signal with a Gaussian background probability density function with mean $N_b$ and standard deviation $\delta N_b$ (see e.g. [7,8] and references therein). Given these assumptions, the probability $p$ that the background fluctuates by chance to the measured value $N_{\mathrm{data}}$ or above is given by

$$p = A \int_0^\infty db\; G(b; N_b, \delta N_b) \sum_{i=N_{\mathrm{data}}}^{\infty} \frac{e^{-b} b^i}{i!},$$

Table 1: Average event trigger efficiency (in %) for events passing various lepton and jet selection criteria described in detail in the indicated sections.

| Trigger | SU1 | SU2 | SU3 | SU4 | SU6 | SU8.1 |
|---|---|---|---|---|---|---|
| | 0-lepton, 4-jet selection [Section 2.1] | | | | | |
| JETS | 44.6 | 51.0 | 33.8 | 7.7 | 51.7 | 48.2 |
| j70_xE70 | 99.7 | 98.7 | 99.5 | 97.2 | 99.6 | 99.7 |
| | 0-lepton, 3-jet selection [Section 2.2] | | | | | |
| JETS | 64.9 | 71.1 | 54.9 | 34.3 | 71.8 | 66.8 |
| j70_xE70 | 100. | 99.8 | 100. | 99.9 | 100. | 100. |
| | 0-lepton, 2-jet selection [Section 2.2] | | | | | |
| JETS | 44.1 | 39.9 | 30.1 | 8.8 | 53.6 | 47.6 |
| j70_xE70 | 100. | 100. | 100. | 99.9 | 100. | 100. |
| | 1-lepton, selection [Section 3] | | | | | |
| JETS | 41.8 | 50.5 | 31.7 | 8.1 | 48.4 | 45.6 |
| j70_xE70 | 99.6 | 99.0 | 98.9 | 95.6 | 98.9 | 99.1 |
| 1LEP (mu20 OR e22i) | 81.2 | 81.0 | 79.9 | 80.3 | 80.4 | 79.5 |
| | OS 2-lepton, selection [Section 4.1] | | | | | |
| JETS | 36.7 | 47.3 | 34.0 | 6.7 | 47.2 | 40.8 |
| j70_xE70 | 99.2 | 100.0 | 98.9 | 94.3 | 99.6 | 100.0 |
| 1LEP (mu20 OR e22i) | 87.0 | 90.0 | 87.5 | 84.8 | 79.6 | 86.4 |
| 2LEP (2mu10 OR 2e15i) | 20.5 | 35.5 | 27.0 | 18.0 | 26.0 | 14.6 |
| | SS 2-lepton, selection [Section 4.2] | | | | | |
| JETS | 39.9 | 48.8 | 29.2 | 1.6 | 46.6 | 34.5 |
| j70_xE70 | 99.3 | 100.0 | 98.9 | 84.1 | 98.3 | 100.0 |
| 1LEP (mu20 OR e22i) | 94.2 | 92.7 | 95.9 | 95.2 | 89.7 | 96.6 |
| 2LEP (2mu10 OR 2e15i) | 32.6 | 41.5 | 32.2 | 25.4 | 25.9 | 31.0 |
| | 3-lepton, selection [Section 5] | | | | | |
| JETS | 43.7 | 60.2 | 40.1 | 17.6 | 46.4 | 48.3 |
| j70_xE70 | 95.6 | 85.4 | 93.5 | 79.8 | 96.4 | 98.3 |
| 1LEP (mu20 OR e22i) | 95.2 | 94.2 | 95.8 | 94.7 | 94.6 | 96.7 |
| 2LEP (2mu10 OR 2e15i) | 49.1 | 60.2 | 51.0 | 44.7 | 47.3 | 53.3 |

where $G(b;N_b,\delta N_b)$ is a Gaussian and the factor

$$A = \left[ \int_0^\infty db\, G(b;N_b,\delta N_b) \sum_{i=0}^{\infty} e^{-b} b^i / i! \right]^{-1}$$

ensures that the function is normalised to unity. If the Gaussian probability density function $G$ is replaced by a Dirac delta function $\delta(b - N_b)$, the estimator $p$ results in a usual Poisson probability.

The probability $p$ is transformed into "standard-deviations", denoted in this note by the symbol $Z_n$, using the formula

$$Z_n = \sqrt{2}\ \mathrm{erf}^{-1}(1 - 2\mathrm{p})$$

The Root [9] library provides functions to calculate $p$ and $Z_n$.

If many different data selections are considered, it becomes more likely that statistical fluctuations would be misinterpreted as new phenomena if the number of selections is not considered in the statistical

Table 2: Number of events surviving subsequent selection cuts as defined in the text for 4-jets analysis normalized to 1fb$^{-1}$ using NLO cross-sections.

| Sample | Cut 1 | Cut 2 | Cut 3 | Cut 4 | Cut 5 | $M_{\text{eff}}$ Cut |
|---|---|---|---|---|---|---|
| SU3 | 9600 | 7563 | 5600 | 5277 | 4311 | 3349 |
| SU1 | 3485 | 2854 | 2004 | 1907 | 1401 | 1229 |
| SU2 | 604 | 369 | 308 | 279 | 169 | 131 |
| SU4 | 79618 | 57803 | 46189 | 42408 | 34966 | 8507 |
| SU6 | 2551 | 2062 | 1468 | 1383 | 1080 | 956 |
| SU8.1 | 3118 | 2540 | 1778 | 1686 | 1448 | 1284 |
| MC@NLO $t\bar{t}$ | 12861 | 8798 | 6421 | 5790 | 4012 | 305 |
| Pythia QCD | 29230 | 7044 | 4667 | 848 | 848 | 13 |
| Alpgen $Z$ | 1626 | 1045 | 732 | 660 | 644 | 162 |
| Alpgen $W$ | 4066 | 2393 | 1654 | 1499 | 1147 | 228 |
| Herwig $WZ$ | 22 | 15 | 9 | 8 | 4 | 1 |
| Total Standard Model | 47805 | 19294 | 13483 | 8806 | 6655 | 708 |
| SU3 $S/B$ | 0.2 | 0.4 | 0.4 | 0.6 | 0.6 | 4.7 |
| $Z_n$ | 0.5 | 1.3 | 1.4 | 2.6 | 2.7 | 13 |
| SU3 eff (excl) | 35.1% | 78.8% | 74.0% | 94.2% | 81.7% | 77.9% |
| SU3 eff (incl) | 35.1% | 27.7% | 20.5% | 19.3% | 15.8% | 12.3% |

procedure. This is known in statistics as the problem of "multiple comparisons". The probability values are therefore corrected for multiple comparisons via a Monte Carlo method. The effect is the reduction of approximately half a unit of $Z_n$ for $Z_n = 3$, decreasing with increasing $Z_n$. In the last section of the note the significance always corresponds to the corrected $Z_n$.

# 2 Zero-lepton mode

A SUSY signal at the LHC is typically dominated by the production of squarks and gluinos. In the $R$-parity conserving case, at the end of each sparticle decay chain one finds an undetected LSPs, which can together generate large $E_{\text{T}}^{\text{miss}}$. The least model-dependent SUSY signature is therefore the search for events with multiple jets and $E_{\text{T}}^{\text{miss}}$. Traditionally searches have been performed requiring at least four jets; the high multiplicity helps to reduce the background from QCD and $W/Z+$jets. Both for this topology and for the leptonic topologies in the following sections we adopt very simple sets of cuts, similar to the ones used in the ATLAS Physics TDR [10]. In addition to the four-jet signatures we have also addressed signatures with lower jet multiplicity. These signatures have more backgrounds, but might be favoured in some SUSY models, and should be more cleanly reconstructed in the detector, because of their less-complex topologies. This may be an advantage in the early phases of the experiment.

## 2.1 Four or more jets in final state

The basic selections applied for this channel are:

1. At least four jets with $p_T > 50$ GeV at least one of which must have $p_T > 100$ GeV; and $E_{\text{T}}^{\text{miss}} > 100$ GeV.

2. $E_{\text{T}}^{\text{miss}} > 0.2 M_{\text{eff}}$.

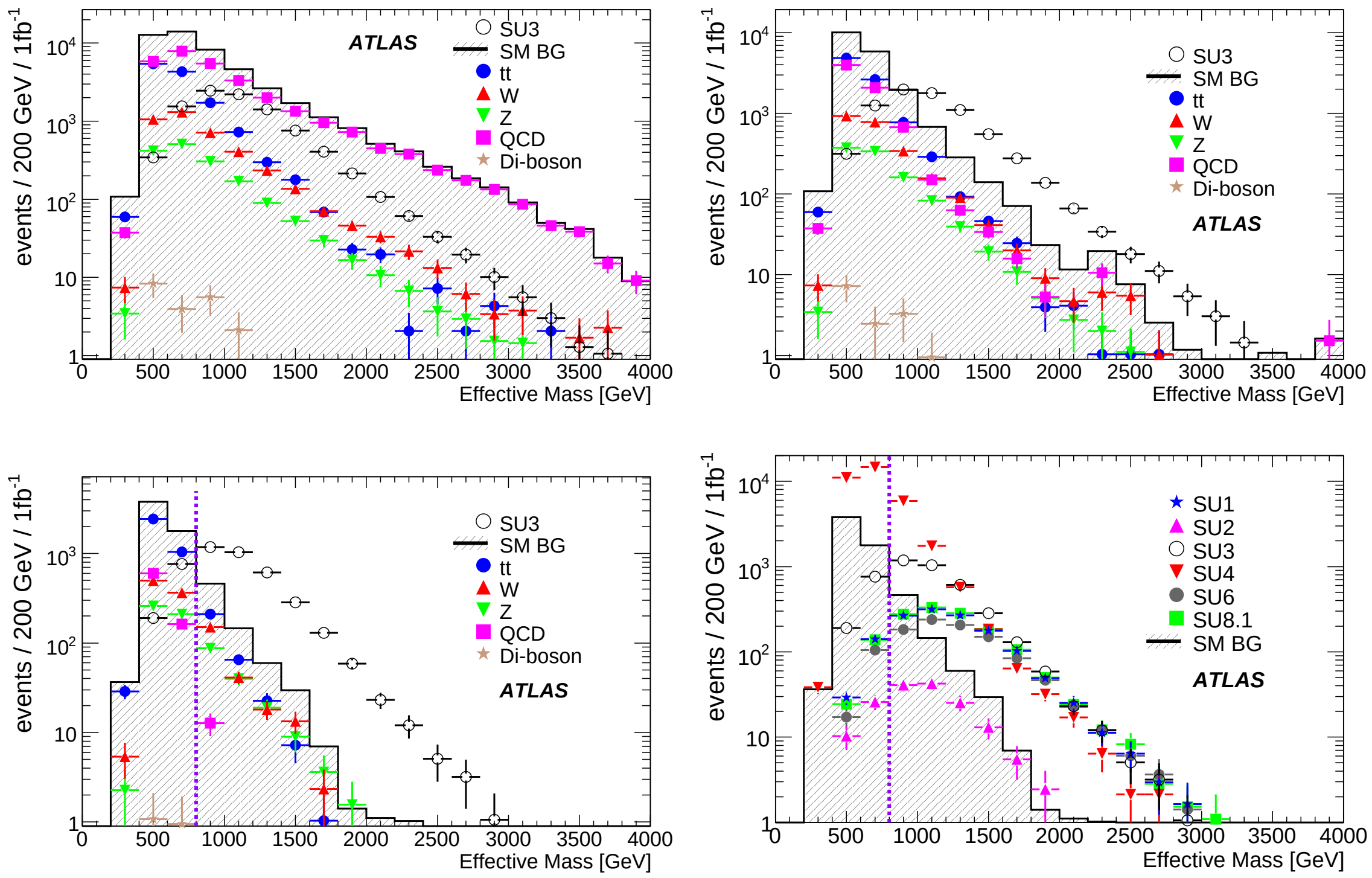


Figure 1: $M_{\text{eff}}$ distribution for events surviving successive selection cuts: cut 1 (top left), cut 2 (top right), and cuts 3-5 (bottom left). The open circles represent point SU3, and the different background contributions are shown according to the legend. The last plot (bottom right) show all of the SUSY benchmark points and the total Standard Model background after cuts 1-5. Open circles represent the SUSY SU3 signal as predicted by Monte Carlo simulation, while the shaded area shows the total Standard Model background.

3. Transverse sphericity, $S_T > 0.2$.
4. $\Delta\phi(\text{jet}_1 - E_{\text{T}}^{\text{miss}}) > 0.2$, $\Delta\phi(\text{jet}_2 - E_{\text{T}}^{\text{miss}}) > 0.2$, $\Delta\phi(\text{jet}_3 - E_{\text{T}}^{\text{miss}}) > 0.2$.
5. Reject events with an $e$ or a $\mu$.
6. $M_{\text{eff}} > 800$ GeV.

Most of the background samples have been filtered at generation level with various requirements on $E_{\text{T}}^{\text{miss}}$ and jet multiplicity. The first cut in the analysis flow applies harder requirements than any of the ones applied at the filter level to minimise the bias to the study from the use of filtered samples.

The main background at this point from QCD events where $E_{\text{T}}^{\text{miss}}$ is produced either by a fluctuation in the measurement of the energy of one or more jets, or by a real neutrino from the decay of a $B$ hadron produced in the fragmentation process. Since the statistical fluctuation on the $E_{\text{T}}^{\text{miss}}$ measurement grow with increasing $M_{\text{eff}}$, the second cut above eliminates the Gaussian part of the $E_{\text{T}}^{\text{miss}}$ measurement fluctuations. In SUSY events the jets are produced from the decay of heavy particles produced approximately at rest, and are thence distributed isotropically in space, whereas for the QCD events the direction of the two partons from the hard scattering provides a privileged direction. The cut on sphericity is intended to exploit this fact. Both for jet mismeasurement and for $b$ decays, the $E_{\text{T}}^{\text{miss}}$ vector will be close the direction of one jets, and so the $\Delta\phi$ cuts are very efficient in reducing the QCD background. The lepton

veto is applied in order to facilitate the combination with analyses requiring leptons, but it is not expected to significantly modify the signal-to-background ratio.

The number of events surviving each of the cuts for an integrated luminosity of $1\,\text{fb}^{-1}$ is shown in Table 2 for all of the considered benchmark points and for the backgrounds. The number of events in the last column includes the effect of the j70xE70 trigger, which, as shown in Table 1, has an efficiency in excess of 97% for all considered signal benchmarks.

The distribution of the final selection variable $M_{\text{eff}}$, for signal and background, is shown in Fig. 1 for point SU3 at different stages of the analysis. The QCD background is dominant after the first cut but is reduced to a similar level to the backgrounds containing real neutrinos by subsequently requiring $E_{\text{T}}^{\text{miss}} > 0.2M_{\text{eff}}$ (cut 2). The cuts on event sphericity and $\Delta\phi$ strongly reduce the QCD background, which becomes concentrated in the region of low $M_{\text{eff}}$. After all cuts $t\bar{t}$ is the dominant background, but there are also significant contributions from $W+$jets and $Z+$jets. The final cut, $M_{\text{eff}} > 800$ GeV, reduces the background to below the level of the signal for all considered benchmark points except for SU2. For this point to be found in the 0-lepton channel, one would have to select larger values of $M_{\text{eff}}$ to enhance the signal-to-background ratio, and a greater integrated luminosity would be required.

The statistical significance $Z_n$ for $1\,\text{fb}^{-1}$ was calculated using the prescription in Section 1.2 including the systematic uncertainty on the background [2, 3]. The significance for point SU3 after each cut is shown in Table 2. The significances $Z_n$ after all cuts are 13 for SU3, 6.3 for SU1, 0.9 for SU2, 25 for SU4, 6.3 for SU6, and 6.5 for SU8.1. Evidently only point SU2, for which the cross section is dominated by direct gaugino production (which is investigated elsewhere in this volume [11]), would not be accessible for the assumed set of cuts, integrated luminosity and level of background understanding.

These numbers should be taken as indicative. The uncertainty on the background used in the calculation is the estimate of what one would obtain using complicated procedures for background evaluation based on a combination of data-driven and Monte Carlo methods. The absolute value of the backgrounds used for this study is derived only from Monte Carlo, and the present uncertainty on this value is much higher. An idea of the robustness of the analysis can be obtained by studying the significance for the benchmark points if the background would be increased by a factor 2. In this case the significance for SU3 would drop to 7.8, and the one for SU6 to approximately 3.1. The significance for SU6 would be, in this situation, dominated by the systematic uncertainty on the background evaluation. Therefore an increase in integrated luminosity would result in an increased reach only if it can be used to reduce the uncertainty on the background evaluation.

## 2.2 Inclusive two-jet and three-jet final states

The analyses based on lower jet-multiplicities are based on very similar requirements to the 4-jet analysis above. The differences are: higher $p_{\text{T}}$ requirements on the remaining jets to cope with the increased QCD background, and a slightly harder $E_{\text{T}}^{\text{miss}}$ cut. The sphericity cut is less relevant in the case of low jet multiplicities and is dropped. For the two (three)-jet analysis the cuts are respectively:

1. At least two (three) jets, the hardest with $p_T > 150$ GeV and the second (and third) with $p_T > 100$ GeV; $E_{\text{T}}^{\text{miss}} > 100$ GeV
2. $E_{\text{T}}^{\text{miss}} > 0.3(0.25)M_{\text{eff}}$.
3. $\Delta\phi(\text{jet}_1 - E_{\text{T}}^{\text{miss}}) > 0.2$, $\Delta\phi(\text{jet}_2 - E_{\text{T}}^{\text{miss}}) > 0.2$, $(\Delta\phi(\text{jet}_3 - E_{\text{T}}^{\text{miss}}) > 0.2)$
4. Reject events with an $e$ or a $\mu$
5. $M_{\text{eff}} > 800$ GeV.

The $M_{\text{eff}}$ variable is different from the one defined in [1] in that only the 2(3) highest $p_T$ jets for the 2–(3–)jet analysis are used.

Since the ALPGEN $W+$jets and $Z+$jets background samples have a filter at generation level requiring 4 jets, samples produced with the PYTHIA generator were used in this case.

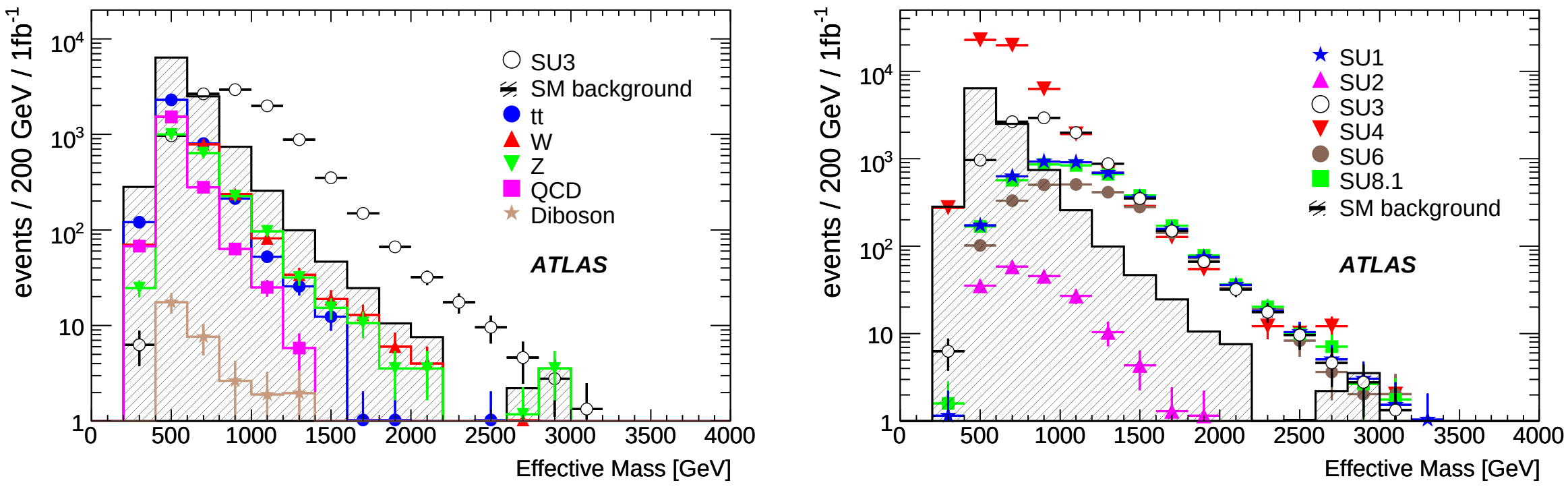


Figure 2: $M_{\text{eff}}$ distribution for the 0-lepton plus 2-jet analysis, after final cuts. Left: The open circles show the SUSY (SU3) signal Monte Carlo prediction, while the total Standard Model background is shown by the shaded histogram. The individual background contributions are shown by the points, as described in the legend. Right: The points show the distribution of the signal for a number of SUn points.

Table 3: Number of events surviving the selection cuts defined in the text for the 2–jet analysis. Entries are normalized to 1 fb$^{-1}$ using next-to-leading-order cross-sections.

| Sample | Cut 1 | Cut 2 | Cut 3 | Cut 4 | $M_{\text{eff}}$ Cut |
|---|---|---|---|---|---|
| SU3 | 18660.7 | 12519.8 | 12217.5 | 10055.2 | 6432.2 |
| SU1 | 7699.9 | 5427.5 | 5318.1 | 3996.8 | 3196.0 |
| SU2 | 642.4 | 319.7 | 301.2 | 185.1 | 90.4 |
| SU4 | 123219 | 64502.4 | 62172.9 | 52108.0 | 9434.4 |
| SU6 | 4483.1 | 3133.5 | 3041.7 | 2418.5 | 1987.0 |
| SU8.1 | 6384.7 | 4482.5 | 4381.8 | 3804.5 | 3067.7 |
| $t\bar{t}$ | 17666.6 | 6273.8 | 5778.6 | 3556.7 | 304.8 |
| QCD | 124513.9 | 7341.7 | 1983.7 | 1983.7 | 107.6 |
| $Z+$jets | 3222.5 | 2192.2 | 2109.5 | 2056.1 | 391.6 |
| $W+$jets | 8887.2 | 4504.5 | 4072.4 | 2775.5 | 395.1 |
| Diboson | 150.4 | 71.2 | 66.0 | 32.1 | 6.8 |
| Standard Model | 154440.5 | 20383.4 | 14010.1 | 10404.1 | 1205.8 |
| SU3 $S/B$ | 0.12 | 0.61 | 0.87 | 0.97 | 5.3 |
| SU3 $S/\sqrt{B}$ | 47.5 | 87.7 | 103.2 | 98.6 | 185.2 |
| SU3 eff (cum) | 67.4% | 45.2% | 44.1% | 36.3% | 23.2% |
| SU3 eff (excl) | 67.4% | 67.1% | 97.6% | 82.3% | 64.0% |

The cut flow for the 2-jet analysis is given in Table 3, and the $M_{\text{eff}}$ distributions before the $M_{\text{eff}}$ cut for the different background contributions and for the different signal points are shown in Figure 2. The

number of events after all cuts includes the effect of the j70xE70 trigger, which, as shown in Table 1, has an efficiency in excess of 99% for all considered SUSY benchmark points. After the $\Delta\phi$ cuts the $t\bar{t}$, $W+$jets, $Z+$jets and QCD all give comparable contributions to the background. After all cuts the surviving events are approximately doubled both for signal and background, as compared to the 4-jet analysis, except for the low-mass SU4 points for which the harder kinematic cuts reduce the signal efficiency.

Assuming the estimated systematic background errors are the same as for the 4–jet case, the estimated significances are 13.3 for SU3, 8.0 for SU1, 17.2 for SU4, 5.5 for SU6, and 7.7 for SU8.1, for 1 fb$^{-1}$ of integrated luminosity. The significance for SU2, for which direct gaugino production is dominant, is less than 1.0. The equivalent numbers for the 3–jet analysis are: 17.0 for SU3, 9.5 for SU1, 25.7 for SU4, 7.3 for SU6 and 9.6 for SU8. Although the signal over background ratio is equivalent or better (for the three-jet topology) than for the 4-jet analysis this is only partially reflected in the significances when the systematic uncertainty is taken into account. This is due to the increased contribution of the QCD background, which has an estimated uncertainty on QCD is of 50%, as compared to 20% for the other backgrounds. Therefore, since the uncertainties of the backgrounds were evaluated for a 4-jet analysis, a dedicated background study would be needed to obtain a correct estimate of the discovery potential in this topology.

The 2-, 3- and 4-jet analyses are based on very similar cuts and therefore have a large overlap in the selected events. About 40% of all 2-jet events are also contained in the 3-jet slection and about 35% in the 4-jet selection. The biggest overlap is for the 3 jet events: about 59% of all 3-jet events are also contained in the 4-jet analysis and about 97% in the 2-jet analysis.

An alternative strategy was explored where cuts 2 and 3 of the $M_{\text{eff}}$ analysis are dropped, and a cut on the $m_{T2}$ variable [12, 13], $m_{T2} > 400$ GeV, is applied as the only discriminating observable. The $m_{T2}$ variable, has the interesting properties that it takes low values for events where either the visible $p_T$ or $E_{\text{T}}^{\text{miss}}$ are small, and in the case of small $\Delta\phi$. It can therefore replace these topological cuts. For semi-invisibly decaying particles $m_{T2}$ is related to the difference in mass between the particles produced in the interactions and their invisible decay products. It can therefore take a larger value for SUSY events than for top or $W$ events. Taking the estimates of the systematic background errors into account, the significances for the 2-jet $m_{T2}$ analysis are: 15.6 for SU3, 11.5 for SU1, 10.9 for SU4, 8.3 for SU6, and 11.1 for SU8.1, somewhat better than the equivalent analysis based on $M_{\text{eff}}$. The most effective strategy will be ultimately defined by how well the systematic uncertainty on the background evaluation can be controlled in the different approaches.

## 3 One-lepton mode

While the 0-lepton mode with multiple jets plus $E_{\text{T}}^{\text{miss}}$ is probably the most generic search mode for SUSY with $R$-parity conservation, it is sensitive to backgrounds from mismeasured QCD multijet events. Requiring one lepton in addition to multiple jets and $E_{\text{T}}^{\text{miss}}$ greatly reduces the potential QCD multijet background; the remaining backgrounds are under better control. Even if $\tau$ decays of gauginos are dominant, leptonic $\tau$ decays provide a significant 1-lepton rate, at least for high masses. It is not surprising, therefore, that the reach in the 1-lepton and 0-lepton modes are comparable.

The cuts in this analysis are similar to those used in the ATLAS Physics TDR [10] but also include a cut on the transverse mass[1], $M_T$, formed from the lepton and $E_{\text{T}}^{\text{miss}}$ which has the role of suppressing the $W+$jets and $t\bar{t}$ backgrounds. The $S_T$ cut is included for historical reasons, but its effectiveness is questionable:

[1]Note the distinction between the transverse mass, $M_T$, which is a function of the momentum of one visible particle and the missing transverse momentum and the *stransverse* mass which is a function of the momenta of two visible particles and the missing transverse momentum.

Table 4: Number of events surviving the selection cuts defined in the text for the 1-lepton analysis. Entries are normalized to 1 $fb^{-1}$ using NLO cross-sections. The last column reports a simple $S/\sqrt{B}$ calculation of the corresponding significance of an observation for the SUSY benchmark points (SUn).

| Sample | Cuts 1–4 | Cut 5 | Cut 6 | Cut 7 | $S/\sqrt{B}$ |
|---|---|---|---|---|---|
| SU1 | 571.7 | 423.0 | 259.9 | 232.3 | 36.0 |
| SU2 | 86.7 | 75.6 | 46.1 | 39.6 | 6.1 |
| SU3 | 995.7 | 767.9 | 450.5 | 363.6 | 56.4 |
| SU4 | 7523.6 | 6260.4 | 2974.4 | 895.8 | 138.9 |
| SU6 | 342.3 | 250.9 | 161.9 | 147.9 | 22.9 |
| SU8.1 | 296.4 | 214.4 | 151.4 | 136.3 | 21.1 |
| $t\bar{t}$ | 2028.5 | 1546.8 | 131.7 | 36.0 | |
| $W$ | 425.2 | 314.8 | 9.9 | 5.4 | |
| $Z$ | 39.0 | 27.3 | 1.7 | 0.2 | |
| Diboson | 7.3 | 5.1 | 0.8 | 0.0 | |
| QCD | 0.0 | 0.0 | 0.0 | 0.0 | |
| Standard Model BG | 2500.1 | 1894.0 | 144.1 | 41.6 | |

1. Exactly one isolated lepton with $p_T > 20$ GeV satisfying the selection criteria described earlier.
2. No additional leptons with $p_T > 10$ GeV. This ensures no overlap with the 0-lepton, 2-lepton, and 3-lepton analyses.
3. At least four jets with $p_T > 50$ GeV at least one of which must have $p_T > 100$ GeV.
4. $E_{\mathrm{T}}^{\mathrm{miss}} > 100$ GeV and $E_{\mathrm{T}}^{\mathrm{miss}} > 0.2M_{\mathrm{eff}}$.
5. Transverse sphericity, $S_T > 0.2$.
6. Transverse mass, $M_T > 100$ GeV.
7. $M_{\mathrm{eff}} > 800$ GeV.

Cuts 1–2 define the 1-lepton analysis, while Cuts 3–4 both reduce the Standard Model backgrounds and ensure compatability with the Standard Model filter cuts. Distributions without these four cuts are not meaningful and so are not shown. Cut 5 reduces the $E_{\mathrm{T}}^{\mathrm{miss}}$ background from mismeasured dijet events; Cut 6 reduces the background from events in which the $E_{\mathrm{T}}^{\mathrm{miss}}$ comes from $W \to \ell\nu$; and Cut 7 selects high-mass final states.

The cut flow table for these cuts is shown in Table 4. The number of events after all cuts includes the effect of the j70xE70 trigger, which, as shown in Table 1, has an efficiency of around 99% for all the benchmark points considered other than the low mass SU4 point, for which the efficiency is still above 95%. Note that the QCD background is reduced to a negligible level by the lepton and $E_{\mathrm{T}}^{\mathrm{miss}}$ cuts as expected. The background after all cuts is dominated by $t\bar{t}$ and $W+$jets, both of which are expected to be better understood than the QCD background. Therefore, while the 1-lepton mode may not have better reach than the 0-lepton mode given the calculated backgrounds, its reach seems more robust against background uncertainties.

The $M_{\mathrm{eff}}$ distribution for point SU3 after each cut is shown in Figure 3. The $E_{\mathrm{T}}^{\mathrm{miss}}$ and $M_{\mathrm{eff}}$ distributions for all the SUn points after all cuts are shown in Figures 4. It is clear from these figures and from Table 4 that the only significant backgrounds to the 1-lepton mode are from $t\bar{t}$ and $W+$jets, as one

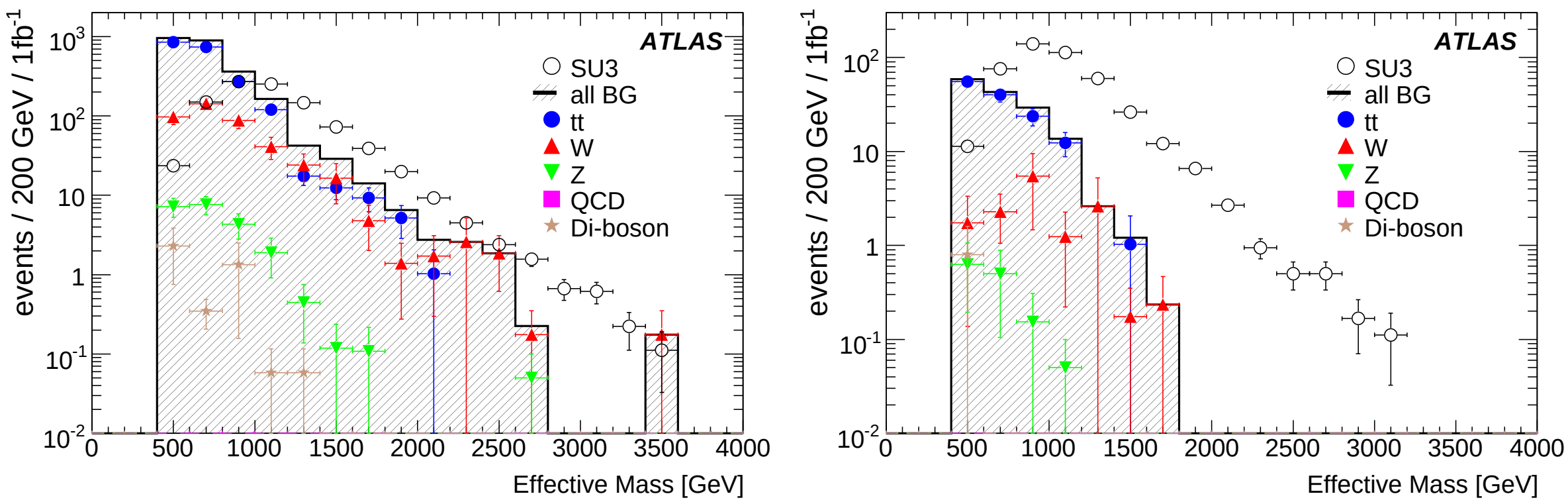


Figure 3: Expected $M_{\text{eff}}$ distributions after Cuts 1–4 (left), and Cut 6 (right) for the 1-lepton analysis. Compare with Table 4.

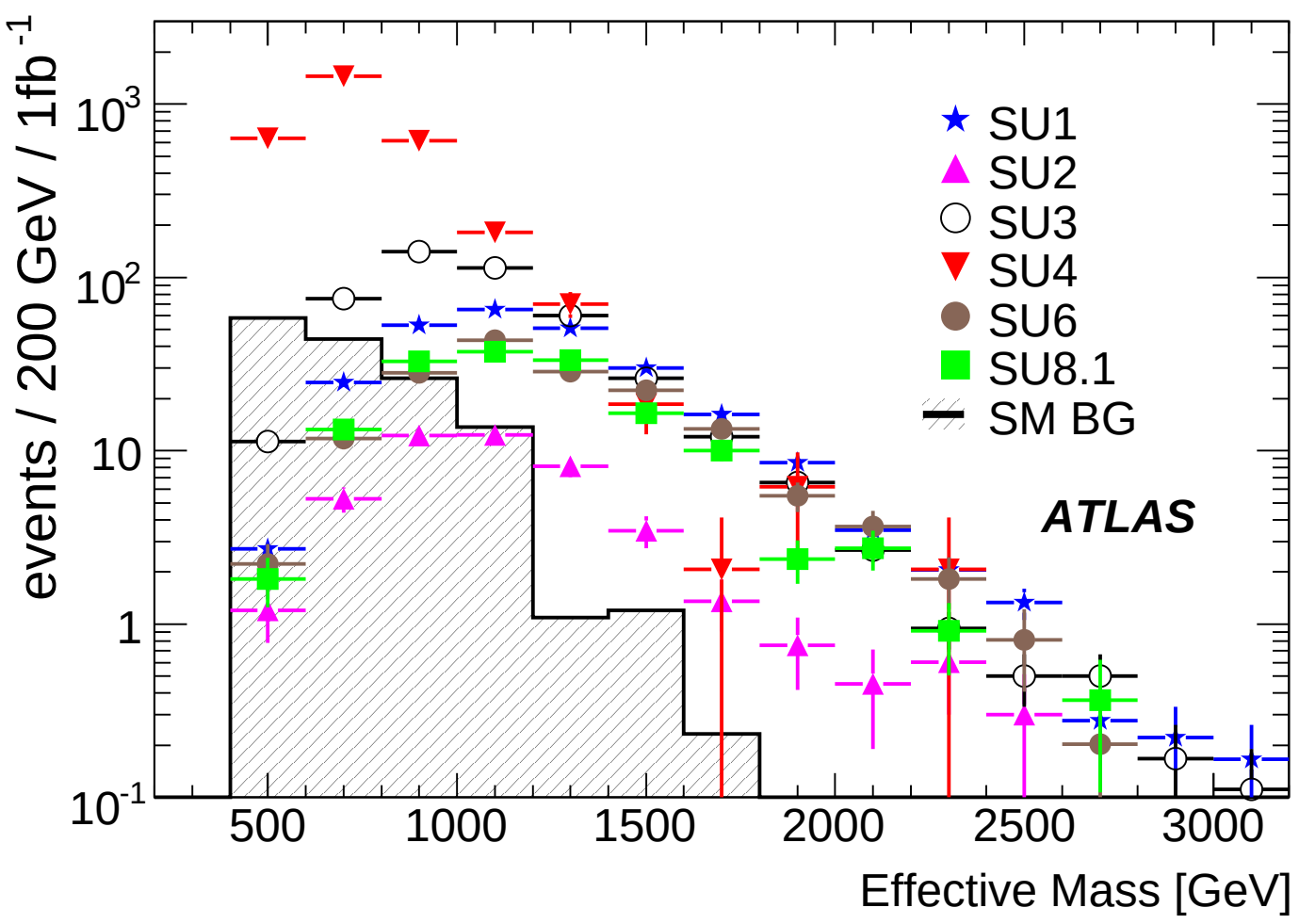


Figure 4: The $M_{\text{eff}}$ distributions for each of the SUn benchmark points, and for the sum of the Standard Model backgrounds with $1\,\text{fb}^{-1}$ for the 1-lepton analysis. All the cuts except on $M_{\text{eff}}$ are applied.

would expect. The estimated error on both of these backgrounds using data-driven methods is $\pm 20\%$ [2]. Given this and the calculated signal and background rates in Table 4, it is evident that all the SUSY points considered except SU2 could be discovered with good significance in the 1-lepton mode. For SU2, the production cross-section is dominated by gaugino pair production, so a different analysis [11] is required.

To make this conclusion more quantitative, the significance $Z_n$ defined in Section 1.2 was calculated. The central value of each background is taken from the current Monte Carlo simulation; the studies of data-driven background estimation [2, 3] provide estimated errors of $\pm 50\%$ for QCD multijet backgrounds and $\pm 20\%$ for $t\bar{t}$, $W+$jets, and all other backgrounds. The results of this calculation are shown in Table 5 for an integrated luminosity of $1\,\text{fb}^{-1}$. Each of these points except SU2 would have $Z_n > 5$ for just $100\,\text{pb}^{-1}$ if the same 20% background uncertainty could be obtained with that luminosity.

The significances, $Z_n$, (Table 5) for $1\,\text{fb}^{-1}$ are much smaller than the $S/\sqrt{B}$ values in Table 4. This reflects the fact that, unlike $S/\sqrt{B}$, the $Z_n$ measure of significance includes the estimated systematic uncertainty on the background. Table 5 also indicates that, provided the relative uncertainties in the background determiations did note increase, harder $M_{\text{eff}}$ cuts would lead to better significances after systematic uncertainties in the background are taken into account.

Table 5: Significance $Z_n$ for the 1-lepton plus 4-jet analysis with 1 fb$^{-1}$ including the systematic uncertainty in the background estimation.

| Sample | $M_{\rm eff} > 400$ GeV | | $M_{\rm eff} > 800$ GeV | | $M_{\rm eff} > 1200$ GeV | |
|---|---|---|---|---|---|---|
| | Events | $Z_n$ | Events | $Z_n$ | Events | $Z_n$ |
| Standard Model BG | 144 | | 42 | | 2 | |
| SU1 | 260 | 7.6 | 232 | 12.3 | 114 | 18.0 |
| SU2 | 46 | 1.5 | 40 | 3.4 | 15 | 6.0 |
| SU3 | 450 | 9.5 | 364 | 16.7 | 110 | 17.7 |
| SU4 | 2974 | 33.7 | 896 | 29.4 | 99 | 16.6 |
| SU6 | 162 | 4.9 | 148 | 8.9 | 76 | 14.2 |
| SU8.1 | 151 | 4.6 | 136 | 8.4 | 66 | 13.1 |

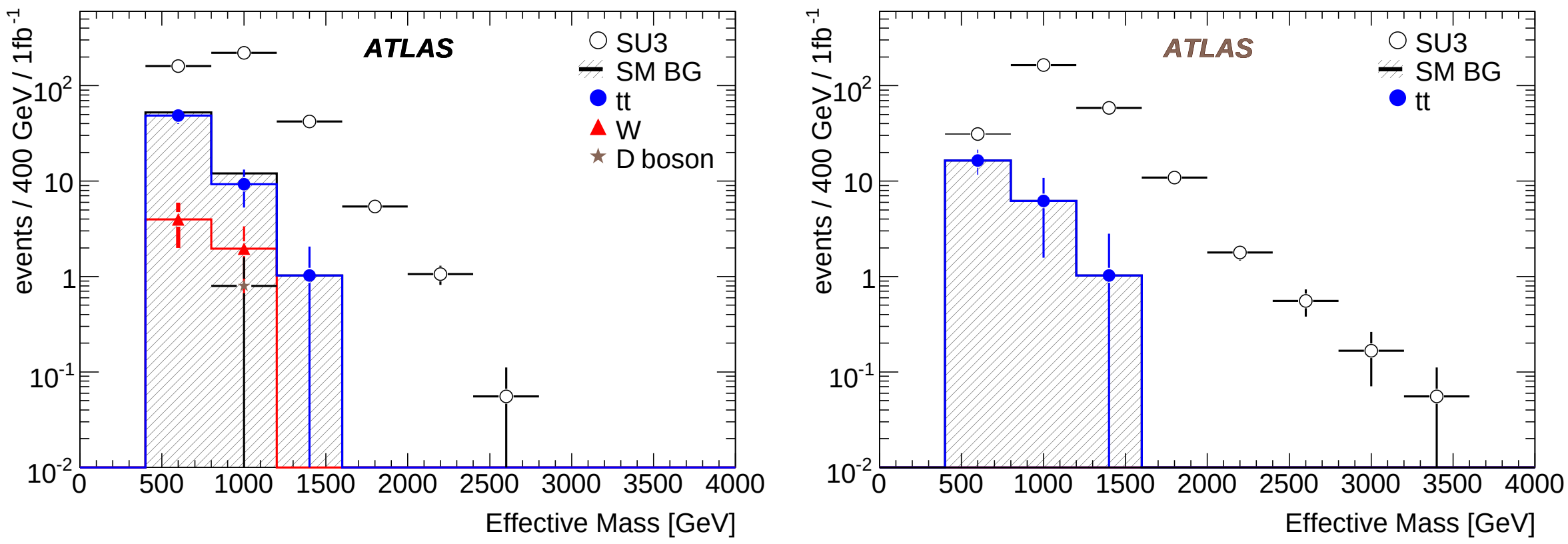


Figure 5: $M_{\rm eff}$ distribution for events with one lepton and: 2 jets (left) or 3 jets (right) after all cuts were applied.

Supersymmetry events need not contain large numbers of jets. For example, in mSUGRA the process

$$\tilde{q}_L + \tilde{q}_R \rightarrow \tilde{\chi}_1^{\pm} q' + \tilde{\chi}_1^0 q$$
$$\quad \hookrightarrow \tilde{\chi}_1^0 \ell^{\pm} \nu$$

can have a large rate and gives one lepton and just two hard jets. An alternative 1-lepton analysis has been performed requiring just two or three jets rather than four. Because a 4-jet selection was applied to the Alpgen samples at the generator level, Pythia was used for the $W +$jets backgrounds. The jet cuts employed are harder: 150 GeV for the leading jet and 100 GeV for the others. The missing energy cut is also harder, $E_{\rm T}^{\rm miss} > \max(100 \text{ GeV}, 0.3M_{\rm eff})$ and $\max(100 \text{ GeV}, 0.25M_{\rm eff})$ for the 2-jet and 3-jet case respectively. The $M_{\rm eff}$ distributions after all cuts are shown in Figure 5. Evidently an analysis requiring a smaller number of jets with harder cuts also can be effective.

Table 6: The number of events surviving the selection cuts defined in the text for the opposite-sign dilepton analysis, normalized to $1\text{fb}^{-1}$ using next-to-leading-order cross-sections. The last two columns give the $S/B$ ratio and the $Z_n$ significance, the latter of which includes the systematic uncertainties on the Standard Model backgrounds.

| Sample | Cuts 1-3 | Cut 4 | $S/B$ | $Z_n$ |
|---|---|---|---|---|
| SU3 | 200.8 | 159.8 | 1.88 | 3.55 |
| SU1 | 91.0 | 72.6 | 0.86 | 1.65 |
| SU2 | 22.5 | 18.8 | 0.22 | 0.43 |
| SU4 | 948.0 | 809.5 | 9.56 | 22.5 |
| $t\bar{t}$ | 111.1 | 81.5 | | |
| $W+$jets | 2.47 | 1.97 | | |
| $Z+$jets | 1.77 | 1.20 | | |
| QCD (J3-J7) | 0 | 0 | | |
| Total Standard Model | 115.34 | 84.67 | | |

# 4 Two-lepton mode

## 4.1 Opposite sign dileptons

Supersymmetry events with two opposite-sign leptons can arise from neutralino decays, especially $\chi_2^0 \to l^\pm l^\mp \chi_1^0$, either directly or through an intermediate slepton. Such dileptons must have the same flavour to avoid inducing $\mu \to e\gamma$ and other lepton-flavour-violating interactions at one loop. By contrast leptons produced from independent decays can give either same-flavour (OSSF) or different-flavour (OSDF) dilepton pairs, again with $\ell \in \{e, \mu\}$.

The opposite-sign dilepton analysis uses the following cuts:

1. Two isolated, opposite-sign leptons with $p_T > 10$ GeV and $|\eta| < 2.5$ which satisfy the cuts described in the introductory SUSY note [1]. Events containing additional leptons were vetoed.

2. At least four jets with $p_T > 50$ GeV at least one of which must have $p_T > 100$ GeV.

3. $E_{\mathrm{T}}^{\mathrm{miss}} > 100$ GeV and $E_{\mathrm{T}}^{\mathrm{miss}} > 0.2 M_{\mathrm{eff}}$.

4. Transverse sphericity, $S_T > 0.2$.

Cut 1 defines the opposite-sign dilepton sample, while Cuts 2 and 3 both suppress the Standard Model backgrounds and provide consistency with the Monte Carlo generator cuts on those backgrounds. After Cuts 1–3 the dominant background by far is $t\bar{t}$, as one would expect. The $S_T$ cut only increases the $S/B$ ratio by about 8% while reducing the signal by 20%.

The signals and backgrounds after the cuts, and the corresponding significances, are shown in Table 6. The number of events after all cuts includes the effect of the j70xE70 trigger, which, as shown in Table 1, has an efficiency of around 99% for all considered signal benchmarks, except for the low mass SU4 point, for which the efficiency is above 95%. The benchmark points SU3 and SU4 both have high discovery potential in the dilepton channel. While SU1 has fairly large dilepton branching ratios, many of the leptons are soft because of the small mass gaps between supersymmetric particles. An improved analysis based low-$p_T$ lepton reconstruction algorithms would help greatly for this point.

It is instructive to see how the significances, $Z_n$, vary with the cut on the leading jet and on $E_{\mathrm{T}}^{\mathrm{miss}}$. This is shown in Figure 6 for each of the points SU1 – SU4. It can be seen that the significance improves with

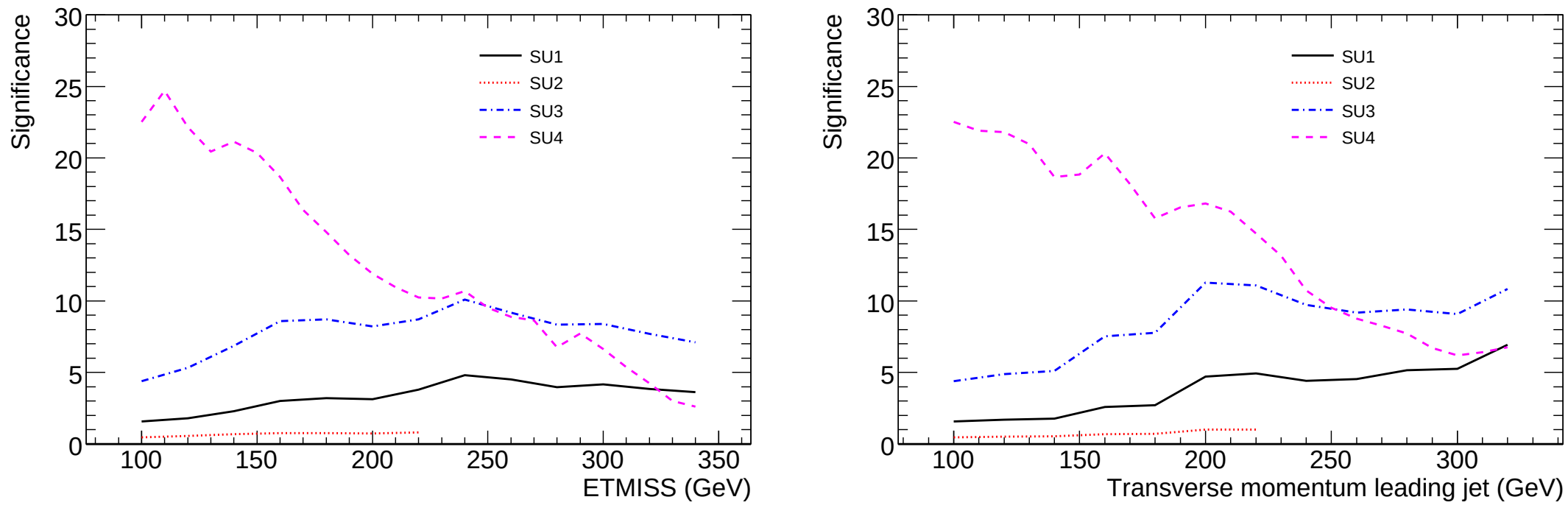


Figure 6: Significance of signal events for the four benchmark points, as a function of the cut on transverse missing energy (left) and the transverse momentum of the leading jet (right), for an integrated luminosity of 1 $\text{fb}^{-1}$.

Table 7: The optimized cuts for each point and corresponding signal, background, and significance. Compare with Table 6.

| Sample | $E_T^{\text{miss}}$ cut | Leading jet cut | signal | background | Significance |
|---|---|---|---|---|---|
| SU1 | 100 GeV | 320 GeV | 37.97 | 6.30 | 6.94 |
| SU2 | 140 GeV | 200 GeV | 13.74 | 22.68 | 1.07 |
| SU3 | 140 GeV | 200 GeV | 125.34 | 22.68 | 11.45 |
| SU4 | 110 GeV | 100 GeV | 772.53 | 66.80 | 24.70 |

harder cuts than those given in the above cut list even for the low-mass point SU4. The optimal cuts for each point and the signal, background and significance are shown in Table 7. Systematic errors of 50% on on all Standard Model backgrounds are included. Of course one should not optimize an analysis for a single point, but the table suggests that, provided the systematic uncertainties on the Standard Model background determinations do not significantly increase, harder cuts would be preferred. Optimization for wider ranges of points is discussed in Section 8.

Observing an non-resonant excess of OSSF dilepton events over OSDF events would be a clear indication of new physics. In SUSY leptonic $\tilde{\chi}_2^0$ decays can produce this excess, and have a characteristic endpoint set by the masses involved. The significance of the difference, calculated as $(N_{OSSF} - N_{OSDF})/\sqrt{N_{OSSF} + N_{OSDF}}$, is shown in Table 8. This significance calculation assumes that the relative $e$ and $\mu$ acceptances are well understood, which is not unreasonable given that all Standard Model processes satisfy $e/\mu/\tau$ universality. For SU1 the combined branching ratio for $\tilde{\chi}_2^0 \to \tilde{\ell}_{L,R}^{\pm}\ell^{\mp}$ is 11.7%, but the acceptance is reduced by the small mass gaps. For SU2 gaugino pair production dominates, so the jet cuts suppress the signal.

## 4.2 Same sign dileptons

In the Standard Model the rate for prompt, isolated, same-sign dileptons is small. Of course some leptons from hadronized heavy or light quarks can also pass the isolation cut and contribute like-sign backgrounds. In SUSY, on the other hand, the gluino is a self-conjugate Majorana fermion, so events containing like-sign dileptons can be common. Thus, same-sign dileptons are a good signature for SUSY and a characteristic feature of it.

The cuts used for the same-sign dilepton analysis are:

Table 8: The number of OSSF and OSDF dilepton events passing the optimized cuts and the corresponding statistical significance for $1\,\text{fb}^{-1}$.

| Sample | $E_T^{\text{miss}}$ cut | Leading jet cut | $N_{OSSF}$ | $N_{OSDF}$ | Significance |
|---|---|---|---|---|---|
| SU1 | 220 GeV | 100 GeV | 90.69 | 58.53 | 2.63 |
| SU2 | 140 GeV | 100 GeV | 31.64 | 29.95 | 0.22 |
| SU3 | 160 GeV | 160 GeV | 93.75 | 38.58 | 4.80 |
| SU4 | 120 GeV | 100 GeV | 392.45 | 281.55 | 4.27 |

Table 9: The number of events surviving the selection cuts (as defined in the text) for the same-sign dileptons analysis, normalized to $1\text{fb}^{-1}$ using next-to-leading-order cross-sections. No background events pass the final cut; the 90% upper limit for $t\bar{t}$ background is given.

| Process | Cuts 1–3 | Cut 4 | $Z_n$ |
|---|---|---|---|
| SU1 | 30.1 | 21.9 | 7.2 |
| SU2 | 13.0 | 6.6 | 1.9 |
| SU3 | 37.9 | 24.9 | 7.7 |
| SU4 | 251.8 | 138.8 | 19.9 |
| SU6 | 18.0 | 13.9 | 4.5 |
| $t\bar{t}$ | 2.1 | $< 2.3$ | |
| $W+\text{jets}$ | 0.7 | 0.0 | |
| $Z+\text{jets}$ | 0.0 | 0.0 | |

1. Exactly 2 same-sign leptons with $p_T > 20$ GeV satisfying the usual isolation and other cuts [1].
2. At least four jets with $p_T > 50$ GeV at least one of which must have $p_T > 100$ GeV.
3. Transverse missing energy $E_T^{\text{miss}} > 100$ GeV.
4. $E_T^{\text{miss}} > 0.2M_{\text{eff}}$.

The first cut defines the same-sign dilepton sample, while Cuts 2-4 suppress the Standard Model backgrounds. Cuts 1 and 2 are used in the Monte Carlo generator filters for some of the backgrounds.

The cut flow table for these cuts is shown in Table 9. The number of events after all cuts includes the effect of the j70xE70 trigger, which, as shown in Table 1, has an efficiency of around 99% for all considered signal benchmarks, except for the low mass SU4 point, for which the efficiency is 84%.

Since the $W+\text{jets}$ and $Z+\text{jets}$ backgrounds have been filtered at the generator level, the results for these are biased until after Cut 3, but evidently they are small. A number of other backgrounds were examined and found to be negligible compared to those listed in the table. None of the Monte Carlo events generated for the Standard Model background determination passed all the cuts. A 90% confidence upper limit of 2.3 Monte Carlo events gives the indicated upper limit on the $t\bar{t}$ background after cut 4. This is used as the estimate of the total background since $t\bar{t}$ is expected to dominate; $b$ jets can produce a second lepton of the same sign and that lepton has a non-negligable probablility of being well-isolated. The assumption of $t\bar{t}$ dominance is consistent with the results after Cut 3 in Table 9. Two possible backgrounds, $W^{\pm}W^{\pm}$ and $t\bar{t}t\bar{t}$, are probably small but have not been studied.

The $E_T^{\text{miss}}$ distributions after the other cuts are shown in Figure 7. While the rates are small, the $S/B$ ratio is good and the signal is distinctive. The $E_T^{\text{miss}}$ cut was varied and the value in the cuts used here

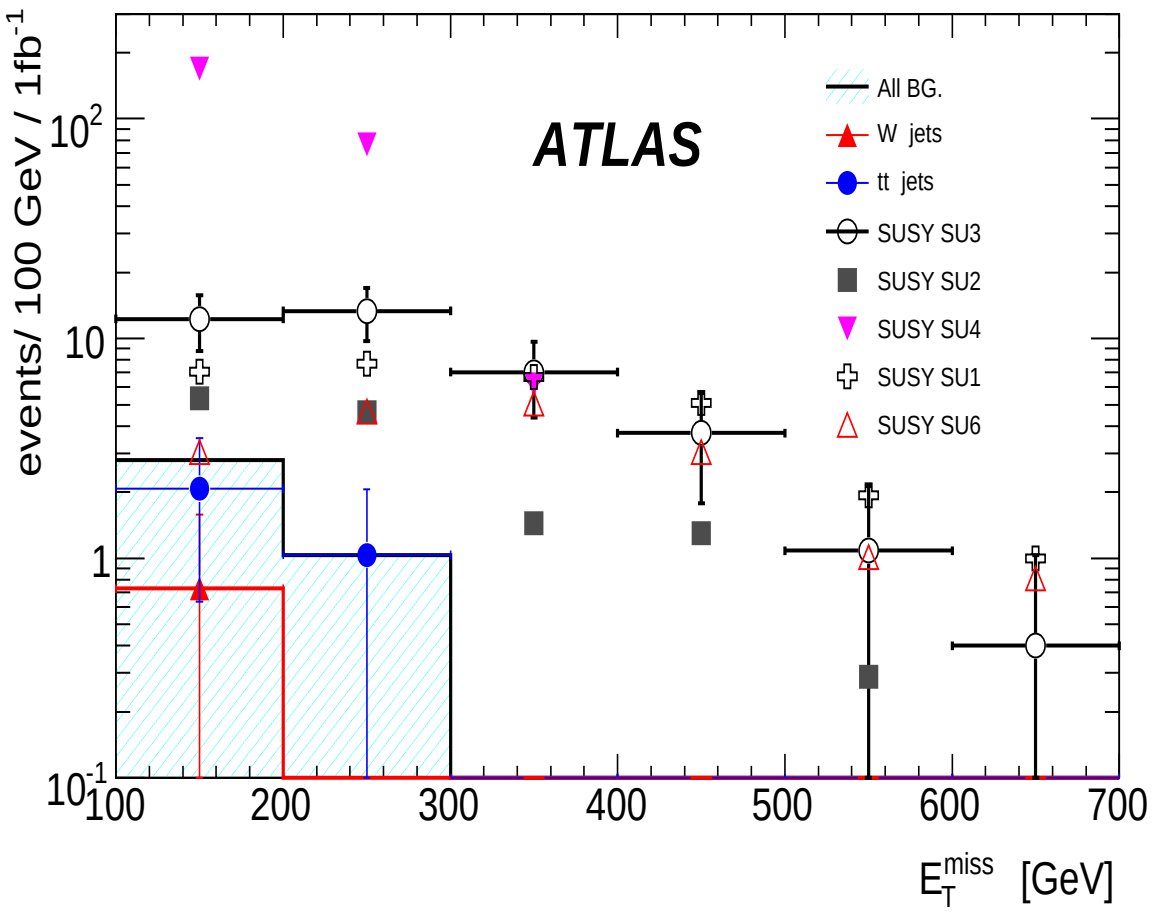


Figure 7: $E_{\mathrm{T}}^{\mathrm{miss}}$ in the same sign dilepton events after all cuts except the $E_{\mathrm{T}}^{\mathrm{miss}}$ cut.

found to be appropriate.

It is clear that the Standard Model same-sign dilepton background is small and is probably dominated by $t\bar{t}$. A data-driven analysis of this background has not yet been done. It is expected that it will be possible to measure the background from processes such as $t \to \ell^+ X$, $\bar{b} \to \mu^+ X$ as a function of the isolation cut and to extrapolate to the cut used here. For $\bar{b} \to e^+ X$, where the $e$ identification cuts impose an implicit isolation cut, it will be necessary to extrapolate from the $\mu$ result using Monte Carlo techniques. In the absence of such studies the significance for the same-sign dilepton analysis has been calculated using the 90% upper limit for $t\bar{t}$ given in Table 9 with the standard systematic uncertainty of $\pm 20\%$. This gives the $Z_n$ values listed in the same table. Although the systematic error on the background is uncertain, certainly SU4 and very likely SU1 and SU3 would be observable with a significance greater than $5\sigma$ with $1\,\mathrm{fb}^{-1}$.

More work on the same-sign dilepton background, and more generally on estimates of leptons from $b$ and $c$ decays passing isolation cuts, is clearly needed.

# 5 Three-lepton mode

The trilepton signal from direct gaugino production [14] is perhaps the best search mode for SUSY at the Tevatron [15, 16]. The corresponding search with ATLAS is described elsewhere in this volume [11]. The analyses discussed here are aimed at trilepton production from all sources, not just from direct production. Two approaches have been followed. The first, the 3-leptons + jet selection makes explicit use of a high-$p_{\mathrm{T}}$ jet, similar to the 1- and 2-lepton analyses described above. The second, the 3-leptons + $E_{\mathrm{T}}^{\mathrm{miss}}$ selection, relies on track isolation cuts to select prompt leptons and is similar to the exclusive analysis [11]. The 1LEP trigger typically gives an efficiency of $\gtrsim 95\%$ for these modes (see Table 1).

## 5.1 Three-lepton + jet analysis

The 3-leptons + jet selection requires:

1. At least three leptons with $p_T > 10$ GeV satisfying the usual identification and isolation cuts [1].

2. At least one jet with $p_T > 200$ GeV.

No $E_{\rm T}^{\rm miss}$ cut is made, so this analysis could be used even if detector problems seriously degraded the $E_{\rm T}^{\rm miss}$ performance. The jet cut is sufficient to suppress the $WZ$ and $W\gamma^*$ backgrounds, so cuts on the invariant mass of opposite-sign same-family dilepton pairs are also not required.

Table 10: The numbers of surviving SUSY and Standard Model events for the benchmark points SU2, SU3 and SU4, as the "3-leptons+jet" inclusive trilepton selection is applied. All numbers are nomalized to 1 fb$^{-1}$ of integrated luminosity.

| Sample | Cut 1 | Cut 2 | $S/B$ | $S/\sqrt{B}$ | $Z_n$ |
|---|---|---|---|---|---|
| SU2 | 35 | 13 | 1.1 | 3.7 | 2.7 |
| SU3 | 139 | 94 | 7.8 | 27.1 | 11.5 |
| SU4 | 1284 | 312 | 26.0 | 90.0 | 24.4 |
| $t\bar{t}$ | 455 | 11 | – | – | – |
| $ZZ$ | 59 | 0 | – | – | – |
| $ZW$ | 193 | 1 | – | – | – |
| $WW$ | 3 | 0 | – | – | – |
| $Z+\gamma$ | 9 | 0 | – | – | – |
| $Zb$ | 656 | 0 | – | – | – |

The cut flow for this selection is shown in Table 10. The number of events after all cuts includes the effect of the 1LEP trigger, which, as shown in Table 1, has an efficiency of around 95% for all of the benchmark points considered. The jet cut (Cut 2) particularly reduces the $ZW$ and $Zb$ backgrounds in which the jets tend to be soft. The dominant background after all cuts is $t\bar{t}$, but there is also a small remaining background from $WZ$. The same table shows the statistical significance $S/\sqrt{B}$ and the significance $Z_n$ including a background uncertainty of 20%. As has already been discussed for the same-sign dilepton selection, the key issue for background determination is the estimation of leptons from $b \to \ell X$ passing the isolation cut in $t\bar{t}$ events. We expect that this could be measured as a function of the isolation cut for $\mu$ and then applied to $e$ using Monte Carlo simulation. Given the large $S/B$ in Table 10, even a 100% background uncertainty would yield $Z_n > 5$ for points SU3 and SU4.

Adding a cut on $E_{\rm T}^{\rm miss}$ to this analysis was investigated. The surviving background events after the trilepton and jet cuts have a wide range of $E_{\rm T}^{\rm miss}$, so a cut to reduce them would also reduce the already rather small signal.

## 5.2 Three-lepton + $E_{\rm T}^{\rm miss}$ analysis

The 3-leptons $+ E_{\rm T}^{\rm miss}$ selection does not require (or veto) jets, so it is sensitive to direct gaugino production as well as to trileptons produced in the decays of squarks and gluinos. The analysis cuts have been somewhat optimized for SU2, for which gaugino pair production dominates. Since the dominant source of trileptons in SUSY includes a decay $\tilde{\chi}_2^0 \to \tilde{\chi}_1^0 \ell^+\ell^-$, at least one OSSF lepton pair is required among the three leptons.

The cuts for this analysis are:

1. $N_\ell \geq 3$ leptons with $p_T > 10$ GeV satisfying the usual identification and isolation cuts [1].

2. At least one OSSF dilepton pair with $M > 20$ GeV to suppress low-mass $\gamma^*$, $J/\psi$, $\Upsilon$, and conversion backgrounds.

3. Lepton track isolation: $p_{T,\rm trk}^{0.2} < 1$ GeV for muons and $< 2$ GeV for electrons, where $p_{T,\rm trk}^{0.2}$ is the maximum $p_T$ of any additional track within a cone $R = 0.2$ around the lepton.

Table 11: Expected event numbers for the 3-leptons + $E_{\mathrm{T}}^{\mathrm{miss}}$ analysis for $1\,\mathrm{fb}^{-1}$ for signal and background processes. The $WW$ and $Z\gamma$ backgrounds are small and so are not listed.

| Process | Cuts 1-2 | Cut 3 | Cut 4 | Cut 5 |
|---|---|---|---|---|
| SU1 | 42.2 | 33.0 | 32.6 | 24.1 |
| SU2 | 29.8 | 24.1 | 21.1 | 17.6 |
| SU3 | 130.1 | 101.2 | 98.6 | 63.9 |
| SU4 | 968.1 | 691.5 | 654.3 | 544.9 |
| SU8.1 | 10.2 | 8.0 | 8.0 | 5.3 |
| $WZ$ | 188.3 | 166.2 | 122.5 | 22.8 |
| $ZZ$ | 55.9 | 46.4 | 10.3 | 1.6 |
| $Zb$ | 582.5 | 221 | 1.3 | 0 |
| $t\bar{t}$ | 283.2 | 59.9 | 56.6 | 47.9 |

Table 12: Number of signal ($S$) and background ($B$) events surviving the 3-leptons + $E_{\mathrm{T}}^{\mathrm{miss}}$ selection and the corresponding values for $S/\sqrt{B}$ and $Z_n$. All numbers are normalized to 1 fb$^{-1}$ .

| | SU1 | SU2 | SU3 | SU4 | SU8 |
|---|---|---|---|---|---|
| $S$ | 24.1 | 17.6 | 63.9 | 544.9 | 5.3 |
| $B$ | | | 73.5 | | |
| $S/\sqrt{B}$ | 2.8 | 2.1 | 7.5 | 63.5 | 0.6 |
| $Z_n$ | 1.3 | 1.0 | 3.5 | 16.4 | 0.3 |

4. $E_{\mathrm{T}}^{\mathrm{miss}} > 30$ GeV.

5. $M < M_Z - 10$ GeV for any OSSF dilepton pair.

Cut 3 provides an additional rejection of leptons from $b$ and $c$ decays beyond the calorimeter isolation cut, while Cut 4 reduces Standard Model backgrounds containing a $Z$.

The significances $S/\sqrt{B}$ and $Z_n$ for this second analysis are shown in Table 12, where the $Z_n$ significance includes the standard 20% background systematic uncertainty. A detailed study of the performance of the lepton isolation cuts is needed to understand the uncertainty on the $t\bar{t}$ background in particular. Since this analysis does not require jets, one might hope that it would be sensitive to the dominant gaugino pair production for SU2, but only SU4 gives a signal with $Z_n > 5$ for $1\,\mathrm{fb}^{-1}$. For all of the benchmark points studied the 3-leptons + jet analysis is more sensitive than the 3-leptons + $E_{\mathrm{T}}^{\mathrm{miss}}$ one.

## 6 Tau mode

SUSY models generically violate $e/\mu/\tau$ universality; $\tau$ decays can even be dominant, especially for $\tan\beta \gg 1$. Hence it is worthwhile to look for signatures involving hadronic $\tau$ decays even though the fake background from jets is much larger than that for $e$ or $\mu$. Leptonic $\tau$ decays are indistinguishable from prompt leptons and are already included in the previous analyses.

The cuts used in this analysis are:

1. At least four jets with $p_T > 50$ GeV and at least one with $p_T > 100$ GeV.

Table 13: The number of signal ($S$) and background ($B$) events after tau selection and corresponding values of significance, normalised to $1\,\text{fb}^{-1}$.

| Sample | $S$ | $B$ | $S/B$ | $S/\sqrt{B}$ | $Z_n$ |
|---|---|---|---|---|---|
| SU3 | 259 | 51 | 5.1 | 36.3 | 12 |
| SU6 | 119 | 51 | 2.3 | 16.7 | 6.8 |

2. $E_{\mathrm{T}}^{\mathrm{miss}} > 100$ GeV.
3. $\Delta\phi(j_i, E_{\mathrm{T}}^{\mathrm{miss}}) > 0.2$ for each of the three leading jets $j_i$, $i = 0, 1, 2$.
4. No isolated leptons using the standard cuts [1].
5. At least one $\tau$ with $p_T > 40$ GeV and $|\eta| < 2.5$ reconstructed by the high $p_T$ $\tau$ algorithm [17] with a likelihood, $L > 4$.
6. $E_{\mathrm{T}}^{\mathrm{miss}} > 0.2M_{\mathrm{eff}}$.
7. $M_T > 100$ GeV, where $M_T$ is calculated using the visible momentum of the hardest $\tau$ and $E_{\mathrm{T}}^{\mathrm{miss}}$.

Cuts 1, 2, and 6 are standard. Cut 3 requires a large $\Delta\phi$ between $E_{\mathrm{T}}^{\mathrm{miss}}$ and the leading jets, thus reducing the background both from mismeasured jets and from $b$ and $c$ decays. Cut 4 makes this analysis disjoint from the 1, 2, and 3-lepton analyses described above. There is still overlap with the 0-lepton analysis described in Section 2. Cut 5 defines the $\tau$ sample; these cuts give an efficiency of $\sim 50\%$ with a purity of $\sim 80\%$ for the SU3 sample. Finally, if the $E_{\mathrm{T}}^{\mathrm{miss}}$ comes from one $W \to \tau\nu$ decay, then $M_T$ used in Cut 7 should satisfy $M_T < m_W$. The applied cuts are a superset of the basic cuts for the inclusive 4-jet 0-lepton analysis, which does not employ a $\tau$ veto and therefore the events selected for this analysis will have an almost complete overlap with the ones selected in the analysis described in the corresponding section. For the same reason, the efficiency of the j70xE70 trigger is expected to be between 97% and $\sim$100% for all the benchmark points, as was the case for the inclusive multi-jet analysis.

The effect of these cuts is indicated graphically in Figure 8. The requirement of a reconstructed $\tau$ (Cut 5) eliminates the QCD background. After the $M_T$ cut the $S/B$ ratio is high. The resulting signal, background, and significance for points SU3 and SU6 are given in Table 13 assuming the usual 20% systematic uncertainty for the background, which is dominated by $t\bar{t}$ with some contribution from $W+$ jets.

The data-driven uncertainty on $\tau$ SUSY backgrounds has not yet been studied. Clearly $\tau$ reconstruction is difficult. However, it should be possible to simulate real $\tau$ backgrounds by selecting backgrounds with reconstructed $e$ and $\mu$ and replacing the leptons with simulated $\tau$ decays. Fake backgrounds can be similarly determined using reconstructed events combined with the measured jet $\to \tau$ fake rate. If the resulting uncertainty on the background is about 20%, as is assumed in Table 13, then both points SU3 and SU6 would be observable in the $\tau$ mode.

## 7 $b$-jet mode

SUSY signals are typically rich in $b$ quarks because the $\tilde{b}$ and $\tilde{t}$ tend to be lighter than first- and second-generation squarks and because Higgsino couplings enhance heavy flavour production. In the benchmark points studied the fractions of events containing $b$ jets range from 14.4% for SU2 to 72.8% for SU4. In QCD events $b$ quarks typically occur at the percent level. Thus, requiring a $b$ quark suppresses the QCD background, which may be difficult to control, just as requiring an $e$ or $\mu$ does.

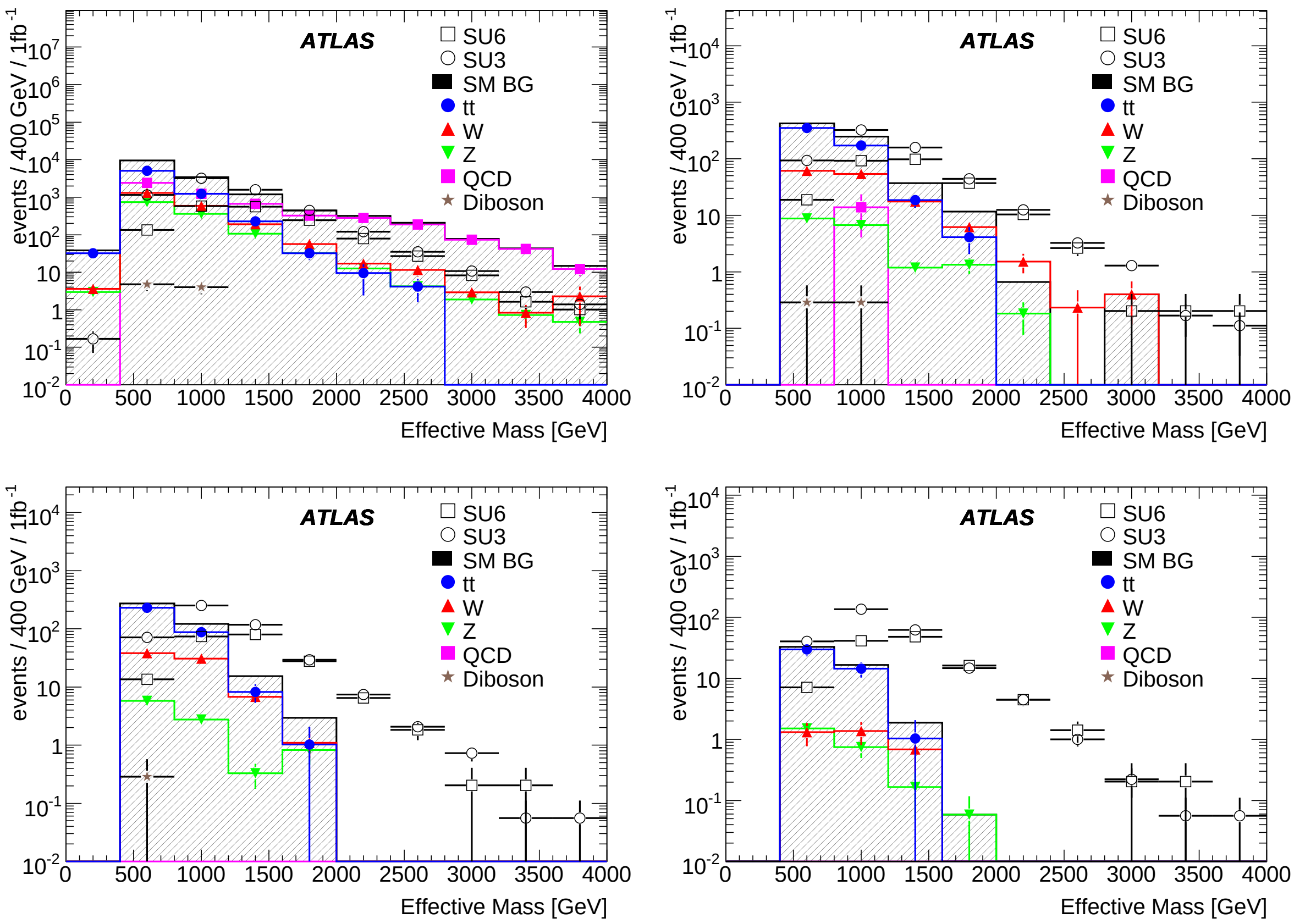


Figure 8: The $M_{\text{eff}}$ distributions for SUSY signals and Standard Model backgrounds in the $\tau$ analysis after Cuts 4, 5, 6, and 7.

In this section an analysis of signatures with $b$ jets is performed for SUSY points SU1, SU3, SU4 and SU6 using full simulation both for the signal and for the Standard Model backgrounds. Isolated leptons may also be present, and all channels with and without leptons are summed. SUSY processes almost always will give $b\bar{b}$ pairs, and this is taken into account. No equivalent analysis was performed in the Physics TDR.

The cuts used in this analysis are as follows:

1. At least 4 jets in the event with $p_{\text{T}} > 50$ GeV.
2. Leading jet $p_{\text{T}} > 100$ GeV.
3. Missing transverse energy, $E_{\text{T}}^{\text{miss}} > 100$ GeV.
4. Missing transverse energy, $E_{\text{T}}^{\text{miss}} > 0.2M_{\text{eff}}$.
5. Transverse sphericity, $S_T > 0.2$.
6. At least 2 jets are tagged as $b$ jets, as described below.
7. $M_{\text{eff}} > 600$, 800, or 1000 GeV.

Note that Cuts 1–3 are also used in Monte Carlo generator filters for some of the background samples. Cut 7 is used to optimize the signal-to-background ratio in the selected events.

Table 14: Number of events surviving selection cuts as defined in the text for the inclusive search with $b$ jets normalized to $1\text{fb}^{-1}$ using NLO cross-sections. The results are shown for three different values of the $M_{\text{eff}}$ cut (Cut 7).

| Sample | Cuts 1–3 | Cut 4 | Cut5 | Cut 6 | Cut 7 | | |
|---|---|---|---|---|---|---|---|
| | | | | | 600 GeV | 800 GeV | 1000 GeV |
| SU1 | 3469 | 2806 | 1994 | 456 | 442 | 375 | 263 |
| SU2 | 608 | 358 | 299 | 170 | 166 | 141 | 87 |
| SU3 | 9357 | 7279 | 5474 | 1158 | 1086 | 818 | 425 |
| SU4 | 79761 | 56697 | 45661 | 16478 | 10204 | 3186 | 926 |
| SU6 | 2557 | 2049 | 1467 | 505 | 495 | 436 | 340 |
| $t\bar{t}$ | 12864 | 8273 | 6117 | 2182 | 836 | 215 | 61 |
| QCD | 29435 | 7402 | 5171 | 740 | 259 | 79 | 5 |
| $W+$jets | 4068 | 2309 | 1600 | 23 | 16 | 7 | 2 |
| $Z+$jets | 1249 | 680 | 432 | 5 | 3 | 1 | 1 |
| Diboson | 22 | 13 | 8 | 2 | 1 | 0 | 0 |
| $B_{SM}$ | 47527 | 18676 | 13328 | 2950 | 1115 | 303 | 69 |

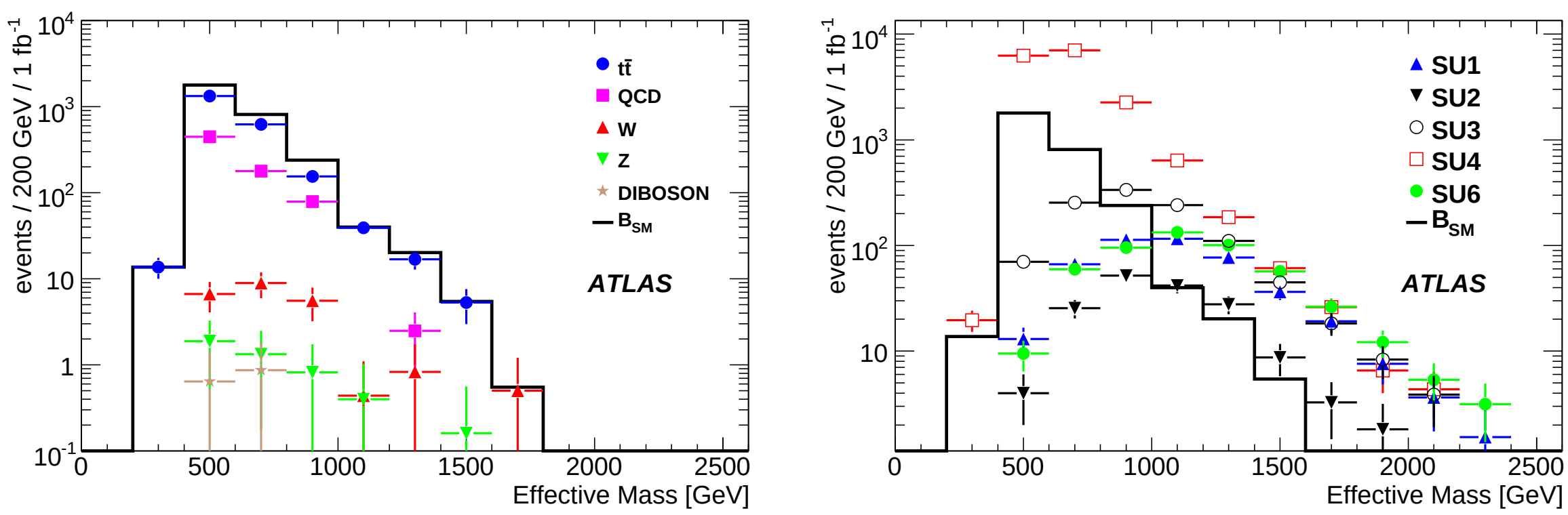


Figure 9: $M_{\text{eff}}$ distributions for $b$-jet analysis. Left: Standard Model backgrounds. Right: SUSY signals with total background.

Jets with $p_T > 20$ GeV are selected as $b$ jets using the default tagging algorithm based on the 3-dimensional impact parameter and secondary vertex detection [18] with a cut weight $> 6.75$, giving a nominal efficiency of 60%. Above about $p_T = 100$ GeV both the efficiency and the light-jet rejection decrease as discussed in the introductory SUSY note [1] and references therein. Naively one would expect that the increase of the $B$ decay length with $\gamma = E_B/M_B \gg 1$ would offset the $1/\gamma$ decrease of the angles since the multiple scattering angular errors also decrease similarly. There is also a substantial dependence of $b$ tagging on the $\eta$ of the jet. Many of the $b$ jets in SUSY events have high $p_T$: the typical $p_T$ of the leading jet in SU3 is about 300 GeV, for which the light-jet rejection during $b$ tagging is $\mathcal{O}(100)$.

Events with zero or more leptons and at least two tagged $b$ jets were combined in a single inclusive analysis. Inevitably this means that there is overlap with the analyses in Sections 2–5. The cut flow is shown in Table 14. After Cut 6 ($N_b \geq 2$) the $t\bar{t}$ background is dominant, as one might expect, but the

Table 15: $S/B$ ratio and signal significance $Z_n$ including systematic effects for the $b$-jet analysis with $0.1\,\text{fb}^{-1}$ and $1\,\text{fb}^{-1}$ with $M_{\text{eff}} > 1000$ GeV.

| | $S/B$ | $Z_n$ for $0.1\,\text{fb}^{-1}$ | $Z_n$ for $1\,\text{fb}^{-1}$ |
|---|---|---|---|
| SU1 | 3.8 | 6.0 | 9.3 |
| SU2 | 1.3 | 2.3 | 5.0 |
| SU3 | 6.2 | 7.5 | 13.0 |
| SU4 | 13.4 | 12.6 | 21.7 |
| SU6 | 4.9 | 7.1 | 11.2 |

QCD background remains substantial. The $b$-tagging performance at high $p_T$ is clearly an important issue for this analysis. As the applied cuts are a superset of the basic cuts for the inclusive 4-jet analyses, the efficiency of the j70xE70 trigger is, as quoted in the corresponding section, between 97% and ∼100% for all the considered benchmark points.

To calculate the significance in this channel an uncertainty of 50% for the QCD background and 20% for the other backgrounds is assumed for $1\,\text{fb}^{-1}$. The uncertainty on the $b$-tagging efficiency of 60% is 5% [2, 18]. This is assumed to be included in the $t\bar{t}$ uncertainty and is ignored for the other, smaller backgrounds.

The hardest effective mass cut, $M_{\text{eff}} > 1000$ GeV, was found to be the most effective, so only results for it are shown here. The resulting significances, $Z_n$, including the above systematic effects, are shown in Table 15, for two luminosities: 0.1 and $\text{fb}^{-1}$, where the same systematic uncertainty is assumed to be the same for both luminosities. All background uncertainties are added linearly. The low-mass point SU4 and perhaps also SU3 could be discovered using this analysis with only $0.1\,\text{fb}^{-1}$ assuming that the background could be understood adequately. All points except SU2 could be discovered with $1\,\text{fb}^{-1}$, for which the background uncertainties are realistic. This analysis seems to be particularly useful compared to some other analyses for point SU6.

# 8 Scans and optimization

The SUSY points studied so far were chosen to give a variety of signatures, but there is no reason to think that they are representative of what might be found at the LHC. This section uses scans over the parameters of several models for SUSY breaking – all with $R$ parity conservation – in order to sample a wider range of possibilities. The goal is to develop one or more search strategies covering as wide a subset of the scanned models as possible. Since each scan includes hundreds of points, this section must rely on ATLFAST [19], the fast parameterized simulation of the ATLAS detector.

Data-driven methods [2,3] will be used to determine the Standard Model backgrounds to the possible SUSY signatures. For $1\,\text{fb}^{-1}$ the estimated errors [2, 3] are typically 50% for QCD jets and 20% for the $W$, $Z$, and $t$ backgrounds. Several approaches were considered to look for an excess above a cut on $M_{\text{eff}}$ or $E_{\text{T}}^{\text{miss}}$ after basic jet and lepton selections. The significance is corrected for multiple cuts as described in Section 1.2. Results are shown here only for the $M_{\text{eff}}$ cut, that yielded best performance. A multivariate optimization using TMVA [20] gave a minor improvement with the available Monte Carlo statistics. This and other cut procedures are still being studied.

The analyses described above in this note have used signal and background cross-sections normalized to next-to-leading-order calculations [1]. This was impractical for scans over many points, each involving many subprocesses. The goal here is not to determine the exact limit or exclusion value but rather to test whether the proposed approaches work for a wide range of models. It was therefore decided to normalize

the signal cross-sections for all scans to the leading-order HERWIG values but to use next-to-leading-order normalizations for the backgrounds. Since next-to-leading-order corrections generally increase cross-sections, the resulting reach estimates are conservative.

## 8.1 SUSY signal samples

It is impossible to scan the 105-dimensional parameter space of the MSSM or even the 19 dimensional subspace with flavour and *CP* conservation and degeneracy of the first two generations. Hence a number of SUSY-breaking models with many fewer parameters were used.

Several of these scans (e.g. the first two mSUGRA scans listed below) ignore dark matter and other existing constraints. Of course any true theory must obey such constraints. It is possible, however, to modify the SUSY breaking model to satisfy the constraints while keeping the basic phenomenology unchanged. One such example, the non-universal-Higgs model (NUHM), is discussed below. Since there is no unique model of SUSY-breaking, all these scans should be viewed only as possible patterns of LHC signatures, not as complete theories.

**mSUGRA fixed grid,** $\tan\beta = 10$**,** $A_0 = 0$**,** $\mu > 0$**:** A $25 \times 25$ grid was made varying $m_0$ from 60 GeV to 2940 GeV in 25 steps of 120 GeV, and $m_{1/2}$ from 30 GeV to 1470 GeV in 25 steps of 60 GeV. SUSY spectra were generated using ISAJET 7.75 [21] with a top quark mass of 175 GeV. Out of the 625 possible points, a spectrum could be successfully generated for 600; the other 25 failed for theoretical reasons. For each good point 20k events were produced using ATLFAST. Constraints other than from direct searches were ignored. While constraints such as the dark-matter relic density constrain specific SUSY-breaking models such as mSUGRA, they are much less restrictive for generic models.

**mSUGRA fixed grid:** $\tan\beta = 50$**,** $A_0 = 0$**,** $\mu < 0$**:** Large $\tan\beta$ increases the mixing of $\tilde{b}_{L,R}$ and $\tilde{\tau}_{L,R}$, leading to enhanced $b$ and $\tau$ production. A grid of $25 \times 25$ points with was generated with $m_0$ varied from 200 to 3000 GeV in steps of 200 GeV and with $m_{1/2}$ varied from 100 to 1500 GeV in steps of 100 GeV. The top mass was fixed at 175 GeV. Constraints other than from direct searches were again ignored.

**mSUGRA random grid with constraints:** In this sample all mSUGRA parameters were varied in two regions[2] previously found [22] to be compatible with dark-matter and other constraints with $\mu > 0$ and $m_t = 175$ GeV.

The mSUGRA parameters were chosen randomly (with $\mu > 0$) and their properties calculated using ISAJET 7.75. All selected points satisfy the LEP Higgs mass limit, $m_h > 114.4$ GeV [23]; the WMAP total dark matter limit, $\Omega h^2 < 0.14$ [24]; within $3\sigma$ the branching ratio limits $B(b \to s\gamma) = (3.55 \pm 0.26) \times 10^{-4}$ [25] within $3\sigma$ and $B(B_s \to \mu^+\mu-) < 1.5 \cdot 10^{-7}$ [26]; and with $\delta a_\mu$ less than the $3\sigma$ upper limit from the muon anomalous magnetic moment measurement $a_\mu = (11659208 \pm 6) \times 10^{-10}$ [27].

**GMSB grid:** $M_{\text{mess}} = 500$ TeV**,** $N_{\text{mess}} = 5$**,** $C_{\text{grav}} = 1$**:** With $N_{\text{mess}} = 5$ the NLSP is a slepton which decays promptly to leptons or $\tau$'s. A fixed grid was made varying $\Lambda$ was varied from 10 TeV to 80 TeV in steps of 10 TeV and $\tan\beta$ from 5 to 40 in steps of 5.

**NUHM grid:** The NUHM model is similar to the mSUGRA model but does not assume that the Higgs masses unify with the squark and slepton ones at the GUT scale. This allows more gaugino/Higgsino mixing at the weak scale and so relaxes the mSUGRA dark matter constraints. The scan uses a step size

[2] The parameters are varied within $\{0 < m_0 < 2$ TeV, $0.5 < m_{1/2} < 1.3$ TeV, $-0.34 < A_0 < 2.4$ TeV, $39 < \tan\beta < 55\}$ and $\{1 < m_0 < 3$ TeV , $m_{1/2} < 0.5$ TeV , $-2.0 < A_0 < 2.0$ TeV, $20 < \tan\beta < 55\}$

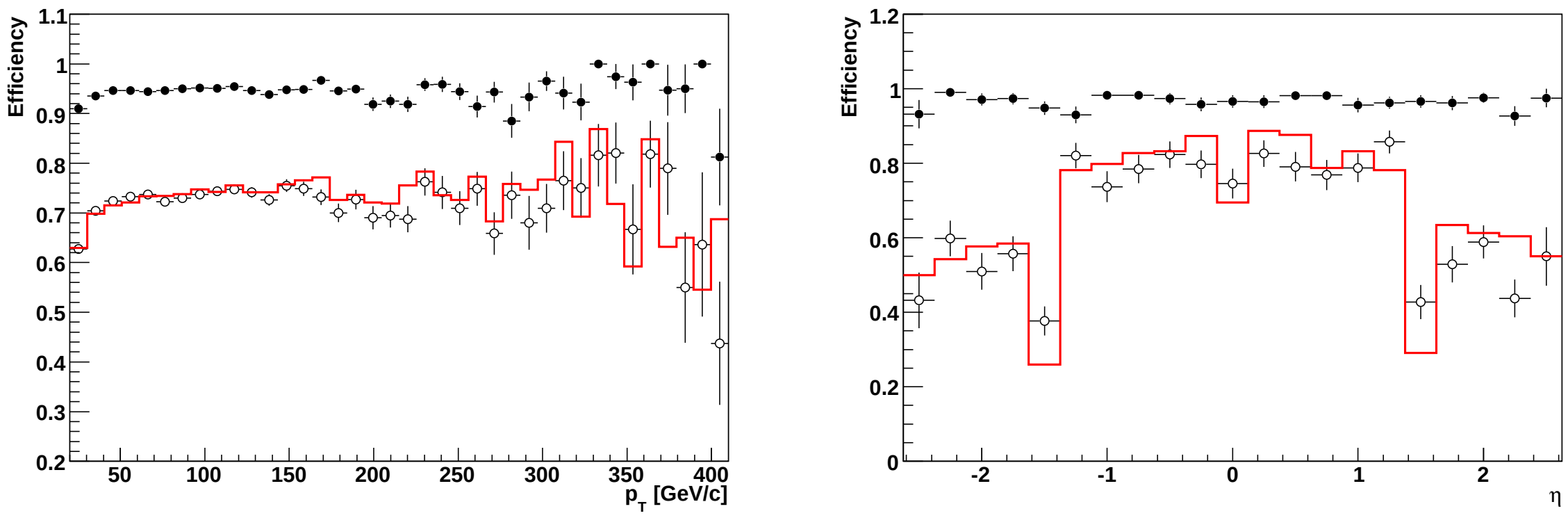


Figure 10: Efficiencies for electrons as a function of $p_T$ for the SU3 sample (left) and $\eta$ for the ALPGEN sample $Z \rightarrow ee$ (right). The solid red line corresponds to the Geant 4 simulation, the solid circles to uncorrected ATLFAST, and the open circles to the corrected version of ATLFAST used for the reach analyses.

of 100 GeV in both $m_0$ and $m_{1/2}$. For each point the values of $\mu$ and $M_A$ at the weak scale are adjusted to give acceptable cold dark matter.

### 8.2 ATLFAST corrections

ATLFAST is a fast parameterized simulation of the ATLAS detector. The version used here is rather idealized. Corrections to the efficiency for $e$ reconstruction were applied as a function of $p_T$ and $\eta$. An example of the effect of these corrections is shown in Figure 10. In addition, the ATLFAST algorithm finding reconstructed cone jets was missing the split-merge step, so jets matched to the same truth jet were combined. With these corrections the ATLFAST and full simulations agree reasonably well. All results shown here use ATLFAST with these corrections.

### 8.3 Discovery reach

The reach plots in this subsection are all based on analyses that require a certain number of jets and leptons ($e$ or $\mu$) and then find an optimal $M_{\text{eff}}$ cut (in steps of 400 GeV) to maximize the significance $Z_n$ corrected for multiple cuts of the signal over the Standard Model background, using background errors estimated from studies of data-driven methods [2,3]. Not all modes were studied because of limited time or Monte Carlo statistics.

The analysis most similar to that in the Physics TDR [10], requires four jets with $p_T > \{100, 50, 50, 50\}$ GeV and $E_{\text{T}}^{\text{miss}} > \max(100\text{ GeV}, 0.2M_{\text{eff}})$. The $5\sigma$ discovery reach for the analyses reqiring zero, one, or two opposite-sign leptons for mSUGRA with $\tan\beta = 10$ are shown in Figure 11. The plot also shows the trilepton reach with just one jet. The 0-lepton mode has the best estimated reach, close to 1.5 TeV for the smaller of $m_{\tilde{g}}$ and $m_{\tilde{q}}$. The 1-lepton estimated reach is somewhat less, but it is more robust against QCD backgrounds which might result from detector problems. Figure 11 also shows that the reach for $\tan\beta = 50$ is similar for the zero- and one-lepton channels. Despite the enhanced $\tau$ decays for $\tan\beta \gg 1$, the one-$\tau$ reach is slightly worse than the reach for zero and one leptons. This reflects the lower efficiency and purity for $\tau$ reconstruction. Compared to $\tau + 4$ jets, the reach for $\tau + 3$ jets is slightly better, while $\tau + 2$ jets is about the same. The curves for the $\tau + 2$-jet and the $\tau + 3$-jet analyses are not shown.

Requiring four jets is not necessarily the best choice. The $5\sigma$ reach contours for the 0-lepton plus $E_{\text{T}}^{\text{miss}}$ and the 1-lepton plus $E_{\text{T}}^{\text{miss}}$ analyses for various jet multiplicities are shown in Figure 12, again for

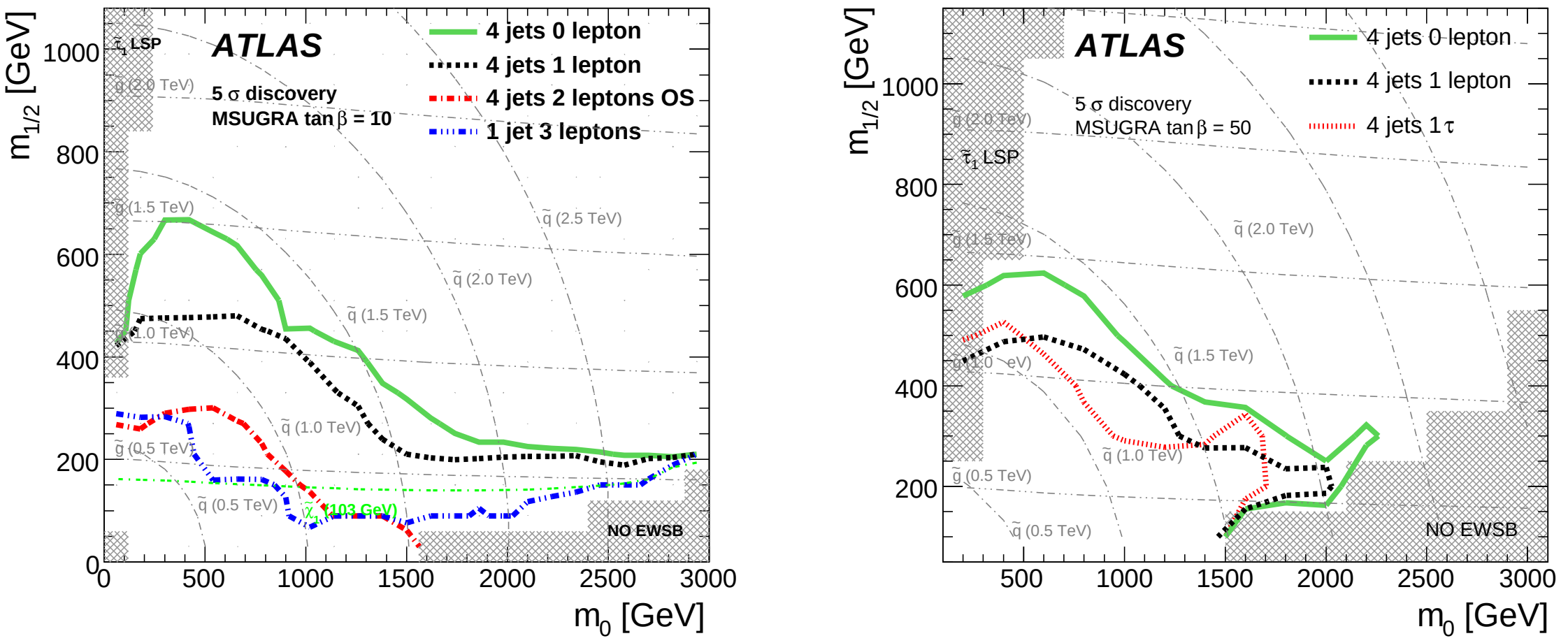


Figure 11: The 1 fb$^{-1}$ $5\sigma$ reach contours for the 4-jet plus $E_{\mathrm{T}}^{\mathrm{miss}}$ analyses with various lepton requirements for mSUGRA as a function of $m_0$ and $m_{1/2}$. Left: $\tan\beta = 10$. Right: $\tan\beta = 50$. The horizontal and curved grey lines indicate gluino and squark mass contours respectively in steps of 500 GeV.

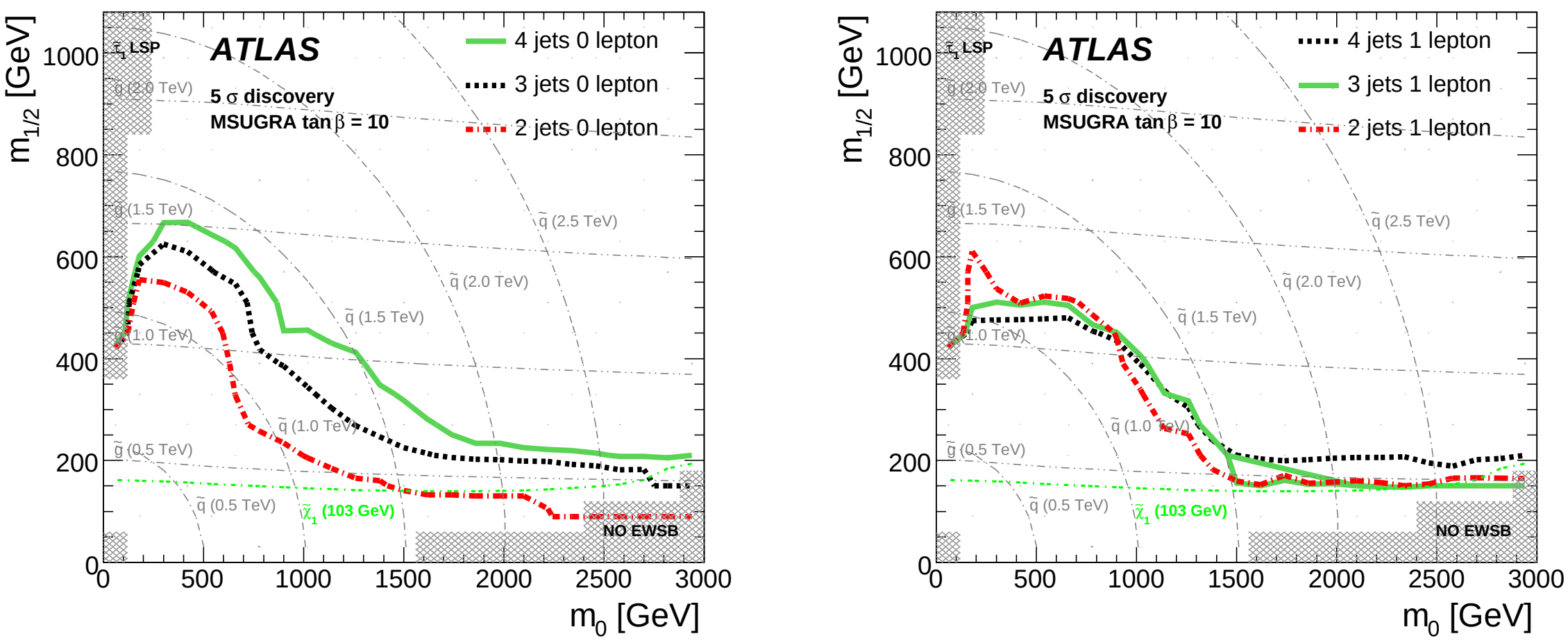


Figure 12: The 1 fb$^{-1}$ $5\sigma$ reach contours for the 0-lepton and 1-lepton plus $E_{\mathrm{T}}^{\mathrm{miss}}$ analyses with various jet requirements as a function of $m_0$ and $m_{1/2}$ for the $\tan\beta = 10$ mSUGRA scan. The horizontal and curved grey lines indicate the gluino and squark masses respectively in steps of 500 GeV.

the $\tan\beta = 10$ mSUGRA scan. For the 0-lepton mode the choice of four jets seems best, while for the 1-lepton mode the 2-jet, 3-jet and 4-jet reaches are all comparable. The reaches (not shown here) for the opposite-sign dilepton plus $E_{\mathrm{T}}^{\mathrm{miss}}$ signature requiring at least 2, 3, or 4 jets are comparable and in all cases are less than the reaches for the 0-lepton and 1-lepton modes. Observing a signal in multiple channels would provide further confidence that the observed excesses were evidence for new physics.

The mSUGRA “random” scan with low-energy constraints samples only a limited range of parameters and hence of gluino and squark masses. The results of this scan are shown as a scatter plot of points in Figure 13 compared to those for the mSUGRA scans. The reach for those mSUGRA points which are compatible with low-energy constraints is comparable to that for generic points. This is not surprising

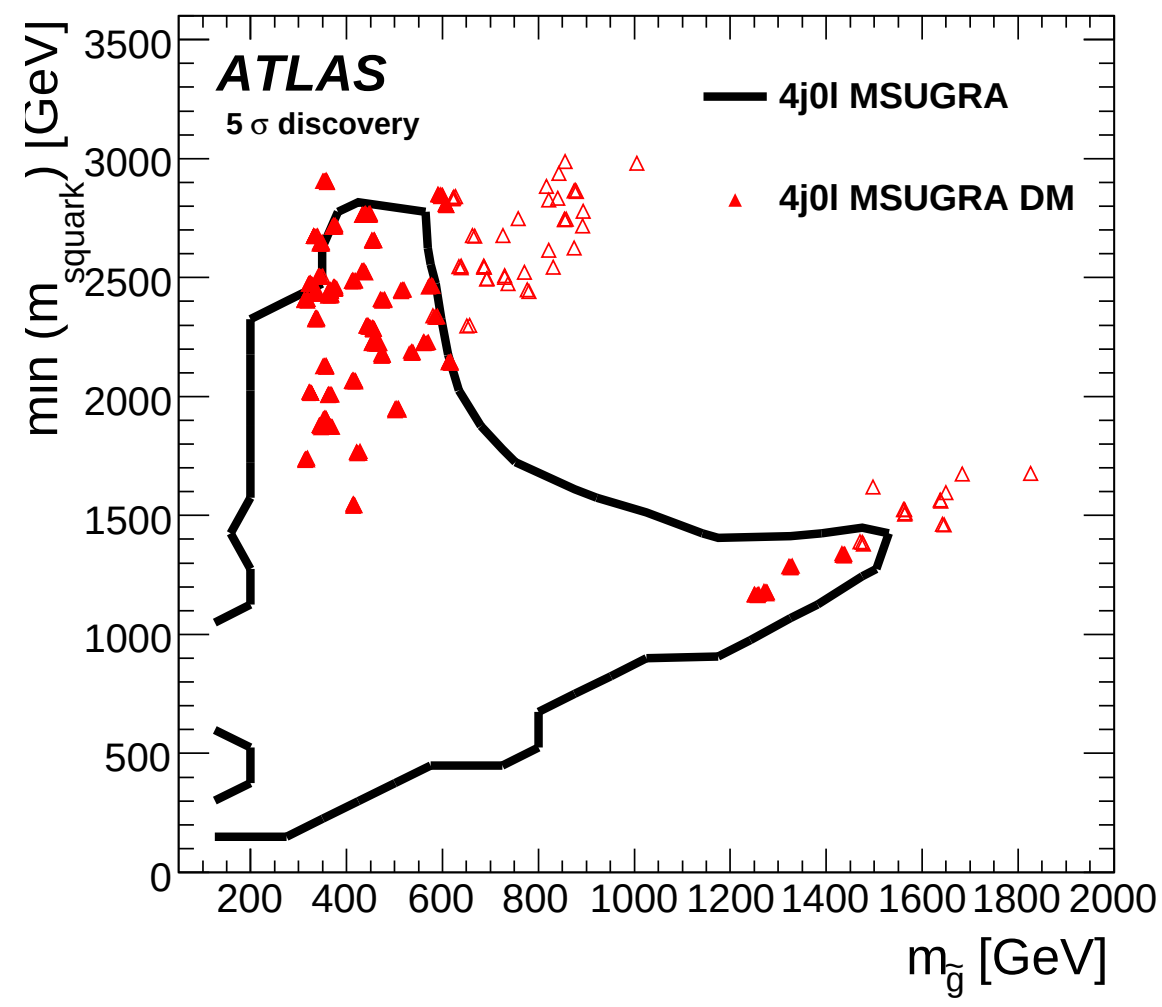


Figure 13: Reach for the “random with constraints” mSUGRA scan plotted in the $m_{\tilde{g}}, m_{\tilde{q}}$ plane. Solid triangles represent points which are observable ($Z_n > 5$) with $1\,\text{fb}^{-1}$, while open triangles show points which are not.

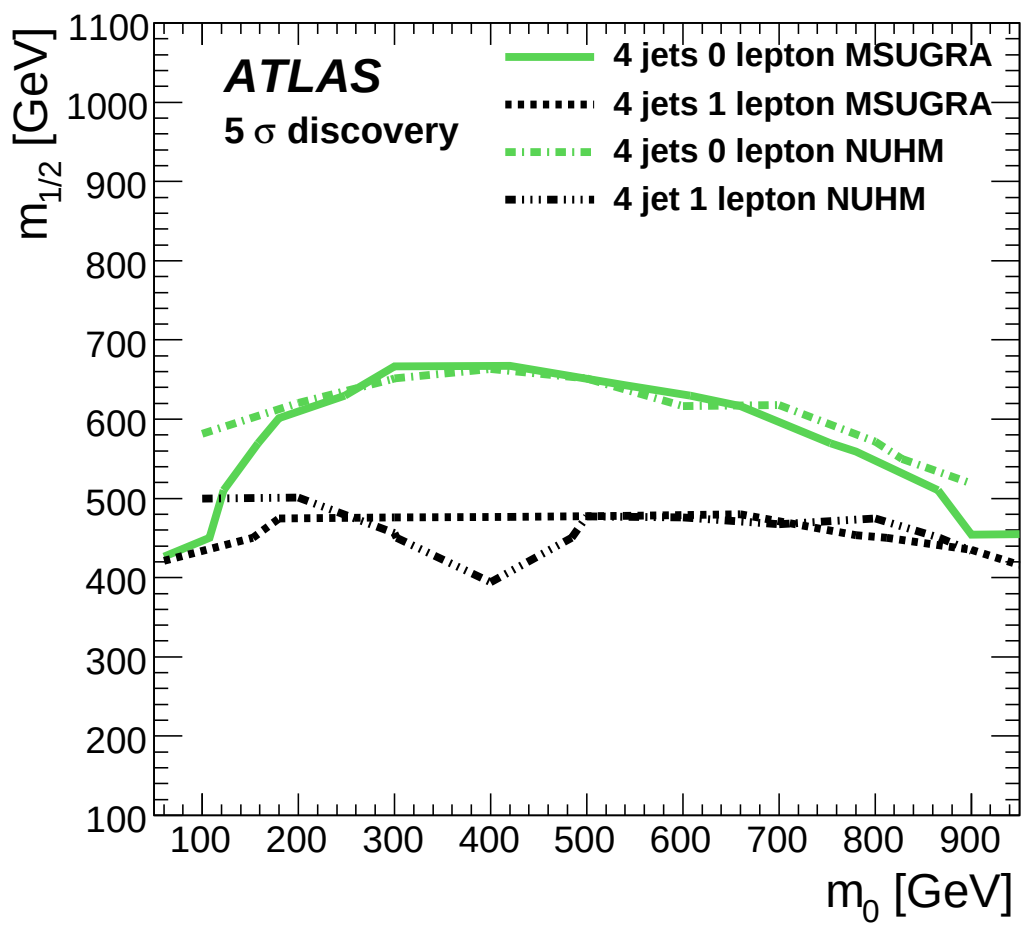


Figure 14: The 1 $\text{fb}^{-1}$ reach for NUHM models with 4 jets, 0 or 1 leptons, and $E_{\text{T}}^{\text{miss}}$. The masses for NUHM are similar to those shown in Figure 11.

given that the SUSY production cross-sections are mainly controlled by the gluino and squark masses, but it adds support to the approach used in this section.

The mSUGRA model is only one possible mechanism for SUSY breaking. The non-universal-Higgs model has qualitatively similar phenomenology but different patterns of masses and decay modes. The reach plots with four jets, zero or one leptons, and $E_{\text{T}}^{\text{miss}}$ for the NUHM are shown in Figure 14. The reach with zero and one leptons is virtually identical to that for mSUGRA. This is as expected: adding some Higgsino mixing allows $\tilde{\chi}_1^0$ annihilation but has a minor effect on the other decays.

Another alternative often considered, Anomaly Mediated SUSY Breaking (AMSB), is not examined here. Previous studies [28] have found an overall reach comparable to mSUGRA with similar assump-

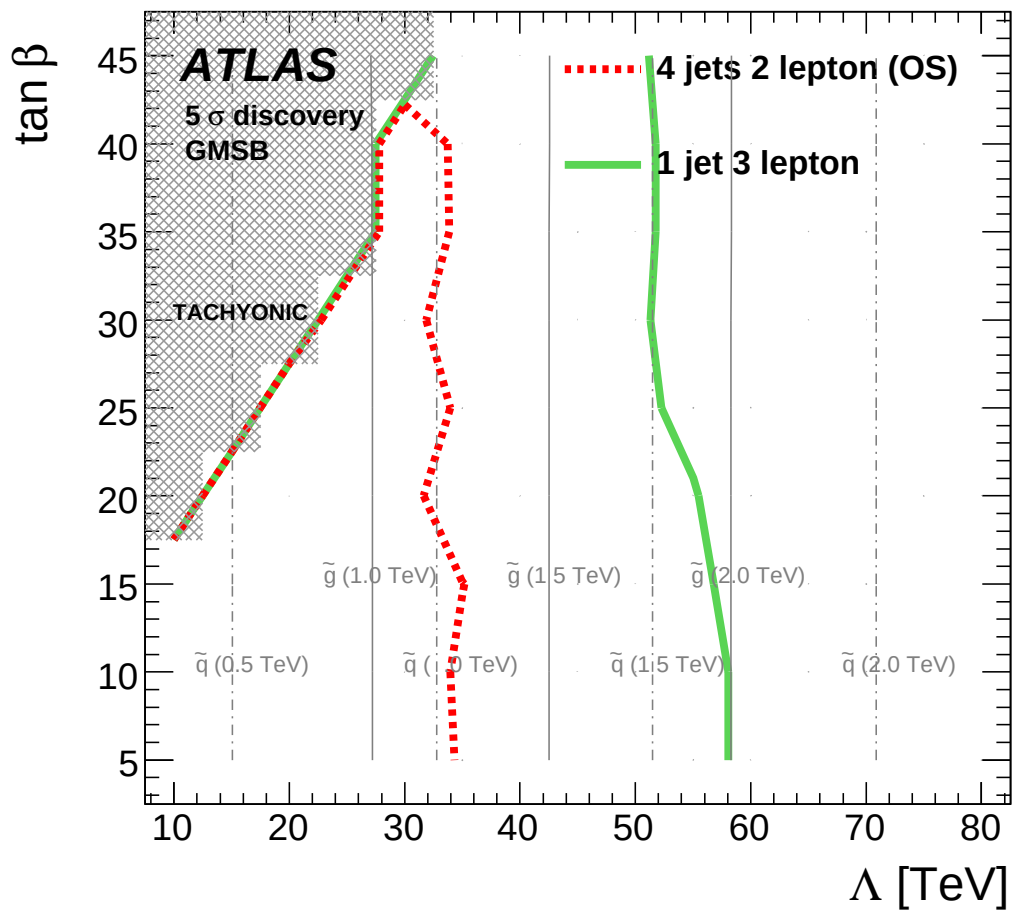


Figure 15: The 1 $\mathrm{fb}^{-1}$ $5\sigma$ reach contours of the 2-lepton and 3-lepton analyses for the GMSB scan. The vertical solid and dashed grey lines indicate the gluino and squark masses respectively in steps of 500 GeV.

tions. The reach in the one-lepton modes is less because the lightest chargino is almost degenerate with the LSP and so does not give visible leptons.

The models considered in the GMSB scan all have at least two leptons or $\tau$'s at the Monte Carlo generator level, so the signatures are easier to distinguish from Standard Model backgrounds. The reach plots for this scan are shown in Figure 15. The reach for three leptons is significantly better than for two leptons and extends well beyond 2 TeV for gluinos for large $\tan\beta$ and is close to 2 TeV for all $\tan\beta$. Special signatures that can result from GMSB models are discussed elsewhere in this volume [4].

## 8.4 Summary

The results of the scans presented in this section together with the full simulation analyses presented earlier indicate that ATLAS should discover signals for $R$-parity conserving SUSY with gluino and squark masses less than $\mathscr{O}(1\ \mathrm{TeV})$ after having accumulated and understood an integrated luminosity of about $1\,\mathrm{fb}^{-1}$. For favorable models the mass reach could be greater. The luminosity required to discover a given SUSY scenario is greater than that estimated previously [29]. The main differences in this analysis are that the uncertainty in the background (derived from data-driven methods) is taken into account and that the signal and background simulations are more realistic. Given the admittedly qualitative naturalness arguments about SUSY masses, it is plausible that SUSY could be found with $1\,\mathrm{fb}^{-1}$ if it exists at the TeV scale. Conversely, if SUSY is not found with $1\,\mathrm{fb}^{-1}$, it might still eventually be discovered at the LHC, but it will be difficult to study in detail.

# Measurements from Supersymmetric Events

**Abstract**
We review the techniques used to reconstruct the decay of supersymmetric particles and measure their properties with ATLAS at the LHC, concentrating on strategies to be applied to a data set of integrated luminosity of 1 $\text{fb}^{-1}$ that can be expected after the first year of operation of the LHC. These techniques are illustrated using several benchmark points chosen in the mSUGRA parameter space, but they are applicable to a broader range of supersymmetric (and other similar) models. The most appropriate methods will be selected and fine-tuned once (and if) signatures consistent with Supersymmetry are established. Supersymmetric cascade decays typically have large transverse missing energy due to the presence of undectected neutralinos, and have characteristic edges and thresholds in the dilepton, dijet and lepton-jet invariant mass distributions. The reconstruction of such edges is the focus of the first part of the paper. The second part of the paper concentrates on the reconstruction of more specific decay channels, involving light stops, staus and Higgs bosons. The final section indicates how sparticle masses and other supersymmetric parameters could be constrained using such measurements.

## 1 Introduction

Supersymmetry (SUSY) can be discovered by the ATLAS experiment at the LHC during the initial running period if some coloured sparticles have masses of the order hundreds of GeV and hence production cross-sections of the order of a few pb. A strategy to establish SUSY discovery is outlined in another paper in this collection [1], while here we concentrate on parameter measurements that can be performed with the early data. The same particle identification conventions and selection criteria as in [2] are used in this paper.

Once a signature consistent with Supersymmetry has been established, the experimental emphasis will move on to measuring the sparticle mass spectrum and constraining the parameters of the model. In the case of R-parity-conserving models, the decay chain of sparticles cannot be completely reconstructed, as sparticles eventually decay into LSPs that can not be detected. For this reason edge positions, rather than mass peaks, are measured in the invariant mass distribution of sparticle decay products. In R-parity-violating models sparticles can have long lifetimes and can be detected by studying their decay in-flight within the detector. These types of signatures are discussed in [3].

A complete coverage of all allowed SUSY models is impossible, so we limit this study to a subset of the models where SUSY breaking is mediated by gravity (mSUGRA), and to the points in parameter space described in [2], however the measurement techniques and fit methods developed can be adapted for many models. During initial data-taking, the error on such measurements will be limited by statistics, making measurements possible only for models with moderate ($\lesssim$ 1 TeV) values of the SUSY mass scale where enough events can be isolated. In this paper we study the cases of a total integrated luminosity of 0.5 $\text{fb}^{-1}$ for the “Low Mass” point (SU4) and of 1 $\text{fb}^{-1}$ for the “Bulk” point (SU3), with the idea of developing the experimental analyses which might be performed after the first year or so of data taking. Some benchmark points require somewhat larger datasets in order to perform kinematic measurements; as an example we show a measurement of the dilepton mass edges for the “Coannihilation” point (SU1) with a dataset corresponding to 18 $\text{fb}^{-1}$.

In sections 3 and 4 we study the decay chain:

$$\tilde{q}_{\text{L}} \rightarrow \tilde{\chi}_2^0 q (\rightarrow \tilde{\ell}^{\pm} \ell^{\mp} q) \rightarrow \tilde{\chi}_1^0 \ell^+ \ell^- q \tag{1}$$

in events containing two opposite-sign isolated electrons or muons, hard jets and missing energy. Kinematic endpoints in the invariant mass spectra of lepton pairs and lepton+jet combinations are fitted and used to derive relations between the masses of sparticles. In the case of first- and second-generation squarks, it will often not be possible to experimentally determine squark flavour, so we define $m_{\tilde{q}_{\rm L}}$ to be the average of the masses of the $\tilde{u}_L$ and $\tilde{d}_L$ squarks, and $m_{\tilde{q}_{\rm R}}$, the average mass of $\tilde{u}_R$ and $\tilde{d}_R$.

Events with tau leptons in the final state are studied in Section 5 and di-tau mass edges in the $\tilde{\chi}_2^0$ decay chain reconstructed. This signature is particularly important in the co-annihilation region where the decay into tau stau pairs is favoured.

In Section 6 we analyse events with two hard jets and missing energy in order to to measure the jet "stransverse mass". This variable is sensitive to the mass of the right-handed squark in events where a pair of squarks are produced, each decaying as:

$$\tilde{q}_{\rm R} \to q\tilde{\chi}_1^0 \tag{2}$$

A kinematical edge depending on the mass of the light stop is reconstructed in Section 7 by exploiting the decay:

$$\tilde{g} \to \tilde{t}_1 t \to \tilde{\chi}_1^{\pm} bt \tag{3}$$

and reconstructing the $tb$ invariant mass.

The reconstruction of the lightest Higgs bosons, produced by the $\tilde{\chi}_2^0$ decay followed by the Higgs decay into a pair of $b$ quarks, is investigated in Section 8. Simulations of the "Higgs" point (SU9) show that if these decays are allowed, then the Standard Model Higgs boson may be initially detected as a SUSY decay product rather than by signatures that involve its production via Standard Model processes.

In Section 9 the parameters measured in sections 3 to 8 are combined to extract information about the SUSY model such as the sparticle mass spectrum and the mSUGRA parameters (under the hypothesis that mSUGRA is realised).

All of these studies use a realistic detector geometry with residual misalignments, and all relevant Standard Model backgrounds are taken into account, as are the trigger efficiencies. The reconstruction of final state objects, the event selection criteria, the strategy used to simulate both signal and background events, and the methods for estimating systematic uncertainties are common across all SUSY analyses and are discussed in the introduction to this chapter [2].

## 2 Measurement of endpoints

The decay chain in Eq. (1) is particularly suited to measure the mass of SUSY particles, as the presence in the final state of charged leptons, missing energy from the escaping neutralino and hadronic jets ensures a large signal to background ratio. Thus, fit results are not very dependent on the precise measurement of the Standard Model background. Although we discuss the reconstruction of edges and thresholds within the mSUGRA framework, the same methodology can be applied to the large variety of SUSY models where the $\tilde{q}_{\rm L}$ decay channel in Eq. (1) is open. In the following we indicate with $\ell$ only electrons and muons (with $\tilde{\ell}$ being their superpartners) while $\tau$ leptons are indicated explicitly.

The endpoint in the di-lepton invariant mass distribution is a function of the masses of the particles involved in the decay. If the sleptons are heavier than the $\tilde{\chi}_2^0$ then the decay proceeds through the three body channel $\tilde{\chi}_2^0 \to \tilde{\chi}_1^0 \ell^+\ell^-$ as in the SU4 model. In this case, the distribution of the invariant mass of the two leptons has a non-triangular shape described in [4, 5] with an endpoint equal to the difference of the mass of the two neutralinos:

$$m_{\ell\ell}^{\rm edge} = m_{\tilde{\chi}_2^0} - m_{\tilde{\chi}_1^0} \tag{4}$$

If at least one of the sleptons is lighter than the $\tilde{\chi}_2^0$ then the two-body decay channel $\tilde{\chi}_2^0 \to \tilde{\ell}^\pm \ell^\mp \to \tilde{\chi}_1^0 \ell^+ \ell^-$ dominates. The distribution of the invariant mass of the two leptons is triangular with an endpoint at:

$$m_{\ell\ell}^{\text{edge}} = m_{\tilde{\chi}_2^0} \sqrt{1 - \left(\frac{m_{\tilde{\ell}}}{m_{\tilde{\chi}_2^0}}\right)^2} \sqrt{1 - \left(\frac{m_{\tilde{\chi}_1^0}}{m_{\tilde{\ell}}}\right)^2}. \tag{5}$$

For the SU3 point, where the $\tilde{\ell}_{\text{R}}$ and the $\tilde{\tau}_1$ are lighter than the $\tilde{\chi}_2^0$, such an endpoint is expected in the $\ell^+\ell^-$ ($\tau^+\tau^-$) distribution for $m_{\ell\ell}^{\text{edge}} = 100.2$ GeV ($m_{\tau\tau}^{\text{edge}} = 98.3$ GeV). For the SU1 point both $\tilde{\ell}_{\text{R}}$ and $\tilde{\ell}_{\text{L}}$ as well as $\tilde{\tau}_1$ and $\tilde{\tau}_2$ are lighter than $\tilde{\chi}_2^0$, resulting in a double triangular distribution for the dilepton invariant mass with two edges.

Measuring the dilepton endpoint allows us to establish a relationship between the masses of the two lightest neutralinos and any sleptons that are lighter than the $\tilde{\chi}_2^0$. For a determination of the masses of all the particles involved in the decay chain Eq. (1), further mass distributions involving a jet are used: $m_{\ell\ell q}$, $m_{\ell\ell q}^{\text{thr}}$, $m_{\ell q(\text{low})}$ and $m_{\ell q(\text{high})}$. Since it is not possible to identify the quark from the $\tilde{q}_{\text{L}}$ decay, we make the assumption that it generates one of the two highest $p_{\text{T}}$ jets in the event, as is normally the case if the $\tilde{q}_{\text{L}}$ is much heavier than the $\tilde{\chi}_2^0$. Hence only the two leading jets are considered. For the $m_{\ell\ell q}$ distribution a maximum value of the distribution is expected so the jet giving the lowest $m_{\ell\ell q}$ value is used. The $m_{\ell\ell q}^{\text{thr}}$ distribution is defined by the additional constraint $m_{\ell\ell} > m_{\ell\ell}^{\text{edge}}/\sqrt{2}$, giving a non-zero threshold value [6, 7]. Since a minimum is sought, the jet giving the highest $m_{\ell\ell q}$ value is used in this distribution. The distributions $m_{\ell q(\text{low})}$ and $m_{\ell q(\text{high})}$ are formed from the lower and higher $m_{lq}$ value of each event using the same jet as for $m_{\ell\ell q}$. Both distributions have well-defined endpoints.

The theoretical values of the kinematic threshold and endpoints listed above can be calculated using the analytical expressions given in [6, 7]. The theoretical positions of the end points for the SU1, SU3 and SU4 models are summarised in Table 1.

Table 1: Value of the end points of the invariant mass distributions for the three benchmark points considered in this section. For SU1 the two endpoints correspond to the two available decay chains of the $\tilde{\chi}_2^0$ involving a right or left slepton.

| Mass Distribution | SU1 end point (GeV) | SU3 end point (GeV) | SU4 end point (GeV) |
|---|---|---|---|
| $m_{\ell\ell}^{\text{edge}}$ | 56.1, 97.9 | 100.2 | 53.6 |
| $m_{\tau\tau}^{\text{edge}}$ | 77.7, 49.8 | 98.3 | 53.6 |
| $m_{\ell\ell q}^{\text{edge}}$ | 611, 611 | 501 | 340 |
| $m_{\ell\ell q}^{\text{thr}}$ | 133, 235 | 249 | 168 |
| $m_{lq(\text{low})}^{\text{max}}$ | 180, 298 | 325 | 240 |
| $m_{lq(\text{high})}^{\text{max}}$ | 604, 581 | 418 | 340 |

Another advantage of the decay chain in 1 is the possibility of estimating both the SUSY combinatorial background and the Standard Model background from the data with high accuracy. The technique, known as *flavour subtraction*, is based on the fact that the signal contains two opposite-sign same-flavour (OSSF) leptons, while the background leptons come from different decay chains, which can be of the same flavour or of different flavour with the same probability. The background thus cancels in the subtraction:

$$N(e^+e^-)/\beta + \beta N(\mu^+\mu^-) - N(e^\pm\mu^\mp) \tag{6}$$

where $\beta = 0.86$ is an efficiency correction factor equal to the ratio of the electron and muon reconstruction efficiencies. The value of $\beta$ is taken from [8, 9], and is assumed in the following to be known with an uncertainty of 10%.

# 3 Dilepton edges

## 3.1 Event Selection

Events with two or three isolated leptons (electrons or muons) with $p_{\mathrm{T}} > 10$ GeV and $|\eta| < 2.5$ are selected. If two leptons are selected, they are required to have opposite signs. If three leptons are present, the two opposite-sign combinations are considered and treated independently in the rest of the analysis.

In order to select SUSY events and reject the Standard Model background it is necessary to require the presence of energetic jets and missing energy. The variables used to discriminate SUSY from the SM background are the transverse missing energy, the transverse momenta of the four leading jets, the ratio between the transverse missing energy and the effective mass, and the transverse sphericity ($S_T$). In order to optimise the cuts on these variables, the value of:

$$S \equiv (N_{\mathrm{OSSF}} - N_{\mathrm{OSDF}})/\sqrt{N_{\mathrm{OSSF}} + N_{\mathrm{OSDF}}} \tag{7}$$

is maximized for each SUSY point, where $N_{\mathrm{OSSF}}$ and $N_{\mathrm{OSDF}}$ are the number of same-flavour and different-flavour lepton pairs respectively.

The $S$ variable can be computed from collider data, since no Monte Carlo information is used. By maximizing the value of $S$ we are maximizing the selection efficiency for signal events while suppressing the Standard Model and the SUSY combinatorial backgrounds.

In order to improve the sensitivity to the signal, only lepton pairs with an invariant mass $m_{\ell\ell} < m_{\ell\ell}^{\mathrm{edge}} + 10$ GeV are considered. Since the true value of the endpoint is *a priori* unknown, this choice implies that the edge has already been observed, and that afterwards the selection cuts are optimised as described here in order to improve the separation between signal and background and the measurement of the endpoint. We are thus focusing here on determining selection cuts that would allow a precise measurement of the endpoint with moderate statistics, rather than on finding the first evidence for an excess of different-flavour lepton pairs or the first evidence for the presence of the edge.

In Table 2 the optimal selection resulting from the scan is shown. For all three points a 2-jet selection is preferred, leaving out cuts on the third and fourth jets, on $S_{\mathrm{T}}$ and on the ratio $E_{\mathrm{T}}^{\mathrm{miss}}/M_{\mathrm{eff}}$. For the "Coannihilation" point (SU1) and the "Bulk" point (SU3) the $S$-value is found to be stable in an interval around the maximum value. For the "Low Mass" point (SU4) the best $S$-value is found for the loosest cut allowed by the available Monte Carlo samples. Hence even looser cuts may be preferred as far as the value of $S$ is concerned. The cuts on $E_{\mathrm{T}}^{\mathrm{miss}}$ and the $p_{\mathrm{T}}$ of leading jet are however required in order to have a high trigger efficiency[1].

The number of signal and background lepton pairs passing the selection cuts is shown in Table 3. All numbers are for 1 fb$^{-1}$. The main Standard Model background is always $t\bar{t}$ accounting for about 95% of the total background. The remaining background events are from W, Z and WW, WZ, ZZ production. The background due to QCD jets is negligible. The fraction of SUSY events in the selected sample with OSSF leptons is 59% for SU1, 77% for SU3, and 80% for SU4.

Table 2: Results of the event-selection optimisation for the S-variable Eq. (7) for signal ($s$) and Standard Model background ($b$) with range limit $m_{\ell\ell} < m_{\ell\ell}^{\rm edge} + 10$ GeV, for SU1, SU3 and SU4 for 1 fb$^{-1}$. The best selection is shown.

| | $p_{\rm T}^{\rm j1}$ | $p_{\rm T}^{\rm j2}$ | $p_{\rm T}^{\rm j3}$ | $p_{\rm T}^{\rm j4}$ | $E_{\rm T}^{\rm miss}$ | $E_{\rm T}^{\rm miss}/M_{\rm eff}$ | $S_{\rm T}$ | $s_{\rm OSSF}$ | $s_{\rm OSDF}$ | $b_{\rm OSSF}$ | $b_{\rm OSDF}$ | $S$ |
|---|---|---|---|---|---|---|---|---|---|---|---|---|
| **SU1** | 200 | 150 | - | - | 120 | - | - | 120 | 64 | 69 | 53 | 5.1 |
| **SU3** | 180 | 100 | - | - | 120 | - | - | 615 | 149 | 93 | 92 | 15.1 |
| **SU4** | 100 | 50 | - | - | 100 | - | - | 3048 | 1574 | 411 | 419 | 19.9 |

Table 3: Number of lepton pairs passing the selection cuts optimized for the SUSY sample SU1 (above), SU3 (middle) and SU4 (below), for 1 fb$^{-1}$ of integrated luminosity. The contribution from $t\bar{t}$ production is indicated separately as it constitutes most of the Standard Model background. The remaining background events are from W, Z and WW, WZ, ZZ production. The background due to QCD jets is negligible.

| Sample | $e^+e^-$ | $\mu^+\mu^-$ | OSSF | OSDF |
|---|---|---|---|---|
| SUSY SU1 | 56 | 88 | 144 | 84 |
| Standard Model ($t\bar{t}$) | 35 (35) | 65 (63) | 101 (99) | 72 (68) |
| SUSY SU3 | 274 | 371 | 645 | 178 |
| Standard Model ($t\bar{t}$) | 76 (75) | 120 (115) | 196 (190) | 172 (165) |
| SUSY SU4 | 1729 | 2670 | 4400 | 2856 |
| Standard Model ($t\bar{t}$) | 392 (377) | 688 (657) | 1081 (1035) | 1104 (1063) |

### 3.2 Reconstruction of the dilepton edge

The distribution of the invariant mass of same-flavour and different-flavour lepton pairs is shown in Fig. 1 for the SUSY benchmark points and backgrounds, after the selection cuts optimized for SU3 (left plot) and SU4 (right plot), and for an integrated luminosity of 1 fb$^{-1}$ and 0.5 fb$^{-1}$ respectively. It can be seen from regions where the signal does not contribute (i.e. for the Standard Model backgrounds and for $m_{\ell\ell} > m_{\ell\ell}^{\rm edge}$ for SUSY) that the different-flavour distributions are similar to the same-flavour backgrounds.

The invariant mass distribution after flavour subtraction is shown in the left plot of Fig. 2 in the presence of the SU3 signal and for an integrated luminosity of 1 fb$^{-1}$. The distribution has been fitted with a triangle smeared with a Gaussian. The value obtained for the endpoint is $(99.7 \pm 1.4 \pm 0.3)$ GeV where the first error is due to statistics and the second is the systematic error on the lepton energy scale and on the $\beta$ parameter [2]. This result is consistent with the true value of 100.2 GeV calculated from Eq. (5).

The right plot of Fig. 2 shows the flavour-subtracted distribution in the presence of the SU4 signal for an integrated luminosity of 0.5 fb$^{-1}$. The fit was performed using the function from [5] which describes the theoretical distribution for the 3-body decay in the limit of large slepton masses, smeared for the experimental resolution. This function vanishes near the endpoint and is a better description of the true distribution for SU4 than the triangle with a sharp edge. The endpoint from the fit is $(52.7 \pm 2.4 \pm$

[1]For this channel, both the lepton triggers and the trigger based on $E_{\rm T}^{\rm miss}$ may be relied upon. The latter are however less efficient, and they would imply different $p_{\rm T}$ thresholds for electrons and muons.

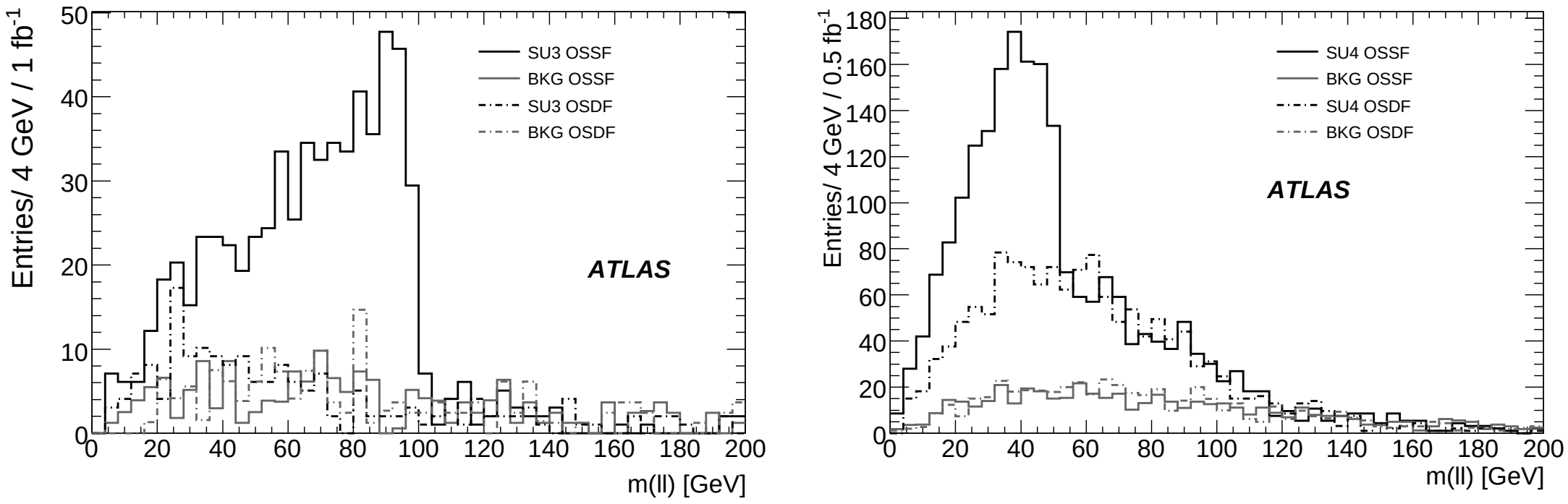


Figure 1: Left: distribution of the invariant mass of same-flavour and different-flavour lepton pairs for the SUSY benchmark points and backgrounds after the cuts optimized from data in presence of the SU3 signal (left), and the SU4 signal (right). The integrated luminosities are 1 $fb^{-1}$ and 0.5 $fb^{-1}$ respectively.

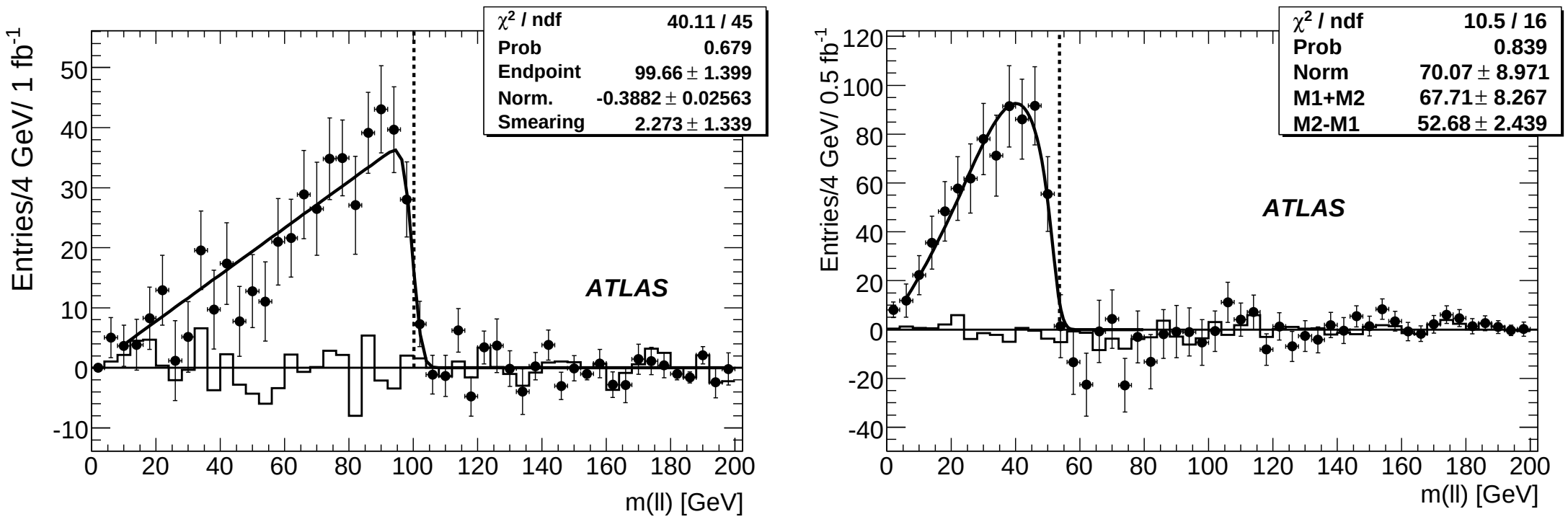


Figure 2: Left: Distribution of invariant mass after flavour subtraction for the SU3 benchmark point with an integrated luminosity of 1 $fb^{-1}$. Right: the same distribution is shown for the SU4 benchmark point and an integrated luminosity of 0.5 $fb^{-1}$. The line histogram is the Standard Model contribution, while the points are the sum of Standard Model and SUSY contributions. The fitting function is superimposed and the expected position of the endpoint is indicated by a dashed line.

0.2) GeV, consistent with the theoretical endpoint of 53.6 GeV.

Since the true distribution will not be known for data, the distribution was also fitted with the smeared triangle expected for the 2-body decay chain. This also gives a good $\chi^2$ with an endpoint of $(49.1 \pm 1.5 \pm 0.2)$ GeV. A larger integrated luminosity will be required to use the shape of the distribution to discriminate between the two-body and the three-body decays.

In Fig. 3 the flavour-subtracted distribution of the dilepton mass is shown for the SU1 point at an integrated luminosity of 1 $fb^{-1}$ (left) and 18 $fb^{-1}$ (right) [2]. While there is already a clear excess of SF-OF entries at 1 $fb^{-1}$ , a very convincing edge structure cannot be located. At 18 $fb^{-1}$ the two edges are visible. A fit function consisting of a double triangle convoluted with a Gaussian, the latter

[2] Only 1 $fb^{-1}$ of simulated Standard Model background was available. To scale the Standard Model contribution to higher luminosities a probability density function for the m(ll) distribution was constructed by fitting a Landau function to the 1 $fb^{-1}$ distribution, assuming statistically identical shapes for $e^+e^-$, $\mu^+\mu^-$ and $e^\pm\mu^\mp$ and normalisation according to a $\beta$ of 0.86. The systematic uncertainty on the endpoint determination from this procedure was estimated to be a small fraction of the statistical uncertainty.

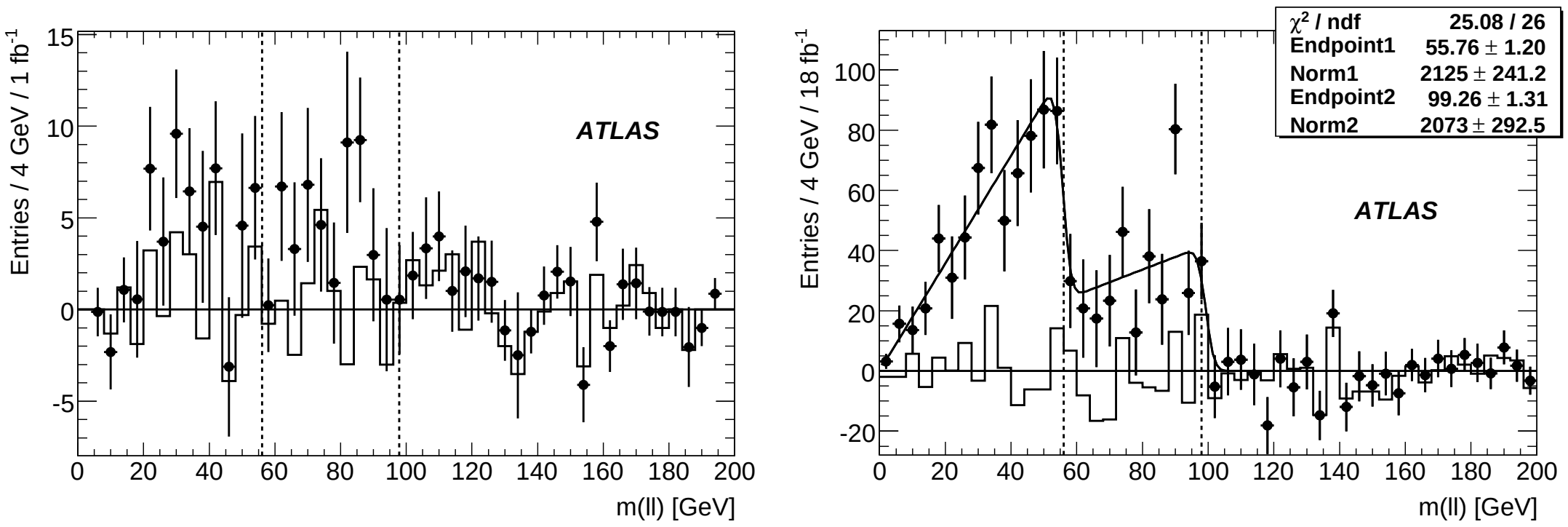


Figure 3: Distribution of invariant mass after flavour subtraction for the SU1 point and for an integrated luminosity of 1 fb$^{-1}$ (left) and 18 fb$^{-1}$ (right). The points with error bars show SUSY plus Standard Model, the solid histogram shows the Standard Model contribution alone. The fitted function is superimposed (right), the vertical lines indicate the theoretical endpoint values.

having a fixed width of 2 GeV, returns endpoint values of $55.8 \pm 1.2 \pm 0.2$ GeV for the lower edge and $99.3 \pm 1.3 \pm 0.3$ GeV for the upper edge, consistent with the true values of 56.1 and 97.9 GeV. As can be seen from Fig. 3 (right) the $m_{\ell\ell}$ distribution also contains a noticeable contribution from the leptonic decay of $Z$ bosons present in SUSY events. Even though the upper edge is located close to the $Z$ mass, adding a $Z$ peak of fixed mass and width to the fit function only affects the endpoints at the 0.2-0.3 GeV level. However the $Z$ peak changes the normalisation of the upper triangle so for considerations of couplings and branching ratios it should be included.

# 4 Leptons+Jets edges

In events selected for the dilepton analysis in the previous section, jets are added to construct further distributions as described in Sect. 2. Additional selection cuts are applied to refine the distributions:

$$m_{\ell\ell} < m_{\ell\ell}^{\text{edge}} + \Delta_1 \qquad \text{(all distributions)} \tag{8}$$

$$m_{\ell\ell q} < m_{\ell\ell q}^{\text{edge}} + \Delta_2 \qquad (m_{\ell q}\ \text{distributions}) \tag{9}$$

Here $m_{\ell\ell}^{\text{edge}}$ and $m_{\ell\ell q}^{\text{edge}}$ refer to experimental values found in this and the previous section. The value $\Delta_1$ is a small number, 10 (3.3) GeV for SU3 (SU4), to account for the fact that the edge stretches slightly beyond the fitted endpoint. One can see from Fig. 1 that this cut should be very effective for SU4, but much less so for SU3. The value $\Delta_2$ serves a similar purpose, but since the determination of $m_{\ell\ell q}^{\text{edge}}$ is less reliable, a looser cut is used, 155 (37) GeV for SU3 (SU4).

The invariant mass distributions $m_{\ell\ell q}$ and $m_{\ell q}$ are shown in Figures 4 and 5 for both the "Bulk" point (SU3) and the "Low Mass" point (SU4) after efficiency-corrected flavour subtraction.

While the Standard Model background causes considerable bin-by-bin fluctuations for integrated luminosities $\lesssim$ 1 fb$^{-1}$, the net contribution of entries beyond the endpoints is mainly due to combinatorics from choosing the wrong jet in a true SUSY event. For the distributions where a clear tail is visible, a straight line is assumed for the background, otherwise it is set to zero. (The statistics box of the plots indicates which background hypothesis is used.) Since the tail is due to SUSY events, it will not be known beforehand and a data-driven approach (not described here) would be required.

The lepton+jet distributions have shapes which depend on the sparticle masses [10]. Depending on the sparticle spectrum, the edge region may contain non-trivial features such as experimentally unde-

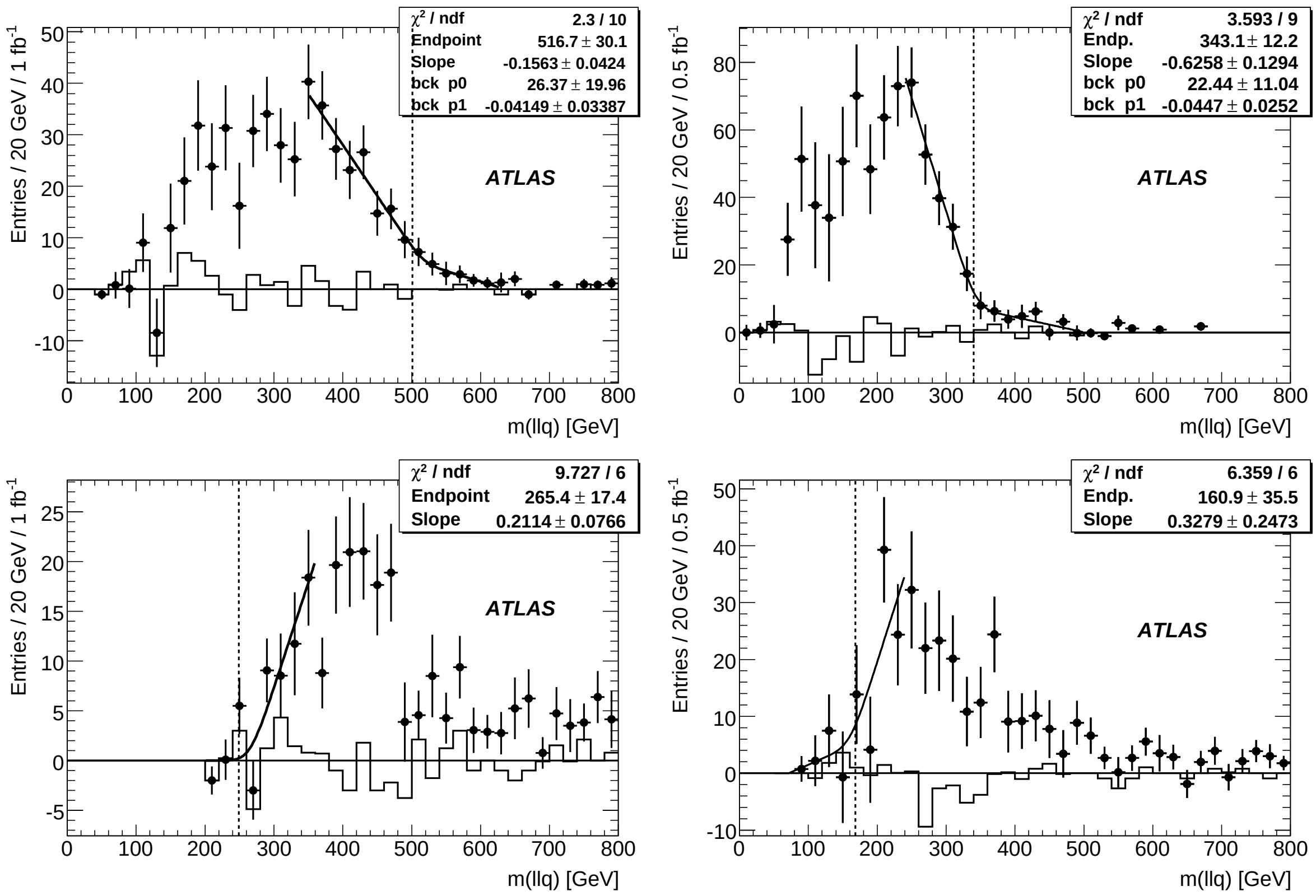


Figure 4: Efficiency-corrected flavour-subtracted distributions of $m_{\ell\ell q}$ (top) and $m_{\ell\ell q}^{\rm thr}$ (bottom) for SU3 (left) for 1 fb$^{-1}$ and SU4 (right) with 0.5 fb$^{-1}$ of integrated luminosity. The points with error bars show SUSY plus Standard Model, the solid histogram shows the Standard Model contribution alone. The fitted function is superimposed, the vertical line indicates the theoretical endpoint value.

tectable 'feet' containing very few events or vertical drops. For all the relevant distributions of two-body scenarios, analytic formulas describing the shape in terms of the sparticle masses are known [11, 12]. With low statistics a straight-line fit is likely to give to give a sufficiently good description in many cases. All the edges and thresholds were fitted with the following formula,

$$f(m) = \frac{1}{\sqrt{2\pi}\sigma} \int \exp\Big(-\frac{(m-m')^2}{2\sigma^2}\Big) \max\{A(m'-m^{\rm EP}),0\}\ {\rm d}m' + \max\{a+bm,0\}\ , \qquad (10)$$

where $m^{\rm EP}$ represents the endpoint (or threshold), $A$ is the slope of the signal distribution, while $a$ and $b$ are the background parameters. The Gaussian smearing gives a smooth transition between the two straight lines and mimics in a simple way the smearing of an edge due to mismeasurement of jet momenta. The smearing parameter $\sigma$ was fixed to 15 GeV. The fitted endpoint values were found not to be very sensitive to the choice of $\sigma$ in the range 0–20 GeV. For $m_{\ell\ell q}$ and the $m_{\ell q}$ distributions the integration range is (0, $m^{\rm EP}$). For the $m_{\ell\ell q}^{\rm thr}$ distribution the lower integration limit is $m^{\rm EP}$ while some 100–200 GeV above the upper fit range is a safe upper integration limit. The endpoints resulting from the fits to the distributions in Figures 4 and 5 are summarised in Table 4.

For the "Bulk" point (SU3) the edges are found to be sufficiently well described by a straight line in Eq. (10) for the signal region, however describing the background region by a straight line results in large systematic uncertainties in the endpoint fit. In particular, the $m_{\ell\ell q}$ distribution does not have a clear edge. Even though the $m_{\ell\ell q}$ cut in Eq. (9) removes a considerable amount of background for the $m_{\ell q({\rm low})}$

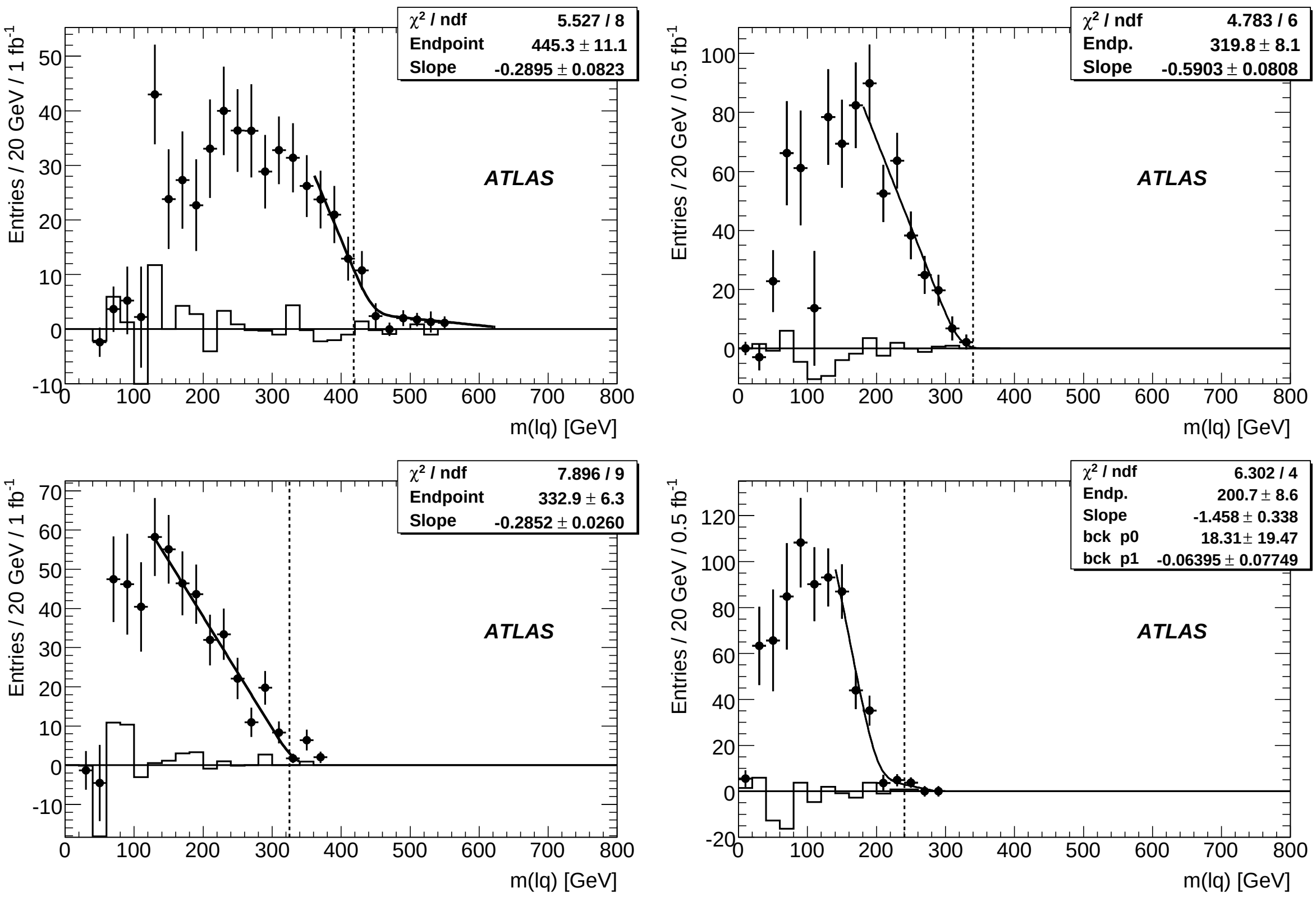


Figure 5: Efficiency-corrected flavour-subtracted distributions of $m_{\ell q(\text{high})}$ (top) and $m_{\ell q(\text{low})}$ (bottom) for SU3 (left) with 1 fb$^{-1}$ and SU4 (right) with 0.5 fb$^{-1}$ of integrated luminosity. The points with error bars show SUSY plus Standard Model, the solid histogram shows the Standard Model contribution alone. The fitted function is superimposed, the vertical line indicates the theoretical endpoint value.

distribution there is still some background left for the $m_{\ell q(\text{high})}$ distribution. A systematic uncertainty is assigned to account for the background estimation for both fits. In case of $m_{\ell q(\text{low})}$ a background-free fit can be made resulting in a few GeV uncertainty. The $m_{\ell\ell q}^{\text{thr}}$ distribution is expected to be concave, so a systematic uncertainty is added in the estimation when fitting the threshold by a straight line fit.

For SU4, the $m_{\ell\ell q}$ and both of the $m_{\ell q}$ distributions have edges which are well described by the fit function Eq. (10). This is confirmed by the dominance of the statistical errors over the systematics ones. The $m_{\ell\ell q}^{\text{thr}}$ fit is more problematic resulting in somewhat larger errors. This could be expected since the contributions from different-family and non-signal same-family peaks in this mass region, whereas they are close to vanishing in the edge regions of the other distributions. Another reason (although probably of less importance at 0.5 fb$^{-1}$ ) is that the threshold edge is concave and only moderately well described by a straight line.

While the fit values of $m_{\ell\ell q}^{\text{edge}}$ and $m_{\ell\ell q}^{\text{thr}}$ are compatible with the theoretical values, the fitted $m_{lq(\text{high})}^{\max}$ and $m_{lq(\text{low})}^{\max}$ are off by $2\sigma$ and $4\sigma$, respectively. This comes from the fact that in SU4 the decay of $\tilde{\chi}_2^0$ is a three-body decay. In such scenarios the $m_{\ell q}$ distributions, and in particular $m_{\ell q(\text{low})}$ are often so sparsely populated towards the high mass values that the endpoints are not experimentally deducible from the edges. Note, however, that there is no hint from the $\chi^2$ of the $m_{\ell q(\text{low})}$ fit that we are in such a situation. This topic is discussed further in Sect. 9.1 where endpoint relations are inverted to give sparticle masses.

Table 4: Endpoint positions for SU3 and SU4, in GeV. The first error is statistical, the second and third are the systematic and the jet energy scale uncertainty, respectively. The theoretical values are also given for ease of comparison to the left of the fitted values. The integrated luminosity assumed is 1 fb$^{-1}$ for SU3 and 0.5 fb$^{-1}$ for SU4.

| Endpoint | SU3 truth | SU3 measured | SU4 truth | SU4 measured |
|---|---|---|---|---|
| $m_{\ell\ell q}^{\mathrm{edge}}$ | 501 | $517 \pm 30 \pm 10 \pm 13$ | 340 | $343 \pm 12 \pm 3 \pm 9$ |
| $m_{\ell\ell q}^{\mathrm{thr}}$ | 249 | $265 \pm 17 \pm 15 \pm 7$ | 168 | $161 \pm 36 \pm 20 \pm 4$ |
| $m_{lq(\mathrm{low})}^{\mathrm{max}}$ | 325 | $333 \pm 6 \pm 6 \pm 8$ | 240 | $201 \pm 9 \pm 3 \pm 5$ |
| $m_{lq(\mathrm{high})}^{\mathrm{max}}$ | 418 | $445 \pm 11 \pm 11 \pm 11$ | 340 | $320 \pm 8 \pm 3 \pm 8$ |

# 5 Tau signatures

## 5.1 Determination of the di-tau endpoint position

The endpoint of the invariant mass distribution from two taus emerging from a $\tilde{\chi}_2^0$ decay,

$$\tilde{\chi}_2^0 \to \tilde{\tau}_1 \tau \to \tilde{\chi}_1^0 \tau^\pm \tau^\mp \ , \tag{11}$$

depends on the masses of the $\tilde{\chi}_2^0$, the $\tilde{\chi}_1^0$ and the $\tilde{\tau}_1$, and therefore can contribute to the determination of SUSY parameters.

Taus play an important role in scenarios like mSUGRA where $\tilde{\chi}_2^0$ is mostly wino, and therefore preferentially couples to *L*-type sfermions. The large *L*-*R* mixing in the stau sector significantly enhances the branching ratio for the decay $\tilde{\chi}_2^0 \to \tilde{\tau}_1^\pm \tau^\mp$ with respect to other leptons. In the scenarios SU1 and SU3 studied here, for example, the branching ratio for the decays into taus is a factor of 10 larger than for decays into electrons or muons. Since in many models (including mSUGRA) the $\tilde{\tau}_1$ is the lightest slepton, for certain values of the SUSY parameters the only allowed two body decay is $\tilde{\chi}_2^0 \to \tilde{\tau}\tau$ as $\tilde{\chi}_2^0 \to \tilde{\chi}_1^0 \mathrm{h}$, $\tilde{\chi}_2^0 \to \tilde{\chi}_1^0 \mathrm{Z}$, and $\tilde{\chi}_2^0 \to \tilde{e}e$ or $\tilde{\mu}\mu$ are kinematically forbidden.

Finally, whereas mass information about $\tilde{\chi}_1^0$ and $\tilde{\chi}_2^0$ can often be obtained more precisely from $\tilde{\chi}_2^0$ decays into electrons and muons (if these decays are open), decays into taus are needed to probe the $\tilde{\tau}$ mass parameters.

In contrast to $\tilde{\chi}_2^0$ decays into electrons or muons, the di-tau invariant mass spectrum does not have a sharp endpoint at the maximum kinematic value. Due to the presence of neutrinos from the tau decays, the $m_{\tau\tau}$ distribution (where $m_{\tau\tau}$ indicates the invariant mass of the visible decay products of the tau pair) falls off smoothly below the maximum value given by either Eq. (4) or Eq. (5). Only hadronic tau decays are considered for tau identification: the tracking-seeded reconstruction algorithm [13] is used to reconstruct taus in the "Coannihilation" point (SU1), while the calorimeter-seeded algorithm [13] is used for the "Bulk" point model (SU3). This choice is motivated by the higher efficiency of the former in reconstructing the low $p_\mathrm{T}$ taus that are present in the SU1 model.

The SU1 point also has a considerably lower cross-section than the SU3 point, so different selection procedures are used to maximize the signal significance. For the SU3 point events are selected with two taus, $E_\mathrm{T}^{\mathrm{miss}} > 230$ GeV, and at least four jets with $p_\mathrm{T}$ greater than 220, 50, 50, 30 GeV respectively. For the SU1 point the cut on $E_\mathrm{T}^{\mathrm{miss}}$ is relaxed to 100 GeV and at least two jets with $p_\mathrm{T}$ greater than 100 and 50 GeV respectively. In addition, an elliptical cut in the space of $E_\mathrm{T}^{\mathrm{miss}}$ and the sum $p_\mathrm{T}(1) + p_\mathrm{T}(2)$ of the two highest $p_\mathrm{T}$ jets is applied to SU1. The semi-axes of the ellipse are 450 GeV for $E_\mathrm{T}^{\mathrm{miss}}$ and 500 GeV for the sum of jet $p_\mathrm{T}$. This cut exploits the anticorrelation between $E_\mathrm{T}^{\mathrm{miss}}$ and $(p_\mathrm{T}(1) + p_\mathrm{T}(2))$ which is different for the signal and the Standard Model background.

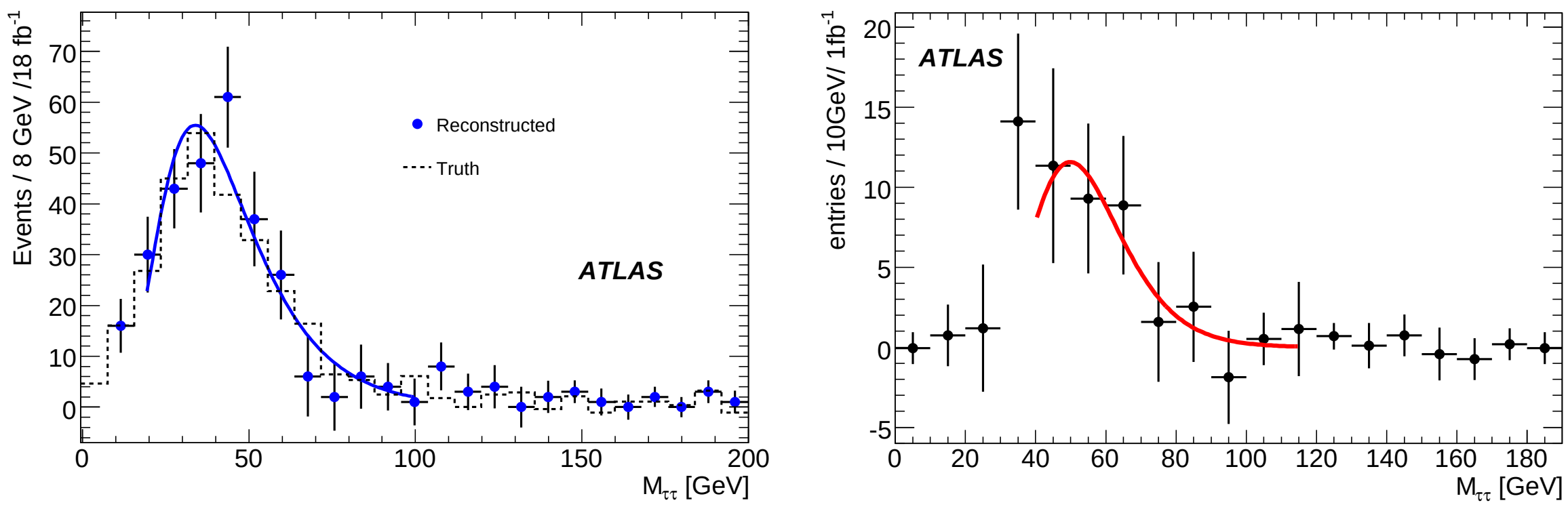


Figure 6: Invariant mass distribution of opposite-sign tau pairs with same-sign tau distribution subtracted, for the SU1 (18 fb$^{-1}$, left) and SU3 scenarios (1 fb$^{-1}$, right). The dashed histogram in the left plot shows the distribution at the generator level, while points show the reconstruction-level distribution.

The invariant mass distribution emerging after the above cuts are applied is shown in Figure 6 for the SU1 and SU3 scenarios, where the corresponding distribution of two taus with the same sign of electric charge is subtracted from tau pairs with opposite sign in order to reduce combinatorial background. This is possible because uncorrelated and fake taus should arise about as often with the same charge as with opposite charges. To decrease the SUSY background further, the two taus arising from the same decay chain are required to have a maximum separation $\Delta R < 2$ in the $\eta$-$\phi$-plane.

The following log normal function with three parameters, inspired by [14], is used to fit the $m_{\tau\tau}$ distribution:

$$f(x) = \frac{p_0}{x} \cdot \exp\left(-\frac{1}{2p_2^2}(\ln(x) - p_1)^2\right) \tag{12}$$

This function does not contain the endpoint position explicitly, but approaches the x-axis asymptotically. The endpoint is then derived from the inflection point $m_{IP}$ of the fit function:

$$m_{IP} = \exp\left(-\frac{1}{2}p_2^2\left(3 - \sqrt{1 + \frac{4}{p_2^2}}\right) + p_1\right) \tag{13}$$

using a Monte Carlo-based calibration procedure.

The inflection point obtained for 14 SU3-like models is plotted against the theoretical endpoint value for each of these models. The SU3-like points are generated using the ATLAS fast simulation program [15] and varying the masses of the $\tilde{\chi}_2^0$, the $\tilde{\tau}_1$ and the $\tilde{\chi}_1^0$ separately, while keeping the other two masses fixed. In Figure 7 the inflection point is plotted as a function of the endpoint $m_{EP}$ and fitted with a straight line, yielding the following calibration function:

$$m_{IP} = (0.47 \pm 0.02)m_{EP} + (15 \pm 2)\ \text{GeV} \tag{14}$$

The covariance between the slope and the axis intercept is $-0.034$ GeV.

The fit using function Eq. (12) for the SU1 and SU3 models is shown in Fig. 6 giving an inflection point at $m_{IP} = 48\pm3$ GeV and $m_{IP} = 62\pm8$ GeV respectively, which translates into endpoints at $m_{EP} = (70 \pm 6.5^{\text{stat}} \pm 5^{\text{syst}})$ GeV (SU1) and $(102 \pm 17^{\text{stat}} \pm 5.5^{\text{syst}})$ GeV (SU3) using the calibration relation in Eq. (14). The systematic uncertainty is dominated by the fitting procedure and is evaluated by changing the binning and fit ranges. The effects of 1% and 5% jet energy scale uncertainties have been tested and found to introduce an additional systematic uncertainty on the endpoint measurement well below 3%, so

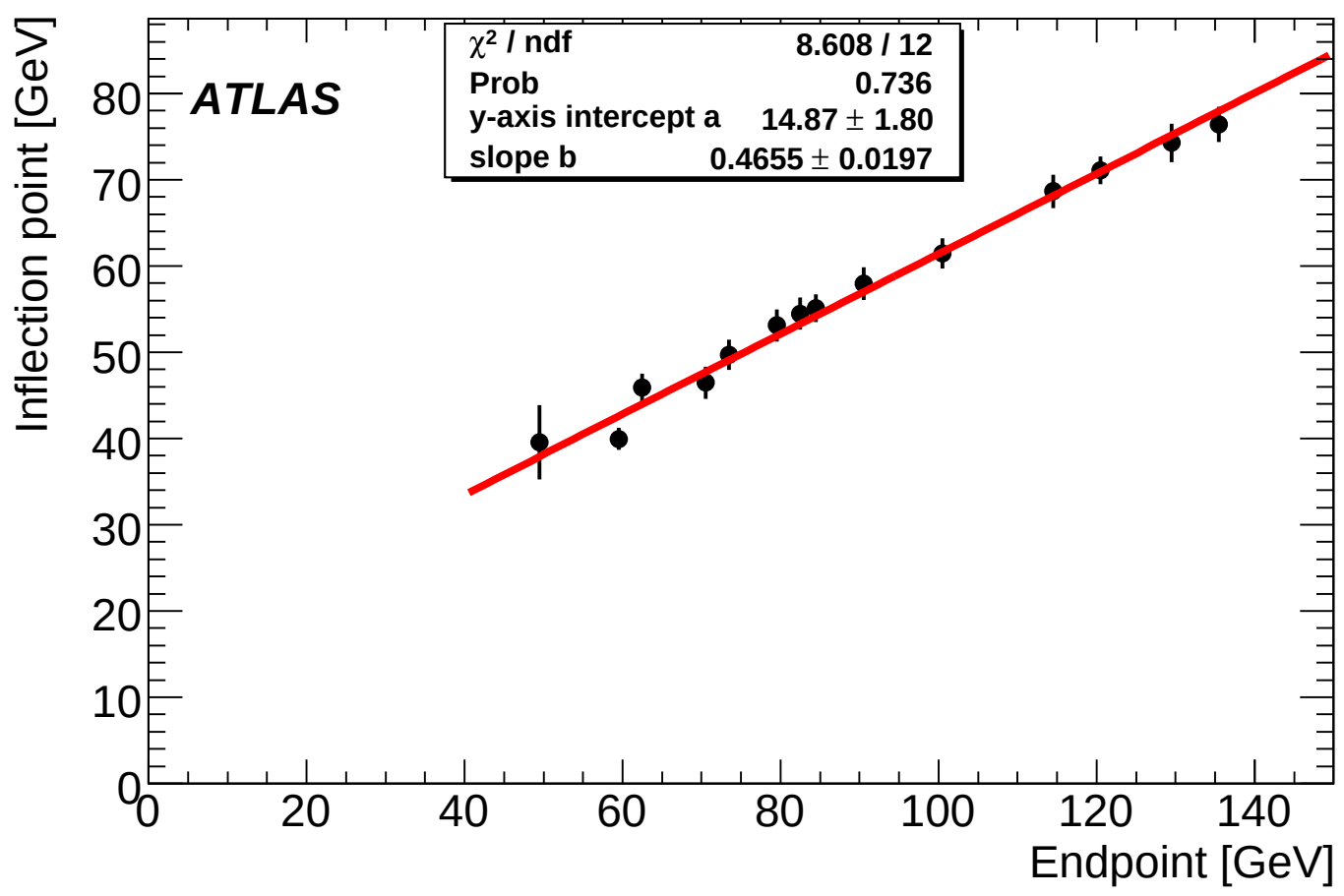


Figure 7: Calibration curve showing the relation between the position of the inflection point (measured and fitted with function Eq. (12) after ATLFAST based detector simulation) and the endpoint (calculated with equation Eq. (5)) of the di-tau mass distribution. The SU3 point is not included.

they are negligible compared to the systematic error introduced by the fitting procedure. The theoretical expectation for the SU1 (SU3) endpoint is 78 GeV (98 GeV). The difference between the theoretical and the extracted endpoint comes from the fitting and fit-calibration procedure rather than from detector effects. As Fig. 6 (left) shows, the generator-level distribution of the visible products is close to the reconstruction-level distribution and after fitting gives an endpoint value of 70 GeV for the SU1 scenario which is the same as for the reconstructed distribution.

## 5.2 Impact of the tau polarization on the di-tau mass spectrum

The method discussed in the previous section assumes a fixed polarization of the two taus from the $\tilde{\chi}_2^0$ and therefore neglects the effect of the polarization on the invariant mass spectra. However, the polarization of the taus from the decay cascade can vary significantly between different SUSY models. Polarization effects on the di-tau mass distribution are studied by simulating samples of events where the polarization of the two taus from the decay Eq. (11) is allowed to vary. The ATLAS fast simulation program [15] is used to simulate a data sample equivalent to 51 fb$^{-1}$ of data.

Parity violation in weak interactions in conjunction with momentum and angular momentum conservation leads to a correlation between the visible tau energy and the polarization of the tau. In case of the decay $\tau- > \nu_\tau \pi^-$, since the pion is a scalar particle the neutrino spin is forced to be parallel to that of the tau and therefore the fixed neutrino helicity determines the neutrino momentum direction. Thus, to conserve momentum, the direction of the pion momentum is forced to be parallel or antiparallel to that of the tau depending on the tau polarization. This leads to the pion getting a boost parallel or antiparallel to the tau momentum resulting in harder and softer pions. This affects the di-tau mass spectrum, as shown in Fig. 8. The curves in the plot are theoretical predictions from [16]. The invariant masses of taus with left chirality (*LL*) are on average smaller than for right (*RR*) taus.

For decays via the vector mesons $\rho$ and $a_1$, the momentum of the vector meson has the same (opposite) direction as in the case of pions for longitudinal (transverse) polarization of the vector meson.

Adding all hadronic tau decay modes finally yields the invariant mass distributions shown in Figure 8 for the chirality options *LL*, *RR* and *LR*/*RL*. The position of the trailing edge is clearly shifted for different polarizations whereas the shape difference calculated in [16] is barely visible after detector

simulation.

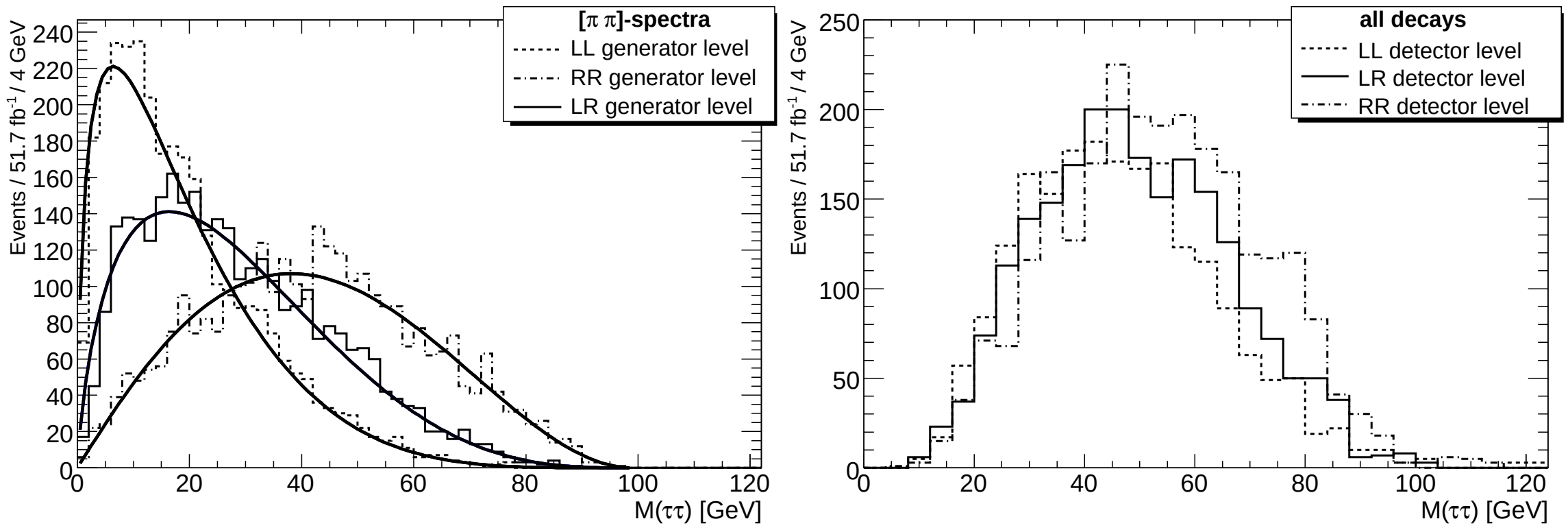


Figure 8: Left: Di-tau invariant mass spectrum for $\tau \rightarrow \pi\nu_\tau$ decays as obtained from Monte Carlo truth information together with the expectation from theory. Right: Di-tau invariant mass spectrum for all hadronic decays after an ATLFAST based detector simulation. Both plots show the mass distributions for the chirality states *LL*, *RR* and *LR*/*RL*.

As a consequence, the inflection point of the distribution is shifted due to polarization effects. The maximum difference in the inflection point measurement $\Delta m_{IPA}(RR-LL)$ has been found to be 7 GeV which is comparable to the statistical error on the position of the inflection point presented in the previous section. Without additional information on the tau polarization we might quickly reach a point where it is not possible to improve the di-tau endpoint measurement. In this case the achievable precision for an integrated luminosity of 1 fb$^{-1}$ for the SU3 model is:

$$m_{\tau\tau}^{\max} = 102 \pm 17^{\text{stat}} \pm 5.5^{\text{syst}} \pm 7^{\text{pol}}$$

The uncertainty due to the polarization effects dominates over the other systematic uncertainties and therefore needs special attention. A study of the different polarization dependencies of the decay kinematics for different decay modes of the tau might be helpful. As the emission direction of vector mesons is opposite for longitudinal and transversal states the net effect is determined by the branching ratio into the two states. It turns out that for the $a_1$ there are as many longitudinal as transverse polarized whereas for the $\rho$ there are more longitudinal vector mesons, leading to different polarization dependencies.

# 6 $\tilde{q}_{\text{R}}$ pair reconstruction

Events where a pair of $\tilde{q}_{\text{R}}$ particles is produced, and where each decays through the process

$$\tilde{q}_{\text{R}} \rightarrow \tilde{\chi}_1^0 q \tag{15}$$

lead to a characteristic signature with two high-$p_{\text{T}}$ jets and large $E_{\text{T}}^{\text{miss}}$ from the escaping neutralinos. $\tilde{q}_{\text{R}}$ pair production represents about 10% (5%) of the total SUSY production cross-section for the "Bulk" point SU3 ("Low Mass" SU4). For both points (and more generally in most of mSUGRA parameter space) the $\tilde{q}_{\text{R}}$ decays almost entirely through the process in Eq. (15).

Events with large $E_{\text{T}}^{\text{miss}}$ and a pair of high-$p_{\text{T}}$ jets are selected by requiring:

- $E_{\text{T}}^{\text{miss}} > \max(200\text{ GeV},\ 0.25 M_{\text{eff}})$ and $M_{\text{eff}} > 500$ GeV
- Two jets with $p_{\text{T}} > \max(200\text{ GeV}, 0.25 M_{\text{eff}})$, $|\eta| < 1$ and $\Delta R > 1$

Table 5: The total event yield and the number of Standard Model background events satisfying selection criteria for $m_{T2}$ reconstruction, signal-to-background ratio and signal statistical significance. Only errors from the detector systematic uncertainties are quoted.

| | Integrated luminosity ($fb^{-1}$) | Event yield | Standard Model | $S/B_{SM}$ | $S/\sqrt{B_{SM}}$ |
|---|---|---|---|---|---|
| SU3 | 1.0 | $282 \pm 20$ | 18 | $14.7 \pm 1.1$ | $62.2 \pm 4.7$ |
| SU4 | 0.5 | $258 \pm 65$ | 9 | $27.7 \pm 7.2$ | $83.0 \pm 21.7$ |

- No additional jet with $p_T > \min(200\ \text{GeV}, 0.15 M_{\text{eff}})$
- No isolated leptons and no jets tagged as $b$ jets
- Transverse sphericity $S_T > 0.2$

These selection cuts are tuned using the information from the event generation to select $\tilde{q}_R$ pair production.

The systematic uncertainty originating from the jet and $E_T^{\text{miss}}$ energy scale and resolution [2] change the event yield by 7% for the case of SU3 and by 25% for the case of SU4. The systematic effect on the energy scale and resolution of leptons is negligible.

The total event yield and the number of Standard Model background events satisfying the selection criteria for 1 $fb^{-1}$ for SU3 and 0.50 $fb^{-1}$ for SU4 are given in Table 5 together with the signal-to-background ratio and the signal statistical significance.

To reconstruct the $\tilde{q}_R$ mass we use the the $m_{T2}$ variable [2, 17, 18], sometimes referred to as "stransverse" mass. This variable uses the kinematic features of the $\tilde{q}_R$ decays to reconstruct $m_{\tilde{q}_R}$[3] making the assumption that $m_{\tilde{\chi}_1^0}$ is known from the measurements in Section 4. The $m_{T2}$ distributions for the SU3 and SU4 points are shown in Figure 9. As is common for SUSY measurements, the distribution is expected to have an edge at $m_{\tilde{q}_R}$ rather than a peak. A linear fit is applied to the right part of the distribution to determine the edge position at $590 \pm 9(\text{stat})^{+13}_{-6}(\text{sys})$ GeV for SU3 and $421 \pm 17(\text{stat})^{+10}_{-3}(\text{sys})$ GeV for SU4. This can be compared to the expected positions of $m_{\tilde{q}_R} = 611$ GeV for SU3 and $m_{\tilde{q}_R} = 406$ GeV for SU4. The systematic error accounts for the choice of the fit limits as well as the jet energy scale systematic.

# 7 Light stop signature

In the "Low Mass" benchmark point (SU4), the SUSY masses are all in the range $m_{\tilde{\chi}_1^0} = 60$ GeV $< m < m_{\tilde{t}_2} = 445$ GeV. The stop $\tilde{t}_1$ is light ($m_{\tilde{t}_1} = 206$ GeV) and always decays by $\tilde{t}_1 \to \tilde{\chi}_1^{\pm} b$. A detailed analysis of the phenomenology of this point can be found in [19].

At this SU4 benchmark point the light stop is produced in the gluino decay Eq. (3) which has a branching ratio of 42%. Associated gluino production with a $\tilde{q}_L$ or $\tilde{q}_R$ followed by the decay in Eq. (3) occurs in $\sim 18\%$ of all SU4 events. In the decay Eq. (3) the final state $tb$ invariant mass distribution has

[3] For the benchmark point considered here, the effect of the SUSY background (i.e. events other than the $\tilde{q}_R$ pair production which pass the event selection) on the position of the edge is small. This has also been shown to be true for other benchmark points [10]. However, the identification of the edge as a measurement of the $\tilde{q}_R$ mass may not hold for all the SUSY parameter space.

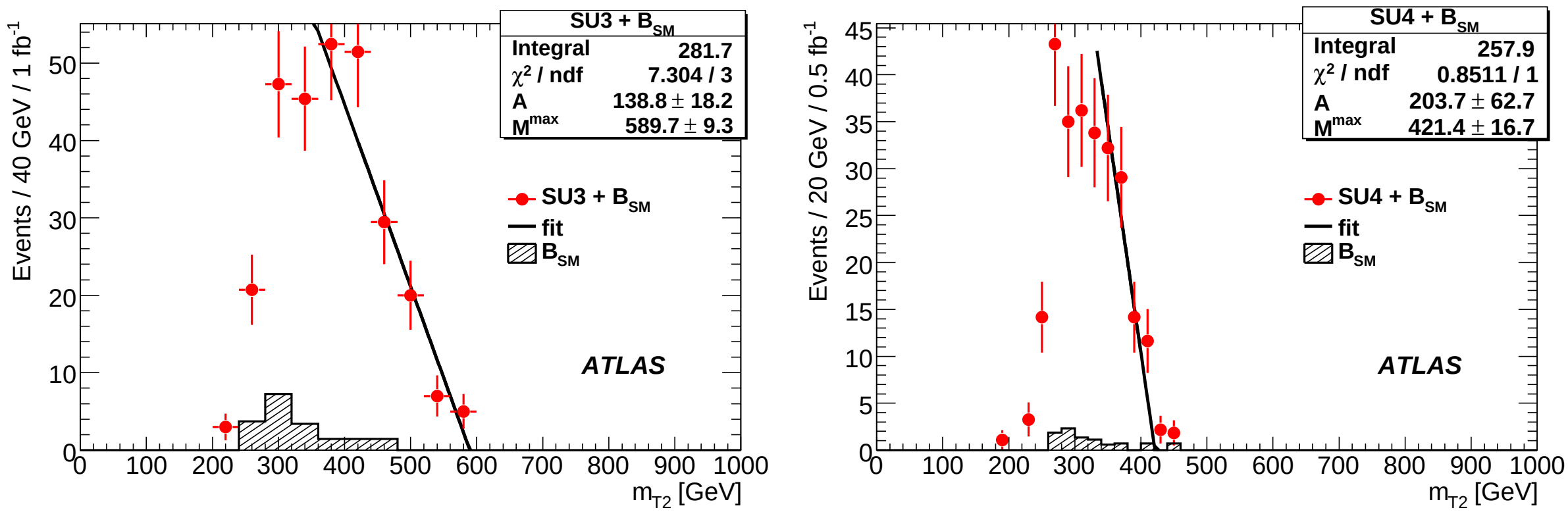


Figure 9: Fit of the sum of the reconstructed $m_{T2}$ distributions in the selected SUSY and the remaining Standard Model background events with 1 fb$^{-1}$ for SU3 and 0.5 fb$^{-1}$ for SU4.

the upper kinematic endpoint:

$$M^{\max}(tb) = \left[ m_t^2 + \frac{m_{\tilde{t}_1}^2 - m_{\tilde{\chi}_1^\pm}^2}{2m_{\tilde{t}_1}^2} \left( (m_{\tilde{g}}^2 - m_{\tilde{t}_1}^2 - m_t^2) + \sqrt{(m_{\tilde{g}}^2 - (m_{\tilde{t}_1} - m_t)^2)(m_{\tilde{g}}^2 - (m_{\tilde{t}_1} + m_t)^2)} \right) \right]^{1/2} . \quad (16)$$

With $m_{\tilde{g}} = 413$ GeV, $m_{\tilde{\chi}_1^\pm} = 113$ GeVand a top mass of 175 GeV, Eq. (16) gives

$$M^{\max}(tb) \sim 300 \text{ GeV}. \quad (17)$$

The other significant decays at this benchmark point which lead to the same final state are:

$$\tilde{g} \rightarrow \tilde{b}_1 b \rightarrow \tilde{\chi}_1^\pm tb, \quad (18)$$
$$\tilde{g} \rightarrow \tilde{b}_1 b \rightarrow \tilde{t}_1 Wb \rightarrow \tilde{\chi}_1^\pm bbW, \quad (19)$$
$$\tilde{g} \rightarrow \tilde{b}_2 b \rightarrow \tilde{\chi}_1^\pm tb, \quad (20)$$
$$\tilde{g} \rightarrow \tilde{b}_2 b \rightarrow \tilde{t}_1 Wb \rightarrow \tilde{\chi}_1^\pm bbW. \quad (21)$$

The final states from the decays Eq. (19) and Eq. (21) are equivalent to the final state from the decay Eq. (3) if the $bW$ invariant mass is close to the top mass. Associated gluino production with left or right squark followed by these decays occurs in 4% (Eq. (18)), 9% (Eq. (19)), 0.1% (Eq. (20)) and 0.9% (Eq. (21)) of all SU4 events. Due to the small mass difference between $\tilde{g}$ and the $\tilde{b}_1$ or $\tilde{b}_2$, the final states Eq. (18) and Eq. (19) can be suppressed by imposing a minimum cut on the $p_{\mathrm{T}}$ of the $b$ jet, while the final states Eq. (20) and Eq. (21) are suppressed because $b$ jets originating from the gluino decay $\tilde{g} \rightarrow \tilde{b}_2 b$ are on the average below the detection threshold.

In order to extract light stop signal from the $\tilde{g}\tilde{q}$ events where gluino decays to stop and top, the final state $tb$ invariant mass distribution in Eq. (3) is reconstructed for top quark decays into hadronic final states only:

$$t \rightarrow Wb \rightarrow qqb \quad (22)$$

making no assumptions about the $\tilde{\chi}_1^\pm$ decay modes which dominantly produce two additional light-quark jets. The hardest jet in the event is assumed to be the light-quark jet originating from the decay of the left or right squark produced in association with the gluino.

We select jets with $p_{\mathrm{T}} > 20$ GeVand $|\eta| < 2.5$. In this range the $b$-tagging efficiency is about 60% [1]. The event selection requires the following:

- At least 5 jets in the event with $p_{\mathrm{T}} > 30$ GeV, where
  - The hardest jet is a light-quark jet with $p_{\mathrm{T}} > 100$ GeV,
  - 2 and only 2 jets are tagged as $b$ jets and they have $p_{\mathrm{T}} > 50$ GeV and
  - At least 2 of the light-quark jets have $p_{\mathrm{T}} > 30$ GeV
- $E_{\mathrm{T}}^{\mathrm{miss}} > 150$ GeV, $M_{\mathrm{eff}} > 400$ GeV, $E_{\mathrm{T}}^{\mathrm{miss}}/M_{\mathrm{eff}} > 0.2$
- $S_T > 0.1$ .

Of the Standard Model backgrounds simulated, only events from our $t\bar{t}$ and QCD samples satisfy these selection criteria. No $W$+jets or $Z$+jets events pass the cuts.

The dominant detector-performance systematic uncertainties come from the jet and $E_{\mathrm{T}}^{\mathrm{miss}}$ energy scale and resolution. Assuming that the uncertainties for 200 pb$^{-1}$ are the same as for 100 pb$^{-1}$, i.e. 10% uncertainties on the jet and $E_{\mathrm{T}}^{\mathrm{miss}}$ energy scale and resolution [20], and including 5% uncertainty on the $b$-tagging efficiency, the resulting systematic uncertainty in the number of selected events would be $\sim 40\%$ for SU4 and $\sim 50\%$ for $t\bar{t}$ .

The top-bottom invariant mass is reconstructed for events satisfying the selection criteria for the light stop search:

- Excluding the hardest jet, all light-quark jets with $p_{\mathrm{T}}^{\mathrm{jet}} > 30$ GeV are combined into dijet pairs.
- All such pairs with invariant mass within the window $|m_{jj} - m_W| < 15$ GeV are combined with each of the two $b$ jets and the $bjj$ combination with invariant mass closest to the top mass is selected.
- The four-vectors of this dijet pair are rescaled such that $m_{jj} = m_W$ and $m_{bjj}$ is recalculated and accepted as top candidate if $|m_{bjj} - m_{top}| < 30$ GeV.
- The same $bjj$ combination is combined with the other $b$ jet and $m_{tb}$ calculated, with the requirement that the angle between top and bottom be $\Delta R(t,b) < 2$.

The $W$ sideband method [21–23] is used to estimate SUSY combinatorial background originating from supersymmetric processes in which jet pairs accidentally have an invariant mass within our $W$-mass window, and so fake $W$ bosons. The sidebands used are the regions of dijet invariant mass 30 GeV below and 30 GeV above our $W$-mass window. The fake $W$ boson contribution to the $m_{tb}$ distribution is evaluated as the average contribution of the jet pairs from the $W$ sidebands after they have been scaled linearly to the $W$ mass zone and the procedure of $m_{tb}$ reconstruction has been repeated.

The numbers of signal and the remaining Standard Model background events, together with the total event yield at 200 pb$^{-1}$, are listed in Table 6. The $m_{tb}$ distribution reconstructed in signal events without the subtraction of the SUSY combinatorial background and the SUSY combinatorial background itself are plotted in Figure 10. The contribution of fake $W$ bosons in SU4 is higher than the number of events remaining after subtraction.

The $m_{tb}$ distributions before and after the subtraction of the SUSY combinatorial background are shown in Figure 10. The background-subtracted distirbution is fitted in order to extract the endpoint. The resulting $m_{tb}$ distribution is fitted with a triangular function smeared with a Gaussian:

$$f(M) = A \int_{-1}^{1} e^{-\frac{(M - M^{\max}\sqrt{\frac{1+x}{2}})^2}{2\sigma^2}} dx + (a + bM),$$

where the kinematic endpoint, $M^{\max}$, and the smearing, $\sigma$, are two of the five fit parameters. The smearing, $\sigma$, models the experimental resolution of the reconstructed $m_{tb}$. The position of the upper kinematic

Table 6: The number of signal and remaining Standard Model background events and the total event yield at 200 pb$^{-1}$. The last row gives the signal-to-background ratio. The errors quoted are from detector sources of systematic uncertainty.

| L = 200 pb$^{-1}$ | Initial selection | $\lvert m_{bjj} - m_t \rvert < 30$ GeV | | $\Delta R(t,b) < 2$ | |
|---|---|---|---|---|---|
| | | without $W$ sub. | with $W$ sub. | without $W$ sub. | with $W$ sub. |
| SU4 | 963 | 537 | 224 | 267 | 120 |
| $t\bar{t}$ | 99 | 28 | 13 | 9 | 4 |
| QCD | 6 | 3 | 2 | 3 | 2 |
| Total | 1068 ± 426 | 568 ± 225 | 239 ± 95 | 279 ± 109 | 126 ± 50 |
| SU4 / ($t\bar{t}$ + QCD) | 9.2 ± 4.1 | 17.3 ± 7.3 | 14.9 ± 6.3 | 22.3 ± 9.1 | 20.0 ± 8.3 |

endpoint obtained from the 5-parameter fit (Figure 10) is $M^{\max} = 297 \pm 9$ GeV with $\sigma = 28 \pm 7$ GeV corresponding to $\sim 10\%$ of the $M^{\max}$ value. The position of the upper kinematic endpoint obtained from the 4 parameters fit is $M^{\max} = 298 \pm 6(\text{stat})^{+16}_{-41}(\text{sys})$ GeV with $\sigma$ set to 10% of $M^{\max}$. The expected value of $m_{tb}$ given by Eq. (17) is 300 GeV.

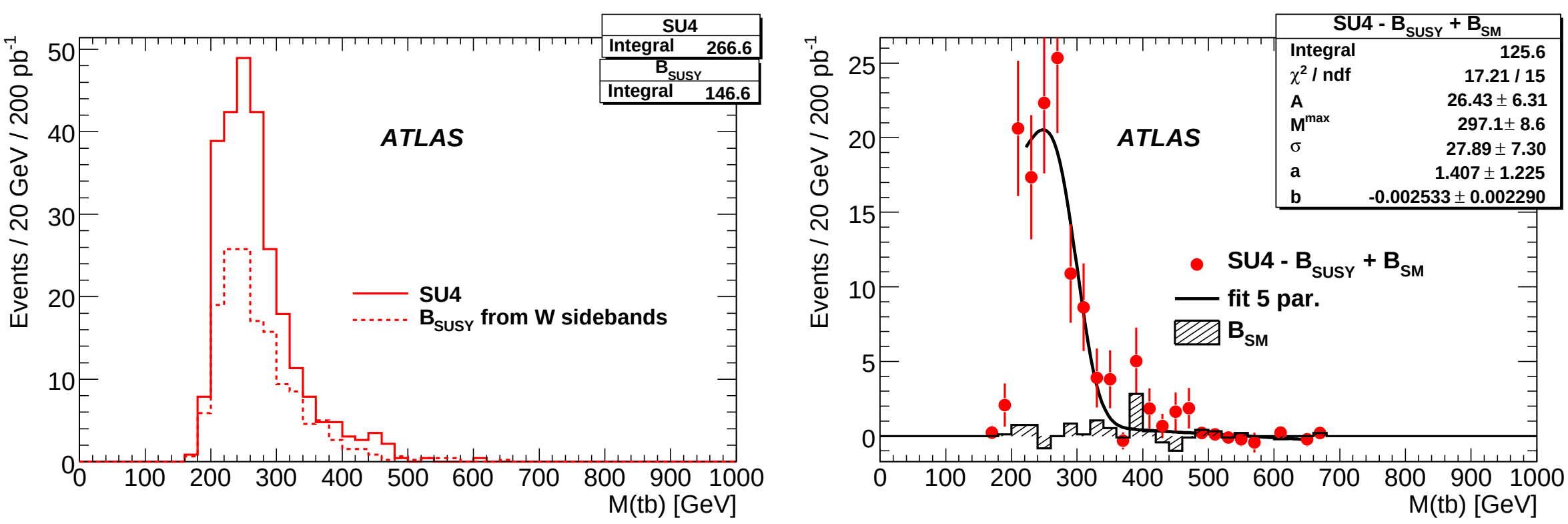


Figure 10: Left: Reconstructed $m_{tb}$ distributions in signal and SUSY combinatorial background events. Right: The 5 parameters fit of the sum of the reconstructed $m_{tb}$ distributions in signal and the remaining Standard Model events after the subtraction of the SUSY combinatorial background; all at 200 pb$^{-1}$.

## 8 Higgs signatures in SUSY events

In the context of supersymmetric models, Higgs bosons at the LHC can be produced in proton collisions either through direct interaction of Standard Model particles, such as gluon-gluon fusion, or through the decay of a supersymmetric particle produced in the initial interaction.

We will consider the possibility of observing the lightest CP-even $h$ boson via the second mechanism. In this case, a missing transverse energy signature, typical of R-parity conserving SUSY scenarios, can be reconstructed in association with the Higgs boson and exploited to reduce the background, making it possible to study the dominant decay channel $h \to b\bar{b}$, which is otherwise hidden by the enormous QCD continuum.

Within mSUGRA the most promising source of Higgs production is the decay of a second-lightest neutralino. Indeed, $\tilde{\chi}_2^0 \rightarrow \tilde{\chi}_1^0 h$, if open, dominates the $\tilde{\chi}_2^0 \rightarrow \tilde{\chi}_1^0 Z$ mode, because the two lightest neutralinos are basically gauginos, so that the higgsino-gaugino-Higgs vertex is enhanced with respect to the higgsino-higgsino-gauge one. However, if the sleptons are lighter than the $\tilde{\chi}_2^0$, the decay channels $\tilde{\ell}^{\pm}\ell^{\mp}$ and $\tilde{\nu}_\ell \bar{\nu}_\ell$ open up, dominating the $\tilde{\chi}_2^0$ width. As a consequence, we expect that mSUGRA points interesting for Higgs searches will not show a clear di-lepton signature. The benchmark point chosen for this analysis is SU9, at which $BR(\tilde{\chi}_2^0 \rightarrow \tilde{\chi}_1^0 h) \sim 87\%$.

Exploiting the capabilities of the ATLAS detector in missing transverse energy measurement and $b$ tagging, the passage of weakly interacting particles may be revealed and a $b\bar{b}$ pair with invariant mass peaking around the Higgs mass can be reconstructed.

Standard Model events with similar signatures which are backgrounds for this analysis, include events with neutrino production causing a genuine $E_{\mathrm{T}}^{\mathrm{miss}}$ signal and QCD events with fake $E_{\mathrm{T}}^{\mathrm{miss}}$ generated by instrumental effects. SUSY events can themselves constitute a background, as they contain many $b$-jet candidates, both true and mistagged. These can be divided into two categories: SUSY cascades without and with production of a Higgs decaying to $b\bar{b}$. In the latter case potential signal events may be incorrectly reconstructed as the selected $b\bar{b}$ pair is not the one coming from the Higgs decay. We will refer to the former type simply as "SUSY background" and to the latter as "combinatorial background".

The following selection cuts are applied:

1. $E_{\mathrm{T}}^{\mathrm{miss}} > 300$ GeV;
2. two light-flavoured jets with $p_{\mathrm{T}} > 100$ GeV;
3. two $b$ jets with $p_{\mathrm{T}} > 50$ GeV;
4. no leptons with $p_{\mathrm{T}} > 10$ GeV.

The first two cuts are typical of SUSY analyses, while the pupose of the last cut is to suppress backgrounds from $t\bar{t}$ and $W$ production.

When three or more $b$ jets with transverse momentum greater than 50 GeV are found in a single event, the second and third leading-$p_{\mathrm{T}}$ $b$-jets are chosen as the candidate Higgs decay pair. This is because an important source of $b$ jets is the decay of a bottom squark to $\tilde{\chi}_2^0$ and $b$ and since $m_{\tilde{b}} - m_{\tilde{\chi}_2^0} \sim 500\text{ GeV} > m_h$ the sbottom daughters get more allowed phase space than the Higgs daughters and thus, in general, higher $p_{\mathrm{T}}$.

In Figure 11 (left) the invariant mass of the selected $b$ jet pairs is shown assuming 10 fb$^{-1}$ of collected luminosity. The shaded histogram corresponds to the sum of the Standard Model backgrounds, the dashed and dotted lines are the SUSY and combinatorial backgrounds respectively. These last two, together with the $t\bar{t}$ production, are the most important backgrounds. The black curve is the result of a least squares fit to a Gaussian function, representing the Higgs resonance, superimposed on a second degree polynomial background. The estimated number of signal and background events is obtained by counting the $b$ pairs with invariant mass inside a $\pm 25$ GeV range around the fitted peak centre. The signal significance, computed in the Gaussian approximation as the number of signal events over the square root of the background, is about 14.

Table 7 summarises the expected event rates after the application of the selection cuts and after the additional mass window request.

SUSY mass spectrum information is reconstructed in the SU9 model by studying the decay:

$$\tilde{q}_{\mathrm{L}} \rightarrow \tilde{\chi}_2^0 q \rightarrow \tilde{\chi}_1^0 h q. \quad (23)$$

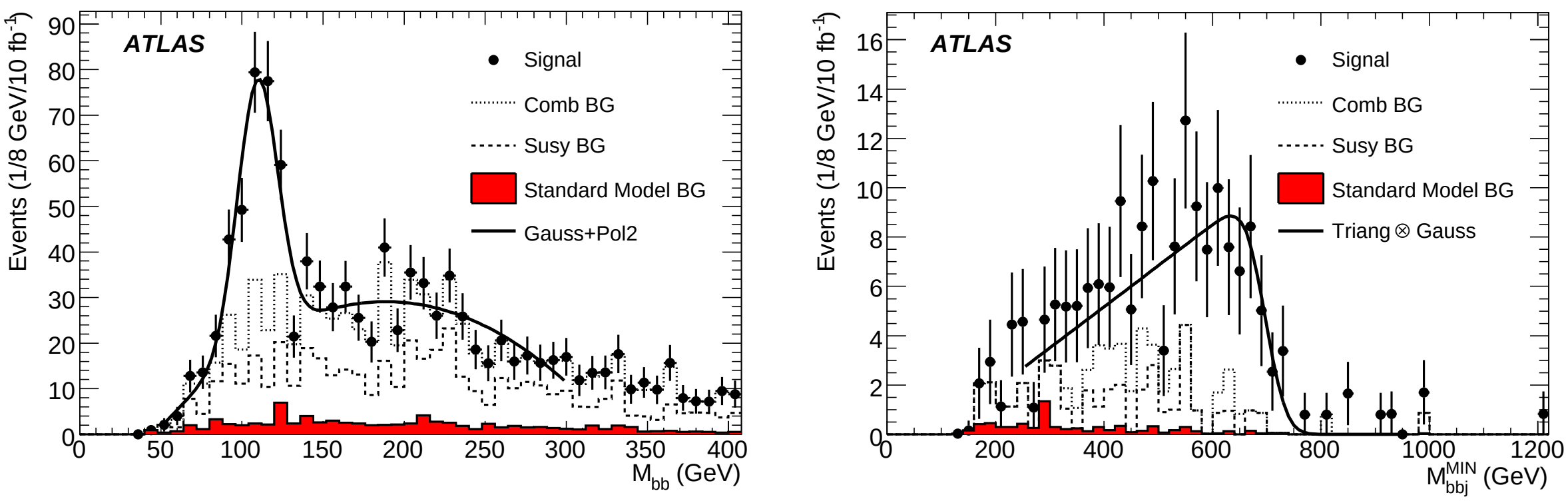


Figure 11: Invariant mass of the selected $b$-jet pairs (left) and invariant mass of the system consisting of the Higgs plus the jet minimising $m_{hq}$ (right) for 10 fb$^{-1}$ of integrated luminosity.

Table 7: Summary of the number of expected SUSY and Standard Model events after the application of the different selection cuts, for 10 fb$^{-1}$ of integrated luminosity.

| **SU9** | Signal | Comb BG | Susy BG |
|---|---|---|---|
| No cuts | 11050 | | 21950 |
| Cut *1, 2, 3* | 356 | 946 | 908 |
| Cut *4* | 230 | 449 | 433 |
| ±25 GeV mass window | 179 | 76 | 76 |

| **Standard Model** | $t\bar{t}$ | $Z$ | $W$ | $b\bar{b}$ |
|---|---|---|---|---|
| Cut *1, 2, 3* | 133 | 12 | 22 | 43 |
| Cut *4* | 53 | 8 | 10 | 21 |
| ±25 GeV mass window | 11 | 2 | 4 | 4 |

As a consequence of two-body kinematics, the invariant mass of the Higgs-quark system shows both a minimum and a maximum value, related to different combinations of the masses of the SUSY particles involved.

The events passing the previous selection cuts, including the mass window cut, are also required to have at least one $b$ jet with $p_T > 100$ GeV. Furthermore, a veto is imposed on additional $b$-tagged jets with $p_T > 50$ GeV. This will result in fewer signal events, but also in a reduced background contamination yielding a clear distribution shape, albeit with lower signal significance. As in Section 4 the quark from the $\tilde{q}_L$ is expected to produce one of the two highest-$p_T$ jets and the two distributions $m_{hq}^{\min}$ and $m_{hq}^{\max}$ are reconstructed using the jet that maximises and minimises the $m_{hq}$ value, respectively. Since the background events will tend to concentrate toward low mass values, the $m_{hq}^{\min}$ (Figure 11, right) is used to determine the mass upper limit $m_{hq,\mathrm{edge}}$. The $m_{hq,\mathrm{threshold}}$ value can determined from the $m_{hq}^{\min}$ distribution.

The mass edge value is obtained by fitting a triangular shape convolved with a Gaussian:

$$m_{hq,\mathrm{edge}} = 695 \pm 15\ (\mathrm{stat}) \pm 3\ (\mathrm{syst}) \pm 35\ (\mathrm{JES}),$$

to be compared to the true value of 732 GeV. The statistical uncertainty is the error on the fitted parameter,

while the systematic comes from the parameter dependence on the fitting boundaries. An additional 5% systematic error is expected from the jet energy scale (JES) uncertainty.

The mass threshold evaluation is more challenging, since background events tend to populate the low mass region. For an integrated luminosity of 10 fb$^{-1}$, it was not possible to fit to the mass distribution. However, high-statistics studies performed with a fast simulation of the ATLAS detector show that a mass threshold value could be extracted from 300 fb$^{-1}$ of collected data, expected after 3 years of LHC running at design luminosity.

# 9 Mass and parameters measurement

This section is devoted to the extraction of SUSY mass spectra and parameters from the measurements described in the previous sections of this note. As an example, the mSUGRA benchmark points SU3 (with a luminosity of 1 fb$^{-1}$) and SU4 (with 0.5 fb$^{-1}$) are chosen to give a flavour of what might be expected in the initial phase of the experiment in case of a rather optimistic SUSY scenario. The small luminosity at this stage results in a limited number of available measurements and rather large uncertainties. In such a situation only models with few parameters can be fitted. In Section 9.1 the masses of the decay chain Eq. (1) are derived from the measurement of the kinematic endpoints described in Section 3 and 4. In Sections 9.2 to 9.4 the parameters of the mSUGRA model are derived instead.

## 9.1 Measurement of masses from SUSY decays

We use the information from the experimentally measured endpoints to extract the masses of the SUSY particles. As described earlier in this paper, in many cases analytic expressions have been deduced for the endpoints expressed in terms of the masses. Examples are shown in Eq. (4) and Eq. (5). If sufficient endpoints are known, then the masses can be deduced. Here we use a numerical $\chi^2$ minimization based on the MINUIT package to extract the SUSY particle masses from a combination of endpoints. We define the $\chi^2$ as

$$\chi^2 = \sum_{k=1}^{n} \frac{(m_k^{\max} - t_k^{\max}(m_{\tilde{\chi}_1^0}, m_{\tilde{\chi}_2^0}, m_{\tilde{\ell}_{\mathrm{R}}}, m_{\tilde{q}_{\mathrm{L}}}))^2}{\sigma_k^2}. \tag{24}$$

For each of the $n$ endpoint measurements, $k$, the quantities $m_k^{\max}$ and $\sigma_k$ denote the fit value and its uncertainty respectively. The $t_k^{max}$ are the theoretical endpoint expressions [6, 10], which contain as parameters the masses of the two lightest neutralinos, the scalar quark $\tilde{q}_{\mathrm{L}}$ , and (for the two-body decay chain only) the scalar lepton $\tilde{\ell}_{\mathrm{R}}$ . As a starting point for the fit the generated masses are used. The masses are constrained to be positive in the fit and the mass hierarchies from the model input are enforced.

With the statistics expected for an integrated luminosity of 1 fb$^{-1}$ (0.5 fb$^{-1}$) for SU3 (SU4) we observe instabilities in the fit. Depending on the fluctuation and precision of the endpoint measurements the fit does not converge. Some of the measured endpoints have shown possible deviations from the generated values of up to several standard deviations. Systematic effects to explain such discrepancies are identified in earlier sections of this article. Significant deviations distort the results and also negatively affect the fit convergence, especially in the presence of degenerate kinematic endpoint equations. We also note large correlations (typically larger than 95%) among the fitted parameters. This is expected, since the endpoints are most sensitive to mass differences. The correlations can also lead to difficulties with convergence and result in larger uncertainties for the fitted masses.

To show a potential result with early data, we quote here the results from converging fits for both the SU3 and SU4 points. We use the endpoints from the lepton+jets edges summarized in Table 4 and the dilepton edge fit from Section 3 ($99.7 \pm 1.4 \pm 0.3$ GeV for SU3 and $52.7 \pm 2.4 \pm 0.3$ GeV for SU4). For the SU3 fit, all dilepton+jets edges are used. In the SU4-fit, we discard the $m_{\ell q(\mathrm{low})}$ measurement as it has been shown not to be very reliable and does not provide additional constraints in three-body

Table 8: Resulting SUSY particle masses and mass differences within SU3 and SU4 from the $\chi^2$ minimization fit using the dilepton and lepton+jets edges. Shown are the measured masses $m_{\text{meas}}$ and mass differences $\Delta m_{\text{meas}}$ followed first by the parabolic errors as returned by MIGRAD and then by the jet energy scale errors. When the measured parameter is anticorrelated with the jet energy scale variation, this is indicated by a $\mp$ sign. The input Monte Carlo masses $m_{\text{MC}}$ and mass differences $\Delta m_{\text{MC}}$ are also shown. The integrated luminosity assumed is 1 fb$^{-1}$ for SU3 and 0.5 fb$^{-1}$ for SU4.

| Observable | SU3 $m_{\text{meas}}$ [GeV] | SU3 $m_{\text{MC}}$ [GeV] | SU4 $m_{\text{meas}}$ [GeV] | SU4 $m_{\text{MC}}$ [GeV] |
|---|---|---|---|---|
| $m_{\tilde{\chi}_1^0}$ | $88 \pm 60 \mp 2$ | 118 | $62 \pm 126 \mp 0.4$ | 60 |
| $m_{\tilde{\chi}_2^0}$ | $189 \pm 60 \mp 2$ | 219 | $115 \pm 126 \mp 0.4$ | 114 |
| $m_{\tilde{q}}$ | $614 \pm 91 \pm 11$ | 634 | $406 \pm 180 \pm 9$ | 416 |
| $m_{\tilde{\ell}}$ | $122 \pm 61 \mp 2$ | 155 | | |
| Observable | SU3 $\Delta m_{\text{meas}}$ [GeV] | SU3 $\Delta m_{\text{MC}}$ [GeV] | SU4 $\Delta m_{\text{meas}}$ [GeV] | SU4 $\Delta m_{\text{MC}}$ [GeV] |
| $m_{\tilde{\chi}_2^0} - m_{\tilde{\chi}_1^0}$ | $100.6 \pm 1.9 \mp 0.0$ | 100.7 | $52.7 \pm 2.4 \mp 0.0$ | 53.6 |
| $m_{\tilde{q}} - m_{\tilde{\chi}_1^0}$ | $526 \pm 34 \pm 13$ | 516.0 | $344 \pm 53 \pm 9$ | 356 |
| $m_{\tilde{\ell}} - m_{\tilde{\chi}_1^0}$ | $34.2 \pm 3.8 \mp 0.1$ | 37.6 | | |

decay scenarios such as SU4. The $m_{\ell q(\text{high})}$ endpoint can be measured more reliably and its kinematic expression only differs from the one of $m_{\ell q(\text{low})}$ by a constant factor. The di-tau edges are not used here.

The masses resulting from the $\chi^2$ fit are shown in Table 8 (upper part). The parabolic errors are the first errors shown in the table and the jet energy scale errors are the second. Asymmetric errors show a large uncertainty on the positive side. The jet energy scale errors are determined by varying all the endpoints along with their fully correlated jet energy scale uncertainties and refitting the masses. The difference in the central values of the fit is taken as the jet energy scale uncertainty for Table 8. One can see that the jet energy scale errors are small compared to the error on the masses, but might be relevant to the mass difference measurements.

We note that a fit with SU3 kinematic assumptions applied to the SU4 endpoints also returns consistent masses. The decision about the mass hierarchy would thus have to be based on additional information from collider data. A possible source is the shape of the dilepton edge as discussed in Section 3.

Besides the masses we can also extract differences of SUSY particle masses. These are more directly related to the endpoints and we expect to be able to determine them more reliably than the masses of individual sparticles. For mass differences, a $\chi^2$ similar to Eq. Eq. (24) is used, where the parameters are written in terms of mass differences to the neutralino $\tilde{\chi}_1^0$. We obtain the results shown in Table 8 (bottom part).

We conclude that a first look at sparticle masses is possible with early data, although with large uncertainties. Appropriate model assumptions and additional information will probably have to be used to constrain the fits.

## 9.2 Observables and fit assumptions

To demonstrate the feasibility of parameter determination with initial data, we show the constraints one would obtain for our benchmark points if one assumed an mSUGRA framework.

The SUSY parameter-fitting package Fittino version 1.4.1 [24] is used, interfaced to a beta version of SPheno3 [25] to perform the theoretical calculations for a given set of parameters.

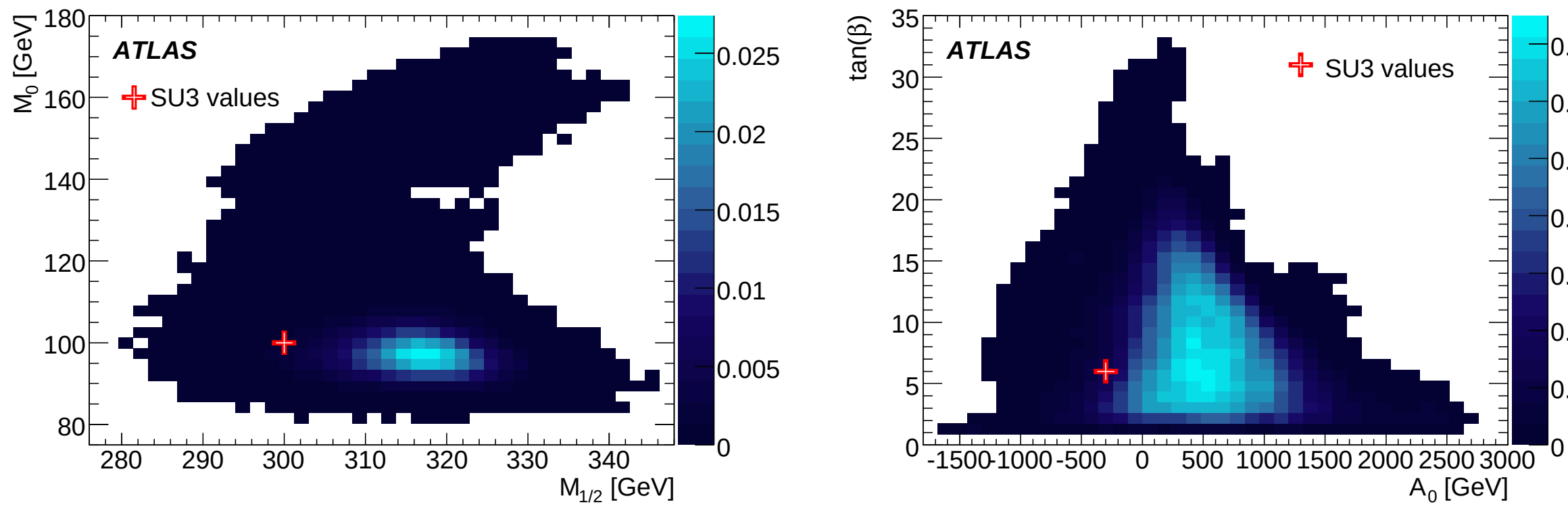


Figure 12: Two-dimensional Markov chain likelihood maps for mSUGRA parameters $M_0$ and $M_{1/2}$ (left) as well as $\tan\beta$ and $A_0$ (right) for sign $\mu = +1$, for benchmark point SU3, with integrated luminosity of 1 fb$^{-1}$. The crosses indicate the actual values of the parameters for that benchmark point.

The fit is given the measurements presented in sections 3, 4 and 6. The lepton and the jet energy scale uncertainties are each considered to be 100% correlated between measurements. Uncertainties on the theoretical predictions are not taken into account. For illustration purposes an additional parameter determination is performed where – following a prescription used in [26] – 1% (0.5%) uncertainty on the theoretical calculation of the pole masses of coloured (un-coloured) sparticles is assumed. No correlations between the theoretical uncertainties on the pole masses are considered.

## 9.3 Markov chain analysis

To obtain a first glimpse of the possible parameter space a Markov chain analysis is performed. With this technique it is possible to efficiently sample from a large-dimensional parameter spaces. This allows us to check whether there are several topologically disconnected parameter regions which are favoured by the given measurements.

Figure 12 shows two-dimensional likelihood maps for $M_0$ and $M_{1/2}$ (left) as well as $\tan\beta$ and $A_0$ (right) for sign $\mu = +1$ obtained for the given set of measurements. The plots demonstrate that for a given sign $\mu$ preferred parameters are found around the true parameter points independent of the starting point. No further preferred regions occur. For $M_0$ and $M_{1/2}$ a clearly preferred region is found around the SU3 values of 100 GeV and 300 GeV, respectively. As expected, given the measurements used, the determination of $\tan\beta$ and $A_0$ is more difficult. Nevertheless, here too the region around the nominal SU3 values is the preferred one.

## 9.4 Parameter determination

In order to determine the derived central values of the parameters and their uncertainties, for each assumption of the sign of $\mu$ a set of 500 toy fits are performed. For each fit the observables are smeared using the full correlation matrix. Simulated annealing followed by a Minuit fit started with the best parameter estimates from simulated annealing is subsequently run using the smeared observables. The four-dimensional distribution of parameters obtained from the toy fits is used to derive the parameter uncertainties and their correlations. Figure 13 shows the one-dimensional projections of parameter distributions for $M_0$, $M_{1/2}$, $\tan\beta$ and $A_0$. The mean and RMS values of the results of the fit are reported in Table 9. As already indicated by the Markov chain analysis $M_0$ and $M_{1/2}$ can be derived reliably with

uncertainties ± 9.3 GeV and ± 6.9 GeV (RMS of the toy fit results), respectively whereas for $\tan\beta$ and $A_0$ only the order of magnitude can be derived from these measurements. The $\chi^2$ distribution of the toy

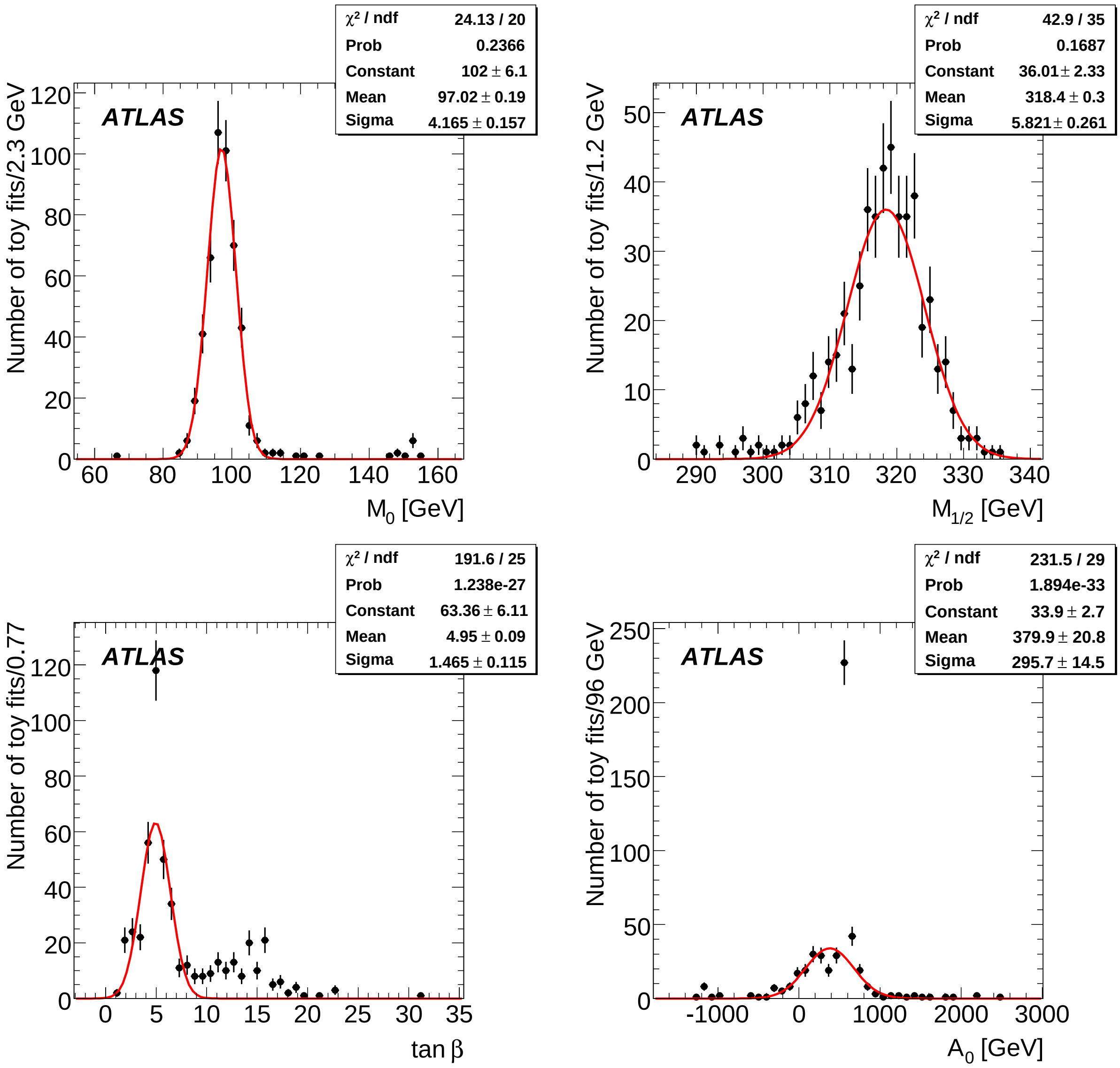


Figure 13: Distributions of the mSUGRA parameters obtained with the fits to pseudo-experiment results.

fits can be used to evaluate the toy fit performance. The observed mean $\chi^2 = 12.6 \pm 0.2$ for sign $\mu$ = +1 is compatible with the expected value of $N_{dof} = 11$. The solutions for the wrong assumption sign $\mu = -1$, also reported in Table 9, cannot however be ruled out as the observed mean $\chi^2 = 15.4 \pm 0.3$ is also acceptable.

## 10 Conclusions

If the supersymmetric partners of quarks and gluons exist at a moderate mass scale ($\lesssim$ 1 TeV) they will be abundantly produced in $pp$ collisions at the LHC centre-of-mass energy of 14 TeV. In this scenario, a few $\text{fb}^{-1}$ of ATLAS data will allow the discovery of the new particles [1], once the commissioning of

Table 9: Results of a fit of the mSUGRA parameters to the observables listed in Sections 3, 4 and 6 for the SU3 point. The mean and RMS of the distribution of the results from the toy fits is reported. The two possible assumptions for the digital parameter $\text{sign}(\mu) = +1$ $\text{sign}(\mu) = \pm 1$ have been used, resulting in different preferred regions for the other parameters. The effect of different assumptions on theoretical uncertainties is also shown.

| Parameter | SU3 value | fitted value | exp. unc. |
|---|---|---|---|
| | $\text{sign}(\mu) = +1$ | | |
| $\tan\beta$ | 6 | 7.4 | 4.6 |
| $M_0$ | 100 GeV | 98.5 GeV | $\pm 9.3$ GeV |
| $M_{1/2}$ | 300 GeV | 317.7 GeV | $\pm 6.9$ GeV |
| $A_0$ | $-300$ GeV | 445 GeV | $\pm 408$ GeV |
| | $\text{sign}(\mu) = -1$ | | |
| $\tan\beta$ | | 13.9 | $\pm 2.8$ |
| $M_0$ | | 104 GeV | $\pm 18$ GeV |
| $M_{1/2}$ | | 309.6 GeV | $\pm 5.9$ GeV |
| $A_0$ | | 489 GeV | $\pm 189$ GeV |

the detector has been completed and the Standard Model backgrounds have been well understood.

The next step after discovery will be to select specific supersymmetric decay chains to measure the properties of the new particles. Here we have focused on those measurements that will be possible using 1 $\text{fb}^{-1}$ of integrated luminosity. Specific benchmarks in parameter space have been used to demonstrate the precision that can be expected from these measurements, but the same (or similar) techniques can be applied to much of the SUSY parameter space accessible with early LHC data.

For the benchmark points considered, the most promising decay chain involves the leptonic decay of the next-to-lightest neutralino ($\tilde{\chi}_2^0 \to \tilde{\chi}_1^0 \ell^+ \ell^-$). The invariant mass of the two leptons shows a clear kinematic maximum (Section 3) which could already be measured with a precision of a few per cent with the limited data set considered. The combination of one or both leptons with the hardest jets in the event would allow observation of several other kinematic minima and maxima (Section 4).

For high values of $\tan\beta$ the decays into taus will be far more abundant than those involving electrons or muons; the excellent performance expected for the identification and measurement of hadronic $\tau$ decays in ATLAS will also allow observation of the dilepton edge in the $\tau^+\tau^-$ invariant mass distribution (Section 5).

The leptonic decays will not be the only channel for early measurements with supersymmetric decays. The $\tilde{q}_R \to q\tilde{\chi}_1^0$ decay can be used to determine the $\tilde{q}_R$ mass (Section 6). The combination of hadronically decaying top quarks and $b$-jets in supersymmetric events is also a promising possibility for low-scale Supersymmetry with decay chains involving scalar top and bottom quarks, as shown by the reconstruction of the edge of the $tb$ invariant mass discussed in Section 7.

If the $\tilde{\chi}_2^0 \to \tilde{\chi}_1^0 h$ decay is open, it will provide a substantial source of Higgs bosons. Since the Standard Model backgrounds can be suppressed by the usual SUSY cuts, it will then become possible to observe the $h \to b\bar{b}$ decay with moderate (5 $\text{fb}^{-1}$) integrated luminosity (Section 8).

The different channels will provide complementary information about the SUSY mass phenomenology. The various measurements will have to be combined to reconstruct the SUSY mass spectrum and attempt to understand the SUSY-breaking mechanism. In Section 9 it is discussed how a selected set

of early studies can be combined to obtain the first measurements of supersymmetric masses and of the parameters of the mSUGRA model. With 1 fb$^{-1}$ the reconstruction of part of the supersymmetric mass spectrum will only be possible for favourable SUSY scenarios and with some assumptions about the decay chains involved. Larger integrated luminosity will help to overcome these limitations, as more measurements become possible and the precision of each increases.

# Multi-Lepton Supersymmetry Searches

**Abstract**
We investigate the potential of the ATLAS detector to discover new physics events containing three leptons and missing transverse momentum. Such final states are predicted in a variety of extensions to the Standard Model. In the context of supersymmetric models, they could result from direct production of gaugino pairs. Using Monte Carlo simulations we present the discovery potential for several benchmark Supersymmetry points. We pay particular attention to the case where all strongly interacting sparticles are heavy. We investigate trigger and reconstruction efficiencies and discuss methods for measuring various systematic uncertainties. A solid discovery is expected with an integrated luminosity of the order of several inverse fb. If coloured particles are heavy direct production of gauginos dominates. In such scenarios, discovery would require about an order of magnitude larger luminosity.

## 1 Introduction

If supersymmetric (or other partner) particle production at the LHC is dominated by particles without colour charge, then one of the most promising discovery channels is in multi-lepton + missing transverse momentum ($E_{\mathrm{T}}^{\mathrm{miss}}$ ) final states with little hadronic activity. In this section we investigate the ability of the ATLAS experiment to discover new physics in events containing three (or more) leptons – either electrons or muons – of which two must have opposite signs but the same flavour (OSSF). We determine the sensitivity which would be obtained for discovery of five benchmark points with an integrated luminosity of 10 fb$^{-1}$.

While an analysis of this channel is clearly sensitive to other models, in this section we use Supersymmetry as our example signature, we assume R-parity conservation, and that the lightest SUSY particle (LSP) is the weakly interacting $\tilde{\chi}_1^0$, which provides the missing transverse energy signal.

In the supersymmetric case, the final states of interest could come from leptonic decay of pairs of heavy gauginos (such as $\tilde{\chi}_2^0$ and $\tilde{\chi}_1^+$) through real or virtual $W^{\pm}$, $Z^0$ or sleptons to leptons and a pair of LSPs. The heavy gauginos may be produced directly, or in the decay of heavier partner particles.

The primary aim is to make the most realistic determination which is possible at this time of the discovery potential in this channel. We also wish to identify the most important Standard Model backgrounds, so that analyses can be prepared to measure them in control regions with the ATLAS data.

Throughout this study we use the inclusive Supersymmetry production Monte Carlo samples described in [1]. The points lie in various regions of mSUGRA parameter space in which the LSP relic density is broadly consistent with the observed cold dark matter density.

The Standard Model backgrounds simulated are listed in Table 1. The most important backgrounds are found to be $t\bar{t}$ , $Zb$ and $ZW$. Fully leptonic $ZW$ events represent one significant source of events containing three leptons and missing energy. Their contribution can be reduced by rejecting OSSF lepton pairs with invariant masses consistent with the $Z$ mass. Leptonic decays of $t\bar{t}$ , and $Zb$ are expected to produce two leptons but can generate a third from leptonic $b$ quark decay. These backgrounds have large cross-sections, but can be reduced by the introduction of stringent cuts on the isolation of the lepton tracks – as we will discuss in Section 3.

A particularly important benchmark point for the trilepton analysis is the point SU2 [1]. This point lies within the 'focus point' region of mSUGRA parameter space which is characterised by very large masses for squarks and sleptons and relatively light gauginos. The heavy squarks and sleptons (see Figure 1 and Table 2 for the SUSY mass hierarchy at SU2), will have very small production cross-sections

Table 1: List of the background samples used, with Monte Carlo generator cross-sections ($\sigma$), next-to-leading-order to leading-order $k$ factors (where known for leading-order generators), average weights ($\langle w \rangle$) and corresponding integrated luminosities. The cross-sections for $WW$, $WZ$, $ZZ$, $Z\gamma$ and $Zb$ are quoted after a filter is applied on the generator output requiring at least one lepton with pseudorapidity, $|\eta| < 2.8$ and transverse momentum, $p_T > 10$ GeV. The diboson ($WW$, $WZ$, $ZZ$) samples have a different cross-section than the diboson MC@NLO samples (found elsewhere in this volume) because they also include contributions from (and interference with) the photon pole.

| Process | $\sigma$ [pb] | $k$ factor | $\langle w \rangle$ | $\int dt \mathscr{L}$ [fb$^{-1}$] |
|---|---|---|---|---|
| $WW$ | 24.5 | 1.67 | 1 | 1.22 |
| $WZ$ | 7.8 | 2.05 | 1 | 2.98 |
| $ZZ$ | 2.1 | 1.88 | 1 | 12.7 |
| $Z\gamma$ | 2.6 | 1.30 | 1 | 2.98 |
| $Zb$ | 154 | 1 | 0.66 | 0.75 |
| $t\bar{t}$ | 450 | - | 0.73 | 0.92 |

at the LHC and make it a difficult region in which to discover SUSY using the analyses described in [2] based on the selection of hadronic jets and missing transverse momentum [3]. However, gaugino and gluino production will still be abundant, so we expect good discovery potential in multi-lepton events. The branching ratios of gauginos for SU2 can be found in Table 3[1].

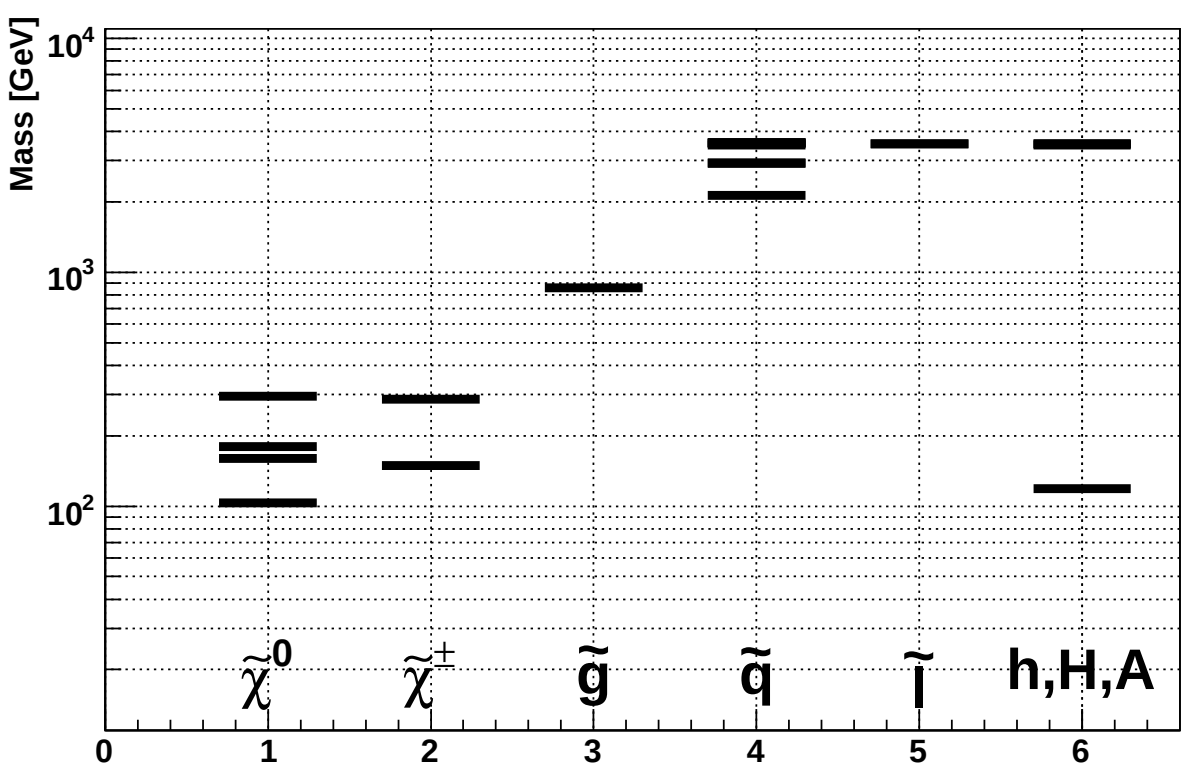


Figure 1: SU2 sparticle mass spectrum.

Table 2: Particle masses for SU2.

| Sparticle | Mass [GeV] | Sparticle | Mass [GeV] |
|---|---|---|---|
| $\tilde{\chi}_1^0$ | 103 | $\tilde{g}$ | 857 |
| $\tilde{\chi}_2^0$ | 160 | $\tilde{u}_L$ | 3563 |
| $\tilde{\chi}_3^0$ | 180 | $\tilde{u}_R$ | 3574 |
| $\tilde{\chi}_4^0$ | 295 | $\tilde{d}_L$ | 3564 |
| $\tilde{\chi}_1^\pm$ | 149 | $\tilde{d}_R$ | 3576 |
| $\tilde{\chi}_2^\pm$ | 287 | $\tilde{b}_1$ | 2925 |
| $\tilde{\ell}_L$ | 3548 | $\tilde{b}_2$ | 3501 |
| $\tilde{\ell}_R$ | 3547 | $\tilde{t}_1$ | 2131 |
| $\tilde{\nu}_L$ | 3546 | $\tilde{t}_2$ | 2935 |

The total cross-section times branching ratio for chargino-neutralino direct pair production and decay to a trilepton final state is 32.6 fb. Table 4 shows the contribution from each $\tilde{\chi}^\pm\ \tilde{\chi}^0$ pair production to a trilepton final state. The contribution from $\tilde{\chi}^\pm\ \tilde{\chi}^\mp$ and $\tilde{\chi}^0\ \tilde{\chi}^0$ pair production to a trilepton final state is not tabulated, but also adds a small contribution to the signal. Exclusive trilepton signal for SU2 is dominated by the pair production $\tilde{\chi}_1^\pm\ \tilde{\chi}_2^0$ followed by $\tilde{\chi}_1^\pm \to \tilde{\chi}_1^0\ \ell\nu$ and $\tilde{\chi}_2^0 \to \tilde{\chi}_1^0\ \ell^+\ell^-$ decays.

The trilepton signal may include contributions both from direct gaugino pair events and from other SUSY events. The latter can lead to trilepton final states when cascades initiated by heavier sparticles decay via gauginos or sleptons. In a search it is not necessary to distinguish between the various contributions to new physics and so both classes of event form part of the signal. However in this study we are particularly interested in the scenario in which all strongly interacting particles are heavy, since in those cases other analyses (requiring jets) will have more difficulty making a discovery. Since we want to be sensitive to SUSY even in this harder case, we define the signal in two different ways:

[1]The masses and branching ratios were calculated using Isajet v7.71 [4] using a top mass of 175 GeV.

Table 3: Some important branching ratios for the benchmark point SU2.

| Sparticle | Decay Mode | B.R. |
|---|---|---|
| $\tilde{\chi}_2^0$ | $\tilde{\chi}_1^0\ell^+\ell^-$ | 7% |
| $\tilde{\chi}_3^0$ | $\tilde{\chi}_1^0\ell^+\ell^-$ | 7% |
| $\tilde{\chi}_4^0$ | $\tilde{\chi}_1^\pm W^\mp$ | 81% |
| | $\tilde{\chi}_3^0 Z$ | 12% |
| $\tilde{\chi}_1^\pm$ | $\tilde{\chi}_1^0\ell\nu$ | 22% |
| $\tilde{\chi}_2^\pm$ | $\tilde{\chi}_2^0 W^\pm$ | 38% |
| | $\tilde{\chi}_3^0 W^\pm$ | 18% |
| | $\tilde{\chi}_1^\pm Z$ | 30% |

Table 4: Leading-order cross-sections and number of trilepton events for integrated luminosity of 10 fb$^{-1}$ for SU2.

| Production | $\sigma$ [fb] | Trilepton events /10 fb$^{-1}$ |
|---|---|---|
| $\tilde{\chi}_1^\pm\tilde{\chi}_2^0$ | 1138.0 | 175 |
| $\tilde{\chi}_1^\pm\tilde{\chi}_3^0$ | 679.3 | 105 |
| $\tilde{\chi}_1^\pm\tilde{\chi}_4^0$ | 51.4 | 6 |
| $\tilde{\chi}_2^\pm\tilde{\chi}_2^0$ | 58.5 | 7 |
| $\tilde{\chi}_2^\pm\tilde{\chi}_3^0$ | 61.6 | 7 |
| $\tilde{\chi}_2^\pm\tilde{\chi}_4^0$ | 310.3 | 26 |
| TOTAL | | 326 |

- "Inclusive SUSY" : Inclusive supersymmetric particle pair production (any sparticles)
- "Direct gaugino" : Direct production of charginos and neutralinos only

Inclusive SUSY represents the signal we would obtain at the benchmark points. Direct gaugino represents a more pessimistic scenario, where coloured sparticles are heavy so have no significant LHC cross-section.

## 2 Lepton selection

The final selection will require three leptons – electrons or muons – which must consist of an OSSF pair and a further third lepton. The initial lepton selection requirements are based on the ATLAS standard criteria. The main definitions are summarised in Table 5, and are briefly discussed below. The relatively low $p_\mathrm{T}$ threshold for electrons and muons of 10 GeV in these analyses is an attempt to increase the number of trilepton events, despite lower lepton reconstruction efficiencies at low $p_\mathrm{T}$ .

- **Muons** must satisfy $p_\mathrm{T} > 10$ GeV and $|\eta| < 2.5$, and tracks found in the muon spectrometer must match inner detector tracks. For each muon spectrometer track only the best matching track in the inner detector is taken. It is required that the $\chi^2$ for the match of the muon track to the points on the track [5] is less than 100. A primary isolation criterion requires less than 10 GeV of transverse energy[2] in the calorimeter in a cone of radius $\Delta R = 0.2$ around the muon[3].

- **Electrons** must satisfy $p_\mathrm{T} > 10$ GeV and $|\eta| < 2.5$. They must satisfy shower shape and isolation requirements as described in [6]. The entire event is rejected if any electron candidate is found in the barrel-endcap transition region (1.37$< |\eta| <$ 1.52) due to the lower electron identification performance in this region. A primary isolation criterion is that electrons must have less than 10 GeV of transverse energy in an annulus of radius $\Delta R$=0.2 surrounding the electron.

Electrons and muons are then subject to further vetoes as follows: electrons and muons within $\Delta R <$ 0.4 of a jet are removed from the event, along with OSSF pairs with invariant mass $M_{OSSF} < 20$ GeV, which are likely to have been produced from photon conversions or hadronic decays.

[2] Transverse energy, $E_T = E\sin\theta$, where $E$ is the energy and $\theta$ is the polar angle relative to the beam direction.

[3] $\Delta R \equiv \sqrt{\Delta\eta^2 + \Delta\phi^2}$ where $\Delta\eta$ is the pseudorapidity difference, with $\eta \equiv -\log\tan(\theta/2)$, and $\Delta\phi$ is the difference in azimuthal angle.

Table 5: Selection criteria for muons, electrons and jets.

| | Muon | Electron | Jet |
|---|---|---|---|
| $p_T$ cut | $> 10$ GeV | $> 10$ GeV | $> 10$ GeV |
| $\eta$ cut | $\lvert\eta\rvert < 2.5$ | $\lvert\eta\rvert < 1.37$ or $1.52 < \lvert\eta\rvert < 2.5$ | $\lvert\eta\rvert < 2.5$ |
| Calorimeter Isolation | $\lvert E\rvert < 10$ GeV in $\Delta R = 0.2$ | $\lvert E\rvert < 10$ GeV in $\Delta R = 0.2$ | - - |

## 2.1 Single lepton selection efficiencies

Searches for SUSY final states with three leptons require both a high lepton reconstruction efficiency and low fake rates. There are few Standard Model processes with similar signatures, but the high cross section backgrounds $t\bar{t}$ and $Zb$ can pass the trilepton requirement as there are both primary leptons in the event and a number of jets, in particular $b$ jets which can introduce secondary leptons. Good isolation criteria are therefore important in order to reduce the fake rate and thus the background. This section presents a study of the lepton efficiency, fake rate and purity for the common object definition which includes a calorimeter based isolation criterion as well as the performance of a track-based alternative.

When collision data are available, the best determination of these quantities will be made from the ATLAS data using the methods described in [7]. Since most of our events of interest will have a similar environment – isolated leptons and little jet activity – we may expect that the values found in those studies should closely match the corresponding quantities in our search. However since those measurements require collision data, in this section we present the efficiencies, fake rates and purities as determined from Monte Carlo information only.

The following isolation criteria have been studied:

- **No isolation cut** - showed as a reference;
- **Calorimeter-based isolation** $E_{\rm cal}^{\Delta R=0.2} < 10$ GeV. $E_{\rm cal}^{\Delta R=0.2}$ is the energy deposited in a cone with $\Delta R = 0.2$ around the lepton candidate;
- **Maximum-$p_{\rm T}$ track-based isolation** $p_{T{\rm track,max}}^{\Delta R=0.2}(\ell) < 2/1$ GeV for $e/\mu$, where $p_{T{\rm track,max}}^{\Delta R=0.2}(\ell)$ is the maximum $p_T$ of any track in a $\Delta R = 0.2$ cone around the lepton;
- **Sum-$p_{\rm T}$ track-based isolation** $p_{T{\rm track},\Sigma}^{\Delta R=0.3}(\ell) < 4$ GeV, where $p_{T{\rm track},\Sigma}^{\Delta R=0.3}$ is the sum of the $p_{\rm T}$ of all tracks above 1 GeV in a $\Delta R = 0.3$ cone around the lepton.

The lepton reconstruction efficiency is defined to be

$$\mathscr{E}_\ell \equiv \frac{n_\ell^{\rm match}}{n_\ell^{\rm MC}} \,, \tag{1}$$

where $n_\ell^{\rm match}$ is the number of *generator-level* leptons matched to a reconstructed candidate and $n_\ell^{\rm MC}$ is the total number of generator-level leptons.

In this study generator-level leptons are defined to be charged leptons ($\ell \in \{e, \mu\}$) which have come from decays of SUSY particles (sleptons and gauginos), Standard Model gauge bosons and tau leptons, but do *not* include leptons from other sources (such as hadronic decays, bremsstrahlung, or photon conversions). No isolation cut has been applied on these generator-level leptons. Generator-level particles are matched to reconstructed candidates which have passed kinematics cuts and object selection according to the definition in Section 2. A match is required to be found within $\Delta R = 0.02$. Figure 2 shows

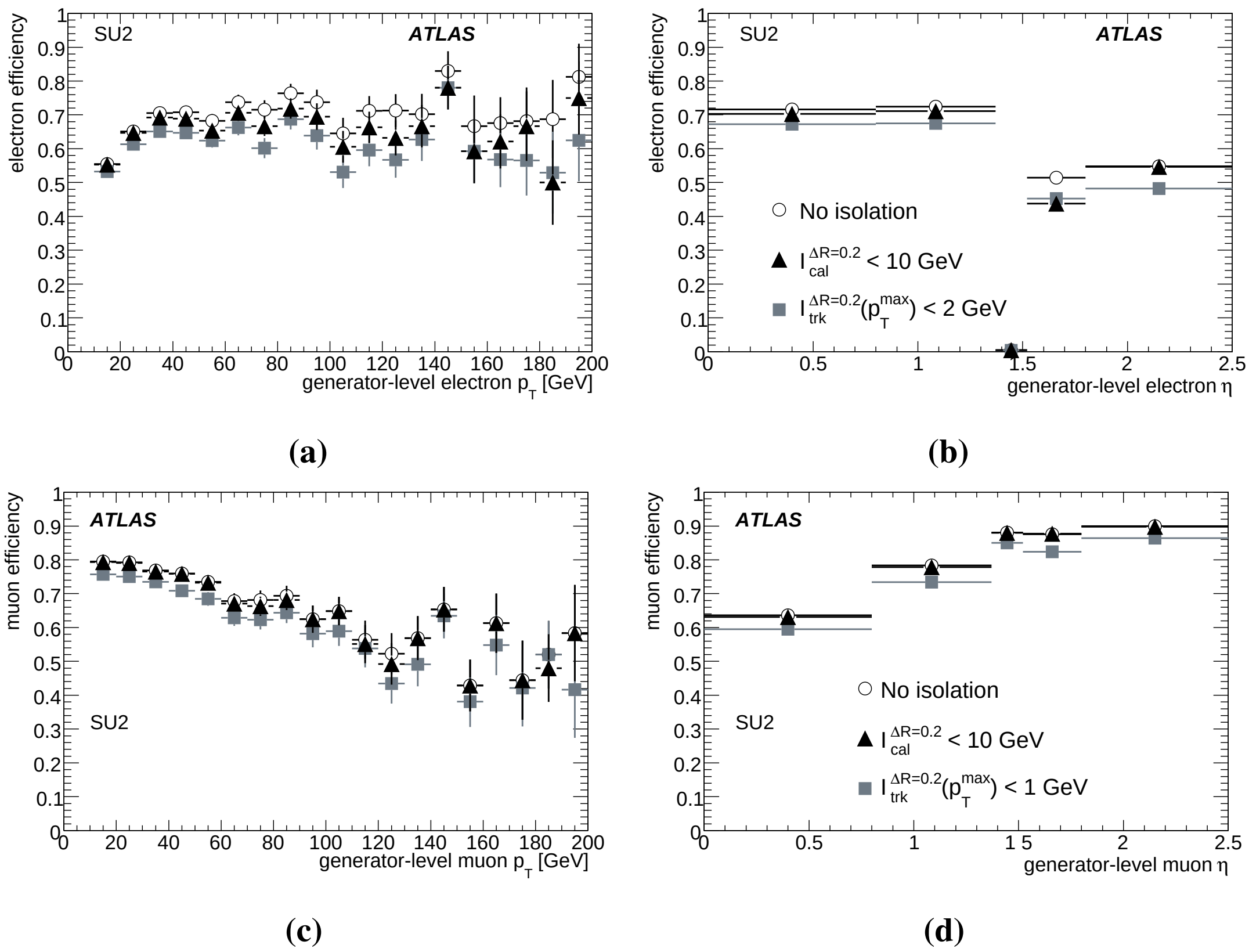


Figure 2: The efficiency for electrons (a,b) and muons (c,d) to be reconstructed and to pass various isolation criteria, for the SU2 sample. The plots are shown as a function of the $p_T$ (a,c) and $\eta$ (b,d) of the matched Monte Carlo lepton.

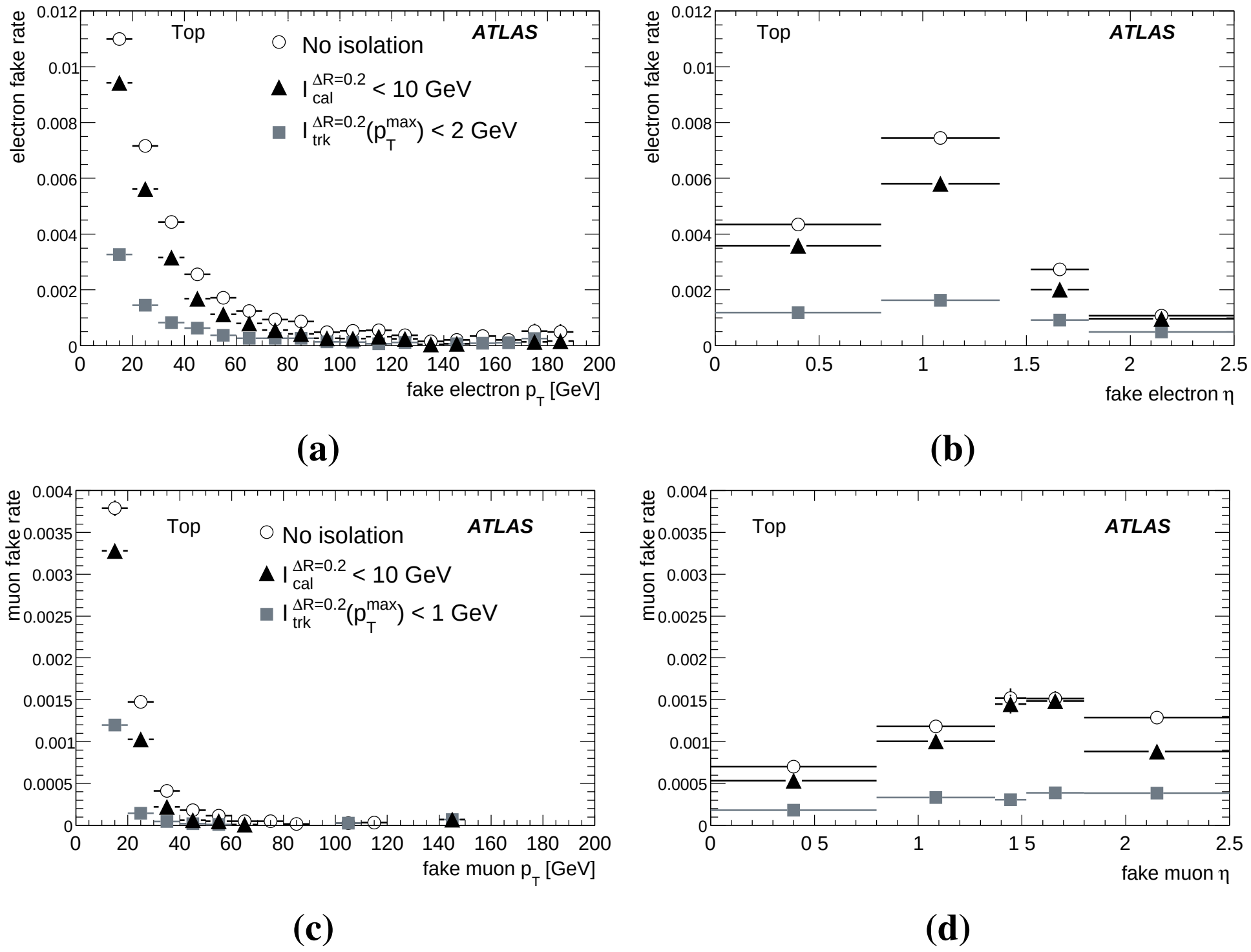


Figure 3: The fake rate for electrons (a,b) and muons (c,d) in the $t\bar{t}$ sample as function of the $p_T$ (a,c) and $\eta$ (b,d) of the fake lepton.

the reconstruction efficiency for electrons and muons from the SU2 sample using different isolation requirements as function of the $p_{\mathrm{T}}$ and $\eta$ of the matched Monte Carlo lepton.

The requirement that leptons should be separated from selected jets by $\Delta R > 0.4$ (referred to as the $\Delta R(\ell,\mathrm{jet}) > 0.4$ cut) is already a strong isolation requirement which reduces the efficiency by $\sim 5\%$ for electrons and $\sim 15\%$ for muons. For the case of $t\bar{t}$ events (not shown in the figures) the effect is even larger as there are more jets in the events.

For the muons, one can see that after introducing the $\Delta R(\ell,\mathrm{jet}) > 0.4$ cut, the efficiency decreases with increasing $p_T$ , while the $|\eta|$ distribution decreases in the central region $|\eta| \lesssim 1.4$. A clear drop in electron efficiency can be found near $|\eta| \approx 1.4$ which is the barrel-endcap transition region.

After applying only the $\Delta R(\ell,\mathrm{jet}) > 0.4$ cut (referred to as “no isolation” in the plots), the total efficiency is $(65.5 \pm 0.5)\%$ and $(75.2 \pm 0.5)\%$ for electrons and muons respectively. The calorimeter-based isolation $E_{\mathrm{cal}}^{\Delta R=0.2} < 10\,\mathrm{GeV}$ reduces the electron efficiency by $\approx 3\%$ to $(63.6 \pm 0.5)\%$ while leaving the muon efficiency almost unchanged. The track-based alternative has similar effect on both lepton flavours causing a loss of about 7 to 8% with respect to the “no isolation” performance.

The fake rate is defined as

$$\mathscr{F}_\ell \equiv \frac{n_\ell^{\mathrm{MC},\overline{\mathrm{match}}}}{n_{\mathrm{jet}}^{\mathrm{MC}}} , \qquad (2)$$

where $n_\ell^{\mathrm{MC},\overline{\mathrm{match}}}$ is the number of reconstructed leptons which are *not* matched to a generator-level lepton and $n_{\mathrm{jet}}^{\mathrm{MC}}$ is the number of jets at the generator level[4].

Only generator-level jets with $|\eta| < 2.5$ are considered. They are rejected if there is an overlap with an electrons within $\Delta R < 0.2$, but in general include photons, hadronic tau jets and $b$ jets. The generator

[4] *i.e.* jets found by running a jet algorithm over the Monte Carlo generator final state, before detector simulation.

jet must have energy, $E > 7$ GeV. Asymmetric $p_{\mathrm{T}}$ cuts are applied to the reconstructed objects compared to their Monte Carlo equivalents. For the efficiency calculation ${p_{\mathrm{T}}}_{\ell}^{\mathrm{MC}} > 10$ compared to $p_{\mathrm{T}\ell} > 5$ GeV, while for the fake rate calculation ${p_{\mathrm{T}}}^{MC} > 5$ and $p_{\mathrm{T}\ell} > 10$ GeV. This reduces the number of mismatches which would be found near the edge of the fiducial region from small mismeasurements of $p_{\mathrm{T}}$ .

Figure 3 shows the fake rate as function of the $p_T$ and $\eta$ of the fake lepton for the $t\bar{t}$ sample (one of the major backgrounds). The $p_{\mathrm{T}}$ distribution is clearly peaked at low values typical for leptons from $b$-jet decays. The $\eta$ distribution follows the detector layout with higher fake rate in the transition region between the barrel and end caps.

The $\Delta R(\ell,\mathrm{jet}) > 0.4$ requirement provides a reduction in the fake rate of about $\approx 80\%$ for muons and $\approx 5\%$ for electrons (as compared to leptons passing the standard selection with $E_{\mathrm{cal}}^{\Delta R=0.2} < 10$). With no isolation other than the $\Delta R(\mu,\mathrm{jet}) > 0.4$ requirement one obtains a fake rate of $(4.1 \pm 0.1) \times 10^{-3}$ and $(1.1 \pm 0.1) \times 10^{-3}$ for electrons and muons respectively. The isolation criteria provide a similar relative suppression of the electron and muon fake: $\approx 20\%$ for $I_{\mathrm{cal}}^{\Delta R=0.2} < 10$ and $\approx 74\%$ for $p_{T\mathrm{track,max}}^{\Delta R=0.2}(\ell) < 2/1$ GeV for $e/\mu$. The effect of the isolation cuts is very similar in the case of the SU2 sample, but the fake rates are almost an order of magnitude lower than in $t\bar{t}$ events.

We define purity by:

$$\mathscr{P}_\ell = \frac{n_\ell^{\mathrm{match}}}{n_\ell} , \quad (3)$$

where $n_\ell$ is the number of reconstructed leptons, and the matching criteria are as described above. The lepton selection described in Section 2 provide samples with a purity of 92% (e) and 97% ($\mu$) with very similar results for both the SU2 and $t\bar{t}$ samples. However after requiring at least 3 leptons in the event there is a clear difference between the two. The purity is almost unchanged for SU2, while, in $t\bar{t}$ trilepton events the purity drops to $\sim 50\%$ for electrons and $\sim 70\%$ for muons.

## 3 Event selection

Events are required to pass either of the two single-lepton triggers, labelled L2_e22i and L2_mu20, which are well-suited for the tri-lepton analysis at $\mathscr{L} = 10^{31-32}\,\mathrm{cm}^{-2}\mathrm{s}^{-1}$. A more detailed discussion of the motivation for using these triggers can be found in Section 5.

Distributions of some important event variables for Standard Model backgrounds and for the benchmark point SU2 (both for inclusive SUSY and direct gaugino) can be found in Figure 4. We show the $p_{\mathrm{T}}$ distributions of reconstructed leptons (electron or muon) and jets. Requiring at least three leptons in the event results in mainly $t\bar{t}$ and $Zb$ Standard Model backgrounds, with larger contributions from dibosons in four lepton events. It can seen in Figures 4a and 4b that $p_{\mathrm{T}}$ distributions of the two hardest leptons in three lepton events are similar in SU2 and $t\bar{t}$ , with $Zb$ adding a significant contribution in the low $p_{\mathrm{T}}$ region. The $p_{\mathrm{T}}$ distributions for the third-hardest leptons are plotted in Figure 4c where it can be seen that whilst $t\bar{t}$ and $Zb$ are the major backgrounds in the low $p_{\mathrm{T}}$ region, there is a large contribution from the dibosons across the entire $p_{\mathrm{T}}$ range. This is because the third leptons from $t\bar{t}$ and $Zb$ are soft leptons from leptonic $b$-quark decay, whereas all the leptons from dibosons are from $Z/W^{\pm}$ boson decays. The $p_{\mathrm{T}}$ distributions of the reconstructed jets are plotted in Figure 4d for events with three or more leptons.

Since an OSSF lepton pair is expected from the $\tilde{\chi}_2^0$ decay, a selection requiring two OSSF leptons is applied (i.e. we require $e^+e^- + \ell$ or $\mu^+\mu^- + \ell$, where $\ell \in e,\mu$)[5]. A further cut requires three or more leptons in the event.

A stringent cut on the isolation of the tracks of the leptons is made next, using the $p_{T\mathrm{track,max}}^{\Delta R=0.2}$ variable described in Section 2.1. The purpose is to reduce backgrounds from bremsstrahlung, hadron decays and

[5]Trilepton events where all three leptons have the Same Sign, (SS, $\ell^\pm\ell^\pm\ell^\pm$, where $\ell \in e,\mu$) have also been investigated, however due to the very low statistics associated with the channel in the signal samples investigated, it is not considered further here.

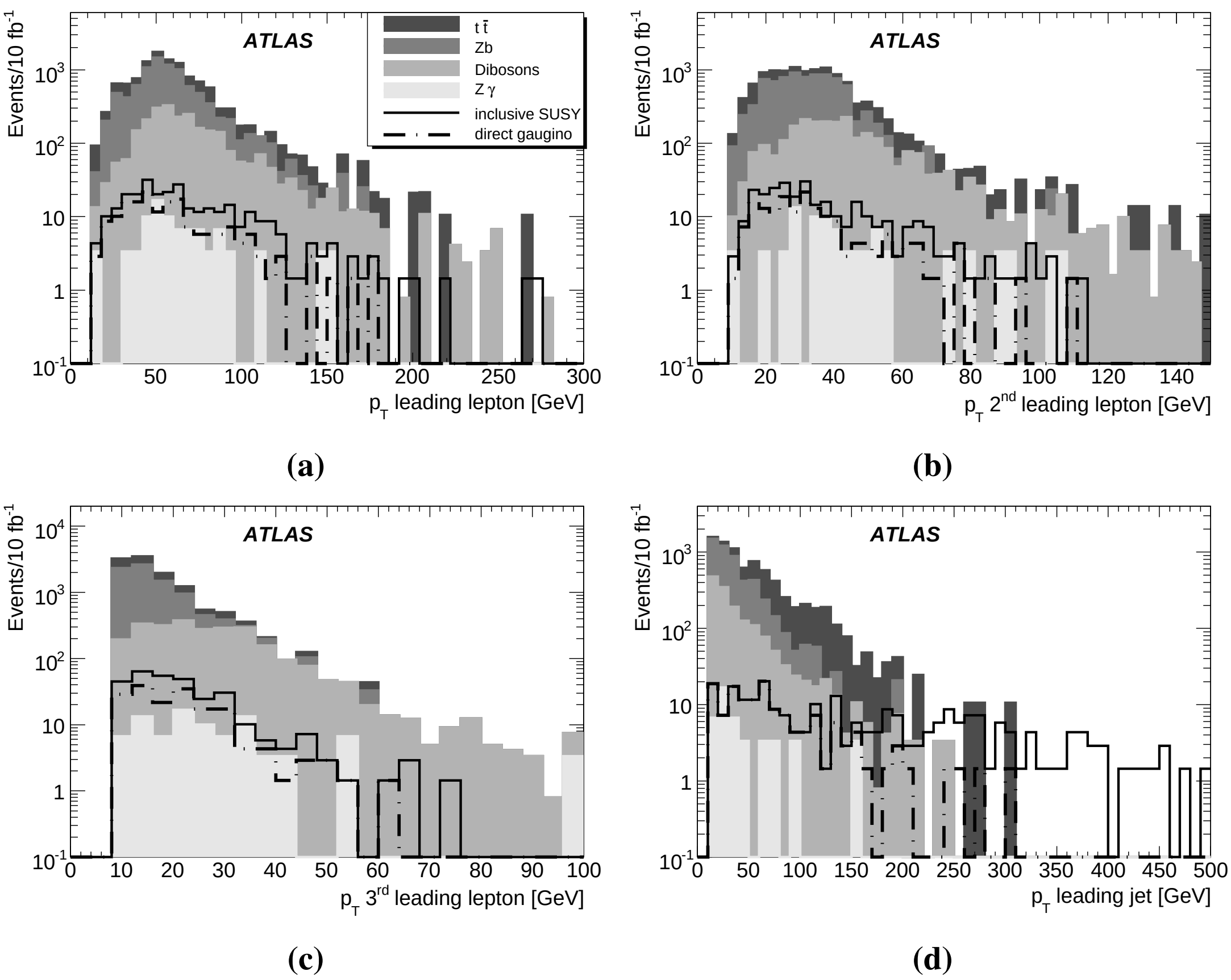


Figure 4: Transverse momentum of the leading three leptons (a-c) and leading-jet $p_{\mathrm{T}}$ (d) after an initial three-lepton requirement has made.

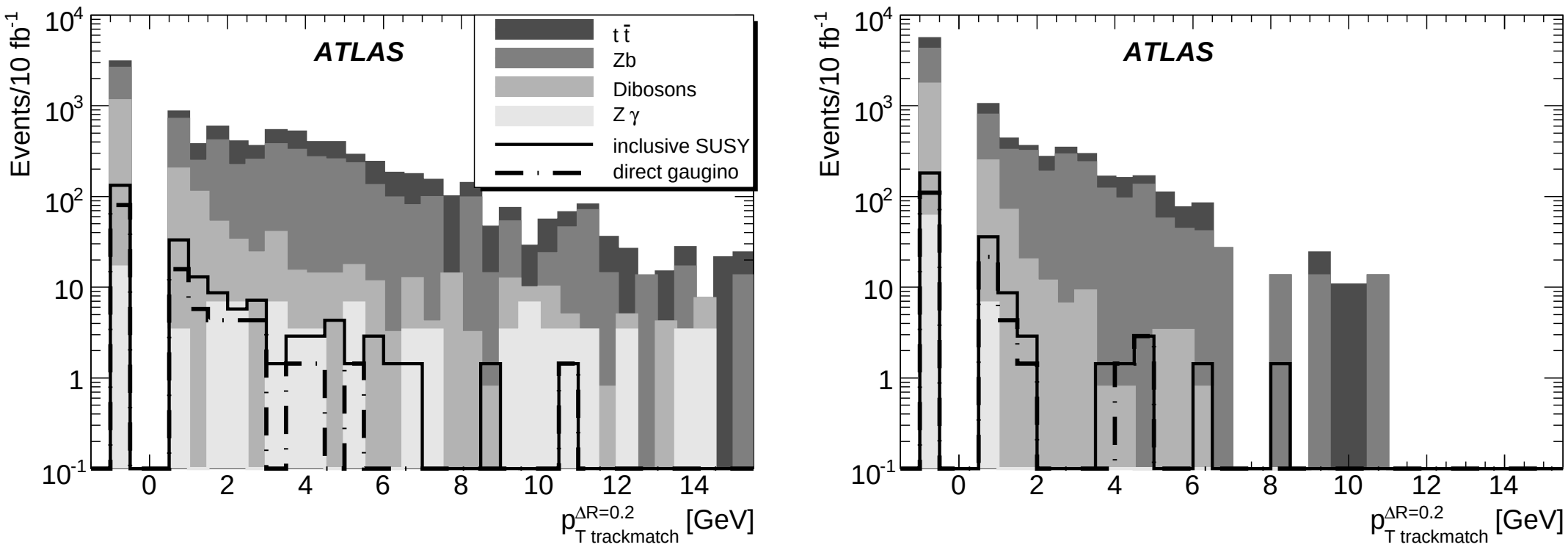


**(a)** **(b)**

Figure 5: $p_{T\text{track,max}}^{\Delta R=0.2}$ for **(a)** electrons and **(b)** muons.

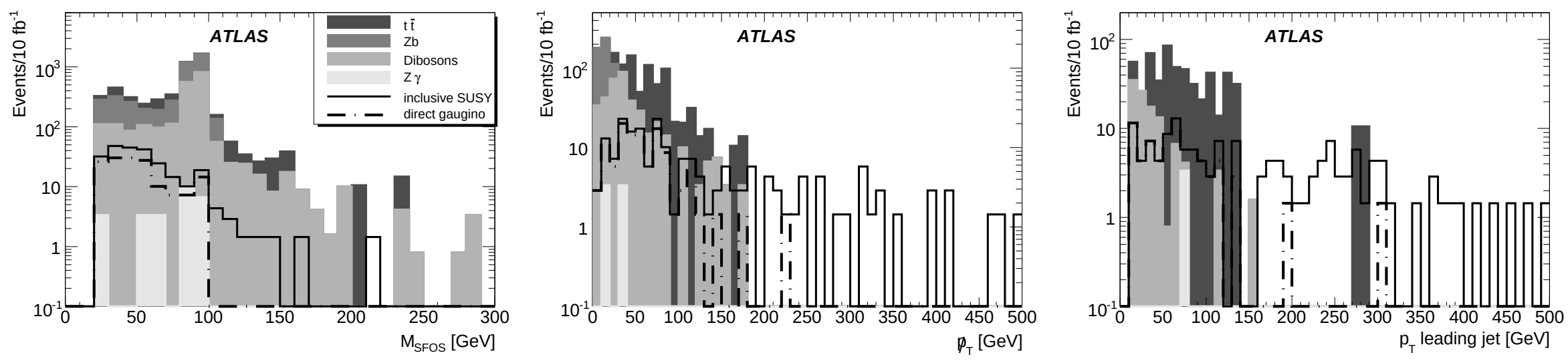


**(a)** **(b)** **(c)**

Figure 6: **(a)** OSSF dilepton invariant mass distribution. **(b)** $E_{\mathrm{T}}^{\mathrm{miss}}$ distribution after the $Z$ mass window cut. **(c)** The $p_{\mathrm{T}}$ distribution of the leading jet after the $E_{\mathrm{T}}^{\mathrm{miss}}$ cut is applied.

photon conversions. The relevant distributions for electrons and muons are plotted in Figure 5a and 5b. The plots contain negative entries when no track is found within $\Delta R$. We require $p_{T\text{track,max}}^{\Delta R=0.2} < 1$ GeV for muons and $< 2$ GeV for electrons. This reduces the number of background events to 23% ($t\bar{t}$ ) and 37% ($Zb$) of their previous levels, whilst keeping 82% (inclusive SUSY), 86% (direct gaugino) of the signal for our SU2 benchmark point.

The diboson and $Zb$ backgrounds will produce the two OSSF leptons mainly from $Z$ decays, which will not be the case for three-body decays of neutralinos. The dilepton invariant mass distribution is shown in Figure 6a, (after applying the cuts already described) and shows the expected peak at the $Z$ mass. A simple way to reduce the these backgrounds is to discard events which have any OSSF dilepton pair with invariant mass in the mass window 81.2 GeV $< M_{OSSF} <$ 101.2 GeV. Note that this exclusion window also offers an excellent control region in which to measure the size of these backgrounds from the data. It is however somewhat model-dependent. Some points in SUSY parameter space preferentially decay through *real* $Z$ bosons, rather than through three-body decays. At those points the OSSF dilepton mass distribution would also be strongly peaked at the $Z$ mass.

A large missing-transverse-momentum cut is used in most SUSY analyses, but in this analysis a smaller cut at 30 GeV is applied since in direct gaugino production the two invisible $\tilde{\chi}_1^0$s are often almost back-to-back in the transverse plane, resulting in a lower overall $E_{\mathrm{T}}^{\mathrm{miss}}$ (Figure 6b). Missing transverse momentum is calculated as described in [8].

Finally there is the possibility of adding a cut can be made on the hadronic activity in the event. This

could be useful in the case where direct gaugino production dominates, since it can be expected to reduce the $t\bar{t}$ background to a greater extent than the signal. Whether this cut is appropriate will depend on the SUSY scenario presented by nature, so analyses both with and without this cut will be necessary.

Selected jets must satisfy $p_{\mathrm{T}} > 10$ GeV and $|\eta| < 2.5$, and are reconstructed based on calorimeter tower signals [9] using a seeded cone algorithm with radius $\Delta R = 0.4$. Jets are initially selected with $p_{\mathrm{T}} > 10$ GeV and $|\eta| < 2.5$. Jets are not considered if they overlap with a reconstructed electron within $\Delta R < 0.2$.

The $p_{\mathrm{T}}$ distribution of the leading jet in the event is plotted in Figure 6c, where it is seen that the direct gaugino production has jets with lower $p_{\mathrm{T}}$ than the Standard Model backgrounds. The level of this cut is chosen at $p_{\mathrm{T}} > 20$ GeV.

The complete event selection is then:

1. At least one pair of OSSF leptons ($e^+e^-$ or $\mu^+\mu^-$)
2. $N_\ell >= 3$ ($\ell \in \{\mathrm{e}, \mu\}$), and where all $\ell_i$ satisfy the requirements in Section 2
3. $p_{T\mathrm{track,max}}^{\Delta R=0.2} < 2$ GeV for electrons, $p_{T\mathrm{track,max}}^{\Delta R=0.2} < 1$ GeV for muons
4. No OSSF dilepton pair has invarint mass in the range 81.2 GeV $< M_{OSSF} <$ 102.2 GeV
5. $E_{\mathrm{T}}^{\mathrm{miss}} > 30$ GeV
6. Optional – no jet with $p_{\mathrm{T}} > 20$ GeV

In Figure 7 we show distributions of the invariant mass of the dilepton pair after all cuts. The distributions have been flavour subtracted; the quantity plotted is

$$n^{\mathrm{OSSF}} - n^{\overline{\mathrm{OSSF}}}\,, \tag{4}$$

where $n^{\mathrm{OSSF}}$ is the number of events in the signal selection (containing OSSF dilepton pairs) as described earlier in this section, and $n^{\overline{\mathrm{OSSF}}}$ is the number of events in which there are three leptons, but no OSSF pair. The non-OSSF events give an indication of the expected background size since many background sources do not necessarily produce OSSF pairs.

# 4 Discovery potential

The numbers of events and the resulting significance at each stage of the analysis are listed in Table 6 for the benchmark point SU2. We use the following definition of signal signficance,

$$\mathscr{S} = \frac{S}{\sqrt{S+B}} \tag{5}$$

where $S$ is the number of signal events and $B$ is the number of background events. After applying the event selection above (including the jet veto), 29 events remain for SU2 inclusive SUSY all of which are direct gaugino, with an expected background of 210 events (mainly $ZW$ production). This corresponds to $\mathscr{S}$ of 1.87 for 10 fb$^{-1}$ . This yields a $5\sigma$ discovery signal after ~80 fb$^{-1}$ of integrated luminsity. With the jet veto selection, $ZW$ is the dominant remaining Standard Model background. Without the jet veto, 177 signal events (95 of them direct gaugino) remain. Statistical significances of 5.94 and 3.34 are found for the inclusive SUSY and direct gaugino signals respectively.

The expected statistical significances for various mSUGRA benchmark points are summarised in Table 7 excluding the jet veto unless indicated otherwise by "SU$x$+JV".

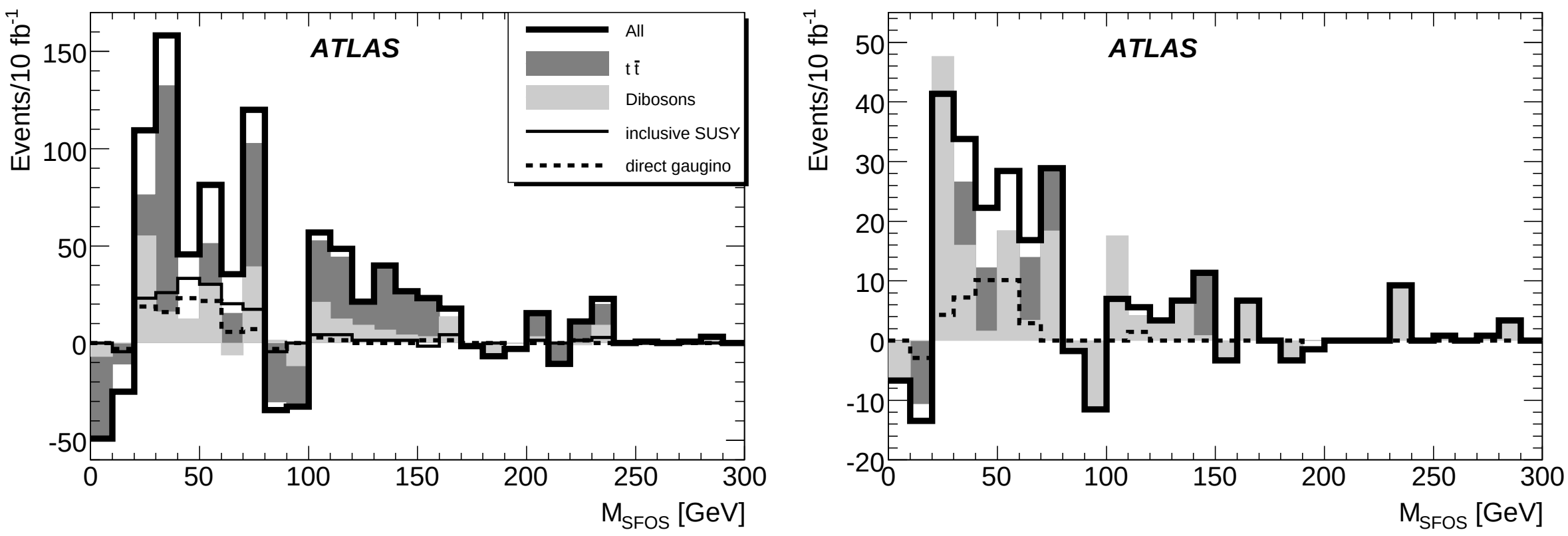


**(a)** **(b)**

Figure 7: Distributions of the OSSF dilepton invariant mass after all selections have been applied **(a)** without the jet veto and **(b)** including a jet veto.

Table 6: Numbers of events and statistical significance as selection is applied for the benchmark point SU2, for integrated luminosity of 10 fb$^{-1}$.

| Kinematic Cut | No Cuts | $N_L >= 2$ | OSSF | $N_L >= 3$ | TrackIsol | $m_{\ell\ell}$ | $E_T^{miss}$ | JetVeto |
|---|---|---|---|---|---|---|---|---|
| SU2 gauginos | 64.0k | 1647 | 1108 | 178 | 153 | 120 | 95 | 29 |
| SU2 other | 7081 | 776 | 353 | 127 | 95 | 85 | 82 | 0 |
| $t\bar{t}$ | 4.41M | 234k | 104k | 2812 | 634 | 507 | 476 | 42 |
| $ZZ$ | 38.2k | 10.4k | 9984 | 580 | 476 | 57 | 13 | 6 |
| $ZW$ | 156k | 17.2k | 14.5k | 1910 | 1682 | 322 | 218 | 154 |
| $WW$ | 400k | 22.7k | 10.7k | 25 | 8 | 8 | 8 | 8 |
| $Z\gamma$ | 32.8k | 7184 | 6970 | 91 | 27 | 7 | 3 | 0 |
| $Zb$ | 1.59M | 57.4k | 559k | 6523 | 2409 | 386 | 0 | 0 |
| inclusive SUSY $\mathscr{S}$ | | 2.60 | 1.74 | 2.76 | 3.36 | 5.31 | **5.94** | **1.87** |
| direct gaugino $\mathscr{S}$ | | 1.77 | 1.32 | 1.61 | 2.09 | 3.20 | **3.34** | **1.87** |

The discovery prospects for inclusive SUSY, reflected in columns titled "SU$x$", are rather encouraging: a 5$\sigma$ discovery can be expected with several fb$^{-1}$ of integrated luminosity[6]. The direct gaugino pair production is highlighted in columns marked as SU2$\chi$ and SU3$\chi$. Both for SU2 and SU3 we see a drop in significance which roughly corresponds to the fraction of the direct gaugino production compared to the total SUSY cross-section. In fact, the drop is somewhat higher, since the $p_T$ spectrum tends to be softer for direct gaugino pair production compared to strong SUSY production with its characteristic decay chains. Here, a discovery can be expected with several tens of fb$^{-1}$.

## 5 Lepton trigger study

Unlike most SUSY searches, in this channel we cannot rely on jet or $E_T^{miss}$ triggers. We have studied a variety of triggers at the second level[7], after the first level trigger has been applied. Other studies [10,11]

[6] Taking into account statistical errors only. Systematic uncertainties are discussed in Section 6.5.

[7] ATLAS has three trigger levels: a first level (L1) hardware trigger, a software-based second level trigger (L2) that examines regions of interest within the decector, and finally a software-based event filter (EF). More details about the electromagnetic [10] and muon [11] triggers may be found elsewhere in this volume.

Table 7: Discovery potential, and integrated luminosity required for $5\sigma$ discovery. The jet veto is only applied in columns headed '+JV'. For a fuller description of the notation, see the text.

| | SU1 | SU2 | SU3 | SU4 | SU8 | SU2$\chi$ | SU3$\chi$ | SU2+JV | SU3+JV |
|---|---|---|---|---|---|---|---|---|---|
| $\mathscr{S}$, 10 fb$^{-1}$ | 7.7 | 5.9 | 17.2 | 69.3 | 1.9 | 3.3 | 1.6 | 1.9 | 1.4 |
| $\int dt\,\mathscr{L}$ for $5\sigma$ | 4.2 | 7.1 | 0.8 | 0.1 | 70.5 | 22.4 | 92.9 | 66.9 | 119.3 |

Table 8: Fraction of events triggered at L2 at three selection stages of the selection for the heavy coloured sparton scenario in SU2 (first block), the direct gaugino production in SU3 (second block), and the inclusive SU3 signal (third block). "$\bigcup$" stands for the OR-combination of L2_e22i and L2_mu20.

| Selection | SU2$\chi$ | | | SU3$\chi$ | | | SU3 incl. | | |
|---|---|---|---|---|---|---|---|---|---|
| Stage | L2_e22i | L2_mu20 | $\bigcup$ | L2_e22i | L2_mu20 | $\bigcup$ | L2_e22i | L2_mu20 | $\bigcup$ |
| OSSF pair | 41% | 54% | 89% | 42% | 54% | 92% | 51% | 51% | 94% |
| OSSF+3$^{\rm rd}\ell$ | 58% | 67% | 93% | 59% | 63% | 95% | 66% | 68% | 98% |
| after all cuts | 57% | 66% | 92% | 58% | 57% | 94% | 66% | 64% | 97% |

demonstrate that objects passing L2 have a high efficiency for passing EF.

As a figure of merit, the fractions of triggered events

- after selecting for an OSSF pair;
- after requiring a further third lepton;
- after all cuts, including the jet veto;

are studied.

We have identified two single-lepton triggers, labelled L2_e22i and L2_mu20, that are well-suited for the tri-lepton analysis at $\mathscr{L} = 10^{31-32}$ cm$^{-2}$s$^{-1}$. At high luminosity it is planned to have corresponding triggers with additional isolation criteria (e22i_tight and mu20i) which will remain unprescaled. The resulting event-triggering efficiencies are listed in Table 8. OR-combined, they have $\approx$92% probability for direct gaugino production for the benchmark point SU2. They have sufficiently high $p_T$ thresholds to be easily studied with leptonic $Z$ decays.

The same trigger set was studied for the direct gaugino and inclusive SUSY for another bechmark point, SU3, that has lighter squarks than SU2. Again the OR-combination of L2_e22i and L2_mu20 shows a good performance of $\sim$92, 95, 94%. Their individual L2 event efficiencies are summarised in the middle block of Table 8.

In addition we have studied the performance of the L2_e22i and L2_mu20 triggers for the tri-lepton analysis, outside of the heavy coloured sparton scenario, for inclusive SU3 production. Again in this case we find L2 event efficiencies of around $\sim$94, 98, 97%.

The efficiency is relatively high for lepton triggers even though the trigger thresholds are fairly high compared to the offline cuts on lepton transverse momenta. The reason for that is simple combinatorics: since the 3 leptons are ordered by their transverse momenta, it is likely that the leading lepton has a high $p_T$ if the third one passes the offline threshold. On the other hand, cases where *all* the three leptons are below $20-30$ GeV but above 10 GeV are unlikely.

We have investigated how the leptonic trigger efficiencies for the inclusive SUSY and direct gaugino for the benchmark point SU2 change as a function of the progressive event selection stages described in Section 3. In both cases, the trigger efficiencies reach their approximate maxima after the three-lepton selection stage. Beyond this cut stage, the efficiency values plateau, indicating that our event

selection requirements should not bias signal trigger efficiencies. It can be concluded that the L2_e22i and L2_mu20 single lepton triggers provide a good performance for the tri-lepton analysis in the early days of ATLAS running at $\mathscr{L} = 10^{31}\,\text{cm}^{-2}\text{s}^{-1}$. Di-leptonic triggers with lower thresholds like 2e12i, 2mu10, and e15i_mu10 can be used to recover events where all three leptons have low transverse momenta around 20 GeV.

# 6 Systematic uncertainties

The dominant sources of systematic uncertainty for this search are different to other SUSY search channels that focus on final states containing jets. For our multi-lepton search the main sources of uncertainty in the backgrounds are described below.

## 6.1 Background rates

The production rates for the majority of background processes such as diboson production and $t\bar{t}$ are known at the parton level at better than next-to-leading order. However, it is generally better to determine the rate of these backgrounds from the ATLAS data themselves, reducing any uncertainty from luminosity, PDFs and other systematics (e.g. from acceptances, efficiencies etc). The background rate measurements should be made in "control regions" in which they dominate. For the $WZ$ background a sensible control region is the region of phase space where the OSSF lepton pair has an invariant mass near the $m_Z$ [12]. For the $t\bar{t}$ background, one would examine single lepton and OSSF dilepton channels [13]. Any statistical uncertainty in the background measurements forms a systematic uncertainty in our analysis. The expected integrated luminosities for cleanly measuring each background and the resulting statistical uncertainties can be found in Table 9.

Any further systematic uncertainties in the background rates are not considered here so as not to double-counting their effects[8]. One must be careful that the extrapolation from the 'control' to the 'measurement' region of parameter space is well understood. The extrapolation for the backgrounds involving the $Z$ peak relies on the $Z$ line-shape, which is theoretically well-known. The extrapolation from dilepton to trilepton final states for $t\bar{t}$ relies on a good knowledge of the rate at which $b$ quarks (from top decays) produce isolated leptons. A method of determining this rate from the ATLAS data is presented in Section 6.3.

Table 9: Expected rates to cleanly select background samples in control regions, and the corresponding statistical uncertainties.

| Background | Statistical uncertainty after 10 fb$^{-1}$ [%] | Reference |
|---|---|---|
| $WW$ | 1.3 | [12] |
| $WZ$ | 2.6 | [12] |
| $ZZ$ | 6.6 | [12] |
| $t\bar{t}$ | $\ll 1$ | [13] |

One can see from Table 6 that, after event selection and jet veto, the $WZ$ and $t\bar{t}$ backgrounds are most significant in the signal region. The systematic uncertainty in measuring the rate of the backgrounds in

[8] In fact, uncertainties on background rates that positively correlate between the selection region and the control region will actually tend to cancel.

the control regions is expected to be of the order of a few percent maximum for the $WZ$ and even smaller for the $t\bar{t}$ (Table 9). Our estimate for the total systematic uncertainty due to statistical fluctuations in the background control sample is therefore dominated by the $WZ$ background. When multiplied by the fraction of the final background that comes from $WZ$, this generates 1.9% (0.8%) uncertainty in the background rate for the analysis with (without) the jet veto.

## 6.2 Trigger and reconstruction efficiencies

Lepton trigger and reconstruction efficiencies can be determined using the "tag-and-probe" method for electrons and muons coming from $Z \to \ell\ell$ [7].

Table 10: Expected systematic uncertainties on background rates from other ATLAS measurements using the tag-and-probe method, and resultant estimated uncertainties for this analysis. The label 'reco' refers to combined reconstruction and selection efficiencies.

| Source | Tag-and-probe [7] | This analysis (10 fb$^{-1}$) | |
|---|---|---|---|
| | ($1\ell^{\pm}$, 1fb$^{-1}$) | $1\ell^{\pm}$ | $3\ell^{\pm}$ |
| $e$ (trigger) | $\ll 1\%$ | } 0.5 %  | } 2.3%  |
| $\mu$ (trigger) | 0.4 % | | |
| $e$ (reco) | 0.5 % | 1 % | |
| $\mu$ (reco) | $\ll 1\%$ | $\ll 1\%$ | |

The precision with which lepton trigger and reconstruction efficiency can be determined has been studied in [7] for an integrated luminosity scaled to 1 fb$^{-1}$. The statistical uncertainty on the trigger efficiency for 1 fb$^{-1}$ is very small ($\ll 1\%$) and is negligible for 10 fb$^{-1}$. The systematic uncertainties in determining the trigger efficiencies are shown in Table 10. The trigger efficiency uncertainty for our analysis, for which the highest-$p_T$ lepton has similar kinematics to the sample used in [7], is therefore estimated to be $\lesssim 0.5\%$ for $\int dt\,\mathscr{L} = 10$ fb$^{-1}$.

Reconstruction and selection efficiencies have also been studied using a similar method [7]. The estimates of the uncertainties in the efficiencies are also contained in Table 10. We have increased the expected uncertainty in the electron reconstruction efficiency from the "tag-and-probe" value of 0.5% to our own estimate of 1% to reflect a somewhat larger uncertainty for our lower $p_T$ electrons.

In the final column of Table 10 we estimate the resulting systematic uncertainty for the trilepton selection efficiency of this analysis. Since we propose using single lepton triggers (Section 5), only the highest $p_T$ lepton contributes, and our event trigger uncertainties are the same as the single-lepton trigger ones. All three leptons are assumed to contribute to reconstruction and selection uncertainties, with equal contributions from electrons and muons. Our resulting uncertainty from combined trigger, reconstruction and selection efficiencies is 2.3%.

For this study we assume that similar uncertainties will apply with and without the jet veto, but we note that there may be larger systematic uncertainties in lepton efficiencies if jets are permitted in the final state.

## 6.3 Lepton fake rates

The $t\bar{t}$ and $Z+b$ backgrounds contribute to the trilepton final state when (along with a lepton pair from dileptonic $t\bar{t}$ decay or $Z$ decay) a $B$ hadron decays leptonically, producing a third isolated lepton. The third lepton requirement reduces each of these backgrounds by a large factor – about two orders of magnitude in each case. About one order of magnitude of reduction can be accounted for by the

branching ratio of $B$ hadrons to leptons, while the other factor of about ten arises from our rejection of non-isolated leptons. We are therefore sensitive to the rate at which leptons from $b$ decays pass the isolation criteria, which we denote $R_{b\to\ell}$ (it should be understood to include leptons both from direct $b \to \ell$ and also from $b \to c \to \ell$).

We hereby propose a method by which $R_{b\to\ell}$ could be measured in the ATLAS data, and estimate the remaining systematic uncertainty after that measurement. The control sample we suggest is semi-leptonic $t\bar{t}$ decays in which a lepton from one (probe) $b$ decay generates a same-sign dilepton pair. To cleanly select the control sample we would require events with $t\bar{t}$ kinematics (*e.g.* consistent invariant masses for $t$ and $W$) and one clean vertex tag from the other (probe) $b$ jet. This sample has the further advantage that the $b$ quarks have the correct kinematical distributions for the background of interest (dileptonic $t\bar{t}$ ). The same-sign dilepton requirement removes contamination from normal dileptonic $t\bar{t}$ decays since the latter will produce opposite-sign lepton pairs.

Dedicated studies [13] suggest that we can expect to cleanly select about $n_{t\bar{t}->\ell} = 5 \times 10^4$ semi-leptonic $t\bar{t}$ events ($e$ and $\mu$ combined) for $\int dt\,\mathscr{L}$=10 fb$^{-1}$. The fractional statistical uncertainty, $\delta$, that we could expect from the proposed measurement of $R_{b\to\ell}$ is then,

$$\delta(R_{b\to\ell}) = \frac{1}{\sqrt{R_{b\to\ell} \times n_{t\bar{t}->\ell}}} \,. \tag{6}$$

From examination of the event record in our Monte Carlo samples, we found $R_{b\to\ell}$ to be approximately $5\times10^{-3}$. If a similar rate were to be found from measurements of the semi-leptonic $t\bar{t}$ events, then the corresponding value for $\delta(R_{b\to\ell})$ would be 6%.

With a jet veto, $t\bar{t}$ forms about 20% of the background after full selection (and $Z+b$ a very small contribution), and there is a resulting $\approx 1.2\%$ uncertainty in the total background. Without the jet veto, the $t\bar{t}$ contribution is much larger (about 66%) and a 4% systematic uncertainty results.

## 6.4 Jet and missing energy scales

The global uncertainty on the jet energy scale is currently conservatively expected to be about 5%, but the true value is difficult to determine without collision data. In events in which the missing transverse momentum is dominated by hadronic activity, the fractional uncertainty in $E_{\mathrm{T}}^{\mathrm{miss}}$ will be of a similar size, since the two measurements will be highly correlated. We therefore determined the effect of a 5% systematic uncertainty in the missing energy when no jet veto is used.

For the analysis which includes a jet veto, we expect missing energy scale uncertainty to be rather smaller than 5%, since the majority of the missing energy will recoil against the (well-measured) leptons. We also assume that in this case the systematic uncertainties in the jet energy scale and the missing energy should also be largely uncorrelated (for the same reason). Measurements of recoil of jets against $Z$ bosons and photons allow the missing energy resolution to be well-determined as a function of jet energy [8]. Any residual uncertainty in the missing energy scale is estimated by us to be about 2% at low hadronic energies (based on systematic differences between methods in [8]).

For the jet-veto analysis, we determined the effects of a 5% variation in the hadronic energy scale and (an independent) 2% variation in the missing energy scale. The results are shown along with the systematic uncertainties from other sources in Table 11.

As well as the uncertainty from the energy scales, there can be some contribution to the uncertainty in the number of events failing the jet veto from our lack of knowledge of initial-state radiation (ISR) in electro-weak processes (such as $WZ$ production). However the ISR spectrum can be readily determined from other electroweak production control regions (as described in [14] for single electroweak gauge boson production). The resulting statistical uncertainties in those control measurements will be small, and the systematic effects tend to cancel with those described in this section, so no additional systematic uncertainty is added here.

### 6.5 Summary of systematic uncertainties

Table 11: Estimates of the dominant uncertainties in the background determination for $\int dt\,\mathscr{L} = 10$ fb$^{-1}$.

| Source | Uncertainty | |
|---|---|---|
| | No jet veto | With jet veto |
| Background production rates | 0.8% | 1.9% |
| Lepton Efficiency | 2.3% | 2.3% |
| Fakes ($R_{b\to\ell}$) | 4.0% | 1.2% |
| Hadronic energy scale | – | 1.8% |
| Missing energy scale | 1.5% | 1.0% |
| *Total systematic* | 4.9% | 3.8% |
| *Statistical* | 3.7% | 6.9% |
| *Statistical + Systematic* | 6.2% | 7.9% |

Our best current estimates of the various sources of uncertainty are summarised in Table 11. One can see that with the full selection, including the jet veto, and if the supporting measurements can be made with the precision expected, then the SUSY-search analysis will be limited by statistical fluctuations in the background samples (about 6.9%) rather than by systematic sources of uncertainty (about 3.8%).

Without the jet veto, the statistical uncertainty is smaller, however the systematic contribution to the total uncertainty increases to 4.9%, largely because we have an increased sensitivity to $R_{b\to\ell}$ since the top background is much larger if no jet veto is applied.

## 7 Conclusions

The ATLAS experiment will have sensitivity to new physics, including supersymmetry, in events with three leptons in association with missing transverse momentum. For most of the SUSY benchmark points studied, a discovery of new physics could be made in this channel with integrated luminosity of several fb$^{-1}$ . Models in which all strongly interacting partner particles are heavy would have smaller cross-sections, but could also be discovered with integrated luminosity of the order of several tens of fb$^{-1}$ .

The major sources of systematic uncertainty were indicated and methods for determining these from data discussed. While the actual sizes of these uncertainties can only be estimated reliably with collision data, estimates based on existing information were presented. In particular, knowledge of the rate at which which $b$ jets lead to seemingly-isolated leptons was found to be an important element in understanding Standard Model backgrounds. A method of measuring this rate from the data using $t\bar{t}$ events was proposed.

# Supersymmetry Signatures with High-$p_T$ Photons or Long-Lived Heavy Particles

**Abstract**
In certain Supersymmetry breaking scenarios, characteristic signatures can be expected which would not necessarily be found in generic SUSY searches for events containing high $p_{\mathrm{T}}$ multi-jets and large missing transverse energy. This paper describes the expected response of the ATLAS detector to four signatures: high-$p_{\mathrm{T}}$ photons which may or may not appear to point back to the primary collision vertex and long-lived charged sleptons and *R* hadrons. Such processes often have the advantage of small Standard Model backgrounds and their observation could provide unique constraints on the different SUSY breaking scenarios. Using these signatures discovery potentials are estimated for either Gauge-Mediated Supersymmetry Breaking or Split-Supersymmetry scenarios. Using Monte Carlo samples of SUSY and background processes corresponding to integrated luminosity of about 1 fb$^{-1}$ we study all aspects of the analysis, including the expected trigger response and offline data reconstruction.

## 1 Introduction

Supersymmetry (SUSY) is one of the most widely investigated theories of physics beyond the Standard Model [1–4]. Searches for signatures of new physics processes predicted within SUSY models are thus central to the physics program of the ATLAS experiment [5], which will be sensitive to the production of SUSY particles with masses up to several TeV [6]. To facilitate the exploitation of early LHC data, a number of preparatory studies have been undertaken by the ATLAS SUSY group to estimate the response of the ATLAS detector to a variety of physics processes and to optimise and test the software which is necessary for the analysis of collider data. As a part of this exercise, the ATLAS SUSY group has performed a number of studies. Techniques to estimate Standard Model backgrounds have been developed [7, 8] and calculations have been made of the discovery potential for generic inclusive SUSY signatures [9], based largely on the minimal SUGRA model [10–14] for which the principal observables are high-$p_{\mathrm{T}}$ jets and missing transverse energy. However, there exist a number of SUSY scenarios which predict specific final state topologies which may not be observed by generic searches, such as prompt photons and long-lived stable massive particles. In this work, the response of the ATLAS detector to such signatures is studied and, where appropriate, calculations are made of discovery potentials for early LHC running for which an integrated luminosity of around 1 fb$^{-1}$ is assumed. Although the work is performed within the framework of SUSY searches, the techniques which have been developed are more generally applicable to searches for new phenomena.

The theoretical models used in this work are based on the minimal Gauge Mediated Supersymmetry Breaking (GMSB) model [6, 15–21] , the Split-SUSY model [22–25] and the gravitino LSP model [26]. A description of the models is given in Section 2.1 while specific choices of parameters are found in the sections devoted to each of the studied signatures. Four signatures are investigated and are described below.

**Two high-$p_{\mathrm{T}}$ photons + $E_{\mathrm{T}}^{\mathrm{miss}}$** : In a GMSB scenario in which the next lightest supersymmetric particle (NLSP) is the $\tilde{\chi}_1^0$ , two high-$p_{\mathrm{T}}$ photons are expected in each signal event each arising from the decay of a $\tilde{\chi}_1^0$ to a $\tilde{G}$ and photon. Such events with two isolated high-$p_{\mathrm{T}}$ photons plus large $E_{\mathrm{T}}^{\mathrm{miss}}$ have small Standard Model backgrounds and a high mass discovery reach is expected even in the early LHC

running. In the high mass regime, however, the production cross-section for SUSY events and the signal is of the same order as the instrumental background which includes misidentified photons arising from electrons or jets, and the irreducible background of radiation from leptons. Earlier experiments have exclusions limits of 93 GeV in $\tilde{\chi}_1^0$ mass and 167 GeV in $\tilde{\chi}_1^{\pm}$ mass [27].

**Non-pointing photons**: In certain GMSB scenarios, the $\tilde{\chi}_1^0$ could be relatively long-lived. When the decay length is comparable to the size of ATLAS inner-detector, high-$p_T$ photons could enter the electromagnetic calorimeter surface at large incident angles with respect to a pointing direction to the beam interaction point. An estimation of how the photon reconstruction and identification efficiency is degraded for such non-pointing photons is essential in any measurement of the $\tilde{\chi}_1^0$ lifetime, which is directly connected to the SUSY breaking scale. Current lower limits on the mass and lifetime are 101 GeV and 5 ns, respectively [28].

**Stable sleptons**: Stable[1] heavy charged sleptons appear in certain regions of parameter space in GMSB scenarios. This signature consists of a penetrating charged track. Since interactions with detectors are ionizations only, the observed tracks will look more like muons, except for their higher energy deposition and longer time of flight than muons. The ATLAS muon system provides a time of flight measurement with an excellent time resolution ($\sigma_{tof} \approx 0.7$ns) [29], which allows for a precise particle mass measurements for slow particles. Owing to the very high LHC bunch crossing rate, the development of an appropriate triggering scheme is critical to the ensuring the detection of such particles. Previous experiments have provided a lower limit of around 105 GeV in slepton mass [30]

**Stable *R* hadrons**: Stable massive supersymmetric hadrons (*R* hadrons) are predicted in Split-SUSY models or in the gravitino LSP scenario of SUGRA models [26]. The signature of *R* hadrons is similar to that of stable sleptons although multiple nuclear interactions before reaching the muon system lead to characteristic event topologies, such as the appearence of high-$p_T$ tracks in the muon system with no matching track in the inner-detector, or the electric charge flipping between the inner-detector and the muon system. Lower mass limits of around 200 GeV have already been established by other experiments [31].

This paper is organized as follows. Section 2 provides an overview of the different models and a description of the Monte Carlo datasets which were used in this work. Sections 3-6 describe the studies of signatures described above. Finally, a conclusion is given in Section 7.

## 2 Overview of the SUSY models and Monte Carlo datasets

The signal Monte Carlo data samples used in this paper are based on the specific supersymetry breaking scenarios. These models are described in Section 2.1. Some technical details and parameter values used in individual datasets are summarized in Section 2.2. Common background samples are used which are described in the introduction to this chapter [32]. In some cases, fast simulation (ATLFAST-I [33]) samples are also used when estimating the background contributions.

### 2.1 SUSY scenarios considered in this paper

- GMSB (Gauge-Mediated Supersymmetry Breaking) scenarios: in GMSB, SUSY breaking which takes place in hidden sector is transmitted to visible MSSM fields through a messenger sector whose mass scale is much below the Planck scale ($M_m \ll M_P$) via the ordinary Standard Model gauge interactions. The gravitino is very light (in general $\ll 1$ GeV) and is always the LSP. In the minimal model of GMSB, all Supersymmetry breaking interactions are determined by a few

[1]The term stable is used throughout this chapter to refer to particles which, although they may possess finite lifetimes, do not decay during their traversal of the ATLAS detector.

parameters. The phenomenologies studied in this paper are based on this minimal model. Discussions of the non-minimal GMSB models are found elsewhere [34, 35].

The squarks, the sleptons, and the gauginos obtain their masses radiatively from the gauge interactions with the massive messengers; their masses therefore depend on the number of messenger generations, $N_5$ (the index 5 comes from the fact that the messenger fields form complete $SU(5)$ representations).

The gaugino masses scale like $N_5$ while the scalar masses scale like $\sqrt{N_5}$. Hence for $N_5 = 1$, the NLSP is the lightest neutralino ($\tilde{\chi}_1^0$ ) which decays into photon and a gravitino ($\tilde{G}$ ). For $N_5 \geq 2$, the NLSP is a charged stau ($\tilde{\tau}_1$). When $\tan\beta$ is not too large, the mass splitting between the $\tilde{\tau}_1$ and the right-handed selectrons, smuons ($\tilde{e}_R, \tilde{\mu}_R$) is small, rendering them co-NLSP's which decay into leptons and gravitinos. In contrast, in the large $\tan\beta$ region, the stau is the sole NLSP.

The effective SUSY breaking order parameter $F_S$, felt by the messengers, may not coincide with the intrinsic underlying SUSY breaking order parameter $F$, which determines the coupling strength. When $F_S$ is smaller, the NLSP decay length becomes longer. The dimensionless factor $C_G = F/F_S(\geq 1)$ is introduced to control the NLSP decay length with all the other parameters fixed. The decay length is proportional to the square of the control parameter, i.e. scales as $C_G^2$.

The effective visible sector SUSY breaking parameter, $\Lambda(= F_S/M_m)$ sets the overall mass scale for all the MSSM superpartners, which scales linearly with $\Lambda$. On the other hand, these masses only depend logarithmically on the messenger scale $M_m$. The MSSM masses are therefore predominantly determined by the scale $\Lambda$.

- Split-SUSY scenario: Stable exotic hadrons feature in a number of SUSY scenarios. Split-SUSY is one such model. Within this approach the hierarchy problem and the fine tuning of the Higgs mass is accepted, or assumed to be set by another, as yet unknown, mechanism. Phenomenologically, within Split-SUSY scenarios the gauginos and higgsinos have light masses of order the weak scale, which are protected by the chiral symmetry while the scalars have a mass scale $m_s$ which can be near the GUT scale. Since gluino decays proceed via internal squark lines, gluinos can be meta-stable. A meta-stable gluino will form a bound state, a so-called $R_{\tilde{g}}$-hadron.

- A gravitino LSP and a stop NLSP scenario: in addition to stable gluinos, meta-stable stops are also features of some SUSY models. Here, the stops are usually the NLSP and decay to a gravitino LSP with gravitational strength interactions. The generic possible candidate for NLSP is the lightest stop $\tilde{t}_1$, which, like the gluino case in split SUSY scenario, would form stable bound states, denoted $R_{\tilde{t}}$.

## 2.2 Parameter values used in this work

The basic feature of the samples used in this paper are similar to those introduced in [32]. However, our studies also investigate sparticles which decay with a measurable decay length, and new techniques have been introduced to allow these to be simulated. As in Ref. [32], ISAJET7.74 [36] was used to generate the sparticle mass spectrum in the context of minimal GMSB scenarios. The leading order (LO) and the next-to-leading order (NLO) cross-sections were assessed independently using PROSPINO2.0 [37–39]. The mass spectrum tables from ISAJET were used by HERWIG/JIMMY [40, 41] to generate the sparticle cascade decays, parton showers, hadronisations, and underlying events. The HERWIG option *pltcut* (lifetime threshold above which the Herwig does not decay the particles) was set to $3.3 \times 10^{-11}$ s, hence sparticles with finite decay length are not decayed at event generator level[2]. In the detector

[2]the value is set in order not to decay $K^0$ and $\Lambda$ at event generator level, but at detector simulation level.

Table 1: Summary of the neutralino NLSP samples. Dataset GMSB1 is a prompt photon decay sample, while dataset GMSB2 and GMSB3 are the non-pointing photon samples. $N_5 = 1, \tan\beta = 5, \mathrm{sgn}(\mu) = +$ are used at each point.

| name | NLO (LO) $\sigma$ [pb] | $\Lambda$ [TeV] | $M_m$ [TeV] | $C_G$ | $c\tau$ [mm] | $M_{\tilde{\chi}_1^0}$ [GeV] |
|---|---|---|---|---|---|---|
| GMSB1 | 7.8 (5.1) | 90 | 500 | 1.0 | 1.1 | 118.8 |
| GMSB2 | 7.8 (5.1) | 90 | 500 | 30.0 | $9.5 \cdot 10^2$ | 118.8 |
| GMSB3 | 7.8 (5.1) | 90 | 500 | 55.0 | $3.2 \cdot 10^3$ | 118.8 |

Table 2: Summary of the slepton NLSP sample. $N_5 = 3, \tan\beta = 5, \mathrm{sgn}(\mu) = +$, and no decay of slepton is assumed.

| name | NLO (LO) $\sigma$ [pb] | $\Lambda$ [TeV] | $M_m$ [TeV] | $M_{\tilde{\tau}_1}$ [GeV] |
|---|---|---|---|---|
| GMSB5 | 21.0 (15.5) | 30 | 250 | 102.3 |

Table 3: $R$-hadron samples. Dataset R-Hadron1 – R-Hadron6 are the $R_{\tilde{g}}$ samples, while dataset R-Hadron7 – R-Hadron9 are the $R_{\tilde{t}}$ samples.

| name | NLO (LO) cross-section [pb] | sparticle | Mass [GeV] |
|---|---|---|---|
| R-Hadron1 | 567 (335) | $\tilde{g}$ | 300 |
| R-Hadron2 | 12.2 (6.9) | $\tilde{g}$ | 600 |
| R-Hadron3 | 0.43 (0.23) | $\tilde{g}$ | 1000 |
| R-Hadron4 | 0.063 (0.033) | $\tilde{g}$ | 1300 |
| R-Hadron5 | 0.011 (0.006) | $\tilde{g}$ | 1600 |
| R-Hadron6 | 0.0014 (0.00075) | $\tilde{g}$ | 2000 |
| R-Hadron7 | 11.4 (7.8) | $\tilde{t}$ | 300 |
| R-Hadron8 | 0.27 (0.18) | $\tilde{t}$ | 600 |
| R-Hadron9 | 0.010 (0.0064) | $\tilde{t}$ | 900 |

simulation phase, the $\tilde{\chi}_1^0$ decaying into $\tilde{G}$ and $\gamma$, decay length was passed to the ATLAS interface to GEANT4 and the sleptons ($\tilde{\tau}_1, \tilde{e}_R$ and $\tilde{\mu}_R$), interaction with detector materials, i.e. ionization loss was implemented.

In $R$-hadrons scenarios, the masses of the stable sparticles ($M_{\tilde{g}}, M_{\tilde{t}}$) fully determine the phenomenology and other supersymmetric parameters (other sparticle masses) do not enter into any calculation. The generated samples are therefore highly model independent, and the only parameter that varies is the mass of the stable sparticle.

Tables 1-2 summarize the properties of the Monte Carlo signal samples used in this note. Tables 1 and 2 describe the GMSB samples with a neutralino or slepton NLSP, respectively, while Table 3 summarizes the $R$-hadron samples.

# 3 Discovery potential of GMSB SUSY with photon signatures

In GMSB models with $N_5 = 1$ and low $\tan\beta$ the lightest neutralino $\widetilde{\chi}_1^0$ is the NLSP and decays to a gravitino $\widetilde{G}$ and, in scenarios where the neutralino is mainly a photino, to a photon. Therefore the standard SUSY decay cascade of squarks and gluinos is extended by the decay $\widetilde{\chi}_1^0 \to \gamma\widetilde{G}$, as shown in Figure 1. Depending on the branching ratios of the squarks and gluinos decaying to various types of SUSY particles, this decay chain may also contain multiple jets. Events with two high energy photons are expected in $pp$ collisions, if the NLSP lifetime is not too long ($C_{\mathrm{grav}} \sim 1$). These photons originate close to the

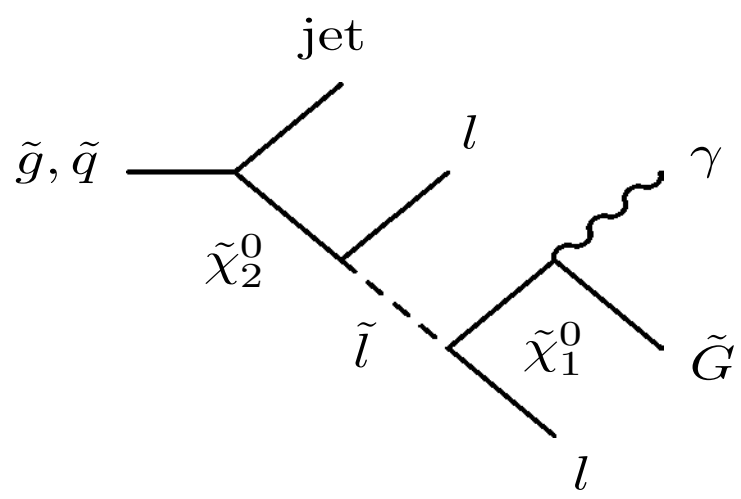


Figure 1: Typical SUSY decay chain for a neutralino NLSP decaying to a photon and a gravitino.

primary interaction vertex ("prompt photons"). The corresponding event signatures and the discovery potential for these models are discussed in this section. The case of large $C_{\text{grav}}$ and therefore long lived neutralinos, resulting in non-pointing photons, is discussed in Section 3.3. For detailed reconstruction and trigger studies we consider the GMSB1 model point as a typical example (see Table 1). For this point the branching ratio of the decay of the lightest neutralino to a photon and a gravitino is $\sim 97\%$, and the total SUSY production cross-section is $\sim 7.8\,\text{pb}$.

In the following we discuss the optimisation of the signal selection, including the trigger selection, the expected background from Standard Model processes and a detailed study of the discovery potential with early data. Since for the latter a fast simulation approach is used, a comparison of fast and full simulation results is also presented.

## 3.1 GMSB1 full simulation studies

### 3.1.1 Signal trigger strategy

The signal events studied here possess the standard SUSY event properties at the LHC: large $E_{\text{T}}^{\text{miss}}$ and multiple jets with high $p_T$. These can be used for triggering the events. In addition, the feature of two high energy photons gives an additional way to trigger on these events independently of $E_{\text{T}}^{\text{miss}}$ and jet triggers. In the following we consider two different trigger menus:

- The first menu is the ATLAS initial menu foreseen for data-taking at a luminosity of $10^{31}\,\text{cm}^{-2}\text{s}^{-1}$ [42]. This menu includes various combinations of jet (J) and $E_{\text{T}}^{\text{miss}}$ triggers (XE) and photon triggers (EM) as shown in Table 4. No prescale values are foreseen for these triggers, whereas the selection of the EM100 trigger is based entirely on the Level-1 (L1) trigger, with no selections envisaged at the Level-2 (L2) trigger or in the event filter. In the following, the trigger efficiency is defined as the ratio of the number of events without any preselection passing a trigger item divided

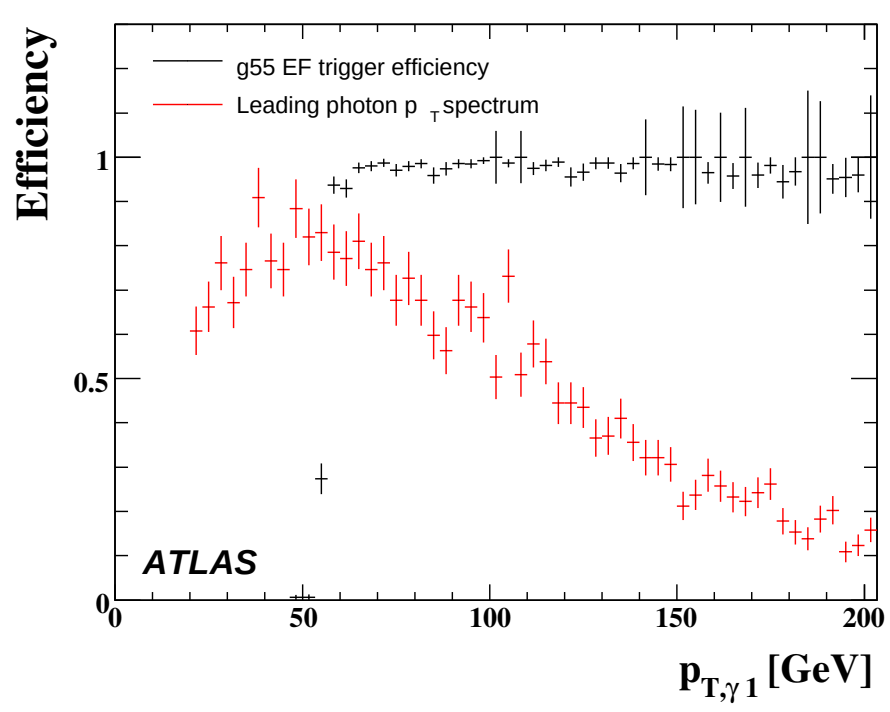


1

Figure 2: The g55 event filter trigger efficiency (see text) for the GMSB1 sample as a function of the reconstructed $p_{\text{T}}$ of the leading photon.

by the total number of events. The total efficiencies for the L1 trigger for the various triggers of this initial menu are summarised in Table 4. It can be seen, that in addition to the standard SUSY triggers based on jets and $E_{\mathrm{T}}^{\mathrm{miss}}$ , the photon triggers have very high efficiencies for selecting GMSB1 events and can be used for initial running.

- The second menu investigated here is the standard ATLAS trigger menu foreseen for the stable running at a luminosity of $\mathscr{L} = 10^{33}\,\mathrm{cm}^{-2}\mathrm{s}^{-1}$. In this menu all three trigger levels are used in the selection. The triggers investigated are combinations of jet (j), $E_{\mathrm{T}}^{\mathrm{miss}}$ (xE) and photon (g) signatures, see Table 5. Again, no prescale values are foreseen for these items. The trigger efficiencies for the signal at L1, L2 and event filter are summarised in Table 5, where the L2 and event filter efficiencies also contain the efficiencies of the previous levels. It can be seen that for both of the planned running phases the photon triggers are as efficient as the $E_{\mathrm{T}}^{\mathrm{miss}}$ and jet triggers for the case of GMSB1 signal events. The efficiency for the g55 photon trigger after the event filter as a function of the reconstructed $p_{\mathrm{T}}$ of the leading photon is shown in Figure 2. As expected, after a steep turn on around 55 GeV a plateau is reached. The integrated efficiency above the threshold of 55 GeV is $\sim 98\%$.

Table 4: Trigger efficiencies and statistical errors for the GMSB1 event sample for ($\mathscr{L} = 10^{31}\,\mathrm{cm}^{-2}\mathrm{s}^{-1}$).

| Trigger item | Efficiency | Trigger item | Efficiency |
|---|---|---|---|
| 2EM13 | 98.71 ± 0.11 | XE70 | 81.39 ± 0.39 |
| EM100 | 83.00 ± 0.38 | J70+XE30 | 93.64 ± 0.25 |
| EM18+XE15 | 98.79 ± 0.12 | 2J42+XE30 | 93.98 ± 0.24 |
| J100 | 91.43 ± 0.28 | 4J23 | 92.27 ± 0.27 |

Table 5: Trigger efficiencies and statistical errors for the GMSB1 event sample for ($\mathscr{L} = 10^{33}\,\mathrm{cm}^{-2}\mathrm{s}^{-1}$).

| Trigger item | L1 | L1+L2 | L1+L2+EF |
|---|---|---|---|
| g55 | 97.18±0.60 | 84.47±1.32 | 80.47±1.44 |
| 2g17i | 71.13±1.65 | 55.07±1.81 | 47.91±1.81 |
| j65+xE70 | 80.66±0.40 | 80.63±0.40 | 69.53±0.46 |
| 3j65 | 83.63±0.37 | 83.55±0.37 | 83.37±0.37 |

In summary it can be said that in the GMSB1 scenario the use of photon triggers is possible for initial running conditions, as well as at a higher luminosity. The efficiencies are as high as for the triggers based on jets and $E_{\mathrm{T}}^{\mathrm{miss}}$ and can thus provide good redundancy.

### 3.1.2 Signal selection

At the benchmark point GMSB1, 48.9% (16.4%) of the signal events have one (two) photons with $p_{\mathrm{T}} > 20$ GeV in the fiducial acceptance ($|\eta| < 2.5$) used for photon identification. For the reconstruction of photons a standard cut-based photon selection [43] is used. This is mainly based on variables using information of the first and second samplings of the electromagnetic calorimeter. The photons are required to be isolated and those located in the transition regions between barrel and endcap calorimeters are excluded. No track veto is applied. After a full GEANT4 simulation of the ATLAS detector, the selection efficiency for photons with $p_{\mathrm{T}} > 20$ GeV is about 65%.

Table 6: ALPGEN Background samples used in this section. The corresponding integrated luminosities is shown.

| Process | | Integrated luminosity ($fb^{-1}$) |
|---|---|---|
| Top | leptonic | $\sim 13.4$ |
| | semi-leptonic | $\sim 3.5$ |
| | hadronic | $\sim 17.1$ |
| Electroweak | $Z \to e^+e^-$ | $\sim 4.8$ |
| + jets | $Z \to \mu^+\mu^-$ | $\sim 8.3$ |
| | $Z \to \tau^+\tau^-$ | $\sim 21.6$ |
| | $Z \to \nu\nu$ | $\sim 9.1$ |
| | $W \to e\nu$ | $\sim 3.4$ |
| | $W \to \mu\nu$ | $\sim 4.7$ |
| | $W \to \tau\nu$ | $\sim 3.9$ |
| QCD | multiple jet production | $\sim 0.03$ |

As background to the signal, events with QCD jets, single gauge boson (W and Z) production and $t\bar{t}$ production are simulated using the ALPGEN generator. The specific processes are listed in Table 6 and the corresponding integrated luminosities are given for each process. In order to separate the signal from the Standard Model background a standard preselection for SUSY-like signatures is first performed:

- At least four jets must be found with $p_T > 50$ GeV ($p_T > 100$ GeV for the leading jet).
- Missing transverse energy $E_T^{miss} > 100$ GeV and $E_T^{miss} > 20\% \cdot M_{eff}$, where the effective mass $M_{eff}$ is defined as the scalar sum of $E_T^{miss}$ and the transverse momenta of the four leading jets.

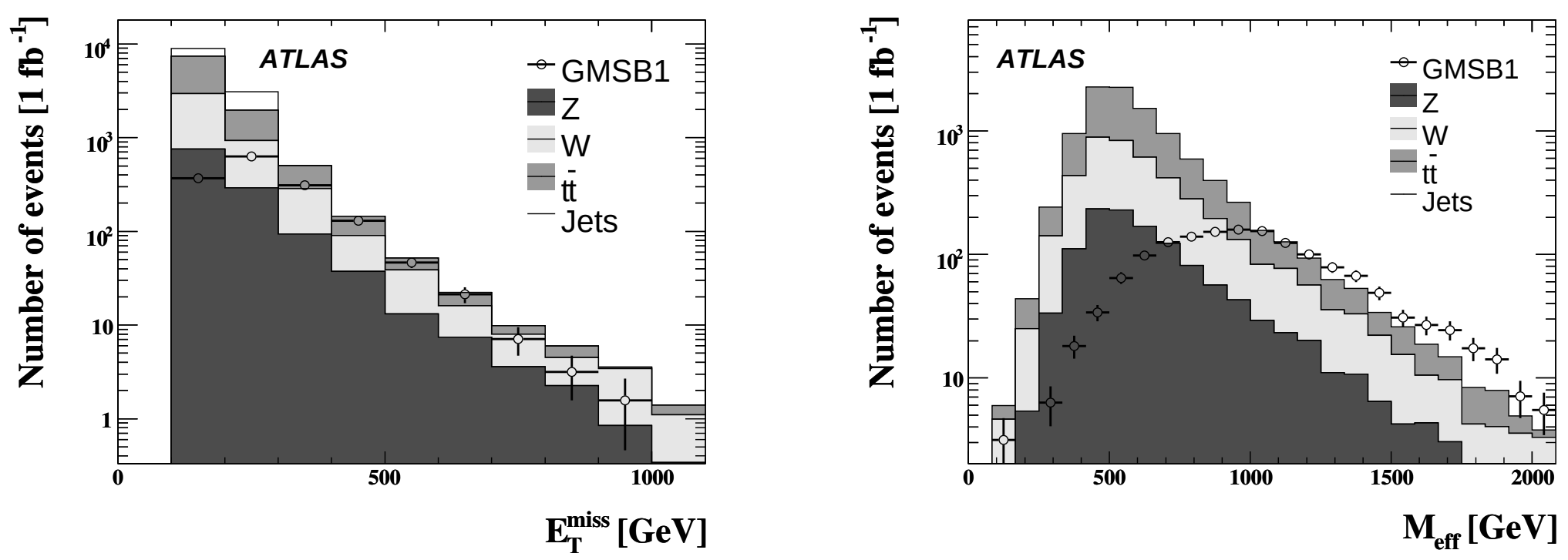


Figure 3: Distributions after preselection for 1 $fb^{-1}$. Left: Missing transverse energy. Right: Effective mass for signal and Standard Model background.

The distributions of $E_T^{miss}$ and $M_{eff}$ after this selection are shown in Figure 3 for the Standard Model background (histograms) and the signal (open symbols) for an integrated luminosity of 1 $fb^{-1}$. No large excess of events is seen over the Standard Model background. However, as shown in Figure 4, a cut on the number of reconstructed photons with $p_T > 20$ GeV and $|\eta| < 2.5$ provides an effective way to further suppress the backgrounds. Figure 3.1.2 shows the $p_T$ distribution of the leading photon after the initial preselection. Table 7 shows the number of selected events for signal ($S$) and background ($B$) after requiring 0, 1 or 2 photons passing the cuts described above and either the g55 or the 2g17i trigger. This combination of triggers has a combined efficiency for the signal of $\sim 85\%$ ($\sim 99\%$) before (after) applying these selection cuts. The number of events is normalized to an integrated luminosity of 1 $fb^{-1}$.

In addition, the signal significance defined as $Sig = S/\sqrt{B}$ is given in the table. In the calculation of $Sig$ it is assumed, that there is at least one background event left. With the requirement of two high-energy photons the selection is mainly free from Standard Model background and the significance becomes very large.

In addition to the selection criteria listed above, checks were made to see weather a better signal significance can be achieved by requiring an opposite sign same flavour (OSSF) lepton pair, which originates in the squark/gluino decay cascade from a $\tilde{\chi}_2^0$ to $\tilde{\chi}_1^0$ decay via a slepton, as depicted in Figure 1. Here, only electrons and muons are accepted as leptons. The requirement of at least one OSSF lepton pair reduces the number of selected signal and background events. The suppression factor for the signal for the combination of one photon and one OSSF pair is larger than for the combination of two photons. Hence, although the background is reduced to a very low level, just using the requirement of two photons gives the largest significance. Requiring an OSSF pair in addition to two photons just reduces the signal, because the background is already very low.

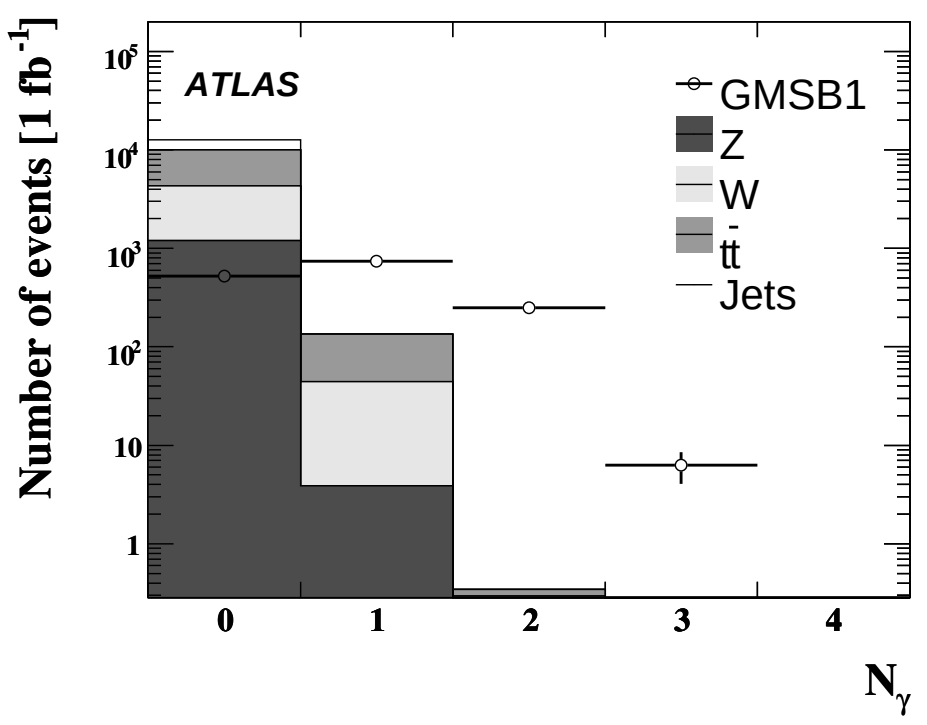


Figure 4: Distributions after preselection for 1 fb$^{-1}$: Number of reconstructed photons with $p_{\mathrm{T}} > 20$ GeV and $|\eta| < 2.5$ (left) and transverse momentum of the leading photon for signal and Standard Model background. (right)

## 3.2 GMSB parameter scan with fast simulation

To investigate the discovery potential of the selection described above, over a wider range of the GMSB parameter space, it is necessary to make use of a fast detector simulation to obtain adequate statistics for signal event reconstruction at various points in the parameter space. The computing requirements of the full simulation make it impractical for this study, so for this part of the analysis the fast simulation package ATLFAST [44] has been used instead. ATLFAST performs no detailed simulation of particle interactions with the detector material, but instead parameterizes the detector response. It has two main features relevant to the analysis discussed here:

- every generated particle is reconstructed.
- there is no distinction between electromagnetic and hadronic calorimeter compartments and the energy of a particle is obtained by smearing the energy of the generated particle with a resolution function. No shower development is simulated.

Note that ATLFAST does not simulate either reconstruction inefficiencies nor particle misidentification for any particles.

Figure 5(a) shows the $M_{\mathrm{eff}}$ distributions of the GMSB1 event sample for full and fast simulation. In the low energy region a small deviation of the fast simulation with respect to the full simulation can be

Table 7: Number of selected signal and background events for 1 fb$^{-1}$ for different cuts on the number of photons and opposite sign same flavour (OSSF) lepton pairs.

| $N_\gamma$ | $N_{OSSF}$ | Signal | $\sum$ Background | Sig | $N_W$ | $N_Z$ | $N_{t\bar{t}}$ |
|---|---|---|---|---|---|---|---|
| 0 | 0 | 1287.4 | 929.6 | 42.3 | 274.4 | 21.0 | 632.8 |
| 0 | 1 | 283.6 | 73.0 | 33.2 | 8.7 | 1.4 | 63.0 |
| 1 | 0 | 902.9 | 51.7 | 126.1 | 19.5 | 2.0 | 30.1 |
| 1 | 1 | 189.1 | 1.4 | 161.4 | 0.2 | 0.0 | 1.2 |
| 2 | 0 | 252.9 | 0.1 | 252.9 | 0.0 | 0.0 | 0.1 |
| 2 | 1 | 37.0 | 0.0 | 37.0 | 0.0 | 0.0 | 0.0 |

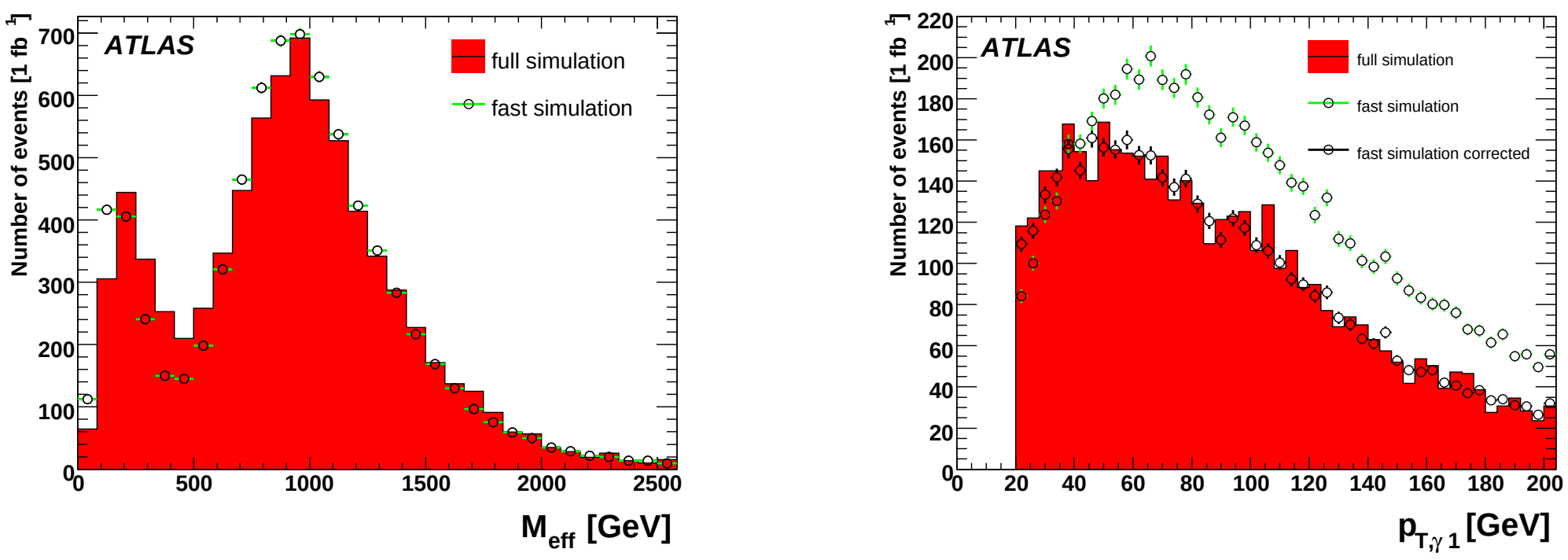


(a) (b)

Figure 5: Signal distributions for full and fast simulation: a) effective mass, b) transverse momentum of the leading photon.

observed, which is due to a slightly higher jet momentum in ATLFAST in the low energy region. However, of more importance for this analysis is the simulation of the photon identification and it is important that the fast simulation provides a reliable modelling. The transverse momentum of the leading photon is shown in Figure 5(b) for full and fast simulation. It can clearly be seen that some significant discrepancies between both simulation approaches exist. This is a result of the photon detection efficiency, which is not included in the fast simulation. The fast simulation assumes a 100% detection efficiency for all truth photons which pass a certain isolation criterion. This effect is taken into account in the analysis by imposing a realistic reconstruction probability on each photon by hand, depending on the transverse momentum of the photon. The corrected $p_{\mathrm{T}}$ distribution is also shown in Figure 5(b).

In Table 8 the numbers of selected signal events for full and fast simulation are shown and good agreement can be observed after each step of the selection. The level of agreement of the fast simulation with the full simulation is of the order of 10% which is considered to be sufficient for a rough estimation of the discovery potential via a scan of the GMSB model parameters using ATLFAST.

Table 8: Number of selected signal events normalised to $\mathscr{L} = 1\,\mathrm{fb}^{-1}$ for full and fast simulation for GMSB1.

| $N_\gamma$ | $N_{OSSF}$ | Signal (full) | Signal (fast) |
|---|---|---|---|
| 0 | 0 | 1287.4 | 1597.7 |
| 1 | 0 | 902.9 | 1029.5 |
| 2 | 0 | 252.9 | 275.3 |

In order to estimate the selection efficiency and the discovery potential of the two photon channel in a more model independent way, a scan of some GMSB model parameters has been performed. The mass spectrum and branching fractions of the different GMSB model points have been calculated using ISASUGRA 7.74 [36]. For each point 12500 signal events have been generated using the HERWIG/JIMMY [40,41]. As discussed above, ATLFAST is used for the detector simulation of the signal with a correction applied to the photon reconstruction efficiency. The estimates for the Standard Model background are taken from the full simulation as described in Section 3.1.2.

For the scan, the SUSY breaking scale parameter $\Lambda$ has been varied from 60 to 200 TeV in steps of 10 TeV and $\tan\beta$ has been varied from 2 to 50 in steps of 2. The other model parameters are fixed to $N_5 = 1$, $M_{\text{mes}} = 500$ TeV, $\text{sgn}\mu = +1$ and $C_{\text{grav}} = 1$, as for the GMSB1 model point. In this part of the parameter space the neutralino is usually the NLSP, which is most often not the case for larger $N_5$ or larger $\tan\beta$. For these regions, other channels need to be used to discover GMSB SUSY. These are discussed in Section 5. The discovery potential for the case of different $C_{\text{grav}}$ and hence non-pointing photons is briefly discussed in Section 3.3.

Figure 6 shows the contour lines where the significance reaches $Sig = 5\sigma$ for the default selection cuts as described above for different integrated luminosities. Since it is assumed that there is 1 background event left, these contour lines represent the lines with 5 signal events. In the regions below and left of the lines a $5\sigma$ discovery can be made with the corresponding amount of data. In the high $\tan\beta$ region above the solid line no sensitivity is quoted for the two photon channel, since in this region the $\tilde{\tau}$ is the NLSP and so no significant excess of photons is expected from the SUSY decay chains. Due to the fact that the SUSY cross section decreases with increasing $\Lambda$, the significance decreases as a function of $\Lambda$ for a given integrated luminosity. In general the discovery potential in most parts of the GMSB model parameter space is high, giving confidence that GMSB SUSY can be discovered in the two photon channel with early data, if it is realised in nature.

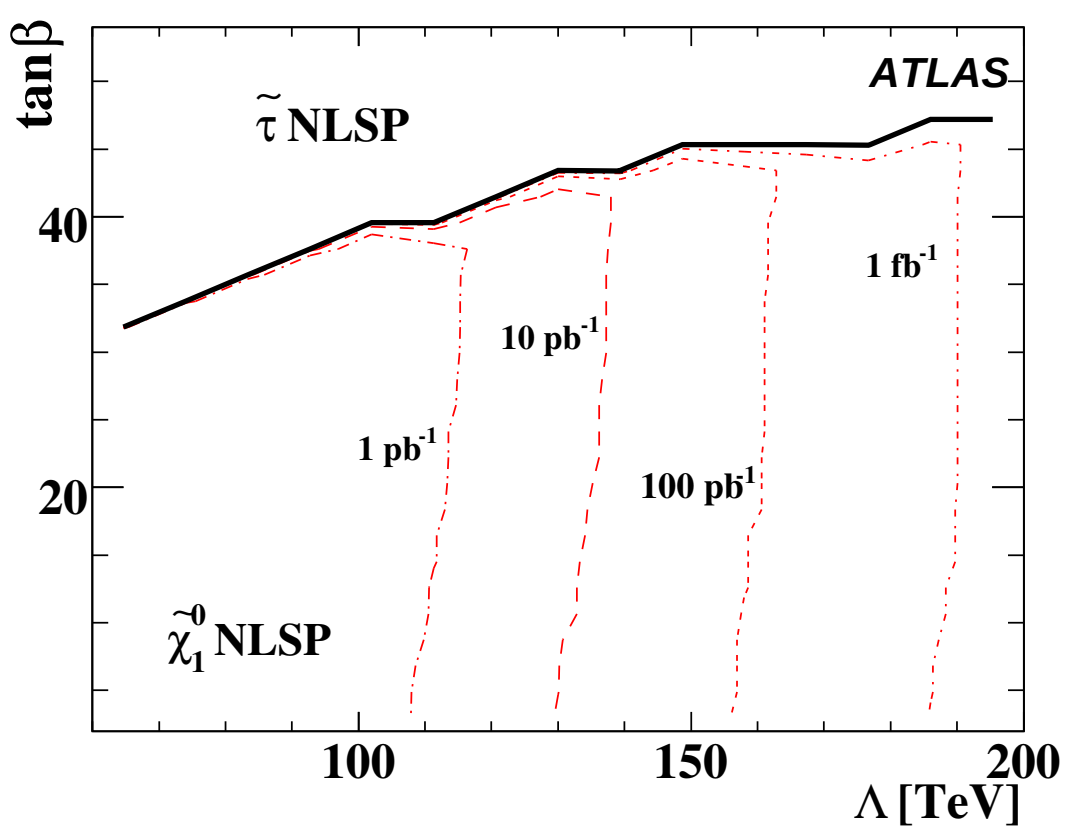


Figure 6: $5\sigma$ discovery potential contour lines for GMSB SUSY in the $\Lambda$ - $\tan\beta$ plane for different integrated luminosities.

### 3.3 GMSB3 (non-pointing photon) full simulation studies

If the gravitino mass parameter $C_{\text{grav}}$ is larger than unity, the NLSP will not decay promptly. In the case where the NLSP is a light neutralino, the resulting photons in the final state will not point back to the interaction point and may therefore be reconstructed and triggered with lower efficiency. The reconstruction efficiency is discussed in greater detail in Section 4.1. Here, the standard photon selection will be used to estimate the discovery potential.

Table 9: Trigger efficiencies and statistical errors for the GMSB3 event sample for ($\mathscr{L} = 10^{33}\,\text{cm}^{-2}\text{s}^{-1}$).

| Trigger item | L1 | L1+L2 | L1+L2+EF |
|---|---|---|---|
| g55 | 90.19±1.08 | 46.04±1.81 | 36.88±1.75 |
| 2g17i | 34.13±1.72 | 17.77±1.39 | 12.87±1.22 |
| j65+xE70 | 80.38±0.56 | 80.24±0.56 | 71.18±0.64 |
| 3j65 | 79.80±0.57 | 79.66±0.57 | 79.62±0.57 |

Table 10: Number of selected signal (GMSB3) and background events for $1\,\text{fb}^{-1}$ for different cuts on the number of photons and opposite sign same flavour (OSSF) lepton pairs.

| $N_\gamma$ | $N_{\text{OSSF}}$ | Signal | $\sum$ Background | Sig | $N_W$ | $N_Z$ | $N_{t\bar{t}}$ |
|---|---|---|---|---|---|---|---|
| 0 | 0 | 825.2 | 929.6 | 27.1 | 274.4 | 21.0 | 632.8 |
| 0 | 1 | 265.2 | 73.0 | 33.2 | 8.7 | 1.4 | 63.0 |
| 1 | 0 | 255.8 | 51.7 | 35.7 | 19.5 | 2.0 | 30.1 |
| 1 | 1 | 68.6 | 1.4 | 58.6 | 0.2 | 0.0 | 1.2 |
| 2 | 0 | 12.5 | 0.1 | 12.5 | 0.0 | 0.0 | 0.1 |
| 2 | 1 | 4.7 | 0.0 | 4.7 | 0.0 | 0.0 | 0.0 |

A suitable model point to study is the GMSB3 point, which has the same parameters as GMSB1, but with $C_{\text{grav}} = 55$. The $\tilde{\chi}_1^0$ decay length in this point is therefore $\gamma\beta c\tau \approx 3$ m. Although only 12.4% (0.6%) of the reconstructed events contain one (two) photons with $p_{\text{T}} > 20$ GeV in the detector acceptance region, this well exceeds the number of background photons. This suggests that one could also use the above defined selection, which is based on the requirement of two hard photons.

Table 9 shows the trigger efficiencies for the same items listed in Table 5. For the L1 trigger, the main source of inefficiency for the g55 trigger is from neutralino decaying to photons outside the inner-detector volume. The larger the $C_{\text{grav}}$ parameter is, the greater is the number of neutralinos that will decay outside the inner-detector. This effect is more pronounced for the 2g17i trigger, which is optimized for the production of both photons within the inner-detector volume. At L2 trigger and at the event filter, cuts are placed on the shape of the electromagnetic showers, which are less efficient for non-pointing photons due to their wider shower shape in the $\eta$ direction compared to prompt photons. The small difference in jet trigger efficiencies between GMSB1 and GMSB3 is again due to the difference in the number of photons produced within the inner-detector volume. Photons in the event are treated as jets up to the event filter, so that GMSB1 has effectively a larger number of jets, compared to the GMSB3 sample, which makes a small but significant difference in jet trigger efficiency.

The resulting numbers of selected events for 1 $\text{fb}^{-1}$ are shown in Table 10. It can clearly be seen that, although the significance, again defined as $Sig = S/\sqrt{B}$, is smaller than in the GMSB1 case, there are enough photons to select a large number of signal events. The difference to the prompt photon case is that with the requirement of an OSSF lepton pair one could obtain a larger significance, which is largest for a combination of one hard photon and one OSSF pair.

### 3.4 Conclusion of GMSB SUSY with photon signatures

In certain regions of the GMSB parameter space the NLSP is a light neutralino, decaying to a gravitino via the emission of hard photons. These photons can be used to efficiently reject the Standard Model background and to discover GMSB SUSY, if it is realized in nature. Attention must be payed to the fact that the photons might not point back to the interaction point leading to losses in reconstruction and

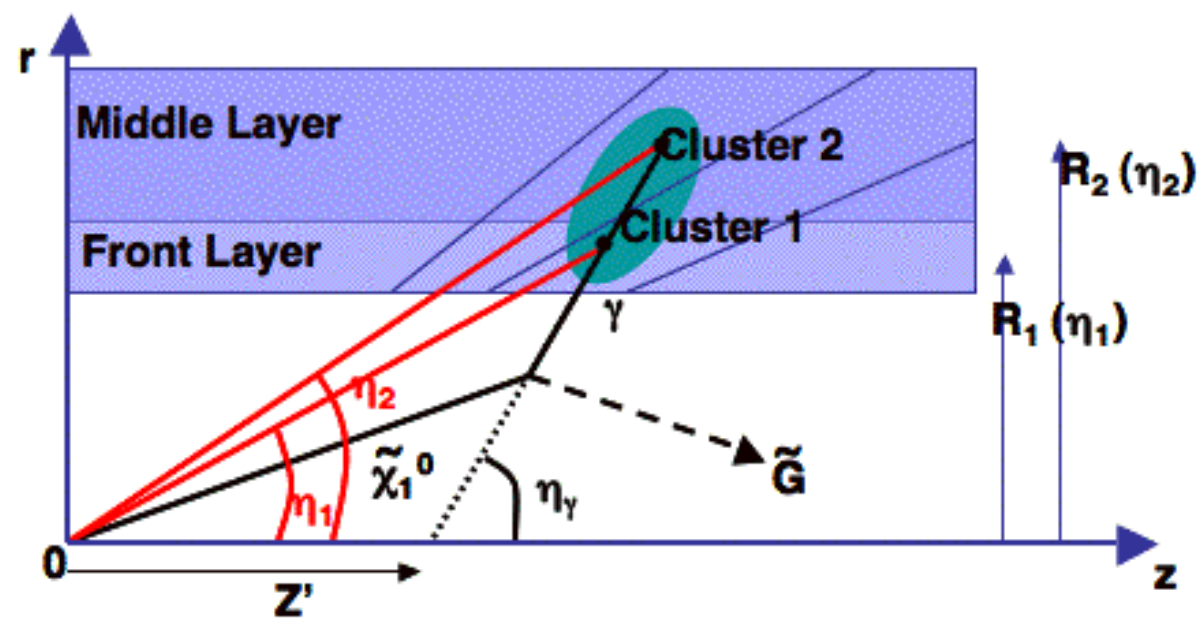


Figure 7: Schematic diagram of a non-pointing photon in the barrel section of the electromagnetic calorimeter. A long lived neutralino can travel a significant distance before decaying into a photon and a gravitino. The gravitino will escape the detector without interacting. A photon produced in this manner can enter the calorimeter at a significantly different angle ($\eta_\gamma$) than a photon produced at the primary vertex ($\eta_1$).

trigger efficiencies.

## 4 Prospects for neutralino lifetime determination

If GMSB SUSY is discovered, the ATLAS calorimeter can be used to first establish whether the neutralino has a long mean lifetime, and then to quantify it. The calorimeter can be used to both measure the direction and the time of the electromagnetic shower. Both the capabilities can be utilised to determine the mean lifetime of the neutralino.

If the neutralino has a significant decay length, a photon can be observed[3] that will not "point-back" to the primary interaction point. This is shown schematically in Figure 7. The neutralino ($\tilde{\chi}_1^0$) travels a significant distance before decaying into a photon ($\gamma$) and a gravitino ($\tilde{G}$). Due to the finite opening angle between the photon and $\tilde{G}$, the path taken by the photon does not extrapolate back to the primary interaction point.

The first sampling layer can measure the $\eta$ position (Cluster 1 in Figure 7) whereas the second sampling layer can measure both $\eta$ and $\phi$ (Cluster 2 in Figure 7). A vector corresponding to the path of the photon can therefore be constructed in the $r-z$ plane. Although we can not measure the exact decay point of the neutralino based on these two measurements, we can extrapolate the path of the photon back to the beam axis, and measure the distance between this point and the primary vertex ($Z'$ in Figure 7).

Since the ATLAS calorimeter has a pointing geometry, if a photon enters the calorimeter at a significant angle, the resulting electromagnetic shower can be spread out over a larger number of calorimeter cells. This wider shower-shape can result in issues for photon reconstruction algorithms and identification criteria. The effects of the reconstruction and identification of the non-pointing photons are discussed in Section 4.1.

If the neutralino has a mean lifetime greater than 0.05 ns[4], the $Z'$ value associated with the photon can be used to establish that the neutralino has a "long" lifetime. Once this observation has been made, the $Z'$ distance can also be used to measure the mean lifetime. This is discussed in Section 4.2.

The neutralino is a massive particle. This means that photons produced from long-lived neutralinos will arrive at the calorimeter later than prompt photons from the primary vertex. A method which uses the calorimeter timing information to calculate the mean neutralino lifetime is discussed in Section 4.3.

[3] As long as the neutralino decays before the calorimeter.

[4] For typical values assumed for the neutralino energy and mass of 200 and 100 GeV respectively, photons with a $Z'$ of at least 1cm from the primary vertex were observed.

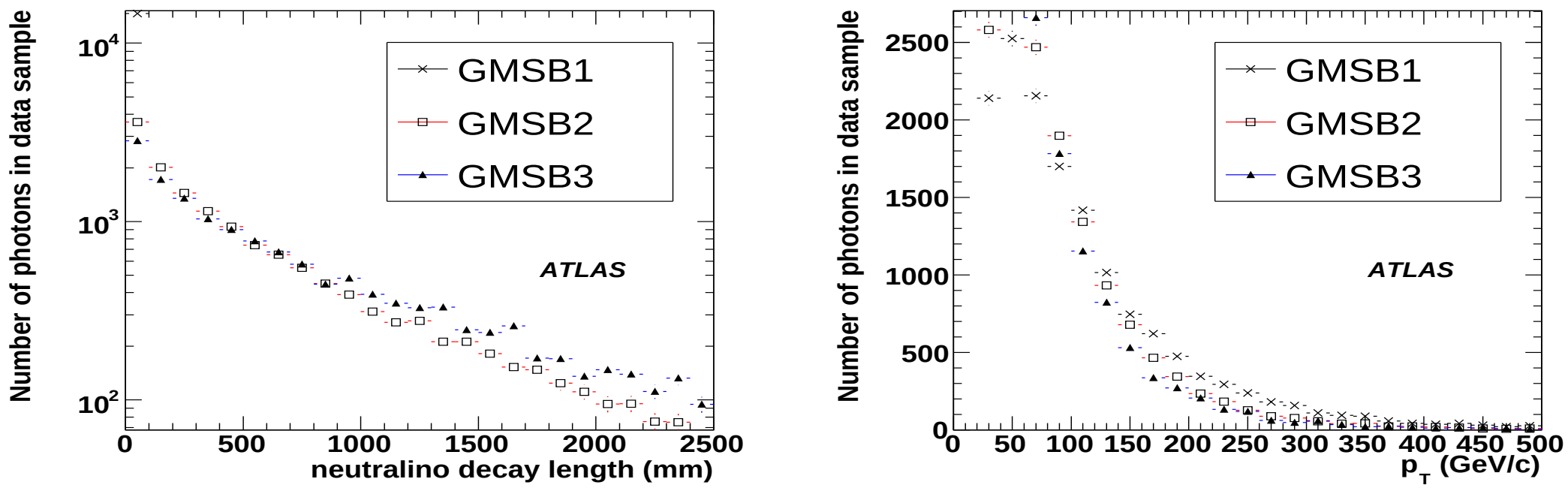


Figure 8: Left: The distributions of the decaylength of the neutralino in the $z$ direction. Right: The transverse momentum of the photons they produce.

## 4.1 Reconstruction and identification of non-pointing photons

### 4.1.1 $\eta$ definitions

Due to the nature of the non-pointing photons, two definitions of $\eta$ are used in this section. The 'truth $\eta$' refers to the $\eta$ from the particle vector from the Monte Carlo event record (shown as $\eta_\gamma$ in Figure 7). The term 'detector $\eta$' refers to the $\eta$ as measured by constructing a vector from (0,0,0) to the barycenter of the electromgnetic shower(shown as $\eta_2$ in Figure 7).

### 4.1.2 Photon reconstruction efficiency

The photon reconstruction efficiency is defined as the fraction of photons, produced from the decay of a neutralino, that are reconstructed as a photon candidate. Using information from the Monte Carlo event record, photons are selected to be used in the efficiency calculation if they satisfy the following criteria:

- originate from a neutralino decay occuring inside the outer envelope of the inner detector,
- $p_T > 20$ GeV,
- $|\text{detector } \eta| < 2.5$.

Photons are declared as successfully reconstructed if a photon candidate is found within $\Delta R < 0.2$ of the position of the truth photon in the calorimeter.

The $z$-component of the decay length and the momentum of the photon in the GMSB samples, as shown in Figure 8 and 8, depend upon the $C_{\text{grav}}$ parameter of the sample.

The overall efficiency for a photon to be reconstructed in the long-lived neutralino data samples, GMSB2 and GMSB3, is $88.3 \pm 0.2\%$ and $83.3 \pm 0.4\%$ respectively. This efficiency includes the reconstruction of photons which have converted to an electron-position pair. This occurs approximately 30% of the time and is dependent on how much material the photon travels through the inner-detector. This means that a photon from a long-lived neutralino will have a smaller probability of converting than a photon produced at the primary vertex. The reconstruction efficiency of a photon which does not convert, is $89.3 \pm 0.2\%$ for the GMSB2 sample and $84.2 \pm 0.5\%$ for the GMSB3 sample.

The difference in overall reconstruction efficiency measured between the GMSB2 and GMSB3 samples is due to the distribution of neutralino lifetimes in the sample, and hence the proportion of photons which are significantly "non-pointing".

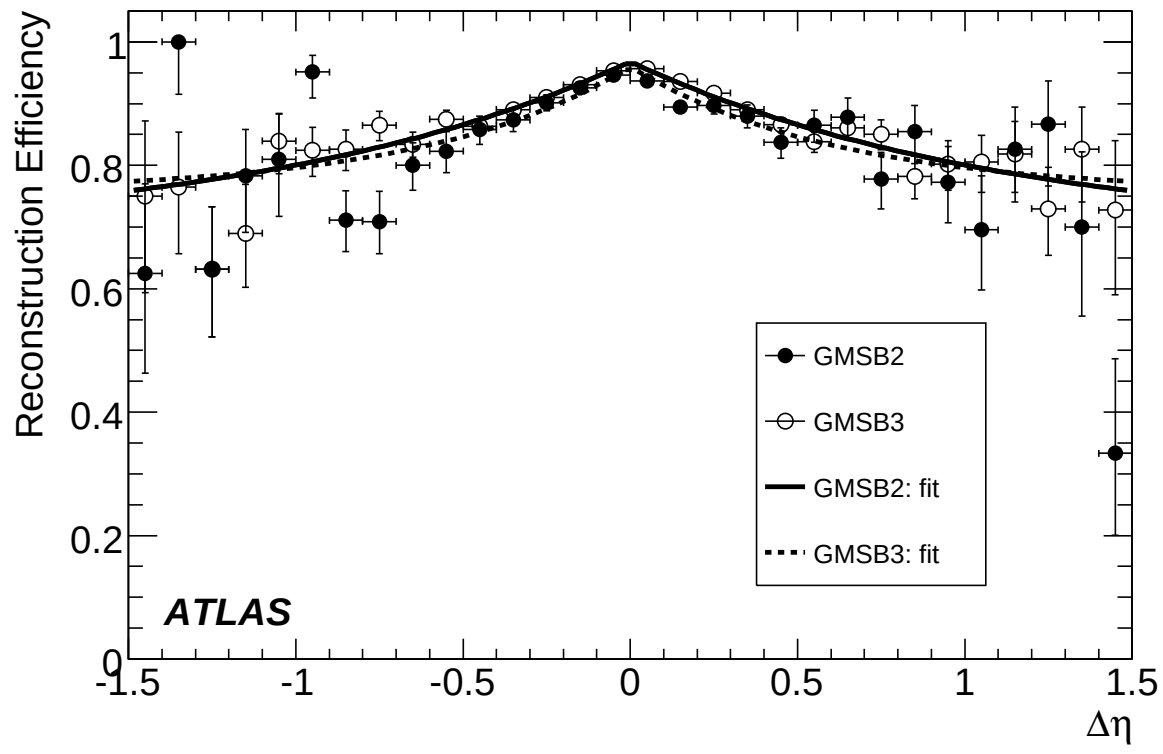


Figure 9: Reconstruction efficiency as a function of $\Delta\eta$ for the GMSB2 and GMSB3 samples.

The reconstruction efficiency is measured independently from the sample parameters as a function the variable $\Delta\eta$ = detector $\eta$ - truth $\eta$ as shown in Figure 9.

There is excellent agreement between the two cases. The efficiency is approximately 90% for $\Delta\eta < 0.25$, and falls steadily to 75% for $\Delta\eta \approx 0.5$. It is clear that the reconstruction efficiency could bias any measured neutralino lifetime distribution. This efficiency distribution is parameterised with an approximation to the top-hat function:

$$R_{\text{eff}}(\Delta\eta) = \frac{b}{1+e^{\frac{|\Delta\eta|+a}{c}}} + d \quad (1)$$

with $a = 4.7(6.9)$, $b = 174(229)$, $c = 0.779(1.16)$ and $d = 0.545(0.351)$ for GMSB2 (GMSB3). The GMSB2 fit result is used in Sections 4.2 and 4.3 (due to greater statistics) to account for any bias in the neutralino lifetime determination due to inefficiency.

### 4.1.3 Study of photon identification efficiency

The dependence of the photon efficiency on the neutralino decay length has been studied for all of the standard photon selection variables [43]. The efficiency for each cut is defined as the fraction of reconstructed photons with $p_T > 20$ GeV and $|\text{detector } \eta| < 2.5$ that pass the standard select requirement for the given variable. Only photons that are successfully identified with true photons that come from the decay of a SUSY particle are used.

Figure 10 shows the efficiency of these cuts as a function of the component of the neutralino decay length parallel to the beam axis. From this figure it can be seen that the hadronic leakage (Had/Em) is independent of the neutralino decay length. Of the cuts forming the standard selection from the second sampling layer of the electromagnetic calorimeter, only the ratio of energy in the $3{\times}3$ / $3{\times}7$ cells (R33) is shown to be flat with respect to the neutralino decay length. The efficiency of the cuts on the ratio of energy in the $3{\times}7$ / $7{\times}7$ cells (R37) and the the lateral width of the shower (weta2) are shown to have a clear dependence on the decay length. These two cuts are removed to form an ‘unbiased photon selection’ in the second sampling layer.

For the cuts forming the standard cut selection in the first sampling layer of the electromagnetic calorimeter, the fraction of energy (f1) and the cuts on the search for a second minima in the first sampling layer (DeltaE and DeltaEmax2) are shown to be relatively stable with respect to the neutralino decay length. There is however, a significant dependence of the efficiency on the decay length for the fraction of energy outside the shower core (fracm), the shower width in three strips (weta1) and the total width of the shower (wtot). These three cuts are removed to form an ‘unbiased photon selection’ in the first sampling layer.

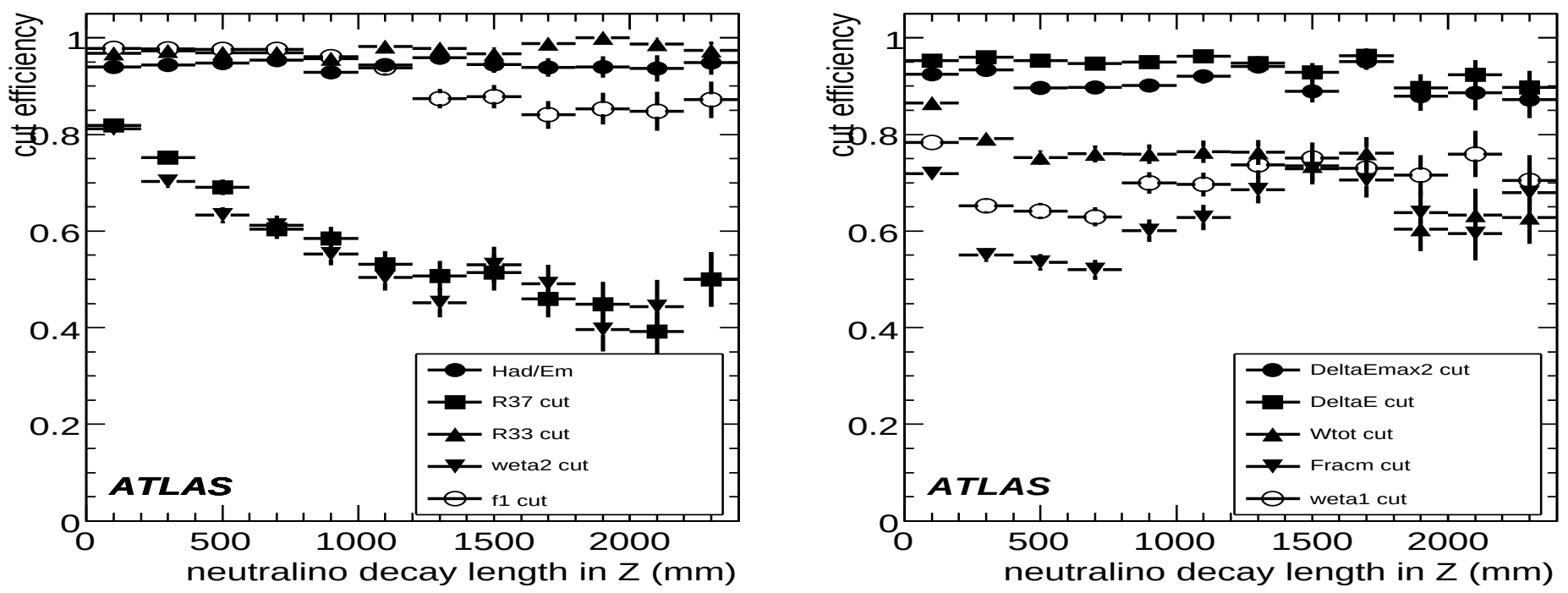


Figure 10: Photon identification cut efficiencies as a function of neutralino decay length in Z

Table 11 shows the effect on the overall efficiency as the five cuts which have been shown to be dependent on the neutralino decay length cuts are excluded from the standard photon selection to form an 'unbiased photon selection'. For simplicity the cuts are seperated according to which calorimeter sampling layer they are based upon.

Table 11: Summary of photons identification efficiencies for the three different signal samples. The hadronic, second sampling and first sampling cuts are applied sequentially.

| Standard photon selection | | | |
|---|---|---|---|
| | hadronic | $2^{nd}$ sampling | $1^{st}$ sampling |
| GMSB1 | $(94.1 \pm 0.2)\%$ | $(75.7 \pm 0.4)\%$ | $(64.1 \pm 0.4)\%$ |
| GMSB2 | $(94.2 \pm 0.1)\%$ | $(56.4 \pm 0.3)\%$ | $(41.9 \pm 0.3)\%$ |
| GMSB3 | $(94.4 \pm 0.3)\%$ | $(49.8 \pm 0.6)\%$ | $(36.1 \pm 0.6)\%$ |

| Unbiased selection | | | |
|---|---|---|---|
| | hadronic | $2^{nd}$ sampling | $1^{st}$ sampling |
| GMSB1 | $(94.1 \pm 0.2)\%$ | $(93.4 \pm 0.2)\%$ | $(85.7 \pm 0.3)\%$ |
| GMSB2 | $(94.2 \pm 0.1)\%$ | $(92.2 \pm 0.1)\%$ | $(82.5 \pm 0.1)\%$ |
| GMSB3 | $(94.4 \pm 0.3)\%$ | $(92.1 \pm 0.3)\%$ | $(80.7 \pm 0.5)\%$ |

The relative effect of loosening cuts on the background is shown in Table 12, which shows the fraction of jets, from a di-jet Monte Carlo data sample, that are reconstructed as photons, that also pass the two different photon selections.

Table 12: The fraction of jets reconstructed as photons and passing all photon criteria. The hadronic, second sampling and first sampling cuts are applied sequentially.

| | Hadronic | $2^{nd}$ sampling | $1^{st}$ sampling |
|---|---|---|---|
| Default photon selection | $(3.4 \pm 0.1)\%$ | $(0.57 \pm 0.06)\%$ | $(0.19 \pm 0.03)\%$ |
| Unbiased selection | $(3.4 \pm 0.1)\%$ | $(2.7 \pm 0.1)\%$ | $(0.70 \pm 0.07)\%$ |

## 4.2 Projected impact-parameter method for neutralino mean lifetime measurement

The distribution of the photon's projected longitudinal impact parameter $Z'$ arising from GMSB signal events can be used to estimate the mean neutralino lifetime.

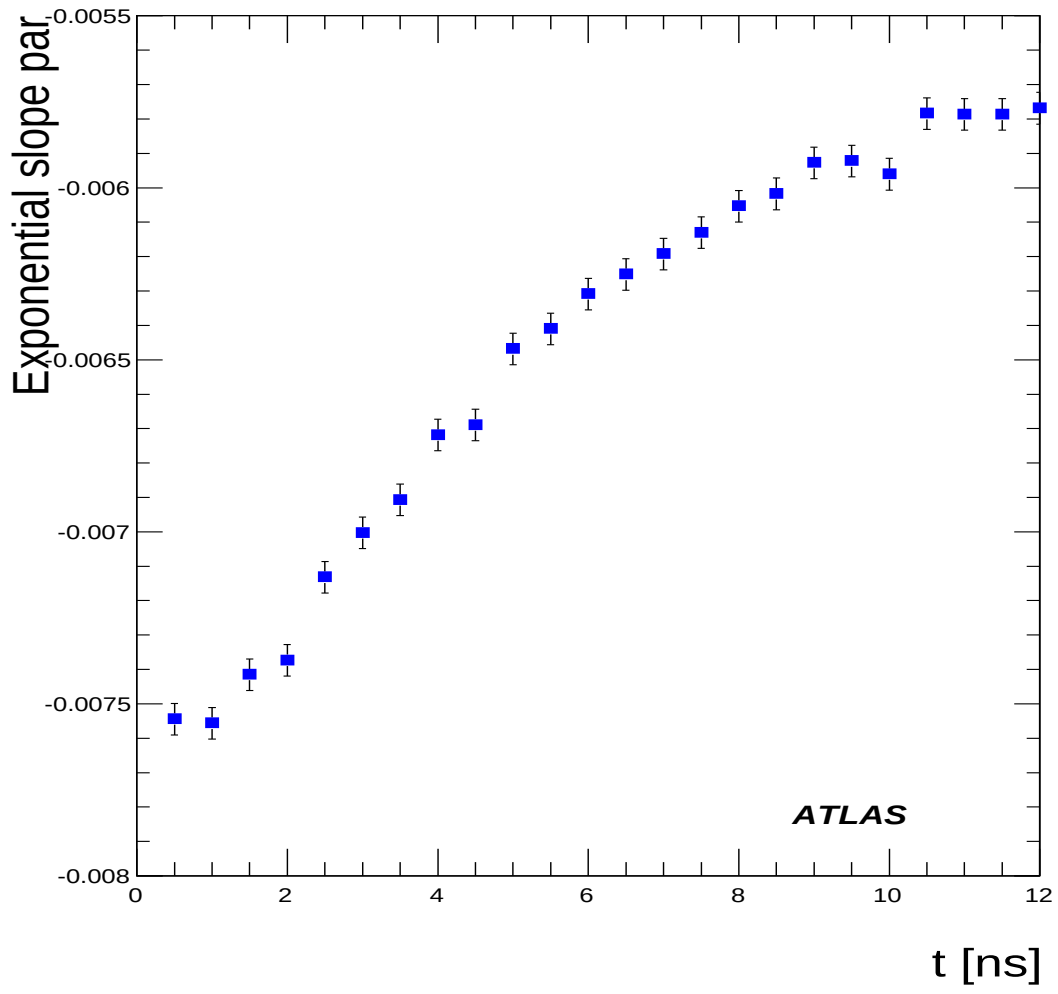


Figure 11: Fitted slope parameters of the projected intersection distributions versus mean neutralino lifetime from the custom-built Monte Carlo simulation for an integrated luminosity of 30 $fb^{-1}$.

A distribution of the intersection points of the $z$-axis with the projected photon path is then created for all reconstructed photons. Each intersection point is corrected for the vertex displacement. An exponential function is fitted to the intersection distribution in order to extract the slope parameter which is sensitive to the mean neutralino lifetime. The range for the fit was chosen to be 50 to 500 mm, to remove any possible vertex effects, and to ensure that the decay occured within the volume of the inner-detector.

Plotting the resulting slope parameters versus the mean neutralino lifetime reveals a clear correlation (Figure 11) between the slope parameter and the neutralino lifetime. These results were obtained using a custom-built Monte Carlo program which provides a detailed parameterisation of the response of the transition radiation tracker and the electromagnetic calorimeter. It is envisaged that a calibration curve such as this, created using full simulation, could be used to determine the mean neutralino lifetime from a measurement of the slope of the $Z'$ distribution. In reality this slope will also be a function of the $\beta$ of the neutralino.

For the GMSB2 data set, a slope of $-4.35(6) \times 10^{-3}$ was measured and for the GMSB3 data set, a slope of $-3.8(2) \times 10^{-3}$ was measured. These slopes were obtained before any photon identification cuts were applied. To estimate the effect of bias due to the reconstruction efficiency, a weight ($= \frac{1}{R_{\text{eff}}(\Delta\eta)}$) is applied to the events, see Section 4.1.2. By comparing the effect of applying this correction on the resultant $Z'$ slope, a $-5 \times 10^{-5}$ systematic error was obtained. Table 13 shows the values obtained when different photon identification cuts are used.

The results from the GMSB2 and GMSB3 samples shown in Table 13 demonstrate very clearly how the slope of the $Z'$ distribution can be affected by photon identification cuts which are based on width measurements of the electromagnetic shower.

### 4.3 Calorimeter timing

A comparison has been made of the timing of the electromagnetic shower in the calorimeter, compared to the lifetime of the generated neutralino (in its rest frame). A Gaussian is fitted to this distribution for different bins of true lifetime, and the resultant mean cluster-time per generated neutralino mean lifetime

Table 13: The slope of the projected-impact-parameter ($Z'$) distributions of the photon, fitted from 50 to 500 mm and measured in two different fully simulated Monte Carlo samples with different photon identification cuts.

| GMSB2: Generated mean lifetime of 3.17 ns | | | |
|---|---|---|---|
| dataset | hadronic | 2[nd] Sampling | 1[st] Sampling |
| default photon selection | $-4.35(6)\times10^{-3}$ | $-7.00(9)\times10^{-3}$ | $-8.3(1)\times10^{-3}$ |
| unbiased selection | $-4.35(6)\times10^{-3}$ | $-4.39(6)\times10^{-3}$ | $-4.54(7)\times10^{-3}$ |
| GMSB3: Generated mean lifetime of 10.7ns | | | |
| dataset | hadronic | 2[nd] Sampling | 1[st] Sampling |
| default photon selection | $-3.8(2)\times10^{-3}$ | $-6.6(2)\times10^{-3}$ | $-7.7(3)\times10^{-3}$ |
| unbiased selection | $-3.8(2)\times10^{-3}$ | $-3.9(2)\times10^{-3}$ | $-4.1(2)\times10^{-3}$ |

Table 14: The mean lifetimes, measured using the calibrated calorimeter time, in two different full simulation Monte Carlo samples with different photon identification cuts. Also shown is the generated mean lifetime of the samples used.

| GMSB2: Generated mean lifetime of 3.17ns | | | |
|---|---|---|---|
| dataset | hadronic | 2[nd] Sampling | 1[st] Sampling |
| default photon selection | $2.9\pm0.2$ ns | $1.1\pm0.07$ ns | $1.33\pm0.05$ ns |
| unbiased selection | $2.9\pm0.2$ ns | $2.9\pm0.2$ ns | $3.0\pm0.2$ ns |
| GMSB3: Generated mean lifetime of 10.7ns | | | |
| dataset | hadronic | 2[nd] Sampling | 1[st] Sampling |
| default photon selection | $9\pm4$ ns | $3.4\pm0.7$ ns | $2.9\pm0.6$ ns |
| unbiased selection | $9\pm4$ ns | $8\pm3$ ns | $19\pm19$ ns |

is plotted in Figure 12. This fit to this plot is used to calibrate the calorimeter time. It has been shown that this method is robust against photon reconstruction or identification efficiency biases. This is because it is independent of the angle of incidence of the photon on the calorimeter. The arrival time of the photon at the calorimeter is a function of the $\beta = v/c$ of the neutralino as well as its lifetime.

Using this calibrated calorimeter time, the neutralino lifetime is plotted for each photon. This distribution has the expected exponential shape modified by acceptance and resolution effects. In order to remove these effects, an exponential is fitted between 0.2 and 1 ns. The mean lifetime of the sample ($\tau$) was calculated from $\tau = \frac{1}{slope}$ where *slope* is the slope of the exponential fitted.

To calculate the effect of bias due to the reconstruction efficiency, a weight ($= \frac{1}{R_{\mathrm{eff}}(\Delta\eta)}$) was applied to the events (see Section 4.1.2). The systematic error on the lifetime determination due to uncertainties on the reconstruction efficiency was determined to be 2%.

To study the effect of the photon identification on this method, the mean lifetime deduced from the slope of the exponential is measured after different identification cuts are applied to the photon sample. These mean lifetimes are shown in Table 14. The large errors on the calculated lifetimes from the GMSB3 sample are due to lack of statistics available for the fit, over the limited range.

In order to obtain estimates for the systematic uncertainty due to the predicted $\beta$ distribution of the GMSB sample (or an error in the timing calibration), the calibration curve (Figure 12) was scaled up and down by 5% corresponding to the difference in mean $\beta$ value between the GMSB2 and GMSB3 samples. The effect on the measured mean lifetime determination corresponds to a systematic error of 0.4(2) ns

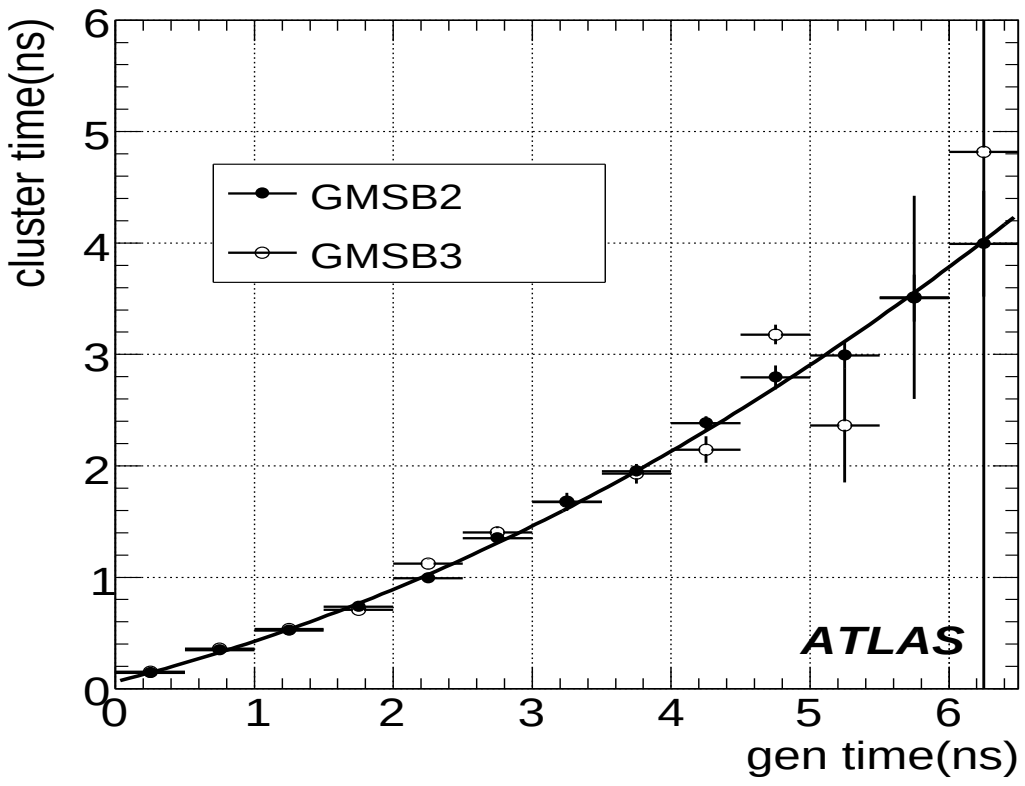


Figure 12: The measured cluster time as a function of the generated neutralino lifetime, for individual neutralinos in the Monte Carlo samples.

for the GMSB2(3) samples.

In order to obtain estimates for the systematic uncertainty in the mean lifetime determination from the choice of fitting region or a global shift in the timing calibration, the fit range of the exponetial distributions was shifted by 100 ps. A systematic error of 1(10) ns was obtained for the GMSB2(3) samples.

The shape of the observed neutralino lifetime distribution is strongly affected by geometric acceptance and event kinematics. To obtain a lifetime measurement unaffected by these issues, a very limited fitting range has been used. A more accurate fit could be achieved by fitting the entire distribution. A full acceptance correction, including model dependent effects, would then be required to relate this distribution to the mean neutralino lifetime. This work, beyond the scope of this publication, should produce a more accurate measurement of the mean neutralino lifetime.

## 4.4 Conclusion of the neutralino lifetime determination

Two independent methods for determining the mean neutralino lifetime have been discussed. The two methods are independent with different issues. For the $Z'$ method of Section 4.2, a study using a simple custom-built Monte Carlo program shows that there is a good correlation between the $Z'$ parameter and the mean neutralino lifetime. However, full simulation of a range of lifetime samples will be required to get the correct parameterisation of the relationship. The calorimeter timing study (Section 4.3), shows there is a good correlation between the calorimeter time and lifetime of the associated neutralino. The largest errors from this studies are due to the limited statistics in the range of the measured neutralino lifetime distribution, used for the exponential fit.

In both methods, one has to take care that the biases introduced by the reconstruction and identification of the photons are both measured and reduced, in order to prevent a distortion of the lifetime distribution.

Both of these methods are signal dependent and rely on simulation calibration, either to produce the $Z'$ calibration curve, or the timing calibration. This is because both the $Z'$ value and the arrival time of the photon in the calorimeter are functions of the $\beta$ of the neutralino as well as its lifetime. One can assume a distribution of $\beta$ from Monte Carlo simulation (as has been done here), but constraints on the both parameters can in principle be achieved by combining the methods. It is proposed that, if non-pointing photons are observed, a multivariate analysis method could be used to combine information from the calorimeter timing and $Z'$ together with information from the primary vertex and cluster positions and energy to place model-independent constraints on the $\beta$ and lifetime of the parent of the non-pointing photon.

# 5 Trigger and reconstruction for searches of long-lived heavy particles

Heavy, charged, long-lived particles are predicted in many models of physics beyond the Standard Model. One example is GMSB where for high $\tan\beta$ the $\tilde{\ell}$ is the NLSP which couples weakly to the gravitino. The signal is a heavy long-lived charged particle with velocity significantly smaller than the speed of light, $\beta < 1$. The momentum (and therefore $\beta$) spectrum of these particles is model dependent. Those which have $\beta$ close to unity are indistinguishable from ordinary muons. Those with $\beta$ significantly lower than 1 could be identified and their mass determined.

In ATLAS event fragments from different parts of the detector are assiged to a particular bunch crossing (BC) using the BC identifier (BCID). The usual assumption is that the particles traverse the detector at nearly at the speed of light ($\beta \approx 1$). Hits from a slower particle may be lost during data collection, or may be marked with the wrong BCID. The implications of low particle speed in the ATLAS trigger and data acquisition design are considered below.

This note does not address the case where the decay length of heavy charged NLSP is such that a significant number of particles will decay inside the tracking volume.

## 5.1 Datasets used

This analysis is based on a data sample of 10,000 events from the CSC production generated with the characteristics of GMSB point 5: $\Lambda = 30$ TeV , $M_m = 250$ TeV, $N_5 = 3$, $\tan\beta = 5$, $\mathrm{sgn}(\mu) = +$, $C_{\mathrm{grav}} = 5000$. At this point the squarks and gluinos have masses around 700 GeV, the neutralino has a mass of 114 GeV and the $\tilde{\tau}$ and $\tilde{\ell}$ have masses of 102 and 100 GeV respectively. The cross-section for this point is 23 pb and the $\tilde{\tau}$ , $\tilde{e}$ and $\tilde{\mu}$ are co-NLSPs and are produced in the decay $\tilde{\chi}^0 \to \tilde{\ell}^{\pm}\ell^{\mp}$. Because of the small mass difference between the neutralino and the slepton, the $\tilde{\ell}$ and lepton are nearly collinear. The $p_{\mathrm{T}}$ and $\beta$ spectra of the sleptons and accompanying leptons are shown in Figure 13.

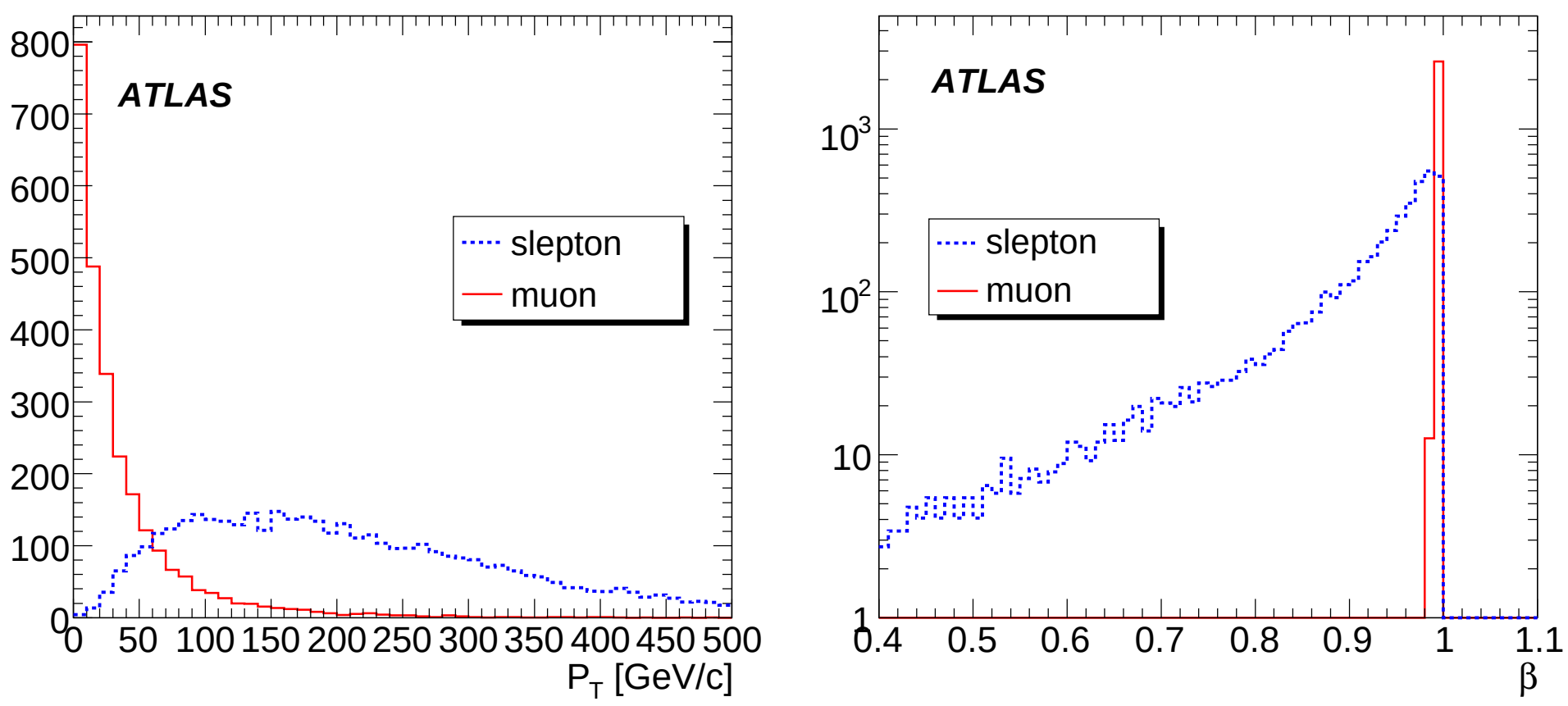


Figure 13: Transverse momentum and velocity spectra for sleptons and accompanying leptons from the GMSB5 sample.

GMSB5 is a single benchmark point and cannot be taken to represent all the possibilities of long-lived particle production. Some issues that impact our ability to discover long-lived new particles, if they exist, depend on the mass and $\beta$ spectrum of these particles. In order to make our study less model dependent we also used for this study additional samples of events containing a single $\tilde{\tau}$ each, generated at different $\beta$ with a uniform $\eta$ distribution between $\eta = -3$ and $\eta = +3$.

Split-SUSY events containing long-lived gluinos with masses of 300 GeV and 1000 GeV were also used to assess the efficiency of the slow particle trigger, as discussed below. The generation and simula-

tion of these Split-SUSY events is described in Section 6.

For the background study we used single muon events from the CSC production. They were produced at constant $p_T$ with a uniform $\eta$ distribution, like the single $\tilde{\tau}$ samples. We used cross-section estimates of 1000 pb (200 pb) for muons with $p_{\mathrm{T}} > 40$ GeV ($> 100$ GeV) inside the ATLAS acceptance of $|\eta| < 2.5$ [45].

The simulation of a long-lived heavy sleptons in the ATLAS detector required a special patch to GEANT4 [46].

## 5.2 Trigger and DAQ issues

When trying to identify slow particles [47, 48] one must pay special attention to the dimensions of ATLAS. Since the detector extends over 20m in length from the interaction point to each detector side and the bunch crossing period is 25ns, this means that particles from three separate bunch crossings can co-exist in the detector at the same time.

As described above, the matching of event fragments from different subdetectors is achieved using the BCID. This is calibrated so that particles originating together at the interaction point and traveling at the speed of light will have the same BCID assigned to them in all detector elements.

When $\beta$ is sufficiently small, the particle will take longer to reach the detectors (especially those far from the interaction point) and hits may be assigned a wrong BCID and thus not be read-out. Figure 14 shows the efficiency with which slepton hits in the muon trigger chambers are associated to the correct bunch crossing as a function of $\beta$. This figure was produced using the single $\tilde{\tau}$ events described above. It can be seen that efficiency drops sharply below $\beta = 0.8(0.7)$ in the endcap (barrel). In order to find particles with $\beta < 0.7$ ($\beta < 0.6$) in the endcap (barrel), ATLAS must collect hits from the following bunch crossing (BC). Fortunately the MDT chambers collect data over a 700 ns interval, and thus hits from many BCs will be present. The RPC and TGC data acquisition can be set up to read out data from $\pm 7$ and $\pm 1$ BCs around the triggered BC respectively. The option to read out the information about extra BC, which was originally intended for debugging, must be switched on during routine ATLAS operation if we want to increase our efficiency for long-lived charged particles.

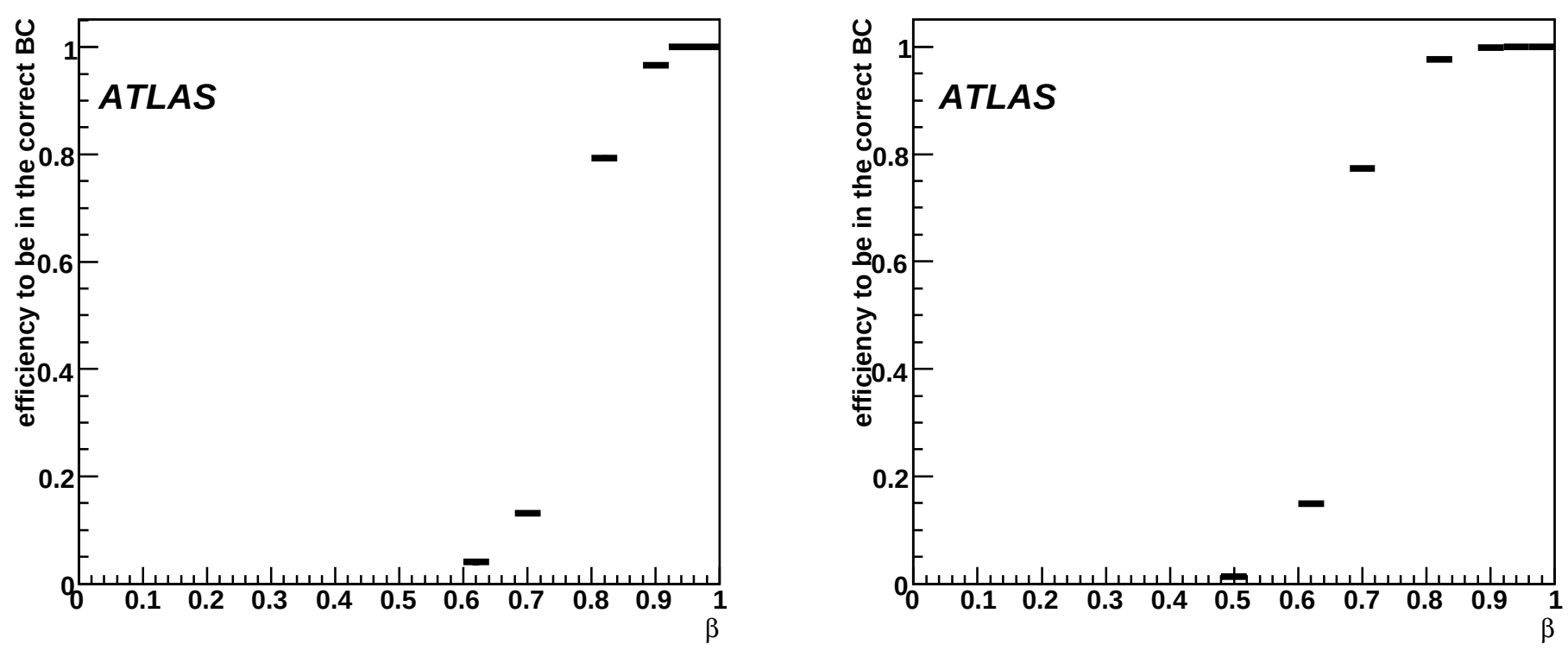


Figure 14: The efficiency, as a function of $\beta$, for all slepton hits in the muon trigger chambers to be included in the same BC with fast particles, for the barrel (right)and the endcap (left).

The hits from a slow particle may fall outside the correct BC, either for all trigger stations (e.g. if the particle is very slow and the trigger was produced by another feature in the event) or hits in the low-$p_{\mathrm{T}}$ stations in the barrel may arrive in the correct BC, but the hits in the outer station may be late. In such a case, even if the $p_{\mathrm{T}}$ of the particle was high, it would produce a low $p_{\mathrm{T}}$ trigger.

Table 15: The L1 trigger efficiencies for GMSB5 simulated events for items from the trigger menu for $\mathscr{L} = 10^{33}\text{cm}^{-}2\text{s}^{-}1$. The description of the trigger items is in [42]

| Trigger label | Description & Efficiency in GMSB5 |
|---|---|
| MU40 | 95% |
| xE200 | 63% |
| EM25i | 46% |

In GMSB5 the two sleptons are produced with different $\beta$'s, and since most of the sleptons have high $\beta$, at least one of the two produced sleptons will have $\beta > 0.7$ in 99% of the events. The level-1 [49] muon trigger efficiency in the correct BC is very high, either from a slepton with high $\beta$, or from one of the accompanying leptons. Table 15 shows the level-1 trigger efficiencies for GMSB5 events, based on the Level-1 thresholds defined in the standard ATLAS menu for a luminosity of $10^{33}$.

Nevertheless, a low $\beta$ slepton, one with good potential for a mass measurement, could arrive to the muon spectrometer, or more likely the outer muon station, in the next BC. In such a case the slow slepton will not be found by the trigger, or be identified as a low $p_{\mathrm{T}}$ muon. In order to identify such a slow particle muon trigger chamber data from the next BC has to be collected.

In Split SUSY, the gluinos are produced directly, and there are few other features in the event. Therefore the *R* hadrons themselves must trigger the event. Since both of the *R* hadrons may be slow, the muon trigger may correspond to the wrong BC for the central parts of the detector. As a result, other event information such as that from the inner detector may be lost in the previous BC. This problem may be solved by also collecting data from the previous BC in the inner detector, but the feasibility of doing this, from the point of view of increased data volume, has not been investigated in this work. Additional data from the calorimeter is not required in order to find long lived heavy charged particles.

## 5.3 A L2 trigger for heavy sleptons

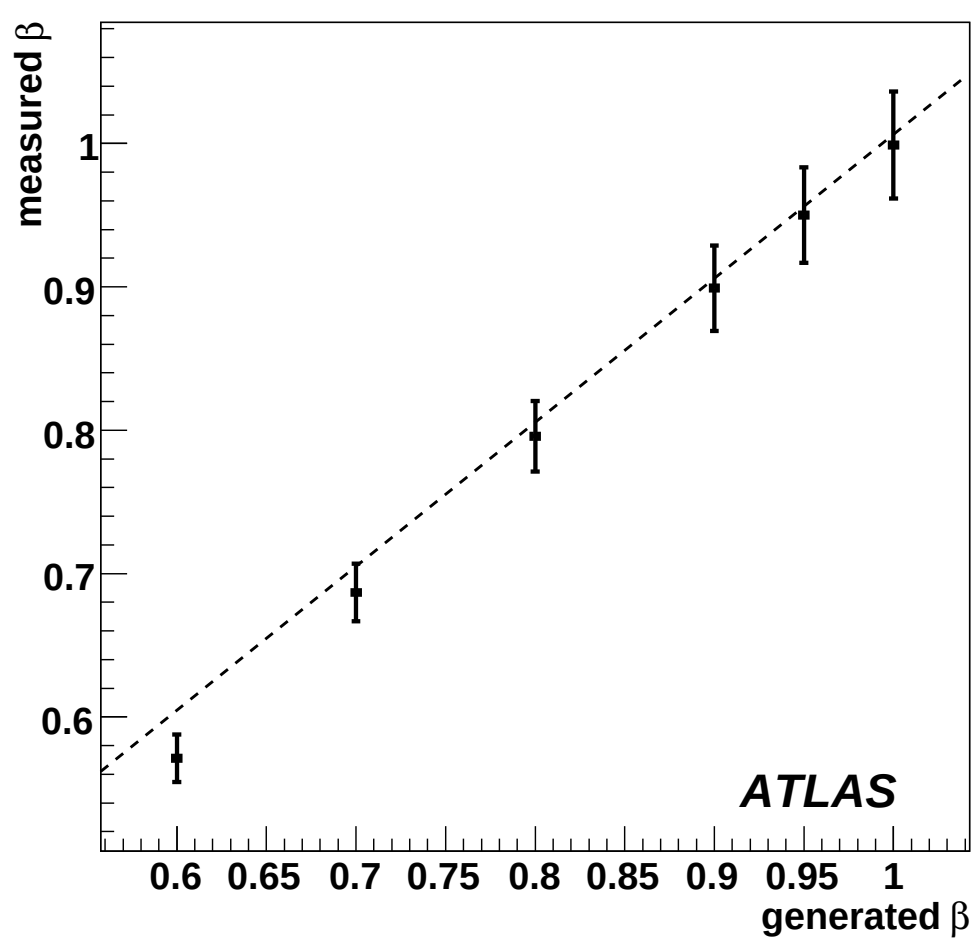


Figure 15: The $\beta$ measured for sleptons in the barrel at L2 for different values of true $\beta$. The error bar represents the fitted sigma of the measured $\beta$ distribution.

At L2 [50], algorithms are activated based on the Region of Interest (RoI) identified at L1. Each algorithm has a reconstruction stage which processes the data from the relevant parts of the subdetectors,

and a hypothesis stage which makes the decision to keep or reject the reconstructed feature. L2 and the Event Filter run in ”trigger chains” which define the order in which algorithms are called and the input and output of each processing stage.

The role of the L2 muon trigger is to confirm muon candidates flagged by the L1 and to give more precise physics quantities of the muon candidate.

The L2 muon selection is performed in two stages. The first stage is performed by the muFast algorithm [51], which starts from a L1 muon RoI and reconstructs the muon in the spectrometer, giving a new $p_{\mathrm{T}}$ estimate. The hypothesis cuts on the estimated $p_{\mathrm{T}}$ are set so that 90% of the muons with $p_{\mathrm{T}}$ above the nominal threshold would pass the selection. The resulting trigger element is then passed to the next algorithm.

Track finding in the inner-detector is performed based on the region of interest found by muFast. The muFast candidate and inner-detector tracks then pass to the next algorithm, muComb, which matches an inner-detector track to the muon spectrometer track and refines the $p_{\mathrm{T}}$ estimate [50].

In the muon barrel, the excellent time resolution (about 3 ns) of the RPC allows measurements of the time of flight (TOF). A method for finding the slepton and measuring its mass at L2 has been developed. We will show that, in the barrel, a slow particle may already be selected effectively at L2. Figure 15 shows the mean value and error of the $\beta$ reconstructed at the L2 as a function of generated $\beta$. This figure was produced using the single $\tilde{\tau}$ events described above.

A selection based only on the $\beta$ measurement would leave too many muons in the sample. At the hypothesis stage, we select using $p_{\mathrm{T}}$(candidate)$>$ 40 GeV, $\beta$(candidate)$<$ 0.97 and $m$(candidate)$>$ 40 GeV. The mass is calculated from the measured $\beta$ and the candidate's $p_T$ and $\eta$ estimated by muFast. Figure 16 show the mass distribution of signal and background resulting from the selection for an integrated luminosity of 500 pb$^{-1}$. It can be seen that the signal to background ratio is already good at the L2.

At a luminosity of $10^{33}$, a slow particle trigger without further trigger selection would produce a rate lower than 1 Hz coming from muons. Further refinement of the slow particle selection using subsequent muon trigger stages can reduce this rate to 0.2 Hz. The measurement of $\beta$ for high $p_{\mathrm{T}}$ muon candidates is

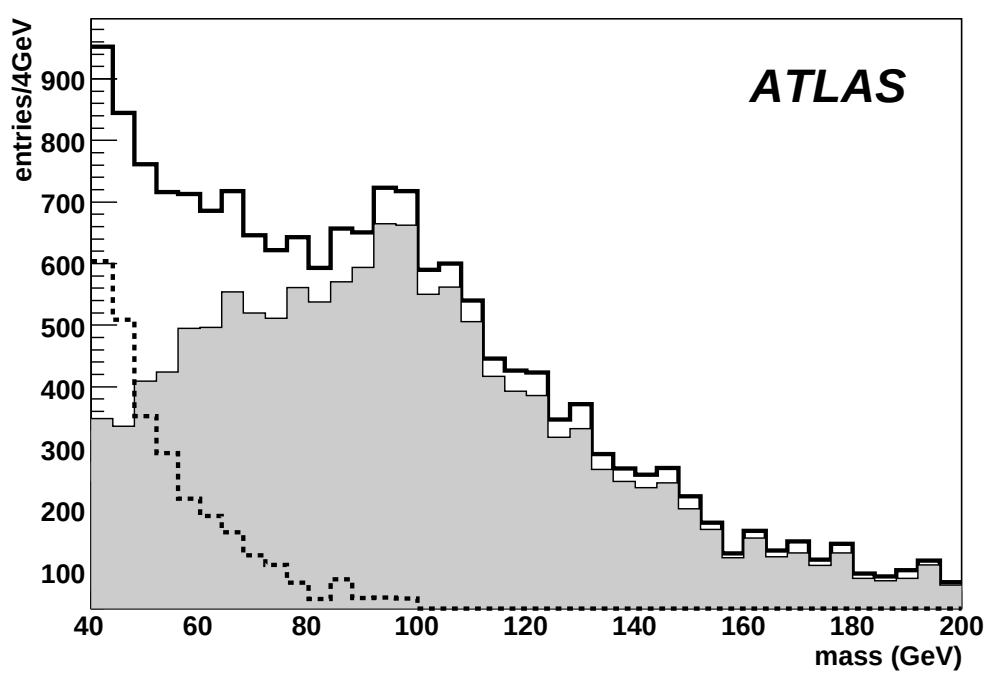


Figure 16: Mass distribution of signal and background resulting from the L2 selection for an integrated luminosity of 500 pb$^{-1}$. The shaded area is the GMSB5 signal, the dashed line is the muon background, and the full line is the sum.

already part of the standard ATLAS L2 program MuFast, and a program to make the selection described above is part of the ATLAS L2 trigger.

Measuring $\beta$ in the muon spectrometer at the second level trigger could be particularly useful for $R$ hadrons, which have a L2 trigger efficiency of about 50% for a mass of 300 GeV. The efficiency loss comes mainly from events where the $R$ hadron is neutral in the inner detector (and undergoes charge exchange before the muon spectrometer), which then fail the second stage of the L2 trigger requiring matching between the candidate found in the muon spectrometer and a track in the inner detector. An-

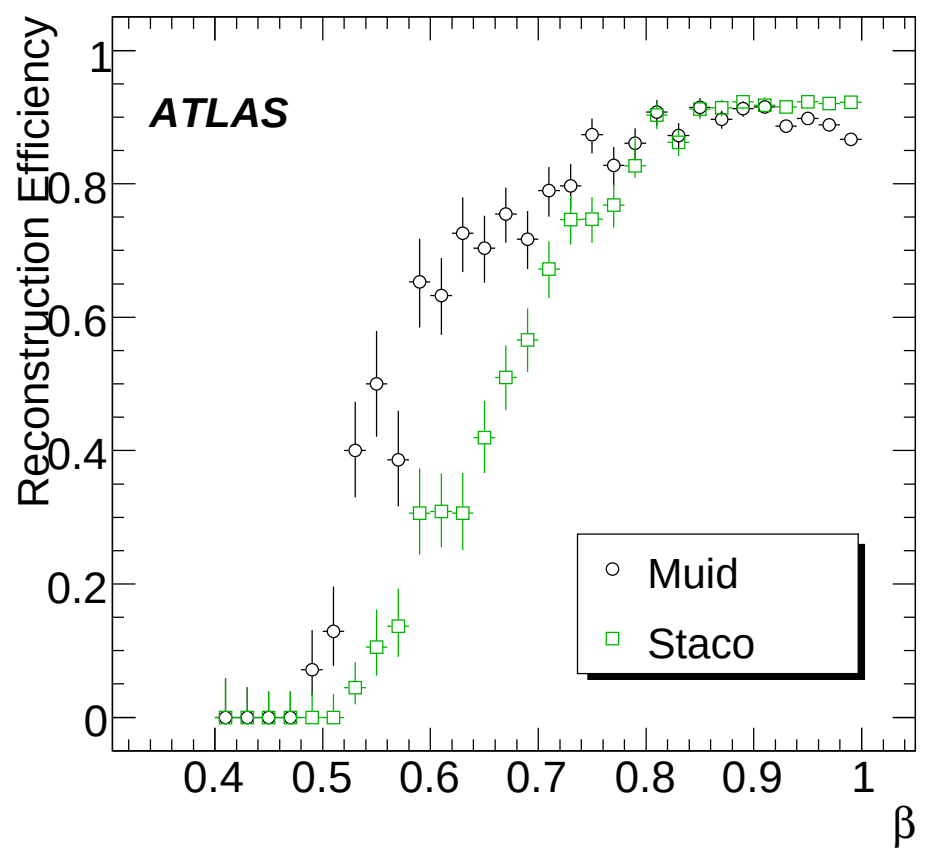


Figure 17: The efficiency to reconstruct sleptons as muons as a function of $\beta$ for two ATLAS muon reconstruction packages.

other source of loss (probably less well simulated) is the assignment of inner detector hits from the charged $R$ hadron to the incorrect bunch crossing.

If the $R$ hadron is identified in the muon spectrometer as a slow particle candidate, it could be accepted without the requirement of a matching inner detector track. The slow particle selection at L2 results in an efficiency of 92% to select events with $R$ hadrons in the barrel. The corresponding muon rate is expected to be completely negligible since all muons have an inner detector track.

The limitations of the L2 slow particle trigger should be noted. Firstly, this selection is performed only in the barrel of the muon spectrometer, where the RPCs are the trigger chambers. The timing information from the endcap is in BC granularity and cannot be used to measure $\beta$. Secondly, a slow particle which does not produce a RoI in the correct BC will not cause the muFast algorithm to be called, and the selection will not be performed. Therefore, slow particles in the endcap, as well as very slow sleptons in events triggered by other objects such as high $p_T$ leptons, can only be identified offline.

The final trigger decision in ATLAS is made in the event filter, which uses algorithms adapted from the offline reconstruction. As will be shown in the next section, the standard muon reconstruction is not efficient for slow particles; therefore many would be rejected at this stage. The combined trigger efficiency of L2 and the event filter for sleptons with velocity $\beta = 0.6$ is 39%.

To avoid the efficiency loss at the event filter, a specific reconstruction algorithm for slow particles, like the one we describe below for reconstruction, is being implemented at the event filter.

Slow particle candidates found at L2 that do not have a matching inner-detector track (such as charge flipped $R$-hadron candidates) should be accepted without further event filter selection. This will increase the combined trigger efficiency of the L2 and the event filter for $R$ hadrons with a mass of 300 GeV from 25% to 92% in the barrel. The muon rate for this selection is completely negligible since all muons have an inner-detector track, but the effect of cavern background on this selection has not been studied yet.

## 5.4 Reconstruction of heavy sleptons with the current muon reconstruction packages

In the standard ATLAS reconstruction [52], stable sleptons are expected to be reconstructed as muons. The efficiency to reconstruct slow sleptons is compromised due to the following issues: muons are reconstructed by forming track segments in the three stations of the Muon Spectrometer. Segments in the $\phi$ direction are found using the trigger chamber data. The data may not be collected if the hits from a slepton are in the next bunch crossing instead of the collision BC. The lack of a $\phi$ segment hinders the reconstruction of the precision segment in $\eta$ using the MDT data. Furthermore, the radii of MDT hits

are distorted by late arrival of the slepton and may not fit well to a segment.

The efficiency of reconstructing the slepton as a muon depends on the reconstruction technique. Figure 17 shows the reconstruction efficiency for sleptons with the two ATLAS combined reconstruction packages, Staco and Muid which are based on the muon standalone packages MuonBoy and MOORE respectively. It can be seen that reconstruction efficiency starts dropping sharply for $\beta < 0.8$. This indicates that special reconstruction techniques will be required for the slow particles. This is discussed in the next subsection.

### 5.5 Identifying heavy long-lived particles at reconstruction

Estimating the velocity and mass in the event reconstruction allows us to identify heavy long lived charged particles, as well as to avoid the efficiency losses suffered when reconstructing them as muons. We do this with the MuGirl package [53], which enables us to select candidates also when the segment reconstruction is imperfect. Offline, the velocity can be also determined using the MDTs, the ATLAS precision muon chambers. In the MDT detectors, the hit position is obtained from the particle drift time. The drift distance is calculated assuming the particle passed the chamber with $\beta = 1$, which is wrong for the slow slepton. Minimizing the $\chi^2$ of fit with respect to the time of arrival to the muon detectors yields an estimate of $\beta$ and of the particle mass. This information is combined with estimates of $\beta$ from the muon trigger chambers. Figure 18 shows the $\beta$ resolution and reconstructed mass of sleptons from GMSB5 obtained from the reconstruction with MuGirl.

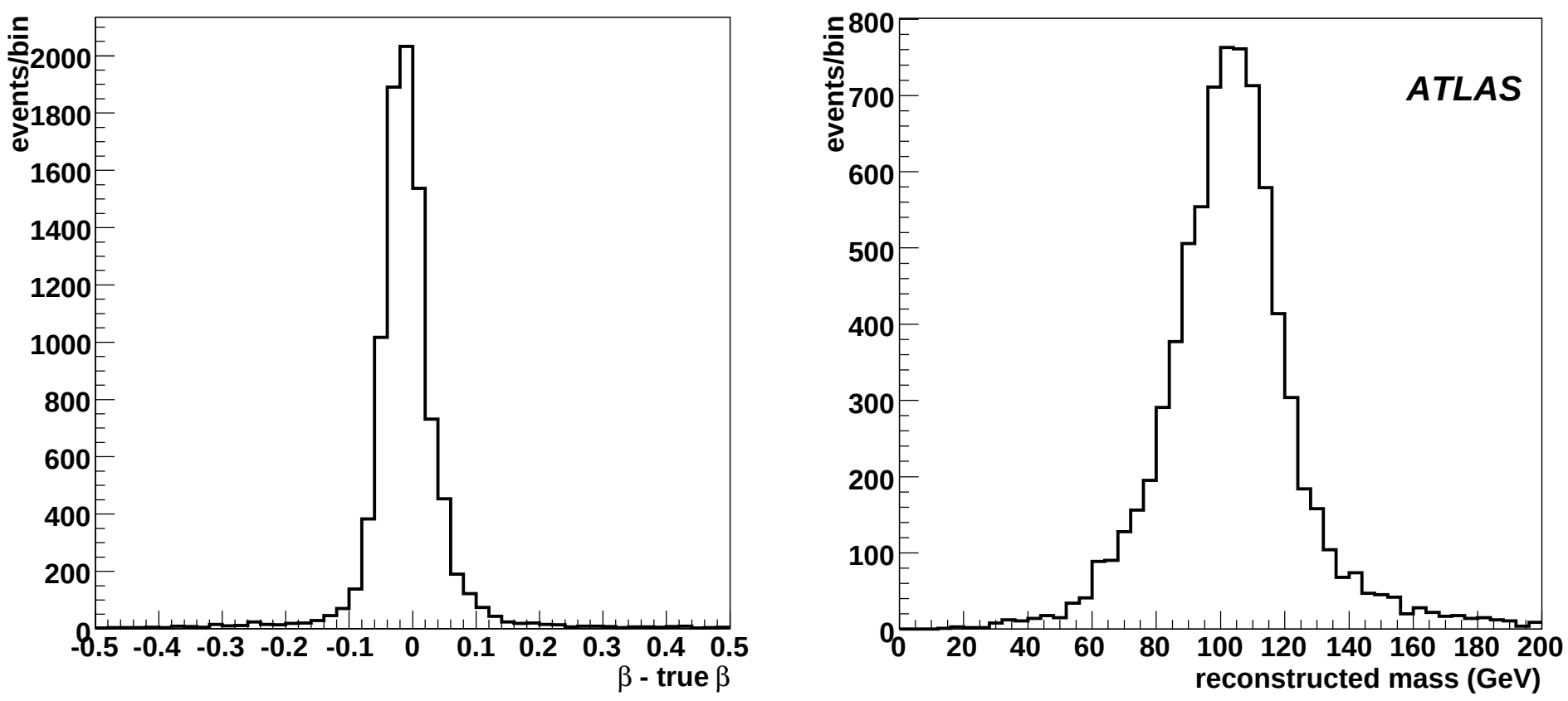


Figure 18: $\beta$ resolution and reconstructed mass for sleptons from the GMSB5 sample.

### 5.6 Conclusion of strategies for the long-lived heavy particle search

Heavy, long-lived charged particles can be discovered in ATLAS, should they exist. However, this cannot be done effectively using the standard ATLAS muon reconstruction tools. Furthermore, much of the discovery must be done before the analysis stage, in the data acquisition, high level trigger and reconstruction. We have added to the standard ATLAS software components which can identify sleptons effectively.

The efficiency to discover slow long-lived charged particles depends on collecting extra data from the bunch crossing following the one in which the interaction occurred. This is most important for data from the muon trigger chambers, where this possibility is included in the data acquisition design.

For the GMSB5 model, discovery would assured with low integrated luminosity, once the MDT and RPC time calibrations are established. The discovery methods are largely independent of the model

characteristics, with discovery for a given integrated luminosity depending primarily on the production cross-section of the slow particles.

# 6 Search strategies for *R* hadrons at ATLAS

## 6.1 Introduction

The presence (or absence) of massive exotic stable hadrons will be an important observable in the search for and quantification of any new physics processes seen at the LHC. Stable exotic coloured particles are predicted in a range of SUSY scenarios (see, for example, Refs. [22–26, 31]). Such particles could be copiously produced at the LHC and sensitivity to particle masses substantially beyond those excluded by earlier collider searches ($\lesssim$200 GeV) [31] could be achieved at ATLAS even with rather modest amounts of integrated luminosity ($\sim 1\text{fb}^{-1}$). This section outlines a strategy for the detection of exotic massive, long-lived hadrons (so-called *R* hadrons) formed from either stable gluinos or stops ($R_{\tilde{g}}$ and $R_{\tilde{t}}$ hadrons, respectively). As described in section 2.1, the $R_{\tilde{g}}$ hadrons ($R_{\tilde{t}}$ hadrons) are considered in the context of a Split-SUSY (stop NLSP/gravitino LSP) scenario. Although this work is performed in the framework of SUSY, the techniques presented here may be used in generic searches for stable heavy exotic hadrons. As in the heavy lepton studies presented in section 4, this work relies on a signature of high-$p_T$ muon-like track, although the distributions presented in this section provide a means of discriminating between lepton and hadron hypotheses for any observed stable massive particle.

This section is organised as follows. First a description is given of the simulation of *R* hadrons at ATLAS, including both the event generation and scattering of *R* hadrons in matter. Final state observables associated with $R_{\tilde{g}}$ and $R_{\tilde{t}}$ hadrons are then presented and it is shown how it may be possible to experimentally distinguish between these two types of particles should a discovery be made of stable massive exotic hadrons. Finally, the discovery potential for $R_{\tilde{g}}$ and $R_{\tilde{t}}$ hadrons is presented.

## 6.2 Physics and detector simulation

### 6.2.1 Event generation

The leading-order event generator PYTHIA [54] was used to produce samples of pair-produced gluino and stop-antistop events for a range of gluino and stop masses, as summarised in Table 16. Production mechanisms of stops and gluinos are illustrated in Figure 19, which shows leading-order Feynman diagrams. The effects of higher orders, which are important for the jet selection used later in section 6.3.3, are computed within PYTHIA using the parton shower technique.

The processes studied were selected such that they provide conservative estimates of likely rates at the LHC, which depend principally on the mass of the heavy object under study and not other free SUSY parameters.

For the gluino generation a Split-SUSY scenario was used in which squarks masses were set to 4 TeV. To ensure the results presented here are as model-independent as possible, only the PYTHIA sub-process $gg \to \tilde{g}\tilde{g}$ was considered whilst neglecting the quark annihilation process $q\bar{q} \to \tilde{g}\tilde{g}$, which is sensitive to the squark mass. The former process is anyway the dominant production channel for gluino masses up to $\sim 1.5$ TeV.

For the stable stop sample, the diagonal production of pairs of the lighter stop state ($\tilde{t_1}$) were assumed and the following sub-processes modelled: $gg \to \tilde{t}_1\tilde{t}_1$ $q\bar{q} \to \tilde{t}_1\tilde{t}_1$. All sparticle masses except that of the $\tilde{t_1}$ quark were set to 4 TeV although, at leading-order, the masses of other sparticles are not relevant for the cross-section calculations [55].

Table 16: Expected number of gluino and stop pair-production events for 1fb$^{-1}$ of integrated luminosity and the equivalent luminosity of signal samples.

| sparticle | **Mass (GeV)** | **Events/fb**$^{-1}$ | $\mathscr{L}$ **(fb**$^{-1}$**)** |
|---|---|---|---|
| $\tilde{g}$ | 300 | $2.69\times10^5$ | $3.72\times10^{-2}$ |
| $\tilde{g}$ | 600 | $4.84\times10^3$ | 2.07 |
| $\tilde{g}$ | 1000 | 138 | 72.5 |
| $\tilde{g}$ | 1300 | 16.4 | 610 |
| $\tilde{g}$ | 1600 | 2.12 | $4.72\times10^3$ |
| $\tilde{g}$ | 2000 | 0.230 | $4.35\times10^4$ |
| $\tilde{t}$ | 300 | $7.82\times10^3$ | 1.12 |
| $\tilde{t}$ | 600 | $1.76\times10^2$ | 35.2 |
| $\tilde{t}$ | 1000 | 6.4 | $1.5\times10^3$ |

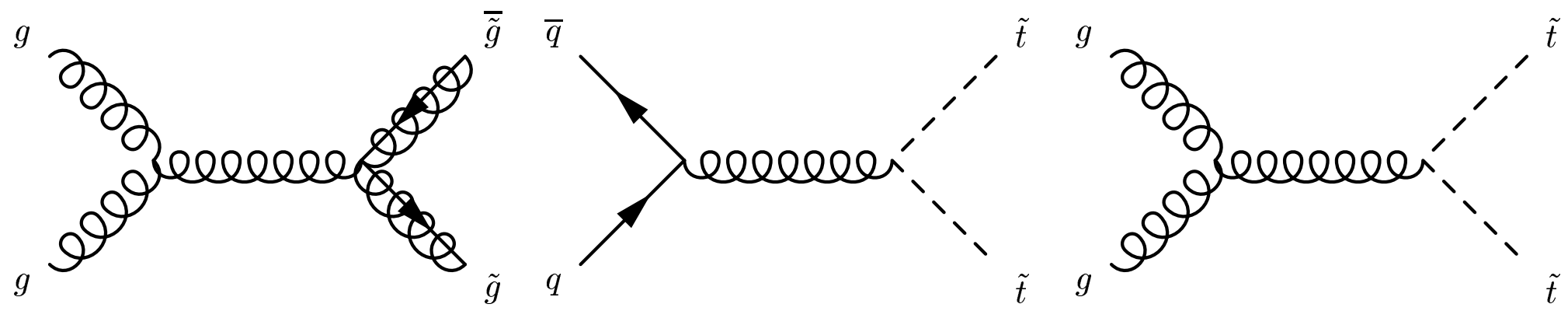


Figure 19: Selection of leading-order processes illustrating the production of gluino and stop particles.

String fragmentation [56] was used to model the momenta of the $R$ hadrons. A Peterson fragmentation function [57],

$$D(z) \propto \frac{1}{z}\left(1-\frac{1}{z}-\frac{\varepsilon_{\tilde{q}\tilde{g}}}{1-z}\right)^{-2}, \tag{2}$$

for which the $\varepsilon$ parameter for the heavy coloured object ($\varepsilon_{\tilde{q}\tilde{g}}$) has a value extrapolated from its value for $b$-quarks ($\varepsilon_{\tilde{q}\tilde{g}}/\varepsilon_b = m_b^2/m_{\tilde{q}\tilde{g}}^2$) [58] was used to model the momentum distribution of the $R$ hadron.

Following hadronisation, and based on calculations of the mass hierarchy of the lowest-lying $R$-hadron states [59], around 55% (40%) of the stable $R_{\tilde{g}}$ ($R_{\tilde{t}}$) hadrons are predicted to have zero electric charge. This difference arises principally due to the possible existence of gluino-gluon states, for which the production probability is set to 10% here, and which are treated as mesons when propagating through matter (Section 6.2.2).

To complement the signal samples, various background samples were used, each corresponding to an integrated luminosity of at least $\sim 1$ fb$^{-1}$. The generated and reconstructed number of events, and the equivalent luminosity of each sample is given in Table 17. As the simulated trigger used for this analysis requires a hard muon-like track, only events which could give rise to a high $p_T$-muon ($p_T > 150$ GeV) were simulated. The following processes were considered. Leading-order 2-to-2 QCD processes, which include all quark flavours except top, and which differ in the values of the internal matrix element cut-offs ($\hat{p}_T$) were generated with PYTHIA, the predictions of which are denoted QCD when discussed in section 6.3.2. In addition, backgrounds arising from diboson (HERWIG [40]) and single boson (PYTHIA) production, denoted electroweak, were produced, again with matrix element cut-offs in order to produce hard muons. A sample of $t\bar{t}$ pair-production events, termed top, was also prepared using the MC@NLO [60] program.

Table 17: Background samples used in this work. The number of generated and reconstructed events are shown along with the equivalent integrated luminosity.Event weighting accounts for the non-integer number of generated and reconstructed $t\bar{t}$ events.

| Sample | Dataset ID | Gen. Events | Rec. Events | $\mathscr{L}(\mathrm{fb}^{-1})$ |
|---|---|---|---|---|
| **QCD**: (PYTHIA) | | | | |
| (140 GeV $< \hat{p}_T <$ 280GeV) | 5013 | $3.125 \times 10^8$ | 2572 | 0.98 |
| (280GeV $< \hat{p}_T <$ 560 GeV) | 5014 | $2.5 \times 10^7$ | 4800 | 1.12 |
| (560GeV $< \hat{p}_T <$ 1120 GeV) | 5015 | $3.5 \times 10^5$ | 738 | 1.01 |
| (1120GeV $< \hat{p}_T <$ 2240GeV) | 5016 | $5 \times 10^4$ | 241 | 9.46 |
| (2240GeV $< \hat{p}_T$ | 5017 | $1 \times 10^4$ | 42 | 442.29 |
| **Electroweak** | | | | |
| ZZ (HERWIG) | 5985 | $2.5 \times 10^4$ | 53 | 9.82 |
| WW (HERWIG) | 5986 | $2 \times 10^4$ | 50 | 1.21 |
| WZ (HERWIG) | 5987 | $1.5 \times 10^4$ | 29 | 2.32 |
| $Z \to \mu\mu$ (PYTHIA) | 5145 | $1.3 \times 10^4$ | 600 | 1.29 |
| $Z \to \tau\tau$ (PYTHIA) | 5106 | $3 \times 10^3$ | 108 | 9.94 |
| $W \to \mu\nu$ (PYTHIA) | 5105 | $3 \times 10^4$ | 600 | 0.94 |
| $W \to \tau\nu$ (PYTHIA) | 5146 | $3 \times 10^4$ | 120 | 7.82 |
| **Top** | | | | |
| $t\bar{t}$: (MC@NLO) | 5200 | $1 \times 10^6$ | 4065.08 | 0.98 |

#### 6.2.2 Simulation of *R*-hadron scattering in matter

A model of *R*-hadron scattering [61,62] recently implemented in Geant4 [63] is used in this work. This is an update of earlier work [59] which, in view of the inherent uncertainties associated with modelling such processes, adopts a pragmatic approach in which the scattering rate is estimated with the geometric cross-section and phase space arguments are used to predict the different 2-to-2 and 2-to-3 reactions. Other approaches to modelling *R*-hadron scattering have been proposed, based on Regge phenomenology [24, 25, 64]. These yield predictions of energy loss and scattering cross-sections which are qualitatively similar to those given by the model used here and any differences would not be expected to change the conclusions of this paper.

The typical energy loss per interaction is predicted to be low (around several GeV [61]) since only the light quarks within the *R* hadron should participate in interactions with matter, leaving the heavy squark or gluino as a spectator. This implies that the fraction of *R* hadrons which would be triggered ($\beta \gtrsim 0.6$, see Section 6.3.1) and which would then be stopped during their traversal of the detector is negligible[5].

In addition to energy loss, another feature of *R*-hadron scattering, which has implications for experimental searches, is the possibility of charge and baryon number exchange. Following repeated scattering $R_{\tilde{g}}$ hadrons and $R_{\tilde{t}}$ hadrons not containing an anti-stop should enter the muon system predominantly as baryons. This is due to the occurence of meson-to-baryon conversion processes for which the inverse reaction is suppressed [59]. Anti-baryons, however, would be expected to quickly annihilate in matter and $R_{\tilde{t}}$ hadrons containing anti-stops would thus largely remain as mesons.

The material budget of the part of the ATLAS detector enclosed by the muon system varies as a function of pseudorapidity between 11 and 21 interaction lengths [5,29], with the calorimeters providing the largest contribution. It is estimated that a $R_{\tilde{g}}$ hadron ($R_{\tilde{t}}$ hadron) will typically undergo 10-20 (7-15)

[5] Although it does not form a part of this work, the possibility of observing the decay of *R* hadrons which would be stopped offers a promising and complementary means of searching for *R* hadrons at the LHC [65,66].

nuclear interactions before reaching the muon system [62]. The difference in scattering rates is due to the smaller number of light valence quarks present in $R_{\tilde{t}}$ hadrons. Owing to the large number of interactions a substantial rate of events is thus expected in which a $R$ hadron appears to possess different values of the electric charge in the inner detector and muon system.

While such topologies represent a challenge for track reconstruction software, they also provide observables useful for the discovery and characterisation of $R$ hadrons, something which is explored in section 6.3. For example, a $R_{\tilde{g}}$ hadron can reverse the sign of its charge. $R_{\tilde{t}}$ hadrons [31] can only show this behaviour in the case where an intermediate neutral state oscillates into its anti-particle [67,68]. Since the expected rates of such are processes are essentially unconstrained by experimental limits on SUSY scenarios, they are not included in the simulation used here. Instead, a conservative, zero oscillation scenario is considered.

## 6.3 Event selection

### 6.3.1 Trigger

The selected level 1 trigger is the *mu*6 trigger [29], which has been considered in previous studies of $R$ hadrons at ATLAS as the most promising trigger for this type of work [69]. This trigger is sensitive to the 'classic' stable massive particle signature of a high transverse momentum muon-like track. The trigger efficiency after the level 2 selection for both $R_{\tilde{g}}$ and $R_{\tilde{t}}$ hadrons falls from around 30% at masses of several hundred GeV to around 20% at 2 TeV. The variation of efficiency with mass is due to the slowness of the $R$ hadron. Here, we consider events in which a $R$-hadron track in the muon system must be associated with the correct bunch crossing [6]. This leads to a rapid fall in efficiency for $\beta \lesssim 0.6$. A gradual decrease in efficiency would therefore naively be expected for increasing $R$-hadron mass. After including requirements that the event filter is passed and the muon-like track is well-reconstructed, there is little mass-dependence in the overall efficiency for $R_{\tilde{g}}$ hadrons (around 10-15%) or $R_{\tilde{t}}$ hadrons (20-30%). This difference arises due to the stringent track cuts in the event filter which, at low masses ($\sim 300$ GeV) reject a substantial proportion of tracks which have reversed the sign of their charge, which, as explained in section 6.2.2, occurs only for $R_{\tilde{g}}$ hadrons. At higher masses, corresponding to higher transverse momentum, the poorer momentum resolution allows more 'charge flippers' to pass. As shown in section 6.3.3 this source of inefficiency has little effect on the discovery potential owing to the large predicted cross-section for low mass $R_{\tilde{g}}$ hadrons. However, future work could involve the development of triggers which do not rely on linked inner detector-muon chamber tracks. Should such a trigger configuration be introduced which is based on the *mu*40 [29] trigger at level 1, this could potentially improve the overall efficiency by a factor of $2-3$ for $\beta \gtrsim 0.6$.

### 6.3.2 *R*-hadron final state observables

Following the trigger selection, reconstructed final state quantities were used to select $R$-hadron events and suppress background. A number of variables are presented which were found to provide discrimination between $R$-hadron and Standard Model background processes. Since observables associated with $R_{\tilde{g}}$ hadrons and $R_{\tilde{t}}$ hadrons are mostly very similar, generally only the $R_{\tilde{g}}$-hadron spectra are presented in this section. Where there is a substantial difference in the spectra of the two particle species, separate distributions are shown.

Figure 20 shows the expected transverse momenta distributions, $\frac{dn}{dp_T}$, of muon-like tracks in $R_{\tilde{g}}$ and $R_{\tilde{t}}$-hadron events, for an integrated luminosity of $1\text{fb}^{-1}$. Distributions from background events are also shown. As expected, the $R$-hadron spectra become harder with larger mass, extending up to $\sim 1$ TeV at the largest mass values, while the background events are comparatively softer.

[6] Section 5 explores possibilities of probing the lower $\beta$ region

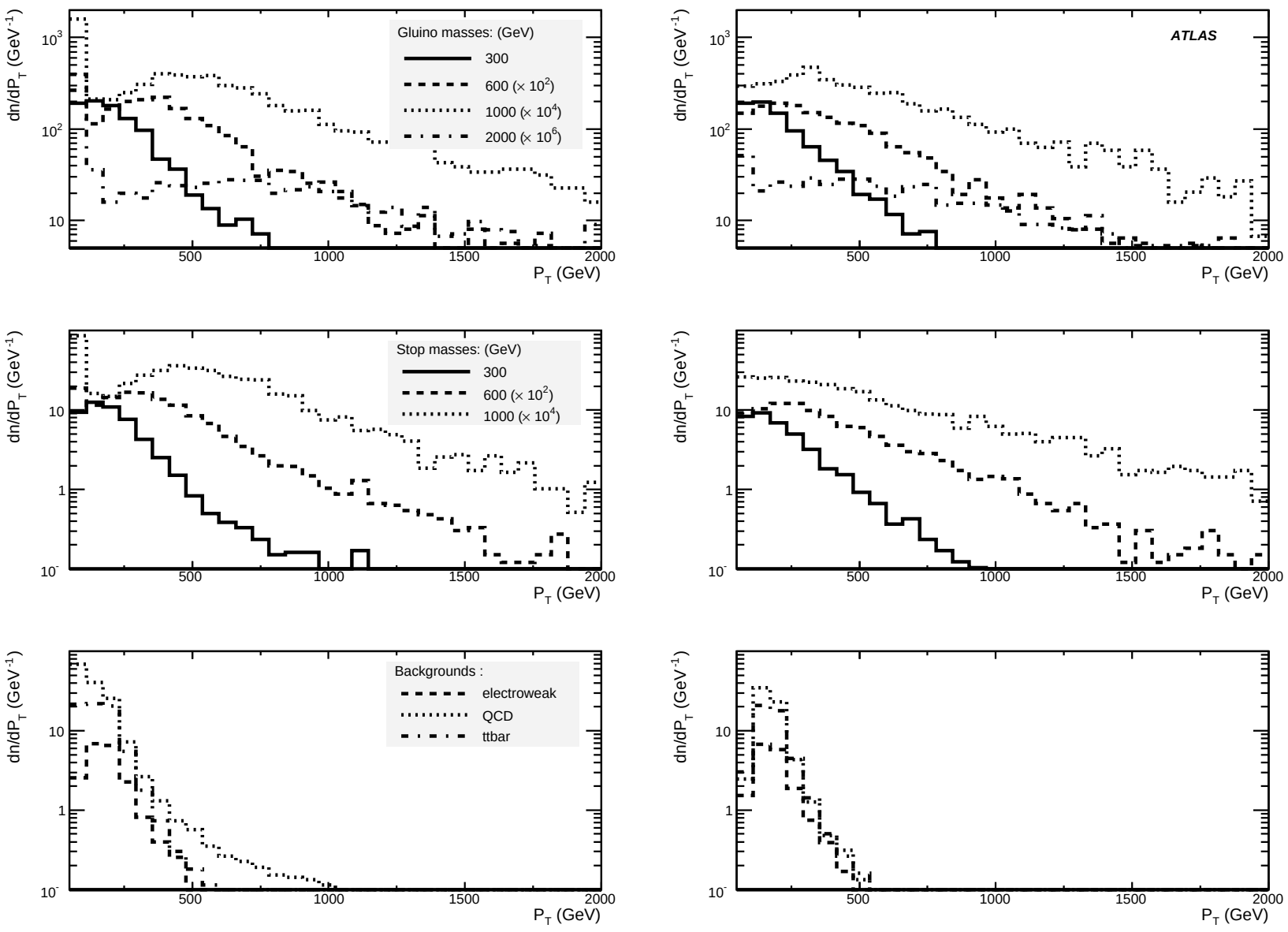


Figure 20: Distributions of transverse momenta $\frac{dn}{dp_T}$ of hard tracks ($p_T > 50$ GeV) as reconstructed in the ID (left) and muon (right) system. The top, middle, and bottom plots show tracks from $R_{\tilde{g}}$, $R_{\tilde{t}}$, and background events, respectively. As labelled, $R$-hadron spectra are scaled according to $R$-hadron mass. The spectra correspond to an equivalent integrated luminosity of 1fb$^{-1}$.

The ratio of high and low threshold $HT/LT$ TRT hit multiplicities is shown in Fig 21 (top left) for $R_{\tilde{g}}$ hadrons of mass 1000 GeV and muon candidates from background events. The different thresholds correspond to low ($\sim$ 200 eV) and high ($\sim$ 6.5 keV) amounts recorded of ionisation energy for a hit. Owing to the large mass and restricted $\beta$ range the simulated $R$-hadron data peak at lower values of $HT/LT$ than the background tracks.

Since $R$ hadrons will typically suffer only several GeV energy loss per interaction through scattering in the calorimeter, it is unlikely they will be associated with a hard calorimeter jet. Figure 21 (top right) shows the distance $R = (\Delta\eta^2 + \Delta\phi^2)^{1/2}$ between a $R_{\tilde{g}}$-hadron track and a jet (defined with the $k_T$ algorithm) with $p_T > 100$ GeV. Clearly for the $R$ hadrons, the spectrum is typically larger than around 1, unlike the background sources which peak at lower values. The QCD background peaks around zero since a large proportion of muons in this sample are produced in the decay of heavy quarks. The top distribution peaks at values around 0.4 reflecting the higher jet multiplicity in such events compared to $R$-hadron events. The distribution is not shown for the electroweak backgrounds owing to the statistical imprecision of the Monte Sample sample (very few events with high $p_T$ jets arise in the selected kinematic region under study here).

In a leading-order picture, $R$ hadrons will be produced approximately back-to-back, unlike a number of background sources, as can be seen in Figure 21 (bottom left), which shows the cosine of the angle between two $R_{\tilde{g}}$-hadron candidates which both leave hard tracks in the muon system $(\cos\Delta\Phi_{\mu,\mu})$; the $R_{\tilde{g}}$-hadron sample peaks strongly at $(\cos\Delta\Phi_{\mu,\mu}) \sim -1$. Figure 21 (bottom right) shows the distribution of the cosine of the angle between hard tracks in the inner detector and muon system $(\cos\Delta\Phi_{ID,\mu})$, which, for pair production events, would be expected to display peaks at $\sim \pm 1$.

As described in section 6.2.2, charge exchange processes can give rise to events in which a linked track is assigned different values of electric charge in the inner detector and muon systems. This is shown in Figure 22 which shows the variable $\frac{q_{ID} p_{T,ID}}{q_\mu p_{T\mu}}$, where $q_{ID}$ ,$q_\mu$, $p_{T,ID}$, and $p_{T\mu}$ are the charge as reconstructed in the inner detector and muon system, and the reconstructed transverse momentum in

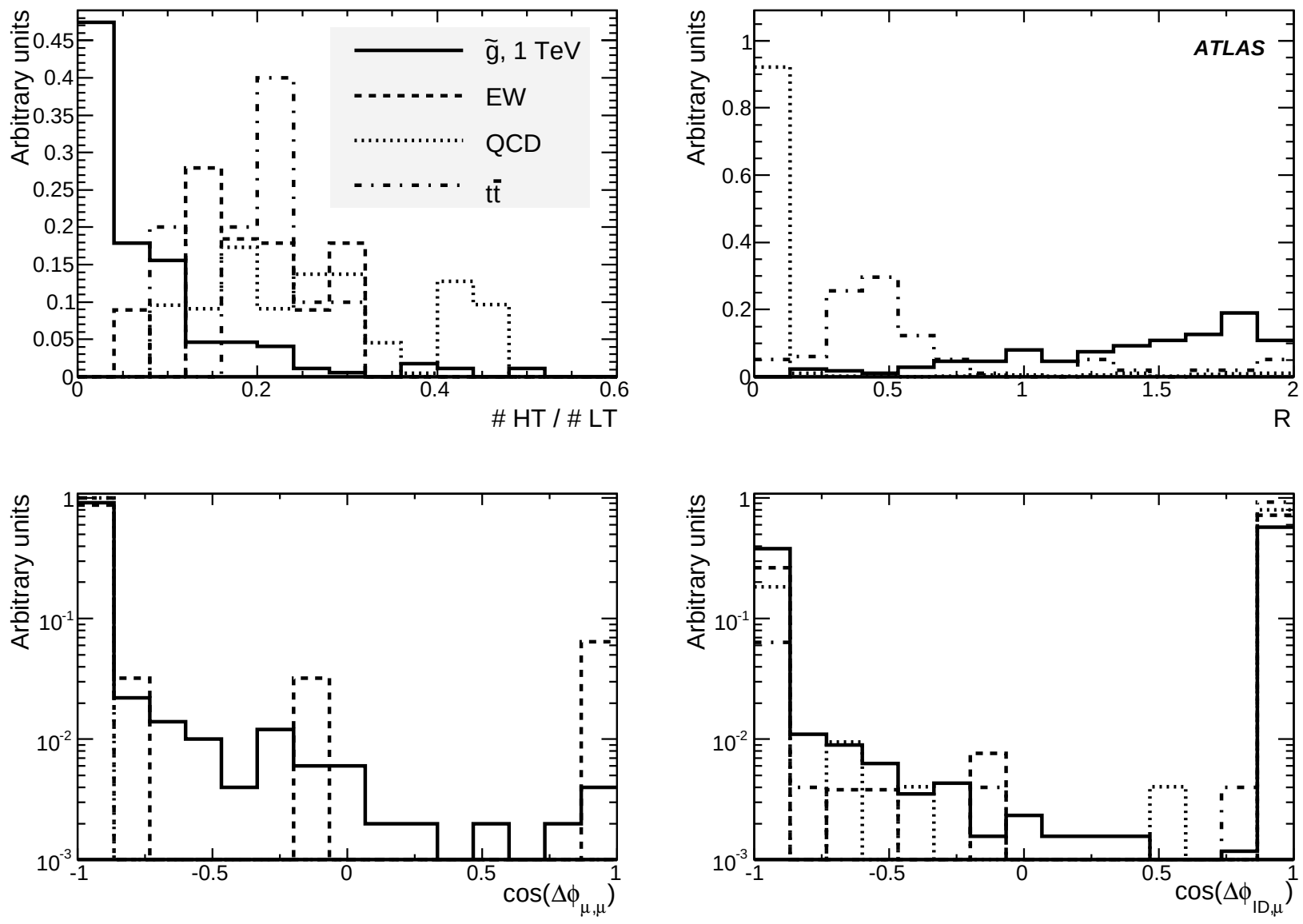


Figure 21: Ratio of the number of high to low threshold hits in the TRT (top left); distance between a $R$-hadron candidate and a jet (top right); cosine of the angle between two high $p_T$ tracks in the muon system (bottom left); and cosine of the angle between high $p_T$ tracks in the ID and muon systems (bottom right). Distributions are shown for $R_{\tilde{g}}$ hadrons of mass 1000 GeV and three background sources.

the inner detector and muon system, respectively. The gluino distributions show a two peak structure, with the peak at negative values of $\frac{q_{ID}p_{T,ID}}{q_\mu p_{T\mu}}$ arising from charge 'flipping' processes. $R_{\tilde{t}}$-hadron spectra indicate a very small rate (several per cent) of candidates with oppositely signed charge in the muon and ID systems which is due to charge misidentification in the muon and ID systems. Both the stop and gluino distributions become broader with increasing mass; this reflects the commensurate increase in $p_T$ and hence degraded resolution of the tracking systems, which has the effect of allowing a greater proportion of charge-'flipping' $R_{\tilde{g}}$ hadrons to satisfy the event filter.

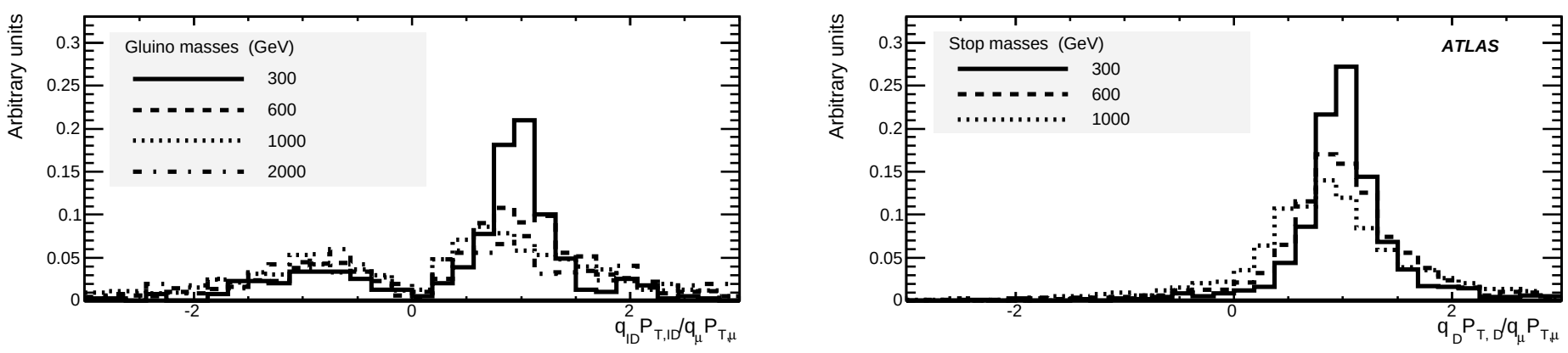


Figure 22: Distributions of $\frac{q_{ID}p_{T,ID}}{q_\mu p_{T\mu}}$ for $R_{\tilde{g}}$ (left) and $R_{\tilde{t}}$ hadrons (right). Predictions for a range of $R$-hadron masses are shown.

### 6.3.3 $R$-hadron selection criteria

Using the information presented in section 6.3.2 $R$-hadron selection criteria were developed following an optimisation procedure [62]. First, no hard muon-like track ($p_T > 250$ GeV) can come within a distance $R < 0.36$ of a hard jet ($p_T > 100$ GeV). Furthermore a candidate $R$ hadron must satisfy at least one of the following conditions listed below. For consistency the same selection is applied both for $R_{\tilde{g}}$ and $R_{\tilde{t}}$ hadrons though criteria 3-4 are only relevant for $R_{\tilde{g}}$ hadrons. However, these have a negligible impact

on the study of the stop discovery potential.

1. The event contains at least one hard muon track with no linked inner detector track. A linked track is defined such that the distance $R = (\Delta\eta^2 + \Delta\phi^2)^{1/2}$ between the measurements in the ID and muon systems is less than 0.1.

2. The event contains two hard back-to-back ID tracks with the TRT hit distribution satisfying $HT/LT <$ 0.05. A back-to-back configuration is defined such that the cosine of the angle between the two muon tracks is less than -0.85.

3. The event contains two hard back-to-back (as defined above) like-sign muon tracks.

4. The event contains at least one hard muon track with a hard matching ID track of opposite charge fulfilling the condition $p_{T,ID} > 0.5 p_{T\mu}$.

Table 18 shows the acceptance numbers and rates for the various samples. It can be seen that for $R$-hadron masses below 1 TeV ATLAS opens up a discovery window with integrated luminosity of the order of 1 fb$^{-1}$. For masses above 1 TeV the rate of signal events is small, and is comparable to the expected background rate, so even discovery would be challenging even with larger data-sets.

Table 18: Number of events selected for the given samples. Background samples not mentioned here are rejected by the selection.

| **Sample** | **Accepted events** | **Rate (Events / fb$^{-1}$)** |
|---|---|---|
| 300 GeV gluino | 235 | $6.44 \times 10^3$ |
| 600 GeV gluino | 551 | $2.70 \times 10^3$ |
| 1000 GeV gluino | 774 | 10.7 |
| 1300 GeV gluino | 732 | 1.20 |
| 1600 GeV gluino | 685 | 0.147 |
| 2000 GeV gluino | 546 | $1.26 \times 10^{-2}$ |
| 300 GeV stop | 78 | 70.0 |
| 600 GeV stop | 134 | 3.9 |
| 1000 GeV stop | 170 | 0.1 |
| J5 | 1 | 0.893 |
| J8 | 1 | $2.26 \times 10^{-3}$ |
| $Z \to \mu\mu$ | 1 | 0.776 |

### 6.4 Conclusion of *R*-hadron search strategies

Stable massive exotic hadrons ($R$ hadrons) are predicted in a number of SUSY scenarios. By exploiting the signature of a hard penetrating particle which may undergo charge exchange in the calorimeter and seemingly does not fall within a jet, ATLAS will be able to discover $R$ hadrons for masses below 1 TeV with relatively low amounts of integrated luminosity ($\sim 1\text{fb}^{-1}$).

## 7 Conclusion

Search strategies at ATLAS have been developed for a range of signatures of new physics processes expected within SUSY models. Studies were made of high transverse momentum photons, which may or

may not have been produced at the primary collision point, and of slow moving stable massive interacting particles (sleptons and *R*-hadrons).

It was shown that with early LHC data, corresponding to an integrated luminosity of around $1\text{fb}^{-1}$, ATLAS opens up a discovery window for those SUSY scenarios giving rise to the aforementioned signatures. Although the studies were performed within the framework of SUSY, the techniques used can be applied to generic searches for physics beyond the Standard Model.

# Exotic Processes

# Dilepton Resonances at High Mass

**Abstract**
We present the discovery potential of a heavy new resonance decaying into a pair of leptons with early LHC data with the ATLAS detector. The dilepton final states are robust channels to analyze because of the simplicity of the event topology. The unprecedented available center-of-mass energy will allow one to probe regions that are inaccessible at previous experiments even with modest amounts of data. After studying the Standard Model predictions and the associated uncertainties one can then look for significant deviations as indication of beyond the Standard Model physics (BSM). The focus of the note is to study the prospects for discovering BSM physics in the dilepton final states with an integrated luminosity ranging from 100 $\text{pb}^{-1}$ to 10 $\text{fb}^{-1}$.

## 1 Introduction

New heavy states forming a narrow resonance decaying into opposite sign dileptons are predicted in many extensions of the Standard Model: grand unified theories, Technicolor, little Higgs models, and models including extra dimensions [1–4]. The discovery of a new heavy resonance would open a new era in our understanding of elementary particles and their interactions. Because of the historic importance of the dilepton channel as a discovery channel and the simplicity of the final state, these channels will be very important to study with early ATLAS data. The strictest direct limits on the existence of heavy neutral particles are from direct searches at the Tevatron [5–7]; the highest excluded mass is currently almost 1 TeV. The LHC will have a center-of-mass energy of 14 TeV which should ultimately increase the search reach for new heavy particles to the 5 - 6 TeV range. Many exotic models can be tested at the LHC, but analyzing all the existing models is impossible. Instead we choose to take a different approach, grouping the early-data analysis by their final state topologies. There have been several other ATLAS studies evaluating the potential for discovery of a heavy resonance [8]. However, this is the first study to include full trigger simulation, misalignments, and data driven methods. Including these experimental issues is important to realistically estimate the analysis potential. We focus on the early data phase of the experiment, defined roughly to include the accumulation of up to 10 $\text{fb}^{-1}$ of ATLAS data.

In the remaining of this introduction, the investigated models are reviewed. In sections 2 and 3, we explore the detector performance concerning the electron, muon and tau reconstruction abilities at high energies and the corresponding trigger efficiencies. In section 4 we investigate the Standard Model predictions and associated uncertainties, as well as the signal cross-section. In section 5, we proceed to search for Exotic resonances.

### 1.1 Models Predicting a $Z'$

Several models [1,3] predict the existence of additional neutral gauge bosons. In particular, grand unified theories, as well as "little Higgs" models, predict their existence as a manifestation of an extended symmetry group. Generically, there are no predictions for the mass of these particles. Since the experimental consequences are very similar in the dilepton final state, we examine only some representative models: the Sequential Standard Model (SSM)[1], the $E_6$ and the Left-Right Symmetric models [9]. The partial width of the $Z'$ boson is given by $\Gamma(Z' \to \ell^+\ell^-) \approx [(g_\ell^R)^2 + (g_\ell^L)^2]\frac{m_{Z'}}{24\pi}$ where $g_\ell^R$ and $g_\ell^L$ are the right and left handed couplings of the charged leptons to the $Z'$ boson and $m_{Z'}$ is the mass of the $Z'$ boson. For

[1]The Sequential Standard Model includes a new heavy gauge boson with exactly the same couplings to the quarks and leptons as the Standard Model $Z$ boson.

| $Z'$ Model | Indirect Searches (GeV) | Direct Searches (GeV) | |
|---|---|---|---|
| | | $e^+e^-$ Colliders | $p^+p^-$ Colliders |
| $Z'_\chi$ | 680 | 781 | 864 |
| $Z'_\psi$ | 481 | 366 | 853 |
| $Z'_\eta$ | 619 | 515 | 933 |
| $Z'_{LRSM}$ | 804 | 518 | – |
| $Z'_{SSM}$ | 1787 | 1018 | 966 |

Table 1: 95% C.L. limits on various $Z'$ models.

the masses and couplings considered here the natural width is typically around 1% of the mass of the resonance.

The strictest limits from direct searches come from the D0 and CDF experiments at the Tevatron [5–7]. Indirect searches have also been undertaken by the LEP experiments [10]. The direct limits range from several hundred GeV to approximately 1 TeV and are shown in Table 1. These limits are not expected to improve much beyond 1 TeV [11]. It should be noted that for models where the $Z'$ couples preferentially to the third generation the limits are lower, therefore we consider it important to look at a lower invariant mass region in this channel.

### 1.2 Randall-Sundrum Graviton

The Randall-Sundrum model [4] addresses the hierarchy problem by adding one extra-dimension linking two branes, the Standard Model brane and the Planck brane. The hierarchy is solved by assuming for the fifth dimension a warped geometry in which the size of the ordinary coordinates decreases exponentially from the Planck scale to the TeV scale. The Randall-Sundrum model predicts the existence of a tower of Kaluza-Klein excitations of the graviton. These should be observable as resonances which decay into lepton pairs at the LHC. The current limits depend on the parameters of the model, and range from several hundred GeV to one TeV [5]. We consider the observability of a Randall-Sundrum graviton decaying into electron pairs. The width of the graviton resonance would be very small. For the parameters considered here it ranges from $10^{-4}$ to a few $10^{-3}$ times the mass.

### 1.3 Technicolor

Strongly interacting theories, like Technicolor and Extended Technicolor, provide a dynamical solution to the problem of Electroweak Symmetry Breaking. Many new technifermions which are bound together by a QCD-like force are predicted. One of the most promising search channels is the dilepton decay of the $\rho_{TC}$ and $\omega_{TC}$. We study the "Technicolor Strawman Model" or TCSM [12, 13] as a benchmark model for generic strongly interacting theories. The most stringent limits on technihadrons in the TCSM framework come from the CDF collaboration, who rules out $\rho_{TC}$ and $\omega_{TC}$ with masses below 280 GeV for a particular choice of the TCSM parameters [14]. The width of the techni-mesons depends on the number of technicolors, but is generally assumed to be small, of the order of a few percent of their mass. More details on the exact values of the parameters considered are discussed in a later section.

## 2 Object Identification and Performance

This section describes the requirements used to select objects for the analyses and summarizes findings on the performance using Monte Carlo simulations of the production and decay of new dilepton resonances.

### 2.1 Electron Identification

The electron identification and performance is described in detail elsewhere [15]; here we summarize the results concerning very high transverse momentum[2] ($p_T$). electron identification and reconstruction. The background to very high $p_T$ electron pairs is expected to be low, therefore only minimal selection criteria need to be applied, in order to maximize efficiency. These minimal criteria are called *loose*. On the other hand, when trying to select very high $p_T$ $\tau$ lepton pairs, where one $\tau$ decays hadronically, a tighter selection on the electron from the other $\tau$ decay is needed.

On top of the minimal requirements that the reconstructed clusters should have an absolute pseudo-rapidity ($\eta$) less than 2.5 and should be associated with a track reconstructed in the inner detector, two electron selections were studied (both described in detail in [15]):

- A *loose* selection based on hadronic leakage and shower shape variables. This selection achieves very high efficiency while maintaining rejection against highly energetic pions with wide showers.
- A *medium* selection, which makes further requirements to obtain better rejection against $\pi^0 \to \gamma\gamma$ background by exploiting the very fine granularity of the first compartment of the electromagnetic calorimeter, and tighter requirements on the associated track.

Figure 1 shows the reconstruction efficiency together with the efficiency of the two selections in a sample $Z' \to e^+e^-$ events with $m_{Z'} = 1$ TeV as a function of transverse momentum and pseudo-rapidity, normalized to truth electrons with $p_T$ greater than 50 GeV and $|\eta|$ smaller than 2.5. The efficiency of the reconstruction is dominated by the efficiency of the cluster to track association and is on average slightly below 80%. It must be noted that this efficiency has improved in the more recent software version used in [15]. The *loose* selection criteria have a relative efficiency close to 1, whereas the *medium* selection leads to an overall average efficiency between 65% and 70%.

The energy resolution for electrons at high $p_T$ is about 1% except in the crack region between the forward and central calorimeters where the resolution is about 5% . The probability to assign the wrong charge to an electron ranges from 1% to at most 5% as the transverse momentum goes from 100 GeV to 1 TeV [16]. For a 1 TeV $Z'$ a dielectron mass resolution of $(0.80 \pm 0.02)\%$ is obtained.

### 2.2 Muon Identification

Here we discuss the requirements used to select muons, as well as a method to extract the identification efficiency from data. The ATLAS detector has an excellent standalone muon spectrometer: muon tracks can be found both in the inner detector and the muon spectrometer. A "combined" muon track consists in matched tracks from both the muon spectrometer and inner detector. We require that a muon, with $p_T \geq 30$ GeV,

- forms a combined track (inner detector and muon spectrometer) with $|\eta| \leq 2.5$,
- has a match $\chi^2 < 100.0$ (5 D.O.F) between the parameters of the inner detector and muon spectrometer tracks.

The muons in the 1 TeV $Z'$ sample have a most probable $p_T$ of about 500 GeV. An efficiency of $(95 \pm 0.2)\%$ with a resolution of approximately 5% is found with this selection. The results are consistent with previous studies [17, 18].

The muon identification efficiency as a function of $p_T$ has been determined using two methods. The first method is the 'tag and probe' method, which has been used successfully at the Tevatron. In this method one uses a 'standard candle' as an in situ calibration point. It involves selecting $Z \to \mu\mu$ events

[2]The transverse momentum is defined as the momentum projected on the plane transverse to the beam axis.

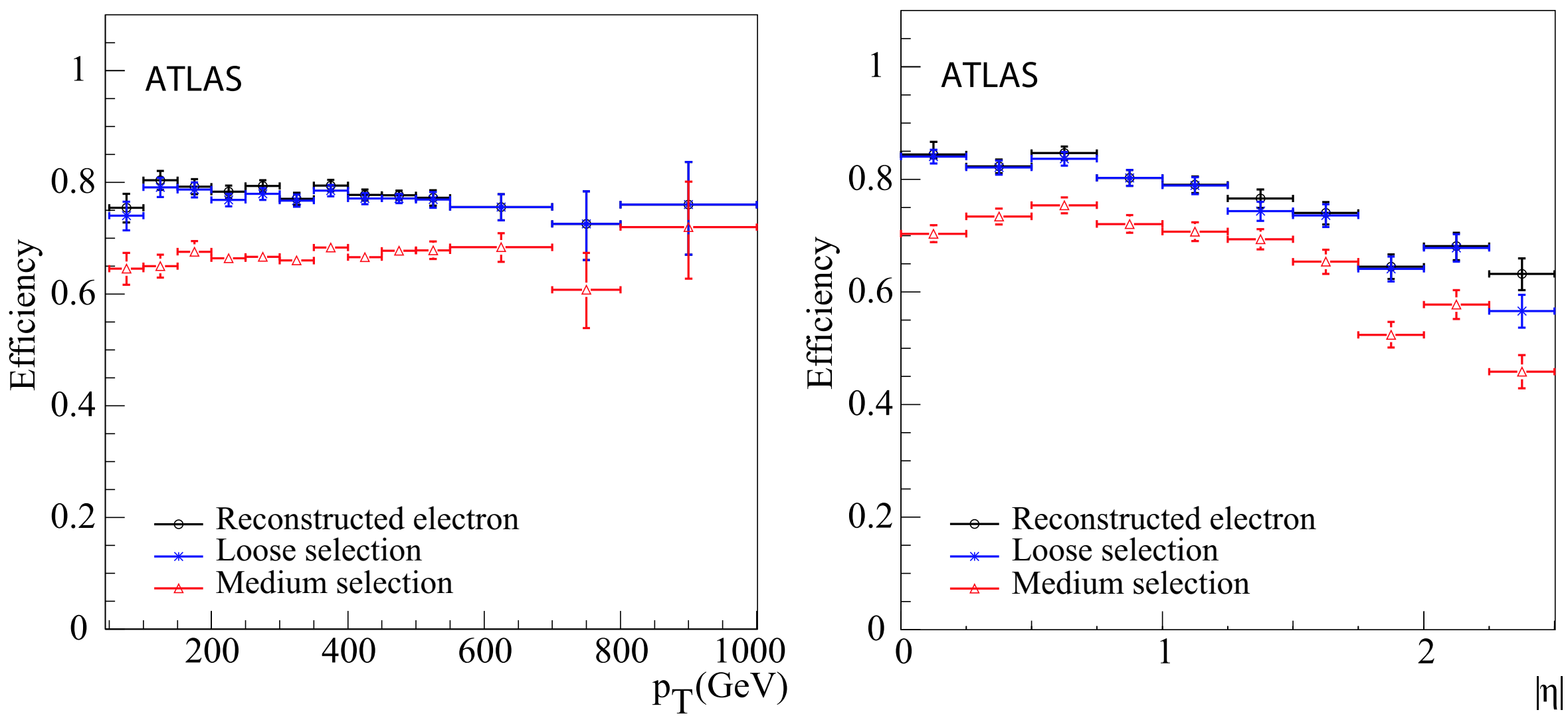


Figure 1: Efficiency of the *loose* and *medium* selection criteria in $Z' \to e^+e^-$ events with $m_{Z'} = 1$ TeV as a function of $p_T$ (left) and $\eta$ (right). The reconstruction efficiency is normalized to truth electrons inside the geometrical acceptance $|\eta| < 2.5$ and with $p_T > 50$ GeV.

and evaluating the reconstruction efficiency from data on these events. One combined muon is used as the tag while an inner detector track is used as a probe track. One can then study how often the probe muon also has a combined track to get an unbiased measurement of the combined muon reconstruction efficiency. The reconstruction efficiency was measured by fitting to the dimuon invariant mass spectrum and finding the fraction of events where the probe track was found as a combined track. A comparison between this tag and probe method and Monte Carlo truth is shown in Fig. 2. The Monte Carlo truth efficiency is determined by counting the number of generated muons with successfully reconstructed inner detector tracks that also have a combined muon track. This study demonstrates that we should be able to use this method to extrapolate into the very high $p_T$ range.

## 2.3 Tau Identification

The algorithm to reconstruct hadronically decaying $\tau$ lepton candidates is described in [19]. It is calorimeter based; it starts from a reconstructed cluster with a transverse energy[3] $E_T > 15$ GeV and then builds identification variables based on information both from the electromagnetic and hadronic calorimeters, as well as the inner tracker. Finally, an electron and a muon veto are applied, which means that hadronically decaying $\tau$ candidates which are matched (which "overlap") with an identified electron or muon are removed.

The reconstruction efficiency, defined as the probability of a true hadronically decaying $\tau$ to be reconstructed as a cluster, and normalized to all true hadronically decaying $\tau$ leptons with $E_T >$15 GeV inside the $\eta$ acceptance, is flat as a function of $\eta$ and $\phi$. The average efficiencies are summarized in Table 2. Efficiencies for electron and muon vetoes are given with respect to all reconstructed $\tau$ leptons.

A likelihood is computed for each $\tau$ candidate. The $\tau$ likelihood combines information from the calorimeter describing the shower shape and tracking information in a multivariate likelihood to maximize the discrimination from background. Detailed studies were done to optimize the $\tau$ lepton efficiency

[3]The transverse energy is defined as the energy multiplied by $\sin\theta$, where $\theta$ is the angle between the beam axis and the direction from the interaction point to the cluster.

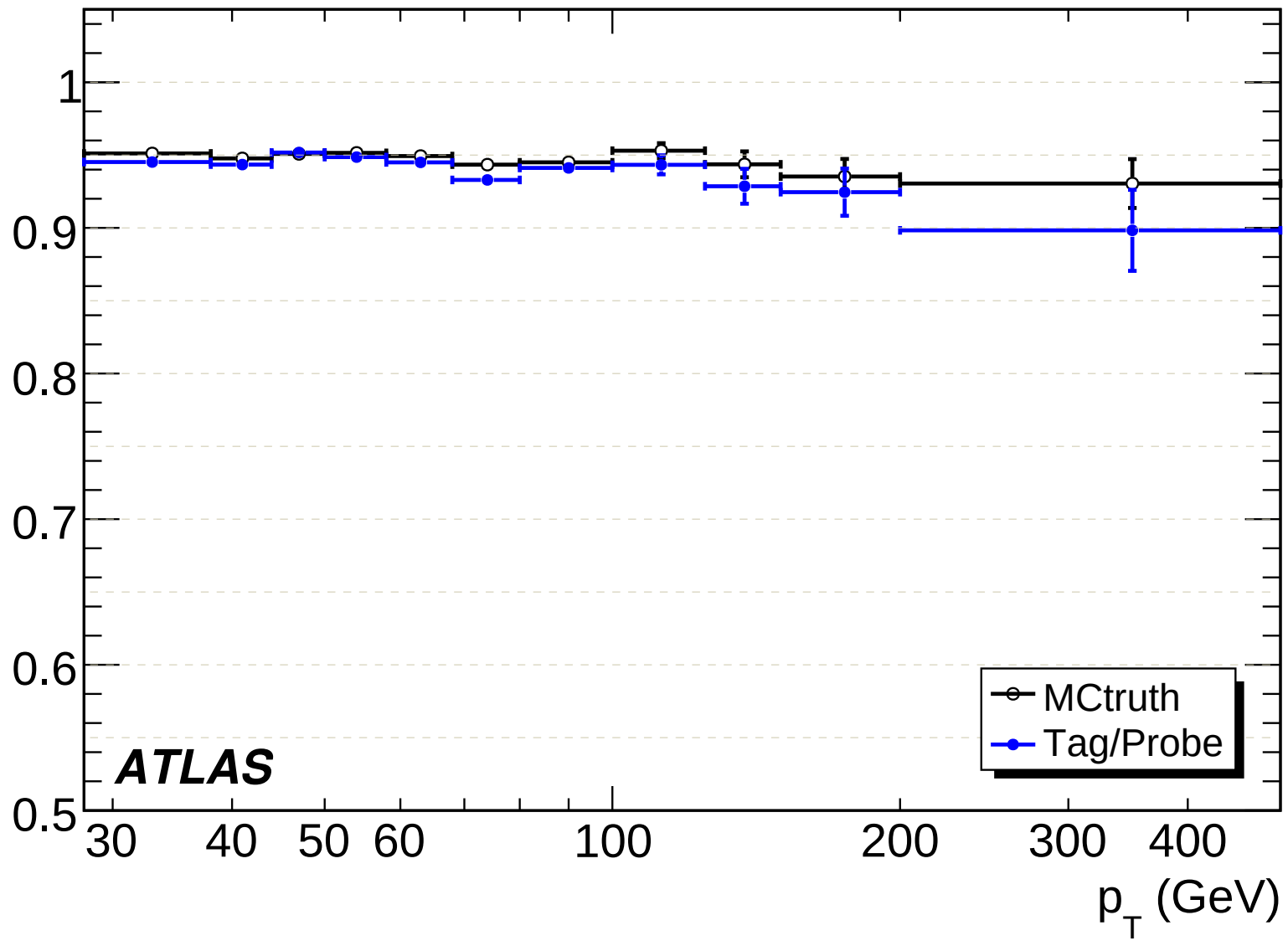


Figure 2: Efficiency of muon reconstruction and identification as a function of $p_T$ from two different methods (see text).

| | |
|---|---|
| Events in $\|\eta\| \leq 2.5$ | $(87.1 \pm 0.1)$ % |
| Events in $\|\eta\| \leq 2.5$ AND $E_T > 15$ GeV | $(85.6 \pm 0.2)$ % |
| Reconstruction | $(98.8 \pm 0.1)$ % |
| Electron veto | $(99.3 \pm 0.1)$ % |
| Muon veto | $(99.9 \pm 0.0)$ % |

Table 2: Reconstruction efficiency, efficiency of e/$\mu$-$\tau$-jet overlap removal for hadronically decaying $\tau$ leptons from $Z'$ boson decays. The efficiencies for kinematic requirements are also given.

| Requirement | Efficiency (%) |
|---|---|
| $E_T > 60$ GeV | $89.8 \pm 0.2$ |
| AND $1 \leq N_{trk} \leq 3$ | $79.2 \pm 0.3$ |
| AND likelihood requirement | $51.0 \pm 0.3$ |

Table 3: Preselection and identification efficiency for $Z' \to \tau\tau$ ($m = 600$ GeV). Efficiency is given with respect to reconstructed hadronically decaying $\tau$ leptons (after removal of overlap with electrons or muons).

and jet rejection for the $Z'$ boson search. The result (shown in Table 3) was to impose a $p_T$-dependent likelihood requirement, a requirement on the number of tracks, and a requirement on the transverse energy.

# 3 Trigger

The aim of the trigger system is to reduce the rate of events flowing through the data acquisition to 200 Hz while maintaining a highly efficient selection for rare signal processes. Even at the initial luminosity of $\mathscr{L} = 10^{31}\ cm^{-2}s^{-1}$, it will be a challenge to keep the trigger highly efficient for all important final states. Several detailed trigger studies were undertaken for the dilepton final state. In this section we summarize those results.

## 3.1 Electron Triggers

There are several proposed triggers which in principle can be used for the dielectron analysis. We studied four triggers: `e55` - requiring one electron with $p_T \geq 60$ GeV, `e22i` - requiring one isolated electron with $p_T \geq 25$ GeV, `2e12` - requiring two electrons with $p_T \geq 15$ GeV, and `2e12i` - requiring two isolated electrons with $p_T \geq 15$ GeV.

Table 4 shows the efficiency at the three ATLAS trigger levels [20]: level 1 (L1), level 2 (L2), and the event filter (EF) for a sample of graviton events. As can be seen in this table, the most efficient triggers are the high $p_T$ triggers that do not require isolation. The low $p_T$ triggers (`2e12`, `2e12i`) will not be considered any further.

| *Signature* | Efficiency (L1/L2/EF) (%) | | | Total Trigger Efficiency (%) |
|---|---|---|---|---|
| `e55` | $99.9 \pm 0.0$ | $95.9 \pm 0.2$ | $94.6 \pm 0.3$ | $90.8 \pm 0.3$ |
| `e22i` | $85.9 \pm 0.3$ | $96.4 \pm 0.4$ | $83.9 \pm 0.3$ | $80.9 \pm 0.4$ |
| `2e12` | $99.9 \pm 0.1$ | $84.9 \pm 0.5$ | $85.5 \pm 0.3$ | $72.6 \pm 0.6$ |
| `2e12i` | $59.1 \pm 0.7$ | $86.1 \pm 0.7$ | $86.2 \pm 0.3$ | $43.9 \pm 0.7$ |

Table 4: Trigger level efficiencies on $G \to e^+e^-$ ($m = 500$GeV) events with respect to *loose* electron offline selection. The last column shows the overall trigger efficiency after all levels.

## 3.2 Muon Triggers

For the dimuon channel we investigated the trigger efficiency for dimuon events using the single muon 20 GeV $p_T$ trigger `mu20` [20] . Results for various signal samples are shown in Table 5. Detailed studies on ways to estimate the trigger efficiency were carried out and presented in [21]. It was found that a tag and probe method, similar to the method described for the offline muon reconstruction, could be used to

| *Sample* | mu20 Efficiency (L1/L2/EF) (%) | | | Total Trigger Efficiency (%) |
|---|---|---|---|---|
| $m = 400$ GeV $\rho_T/\omega_T$ | $97.6 \pm 0.1$ | $98.8 \pm 0.1$ | $99.5 \pm 0.1$ | $96.0 \pm 0.1$ |
| $m = 600$ GeV $\rho_T/\omega_T$ | $98.1 \pm 0.1$ | $98.5 \pm 0.1$ | $99.2 \pm 0.1$ | $95.9 \pm 0.1$ |
| $m = 800$ GeV $\rho_T/\omega_T$ | $97.6 \pm 0.1$ | $98.7 \pm 0.1$ | $99.2 \pm 0.1$ | $95.6 \pm 0.1$ |
| $m = 1$ TeV $\rho_T/\omega_T$ | $97.6 \pm 0.1$ | $98.7 \pm 0.1$ | $99.2 \pm 0.1$ | $95.6 \pm 0.1$ |
| $m = 1$ TeV $Z'_\chi$ | $97.8 \pm 0.1$ | $98.9 \pm 0.1$ | $99.5 \pm 0.0$ | $96.3 \pm 0.1$ |
| $m = 2$ TeV $Z'_{SSM}$ | $97.6 \pm 0.1$ | $98.7 \pm 0.1$ | $98.9 \pm 0.1$ | $95.3 \pm 0.2$ |

Table 5: Simulated trigger level efficiencies of dimuon resonance samples with respect to offline selection. The last column shows the overall trigger efficiency after all levels.

extrapolate $Z \rightarrow \mu\mu$ results to high $p_T$. In addition, efficiencies were obtained using orthogonal triggers, giving a sample which was minimally biased with respect to the muon triggers. It was found in [21] that these estimates agreed with both the tag and probe method and the results from the simulated samples shown here. As can be seen from Table 5 the single muon triggers are highly efficient for any of our signal samples with a total trigger efficiency around 95%.

### 3.3 Triggers for Taus

The $\tau$ lepton decays to hadronic states in 65% of the cases, and the rest of the time to lighter leptons ($e$ or $\mu$). In our studies of ditau final states we select events triggered with a single lepton ($e/\mu$) trigger. Thus, we consider two true final states, which we denote $e\tau_h$ and $\mu\tau_h$.

For the $e\tau_h$ channel we consider two triggers: e22i and e55, as studied in section 3.1. The $\mu\tau_h$ events are selected using the mu20 trigger already used in section 3.2. Note that the efficiencies shown here are lower than for the dimuon or dielectron channel. This arises because there is only one electron or muon in the final state considered here while in the dielectron or dimuon final state either electron/muon can satisfy trigger requirements. Table 6 summarizes the trigger efficiencies.

| *Signature* | Efficiency (L1/L2/EF) (%) | | | Total Trigger Efficiency (%) |
|---|---|---|---|---|
| e22i | 85.2 ±0.5 | 89.8 ±0.4 | 90.2 ±0.3 | 69.1 ±0.6 |
| e55 | 90.0 ±0.3 | 74.2 ±0.4 | 78.7 ±0.6 | 52.6 ±0.8 |
| e22i or e55 | 96.7 ±0.1 | 88.7 ±0.4 | 88.9 ±0.3 | 75.5 ±0.5 |
| mu20 | 79.8 ±0.6 | 90.7 ±0.4 | 97.5 ±0.4 | 70.6 ±0.5 |

Table 6: Trigger level efficiencies for different triggers for $\tau$ leptons from $m = 600$ GeV $Z'$ bosons decaying to $e\tau_h$ and $\mu\tau_h$ final states with respect to offline selection. The last column shows the overall trigger efficiency after all levels.

## 4 Standard Model Predictions and Other Sources of Systematic Uncertainties

In this section, we investigate the main background sources and we show the dominance of the neutral Drell-Yan process. Then we investigate the uncertainties in the Standard Model predictions for Drell-Yan production, as well as for an extra neutral gauge boson. Finally, we discuss the experimental sources of uncertainties, including a dedicated study of the effect of the muon spectrometer alignment.

### 4.1 Background Sources

The neutral Drell-Yan (DY) process constitutes the irreducible background in the search for new heavy dilepton resonances.

The dielectron reducible background results from events in which one or two electrons come from the jet→electron or photon→electron contamination. In addition, true isolated electrons can produced by $W \to e\nu$ or $Z \to ee$ decays. By combining these effects in the dielectron case, one can list the reducible background sources: inclusive jets, $W$+jets, $W$+photon, $Z$+jets, $Z$+photon, photon+jet and photon+photon. For a first estimation of these backgrounds, we have used the event generator PYTHIA [22] to compute the differential cross-sections as a function of the invariant mass of the object pair. The results are shown on the left of Fig. 3. The neutral Drell-Yan process has a much lower cross-section than most of the backgrounds. For each electron-candidate leg originating from a jet (photon), we then apply a rejection factor[4] of $R_{e-jet} = 4 \times 10^3$ ($R_{e-\gamma} = 10$) [15]. We apply an additional requirement to take into account the geometrical acceptance in which the electrons are identified, i.e. $|\eta| < 2.5$, and require at least one object with $p_T \geq 65$ GeV. The resulting differential cross-sections are shown on the right of Fig. 3. One can see that each contribution represents at most 25% of the neutral Drell-Yan process. The sum of all contributions does not exceed 30%. Both the transverse momentum and the rapidity requirements play an important role in reducing the QCD-jet background because it is produced mainly in the $t$-channel resulting in jets with high rapidities. Further reduction of these backgrounds may be obtained by requiring opposite electric charges.

The $WW$, $WZ$, $ZZ$ and top pair processes can also produce two opposite sign electrons. Whereas the cross-sections of the diboson processes are of the same order as the smallest backgrounds above, the top pair cross-section is not negligible. As the topology of the events is not as simple as in the above backgrounds, no conclusion can be drawn without a full simulation study. Using a sample of fully leptonic and semi-leptonic $t\bar{t}$ events, it was checked that the $t\bar{t}$ background was of the order of 10% of the Drell-Yan contribution for di-object masses above 500 GeV after applying electron identification and the same rejection factor as above to the most energetic jet.

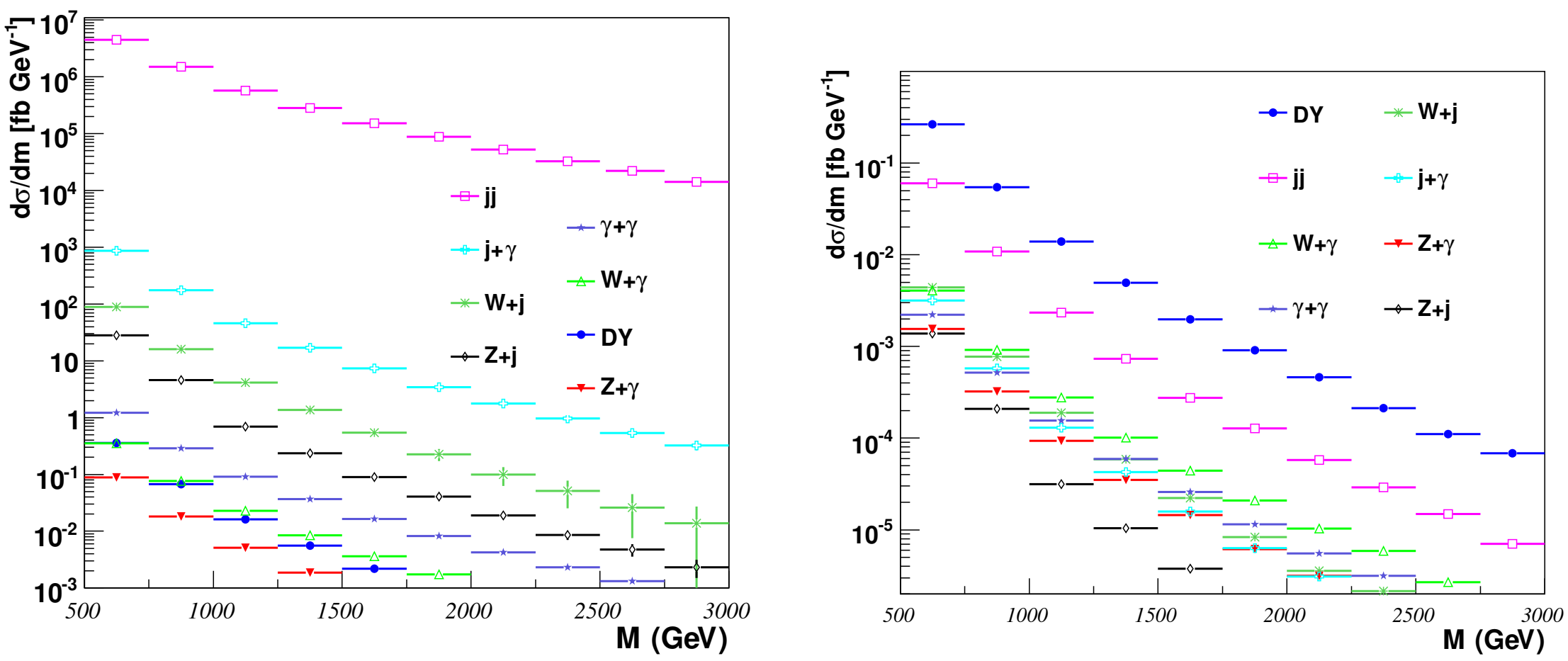


Figure 3: Background contribution to the $e^+e^-$ invariant mass spectrum: before selection requirements (left) and after selection requirements (right).

[4]This number is given for the *medium* selection while the *loose* selection was used here. However, this corresponds to an efficiency of $(80.6 \pm 0.2)\%$, which is higher than our average efficiency using *loose* criteria, due to the better performance of the more recent release used in [15].

The rejection factors $\mu$-jet and $\mu$-photon are higher than the ones corresponding to the electrons and the resulting reducible backgrounds are lower.

All these assumptions will have to be checked with the real data. One can use for instance electron-muon or same charge samples, in which no signal is expected.

In the following, only the neutral Drell-Yan is considered as a source of background in the dilepton channel. The ditau case is treated later.

## 4.2 Controlling the Dilepton Cross-Section

The background estimations in the last section were performed with the PYTHIA event generator which uses tree-level calculations of the cross-sections. The tree-level dilepton cross-sections are subject to large higher order electroweak and QCD corrections. These are known at least to next-to-leading order (NLO) of perturbation theory, not only for the Standard Model Drell-Yan process, but also for a number of new physics processes. They have the additional benefit of reducing the uncertainty induced by the *a priori* unknown renormalization and factorization scales $\mu_{R,F}$. In the following, we discuss in detail the various known radiative corrections and the remaining theoretical uncertainties, focusing on the Standard Model Drell-Yan process and the corrections to the tree-level cross-section.

### 4.2.1 NLO Electroweak Corrections

The electroweak corrections to the Drell-Yan process are known to NLO in the fine-structure constant $\alpha$ [23, 24]. Initial-state photon radiation must be factorized into the parton density functions (PDFs), which in principle modifies the DGLAP evolution of quarks and gluons, but has in practice little effect on the quality of the global fit [25]. Only at very large $x$ and $\mu_F^2$ can the correction become of the order of 1%. Multiple initial-state photon emission can also be resummed, leading to a 0.3% modification of the cross-section [26], or matched to parton showers [27]. The remaining initial-state QED contributions are also small, whereas the photon radiation emitted by the final state leptons can have a significant impact on their mass ($M$) and transverse momentum spectra as well as the forward-backward asymmetry $A_{\rm FB}$ [28].

In the vector-boson resonance region(s) these and the universal parts of the weak corrections, which can amount to +80 (+40) % for muon (electron) pairs below and −18 (−10) % above the resonance, can be taken into account by using a running value of $\alpha(M^2)$ or, more generally, effective vector and axial vector couplings in the Effective Born Approximation. The corrections are then reduced to +6 (+2) % for muon (electron) pairs below and +1 ($< +1$) % above the resonance. While the presence of new physics can modify the running of the weak parameters, the QED corrections remain unaffected.

The electroweak corrections coming from non-factorisable box diagrams with double-boson exchange are small in the $Z$ (and $Z'$) resonance region(s), but they can be quite large away from these resonances (−4 to −16 % for electron pairs, −12 to −38 % for muon pairs of invariant mass 300 GeV to 2 TeV at the LHC, see Fig. 4).

### 4.2.2 NLO QCD Corrections

The QCD corrections to the Standard Model Drell-Yan process are known at NLO [29] and next-to-next-to-leading order (NNLO) [30, 31] in the strong coupling constant $\alpha_s$. The latter include in principle non-factorisable corrections through $qq$ and $gg$ initial states, which remain, however, smaller than 1% in practice, even at small values of $x$, where the gluon density is large. The effects of multiple soft-gluon radiation have been resummed simultaneously in the low-$p_T$ and high-mass (above 500 GeV) regions at next-to-leading logarithmic (NLL) accuracy not only for Standard Model $Z$-bosons [32], but also for $Z'$ bosons [33]. They were shown to be in good agreement with the NNLO result as well as the one

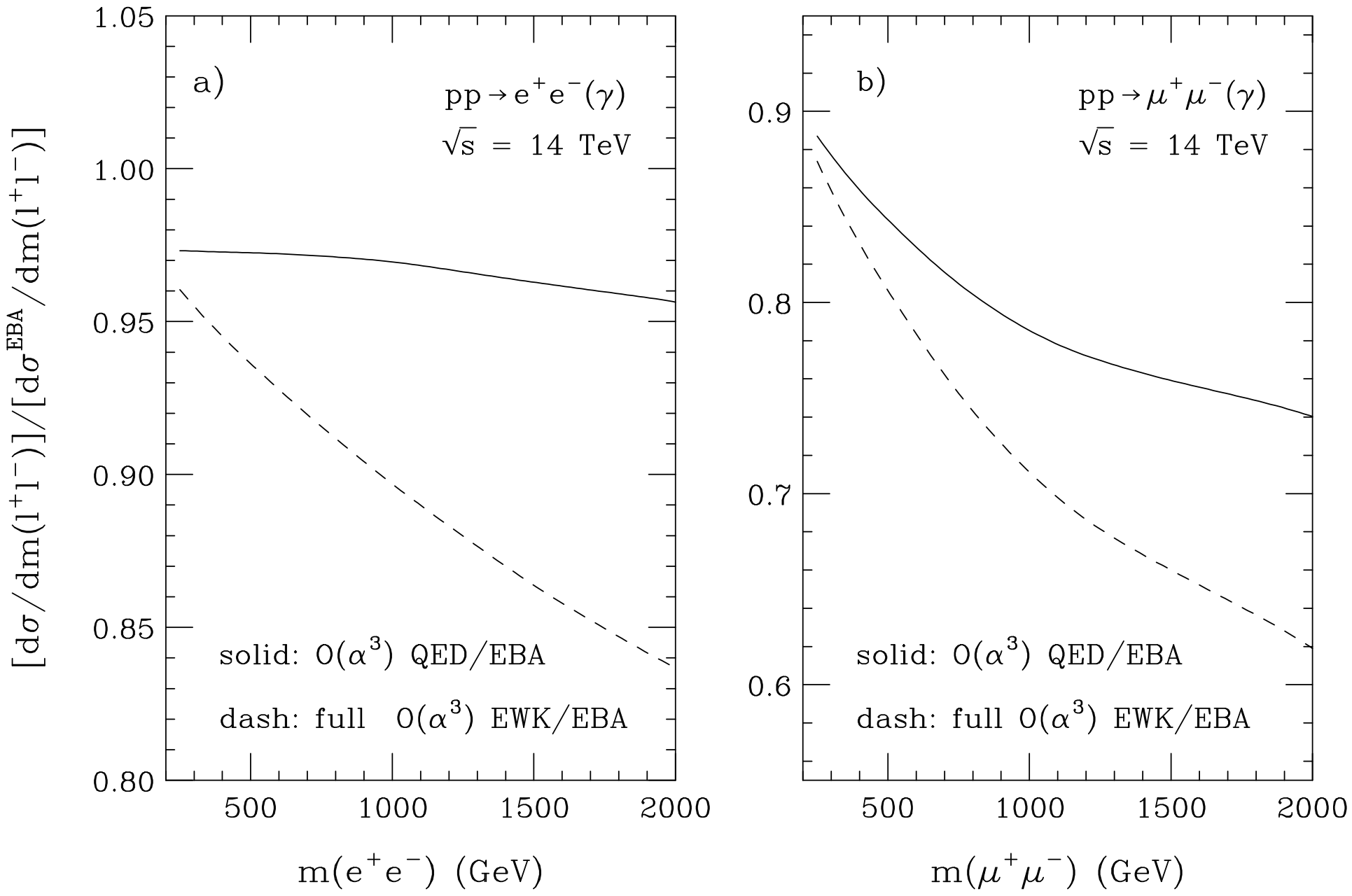


Figure 4: NLO electroweak corrections in the high-mass region for Standard Model electron and muon pair production at the LHC [23]. In the presence of a new resonance, these relative corrections would be largely reduced (see text).

obtained by matching NLO QCD to parton showers in MC@NLO [33, 34]. In contrast, the matching of tree-level matrix elements to parton showers in PYTHIA [22] requires the *ad hoc* application of a (slightly) mass-dependent correction ($K$) factor and leads to an unsatisfactory description of the $p_T$ spectrum. For resonant spin-2 graviton production, which involves not only color-triplet quark, but also color-octet gluon initial states, the NLO QCD corrections are substantially larger ($K \simeq 1.6$) [35, 36] than those for Standard Model or extra neutral gauge bosons ($K \simeq 1.26$, see Fig. 5). In this case, the matching of matrix elements to parton showers has only been performed at the tree-level [37], and resummation has only been performed in the low-$p_T$ region [36].

While the NLO total cross-sections for vector bosons and gravitons still change substantially when the renormalization and factorization scales are varied simultaneously around the resonance mass $M$ by a factor of two ($\pm 9\%$) [33, 36], the scale uncertainty is reduced to the percent-level at NNLO [30, 31] or, alternatively, to $+6$ and $-3$ % after joint resummation at the NLL order [33].

The theoretical uncertainty coming from different parameterizations of parton densities is estimated in Fig. 5 [33] for invariant masses above 500 GeV. Since the invariant mass of the lepton pair is correlated with the momentum fractions of the partons in the external protons, the normalized mass spectra (left) are indicative of the different shapes of the quark and gluon densities in the CTEQ6M[5] parameterization [38]. The latter also influence the transverse-momentum spectra (right). The shaded bands show the uncertainty induced by variations, added in quadrature, along the 20 independent directions that span the 90% confidence level of the data sets entering the CTEQ6 global fit [39]. With about $\pm 5\%$ at 1 TeV ($\pm 11\%$ at 3 TeV), the PDF uncertainty is slightly larger than the scale uncertainty [33, 40].

[5] CTEQ collaboration has recently proposed new sets of PDFs. Using them, both the central values and uncertainties may change by several percents.

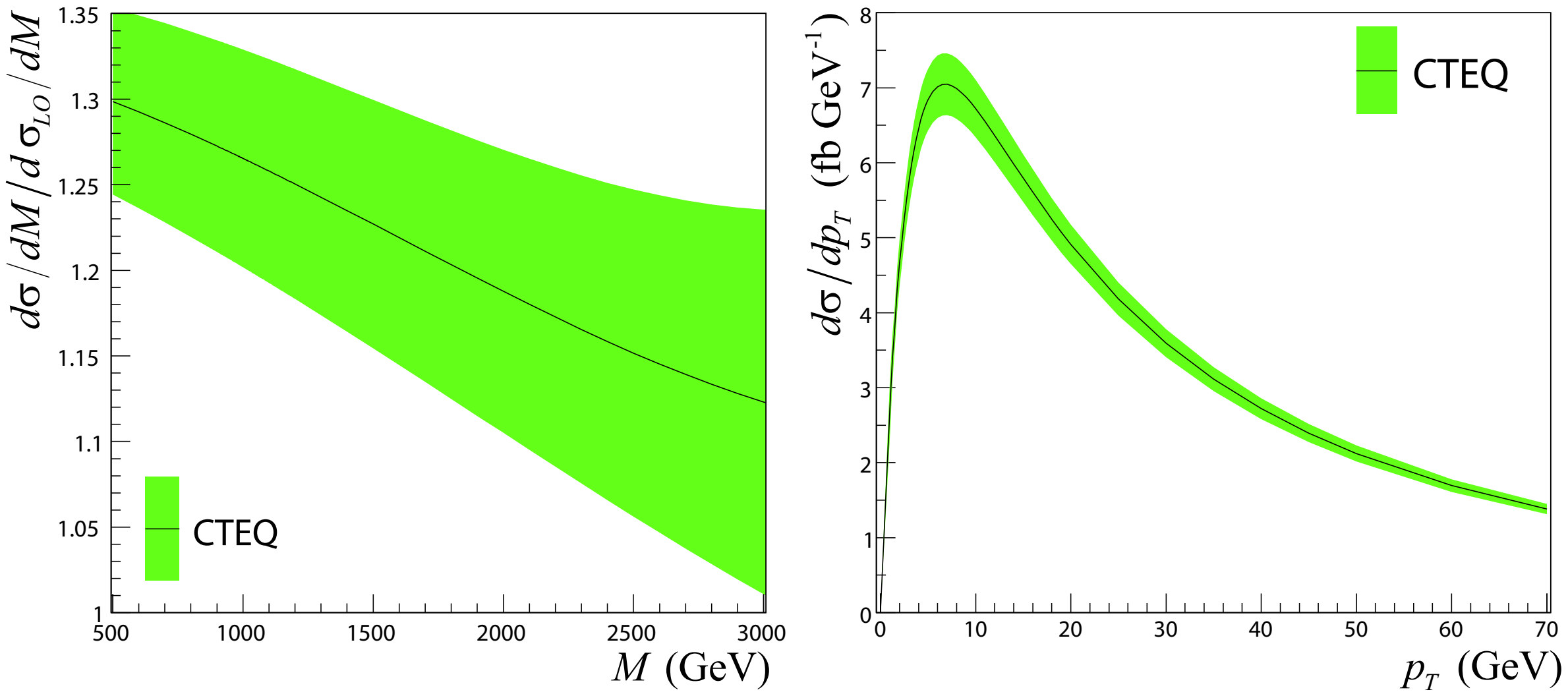


Figure 5: Mass (left) and transverse-momentum (right) spectra after matching the NLO QCD corrections to joint resummation with CTEQ6M parton densities. The mass spectra have been normalized to the LO QCD prediction using CTEQ6L parton densities. The shaded bands indicate the deviations allowed by the up and down variations along the 20 independent directions that span the 90% confidence level of the data sets entering the CTEQ6 global fit.

The uncertainty at low transverse momenta coming from non-perturbative effects in the PDFs is usually parameterized with a Gaussian form factor describing the intrinsic transverse momentum of partons in the proton. Three different parameterizations of this form factor have been proposed [41–43]. In all three cases the transverse-momentum distribution is changed by less than $+3$ and $-6$ % for $p_T >$ 5 GeV [33].

Combining the three contributions (from the scales, the PDFs and the non-perturbative form factor), the total theoretical QCD uncertainty is $\pm 8.5\%$ at 1 TeV, $\pm 14\%$ at 3 TeV. It must be noted that these uncertainties are common to the signal (heavy resonance) and background (Standard Model Drell-Yan).

### 4.3 Effect of Muon Spectrometer Misalignment

At large $p_T$ ($\geq$ 100 GeV), an important contribution to the muon momentum resolution is the alignment of the muon spectrometer. In the early data period, the resolution is expected to be dominated by the alignment. The ultimate goal of the alignment system is to determine the position of the chambers in the muon spectrometer to about 40 $\mu$m and $\sigma_{rot}(mrad) = 0.5\sigma_{trans}(mm)$.

A detailed study was carried out in order to determine the effect of possible larger uncertainties in the position of the chambers to the $Z'$ search. For the analysis, in addition to the ideal case of no misalignment at all, we have chosen 7 different hypotheses of misalignment: $(40\mu m, 20\mu rad)$, corresponding to the target value of the alignment system, (100 $\mu m$, 50 $\mu rad$), (200 $\mu m$, 100 $\mu rad$), (300 $\mu m$, 150 $\mu rad$), (500 $\mu m$, 250 $\mu rad$), (700 $\mu m$, 350 $\mu rad$) and (1000 $\mu m$, 500 $\mu rad$). In the last two cases, the alignment resolution is of the same order or greater than the track sagitta we want to measure.

As shown in Fig. 6, the dominant effect leading to a wash out of the signal is the resolution loss[6]. The

[6]The slight excess of events around 600 GeV is due to mis-reconstructed $Z$ bosons. In later versions of the software, this effect is not present anymore.

loss of resolution due to misalignment will deteriorate our ability to determine the charge of the muon. This was also studied as a function of the misalignment and is summarized in Table 7.

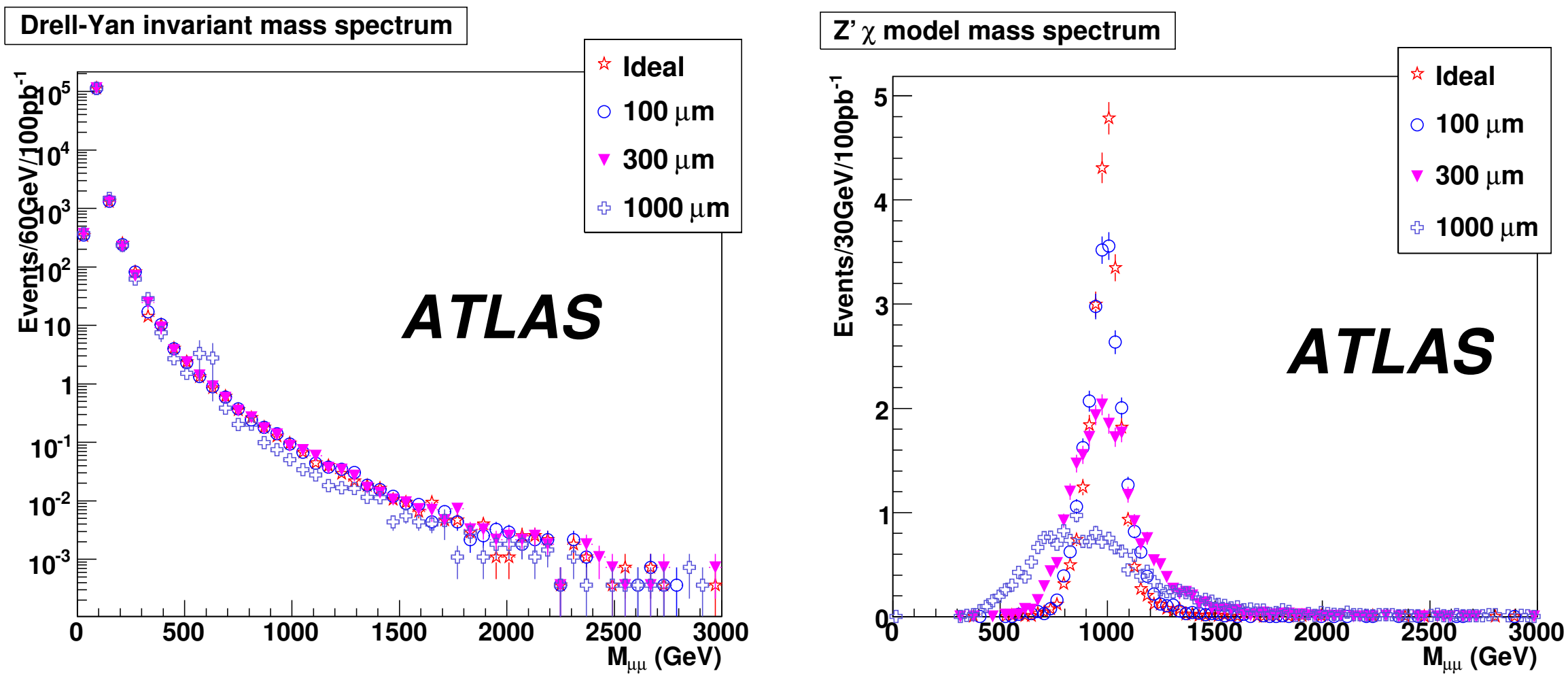


Figure 6: Left: reconstructed invariant mass distribution of Drell-Yan events for different misalignment hypotheses. The numbers corresponds to an integrated luminosity of 100 pb$^{-1}$. Right: reconstructed invariant mass of the $Z'_\chi$ model for the seven misalignment scenarios.

| Misalignment ($\mu m$) | Ideal | 40 | 100 | 200 | 300 | 500 | 700 | 1000 |
|---|---|---|---|---|---|---|---|---|
| Relative efficiency | 0.984 | 0.984 | 0.984 | 0.98 | 0.973 | 0.948 | 0.918 | 0.877 |

Table 7: Loss in signal efficiency due to the charge misidentification for seven misalignment hypotheses.

### 4.4 Other Systematic Uncertainties

Additional experimental systematic uncertainties must be taken into account, listed as follows:

- the uncertainty in the efficiency of object identification was assumed to be 5% for muons, 1% for electrons, and 5% for $\tau$ leptons;
- the uncertainty in the energy scale was assumed to be 1% for muons, 1% for electrons, and 5% for $\tau$ leptons;
- the uncertainty in the resolution of the objects is as follows: $\sigma(\frac{1}{p_T}) = \frac{0.011}{p_T} \oplus 0.00017$ for muons, 20 % for electrons, and 45% for $\tau$ leptons.
- the uncertainty in the luminosity was assumed to be 20% with an integrated luminosity of 100 pb$^{-1}$ of data and 3% for 10 fb$^{-1}$.

The effect of all the above on the discovery potential is discussed in the next section for individual channels.

# 5 Search for Exotic Physics

In this section we present the discovery potential for several resonant signatures in the early running of ATLAS. We focus on the reach with an integrated luminosity of up to 10 $\text{fb}^{-1}$ of data.

The statistical significance of an expected signal can be evaluated in several ways. The simplest approach, "number counting" is based on the expected rate of events for the signal and background processes. From these rates, and assuming Poisson statistics, one can determine the probability that background fluctuations produce a signal-like result according to some estimator; e.g. the likelihood ratio. In the "shape analysis" approach, a detailed knowledge of the expected spectrum of the signal and background for one observable (like the invariant mass distribution for example) can be used to improve the sensitivity of the search by treating each mass bin as an independent search channel, and combining them accordingly.

The resulting sensitivity is in general higher in the shape analysis than the estimation given in the number counting approach. In the shape analysis, the data is fitted or compared to two models: a background-only model and a signal-plus-background model. These are also called "null hypothesis", noted H0 and "test hypothesis", noted H1, respectively. The input signal and background shapes are given to the fitting algorithms either as histograms in the non-parameterized approach [44] or as functions in the parameterized approach. For each of the models, a likelihood or a $\chi^2$ distribution is computed and the log of the ratio of the two likelihoods (LLR) or the difference of two $\chi^2$s are estimated and used to compute the confidence levels. Either $CL_b = CL_{HO}$ alone, or $CL_s = CL_{H1}/CL_{H0}$ (in the "modified frequentist approach" [44]) can then be used to compute the significance $S$:

$$S = \sqrt{2} \times Erf^{-1}(1 - CL_b) \quad \text{or} \quad S = \sqrt{2} \times Erf^{-1}(1 - \frac{1}{CL_s}) \tag{1}$$

in the *double tail* convention[7].

A convenient way to compute the LLR is to use the Fast Fourier Transform (FFT) method presented in [45]. The advantage of this method is that it does not require the generation of millions of pseudo-experiments needed for high significances and which can be time consuming. The sources of systematic uncertainties can then be incorporated as nuisance parameters.

The above methods have been used to investigate the discovery potential of the $Z'$ boson in the dilepton ($e$, $\mu$, $\tau$) channels, of the graviton in the dielectron channel, and of Technicolor in the dimuon channel. This is presented in the following sections.

## 5.1 Background Estimation

As discussed in the previous section, neutral Drell-Yan production of lepton pairs is expected to be the dominant background for all the analyses (but $\tau^+\tau^-$) and other contribution will be neglected here. Since different techniques are used to estimate the signal significance we also treat the Drell-Yan background in a few different but entirely consistent ways:

- In the "number counting" approach, we simply count the expected number of events under the resonance peak from various background sources, including the Drell-Yan process.
- In the non-parameterized $CL_s$ method, we use the number and shape of the mass distribution by producing a histogram for the background.
- For several analyses, we perform a fit to the Drell-Yan background parameterizing the shape which allows to estimate the number of background events and extrapolate it to higher masses.

[7] In this convention, $1 - CL_b$ has to be lower than $2.87 \times 10^{-7}$ to correspond to a $5\sigma$ significance.

Each of these methods produces a complementary and consistent approach to estimating the main background. When the Drell-Yan fit is needed, we parametrize the shape of the background by the formula $ae^{-bM^c}$, where $M$ is the invariant mass of the lepton pair and $a, b, c$ are parameters of the fit. Fits to the Drell-Yan spectrum presented in section 4.2 suggest that the parameterization $exp(-2.2M^{0.3})$ used by [46] describes the background shape well. It is this one which is used in the $Z' \to \mu\mu$, $G \to ee$ and technicolor analyses. In the $Z' \to ee$ analysis, these parameters are allowed to vary in the individual ensemble tests. The fit to the entire spectrum letting all parameters float is consistent with this prescription.

## 5.2 $Z' \to ee$ Using a Parameterized Fit Approach

### 5.2.1 Event Selection

The selection of events with two electrons coming from a $Z'$ has been studied in samples of fully simulated $Z'_\chi \to e^+e^-$ events with $Z'$ boson masses of 1, 2 and 3 TeV, corresponding respectively to integrated luminosities of 21 fb$^{-1}$, 204 fb$^{-1}$ and 2392 fb$^{-1}$.

The first requirement is that the two highest $p_T$ clusters in the event be in the geometrical acceptance. The next requirement is that these clusters be associated with a track; its efficiency is 67% at 1 TeV and decreases for higher masses. The third requirement is that these two reconstructed electron candidates be identified as *loose* electrons. The relative efficiency of such a selection is at least 94% and increases with invariant mass. The trigger studies have been normalized to events with two *loose* electrons. As shown in section 3.1, the highest trigger efficiency is obtained with a non-isolated single electron trigger (`e55`). Its efficiency is 90.8% per event. The last requirement is that the two electrons have opposite electric charges. The requirement flow is presented in Table 8, where the events are counted in a window of $\pm 4\ \Gamma_{Z'}$ around the center of the resonance. Although the opposite charge requirement is optional in the absence of a large background, especially at very high invariant mass, it allows to have a control sample (made of same sign dielectrons) for the background. The resulting overall efficiency is 48% at $m = 1$ TeV, 42% at $m = 2$ TeV and about 34% at $m = 3$ TeV.

| Selection | Signal at 1 TeV | DY at 1 TeV | Signal at 2 TeV | DY at 2 TeV | Signal at 3 TeV | DY at 3 TeV |
|---|---|---|---|---|---|---|
| | 347. | 3.56 | 14.7 | 0.16 | 1.22 | 0.015 |
| 2 generated $e^\pm$, $\lvert\eta\rvert < 2.5$ | 299. | 3.07 | 13.7 | 0.15 | 1.16 | 0.013 |
| 2 clusters with a track | 201. | 2.06 | 8.0 | 0.09 | 0.62 | 0.009 |
| 2 *loose* electrons | 190. | 1.96 | 7.2 | 0.08 | 0.52 | 0.008 |
| At least one $p_T > 65$ GeV | 190. | 1.96 | 7.2 | 0.08 | 0.52 | 0.008 |
| Event triggered | 173. | 1.77 | 6.6 | 0.07 | 0.47 | 0.007 |
| 2 opposite charges | 166. | 1.70 | 6.2 | 0.07 | 0.43 | 0.007 |

Table 8: Requirement flow table for the $Z' \to e^+e^-$ analysis: cross-sections in fb. The events are counted in a window of $\pm 4\ \Gamma_{Z'}$ around the resonance.

The above efficiencies, normalized to events in the geometrical acceptance ($|\eta| < 2.5$), are shown in Fig. 7 (left) as a function of the invariant mass of the electrons. They do not depend on the model used to generate the $Z'$ samples. Only the requirement that the two electrons be in the geometrical acceptance depend on the model. Indeed, the relative proportions of initial quark flavors depend on the couplings of the $Z'$ to the quarks. The PDF of the up quarks being harder than that of the down quarks, $Z'$ produced by a $u\bar{u}$ pair tend to be slightly more boosted, and therefore the electrons stemming from their decay tend to be produced at slightly higher pseudo-rapidities. This effect is visible in Fig. 7 (right) showing

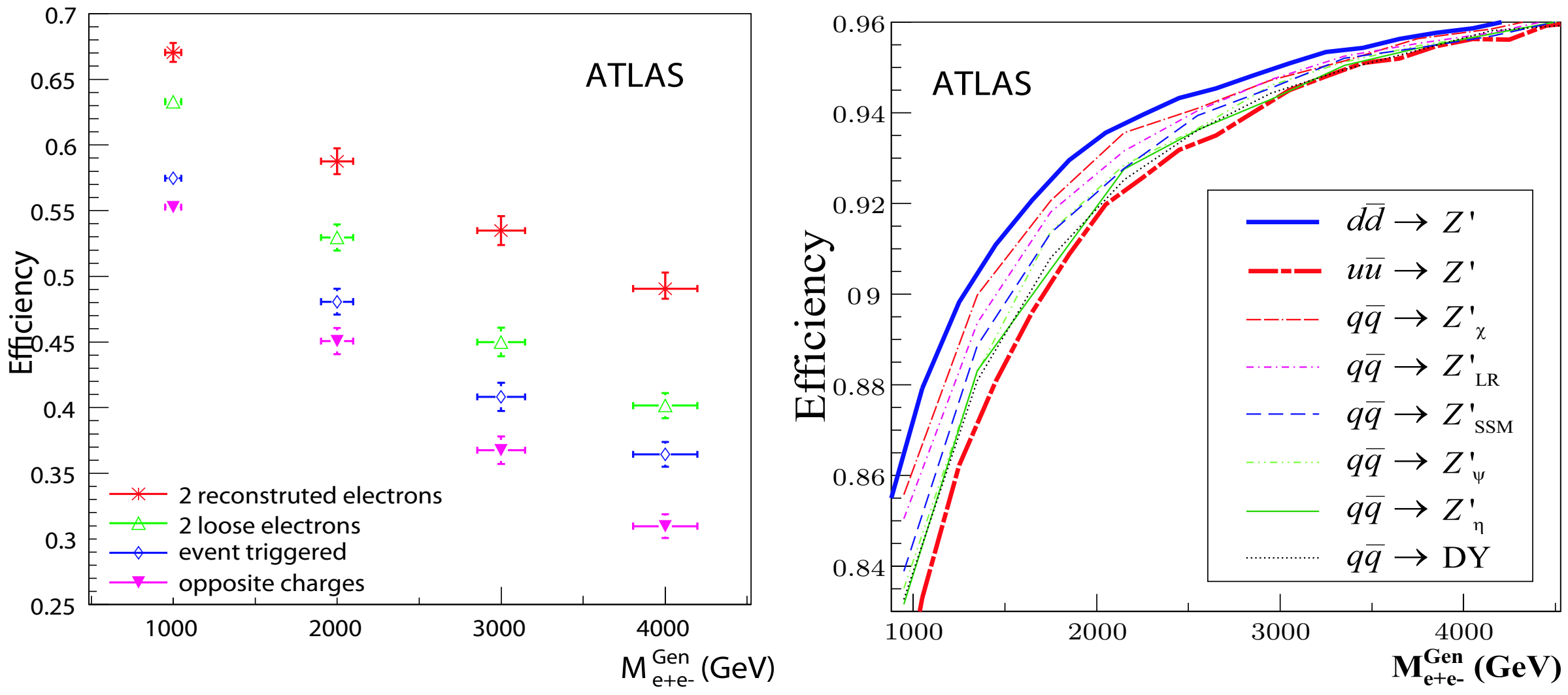


Figure 7: $Z'_\chi \to e^+e^-$ selection efficiency as a function of the generated invariant mass. Left: all selections, normalized to events in the geometrical acceptance; right: $|\eta| < 2.5$ criteria for $u\bar{u}$ and $d\bar{d}$ events separately and for different $Z'$ models (generator level).

the efficiency of the $|\eta|$ selection for $u\bar{u}$ and $d\bar{d}$ events separately, and for a number of benchmark $Z'$ models: the Sequential Standard Model ($Z'_{SSM}$), the E$_6$ models $Z'_\psi$, $Z'_\chi$, $Z'_\eta$, and the left-right symmetric model ($Z'_{LR}$). It is therefore possible to generalize the efficiencies that have been measured in the fully simulated samples to models which haven't been simulated as well as to intermediate masses.

### 5.2.2 Discovery Potential

**Modeling of the dilepton invariant mass spectrum.** In order to compute the significance for several $Z'$ models, a parameterization of the mass spectrum of the signal and of the background has been used. The differential cross-section can be factorized with a good precision in a parton-level term $\frac{d\hat{\sigma}}{dm}$ and a PDF-dependent term $G_{PDF}(m)$:

$$\frac{d\sigma}{dm}(m) = \frac{d\hat{\sigma}}{dm}(m) \times G_{PDF}(m) \quad (2)$$

Using this factorization, one can write:

$$\left.\frac{d\sigma}{dm}\right|_{\mathrm{DY}}(m) = \frac{1}{m^2} \times G_{PDF}(m) \quad (3)$$

$$\begin{aligned} \left.\frac{d\sigma}{dm}\right|_{\mathrm{Signal}}(m) &= \frac{1}{m^2} \times G_{PDF}(m) \\ &+ \mathscr{A}_{\mathrm{peak}} \times \frac{\Gamma^2_{Z'}}{m^2_{Z'}} \frac{m^2}{(m^2 - m^2_{Z'})^2 + m^2_{Z'}\Gamma^2_{Z'}} \times G_{PDF}(m) \\ &+ \mathscr{A}_{\mathrm{interf}} \times \frac{\Gamma^2_{Z'}}{m^2_{Z'}} \frac{m^2 - m^2_{Z'}}{(m^2 - m^2_{Z'})^2 + m^2_{Z'}\Gamma^2_{Z'}} \times G_{PDF}(m) \end{aligned} \quad (4)$$

where $\mathscr{A}_{\mathrm{peak}}$ is the amplitude of the $Z'$ process and $\mathscr{A}_{\mathrm{interf}}$ is the amplitude of the interference $Z'/Z$ and $Z'/\gamma$, both normalized to the Drell-Yan process. This parameterization only depends on four parameters:

$m_{Z'}$, $\Gamma_{Z'}$, $\mathscr{A}_{\text{peak}}$ and $\mathscr{A}_{\text{interf}}$. The differential cross-section is then multiplied by the appropriate $K$-factor (see section 4.2). The detector performance is accounted for as follows: the differential cross-section is multiplied by the efficiency computed above and convoluted by the invariant mass resolution (see section 2.1). The agreement between this parameterization and the full simulation is shown in Fig. 8 (left) for a $Z'_\chi$ at 1 TeV.

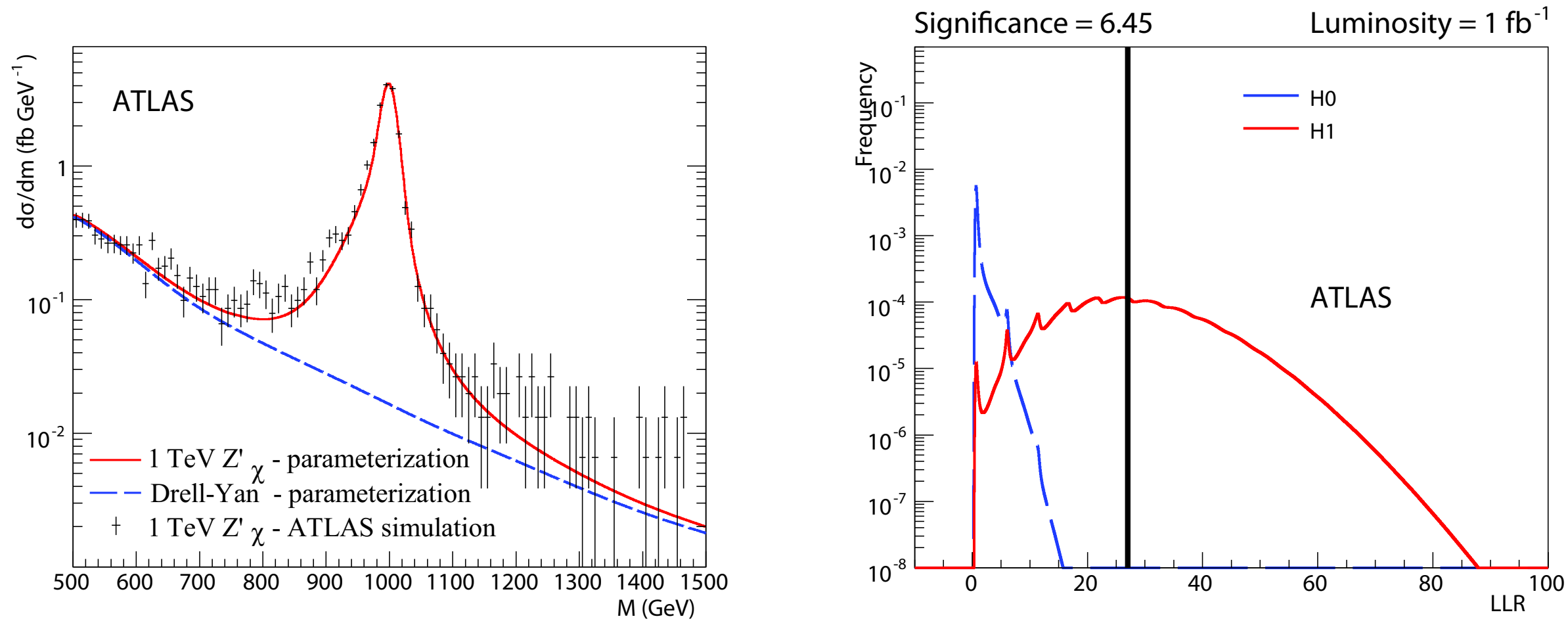


Figure 8: Left: mass spectrum for a $m = 1$ TeV $Z'_\chi \to e^+e^-$ obtained with ATLAS full simulation (histogram) and the parameterization (solid line). The dashed line corresponds to the parameterization of the Drell-Yan process (irreducible background). Right: Log-likelihood ratio densities with 1 fb$^{-1}$ for a $m = 2$ TeV $Z'_\chi$ for the signal and background hypotheses. The vertical line is the median experiment in the H1 hypothesis.

**Results** Using the parameterization presented above to generate mass spectra for signal ($\gamma/Z/Z' \to e^+e^-$) and background ($\gamma/Z \to e^+e^-$), one can compute the distributions of the log-likelihood ratio of the signal (H1) and background (H0) hypotheses.
Figure 8 (right) shows the LLR distributions obtained for a 2 TeV $Z'_\chi$ with 1 fb$^{-1}$ as well as the median signal experiment used to calculate $CL_s$. The FFT method [45] was used in the computation of the LLR distributions. It is important to note that the mass window used to perform the analysis does not affect the result.

Figure 9 (left) shows the integrated luminosity needed for a $5\sigma$ discovery of the usual benchmark $Z'$ models as a function of the $Z'$ mass. Only statistical uncertainties were taken into account. The systematic uncertainties are discussed in the next paragraph. A fixed mass window of [500 GeV – 4 TeV] was used to compute the significance. Roughly speaking, less than 100 pb$^{-1}$ are needed to discover a 1 TeV $Z'$, about 1 fb$^{-1}$ are needed to discover a 2 TeV $Z'$, and about 10 fb$^{-1}$ are needed to discover a 3 TeV $Z'$.

**Systematic Uncertainties** The sources of systematic uncertainties were listed in section 4. Since the main background is the Drell-Yan process, the systematic uncertainties from both the efficiencies and the theoretical predictions on the cross-section will affect the number of signal and background events in the same way, and can be added in quadrature. The uncertainties in the event selection efficiency mainly come from the electron identification and the geometrical acceptance. The former amounts to $2 \times \pm 1\% = \pm 2\%$ for two electrons. Taking the extreme efficiencies for pure $u\bar{u}$ and $d\bar{d}$ events as a conservative estimate, the latter goes from $\pm 3$ to $\pm 0.5\%$. Overall, this represents a systematic uncertainty

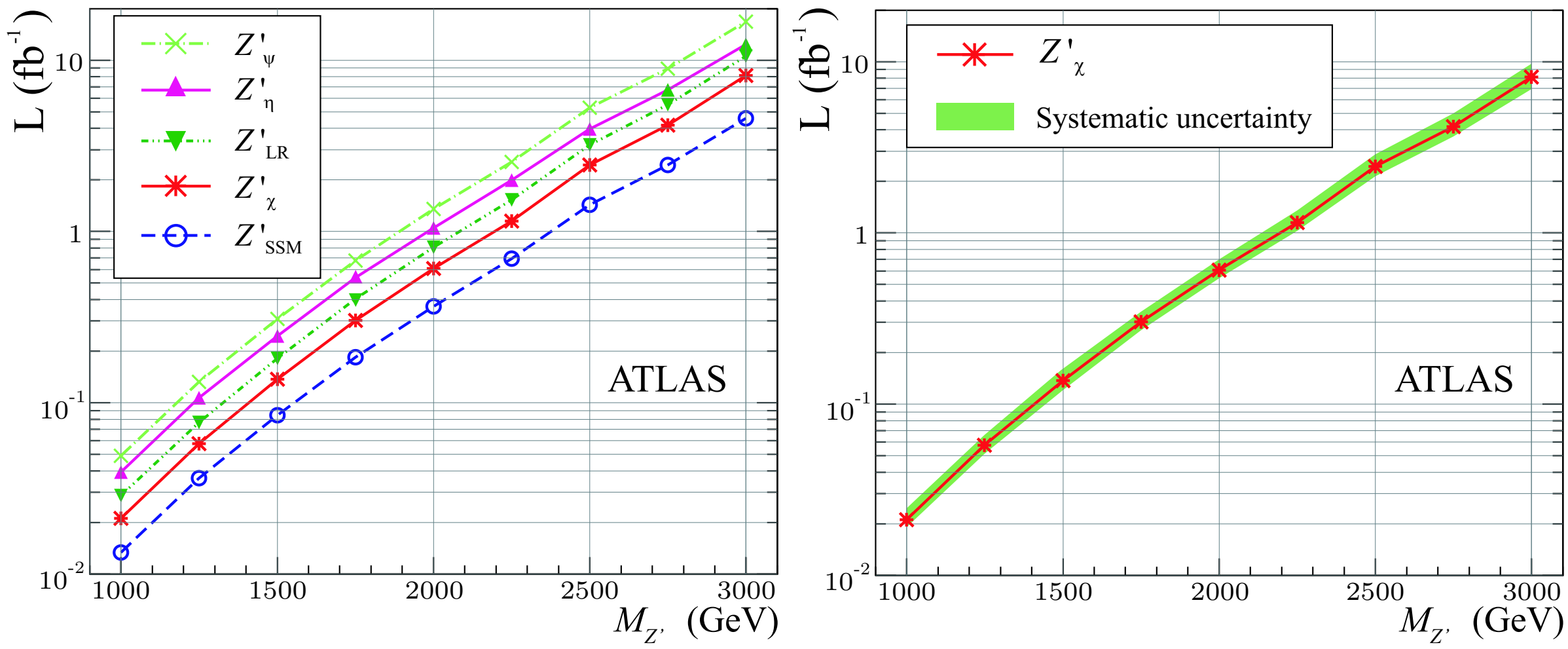


Figure 9: Integrated luminosity needed for a $5\sigma$ discovery of $Z' \to e^+e^-$ as a function of the $Z'$ mass. Left: for various benchmark models with statistical uncertainties only; right: for the $Z'_\chi$ with systematic uncertainties included.

of $\pm 3.6\%$ to $\pm 0.6\%$ from the event selection. This is small as compared to the theoretical uncertainties, which range from $\pm 8.5\%$ to $\pm 14\%$. The effect of these combined uncertainties on the luminosity needed to discover 1, 2 and 3 TeV $Z'$s is $^{+9}_{-10}\%$, $^{+14}_{-10}\%$, $^{+15}_{-13}\%$ (respectively).

The uncertainty in backgrounds other than the Drell-Yan process is another type of uncertainty. However, given that the Drell-Yan contribution is at the level of about 1% of the signal, any variation of the level of non-Drell-Yan background, which is more than ten times smaller, is negligible.

The uncertainty in the electron energy resolution is another type of uncertainty. In addition to the expected uncertainties in the energy resolution as measured in the calorimeter (see section 4), we have conservatively assumed that there was no increase in precision on the measured dielectron invariant mass coming from the angle measurement provided by the tracker. In this case, the resolution of invariant mass increases from about 1% (see section 2.1) to about 1.5%. The effect of these uncertainties on the luminosity needed for a discovery is $^{+5}_{-2}\%$, independent of the Z$'$ mass.

The last type of uncertainty which has been considered is the electron energy scale. When varied within the expected uncertainties, the discovery luminosity varies by $^{+2.5}_{-0}\%$, independent of the Z$'$ mass.

Combining all the above systematic uncertainties, the luminosity needed to discover, for example, a $Z'_\chi$ is shown in Fig. 9 (right). It must be noted that the systematic effect coming from the fact that we do not know a priori the mass of the signal was not taken into account. This is adressed separately in appendix A.

### 5.3 $Z' \to \mu\mu$ Using a Parameterized Fit Approach

The dimuon channel represents an important complement to the dielectron channel. Although the resolution is expected to be up to an order of magnitude worse in the kinematic regime of interest, reducible backgrounds are expected to be considerably lower as discussed in Section 4.1. This feature makes the dimuon channel competitive, especially with early data where the design background rejection may not be achieved. In this section we consider two signal models decaying into dimuons - the $Z'_{SSM}$ and the $Z'_\chi$ boson.

### 5.3.1 Event Selection

To select events from the $Z' \to \mu\mu$ process we require two muons of opposite charge. The muons are required to fulfill the muon identification criteria studied in Section 2.2, including $p_T \geq 30$ GeV and $|\eta| \leq 2.5$ . Events are triggered using the mu20 trigger described in Section 3.2. As seen in Section 4.1, this should select a sample which consists mainly of $Z/\gamma \to \mu\mu$ with limited contamination from other sources of the order of a few percent. Table 9 indicates the effects of the various requirements on both the signal and background samples.

| Sample | $Z'_{SSM}$ (1 TeV) | $Z'_{\chi}$ (1 TeV) | Drell-Yan |
|---|---|---|---|
| Generated | 508.6 | 380.6 | 13.5 |
| $\|\eta\| \leq 2.5$ | 366.8 | 271.5 | 10.8 |
| $p_T \geq 30$ GeV | 364.0 | 270.1 | 10.7 |
| Muon identification | 342.3 | 256.0 | 10.0 |
| Trigger | 325.2 | 243.2 | 9.5 |
| Opposite charge | 324.8 | 243.0 | 9.5 |

Table 9: Selection requirement flow for the $Z' \to \mu\mu$ analysis - cross-sections in fb. Events are counted in a mass window of $\pm 50$ GeV of the resonance mass (signal) and for $m_{\mu\mu} > 800$ GeV (background).

### 5.3.2 Discovery Potential

To evaluate the discovery potential, we use the FFT method [45], as in section 5.2. The amount of data required to discover a $Z'$ boson is computed from the log-likelihood ratio (LLR) of the signal (H1) and background (H0) hypotheses. Figure 10 shows the $1 - CL_b$ obtained as a function of the integrated luminosity for the two studied $Z'$ boson models at $m = 1$ TeV. The largest expected systematic uncertainty (from misalignment of the muon spectrometer) is shown separately. One can see that the amount of luminosity needed for a $5\sigma$ discovery ranges from 20 to 40 pb$^{-1}$, which is competitive with the dielectron channel.

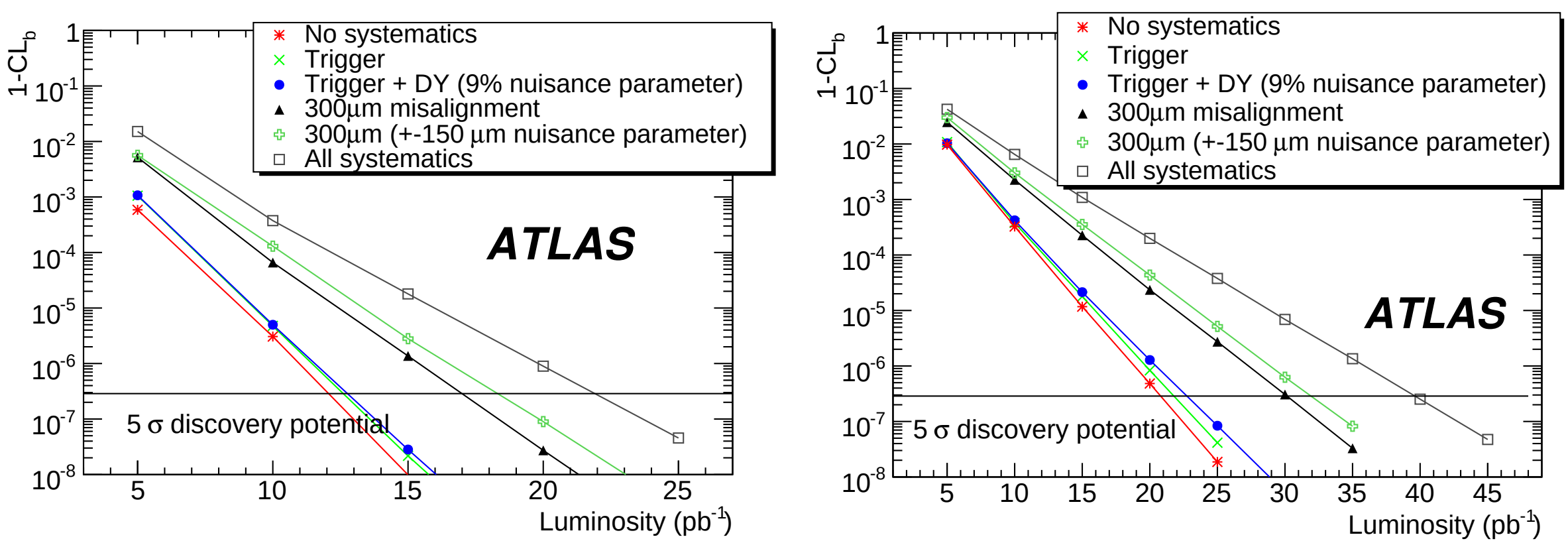


Figure 10: Results of the FFT computation of $1 - CL_b$ for $m = 1$ TeV $Z'_{SSM}$ (left) and $Z'_{\chi}$ (right) bosons. The horizontal line indicates the $1 - CL_b$ value corresponding to $5\sigma$.

**Systematic Uncertainties** Section 4 describes the systematic uncertainties that were considered. As can be seen from Fig. 10 the effect of the nominal systematic uncertainties is modest in this channel. The largest theoretical uncertainty entering this study is the knowledge of the Standard Model Drell-Yan cross-section. In the dimuon channel, the largest experimental uncertainty is the resolution for high $p_T$ muons which will be initially dominated by the alignment of the muon spectrometer. As already discussed, the nominal alignment precision may not be achievable with the integrated luminosities presented here and hence could significantly alter the conclusions. Figure 10 shows that the integrated luminosity needed to reach $5\sigma$ increases from 13 to 20 pb$^{-1}$ if the muon spectrometer is aligned with a precision of 300 $\mu m$. This takes into account an uncertainty of 150 $\mu m$ on the alignment precision estimate, which will have to be measured in data (e.g. from the $Z \to \mu\mu$ sample) and which is treated as a nuisance parameter in the sensitivity computation.

## 5.4 $Z' \to \tau\tau$ Using a Number Counting Approach

The ditau signature is an important component to the high mass resonance search. In particular, there are models in which a hypothetical new resonance couples preferentially to the third generation [47]. For these models the branching ratios are such that the dielectron and dimuon channels are not viable - hence it is critical that we consider all possible channels including ditaus. In this section we discuss the discovery potential for such a resonance. Because of finite resources we restrict ourselves to the process $Z' \to \tau\tau$ with a single mass point $m = 600$ GeV although much of the discussion generalizes to a generic ditau resonance search. The ditau final state can be divided into three final states: hadron-hadron (where both $\tau$ leptons decay hadronically), hadron-lepton (where one $\tau$ lepton decays hadronically and one decays leptonically), and lepton-lepton (where both decay leptonically). Here we consider the hadron-lepton ($h - \ell$) final state. The possibility of observing the hadron-hadron final state using a hadronic $\tau$ trigger will be examined later.

### 5.4.1 Event Selection

To select events in the hadron-lepton final state, we select events with a "hadronic $\tau$" candidate, a charged lepton (muon or electron), and missing transverse energy [8] ($\not\!\!E_T$). As opposed to the dielectron or dimuon channel, the backgrounds to the ditau channel are considerably larger and include Drell-Yan production, $W$+jets, $t\bar{t}$ and dijet events. After the initial object selection several additional requirements are needed to maximize the expected signal significance.

We consider hadronic $\tau$ candidates with $p_T > 60$ GeV and impose a requirement on the likelihood as a function the $\tau$ transverse energy as described in Section 2.3. Candidates which overlap with an electron or muon are removed.

For electron candidates we require *medium* electron selection criteria in this analysis (see Section 2.1). The initial muon selection is the same as described in Section 2.2. Since this channel only requires one high $p_T$ lepton the backgrounds are considerably higher than for the dielectron or dimuon final states. To address this we impose additional requirements on the isolation of the lepton. The isolation requirement imposed on electron candidates is $\sum E_{T_{EM}}^{\Delta R<0.2}/p_T < 0.1$ where $\sum E_{T_{EM}}^{\Delta R<0.2}$ is the sum of the energy deposits in the electromagnetic calorimeter within a cone of $\Delta R = 0.2$ from the location in $\eta$-$\phi$ of the electron, less the electron candidate energy. Isolated electrons are required to have $p_T > 27$ GeV. We impose an isolation requirement similar to that of electrons on muon candidates: $\sum E_{T_{EM}}^{\Delta R<0.2} < 0.1$. For isolated muons we require that the $\chi^2$ lie between 0 and 20 and to be considered by the analysis muons must have $p_T > 22$ GeV.

[8] The missing transverse energy was reconstructed using the cell based algorithm described in [48].

| Selection | Signal | $t\bar{t}$ | Drell-Yan | Multijet | $W$+jet |
|---|---|---|---|---|---|
| Trigger | 1356. | 213600. | 2.3950 $10^7$ | 4.19000 $10^6$ | 6.69400 $10^6$ |
| Lepton | 905. | 150900. | 1.2600 $10^7$ | 1.08230 $10^6$ | 120400. |
| $\tau$ selection | 368. | 7818. | 145680 | 40080 | 4587. |
| Opposite charge | 315. | 2498. | 5306 | 23240 | 771. |
| $\not{E}_T$>30 GeV | 270. | 2040. | 2562 | 835 | 162. |
| $m_T < 35$ GeV | 203.2 | 302.4 | 388.0 | 436.4 | 83.8 |
| $p_T^{tot} < 70$ GeV | 155.0 | 106.7 | 331.5 | 221.6 | 28.4 |
| $m_{vis} > 300$ GeV | 132.5 | 26.2 | 105.6 | 33.8 | 15.0 |
| $\cos\Delta\phi_{\ell h} > -0.99$ | 13.3 | 2.1 | 5.5 | 2.3 | 2.7 |

Table 10: Requirement flow table for the $m = 600$ GeV $Z'_{SSM} \rightarrow \tau\tau \rightarrow \ell h$ analysis - cross-sections given in fb. The Drell-Yan process includes all flavors of leptons ($e^+e^-$, $\mu^+\mu^-$, $\tau^+\tau^-$) with an invariant mass of at least 60 GeV.

After making the $\tau$ candidate selection we make several further requirements to maximize the signal significance. First, we require that $\not{E}_T \geq 30$ GeV. To greatly help with the rejection of the $t\bar{t}$ backgrounds we employ a requirement on the total event $p_T$ which is defined as the sum of $\not{E}_T$and the vector sum of the hadronic $\tau$ with the lepton transverse momentum. We require $p_T^{tot} < 70$ GeV.

The transverse mass of the event is determined by using the lepton kinematics and the event $\not{E}_T$. Defining a four-vector for the missing energy: $\underline{\not{p}_T} = (\not{E}_{T_x}, \not{E}_{T_y}, 0, |\not{E}_T|)$, the transverse mass is calculated as:

$$m_T = \sqrt{2 p_{T,\ell} \not{p}_T (1 - \cos\Delta\phi_{\ell,\not{p}_T})}.$$

We require that $m_T < 35$ GeV.

In the case of the lepton-hadron channel one cannot simply reconstruct the invariant mass of the resonance as energy is taken away from the event by the neutrinos. However, two quantities can be constructed

- A visible mass variable is calculated as defined by CDF [49] using the hadronic $\tau$ and the lepton four-vector information:
$$m_{vis} = \sqrt{(\underline{p_\ell} + \underline{p_h} + \underline{\not{p}_T})^2}$$
- The collinear approximation is used to build up the event-by-event invariant mass. The fraction of the $\tau$ momentum carried by the visible decay daughters, $x_\ell$ and $x_h$, are calculated with the following formulas:
$$x_\ell = \frac{p_{x,\ell} p_{y,h} - p_{x,h} p_{y,\ell}}{p_{y,h} p_{x,\ell} + p_{y,h} \not{p}_x - p_{x,h} p_{y,\ell} - p_{x,h} \not{p}_y},\ x_h = \frac{p_{x,\ell} p_{y,h} - p_{x,h} p_{y,\ell}}{p_{y,h} p_{x,\ell} + p_{x,\ell} \not{p}_y - p_{x,h} p_{y,\ell} - p_{y,\ell} \not{p}_x}.$$
The reconstructed mass is then calculated as $m_{\tau\tau} = \frac{m_{\ell,h}}{\sqrt{x_\ell x_h}}$.

To greatly help the background rejection and to restrict our search to the region of interest we require $m_{vis} > 300$ GeV. Since the collinear approximation breaks down when the two $\tau$ leptons are back-to-back, we impose the requirement that $\cos\Delta\phi_{\ell h} > -0.99$. Of course, since a very heavy particle tends to be produced at rest, the decay objects are mostly back-to-back, leading to a highly inefficient mass reconstruction.

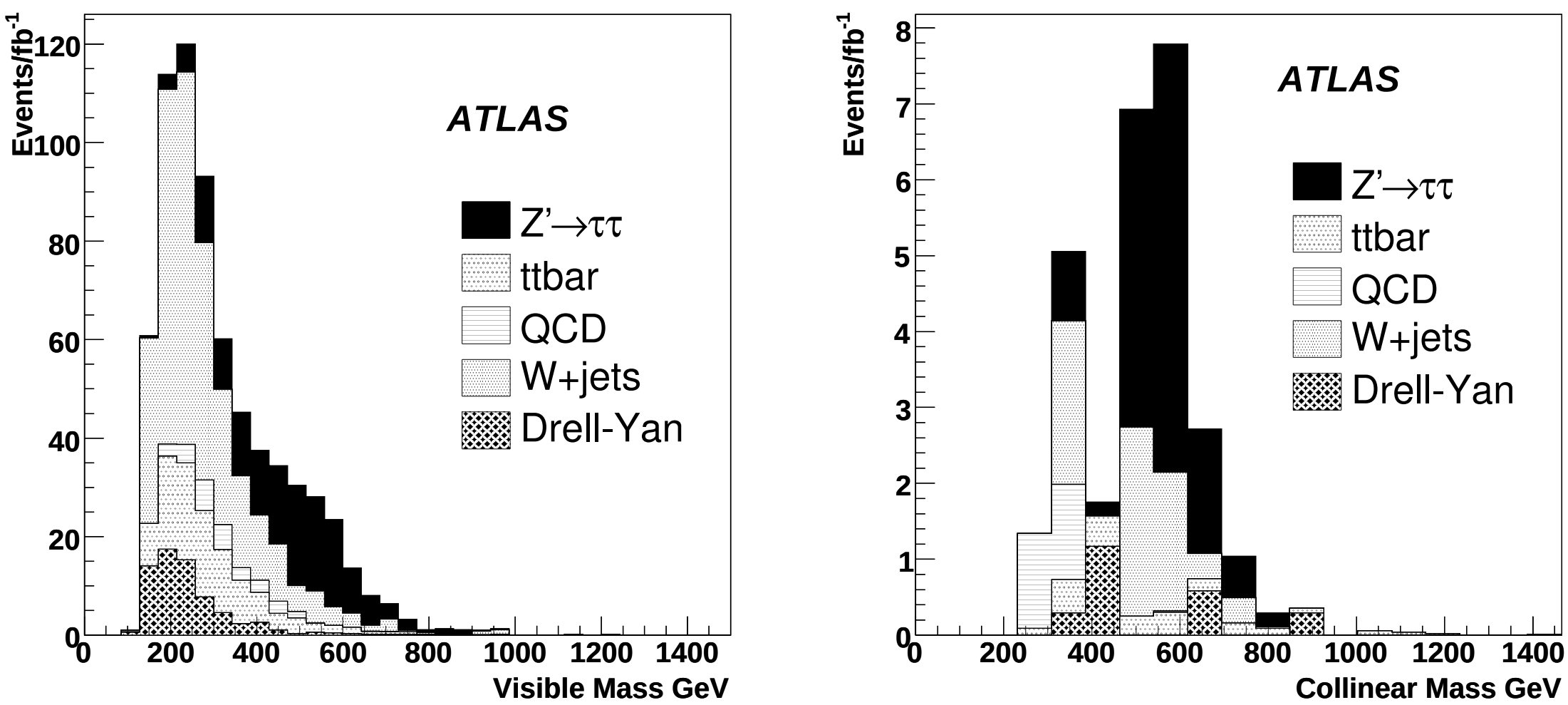


Figure 11: Left: the visible mass distribution in the $Z' \to \tau\tau \to \ell h$ analysis for signal and background processes (1 fb$^{-1}$ of data is assumed). Right: the reconstructed invariant mass obtained using the collinear approximation.

### 5.4.2 Discovery Potential for 1 fb$^{-1}$ of Data

Table 10 shows the effect of the various selection requirements for the signal as well as all background processes considered. Distributions of the visible and reconstructed masses for signal and background are shown in Fig. 11. Here we assume a 600 GeV $Z'$ and the SSM cross-section. In 1 fb$^{-1}$ of ATLAS data we estimate 132. signal events and 181. background events after imposing the event selection up to the requirement on visible mass. Using $S/\sqrt{B}$ we estimate the signal significance to be 9.9. The collinear approximation breaks down when the two $\tau$ leptons are back-to-back, so that even a loose requirement (such as $\cos\Delta\phi_{\ell h} > -0.99$) reduces the signal by a large factor. Hence, we expect that the search will proceed by looking at the visible mass. If a significant excess over background is seen, the collinear approximation will then be used to help establish the presence of a new resonance.

**Systematic uncertainties** The systematic uncertainties that were considered are described in Section 4. For an analysis of 1 fb$^{-1}$ of data the dominant systematic source on the signal, just over $\pm 18\%$, comes from the uncertainty in the luminosity. The second most dominant systematic, the hadronic $\tau$ energy scale, affects the signal at the $\pm 10\%$ level. Summing in quadrature the effect of all systematic uncertainties on the signal Monte Carlo sample results in a total systematic uncertainty of about $\pm 20\%$. The current Monte Carlo samples available for the backgrounds to the ditau analysis are statistically limited and hence prevent a rigorous evaluation of the systematics at this time. As a conservative estimate, we assume that the total systematic uncertainty in the backgrounds is identical to that observed in the signal Monte Carlo. This is a conservative estimate because the majority of the backgrounds in the data have very large cross-sections (dijets, W+jets, etc.) and in principle the evaluation of systematic uncertainties there should be less sensitive to statistical fluctuations than for the signal events. Summing these systematic uncertainties in quadrature and using the formula $S/\sqrt{B+\delta B^2}$ gives a significance of 3.4 in 1 fb$^{-1}$.

### 5.5 $G \to e^+e^-$ in a Parameterized Fit Approach

In this section we present a sensitivity study for the Randall-Sundrum $G \to ee$ final state. In this channel, it is assumed that there is no interference between the $G$ and the dilepton background. Table 11 shows the parameters of the different $G$ samples used in this analysis. $\Gamma_G$ is the simulated graviton resonance width and $\sigma_m$ stands for the width of the observed resonance after convolution with detector resolutions. For $k/\bar{M}_{pl} < 0.06$ the resonance is narrow compared to the experimental resolution.

The main Standard Model background is neutral Drell-Yan production. Other backgrounds such as dijets with both jets misidentified as electrons are expected to be small and neglected at this time.

| Model Parameters | | $\Gamma_G$ | $\sigma_m$ | $\sigma \cdot BR(G \to e^+e^-)$ |
|---|---|---|---|---|
| $m_G$ | $k/\bar{M}_{pl}$ | [GeV] | [GeV] | [fb] |
| 500 GeV | 0.01 | 0.08 | 4.6 | 187.4 |
| 750 GeV | 0.01 | 0.10 | 6.4 | 27.7 |
| 1.0 TeV | 0.02 | 0.57 | 7.9 | 26.0 |
| 1.2 TeV | 0.03 | 1.62 | 10.3 | 22.4 |
| 1.3 TeV | 0.04 | 2.98 | 11.4 | 25.3 |
| 1.4 TeV | 0.05 | 5.02 | 13.1 | 26.8 |

Table 11: Parameters of the $G \to ee$ samples used: natural width ($\Gamma_G$), Gaussian width after detector effects ($\sigma_m$) and leading order cross-section.

#### 5.5.1 Event Selection

In reconstructing the resonance mass, we require a pair of electrons – we do not make any charge requirements – with $p_T \geq 65$ GeV using the *loose* electron selection criteria described in Section 2.1. We require that the events pass the e55 single electron trigger (see section 3.1). Finally we require that the two electrons are roughly back-to-back in $\phi$ with $\cos\Delta\phi_{ee} < 0$ between the two electrons. Table 12 shows the remaining cross-section at each stage of the selection and the total eficiency for different mass points. The efficiency decreases at high graviton masses, due to the track match requirement, which is consistent with the $Z'$ boson analysis (see section 5.2). Table 13 shows the same requirement flow for the Drell-Yan.

The Drell-Yan background distribution after this event selection is shown in Fig. 12 along with signal at $m_G = 1$ TeV and coupling $k/\bar{M}_{pl} = 0.02$. The exponential described in Section 5.1 has been used to model the shape of the background.

#### 5.5.2 Discovery Potential

We search for an excess of events in the mass range from 300 GeV up to 2 TeV and study the signal sensitivity by use of "extended maximum likelihood" fitting. We consider two hypotheses. The null hypothesis, H0, is the hypothesis that the data are described by the Standard Model. The test hypothesis, H1, is that the data are described by the sum of the background and a narrow Gaussian resonance.

To investigate the potential for discovery pseudo-experiments are generated from both the null and test hypothesis. Each pseudo-experiment is fit twice. The first fit assumes the data are described by

| Selection / Sample | 500 GeV | 750 GeV | 1.0 TeV | 1.2 TeV | 1.3 TeV | 1.4 TeV |
|---|---|---|---|---|---|---|
| Generated | 187.4 | 27.7 | 26.0 | 22.4 | 25.3 | 26.8 |
| Acceptance | 172.4 | 25.9 | 24.7 | 21.2 | 24.0 | 25.4 |
| Trigger | 168.7 | 25.0 | 22.6 | 19.1 | 21.4 | 22.3 |
| Electron Id. | 127.9 | 18.3 | 16.4 | 12.8 | 14.6 | 14.7 |
| $p_T \geq 65$ GeV | 125.7 | 18.2 | 16.3 | 12.7 | 14.5 | 14.6 |
| $\cos\Delta\phi_{ee} < 0$ | 123.0 | 17.8 | 16.0 | 12.6 | 14.3 | 14.4 |
| Selection efficiency (%) | $65.6 \pm 1.1$ | $64.4 \pm 1.1$ | $61.7 \pm 1.1$ | $56.3 \pm 1.1$ | $56.4 \pm 1.1$ | $53.9 \pm 1.1$ |

Table 12: Requirement flow for the $G \rightarrow ee$ analysis. The remaining cross-section (in fb) is given at each stage. The mass window is chosen as $\pm 4\sigma_m$ around the signal peak.

| Selection/Sample | 500 GeV | 750 GeV | 1.0 TeV | 1.2 TeV | 1.3 TeV | 1.4 TeV |
|---|---|---|---|---|---|---|
| Generated | 20.33 | 4.91 | 1.43 | 0.90 | 0.51 | 0.51 |
| Acceptance | 18.53 | 4.50 | 1.36 | 0.87 | 0.48 | 0.49 |
| Trigger | 18.45 | 4.25 | 1.16 | 0.80 | 0.45 | 0.44 |
| Electron Id. | 14.13 | 3.18 | 0.88 | 0.58 | 0.38 | 0.33 |
| $p_T \geq 65$ GeV | 13.85 | 3.15 | 0.88 | 0.57 | 0.38 | 0.33 |
| $cos\Delta\phi_{ee} < 0$ | 13.41 | 3.09 | 0.85 | 0.56 | 0.36 | 0.33 |

Table 13: Remaining Drell-Yan cross-section (in fb) at each stage of the $G \rightarrow ee$ analysis. The mass window is chosen as $\pm 4\sigma_m$ around the signal peak.

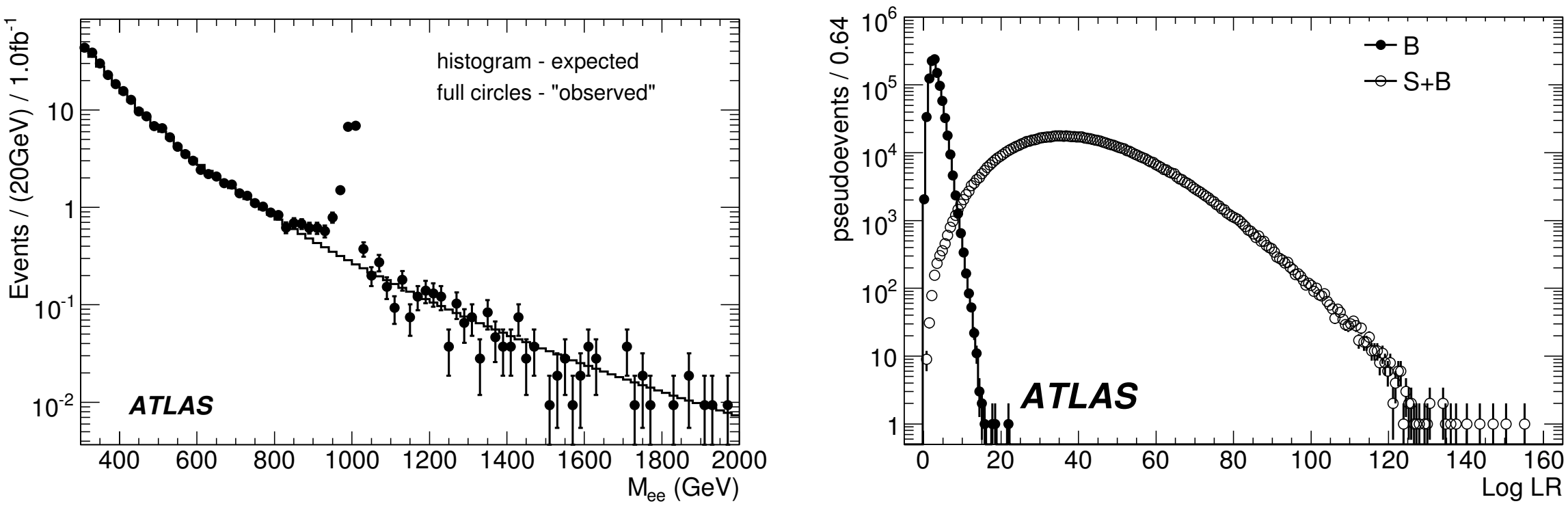


Figure 12: Left: expected (histogram) and "observed" (filled circles) Drell-Yan spectrum from full simulation. The observed distribution includes a graviton with mass of 1 TeV and coupling $k/\bar{M}_{pl} = 0.02$. Note that for the purposes of this plot the vertical axis has been rescaled: the error bars correspond to an integrated luminosity of 100 fb$^{-1}$. Right: Log likelihood ratio curves for one million pseudo-experiments generated with background only (filled circles), and signal plus background (empty circles) for the same $m = 1$ TeV signal point.

the Standard Model using the function described in Section 5.1. The second fit assumes the data are described by the sum of a Gaussian and the shape describing the Drell-Yan background. During this second fit the mean of the Gaussian is allowed to float throughout the entire mass region considered, and the width is fixed to the detector resolution.

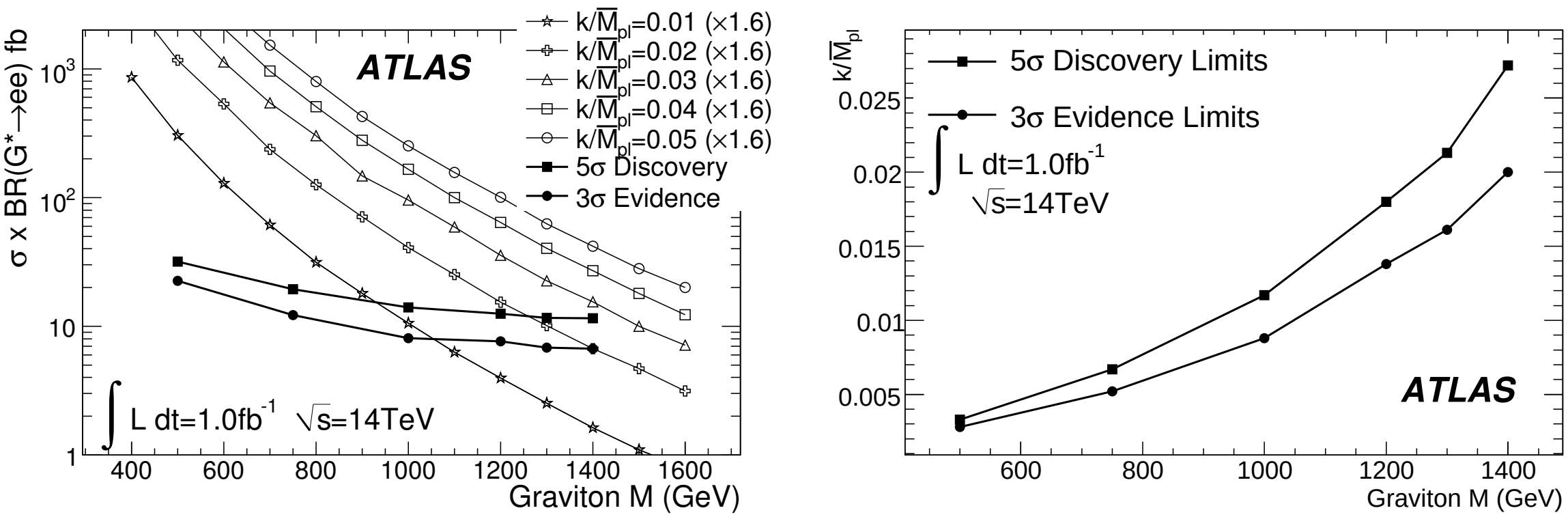


Figure 13: 5$\sigma$ discovery potential (full squares) as a function of the graviton mass. The 3$\sigma$ evidence potential is also shown (full circles). Left: shown with cross-sections as calculated by PYTHIA (LO) and multiplied by a $K$ factor of 1.6 for several values of the coupling; right: dependence of the discovery potential on the coupling.

We can then compare the likelihood of the signal and background hypotheses. The distribution of the logarithm of the likelihood ratio between H0 and H1 is constructed, and shown for one signal point in Fig. 12. Based on this, we calculate the average expected discovery potential from the fraction of the likelihood ratio distribution for background-only pseudo-experiments that extends beyond the mean of the distribution for signal plus background experiments. Figure 13 shows the 5$\sigma$ discovery and 3$\sigma$ evidence reach in cross-section and $k/\bar{M}_{pl}$ coupling constant as a function of graviton mass, estimated

for an integrated luminosity of 1 $fb^{-1}$.

The LO cross-sections are multiplied by the $K$-factors discussed in section 4.2.2 for both signal and Drell-Yan background. Various sources of systematic uncertainties for signal and background are considered in the evaluation of the experimental sensitivity, including luminosity, energy scale, energy resolution, electron identification efficiency and Drell-Yan background uncertainties as listed in section 4.4. The combined effect of the systematic uncertainties is to increase the amount of integrated luminosity needed for discovery between 10 and 15 percent for the different parameter sets.

## 5.6 Technicolor Using a Non-Parameterized Approach

Topcolor-assisted Technicolor models with walking gauge coupling predict new technihadron states that would be copiously produced at the LHC. The lowest mass states are the scalar technipions ($\pi_T^{\pm,0}$) and the vector technirho and techniomega ($\rho_T^{\pm,0}$ and $\omega_T^0$). The vector mesons decay into a gauge boson plus technipion ($\gamma\pi_T$, $W\pi_T$ or $Z\pi_T$,) and fermion-antifermion pairs. This analysis searches for the decays $\rho_T \to \mu^+\mu^-$ and $\omega_T \to \mu^+\mu^-$. The dimuon mode has a lower branching fraction than the modes involving technipions but the signal is clean, straightforward to trigger on, and can be readily observed with early ATLAS data.

The particular model studied here is the "Technicolor Strawman Model" or TCSM [12, 13]. In the TCSM, it is expected that techni-isospin is an approximate good symmetry and therefore the isotriplet $\rho_T$ and isosinglet $\omega_T$ will be nearly degenerate. We will assume for what follows that $m_{\rho_T} = m_{\omega_T}$. The technipions are also expected to be nearly degenerate. In the TCSM, the technipion masses are generically not small. In particular, if $m_{\pi_T} > m_{\rho_T}/2$ the decays of the $\rho_T$ and $\omega_T$ to technipions would be kinematically forbidden [50]. The dimuon rate is expected to come dominantly from the $\omega_T$ with a smaller contribution from the $\rho_T$.

The event selection is summarized in Table 14. The technivector meson natural widths are less than a GeV, so the observed width $\sigma(m)$ is entirely due to detector resolution.

In principle, the best search sensitivity is not obtained by examining the entire dimuon mass distribution for a bump all at once but by using an optimized mass window that maximizes the signal significance for a given assumed signal mass. A prescription for the optimal window size is taken from an analytic calculation in Ref. [51]. Assuming a narrow Gaussian peak on a linear background, the optimal window was found to be $\pm 1.4\sigma$ about the peak mass. Since we are not really in the narrow resonance regime, we did a study using full-simulation ATLAS Monte Carlo for a Technicolor signal on a Drell-Yan background. Taking $S/\sqrt{B}$ as our measure of significance, Fig. 14 (left) shows that a window size of $\pm \sim 1.5\sigma$ or a bit larger is optimal. For this study, a window size of $\pm 1.5\sigma$ about the peak mass is used.

Figure 14 (right) shows the integrated luminosity necessary to observe either $3\sigma$ evidence or a $5\sigma$ discovery, using the modified frequentist approach [44], of technimesons in this channel. The systematic uncertainties summarized in section 4.4 were included in this calculation of technimeson search sensitivity. It should be noted that the integrated luminosity needed for $5\sigma$ discovery will be affected by the level of misalignment of the muon spectrometer. The contours in Fig. 14 were computed assuming the level of alignment we expect to achieve. The studies in sections 4.3 and 5.3 show that for an initial precision of 300 $\mu$m with an uncertainty of 150 $\mu$m the amount of data needed to reach $5\sigma$ would increase by approximately 50%.

| $m_{\rho_T,\omega_T}$ (GeV) | 400 | 600 | 800 | 1000 |
|---|---|---|---|---|
| Peak mass (GeV) | 403 | 603 | 804 | 1004 |
| $\sigma(m)$ (GeV) | 13 | 22 | 34 | 46 |
| Requirement | | | | |
| Generated | 201 | 60.8 | 23.0 | 10.1 |
| $\|\eta\| < 2.5$ | 116 | 39.8 | 15.8 | 7.3 |
| $p_T > 30$ GeV | 114 | 39.5 | 15.7 | 7.2 |
| L1_MU20 | 112 | 38.7 | 15.3 | 7.0 |
| L2_mu20 | 110 | 38.0 | 15.1 | 6.9 |
| EF_mu20 | 109 | 37.5 | 14.9 | 6.8 |
| Match $\chi^2 < 100$ | 104 | 35.7 | 14.0 | 6.4 |
| Opposite charge | 104 | 35.7 | 14.0 | 6.4 |
| Mass window | 78.2 | 26.3 | 10.3 | 4.7 |
| Drell-Yan background | 46.9 | 14.1 | 6.1 | 2.8 |
| Selection efficiency (%) | $38.9 \pm 0.5$ | $43.2 \pm 0.5$ | $44.8 \pm 0.5$ | $46.8 \pm 0.5$ |

Table 14: Selection requirement flow for the analysis - cross-section in fb.

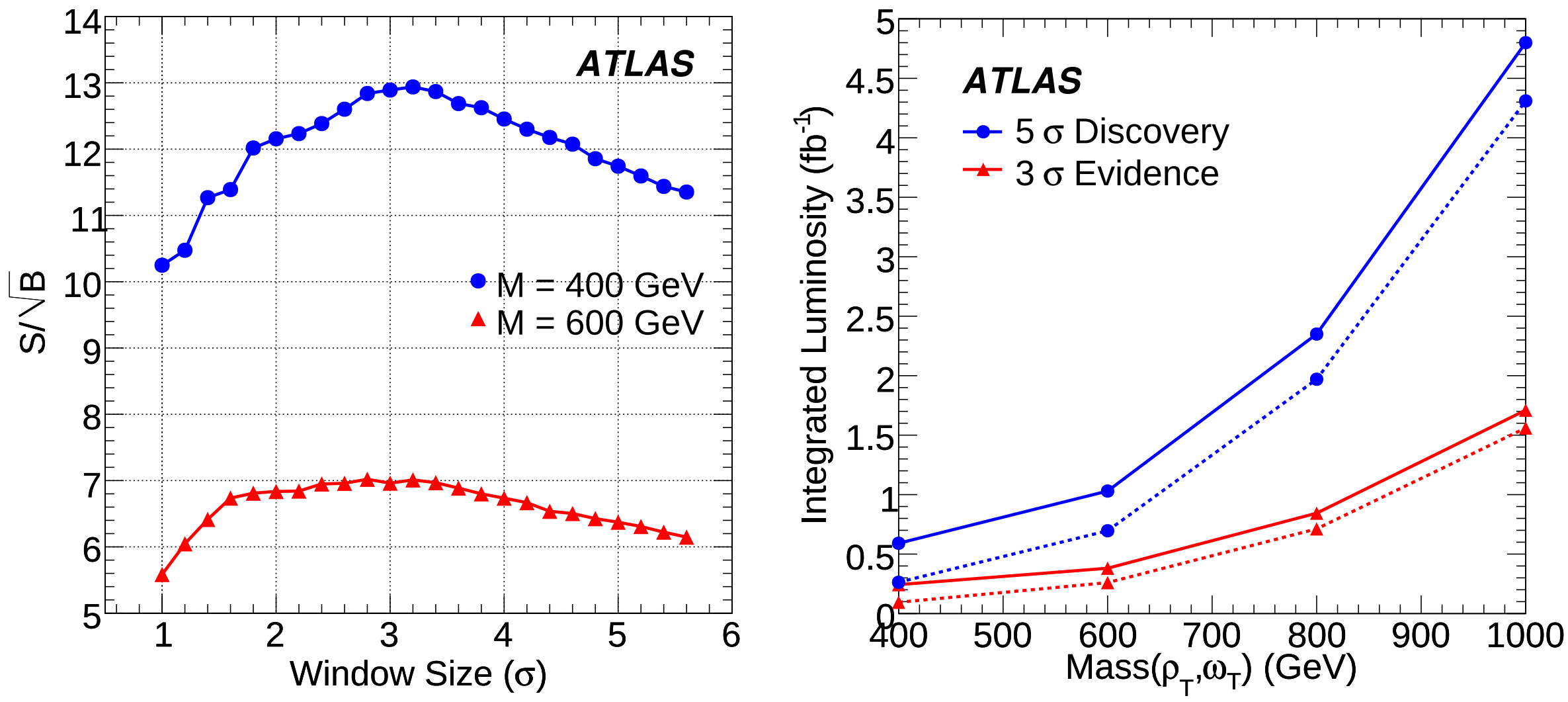


Figure 14: Left: for two different $\rho_T, \omega_T$ signal masses, $S/\sqrt{B}$ is plotted as a function of mass-window size for windows centered on the peak mass. Right: integrated luminosity needed for $3\sigma$ evidence or $5\sigma$ discovery as a function of $\rho_T, \omega_T$ mass. The dashed lines include only statistical uncertainties while the solid lines contain the systematic uncertainties as well.

# 6 Summary and Conclusions

Several models which lead to resonances in the dilepton final state have been studied. Various systematic studies have been undertaken which estimate the effect of uncertainties from both theoretical knowledge of Standard Model processes as well as expected and assumed early detector performance. Data-driven methods have been developed to evaluate efficiencies, backgrounds, and uncertainties. It has been shown that even with early data the discovery potential can be dramatically increased from current limits. The discovery potential with an integrated luminosity of 10 $\text{fb}^{-1}$ depends on the particular model and varies in the $m = 1.0$ to 3.5 TeV range. It should be noted that resonance masses above 1 TeV which are unreachable by the Tevatron experiments could be discovered with 100 $\text{pb}^{-1}$ of data already.

# A Effect of the Unknown Location and Rate

When estimating the significance of a local excess of events, the size of the region considered and uncertainties in the shape of the background can significantly reduce the sensitivity of the search. This appendix presents an assessment of the size of this effect for the $Z'$ boson to dilepton searches. If an excess is found in the dilepton invariant mass, its significance needs to be evaluated in a way that takes into account the possibility of background fluctuations of different masses, cross-sections and widths. One possible way to do this is through the use of maximum likelihood fits, where these quantities are free parameters.

To estimate the effect on the sensitivity of the unknown rate and location of a dilepton resonance, the decay $Z'_{SSM} \to ee$ and $Z'_{SSM} \to \mu\mu$ were both generated for 16 true $Z'$ masses between 1 and 4 TeV (evenly spaced every 200 GeV), with a lower cut on the true dilepton mass of 0.5 TeV in all cases. Each sample was simulated and reconstructed using fast simulation, and events were required to have two back-to-back ($\Delta\phi > 2.9$) leptons of opposite charge with $p_T > 20$ GeV and within $|\eta| < 2.5$. For an estimation of the expected background, Standard Model Drell-Yan production was used.

The dilepton resonance was modeled using an ad-hoc parameterization that models appropriately the shapes of both the $Z' \to ee$ and $Z' \to \mu\mu$ modes, consisting of a product between a Breit-Wigner and a Landau distribution with a common mean, and where the width of the Landau was parameterized as a function of the width of the Breit-Wigner[9]. The common mean, the width parameter and the amplitude of the signal are allowed to float in the fits.

Figure 15 shows the likelihood ratio distributions for an $m = 3$ TeV $Z'_{SSM} \to ee$ fit-based significance, where the signal rate, the peak's width and the mean mass all float in the fit, corresponding to an integrated luminosity of 4 $\text{fb}^{-1}$. The distributions of the log-likelihood ratio for fits to H0 pseudo-experiments and for fits to H1 pseudo-experiments are shown. The fraction $p$ of the H0 distribution that has a likelihood ratio larger than the mean of the H1 distribution is shaded. The value of $p$ is then transformed into a significance following the convention under which $p = 2.87 \times 10^{-7}$ corresponds to $5\sigma$ (see section 5). The fraction shown in the plot corresponds to a significance of $4.29\sigma$.

Several million pseudo-experiments were generated and fit, covering different masses and luminosities. Figure 16 shows the significance for different approaches in the case of an $m = 3$ TeV $Z'_{SSM}$ for both the dielectron (left) and the dimuon (right) cases. The plots compare the significance as obtained from number counting (circles), fixed mass fits (dots) and floating mass fits (squares). The floating-mass significances are on average 20% lower than the fixed-mass calculations for $Z' \to ee$, and about 15% lower in the dimuon case (in obtaining these numbers, we exclude the region below 2.25 $\text{fb}^{-1}$, which is affected by low statistics effects).

[9] The best motivated shape is a Breit-Wigner convoluted with a Gaussian resolution. Unfortunately, the convolution fit is very time consuming and for this study millions of fits were performed. Empirically the combination of a Breit-Wigner and a Landau were found to give essentially identical results.

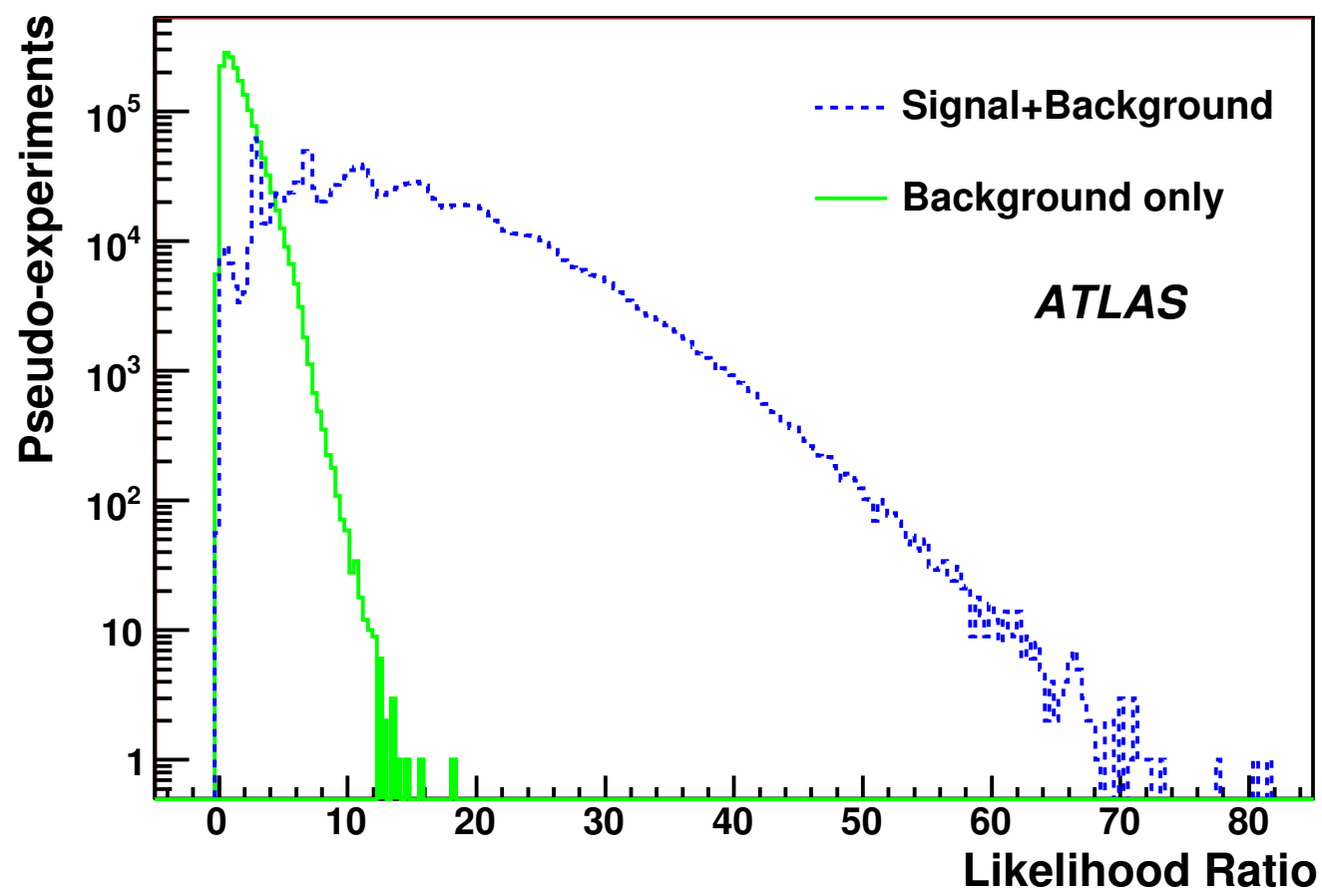


Figure 15: Likelihood ratio distribution for an $m = 3$ TeV $Z'_{SSM} \to ee$; the distribution on the left corresponds to background-only pseudo-experiments; the one on the right, to signal plus background.

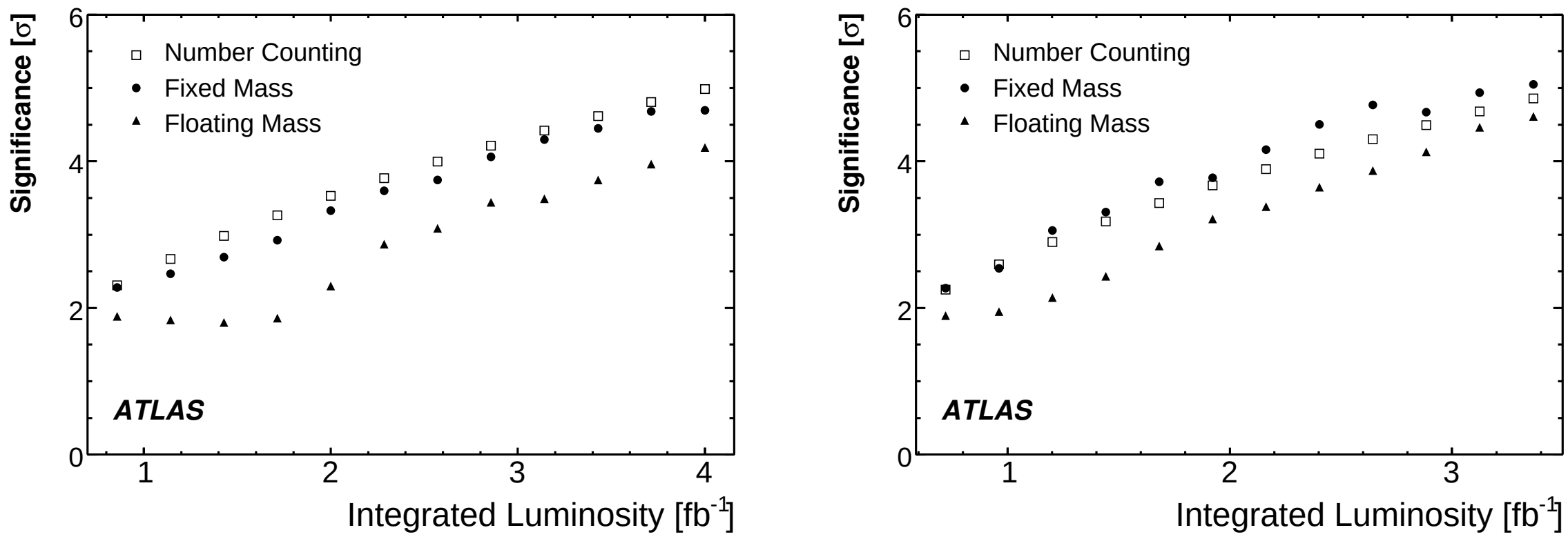


Figure 16: Comparison of the fit-based significance for fixed-mass (dots) and floating-mass (squares) fits for both cases, $Z' \to ee$ (left) and $Z' \to \mu\mu$ (right). Circles show the estimation from number counting.

# Lepton plus Missing Transverse Energy Signals at High Mass

**Abstract**
The prospects for the discovery of heavy lepton-neutrino resonances with the ATLAS detector are evaluated using full detector simulation. The performance of large missing transverse momentum measurement is studied. Its impact on the lepton-neutrino transverse mass reconstruction, and on the backgrounds rejection, is then discussed. As benchmark, the sensitivity to a Standard Model like $W'$ is evaluated. Emphasis is put on the discovery potential of ATLAS with early data, namely with an integrated luminosity of 10 pb$^{-1}$ to 10 fb$^{-1}$.

## 1 Introduction

The Standard Model of particle physics has been able to predict or describe, within errors, almost all measurements performed within its domain. However, several fundamental questions remain unresolved. Its mechanism for electroweak symmetry breaking has not been experimentally confirmed. The model parameters still lack a theoretical explanation. There are indications, therefore, that the Standard Model is not a fundamental theory, but a good approximation of nature at the energy ranges that have been so far accessible to experiment. Thus, the search for physics beyond the Standard Model is an important part of the ATLAS physics program. In this document, a study is presented of the potential for the search of final states comprised of one electron or muon (*lepton*, in what follows) plus missing transverse energy.

A large variety of theories beyond the Standard Model, predict additional gauge bosons. Any charged, spin 1 gauge boson which is not included in the Standard Model is called $W'$ boson and according to several predictions there is at least one $W'$ boson detectable at the LHC. These theories and models which predict new charged gauge bosons range from the Grand Unified Theories [1–3], the various Left-Right Symmetric Models [1, 4–10], Kaluza-Klein theories [11–15], Little Higgs models [16–18], dynamical symmetry breaking models [19] and even models inspired from superstrings [20–22]. As an example, the 45 decompositions of the SO(10) gauge group, which is a candidate for large GUT symmetries, under the $SU(3)_C \times SU(2)_L \times SU(2)_R \times U(1)_{B-L}$ gives rise to a (1,1,3,0) triplet coming from the $SU(2)_R$ group. That is, a triplet of right-handed $W^{\pm,0}$ fields, which carry weak (V+A) interactions. A theoretical model, based on the gauge group $SU(3)_C \times SU(2)_L \times SU(2)_R \times U(1)_{B-L}$ which is called a Left-Right Symmetric Model (LRSM), after spontaneous symmetry breaking, predicts a right-handed $W_R$ gauge boson mixes with the left-handed $W_L$ boson of the Standard Model. The $W_R$ gauge boson is a very attractive $W'$ boson candidate. The search for these particles is an important part of the studies for new physics to be performed at LHC. Studies presented here are based on predictions of a "Standard Model-like" $W'$ boson from so-called extended gauge models [23]. This $W'$ boson has Standard Model-like couplings to fermions and its decays to $WZ$ bosons are suppressed.

The D0 experiment, at Fermilab, has set the present lower limit for the $W'$ boson mass [24] to $m_{W'} > 1$ TeV at 95% C.L. The LHC, with a centre-of-mass energy of 14 TeV, is expected to increase the search reach even at early stages of data taking. Other ATLAS studies have evaluated the potential for discovery of $W' \to \ell\nu_\ell$ where $\ell = \mu, e$ [25]. This study is based on the most recent realistic detector description, including a complete simulation of the trigger chain.

The remaining of this paper is organized as follows. The Monte Carlo samples which were used are summarized in section 2. Section 3 discusses the expected performance on lepton reconstruction, as well as on missing transverse energy. Section 4 briefly describes the triggers which were used in this study. The event selection is discussed in section 5. The discovery potential of a $W'$ with Standard Model couplings is assessed in section 7, after examining the systematic uncertainties in section 6.

## 2 Monte Carlo Samples

Table 1 summarizes the samples used in this study; a detailed account of the procedures, generators and settings used is given in [26]. Signal samples for masses other than 1 and 2 TeV were produced locally and validated against central production samples.

For the signal, samples of $W' \to \ell\nu$ events were generated with PYTHIA v6.403 [27], based on the leading order cross sections and the parton distribution functions CTEQ6 [28], where $\ell$ can be any type of lepton ($\tau$ included), for true $W'$ boson masses ranging from 1 to 4 TeV.

The main background for a $W'$-type state is the high-mass tail of Standard Model $W$ boson production; in order to provide enough background to study also the higher $W'$ boson masses, two samples of Standard Model $W$ boson events were produced, with different requirements on the true invariant mass of the $W$ boson: one with 200 GeV $< m_W <$ 500 GeV, and one with $m_W >$ 500 GeV. In these studies, the alignment and calibration of the detector is assumed to be well described in the reconstruction algorithms.

| Process | Generator | $\sigma \times BR$ [fb] | Comments | Events |
|---|---|---|---|---|
| 1 TeV $W' \to \ell\nu$ | PYTHIA | 9430. | | 30K |
| 1.5 TeV $W' \to \ell\nu$ | PYTHIA | 1786. | | 2.8K |
| 2 TeV $W' \to \ell\nu$ | PYTHIA | 437. | | 30K |
| 2.5 TeV $W' \to \ell\nu$ | PYTHIA | 146. | | 2K |
| 3 TeV $W' \to \ell\nu$ | PYTHIA | 54. | | 10K |
| 3.5 TeV $W' \to \ell\nu$ | PYTHIA | 20. | | 2.8K |
| Standard Model $W \to \ell\nu$ | PYTHIA | 18721.1 | 200 GeV$< m_W <$500 GeV | 20K |
| Standard Model $W \to \ell\nu$ | PYTHIA | 708.26 | $m_W >$500 GeV | 20K |
| $t\bar{t}$ | MC@NLO | 452000 | | 340K |
| Dijet J0 | PYTHIA | $1.76 \times 10^{13}$ | $\hat{p_T} = 8 - 17$ GeV | 380K |
| Dijet J1 | PYTHIA | $1.38 \times 10^{12}$ | $\hat{p_T} = 17 - 35$ GeV | 380K |
| Dijet J2 | PYTHIA | $9.33 \times 10^{10}$ | $\hat{p_T} = 35 - 70$ GeV | 390K |
| Dijet J3 | PYTHIA | $5.88 \times 10^{9}$ | $\hat{p_T} = 70 - 140$ GeV | 380K |
| Dijet J4 | PYTHIA | $3.08 \times 10^{8}$ | $\hat{p_T} = 140 - 280$ GeV | 390K |
| Dijet J5 | PYTHIA | $1.25 \times 10^{7}$ | $\hat{p_T} = 280 - 560$ GeV | 370K |
| Dijet J6 | PYTHIA | $3.60 \times 10^{5}$ | $\hat{p_T} = 560 - 1120$ GeV | 380K |
| Dijet J7 | PYTHIA | $5.71 \times 10^{3}$ | $\hat{p_T} = 1120 - 2240$ GeV | 430K |

Table 1: Monte Carlo samples used for the study of $W'$ bosons. $\hat{p_T}$ represents the transverse momentum of the partons in their rest frame. The $t\bar{t}$ sample includes only fully leptonic and semi-leptonic channels.

## 3 Reconstruction Performance

### 3.1 Muon Reconstruction

Muon reconstruction in ATLAS uses all main detector subsystems. The Muon Spectrometer (MS) is designed to provide efficient and precise stand-alone momentum measurement for muons of transverse momentum up to O($p_T$= 1 TeV). During the back-tracking of the muon to the production vertex, energy loss fluctuations can be measured with the use of the calorimeters, which can also provide independent

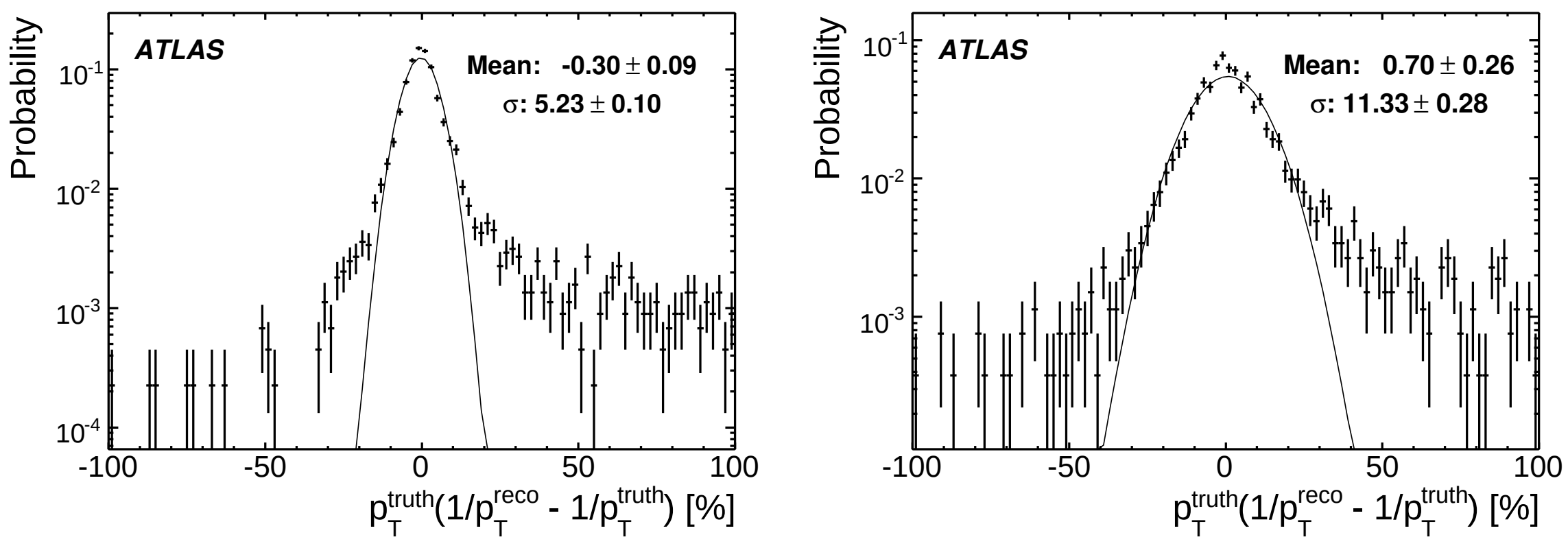


Figure 1: Inverse $p_T$ resolution for muons from $W'$ boson decays; left: $p_T < 400$ GeV (muons from $m = 1$ TeV $W'$ bosons), right: $p_T > 800$ GeV (muons from $m = 2$ TeV $W'$ bosons).

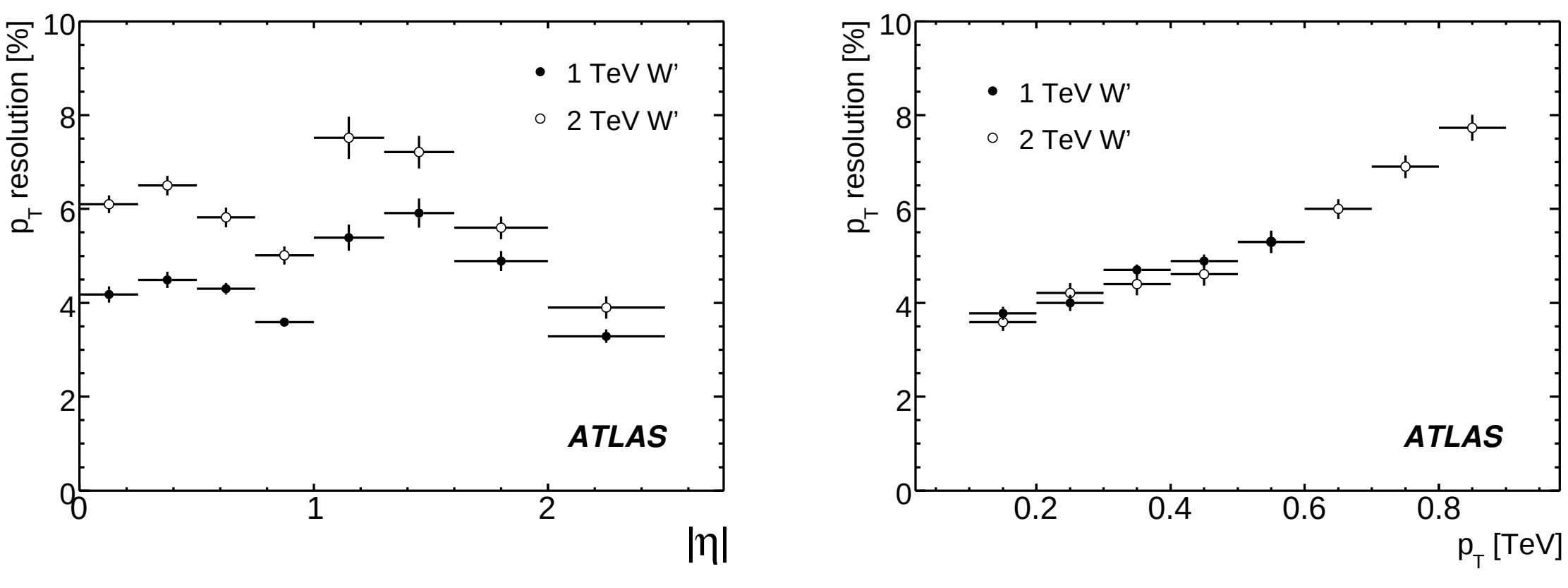


Figure 2: Muon transverse momentum resolution as a function of $\eta$ (left) and $p_T$ (right) in $W'$ boson decays. Filled circles represent muons from $m = 1$ TeV $W'$ bosons, while open circles correspond to muons from $m = 2$ TeV $W'$ bosons.

muon tagging to increase the identification efficiency. For optimum performance in momentum resolution, the MS information is combined with the track information obtained in the Inner Detector (ID). A full description of the algorithms for muon performance and identification can be found in [29] and [30].

Figure 1 shows the inverse transverse momentum $(1/p_T)$ resolution for muons from decays of a $W'$ boson for muons below $p_T = 400$ GeV and above $p_T = 800$ GeV. Especially relevant for the analysis are the negative tails in these plots, since they correspond to reconstructed muon candidates that have a $p_T$ larger than that of the true particle. The relative contribution of these tails can be assessed by the fraction of muon candidates separated by more than $2\sigma$ from the mean of the distribution (which in both cases is consistent with zero, as it should). The fraction in that negative tail is $(4.9 \pm 0.3)\%$ for muons with $p_T < 400$ GeV, and $(3.8 \pm 0.4)\%$ for $p_T > 800$ GeV (to be compared with 2.275% for a gaussian distribution). The transverse momentum resolution achieved is shown as a function of the pseudo-rapidity $\eta$ and $p_T$ in Fig. 2. On average a resolution of 4.5 and 5.5% is recorded for muons from $W'$ bosons of $m = 1$ TeV and 2 TeV respectively.

Figures 3 and 4 show the efficiency for combined muon reconstruction as a function of pseudo-

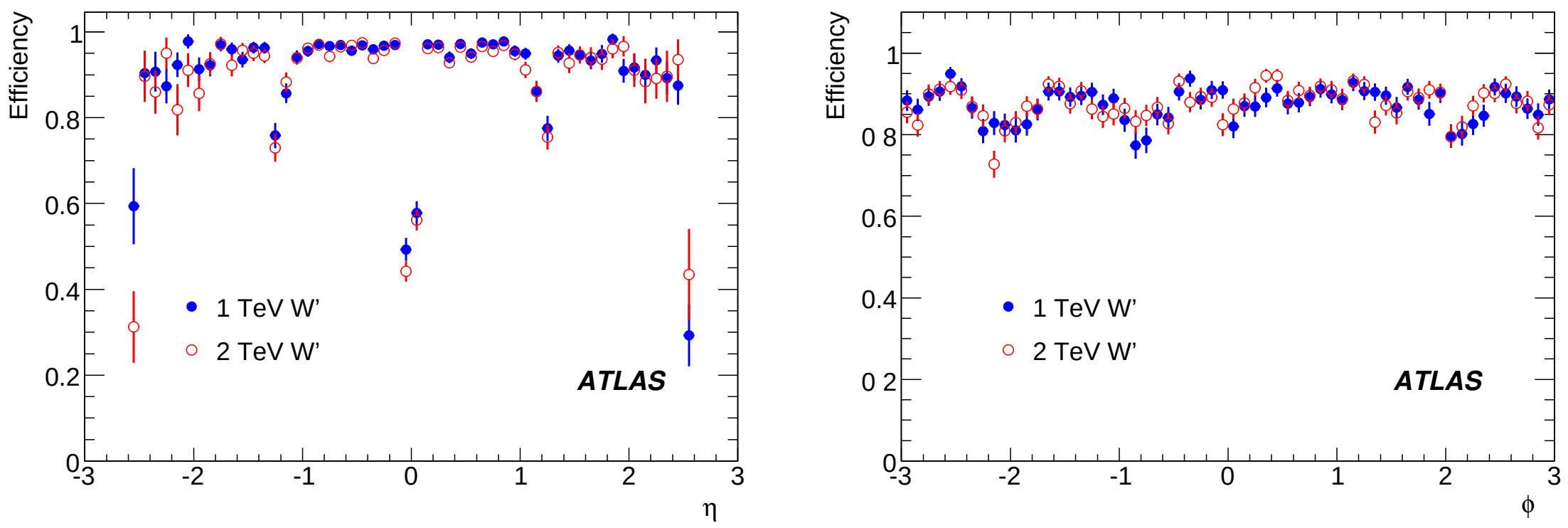


Figure 3: Combined muon reconstruction efficiency as a function of $\eta$ and $\phi$ for muons from $W'$ boson decays.

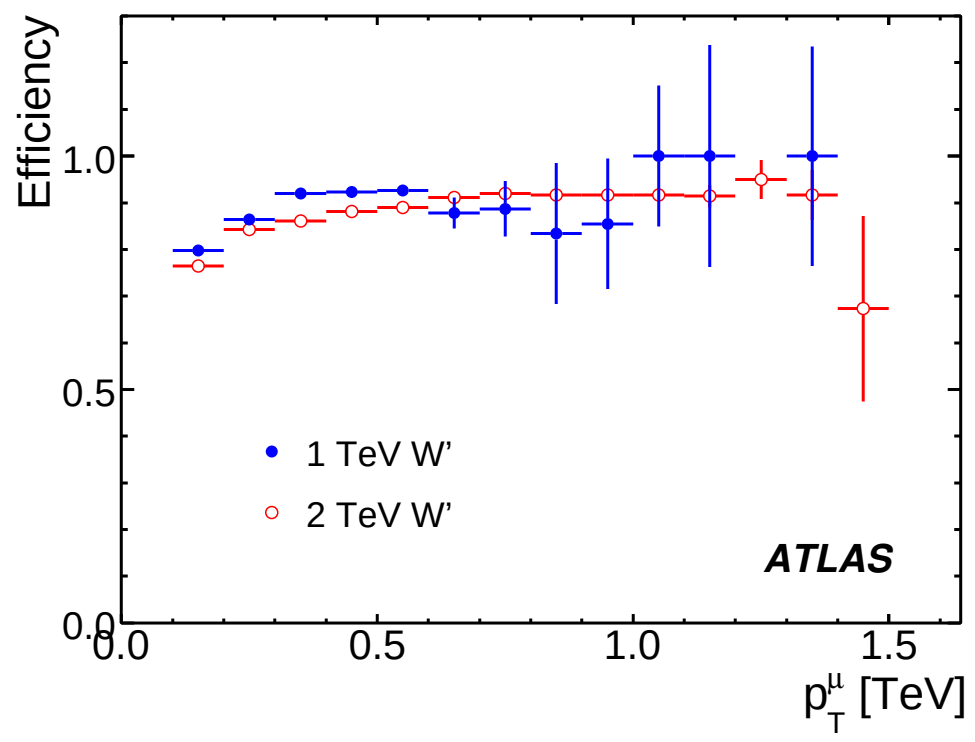


Figure 4: Combined muon reconstruction efficiency as a function of $p_T$ for muons from $W'$ boson decays.

rapidity ($\eta$), azimuthal angle ($\phi$) and transverse momentum ($p_T$) for muons from fully simulated $W'$ boson decays. An overall efficiency of 93.6% and 92.4% is measured for $m = 1$ TeV and 2 TeV $W'$ boson samples, respectively. The regions with lower efficiency in muon reconstruction are observed, as expected, in the middle plane ($\eta = 0$) and in the transition regions between the barrel and the end-cap sections of the MS (at $|\eta| \sim 1.2$). The regions with low efficiency in $\phi$ correspond to the feet of the detector ($\phi \simeq -2, -1$) and to passages for services.

One important issue in this study concerns the background that can rise from badly reconstructed muons. Their momentum being wrongly estimated upwards, can cause both the presence of a high $p_T$ muon and, correspondingly, large missing transverse energy. In the definition of the muon, extra quality criteria may be imposed in order to diminish this probability. In this study, the following mild requirements are adopted:

- A matching $\chi^2$, between MS and ID tracks, smaller than 100 (further discussed in [30]).
- An impact parameter in the $z$-axis (i.e. the beam axis) smaller than 200 mm.
- An impact parameter significance in the transverse (R-$\phi$) plane smaller than 10.

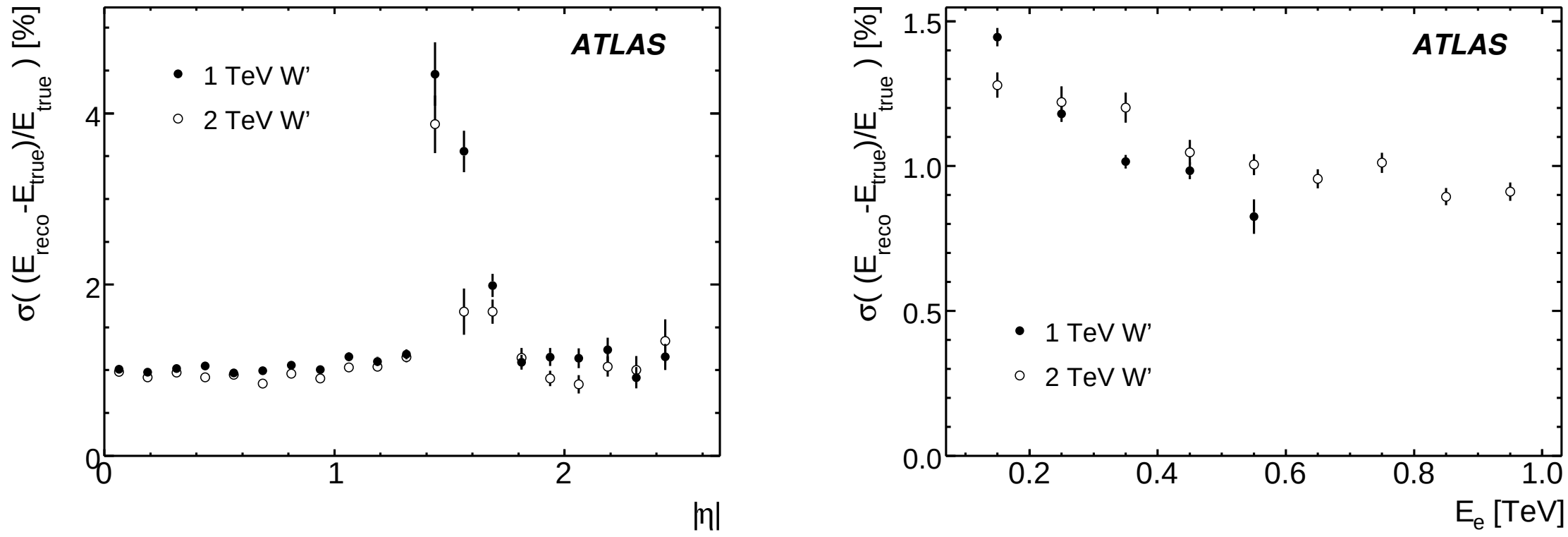


Figure 5: Electron energy resolution as a function of pseudo-rapidity (left) and energy (right) in $W'$ boson decays. Filled circles represent electrons from $m = 1$ TeV $W'$ bosons, while open circles correspond to $m = 2$ TeV $W'$ bosons.

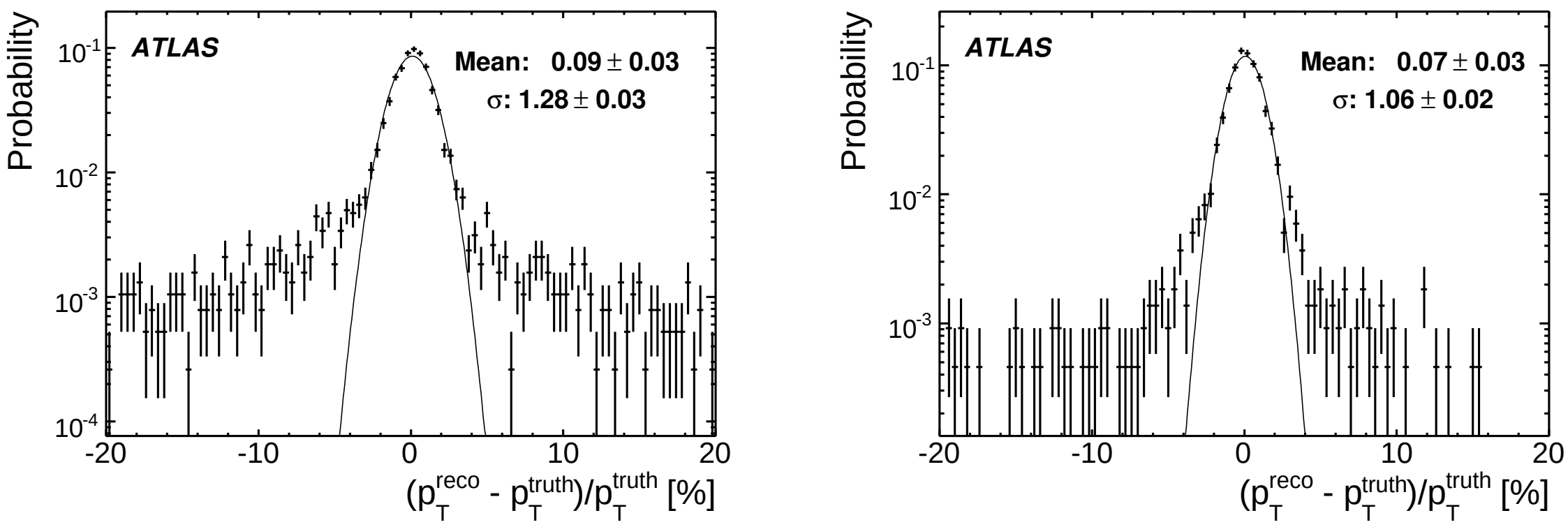


Figure 6: Electron $p_T$ resolution in $W'$ boson decays; left: $p_T < 400$ GeV (muons from $m = 1$ TeV $W'$ bosons), right: $p_T > 800$ GeV (muons from $m = 2$ TeV $W'$ bosons).

## 3.2 Electron Reconstruction

Electron candidates are built starting from clusters of calorimeter cell energy depositions, which are matched to a track from the inner detector. Electron identification and reconstruction are described in detail in [31] and [32], where three standard selections were developed to be used in physics searches. The present study uses the *medium* set of selection requirements, which consists in several requirements on the clusters used (size, containment, association with a track, shower shapes and quality of the track match).

Figure 5 shows the electron energy resolution (in percentage) as a function of pseudo-rapidity ($|\eta|$) and true energy. The average energy resolution for electrons in this energy range is close to 1%, and is worse in the transition region between the two calorimeter systems. Figure 6 shows the relative difference between reconstructed and true transverse momenta of isolated electrons with true $p_T$ lower than 400 GeV and higher than 800 GeV. The fractions of events in the upper tails (more than $2\sigma$ over the fitted mean) are $(11.8 \pm 0.6)\%$ and $(5.3 \pm 0.5)\%$, respectively. These non-Gaussian tails are due to the amount of material in the inner detector and are, therefore, $\eta$-dependent.

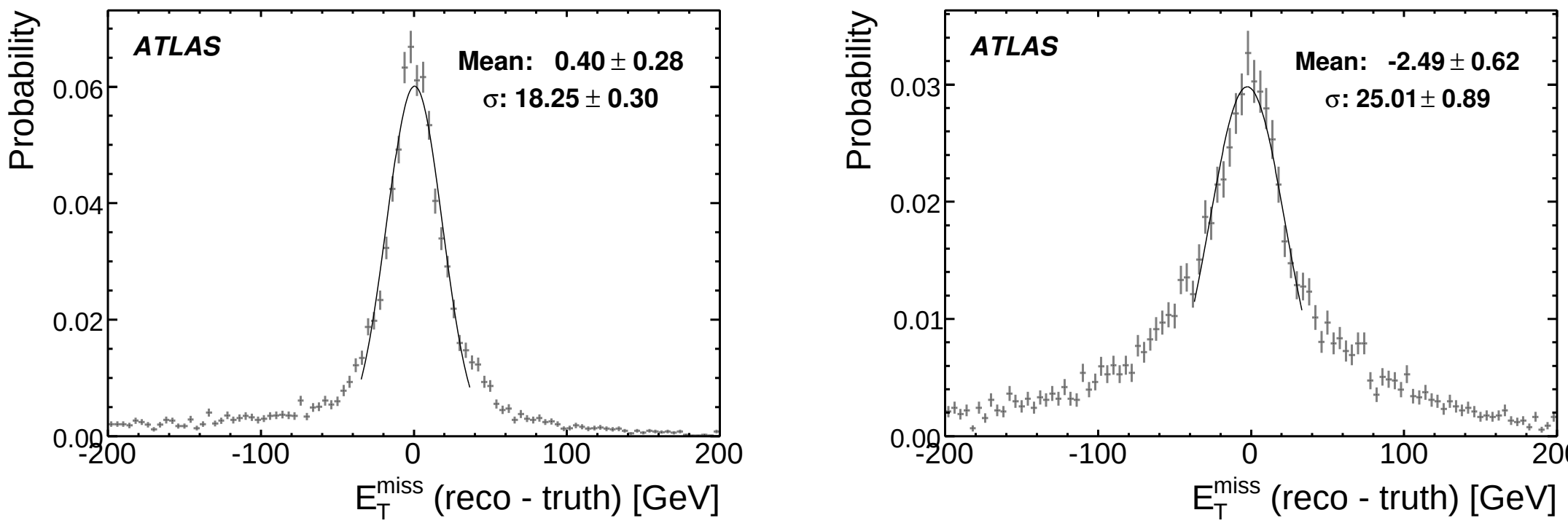


Figure 7: $\not{E}_T$ resolution in muonic $W'$ boson decays. $m_{W'}$ = 1 TeV (left) and 2 TeV (right).

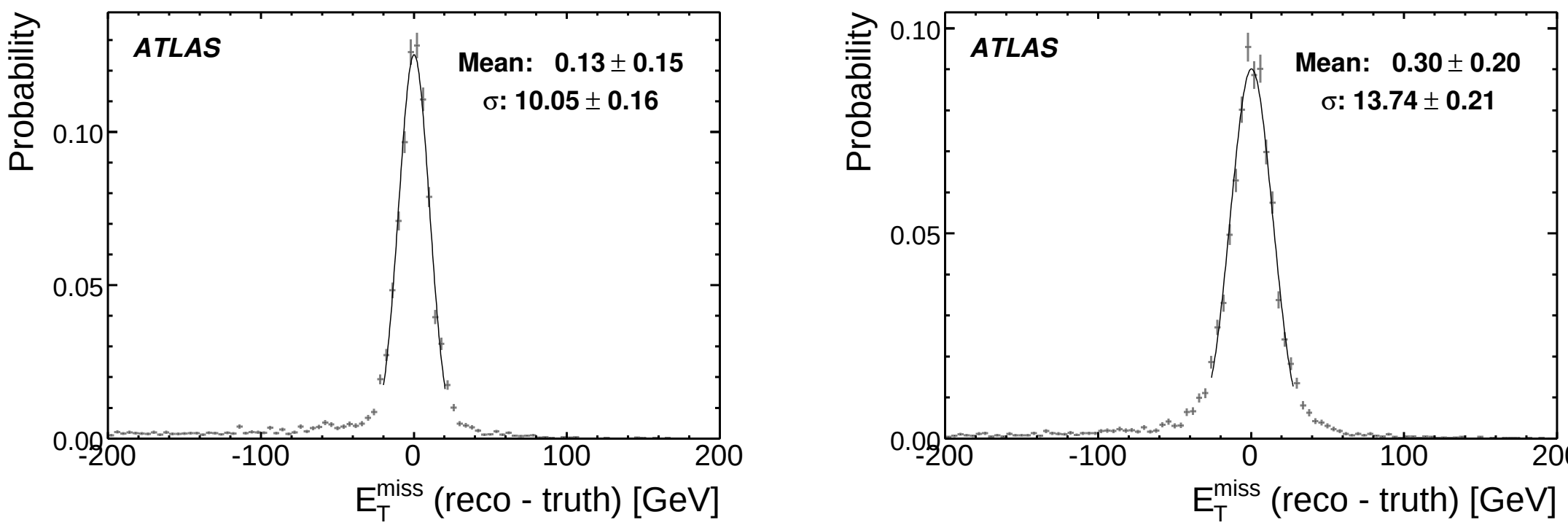


Figure 8: $\not{E}_T$ resolution in $W'$ boson decays to electrons. $m_{W'}$ = 1 TeV (left) and 2 TeV (right).

### 3.3 Reconstruction of the Missing Transverse Energy

The final state under consideration includes a neutrino, whose momentum information can be inferred only partially from the energy imbalance in the detector (since the total transverse momentum of the event has to add up to zero). The reconstruction of the missing transverse energy ($\not{E}_T$) in ATLAS is described in detail in [33].

The resolution of $\not{E}_T$ reconstruction in $W'$ boson events containing muons can be seen in Fig. 7. An average resolution of about 18 GeV is observed for $m_{W'} = 1$ TeV (25 GeV for $m_{W'} = 2$ TeV). In the case of $m_{W'} = 2$ TeV the non-Gaussian tails in the resolution are more pronounced, and come from the degraded performance of muon reconstruction at high $p_T$.

Figure 8 shows the $\not{E}_T$ resolution for events that contain one high-$p_T$ electron from a $W'$ boson decay. The left plot corresponds to the $m$ =1 TeV $W'$ boson, and the right plot to 2 TeV; the resolutions are around 10 and 14 GeV, respectively. These values agree well with the expected $\not{E}_T$ resolution from the mean of the scalar sum of transverse energy ($< \sum E_T >$) in each case; for the $m = 1$ TeV sample, $< \sum E_T >$ for the selected events is 439 GeV, which yields an estimated $\sigma(\not{E}_T) \sim 0.5\sqrt{\sum E_T} = 10.5$ GeV; for $m = 2$ TeV, the expected value (based on $< \sum E_T >$) is 13.3 GeV.

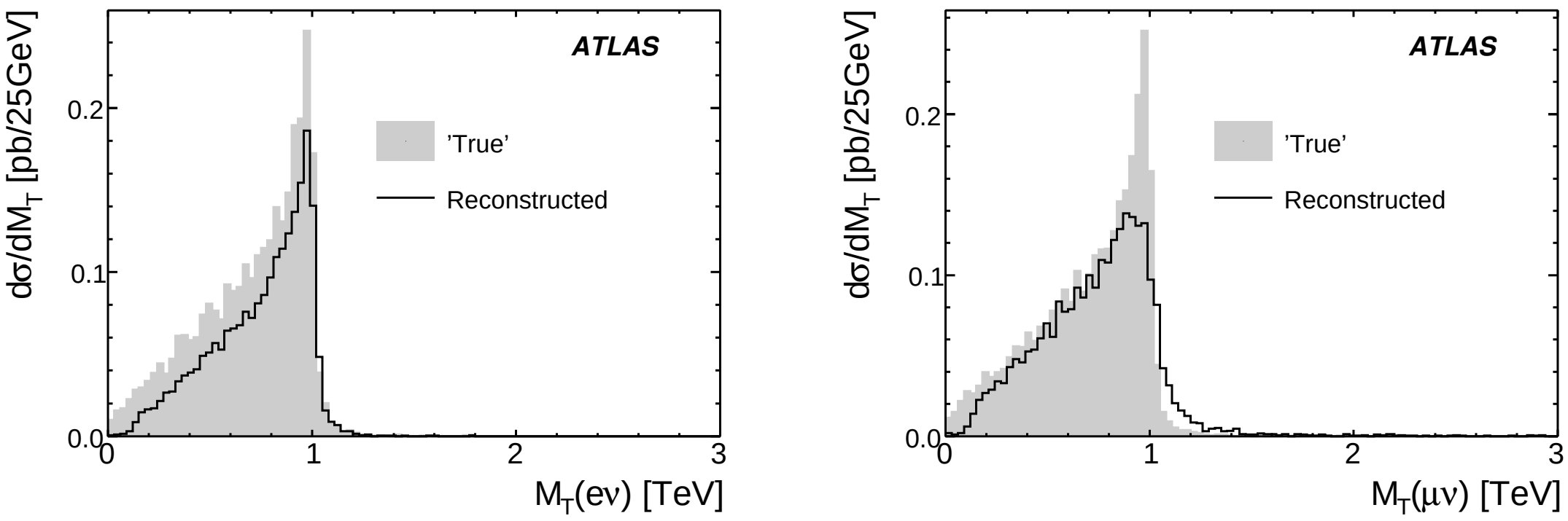


Figure 9: Transverse mass distribution for $m = 1$ TeV $W'$ bosons, as obtained from the true particles' momenta (filled histograms), and from reconstructed information after basic selection (black outline). Left: electron mode; right: muon mode.

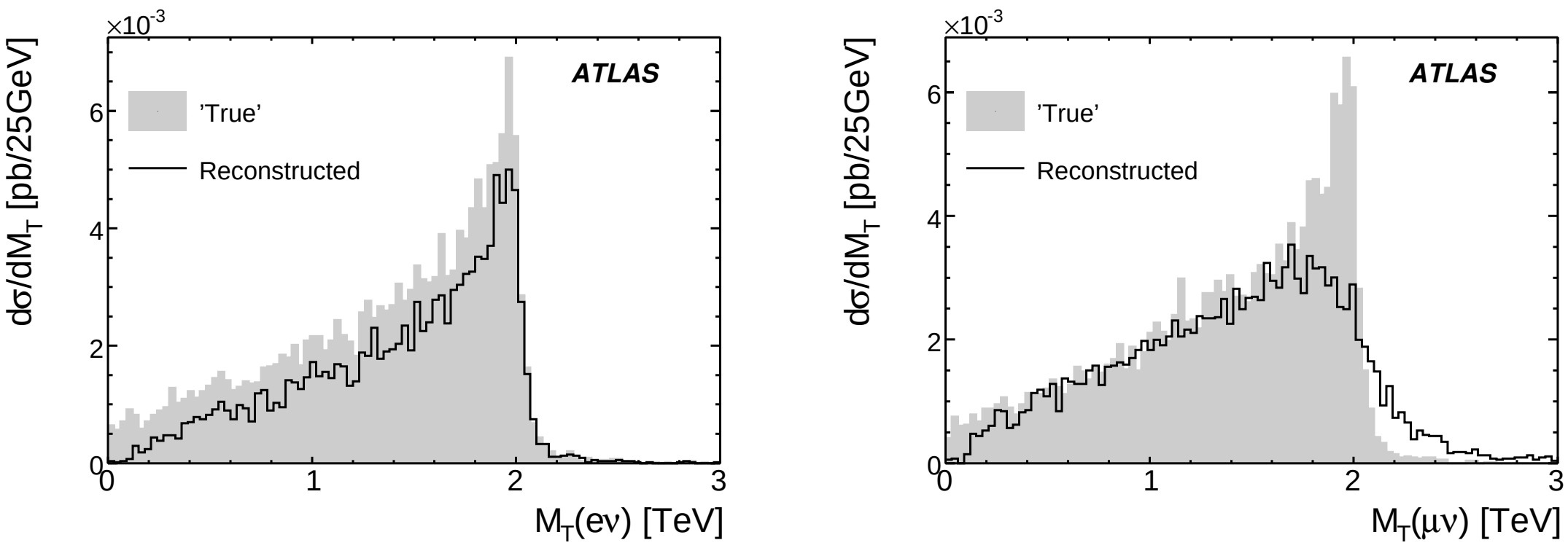


Figure 10: As Fig. 9, for $m = 2$ TeV $W'$ bosons, filled: from true information; outline: reconstructed transverse mass.

### 3.4 Transverse Mass Reconstruction

In the $W'$ boson search, the transverse momentum $p_T$ of the single lepton in the event and the missing transverse energy $\not{E}_T$ are combined to obtain the *transverse mass* as follows:

$$m_T = \sqrt{2 p_T \not{E}_T (1 - cos\Delta\phi_{\ell,\not{E}_T})} \tag{1}$$

where $\Delta\phi_{\ell,\not{E}_T}$ is the angle between the momentum of the lepton and the missing momentum, in the transverse plane. Figures 9 and 10 show the transverse mass distributions for $m = 1$ and 2 TeV signals, respectively, as obtained from truth information (light gray filled histograms) and the degradation due to detector resolution and efficiency (black hollow histograms). As can be expected from Figs. 2 and 5, the shape of the transverse mass spectrum has a larger distortion in the muon channel than in the electron channel, with larger tails for higher $W'$ boson masses. On the other hand, the reconstruction efficiency is higher in the muon channel (over 86% for each mass) than in the electron channel (about 72%).

Figures 11 and 12 show the distribution of the difference between the "true" transverse mass (i.e. as obtained from the true momenta of the lepton and the neutrino) and its reconstructed value, for electron and muon modes, and for $m = 1$ and 2 TeV $W'$ boson masses. In Fig. 11, single Gaussian fits are shown; a fitted width of about 12 GeV is obtained for the electron channel, while the muon channel, besides having much larger non-Gaussian tails, has a fitted width of about 23 GeV.

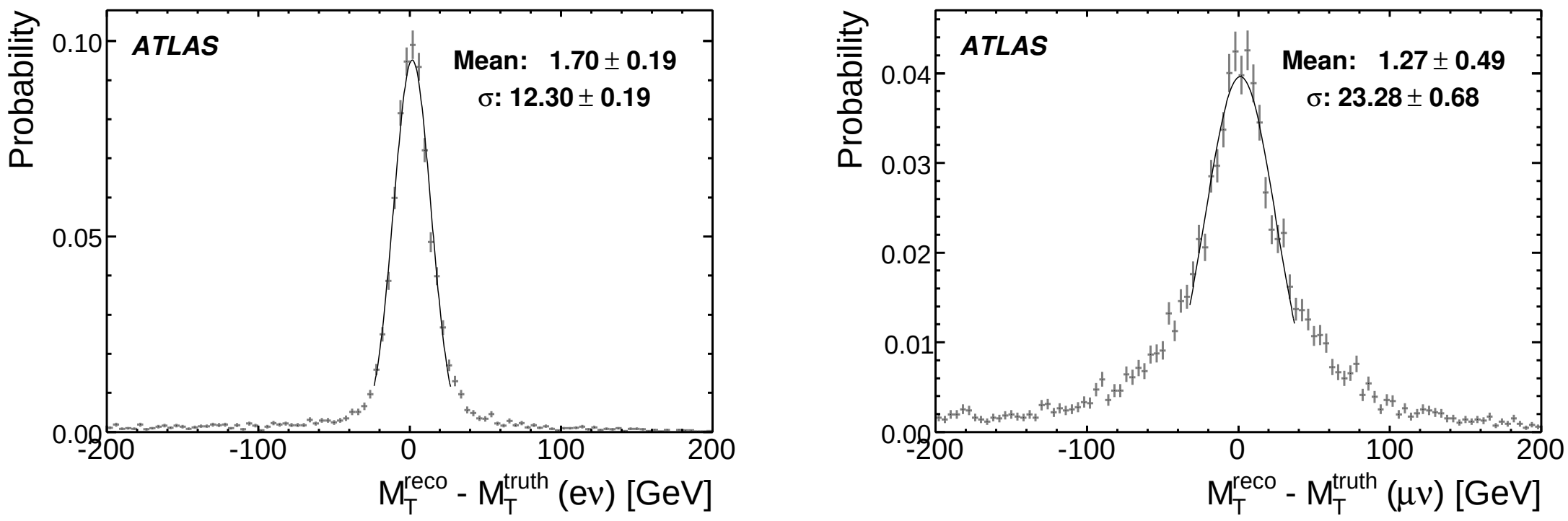


Figure 11: Distribution of the event-by-event difference between the reconstructed and "true" transverse mass for the electron and muon channel, for $m = 1$ TeV $W'$ bosons.

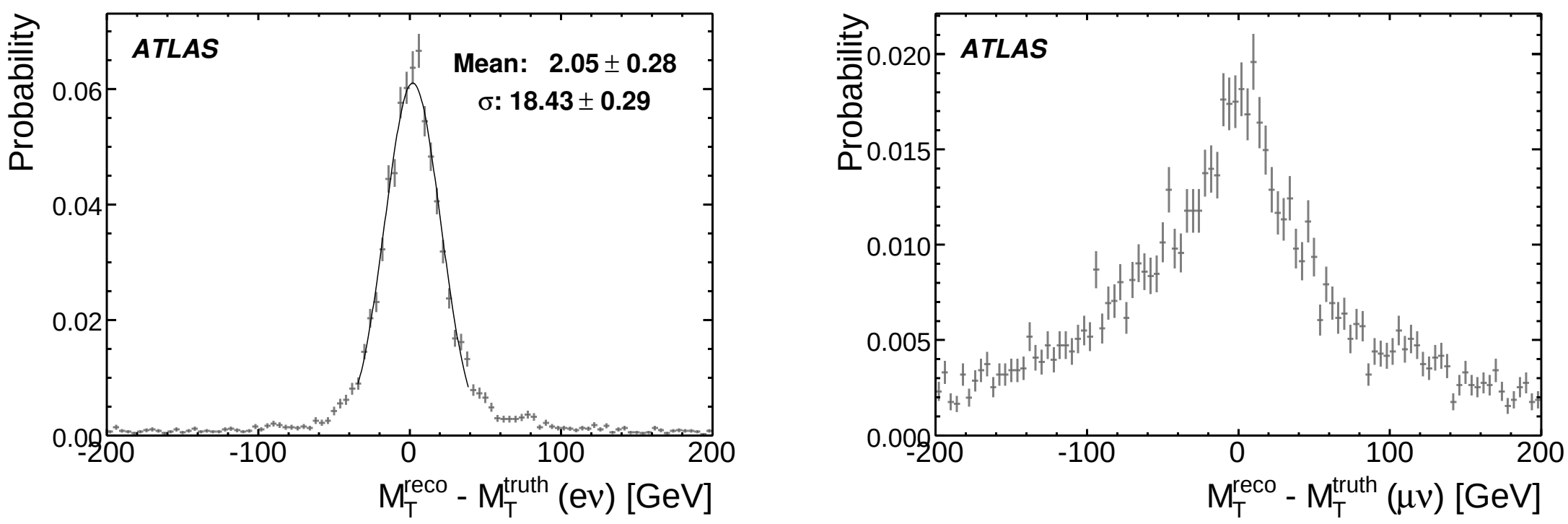


Figure 12: As Fig. 11, for $m = 2$ TeV $W'$ bosons.

Figure 12 shows the corresponding comparison for a 2 TeV signal; however, in this case, the muon channel (on the right) has a stronger non-Gaussian character, which is why no fit was performed. The quadratic mean of the distribution is about 84 GeV.

# 4 Trigger

The ATLAS trigger [34] has three levels: events passed by the L1 (level 1) hardware trigger are partially reconstructed in L2 (level 2) processors and, if accepted there, are fully processed in the EF (event filter) processor farm. Only events accepted by the EF (and thus also by L1 and L2) are recorded for later reconstruction and analysis.

Trigger rates are estimated in separate studies of the electron [35] and muon [36] trigger systems. However, this studies were performed with a more recent version of the software than the one used here[1]. Therefore, we have measured some of the rates directly in simulation using the dijet and top samples described earlier in this note. We additionally measured L1 rates and efficiencies for single-electron and single-muon triggers with thresholds higher than those defined in the simulated trigger menu. The errors we assign to our rate estimates are purely statistical.

[1]Especially for electrons, the trigger menu and algorithms in the simulated samples are quite different from those in the above notes which are much closer to those expected to be used during actual data acquisition.

### 4.1 Electron trigger

For a single electron trigger with an $E_T$ threshold of 100 GeV, we measure a L1 rate of $14 \pm 1$ Hz at an instantaneous luminosity of $10^{32}$ cm$^{-2}$s$^{-1}$, similar to the electron trigger study estimate of 10 Hz. The efficiency to trigger on $W' \rightarrow e\nu$ events for $|\eta| < 2.5$ is 98% for a mass of either 1 or 2 TeV. If the threshold is raised to 250 GeV, we measure a rate of $25 \pm 4$ Hz at $10^{33}$ cm$^{-2}$s$^{-1}$ and an efficiency of 96% for the 2 TeV mass.

Loose requirements in L2 and EF can further reduce these rates with a moderate degradation of the efficiency. For definiteness in the calculations in the following sections, we assume that a trigger efficiency (applied after all requirements) of $0.90 \pm 0.10$ is achieved with an acceptable rate for all $W'$ boson masses.

### 4.2 Muon trigger

The trigger menu and algorithms in the simulation samples are similar to those in the muon trigger study and those expected for data acquisition. In contrast to the electron case, lower thresholds can be applied thanks to the lower fake rates. A significant decrease in rate is then obtained thanks to an improved measurement of $p_T$ at each level. Applying a threshold of 20 GeV at each trigger level, we obtain an EF rate of $20 \pm 10$ Hz for an instantaneous luminosity of $10^{32}$ cm$^{-2}$s$^{-1}$, consistent with the muon study prediction of 13 Hz. We measure a $W' \rightarrow \mu\nu$ trigger efficiency for $|\eta| < 2.5$ of 74% for $m = 1$ TeV and 73% at 2 TeV. At $10^{33}$ cm$^{-2}$s$^{-1}$, we apply a $p_T$ threshold of 40 GeV, the maximum L1 value, and obtain a trigger rate of $4.1 \pm 0.7$ Hz close to the 5.6 Hz obtained in the trigger study. The corresponding trigger efficiency for the $m = 2$ TeV $W'$ boson is 69%.

It should be noted that most of the efficiency loss comes from holes in the coverage of the muon system, where the reconstruction is also inefficient.

## 5 Event Selection

The decay $W' \rightarrow \ell\nu$ provides a rather clean signature consisting of a high-energy isolated lepton and large missing transverse energy. The largest backgrounds are the high-$p_T$ tail of the $W \rightarrow \ell\nu$ decays and $t\bar{t}$ production. Both these final states are accompanied by significant jet activity, but contain also leptons that are as isolated as those expected from $W' \rightarrow \ell\nu$ decays.

A potentially dangerous background is the one arising from fake leptons; since this issue is more likely to be significant for electrons than for muons, the backgrounds will be presented separately for $W' \rightarrow e\nu$ and $W' \rightarrow \mu\nu$ final states.

### 5.1 Event Preselection

In addition to the electron and muon identification criteria described above, events are required to have:

- Only one reconstructed lepton with $p_T > 50$ GeV within $|\eta| < 2.5$.
- Missing transverse energy $\not\!\!E_T > 50$ GeV.

Figure 13 shows, on top, differential cross-section as a function of the lepton $p_T$ for the $m = 1$ TeV and 2 TeV signal samples, Standard Model $W$ boson, $t\bar{t}$ and dijet production. The dashed vertical line shows the requirement value (50 GeV). The bottom plots in Fig. 13 show the $\not\!\!E_T$ distributions for the same processes after requiring only one lepton with $p_T > 50$GeV; again, the requirement value (at 50 GeV) is shown with the dashed vertical line. This selection provides a relatively clean signal in the high transverse mass region, as shown in Fig. 14, which shows the differential cross-section as a function

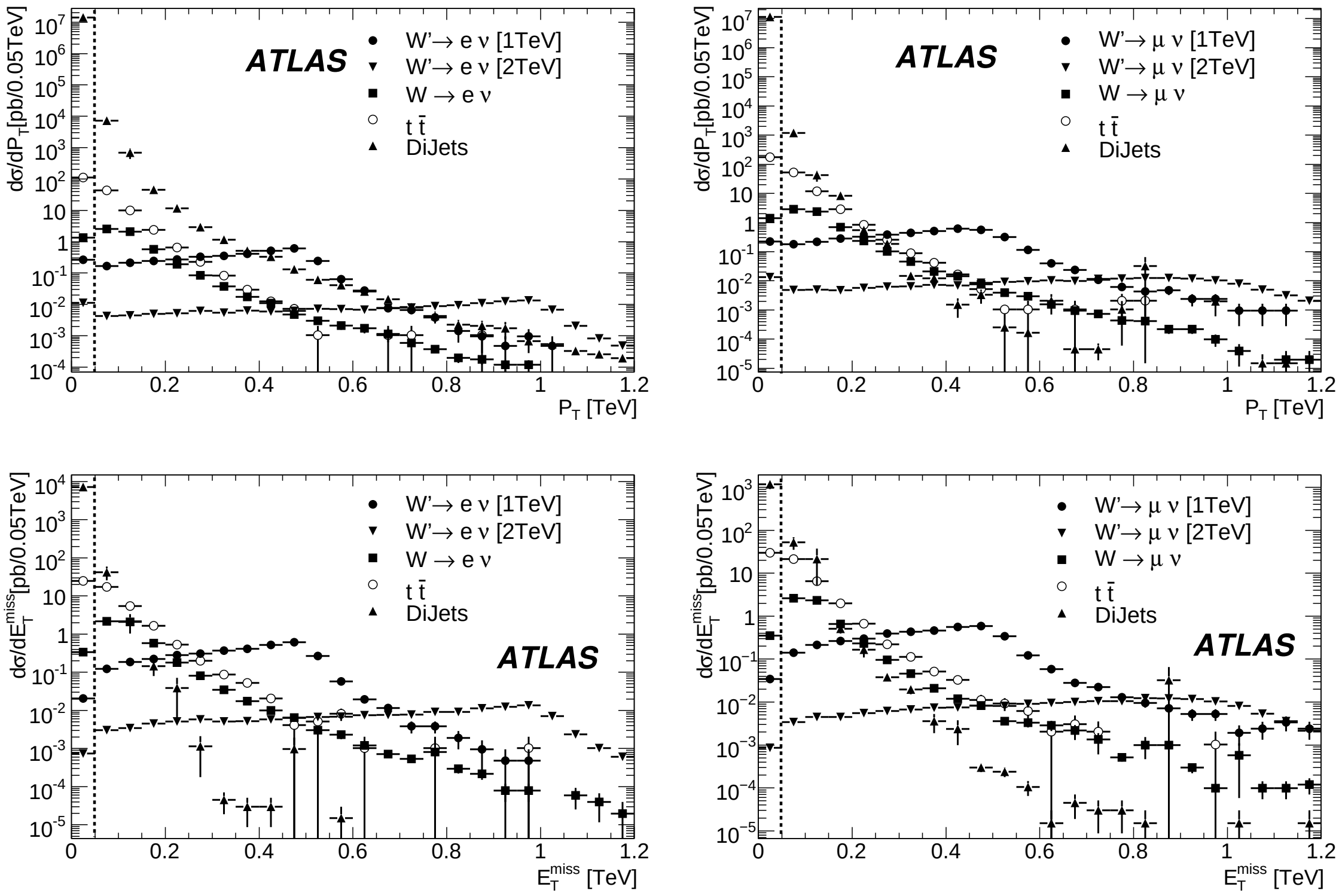


Figure 13: Top: leading lepton $p_T$ distributions (left: electron events, right: muon events). Bottom: $\not{E}_T$ distribution of events with only one reconstructed lepton with $p_T > 50$ GeV (left: electron events, right: muon events).

of the transverse mass after the requirements on $p_T$ and $\not{E}_T$. The background can be further rejected by exploiting additional observables, described in next sections: lepton isolation, lepton fraction and jet veto criteria.

## 5.2 Background Rejection

After the kinematic requirements are applied, the $t\bar{t}$ and dijets backgrounds are still larger than the high-mass tail of the Standard Model $W$ boson close to the threshold value on the lepton $p_T$ and on $\not{E}_T$. Since the uncertainties on the rate of these backgrounds are large, it is desirable to bring them below the irreducible background from $W$ bosons. To achieve this, additional requirements are imposed on lepton isolation and on the lepton fraction, described below. A simpler selection strategy, based on a jet veto, is also explored, since it could prove useful during the first stages of data taking.

### 5.2.1 Lepton Isolation

As the lepton from a $W'$ boson decay is expected to be isolated, only events without high energy tracks around the lepton trajectory are accepted. The tracking isolation is done by requiring that the sum of the $p_T$ of tracks in a $\Delta R$-cone around the lepton be below a threshold; $\Delta R$ is defined as

$$\Delta R \equiv \sqrt{(\Delta\phi)^2 + (\Delta\eta)^2},$$

where $\Delta\phi$ and $\Delta\eta$ are the distances in azimuthal angle and in pseudo-rapidity, respectively, from the lepton under consideration.

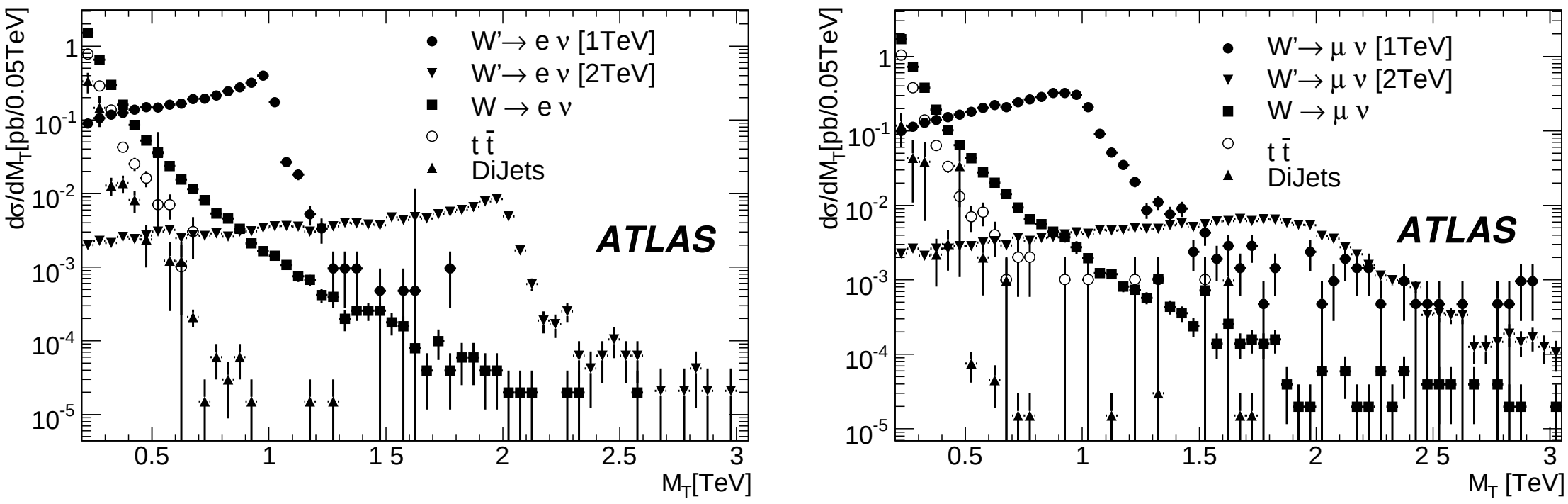


Figure 14: Transverse mass spectrum after the basic kinematic requirements for background and signal ($m_{W'} = 1$ and 2 TeV). Left: electron mode; right: muon mode.

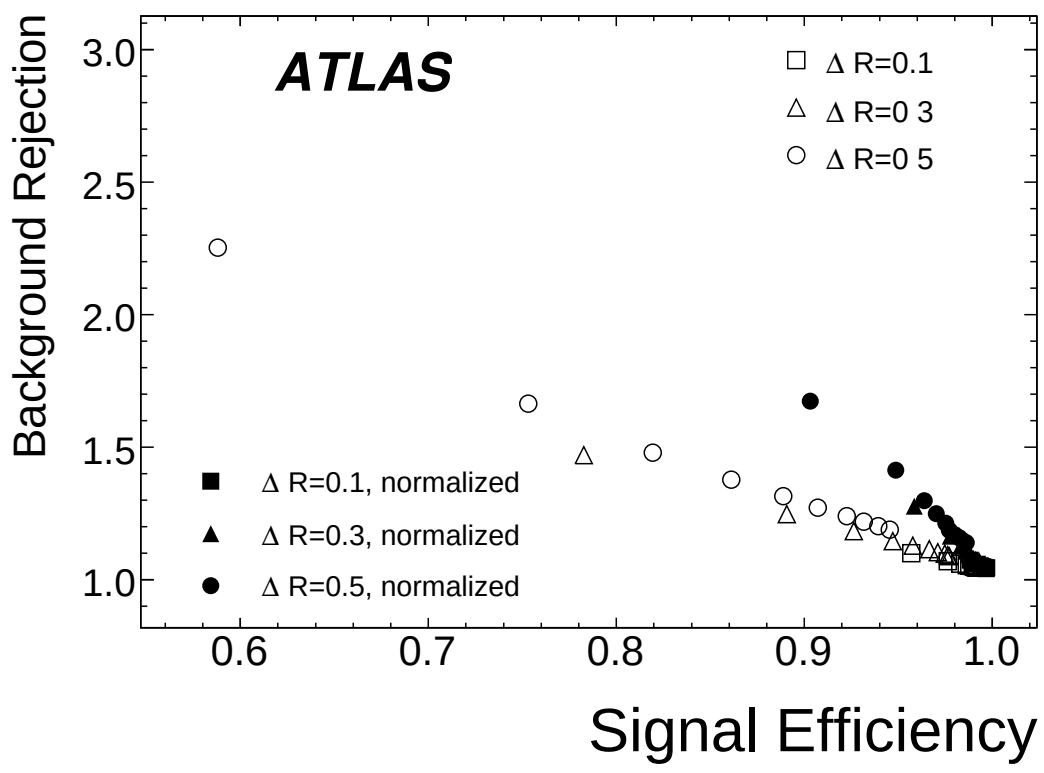


Figure 15: $t\bar{t}$ background rejection and signal efficiency for different requirement values on the $\sum p_T$ (open markers) and $(\sum p_T)/p_{T\text{lepton}}$ (filled markers), for the muon channel. Each marker type corresponds to a different value for $\Delta R$, from 0.1 to 0.5.

Calorimeter isolation was also explored (the calorimetric energy deposited within the volume between two $\Delta R$-cones is required to be below a threshold).

Besides requiring a maximum value of $\sum p_{T\text{tracks}}$ (of 10 GeV to 1 GeV), the use of a *normalized* isolation requirement was also explored, in which the requirement is applied to the $\sum p_{T\text{tracks}}/p_{T\text{lepton}}$ ratio. This ratio is required to be smaller than 0.1 to 0.01. Five different $\Delta R$ values were used in both cases; as shown in Fig. 15 for muons, the normalized isolation selection achieves a higher $t\bar{t}$ rejection for the same efficiencies. The efficiencies and rejections achieved for electrons are similar.

The calorimeter energy difference in two cones is not only of use in the electron case, but also in the muon one. High $p_T$ muons coming from $W'$ boson decays can also radiate a lot inside the material preceding the MS. This radiation appears as energy depositions close to the muon in the calorimeters. As can be seen in Fig. 16 (right), the energy deposition in a cone of $\Delta R < 0.1$ around the muon is much higher, around 30 GeV on average, than the deposition in a cone of $\Delta R < 0.5$ when the inner cone is subtracted (in this case the average is aboout 7 GeV). Moreover, in Fig. 16 (left) it is shown that in the majority of the cases, a high reconstructed energy deposition indicates the existence of high final state radiation. Therefore, the energy deposition in an inner cone (e.g. $\Delta R < 0.1$) must be subtracted also in the case of muons when isolation criteria based on calorimetry are applied. Also on track based isolation criteria an inner cone containing the muon track itself must be subtracted. In this case however, the inner cone can be much narrower, since it only needs to be able to exclude the track associated with the

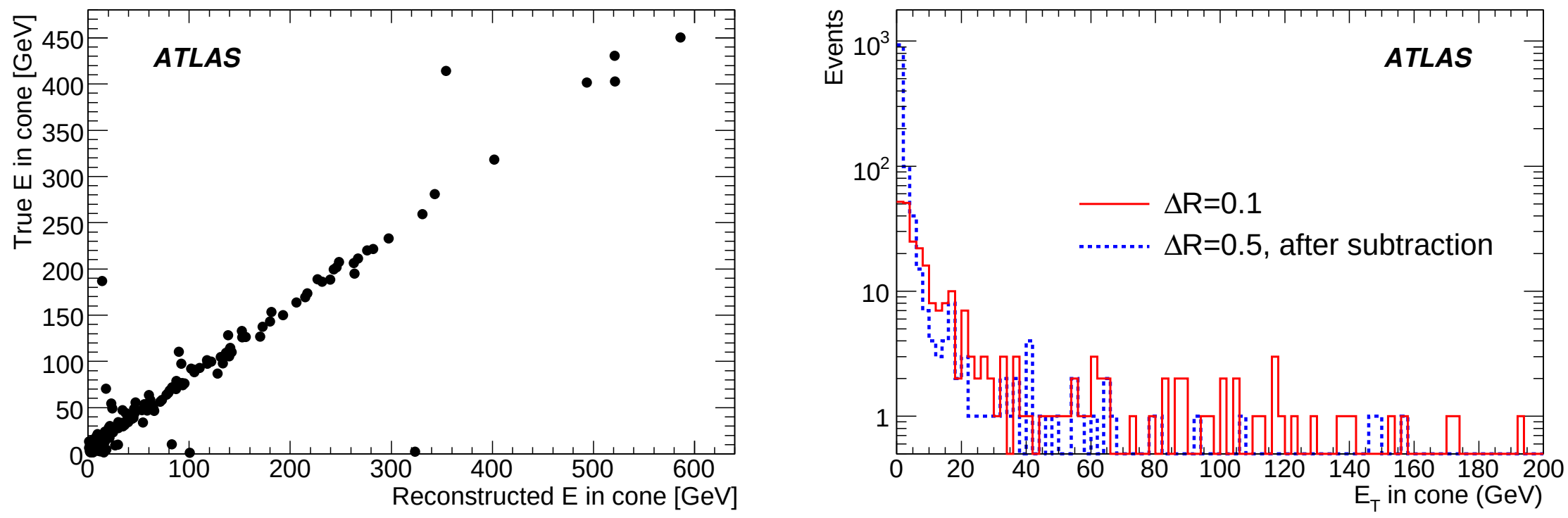


Figure 16: Left: the true energy deposition as a function of the reconstructed one in a cone of $\Delta R$=0.1 for muons coming from decays of $m = 2$ TeV $W'$ bosons. Right: the solid histogram shows the energy recorded in a cone of $\Delta R$=0.1 around the muon. The dashed histogram shows the energy recorded in a cone of $\Delta R$=0.5 after the subtraction of the inner cone deposition.

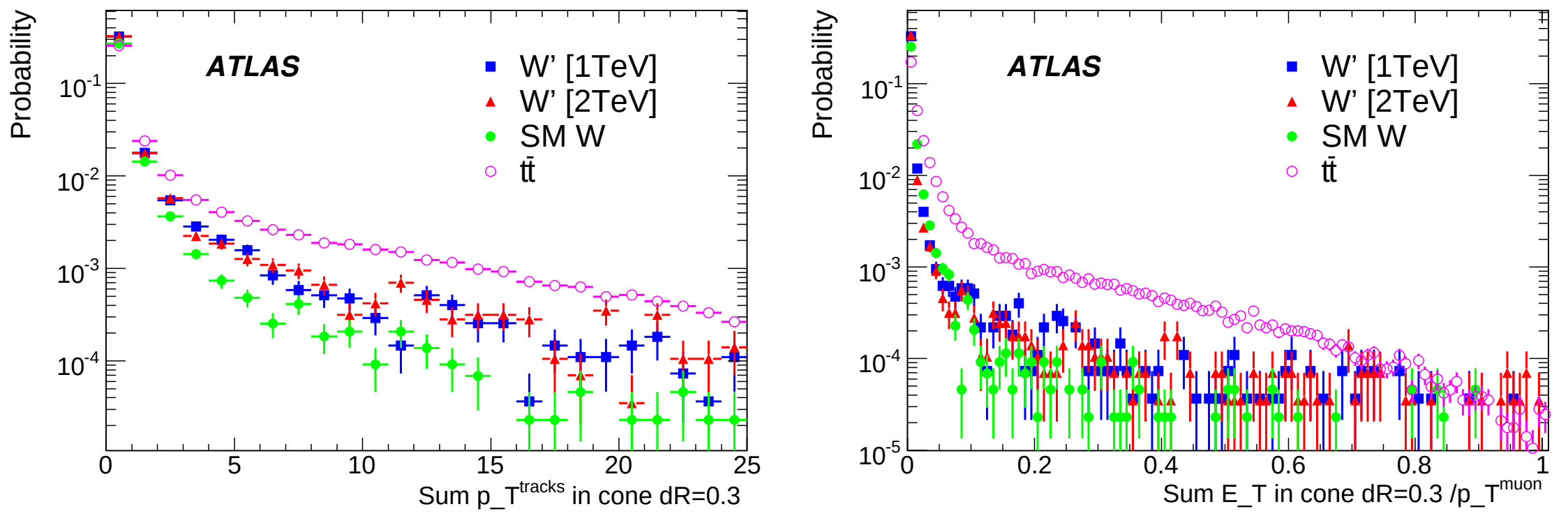


Figure 17: Left: distribution of an absolute track based isolation variable for muons. Right: distribution of a relative calorimetry based isolation variable for muons. In both cases the inner cone of $\Delta R$=0.1 is subtracted.

lepton under consideration. Figure 17 shows the distributions of the isolation energy for different event categories. For these plots, muons with $p_T > 20$ GeV are considered.

For the analysis, a loose requirement of 0.05 is used on the normalized track-based isolation for both channels (electron and muon), and no requirement on the calorimeter-based isolation is applied. Tracks are included in the sum if $0.02 < \Delta R(\text{track}, \text{lepton}) < 0.3$. This requirement keeps about 99% of the signal for both masses ($m_{W'} = 1$ and 2 TeV), rejects about 10% of the $t\bar{t}$ events left after the basic selection and rejects over 99% of the dijet background.

### 5.2.2 Lepton Fraction

Another variable that can be used to reduce the dijet and $t\bar{t}$ backgrounds is the "lepton fraction" of the event, which can be expressed as $\sum p_T^{leptons}/(\sum p_T^{leptons} + \sum E_T)$, where the scalar sum on the lepton $p_T$ sums over $\not{E}_T$ as well. Essentially this variable measures the fraction of energy that can be attributed to leptons (including neutrinos, which are assumed to be the main contribution to $\not{E}_T$) in an event. Here, out of the visible leptons, only the most energetic one is included in the sum (its $p_T$ is added to the $\not{E}_T$ to form $\sum p_T^{leptons}$). The distribution of this variable is shown for different event categories in Fig. 18 (left). As expected, it shows a much lower value for $t\bar{t}$ events (in pink) than for the rest of the samples used

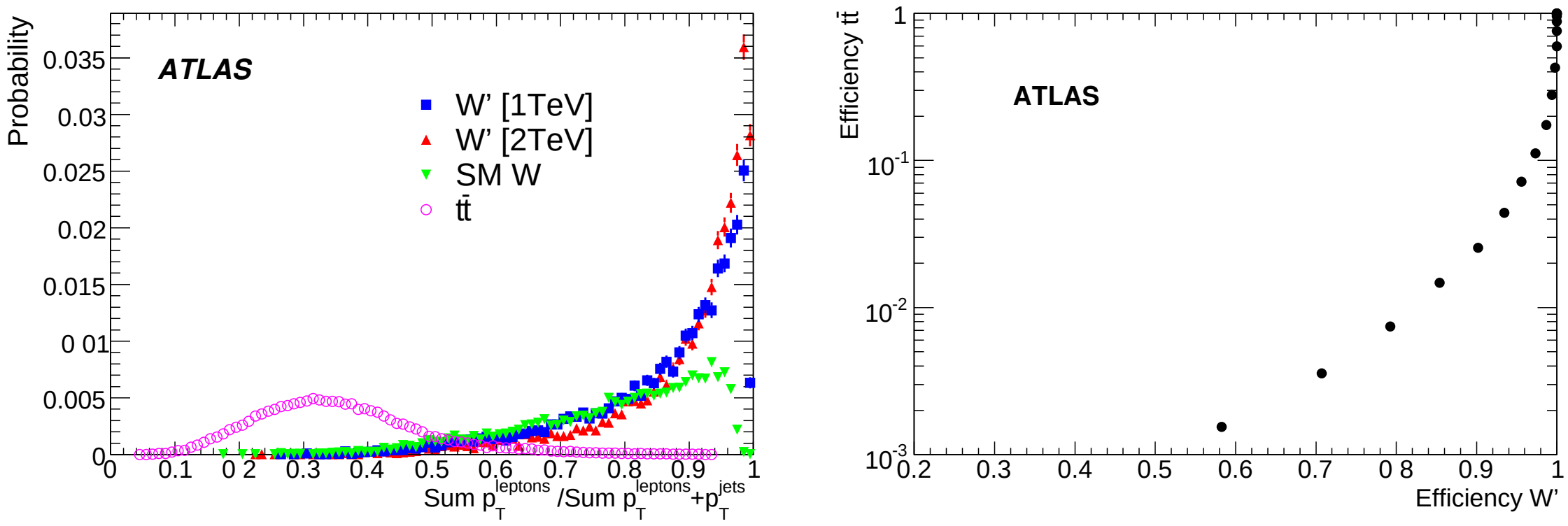


Figure 18: Left: distribution of the lepton fraction variable (see text) for different event categories. Right: signal efficiency versus $t\bar{t}$ efficiency for different requirement values on the lepton fraction variable.

($W'$ boson signals and Standard Model $W$ bosons). The efficiency for signal versus $t\bar{t}$ events for different values of the variable, is shown in Fig. 18 (right). A requirement at 0.5 results in a signal efficiency of $\sim 96\%$ in both channels and a rejection factor of $\sim$45 against the $t\bar{t}$ background, and it suppresses all the remaining dijet events. This value will be used subsequently.

### 5.2.3 Jet Veto and Jet Multiplicity Requirements

A selection procedure based solely on veto-ing events with high jet activity could provide an alternative way to extract a signal in this search. Several requirements on jet activity were explored; in some, events are rejected if they include any jet over an energy threshold, in others, jet multiplicity information is used. The jet veto was applied just after the basic selection (i.e., lepton identification, $p_T$ and $\not{E}_T$ requirements).

Figure 19 shows the distribution of the $p_T$ of the leading jet after the basic selection; the distribution on the left corresponds to the electron channel and the one on the right to the muon channel. Tables 2 and 3 show the expected rates for several jet veto criteria .

Figure 20 shows how after a 200 GeV jet veto requirement (and without isolation or lepton fraction requirements), most of the $t\bar{t}$ and dijet background is rejected, and the signal to background ratio is good for high transverse mass values. Although the signal is reduced by between 5 and 10% with respect to selecting on isolation and lepton fraction, a jet veto requirement may be a good tool if the calibration of the $\sum E_T$ (used to compute the lepton fraction) is not well understood in early data. However, in what follows, this requirement is not used.

## 5.3 Event Selection Results

Figure 21 shows the expected transverse momentum spectra for signal and background for both channels after all requirements (preselection, isolation, and lepton fraction). The selection requirement flow is shown in Tables 4 and 5. The transverse mass requirement has been chosen by minimizing the luminosity needed to get a $5\sigma$ excess. The initial cross-sections for the $W'$ boson signals and for the high mass $W$ boson tail include the $K$-factor obtained in section 6.1.

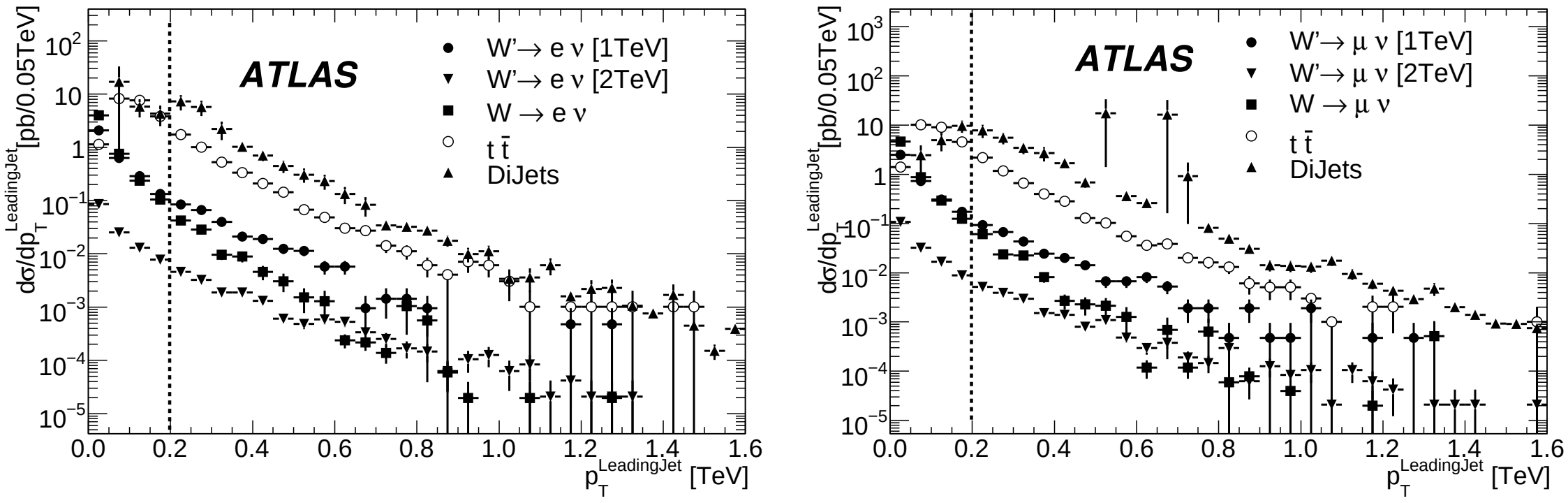


Figure 19: Distributions for the $p_T$ of the leading jet after the basic selection. Left: electron selection. Right: muon selection.

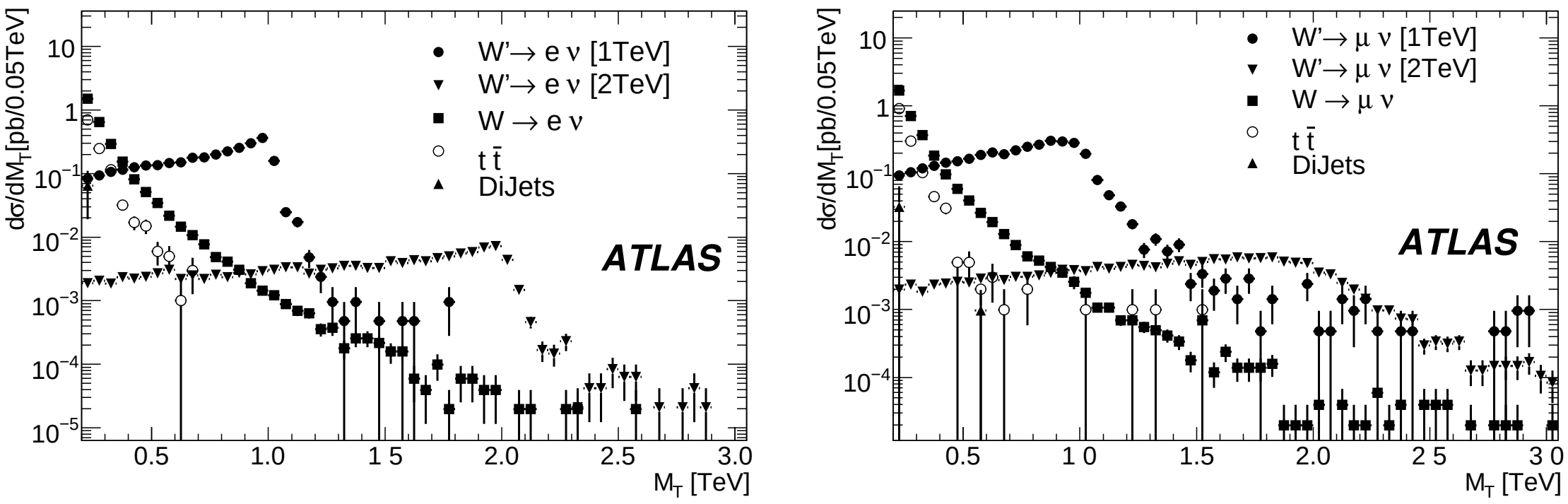


Figure 20: $m_T$ spectrum after preselection requirements and a jet veto of $E_T < 200$ GeV. Left: events with a high-$p_T$ electron; right: events with a high $p_T$ muon.

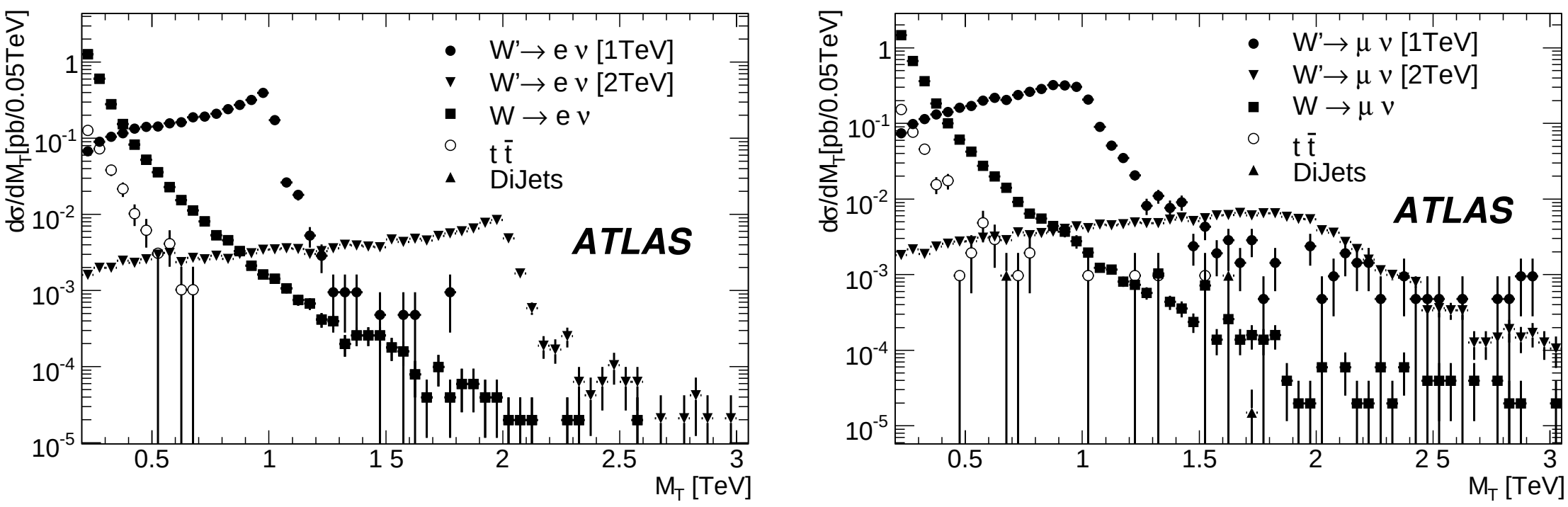


Figure 21: Expected transverse mass spectra after all requirements. Left: electron channel; right: muon channel.

| Requirement | $\sigma$ [pb] | | | | |
|---|---|---|---|---|---|
| | $W'$ 1TeV | $W'$ 2TeV | $W$ | $t\bar{t}$ | Dijets |
| No jets with $p_T > 100\,\text{GeV}$ | 2.71(4) | 0.112(2) | 4.74(5) | 7.07(7) | 17±16 |
| No jets with $p_T > 200\,\text{GeV}$ | 3.13(4) | 0.132(2) | 5.09(5) | 15.7(1) | 27±16 |
| No jets with $p_T > 500\,\text{GeV}$ | 3.38(4) | 0.146(2) | 5.18(5) | 18.7(1) | 44±17 |
| Less than 4 jets with $p_T > 40\,\text{GeV}$ | 3.38(4) | 0.148(2) | 5.18(5) | 14.0(1) | 43±17 |
| Less than 3 jets with $p_T > 100\,\text{GeV}$ | 3.39(4) | 0.148(2) | 5.18(5) | 17.8(1) | 44±17 |
| Less than 2 jets with $p_T > 200\,\text{GeV}$ | 3.38(4) | 0.148(2) | 5.18(5) | 18.4(1) | 44±17 |
| 200 GeV veto, $m_T > 0.7\,\text{TeV}$ | 1.73(3) | | 0.0290(8) | – | – |
| 200 GeV veto, $m_T > 1.4\,\text{TeV}$ | | 0.066(1) | 0.0013(1) | – | – |

Table 2: Cross-sections for signal and backgrounds for dijets, $t\bar{t}$, $W$ and $W'$ boson samples for different requirements on jet content for the electron channel. The number in brackets is the error on the least significant digit.

| Requirement | $\sigma$ [pb] | | | | |
|---|---|---|---|---|---|
| | $W'$ 1TeV | $W'$ 2TeV | $W$ | $t\bar{t}$ | Dijets |
| No jets with $p_T > 100\,\text{GeV}$ | 3.22(4) | 0.141(2) | 5.50(5) | 8.77(8) | 2(1) |
| No jets with $p_T > 200\,\text{GeV}$ | 3.70(4) | 0.166(2) | 5.92(5) | 19.1(1) | 17(4) |
| No jets with $p_T > 500\,\text{GeV}$ | 3.96(4) | 0.182(2) | 6.04(5) | 22.7(1) | 39(5) |
| Less than 4 jets with $p_T > 40\,\text{GeV}$ | 3.98(4) | 0.184(2) | 6.03(5) | 16.9(1) | 53(5) |
| Less than 3 jets with $p_T > 100\,\text{GeV}$ | 3.98(4) | 0.185(2) | 6.04(5) | 21.5(1) | 73(5) |
| Less than 2 jets with $p_T > 200\,\text{GeV}$ | 3.98(4) | 0.185(2) | 6.04(5) | 22.3(1) | 73(5) |
| 200 GeV veto, $m_T > 0.7 TeV$ | 2.07(3) | | 0.040(1) | 0.005(2) | – |
| 200 GeV veto, $m_T > 1.4 TeV$ | | 0.084(1) | 0.0033(8) | 0.0008(8) | – |

Table 3: Cross-sections for signal and backgrounds for dijets, $t\bar{t}$, $W$ and $W'$ boson samples for different requirements on jet content for the muon channel. The number in brackets is the error on the least significant digit.

# 6 Systematic Uncertainties

## 6.1 Generator-level Systematic Uncertainties

The input for the full simulation studies described in earlier sections was obtained by generating $W'$ boson events using PYTHIA [27]. Events in the high-mass tail of the $W$ boson were generated using PYTHIA as well. Both use the default PYTHIA parton distribution functions (PDFs), CTEQ6l, the CTEQ6 [37] LO (leading-order) fit with NLO (next-to-leading-order) $\alpha_S$. Here we report on generator-level studies which examine the effects of making use of the NLO matrix elements and varying the PDFs.

### 6.1.1 Higher Orders

To evaluate contributions from higher order diagrams, we used MC@NLO [38] input to the HERWIG [39] event generator. Both $W$ and $W'$ boson events were generated using the $W$ boson production process with the $W$ boson mass set to the $W'$ boson value for the latter. The $W'$ boson widths were set to the values calculated by PYTHIA. The masses and widths used are listed in Table 6. Both MC@NLO and HERWIG were run using the default HERWIG PDFs, MRST2004nlo, the MRST 2004 fit using the standard $\overline{\text{MS}}$ scheme at NLO [40].

One million events were generated for each generator at each of the masses. The cross-section is

| Requirement | $\sigma$ [pb] | | | | |
|---|---|---|---|---|---|
| | $W'$ (1 TeV) | $W'$ (2 TeV) | $W$ tail | $t\bar{t}$ | Dijets[1-7] |
| (No requirement) | 4.99 | 0.231 | 10.28 | 452 | $1.91 \times 10^{10}$ |
| Preselection | 3.67±0.04 | 0.160±0.002 | 6.80±0.06 | 150.57±0.40 | $(13.6 \pm 0.2) \times 10^6$ |
| $p_T > 50$ GeV | 3.43±0.04 | 0.150±0.002 | 5.53±0.05 | 51.13±0.23 | $(7.23 \pm 0.6) \times 10^3$ |
| $\not{E}_T > 50$ GeV | 3.40±0.04 | 0.149±0.002 | 5.19±0.05 | 25.78±0.16 | 45.33±16.65 |
| Isolation | 3.36±0.04 | 0.148±0.002 | 5.01±0.05 | 23.30±0.16 | 0.65±0.13 |
| Lepton fraction | 3.25±0.04 | 0.145±0.002 | 4.10±0.04 | 0.50±0.02 | – |
| $m_T > 700$ GeV | 1.86±0.03 | | 0.0317±0.0008 | 0 | – |
| $m_T > 1400$ GeV | | 0.0740±0.001 | 0.0014±0.0002 | 0 | – |

Table 4: Cross-section for signal and backgrounds after each requirement. Electron mode.

| Requirement | $\sigma$ [pb] | | | | |
|---|---|---|---|---|---|
| | $W'$ (1 TeV) | $W'$ (2 TeV) | $W$ tail | $t\bar{t}$ | Dijets[1-7] |
| (No requirement) | 4.99 | 0.231 | 10.28 | 452 | $1.91 \times 10^{10}$ |
| Preselection | 4.28±0.05 | 0.199±0.002 | 7.77±0.06 | 205.30±0.46 | $(11.2 \pm 0.19) \times 10^6$ |
| $p_T > 50$ GeV | 4.03±0.04 | 0.187±0.002 | 6.40±0.06 | 61.71±0.25 | $(1.24 \pm 0.26) \times 10^3$ |
| $\not{E}_T > 50$ GeV | 4.00±0.04 | 0.186±0.002 | 6.04±0.05 | 31.34±0.18 | 74.32±23.28 |
| Isolation | 3.95±0.04 | 0.185±0.002 | 5.99±0.05 | 28.70±0.17 | 1.00±0.82 |
| Lepton fraction | 3.81±0.04 | 0.181±0.002 | 4.85±0.05 | 0.64±0.03 | $(1.96 \pm 1.38) \times 10^{-3}$ |
| $m_T > 700$ GeV | 2.20±0.03 | | 0.043±0.002 | 0.007±0.003 | 0.001±0.001 |
| $m_T > 1400$ GeV | | 0.094±0.0001 | 0.0031±0.0006 | 0.001±0.001 | 0.001±0.001 |

Table 5: Cross-section for signal and backgrounds after each requirement. Muon mode.

calculated for transverse mass above 70% of the $W'$ boson mass, i.e. above the values listed in Table 6.

We define the $K$-factor to be the ratio of the MC@NLO cross-section to that from PYTHIA. These are shown as functions of $\eta$ in Fig. 22.

Integrals of the $W'$ boson and $W$ boson tail differential cross-sections are given in Table 7. The NLO predictions are 30-40% higher than those from PYTHIA, with little change with the variations in scale. Although the NLO/LO cross-section and acceptance ratios are of order 40%, the uncertainties on the NLO values are expected to be significantly smaller. Also, the QED corrections are partially included through PHOTOS [41] for FSR, and should have a small impact on the measurements in case of the observation of a signal.

### 6.1.2 Parton Distribution Functions

The LHC will take data in a new energy regime and so we expect significant uncertainty in signal and background predictions due to our uncertainty in knowledge of the PDFs.

The CTEQ6.1 fits include 40 error PDFs corresponding to the two limits on each of 20 eigenvectors.

| M (GeV) | $\Gamma$ (GeV) | Minimum $m_T$ (GeV) |
|---|---|---|
| 1000 | 34.739 | 700 |
| 2000 | 70.540 | 1400 |
| 3000 | 106.390 | 2100 |

Table 6: Masses and widths used as input to MC@NLO/HERWIG generation of $W'$ boson events. The third column gives the lower limit for the masses used to calculate cross-sections.

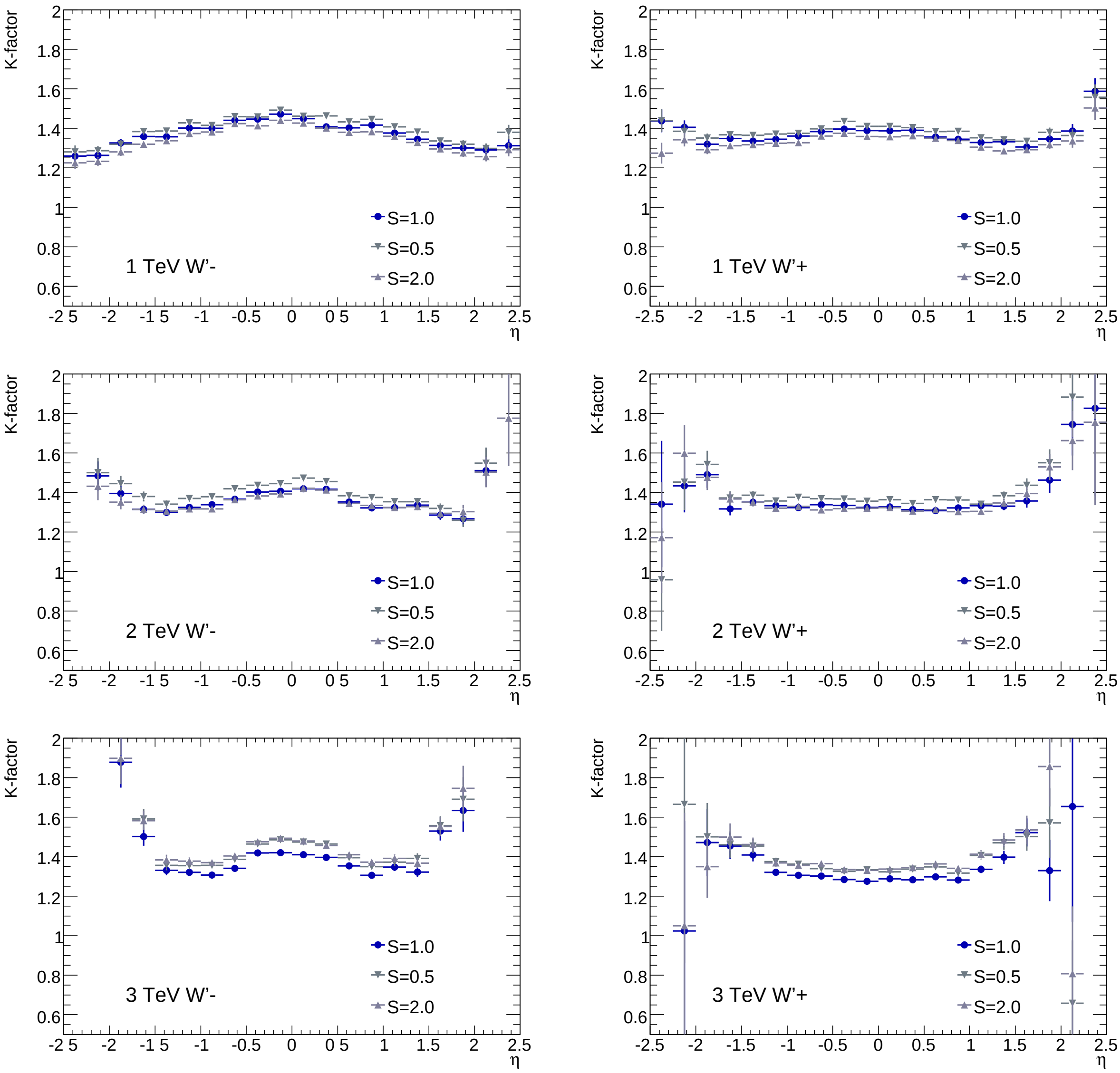


Figure 22: $W'$ boson $K$-factors (ratios of MC@NLO and PYTHIA cross-sections) as functions of $\eta$ for positive (left) and negative (right) charge for masses of 1 (top), 2 (middle) and 3 TeV (bottom). $S$ is the common scale factor. The errors are statistical.

| Process | Min. $m_T$ | PYTHIA $\sigma$ (fb) | NLO $\sigma$ (fb) | $K$-factor | S=0.5 | S=2.0 |
|---|---|---|---|---|---|---|
| $W'(m=1$ TeV)+ | 700 | 534. (1) | 742. (1) | 1.389 (4) | 1.8% (2) | -1.8% (2) |
| $W'(m=1$ TeV)- | 700 | 1204. (1) | 1644. (2) | 1.365 (3) | 1.7% (2) | -1.8% (2) |
| $W'(m=2$ TeV)+ | 1400 | 62.6 (1) | 83.0 (1) | 1.327 (3) | 2.7% (2) | -1.6% (2) |
| $W'(m=2$ TeV)- | 1400 | 20.3 (6) | 27.7 (4) | 1.362 (4) | 3.0% (2) | -1.4% (2) |
| $W'(m=3$ TeV)+ | 2100 | 6.73 (1) | 8.69 (1) | 1.292 (3) | 3.7% (2) | 4.4% (2) |
| $W'(m=3$ TeV)- | 2100 | 1.791 (6) | 2.540 (4) | 1.370 (5) | 3.7% (2) | 4.4% (2) |
| $W+$ | 700 | 20.22 (7) | 27.66 (8) | 1.368 (6) | 2.2% (4) | -0.6% (4) |
| $W$- | 700 | 8.93 (5) | 12.56 (4) | 1.407 (9) | 2.6% (5) | -0.8% (5) |
| $W+$ | 1400 | 1.042 (4) | 1.424 (4) | 1.366 (7) | 2.2% (5) | -1.5% (5) |
| $W$- | 1400 | 0.354 (2) | 0.499 (2) | 1.41 (1) | 1.8% (6) | -1.6% (5) |
| $W+$ | 2100 | 0.1231 (3) | 0.1657 (3) | 1.346 (4) | 3.0% (3) | 2.4% (3) |
| $W$- | 2100 | 0.0346 (1) | 0.0492 (1) | 1.421 (6) | 3.1% (3) | 2.5% (3) |

Table 7: Integrated $W'$ boson and $W$ boson tail cross-sections for PYTHIA and MC@NLO with common scale factor S=1. Integral is over the full $\eta$ range $-2.5 < \eta < 2.5$. The listed $K$-factors are the ratios of the integrated MC@NLO and PYTHIA cross-sections. The last two columns give the change in the MC@NLO cross-section when the common scale factor is changed by a factor of two. The statistical error in the last digit of each calculated quantity is shown in parentheses.

These can be used to estimate the uncertainty in predictions obtained with the fit. Figure 23 shows the PYTHIA prediction for the $m = 1$ TeV $W'$ boson differential cross-section as a function of $\eta$ for the CTEQ6.1 central value and each of the 40 error sets. Events are required to have transverse mass above the threshold in Table 6. The difference in shape between the positively and negatively charged bosons is a consequence of the parton distribution functions since $W'^{+}$ are from $u\bar{d}$ fusion and $W'^{-}$ from $d\bar{u}$ fusion.

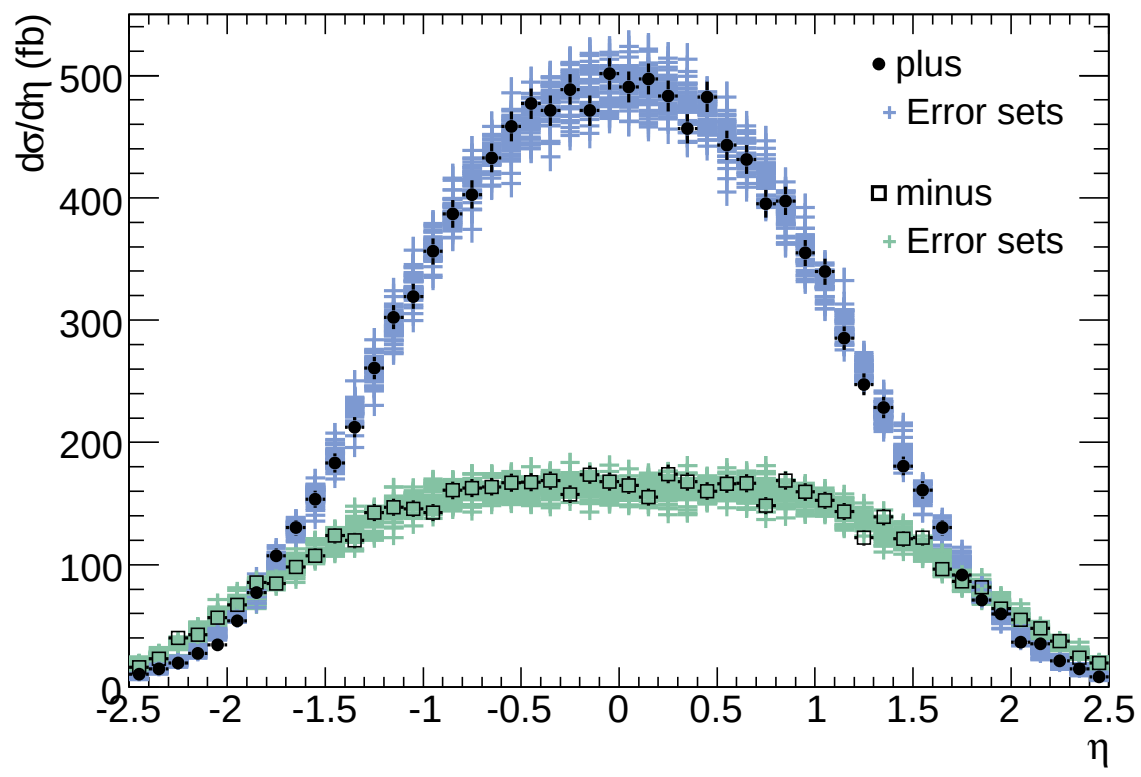


Figure 23: Muon $\eta$ distributions for positively- and negatively-charged $m = 1$ TeV $W'$ bosons using the CTEQ 6.1 PDF central value (black) and 40 error sets.

We calculated cross sections for $W'$ boson production with mass of 1 TeV using the CTEQ6.1 central value and error PDFs by integrating over the full $\eta$ range ($|\eta| < 2.5$) in Fig. 23. To estimate the overall uncertainty, the positive and negative deviations for each eigenvector were summed separately in quadrature for each charge sign. Where both deviations for an eigenvector had the same sign, only the

larger magnitude was included in the sums. Table 8 shows the results.

| Process | Min. $m_T$ | $W$+ | $W$- |
|---|---|---|---|
| $W'$ ($m = 1$ TeV) | 700 | -4.1% (5), +8.2% (5) | -11.1% (7), +3.5% (8) |

Table 8: CTEQ6.1 combined error set deviations for $W'$ boson cross-sections. The statistical error on the last digit is shown in parentheses.

Combining all the above, we assign a common $K$-factor of 1.37 for all masses and charges and assign an 8% uncertainty on this factor.

## 6.2 Instrumental Uncertainties

Detector related uncertainties for these studies can be divided in two categories: the ones related to the reconstruction of the leptons and the ones corresponding to the global event activity as the $\not\!\!E_T$ and the jet characteristics. However, the lepton reconstruction uncertainties can be the dominant factor in the $\not\!\!E_T$ resolution.

### 6.2.1 Lepton Reconstruction

Three main contributions can be identified in this category. The efficiency of lepton identification, as well as the fake rates associated with this, the $p_T$ or $E_T$ scale and its measurement resolution.

Systematic errors on the momentum scale of the muons can arise for instance due to the non-perfect knowledge of the magnetic field. To take into account such effects, a variation of $\pm 1\%$ is applied to the $p_T$ of the reconstructed muons. Positive and negative variations are considered separately. In a similar way but for energy, a variation of $\pm 0.5\%$ was made for electrons.

An incomplete understanding of the material distributions inside the detector as well as possible misalignments in the MS can lead to an additional smearing of the momentum measurement resolution of muons. To evaluate the impact of such contributions on the analysis, a smearing, based on early calibrations of $\sigma(1/p_T) = 0.011/p_T \oplus 0.00017$ is applied. The first term enhances the Coulomb scattering smearing, while the second enhances the alignment contribution, and is the crucial factor in this study.

For the energy measurement resolution for electrons, the total $\sigma(E_T)$ is smeared by $0.0073 \times E_T$, which enhances the constant term only.

Lepton identification efficiency is obviously important for this analysis. The identification efficiency can be estimated from the data, using the tag-and-probe method described in [42] for muons in the region $20 < p_T < 50$ GeV and extrapolated to higher $p_T$ using simulated data. A value of $\pm 5\%$ has been chosen for the evaluation of this uncertainty, corresponding to the early running period of integrated luminosities $\mathscr{L} < 100$ pb$^{-1}$. In the case of electrons a $\pm 1\%$ variation has been applied.

### 6.2.2 Jet Reconstruction

An uncertainty on the jet energy scale of $\pm 7\%$ was imposed, together with an uncertainty on its resolution of $\sigma(E_T) = 0.45 \times \sqrt{E_T} \oplus 5\%$.

### 6.2.3 Missing Energy

If jets or leptons are systematically shifted, then missing transverse energy should be systematically shifted in a known direction. Based on the jet and leptons performance, the missing energy is shifted as follows:

- $\not{E}_{T\,(shifted)}(x) = \not{E}_T(x) + E^{lepton/jet}(x) - E^{lepton/jet}_{shifted}(x)$
- $\not{E}_{T\,(shifted)}(y) = \not{E}_T(y) + E^{lepton/jet}(y) - E^{lepton/jet}_{shifted}(y)$
- $\sum E_{T(shifted)} = \sum E_T + E^{lepton/jet}_T - E^{lepton/jet}_{T(shifted)}$

In the case of muons, momentum is used instead of energy.

### 6.2.4 Summary of Experimental Systematic Uncertainties

The effects of the experimental uncertainties are summarized in Tables 9 and 10. In the high-$p_T$ range

| | electrons | | muons | |
|---|---|---|---|---|
| Description of systematic | $\delta_s$ [%] | $\delta_b$ [%] | $\delta_s$ [%] | $\delta_b$ [%] |
| Lepton energy scale + | +0.8 | +1.8 | +1.2 | +4.6 |
| Lepton energy scale - | -0.7 | -2.1 | -1.2 | -4.4 |
| Lepton energy resolution | +0.1 | +0.2 | -1.0 | +3.7 |
| Lepton identification efficiency + | +1.0 | +1.0 | +5. | +5. |
| Lepton identification efficiency - | -1.0 | -1.0 | -5. | -5. |
| Jet energy scale + | +0.1 | -0.2 | -0.1 | +0.1 |
| Jet energy scale - | +0.1 | -0.2 | +0.1 | +0.7 |
| Jet energy resolution | +0.0 | +0.1 | -0.1 | +0.3 |
| Luminosity | ±3. | | ±3. | |

Table 9: Effect of the detector systematics in percentage for $m_{W'} = 1$ TeV. $\delta_s$ is the uncertainty on the signal, $\delta_b$ is the uncertainty on the background.

| | electrons | | muons | |
|---|---|---|---|---|
| Description of systematic | $\delta_s$ [%] | $\delta_b$ [%] | $\delta_s$ [%] | $\delta_b$ [%] |
| Lepton energy scale + | +0.7 | +1.2 | +1.5 | +3.4 |
| Lepton energy scale - | -0.4 | -3.7 | -1.7 | -2.5 |
| Lepton energy resolution | -0.03 | 0.0 | -4.2 | +6.8 |
| Lepton identification efficiency + | +1.0 | +1.0 | +5. | +5. |
| Lepton identification efficiency - | -1.0 | -1.0 | -5. | -5. |
| Jet energy scale + | +0.1 | 1.2 | +0.1 | +0.8 |
| Jet energy scale - | -0.1 | 0 | -0.3 | -0.1 |
| Jet energy resolution | -0.1 | 0 | +0.1 | -0.1 |
| Luminosity | ±3. | | ±3. | |

Table 10: Effect of the detector systematics in percentage for $m_{W'} = 2$ TeV. $\delta_s$ is the uncertainty on the signal, $\delta_b$ is the uncertainty on the background.

under consideration, the systematic uncertainties on the quality of single lepton reconstruction have a stronger effect on the muon channel, for which there are comparable contributions from energy scale, resolution and identification efficiency (with the resolution uncertainty becoming more important for a higher $W'$ mass); out of these three, the energy scale uncertainty dominates in the electron channel for $m_{W'} = 1$ TeV, but becomes less important for $m_{W'} = 2$ TeV. Jet uncertainties do not play a strong role on either channel.

# 7 Discovery Potential

In order to assess the ATLAS discovery potential in the search for a $W' \to \ell + \not\!\!E_T$ signal, the luminosity needed for a $5\sigma$ excess is obtained as a function of the mass of the $W'$ boson.

The significance is obtained from the expected number of signal and background events in the region $m_T > 0.7 m_{W'}$, where $m_{W'}$ is the mass of the hypothesized $W'$ boson. Calling these expected numbers $s$ and $b$, respectively, the significance $S$ is obtained as

$$S = \sqrt{2((s+b)\ln(1+s/b)-s)}$$

which gives a good approximation to the likelihood-ratio based significance in the low statistics regime. Figure 24 shows the expected integrated luminosity needed for a 5-sigma excess as a function of the mass of the $W'$ boson.

Higher order corrections for $W'/W \to \ell\nu$ processes are taken into account as stated in section 6.1. Systematic uncertainties listed in Tables 9 and 10 are taken into account by increasing the expected background by the sum in quadrature of its positive expected variations, and by reducing the signal by the sum in quadrature of its expected negative variations; this assumes no correlations of the expected signal and background expectations and, as a result, produces a conservative estimate.

For comparison, the integrated luminosity values for a $5\sigma$ significance were also obtained taking into account the shape of the signal and background $m_T$ distributions. This was done using a technique in which, instead of an ensemble of Monte Carlo pseudo-experiments [43], a Fast Fourier Transform (FFT) is used to calculate the experimental estimator distributions [44]. This method allows a fast determination of the probability that background fluctuations produce a signal-like result, but it depends on the assumption that both the location of the signal and its shape are well known. Treating each bin of the transverse mass distribution as an independent search channel, and combining them accordingly, the resulting sensitivity is in general higher than the estimation given in the number counting approach. With this method, the luminosity required for a $5\sigma$ effect was reduced between 20 and 35% with respect to the values shown in Fig. 24.

Even for very low integrated luminosities (of the order of picobarns), a $W'$ boson with a mass above the current experimental limits could be found with a significance in excess of $5\sigma$, while, with 1 $\text{fb}^{-1}$, masses of the order of 3 TeV can be reached. As an illustration, Figs 25 and 26 show Monte Carlo outcomes of pseudo-experiments corresponding to 10 $\text{pb}^{-1}$ and 100 $\text{pb}^{-1}$, respectively, for both channels. The solid line histograms depict the expected background, those in dotted lines the $m = 1$ TeV $W'$ boson signal and the dashed-dotted line histograms show possible $m = 2$ TeV $W'$ boson signals.

# 8 Summary and Conclusion

The potential for the ATLAS experiment to reconstruct and identify the decay of a heavy, charged gauge boson into a lepton and a neutrino has been studied. Various systematic and theoretical uncertainties have been considered, as well as plausible estimations of our uncertainties about the performance of the detector in the early stages of data taking. These studies show that, even with integrated luminosities as low as 10 $\text{pb}^{-1}$ of data, it would be possible to discover this type of bosons, should they exist not far beyond the current experimental limits and have Standard Model like couplings. With an integrated luminosity of a few $\text{fb}^{-1}$, ATLAS has the potential to discover these particles for masses up to 4 TeV.

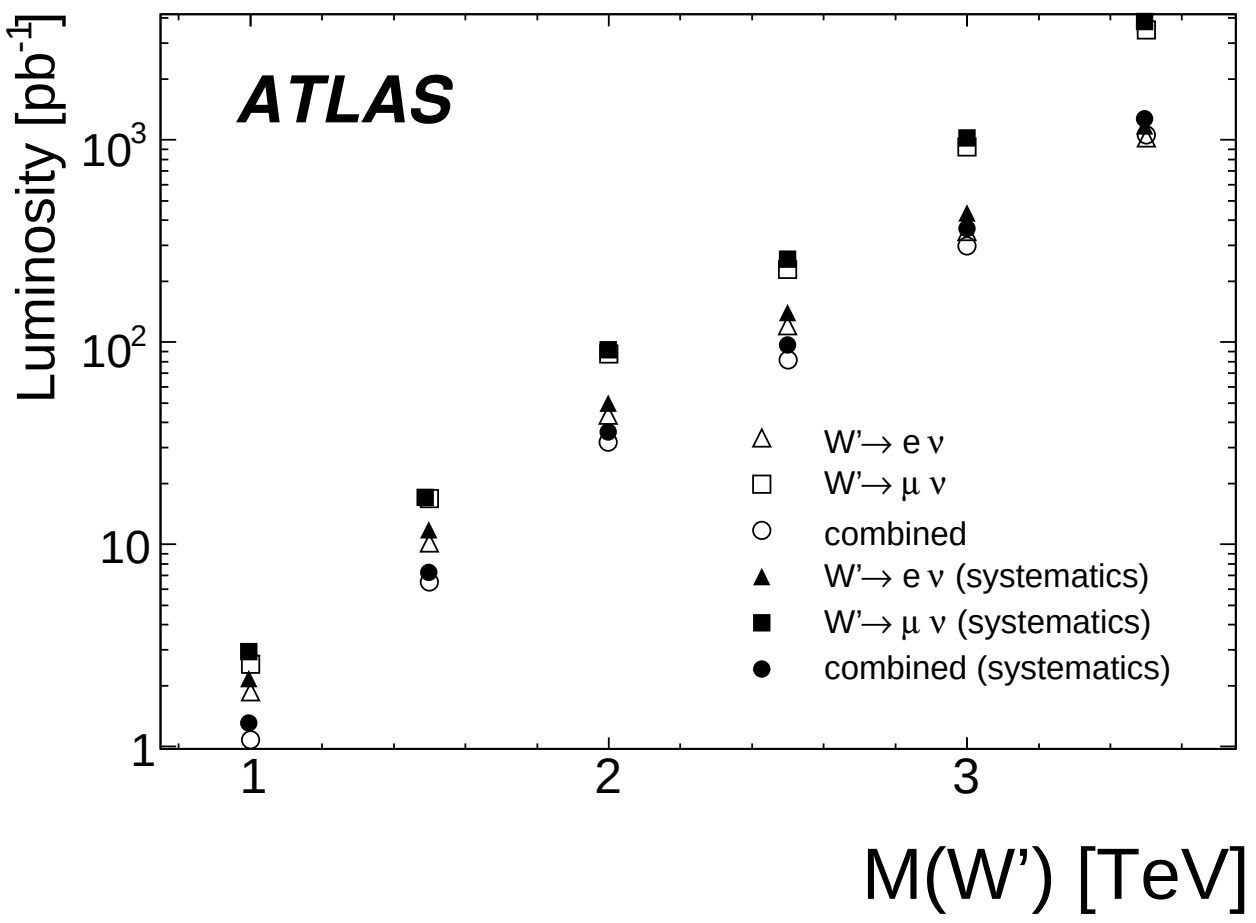


Figure 24: Integrated luminosity needed to have a $5\sigma$ discovery as a function of the mass of the $W'$ bosons; triangles correspond to the $e\nu$ search, squares to $\mu\nu$, circles to the combined search. Filled markers include the effect of systematic uncertainties.

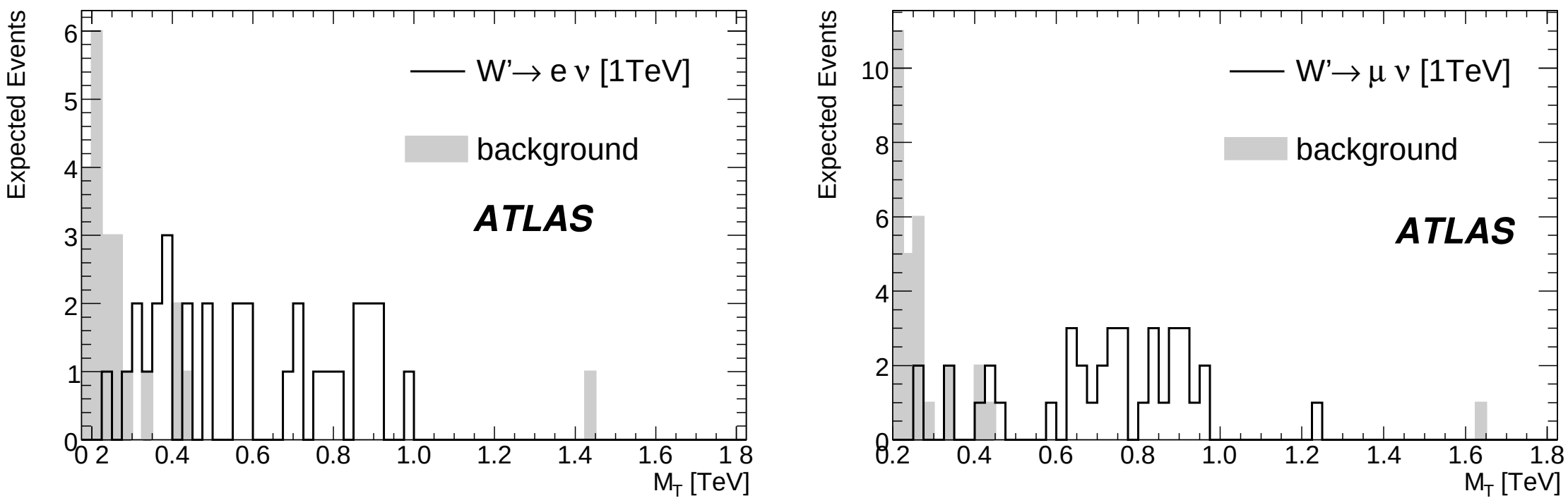


Figure 25: Monte Carlo pseudo-experiment for 10 pb$^{-1}$. Left: electron channel; right: muon channel.

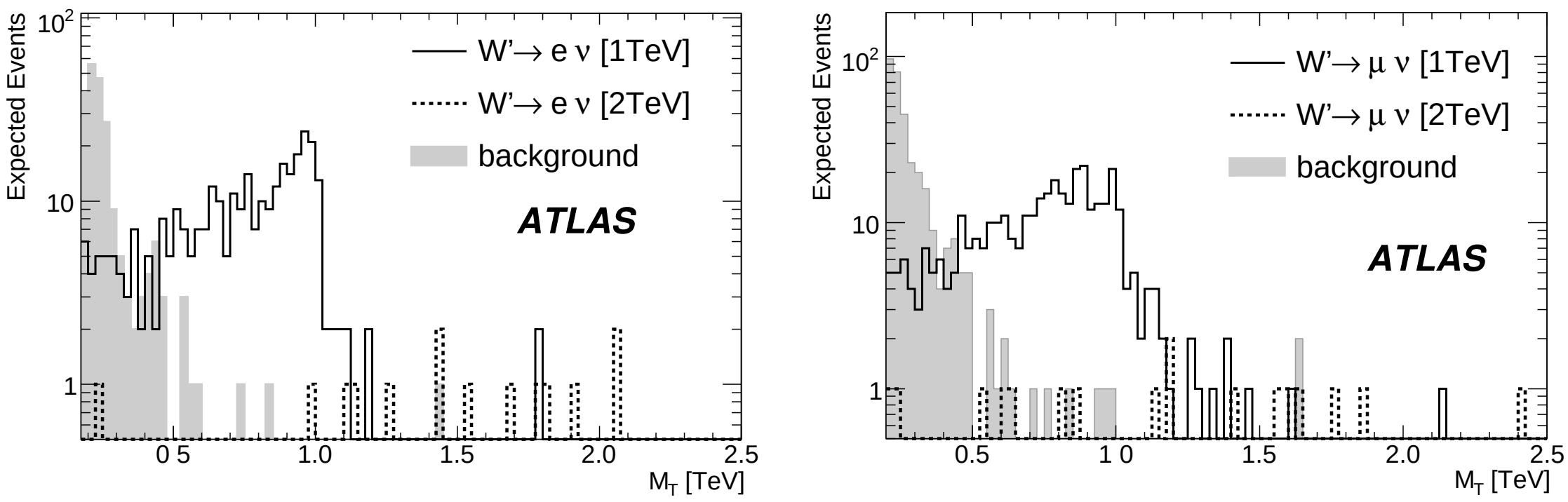


Figure 26: Monte Carlo pseudo-experiment for 100 pb$^{-1}$. Left: electron channel; right: muon channel.

# Search for Leptoquark Pairs and Majorana Neutrinos from Right-Handed $W$ Boson Decays in Dilepton-Jets Final States

**Abstract**

Final states with high-$p_T$ leptons and jets are predicted by many Beyond the Standard Model scenarios. Two prominent models are used here as guides to understanding the event topologies: the scalar leptoquarks and the Left-Right Symmetry. In contrast to many SUSY signatures, their topologies rarely contain missing energy. Their discovery potential with early ATLAS data, corresponding to an integrated luminosity of a few hundred inverse picobarns, is discussed.

## 1 Introduction

Grand Unification has inspired many extensions of the Standard Model. Such models introduce new, usually very heavy particles, and previous searches for Grand Unification Theory (GUT) signatures have placed limits on masses and interaction strengths of the new particles. The LHC will probe new regions of parameter space, allowing for a direct search for these particles. Decays characterized by final states with two highly energetic leptons, two jets and no missing transverse energy are studied in this note. The models for new physics considered for this note are described below. The simulation of signal and background processes is described in section 2. In section 4 the baseline selection that is used for all analyses is explained. After the trigger requirements are given (section 3), section 5 details the specifics of each of the analyses. The systematics are described in section 6 and the final sensitivity estimates are given in section 7.

### 1.1 Leptoquarks

The experimentally observed symmetry between leptons and quarks has motivated the search for leptoquarks (LQ), hypothetical bosons carrying both quark and lepton quantum numbers, as well as fractional electric charge [1–5]. Leptoquarks could, in principle, decay into any combination of a lepton and a quark. Experimental limits on lepton number violation, flavor-changing neutral currents, and proton decay favour three generations of leptoquarks. In such a scenario, each leptoquark couples to a lepton and a quark from the same Standard Model generation [6]. Leptoquarks can either be produced in pairs by the strong interaction or in association with a lepton via the leptoquark-quark-lepton coupling. Figure 1 shows the Feynman diagrams for leptoquark production processes accessible at the LHC.

This note describes the search for leptoquarks decaying to either an electron and a quark or a muon and a quark. The branching ratio of a leptoquark to a charged lepton and a quark is denoted as $\beta$. Decays to neutrinos are not considered, and events are not explicitly selected based on the flavor of the quark. The experiments at the Tevatron have searched for first (decaying to $eq$), second (decaying to $\mu q$), and third (decaying to $\tau q$) generation scalar leptoquarks. For $\beta = \mathscr{B}(LQ \to \ell^{\pm} q) = 1$, the DØ [7] and CDF [8] collaborations have set 95%CL limits for first generation scalar leptoquarks of $m_{\mathrm{LQ}_1} > 256$ GeV and $m_{\mathrm{LQ}_1} > 236$ GeV, respectively. These limits are based on integrated $p\bar{p}$ luminosities of approximately 250 pb$^{-1}$ and 200 pb$^{-1}$. The results for second generation leptoquarks, $m_{\mathrm{LQ}_2} > 251$ GeV and $m_{\mathrm{LQ}_2} > 226$ GeV, were obtained with 300 pb$^{-1}$ and 200 pb$^{-1}$ by the DØ [9] and CDF [10] experiments, respectively.

The Tevatron exclusion limits are expected to reach 300-350 GeV in the near future.

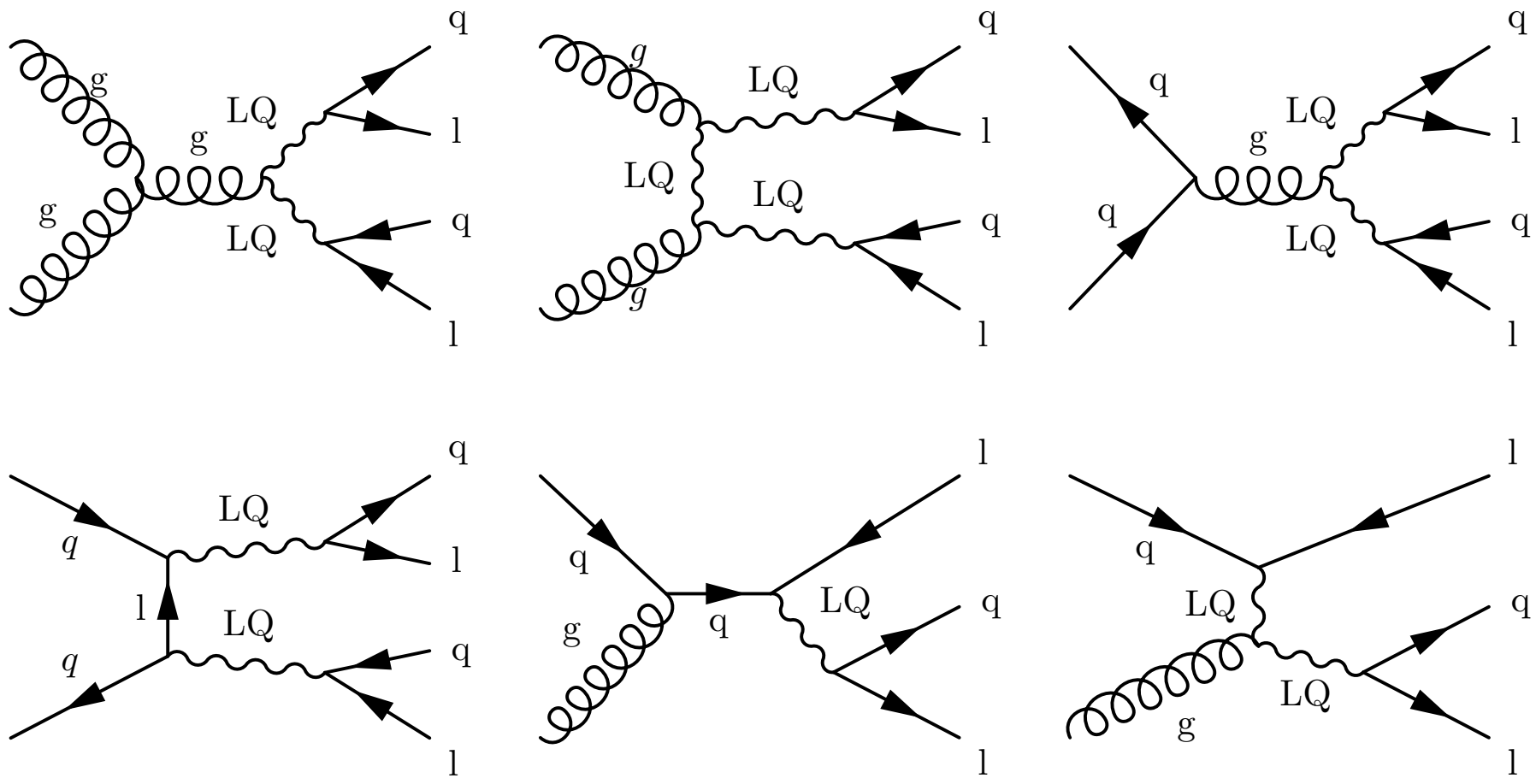


Figure 1: Feynman diagrams for leptoquark production.

## 1.2 Left-Right Symmetry

Left-Right Symmetric Models (LRSMs) of the weak interaction address two important topics: the nonzero masses of the three known left-handed neutrinos [11] and baryogenesis. LRSMs conserve parity at high energies by introducing three new heavy right-handed Majorana neutrinos $N_e$, $N_\mu$ and $N_\tau$. The smallest gauge group that implements an LRSM is $SU(2)_L \times SU(2)_R \times U(1)_{B-L}$. At low energies, the left-right symmetry is broken and parity is violated. The Majorana nature of the new heavy neutrinos explains the masses of the three left-handed neutrinos through the see-saw mechanism [12]. The lepton number $L$ could be violated in processes that involve the Majorana neutrinos. This opens a window to the very attractive theoretical scenario for baryogenesis via leptogenesis, where baryon and lepton numbers $B$ and $L$ are violated but $B-L$ is conserved.

In addition to the Majorana neutrinos, most general LRSMs also introduce the new intermediate vector bosons $W_R$ and $Z'$, Higgs bosons, and a left-right mixing parameter. The most restrictive lower limit on the mass of the $W_R$ boson comes from the $K_L - K_S$ mass difference which requires $m_{W_R} >$ 1.6 TeV. This lower limit is subject to large corrections from higher-order QCD effects. Heavy right-handed Majorana neutrinos with masses of about a few hundred GeV would be consistent with the data from supernova SN1987A. Such heavy neutrinos would allow for a $W_R$ boson at the TeV mass scale. This scenario would also be consistent with LEP data on the invisible width of the $Z$ boson. Present experimental data on neutral currents imply a lower limit on the mass of a $Z'$ boson of approximately 400 GeV. Recent direct searches [13] for the $W_R$ boson at DØ give a lower mass limit of 739 GeV and 768 GeV, assuming the $W_R$ boson could decay to both lepton pairs and quark pairs, or only to quark pairs, respectively. However, heavy Majorana neutrinos decaying to a lepton and a pair of quarks (detected as jets) were not searched for in those analyses.

The new intermediate vector bosons $W_R$ and $Z'$ would be produced at the LHC via the Drell-Yan (DY) process like Standard Model $W$ and $Z$ bosons. Their decays would be a source of new Majorana neutrinos. The Feynman diagram for $W_R$ boson production and its subsequent decay to a Majorana neutrino is shown in Fig. 2. This note describes an analysis of $W_R$ boson production and its decays $W_R \rightarrow eN_e$ and $W_R \rightarrow \mu N_\mu$, followed by the decays $N_e \rightarrow eq'\bar{q}$ and $N_\mu \rightarrow \mu q'\bar{q}$, which can be detected in final states with (at least) two leptons and two jets.

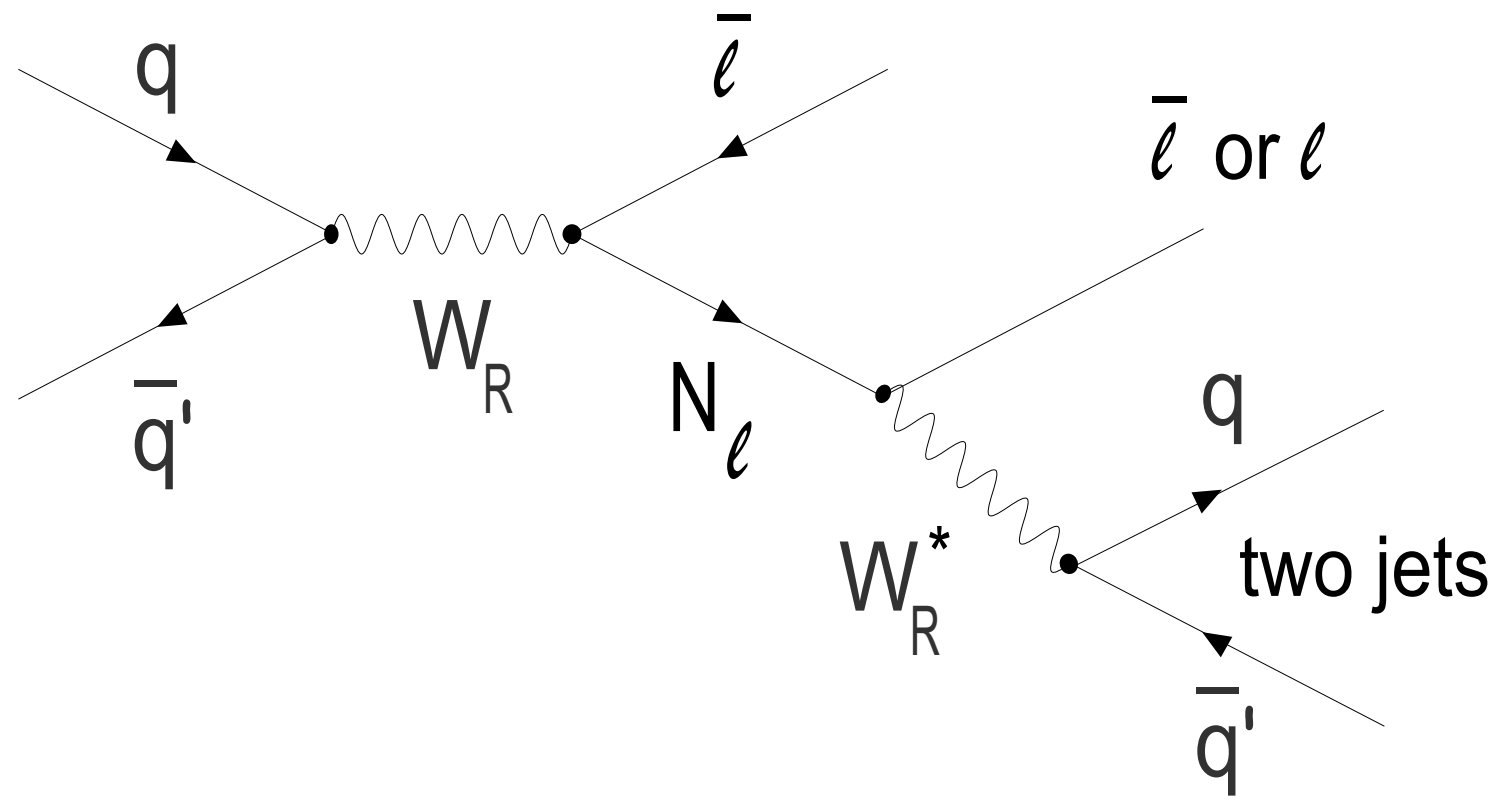


Figure 2: Feynman diagram for $W_R$ boson production and its decay to a Majorana neutrino $N_\ell$.

| $m_{LQ}$ in GeV | $\sigma(pp \to LQLQ)$ (NLO) in pb |
|---|---|
| 300 | $10.1 \pm 1.5$ |
| 400 | $2.24 \pm 0.38$ |
| 600 | $0.225 \pm 0.048$ |
| 800 | $0.0378 \pm 0.0105$ |

Table 1: NLO cross-sections for scalar leptoquark pair production at the LHC [16].

# 2 Simulation of Physics Processes

## 2.1 Leptoquarks

The signals have been studied using samples of first generation (1st gen.) and second generation (2nd gen.) scalar leptoquarks simulated with the Monte Carlo (MC) generator PYTHIA [14] and using the CTEQ6L1 parameterization [15] of the parton density functions (PDFs). A leptoquark-lepton-quark coupling $\lambda = 0.8$ was used in the event generation leading to a natural width of the leptoquarks of 0.63 GeV and 1.3 GeV for leptoquark masses of 400 GeV and 800 GeV respectively. The next to leading order (NLO) cross-sections for leptoquark pair production at 14 TeV $pp$ centre-of-mass energy were taken from Ref. [16] and are shown in Table 1 for the four simulated leptoquark masses.

## 2.2 Left-Right Symmetry

Studies of the discovery potential for $W_R$ bosons and the Majorana neutrinos, $N_e$ and $N_\mu$ produced in their decays, were performed using datasets simulated with the MC generator PYTHIA according to a particular implementation [17] of an LRSM described in [18]. The Standard Model axial and vector couplings, the CKM matrix for the quark sector, no mixing between the new and Standard Model intermediate vector bosons, and phase space isotropic decays of Majorana neutrinos are assumed for the right-handed sector in this model. The products of leading-order production cross-sections $\sigma(pp \to W_R X)$ and branching fractions to studied final states $W_R \to \ell N_\ell \to \ell\ell jj$ are 24.8 pb for $m_{W_R} = 1800$ GeV, $m_{N_e} = m_{N_\mu} = 300$ GeV and 47.0 pb for $m_{W_R} = 1500$ GeV, $m_{N_e} = m_{N_\mu} = 500$ GeV. In the rest of this note, these samples are referred to as LRSM_18_3 and LRSM_15_5, respectively. The Majorana nature of the new heavy neutrinos allows for same-sign and opposite-sign dileptons.

### 2.3 Background Processes

The main sources of background for the analyses presented here are $t\bar{t}$ and inclusive $Z/\gamma^*$ production processes. Multijet production, where two jets are misidentified as leptons, represents another background. In addition, minor contributions arise from diboson production. Other potential background sources, such as single-top production, were also studied. Their contribution was found to be insignificant.

- $Z/\gamma^*$ background was studied using a combination of two MC samples with generator-level dilepton invariant mass preselections of $m_{\ell\ell} > 60$ GeV and $m_{\ell\ell} > 150$ GeV, the latter sample corresponding to a much larger integrated luminosity than the former. The samples were normalized to the given luminosity using their partial cross-sections and the NLO estimate $\sigma(pp \to Z) \times \mathscr{B}(Z \to \ell^+\ell^-) = 2032$ pb, obtained with the MC generator FEWZ [19, 20]. A lepton filter was applied at the event generation, requiring at least one electron or muon with transverse momentum greater than 10 GeV and absolute pseudo-rapidity smaller than 2.7, resulting in an effective cross-section of 1808 pb.

  For logistical reasons, the sample with the lower mass preselection was generated using the MC generator PYTHIA [14], and the sample with higher mass preselection was generated using HERWIG [21]. In both cases, the CTEQ6L1 [15] parton distribution functions were used. The consistency between the two samples was verified at high dilepton masses.

- $t\bar{t}$ background was simulated using the MC generator MC@NLO [22] using the CTEQ6M [15] parton distribution functions. It was normalized to the given integrated luminosity using a production cross-section of 833 pb estimated to the next-to-leading order (NLO+NLL) [23]. In addition, a lepton filter was applied, requiring at least one electron or muon with transverse momentum greater than 1 GeV, which resulted in an effective cross-section of 450 pb.

- The diboson samples were generated using HERWIG with a generator-level preselection on the invariant mass of $Z/\gamma^* > 20$ GeV. With this requirement, the NLO partial cross-sections for $WW$, $WZ$ and $ZZ$ boson pair production processes were numerically estimated (using MC@NLO) to be 117.6 pb, 56.4 pb, 17.8 pb, respectively. The CTEQ6L1 parton distribution functions were used for event generation. Again, a lepton filter was applied, with a transverse momentum threshold of 10 GeV and a maximum absolute pseudo-rapidity of 2.8. This resulted in a total effective cross-section of 60.9 pb.

- The multijet background was simulated using PYTHIA with the CTEQ6L1 structure functions. The normalization was based on PYTHIA cross-section estimates. The statistics of these samples are very limited, such that no reliable estimate of this background could be made at this time.

## 3 Trigger Requirements

The trigger system [24] of the ATLAS experiment has three levels, L1, L2 and the Event Filter (EF). To ensure high overall trigger efficiencies, our analyses rely on single lepton trigger streams with relatively low thresholds. The dielectron analyses rely on the single electron-based trigger called `e55` which has a threshold of around 60 GeV [24]. When selected events fail this trigger, the analyses rely on the lower-threshold (about 25 GeV) single electron trigger called `e22i` [24] in which the electron is required to be isolated. A single muon trigger with threshold about 20 GeV (`mu20` [24]) is used in the dimuon analyses.

Final states studied in this note always contain two high-$p_T$ leptons. While the baseline selection described in section 4 requires two leptons with $p_T > 20$ GeV, most signal events contain at least one lepton with significantly higher $p_T$. As a result, the overall trigger efficiency for events that satisfy all

analysis selection criteria (section 5) exceeds 95%. The trigger efficiencies for signal MC events that satisfy all selection criteria are shown in Table 2.

| Process | L1 | L2 | EF | L1*L2*EF |
|---|---|---|---|---|
| 1st gen. leptoquarks $m_{LQ} = 400$ GeV | 100.0% | 99.4% | 97.6% | 97.0% |
| 2nd gen. leptoquarks $m_{LQ} = 400$ GeV | 97.7% | 99.1% | 99.7% | 96.5% |
| LRSM (ee) $m_{W_R} = 1800$ GeV, $m_{N_e} = 300$ GeV | 100.0% | 99.2% | 97.2% | 96.4% |
| LRSM ($\mu\mu$) $m_{W_R} = 1800$ GeV, $m_{N_\mu} = 300$ GeV | 96.8% | 98.7% | 98.9% | 94.5% |

Table 2: Overall trigger efficiencies for signal events that satisfy all selection criteria.

## 4 Baseline Event Selection

The baseline event selection, common for all analyses presented in this note, requires two leptons and two jets. All analyses use the same selection criteria for signal electron, muon, and jet candidates. The baseline selection criteria for these reconstructed objects are summarized below. Performance studies are described elsewhere [25–28].

Electron candidates are identified as energy clusters reconstructed in the liquid argon electromagnetic calorimeter that match tracks reconstructed in the inner tracking detector and satisfy the *medium* electron identification requirements [25].

Muon candidates are identified as tracks reconstructed in the muon spectrometer [26] that, when extrapolated to the beam axis, match a track reconstructed in the inner detector, and satisfy relative isolation energy requirements $E_T^{iso}/p_T^{\mu} \leq 0.3$. $p_T^{\mu}$ is the muon candidate's transverse momentum and $E_T^{iso}$ is the energy detected in the calorimeters in a cone of $\Delta$R=$\sqrt{\Delta\eta^2+\Delta\phi^2}$=0.2 around the muon candidate's reconstructed trajectory, corrected for the expected energy deposition by a muon.

Jets are identified as energy clusters reconstructed in the calorimeters using a $\Delta$R=0.4 cone algorithm [27]. $\Delta R$ between a jet and any electron candidate (as defined above) must be larger than 0.1. This veto is imposed to avoid electrons being misidentified as jets. It is applied in all analyses, regardless of whether electrons are explicitly considered in the final states or not. The jet energy scale calibration is performed using full MC simulation and requires that the average reconstructed jet energy agrees with the average energy of the jets reconstructed with the Monte Carlo truth particles. The same jet reconstruction algorithm, with cone size $\Delta R = 0.4$, is used for both reconstruction and calibration.

All objects are required to have $p_T \geq 20$ GeV, the leptons must have an absolute pseudo-rapidity $|\eta|$ smaller than 2.5 and jets must have $|\eta| \leq 4.5$.

To suppress contributions from Drell-Yan backgrounds, the dilepton invariant mass is required to be at least 70 GeV. Tighter analysis-specific requirements are later applied to this and other variables in order to achieve the best sensitivities in individual studies, as described in the following section.

## 5 Individual Analyses

### 5.1 Search for Leptoquark Pair Production

Following the baseline object identification criteria described above, the leptoquark pair analyses require events to have at least two oppositely charged leptons of the same flavour and at least two jets. Signal sensitivity and discovery potential are estimated using a sliding mass window algorithm: only events in the mass region around the assumed mass of the leptoquark are analyzed.

For large leptoquark masses, signal leptons and jets have, on average, larger transverse momenta than background particles. The following kinematic quantities are used to separate the signal from backgrounds: the transverse momentum of the leptons ($p_T$ ), the scalar sum of the transverse momenta of the two most energetic jets and leptons ($S_T = \sum|\vec{p}_T|_{jet} + \sum|\vec{p}_T|_{lep}$), the dilepton invariant mass ($m_{\ell\ell}$), and lepton-jet invariant mass. The lepton-jet invariant mass represents the mass of the leptoquark if the correct lepton-jet combination is chosen. Since there are two leptons and two jets there are two possible combinations, and we choose the combination which gives the smallest difference between the masses of the first and second leptoquark candidates.

| Physics sample | Before selection | Baseline selection | $S_T \geq$ 490 GeV | $m_{ee} \geq$ 120 GeV | $m^1_{lj}$ - $m^2_{lj}$ window (GeV) [320-480] - [320-480] | [700-900] - [700-900] |
|---|---|---|---|---|---|---|
| $LQ$ ($m = 400$ GeV) | 2.24 | 1.12 | 1.07 | 1.00 | 0.534 | - |
| $LQ$ ($m = 800$ GeV) | 0.0378 | 0.0177 | 0.0177 | 0.0174 | - | 0.0075 |
| $Z/\gamma^* \geq 60$ GeV | 1808. | 49.77 | 0.722 | 0.0664 | 0.0036 | 0.00045 |
| $t\bar{t}$ | 450. | 3.23 | 0.298 | 0.215 | 0.0144 | $< 0.0012$ |
| Vector Boson pairs | 60.9 | 0.610 | 0.0174 | 0.00384 | $< 0.002$ | $< 0.0014$ |
| Multijet | $10^8$ | 20.51 | 0.229 | 0.184 | 0.0 | 0.0 |

Table 3: 1st generation leptoquark analysis. Partial cross-sections (pb) that survive selection criteria. The upper limits are given at 68% confidence level.

| Physics sample | Before selection | Baseline selection | $p_T^{\mu}\geq$60 GeV $p_T^{jet}\geq$25 GeV | $S_T \geq$ 600 GeV | $m_{\mu\mu} \geq$ 110 GeV | $m_{lj}$ window (GeV) [300-500] | [600-1000] |
|---|---|---|---|---|---|---|---|
| LQ (400 GeV) | 2.24 | 1.70 | 1.53 | 1.27 | 1.23 | 0.974 | - |
| LQ (800 GeV) | 0.0378 | 0.0313 | 0.0306 | 0.0304 | 0.030 | - | 0.0217 |
| $Z/\gamma^* \geq$60 GeV | 1808. | 79.99 | 2.975 | 0.338 | 0.0611 | 0.021 | 0.014 |
| $t\bar{t}$ | 450. | 4.17 | 0.698 | 0.0791 | 0.0758 | 0.0271 | 0.0065 |
| VB pairs | 60.9 | 0.876 | 0.0654 | 0.00864 | 0.00316 | 0.00185 | 0.00076 |
| Multijet | $10^8$ | 0.0 | 0.0 | 0.0 | 0.0 | 0.0 | 0.0 |

Table 4: 2nd generation leptoquark analysis. Partial cross-sections (pb) that survive selection criteria.

In both channels, the values of these selection criteria are optimized[1] to achieve discovery with $5\sigma$ significance at the lowest luminosity possible. Tables 3 and 4 show the values of the selection criteria and resulting signal and background cross-sections for 1st and 2nd generation channels, respectively. One important difference between the two channels is the background due to jets being misidentified as electrons. This background can be significantly reduced by requiring both reconstructed jet-electron masses, ($m^1_{lj}$, $m^2_{lj}$), to be close to the tested leptoquark mass. However, such a selection in the 2nd generation analysis would significantly reduce the signal efficiency, especially for larger leptoquark masses. Therefore, a less strigent selection is applied, and only the average of the two muon-jet masses ($m^{av}_{lj}$) is required to be near the tested leptoquark mass.

Figure 3 shows the $S_T$ variable distribution with $m_{LQ} = 400$ GeV, along with the main backgrounds, Drell-Yan and $t\bar{t}$ production, after baseline selection plus, for the 2nd generation case, the requirements

[1] At this stage, only statistical uncertainties are taken into account.

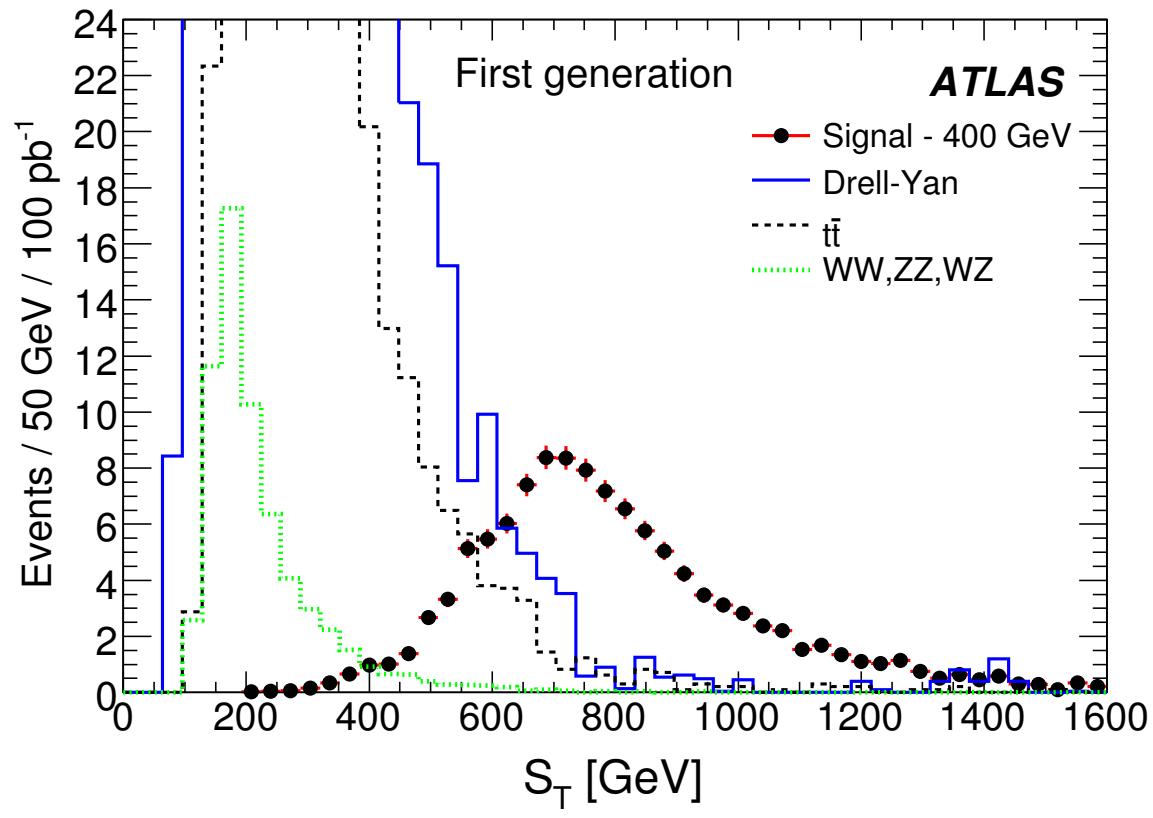


Figure 3: $S_T$ in leptoquark MC events ($m_{LQ} = 400$ GeV) after baseline selection. Left: 1st generation, right: 2nd generation with the additional requirements $p_T^{\mu} > 60$ GeV and $p_T^{jet} > 25$ GeV.

$p_T^{\mu} > 60$ GeV and $p_T^{jet} > 25$ GeV.

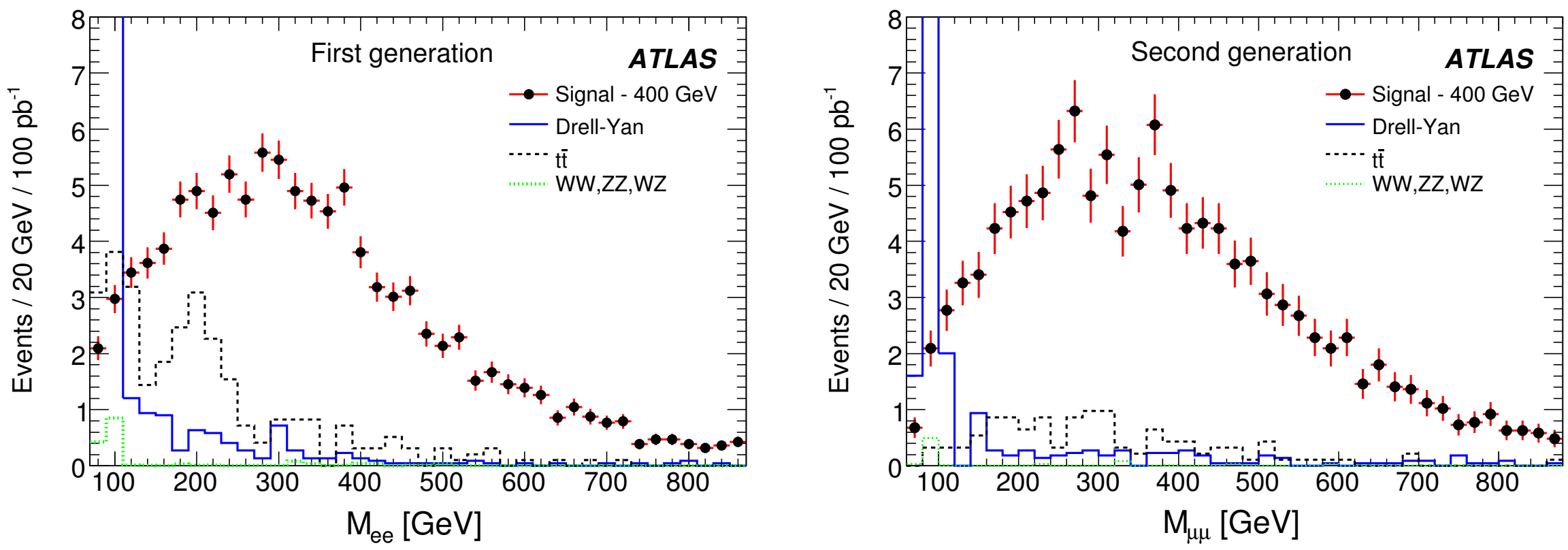


Figure 4: $m_{\ell\ell}$ of the selected lepton pair after $S_T$ selection in leptoquark 1st generation (left) and 2nd generation (right) events ($m_{LQ} = 400$ GeV).

The dilepton mass distribution after the $S_T$ selection is shown in Figure 4.

Figures 5 and 6 show the reconstructed invariant mass of leptoquark candidates ($m_{LQ}$=400 GeV) in signal events and the main backgrounds, Drell-Yan and $t\bar{t}$ production, after the subsequent selections on dimuon mass and $S_T$. Due to gluon radiation, quarks produced in the decays of heavy particles are not equivalent to standard jets. This shifts the peak of the jet energy resolution function towards smaller energies and results in a low-mass shoulder in the distribution of reconstructed masses of heavy particles. Figure 5 shows two entries per event corresponding to the two reconstructed electron-jet objects obtained by adding x and y mass projections of ($m_{lj}^1$, $m_{lj}^2$) on a common axis, $m_{lj}$.

All figures show predicted distributions for an integrated luminosity of 100 pb$^{-1}$.

The trigger efficiency is not included in the plots and tables shown in this section. However events satisfying all selection criteria would trigger with an efficiency exceeding 95%, as discussed in Section 3.

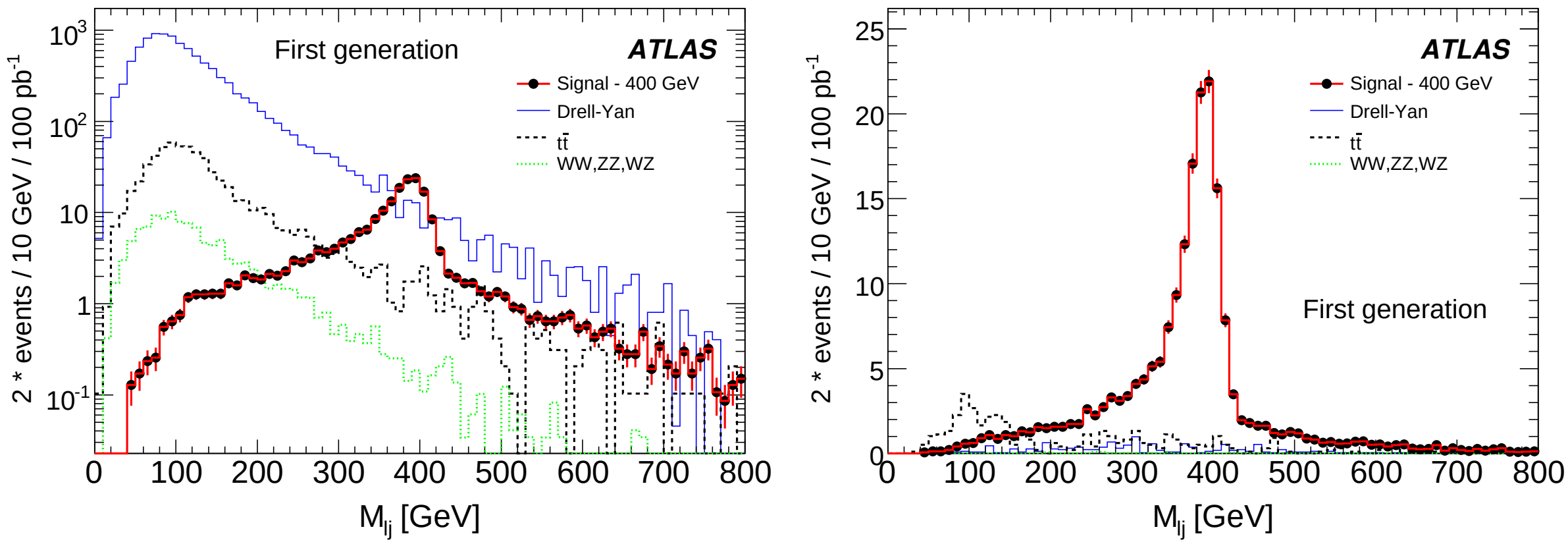


Figure 5: Reconstructed electron-jet invariant mass in the 1st generation leptoquark ($m_{LQ}$=400 GeV) analysis for signal and background MC events after baseline selection (left) and after all selection criteria (right). All distributions are given for 100 pb$^{-1}$ of integrated luminosity.

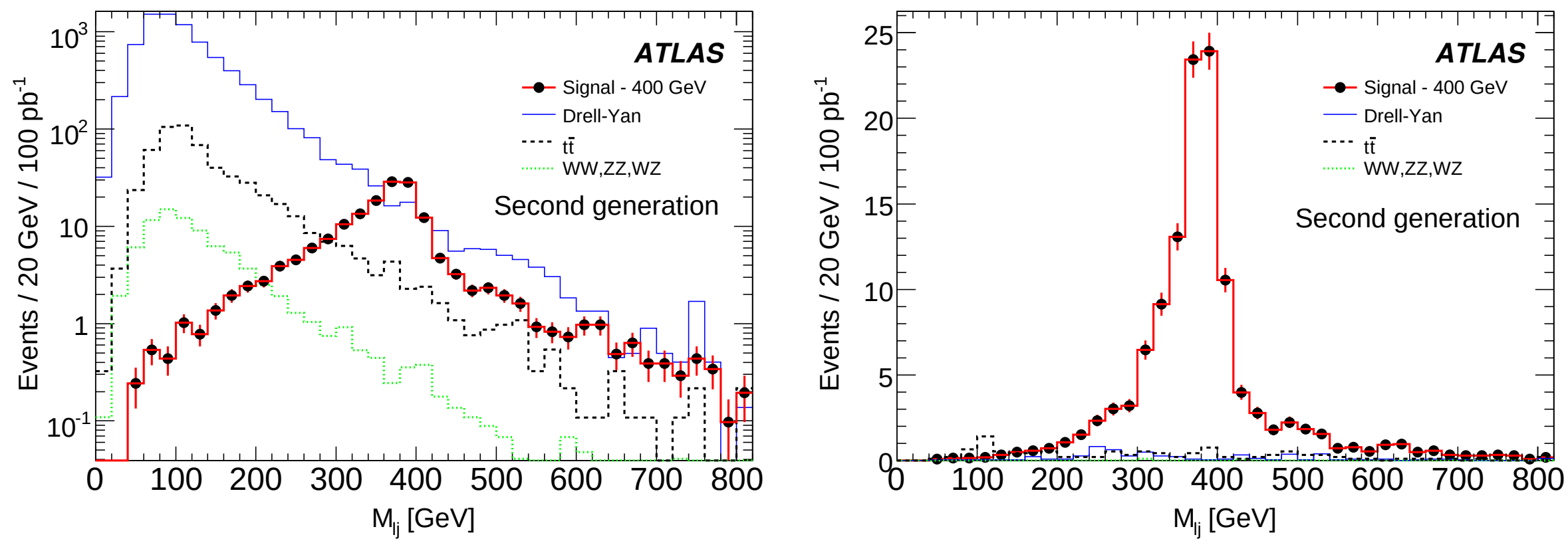


Figure 6: Reconstructed muon-jet invariant mass for 2nd generation leptoquarks ($m_{LQ}$ = 400 GeV) in signal and background MC events after baseline selection (left) and after all selection criteria (right). All distributions are given for 100 pb$^{-1}$ of integrated luminosity.

## 5.2 Search for New Particles from Left-Right Symmetric Models

Signal event candidates are required to contain (at least) two electron or muon candidates and two or more jets that pass the baseline selection criteria. As previously described, the minimum separation between a jet and an electron candidate $\Delta R \geq 0.1$ is required. The two leading $p_T$ lepton candidates and the two leading $p_T$ jets are assumed to be the decay products of the $W_R$ boson. The signal jet candidates are combined with each signal lepton, and the combination that gives the smallest invariant mass is considered as the heavy neutrino ($N_\ell$ in Fig. 2). This assignment is correct in more than 99% of signal MC events. The other lepton is assumed to come directly from the decay of the $W_R$ boson.

When the $W_R$ boson is at least twice as heavy as the Majorana neutrino, the daughter lepton from the neutrino's decay often begins to partially merge with one of the daughter jets. In the dielectron analysis, when the separation between this lepton and a signal jet candidate is in the range $0.1 \leq \Delta R \leq 0.4$, using all three reconstructed objects to estimate the invariant mass of the neutrino would often result in double-counting. To solve this problem, signal event candidates in the dielectron analysis are divided into two groups. When the separation is outside the discussed range, *i.e.* $\Delta R > 0.4$, all three objects are used. However, when the separation is in the critical range, *i.e.* $0.1 \leq \Delta R \leq 0.4$, only jets are used to estimate

the mass of the $N_e$ neutrino. It must be noted that this procedure has little effect on the $N_e$ neutrino mass resolution, because it is dominated by the resolution on the jets energy. The fraction of events falling in the critical range depends on the ratio of $W_R$ boson and Majorana neutrino masses, and increases with this ratio, being 8% for the 1500 GeV to 500 GeV ratio and 26% for the 1800 GeV to 300 GeV ratio considered here. No such problem exists in the dimuon analysis because muon reconstruction is possible even when the reconstructed trajectory's projection into the calorimeters randomly coincides with jet activity. The mass of the Majorana neutrino can be reconstructed with a relative resolution of about 6%, and the mass of the $W_R$ boson can be reconstructed with a relative resolution of 5% to 8%; better resolution on the latter is achieved in the dielectron analyses because the muon spectrometer resolution is degraded at high transverse momenta.

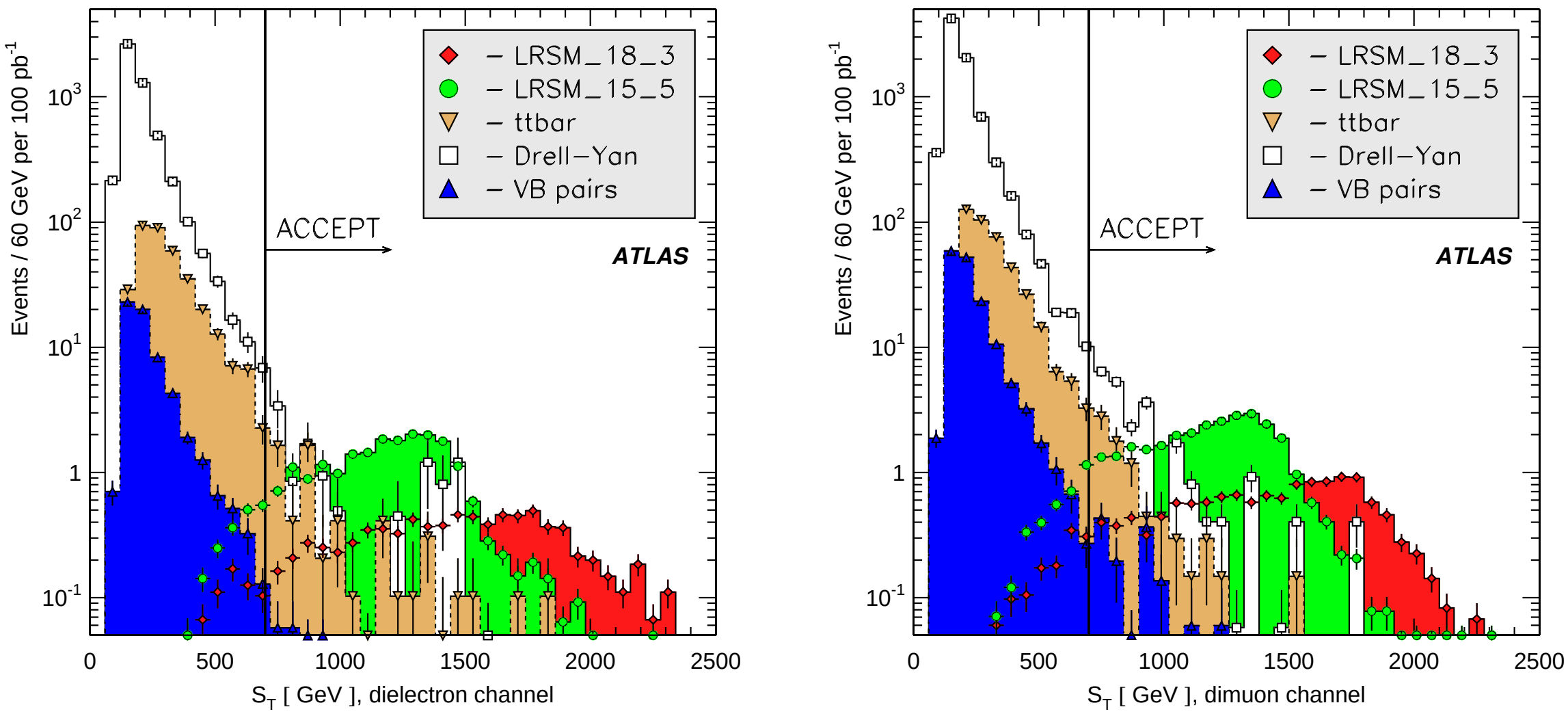


Figure 7: LRSM analysis. $S_T$ distributions for signals and backgrounds normalized to 100 pb$^{-1}$ of integrated luminosity after baseline selection in dielectron (left) and dimuon (right) analyses. Vertical lines indicate the region used in the analysis.

While the main background sources in LRSM analyses are $t\bar{t}$, $Z/\gamma^*$, and vector boson pair production processes, multijets were also identified as a source of potentially dangerous background in the dielectron analysis. The distributions of the scalar sum of signal object candidates' transverse momenta $S_T$, and the reconstructed dilepton invariant mass $m_{\ell\ell}$ for signal and background events, normalized to an integrated luminosity of 100 pb$^{-1}$, are shown in Figs. 7 and 8.

The choice of the selection criteria $S_T \geq 700$ GeV and $m_{\ell\ell} \geq 300$ GeV is made in order to maintain good efficiency not only for mass values used in this study, but also for signals with $m_{W_R} \geq 1000$ GeV.

Partial cross-sections for signal and background processes passing the selection criteria are shown in Tables 5 and 6. Some remarks are in order concerning the selection criteria's efficiencies. First, the dimuon channel is more efficient than the dielectron channel. This is due to the jet-electron merging discussed previously. This issue becomes especially important for a larger ratio of masses $m_{W_R}/m_{N_e}$. However, for a very heavy $W_R$ boson, the dielectron channel could become more significant because the $W_R$ boson mass resolution does not become as wide in the dielectron channel as it does in the dimuon channel. Also, because of its heavy mass, the potential to discover the $W_R$ boson and the heavy neutrino together is much better than in the inclusive search for the new heavy neutrino (assuming the same production mechanism) because of backgrounds.

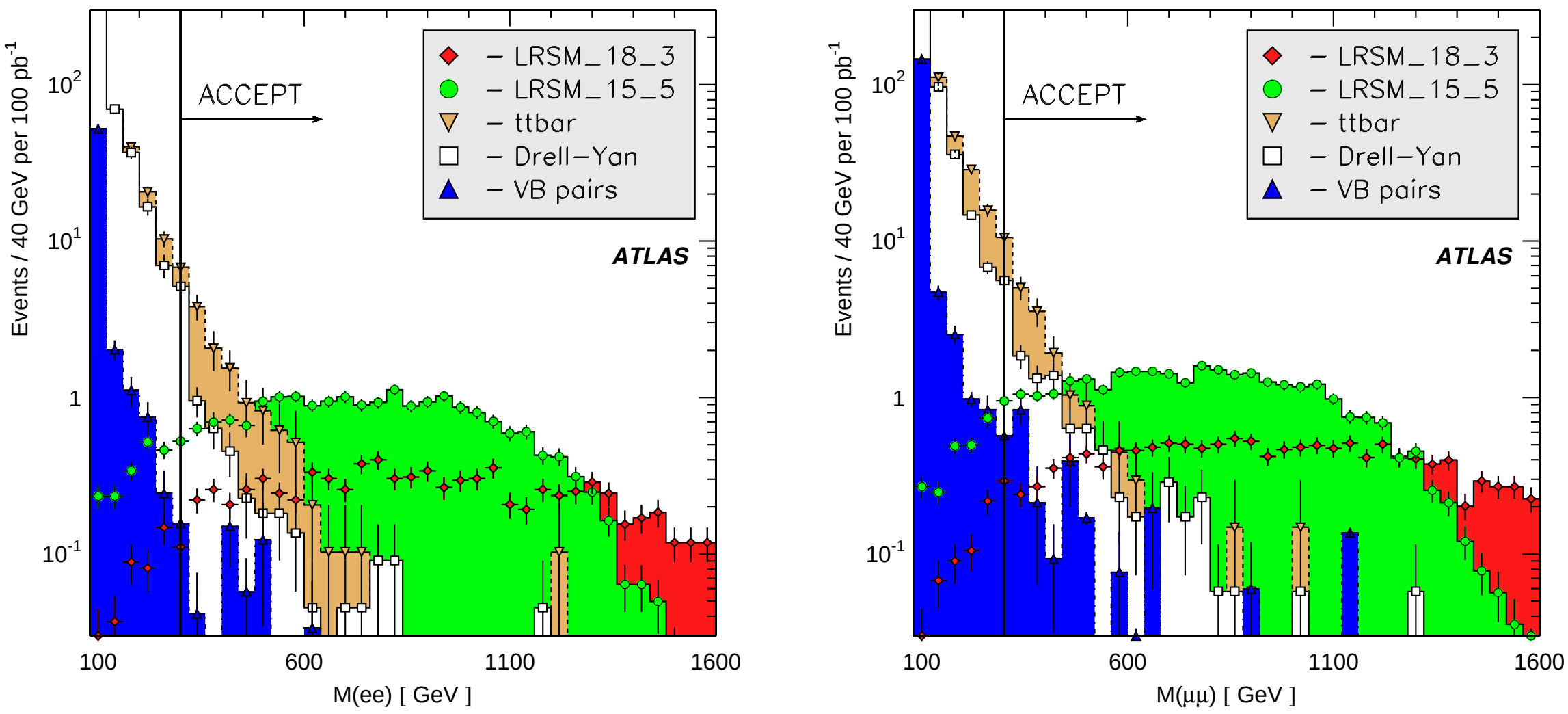


Figure 8: LRSM analysis. The distributions of $m_{\ell\ell}$ for signals and backgrounds normalized to 100 pb$^{-1}$ of integrated luminosity after baseline selection in dielectron (left) and dimuon (right) analyses. Vertical lines indicate the region used in the analysis.

| Physics sample | Before selection | Baseline selection | $m_{ejj}$ $\geq$ 100 GeV | $m_{eejj}$ $\geq$ 1000 GeV | $m_{ee}$ $\geq$ 300 GeV | $S_T$ $\geq$ 700 GeV |
|---|---|---|---|---|---|---|
| LRSM_18_3 | 0.248 | 0.0882 | 0.0882 | 0.0861 | 0.0828 | 0.0786 |
| LRSM_15_5 | 0.470 | 0.220 | 0.220 | 0.215 | 0.196 | 0.184 |
| $Z/\gamma^*, m \geq 60$ GeV | 1808. | 49.77 | 43.36 | 0.801 | 0.0132 | 0.0064 |
| $t\bar{t}$ | 450. | 3.23 | 3.13 | 0.215 | 0.0422 | 0.0165 |
| VB pairs | 60.9 | 0.610 | 0.522 | 0.0160 | 0.0016 | 0.0002 |
| Multijet | $10^8$ | 20.51 | 19.67 | 0.0490 | 0.0444 | 0.0444 |

Table 5: LRSM dielectron analysis. Partial cross-sections (pb) that survive the selection criteria.

Figures 9 and 10 show the distributions of the reconstructed invariant masses of the heavy neutrino and $W_R$ boson candidates for signal and background MC samples before and after the selection criteria are applied. All distributions are normalized to 100 pb$^{-1}$ of integrated luminosity. It should be remarked that the trigger efficiency is not included in the plots and tables shown in this section. However, events satisfying all selection criteria would trigger with an efficiency exceeding 95%, as discussed in Section 3.

Background contributions to signal invariant mass spectra could also arise from jets that are misidentified as signal electrons. In principle, such misidentified jets are efficiently suppressed because at least two signal electron candidates are required, but at present this background remains poorly understood because larger statistics of multijet MC, or better, real data, would be necessary to evaluate its contribution reliably. If needed, a better suppression of events with multijets that are misidentified as electrons is possible by applying a more sophisticated isolation energy requirement. The multijet background does not pose a problem in the dimuon analysis, where estimates of the misidentification rate predict a vanishing contribution from multijet to dimuon events.

| Physics sample | Before selection | Baseline selection | $m_{\mu jj}$ $\geq 100$ GeV | $m_{\mu\mu jj}$ $\geq 1000$ GeV | $m_{\mu\mu}$ $\geq 300$ GeV | $S_T$ $\geq 700$ GeV |
|---|---|---|---|---|---|---|
| LRSM_18_3 | 0.248 | 0.145 | 0.145 | 0.141 | 0.136 | 0.128 |
| LRSM_15_5 | 0.470 | 0.328 | 0.328 | 0.319 | 0.295 | 0.274 |
| $Z/\gamma^*, m \geq 60$ GeV | 1808. | 79.99 | 69.13 | 1.46 | 0.0231 | 0.0127 |
| $t\bar{t}$ | 450. | 4.17 | 4.11 | 0.275 | 0.0527 | 0.0161 |
| VB pairs | 60.9 | 0.876 | 0.824 | 0.0257 | 0.0047 | 0.0015 |
| Multijet | $10^8$ | 0.0 | 0.0 | 0.0 | 0.0 | 0.0 |

Table 6: LRSM dimuon analysis. Partial cross-sections (pb) that survive selection criteria.

Finally, the analyses described in this note do not discriminate between same-sign and opposite-sign dileptons. Same-sign dileptons, however, are a very important signature of Majorana neutrinos, which, being their own anti-particles, could decay to a lepton of either charge. The background contribution to same-sign dileptons is much smaller than to opposite-sign dileptons. Of course, both channels would have to be studied if the discovery is made. The studies of charge misidentification performed in the framework of the presented analyses, predict a rate as high as 5% for high-$p_T$ leptons which is strongly $\eta$-dependent.

## 6 Systematic Uncertainties

The following sources of systematic uncertainties have been considered in the described analyses:

- 20% uncertainty was assumed on the integrated luminosity.
- In the dielectron analyses, 1% was used for the uncertainty in overall trigger efficiency.
- For electron identification and reconstruction efficiency, an uncertainty of 1% was assumed.
- For muon identification, including trigger and reconstruction efficiencies, an uncertainty of 5% was assumed.
- The uncertainty on the electron energy scale was assumed to be $\pm 1\%$.
- The uncertainty on the muon momentum scale was assumed to be $\pm 1\%$.
- The uncertainty on the jet energy scale was estimated by changing the energies of all jets simultaneously by $\pm 10\%$ and $\pm 20\%$, for $|\eta_{jet}| \leq 3.2$ and $|\eta_{jet}| > 3.2$, respectively.
- The 20% uncertainty in electron $p_T$ resolution was estimated using a Gaussian smearing of $p_T$ with a relative width of $0.66 * (0.10/\sqrt{p_T} \bigoplus 0.007)$, where $p_T$ is in GeV.
- The uncertainty due to muon $1/p_T$ resolution was estimated using a Gaussian smearing of $1/p_T$ with a width of $0.011/p_T \bigoplus 0.00017$, where $p_T$ is in GeV.
- The uncertainty due to jet energy resolution was estimated using a Gaussian smearing of jet energies in such a way that the relative jet energy resolution widens from $0.60/\sqrt{E} \bigoplus 0.05$ to $0.75/\sqrt{E} \bigoplus 0.07$ for $|\eta_{jet}| \leq 3.2$, and from $0.90/\sqrt{E} \bigoplus 0.07$ to $1.10/\sqrt{E} \bigoplus 0.10$ for $|\eta_{jet}| > 3.2$, where $E$ is in GeV.

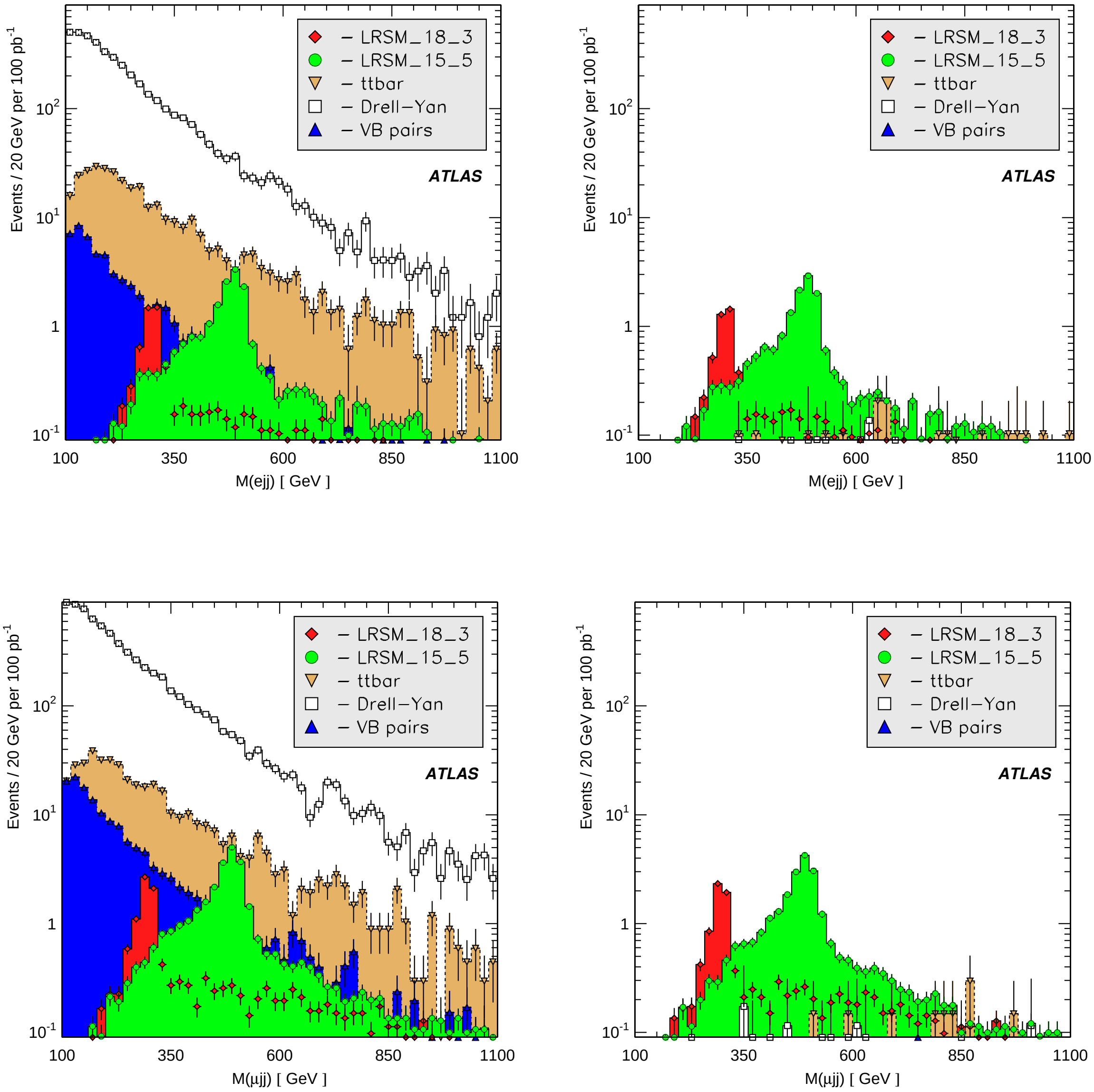


Figure 9: LRSM analysis. The distributions of the reconstructed invariant masses for $N_e$ (top) and $N_\mu$ (bottom) candidates in background and signal (LRSM_18_3 and LRSM_15_5) events before (left) and after (right) background suppression is performed in dielectron and dimuon analyses. All distributions are normalized to 100 pb$^{-1}$ of integrated luminosity. LRSM_15_5 and LRSM_18_3 refer to two sets of LRSM mass hypotheses. See the text for more information.

- Statistical uncertainties on the number of background MC events were considered as systematic uncertainties on the number of background events.
- The systematic uncertainty on the leptoquark cross-section (NLO) [16] was calculated by taking the 40 PDF CTEQ6M tables (two per eigenvector of PDF variations, provided by the CTEQ group for calculating uncertainties [15]), recalculating the leptoquark cross-section with each of these tables, and taking the largest difference of the two variations for each of the 20 eigenvectors to

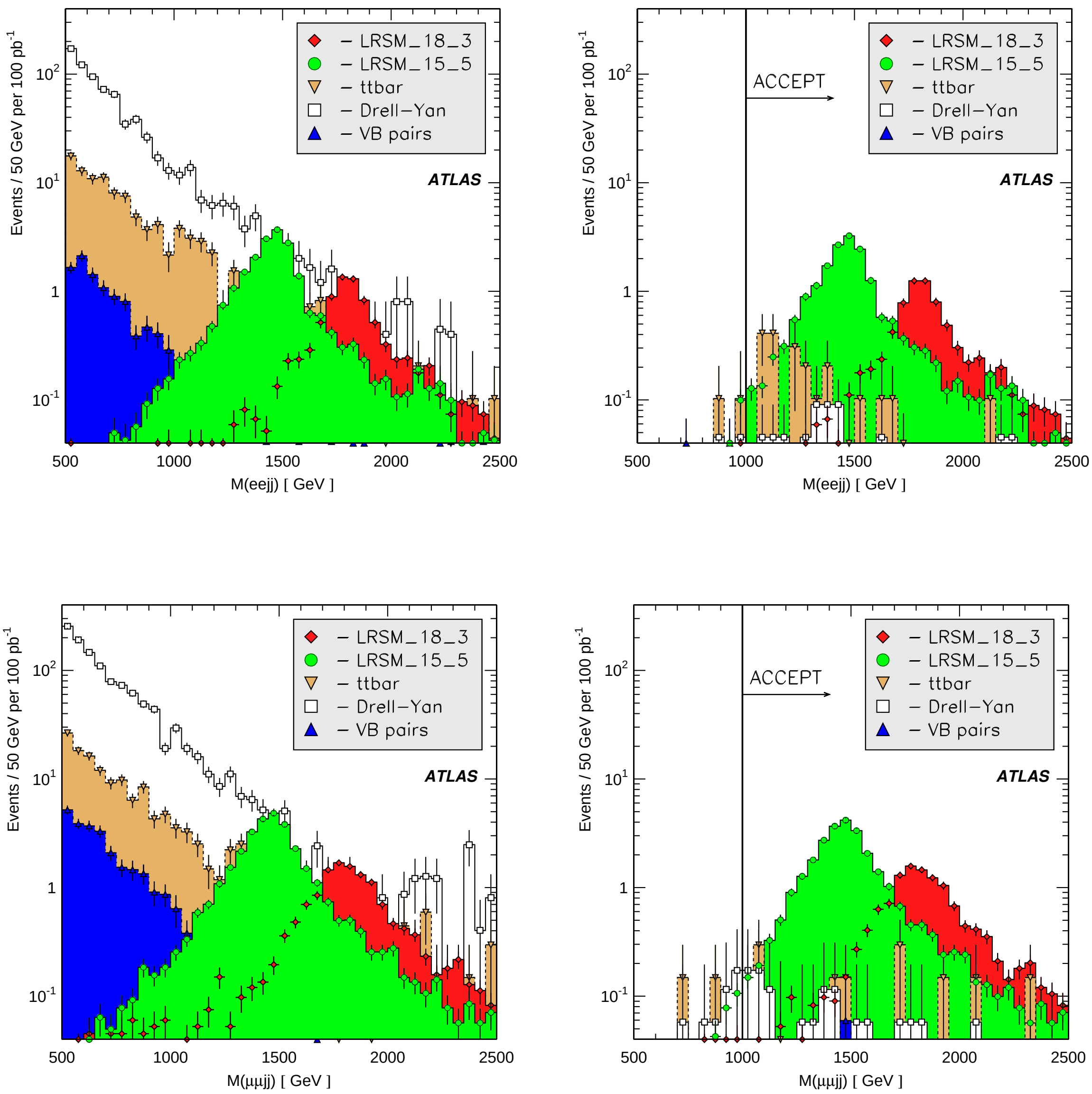


Figure 10: LRSM analysis. The distributions of the reconstructed invariant masses for $W_R \to eN_e$ (top) and $W_R \to \mu N_\mu$ (bottom) candidates in background and signal (LRSM_18_3 and LRSM_15_5) events before (left) and after (right) background suppression is performed in dielectron and dimuon analyses. All distributions are normalized to 100 pb$^{-1}$ of integrated luminosity. Notice that the invariant mass of the $W_R$ boson is shown *before* the requirement $m_{\ell\ell jj} \geq 1000$ GeV is imposed. This variable is strongly correlated with the background-suppressing variables $S_T$ and $m_{\ell\ell}$. LRSM_15_5 and LRSM_18_3 refer to two sets of LRSM mass hypotheses. See the text for more information.

the cross-section calculated with the standard CTEQ6M table. The estimate shown is the sum in quadrature of these 20 differences and the relative difference in cross-section obtained by varying renormalization and factorization scales by a factor of 2. The systematic uncertainty is between 15% and 28% for the tested leptoquark masses.

- The uncertainty of the jet modeling in $Z/\gamma^*$ events was estimated by comparing the background

predictions obtained using MC samples produced with PYTHIA to MC samples produced with ALPGEN. For the leptoquark pair analysis, this results in an uncertainty of about 30% on the background from $Z/\gamma^*$ events.

- Background cross-sections for $t\bar{t}$ and $Z/\gamma^*$ processes were assumed to have uncertainties of 12% and 10%, respectively.

Systematic uncertainties affect both signal and background efficiencies, however the significance computation (next section) is mainly affected by the uncertainty on the background. The dominant systematic effects on the background are due to the uncertainties in integrated luminosity (20%), the jet energy scale (16%-35%), jet energy resolution (6%-28%), and the limited statistics of background MC samples (15%-30%). Possible other sources of systematic uncertainties such as initial and final state radiation modeling, or pile-up, were not evaluated. The total systematic uncertainties for signals and backgrounds are summarized in Table 7.

| analysis | effect on signal events | | effect on background events | |
|---|---|---|---|---|
| | 1st gen. | 2nd gen. | 1st gen. | 2nd gen. |
| leptoquark | ±27% | ±29% | ±53% | ±51% |
| LRSM | ±23% | ±25% | ±45% | ±40% |

Table 7: Summary of total systematic uncertainties (%) for 100 pb$^{-1}$ luminosity.

# 7 Results

The program $S_{cp}$ [29] is used to calculate the significances of possible observations of the signals studied in this note. The significance is defined in units of Gaussian standard deviations, corresponding to the (one-sided) probability of observing a certain number of events exceeding the MC-predicted background $N_b$ at a given integrated luminosity. This probability is usually referred to as $CL_b(N)$, where $N$ is the number of observed events. We report the $5\sigma$ discovery potential evaluated in terms of $CL_b(N_s+N_b)$, where $N_s$ is the expected number of signal events. Systematic uncertainties in the number of background events were also included in the significance calculations. For second generation leptoquarks, the signal selection was optimized at each mass point to minimize the cross-section times branching ratio needed to reach a $5\sigma$ discovery, while for all other analyses the selection cuts presented in earlier sections were used.

The overall reconstruction and trigger efficiencies discussed earlier are used to estimate ATLAS' sensitivity and discovery potential for the studied final states below. These estimates include the trigger efficiency for signal and background events, as discussed in Section 5, Table 2.

## 7.1 Leptoquarks

The integrated luminosities needed for a $5\sigma$ discovery of the 1st and 2nd generation scalar leptoquark signals are shown in Table 8 as function of leptoquark mass, assuming $\beta = 1$. Also, Fig. 11 predicts the integrated luminosities needed for a 400 GeV leptoquark mass discovery, with various values of $\beta^2$, at a $5\sigma$ level.

Finally, Fig. 12 shows the minimum $\beta^2$ that can be probed with ATLAS with 100 pb$^{-1}$ of integrated luminosity as a function of leptoquark mass. Lighter leptoquark masses can be probed with a smaller $\beta$ because of their larger cross-section. It is evident from this figure that ATLAS is sensitive to leptoquark

| Leptoquark mass | Expected luminosity needed for a $5\sigma$ discovery | |
|---|---|---|
| | 1st gen. | 2nd gen. |
| 300 GeV | 2.8 $pb^{-1}$ | 1.6 $pb^{-1}$ |
| 400 GeV | 11.8 $pb^{-1}$ | 7.7 $pb^{-1}$ |
| 600 GeV | 123 $pb^{-1}$ | 103 $pb^{-1}$ |
| 800 GeV | 1094 $pb^{-1}$ | 664 $pb^{-1}$ |

Table 8: The integrated luminosities needed for a $5\sigma$ discovery of 1st and 2nd gen. scalar leptoquarks for different mass hypotheses.

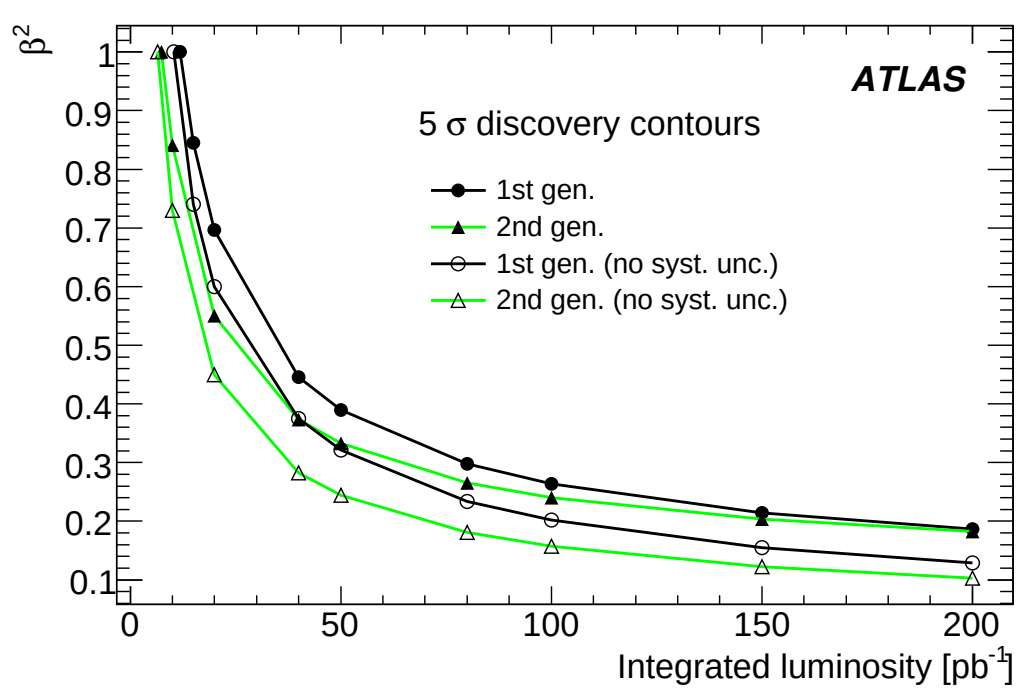


Figure 11: $5\sigma$ discovery potential for 1st and 2nd gen. $m = 400$ GeV scalar leptoquarks versus $\beta^2$ with and without background systematic uncertainty included.

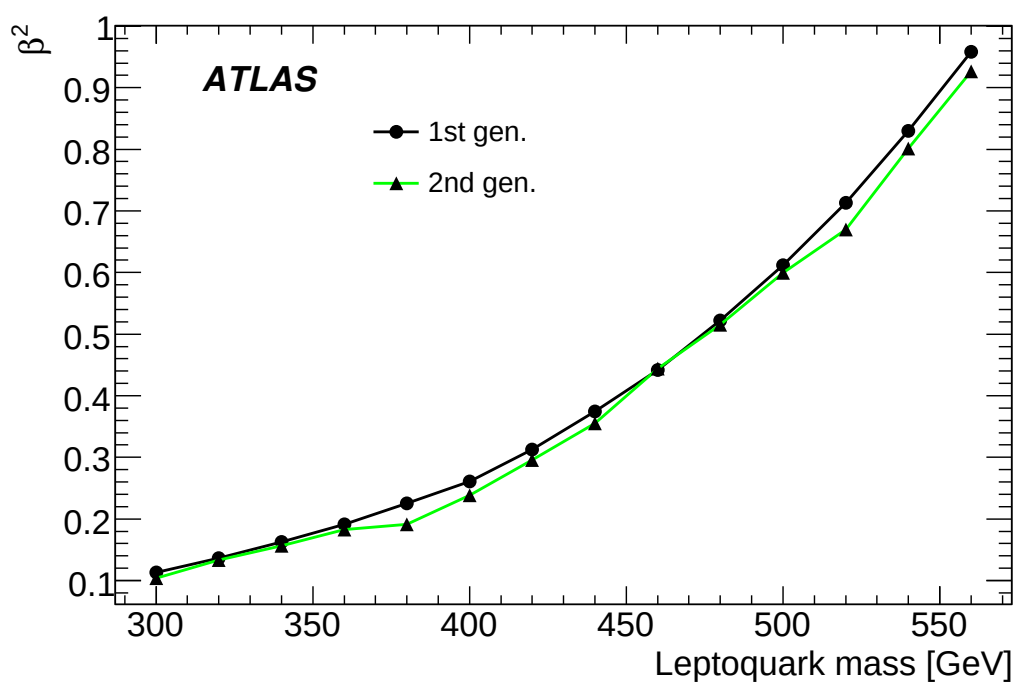


Figure 12: Minimum $\beta^2$ of scalar leptoquarks versus leptoquark mass for 100 $pb^{-1}$ of integrated luminosity at $5\sigma$ (background systematic uncertainty included.)

masses of about 565 GeV and 575 GeV for 1st gen. and 2nd gen., respectively, at the given integrated luminosity, provided leptoquarks always decay into charged leptons and quarks.

### 7.2 Left-Right Symmetry

The significances of studied signals versus integrated luminosity are shown in Fig. 13. Figure 14 shows the product of signal cross-section and dilepton branching fraction versus the integrated luminosity necessary for a $5\sigma$ discovery.

The overall relative systematic uncertainties on Drell-Yan and $t\bar{t}$ backgrounds are approximately 45% and 40% in the dielectron and dimuon analyses, respectively. These estimates are dominated by contributions from jet reconstruction, uncertainty in integrated luminosity and insufficient MC statistics. Currently, multijet background is poorly understood and is not included in the presented sensitivity estimates for the dielectron channel.

## 8 Summary and Conclusions

Studies of final states with two leptons and multiple jets have been discussed, considering both electrons and muons. The early-data discovery potential for Beyond the Standard Model physics predicted by two prominent GUT-inspired models has been investigated.

Both 1st and 2nd generation scalar leptoquark pair production could be discovered with less than 100 $pb^{-1}$ of integrated luminosity, provided that the mass of the leptoquarks is smaller than 500 GeV and the branching ratio into a charged lepton and a quark is 100%.

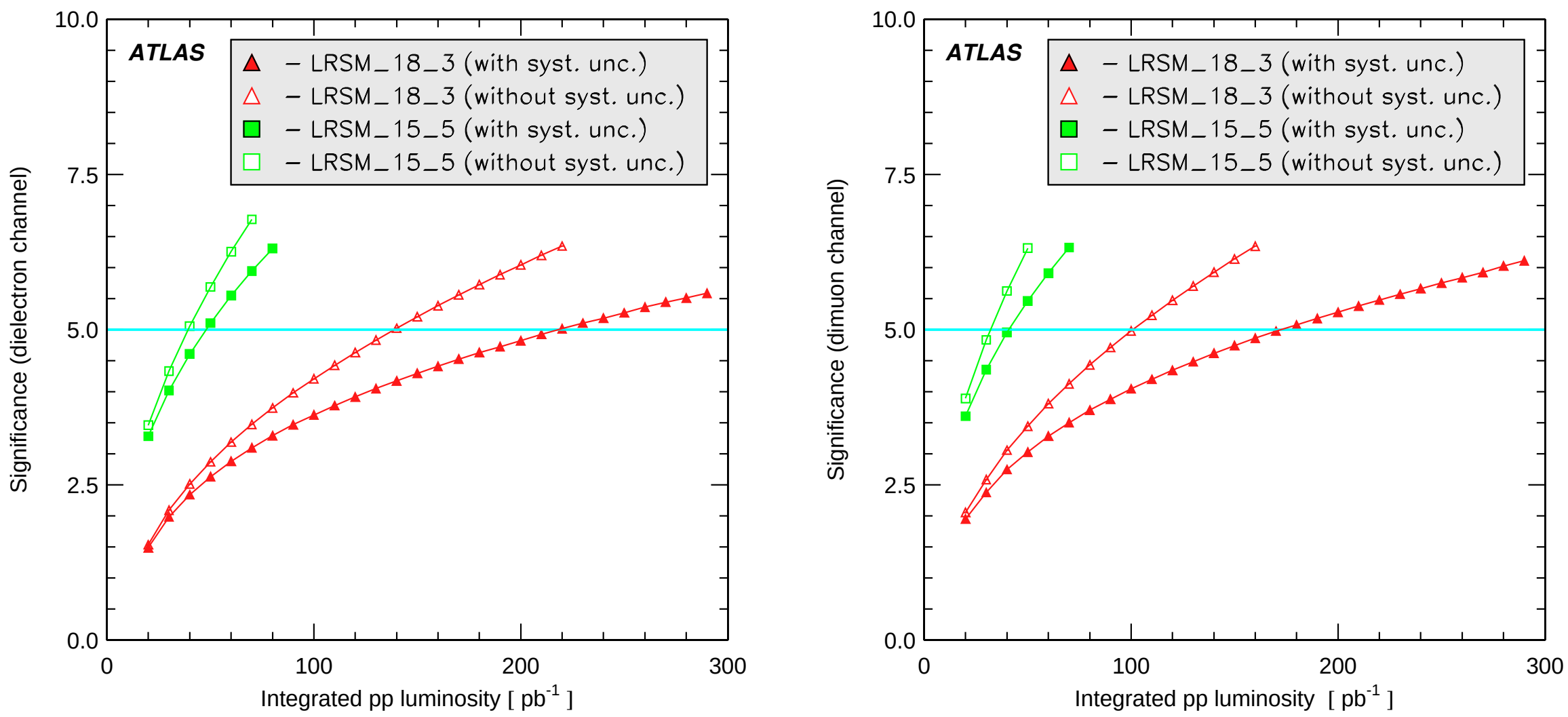


Figure 13: LRSM analysis. Expected signal significances versus integrated luminosity for $N_e$, $N_\mu$ neutrino and $W_R$ boson mass hypotheses, according to signal MC samples LRSM_18_3 and LRSM_15_5. Open symbols show sensitivities without systematic uncertainties. Sensitivities shown with closed symbols include an overall relative uncertainty of 45% (40%) estimated for background contributions in the dielectron (dimuon) analysis. LRSM_15_5 and LRSM_18_3 refer to two sets of LRSM mass hypotheses. See the text for more information.

Two LRSM mass points ($m_{W_R} = 1.8$ TeV,$m_{N_\ell} = 300$ GeV and $m_{W_R} = 1.5$ TeV,$m_{N_\ell} = 500$ GeV) for the right-handed $W_R$ boson and Majorana neutrinos $N_\ell$ have been studied in the dielectron and dimuon channels. It was found that discovery of these new particles at these mass points would require integrated luminosities of 150 pb$^{-1}$ and 40 pb$^{-1}$, respectively.

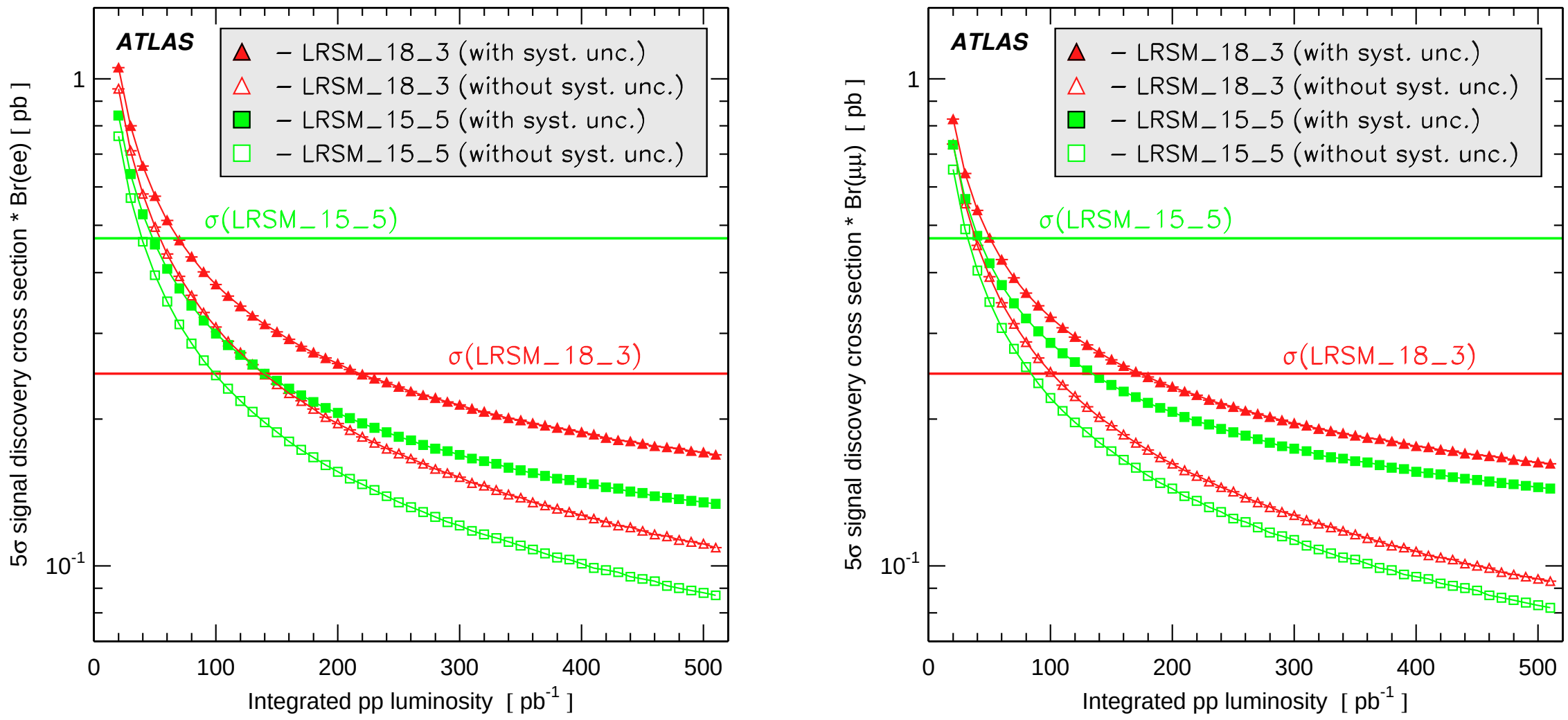


Figure 14: LRSM analysis. The product of signal cross-section and branching fraction to dielectron and dimuon final states versus integrated luminosity necessary for a $5\sigma$ discovery. $N_e$, $N_\mu$ neutrino and $W_R$ boson mass hypotheses are for signal MC samples LRSM_18_3 and LRSM_15_5. Horizontal lines indicate nominal cross-sections for two signal MC samples, according to the LRSM implementation in the MC simulation. Open symbols show discovery potentials without systematic uncertainties. Discovery potentials shown with closed symbols include an overall relative uncertainty of 45% (40%) assumed for the background contribution in the dielectron (dimuon) analysis. LRSM_15_5 and LRSM_18_3 refer to two sets of LRSM mass hypotheses. See the text for more information.

# Vector Boson Scattering at High Mass

**Abstract**

In the absence of a light Higgs boson, the mechanism of electroweak symmetry breaking will be best studied in processes of vector boson scattering at high mass. Various models predict resonances in this channel. Here, we investigate $WW$ scalar and vector resonances, $WZ$ vector resonances and a $ZZ$ scalar resonance over a range of diboson centre-of-mass energies. Particular attention is paid to the application of forward jet tagging and to the reconstruction of dijet pairs with low opening angle resulting from the decay of highly boosted vector bosons. The performances of different jet algorithms are compared. We find that resonances in vector boson scattering can be discovered with a few tens of inverse femtobarns of integrated luminosity.

## 1 Introduction

In the absence of a light Higgs boson, an alternative scenario to the Standard Model, Supersymmetry, or Little Higgs models must be invoked. In particular, Electroweak Symmetry Breaking (EWSB) could result from a strong coupling interaction. Here, we will make no assumptions about the underlying dynamics of EWSB; we treat the Standard Model as a low energy effective theory, and evaluate the potential for measuring vector boson scattering. In the Standard Model, perturbative unitarity is violated [1] in vector boson scattering at high energy for a Higgs mass $m_H > 870$ GeV or, if there is no Higgs ($m_H \to \infty$), for a centre-of-mass energy above a critical value of around 1.7 TeV. The only way to avoid a light Higgs boson is therefore to presume new physics at high energy [2], possibly in the form of vector boson pair resonances. Such resonances are predicted in many models such as QCD-like technicolour models with the required Goldstone bosons resulting from chiral symmetry breaking [3]; Higgsless extra dimension models [4], where Kaluza-Klein states of gauge bosons are exchanged in the $s$-channel [5]; as well as in models with extra vector bosons, from GUT or from strong interaction (BESS models [6]) mixing with the Standard Model vector bosons. The present search for resonances in vector boson scattering can be considered generic and may be interpreted in terms of any of these models.

### 1.1 The Chiral Lagrangian Model

The Chiral Lagrangian (ChL) model is an effective theory valid up to $4\pi v \sim 3$ TeV, where $v = 246$ GeV is the vacuum expectation value of the Standard Model Higgs field. It can provide a description of longitudinal gauge boson scattering at the TeV scale when no light scalar Higgs boson is present. Electroweak symmetry breaking is realised non-linearly. A set of dimension-4 effective operators describe the low energy interactions (see for example [7]). Since, at the LHC, vector boson scattering can occur at the TeV energy scale where the interaction becomes strong, it is necessary to unitarise the scattering amplitudes. One popular unitarisation prescription is the so-called Padé prescription, or Inverse Amplitude Method [8]. This is based on meson scattering in QCD, where it gives an excellent description [9], reproducing observed resonances. Among the terms of the Lagrangian which describe vector boson scattering, under some basic assumptions (custodial symmetry and CP conservation), only 2 parameters (namely $\alpha_4$ and $\alpha_5$) are important for this process. Depending on the values of these two parameters, one can obtain Higgs-like scalar resonances and/or technicolour-like vector resonances [10]. The resulting properly-unitarised amplitudes for vector boson scattering may therefore give information in a higher energy range. They yield poles for certain values of $\alpha_4$ and $\alpha_5$ that can be interpreted as resonances, as shown in Fig. 1. The hashed region in the figure is forbidden by causality arguments [11].

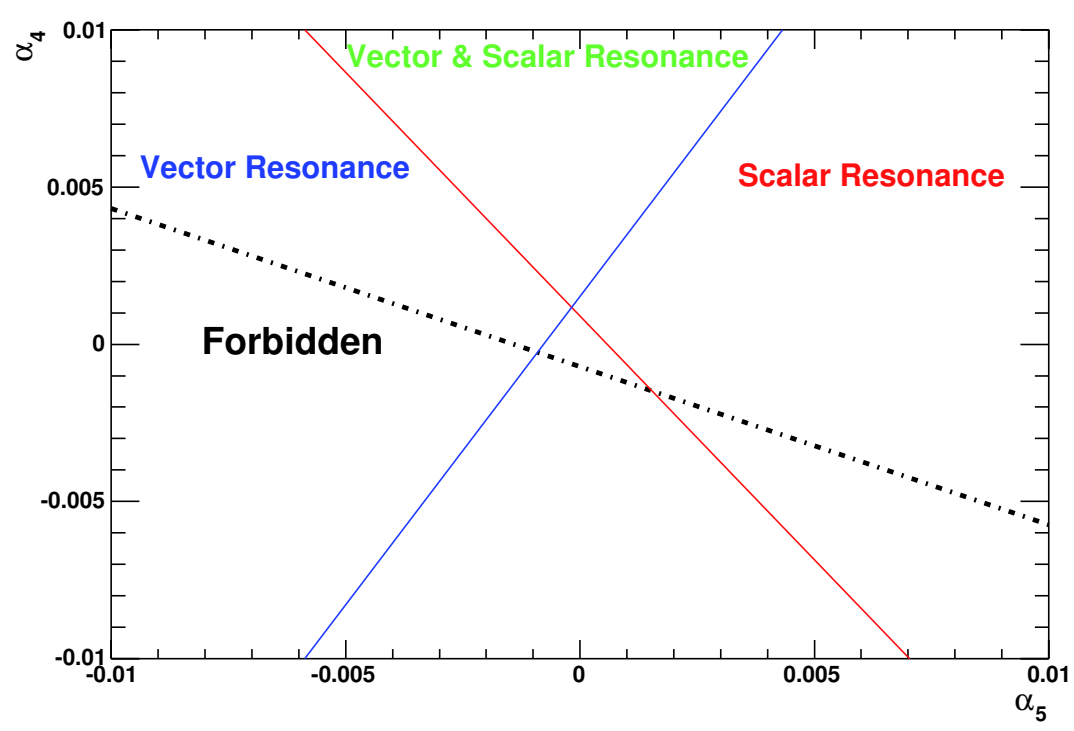


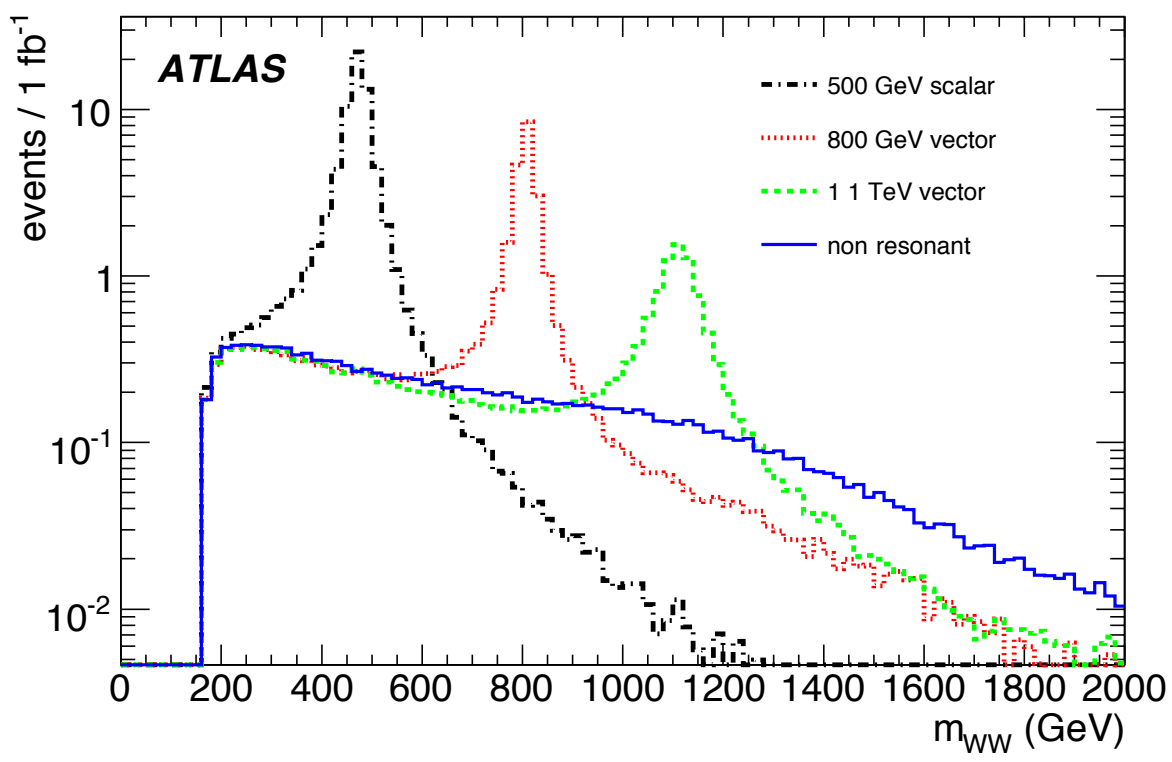


Figure 1: Left: regions in the $(\alpha_5, \alpha_4)$ parameter space indicating which values exhibit vector and/or scalar resonances in the Padé unitarisation scheme. Right: number of events per fb$^{-1}$ as a function of the di-boson invariant mass for different resonance masses studied here.

Other unitarisation procedures are possible, such as the K-matrix method [12] or the $N/D$ method [13]. In general, resonances are not necessarily produced. In non-resonant cases, it remains vital to measure the vector boson scattering cross section, but high luminosity and a very good understanding of backgrounds will be required in order to measure the regularisation of the cross section.

## 1.2 Characteristic Signatures of Vector Boson Scattering

Discovery of the physics signals studied here will, in general, require high integrated luminosity. It will require also extremely large samples of simulated backgrounds, fine tuning of all reconstruction algorithms, and a good understanding of the detector performance, which will only gradually develop after the first few years of LHC running. The main purpose of this note is not, therefore, to evaluate with precision the discovery potential of ChL resonances, but to establish a strategy for the search of this important signal. The main emphasis will be put on those aspects most particular to the high mass vector boson scattering process; that is, the reconstruction of hadronically decaying vector bosons at high $p_T$ and the reconstruction of the high rapidity tag jets.

The decay of a high mass ChL resonance will produce two highly boosted vector bosons in the central rapidity region of the detector. For transverse momenta greater than about 250 GeV, a hadronically decaying vector boson will be seen as one single wide and heavy jet. Methods of distinguishing such jets from single-parton jets will be investigated with different jet algorithms.

A characteristic signature of vector boson scattering is the presence of two high rapidity and high energy “tag” jets [14], arising from the quarks which radiate the incoming vector bosons. The process can thus be efficiently distinguished from contributions to the production of (mostly transversely polarised) final state vector bosons due to bremsstrahlung of these vector bosons from the quarks. In that case, the accompanying jets are softer and more central. A further component of the signature is the suppression of QCD radiation in the rapidity interval between the tag jets due to the fact that no colour is exchanged between the protons in these processes [15]. This characteristic feature allows for efficient use of central jet veto to suppress backgrounds.

The high QCD background at the LHC naturally leads us to focus on “semi-leptonic” vector boson events; that is, those events when one $W$ or $Z$ boson decays leptonically, and the other decays hadronically. These channels represent the best compromise in that there is only at most one neutrino, so the diboson mass may be reconstructed with reasonable resolution, and the backgrounds can be reduced to a manageable level by the requirement of leptons and/or missing transverse energy ($\not\!\!E_T$). Fully-leptonic

events are also useful in cases where clear resonances are present, where a kinematic edge may be visible and the backgrounds may be reduced even further. The case of resonant $ZZ \to \ell^+\ell^- \nu\bar{\nu}$ can also lead to a clean signature. Fully hadronic events may be useable at very high diboson energies, but this possibility is not considered further here. Thus, the study of vector boson scattering events will also require a good understanding of detector performance for electrons, muons and $\not{E}_T$. Although many ATLAS analyses will depend on the reconstruction of these objects, the quality of such reconstruction is evaluated here for the case of high energy leptons.

The note is organised as follows. In the next section (2) we describe the Monte Carlo simulations and the samples used. Next, the trigger is discussed (Section 3), then the detector performance with particular focus on the challenges of this analysis (Section 4). After this, the event selections, efficiencies and purities for the various final states are given (Section 5). An attempt to evaluate the expected sensitivity is made (Section 6), and the systematic uncertainties are discussed (Section 7), before a final summary and conclusion.

# 2 Signals and Background Simulation

## 2.1 Definition of Signal

In order to have a gauge invariant set of diagrams for the background, in spite of a Higgsless scenario, a low mass Higgs will be assumed. A resonance *signal* will be defined here as an excess of events in the resonance mass region over the number expected from the Standard Model continuum when the Higgs boson mass is set at 100 GeV. This ensures that longitudinal vector boson scattering will contribute negligibly to the process. This definition follows the prescription of [16]. We note that measurement of even a continuum cross section for this process at such high energies would be of great importance, but will not be considered here as it should require high luminosity and a very good understanding of backgrounds.

## 2.2 Overview of Generators

The Monte Carlo (MC) generators used in the main analysis are as follows.

- PYTHIA [17] version 6.4.0.3 was used for the signal, with the CTEQ6L parton distribution function and the renormalisation and factorisation scale $Q^2 = m_W^2$. The hard process was modified to include new vector boson scattering amplitudes (see below).
- MADGRAPH [18], version 3.95, with PYTHIA for parton shower, hadronisation and underlying event, was used for $W$+jets and $Z$+jets backgrounds. The default values of fixed renormalisation and factorisation scales of $Q^2 = m_Z^2$ were set and CTEQ6L1 parton distribution functions were used.
- MC@NLO [19], with HERWIG [20], for parton shower and hadronisation and JIMMY [21,22] for underlying event, was used for $t\bar{t}$ background.

The underlying event samples were tuned to data from previous experiments [22]. All samples use PHOTOS [23] to simulate final state radiation. WHIZARD [24] and ALPGEN [25, 26] are also used for some generator level comparisons. WHIZARD uses PYTHIA for parton showering, hadronisation, and underlying event. ALPGEN uses HERWIG/JIMMY.

The different choice of scales for MADGRAPH and PYTHIA is not ideal, but retained for historical reasons since large samples were generated with these choices. However, studies showed that the

| **Sample name** | **Generator** | $\sigma \times Br$, fb |
|---|---|---|
| $qqWZ \to qqjj\ell\ell$, $m = 500$ GeV | PYTHIA-73 | 25.2 |
| $qqWZ \to qq\ell\nu jj$, $m = 500$ GeV | PYTHIA-73 | 83.9 |
| $qqWZ \to qq\ell\nu\ell\ell$, $m = 500$ GeV | PYTHIA-73 | 8.0 |
| $qqWZ \to qqjj\ell\ell$, $m = 800$ GeV | PYTHIA-ChL | 10.5 |
| $qqWZ \to qq\ell\nu jj$, $m = 800$ GeV | PYTHIA-ChL | 35.2 |
| $qqWZ \to qq\ell\nu\ell\ell$, $m = 800$ GeV | PYTHIA-ChL | 3.4 |
| $qqWZ \to qqjj\ell\ell$, $m = 1.1$ TeV | PYTHIA-ChL | 3.7 |
| $qqWZ \to qq\ell\nu jj$, $m = 1.1$ TeV | PYTHIA-ChL | 12.3 |
| $qqWZ \to qq\ell\nu\ell\ell$, $m = 1.1$ TeV | PYTHIA-ChL | 1.18 |
| $qqWW \to qq\ell\nu jj$, $m = 499$ GeV (s) | PYTHIA-ChL | 66.5 |
| $qqWW \to qq\ell\nu jj$, $m = 821$ GeV (s) | PYTHIA-ChL | 27.5 |
| $qqWW \to qq\ell\nu jj$, $m = 1134$ GeV (s) | PYTHIA-ChL | 17.0 |
| $qqWW \to qq\ell\nu jj$, $m = 808$ GeV (v) | PYTHIA-ChL | 29.8 |
| $qqWW \to qq\ell\nu jj$, $m = 1115$ GeV (v) | PYTHIA-ChL | 17.9 |
| $qqWW \to qq\ell\nu jj$, non-resonant | PYTHIA-ChL | 10.0 |
| $qqZZ \to qq\nu\nu\ell\ell$, $m = 500$ GeV | PYTHIA-ChL | 4.0 |
| $jjWZ \to jj\ell\nu\ell\ell$, background | MADGRAPH | 96 |
| $jjZZ \to jj\nu\nu\ell\ell$, background | MADGRAPH | 45.5 |
| | | $\sigma$ (no Br), pb |
| $W^+$ + 4 jets | MADGRAPH | $165 \pm 0.1$ |
| $Z$ + 4 jets | MADGRAPH | $87 \pm 0.7$ |
| $W^+$ + 3 jets | MADGRAPH | $6.2 \pm 0.02$ |
| $Z$ + 3 jets | MADGRAPH | $3.8 \pm 0.02$ |
| $t\bar{t}$ | MC@NLO | $833 \pm 100$ |

Table 1: Data samples and generators used in the present study

major effect is on the cross section rather than on event shapes, and the cross section normalisation is determined independently as described below.

Further details specific to the samples are given below.

## 2.3 List of samples

Table 1 lists the Monte Carlo samples, produced with full detector simulation, used in the present analysis.

The first set of samples represents different reference cases of vector boson scattering signals:

- PYTHIA-73: For the samples labelled "PYTHIA-73", the process 73 (longitudinal $WZ$ scattering) was selected, with MSTP(46)=5 (QCD-like model of [27] with Padé unitarisation). All other switches were left as default. This is meant to represent a generic narrow $WZ$ resonance.

- PYTHIA-ChL: datasets with generator labelled "PYTHIA-ChL" in the table use a modified version of PYTHIA routine PYSGHG. The modification involves replacing the scattering amplitudes calculated for processes 73–77 by those given by Dobado et al [10] with parameters $a_4$ and $a_5$. These parameters were chosen so as to produce a vector or scalar (indicated by a (v) or an (s) in the table) resonance at the desired mass, or signal with no resonances at all. Note that only vector $WZ$ and

scalar $ZZ$ resonances are possible, but both scalar and vector $WW$ resonances can be produced. A continuum sample was also generated using this model.

Background samples include events with two vector bosons and two jets in the final state, arising from gluon or electroweak vector boson exchange between incoming quarks. The vector bosons are here mostly transverse and emitted more centrally than in the case of longitudinal vector boson pair scattering.

- $jjWZ$ final state, where $j$ is a quark or gluon: The decays of the vector bosons are performed in PYTHIA. Note that the semi-leptonic cases are already included in samples $W$+jets and $Z$+ jets (see below). Only the purely leptonic cases make use of this background.
- The background process: $\ell\ell\nu\nu$ with a pair of jets (quark or gluons). The cross section shown in Table 1 is for non-hadronic decay of the $ZZ$'s, with a filter requiring two leptons with $p_T > 5$ GeV and $|\eta| < 2.8$.
- $W/Z+3$ jets and $W/Z+4$ jets: they constitute backgrounds for the cases of high mass and lower mass resonances respectively since, in the former case, we expect that most of the vector bosons which decay hadronically will be reconstructed as a single jet. A correction factor of 1.38 is applied to the $W^+$+jets cross sections to account for the $W^-$+j process. These datasets include all tree level diagrams leading to $W$+4j, $Z$+4j, $W$+3j and $Z$+3j, with the vector bosons decaying leptonically, including all QCD and electroweak contributions. To keep the cross section manageable, preselection cuts were applied at MADGRAPH level. For the $W,Z+4$ jets case, we tag the highest rapidity jet (fjet), backward jet (bjet) and 2 central jets by requiring that $|\eta_{fjet}| > 1.5$, $|\eta_{bjet}| > 1.5$, that the forward and backward jet candidates be on different hemispheres: $\eta_{fjet}\eta_{bjet} < 0$, that at least one forward jet have energy $E > 300$ GeV, and that the invariant mass of the combined forward jets be $m_{jj} > 250$ GeV. We further require $p_T$ of at least one of the central jets to be $p_T^j > 50$ GeV, the $p_T$ of the vectorial addition of the central jets to be $p_T^{jj} > 60$ GeV and the invariant mass of the combined central jets $m_{jj} > 60$ GeV. We note that the forward jet preselection suppresses this background by a factor of 3.5.

  For the case $W,Z+3$ jets, we add the requirements: $p_T$ of the $W$ or $Z$ boson $p_T > 200$ GeV, $|\eta_{W/Z}| < 2$, and $p_T$ of one jet (central) $p_T^j > 200$GeV, $|\eta_j| < 2$.

Additional samples were produced with fast detector simulation to improve background statistics.

## 2.4 Comparative studies of generators

### 2.4.1 Parton shower matching to matrix elements

Here, MADGRAPH was used to generate the $W+$jets background. A better evaluation of this background would be obtained using a generator for which $W$+n partons, n=0, 1, 2, 3 or 4 inclusive, are combined in a manner which avoids double counting of jets produced by the parton shower in PYTHIA. ALPGEN is one such generator (and in fact such matching is now implemented in more recent versions of MADGRAPH). However, due to time constraints, and in order to have a manageable size of background samples, it was not practical to use this technique. In order to validate the use of MADGRAPH, a comparison was made of the $W+4$jets sample with an appropriate ALPGEN sample, with same analysis cuts applied. The ALPGEN samples are not used in the final analysis since they lack sufficient statistics.

Distributions of the vector bosons and jets were compared. As an example, the distributions for the forward jets are shown in Fig. 2. The overall conclusion is that the shapes of the distributions are in reasonable agreement. Therefore, neglecting the effect of parton shower double counting does not significantly affect the event topology. To the extent that such an error is made, the tag jets in the

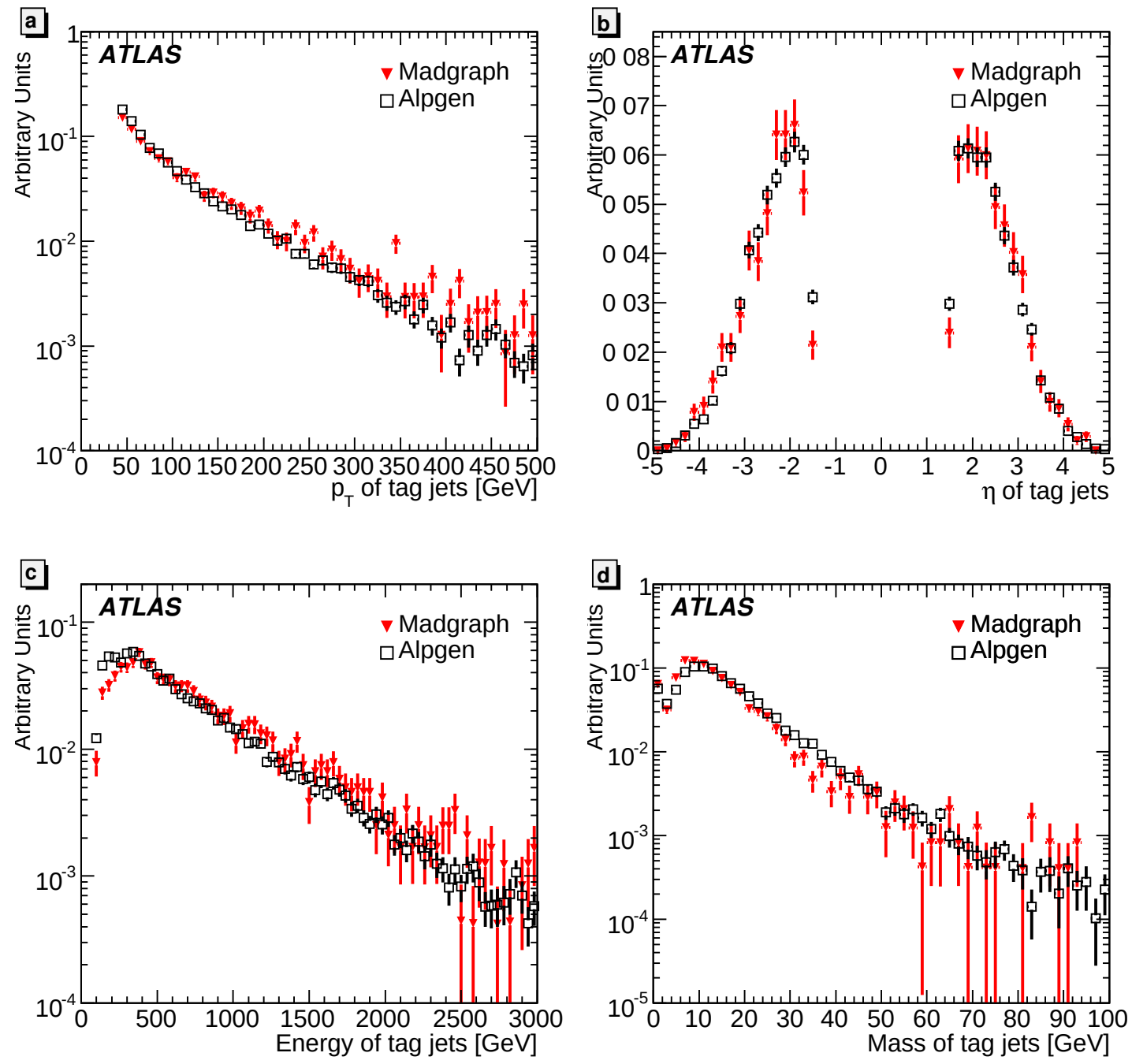


Figure 2: Distributions of the forward jets in the $W+4$jet background for MADGRAPH (red) and ALPGEN (black) samples (area normalised). The error bars show the statistical error in each sample.

ALPGEN sample have a lower energy (leading to a depletion with respect to the MADGRAPH samples at high energies of a few %) and so the backgrounds in this analysis can be considered to be conservatively overestimated. The sensitivity of such backgrounds to the scales has been discussed, for example, in [28, 29]. The difference in $Q^2$ scale of the two samples (ALPGEN uses $Q^2 = m_W^2 + p_T(W)^2$) leads to about a factor two discrepancy in cross section. This was confirmed by running MADGRAPH on a small sample with the same scale as ALPGEN, yielding cross sections smaller by factors 2.05 and 1.77 for the QCD and QED processes respectively. These factors will be applied in the present analysis.

### 2.4.2 Effective $W$ Approximation

WHIZARD [24] is a relatively new event generator originally developed for the ILC. It is able to calculate the full $2 \rightarrow 6$ matrix element needed for the vector boson scattering processes and it implements the Chiral Lagrangian model with the K-Matrix unitarisation scheme [27] which does not lead to vector boson resonances. Further unitarisation schemes in the form of arbitrary resonances are planned.

The generator does not assume an effective $W$ approximation, whereby the bosons emitted from the quarks are treated as partons, allowing vector boson scattering diagrams to form a gauge-invariant subset. This approximation is made in the PYTHIA signal samples, and might be expected to particularly affect the tag jet kinematics. Comparisons between the tag jet distributions from WHIZARD and PYTHIA are shown in Fig. 3. The two samples are not strictly comparable since here, WHIZARD simulates the $2 \rightarrow 4$ processes including non-scattering electroweak diagrams and applies K-matrix unitarisation. Although there are differences (e.g. the tag jets from WHIZARD are somewhat harder than those from PYTHIA) they do not strongly depend upon the vector boson centre-of-mass, and thus the effective $W$ approximation is unlikely to be the culprit. The harder tag jets in WHIZARD mean that if the signal looks more like WHIZARD than PYTHIA, it would be more likely to pass the selection cuts, thus improving the sensitivity. The potential size of the effect was investigated and estimated to be at the few per cent level.

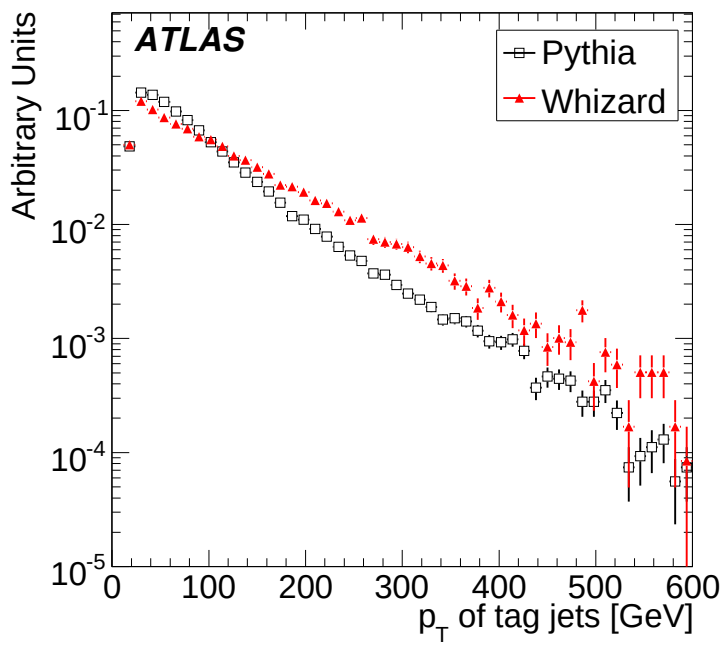


(a)

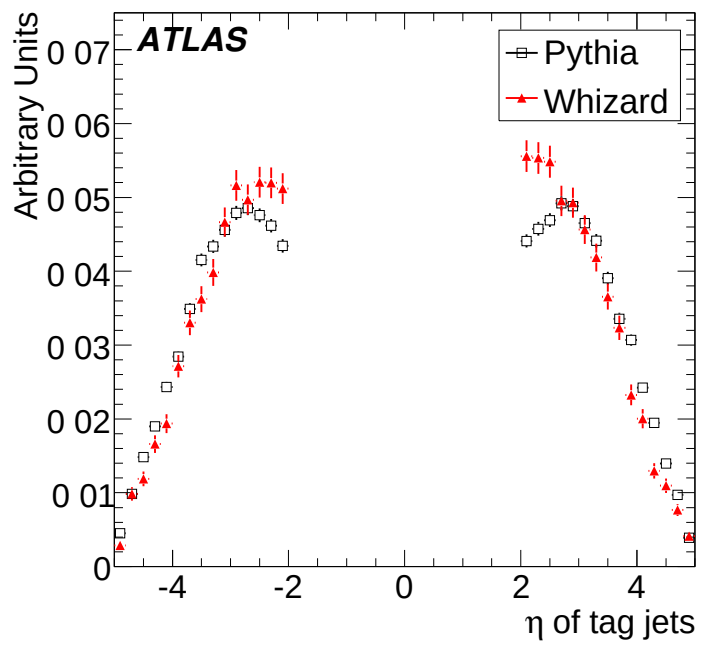


(b)

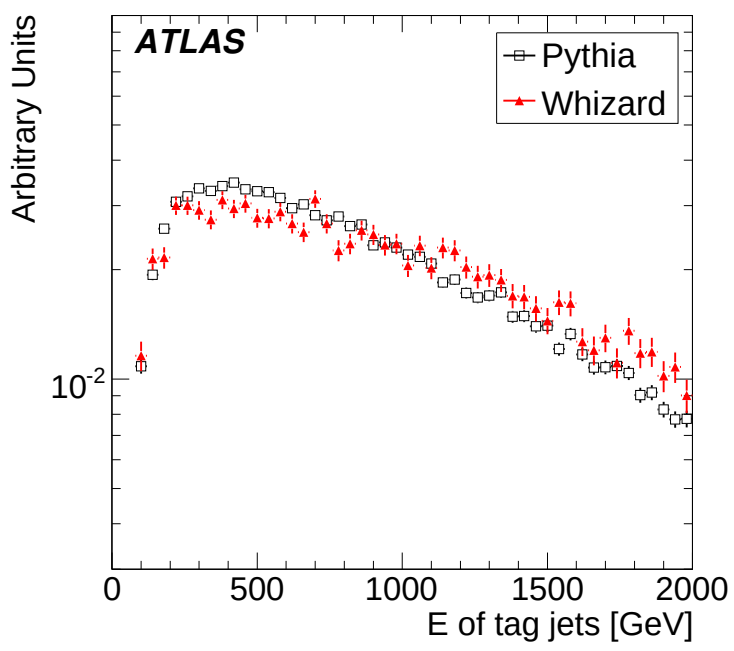


(c)

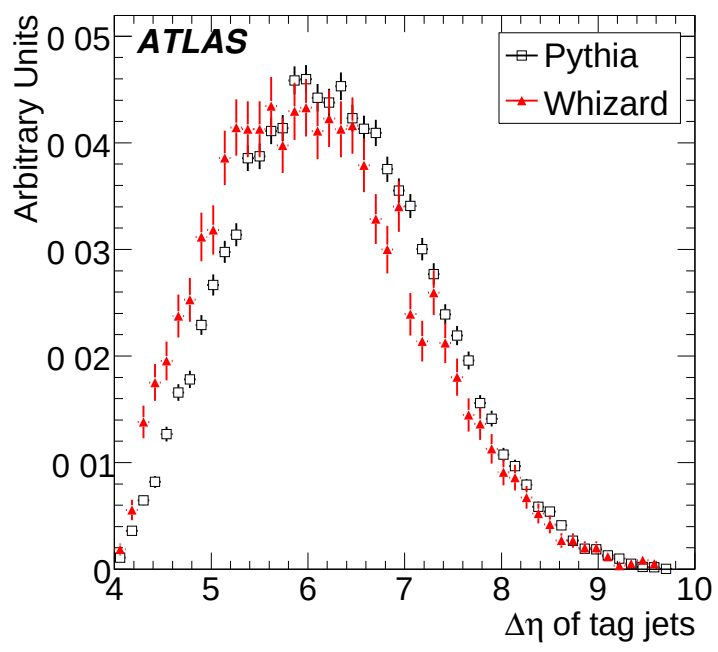


(d)

Figure 3: Differences between WHIZARD (red) and PYTHIA (black) for vanishing anomalous couplings for tag jet distributions of: transverse momentum (a), pseudo rapidity (b), energy (c) and pseudo rapidity difference (d).

## 3 Trigger

As a first step in the analysis, it is important to evaluate the efficiency of the basic trigger menus for a luminosity of $10^{33}$ cm$^{-2}$s$^{-1}$. The triggers chosen were based on an early menu [30] as an example, and the real physics menu is likely to be very different. However, since the signal is at relatively high $p_T$, and triggering on vector bosons is a high priority, this is not likely to have a large impact. To evaluate this efficiency, we apply the following cuts: for electrons and muons we require single leptons to have $p_T$ greater than the value corresponding to the threshold dictated by the trigger signature and $|\eta| < 2.5$; similarly for jets, but with a pseudorapidity cut of $|\eta| < 3.2$. This is necessary because trigger signatures for forward jets exist separately, but unfortunately that trigger information was not available in the simulation version used for this study. The trigger efficiency is defined as the number of times the trigger passed (with the corresponding cuts applied) divided by the number of truth events in the samples (with the same cuts applied). In Table 2 we present a detailed list of efficiencies for the signals $qqWZ \to qqjj\ell\ell$ ($m = 1.1$ TeV) in the left column and $qqWW \to qqjj\ell\nu$ (non-resonant) in the right[1].

The poor efficiency of the `e25i` (see Fig. 4, left) and `2e12i` triggers is understood to be due to the isolation criterion, which was not optimised for high energy electrons.

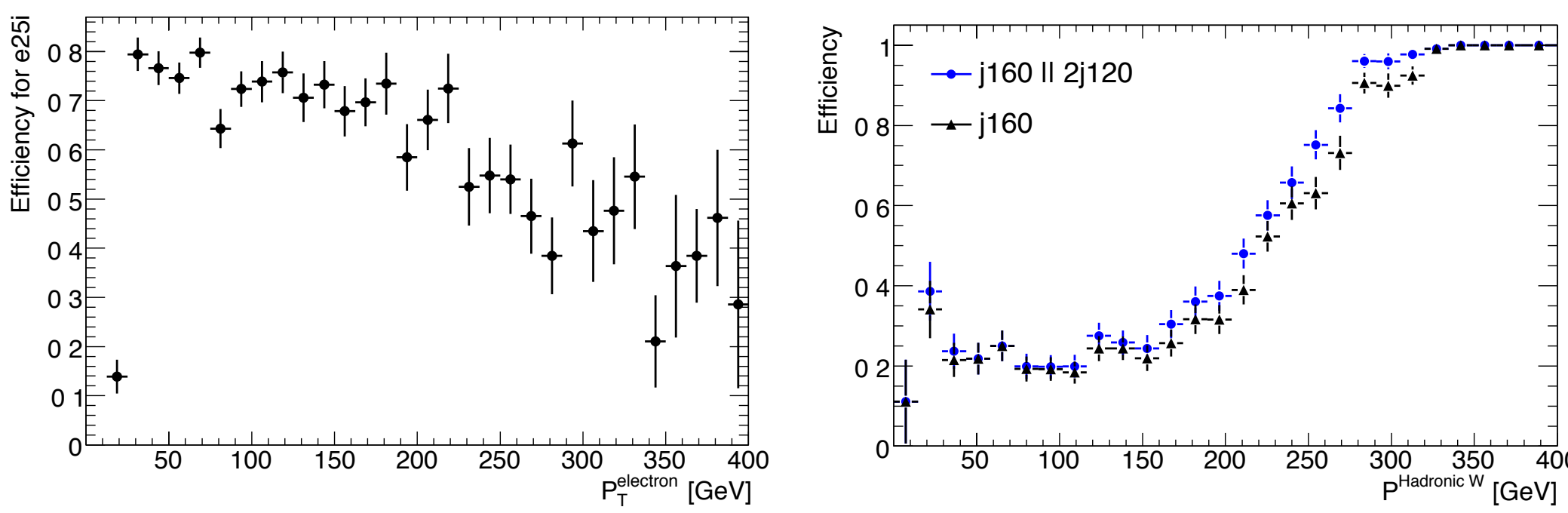


Figure 4: Trigger efficiencies computed with the $WW$ continuum signal. Left: efficiency of the `e25i` trigger as a function of the $p_T$ of electrons from the true leptonically-decaying $W$ boson. Right: efficiency of the `j160` trigger (black triangles) as a function of the $p_T$ of the true hadronically-decaying $W$ boson. Also shown with blue circles is the efficiency when the `j160` and `2j120` triggers are logically OR'ed.

It is worth mentioning that the efficiency for the `2j120` trigger (Fig. 4, right), which requires two jets with $p_T > 120$ GeV suffers partly from the fact that the two jets from the vector boson decay are merged due to the boost as described in Section 4.1. It is also significantly higher for events with true electrons than for those with true muons, probably because the electrons themselves are also reconstructed as jets in the calorimeter.

Finally, various combinations of the trigger signatures might be explored in the future to improve the efficiency. For instance, the `e60` trigger might be used in conjunction with the `e25i` to compensate the low efficiency of the latter for high-momentum electrons (Fig. 4). Likewise, the `2j120` trigger might be used together with `j160`, since the efficiency of the latter drops significantly when the hadronically-decaying vector boson has $p_T < 300$ GeV and decays into two distinctly resolvable jets.

[1] Triggers 2j120 and j160 have been removed from the menu as they are expected to give too high a rate. However, forward jet triggering will be available, and more recent developments in the electron trigger has resulted in improved efficiency.

| | $WZ$ signal | | $WW$ signal | |
|---|---|---|---|---|
| **Trigger Signature** | **Cut Loss** | **Efficiency** | **Cut Loss** | **Efficiency** |
| *Electrons* | | | | |
| 2e12i | 13% | 36% | > 99% | — |
| e25i | 1% | 78% | 11% | 65% |
| e60 | 5% | 82% | 29% | 73% |
| *Muons* | | | | |
| mu6 | 5% | 95% | 6% | 80% |
| mu20 | 5% | 92% | 9% | 73% |
| *Jets* | | | | |
| 2j120 | 67% | 73% | 50% | 80% |
| j160 | 34% | 96% | 30% | 86% |

Table 2: Table of high level trigger efficiencies for $qqWZ \to qqjj\ell\ell$ ($m = 1.1$ TeV) and $qqWW \to qqjj\ell\nu$ (non-resonant). The "Cut Loss" columns indicate the fraction of true events that would be lost by applying the $p_T$ requirements of each trigger signature on the true electrons, muons and jets. Since such events are unlikely to satisfy the trigger conditions, they are not taken into account when the trigger efficiencies are evaluated.

# 4 Reconstruction Challenges

In this section, we focus on those parts of the reconstruction which are most particular to vector boson fusion at high masses. We discuss the following:

- Reconstruction of hadronically-decaying vector bosons. In our regime these typically have high $p_T$ and the decay products are very collimated. We discuss two alternative methods, using $k_\perp$ jets and subjets, and using cone jets with different radii.
- Leptonically decaying vector bosons. These require good lepton and $\not{E}_T$ measurement, but the challenges here are not unique to these channels.
- Forward 'tag' jets. Measuring jets close to the edge of the detector rapidity acceptance is a challenge in common with low mass Higgs searches in vector boson fusion.
- Central jet veto. Since the vector boson scattering process involves no colour exchange between the protons, a suppression of QCD radiation is expected. This can be used to distinguish between signal and background, but is sensitive to underlying event and pile-up.
- Top veto. $t\bar{t}$ production is a major background for the channels which do not contain leptonic $Z$ boson decays. A large fraction of this is removed by explicitly rejecting events containing top candidates.

## 4.1 Hadronic Vector Boson Identification

At lower masses and $p_T$, the hadronically-decaying vector bosons are identified as dijet pairs. However, for events where a hadronically-decaying vector boson is highly boosted, the decay products are often collimated into a single jet. Cuts such as a dijet invariant mass window are no longer applicable in this scenario, but a single jet mass cut can be used.

The single jet mass is defined as the invariant mass evaluated from the 4-vectors of the constituents of the jet. In the ATLAS detector, these constituents are at present calorimeter objects, either topologically defined clusters with some local hadronic calibration, called here *topoclusters*, or calorimeter towers. For jets containing the decay products of a boosted vector boson, this single jet mass is near the mass of the parent boson. For light quark and gluon jets this mass is generally much lower. Since the background to hadronic vector boson identification is so severe, further cuts may be applied on the subjet structure of the candidate jet.

In addition, the transition between the dijet and single jet case as $p_T$ increases needs to be dealt with. Two methods are used, as follows;

1. Dynamically select the appropriate method. To do this, we first look at the highest $p_T$ jet. If this passes the mass window cut, then the single jet selection is applied, as described below. If it does not, then combinations of jet pairs in the event are considered. The vector boson is still expected to be the highest $p_T$ hadronic system, and so the $p_T$ of all jet pairs is evaluated, and the highest $p_T$ pair is taken to be the vector boson candidate. A mass window cut (dependent upon the jet algorithm) is then applied to this pair. Thus a single analysis can be used to scan the data for signs of resonances without bias.

2. When the single jet and jet pair cases yield very different signal to background ratios, it is preferable to choose a priori which mass region is being investigated, and to use the single jet reconstruction for high masses and the dijet reconstruction for low masses. This approach is used in the cone algorithm analysis. For the $m = 800$ GeV resonance, both the dijet and single jet approaches are tried independently.

### 4.1.1 $K_\perp$ Algorithm

The $k_\perp$ algorithm is run with an $R$-parameter (which determines the "jet size") of 0.6 on calibrated topoclusters. The algorithm [31] merges pairs of constituents.

The $k_\perp$ analysis uses the dynamic selection technique described above to decide whether to use a dijet or a single jet for the vector boson candidate. The fraction of vector bosons reconstructed as a single jet, as a function of $p_T$ of the vector boson candidate, is given in Fig. 5. The transition between dijet and single jet takes place between $p_T = 200$ and 300 GeV for this algorithm.

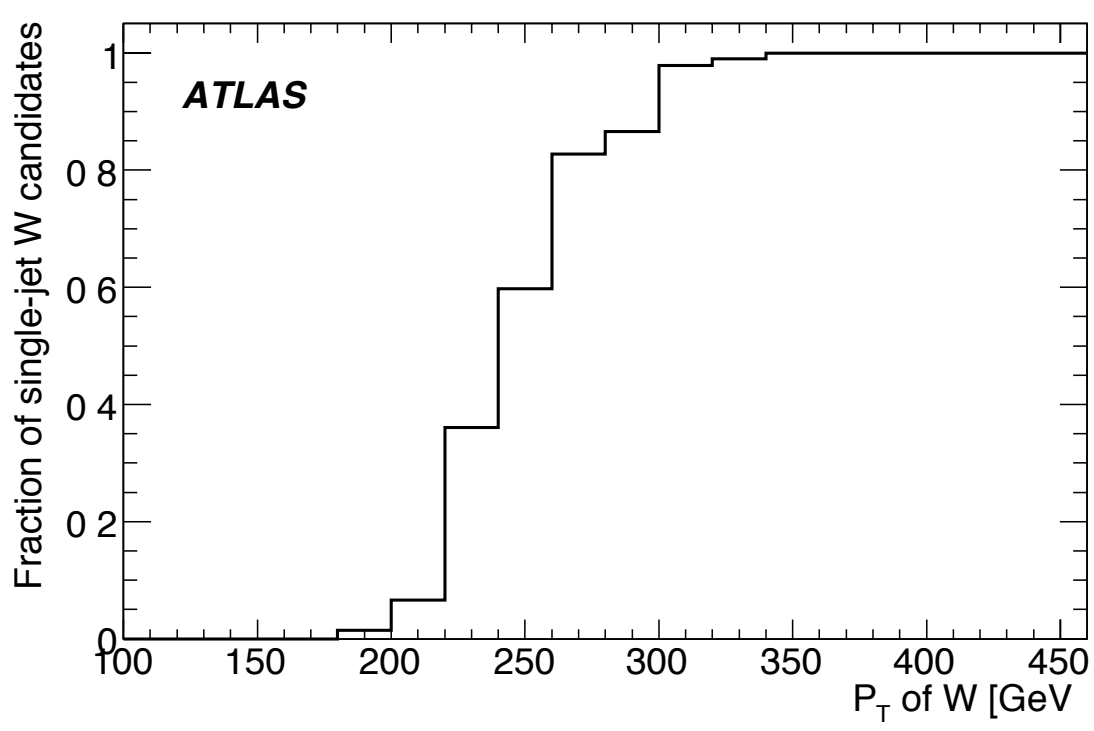


Figure 5: Fraction of $W$ boson candidates reconstructed from a single jet, as a function of the transverse momentum of the reconstructed vector boson, for the $WW$ $m = 1.1$ TeV signal sample.

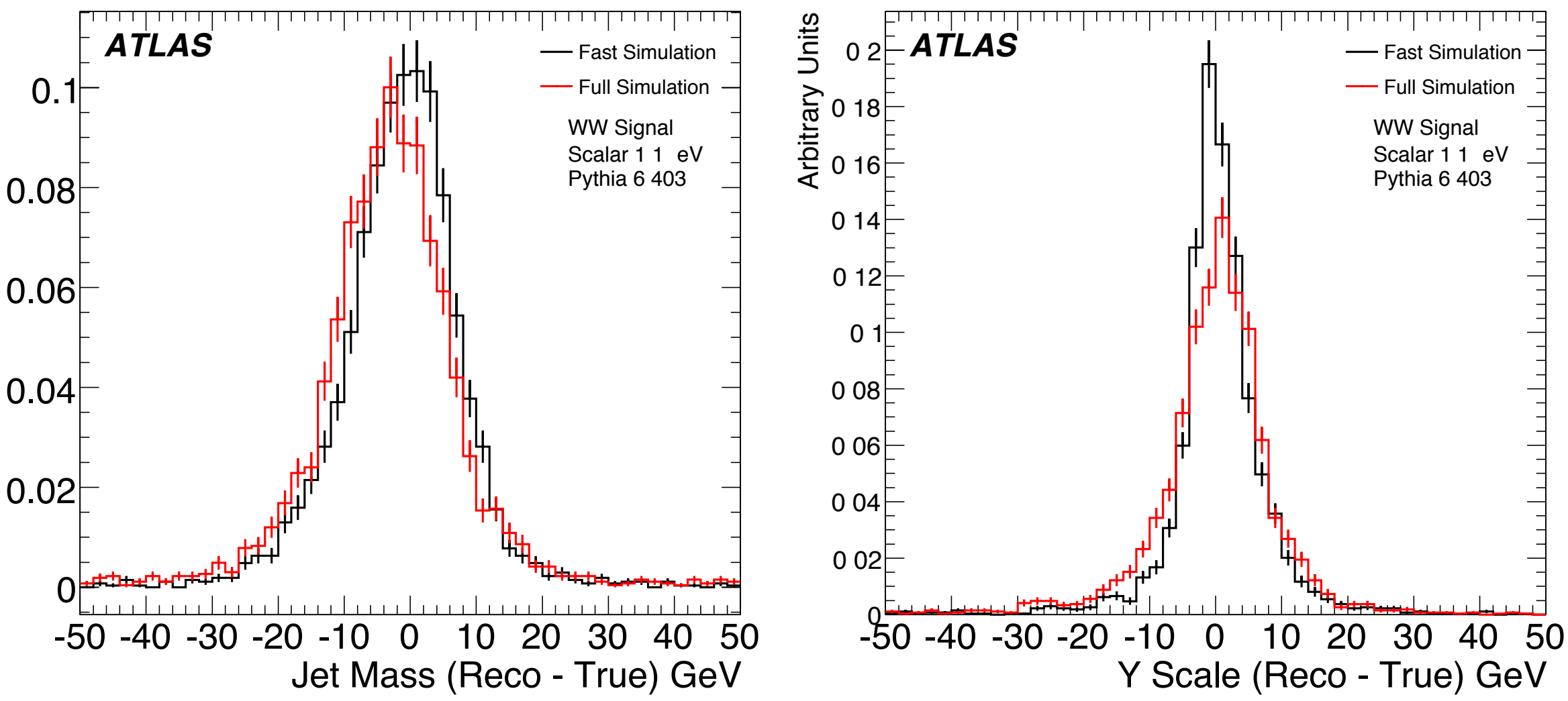


Figure 6: Single jet mass residuals (left) and Y scale residuals (right) from different detector simulations, using the $k_\perp$ algorithm. The truth is defined by running the jet algorithm on the hadronic final state of the MC generator.

### Single jet mode

The resolution of the single jet mass for the $k_\perp$ algorithm has been evaluated for both detector simulations (full and fast) for several samples. For the sample with a $WW$ resonance at $m = 1.1$ TeV (Fig. 6 left) for example, the $W$ boson singlet jet mass resolution was found to be $7.4 \pm 0.2\%$ GeV from full and fast detector simulation.

A mass cut around the window from $m = 68.4$ GeV to 97.2 GeV is applied to $W$ boson candidates, and from $m = 68.7$ GeV to 106.3 GeV for hadronic $Z$ boson candidates reconstructed in the single jet mode. These mass windows are determined by considering the resolution, the tails, and the background contamination.

The $k_\perp$ merging is intrinsically ordered in scale, making the final merging the hardest. The algorithm provides a $y$ value for this final merging, which is a measure of the highest scale at which a jet can be resolved into two subjets. The $y$ value can be converted into a "Y scale" in GeV using the relation Y scale $= E_T \times \sqrt{y}$, where $E_T$ is the jet transverse energy. This Y scale is expected to be $\mathcal{O}(m_V/2)$ (where $m_V$ is the mass of the vector boson) for boosted vector boson jets, and much lower than $E_T$ for light jets [32]. At the truth and fast simulation levels this variable has been shown to have discriminating power even after a single jet mass cut [32–35]. The resolution of the ATLAS detector for this variable is presented in Fig. 6 (right). The resolutions, for the same sample as above, are $12.3 \pm 0.3\%$ and $8.8 \pm 0.2\%$ with full and fast detector simulation, respectively.

Based on the resolution, the tails, and the background contamination, a Y scale cut around the window from 30 GeV to 100 GeV is applied to $W$ and $Z$ boson candidates reconstructed in the single jet mode. To evaluate the benefit of cutting on Y scale, a sample of single jet vector boson candidates is selected in signal and background by applying a $p_T > 300$ GeV cut, motivated by Fig. 5, and a mass window cut. Starting from this sample, the efficiency of the Y scale cut is given in Table 3 for full and fast simulation. The numbers suggest that for the $W$+jets background, an additional rejection factor of approximately 2 is provided by the Y scale cut even after a single jet mass cut has been applied. This is achieved with a signal efficiency of approximately 80%.

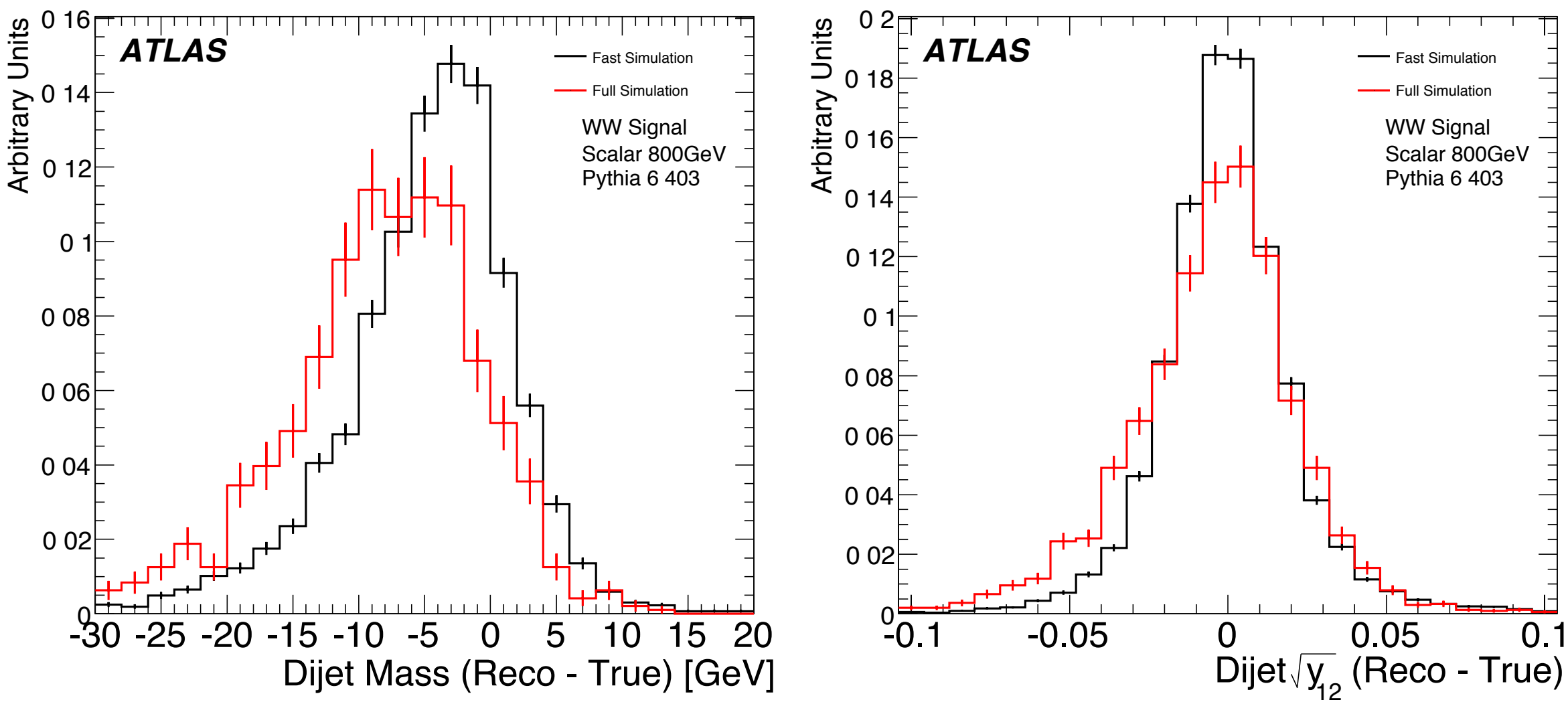


Figure 7: Dijet mass residuals (left) and $y$ residuals (right) from different detector simulations, using the $k_\perp$ algorithm. The truth is defined by running the jet algorithm on the hadronic final state of the MC generator.

| | 1.1 TeV Vector Resonance | $W$+4 jets | $t\bar{t}$ |
|---|---|---|---|
| Jet Mass | 68% (67%) | 14% (14%) | 28% (28%) |
| Y Scale | 77% (84%) | 29% (40%) | 63% (70%) |

Table 3: Efficiency of the jet mass cut and of the Y scale cut in the one-jet case for full (fast) simulation. The Y scale cut is applied after the mass cut.

| | 800 GeV Scalar Resonance | $W$+4 jets | $t\bar{t}$ |
|---|---|---|---|
| Jet Mass | 17% (20%) | 6% ( 7%) | 14% (14%) |
| Y Scale | 79% (83%) | 48% (49%) | 84% (82%) |

Table 4: Efficiency of the jet mass cut and of the Y scale cut in the two-jet case for full (fast) simulation. The Y scale cut is applied after the mass cut.

**Dijet mode**

A mass cut around the window from $m = 62$ GeV to 94 GeV is applied to $W$ boson candidates, and from $m = 66.6$ GeV to 106.2 GeV for hadronic $Z$ boson candidates reconstructed in the dijet mode.

A variable analogous to the $y$ may be calculated, using the $p_T$ of the softer jet relative to the harder one. This variable is required to be in the range $0.1 < \sqrt{y} < 0.45$. The efficiency is shown in Table 4.

The mass and $y$ windows are again determined by considering the resolution, the tails, and the background contamination.

The resolutions of the dijet mass and the $y$ variable for dijet vector boson candidates are shown in Fig. 7. They are found to be approximately 5% for the mass and for the $y$ variable, and are comparable in fast and full simulation.

### 4.1.2 Cone Algorithm

The problem of the two jets from a boosted hadronically decaying vector boson merging into a single jet has also been studied for jets reconstructed using the Cone Algorithm. With this algorithm, jet reconstruction starts from *seeds* i.e. constituents (clusters) with $p_T > 1$ GeV. The algorithm collects all constituents around a seed within $\Delta R = \sqrt{(\Delta\eta)^2 + \Delta\phi)^2} < R_0$ (where $R_0$ can be, for instance 0.4) and adds their momenta vectorially. Then it repeats the procedure over the collection around the direction of the sum, and computes a new sum. It continues repeating this operation until the resulting sum direction is stable.

Single jet hadronic $W$ boson candidates are identified with the highest $p_T$ object in the central region, after having removed overlaps with all electrons in the event within a $\Delta R$ of 0.1. A mass cut in a window around the reconstructed $W$ boson mass is applied.

Figure 8 shows an example of $W$ boson reconstruction using the cone algorithm for the jet-pair case ($m = 500$ GeV resonance) and single jet case ($m = 800$ GeV resonance). A cone size of 0.8 is used for selecting a single jet $W$ boson and 0.4 for the case of a jet pair. The low mass tail is due to events where the two jets from the $W$ are well separated. There is a small difference in the $W$ boson mass peak reconstruction for the two cases. The jets chosen for this selection have a minimum $p_T$ cut of 20 GeV, and those overlapping with electrons have been removed.

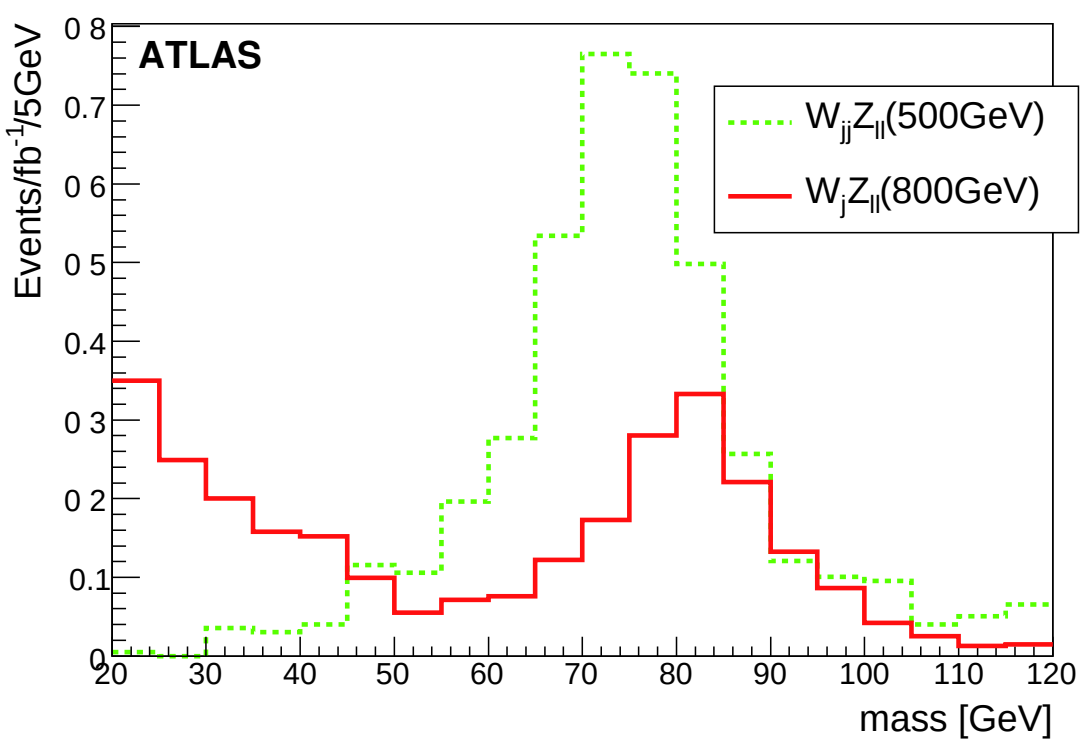


Figure 8: Reconstructed $W$ boson for cases where it forms two separated jets (500 GeV) and a single jet (800 GeV). The samples used are the $m = 500$ GeV resonance (in green) and $m = 800$ GeV resonance (in red).

The exploration of the substructure of a wide jet (typically of size 0.8) is done by searching for 2 narrow jets (size $\sim 0.2$) fitting inside the big jet. Various variables can then be studied, among which

| | 1.1 TeV Vector Resonance | $W$+4jets (QCD) | $t\bar{t}$ | $Z$+3jets (QED) |
|---|---|---|---|---|
| small jets cut | 76.3% | 15.2% | 38.6% | 13.8% |

Table 5: Comparison of efficiencies for the jet sub-structure selection for a typical signal and backgrounds. Efficiencies are relative to the selection of a single jet as described in the Y scale method (Table 3). For the subjet selection, we require 2 small jets with $p_T > 15$ GeV and invariant mass $> 60$ GeV (see text).

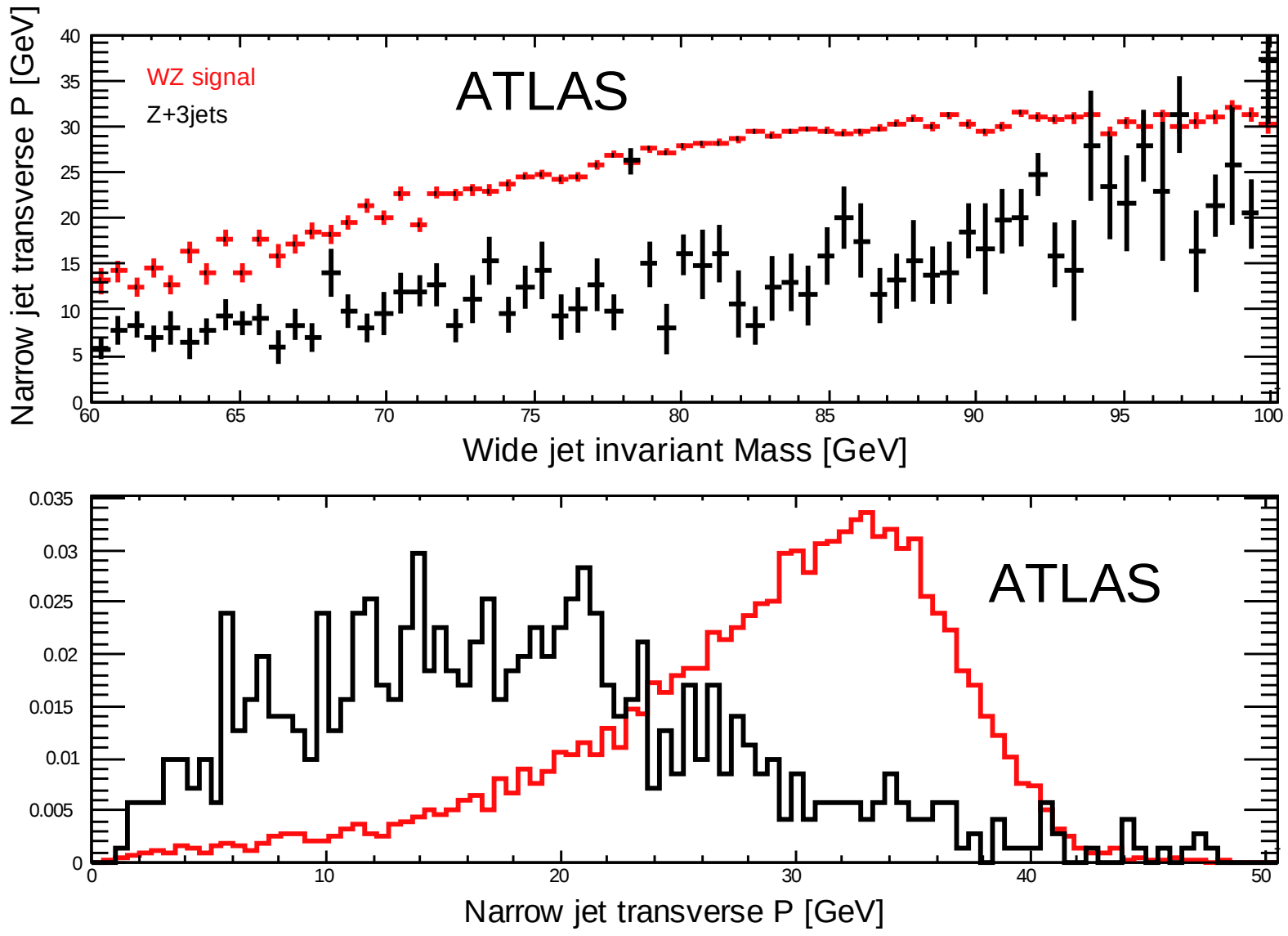


Figure 9: Profile histogram of the momentum of the narrow jet orthogonal to the wide jet direction vs the invariant mass of the wide jet, for $W$ boson hadronic decay of the resonance signal $qqW_{jj}Z_{\ell\ell}$ of $m =$ 1.1 TeV (red) and for $Z$+3 jets sample (black). Lower graph : normalized distributions of narrow jet orthogonal momentum.

are the energy ratio of the narrow jets, their invariant mass, the distance $\Delta R$ between the leading narrow jet and the wide jet, or the momentum component of this narrow jet transverse to the wide jet direction. The discriminating power is illustrated for the $WZ \to \ell\ell jj$ channel (1.1 TeV resonance) and its principal background in Fig. 9 which shows the latter variable (called here 'p transverse') versus the invariant mass reconstructed from two narrow jets. Cutting in the ($p_T$, invariant mass) plane gives results comparable to those obtained with the Y scale method above, as illustrated in Table 5. Similarly, Fig. 10 shows the $p_T$ versus $\Delta R$ between the leading narrow jet and the wide jet.

## 4.2 Leptonic Vector Boson Identification

### 4.2.1 Lepton Reconstruction Efficiencies

All the signals studied in this note involve at least one leptonic vector boson decay. Electrons and muons are selected using standard ATLAS criteria [36, 37] for the case of a resonance $m(WZ)$ of 800 MeV. Figure 11 shows the efficiency for $W$ boson daughter leptons, where the trigger consition has not been applied. The results for different electron selection criteria are given. The loss of efficiency occurs in the forward regions, near the limits of the tracking detectors and at $p_T$ values close to the applied cut. The efficiencies for the leptonic $Z$ boson channels have been found to be similar.

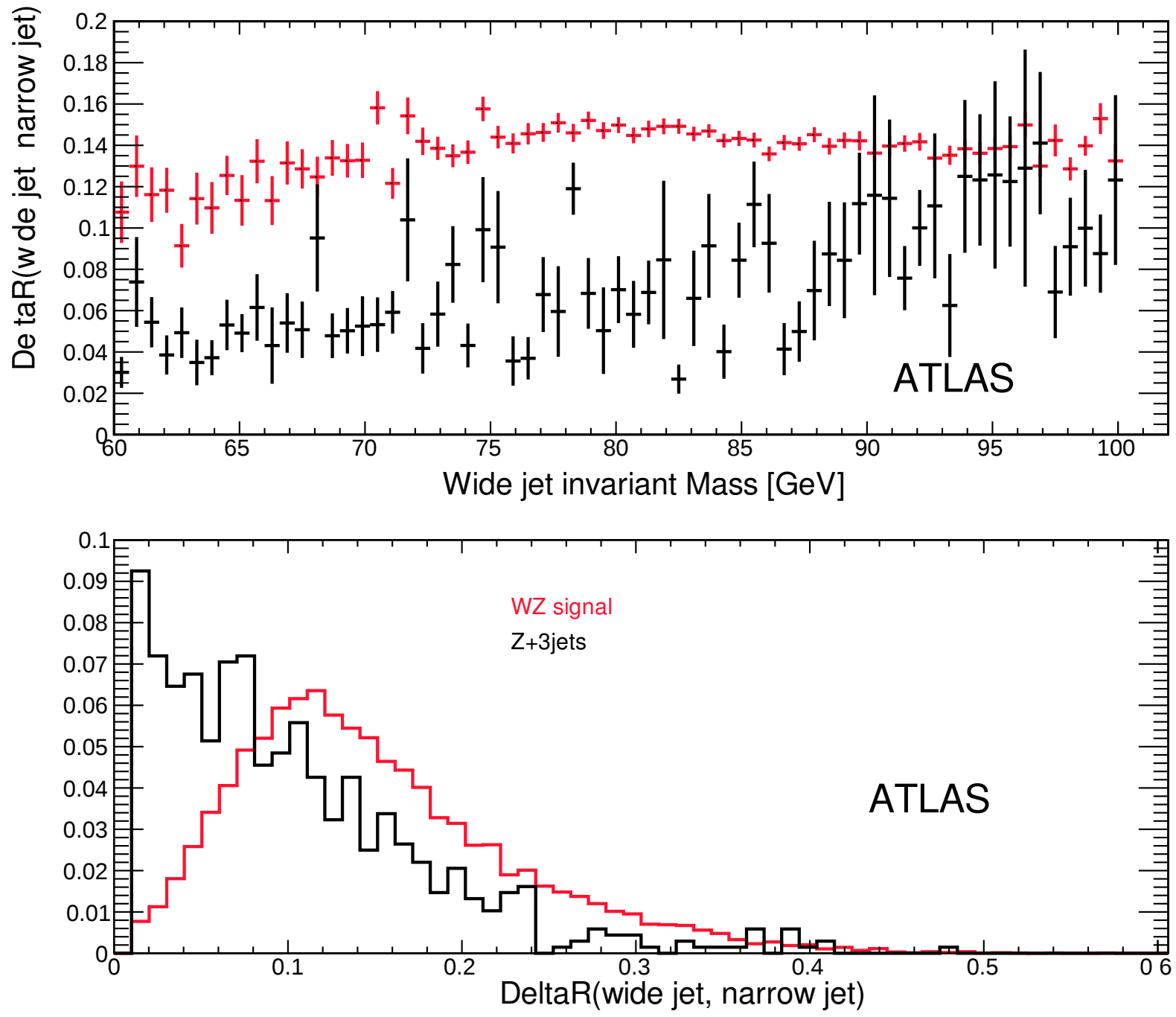


Figure 10: Profile histogram of the distance (narrow jets, wide jet) versus the invariant mass of the wide jet, for $W$ boson hadronic decay of the resonance signal $qqW_{jj}Z_{\ell\ell}$ of $m = 1.1$ TeV (red) and for $Z$+3 jets sample (black). Lower graph : normalized distributions of distance (narrow jets, wide jet).

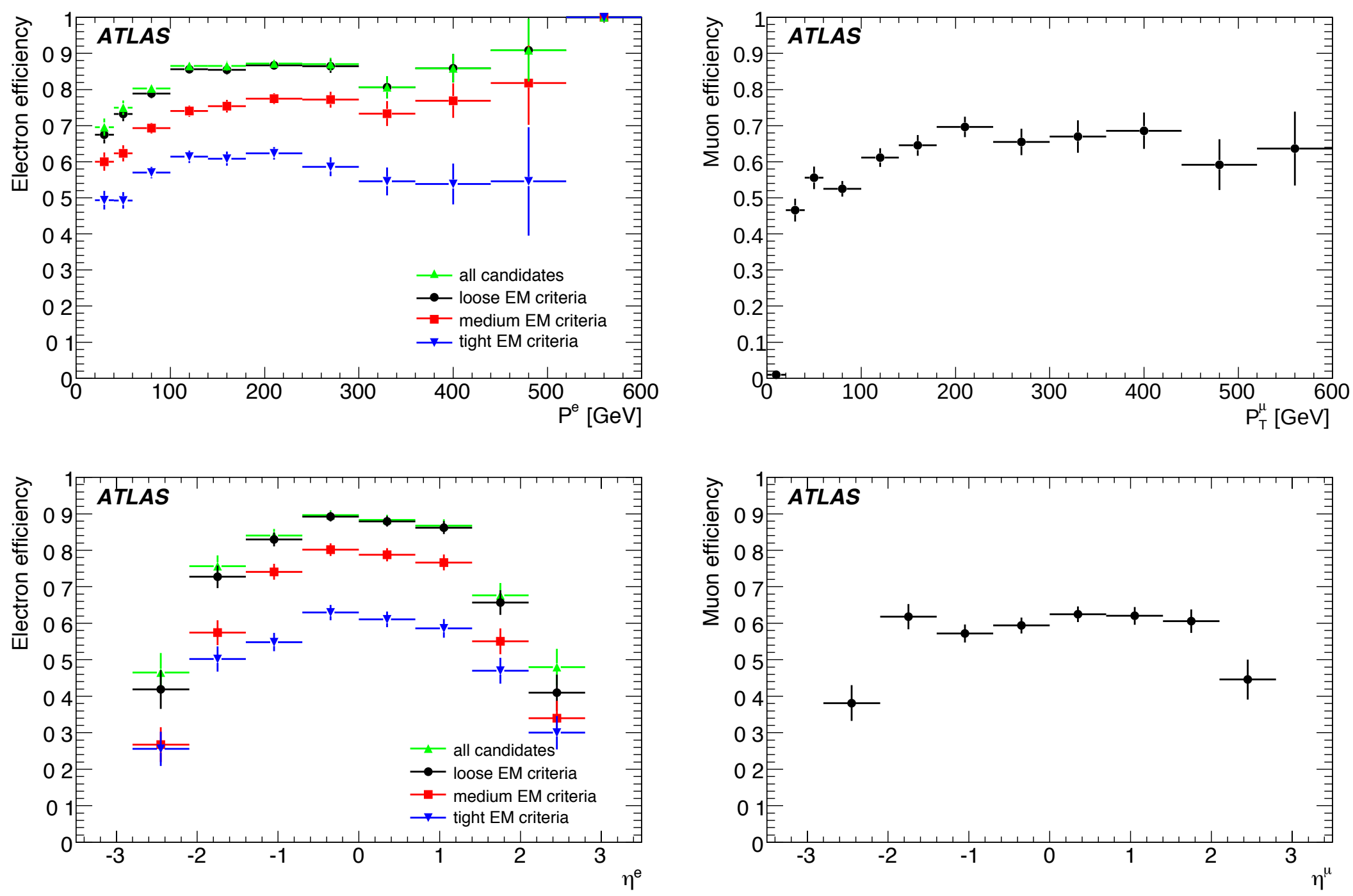


Figure 11: Efficiency of reconstructing and identifying $W$-daughter electrons (left) and muons (right) as functions of true lepton momentum (top) and pseudo-rapidity (bottom). The electron plots show the efficiency for 4 different electron selection criteria: All candidate objects (green), *loose* (black circles), *medium* (red squares), and *tight* (blue triangles).

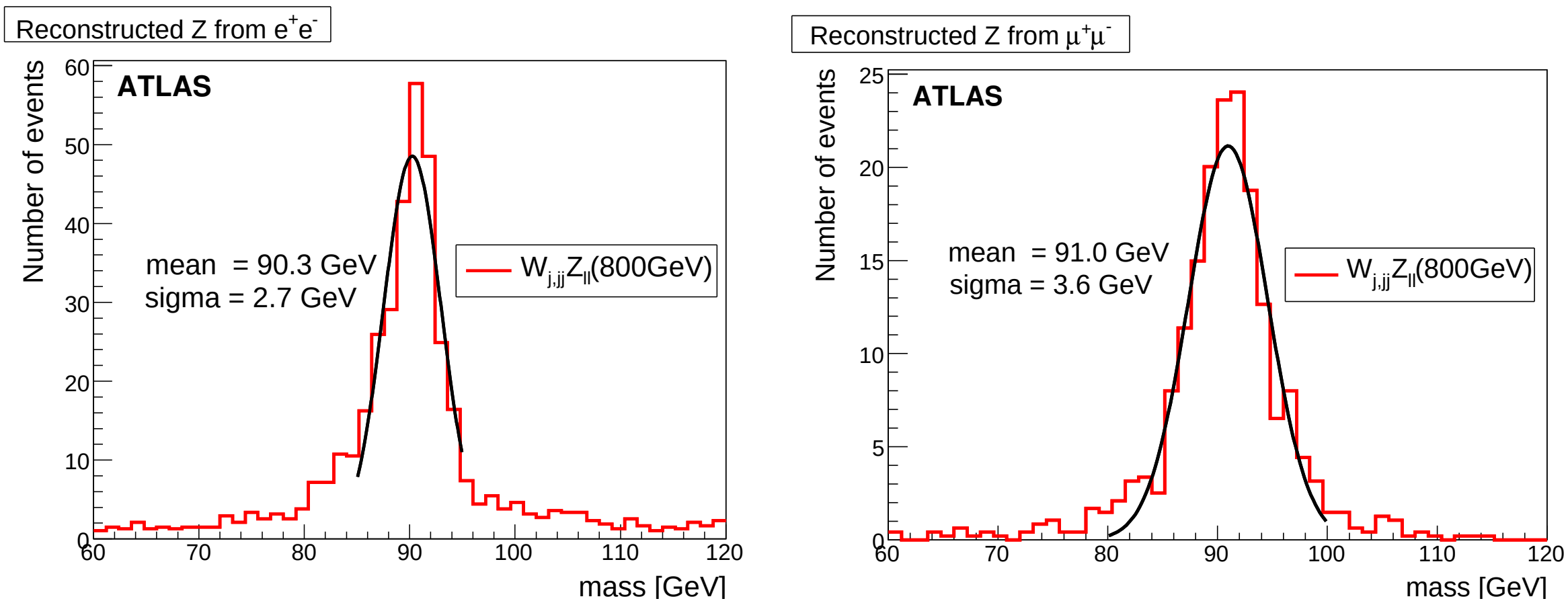


Figure 12: Reconstructed $Z$ boson from electron pairs (left) and muon pairs (right).

### 4.2.2 Leptonic $Z$ Boson Reconstruction

The $Z$ boson candidates are reconstructed from pairs of $e^+e^-$ or $\mu^+\mu^-$. In the electron case, the mass resolution is about 2.7 GeV as is shown in Fig. 12 left, suggesting a mass window selection between $m = 85$ GeV and 97 GeV for $m_{ee}$. In the case of muons, the resolution for the $Z$ mass reconstruction is 3.6 GeV (see Fig. 12 right), so the mass requirement is loosened to be between $m = 83$ GeV and 99 GeV. Furthermore, to reduce the backgrounds (particularly the background from $t\bar{t}$ events), the $p_T$ of one of the leptons is required to be $p_T > 50$ GeV, and that of the other $p_T > 35$ GeV. In the unlikely case that more than one combination of leptons satisfy all these requirements, we choose the composite $Z_{\ell^+\ell^-}$ with the mass closest to the actual $Z$ boson mass.

### 4.2.3 Leptonic $W$ Boson Reconstruction

For the signal, after reconstruction of the hadronic vector boson candidate, the highest $p_T$ lepton corresponds to the lepton from the $W$ boson decay in 96% of cases. Attributing the missing momentum to the neutrino, and taking the nominal $W$ boson mass ($m_W = 80.42$ GeV) as a constraint, a quadratic equation is obtained for the $z$-component of the neutrino's momentum. The $z$ component is required in order to reconstruct the diboson mass in the final analysis. Only events for which at least one real solution exists for this quadratic equation are retained. When there are two possible solutions, one is chosen at random to avoid kinematic bias on the resonance mass. Options such as selecting the reconstructed $W$ boson which is more central have also been considered, and little difference was found in the purity of the reconstruction. In the fully leptonic case, the leptonic $Z$ boson is reconstructed and its daughter leptons removed before applying the above procedure.

## 4.3 Tagged Forward Jets

One of the well known characteristic features of vector boson scattering is the presence of high energy forward jets [14], resulting from the primary quarks from which the vector bosons have radiated (see Fig. 13). Such forward jets are expected to be much less prominent in processes involving gluon or electroweak boson exchange with bremsstrahlung of vector bosons. In the latter case, these vector bosons are mostly transverse and have a harder $p_T$ spectrum than in $W_LW_L$ scattering. Correspondingly, the outgoing primary quarks have a harder $p_T$ and are therefore less forward.

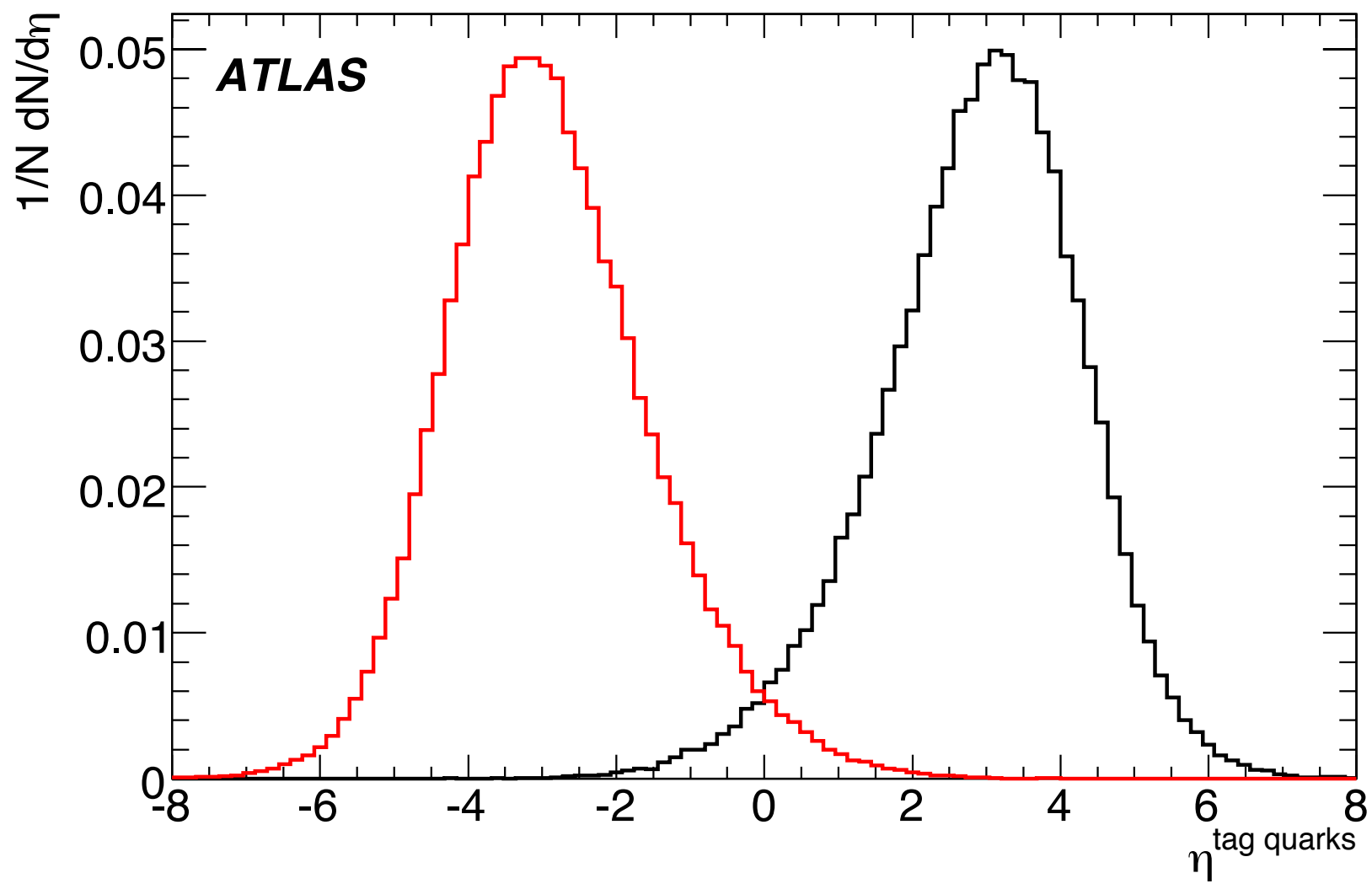


Figure 13: Pseudo-rapidities of the forward and backward quarks in signal events after they radiate the vector bosons, obtained from PYTHIA before any showering, fragmentation, etc.

Many different strategies are possible for implementing a tag-jet selection. A number of these were compared, and the best rejection factors for a given efficiency were obtained as follows:

1. Require two jets with
   - $|\eta(jet)| > \eta_{\rm cut}$ and $p_T(jet) > p_{T{\rm cut}}$
   - opposite signed rapidity
   - at least one of them has an energy greater than a critical value $E_{\rm cut}$
2. If more than one jet with the same sign rapidity satisfies the above cuts, choose the most energetic, labelled FJ1. The next one is labelled FJ2.
   - Require the tag-jet with the opposite sign of rapidity to satisfy $\Delta\eta(FJ1,FJ2) > \Delta\eta_{\rm cut}$ and $E(FJ2) > E_{2{\rm cut}}$

In addition a dijet mass cut is currently applied in the cone algorithm analyses. The specific values of the cuts in each case are to be optimised depending upon the kinematic region under study.

### 4.4 Central Jet Veto

A useful analysis strategy to suppress backgrounds such as $t\bar{t}$ is to apply a central jet veto [15, 38, 39]. For vector boson scattering, one expects little QCD radiation in the central region since only colourless electroweak vector bosons are produced and the forward jets are not colour connected. Given the forward jet cut definition, we unambiguously define the central region of the event as the $\eta$ region between them. The central jet veto then simply requires that no other high $p_T$ jet (here taken as $p_T > 30$ GeV) other than those resulting from the hadronically decaying vector boson lie in the central region.

Specifically, in the analyses where it is applied, the central jet veto rejects events if there are any additional jets with a chosen maximum value for $|\eta|$ and minimum value for $p_T$.

### 4.5 Top Quark Rejection

While the final states from a leptonically decaying $Z$ boson are mostly free of background from top processes, $t\bar{t}$ and $tW$ events form an important source of background for the $WW$ signals. To suppress them, events can be vetoed if a reconstructed $W$ boson candidate, combined with another jet in the event (excluding those overlapping with an identified electron or within $\Delta R < 0.8$ of a $W$ candidate), leads to an invariant mass close to that of the top quark [32]. A typical mass window is $130 < m_t < 240$ GeV.

In a future analysis it is likely that this cut can be improved using $b$-tag information and better jet mass reconstruction, but this has not been investigated here.

## 5 Event Selection

Using the tools outlined in the previous section, we now characterise the samples and outline the specific cuts applied for each final state considered.

### 5.1 $W^+W^- \to \ell^\pm \nu\, jj$ and $W^\pm Z \to \ell^\pm \nu jj$

The hadronic vector boson candidates and the tag jets are obtained using the $k_\perp$ algorithm as discussed in Section 4.1. The leptonic $W$ is identified as described in Section 4.2.3. Both vector boson candidates are required to have $p_T > 200$ GeV, $|\eta| < 2$. Tag jet cuts are made as described in Section 4.3, with $p_{T\mathrm{cut}} = 10$ GeV, $E_{\mathrm{cut}} = E_{2\mathrm{cut}} = 300$ GeV and $\Delta\eta_{\mathrm{cut}} = 5$. The top veto (Section 4.5) and the the central jet veto (Section 4.4) are applied.

The kinematic distributions for the $WW$ channel are shown in Fig. 14. Note that these are significantly biased by the generator-level cuts. The selection efficiencies of the cuts on $WW$ events from four example scenarios are summarised in Table 6, along with the efficiency of the combined trigger selection described in Section 3. No significant differences are observed between the scalar and vector resonances, nor between the $WZ$ and $WW$ channels, except for the $m = 500$ GeV-resonant samples, where the $p_T$ requirements are found to be less efficient for the $WZ$ sample. The QCD-like model of [27] tends to predict softer vector bosons.

Due to the small background statistics available with full simulation, the full-simulation signal samples are used together with fast-simulation background samples in obtaining the final results. The modelling of the kinematics is good, as shown in Fig. 14 [2]. However, in general the efficiency for selecting both signal and background is higher in fast simulation by about 25% compared to full simulation. To account for this, a constant scaling factor is applied to the fast-simulation samples in estimating the significance of the signal over the background (Section 6).

Figure 15 shows a comparison of the final $WW$ mass for the signal samples using fast and full simulation.

The final $WW$ mass spectra obtained using this analysis are shown in Fig. 16. The backgrounds shown have been obtained from the fast-simulation samples and the above mentioned scaling factor has been applied.

### 5.2 $W^\pm Z \to jj\, \ell^+\ell^-$

This channel benefits from a very good resolution on the $Z$ boson leptonic reconstruction, which allows good suppression of the $t\bar{t}$ background.

[2] To achieve this agreement it was necessary to correct the lepton-finding efficiency in the fast simulation by fitting a function to the efficiency from fast and full simulation as a function of lepton $p_T$ and correcting the fast simulation by the ratio of the functions.

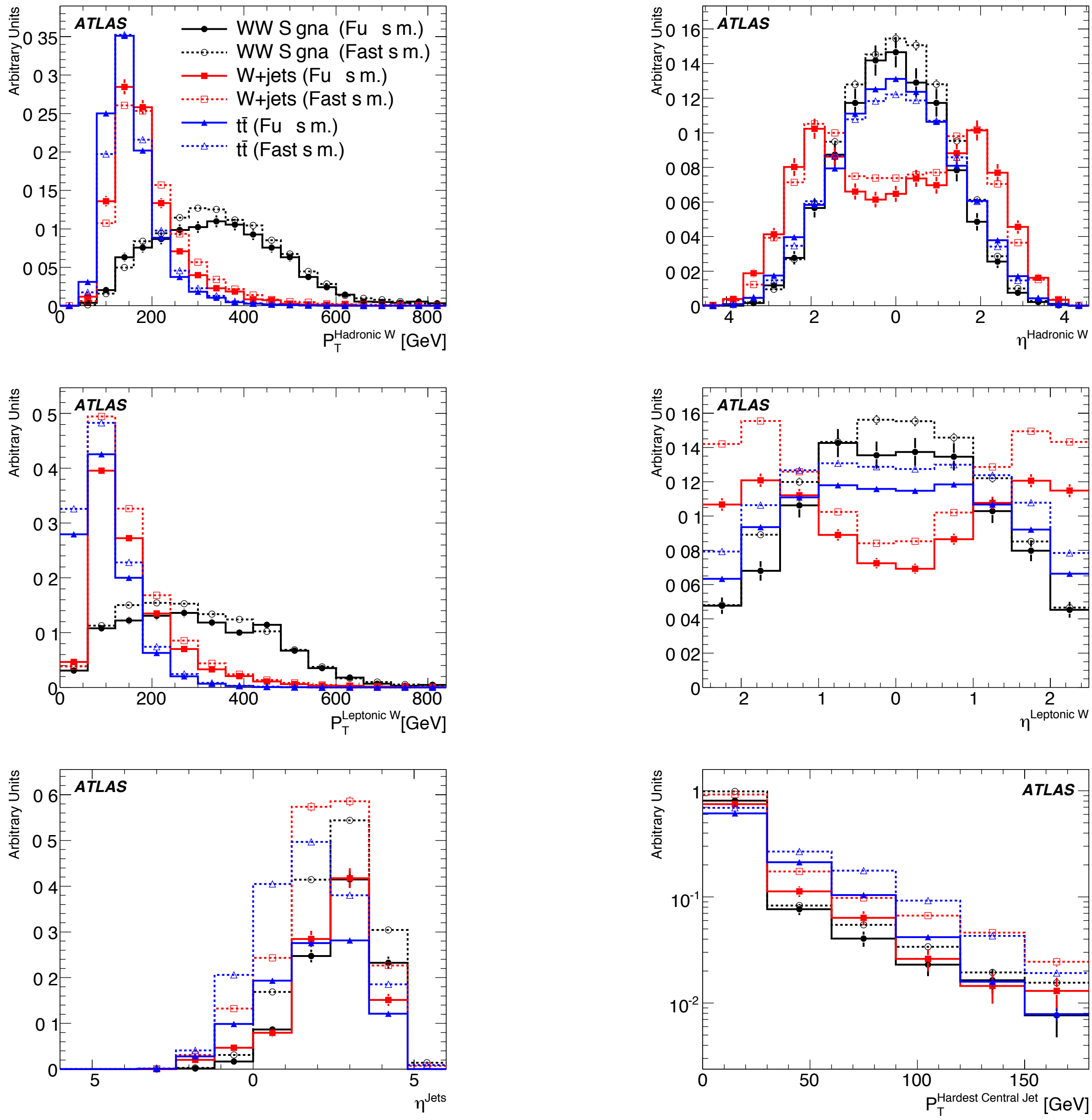


Figure 14: Kinematic distributions for the generated signal and backgrounds ($t\bar{t}$ and $W$+4 jets QCD) in the 1.1 TeV $W^+W^- \rightarrow \ell^\pm \nu\ j(j)$ channel. The top two plots show the $p_T$ and $\eta$ for the hadronic $W$ boson candidate. The middle two show the same variables for the leptonic $W$ boson candidate. The bottom two plots show the $\eta$ distribution of all jets which are at higher rapidity than the $W$ boson candidates, and the $p_T$ distribution of the highest-$p_T$ central jet. In each plot, the full-simulation histograms (solid lines) have been normalised to unit area and the fast-simulation histograms (dashed lines) have been normalised to the same cross section as their full-simulation counterparts.

| Cut | $m = 500$ GeV Scalar Resonance | | $m = 800$ GeV Scalar Resonance | | $m = 1.1$ TeV Vector Resonance | |
|---|---|---|---|---|---|---|
| | Efficiency (%) | $\sigma$ (fb) | Efficiency (%) | $\sigma$ (fb) | Efficiency (%) | $\sigma$ (fb) |
| **Starting sample** | – | 66 | – | 28 | – | 18 |
| $\equiv 1$ Hadronic $W$ | $32.1 \pm 0.5$ ( 34) | 21 ( 23) | $40.0 \pm 0.5$ ( 45) | 11 ( 13) | $39.5 \pm 0.7$ ( 43) | 7.1 (7.8) |
| $\equiv 1$ Leptonic $W$ | $45.4 \pm 0.9$ ( 54) | 9.6 ( 12) | $48.5 \pm 0.8$ ( 57) | 5.4 (7.1) | $48.8 \pm 1.2$ ( 55) | 3.5 (4.3) |
| $p_T$ (Had. $W$) $> 200$ GeV | $57.6 \pm 1.3$ ( 69) | 5.5 (8.5) | $88.2 \pm 0.7$ ( 90) | 4.8 (6.4) | $86.6 \pm 1.1$ ( 88) | 3.0 (3.8) |
| $\|\eta\|$ (Had. $W$) $< 2$ | $91.9 \pm 0.9$ ( 93) | 5.1 (7.9) | $95.3 \pm 0.5$ ( 95) | 4.6 (6.1) | $93.4 \pm 0.9$ ( 92) | 2.8 (3.5) |
| $p_T$ (Lep. $W$) $> 200$ GeV | $43.8 \pm 1.8$ ( 42) | 2.2 (3.3) | $91.3 \pm 0.7$ ( 89) | 4.2 (5.4) | $92.4 \pm 1.0$ ( 89) | 2.6 (3.1) |
| $\|\eta\|$ (Lep. $W$) $< 2$ | $95.5 \pm 1.1$ ( 94) | 2.1 (3.1) | $95.3 \pm 0.6$ ( 95) | 4.0 (5.1) | $92.8 \pm 1.0$ ( 93) | 2.4 (2.9) |
| $\equiv 2$ tag jets | $32.0 \pm 2.6$ ( 37) | 0.7 (1.1) | $42.4 \pm 1.3$ ( 49) | 1.7 (2.5) | $43.7 \pm 2.0$ ( 55) | 1.1 (1.6) |
| $\equiv 0$ top candidates | $50.0 \pm 5.0$ ( 40) | 0.3 (0.5) | $52.0 \pm 2.1$ ( 41) | 0.9 (1.0) | $51.4 \pm 3.0$ ( 44) | 0.5 (0.7) |
| Central jet veto | $100.0 \pm 0.0$ ( 98) | 0.3 (0.4) | $96.7 \pm 1.0$ ( 97) | 0.8 (1.0) | $91.6 \pm 2.3$ ( 93) | 0.5 (0.7) |
| Trigger efficiency | $96 \pm 3$ | 0.3 (0.4) | $98 \pm 1$ | 0.8 (1.0) | $98 \pm 1$ | 0.5 (0.7) |

| Cut | Non-resonant Signal | | $t\bar{t}$ Background | | $W$+jets Backgrounds | |
|---|---|---|---|---|---|---|
| | Efficiency (%) | $\sigma$ (fb) | Efficiency (%) | $\sigma$ (fb) | Efficiency (%) | $\sigma$ (fb) |
| **Starting sample** | – | 10 | – | 450000 | – | 21365 |
| $\equiv 1$ Hadronic $W$ | $38.0 \pm 0.7$ ( 41) | 3.8 (4.1) | $18.9 \pm 0.1$ (19) | 85000 ( 84000) | $8.3 \pm 0.1$ (9) | 1760 ( 1820) |
| $\equiv 1$ Leptonic $W$ | $48.2 \pm 1.1$ ( 55) | 1.8 (2.3) | $22.1 \pm 0.2$ (29) | 19000 ( 25000) | $23.3 \pm 0.7$ (31) | 410 ( 570) |
| $p_T$ (Had. $W$) $> 200$ GeV | $82.1 \pm 1.3$ ( 86) | 1.5 (1.9) | $16.8 \pm 0.4$ (20) | 3200 ( 5000) | $34.4 \pm 1.7$ (43) | 140 ( 240) |
| $\|\eta\|$ (Had. $W$) $< 2$ | $94.4 \pm 0.8$ ( 94) | 1.4 (1.8) | $90.3 \pm 0.7$ (90) | 2900 ( 4500) | $80.1 \pm 2.4$ (77) | 110 ( 190) |
| $p_T$ (Lep. $W$) $> 200$ GeV | $90.4 \pm 1.1$ ( 87) | 1.3 (1.6) | $34.5 \pm 1.3$ (29) | 990 ( 1300) | $48.5 \pm 3.3$ (40) | 55 ( 75) |
| $\|\eta\|$ (Lep. $W$) $< 2$ | $96.0 \pm 0.8$ ( 96) | 1.2 (1.5) | $94.6 \pm 1.0$ (90) | 930 ( 1200) | $80.4 \pm 3.9$ (79) | 44 ( 59) |
| $\equiv 2$ tag jets | $45.1 \pm 2.0$ ( 54) | 0.6 (0.8) | $8.1 \pm 1.3$ (10) | 76 ( 120) | $13.9 \pm 3.5$ (22) | 6 ( 13) |
| $\equiv 0$ top candidates | $56.5 \pm 3.0$ ( 47) | 0.3 (0.4) | $7.9 \pm 4.4$ ( 2) | 5 ( 2) | $60.5 \pm 13.1$ (23) | 4 ( 3) |
| Central jet veto | $91.1 \pm 2.3$ ( 94) | 0.3 (0.4) | $< 50$ ($< 25$) | $< 5$ ( $< 1$) | $84.9 \pm 13.7$ (91) | 3 ( 3) |
| Trigger efficiency | $98 \pm 1$ | 0.3 (0.4) | $\sim 100$ | $< 5$ ( $< 1$) | $82 \pm 16$ | 3 ( 3) |

Table 6: Efficiencies of the cuts for four different $qqWW \rightarrow qq\ell\nu qq$ signal samples and the backgrounds. The trigger efficiency row shows the efficiency of the logical OR of `mu20i`, `e25i` and `jet160` signatures on the samples after all the cuts have been consecutively applied. The numbers in brackets are from the fast simulation.

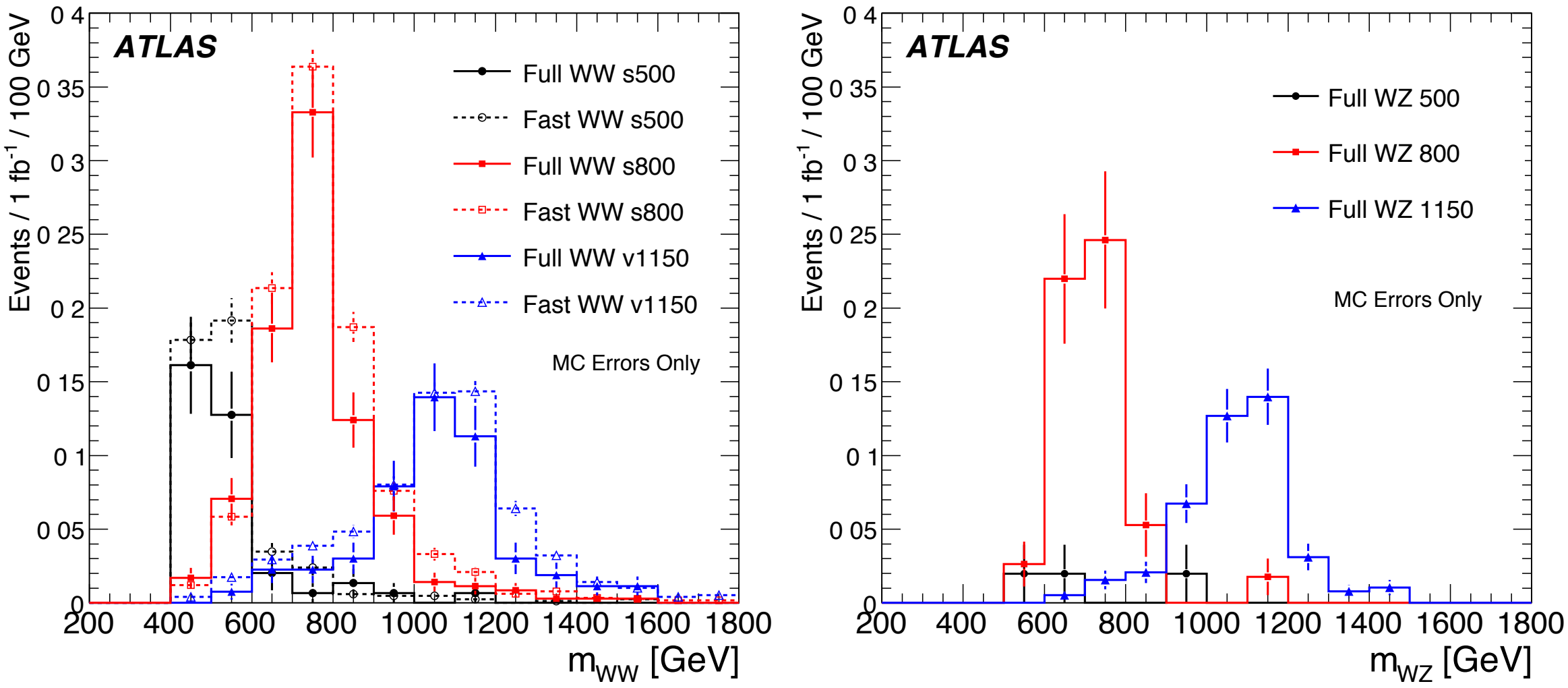


Figure 15: $WW$ (left) and $WZ$ (right) invariant mass spectra in the $\ell\nu j(j)$ semileptonic channel for the three resonant signal samples. The $WW$ plot shows a comparison of the fast- and full-simulation results.

For the $m = 1.1$ TeV $WZ$ resonance, only the case of a single heavy jet from the $W$ boson decay will be considered as it constitutes the majority of the events. For the $m = 800$ GeV resonance, not all $W$ bosons are boosted sufficiently to produce a single jet. We therefore consider separately the cases of a $W$ boson from a single heavy jet and from a jet pair. Finally, for the $m = 500$ GeV resonance, we only consider the jet pair case. In this section, the cone algorithm will be used and compared with an analysis using the $k_\perp$ jet algorithm.

### 5.2.1 $W$ boson from a single jet

The main backgrounds will here be $Z$+ 3 jets and $t\bar{t}$.

Table 7 shows the cut flow for the electron-based and the muon-based analyses for the ChL $WZ$ resonances of mass $m = 1.1$ TeV and $m = 800$ GeV. The $m = 500$ GeV case is not considered here since the $W$ and $Z$ bosons will not be sufficiently boosted, in general, to produce a single jet. The $Z \to e^+e^-$ and $Z \to \mu^+\mu^-$ selections are shown, which correspond each to about 50% of the sample events. After applying electron quality cuts (*medium* electrons) we select the two highest $p_T$ leptons which should satisfy respectively: $p_T(e,\mu) > 50$ GeV and $p_T(e,\mu) > 35$ GeV. The low efficiency of the lepton pair cut is approximately consistent with the expected selection efficiency per lepton, as shown in Fig. 11, as well as the detector acceptance. As can be seen in Fig. 17, the $p_T$ cut suppresses mostly the $t\bar{t}$ background. A leptonic $Z_{e^+e^-}$ or $Z_{\mu^+\mu^-}$ is afterwards reconstructed as described in Section 4.2.2, almost eliminating completely this background. The efficiency of this cut is somewhat poorer for the signal than for the background because most of the background events have $Z$ bosons of relatively low $p_T$, with different lepton pair energies and opening angle.

Using the cone algorithm, size 0.8, the hadronic $W$ boson candidate is identified as a heavy single jet having a mass between 70 and 100 GeV, and separated in azimuthal angle from the $Z$ boson candidate by $\Delta\phi(W,Z) > 2$, as described in Section 4.1 (see Fig. 18). At this stage, considering that the fraction of single jet $W$ bosons becomes important for $p_T > 250$ GeV (see Fig. 5), and in order to be consistent with the preselection cuts on the $Z$+3 jets background, we apply the following cuts to the reconstructed

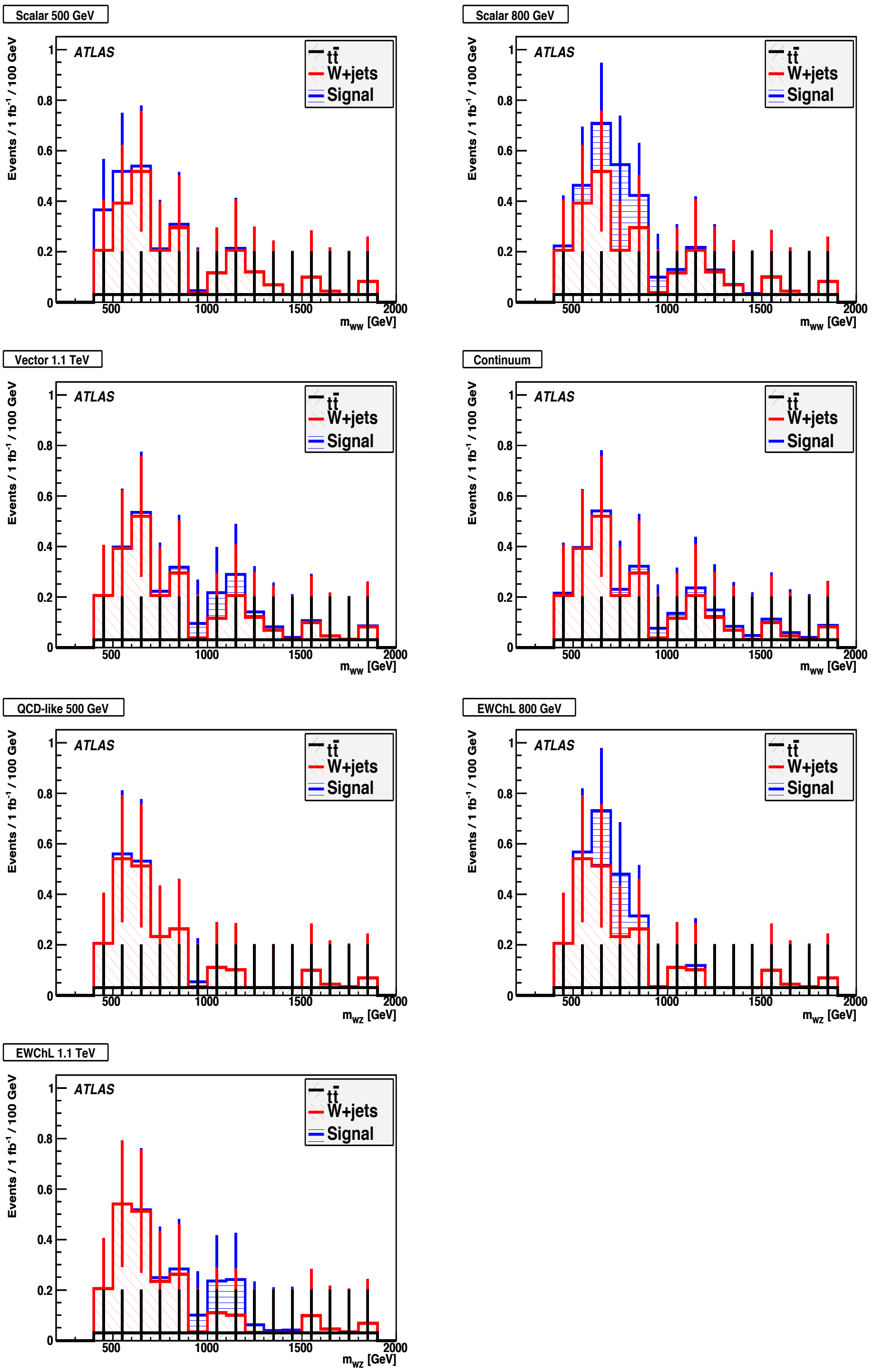


Figure 16: $WW$ (top 4) and $WZ$ (bottom 3) invariant mass spectra in the $\ell\nu\, j(j)$ semileptonic channel, showing the total $W$+jets and $t\bar{t}$ backgrounds and the signal for the three resonant signal samples and the continuum sample. The error bars reflect the uncertainty from the Monte Carlo statistics.

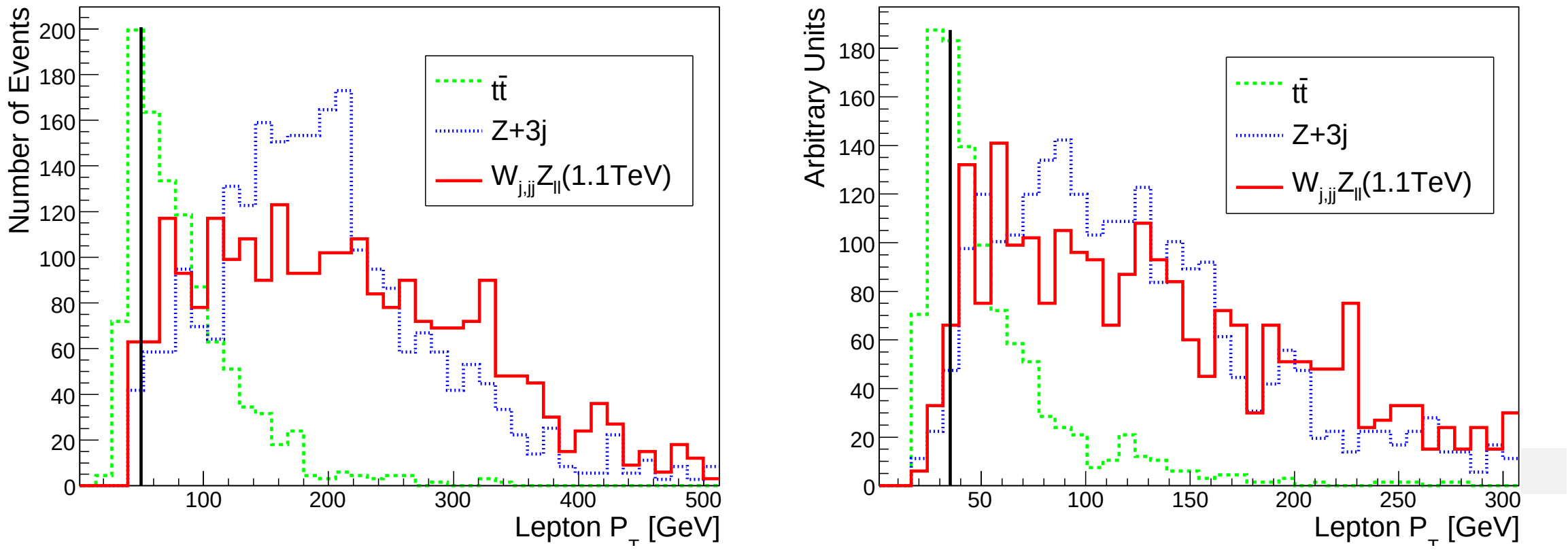


Figure 17: $p_T$ of the highest $p_T$ and second highest $p_T$ electrons from reconstructed $Z$ bosons in the $m =$ 1.1 TeV resonance sample. Distributions are arbitrarily normalised. The line indicates the cut value.

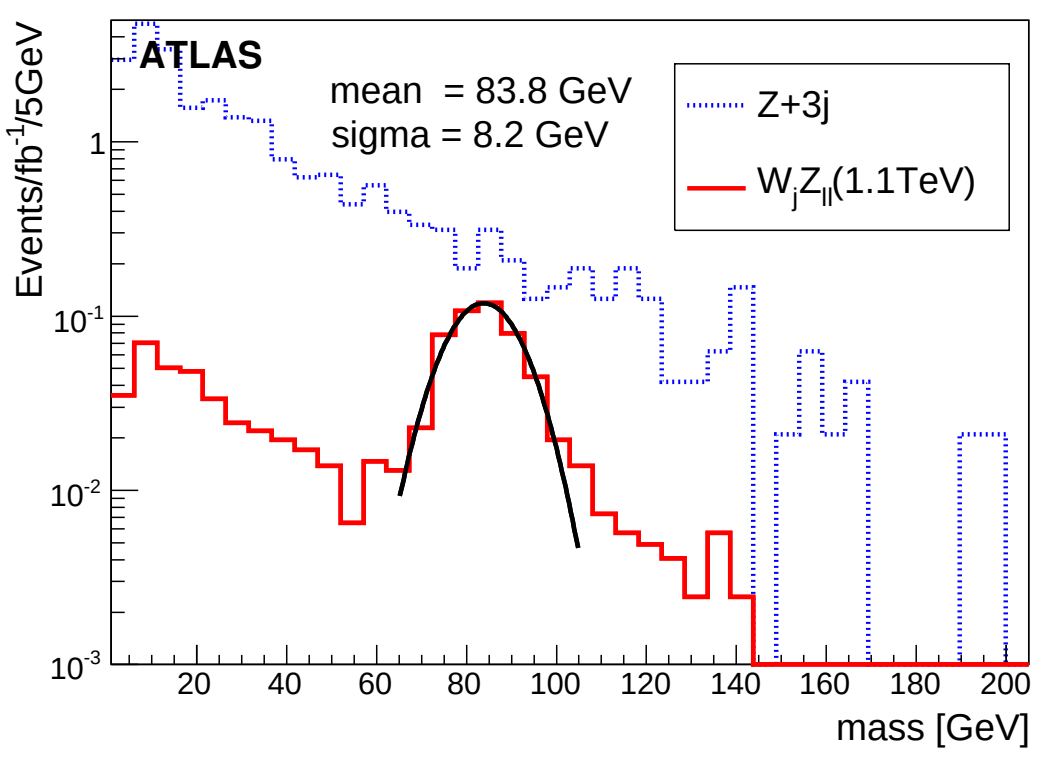


Figure 18: Mass of the heavy jet for the $m =$ 1.1 TeV ChL resonance and corresponding backgrounds. No $t\bar{t}$ event is left.

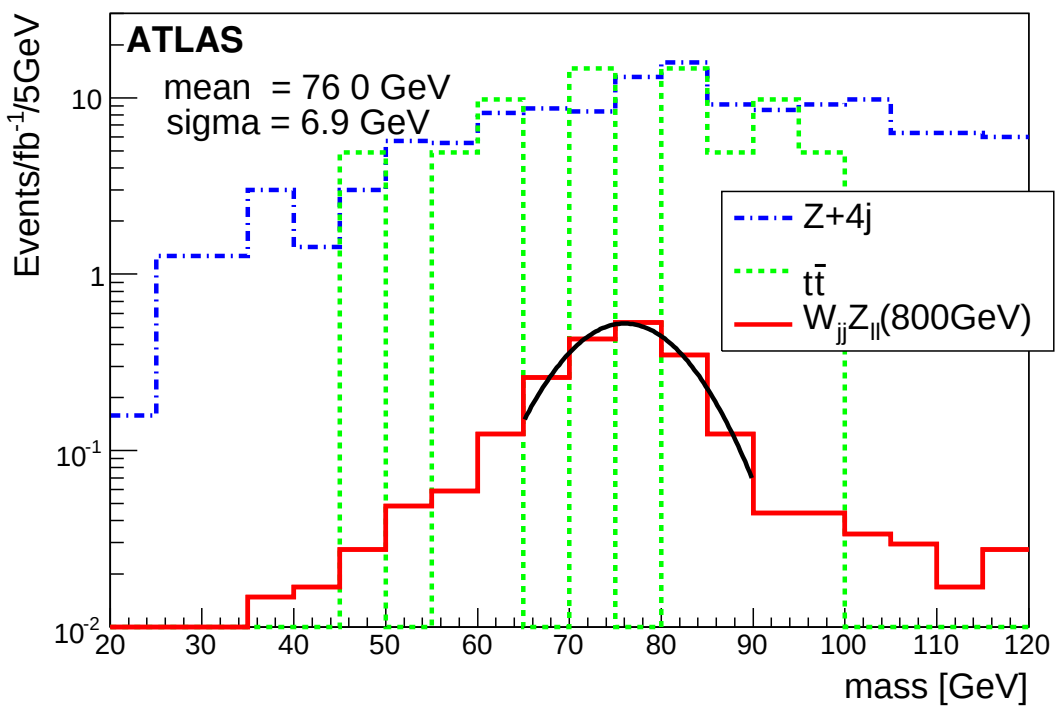


Figure 19: Reconstructed $W$ boson mass from a jet pair for the $m = 800$ GeV resonance.

$W$ and $Z$ bosons: $p_T^{W,Z} > 250$ GeV and $|\eta^{W,Z}| < 2.0$.

After a forward jet selection, (see Section 4.3, $p_{T\text{cut}} = 20$ GeV, $E_{\text{cut}} = E_{2\text{cut}} = 300$ GeV, $\eta_{\text{cut}} = 1.5$, $|\eta_{f\,jet}| > \eta_{\text{central jet}}$, $\Delta\eta_{\text{cut}} = 4.5$), the invariant mass of these two jets is required to be greater than 700 GeV. Note that the efficiency of the forward jet cuts appearing in Table 7 appears artificially good for the background because a preselection was already applied.

A central jet veto was found to be unnecessary, as no $t\bar{t}$ event survived the selection. Because of the lack of statistics for the $t\bar{t}$ sample, it is not possible to exclude completely a contribution from this background. The normalisation factor is 4.9, meaning that $t\bar{t}$ is excluded, over the whole mass range, at the level of 11.3 fb at 90% C.L. To have an estimate of the efficiency of the last two cuts at rejecting this background, the mass window for the cut on the $Z$ boson mass was loosened: $60 < m_Z < 120$ GeV, allowing 44 events (215 fb) to pass for the $Z \to ee$ channel and 38 events (185 fb) for the $Z \to \mu\mu$ channel. The $W$ boson mass cut alone is found to have an efficiency of 12% and the forward jet cut alone lets no event survive. Assuming that the cuts are independent, the overall efficiency of the heavy jet mass cut and forward jet tagging combined is higher than 0.15%. The exclusion limit at 95% C.L. (1.64 $\sigma$) for the $t\bar{t}$ background is shown in Table 7 and it will be assumed that this is negligible in the mass window of

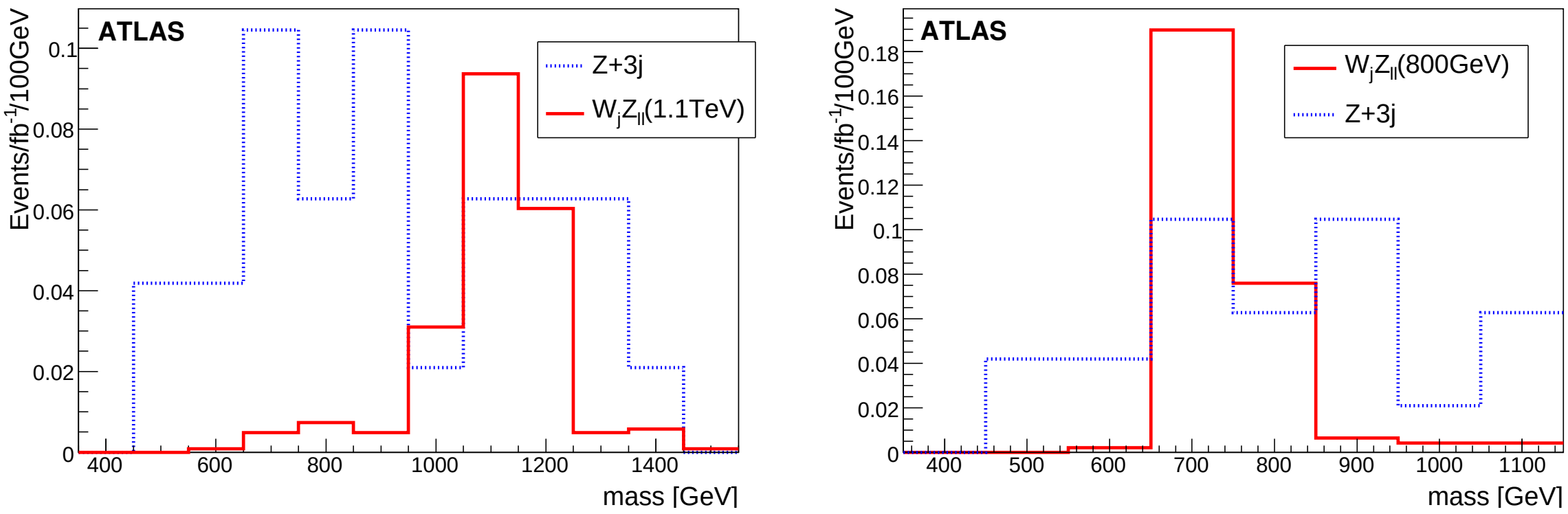


Figure 20: Reconstruction of ChL resonance at $m = 1.1$ TeV (left) and $m = 800$ GeV (right) in the channel $qqW_jZ_{\ell\ell}$ (with $\ell = e, \mu$), where a single jet cone 0.8 has been used to reconstruct the $W$.

the resonance. The $Z$+4 jets background was not included here because it may be double-counting with $Z$+3 jets with parton shower. In order to evaluate the level of this background, an average over the high mass region was taken because of the relatively poor Monte Carlo statistics, yielding about 0.03 fb/100 GeV.

For the $m = 1.1$ TeV case, it was found that the trigger efficiency, based on the OR of `e60, mu20` and `j160`, was 100% at the end of the selection.

Figure 20 shows the resonance mass resulting when the $Z$ boson has been reconstructed from electrons or muons and the $W$ boson from a single jet of size 0.8.

### 5.2.2 *W* boson from a jet pair

As above, after applying electron quality cuts, the lepton transverse momenta are required to satisfy $p_T(e_1, \mu_1) > 50$ GeV and $p_T(e_2, \mu_2) > 35$ GeV, and a $Z_{e^+e^-}$ ($Z_{\mu^+\mu^-}$) boson having a mass between 85 and 97 GeV (83 and 99 GeV) is then reconstructed. Considering all pairs of jets with $p_T > 30$ GeV in the central region ($|\eta| < 3.0$) not overlapping with the electron jets from the $Z$ decay, the one yielding an invariant mass closest to the mass of a $W$ boson will be the $W$ boson candidate (see Fig. 19). The low efficiency of this cut can be explained in part by the fact that a good fraction of events are constituted of a single jet $W$ boson. Forward and backward jet selection proceeds as in 5.2.1. A central jet veto is also applied: we exclude events with an extra jet, having a $p_T > 30$ GeV, not corresponding to the jets from the $W$ boson or the forward and backward jets and we require the $W$ and $Z$ directions to be in the central region $|\eta| < 2$. Figure 21 show the resulting reconstructed resonance masses. Table 7 summarizes the cut flow for this analysis. Here, by using the technique of widening the Z boson mass window as in Sect. 5.2.1, it is estimated that the $t\bar{t}$ background could be approximately 0.13 fb and it will be assumed that this is negligible in the mass window of the resonance.

### 5.2.3 Comparison to $k_\perp$ analysis

This channel was also studied with the analysis techniques described in Section 5.1, with the $k_\perp$ algorithm but using the leptonic $Z$ boson identification (Section 4.2.2) instead of the leptonic $W$ boson identification (Section 4.2.3). The hadronically-decaying $W$ boson is reconstructed dynamically from one or two jets (Section 4.1).

The signal mass distributions for fast and full simulation are shown in Fig. 22. The final mass distributions are shown in Fig. 23, and are comparable to the results from the cone algorithm method for

| | $m = \mathbf{1.1\ TeV}$ | | $m = \mathbf{800\ GeV}$ | | $m = \mathbf{500\ GeV}$ | | $Z+3j$ | | $Z+4j$ | | $t\bar{t}$ | |
|---|---|---|---|---|---|---|---|---|---|---|---|---|
| | $\sigma$ (fb) | eff. | $\sigma$ (fb) | eff. | $\sigma$ (fb) | eff. | $\sigma$ (fb) | eff. | $\sigma$ (fb) | eff. | $\sigma$ (fb) | eff. |
| $Z \to e^+e^-$ | | | | | | | | | | | | |
| $p_T(e1) > 50$ GeV, $p_T(e2) > 35$ GeV | 0.79 | 22% | 2.14 | 20% | 4.15 | 16% | 22.2 | 20% | 195 | 7.9% | 1055 | 29% |
| 85 GeV $< m_Z <$ 97 GeV | 0.63 | 80% | 1.69 | 79% | 3.34 | 80% | 18.5 | 87% | 176 | 90% | 39 | 3.7% |
| $Z \to \mu^+\mu^-$ | | | | | | | | | | | | |
| $p_T(\mu_1) > 50$ GeV, $p_T(\mu_2) > 35$ GeV | 0.60 | 16% | 1.67 | 16% | 3.11 | 12.4% | 17.2 | 16% | 170 | 7.0% | 821 | 22% |
| 83 GeV $< m_Z <$ 99 GeV | 0.48 | 81% | 1.40 | 84% | 2.68 | 86% | 15.8 | 90% | 163 | 95% | 64 | 7.8% |
| $Z \to e^+e^-$ **and** $Z \to \mu^+\mu^-$ | | | | | | | | | | | | |
| $W_jZ_{ll}$ | | | | | | | | | | | | |
| Heavy jet mass $W \to j$ | 0.57 | 51% | 0.75 | 24% | – | – | 2.99 | 8.7% | – | – | 0 | 0% |
| Forward jet tagging | 0.22 | 39% | 0.29 | 39% | – | – | 0.67 | 22% | – | – | $< 0.25$ | – |
| $W_{jj}Z_{ll}$ | | | | | | | | | | | | |
| 65 GeV $< m_{jj} <$ 90 GeV and $\Delta\phi(W_{jj}, Z) > 2.0$ | – | – | 1.51 | 25% | 2.21 | 37% | – | – | 37.6 | 11% | 9.8 | 9.5% |
| Forward jet tagging | – | – | 0.62 | 41% | 0.68 | 31% | – | – | 9.57 | 25% | – | – |
| Central jet veto | – | – | 0.29 | 47% | 0.32 | 47% | – | – | 4.85 | 51% | – | – |

Table 7: Cut flow for the $W_{jj}Z_{\ell\ell}$, $m = 1.1$ TeV, 800 and 500 GeV signals. For each process, the cross section (fb) surviving the successive application of the cuts is shown, as well as the efficiency of each cut. The upper limit for $t\bar{t}$ in the last lines is for 95% C.L.

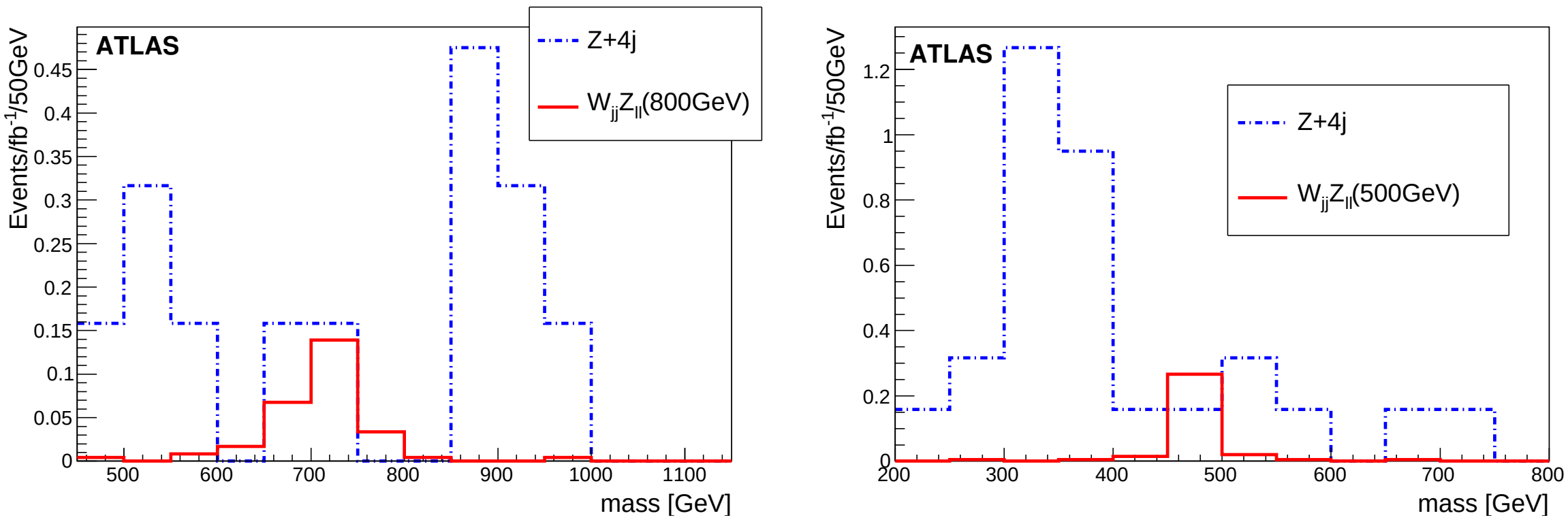


Figure 21: Reconstructed ChL resonance at $m = 800$ GeV (left) and $m = 500$ GeV (right) in the channel $qqW_{jj}Z_{\ell\ell}$ (with $\ell = e,\mu$) where two jets of cone size 0.4 have been used to reconstruct the $W$ boson. No $t\bar{t}$ events survive the selection.

| | $W_{\ell\nu}Z_{\ell\ell}$ ($m =$ **500 GeV**) | | $W_{\ell\nu}Z_{\ell\ell}$ ($m =$ **1.1 TeV**) | | $W_{\ell\nu}Z_{\ell\ell}jj(SM)$ | |
|---|---|---|---|---|---|---|
| | $\sigma$ (fb) | eff. | $\sigma$ (fb) | eff. | $\sigma$ (fb) | eff. |
| $Z_{ee}$ | 1.47 | 18% | 0.23 | 20% | 20.7 | 16% |
| $Z_{\mu\mu}$ | 1.09 | 14% | 0.18 | 15% | 16.7 | 13% |
| $W$ reconstruction | 1.43 | 56% | 0.25 | 61% | 18.9 | 51% |
| Forward jet tagging | 0.63 | 44% | 0.14 | 56% | 1.6 | 8.5% |

Table 8: Cut flow for the $W_{\ell\nu}Z_{\ell\ell}$ ($m = 500$ GeV and 1.1 TeV) signals. All the cuts are described in detail in this section.

the 1.1 TeV case which is not sensitive to the cut on the $p_T$ of the VB's of 200 GeV.

### 5.3 $W^{\pm}Z \to \ell^{\pm}\nu\ \ell^{+}\ell^{-}$

This purely leptonic channel consists of four different signatures: $W_{\ell^{\pm}\nu}Z_{\ell^{\pm}\ell^{\mp}}$ with $\ell = e,\mu$. The main background will be $WZjj$ production from the Standard Model. The analysis starts by identifying leptonic $Z_{e^+e^-}$ ($Z_{\mu^+\mu^-}$) bosons as described in Section 4.2.2, after requiring two leptons with $p_T$ greater than 50 and 35 GeV.

As a second step, we proceed to reconstruct the $W$ boson from the highest $p_T$ lepton among those remaining in the event, if there is one, and the measured missing transverse energy, as described in Section 4.2.3. The solution which yields the highest $p_T$ $W$ boson is kept.

The forward and backward jet selection follows the prescription of the Section 4.3 ($p_{T\text{cut}} = 20$ GeV, $E_{\text{cut}} = E_{2\text{cut}} = 300$ GeV, $\eta_{\text{cut}} = 1.5$, $|\eta_{fjet}| > \eta_{\text{central jet}}$), $\Delta\eta_{\text{cut}} = 4.5$).

In Table 8 we present the cut flow of the reconstruction of the resonances for 1.1 TeV and 500 GeV. Also in Fig. 24 and Fig. 25 we present the reconstructed resonance and the background $WZjj$ for the same resonance mass.

### 5.4 $ZZ \to \nu\nu\ \ell^{+}\ell^{-}$

This scalar resonance can be interpreted as a Standard Model Higgs boson produced by vector boson fusion. At leading order, the cross section times branching ratio would be 6 fb, compared to 4 fb obtained for the ChL model. This signal is characterised by a leptonic $Z$ boson accompanied by large $\not{E}_T$, yielding a large transverse mass. The backgrounds considered are: $ZZjj \to \ell\ell\nu\nu jj$ and $WZjj \to \ell\nu\ell\ell jj$. Other

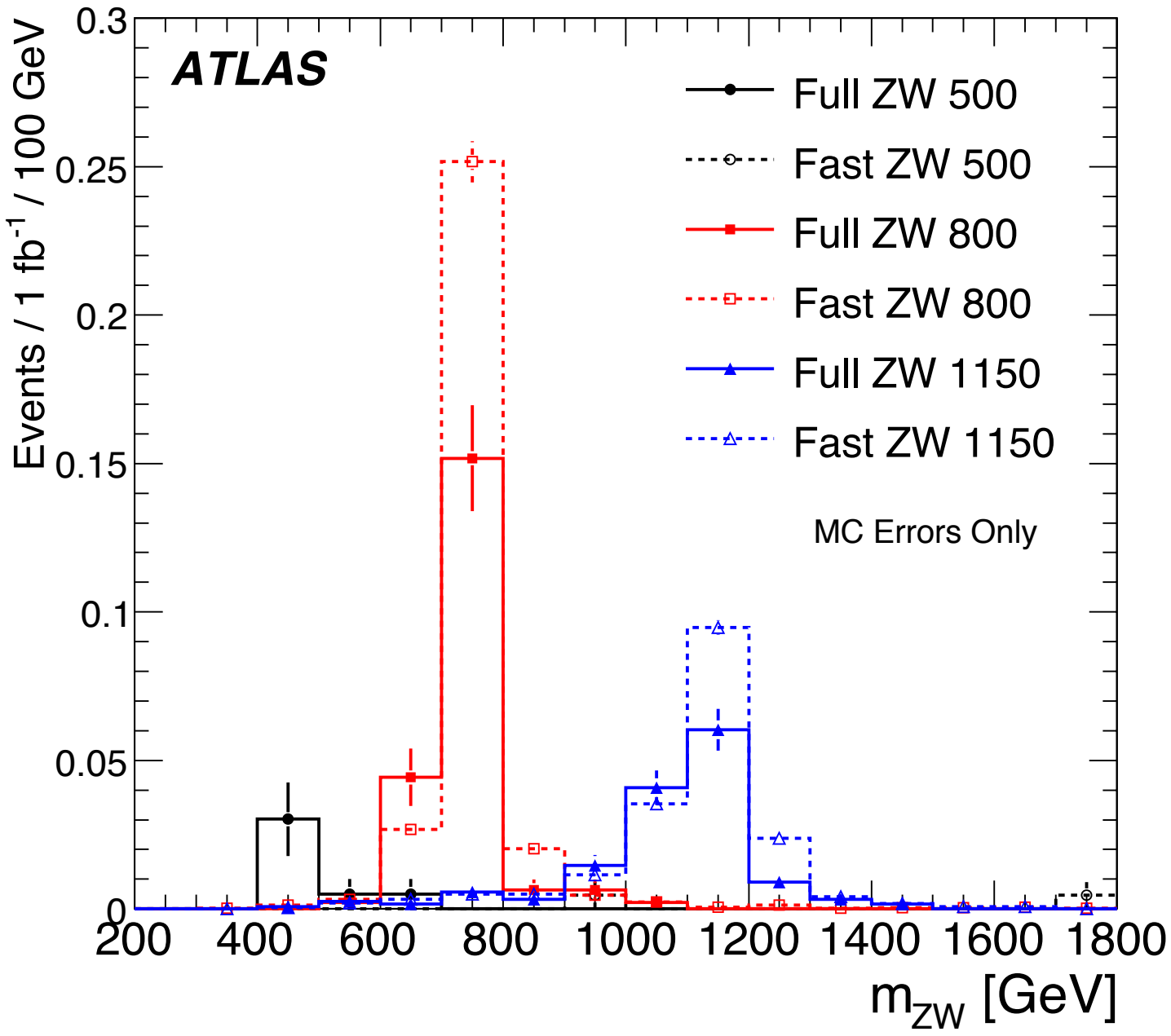


Figure 22: $WZ$ invariant mass spectrum in the $\ell^+\ell^- j(j)$ channel for the three resonant signal samples obtained using $k_\perp$ algorithm approach. Dotted lines indicate the fast simulation results.

| | $Z_{\nu\nu}Z_{\ell\ell}qq$ ($m = $ **500GeV**) | | $W_{\ell\nu}Z_{\ell\ell}jj$ **(SM)** | | $Z_{\nu\nu}Z_{\ell\ell}jj$ **(SM)** | |
|---|---|---|---|---|---|---|
| | $\sigma$ (fb) | eff. | $\sigma$ (fb) | eff. | $\sigma$ (fb) | eff. |
| $Z_{ee}$ | 0.72 | 17.6% | 20.78 | 22% | 9.1 | 20% |
| $Z_{\mu\mu}$ | 0.58 | 15% | 16.7 | 17% | 6.6 | 15% |
| Forward jet tagging | 0.58 | 45% | 3.2 | 8.6% | 0.47 | 3% |
| $\not{E}_T > 150$ GeV | 0.44 | 75% | 0.46 | 14% | 0.12 | 26% |

Table 9: Cut flow for the $Z_{\nu\nu}Z_{\ell\ell}qq$ ($m = 500$ GeV) signal. All the cuts are described in detail in this section.

background can result from $Z$+jets production, where the tail of the missing transverse energy distribution can fake a signal.

After selecting the leptonically decaying $Z_{e^+e^-}$ ($Z_{\mu^+\mu^-}$) boson as usual, with mass between 85 and 97 GeV (83 and 99 GeV), a minimum $\not{E}_T$ of 150 GeV is required. For this high value of $\not{E}_T$, $Z$+jets background is expected to be negligible for a Standard Model Higgs boson signal [39]. The forward jet selection is applied (see Section 4.3, $p_{T\text{cut}} = 20$ GeV, $E_{\text{cut}} = E_{2\text{cut}} = 300$ GeV, $\eta_{\text{cut}} = 1.5$, $\Delta\eta_{\text{cut}} = 4.5$).

The transverse mass, defined as:

$$m_T^2 = (\sqrt{p_T(Z)^2 + m_Z^2} + \not{E}_T)^2 - (\vec{p}_T(Z) + \vec{\not{p}}_T)^2 \tag{1}$$

is shown in Fig. 26 and the cut flow can be found in Table 9.

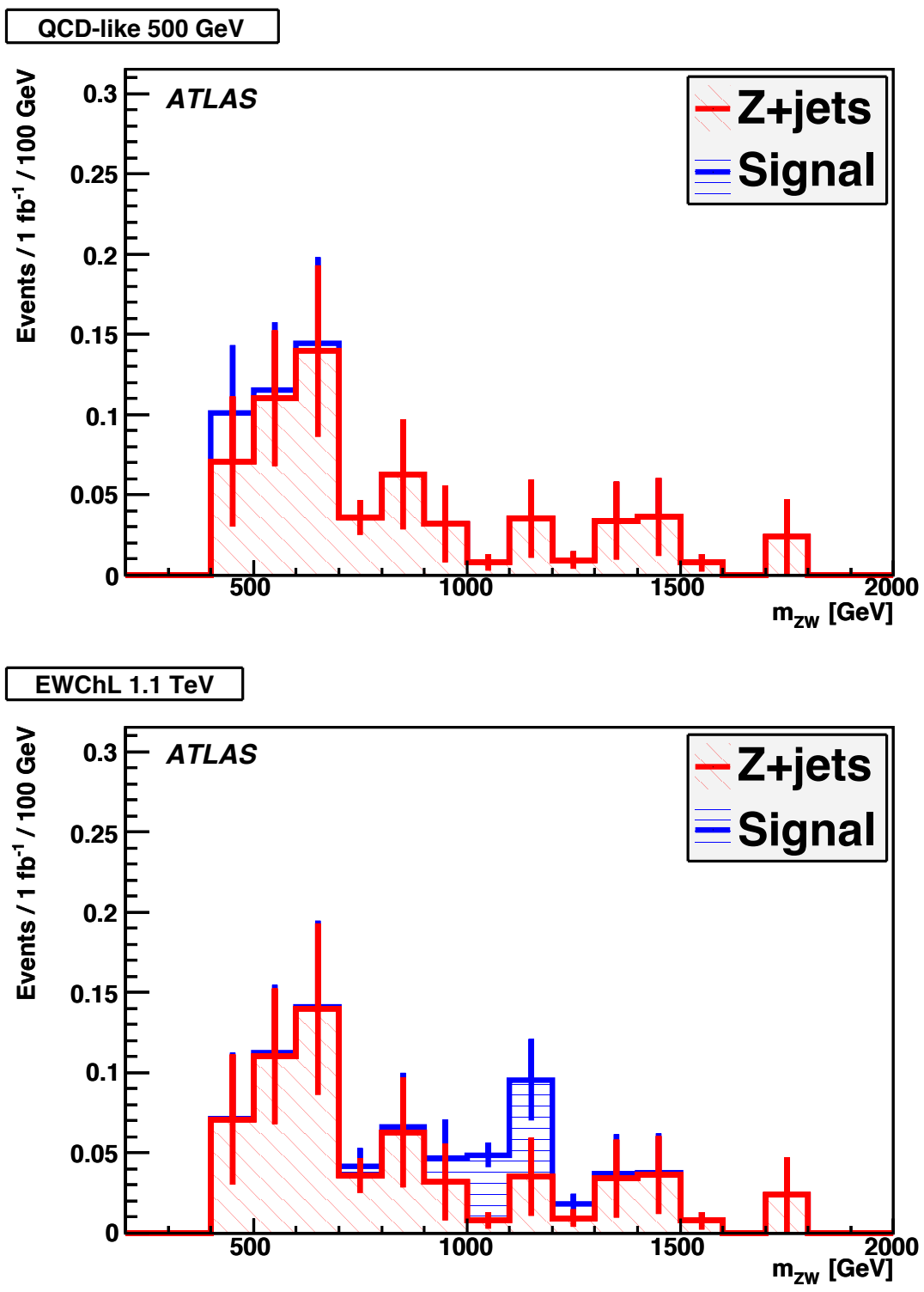


Figure 23: $WZ$ invariant mass spectrum in the $\ell^+\ell^- j(j)$ channel for the three resonant signal samples obtained using $k_\perp$ algorithm approach. $Z$+jet histogram (in red) represents a direct sum of all $Z$+3 or 4 jets backgrounds with no matching, and hence is a conservative estimate. The background from $t\bar{t}$ events has been found to be negligible.

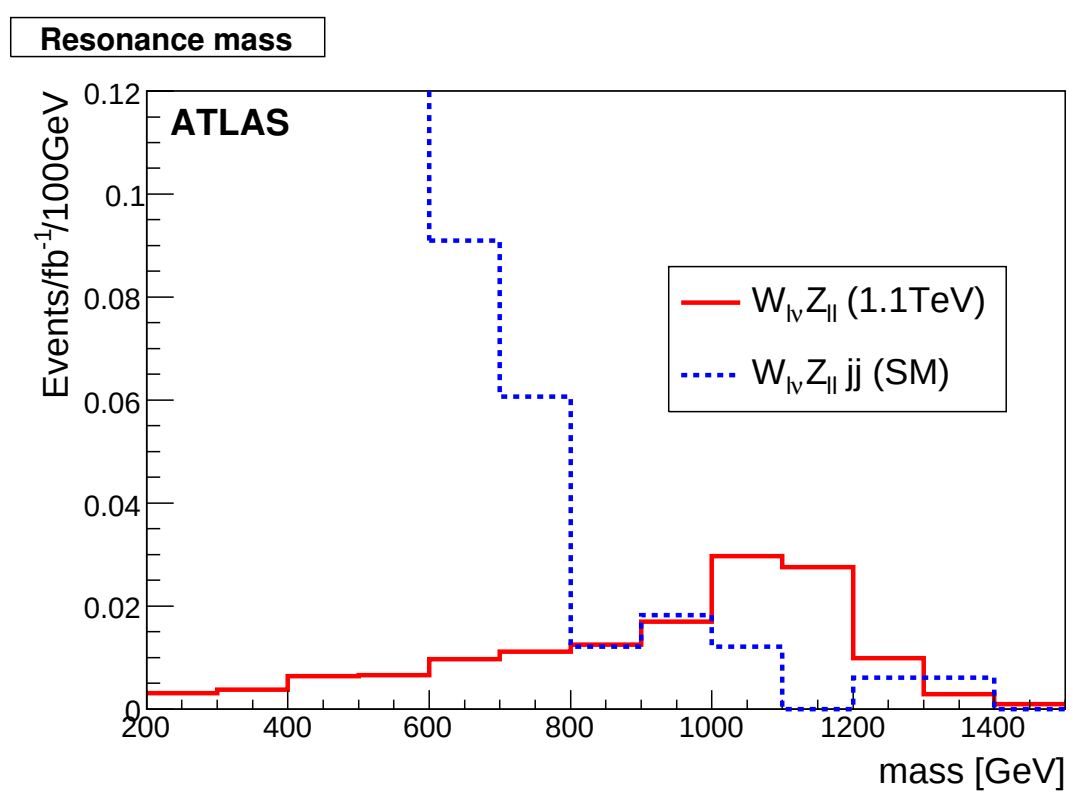


Figure 24: Full reconstruction of ChL resonance $m \sim 1.1$ TeV ($W_{\ell^\pm\nu} Z_{\ell^\pm\ell^\mp}$).

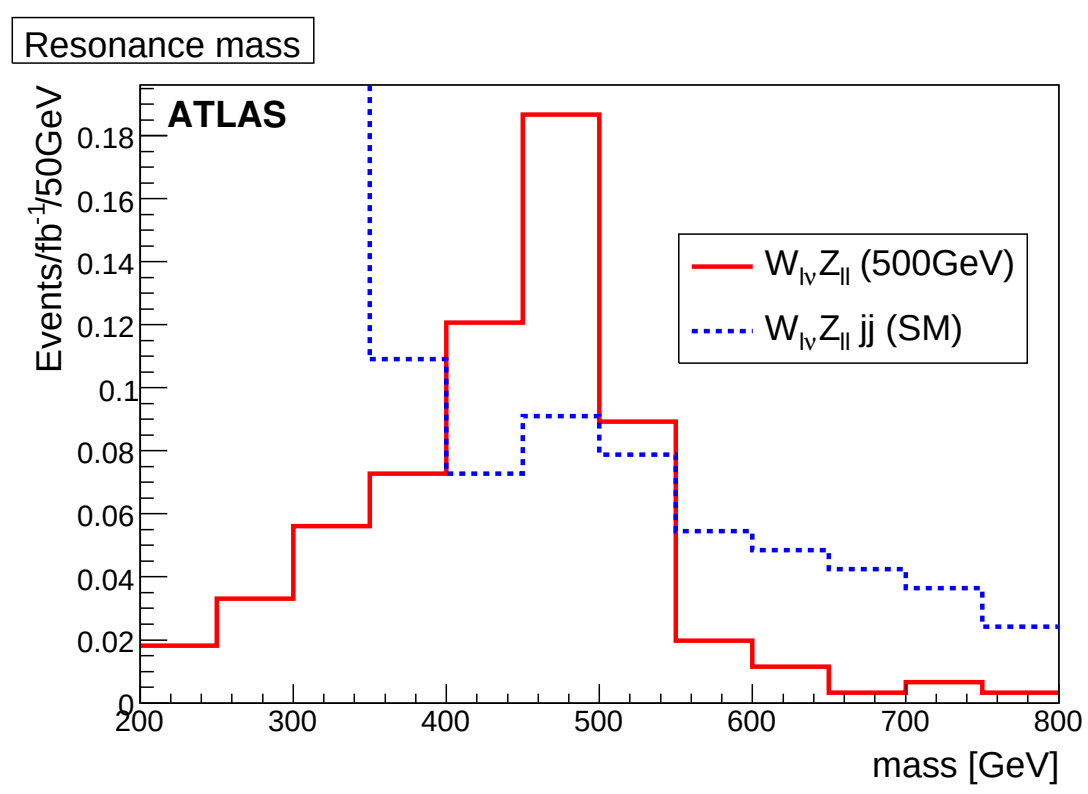


Figure 25: Full reconstruction of QCD-like resonance $m \sim 500$ GeV ($W_{\ell^\pm\nu} Z_{\ell^\pm\ell^\mp}$).

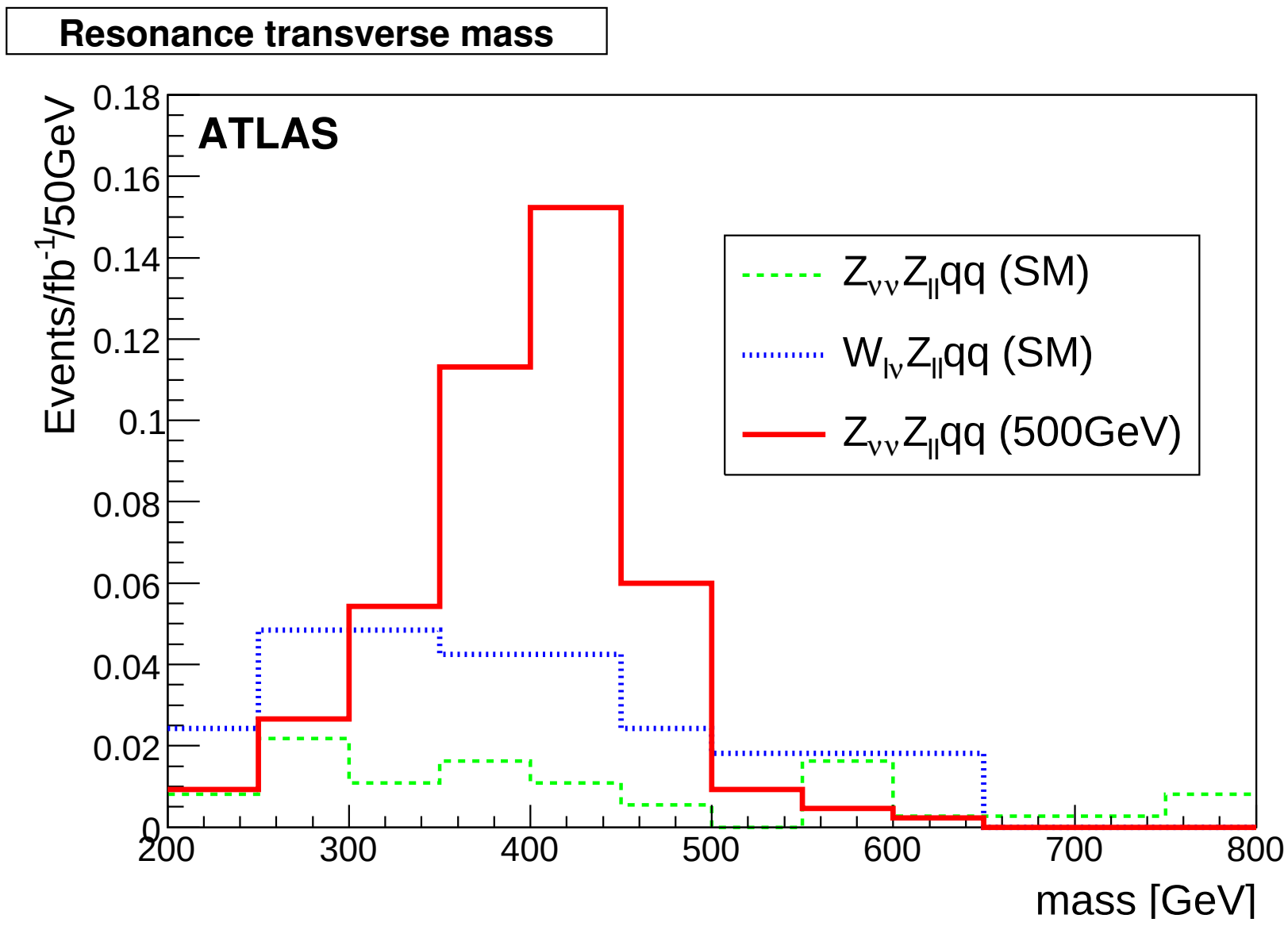


Figure 26: Transverse mass of the $m = 500$ GeV resonance $Z_{\nu\nu}Z_{\ell\ell}$.

## 6 Results

The significance of the signals and the luminosity required for a possible discovery is estimated here. From the reconstructed resonance mass distributions in Section 5 one can evaluate the size of the signal and background in the resonance mass window. Table 10 summarises the approximate cross sections expected after the analyses described above. The table also gives the luminosity required to observe a significant excess over the background, showing the uncertainty from MC statistics only, and the significance of a signal for an integrated luminosity of 100 fb$^{-1}$. Because of the large statistical and systematic uncertainties (see Section 7), the numbers given here must be taken as an only an approximate indication of the reach of the LHC for such resonances.

The significance is calculated as

$$\text{significance} = \sqrt{2((S+B)\ln(1+S/B)-S)} \quad , \tag{2}$$

where S (B) is the number of expected signal (background) events in the signal *peak* region, which is defined as the three consecutive bins (of size given in the figures, chosen to represent the resolution), with the highest total number of signal events. The background is averaged over this region.

In the $WW$ case, only the semileptonic channel is accessible. Thus, as shown in Table 10, around 25 fb$^{-1}$ is needed to start seeing indications of a resonance even in the most optimistic case studied here, and around 70 fb$^{-1}$ is needed for a discovery.

In the continuum case, the “signal” is spread over an extended mass region with a total of about 0.3 events expected for each fb$^{-1}$ in the mass range $m = 400$–$1900$ GeV, compared to about 2.5 background events. Measuring this cross-section with any accuracy using the techniques developed here would require an integrated luminosity of several hundred fb$^{-1}$.

Since the mass windows for hadronic $W$ and $Z$ boson decays overlap, in practice these scenarios can probably not be distinguished in this channel, and a combined analysis would in reality have to be performed, which would then be compared to those channels containing a leptonic $Z$ boson decay.

For each of the $WZ$ resonances, results of the different channels, $W_{\ell\nu}Z_{\ell\ell}$, $W_{jj}Z_{\ell\ell}$ and $W_{\ell\nu}Z_{jj}$ can, in principle, be combined. From Table 10, one can conclude that for two of three mass regions, $m =$

| Process | Cross section (fb) | | Luminosity ($fb^{-1}$) | | Significance |
|---|---|---|---|---|---|
| | signal | background | for $3\sigma$ | for $5\sigma$ | for 100 $fb^{-1}$ |
| $WW/WZ \to \ell\nu\, jj$, $m = 500$ GeV | $0.31 \pm 0.05$ | $0.79 \pm 0.26$ | 85 | 235 | $3.3 \pm 0.7$ |
| $WW/WZ \to \ell\nu\, jj$, $m = 800$ GeV | $0.65 \pm 0.04$ | $0.87 \pm 0.28$ | 20 | 60 | $6.3 \pm 0.9$ |
| $WW/WZ \to \ell\nu\, jj$, $m = 1.1$ TeV | $0.24 \pm 0.03$ | $0.46 \pm 0.25$ | 85 | 230 | $3.3 \pm 0.8$ |
| $W_{jj}Z_{\ell\ell}$, $m = 500$ GeV | $0.28 \pm 0.04$ | $0.20 \pm 0.18$ | 30 | 90 | $5.3 \pm 1.9$ |
| $W_{\ell\nu}Z_{\ell\ell}$, $m = 500$ GeV | $0.40 \pm 0.03$ | $0.25 \pm 0.03$ | 20 | 55 | $6.6 \pm 0.5$ |
| $W_{jj}Z_{\ell\ell}$, $m = 800$ GeV | $0.24 \pm 0.02$ | $0.30 \pm 0.22$ | 60 | 160 | $3.9 \pm 1.2$ |
| $W_{j}Z_{\ell\ell}$, $m = 800$ GeV | $0.27 \pm 0.02 \pm 0.05$ | $0.23 \pm 0.07 \pm 0.05$ | 38 | 105 | $4.9 \pm 1.1$ |
| $W_{j}Z_{\ell\ell}$, $m = 1.1$ TeV | $0.19 \pm 0.01 \pm 0.04$ | $0.22 \pm 0.07 \pm 0.05$ | 68 | 191 | $3.6 \pm 1.0$ |
| $W_{\ell\nu}Z_{\ell\ell}$, $m = 1.1$ TeV | $0.070 \pm 0.004$ | $0.020 \pm 0.009$ | 70 | 200 | $3.6 \pm 0.5$ |
| $Z_{\nu\nu}Z_{\ell\ell}$, $m = 500$ GeV | $0.32 \pm 0.02$ | $0.15 \pm 0.03$ | 20 | 60 | $6.6 \pm 0.6$ |

Table 10: Approximate signal and background cross sections expected after the analyses. An approximate value of the luminosity required for $3\sigma$ and $5\sigma$ significance, and the expected significance for 100 $fb^{-1}$ are shown. The uncertainties, when given, are due to Monte Carlo statistics only.

500 GeV and 800 GeV, a chiral Lagrangian vector resonance can be discovered with less than 100 $fb^{-1}$. The expectations with the alternative $k_\perp$ analysis described in Section 5.2.3 are not far from the values in Table 10. As an example, the integrated luminosity needed for $3\sigma$ observation of the $m = 800$ GeV signal is 63 $fb^{-1}$, and of the $m = 1.1$ TeV signal is 81 $fb^{-1}$.

A scalar resonance at $m = 500$ GeV will require about 60 $fb^{-1}$ to be seen in the $ZZ \to \nu\nu\ell\ell$ channel.

# 7 Systematic Uncertainties

A number of large systematic uncertainties affect the signals studied here. Because of the small cross sections and the important backgrounds, it is difficult to estimate them with precision from Monte Carlo simulations. Data driven tests will be required to understand better the systematic effects. Some discussion of the most significant effects is given here.

## 7.1 Background Cross sections

As was discussed in Section 2, the renormalisation and factorisation scales, $Q^2$, can affect the cross section by as much as a factor of two. This is especially true at high centre of mass energies, where the degree of virtuality of partons and choice of scale for $\alpha_s$ are quite critical [28, 29]. At present this represents a theoretical uncertainty on the current sensitivity estimate. While the predictions may improve in future, in an eventual analysis, the backgrounds would have to be measured from data and the eventual size of the associated systematic uncertainty has not been studied here.

Another consideration is that for the analyses which dynamically move between the dijet and single jet reconstruction technique for the hadronically decaying vector boson (see Section 4.1), to evaluate the background with the samples available both $W$+3 jet and $W$+4 jet samples must be used. This implies some double-counting due to the lack of parton-shower matching in these samples, and so the background will be overestimated. This is in addition to the fact that as shown in Section 2.4.1, the MADGRAPH samples used overestimate slightly the energy of the tag jets. The effect is expected to be at the few per cent level.

## 7.2 Signal Cross sections

The PYTHIA signal generation produces softer tag jets than the more exact WHIZARD MC, and thus the signal efficiencies are likely to be underestimated.

## 7.3 Monte Carlo Statistics

We are limited by the very large size of the background samples required. Fast MC simulation was shown to be in good agreement with full simulation and was used to evaluate $t\bar{t}$ background for the $WW$ signals. In some cases, only upper limits on the backgrounds can be given, although it is expected that these limits are very conservative. Again, this represents a systematic uncertainty on the current sensitivity estimates, but will not be present in a final data analysis, assuming sufficient simulated data will eventually be available.

## 7.4 Pile-up and Underlying Event

Pile-up and underlying event are separate effects which have potentially similar and crucial impact on the efficiency of the forward jet cuts, the central jet veto and the top veto, as well as on the jet mass resolution.

Of particular concern is the fact that the top veto in the $WW$ analysis uses jets down to $p_T = 10\,\text{GeV}$, expected to be strongly affected by these effects [40]. Simply raising the cut to 20 GeV admits significantly larger background. Some of this can be removed for the higher mass resonances by raising the $p_T$ cut on the vector boson. However, a more promising approach is likely to be to exploit $b$-tagging and improved jet mass reconstruction to improve the veto.

### 7.4.1 Pile-up

Fully simulated samples with pile-up at low luminosity ($10^{33}\text{cm}^{-2}\text{s}^{-1}$) were available, but with much lower statistics: we restricted the analysis to one signal sample and one background sample. However pile-up effects should be approximately independent of the underlying physics sample, and we assume we can safely generalise the results obtained here.

We compared the same events of the $W_{jj}Z_{\ell\ell}$ at $m = 1.1\,\text{TeV}$ sample reconstructed with and without pile-up simulation. This allows computation of the fraction of events with pile-up having tagged forward jets with respect to corresponding non-pile-up events which fail the tagged jet criterion, thus defining a ‘fake’ rate. The reciprocal fraction defines a ‘miss’ rate. The effect of pile-up increases with increasing jet radius and decreasing energy threshold, as would be expected. We found that both ‘fake’ and ‘miss’ effects are essentially due to the degradation of energy resolution in presence of pile-up and that their combination contributes to an uncertainty on the efficiency of the order of 5%.

### 7.4.2 Underlying Event

Current simulations use underlying event models tuned to Tevatron and other data [22], but there is a large extrapolation needed to 14 TeV. The underlying event would be have to be measured in LHC data, and its level is not currently known.

## 7.5 Other Systematic Effects

Systematic effects, such as uncertainties in the luminosity, in efficiencies and resolutions, jet energy scale, etc. are of the order of a few percent and will therefore be completely dominated by the above effects and by statistical uncertainties.

## 8 Summary and Conclusion

The Chiral Lagrangian model with Padé unitarisation provides a framework for studying vector boson scattering at high mass, in case a light Higgs boson is not found at the LHC in the first years of running. With full detector simulation, the search for vector and scalar resonances of masses $m = 500$, 800 and 1100 GeV is studied. To suppress the very high backgrounds from $W$+jets and $Z$+jets to acceptable levels requires special techniques investigated here. In particular, at these high masses, hadronic vector boson decay results in a single jet. The reconstruction of the jet mass is found to be generally quite efficient at rejecting QCD jets. The $k_\perp$ and the cone algorithms can be applied to this heavy jet to resolve it into two light jets, suppressing further the background. Other conventional techniques for the study of vector boson fusion are also found essential for the present analysis: forward jet tagging, central jet veto and top-jet veto.

The cut-based analysis presented here is performed with realistic simulation and reconstruction of leptons and jets. Improvements can be expected by more sophisticated analysis and, with real data and a good understanding of the detector, further gains can be achieved by improvements in the reconstruction efficiencies.

The discovery of resonances in vector boson scattering at high mass will take a few tens of fb$^{-1}$, but the different decay channels of the vector boson pairs allow a cross-check of the presence of a resonance. These results can be considered generic of vector boson scattering and can therefore be interpreted in terms of other theoretical models with possibly different cross-sections.

# Discovery Reach for Black Hole Production

**Abstract**

Models with extra space dimensions, in which our Universe exists on a 4-dimensional brane embedded in a higher dimensional bulk space-time, offer a new way to address outstanding problems in and beyond the Standard Model. In such models the Planck scale in the bulk can be of the order of the electroweak symmetry breaking scale. This allows the coupling strength of gravity to increase to a size similar to the other interactions, opening the way to the unification of gravity and the gauge interactions. The increased strength of gravity in the bulk space-time means that quantum gravity effects would be observable in the TeV energy range reachable by the LHC. The most spectacular phenomenon would be the production of black holes, which would decay semi-classically by Hawking radiation emitting high energy particles. In this note, we discuss the potential for the ATLAS experiment to discover such black holes in the early data (1–1000 $\text{pb}^{-1}$)

## 1 Introduction

In this study we simulate the search for black holes in the first $100\,\text{pb}^{-1}$ of LHC data with the ATLAS detector and software framework.

The document's structure is as follows: Section 2 gives an overview of the extra dimension models, present limits on the size of the extra dimensions and a discussion of black hole production and decay. In Section 3 the Monte Carlo simulation samples are described. Section 4 presents basic event properties, and is followed by Sections 5 and 6 dealing with the triggering and analysis selection, respectively. The expected systematic uncertainties are given in Section 7. Finally, the extraction of model parameters, especially of black hole properties, is covered in Section 8. A summary is given in Section 9.

## 2 Theory

### 2.1 Theoretical Motivation

The electroweak energy scale and the Planck scale, at which gravitational interactions become strong, differ by about sixteen orders of magnitude. This large difference between the scales of the two fundamental interactions is known as the hierarchy problem. Explaining the hierarchy problem is one of the outstanding challenges in particle physics.

Arkani-Hamed, Dimopoulos and Dvali (ADD) [1–3], and Randall and Sundrum (RS) [4, 5] have pioneered approaches to solving the hierarchy problem by using extra-dimensional space. The hierarchy is generated by the geometry of the additional spatial dimensions. ADD models postulate additional flat extra dimensions, while RS models invoke a single warped extra dimension. The observed weakness of gravity is thus due to the gravitational field being allowed to expand into the higher-dimensional space (bulk), while the Standard Model particles are confined to our familiar three-dimensional space (3-brane). Extra-dimensional models can also be motivated by string theory.

In extra-dimensional models, the $D$-dimensional Planck scale $M_D$ is the fundamental scale from which the Planck scale $M_{\text{Pl}} = 1.22 \times 10^{19}$ GeV in four dimensions is derived[1]. The relationship between

[1]Several conventions exist for the $D$-dimensional Planck scale in the ADD model. We denote by $M_D$ the parameter defined by Giudice, Rattazzi and Wells [6] and used by the PDG [7]: $M_D^{D-2} = (2\pi)^{D-4}/(8\pi G_D)$, where $G_D$ is the $D$-dimensional Newton gravity constant. In an alternative convention given by Dimopoulos and Landsberg [8] the $D$-dimensional Planck scale

the two scales is determined by the volume of the extra dimensions in ADD models or by the warp factor in RS models. For large extra dimensions or a strongly warped extra dimension, the fundamental scale of gravity can be as low as the electroweak scale. If the Planck scale is low enough, black holes could be produced at the Large Hadron Collider (LHC) [9, 10]. Detecting them will not only test general relativity and probe extra dimensions, but would also teach us about quantum gravity.

## 2.2 Experimental Limits

Assuming that low-scale gravity is due to the existence of extra dimensions[2], most experimental searches for unusual low-scale gravity effects have focused on detecting evidence for extra dimensions. Current experimental limits allow the fundamental scale of gravity to be as low as about 1 TeV. In testing the ADD models and deriving limits in these models, the compactification radius of all extra dimensions is assumed to be the same. ADD models have been tested at length scales comparable to the radius of the compactified (i.e. curled-up) dimensions $R$. Were the effective number of large extra dimensions to be $n = D-4$, the inverse-square law would smoothly change from the $1/r^2$ form for $r \gg R$ to a $1/r^{2+n}$ form for $r \ll R$. Searches have been performed and constraints on the Planck scale have been set by tabletop and particle accelerator experiments, astrophysical observations, cosmic-ray measurements and cosmological considerations. Direct searches for black holes at collider experiments have not yet been performed. The only direct limits on black hole production in high energy interactions were obtained using cosmic-ray data. Table top experiments lead to an upper bound of $R \leq 44$ µm, at the 95% confidence level [12]. The LEP bounds obtained with direct searches vary from 1.5 TeV for $n = 2$ extra dimensions to 0.75 TeV for 5 extra dimensions [13]. The latest direct search result from the CDF collaboration has set lower bounds with 1.1 fb$^{-1}$ of Run II data on $M_D$ of 1.33 TeV for $n = 2$ to 0.88 TeV for $n = 6$ [14]. All four LEP experiments combined set a lower limit on $M_D$ of 1.2 TeV for positive interference, or 1.1 TeV for negative interference between Standard Model diagrams and graviton exchange [15, 16]. Indirect searches by the DØ [17] collaboration set lower limits around 1.28 TeV[3]. Astrophysics places the most stringent lower limits on $M_D$ in ADD models which however fall sharply with increasing number of extra dimension [18–24]. Considerations of neutron star dynamics imposes the strongest constraints: $M_D > 1760$, 77, 9 and 2 TeV for $n = 2$, 3, 4 and 5 extra dimensions, respectively [18]. One should note that all astrophysical and cosmological constraints are based on a number of assumptions, whose uncertainties are not included in the limit derivations, so the results are reliable only as order of magnitude estimates. Ultra high-energy cosmic-ray particles, through their interaction with the Earth's atmosphere, offer a complementary probe of extra dimensions. Cosmic-rays interact with the atmosphere and earth's crust with centre-of-mass energies of the order of 100 TeV. The particles can produce black holes deep in the atmosphere, leading to quasi-horizontal giant air showers. So far a lower bound on $M_D$ in the ADD model, ranging from 1.0 to 1.4 TeV for scenarios with 4 to 7 extra dimensions has been set at 95% confidence level [25]. It is expected that the Pierre Auger Observatory will be able to set more stringent limits during the first five years of operation; the estimates place $M_D \gtrsim 3$ TeV for $n \geq 4$ [26].

## 2.3 Working Model

Our working model for black holes uses the black disk cross-section, which depends only on the horizon radius. The $(4+n)$-dimensional Myers-Perry solution [27], similar to the 4-dimensional Schwarzschild radius, is chosen for the horizon radius $r_{\mathrm{h}}$. It depends only on the number of dimensions and the Planck

$M_{\mathrm{DL}}$ is defined via $M_{\mathrm{DL}}^{D-2} = 1/G_D$.

[2]See Ref. [11] for a model of TeV gravity in four dimensions.

[3]This lower limit uses the Hewett approach for the calculation of $M_D$ and implies a positive interference term $\lambda$ with the Standard Model diagrams.

scale. The classical black hole cross-section at the parton level is

$$\hat{\sigma}_{ab\to \mathrm{BH}} = \pi r_{\mathrm{h}}^2, \tag{1}$$

where $a$ and $b$ are the parton types. In most cases, we work with initial black hole masses at least five times higher than the Planck scale at which the expression for the cross section should be valid.

The total cross-section is obtained by convoluting the parton-level cross-section with the parton distribution functions (PDFs), integrating over the phase space, and summing over the parton types.

Throughout this study we use the CTEQ6L1 (leading order with leading order $\alpha_s$) parton distribution functions [28] within the LHAPDF framework [29]. The momentum scale for the PDFs is set equal to the black hole mass for convenience.

The transition from the parton-level to the hadron-level cross-section is based on a factorisation ansatz. The validity of this formula for the energy region above the Planck scale is unclear. Even if factorisation is valid, the extrapolation of the parton distribution functions into this transplanckian region based on Standard Model evolution from present energies is questionable, since the evolution equations neglect gravity and possible KK states in the proton.

The details of horizon formation, and the balding and spin-down phases have been ignored[4]. The important effects of angular momentum in the production and decay of the black hole in extra dimensions are not accounted for in the Monte Carlo event generator. The black holes are considered as $D$-dimensional Schwarzschild solutions. Only the Hawking evaporation phase is generated by the simulation.

We can view the Hawking evaporation phase as consisting of two parts: determination of the particle species and assigning energy to the decay products. A particle species is selected randomly with a probability determined by its number of degrees of freedom and the ratio of emissivities[5]. The degrees of freedom take into account polarisation, charge and colour. The emitted charge is chosen such that the magnitude of the black hole charge decreases. All Standard Model particles are considered, including a Higgs boson[6]. The particles are treated as massless, including the gauge bosons and heavy quarks.

Gravitons have not been included in the simulation, which is another drawback of the current model. Because the graviton lives in the bulk, the number of degrees of freedom of the graviton becomes significant for high numbers of dimensions. In addition, the graviton emissivity is highly enhanced as the space-time dimensionality increases. Therefore the black hole may lose a significant fraction of its mass into the bulk, resulting in missing transverse energy.

The energy assignment to the decay particles in the Hawking evaporation phase has been implemented as follows. The particle species selected by the model described above is given an energy randomly according to its extra-dimensional decay spectrum. A different decay spectrum is used for scalars, fermions and vector bosons, i.e. the spin statistics factor is taken into account. Grey-body spectra are used without approximations [30]. The grey-body factors depend on the number of dimensions. The Hawking temperature is updated after each decay. It is assumed the decay is quasi-stationary in the sense that the black hole has time to come into equilibrium at each new temperature before the next particle is emitted. The energy of the particle given by the spectrum must be constrained to conserve energy and momentum at each step.

The evaporation phase ends when the chosen energy for the emitted particle is ruled out by the kinematics of a two-body decay. At this point an isotropic two-body phase-space decay is performed. In our simulation, the decay is performed totally to Standard Model particles and no stable exotic remnants survive.

---

[4]The main effects of these are to reduce the black hole production cross-section, possibly by as much as a few orders of magnitude. Work is underway to try to estimate the magnitude of this.

[5]Here emissivity is the fractional emission rate per degree of freedom. Mathematically it is ratio of $dE/dt$ for different species, or roughly speaking the area under the black body distribution for a particular type of particle.

[6]Including a scalar Higgs boson is not significant since it has only one degree of freedom.

| Name | Description | Value |
|---|---|---|
| MINMSS | Minimum mass of black holes | 5 TeV |
| MAXMSS | Maximum mass of black holes | 14 TeV |
| MPLNCK | Planck scale | 1 TeV |
| MSSDEF | Convention for Planck scale | 2 |
| TOTDIM | Total number of dimensions | 6 |
| NBODY | Number of particles in remnant decay | 2 |
| GTSCA | Black hole mass used as PDF momentum scale | True |
| TIMVAR | Allow $T_H$ to change with time | True |
| MSSDEC | Use all Standard Model particles as decay products | True |
| GRYBDY | Include grey-body effects | True |
| KINCUT | Use a kinematic cut-off on the decay | True |

Table 1: Default parameters used in the CHARYBDIS generator.

Baryon number, colour and electric charge are conserved in the black hole production and decay in this model. Missing transverse energy in the generator comes only from the neutrinos, while in reality missing transverse energy is also possible due to the lost energy in inelastic production, graviton emission, a non-detectable black hole remnant and the possibility that the black hole can leave the Standard Model brane. For the black holes we consider, only a small amount of energy, on average, is lost due to neutrinos. If gravitons were considered, the average energy loss would be approximately 9% [31].

# 3 Monte Carlo Simulations

## 3.1 Production of Signal and Background Events

The event generator CHARYBDIS [32, 33] version 1.003 was used within the ATLAS software framework to generate Monte Carlo signal samples. It was interfaced via the Les Houches accord [29] to HERWIG [34, 35] which provides the parton evolution and hadronisation, as well as Standard Model particle decays.

Table 1 shows the default CHARYBDIS parameters used, for which approximately 25000 events were generated. Three other black hole signal samples were generated with variations in the number of dimensions and in the black hole minimum mass. In all simulations, the parameter MSSDEF was set equal to 2, setting the Planck scale MPLNCK to be the $D$-dimensional Planck scale $M_{\mathrm{DL}}$ in the convention of Ref. [8]. The above samples subsequently underwent the full ATLAS detector simulation and reconstruction. Fast simulation using ATLFAST [36] was employed to widen the range of signal samples studied, enabling investigation of the many theoretical uncertainties modelled by generator parameter switches (see Table 2).

Black holes decay democratically to all particles of the Standard Model, so few Standard Model processes should produce the same particle spectrum. Black hole decays are characterised by a number of high energy and transverse momentum objects, so the primary Standard Model backgrounds are states with high multiplicity or high energy jets. The predominant backgrounds to our signal are described below and their datasets and cross-sections are listed in Table 3. Sizeable samples are required due to their large cross-sections at the LHC.

- $t\bar{t}$ leptonic and hadronic decay modes. This process yields the largest contribution to the background due to its large cross-section at the LHC and the large branching ratio to hadronic final

| $n$ | $m_{\mathrm{BH}}$ ( TeV) | $\sigma$ ( pb) | Note |
|---|---|---|---|
| 2 | 5-14 | 40.7 | |
| 2 | 8-14 | 0.34 | |
| 4 | 5-14 | 24.3 | |
| 7 | 5-14 | 22.3 | |
| 2 | 5-14 | 6.4 | MPLNCK=2 |
| 3 | 5-14 | 28.5 | |
| 5 | 5-14 | 22.7 | |
| 2 | 5-14 | 40.7 | KINCUT=0 |
| 7 | 5-14 | 22.3 | KINCUT=0 |
| 2 | 5-14 | 40.7 | TIMEVAR=0 |
| 7 | 5-14 | 22.3 | TIMEVAR=0 |
| 2 | 5-14 | 40.7 | NBODY=4 |
| 7 | 5-14 | 22.3 | NBODY=4 |

Table 2: Monte Carlo datasets and their respective cross-sections used in this analysis. The first four samples were simulated using both full and fast simulations; the lower nine samples were simulated using the fast simulation ATLFAST. The final column shows the CHARYBDIS parameter that was changed with respect to the reference set shown in Table 1.

states. The matrix element calculation is done with MC@NLO [37] and HERWIG is used to perform the parton shower evolution, parton decay and their hadronisation.

- QCD dijet production. The requirements placed on the hadronic part of the signal events reduces the contribution from low-$p_T$QCD jets. This background is generated using PYTHIA 6.4 [38]. Note that the complete QCD inclusive jet production is not fully modelled by the PYTHIA dijet simulation due to the lack of higher-order QCD contributions. Very low-statistics samples of multijet samples generated by ALPGEN [39] were also used.
- $W \to \ell\nu$ + jets production. These backgrounds, though coming from the hard process, have cross-sections that rapidly become small compared to the signal as more jets are added. Vector boson plus jets samples were generated using ALPGEN.
- $Z \to \ell\ell$ + jets production.
- $\gamma(\gamma)$ + jets production.

### 3.2 Detector Simulation

The detector simulation and reconstruction of both signal and background Monte Carlo events were performed within the ATLAS offline framework.

Fast simulation (ATLFAST) was used to widen the range of signal samples studied. The primary advantage of this is one of processing rate: since no detector interactions are modelled it requires less than one second per event. In contrast, full simulation requires approximately 15 minutes for typical Standard Model events, and over 30 minutes per black hole event. Despite this advantage, there are drawbacks: the fast simulation does not include a complete treatment of lepton isolation and misidentification nor of photon conversion.

The same generator signal samples were passed through the full and fast simulations in order to understand the differences in black hole events. Variables ranging from simple multiplicities to event

| Process | $\sigma$ ( pb) |
|---|---|
| Semi/Fully Leptonic $t\bar{t}$ | 463 |
| Hadronic $t\bar{t}$ | 370 |
| QCD dijets | $12.84 \times 10^3$ |
| $W \to e\nu_e + \text{jets}$ | 281 |
| $W \to \mu\nu_\mu + \text{jets}$ | 279 |
| $Z \to ee + \text{jets}$ | 25.8 |
| $Z \to \mu\mu + \text{jets}$ | 26.0 |
| $\gamma + \text{jets}$ | $5.00 \times 10^3$ |
| $\gamma\gamma + \text{jets}$ | 67.6 |

Table 3: Background Monte Carlo datasets and their respective branching ratio times cross-sections.

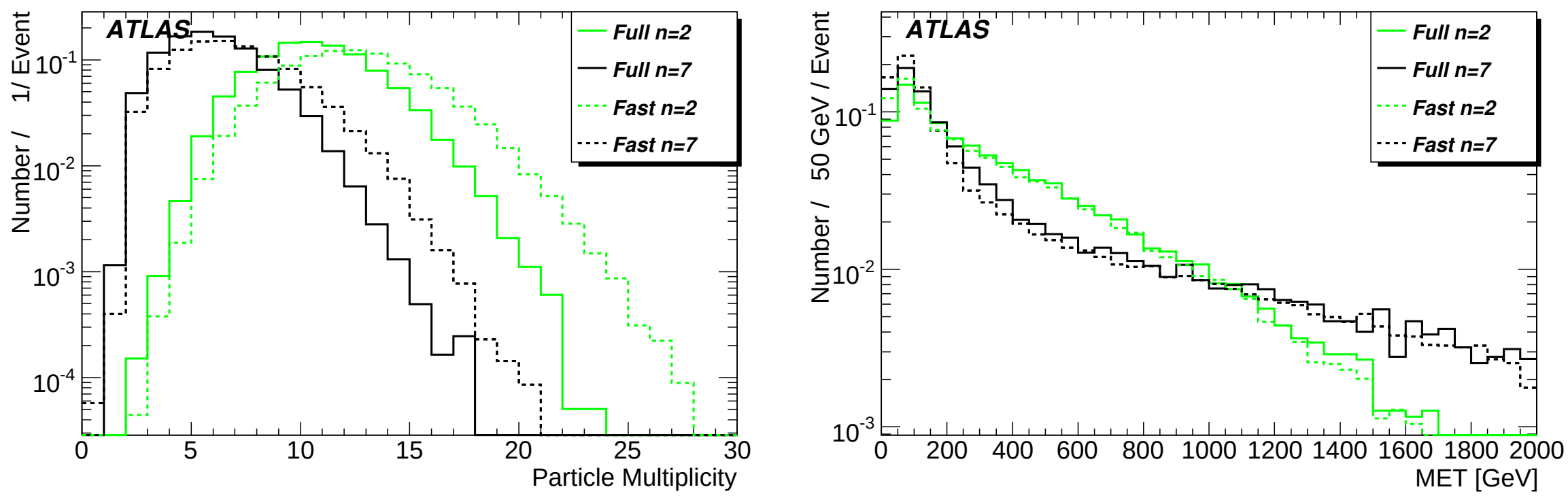


Figure 1: Total particle multiplicity and missing transverse energy distributions for signal samples from fast and full simulation.

shapes were compared. Sample distributions of particle multiplicity and $\not{E}_T$ are shown in Figure 1. The multiplicity difference is due to differing default jet algorithms; we use a cone algorithm with radius $\Delta R = 0.4$ in the full simulation, whereas ATLFAST uses a $k_\perp$ algorithm. Using the same algorithm for both samples gives very close agreement, nonetheless this discrepancy will have an effect in analyses dependent purely upon multiplicity information. All other variables investigated showed concordant results.

## 4 Event Properties

The high mass scale, and the thermal nature of the decay process, result in black hole events being characterised by a large number of high-$p_T$ final state particles, including all the Standard Model fields. Graviton emission is also expected, but is not simulated in CHARYBDIS. Of the final state particles, the detector can measure jets, electrons, muons and photons well, and will be able to reconstruct some of the $Z$ and $W$ bosons. The missing transverse energy, produced mainly by neutrino and graviton emission, can also be measured. In this section, the data sample with two extra dimensions and black hole masses above 5 TeV is used as the reference signal sample.

A key feature of black hole decays is that the Hawking temperature is higher for larger $n$, for a given black hole mass. A higher temperature produces higher energy emissions, with the consequence that the

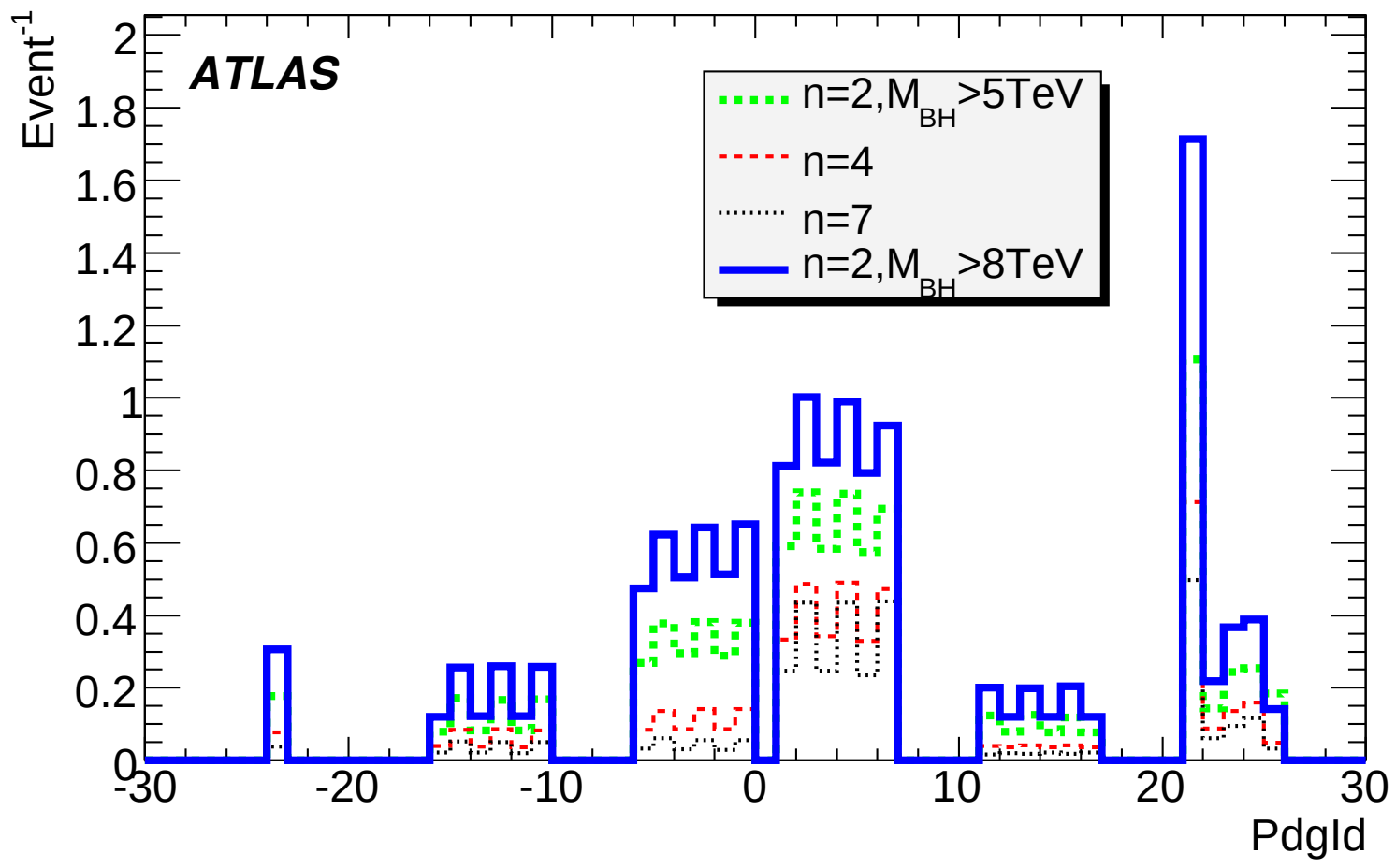


Figure 2: PDG code of particles emitted from black hole decay for a minimum black hole mass of 5 TeV and $n = 2$, 4 and 7 and for a minimum black hole mass of 8 TeV and $n = 2$ ($|PdgId| = 1 - 6$ are quarks, $11 - 16$ leptons, and $21 - 25$ gauge and Higgs bosons). The vertical axis shows multiplicity per black hole decay.

energy is shared between fewer particles. This has a significant effect on the multiplicity and event shape distributions. Similarly, the samples with a higher black hole low-mass cutoff produce more high energy final state particles.

### 4.1 Particle Types and Multiplicities

Figure 2 shows the types of particles produced directly by black hole decay. The vertical axis shows the average number of particles per black hole decay. From this figure, we see that a heavier black hole has more decay products. The particle-antiparticle balance is broken by the initial state of two protons colliding. Moreover, due to conservation of energy and momentum, colour connection etc., a perfect democratic decay cannot be achieved, e.g., the number of top quarks is smaller than that of ligher quarks. The possibility of identifying fermions and bosons and determining their branching ratios in black hole decays was studied in [40].

Figure 3 shows $p_T$ and pseudorapidity ($\eta$) distributions of particles produced directly from black hole decays. As expected, the shape depends little on particle type. Figure 4 shows the reconstructed multiplicity of final-state jets, leptons and photons. Four signal samples are shown for $n = 2$, 4 and 7 with a minimum black hole mass of 5 TeV, and for $n = 2$ with a minimum black hole mass of 8 TeV. The figure also compares the reference signal to the backgrounds. The multiplicity in the signal falls as $n$ rises, because the black holes decay at a higher temperature.

### 4.2 Event Shape

At first sight, one would expect black hole events to be very different from the background in event shape variables [38, 41, 42] such as sphericity, because of the high multiplicity thermal decay. However, the event shape of the black hole events varies considerably with $n$, making such variables less useful than could be hoped. Though the background distributions show less variation, when these are scaled by their large cross sections, there is a large degree of overlap, disfavouring their use as a cut variable. Additionally, our ignorance of the decay modes of the final black hole remnant introduces a significant

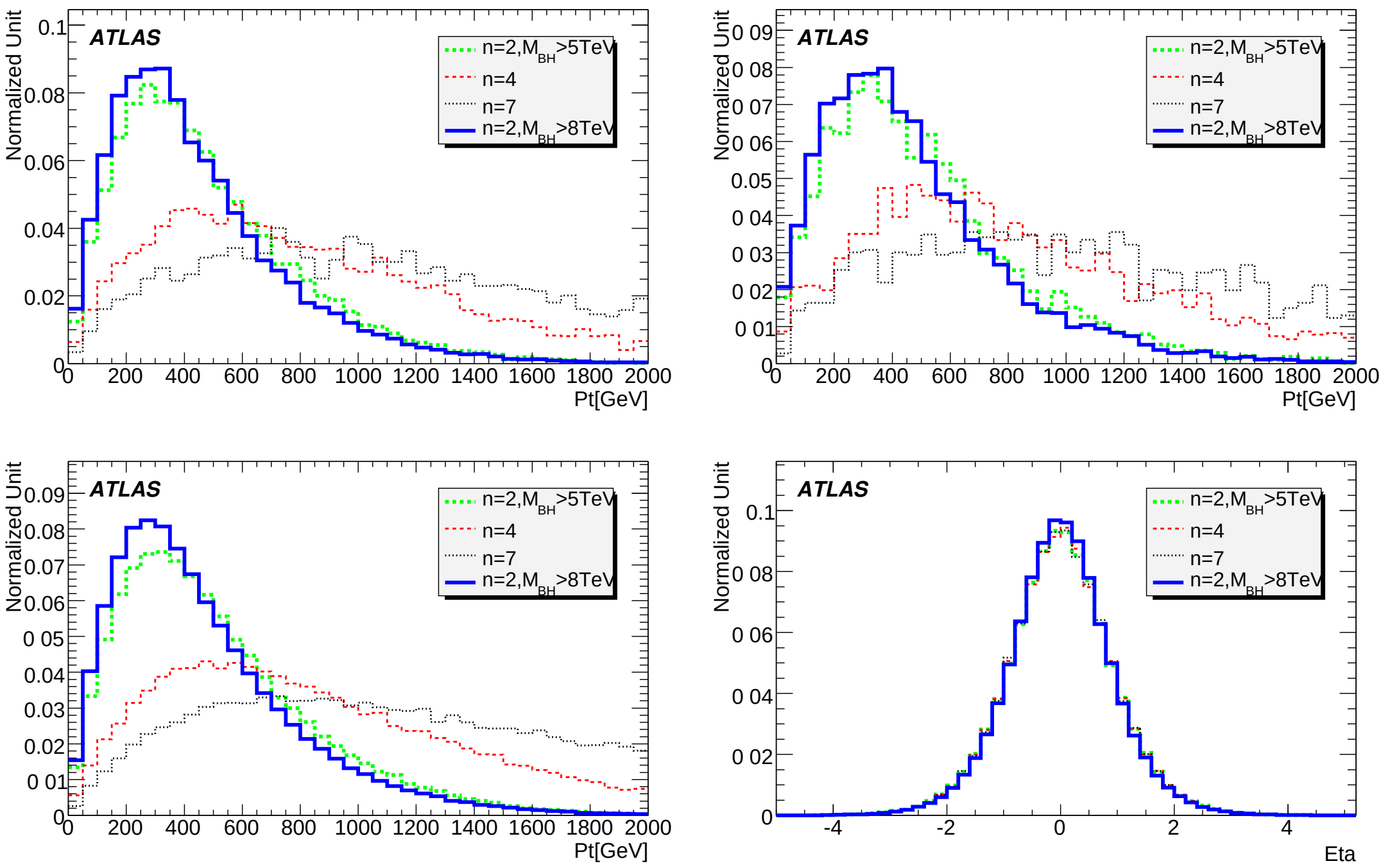


Figure 3: Generator $p_T$ distributions (top row): leptons (left) and $Z$ bosons (right) emitted from the black hole. The bottom row shows $p_T$ and $\eta$ spectra for all particles emitted from the black hole.

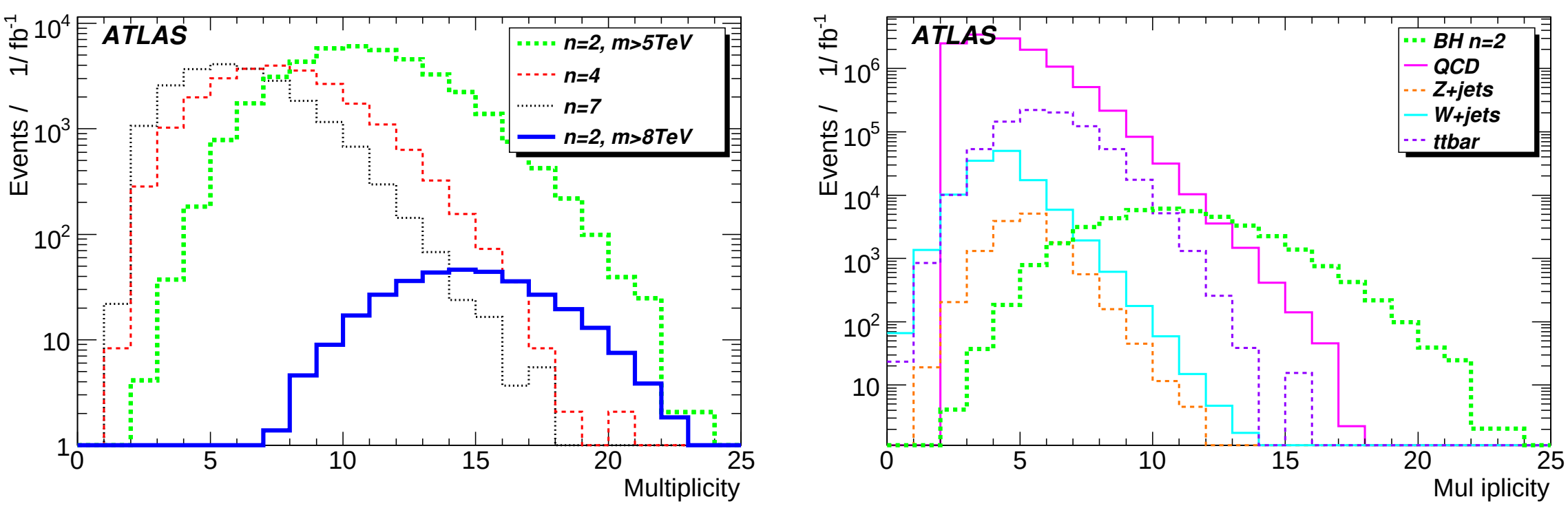


Figure 4: Multiplicities of reconstructed objects for (left) black hole samples and (right) backgrounds. They are normalised to the integrated luminosity of 1 $fb^{-1}$.

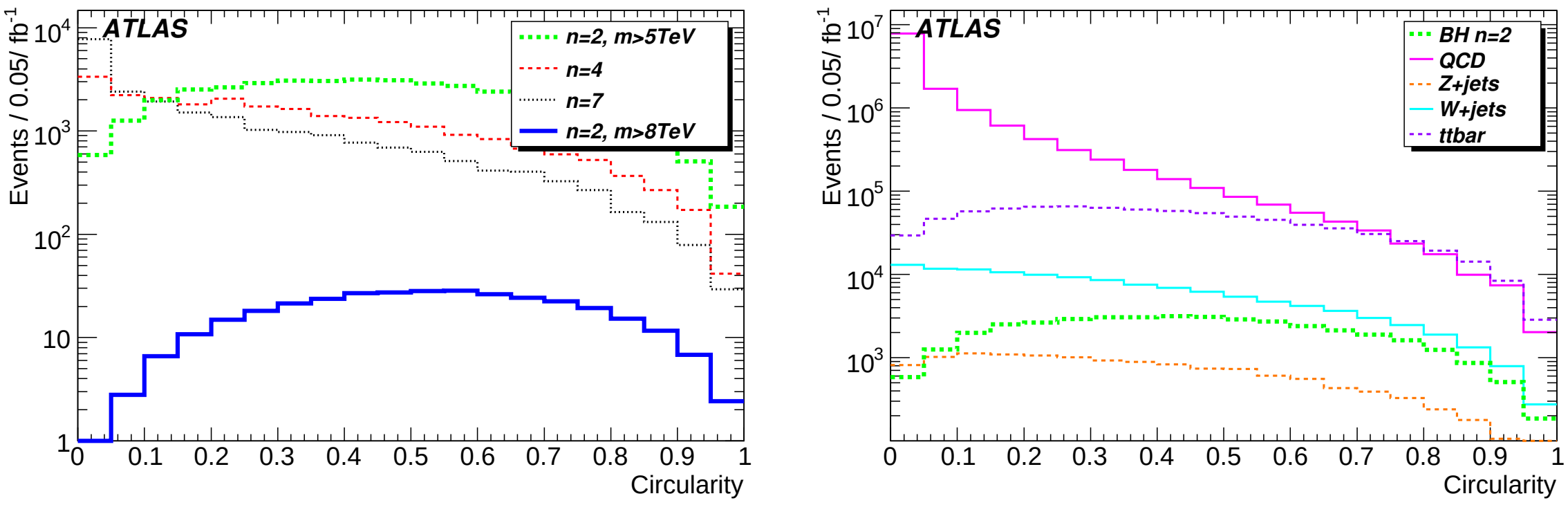


Figure 5: Circularity calculated from reconstructed objects for (left) black hole samples and (right) backgrounds.

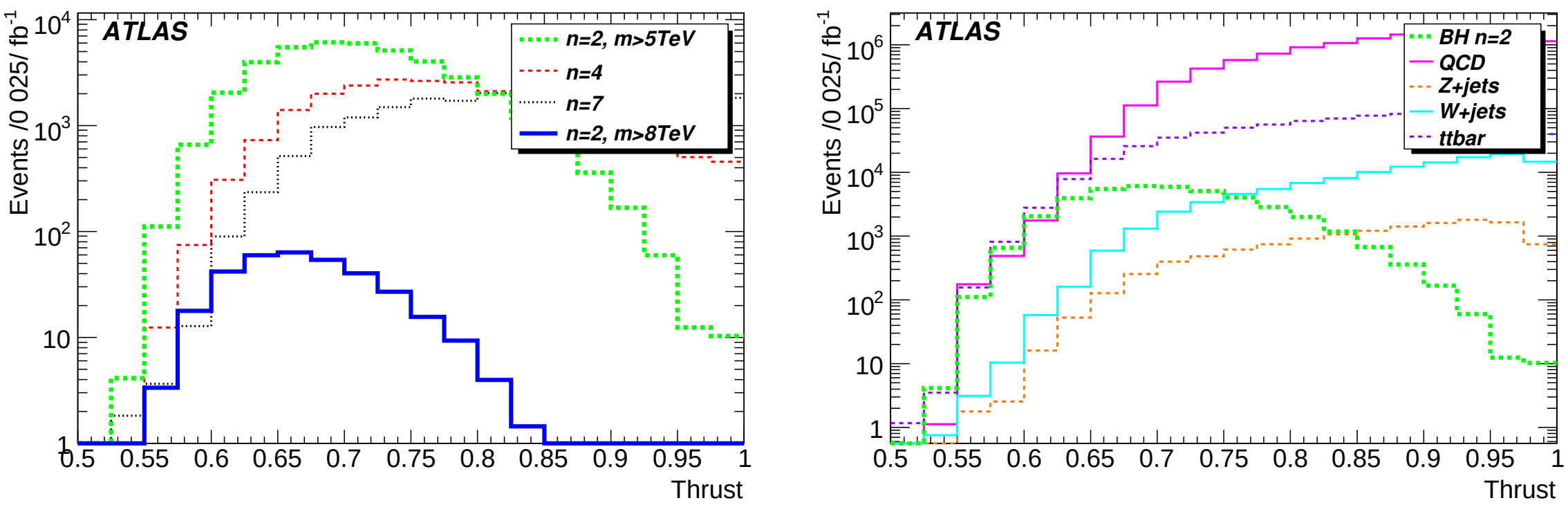


Figure 6: Thrust calculated from reconstructed objects for (left) black hole samples and (right) QCD dijet and $t\bar{t}$ backgrounds.

systematic effect. In our version of CHARYBDIS, once the mass of the black hole has dropped below the Planck scale, the remnant decays to either 2 or 4 bodies. We have selected the two-body option for our standard samples. This means that at high $n$, where events can reach this stage after few emissions, the circularity of the events is reduced, and the thrust increased.

The distinguishing power between signal and backgrounds of a selection of event shape variables was studied; Figure 5 shows the circularity distribution for the same samples as Figure 4; similarly Figs. 6 and 7 show their thrust distributions, sphericity and aplanarity. The expected bias towards more "jet-like" events is clearly seen at high $n$. For this reason, we choose not to use event shape variables as a discriminant in this analysis.

# 5 Trigger

The ATLAS trigger and data-acquisition system consists of three levels (L1, L2, EF) of online event selection [43]. Each subsequent trigger level refines the decisions made at the previous level and may apply additional selection criteria. The ATLAS trigger is described in detail in ref. [44].

(a) Thrust major

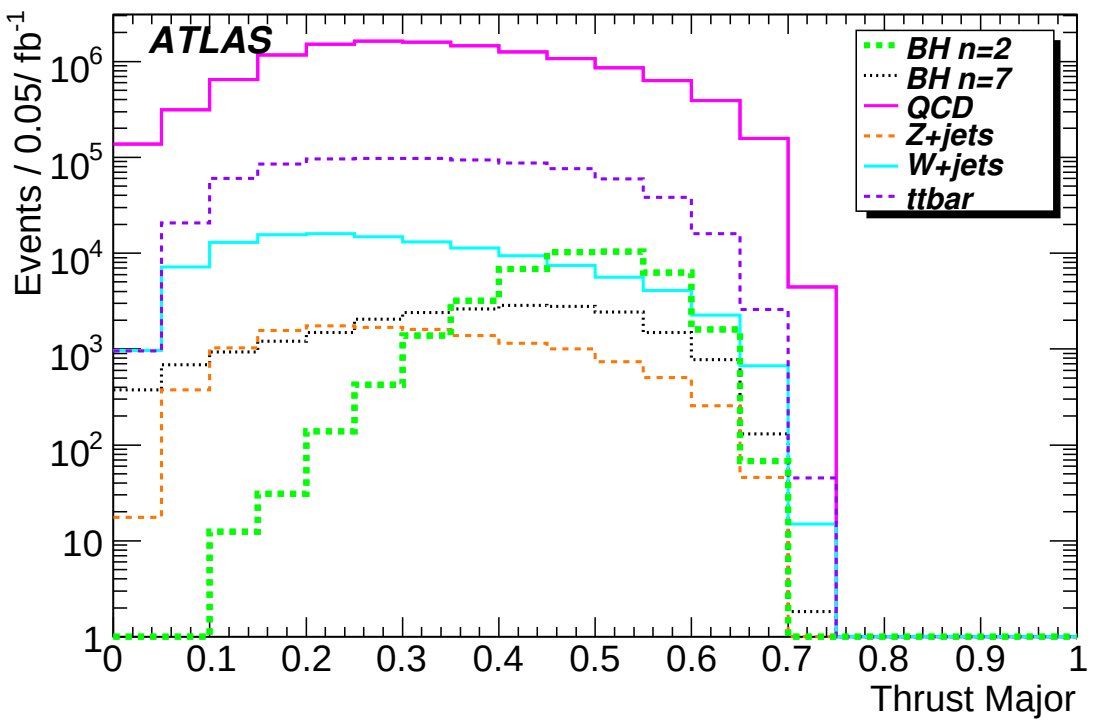


(b) Thrust minor

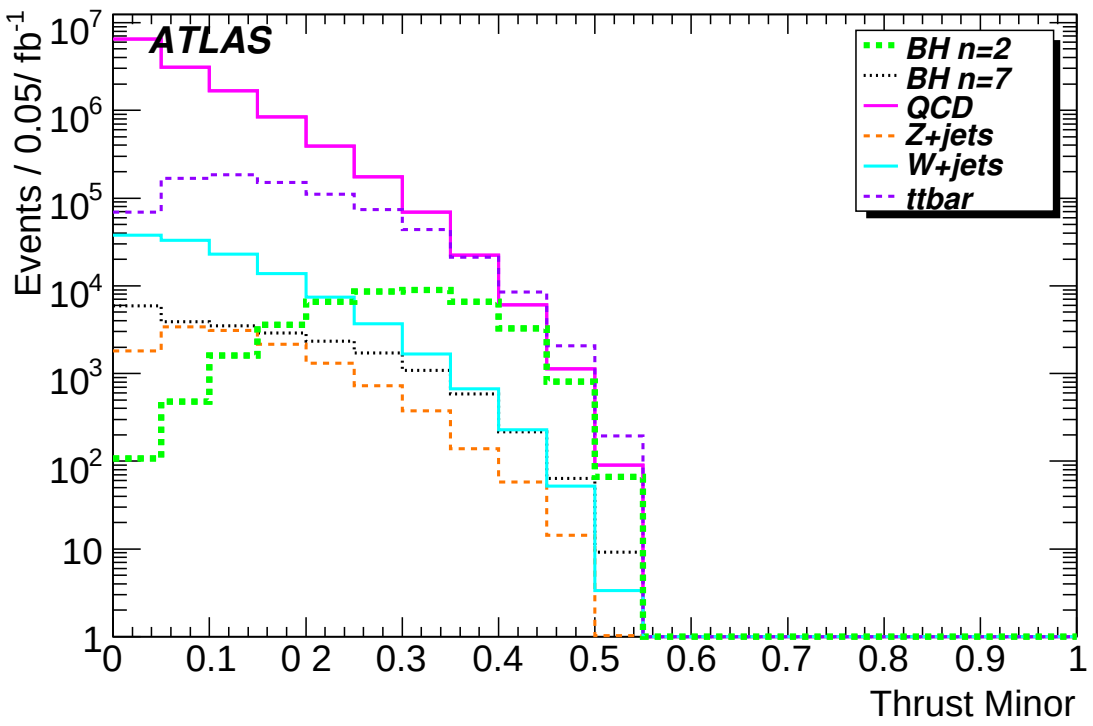


(c) Sphericity

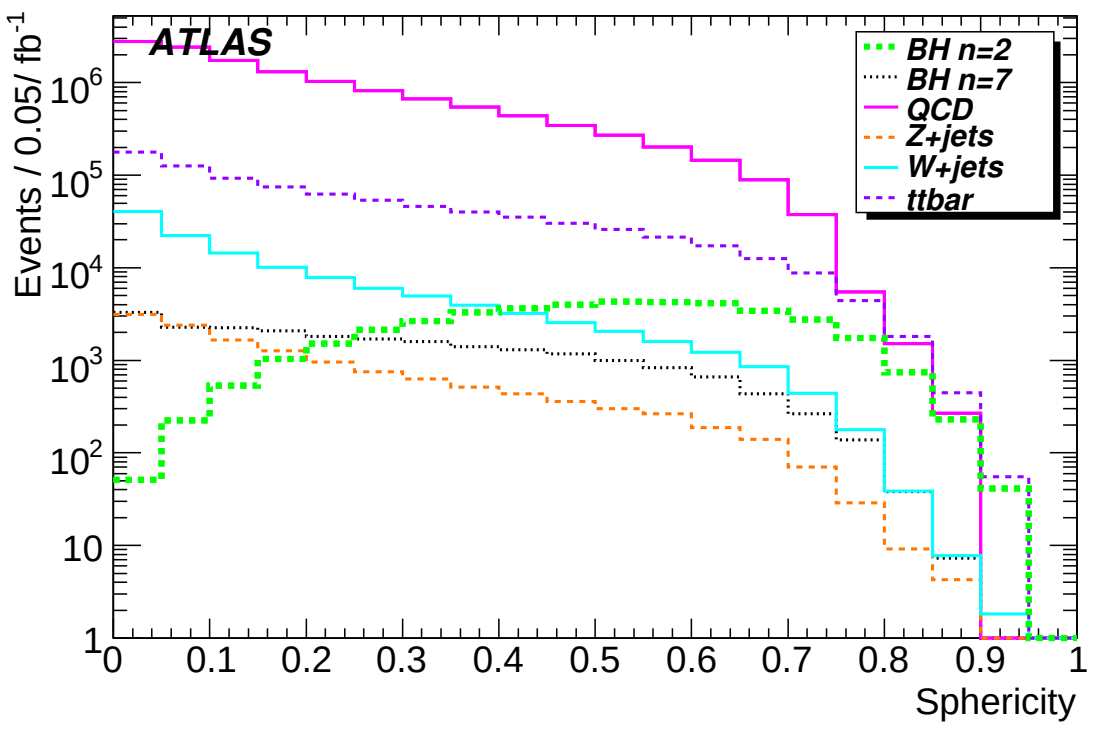


(d) Aplanarity

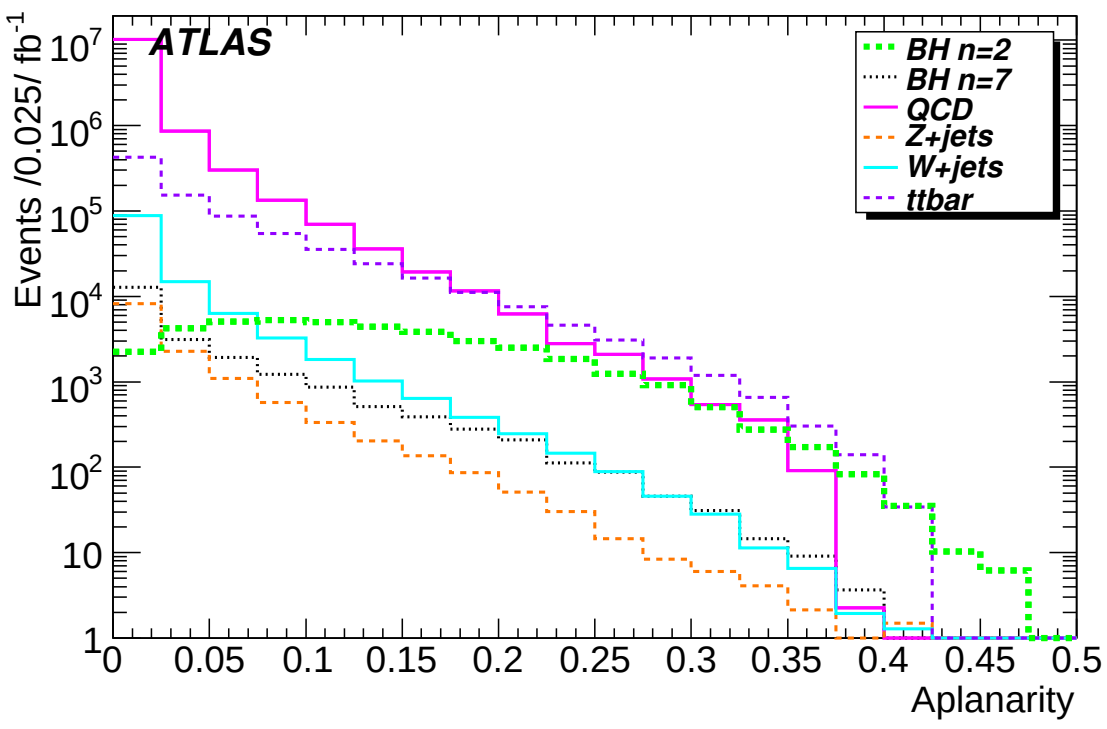


Figure 7: Different event shape variables for black hole samples and backgrounds. They are normalised to the integrated luminosity of 1 $\mathrm{fb}^{-1}$.

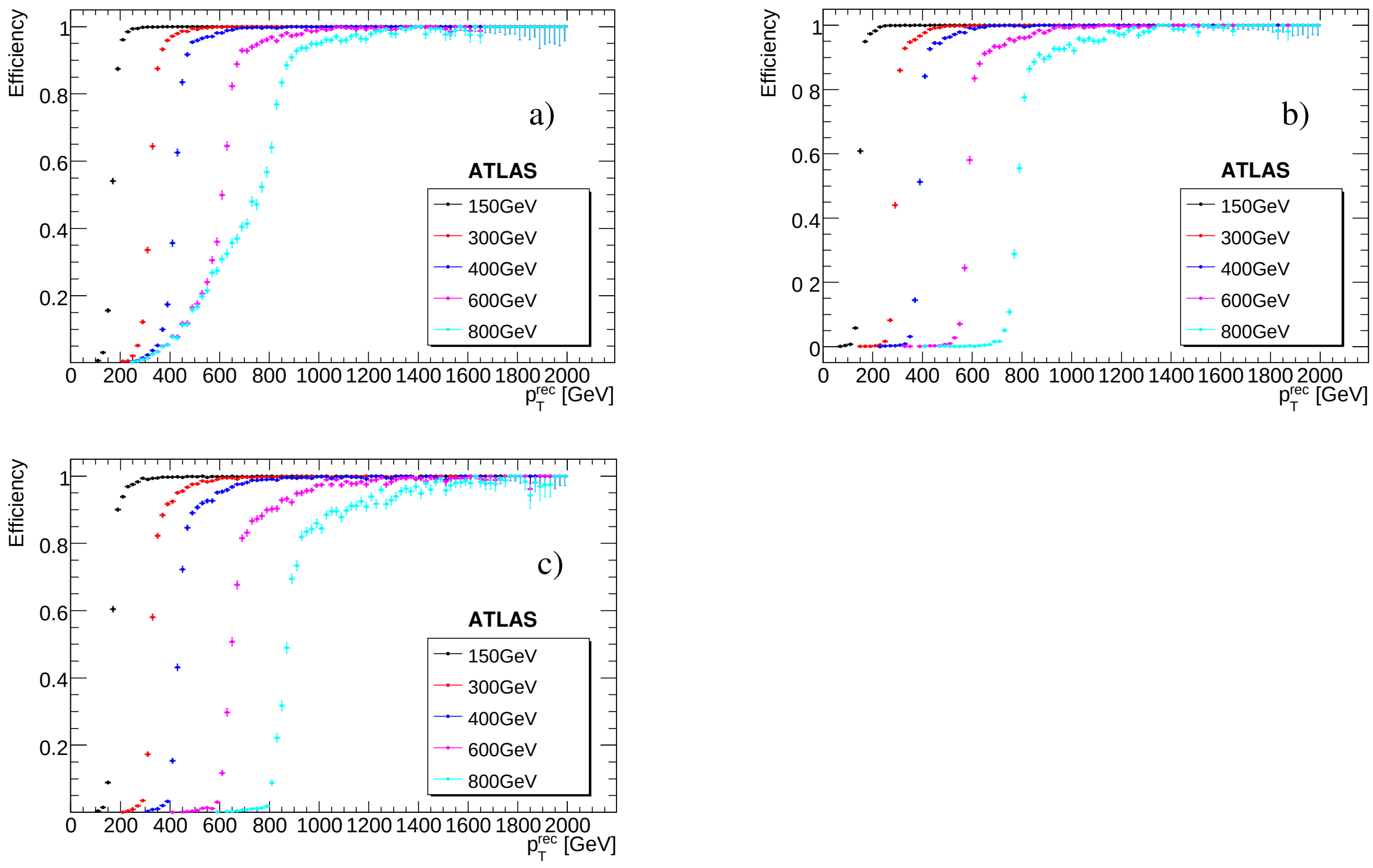


Figure 8: Simulated jet trigger efficiencies as functions of the offline reconstructed jet $p_T$for a) L1, b) L2 and c) EF. The efficiencies are determined for different $p_T$-thresholds: 150 GeV (black), 300 GeV (red), 400 GeV (blue), 600 GeV (magenta) and 800 GeV (cyan).

## 5.1 Triggering on Black Holes

Each black hole produces multiple decay products, including hadronic jets, leptons and photons, as described in Section 4. The jets typically carry a dominant fraction of the visible decay energy and hence provide the best option for triggering black hole events.

The response of the jet trigger, as simulated in the current version of the ATLAS detector simulation, is demonstrated in Figure 8. The plots show the trigger efficiencies for various $p_T$-thresholds as functions of the jet $p_T$ reconstructed offline. For these plots, a match between the jet reconstructed offline and at the respective trigger level is required. The matching consists of searching for the closest offline jet in $\Delta R = \sqrt{(\Delta\eta)^2 + (\Delta\phi)^2}$, where $\Delta\eta$ and $\Delta\phi$ are the distances between the reconstructed jet and the trigger jet in pseudorapidity $\eta$ and azimuth $\phi$, respectively. To avoid incorrect matching for L1 jets, a modified criterion is applied: the L1 jet closest in energy to the reconstructed jet is chosen among the jets found within the $\Delta R = 0.5$ distance around the reconstructed jet. The shape of the L1 efficiency distribution for the 800 GeV threshold is due to the saturation of the L1 trigger tower energies at 255 GeV. Events in which the transverse energy in one trigger tower exceeds 255 GeV are automatically accepted, as larger values fill up the memory of the L1 trigger analog-to-digital converters.

The efficiency at each trigger level is determined independently of the decisions at the other levels. Were the trigger chain to have the same threshold on all levels, the total efficiency would be the convolution of the respective functions. The L2 algorithms are based on regions of interest provided by L1, hence it is not possible to determine their efficiency completely independently of the L1 decisions. The L2 algorithms were run on all L1 jet RoI starting from the lowest L1 jet $p_T$threshold of 35 GeV. This is much lower than the thresholds studied, making the L2 decision (shown in Figure 8b) virtually

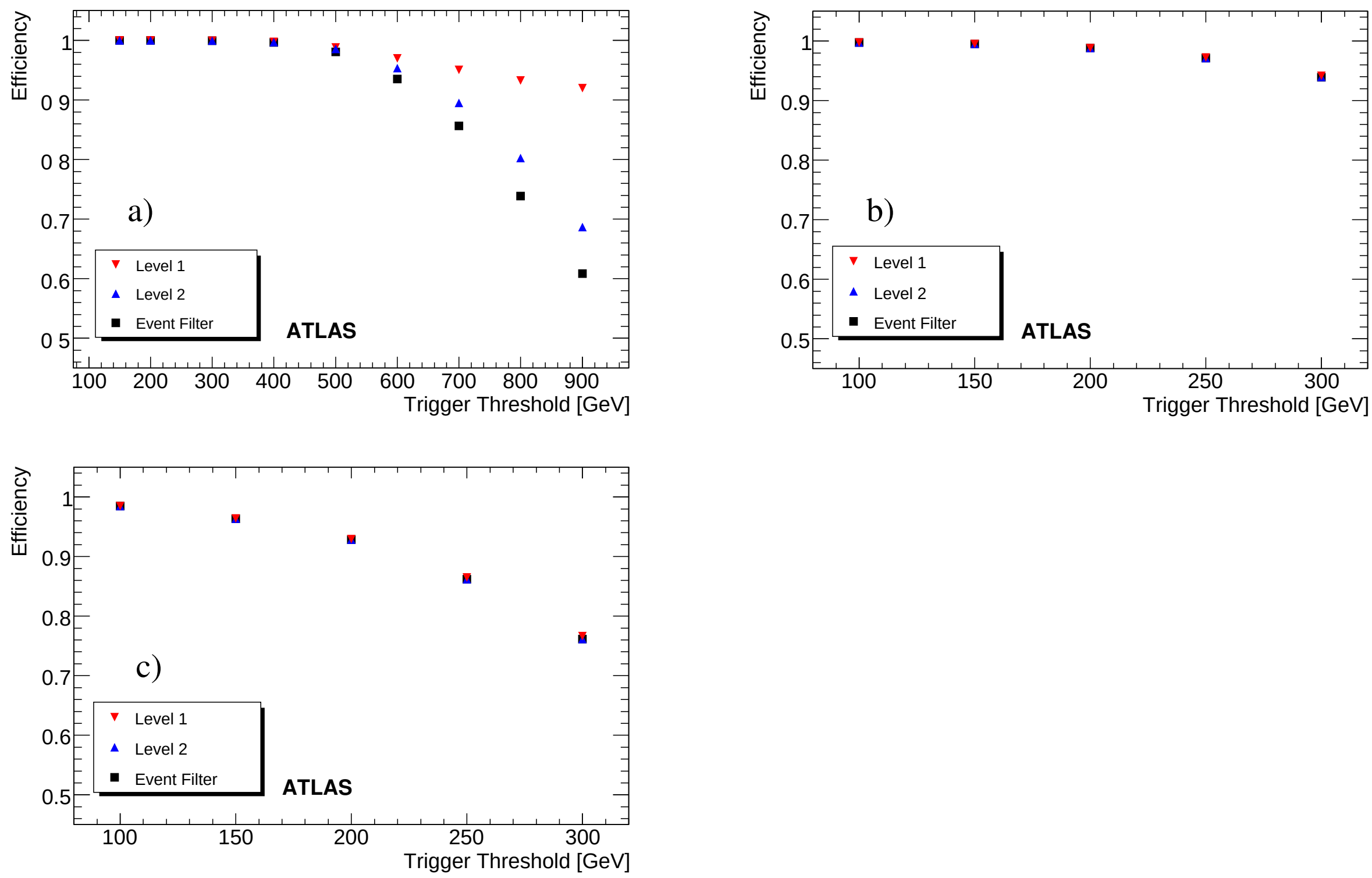


Figure 9: Simulated jet trigger efficiencies for black hole events from the signal sample with $n = 2$ and $m > 5$ TeV as functions of the jet $p_T$threshold for a) single-jet trigger, b) 3-jet trigger and c) 4-jet trigger. The efficiencies are determined for L1 (black), L2 (red) and EF (blue).

independent of the L1 efficiency.

The total trigger efficiencies are listed in Table 4 for three signal samples and demonstrated in Figure 9 for the signal sample with $n = 2$ and $m > 5$ TeV. The highest efficiency is provided by the single-jet trigger, which we consider to be the master trigger for the black hole events. The presence of multiple high-$p_T$ jets per event, each of which is likely to pass the trigger, results in very high total efficiencies. Setting this trigger threshold at 400 GeV will provide greater than 99% efficiency at all trigger levels. The Standard Model process rate at this threshold is expected to be less than 0.1 Hz at an instantaneous luminosity of $10^{31}$ cm$^{-2}$s$^{-1}$, which should allow this trigger to run at this threshold without prescaling for the first few years of LHC data taking. The rate of black hole events is expected to be less than 5 mHz at the $10^{31}$ cm$^{-2}$s$^{-1}$ luminosity. For the start-up running at the luminosity of $10^{31}$ cm$^{-2}$s$^{-1}$, it is planned to set the highest threshold for the single-jet trigger at 120 GeV, guaranteeing an efficiency of almost 100% for black hole events.

Alternatively, a trigger based on the scalar sum of transverse energies of all recorded decay products ("sum-$E_T$ trigger") can be used. No simulation of this trigger is available in the samples used in this study. Looking at this sum in the offline reconstruction suggests that this trigger would collect nearly 100% of black hole events for Planck scales above 1 TeV. It is foreseen to run this trigger in the start-up data taking, unprescaled at the threshold of 650 GeV.

Based on experience from previous collider experiments, one may expect detector hardware problems at the beginning of data taking. In particular, noisy channels in the calorimeter or trigger electronics may cause high trigger rates for the single-jet trigger and for the sum-$E_T$ trigger, such that even the highest threshold triggers have to be prescaled. In such cases, a multijet (3- or 4-jet) trigger is considered for use

a) CHARYBDIS: $n = 2, m > 5$ TeV

| Trigger | L1 | L2 | EF |
|---|---|---|---|
| j100 | 1 | 1 | 1 |
| j400 | 0.997 | 0.997 | 0.997 |
| 3j100 | 0.998 | 0.998 | 0.998 |
| 3j250 | 0.972 | 0.971 | 0.971 |
| 4j100 | 0.985 | 0.985 | 0.985 |
| 4j250 | 0.865 | 0.862 | 0.862 |

b) CHARYBDIS: $n = 4, m > 5$ TeV

| Trigger | L1 | L2 | EF |
|---|---|---|---|
| j100 | 1 | 1 | 1 |
| j400 | 0.997 | 0.997 | 0.996 |
| 3j100 | 0.952 | 0.952 | 0.952 |
| 3j250 | 0.886 | 0.885 | 0.885 |
| 4j100 | 0.807 | 0.806 | 0.806 |
| 4j250 | 0.612 | 0.607 | 0.607 |

c) CHARYBDIS: $n = 7, m > 5$ TeV

| Trigger | L1 | L2 | EF |
|---|---|---|---|
| j100 | 1 | 1 | 1 |
| j400 | 0.990 | 0.987 | 0.985 |
| 3j100 | 0.807 | 0.806 | 0.805 |
| 3j250 | 0.710 | 0.704 | 0.704 |
| 4j100 | 0.525 | 0.522 | 0.522 |
| 4j250 | 0.343 | 0.341 | 0.341 |

Table 4: Simulated jet trigger efficiencies for black hole events as functions of the jet-$p_T$threshold for different simulation samples.

until the detector problems are resolved. The efficiencies of such triggers are listed in Table 4.

In the present study, the minimum mass of a black hole is set at 5 TeV or more in order to be safely above the Planck scale. At lower masses one may expect an increased rate of dijet events described by a contact interaction. The single-jet trigger or the sum-$E_T$ trigger at the thresholds considered above are well suited for detecting such signatures. Such events may not, however, be selected by multijet triggers.

The trigger efficiencies, studied here in the simulation, have to be determined from data. An unbiased determination requires an "orthogonal" trigger, e.g. a trigger based on fully independent information from that used by the master trigger. A muon trigger which is based solely on signals in the muon detector should be well suited for such studies.

# 6 Signal Selection and Background Rejection

## 6.1 Event Selection

Since all types of Standard Model particles are produced from black hole decay, we make full use of particle identification information (PID) from our detectors. First we select muons, electrons, photons and jets, which are called *objects* in this section. Table 5 shows the details of their selection criteria.

The identification of objects is sometimes ambiguous: e.g., an electron could be simultaneously reconstructed as a jet. To resolve this, we apply PID to each object, selecting muons, electrons, photons and jets in that order of priority. Once an object passes the PID criteria in a given category, any remaining ambiguous assignments are removed if they match the chosen object within a $\Delta R$ of less than 0.1.

Next we select black hole events using these objects as described below. Then we reconstruct a black hole from all the identified objects for the selected event. The mass of the black hole in an event is calculated from the four-momenta of the reconstructed final state objects and missing $E_T$, which is

included in the calculation to improve the reconstructed mass resolution:

$$p_{\rm BH} = \sum_{i=\rm objects} p_i + (\not{E}_T, \not{E}_{T_x}, \not{E}_{T_y}, 0) \ , \tag{2}$$

$$m_{\rm BH} = \sqrt{p_{\rm BH}^2} \ . \tag{3}$$

We present two methods to select black hole events. One is based on the scalar summation of $p_T$ and the other on the multiplicity of high-$p_T$ objects. Both make use of the characteristic of a black hole having large mass. After that, we require a high-$p_T$ lepton to reject backgrounds further.

Figure 10 shows the scalar summation of the $p_T$ of each object, $\sum|p_T|$, which demonstrates good background discrimination and high signal efficiency for all black hole samples. We require $\sum|p_T|$ to be larger than 2.5 TeV to reject backgrounds. This requirement is relatively unaffected by changes in the model, in particular by changes to the number of extra dimensions $n$. Figure 11 shows $m_{\rm BH}$ distributions after this requirement. The QCD dijet background is already well suppressed, but we also investigated the effect of a further selection, requiring a lepton with a $p_T > 50$ GeV. This resulted in the QCD dijet background being rejected by a factor greater than $10^6$ as shown in Table 6, which summarises the event numbers for an integrated luminosity of 1 fb$^{-1}$. Though the high statistics QCD samples used were generated with PYTHIA, a leading order generator, there were also $p_T$-sliced small ALPGEN multijet samples available. When investigated using the $\sum|p_T|$ and lepton cut method, a very similar, marginally lower number of background events was predicted according to the very limited statistics available. Larger scale studies would be needed to conclude anything more concrete. Poisson confidence limits are used for samples where fewer than 20 events passed the requirements. Signal cross-section errors are statistical only; the theoretical uncertainties are large as discussed in Section 2. [7]

An alternative selection procedure was also used. Figure 12 shows the $p_T$ distributions of the leading, 2nd-, 3rd- and 4th-leading objects out of all the selected objects. The 4th-leading object still has larger $p_T$ in the signal events than in the background events. We require the number of objects with $p_T > 200$ GeV to be equal to or greater than four. Figure 14 (left) shows $m_{\rm BH}$ distributions after this requirement. Since QCD processes still remain large, a lepton requirement is again used to decrease it. Figure 13 shows the distribution of the highest $p_T$ lepton (muon or electron). As expected, the number of leptons from QCD processes is small. Requiring the number of leptons (muons or electrons) with $p_T > 200$ GeV to be equal to or greater than one results in the $m_{\rm BH}$ distributions shown in Figure 14(right).

The shape of the background in the region of high $m_{\rm BH}$ was fitted with a gaussian plus an asymmetric gaussian (Figure 15) and that function is used to estimate the number of background events.

CHARYBDIS does not include graviton emission. In practice this, and the energy lost in gravitational interactions during the balding phase, would be another source of $\not{E}_T$. Consequently we expect

[7] In the case of two hadronic subsamples ($t\bar{t}$ and dijets) where very few events passed the $\sum|p_T|$ requirement, the lepton requirement rejection factor was applied to the $\sum|p_T|$ requirement's Poisson bound to estimate the background distribution error.

| (a) muon | (b) electron |
|---|---|
| $|\eta| < 2.5$, $p_T > 15$ GeV<br>Central track match ($0 \le \chi^2 < 100$)<br>Isolation $E_{T,\rm cone0.2} < \min(100, 0.2p_T + 20)$ GeV | $|\eta| < 2.5$ except for 1.00< $|\eta|$ <1.15, 1.37< $|\eta|$ <1.52<br>$p_T > 15$ GeV<br>*medium* selection [45] |
| **(c) photon** | **(d) jet** |
| $|\eta| < 2.5$, $p_T > 15$ GeV<br>*tight* selection [45]<br>Isolation $E_{T,\rm cone0.2} < 0.2p_T + 20$ GeV | Cone algorithm ($R = 0.4$) based on calorimeter towers<br>$|\eta| < 2.5$, $p_T > 20$ GeV |

Table 5: Particle selection

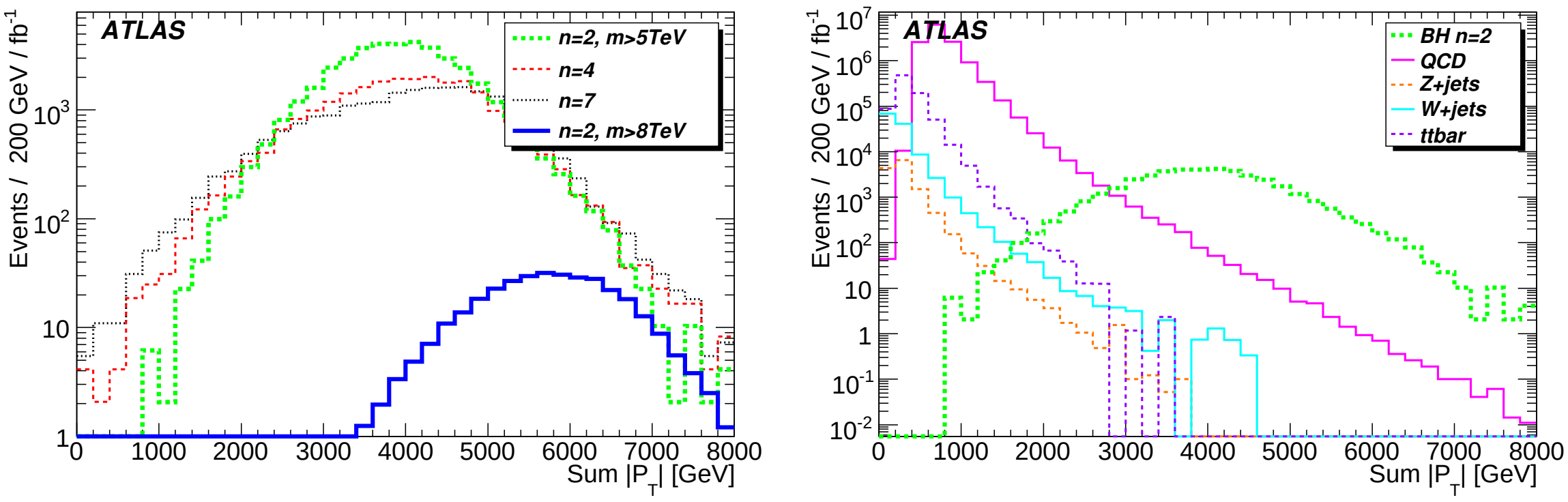

Figure 10: $\sum |p_T|$ distributions for (left) black hole samples and (right) backgrounds (QCD dijet, $t\bar{t}$ and vector boson plus jets), along with one signal sample for reference. They are normalised to an integrated luminosity of 1 fb$^{-1}$.

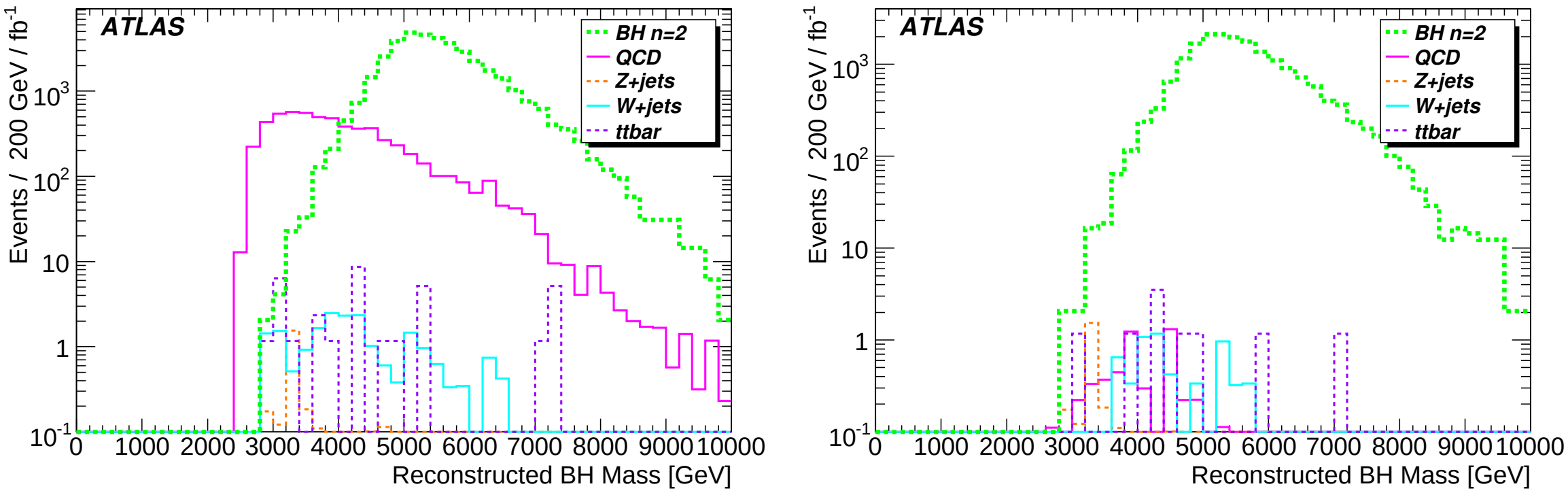

Figure 11: Black hole mass distribution with a requirement $\sum |p_T|$ >2.5 TeV (left), and black hole mass distribution with an additional requirement on the lepton-$p_T$ of $p_T > 50$ GeV (right). The signal sample with $n = 2$ and $m > 5$ TeV and backgrounds are shown.

| Dataset | Before selection (fb) | $\sum \lvert p_T \rvert > 2.5$ TeV (fb) | After requiring a lepton (fb) | acceptance |
|---|---|---|---|---|
| $n = 2, m > 5$ TeV | $40.7 \pm 0.1 \times 10^3$ | $39.2 \pm 0.3 \times 10^3$ | $18.6 \pm 0.2 \times 10^3$ | 0.46 |
| $n = 4, m > 5$ TeV | $24.3 \pm 0.1 \times 10^3$ | $22.6 \pm 0.2 \times 10^3$ | $6668 \pm 83$ | 0.27 |
| $n = 7, m > 5$ TeV | $22.3 \pm 0.1 \times 10^3$ | $20.1 \pm 0.2 \times 10^3$ | $3574 \pm 60$ | 0.17 |
| $n = 2, m > 8$ TeV | $338.2 \pm 1$ | $338.1 \pm 2.5$ | $212 \pm 16$ | 0.63 |
| $t\bar{t}$ | $833 \pm 100 \times 10^3$ | $23.6^{+12.2}_{-6.7}$ | $8.2^{+2.43}_{-2.43}$ | $9.8 \times 10^{-6}$ |
| QCD dijets | $12.8 \pm 3.7 \times 10^6$ | $5899^{+1773}_{-1771}$ | $5.37^{+3.25}_{-2.02}$ | $4.3 \times 10^{-7}$ |
| $W_{\ell\nu} + \geq 2$ jets | $1.9 \pm 0.04 \times 10^6$ | $12.3^{+9.0}_{-1.8}$ | $4.67^{+8.75}_{-0.93}$ | $2.4 \times 10^{-6}$ |
| $Z_{\ell\ell} + \geq 3$ jets | $51.8 \pm 1 \times 10^3$ | $2.75^{+2.02}_{-2.01}$ | $2.57^{+0.95}_{-0.64}$ | $5.0 \times 10^{-5}$ |

Table 6: Acceptance for each signal and background dataset in fb after requiring $\sum |p_T|$ >2.5 TeV, and a lepton with $p_T > 50$ GeV.

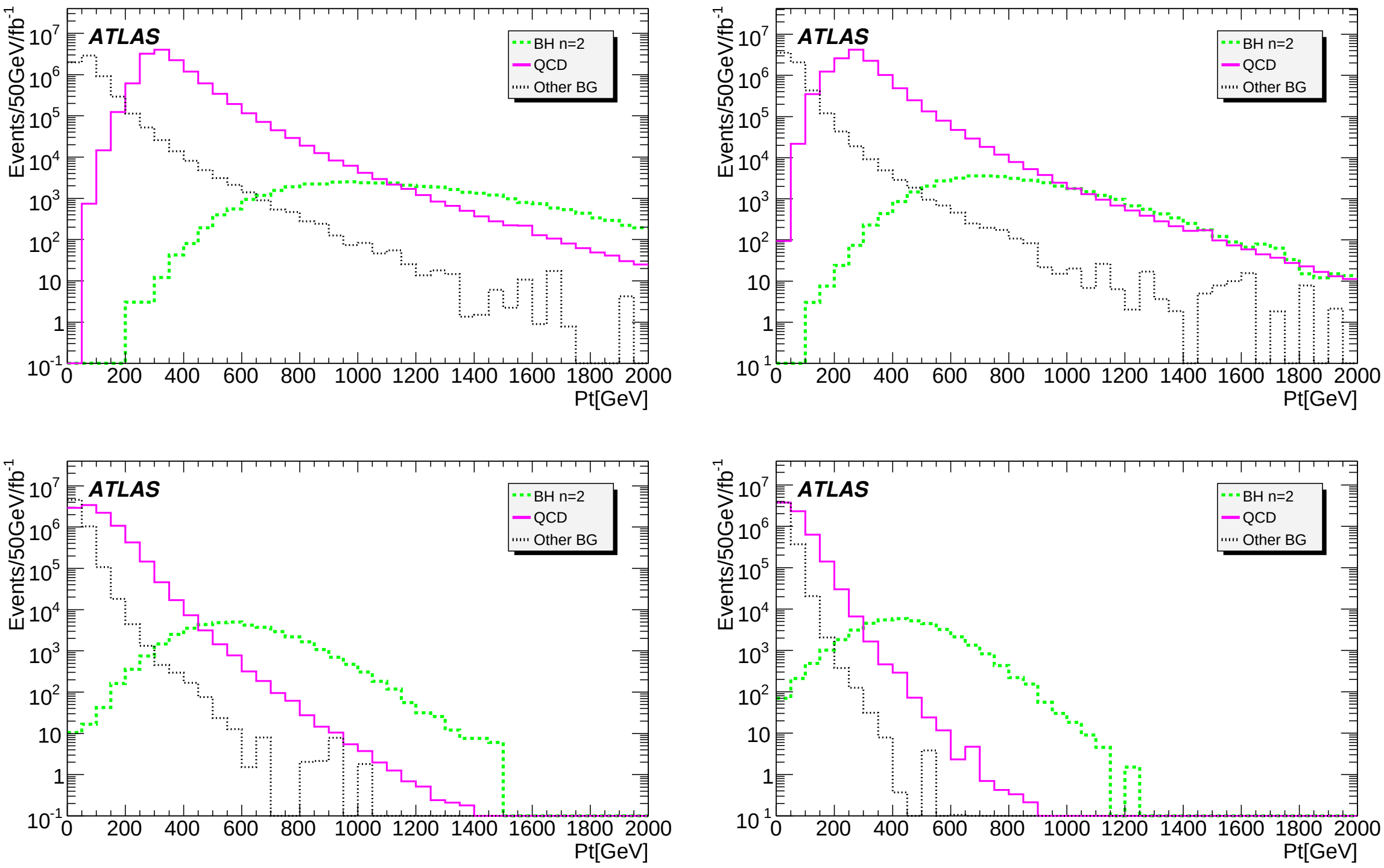

Figure 12: $p_T$ distributions of leading (top left), 2nd- (top right), 3rd- (bottom left) and 4th-leading (bottom right) objects out of all the selected objects for the signal sample with $n = 2$ and $m > 5$ TeV and backgrounds (see Table 7).

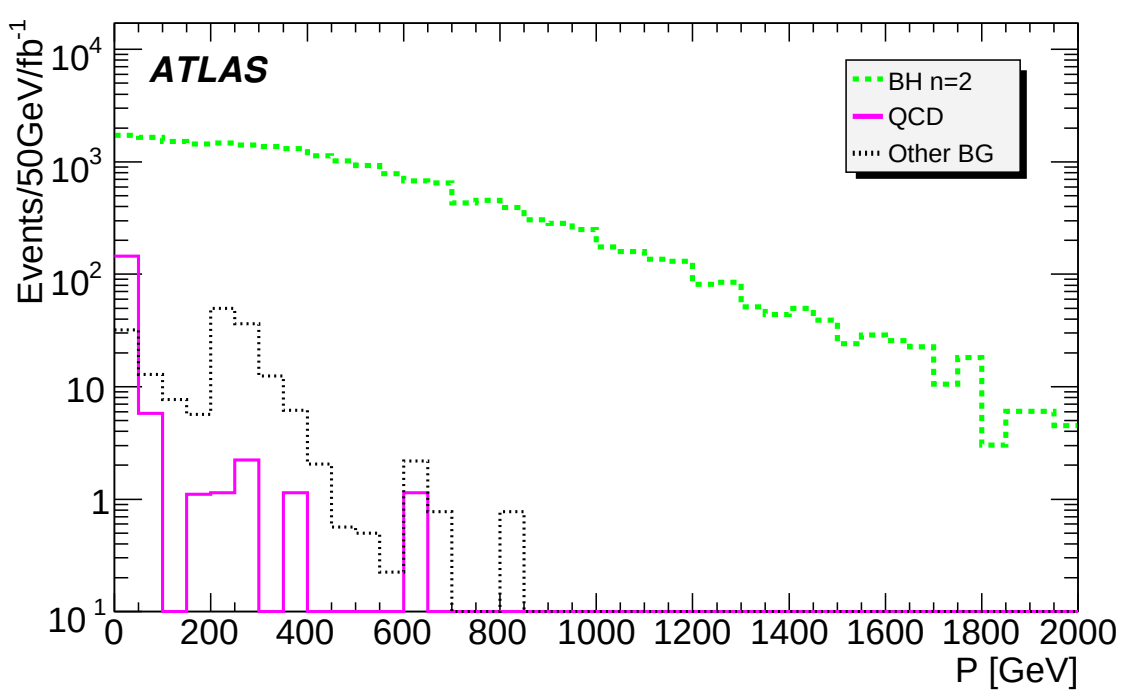

Figure 13: $p_T$ distributions of the leading lepton (electron or muon) after requiring the number of objects (electron, muon, photon or jet) with $p_T > 200$ GeV to be larger than 3 for the signal sample with $n = 2$ and $m > 5$ TeV and backgrounds (see Table 7).

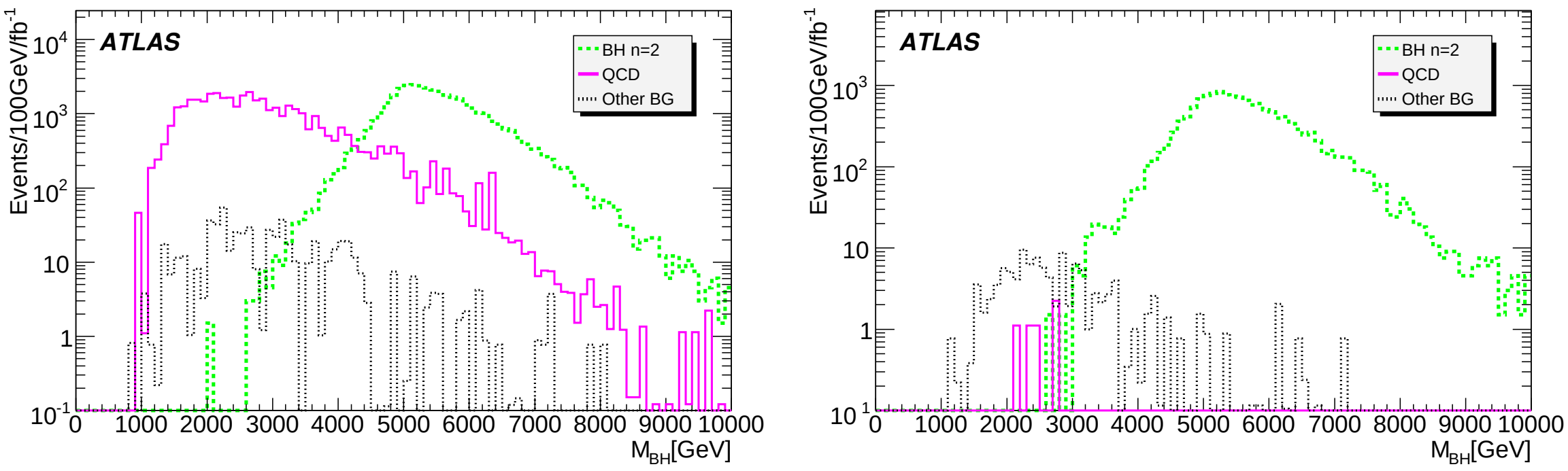

Figure 14: Black hole mass distribution for the signal sample with $n = 2$ and $m > 5$ TeV and backgrounds (see Table 7) after multiplicity requirement of at least 4 objects with $p_T > 200$ GeV (left plot) and an additional requirement of a lepton (electron or muon) with $p_T > 200$ GeV (right plot).

| Dataset | Before selection (fb) | After multi-object requirement (fb) | After lepton requirement (fb) | Acceptance |
|---|---|---|---|---|
| $n = 2, m > 5$ TeV | $40.7 \times 10^3$ | $38.9 \pm 0.4 \times 10^3$ | $14.0 \pm 0.2 \times 10^3$ | 0.34 |
| $n = 4, m > 5$ TeV | $24.3 \times 10^3$ | $17.9 \pm 0.3 \times 10^3$ | $4521 \pm 126$ | 0.19 |
| $n = 7, m > 5$ TeV | $22.3 \times 10^3$ | $9953 \pm 185$ | $1956 \pm 82$ | 0.087 |
| $n = 2, m > 8$ TeV | 338 | $338 \pm 4$ | $164 \pm 3$ | 0.49 |
| $t\bar{t}$ | $833 \times 10^3$ | $129 \pm 27$ | $36^{+12}_{-9}$ | $4.3 \times 10^{-5}$ |
| QCD dijets | $12.8 \times 10^6$ | $38.9 \pm 1.9 \times 10^3$ | $6^{+107}_{-3}$ | $5.6 \times 10^{-7}$ |
| W+jets | $560 \times 10^3$ | $99^{+28}_{-22}$ | $56^{+24}_{-13}$ | $1 \times 10^{-3}$ |
| Z+jets | $51.8 \times 10^3$ | $29^{+90}_{-4}$ | $19^{+90}_{-3}$ | $4 \times 10^{-4}$ |
| $\gamma(\gamma)$+jets | $5.1 \times 10^6$ | $285^{+87}_{-76}$ | $0^{+40}_{-0}$ | $< 10^{-5}$ |

Table 7: Acceptance of the 4-object requirements for each dataset in fb. 90% confidence limits are used when no events passed the requirements.

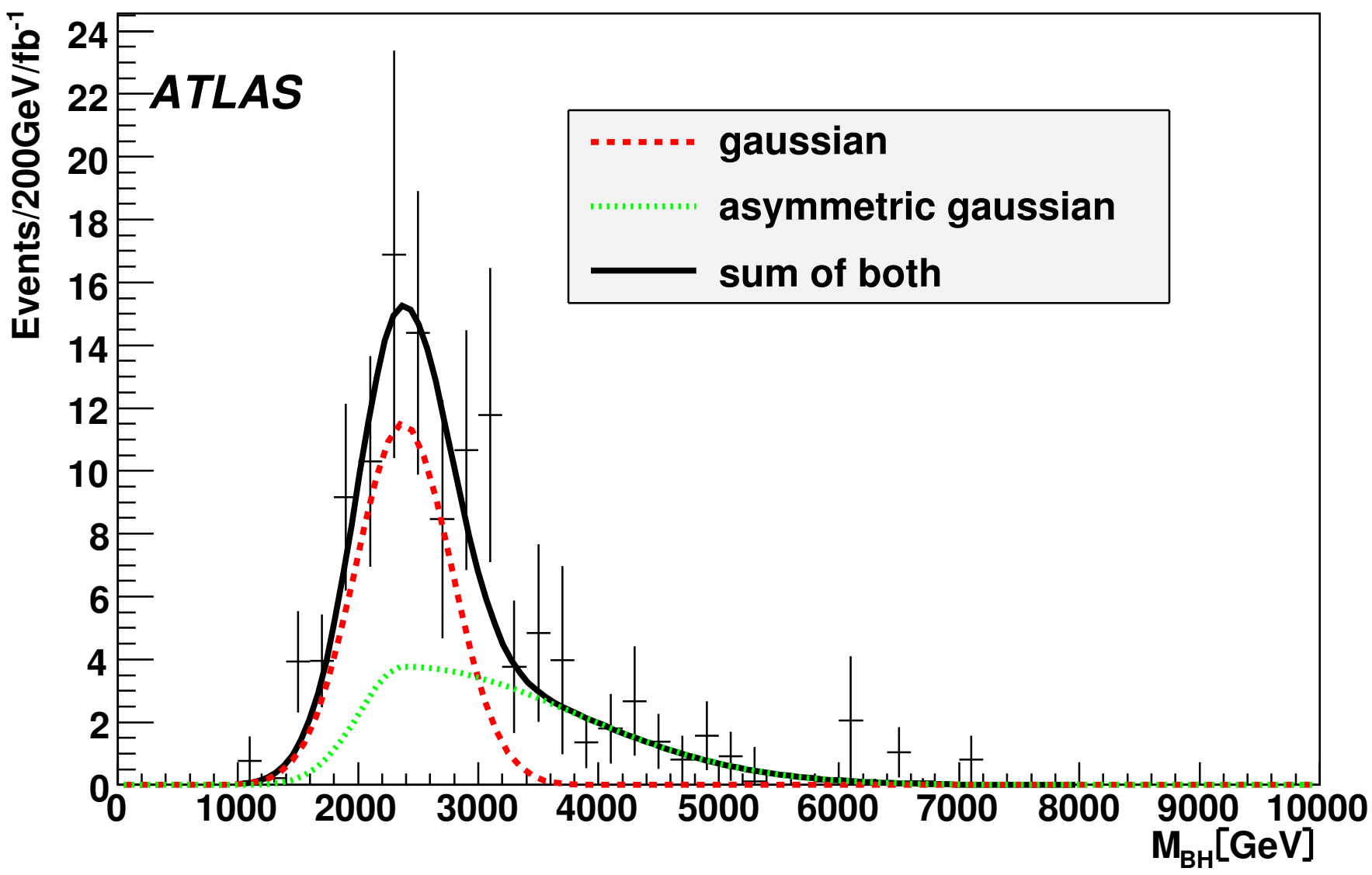

Figure 15: The background shape after the 4-object and lepton requirements is shown (data points). The points were fitted by the sum (black line) of a gaussian (red line) and an asymmetric gaussian (green line).

CHARYBDIS to underestimate this for black hole events. Nonetheless, each of the black hole samples studied in this analysis often have very wide distributions of $\not\!\!E_T$, with tails extending out to several TeV. This property of models with black holes is most unusual and hard to reproduce in other new physics scenarios, and should make it possible to distinguish between Black Holes and the majority of SUSY models for example.

A requirement on $\not\!\!E_T$ above $\sim 500-600$ GeV was studied as an alternative to a lepton requirement for black hole signal selection. Figure 16 shows the potential of this method, and contrasts these models with three common supersymmetric models of different cross-section and mass scale. Despite the possibility of early evidence for the presence of black holes, there are disadvantages to relying on such a selection. Firstly, our ability to reconstruct the black hole's mass is aided by limiting $\not\!\!E_T$ to be under 100 GeV (Figure 22). Such a signal from high $\not\!\!E_T$ events would be dominated by those events reconstructed most poorly, limiting their use for cross-section measurement and discovery. The theoretical uncertainties are large and difficult to quantify, and finally there are experimental difficulties in calibrating and accurately measuring this variable across a wide energy range.

## 6.2 Discovery Reach

Making a robust discovery potential for black hole events is difficult, because the semi classical assumptions used to model them are only valid well above the Planck scale. Close to the Planck scale, events may occur due to gravitational effects with lower multiplicities, but without the signatures anticipated by our event selections. As the energy rises above the threshold needed for black hole creation, our requirements should become more efficient. Lack of theoretical understanding makes it impossible to model this threshold region.

To account for this, we impose a lower requirement on the true mass of black holes created in our simulated samples, $BH_{thresh}$, normally set at 5 TeV, and we do not attempt to account for any additional signal from lower masses. In order to estimate the discovery potential, two methods have been consid-

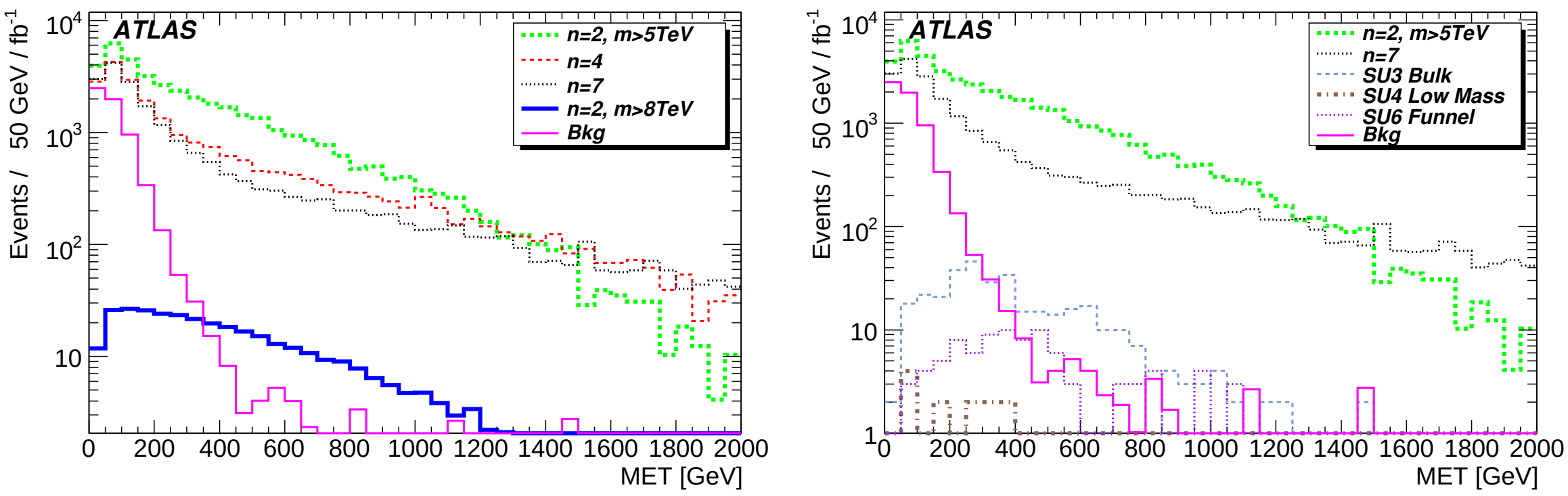

Figure 16: The left hand plot shows the missing transverse energy distributions after a $\sum |p_T| > 2.5$ TeV requirement. A requirement of $\not{E}_T > 500$ GeV would leave negligible background and a large number of signal events for all samples. The right hand plot compares two black hole samples with three supersymmetric models with a range of mass scales; the two classes of models can easily be distinguished by their differing cross-sections and the extent of the $\not{E}_T$ tail.

ered:

1. we keep our signal selection requirements constant, and increase the value of $BH_{thresh}$. Since the analysis requirements are unchanged, the background remains constant, while the signal drops as the production of events occurs at higher mass. We then evaluate the luminosity required to detect a minimum of 10 signal events, with $S/\sqrt{B} > 5$, assuming the production cross-section is as high as predicted. Such a study is shown in Figure 17, using the $\Sigma|p_T|$ and lepton requirements. This method produces conservative limits, taking some account of the uncertainty in the production cross-section near the threshold.

2. We keep the production model unchanged with $BH_{thresh} = 5$ TeV, but apply an additional requirement on the reconstructed black hole mass. This requirement reduces substantially background events, while allowing the higher mass signal to pass unchanged. This is less conservative, since it allows black hole signal events to be produced at low mass, but to migrate above the reconstructed mass requirement because of the detector mass resolution, hence increasing the signal. As before, we use the nominal value of the production cross-section, and evaluate the luminosity needed to meet our discovery criteria, this time as a function of reconstructed mass. A study using this method is shown in Figure 18 using the 4-object and lepton requirements.

The two approaches are complementary and illustrate the uncertainties in different ways. We observe that the search reach is limited eventually at high mass by the falling production cross-section, reflecting the falling parton luminosity and the limited energy of the LHC. We conclude that, if the semi classical cross-section estimates are valid, black holes can be discovered above a 5 TeV threshold with a few pb$^{-1}$ of data, while 1 fb$^{-1}$ would allow a discovery to be made even if the production threshold was at 8 TeV.

# 7 Systematic Uncertainties

## 7.1 Signal uncertainties

We have investigated the systematic uncertainties using fast simulation runs, having checked that the full and fast simulations agree well for this purpose.

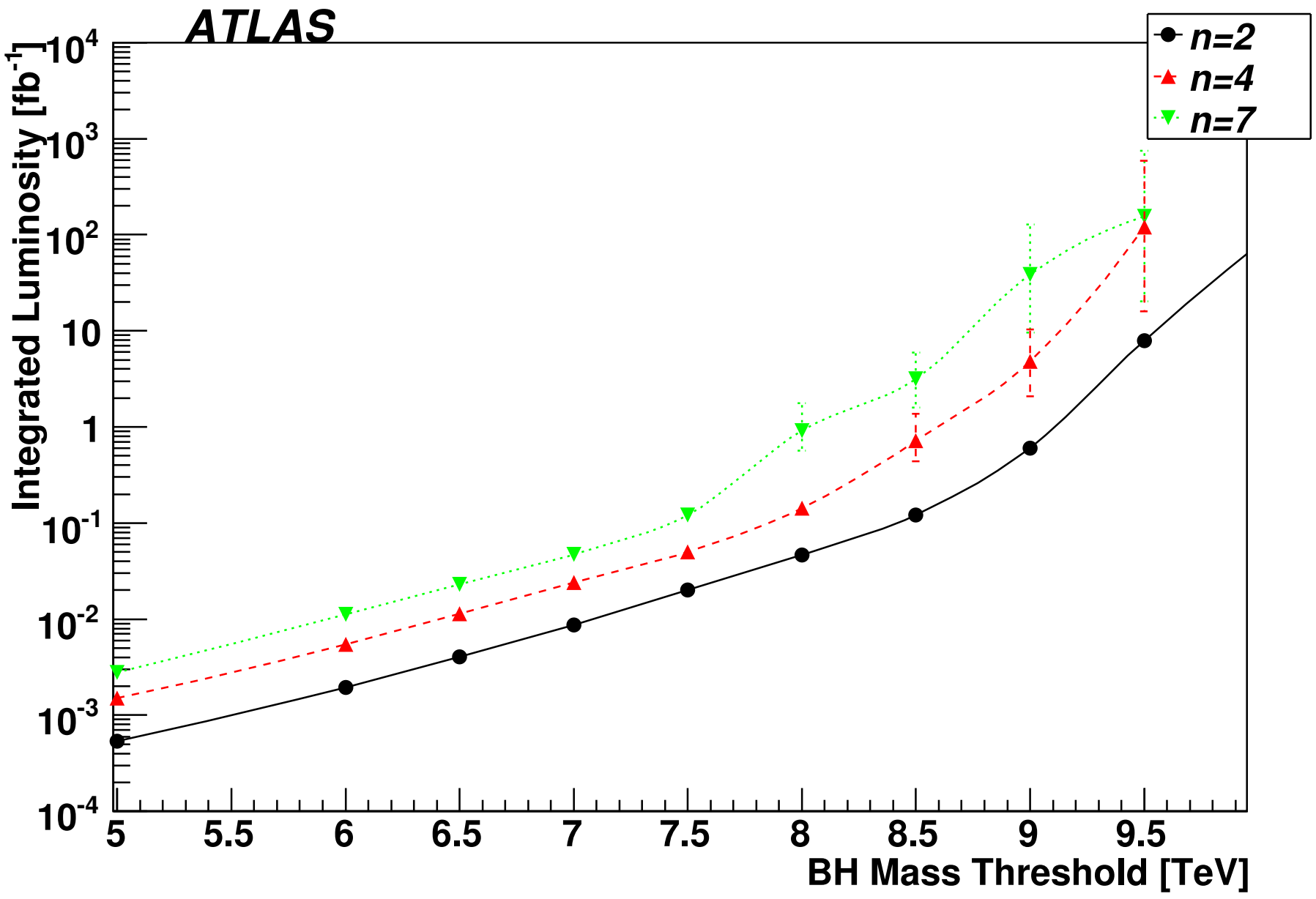

Figure 17: Discovery potential using $\sum |p_T|$ and lepton selections: required luminosity as a function of black hole mass threshold. Error bars reflect statistical uncertainties only.

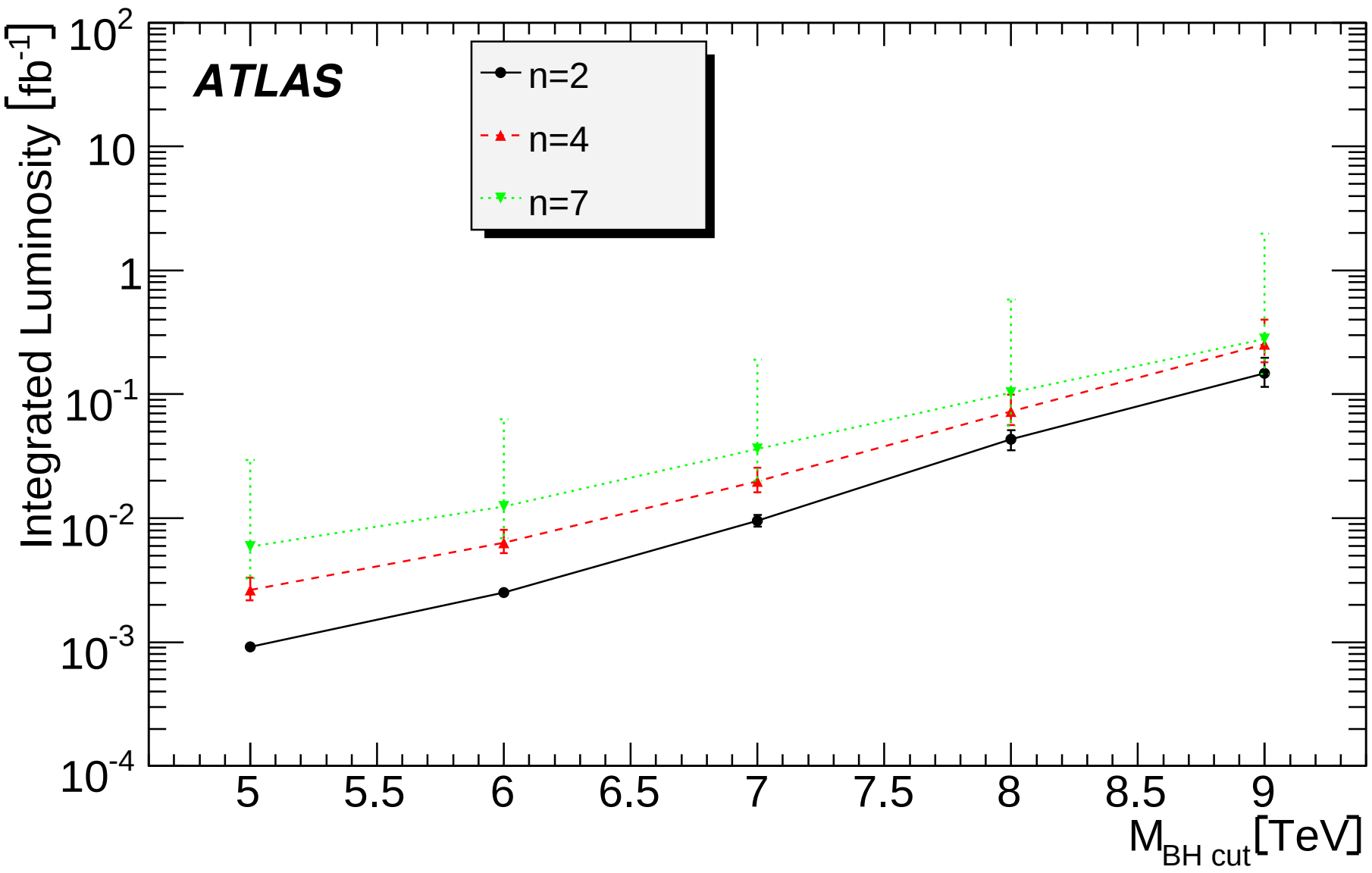

Figure 18: Discovery potential for black holes using four-object and lepton requirements. The required luminosity is shown as a function of the requirement on the reconstructed black hole mass. The error bars correspond to experimental systematic uncertainties. (See text for constraints.)

There are a number of theoretical parameters associated with CHARYBDIS which can generate systematic errors in the estimates of the acceptance for signal events. These are:

- The kinematic cutoff. This parameter is normally true, and causes the generator to end thermal emission if an unphysical emission is randomly selected. The generator moves immediately to the final remnant decay phase. This approximation deteriorates at high numbers of extra dimensions because of the high temperature and emitted particle energies. We have investigated the alternative, where a new emission is selected until a physical one is chosen. In this case, thermal emission will continue until the black hole mass falls below $M_{\mathrm{DL}}$.
- Temperature variation. The Hawking temperature of the black hole is normally allowed to increase as its mass decreases, as expected if the black hole has time to equilibrate between decays. We have investigated the alternative of keeping the temperature fixed at the initial value, as would be the case if the black hole decayed very quickly or "suddenly" (see Table 8).
- Number of extra dimensions. In addition to our full simulation samples with $n = 2$, 4 and 7, we have simulated $n = 3$ and 5 with the fast simulation. As noted above, the events become more jet-like at high $n$ and the particle multiplicity drops (see Figure 20), due to the increased Hawking temperature. Our signal selection remains robust, as shown in Table 8.
- Planck scale. We have investigated changing the Planck scale from its default value of $M_{\mathrm{DL}} =$ 1 TeV to 2 TeV. We note that, since the model is only valid for black hole masses much larger than the Planck scale, this scenario is not well modelled in the range of masses accessible at the LHC.
- Remnant decay. We have investigated changing the remnant decay model from a two-body to a four-body mode (see Figure 20).

Figure 19 shows the effect of changing the kinematic cutoff on the particle multiplicity and $\sum |p_T|$ distributions. Since the black hole is forced to decay thermally until it falls below $M_{\mathrm{DL}}$, the multiplicity is higher, and the events have lower energy emissions. The total energy remains constant, however so the $\sum |p_T|$ distribution is relatively stable.

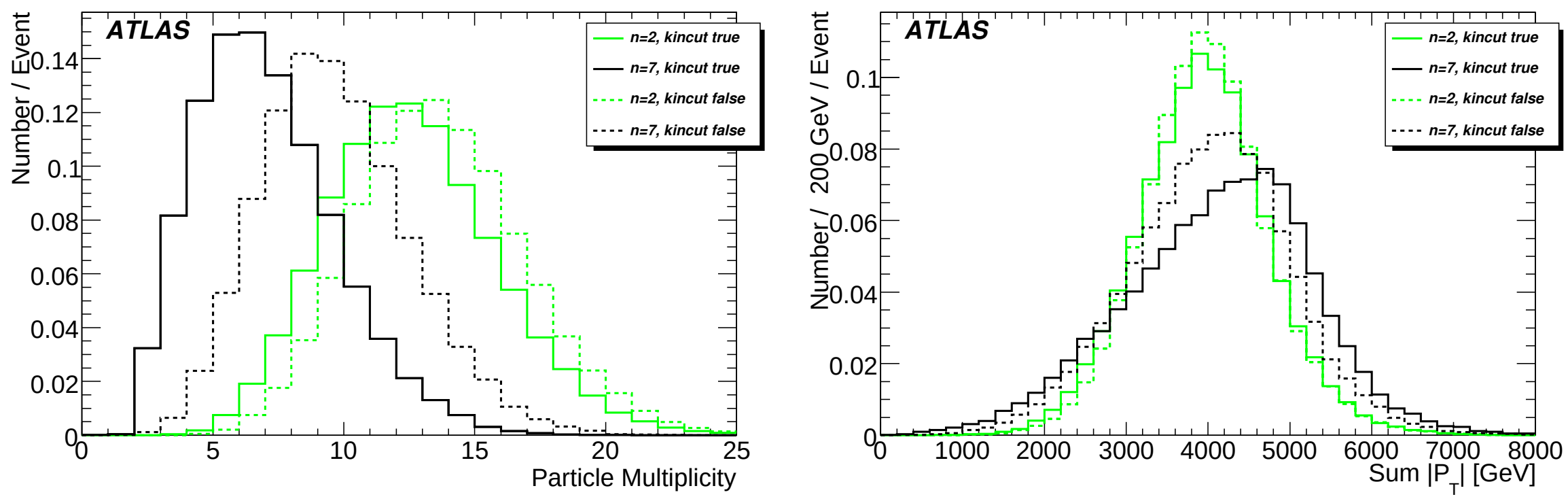

Figure 19: Particle multiplicity and $\sum |p_T|$ distributions, showing the variation when the kinematic cut-off parameter is changed.

The acceptance of the signal for various parameter choices are shown in Table 8. Compared to the standard full simulation, the largest effect is observed for high $n$, when the kinematic cut-off parameter is changed. This is as expected, since this parameter has a large effect on the evolution of the black hole during its decay. Similarly changing the remnant decay from 2 to 4 bodies has a large effect at high $n$; the effect of this, and of changing number of extra dimensions is shown in Figure 20.

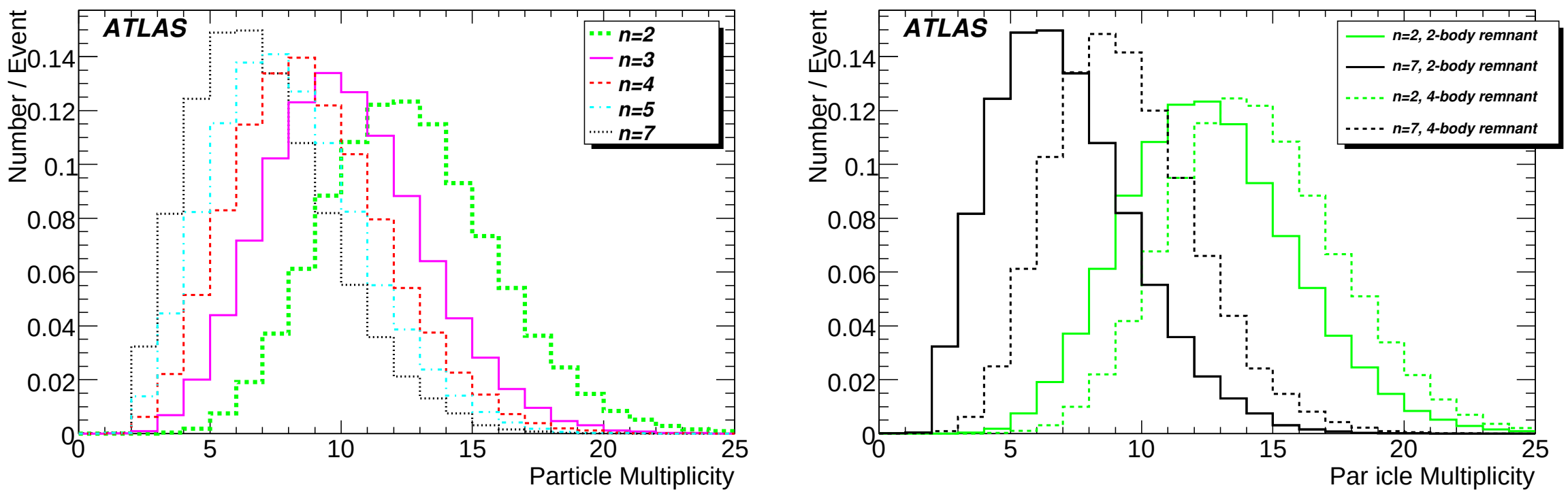

Figure 20: Particle multiplicity distributions, showing the variation produced by a change in the number of extra dimensions, or the number of particles produced by remnant decay.

| $n$ | Full Sim | Fast Sim | Kin. Cut off | $T_H$-variation off | 4-body remnant |
|---|---|---|---|---|---|
| 2 | 45.8 | 42.9 | 47.2 | 48.7 | 47.9 |
| 3 | - | 33.2 | - | - | - |
| 4 | 27.4 | 26.6 | - | - | - |
| 5 | - | 21.7 | - | - | - |
| 7 | 16.1 | 15.9 | 29.2 | 16.6 | 27.4 |

Table 8: Signal acceptance (%) for different model assumptions.

### 7.2 Uncertainties on detector performance

Two kinds of systematic uncertainties on detector performance were studied. One is an uncertainty on lepton identification efficiency. To estimate this effect, we loosened and tightened particle identification selections, by changing the hadronic leakage for electrons and the isolation cut for muons. The changes in signal efficiencies are around 2%.

The effect of a 5% error in the jet energy scale (JES) was also considered. The effect of these uncertainties on the discovery potential is shown in Figure 18.

## 8 Results

### 8.1 Search Reach for Black Hole Production at the LHC

The studies presented show that the ATLAS detector is capable of discovering the production of black holes up to the kinematic limit of the LHC, assuming that the signal is correctly modelled. This conclusion is largely based on the predicted huge production cross-section, and the small background from QCD dijets at very high-$p_T$, especially when the presence of a high energy lepton is required. However, both of these assumptions are suspect. The high production cross-section is subject to considerable discussion in the literature, as discussed in Section 2. Moreover, until the LHC has measured the QCD cross-section at 14 TeV, we cannot be certain of the tails of the QCD distributions. The Monte Carlo simulations of these tails are working at the limit of their validity, given the high energies and large multiplicities involved.

For these reasons, we prefer not to place too much weight on the detailed search reach limits. Instead, we confine ourselves to the statement that, with current understanding, the black hole signature considered should be clearly visible if it exists.

### 8.2 Determination of Model Parameters

We have considered the possibility of extracting model parameters from the data, should a signal be observed. There are two key parameters: the Planck scale $M_D$ (or $M_{\mathrm{DL}}$ depending on the convention) and the number of extra dimensions $n$. In Ref. [46, 47] a method was proposed to extract $M_D$ from the cross-section data, which fixes the Planck scale (within the model assumptions), and from events with high energy emissions.

The Hawking temperature $T_H$ of the black hole depends on $n$. If we detect events with emissions near $m_{\mathrm{BH}}/2$, the energy of those emissions is a measure of the initial $T_H$. Hence, over the sample of black holes, the probability of such emissions is a measure of the characteristic temperature, and can be used to extract $n$. This method was first put forward in Ref. [46], and here is made compatible with the need for background rejection requirements. The requirements described in Section 6 are not appropriate: the lepton requirement biases the selected events in favour of final states with many particles, and hence against those events with a single high energy emission. A suitable requirement was found to be $\sum |p_T| > 3.5$ TeV; this removes the background without biasing the signal events selected.

Figure 21 shows the probability of a hard emission for two samples, compared to the predictions. The method requires accurate mass resolution, and so an additional requirement on $\not{E}_T < 100$ GeV is applied. The addition of this requirement lowers the efficiency noticeably, but does improve the black hole mass reconstruction; details of the resolution and efficiency can be found in Figure 22. The data are consistent with the expected value of $n$, but due to this reduction in signal efficiency, more data would be required to make a definitive measurement. It should be noted that this measurement requires the Planck scale to be known. If this cannot be determined from the production cross-section, it is likely that the threshold behaviour near the Planck scale would provide an indication of its value. At present, no theoretical model exists to allow us to make predictions in this region.

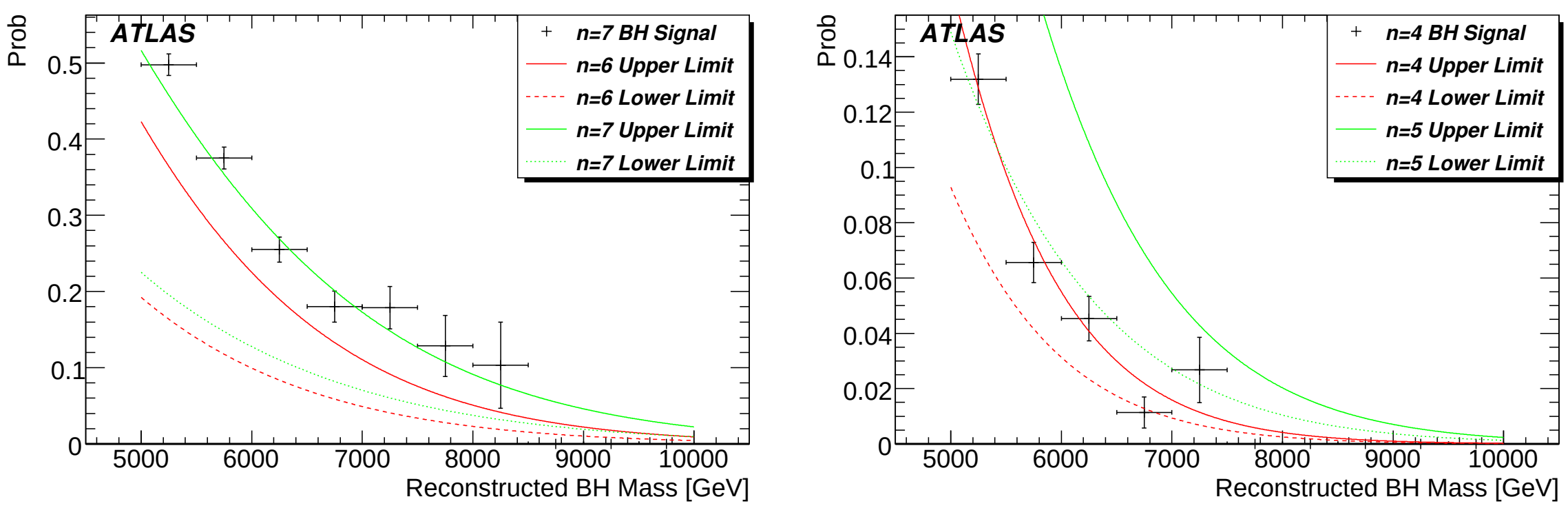

Figure 21: The probability of a hard emission near $m_{\mathrm{BH}}/2$ for $n = 7$ and $n = 4$. The bands show the expected range for $n = 7$ and $n = 6$, and for $n = 5$ and $n = 4$, respectively, for a luminosity of approximately 0.75 $\mathrm{fb}^{-1}$.

## 9 Summary

The search for black holes in the first $100\,\mathrm{pb}^{-1}$ of LHC data with the ATLAS detector and software framework was simulated.

| | | Normalisation | Mean (GeV) | Resolution (GeV) |
|---|---|---|---|---|
| Without | Narrow | $1018 \pm 26$ | $-217 \pm 5$ | $276 \pm 9$ |
| $\not{E}_T$ requirement | Wide | $276 \pm 30$ | $-148 \pm 9$ | $722 \pm 13$ |
| With | Narrow | $318 \pm 12$ | $-116 \pm 8$ | $215 \pm 9$ |
| $\not{E}_T$ requirement | Wide | $108 \pm 7$ | $118 \pm 18$ | $635 \pm 16$ |

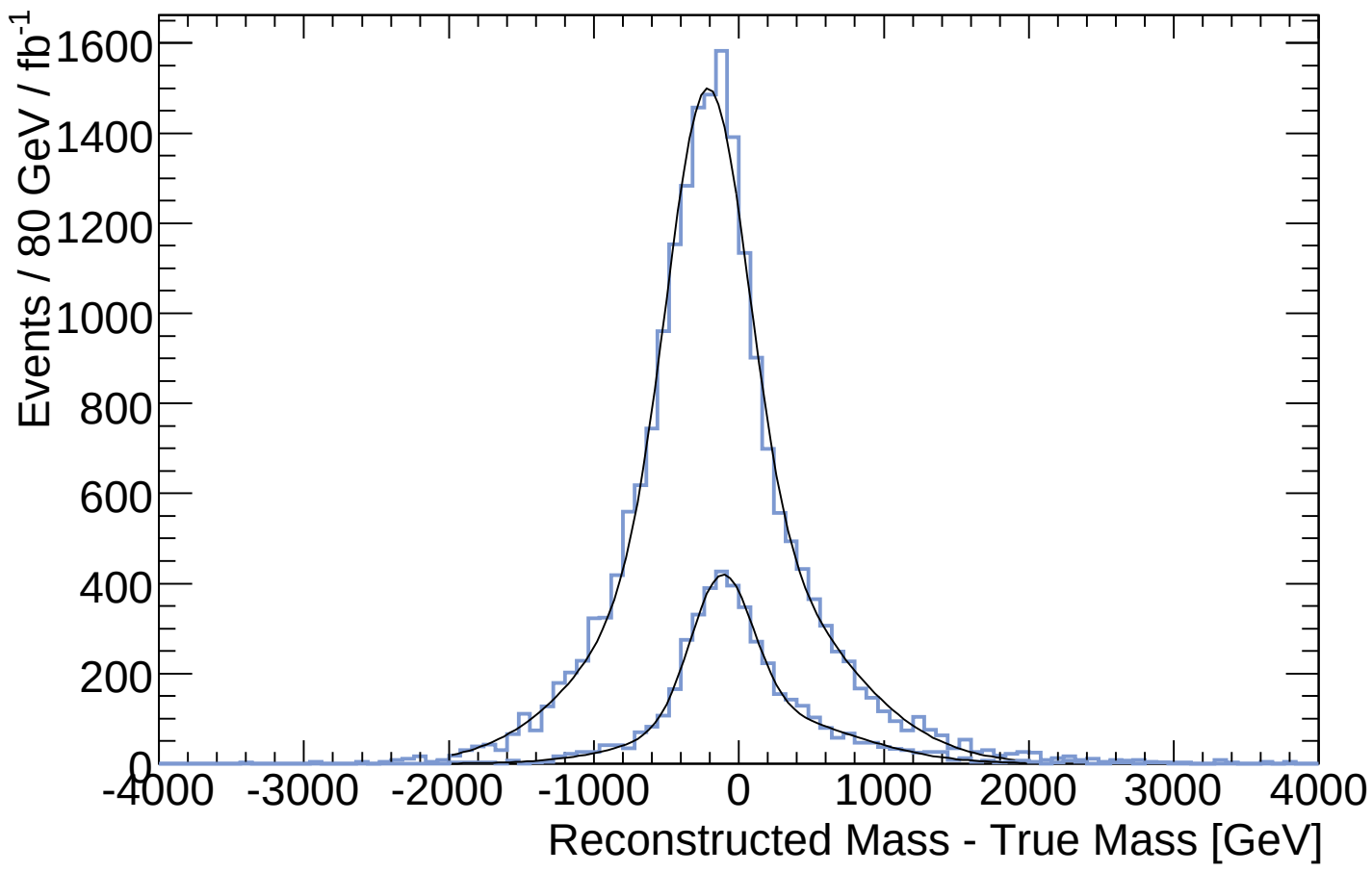

Figure 22: Black hole mass resolution distributions and their fits to double Gaussian functions, using a $\sum |p_T|$ and lepton requirement (1 $\mathrm{fb}^{-1}$). The upper curve is without a $\not{E}_T$cut. The lower curve has an additional requirement on $\not{E}_T < 100$ GeV.

We summarised the current experimental limits on black hole production and studied, with the help of the black hole event generator CHARYBDIS and Standard Model Monte Carlo data sets, the basic event properties, trigger and selection efficiencies, theoretical and experimental uncertainties of black hole production at the LHC for a flat ADD extra dimension scenario with the Planck scale $M_{\mathrm{DL}} = 1$ TeV. We have explored the uncertainties inherent in the theoretical modelling and our understanding of the detector. We conclude that, if the semi-classical cross section estimates are valid, black holes above a 5 TeV threshold can be discovered with a few $\mathrm{pb}^{-1}$ of data, while 1 $\mathrm{fb}^{-1}$ would allow a discovery to be made even if the production threshold was 8 TeV.